\documentclass{article}

\usepackage{graphicx}
\usepackage{amsmath}
\usepackage{amssymb}
\usepackage{amsthm}
\usepackage{graphicx}
\usepackage[margin=0.75in]{geometry}
\usepackage{float}
\usepackage{algorithm}
\usepackage{algorithmic}
\usepackage{caption}
\usepackage{subcaption}

\floatname{algorithm}{Procedure}

\theoremstyle{theorem}
\newtheorem{theorem}{Theorem}
\newtheorem{lemma}{Lemma}
 
\theoremstyle{definition}

\newcommand{\ie}{i.e., }
\newcommand{\eg}{e.g., }	
\newcommand{\anagram}[2]{\texttt{#1} $\rightarrow$ \texttt{#2}}
\newcommand{\Int}{{\mathbb{Z}}}
\renewcommand{\vec}[1]{\ensuremath{\boldsymbol{#1}}}
\newcommand{\ba}[1]{\begin{array}{#1}}
\newcommand{\ea}{\end{array}}
\newcommand{\bmat}{\left[\ba{rrrrrrrrrr}}
\newcommand{\emat}{\ea\right]}

\graphicspath{{./figures/}} 

\title{Star Anagram Detection and Classification}
\author{Jason Parker\footnote{US Air Force Research Laboratory, Wright Patterson AFB, USA}\\
Dan Barker\footnote{Freedom From Religion Foundation, Madison, WI, USA}}
	
\begin{document}

\maketitle

\begin{abstract}
A star anagram is a rearrangement of the letters of one word to produce another word where no letter retains its original neighbors.
These maximally shuffled anagrams are rare, comprising only about 5.7\% of anagrams in English. 
They can also be depicted as unicursal polygons with varying forms, including the eponymous stars.
We develop automated methods for detecting stars among other anagrams and for classifying them based on their polygon's degree of both rotational and reflective symmetry.
Next, we explore several properties of star anagrams including proofs for two results about the edge lengths of perfect, \ie maximally symmetric, stars leveraging perhaps surprising connections to modular arithmetic and the celebrated Chinese Remainder Theorem.
Finally, we conduct an exhaustive search of English for star anagrams and provide numerical results about their clustering into common shapes along with examples of geometrically noteworthy stars.

\end{abstract}

%%%%%%%%%%%%%%%%%%%%%%%%%%%%%%%%
\section{Introduction}

\subsection{Motivation}
An \emph{anagram} is a word or phrase formed by rearranging the letters of another word or phrase. 
In this article, we use the word anagram to refer only to a pair of single words that are anagrams of each other.
We consider the pair to be ordered and view the second word as a rearrangement of the letters in the first word, e.g., \anagram{EARTH}{HEART}. 
Notice that \anagram{EARTH}{HEART} is a simple rotation. 
All letters keep their original neighbors; nothing has been shuffled.

\begin{figure}[hbt]
	\centering
		\includegraphics[width=0.6\textwidth]{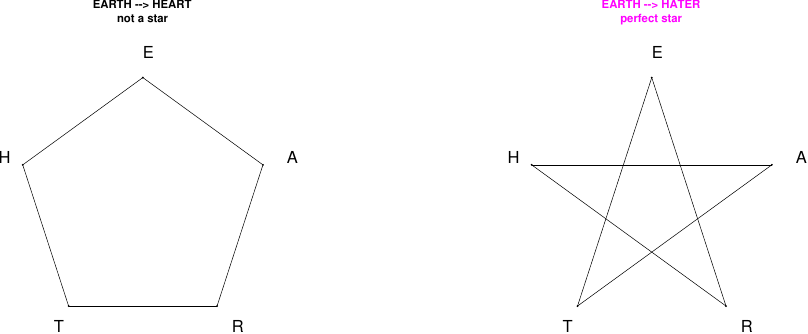}
	\caption{(Left) An anagram that is not a star. (Right) An example star anagram.}
	\label{fig:starExample}
\end{figure}

A \emph{star anagram} is a special class of anagrams in which the letters have been maximally shuffled: no letter in the second word is adjacent to one of its original neighbors, counting the first and last letters as neighbors.
An example is \anagram{EARTH}{HATER}. 
The name \emph{star} anagram derives from an interesting geometric property of these anagrams. 
In particular, if we arrange the letters of the first word in a circle and trace the path formed by the rearranged second word, we obtain a star as on the right in Figure \ref{fig:starExample}. 
Most anagrams, like our earlier example \anagram{EARTH}{HEART}, do not produce a star shape, as we see on the left of the Figure. 

\begin{figure}[tbh]
	\centering
		\includegraphics[width=\textwidth]{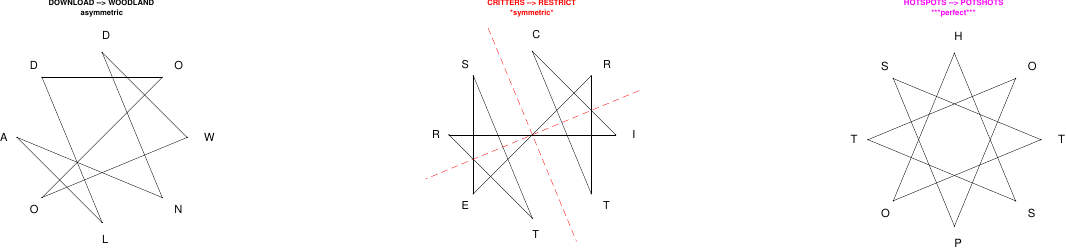}
	\caption{Length $8$ examples of the three star anagrams classes.}
	\label{fig:classExamples}
\end{figure}

An interesting subset of star anagrams are \emph{symmetric}. They can be folded perfectly along some dividing line (reflective symmetry) or rotated less than one full turn and look the same (rotational symmetry). 
Stars which lack this symmetry are \emph{asymmetric}. 
An even rarer type of star anagram has all edges of the same length, which we denote as \emph{perfect}.
Note that perfect stars are both reflexively and rotationally symmetric, although we place them into a special class. 
Figure \ref{fig:classExamples} provides examples of all three classes of star anagrams for words of length $8$. 

Barker \cite{barker} originally coined the term star anagram and introduced it to the first author. 
Prior to the work described in this article, Barker searched for star anagrams without automated tools. 
While intellectually rewarding (because it challenges you to turn one dimension into two), this approach makes it difficult to identify large groups of star anagrams, particularly among longer words and those with repeated letters. 

\subsection{Contribution}
The primary contribution of this paper is a numerically inexpensive method for automatically detecting star anagrams and classifying them based on their degree of rotational and reflective symmetry, including a simple test for perfection. 
All of these methods rely on simple operations computed from the edge lengths of the anagram's representation as a unicursal polygon. 
We also use the Chinese Remainder Theorem to prove that perfect stars must have edge lengths that are coprime with their word length, a result already well known in the study of star polygons. 

A star anagram's polygon changes based on the ordering of the words. 
We prove that reversing the order of the anagram preserves both starriness and perfection.
A surprising result on the edge length of reversed perfect stars is also provided, demonstrating that the edge lengths of a perfect star and its reversed star are modular inverses in the parlance of number theory. 

Finally, we conduct a detailed numerical study of the star anagrams in English. 
First, all star anagrams in a large database of English words are detected and classified.
We then provide numerical results on the clustering of these star anagrams into common shapes and their distribution across word lengths.
An Appendix provides a complete set of figures depicting all star anagrams detected in English. 
We also discuss the initially surprising notion of  \emph{autostars}, which are words that can be star anagrams of themselves. 
An exhaustive search of autostars provides interesting examples of polygon shapes that do not appear among normal star anagrams in the English language.

\subsection{Outline}
The remainder of this paper is organized as follows. 
Section \ref{sec:detection} describes our approach for detecting star anagrams, and Section \ref{sec:classification} describes our approach for star anagram classification. 
In Section \ref{sec:remarks}, we prove several properties of star anagrams, discuss clustering of stars into common shapes, and introduce autostars.  
Section \ref{sec:search} presents the numerical results for our search of English words for star anagrams. 
Finally, Section \ref{sec:conclusions} provides concluding remarks and possible future work. 

\subsection{Notation}
Throughout this article we will use 
bold face capital letters for matrices (\eg $\vec{A}$), 
bold face lower case letters for vectors (\eg $\vec{p}$), 
and non-bold letters for scalars (\eg $N$).  
We will denote the set of integers as $\Int$.
A length $N$ vector $\vec{p}$ of integers will be written as $\vec{p} \in \Int^N$, with the $n^{th}$ entry denoted as $p_n$. 
Similarly, a matrix with $M$ rows and $N$ columns will be denoted as $\vec{A} \in \Int^{M\times N}$, with the scalar entry in the $m^{th}$ row and $n^{th}$ column denoted as $a_{mn}$.  
Note that we use $0$ based indexing, \eg numbering the columns from $0 \ldots N-1$, throughout this article. 

We use $N!$ for the factorial of a scalar $N$, and the magnitude of a scalar $p$ will be given as $|p|$. 
The  modulo $N$ operation (\ie remainder after division by $N$) for a scalar integer $K$ will be denoted as $K \bmod N$.
We say that $a \equiv b \pmod{N}$ if $a \bmod N = b \bmod N$. 
Finally, we say that two integers $N$ and $L$ are \emph{coprime} if they have no common positive divisor other than $1$.

%%%%%%%%%%%%%%%%%%%%%%%%%%%%%%%%
\section{Star Anagram Detection}
\label{sec:detection}
Generating a comprehensive list of all anagrams in a given set of words is straightforward, \eg by exhaustively comparing the sorted letters of equal length words. 
Omitting those details, we will focus on an algorithmic approach for detecting whether a given anagram is a star. 
Before describing this approach, we need to explain how to think of anagrams as paths. 

\subsection{Anagrams as Paths}
We number the letters of any  length $N$ word with the integer values $0$ to $N-1$. 
Any rearrangement of these letters can be viewed as traversing a \emph{path} that connects the letters in the specified order. 
We will represent such a path as a vector $\vec{p} \in \Int^N$ with entries $\{p_n\}_{n=0}^{N-1}$.
For our example \anagram{EARTH}{HATER}, we obtain the path $\vec{p} = [4,1,3,0,2]$. 
Note that for a given word of length $N$ there are $N!$ possible paths, including the original word and many nonsense arrangements.
   
In our analysis of star anagrams, we think of the letters of the original word as nodes arranged uniformly around a circle or ring. 
The path is drawn as a series of line segments connecting the nodes in the specified order, including a segment from node $p_{N-1}$ back to node $p_0$ to close the figure. 
These shapes produced by a continuous path are known in geometry as \emph{unicursal polygons} and have been widely studied.
Indeed, \emph{star polygons}, which are the unicursal polygons produced by our star anagrams, have been studied since at least the fourteenth century~\cite[Section 2.8]{coxeter}. 
In the sequel, we will see that star anagram detection and classification can be done entirely by looking at the properties of anagram paths.    

\subsection{Identifying All Possible Paths}
Our simple example \anagram{EARTH}{HATER} conveniently hid a complication. 
In particular, the path for an anagram with repeated letters is not unique. 
The nodes of the repeated letters can be swapped in the path without changing the resulting word. 
For an anagram with $R$ repeated letters each of which appears $w_r$ times, the number of possible paths $P$ will be $P = \prod_{r=0}^{R-1} w_r!$, which can be large.

\begin{figure}[tbh]
	\centering
		\includegraphics[width=\textwidth]{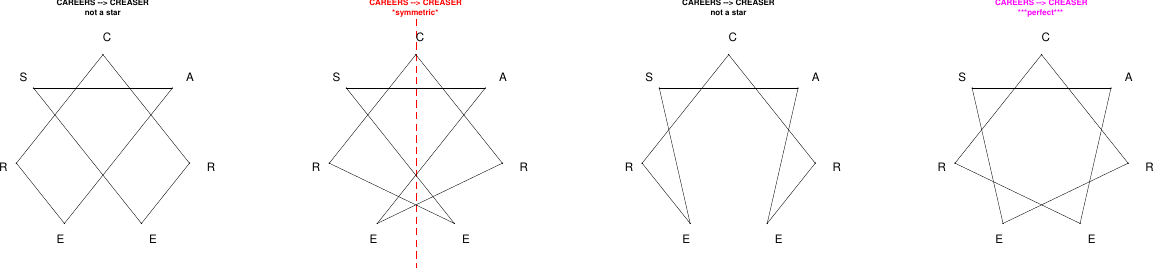}
	\caption{The $4$ possible paths for the perfect star anagram \anagram{CAREERS}{CREASER}.}
	\label{fig:allPathExample}
\end{figure}

For example, consider \anagram{CAREERS}{CREASER}. 
The two ``e'' and ``r'' letters can both be visited in either order in the path, leading to $P = (2! \cdot 2!) = 4$ possible paths as shown in Figure \ref{fig:allPathExample}. 
Only the fourth path reveals this anagram to be a perfect star. 

Our solution to this problem is straightforward. 
We simply generate all possible paths for each anagram using an exhaustive recursive enumeration and evaluate each path. 
We then select a single path $\vec{p}^*$ out of the set of $P$ possible paths $\{\vec{p}_i\}_{i=0}^{P-1}$ to represent the anagram. 
A perfect star path is selected if found, followed in preference by a symmetric star path, an asymmetric star path, and finally a non-star path.
If multiple paths are found within the preferred class, we select one arbitrarily.
Multiple symmetric star paths are handled differently, which we describe after discussing classification. 

\subsection{Step Sizes}
Now that we have identified the set of paths to test for a given anagram, we turn our attention to computing the steps around the circle represented by a given path. 
First, given a path $\vec{p}$, we define the path differences $\vec{d} \in \Int^N$ with entries $\{d_n\}_{n=0}^{N-1}$ given by
\begin{align}
\label{eq:d}
d_n = p_{n+1} - p_n,
\end{align}
where for convenience we define $p_{N} = p_0$. 
We would like to use these differences to analyze the geometric properties of the path around the circle.

However, these raw differences of the node locations include ambiguities that make direct analysis difficult.
Notice that $d_n$ can take on values from $-(N-1)$ to $N-1$, yielding $2N - 2$ possible values\footnote{Notice $d_n \ne 0$ since all the $\{p_n\}_{n=0}^{N-1}$ are distinct.}.    
Clearly these values are redundant, since starting from a given node there are only $N-1$ possible \emph{steps} to the next node. 
Our goal will be to map these path differences to unambiguous steps. 

If the next node is $k_1$ steps away around the circle in the clockwise direction, then it will be $k_2 = N - k_1$ steps away in the counter-clockwise direction. 
Each of the other nodes can thus always be reached by two complementary steps with sizes satisfying $k_1 + k_2 = N$. 
To avoid this ambiguity, we will use the smaller length for our steps.
This choice also allows us to define the length of each edge as the magnitude of the corresponding step.
We will also define a clockwise step as positive and a counter-clockwise step as negative.
Finally, when $N$ is even, the clockwise and counter-clockwise steps directly across the circle will both have length $N/2$ with opposite signs. 
This ambiguity corresponds exactly to the $\pm 180^\circ$ ambiguity when measuring angles. 
To avoid this issue, we will simply define a step directly across the circle as positive $N/2$. 

\begin{figure}[hbt]
	\centering
		\includegraphics[width=0.6\textwidth]{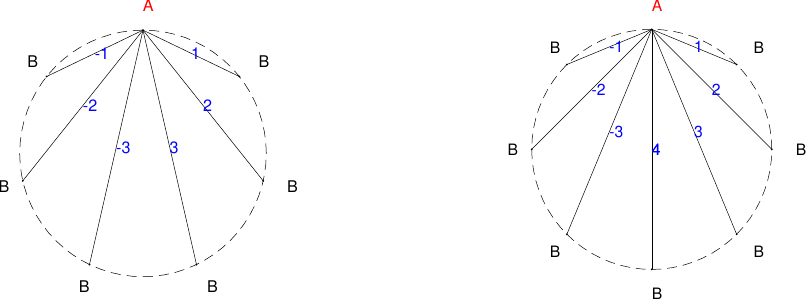}
	\caption{All possible steps from the red ``A'' node for $N=7$ (left) and $N=8$ (right).}
	\label{fig:stepSizes}
\end{figure}

\begin{figure}[hbt]
	\centering
		\includegraphics[width=0.6\textwidth]{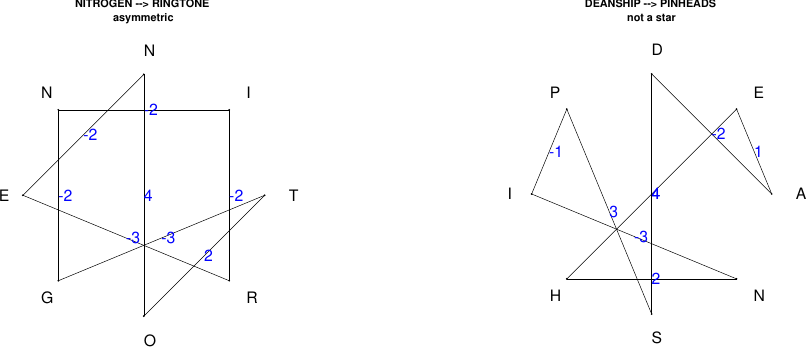}
	\caption{Two example anagrams with the steps $\{s_n\}_{n=0}^{N-1}$ labeled in blue.}
	\label{fig:stepExamples}
\end{figure}

Putting all of these definitions together, we arrive at a unique set of $N-1$ possible steps $s$ from a given node that satisfy $-N/2 < s \le N/2$, with $s=N/2$ only allowed for even $N$ and $s \ne 0$.
For a path $\vec{p}$ with path differences $\vec{d}$ we can compute the vector of corresponding steps $\vec{s} \in \Int^N$ with entries $\{s_n\}_{n=0}^{N-1}$ as
\begin{align}
\label{eq:s}
s_n &=
	\begin{cases}
	d_n, & |d_n| < \frac{N}{2}\\
	\frac{N}{2}, & |d_n| =  \frac{N}{2}\\
	d_n - N, & d_n >  \frac{N}{2}\\
	d_n + N, & d_n <  -\frac{N}{2}.
	\end{cases} 
\end{align}
Figure \ref{fig:stepSizes} illustrates these steps from the top node of the circle for $N=7$ and $N=8$.  
As an example, for the $N=8$ asymmetric star \anagram{NITROGEN}{RINGTONE} we obtain
\begin{align*}
\begin{tabular}{lllrrrrrrrrl}
$\vec{p}$ & = & {[} & 3  & 1  & 7  & 5  & 2 & 4  & 0  & 6  & {]} \\
$\vec{d}$ & = & {[} & -2 & 6  & -2 & -3 & 2 & -4 & 6  & -3 & {]} \\
$\vec{s}$ & = & {[} & -2 & -2 & -2 & -3 & 2 & 4  & -2 & -3 & {].}
\end{tabular}
\end{align*}
This star is shown on the left of Figure \ref{fig:stepExamples} with the steps labeled. 

\begin{lemma}[Steps]
The steps $\{s_n\}_{n=0}^{N-1}$ for a path $\vec{p} \in \Int^N$ satisfy
\begin{align}
	\label{eq:modSteps}
	p_{n+1} &= \left(p_n + s_n\right) \bmod N.
\end{align}
\end{lemma}

\begin{proof}
Examining \eqref{eq:s}, we see that $s_n \equiv d_n \pmod{N}$.
Notice also that $p_n = p_n \bmod N$. 
Using \eqref{eq:d}, we combine these facts to obtain $p_{n+1} = p_{n+1} \bmod N = (p_n + d_n) \bmod N = (p_n + s_n) \bmod N$.
\end{proof}
\noindent
This relationship will be useful for proving properties of edge lengths in the sequel.

\subsection{Detecting a Star Path}
Our remaining task for this section is to determine if a given path corresponds to a star anagram. 

\begin{theorem}[Star Detection]
A path $\vec{p} \in \Int^N$ is a star anagram path if and only if the path's steps satisfy
$|s_{n}| \ne 1$ for all $n \in \left\{0,1,\ldots N-1\right\}$.
\end{theorem}

\begin{proof}
Recall that an anagram is a star if no letter in the new word retains its original neighbors. 
By \eqref{eq:modSteps}, this condition occurs exactly when no step is to the nearest clockwise neighbor ($s_n = 1$) or the nearest counter-clockwise neighbor ($s_n = -1)$.
Recalling that $d_{N-1} = p_{0} - p_{N-1}$, we see that testing the $N$ steps captures all pairs of possible former neighbors.
\end{proof}
\noindent
Notice that this check is easily performed for each of the $P$ possible paths for each anagram, requiring only a few operations on $N$ scalar values.

%%%%%%%%%%%%%%%%%%%%%%%%%%%%%%%%
\section{Star Anagram Classification}
\label{sec:classification}
Now that we have a reliable method for detecting star anagrams, we turn our attention to classification. 
We start with the simpler test for perfection. 

\subsection{Identifying Perfection}
Recall that a perfect star anagram has all edges of the same length. 
Since the edge lengths are given by the magnitudes of the steps $\vec{s}$, we arrive immediately at our test for perfection.
\begin{theorem}[Perfection Test]
A path $\vec{p} \in \Int^N$ is a perfect star path if and only if the path's steps satisfy
$s_{n} = S$ for all $n \in \left\{0,1,\ldots N-1\right\}$ for a constant $S$ satisfying $|S| = L > 1$.
\end{theorem}

\begin{proof}
By definition, a perfect star path must satisfy $|s_n| = L$ for a constant $L > 1$. We see that the steps must have the same sign, since two consecutive steps with equal magnitude and opposed signs would cause the path to repeat the previous node.
\end{proof}
\noindent

To see this test in action, consider our earlier example \anagram{EARTH}{HATER}. We obtain
\begin{align*}
\begin{tabular}{llrrrrrrl}
$\vec{p}$ & = & {[} & 4  & 1 & 3  & 0 & 2 & {]} \\
$\vec{d}$ & = & {[} & -3 & 2 & -3 & 2 & 2 & {]} \\
$\vec{s}$ & = & {[} & 2  & 2 & 2  & 2 & 2 & {]}.
\end{tabular}
\end{align*}
This star, like all length $5$ stars, is a perfect pentagram with $L = 2$. 
Indeed, this characteristic shape was the origin of the name star anagram. 

\subsection{Symmetry and Edges}
Finally, our last task is to check if an anagram that is known to be a non-perfect star is symmetric. 
The key will again be to leverage our steps $\vec{s}$. 
However, we will find it convenient to work with a representation of the steps that makes the $2$ edges associated with each node easier to analyze. 

We define the matrix of edges $\vec{E} \in \Int^{2 \times N}$ such that $e_{1n}$ and $e_{2n}$ are the two steps taken from  the letter labeled $n$ to the two letters connected to that node along the path. 
One of these steps is part of the path and corresponds exactly to $s_k$ where $k$ is the index for which $p_{k} = n$. 
The other is a step backwards along the path and will be $-s_{k-1}$ for the same $k$, where we define $s_{-1} = s_{N-1}$ for convenience. 
We also sort the entries such that $e_{1n} \le e_{2n}$. 
Finally, we set steps across the circle with magnitude $N/2$ to $0$ in the edge matrix to avoid ambiguities when dealing with negated values in the edge matrix.

An example will likely be helpful for the reader. Consider the non-star anagram \anagram{DEANSHIP}{PINHEADS} shown on the right in Figure \ref{fig:stepExamples}.
We obtain the steps for this anagram as
\begin{align*}
\begin{tabular}{lllrrrrrrrrl}
$\vec{p}$ & = & {[} & 7  & 6  & 3 & 5  & 1 & 2  & 0 & 4 & {]} \\
$\vec{d}$ & = & {[} & -1 & -3 & 2 & -4 & 1 & -2 & 4 & 3 & {]} \\
$\vec{s}$ & = & {[} & -1 & -3 & 2 & 4  & 1 & -2 & 4 & 3 & {],}
\end{tabular}
\end{align*}
where $s_0 = -1$ and $s_4 = 1$ reveal the anagram to be a non-star. 
We obtain the edges $\vec{E}$ for this anagram as
\begin{align*}
	\vec{E} &= 
	\bmat 
		0 & 0 & -2 & 2 & 0 & -2 & -3 & -3\\
		2 & 1 & -1 & 3 & 3 & 0 & 1 & -1\\
	\emat.
\end{align*}

\begin{lemma}
Two paths produce identical shapes if and only if they have identical edge matrices.
\end{lemma}

\begin{proof}
By construction, the edge matrix encodes the connections between nodes in our unicursal polygons. 
Sorting the entries within each column ensures a unique representation regardless of the direction or starting point of the paths.
\end{proof}

We say that two shapes are \emph{equivalent} if they are identical up to a rotation in the plane. 
Noticing that rotating a shape does not change the lengths or signs of the edges, we see that rotating a shape in the plane is precisely equivalent to performing a circular shift of the columns in the edge matrix, proving the following.

\begin{lemma}
\label{lem:edgeRotate}
Two shapes are equivalent if and only if their edge matrices are equal up to a circular shift of the columns.
\end{lemma}

\subsection{Identifying Rotational Symmetry}
With the edge matrix $\vec{E}$ defined, we are ready to classify star anagram symmetry. 
We begin with the simpler case of rotational symmetry. 

Consider rotating the shape created by drawing the star anagram's path completely through $360^\circ$. 
If throughout this single rotation, the anagram appears identical to its original position $O_{rot}$ times, then we say that the anagram has rotational symmetry of order $O_{rot}$. 
Clearly, we have that $O_{rot} \le N$.

Recalling Lemma \ref{lem:edgeRotate}, we can test for rotational symmetry by looking for periodicity in the columns of $\vec{E}$. 
This condition can be easily checked by circularly shifting the columns of $\vec{E}$ and testing for equality with the original matrix.
We immediately see that if the smallest circular shift that produces the original matrix $\vec{E}$ is a shift of $K$, then the star anagram's shape will repeat $N/K$ times through a complete rotation. 
Therefore, the anagram has rotational symmetry of order $O_{rot} = N/K$, with $O_{rot} = N/N = 1$ if it lacks rotational symmetry. 
We summarize these results in the following theorem. 

\begin{theorem}[Rotational Symmetry Test]
A star anagram has rotational symmetry $O_{rot} = N/K$, where $K$ is the smallest circular shift that recovers the original edge matrix.
\end{theorem}

Notice that all the columns of $\vec{E}$ are identical for perfect star, yielding $O_{rot} = N$.
Indeed, we note that order $N$ rotational symmetry is an alternative test for perfection of a star anagram, excluding the non-star case with all $|s_n| =1$ that produces a regular $N$-sided polygon. 

\begin{figure}[tbh]
	\centering
		\includegraphics[width=\textwidth]{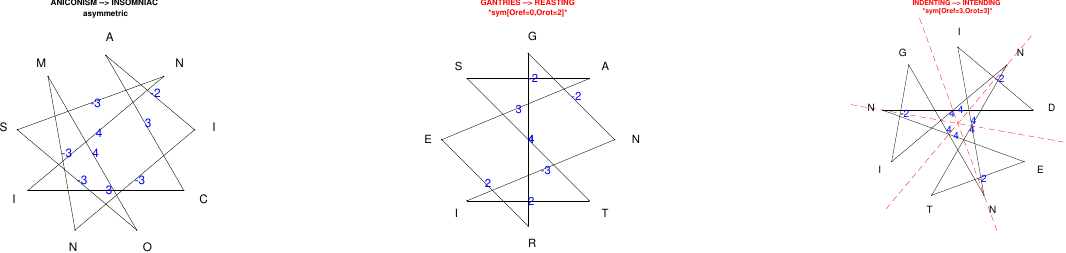}
	\caption{Examples of star anagrams with $O_{rot} = 1$ (left), $O_{rot} = 2$ (middle), and $O_{rot} = 3$ (right) with the $\{s_n\}_{n=0}^{N-1}$ labeled in blue.}
	\label{fig:rotExamples}
\end{figure}

Figure \ref{fig:rotExamples} provides examples of star anagrams with rotational symmetry of order one through three identified with this test. 
Other than perfect stars, we are unaware of a star anagram in English with $O_{rot} > 3$.  
Note that \anagram{INDENTING}{INTENDING} also has reflective symmetry, which we will discuss next.

\subsection{Identifying Reflective Symmetry}
A star anagram is said to be reflectively symmetric if the anagram's shape can be symmetrically folded across one or more dividing lines passing through the shape's center. 
We define the number of such lines to be the order of reflective symmetry $O_{ref}$.
An asymmetric star anagram thus has $O_{ref} = 0$. 

\begin{figure}[tbh]
	\centering
		\includegraphics[width=\textwidth]{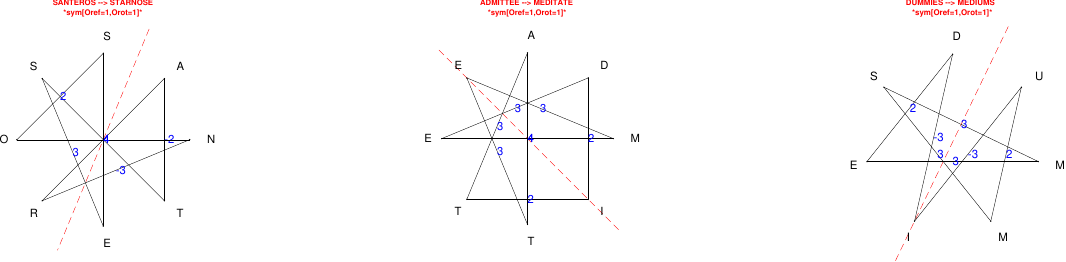}
	\caption{Examples of the $3$ possible reflective symmetry conditions with the steps $\{s_n\}_{n=0}^{N-1}$ labeled in blue. (Left) $N$ even, between nodes. (Middle) $N$ even, through nodes. (Right) $N$ odd, passing through one node.}
	\label{fig:symExamples}
\end{figure}

Such a line of symmetry can either pass through two nodes on the circle, pass between two nodes as it enters/leaves the circle, or pass through a node on one side and between two nodes on the other. 
Considering the geometry of these figures, the first and last cases can only happen when $N$ is even, while the middle case always occurs when $N$ is odd. 
Figure \ref{fig:symExamples} provides examples of these $3$ conditions to make the ideas concrete. 
Notice that all perfect stars satisfy $O_{ref} = N$, with a line of symmetry passing through every node (or pair of nodes for even $N$) and another $N/2$ lines of symmetry passing between nodes for even $N$.
Similar to the rotational case, order $N$ reflective symmetry is an alternative test for perfection, again excluding the regular polygons produced with all $|s_n| =1$. 

Given the path $\vec{p}$ for a non-perfect star with even $N$, we need to look for lines of symmetry that pass either through or between nodes, as in the first two panels of Figure \ref{fig:symExamples}. 
We start with the more intuitive case of a line of symmetry that passes between nodes.
Examining the figure, the condition we want to test is that paired nodes walking around the circle in opposite directions have edges with equal magnitudes and opposite signs, \ie that are mirror images of each other. 

In terms of the matrix $\vec{E}$, this situation will occur when the columns can be split into two groups with the second group in reverse order and negated. 
Note that the columns must be re-sorted after negation and reversing their order to test for equality.
We refer to columns that can be split into two such groups as a \emph{negated palindrome}. 
We can test our anagram for each of the $N/2$ possible lines of symmetry by circularly shifting $\vec{E}$ so that each of the first $N/2$ nodes are in the zeroth column and checking for this negated symmetry in the columns. 

Testing for the second case in Figure \ref{fig:symExamples} for even $N$ is similar, except we have to handle the two nodes that the line of symmetry passes through differently. 
Rather than appearing as a pair of mirrored nodes, these nodes need to satisfy $e_{1n} = -e_{2n}$ so that the symmetry of the two edges ``departing'' from the node is maintained.
The process for testing a star anagram with $N$ odd is also similar, except each of the possible $N$ lines of symmetry to be checked passes through exactly one node. 
We summarize these arguments with the following theorem.

\begin{theorem}[Reflective Symmetry Test]
A star anagram path has reflective symmetry order $0 \le O_{ref} \le N$ corresponding to the number of lines of reflective symmetry. 
Each possible line of symmetry is a line of reflective symmetry if and only if the edge matrix $\vec{E}$ satisfies $e_{1n} = -e_{2n}$ for each node $n$ intersected by the line of symmetry, with the remaining columns forming a negated palindrome.
\begin{itemize}
\item For $N$ even, there are $N/2$ possible lines of symmetry passing through opposing nodes and $N/2$ possible lines of symmetry passing between pairs of nodes.
\item For $N$ odd, each possible line of symmetry passes through exactly one node.
\end{itemize}
\end{theorem}

\subsection{Symmetry and Multiple Paths}
We consider a star to be symmetric if it has rotational ($O_{rot} > 1$) or reflective ($O_{ref} > 0$) symmetry. 
Recall that we test each of the $P$ possible paths $\{\vec{p}_i\}_{i=0}^{P-1}$ for an anagram.
We still need to address how we choose the path to represent the star, denoted $\vec{p}^* \in \{\vec{p}_i\}_{i=0}^{P-1}$,  when multiple paths for a given non-perfect star anagram are symmetric.

\begin{figure}[tbh]
	\centering
		\includegraphics[width=\textwidth]{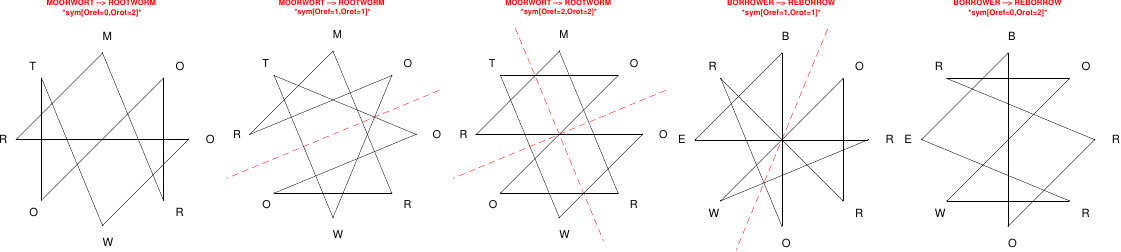}
	\caption{(Left $3$ panes) The $3$ symmetric star paths for the anagram \anagram{MOORWORT}{ROOTWORM}. (Right $2$ panes) The $2$ symmetric star paths for the anagram \anagram{BORROWER}{REBORROW}.}
	\label{fig:moorwort-borrower}
\end{figure}

In most cases, one of the paths will have the sum $O_{rot} + O_{ref}$ greater than the sum for all other paths. 
In these cases, we simply take this strictly more symmetric, or \emph{dominant}, path for the anagram as our selection $\vec{p}^*$ and report the order of its symmetries. 
An interesting example is the symmetric star \anagram{MOORWORT}{ROOTWORM}.
The $3$ symmetric paths of its $12$ possible paths are shown in Figure \ref{fig:moorwort-borrower}.
The first path is only rotationally symmetric, the second is only reflectively symmetric, and the third path, which we select as $\vec{p}^*$, is both. 
There are also a handful of cases that have no clearly dominant path. 
As an example, consider the symmetric star \anagram{BORROWER}{REBORROW}.
Both of its symmetric paths shown in Figure \ref{fig:moorwort-borrower} satisfy $O_{rot} + O_{ref} = 2$, and we simply select one arbitrarily to represent the anagram.

\subsection{Summary of Star Anagram Detection and Classification}
We briefly summarize our overall approach to star anagram detection and classification.
First, for ease of discussion, we consider ``non-star'' as a fourth class, rather than describing detection as a separate process.
Thus, we consider placing a given anagram into one of four classes: non-star, asymmetric star, symmetric star, and perfect star. 
We also consider the classes to be ordered as given so that we can apply comparison operators between classes when describing our procedure. 

With these ordered classes in mind, we first generate the set of all possible paths for the anagram. 
Next, for each path, we determine the class and symmetry orders $(O_{rot},O_{ref})$ as described in Procedure \ref{pro:classPath}.
Looking across these results, we select the path that gives the ``best'' or greatest class given our ordering above. 
Ties among paths of the same class are broken by maximizing the sum $O_{rot} + O_{ref}$.
For paths with the same class and sum of symmetry orders, we simply select the last one in our enumerated list.
This overall process is summarized in Procedure \ref{pro:classAnagram}. 
 
%%%%%%%%%%%%%%%%%%%%%%%%%%%
%Classify a path
\begin{algorithm}
\caption{Classify a path $\vec{p}$}
\begin{algorithmic} 
\label{pro:classPath}
\REQUIRE $\vec{p}$
\ENSURE $\textrm{class} \in \{\textbf{non-star}, \textbf{asymmetric}, \textbf{symmetric}, \textbf{perfect} \}$, $O_{rot}$, $O_{ref}$ 
\STATE \textbf{Initialize:}
\STATE $N \leftarrow \textrm{length of } \vec{p}$
\STATE $O_{rot} \leftarrow -1$; $O_{ref} \leftarrow -1$   \COMMENT{Not determined}
\STATE
\STATE \textbf{Classify:}
\STATE Compute $\vec{s}$ \COMMENT {Compute steps}
\IF{$|s_n| = 1$  for any $n \in \{0,1,\ldots N-1\}$}
	\STATE $\textrm{class} \leftarrow \textbf{non-star}$
	\RETURN \COMMENT{Not a star. Stop.}
\ENDIF
	
\IF{$|s_n| = L,\quad \forall n \in \left\{0,1,\ldots N-1\right\},$}
	\STATE $\textrm{class} \leftarrow \textbf{perfect}$
	\STATE $O_{rot} \leftarrow N$; $O_{ref} \leftarrow N$
	\RETURN \COMMENT{Perfect star found. Stop.}
\ENDIF

\STATE Compute $\vec{E}$ for $\vec{p}$ \COMMENT{Compute edges}
\STATE Compute $O_{rot},O_{ref}$ for $\vec{E}$ \COMMENT{Test for symmetry}
	
\IF{$O_{rot} + O_{ref} > 1$}
	\STATE $\textrm{class} \leftarrow \textbf{symmetric}$
	\ELSE
	\STATE $\textrm{class} \leftarrow \textbf{asymmetric}$
\ENDIF
\end{algorithmic}
\end{algorithm}
%%%%%%%%%%%%%%%%%%%%%%%%%%%

%%%%%%%%%%%%%%%%%%%%%%%%%%%
%Classify an anagram
\begin{algorithm}
\caption{Classify the anagram $w_1 \rightarrow w_2$}
\begin{algorithmic} 
\label{pro:classAnagram}
\REQUIRE $w_1$, $w_2$
\ENSURE $\textrm{class}^* \in \{\textbf{non-star}, \textbf{asymmetric}, \textbf{symmetric}, \textbf{perfect} \}$, $\vec{p}^*$, $O^*_{rot}$, $O^*_{ref}$ 
\STATE \textbf{Initialize:}
\STATE $\textrm{class}^* \leftarrow  \textbf{non-star}$ \COMMENT{Define: $\textbf{non-star}<\textbf{asymmetric}< \textbf{symmetric}< \textbf{perfect}$}
\STATE $\vec{p}^* \leftarrow \vec{p}_1$
\STATE $O^*_{rot} \leftarrow -1$; $O^*_{ref} \leftarrow -1$   \COMMENT{Not determined}
\STATE Compute all paths $\{\vec{p}_i\}_{i=0}^{P-1}$ for $w_1 \rightarrow w_2$
\STATE

\STATE \textbf{Test each path:}

\FOR {$i=0$ \TO $P-1$} 
	
	\STATE Compute ($\textrm{class}^i, O_{rot}^i, O_{ref}^i$) for $\vec{p}_i$ \COMMENT{Using Procedure \ref{pro:classPath}}
	
	\IF{$\textrm{class}^i > \textrm{class}^*$ \OR ($\textrm{class}^i = \textrm{class}^*$ \AND 
		$O_{rot}^i + O_{ref}^i \ge O_{rot}^* + O_{ref}^*$)}
			\STATE $\textrm{class}^* \leftarrow \textrm{class}^i$
			\STATE $\vec{p}^* \leftarrow \vec{p}^i$
			\STATE $O_{rot}^* \leftarrow O_{rot}^i$;  $O_{ref}^* \leftarrow O_{ref}^i$
	\ENDIF
		
\ENDFOR
\end{algorithmic}
\end{algorithm}
%%%%%%%%%%%%%%%%%%%%%%%%%%%

%%%%%%%%%%%%%%%%%%%%%%%%%%%%%%%%
\section{Remarks on Star Anagrams}
\label{sec:remarks}
Now that we can detect and classify star anagrams, we make some remarks about interesting properties of certain star anagram classes. 

\subsection{Perfect Star Edge Lengths}
Recall that a perfect star is one for which $|s_n| = L$ for all $n$. 
One naturally asks: what edge lengths $L$ are possible for perfect stars with a given word length $N \ge 5$? 

First, we note that $L \ge 2$, since $L = 1$ does not produce a star path. 
Next, recalling the definition of the steps $\vec{s}$, we see that\footnote{We will soon see that $L=N/2$ is not a valid edge length for a perfect star; hence the strict inequality here.} $L < N/2$. 
As noted previously, one can easily show that the steps $s_n $ for a perfect star all have the same sign. 
Since reversing the path for a perfect star will produce the same shape, we can without loss of generality consider only positive valued steps with $s_n = L$ for all $n$. 
Furthermore, since a perfect star path can be rotated to start with any node, we can without loss of generality consider paths that begin with $p_0 = 0$. 

Our task is thus reduced to finding the constant step sizes $2 \le L < N/2$ such that a path with these steps and $p_0 = 0$ forms a perfect star.
Incrementing in constant steps of size $s_n = L$ and recalling equation \ref{eq:modSteps}, we obtain for $1 \le n \le N-1$
\begin{align*}
p_n &= \left(p_{n-1} + L\right) \bmod N = \left(p_0 + \sum_{i=0}^{n-1} L\right) \bmod N = q_n \bmod N,
\end{align*}
where we define $q_n = nL$ for $n = 0, 1, \ldots N-1$. 
Notice that we also have $p_0 = q_0 \bmod N$, since $q_0 = 0$.  
We also state a simplified version of the Chinese Remainder Theorem~\cite[Theorem 4.13]{rosen} that will be useful in our proof.

\begin{theorem}[Simplified Chinese Remainder Theorem]
\label{thm:CRT}
For coprime integers $A,B > 0$, each integer $C \ge 0$ with $C < AB$ can be uniquely identified from the remainders $C \bmod A$ and $C \bmod B$. 
\end{theorem}

With these ideas in mind, we can arrive at the following.

\begin{theorem}[Valid Edge Lengths for Perfect Stars]
\label{thm:starEdges}
A perfect star path with length $N$ and constant edge length $L$ exists if and only if $L$ and $N$ are coprime with $2 \le L < N/2$ and $N \ge 5$.
\end{theorem}

\begin{proof}
First, we show that an $L$ that is not coprime with $N$ cannot be a valid edge length for a perfect star. 
Assume $L$ and $N$ are \emph{not} coprime.
Then there exists integers $a,b \ge 2$ and $c \ge 1$ such that $N = ab$ and $L = ac$.
We see that $p_{b} = bL \bmod N = bac \bmod N = Nc \bmod N = 0$.
Since $b < N - 1$, this implies that the path revisits node $0$ before visiting all the other nodes. 
Thus, this is not a valid path for a star anagram, and $L$ is not a valid edge length for a perfect star.
Note that this also excludes $L = N/2$ as a valid edge length, since $N/2$ and $N$ are not coprime. 

Now assume that $L$ and $N$ are coprime. 
Notice that $q_n < NL$ for all $n$.
Since $N$ and $L$ are coprime, Theorem \ref{thm:CRT} requires that any integer $q_n < NL$ be uniquely identifiable from the values $q_n \bmod L$ and $q_n \bmod N$.
Since, $q_n \bmod L = nL \bmod L = 0$ for all $n$, this requires that $q_n \bmod N$ is distinct for each $n$. 
But these $N$ values between $0$ and $N-1$ are precisely the elements $\{p_n\}_{n=0}^{N-1}$, which shows that this path visits each node exactly once. 
Thus, this path forms a valid perfect star path with $s_n = L$. 
\end{proof}
\noindent We note that this result is already well known in the study of star polygons, \eg \cite[Section 2.8]{coxeter}.

Notice the interesting consequence that there are no perfect stars for words of length $N=6$. 
We note that this result is perhaps ironic given that $6$ is a perfect integer\footnote{The integer $6$ is prefect in the sense that $6 = 1 + 2 + 3$. In other words, $6$ is equal to the sum of its factors.}. 
While perfect stars of any other length are possible, we will see below that perfect star anagrams for large $N$ were not found in our search over the English language. 

\subsection{Commutativity}
We defined an anagram to be an ordered pair.
A natural question arises: if \anagram{$w_0$}{$w_1$} is a star, does it follow that \anagram{$w_1$}{$w_0$} is also a star? 
Put another way, is starriness commutative? 
We shall see that this is indeed the case. 

Let $\vec{p}$ with entries $\{p_n\}_{n=0}^{N-1}$ be a path for \anagram{$w_0$}{$w_1$}. 
We wish to construct the reversed path\footnote{Note that each path for an anagram with repeated letters has a corresponding reversed path.} $\vec{\bar{p}}$ with entries $\{\bar{p}_n\}_{n=0}^{N-1}$ for \anagram{$w_1$}{$w_0$}. 
If we label each letter with its position in both words, we notice that the two sets of values exchange roles as the indices and node locations for the path and its reversed path.
An example may be helpful;
consider our familiar $N=5$ star with the paths labeled in both directions
\begin{align*}
\begin{tabular}{ccccccccccc}
 &  &  &  &  &   & $p_0$ & $p_1$ & $p_2$ & $p_3$  & $p_4$ \\
0 & 1 & 2 & 3 & 4 & $\rightarrow$  & 4 & 1 & 3 & 0 & 2 \\
E & A & R & T & H & & H & A & T & E & R \\
3 & 1 & 4 & 2 & 0 & $\leftarrow$  & 0 & 1 & 2 & 3 & 4\\
$\bar{p}_0$ & $\bar{p}_1$ & $\bar{p}_2$ & $\bar{p}_3$  & $\bar{p}_4$. &  &  &  &  &  & 
\end{tabular}
\end{align*}
Notice that the pair of values above and below each letter remains the same in both words. 
In general, we see that the two paths satisfy
\begin{align}
p_{\bar{p}_n} &= n \label{eq:pbar}\\
\bar{p}_{p_n} &= n \label{eq:pbar2}
\end{align} 
for $n = 0,1,2,\ldots N-1$. 
This relationship will be useful shortly when we examine perfect stars. 
We first state a result on general star anagrams. 

\begin{theorem}[Starriness Commutes]
If \anagram{$w_0$}{$w_1$}  is a star anagram with star path $\vec{p}$, then \anagram{$w_1$}{$w_0$}  is a star anagram with the reversed path $\vec{\bar{p}}$. 
\end{theorem}

\begin{proof}
Suppose that $\vec{\bar{p}}$ is not a star path. 
This requires a pair of letters that are adjacent in $w_0$ to also be adjacent in $w_1$.
However, this contradicts the starriness of $\vec{p}$, and we conclude that $\vec{\bar{p}}$ is a star path for \anagram{$w_1$}{$w_0$}. 
\end{proof}

\begin{figure}[hbt]
	\centering
		\includegraphics[width=0.6\textwidth]{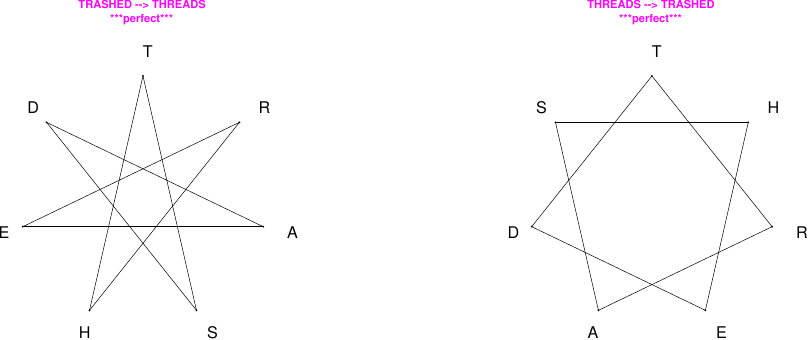}
	\caption{An example perfect star with $S=-3$ (left) and the corresponding reversed perfect star (right) with $\bar{S} = 2$. As expected from Theorem \ref{thm:perfectionCommutes}, $\bar{S}S \bmod N = 2(-3) \bmod 7 = 1$.}
	\label{fig:threads}
\end{figure}

Next we turn our attention to perfection, which also commutes. 
However, the edge length of the reversed star is often not the same.
An example is provided in Figure \ref{fig:threads} with $L=3$ for the perfect star and $L=2$ for the reversed perfect star.  
The following theorem provides the exact relationship between the edge lengths in terms of the possibly negative constant step size $S$, recalling that $L = |S|$ for a perfect star.  

\begin{theorem}[Perfection Commutes]
\label{thm:perfectionCommutes}
If \anagram{$w_0$}{$w_1$}  is a length $N$ perfect star anagram with path $\vec{p}$ and constant step size $S$, 
then \anagram{$w_1$}{$w_0$} is a prefect star with the reversed path $\vec{\bar{p}}$ and step size 
$\bar{S}$ satisfying $\bar{S}S \bmod N = 1$.
\end{theorem}

\begin{proof}
From (\ref{eq:modSteps}), $\vec{p}$ satisfies 
$p_{n} = (p_{0} + nS) \bmod N$ for $n = 0, 1, 2, \ldots N-1$. 
Since this a star path, there must exist a unique $2 \le K < N$ such that 
$p_K = (p_{0} + KS) \bmod N = (1 + p_{0}) \bmod N$, \ie the path must visit the node $1$ greater than $p_0$ exactly once and not on the step $p_1$.
Adding the integer $p_{n} - p_{0}$ to both sides, we obtain for each $n$ and our constant $K$ the relationship
$(1 + p_{n}) \bmod N = (p_{n} + KS) \bmod N = p_{\left(n + K \bmod N\right)}$,
where the last equality follows by noticing that $p_0 = (p_0 + NS) \bmod N$ and the properties of modular arithmetic. 
Note that $p_n = 0$ for some $n$, and hence $KS \bmod N = 1$.

Consider now the path differences for the reversed path $\{\bar{d}_n\}_{n=0}^{N-1}$.
By definition we have 
$\bar{d}_n = \bar{p}_{n+1} - \bar{p}_{n}$. 
Recalling that we defined $p_N = p_0$ for notational convenience, we can more precisely write this as
$\bar{d}_n = \bar{p}_{\left(n+1\bmod N\right)} - \bar{p}_{n}$. 
For a given $n$, recalling equation \ref{eq:pbar}, there is a unique $n^*$ such that $p_{n^*} = n$.
From our results above, we have that
$(n+1) \bmod N  = (p_{n^*} + 1) \bmod N = p_{\left(n^* + K \bmod N\right)}$.
Putting these together and applying (\ref{eq:pbar2}), this allows us to obtain
$\bar{d}_n = \bar{p}_{p_{\left(n^* + K \bmod N\right)}} - \bar{p}_{p_{n^*}} = (n^* + K) \bmod N - n^*$.
Since $0 \le n^*,K < N$, we can easily see that $\bar{d}_n \in \{K,K-N\}$ for $n=0,1,2,\ldots N-1$. 

Recalling the definition of $\bar{s}_n$ from (\ref{eq:s}), we can see that $\bar{s}_n = K$ for all $n$ if $K \le N/2$ or $\bar{s}_n = K-N$ for all $n$ if $K > N/2$. 
Thus, $\vec{\bar{p}}$ has a constant step step size $\bar{S} \in \{K,K-N\}$ and is a perfect star path.
Recalling $KS \bmod N = 1$, and adding $-NS$ if $\bar{S} = K - N$, we have $\bar{S}S \bmod N = 1$.
\end{proof}

We note that $\bar{S}$ is the \emph{modular inverse} of $S$ in the parlance of number theory, which exists uniquely modulo $N$ precisely when $N$ and $S$ are coprime~\cite[Section 4.2]{rosen}. 
Our perfect step size $S$ is of course coprime with $N$ by Theorem \ref{thm:starEdges}. 

\begin{figure}[hbt]
	\centering
		\includegraphics[width=0.6\textwidth]{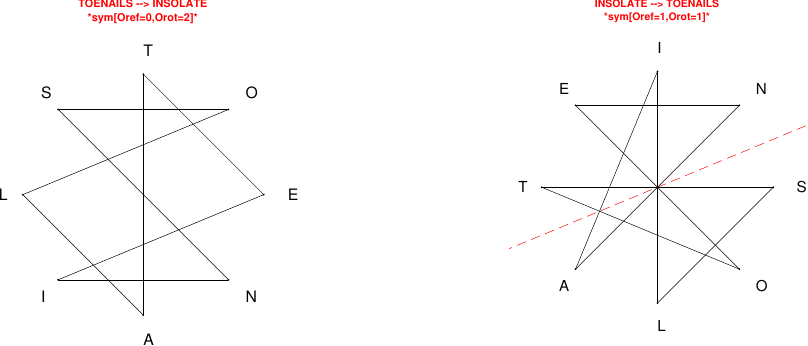}
	\caption{An example symmetric star with a reversed path that differs in both symmetry orders.}
	\label{fig:toenails}
\end{figure}

Finally, we address the reversed paths of symmetric stars. It is not the case that reversing the path preserves rotational or reflective symmetry.
A counter example is the symmetric star \anagram{TOENAILS}{INSOLATE} shown in Figure \ref{fig:toenails}.
However, we have verified by brute force that the reversed star of every symmetric star has some form of symmetry, which suggests the conjecture that general symmetry commutes. 
We defer a rigorous analysis and possible proof of this conjecture to future work. 

\subsection{Star Clusters}
Recall that two shapes are equivalent if they are identical after a rotation in the plane. 
We will refer to a set of star anagrams with equivalent polygons as a \emph{star cluster}. 
From Theorem \ref{thm:starEdges}, we know that the number of  possible perfect star polygons for length $N$ is simply the number of integers less than $N/2$ that are coprime with $N$.
The total number of possible star polygons was calculated in \cite{gordon}, specializing earlier work in \cite{golomb} on the enumeration of polygons in general. 

We provide the number of possible star polygons of each class for several lengths $N$ in Table \ref{tab:shapeCounts}. 
The counts in the table were computed by brute force enumeration and match the theoretical results for the perfect and total columns.
We are not aware of a closed form result for the number of possible symmetric star polygons and leave such a result for future work. 

\begin{table}[h]
\centering
\caption{Number of possible unique star polygons, up to a rotation, of each class for several word lengths $N$. The perfect column is consistent with Theorem \ref{thm:starEdges}, and the total column matches the results in \cite{gordon}.}

\begin{tabular}{crrrr}
\hline
\multicolumn{1}{}{} & \multicolumn{4}{c}{Unique star polygons up to a rotation}                                                                                         \\ \hline
$N$                    & \multicolumn{1}{c}{Asymmetric} & \multicolumn{1}{c}{Symmetric} & \multicolumn{1}{c}{Perfect} & \multicolumn{1}{c}{Total} \\ \hline
5                      & \multicolumn{1}{r}{0}          & \multicolumn{1}{r}{0}         & \multicolumn{1}{r}{1}       & 1                          \\ \hline
6                      & \multicolumn{1}{r}{0}          & \multicolumn{1}{r}{1}         & \multicolumn{1}{r}{0}       & 1                          \\ \hline
7                      & \multicolumn{1}{r}{0}          & \multicolumn{1}{r}{3}         & \multicolumn{1}{r}{2}       & 5                          \\ \hline
8                      & \multicolumn{1}{r}{12}         & \multicolumn{1}{r}{14}        & \multicolumn{1}{r}{1}       & 27                         \\ \hline
9                      & \multicolumn{1}{r}{126}        & \multicolumn{1}{r}{47}        & \multicolumn{1}{r}{2}       & 175                        \\ \hline
10                     & \multicolumn{1}{r}{1354}       & \multicolumn{1}{r}{178}       & \multicolumn{1}{r}{1}       & 1533                       \\ \hline
\end{tabular}

\label{tab:shapeCounts}
\end{table}

\subsection{Autostars}
For a word with no repeated letters, the only path that produces the original word will be a regular polygon, making a simple trip around the circle. 
This does not meet our definition of an anagram, since the letters are not re-arranged. 
We might denote this trivial case as a \emph{self anagram}. 
However, a word with repeated letters admits multiple paths around the circle that produce the original word. 
It turns out that some words with repeated letters are thus star anagrams of themselves. 
We refer to these words as \emph{autostars.} 
In the following section, we will see that autostars are quite common in English and provide interesting star shapes not observed among normal star anagrams. 

%%%%%%%%%%%%%%%%%%%%%%%%%%%%%%%%
\section{Star Search}
\label{sec:search}
We now turn our attention toward detecting and classifying all star anagrams in a set of $501,603$ English words\footnote{We used the Spell Checker Oriented Word List (SCOWL) \cite{scowl} Version 2020.12.07 with United States spelling including British variations, ``insane'' (95) size, common spelling variants to level $3$, and included ``hacker'' words in the list. 
We also note that accent marks were suppressed for anagram search purposes. }. 
Particularly for large $N$, this extensive word list includes many jargon and highly obscure words.

Our code first scans this word list and produces an exhaustive enumeration of all possible anagrams, yielding $108,716$ total anagrams.
Each anagram is then processed using the methods in Section \ref{sec:classification} to classify it as a non-star, asymmetric star, symmetric star, or perfect star.
The resulting collection of star anagrams is organized first by length $N$, then by class, and finally by shape into star clusters. 
This process produced a total of $6,204$ star anagrams ranging in length from $N=5$ to $N=21$. 
These results included $1,614$ asymmetric stars,  $3,335$ symmetric stars, and $1,255$ perfect stars. 
Thus, about $5.7\%$ of English anagrams are stars. 

\begin{figure}[hbt]
	\centering
		\includegraphics[width=0.5\textwidth]{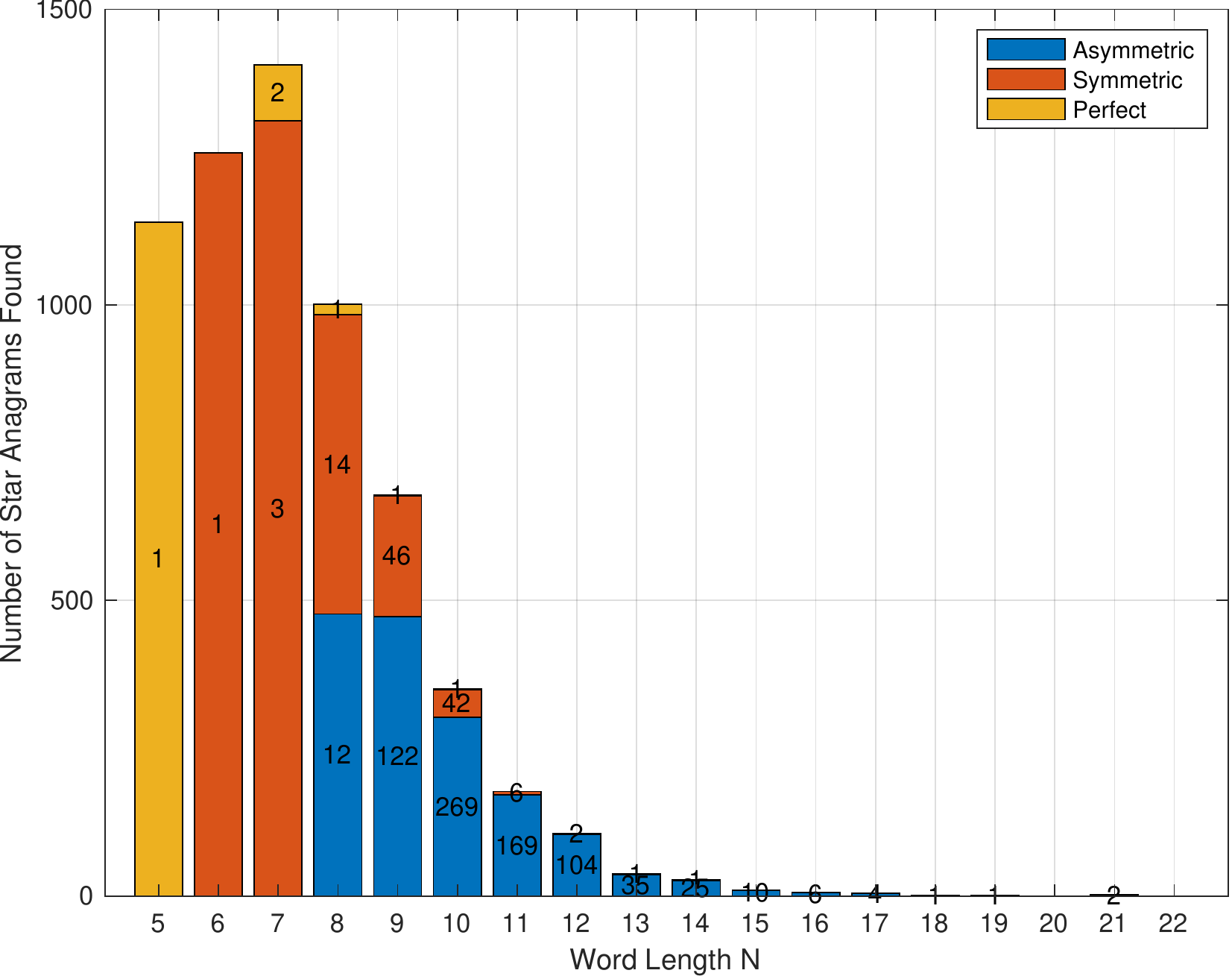}
	\caption{The number of each class of star anagram found for each value of word length $N$. The black number on each bar indicates how many unique shapes were found for that class and word length.}
	\label{fig:frequency}
\end{figure}

Figure \ref{fig:frequency} summarizes the results.
Each vertical bar shows the number of stars found for a given value of $N$, broken into up to $3$ pieces to show the split between asymmetric, symmetric, and perfect stars for that word length.
Only lengths $8$, $9$, and $10$ have stars from all three classes, and hence only those bars are split into $3$ pieces. 
The black number on each bar piece indicates the number of star clusters found for that class and word length. 
Comparison with Table \ref{tab:shapeCounts} reveals that all possible shapes are only observed for $N < 9$. 

\begin{figure}[tbh]
	\centering
		\includegraphics[width=\textwidth]{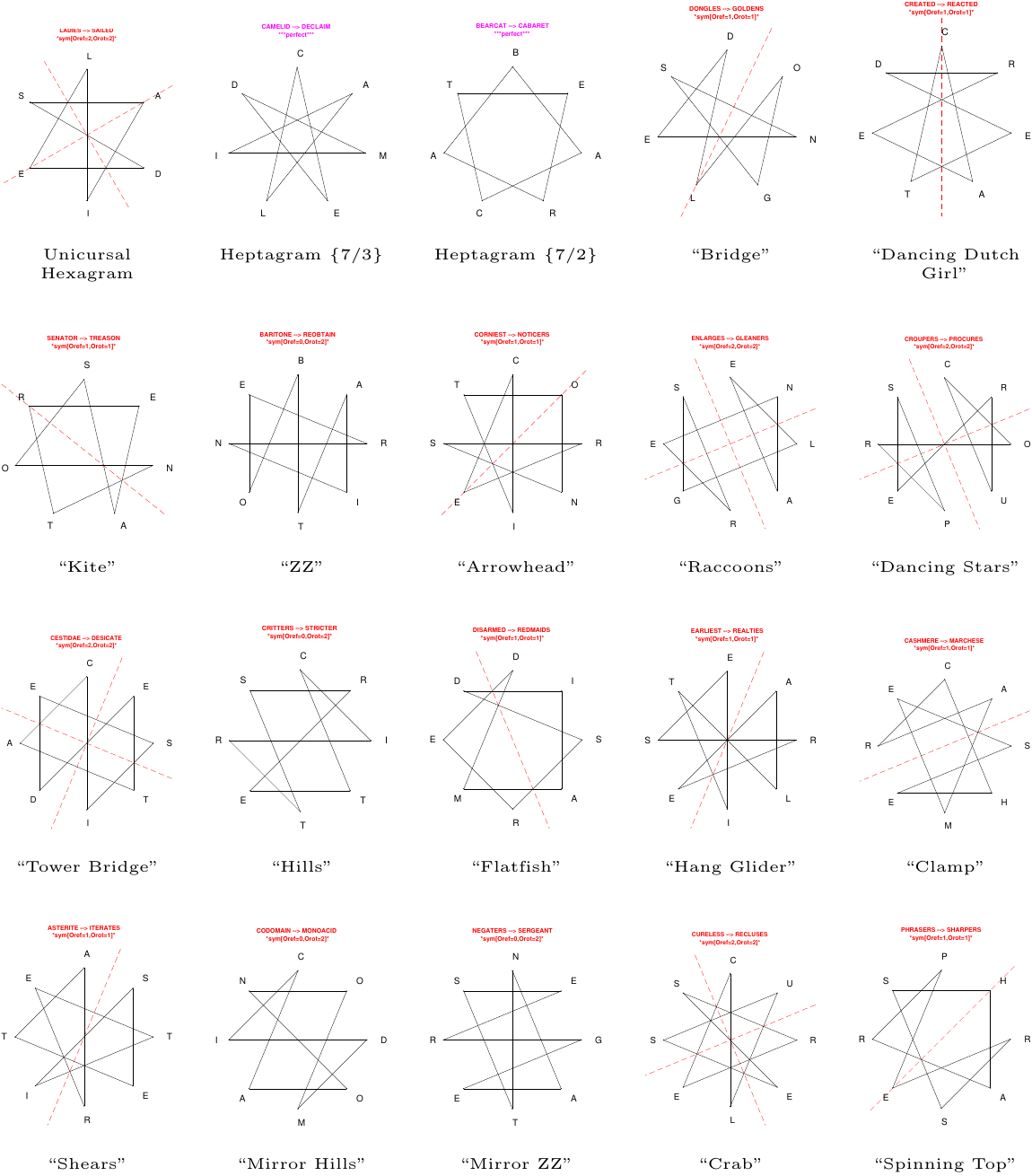}
	\caption{Examples of $20$ of the $21$ unique shapes for symmetric and perfect stars of length $N \in \{6,7,8\}$ along with our nicknames or common names for these shapes. See Figure \ref{fig:classExamples} for an example Octagram \{8/3\}.}
	\label{fig:shapes678}
\end{figure}

\begin{figure}[tbh]
	\centering
		\includegraphics[width=\textwidth]{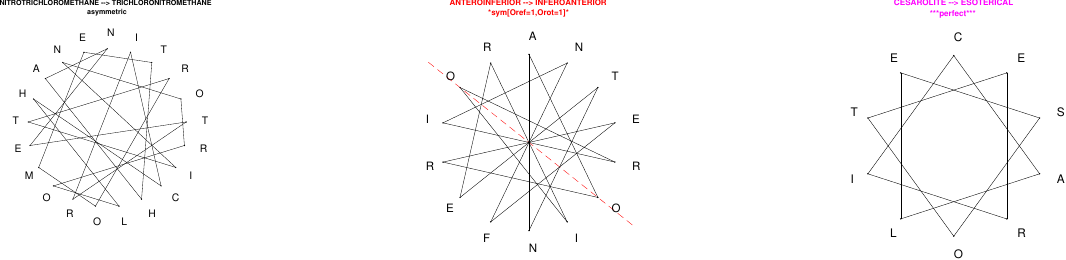}
	\caption{Examples of the longest stars found for each class. (Left) Asymmetric star with $N=21$. (Middle) Symmetric star with $N=14$. (Right) Perfect star with $N=10$.}
	\label{fig:longStars}
\end{figure}

We provide examples from each of the symmetric and perfect clusters for $N \in \{6,7,8\}$ in Figure \ref{fig:shapes678}. 
We also include either common names for these shapes or our own nicknames.
The perfect stars are denoted with their key parameters written as $\{N/L\}$, a convention known as \emph{Schläfli notation}~\cite[Section 2.8]{coxeter}. 
Examples of the longest stars found for each class are shown in Figure \ref{fig:longStars}.
Star anagrams become exceedingly rare for large $N$. 
Indeed, we found only $25$ stars with $N > 14$. 
A complete set of figures depicting all the star anagrams found in our search is provided in Appendix \ref{ap:library}.

\subsection{Autostar Results}
Words that are autostars are surprisingly common, with $9,948$ examples found in our search over the English language. 
For very long words with repeated letters, the number of possible paths to search becomes computationally prohibitive. 
For example, the word SUPERCALIFRAGILISTICEXPIALIDOCIOUS has $209,018,880$ possible autostar paths. 
To avoid this problem, we excluded words with more than $3,000,000$ possible paths from our search for autostars. 
Only $21$ words were eliminated by this criteria from our list of $501,603$ English words. 

\begin{figure}[tbh]
	\centering
		\includegraphics[width=\textwidth]{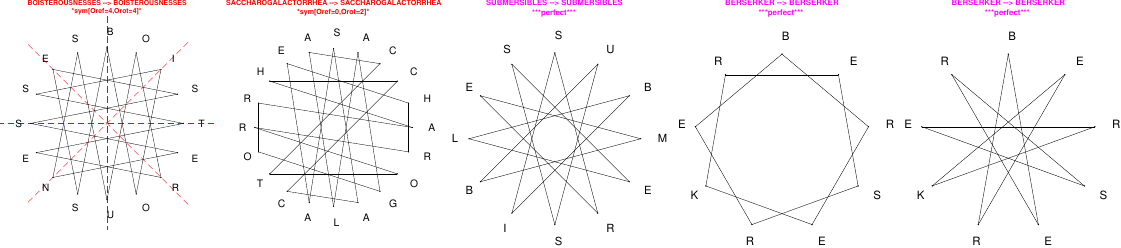}
	\caption{Example autostar results including a symmetric autostar with $O_{ref} = 4$, a symmetric autostar with $N=20$, a perfect autostar with $N=12$, and a perfect autostar with two possible edge lengths.}
	\label{fig:autostars}
\end{figure}

Autostars can of course be symmetric. We found $3,542$ symmetric autostars. 
Two interesting examples are shown in Figure \ref{fig:autostars}. 
The first example in the figure is particularly interesting, since it has rotational symmetry of order $4$ without being perfect. 
We are not aware of any regular stars with this property.
The longest symmetric autostar we found has length $N=20$ and appears in the Figure's second pane.
We also found perfect autostars with length $12$, exceeding the maximum length $10$ we found among regular stars.
The third pane of Figure \ref{fig:autostars} provides an example.
Finally, the last two panes of Figure \ref{fig:autostars} show two perfect paths for the autostar \anagram{BERSERKER}{BERSERKER} with different edge lengths $L$. We did not find such an example among regular stars. 

%%%%%%%%%%%%%%%%%%%%%%%%%%%%%%%%
\section{Conclusions}
\label{sec:conclusions}
We have developed an efficient algorithmic approach for detecting and classifying star anagrams and proved several interesting properties relating to their edge lengths and reversability.
While we proved that starriness and perfection are preserved by reversing a star anagram, we leave for possible future work a proof that symmetry is preserved after reversal. 

Results from a detailed numerical study of all star anagrams in a large database of English words were also presented, including geometrically interesting examples of autostars with properties not found among normal star anagrams. 
Results on the clustering of star anagrams into common shapes were also provided. 
Future work could include a rigorous analysis of the number of possible shapes of symmetric stars.
Finally, a numerical study of star anagrams in other languages, including those with different alphabets, might yield interesting geometric results with different distributions across symmetry orders and word lengths than those found in English.

%%%%%%%%%%%%%%%%%%%%%%%%%%%%%%%%
%%%%%%%%%%%%%%%%%%%%%%%%%%%%%%%%
%%%%%%%%%%%%%%%%%%%%%%%%%%%%%%%%
\section*{Declaration of competing interest}
The authors declare that they have no known competing financial interests or personal relationships that could have appeared to influence the work reported in this paper.

\section*{Data availability}
Data is publicly available.

\section*{Acknowledgment}
The authors conducted this work independently without sponsorship. 
The views in this article are not those of the U.S. government, the U.S. Air Force Research Laboratory, or the Freedom From Religion Foundation. 

\bibliographystyle{elsarticle-num} 
\bibliography{starAnagrams}

%The library
\appendix
\clearpage
\section{Star Anagram Library}
\label{ap:library}
The following sections provide a complete enumeration of all star anagrams found in the SCOWL word list using the detection and classification algorithms described above. 

%%%%%%%%%%%%%%%%%%
\clearpage
\subsection{Star Anagrams $N = 21$}
For $N=21$, only two asymmetric star anagrams were detected.

\begin{figure}[H]
\centering
\begin{subfigure}[T]{0.19\textwidth}
\centering
\includegraphics[width=\textwidth]{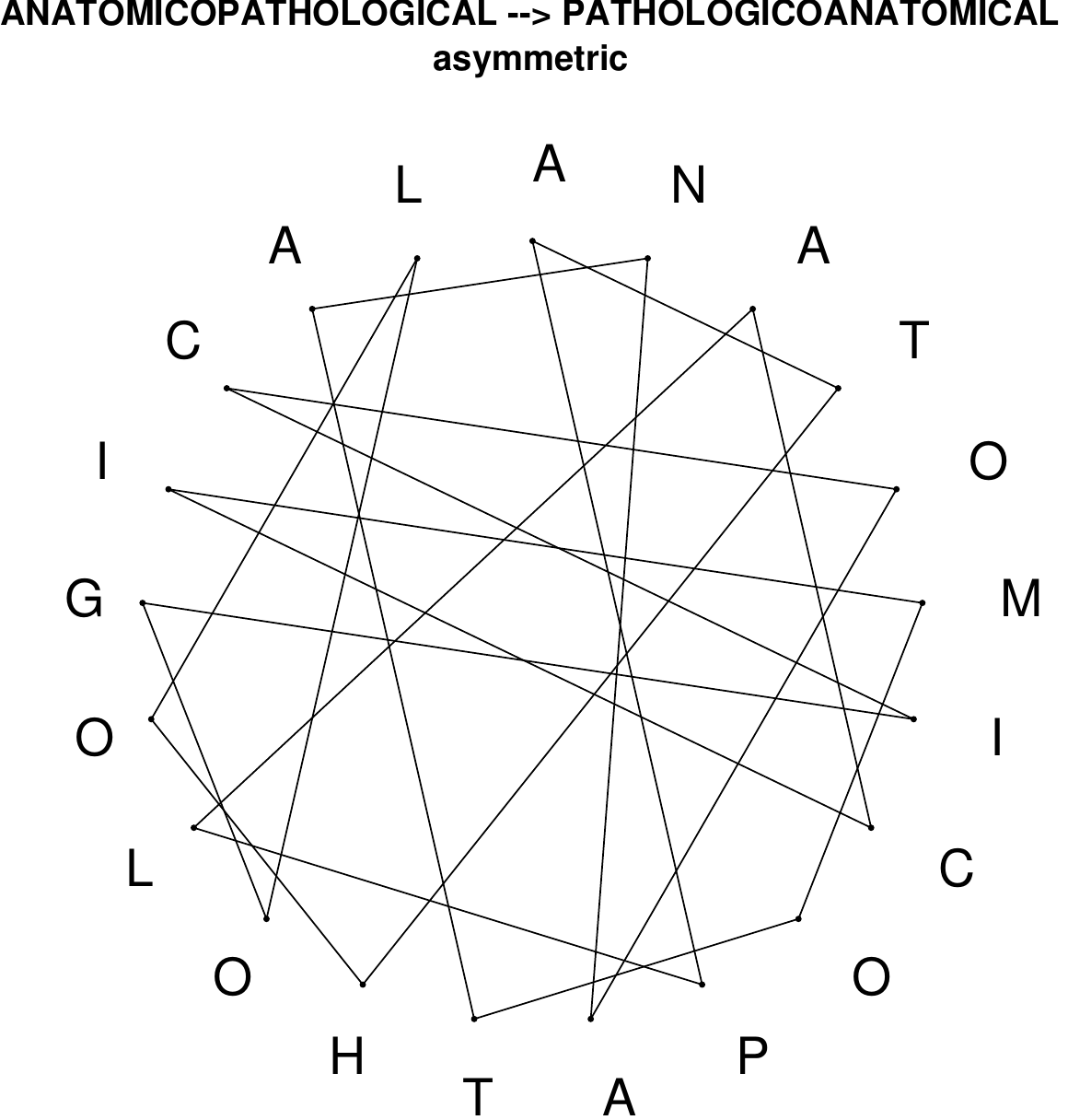}
\end{subfigure}
\hfill
\begin{subfigure}[T]{0.19\textwidth}
\centering
\includegraphics[width=\textwidth]{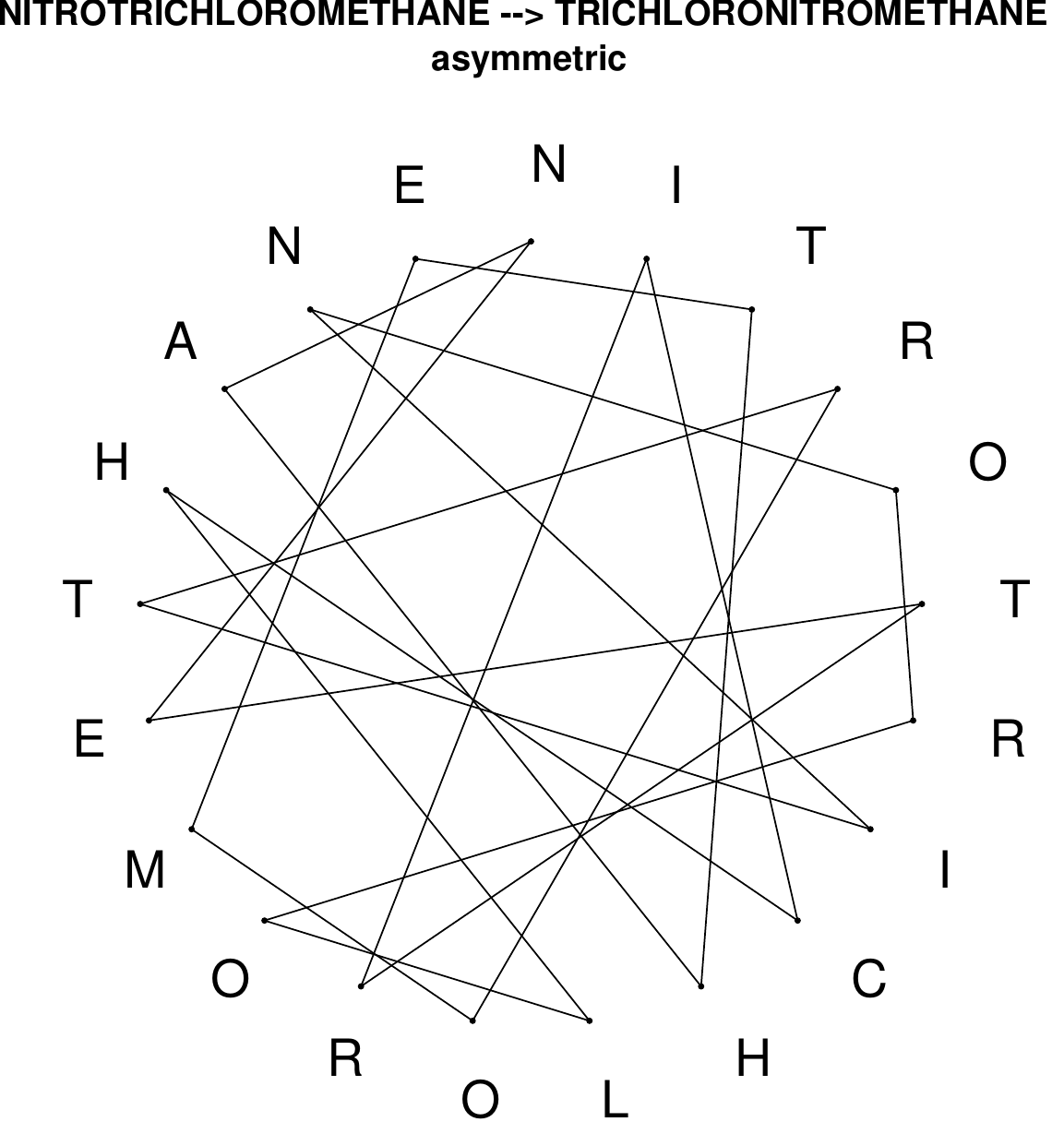}
\end{subfigure}
\hfill
\end{figure}

%%%%%%%%%%%%%%%%%%
\clearpage
\subsection{Star Anagrams $N = 20$}
For $N=20$, we did not find any star anagrams.

%%%%%%%%%%%%%%%%%%
\clearpage
\subsection{Star Anagrams $N = 19$}
For $N=19$, we found a single asymmetric star anagram.

\begin{figure}[H]
\centering
\begin{subfigure}[T]{0.19\textwidth}
\centering
\includegraphics[width=\textwidth]{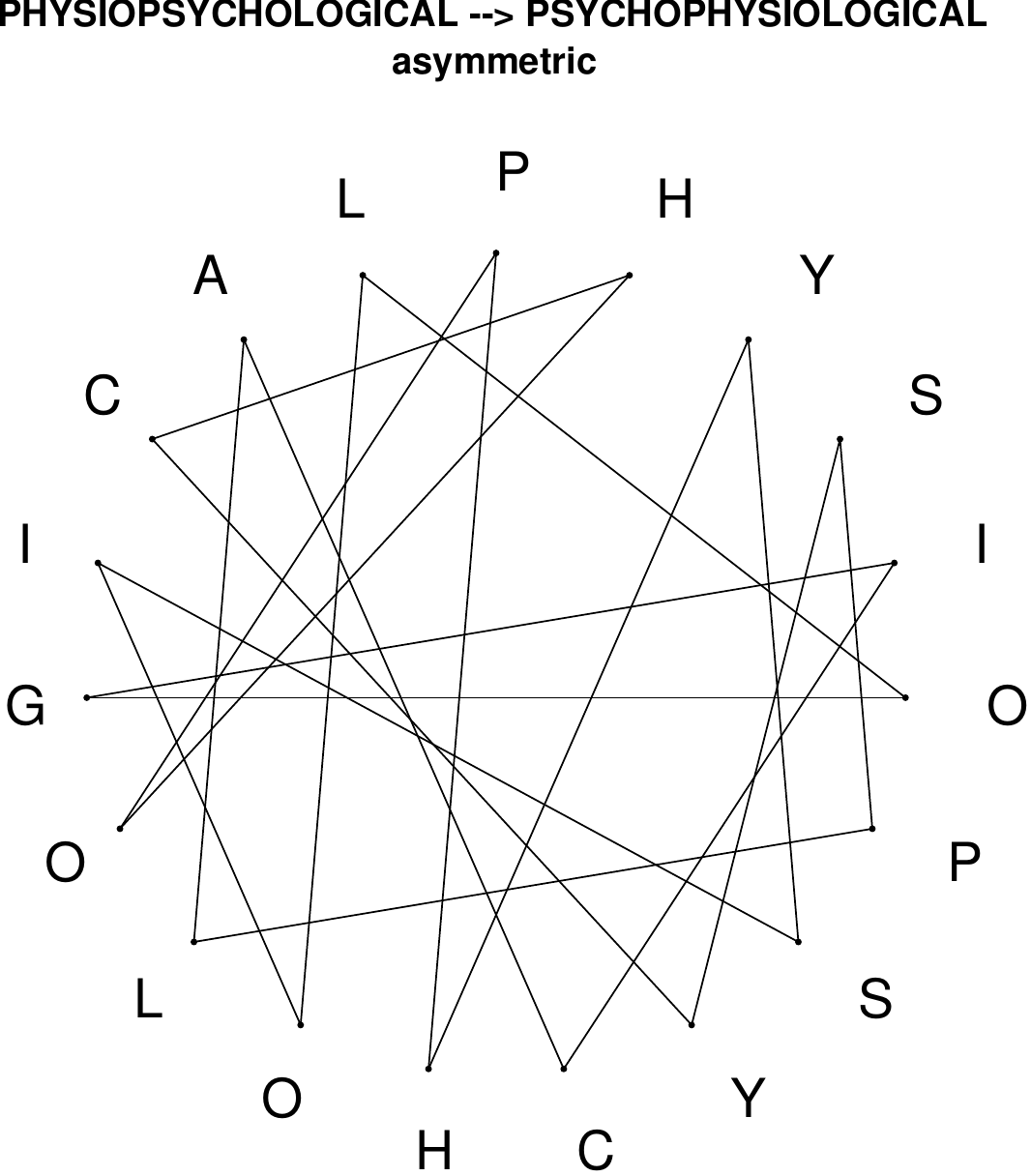}
\end{subfigure}
\hfill
\end{figure}

%%%%%%%%%%%%%%%%%%
\clearpage
\subsection{Star Anagrams $N = 18$}
For $N=18$, we found a single asymmetric star anagram.

\begin{figure}[H]
\centering
\begin{subfigure}[T]{0.19\textwidth}
\centering
\includegraphics[width=\textwidth]{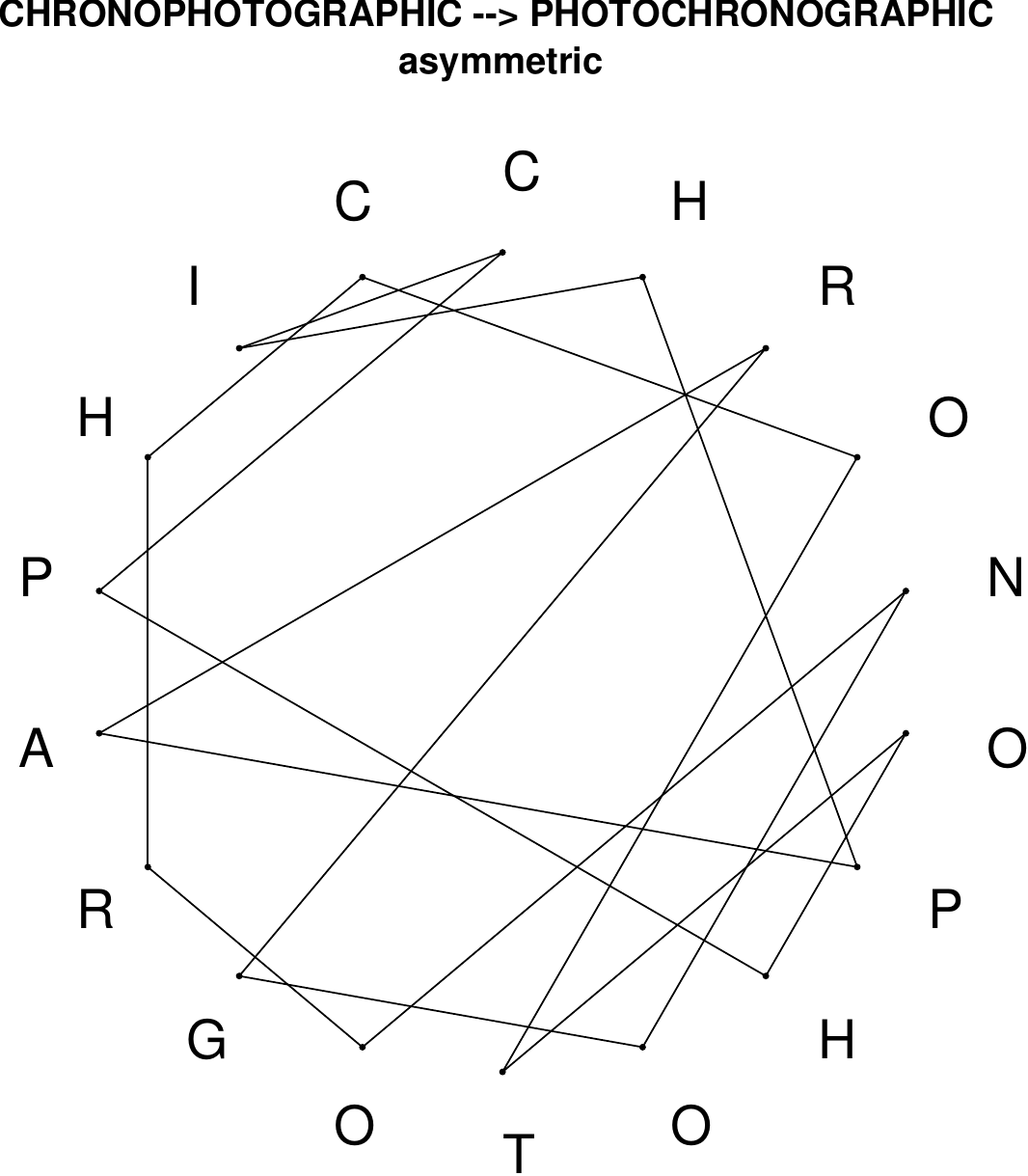}
\end{subfigure}
\hfill
\end{figure}

%%%%%%%%%%%%%%%%%%
\clearpage
\subsection{Star Anagrams $N = 17$}
For $N=17$, we found a handful of asymmetric star anagrams.

\begin{figure}[H]
\centering
\begin{subfigure}[T]{0.19\textwidth}
\centering
\includegraphics[width=\textwidth]{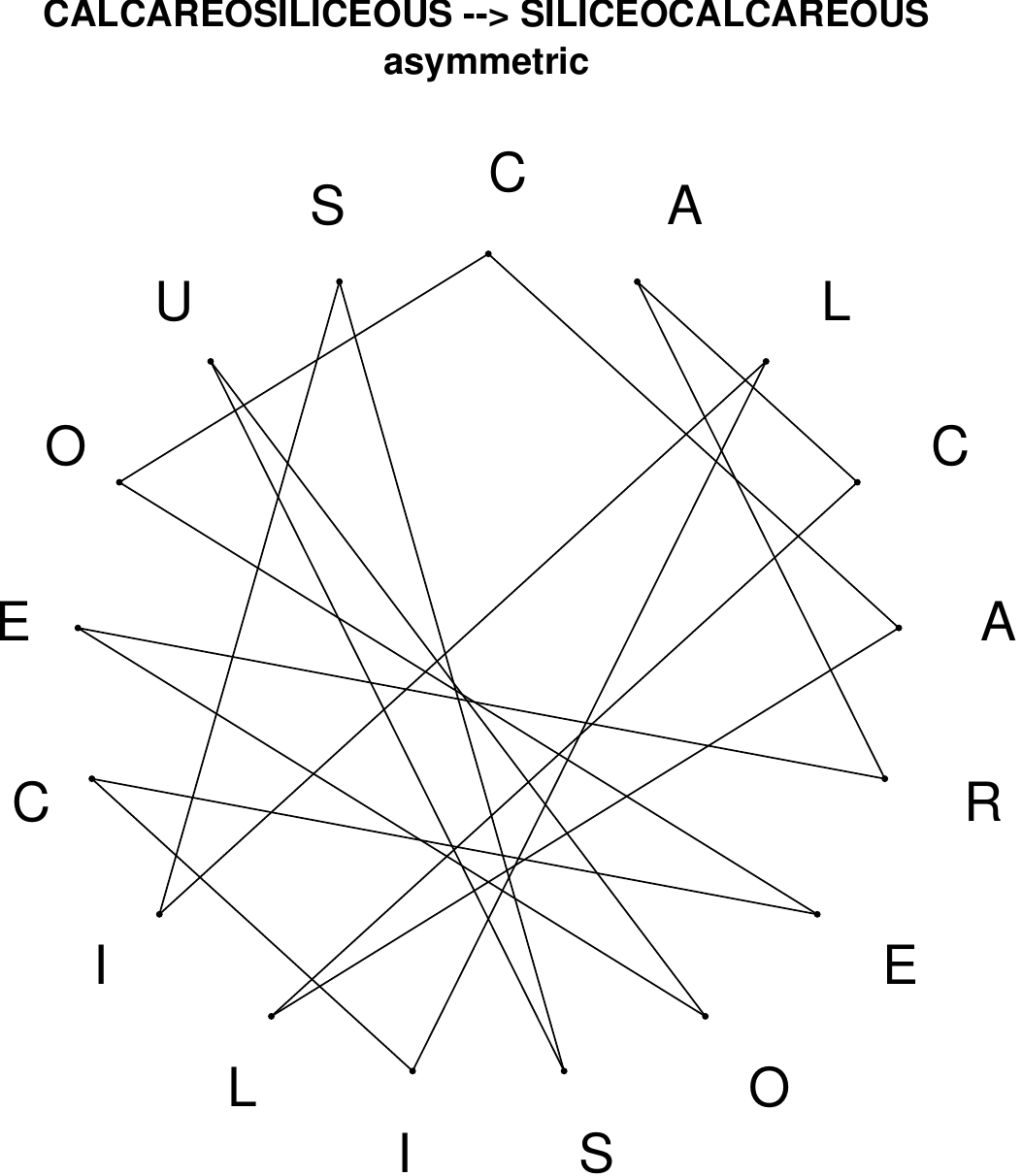}
\end{subfigure}
\hfill
\begin{subfigure}[T]{0.19\textwidth}
\centering
\includegraphics[width=\textwidth]{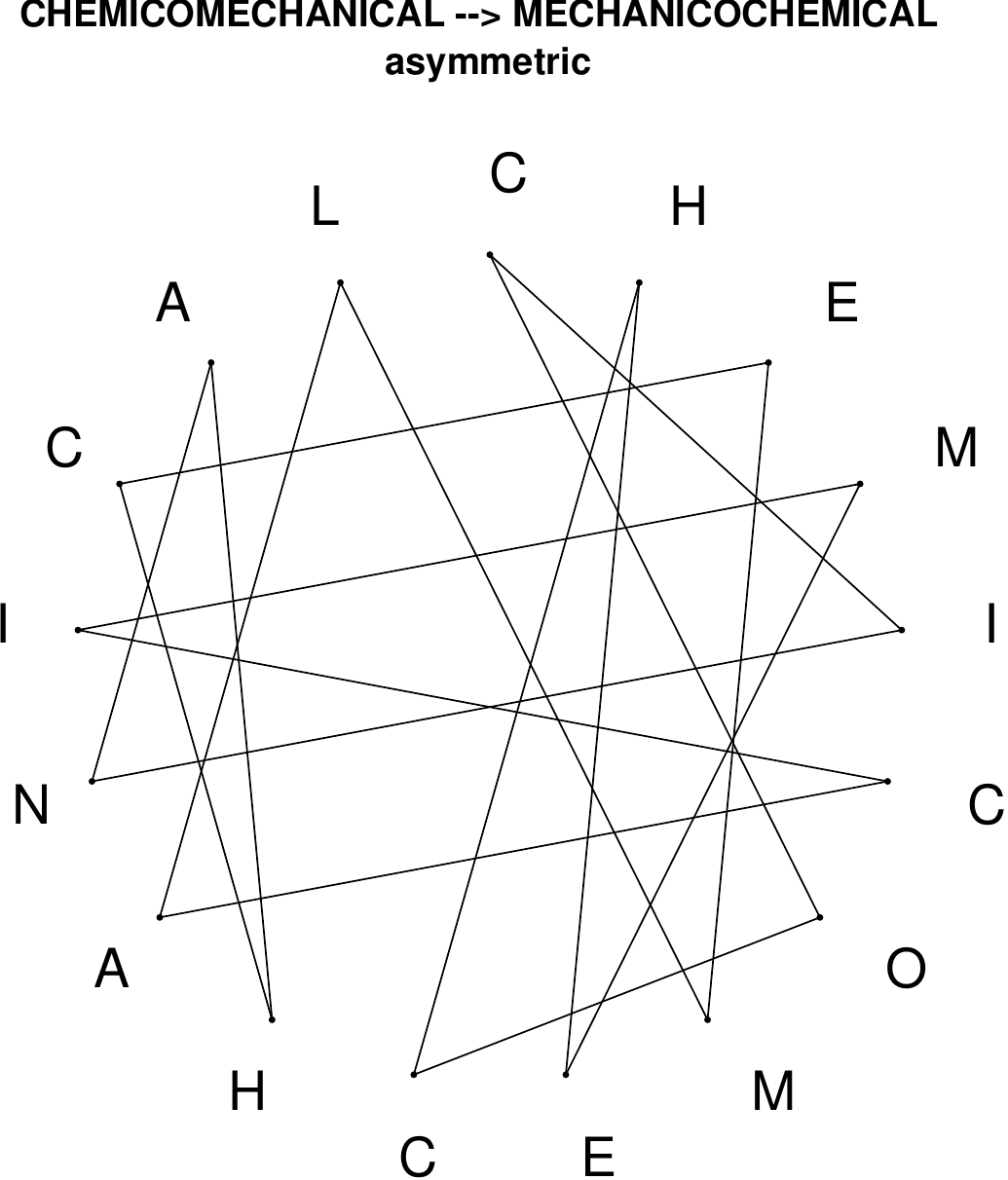}
\end{subfigure}
\hfill
\begin{subfigure}[T]{0.19\textwidth}
\centering
\includegraphics[width=\textwidth]{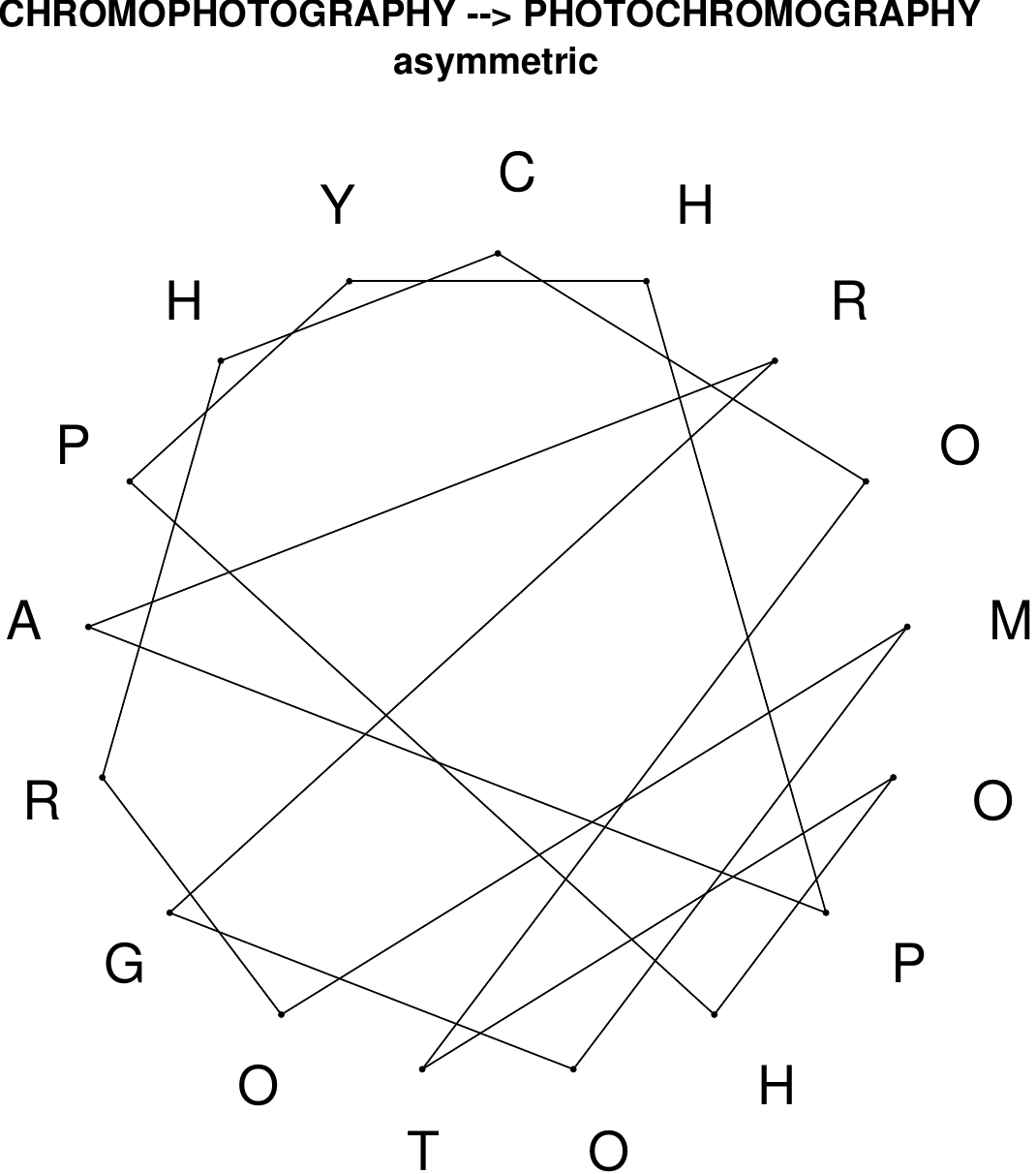}
\end{subfigure}
\hfill
\begin{subfigure}[T]{0.19\textwidth}
\centering
\includegraphics[width=\textwidth]{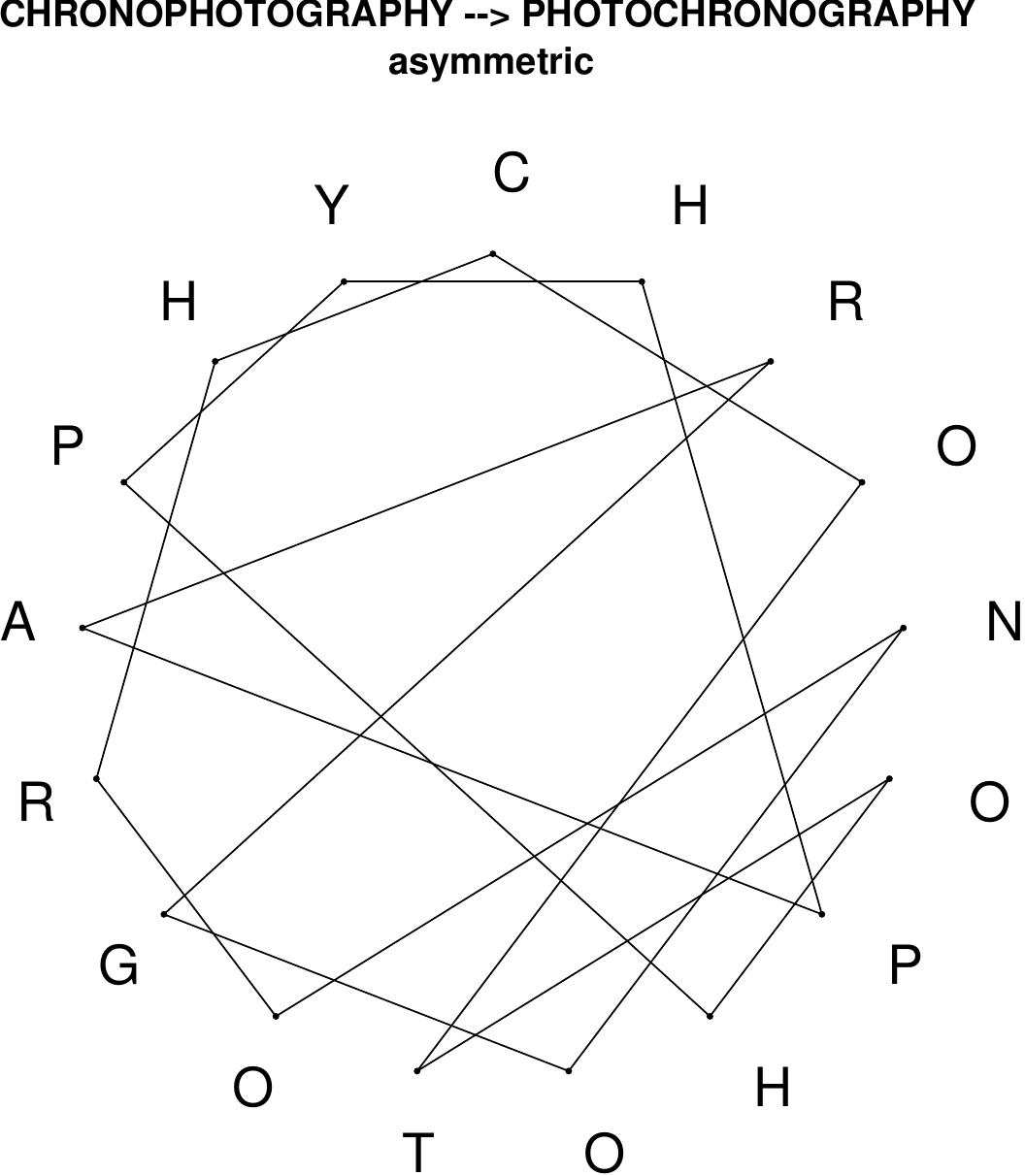}
\end{subfigure}
\hfill
\begin{subfigure}[T]{0.19\textwidth}
\centering
\includegraphics[width=\textwidth]{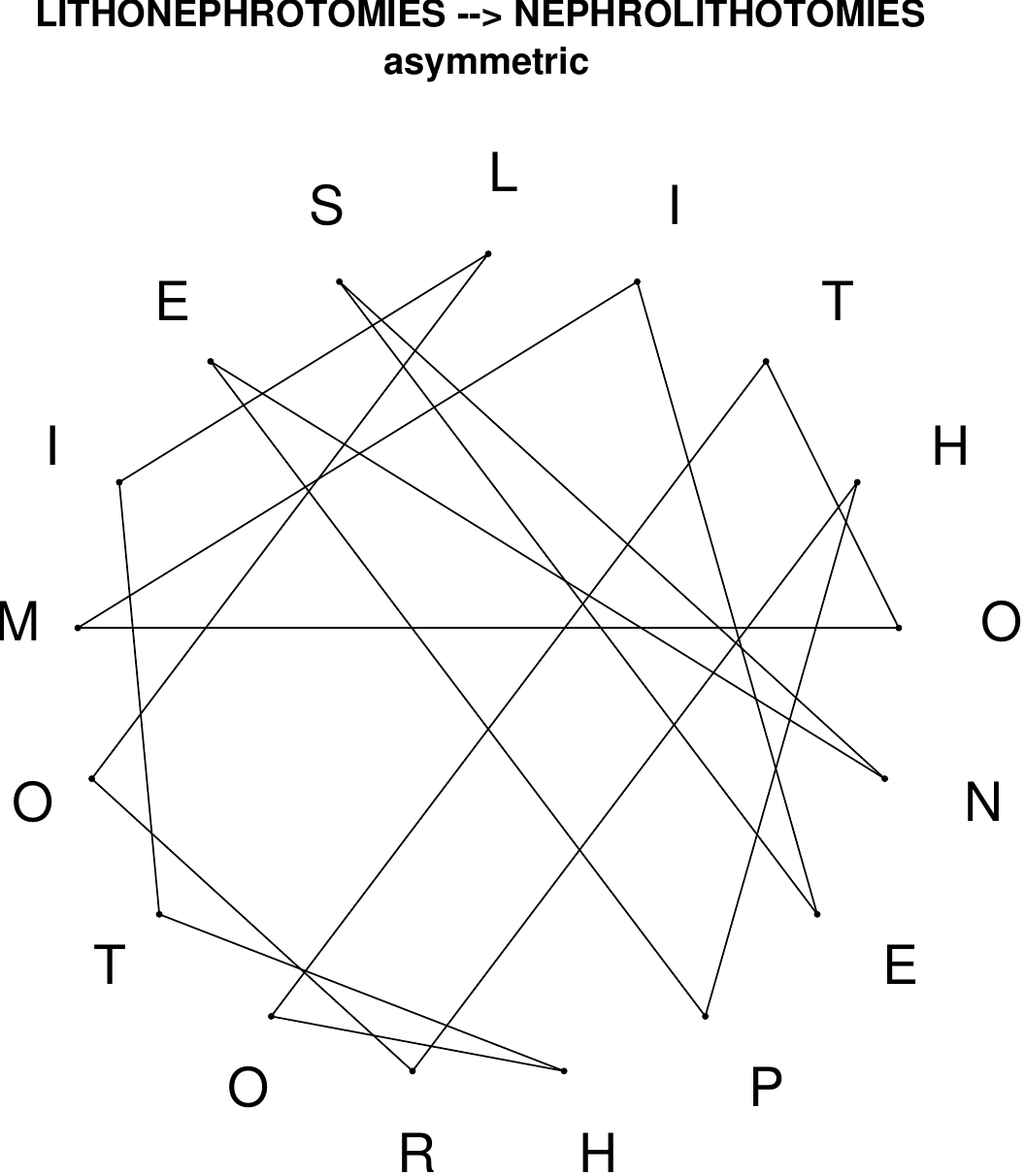}
\end{subfigure}
\end{figure}

%%%%%%%%%%%%%%%%%%
\clearpage
\subsection{Star Anagrams $N = 16$}
For $N=16$, we found a handful of asymmetric star anagrams.

\begin{figure}[H]
\centering
\begin{subfigure}[T]{0.19\textwidth}
\centering
\includegraphics[width=\textwidth]{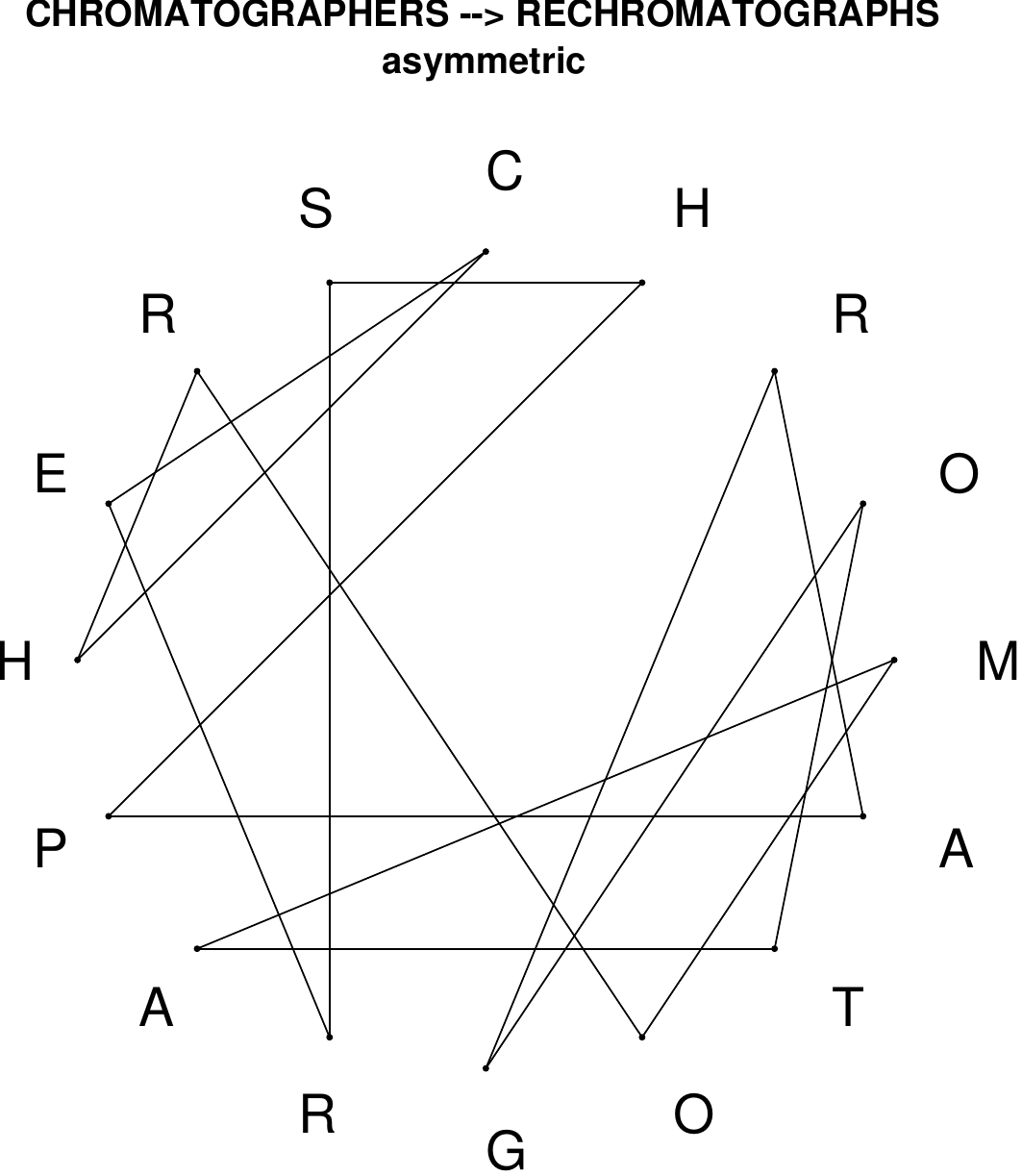}
\end{subfigure}
\hfill
\begin{subfigure}[T]{0.19\textwidth}
\centering
\includegraphics[width=\textwidth]{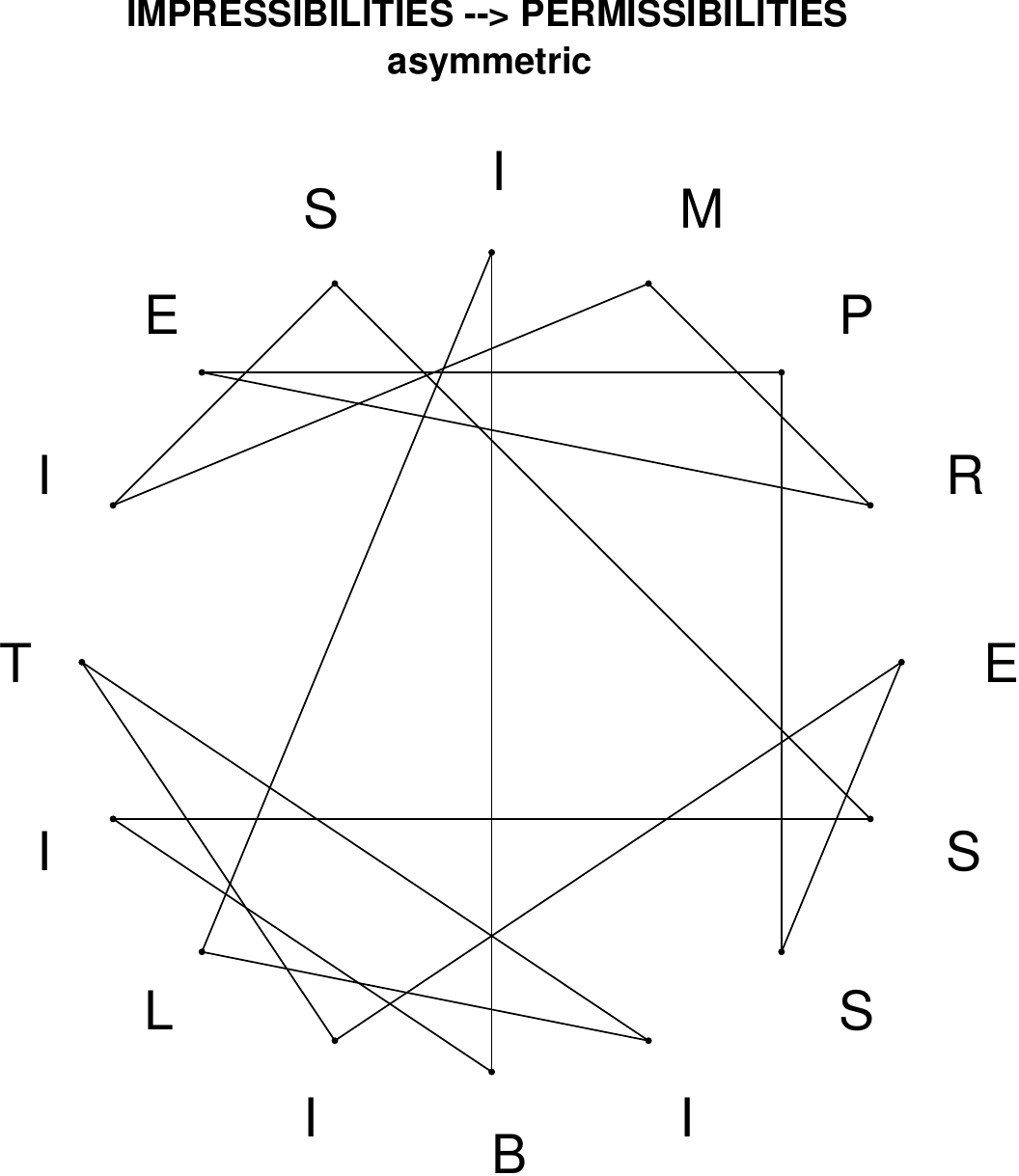}
\end{subfigure}
\hfill
\begin{subfigure}[T]{0.19\textwidth}
\centering
\includegraphics[width=\textwidth]{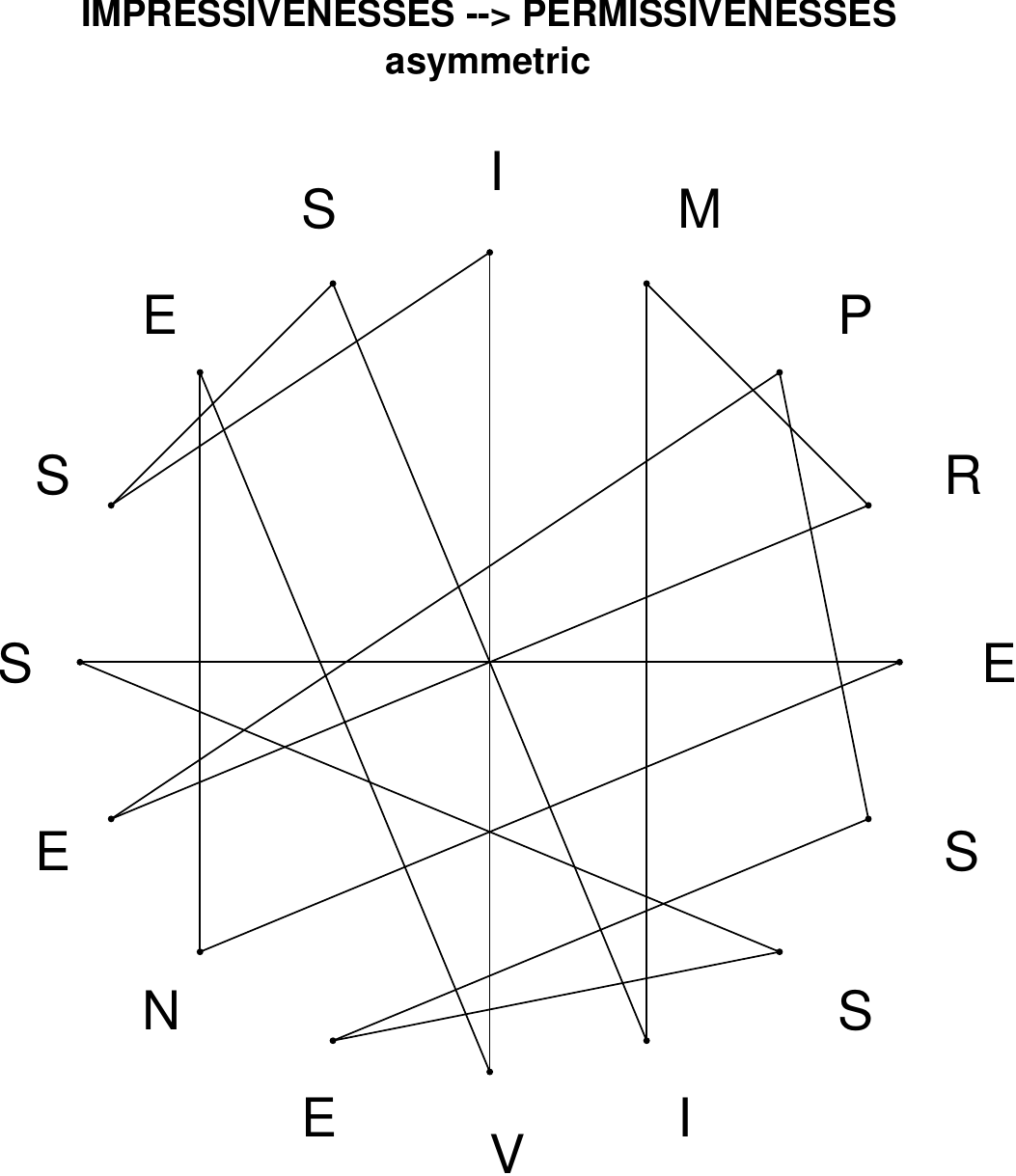}
\end{subfigure}
\hfill
\begin{subfigure}[T]{0.19\textwidth}
\centering
\includegraphics[width=\textwidth]{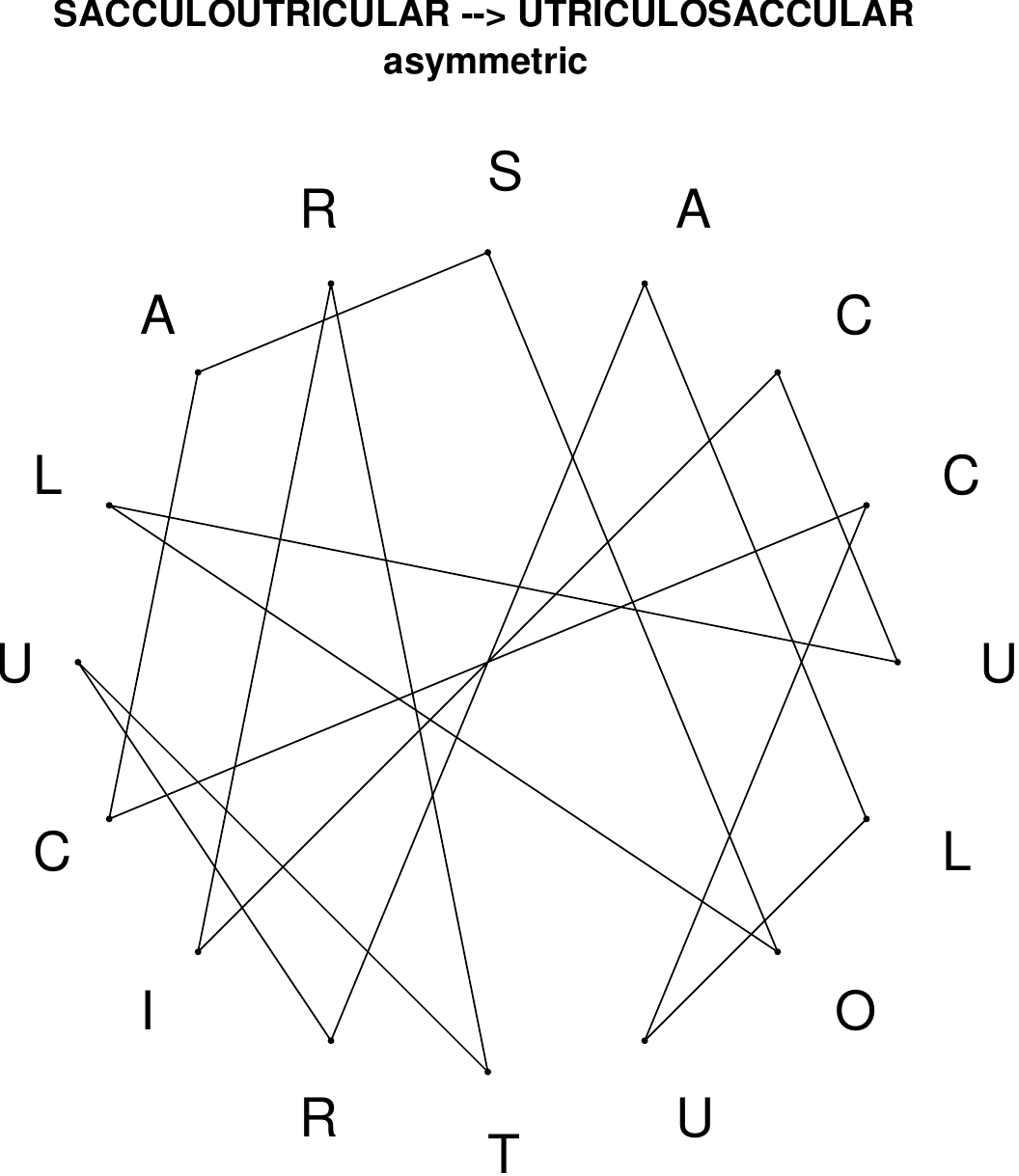}
\end{subfigure}
\hfill
\begin{subfigure}[T]{0.19\textwidth}
\centering
\includegraphics[width=\textwidth]{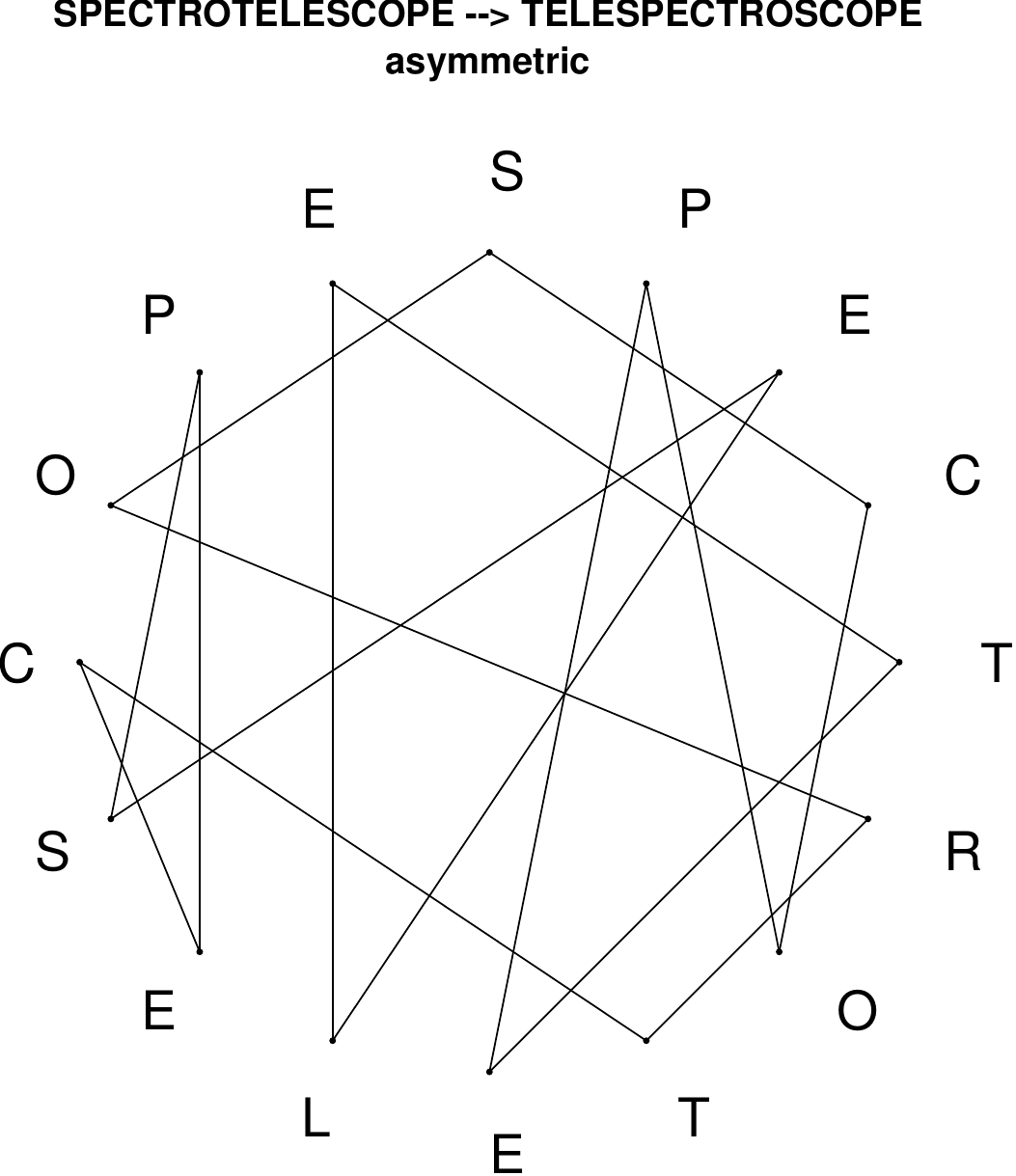}
\end{subfigure}
\end{figure}

\begin{figure}[H]
\centering
\begin{subfigure}[T]{0.19\textwidth}
\centering
\includegraphics[width=\textwidth]{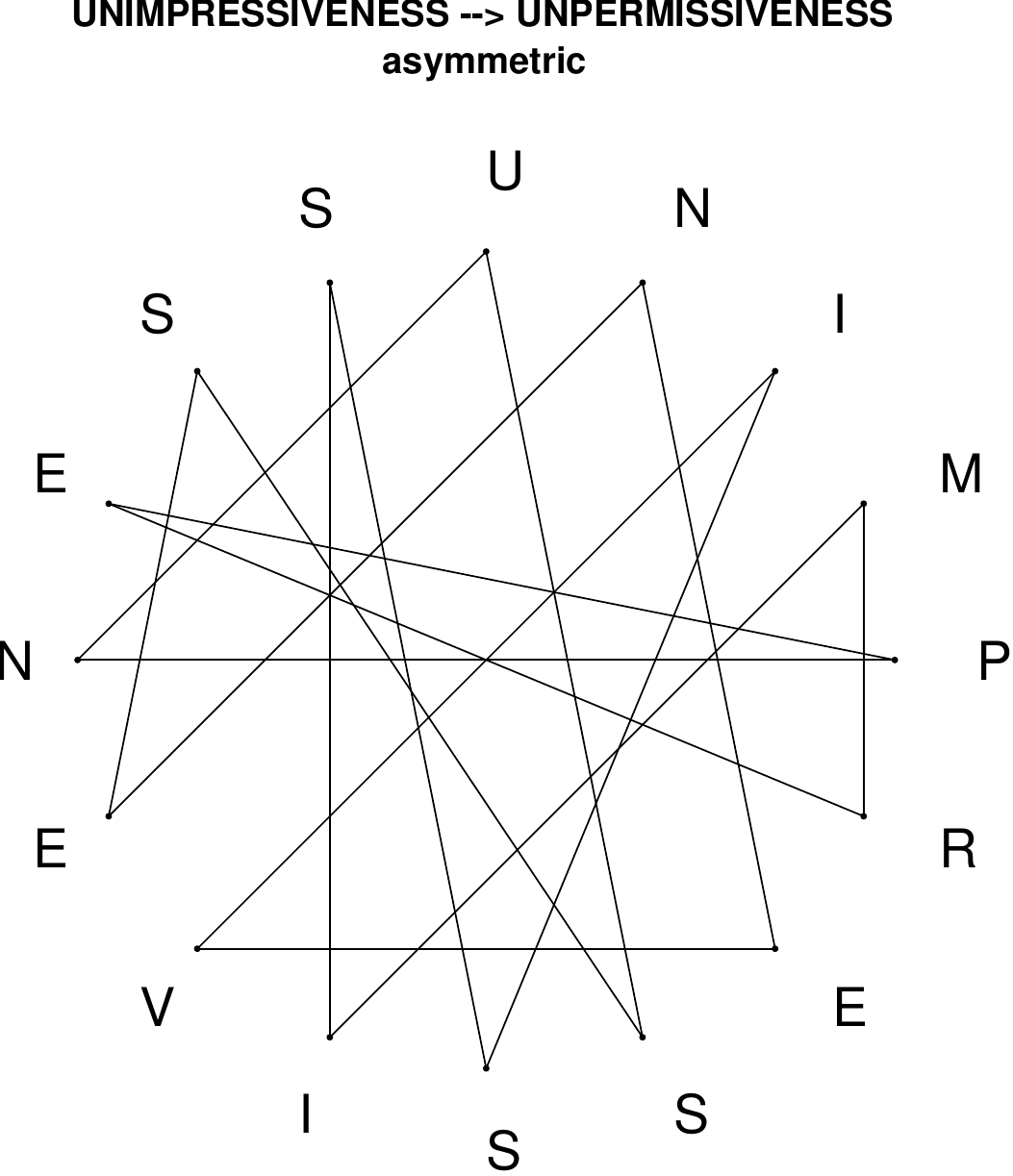}
\end{subfigure}
\hfill
\end{figure}

%%%%%%%%%%%%%%%%%%
\clearpage
\subsection{Star Anagrams $N = 15$}
For $N=15$, we found a handful of asymmetric star anagrams.

\begin{figure}[H]
\centering
\begin{subfigure}[T]{0.19\textwidth}
\centering
\includegraphics[width=\textwidth]{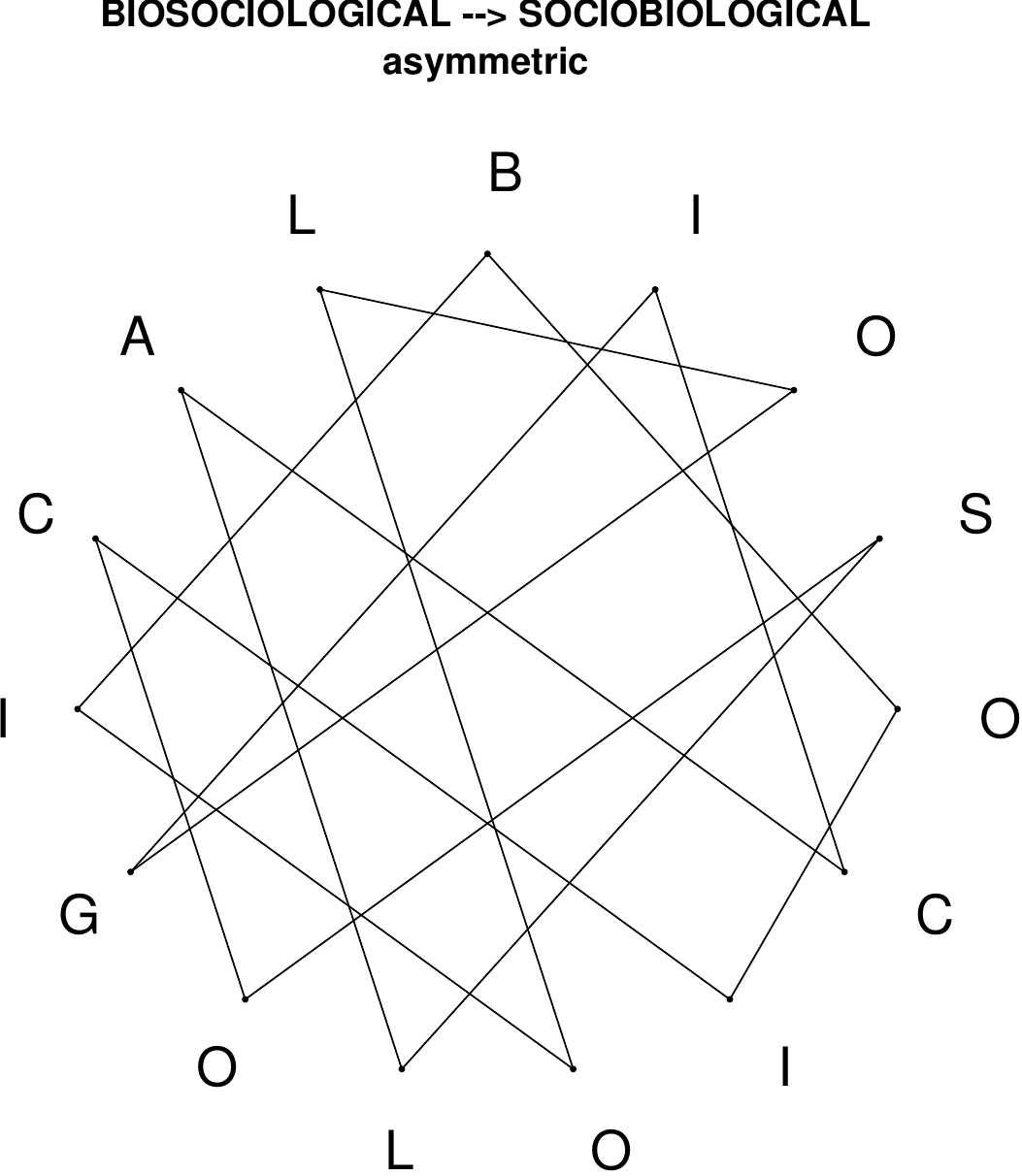}
\end{subfigure}
\hfill
\begin{subfigure}[T]{0.19\textwidth}
\centering
\includegraphics[width=\textwidth]{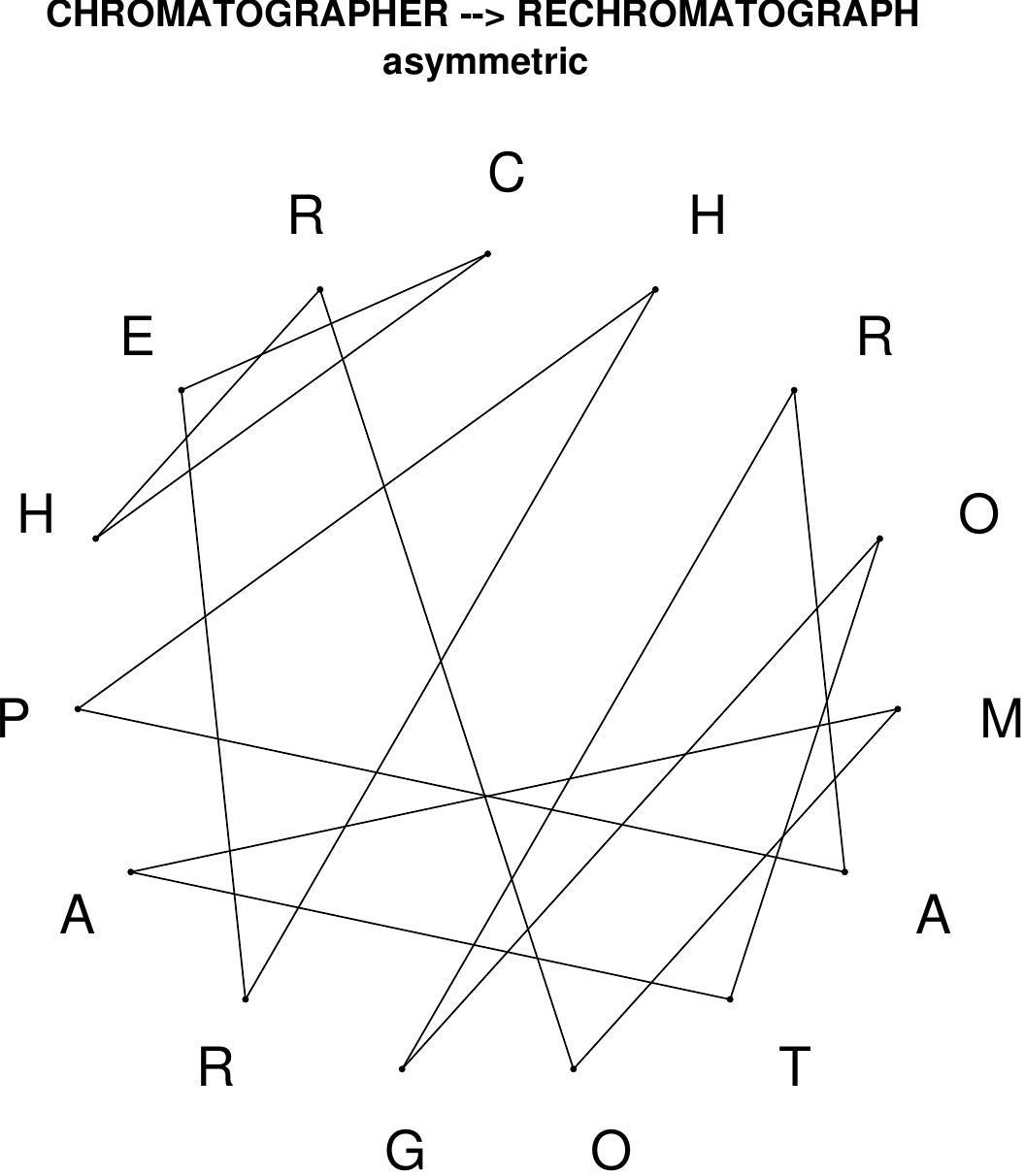}
\end{subfigure}
\hfill
\begin{subfigure}[T]{0.19\textwidth}
\centering
\includegraphics[width=\textwidth]{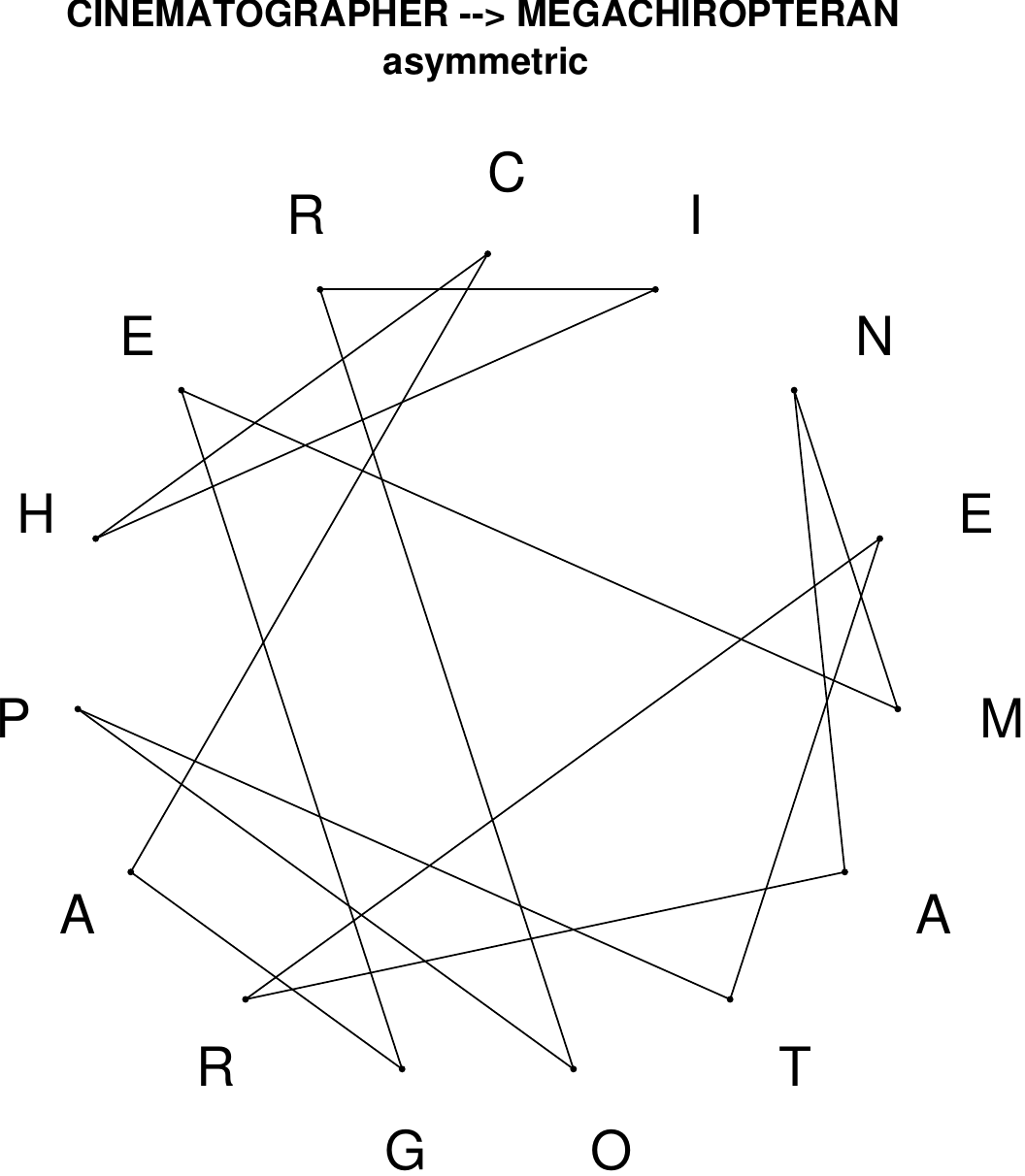}
\end{subfigure}
\hfill
\begin{subfigure}[T]{0.19\textwidth}
\centering
\includegraphics[width=\textwidth]{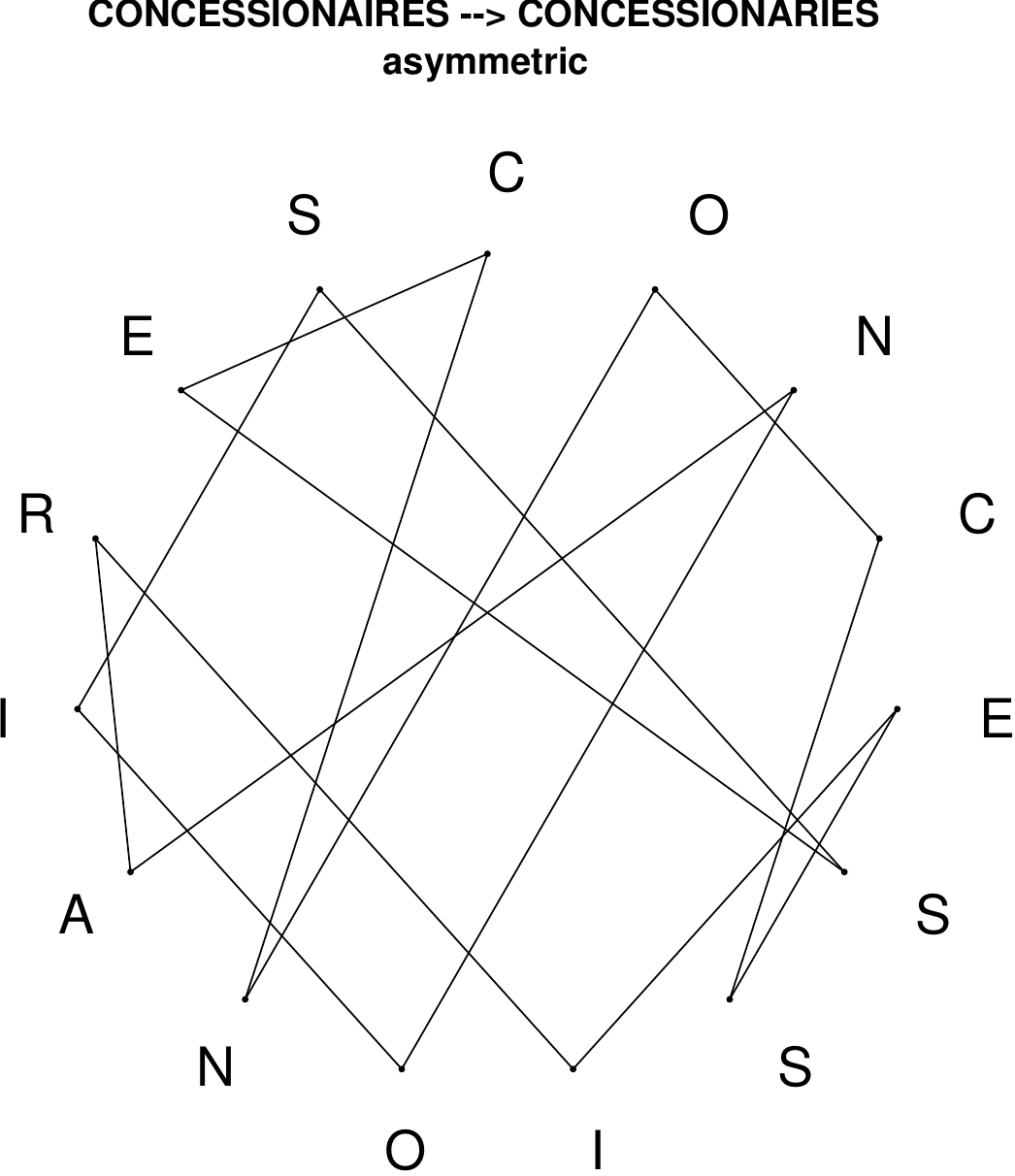}
\end{subfigure}
\hfill
\begin{subfigure}[T]{0.19\textwidth}
\centering
\includegraphics[width=\textwidth]{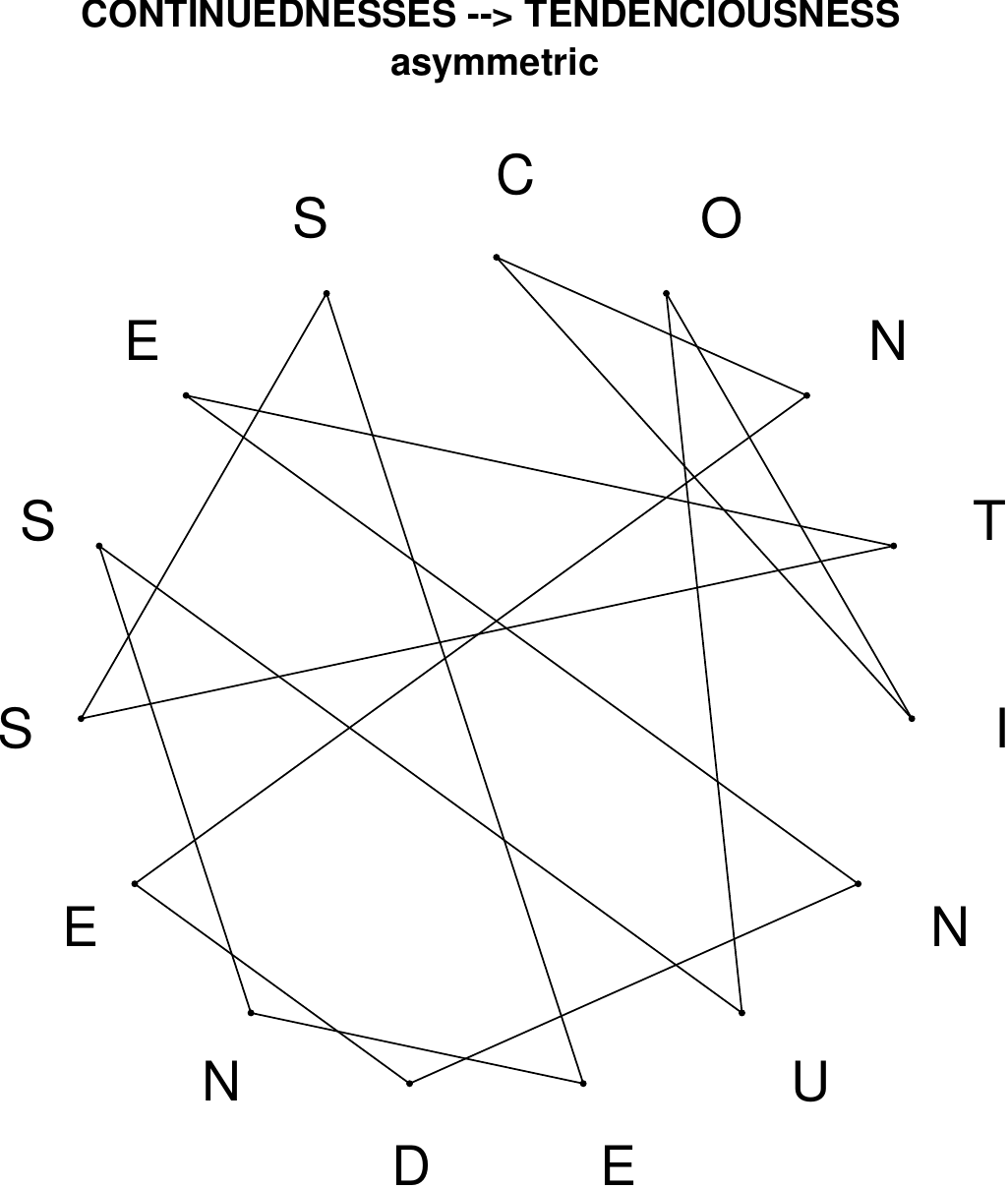}
\end{subfigure}
\end{figure}

\begin{figure}[H]
\centering
\begin{subfigure}[T]{0.19\textwidth}
\centering
\includegraphics[width=\textwidth]{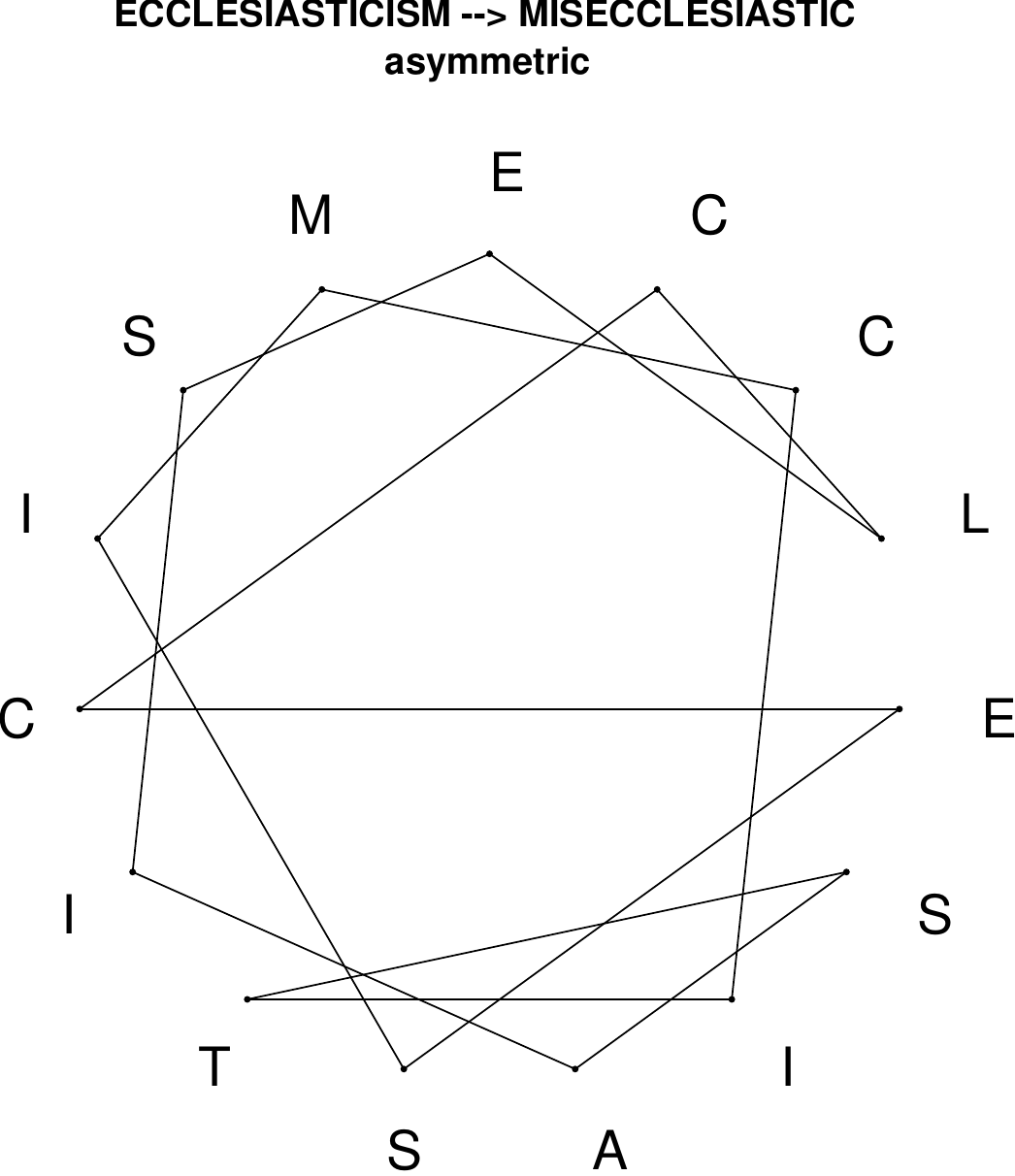}
\end{subfigure}
\hfill
\begin{subfigure}[T]{0.19\textwidth}
\centering
\includegraphics[width=\textwidth]{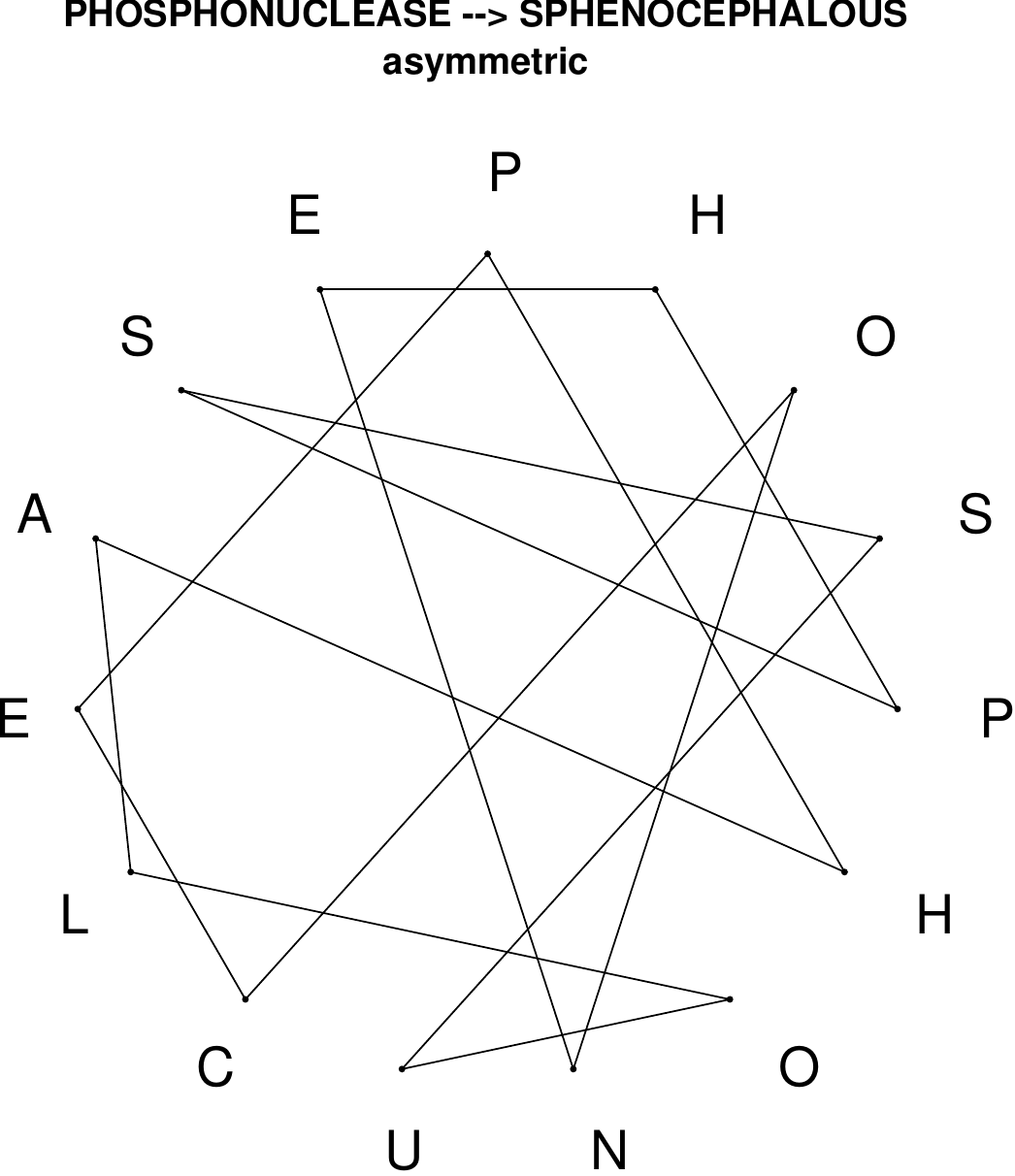}
\end{subfigure}
\hfill
\begin{subfigure}[T]{0.19\textwidth}
\centering
\includegraphics[width=\textwidth]{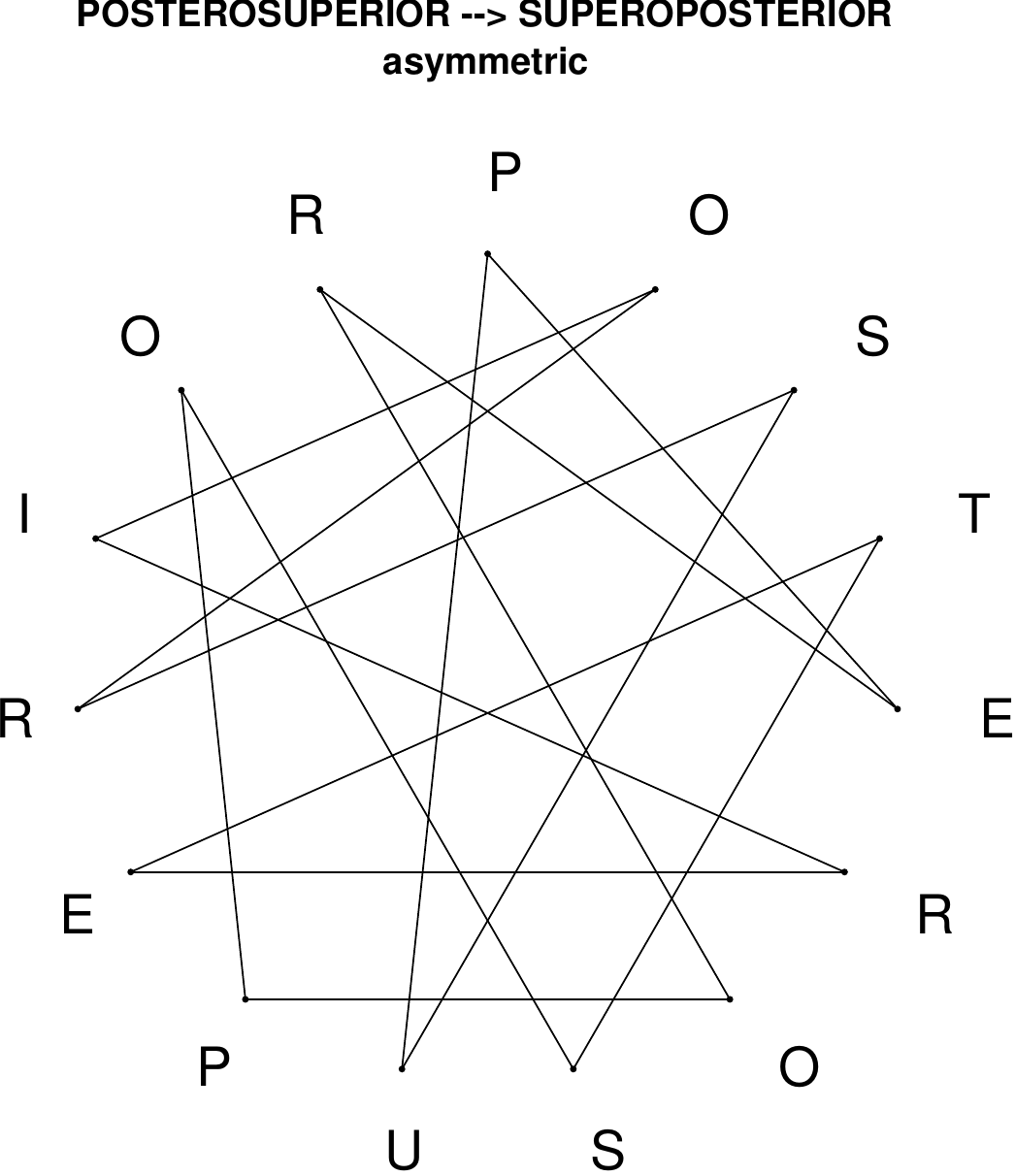}
\end{subfigure}
\hfill
\begin{subfigure}[T]{0.19\textwidth}
\centering
\includegraphics[width=\textwidth]{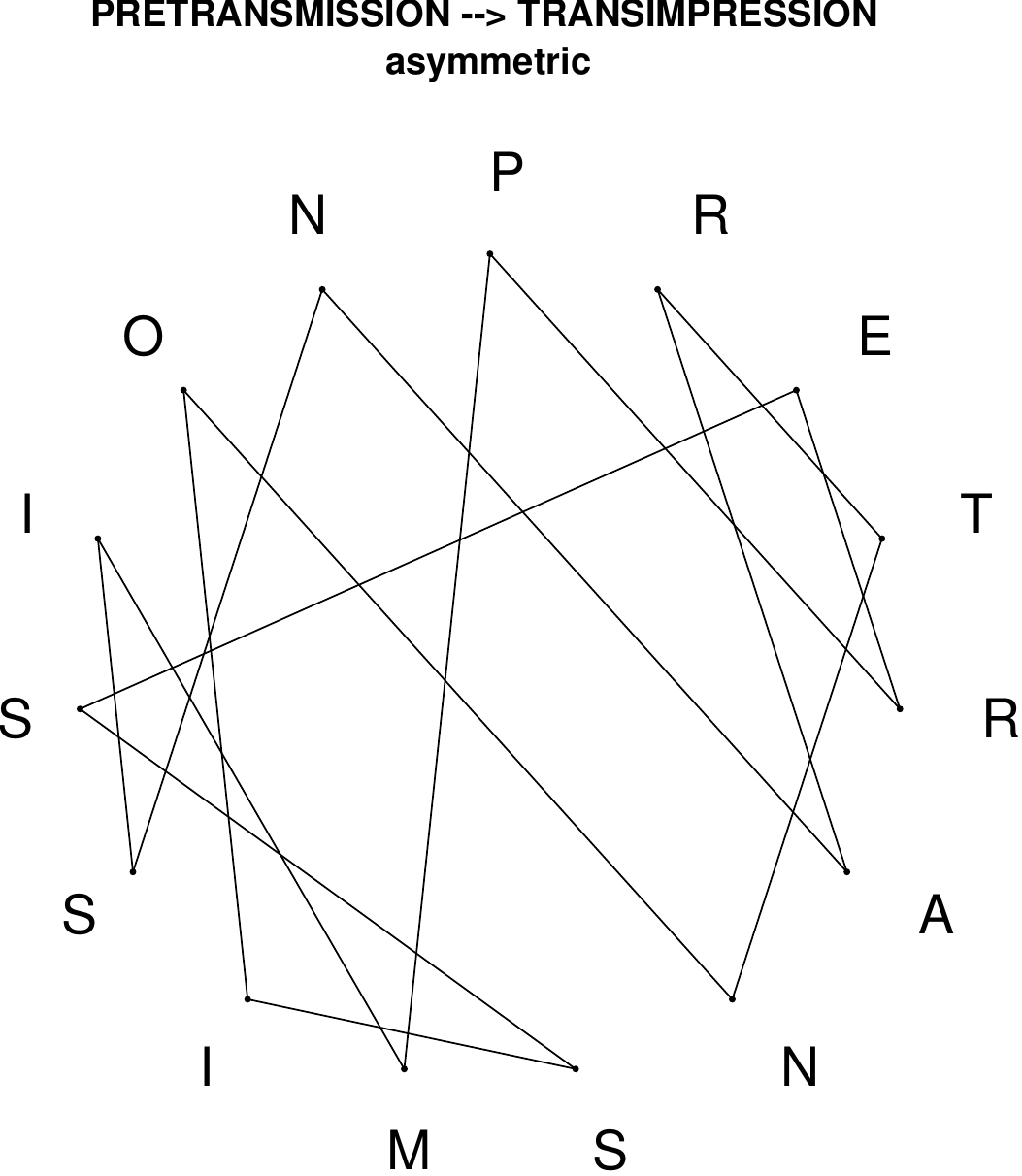}
\end{subfigure}
\hfill
\begin{subfigure}[T]{0.19\textwidth}
\centering
\includegraphics[width=\textwidth]{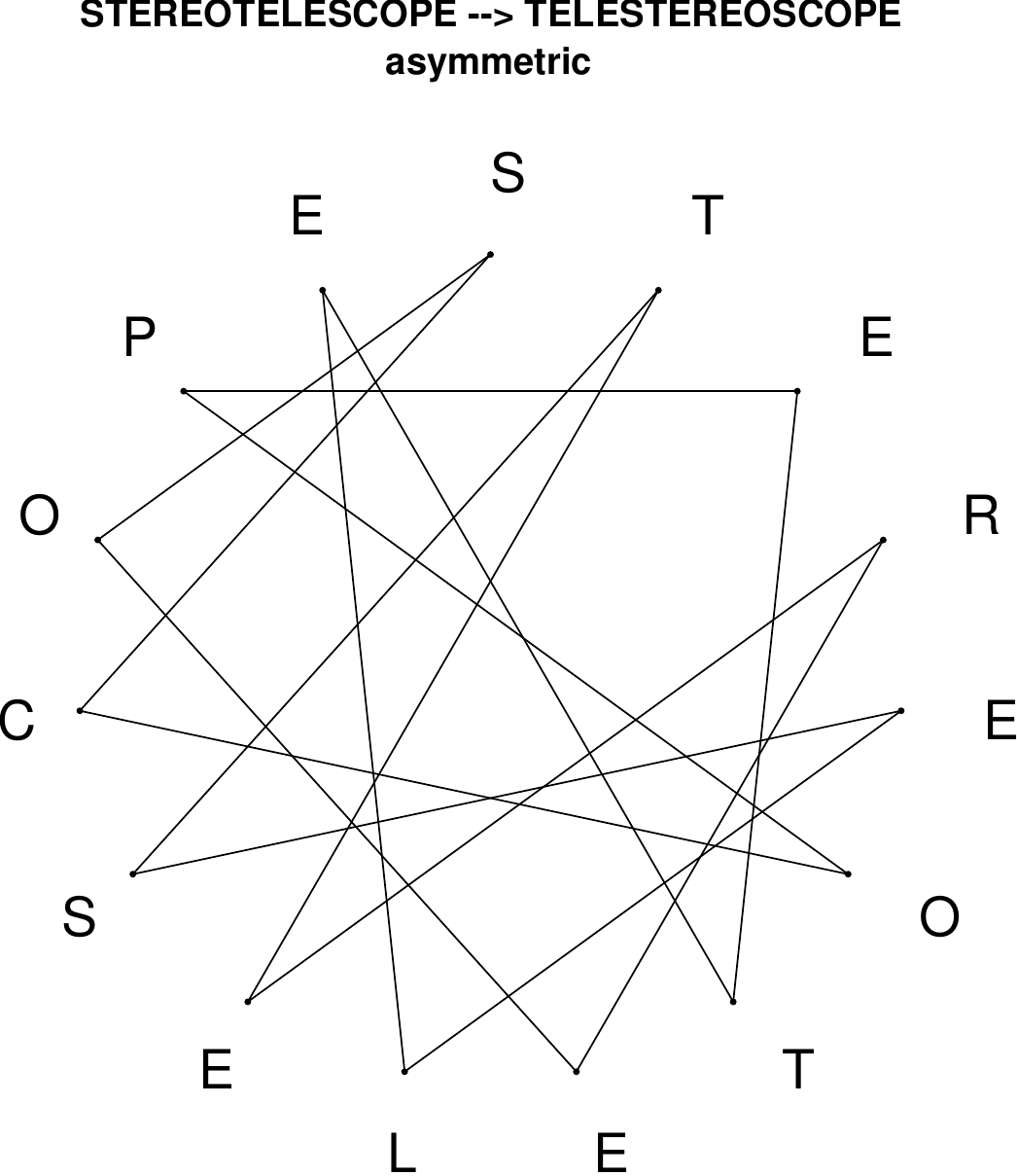}
\end{subfigure}
\end{figure}

%%%%%%%%%%%%%%%%%%
\clearpage
\subsection{Star Anagrams $N = 14$}
For $N=14$, we found a few dozen asymmetric star anagrams. We also see our first and longest symmetric star. 

\subsubsection{Symmetric Stars $N=14$}

\begin{figure}[H]
\centering
\begin{subfigure}[T]{0.19\textwidth}
\centering
\includegraphics[width=\textwidth]{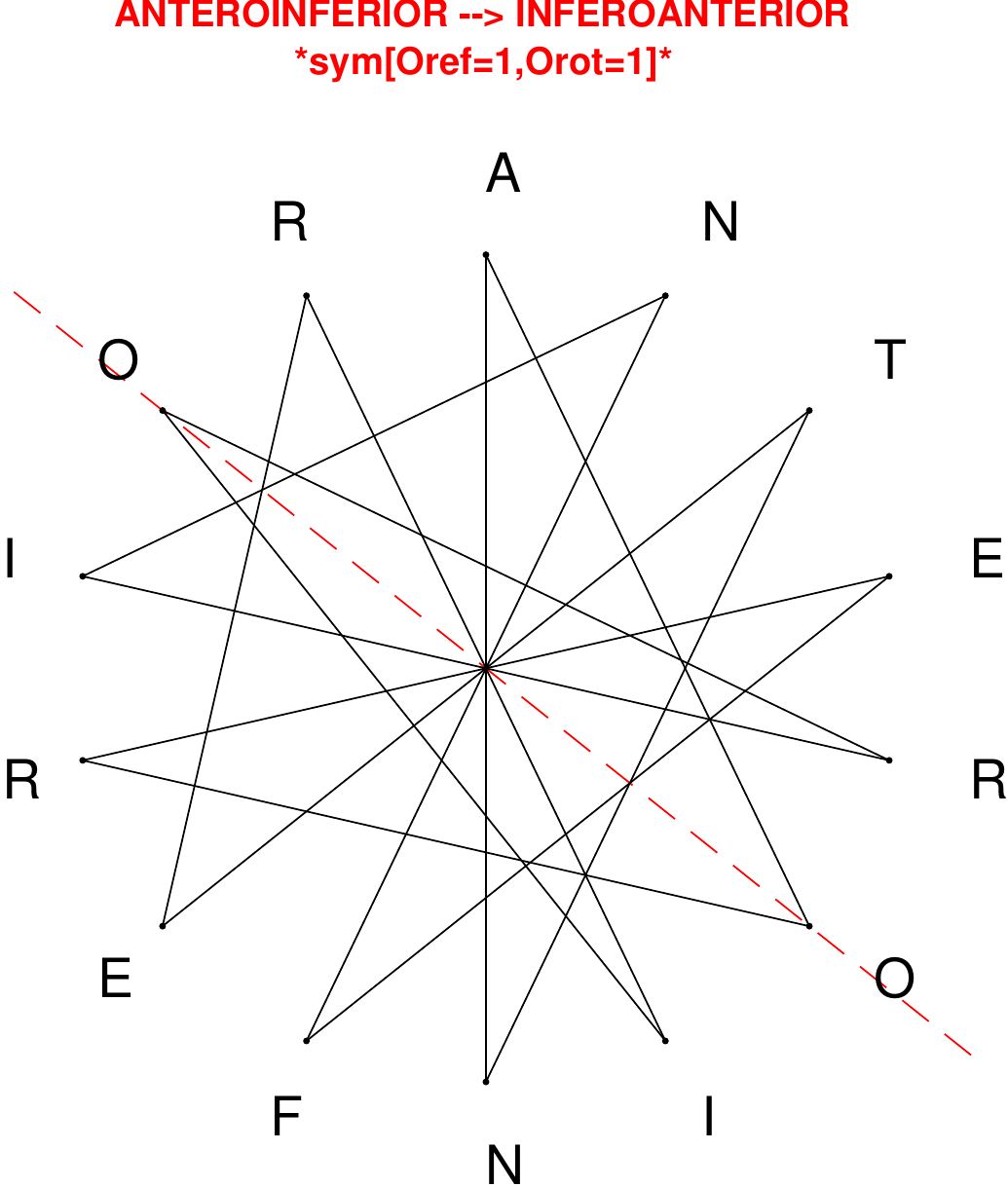}
\end{subfigure}
\hfill
\end{figure}

\subsubsection{Asymmetric Stars $N=14$}

\begin{figure}[H]
\centering
\begin{subfigure}[T]{0.19\textwidth}
\centering
\includegraphics[width=\textwidth]{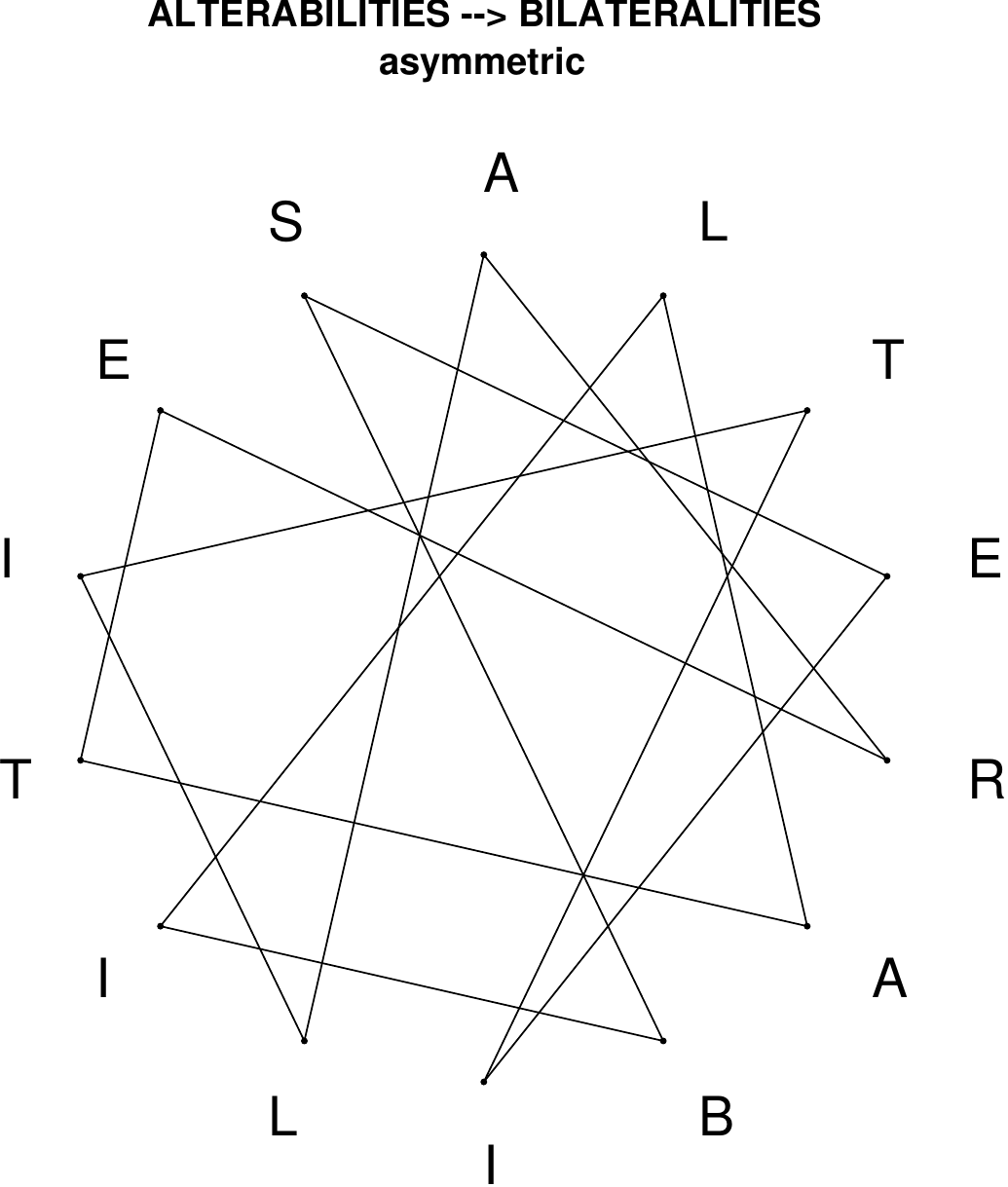}
\end{subfigure}
\hfill
\begin{subfigure}[T]{0.19\textwidth}
\centering
\includegraphics[width=\textwidth]{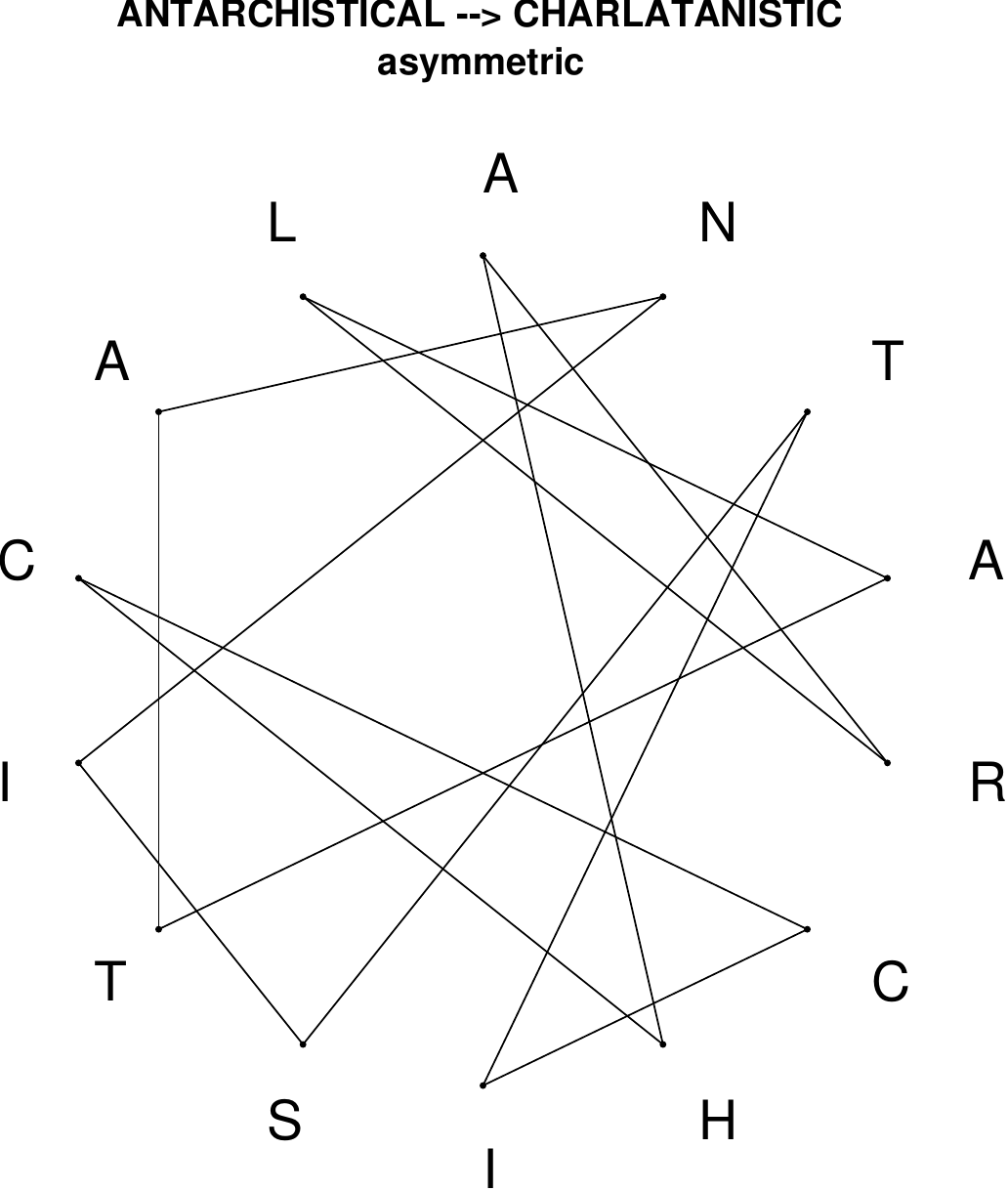}
\end{subfigure}
\hfill
\begin{subfigure}[T]{0.19\textwidth}
\centering
\includegraphics[width=\textwidth]{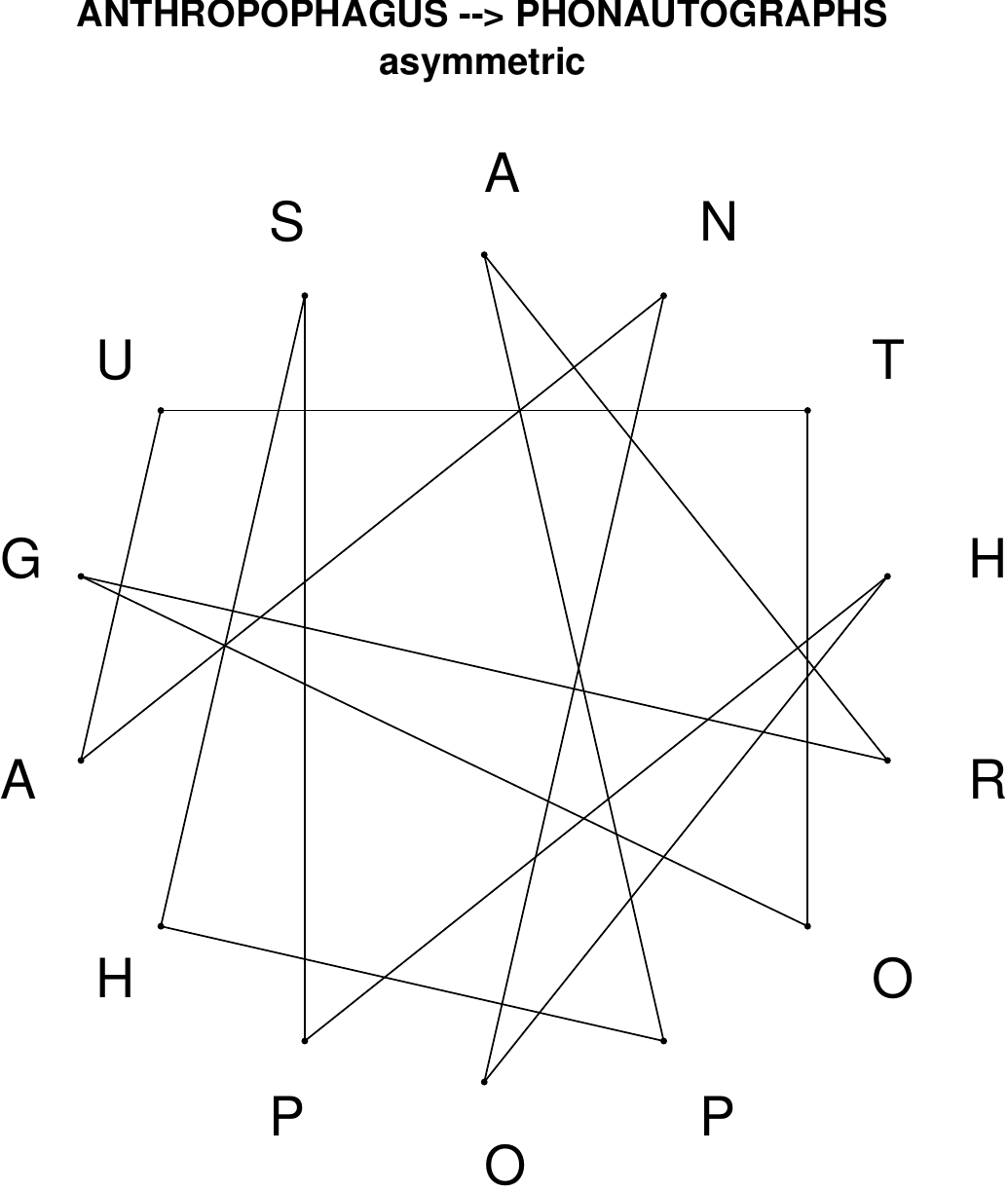}
\end{subfigure}
\hfill
\begin{subfigure}[T]{0.19\textwidth}
\centering
\includegraphics[width=\textwidth]{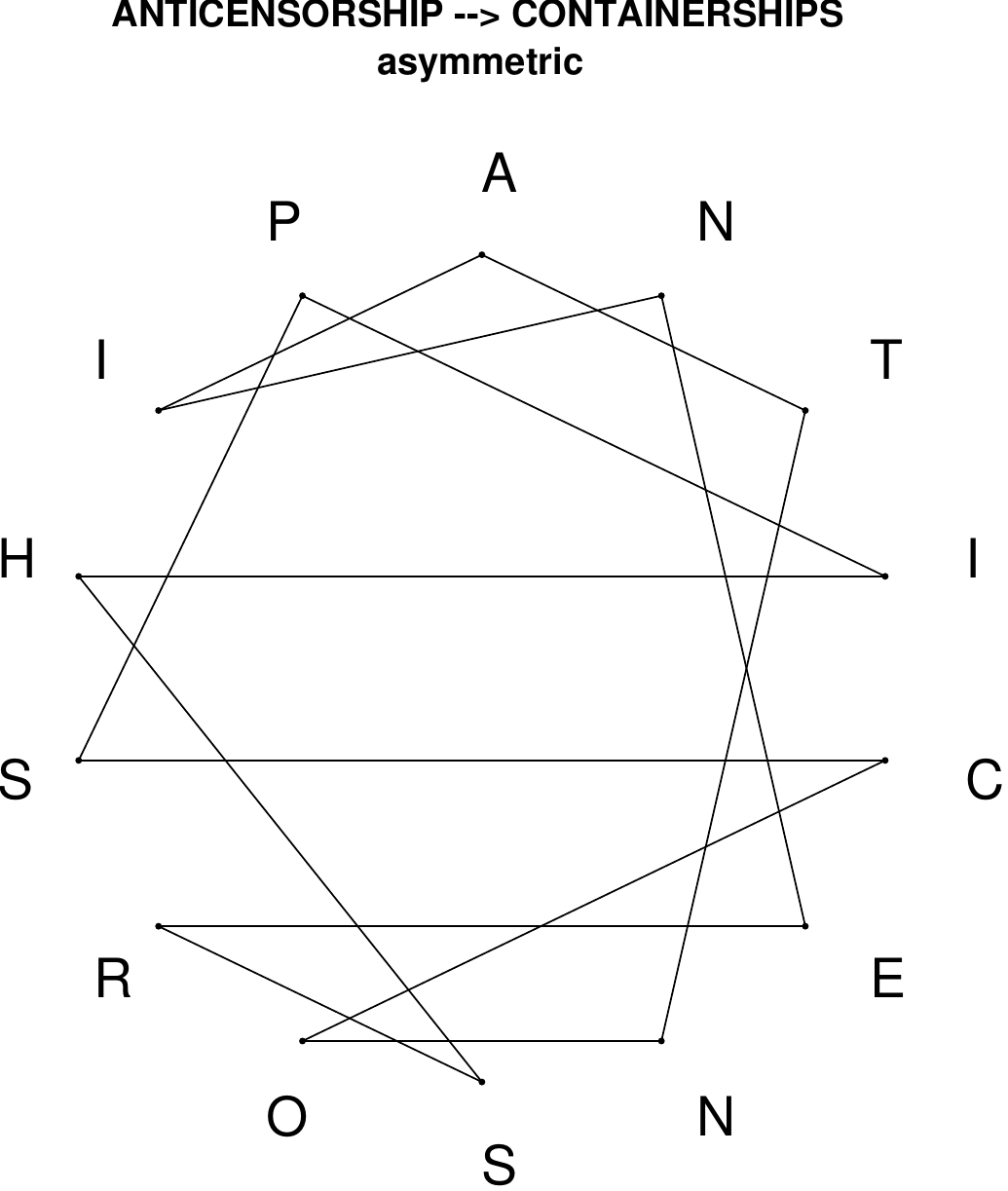}
\end{subfigure}
\hfill
\begin{subfigure}[T]{0.19\textwidth}
\centering
\includegraphics[width=\textwidth]{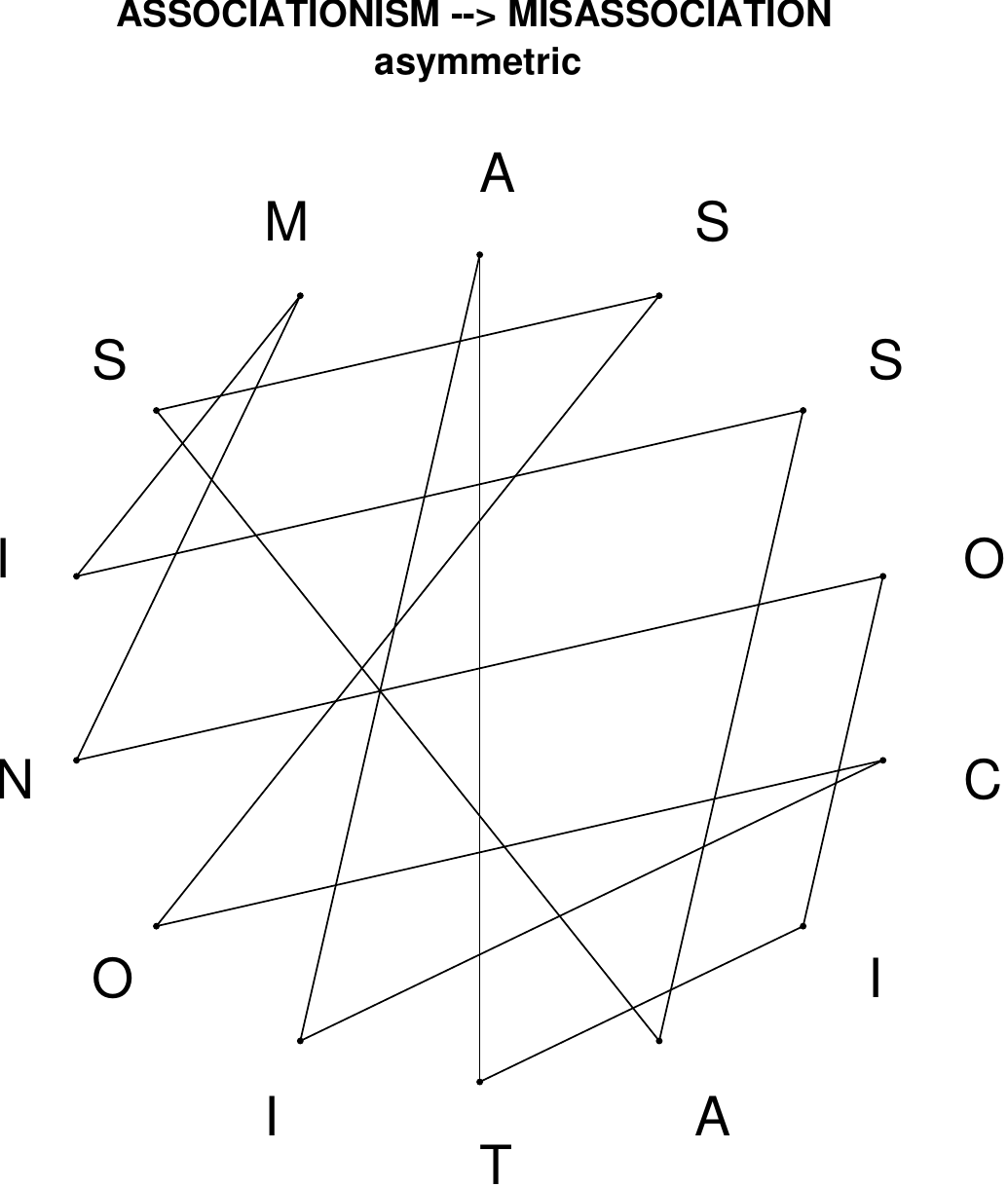}
\end{subfigure}
\end{figure}

\begin{figure}[H]
\centering
\begin{subfigure}[T]{0.19\textwidth}
\centering
\includegraphics[width=\textwidth]{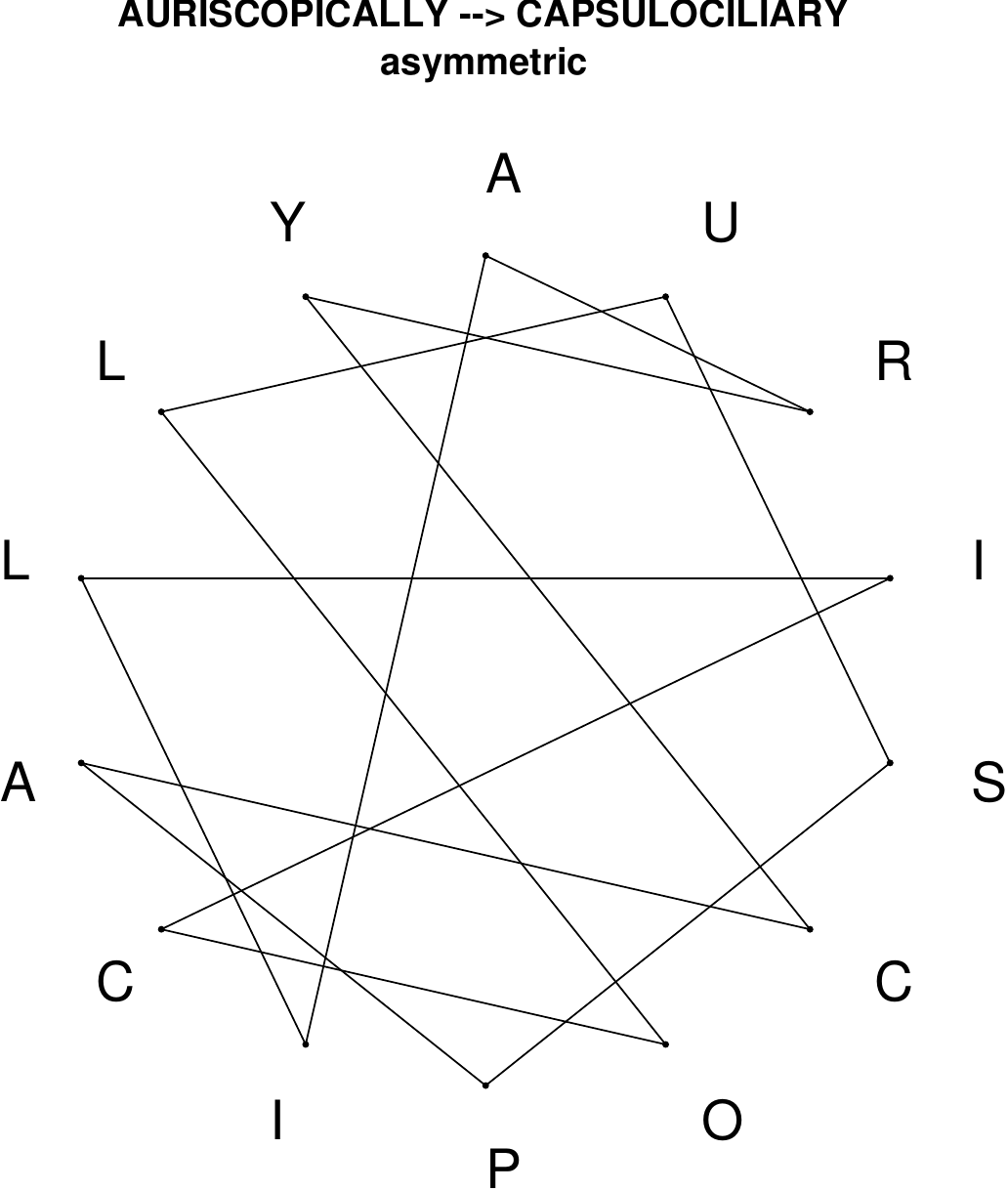}
\end{subfigure}
\hfill
\begin{subfigure}[T]{0.19\textwidth}
\centering
\includegraphics[width=\textwidth]{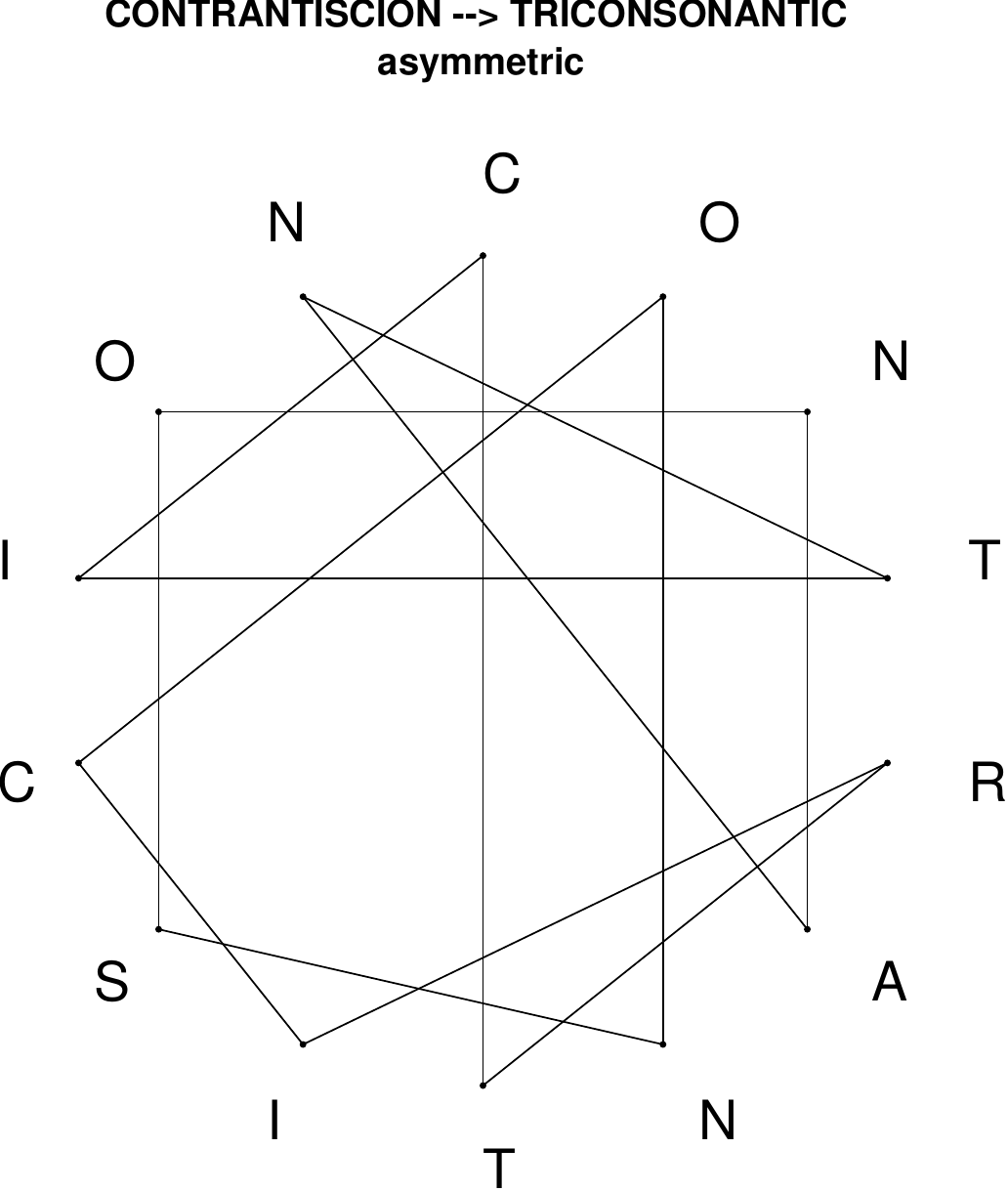}
\end{subfigure}
\hfill
\begin{subfigure}[T]{0.19\textwidth}
\centering
\includegraphics[width=\textwidth]{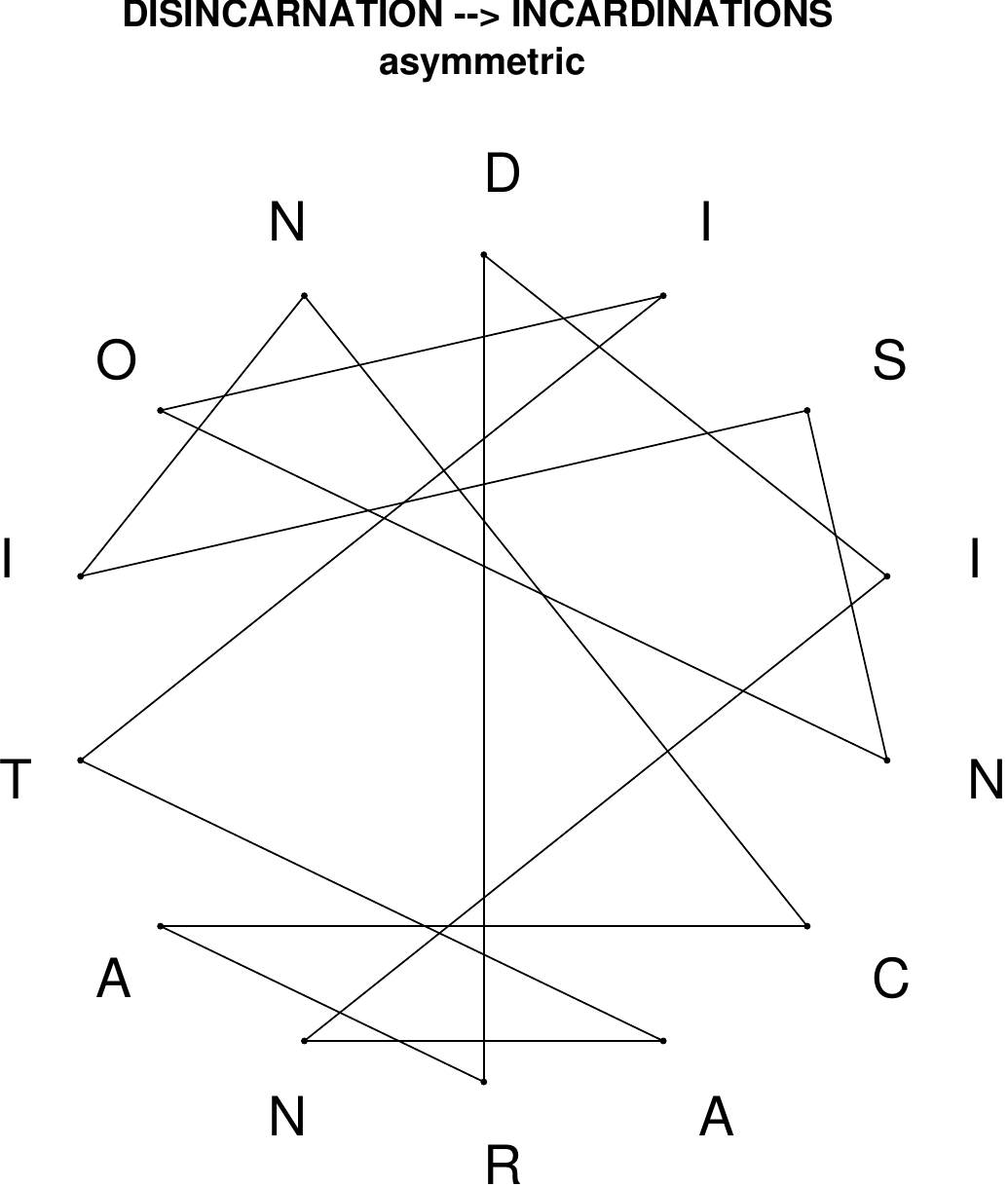}
\end{subfigure}
\hfill
\begin{subfigure}[T]{0.19\textwidth}
\centering
\includegraphics[width=\textwidth]{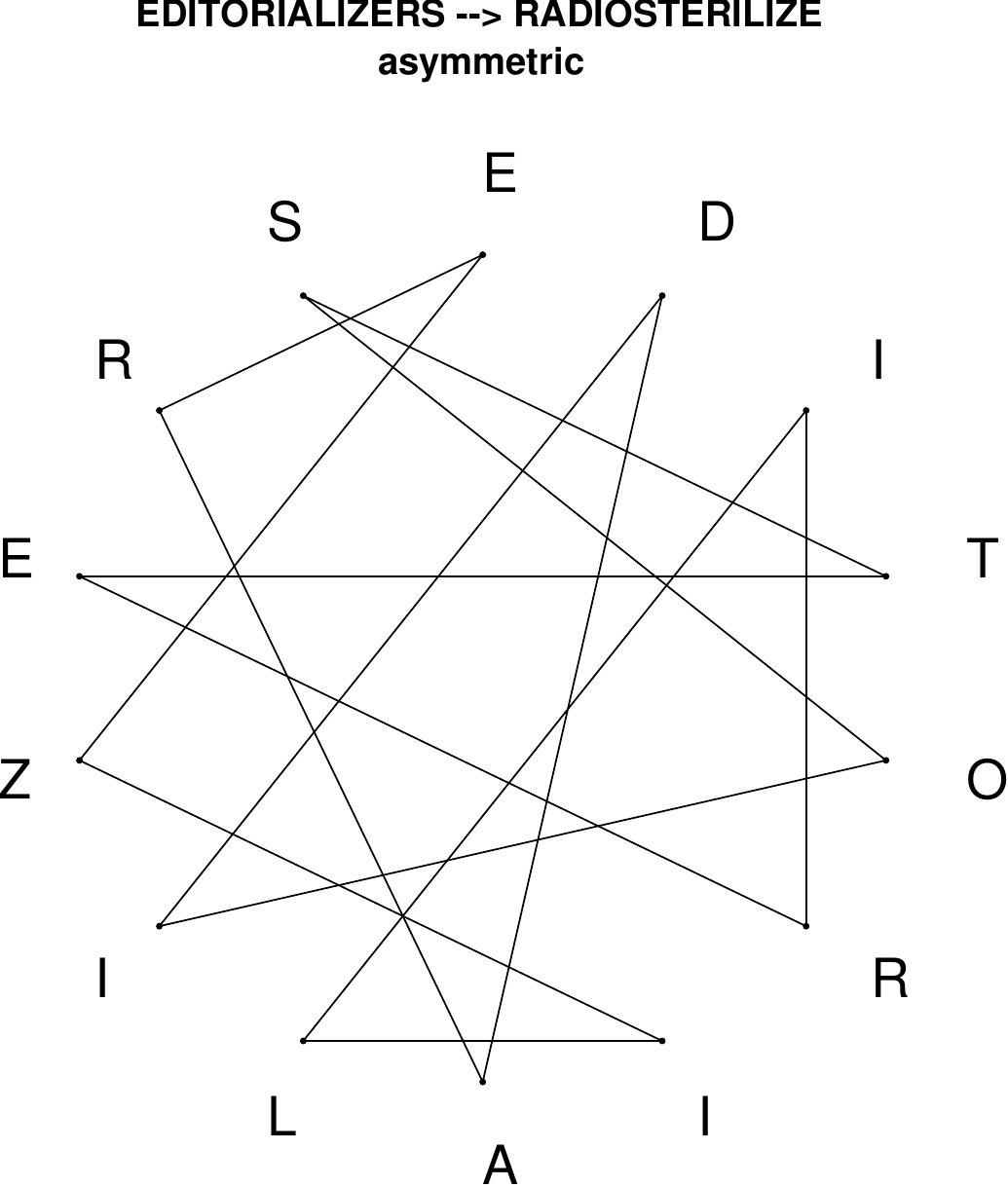}
\end{subfigure}
\hfill
\begin{subfigure}[T]{0.19\textwidth}
\centering
\includegraphics[width=\textwidth]{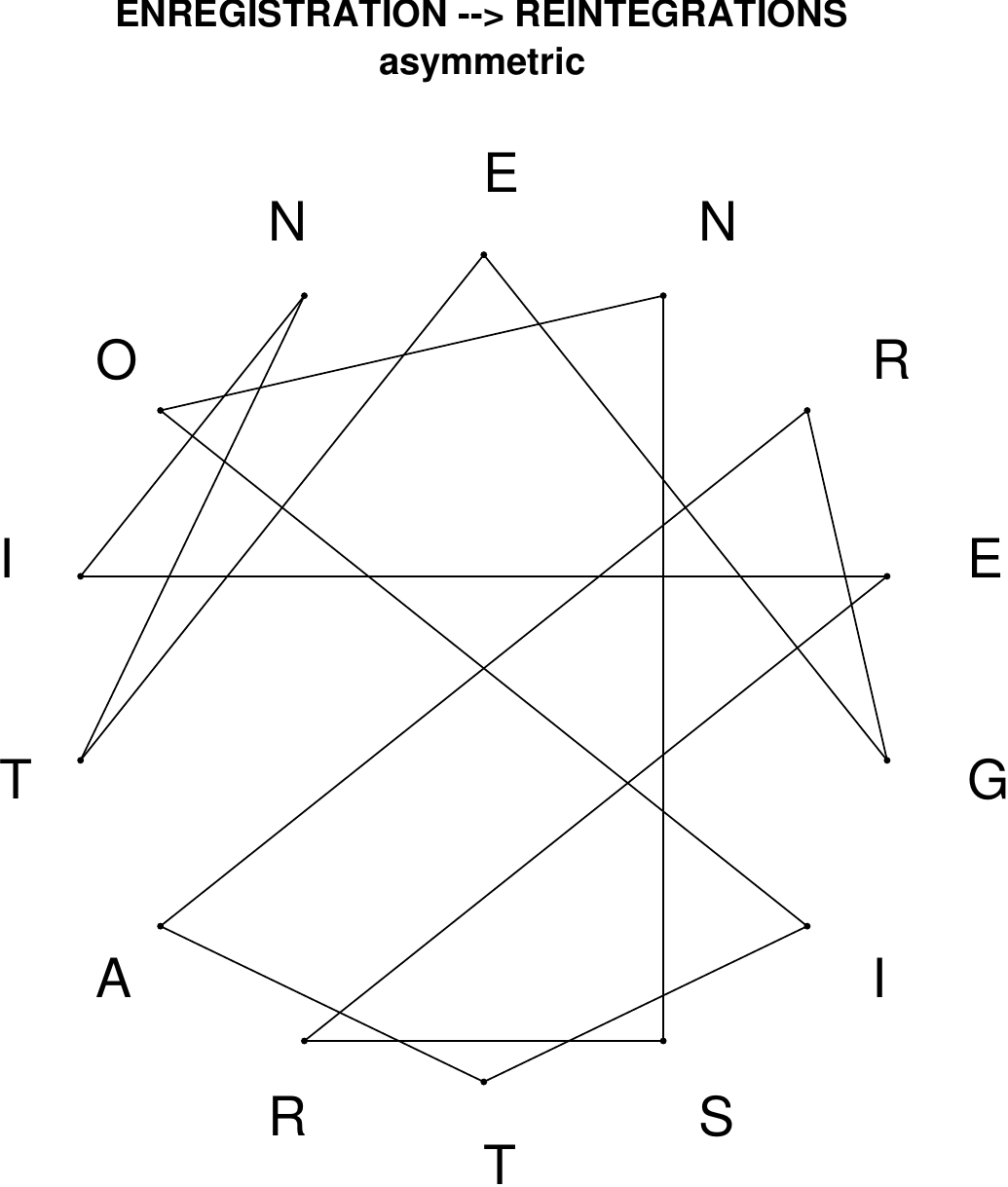}
\end{subfigure}
\end{figure}

\begin{figure}[H]
\centering
\begin{subfigure}[T]{0.19\textwidth}
\centering
\includegraphics[width=\textwidth]{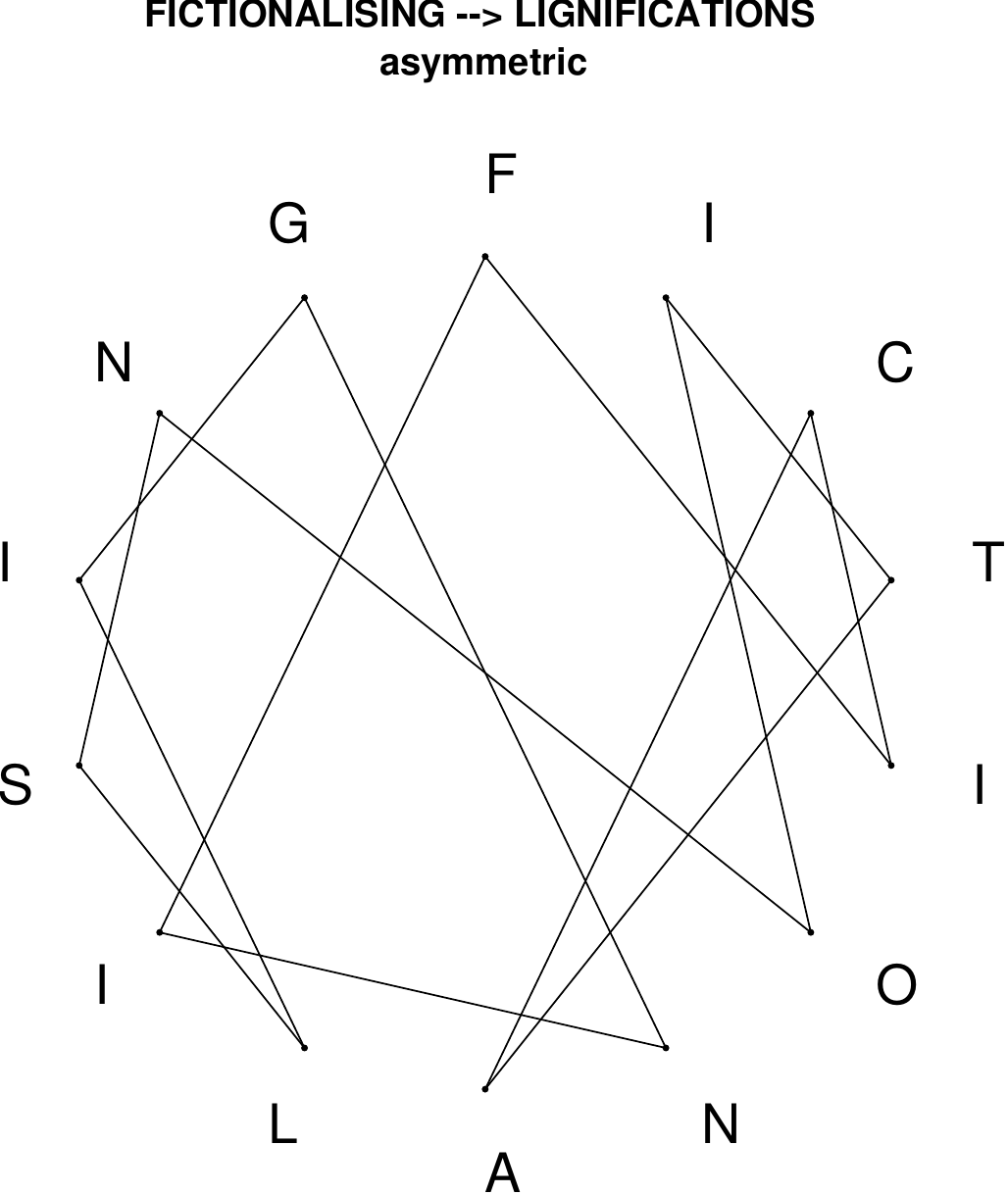}
\end{subfigure}
\hfill
\begin{subfigure}[T]{0.19\textwidth}
\centering
\includegraphics[width=\textwidth]{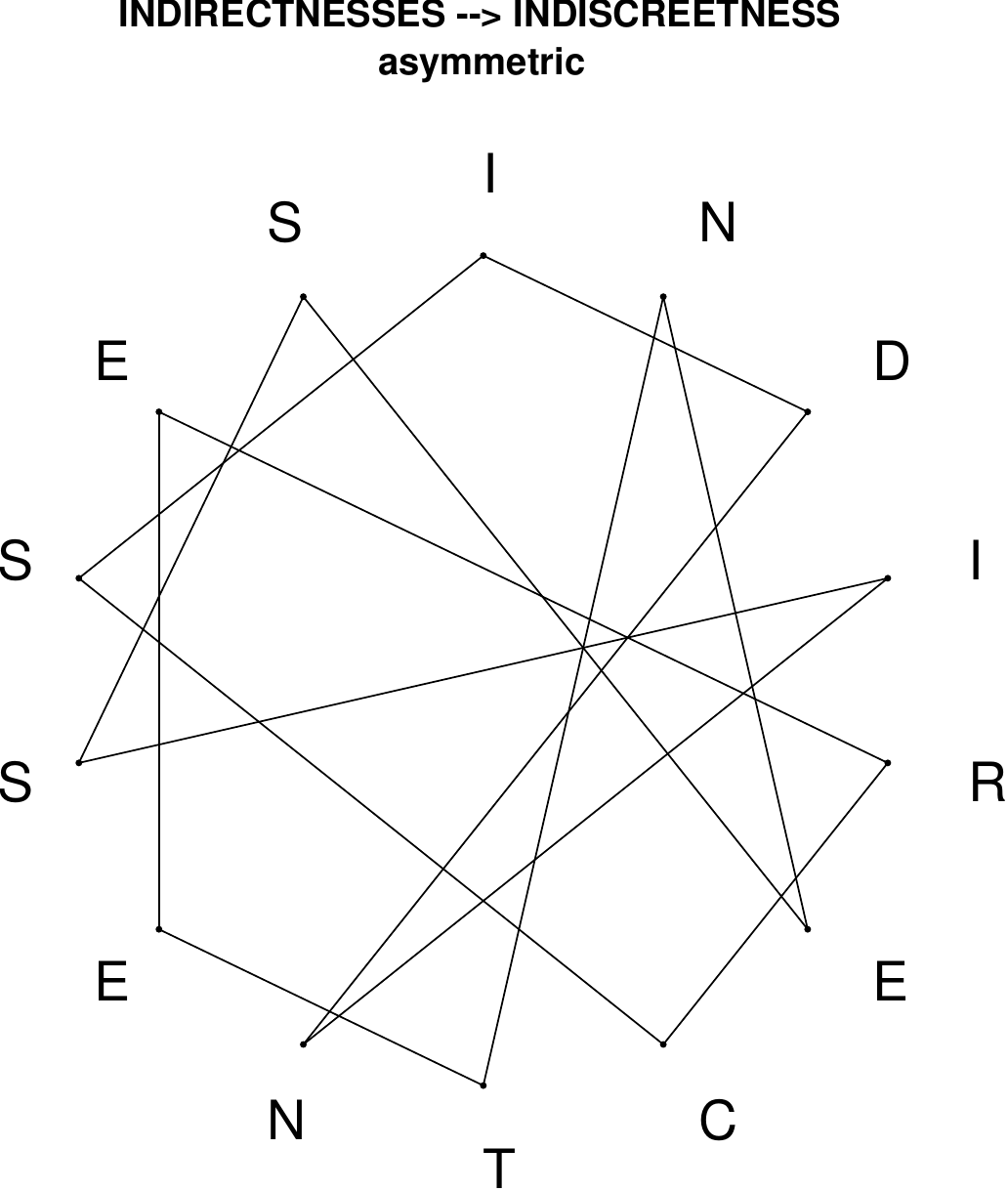}
\end{subfigure}
\hfill
\begin{subfigure}[T]{0.19\textwidth}
\centering
\includegraphics[width=\textwidth]{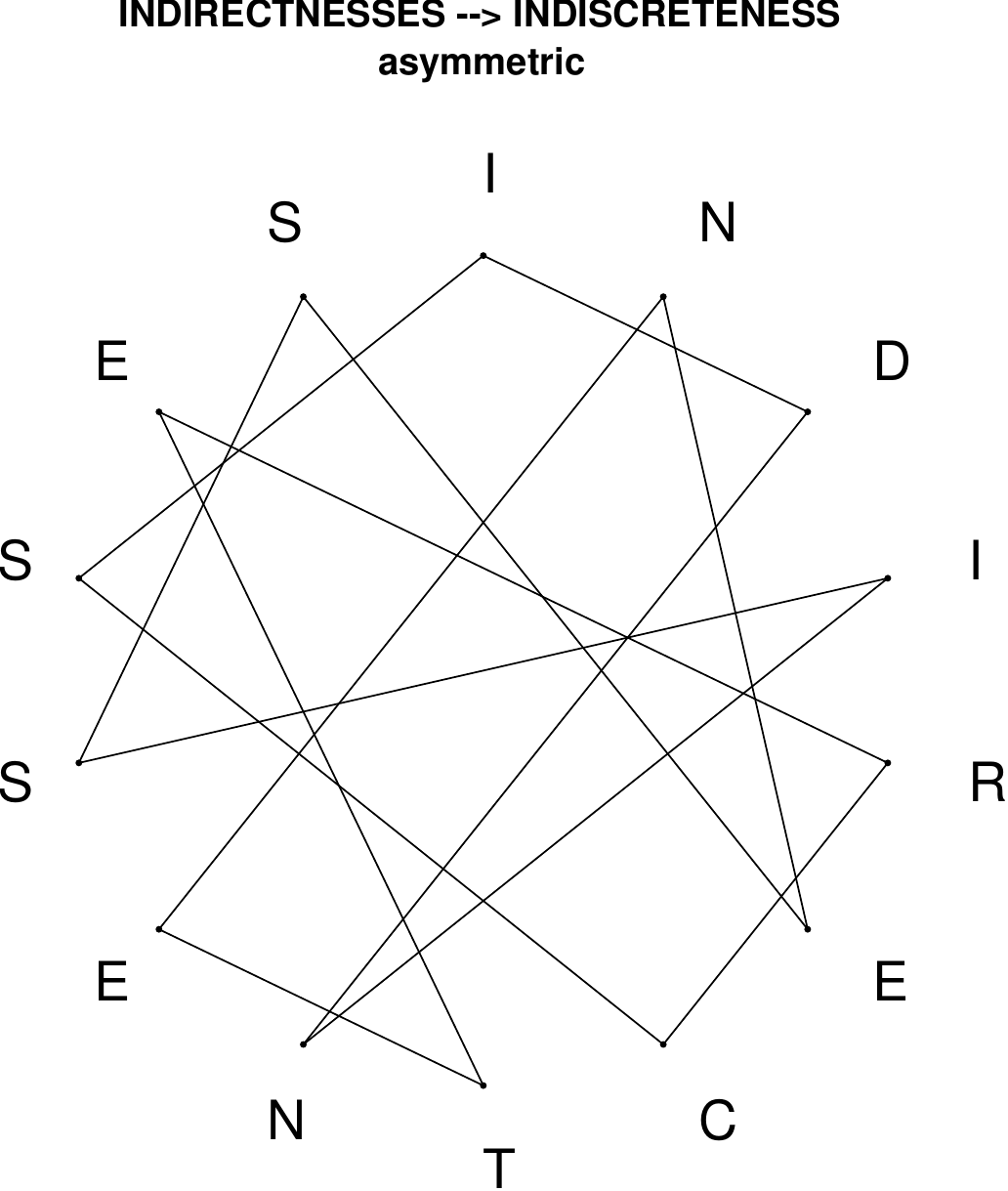}
\end{subfigure}
\hfill
\begin{subfigure}[T]{0.19\textwidth}
\centering
\includegraphics[width=\textwidth]{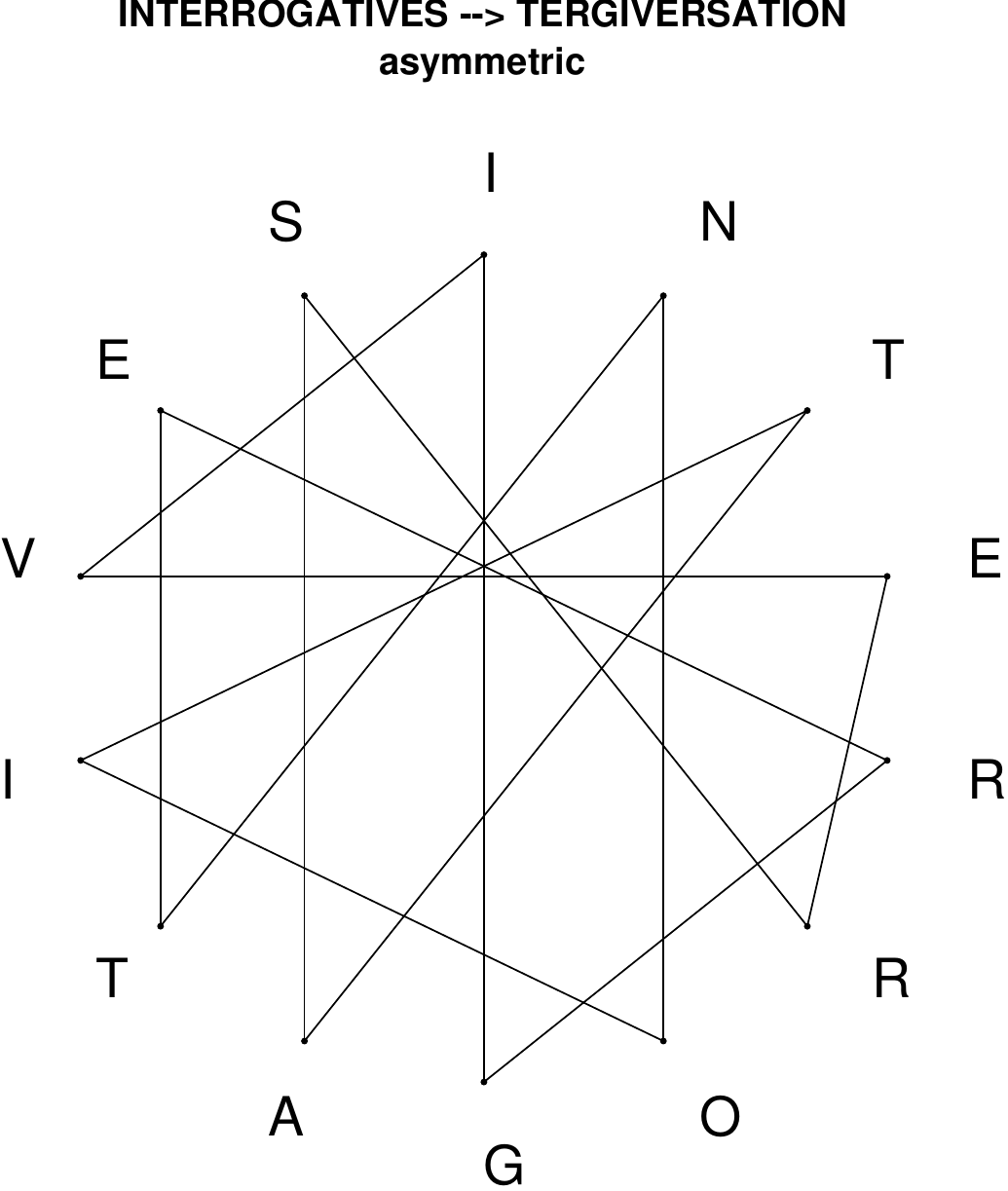}
\end{subfigure}
\hfill
\begin{subfigure}[T]{0.19\textwidth}
\centering
\includegraphics[width=\textwidth]{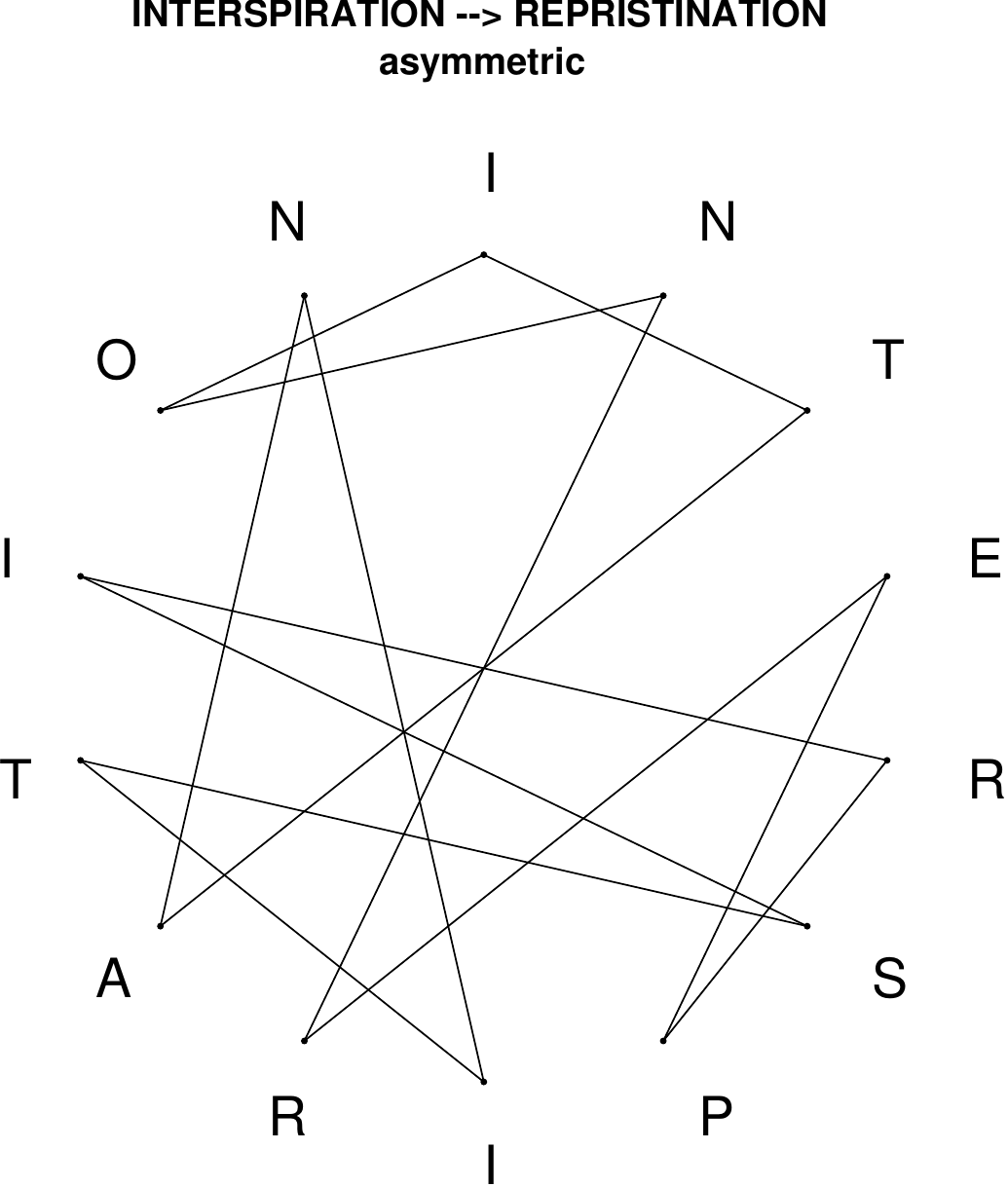}
\end{subfigure}
\end{figure}

\begin{figure}[H]
\centering
\begin{subfigure}[T]{0.19\textwidth}
\centering
\includegraphics[width=\textwidth]{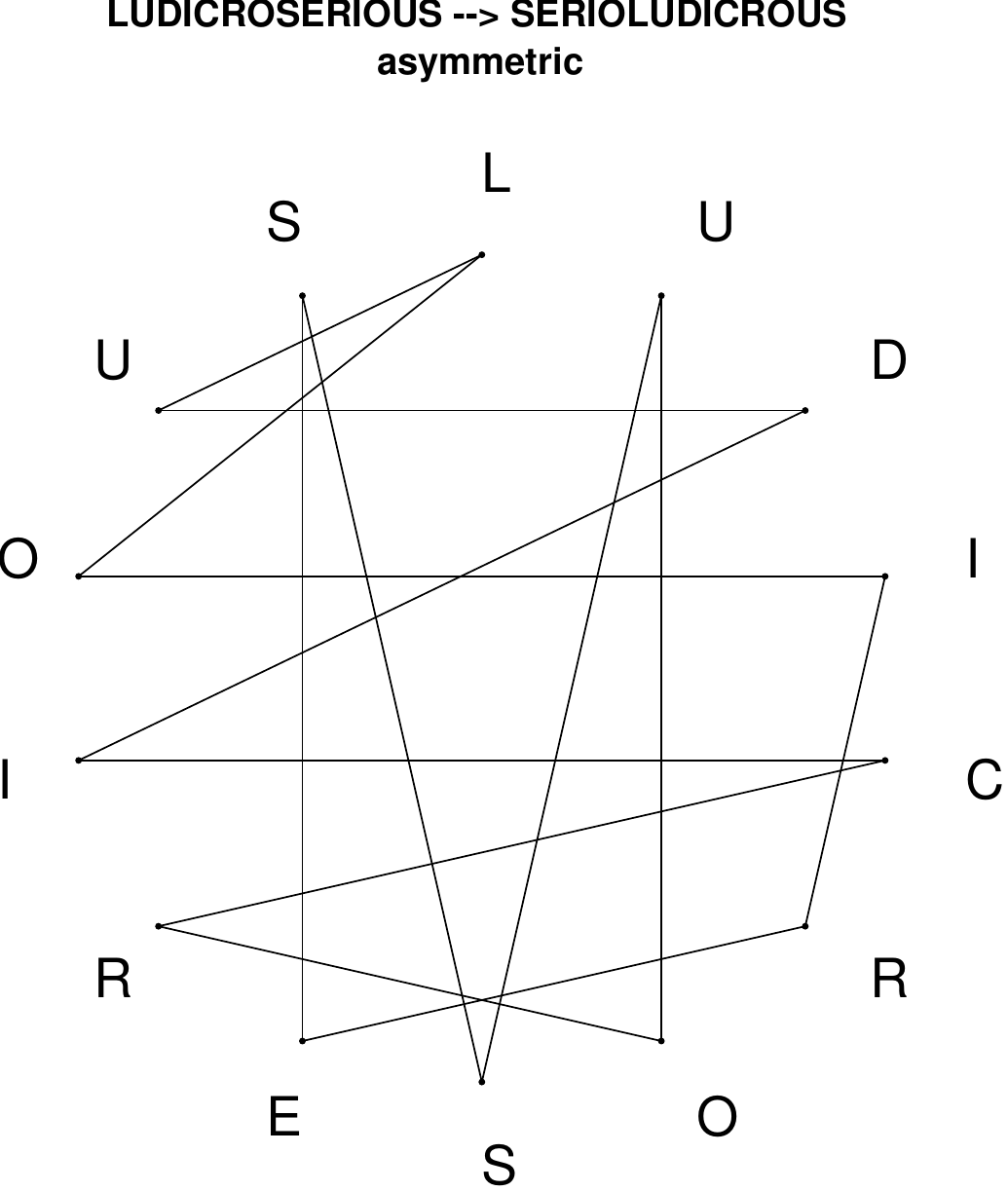}
\end{subfigure}
\hfill
\begin{subfigure}[T]{0.19\textwidth}
\centering
\includegraphics[width=\textwidth]{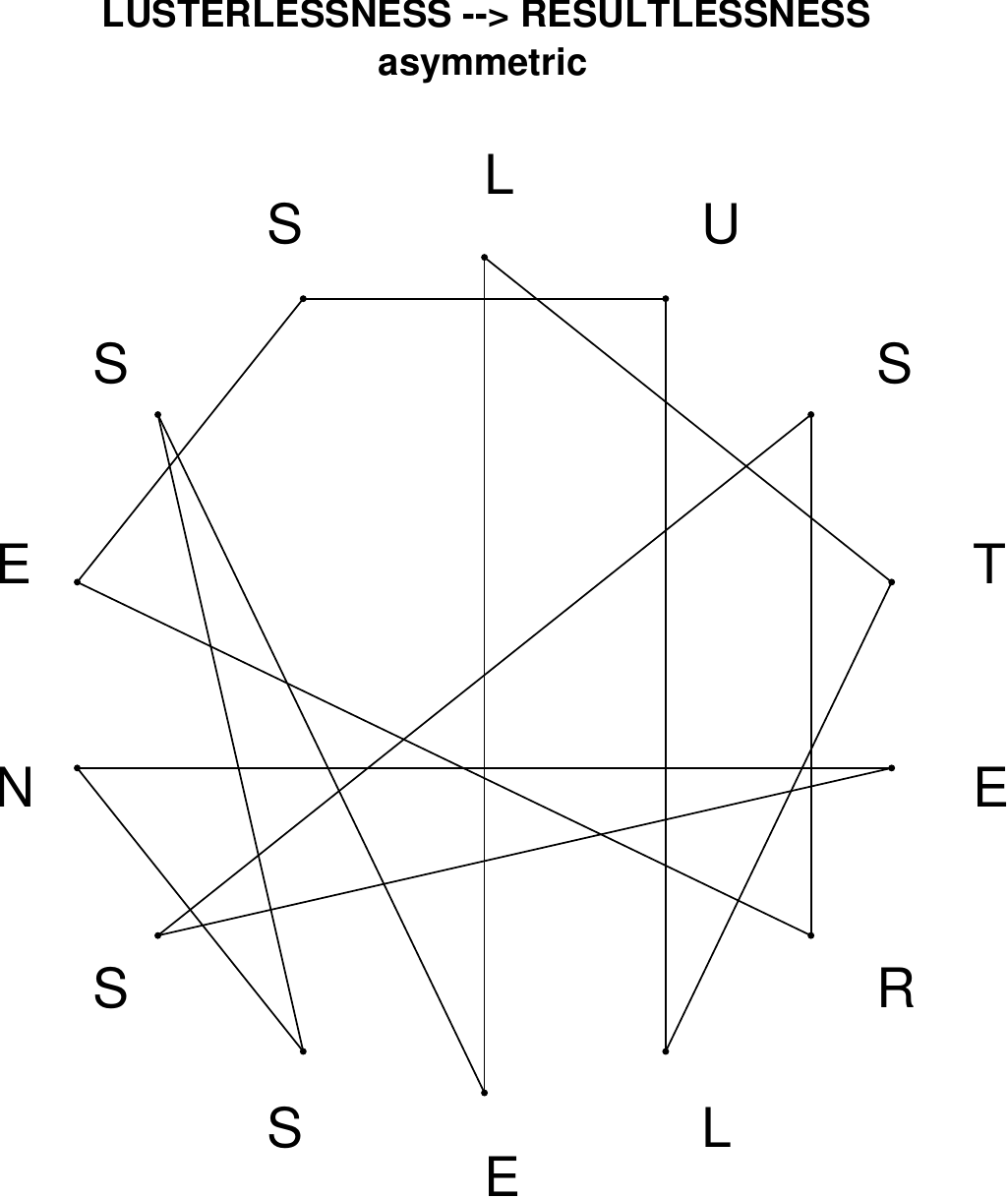}
\end{subfigure}
\hfill
\begin{subfigure}[T]{0.19\textwidth}
\centering
\includegraphics[width=\textwidth]{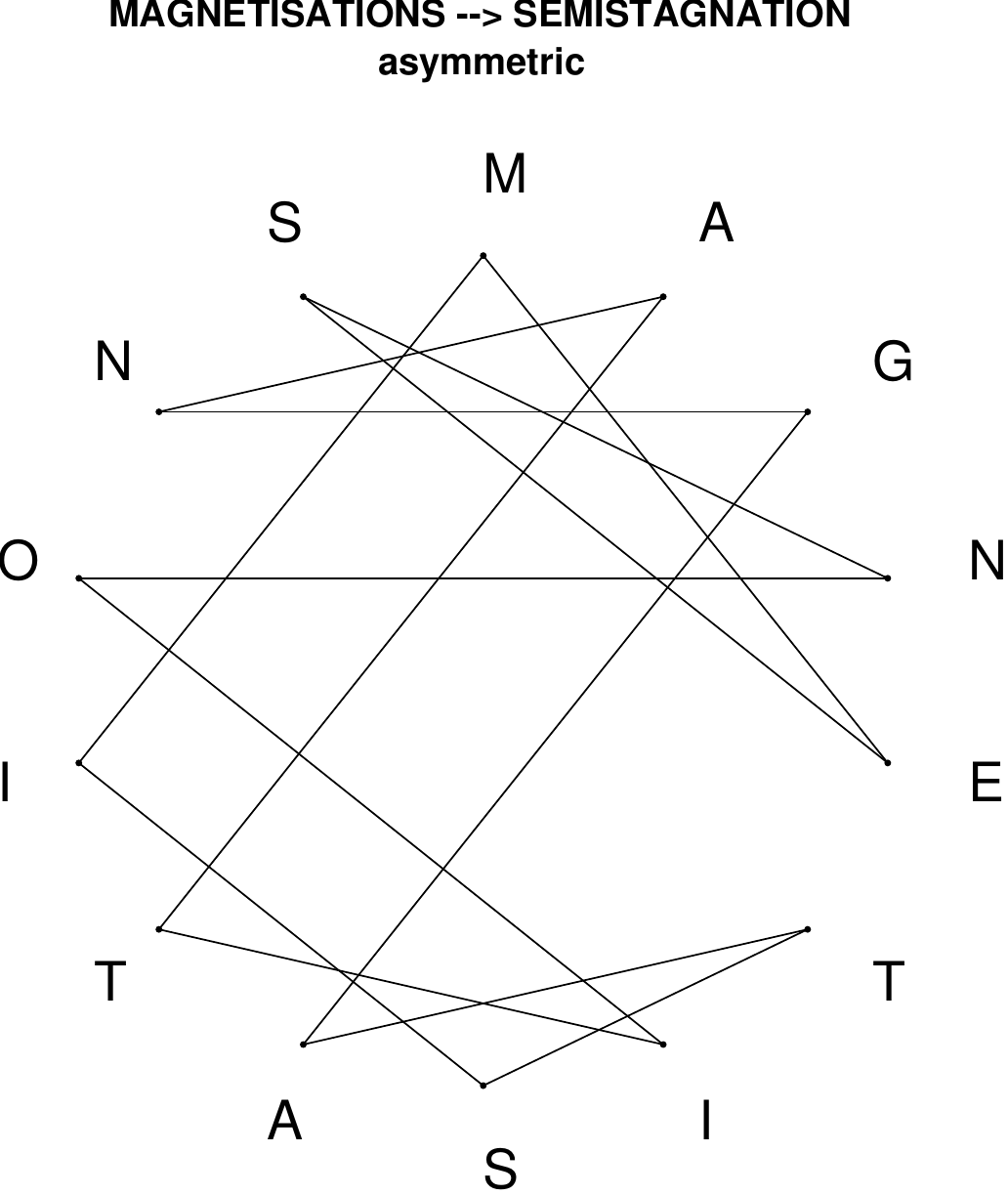}
\end{subfigure}
\hfill
\begin{subfigure}[T]{0.19\textwidth}
\centering
\includegraphics[width=\textwidth]{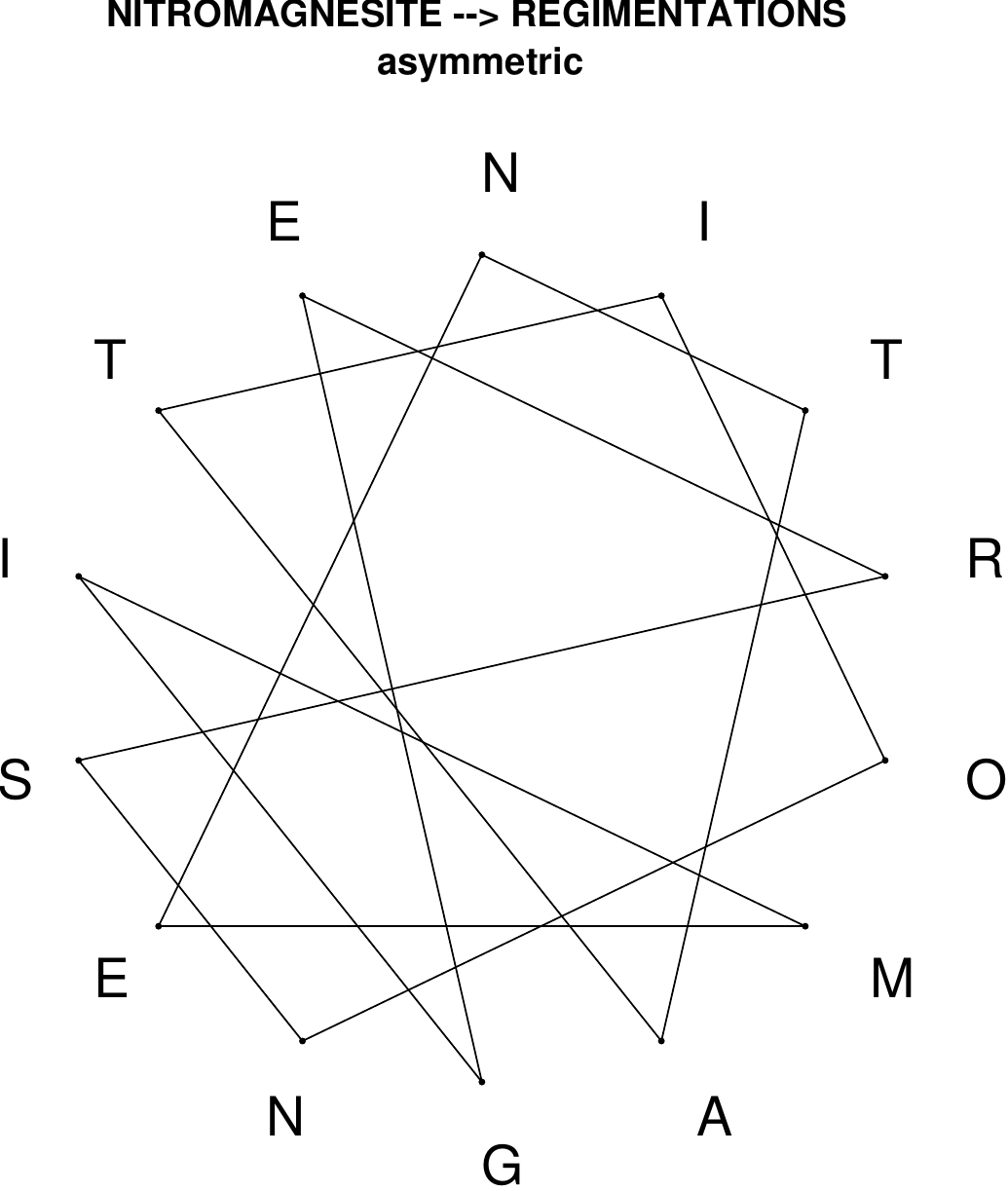}
\end{subfigure}
\hfill
\begin{subfigure}[T]{0.19\textwidth}
\centering
\includegraphics[width=\textwidth]{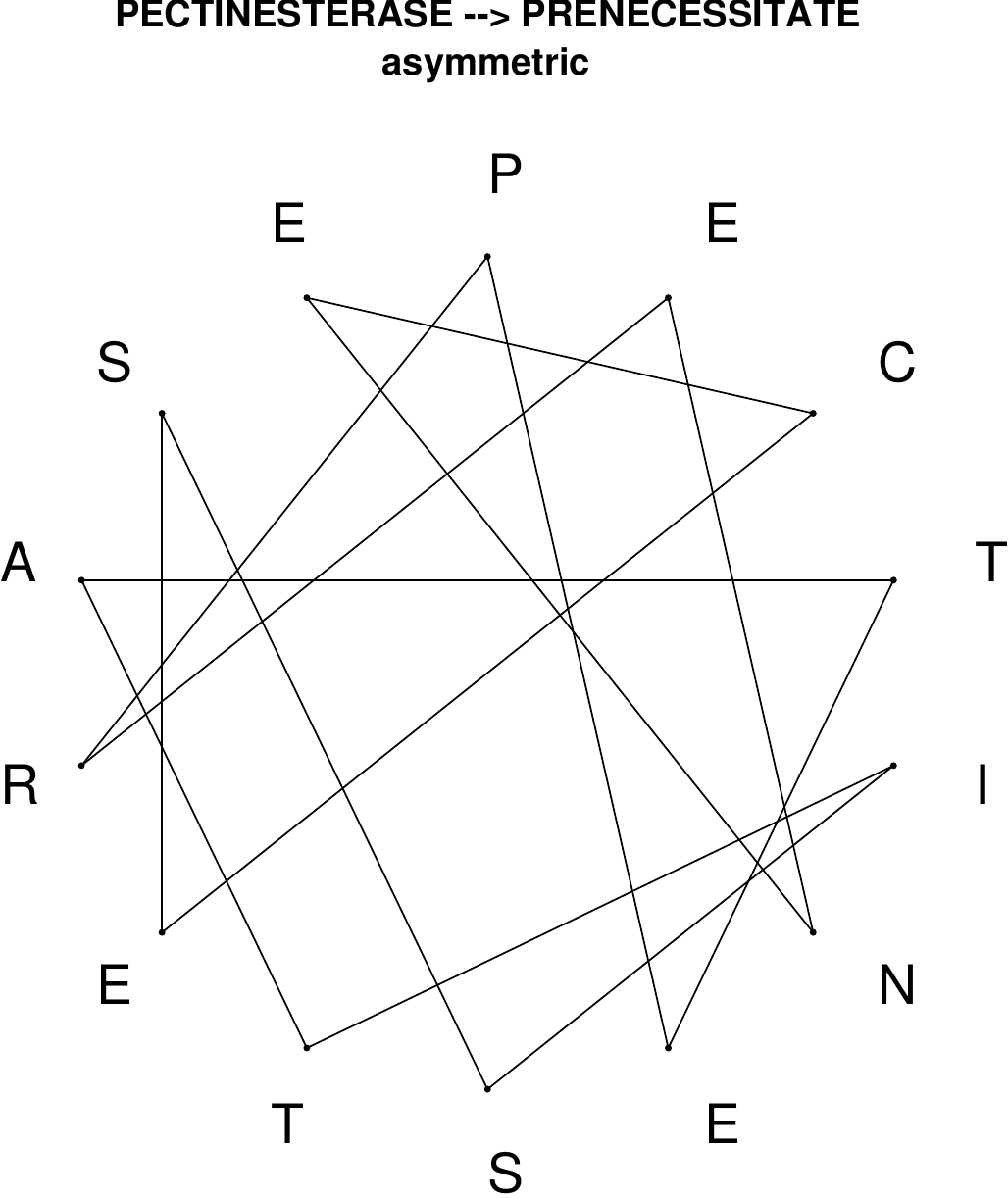}
\end{subfigure}
\end{figure}

\begin{figure}[H]
\centering
\begin{subfigure}[T]{0.19\textwidth}
\centering
\includegraphics[width=\textwidth]{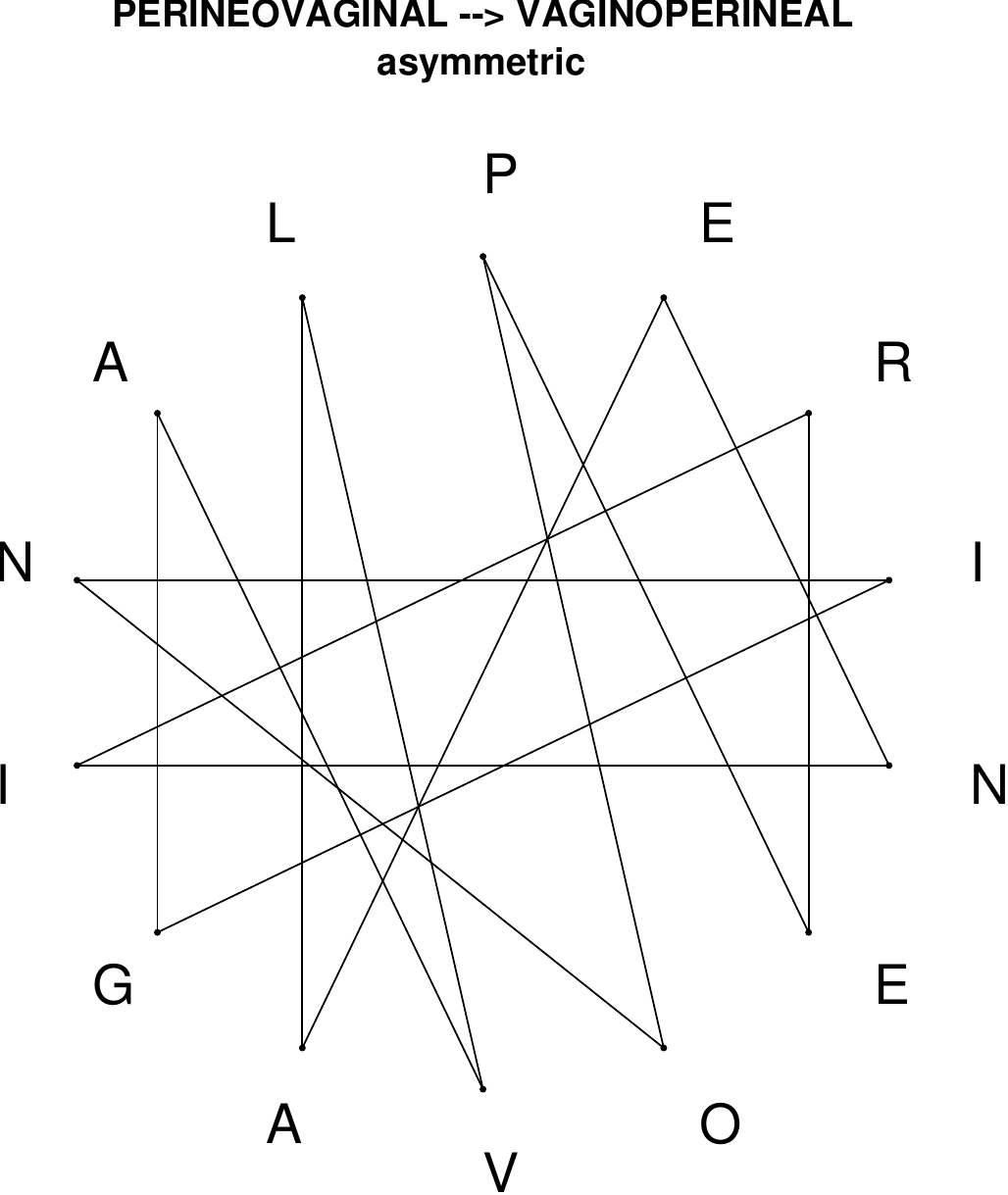}
\end{subfigure}
\hfill
\begin{subfigure}[T]{0.19\textwidth}
\centering
\includegraphics[width=\textwidth]{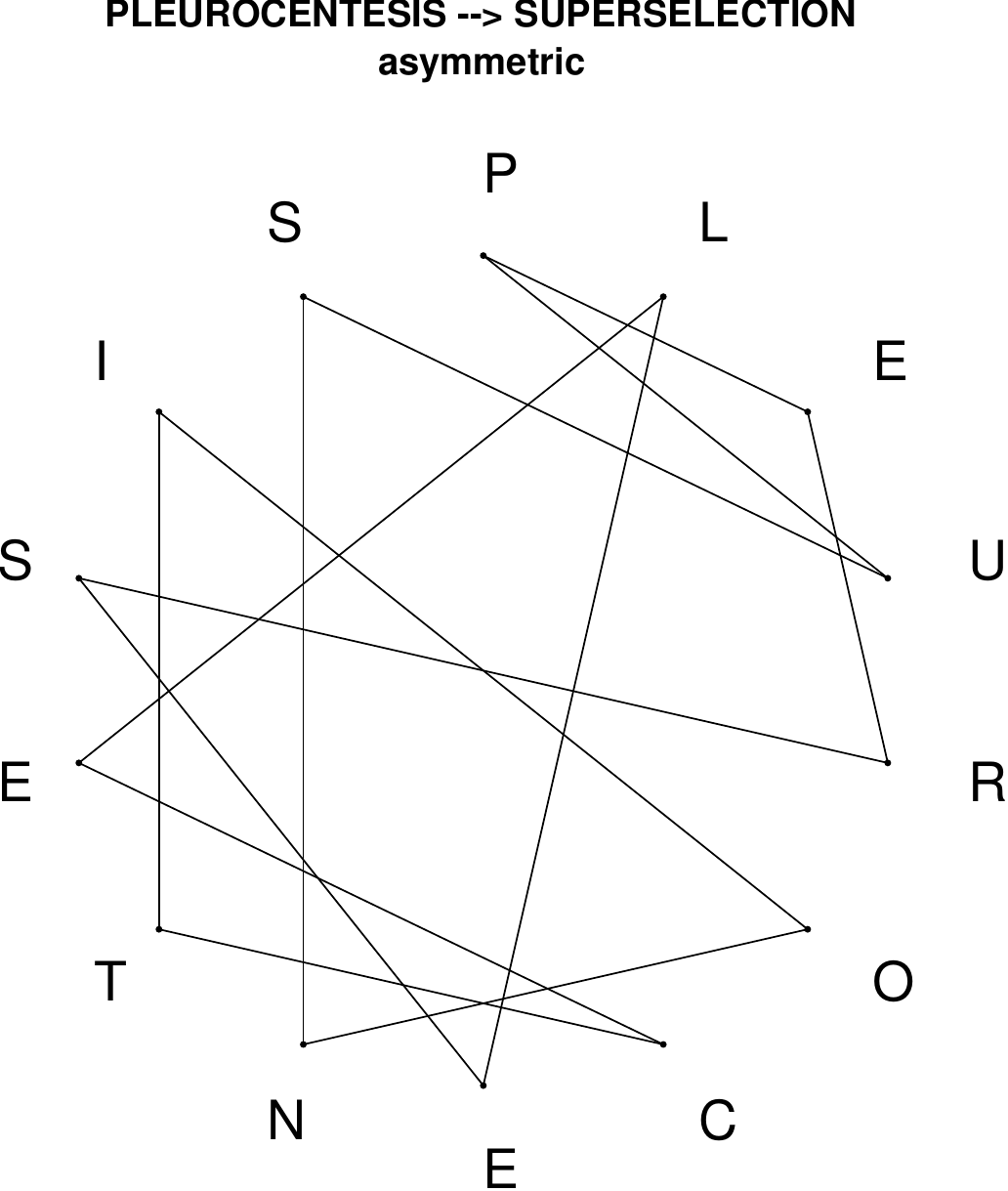}
\end{subfigure}
\hfill
\begin{subfigure}[T]{0.19\textwidth}
\centering
\includegraphics[width=\textwidth]{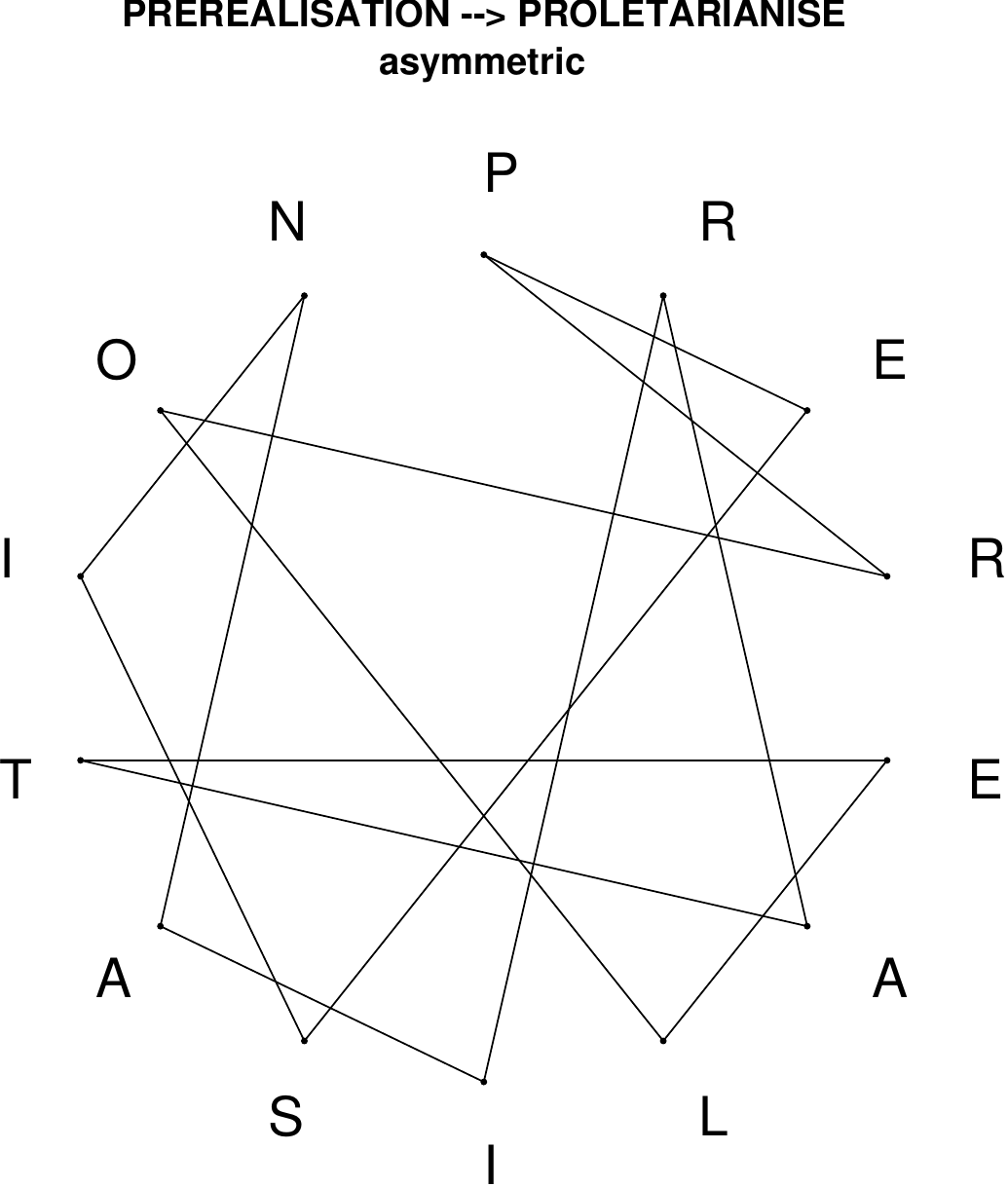}
\end{subfigure}
\hfill
\begin{subfigure}[T]{0.19\textwidth}
\centering
\includegraphics[width=\textwidth]{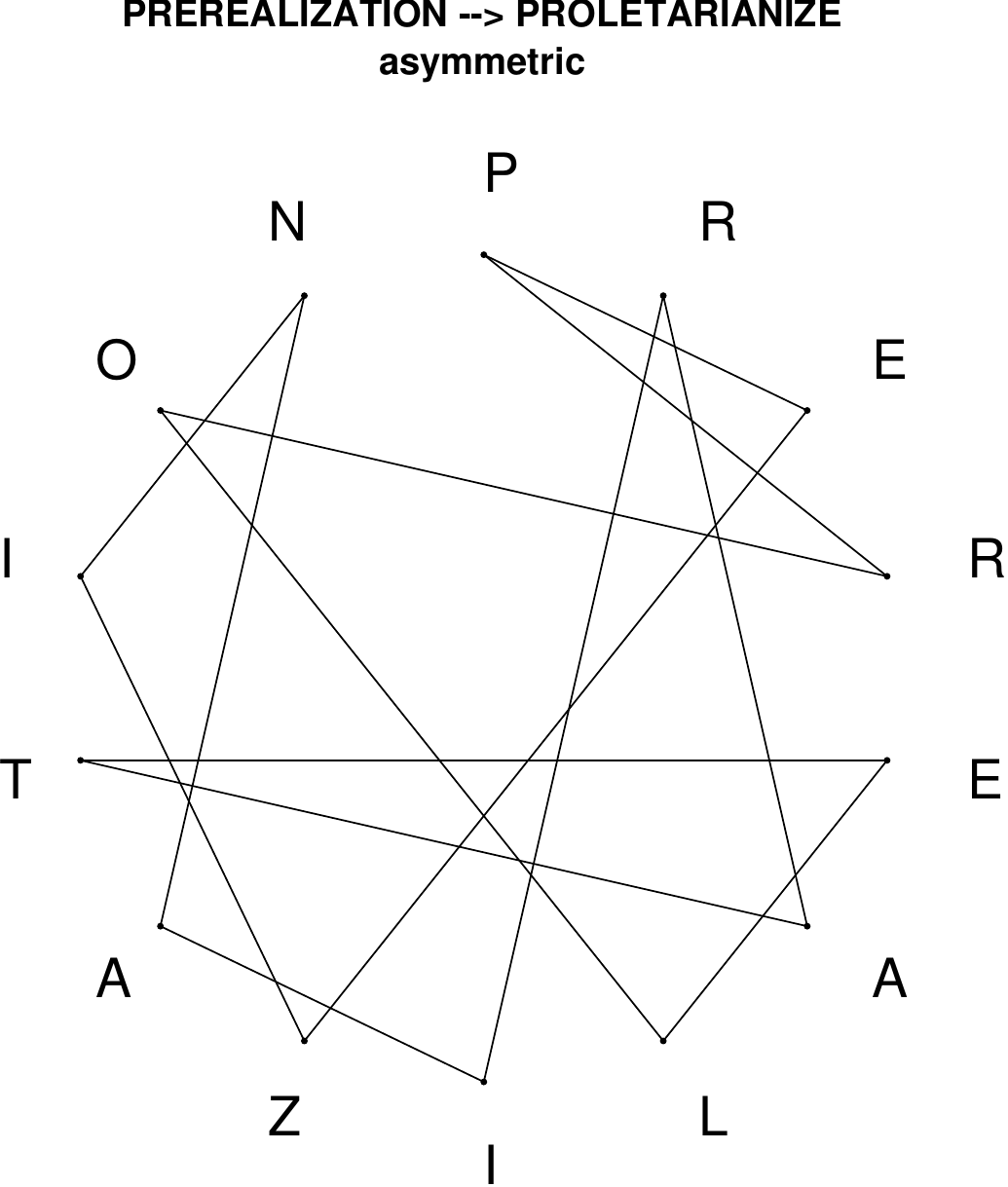}
\end{subfigure}
\hfill
\begin{subfigure}[T]{0.19\textwidth}
\centering
\includegraphics[width=\textwidth]{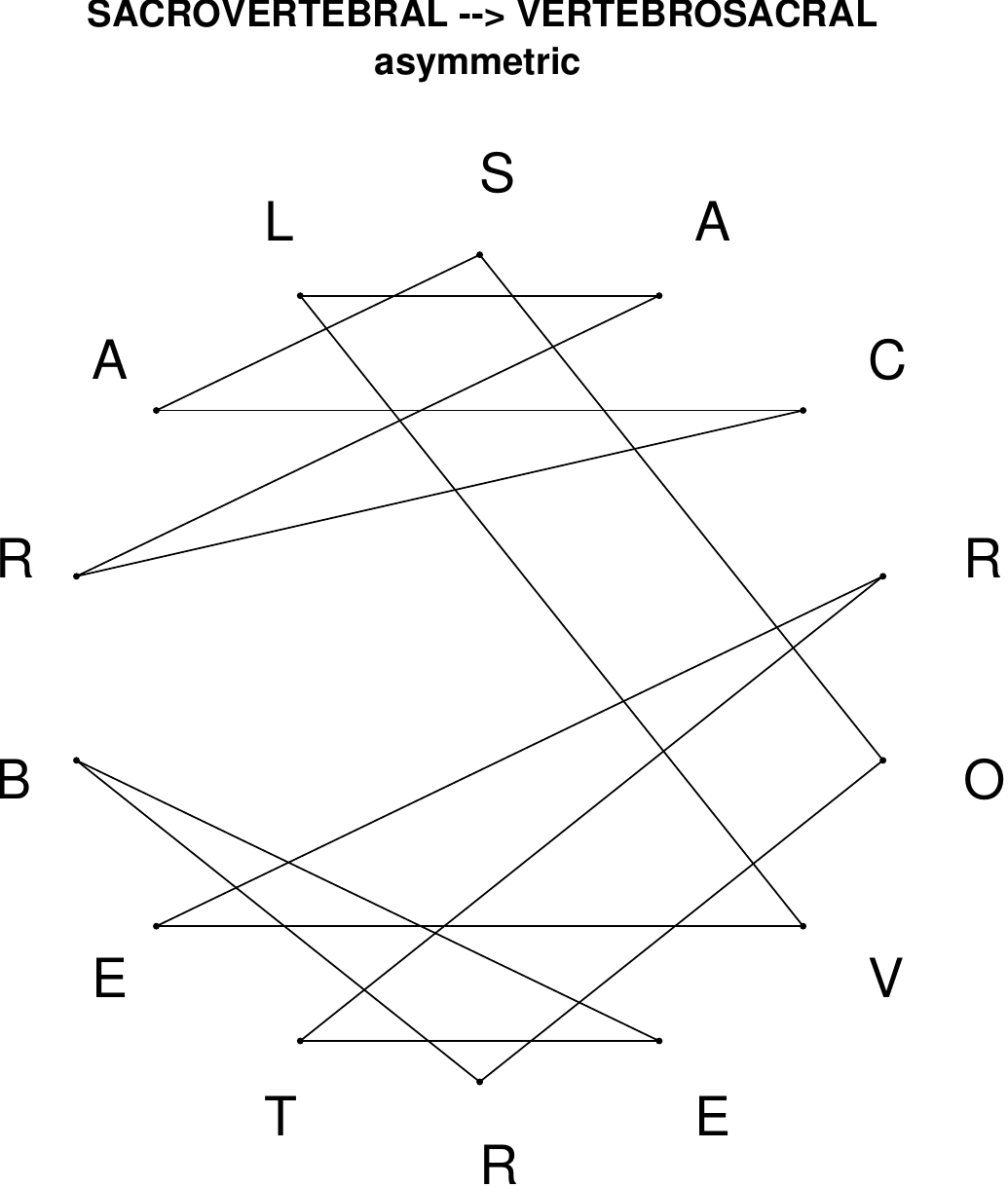}
\end{subfigure}
\end{figure}

\begin{figure}[H]
\centering
\begin{subfigure}[T]{0.19\textwidth}
\centering
\includegraphics[width=\textwidth]{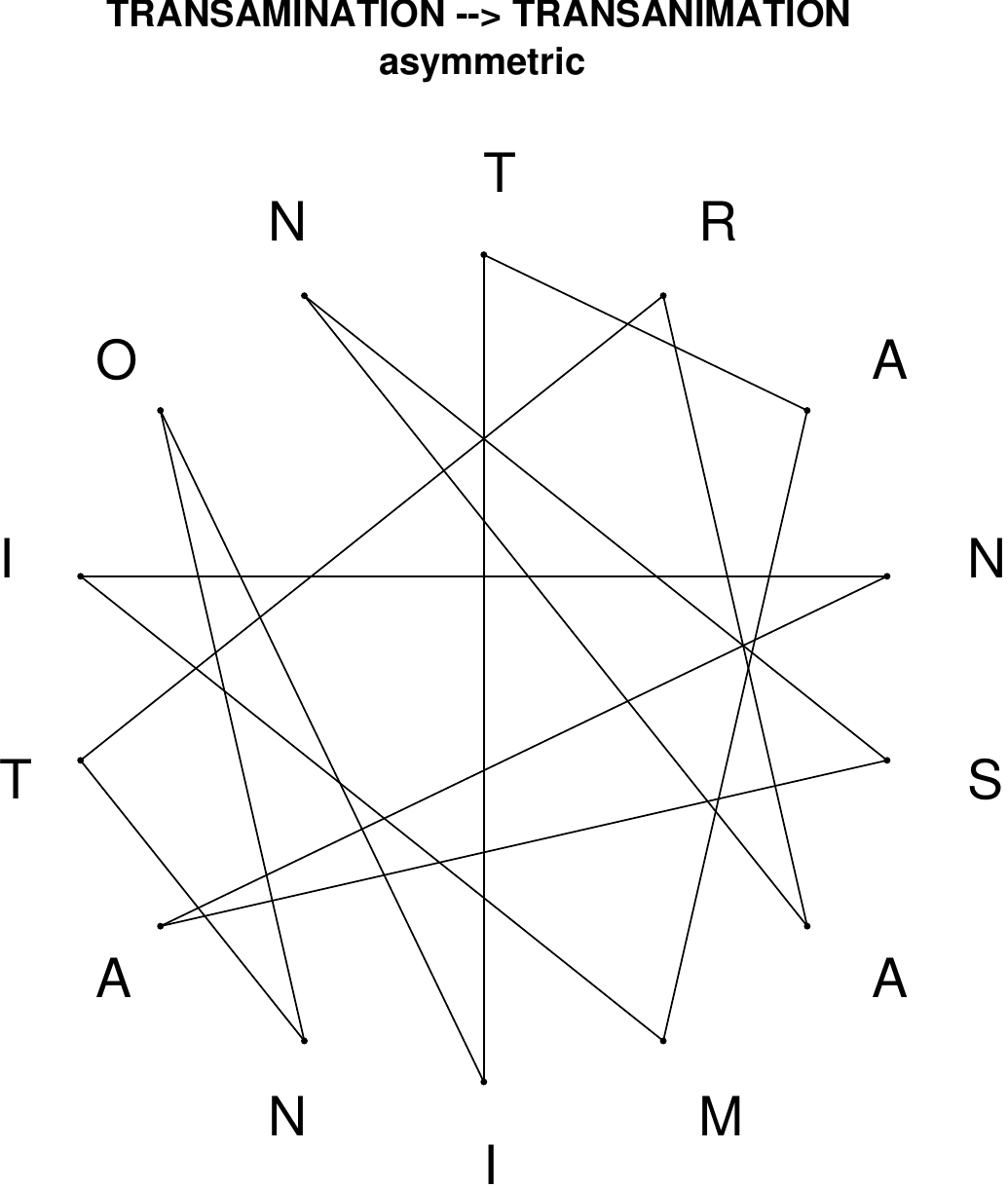}
\end{subfigure}
\hfill
\end{figure}

%%%%%%%%%%%%%%%%%%
\clearpage
\subsection{Star Anagrams $N = 13$}
For $N=13$, we found a single symmetric star among a few dozen asymmetric stars. 

\subsubsection{Symmetric Stars $N=13$}

\begin{figure}[H]
\centering
\begin{subfigure}[T]{0.19\textwidth}
\centering
\includegraphics[width=\textwidth]{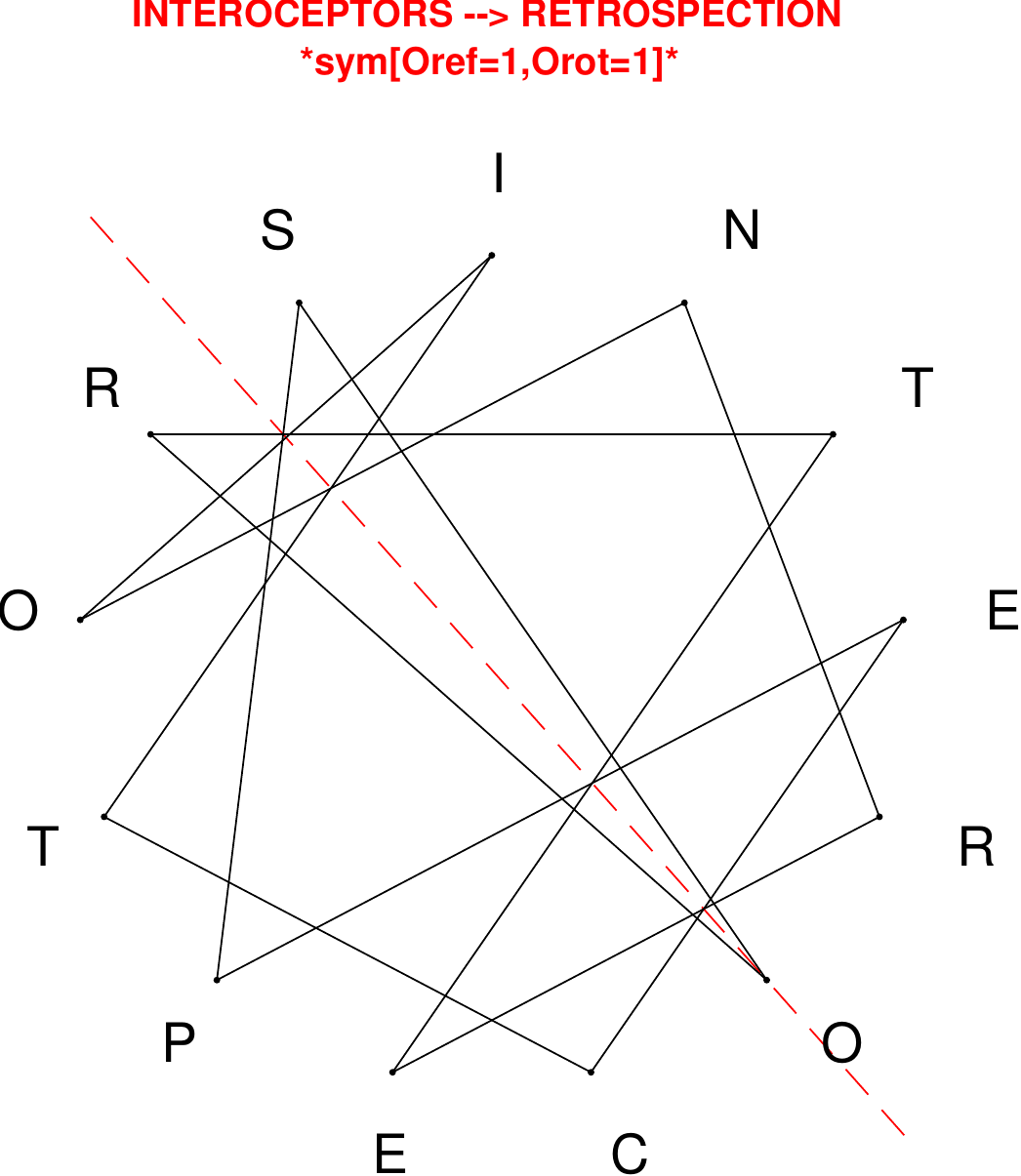}
\end{subfigure}
\hfill
\end{figure}

\subsubsection{Asymmetric Stars $N=13$}

\begin{figure}[H]
\centering
\begin{subfigure}[T]{0.19\textwidth}
\centering
\includegraphics[width=\textwidth]{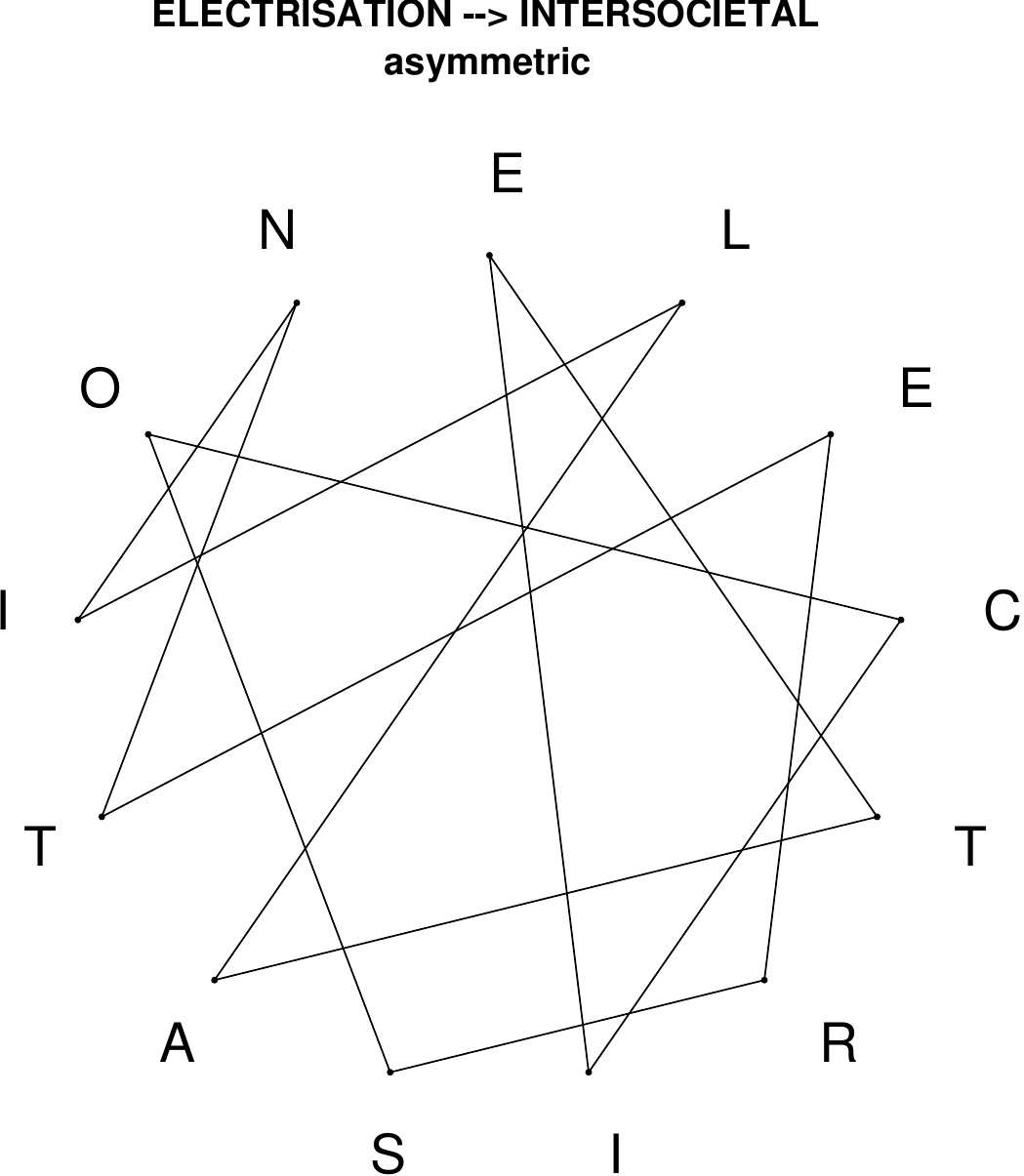}
\end{subfigure}
\hfill
\begin{subfigure}[T]{0.19\textwidth}
\centering
\includegraphics[width=\textwidth]{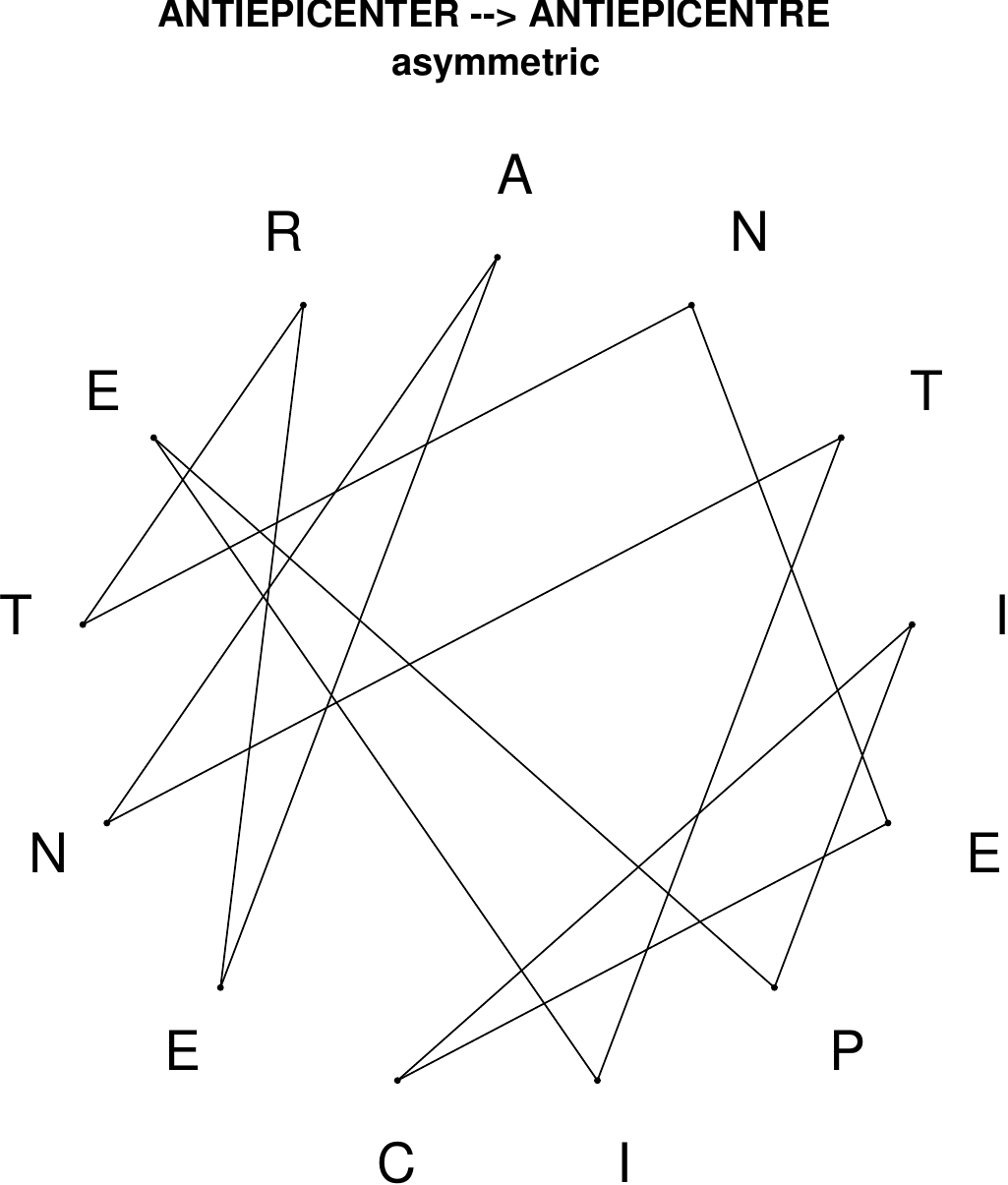}
\end{subfigure}
\hfill
\begin{subfigure}[T]{0.19\textwidth}
\centering
\includegraphics[width=\textwidth]{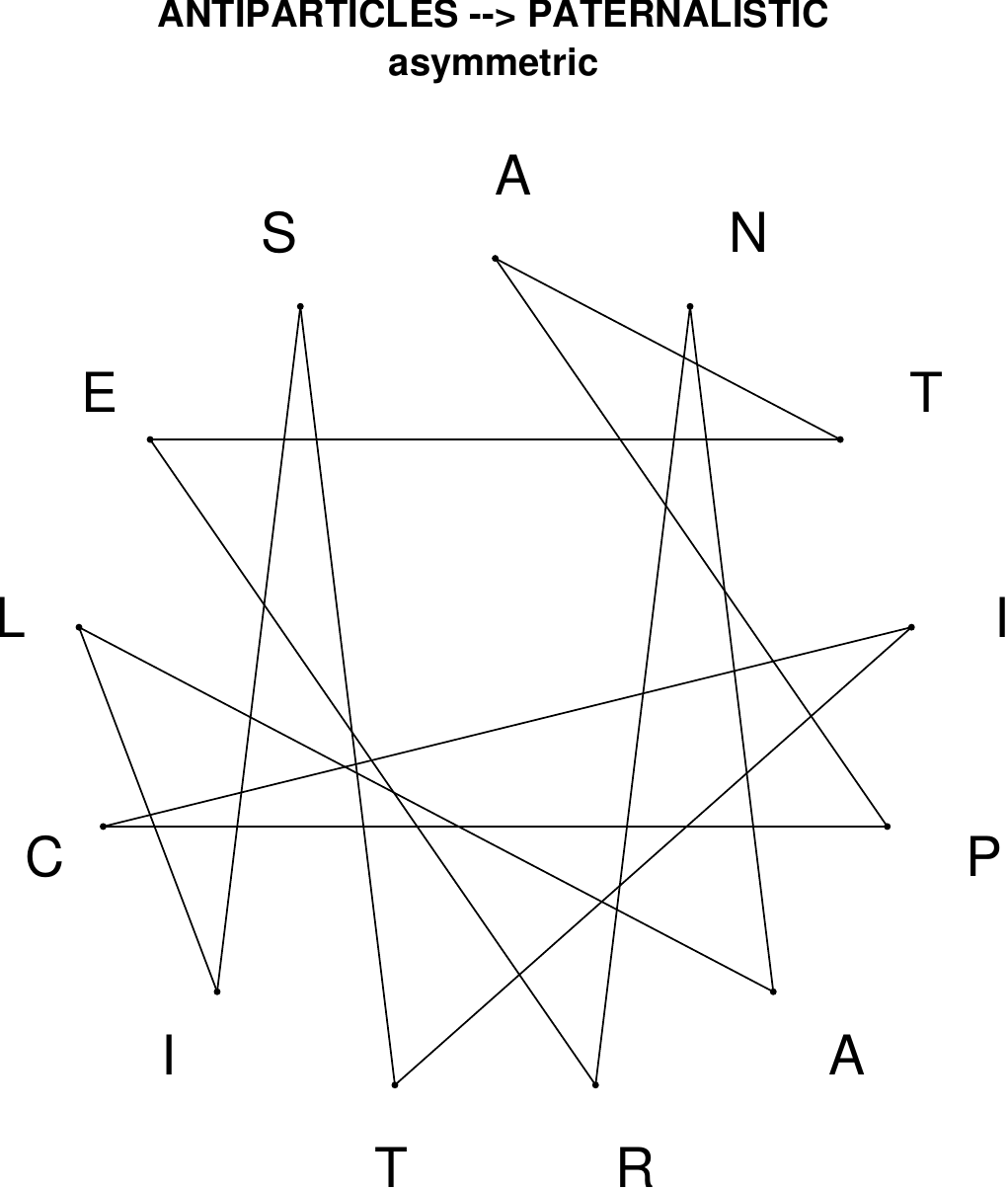}
\end{subfigure}
\hfill
\begin{subfigure}[T]{0.19\textwidth}
\centering
\includegraphics[width=\textwidth]{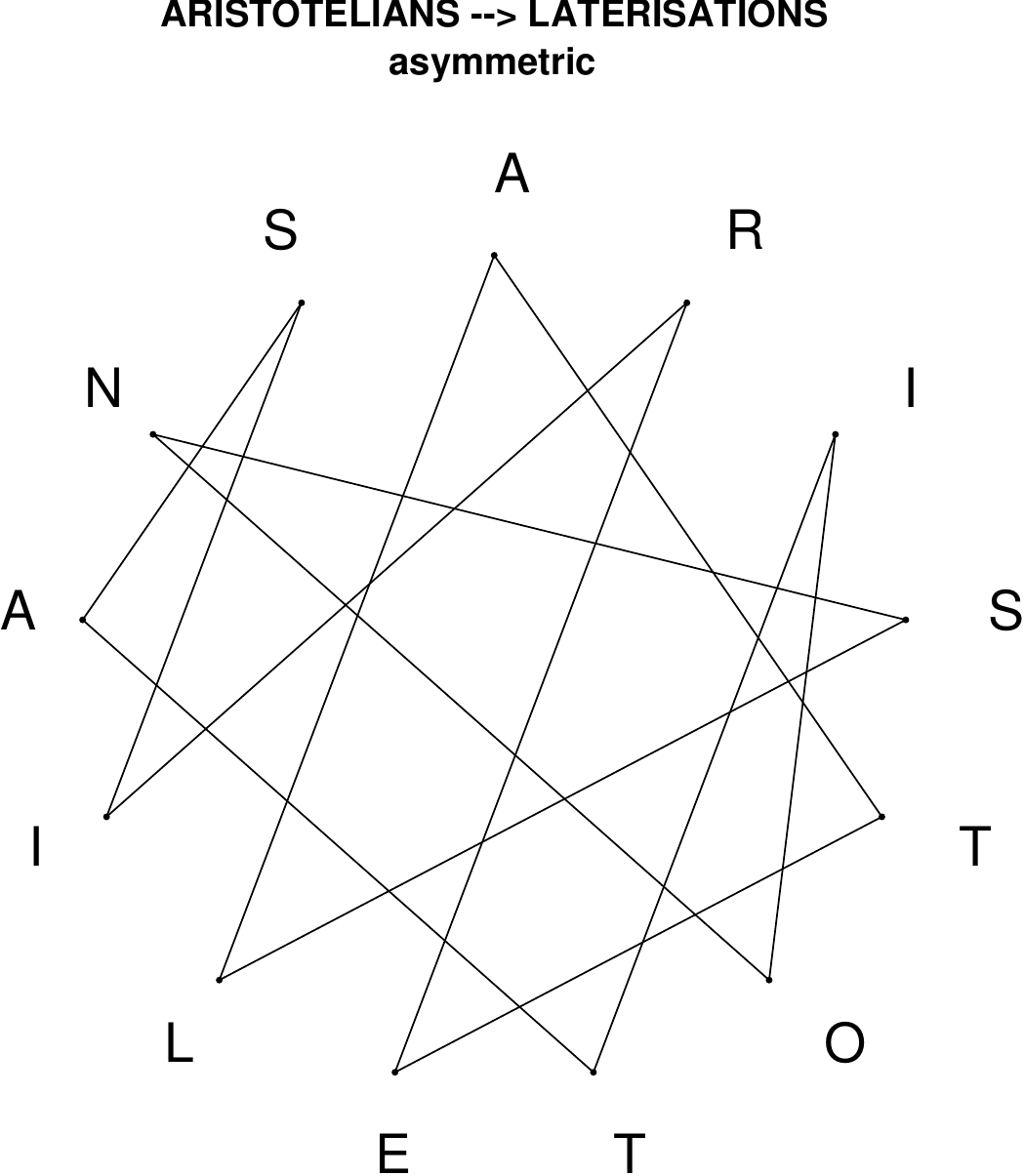}
\end{subfigure}
\hfill
\begin{subfigure}[T]{0.19\textwidth}
\centering
\includegraphics[width=\textwidth]{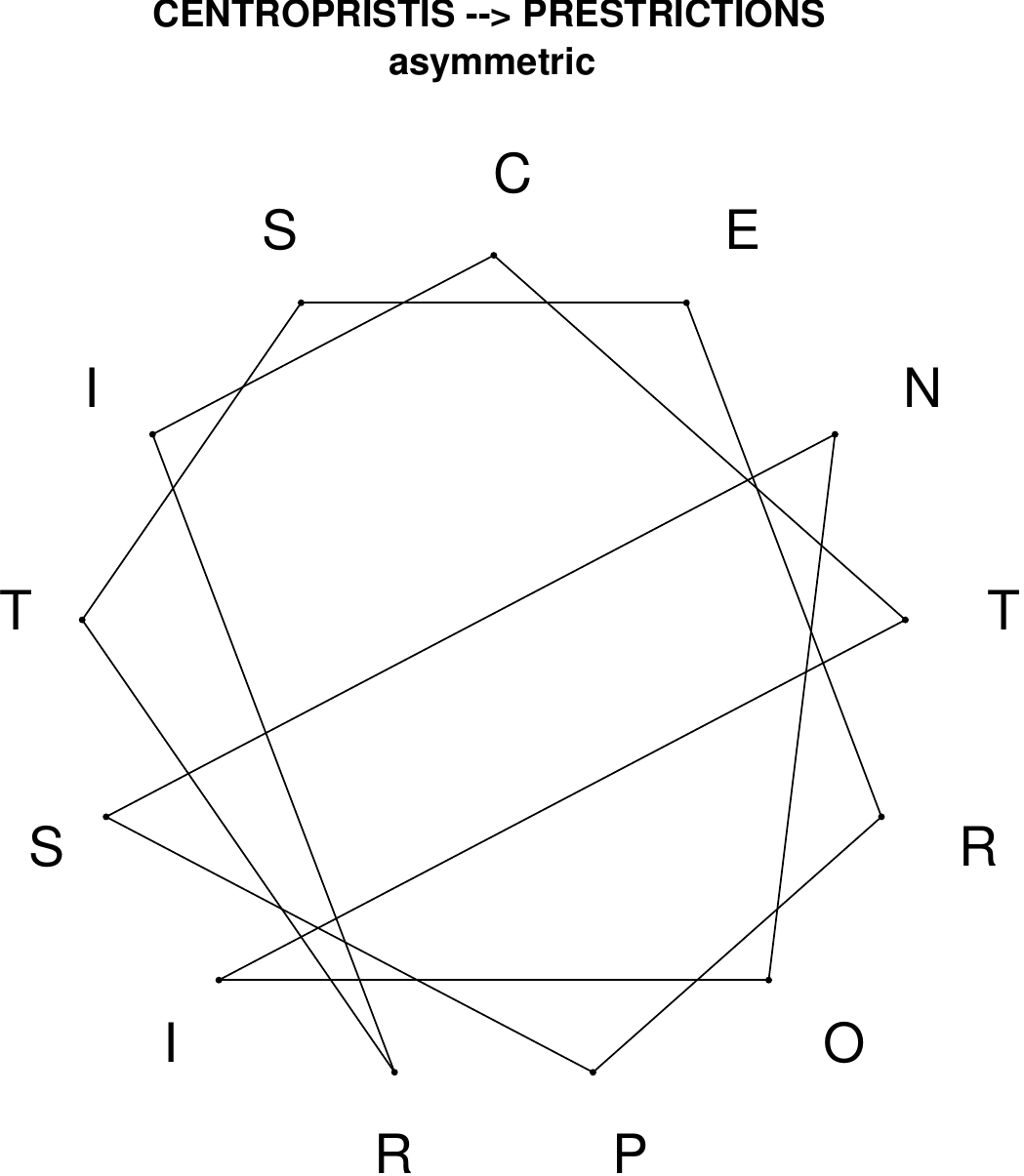}
\end{subfigure}
\end{figure}

\begin{figure}[H]
\centering
\begin{subfigure}[T]{0.19\textwidth}
\centering
\includegraphics[width=\textwidth]{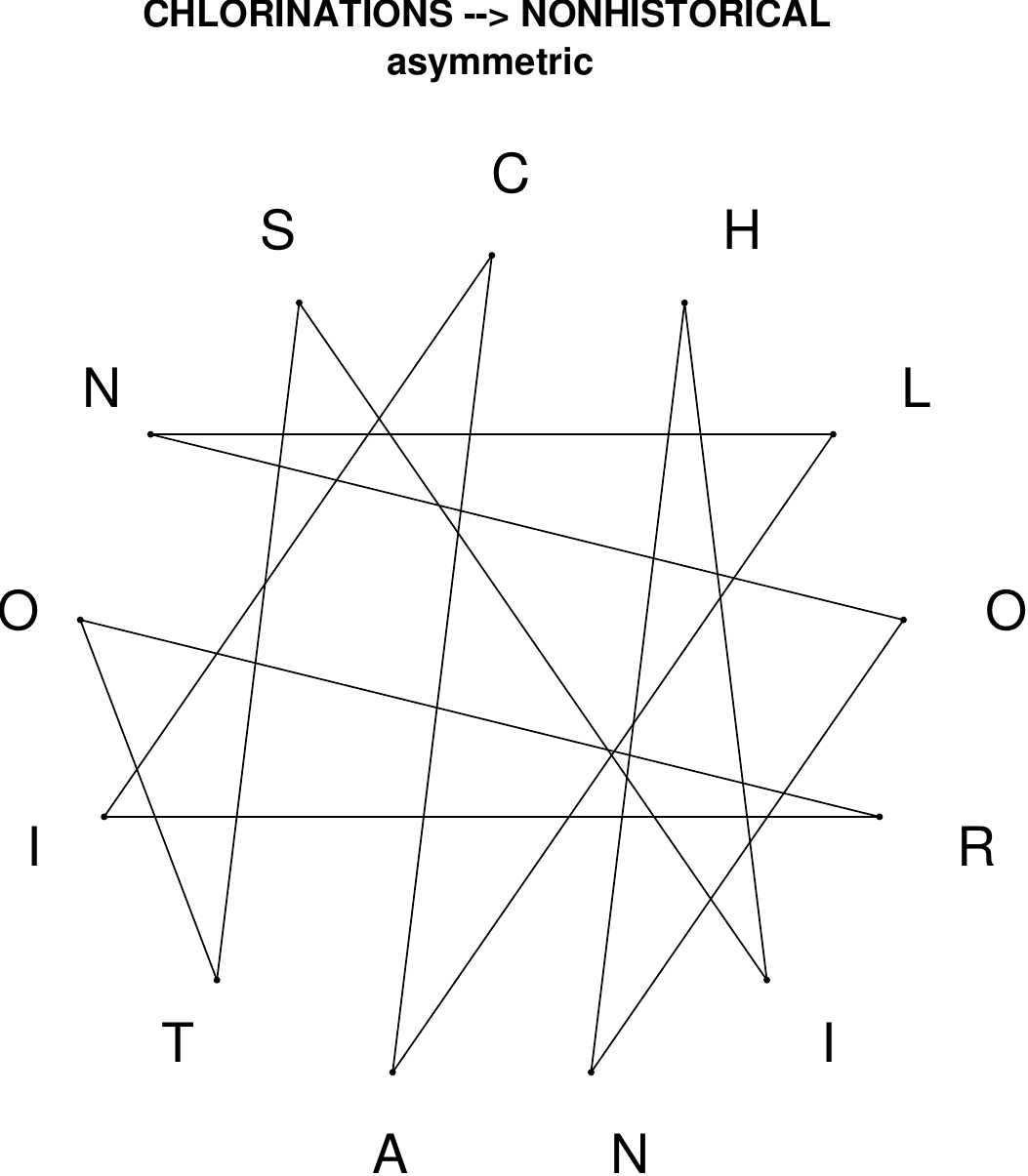}
\end{subfigure}
\hfill
\begin{subfigure}[T]{0.19\textwidth}
\centering
\includegraphics[width=\textwidth]{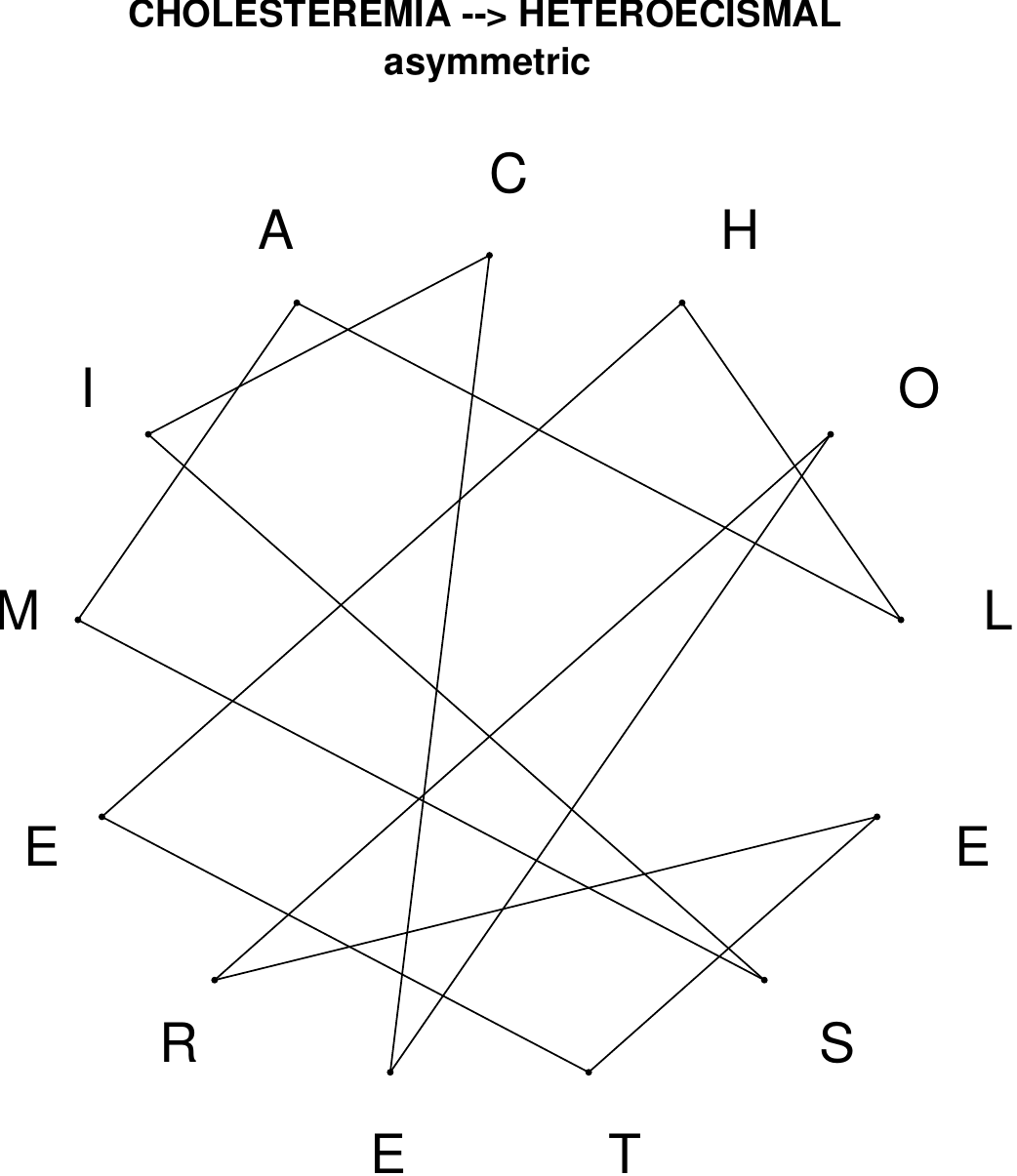}
\end{subfigure}
\hfill
\begin{subfigure}[T]{0.19\textwidth}
\centering
\includegraphics[width=\textwidth]{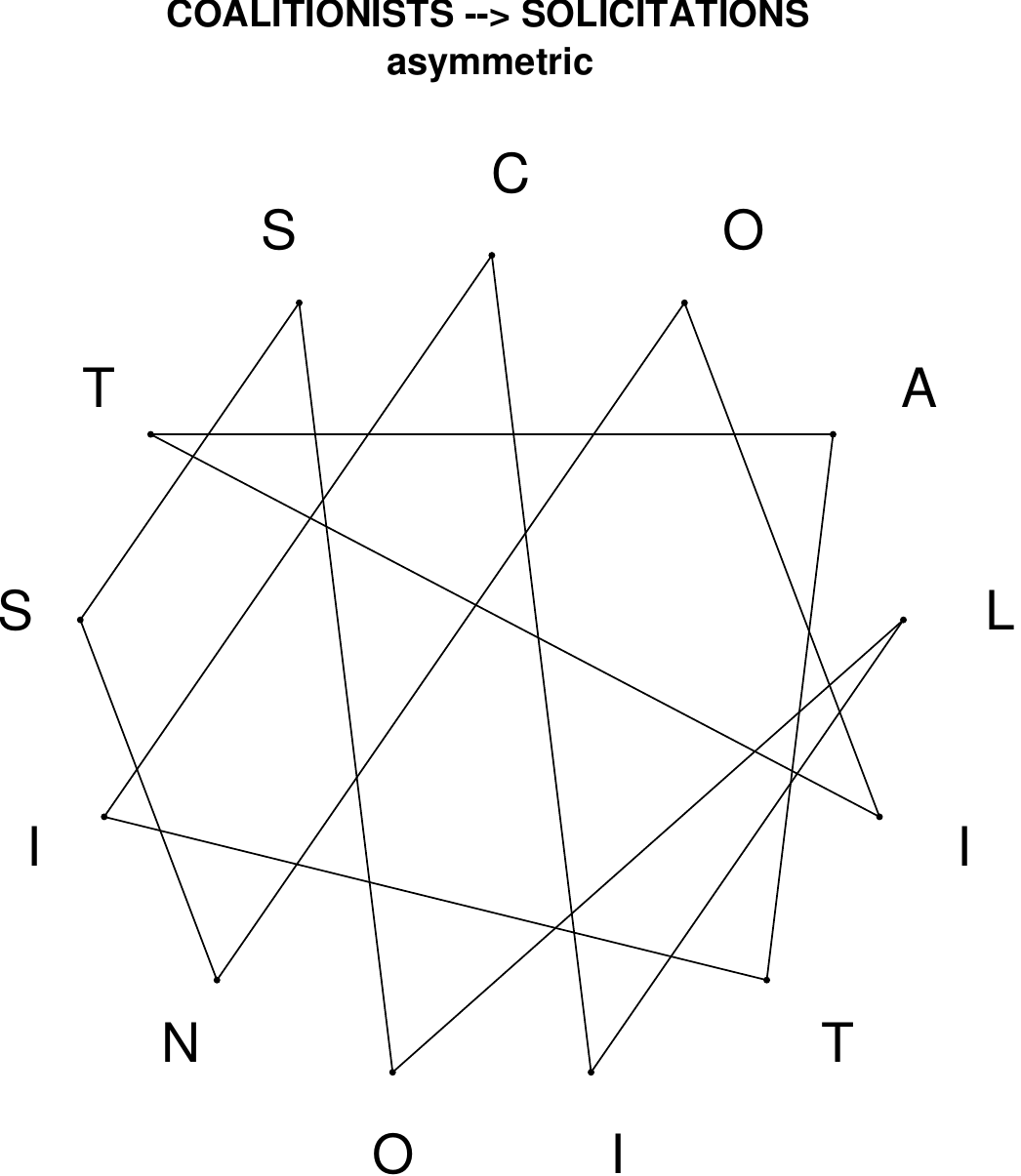}
\end{subfigure}
\hfill
\begin{subfigure}[T]{0.19\textwidth}
\centering
\includegraphics[width=\textwidth]{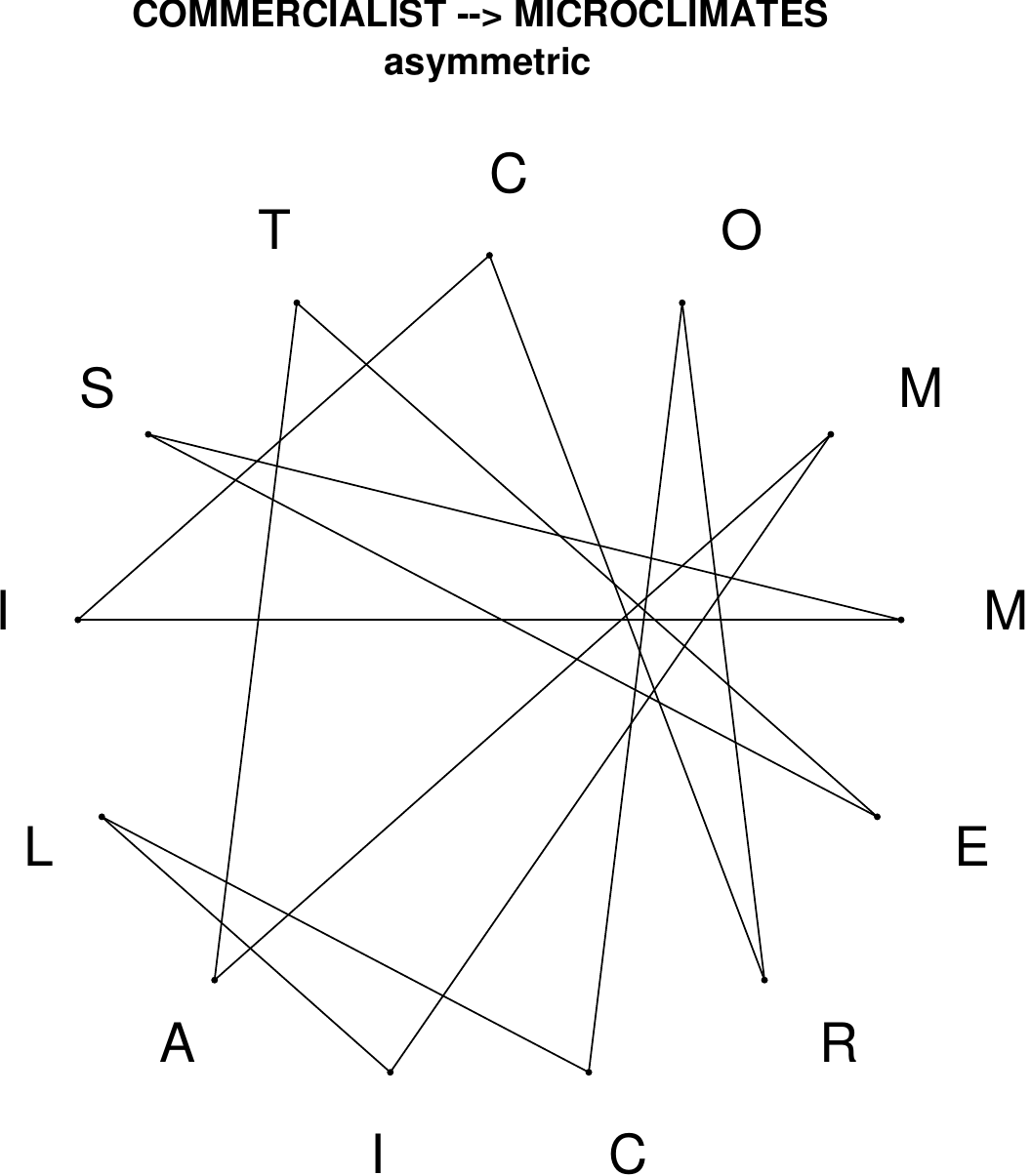}
\end{subfigure}
\hfill
\begin{subfigure}[T]{0.19\textwidth}
\centering
\includegraphics[width=\textwidth]{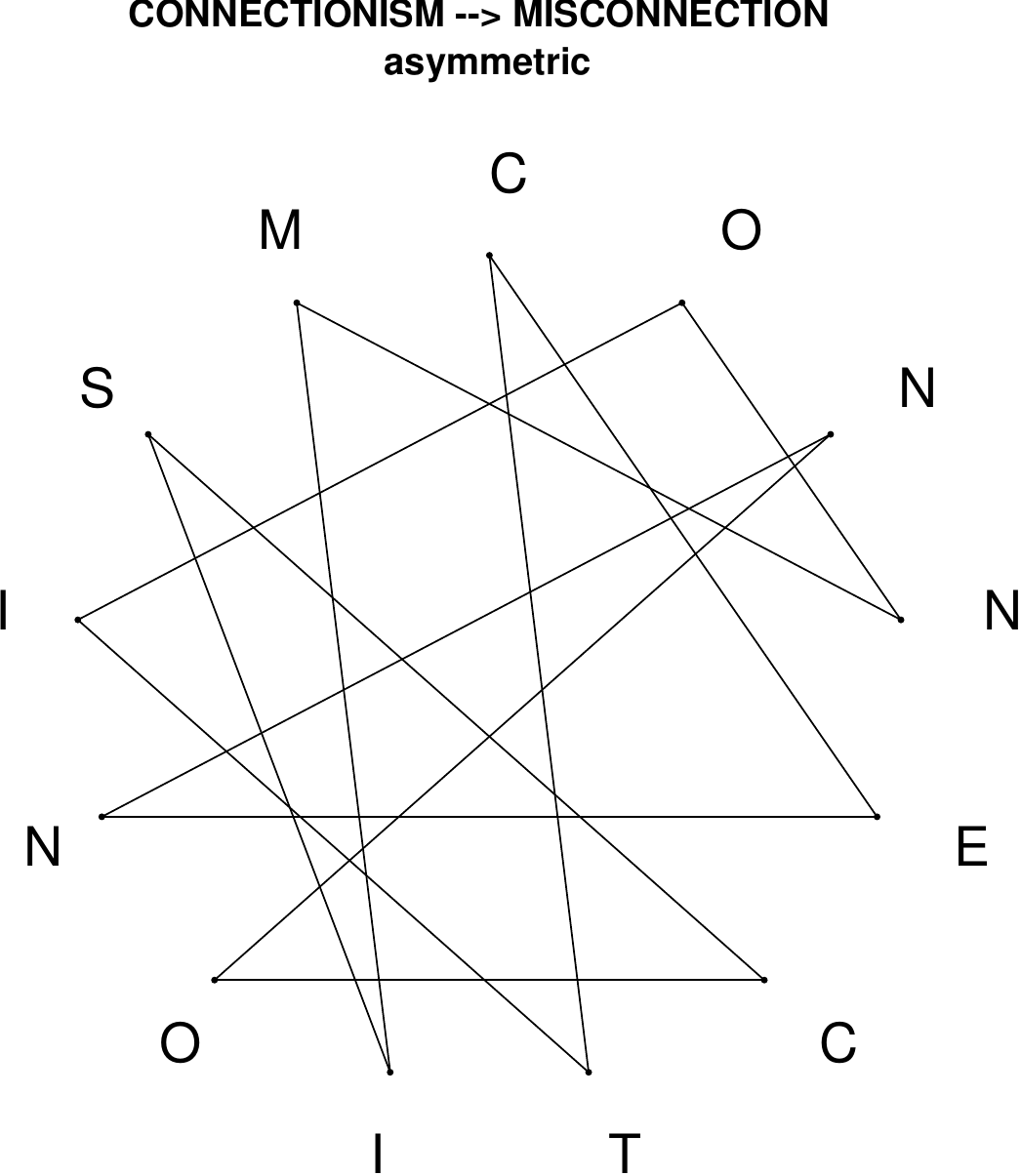}
\end{subfigure}
\end{figure}

\begin{figure}[H]
\centering
\begin{subfigure}[T]{0.19\textwidth}
\centering
\includegraphics[width=\textwidth]{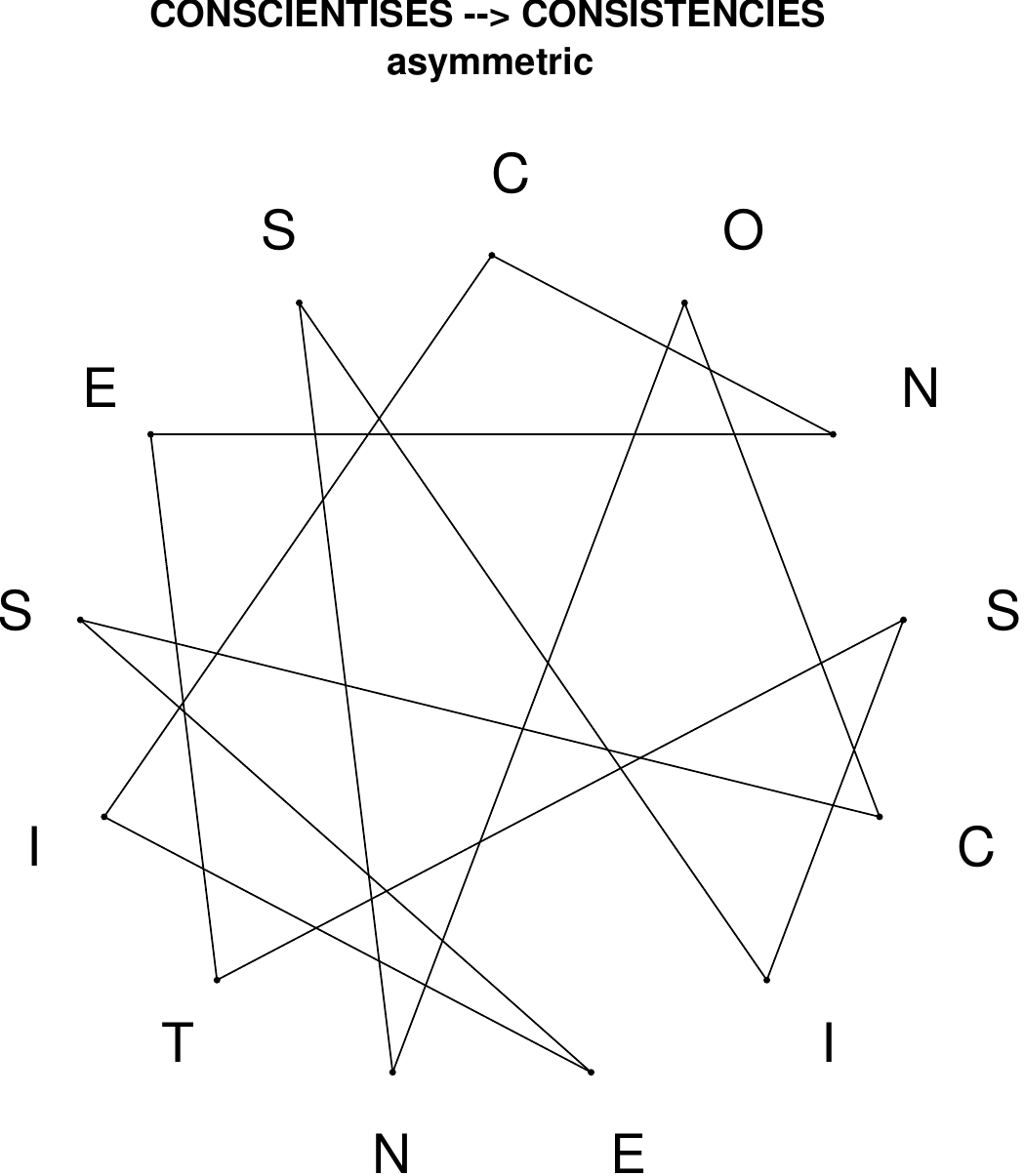}
\end{subfigure}
\hfill
\begin{subfigure}[T]{0.19\textwidth}
\centering
\includegraphics[width=\textwidth]{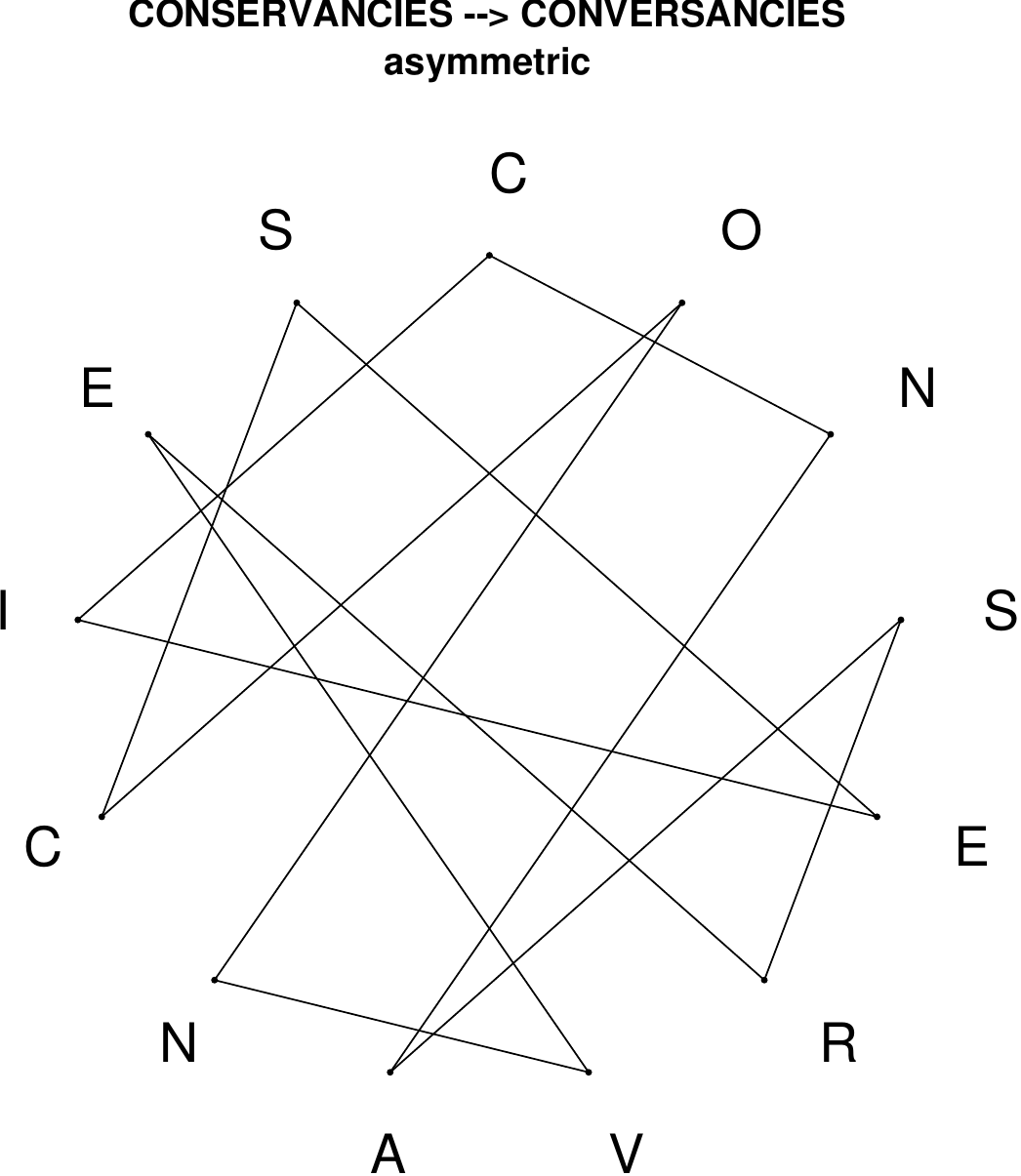}
\end{subfigure}
\hfill
\begin{subfigure}[T]{0.19\textwidth}
\centering
\includegraphics[width=\textwidth]{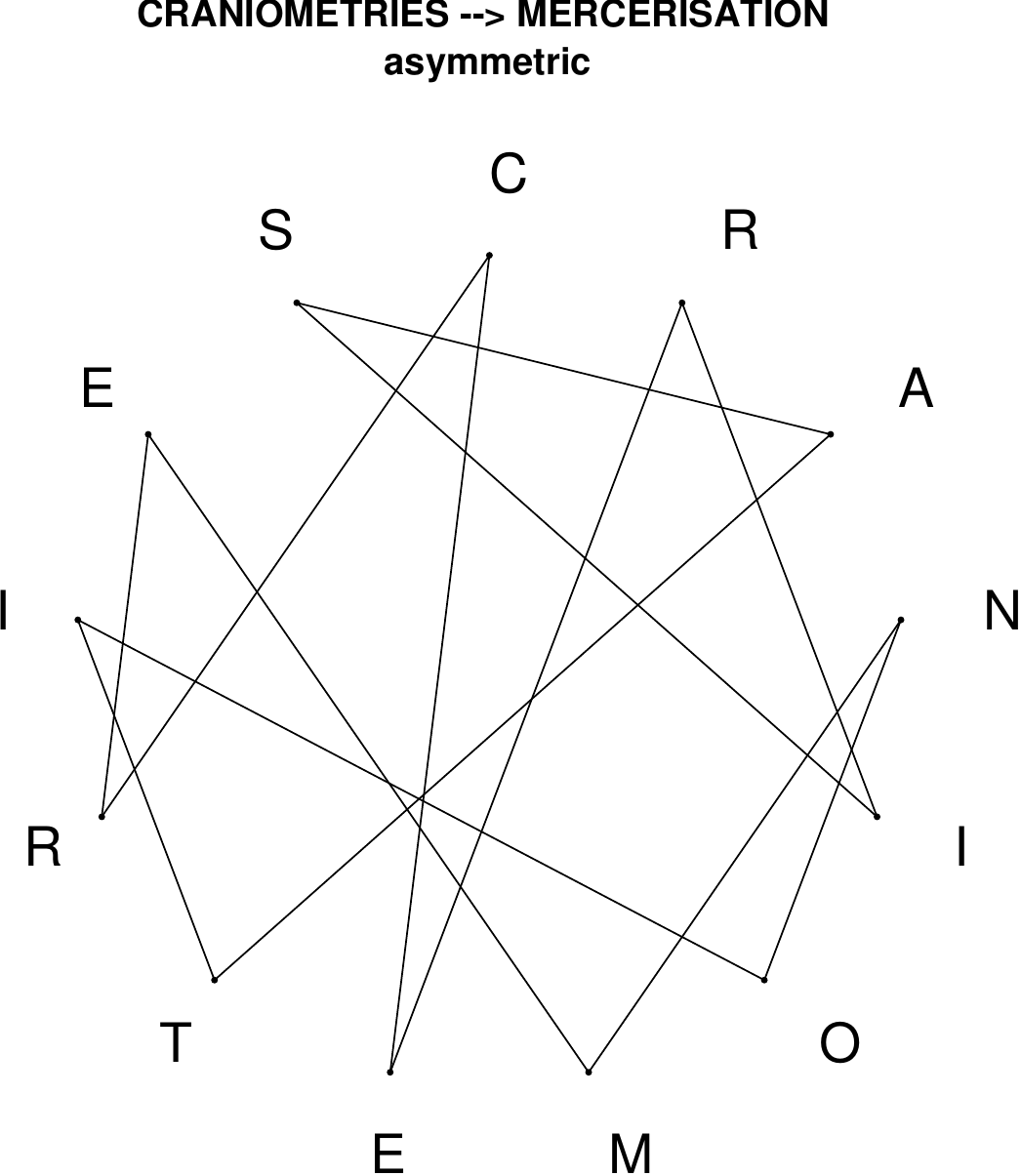}
\end{subfigure}
\hfill
\begin{subfigure}[T]{0.19\textwidth}
\centering
\includegraphics[width=\textwidth]{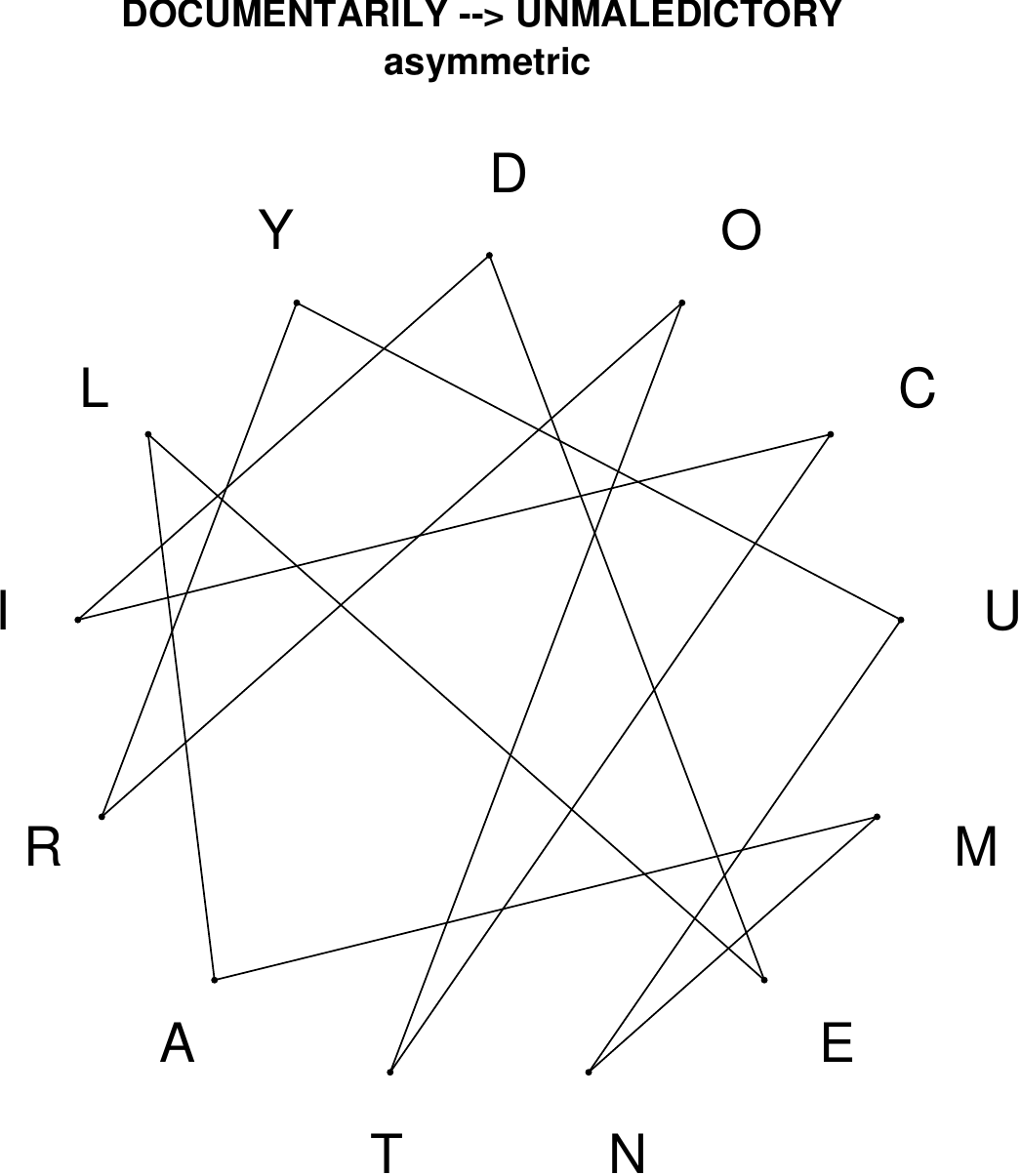}
\end{subfigure}
\hfill
\begin{subfigure}[T]{0.19\textwidth}
\centering
\includegraphics[width=\textwidth]{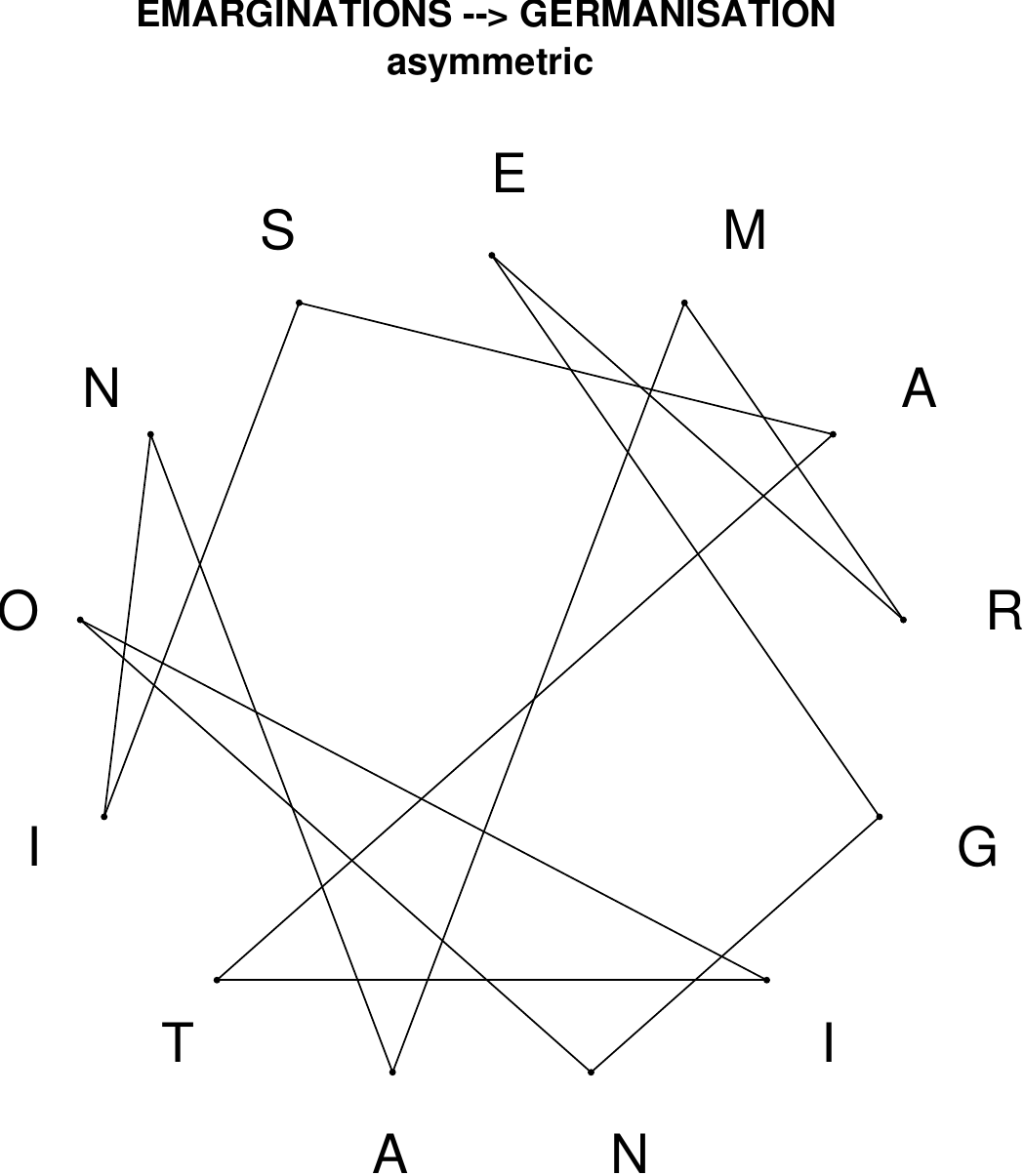}
\end{subfigure}
\end{figure}

\begin{figure}[H]
\centering
\begin{subfigure}[T]{0.19\textwidth}
\centering
\includegraphics[width=\textwidth]{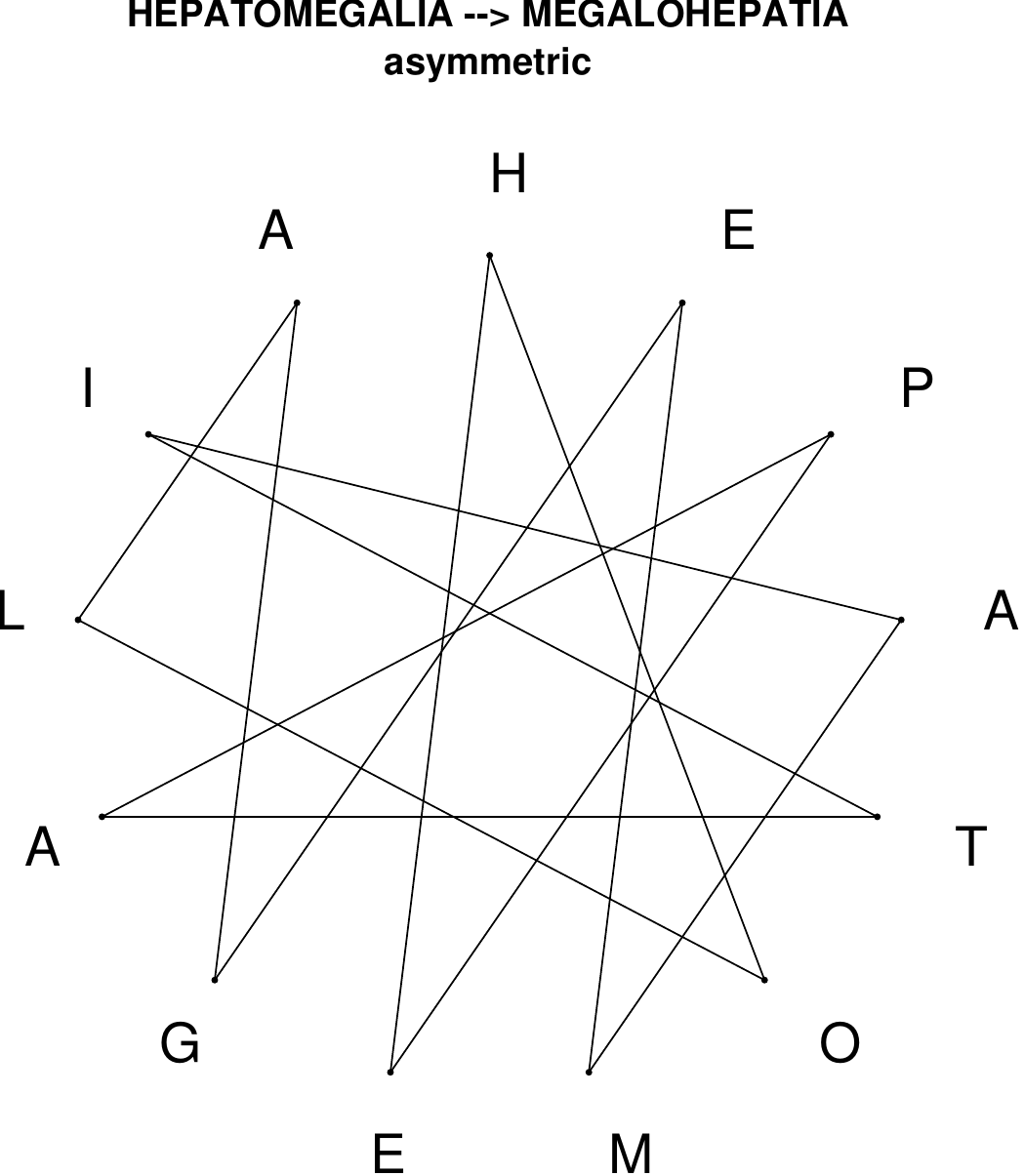}
\end{subfigure}
\hfill
\begin{subfigure}[T]{0.19\textwidth}
\centering
\includegraphics[width=\textwidth]{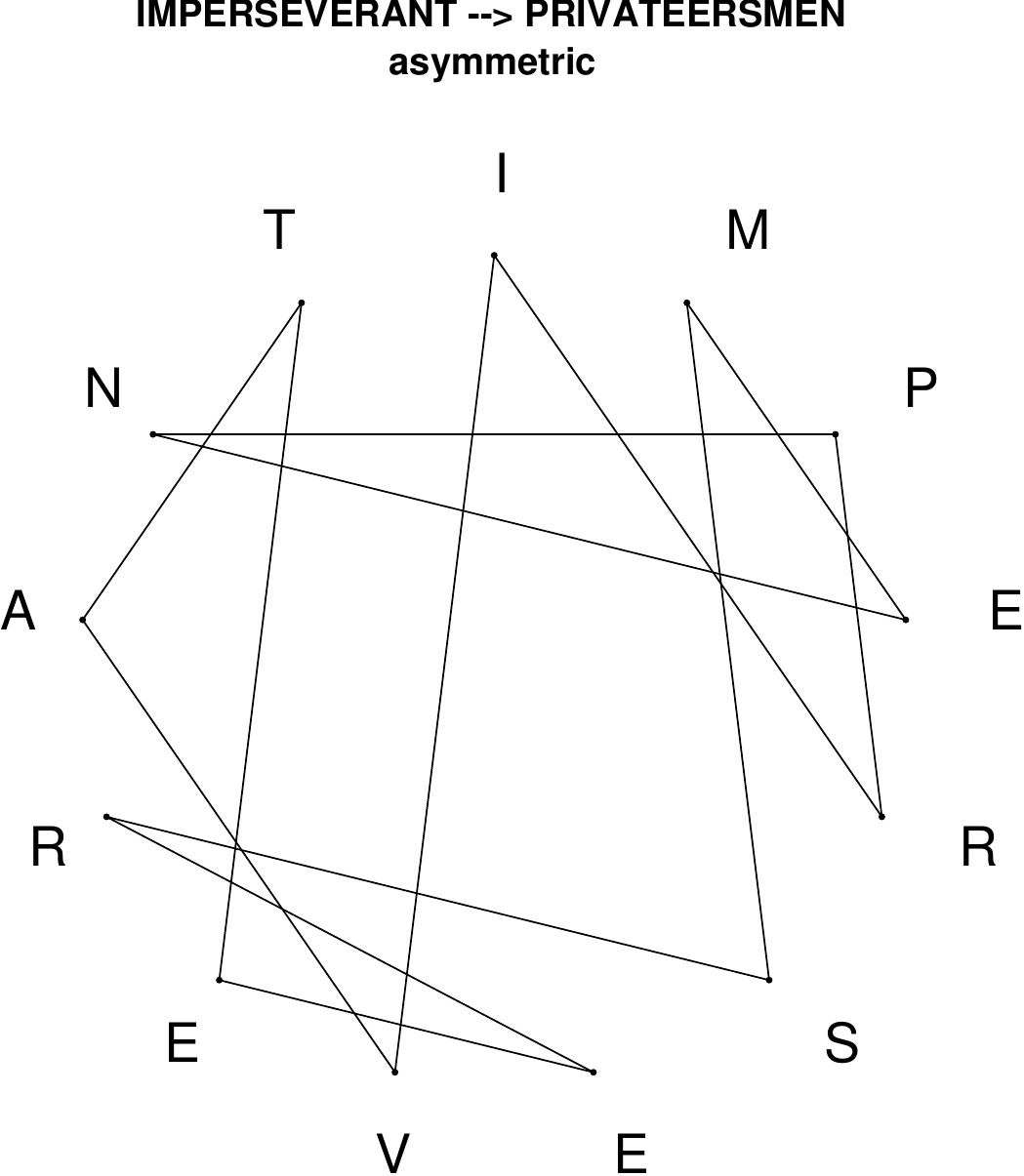}
\end{subfigure}
\hfill
\begin{subfigure}[T]{0.19\textwidth}
\centering
\includegraphics[width=\textwidth]{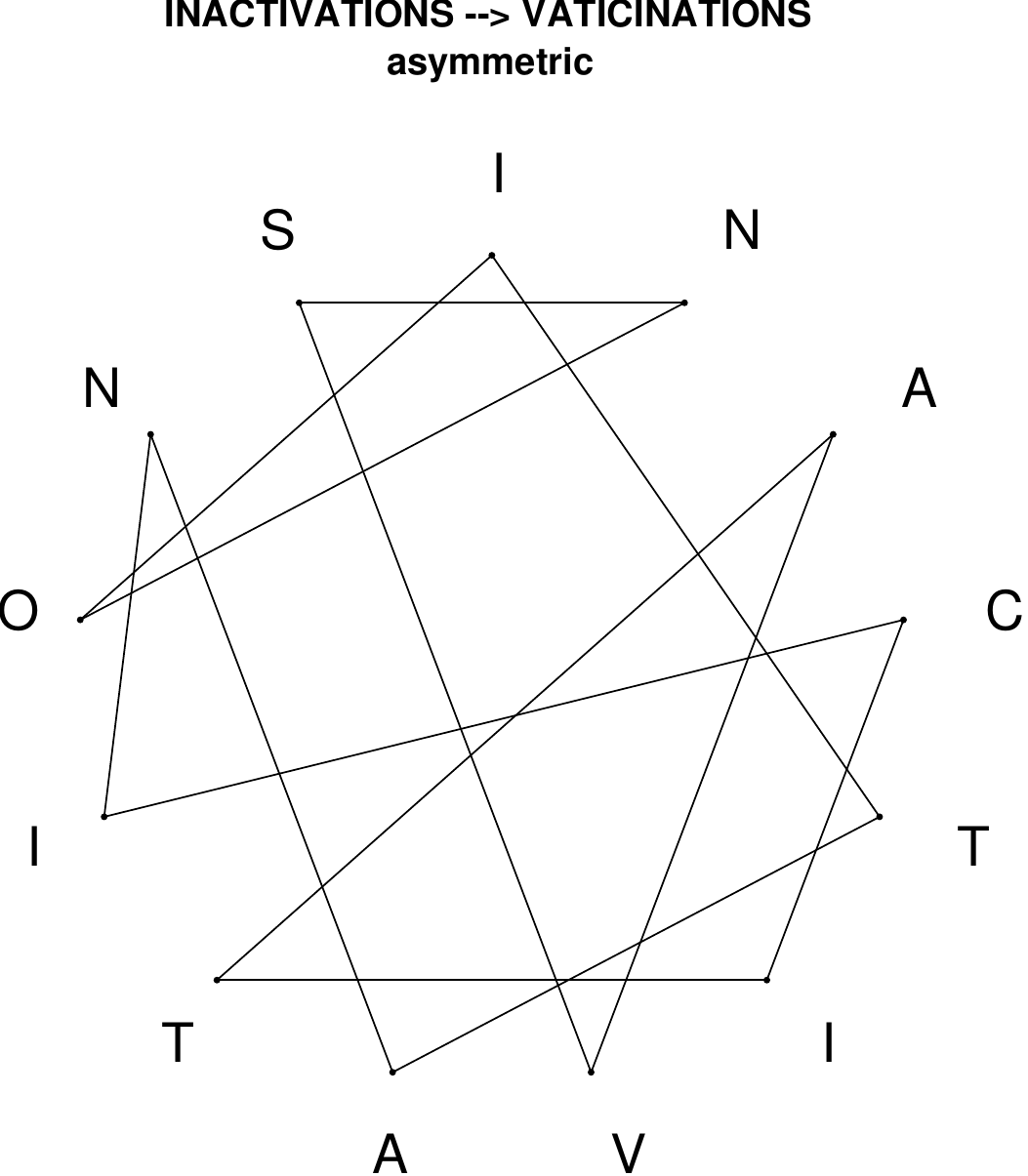}
\end{subfigure}
\hfill
\begin{subfigure}[T]{0.19\textwidth}
\centering
\includegraphics[width=\textwidth]{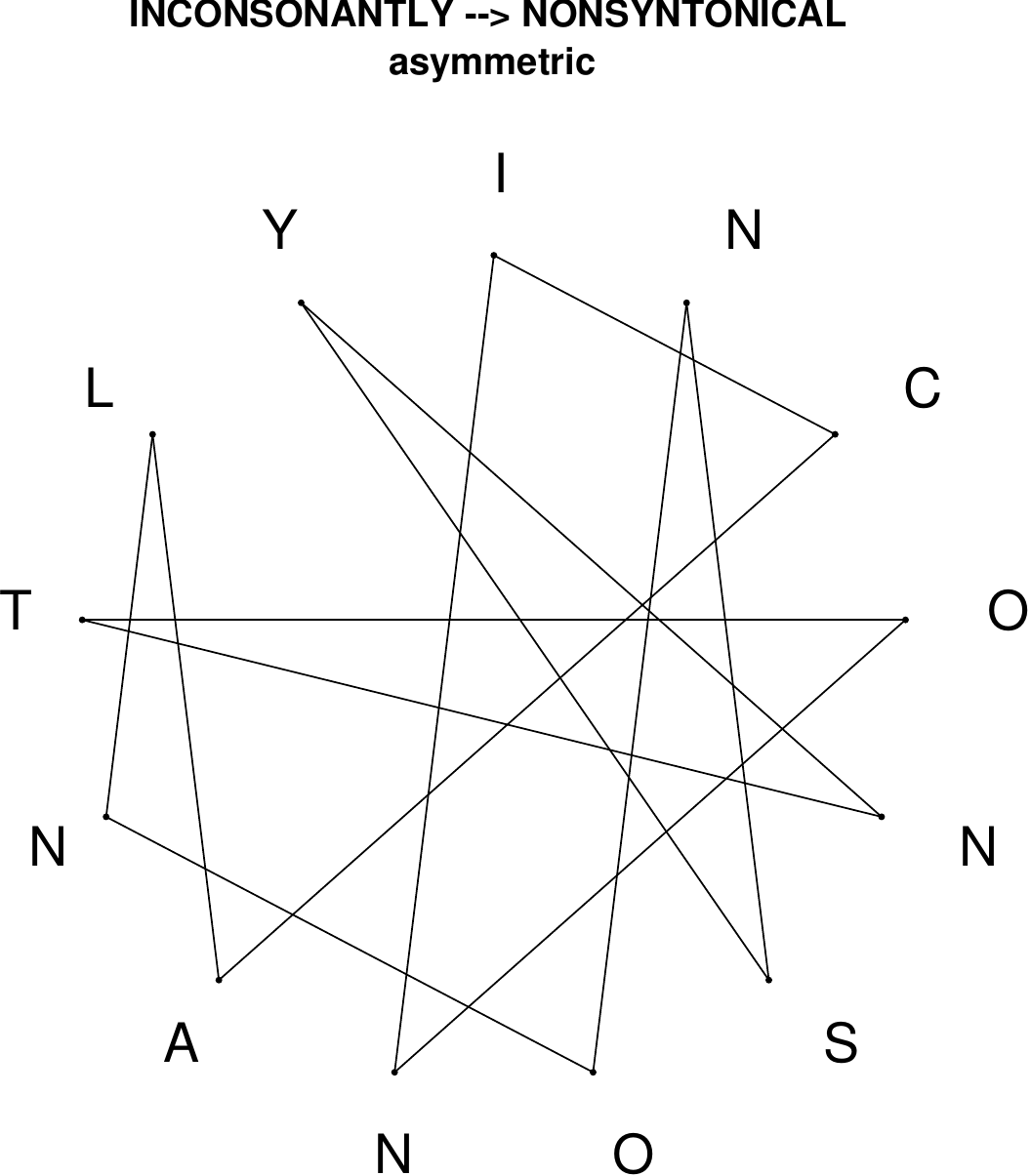}
\end{subfigure}
\hfill
\begin{subfigure}[T]{0.19\textwidth}
\centering
\includegraphics[width=\textwidth]{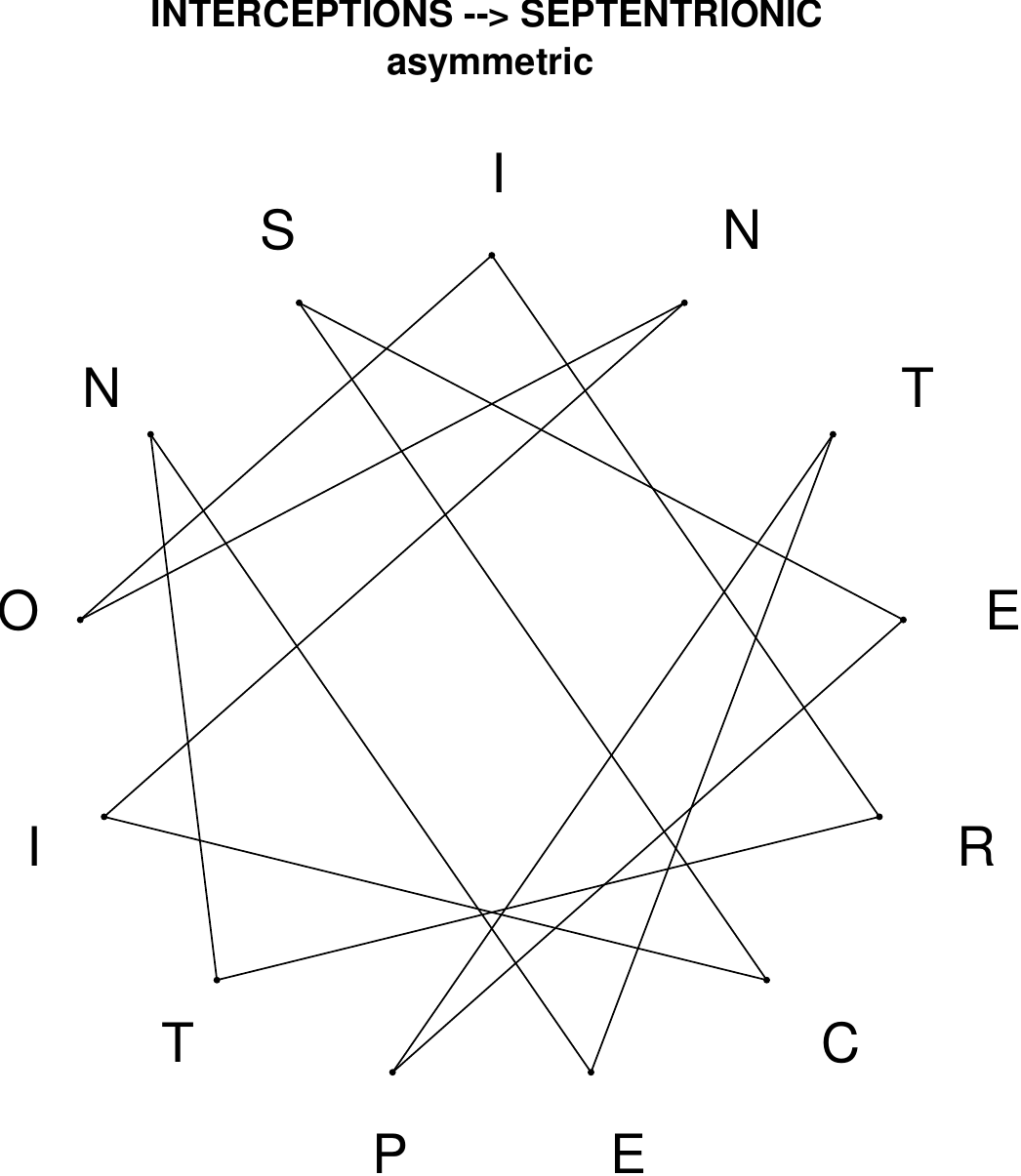}
\end{subfigure}
\end{figure}

\begin{figure}[H]
\centering
\begin{subfigure}[T]{0.19\textwidth}
\centering
\includegraphics[width=\textwidth]{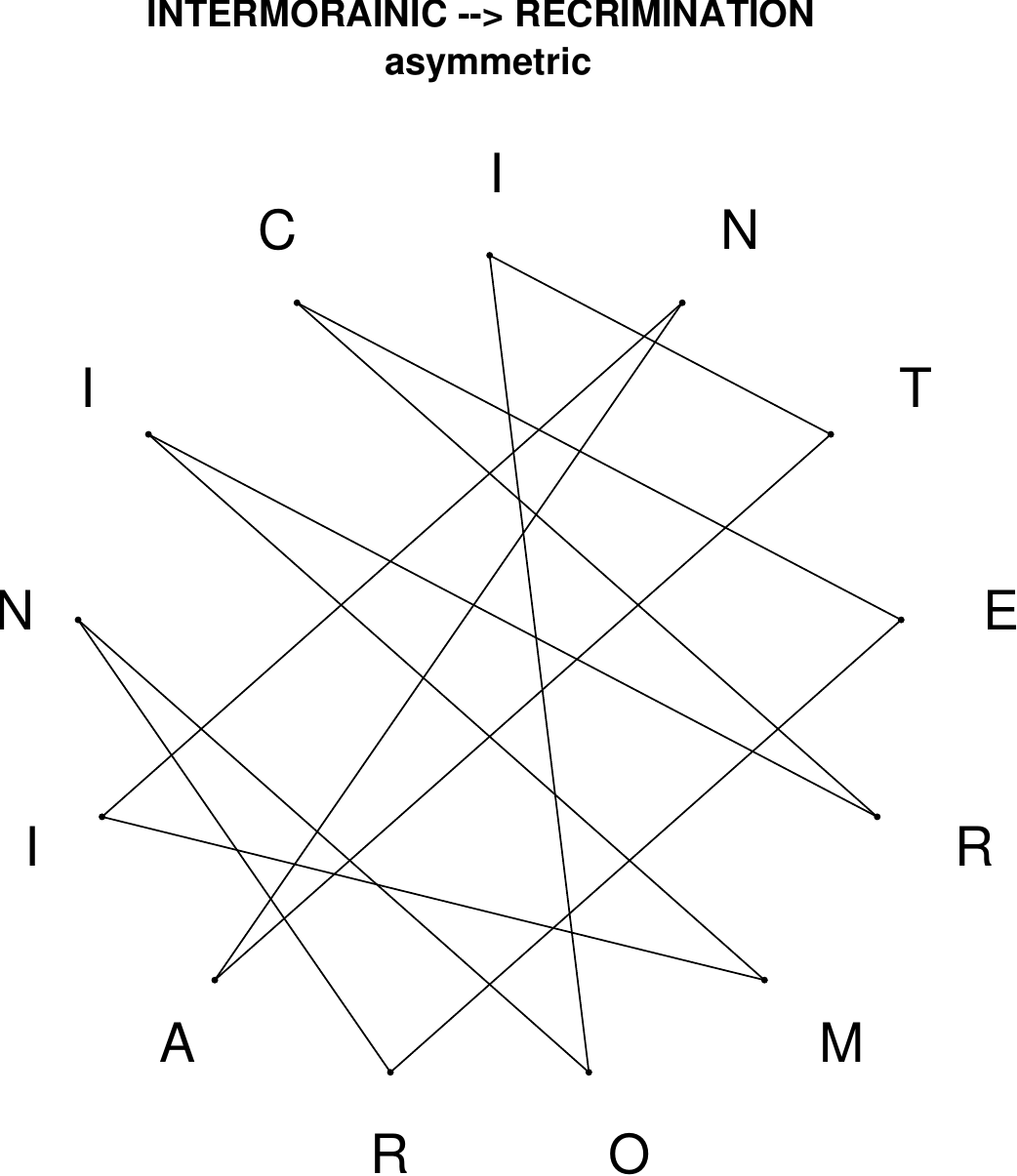}
\end{subfigure}
\hfill
\begin{subfigure}[T]{0.19\textwidth}
\centering
\includegraphics[width=\textwidth]{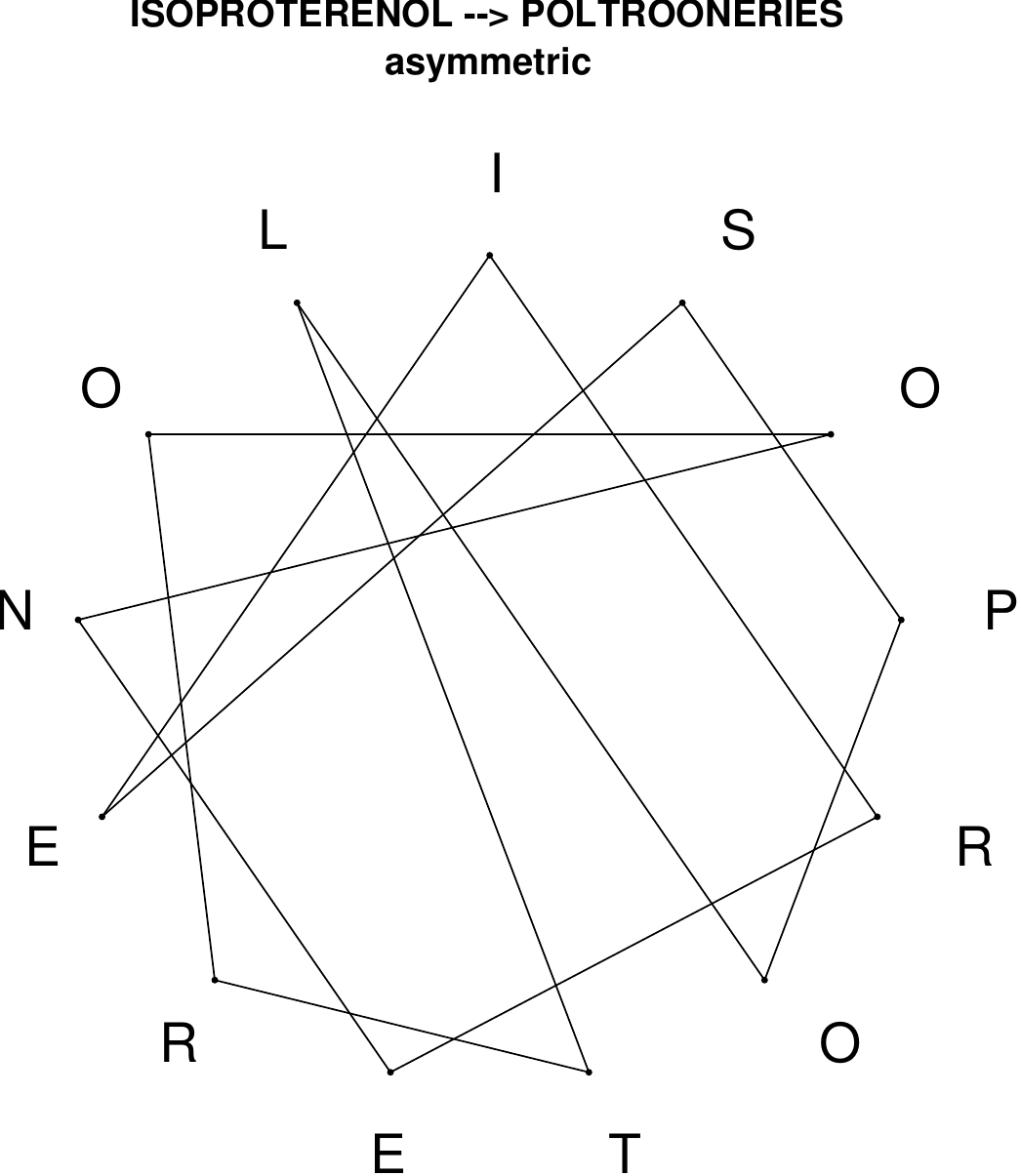}
\end{subfigure}
\hfill
\begin{subfigure}[T]{0.19\textwidth}
\centering
\includegraphics[width=\textwidth]{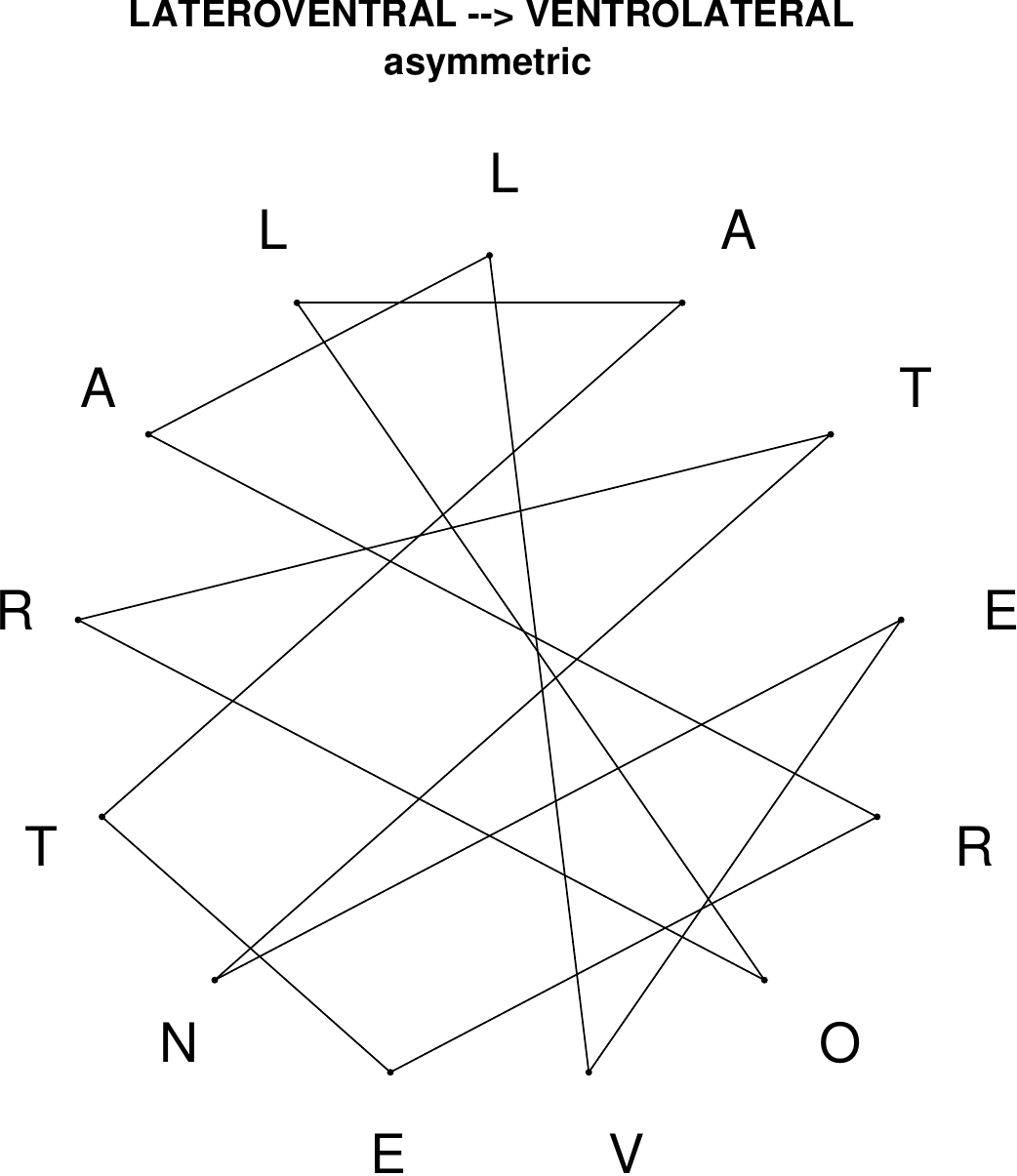}
\end{subfigure}
\hfill
\begin{subfigure}[T]{0.19\textwidth}
\centering
\includegraphics[width=\textwidth]{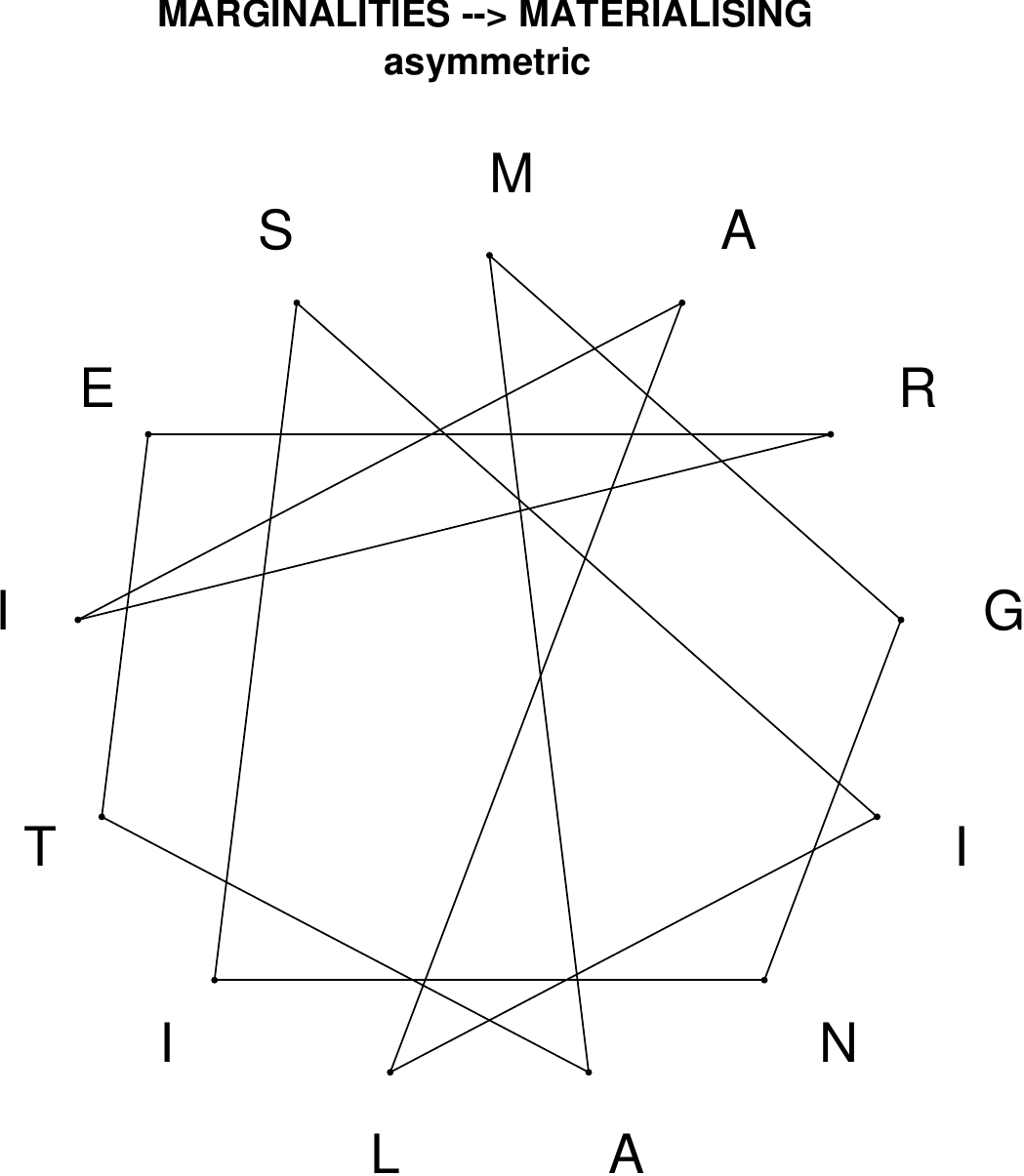}
\end{subfigure}
\hfill
\begin{subfigure}[T]{0.19\textwidth}
\centering
\includegraphics[width=\textwidth]{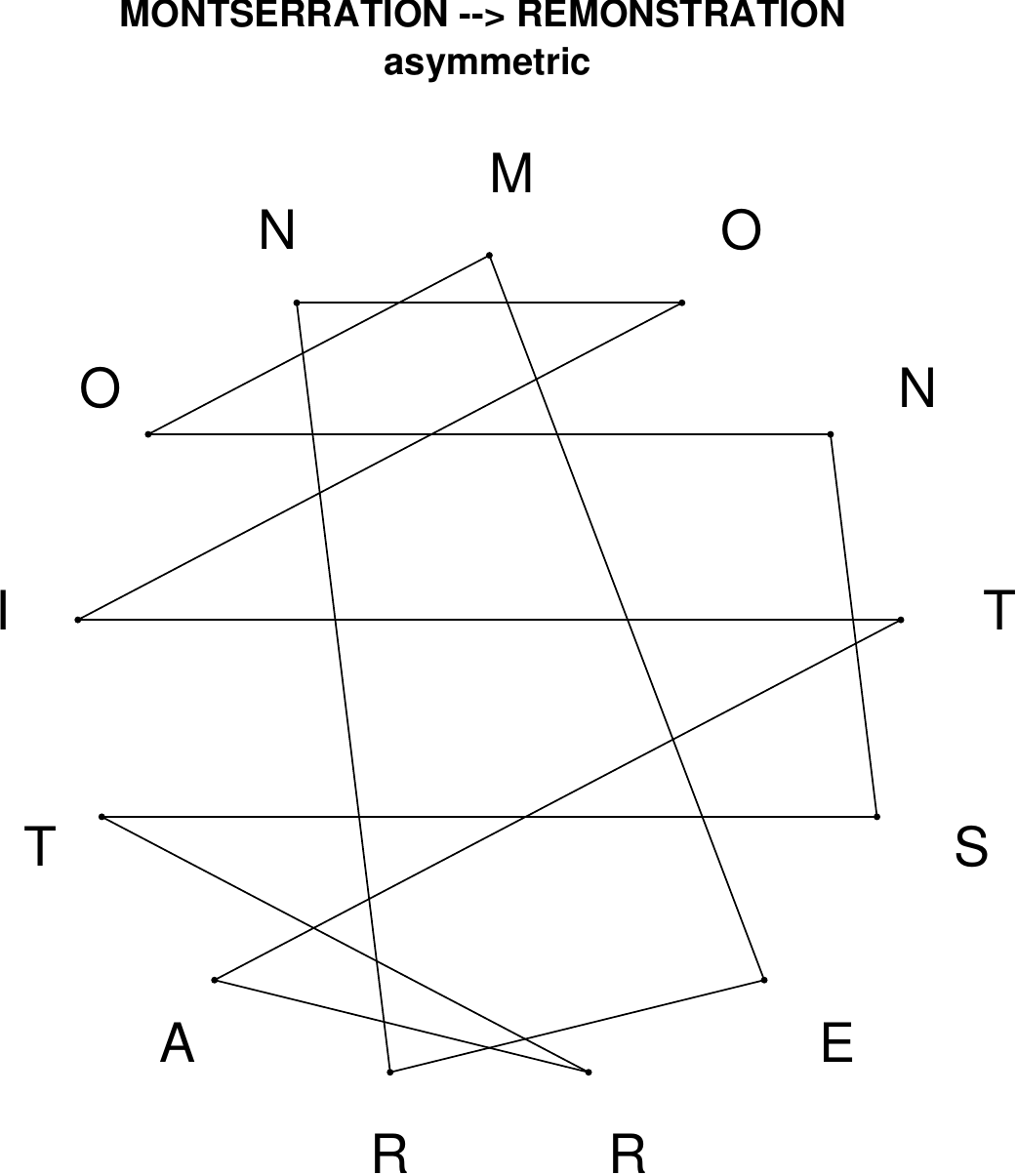}
\end{subfigure}
\end{figure}

\begin{figure}[H]
\centering
\begin{subfigure}[T]{0.19\textwidth}
\centering
\includegraphics[width=\textwidth]{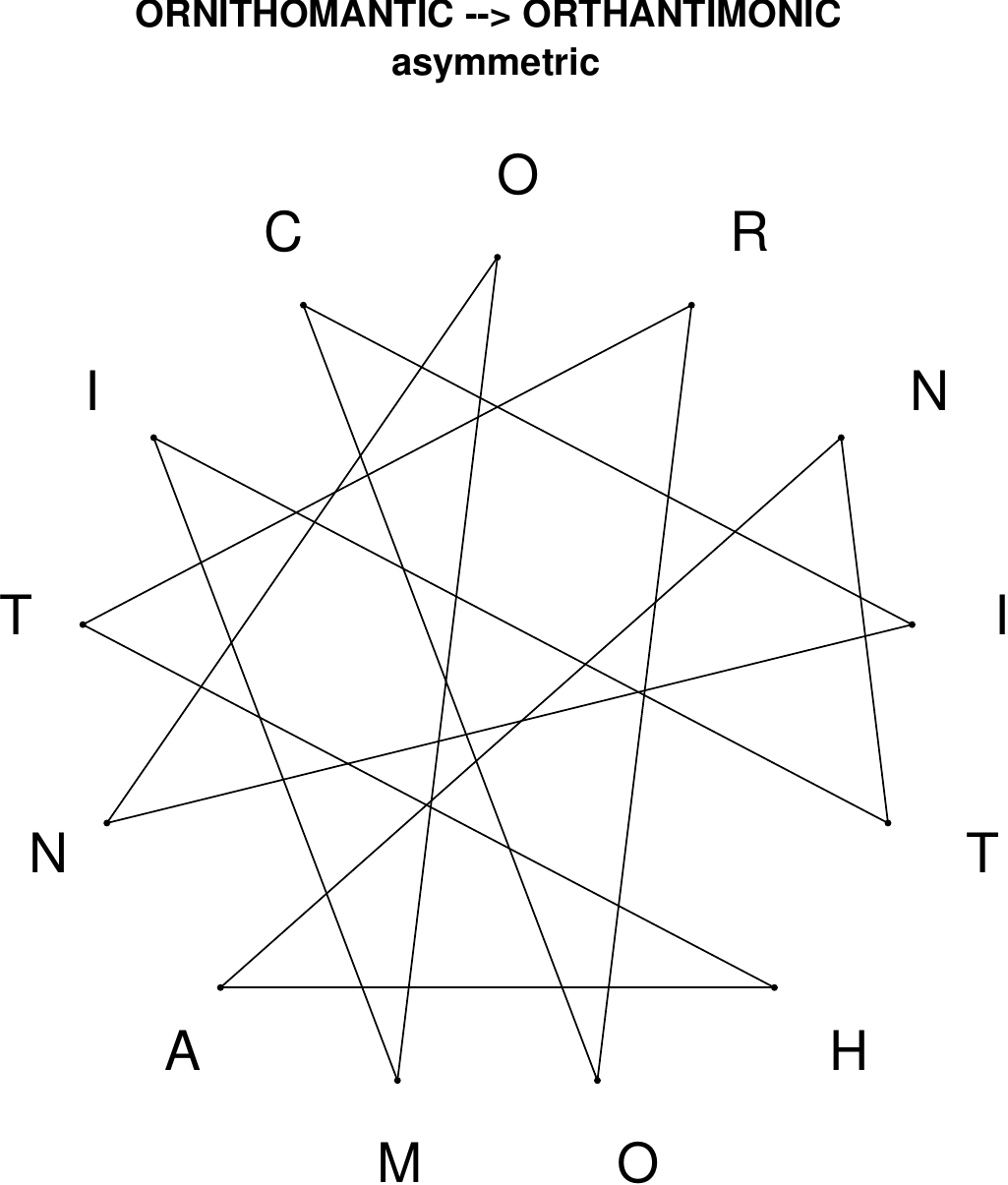}
\end{subfigure}
\hfill
\begin{subfigure}[T]{0.19\textwidth}
\centering
\includegraphics[width=\textwidth]{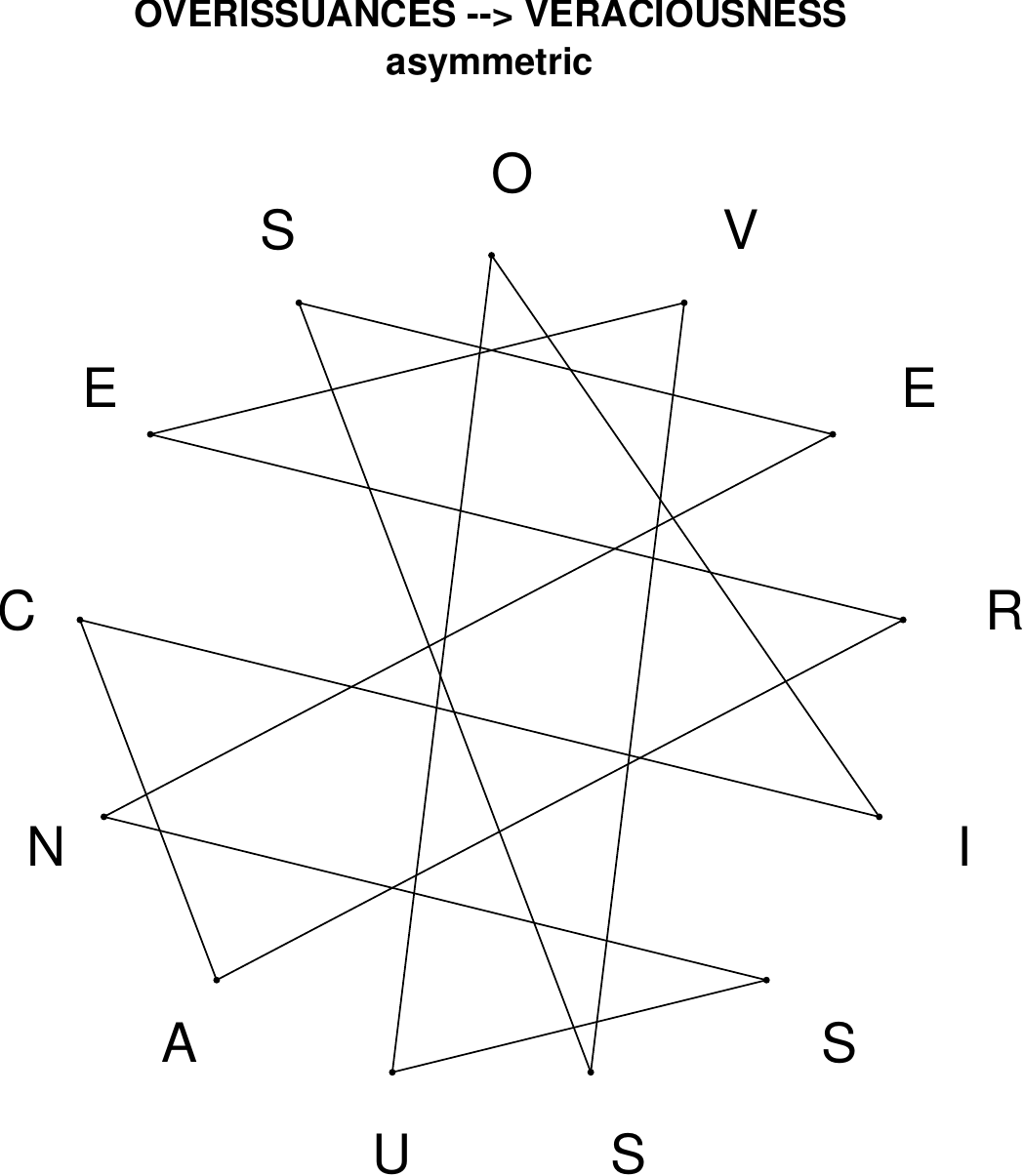}
\end{subfigure}
\hfill
\begin{subfigure}[T]{0.19\textwidth}
\centering
\includegraphics[width=\textwidth]{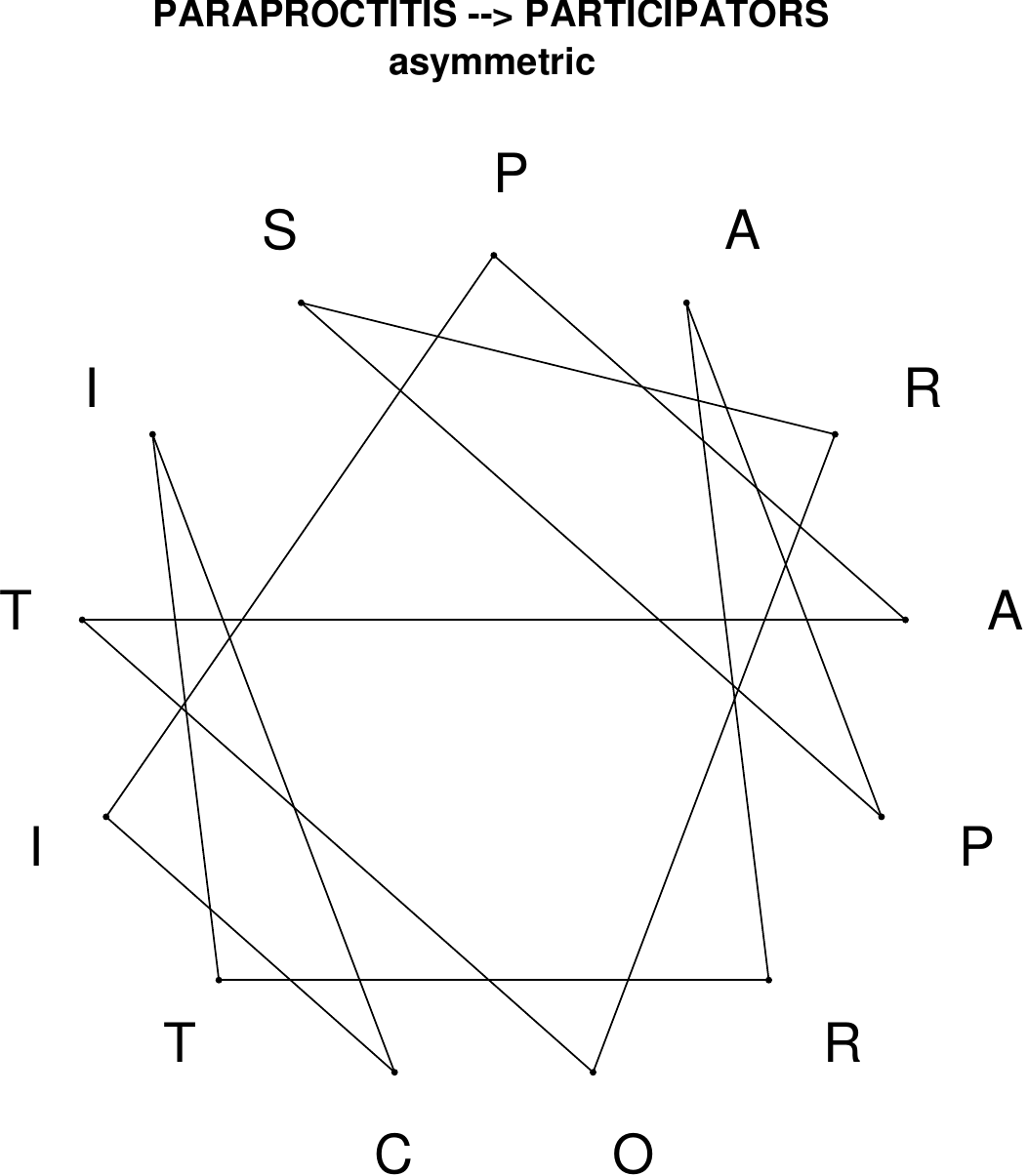}
\end{subfigure}
\hfill
\begin{subfigure}[T]{0.19\textwidth}
\centering
\includegraphics[width=\textwidth]{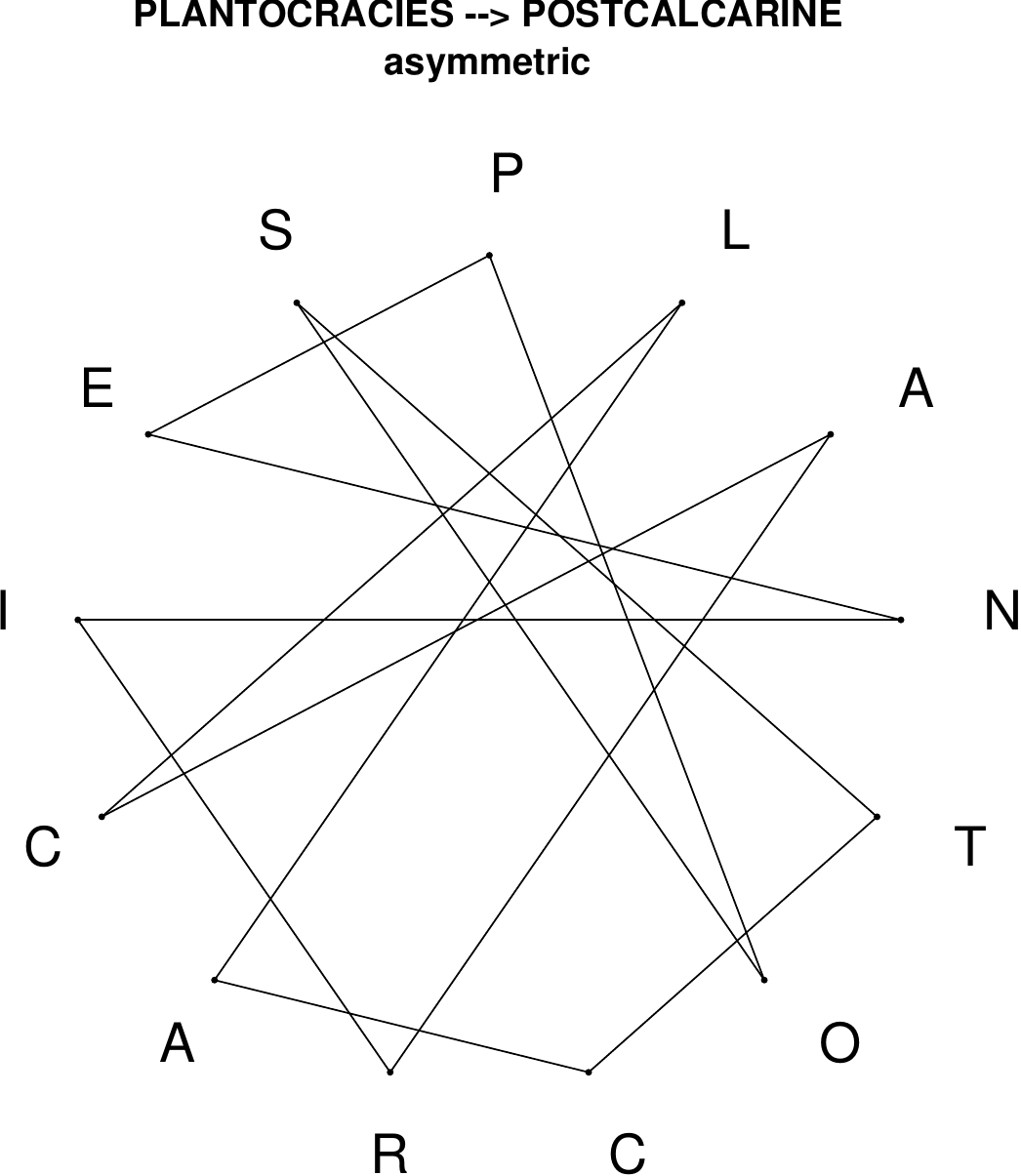}
\end{subfigure}
\hfill
\begin{subfigure}[T]{0.19\textwidth}
\centering
\includegraphics[width=\textwidth]{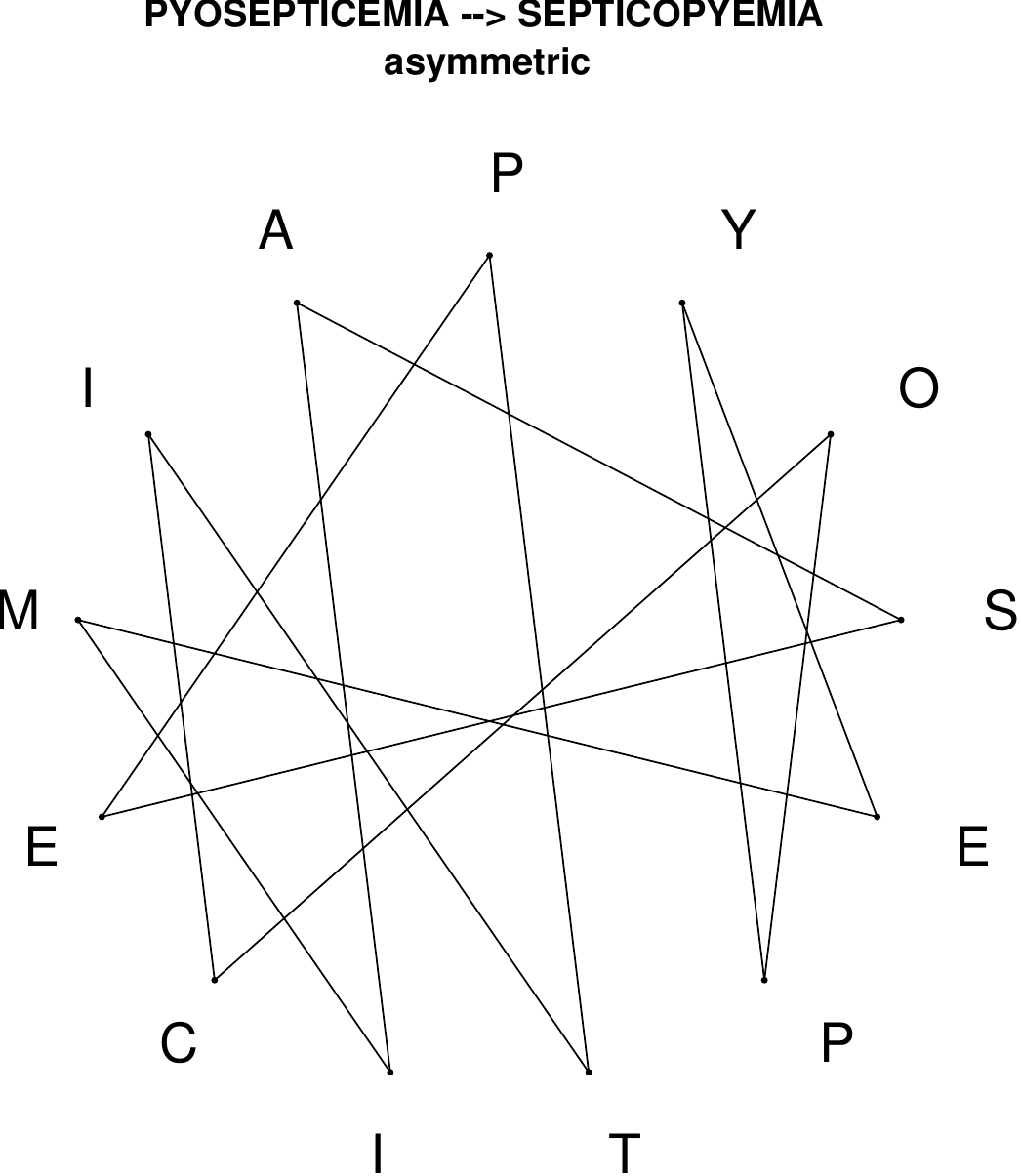}
\end{subfigure}
\end{figure}

\begin{figure}[H]
\centering
\begin{subfigure}[T]{0.19\textwidth}
\centering
\includegraphics[width=\textwidth]{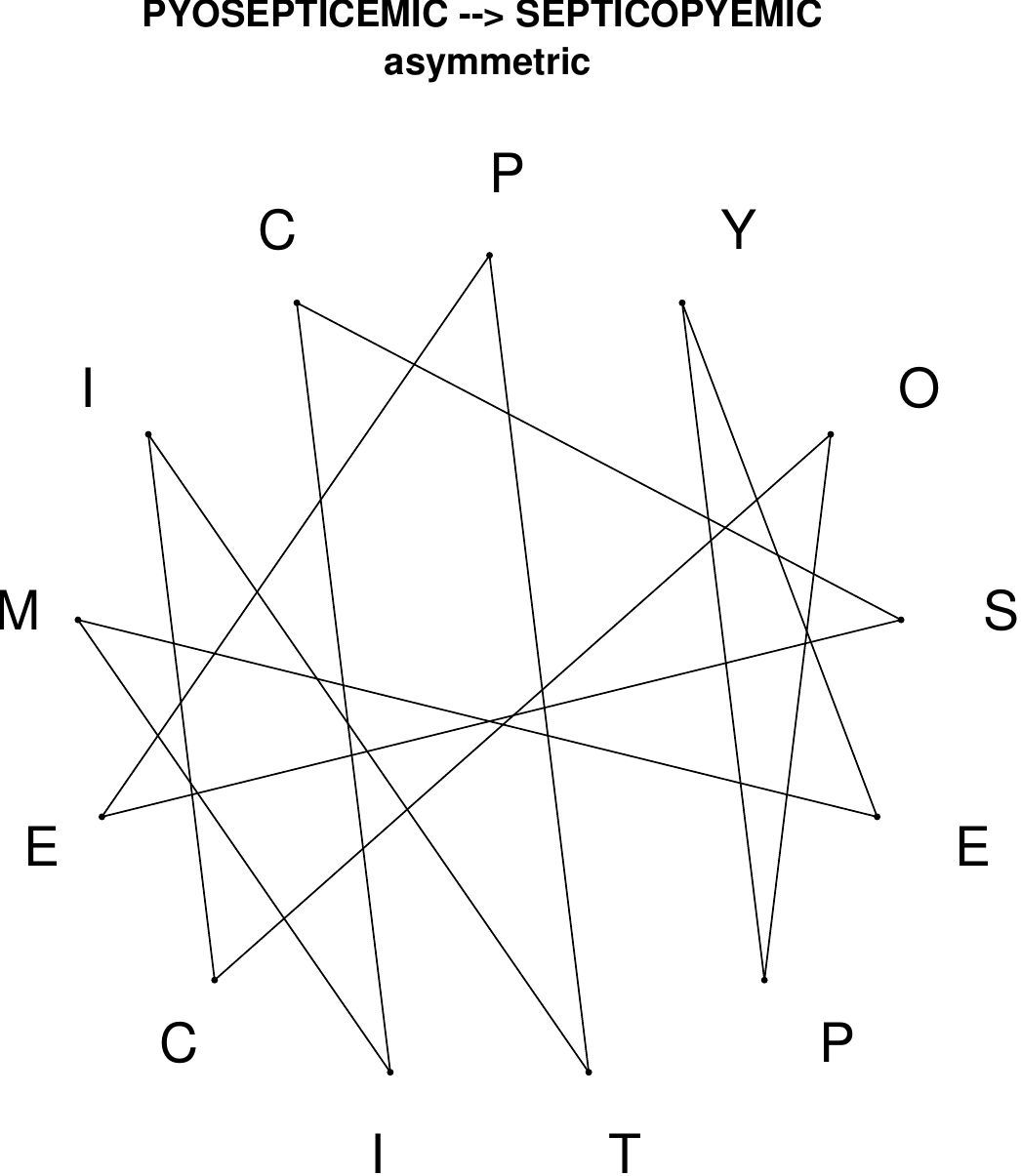}
\end{subfigure}
\hfill
\begin{subfigure}[T]{0.19\textwidth}
\centering
\includegraphics[width=\textwidth]{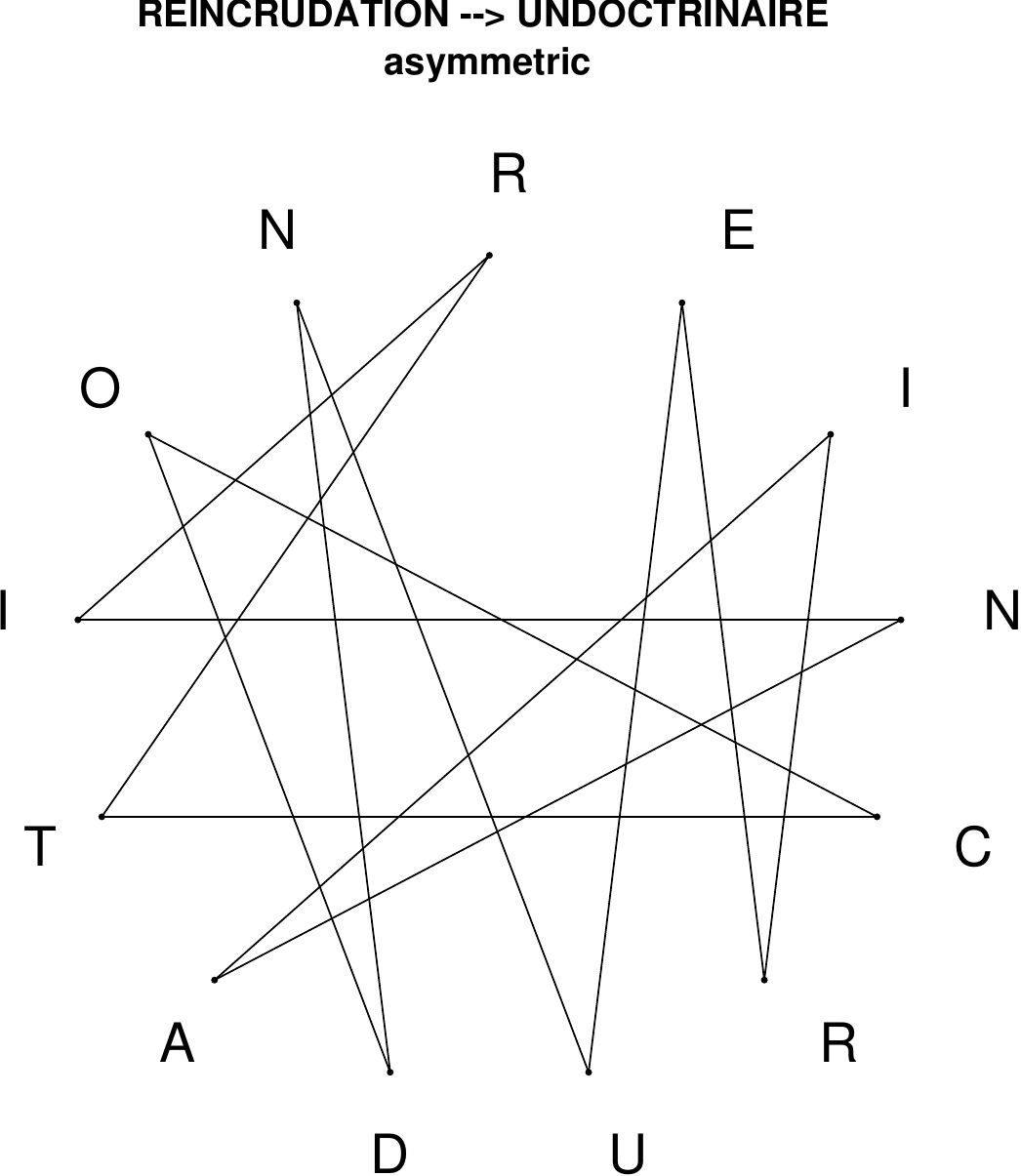}
\end{subfigure}
\hfill
\begin{subfigure}[T]{0.19\textwidth}
\centering
\includegraphics[width=\textwidth]{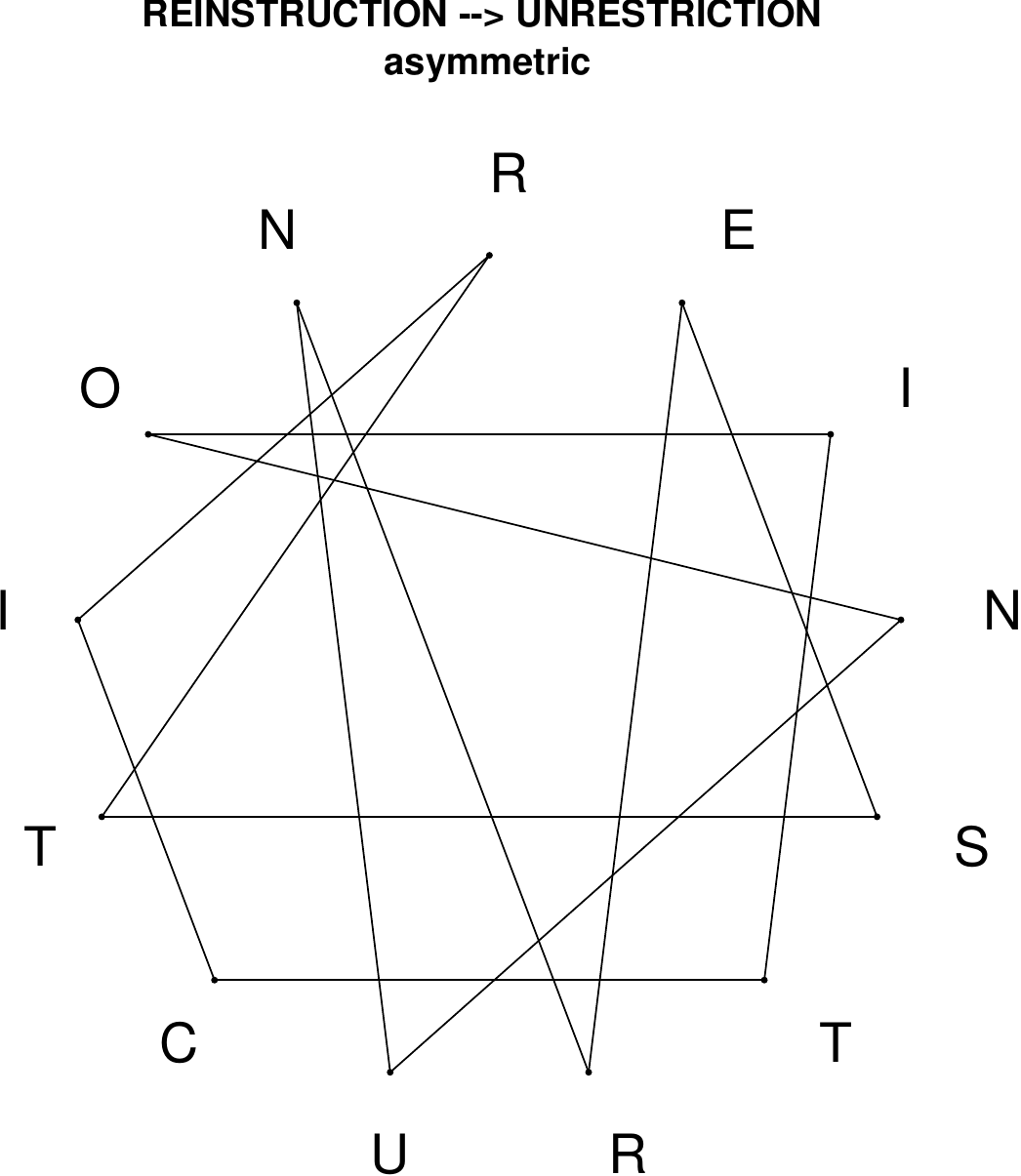}
\end{subfigure}
\hfill
\begin{subfigure}[T]{0.19\textwidth}
\centering
\includegraphics[width=\textwidth]{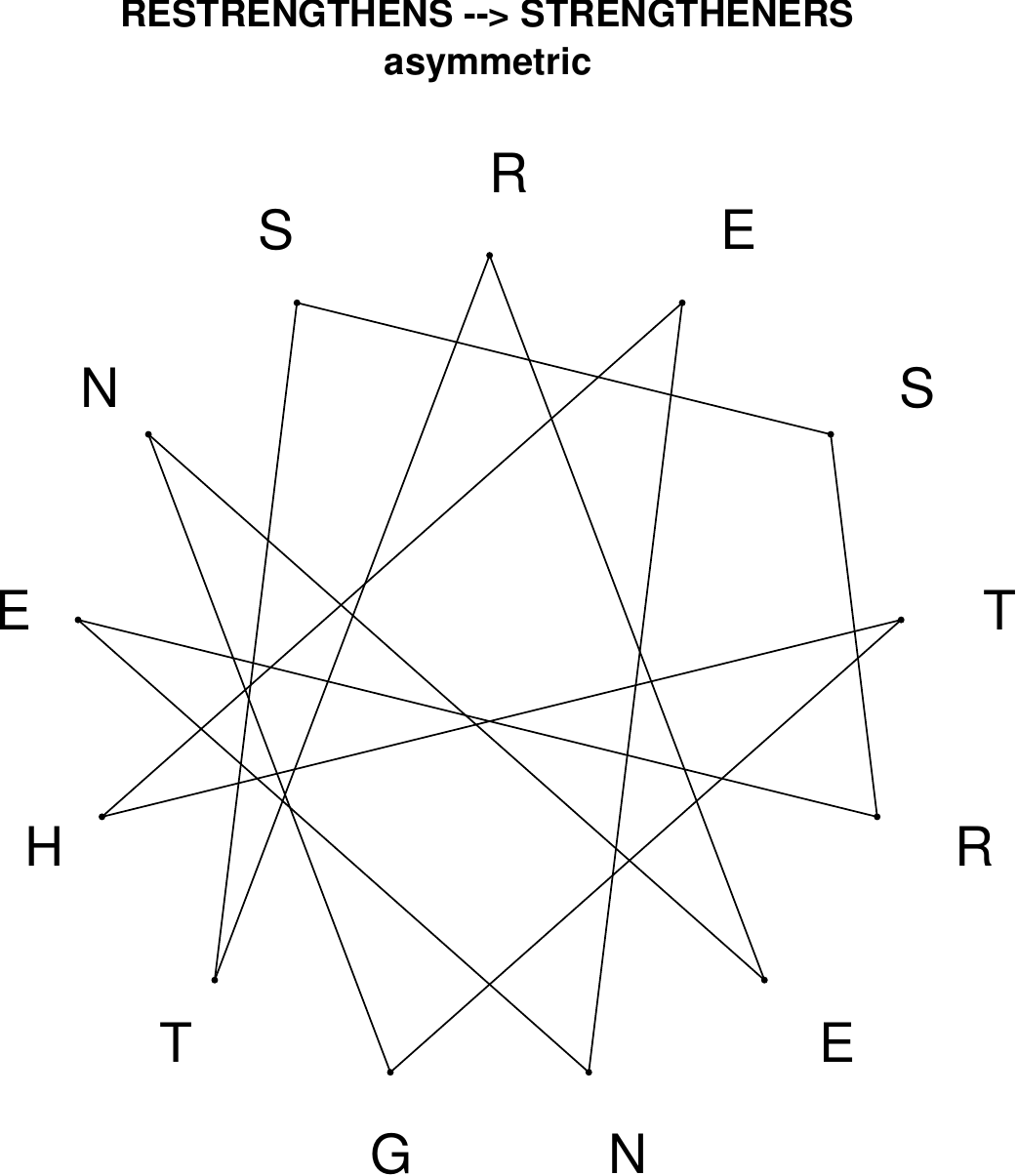}
\end{subfigure}
\hfill
\begin{subfigure}[T]{0.19\textwidth}
\centering
\includegraphics[width=\textwidth]{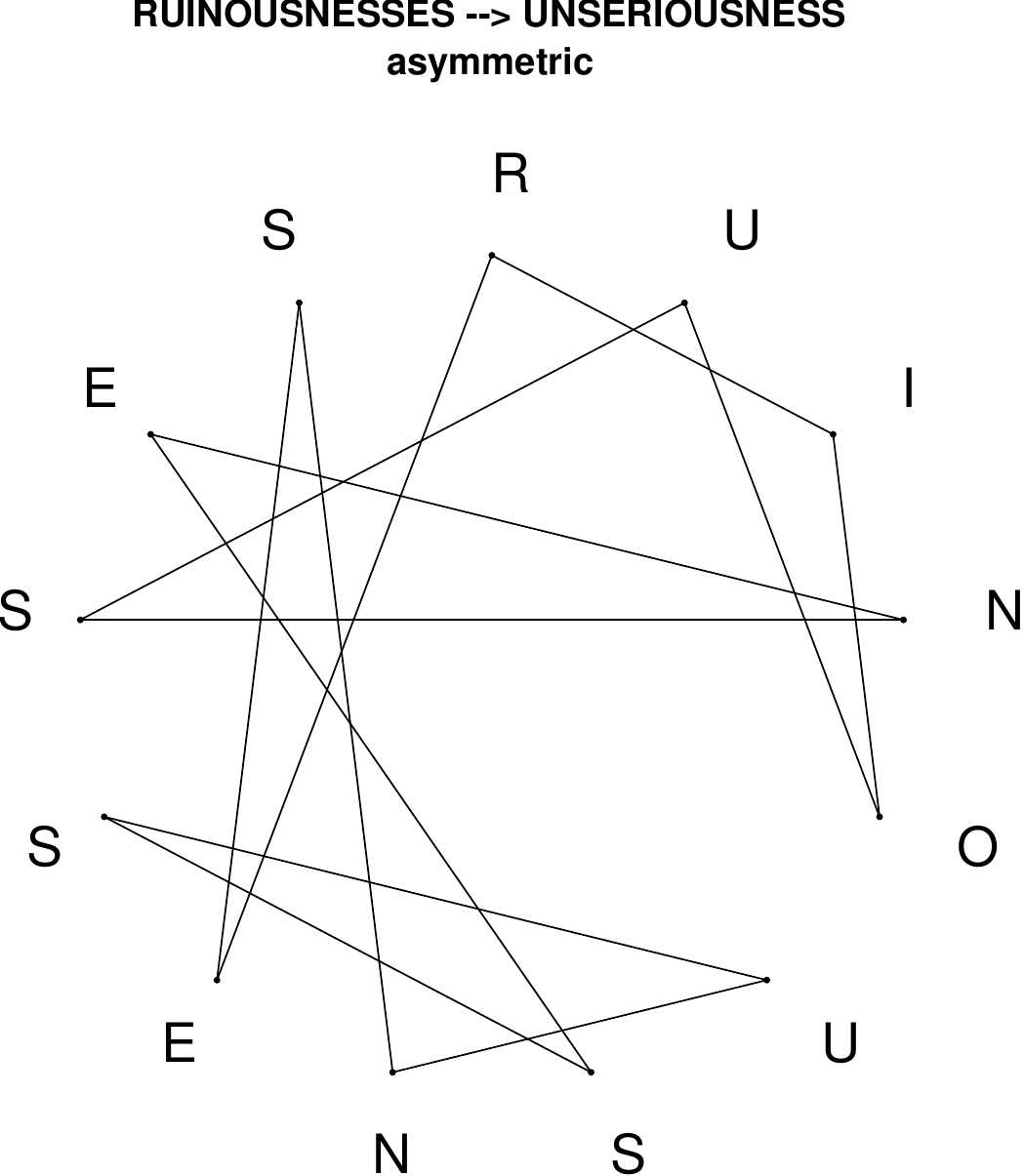}
\end{subfigure}
\end{figure}

\begin{figure}[H]
\centering
\begin{subfigure}[T]{0.19\textwidth}
\centering
\includegraphics[width=\textwidth]{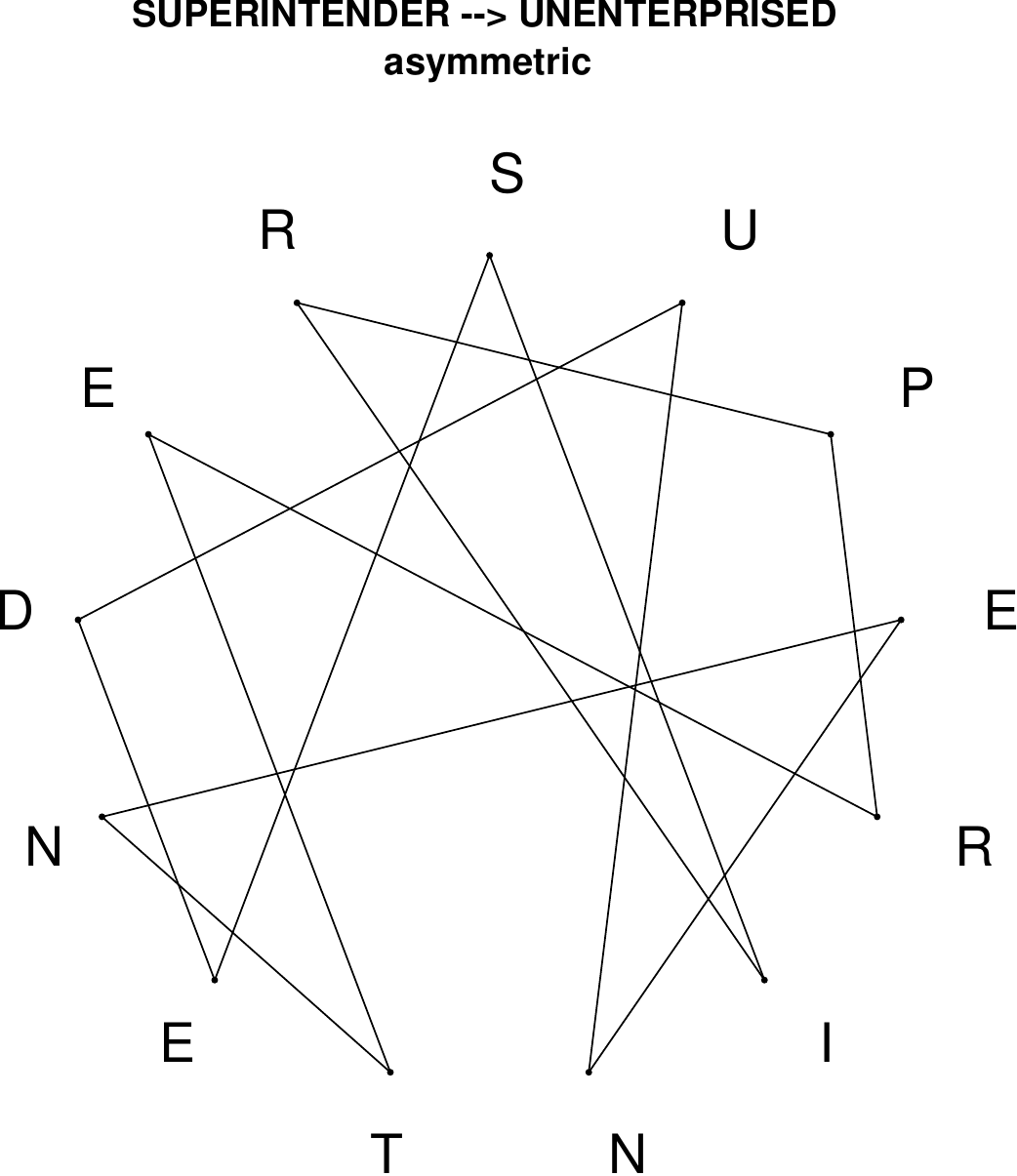}
\end{subfigure}
\hfill
\end{figure}

%%%%%%%%%%%%%%%%%%
\clearpage
\subsection{Star Anagrams $N = 12$}
For $N=12$, we found two symmetric stars among a larger group of asymmetric stars. 

\subsubsection{Symmetric Stars $N=12$}

\begin{figure}[H]
\centering
\begin{subfigure}[T]{0.19\textwidth}
\centering
\includegraphics[width=\textwidth]{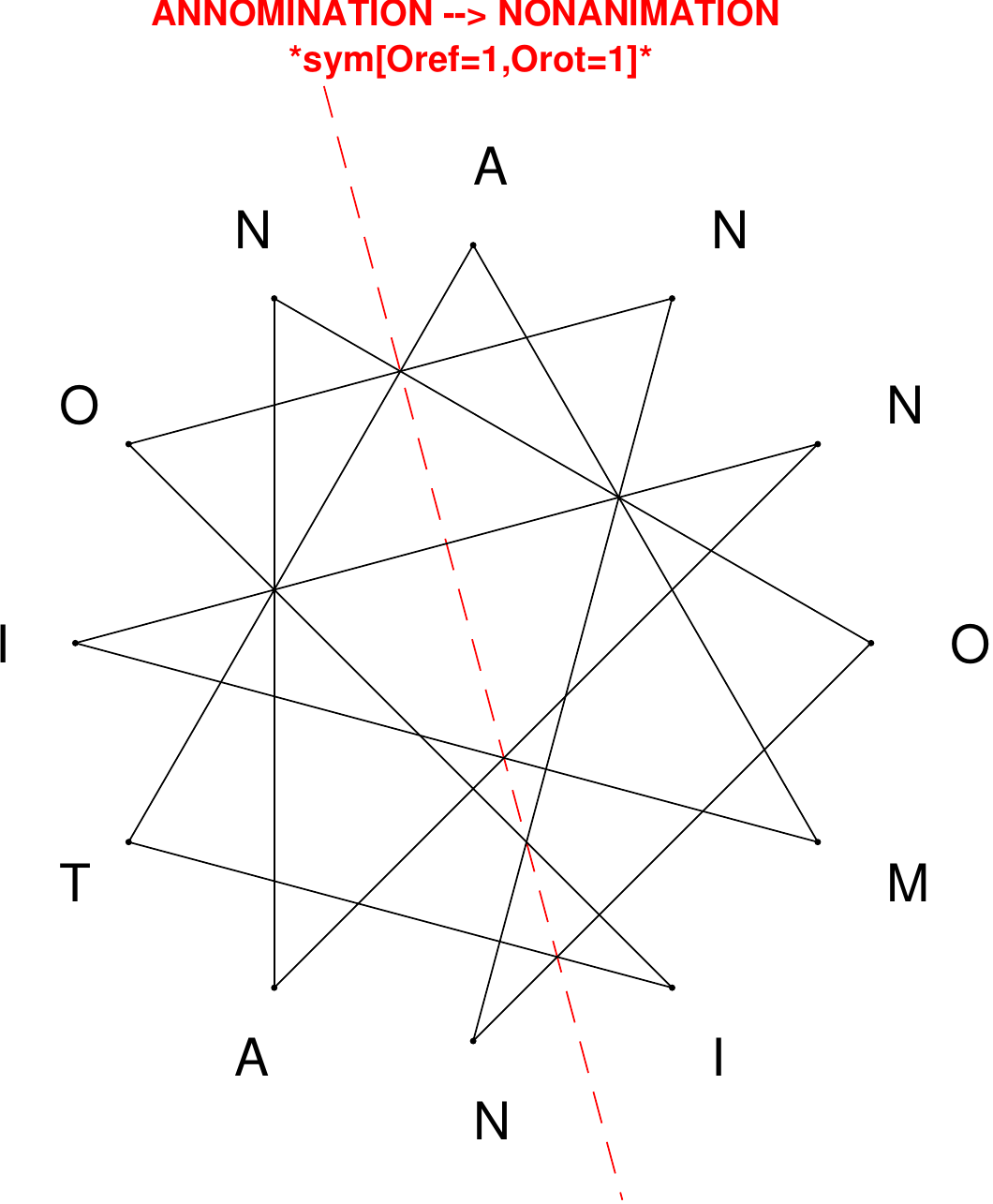}
\end{subfigure}
\hfill
\begin{subfigure}[T]{0.19\textwidth}
\centering
\includegraphics[width=\textwidth]{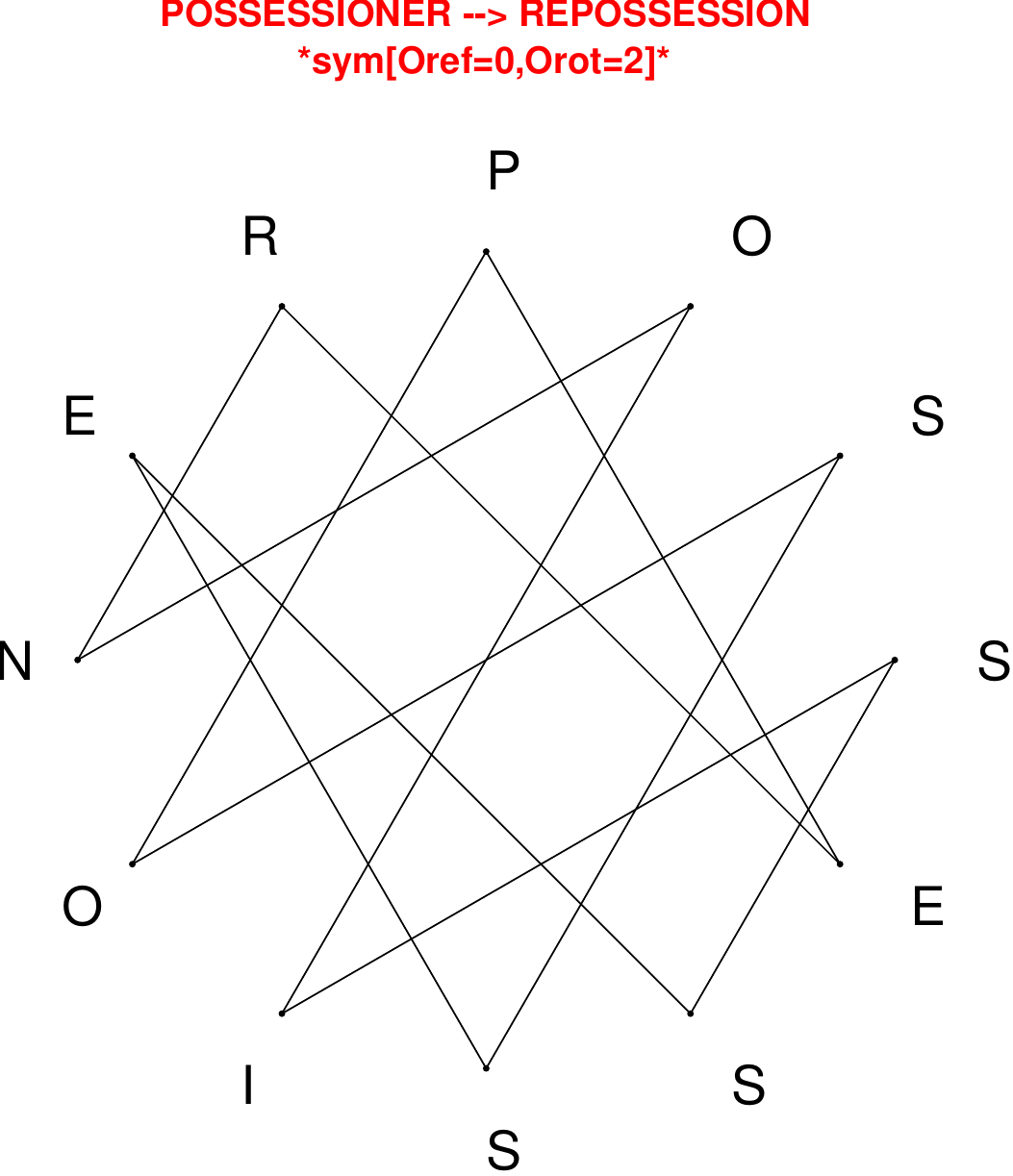}
\end{subfigure}
\hfill
\end{figure}

\subsubsection{Asymmetric Stars $N=12$}

\begin{figure}[H]
\centering
\begin{subfigure}[T]{0.19\textwidth}
\centering
\includegraphics[width=\textwidth]{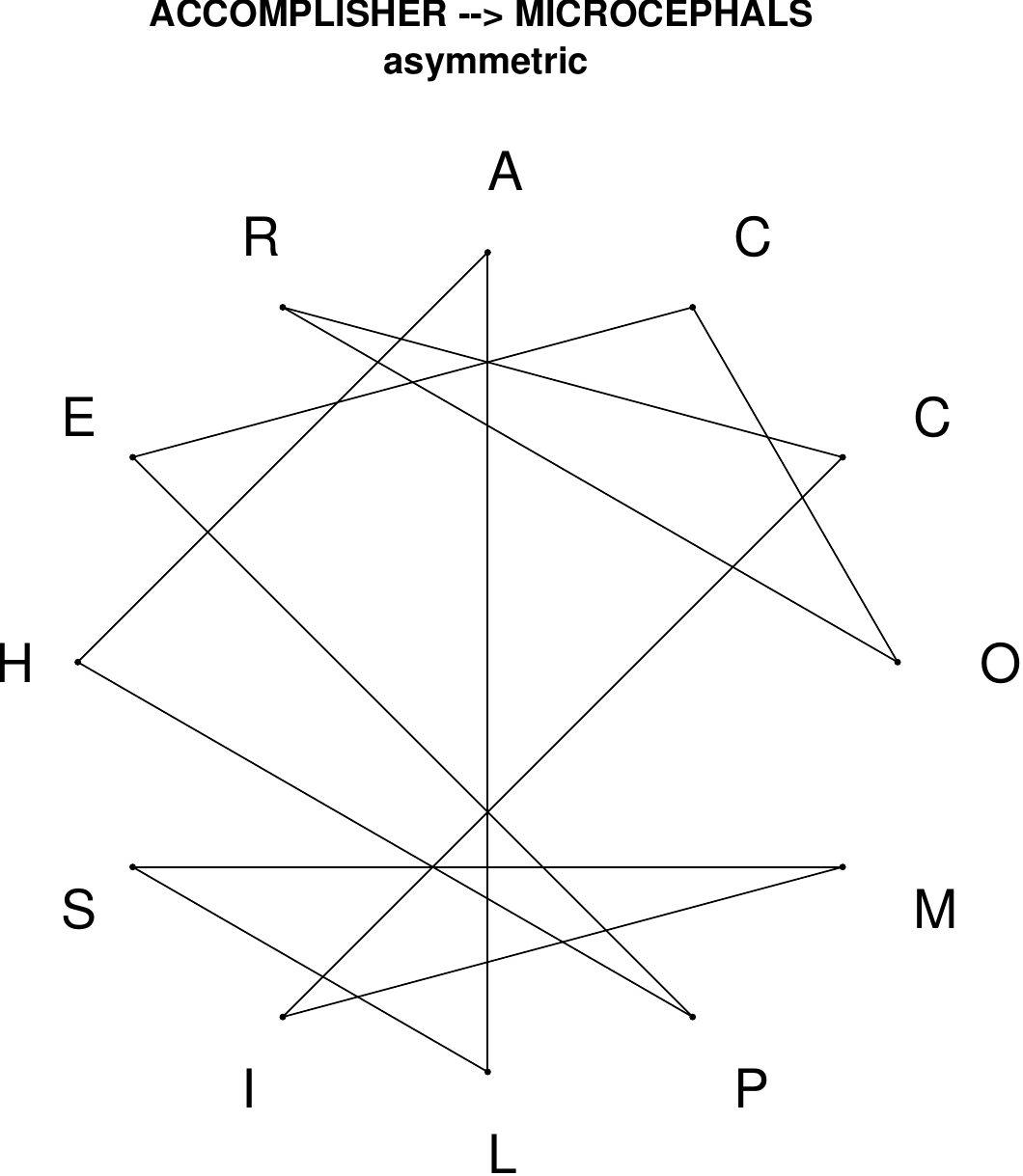}
\end{subfigure}
\hfill
\begin{subfigure}[T]{0.19\textwidth}
\centering
\includegraphics[width=\textwidth]{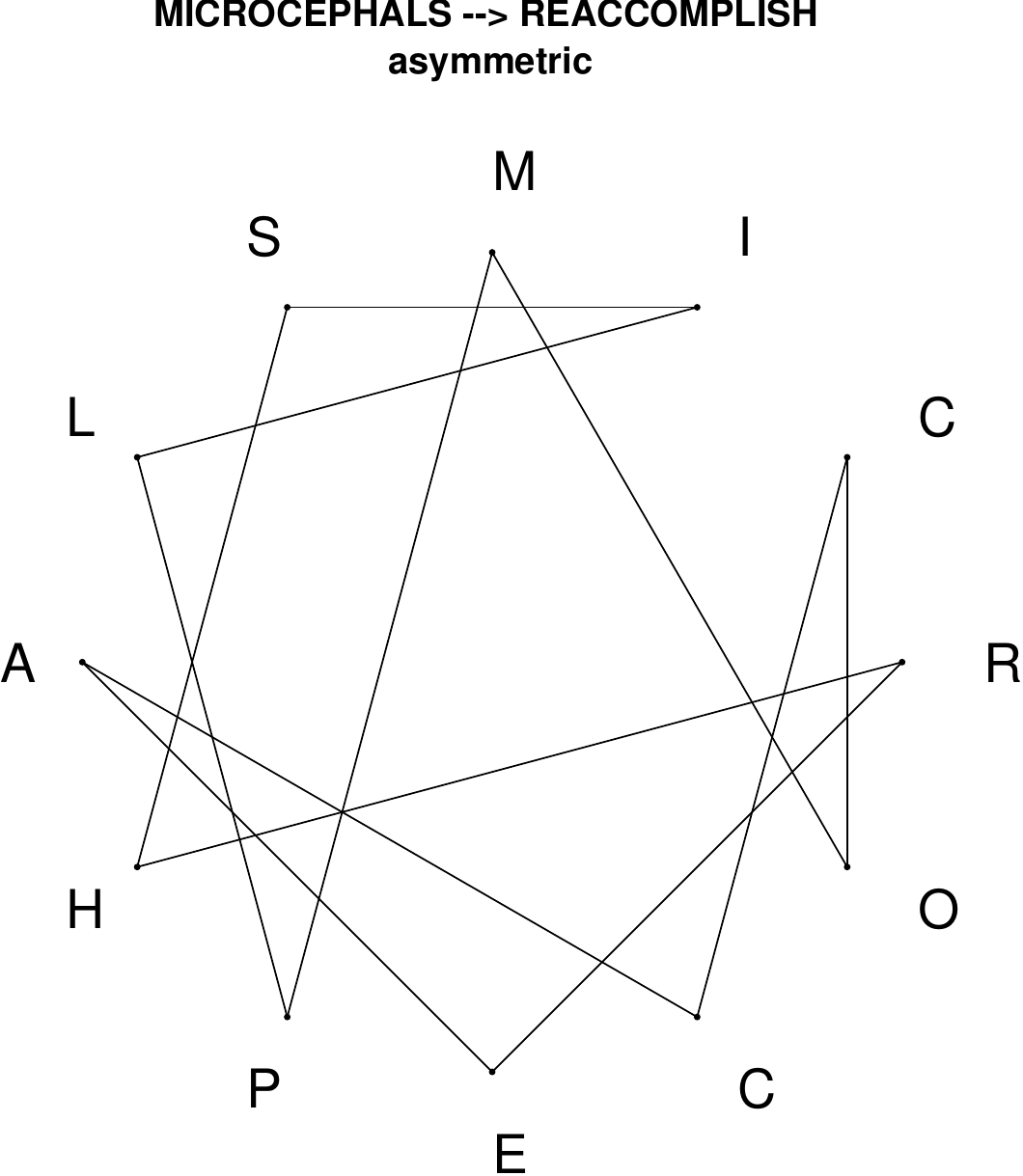}
\end{subfigure}
\hfill
\begin{subfigure}[T]{0.19\textwidth}
\centering
\includegraphics[width=\textwidth]{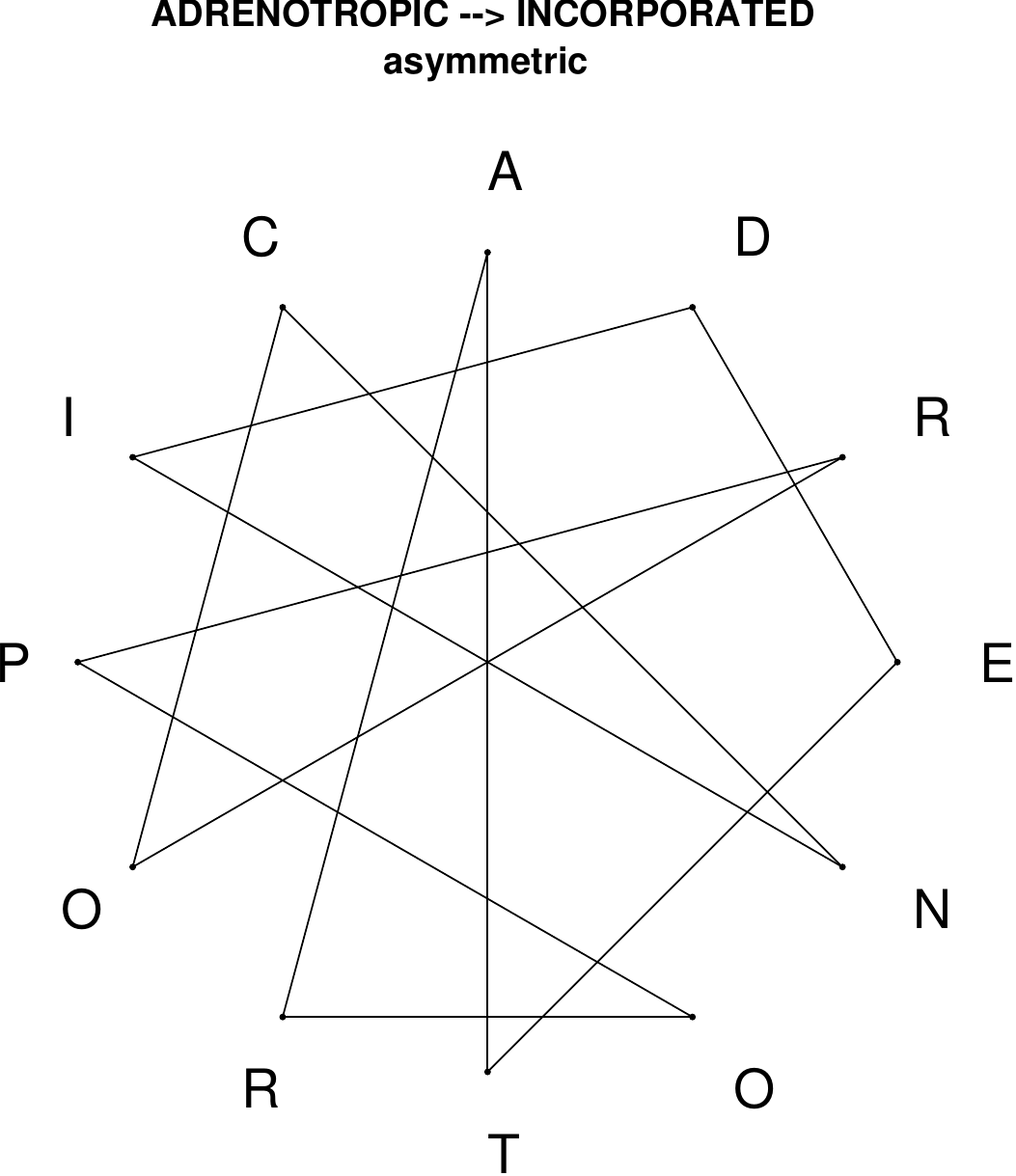}
\end{subfigure}
\hfill
\begin{subfigure}[T]{0.19\textwidth}
\centering
\includegraphics[width=\textwidth]{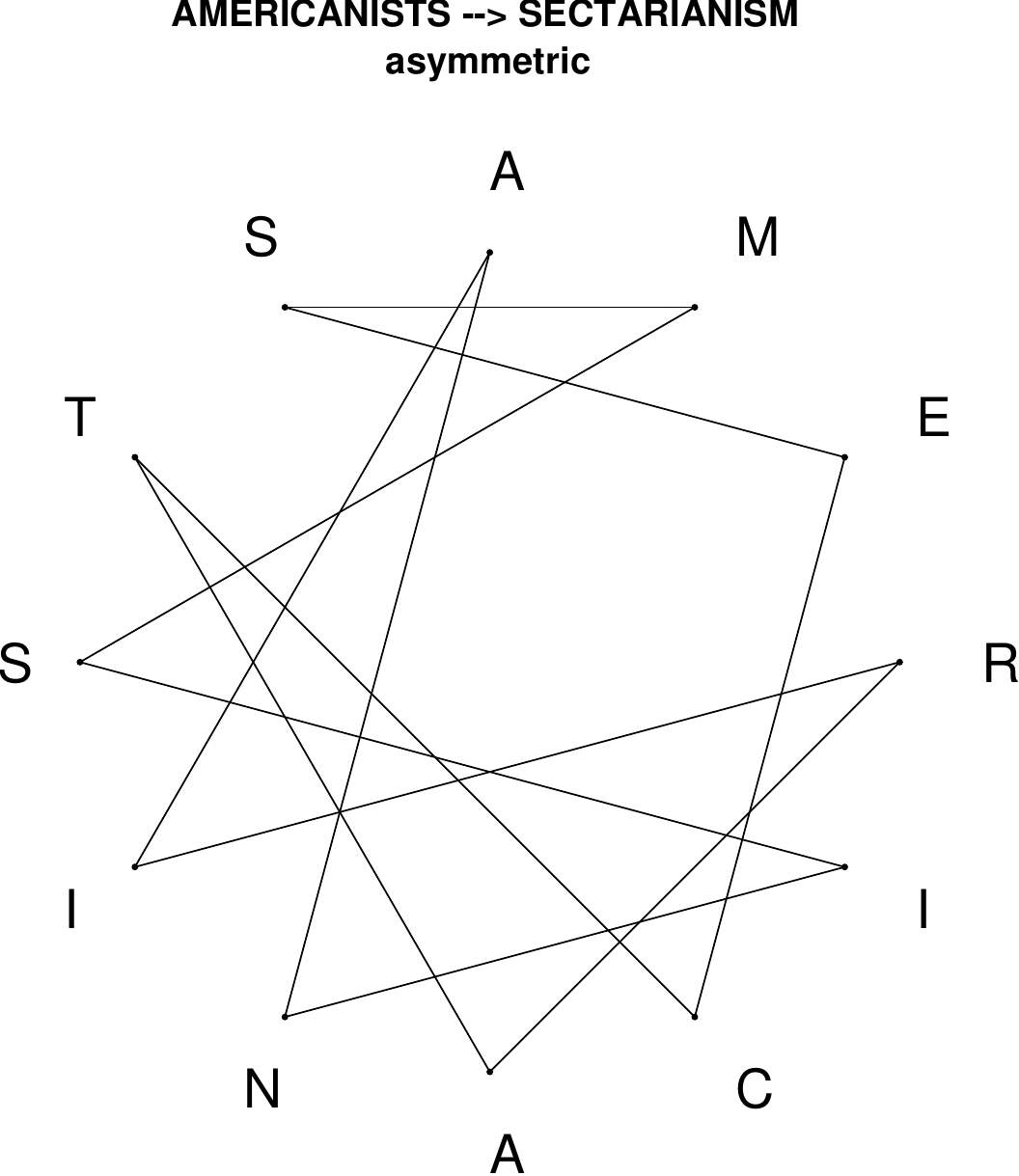}
\end{subfigure}
\hfill
\begin{subfigure}[T]{0.19\textwidth}
\centering
\includegraphics[width=\textwidth]{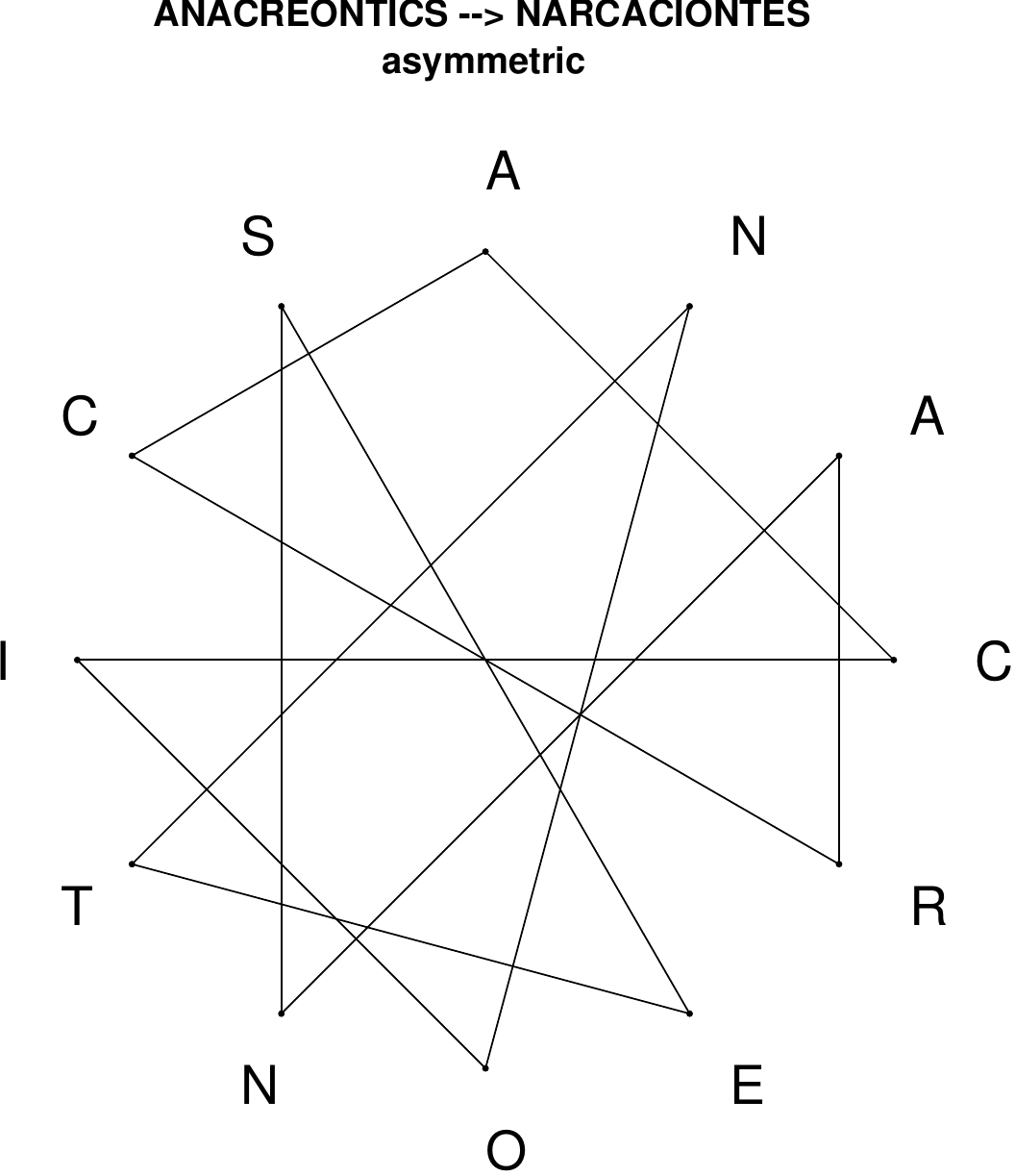}
\end{subfigure}
\end{figure}

\begin{figure}[H]
\centering
\begin{subfigure}[T]{0.19\textwidth}
\centering
\includegraphics[width=\textwidth]{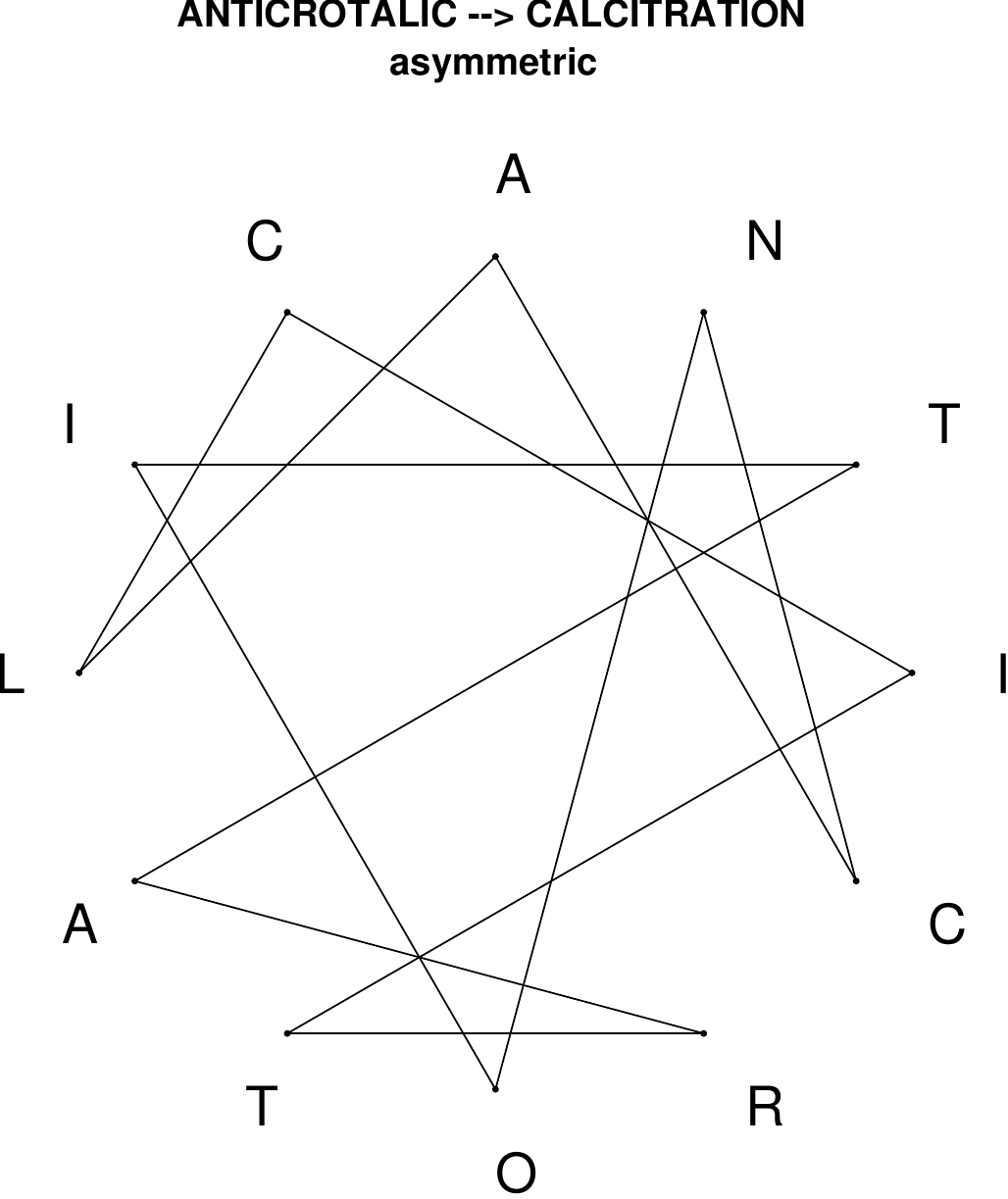}
\end{subfigure}
\hfill
\begin{subfigure}[T]{0.19\textwidth}
\centering
\includegraphics[width=\textwidth]{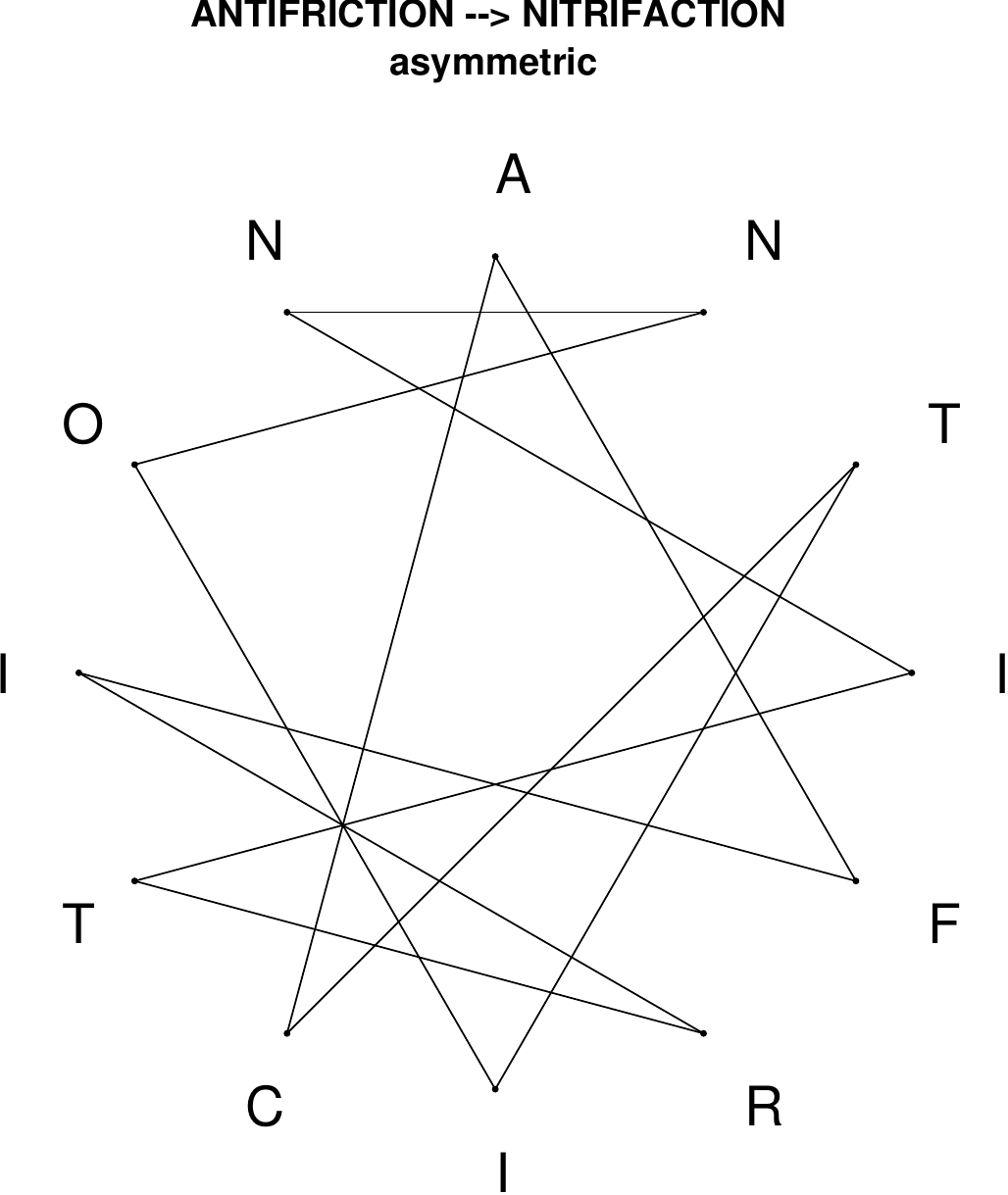}
\end{subfigure}
\hfill
\begin{subfigure}[T]{0.19\textwidth}
\centering
\includegraphics[width=\textwidth]{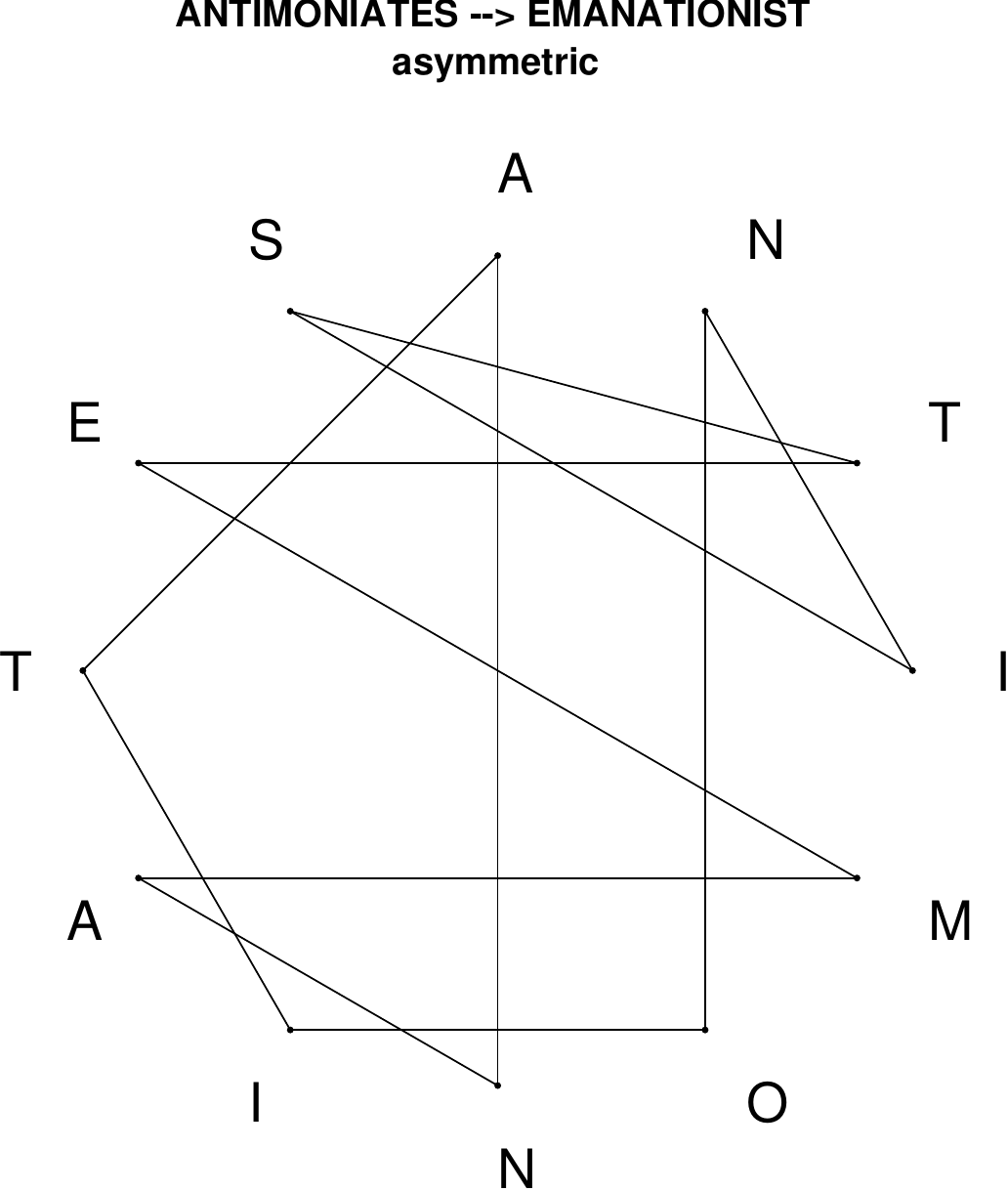}
\end{subfigure}
\hfill
\begin{subfigure}[T]{0.19\textwidth}
\centering
\includegraphics[width=\textwidth]{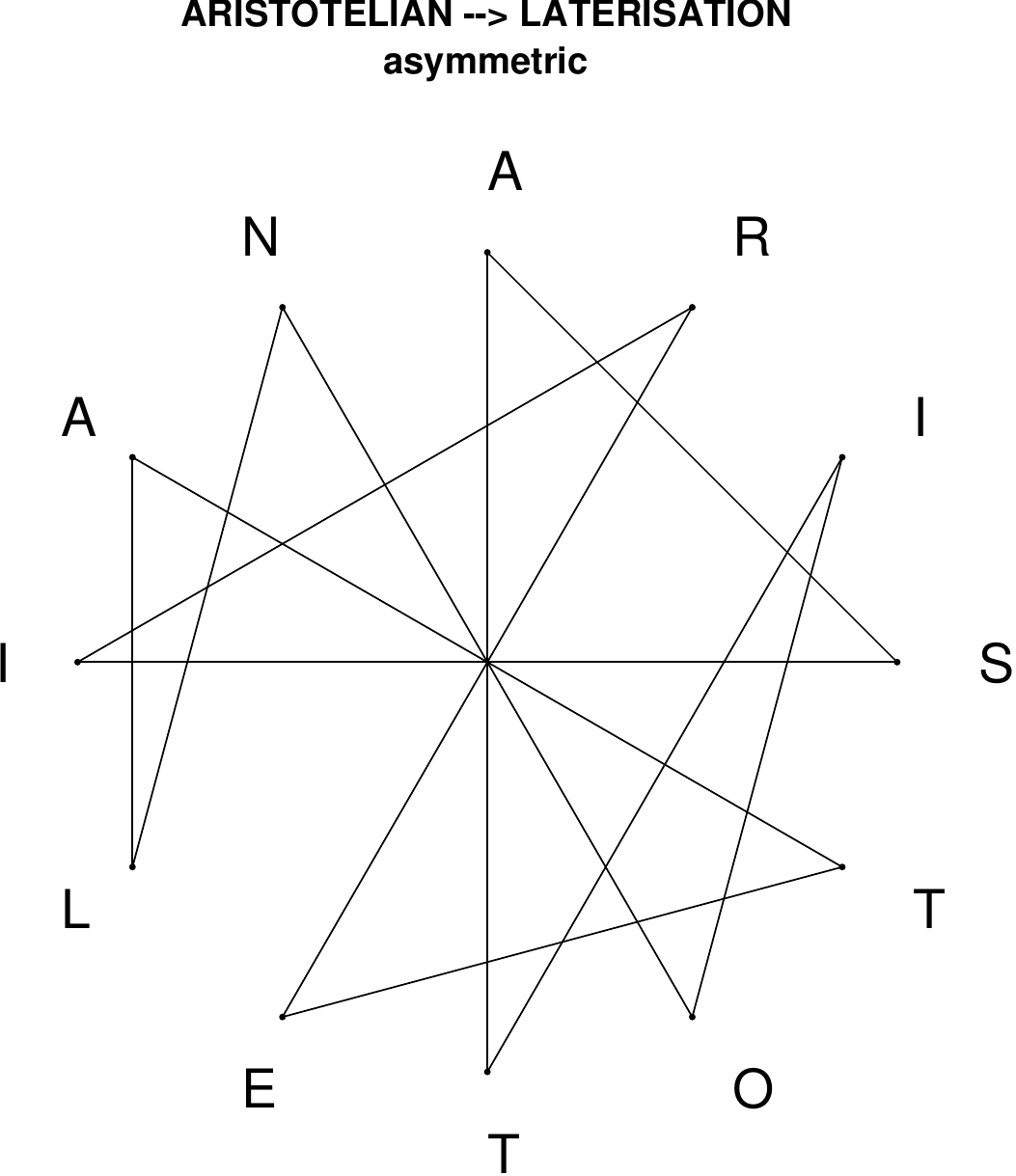}
\end{subfigure}
\hfill
\begin{subfigure}[T]{0.19\textwidth}
\centering
\includegraphics[width=\textwidth]{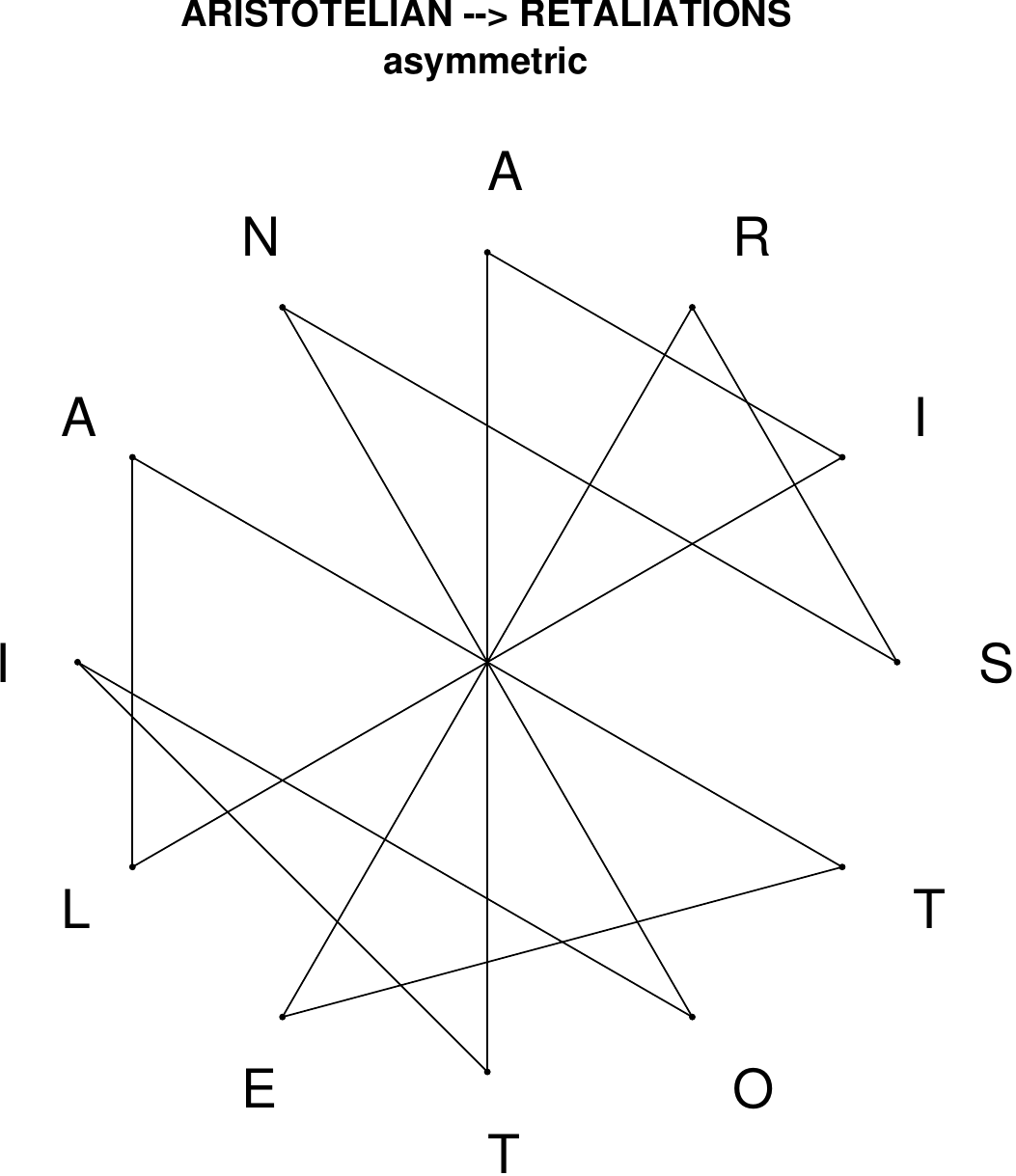}
\end{subfigure}
\end{figure}

\begin{figure}[H]
\centering
\begin{subfigure}[T]{0.19\textwidth}
\centering
\includegraphics[width=\textwidth]{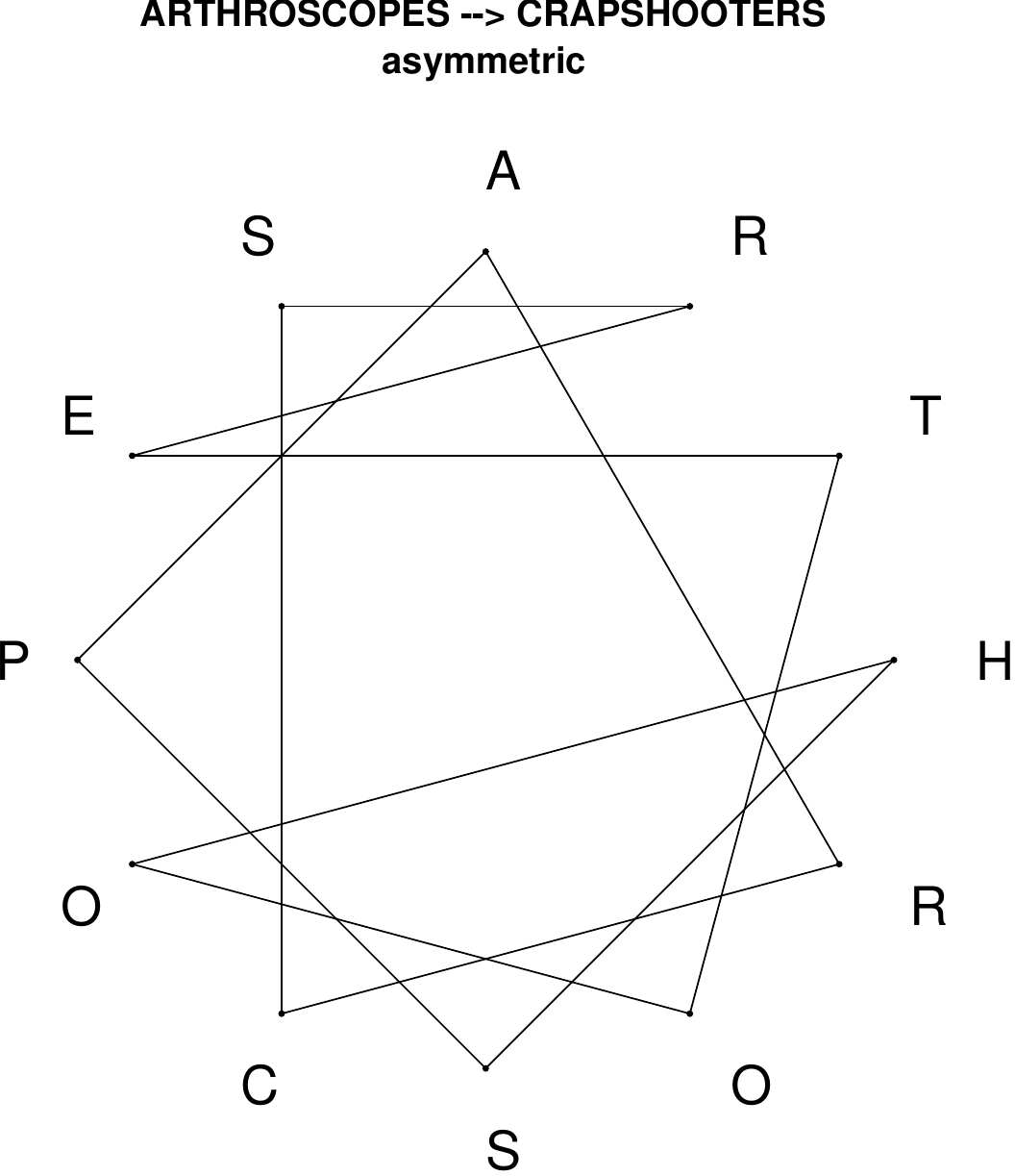}
\end{subfigure}
\hfill
\begin{subfigure}[T]{0.19\textwidth}
\centering
\includegraphics[width=\textwidth]{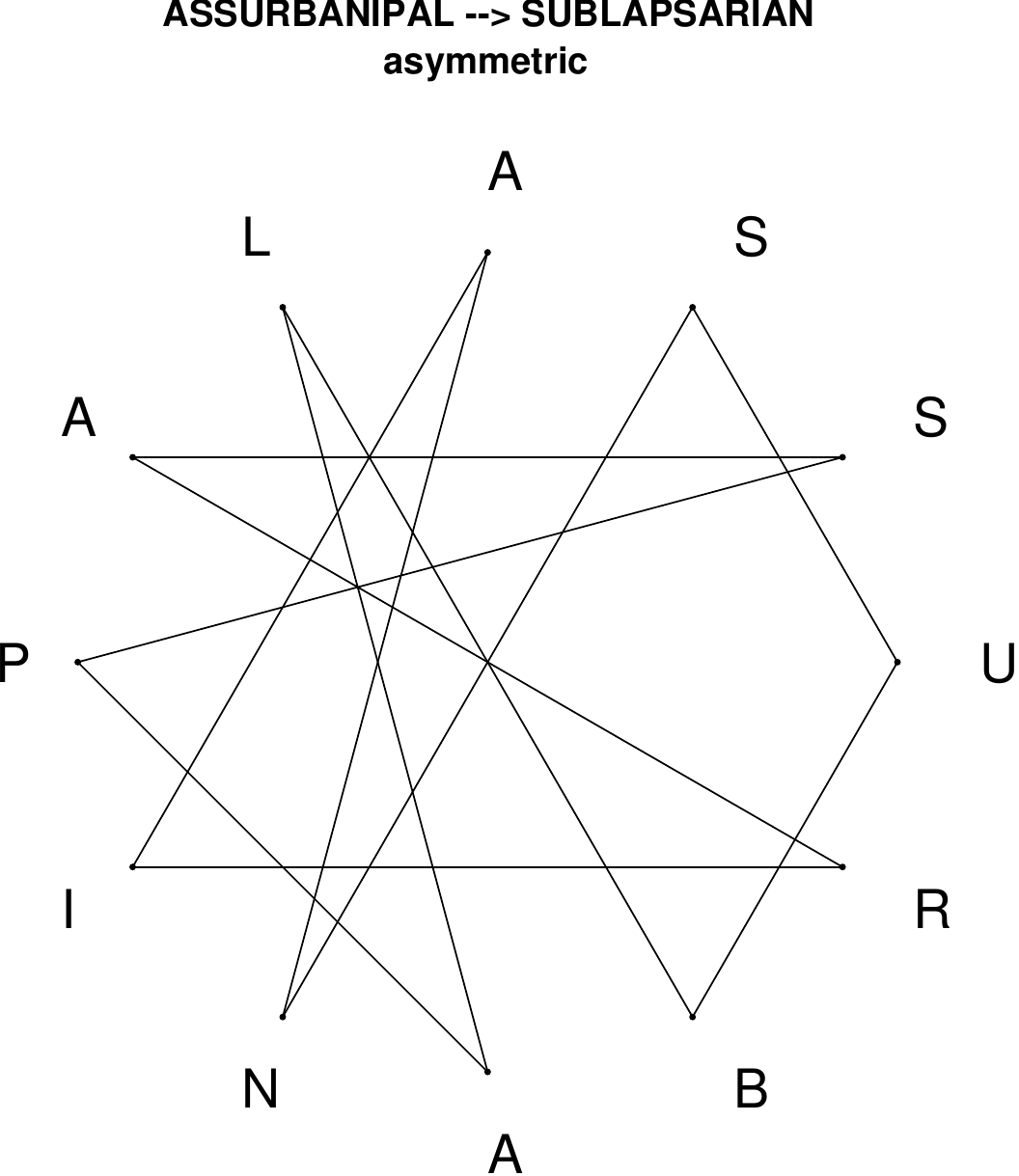}
\end{subfigure}
\hfill
\begin{subfigure}[T]{0.19\textwidth}
\centering
\includegraphics[width=\textwidth]{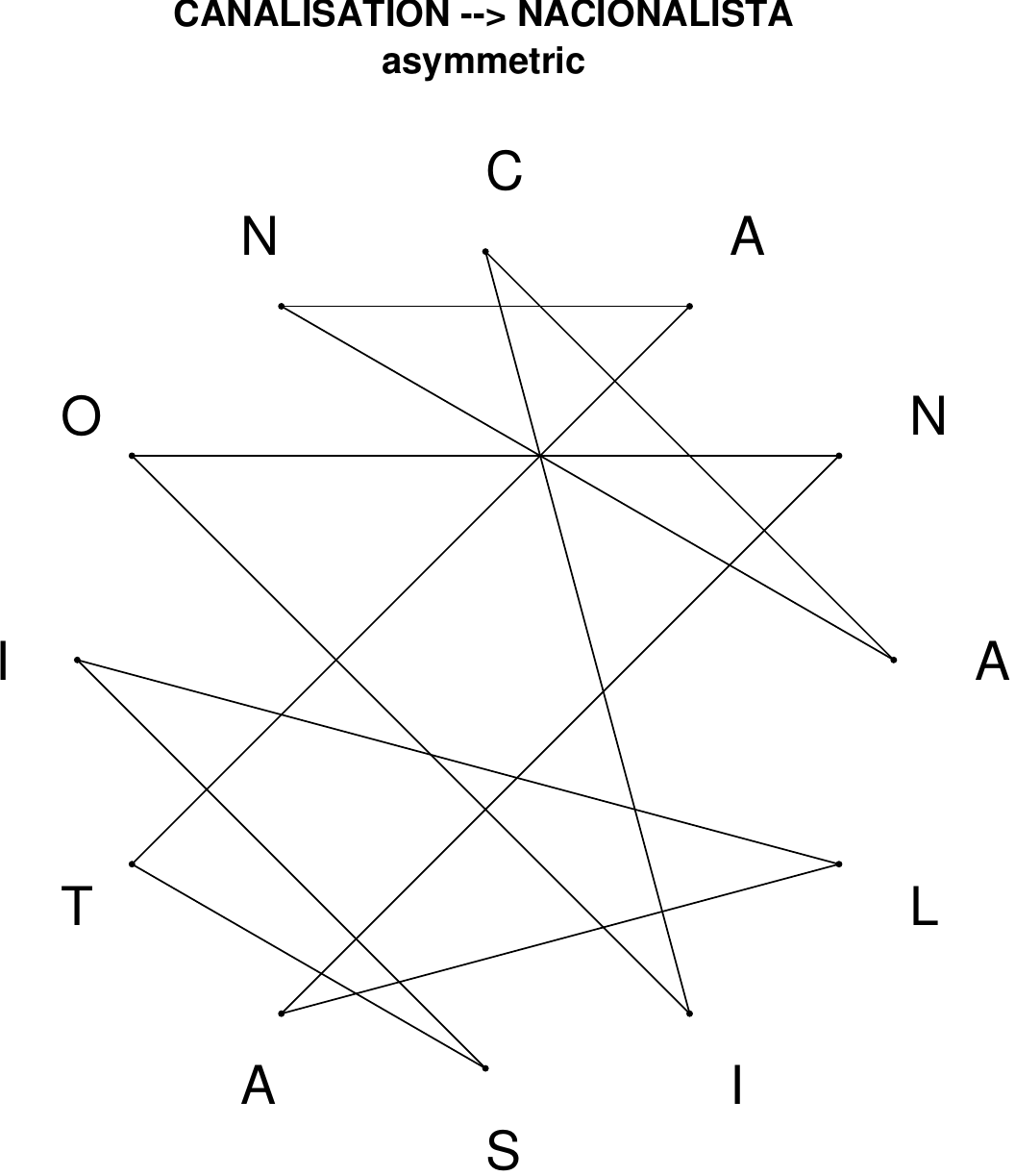}
\end{subfigure}
\hfill
\begin{subfigure}[T]{0.19\textwidth}
\centering
\includegraphics[width=\textwidth]{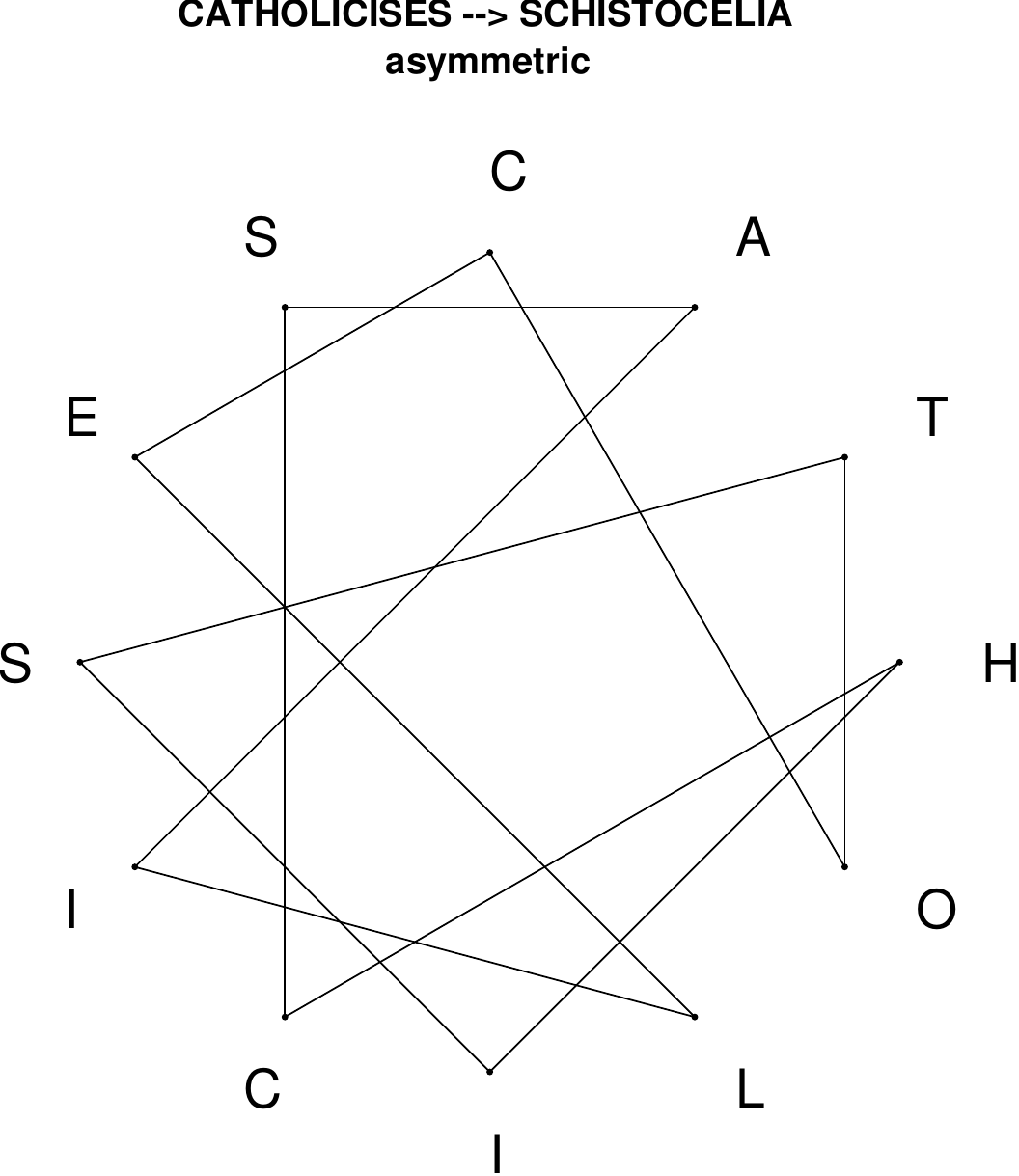}
\end{subfigure}
\hfill
\begin{subfigure}[T]{0.19\textwidth}
\centering
\includegraphics[width=\textwidth]{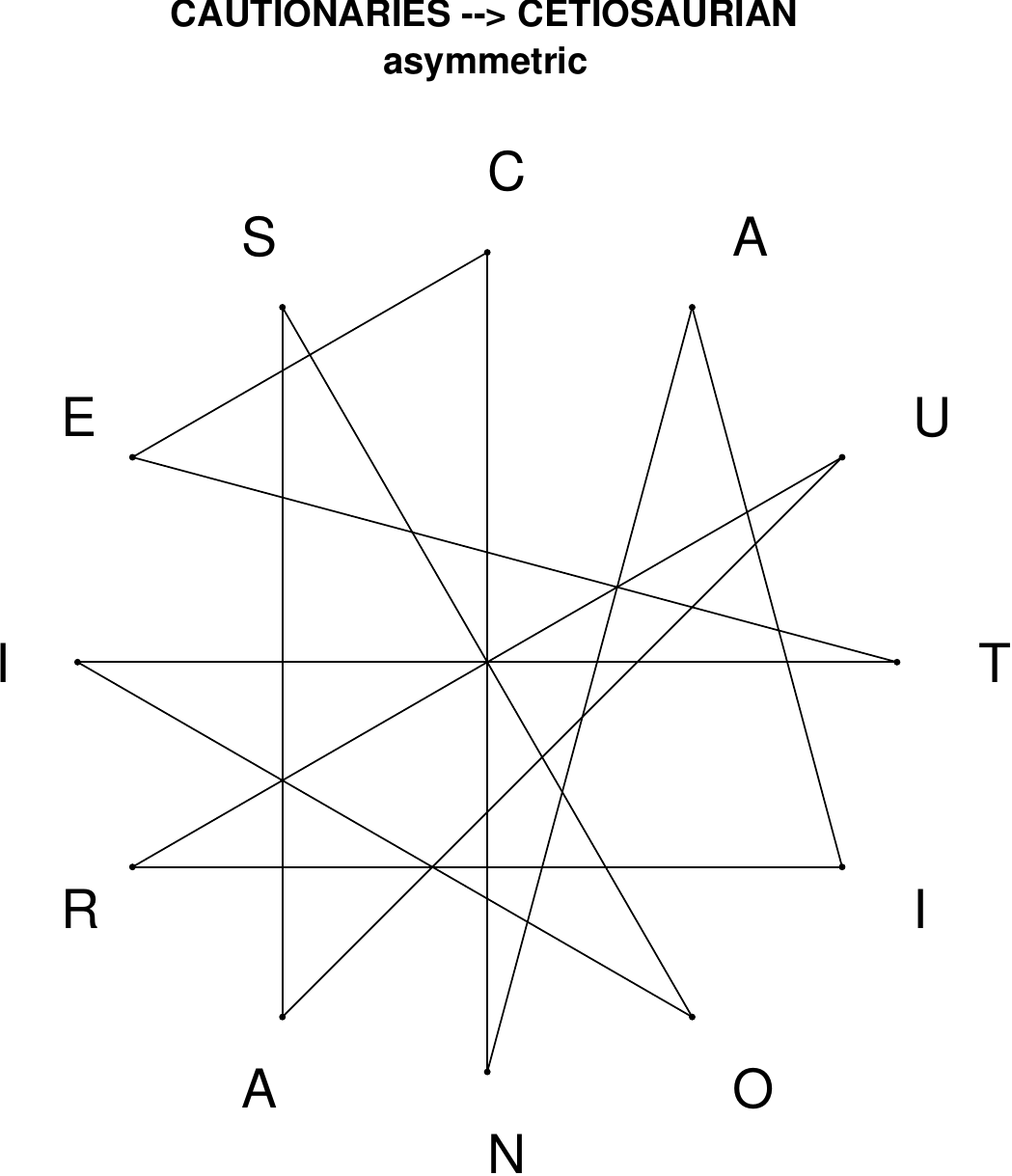}
\end{subfigure}
\end{figure}

\begin{figure}[H]
\centering
\begin{subfigure}[T]{0.19\textwidth}
\centering
\includegraphics[width=\textwidth]{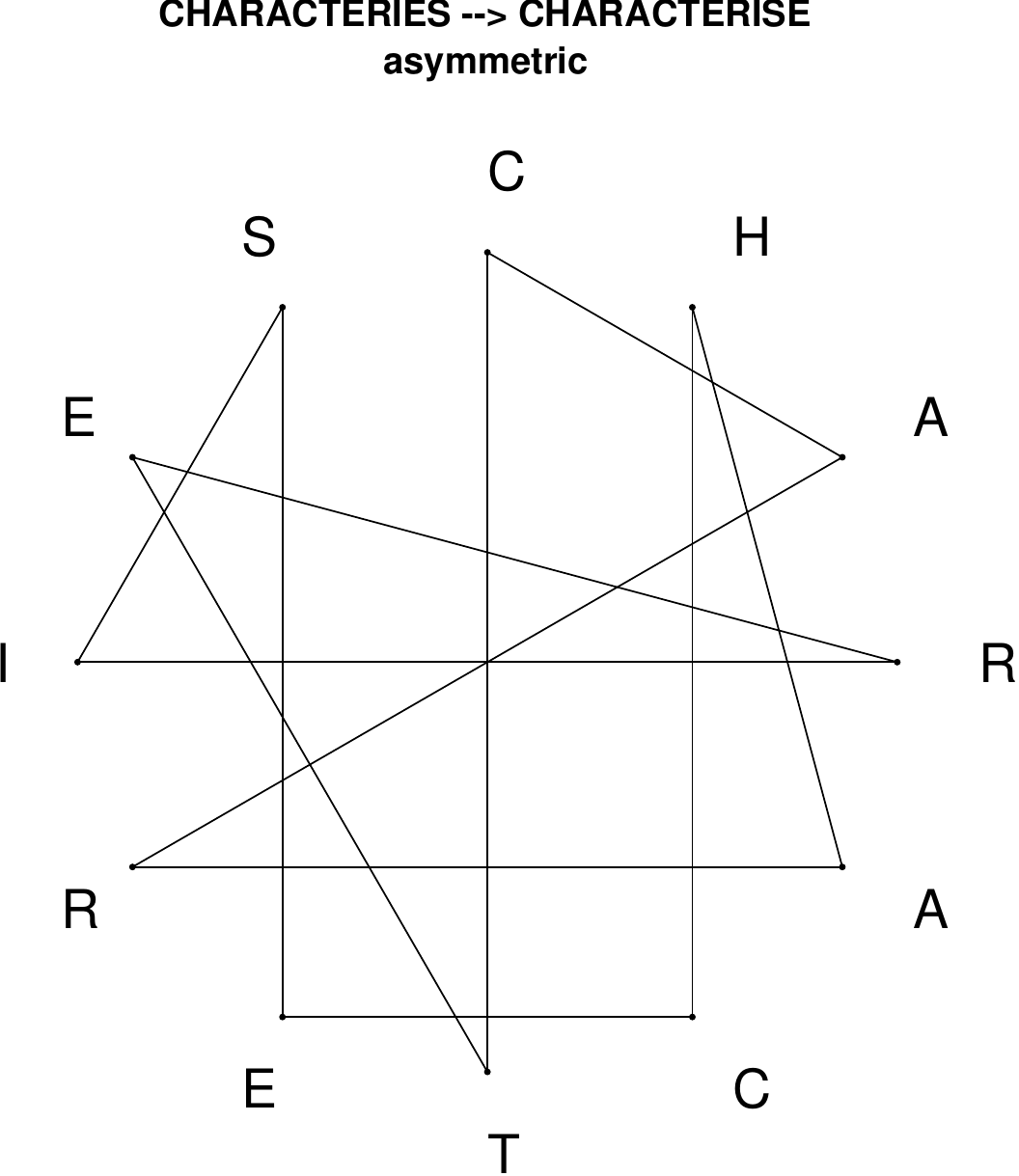}
\end{subfigure}
\hfill
\begin{subfigure}[T]{0.19\textwidth}
\centering
\includegraphics[width=\textwidth]{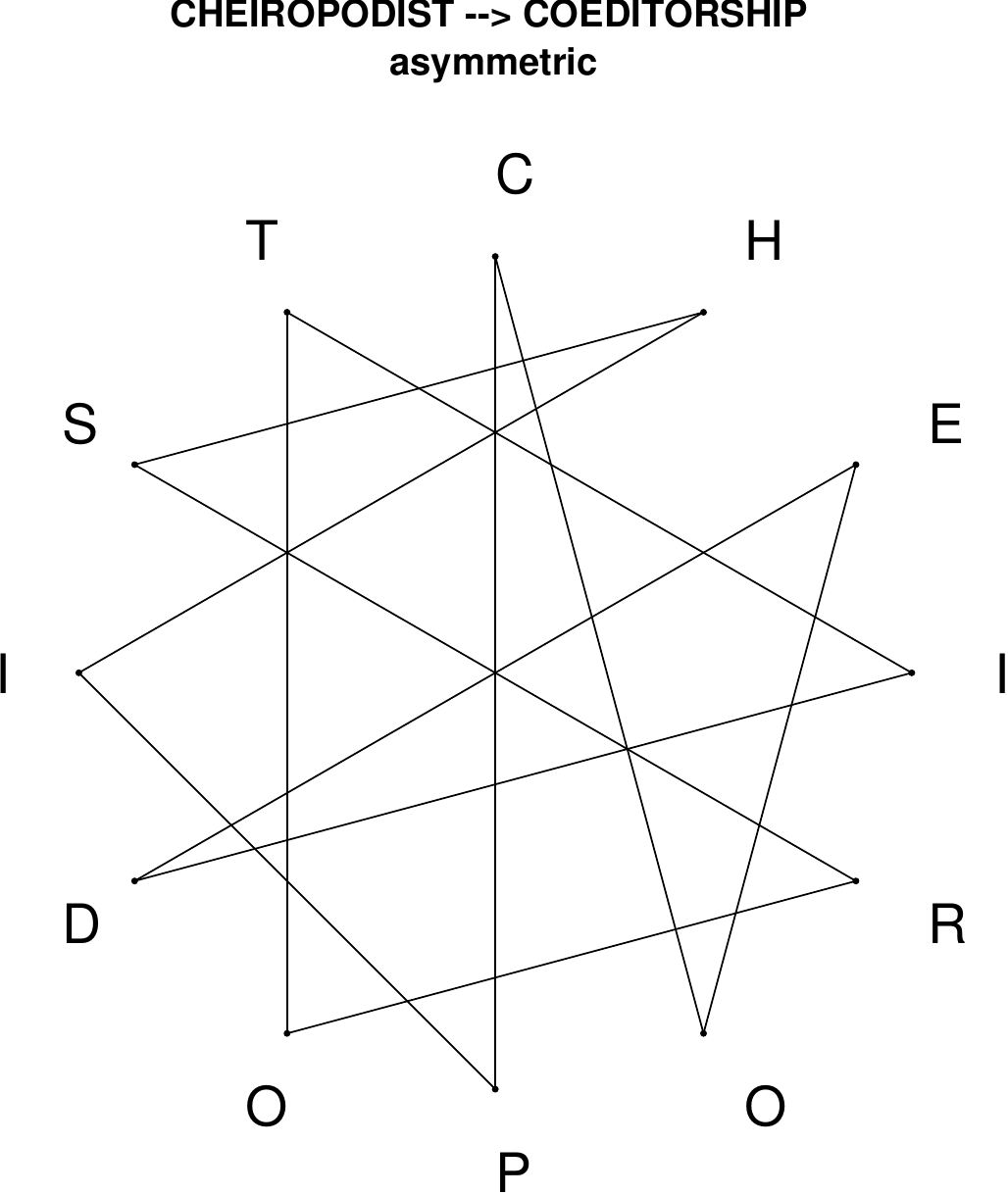}
\end{subfigure}
\hfill
\begin{subfigure}[T]{0.19\textwidth}
\centering
\includegraphics[width=\textwidth]{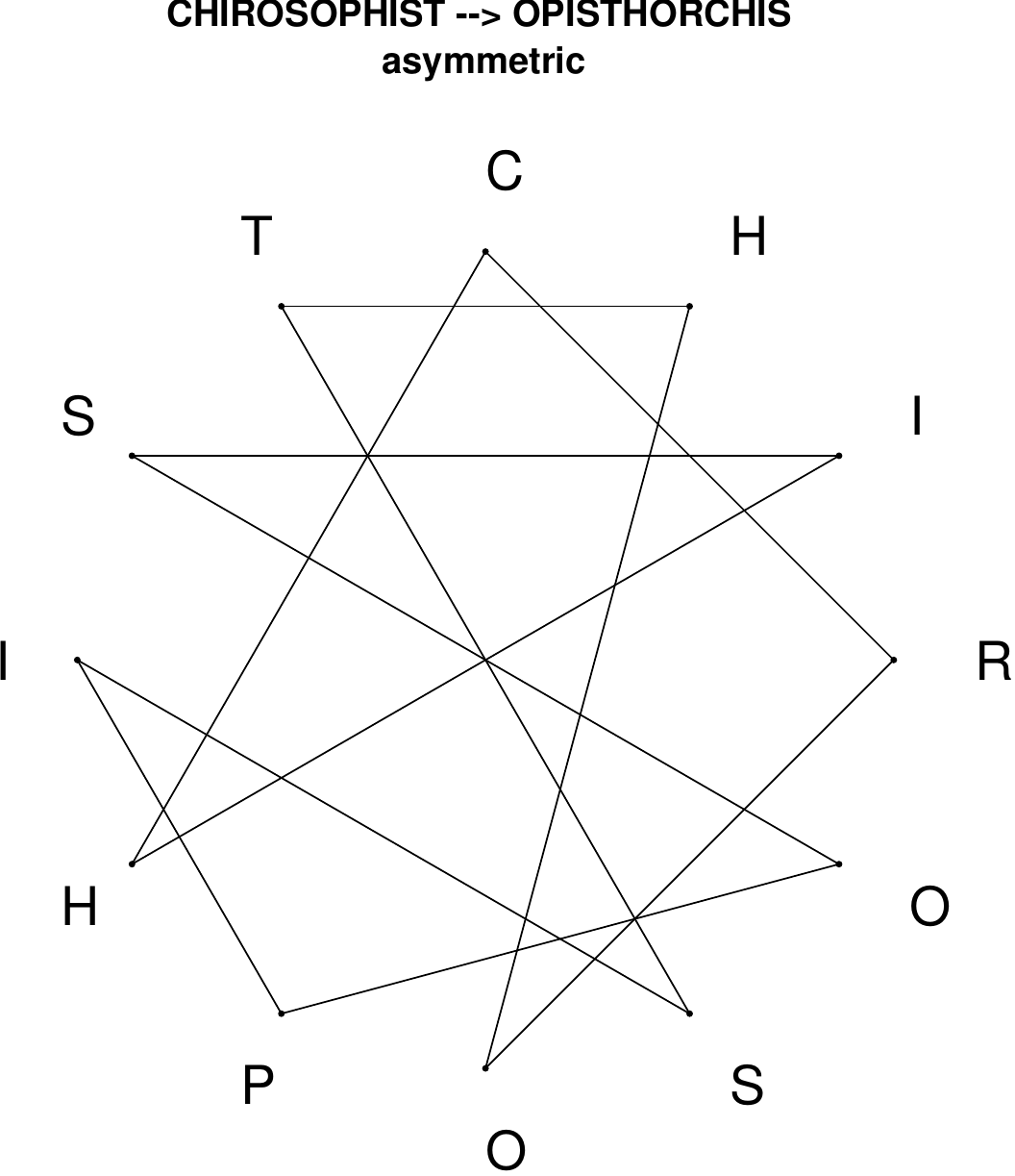}
\end{subfigure}
\hfill
\begin{subfigure}[T]{0.19\textwidth}
\centering
\includegraphics[width=\textwidth]{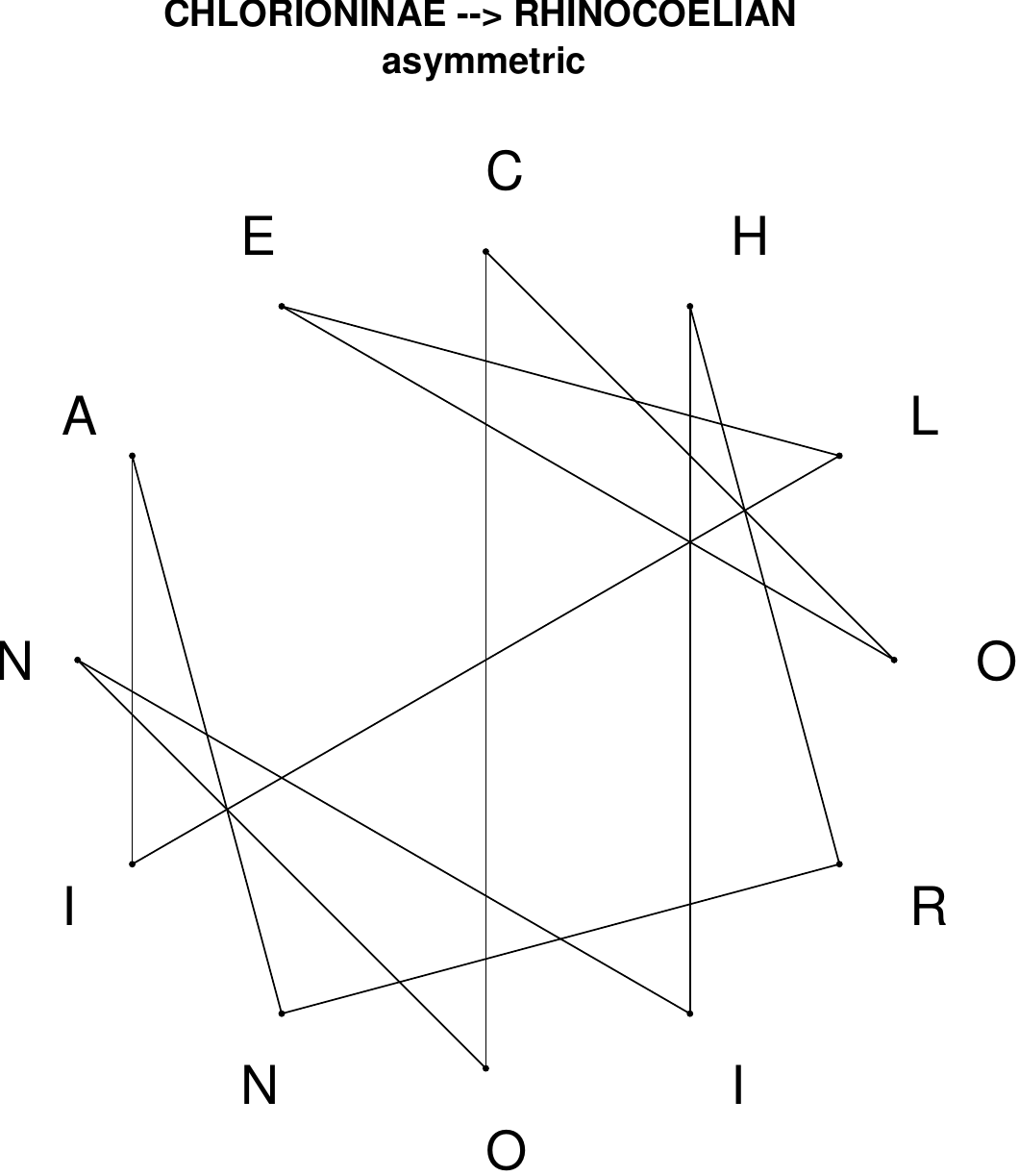}
\end{subfigure}
\hfill
\begin{subfigure}[T]{0.19\textwidth}
\centering
\includegraphics[width=\textwidth]{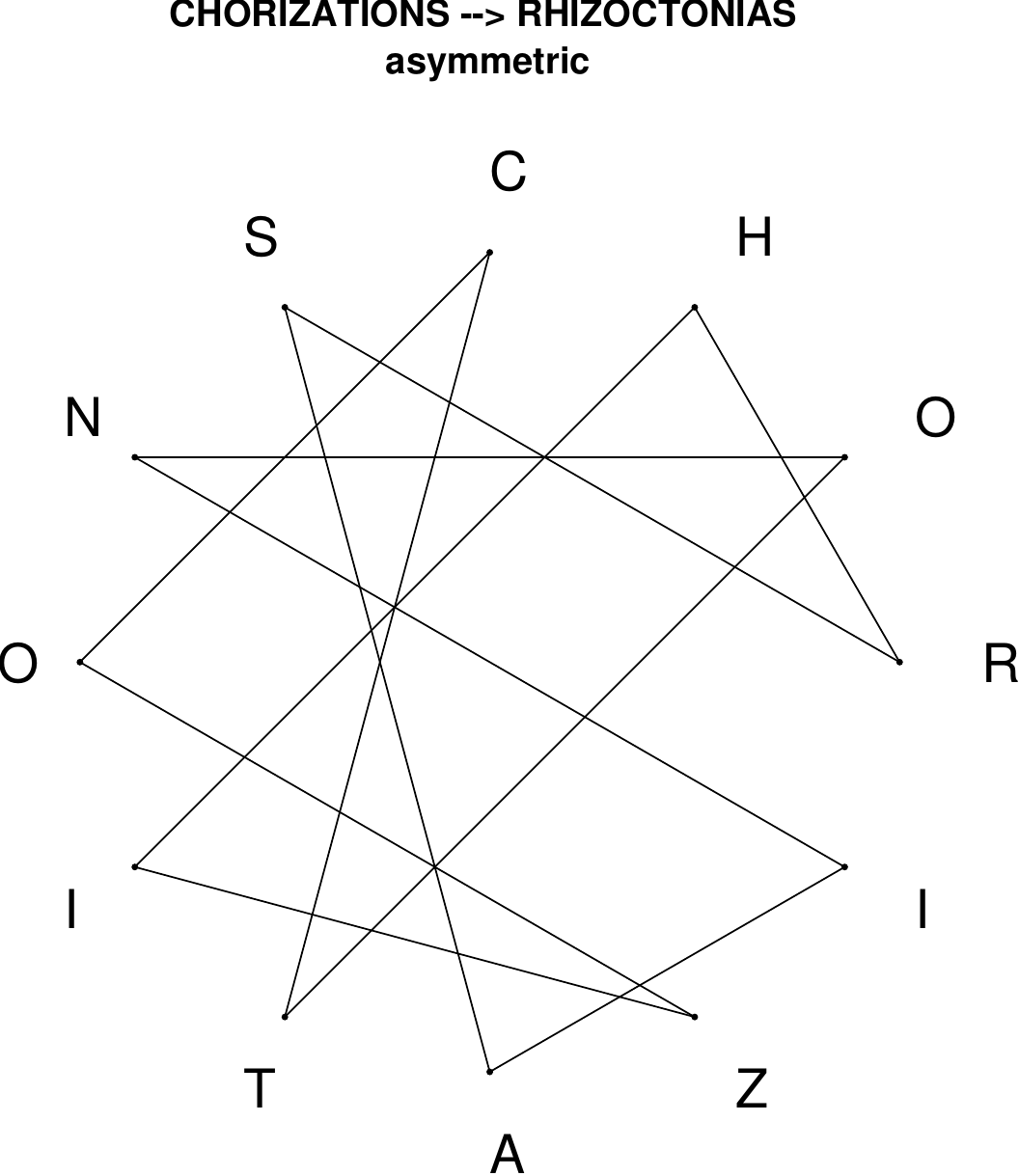}
\end{subfigure}
\end{figure}

\begin{figure}[H]
\centering
\begin{subfigure}[T]{0.19\textwidth}
\centering
\includegraphics[width=\textwidth]{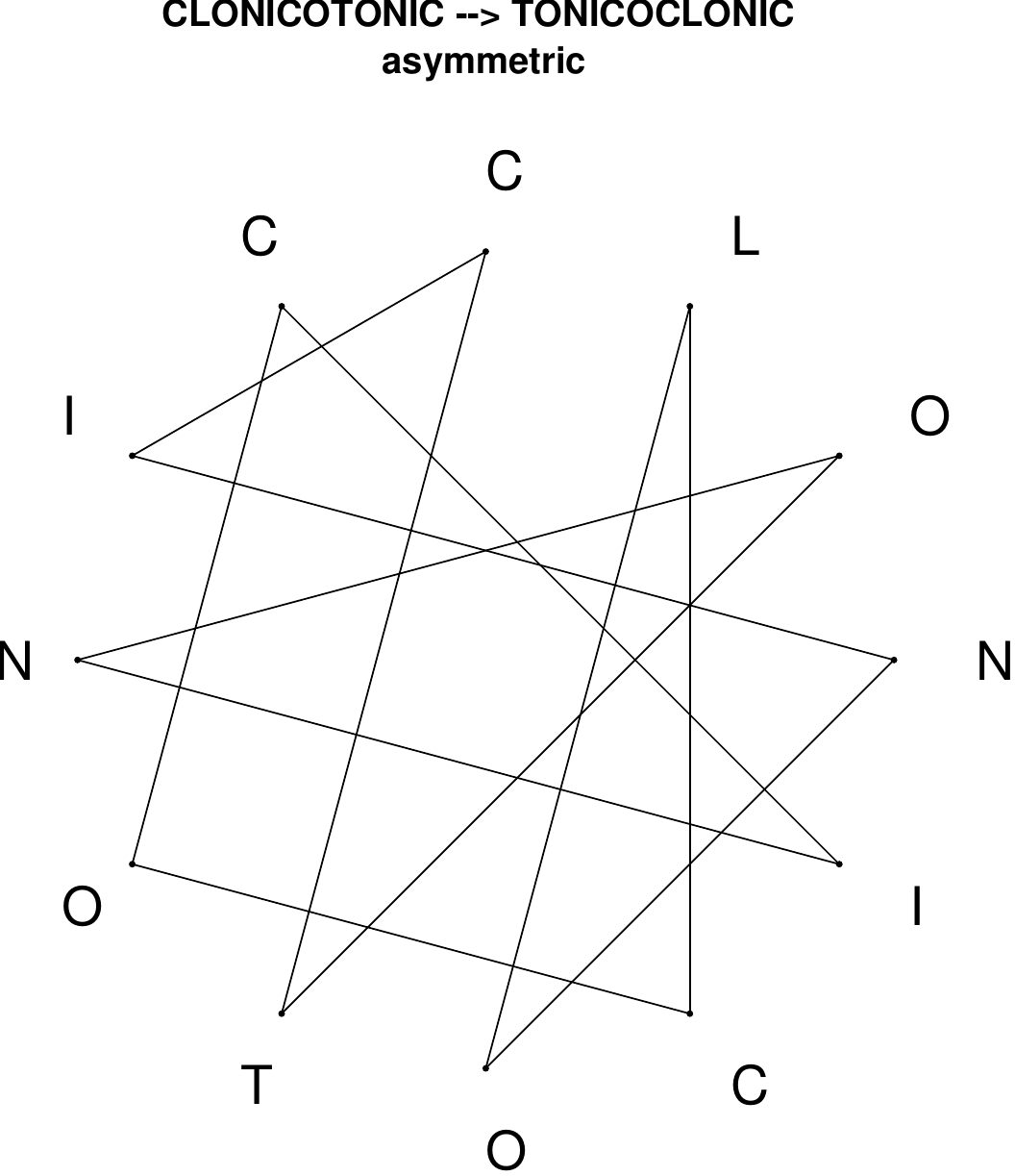}
\end{subfigure}
\hfill
\begin{subfigure}[T]{0.19\textwidth}
\centering
\includegraphics[width=\textwidth]{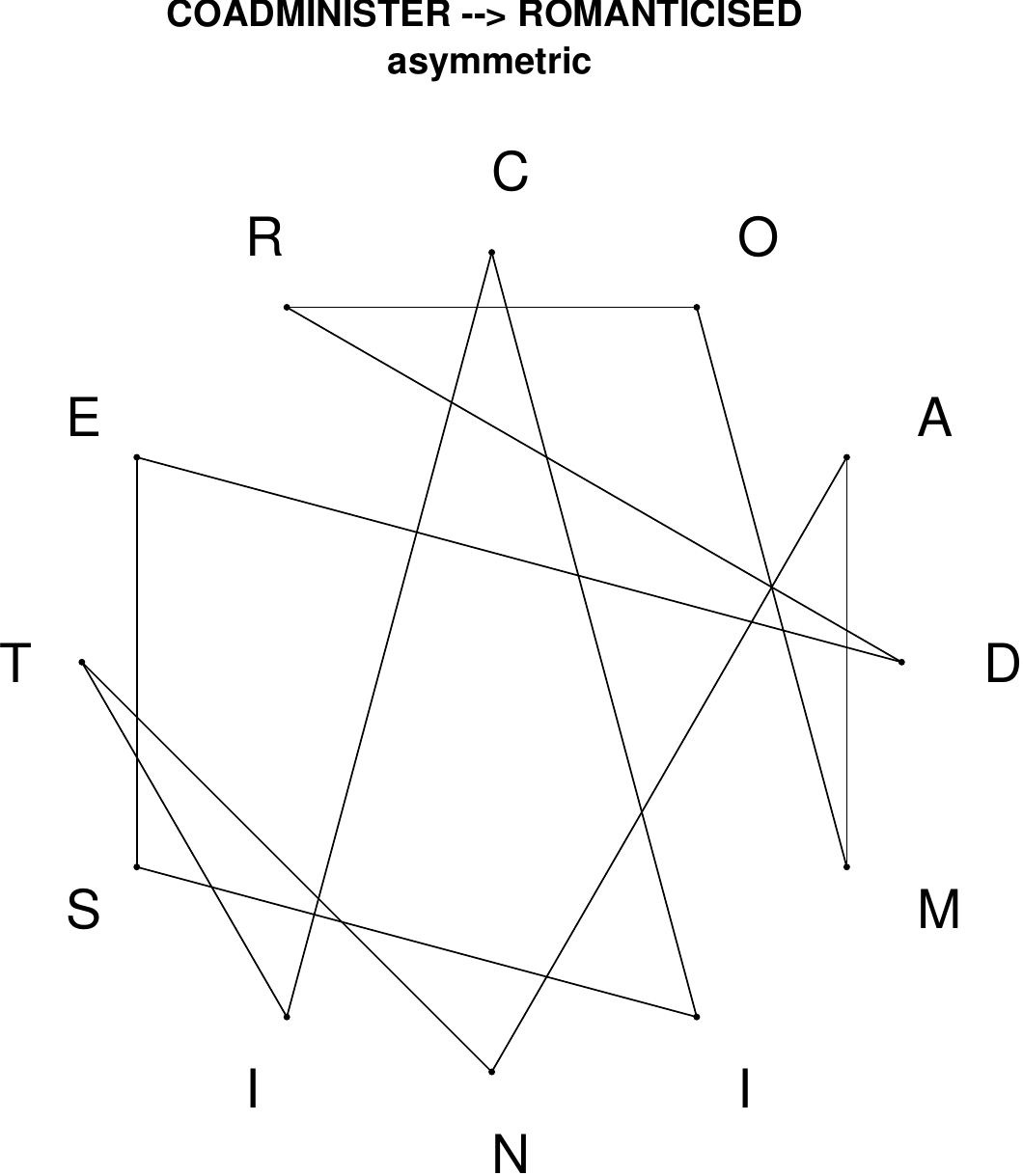}
\end{subfigure}
\hfill
\begin{subfigure}[T]{0.19\textwidth}
\centering
\includegraphics[width=\textwidth]{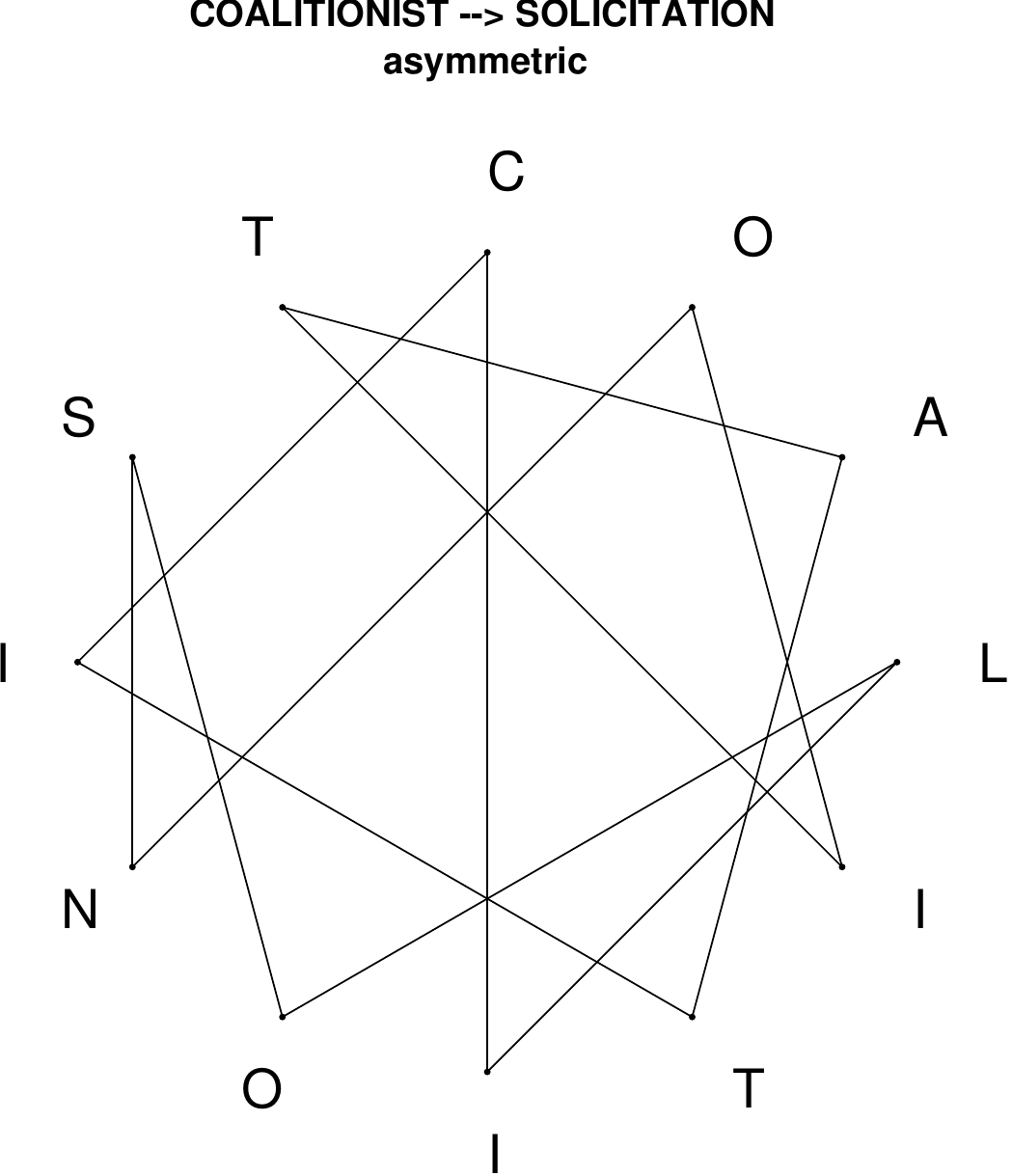}
\end{subfigure}
\hfill
\begin{subfigure}[T]{0.19\textwidth}
\centering
\includegraphics[width=\textwidth]{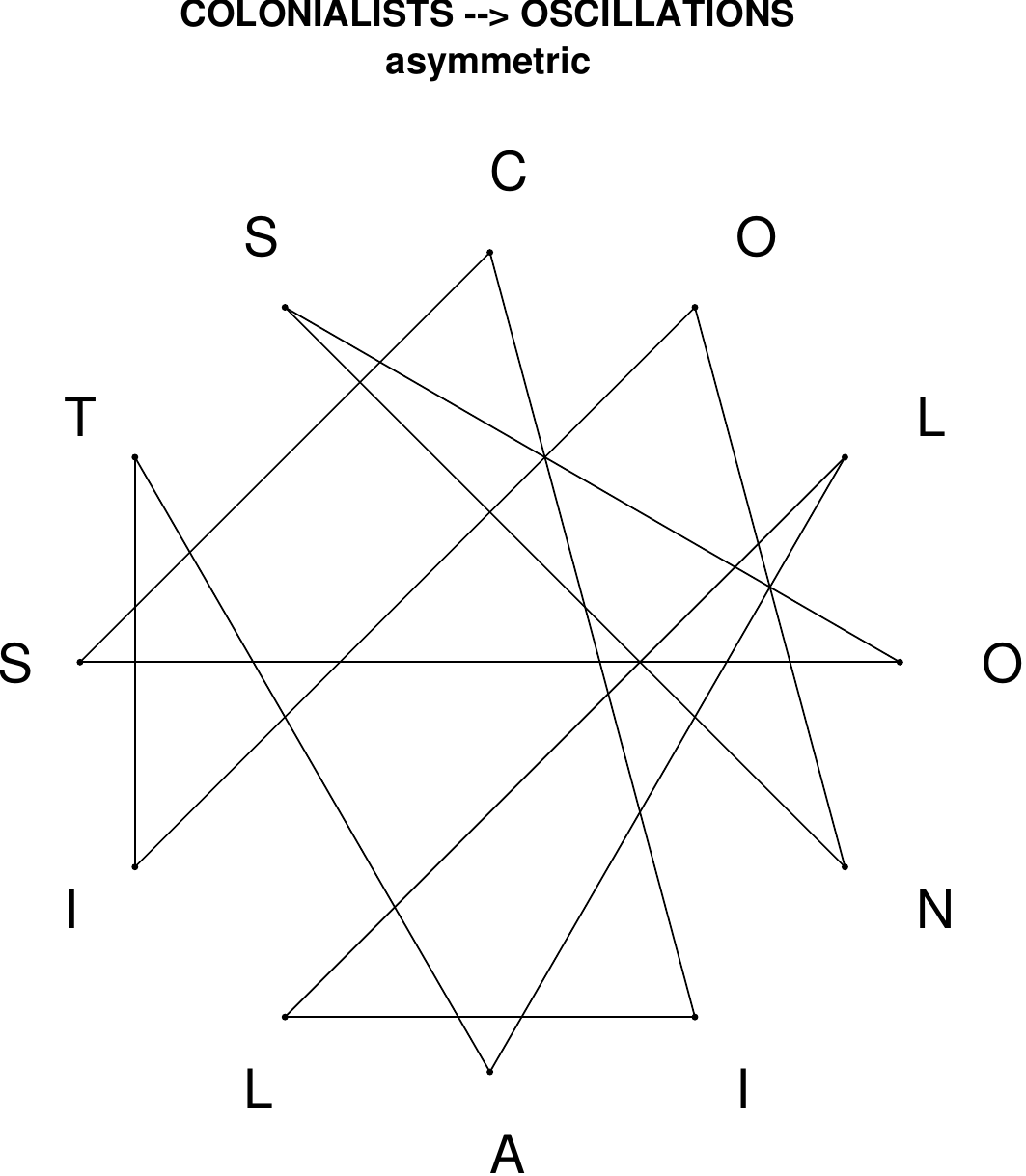}
\end{subfigure}
\hfill
\begin{subfigure}[T]{0.19\textwidth}
\centering
\includegraphics[width=\textwidth]{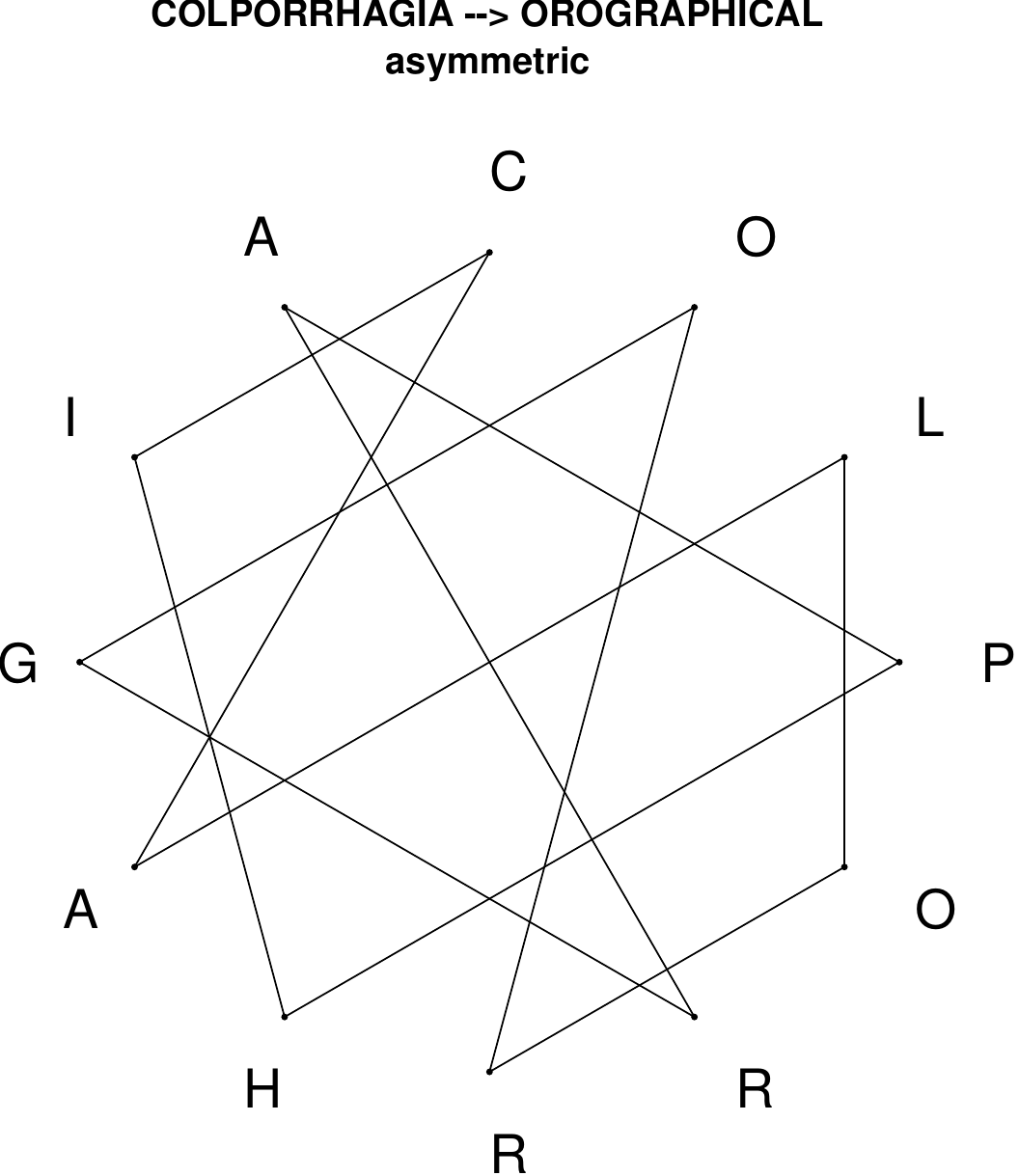}
\end{subfigure}
\end{figure}

\begin{figure}[H]
\centering
\begin{subfigure}[T]{0.19\textwidth}
\centering
\includegraphics[width=\textwidth]{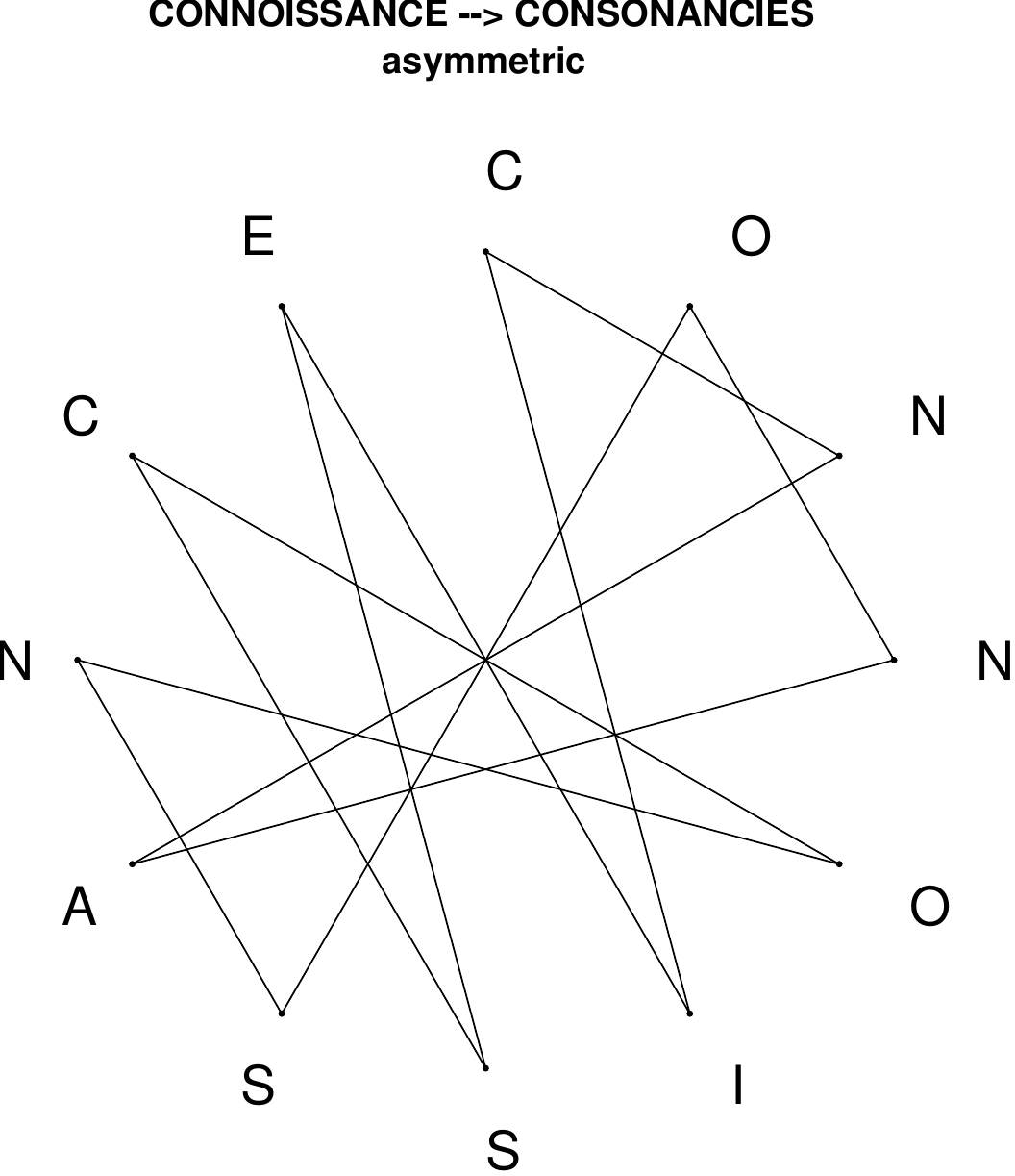}
\end{subfigure}
\hfill
\begin{subfigure}[T]{0.19\textwidth}
\centering
\includegraphics[width=\textwidth]{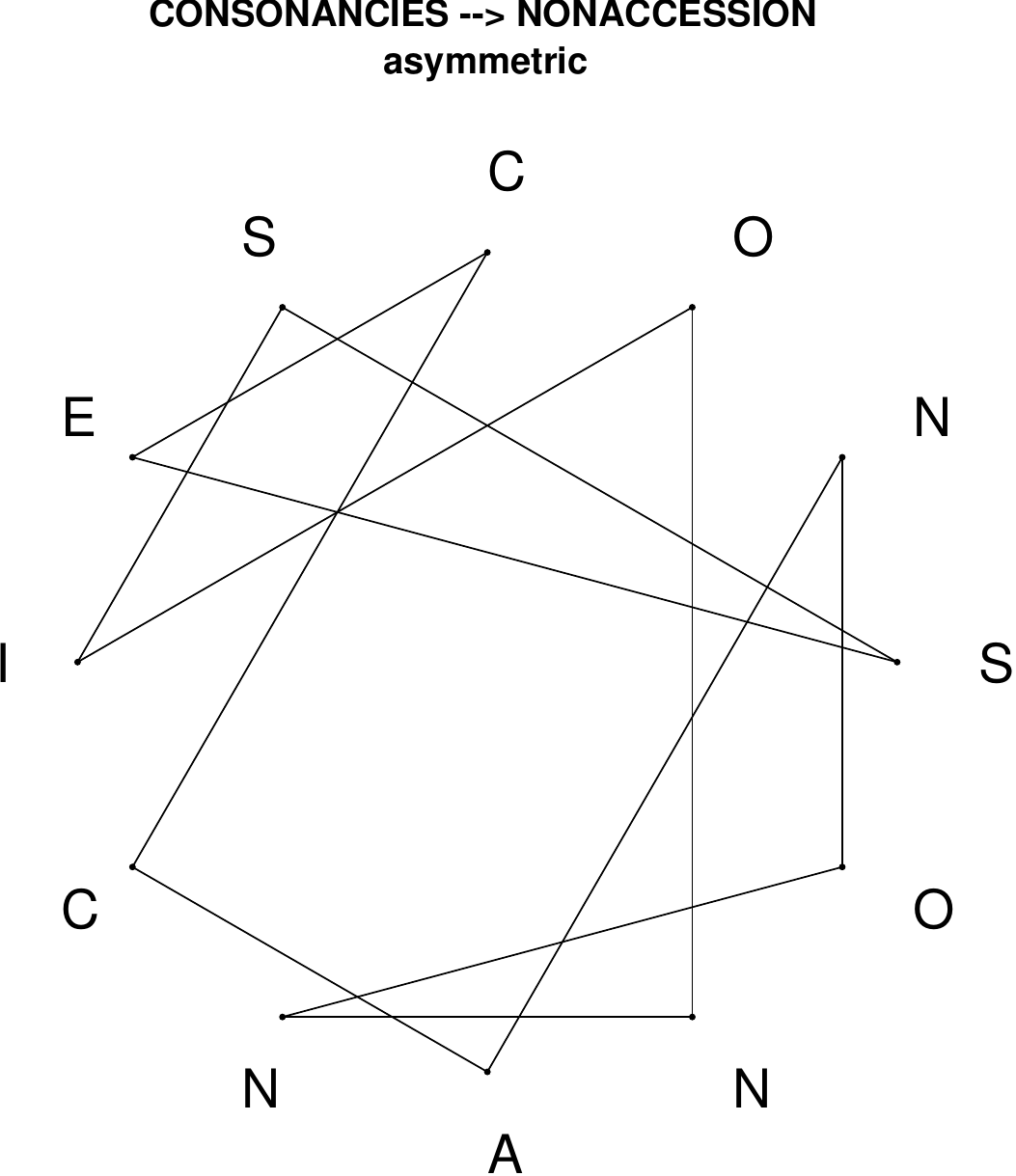}
\end{subfigure}
\hfill
\begin{subfigure}[T]{0.19\textwidth}
\centering
\includegraphics[width=\textwidth]{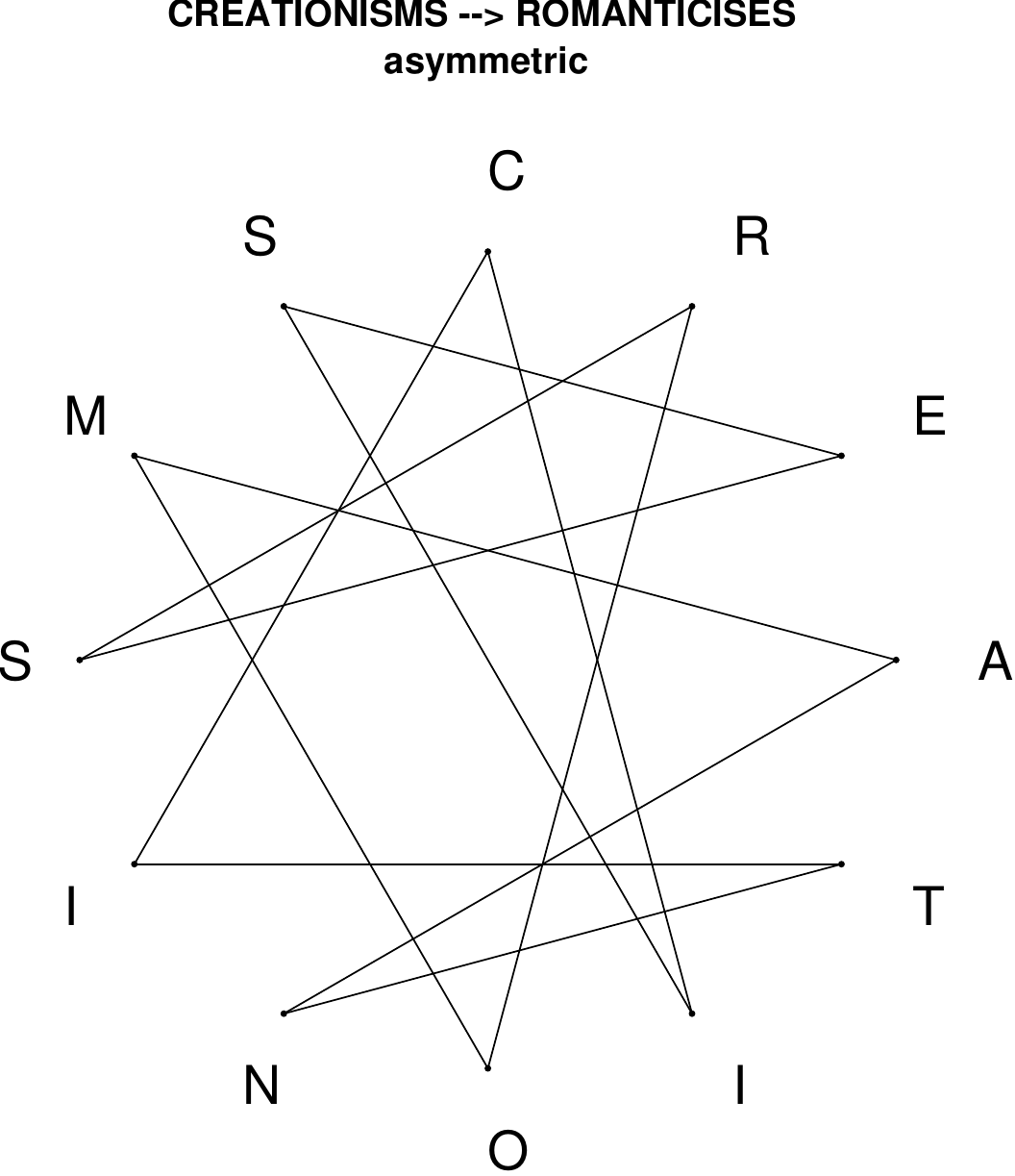}
\end{subfigure}
\hfill
\begin{subfigure}[T]{0.19\textwidth}
\centering
\includegraphics[width=\textwidth]{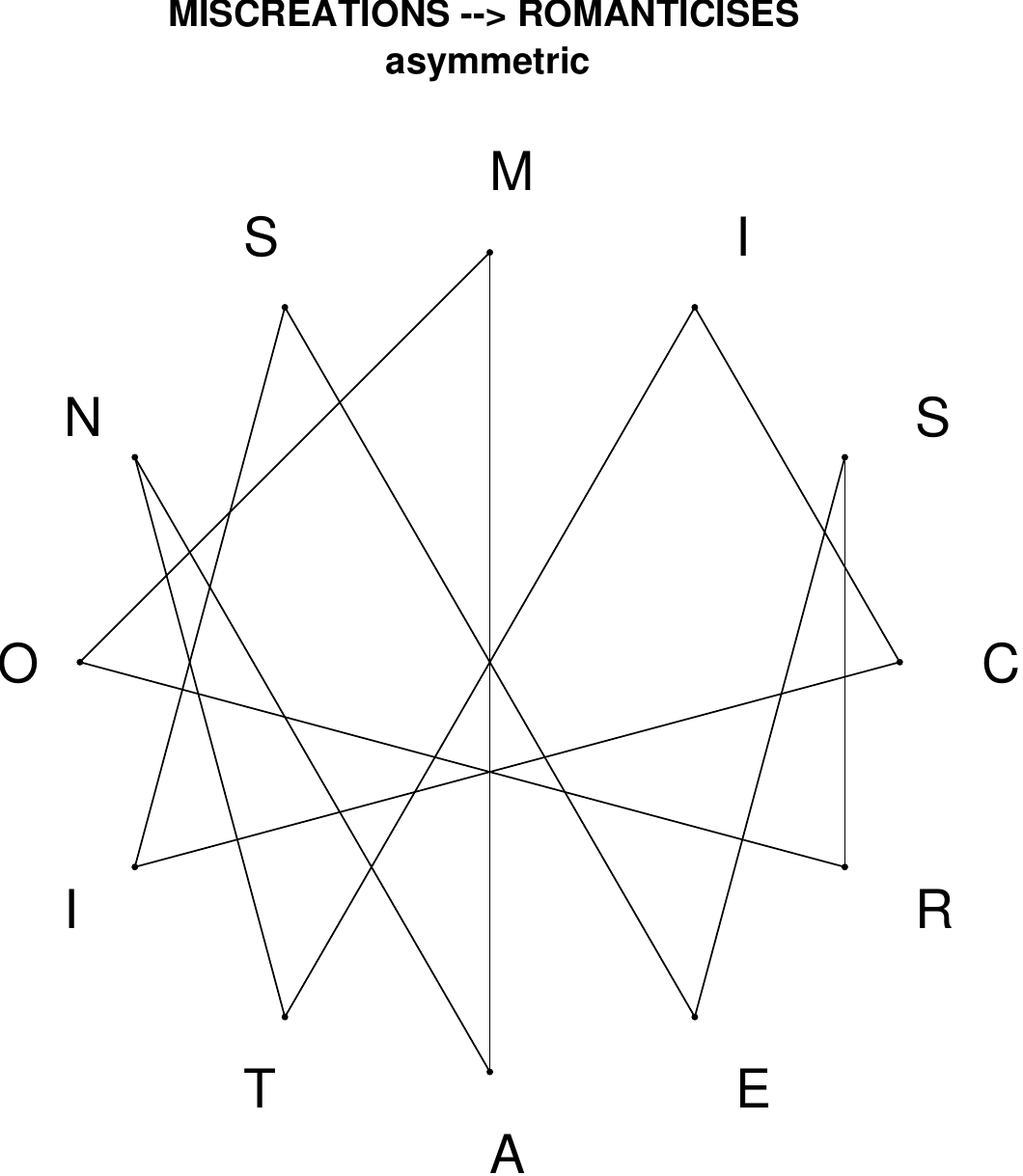}
\end{subfigure}
\hfill
\begin{subfigure}[T]{0.19\textwidth}
\centering
\includegraphics[width=\textwidth]{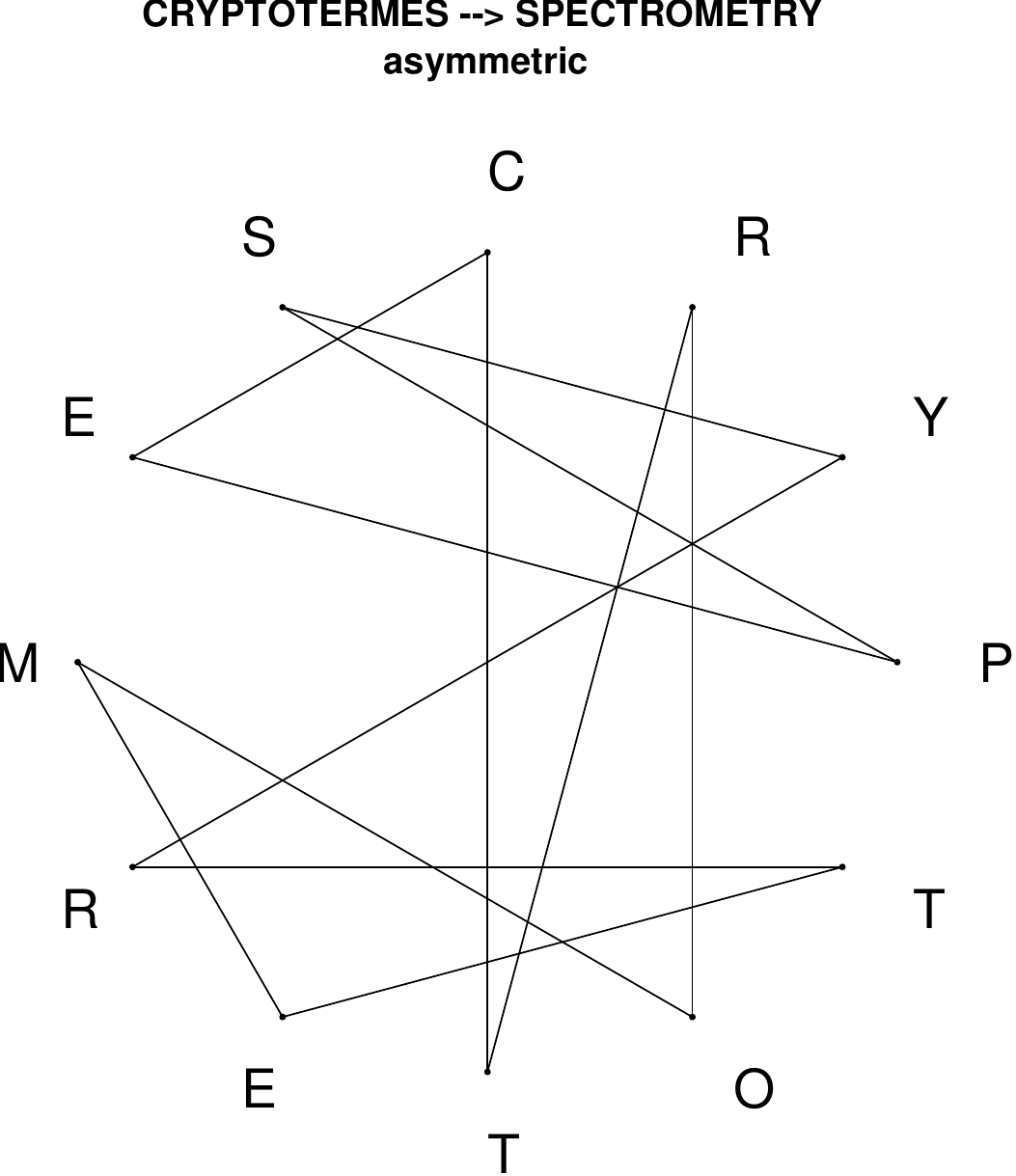}
\end{subfigure}
\end{figure}

\begin{figure}[H]
\centering
\begin{subfigure}[T]{0.19\textwidth}
\centering
\includegraphics[width=\textwidth]{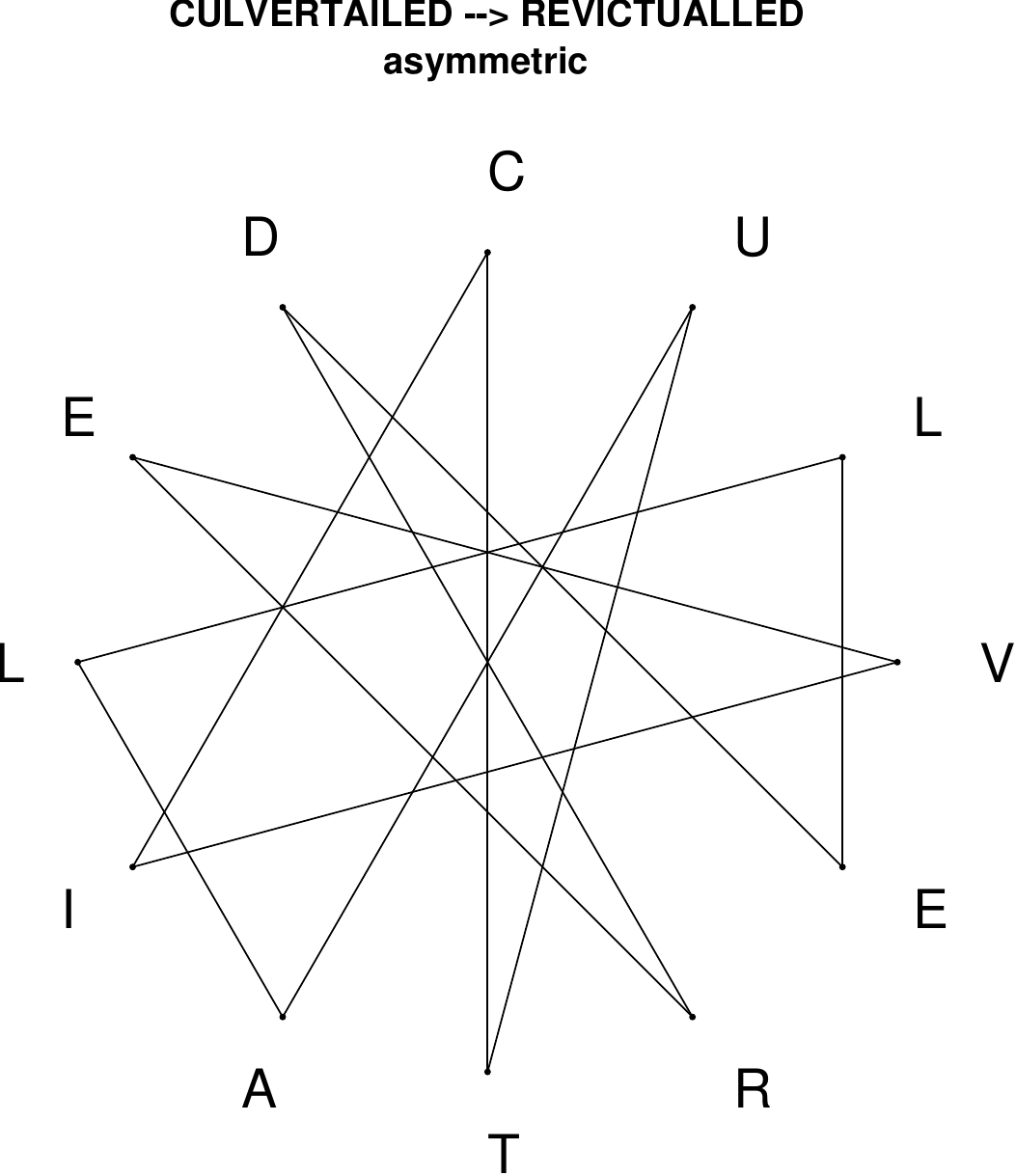}
\end{subfigure}
\hfill
\begin{subfigure}[T]{0.19\textwidth}
\centering
\includegraphics[width=\textwidth]{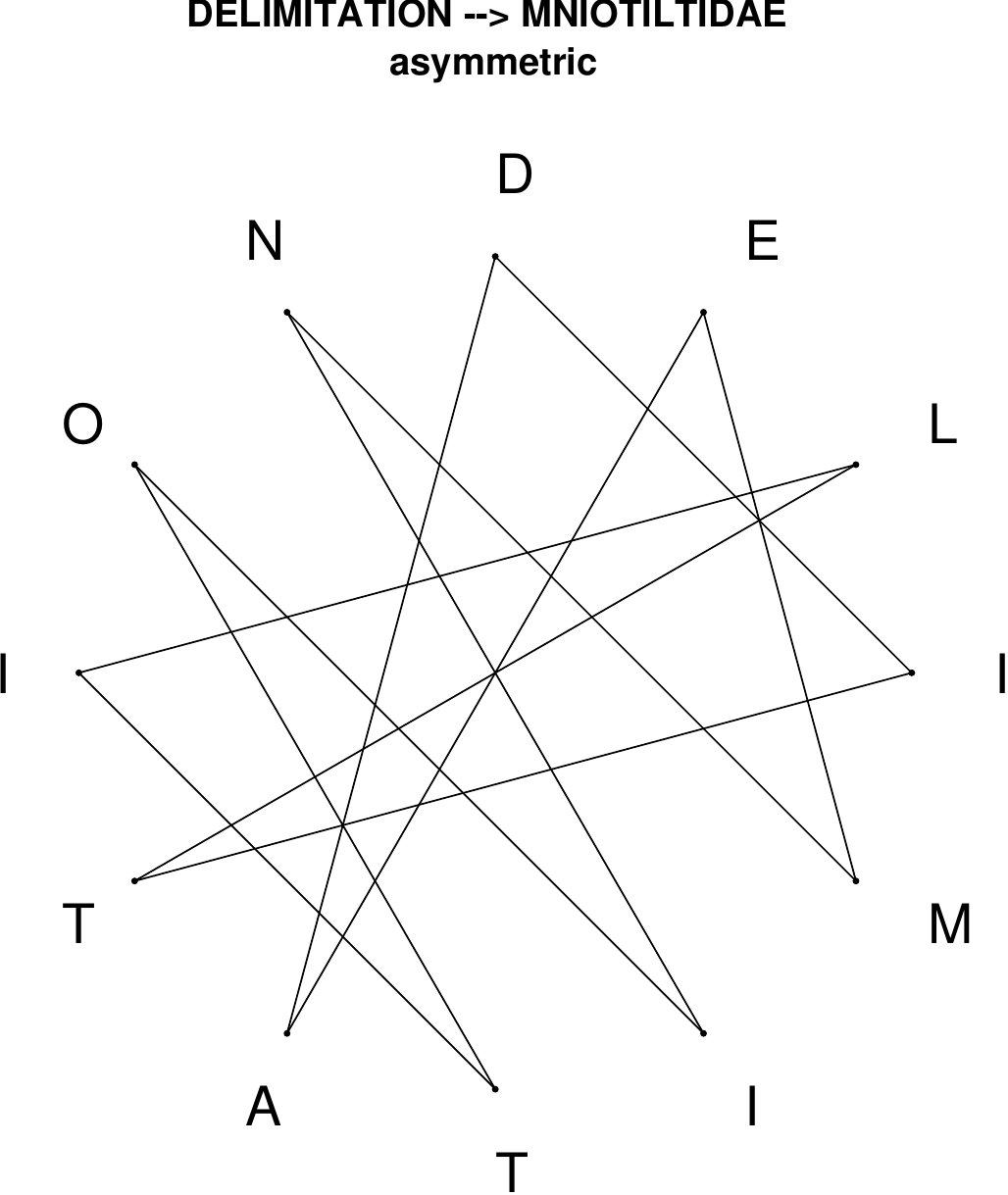}
\end{subfigure}
\hfill
\begin{subfigure}[T]{0.19\textwidth}
\centering
\includegraphics[width=\textwidth]{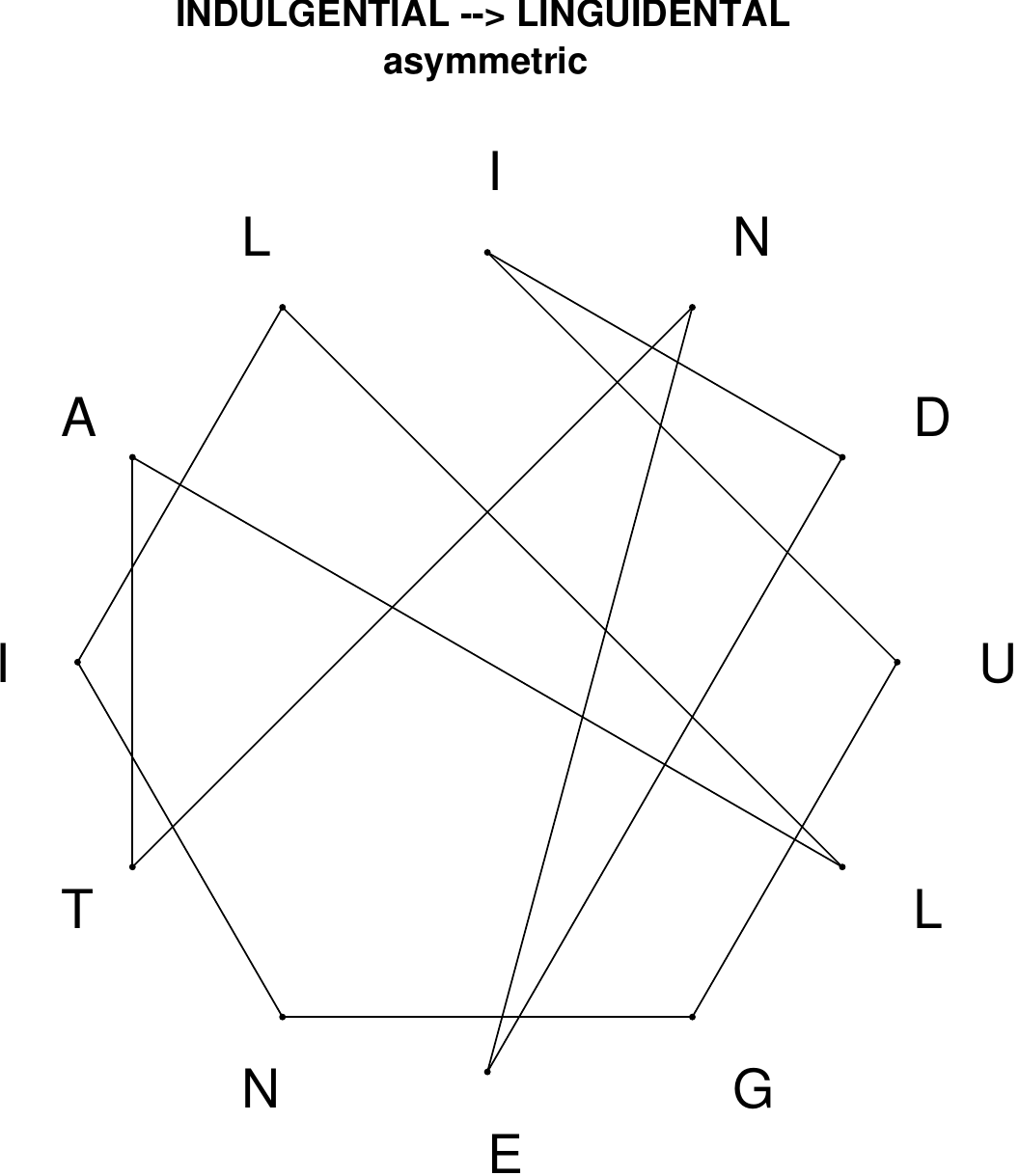}
\end{subfigure}
\hfill
\begin{subfigure}[T]{0.19\textwidth}
\centering
\includegraphics[width=\textwidth]{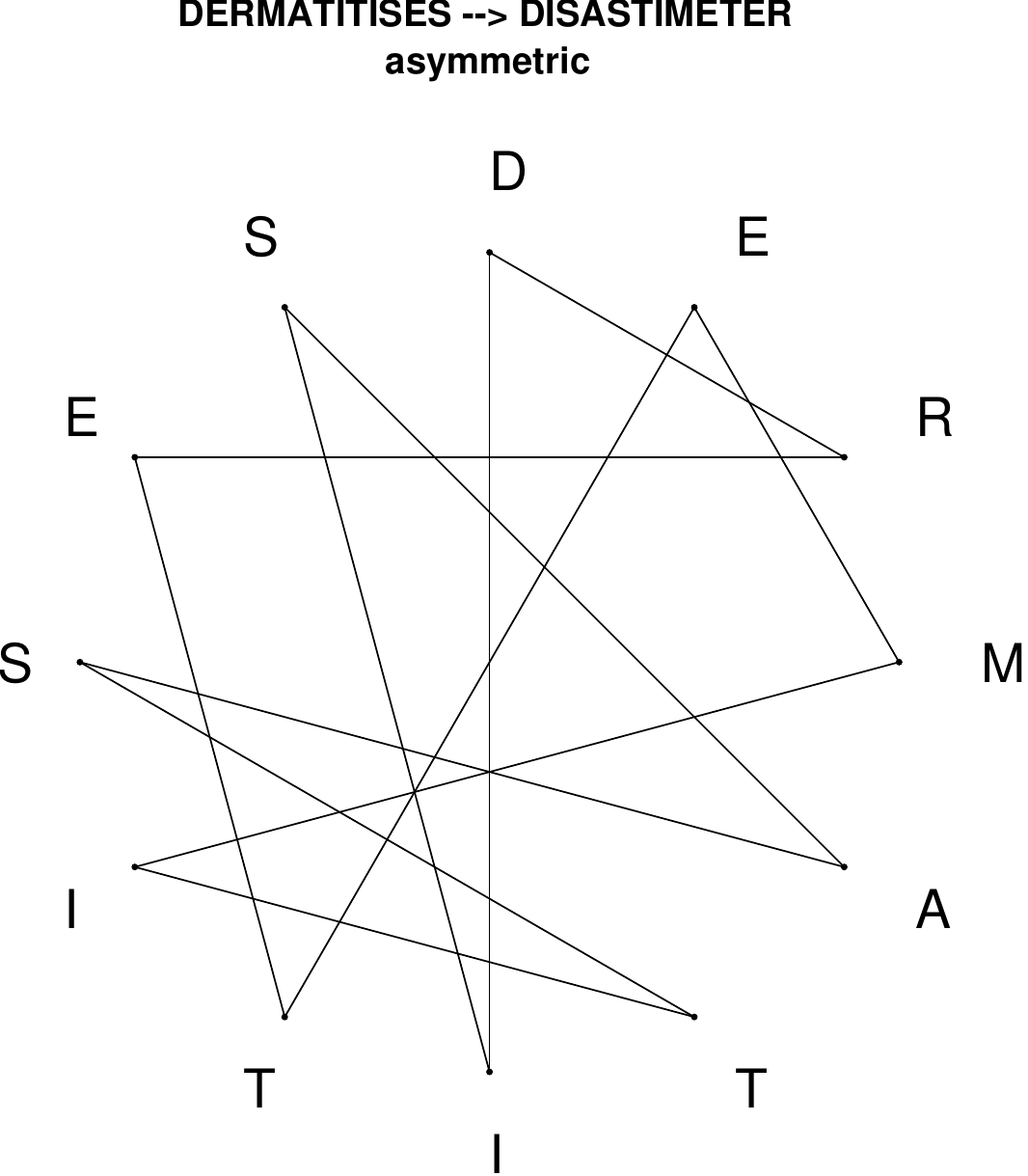}
\end{subfigure}
\hfill
\begin{subfigure}[T]{0.19\textwidth}
\centering
\includegraphics[width=\textwidth]{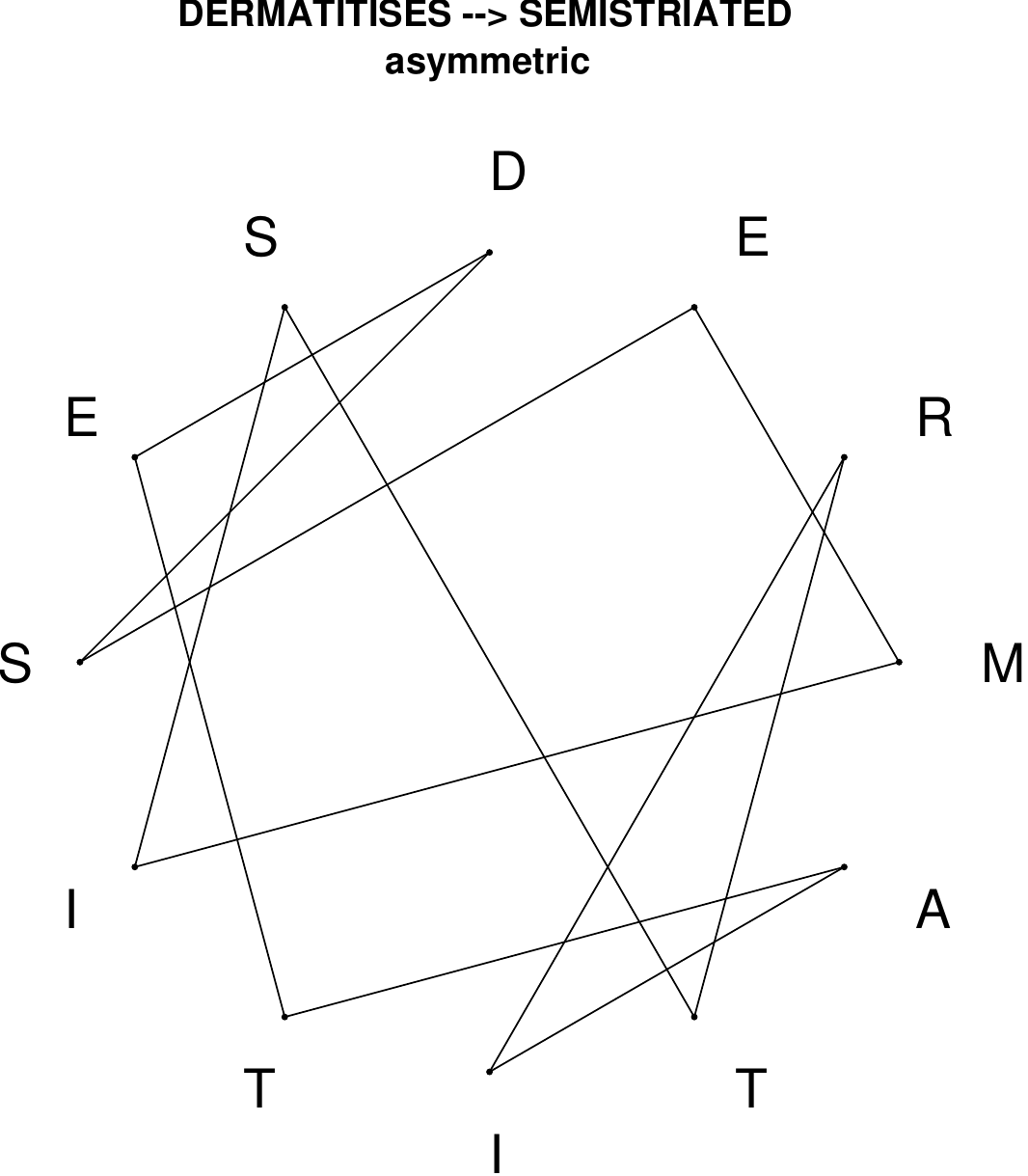}
\end{subfigure}
\end{figure}

\begin{figure}[H]
\centering
\begin{subfigure}[T]{0.19\textwidth}
\centering
\includegraphics[width=\textwidth]{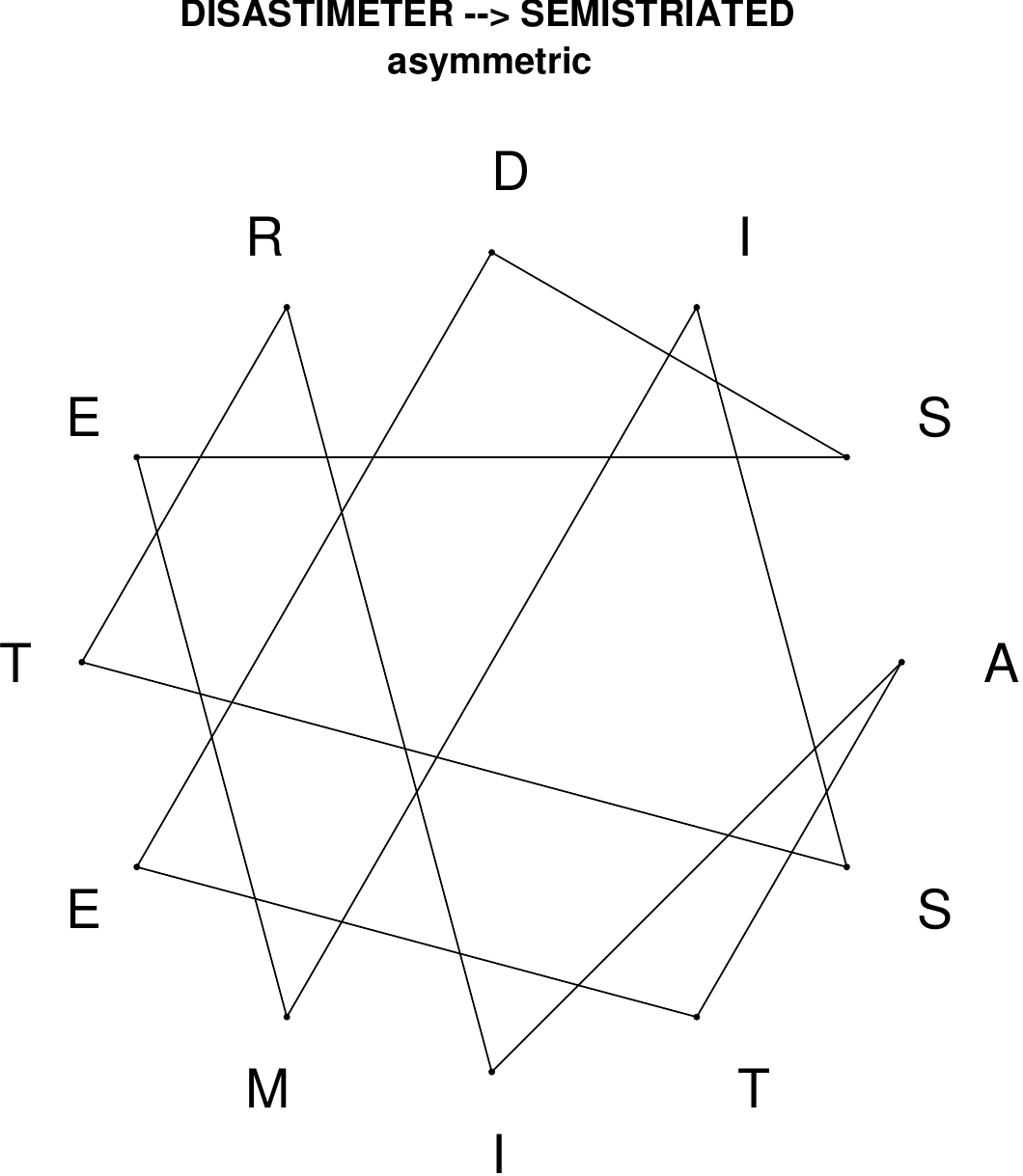}
\end{subfigure}
\hfill
\begin{subfigure}[T]{0.19\textwidth}
\centering
\includegraphics[width=\textwidth]{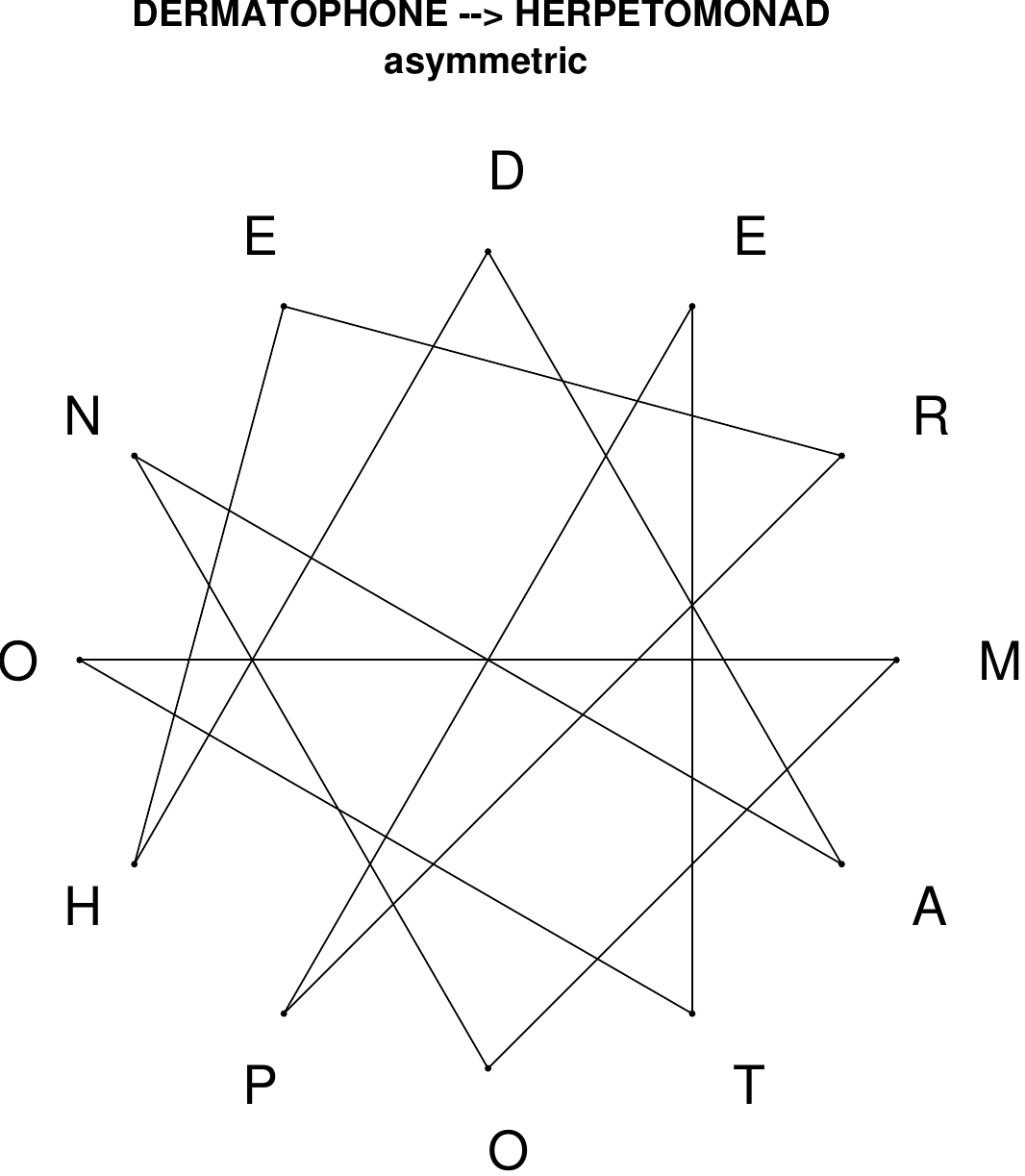}
\end{subfigure}
\hfill
\begin{subfigure}[T]{0.19\textwidth}
\centering
\includegraphics[width=\textwidth]{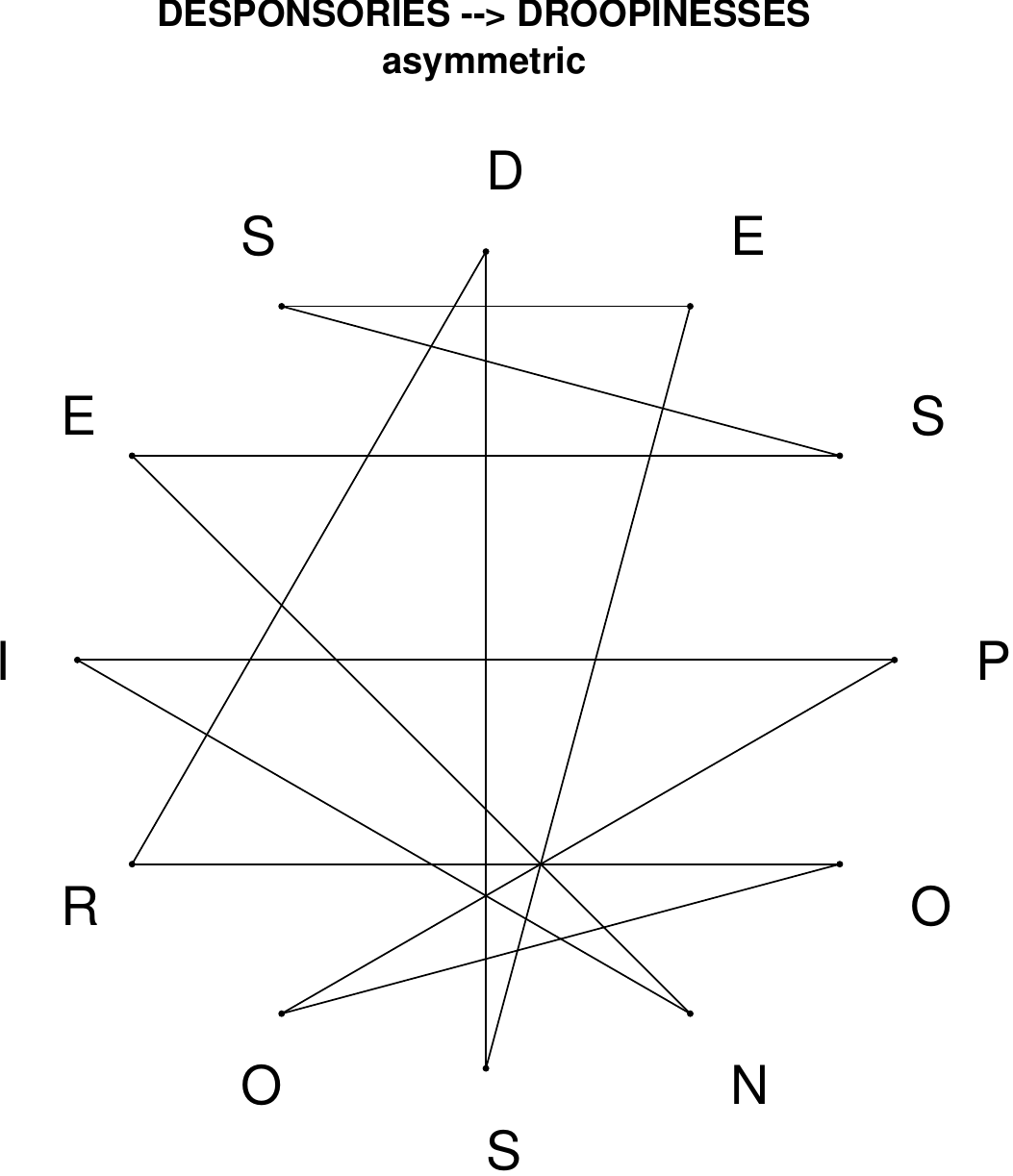}
\end{subfigure}
\hfill
\begin{subfigure}[T]{0.19\textwidth}
\centering
\includegraphics[width=\textwidth]{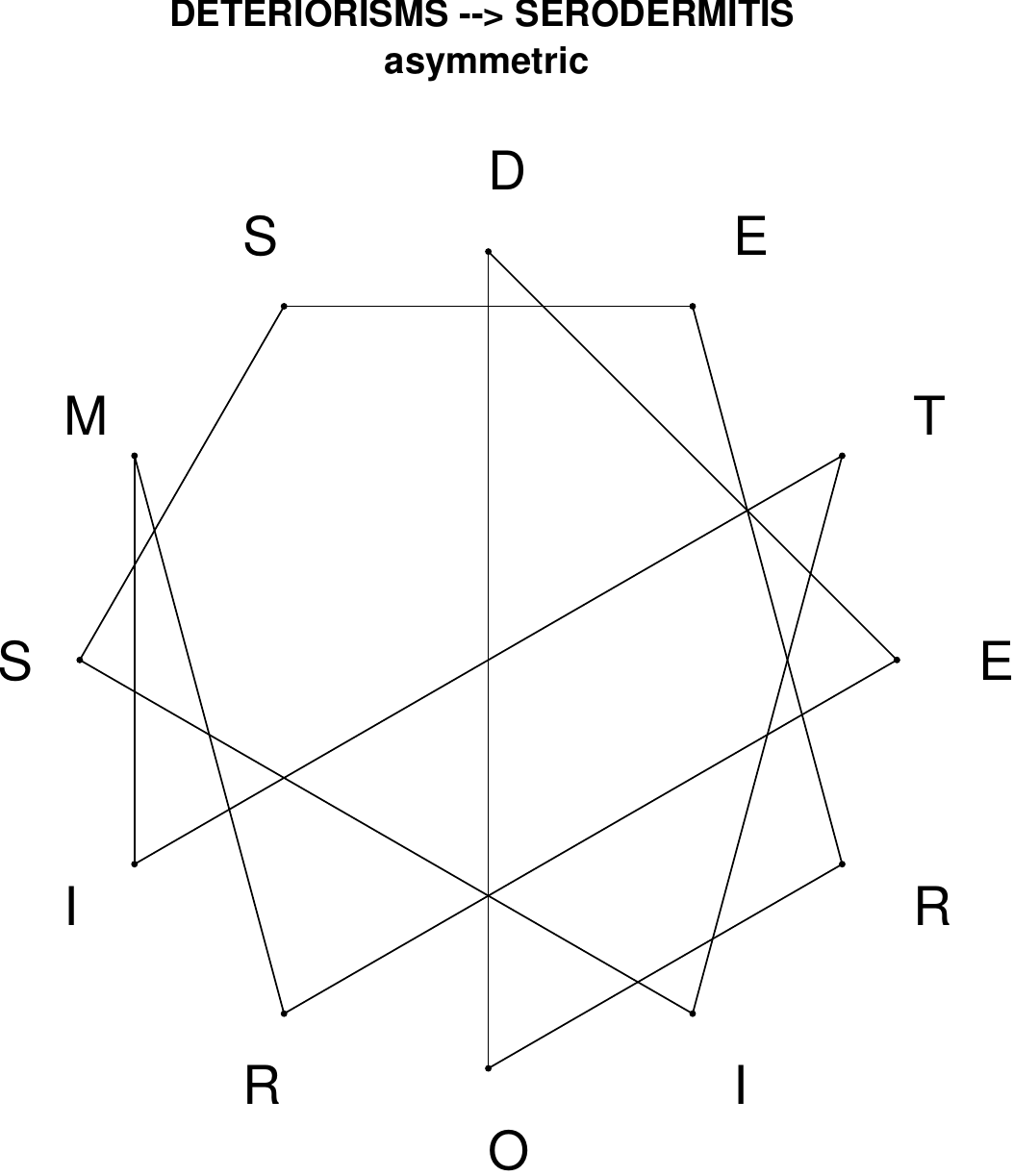}
\end{subfigure}
\hfill
\begin{subfigure}[T]{0.19\textwidth}
\centering
\includegraphics[width=\textwidth]{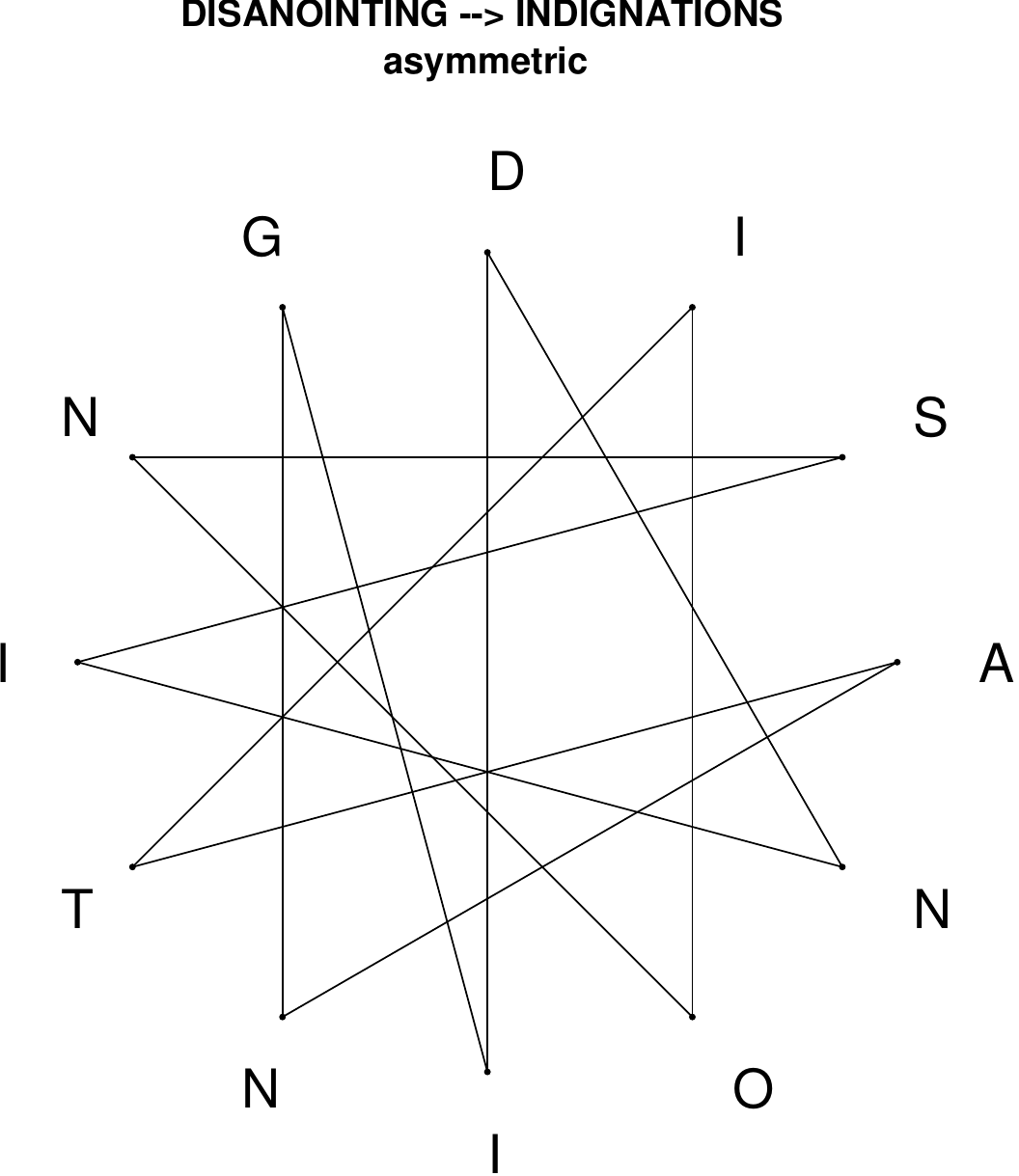}
\end{subfigure}
\end{figure}

\begin{figure}[H]
\centering
\begin{subfigure}[T]{0.19\textwidth}
\centering
\includegraphics[width=\textwidth]{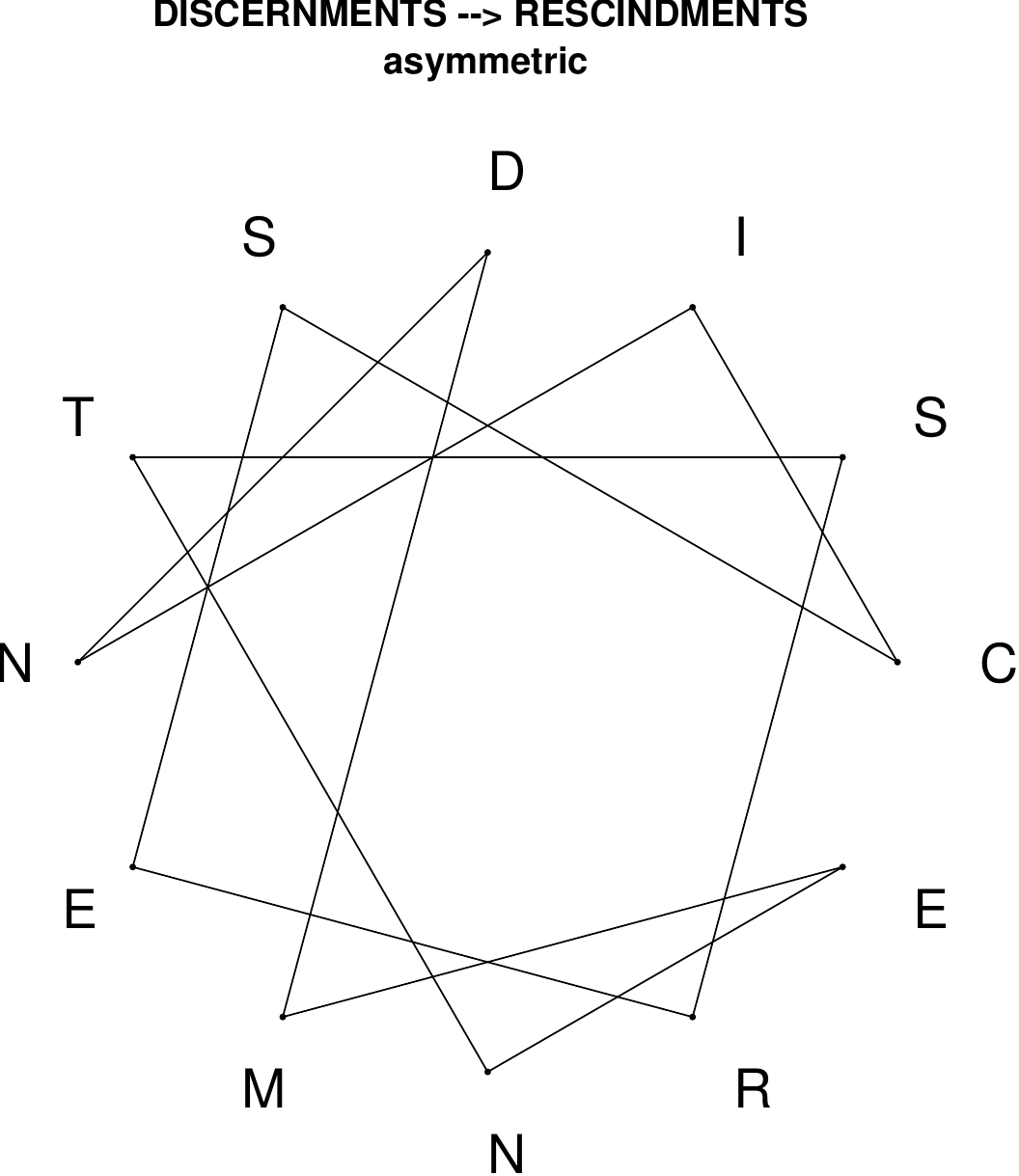}
\end{subfigure}
\hfill
\begin{subfigure}[T]{0.19\textwidth}
\centering
\includegraphics[width=\textwidth]{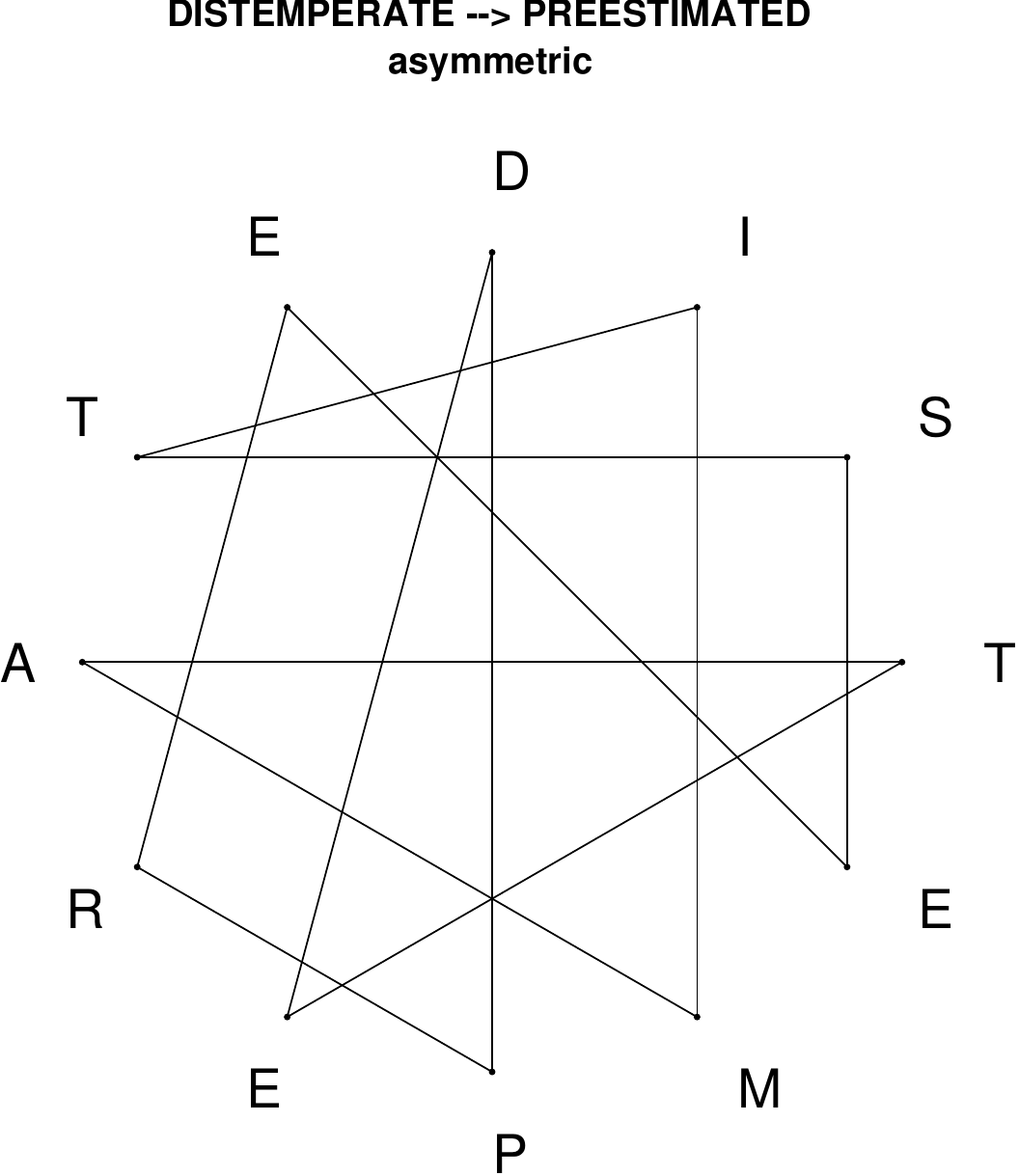}
\end{subfigure}
\hfill
\begin{subfigure}[T]{0.19\textwidth}
\centering
\includegraphics[width=\textwidth]{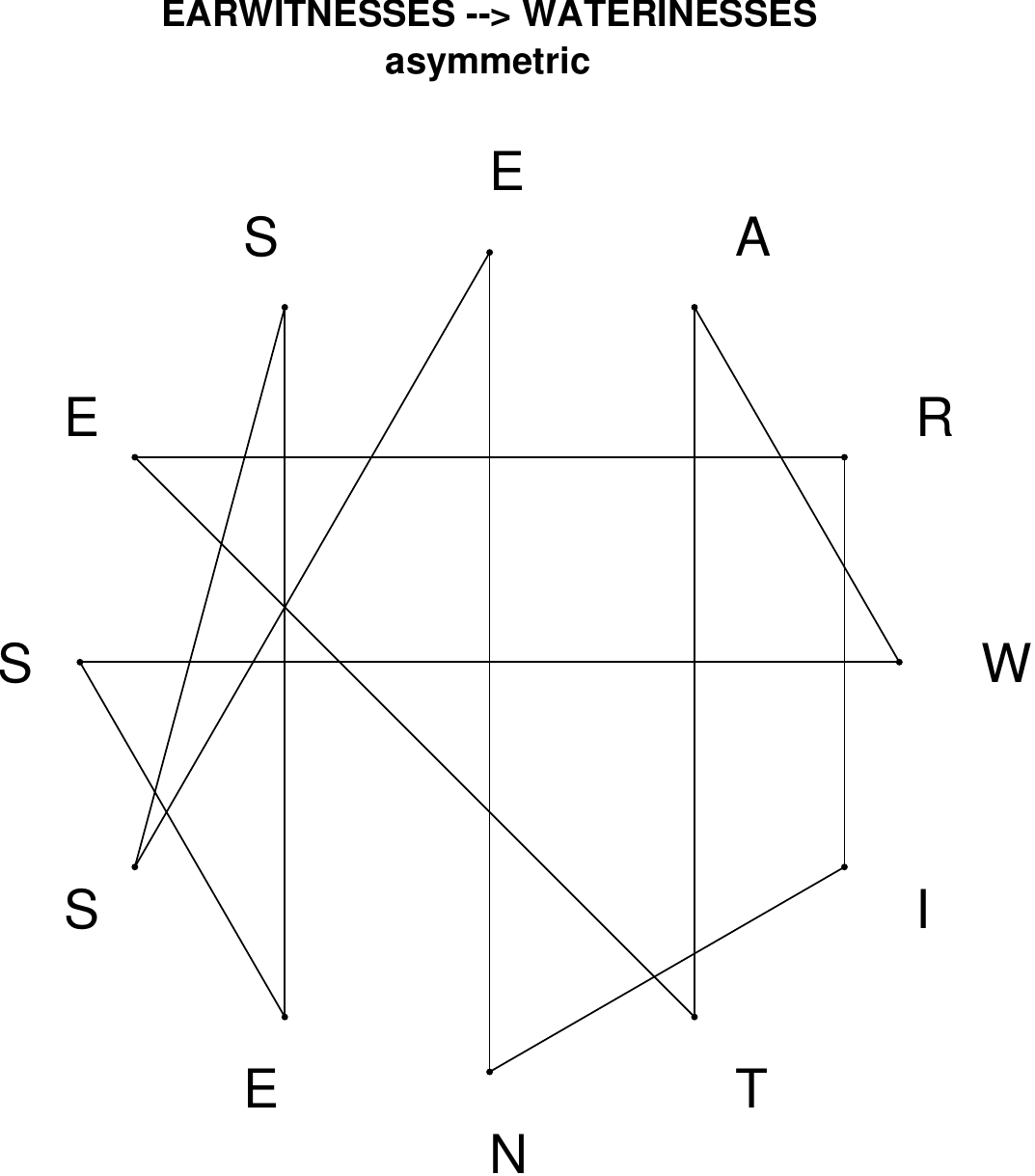}
\end{subfigure}
\hfill
\begin{subfigure}[T]{0.19\textwidth}
\centering
\includegraphics[width=\textwidth]{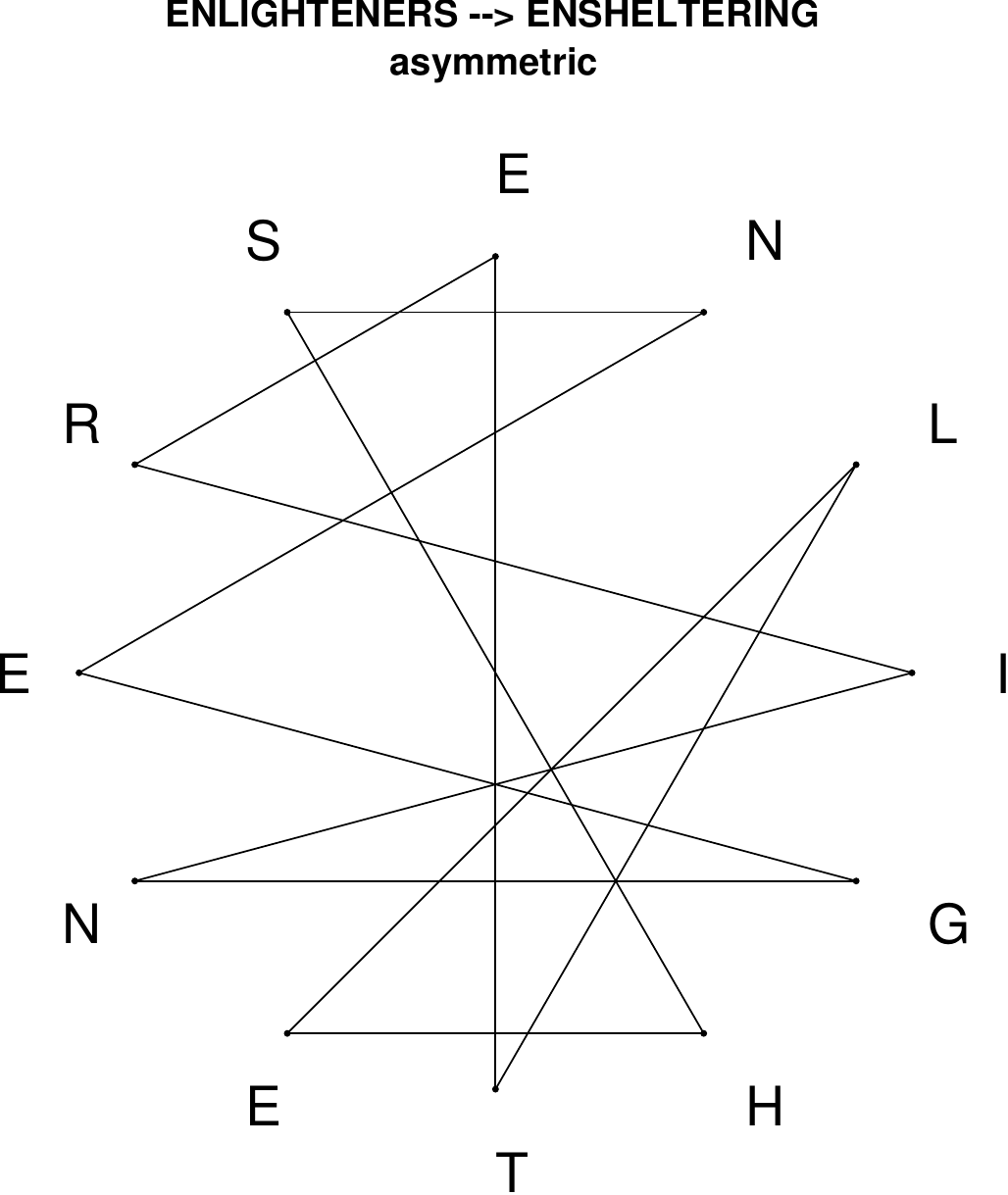}
\end{subfigure}
\hfill
\begin{subfigure}[T]{0.19\textwidth}
\centering
\includegraphics[width=\textwidth]{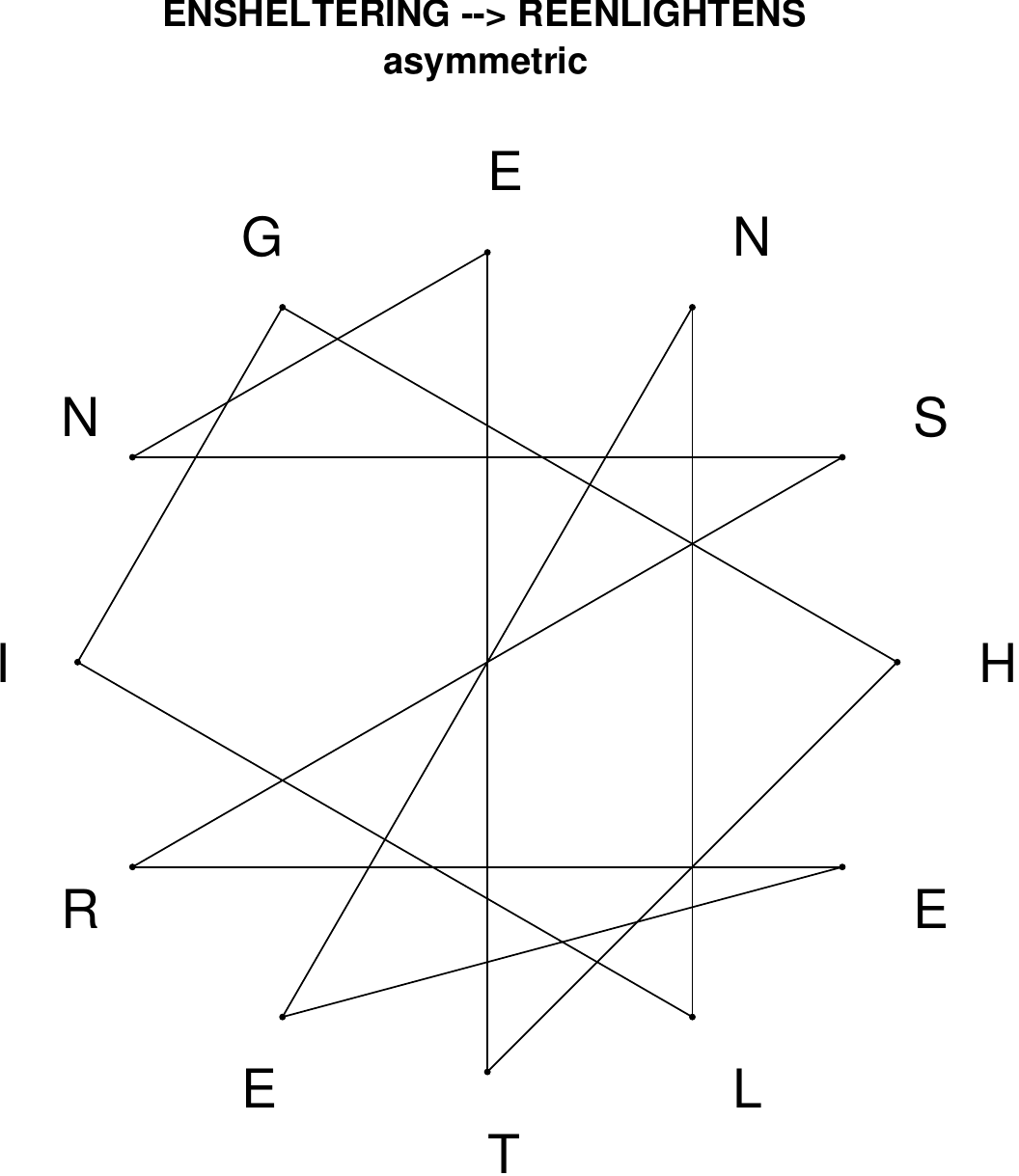}
\end{subfigure}
\end{figure}

\begin{figure}[H]
\centering
\begin{subfigure}[T]{0.19\textwidth}
\centering
\includegraphics[width=\textwidth]{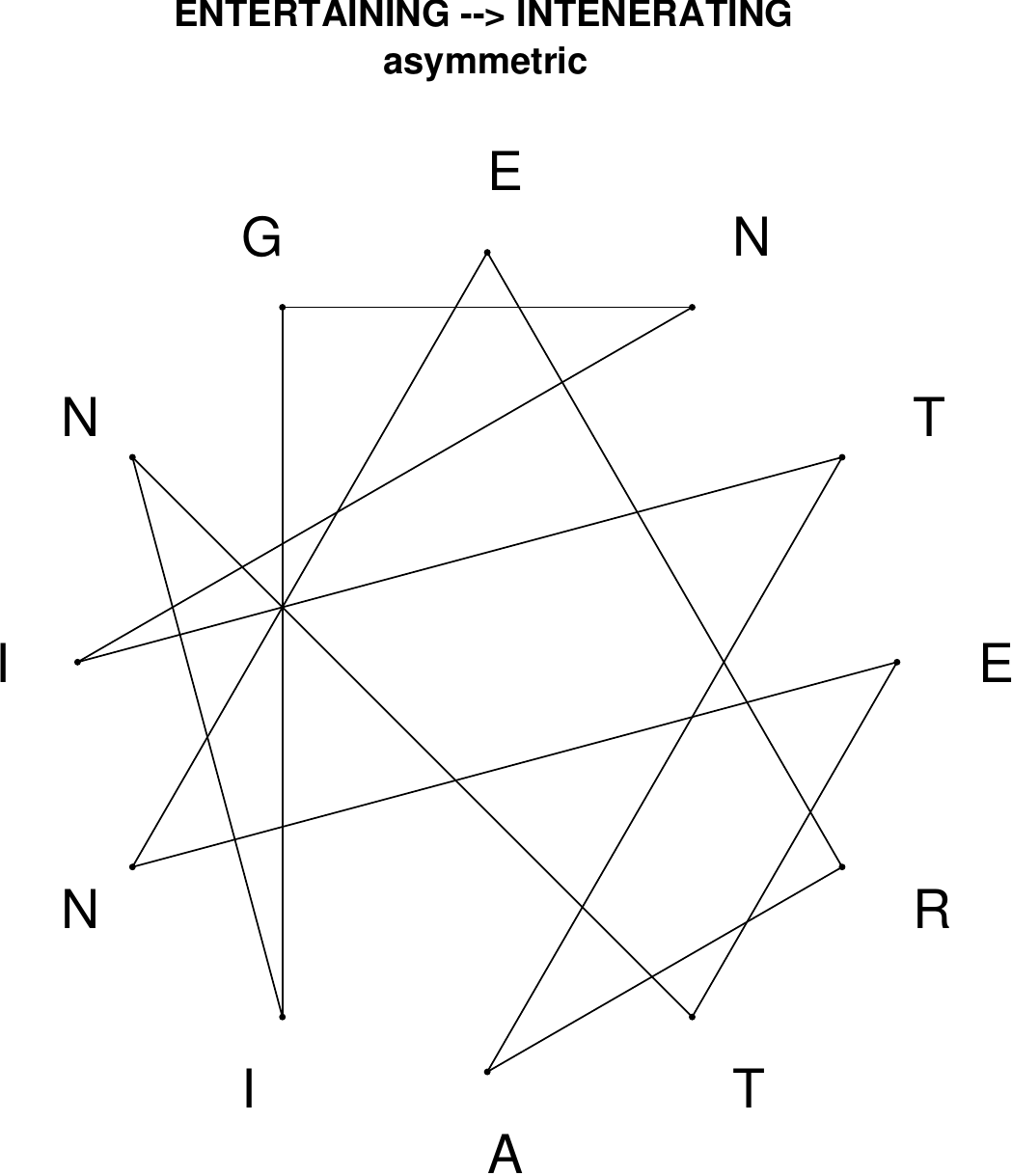}
\end{subfigure}
\hfill
\begin{subfigure}[T]{0.19\textwidth}
\centering
\includegraphics[width=\textwidth]{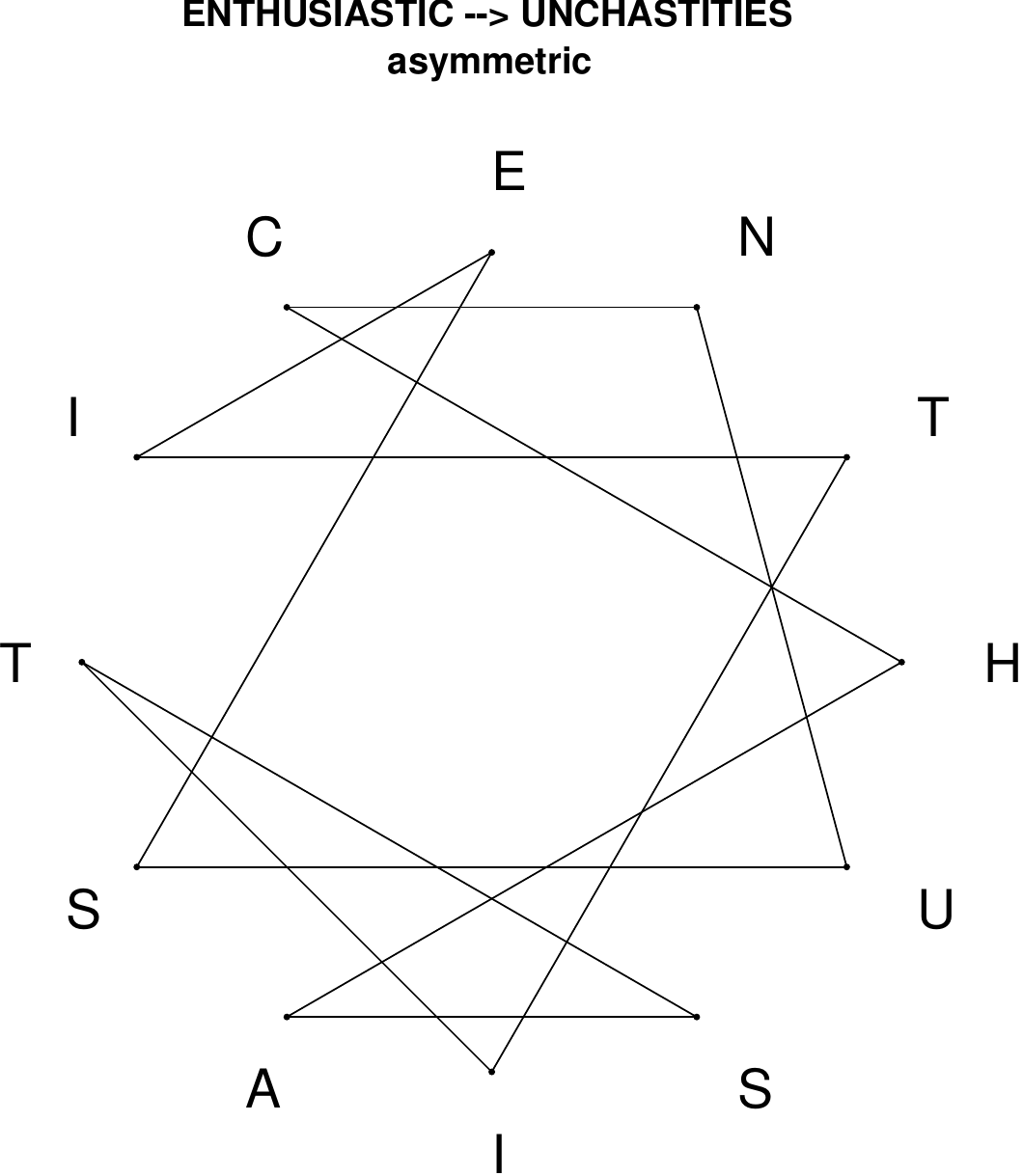}
\end{subfigure}
\hfill
\begin{subfigure}[T]{0.19\textwidth}
\centering
\includegraphics[width=\textwidth]{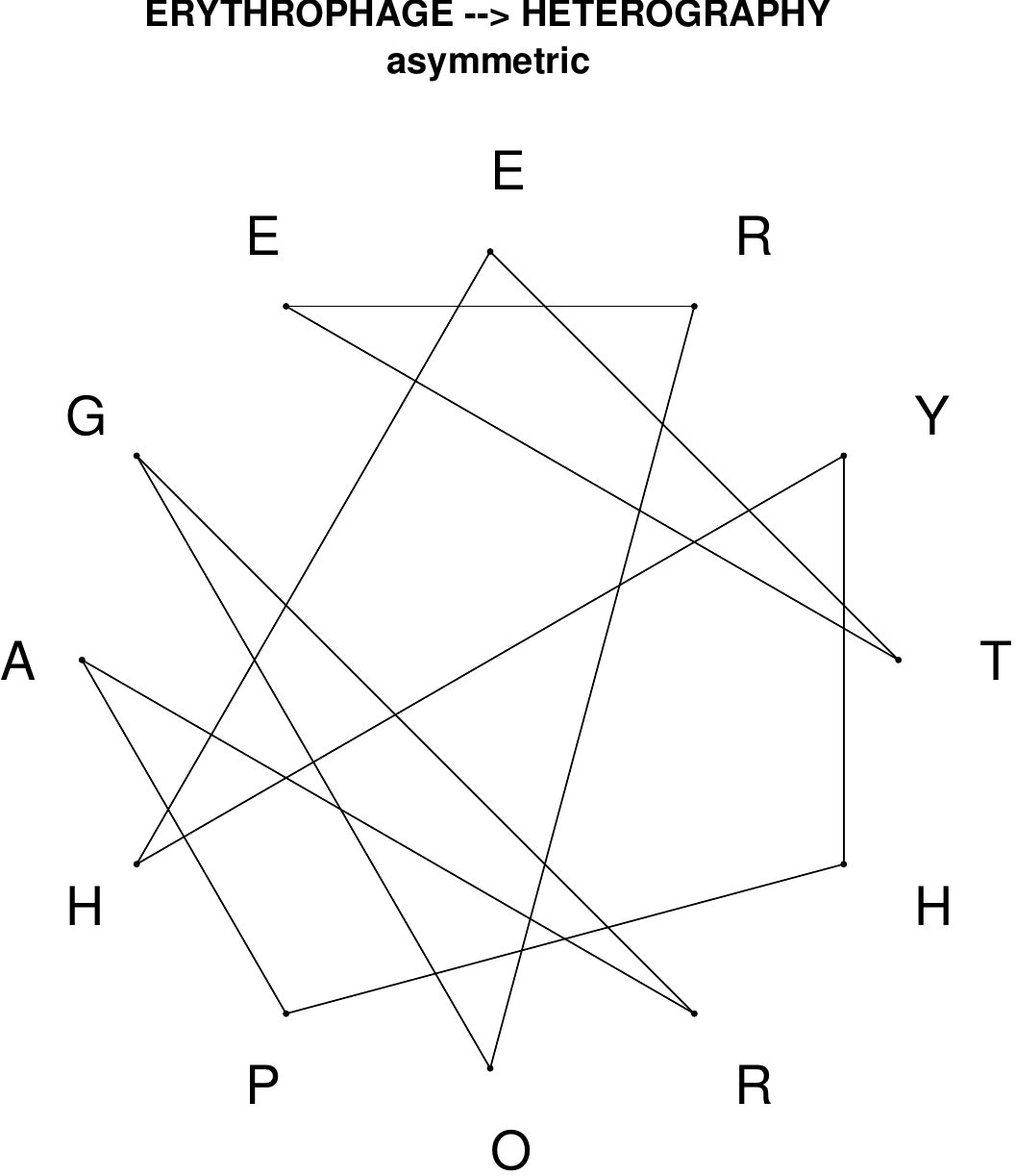}
\end{subfigure}
\hfill
\begin{subfigure}[T]{0.19\textwidth}
\centering
\includegraphics[width=\textwidth]{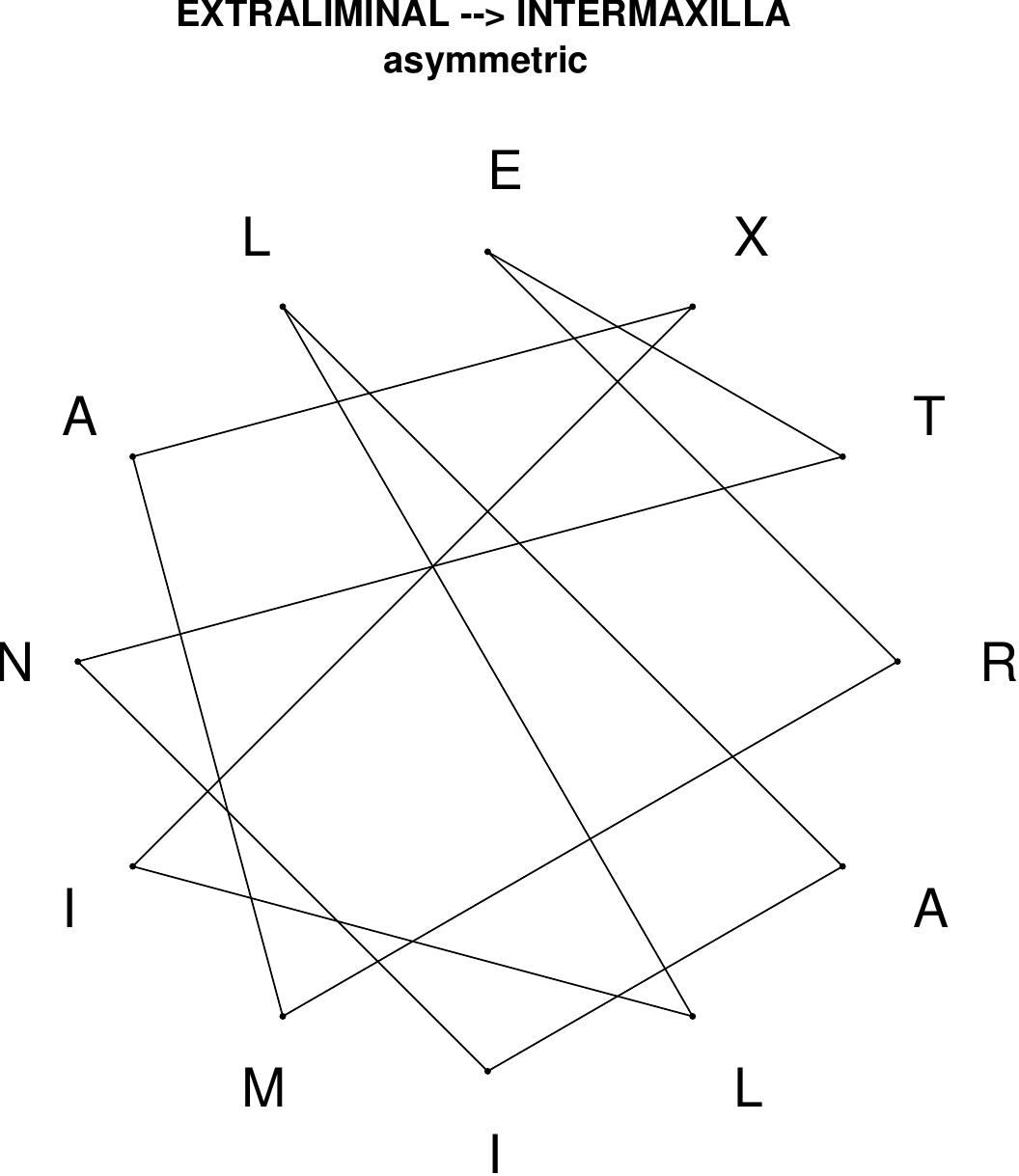}
\end{subfigure}
\hfill
\begin{subfigure}[T]{0.19\textwidth}
\centering
\includegraphics[width=\textwidth]{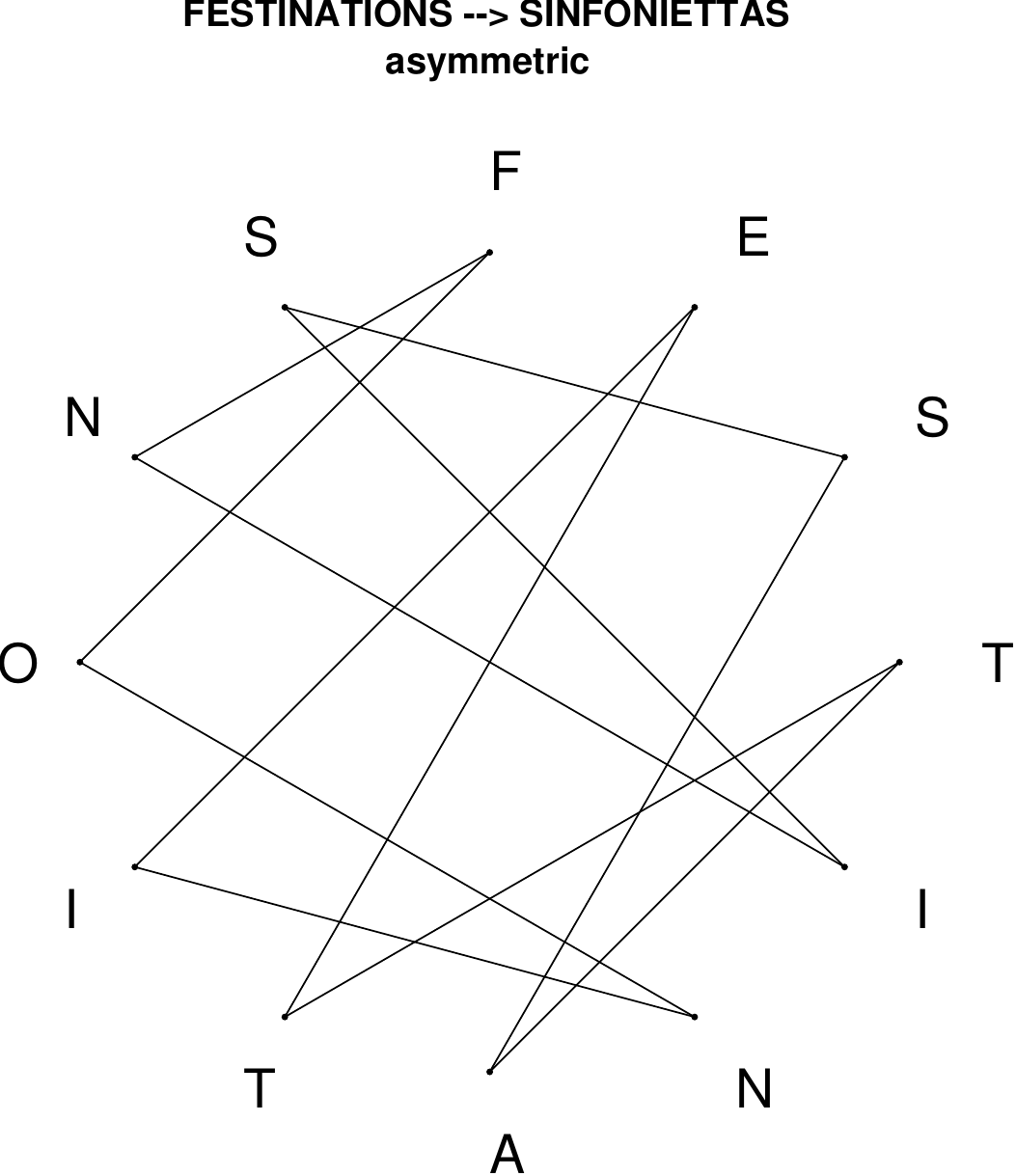}
\end{subfigure}
\end{figure}

\begin{figure}[H]
\centering
\begin{subfigure}[T]{0.19\textwidth}
\centering
\includegraphics[width=\textwidth]{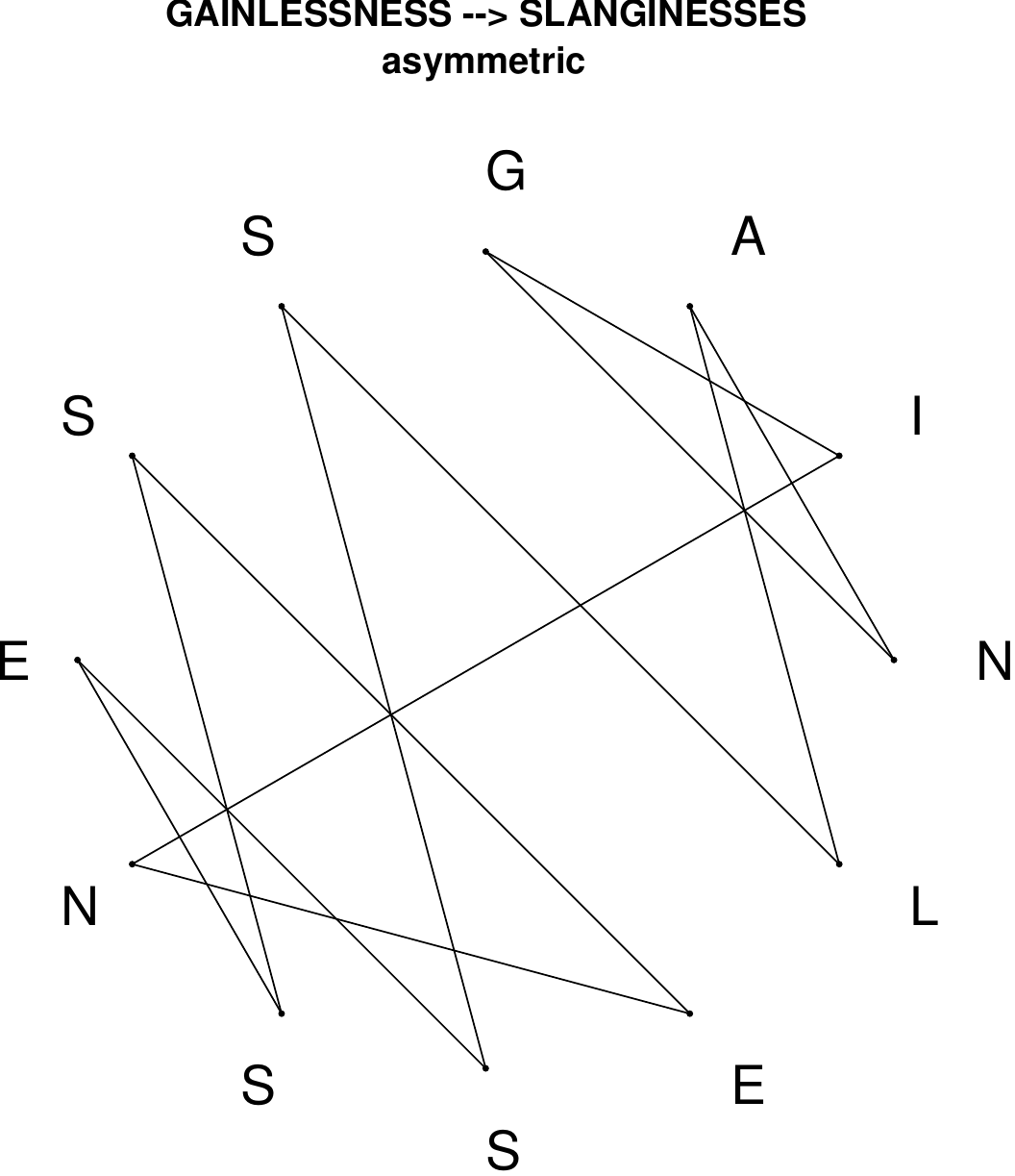}
\end{subfigure}
\hfill
\begin{subfigure}[T]{0.19\textwidth}
\centering
\includegraphics[width=\textwidth]{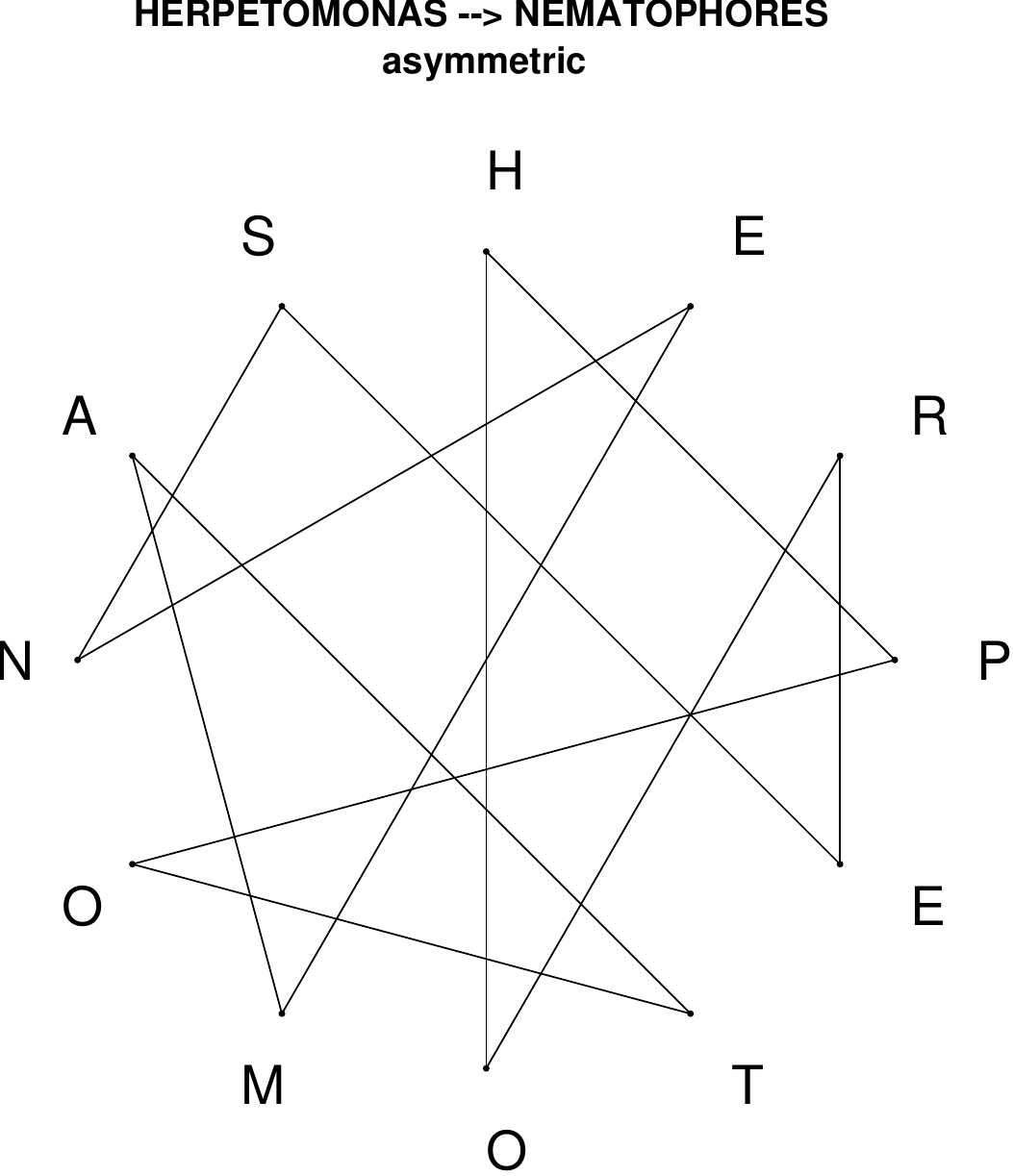}
\end{subfigure}
\hfill
\begin{subfigure}[T]{0.19\textwidth}
\centering
\includegraphics[width=\textwidth]{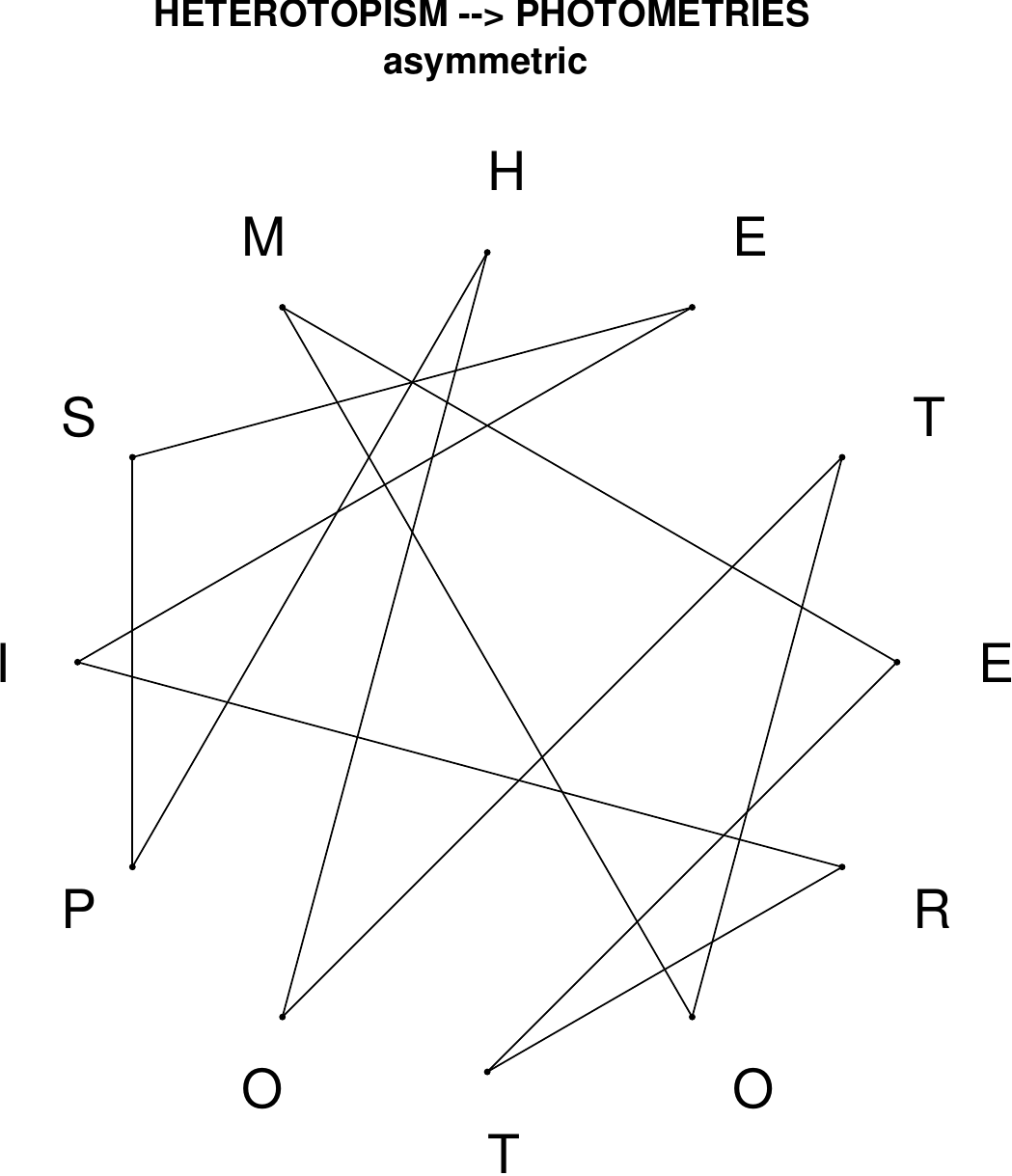}
\end{subfigure}
\hfill
\begin{subfigure}[T]{0.19\textwidth}
\centering
\includegraphics[width=\textwidth]{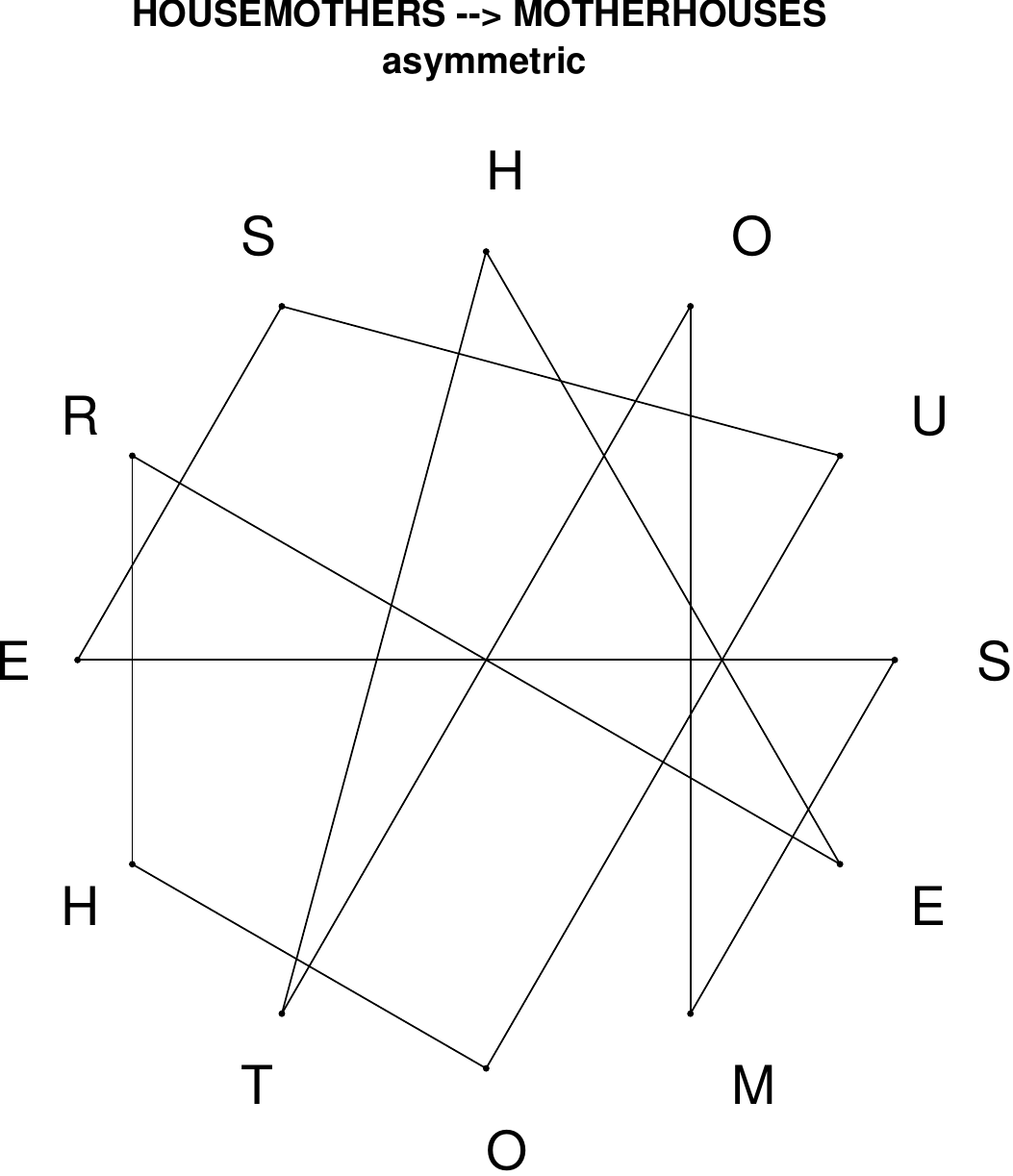}
\end{subfigure}
\hfill
\begin{subfigure}[T]{0.19\textwidth}
\centering
\includegraphics[width=\textwidth]{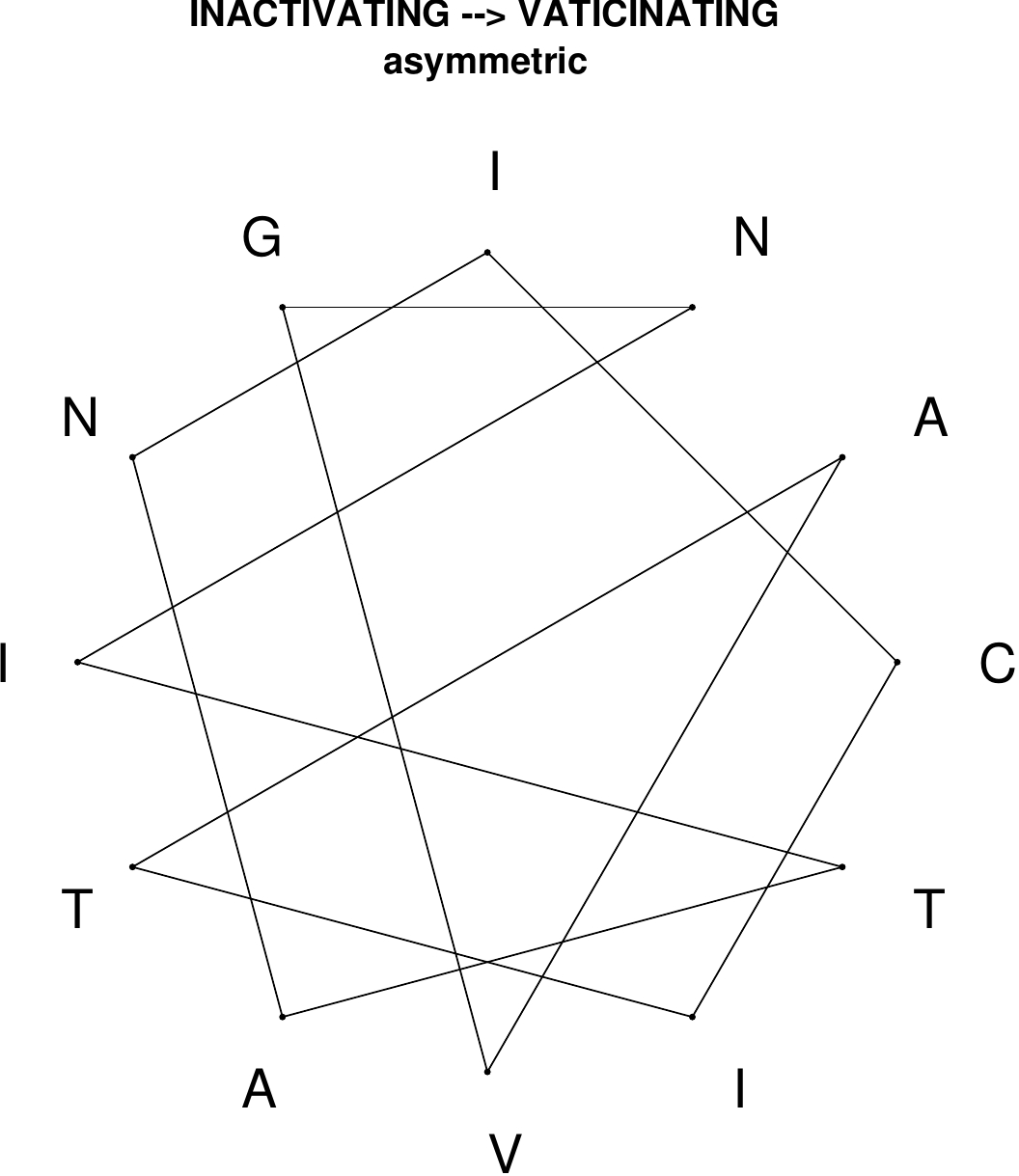}
\end{subfigure}
\end{figure}

\begin{figure}[H]
\centering
\begin{subfigure}[T]{0.19\textwidth}
\centering
\includegraphics[width=\textwidth]{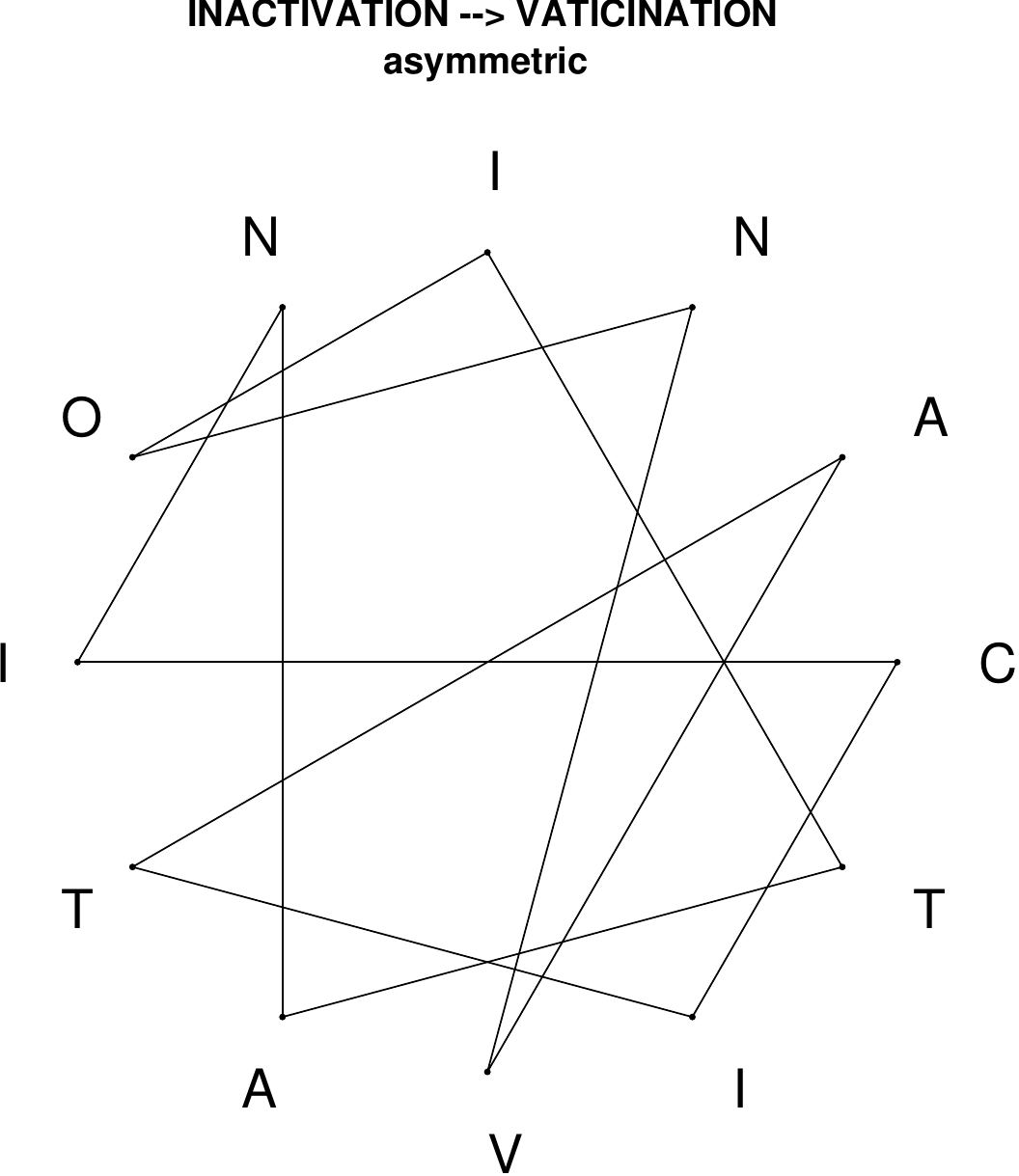}
\end{subfigure}
\hfill
\begin{subfigure}[T]{0.19\textwidth}
\centering
\includegraphics[width=\textwidth]{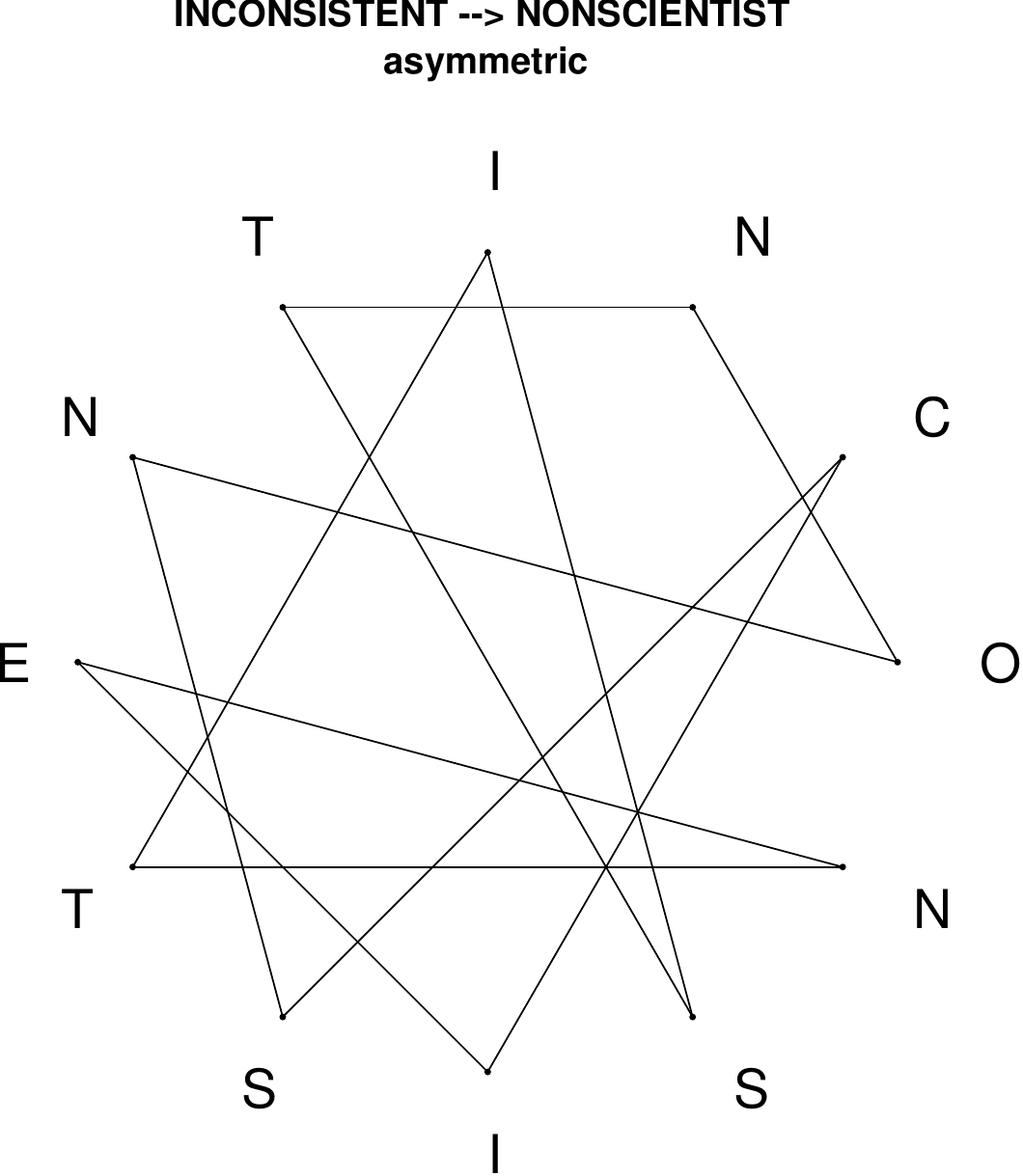}
\end{subfigure}
\hfill
\begin{subfigure}[T]{0.19\textwidth}
\centering
\includegraphics[width=\textwidth]{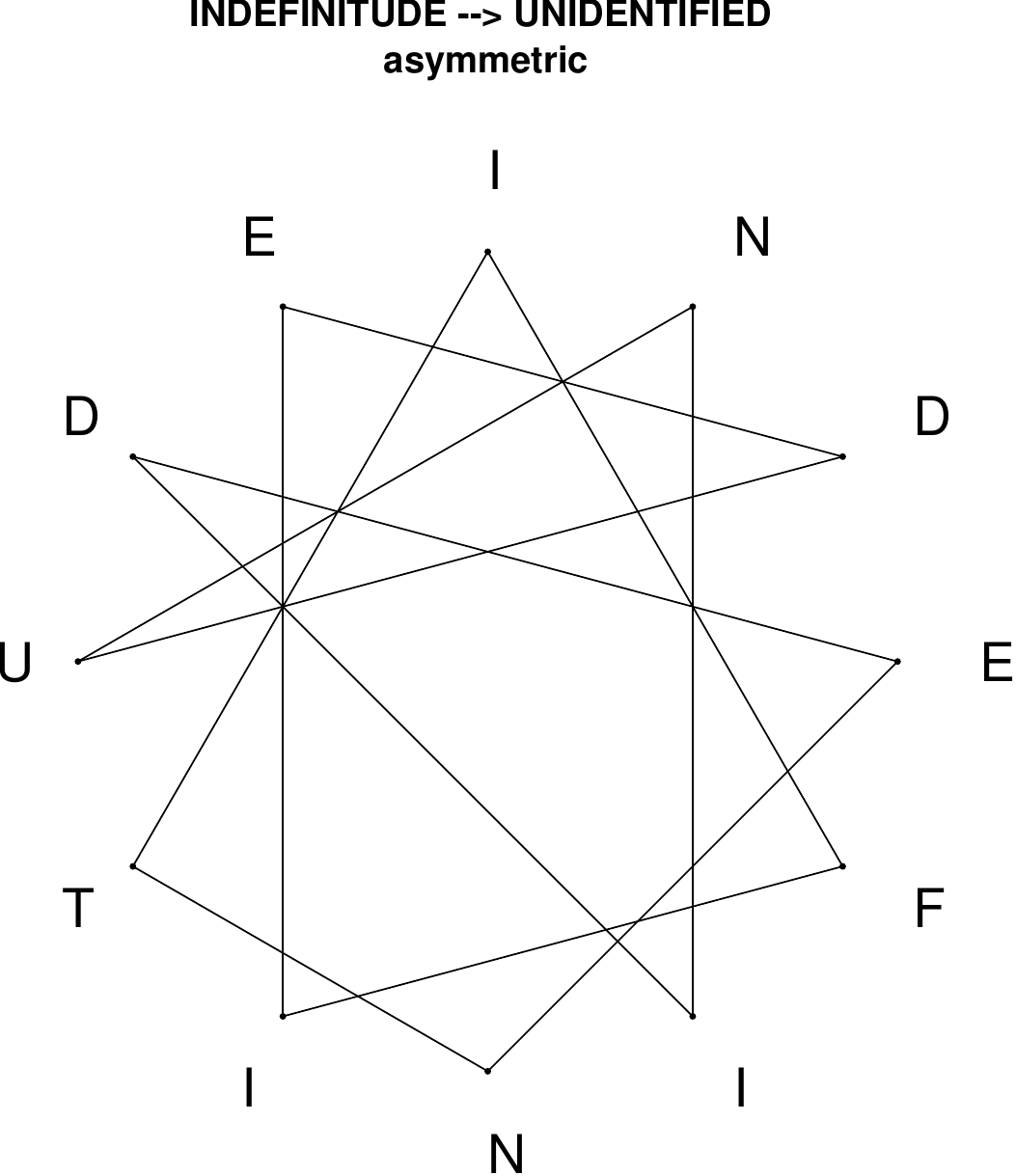}
\end{subfigure}
\hfill
\begin{subfigure}[T]{0.19\textwidth}
\centering
\includegraphics[width=\textwidth]{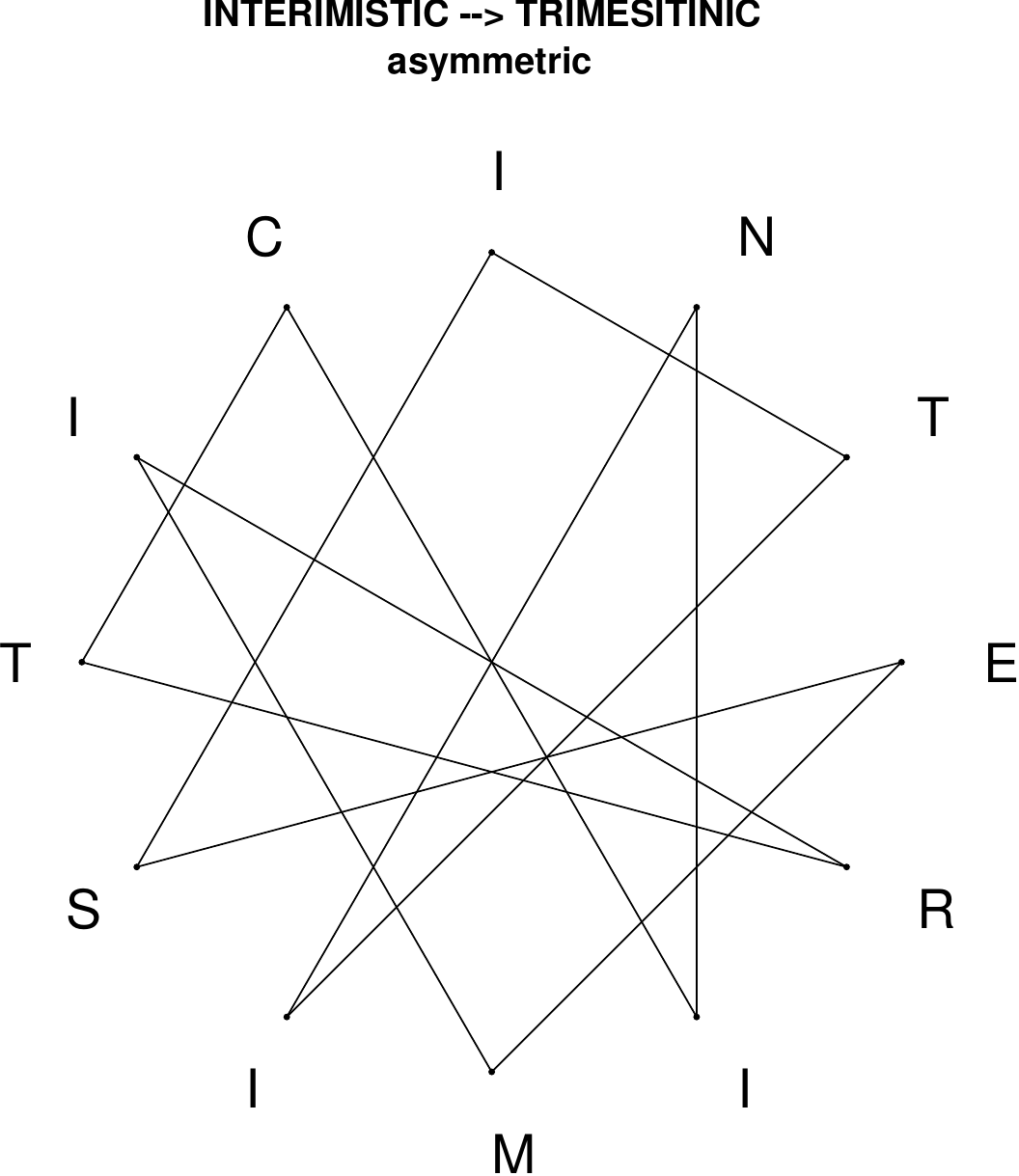}
\end{subfigure}
\hfill
\begin{subfigure}[T]{0.19\textwidth}
\centering
\includegraphics[width=\textwidth]{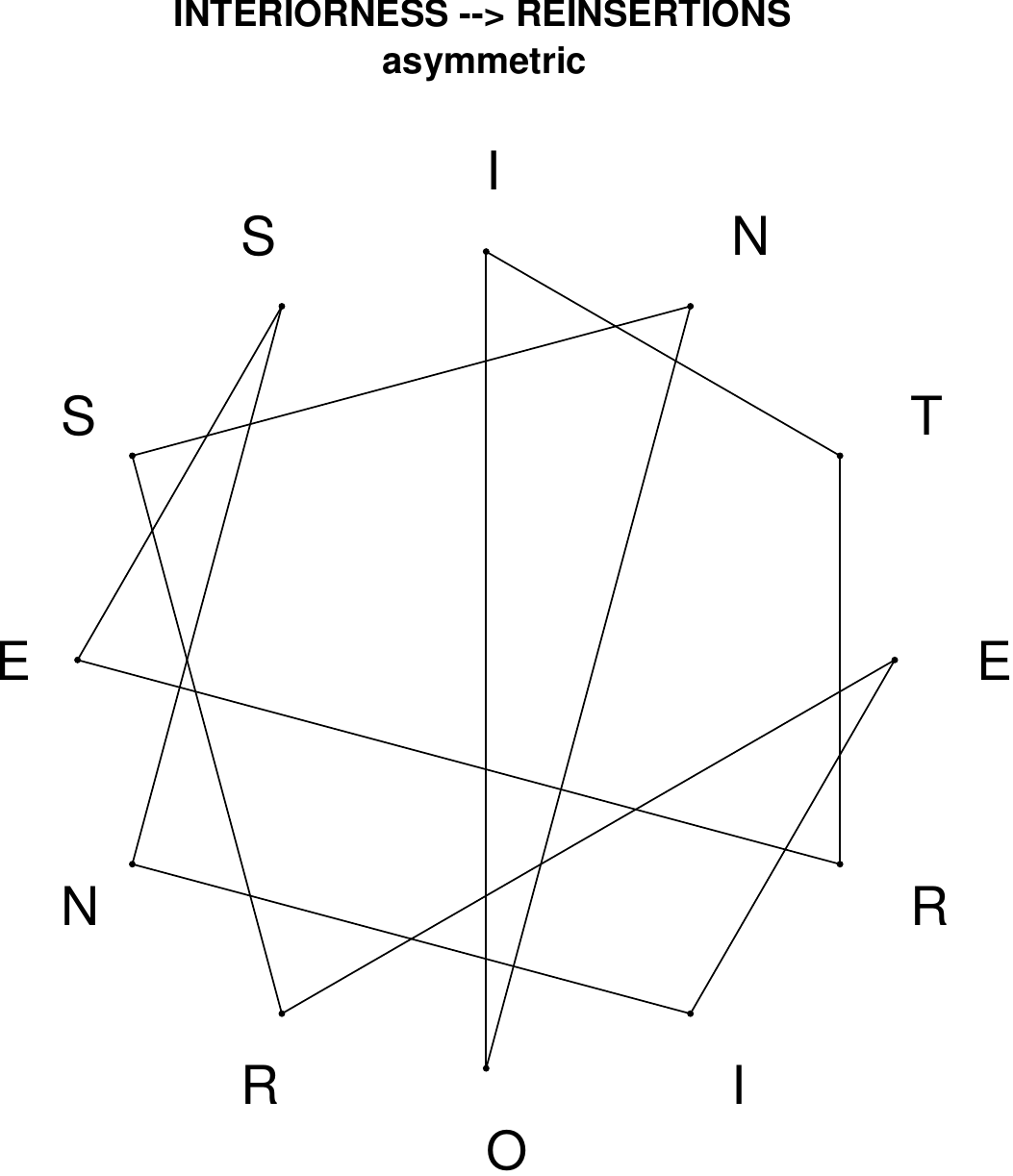}
\end{subfigure}
\end{figure}

\begin{figure}[H]
\centering
\begin{subfigure}[T]{0.19\textwidth}
\centering
\includegraphics[width=\textwidth]{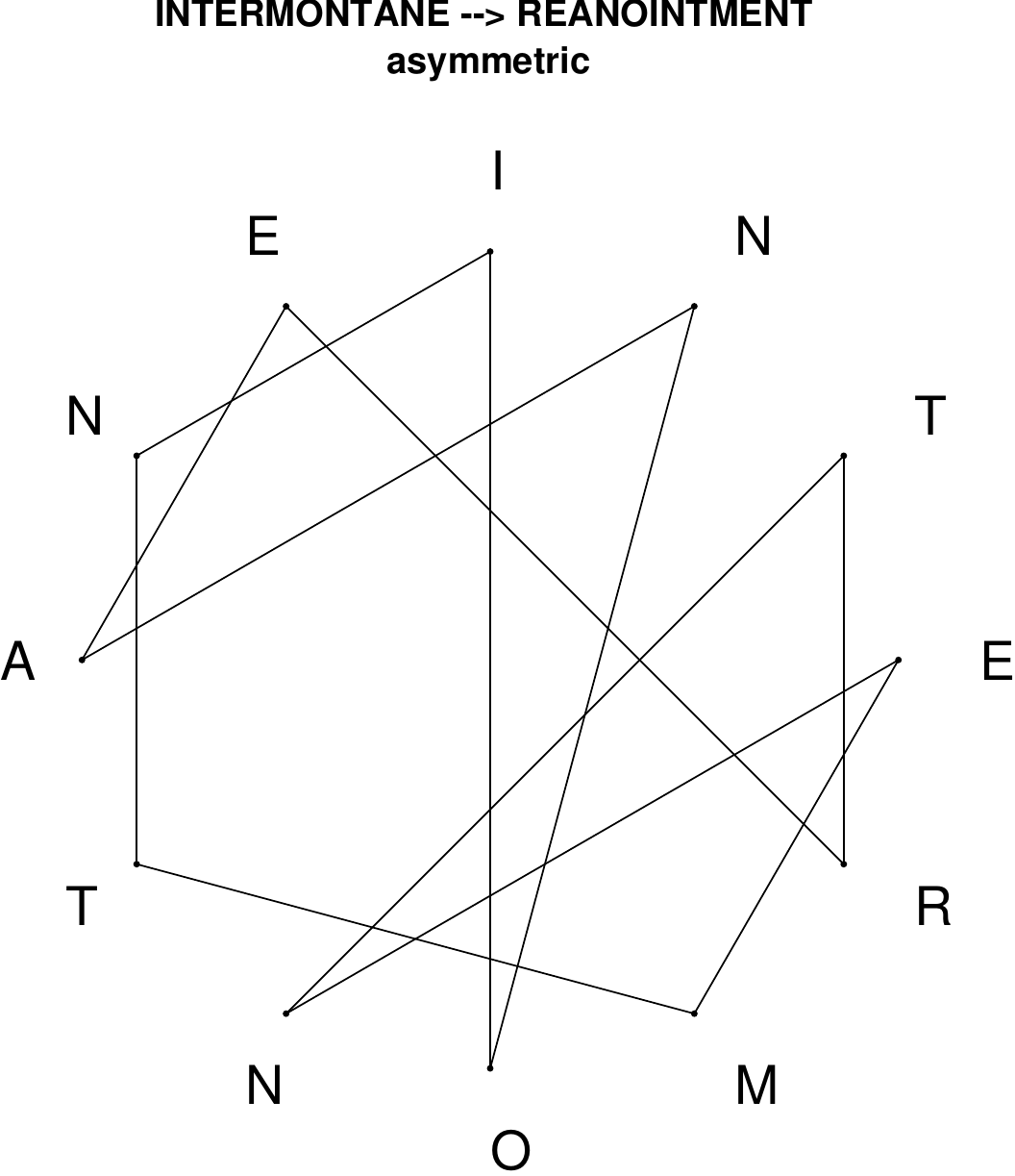}
\end{subfigure}
\hfill
\begin{subfigure}[T]{0.19\textwidth}
\centering
\includegraphics[width=\textwidth]{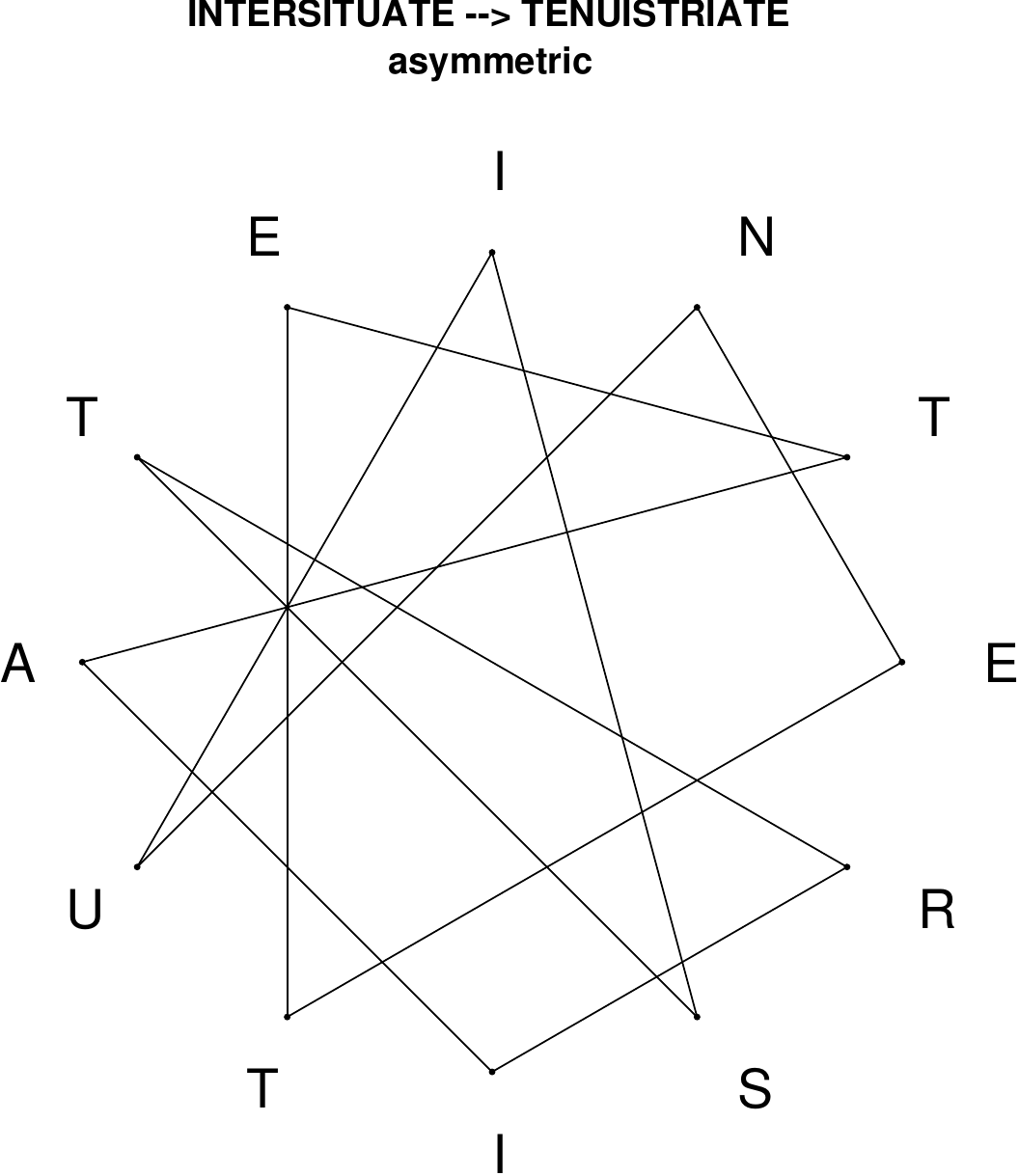}
\end{subfigure}
\hfill
\begin{subfigure}[T]{0.19\textwidth}
\centering
\includegraphics[width=\textwidth]{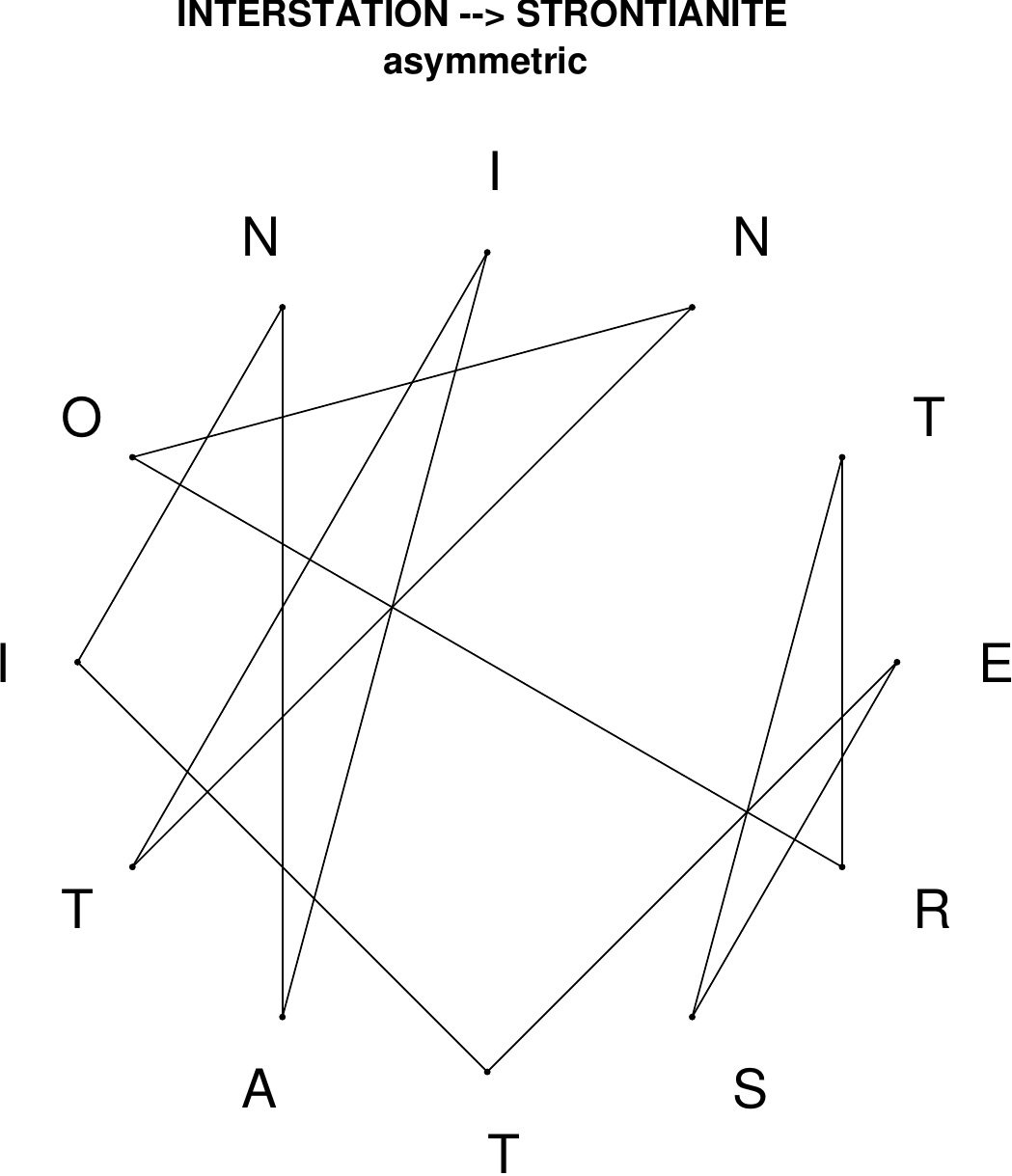}
\end{subfigure}
\hfill
\begin{subfigure}[T]{0.19\textwidth}
\centering
\includegraphics[width=\textwidth]{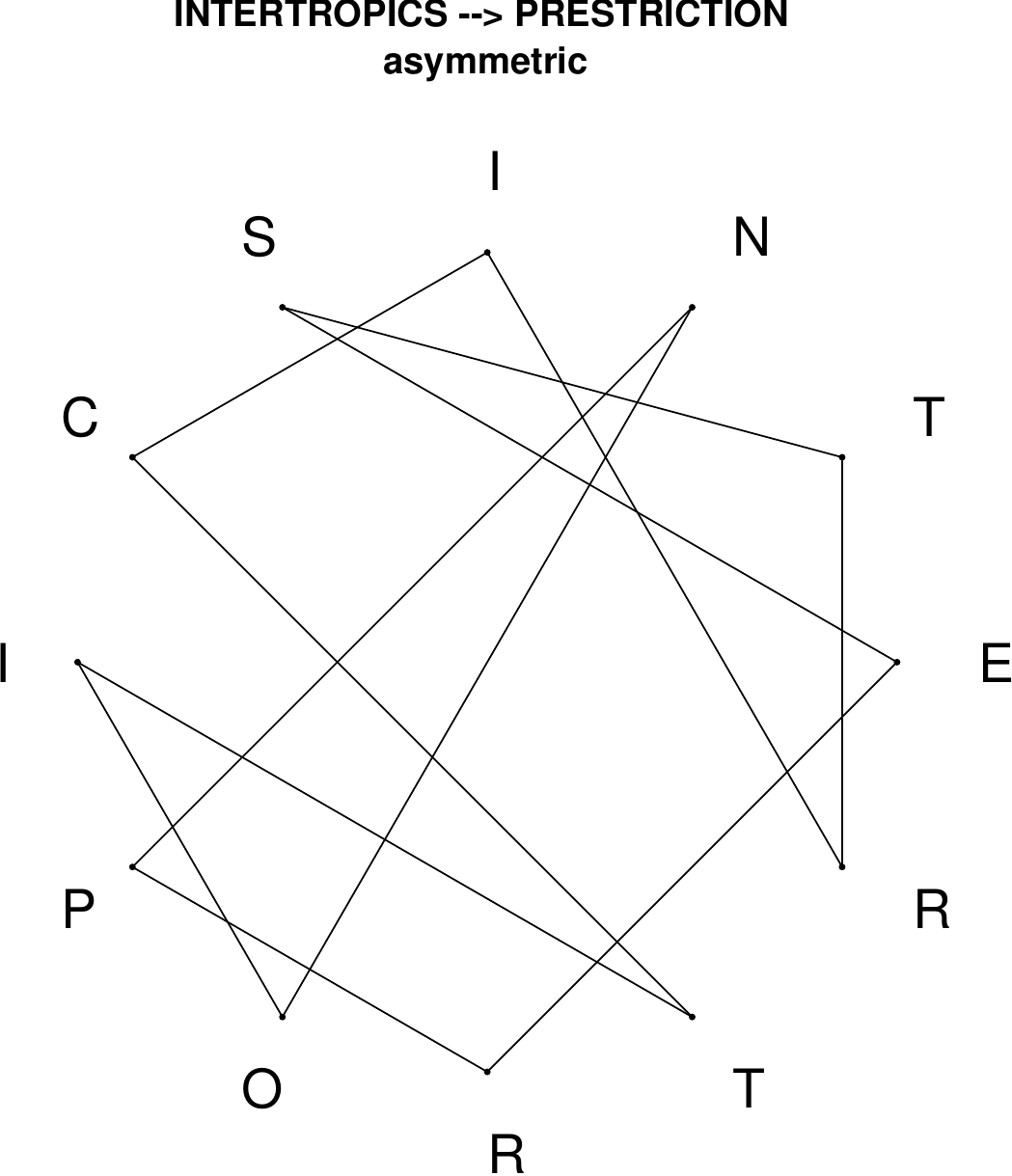}
\end{subfigure}
\hfill
\begin{subfigure}[T]{0.19\textwidth}
\centering
\includegraphics[width=\textwidth]{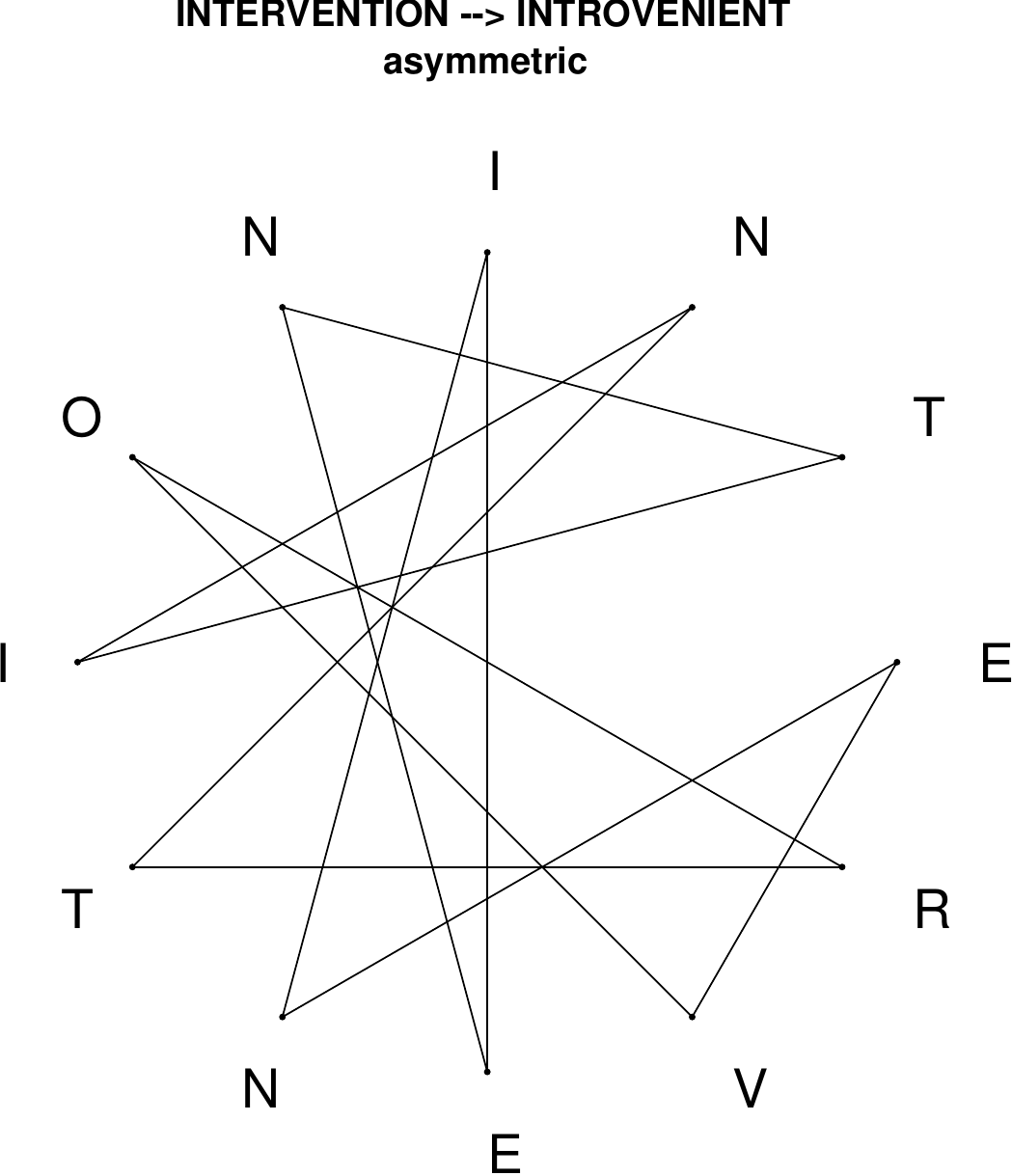}
\end{subfigure}
\end{figure}

\begin{figure}[H]
\centering
\begin{subfigure}[T]{0.19\textwidth}
\centering
\includegraphics[width=\textwidth]{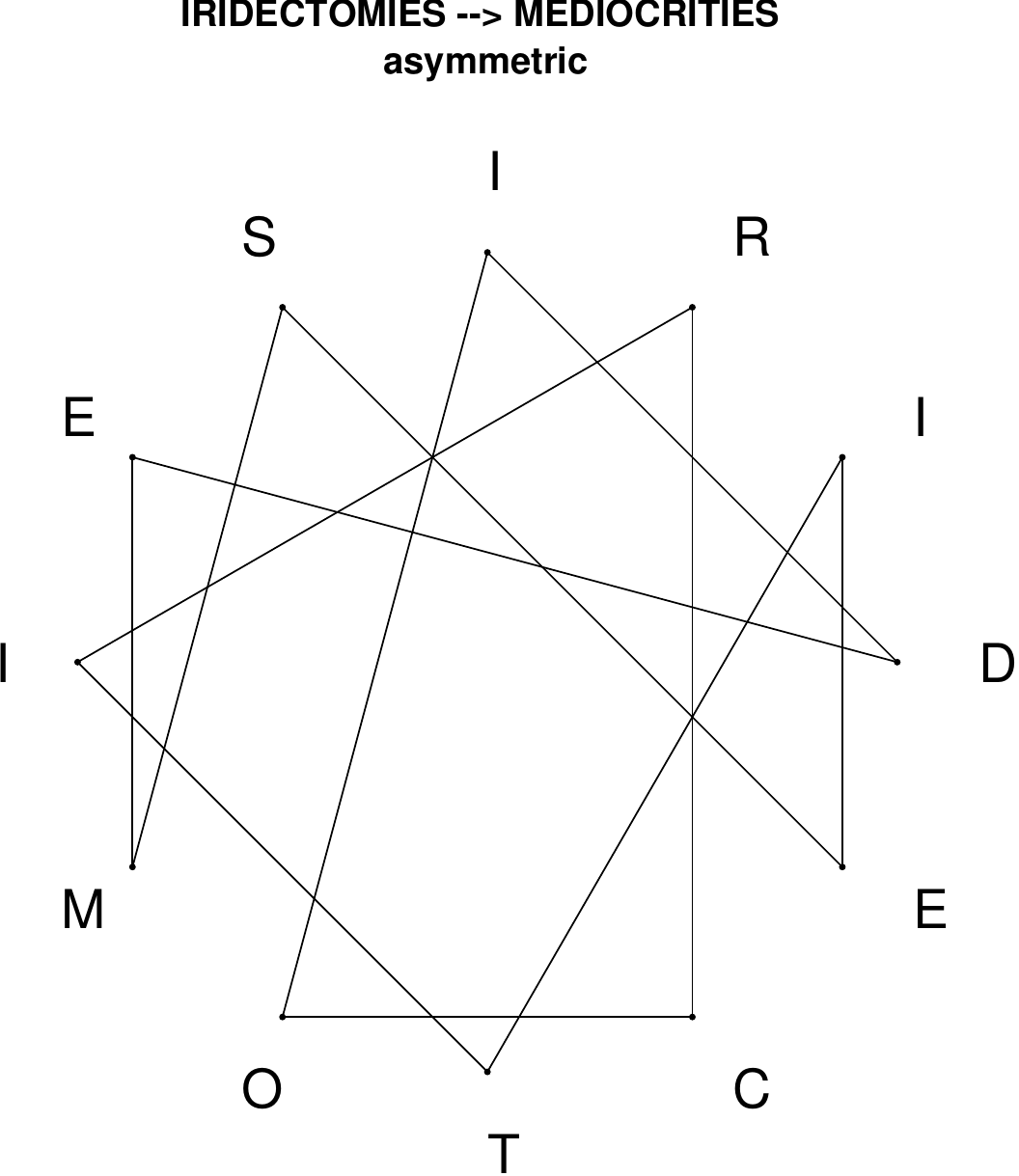}
\end{subfigure}
\hfill
\begin{subfigure}[T]{0.19\textwidth}
\centering
\includegraphics[width=\textwidth]{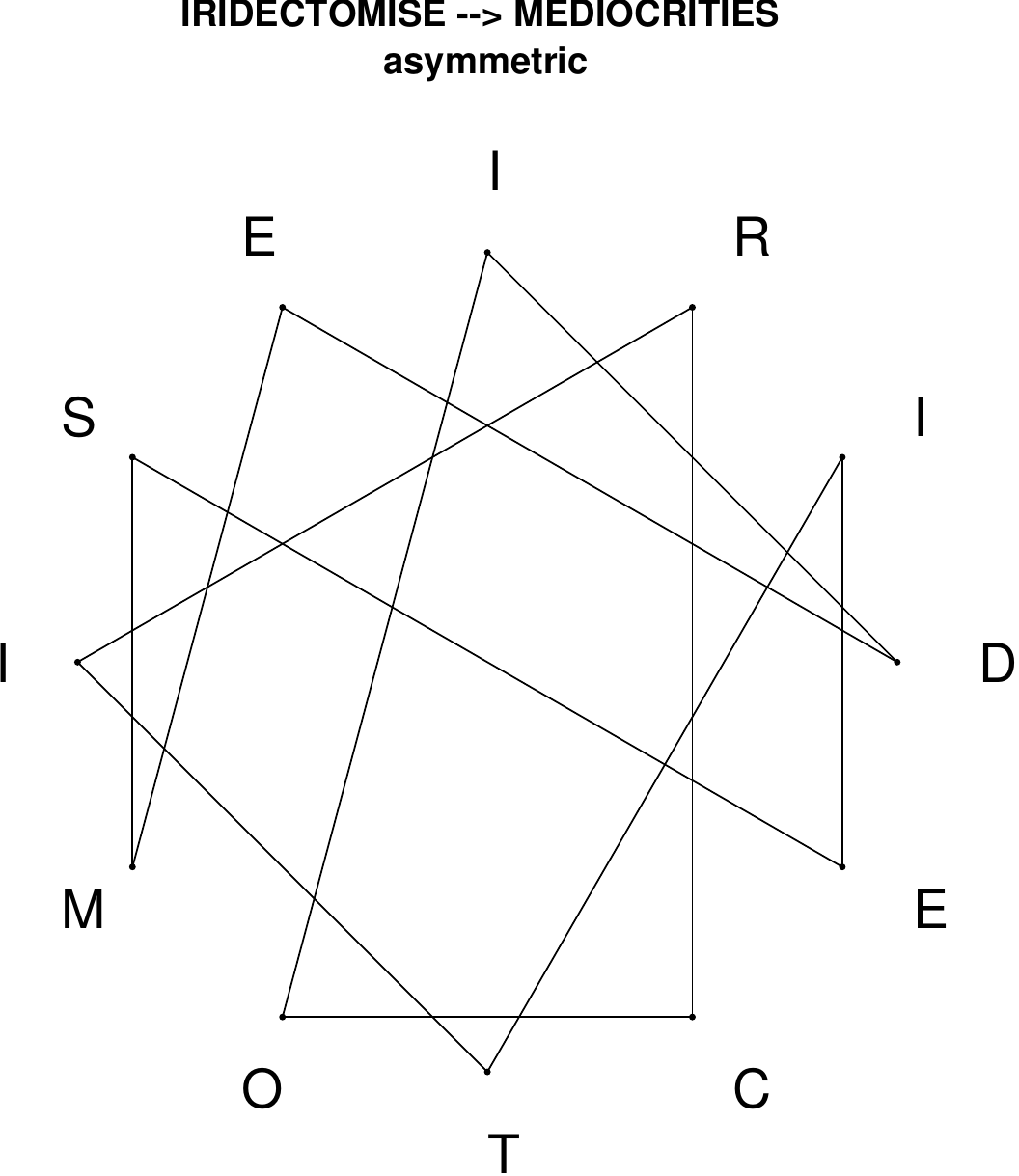}
\end{subfigure}
\hfill
\begin{subfigure}[T]{0.19\textwidth}
\centering
\includegraphics[width=\textwidth]{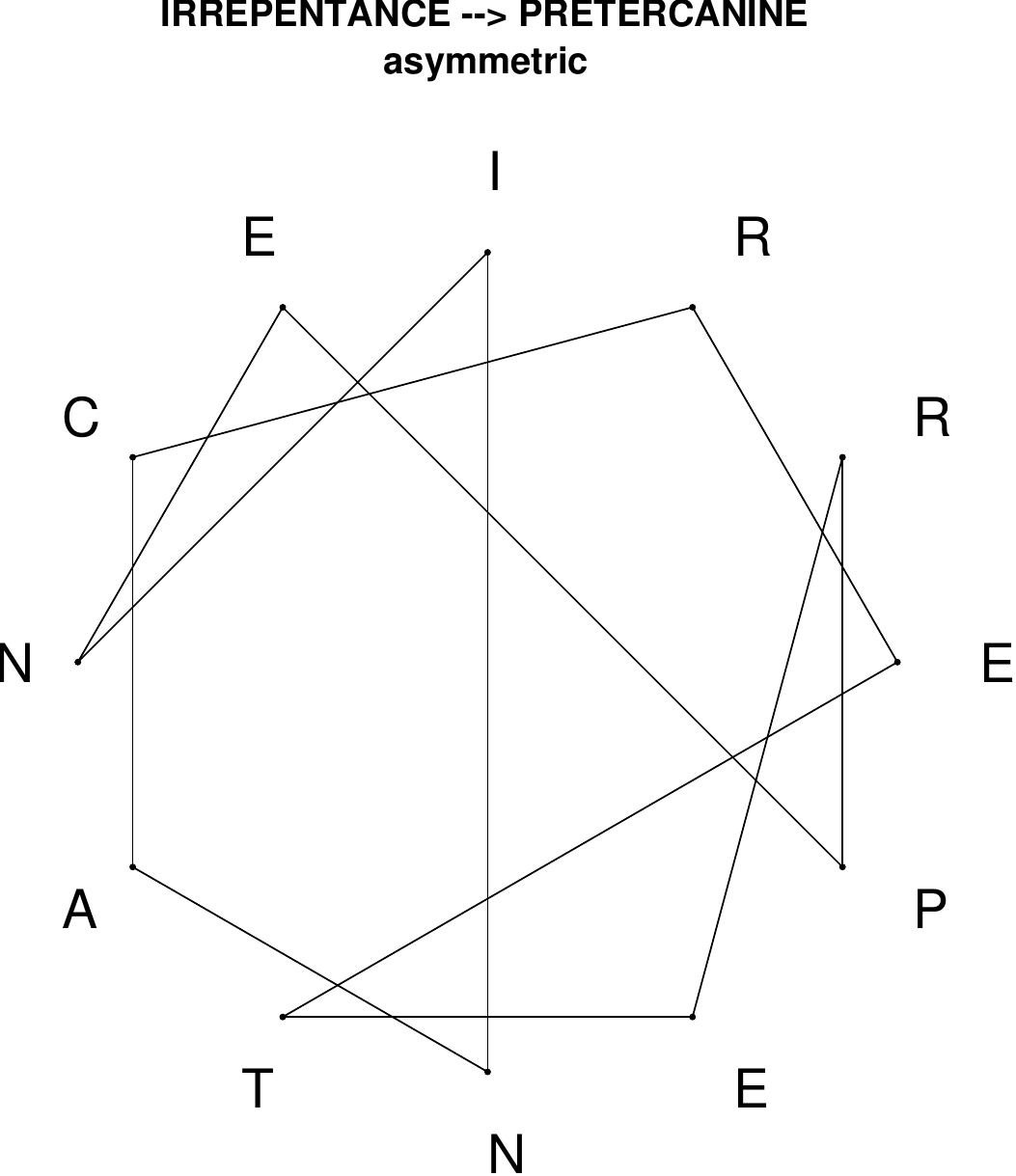}
\end{subfigure}
\hfill
\begin{subfigure}[T]{0.19\textwidth}
\centering
\includegraphics[width=\textwidth]{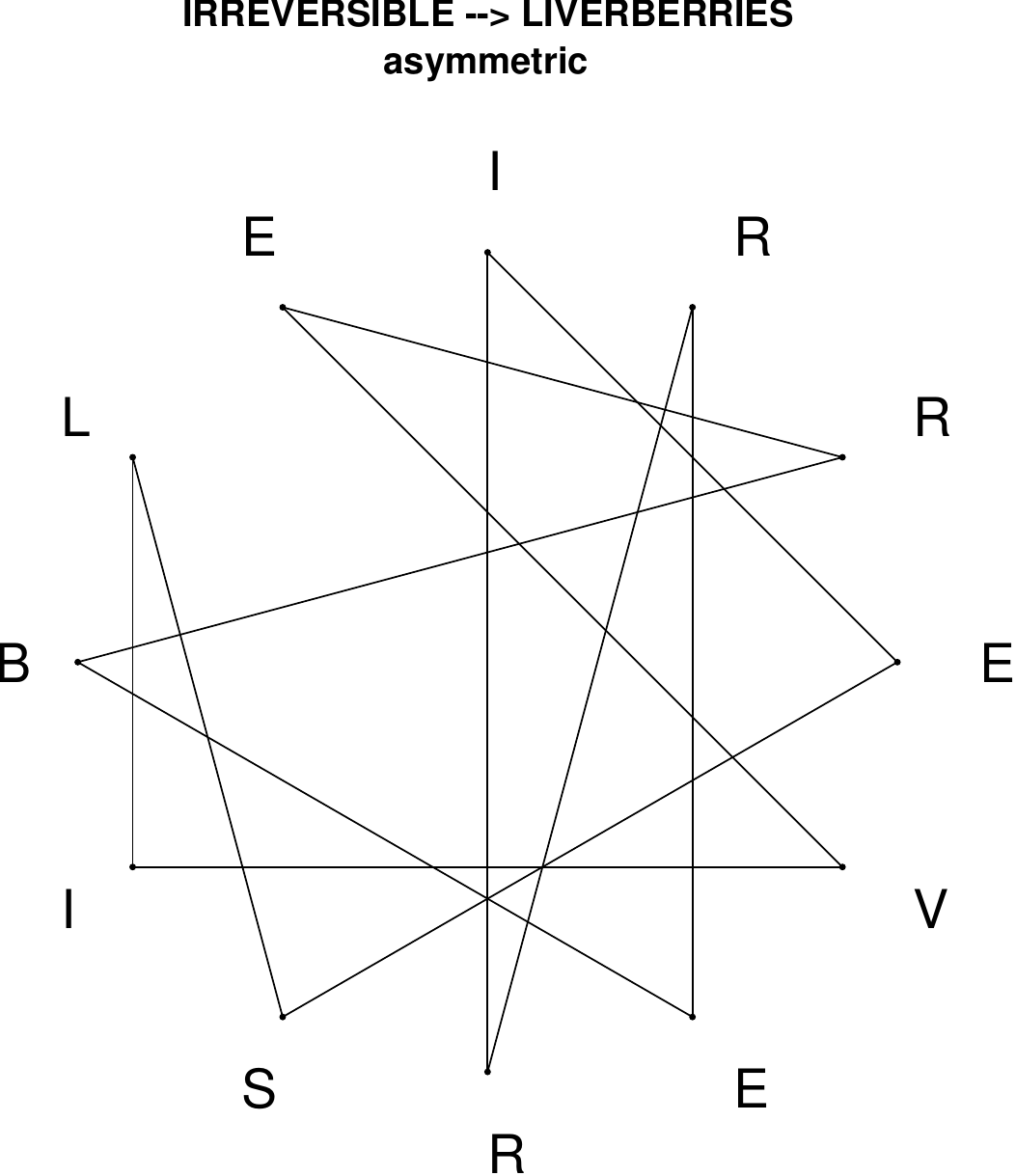}
\end{subfigure}
\hfill
\begin{subfigure}[T]{0.19\textwidth}
\centering
\includegraphics[width=\textwidth]{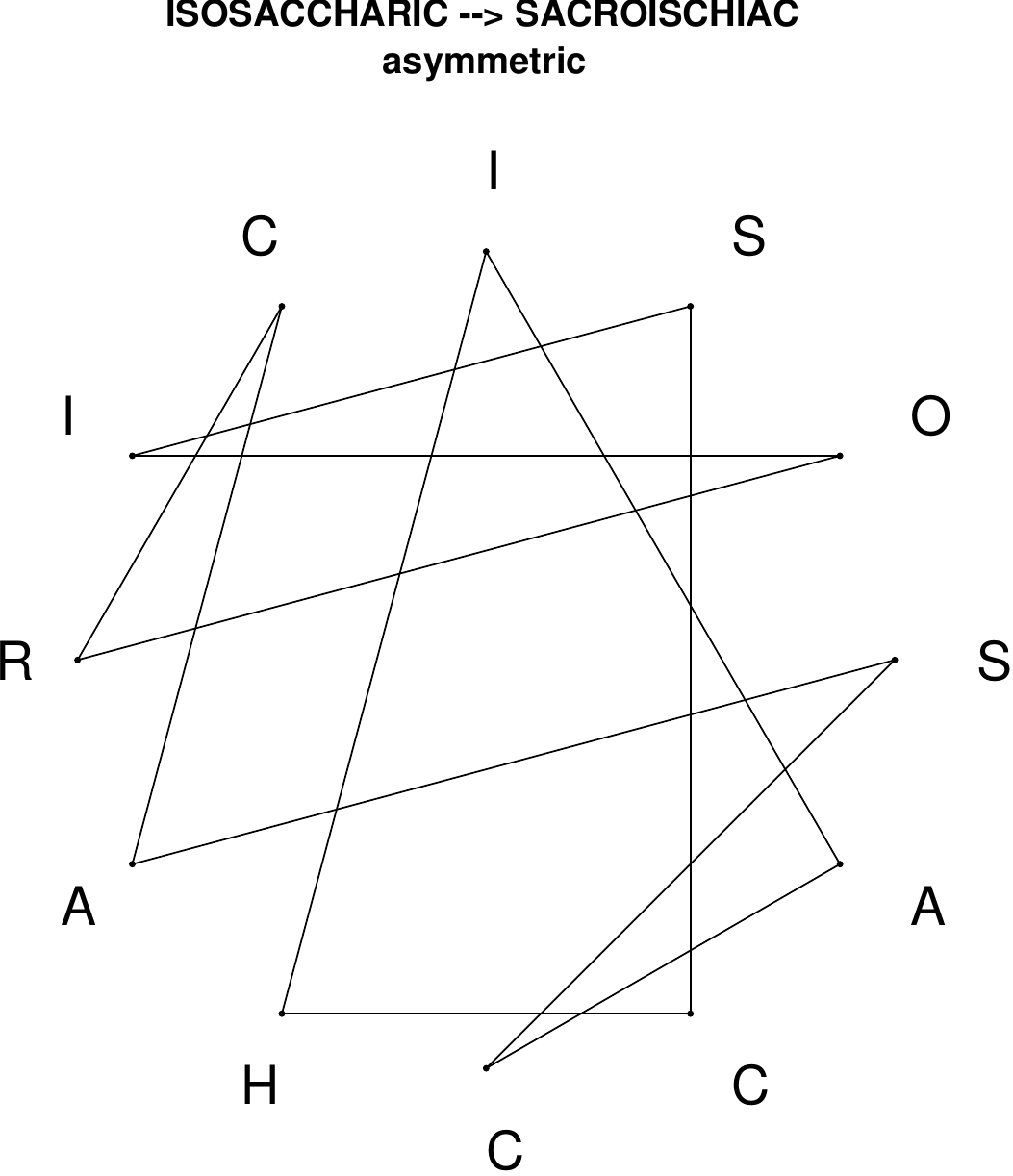}
\end{subfigure}
\end{figure}

\begin{figure}[H]
\centering
\begin{subfigure}[T]{0.19\textwidth}
\centering
\includegraphics[width=\textwidth]{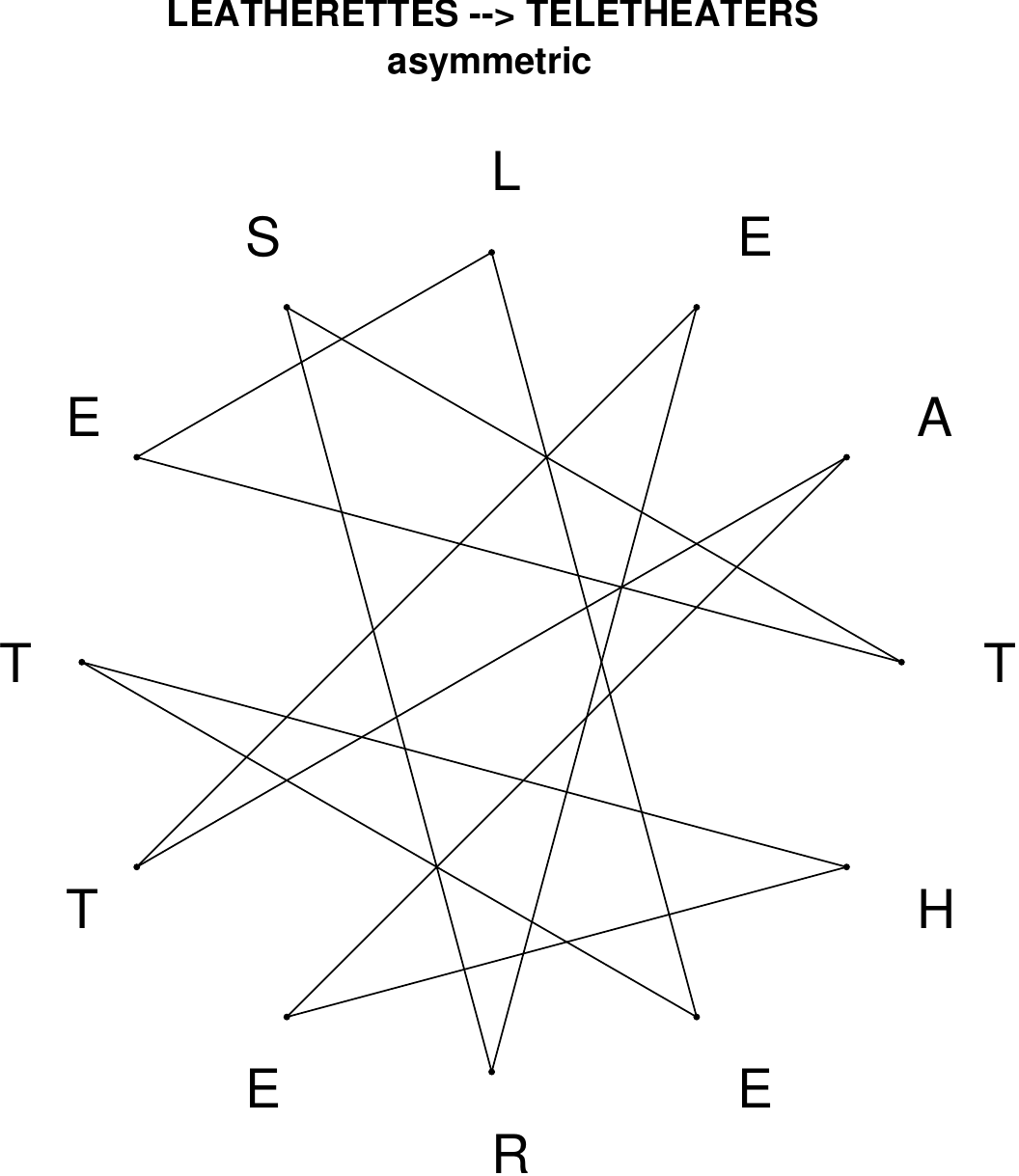}
\end{subfigure}
\hfill
\begin{subfigure}[T]{0.19\textwidth}
\centering
\includegraphics[width=\textwidth]{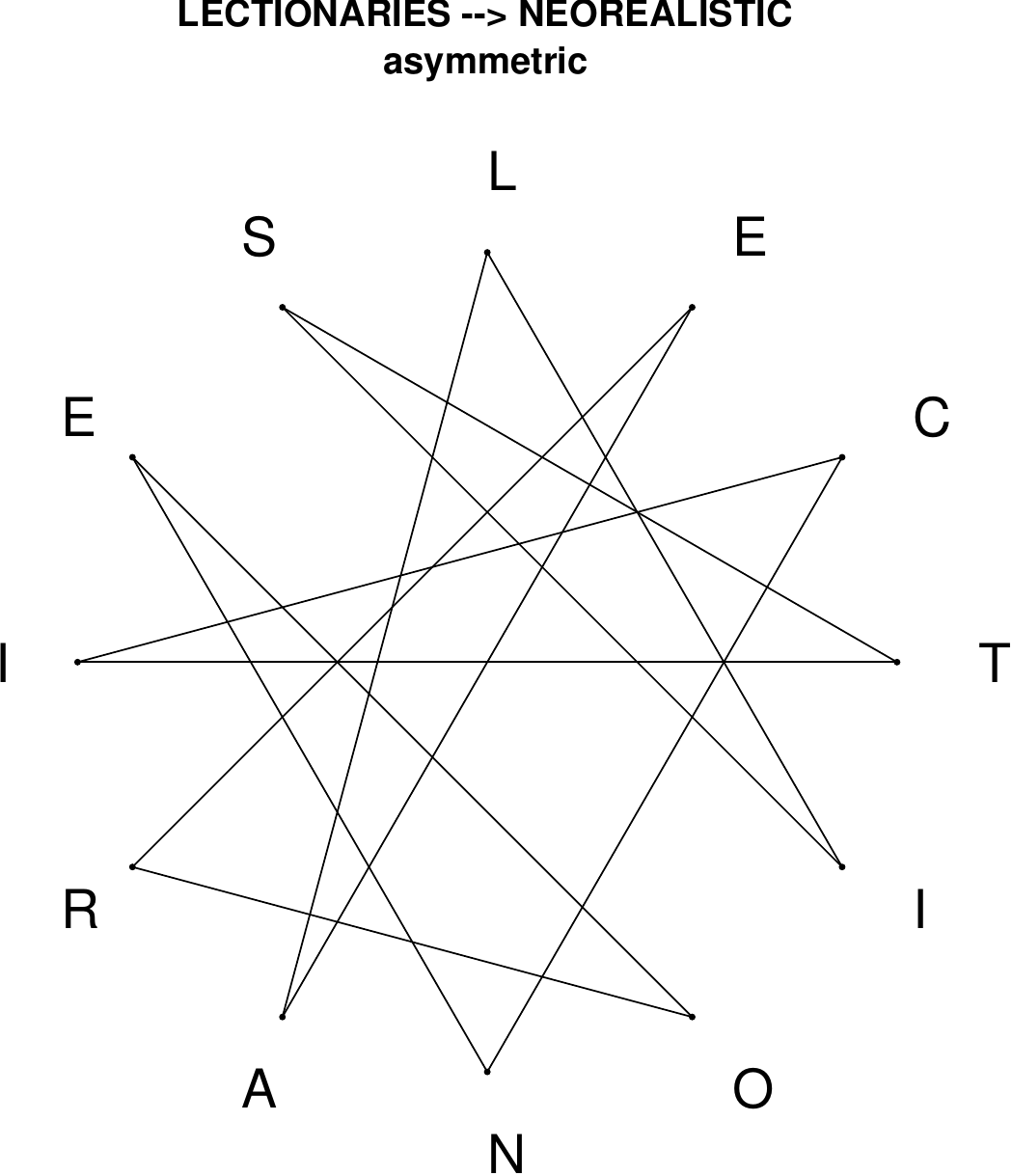}
\end{subfigure}
\hfill
\begin{subfigure}[T]{0.19\textwidth}
\centering
\includegraphics[width=\textwidth]{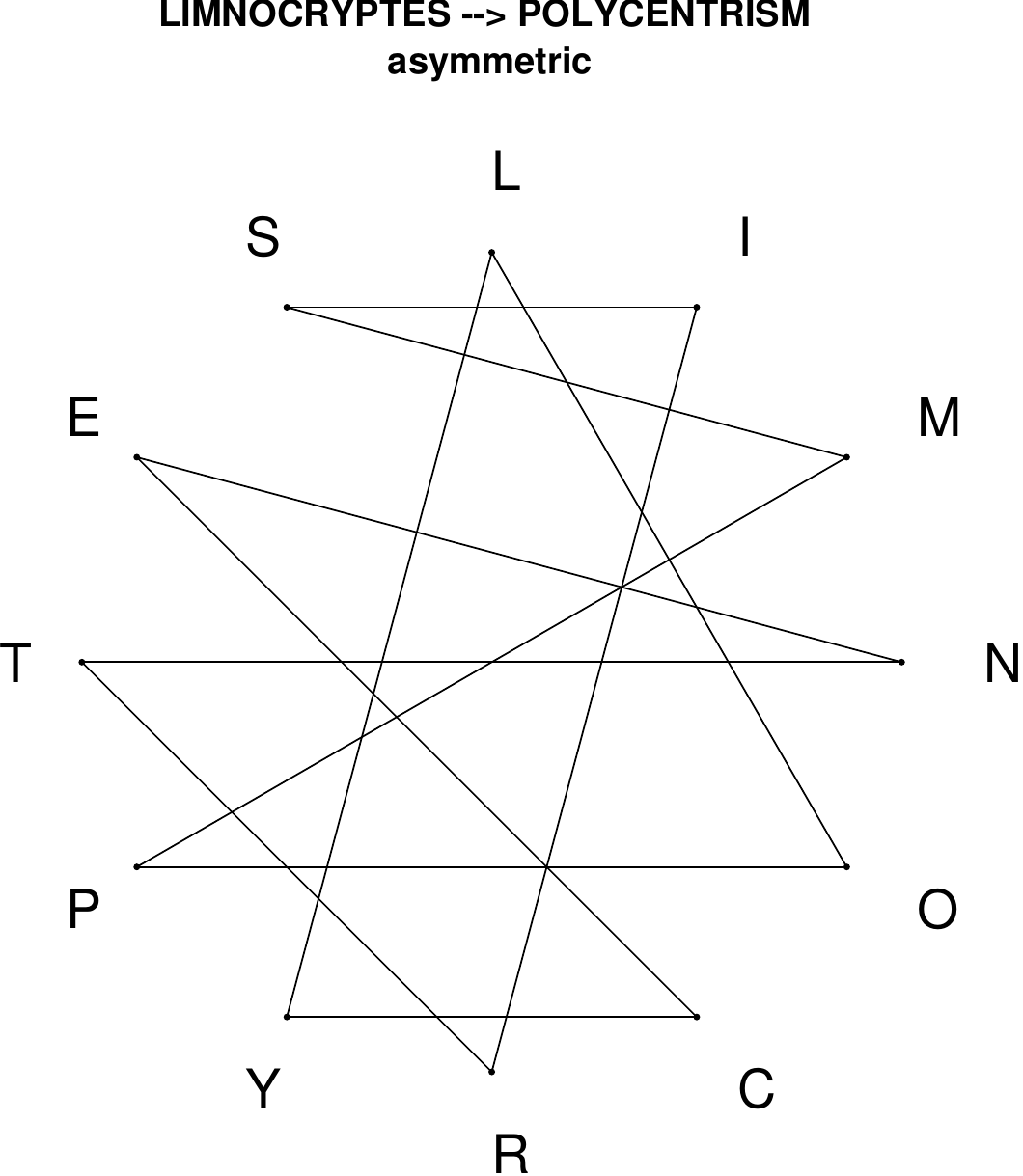}
\end{subfigure}
\hfill
\begin{subfigure}[T]{0.19\textwidth}
\centering
\includegraphics[width=\textwidth]{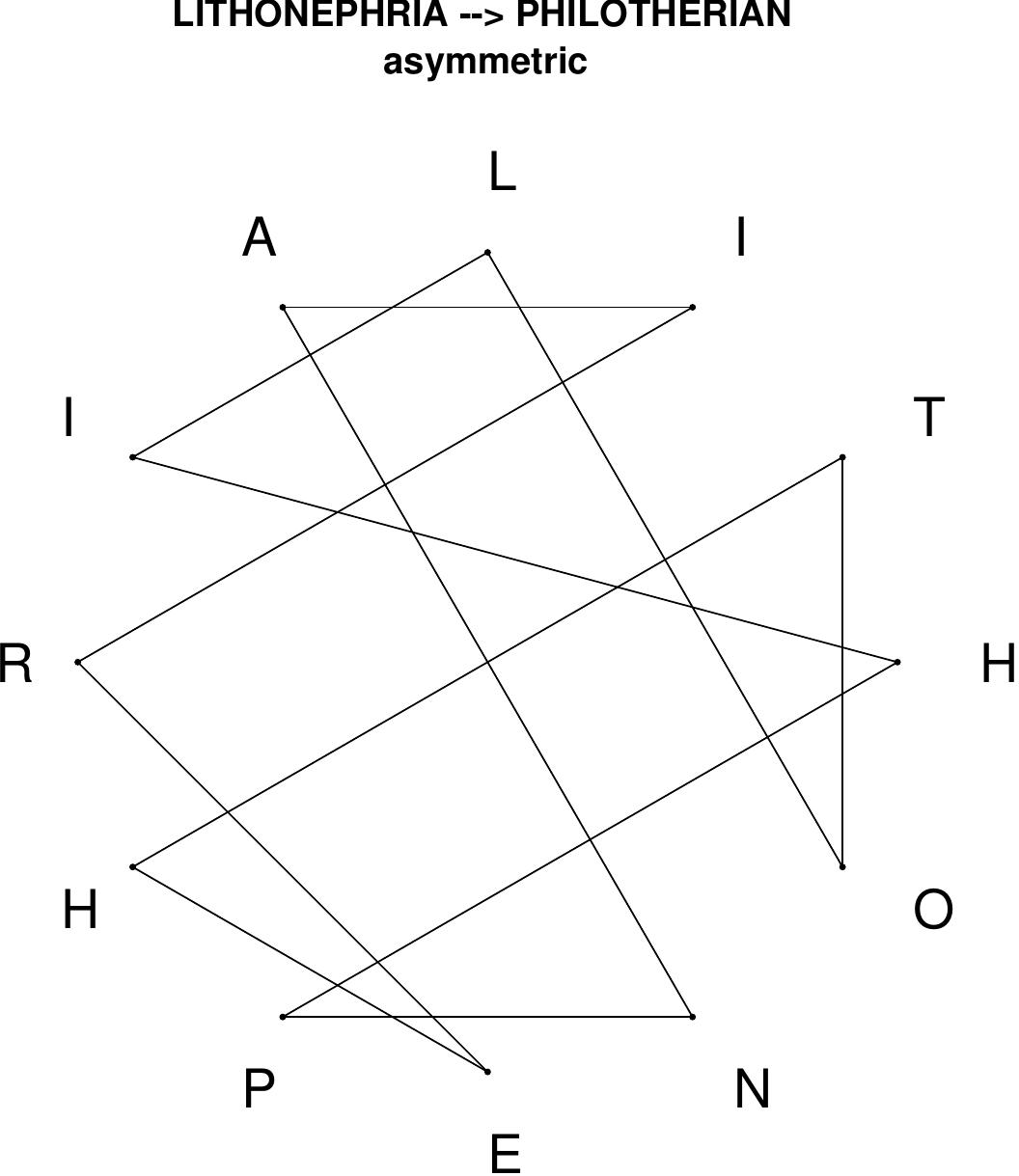}
\end{subfigure}
\hfill
\begin{subfigure}[T]{0.19\textwidth}
\centering
\includegraphics[width=\textwidth]{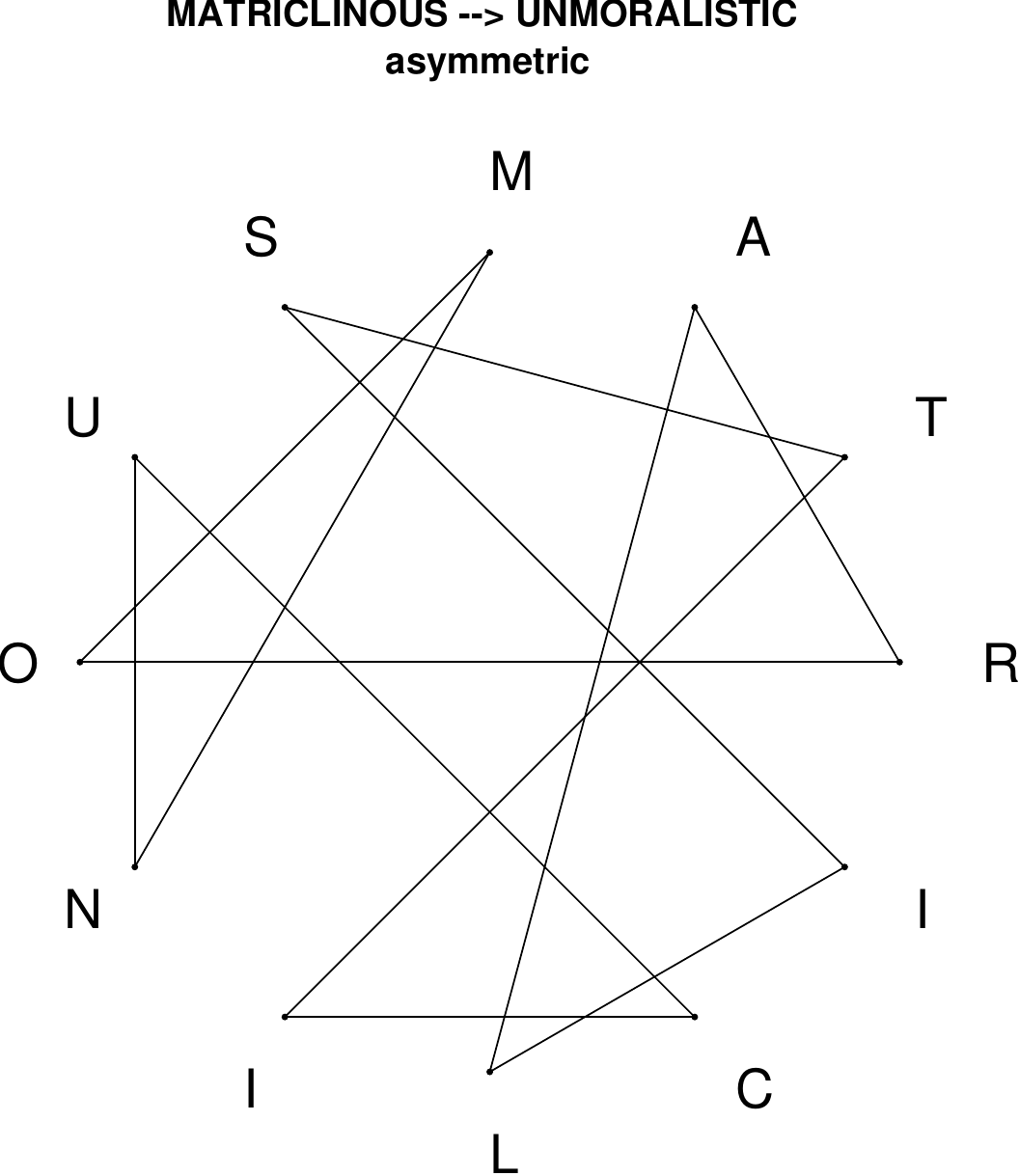}
\end{subfigure}
\end{figure}

\begin{figure}[H]
\centering
\begin{subfigure}[T]{0.19\textwidth}
\centering
\includegraphics[width=\textwidth]{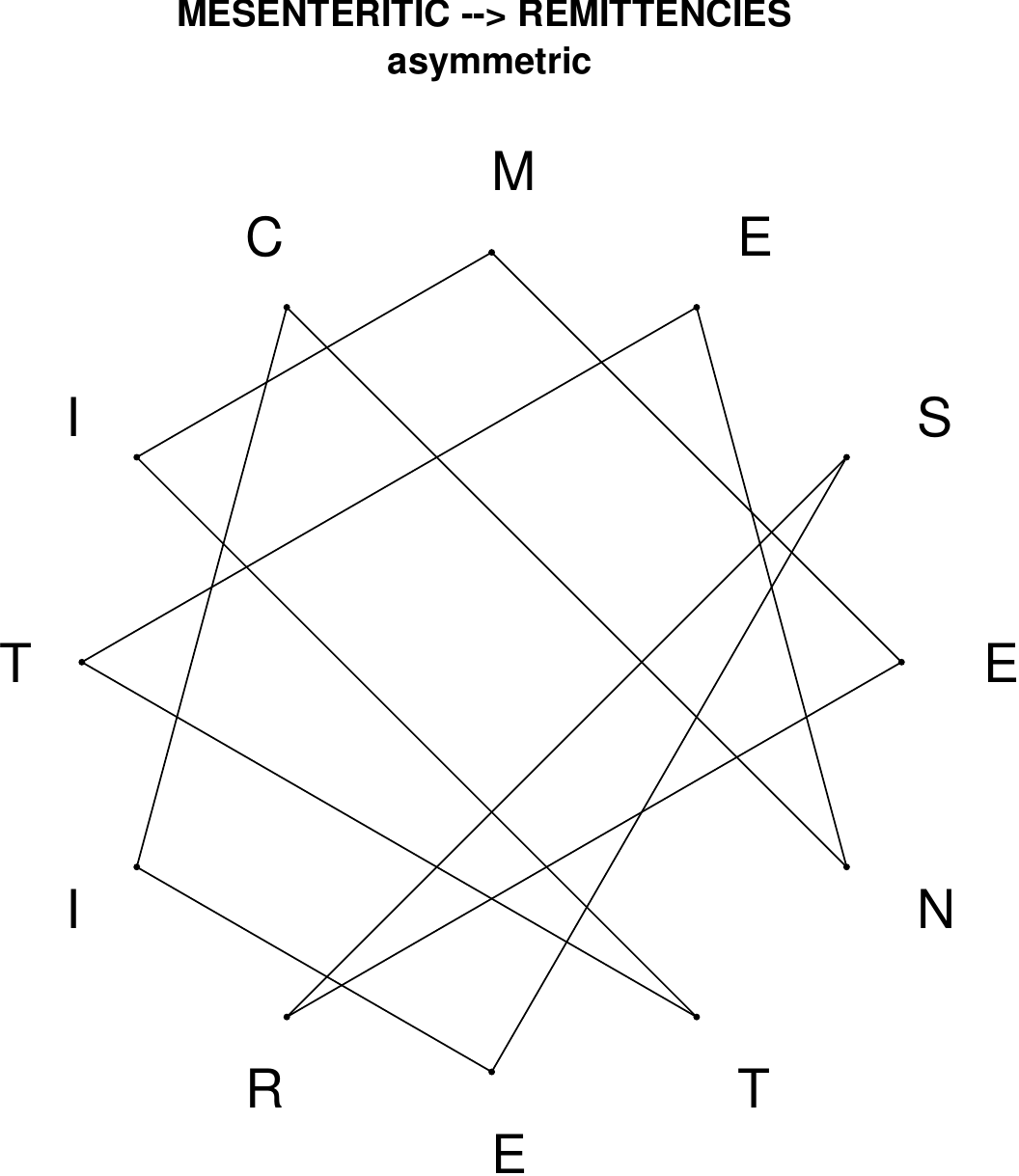}
\end{subfigure}
\hfill
\begin{subfigure}[T]{0.19\textwidth}
\centering
\includegraphics[width=\textwidth]{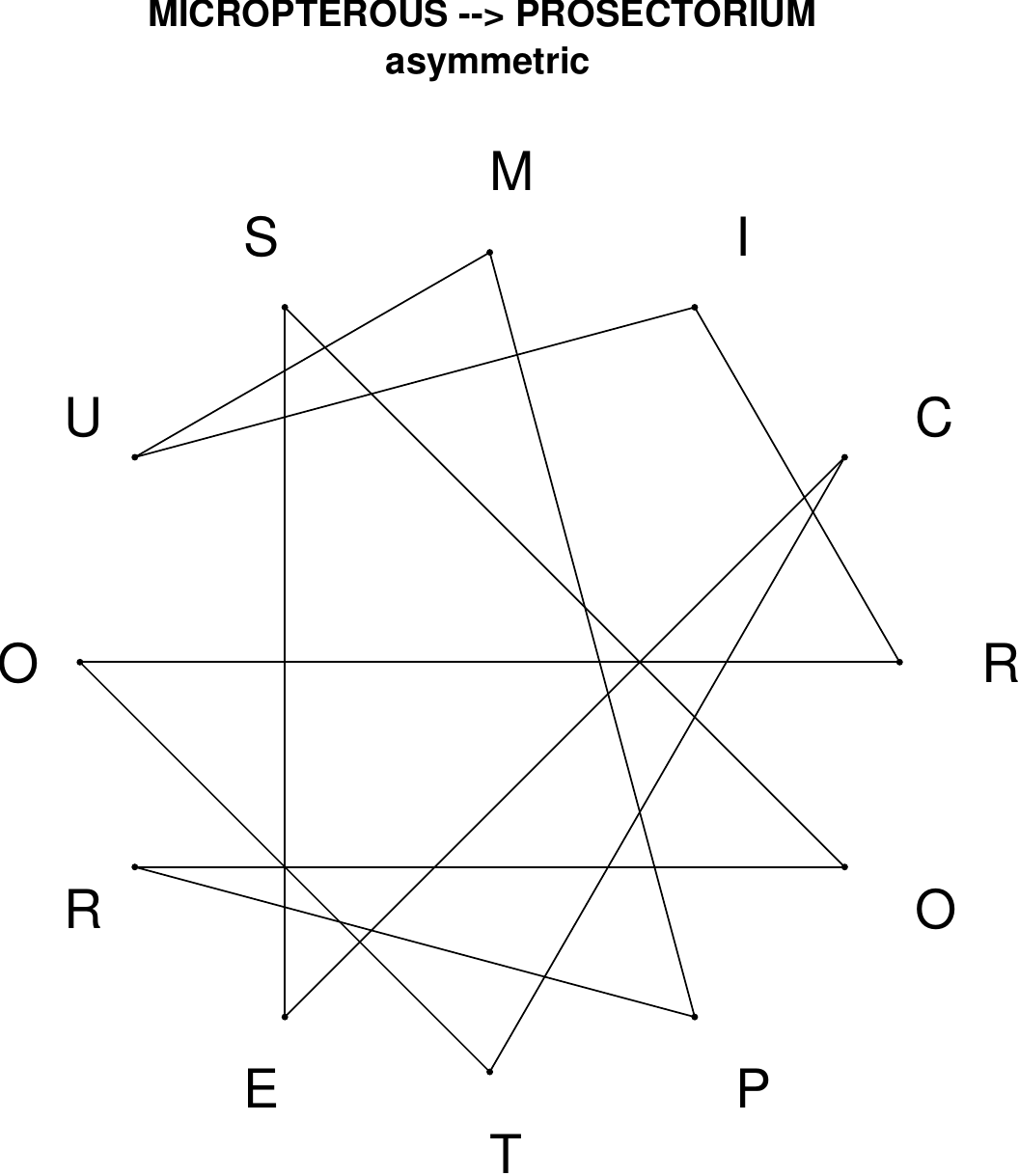}
\end{subfigure}
\hfill
\begin{subfigure}[T]{0.19\textwidth}
\centering
\includegraphics[width=\textwidth]{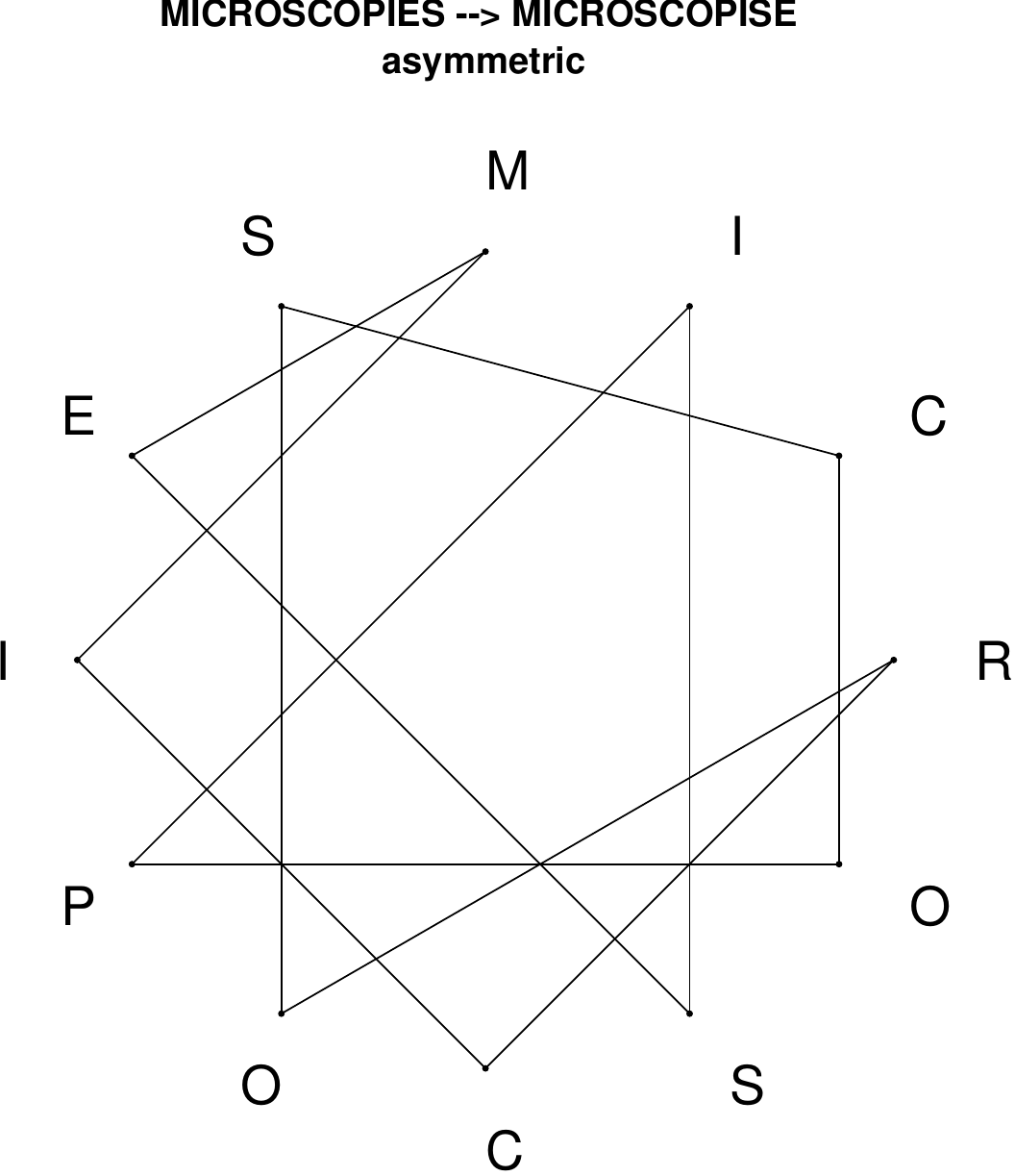}
\end{subfigure}
\hfill
\begin{subfigure}[T]{0.19\textwidth}
\centering
\includegraphics[width=\textwidth]{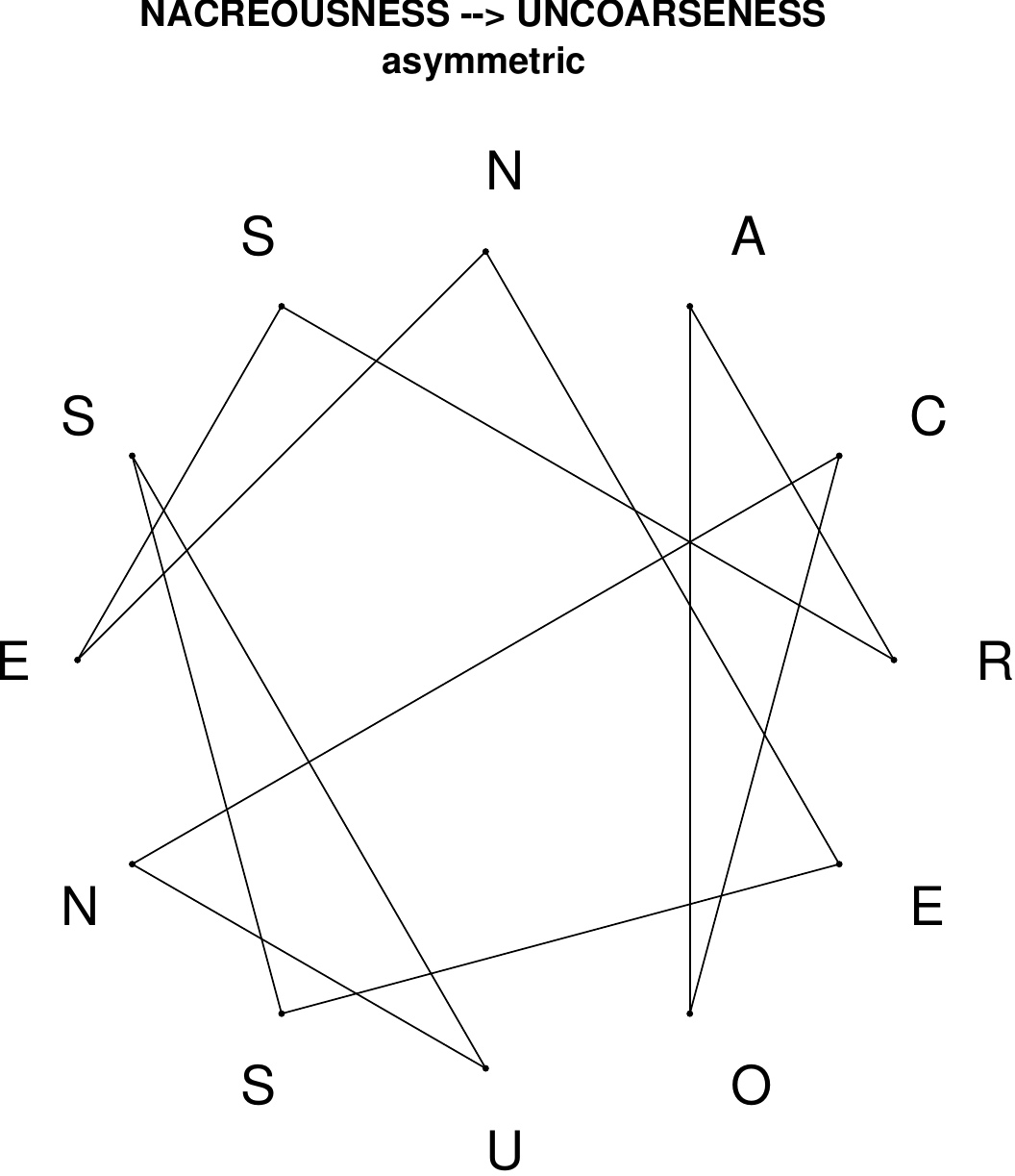}
\end{subfigure}
\hfill
\begin{subfigure}[T]{0.19\textwidth}
\centering
\includegraphics[width=\textwidth]{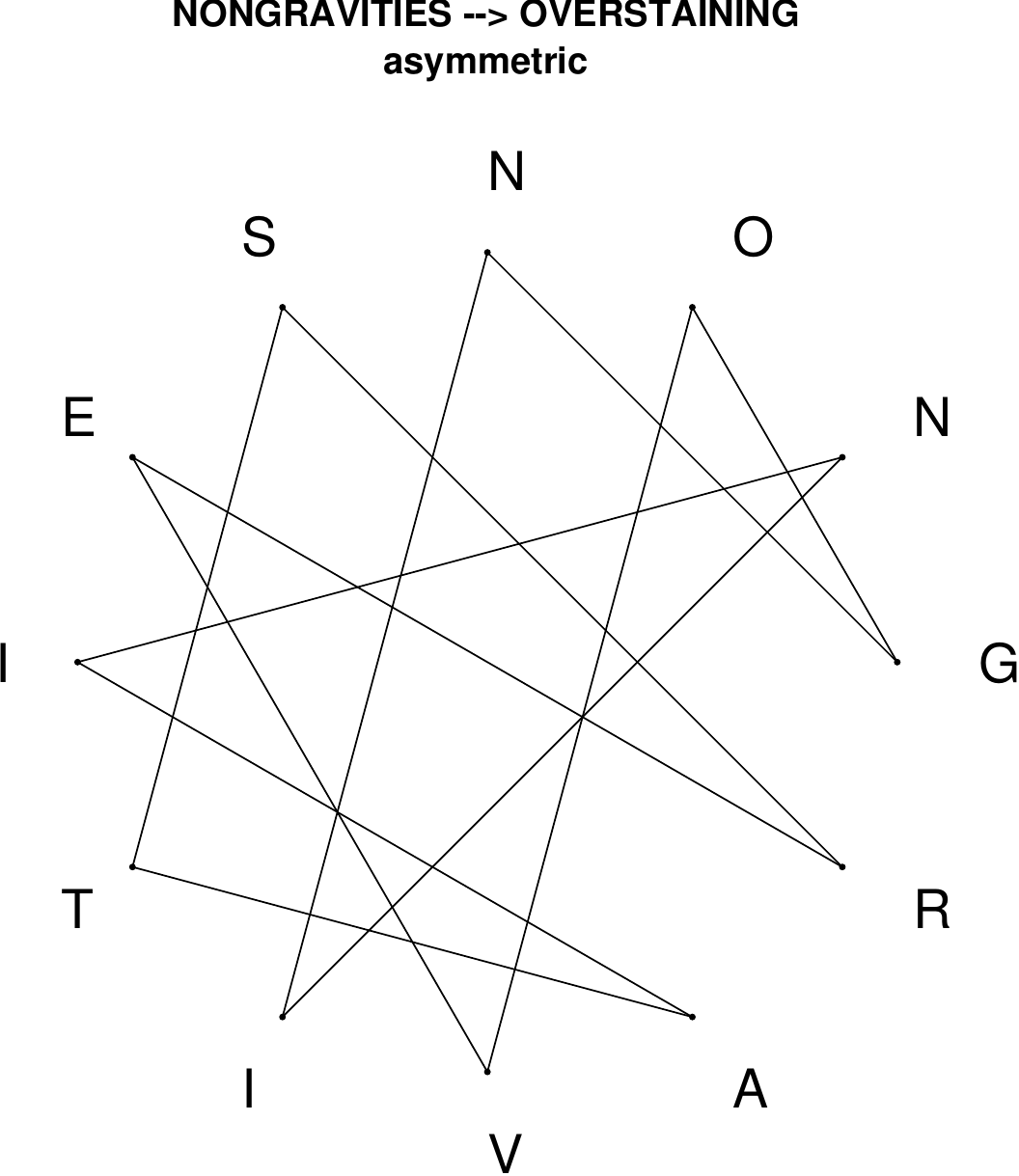}
\end{subfigure}
\end{figure}

\begin{figure}[H]
\centering
\begin{subfigure}[T]{0.19\textwidth}
\centering
\includegraphics[width=\textwidth]{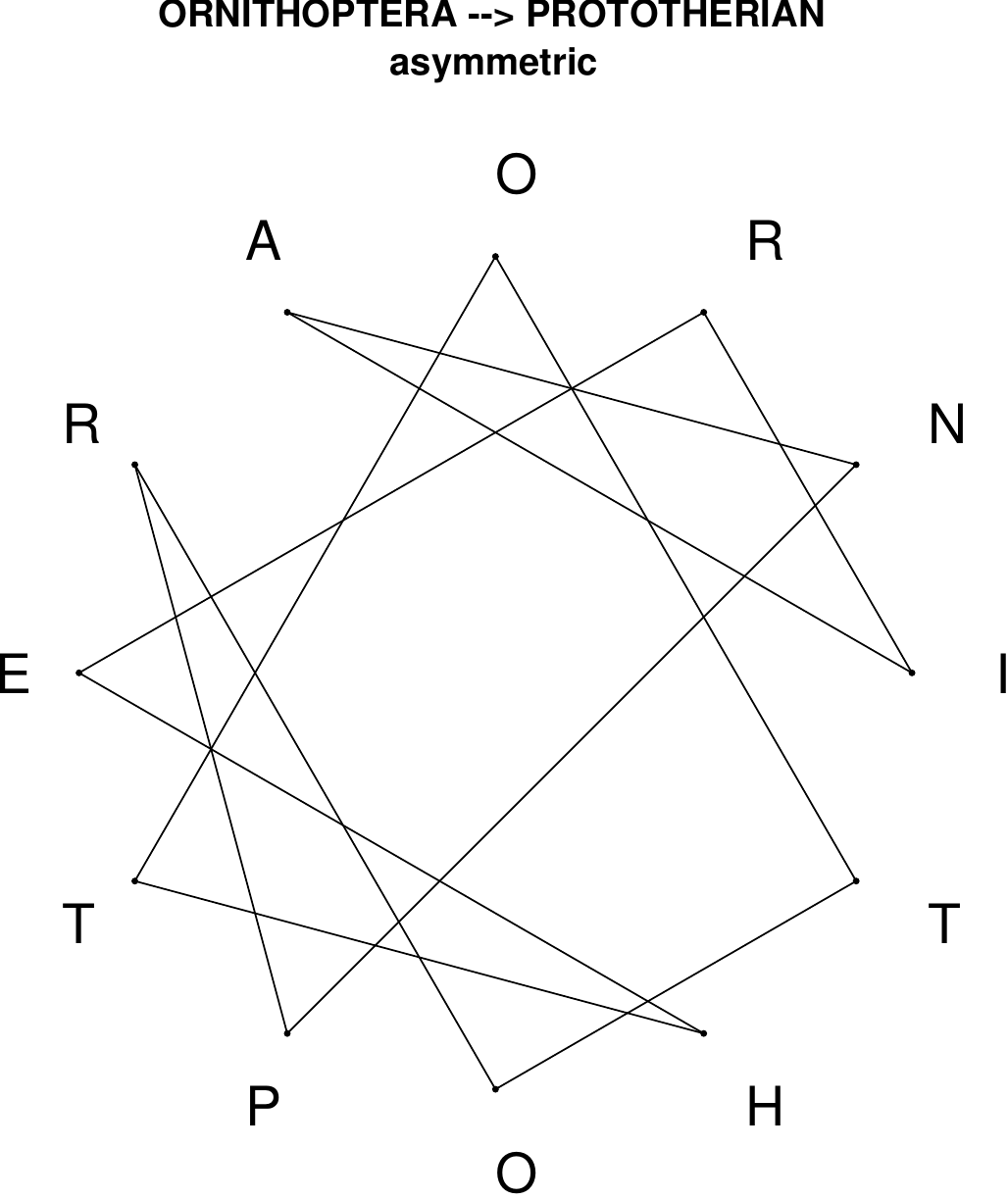}
\end{subfigure}
\hfill
\begin{subfigure}[T]{0.19\textwidth}
\centering
\includegraphics[width=\textwidth]{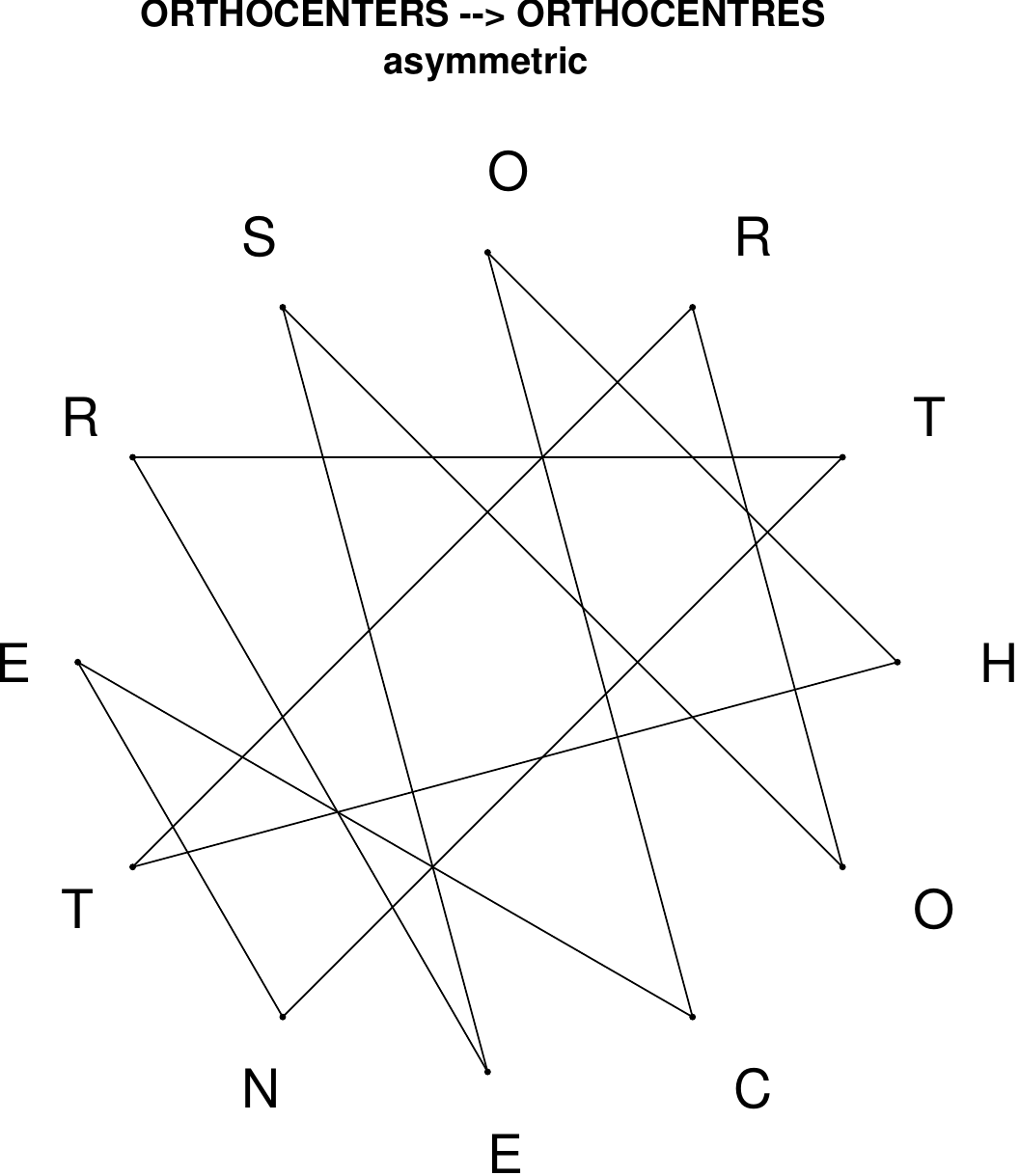}
\end{subfigure}
\hfill
\begin{subfigure}[T]{0.19\textwidth}
\centering
\includegraphics[width=\textwidth]{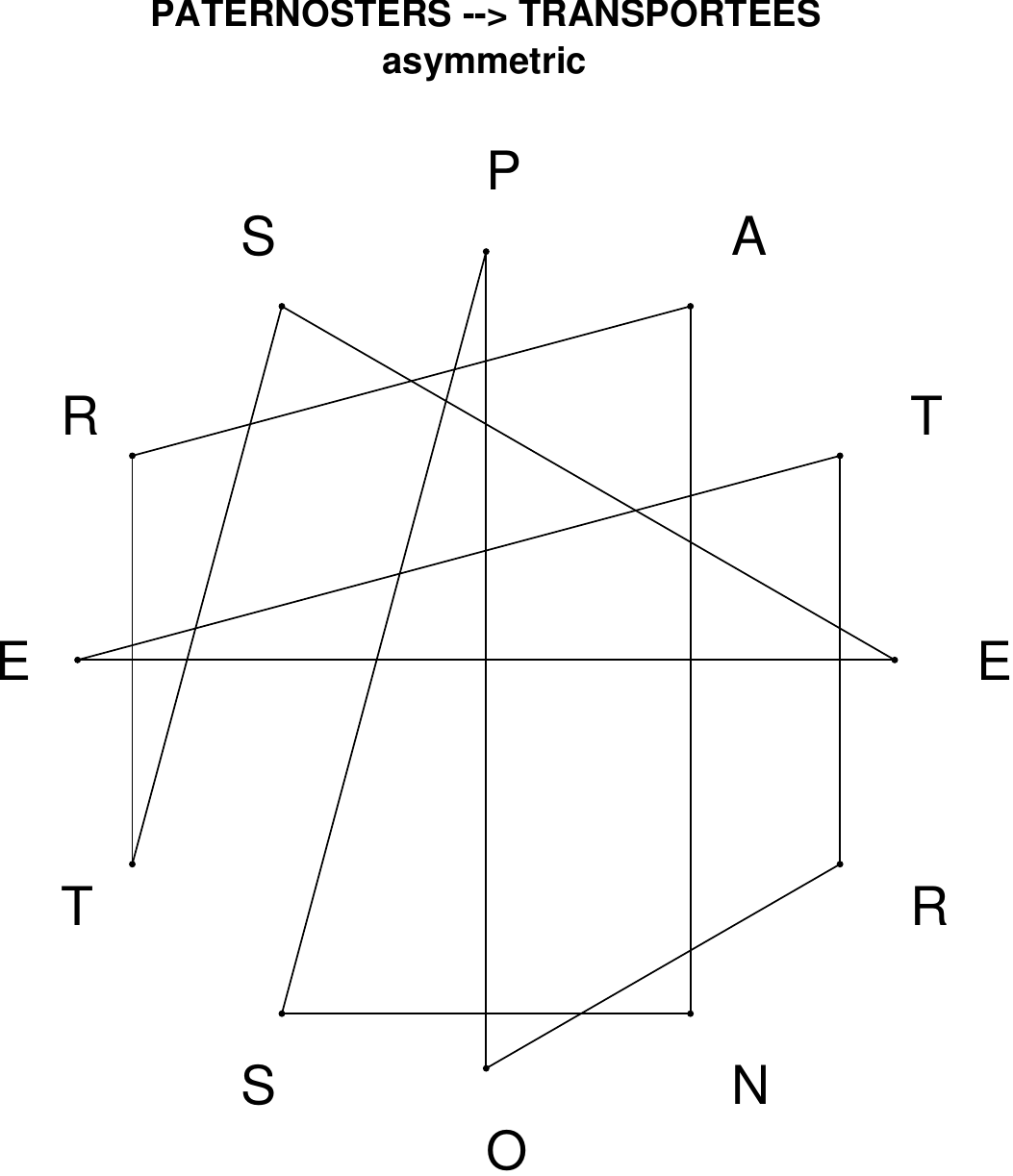}
\end{subfigure}
\hfill
\begin{subfigure}[T]{0.19\textwidth}
\centering
\includegraphics[width=\textwidth]{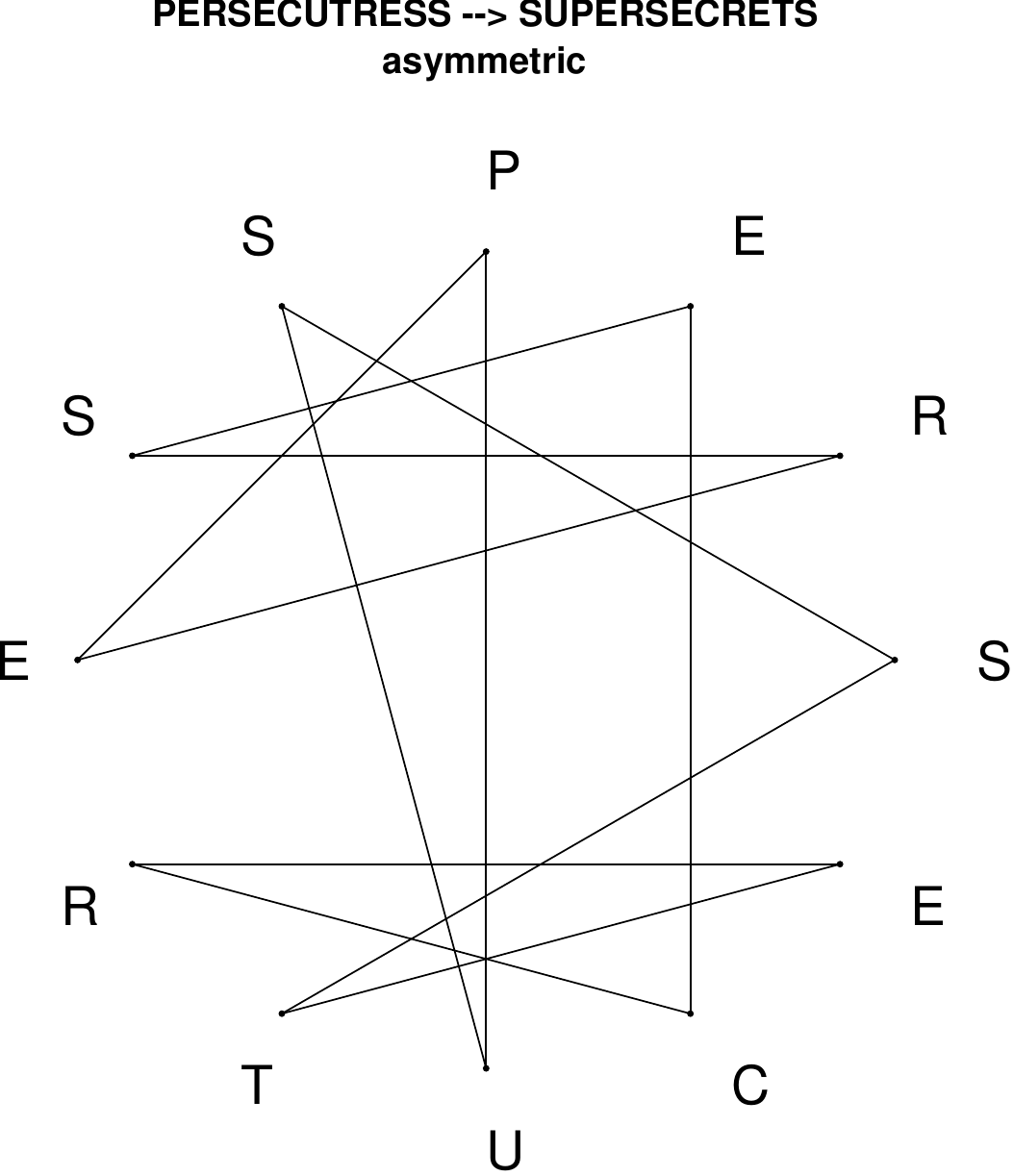}
\end{subfigure}
\hfill
\begin{subfigure}[T]{0.19\textwidth}
\centering
\includegraphics[width=\textwidth]{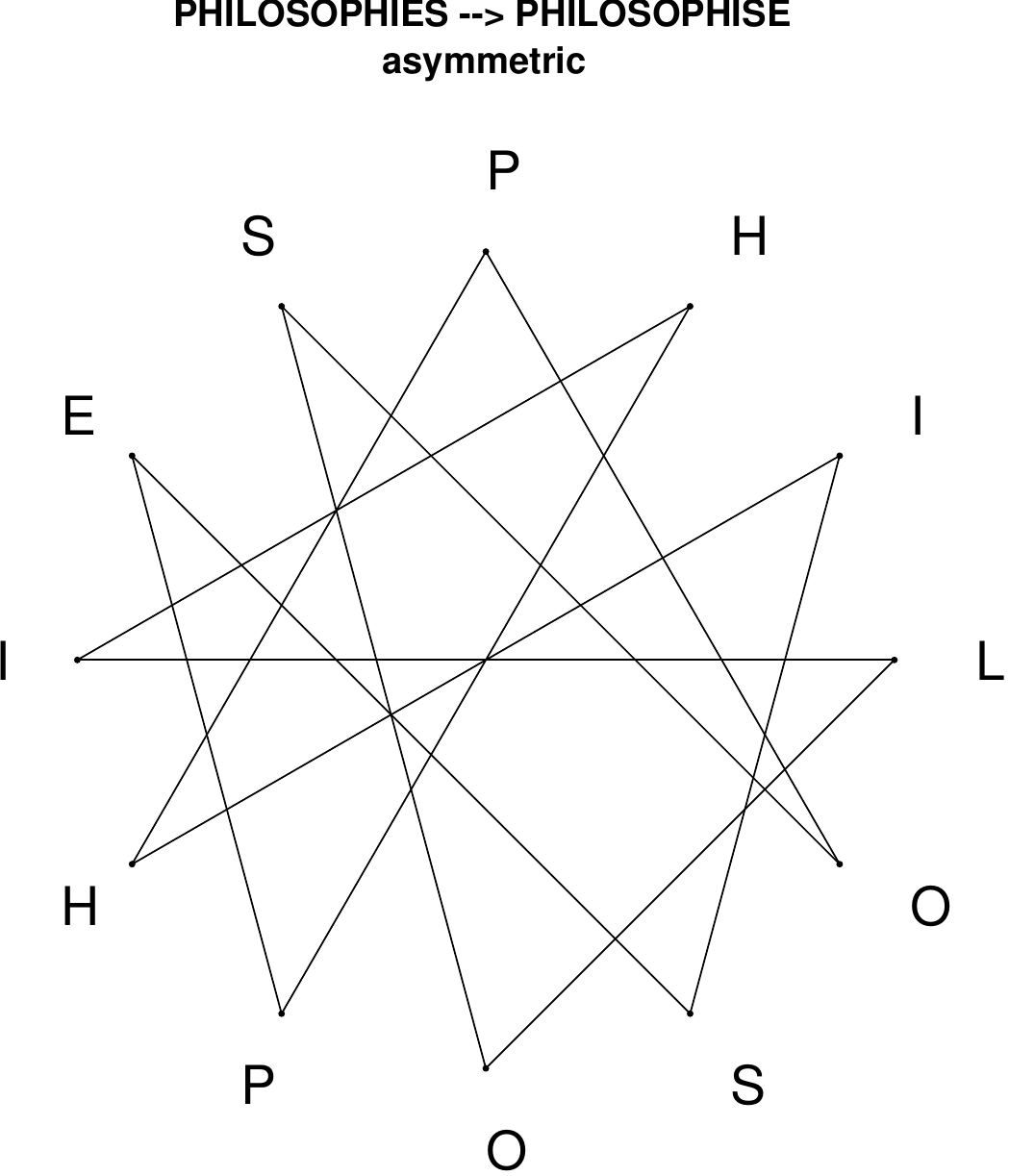}
\end{subfigure}
\end{figure}

\begin{figure}[H]
\centering
\begin{subfigure}[T]{0.19\textwidth}
\centering
\includegraphics[width=\textwidth]{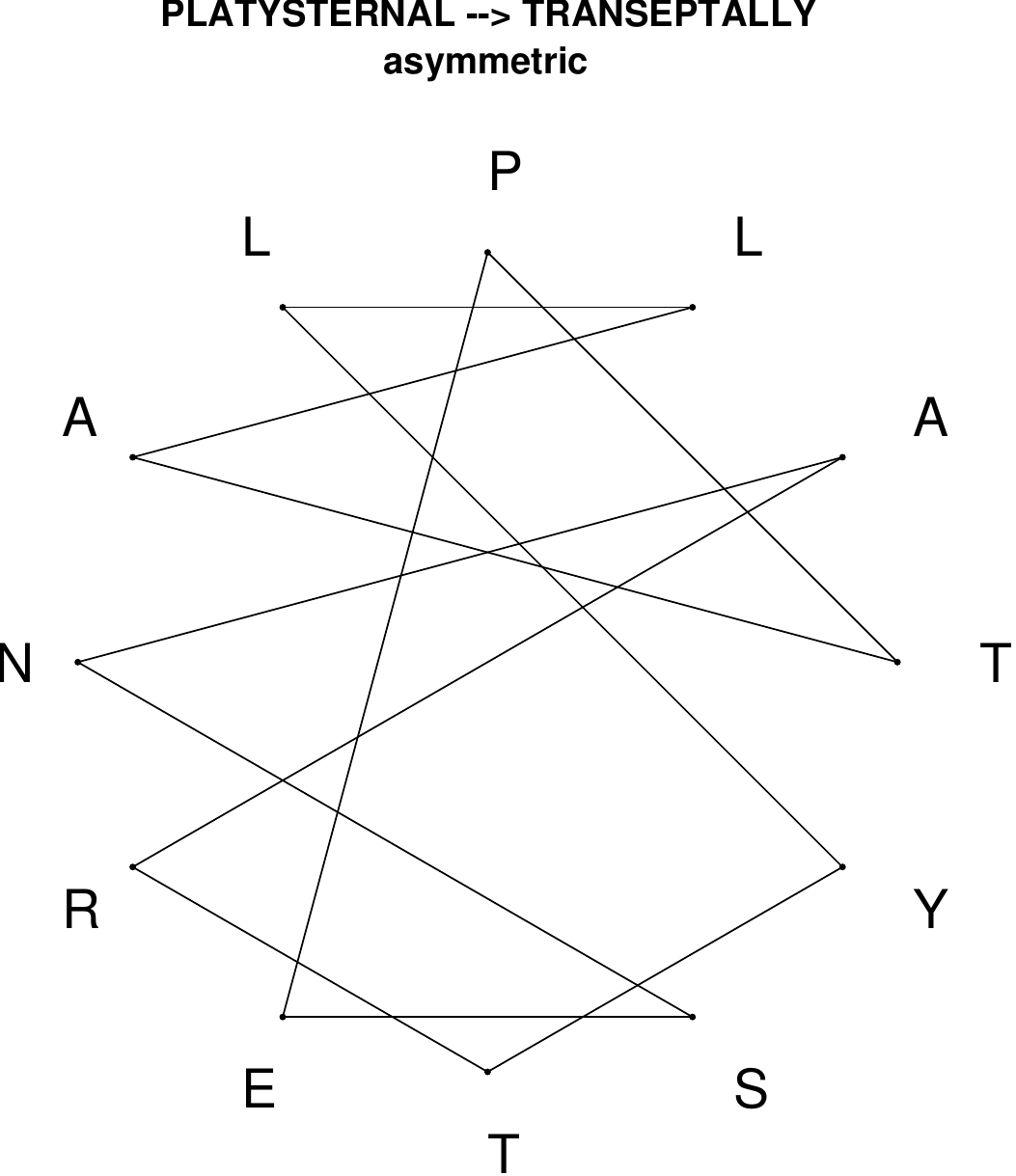}
\end{subfigure}
\hfill
\begin{subfigure}[T]{0.19\textwidth}
\centering
\includegraphics[width=\textwidth]{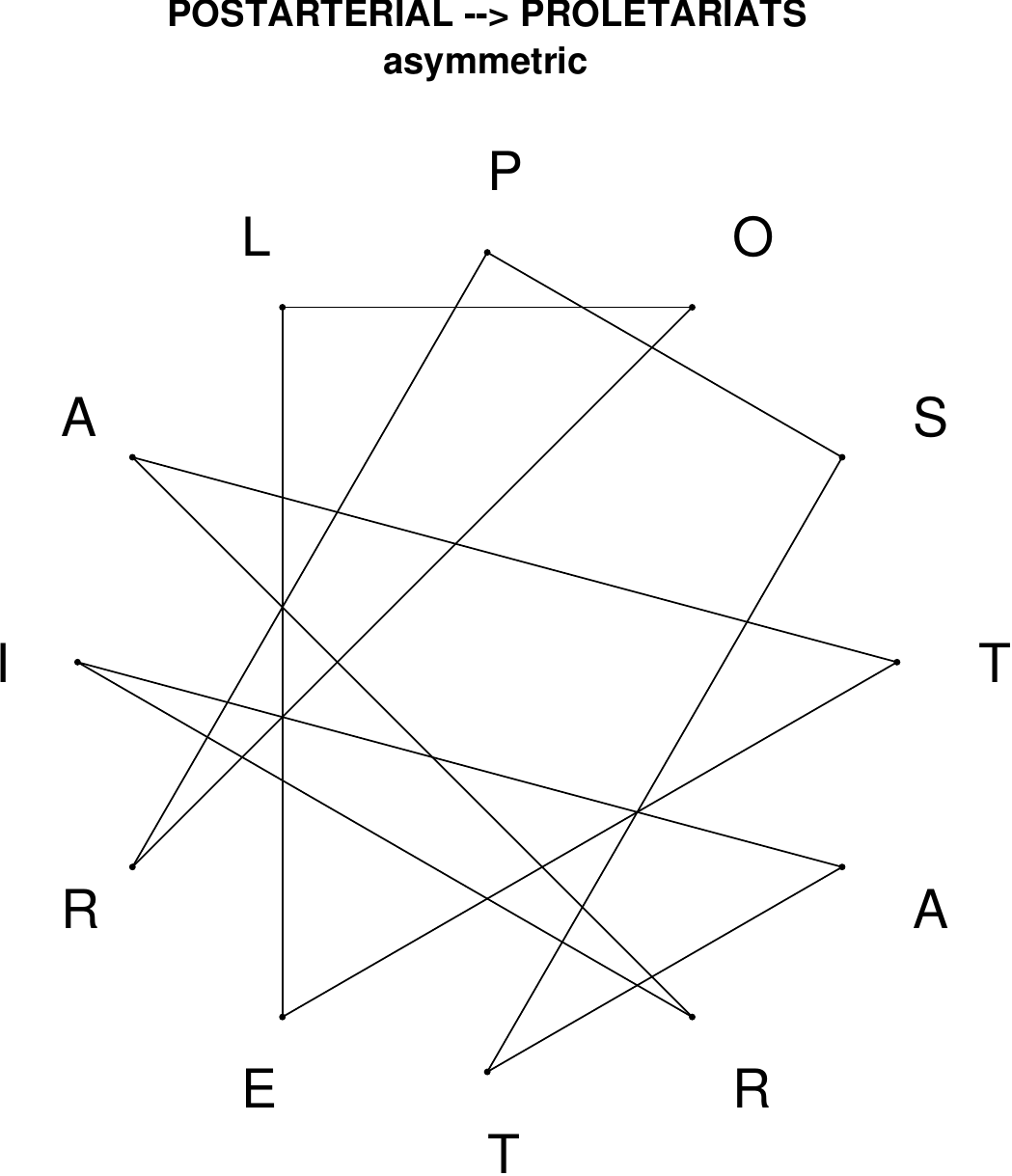}
\end{subfigure}
\hfill
\begin{subfigure}[T]{0.19\textwidth}
\centering
\includegraphics[width=\textwidth]{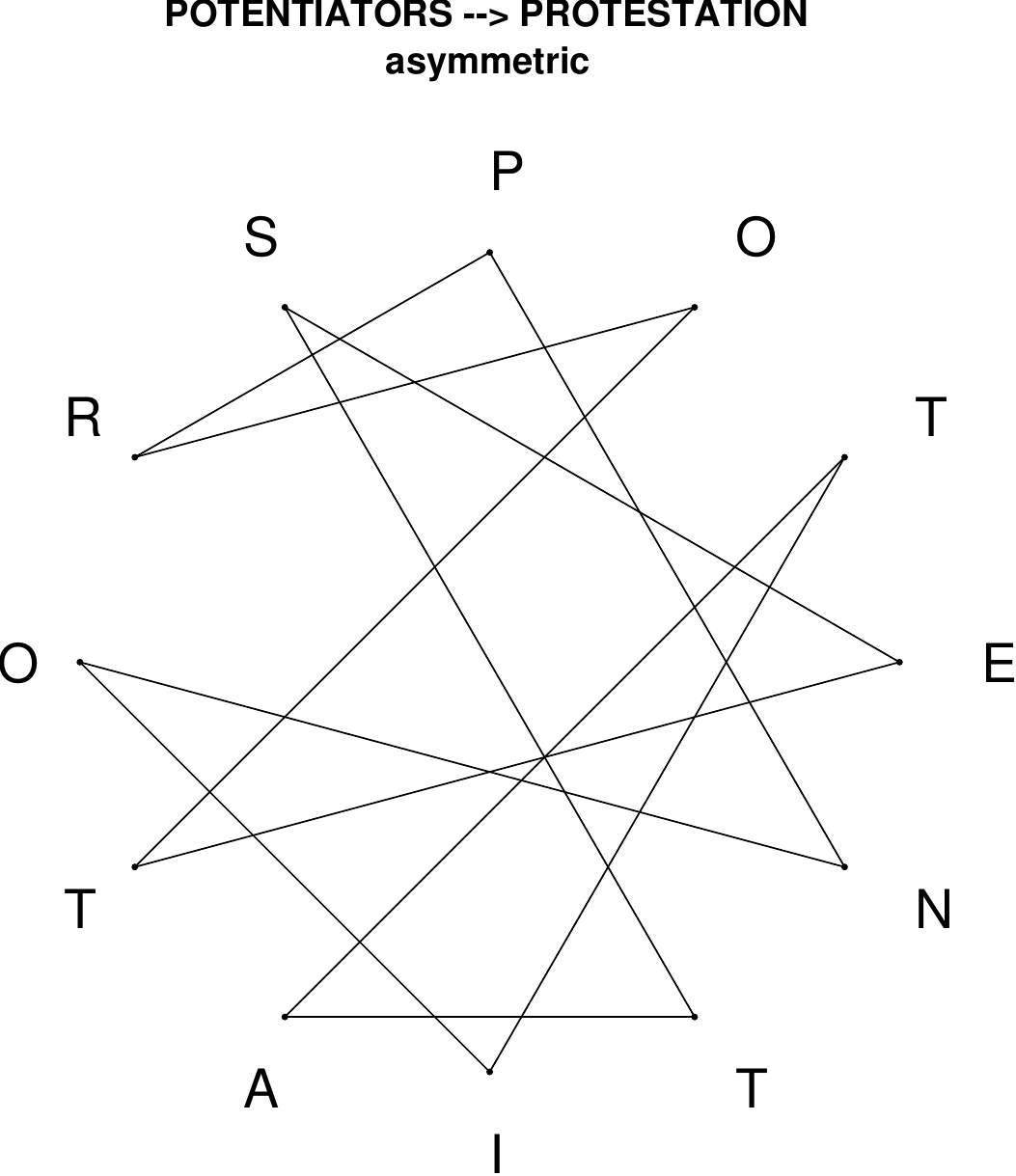}
\end{subfigure}
\hfill
\begin{subfigure}[T]{0.19\textwidth}
\centering
\includegraphics[width=\textwidth]{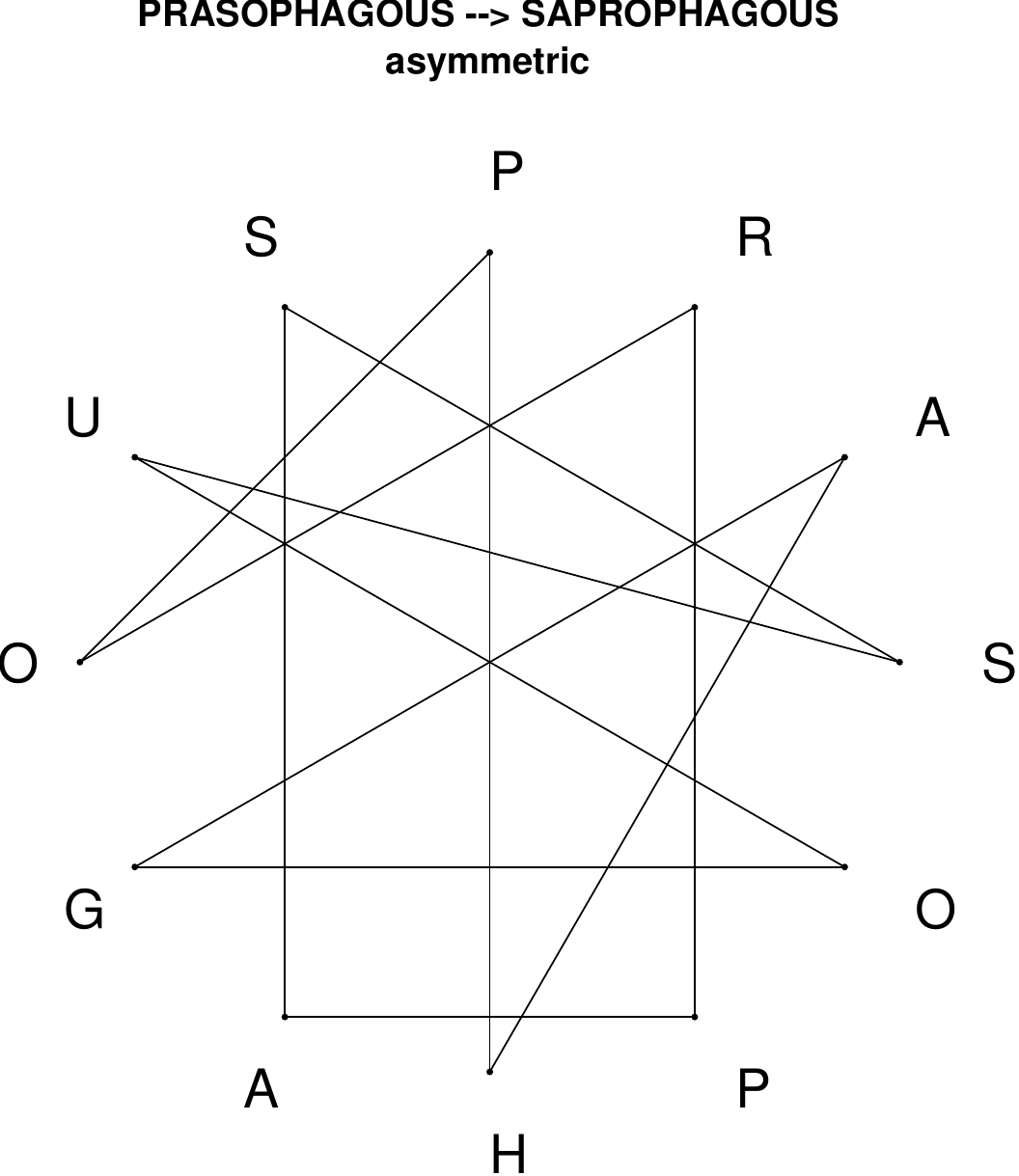}
\end{subfigure}
\hfill
\begin{subfigure}[T]{0.19\textwidth}
\centering
\includegraphics[width=\textwidth]{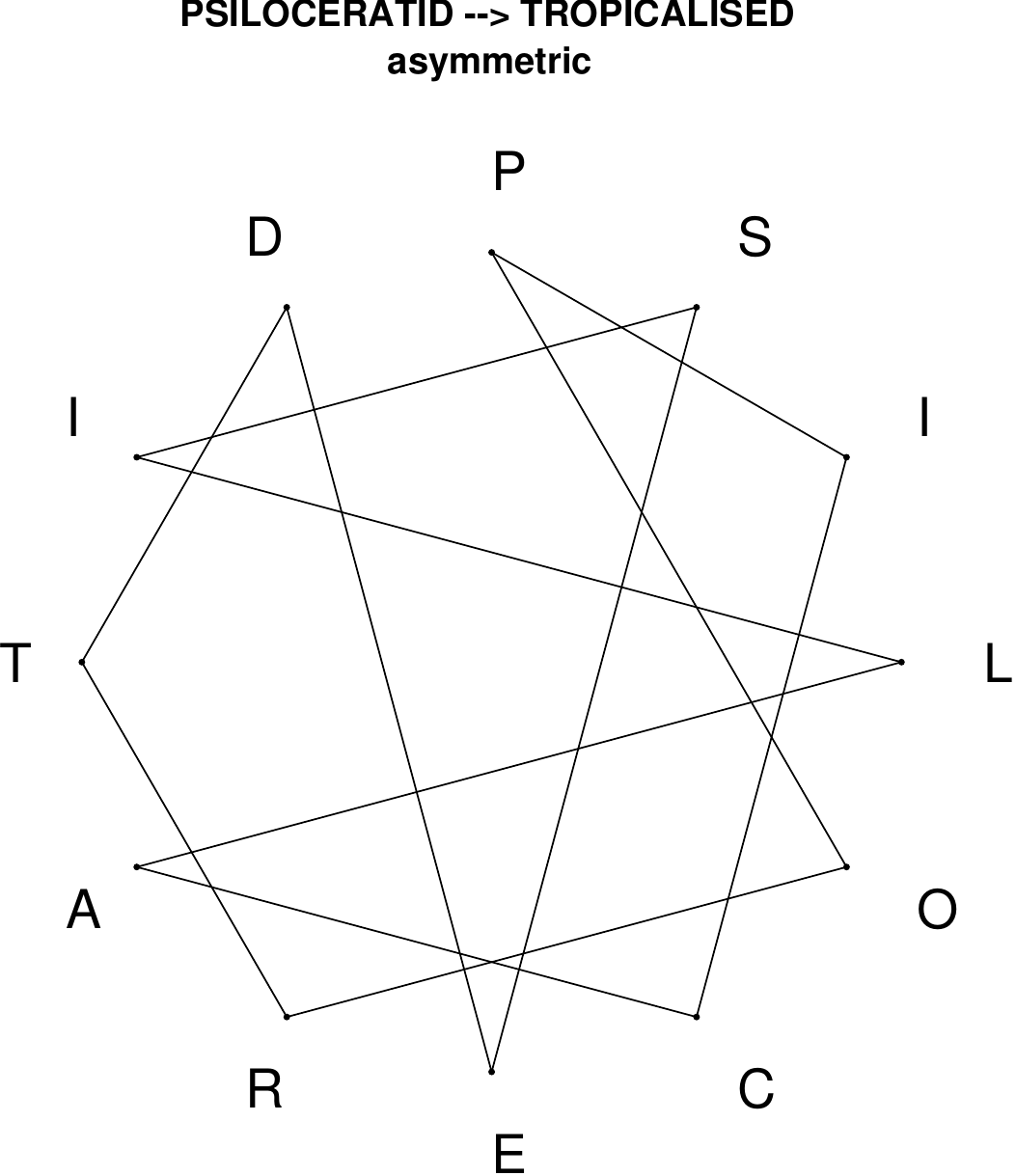}
\end{subfigure}
\end{figure}

\begin{figure}[H]
\centering
\begin{subfigure}[T]{0.19\textwidth}
\centering
\includegraphics[width=\textwidth]{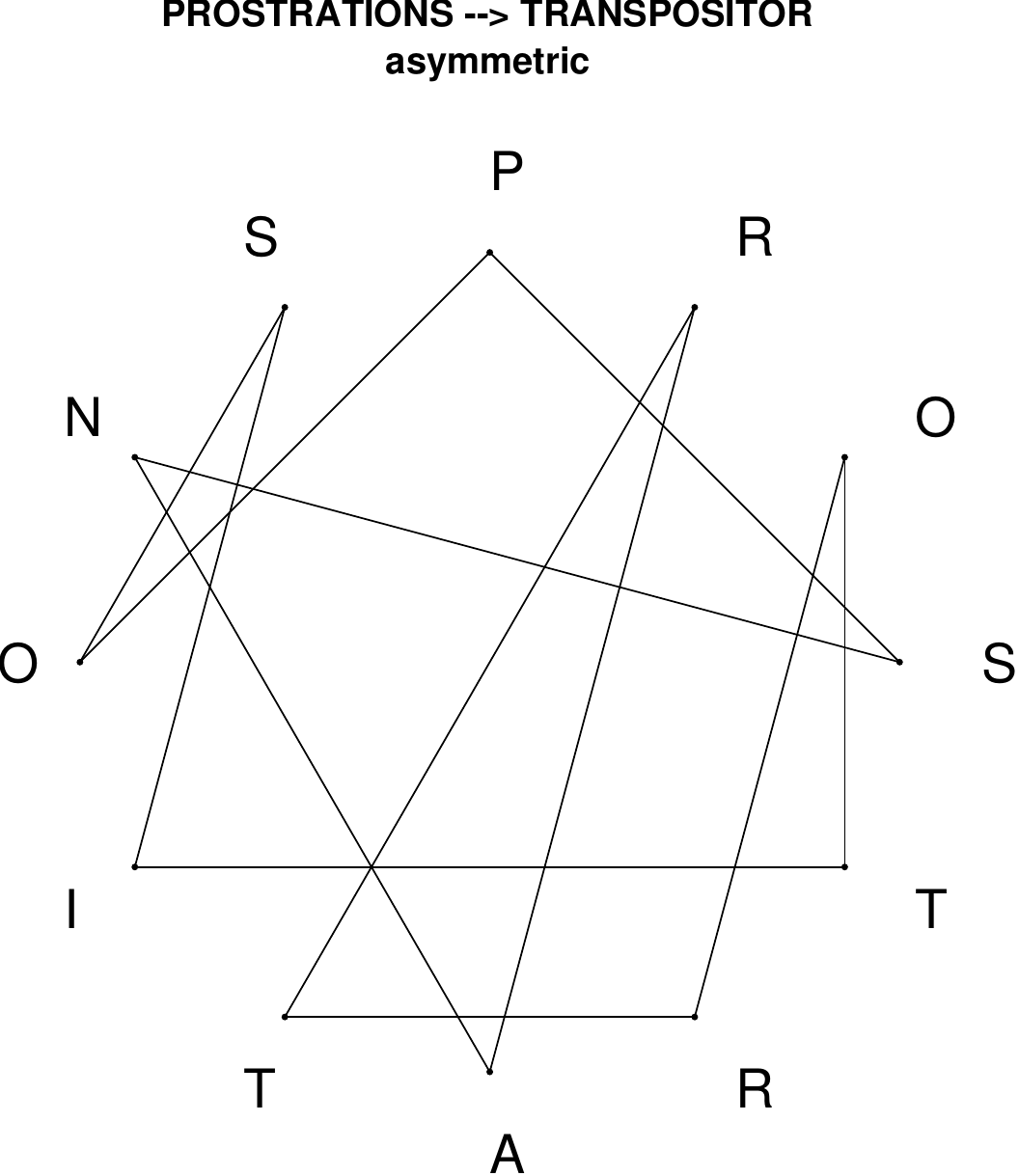}
\end{subfigure}
\hfill
\begin{subfigure}[T]{0.19\textwidth}
\centering
\includegraphics[width=\textwidth]{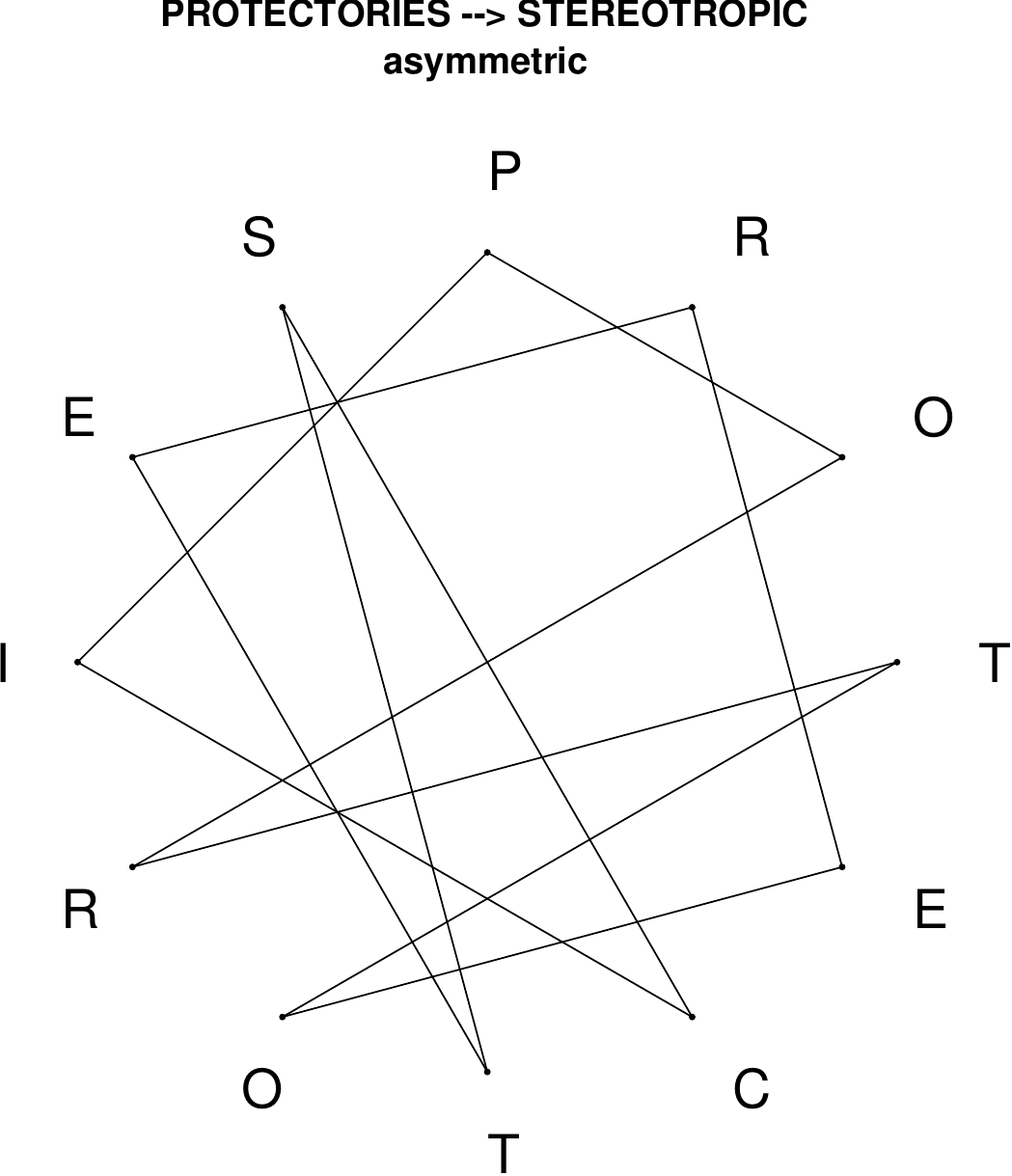}
\end{subfigure}
\hfill
\begin{subfigure}[T]{0.19\textwidth}
\centering
\includegraphics[width=\textwidth]{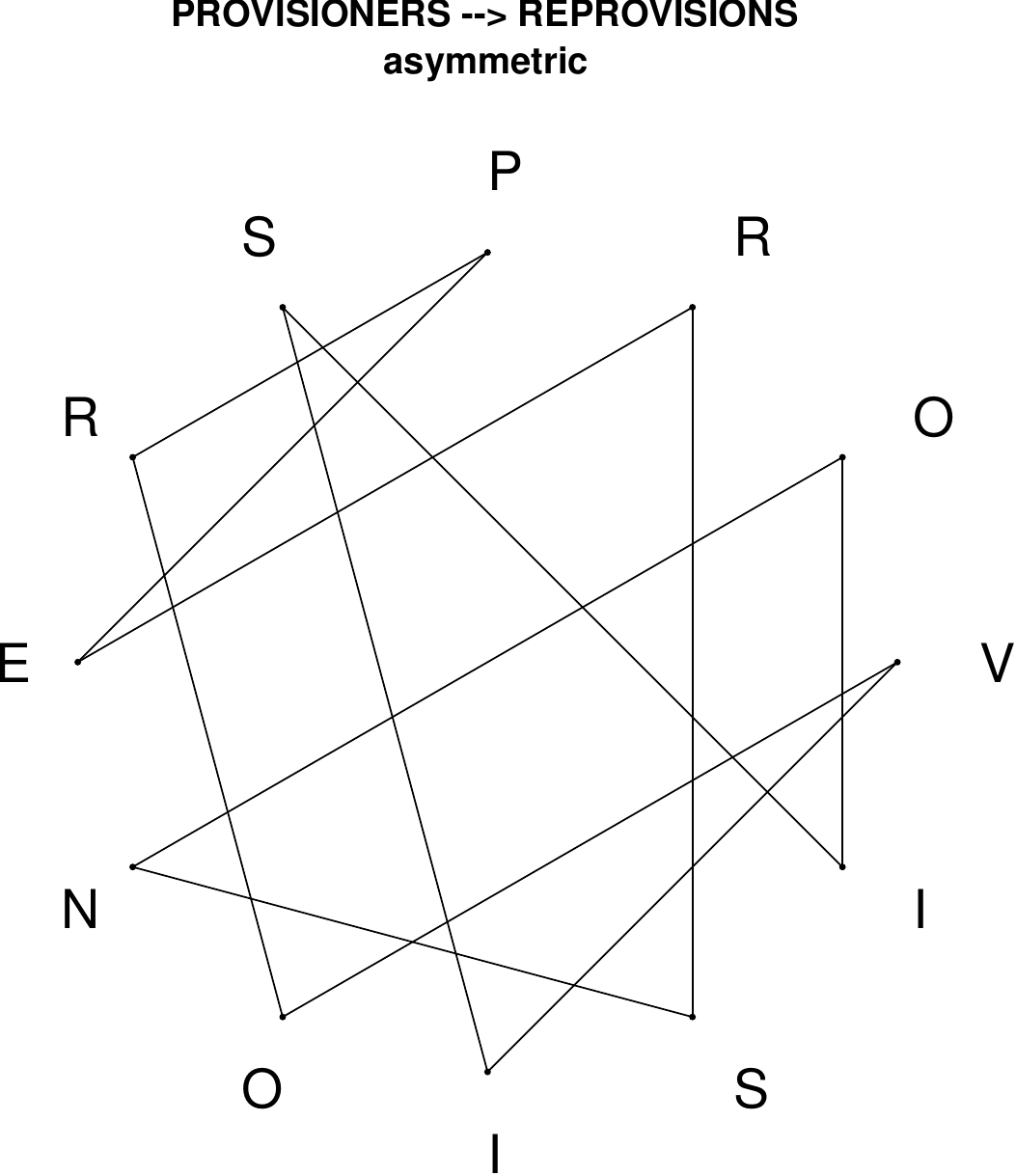}
\end{subfigure}
\hfill
\begin{subfigure}[T]{0.19\textwidth}
\centering
\includegraphics[width=\textwidth]{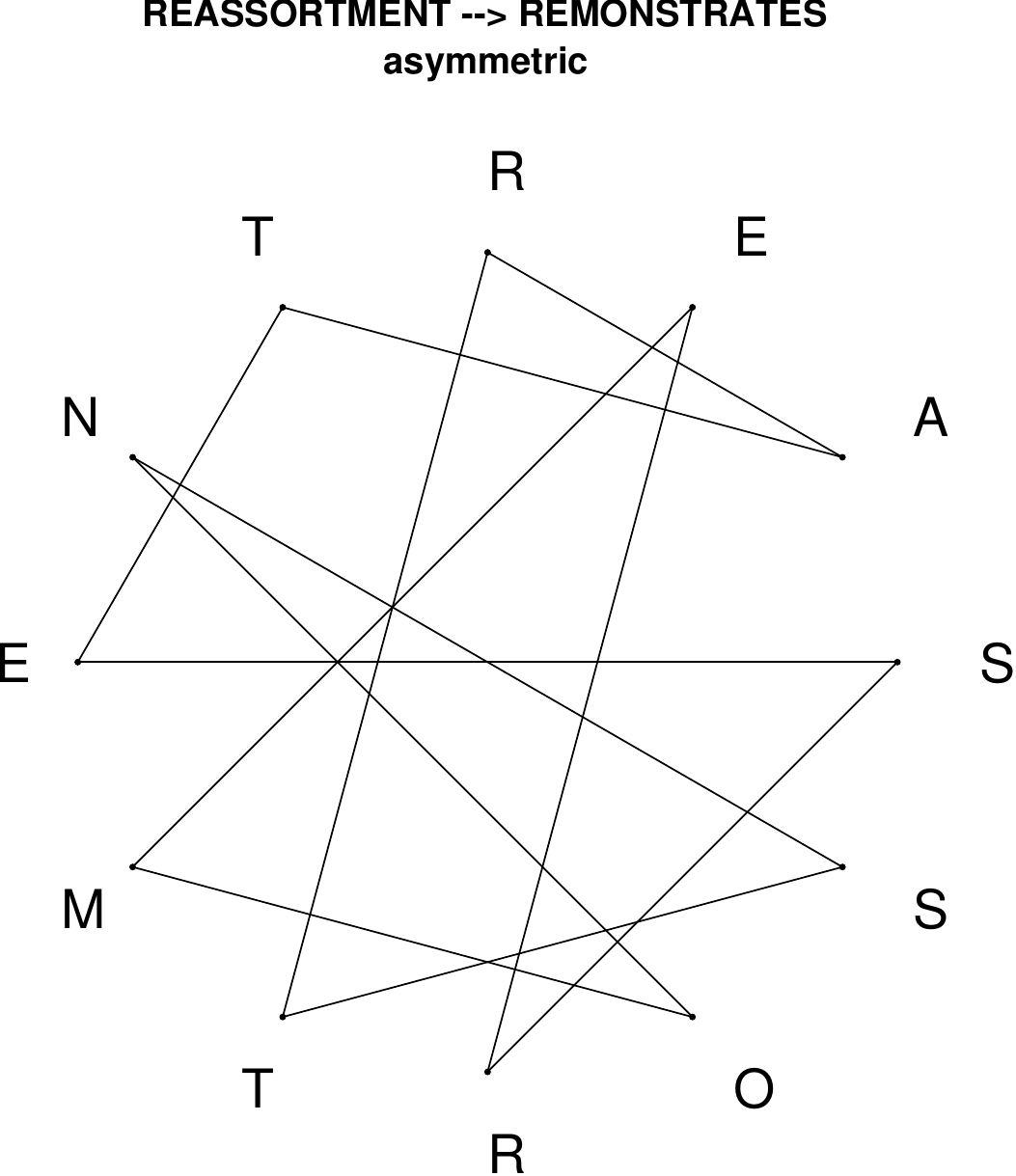}
\end{subfigure}
\hfill
\begin{subfigure}[T]{0.19\textwidth}
\centering
\includegraphics[width=\textwidth]{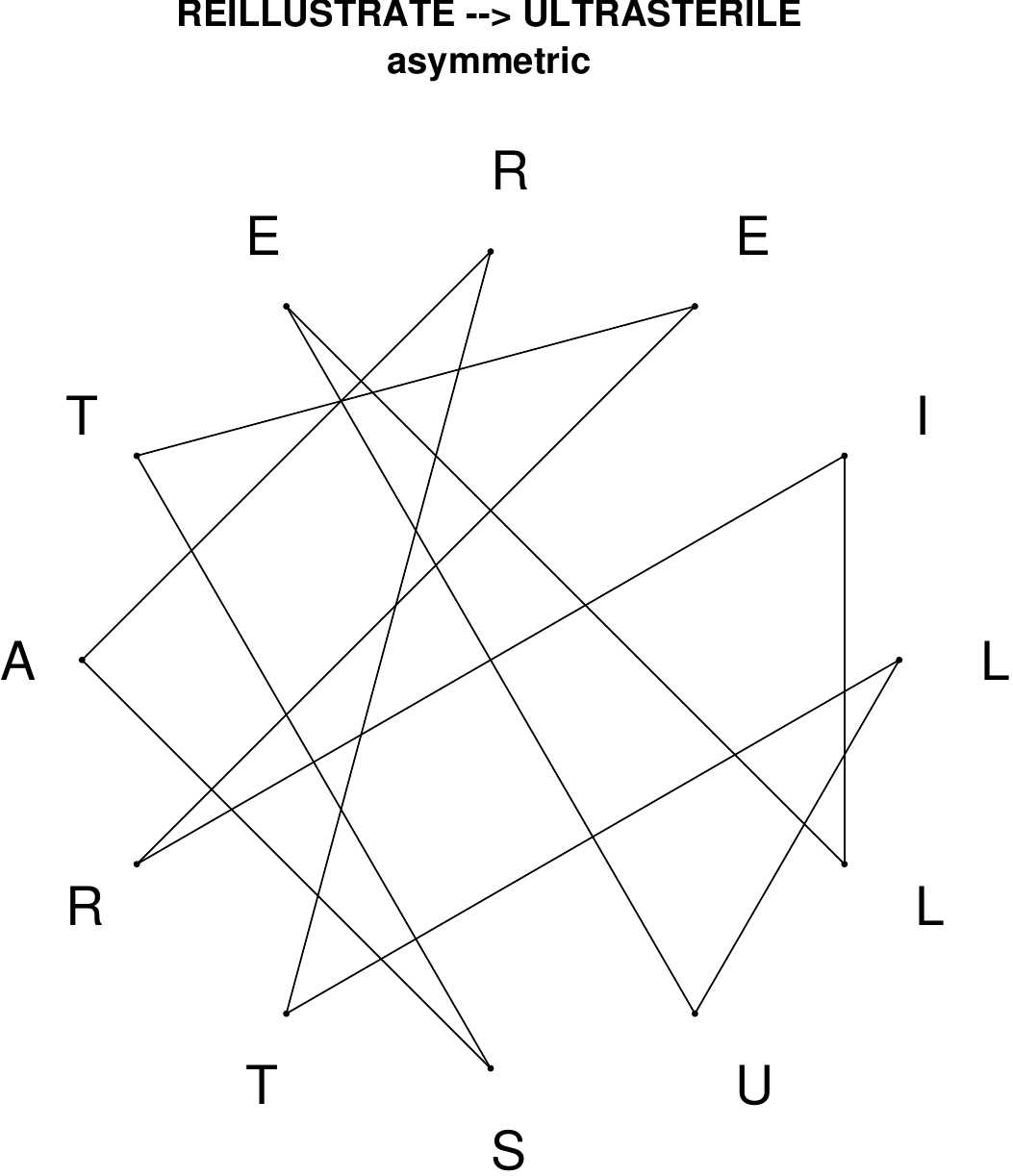}
\end{subfigure}
\end{figure}

\begin{figure}[H]
\centering
\begin{subfigure}[T]{0.19\textwidth}
\centering
\includegraphics[width=\textwidth]{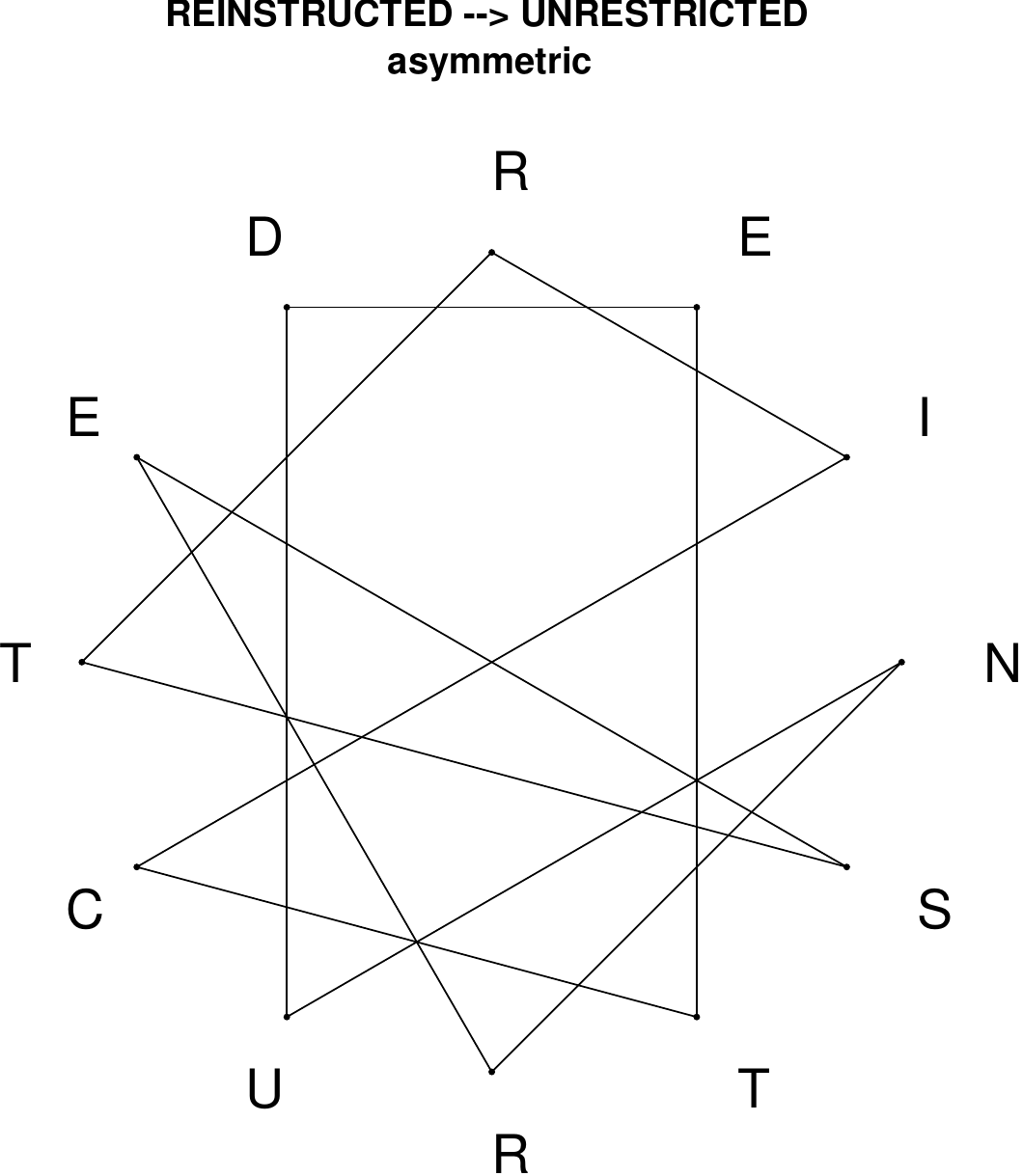}
\end{subfigure}
\hfill
\begin{subfigure}[T]{0.19\textwidth}
\centering
\includegraphics[width=\textwidth]{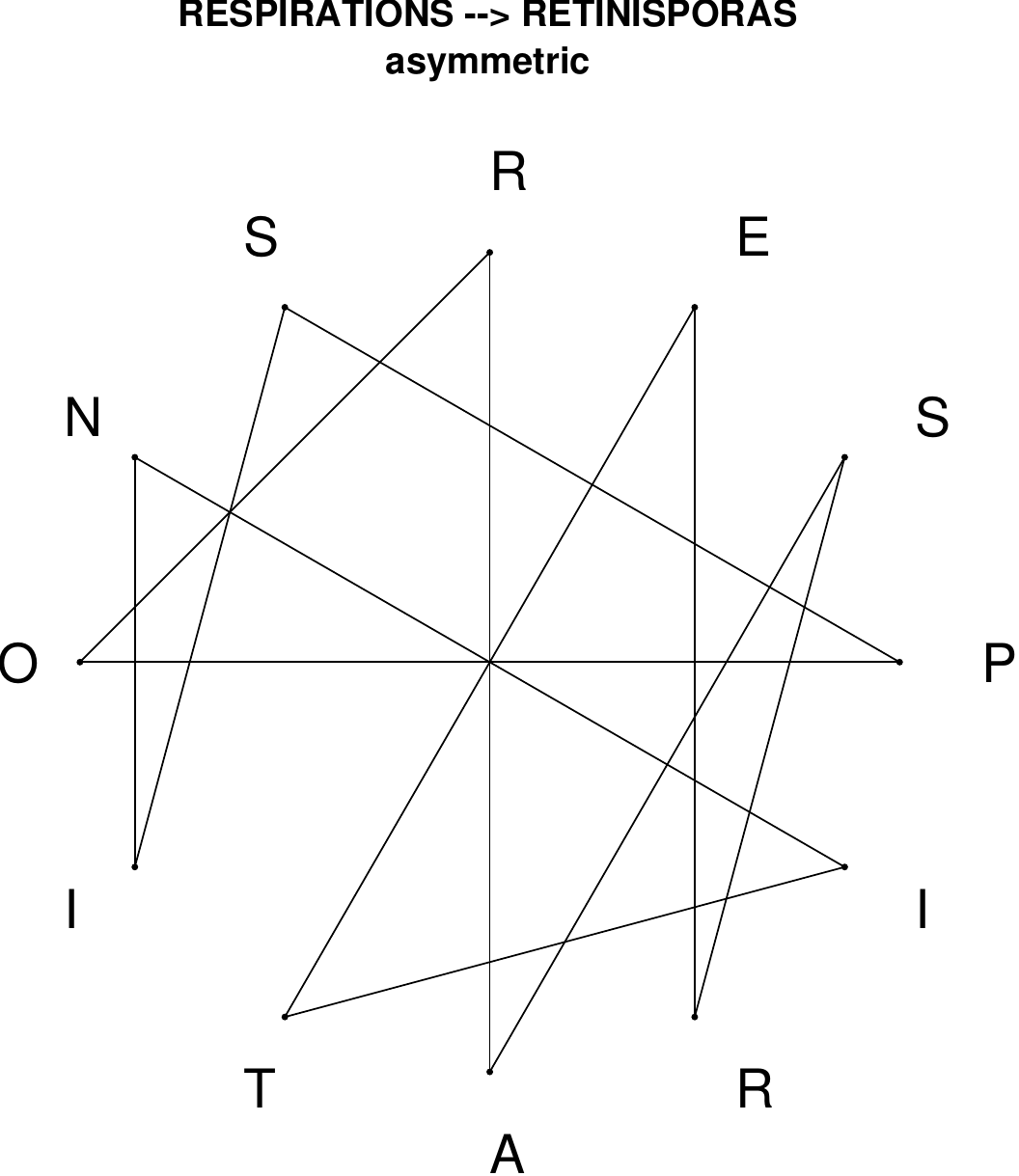}
\end{subfigure}
\hfill
\begin{subfigure}[T]{0.19\textwidth}
\centering
\includegraphics[width=\textwidth]{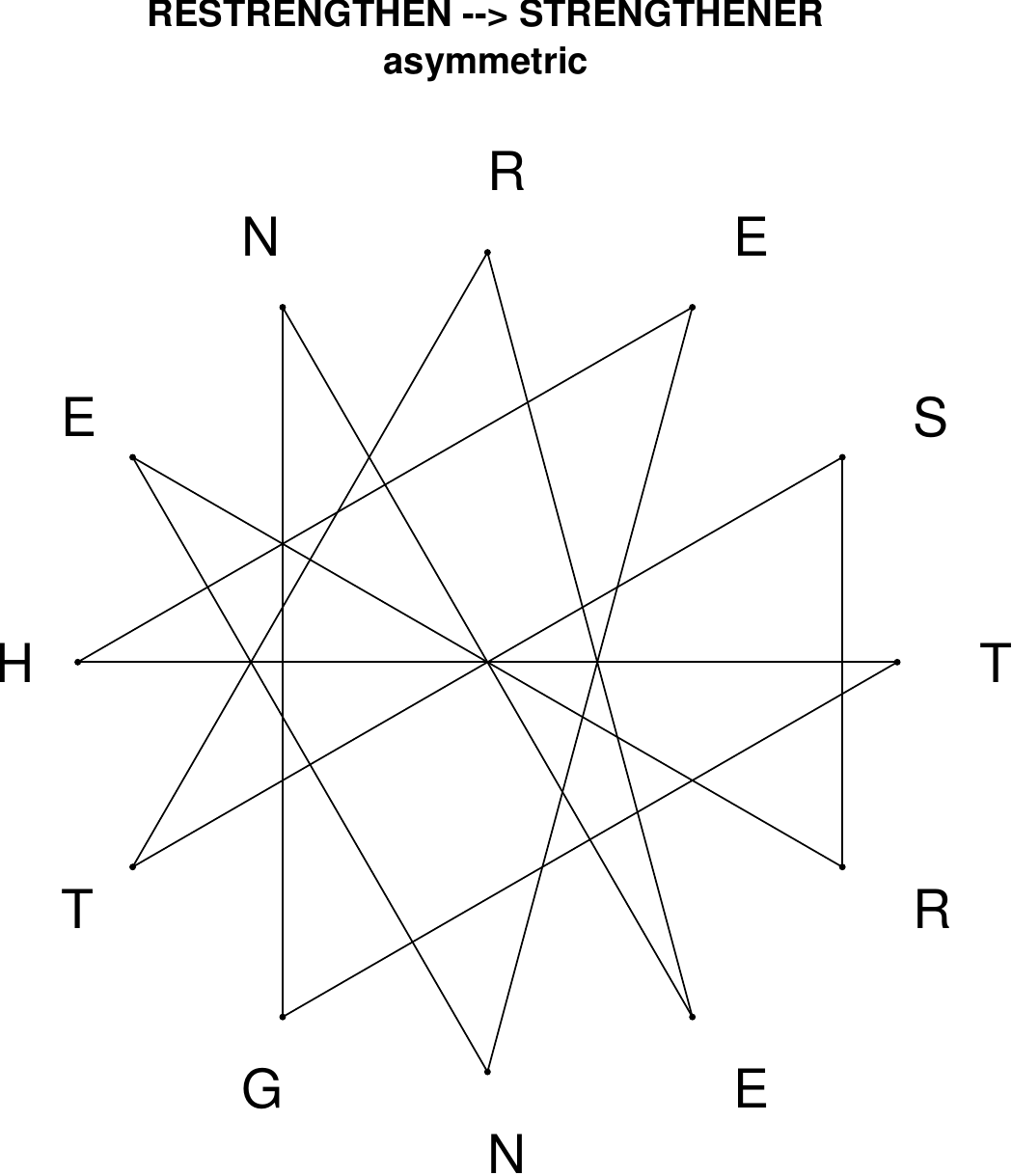}
\end{subfigure}
\hfill
\begin{subfigure}[T]{0.19\textwidth}
\centering
\includegraphics[width=\textwidth]{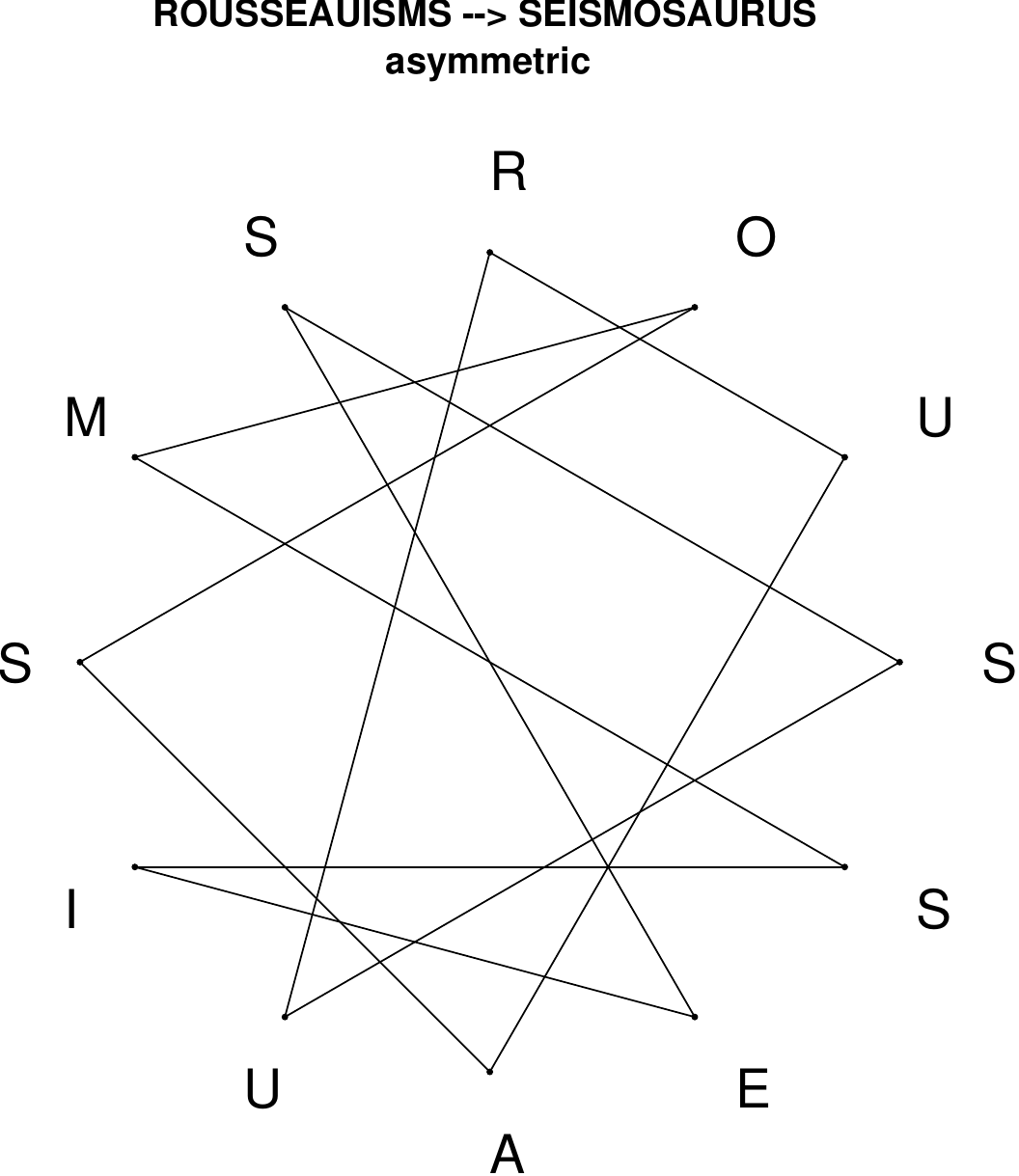}
\end{subfigure}
\hfill
\begin{subfigure}[T]{0.19\textwidth}
\centering
\includegraphics[width=\textwidth]{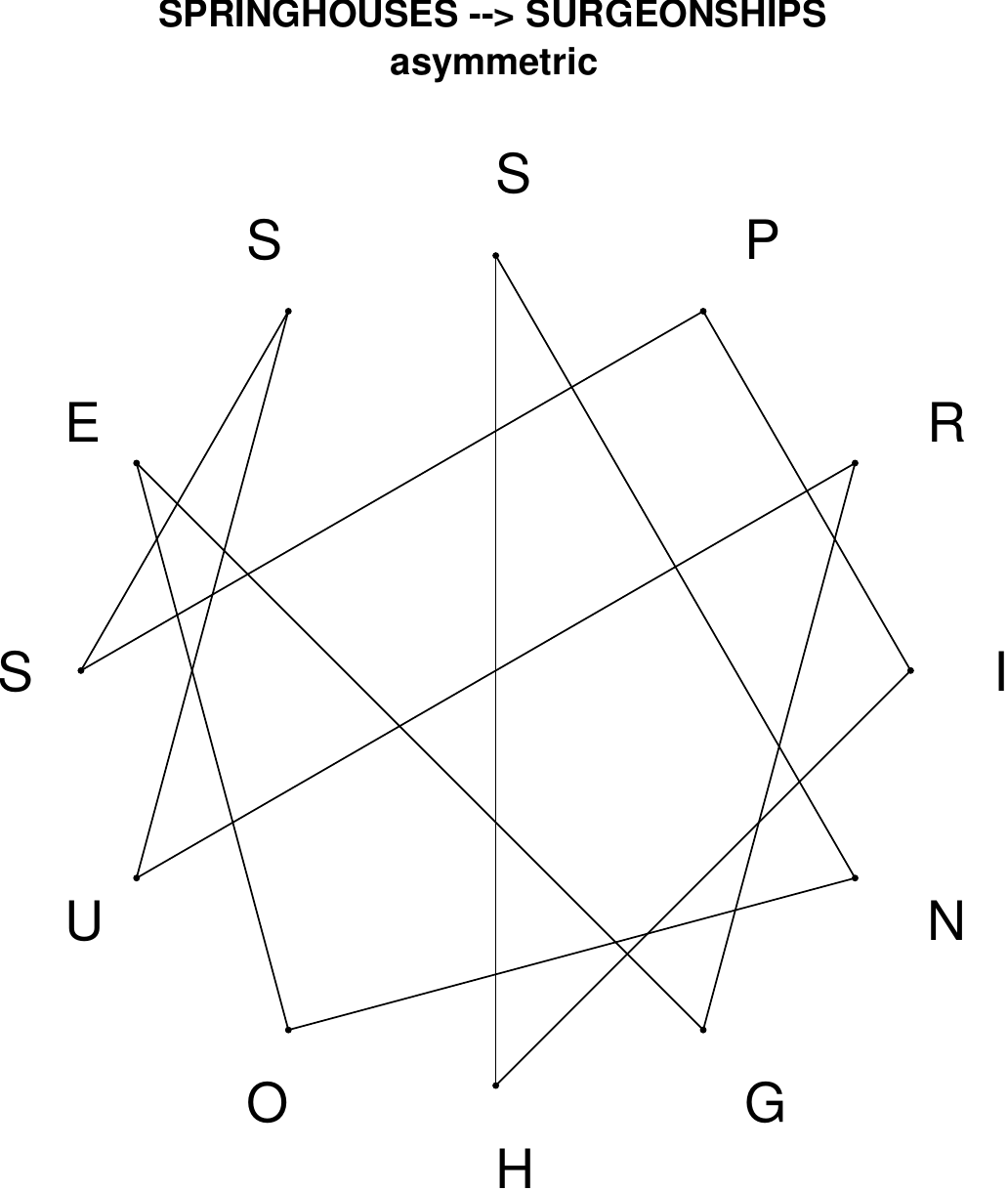}
\end{subfigure}
\end{figure}

\begin{figure}[H]
\centering
\begin{subfigure}[T]{0.19\textwidth}
\centering
\includegraphics[width=\textwidth]{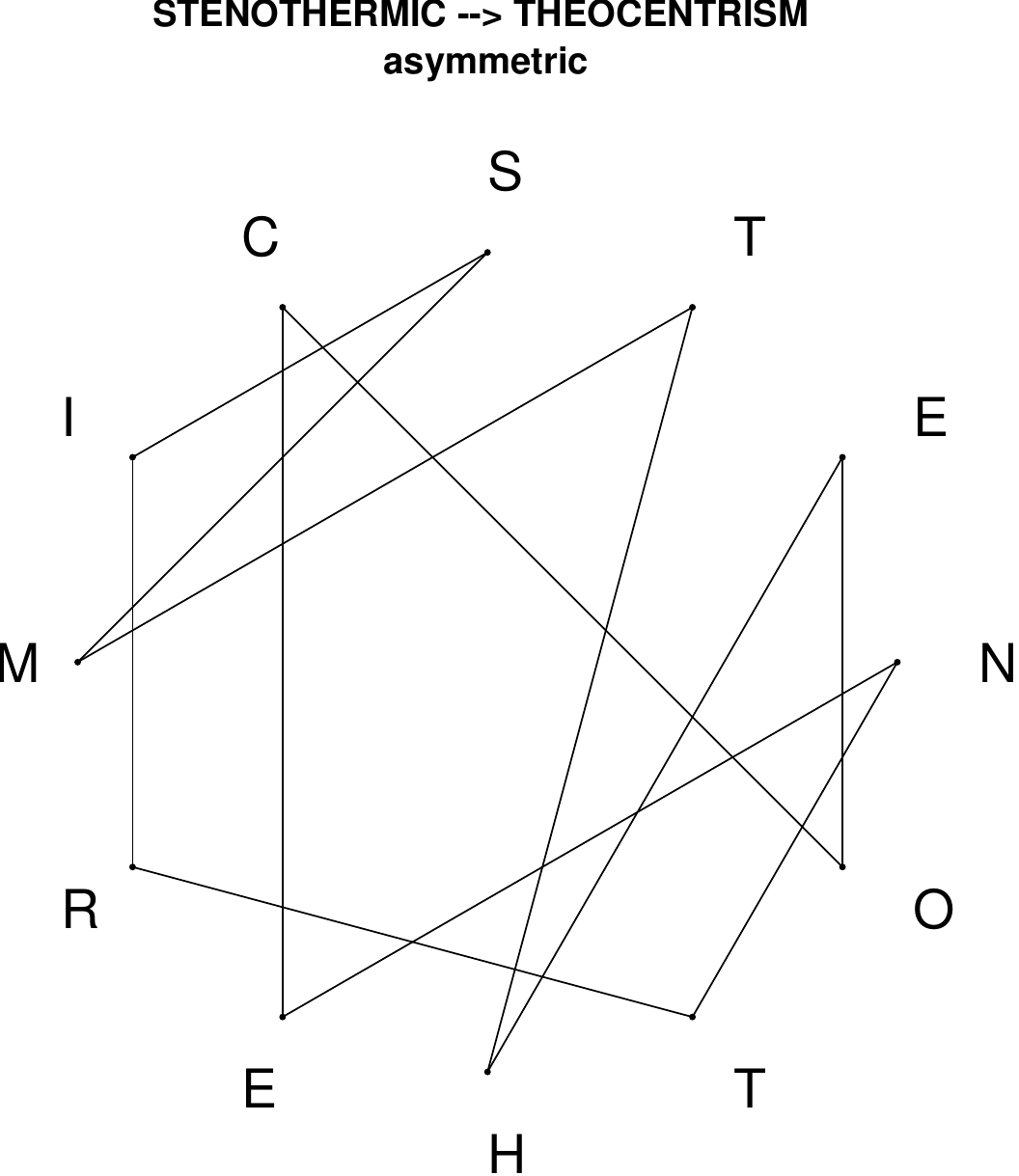}
\end{subfigure}
\hfill
\begin{subfigure}[T]{0.19\textwidth}
\centering
\includegraphics[width=\textwidth]{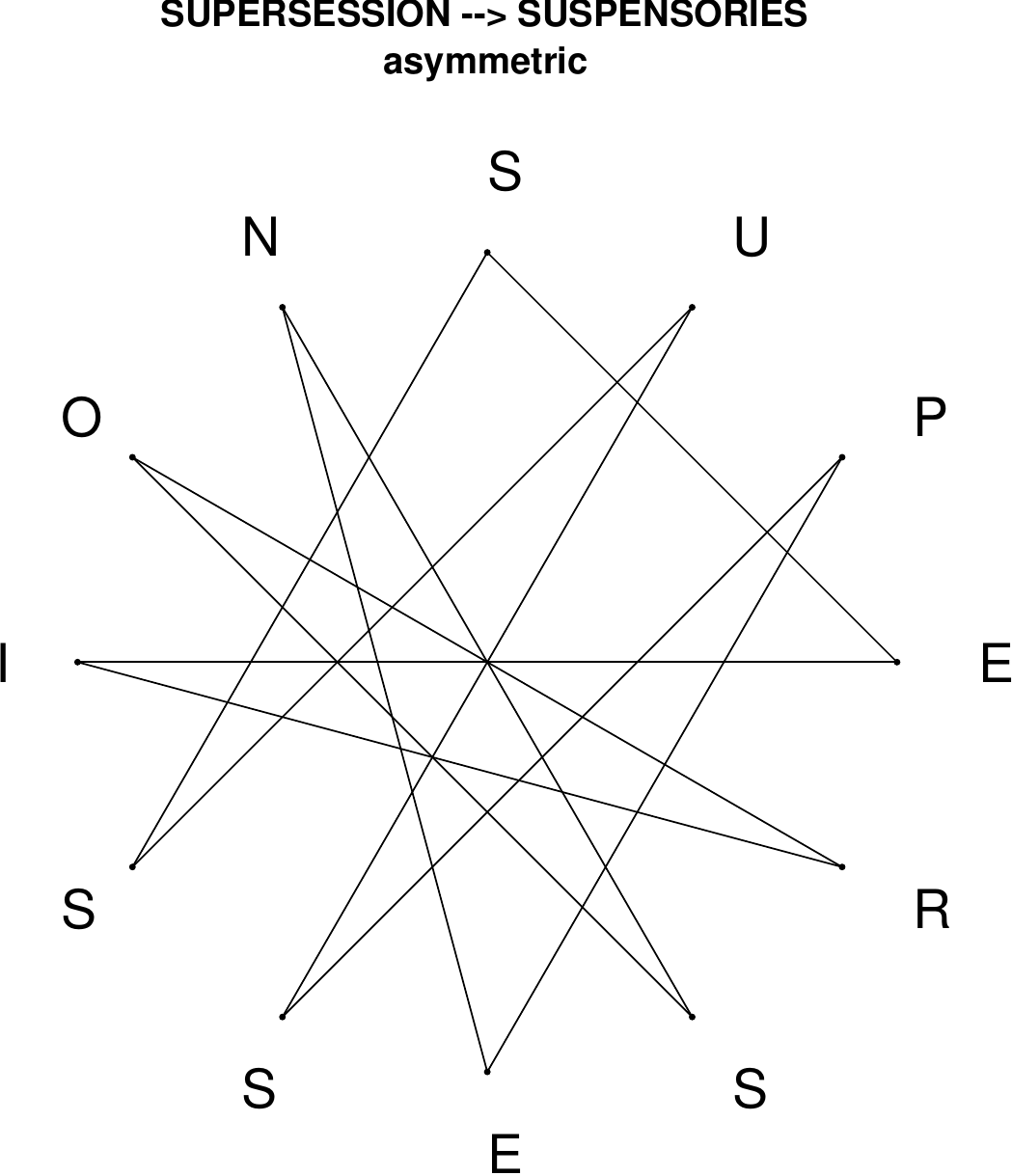}
\end{subfigure}
\hfill
\begin{subfigure}[T]{0.19\textwidth}
\centering
\includegraphics[width=\textwidth]{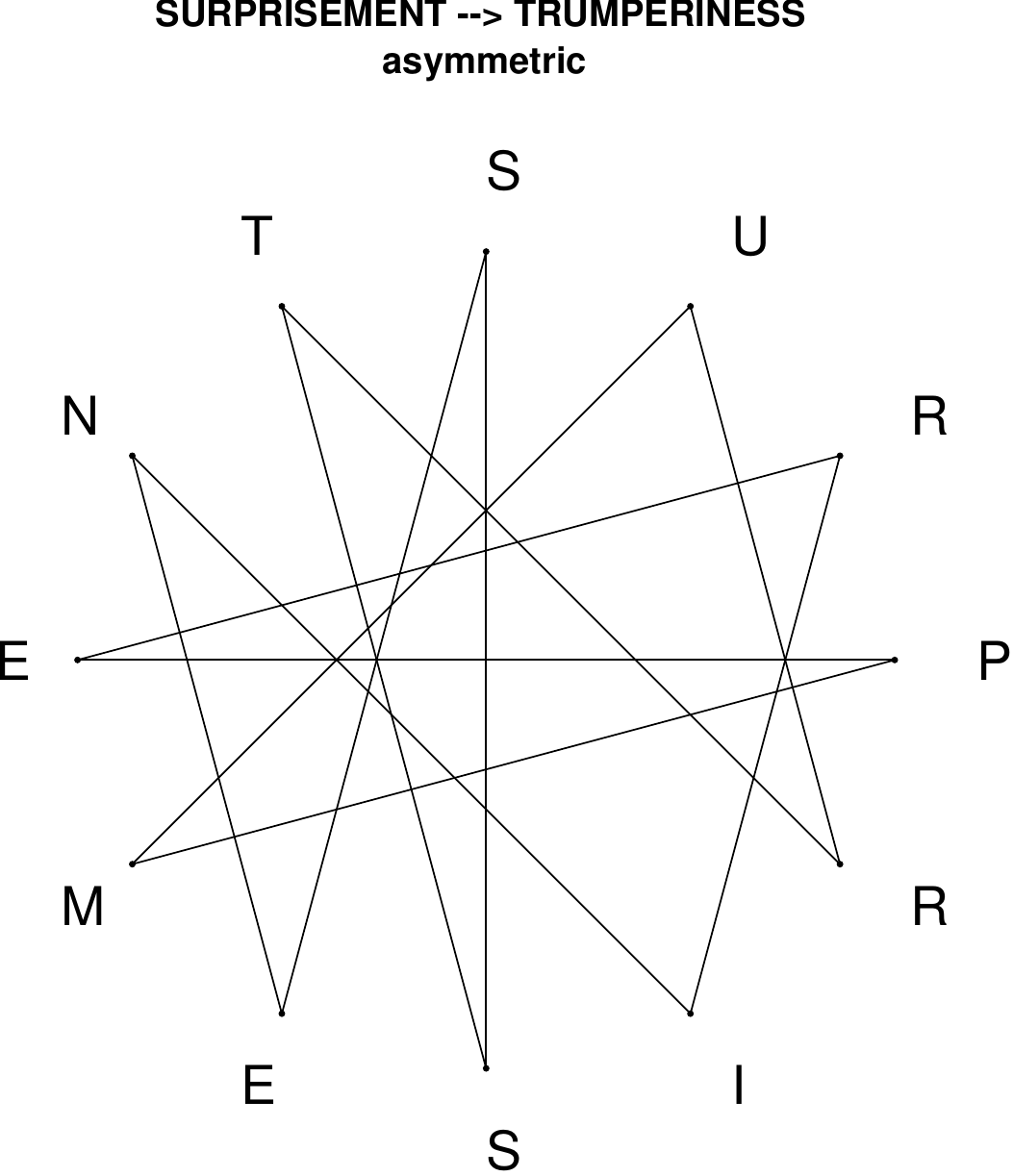}
\end{subfigure}
\hfill
\begin{subfigure}[T]{0.19\textwidth}
\centering
\includegraphics[width=\textwidth]{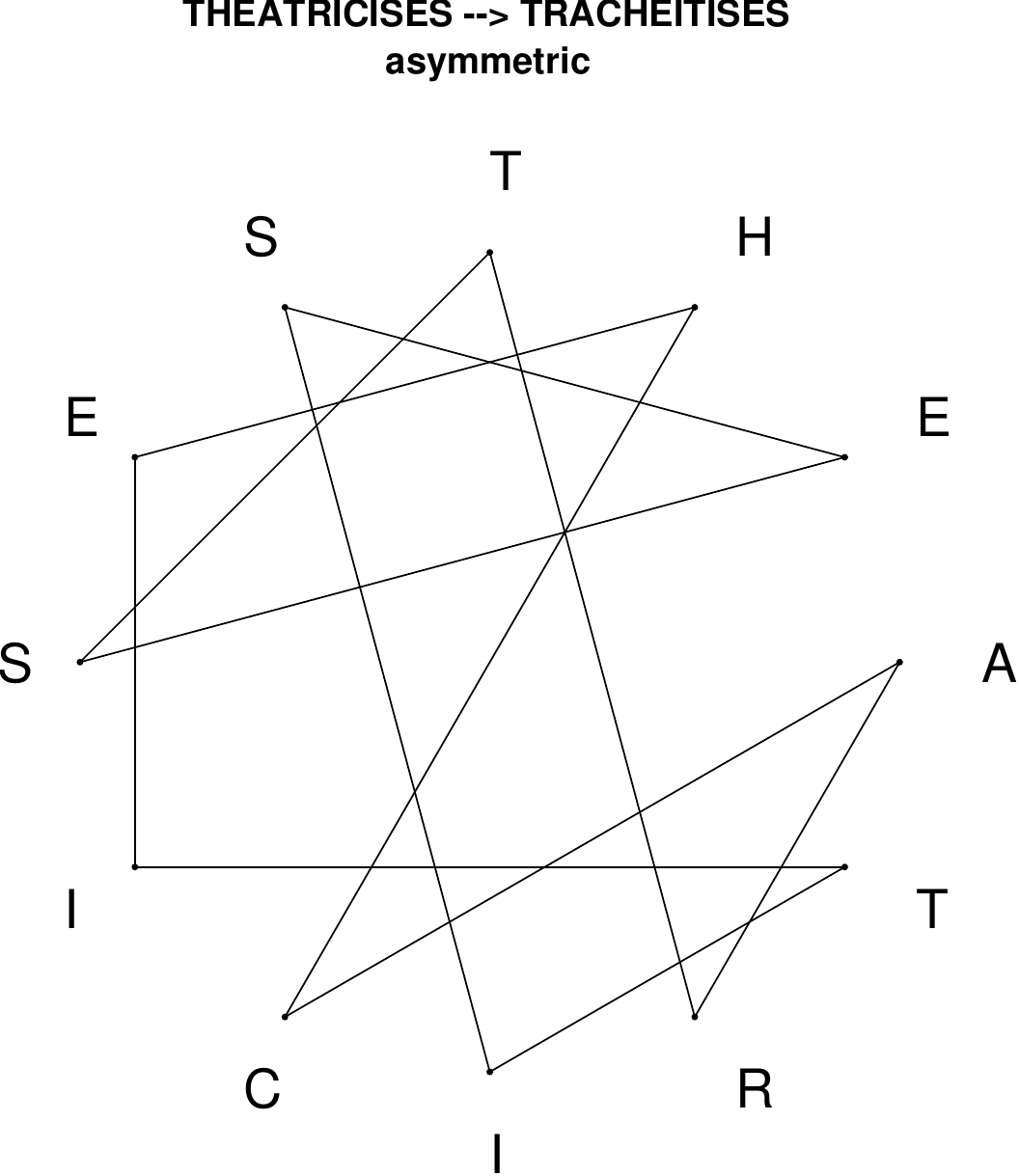}
\end{subfigure}
\hfill
\end{figure}

%%%%%%%%%%%%%%%%%%
\clearpage
\subsection{Star Anagrams $N = 11$}
For $N=11$, we found a handful of symmetric stars among a much larger group of asymmetric stars. 

\subsubsection{Symmetric Stars $N=11$}

\begin{figure}[H]
\centering
\begin{subfigure}[T]{0.19\textwidth}
\centering
\includegraphics[width=\textwidth]{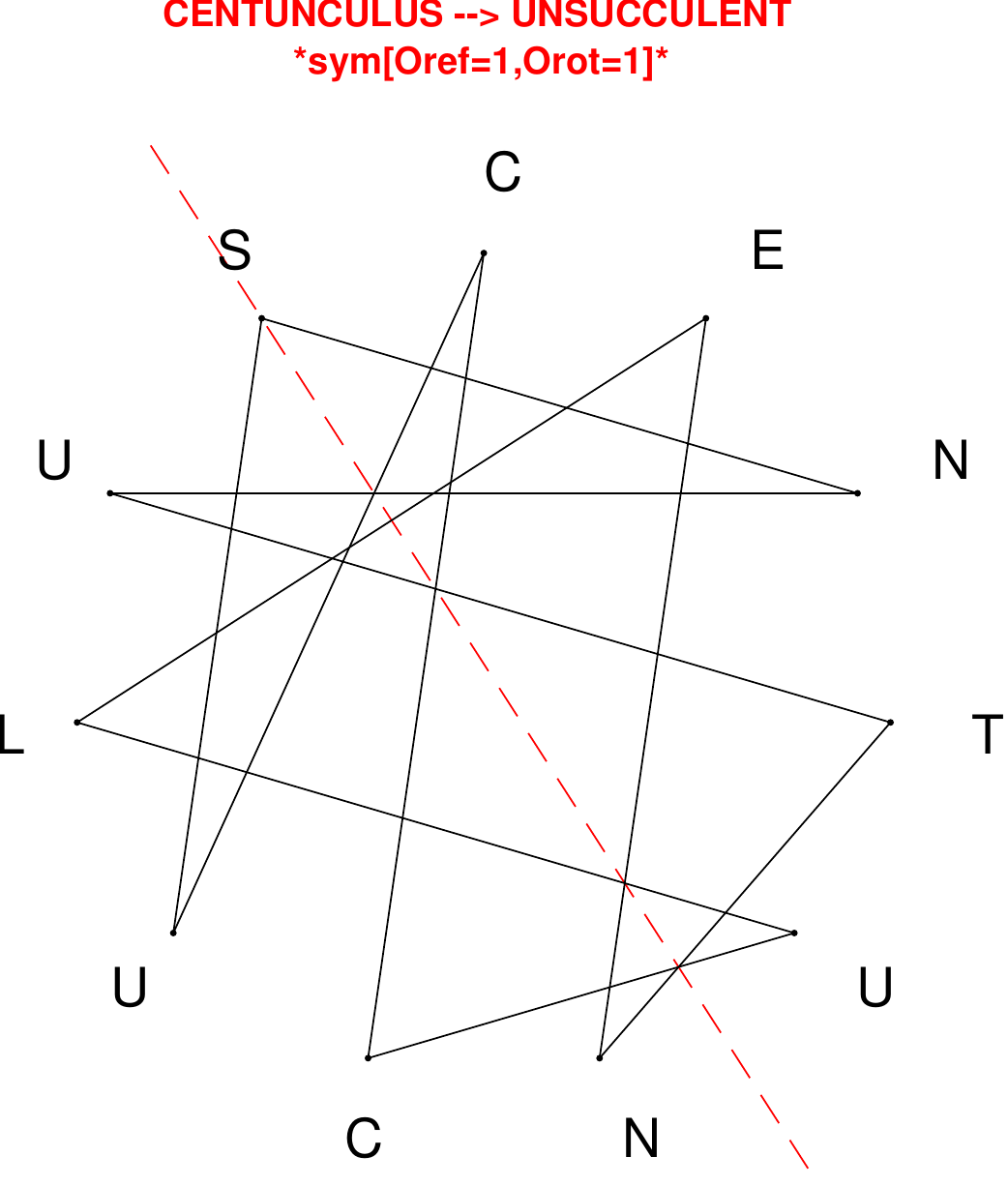}
\end{subfigure}
\hfill
\begin{subfigure}[T]{0.19\textwidth}
\centering
\includegraphics[width=\textwidth]{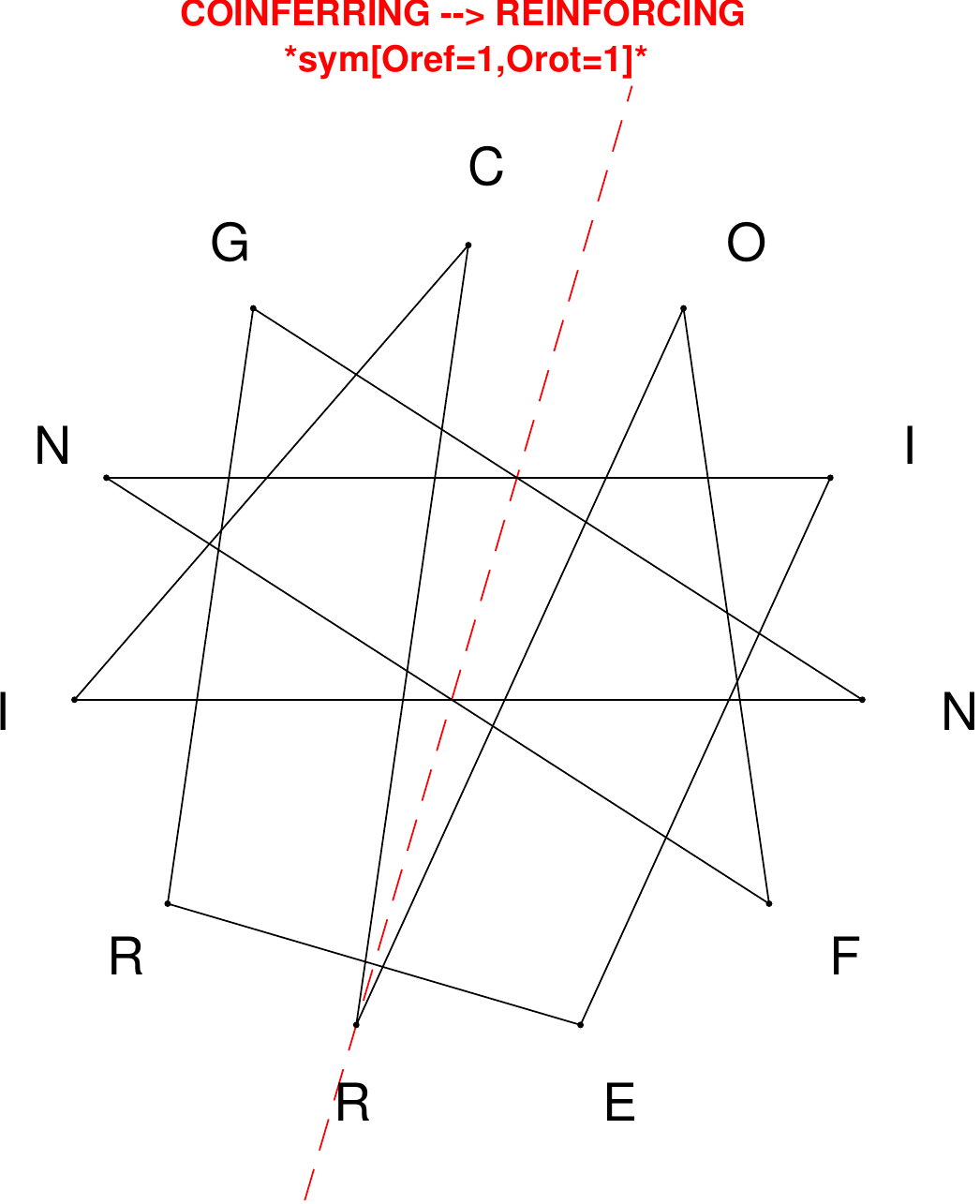}
\end{subfigure}
\hfill
\begin{subfigure}[T]{0.19\textwidth}
\centering
\includegraphics[width=\textwidth]{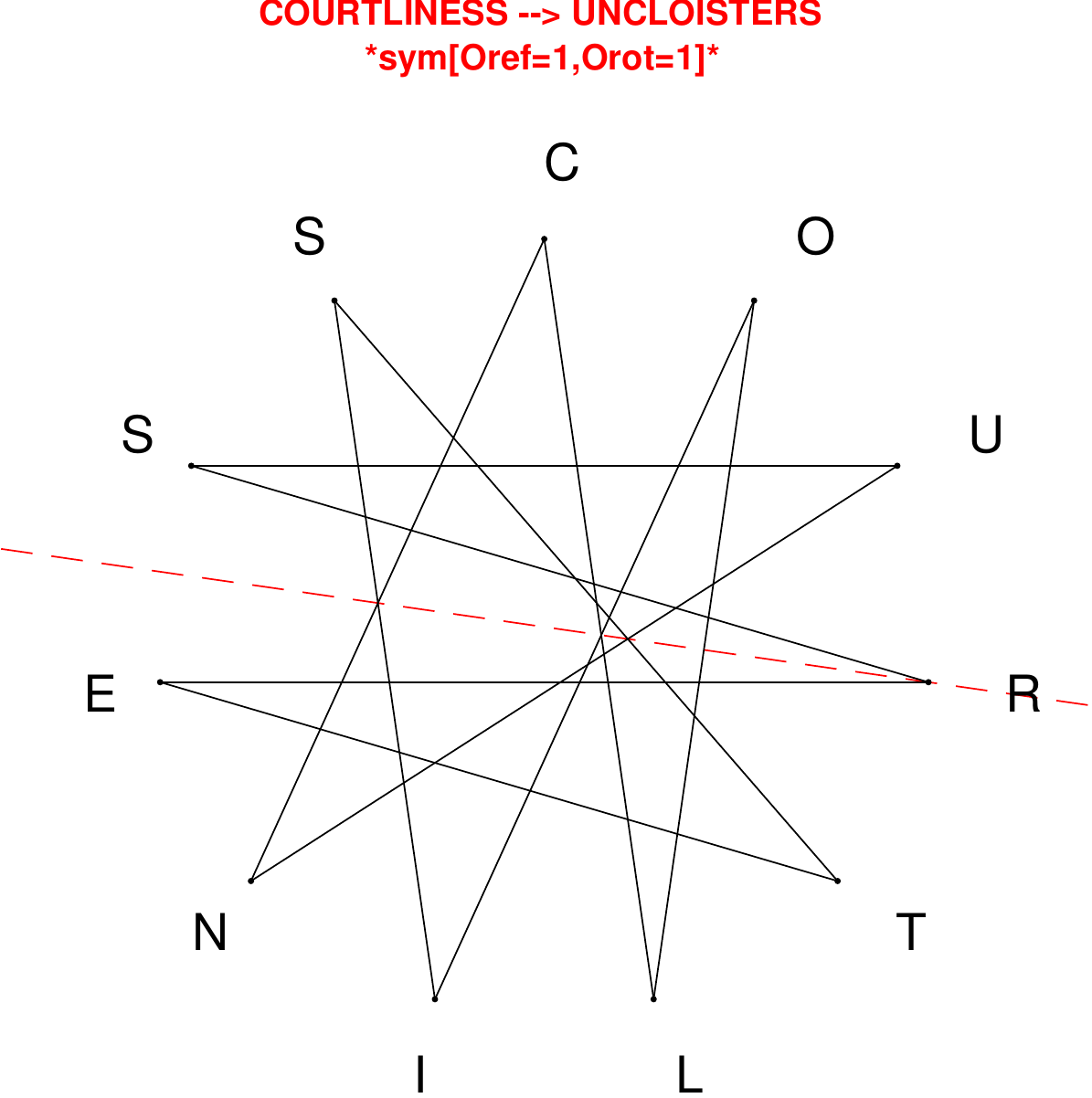}
\end{subfigure}
\hfill
\begin{subfigure}[T]{0.19\textwidth}
\centering
\includegraphics[width=\textwidth]{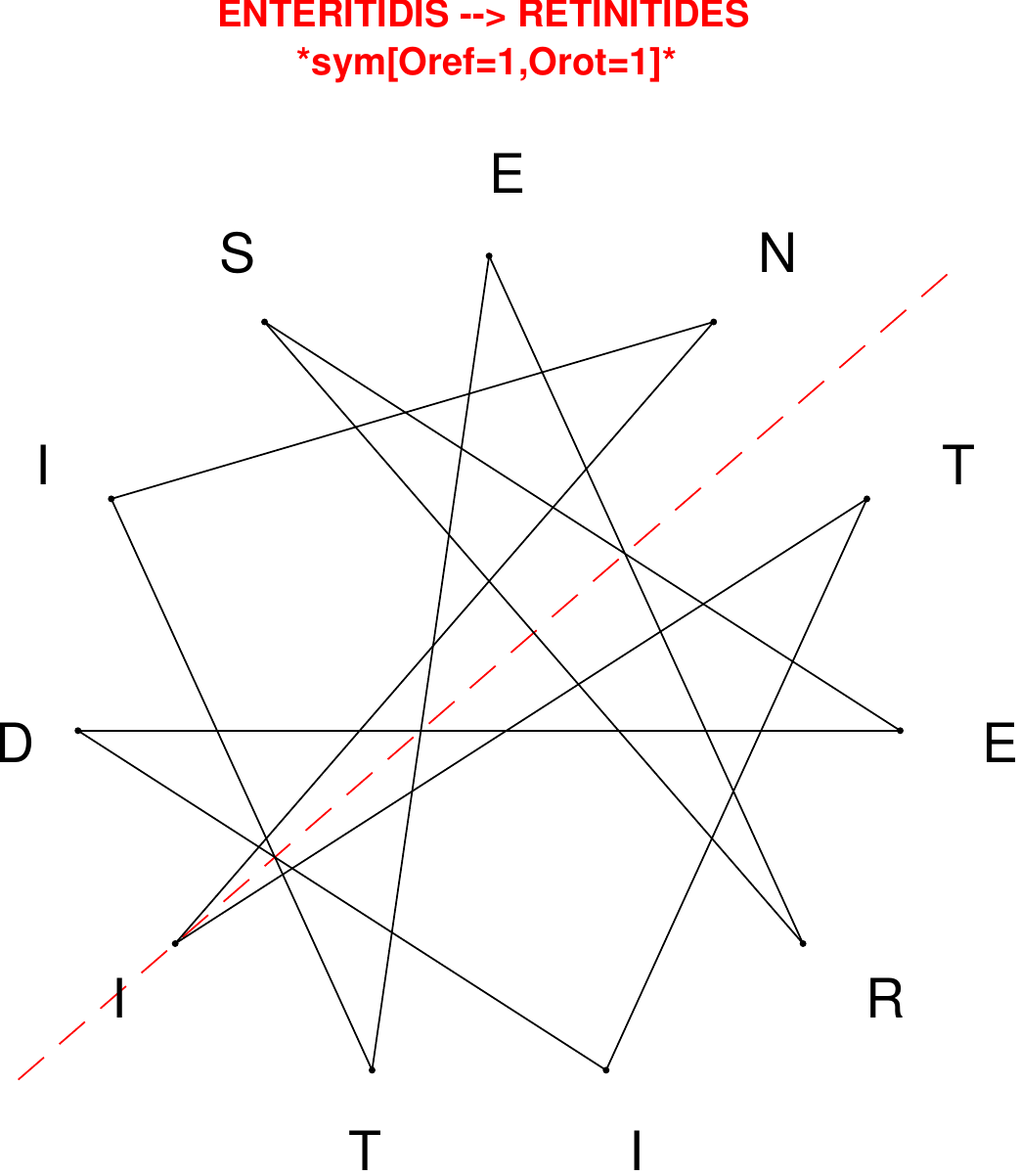}
\end{subfigure}
\hfill
\begin{subfigure}[T]{0.19\textwidth}
\centering
\includegraphics[width=\textwidth]{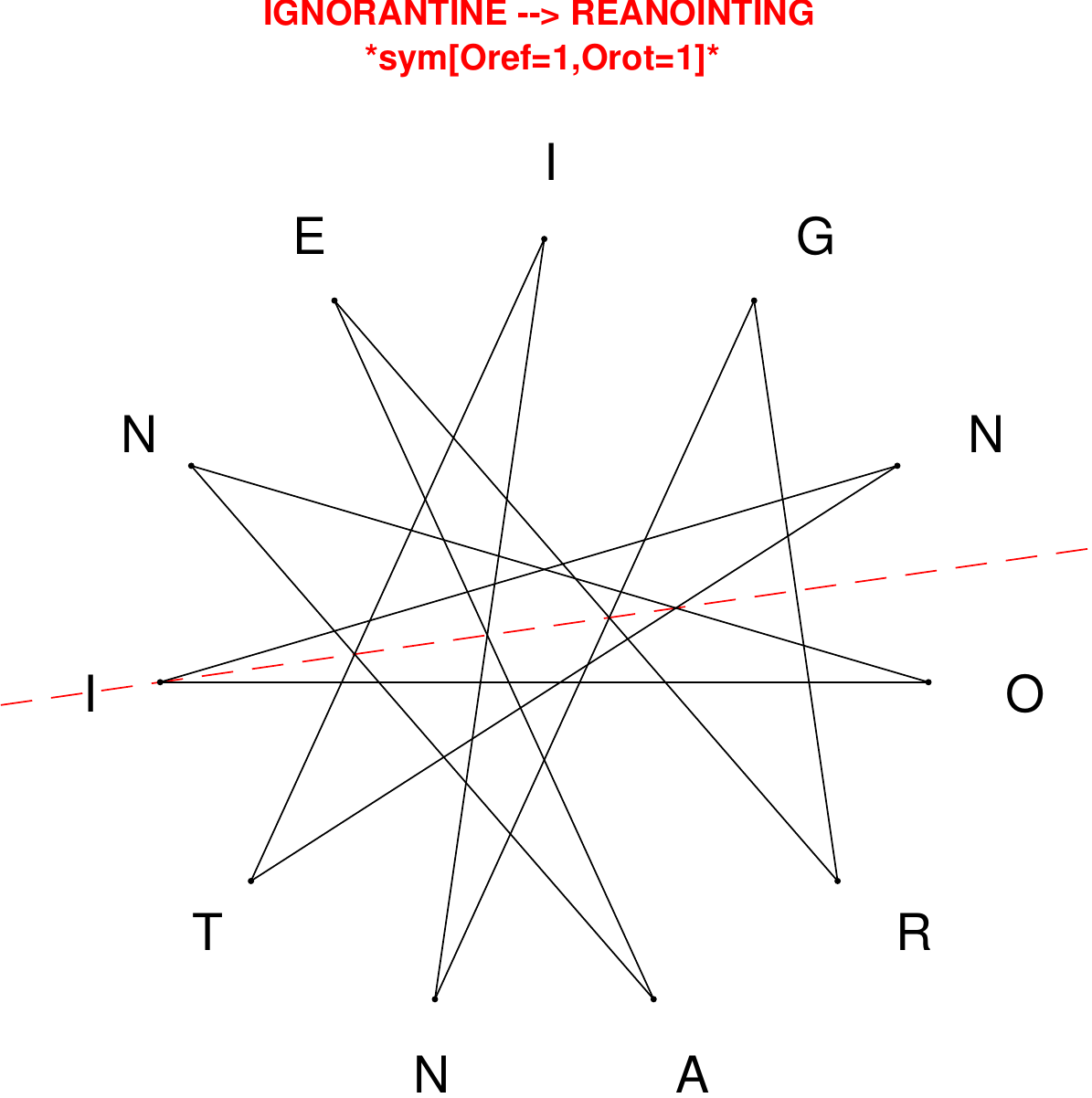}
\end{subfigure}
\end{figure}

\begin{figure}[H]
\centering
\begin{subfigure}[T]{0.19\textwidth}
\centering
\includegraphics[width=\textwidth]{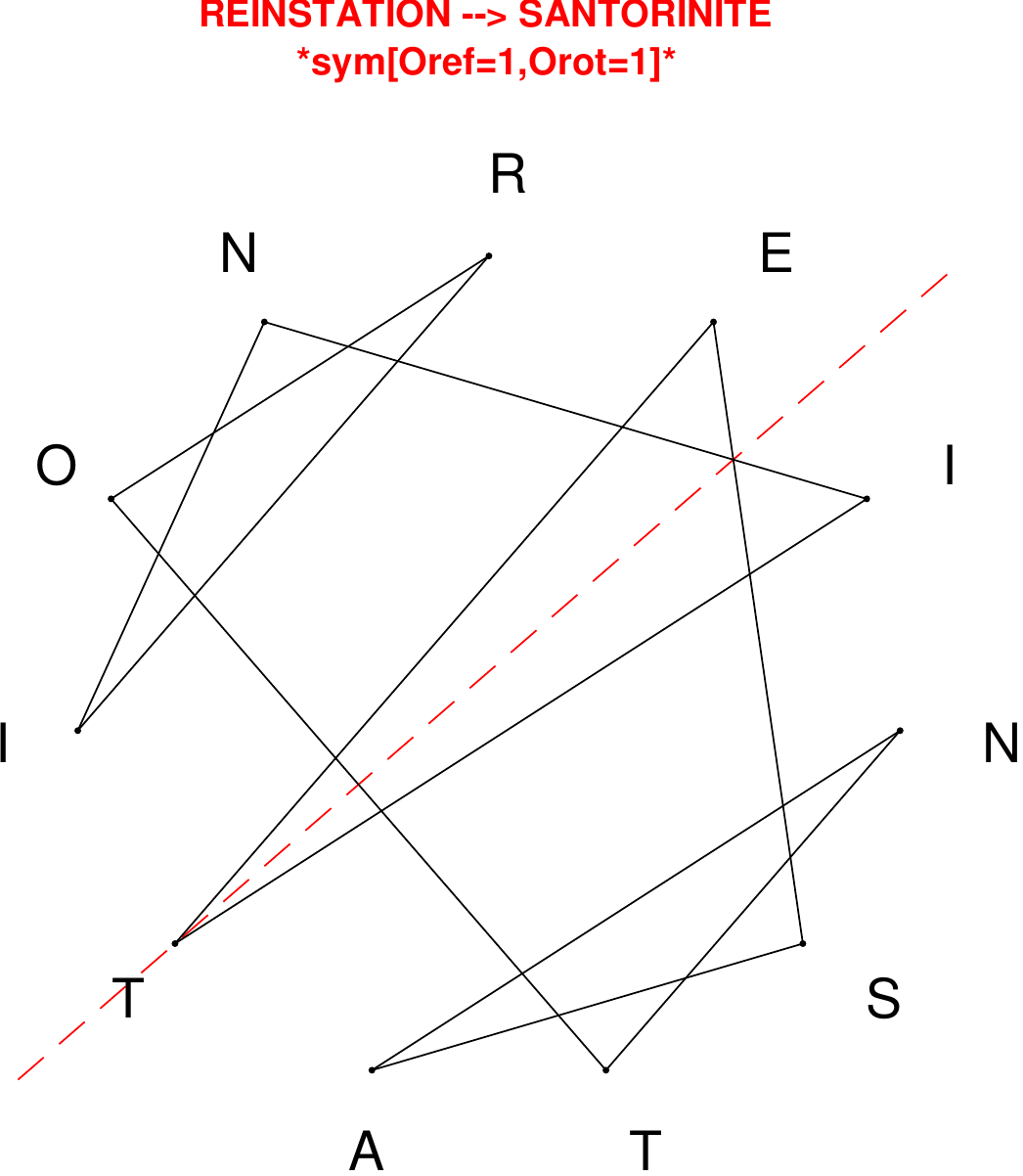}
\end{subfigure}
\hfill
\end{figure}

\subsubsection{Asymmetric Stars $N=11$}

\begin{figure}[H]
\centering
\begin{subfigure}[T]{0.19\textwidth}
\centering
\includegraphics[width=\textwidth]{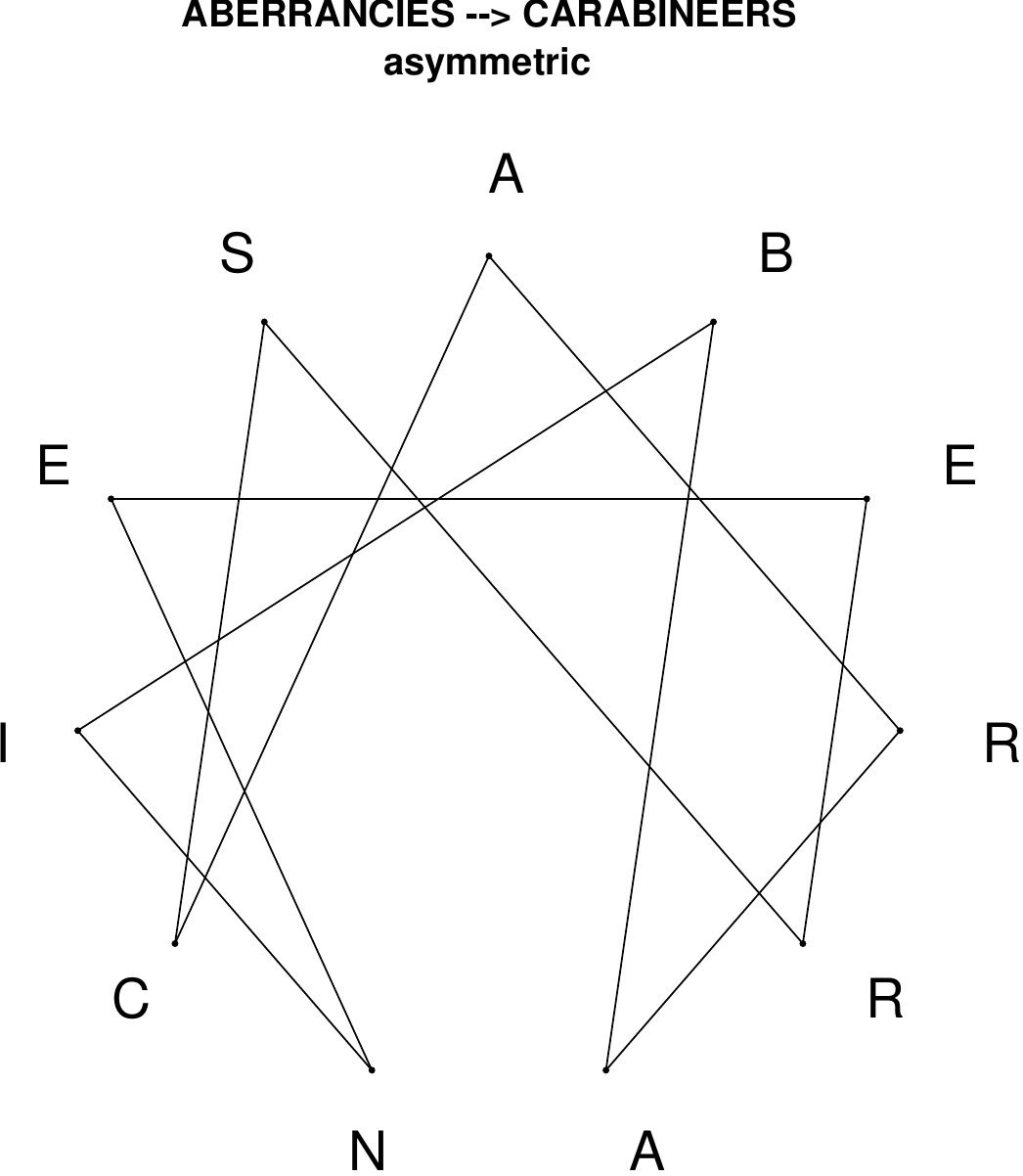}
\end{subfigure}
\hfill
\begin{subfigure}[T]{0.19\textwidth}
\centering
\includegraphics[width=\textwidth]{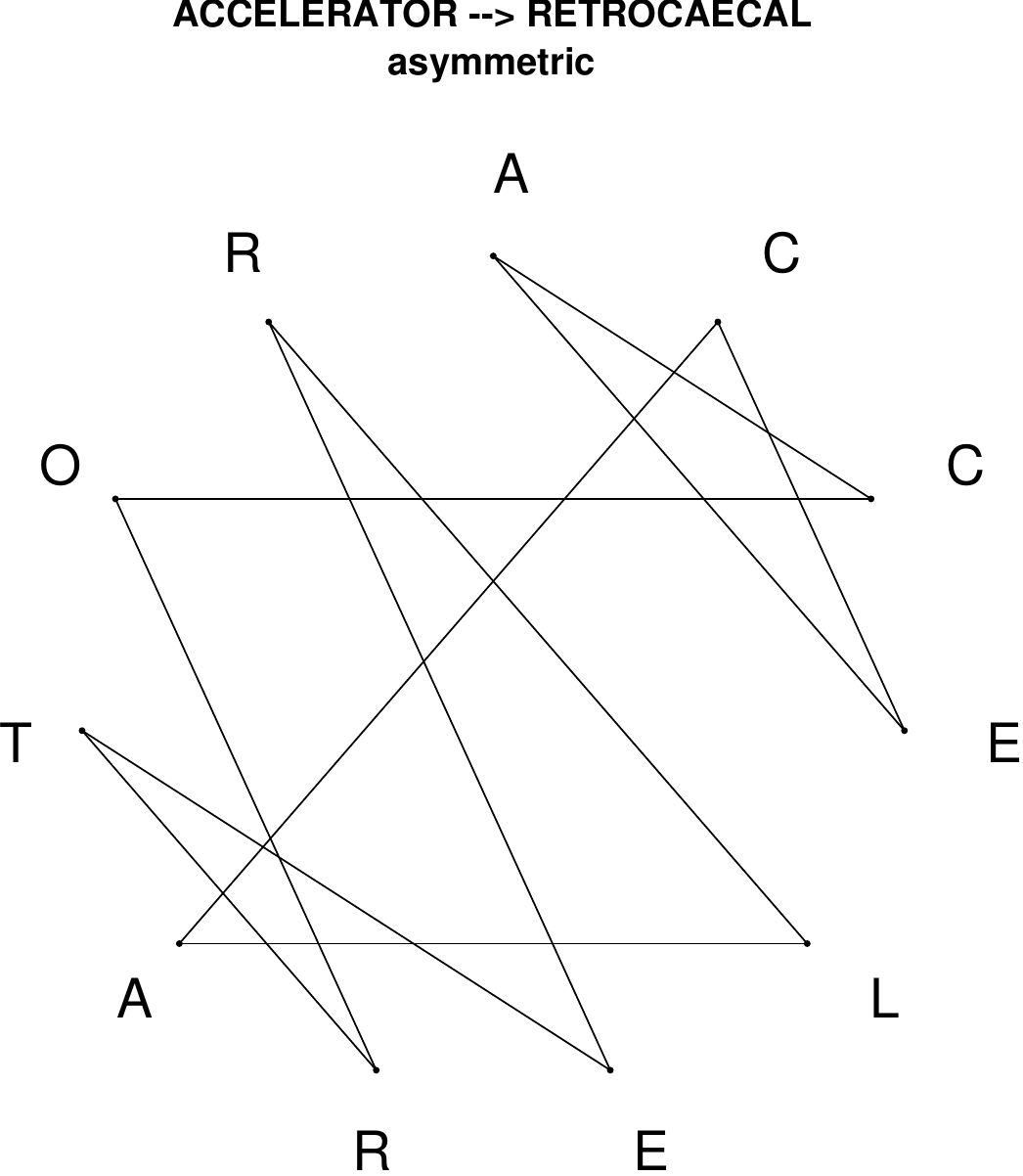}
\end{subfigure}
\hfill
\begin{subfigure}[T]{0.19\textwidth}
\centering
\includegraphics[width=\textwidth]{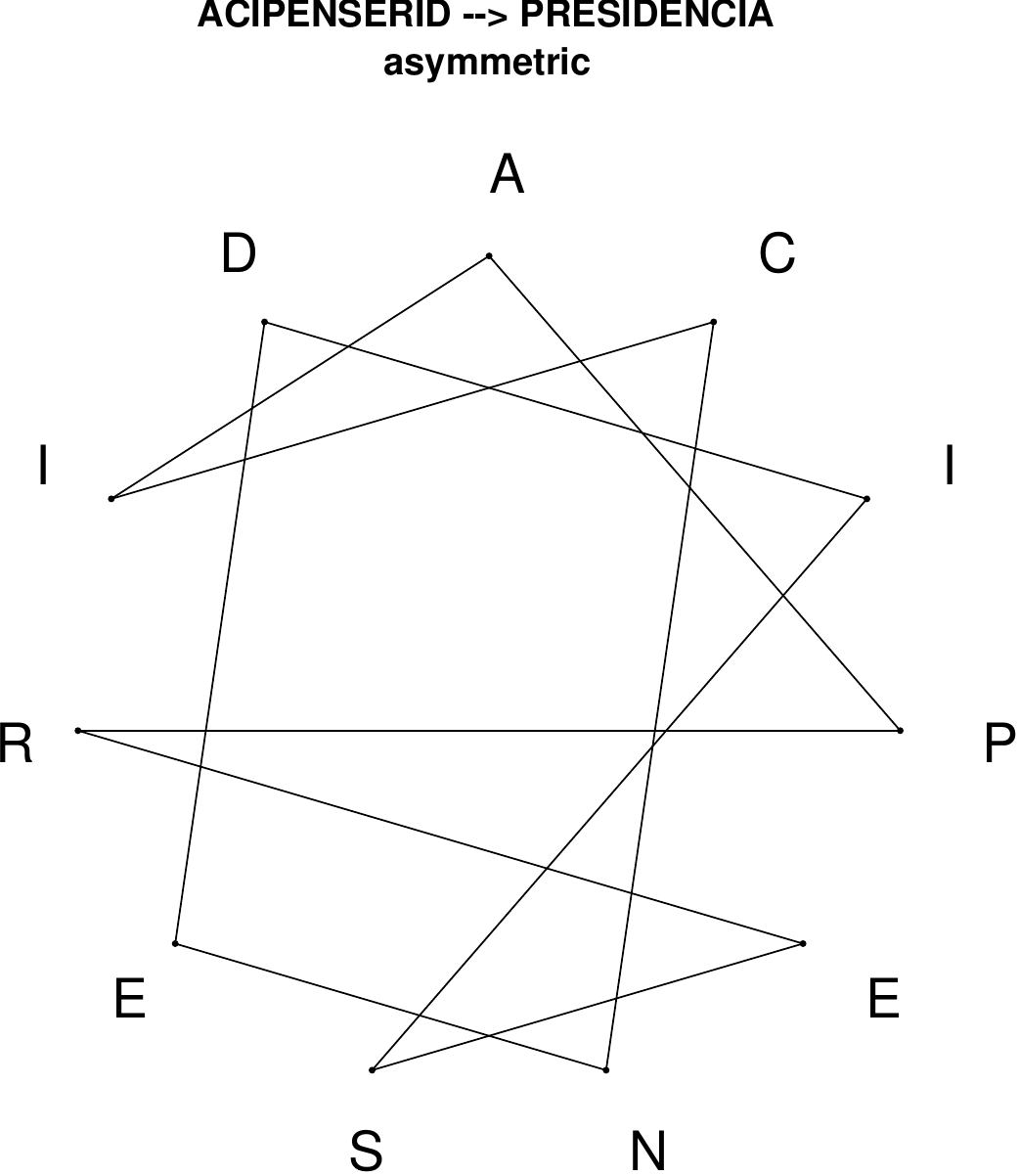}
\end{subfigure}
\hfill
\begin{subfigure}[T]{0.19\textwidth}
\centering
\includegraphics[width=\textwidth]{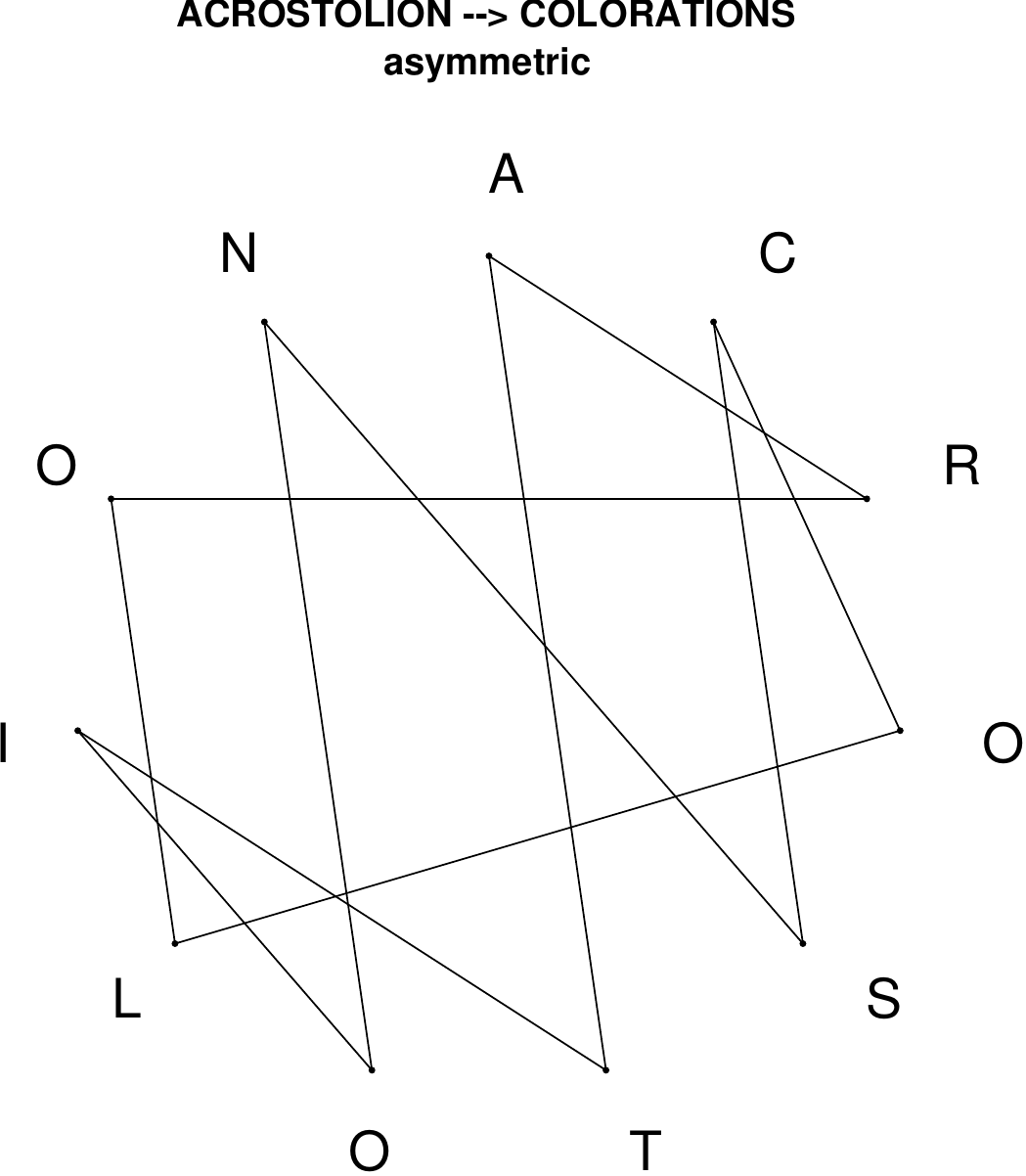}
\end{subfigure}
\hfill
\begin{subfigure}[T]{0.19\textwidth}
\centering
\includegraphics[width=\textwidth]{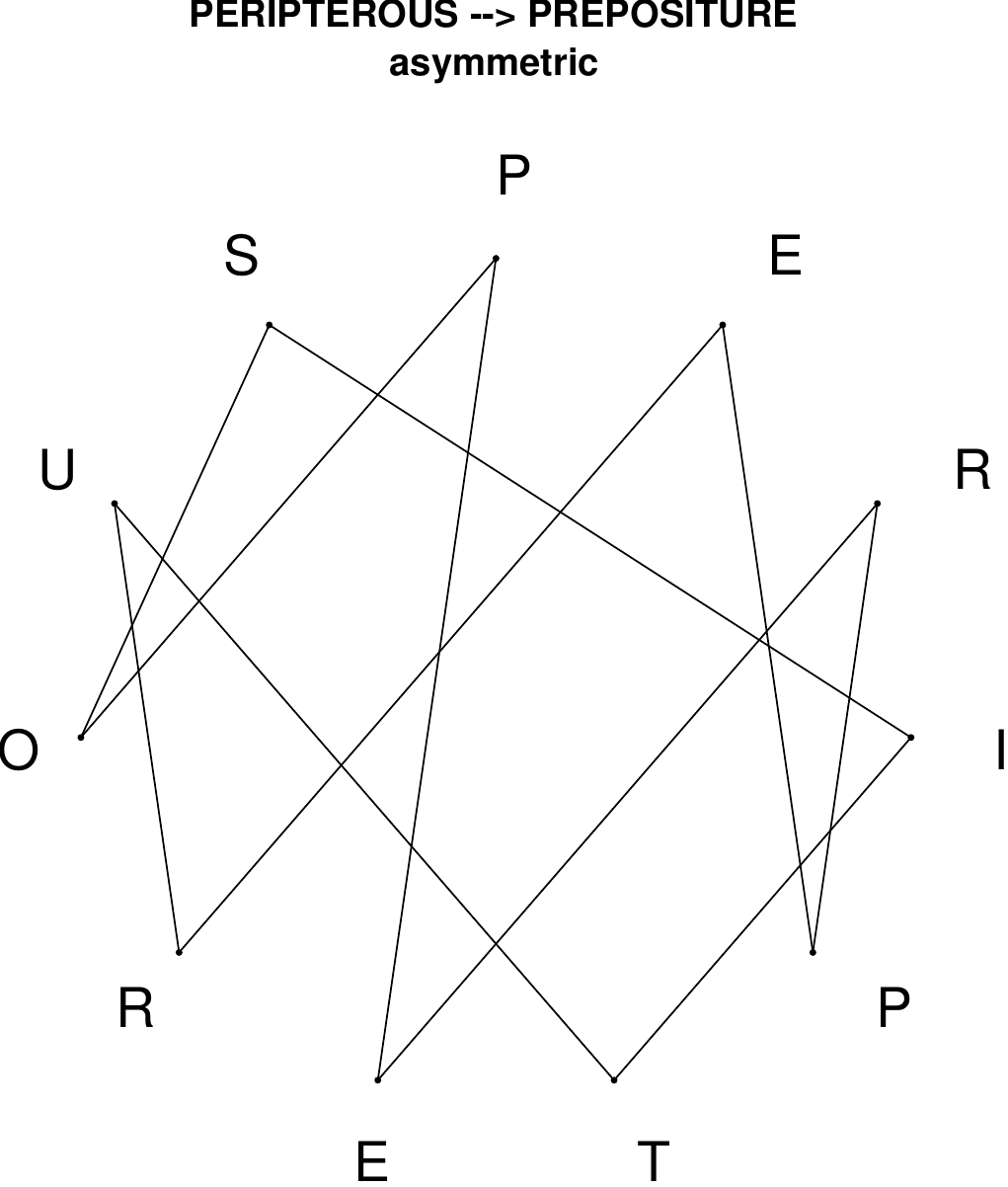}
\end{subfigure}
\end{figure}

\begin{figure}[H]
\centering
\begin{subfigure}[T]{0.19\textwidth}
\centering
\includegraphics[width=\textwidth]{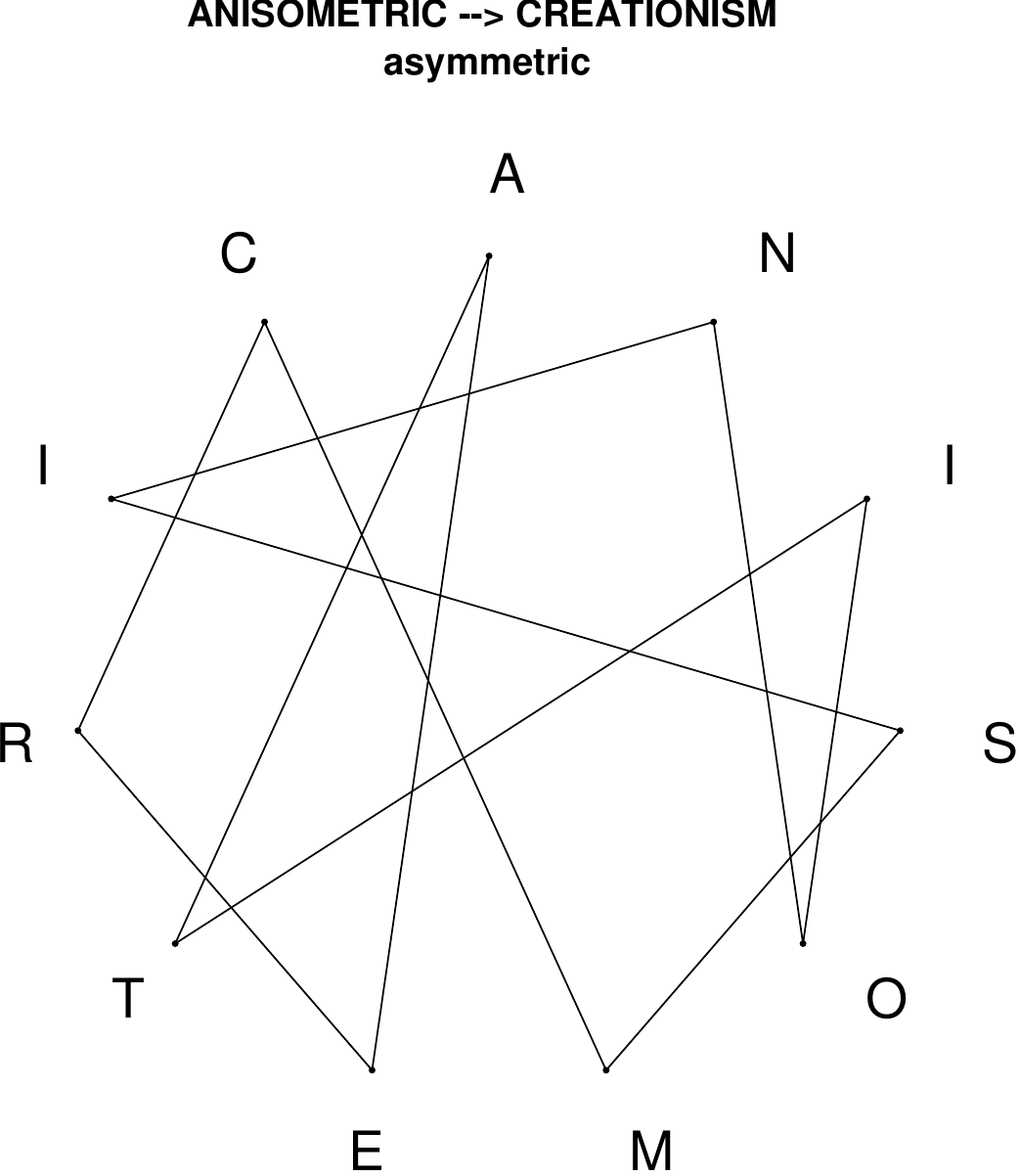}
\end{subfigure}
\hfill
\begin{subfigure}[T]{0.19\textwidth}
\centering
\includegraphics[width=\textwidth]{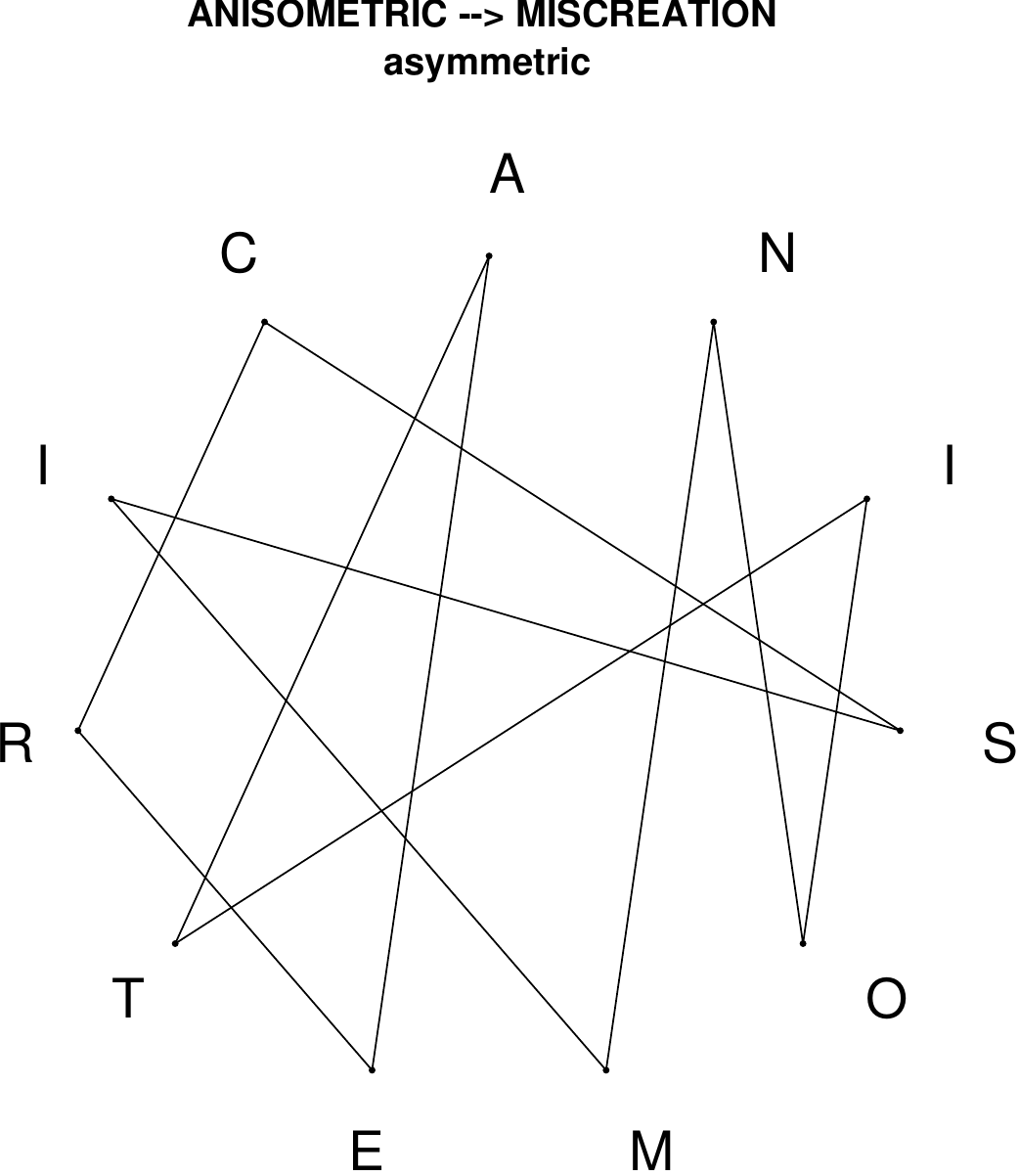}
\end{subfigure}
\hfill
\begin{subfigure}[T]{0.19\textwidth}
\centering
\includegraphics[width=\textwidth]{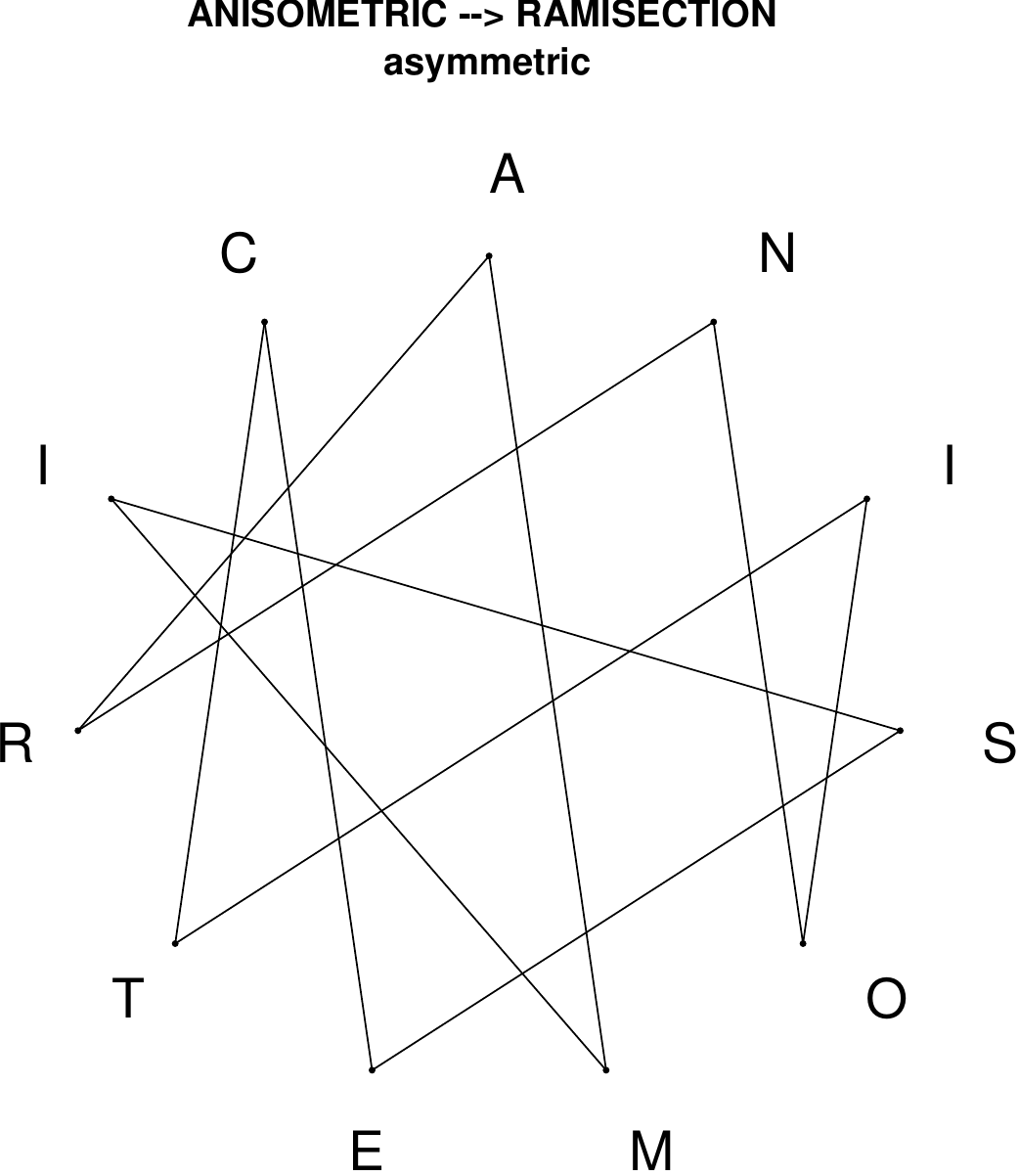}
\end{subfigure}
\hfill
\begin{subfigure}[T]{0.19\textwidth}
\centering
\includegraphics[width=\textwidth]{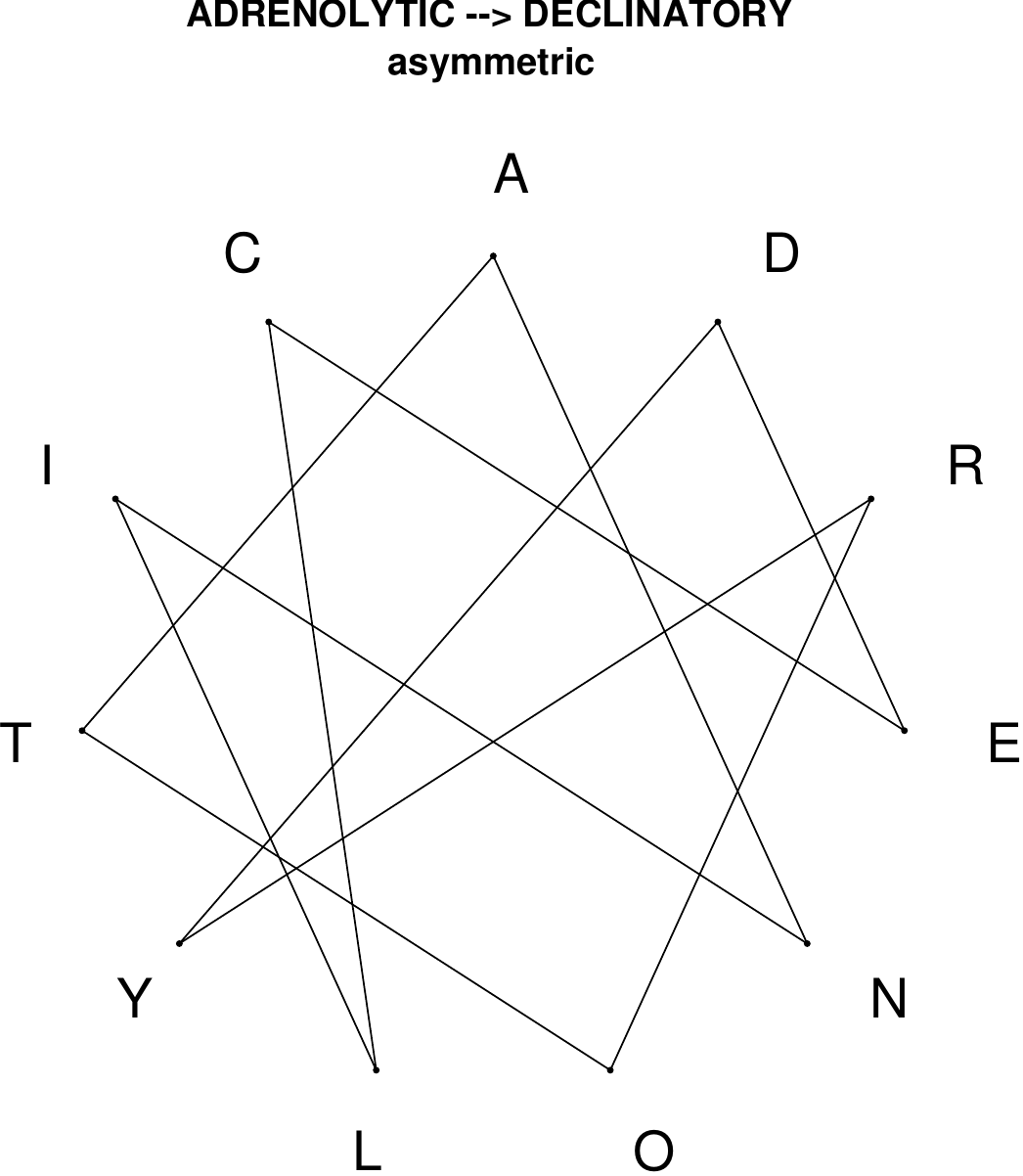}
\end{subfigure}
\hfill
\begin{subfigure}[T]{0.19\textwidth}
\centering
\includegraphics[width=\textwidth]{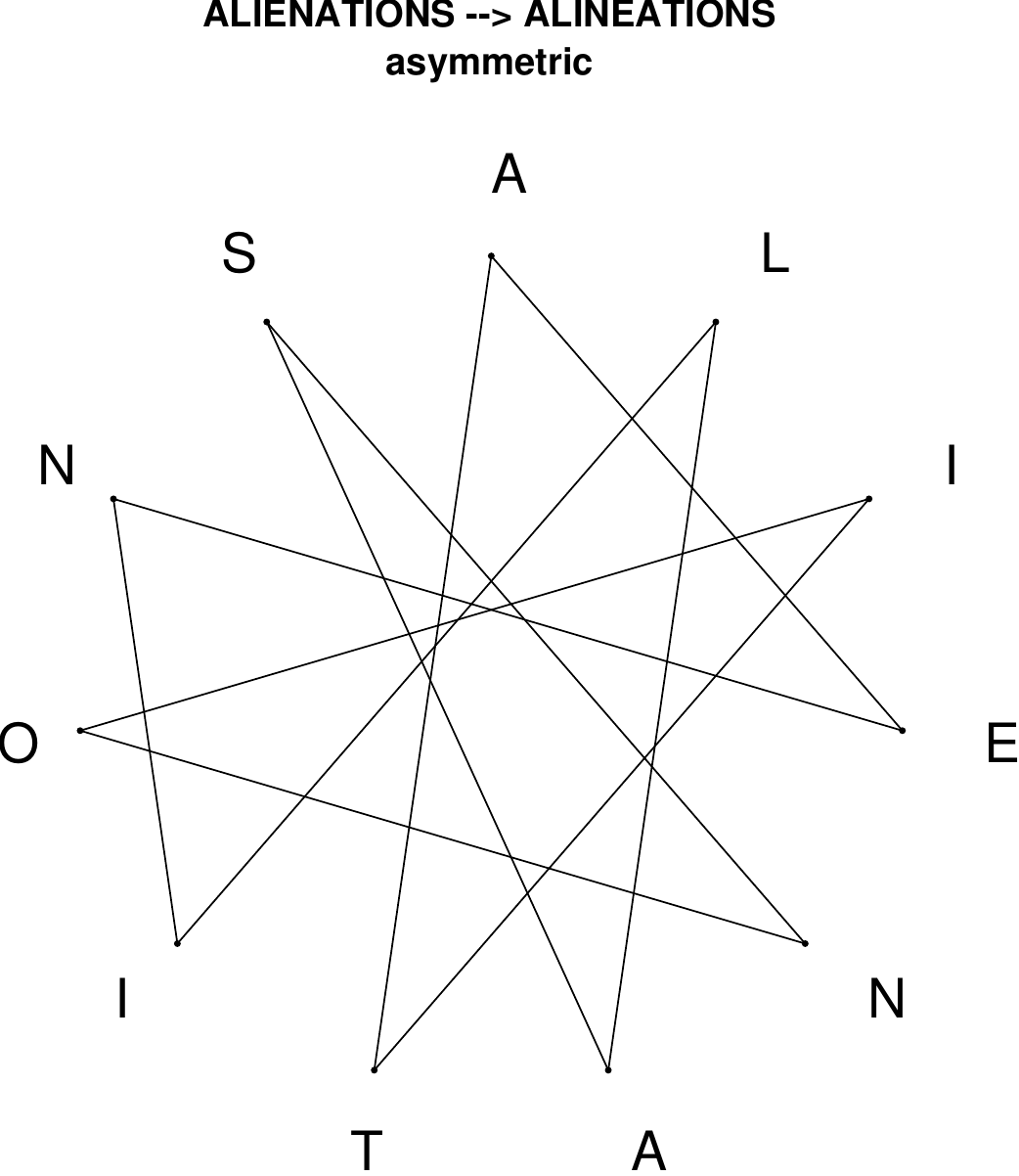}
\end{subfigure}
\end{figure}

\begin{figure}[H]
\centering
\begin{subfigure}[T]{0.19\textwidth}
\centering
\includegraphics[width=\textwidth]{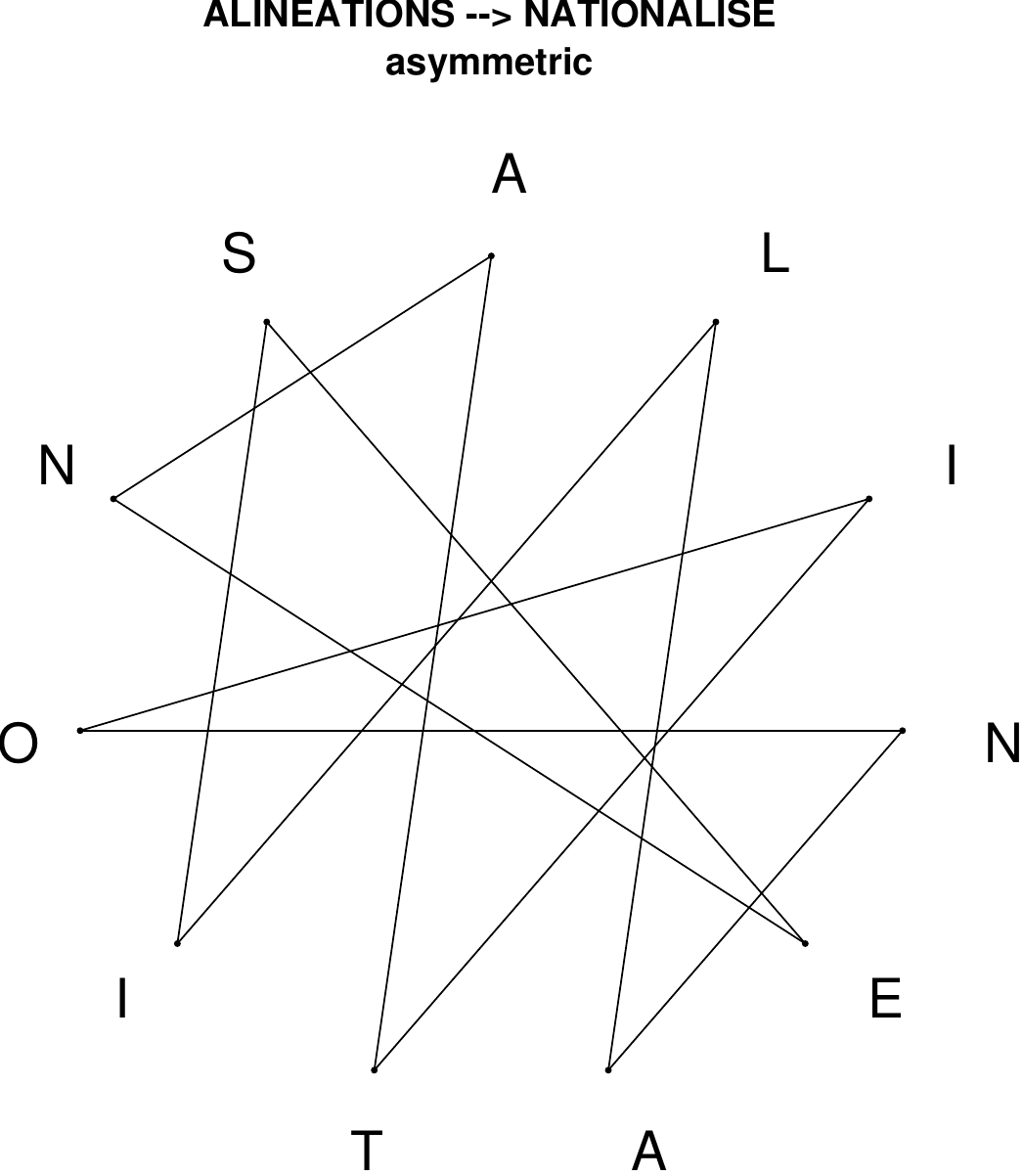}
\end{subfigure}
\hfill
\begin{subfigure}[T]{0.19\textwidth}
\centering
\includegraphics[width=\textwidth]{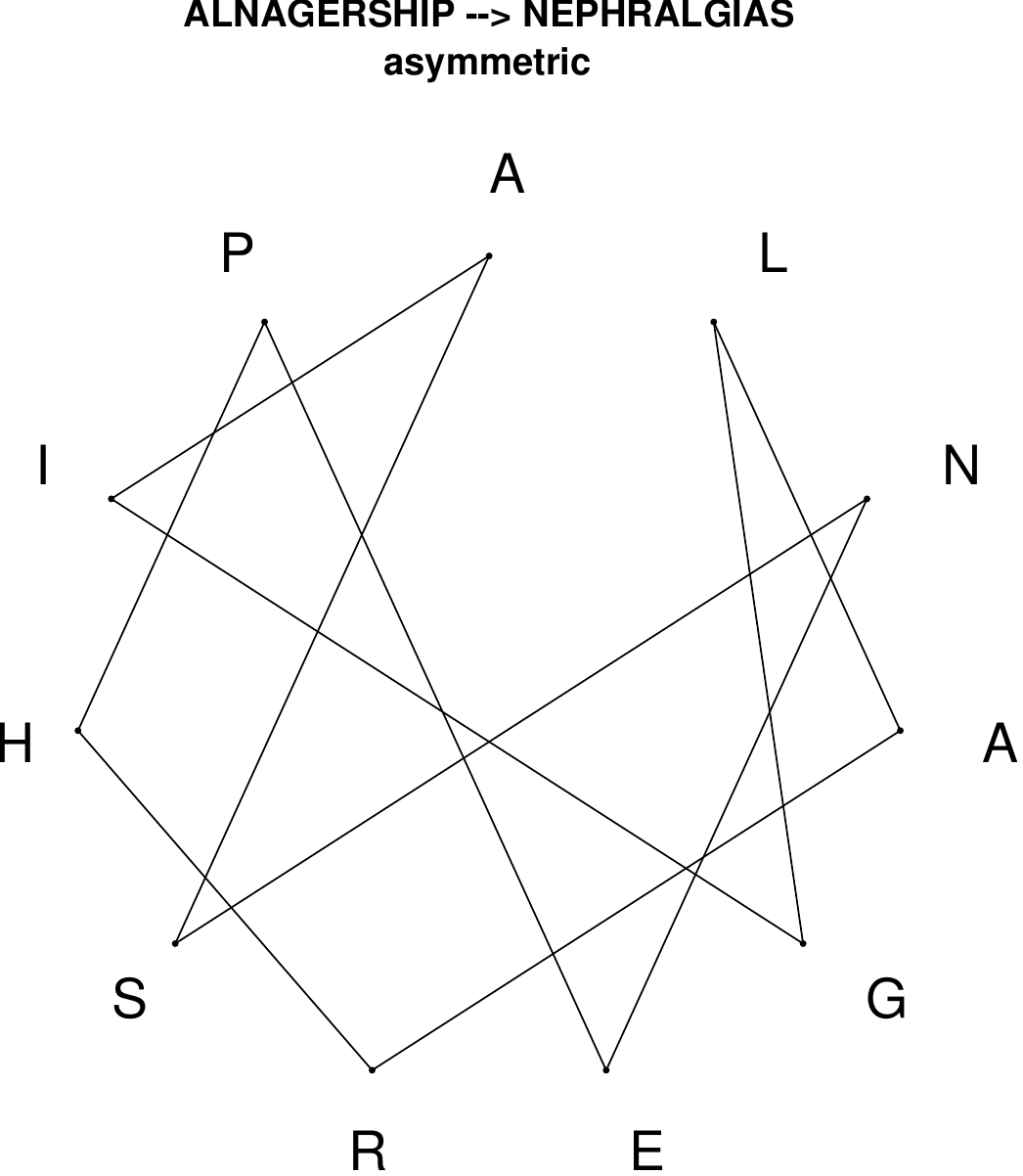}
\end{subfigure}
\hfill
\begin{subfigure}[T]{0.19\textwidth}
\centering
\includegraphics[width=\textwidth]{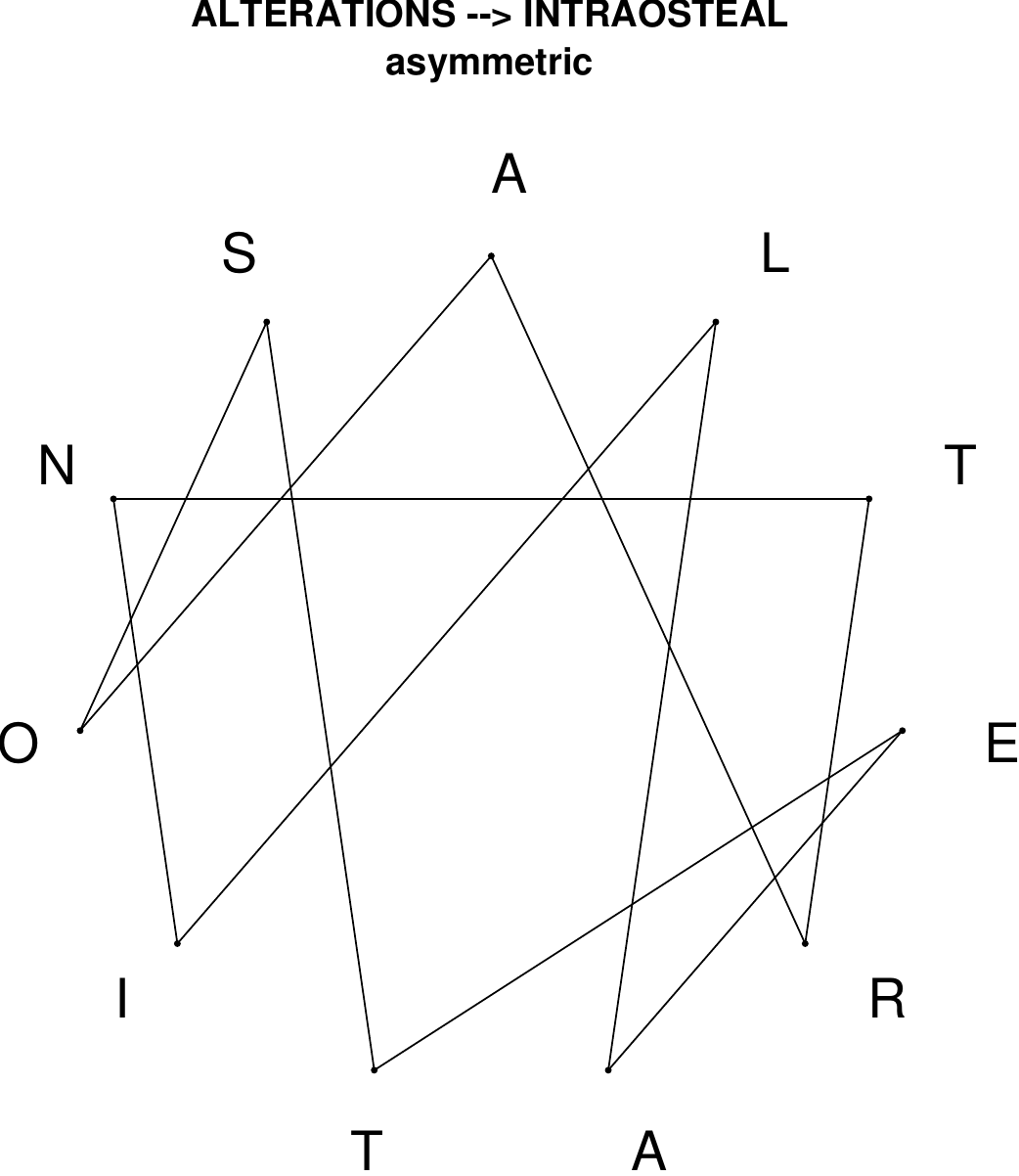}
\end{subfigure}
\hfill
\begin{subfigure}[T]{0.19\textwidth}
\centering
\includegraphics[width=\textwidth]{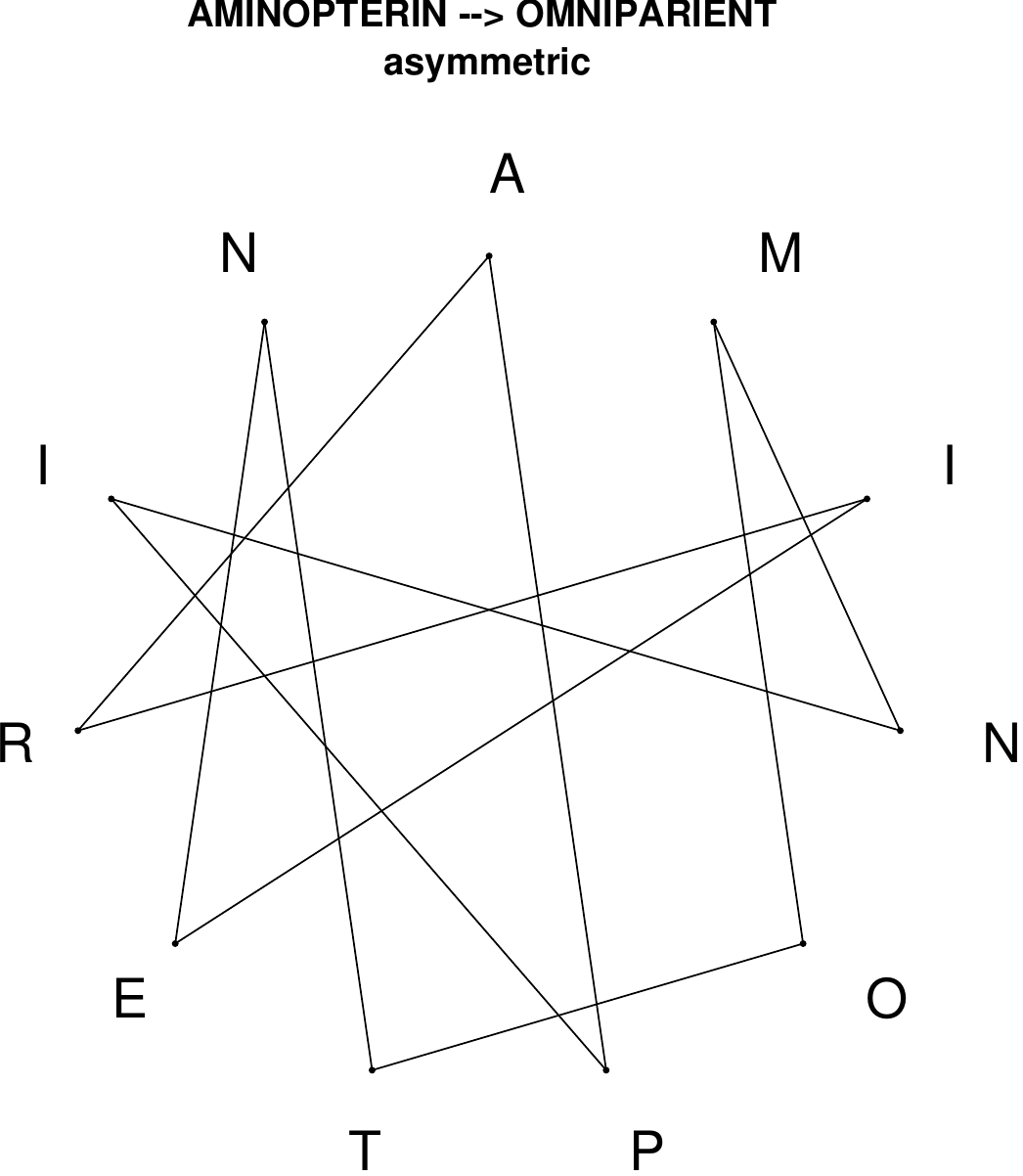}
\end{subfigure}
\hfill
\begin{subfigure}[T]{0.19\textwidth}
\centering
\includegraphics[width=\textwidth]{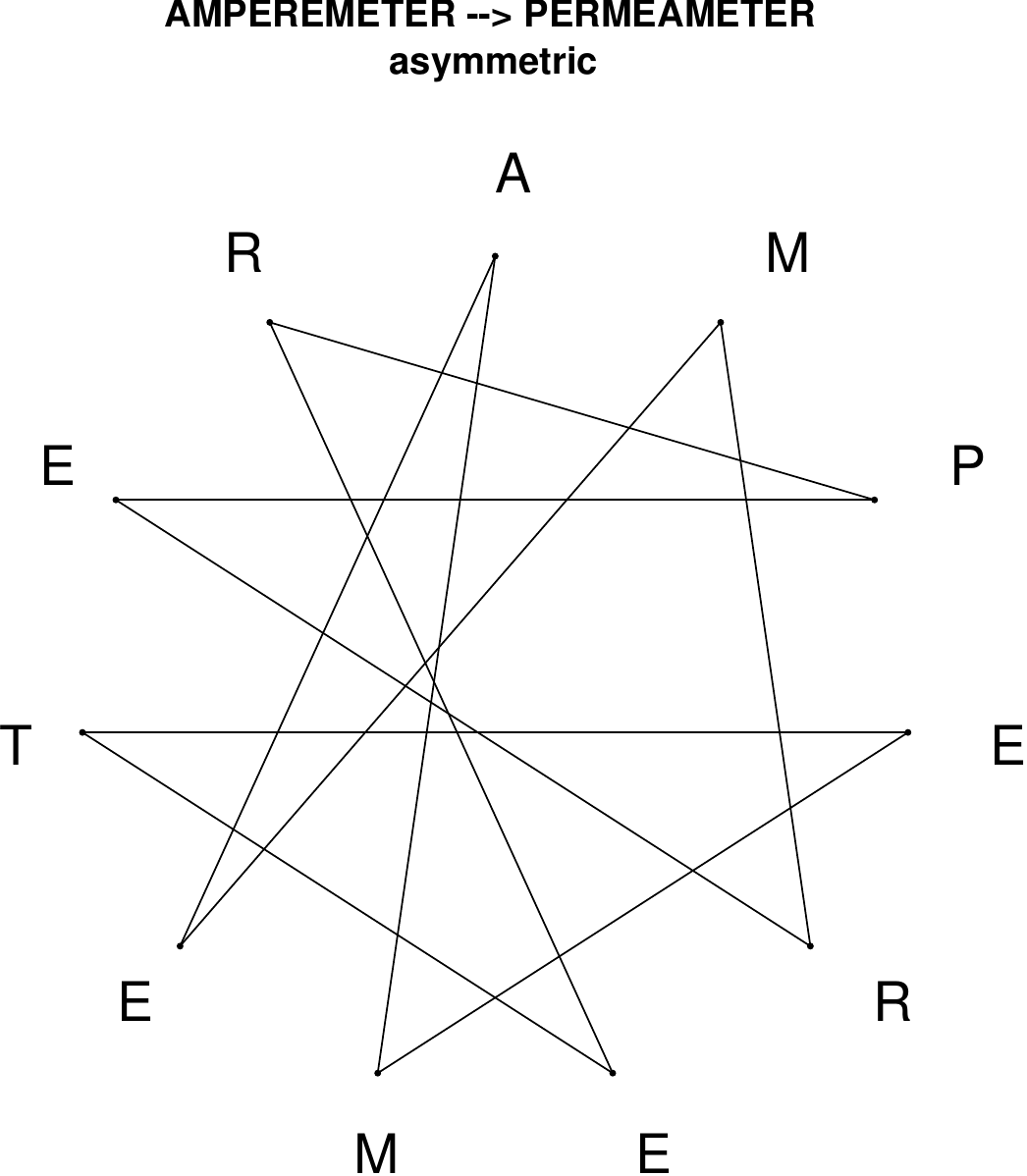}
\end{subfigure}
\end{figure}

\begin{figure}[H]
\centering
\begin{subfigure}[T]{0.19\textwidth}
\centering
\includegraphics[width=\textwidth]{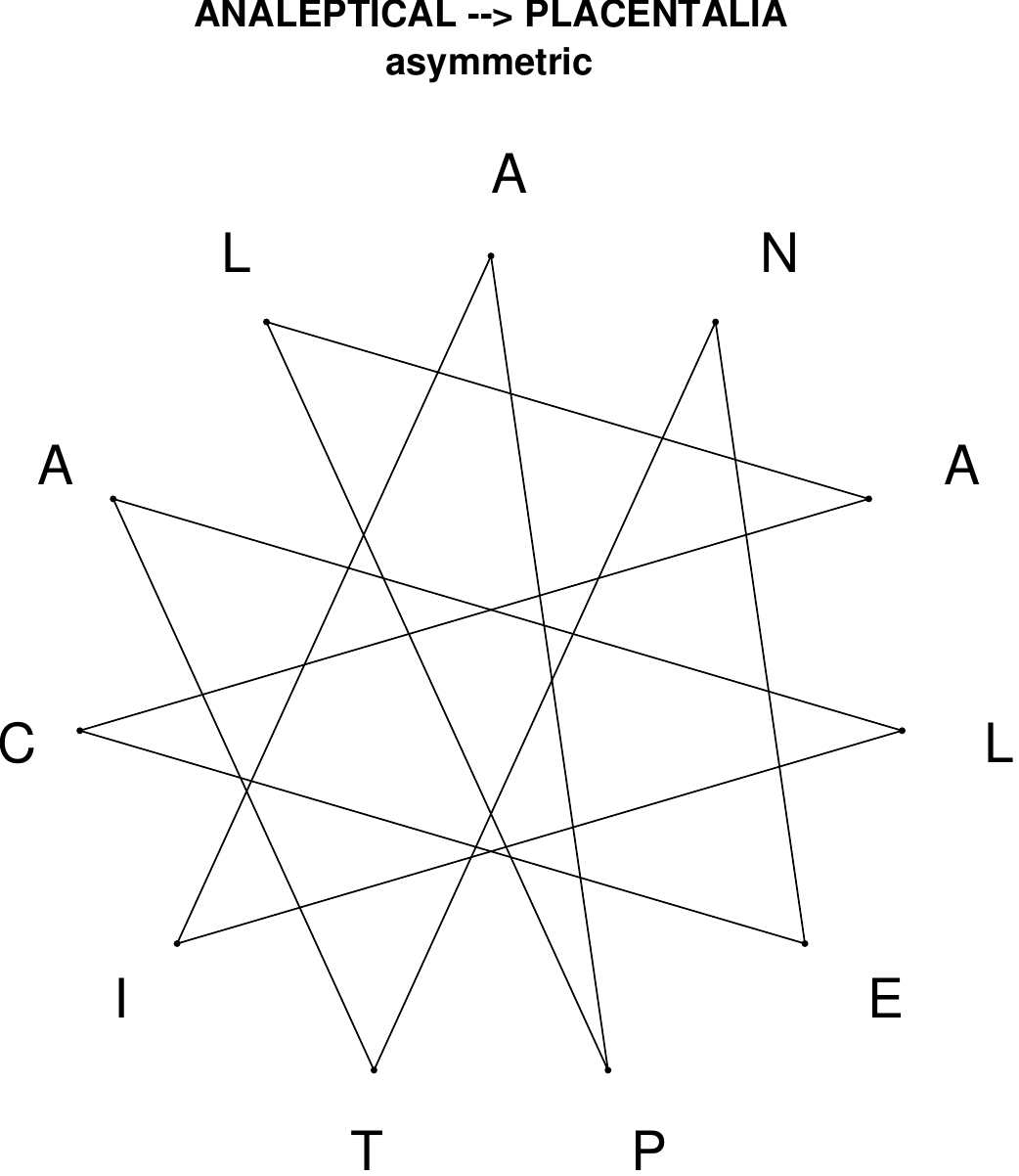}
\end{subfigure}
\hfill
\begin{subfigure}[T]{0.19\textwidth}
\centering
\includegraphics[width=\textwidth]{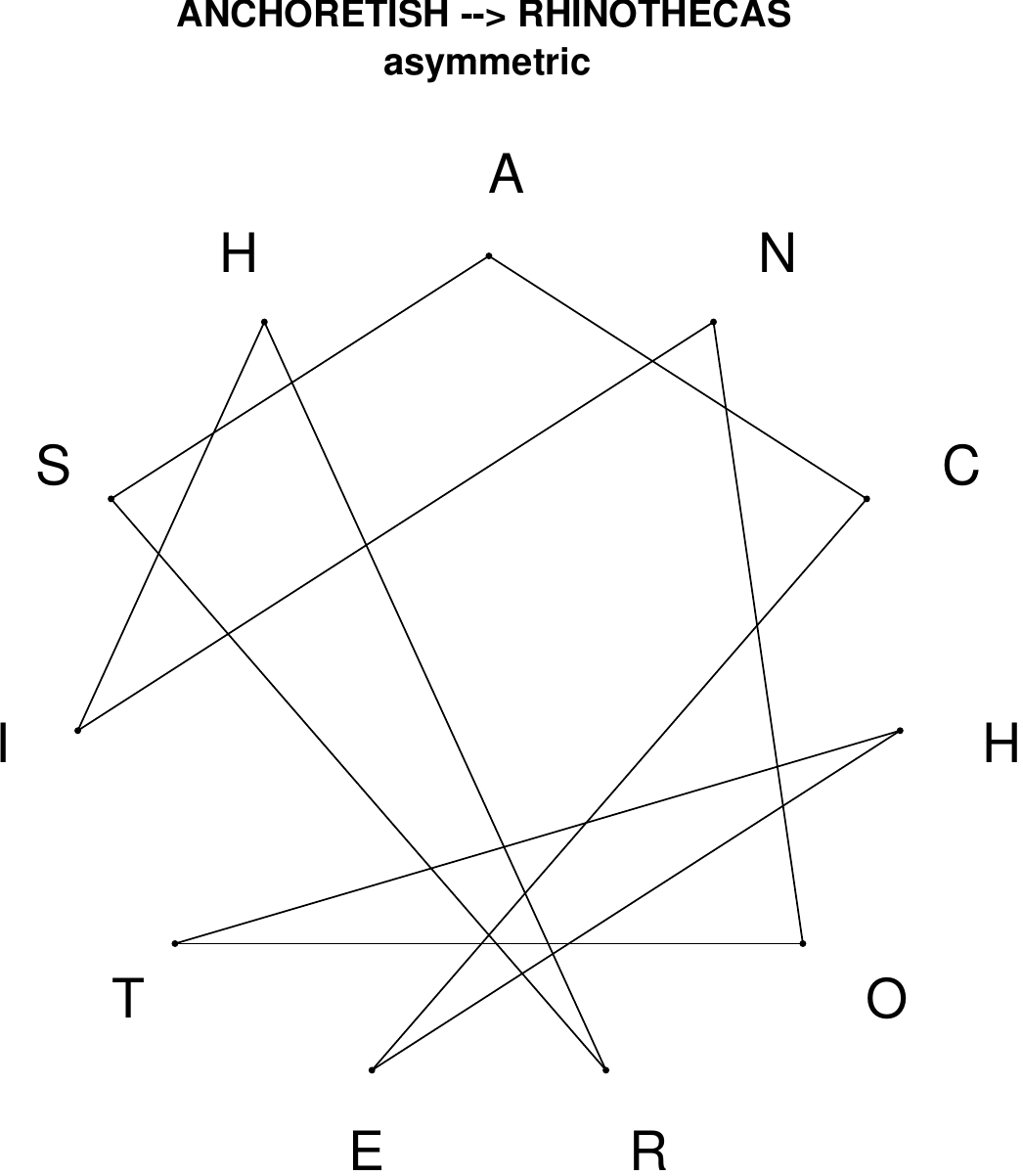}
\end{subfigure}
\hfill
\begin{subfigure}[T]{0.19\textwidth}
\centering
\includegraphics[width=\textwidth]{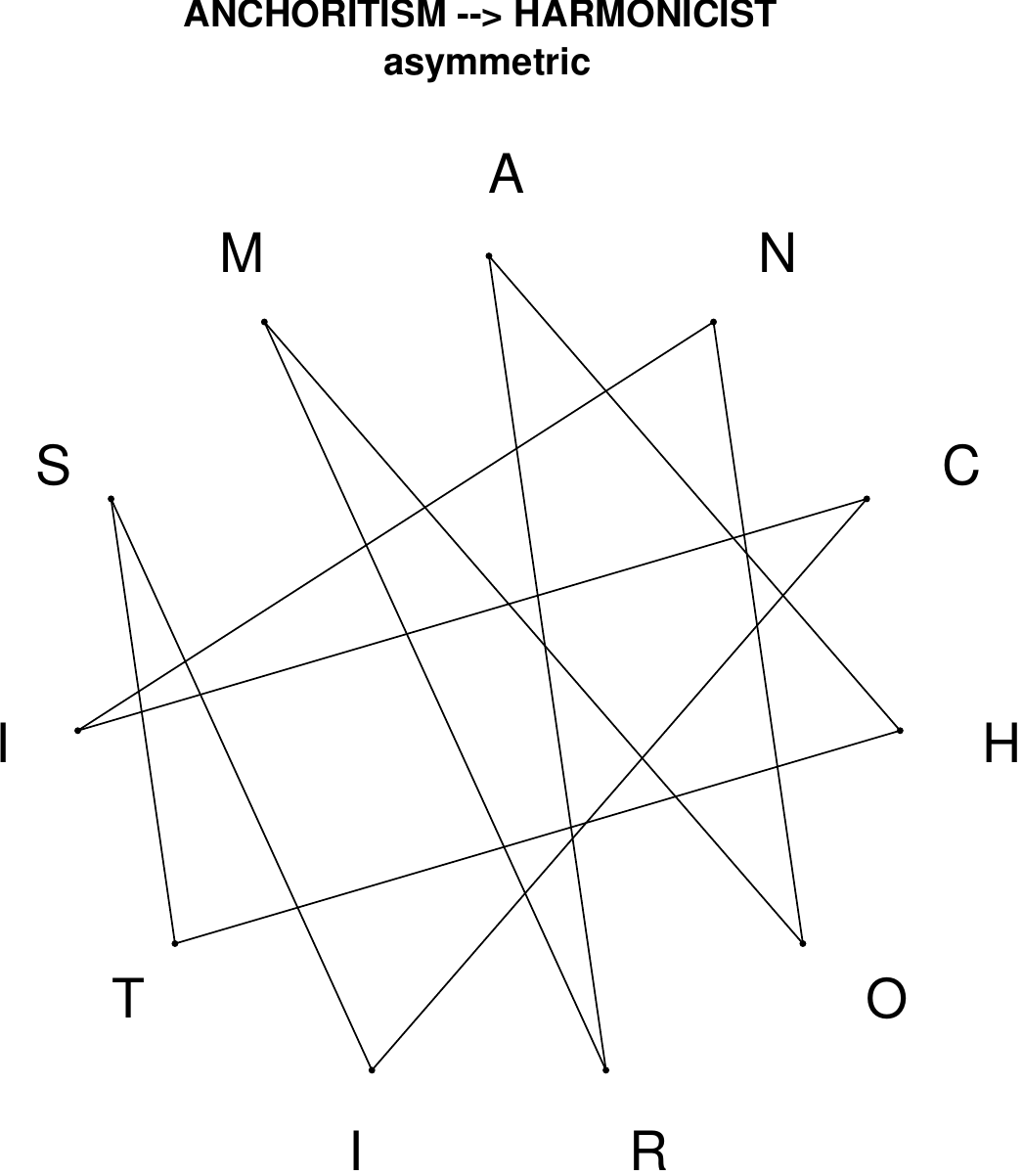}
\end{subfigure}
\hfill
\begin{subfigure}[T]{0.19\textwidth}
\centering
\includegraphics[width=\textwidth]{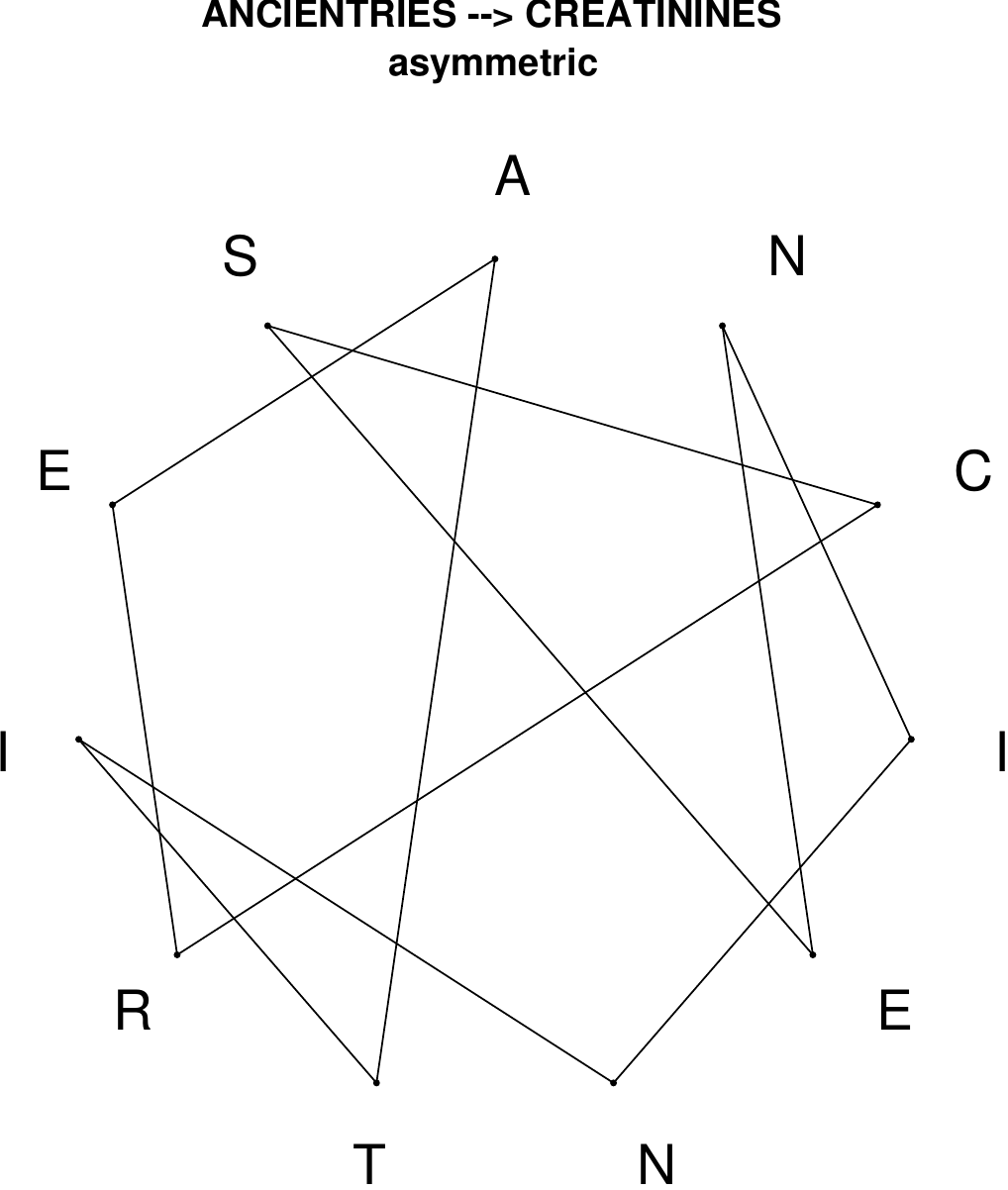}
\end{subfigure}
\hfill
\begin{subfigure}[T]{0.19\textwidth}
\centering
\includegraphics[width=\textwidth]{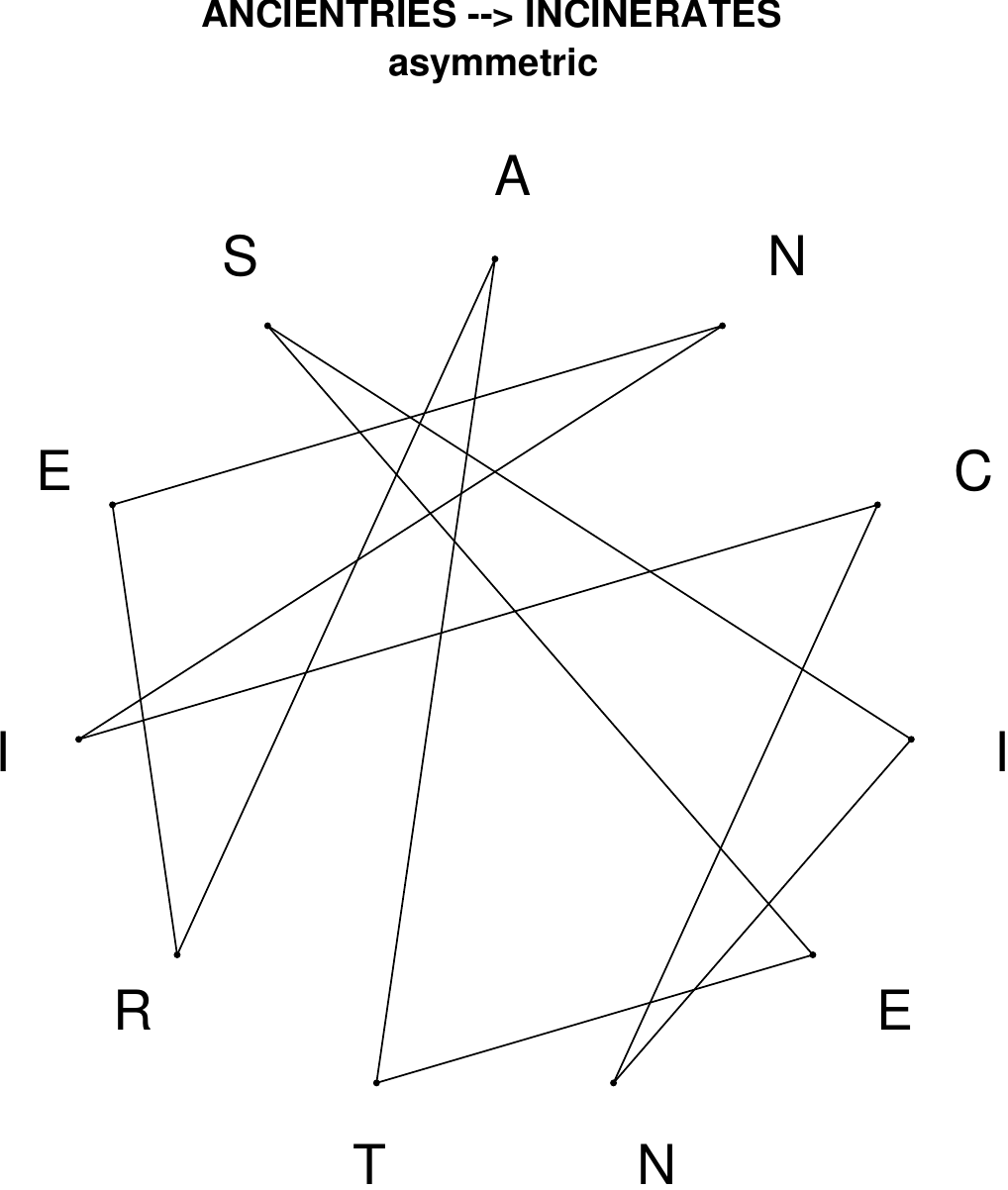}
\end{subfigure}
\end{figure}

\begin{figure}[H]
\centering
\begin{subfigure}[T]{0.19\textwidth}
\centering
\includegraphics[width=\textwidth]{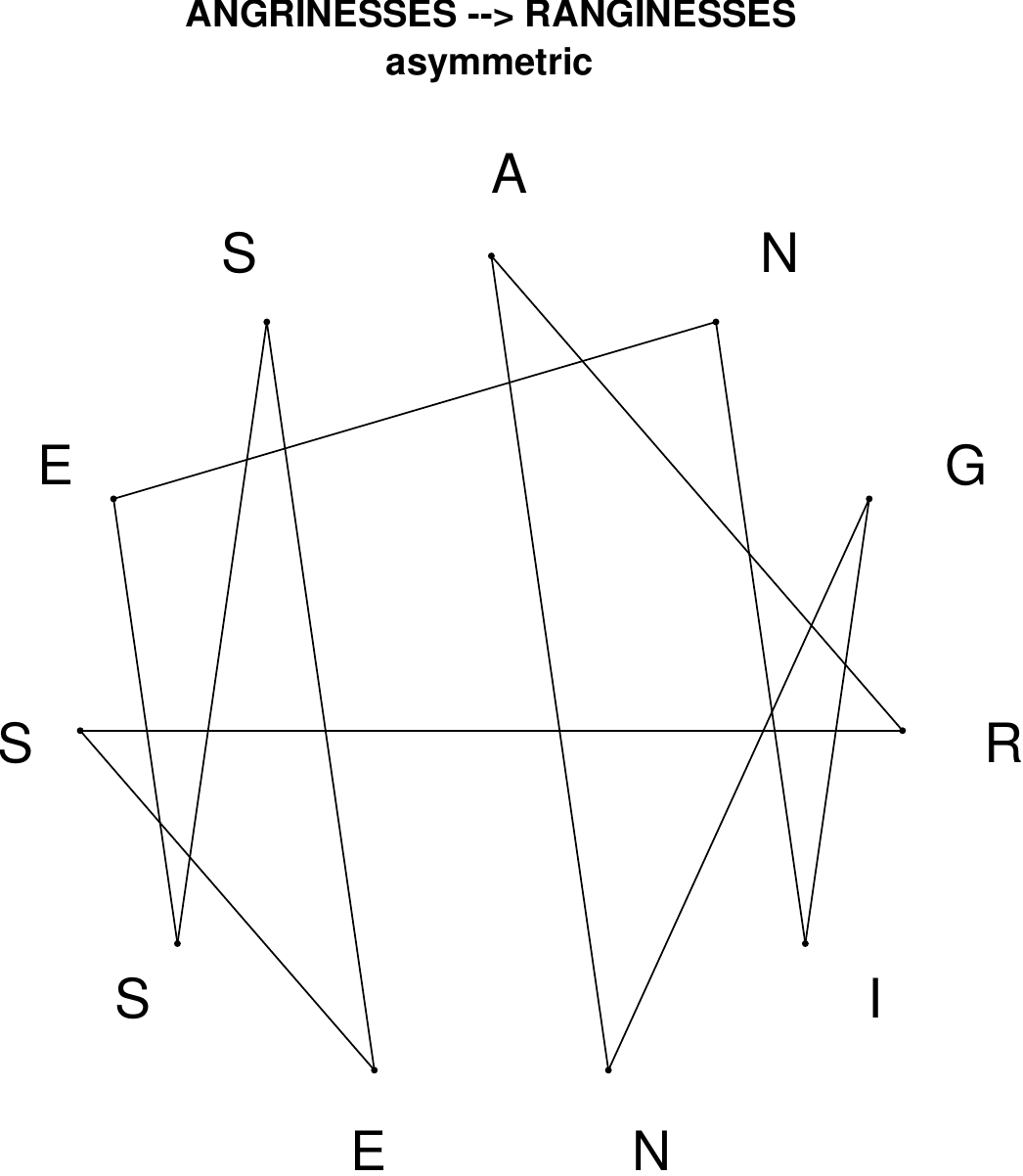}
\end{subfigure}
\hfill
\begin{subfigure}[T]{0.19\textwidth}
\centering
\includegraphics[width=\textwidth]{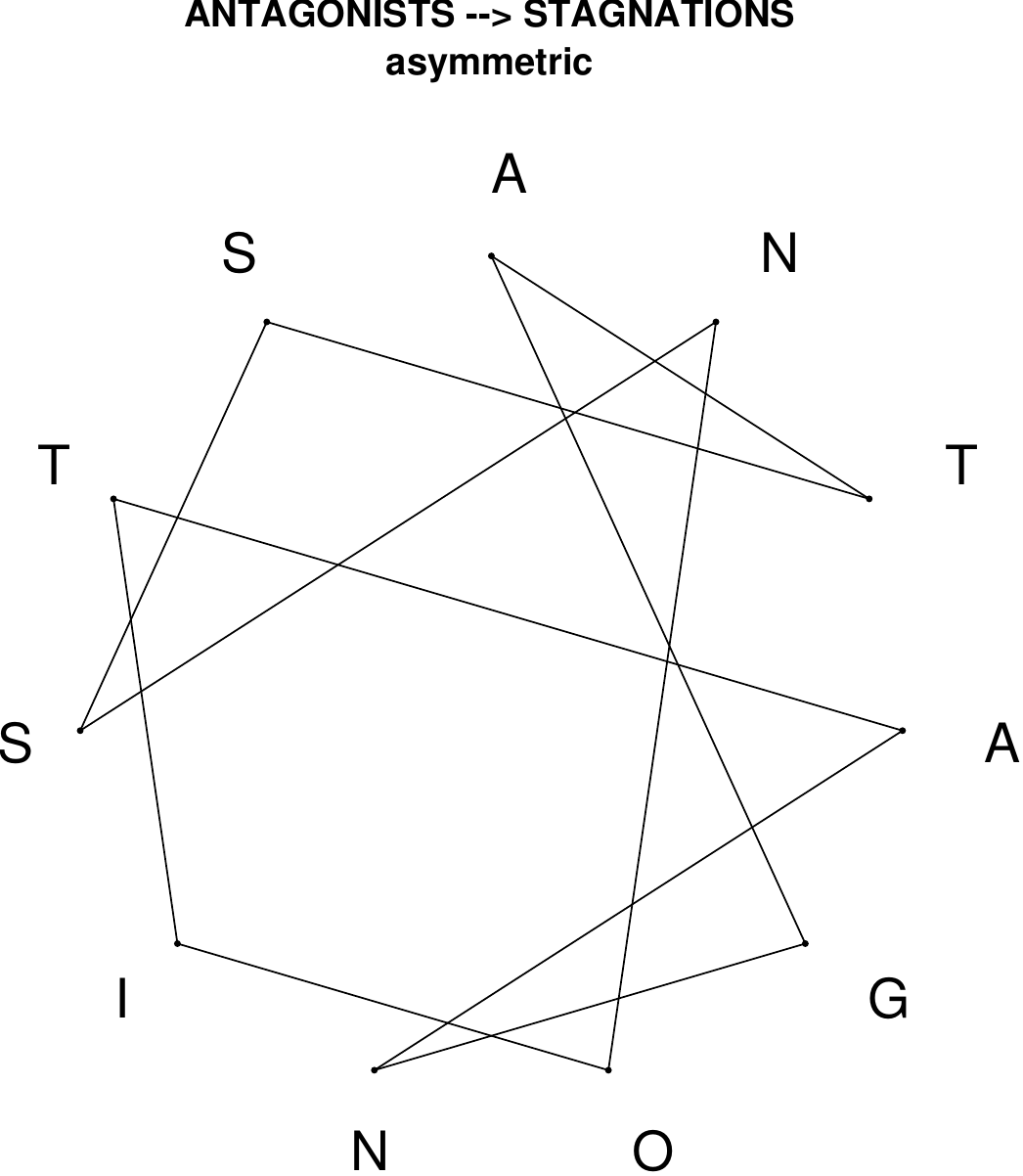}
\end{subfigure}
\hfill
\begin{subfigure}[T]{0.19\textwidth}
\centering
\includegraphics[width=\textwidth]{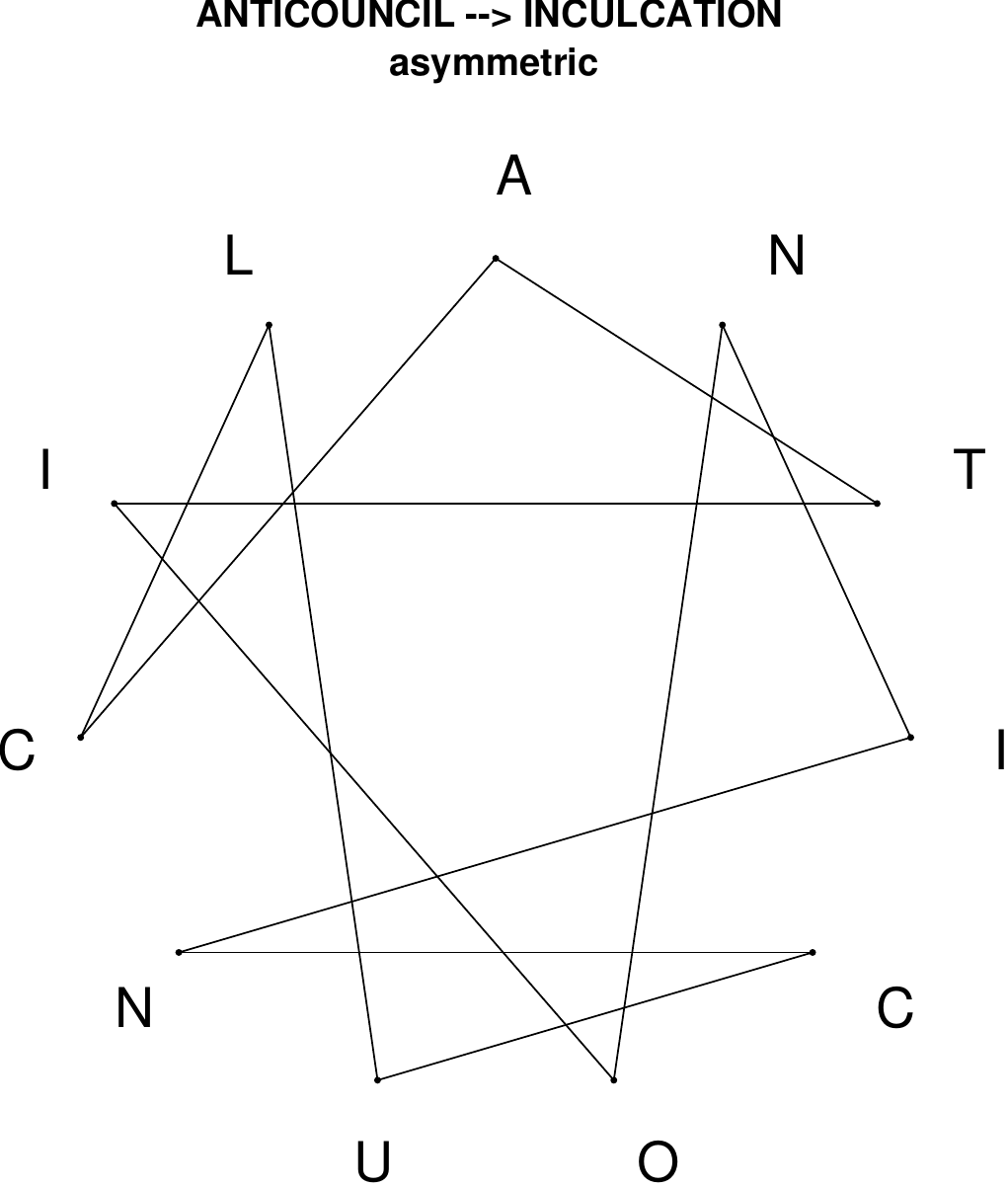}
\end{subfigure}
\hfill
\begin{subfigure}[T]{0.19\textwidth}
\centering
\includegraphics[width=\textwidth]{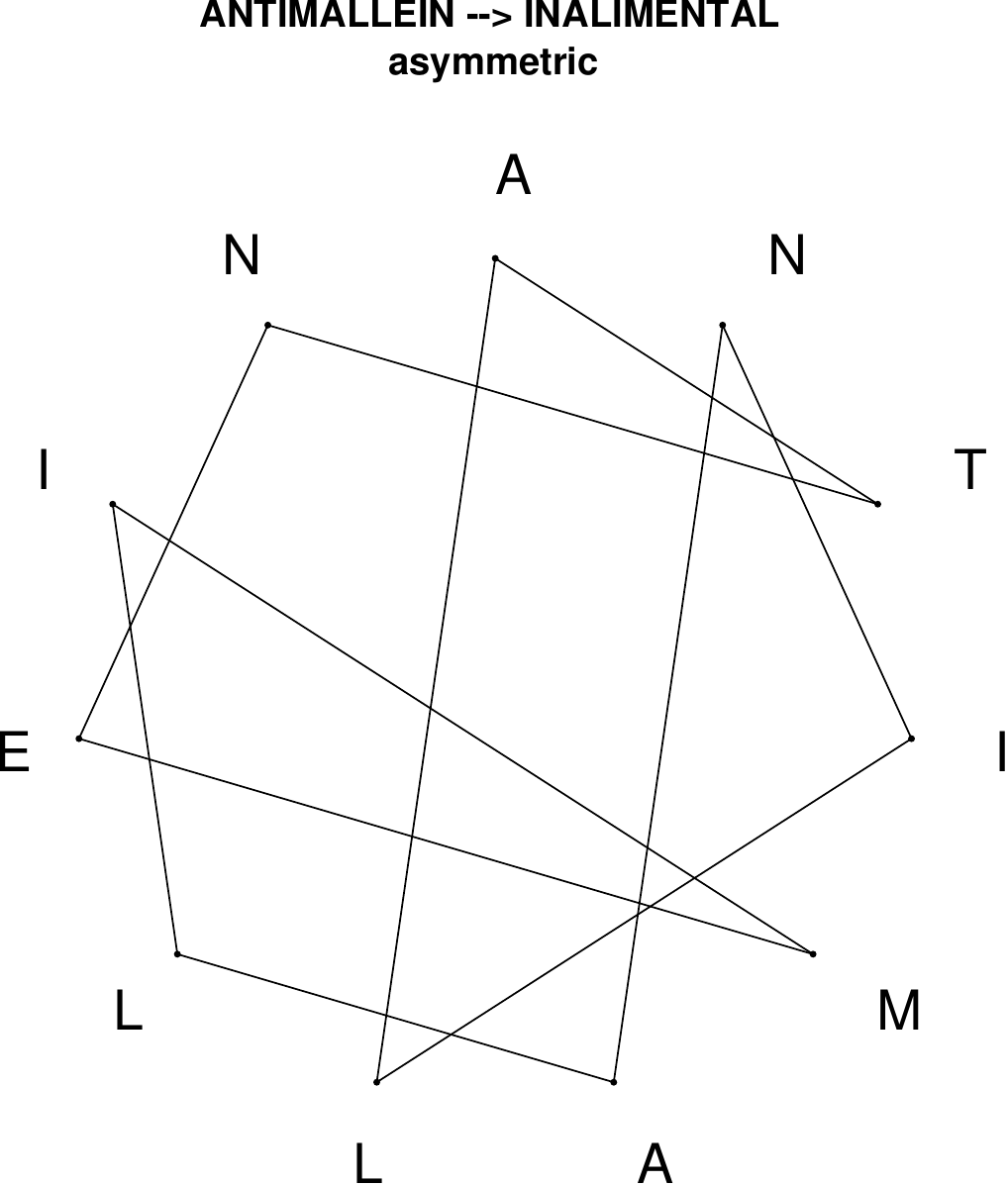}
\end{subfigure}
\hfill
\begin{subfigure}[T]{0.19\textwidth}
\centering
\includegraphics[width=\textwidth]{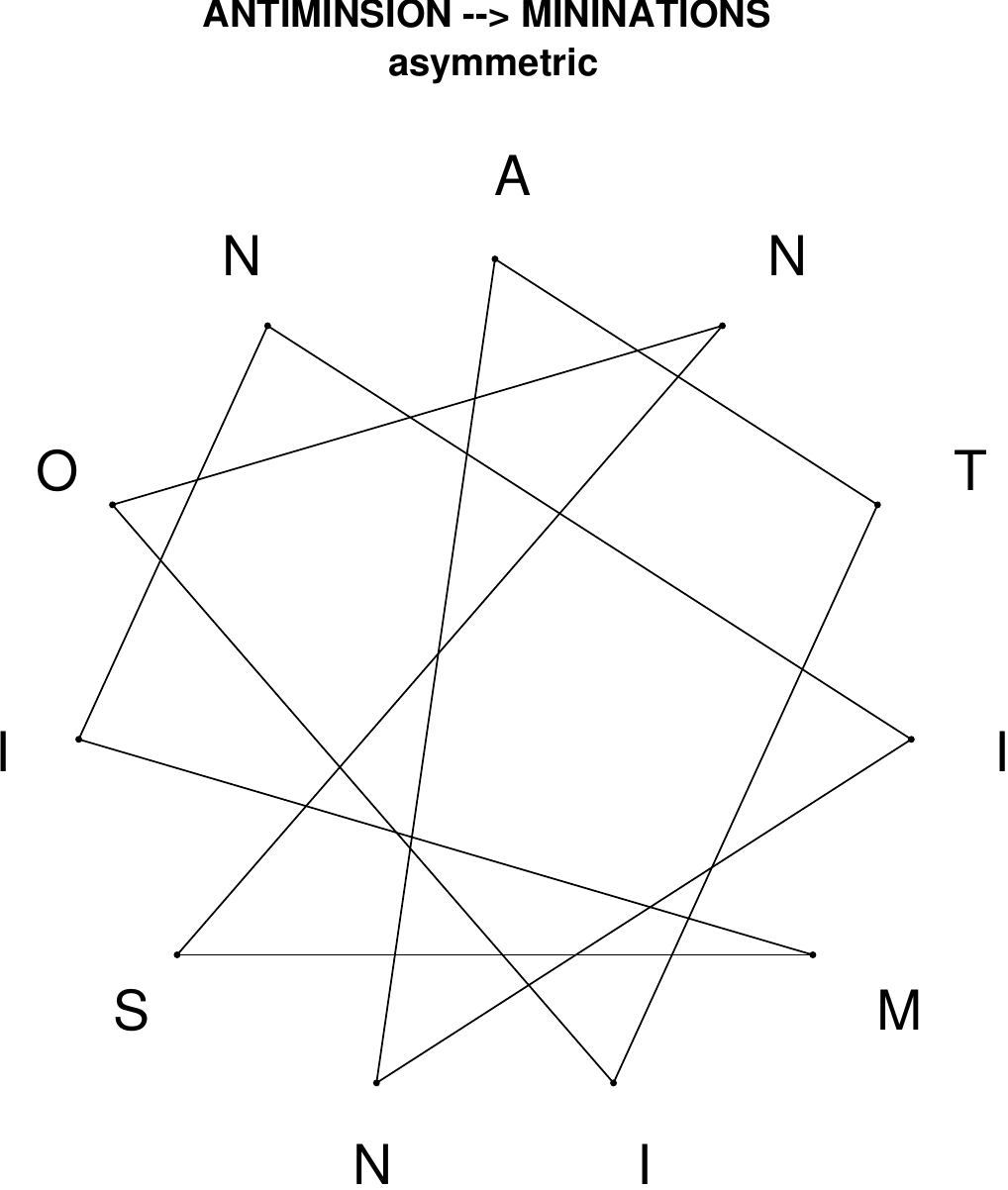}
\end{subfigure}
\end{figure}

\begin{figure}[H]
\centering
\begin{subfigure}[T]{0.19\textwidth}
\centering
\includegraphics[width=\textwidth]{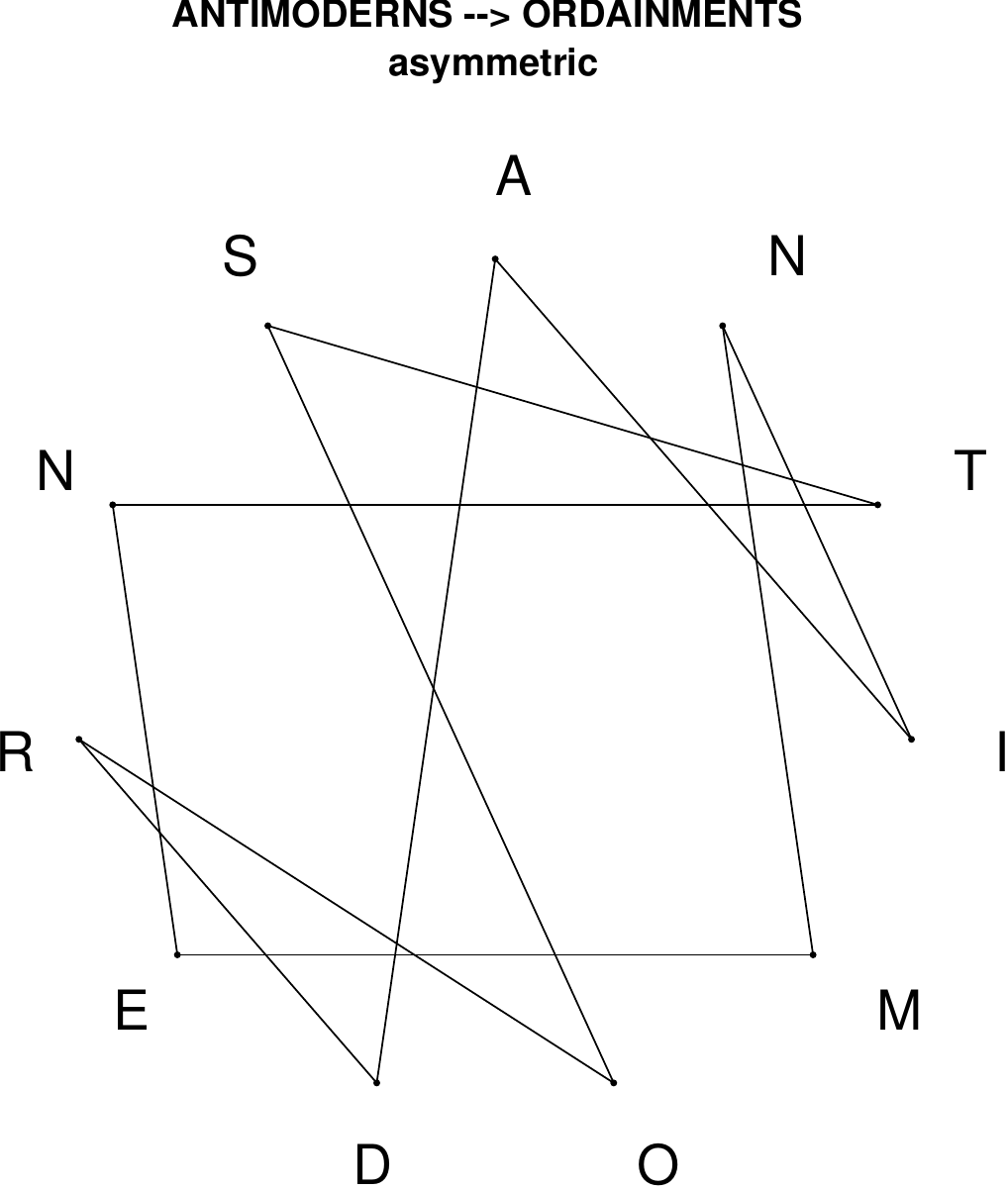}
\end{subfigure}
\hfill
\begin{subfigure}[T]{0.19\textwidth}
\centering
\includegraphics[width=\textwidth]{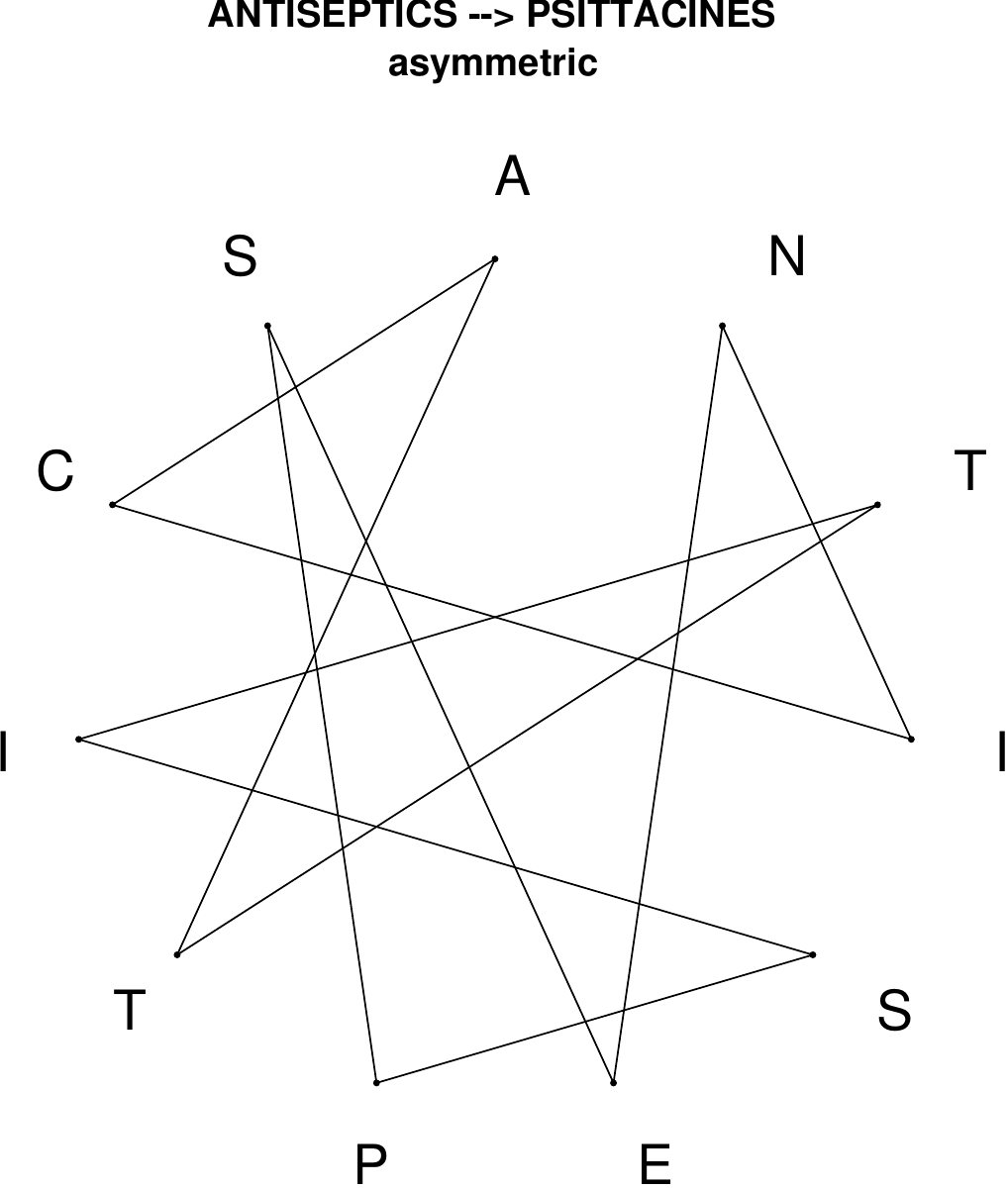}
\end{subfigure}
\hfill
\begin{subfigure}[T]{0.19\textwidth}
\centering
\includegraphics[width=\textwidth]{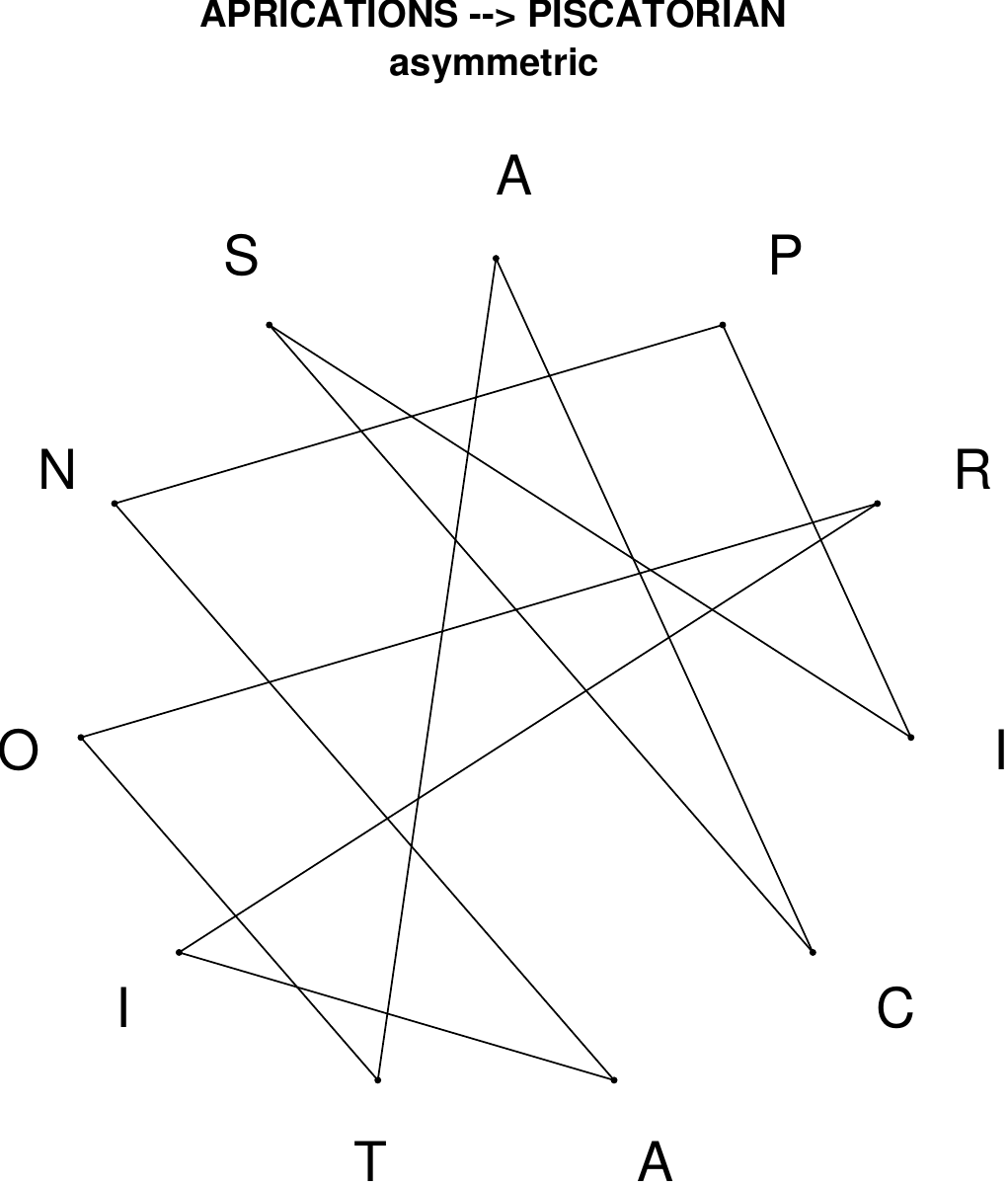}
\end{subfigure}
\hfill
\begin{subfigure}[T]{0.19\textwidth}
\centering
\includegraphics[width=\textwidth]{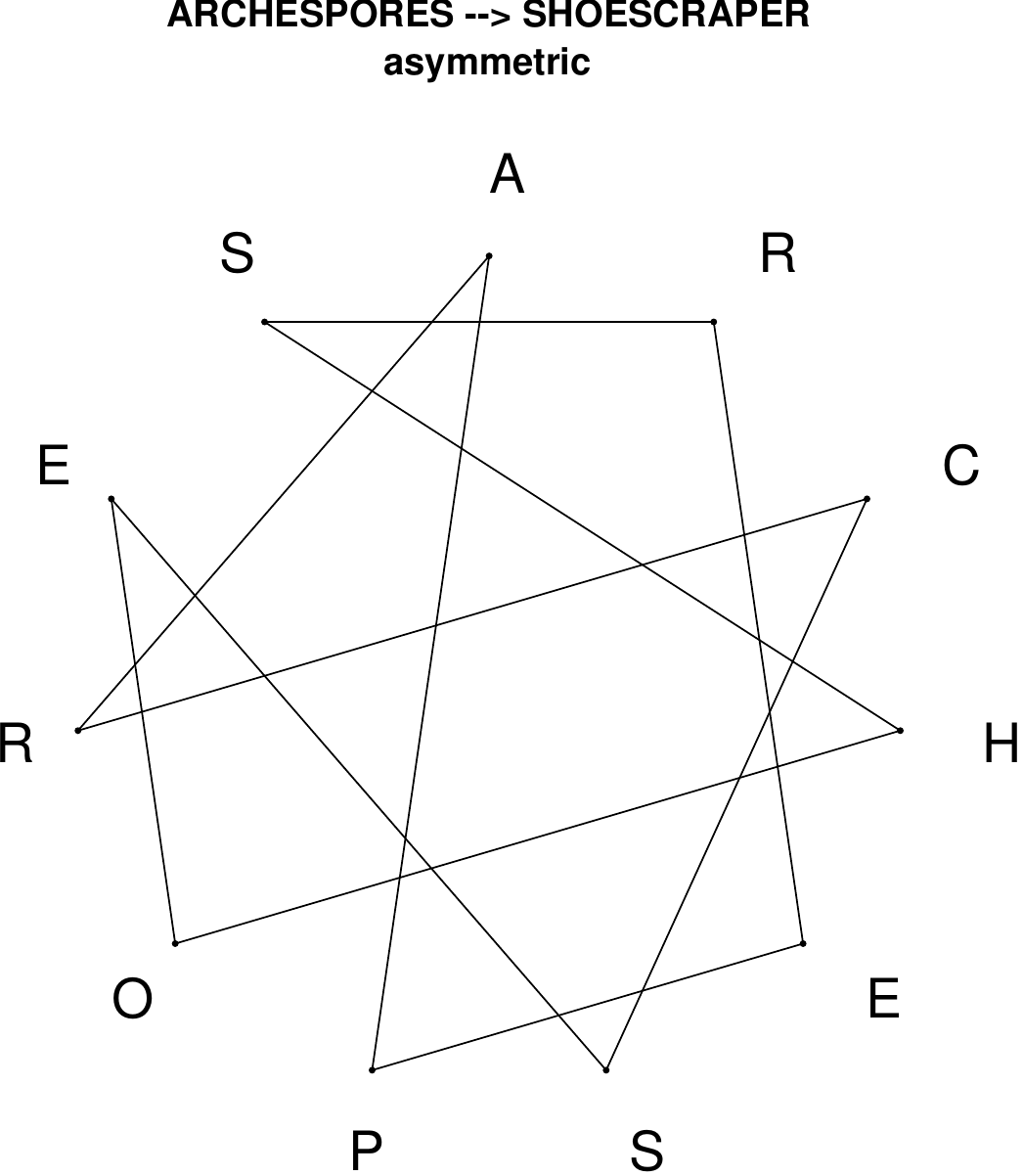}
\end{subfigure}
\hfill
\begin{subfigure}[T]{0.19\textwidth}
\centering
\includegraphics[width=\textwidth]{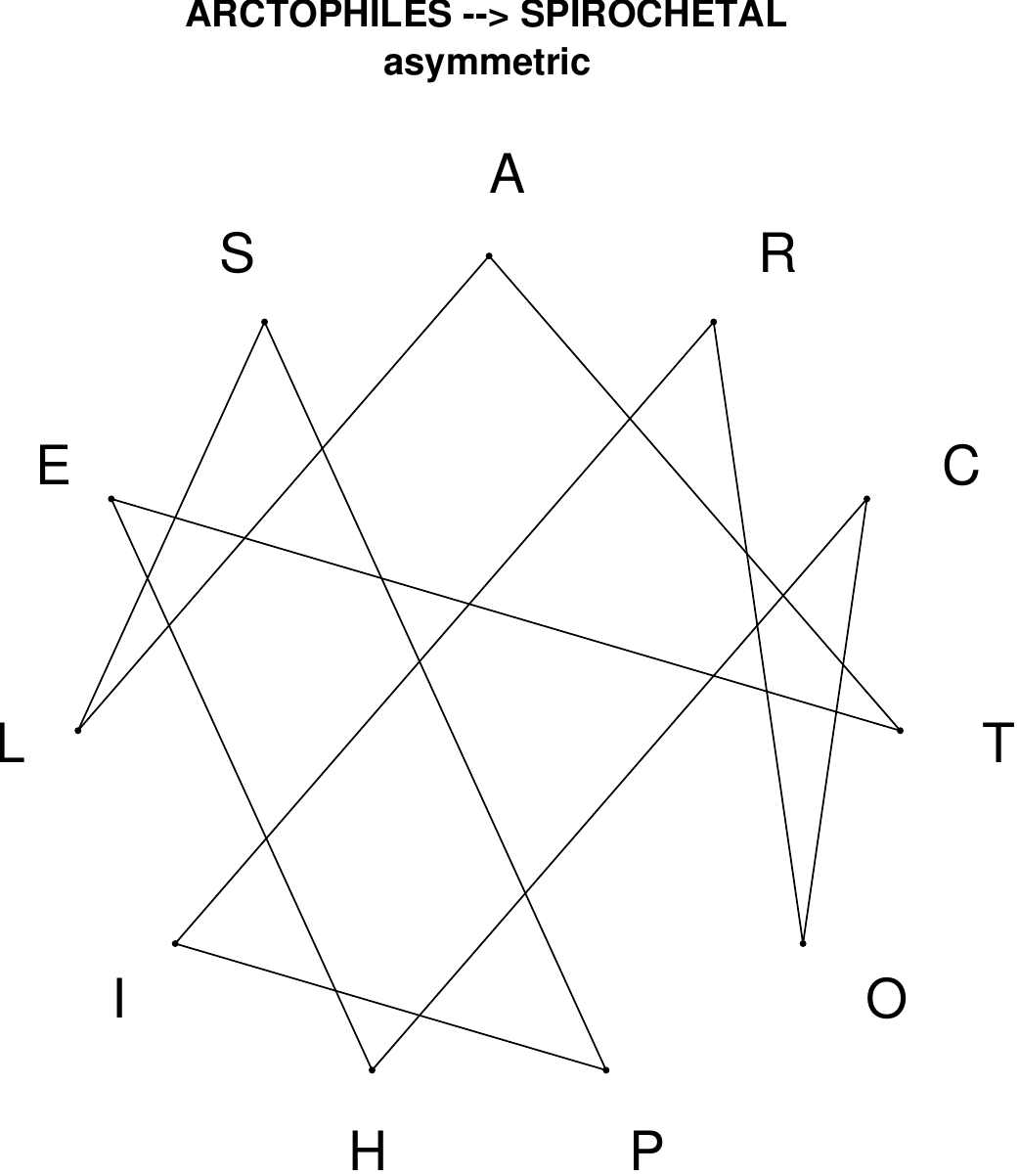}
\end{subfigure}
\end{figure}

\begin{figure}[H]
\centering
\begin{subfigure}[T]{0.19\textwidth}
\centering
\includegraphics[width=\textwidth]{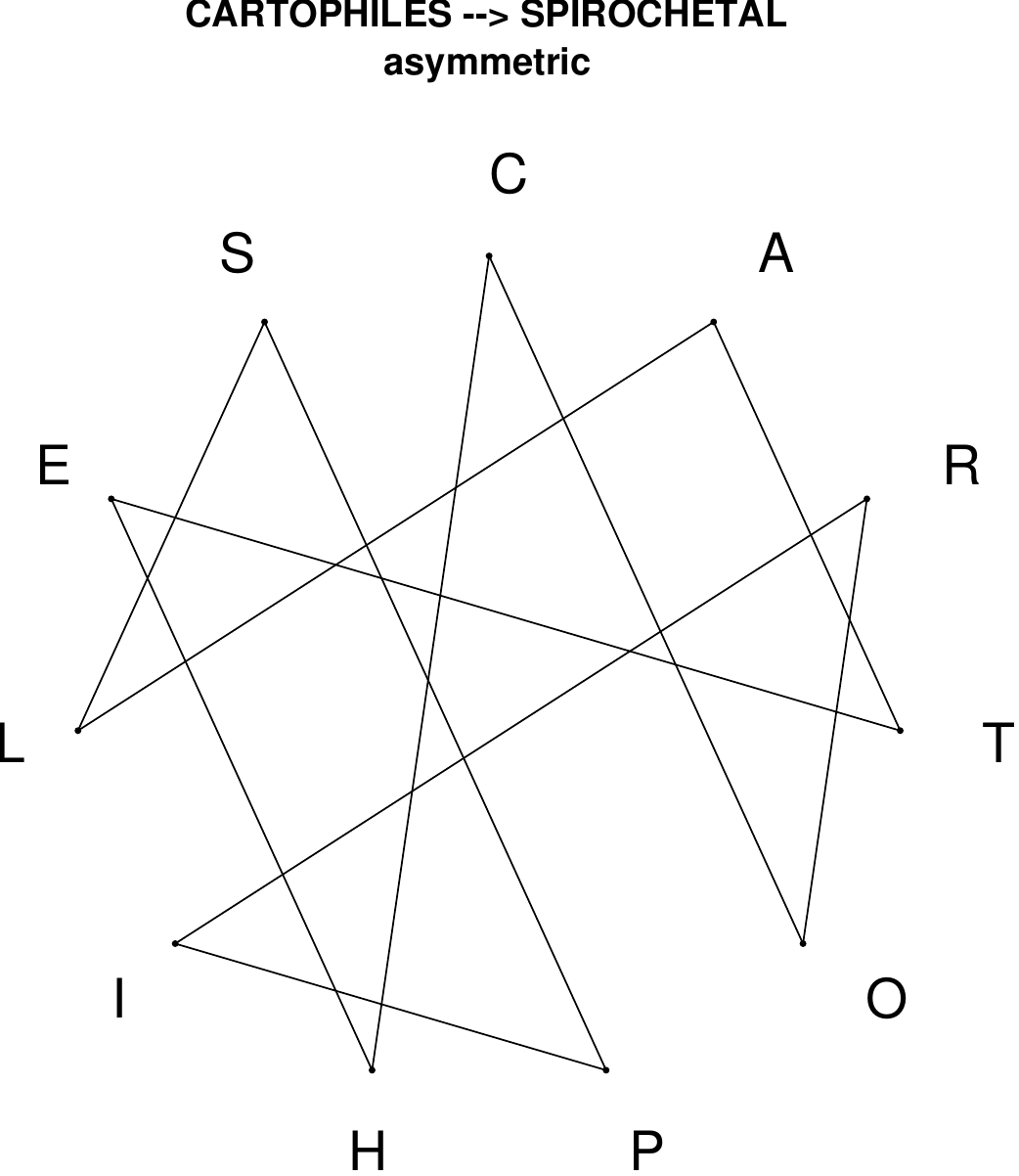}
\end{subfigure}
\hfill
\begin{subfigure}[T]{0.19\textwidth}
\centering
\includegraphics[width=\textwidth]{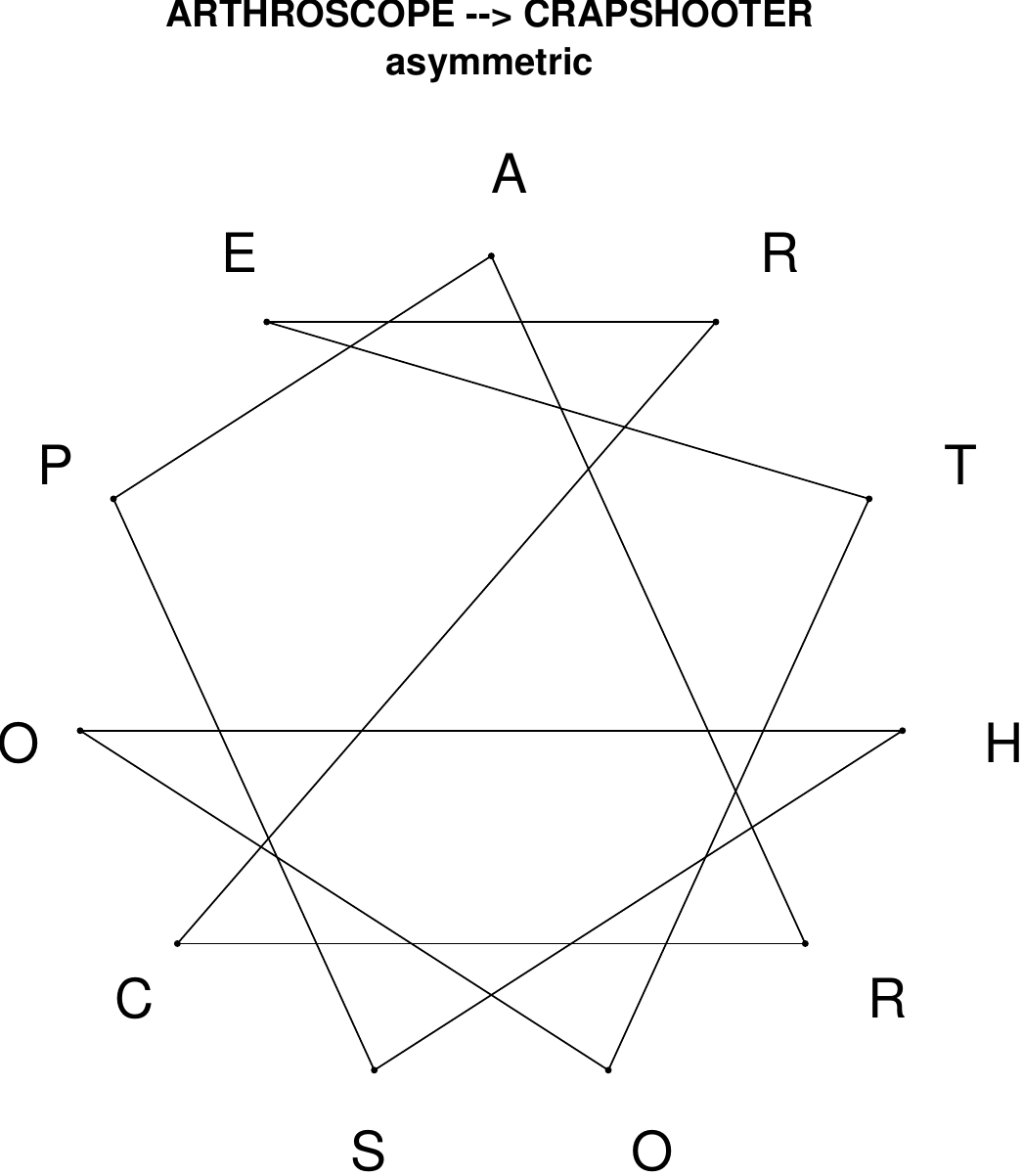}
\end{subfigure}
\hfill
\begin{subfigure}[T]{0.19\textwidth}
\centering
\includegraphics[width=\textwidth]{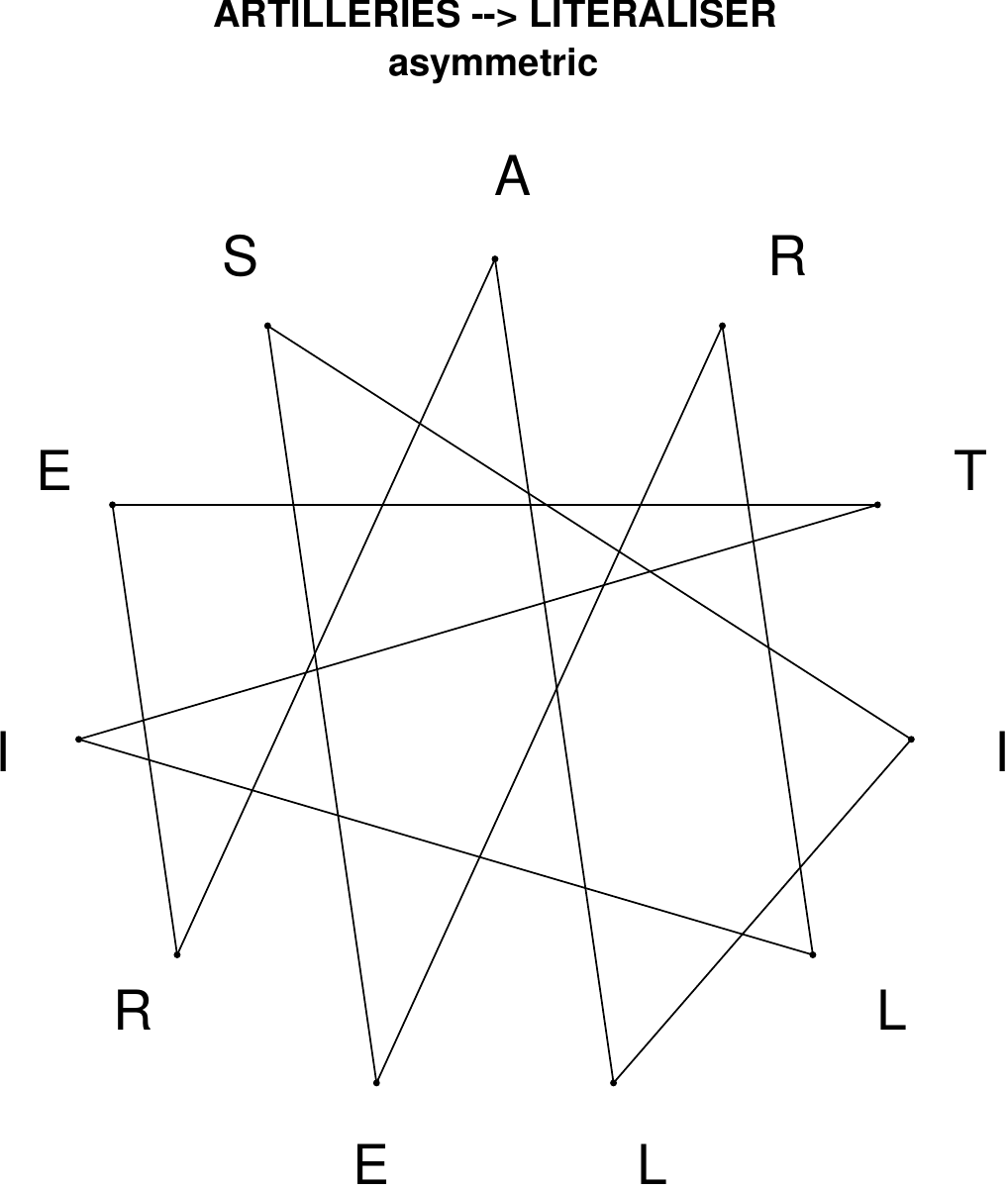}
\end{subfigure}
\hfill
\begin{subfigure}[T]{0.19\textwidth}
\centering
\includegraphics[width=\textwidth]{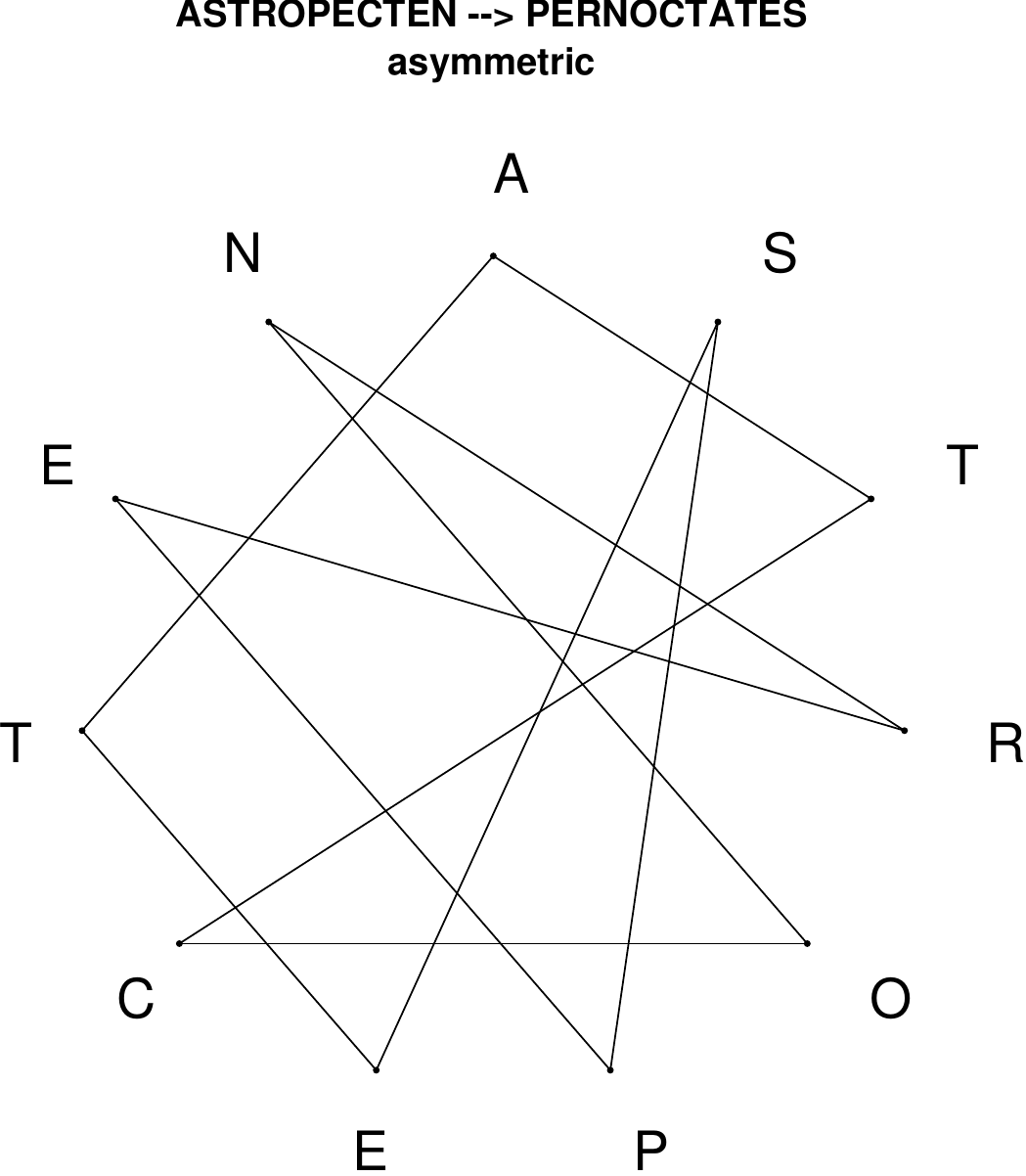}
\end{subfigure}
\hfill
\begin{subfigure}[T]{0.19\textwidth}
\centering
\includegraphics[width=\textwidth]{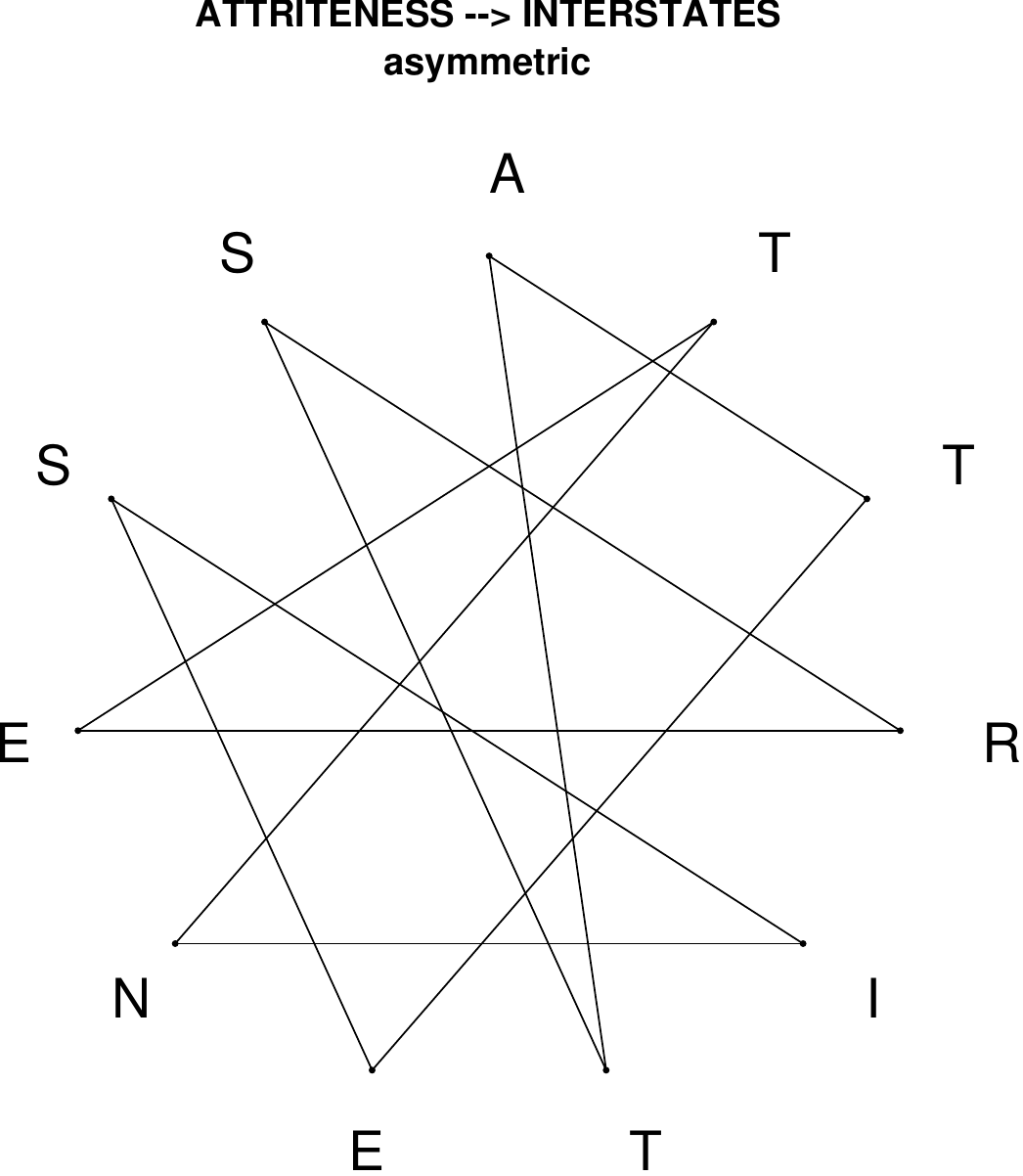}
\end{subfigure}
\end{figure}

\begin{figure}[H]
\centering
\begin{subfigure}[T]{0.19\textwidth}
\centering
\includegraphics[width=\textwidth]{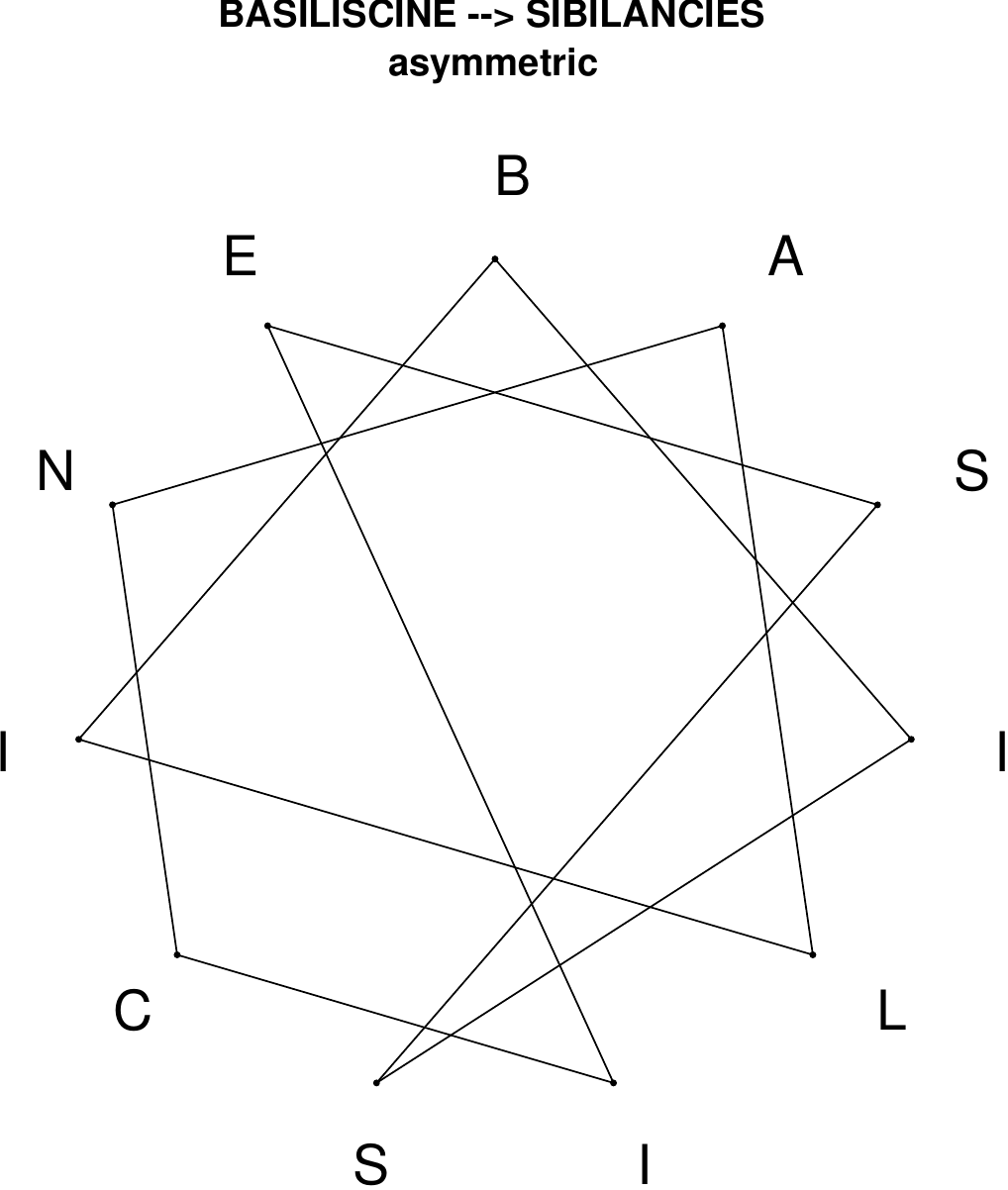}
\end{subfigure}
\hfill
\begin{subfigure}[T]{0.19\textwidth}
\centering
\includegraphics[width=\textwidth]{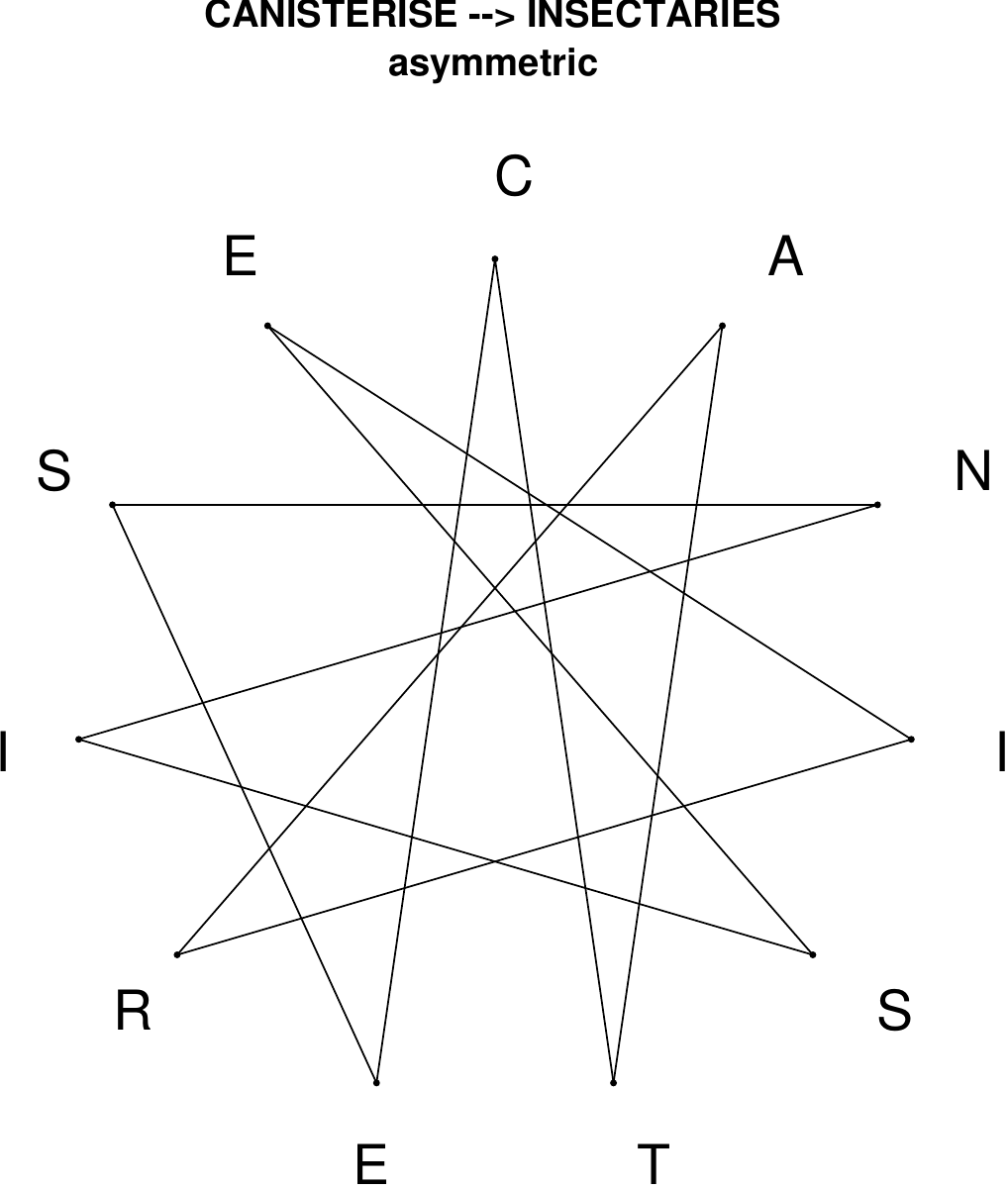}
\end{subfigure}
\hfill
\begin{subfigure}[T]{0.19\textwidth}
\centering
\includegraphics[width=\textwidth]{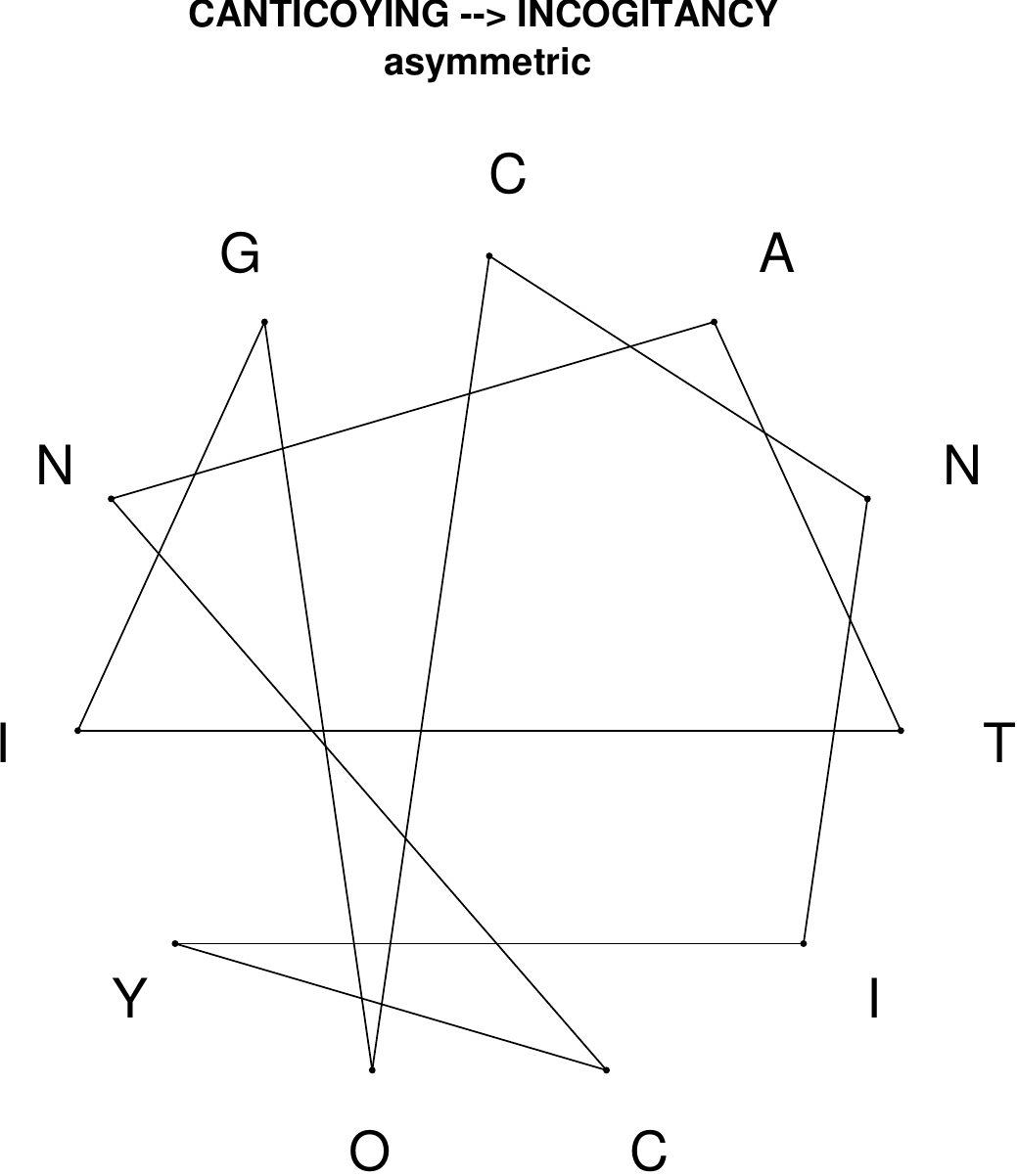}
\end{subfigure}
\hfill
\begin{subfigure}[T]{0.19\textwidth}
\centering
\includegraphics[width=\textwidth]{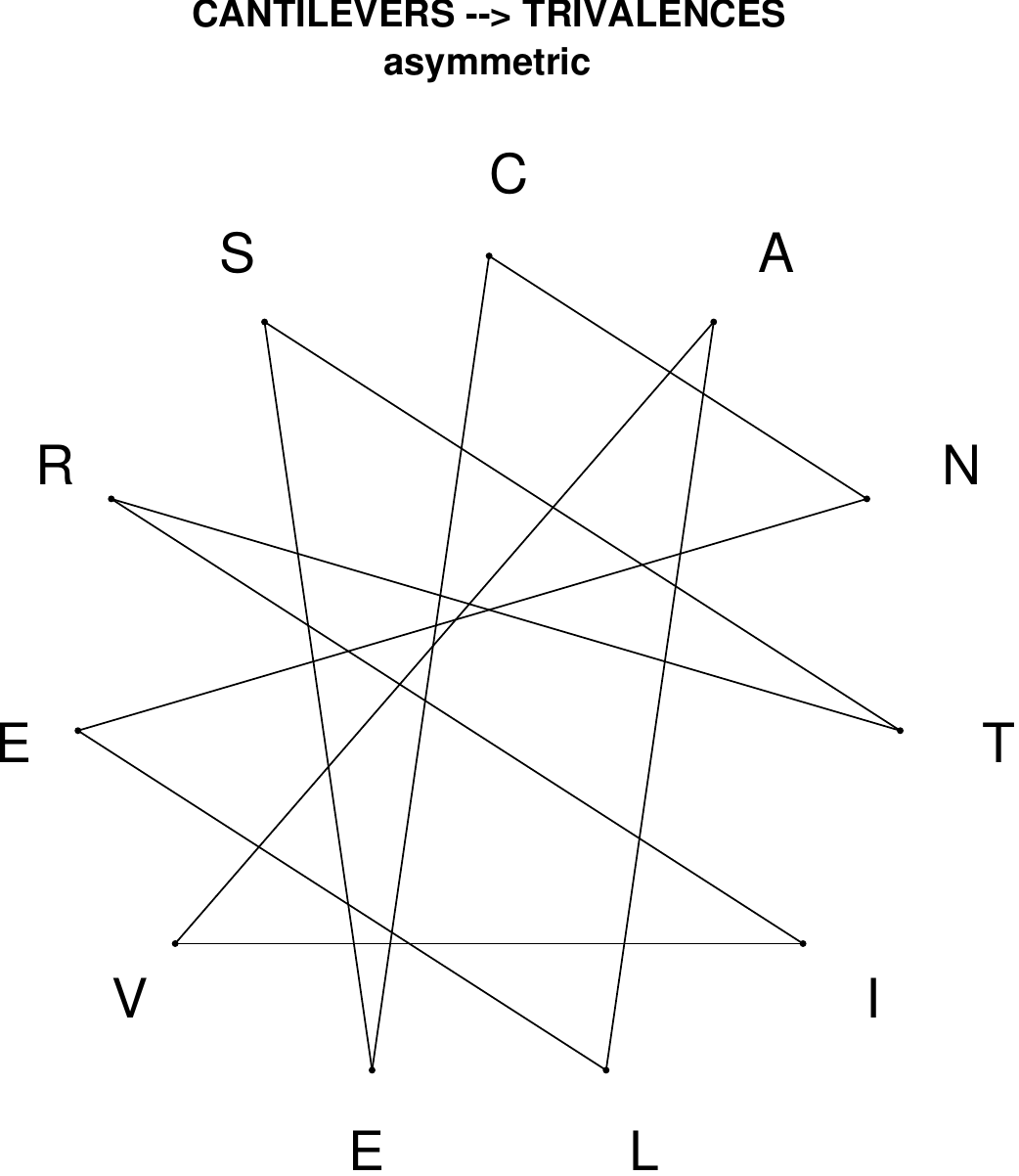}
\end{subfigure}
\hfill
\begin{subfigure}[T]{0.19\textwidth}
\centering
\includegraphics[width=\textwidth]{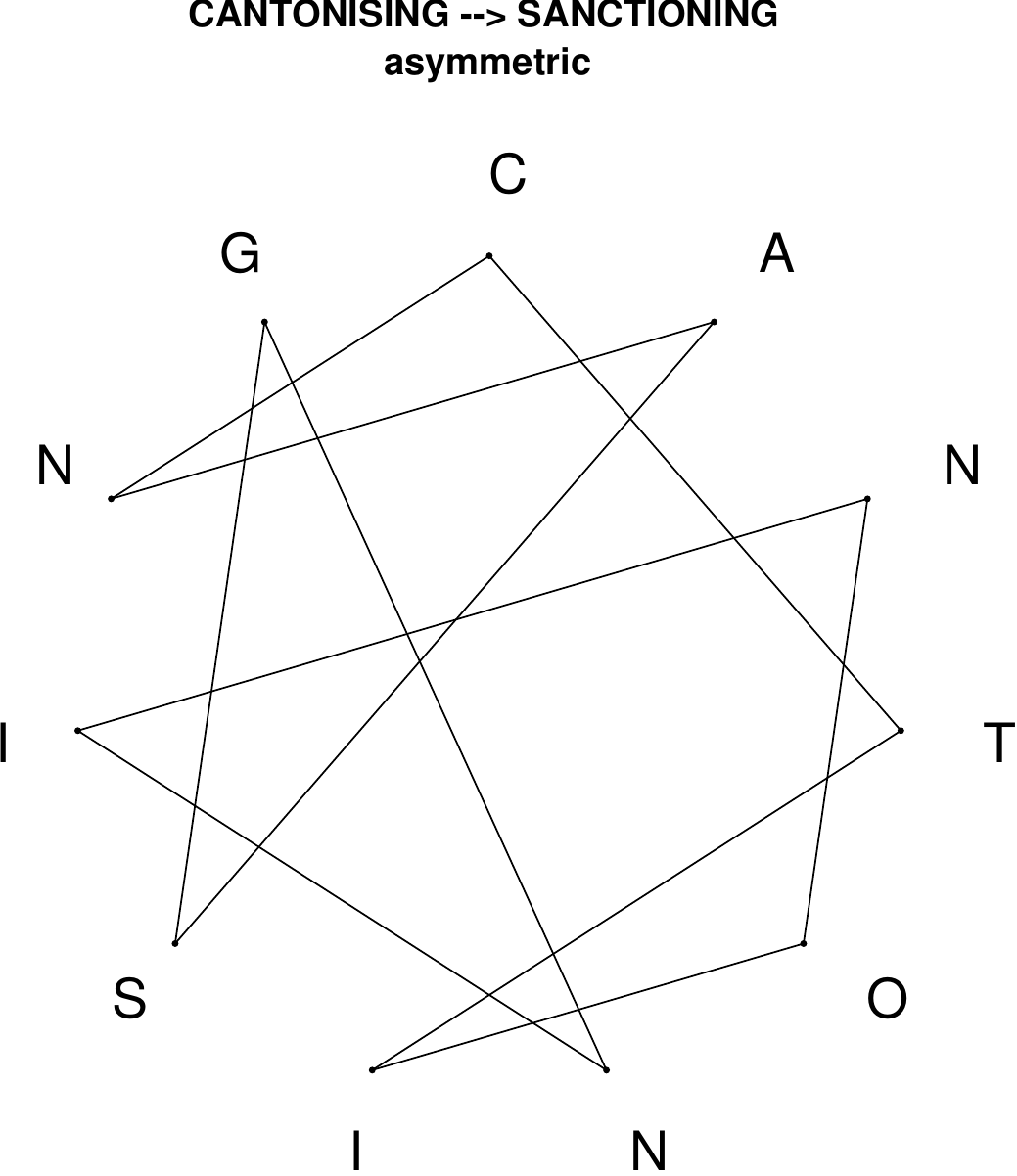}
\end{subfigure}
\end{figure}

\begin{figure}[H]
\centering
\begin{subfigure}[T]{0.19\textwidth}
\centering
\includegraphics[width=\textwidth]{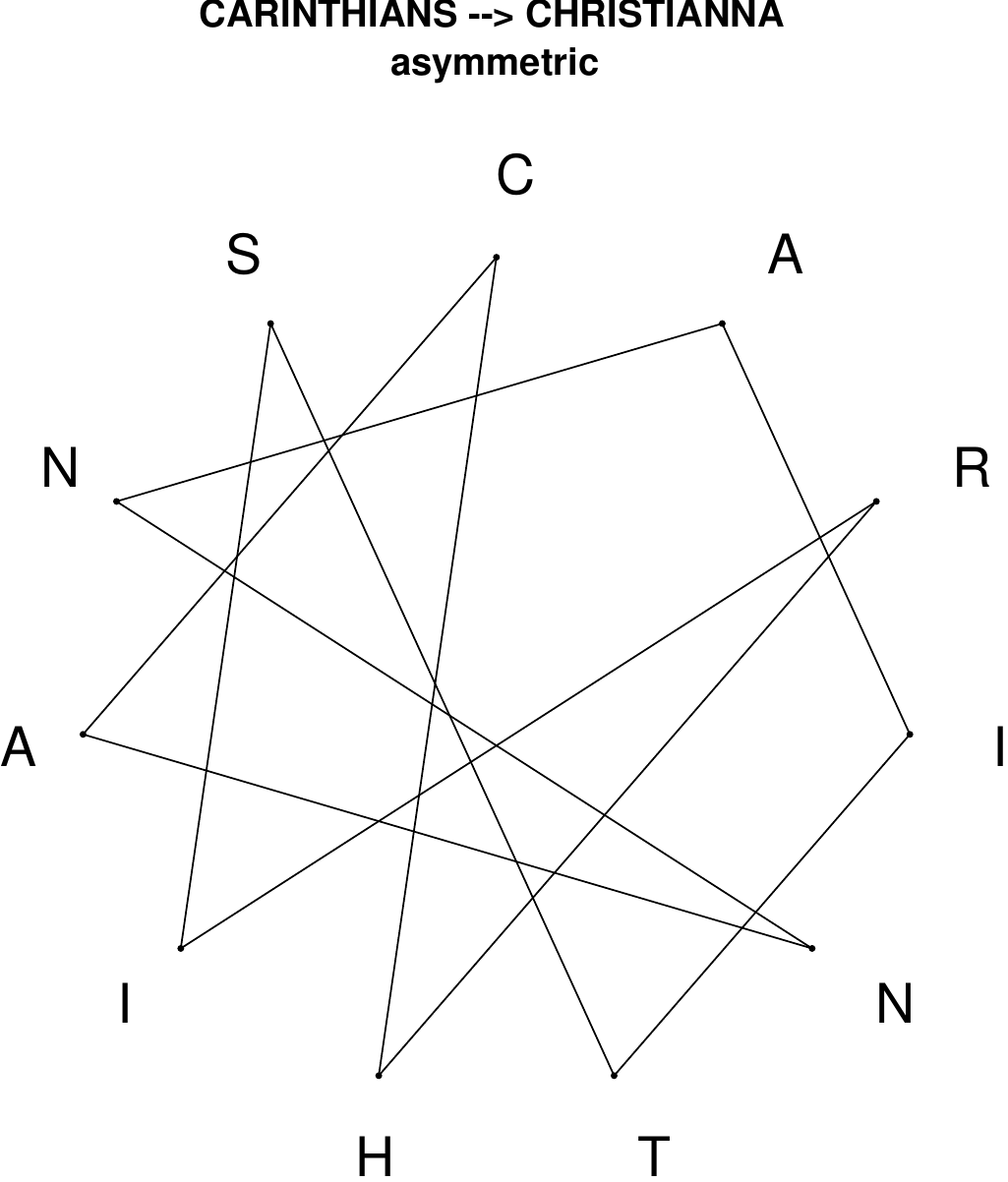}
\end{subfigure}
\hfill
\begin{subfigure}[T]{0.19\textwidth}
\centering
\includegraphics[width=\textwidth]{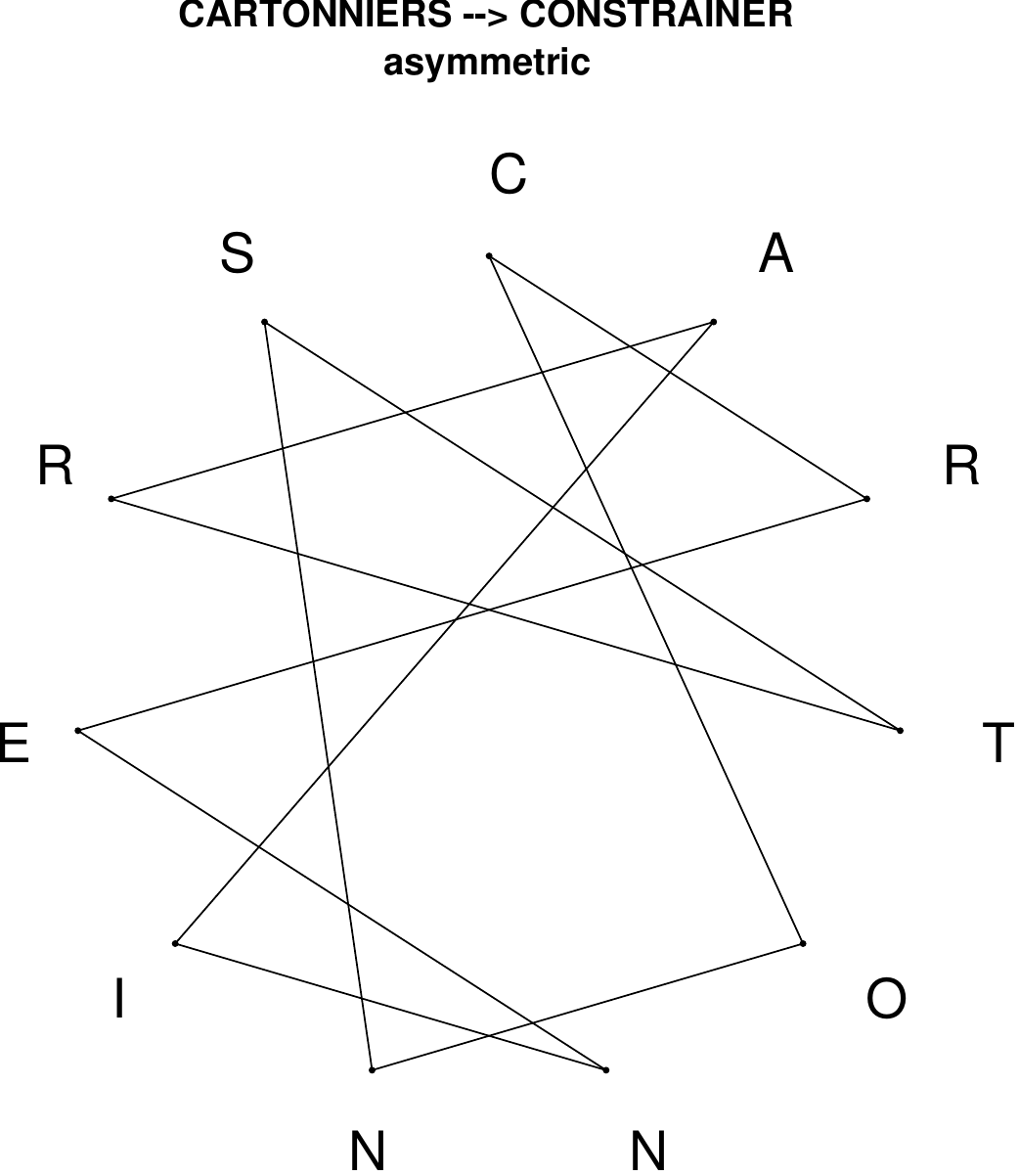}
\end{subfigure}
\hfill
\begin{subfigure}[T]{0.19\textwidth}
\centering
\includegraphics[width=\textwidth]{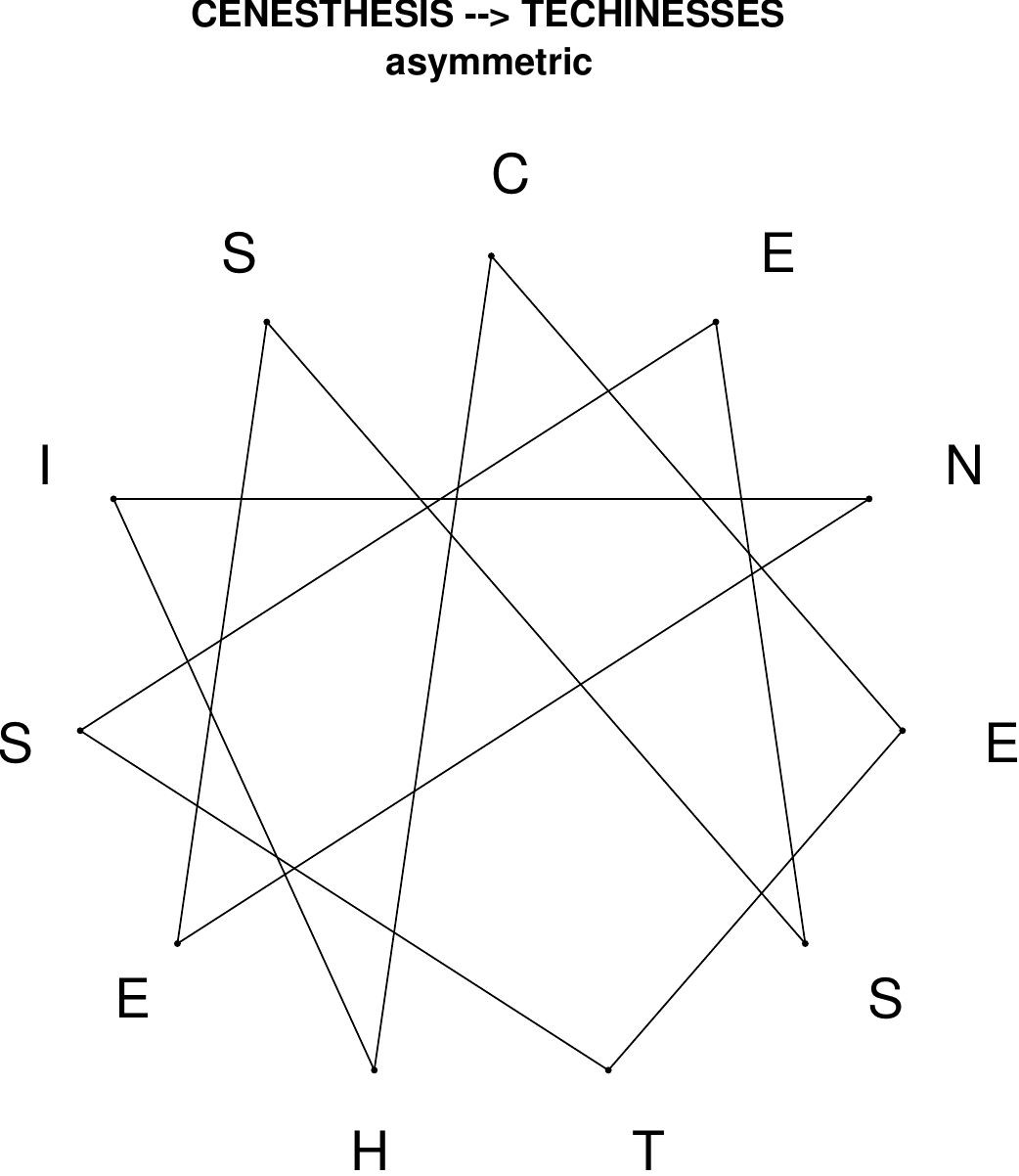}
\end{subfigure}
\hfill
\begin{subfigure}[T]{0.19\textwidth}
\centering
\includegraphics[width=\textwidth]{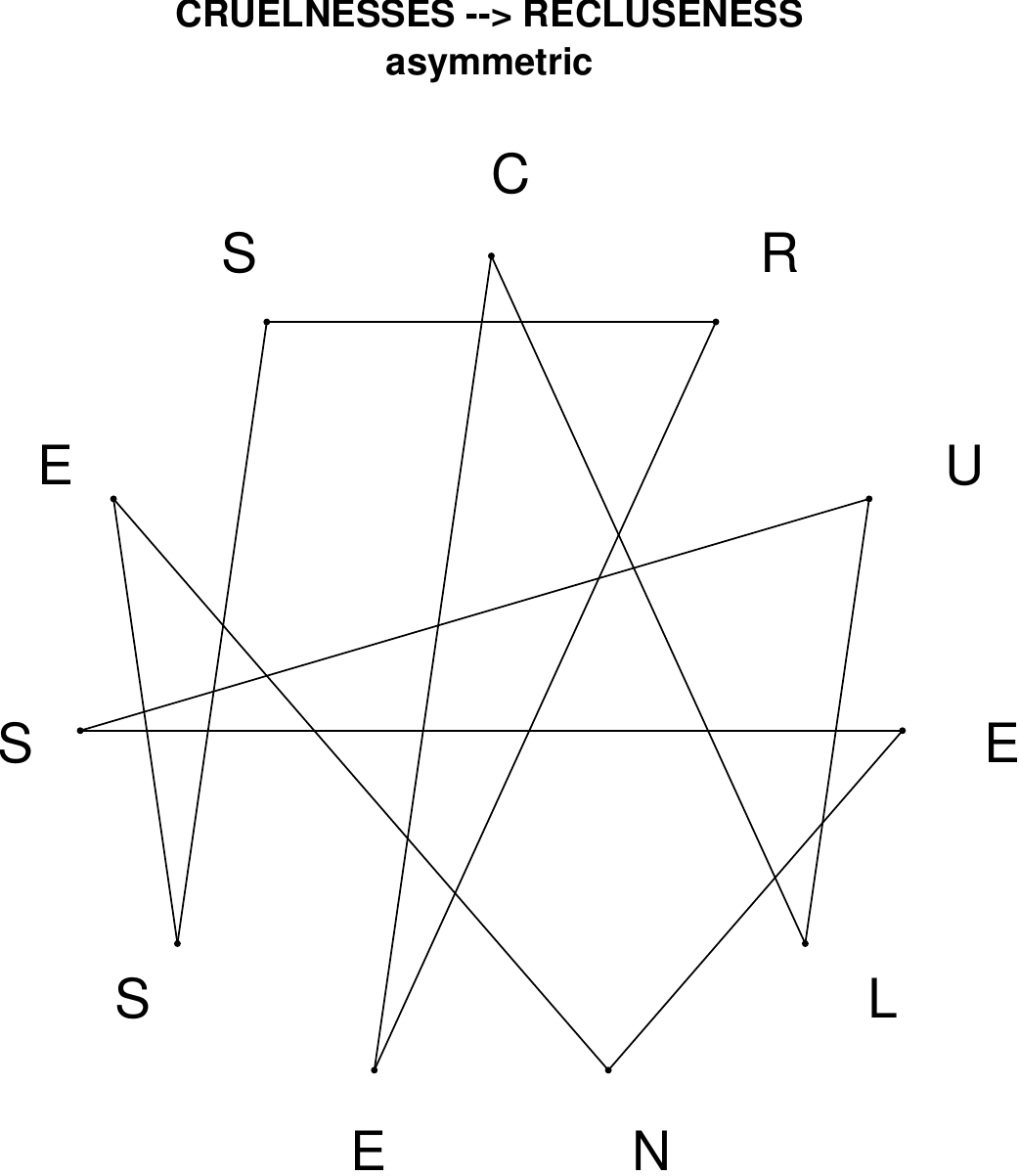}
\end{subfigure}
\hfill
\begin{subfigure}[T]{0.19\textwidth}
\centering
\includegraphics[width=\textwidth]{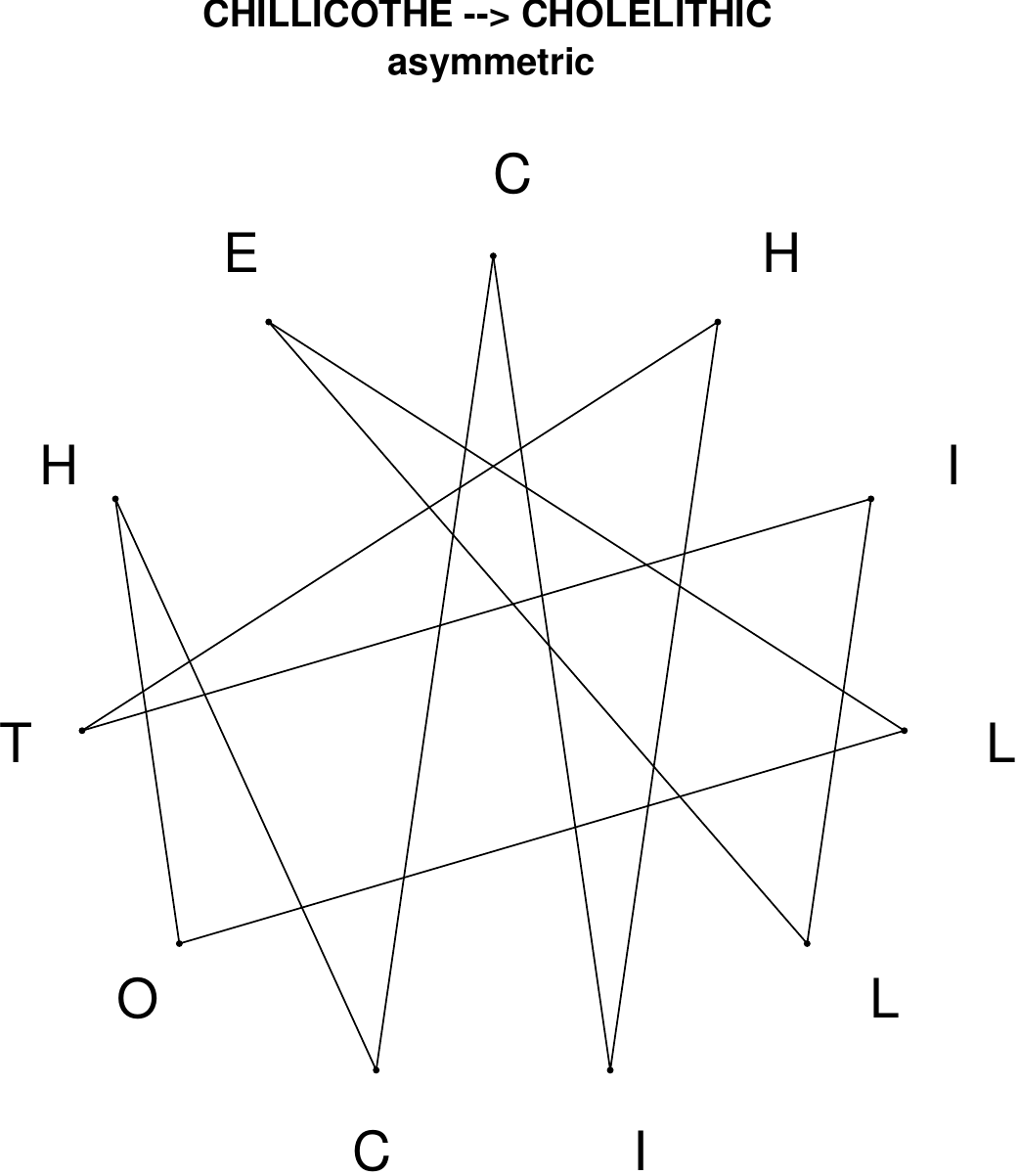}
\end{subfigure}
\end{figure}

\begin{figure}[H]
\centering
\begin{subfigure}[T]{0.19\textwidth}
\centering
\includegraphics[width=\textwidth]{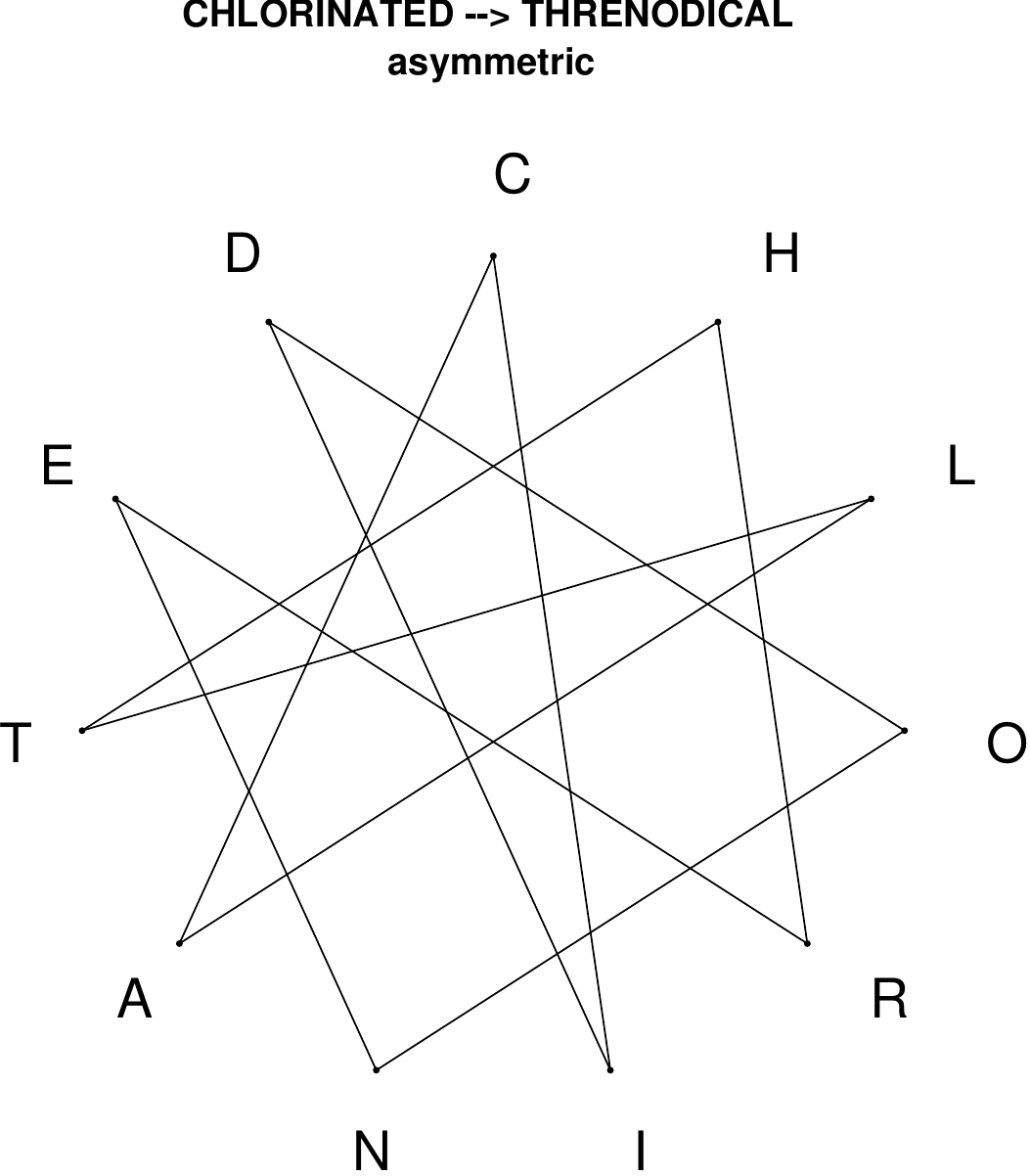}
\end{subfigure}
\hfill
\begin{subfigure}[T]{0.19\textwidth}
\centering
\includegraphics[width=\textwidth]{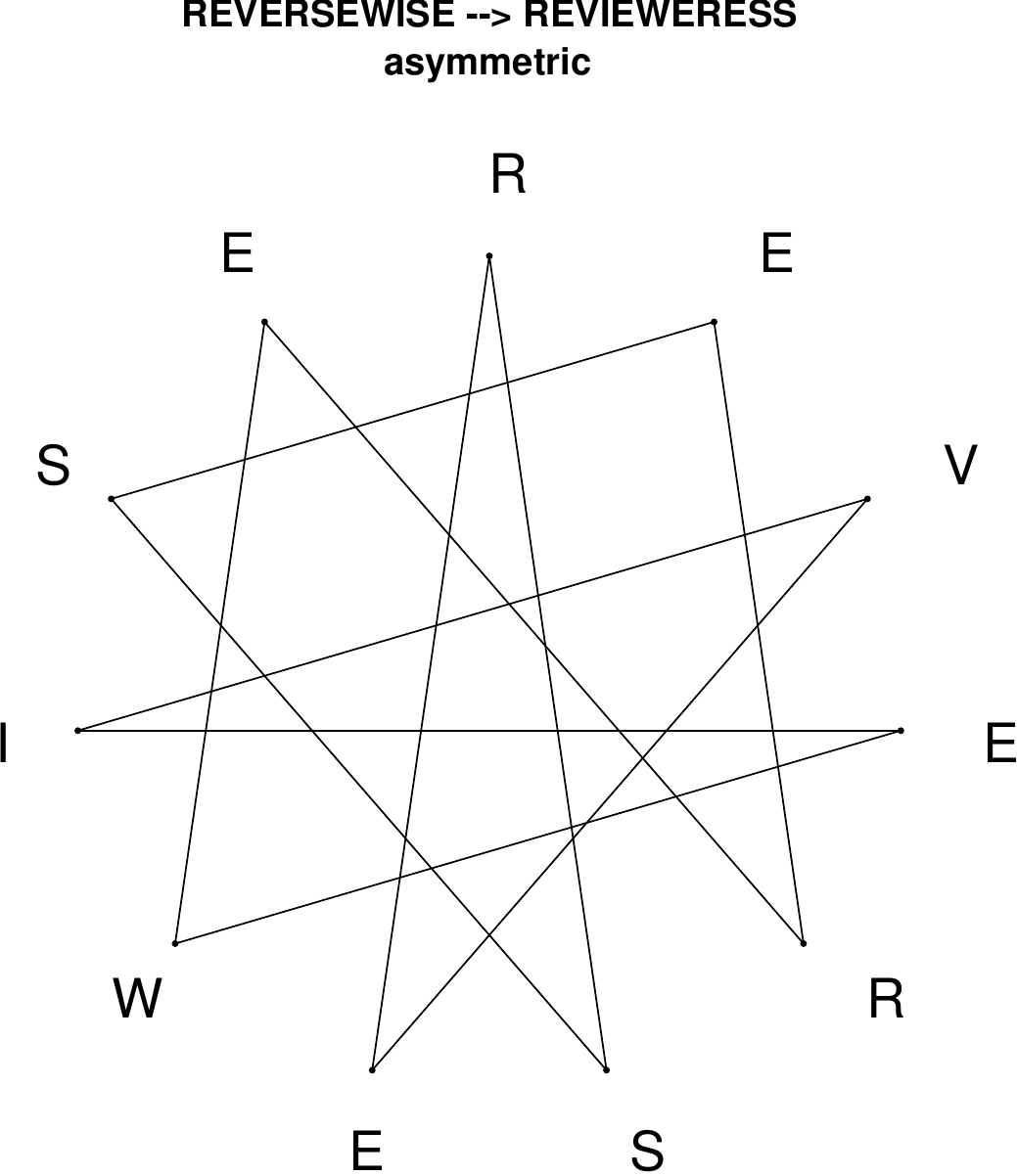}
\end{subfigure}
\hfill
\begin{subfigure}[T]{0.19\textwidth}
\centering
\includegraphics[width=\textwidth]{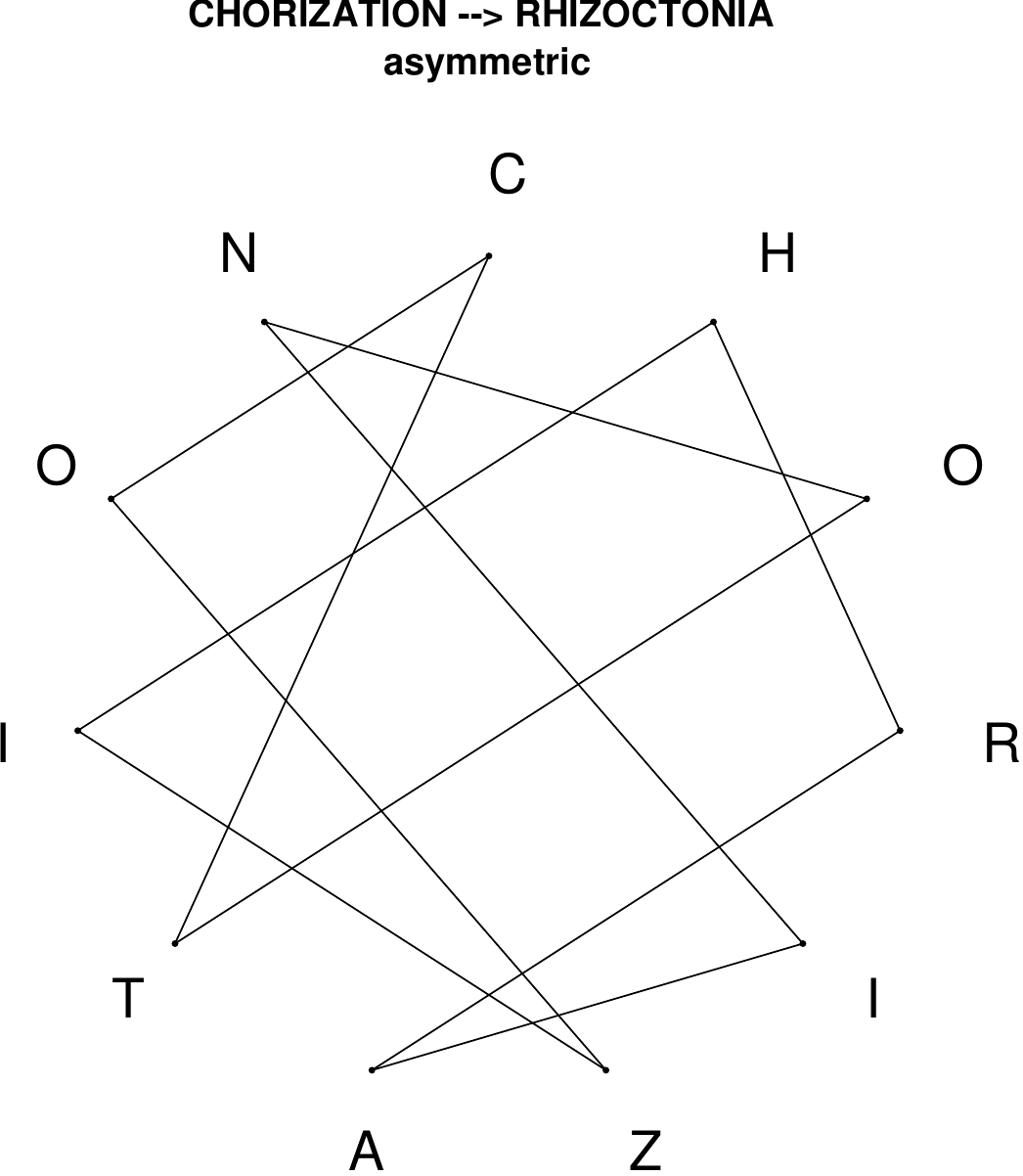}
\end{subfigure}
\hfill
\begin{subfigure}[T]{0.19\textwidth}
\centering
\includegraphics[width=\textwidth]{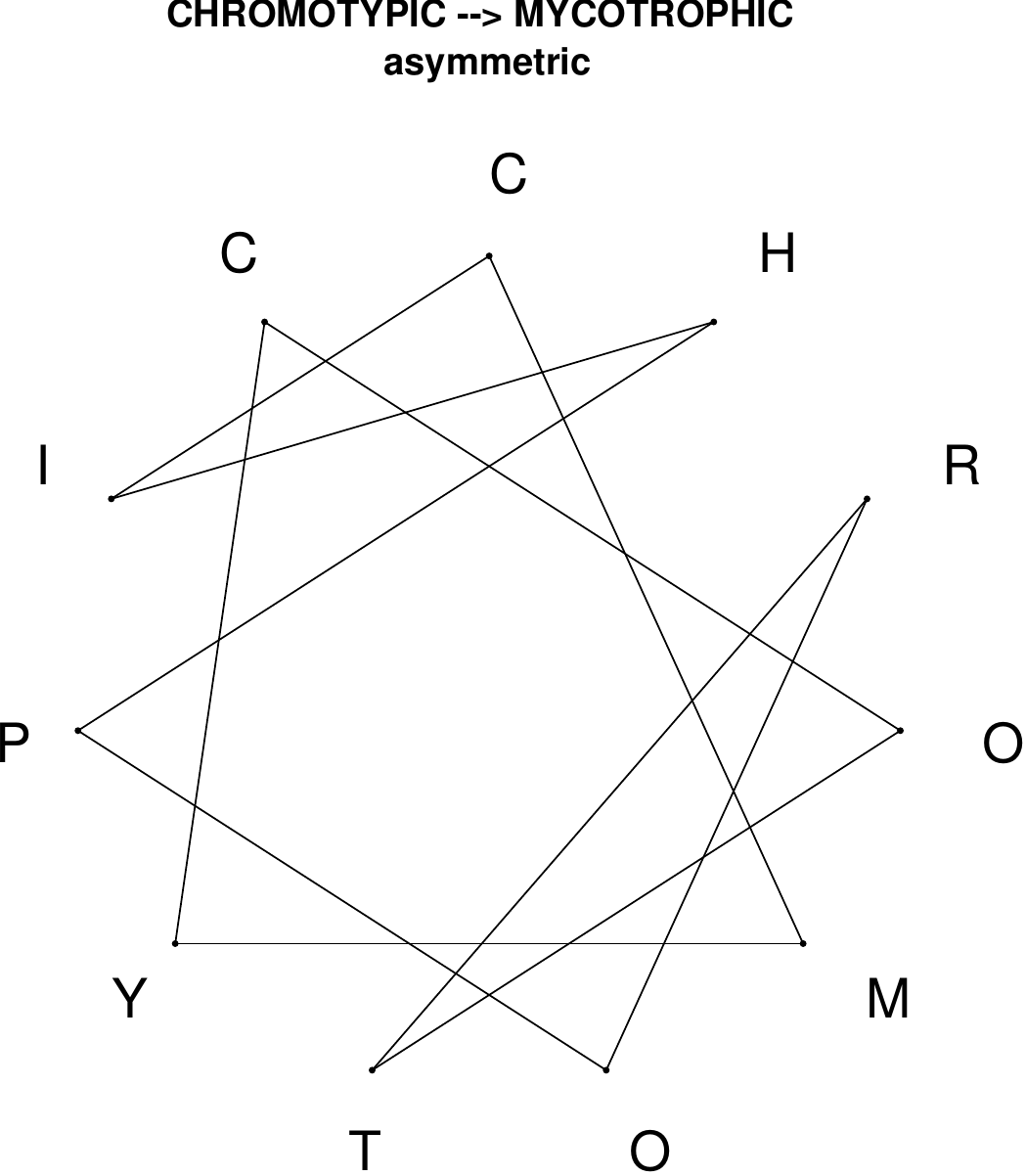}
\end{subfigure}
\hfill
\begin{subfigure}[T]{0.19\textwidth}
\centering
\includegraphics[width=\textwidth]{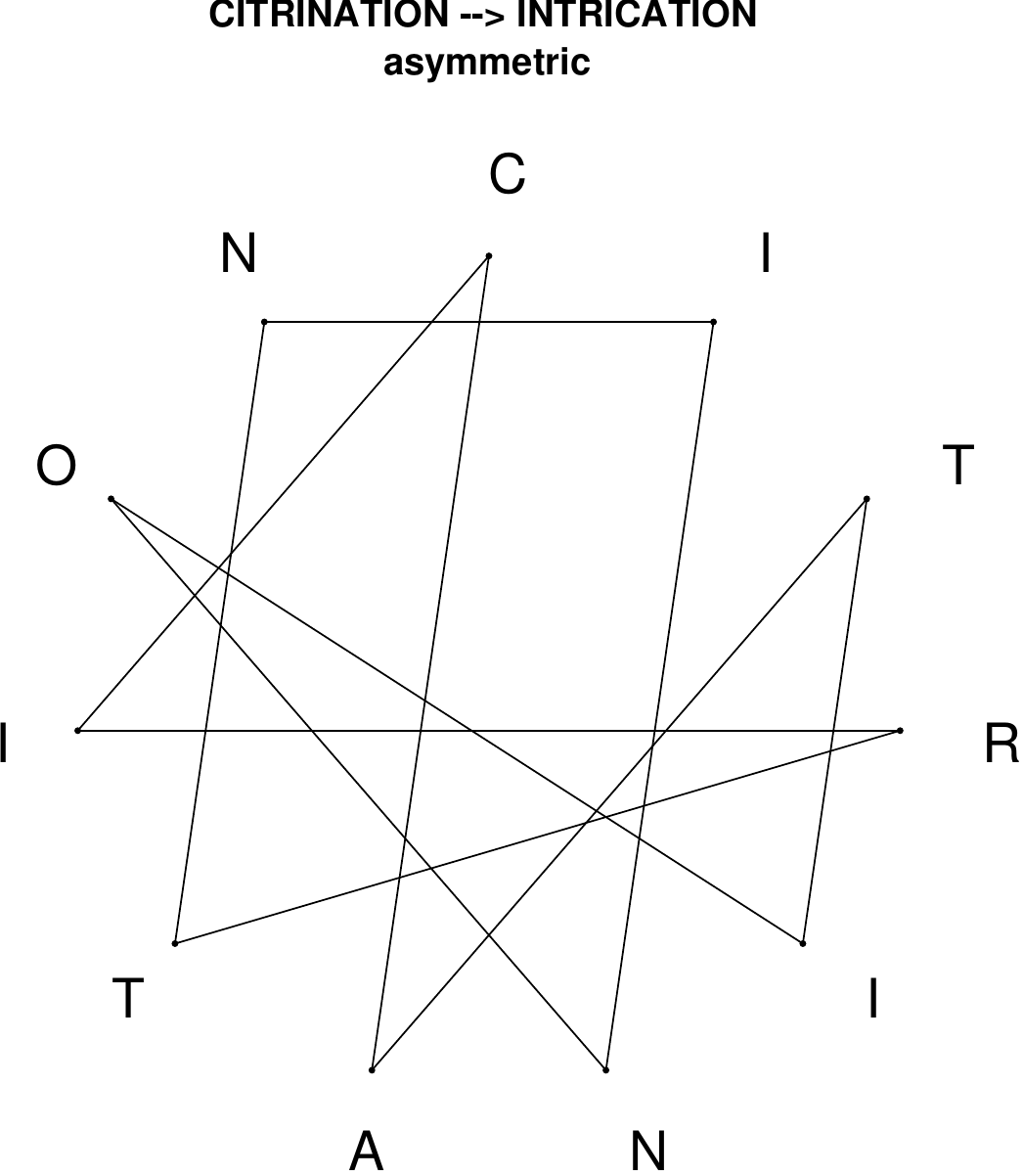}
\end{subfigure}
\end{figure}

\begin{figure}[H]
\centering
\begin{subfigure}[T]{0.19\textwidth}
\centering
\includegraphics[width=\textwidth]{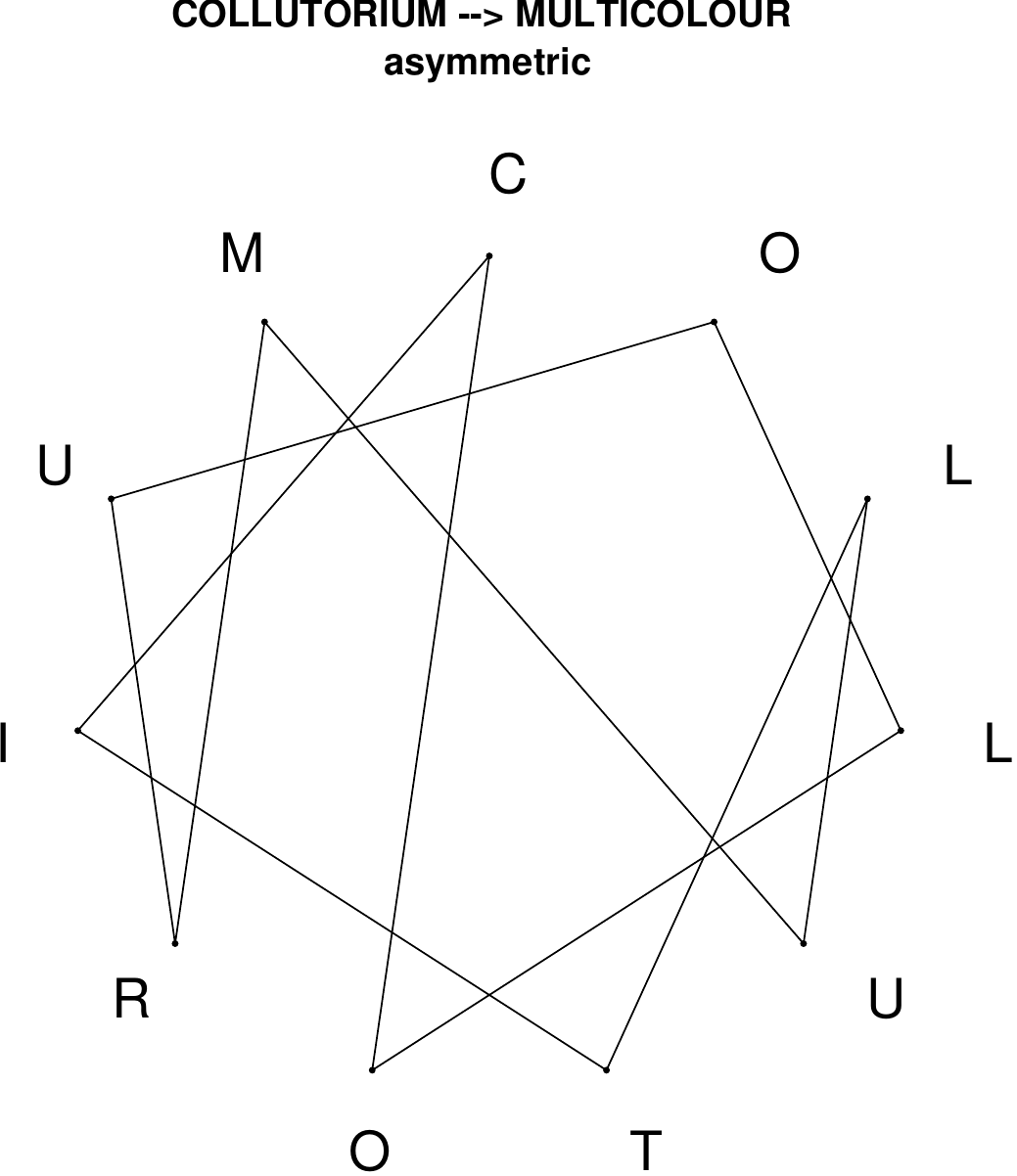}
\end{subfigure}
\hfill
\begin{subfigure}[T]{0.19\textwidth}
\centering
\includegraphics[width=\textwidth]{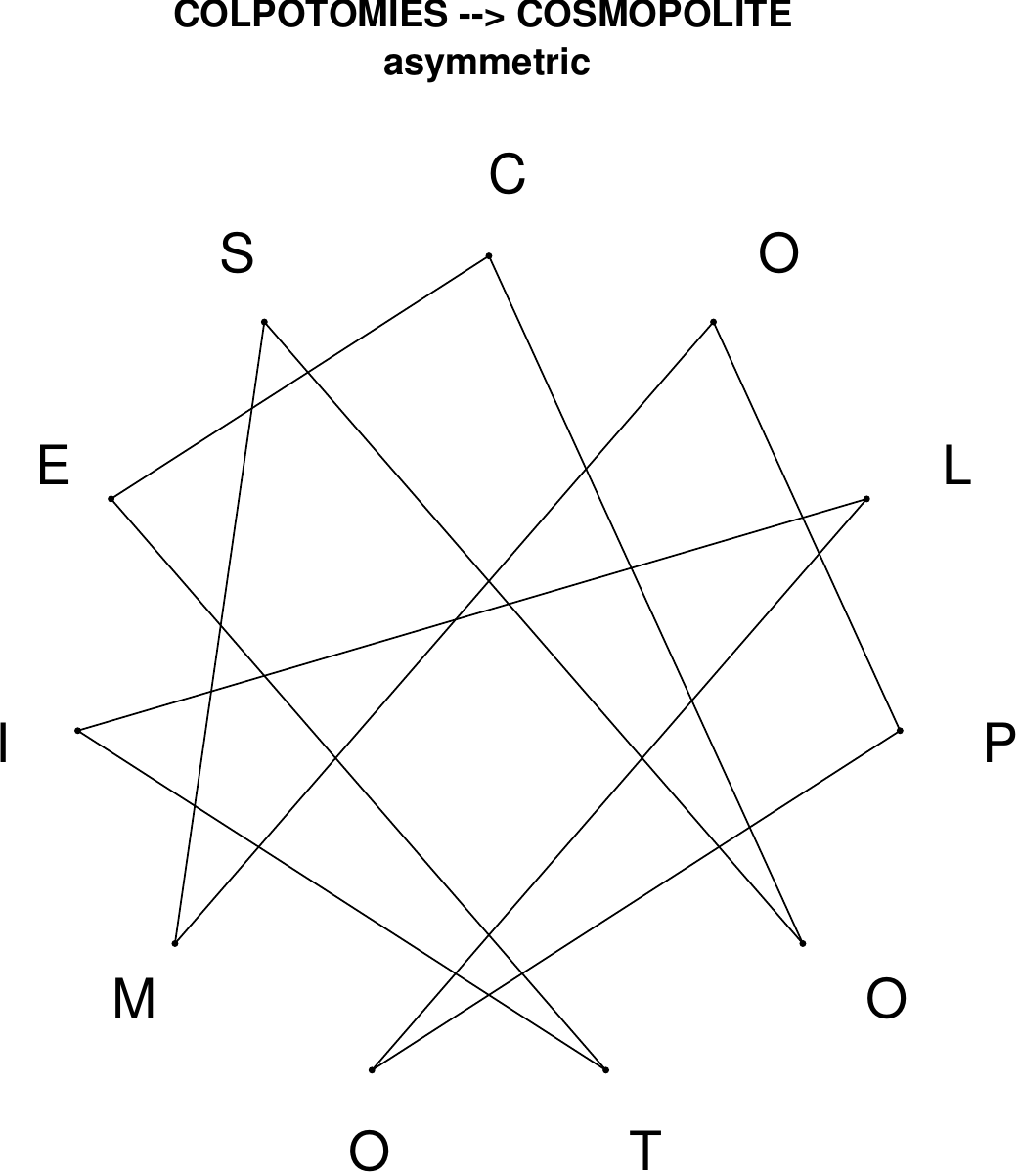}
\end{subfigure}
\hfill
\begin{subfigure}[T]{0.19\textwidth}
\centering
\includegraphics[width=\textwidth]{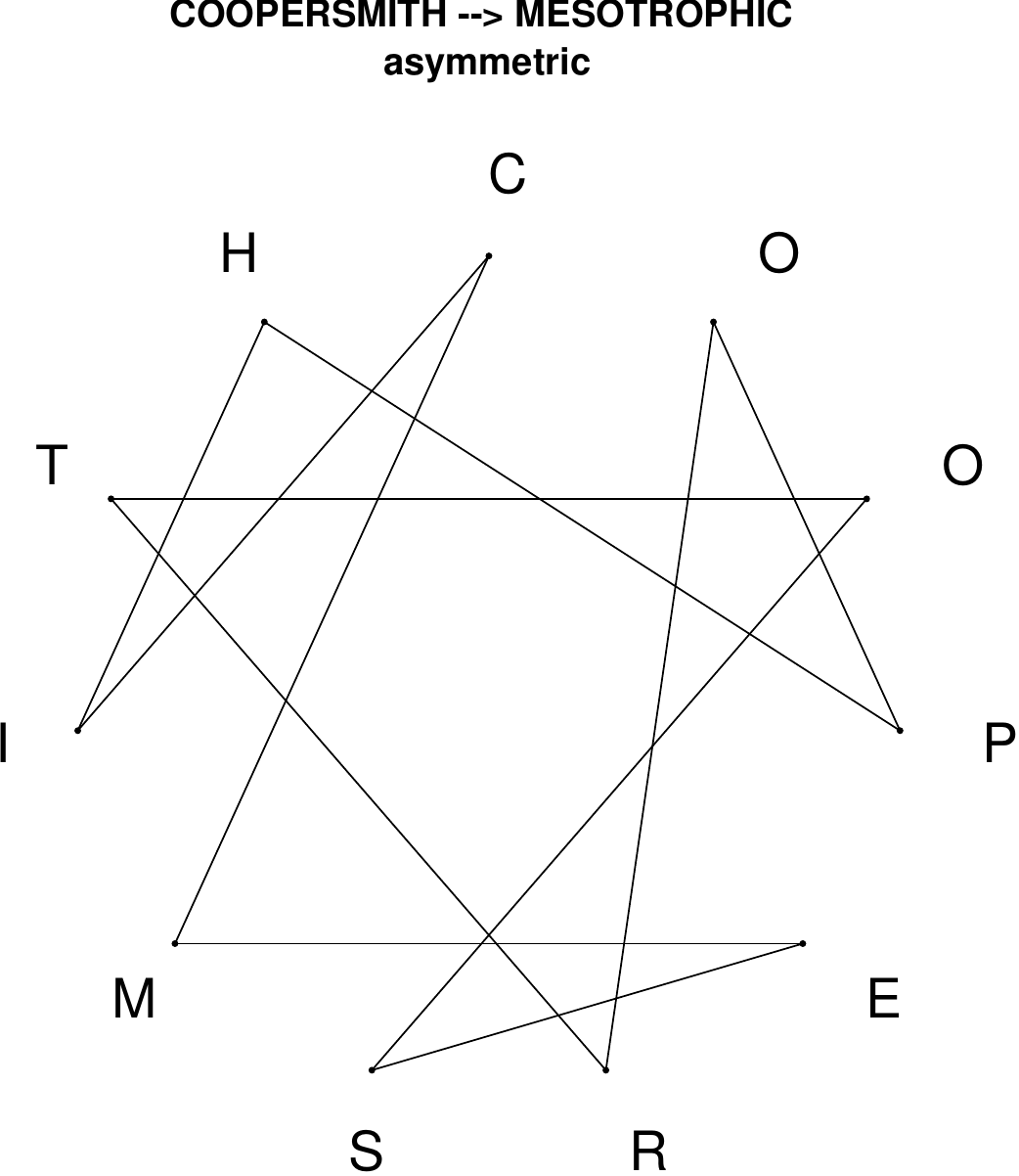}
\end{subfigure}
\hfill
\begin{subfigure}[T]{0.19\textwidth}
\centering
\includegraphics[width=\textwidth]{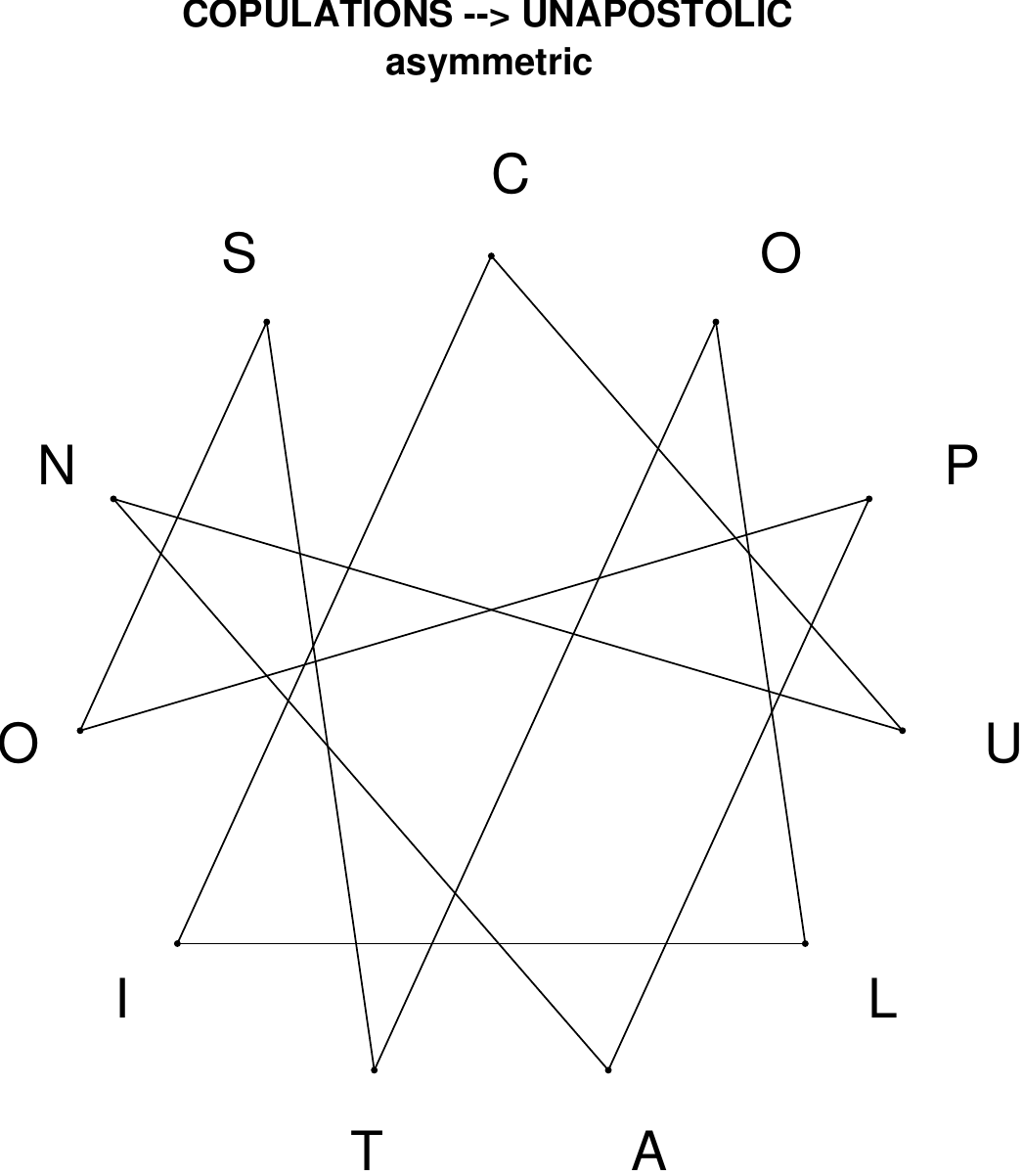}
\end{subfigure}
\hfill
\begin{subfigure}[T]{0.19\textwidth}
\centering
\includegraphics[width=\textwidth]{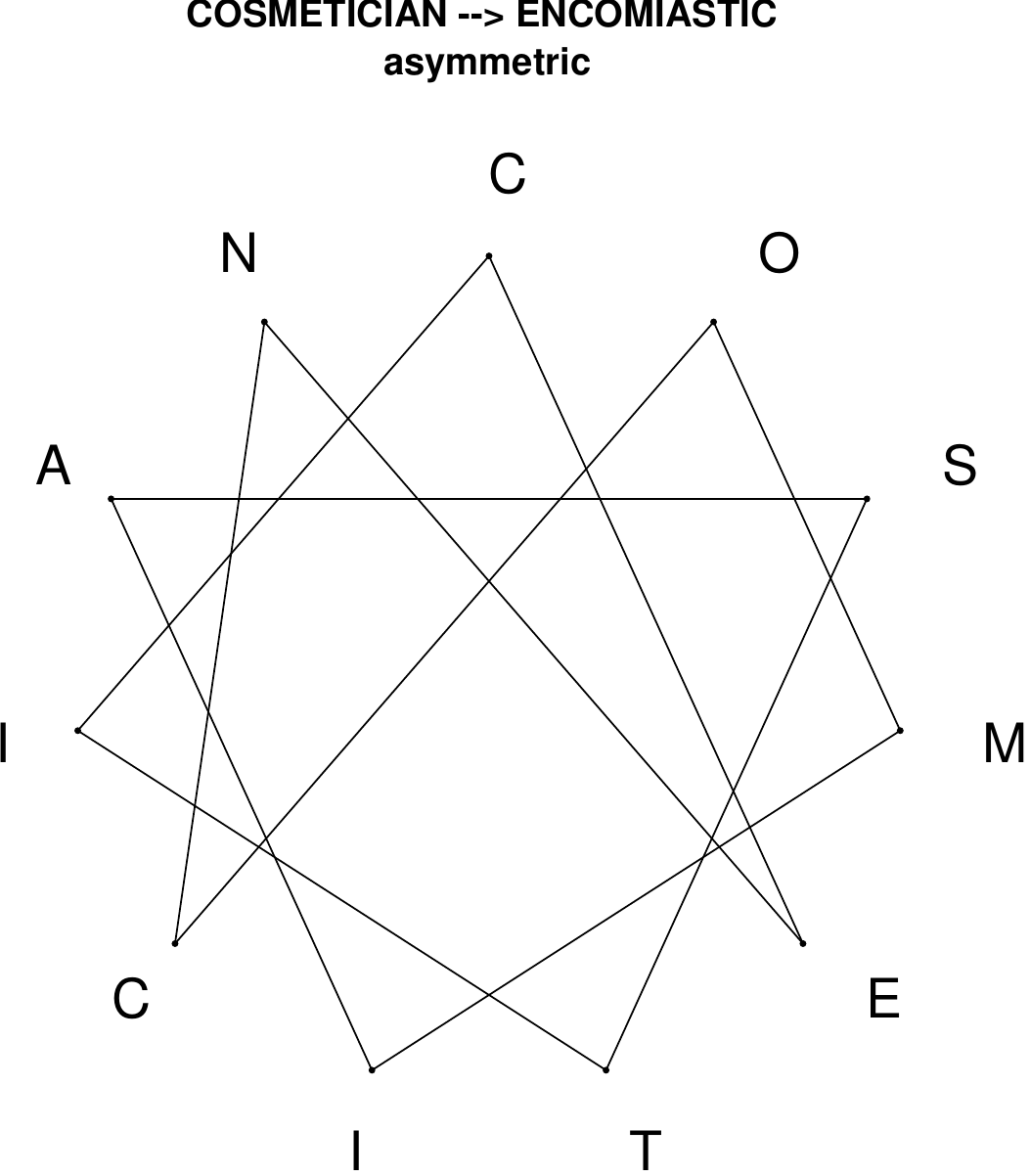}
\end{subfigure}
\end{figure}

\begin{figure}[H]
\centering
\begin{subfigure}[T]{0.19\textwidth}
\centering
\includegraphics[width=\textwidth]{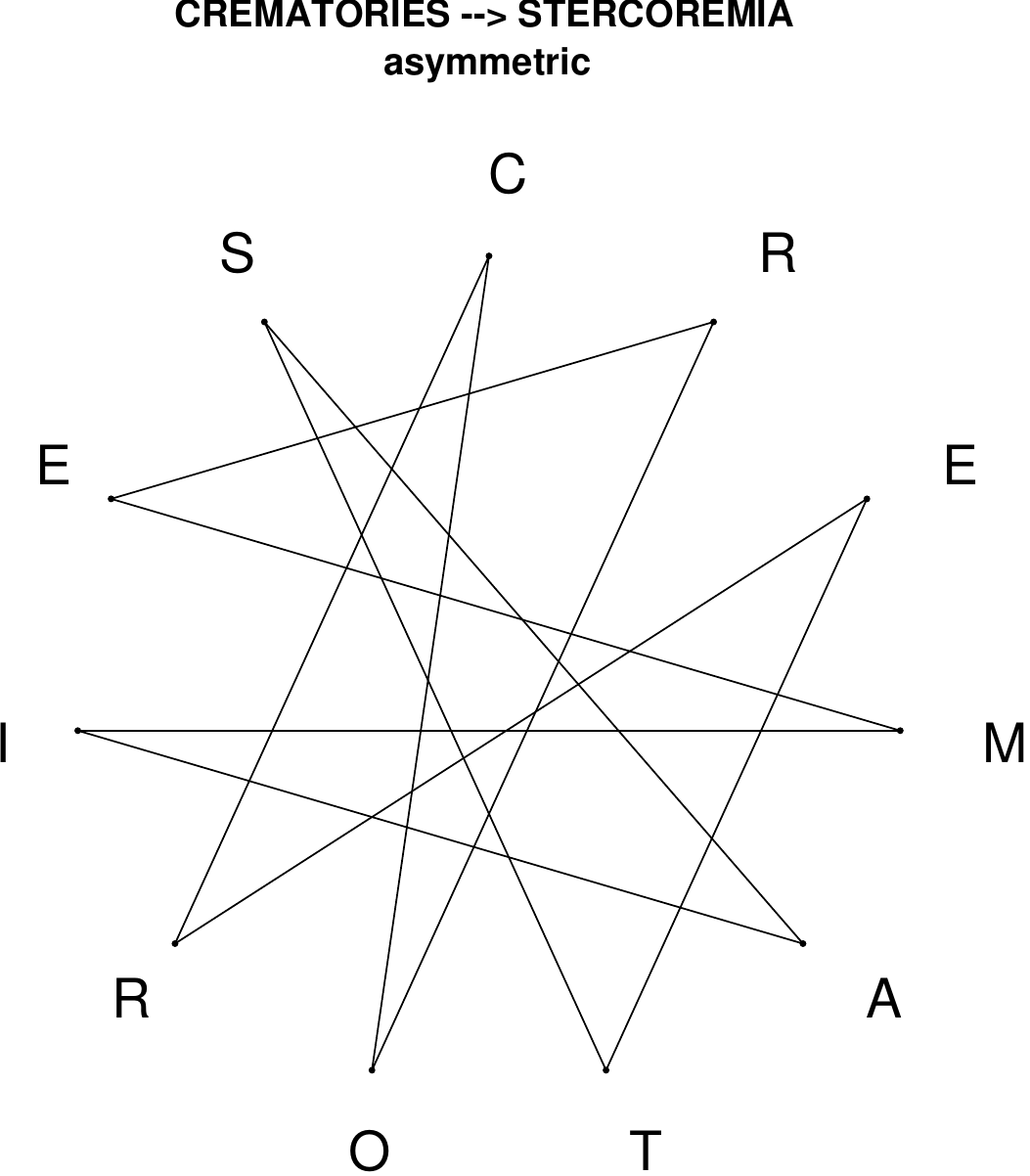}
\end{subfigure}
\hfill
\begin{subfigure}[T]{0.19\textwidth}
\centering
\includegraphics[width=\textwidth]{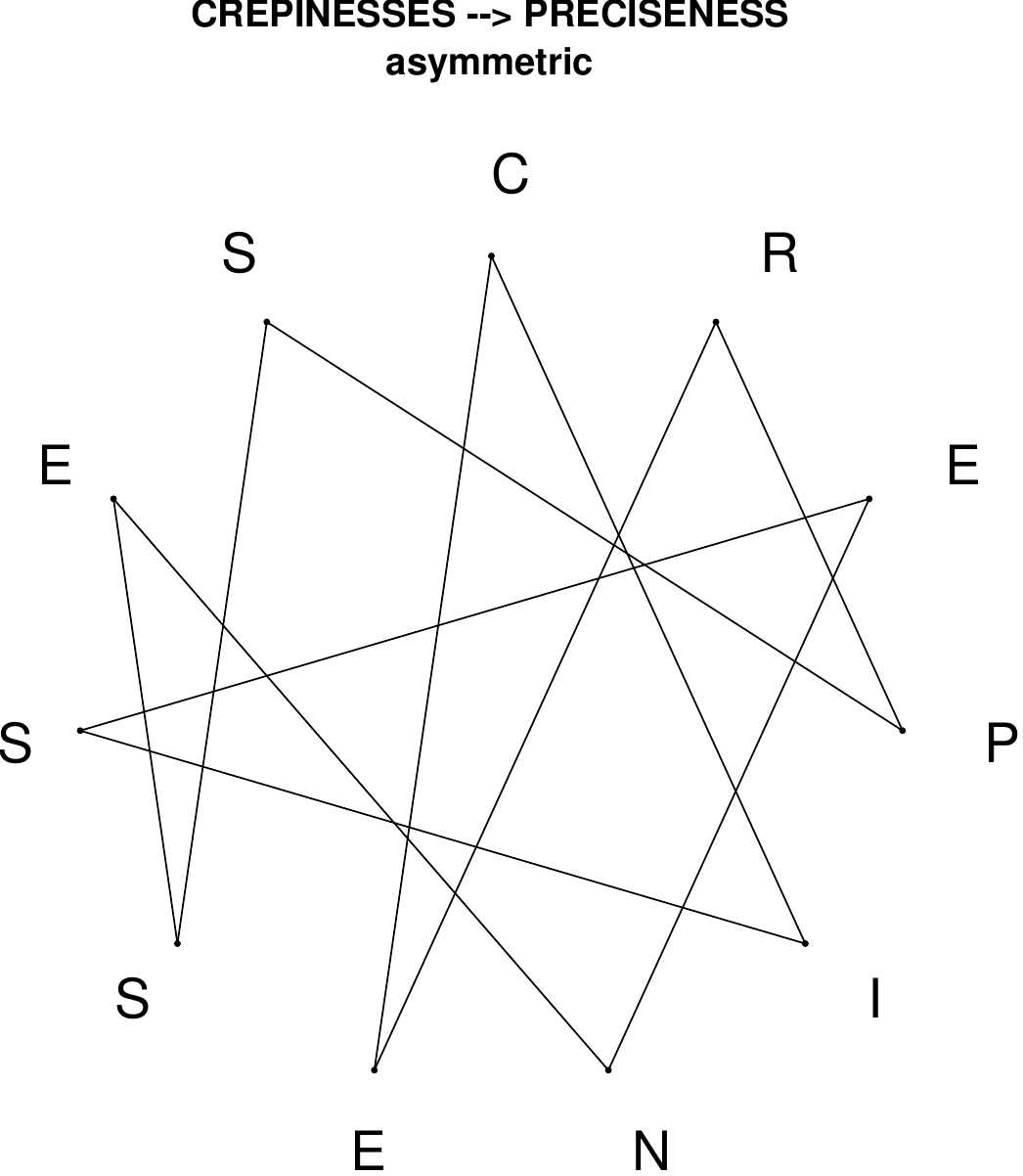}
\end{subfigure}
\hfill
\begin{subfigure}[T]{0.19\textwidth}
\centering
\includegraphics[width=\textwidth]{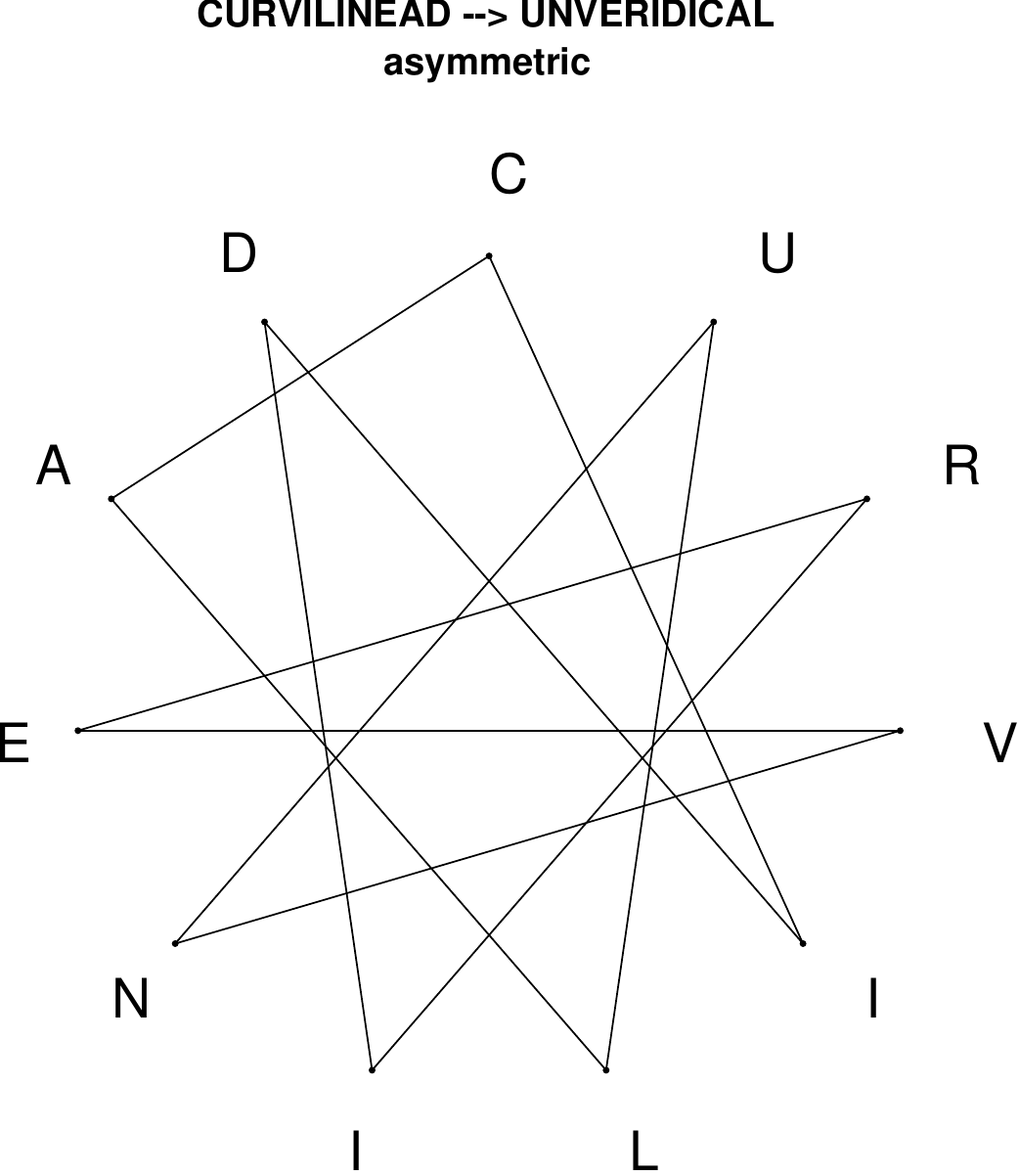}
\end{subfigure}
\hfill
\begin{subfigure}[T]{0.19\textwidth}
\centering
\includegraphics[width=\textwidth]{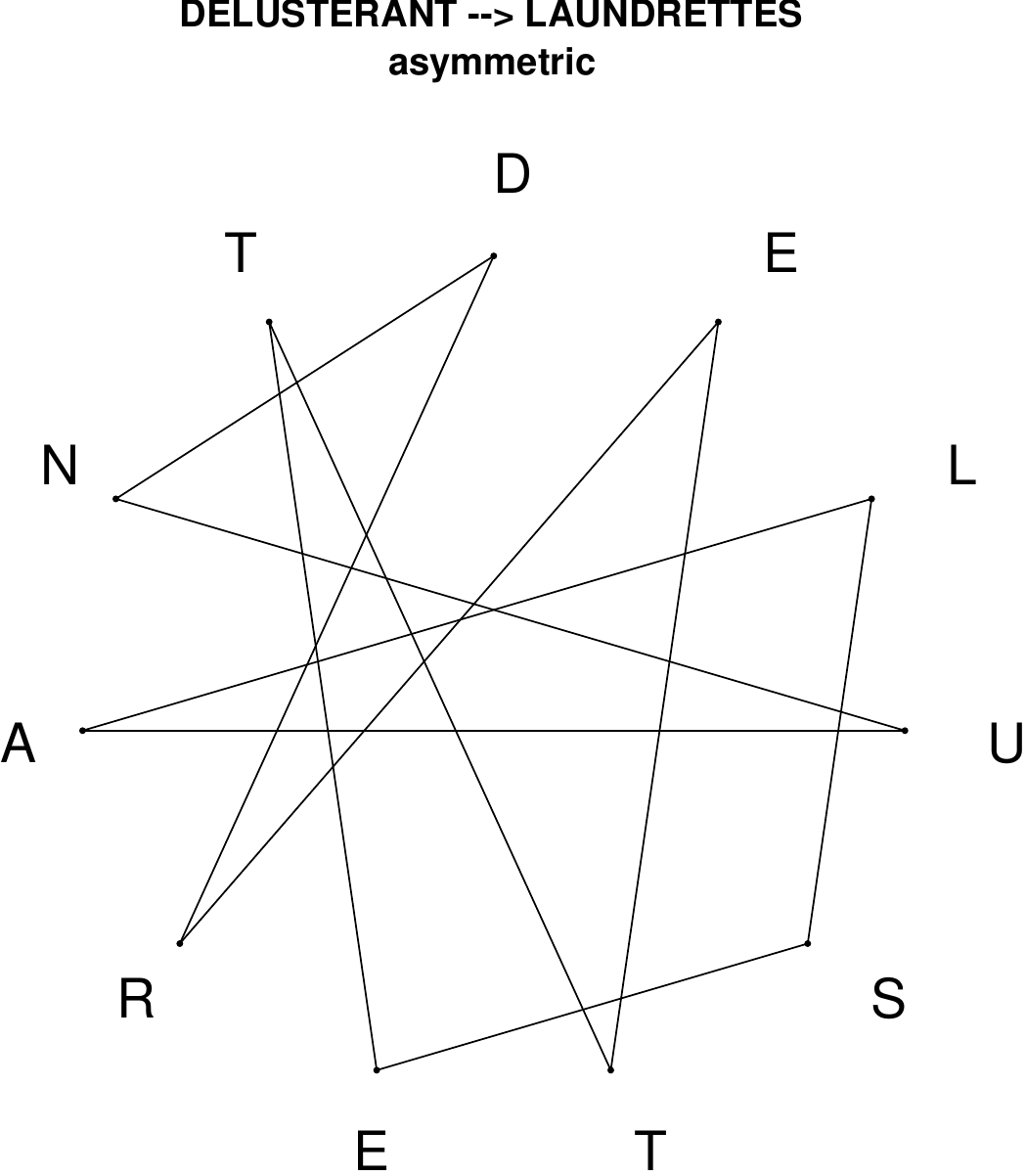}
\end{subfigure}
\hfill
\begin{subfigure}[T]{0.19\textwidth}
\centering
\includegraphics[width=\textwidth]{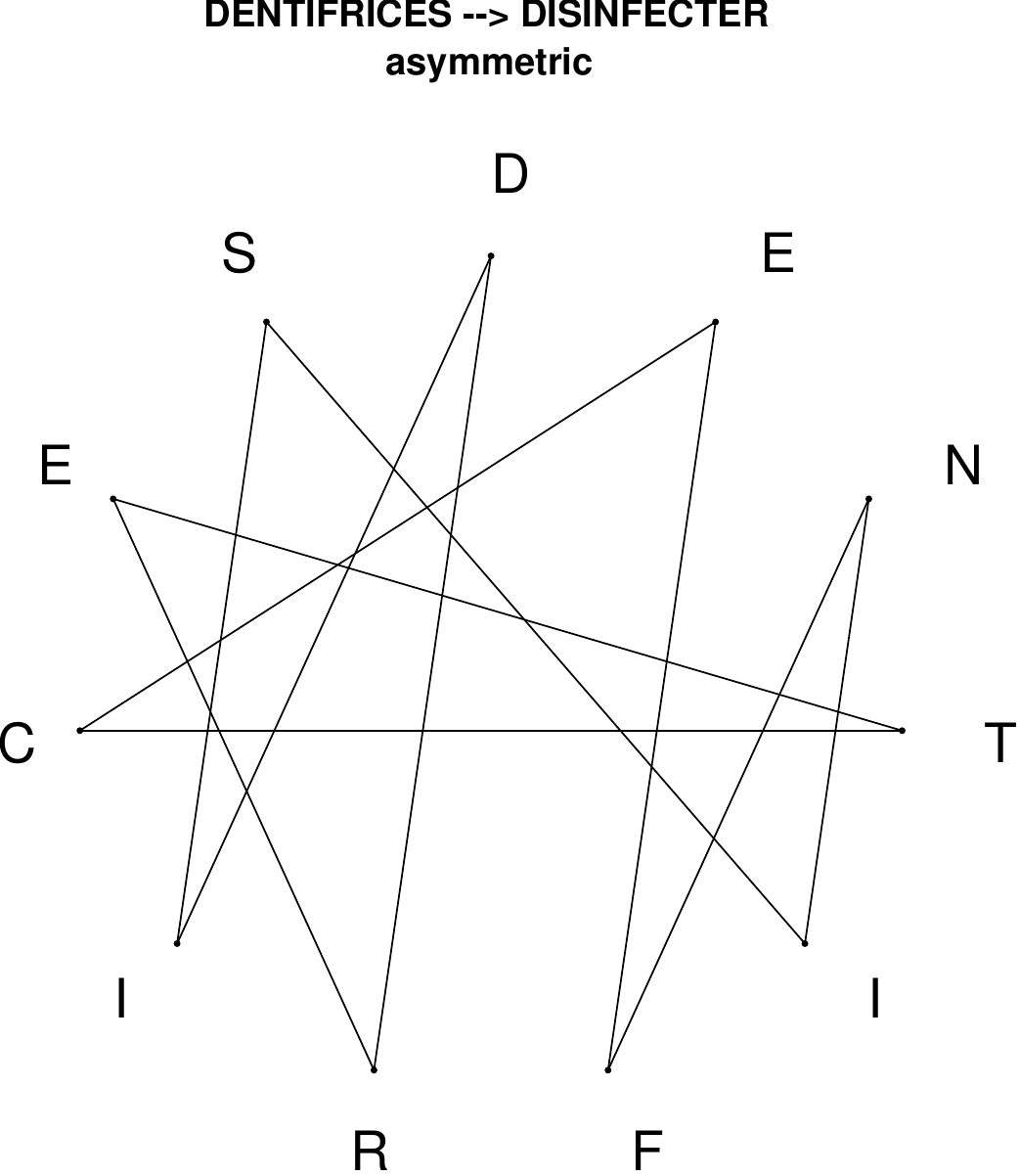}
\end{subfigure}
\end{figure}

\begin{figure}[H]
\centering
\begin{subfigure}[T]{0.19\textwidth}
\centering
\includegraphics[width=\textwidth]{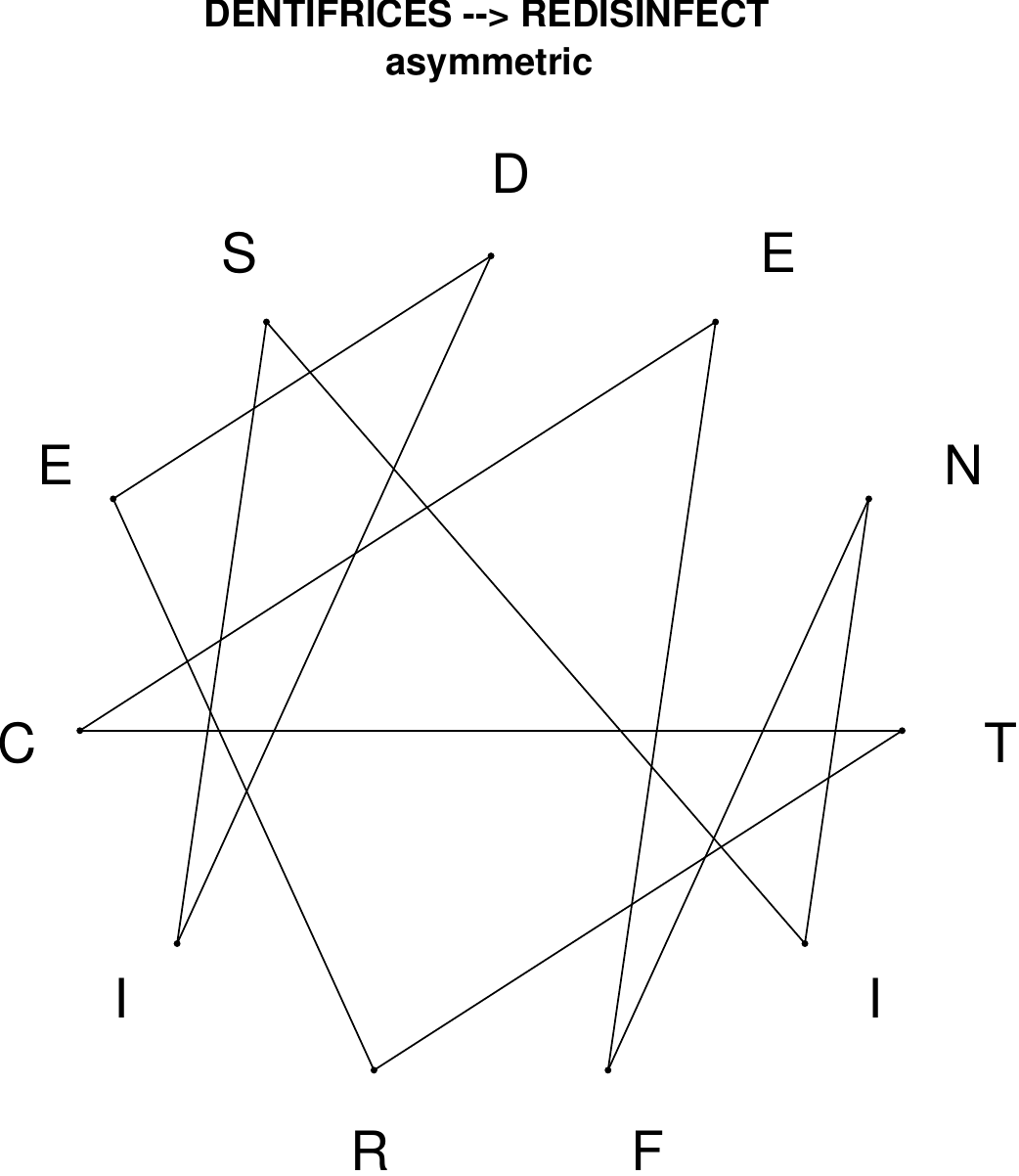}
\end{subfigure}
\hfill
\begin{subfigure}[T]{0.19\textwidth}
\centering
\includegraphics[width=\textwidth]{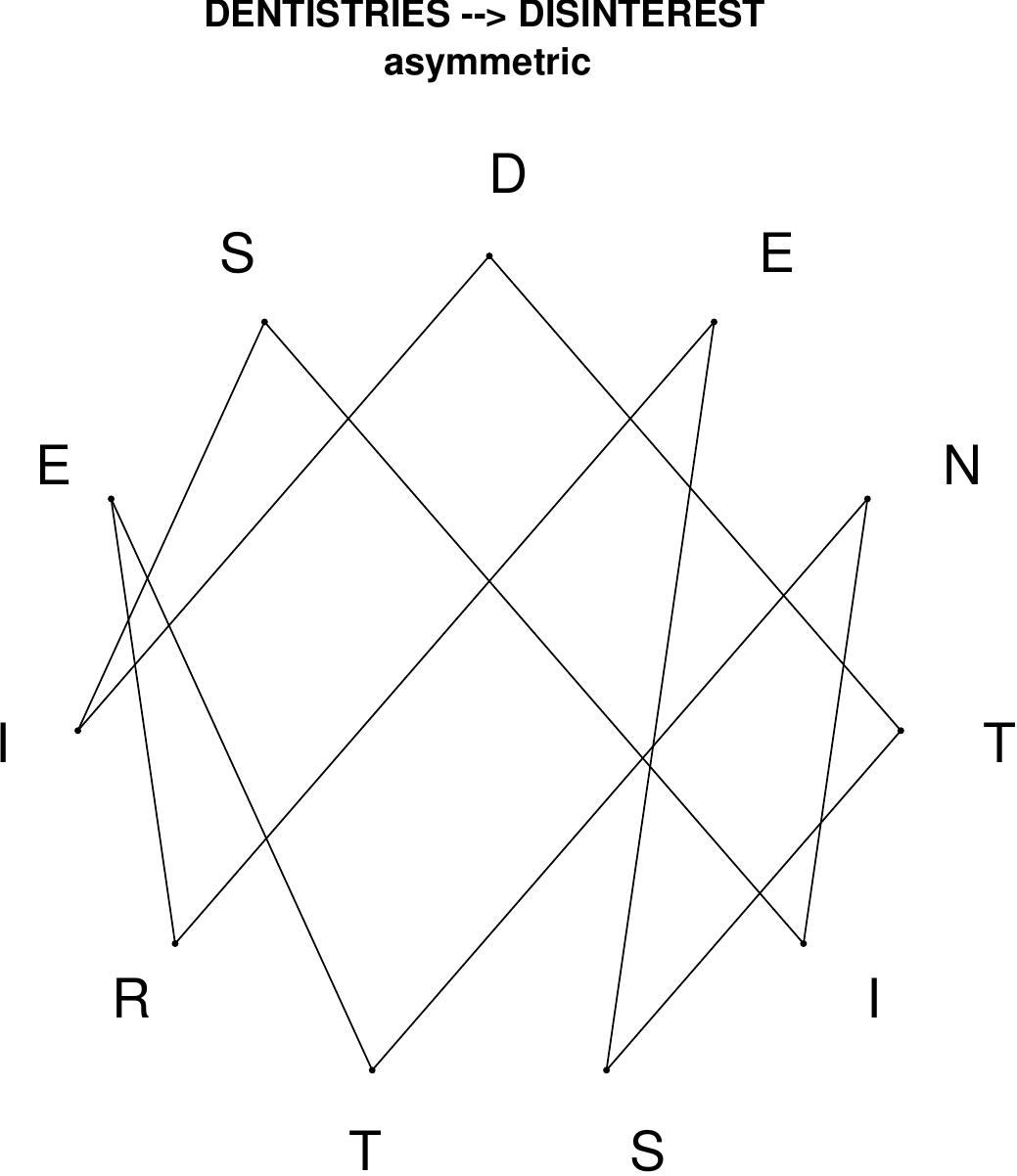}
\end{subfigure}
\hfill
\begin{subfigure}[T]{0.19\textwidth}
\centering
\includegraphics[width=\textwidth]{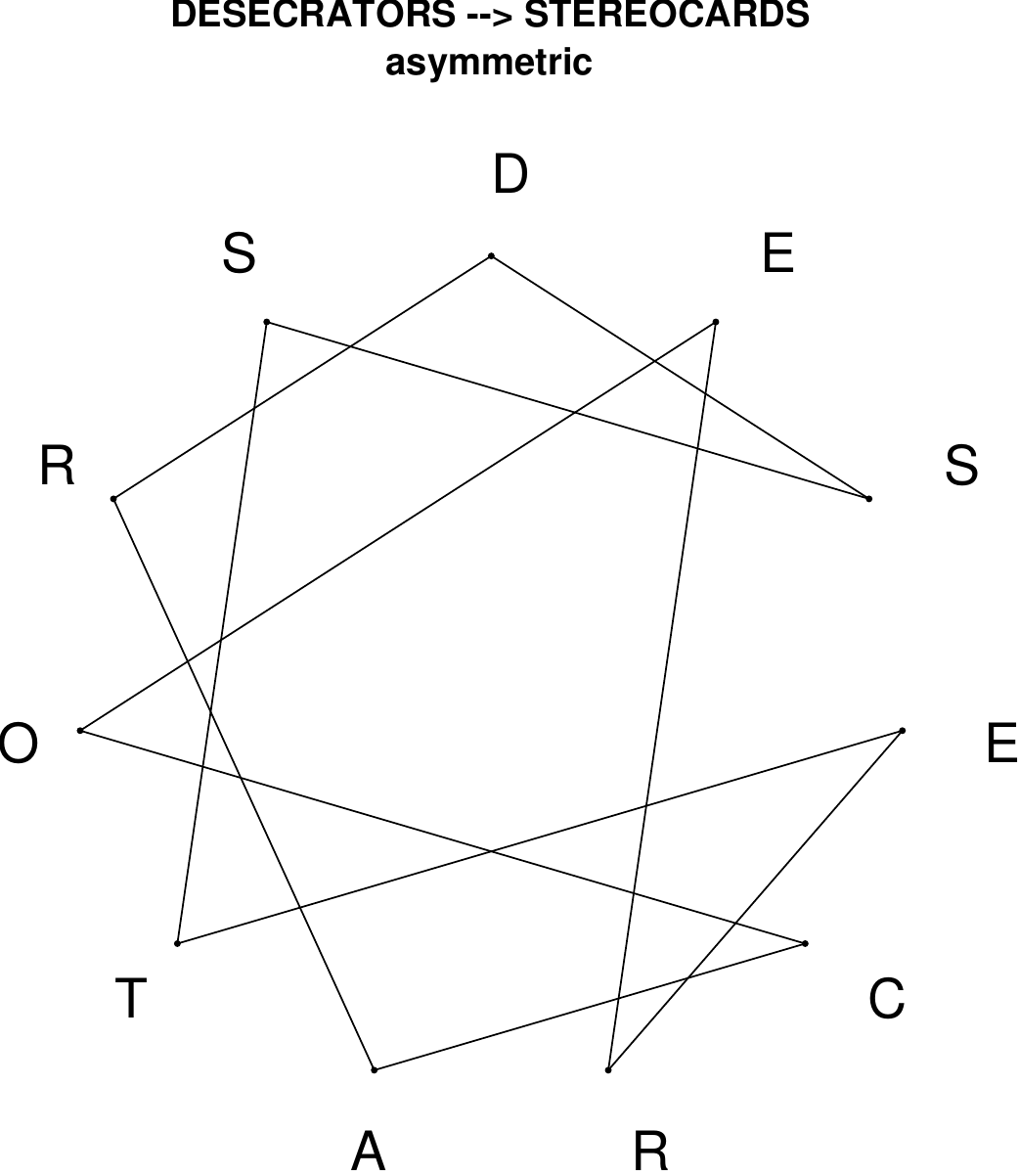}
\end{subfigure}
\hfill
\begin{subfigure}[T]{0.19\textwidth}
\centering
\includegraphics[width=\textwidth]{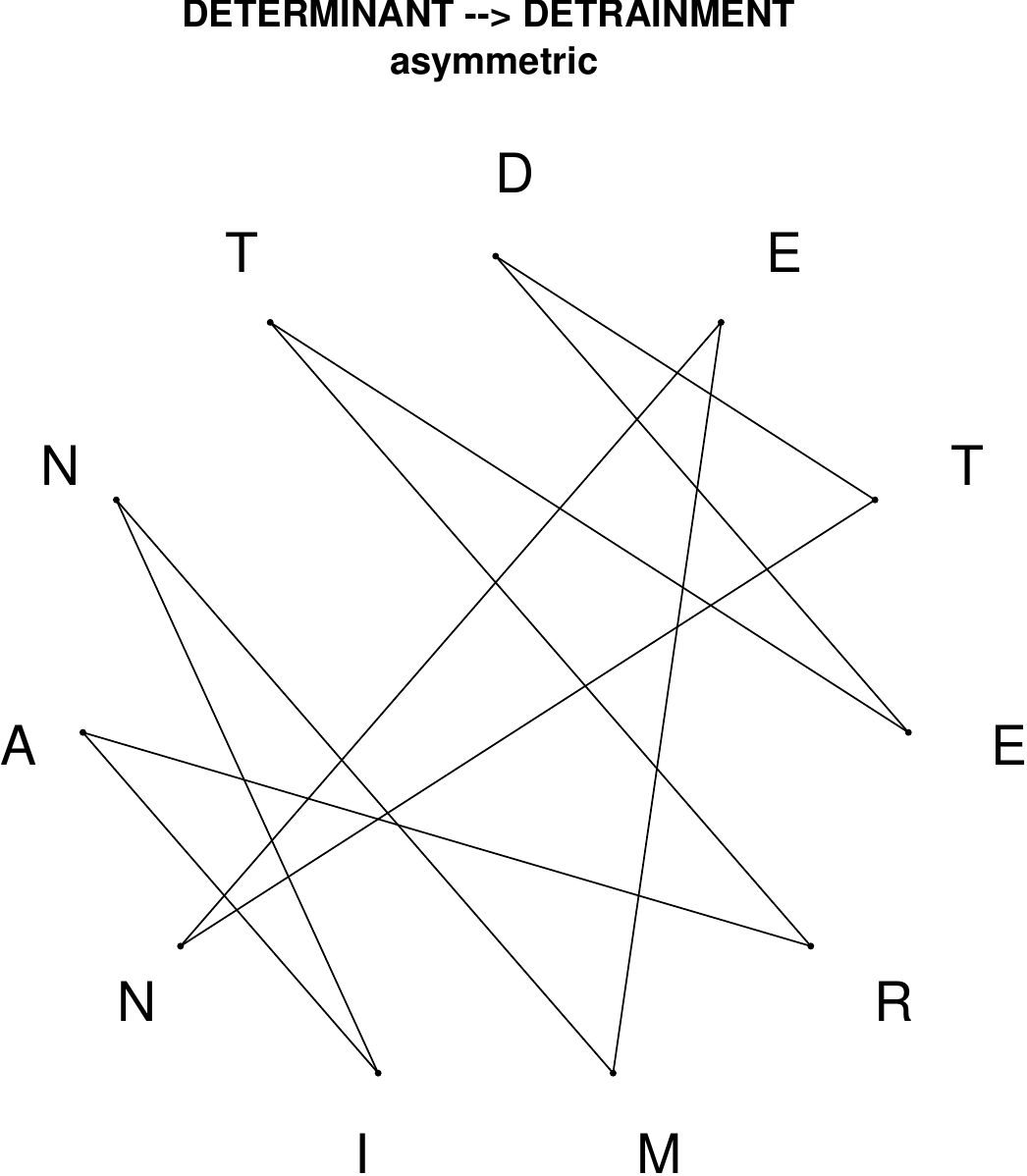}
\end{subfigure}
\hfill
\begin{subfigure}[T]{0.19\textwidth}
\centering
\includegraphics[width=\textwidth]{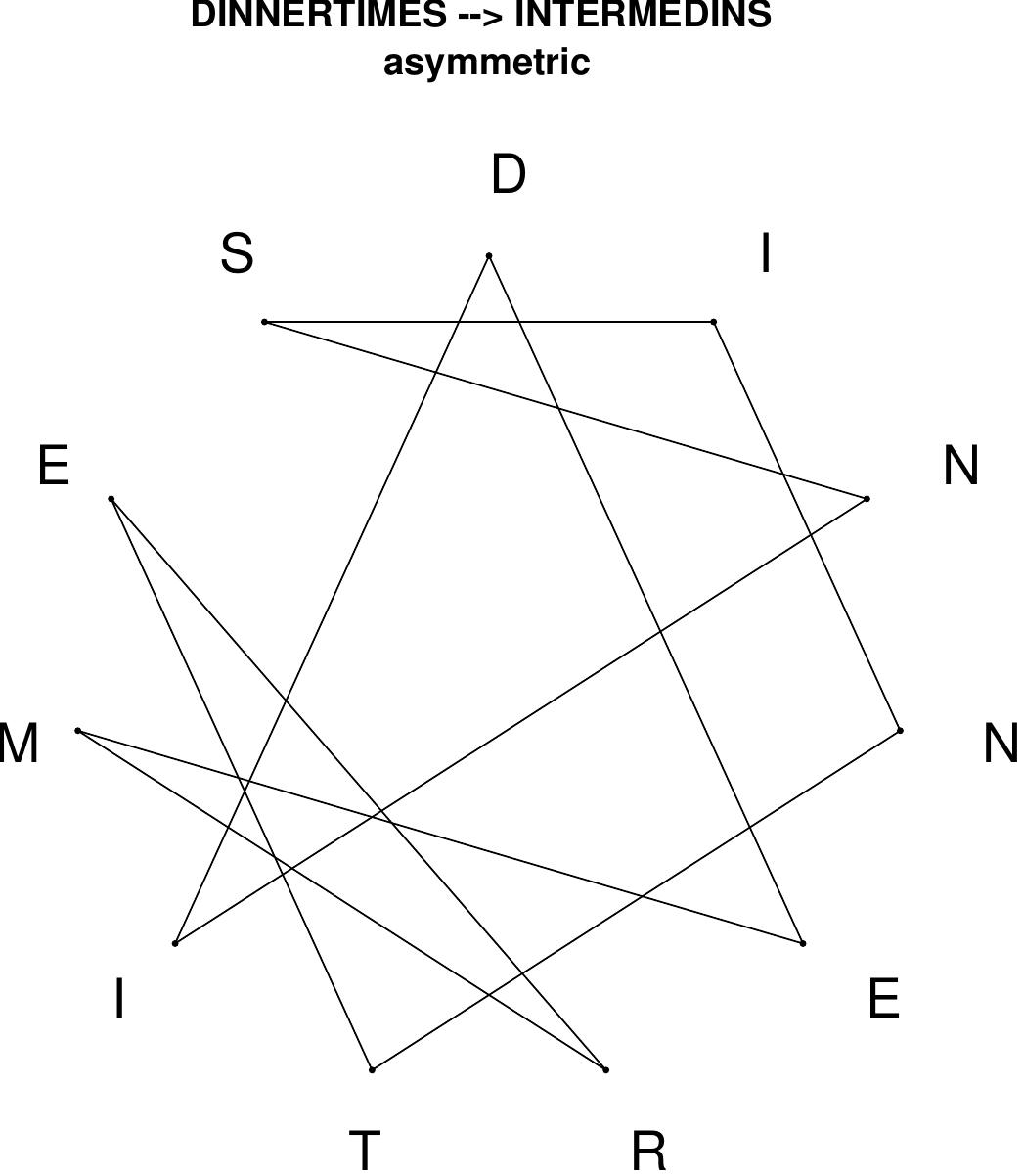}
\end{subfigure}
\end{figure}

\begin{figure}[H]
\centering
\begin{subfigure}[T]{0.19\textwidth}
\centering
\includegraphics[width=\textwidth]{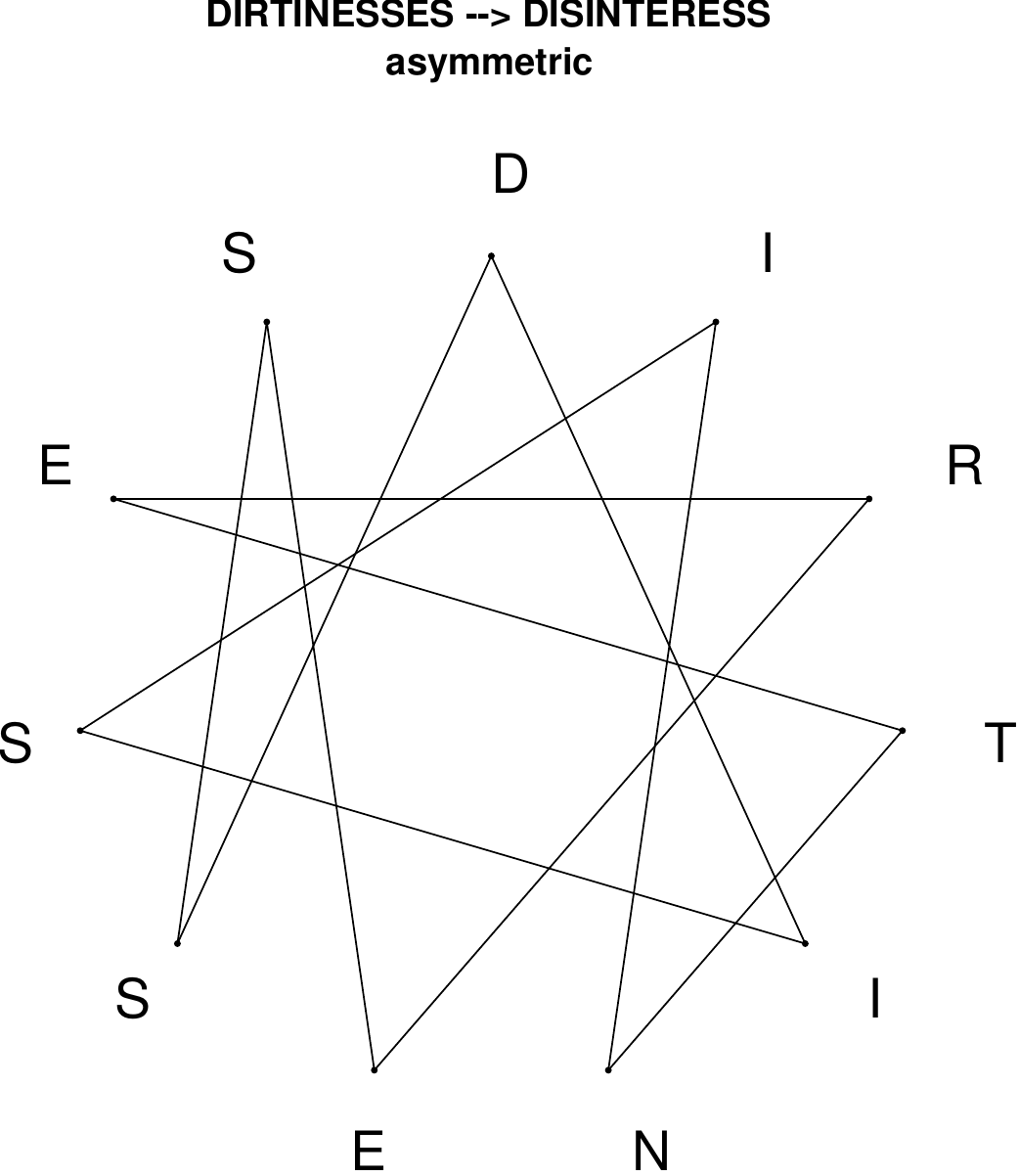}
\end{subfigure}
\hfill
\begin{subfigure}[T]{0.19\textwidth}
\centering
\includegraphics[width=\textwidth]{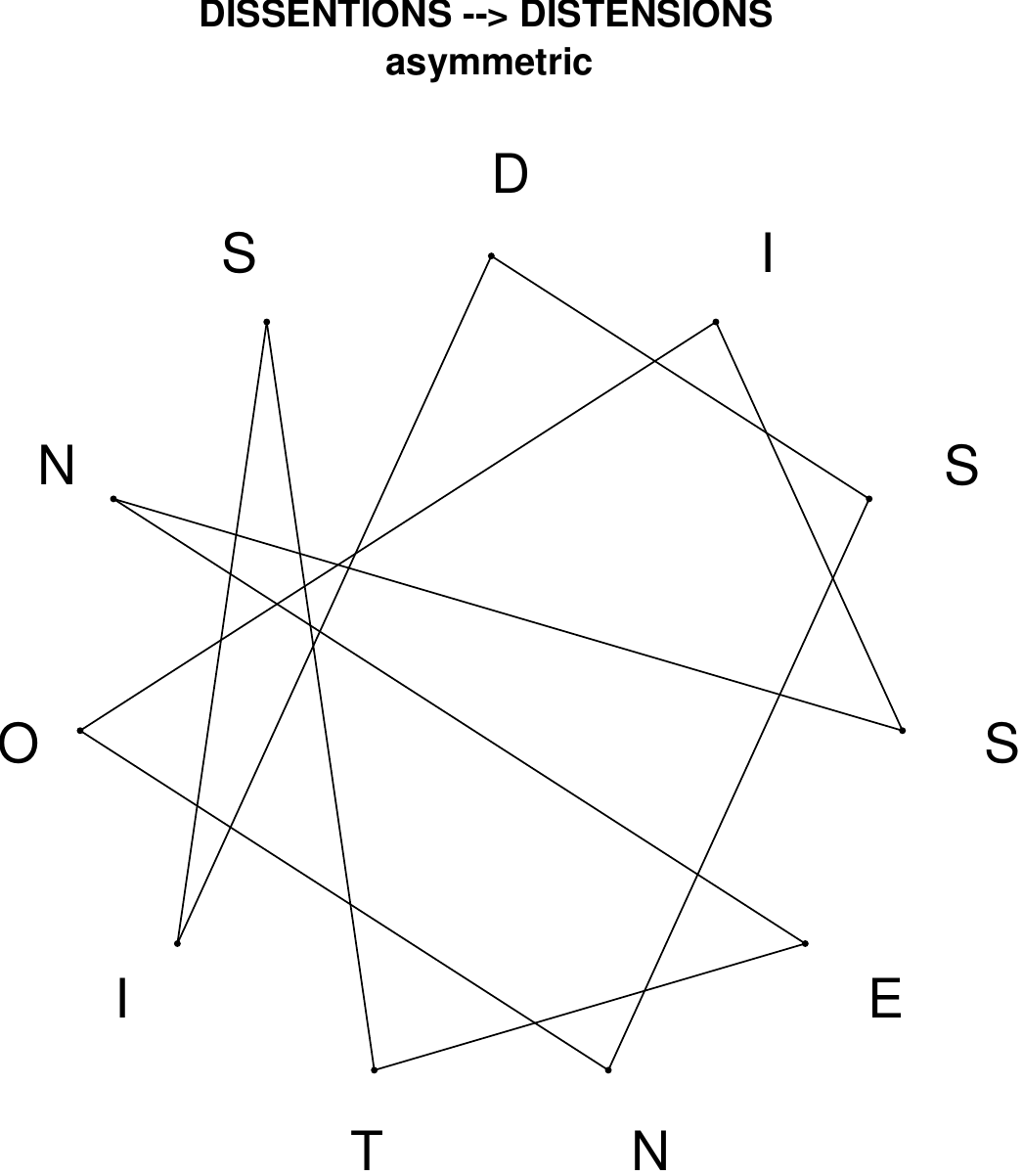}
\end{subfigure}
\hfill
\begin{subfigure}[T]{0.19\textwidth}
\centering
\includegraphics[width=\textwidth]{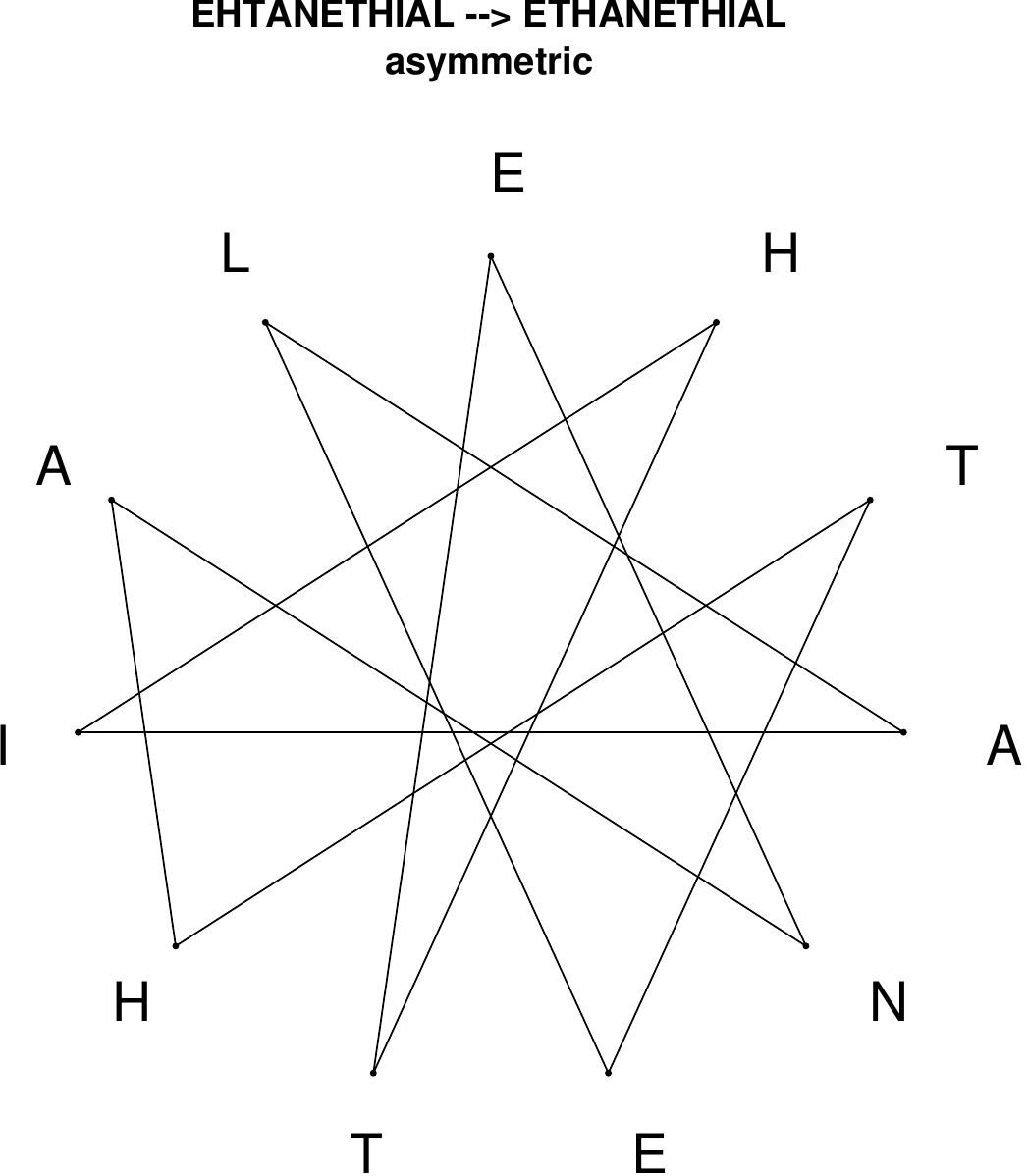}
\end{subfigure}
\hfill
\begin{subfigure}[T]{0.19\textwidth}
\centering
\includegraphics[width=\textwidth]{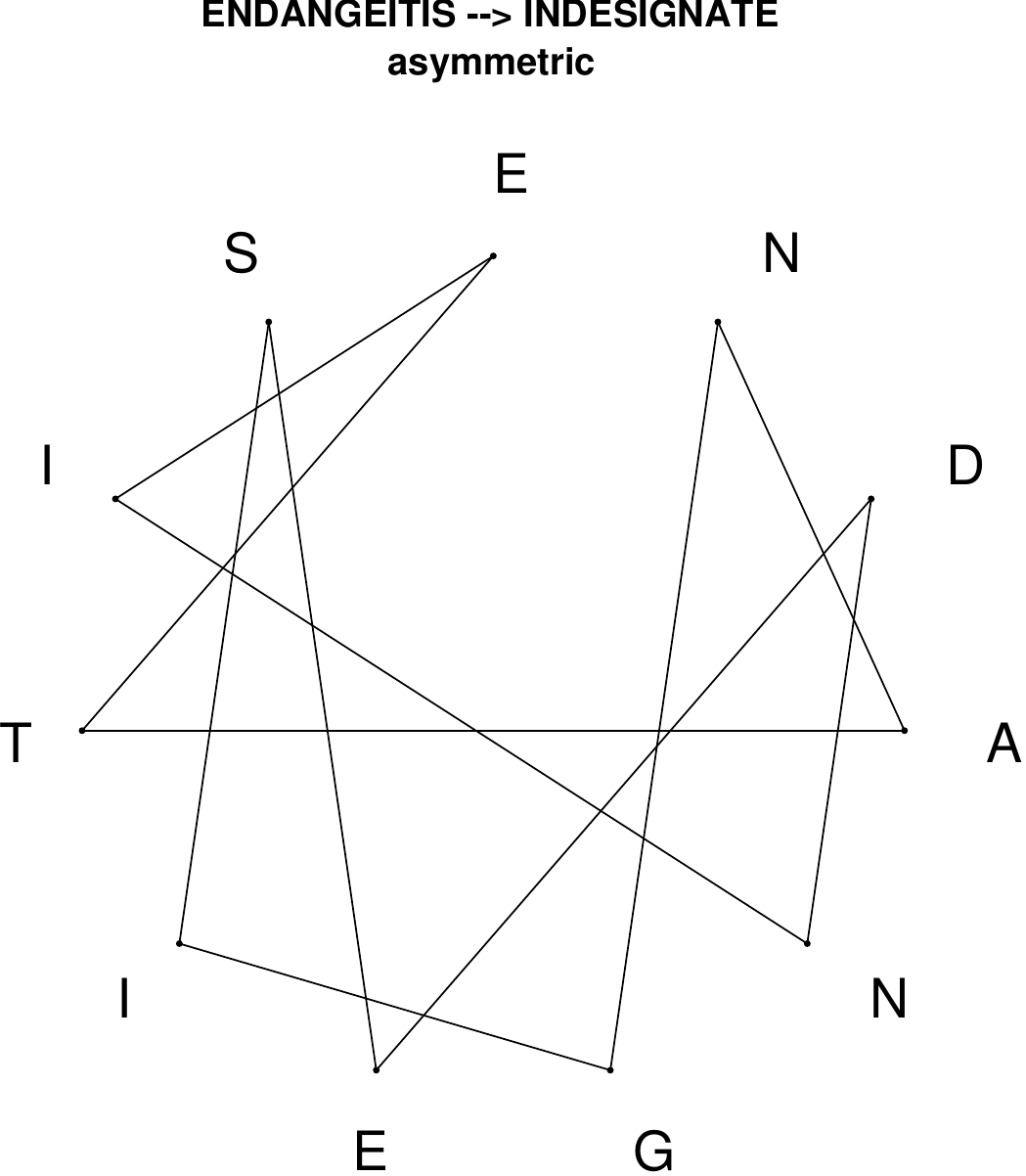}
\end{subfigure}
\hfill
\begin{subfigure}[T]{0.19\textwidth}
\centering
\includegraphics[width=\textwidth]{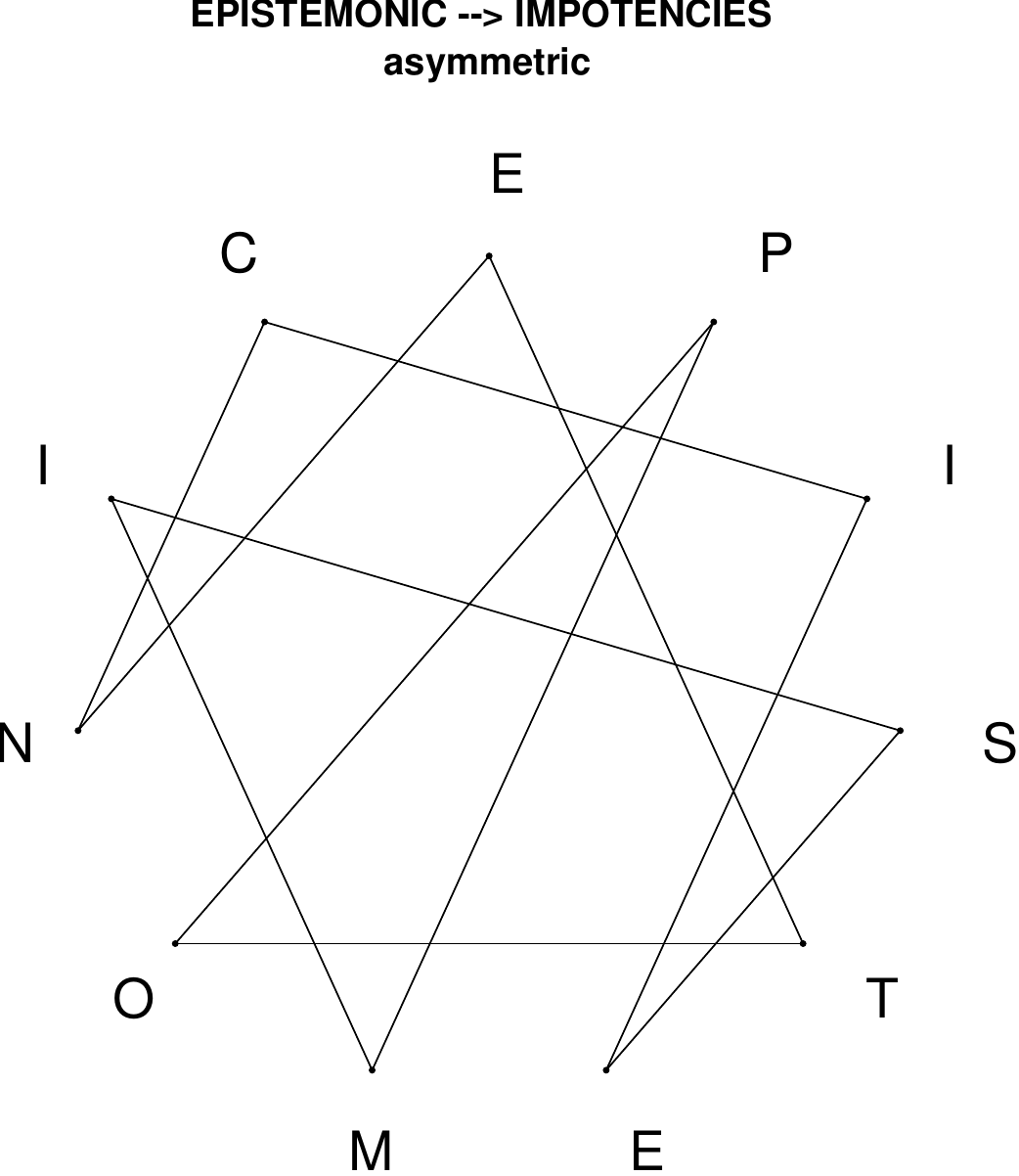}
\end{subfigure}
\end{figure}

\begin{figure}[H]
\centering
\begin{subfigure}[T]{0.19\textwidth}
\centering
\includegraphics[width=\textwidth]{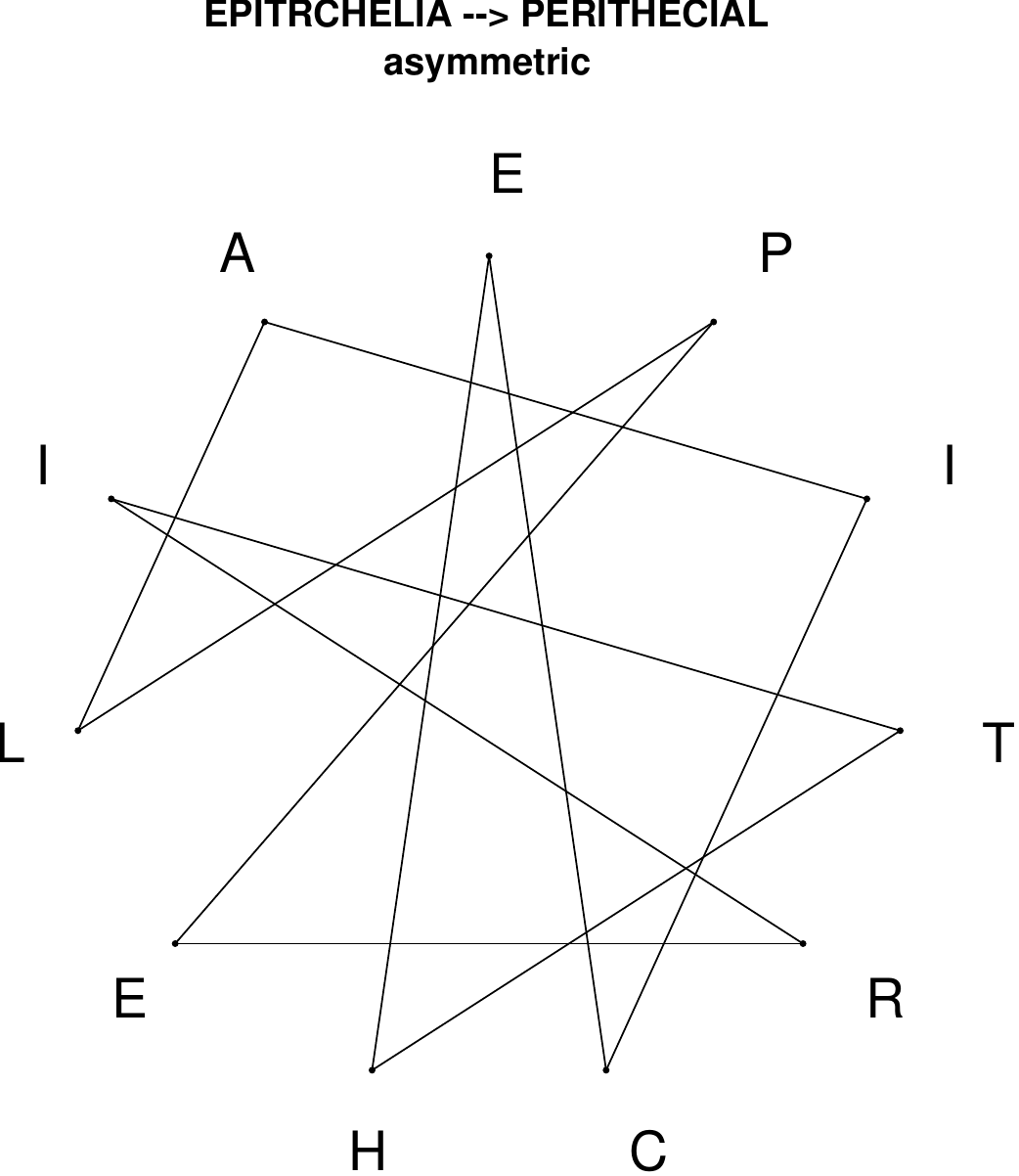}
\end{subfigure}
\hfill
\begin{subfigure}[T]{0.19\textwidth}
\centering
\includegraphics[width=\textwidth]{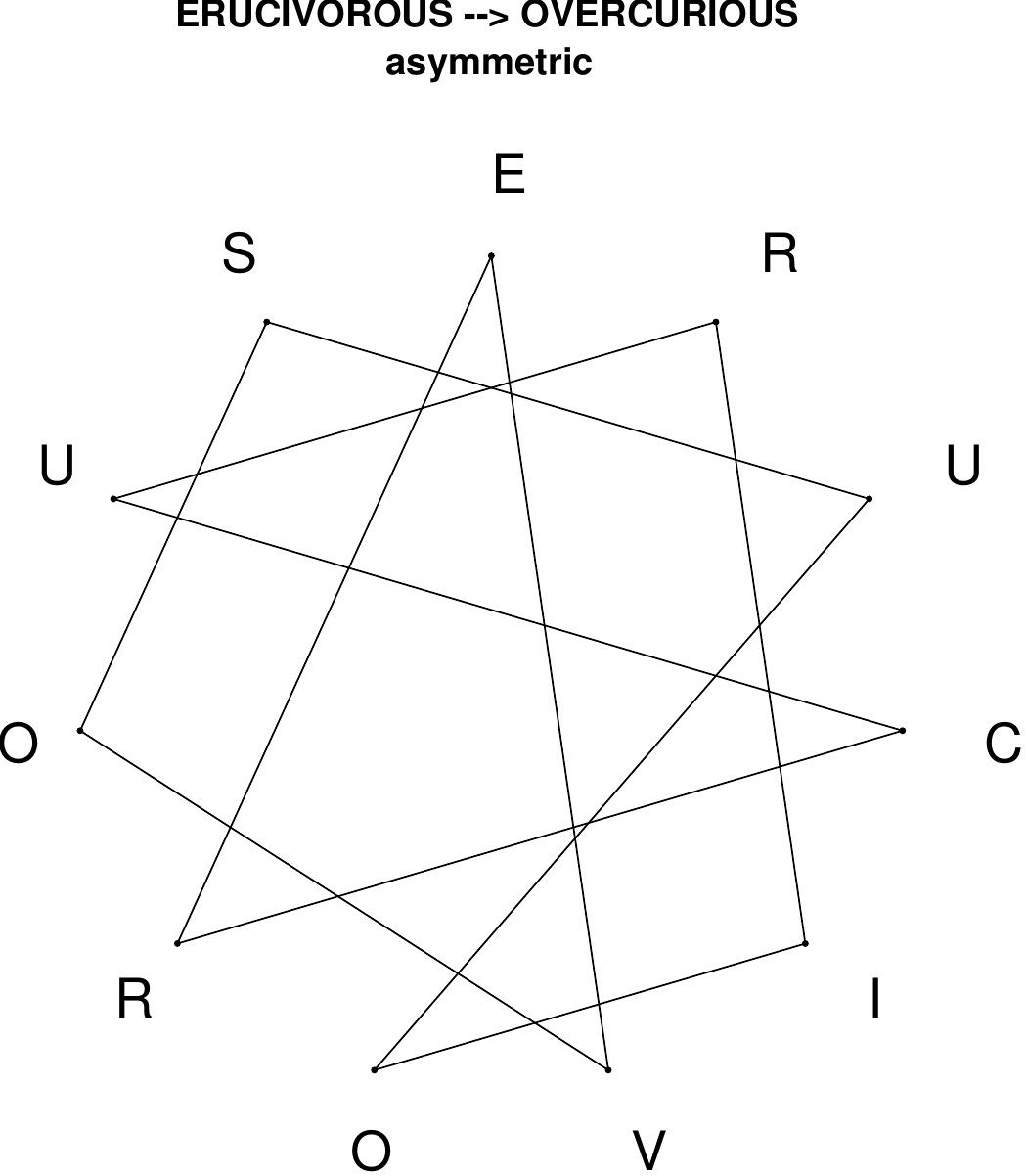}
\end{subfigure}
\hfill
\begin{subfigure}[T]{0.19\textwidth}
\centering
\includegraphics[width=\textwidth]{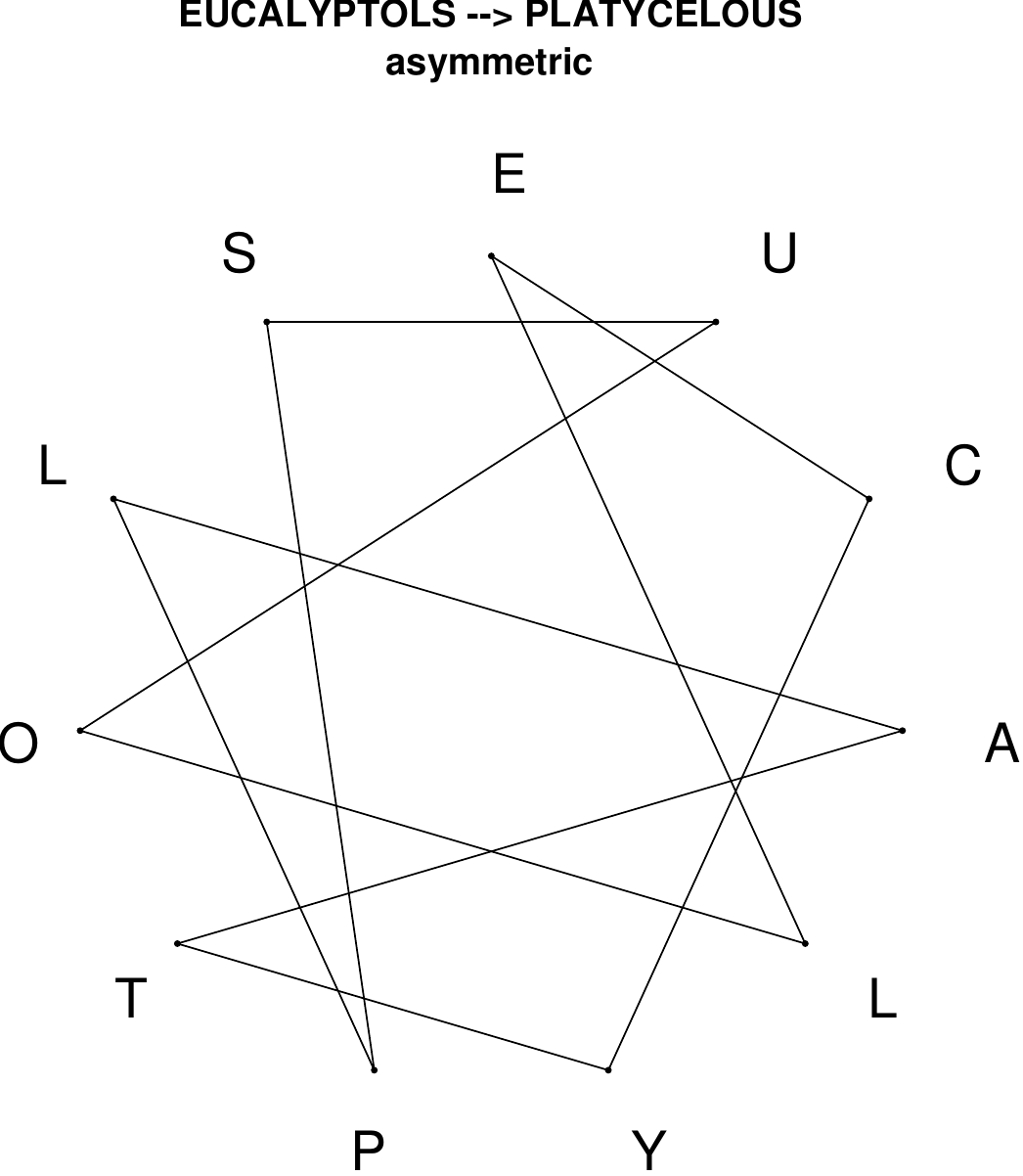}
\end{subfigure}
\hfill
\begin{subfigure}[T]{0.19\textwidth}
\centering
\includegraphics[width=\textwidth]{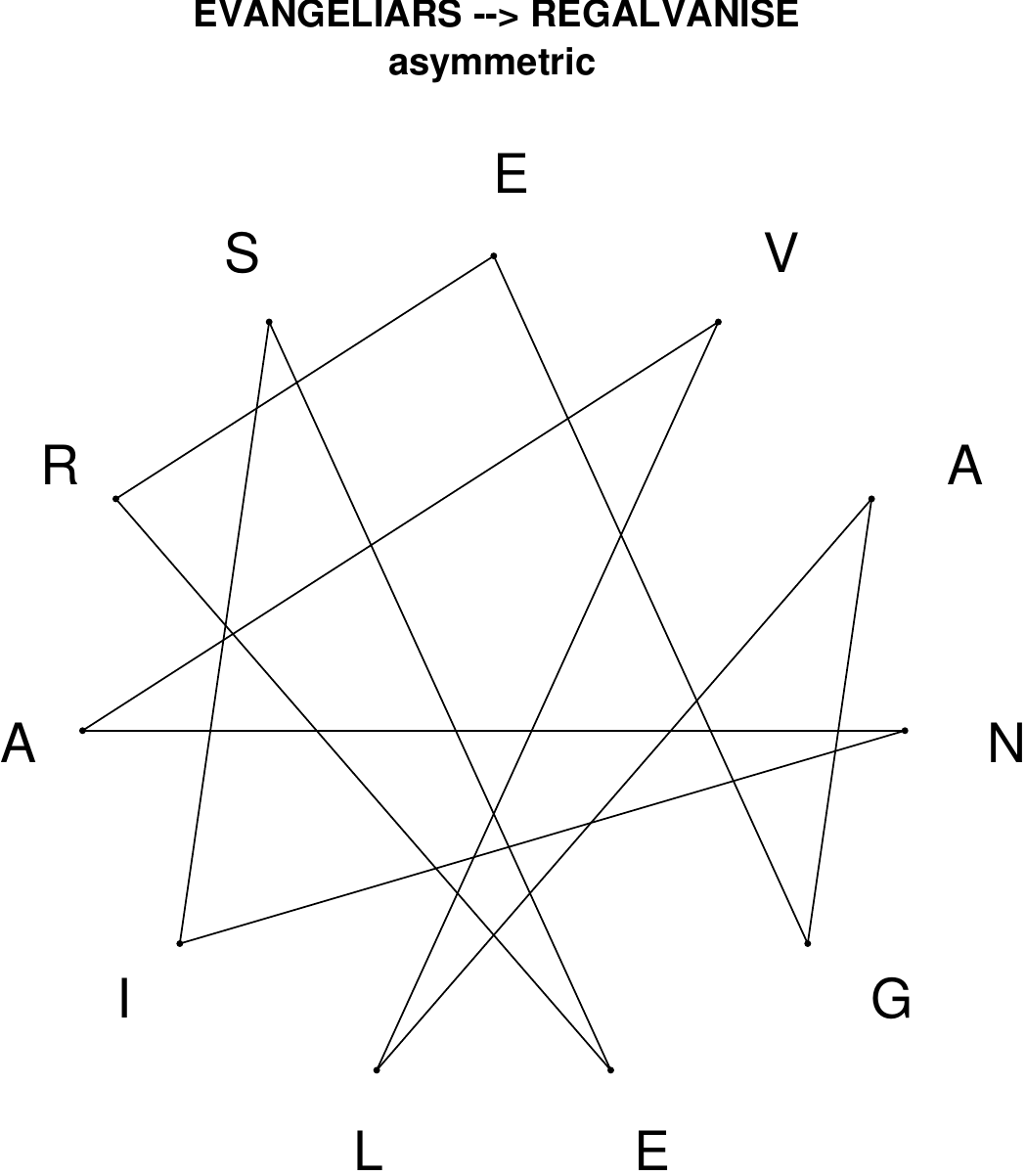}
\end{subfigure}
\hfill
\begin{subfigure}[T]{0.19\textwidth}
\centering
\includegraphics[width=\textwidth]{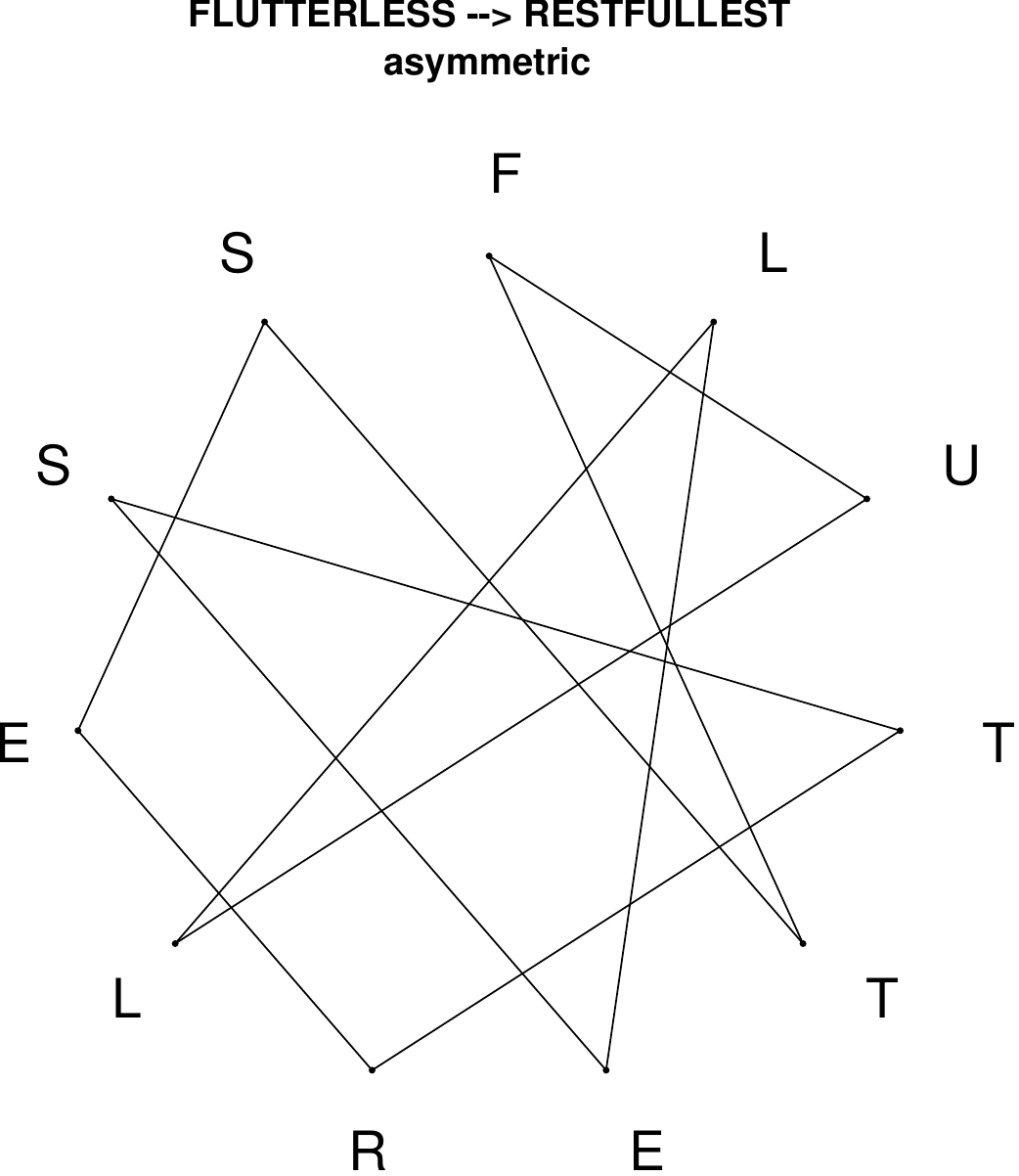}
\end{subfigure}
\end{figure}

\begin{figure}[H]
\centering
\begin{subfigure}[T]{0.19\textwidth}
\centering
\includegraphics[width=\textwidth]{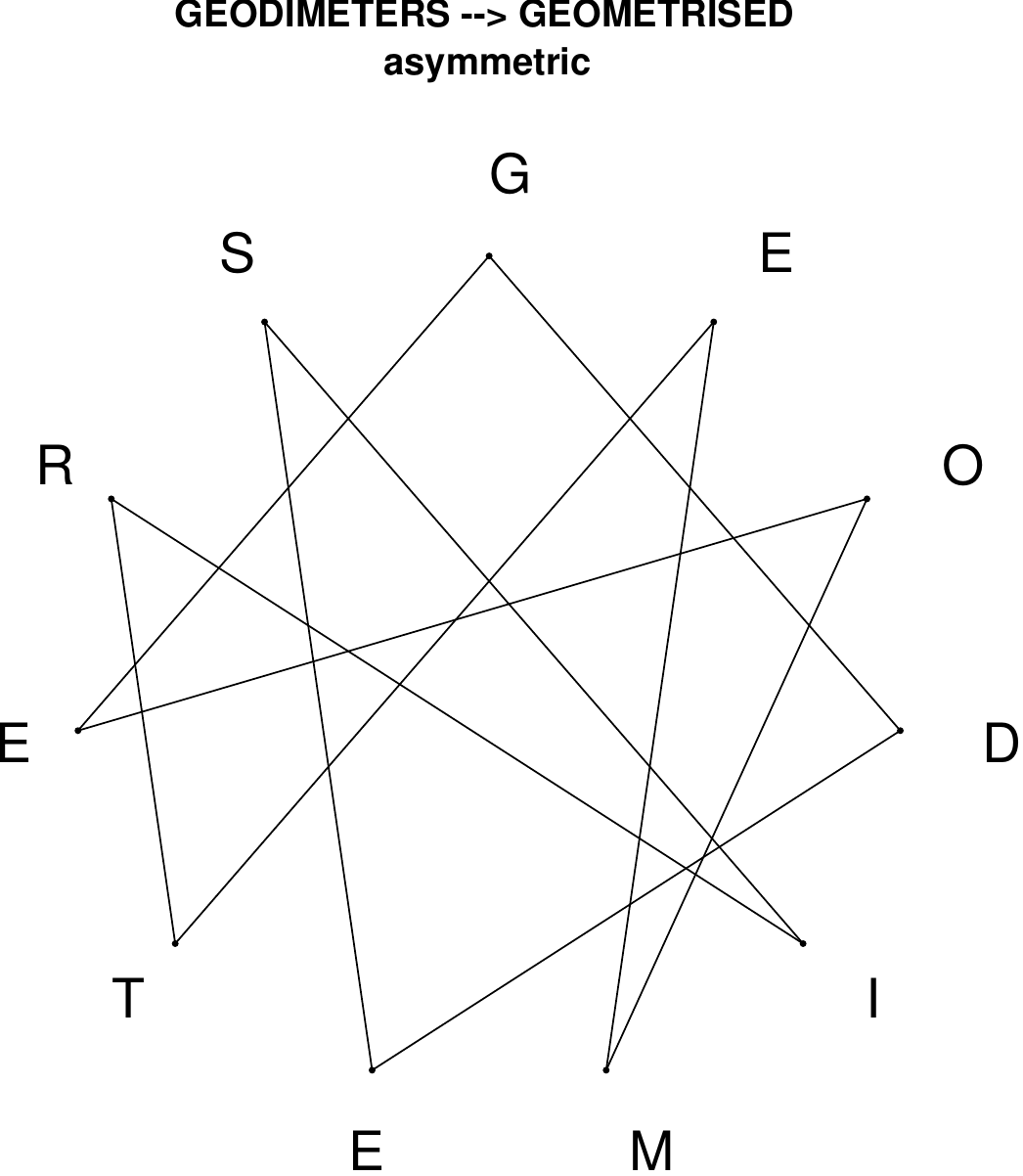}
\end{subfigure}
\hfill
\begin{subfigure}[T]{0.19\textwidth}
\centering
\includegraphics[width=\textwidth]{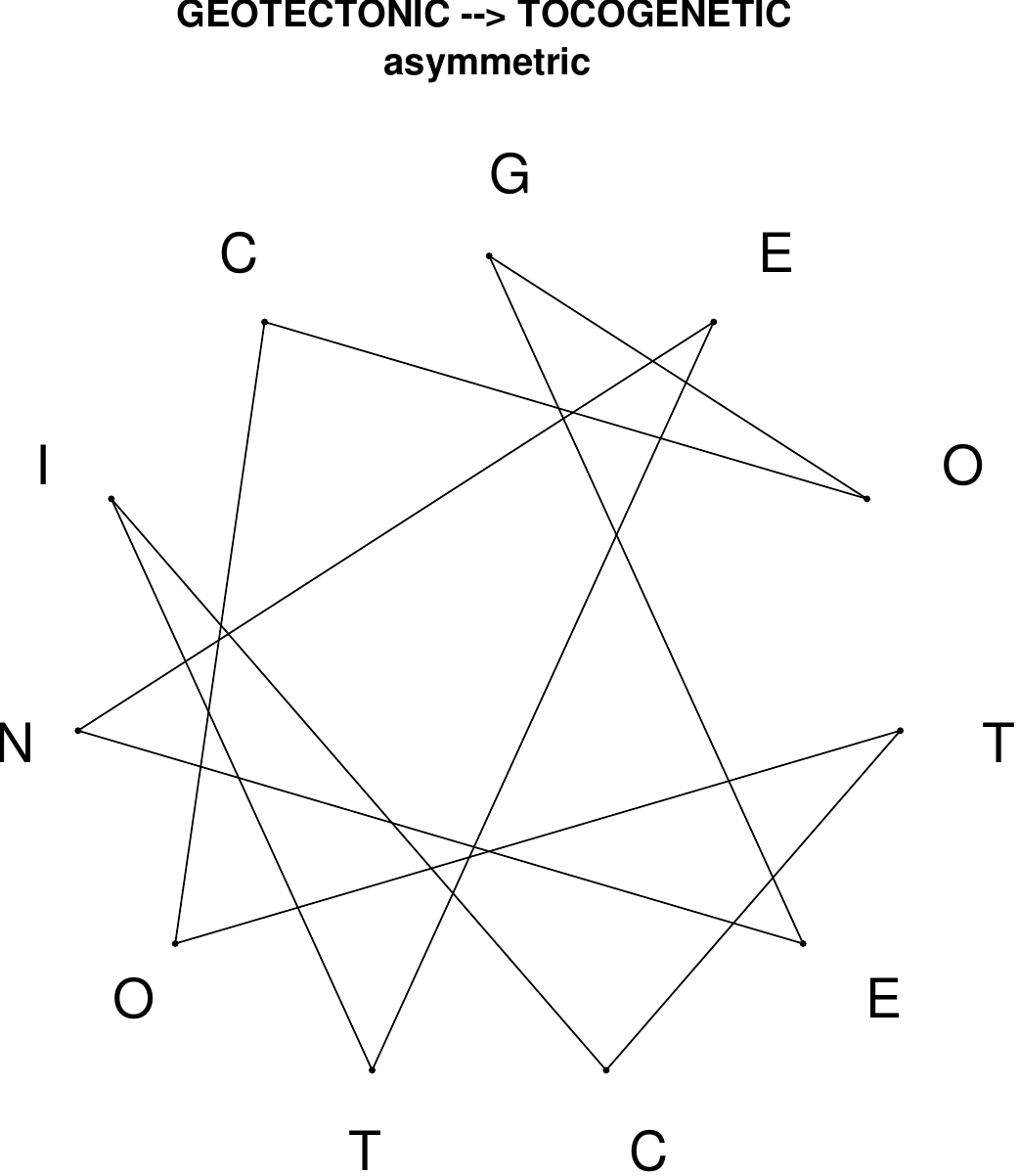}
\end{subfigure}
\hfill
\begin{subfigure}[T]{0.19\textwidth}
\centering
\includegraphics[width=\textwidth]{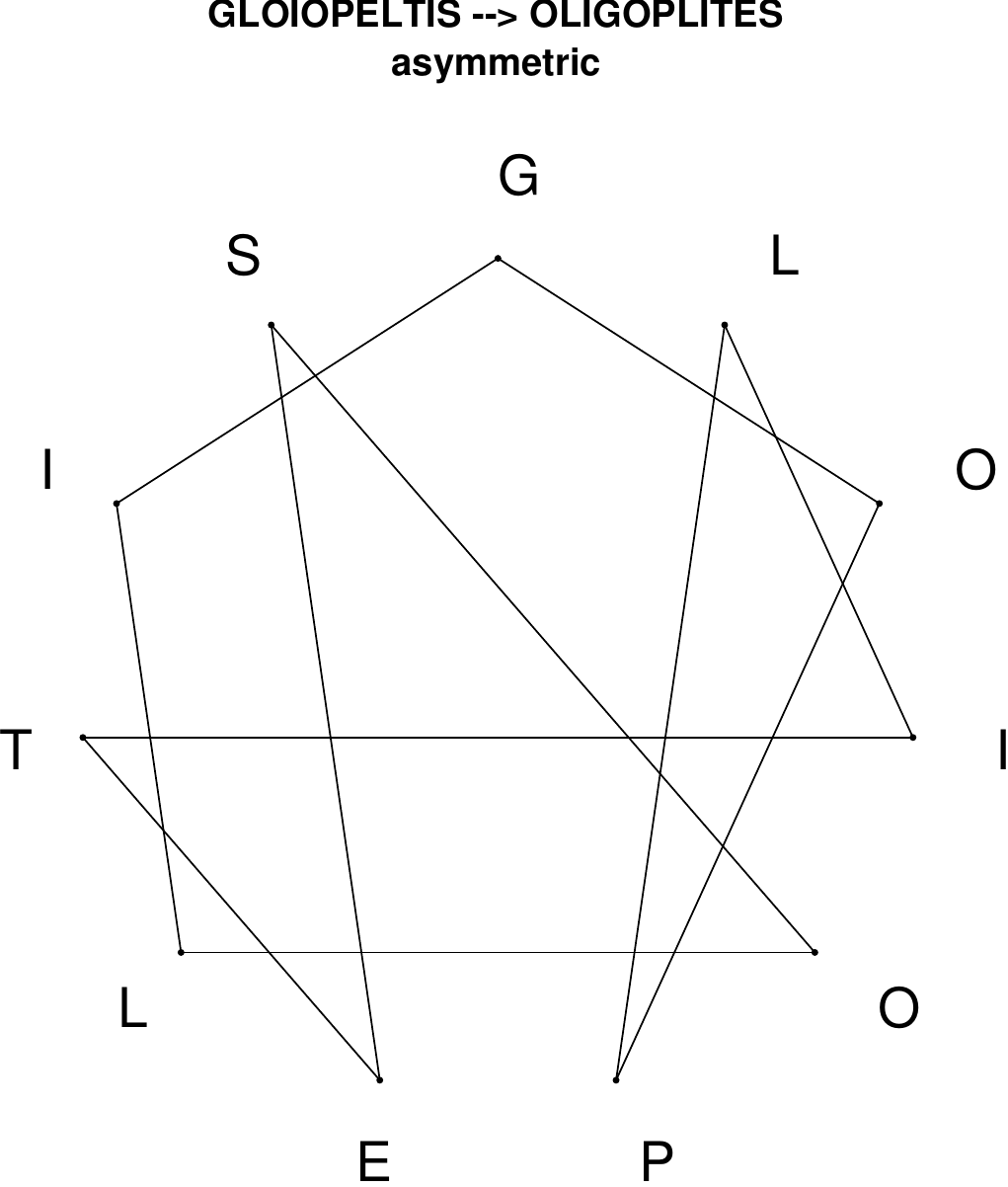}
\end{subfigure}
\hfill
\begin{subfigure}[T]{0.19\textwidth}
\centering
\includegraphics[width=\textwidth]{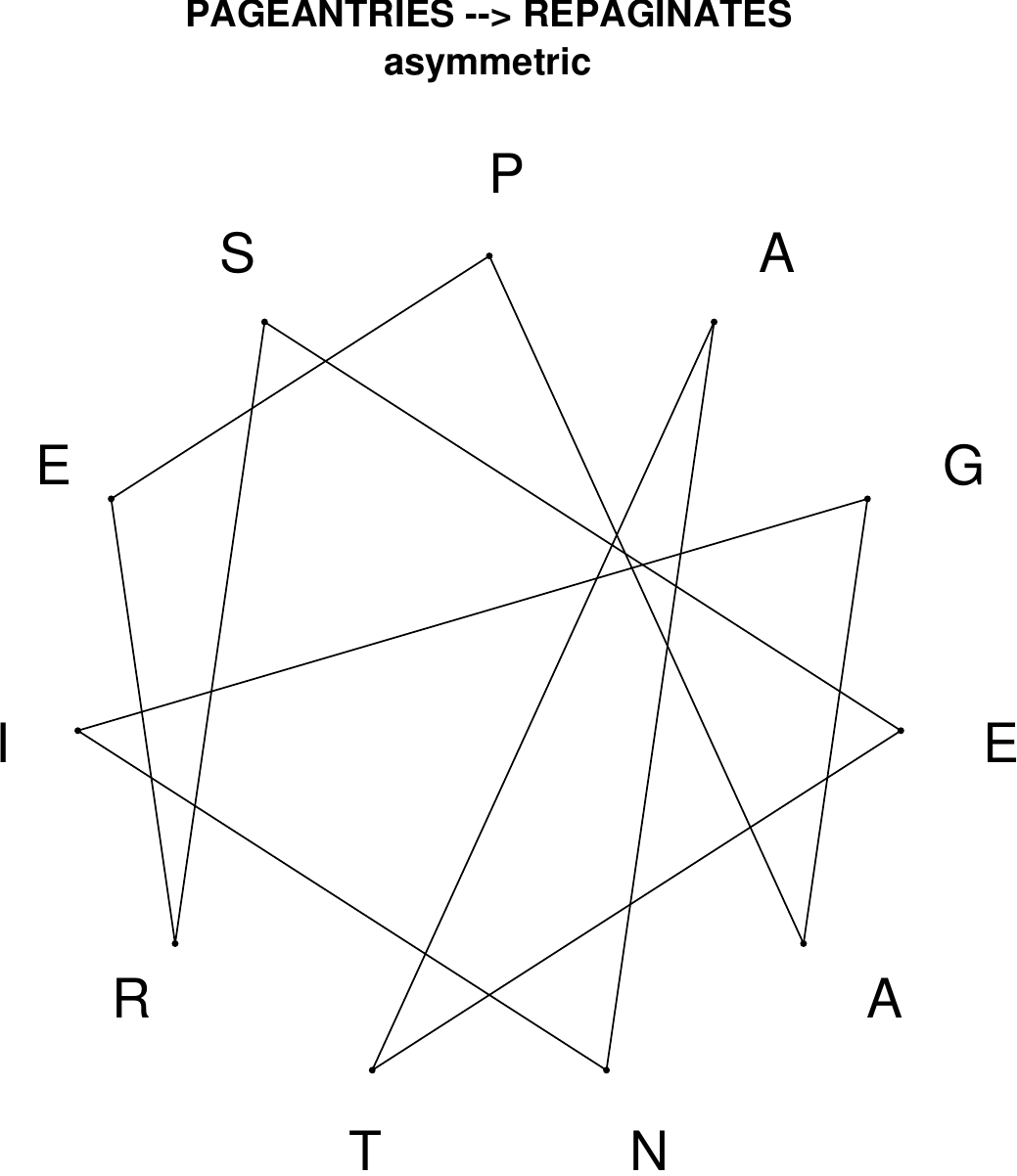}
\end{subfigure}
\hfill
\begin{subfigure}[T]{0.19\textwidth}
\centering
\includegraphics[width=\textwidth]{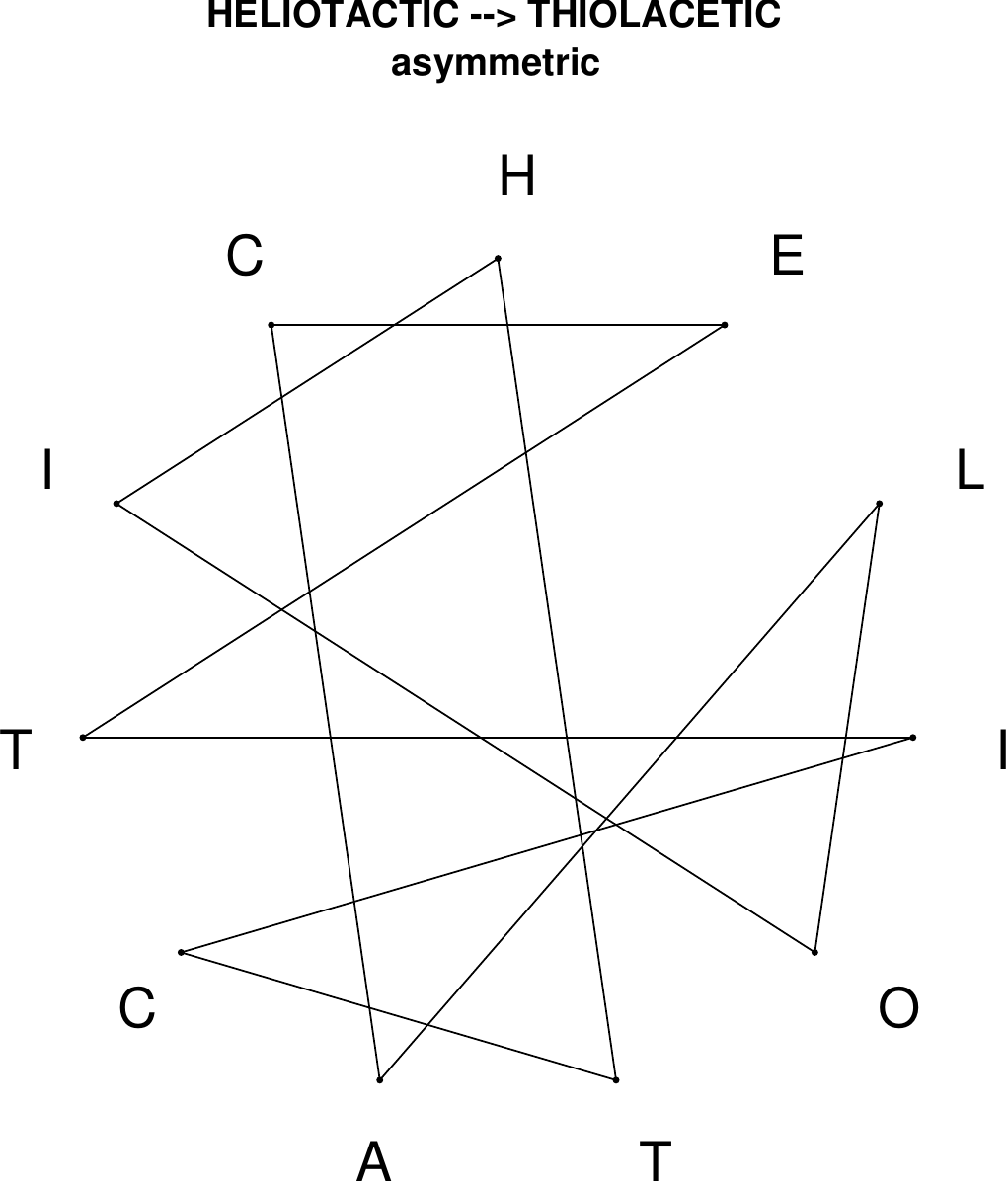}
\end{subfigure}
\end{figure}

\begin{figure}[H]
\centering
\begin{subfigure}[T]{0.19\textwidth}
\centering
\includegraphics[width=\textwidth]{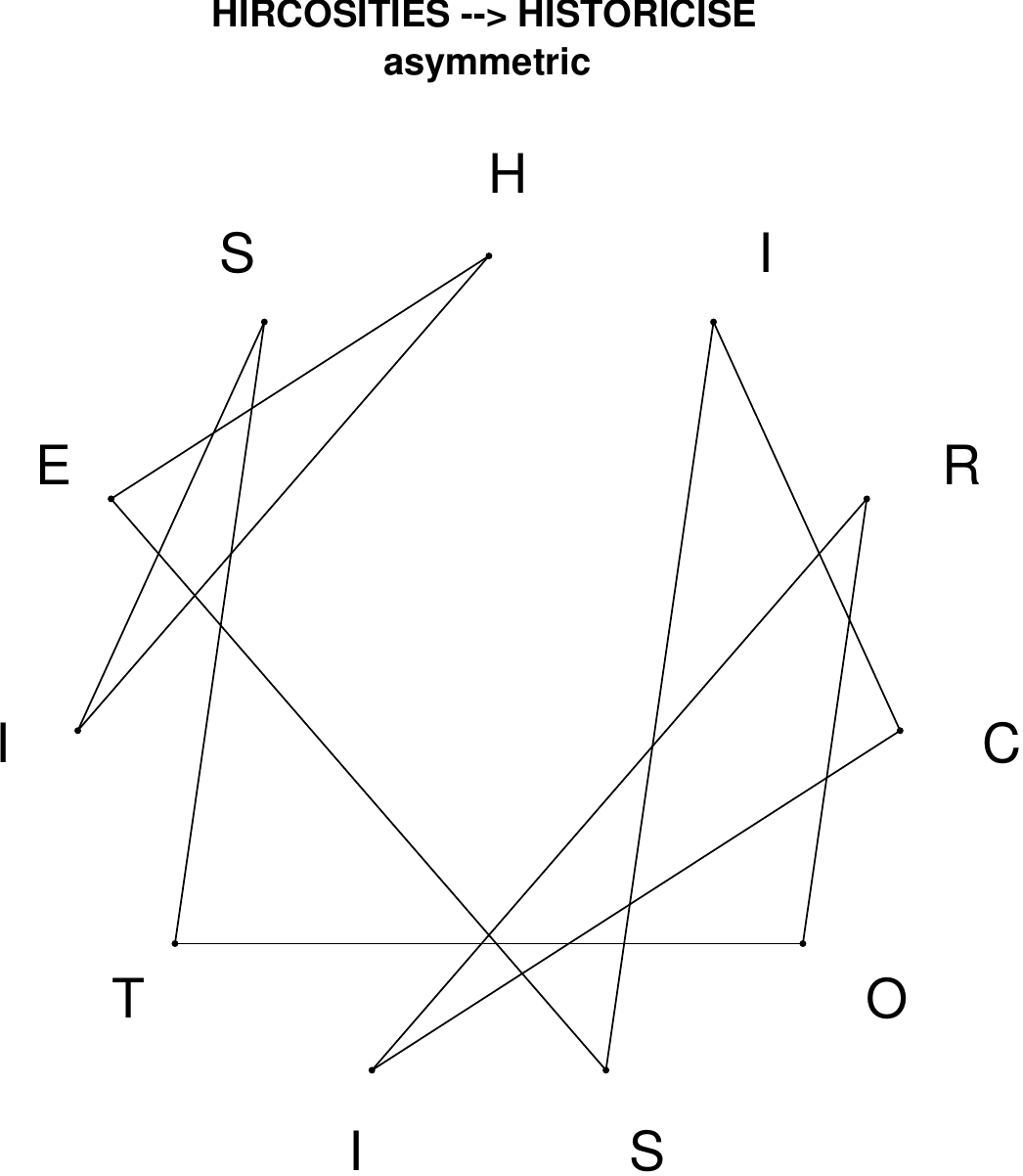}
\end{subfigure}
\hfill
\begin{subfigure}[T]{0.19\textwidth}
\centering
\includegraphics[width=\textwidth]{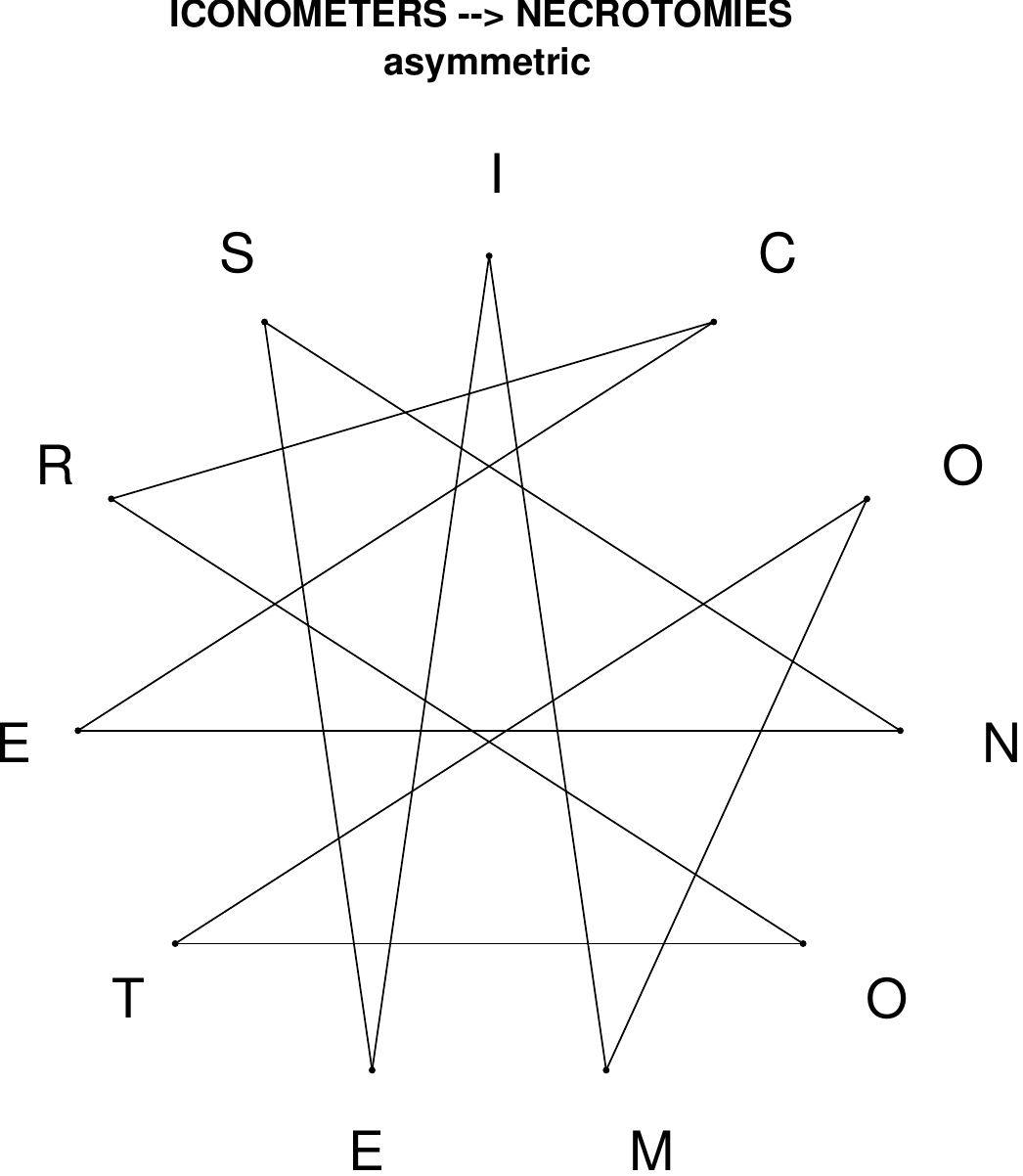}
\end{subfigure}
\hfill
\begin{subfigure}[T]{0.19\textwidth}
\centering
\includegraphics[width=\textwidth]{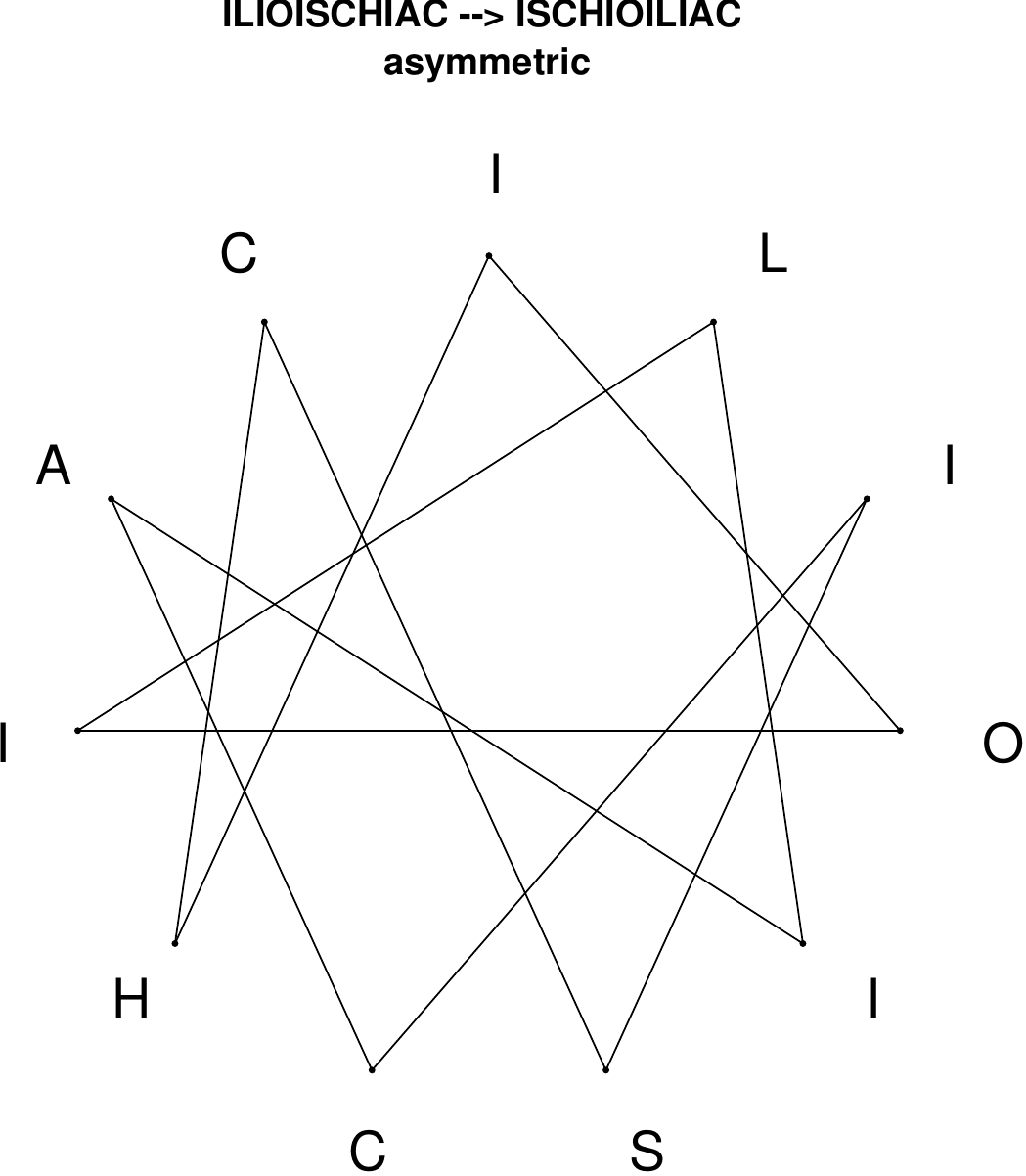}
\end{subfigure}
\hfill
\begin{subfigure}[T]{0.19\textwidth}
\centering
\includegraphics[width=\textwidth]{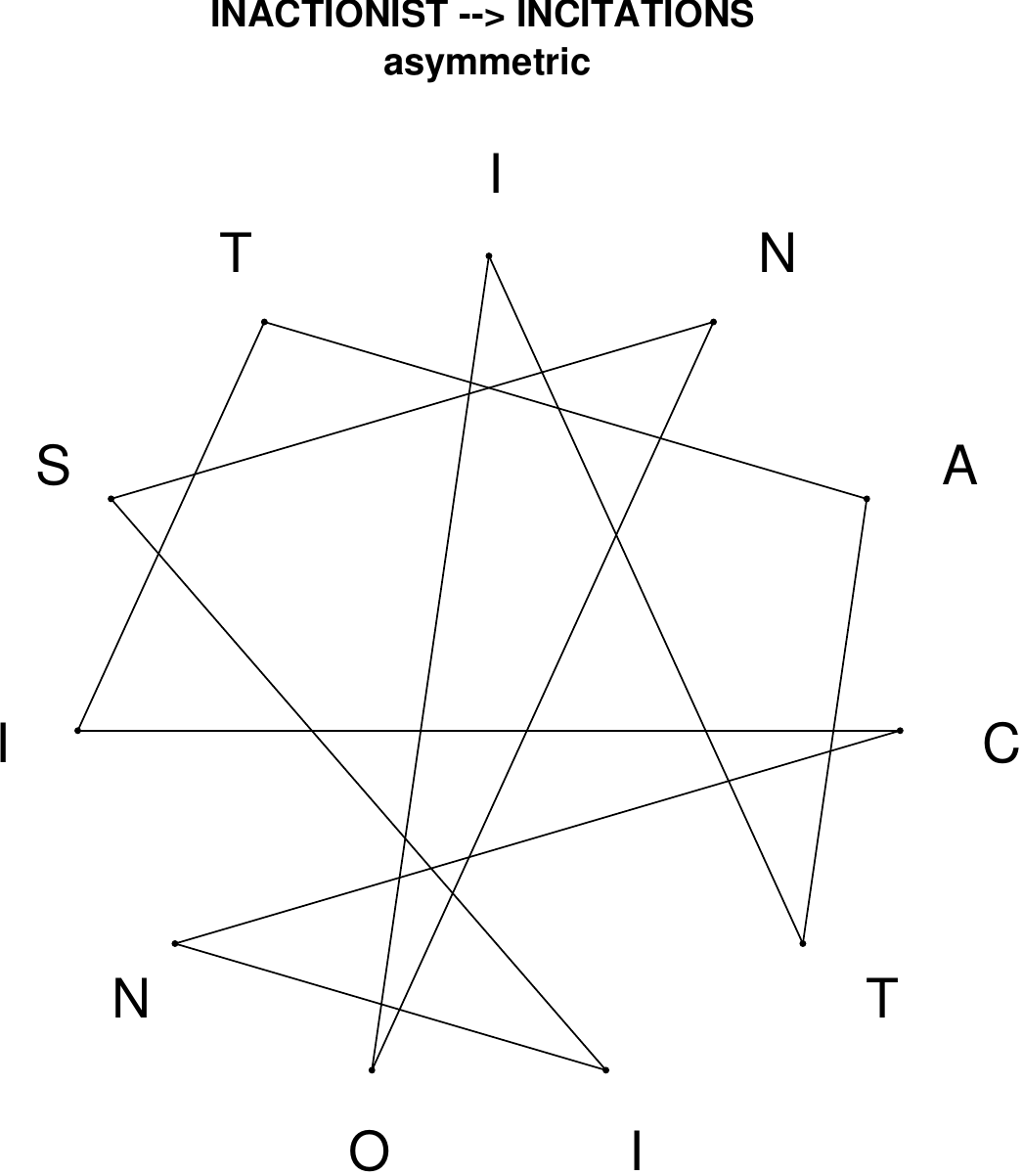}
\end{subfigure}
\hfill
\begin{subfigure}[T]{0.19\textwidth}
\centering
\includegraphics[width=\textwidth]{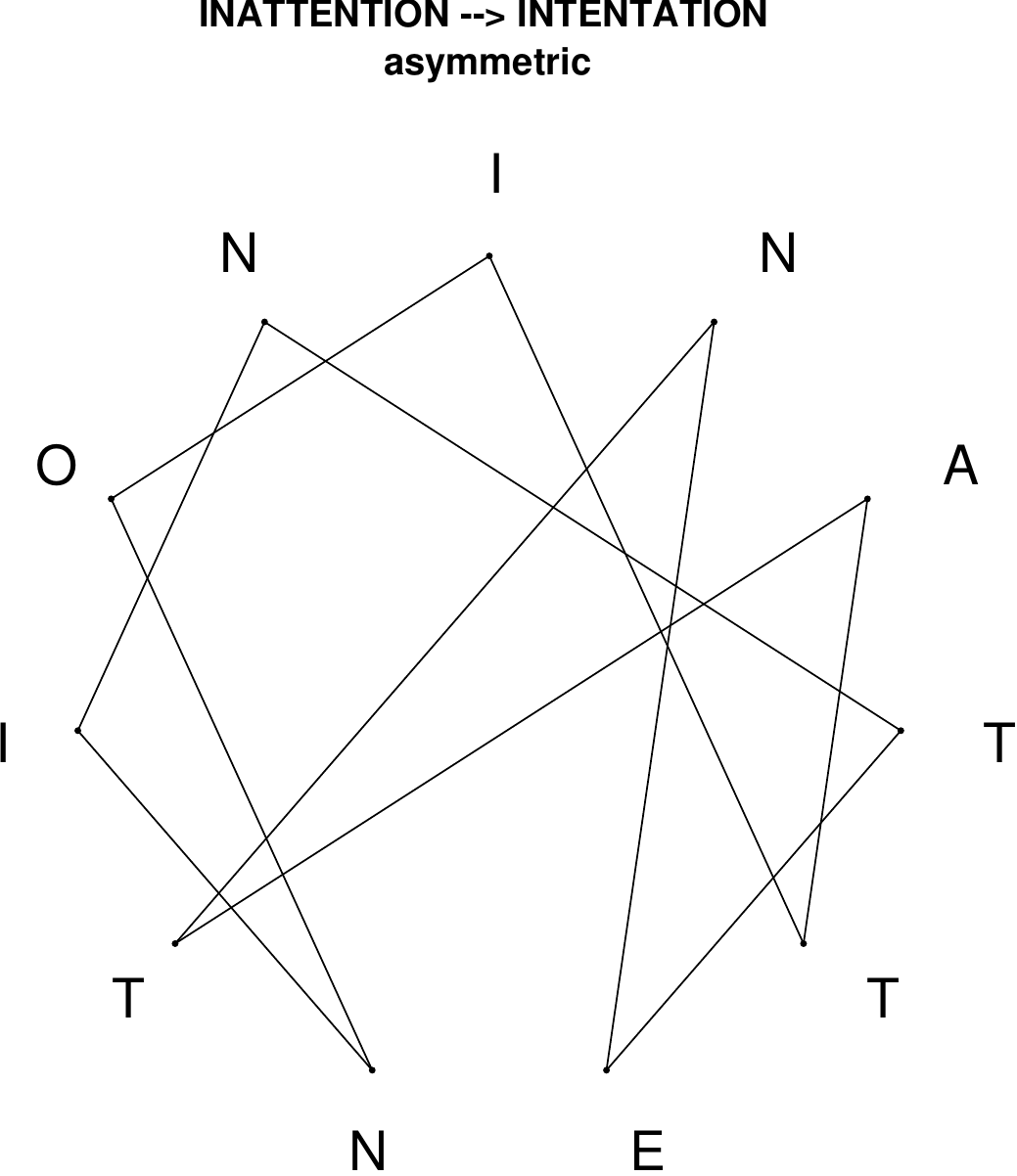}
\end{subfigure}
\end{figure}

\begin{figure}[H]
\centering
\begin{subfigure}[T]{0.19\textwidth}
\centering
\includegraphics[width=\textwidth]{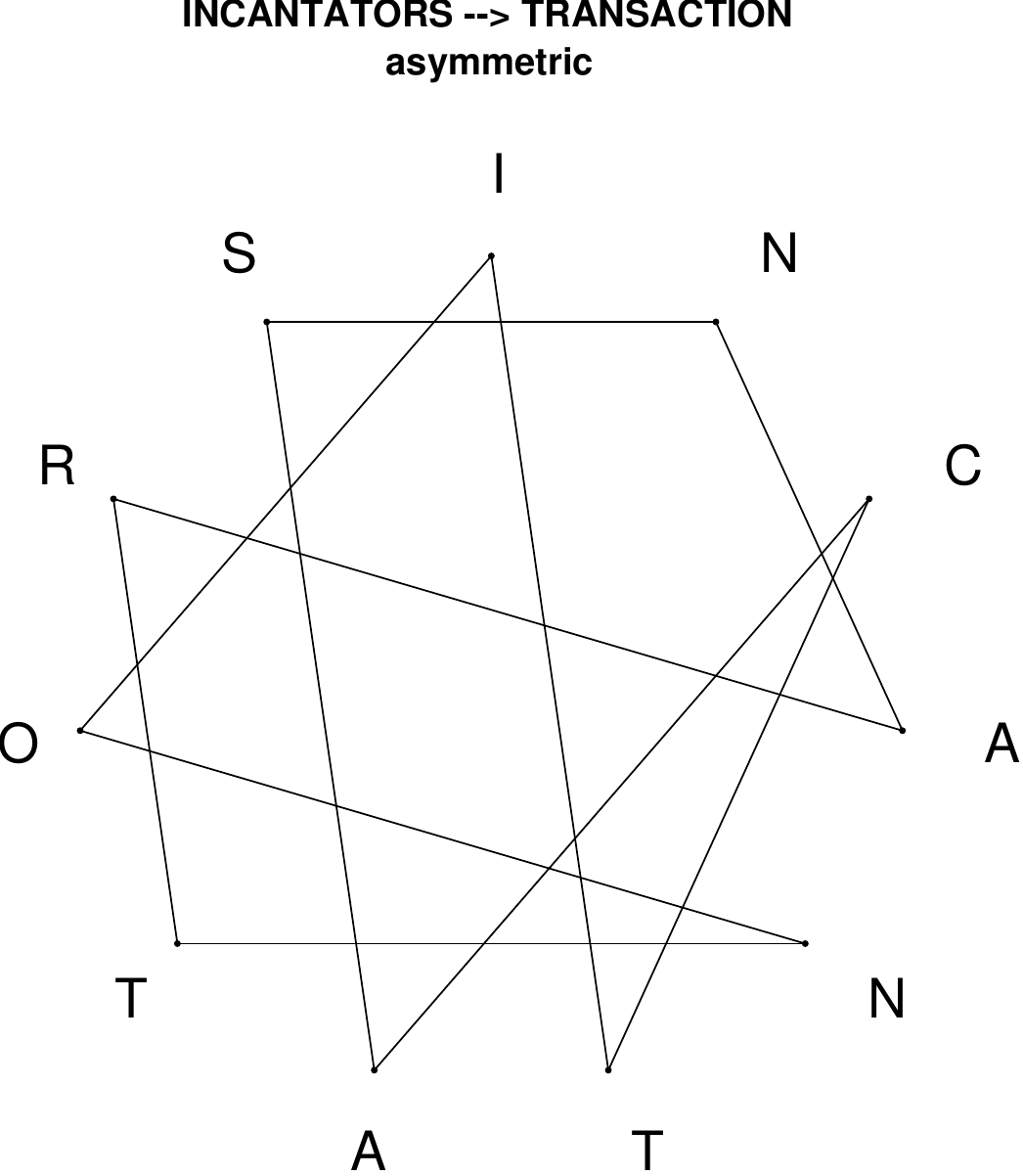}
\end{subfigure}
\hfill
\begin{subfigure}[T]{0.19\textwidth}
\centering
\includegraphics[width=\textwidth]{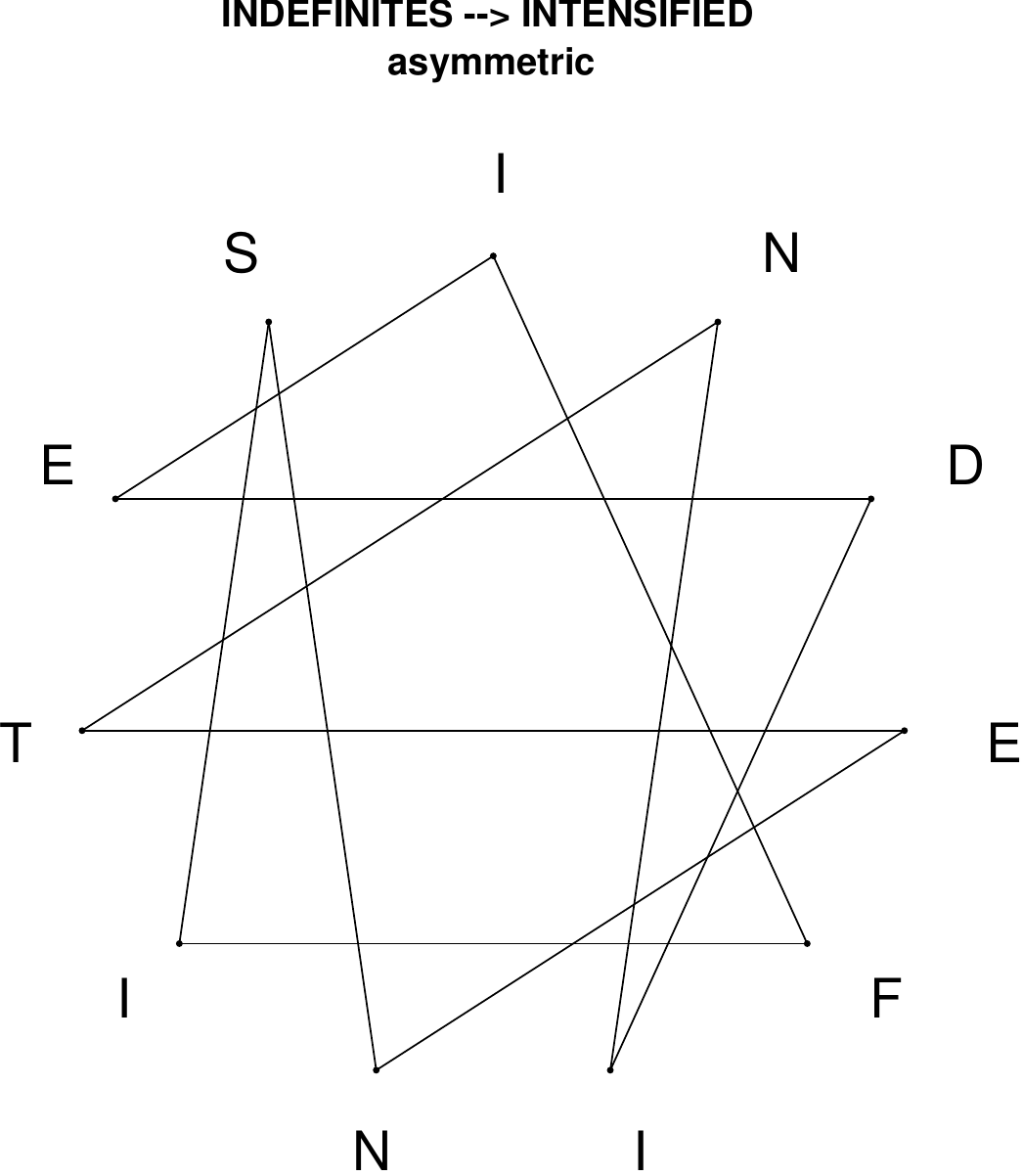}
\end{subfigure}
\hfill
\begin{subfigure}[T]{0.19\textwidth}
\centering
\includegraphics[width=\textwidth]{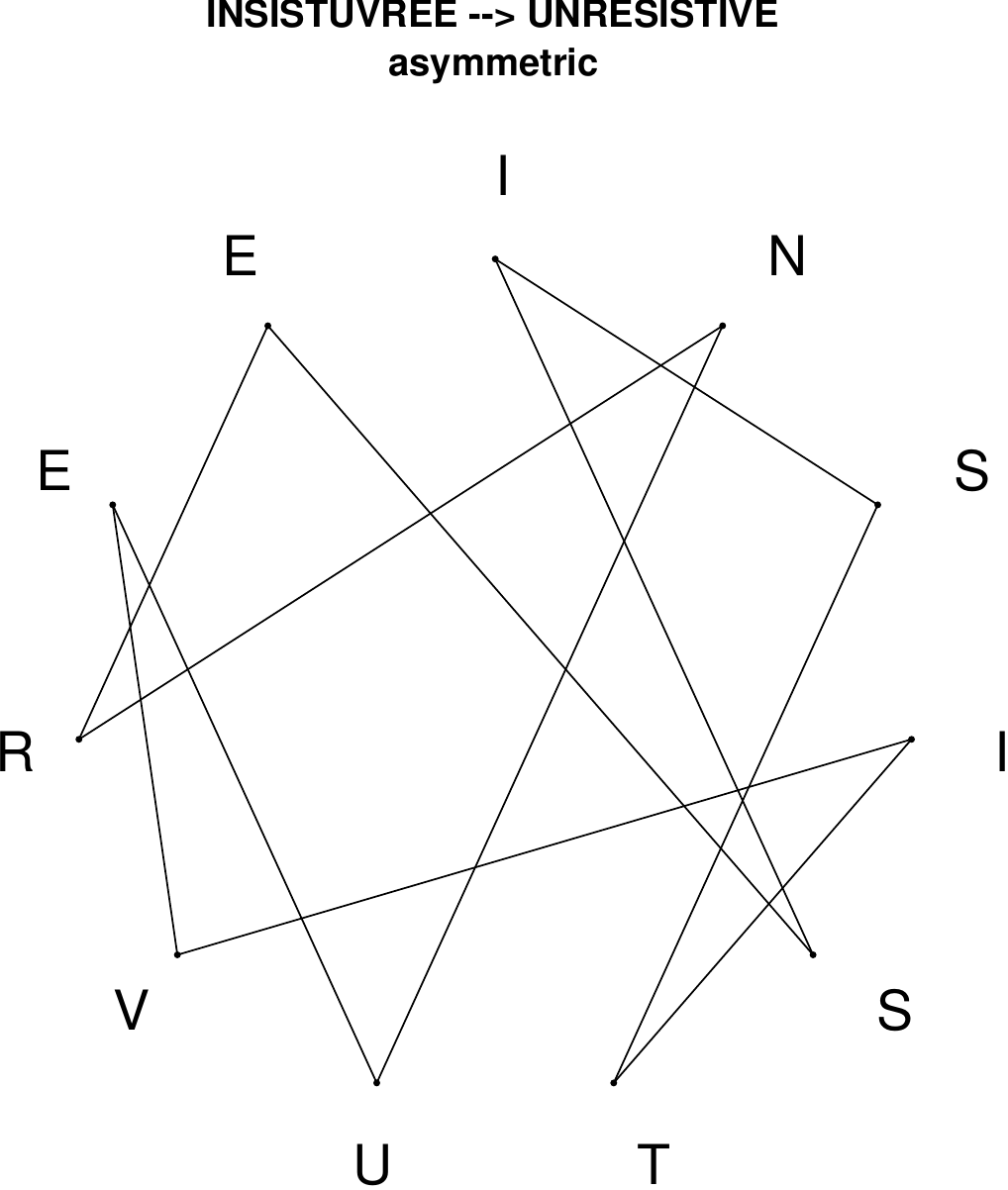}
\end{subfigure}
\hfill
\begin{subfigure}[T]{0.19\textwidth}
\centering
\includegraphics[width=\textwidth]{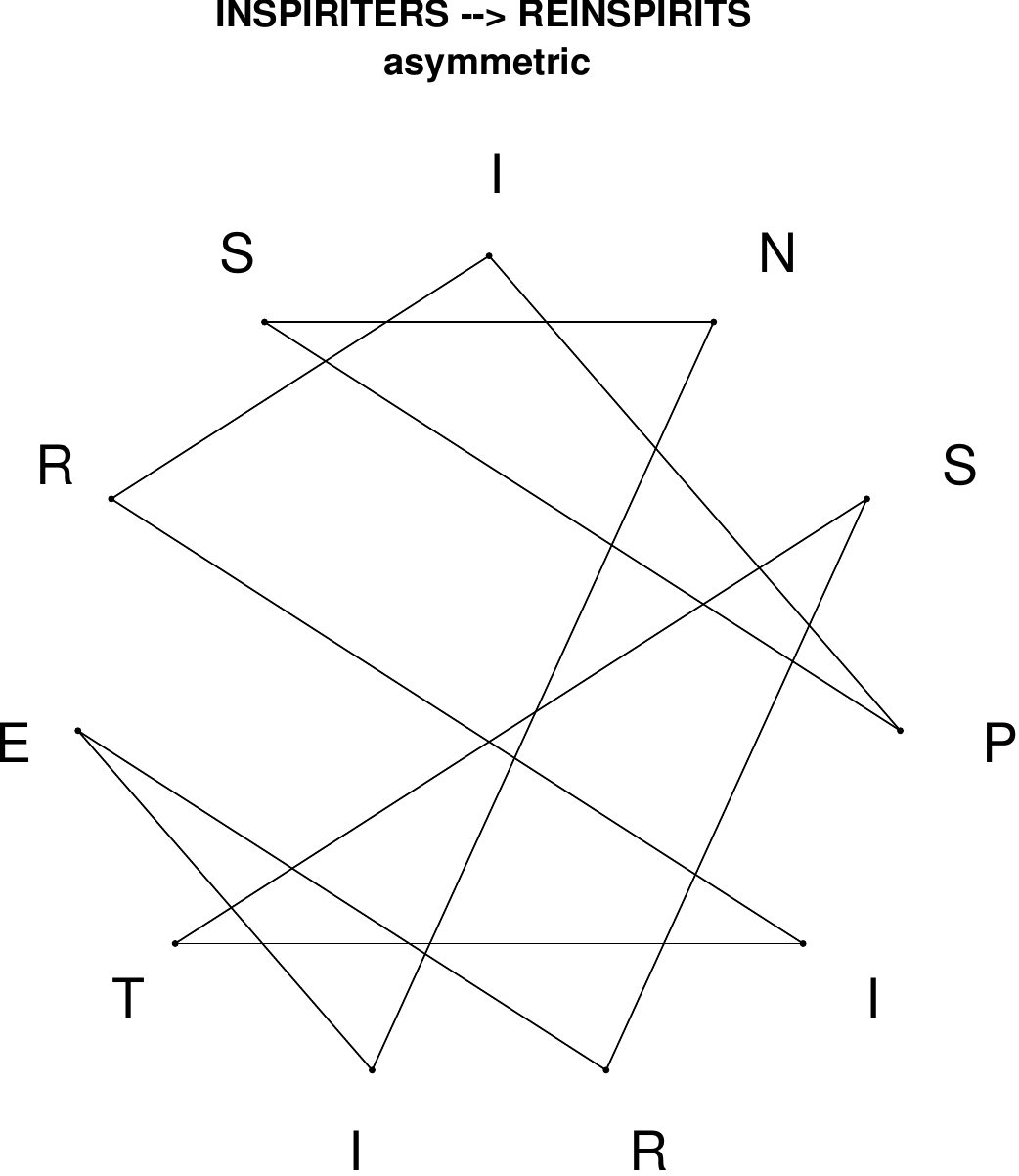}
\end{subfigure}
\hfill
\begin{subfigure}[T]{0.19\textwidth}
\centering
\includegraphics[width=\textwidth]{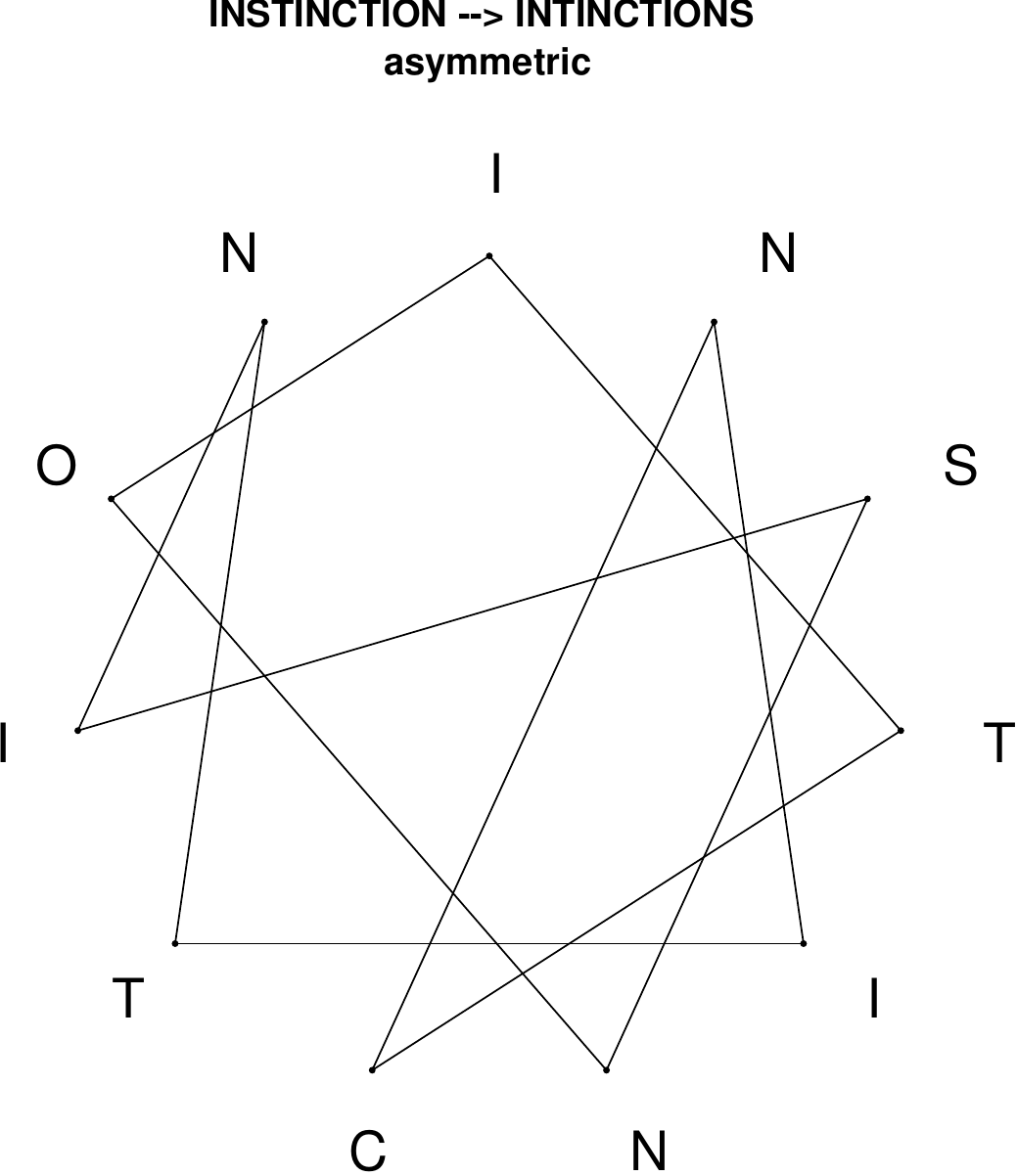}
\end{subfigure}
\end{figure}

\begin{figure}[H]
\centering
\begin{subfigure}[T]{0.19\textwidth}
\centering
\includegraphics[width=\textwidth]{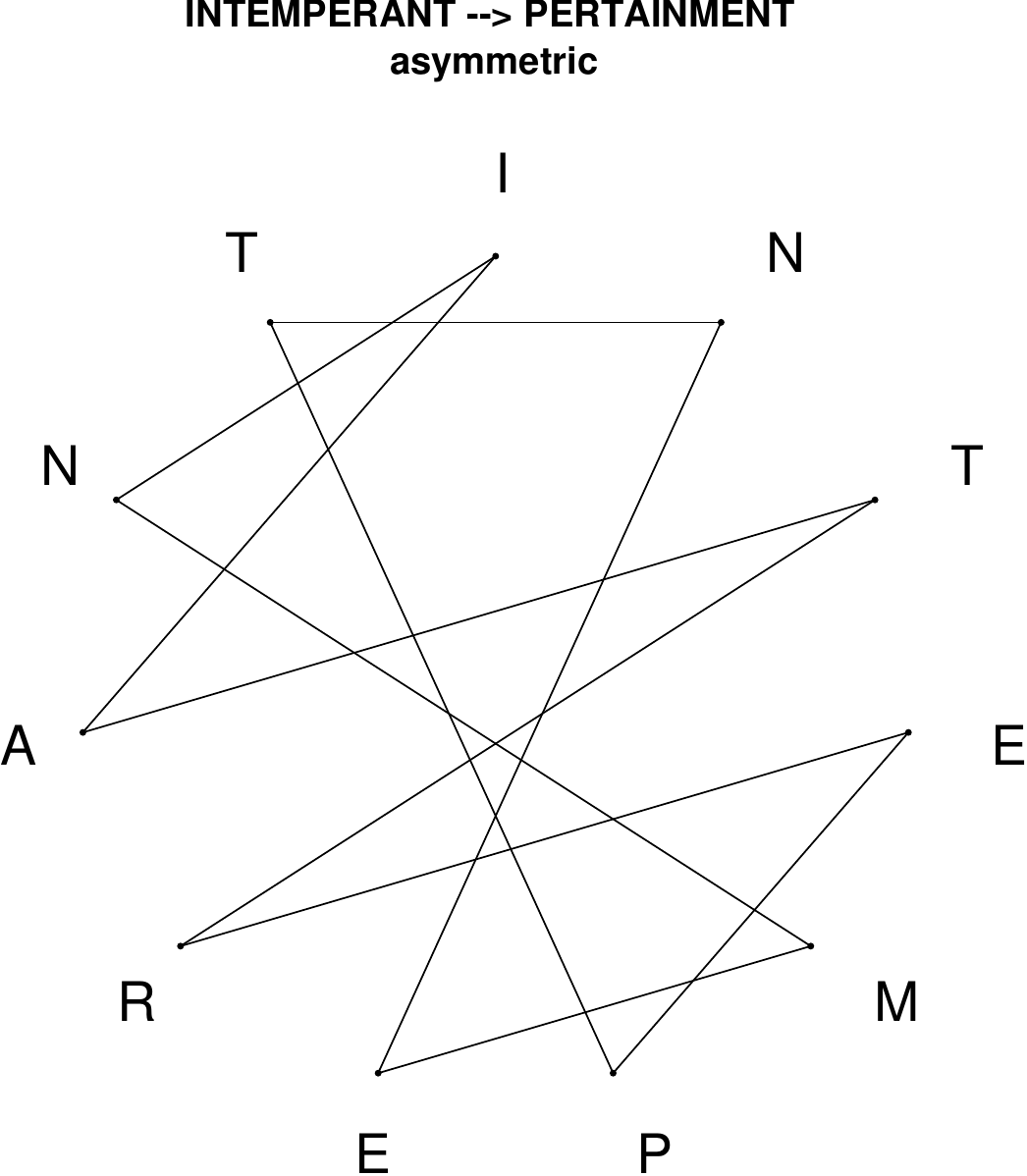}
\end{subfigure}
\hfill
\begin{subfigure}[T]{0.19\textwidth}
\centering
\includegraphics[width=\textwidth]{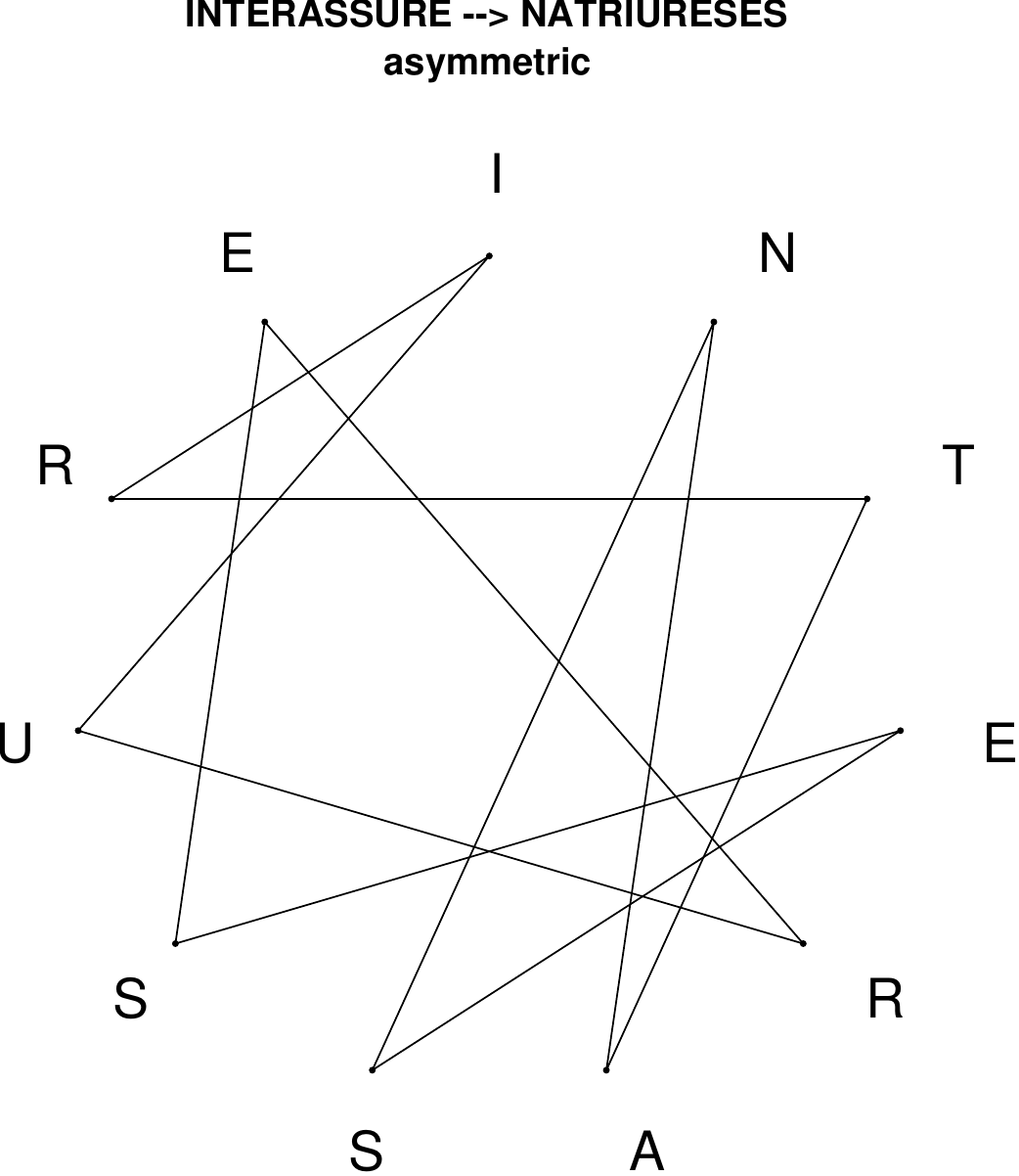}
\end{subfigure}
\hfill
\begin{subfigure}[T]{0.19\textwidth}
\centering
\includegraphics[width=\textwidth]{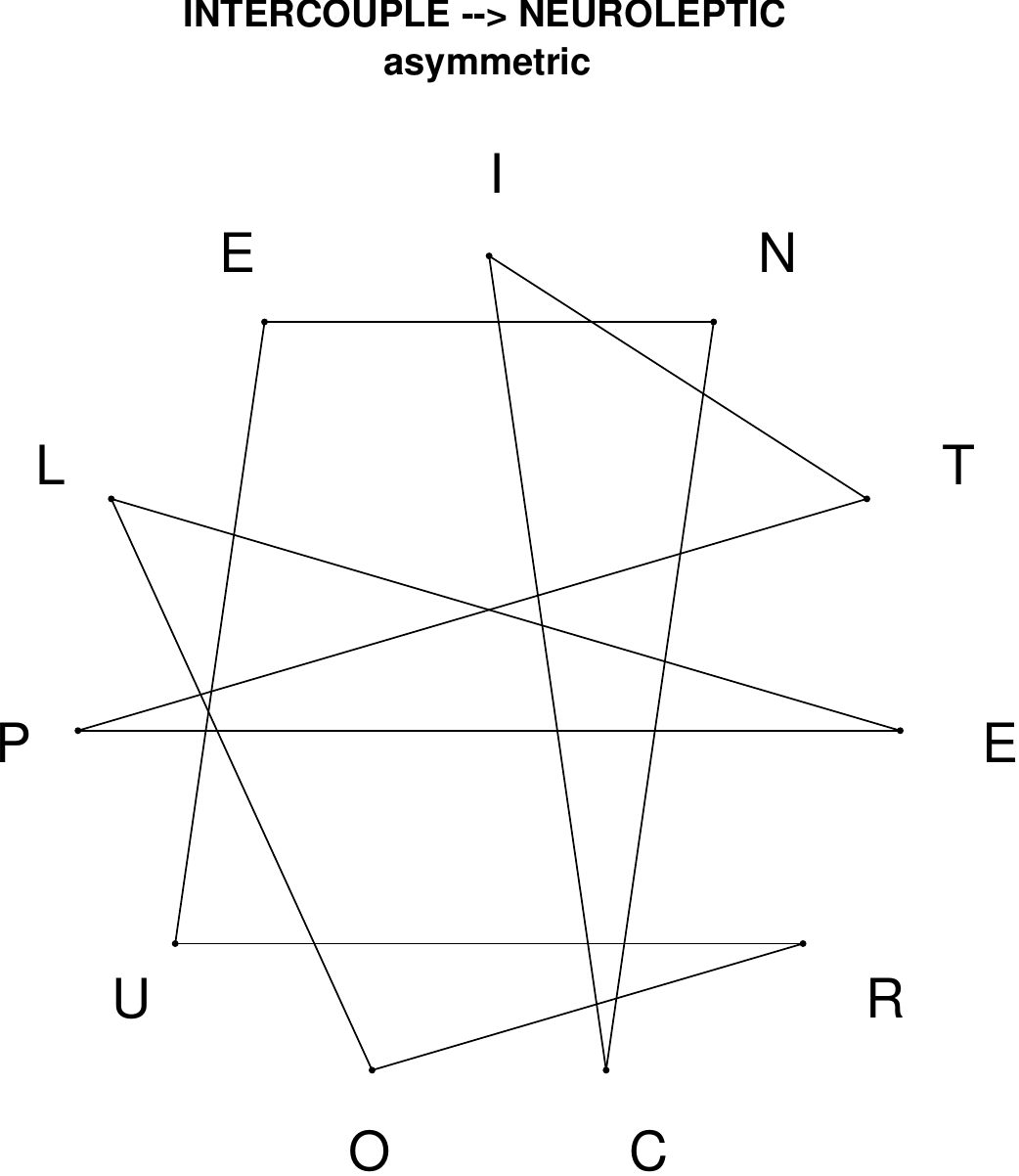}
\end{subfigure}
\hfill
\begin{subfigure}[T]{0.19\textwidth}
\centering
\includegraphics[width=\textwidth]{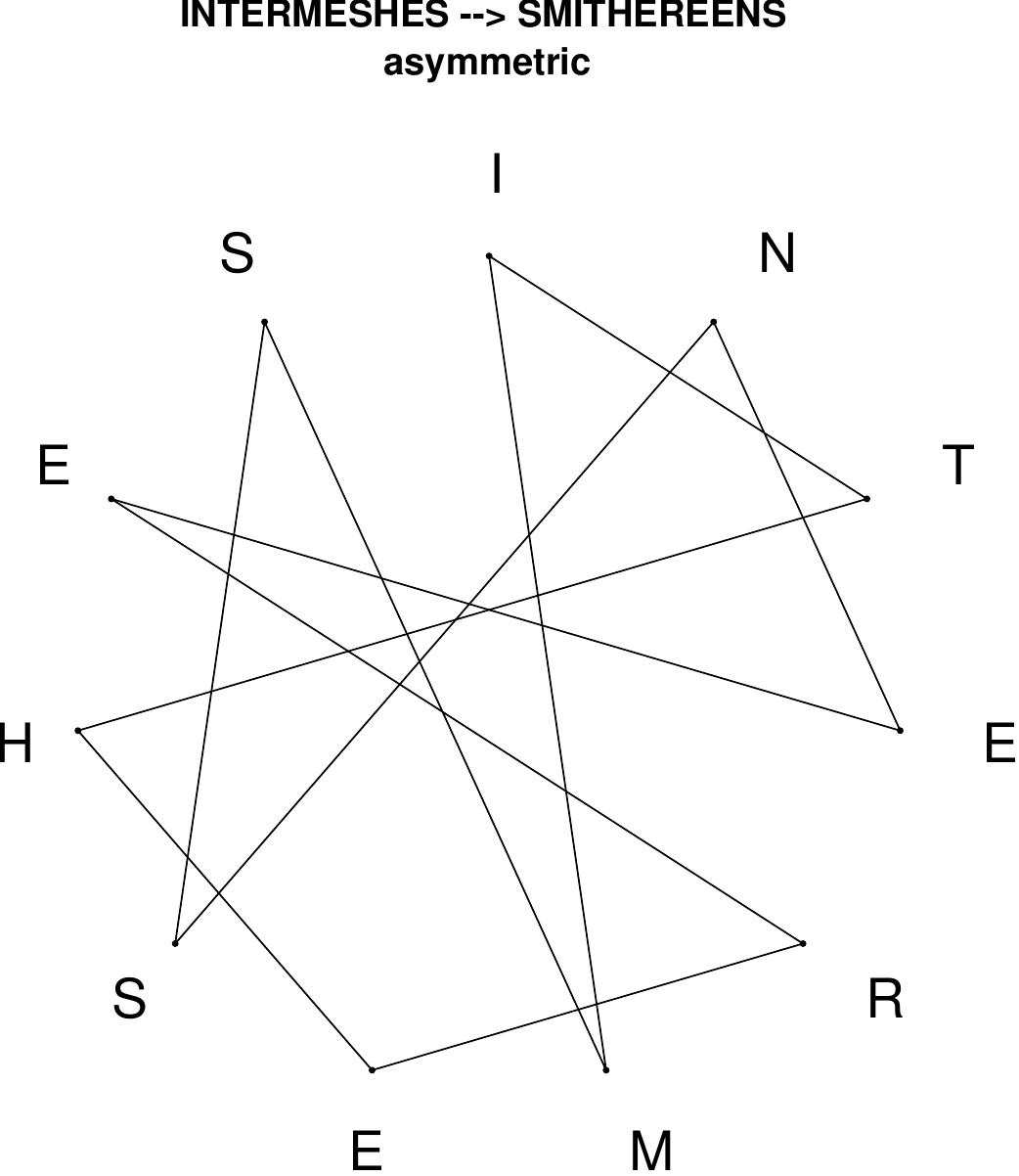}
\end{subfigure}
\hfill
\begin{subfigure}[T]{0.19\textwidth}
\centering
\includegraphics[width=\textwidth]{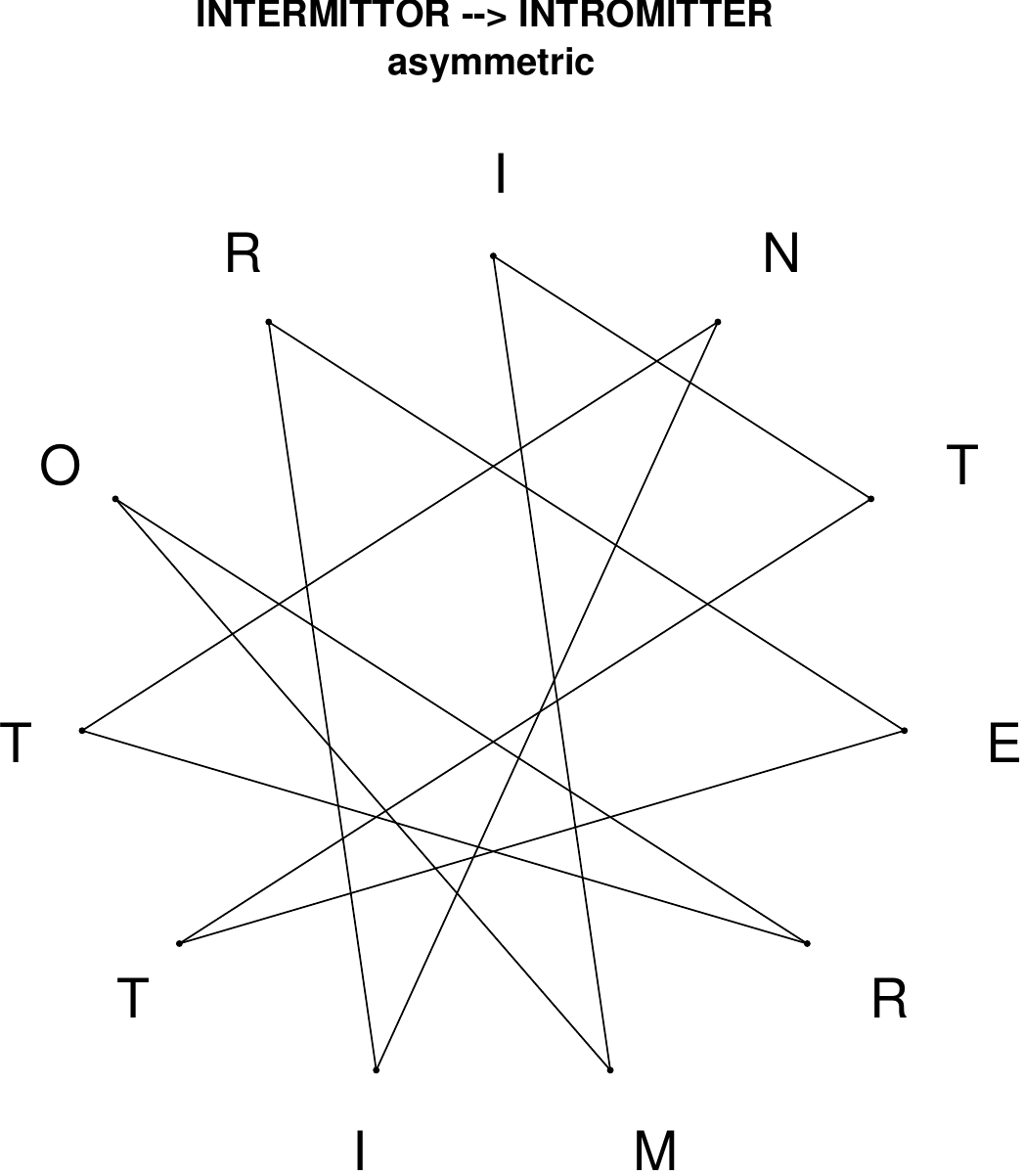}
\end{subfigure}
\end{figure}

\begin{figure}[H]
\centering
\begin{subfigure}[T]{0.19\textwidth}
\centering
\includegraphics[width=\textwidth]{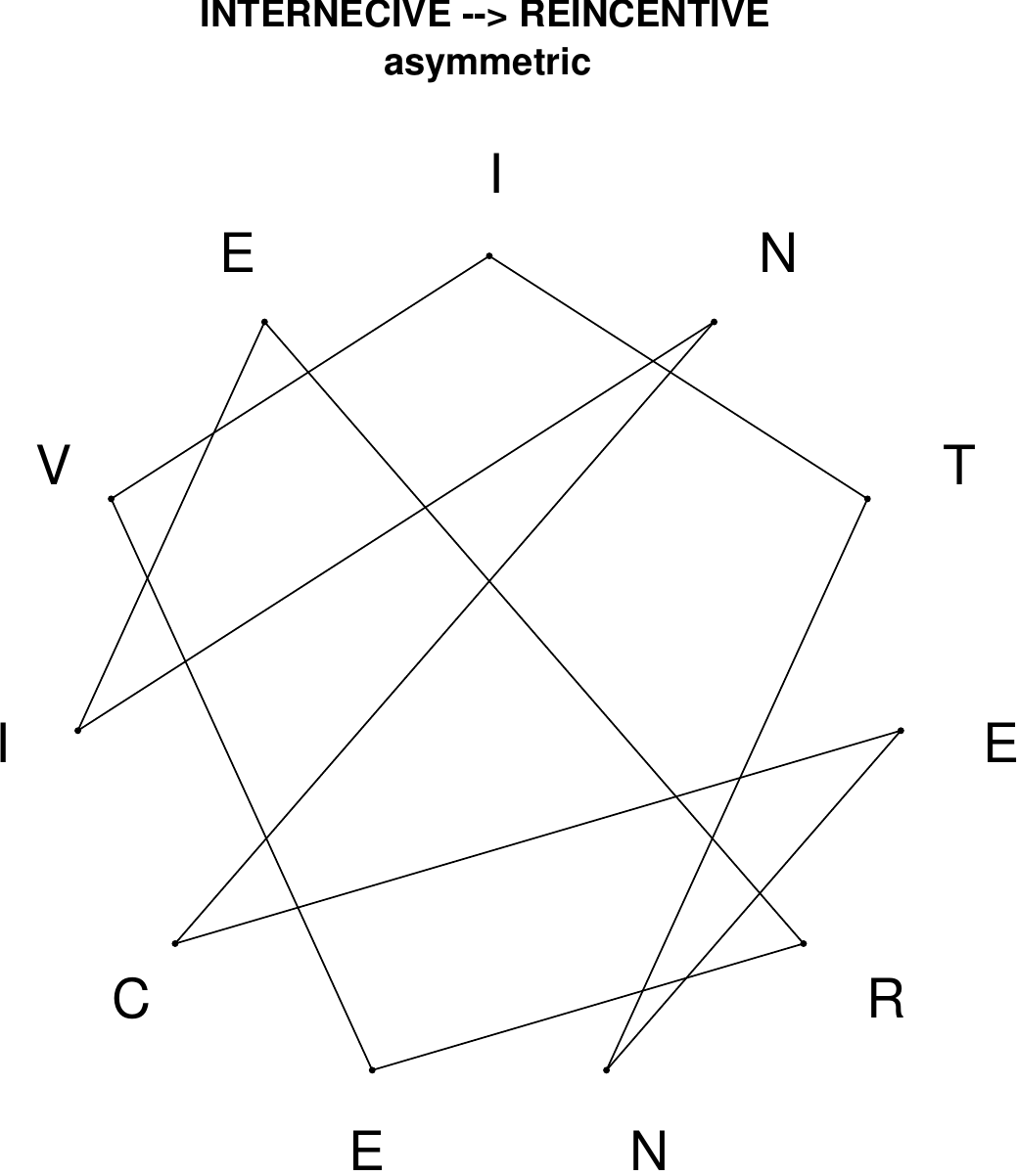}
\end{subfigure}
\hfill
\begin{subfigure}[T]{0.19\textwidth}
\centering
\includegraphics[width=\textwidth]{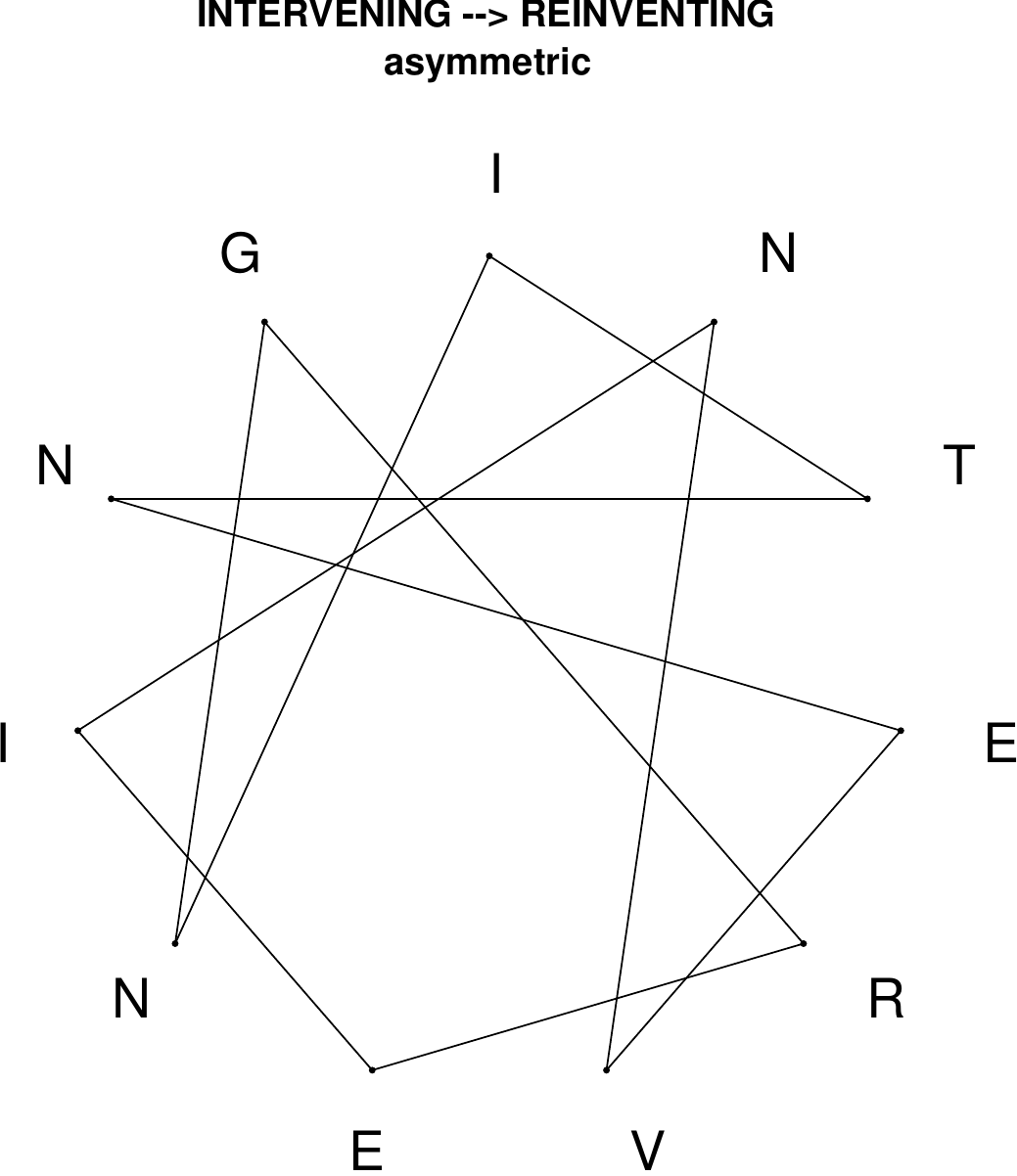}
\end{subfigure}
\hfill
\begin{subfigure}[T]{0.19\textwidth}
\centering
\includegraphics[width=\textwidth]{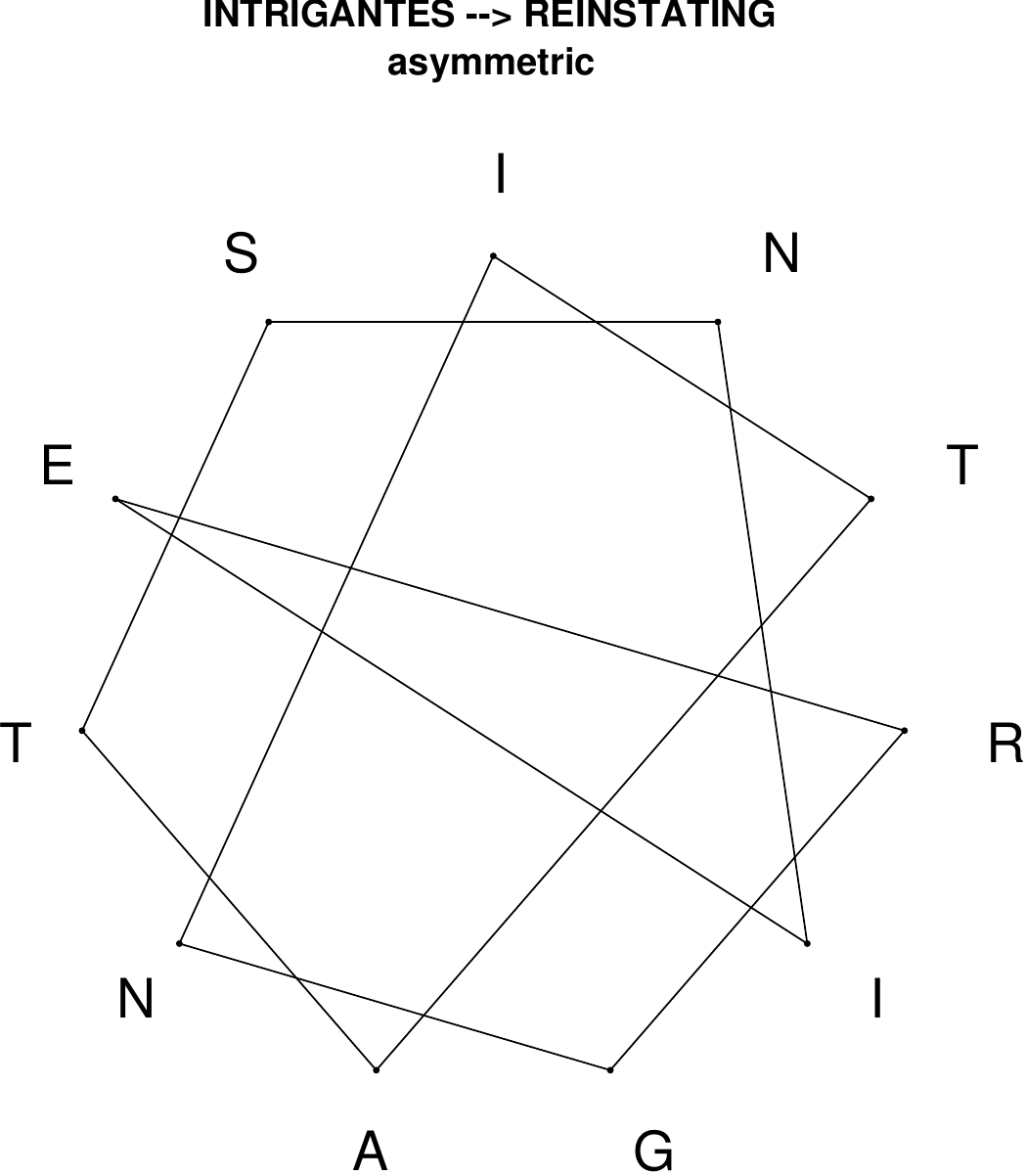}
\end{subfigure}
\hfill
\begin{subfigure}[T]{0.19\textwidth}
\centering
\includegraphics[width=\textwidth]{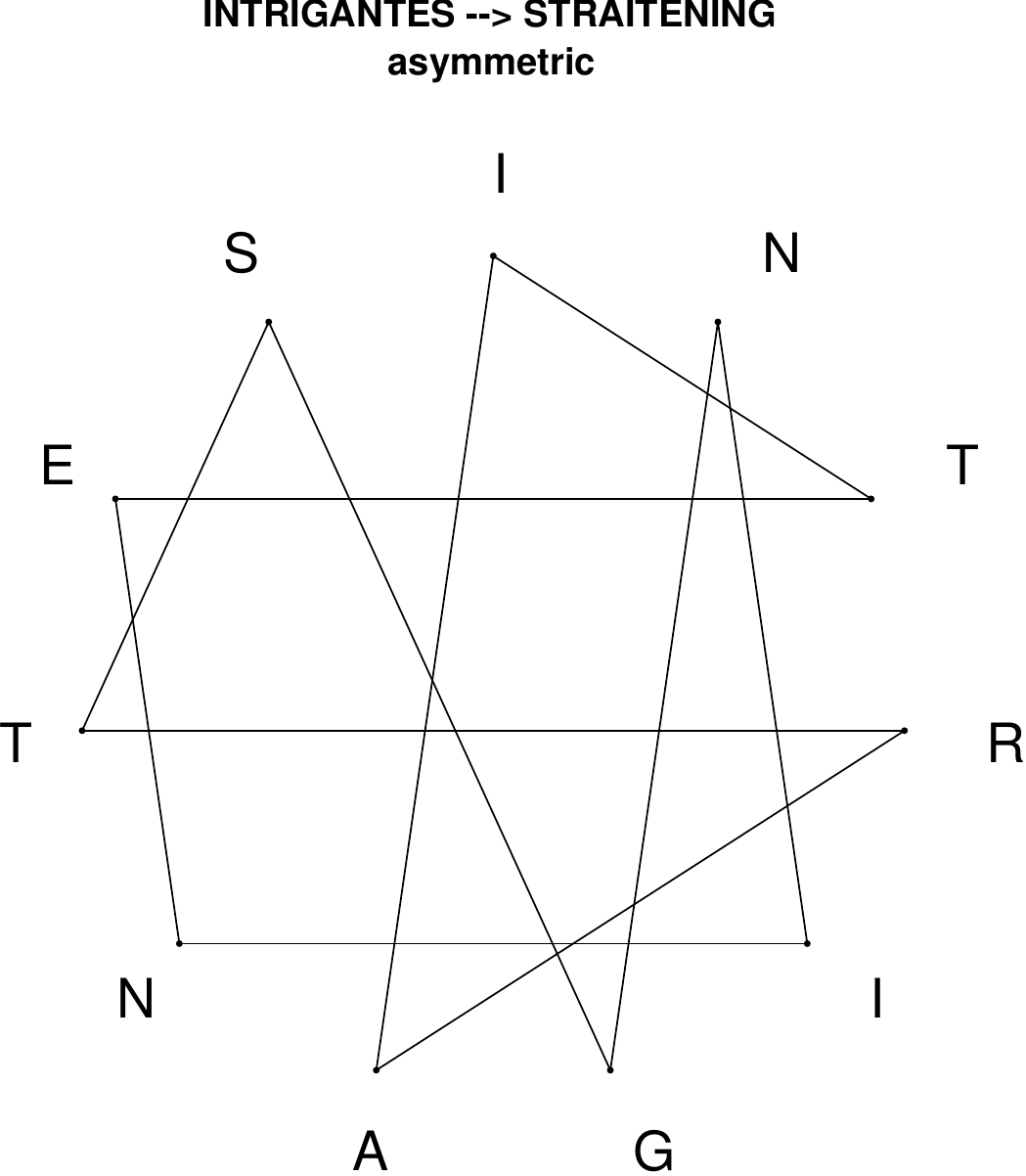}
\end{subfigure}
\hfill
\begin{subfigure}[T]{0.19\textwidth}
\centering
\includegraphics[width=\textwidth]{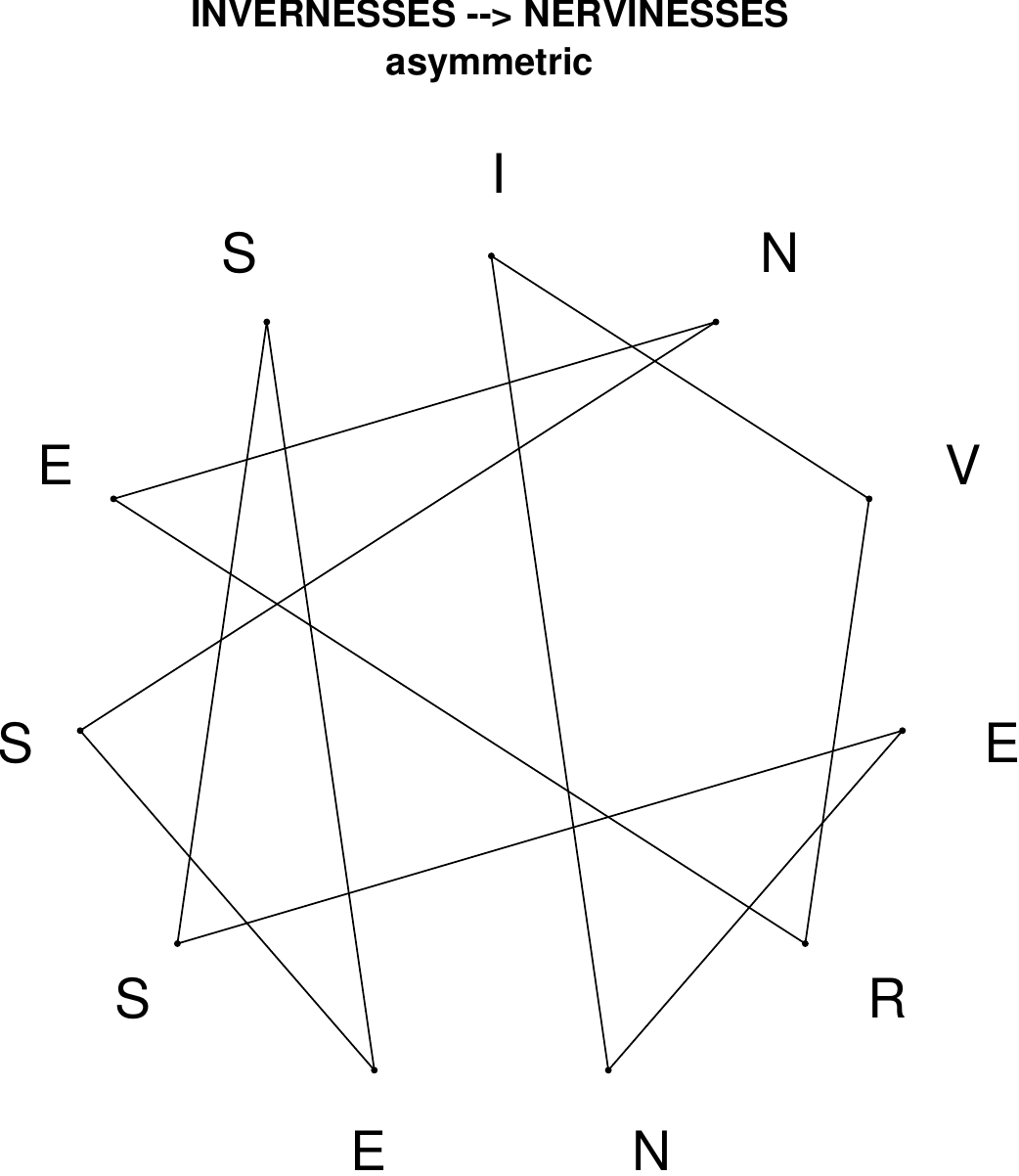}
\end{subfigure}
\end{figure}

\begin{figure}[H]
\centering
\begin{subfigure}[T]{0.19\textwidth}
\centering
\includegraphics[width=\textwidth]{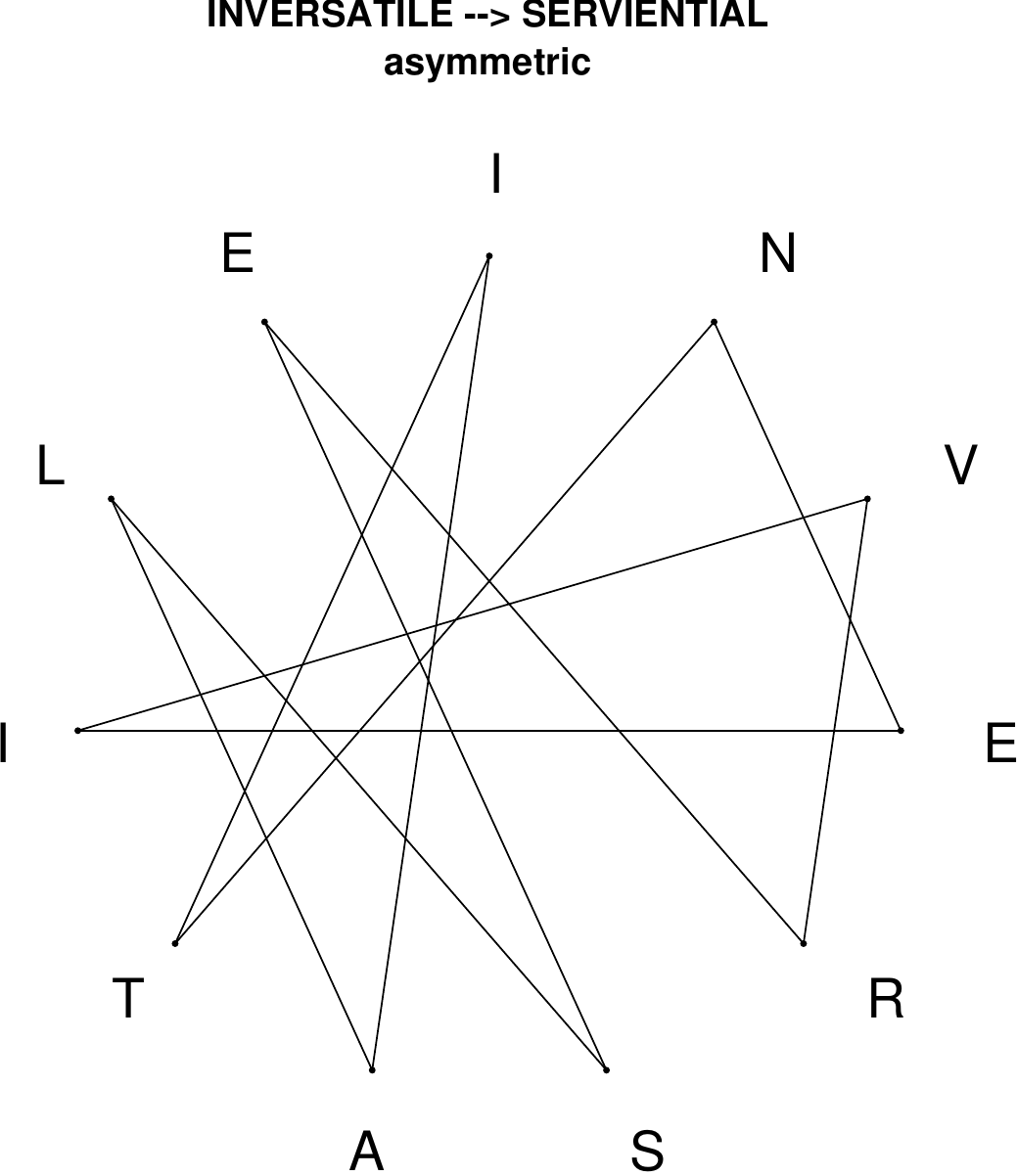}
\end{subfigure}
\hfill
\begin{subfigure}[T]{0.19\textwidth}
\centering
\includegraphics[width=\textwidth]{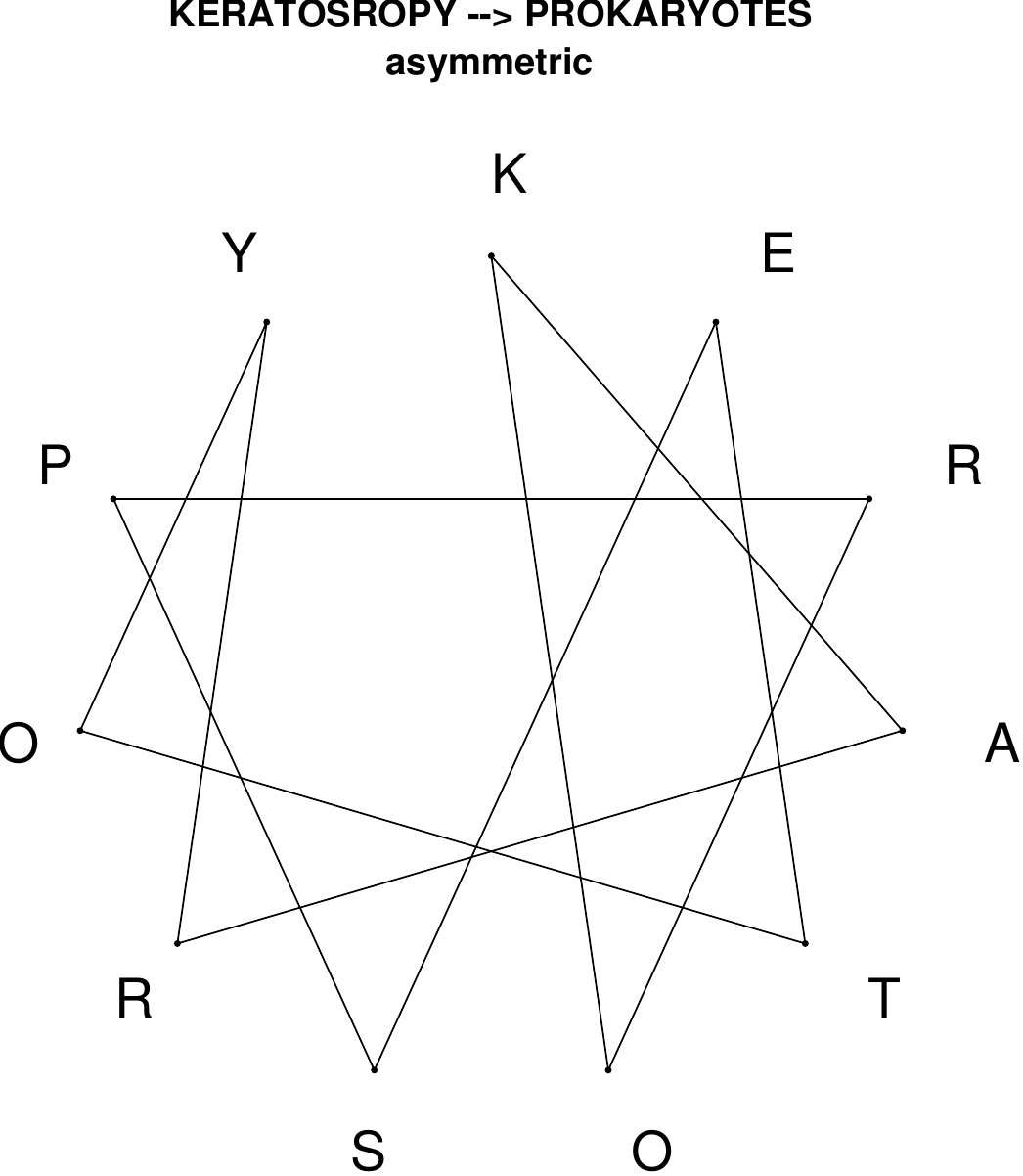}
\end{subfigure}
\hfill
\begin{subfigure}[T]{0.19\textwidth}
\centering
\includegraphics[width=\textwidth]{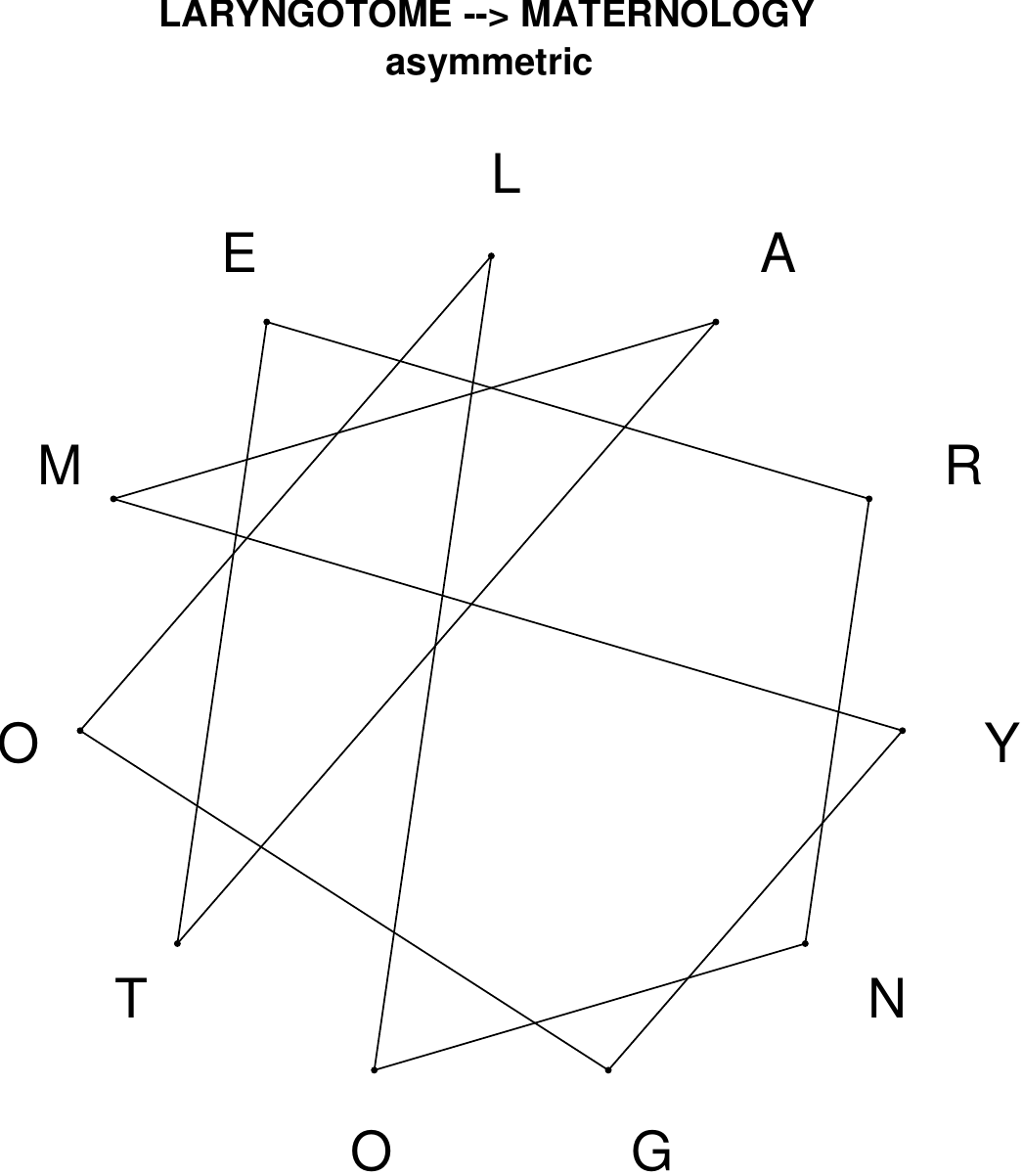}
\end{subfigure}
\hfill
\begin{subfigure}[T]{0.19\textwidth}
\centering
\includegraphics[width=\textwidth]{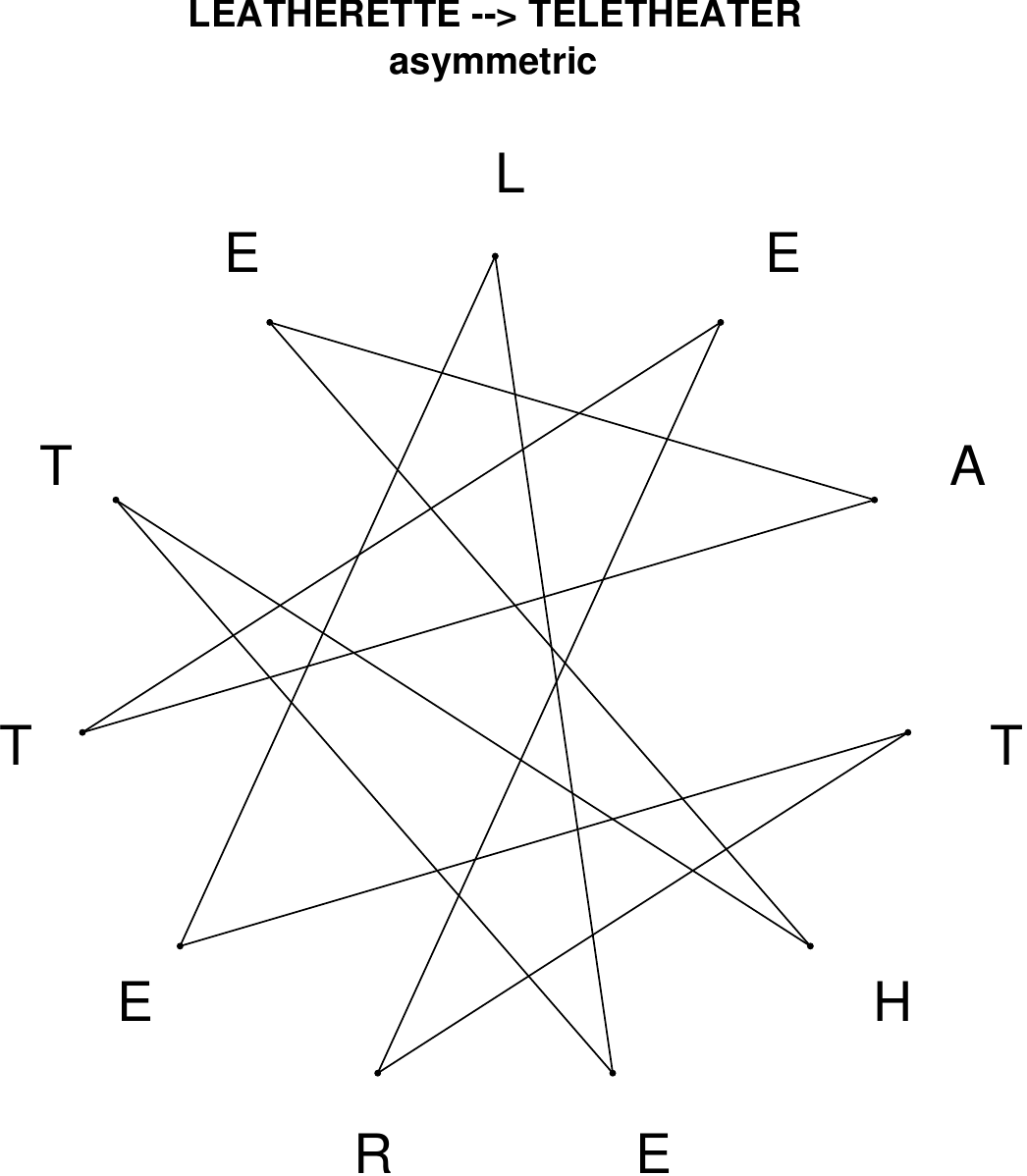}
\end{subfigure}
\hfill
\begin{subfigure}[T]{0.19\textwidth}
\centering
\includegraphics[width=\textwidth]{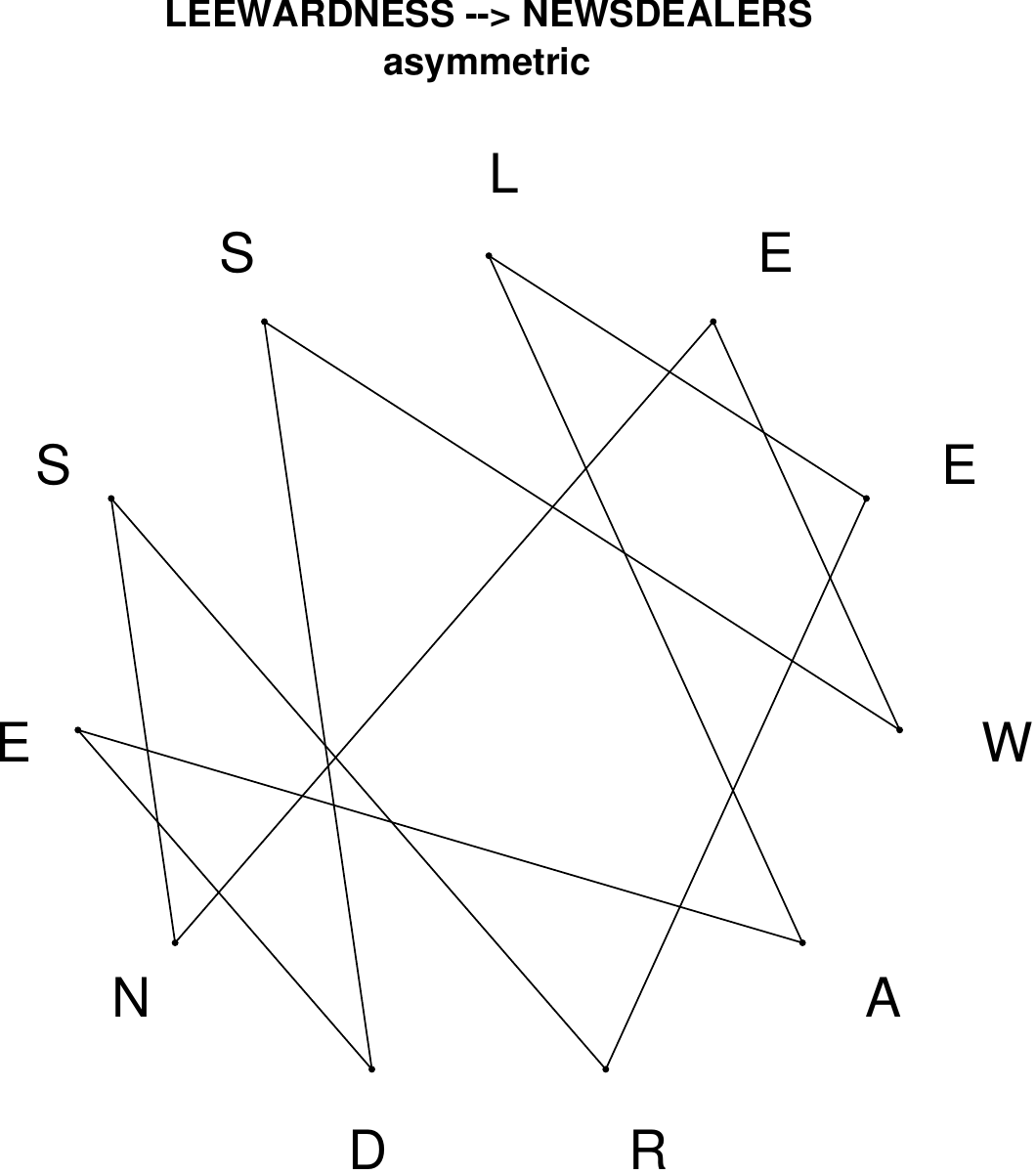}
\end{subfigure}
\end{figure}

\begin{figure}[H]
\centering
\begin{subfigure}[T]{0.19\textwidth}
\centering
\includegraphics[width=\textwidth]{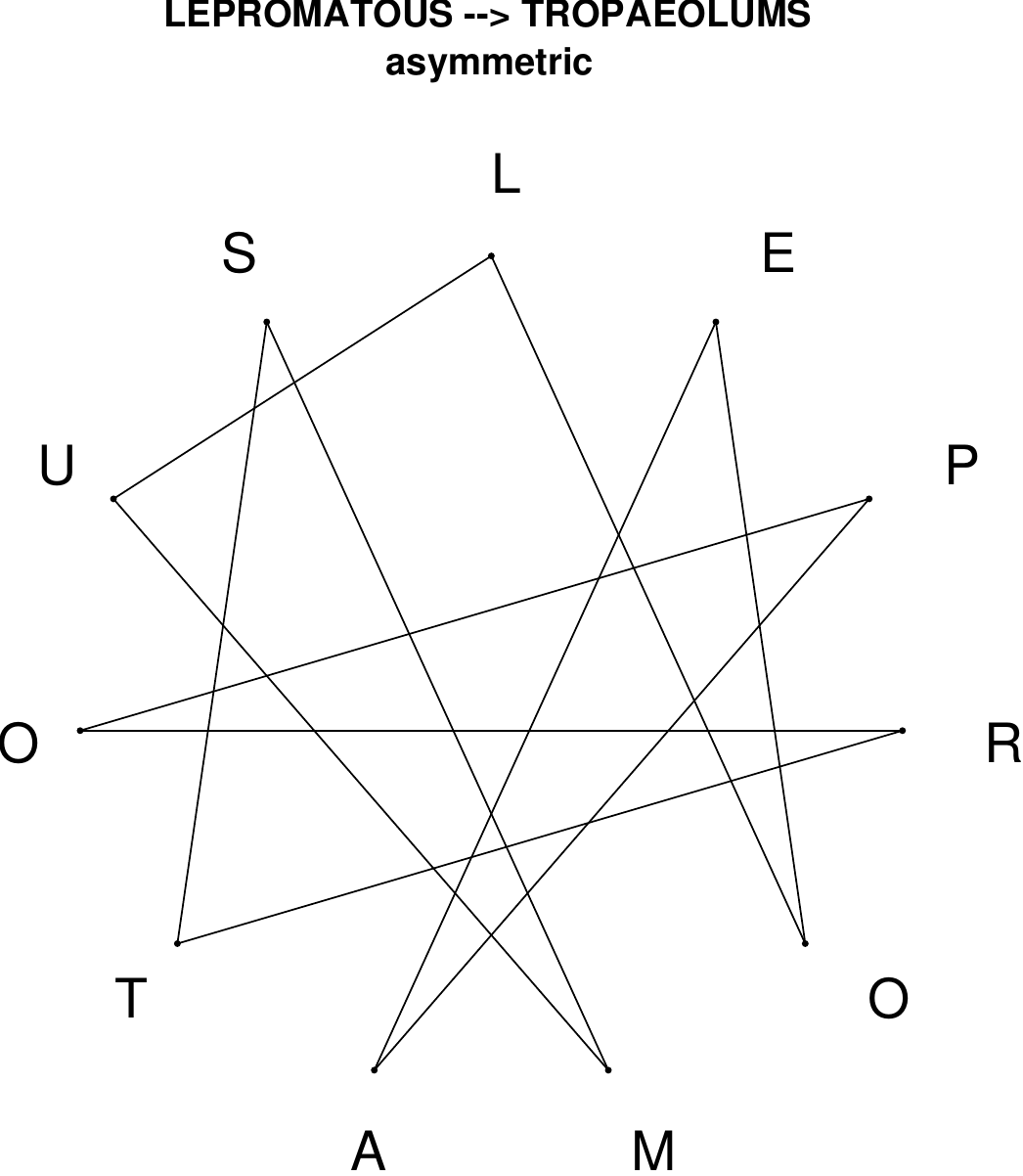}
\end{subfigure}
\hfill
\begin{subfigure}[T]{0.19\textwidth}
\centering
\includegraphics[width=\textwidth]{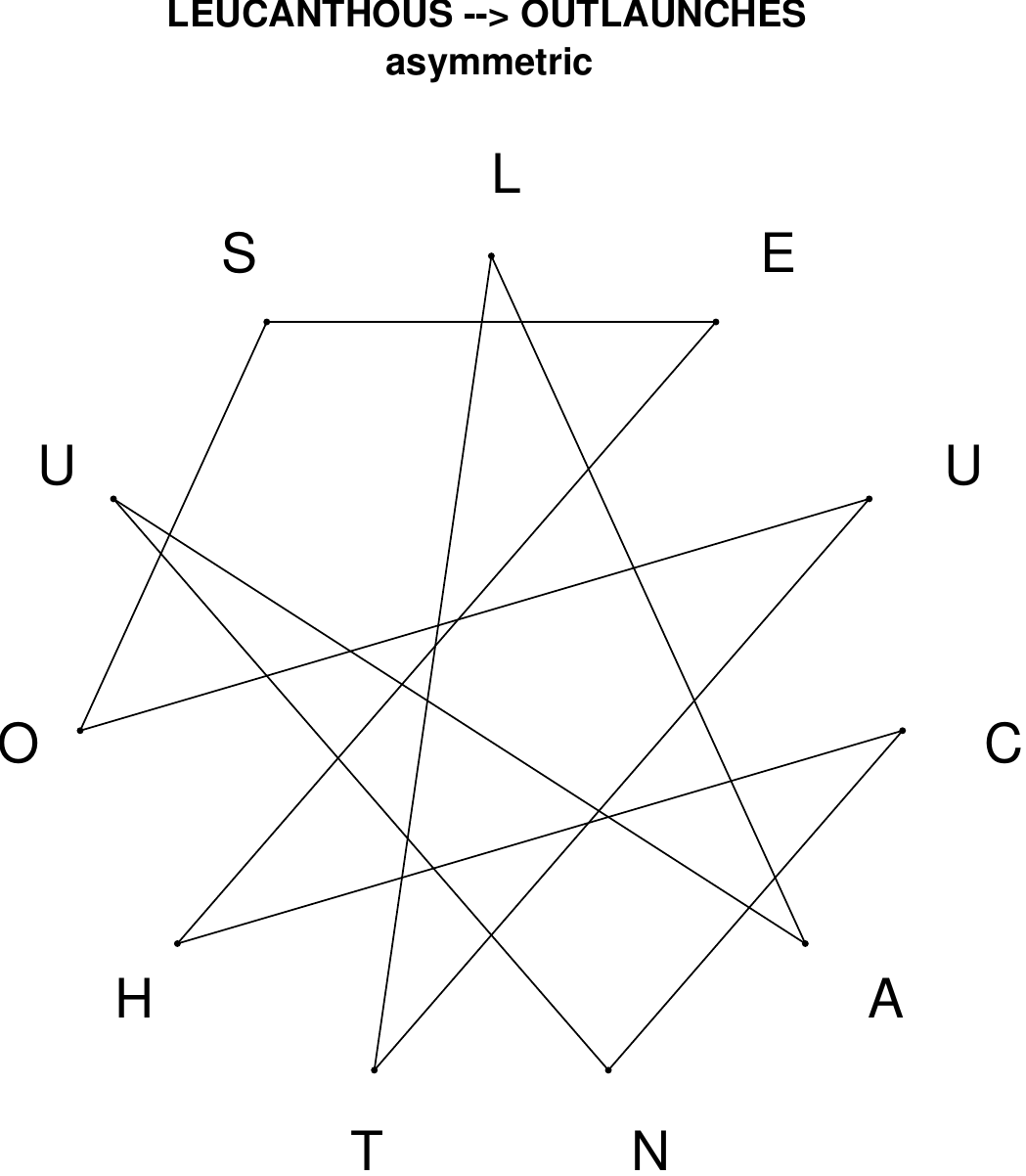}
\end{subfigure}
\hfill
\begin{subfigure}[T]{0.19\textwidth}
\centering
\includegraphics[width=\textwidth]{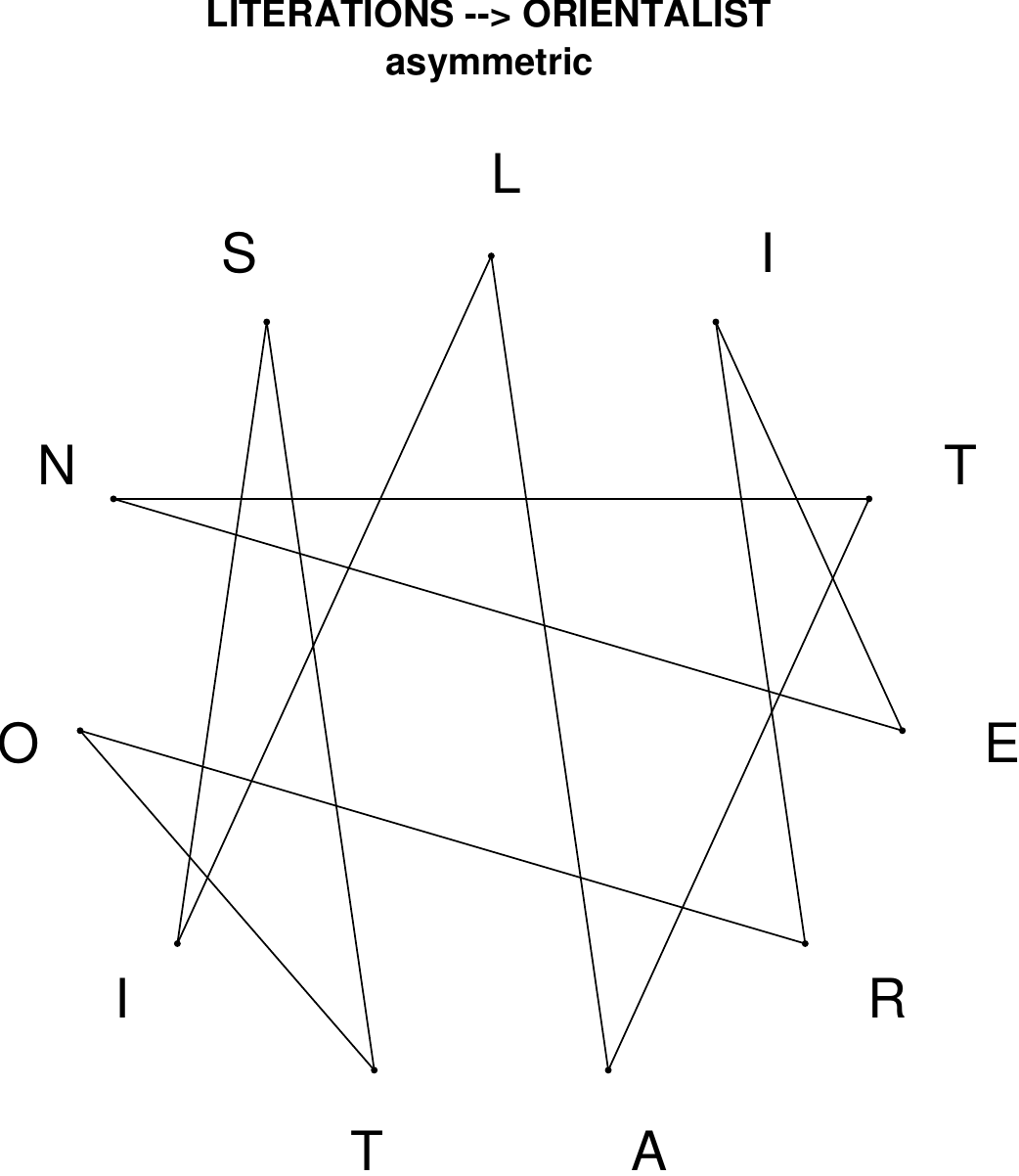}
\end{subfigure}
\hfill
\begin{subfigure}[T]{0.19\textwidth}
\centering
\includegraphics[width=\textwidth]{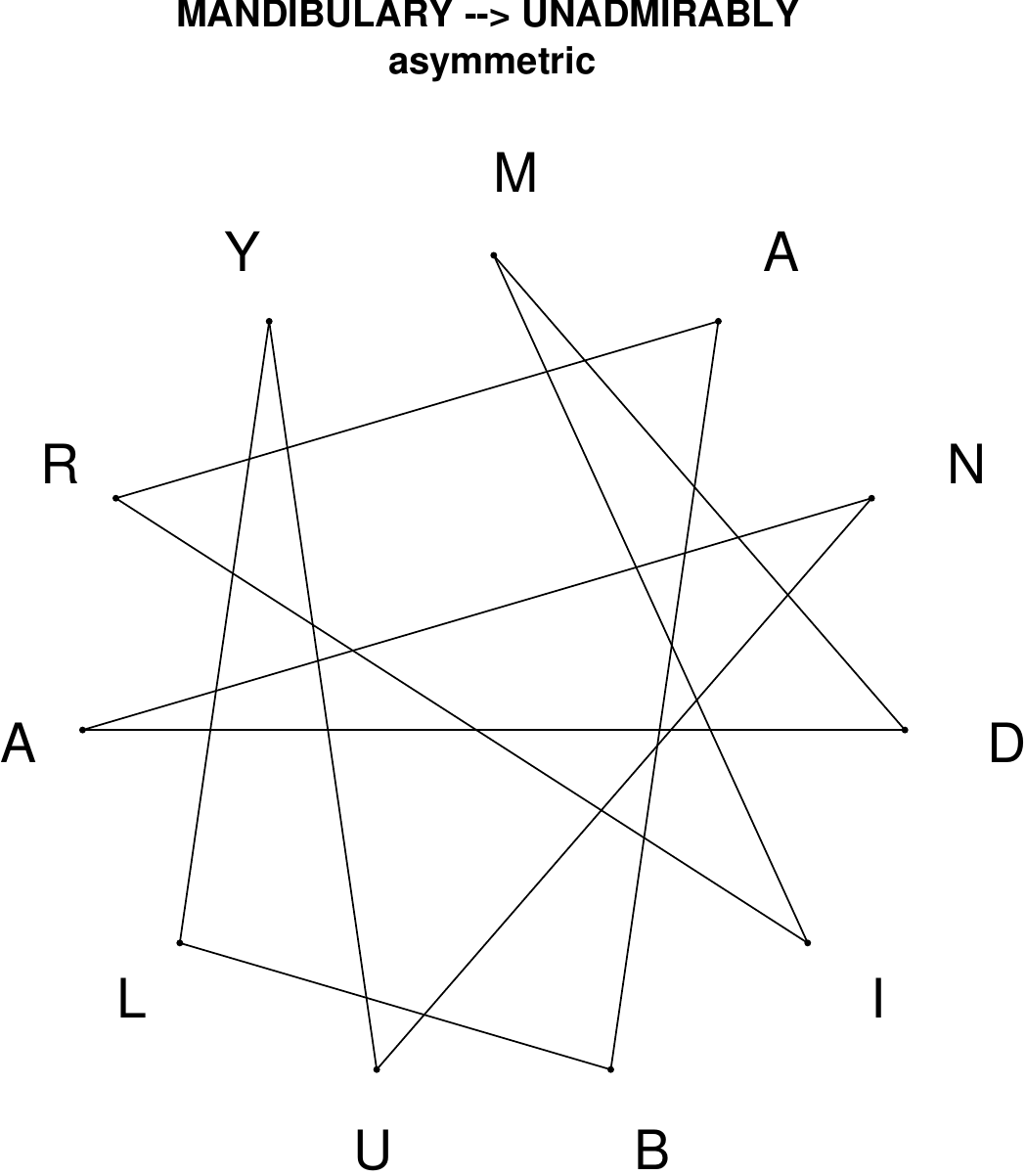}
\end{subfigure}
\hfill
\begin{subfigure}[T]{0.19\textwidth}
\centering
\includegraphics[width=\textwidth]{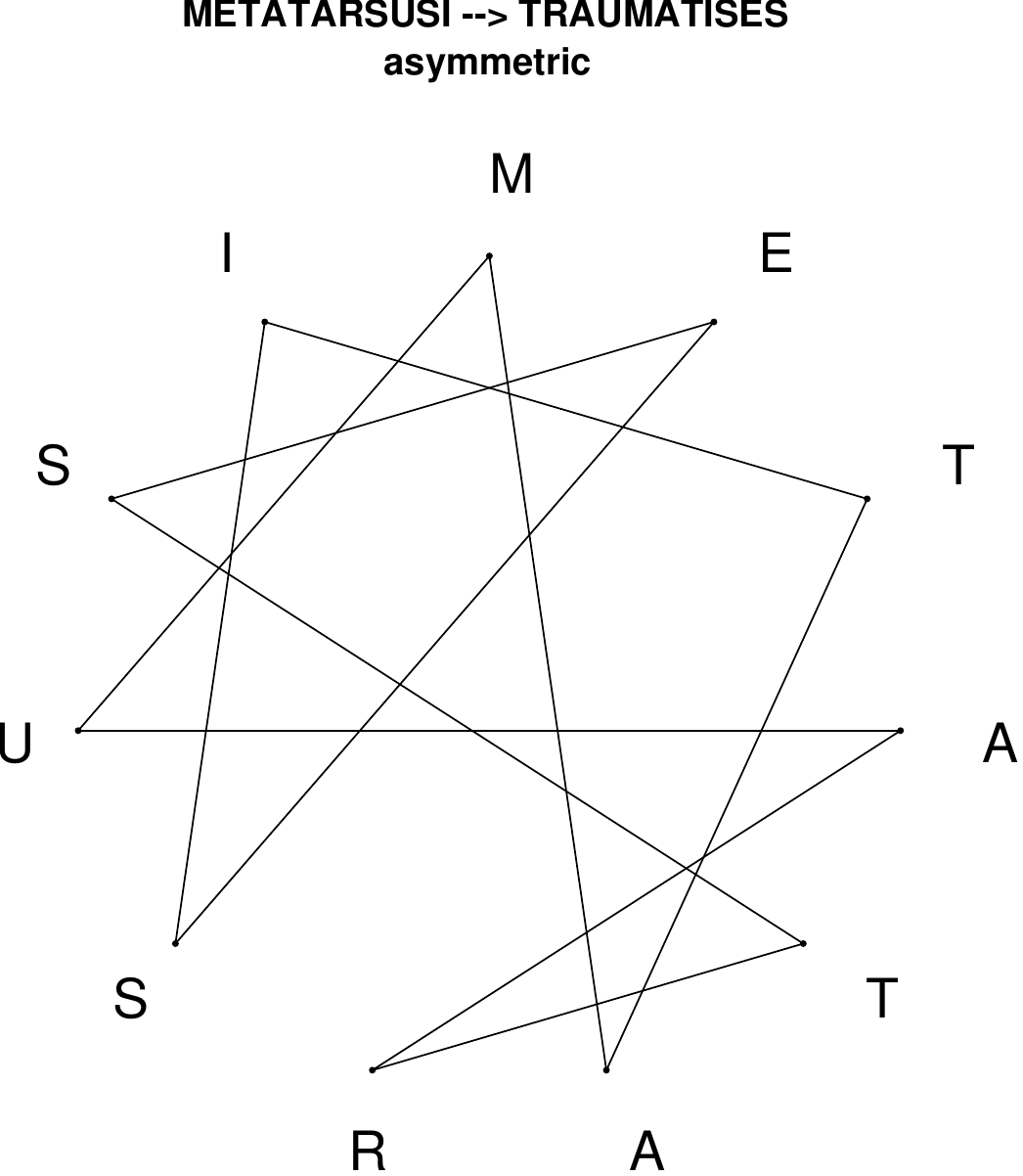}
\end{subfigure}
\end{figure}

\begin{figure}[H]
\centering
\begin{subfigure}[T]{0.19\textwidth}
\centering
\includegraphics[width=\textwidth]{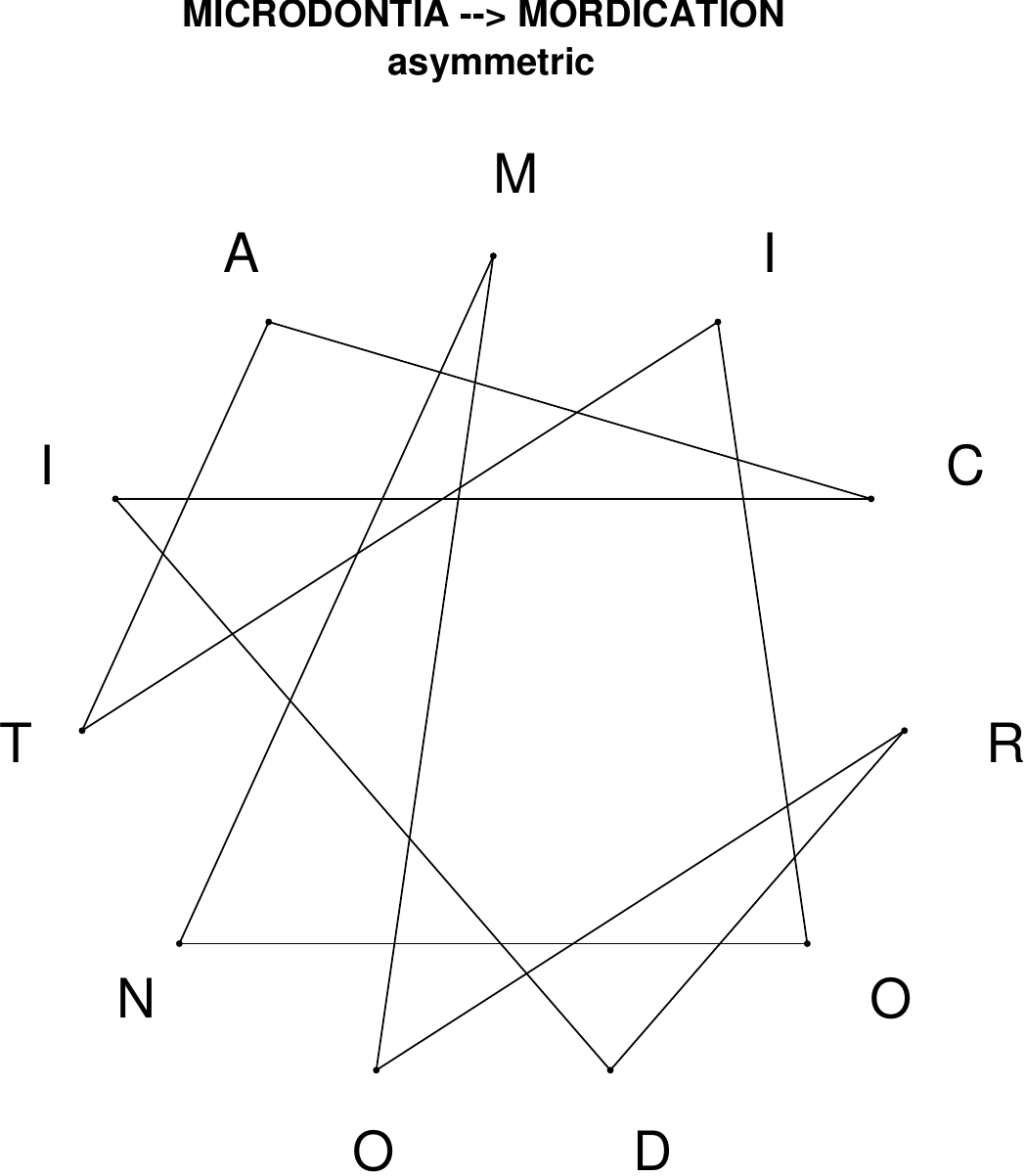}
\end{subfigure}
\hfill
\begin{subfigure}[T]{0.19\textwidth}
\centering
\includegraphics[width=\textwidth]{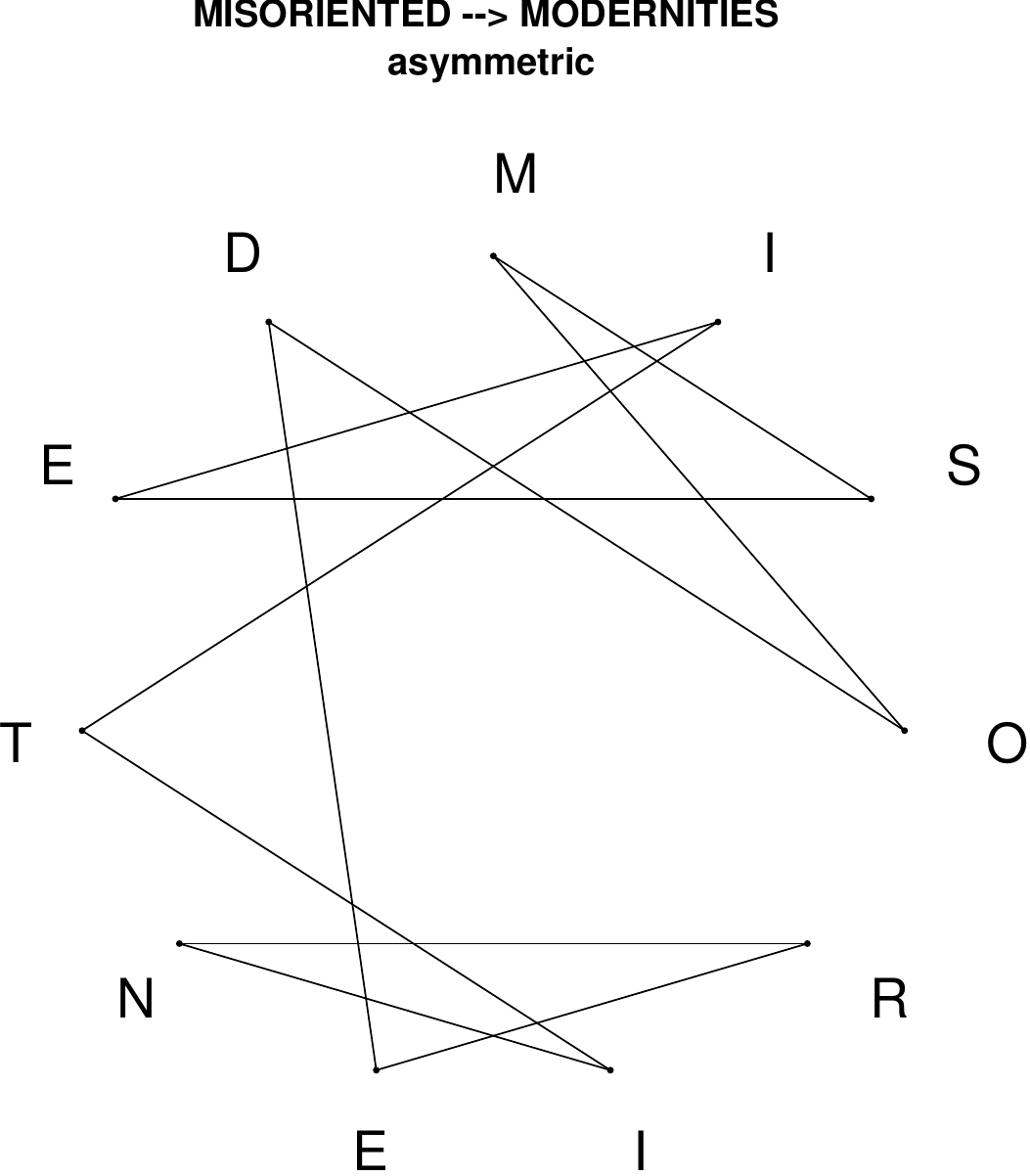}
\end{subfigure}
\hfill
\begin{subfigure}[T]{0.19\textwidth}
\centering
\includegraphics[width=\textwidth]{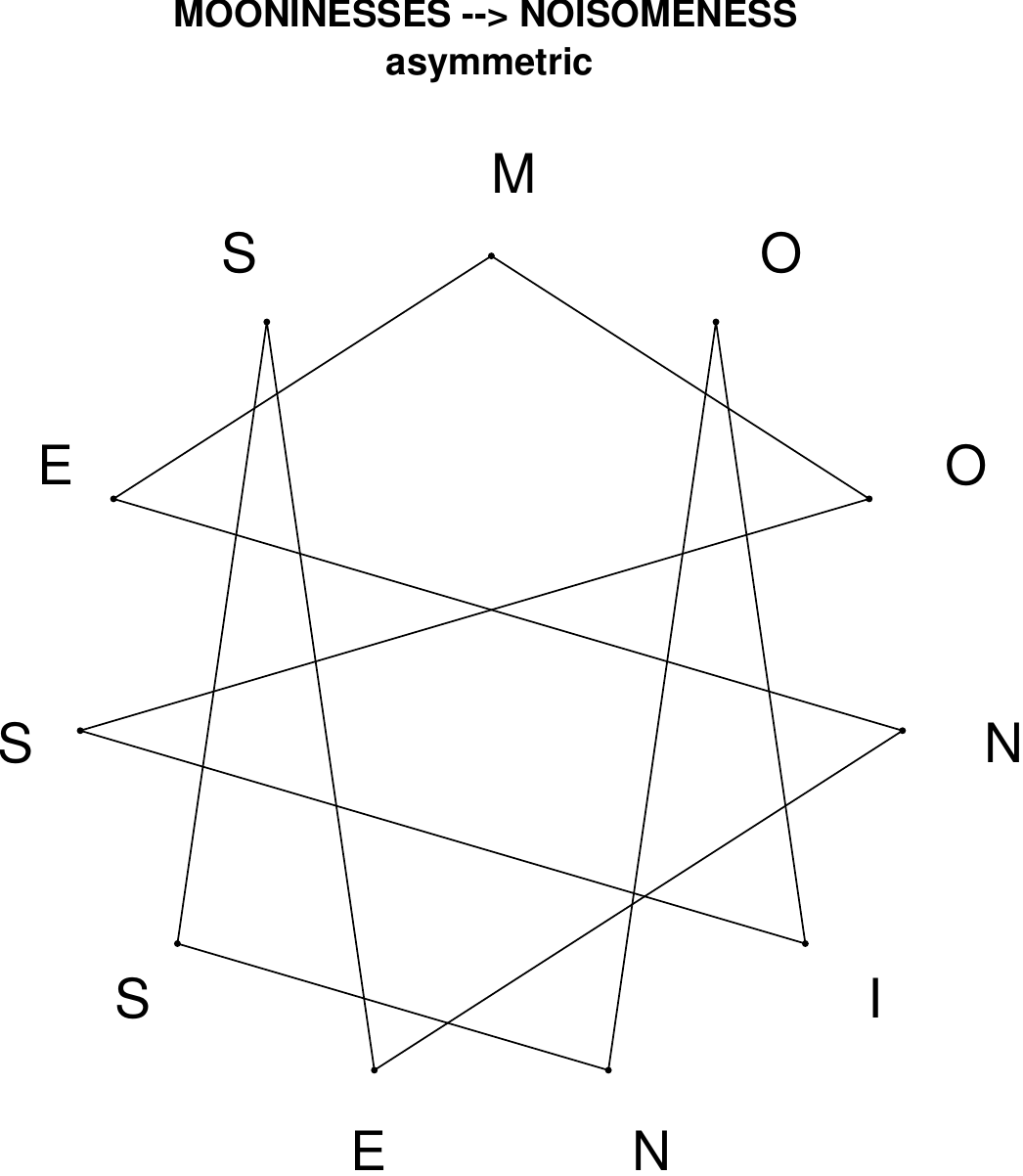}
\end{subfigure}
\hfill
\begin{subfigure}[T]{0.19\textwidth}
\centering
\includegraphics[width=\textwidth]{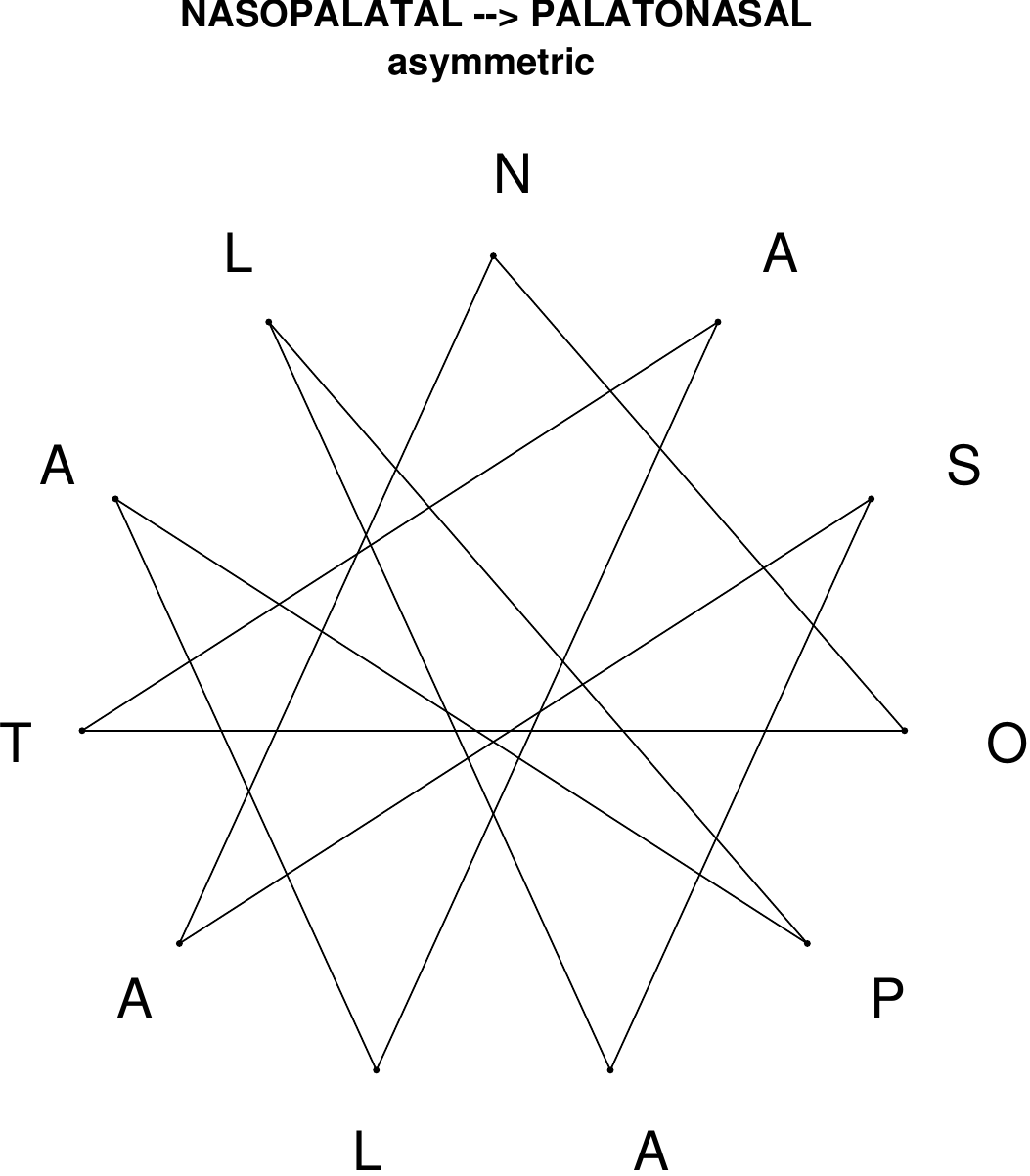}
\end{subfigure}
\hfill
\begin{subfigure}[T]{0.19\textwidth}
\centering
\includegraphics[width=\textwidth]{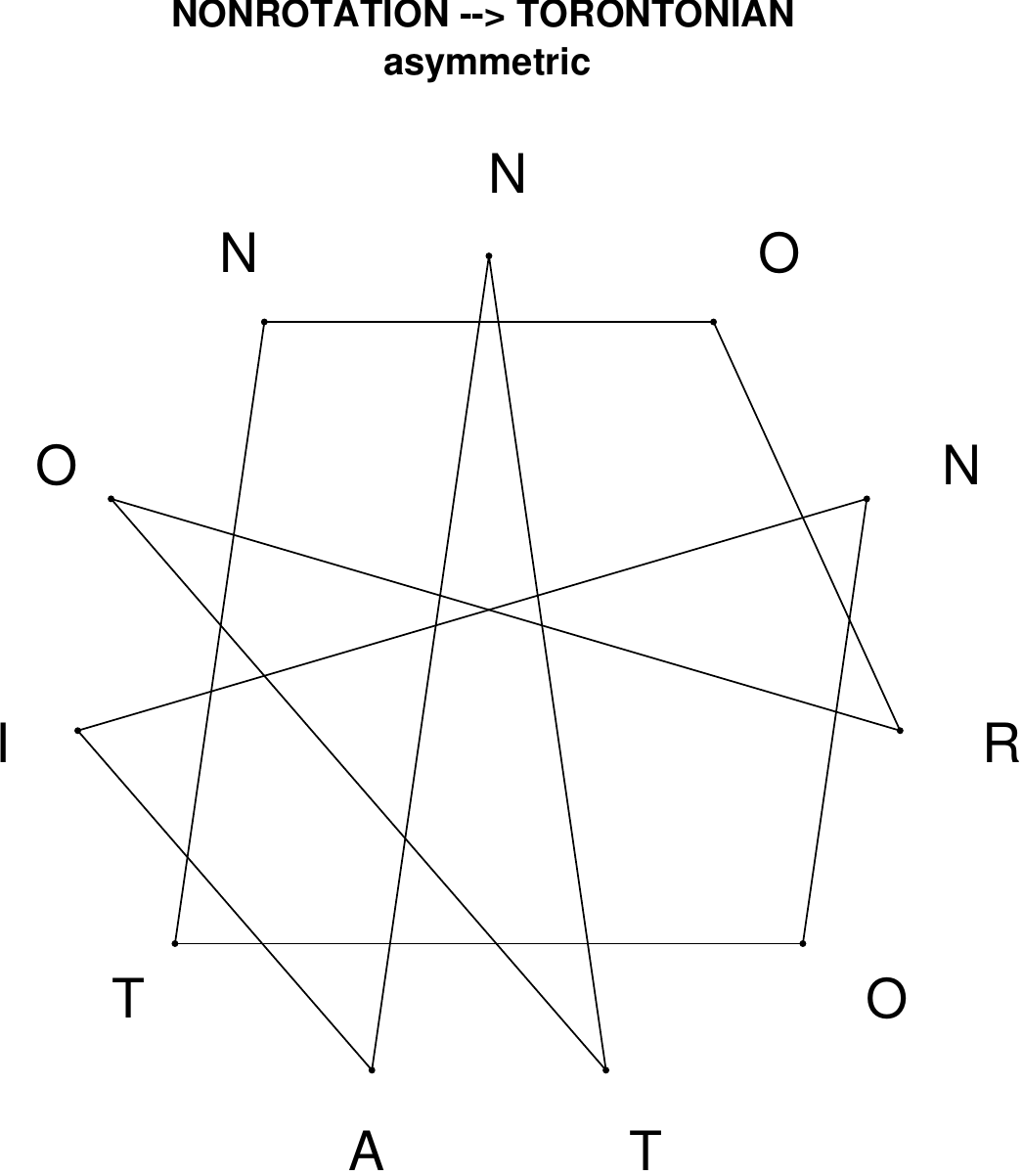}
\end{subfigure}
\end{figure}

\begin{figure}[H]
\centering
\begin{subfigure}[T]{0.19\textwidth}
\centering
\includegraphics[width=\textwidth]{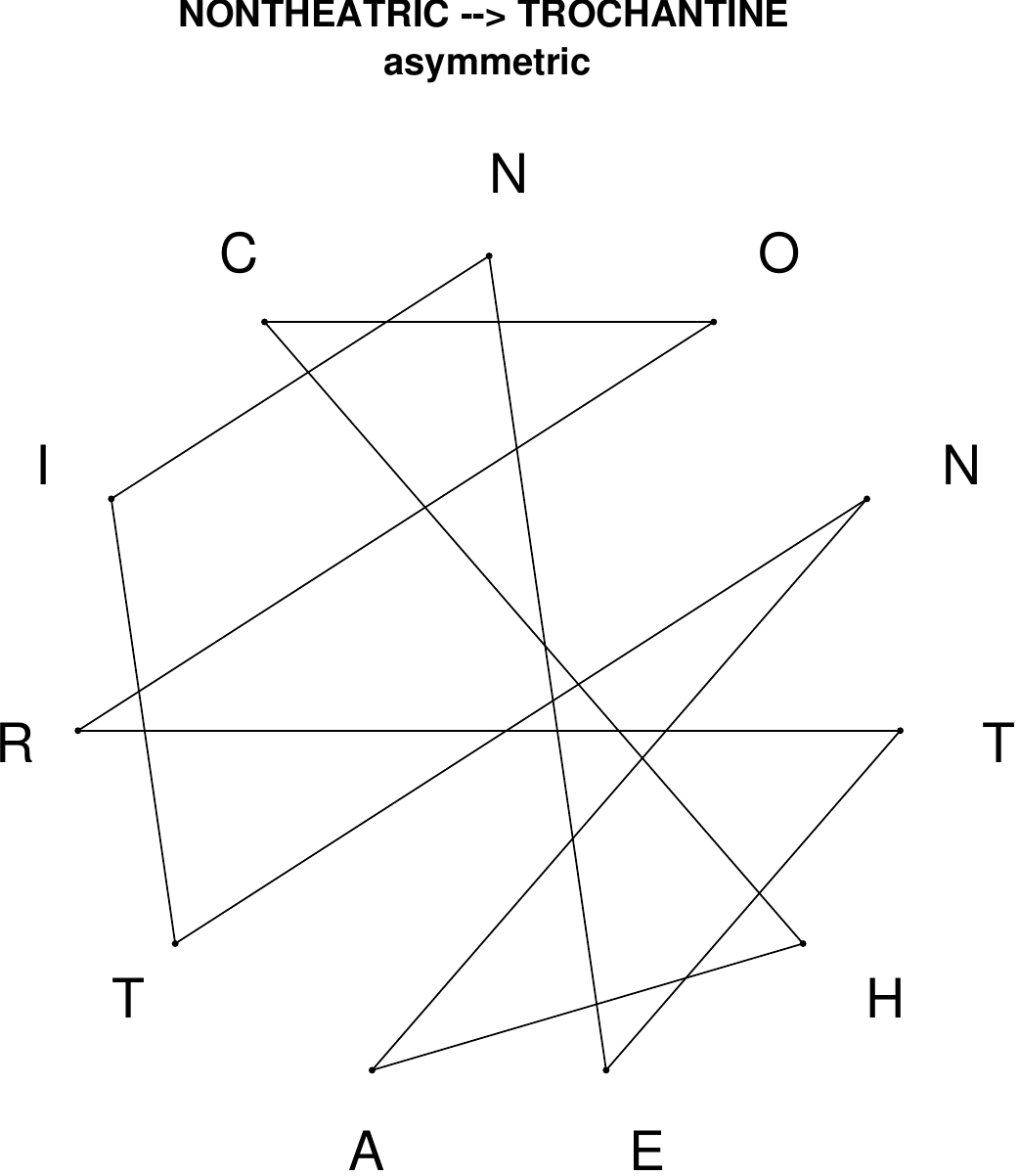}
\end{subfigure}
\hfill
\begin{subfigure}[T]{0.19\textwidth}
\centering
\includegraphics[width=\textwidth]{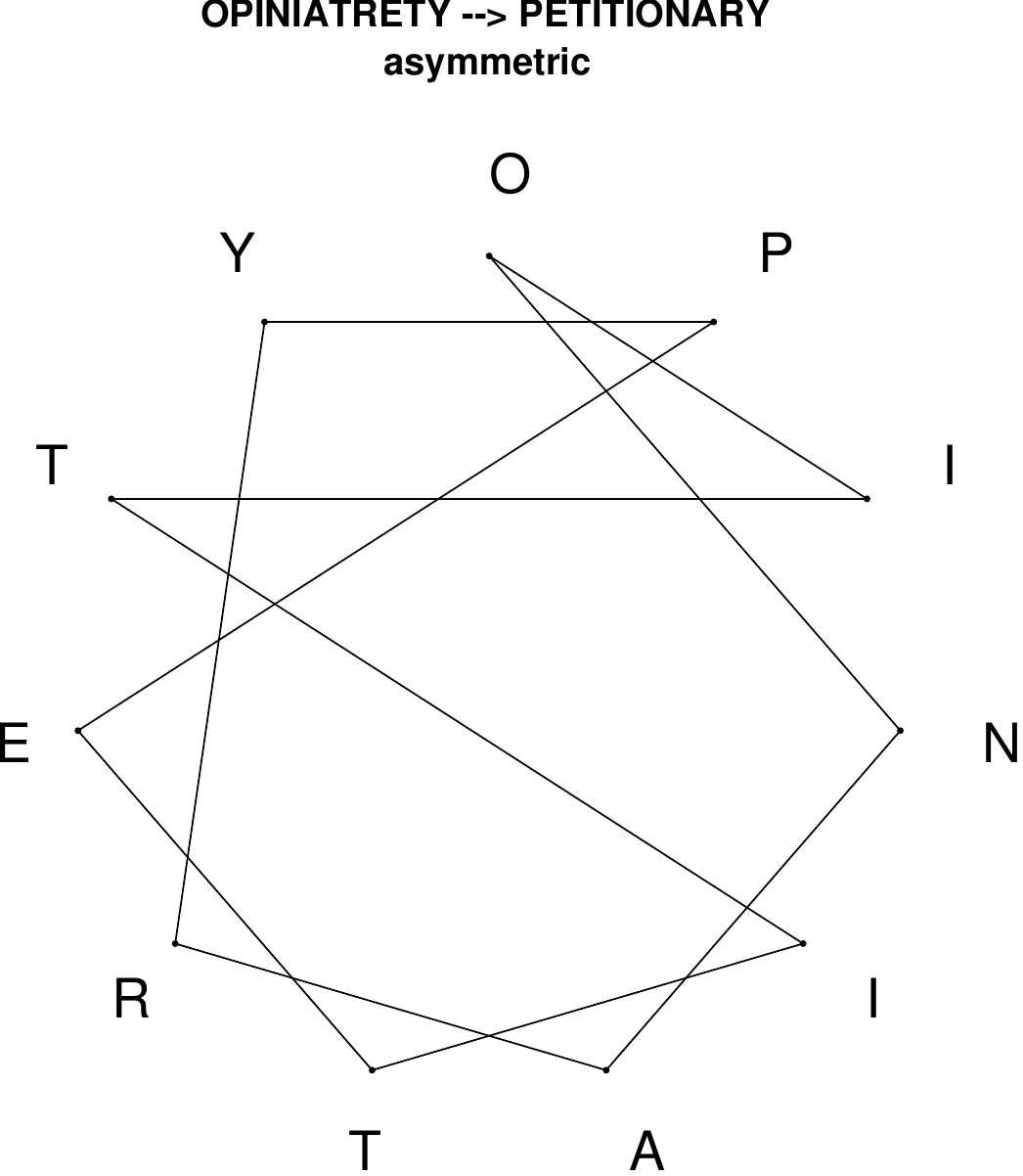}
\end{subfigure}
\hfill
\begin{subfigure}[T]{0.19\textwidth}
\centering
\includegraphics[width=\textwidth]{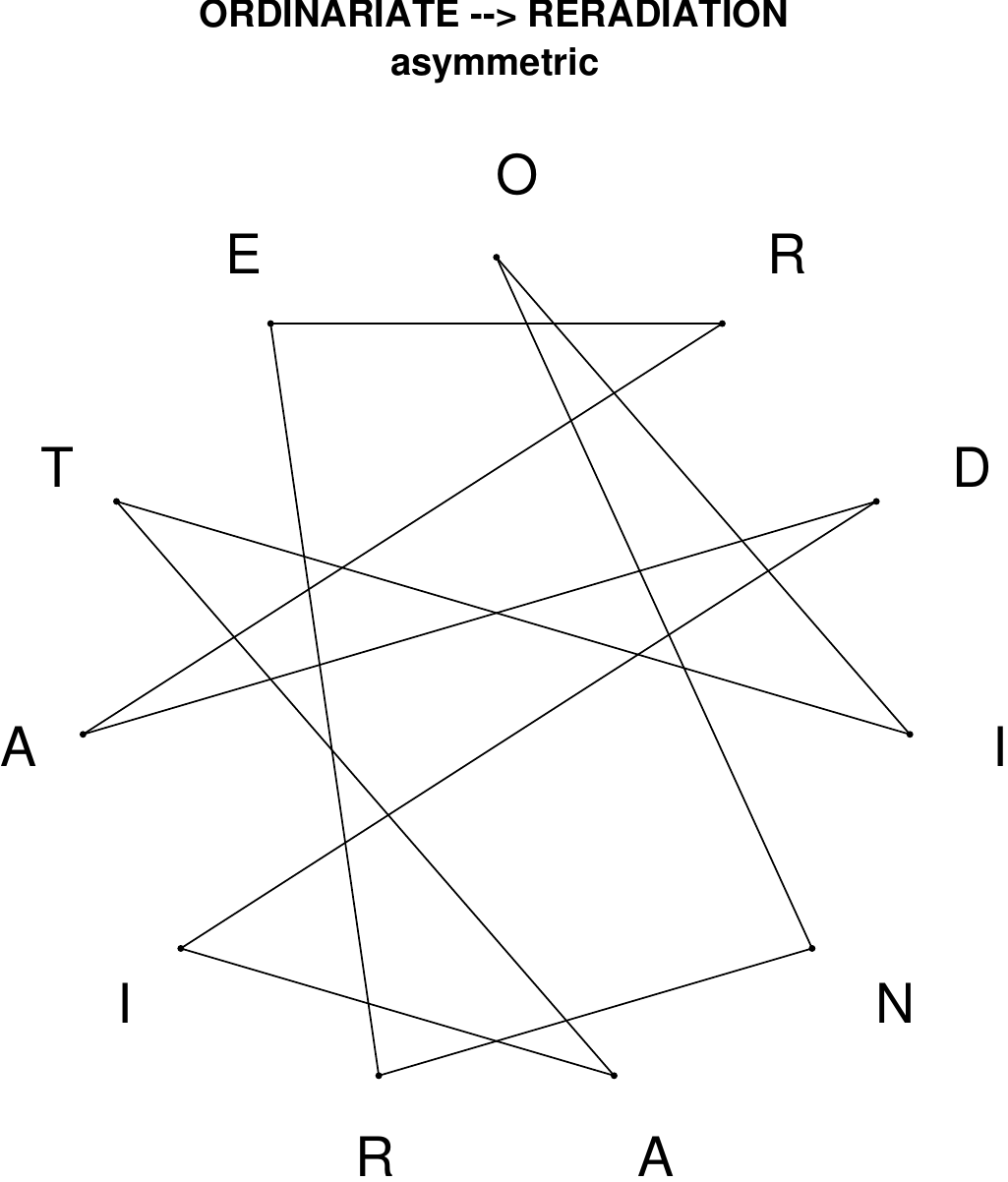}
\end{subfigure}
\hfill
\begin{subfigure}[T]{0.19\textwidth}
\centering
\includegraphics[width=\textwidth]{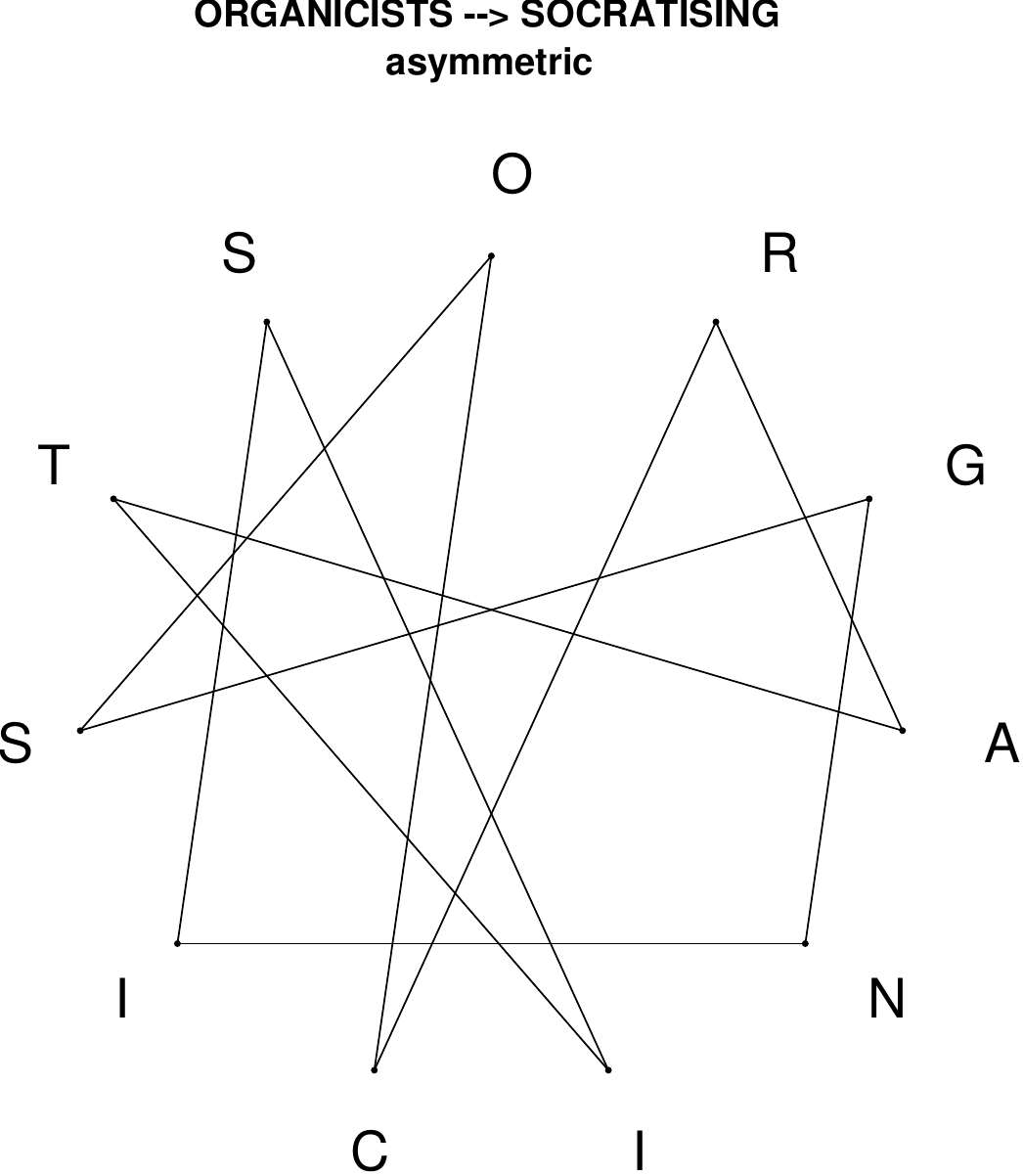}
\end{subfigure}
\hfill
\begin{subfigure}[T]{0.19\textwidth}
\centering
\includegraphics[width=\textwidth]{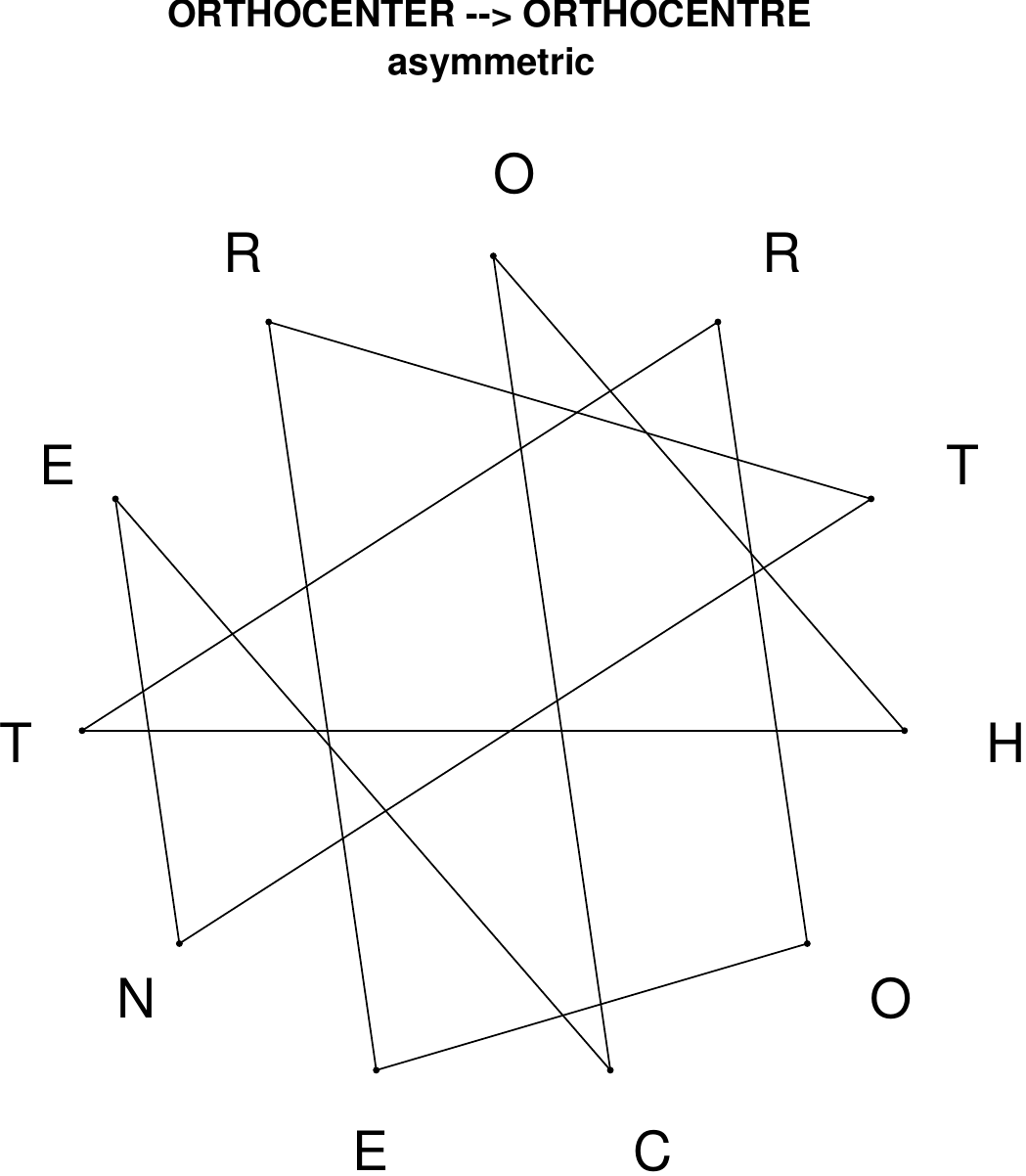}
\end{subfigure}
\end{figure}

\begin{figure}[H]
\centering
\begin{subfigure}[T]{0.19\textwidth}
\centering
\includegraphics[width=\textwidth]{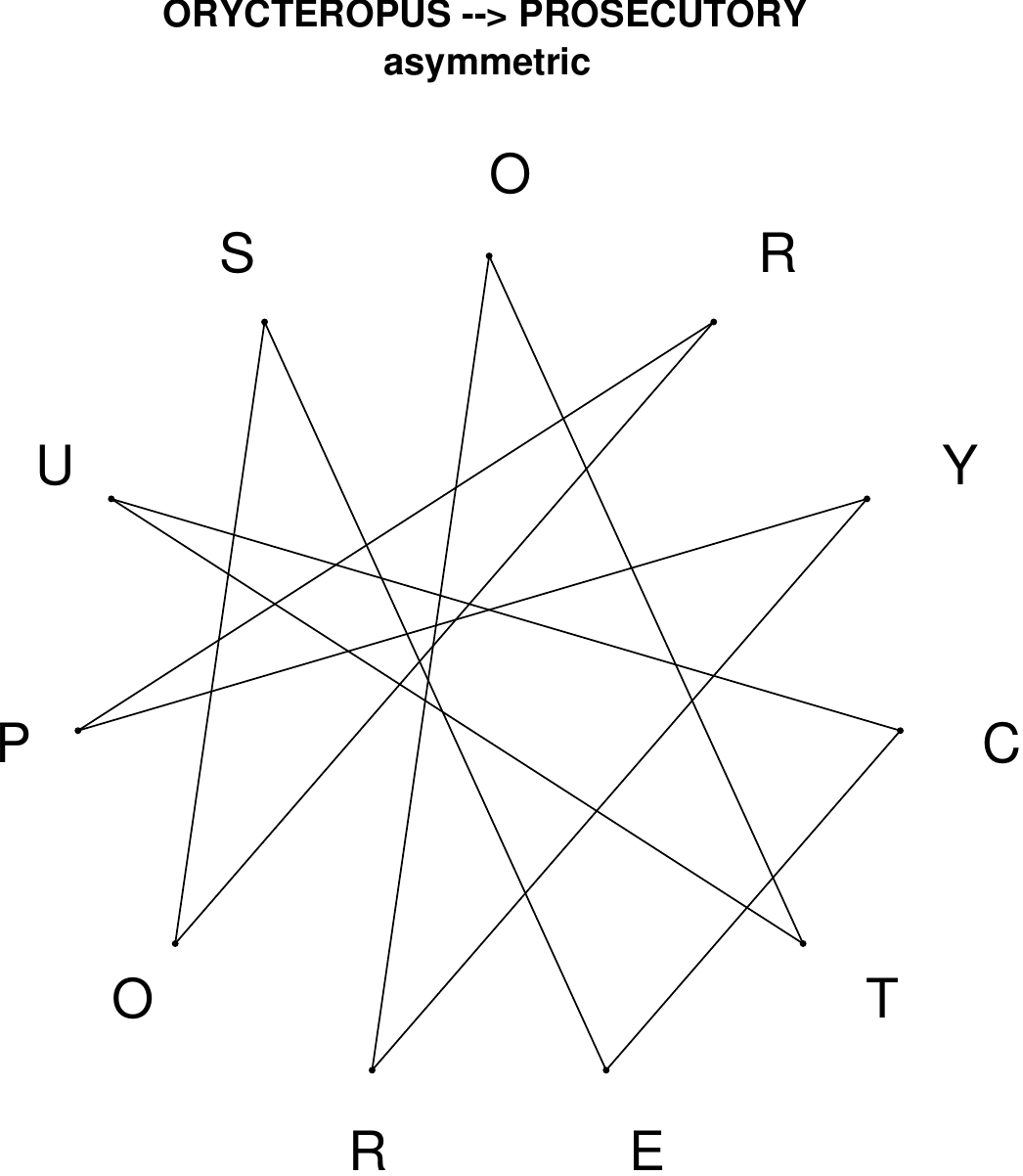}
\end{subfigure}
\hfill
\begin{subfigure}[T]{0.19\textwidth}
\centering
\includegraphics[width=\textwidth]{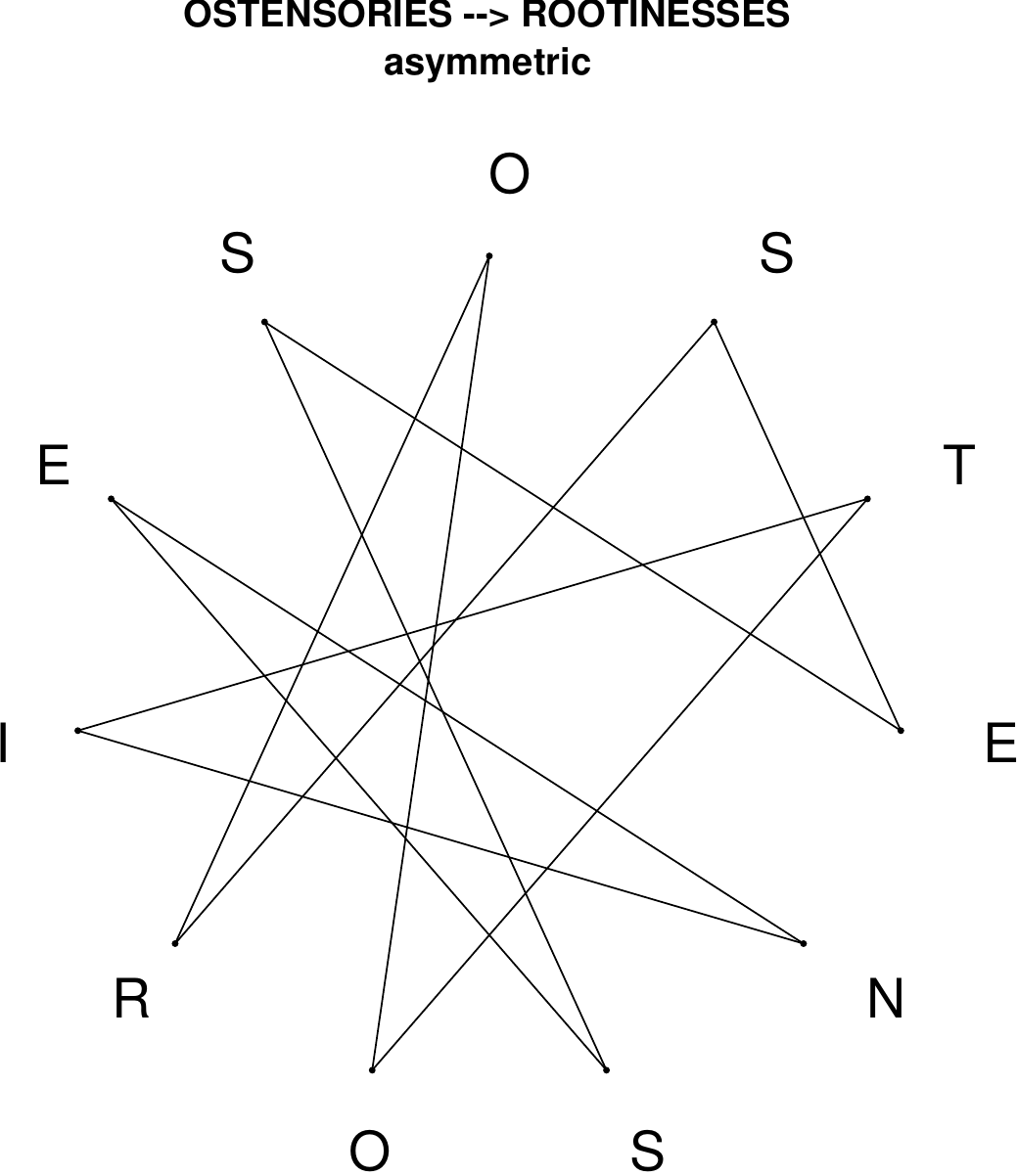}
\end{subfigure}
\hfill
\begin{subfigure}[T]{0.19\textwidth}
\centering
\includegraphics[width=\textwidth]{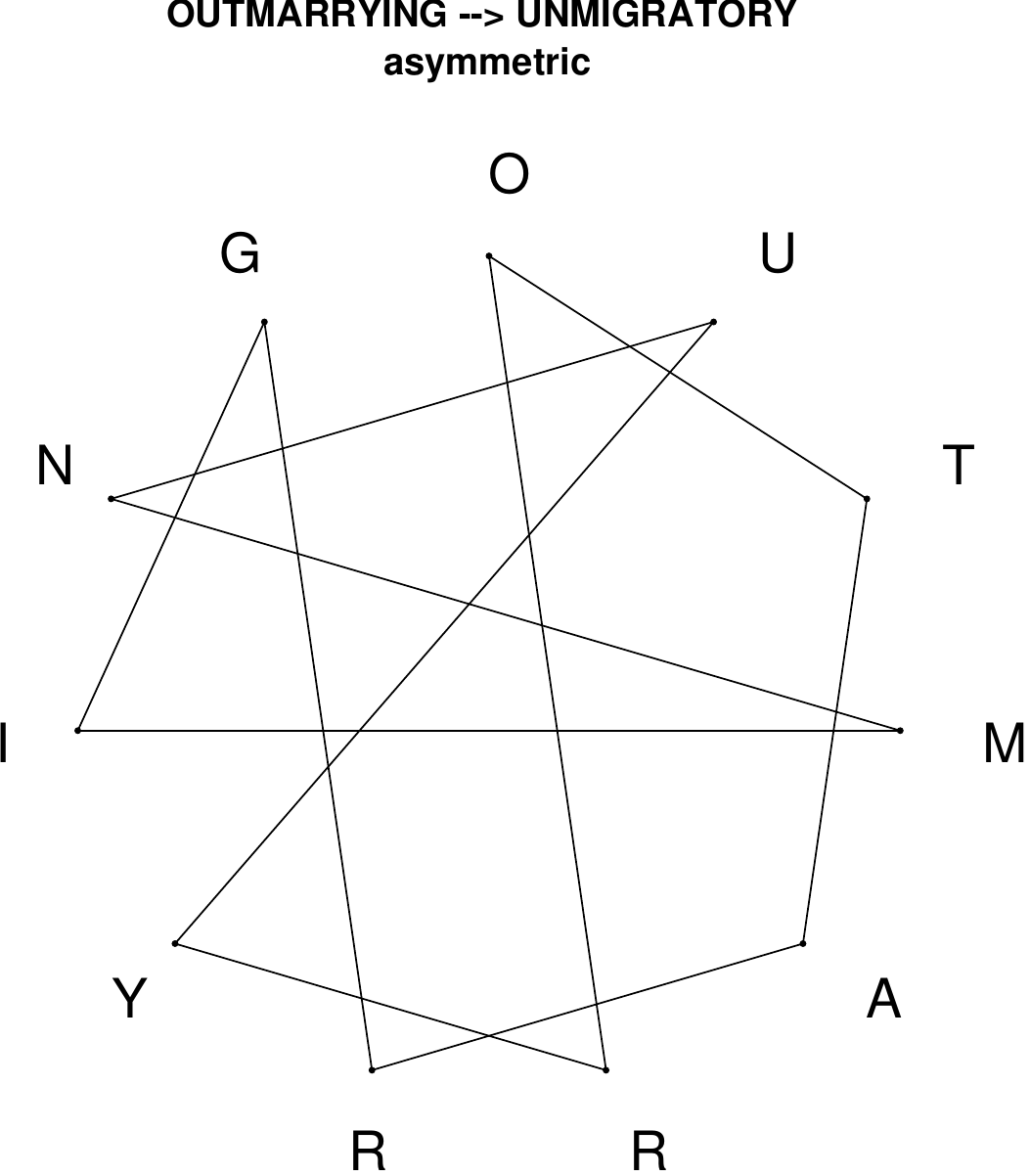}
\end{subfigure}
\hfill
\begin{subfigure}[T]{0.19\textwidth}
\centering
\includegraphics[width=\textwidth]{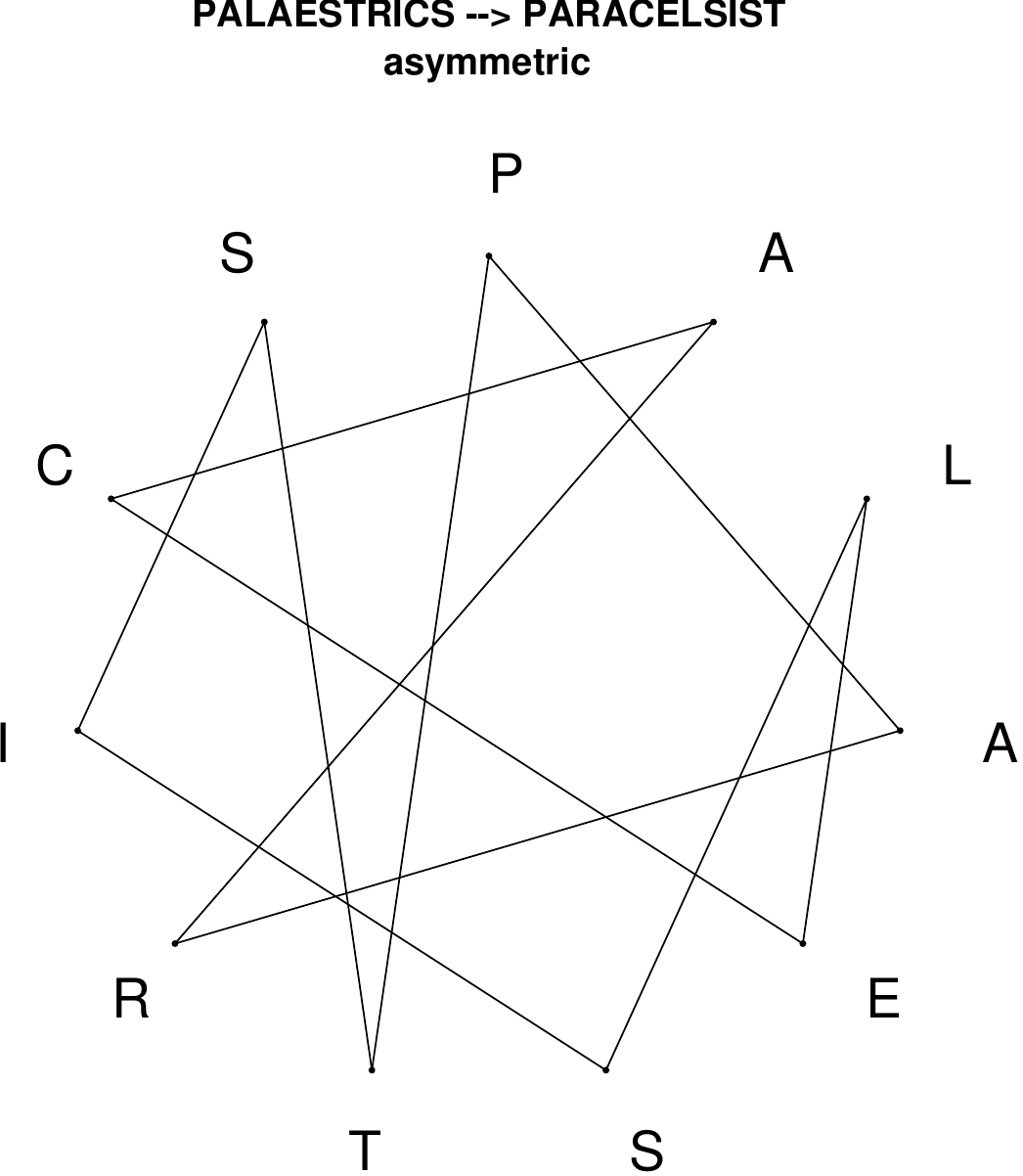}
\end{subfigure}
\hfill
\begin{subfigure}[T]{0.19\textwidth}
\centering
\includegraphics[width=\textwidth]{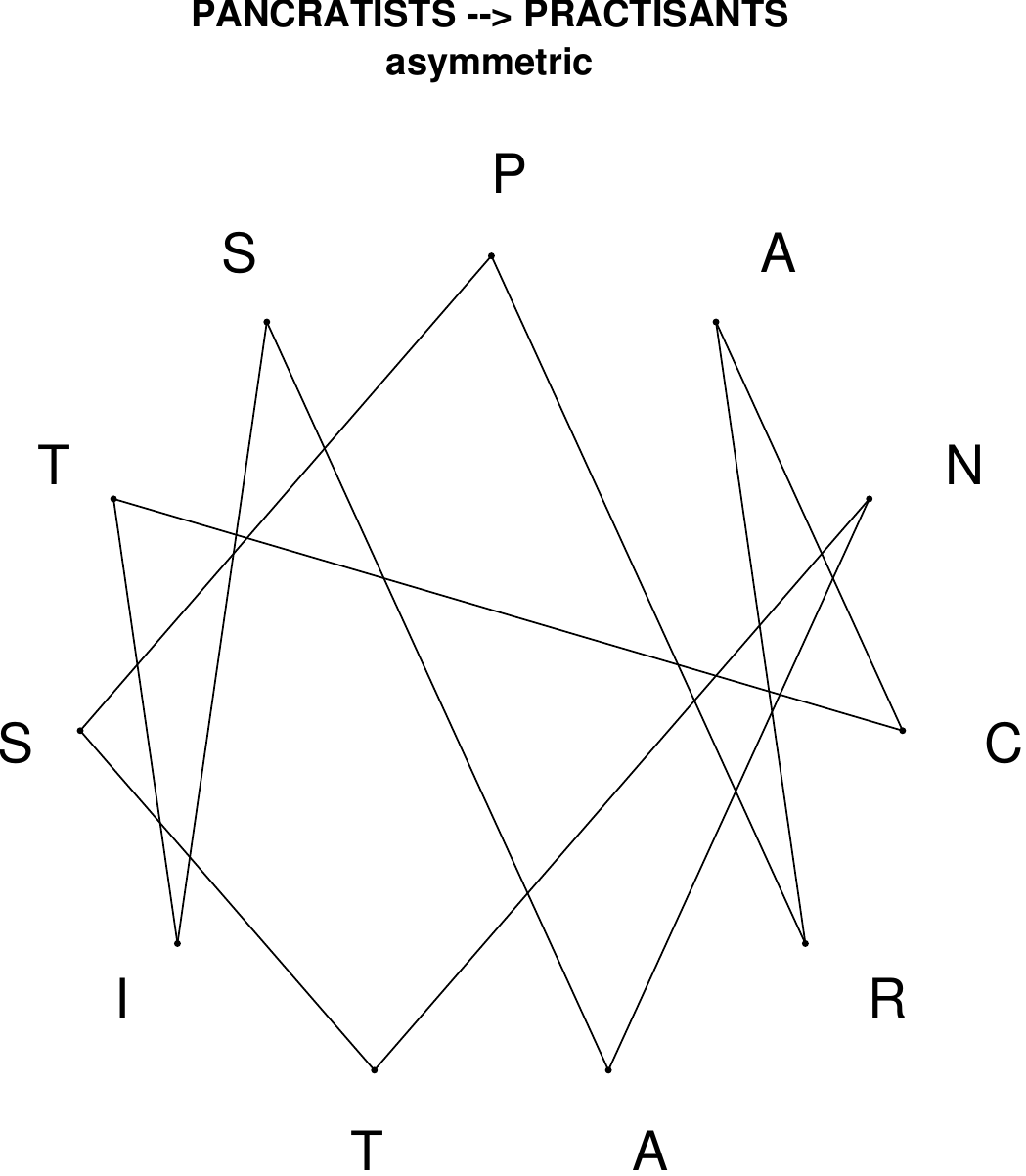}
\end{subfigure}
\end{figure}

\begin{figure}[H]
\centering
\begin{subfigure}[T]{0.19\textwidth}
\centering
\includegraphics[width=\textwidth]{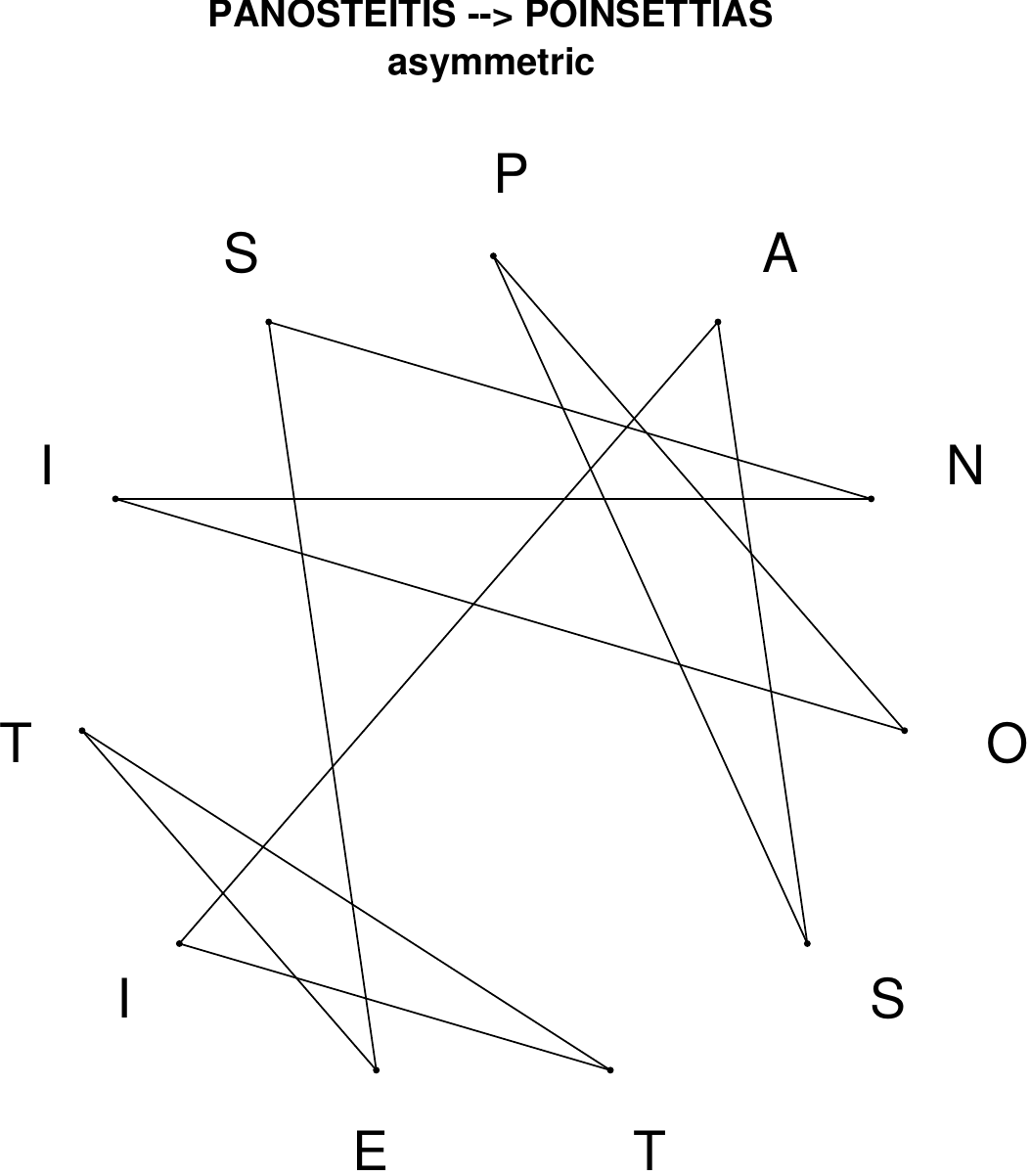}
\end{subfigure}
\hfill
\begin{subfigure}[T]{0.19\textwidth}
\centering
\includegraphics[width=\textwidth]{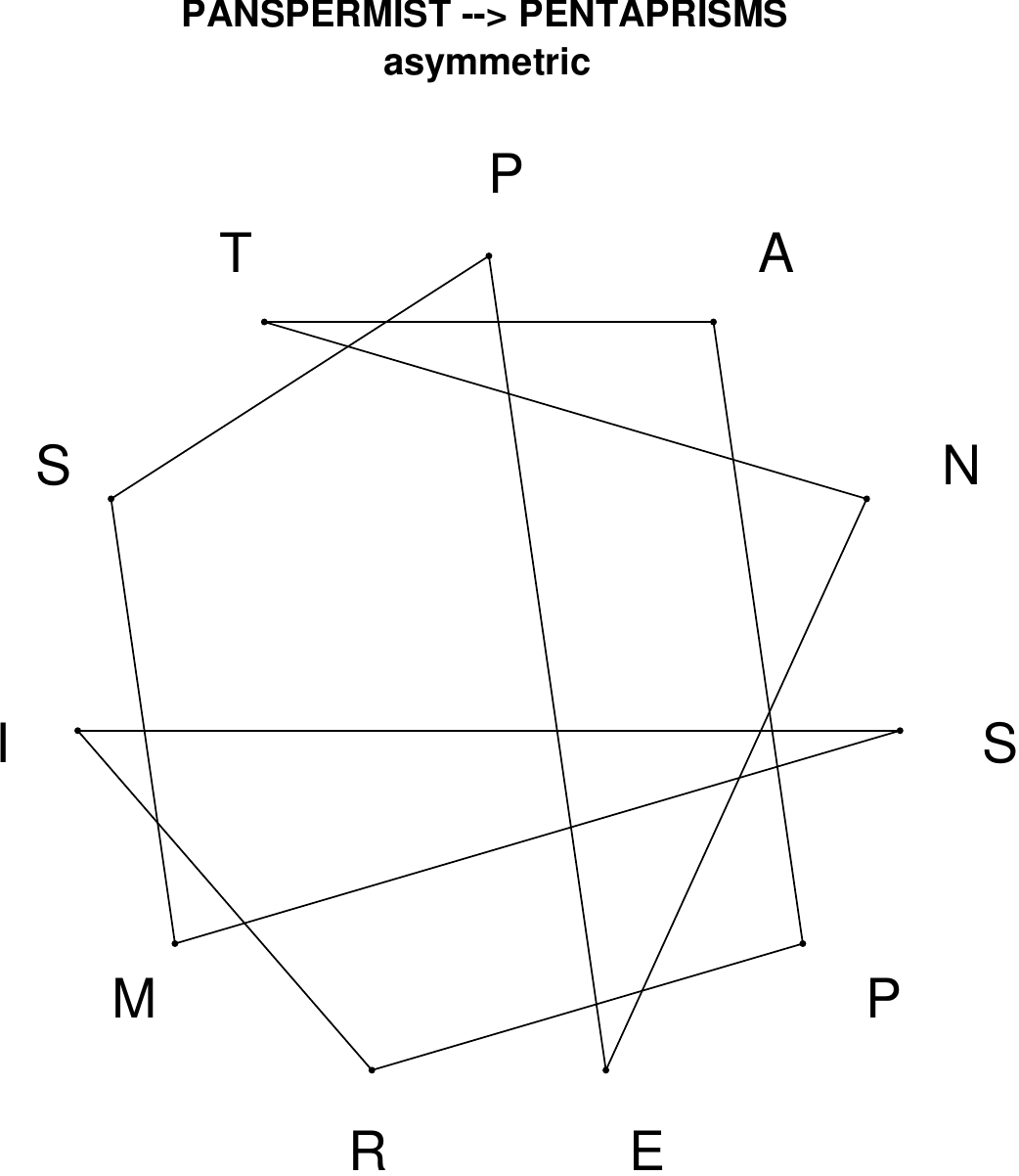}
\end{subfigure}
\hfill
\begin{subfigure}[T]{0.19\textwidth}
\centering
\includegraphics[width=\textwidth]{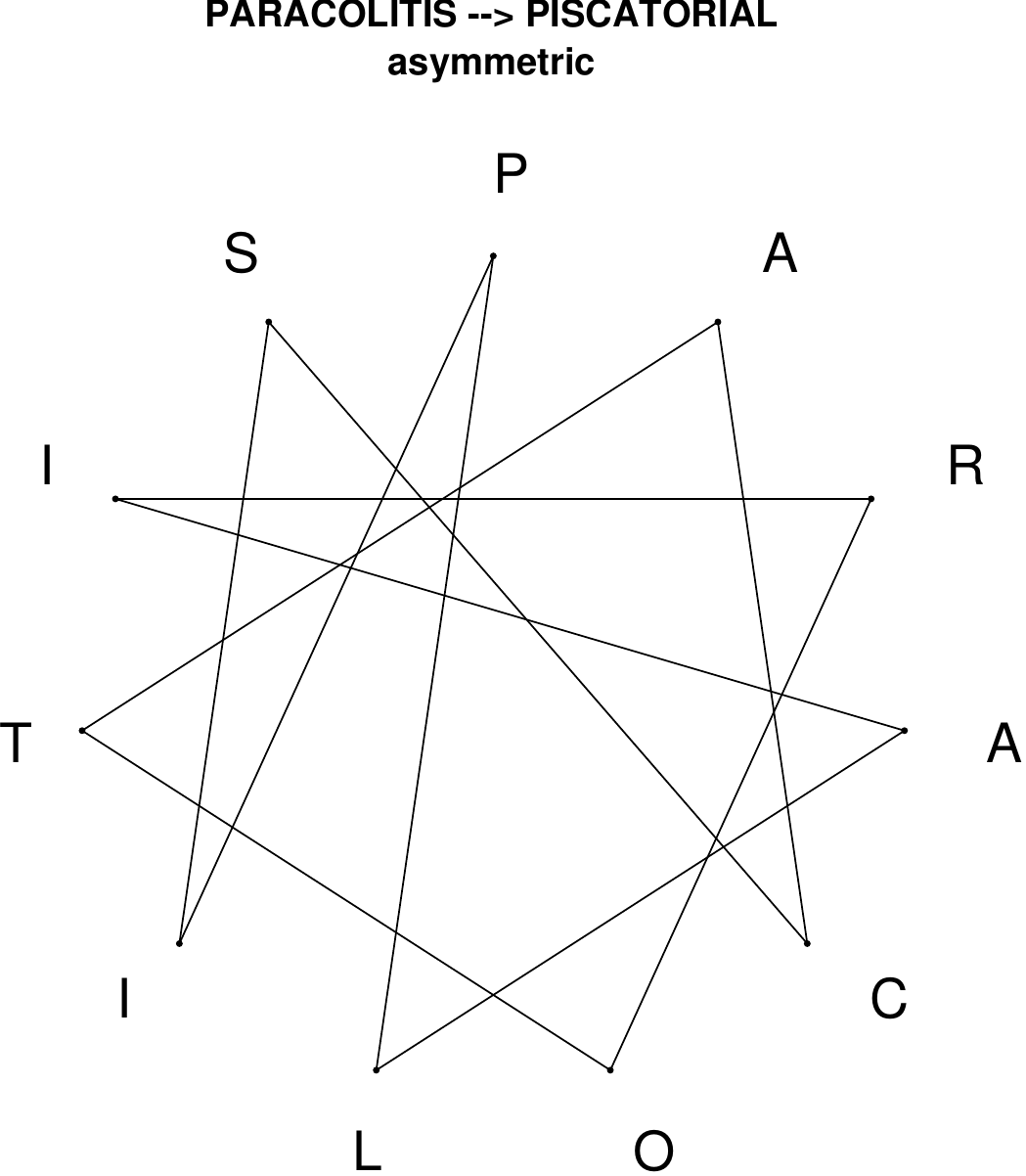}
\end{subfigure}
\hfill
\begin{subfigure}[T]{0.19\textwidth}
\centering
\includegraphics[width=\textwidth]{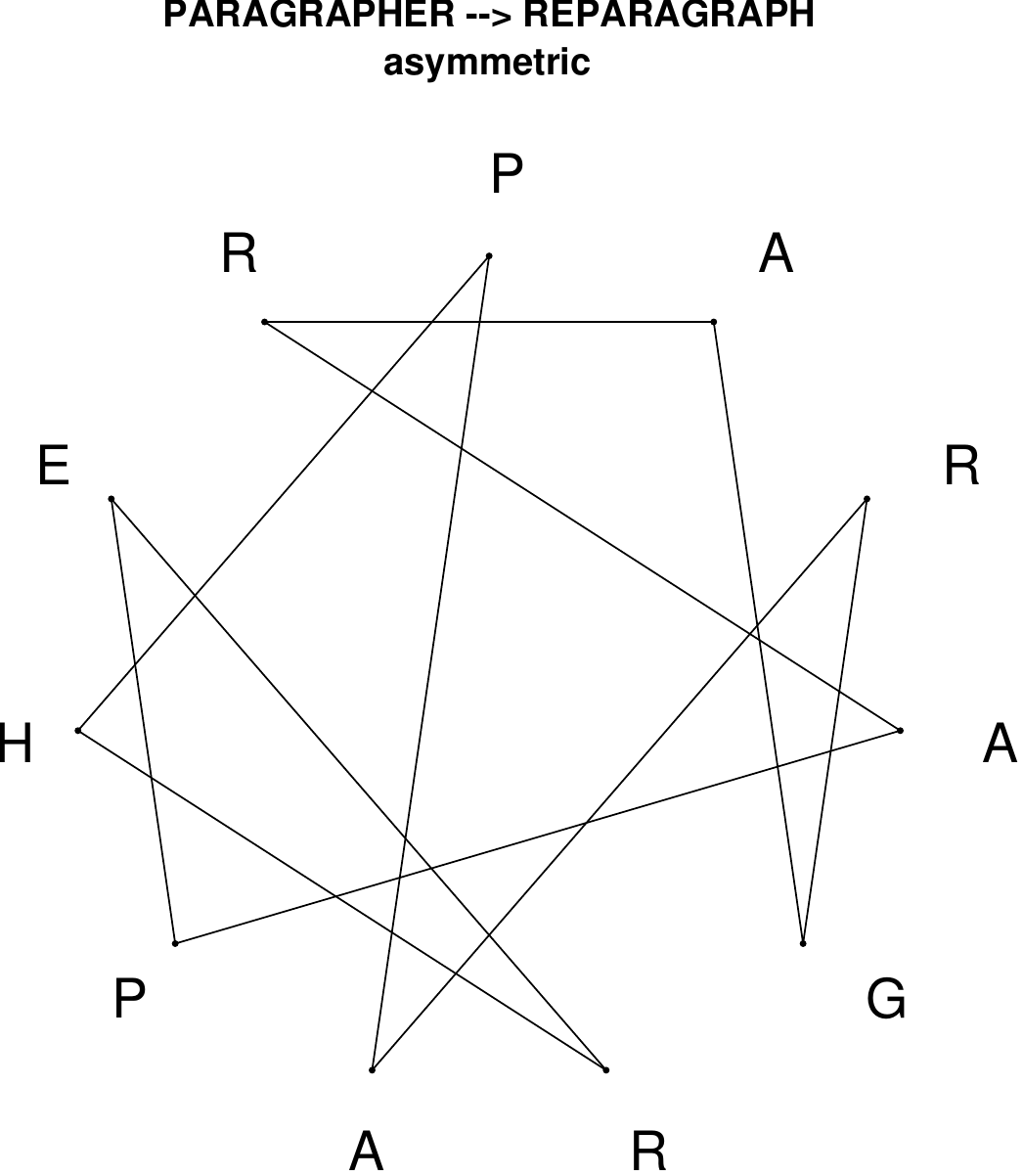}
\end{subfigure}
\hfill
\begin{subfigure}[T]{0.19\textwidth}
\centering
\includegraphics[width=\textwidth]{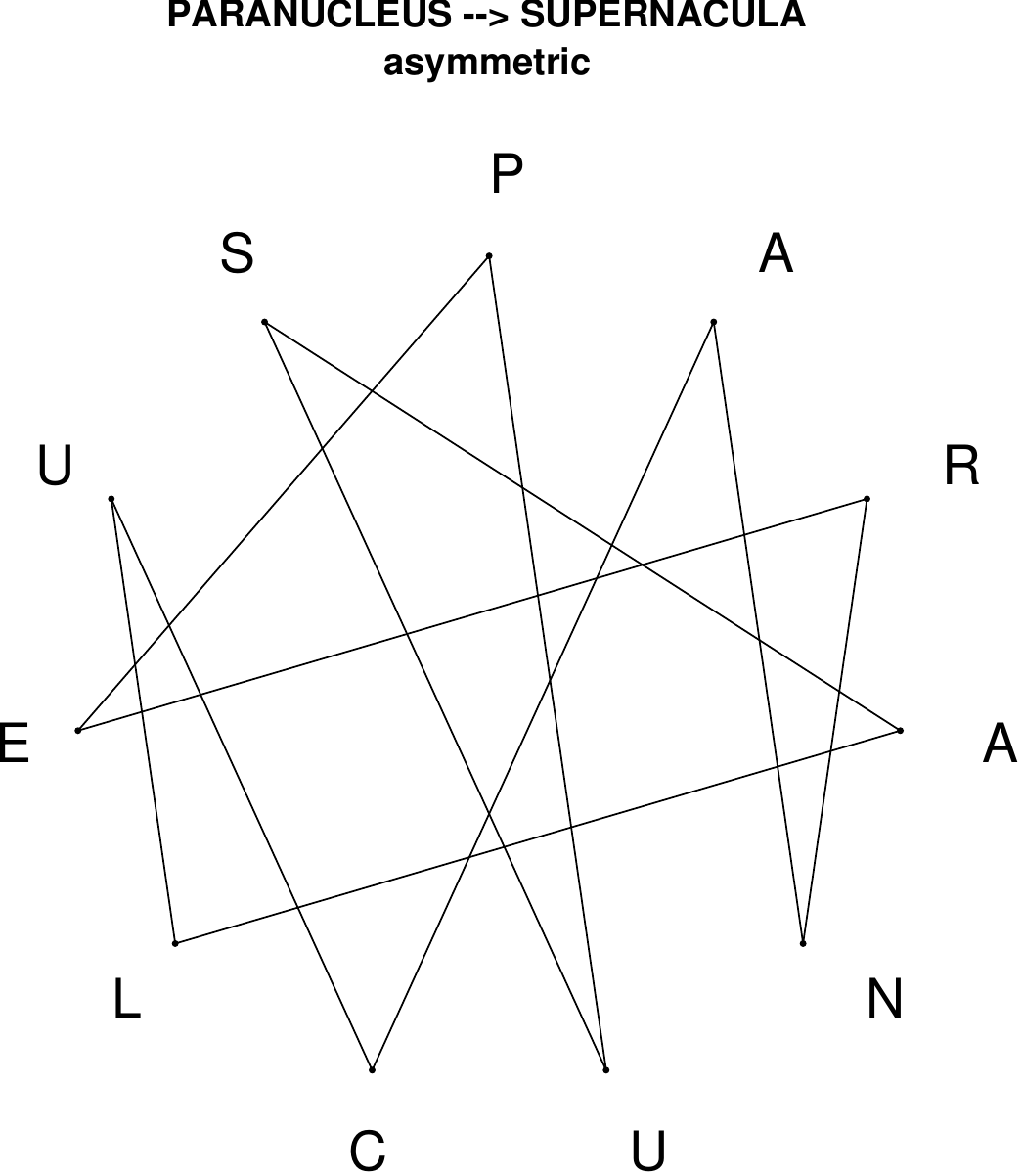}
\end{subfigure}
\end{figure}

\begin{figure}[H]
\centering
\begin{subfigure}[T]{0.19\textwidth}
\centering
\includegraphics[width=\textwidth]{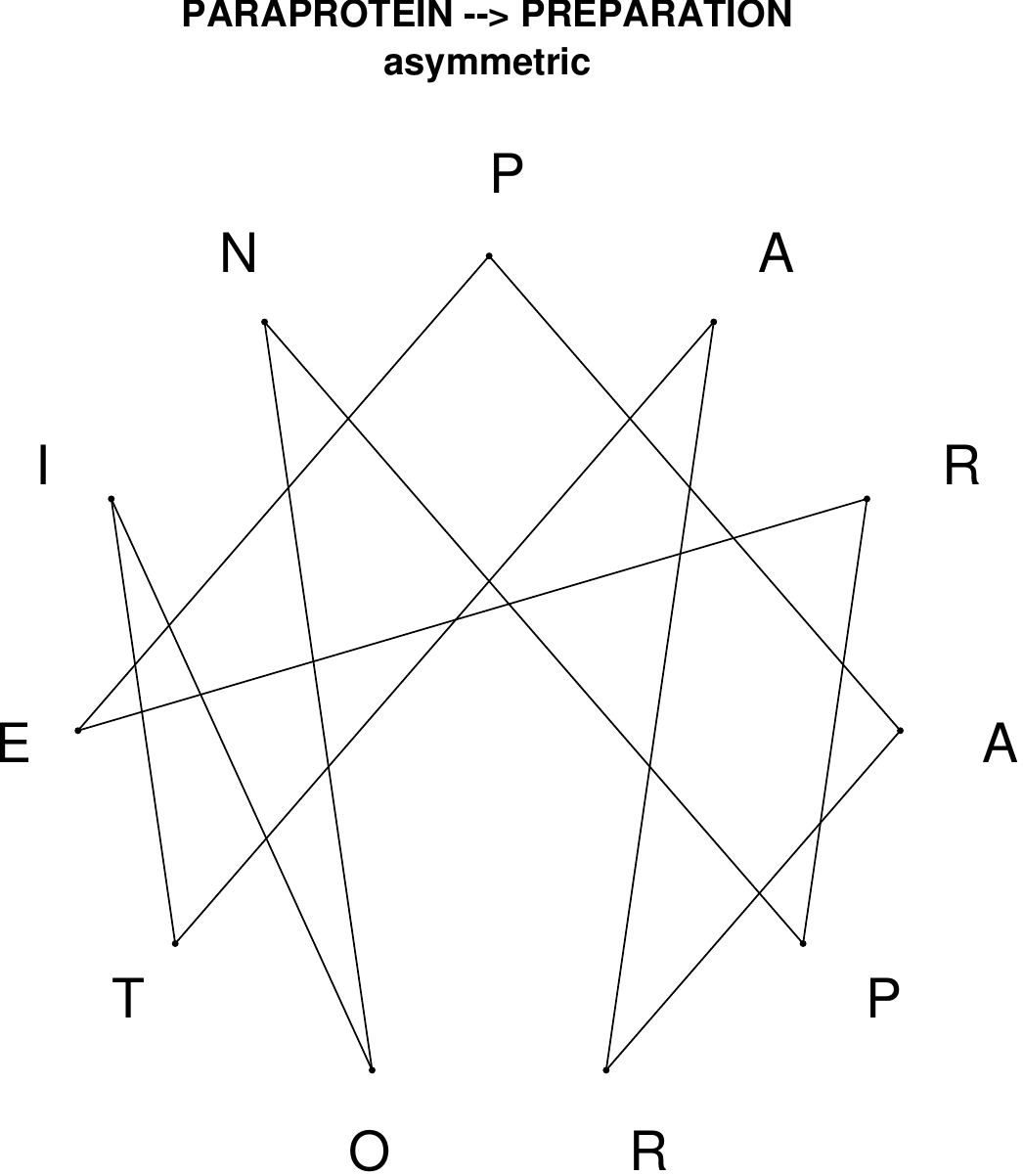}
\end{subfigure}
\hfill
\begin{subfigure}[T]{0.19\textwidth}
\centering
\includegraphics[width=\textwidth]{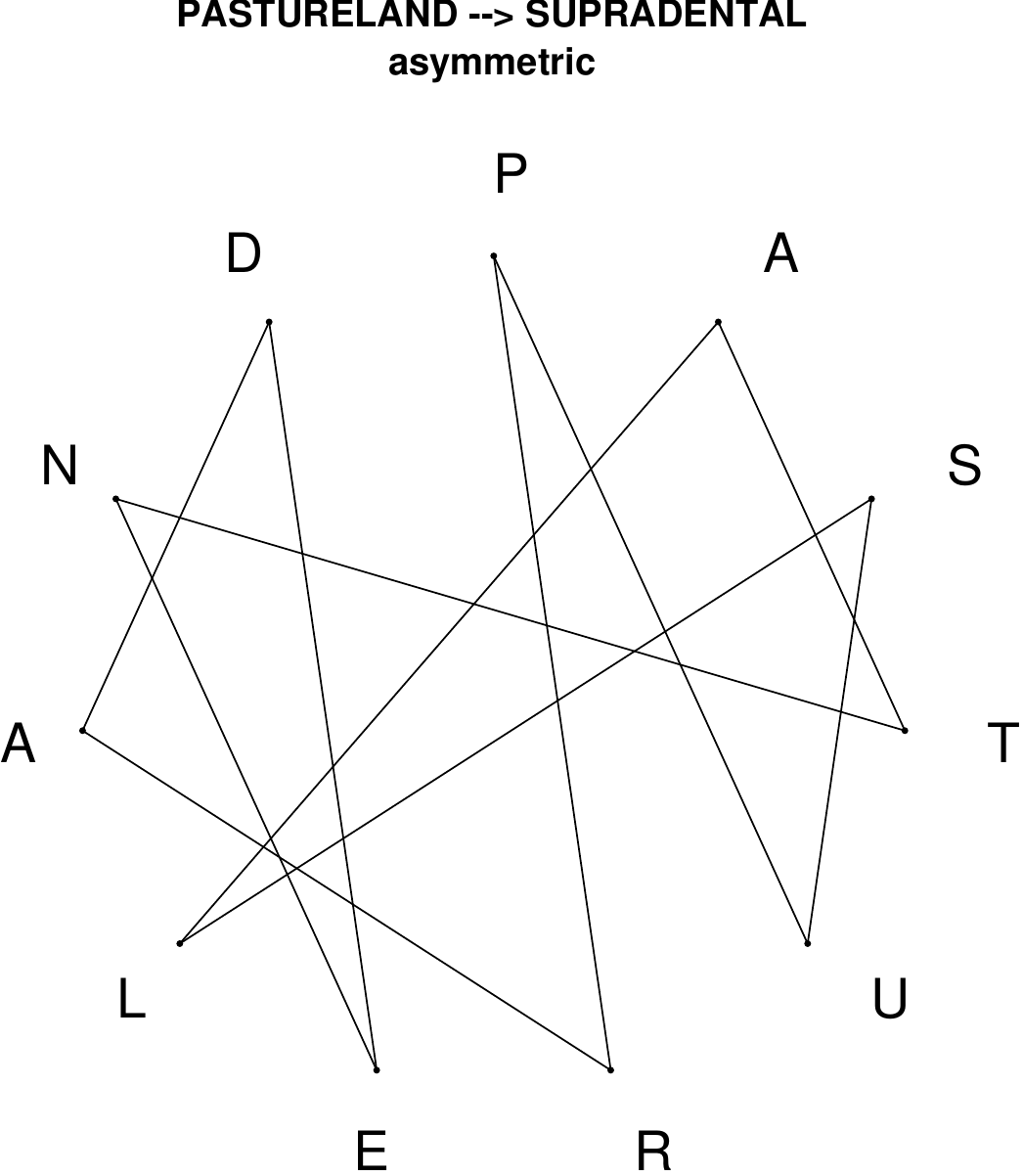}
\end{subfigure}
\hfill
\begin{subfigure}[T]{0.19\textwidth}
\centering
\includegraphics[width=\textwidth]{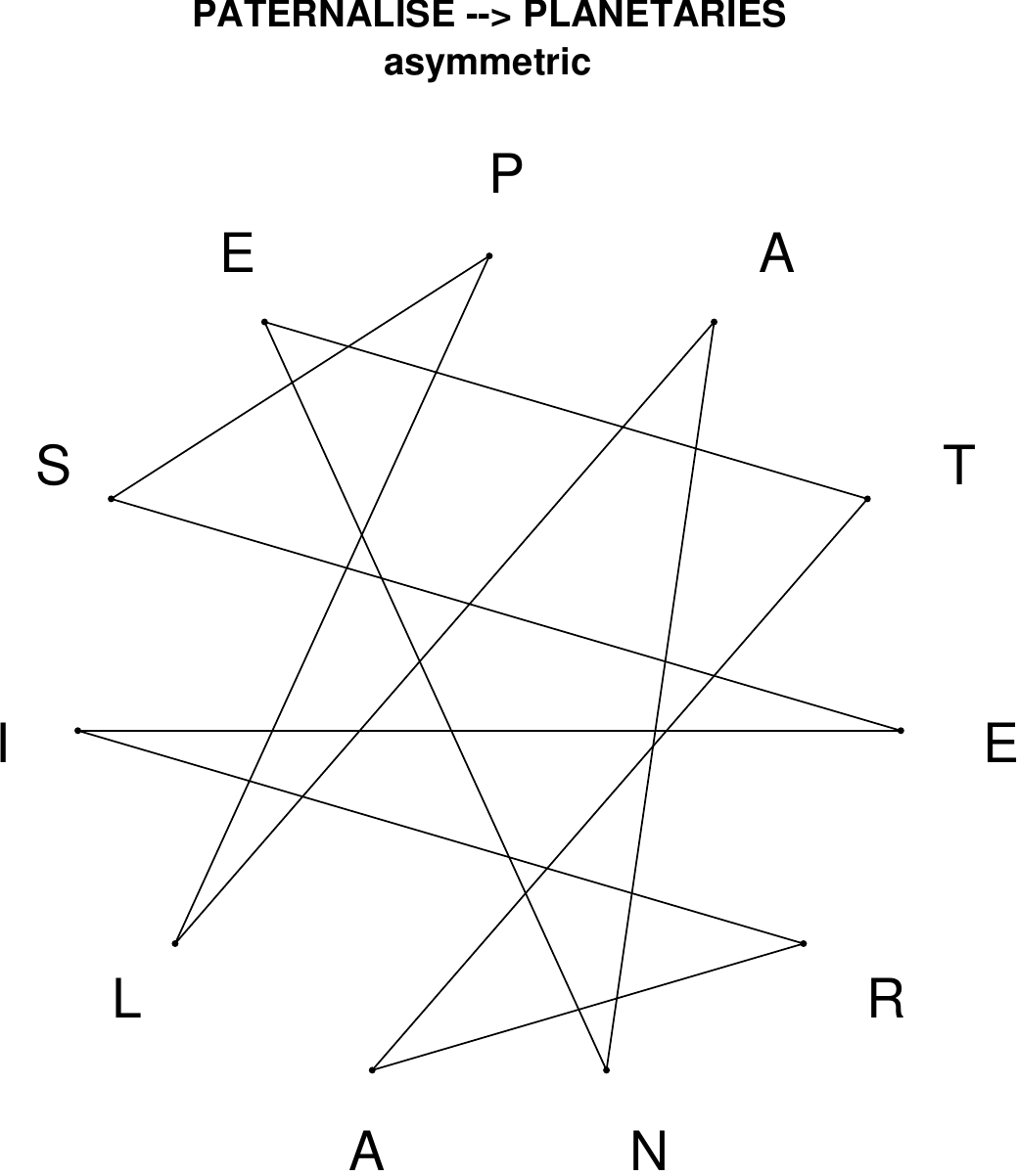}
\end{subfigure}
\hfill
\begin{subfigure}[T]{0.19\textwidth}
\centering
\includegraphics[width=\textwidth]{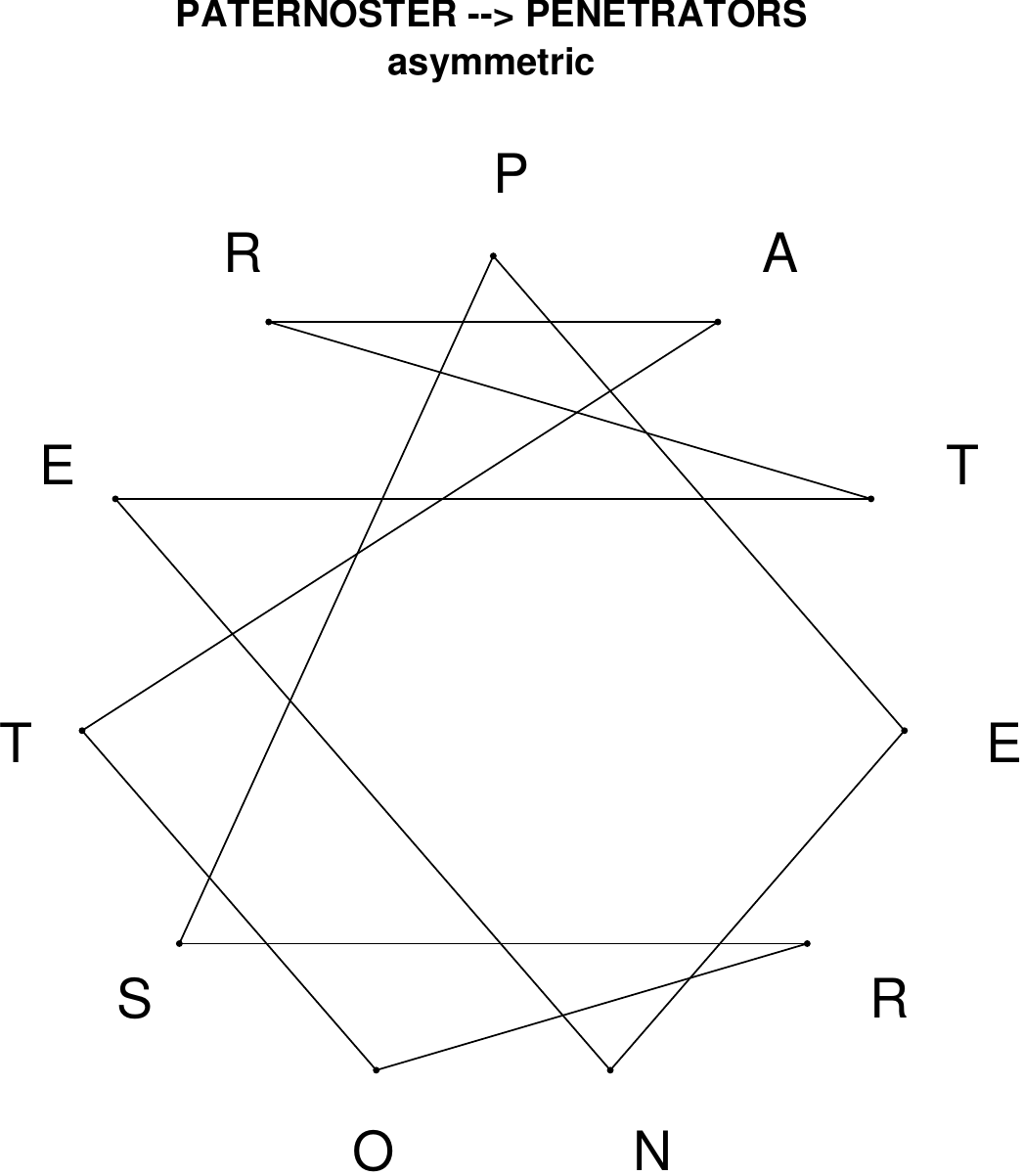}
\end{subfigure}
\hfill
\begin{subfigure}[T]{0.19\textwidth}
\centering
\includegraphics[width=\textwidth]{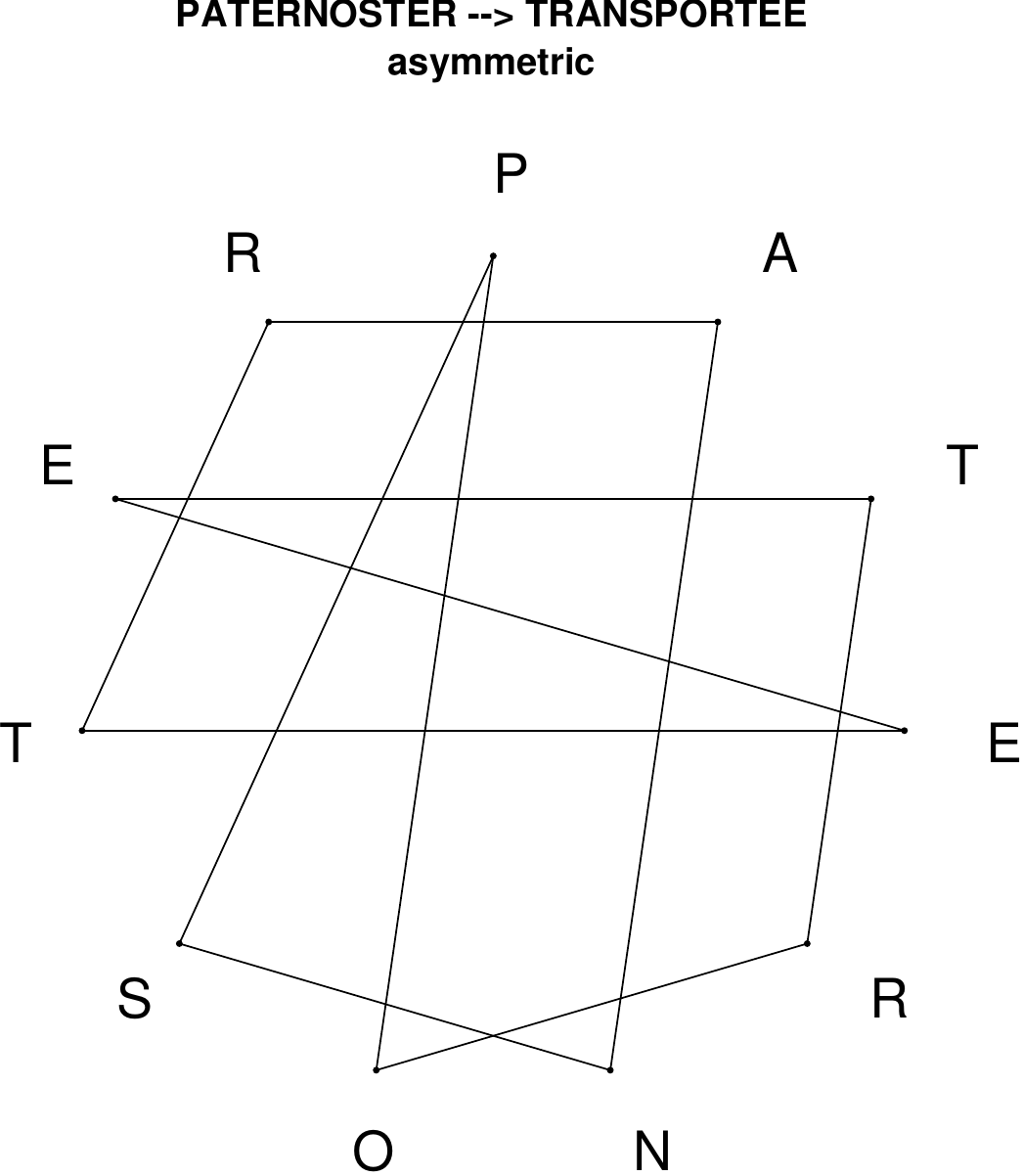}
\end{subfigure}
\end{figure}

\begin{figure}[H]
\centering
\begin{subfigure}[T]{0.19\textwidth}
\centering
\includegraphics[width=\textwidth]{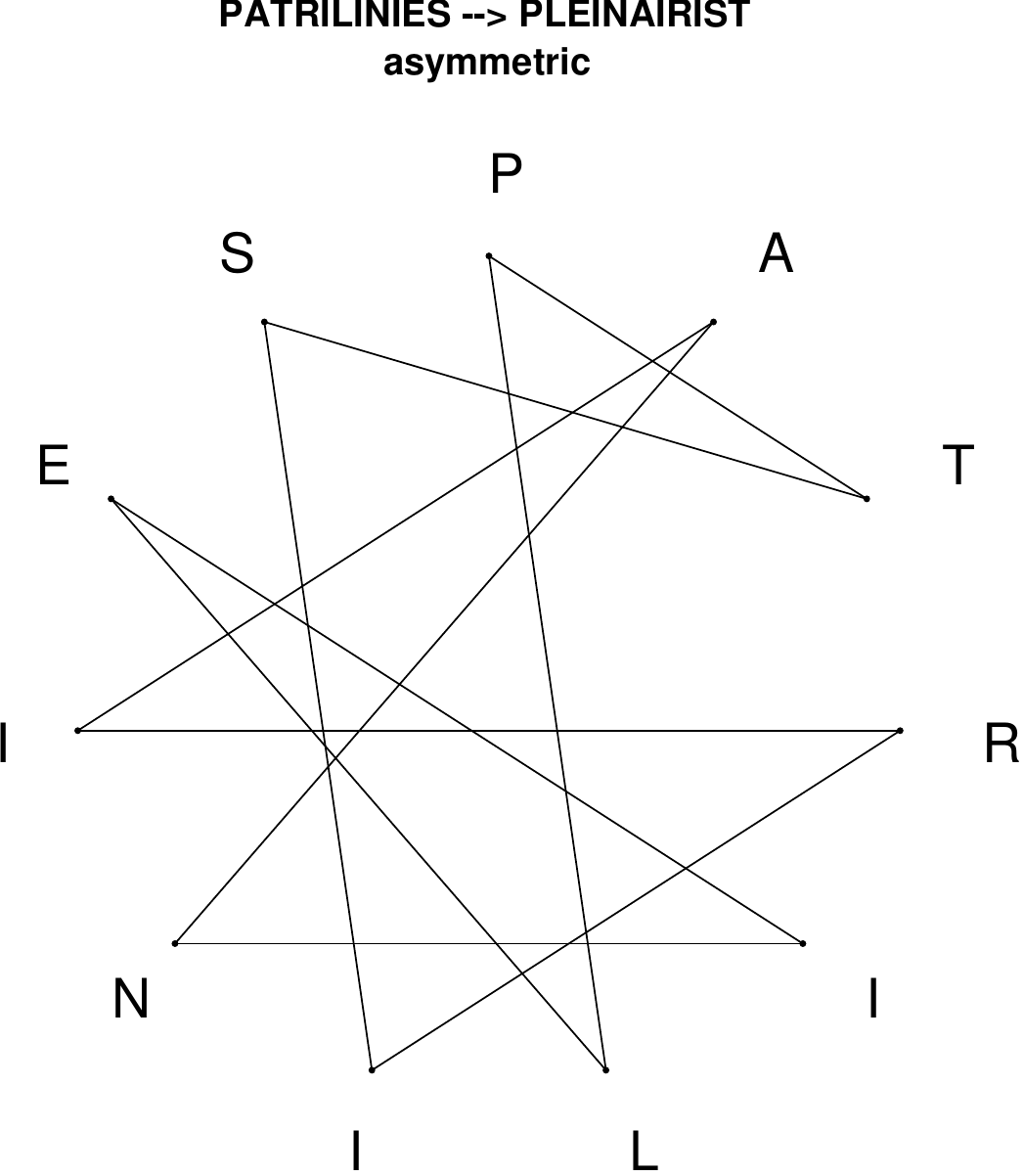}
\end{subfigure}
\hfill
\begin{subfigure}[T]{0.19\textwidth}
\centering
\includegraphics[width=\textwidth]{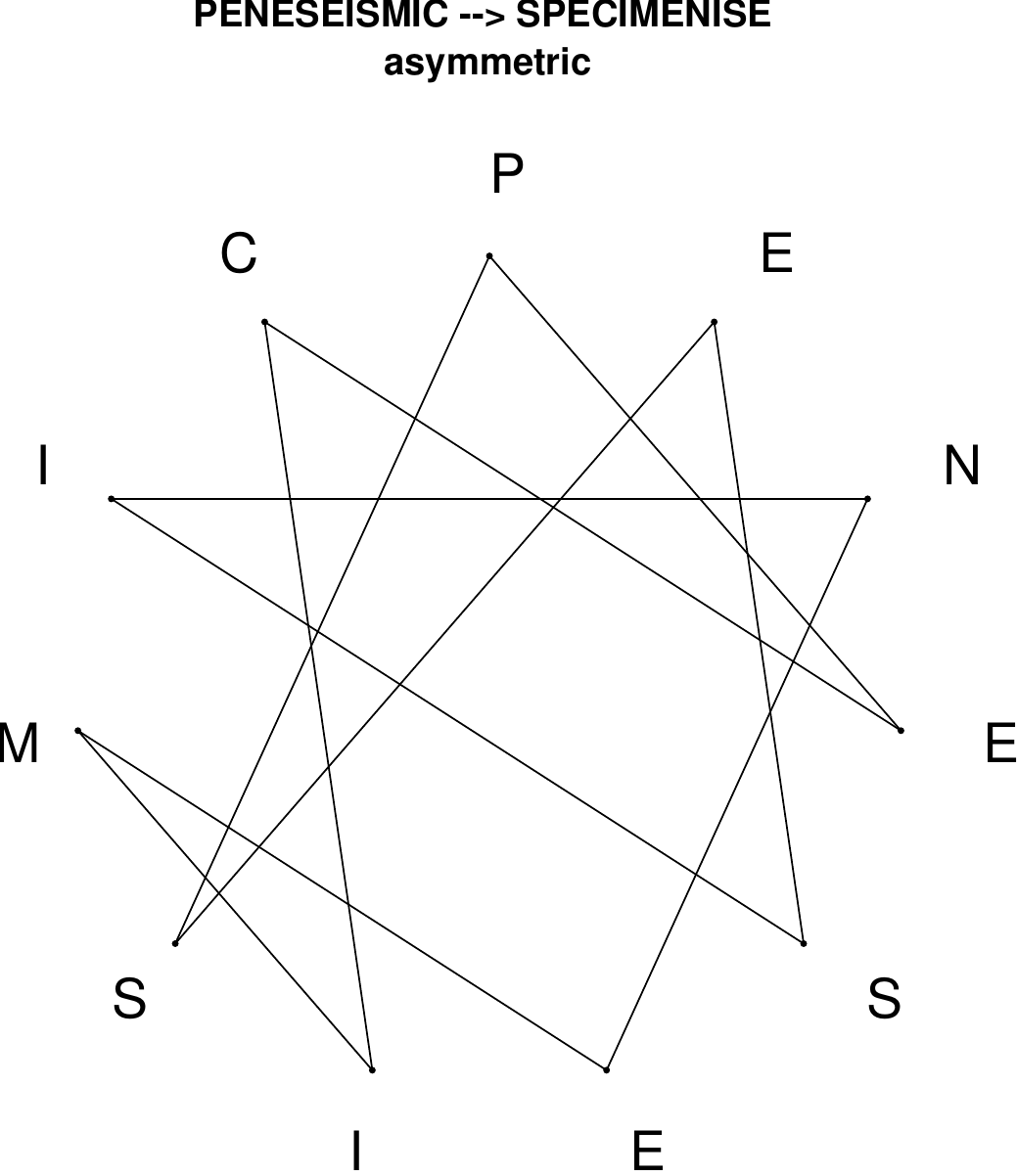}
\end{subfigure}
\hfill
\begin{subfigure}[T]{0.19\textwidth}
\centering
\includegraphics[width=\textwidth]{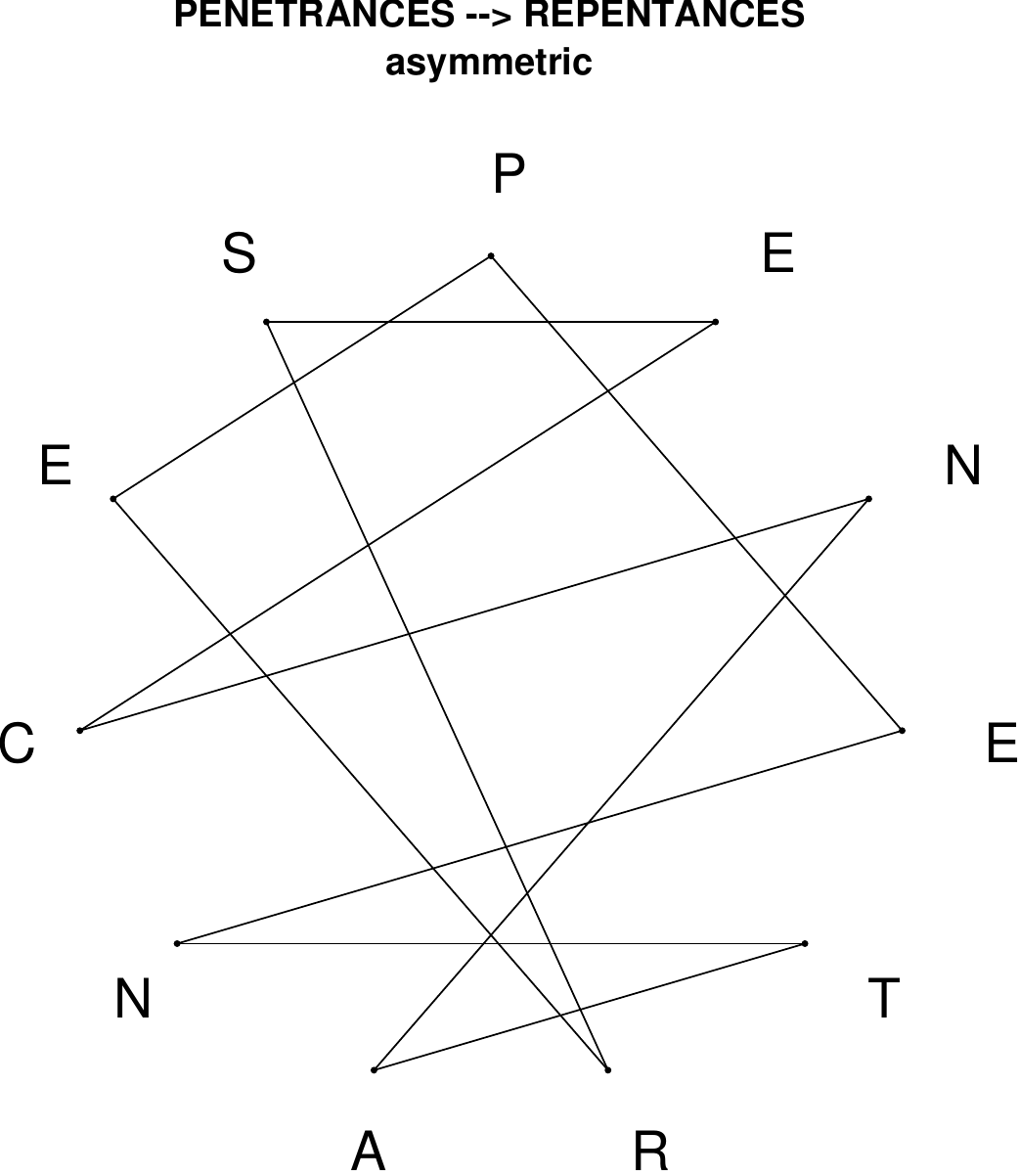}
\end{subfigure}
\hfill
\begin{subfigure}[T]{0.19\textwidth}
\centering
\includegraphics[width=\textwidth]{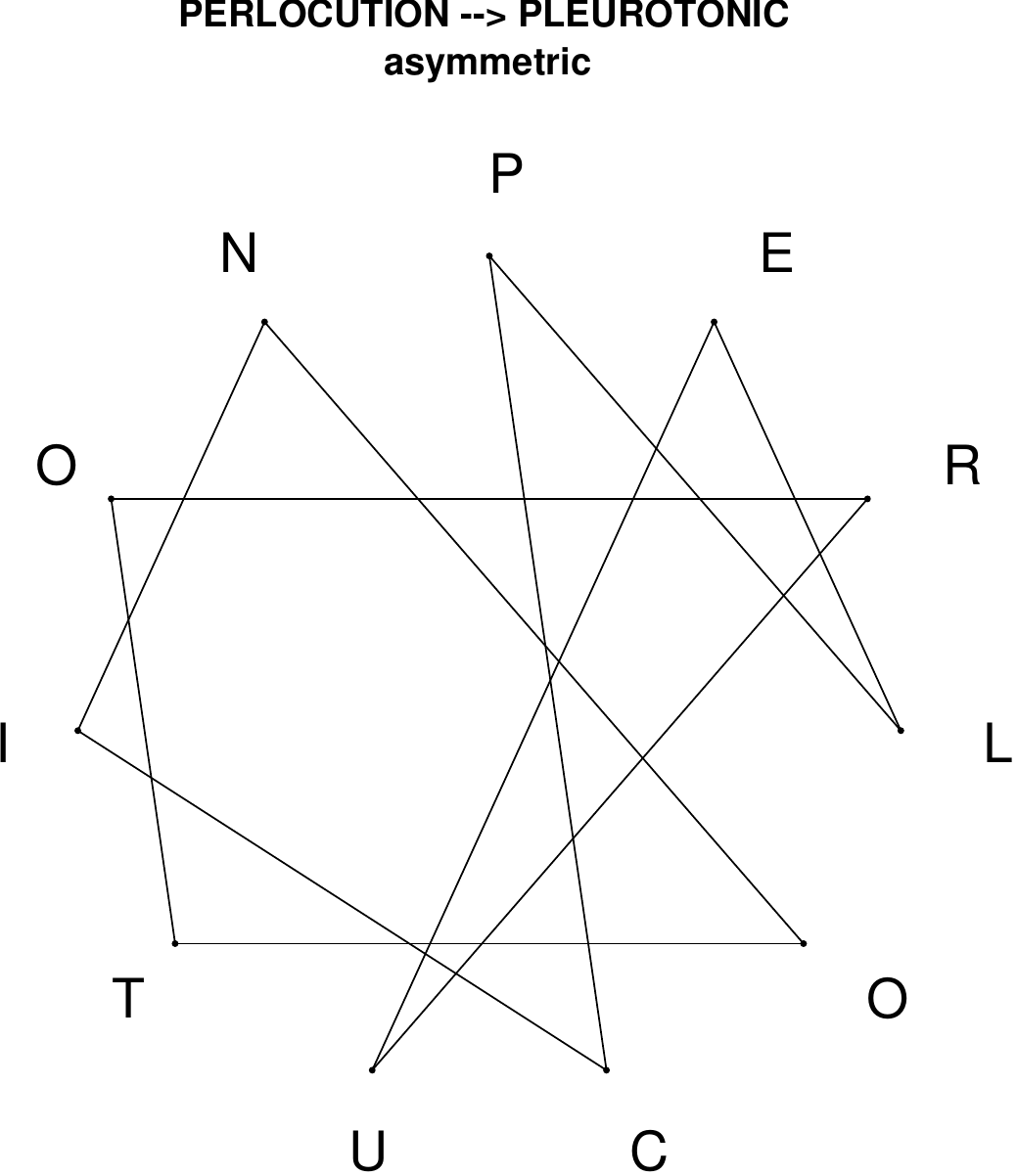}
\end{subfigure}
\hfill
\begin{subfigure}[T]{0.19\textwidth}
\centering
\includegraphics[width=\textwidth]{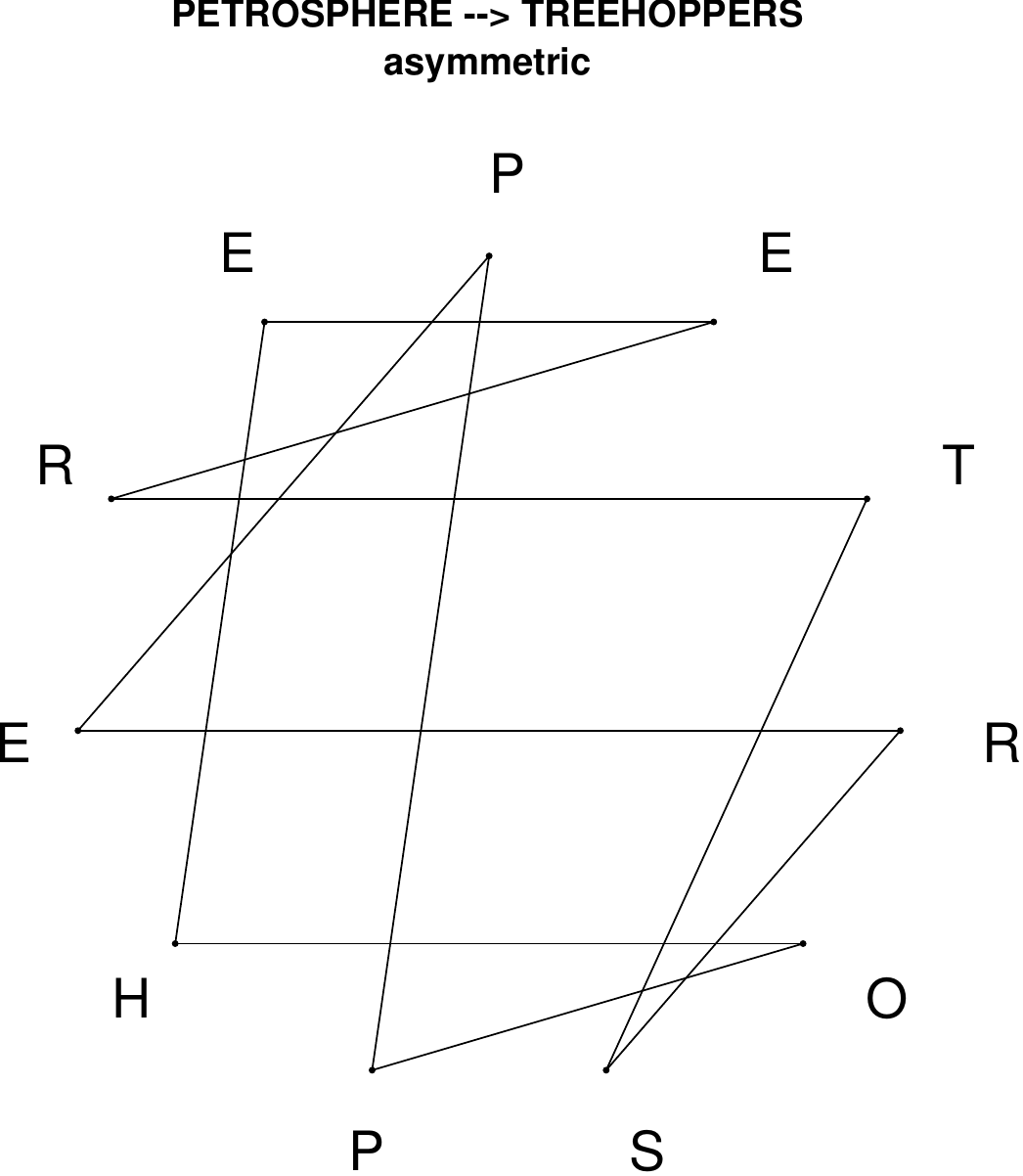}
\end{subfigure}
\end{figure}

\begin{figure}[H]
\centering
\begin{subfigure}[T]{0.19\textwidth}
\centering
\includegraphics[width=\textwidth]{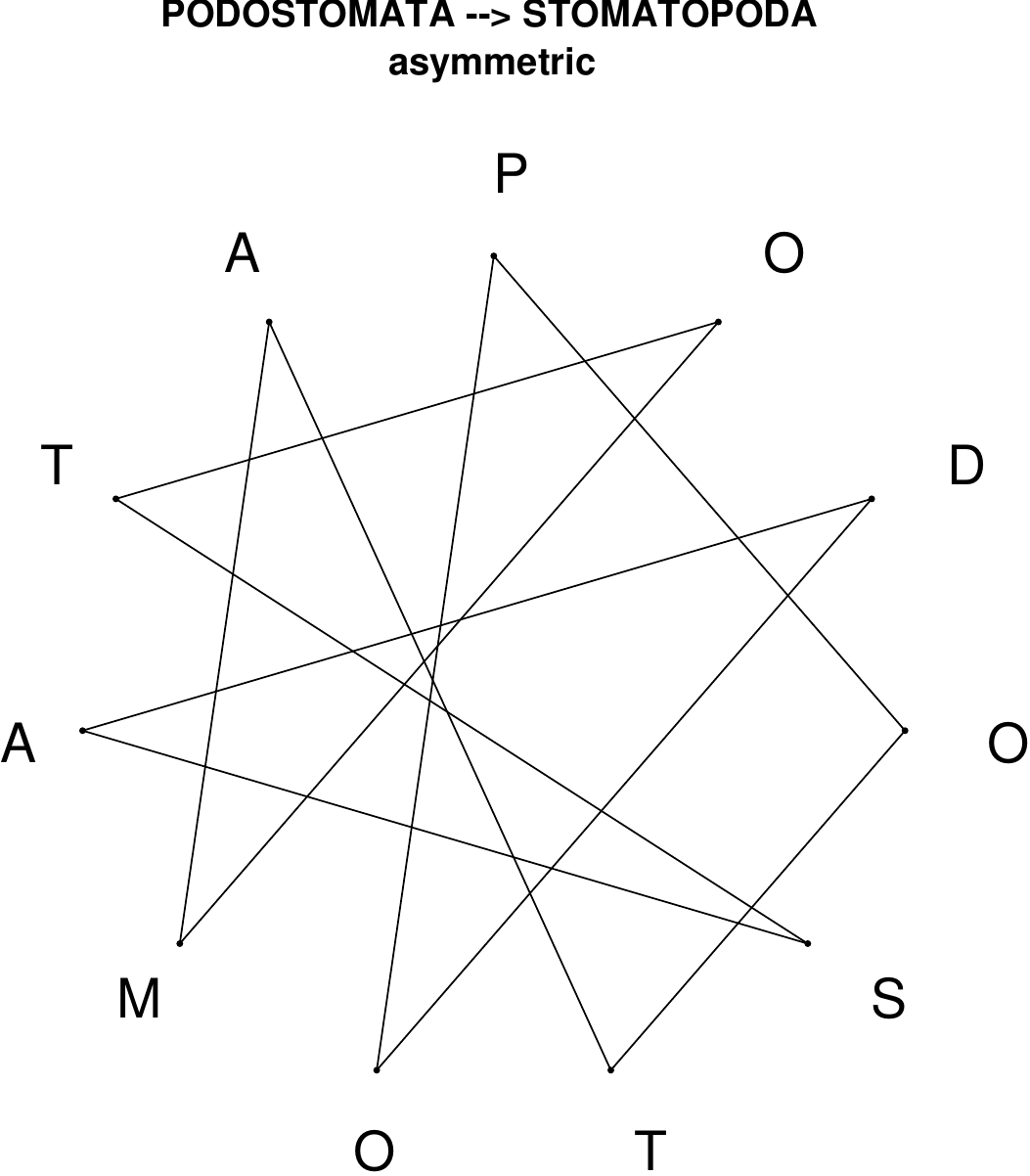}
\end{subfigure}
\hfill
\begin{subfigure}[T]{0.19\textwidth}
\centering
\includegraphics[width=\textwidth]{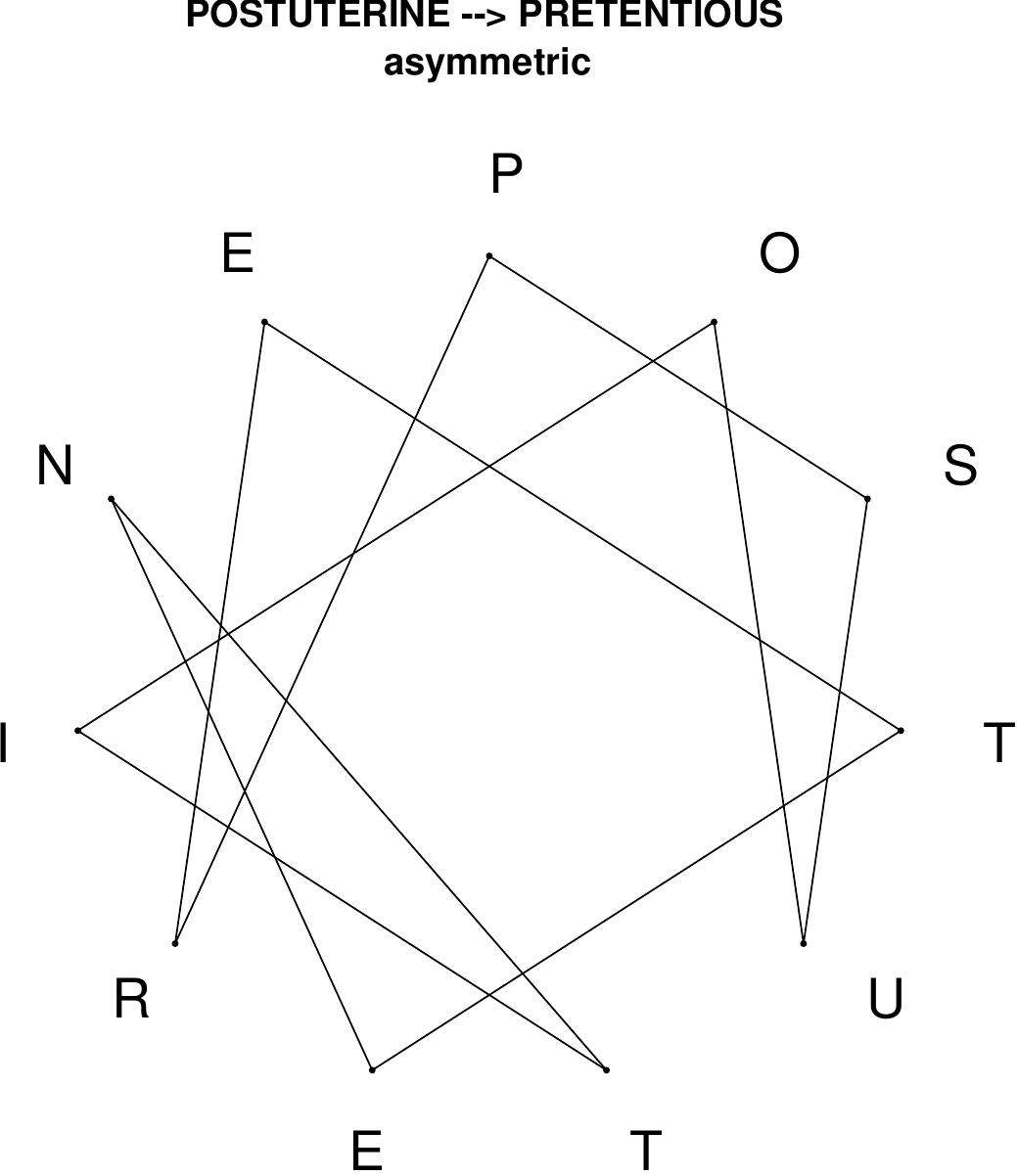}
\end{subfigure}
\hfill
\begin{subfigure}[T]{0.19\textwidth}
\centering
\includegraphics[width=\textwidth]{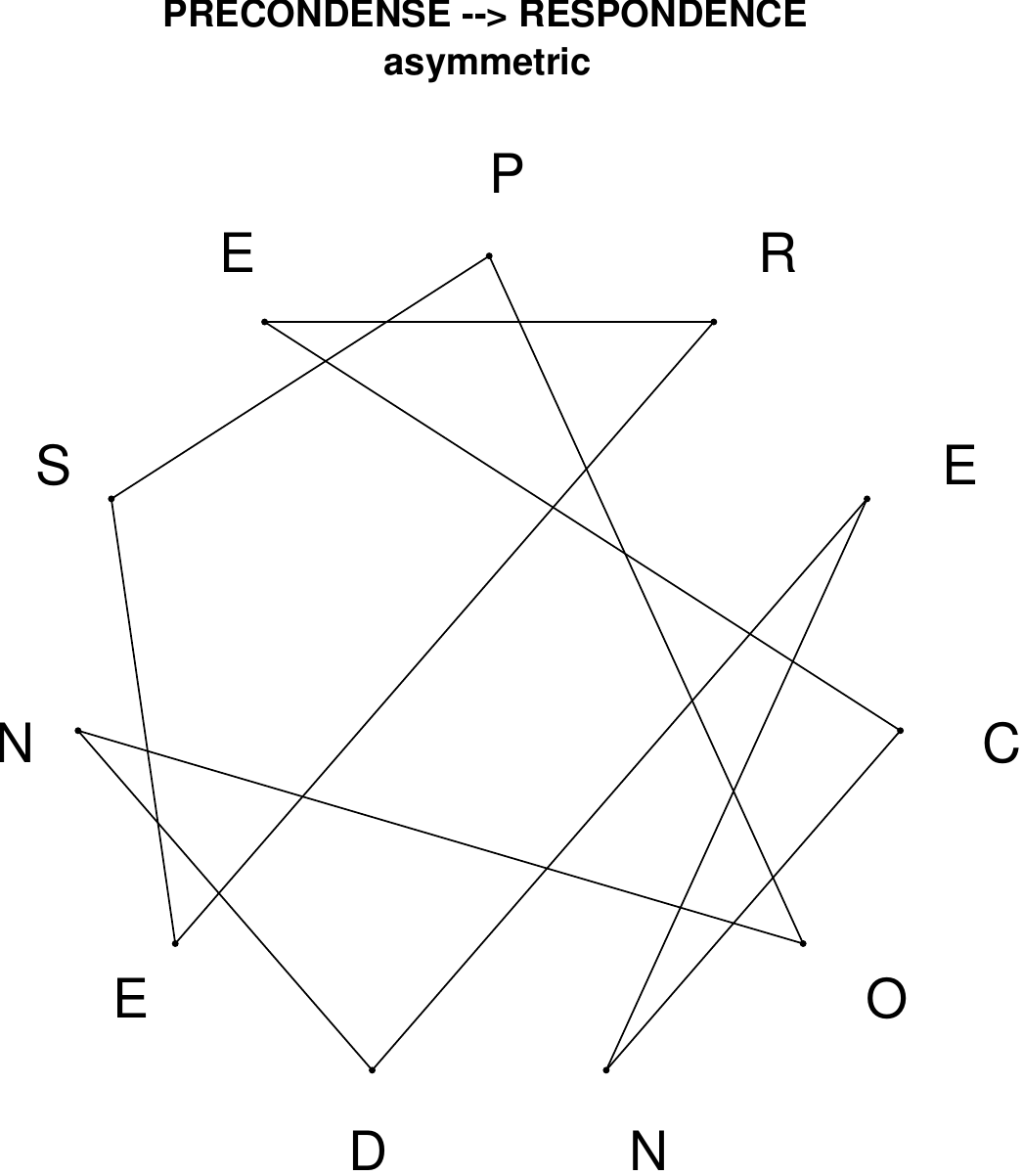}
\end{subfigure}
\hfill
\begin{subfigure}[T]{0.19\textwidth}
\centering
\includegraphics[width=\textwidth]{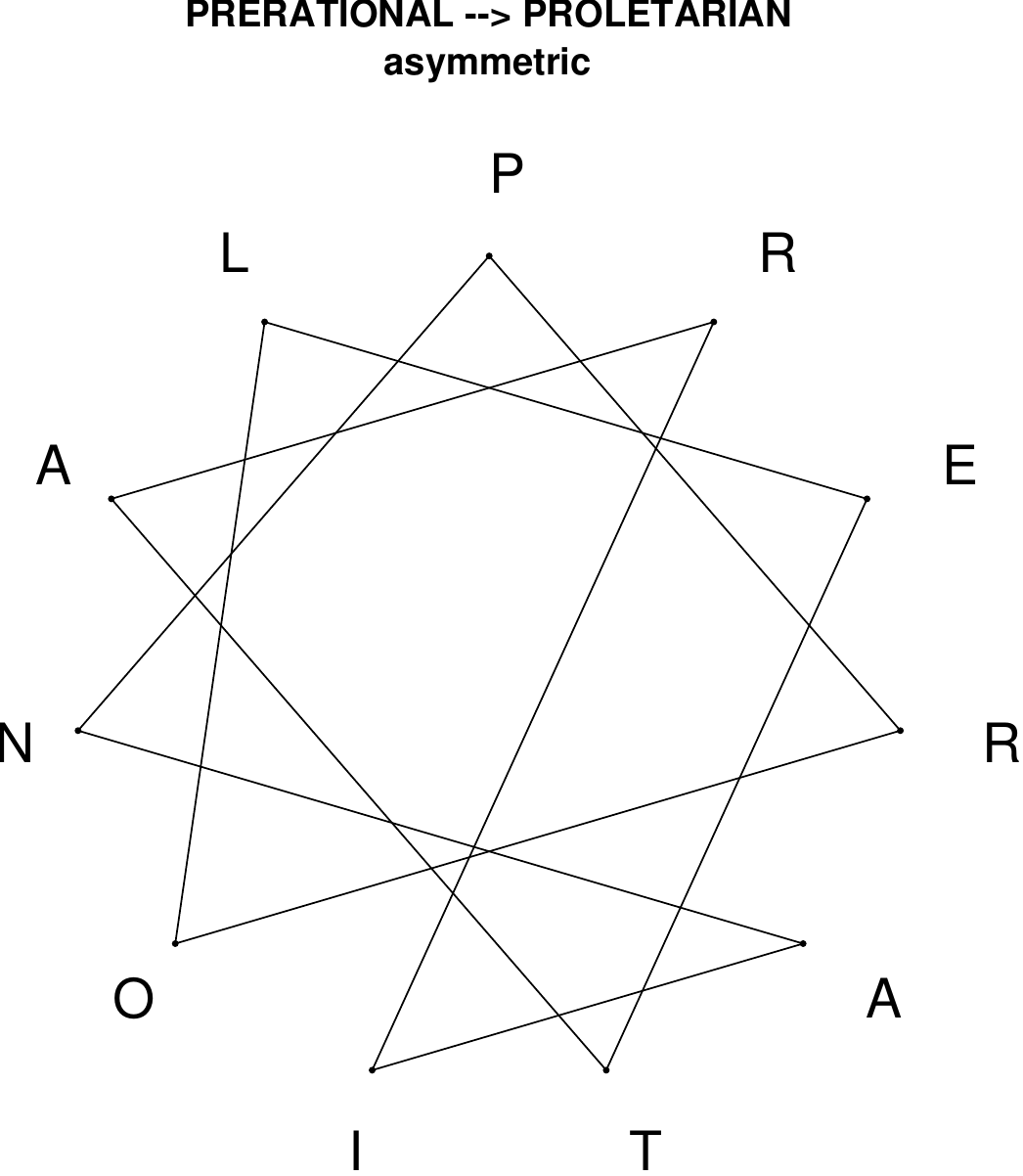}
\end{subfigure}
\hfill
\begin{subfigure}[T]{0.19\textwidth}
\centering
\includegraphics[width=\textwidth]{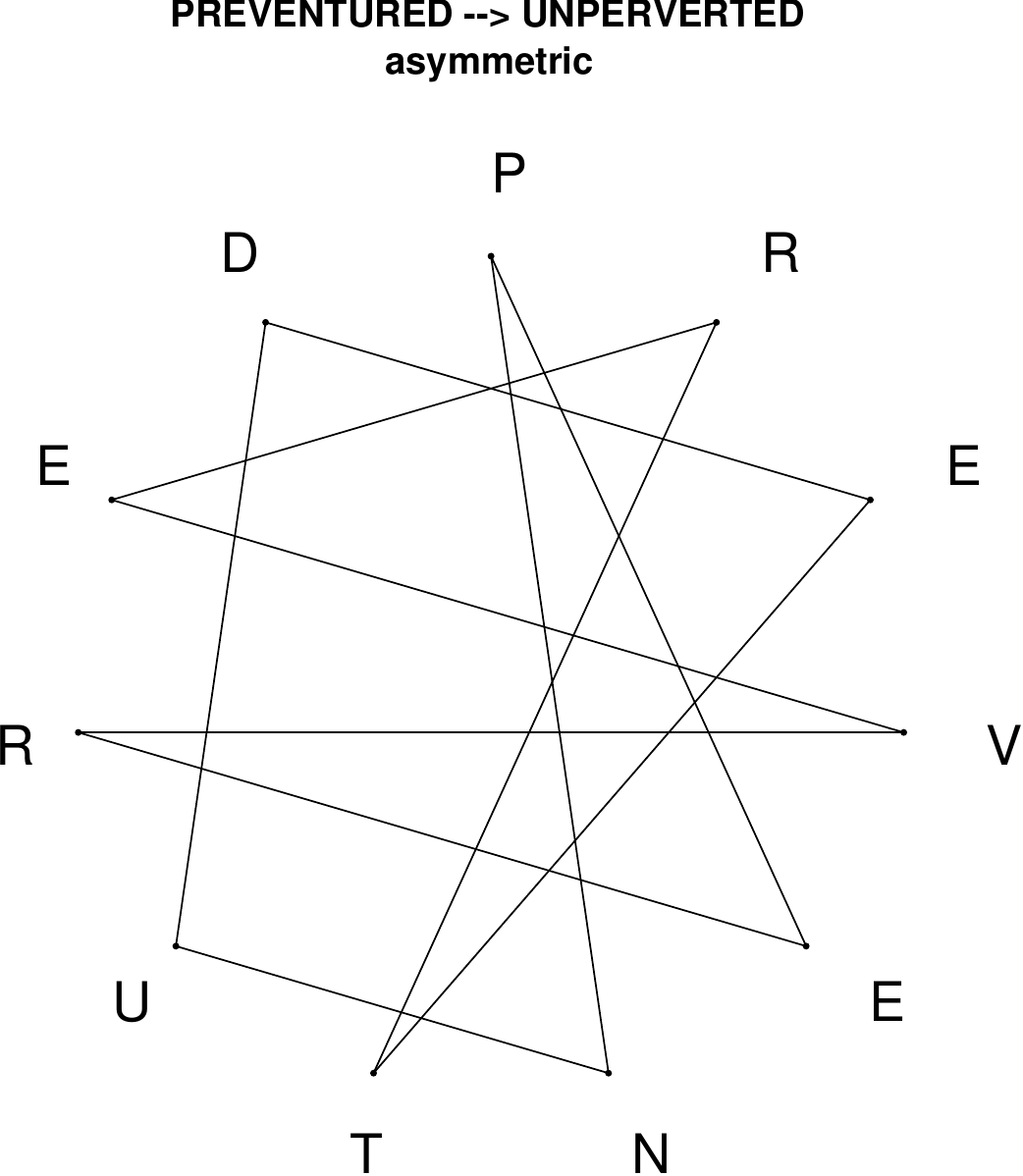}
\end{subfigure}
\end{figure}

\begin{figure}[H]
\centering
\begin{subfigure}[T]{0.19\textwidth}
\centering
\includegraphics[width=\textwidth]{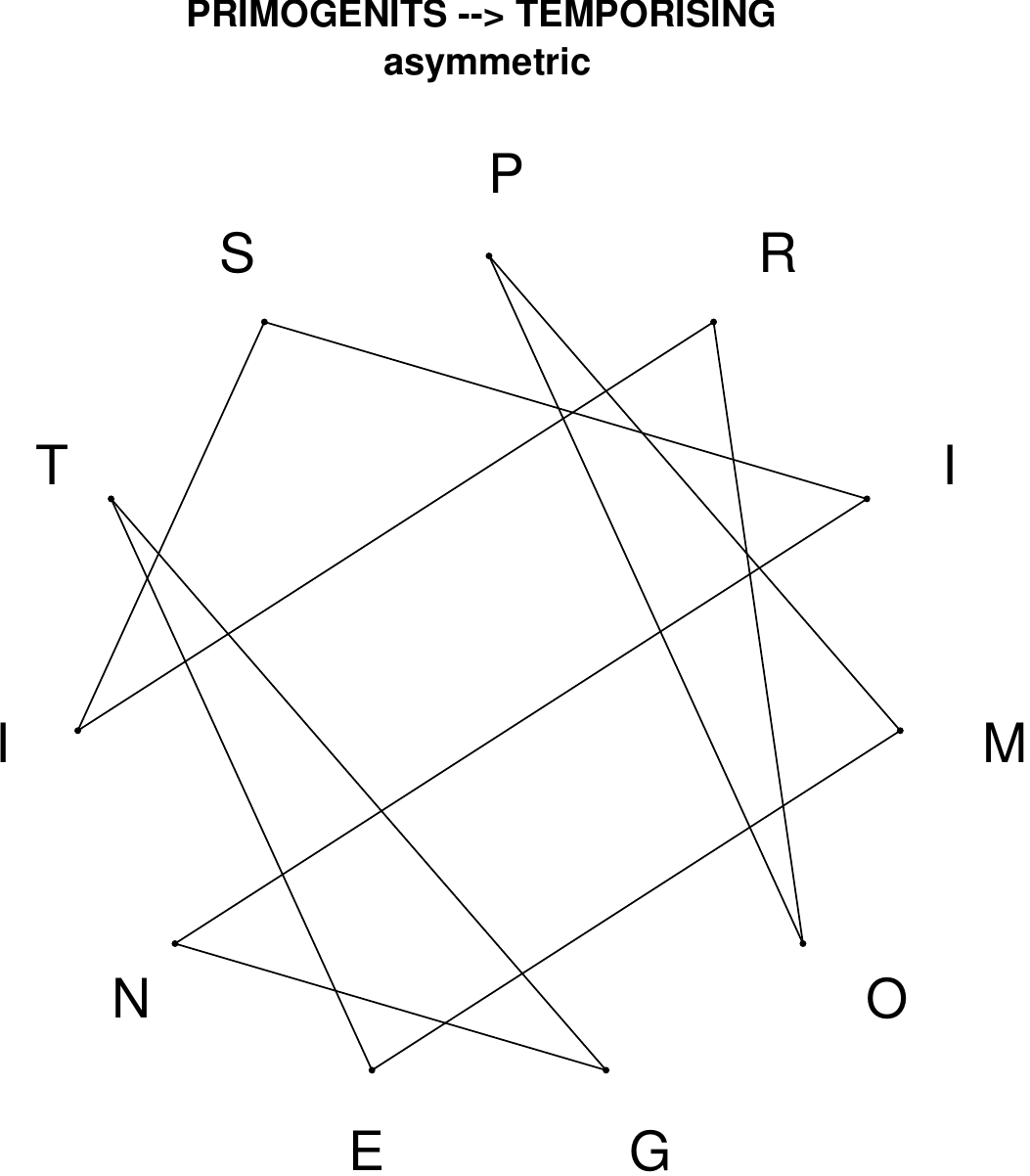}
\end{subfigure}
\hfill
\begin{subfigure}[T]{0.19\textwidth}
\centering
\includegraphics[width=\textwidth]{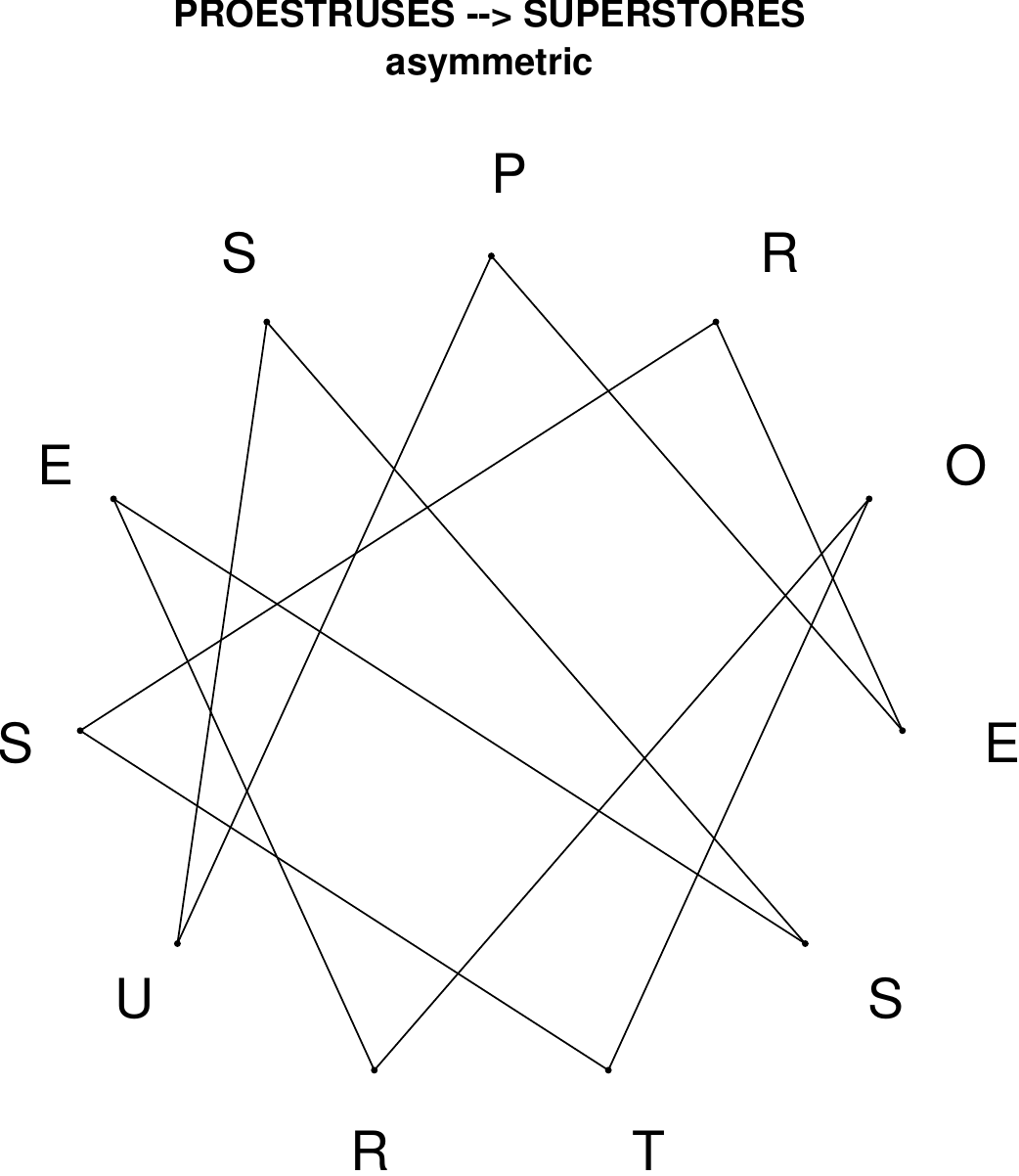}
\end{subfigure}
\hfill
\begin{subfigure}[T]{0.19\textwidth}
\centering
\includegraphics[width=\textwidth]{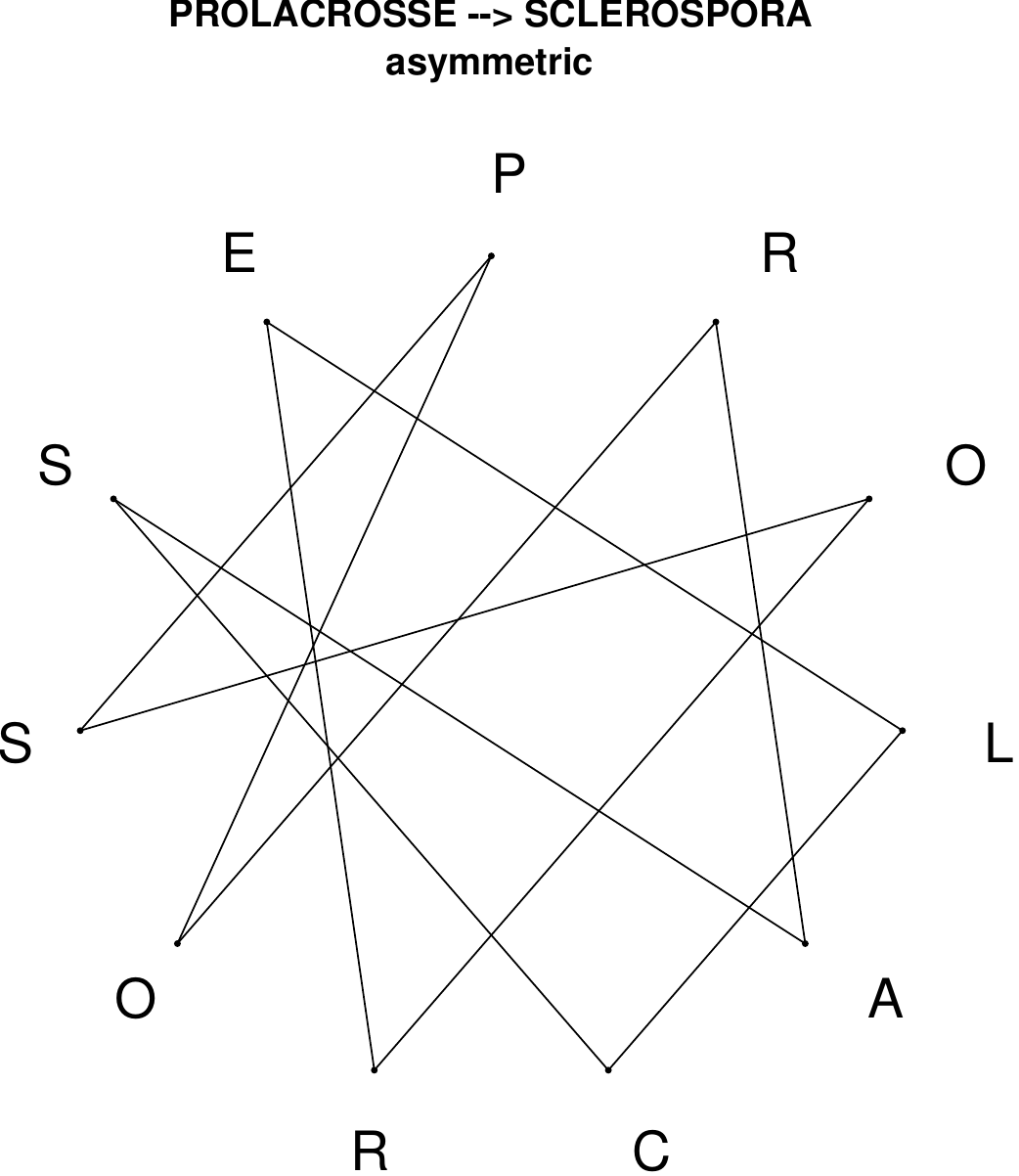}
\end{subfigure}
\hfill
\begin{subfigure}[T]{0.19\textwidth}
\centering
\includegraphics[width=\textwidth]{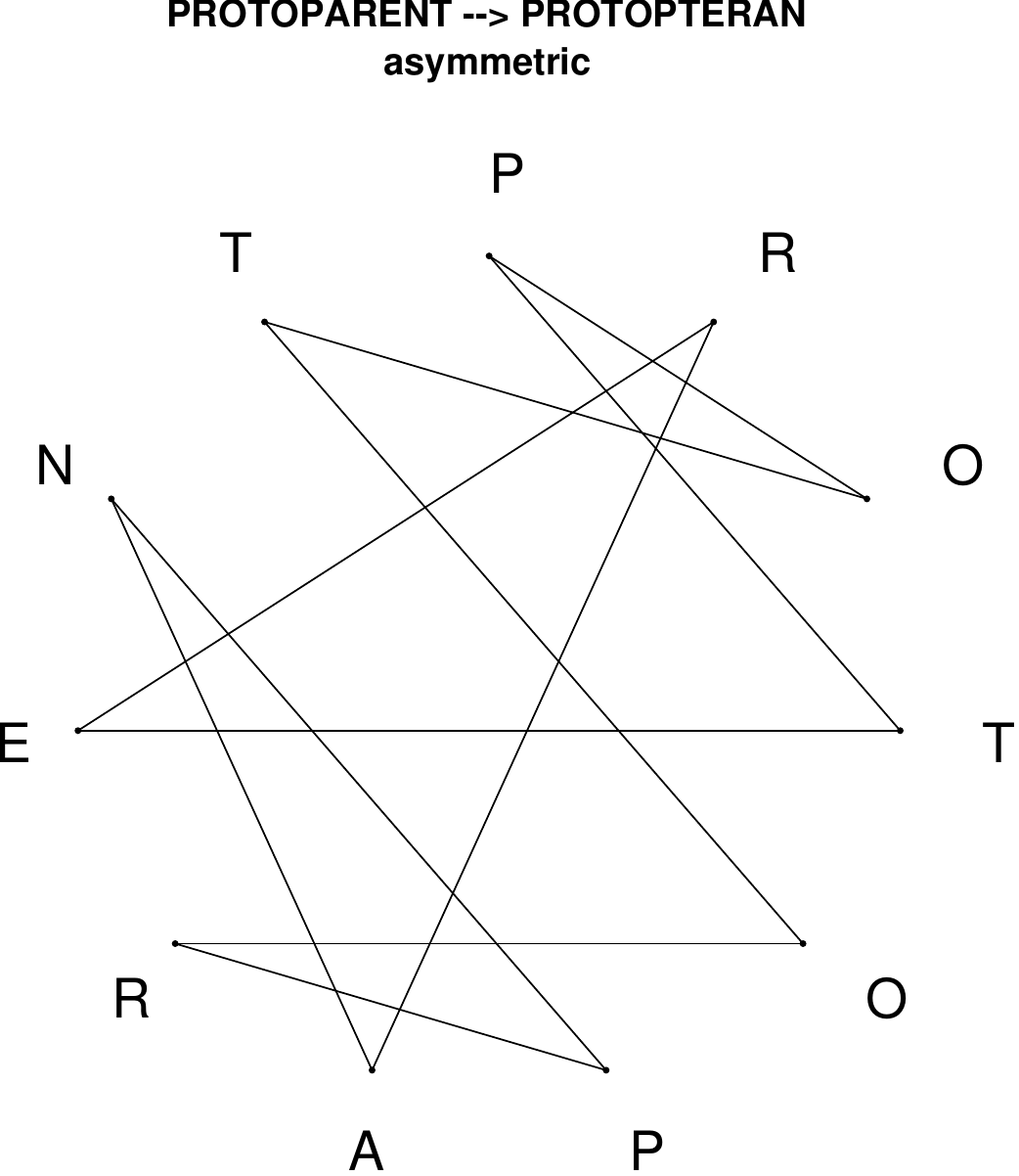}
\end{subfigure}
\hfill
\begin{subfigure}[T]{0.19\textwidth}
\centering
\includegraphics[width=\textwidth]{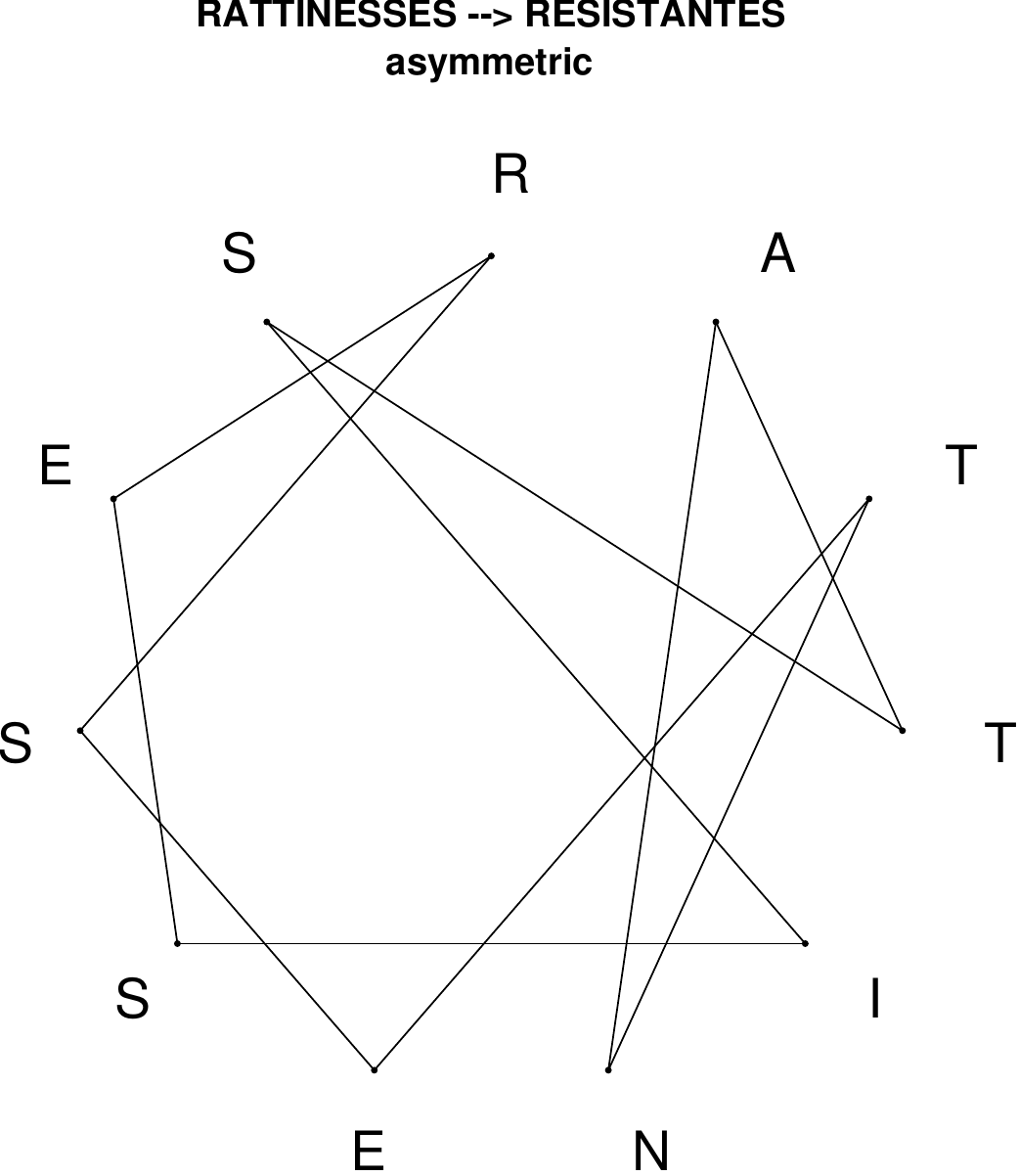}
\end{subfigure}
\end{figure}

\begin{figure}[H]
\centering
\begin{subfigure}[T]{0.19\textwidth}
\centering
\includegraphics[width=\textwidth]{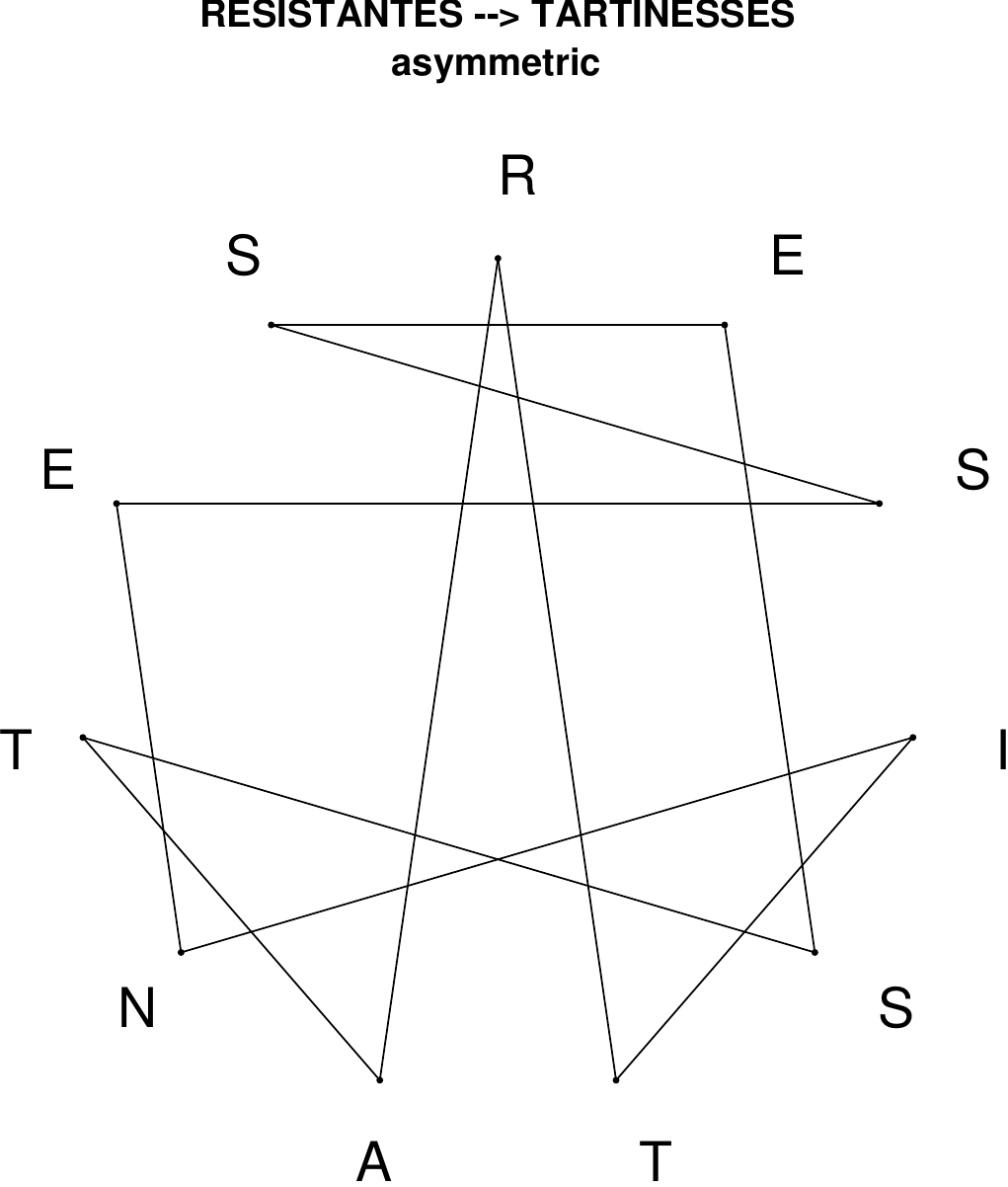}
\end{subfigure}
\hfill
\begin{subfigure}[T]{0.19\textwidth}
\centering
\includegraphics[width=\textwidth]{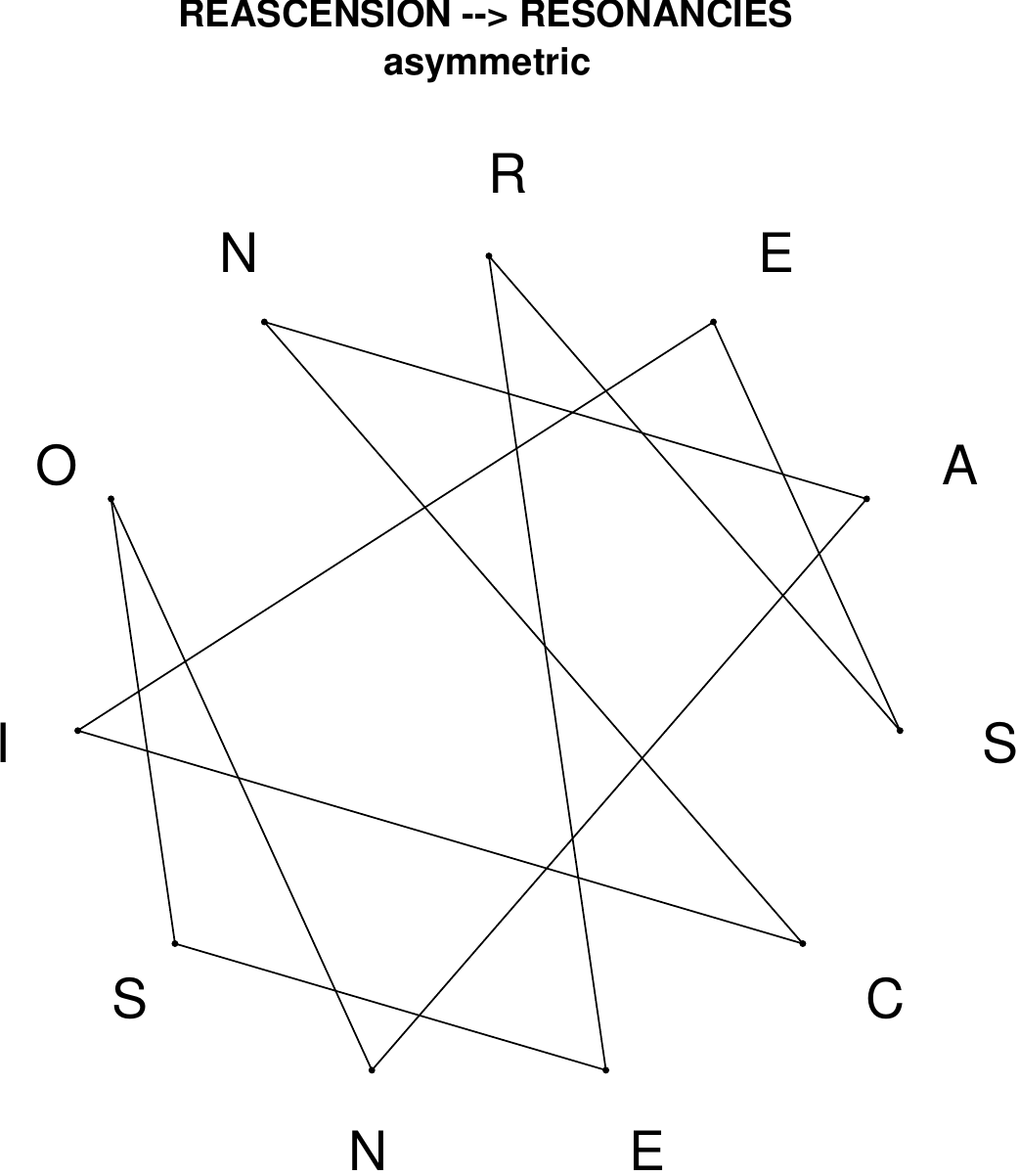}
\end{subfigure}
\hfill
\begin{subfigure}[T]{0.19\textwidth}
\centering
\includegraphics[width=\textwidth]{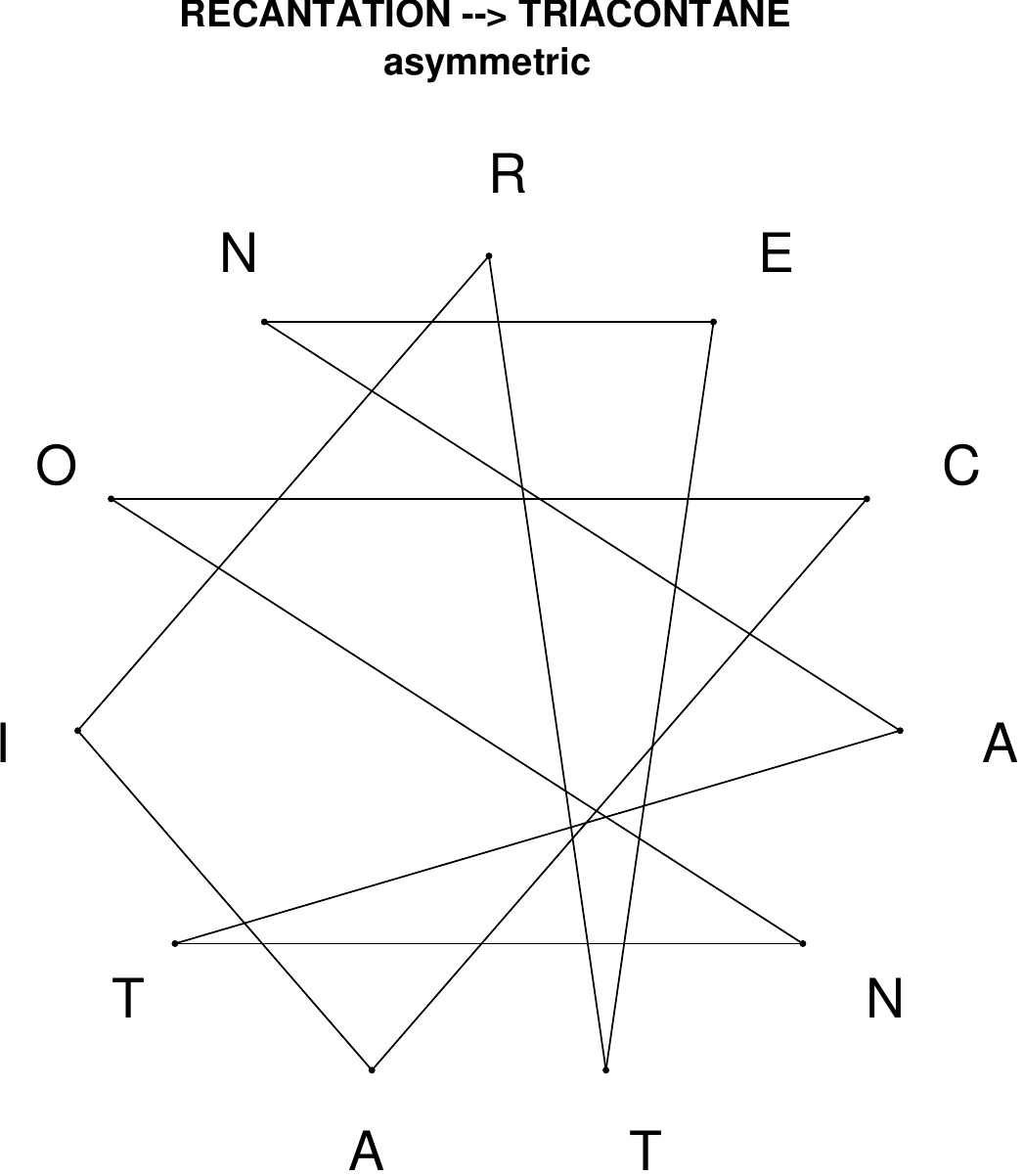}
\end{subfigure}
\hfill
\begin{subfigure}[T]{0.19\textwidth}
\centering
\includegraphics[width=\textwidth]{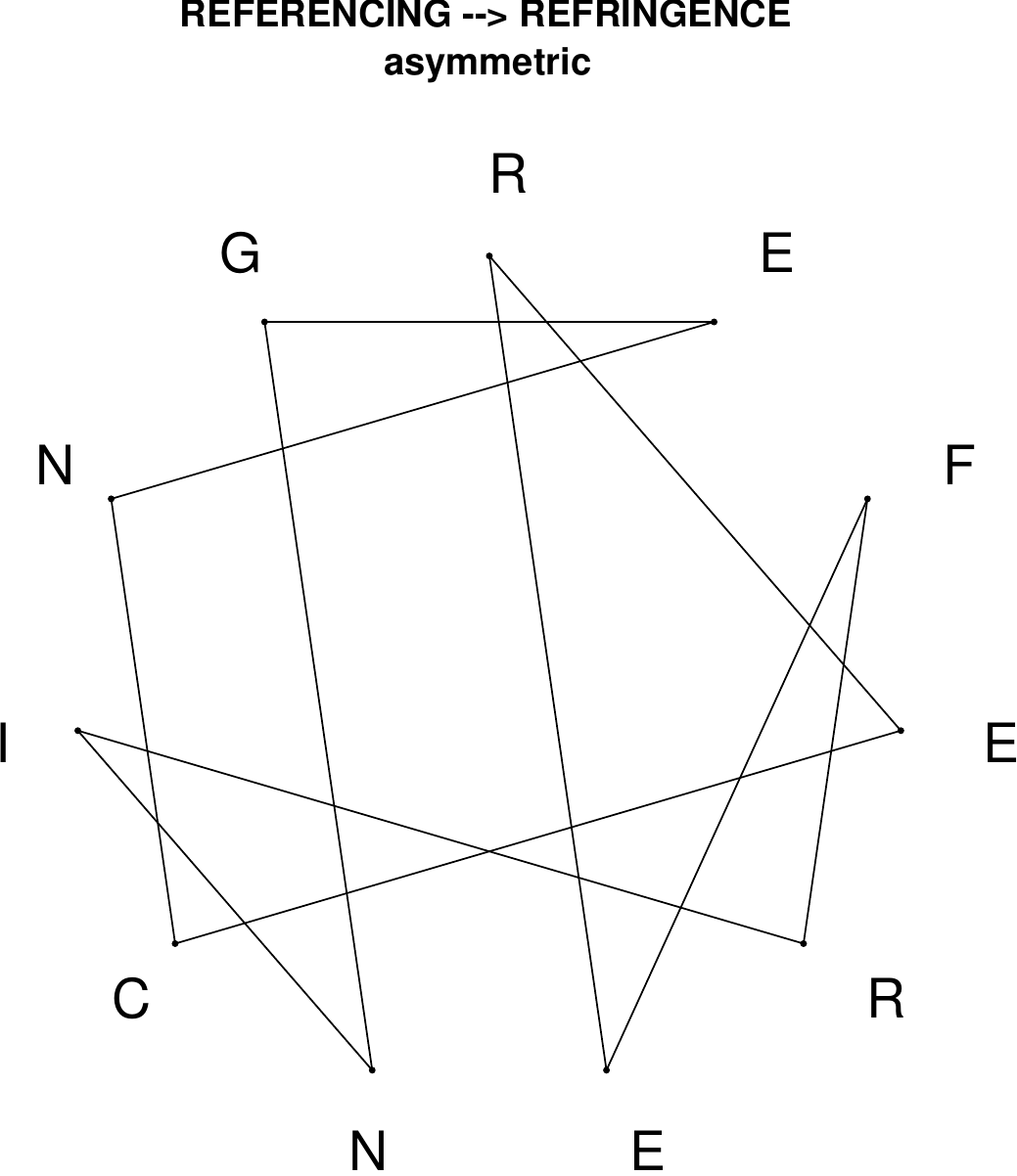}
\end{subfigure}
\hfill
\begin{subfigure}[T]{0.19\textwidth}
\centering
\includegraphics[width=\textwidth]{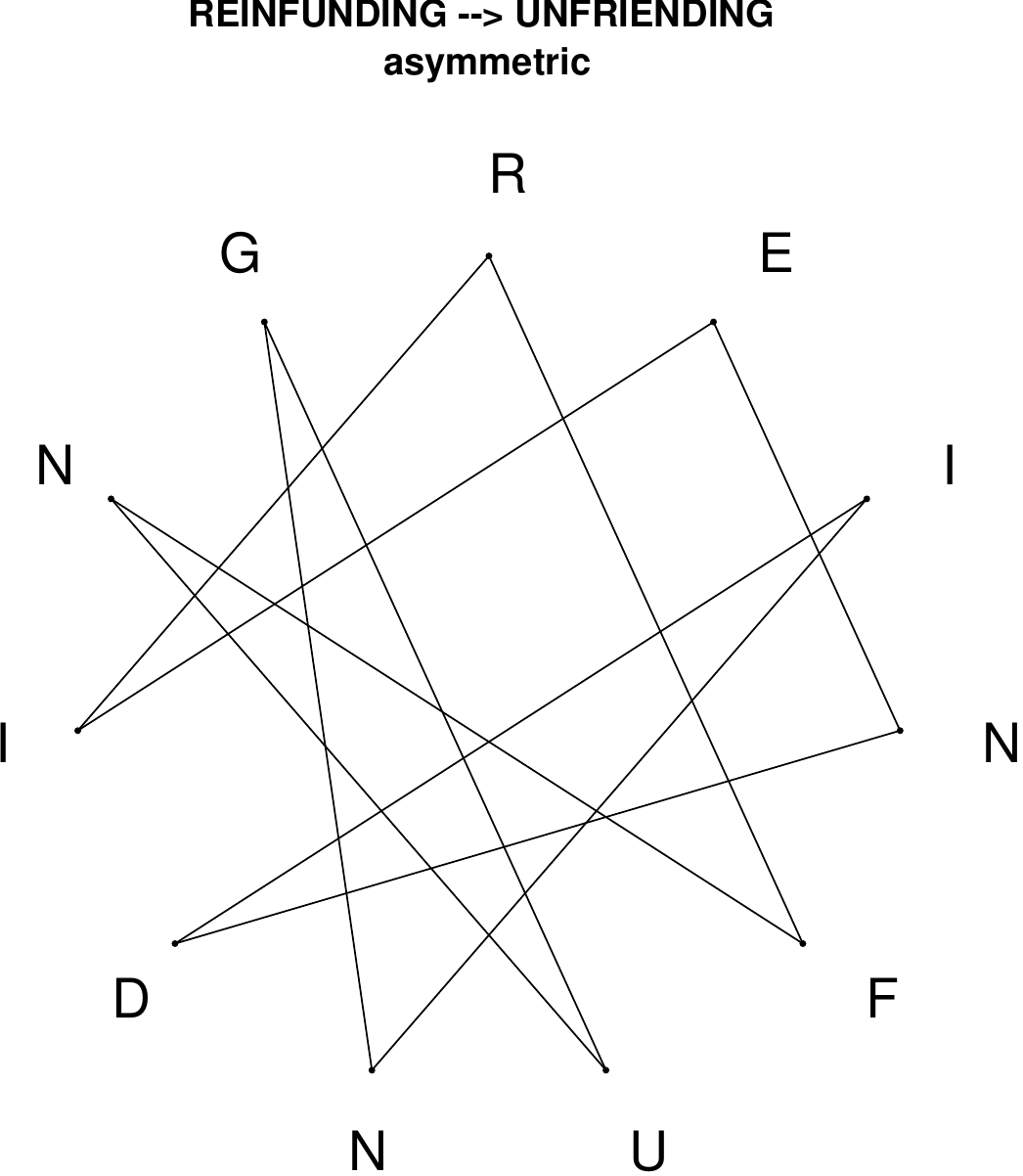}
\end{subfigure}
\end{figure}

\begin{figure}[H]
\centering
\begin{subfigure}[T]{0.19\textwidth}
\centering
\includegraphics[width=\textwidth]{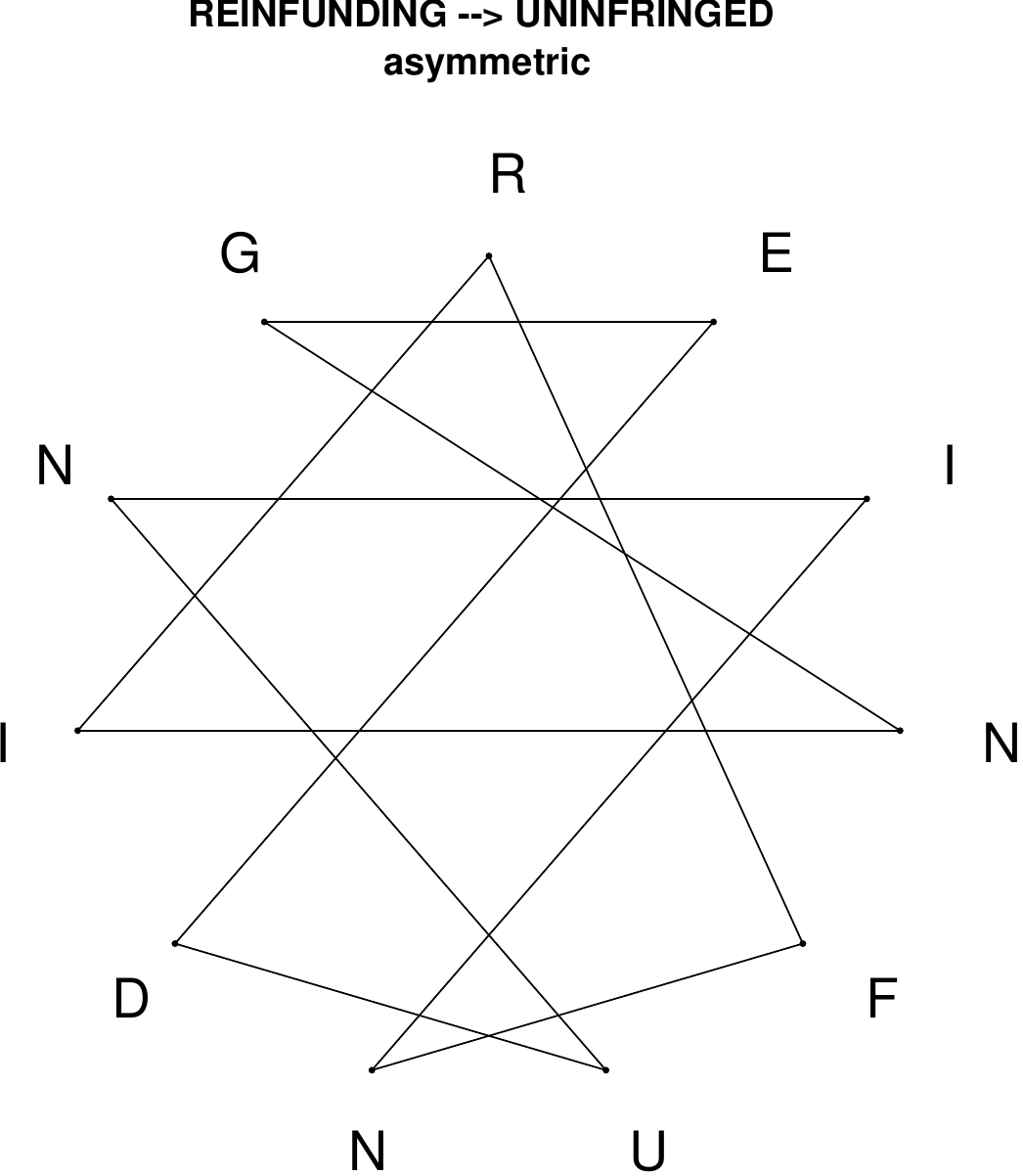}
\end{subfigure}
\hfill
\begin{subfigure}[T]{0.19\textwidth}
\centering
\includegraphics[width=\textwidth]{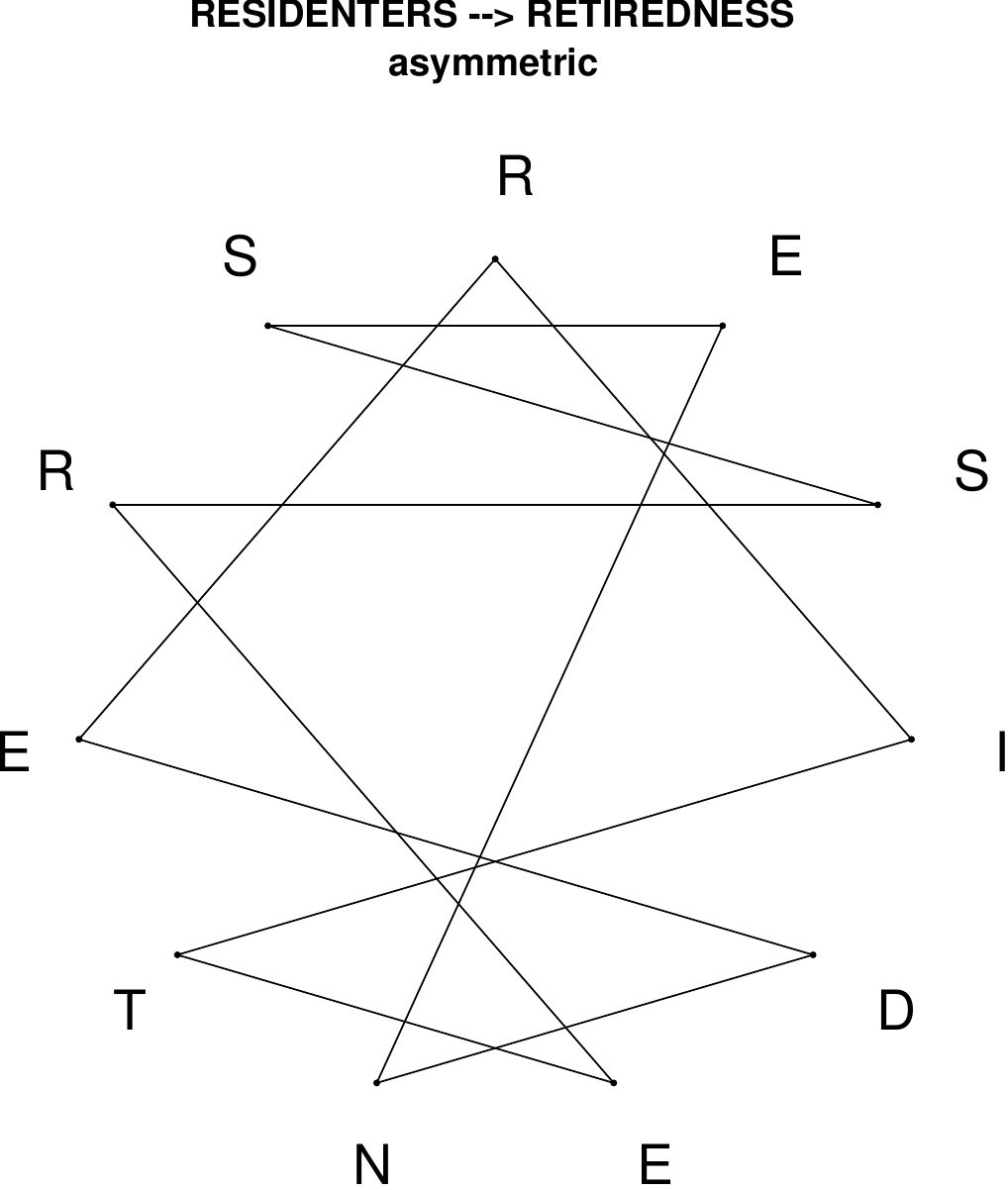}
\end{subfigure}
\hfill
\begin{subfigure}[T]{0.19\textwidth}
\centering
\includegraphics[width=\textwidth]{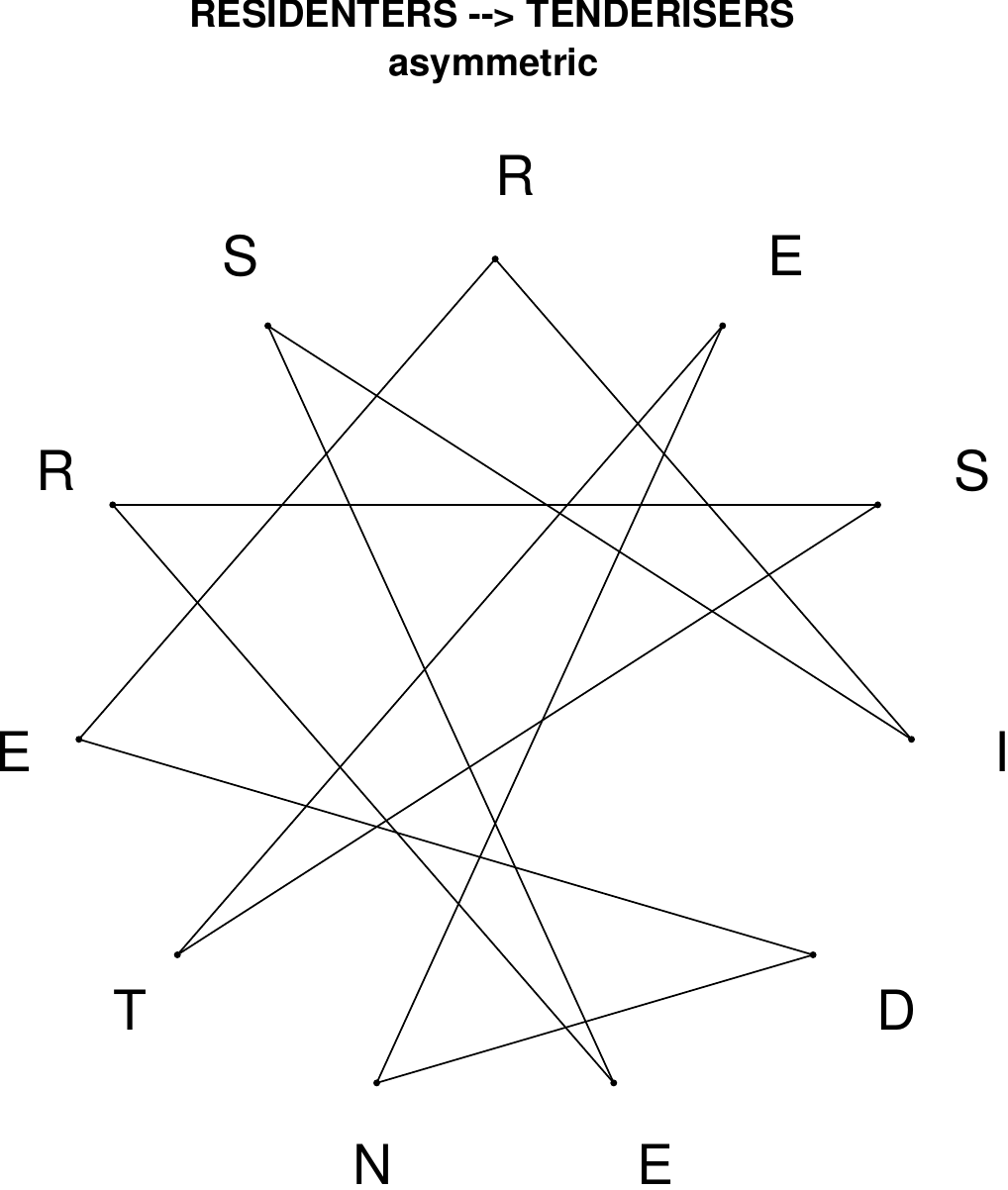}
\end{subfigure}
\hfill
\begin{subfigure}[T]{0.19\textwidth}
\centering
\includegraphics[width=\textwidth]{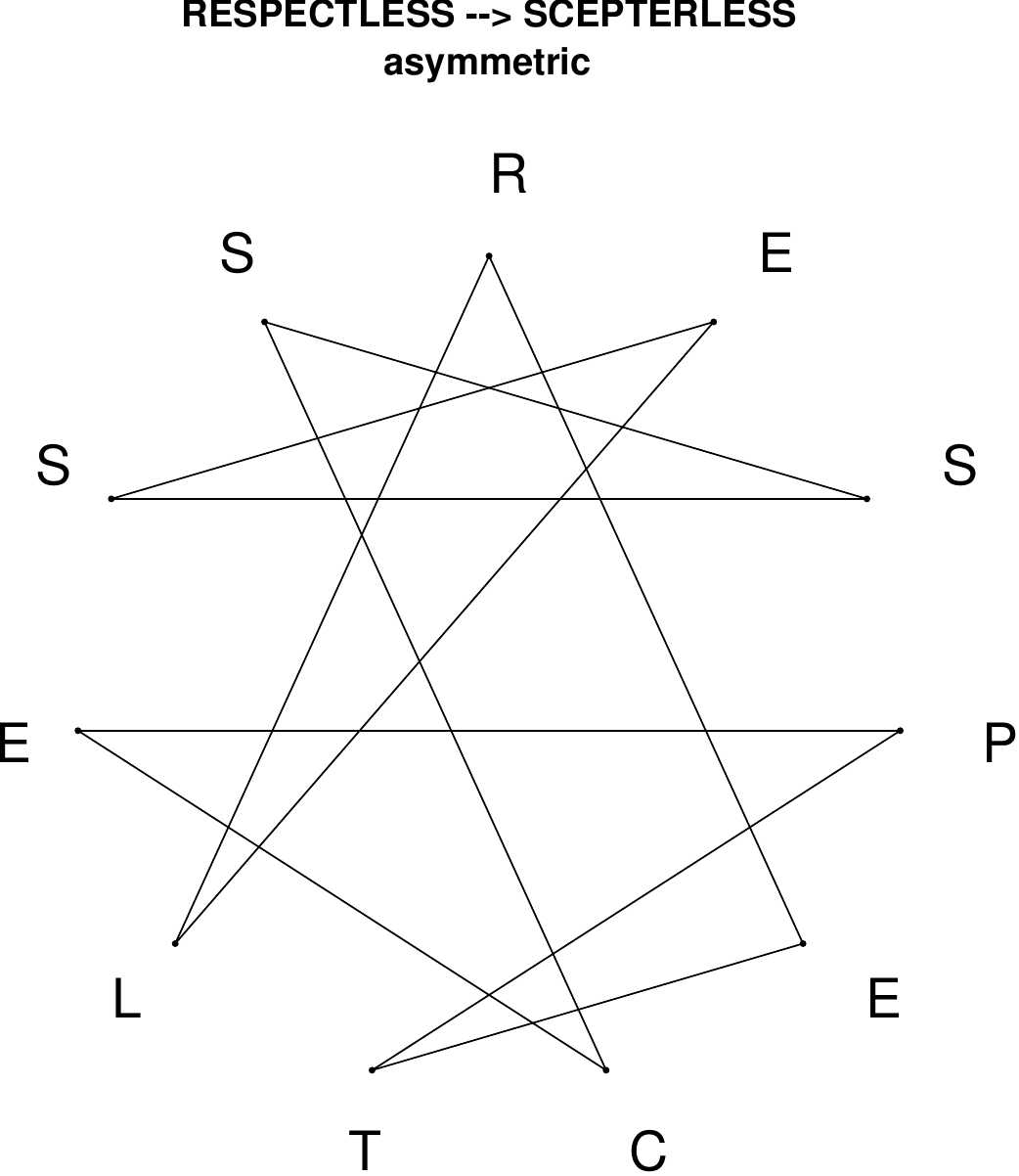}
\end{subfigure}
\hfill
\begin{subfigure}[T]{0.19\textwidth}
\centering
\includegraphics[width=\textwidth]{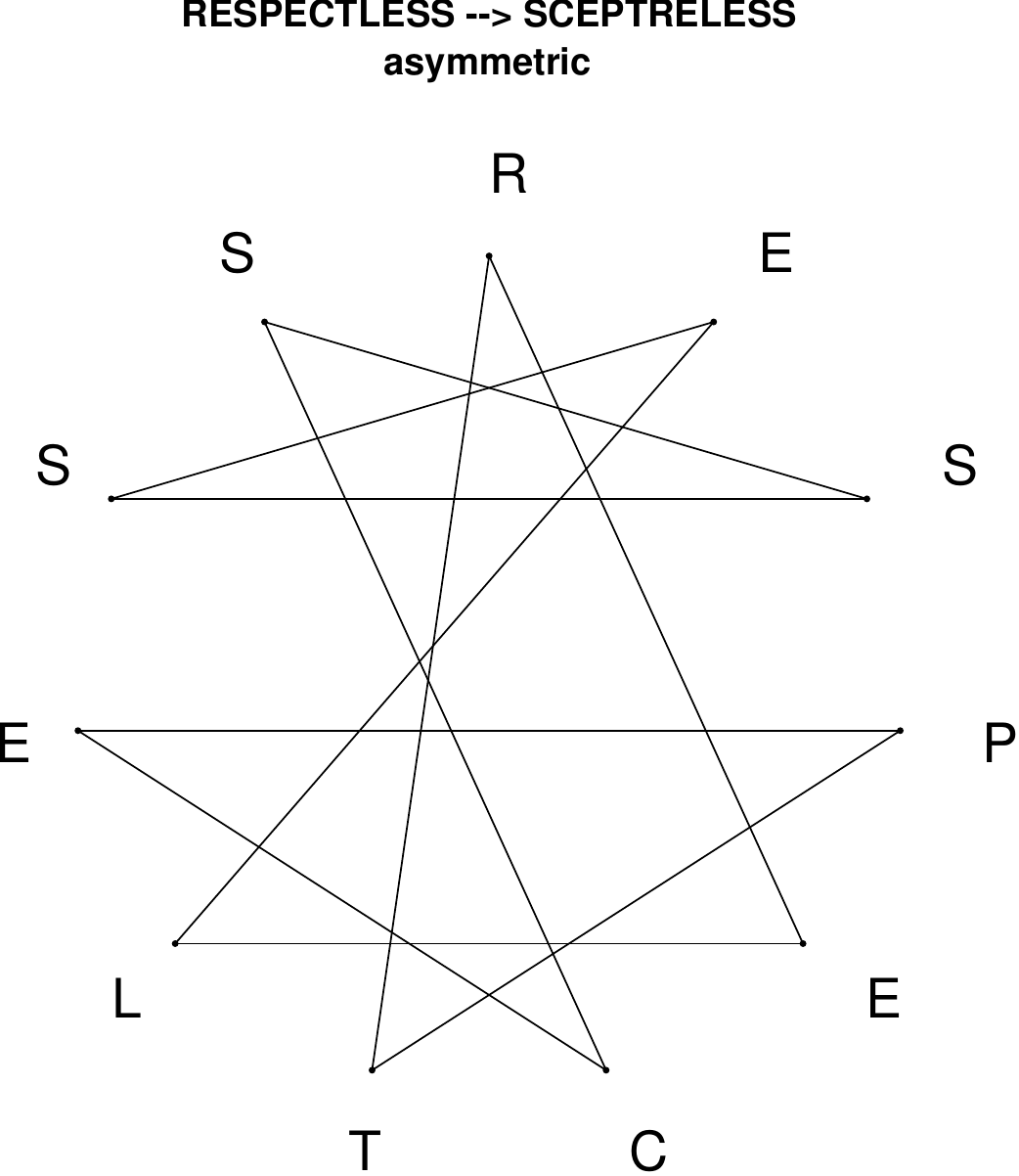}
\end{subfigure}
\end{figure}

\begin{figure}[H]
\centering
\begin{subfigure}[T]{0.19\textwidth}
\centering
\includegraphics[width=\textwidth]{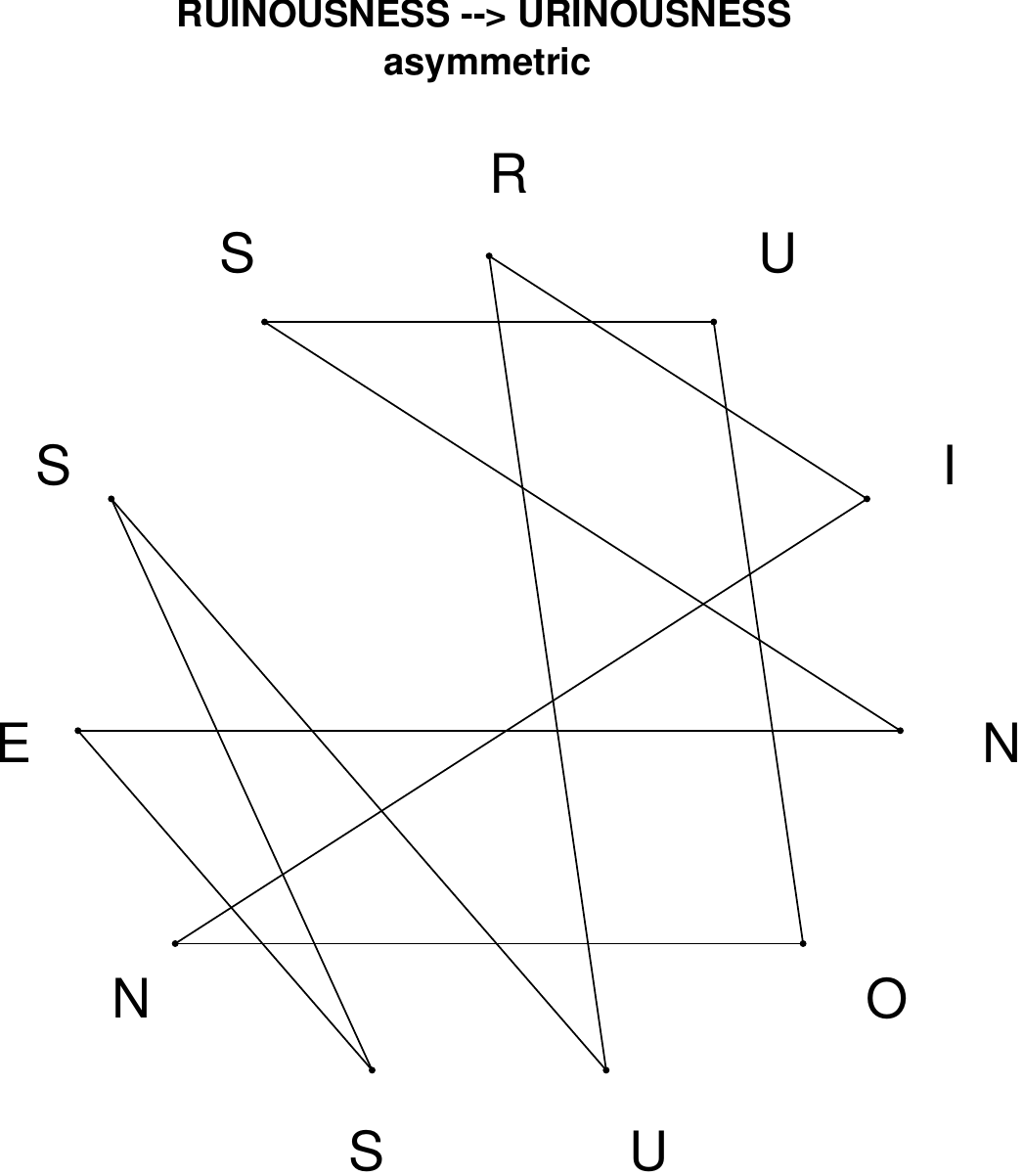}
\end{subfigure}
\hfill
\begin{subfigure}[T]{0.19\textwidth}
\centering
\includegraphics[width=\textwidth]{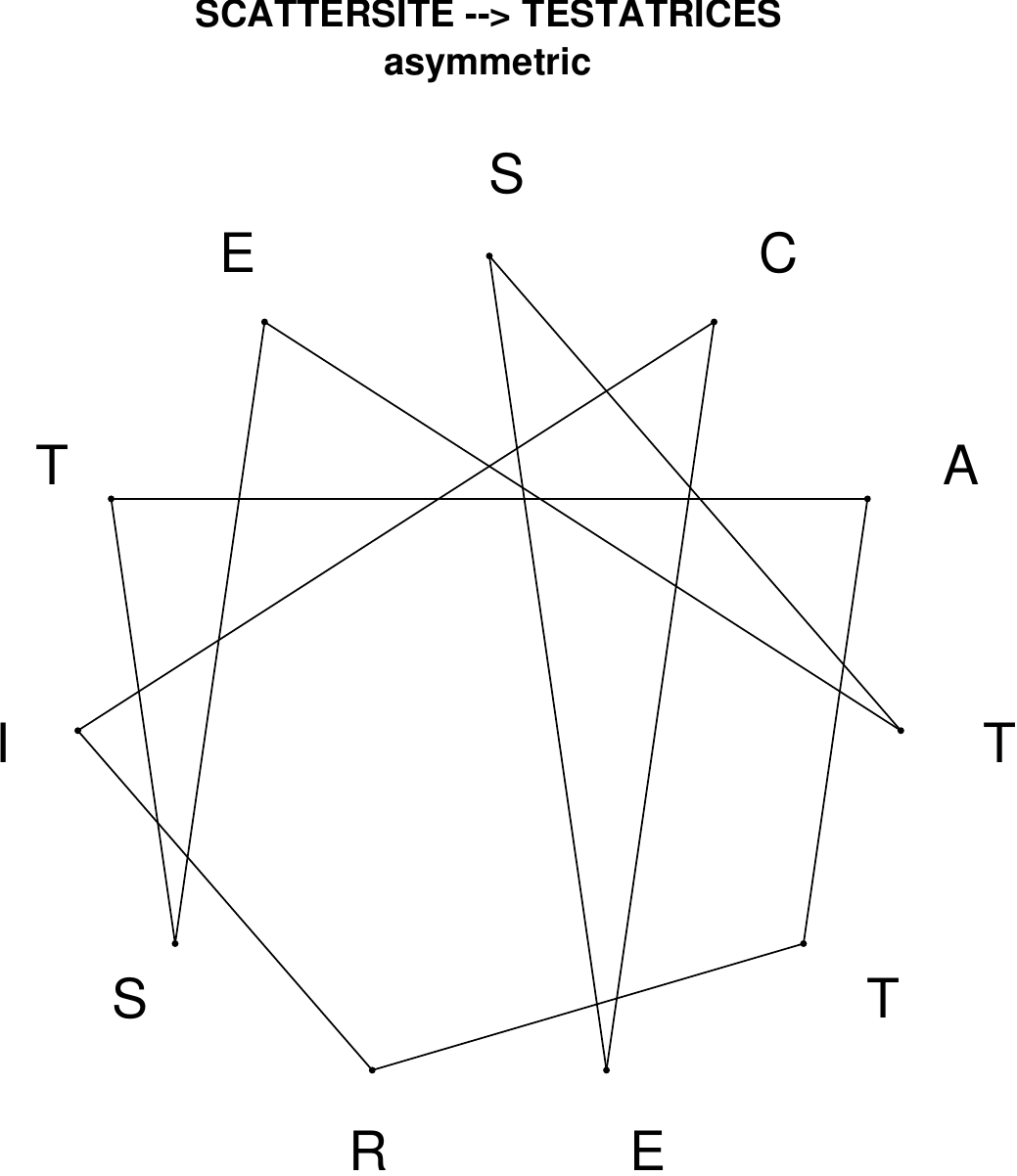}
\end{subfigure}
\hfill
\begin{subfigure}[T]{0.19\textwidth}
\centering
\includegraphics[width=\textwidth]{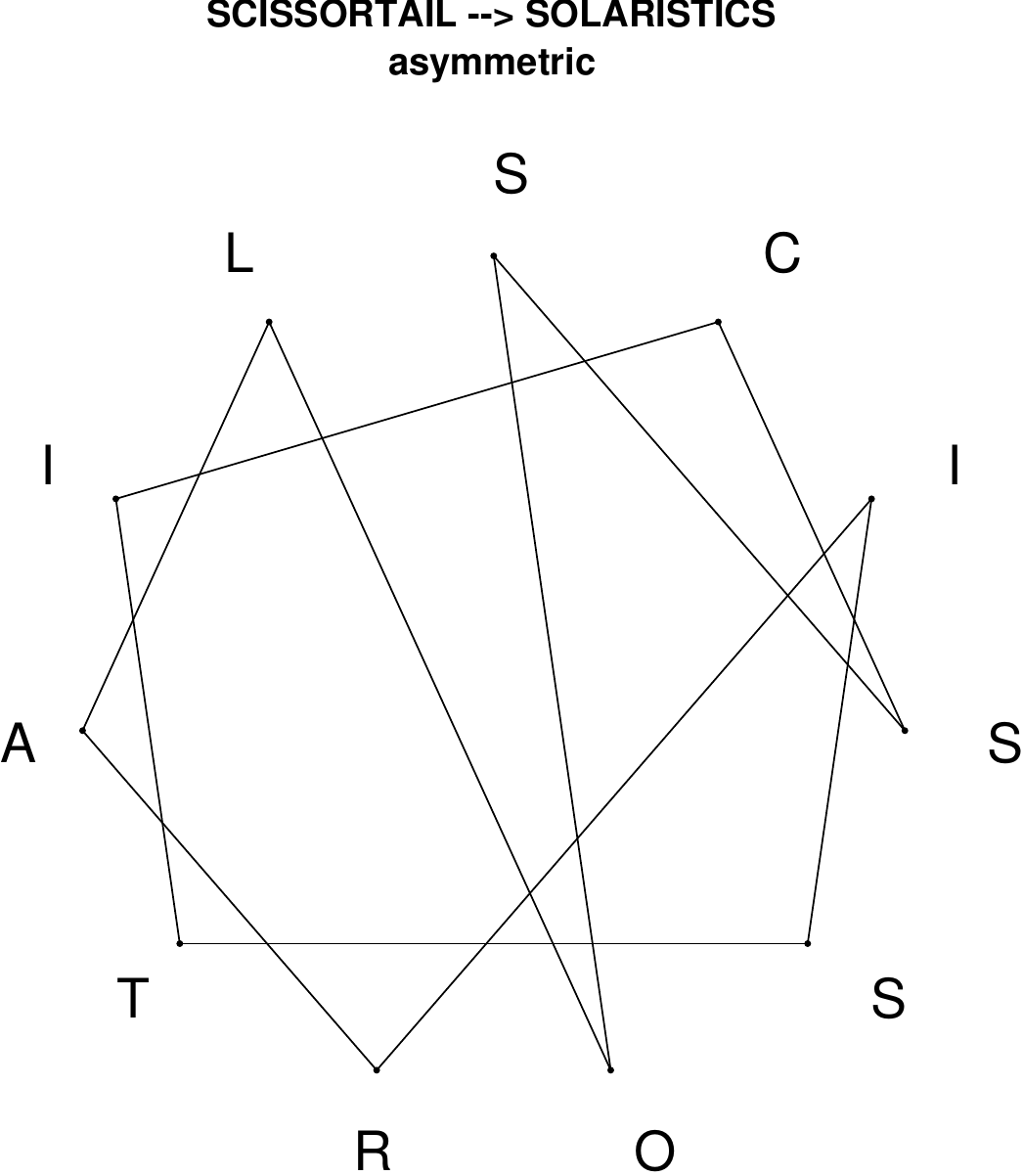}
\end{subfigure}
\hfill
\begin{subfigure}[T]{0.19\textwidth}
\centering
\includegraphics[width=\textwidth]{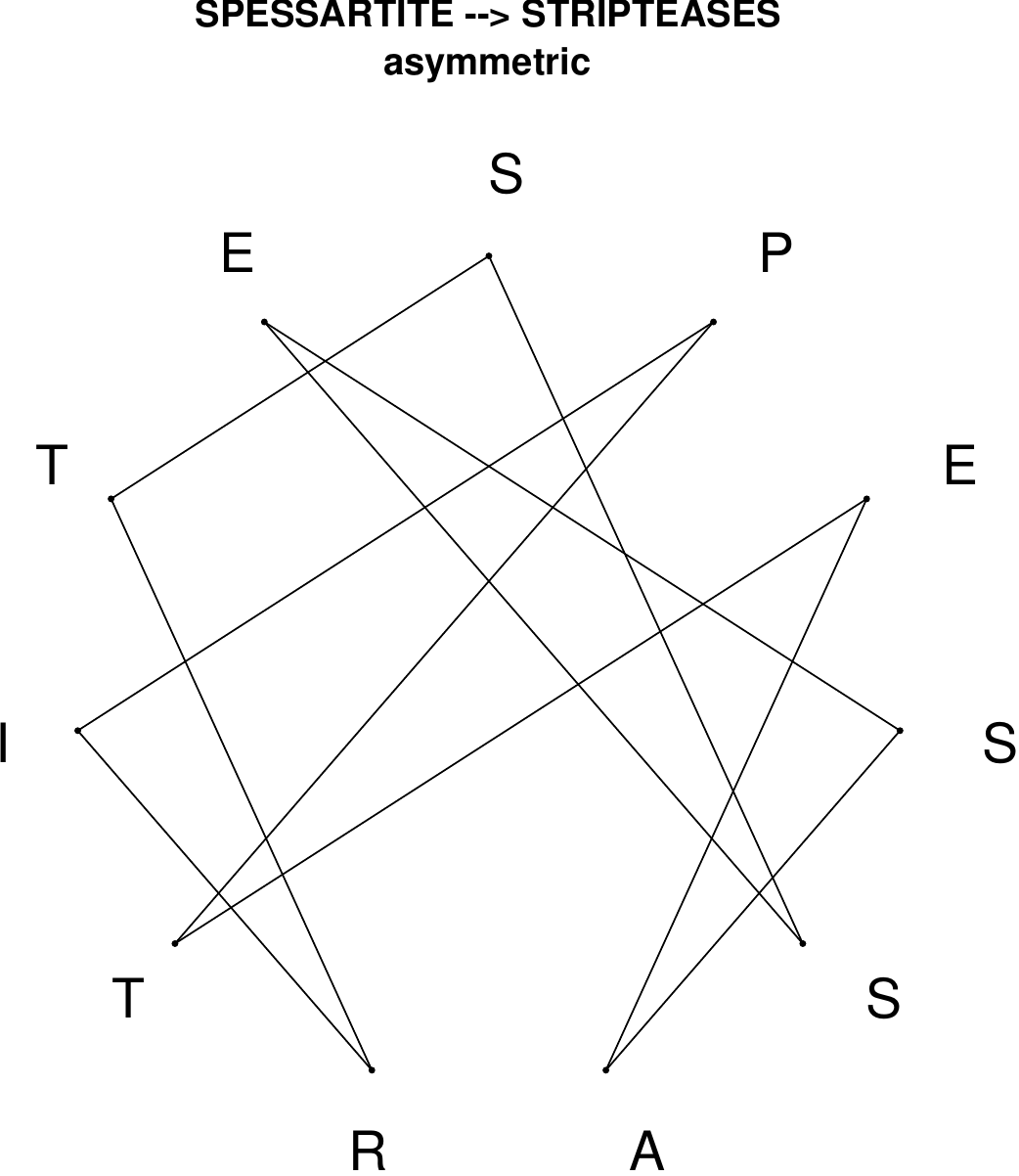}
\end{subfigure}
\hfill
\begin{subfigure}[T]{0.19\textwidth}
\centering
\includegraphics[width=\textwidth]{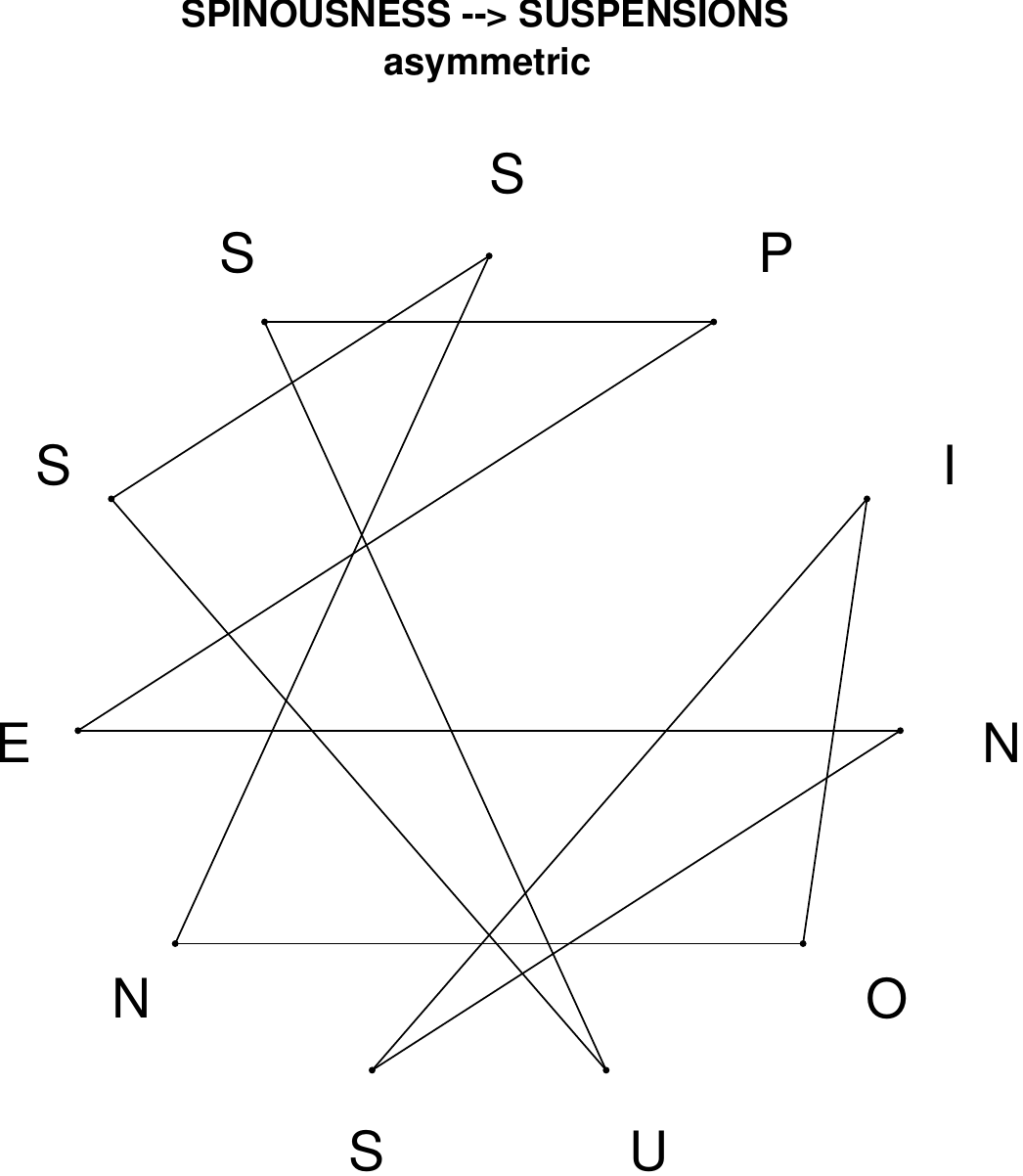}
\end{subfigure}
\end{figure}

\begin{figure}[H]
\centering
\begin{subfigure}[T]{0.19\textwidth}
\centering
\includegraphics[width=\textwidth]{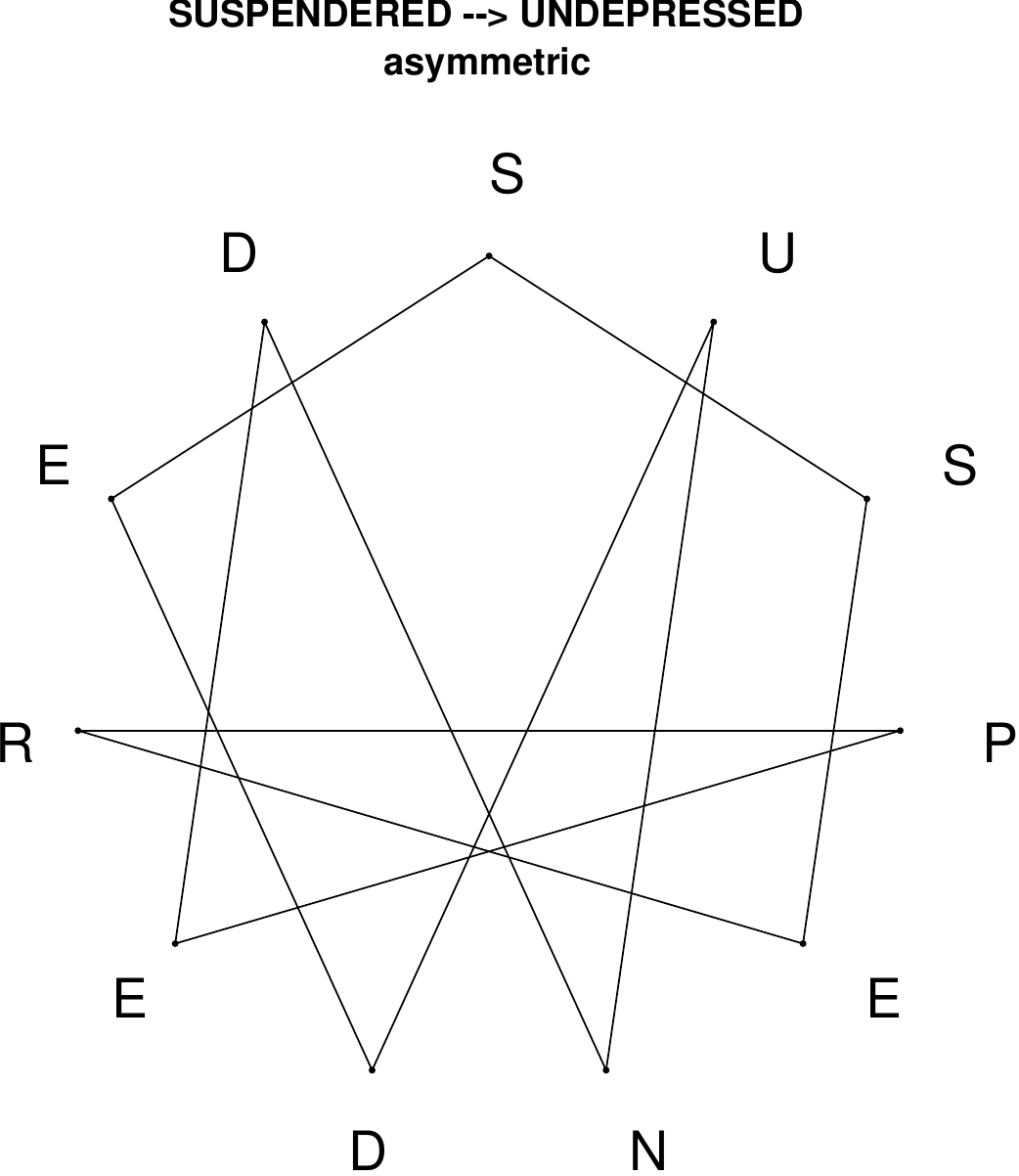}
\end{subfigure}
\hfill
\begin{subfigure}[T]{0.19\textwidth}
\centering
\includegraphics[width=\textwidth]{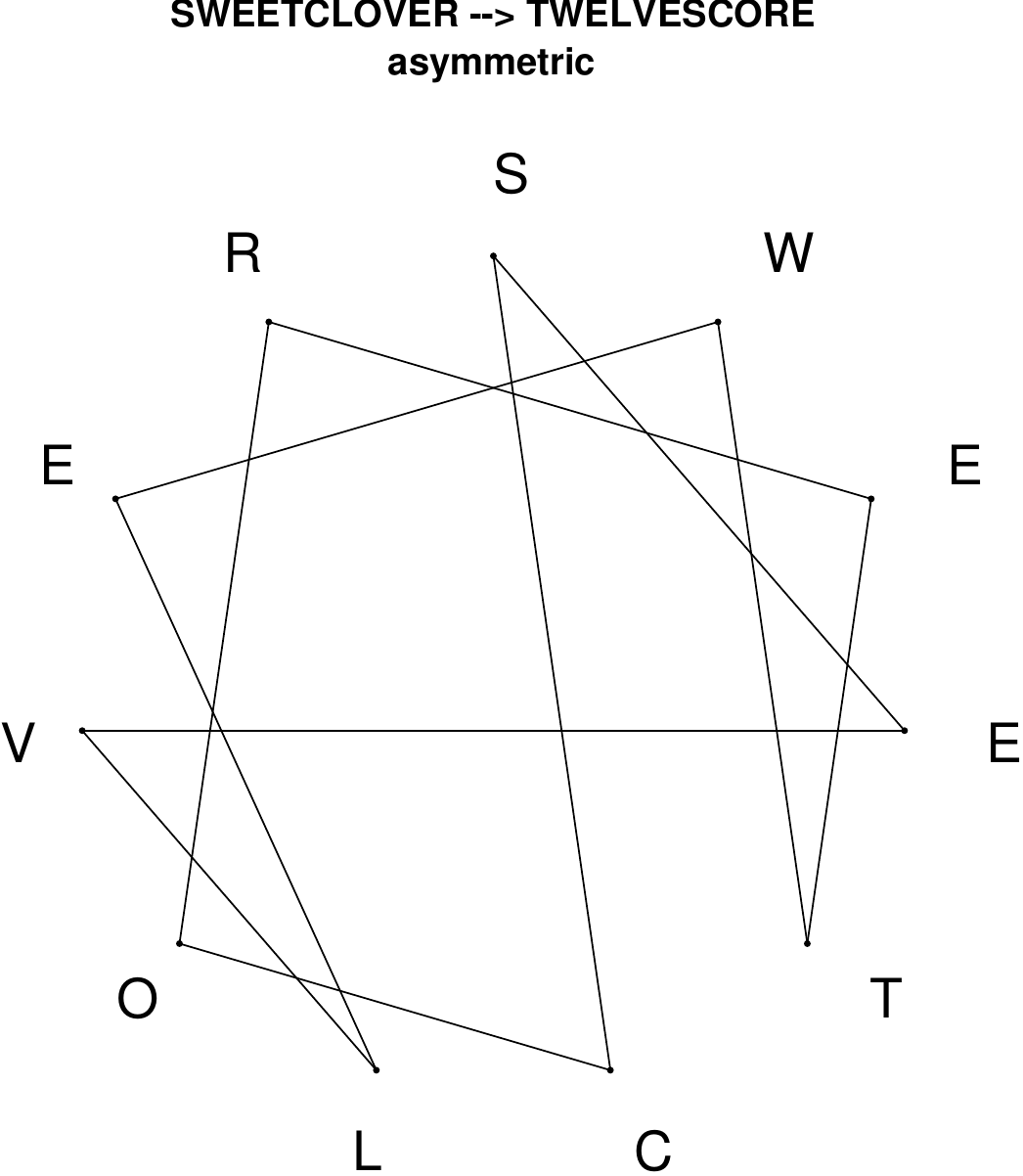}
\end{subfigure}
\hfill
\begin{subfigure}[T]{0.19\textwidth}
\centering
\includegraphics[width=\textwidth]{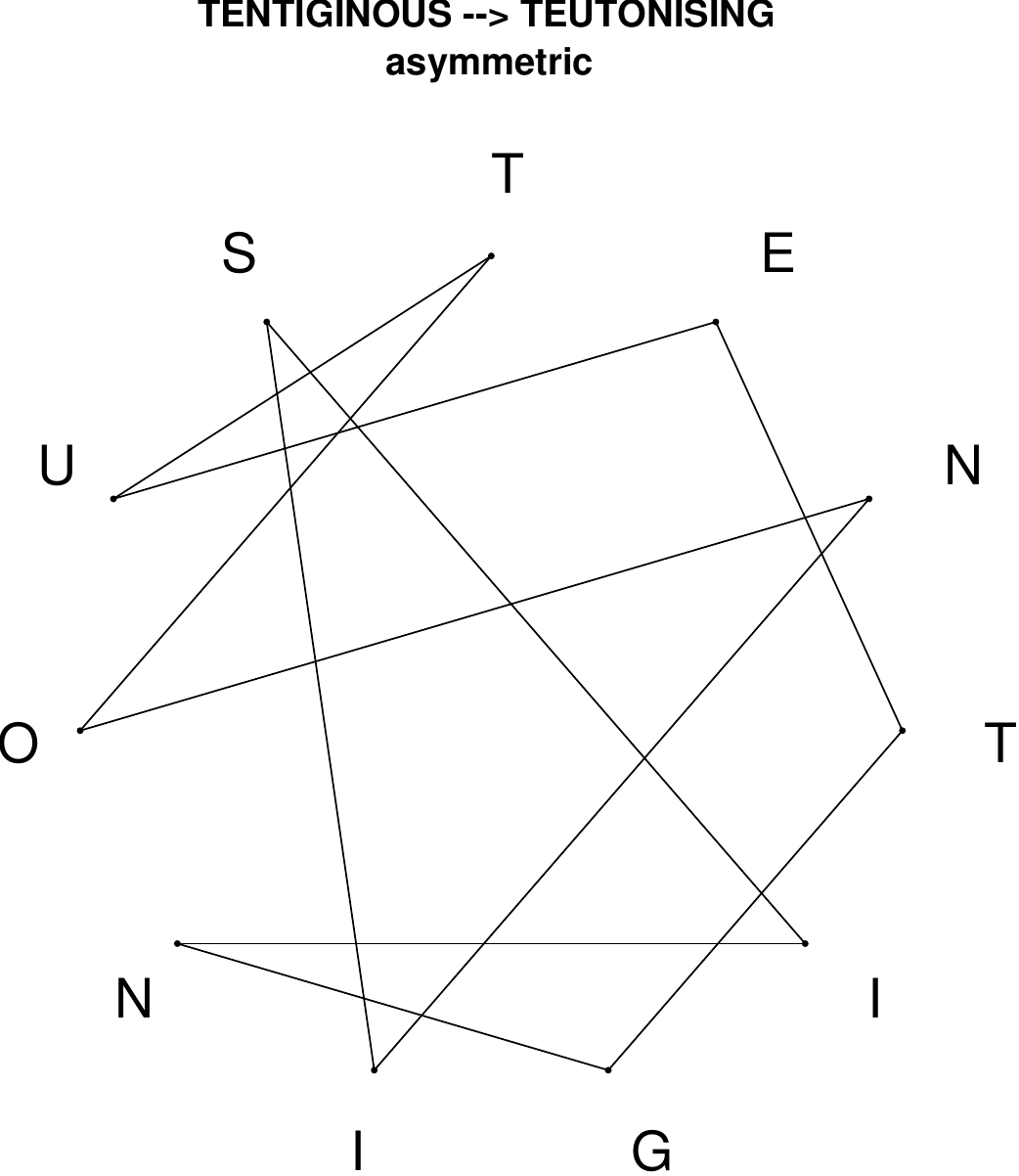}
\end{subfigure}
\hfill
\begin{subfigure}[T]{0.19\textwidth}
\centering
\includegraphics[width=\textwidth]{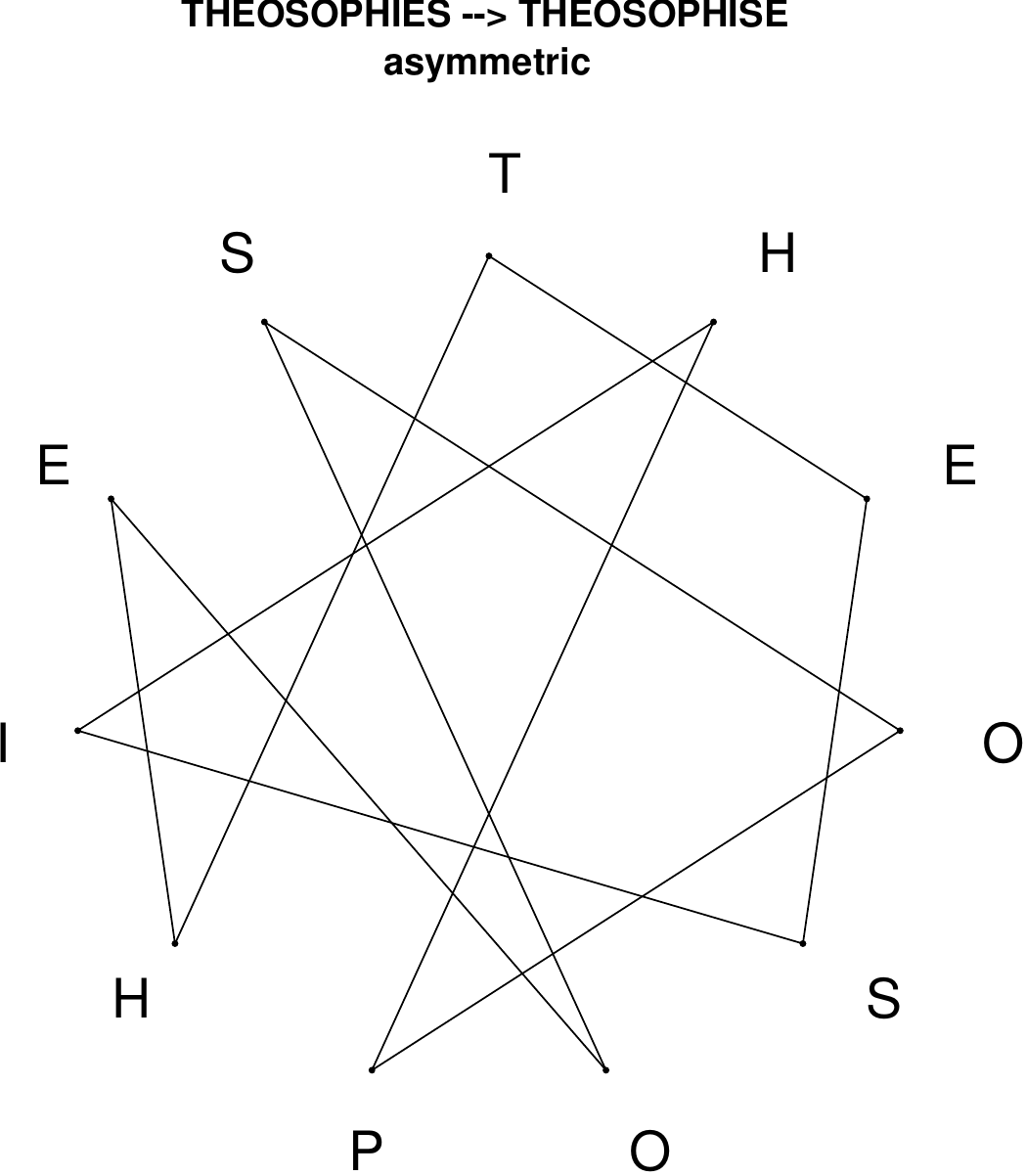}
\end{subfigure}
\hfill
\begin{subfigure}[T]{0.19\textwidth}
\centering
\includegraphics[width=\textwidth]{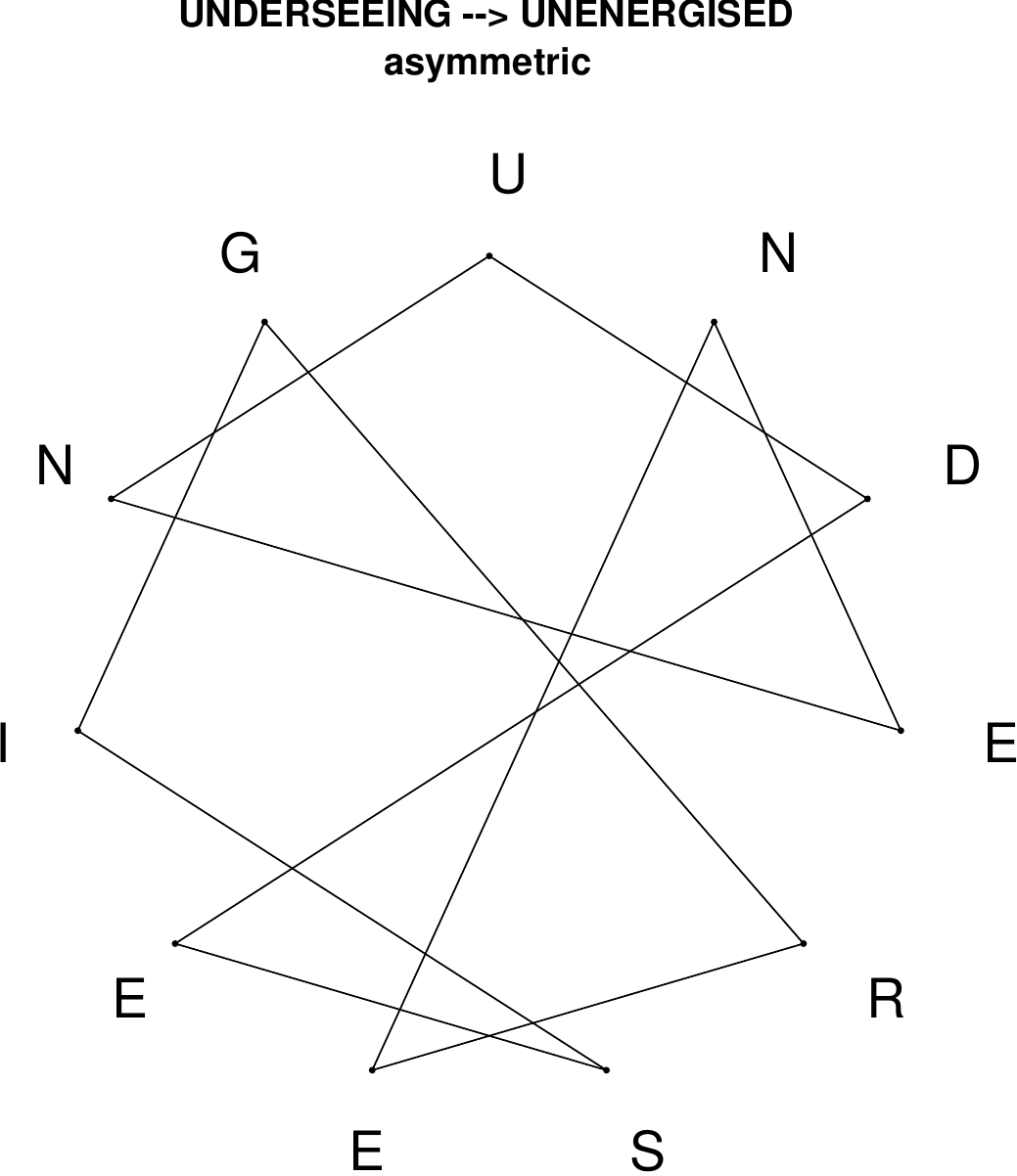}
\end{subfigure}
\end{figure}

\begin{figure}[H]
\centering
\begin{subfigure}[T]{0.19\textwidth}
\centering
\includegraphics[width=\textwidth]{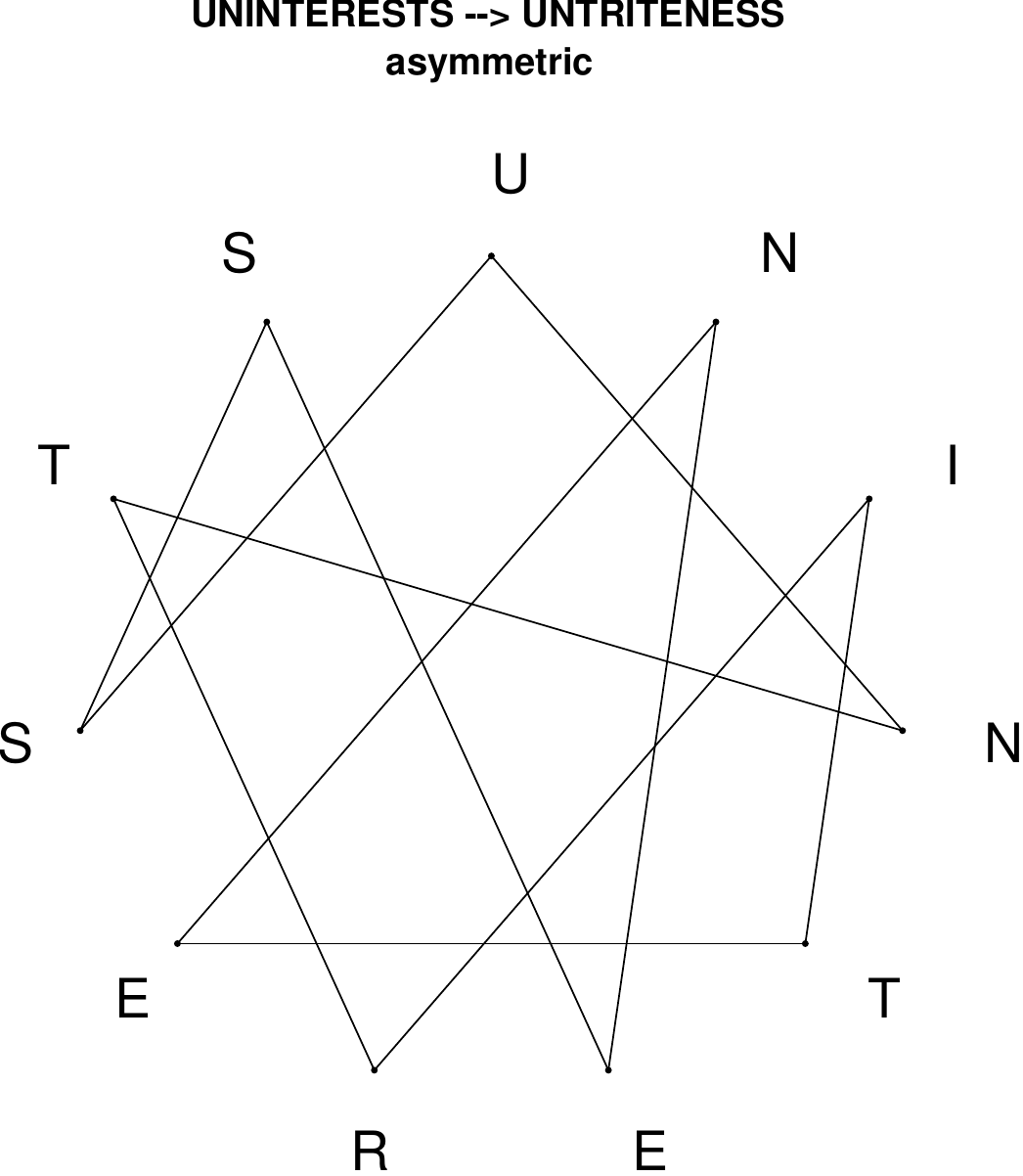}
\end{subfigure}
\hfill
\end{figure}

%%%%%%%%%%%%%%%%%%
\clearpage
\subsection{Star Anagrams $N = 10$}
For $N=10$, we found our first and longest perfect star along with a small group of symmetric stars and a large group of asymmetric stars. As mentioned above, $N=10$ is the first word length for which we found stars from all three classes. 

\subsubsection{Perfect Stars $N=10$}

\begin{figure}[H]
\centering
\begin{subfigure}[T]{0.19\textwidth}
\centering
\includegraphics[width=\textwidth]{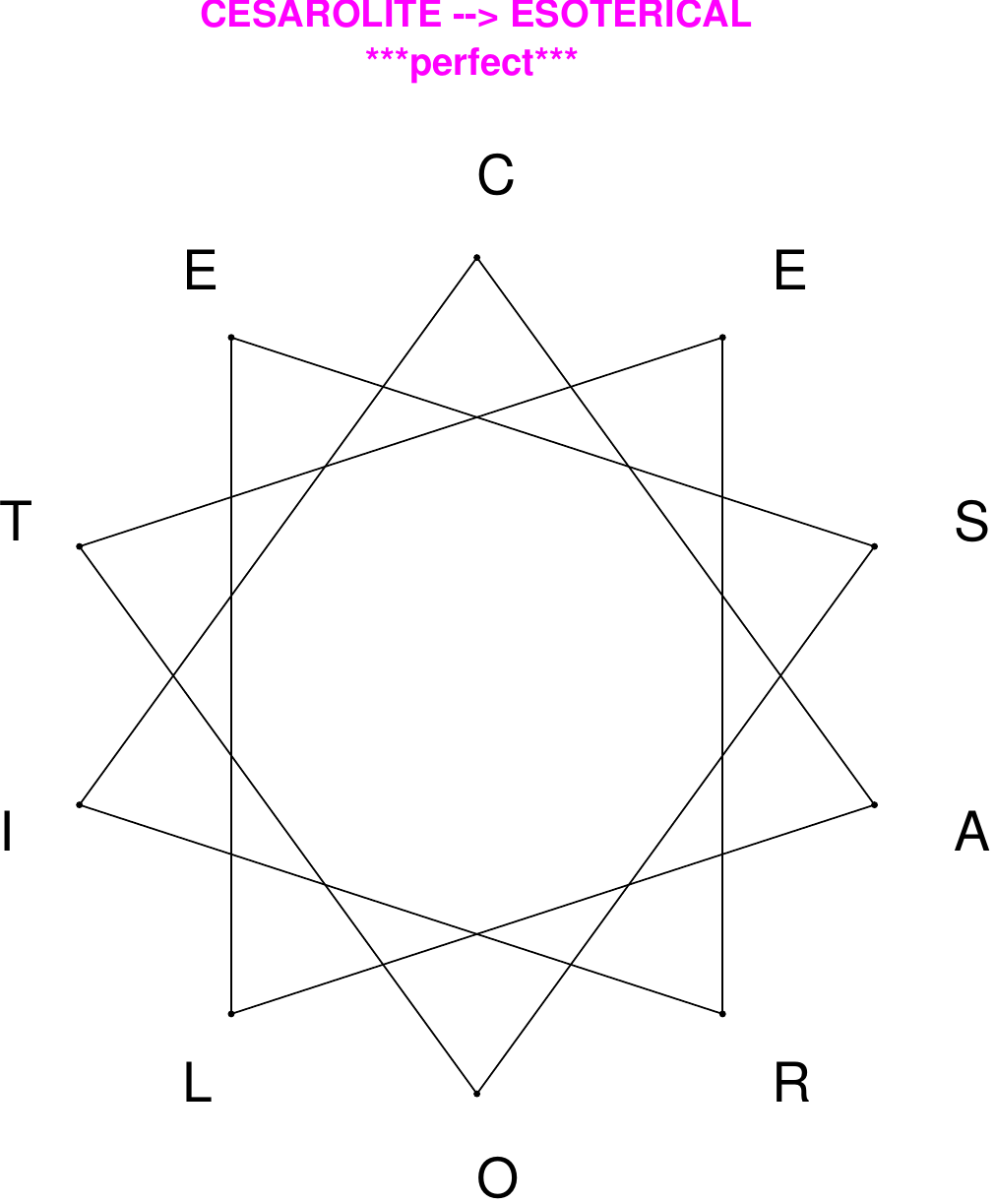}
\end{subfigure}
\hfill
\end{figure}

\subsubsection{Symmetric Stars $N=10$}

\begin{figure}[H]
\centering
\begin{subfigure}[T]{0.19\textwidth}
\centering
\includegraphics[width=\textwidth]{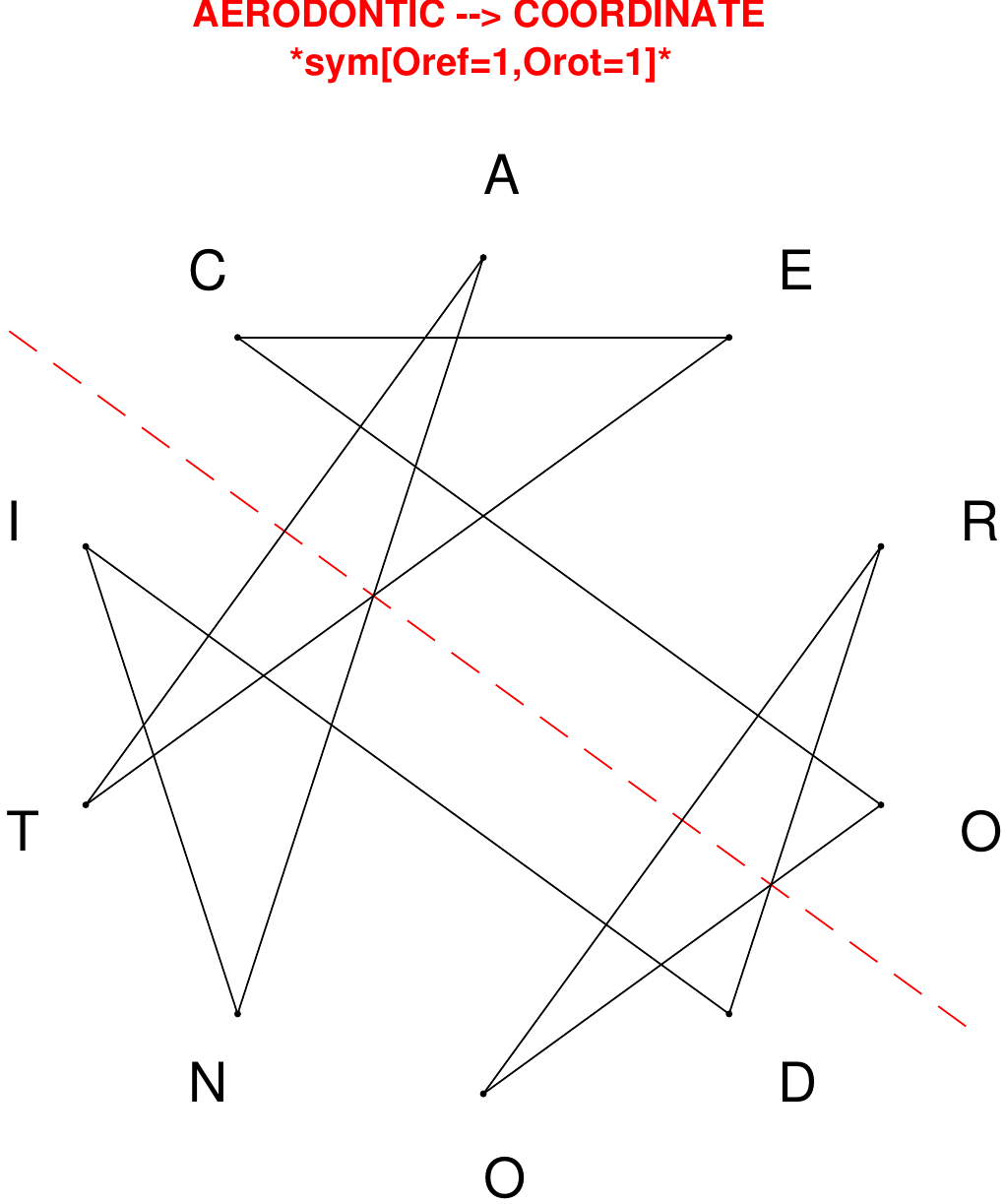}
\end{subfigure}
\hfill
\begin{subfigure}[T]{0.19\textwidth}
\centering
\includegraphics[width=\textwidth]{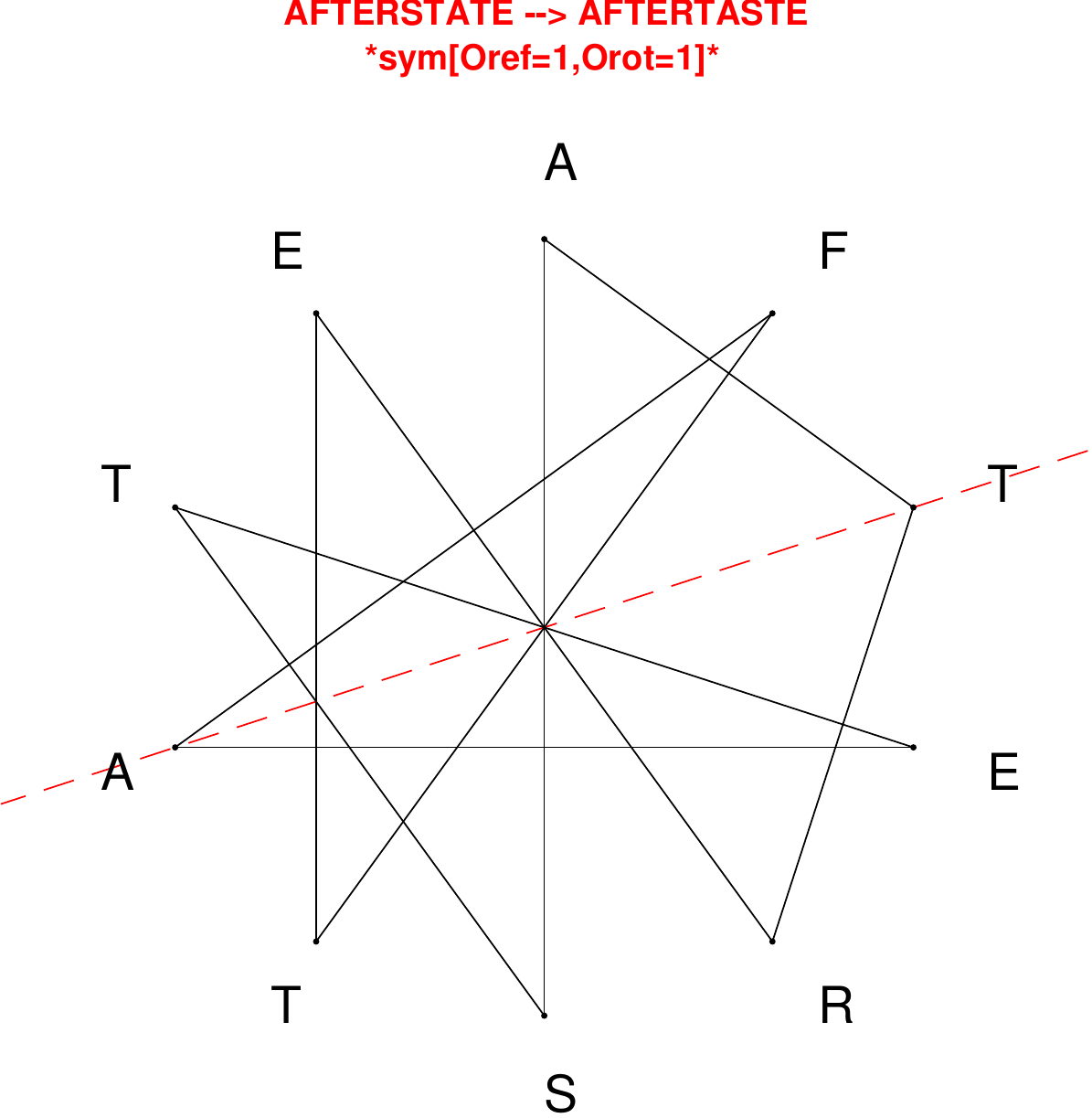}
\end{subfigure}
\hfill
\begin{subfigure}[T]{0.19\textwidth}
\centering
\includegraphics[width=\textwidth]{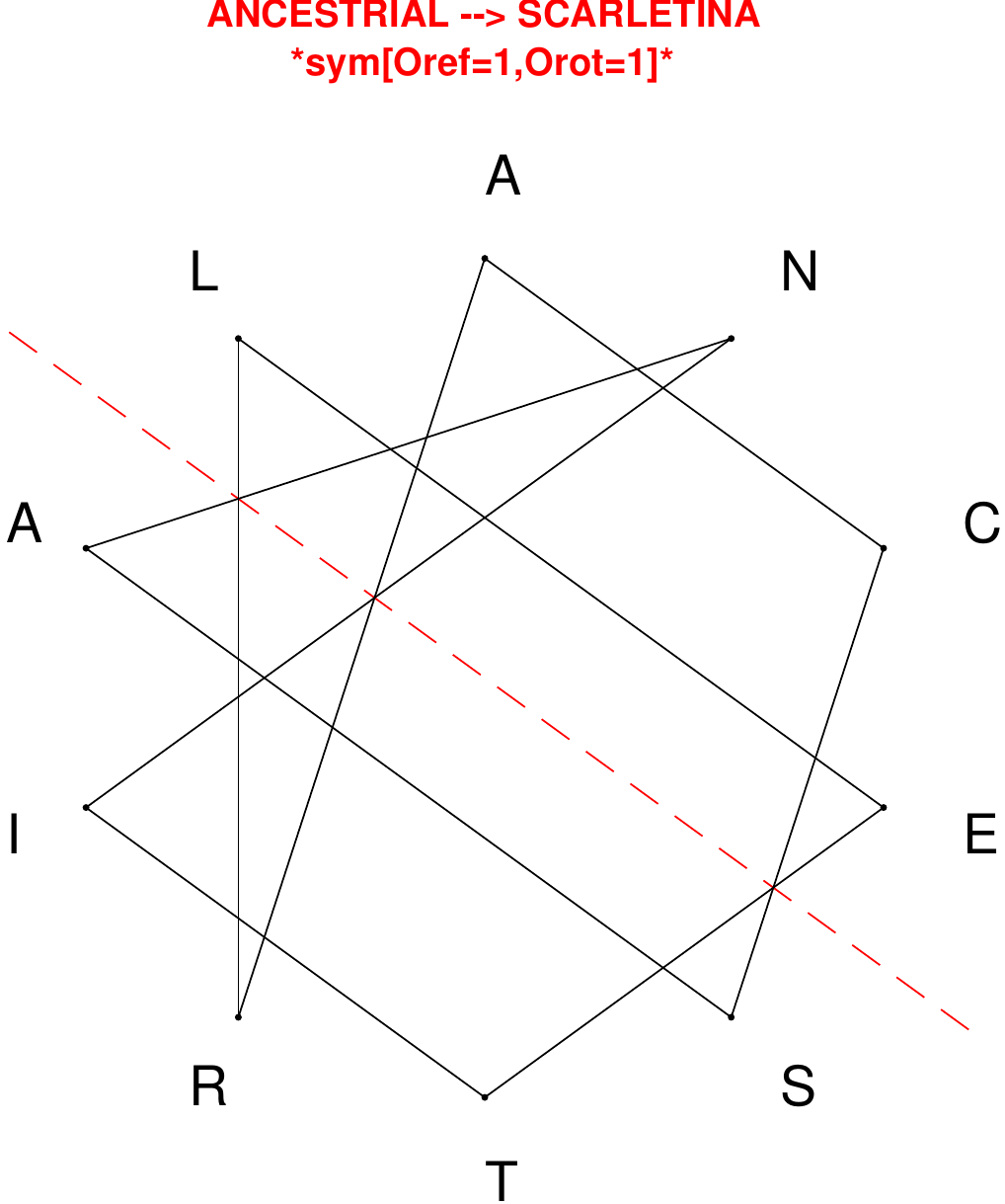}
\end{subfigure}
\hfill
\begin{subfigure}[T]{0.19\textwidth}
\centering
\includegraphics[width=\textwidth]{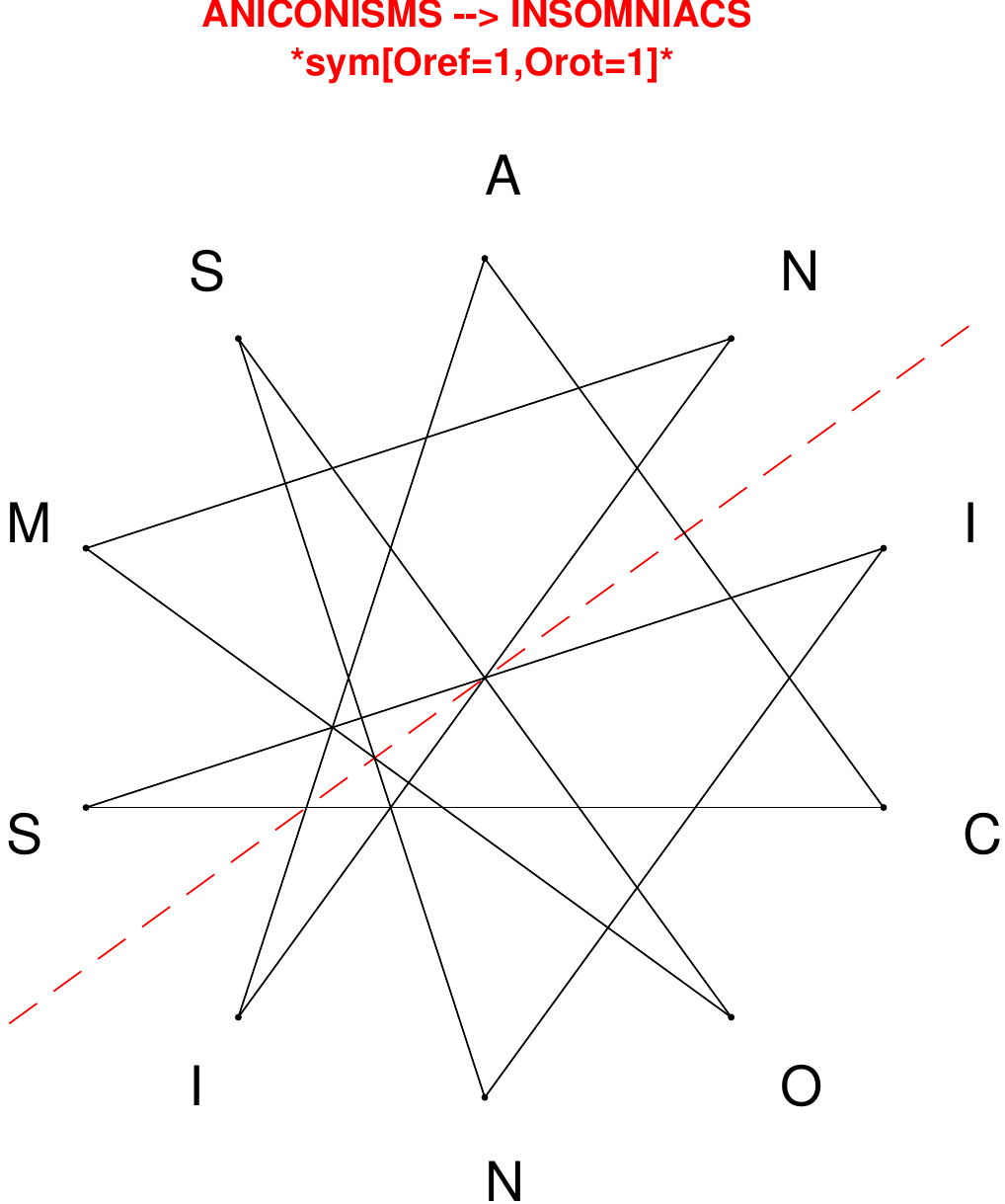}
\end{subfigure}
\hfill
\begin{subfigure}[T]{0.19\textwidth}
\centering
\includegraphics[width=\textwidth]{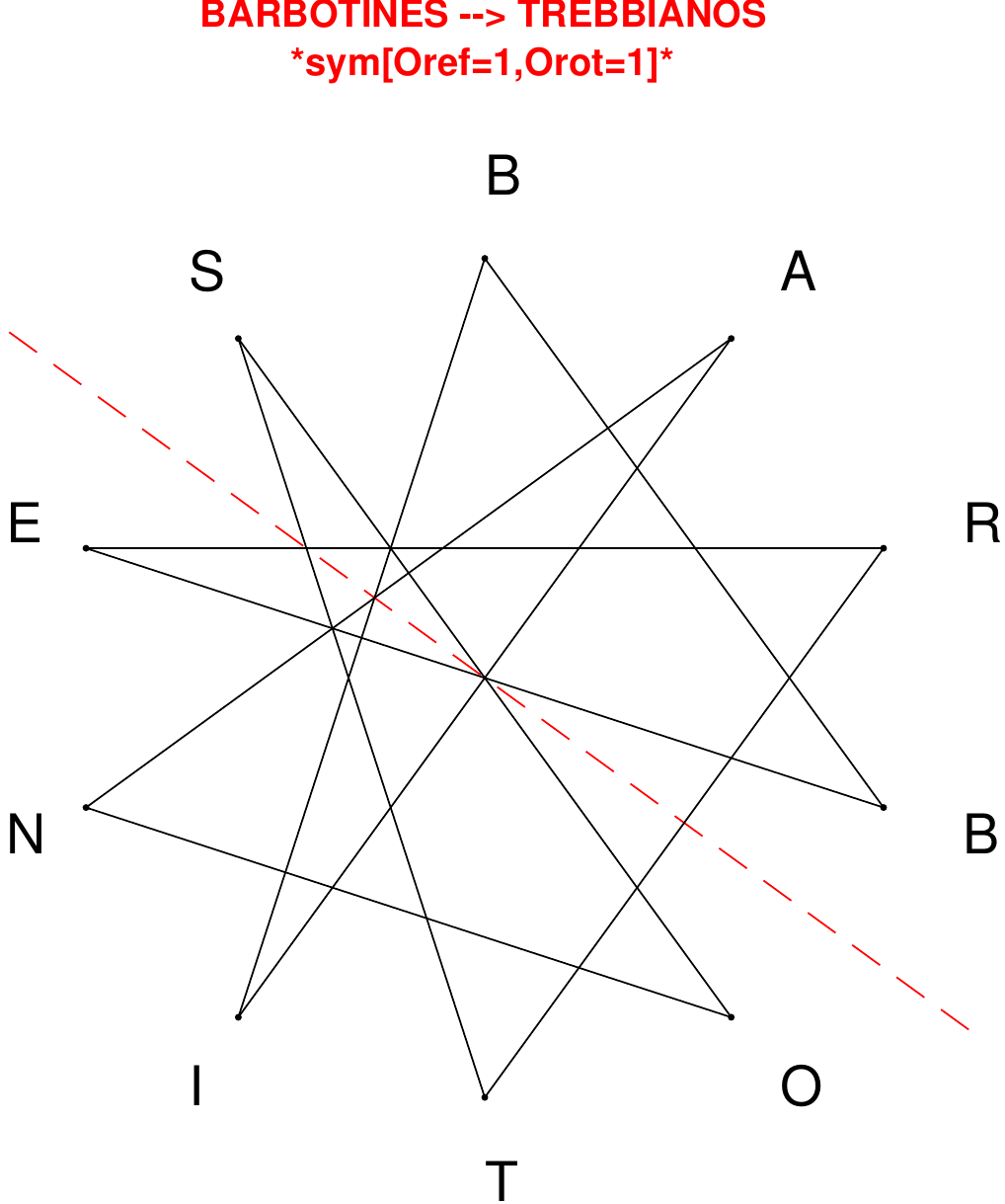}
\end{subfigure}
\end{figure}

\begin{figure}[H]
\centering
\begin{subfigure}[T]{0.19\textwidth}
\centering
\includegraphics[width=\textwidth]{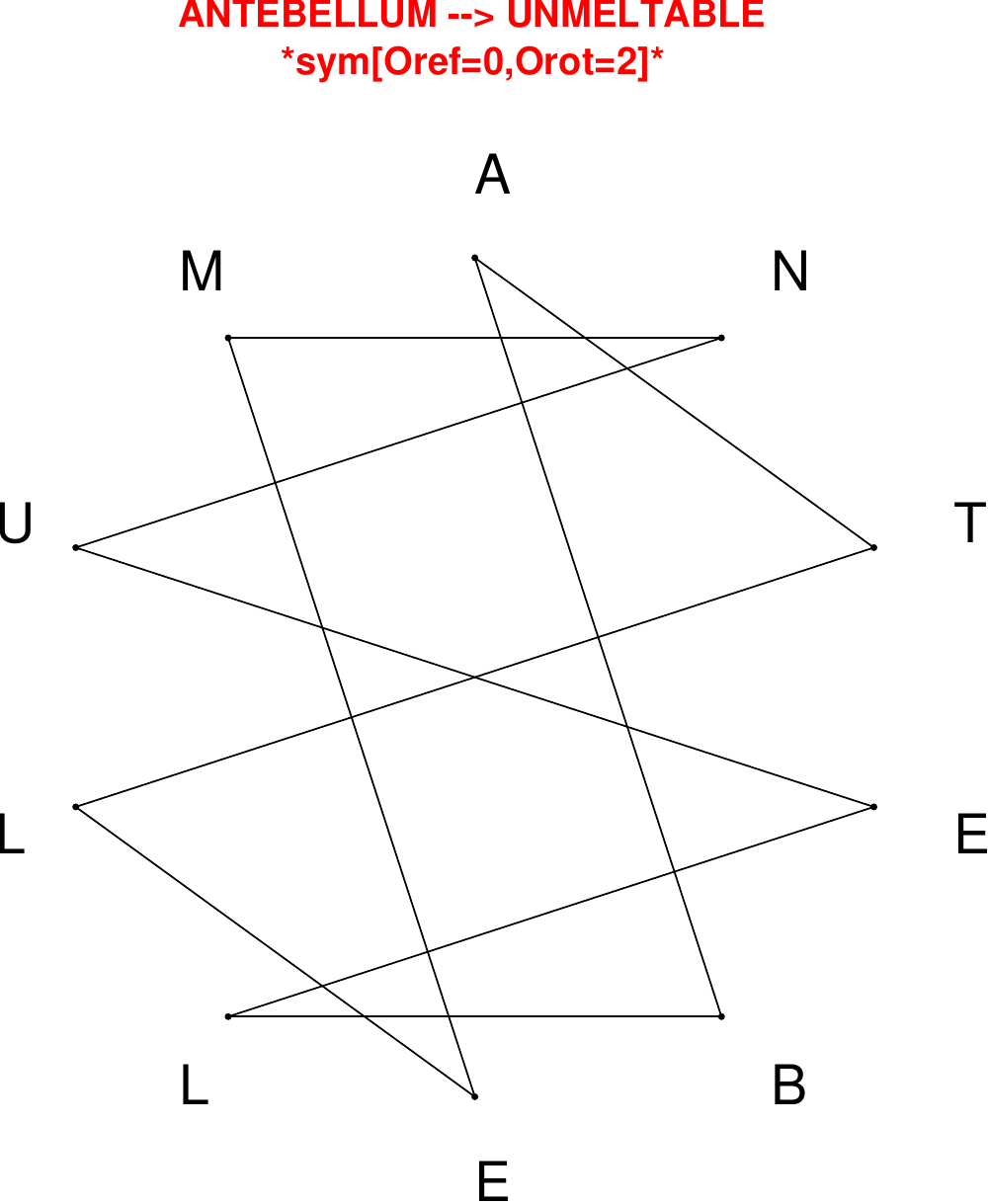}
\end{subfigure}
\hfill
\begin{subfigure}[T]{0.19\textwidth}
\centering
\includegraphics[width=\textwidth]{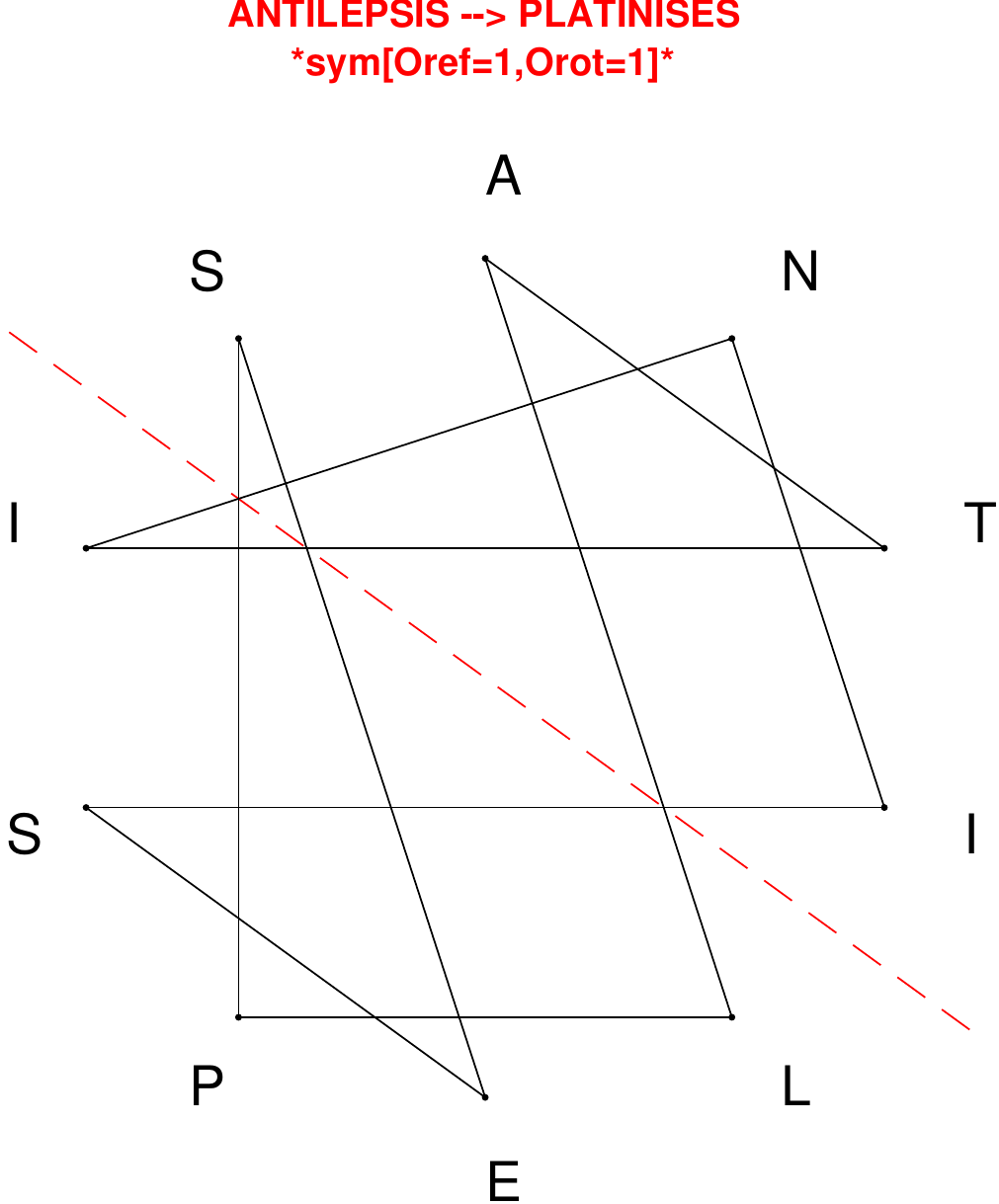}
\end{subfigure}
\hfill
\begin{subfigure}[T]{0.19\textwidth}
\centering
\includegraphics[width=\textwidth]{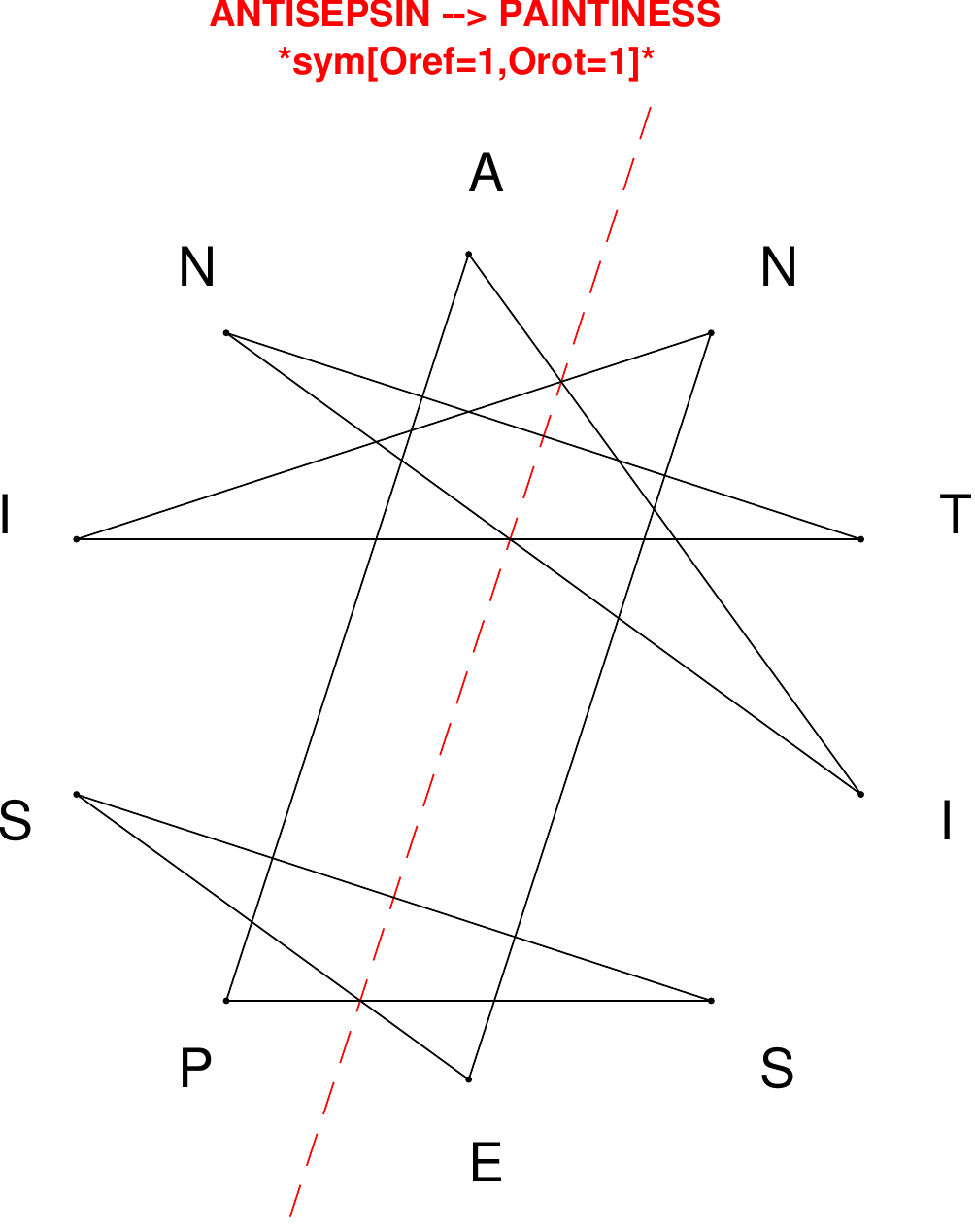}
\end{subfigure}
\hfill
\begin{subfigure}[T]{0.19\textwidth}
\centering
\includegraphics[width=\textwidth]{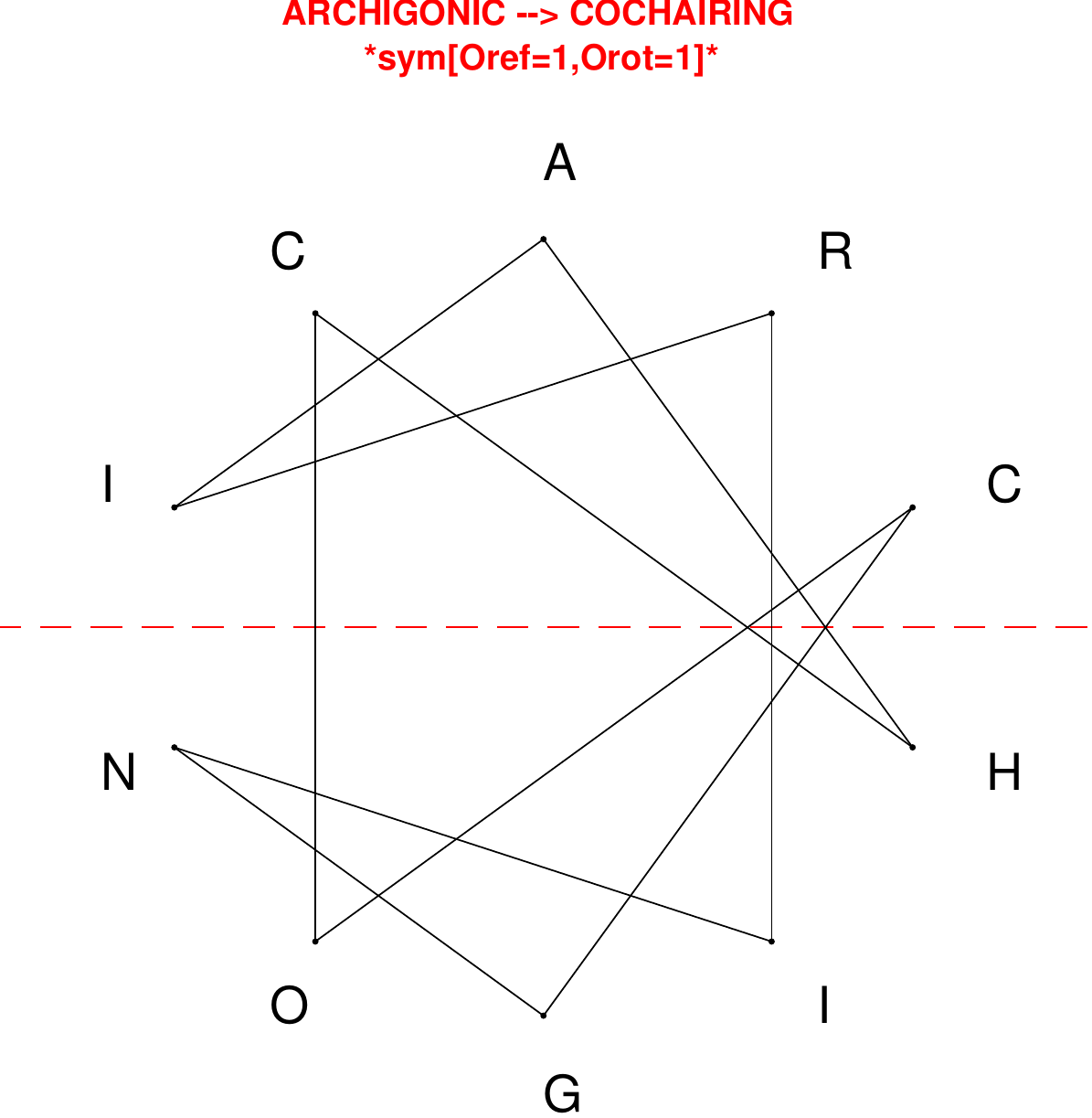}
\end{subfigure}
\hfill
\begin{subfigure}[T]{0.19\textwidth}
\centering
\includegraphics[width=\textwidth]{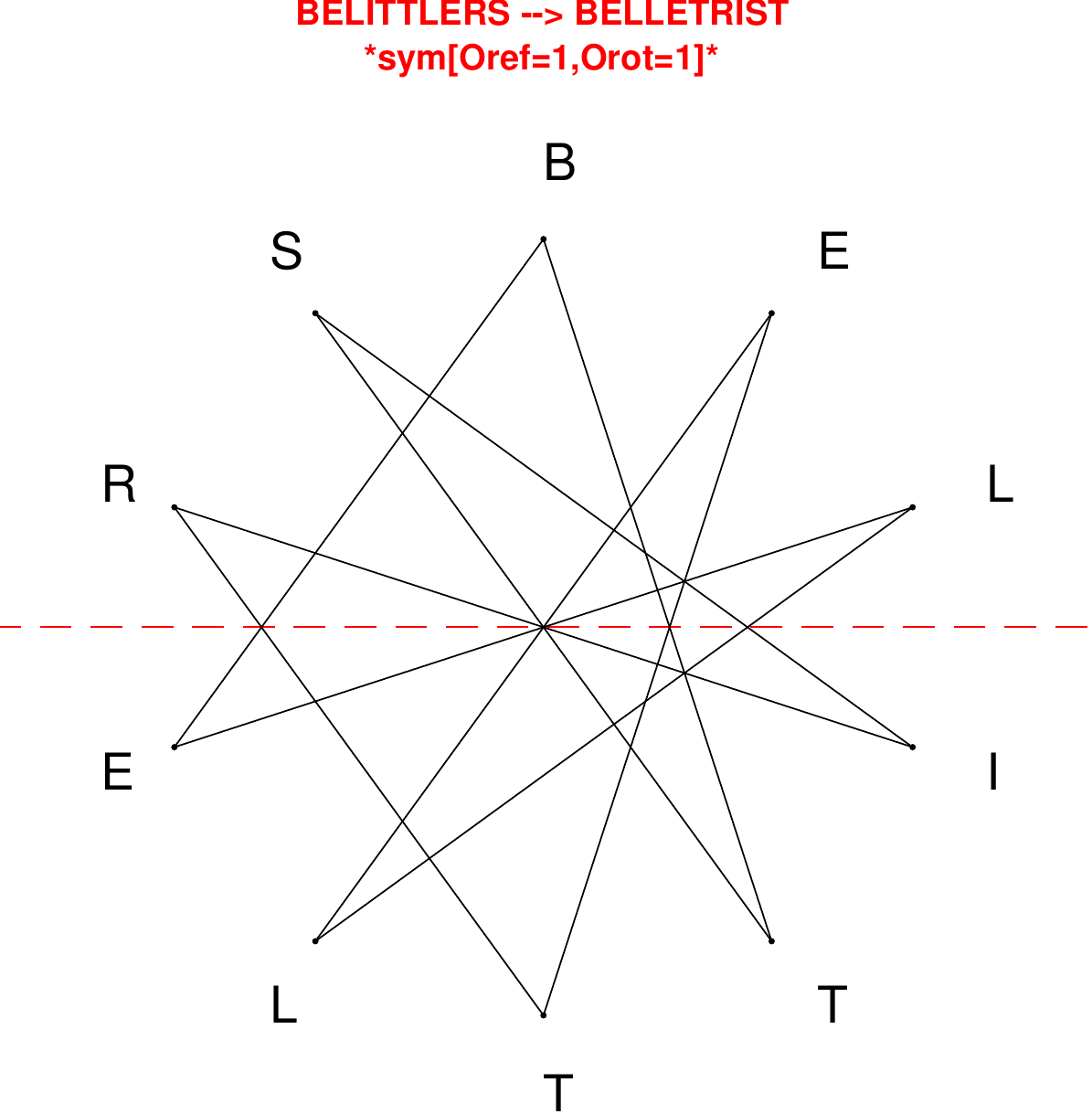}
\end{subfigure}
\end{figure}

\begin{figure}[H]
\centering
\begin{subfigure}[T]{0.19\textwidth}
\centering
\includegraphics[width=\textwidth]{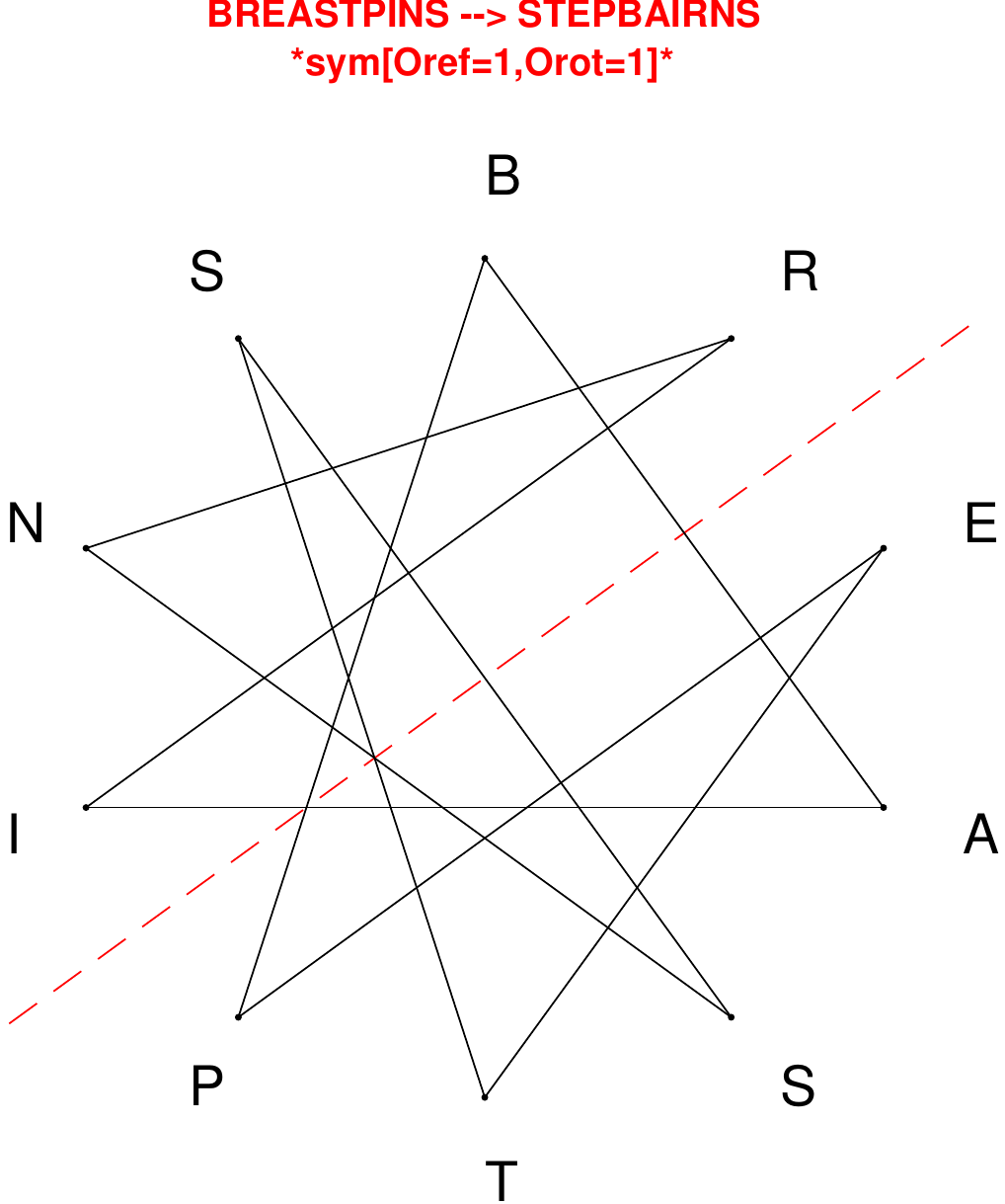}
\end{subfigure}
\hfill
\begin{subfigure}[T]{0.19\textwidth}
\centering
\includegraphics[width=\textwidth]{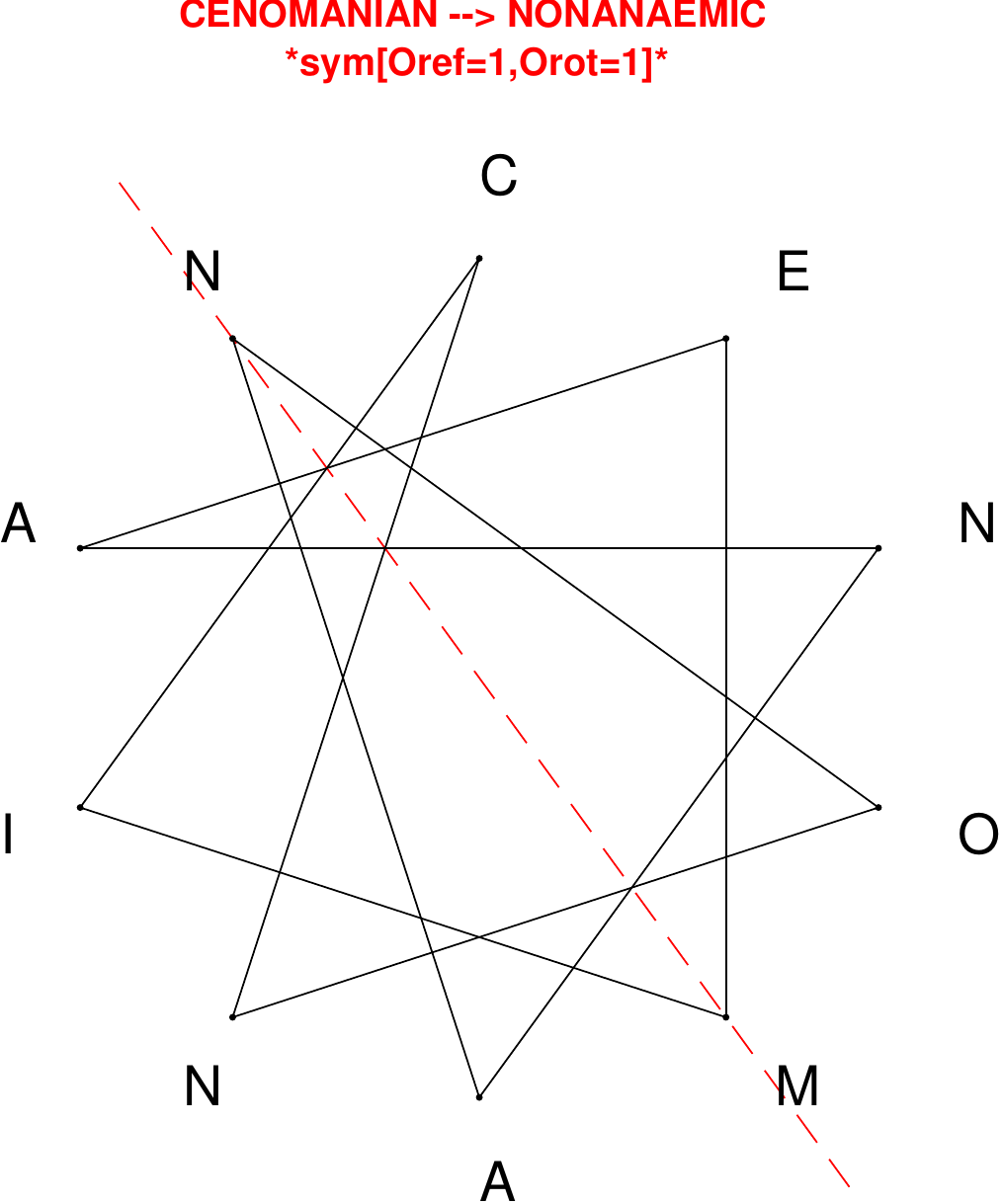}
\end{subfigure}
\hfill
\begin{subfigure}[T]{0.19\textwidth}
\centering
\includegraphics[width=\textwidth]{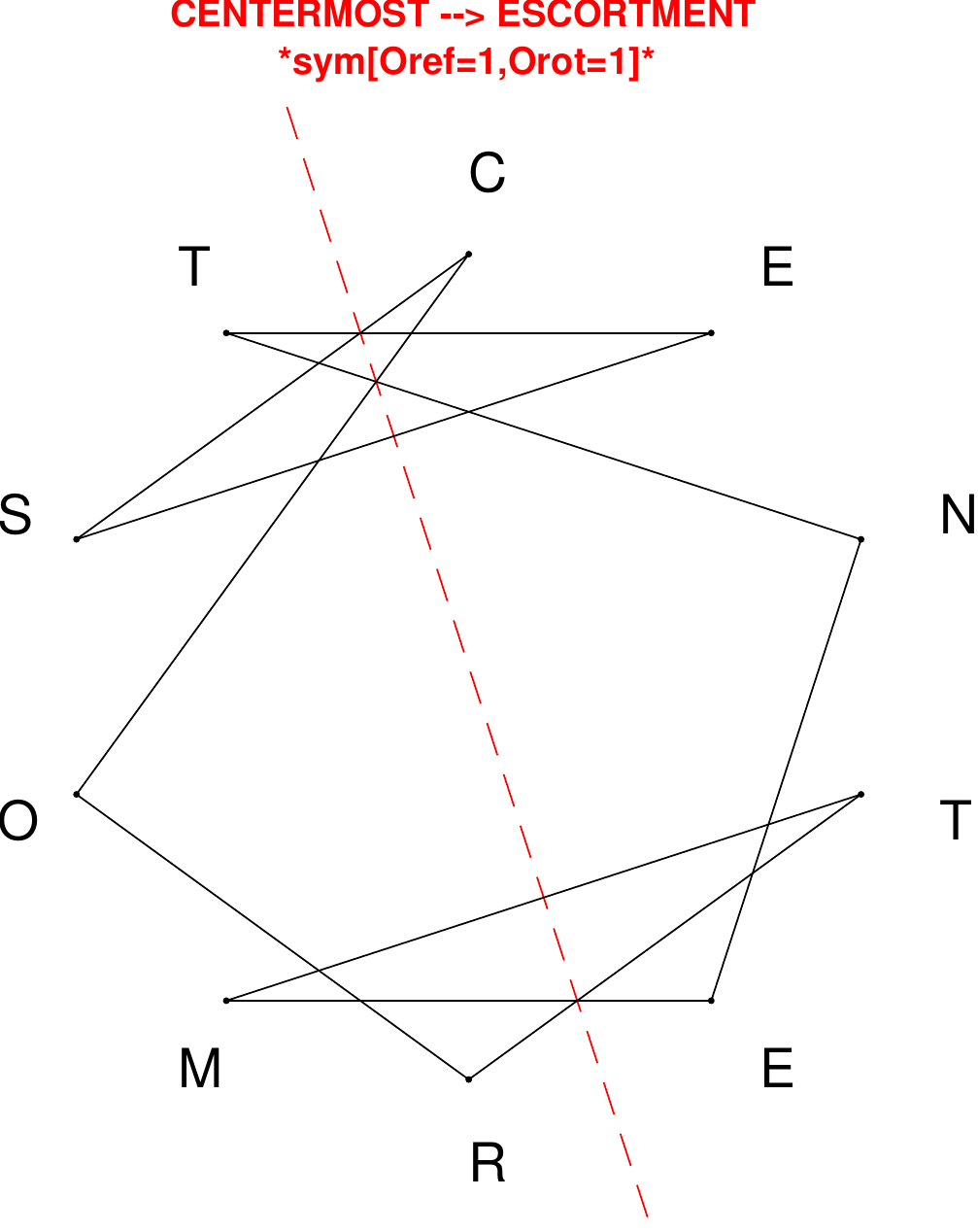}
\end{subfigure}
\hfill
\begin{subfigure}[T]{0.19\textwidth}
\centering
\includegraphics[width=\textwidth]{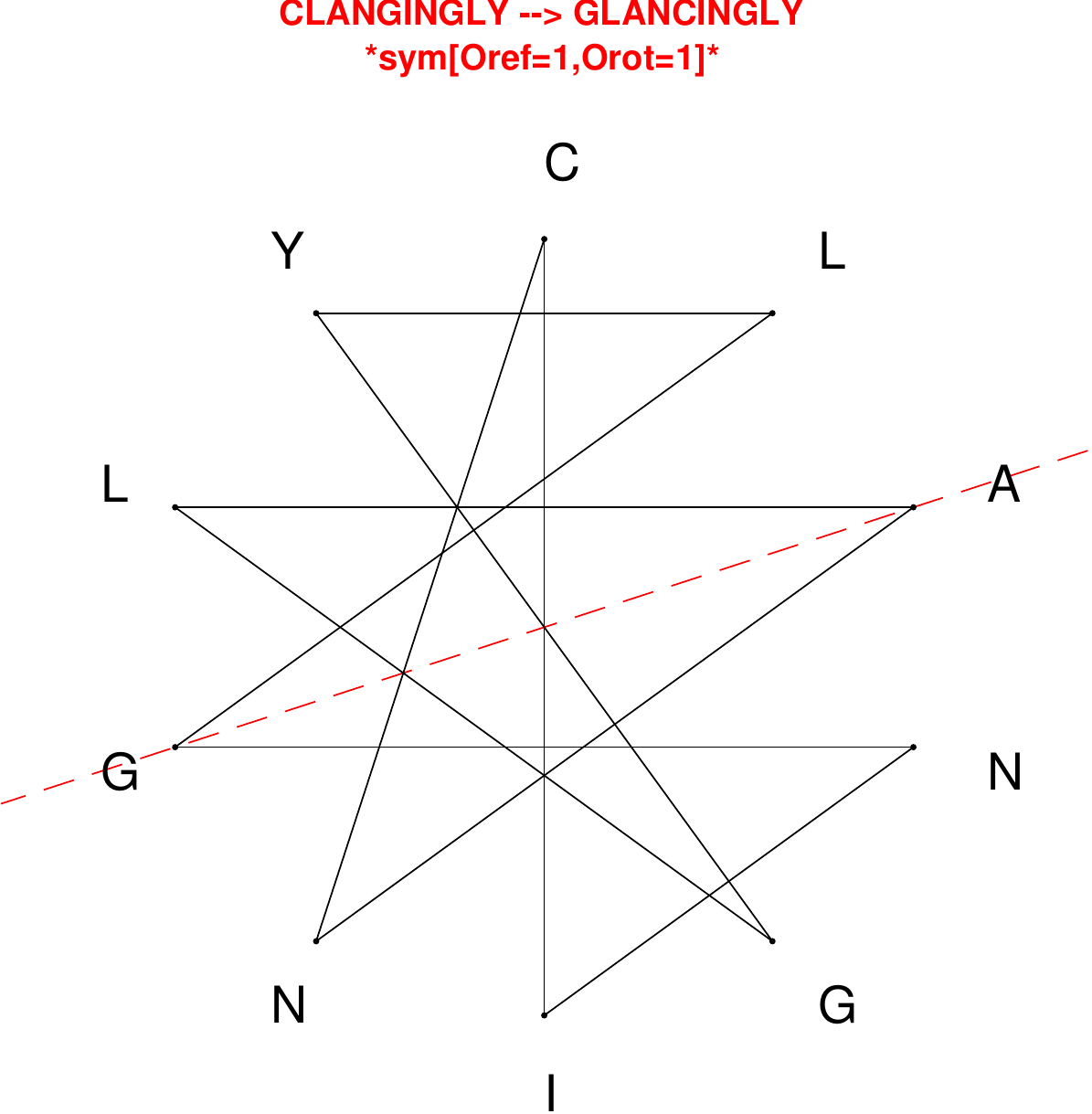}
\end{subfigure}
\hfill
\begin{subfigure}[T]{0.19\textwidth}
\centering
\includegraphics[width=\textwidth]{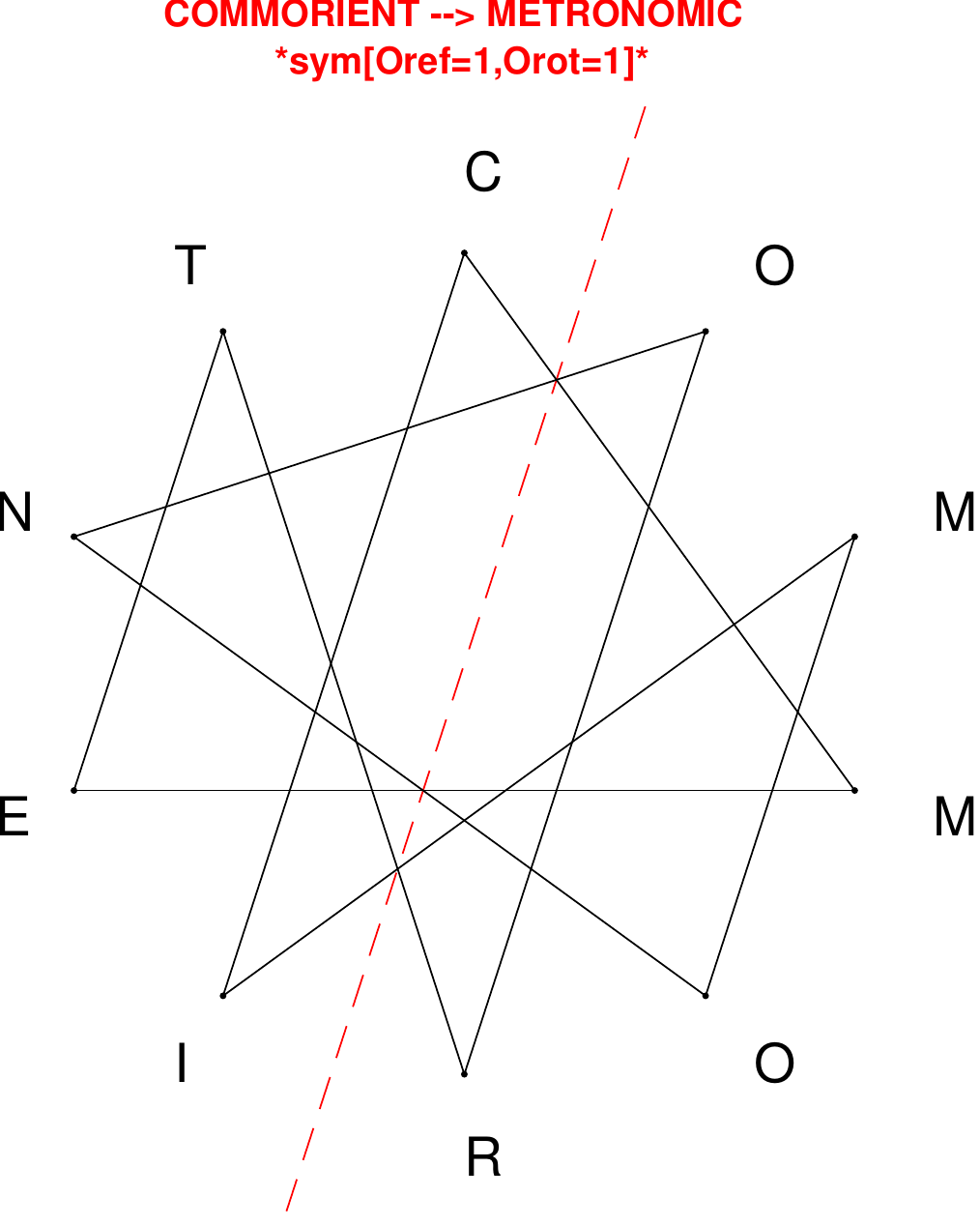}
\end{subfigure}
\end{figure}

\begin{figure}[H]
\centering
\begin{subfigure}[T]{0.19\textwidth}
\centering
\includegraphics[width=\textwidth]{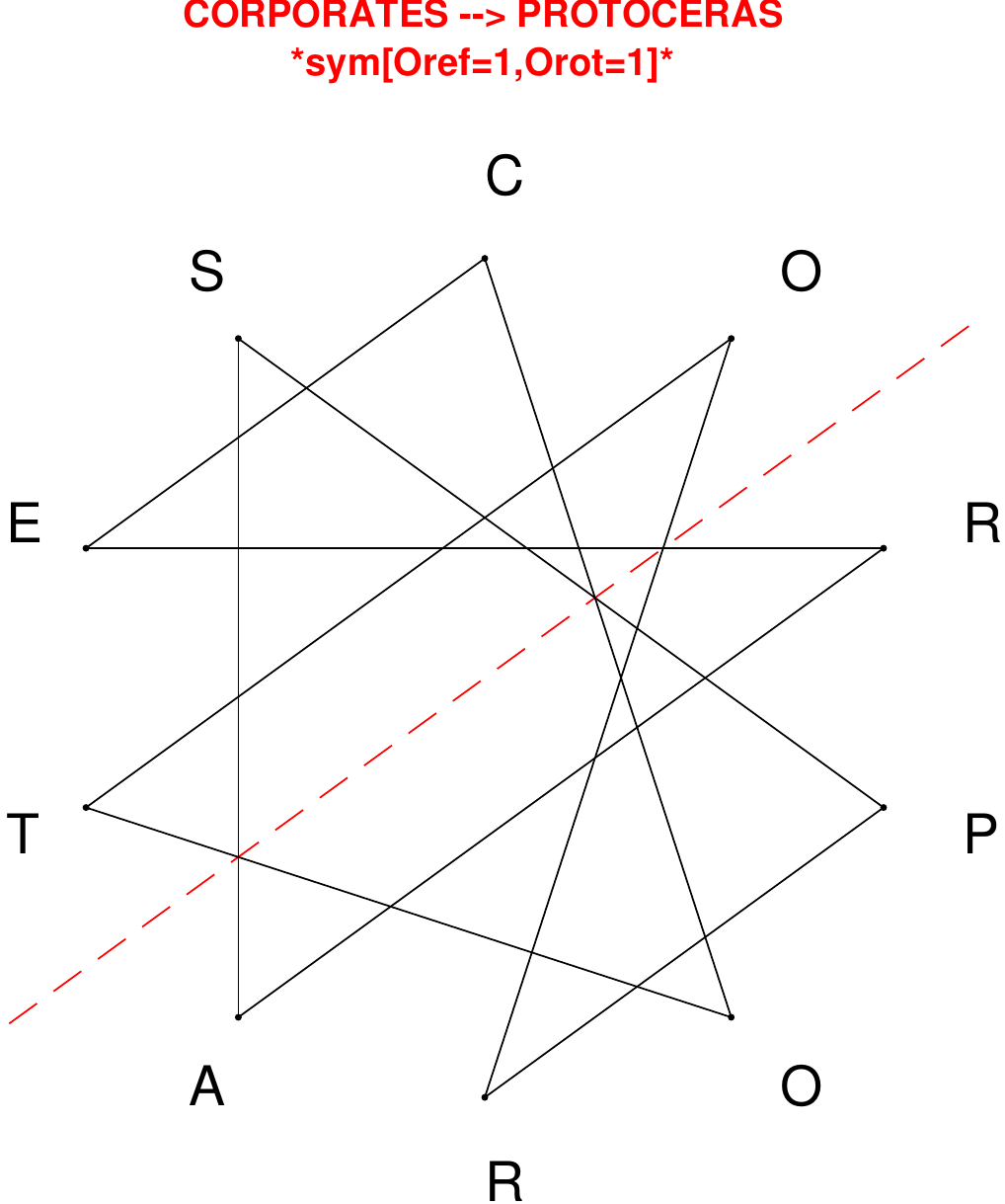}
\end{subfigure}
\hfill
\begin{subfigure}[T]{0.19\textwidth}
\centering
\includegraphics[width=\textwidth]{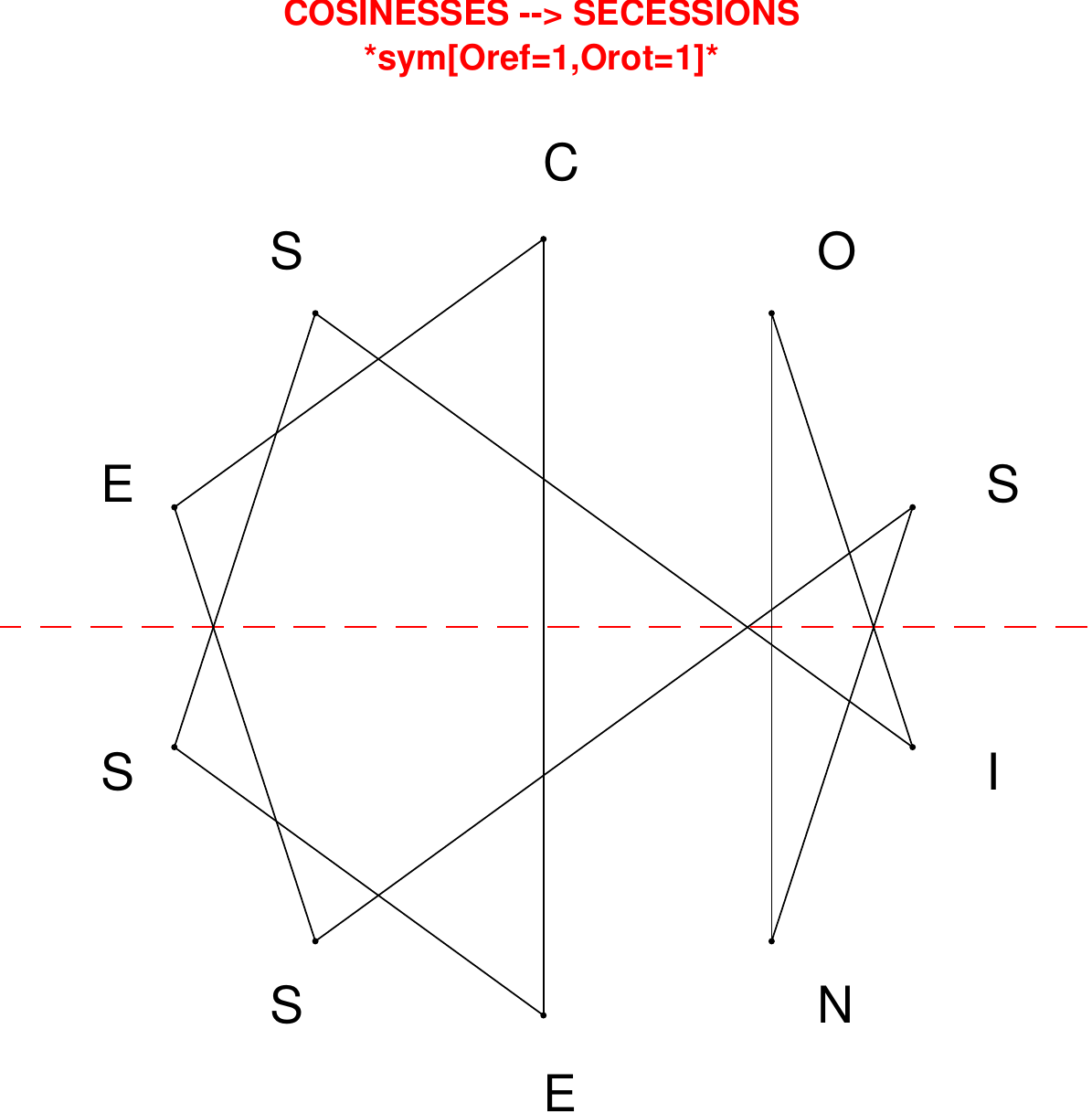}
\end{subfigure}
\hfill
\begin{subfigure}[T]{0.19\textwidth}
\centering
\includegraphics[width=\textwidth]{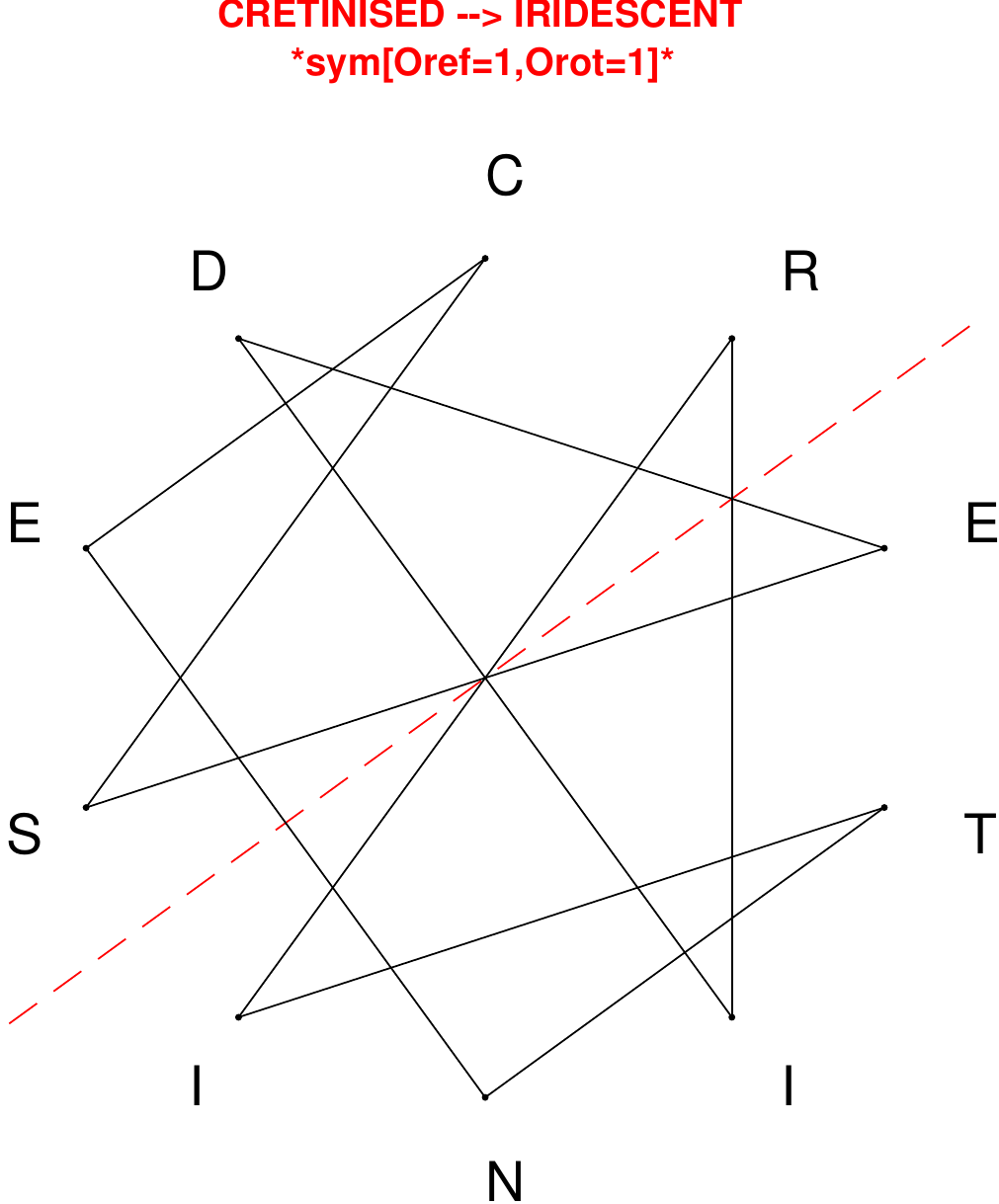}
\end{subfigure}
\hfill
\begin{subfigure}[T]{0.19\textwidth}
\centering
\includegraphics[width=\textwidth]{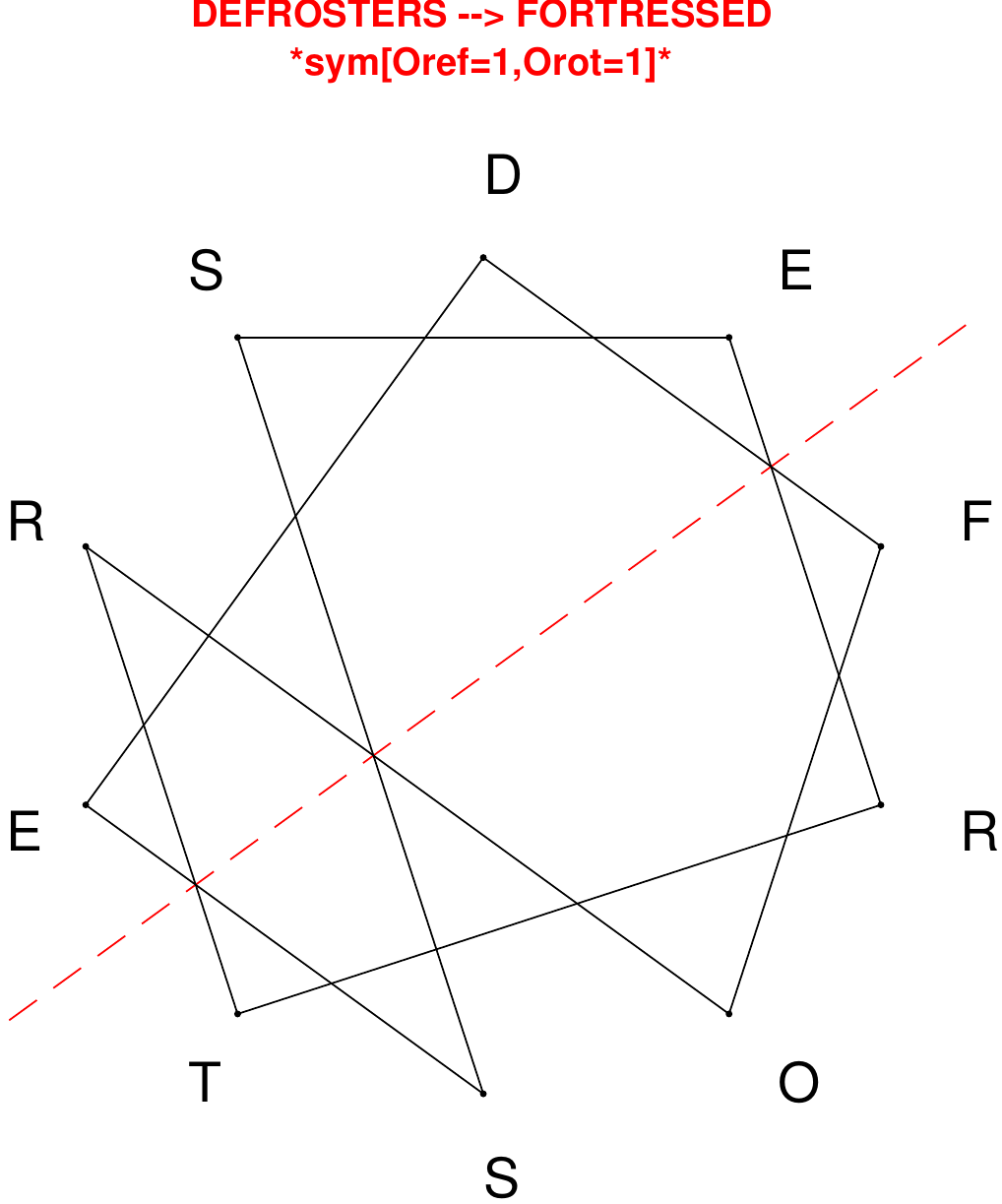}
\end{subfigure}
\hfill
\begin{subfigure}[T]{0.19\textwidth}
\centering
\includegraphics[width=\textwidth]{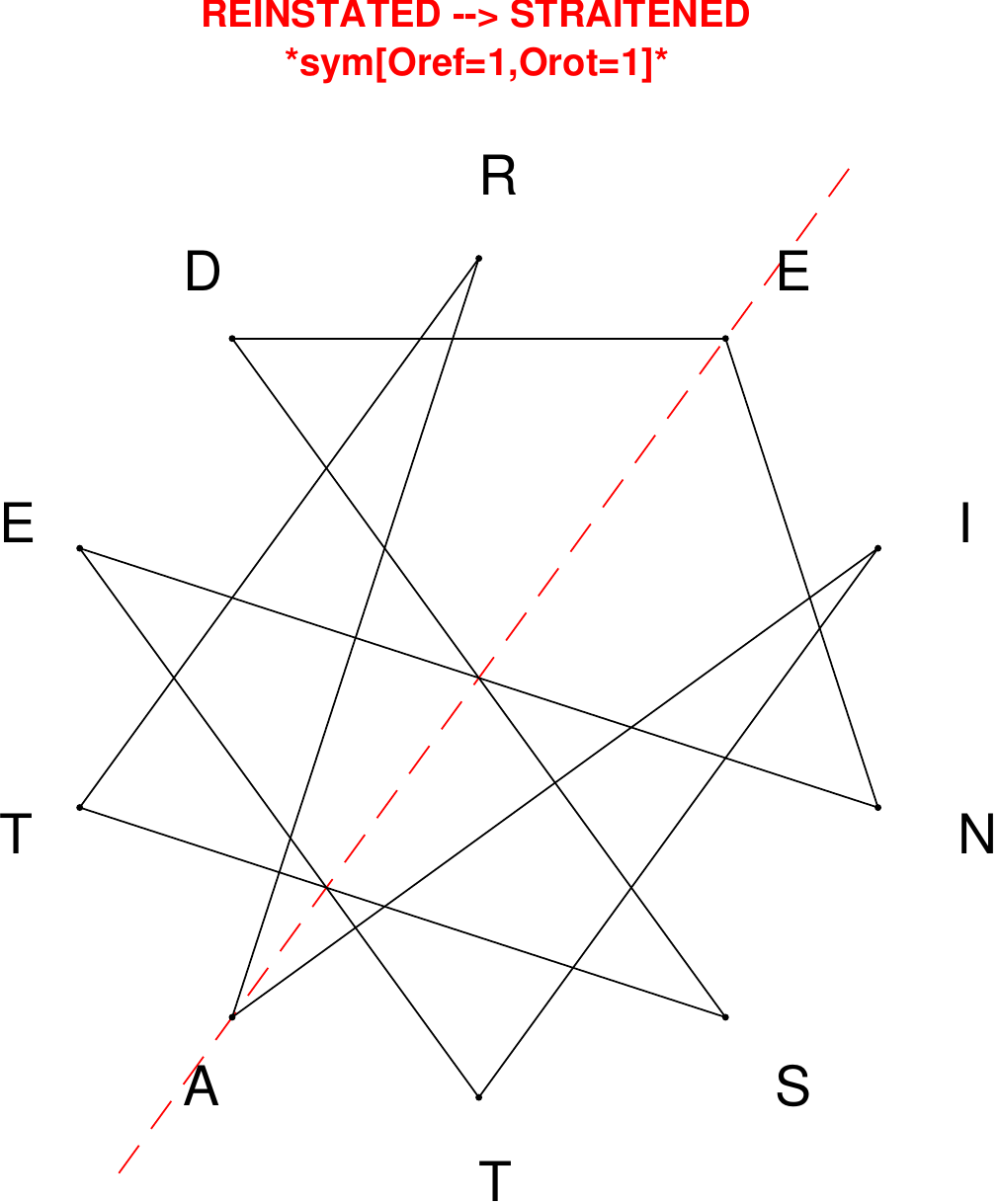}
\end{subfigure}
\end{figure}

\begin{figure}[H]
\centering
\begin{subfigure}[T]{0.19\textwidth}
\centering
\includegraphics[width=\textwidth]{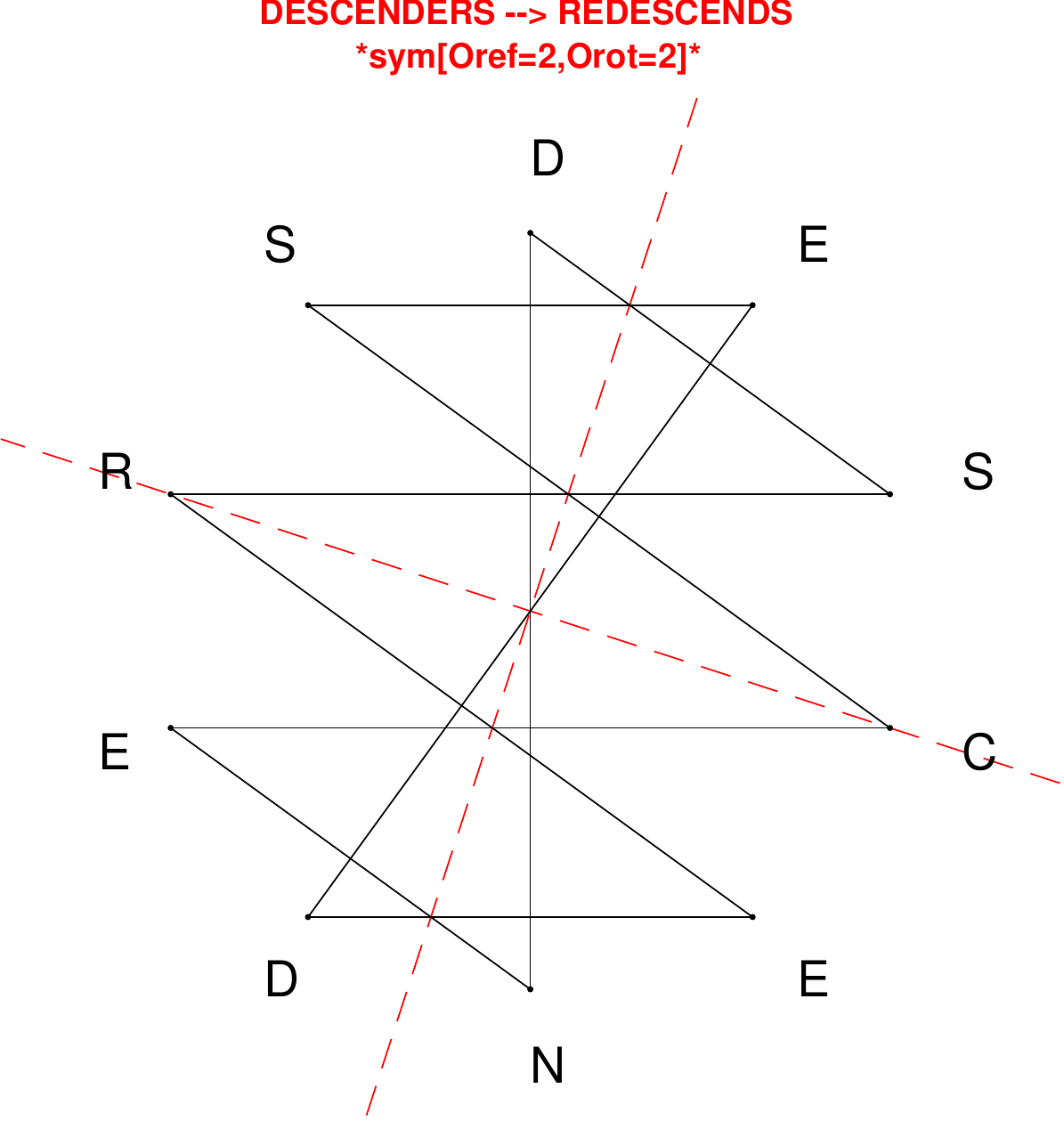}
\end{subfigure}
\hfill
\begin{subfigure}[T]{0.19\textwidth}
\centering
\includegraphics[width=\textwidth]{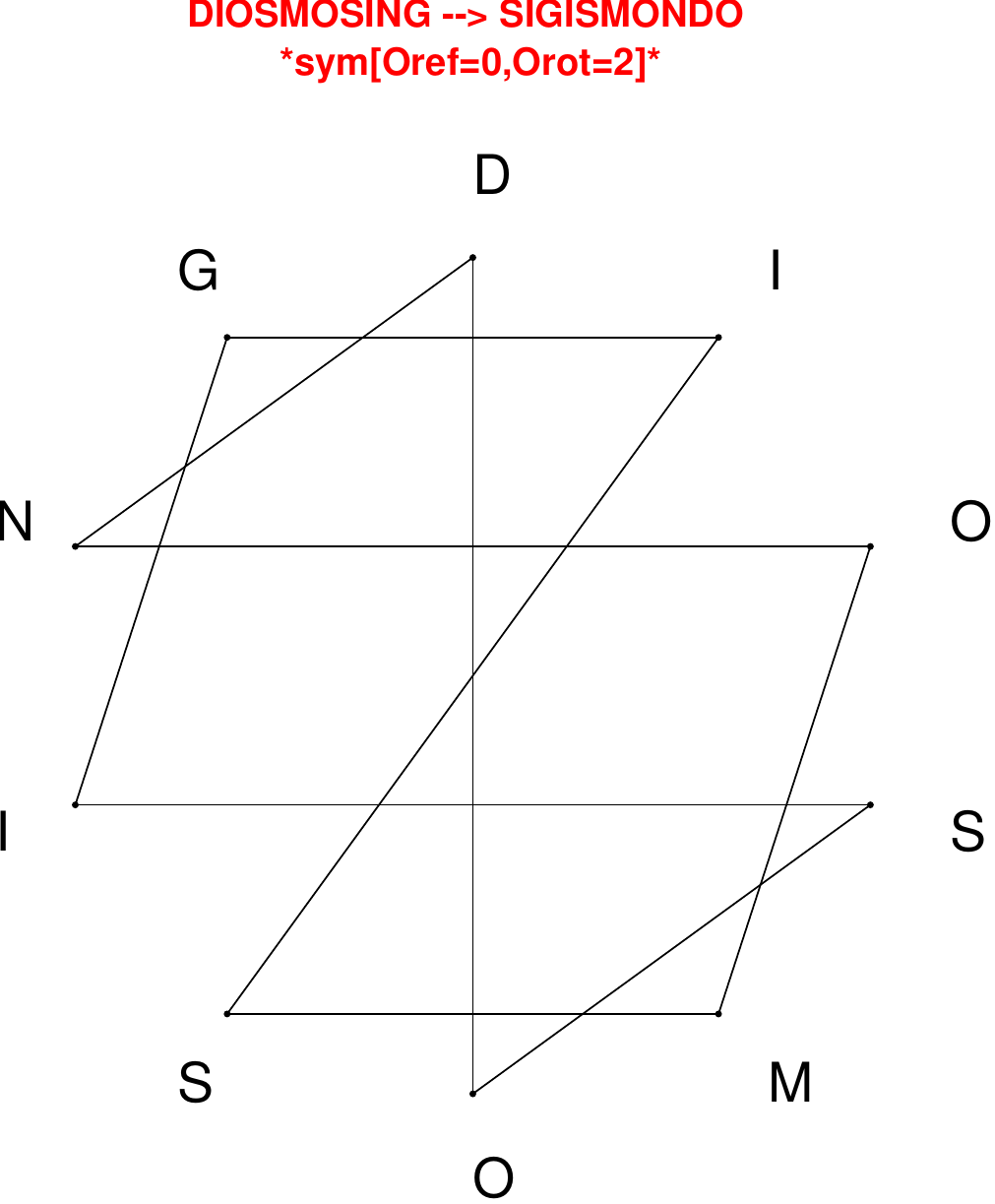}
\end{subfigure}
\hfill
\begin{subfigure}[T]{0.19\textwidth}
\centering
\includegraphics[width=\textwidth]{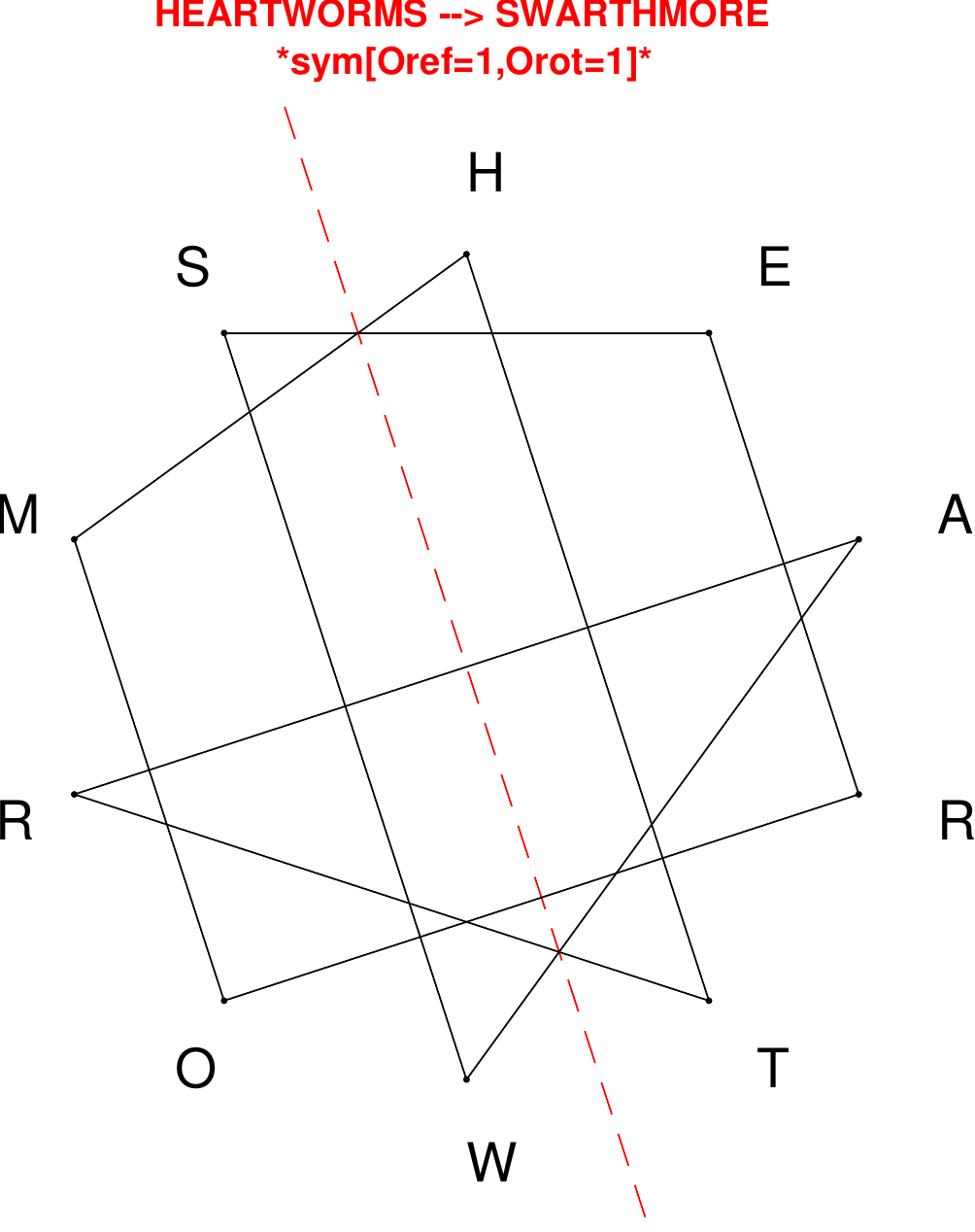}
\end{subfigure}
\hfill
\begin{subfigure}[T]{0.19\textwidth}
\centering
\includegraphics[width=\textwidth]{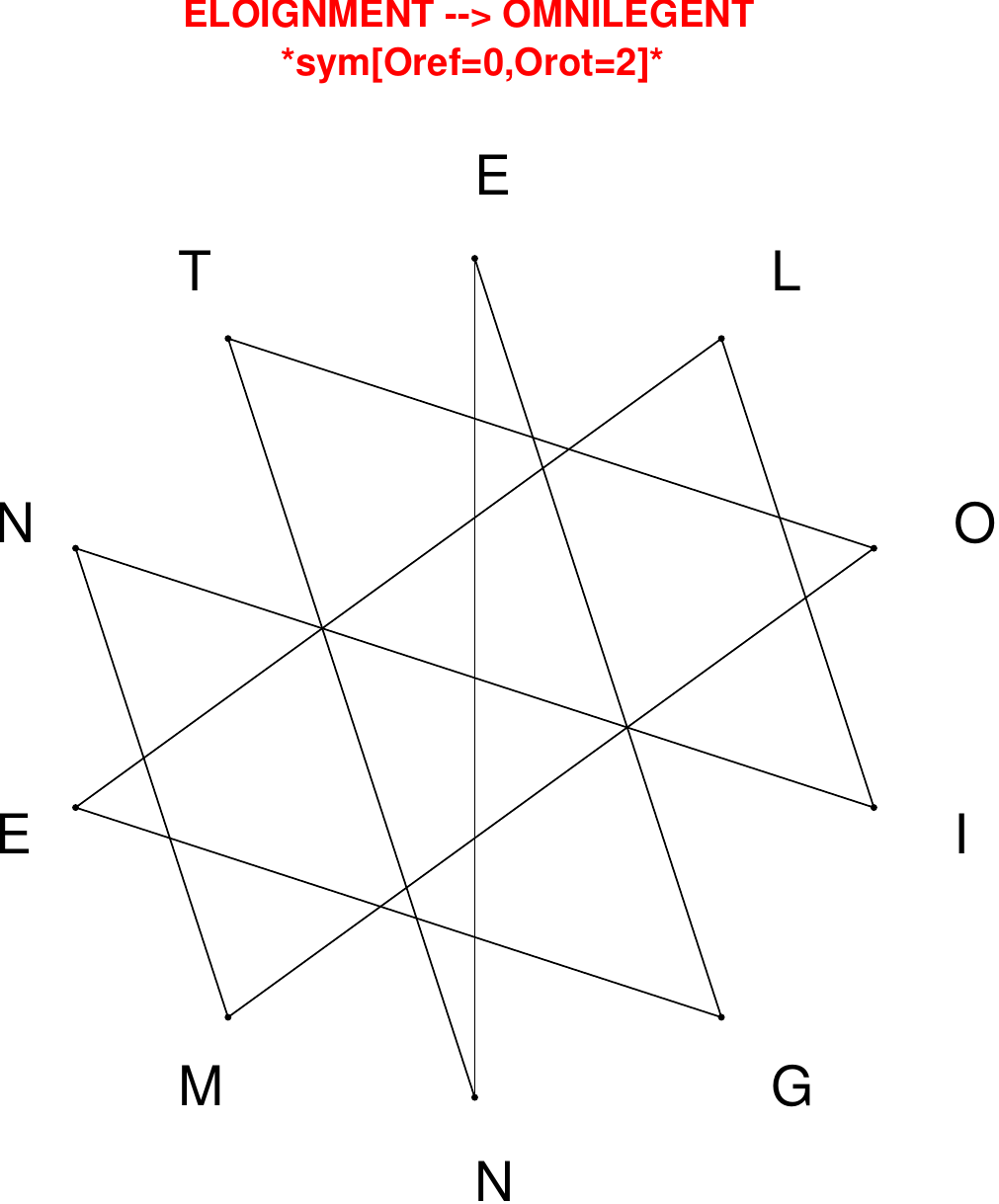}
\end{subfigure}
\hfill
\begin{subfigure}[T]{0.19\textwidth}
\centering
\includegraphics[width=\textwidth]{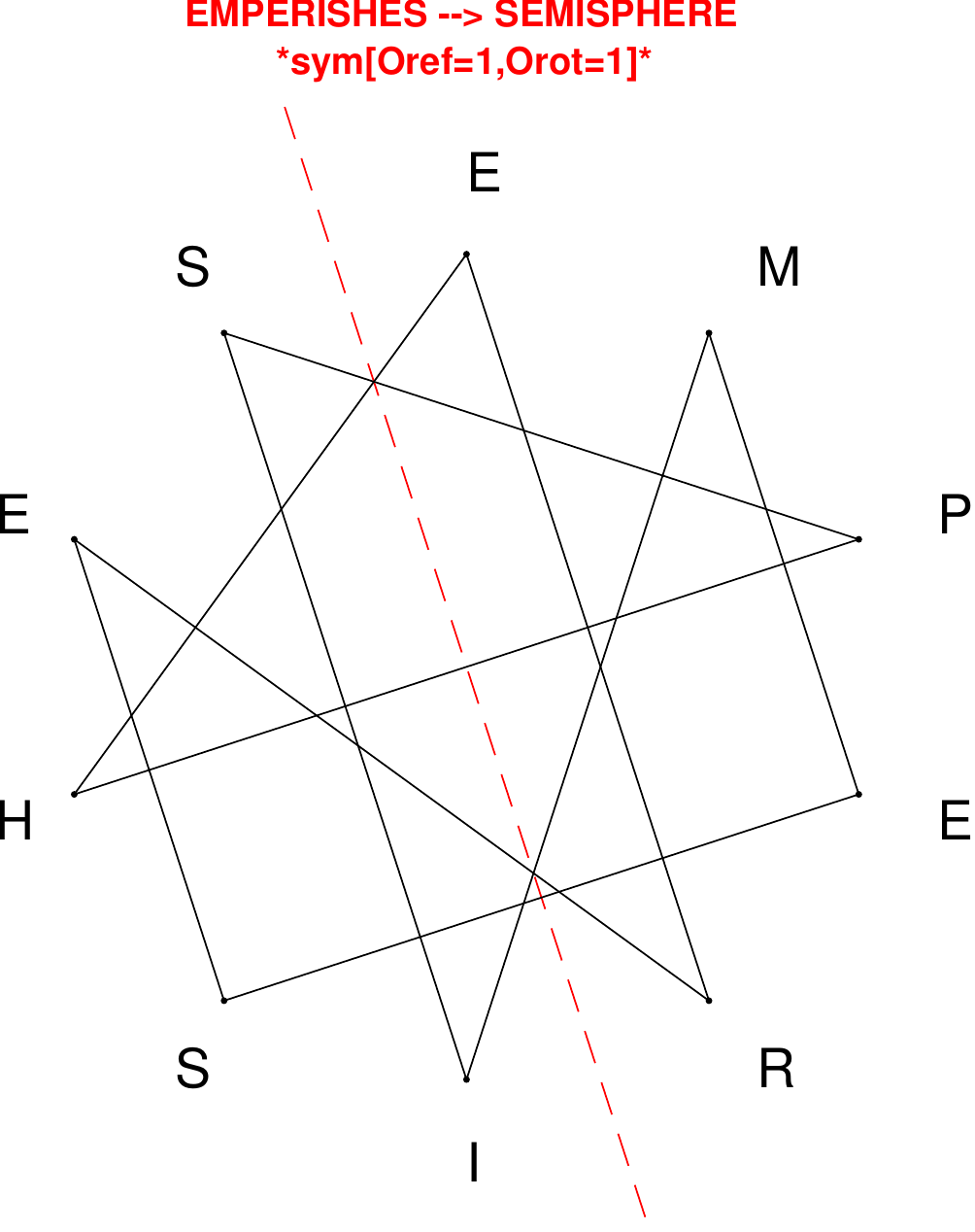}
\end{subfigure}
\end{figure}

\begin{figure}[H]
\centering
\begin{subfigure}[T]{0.19\textwidth}
\centering
\includegraphics[width=\textwidth]{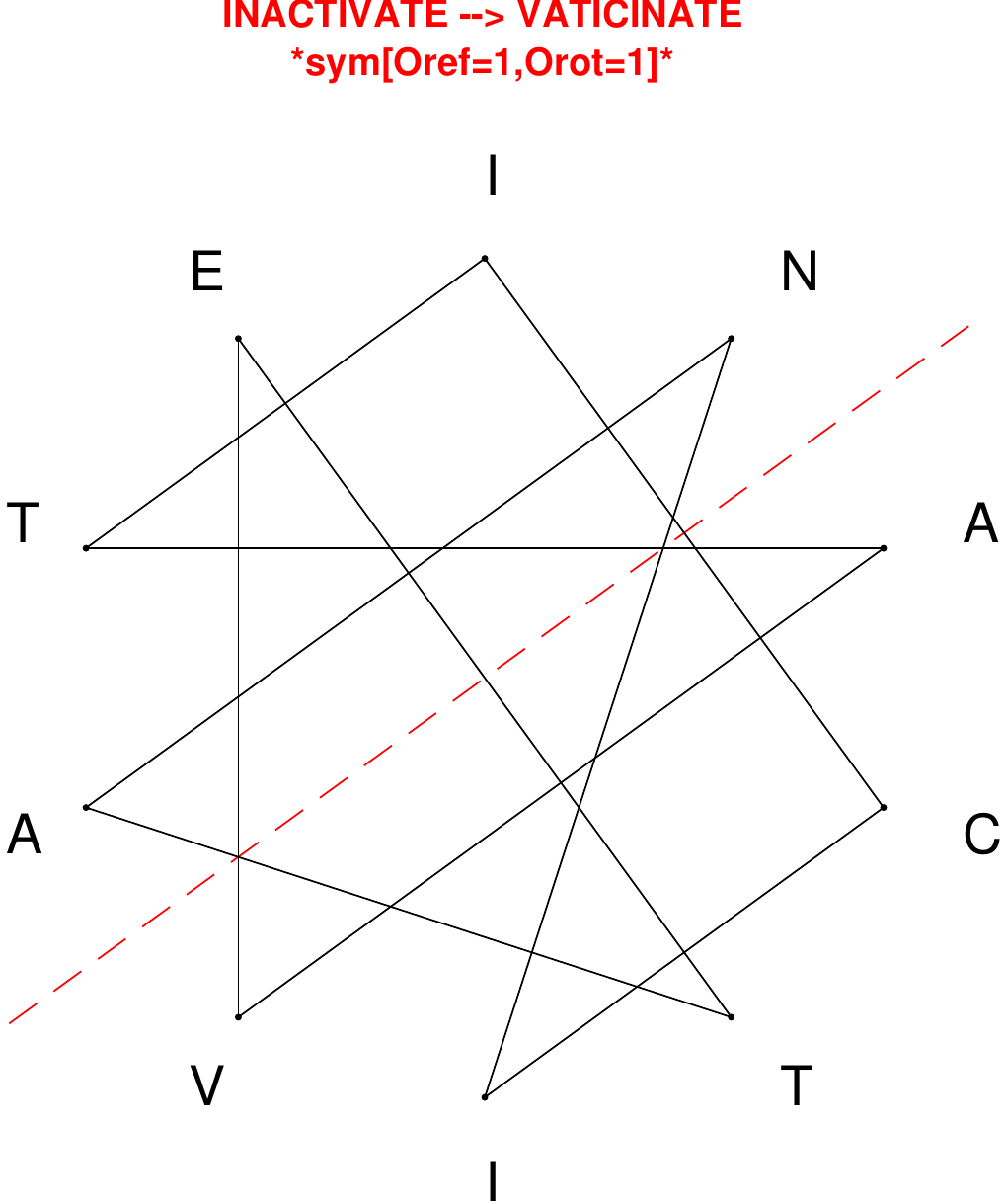}
\end{subfigure}
\hfill
\begin{subfigure}[T]{0.19\textwidth}
\centering
\includegraphics[width=\textwidth]{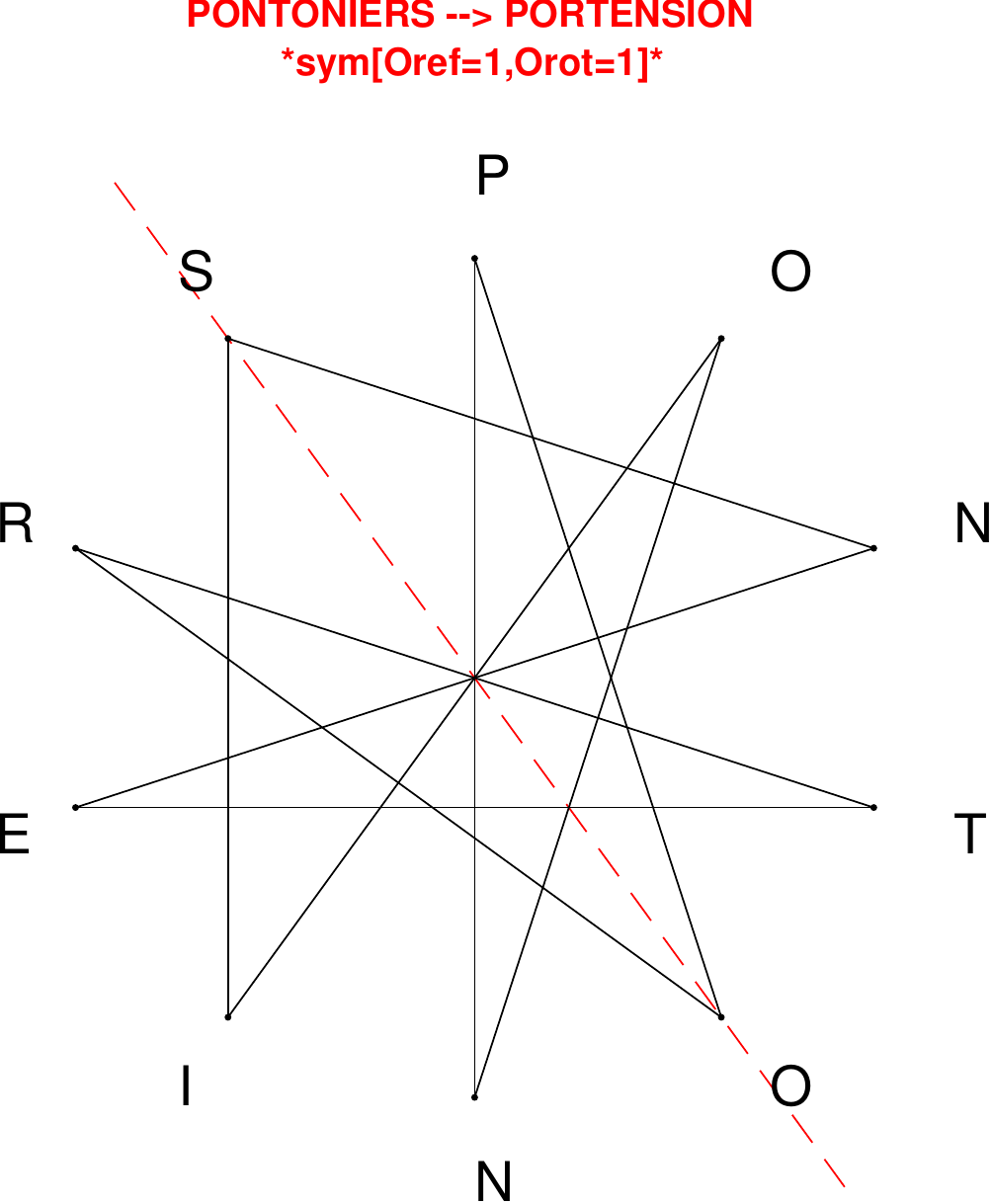}
\end{subfigure}
\hfill
\begin{subfigure}[T]{0.19\textwidth}
\centering
\includegraphics[width=\textwidth]{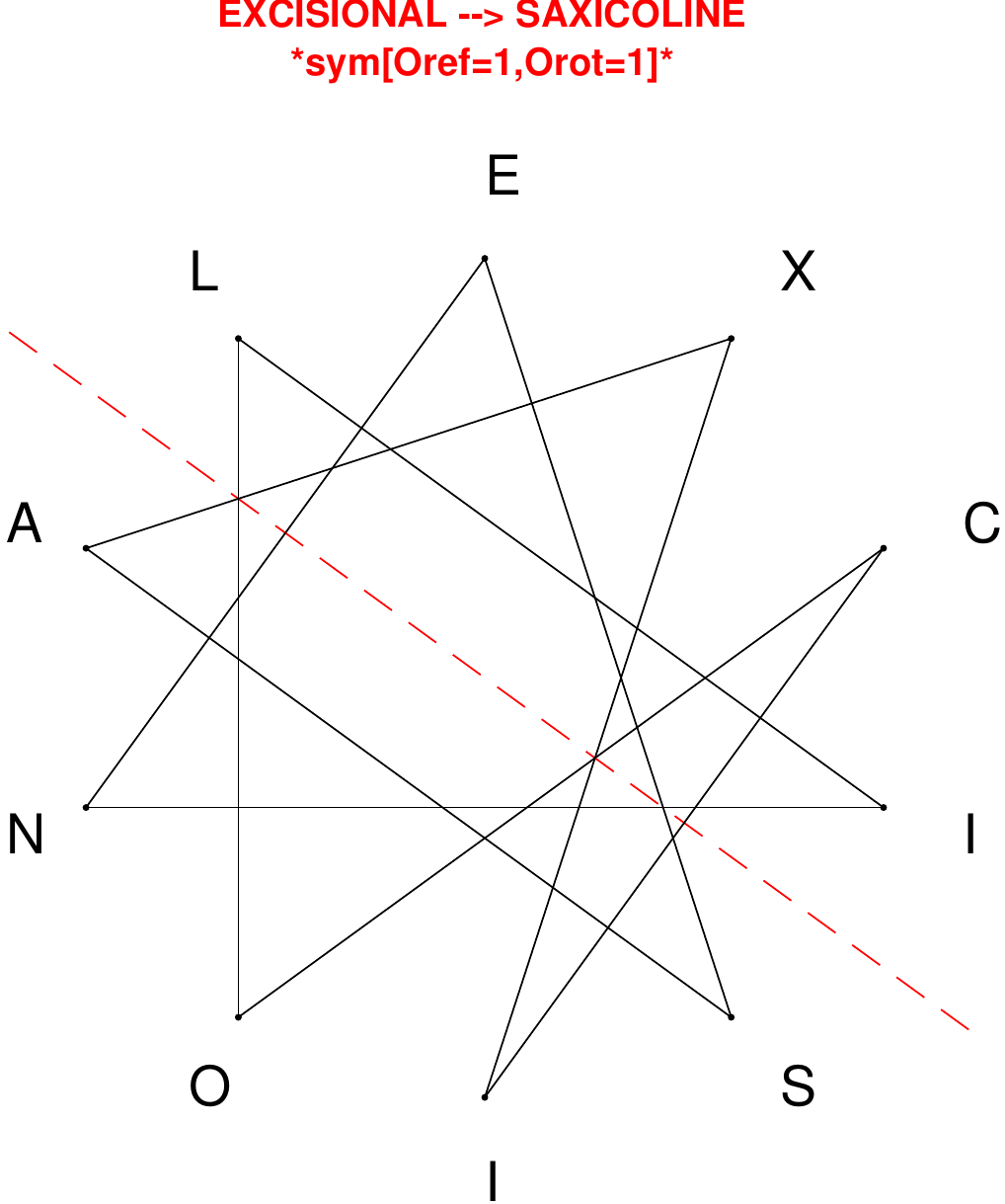}
\end{subfigure}
\hfill
\begin{subfigure}[T]{0.19\textwidth}
\centering
\includegraphics[width=\textwidth]{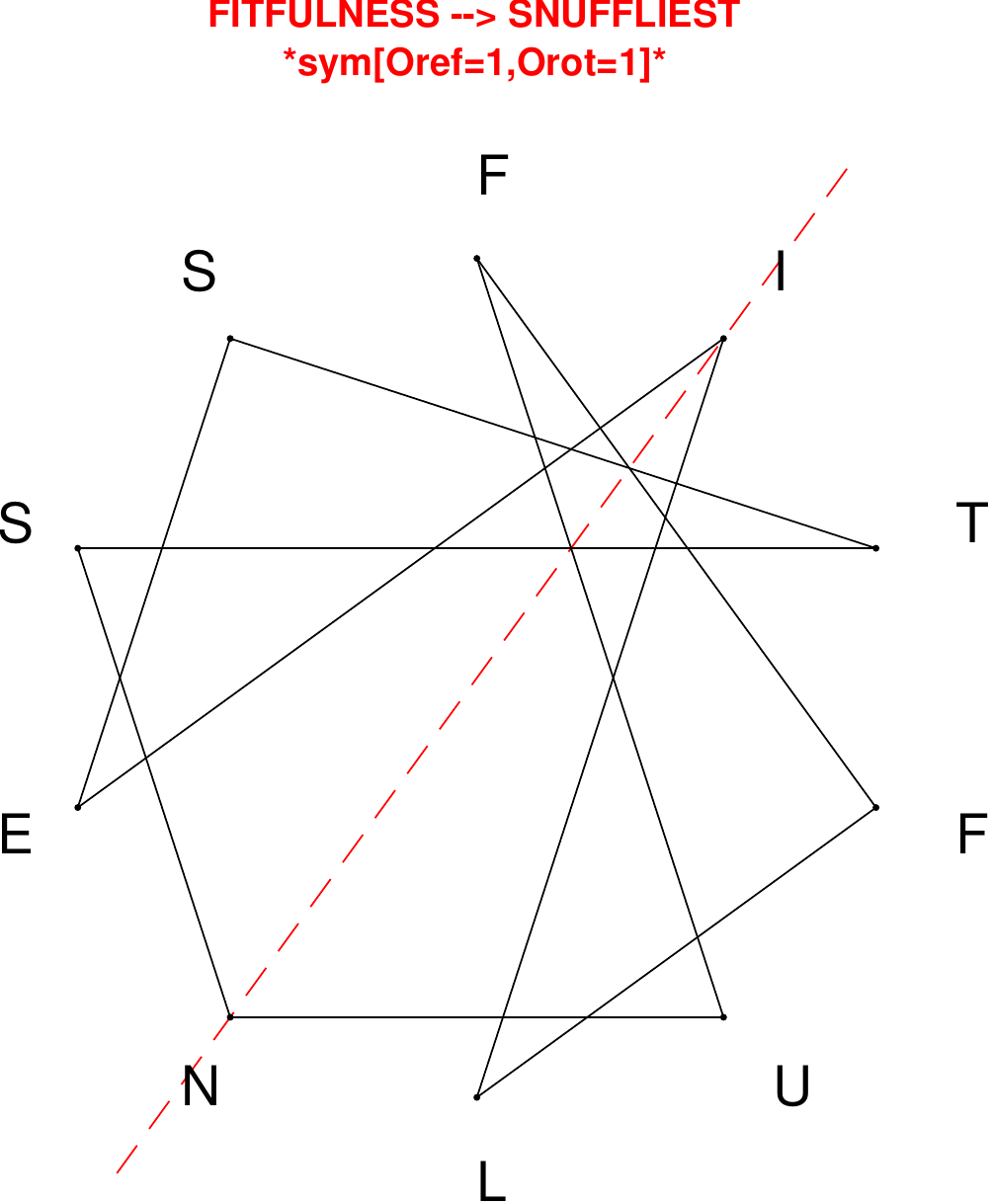}
\end{subfigure}
\hfill
\begin{subfigure}[T]{0.19\textwidth}
\centering
\includegraphics[width=\textwidth]{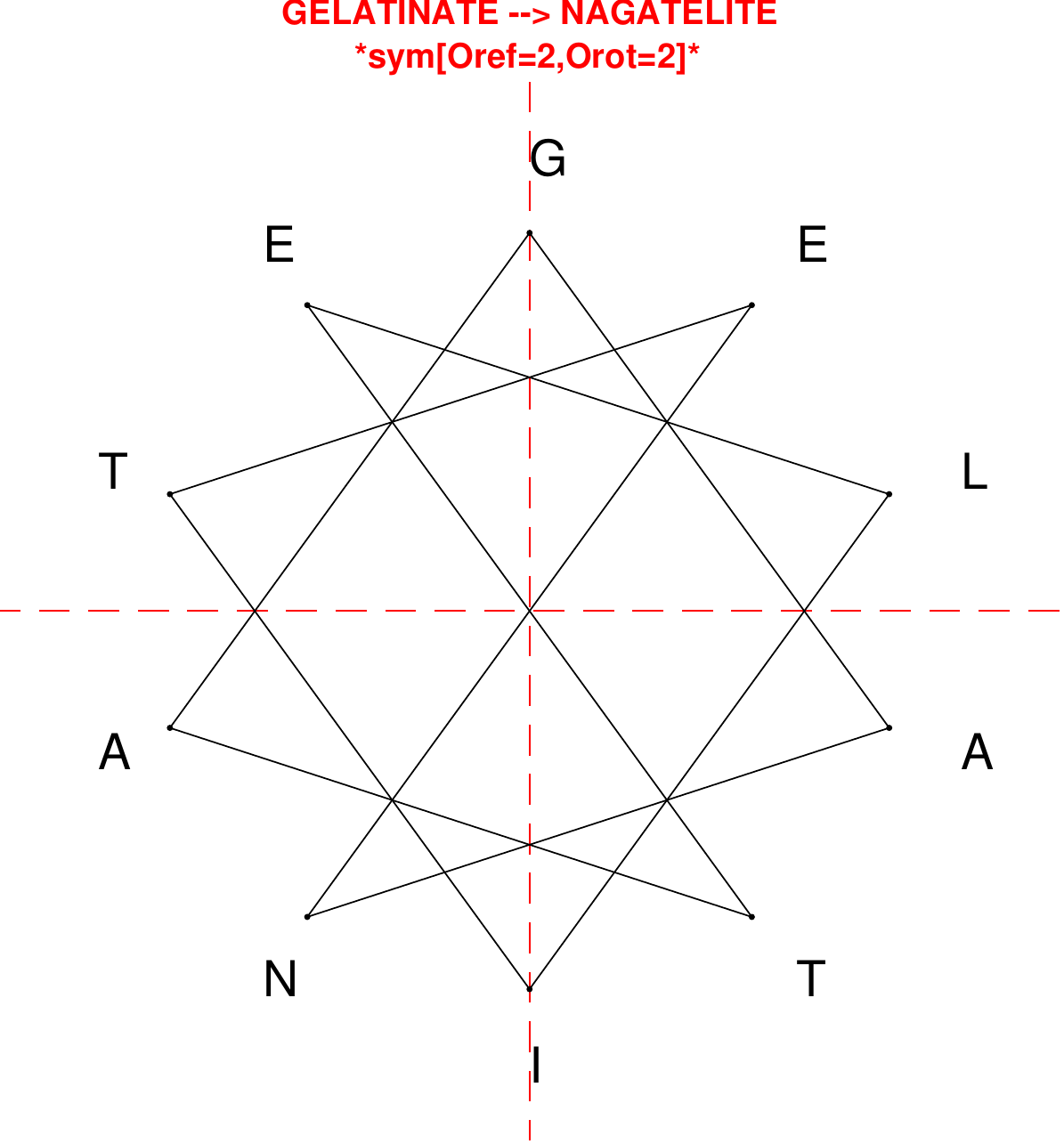}
\end{subfigure}
\end{figure}

\begin{figure}[H]
\centering
\begin{subfigure}[T]{0.19\textwidth}
\centering
\includegraphics[width=\textwidth]{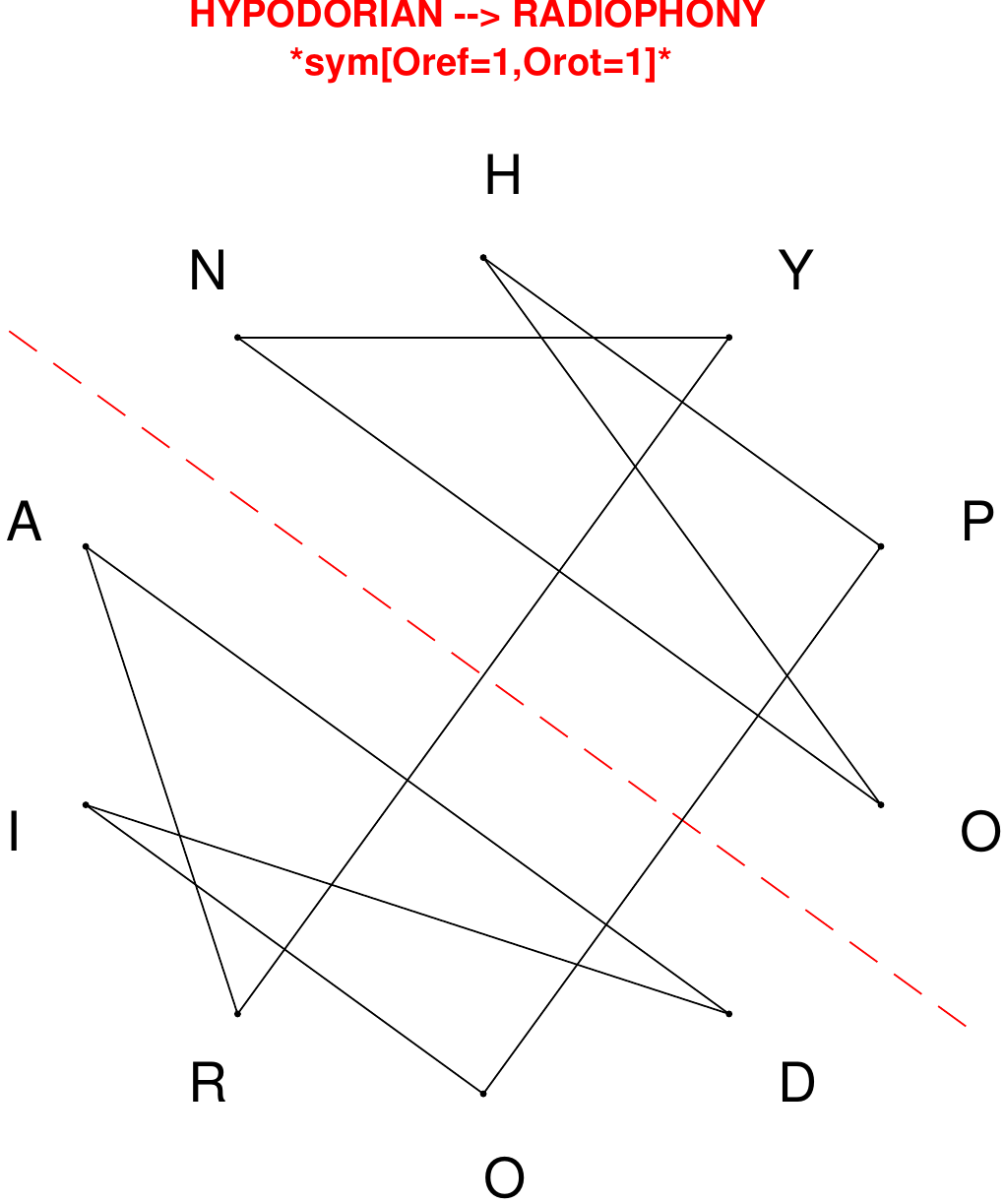}
\end{subfigure}
\hfill
\begin{subfigure}[T]{0.19\textwidth}
\centering
\includegraphics[width=\textwidth]{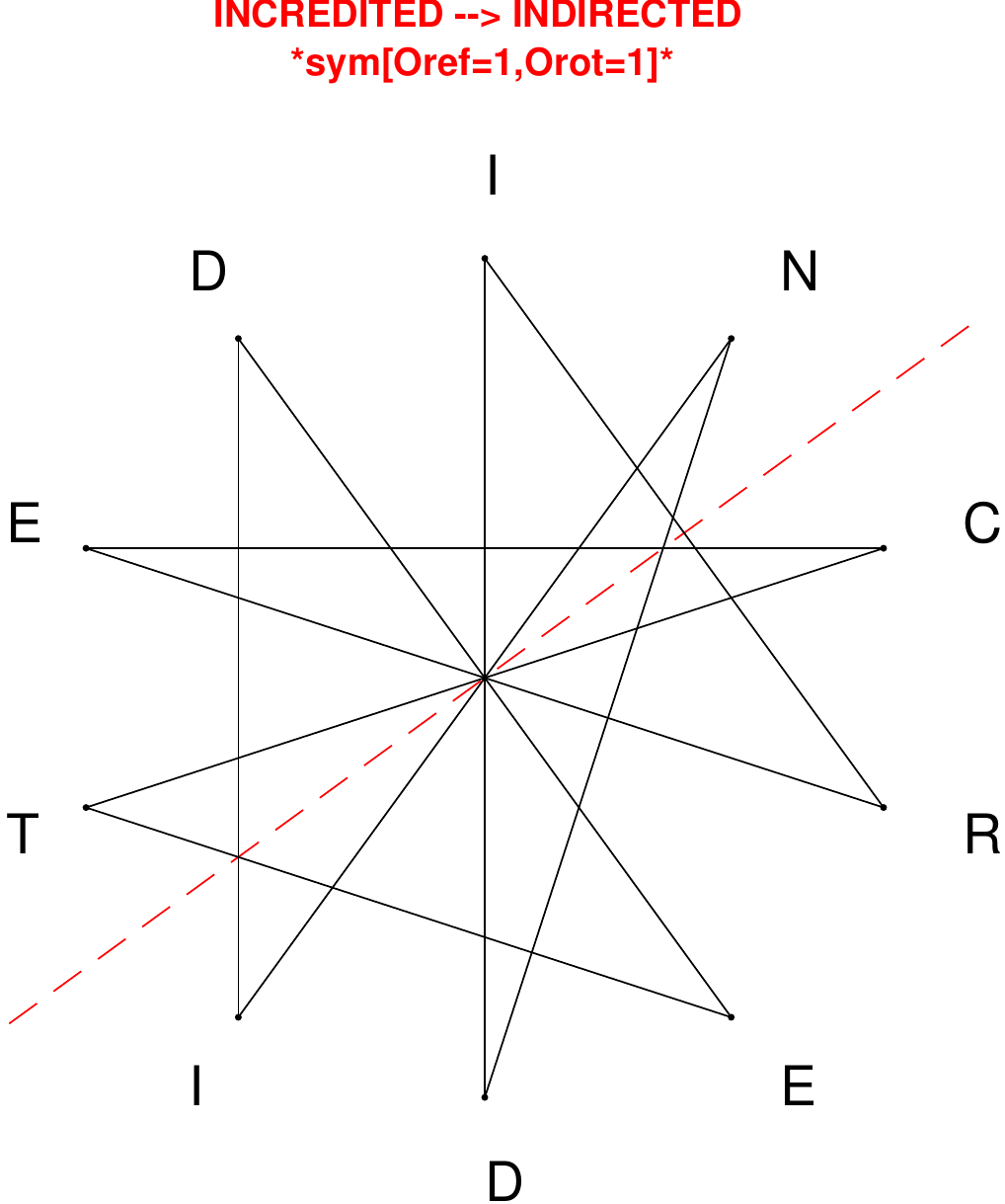}
\end{subfigure}
\hfill
\begin{subfigure}[T]{0.19\textwidth}
\centering
\includegraphics[width=\textwidth]{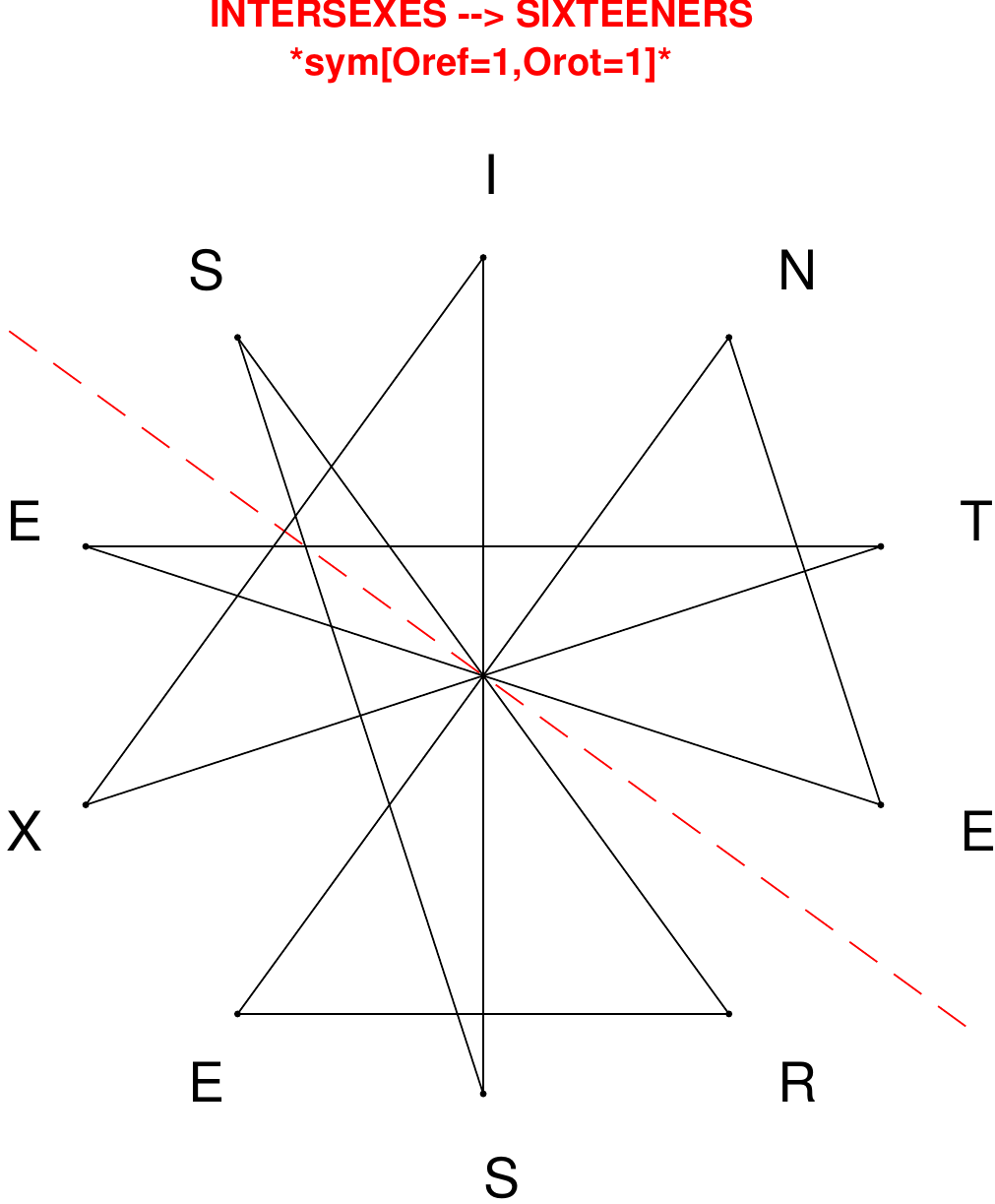}
\end{subfigure}
\hfill
\begin{subfigure}[T]{0.19\textwidth}
\centering
\includegraphics[width=\textwidth]{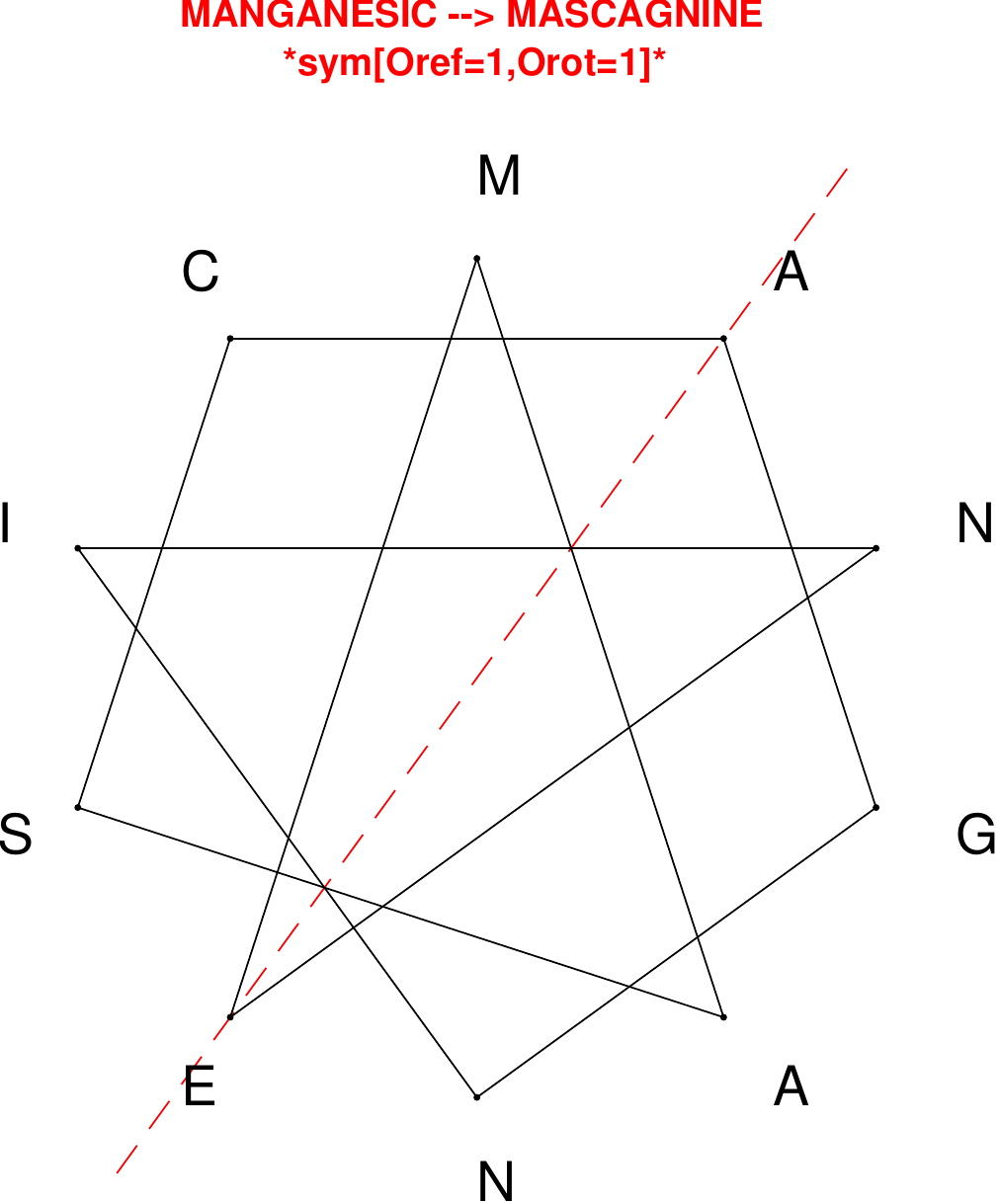}
\end{subfigure}
\hfill
\begin{subfigure}[T]{0.19\textwidth}
\centering
\includegraphics[width=\textwidth]{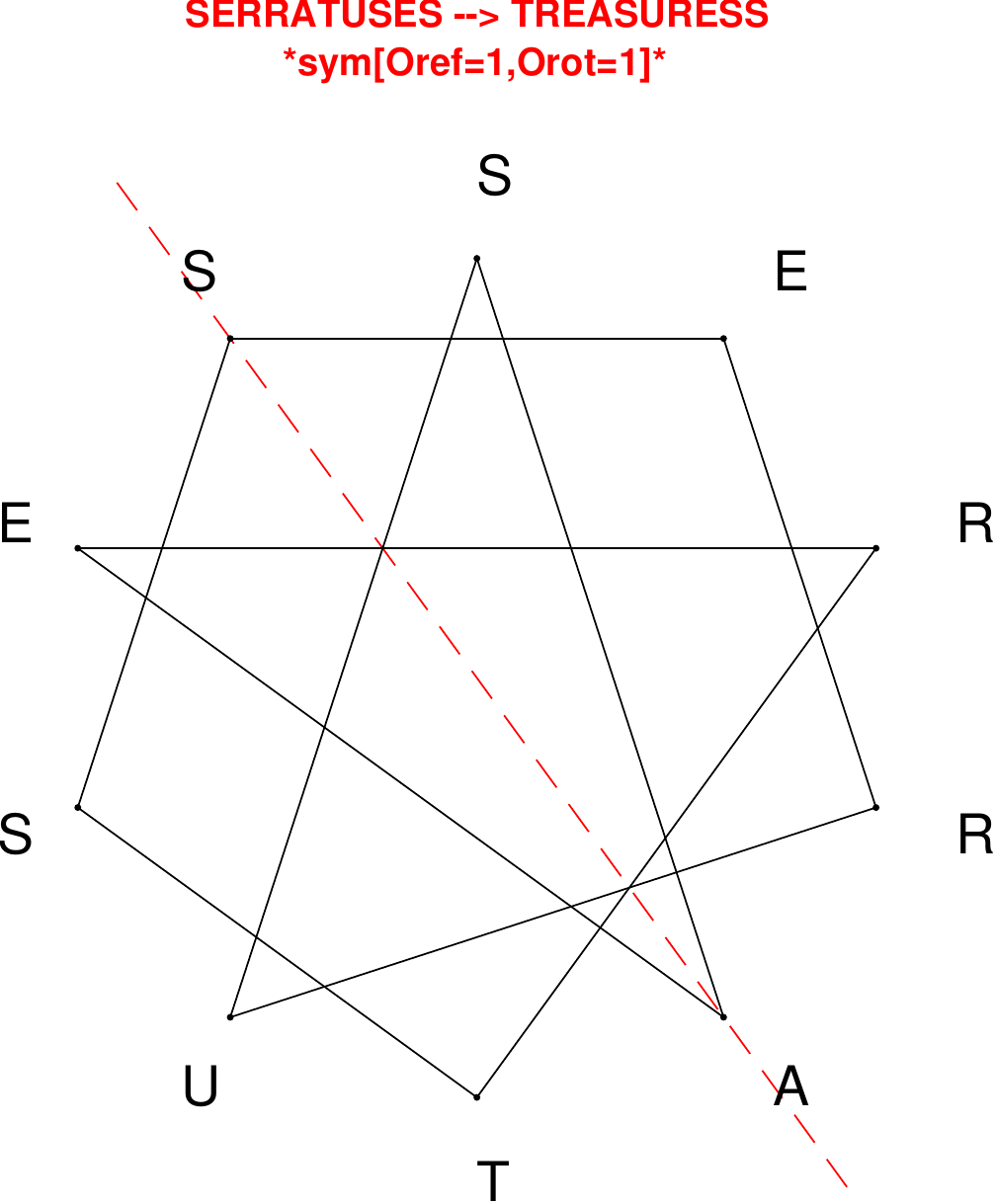}
\end{subfigure}
\end{figure}

\begin{figure}[H]
\centering
\begin{subfigure}[T]{0.19\textwidth}
\centering
\includegraphics[width=\textwidth]{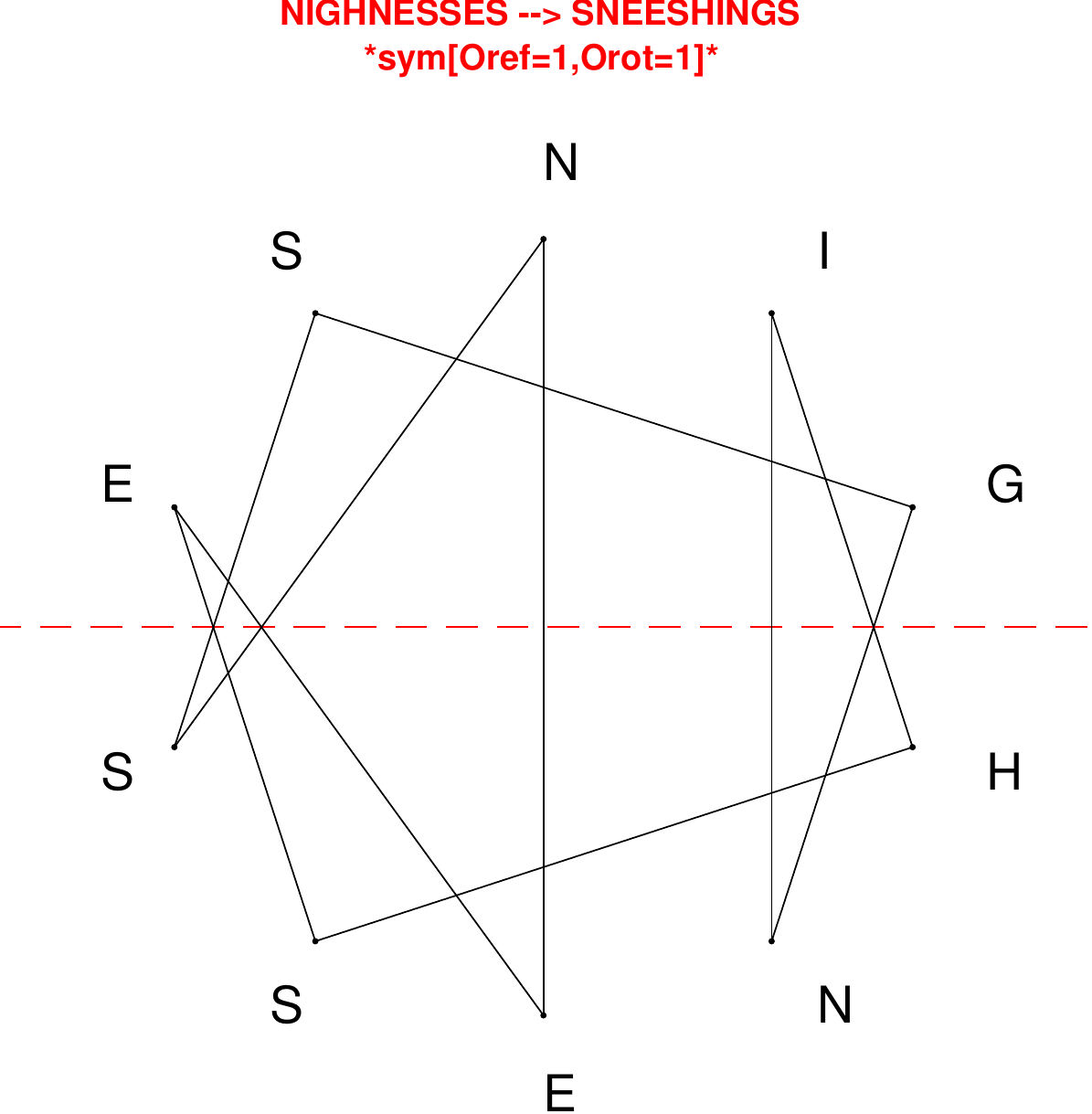}
\end{subfigure}
\hfill
\begin{subfigure}[T]{0.19\textwidth}
\centering
\includegraphics[width=\textwidth]{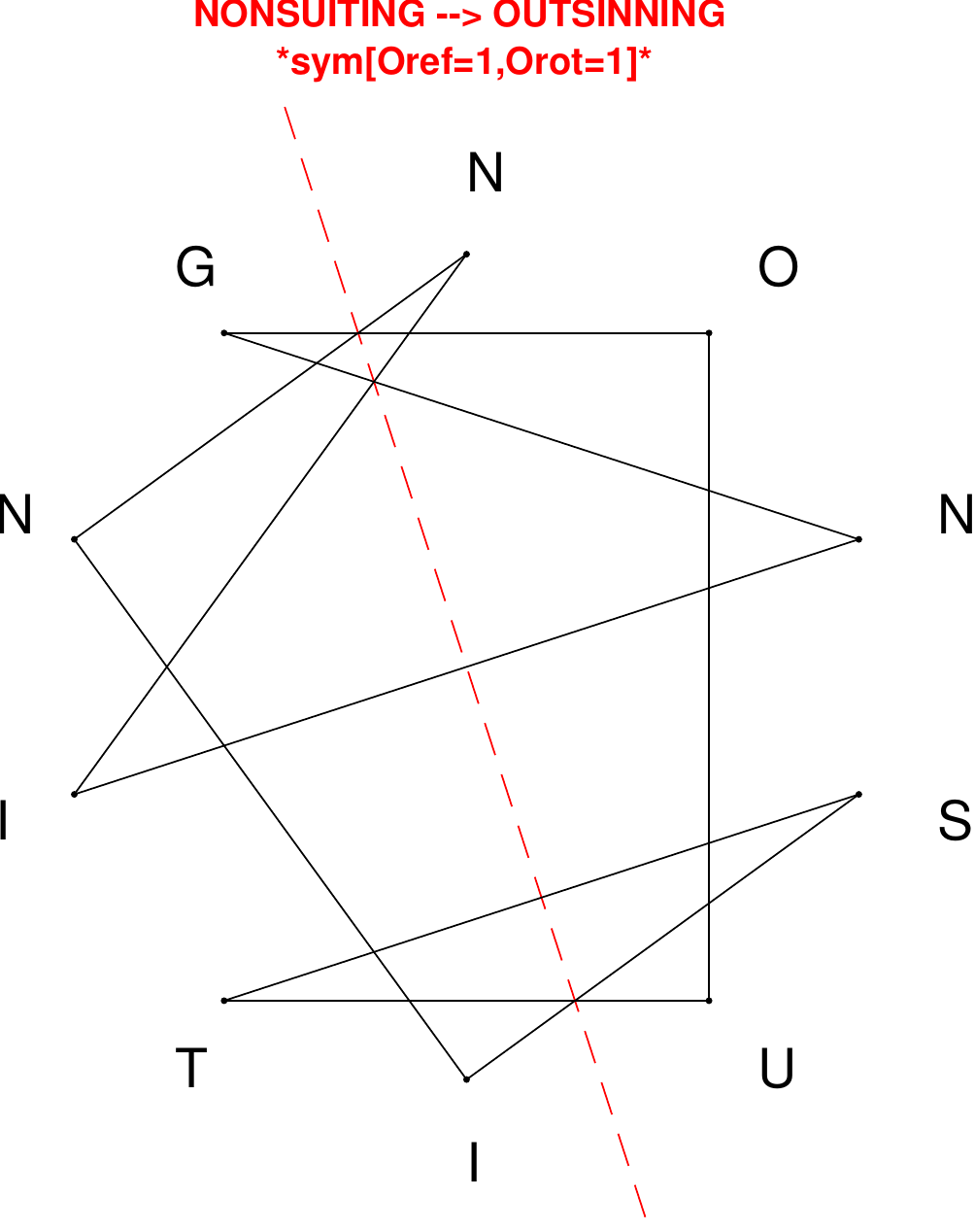}
\end{subfigure}
\hfill
\begin{subfigure}[T]{0.19\textwidth}
\centering
\includegraphics[width=\textwidth]{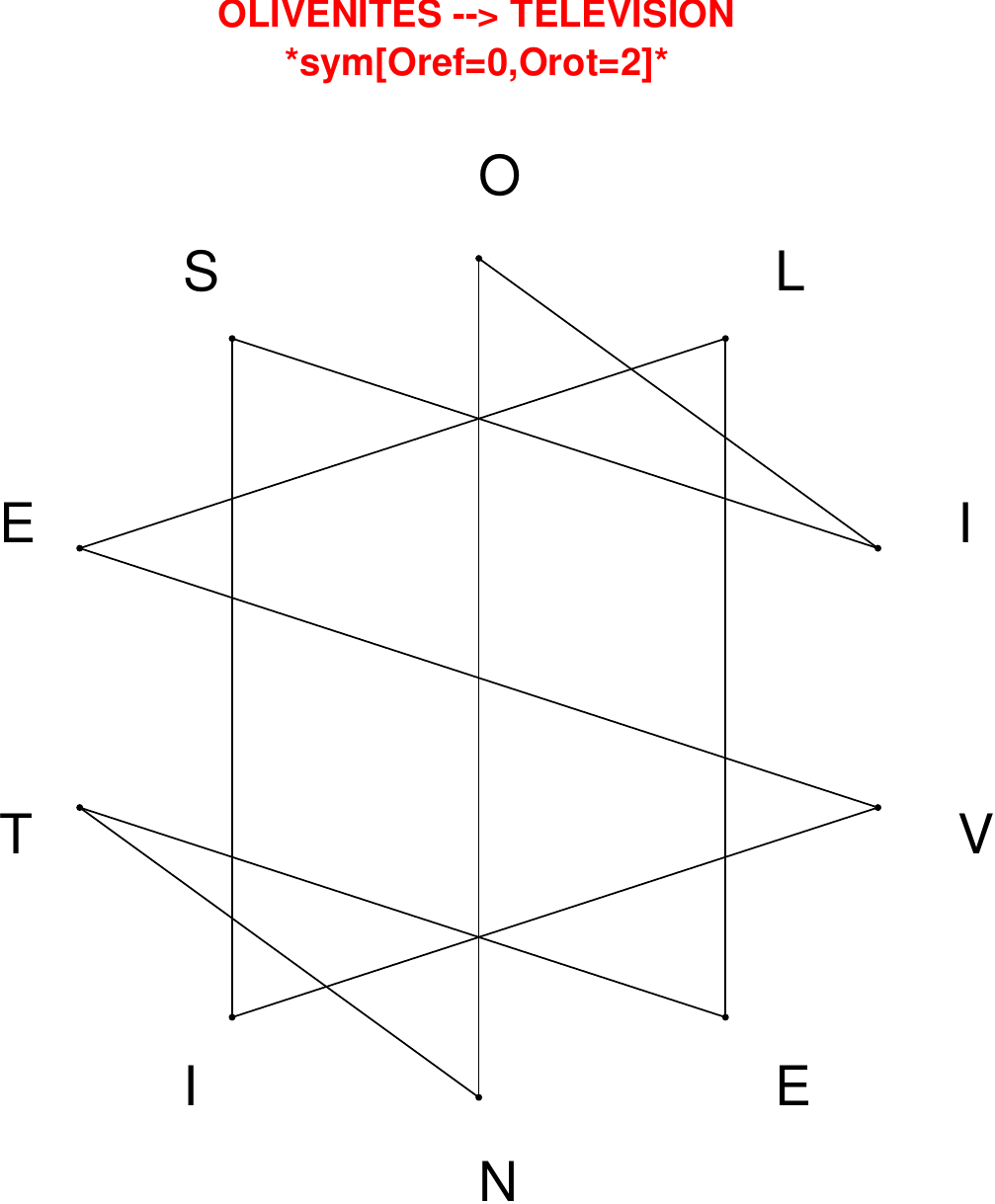}
\end{subfigure}
\hfill
\begin{subfigure}[T]{0.19\textwidth}
\centering
\includegraphics[width=\textwidth]{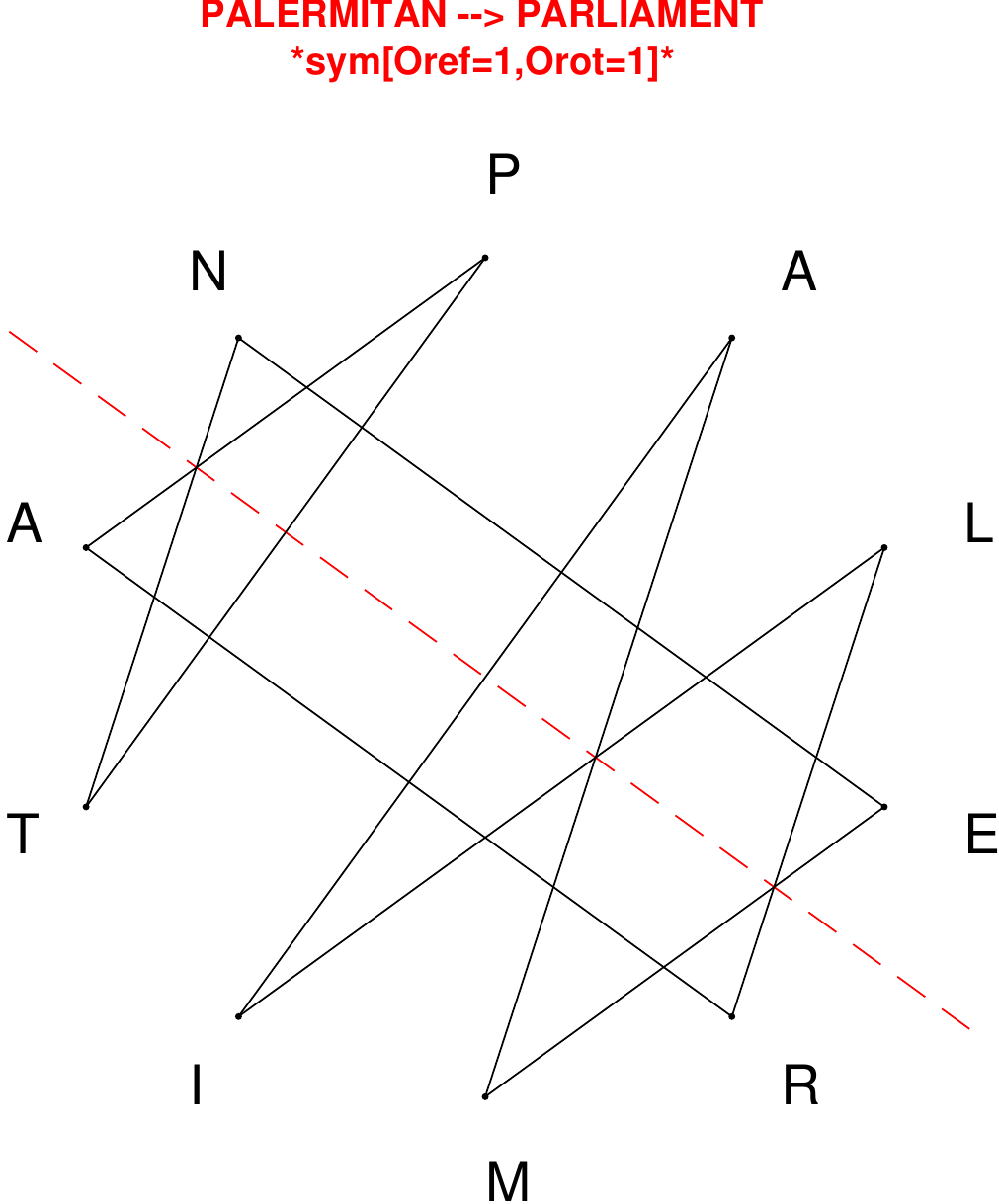}
\end{subfigure}
\hfill
\begin{subfigure}[T]{0.19\textwidth}
\centering
\includegraphics[width=\textwidth]{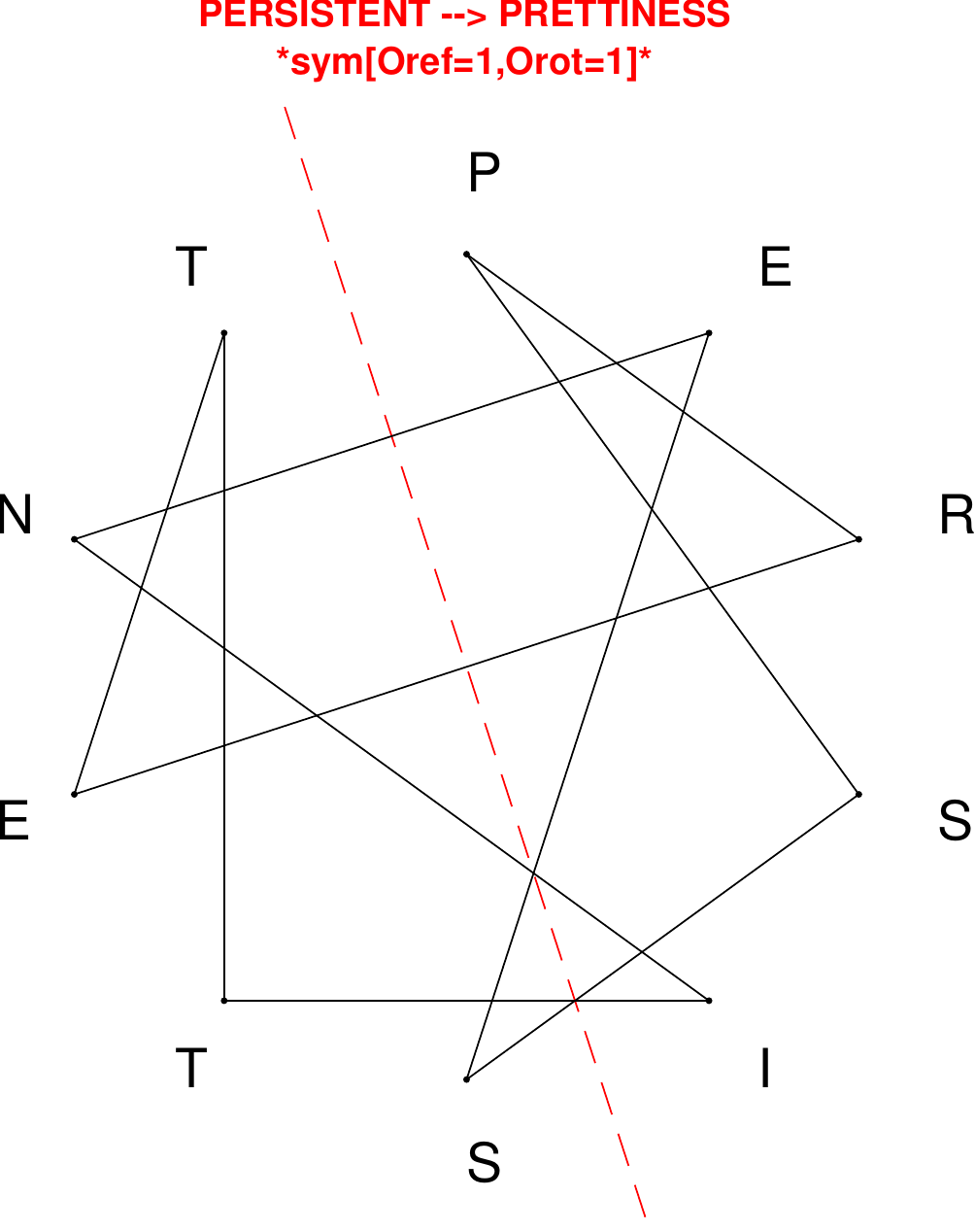}
\end{subfigure}
\end{figure}

\begin{figure}[H]
\centering
\begin{subfigure}[T]{0.19\textwidth}
\centering
\includegraphics[width=\textwidth]{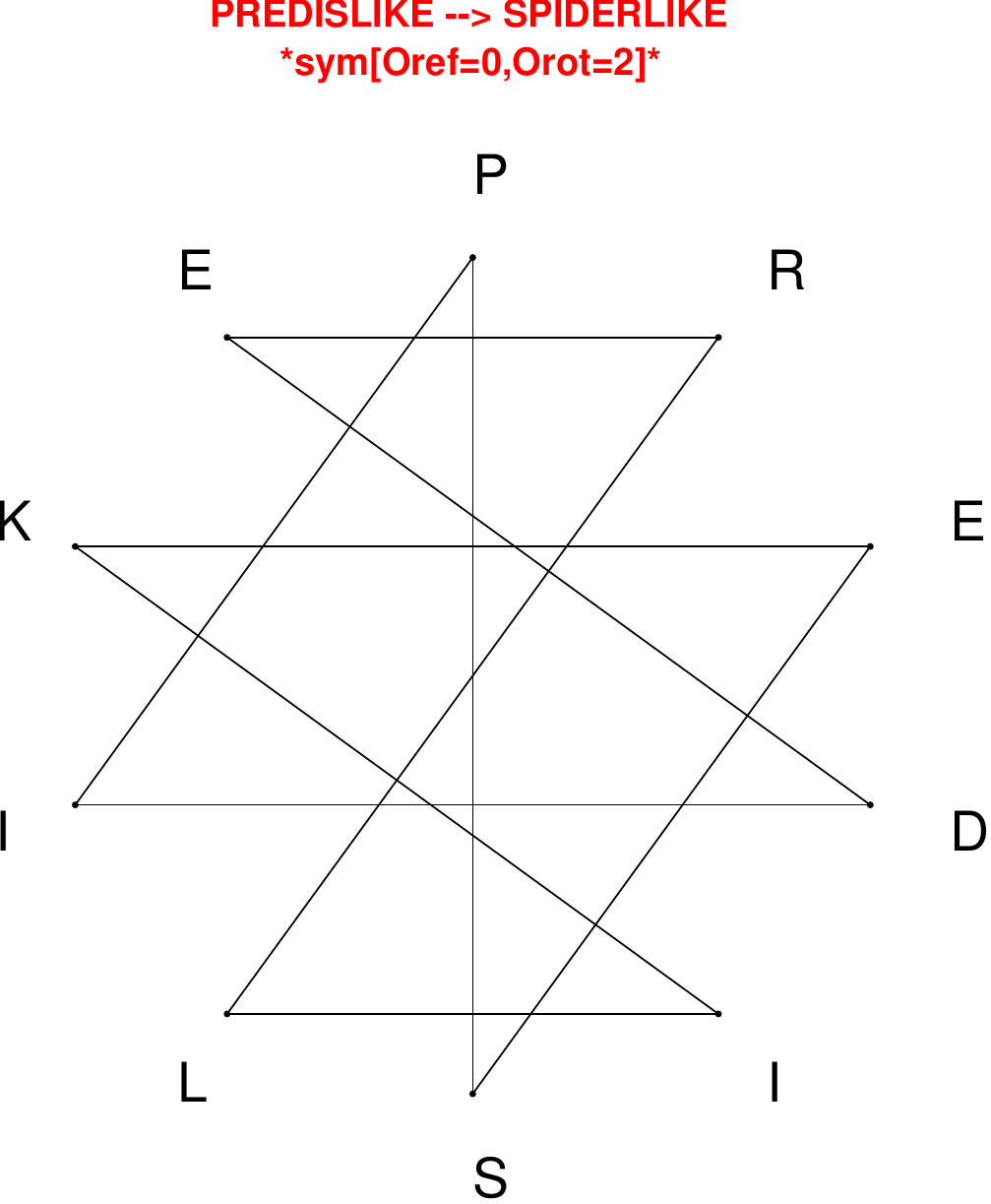}
\end{subfigure}
\hfill
\begin{subfigure}[T]{0.19\textwidth}
\centering
\includegraphics[width=\textwidth]{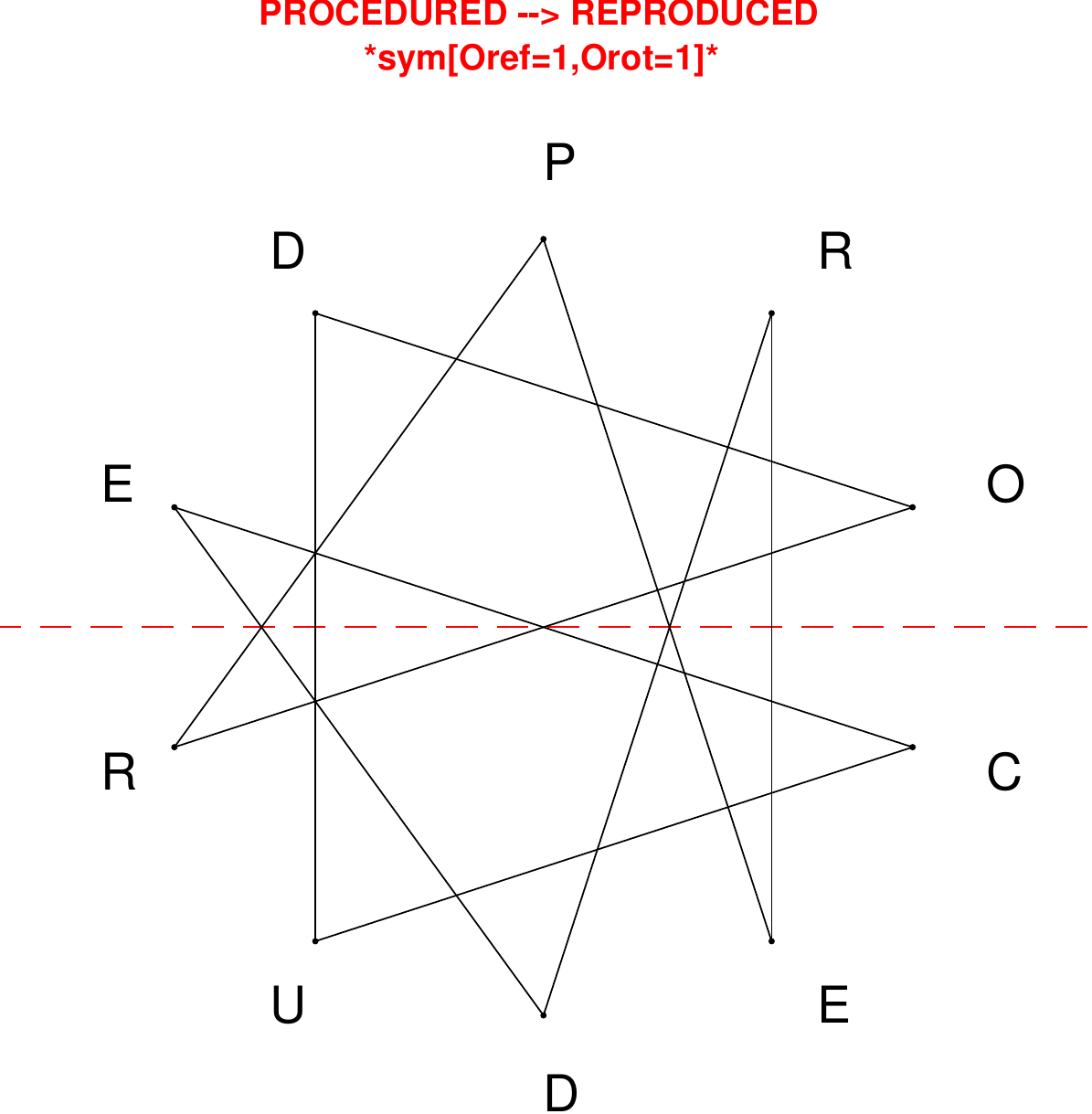}
\end{subfigure}
\hfill
\begin{subfigure}[T]{0.19\textwidth}
\centering
\includegraphics[width=\textwidth]{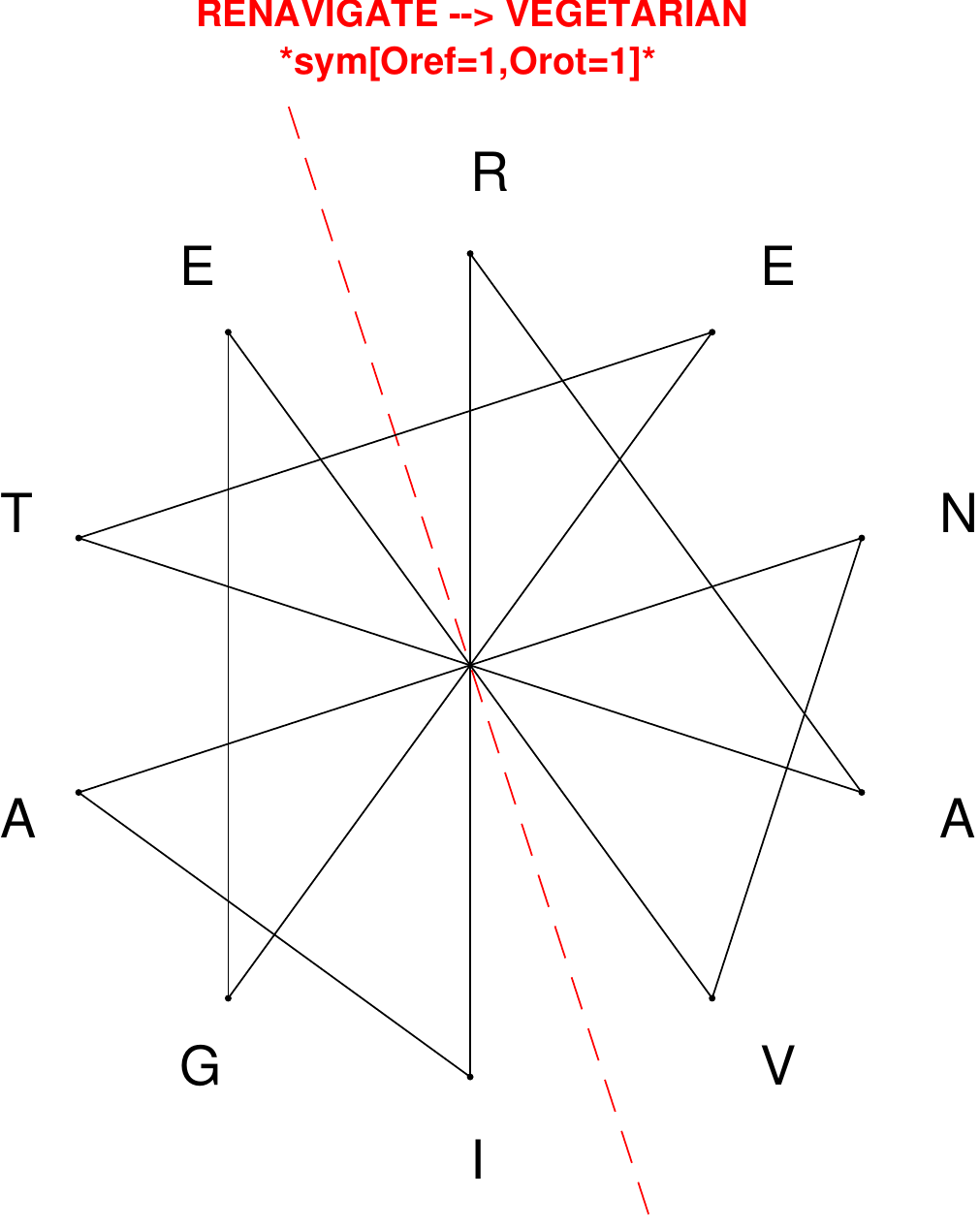}
\end{subfigure}
\hfill
\begin{subfigure}[T]{0.19\textwidth}
\centering
\includegraphics[width=\textwidth]{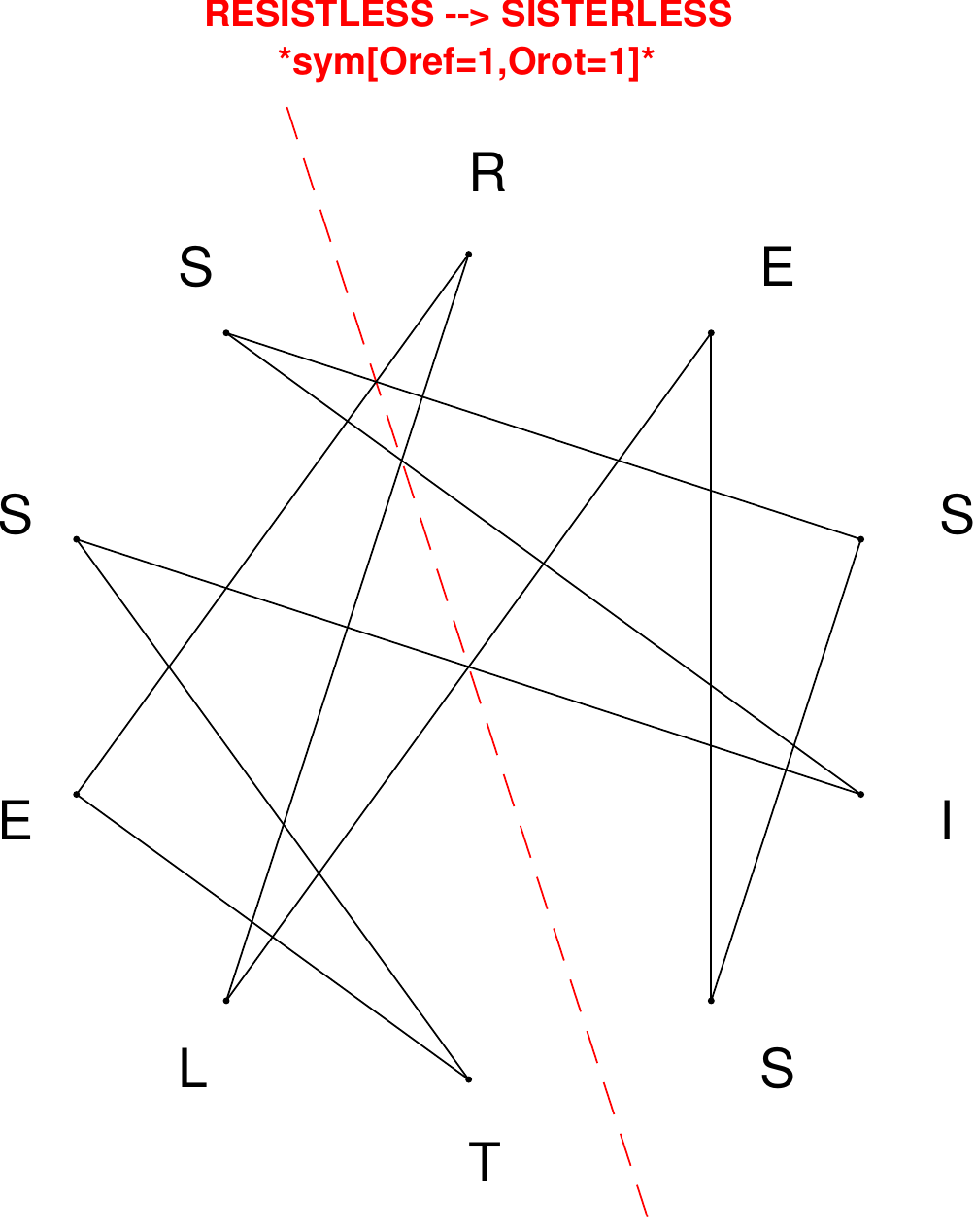}
\end{subfigure}
\hfill
\begin{subfigure}[T]{0.19\textwidth}
\centering
\includegraphics[width=\textwidth]{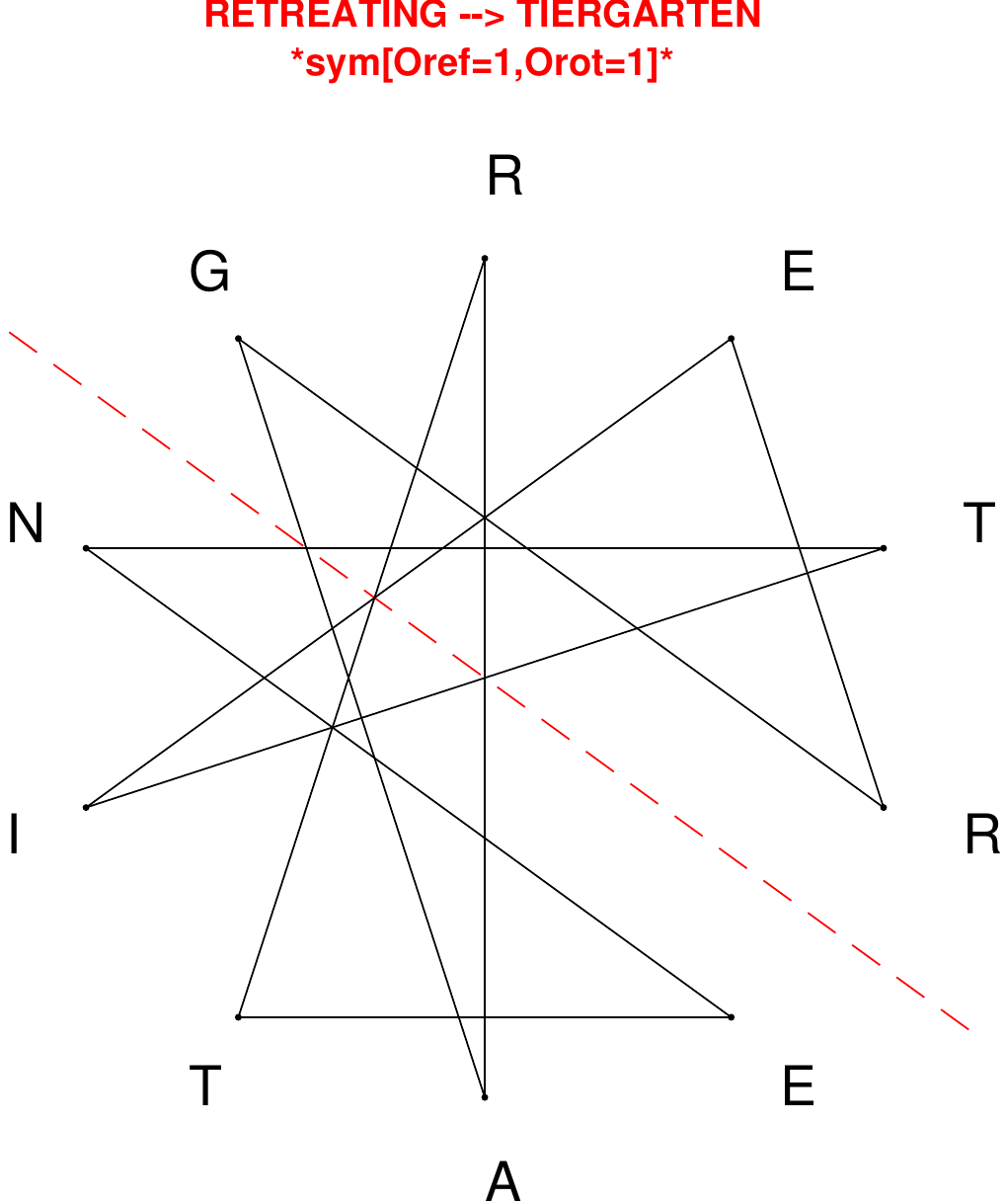}
\end{subfigure}
\end{figure}

\begin{figure}[H]
\centering
\begin{subfigure}[T]{0.19\textwidth}
\centering
\includegraphics[width=\textwidth]{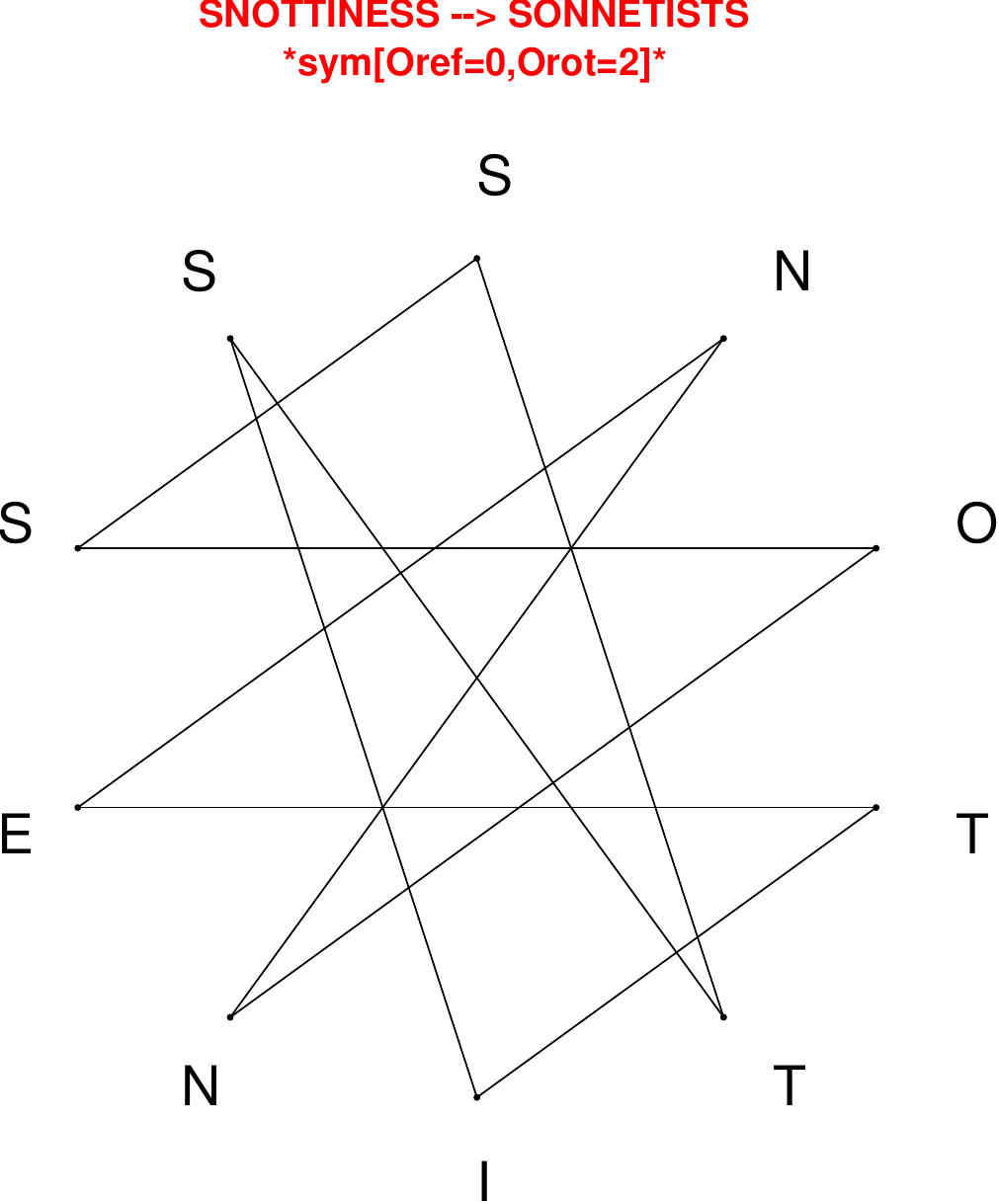}
\end{subfigure}
\hfill
\begin{subfigure}[T]{0.19\textwidth}
\centering
\includegraphics[width=\textwidth]{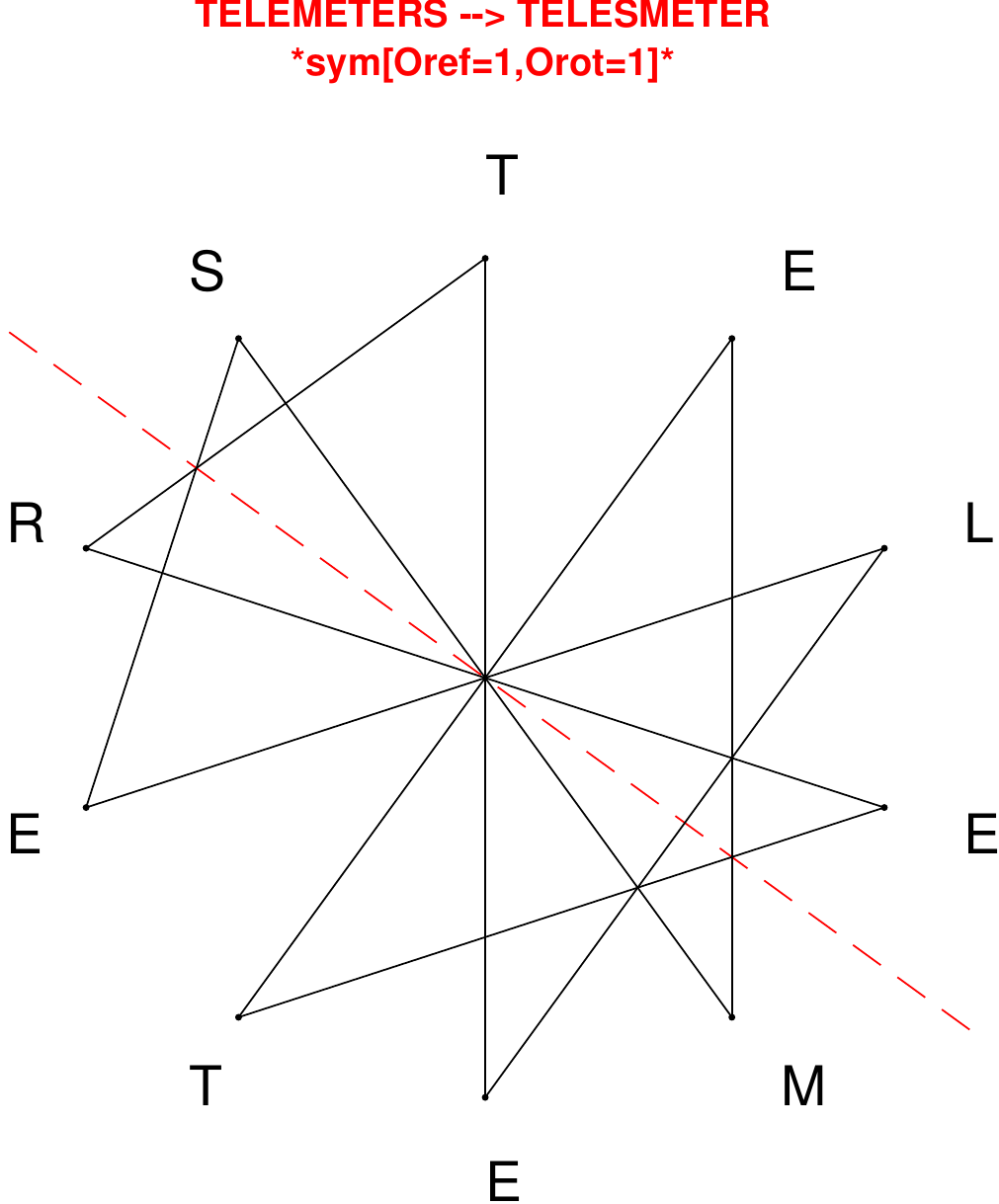}
\end{subfigure}
\hfill
\end{figure}

\subsubsection{Asymmetric Stars $N=10$}

\begin{figure}[H]
\centering
\begin{subfigure}[T]{0.19\textwidth}
\centering
\includegraphics[width=\textwidth]{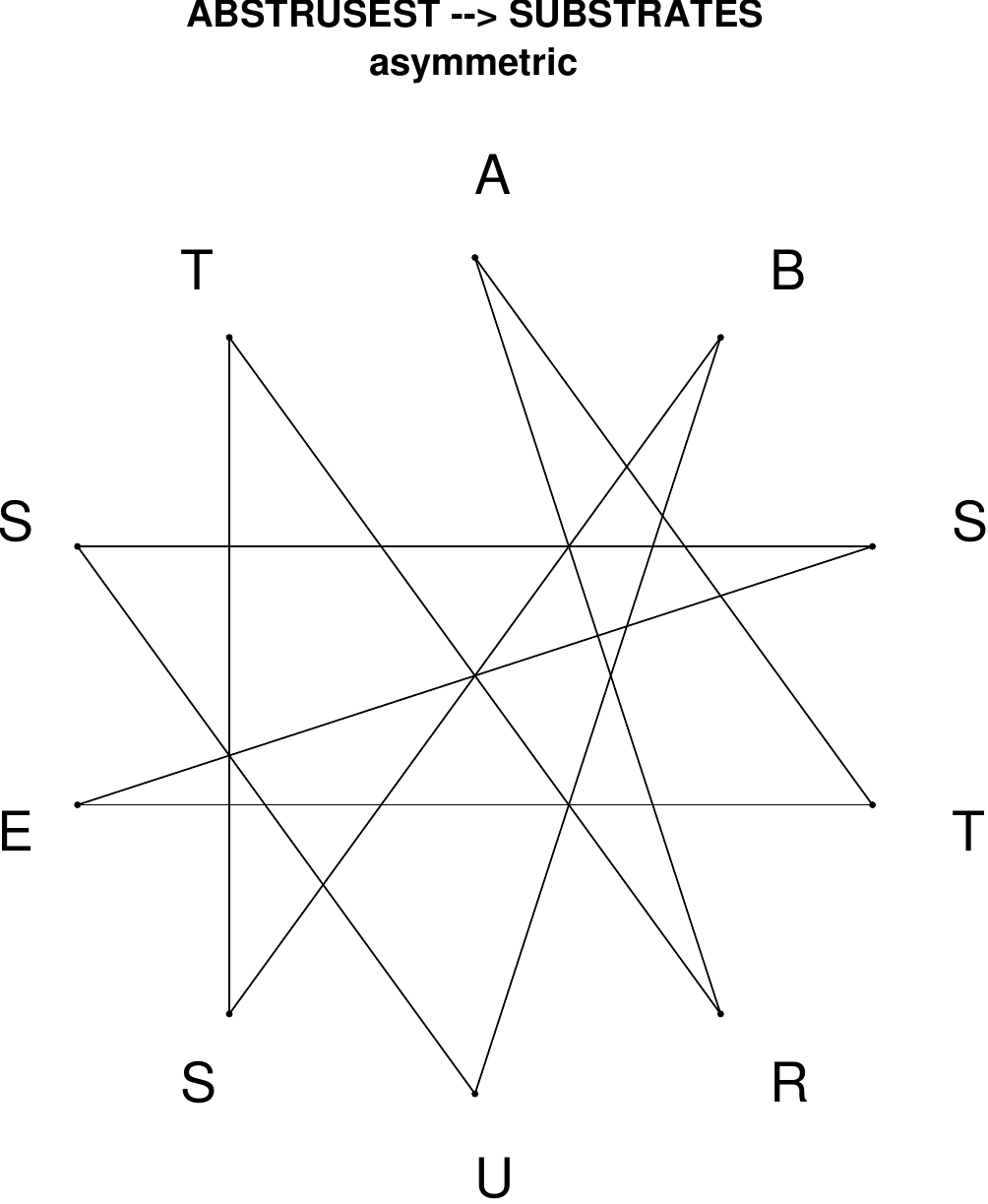}
\end{subfigure}
\hfill
\begin{subfigure}[T]{0.19\textwidth}
\centering
\includegraphics[width=\textwidth]{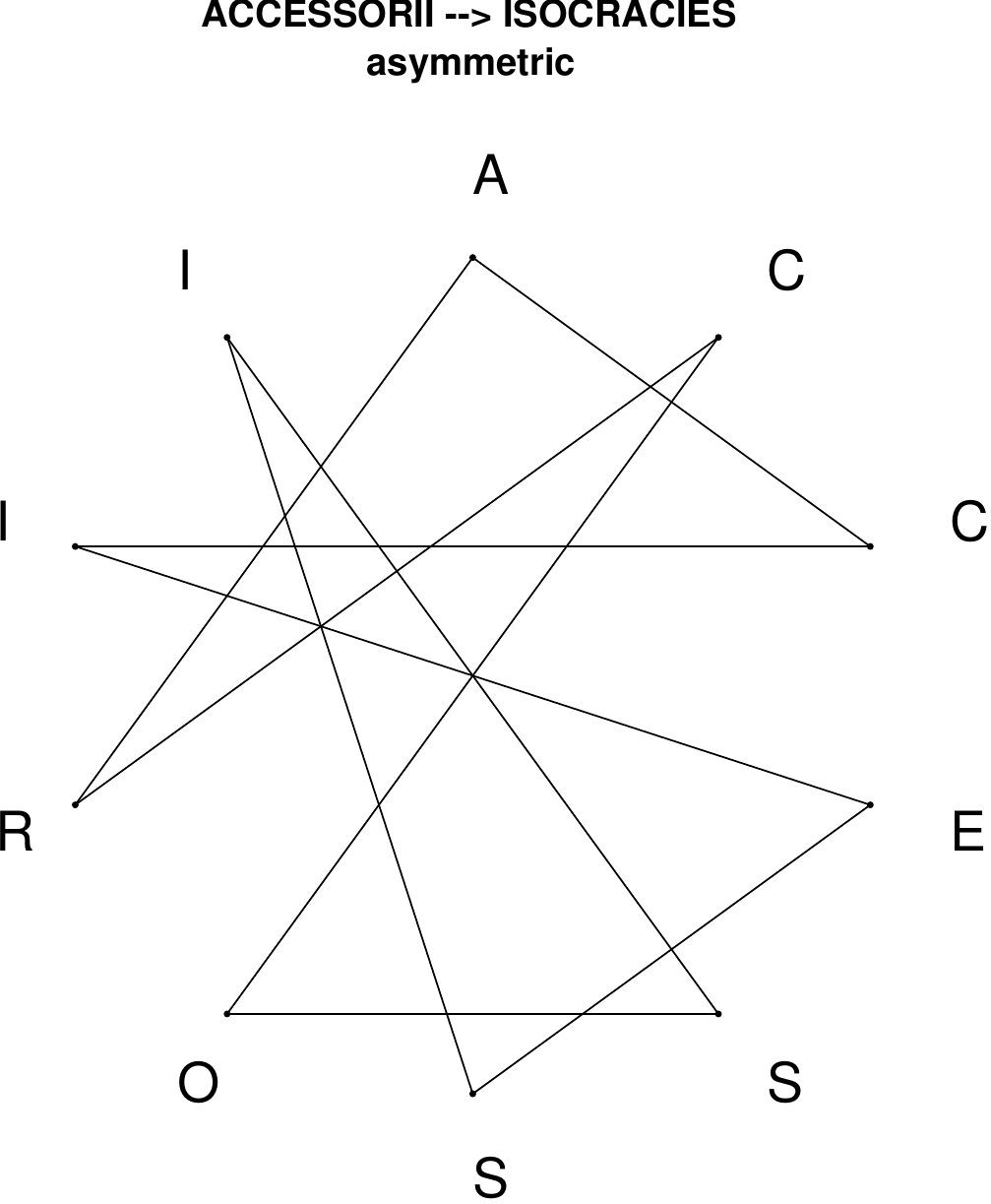}
\end{subfigure}
\hfill
\begin{subfigure}[T]{0.19\textwidth}
\centering
\includegraphics[width=\textwidth]{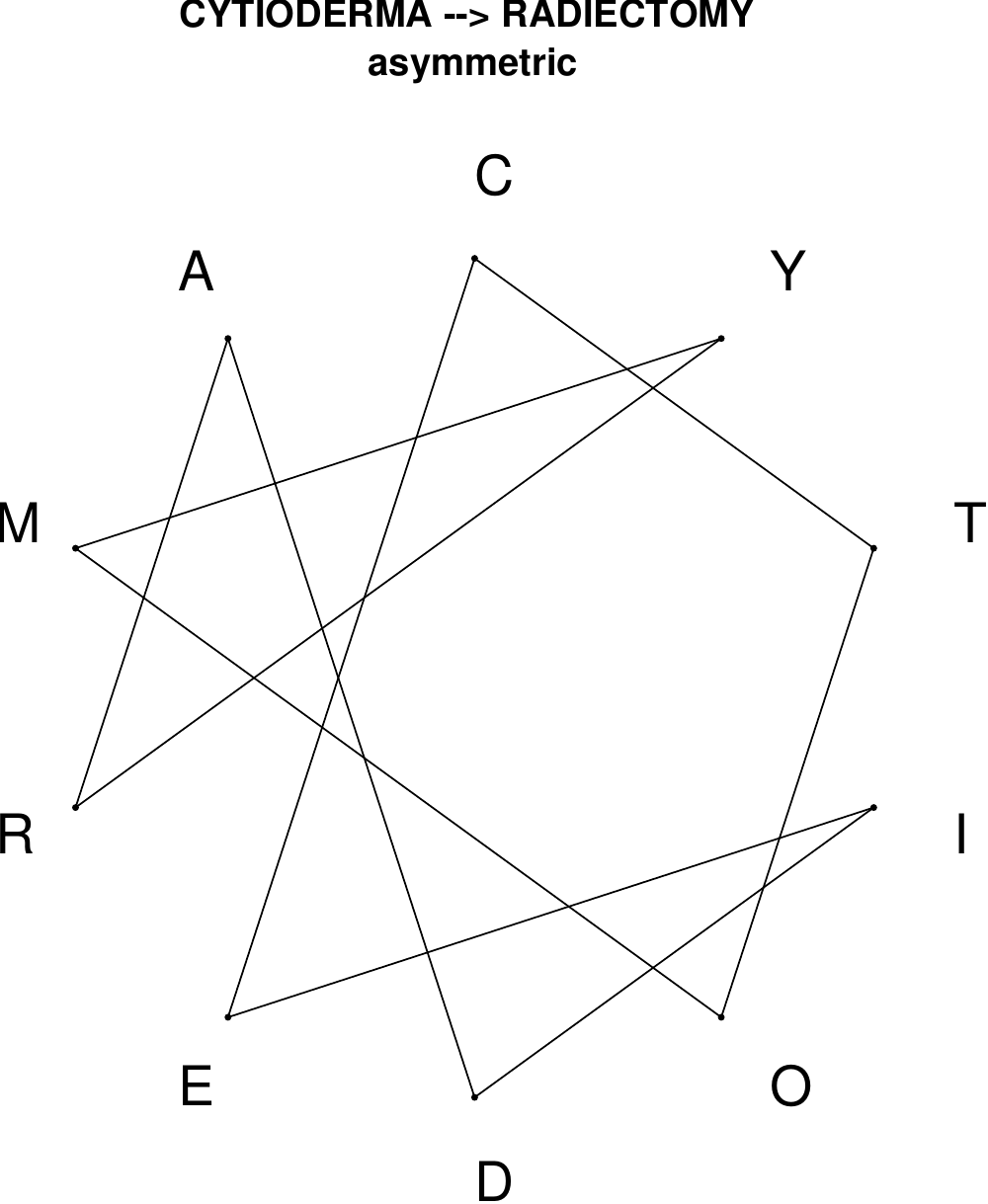}
\end{subfigure}
\hfill
\begin{subfigure}[T]{0.19\textwidth}
\centering
\includegraphics[width=\textwidth]{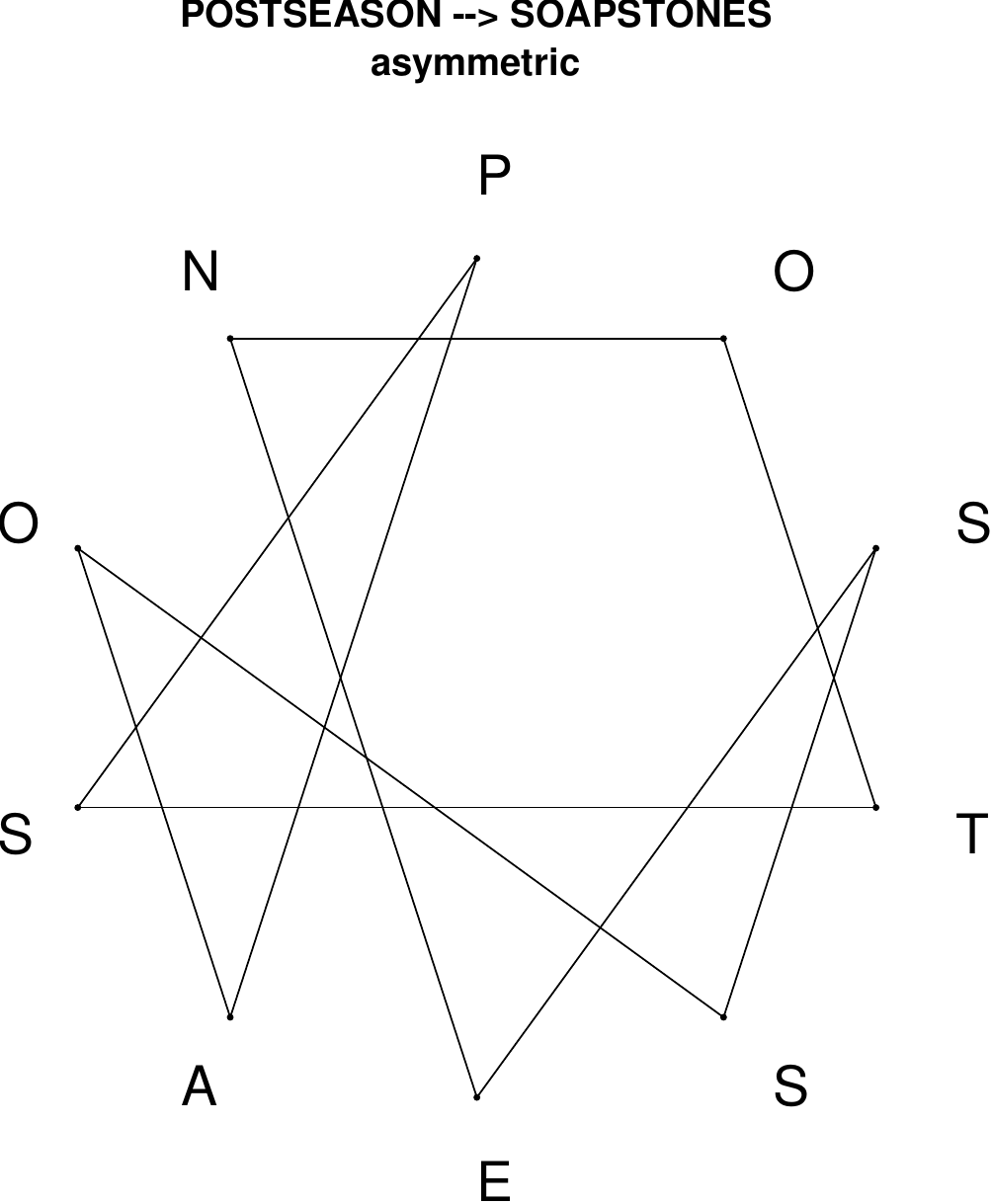}
\end{subfigure}
\hfill
\begin{subfigure}[T]{0.19\textwidth}
\centering
\includegraphics[width=\textwidth]{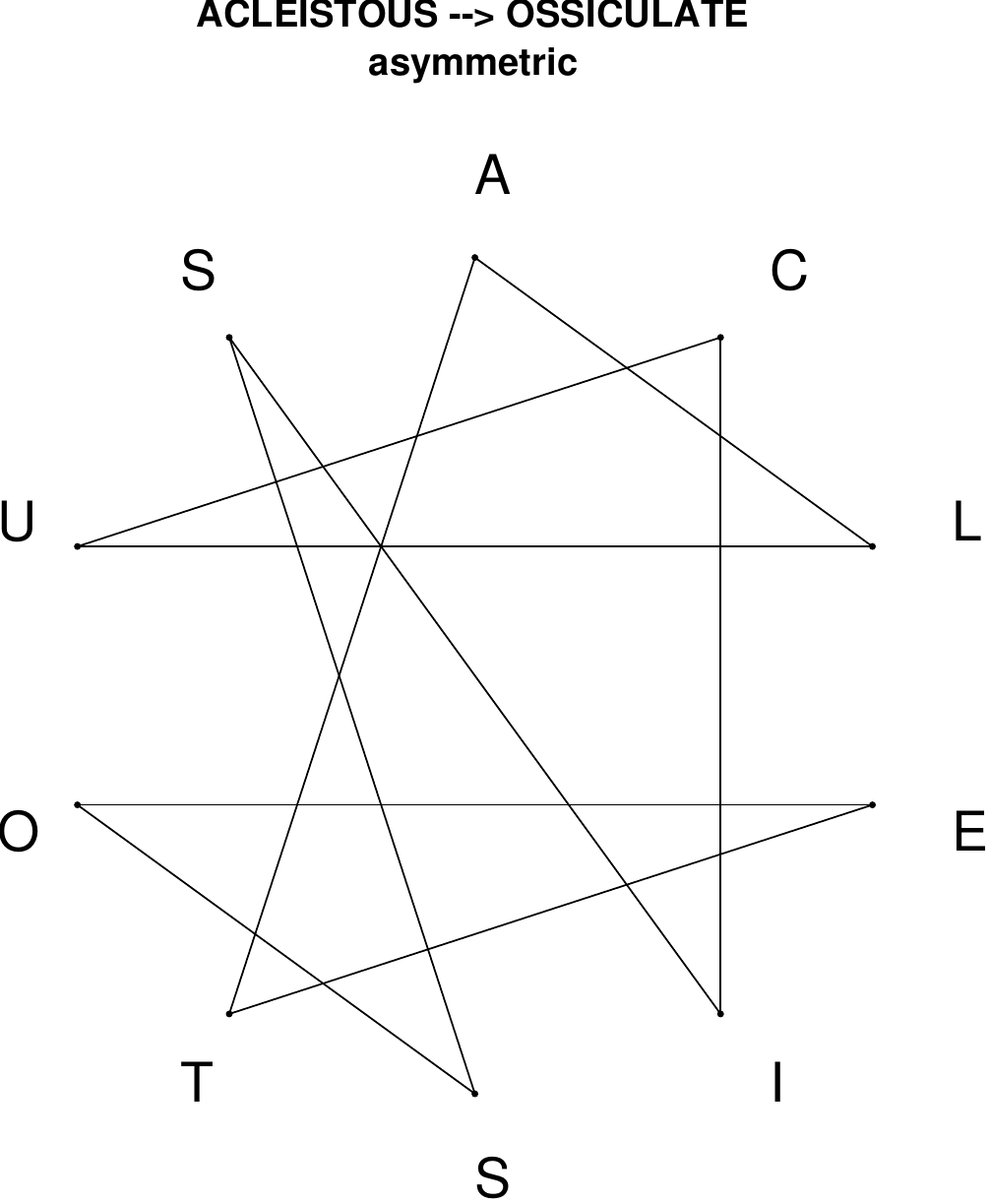}
\end{subfigure}
\end{figure}

\begin{figure}[H]
\centering
\begin{subfigure}[T]{0.19\textwidth}
\centering
\includegraphics[width=\textwidth]{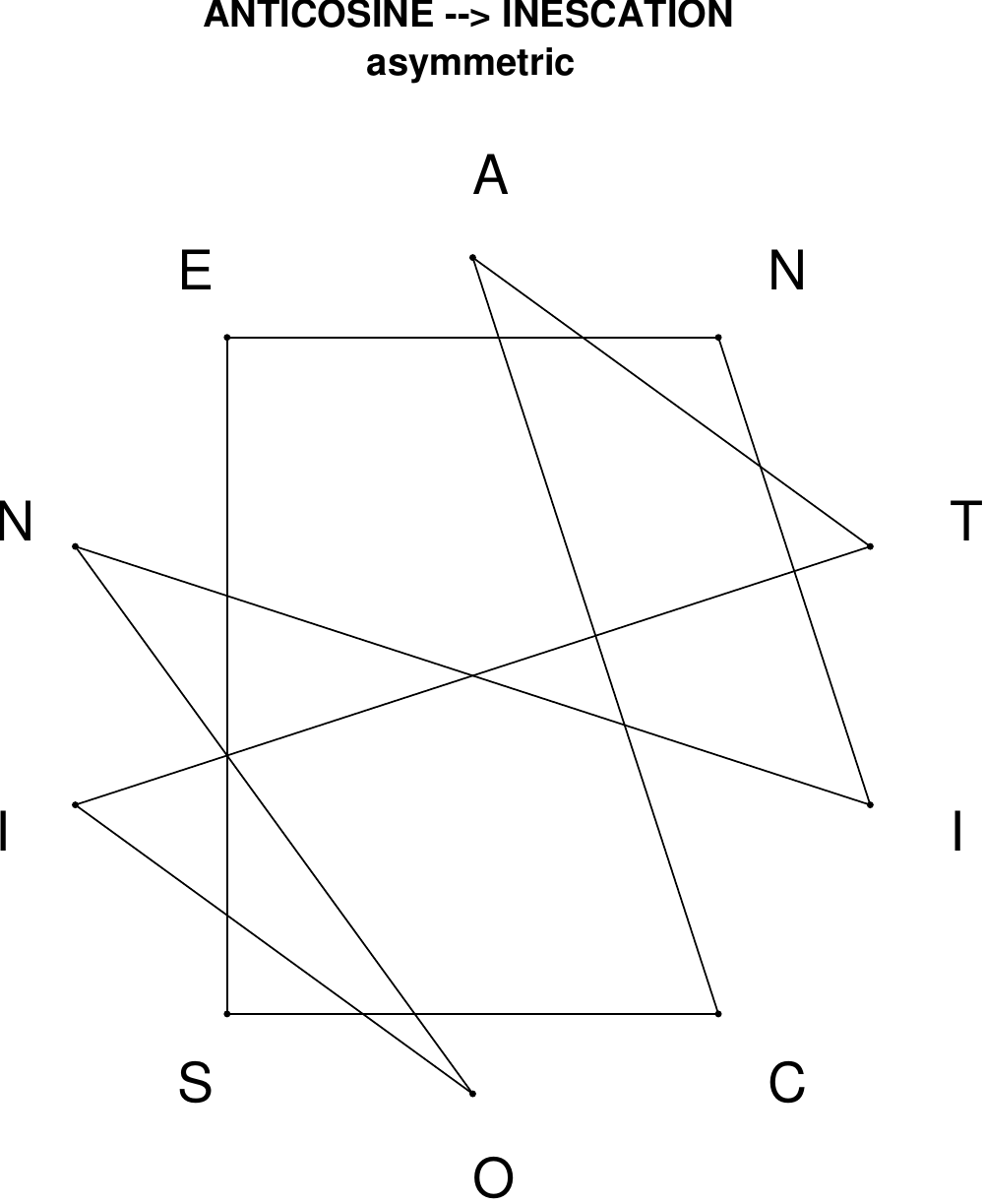}
\end{subfigure}
\hfill
\begin{subfigure}[T]{0.19\textwidth}
\centering
\includegraphics[width=\textwidth]{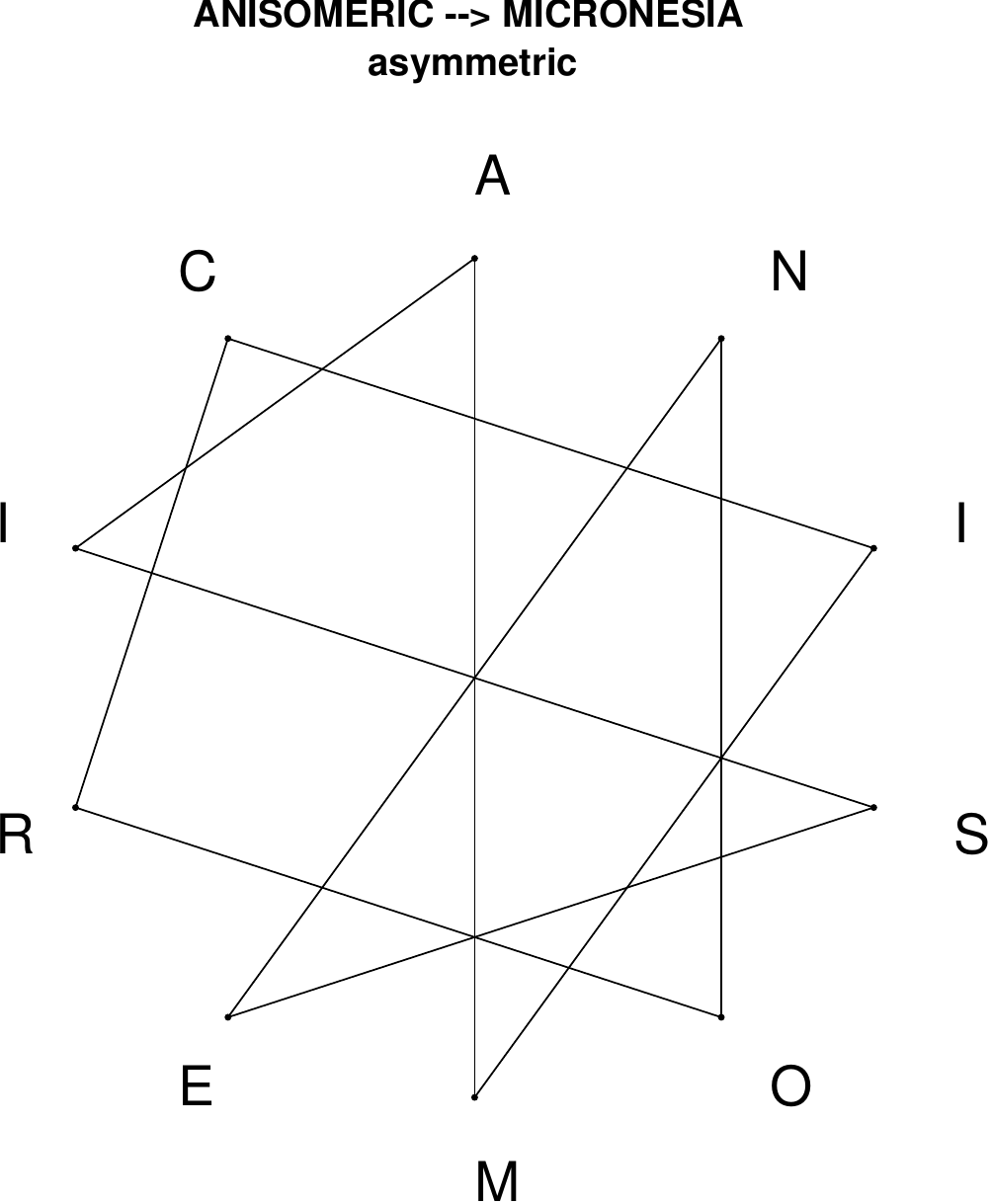}
\end{subfigure}
\hfill
\begin{subfigure}[T]{0.19\textwidth}
\centering
\includegraphics[width=\textwidth]{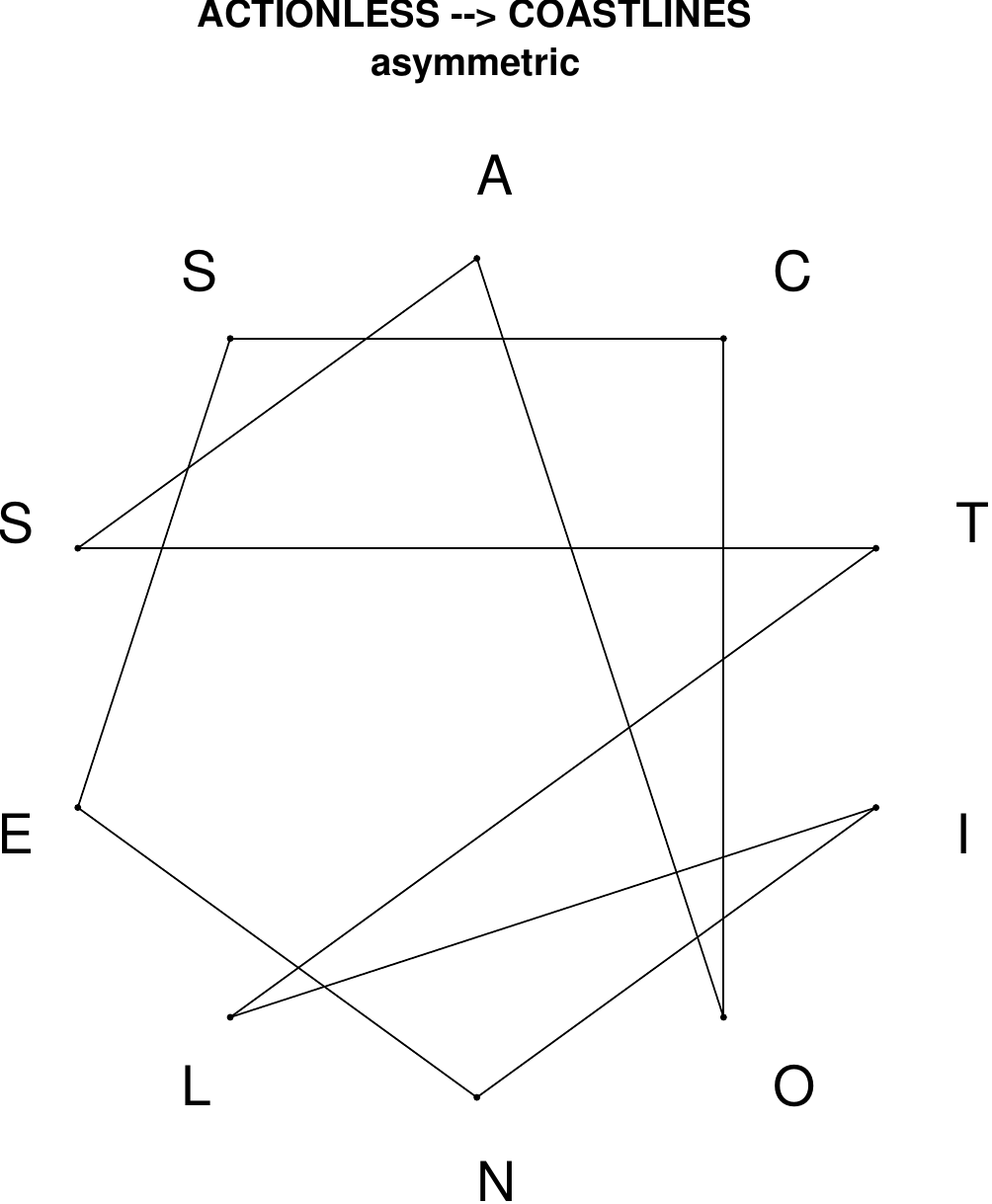}
\end{subfigure}
\hfill
\begin{subfigure}[T]{0.19\textwidth}
\centering
\includegraphics[width=\textwidth]{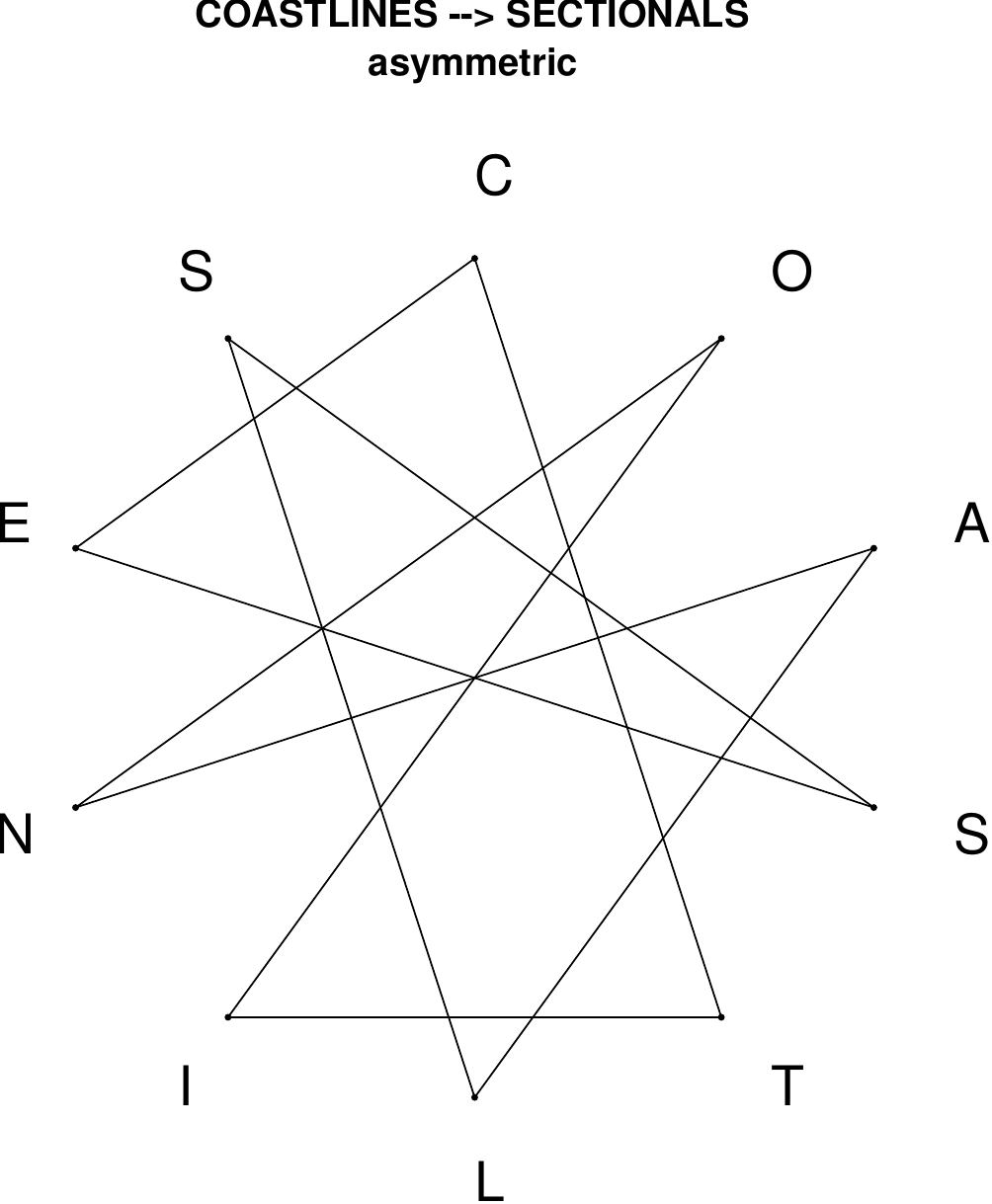}
\end{subfigure}
\hfill
\begin{subfigure}[T]{0.19\textwidth}
\centering
\includegraphics[width=\textwidth]{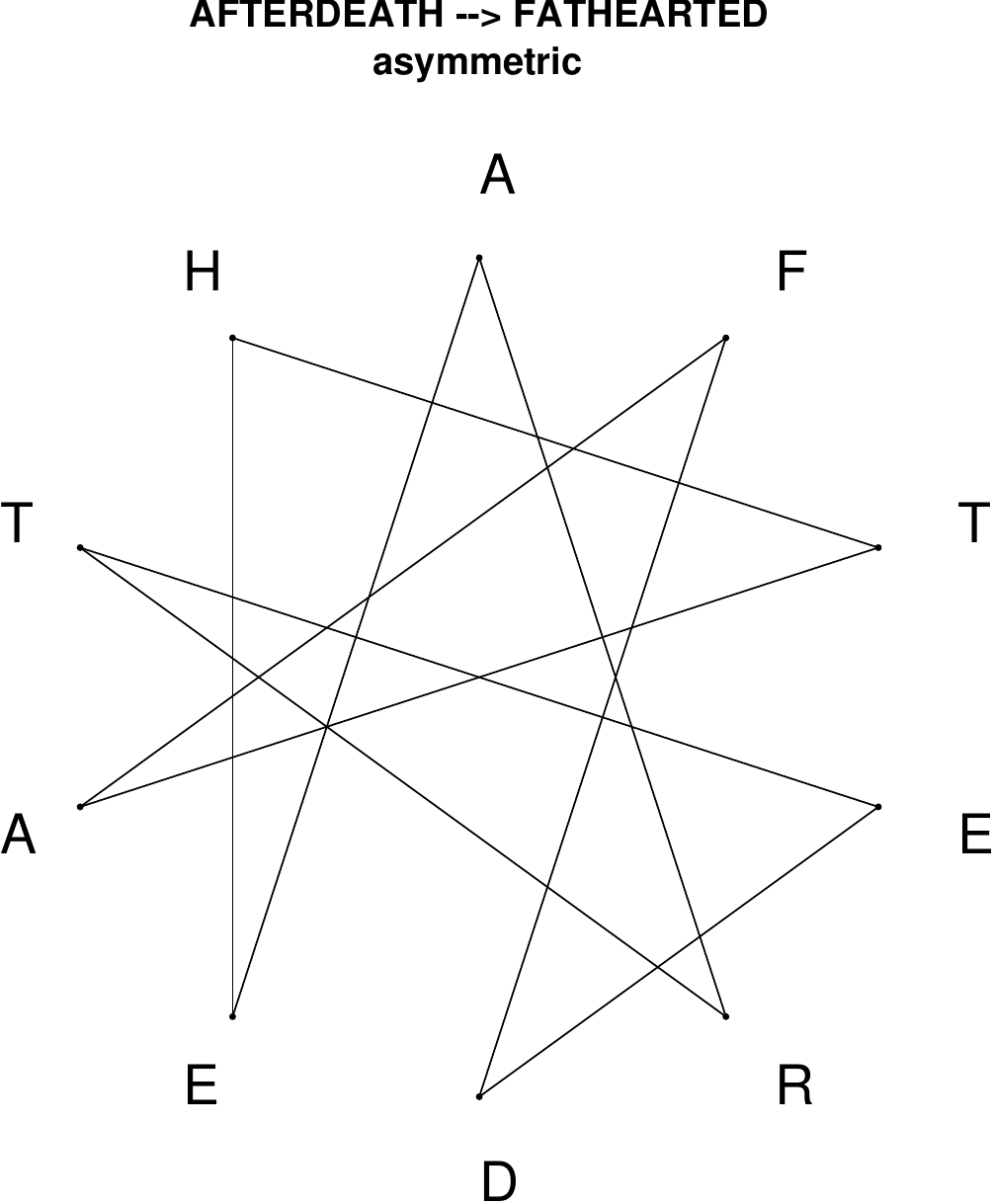}
\end{subfigure}
\end{figure}

\begin{figure}[H]
\centering
\begin{subfigure}[T]{0.19\textwidth}
\centering
\includegraphics[width=\textwidth]{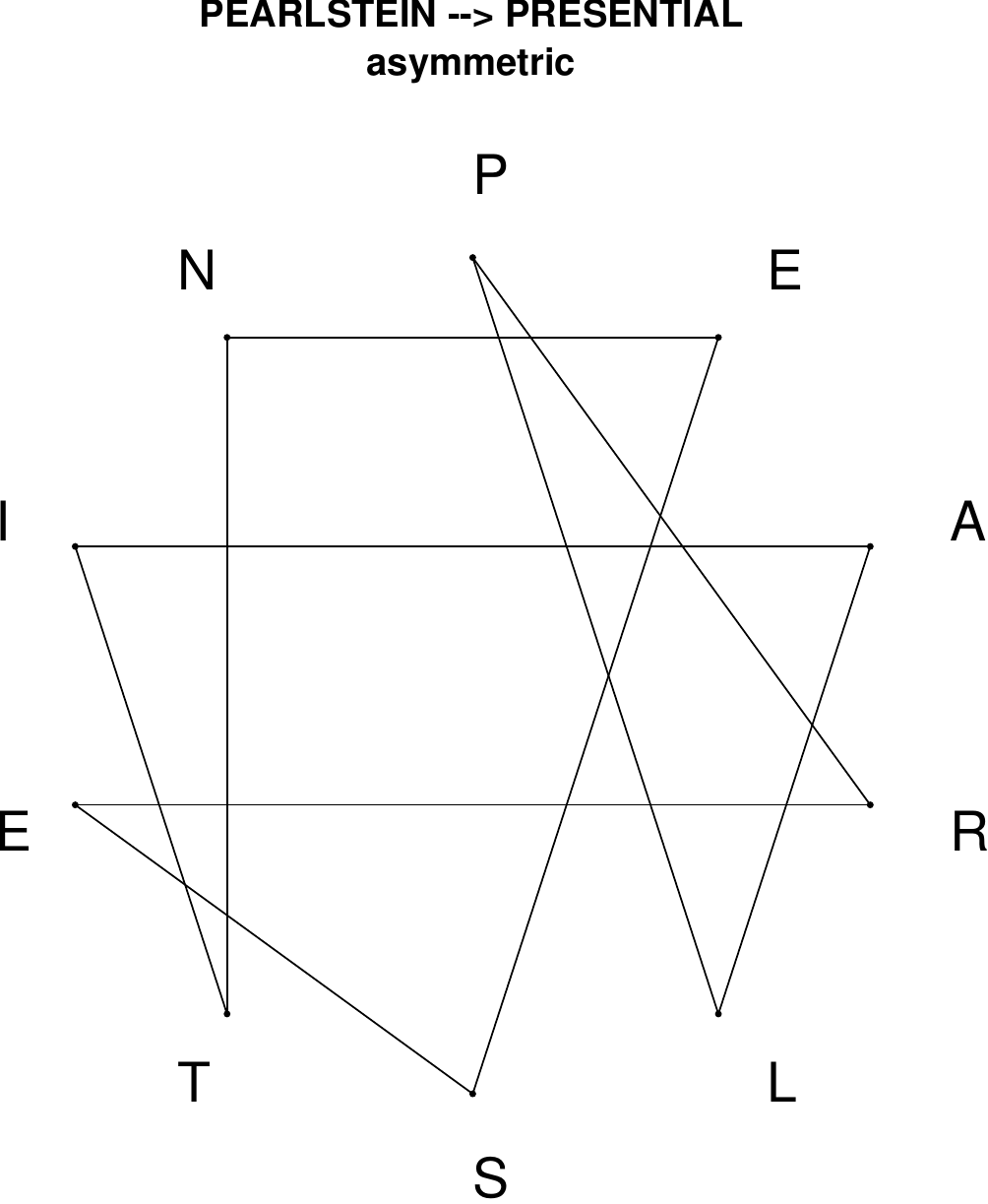}
\end{subfigure}
\hfill
\begin{subfigure}[T]{0.19\textwidth}
\centering
\includegraphics[width=\textwidth]{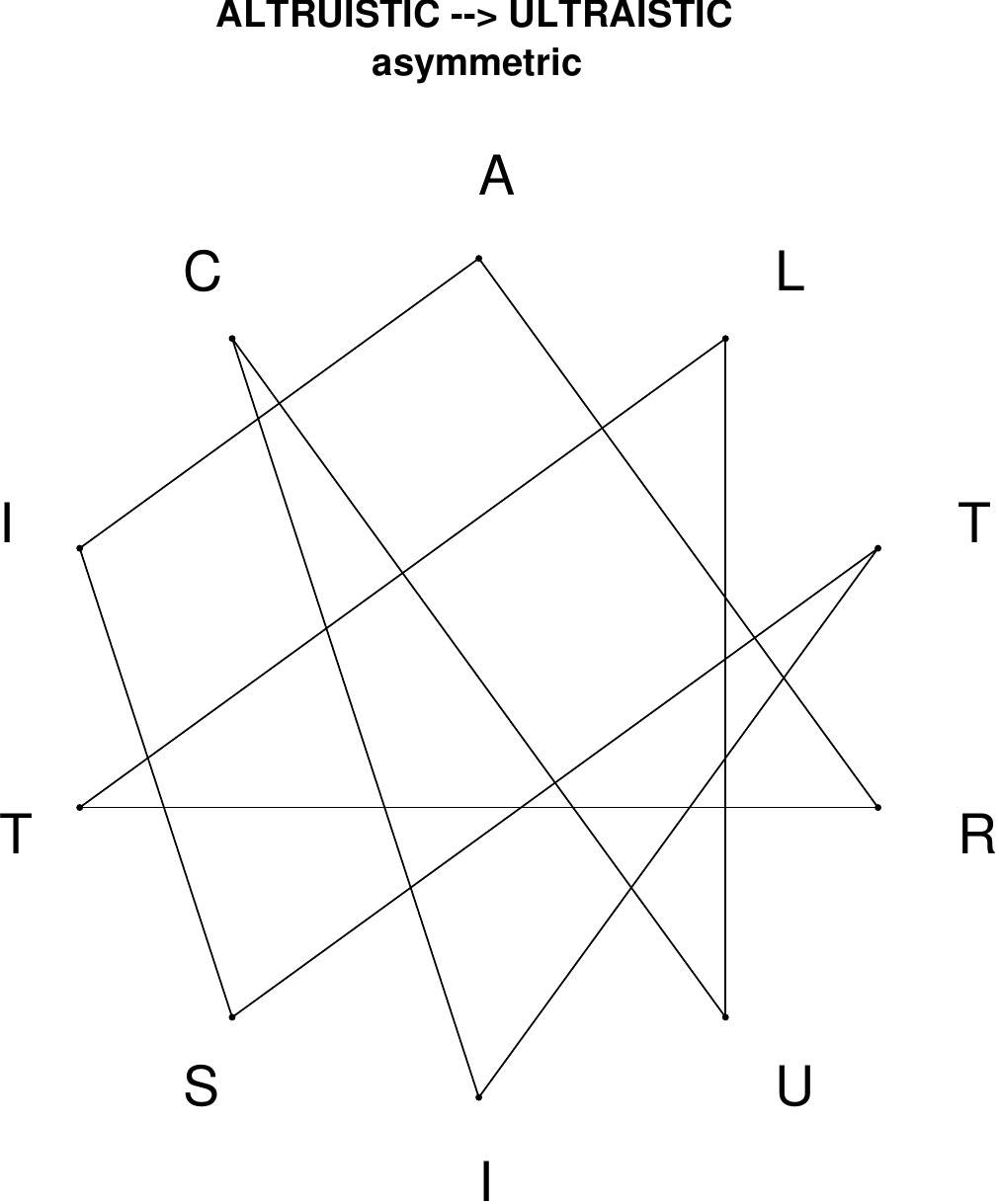}
\end{subfigure}
\hfill
\begin{subfigure}[T]{0.19\textwidth}
\centering
\includegraphics[width=\textwidth]{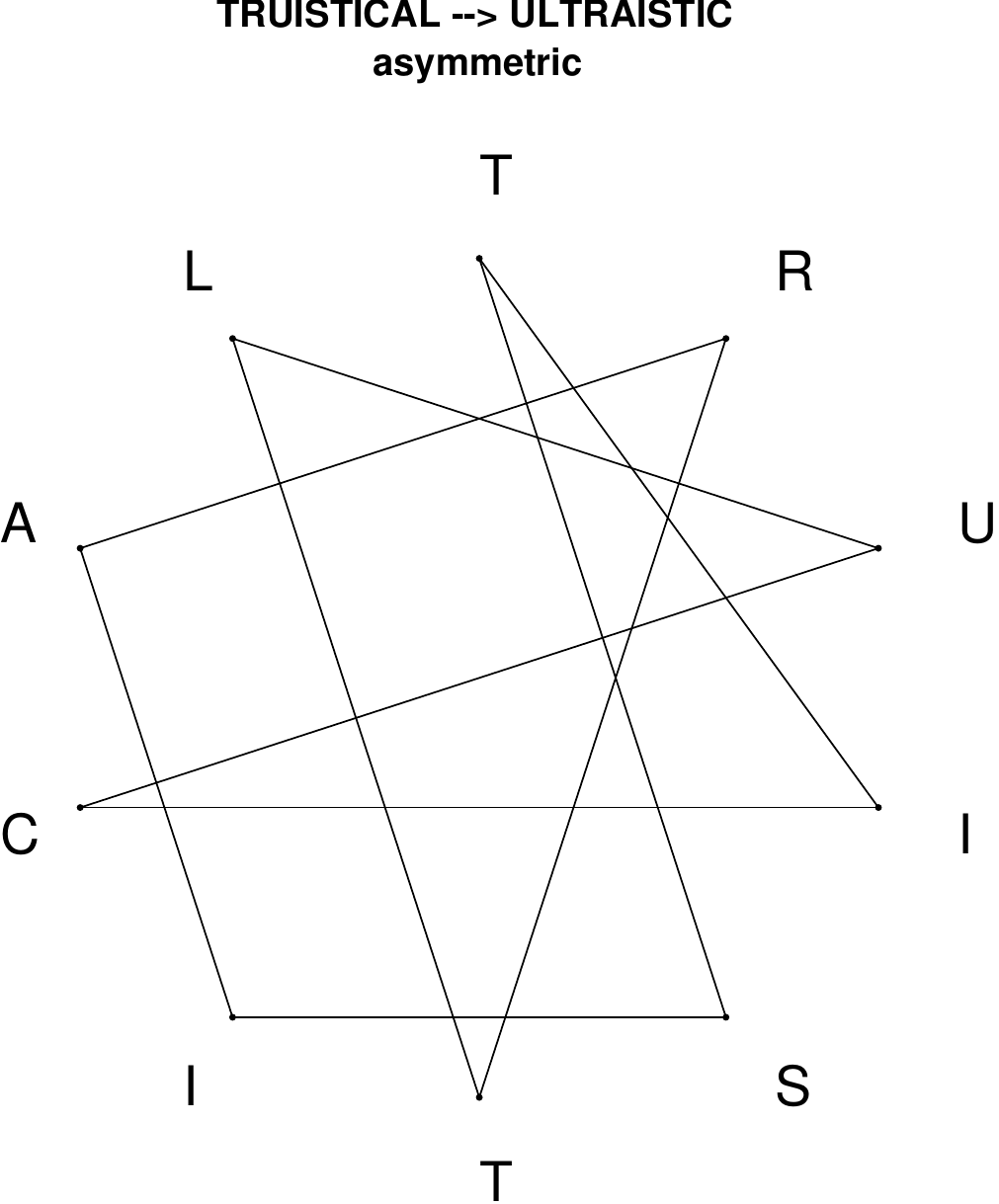}
\end{subfigure}
\hfill
\begin{subfigure}[T]{0.19\textwidth}
\centering
\includegraphics[width=\textwidth]{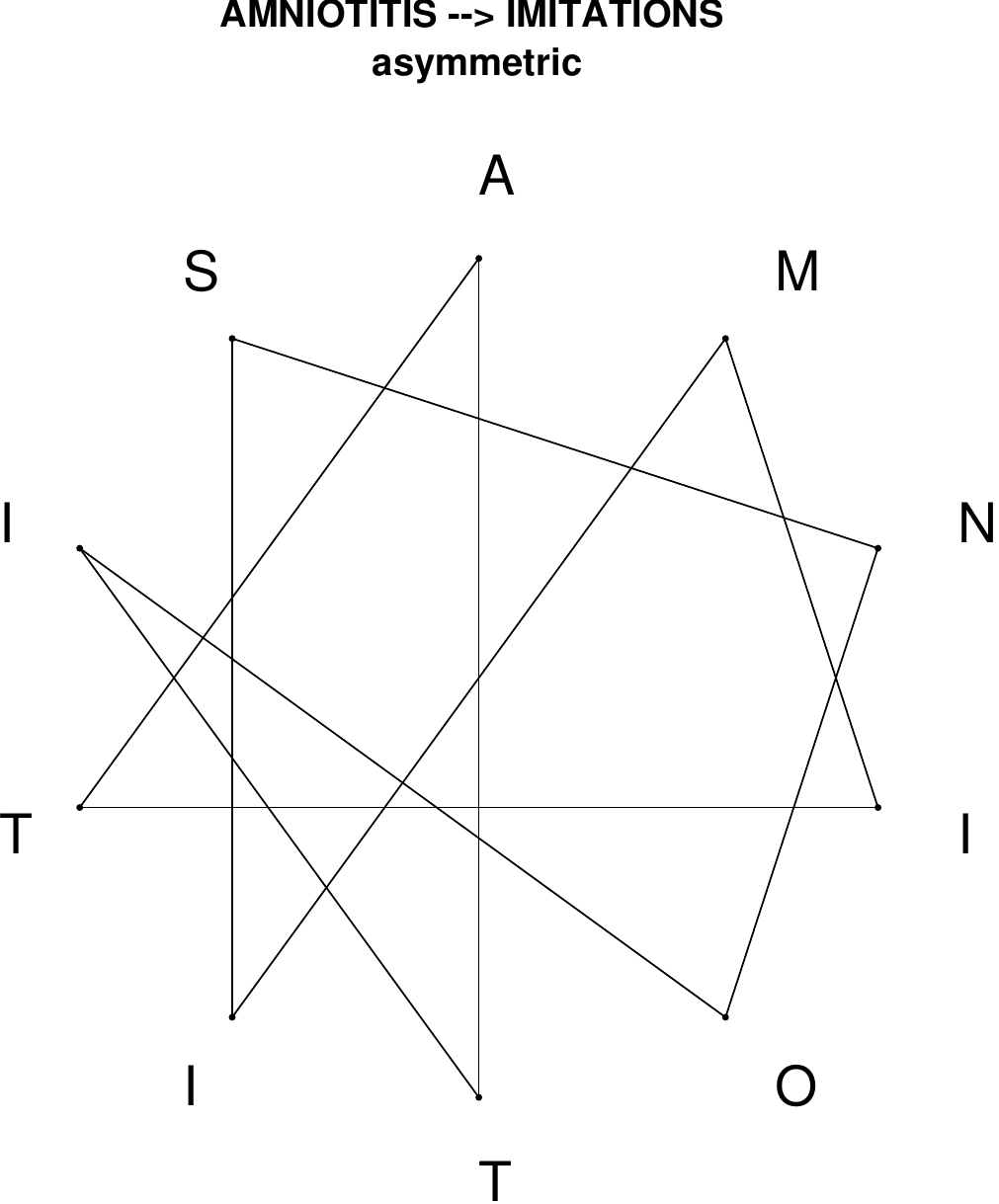}
\end{subfigure}
\hfill
\begin{subfigure}[T]{0.19\textwidth}
\centering
\includegraphics[width=\textwidth]{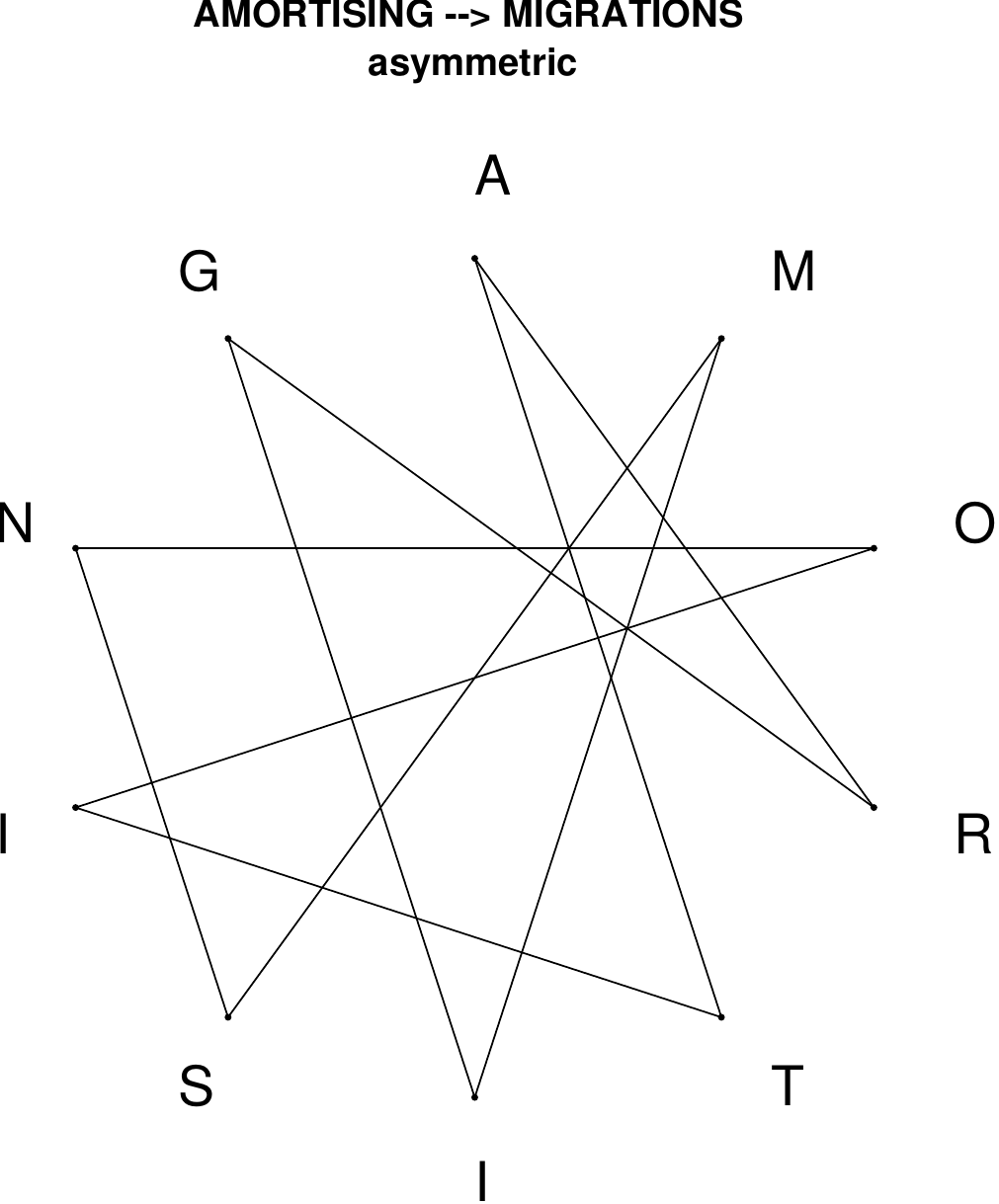}
\end{subfigure}
\end{figure}

\begin{figure}[H]
\centering
\begin{subfigure}[T]{0.19\textwidth}
\centering
\includegraphics[width=\textwidth]{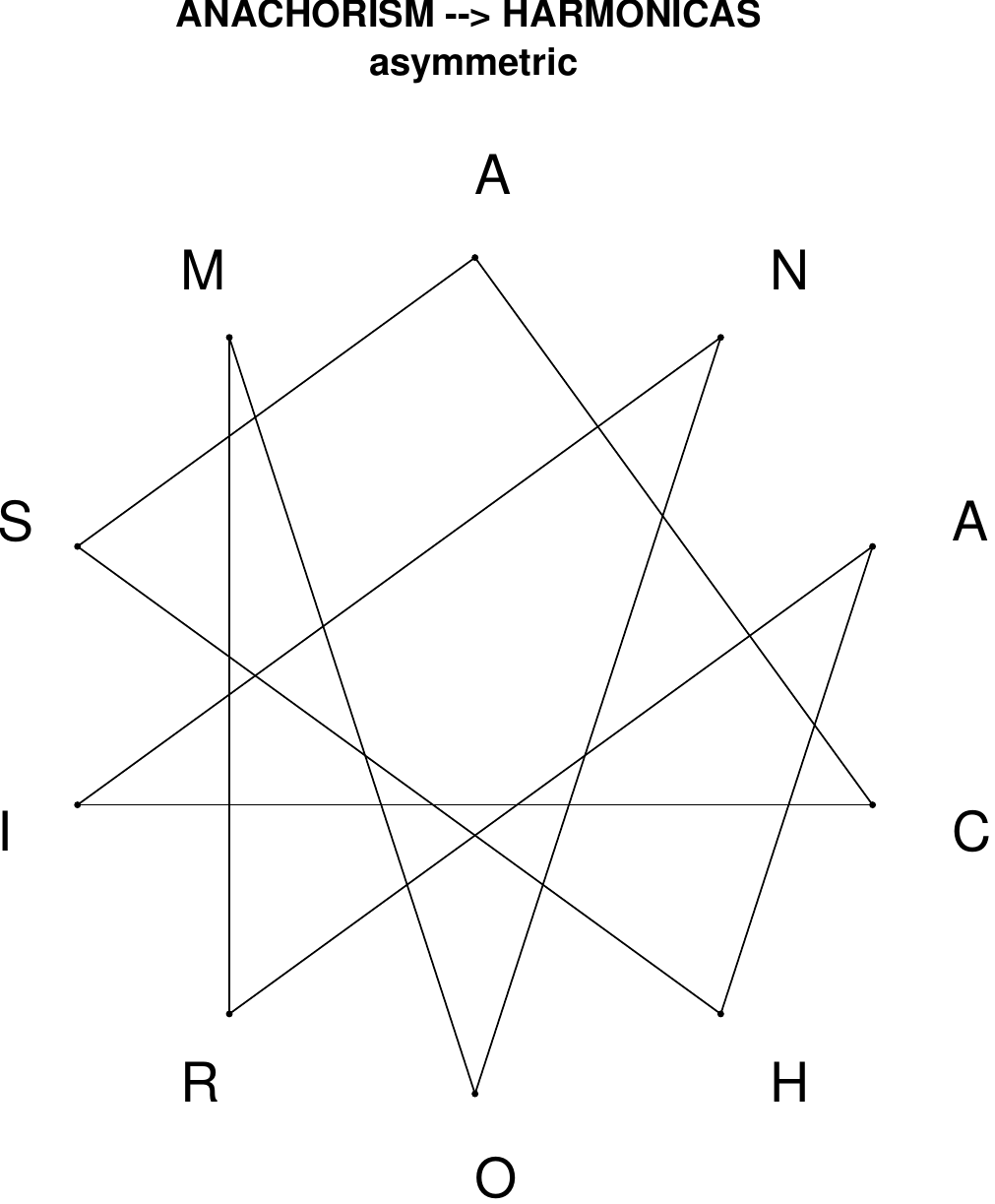}
\end{subfigure}
\hfill
\begin{subfigure}[T]{0.19\textwidth}
\centering
\includegraphics[width=\textwidth]{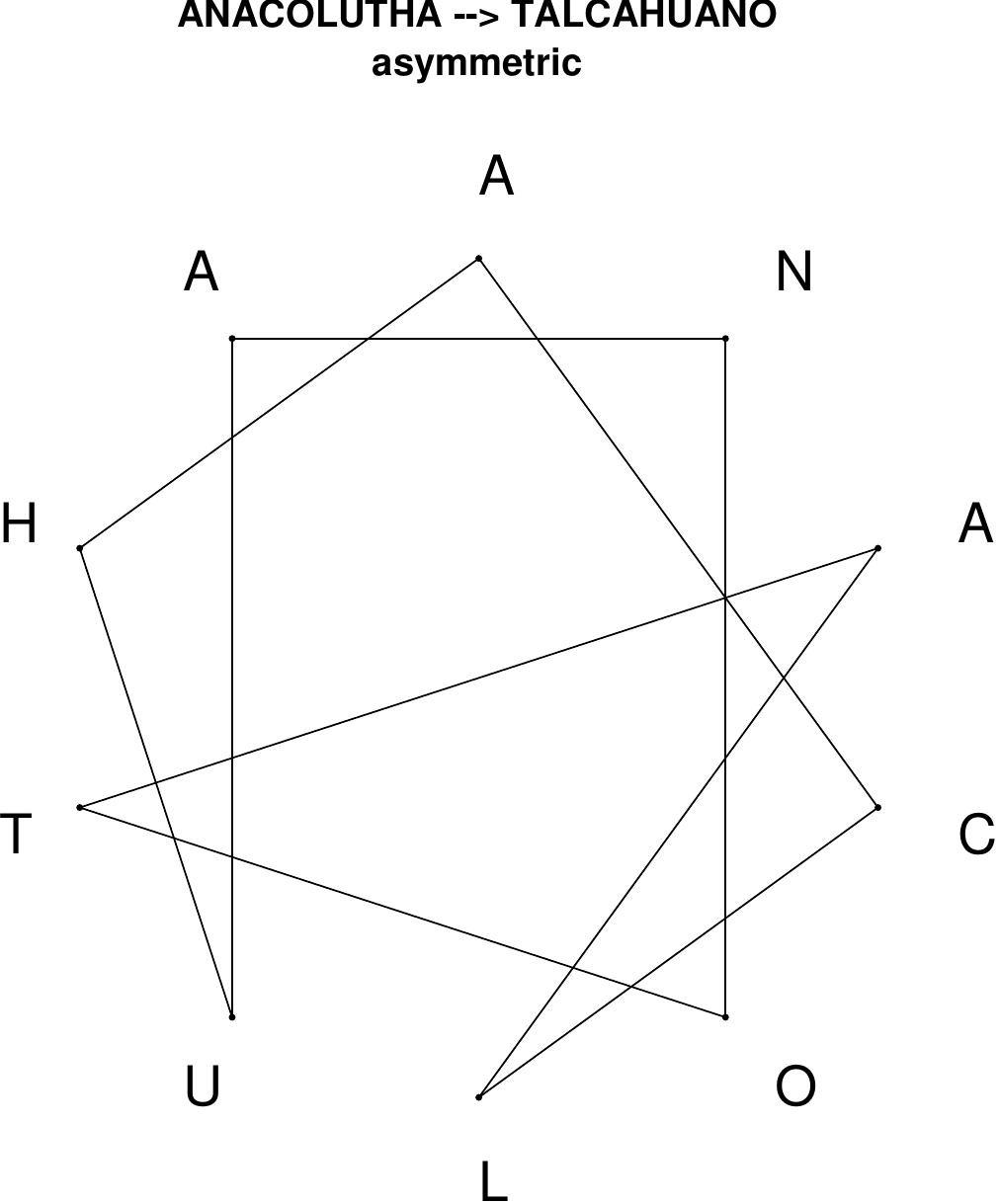}
\end{subfigure}
\hfill
\begin{subfigure}[T]{0.19\textwidth}
\centering
\includegraphics[width=\textwidth]{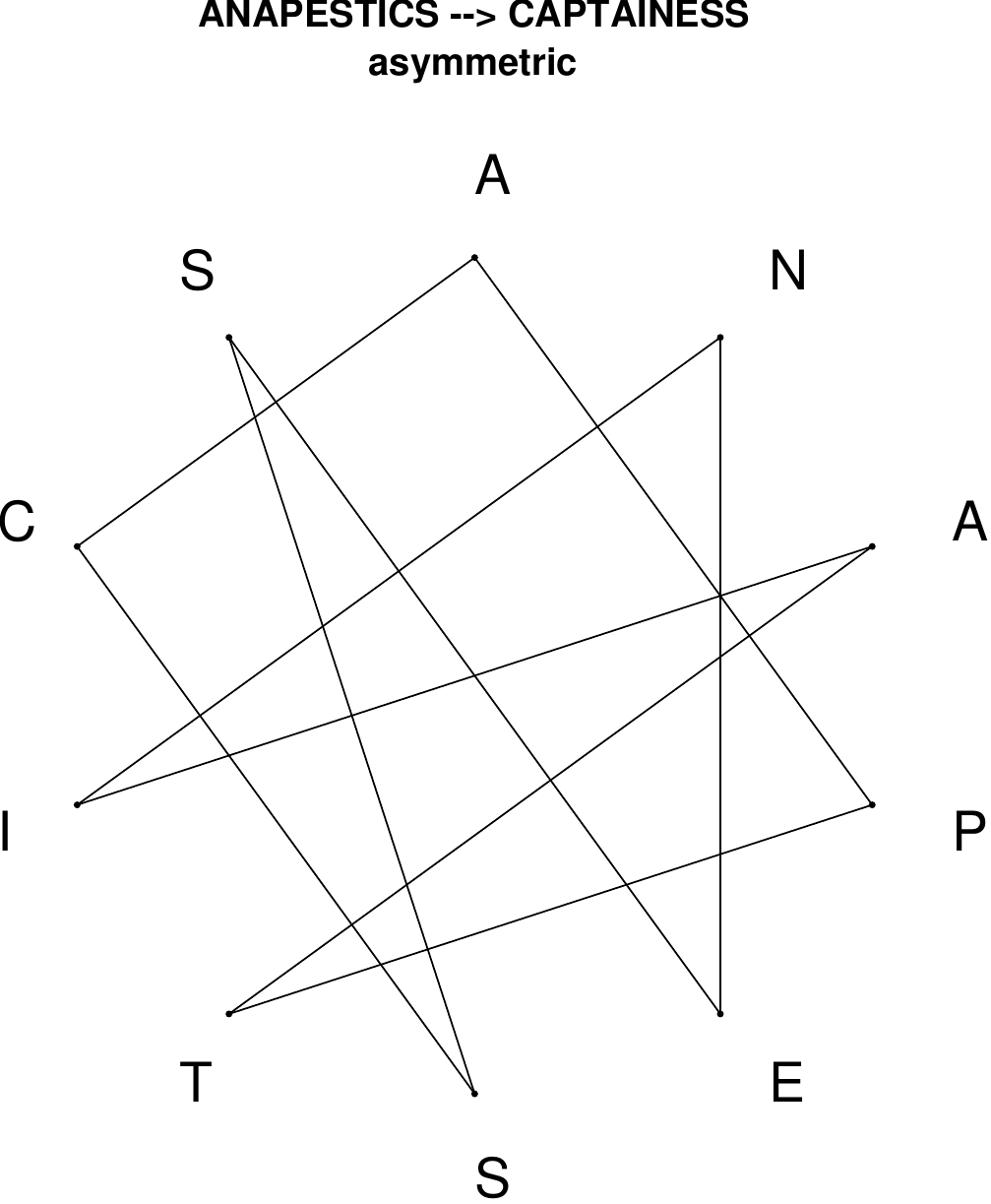}
\end{subfigure}
\hfill
\begin{subfigure}[T]{0.19\textwidth}
\centering
\includegraphics[width=\textwidth]{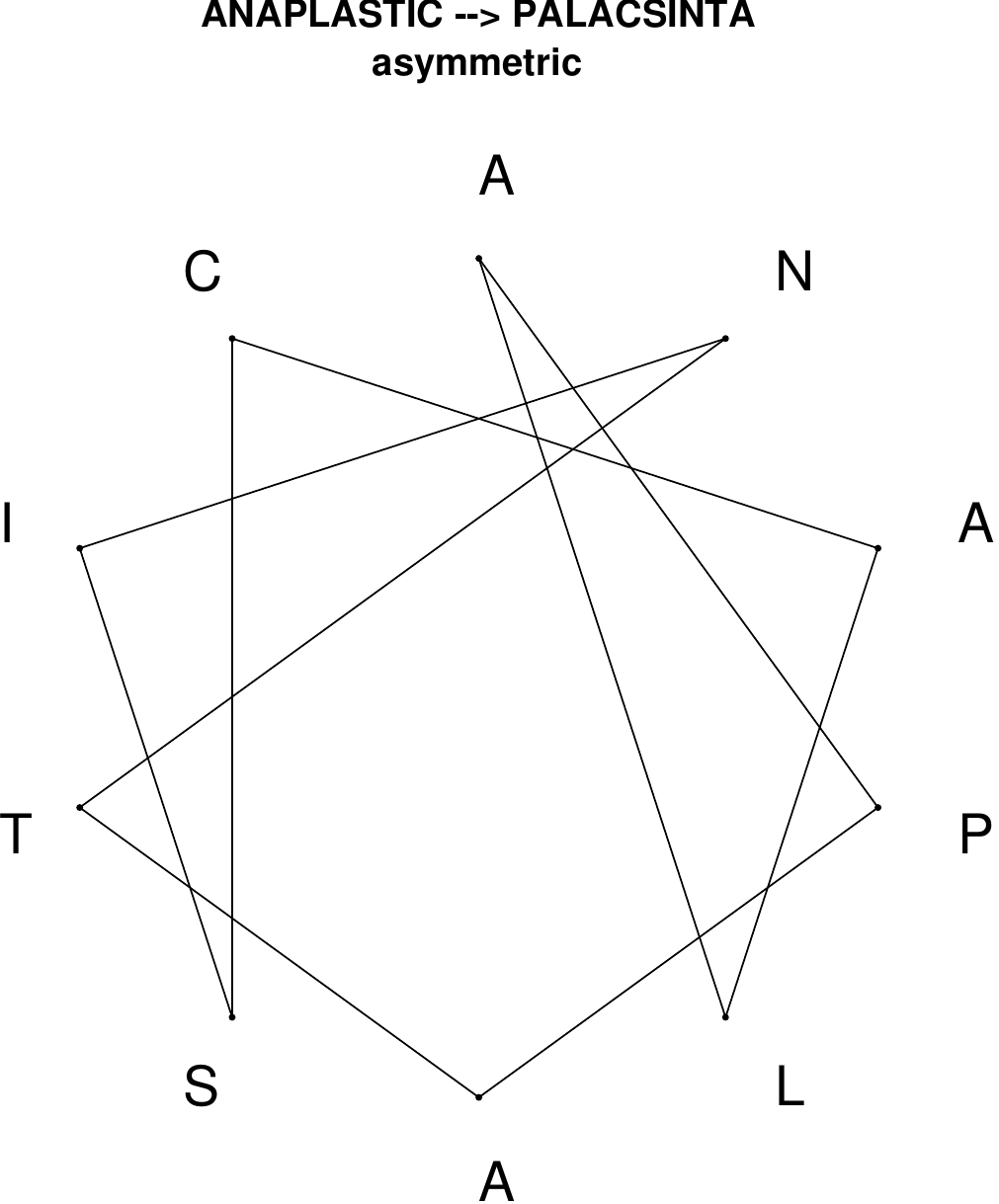}
\end{subfigure}
\hfill
\begin{subfigure}[T]{0.19\textwidth}
\centering
\includegraphics[width=\textwidth]{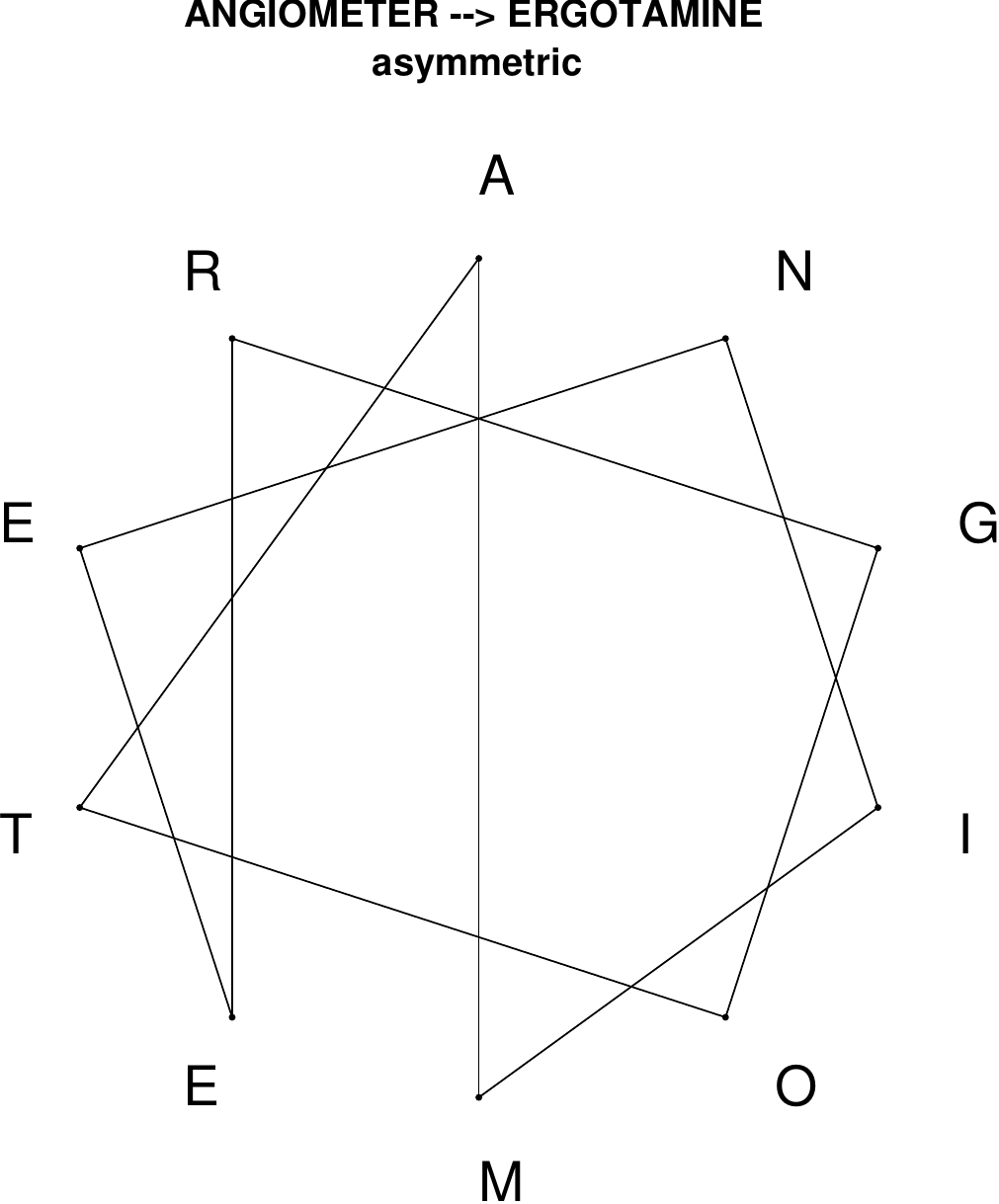}
\end{subfigure}
\end{figure}

\begin{figure}[H]
\centering
\begin{subfigure}[T]{0.19\textwidth}
\centering
\includegraphics[width=\textwidth]{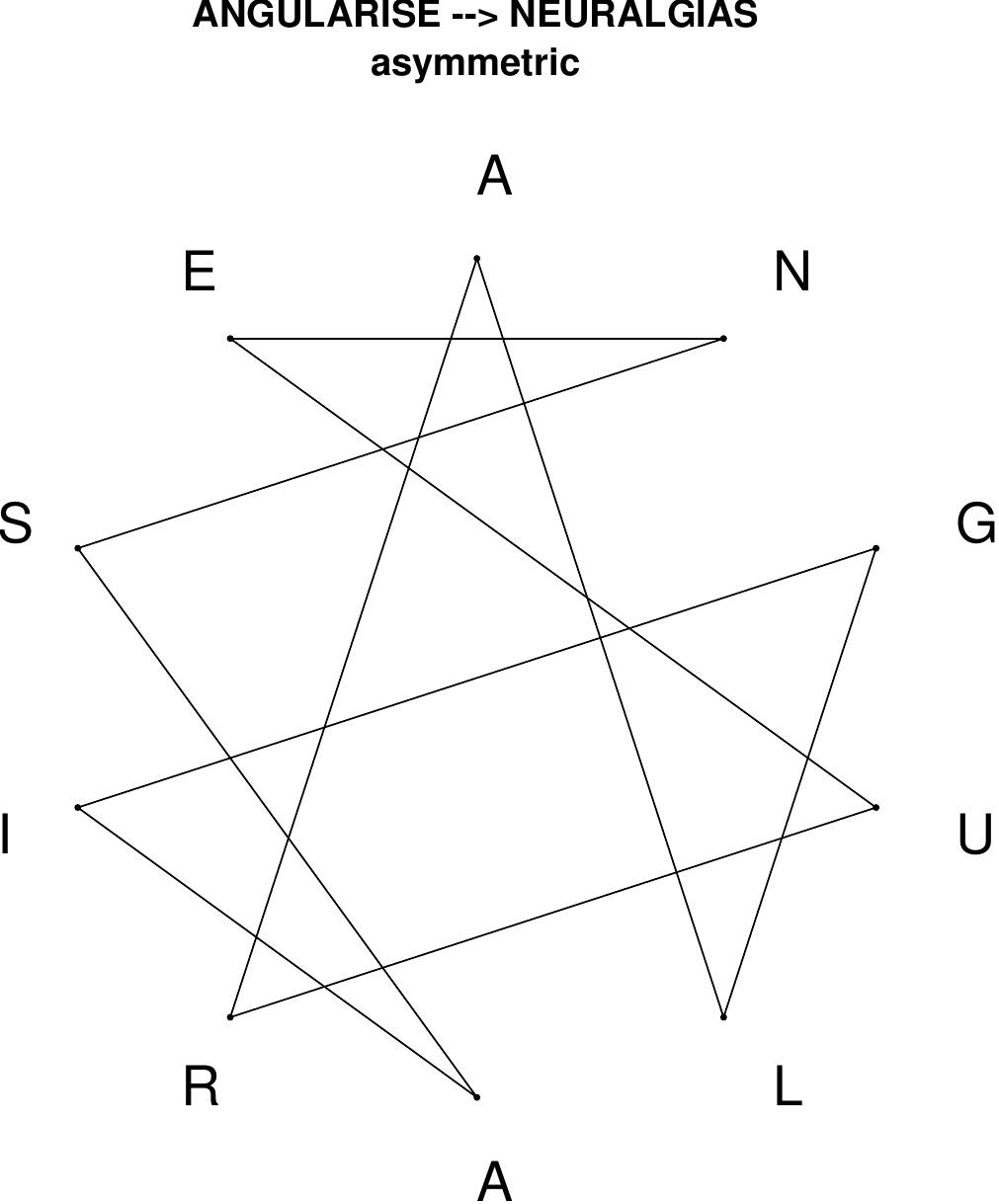}
\end{subfigure}
\hfill
\begin{subfigure}[T]{0.19\textwidth}
\centering
\includegraphics[width=\textwidth]{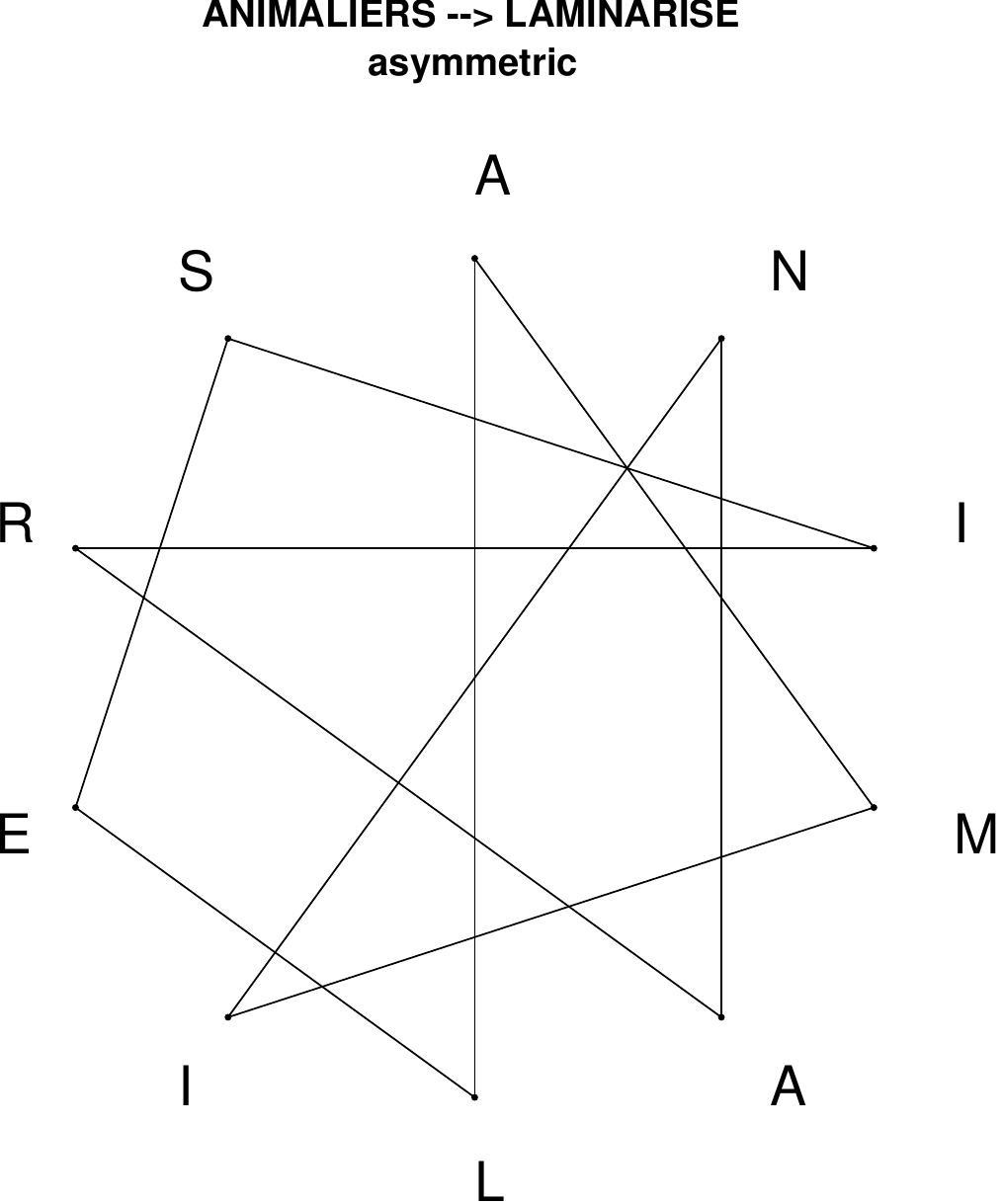}
\end{subfigure}
\hfill
\begin{subfigure}[T]{0.19\textwidth}
\centering
\includegraphics[width=\textwidth]{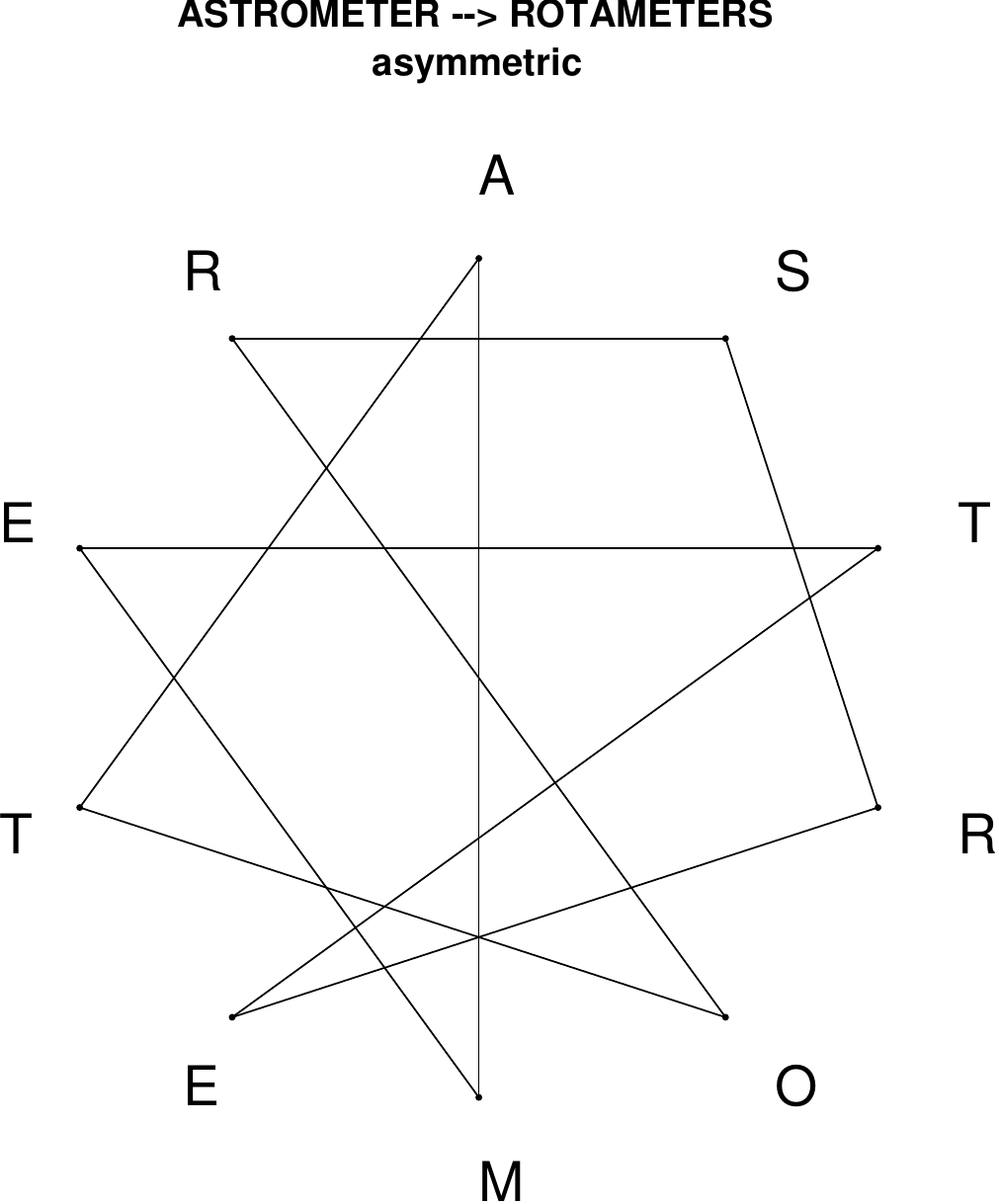}
\end{subfigure}
\hfill
\begin{subfigure}[T]{0.19\textwidth}
\centering
\includegraphics[width=\textwidth]{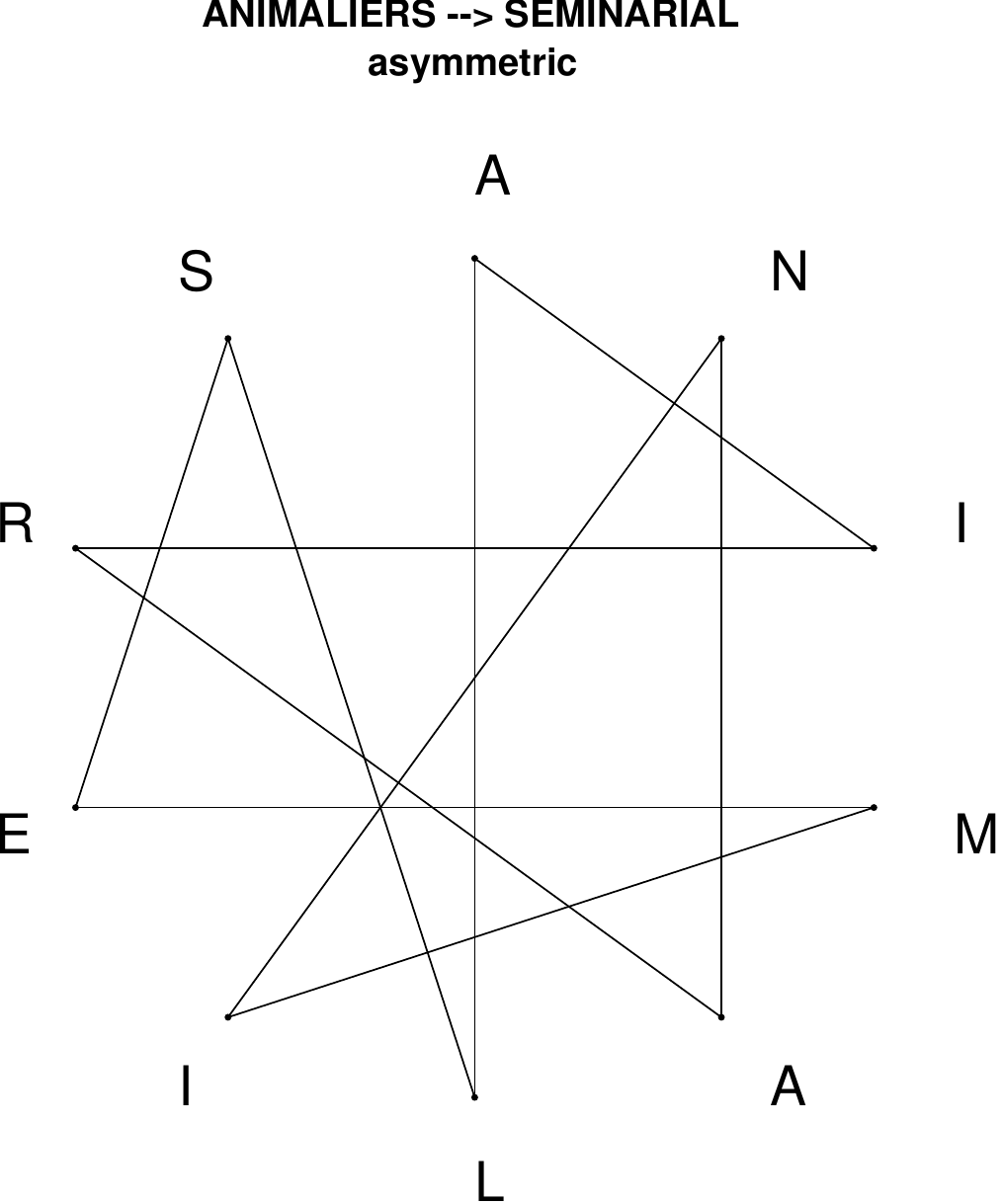}
\end{subfigure}
\hfill
\begin{subfigure}[T]{0.19\textwidth}
\centering
\includegraphics[width=\textwidth]{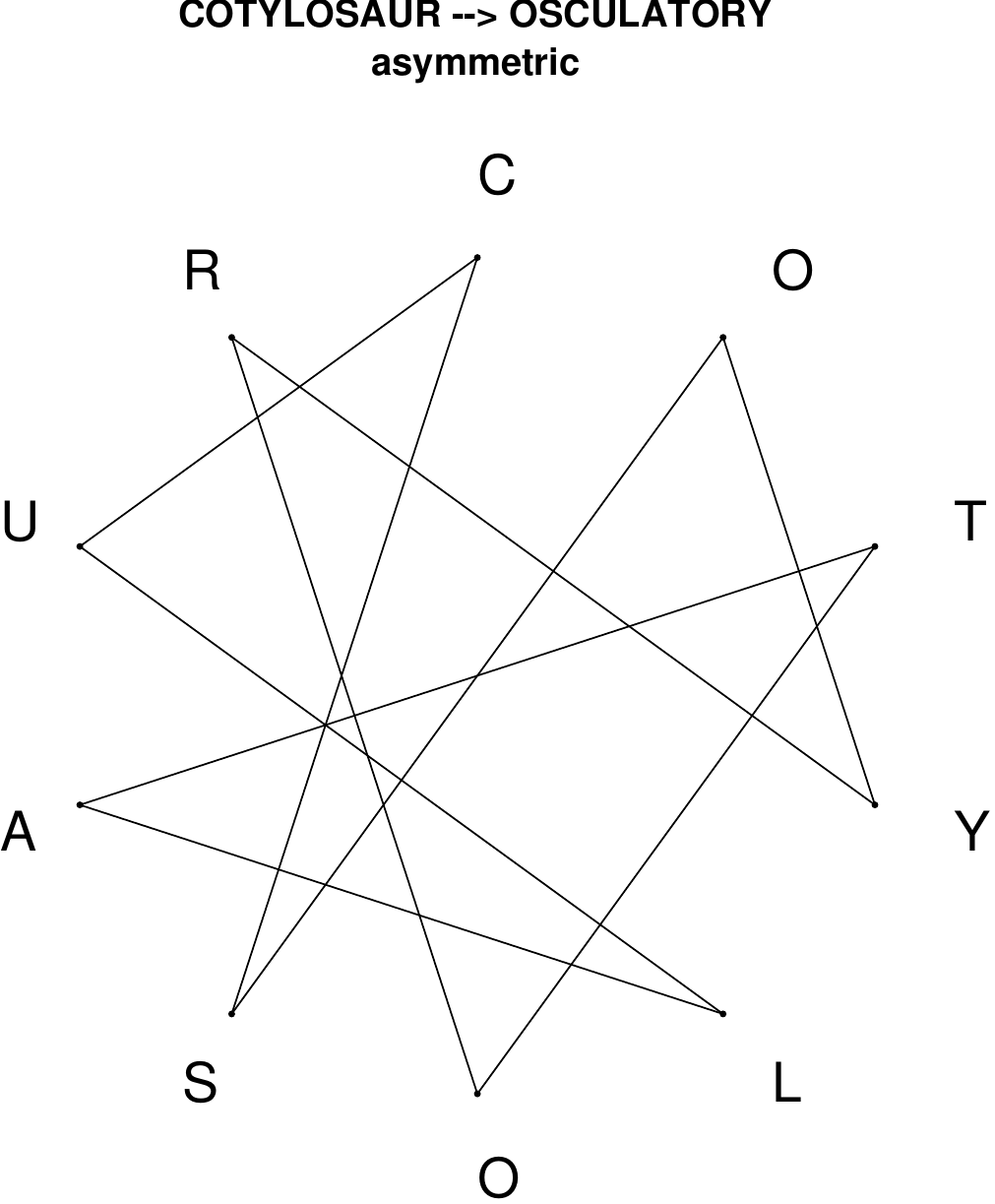}
\end{subfigure}
\end{figure}

\begin{figure}[H]
\centering
\begin{subfigure}[T]{0.19\textwidth}
\centering
\includegraphics[width=\textwidth]{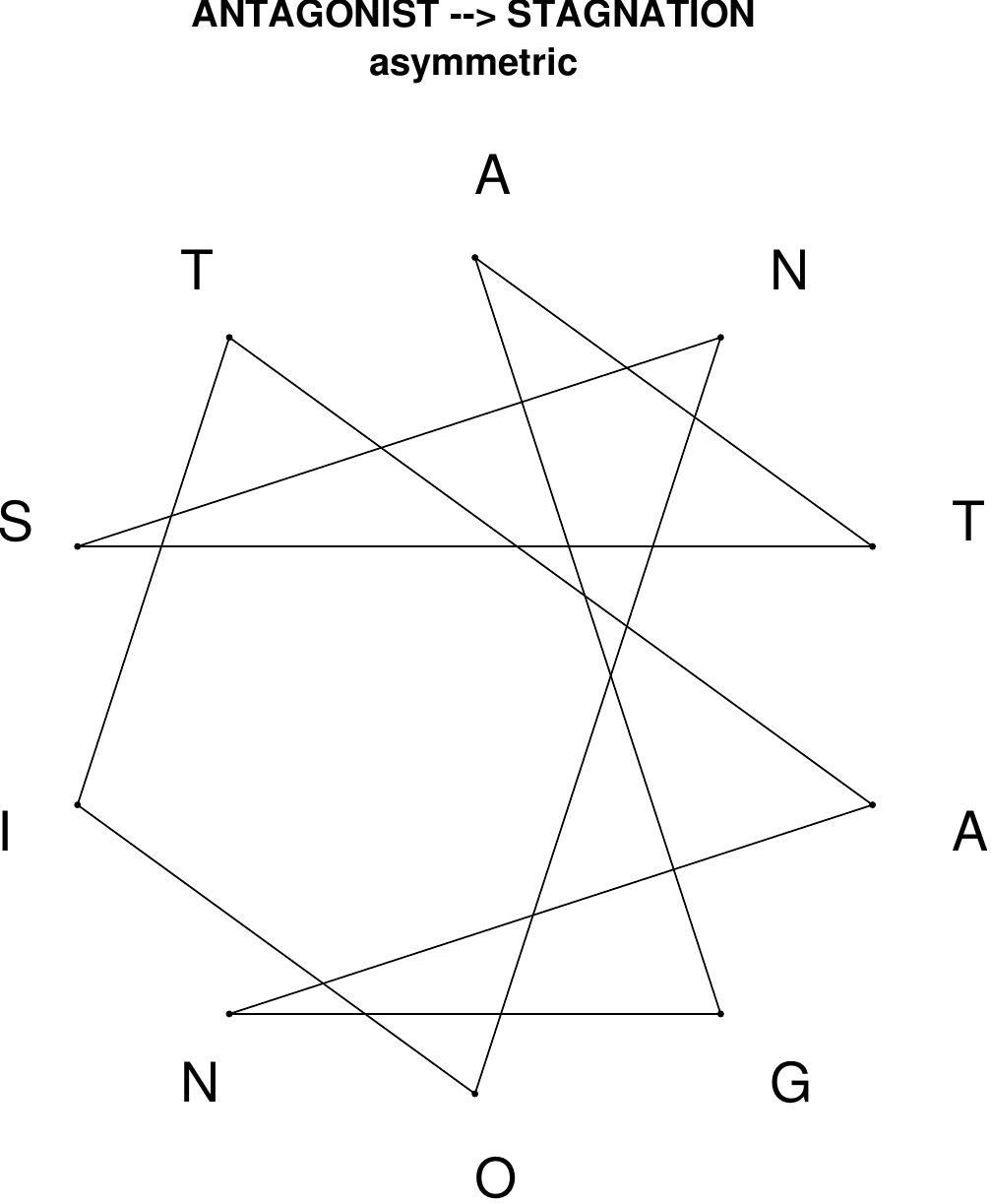}
\end{subfigure}
\hfill
\begin{subfigure}[T]{0.19\textwidth}
\centering
\includegraphics[width=\textwidth]{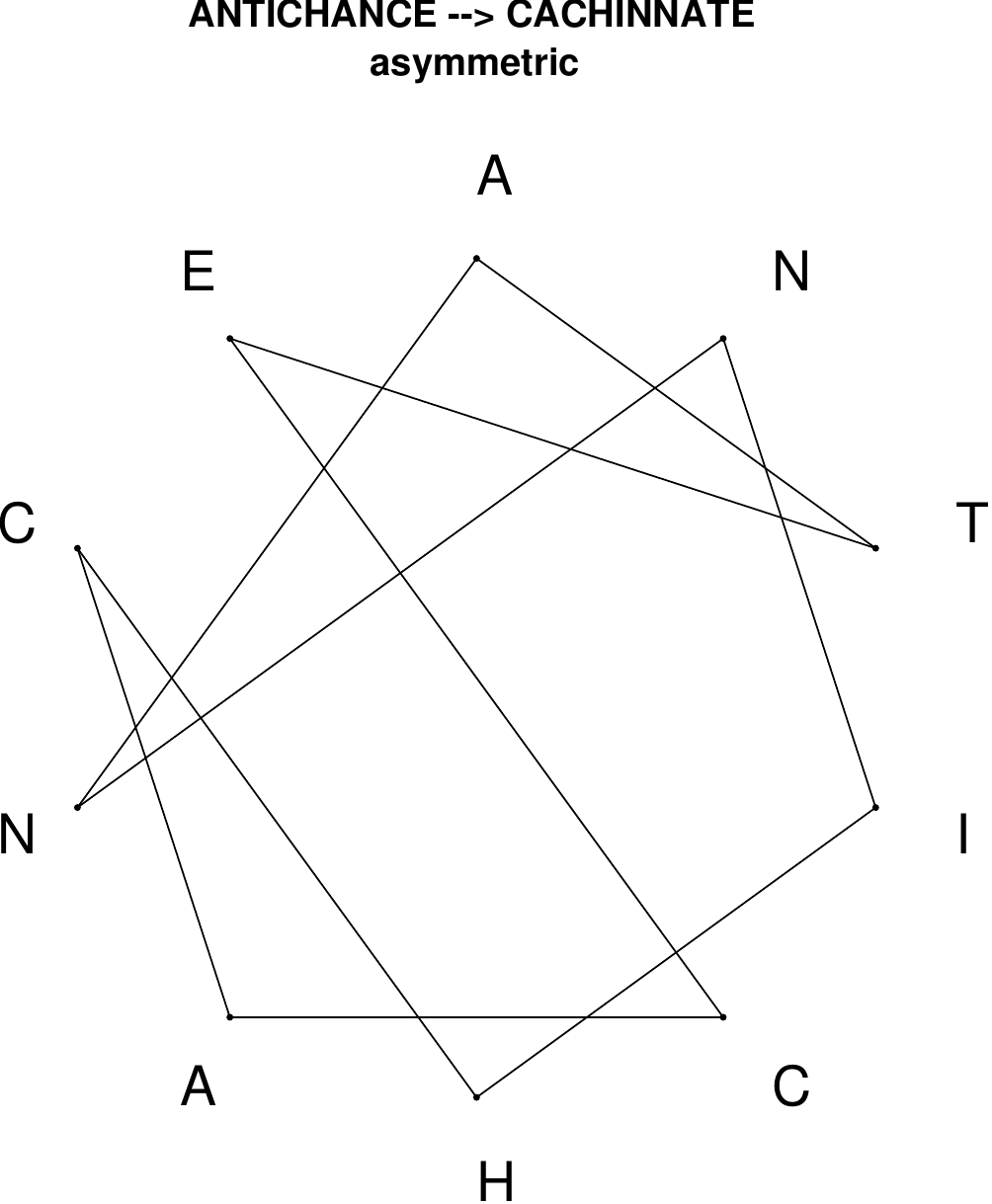}
\end{subfigure}
\hfill
\begin{subfigure}[T]{0.19\textwidth}
\centering
\includegraphics[width=\textwidth]{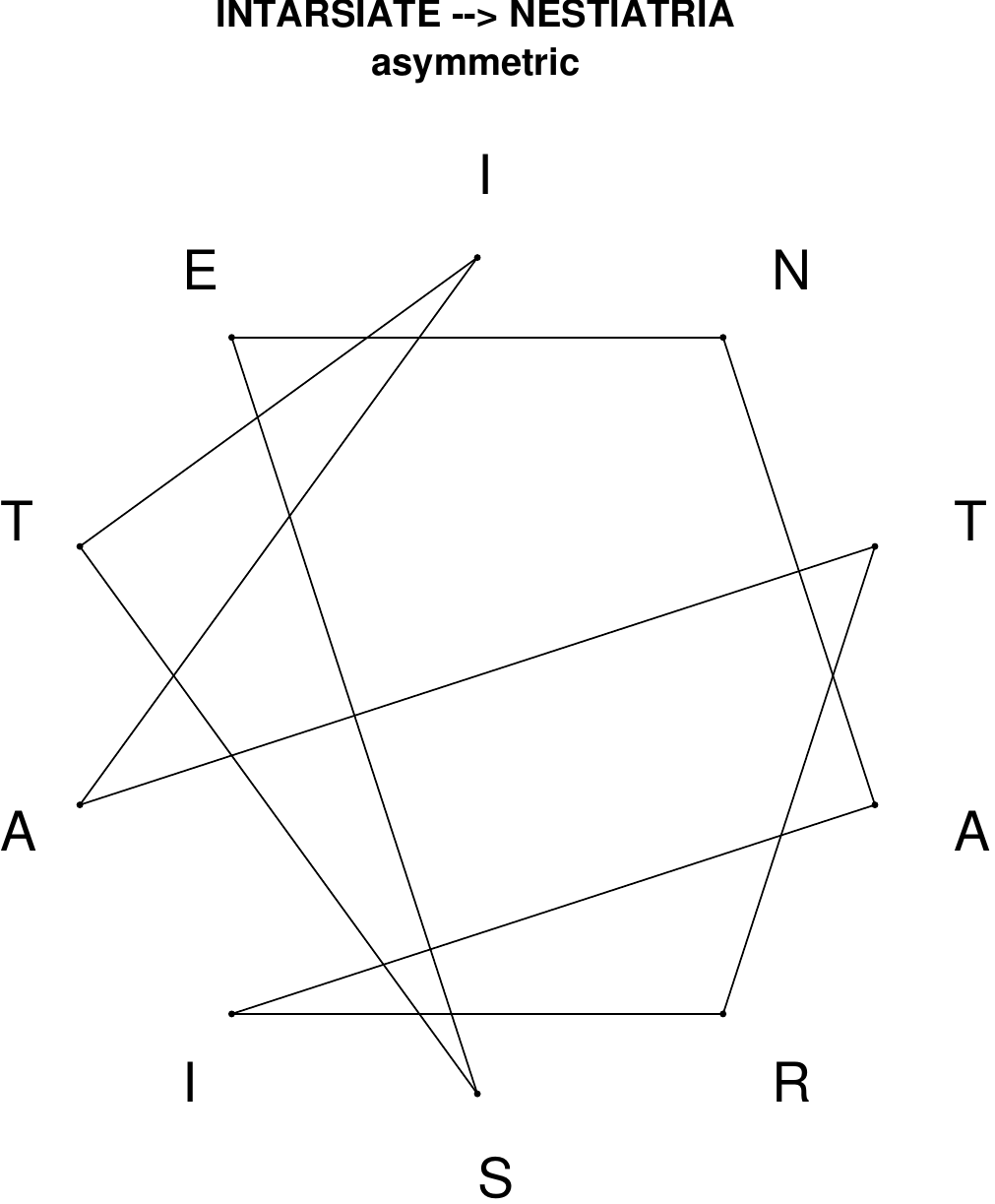}
\end{subfigure}
\hfill
\begin{subfigure}[T]{0.19\textwidth}
\centering
\includegraphics[width=\textwidth]{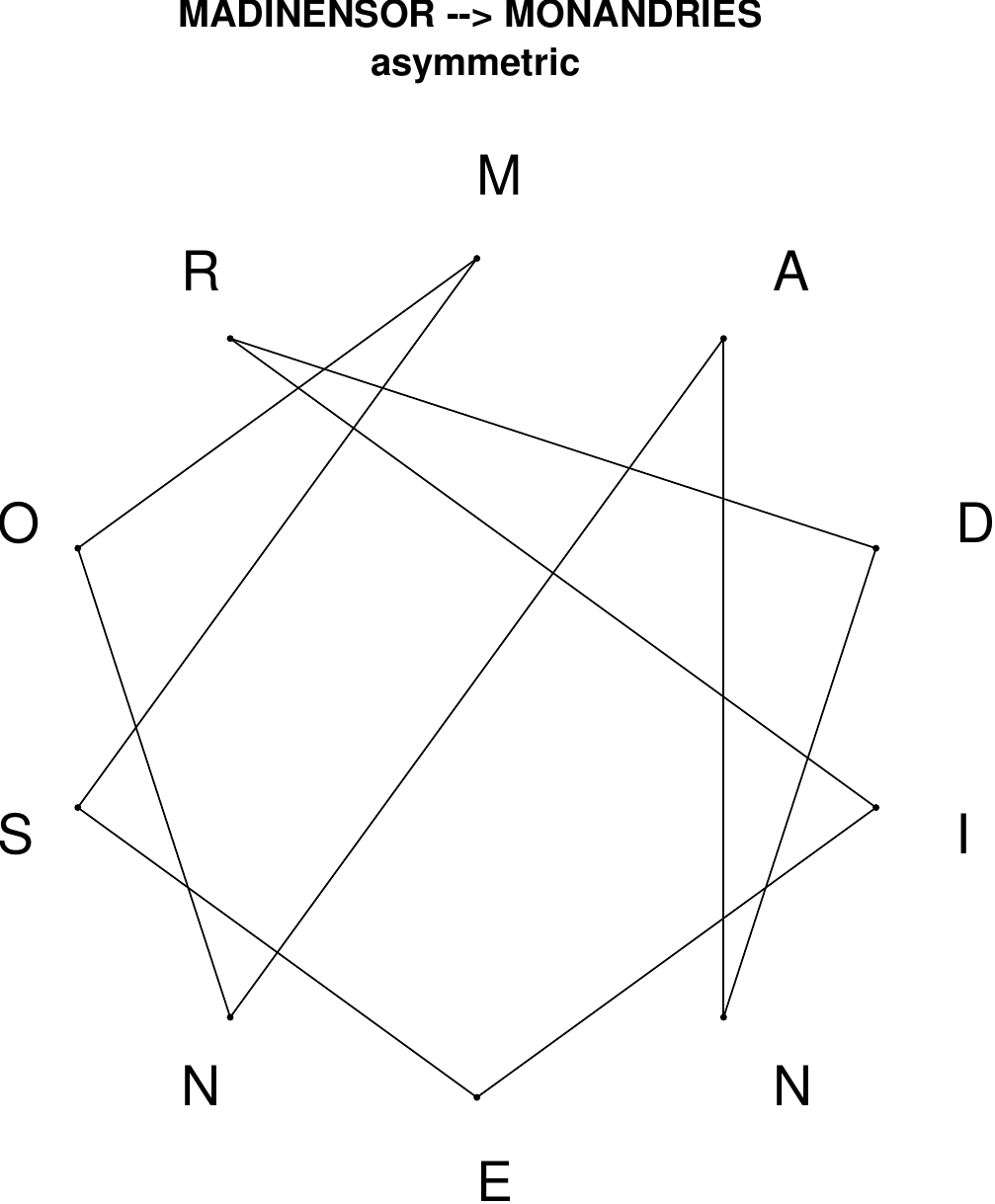}
\end{subfigure}
\hfill
\begin{subfigure}[T]{0.19\textwidth}
\centering
\includegraphics[width=\textwidth]{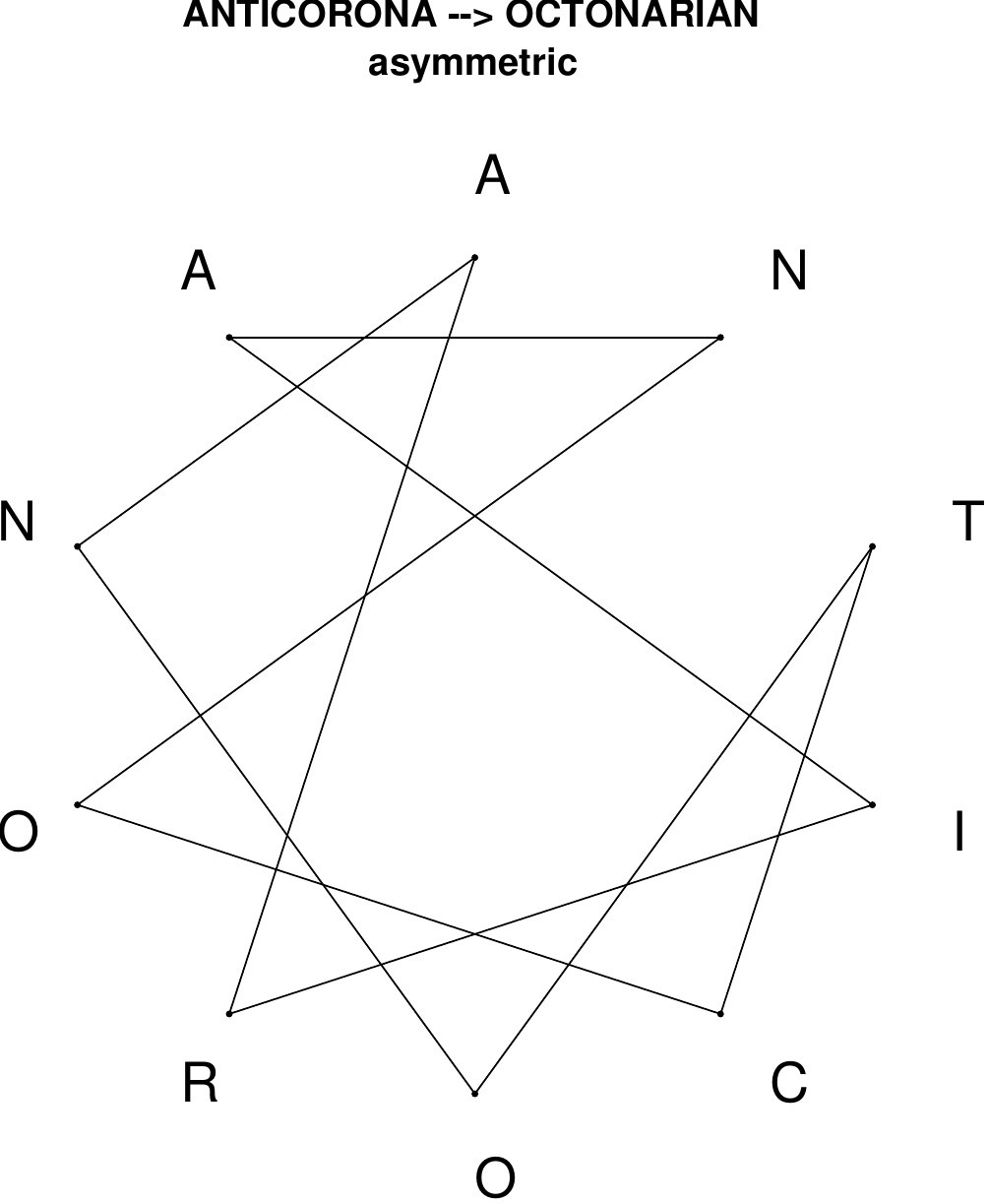}
\end{subfigure}
\end{figure}

\begin{figure}[H]
\centering
\begin{subfigure}[T]{0.19\textwidth}
\centering
\includegraphics[width=\textwidth]{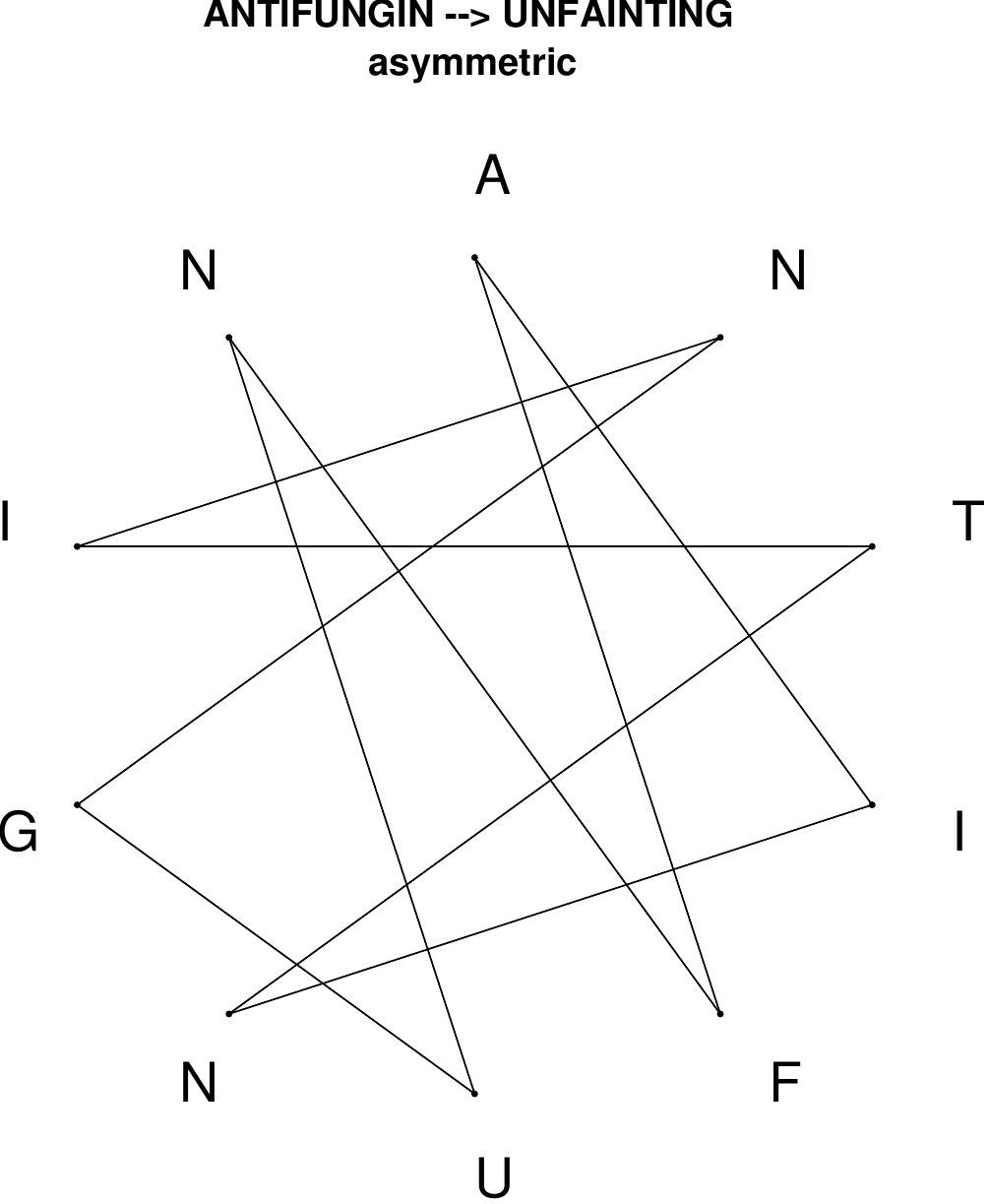}
\end{subfigure}
\hfill
\begin{subfigure}[T]{0.19\textwidth}
\centering
\includegraphics[width=\textwidth]{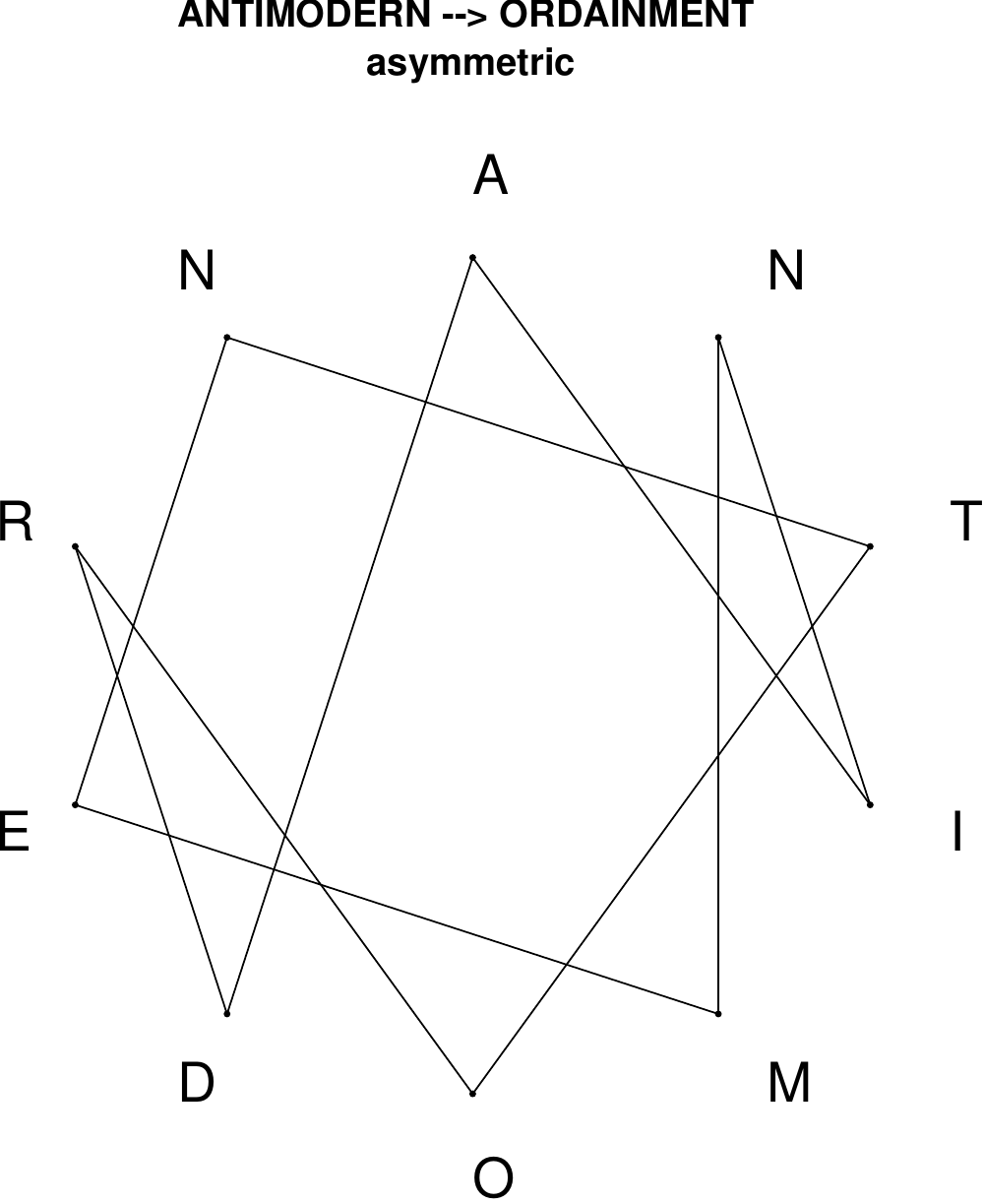}
\end{subfigure}
\hfill
\begin{subfigure}[T]{0.19\textwidth}
\centering
\includegraphics[width=\textwidth]{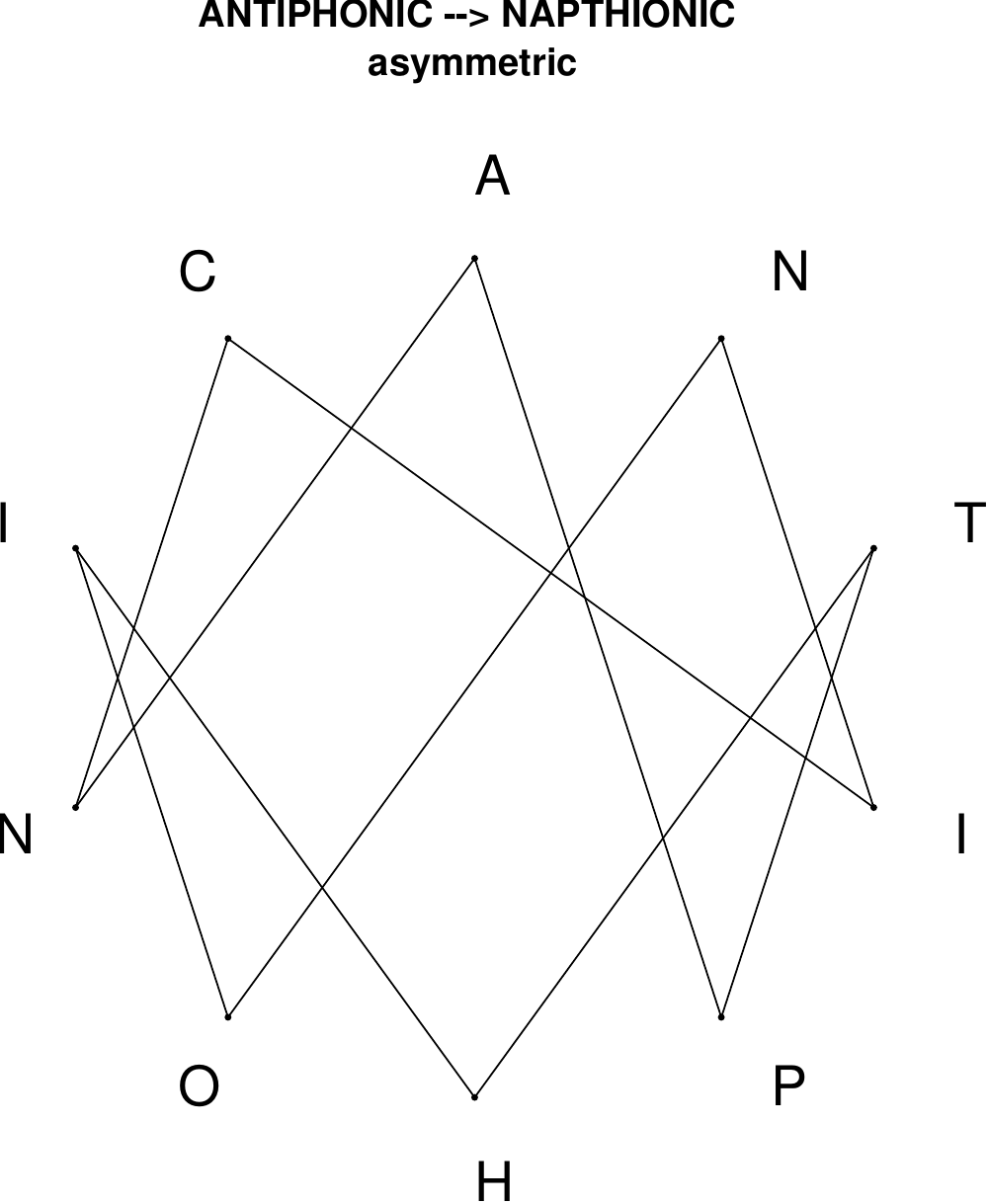}
\end{subfigure}
\hfill
\begin{subfigure}[T]{0.19\textwidth}
\centering
\includegraphics[width=\textwidth]{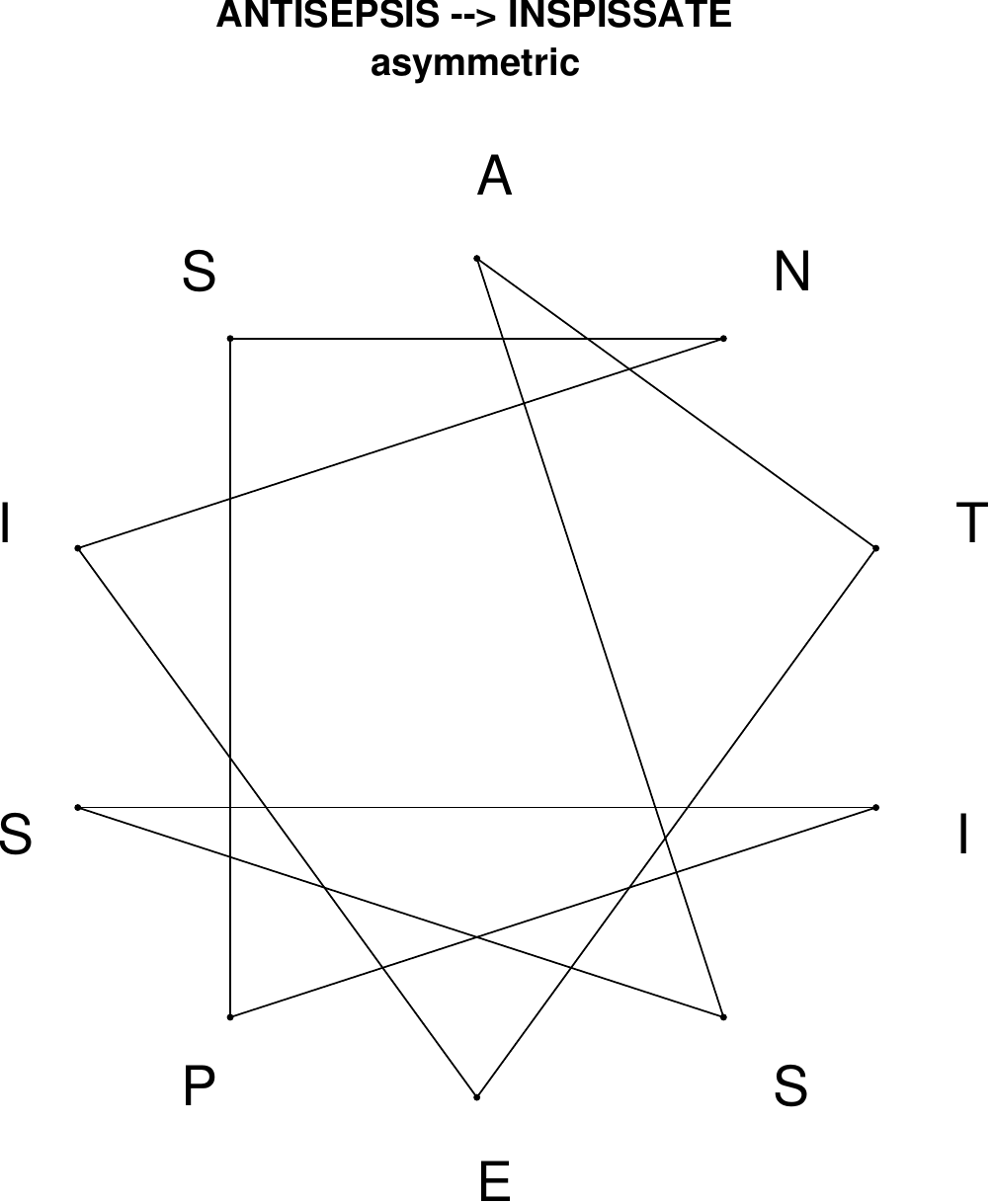}
\end{subfigure}
\hfill
\begin{subfigure}[T]{0.19\textwidth}
\centering
\includegraphics[width=\textwidth]{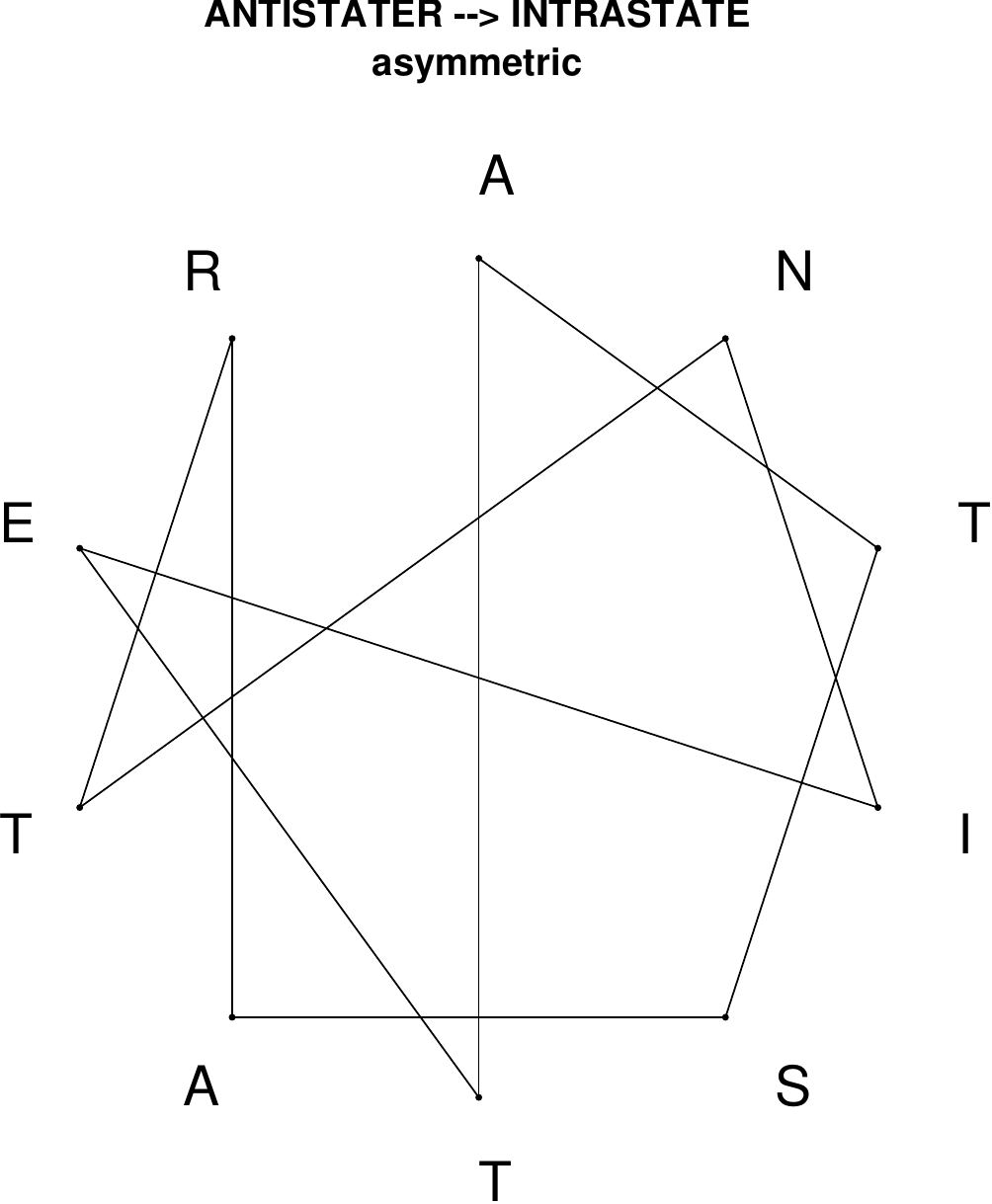}
\end{subfigure}
\end{figure}

\begin{figure}[H]
\centering
\begin{subfigure}[T]{0.19\textwidth}
\centering
\includegraphics[width=\textwidth]{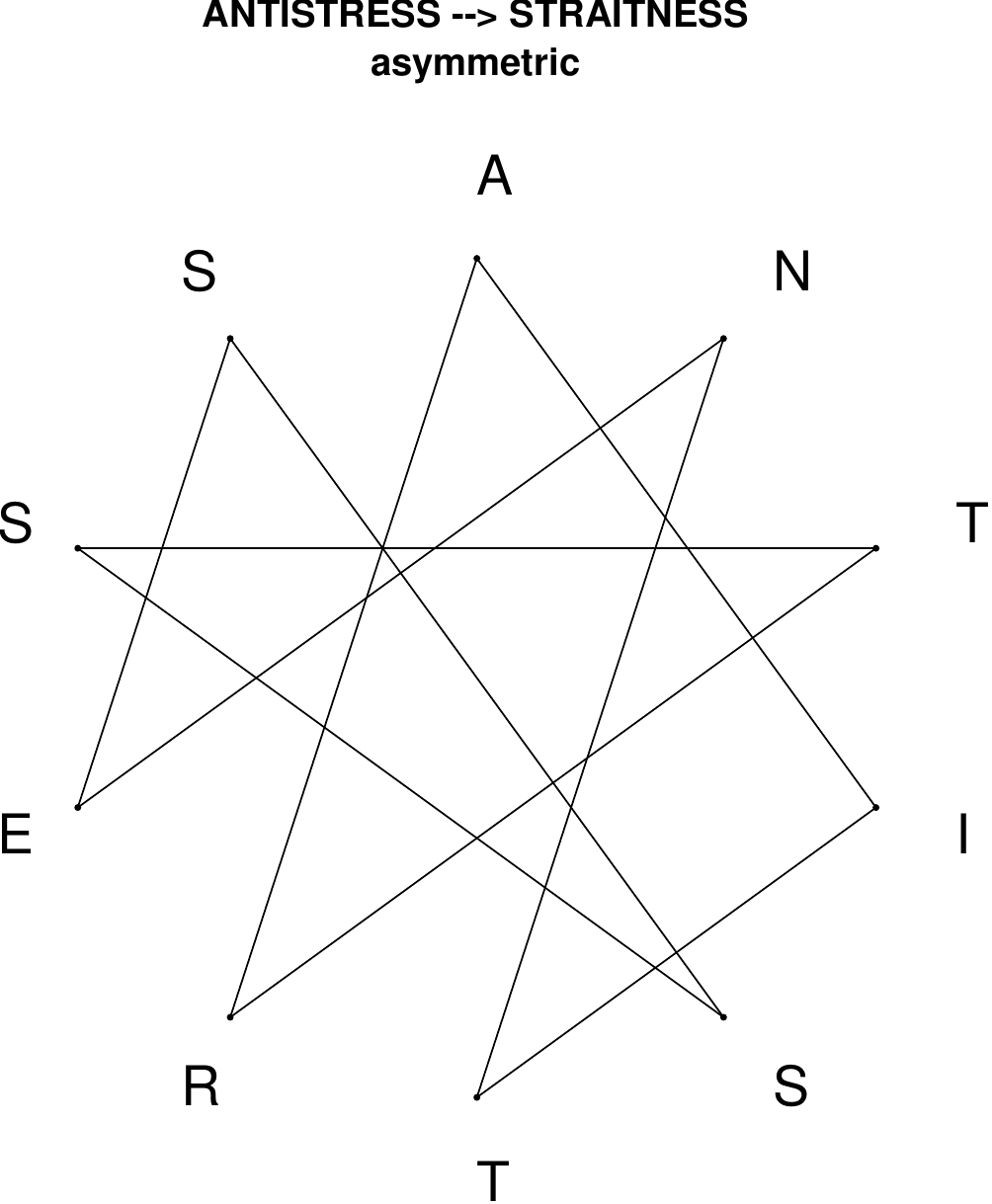}
\end{subfigure}
\hfill
\begin{subfigure}[T]{0.19\textwidth}
\centering
\includegraphics[width=\textwidth]{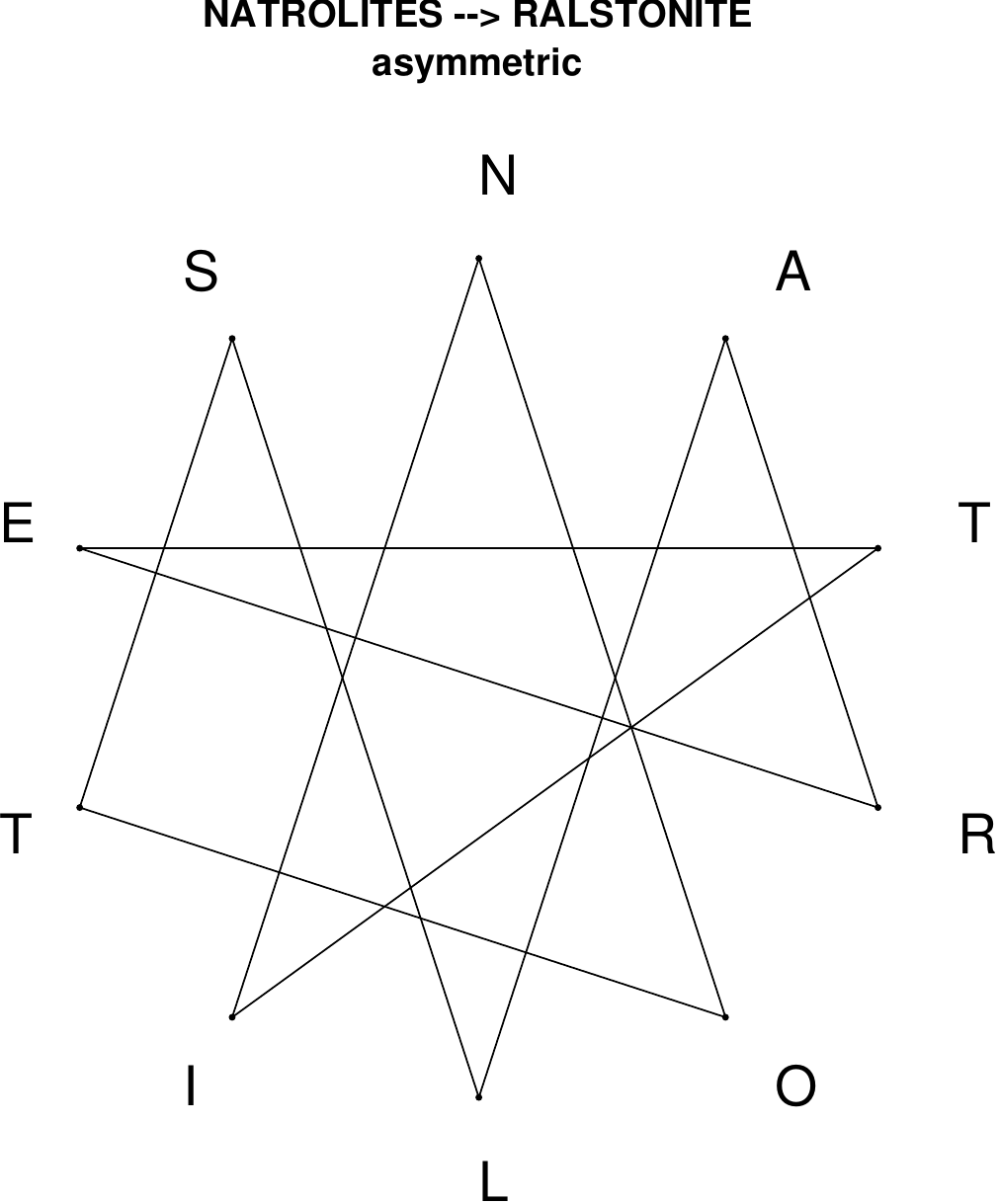}
\end{subfigure}
\hfill
\begin{subfigure}[T]{0.19\textwidth}
\centering
\includegraphics[width=\textwidth]{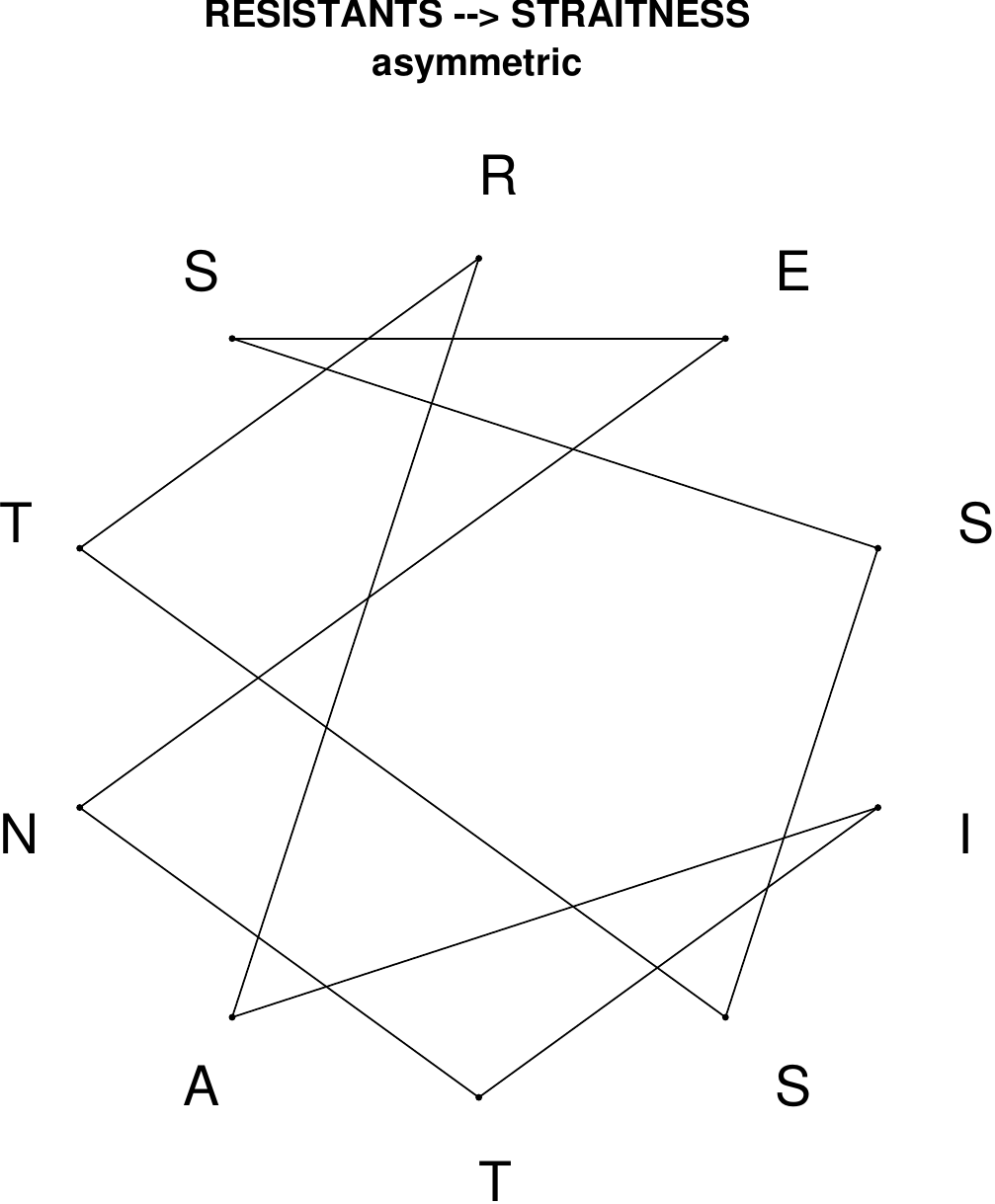}
\end{subfigure}
\hfill
\begin{subfigure}[T]{0.19\textwidth}
\centering
\includegraphics[width=\textwidth]{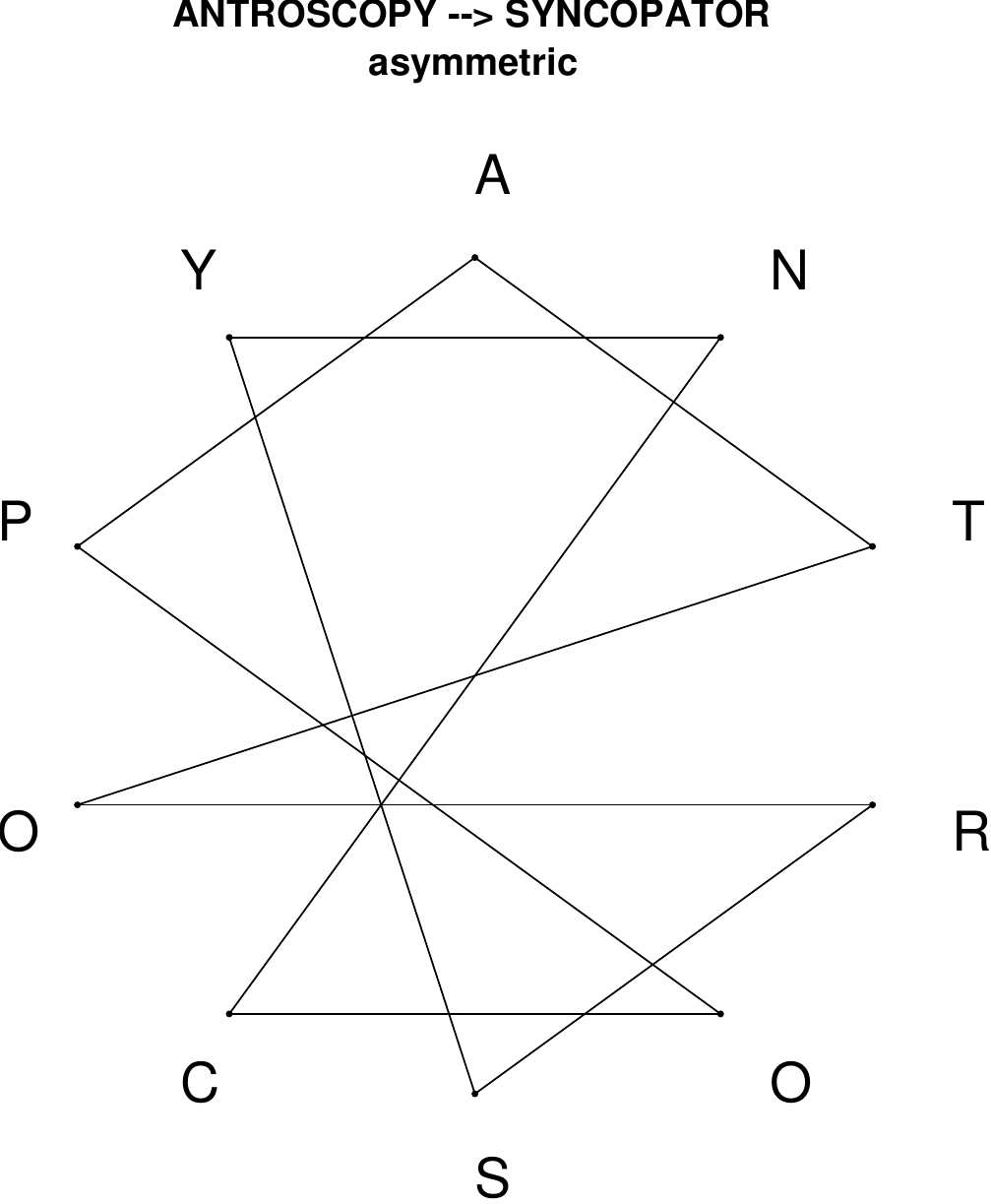}
\end{subfigure}
\hfill
\begin{subfigure}[T]{0.19\textwidth}
\centering
\includegraphics[width=\textwidth]{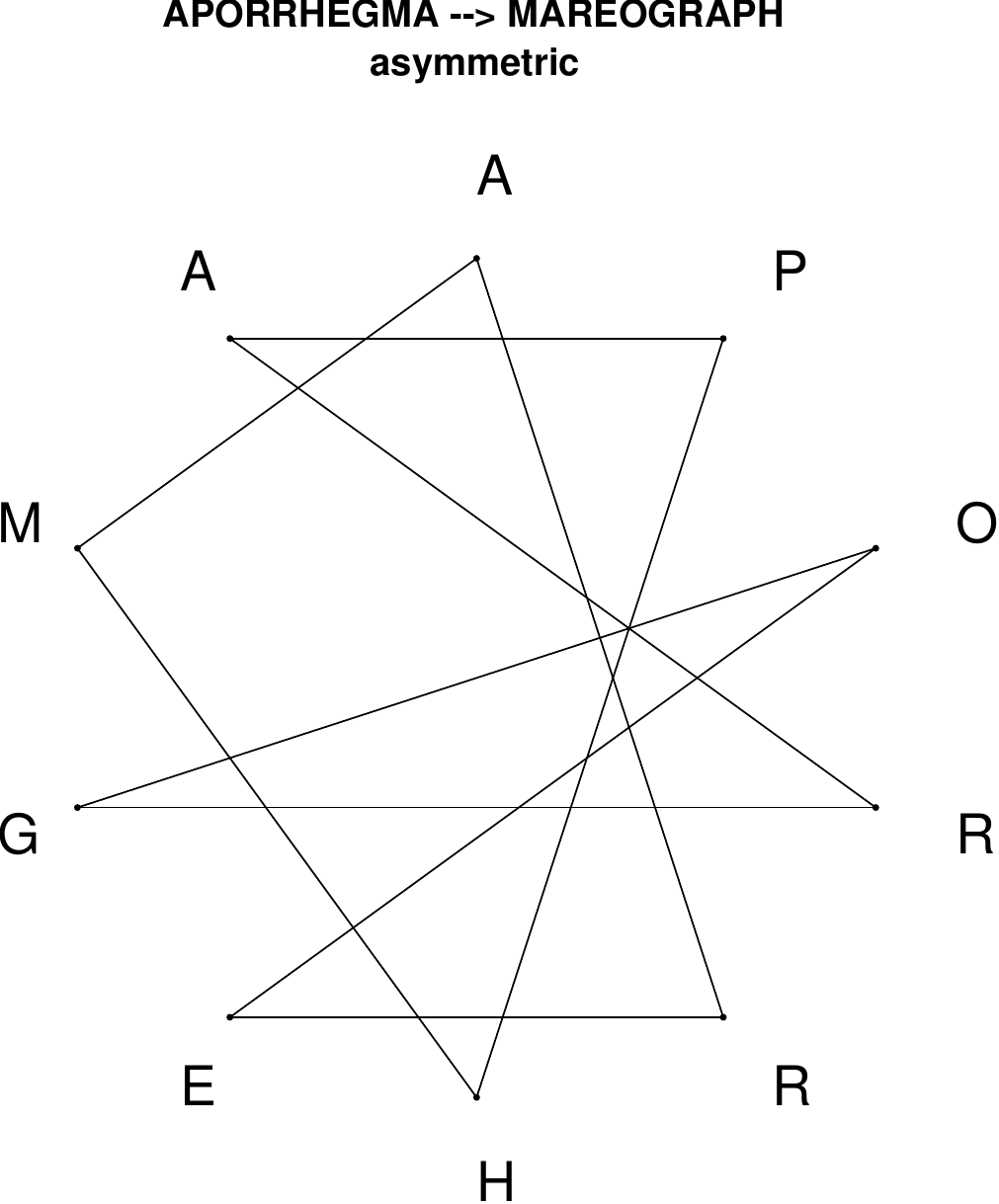}
\end{subfigure}
\end{figure}

\begin{figure}[H]
\centering
\begin{subfigure}[T]{0.19\textwidth}
\centering
\includegraphics[width=\textwidth]{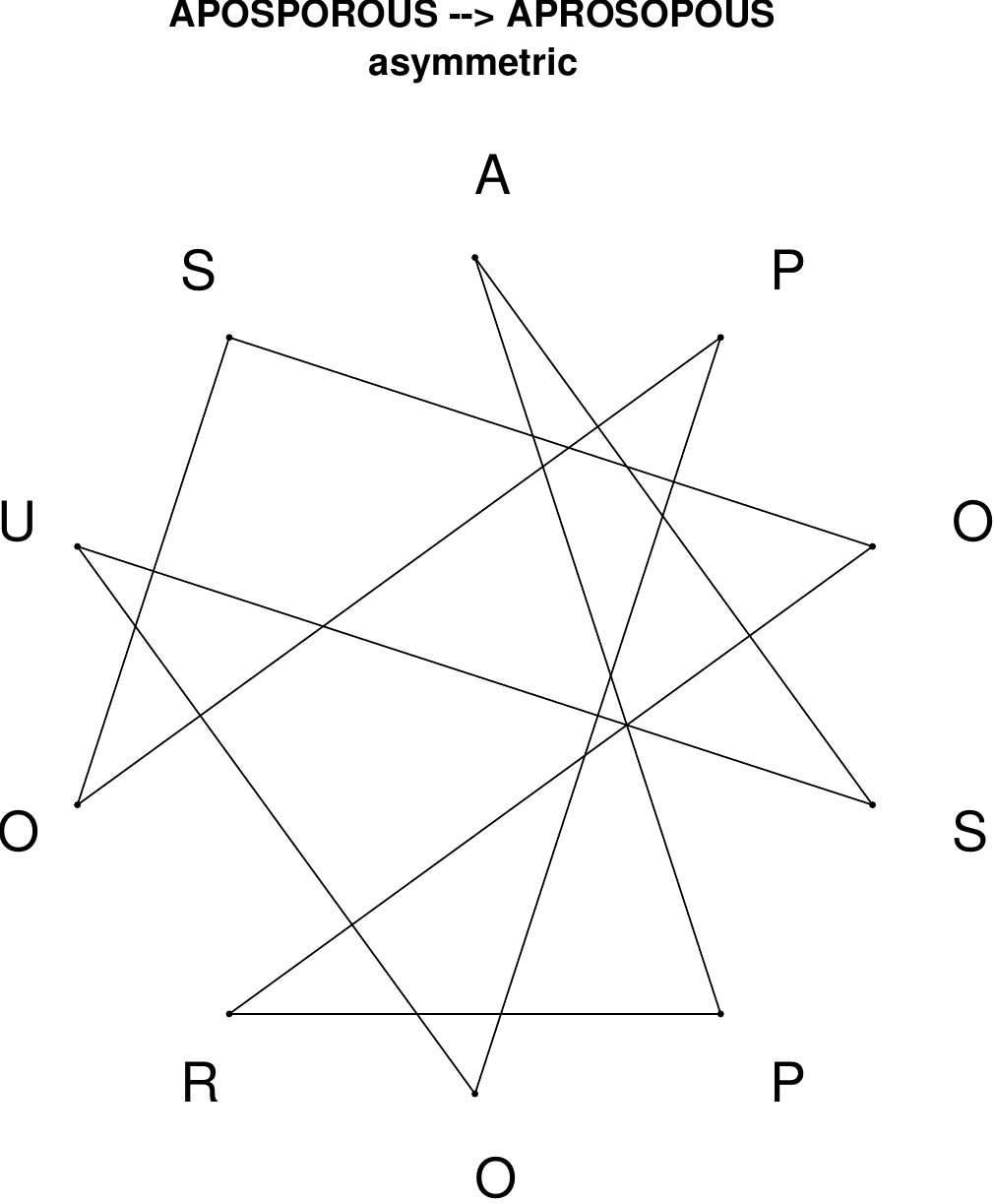}
\end{subfigure}
\hfill
\begin{subfigure}[T]{0.19\textwidth}
\centering
\includegraphics[width=\textwidth]{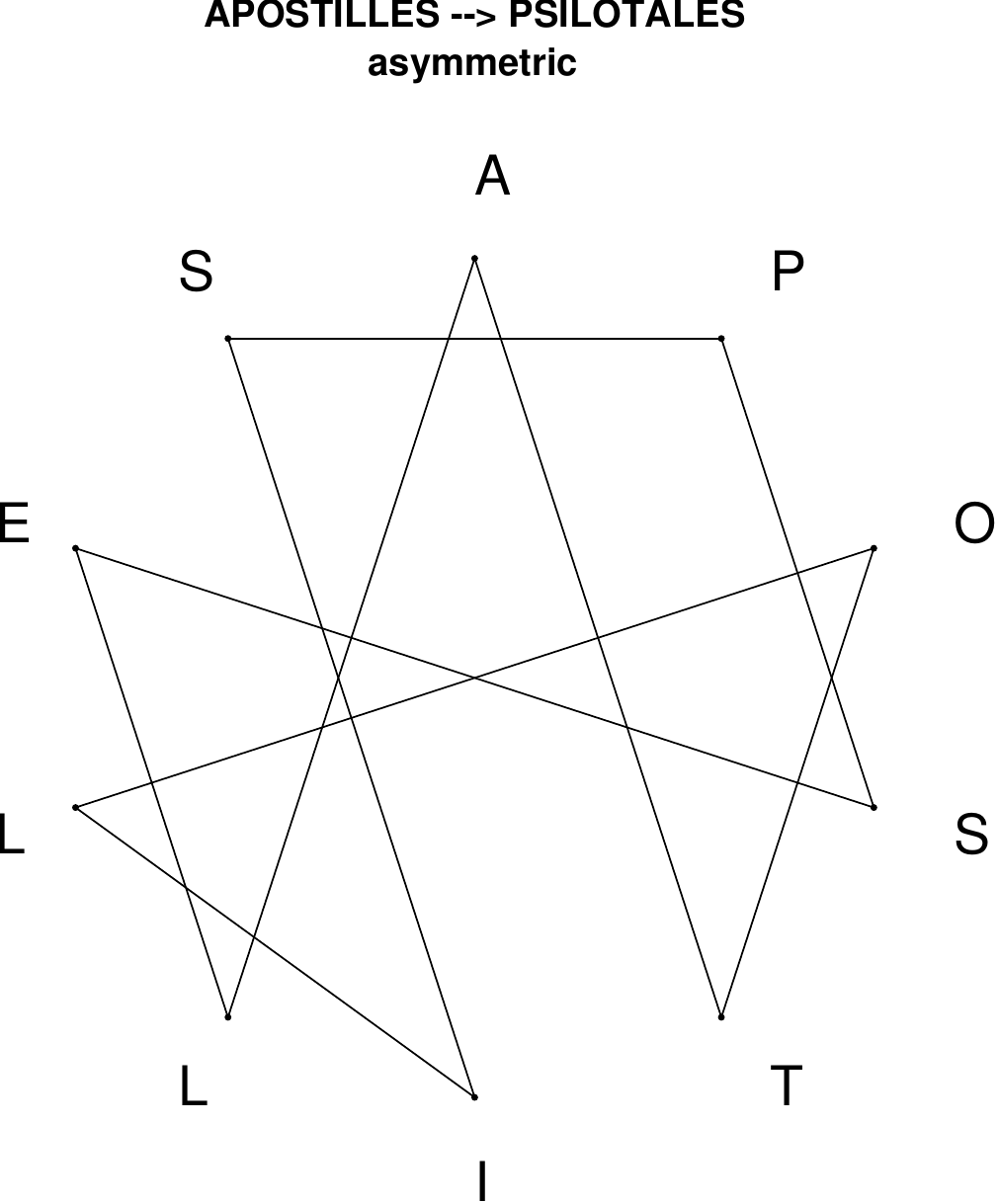}
\end{subfigure}
\hfill
\begin{subfigure}[T]{0.19\textwidth}
\centering
\includegraphics[width=\textwidth]{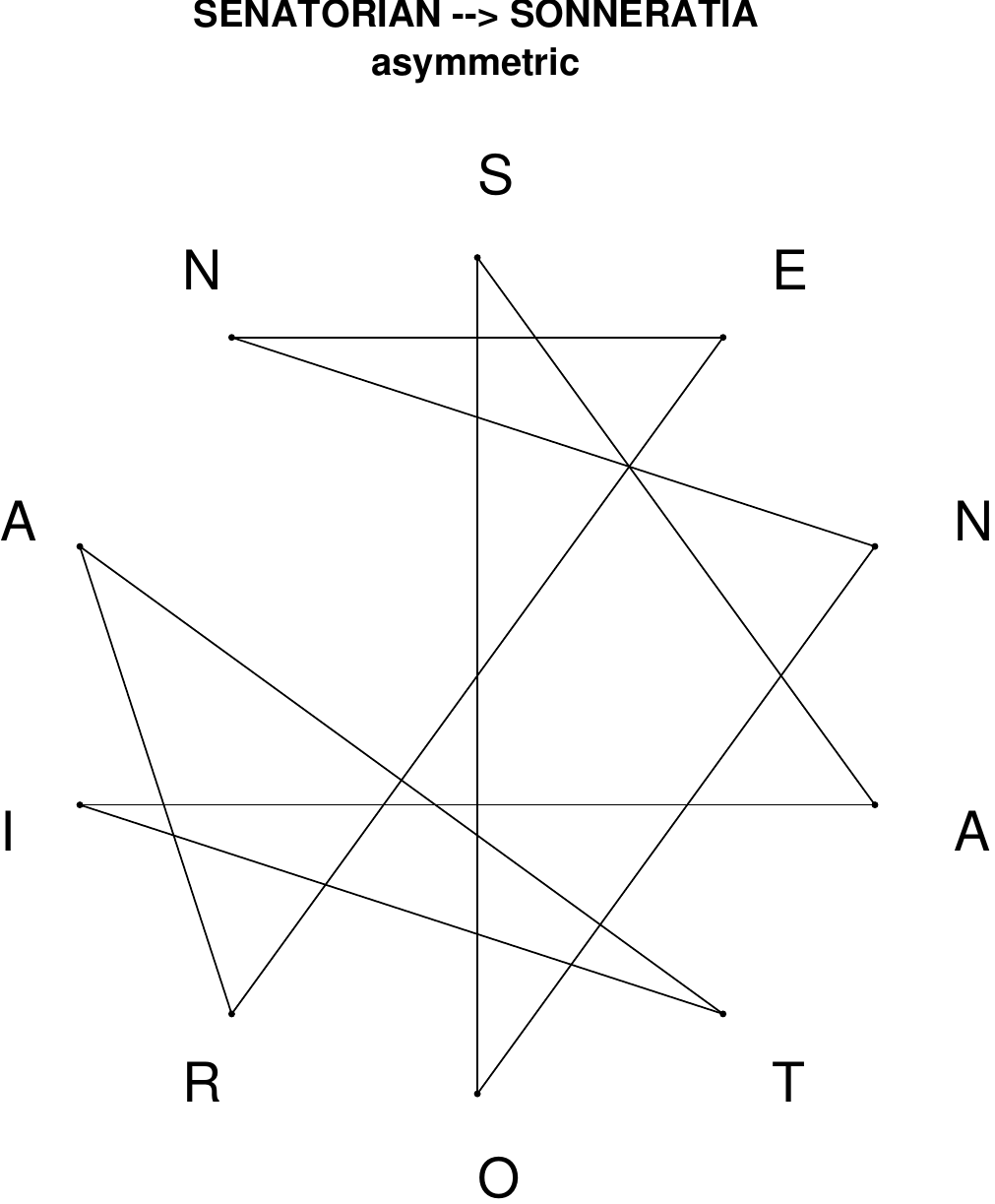}
\end{subfigure}
\hfill
\begin{subfigure}[T]{0.19\textwidth}
\centering
\includegraphics[width=\textwidth]{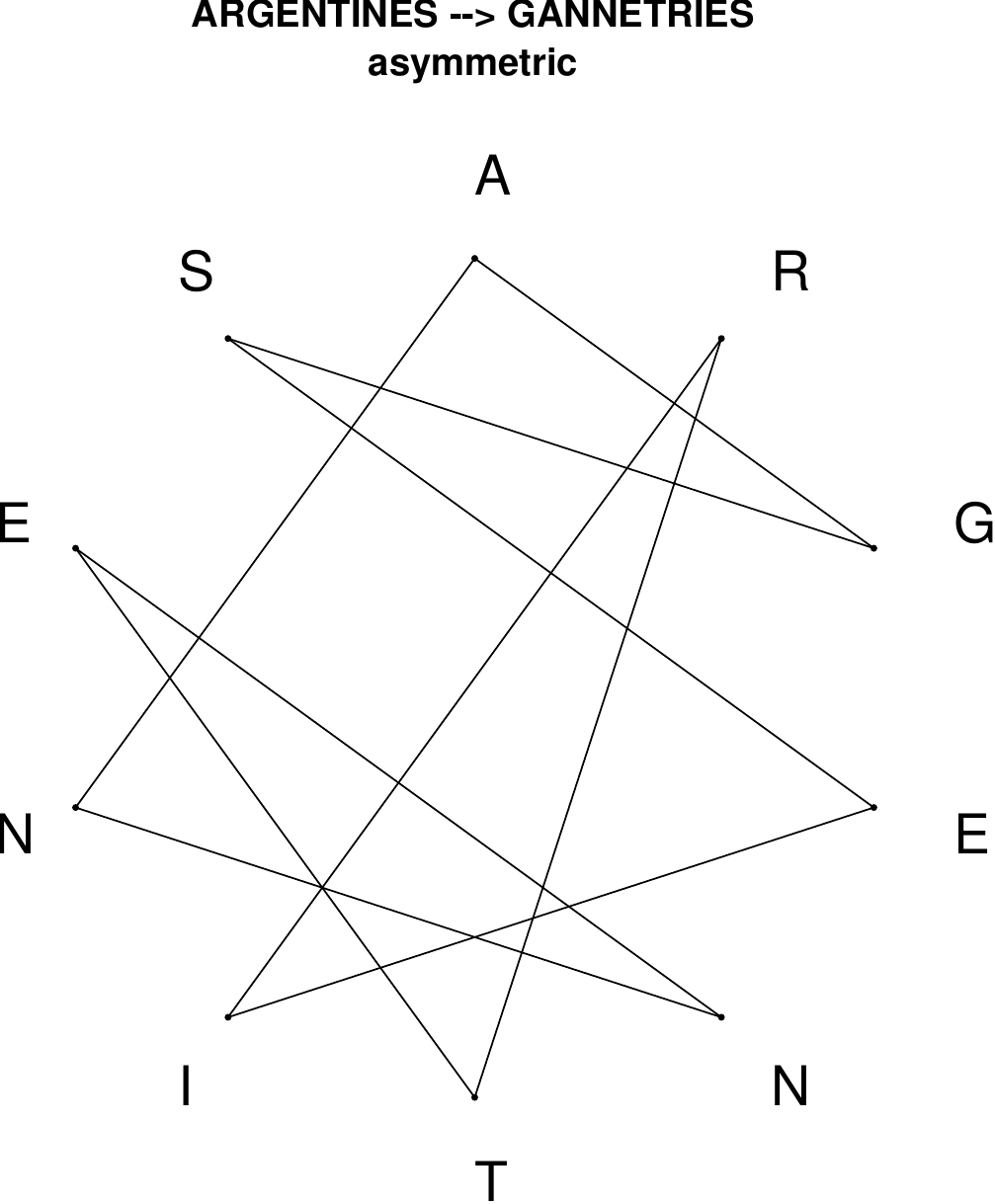}
\end{subfigure}
\hfill
\begin{subfigure}[T]{0.19\textwidth}
\centering
\includegraphics[width=\textwidth]{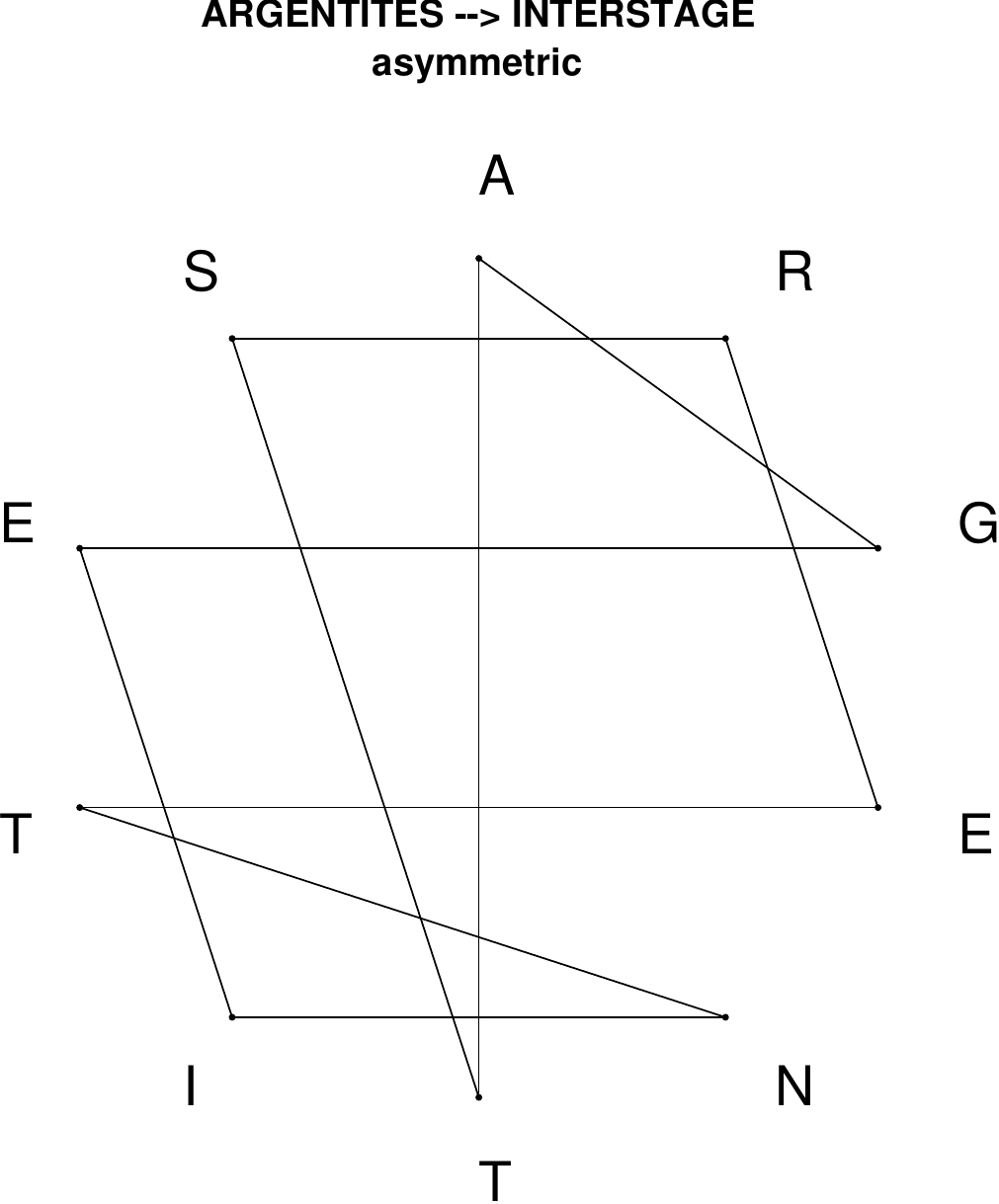}
\end{subfigure}
\end{figure}

\begin{figure}[H]
\centering
\begin{subfigure}[T]{0.19\textwidth}
\centering
\includegraphics[width=\textwidth]{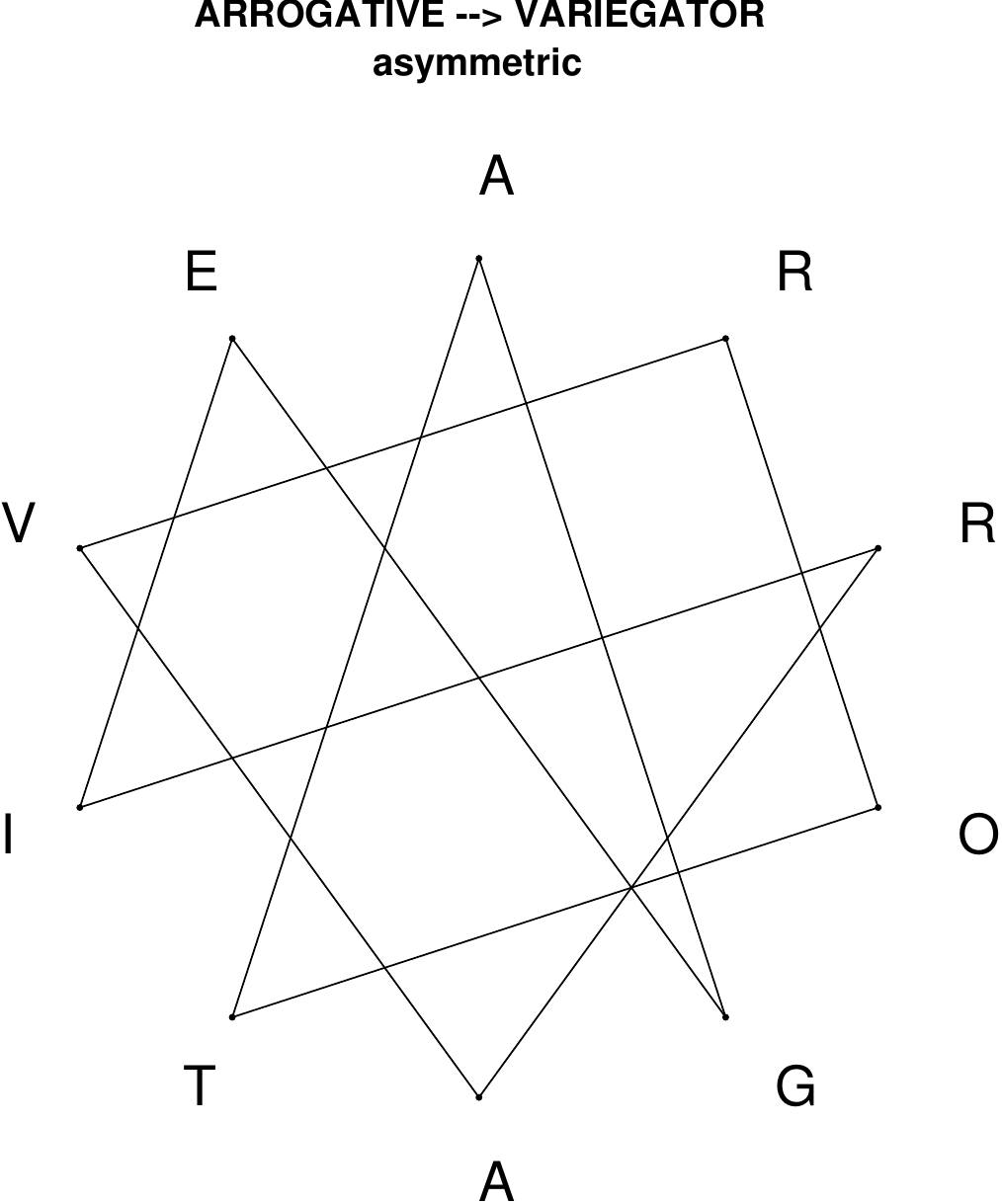}
\end{subfigure}
\hfill
\begin{subfigure}[T]{0.19\textwidth}
\centering
\includegraphics[width=\textwidth]{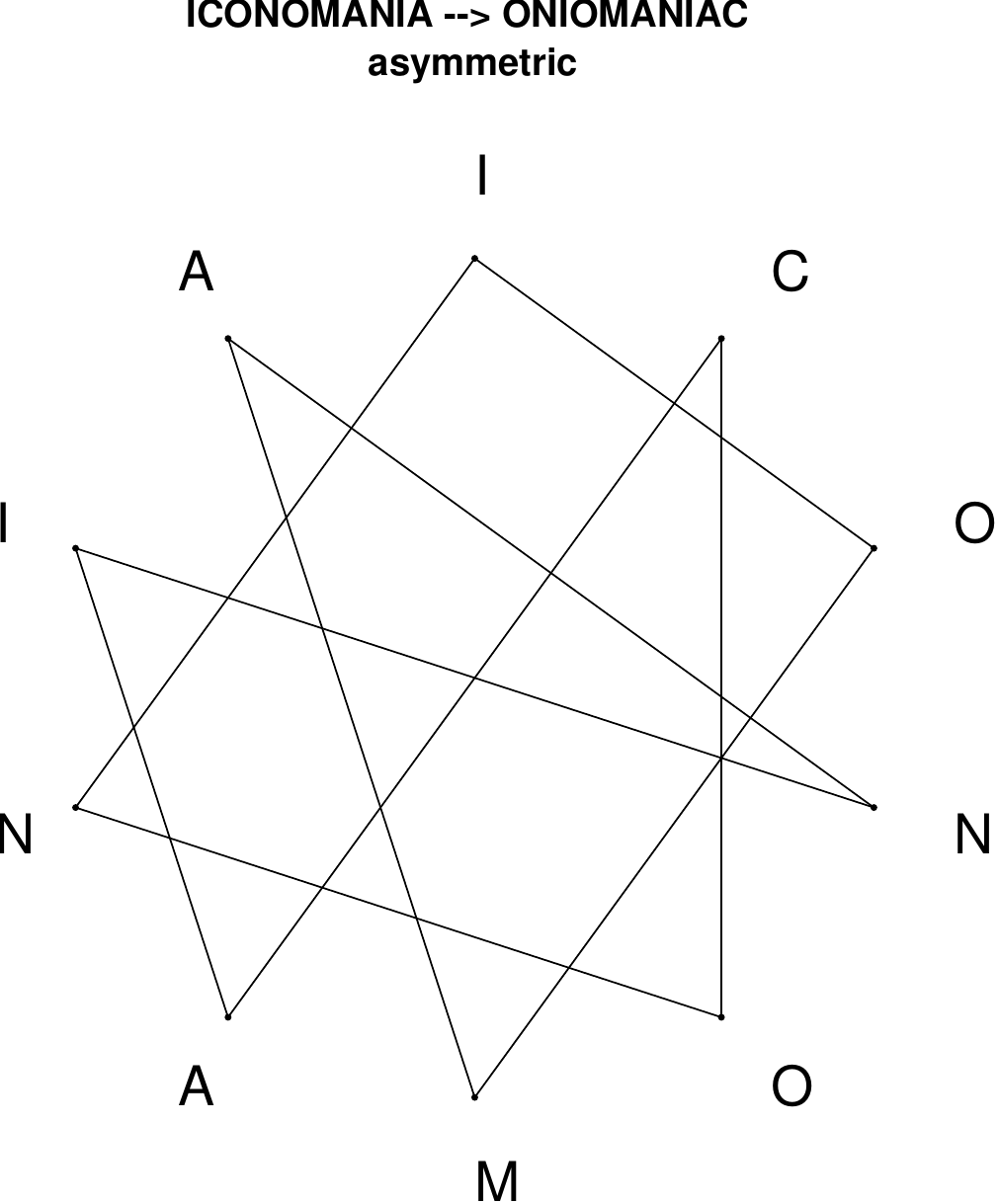}
\end{subfigure}
\hfill
\begin{subfigure}[T]{0.19\textwidth}
\centering
\includegraphics[width=\textwidth]{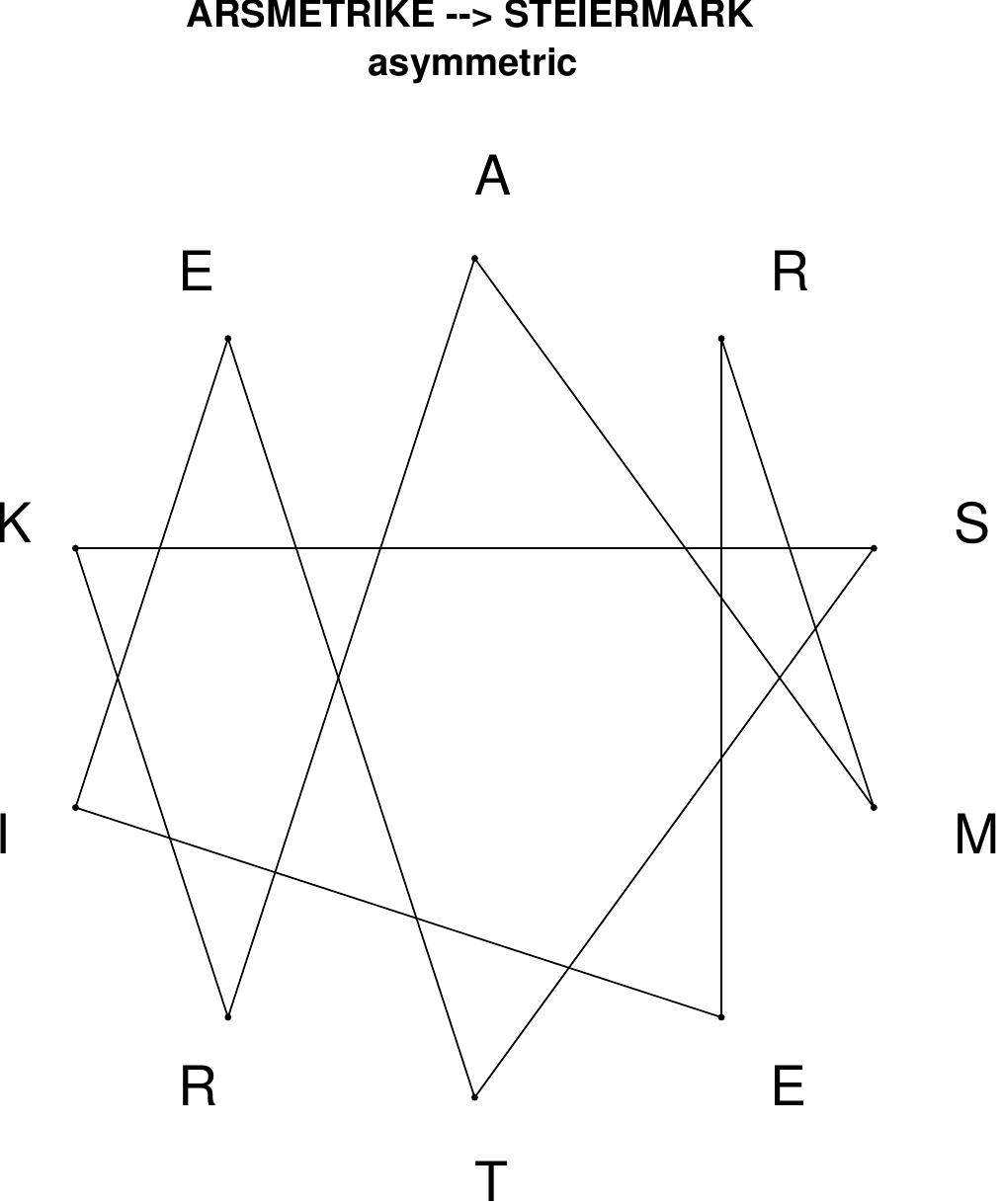}
\end{subfigure}
\hfill
\begin{subfigure}[T]{0.19\textwidth}
\centering
\includegraphics[width=\textwidth]{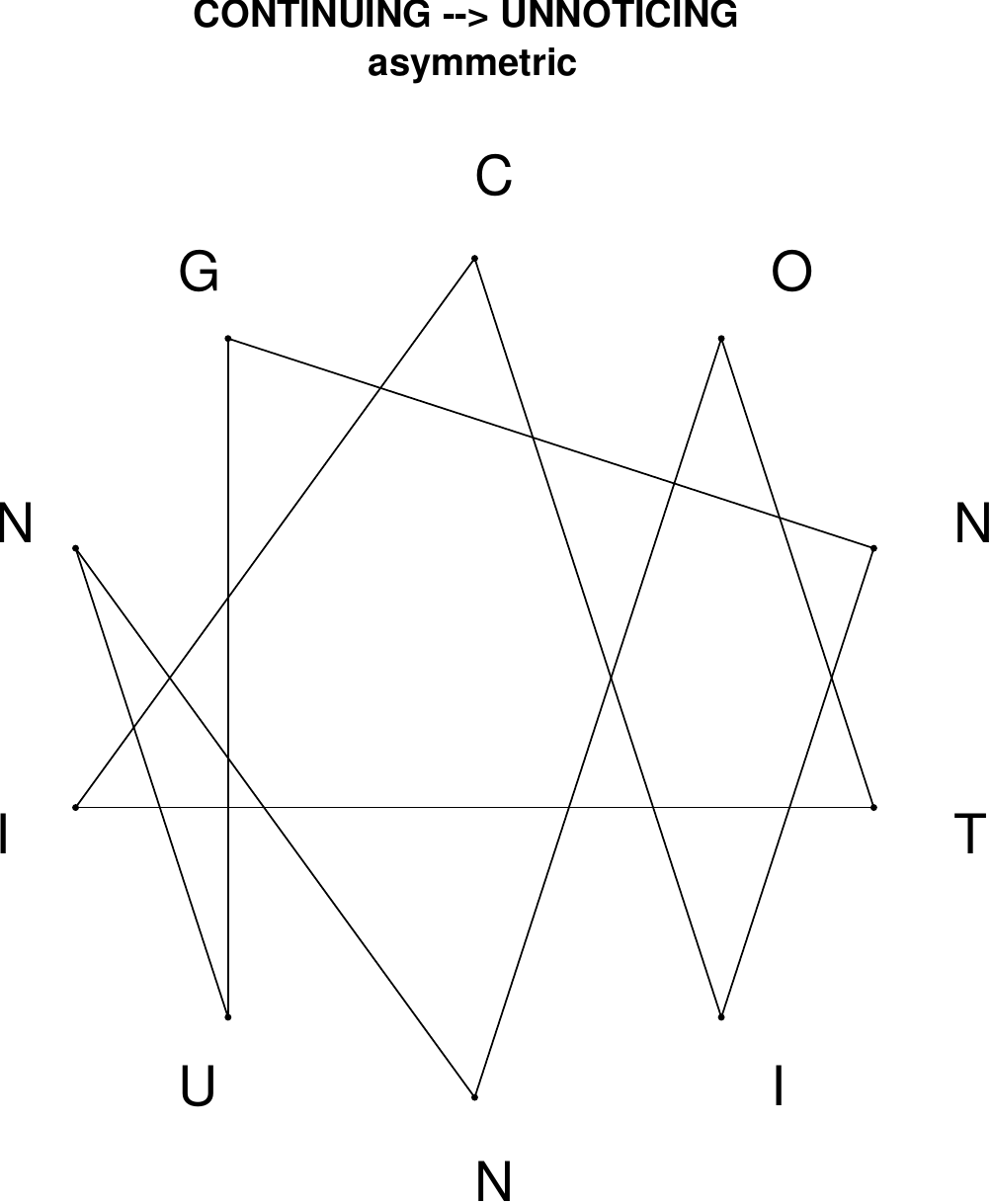}
\end{subfigure}
\hfill
\begin{subfigure}[T]{0.19\textwidth}
\centering
\includegraphics[width=\textwidth]{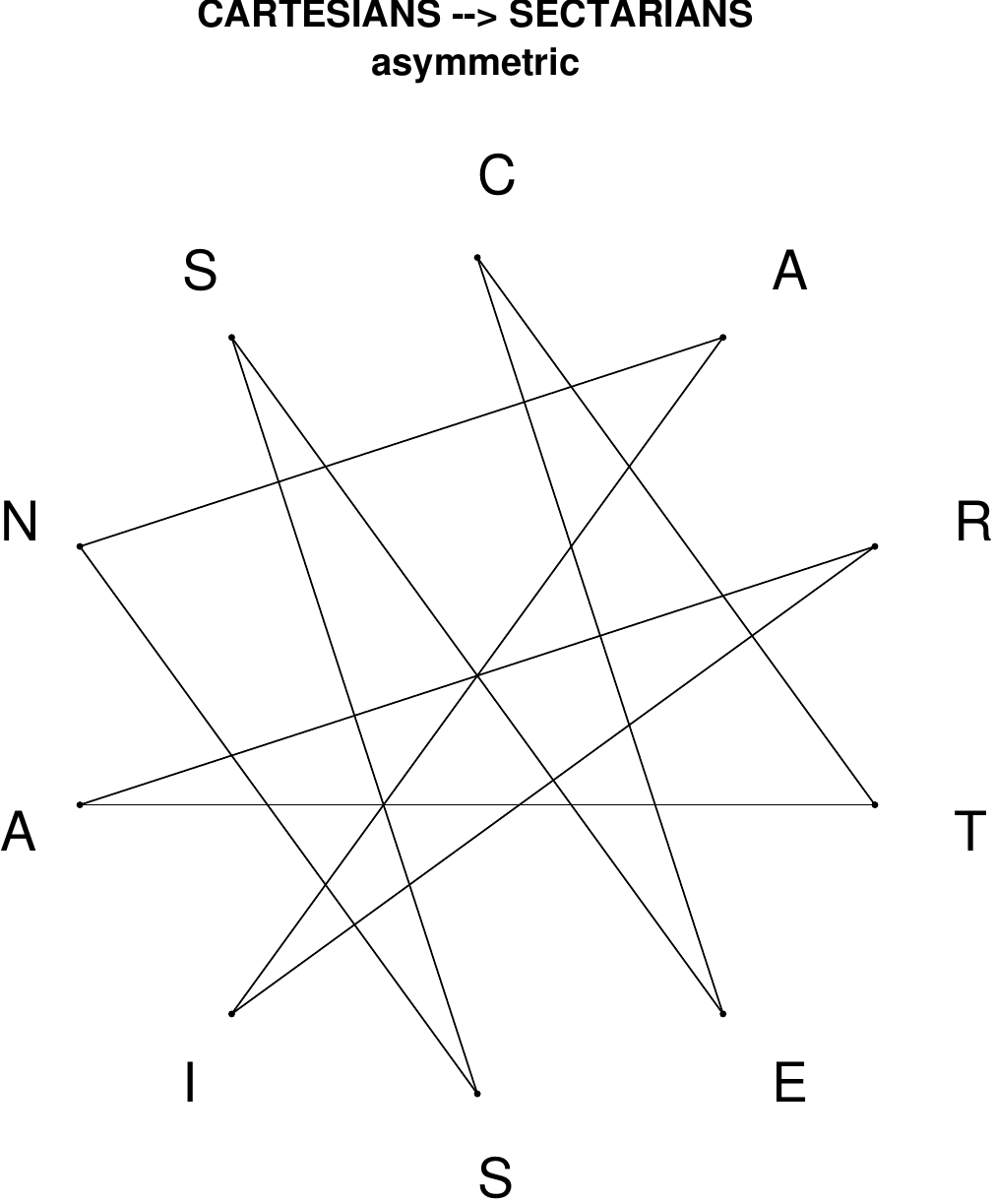}
\end{subfigure}
\end{figure}

\begin{figure}[H]
\centering
\begin{subfigure}[T]{0.19\textwidth}
\centering
\includegraphics[width=\textwidth]{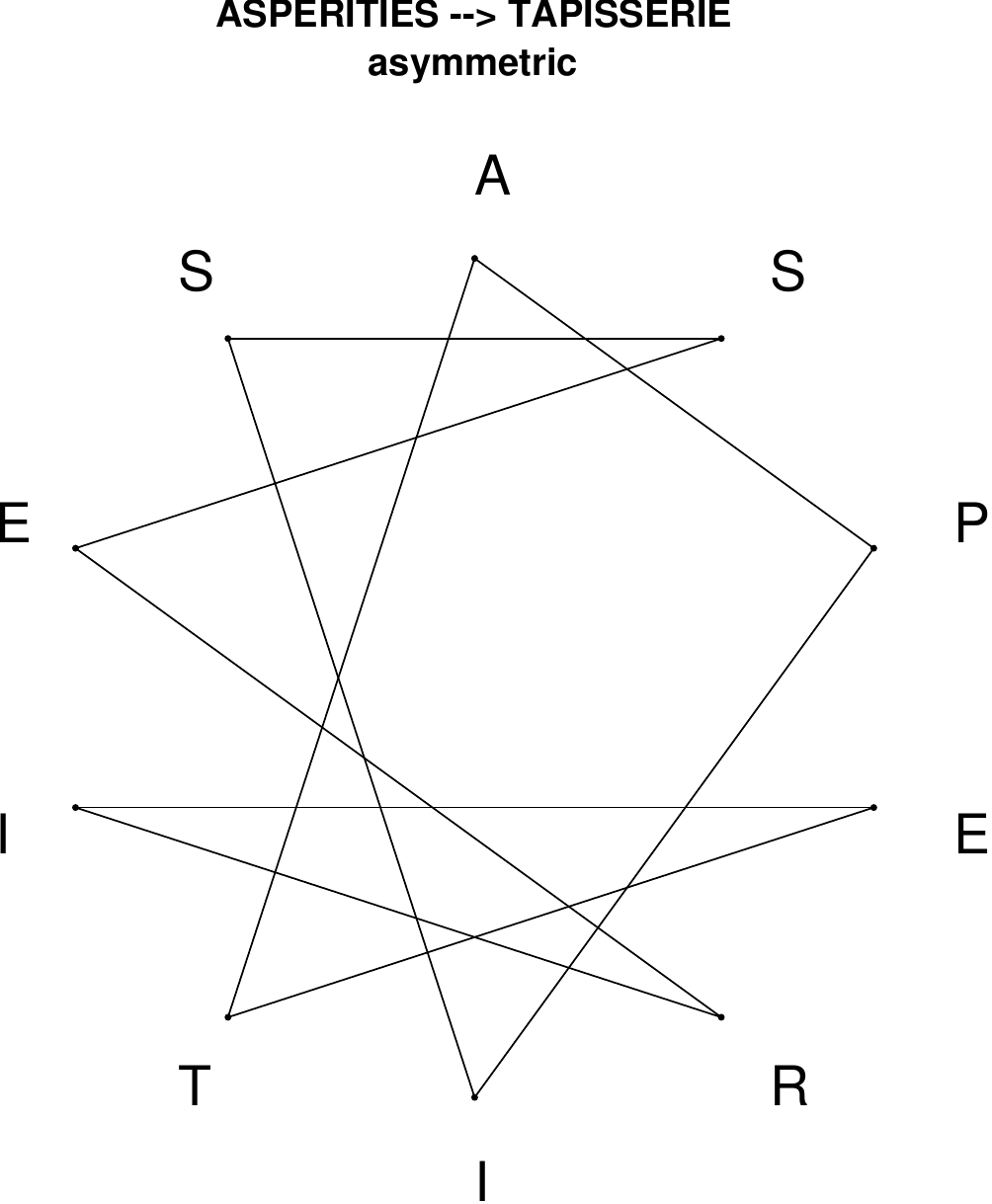}
\end{subfigure}
\hfill
\begin{subfigure}[T]{0.19\textwidth}
\centering
\includegraphics[width=\textwidth]{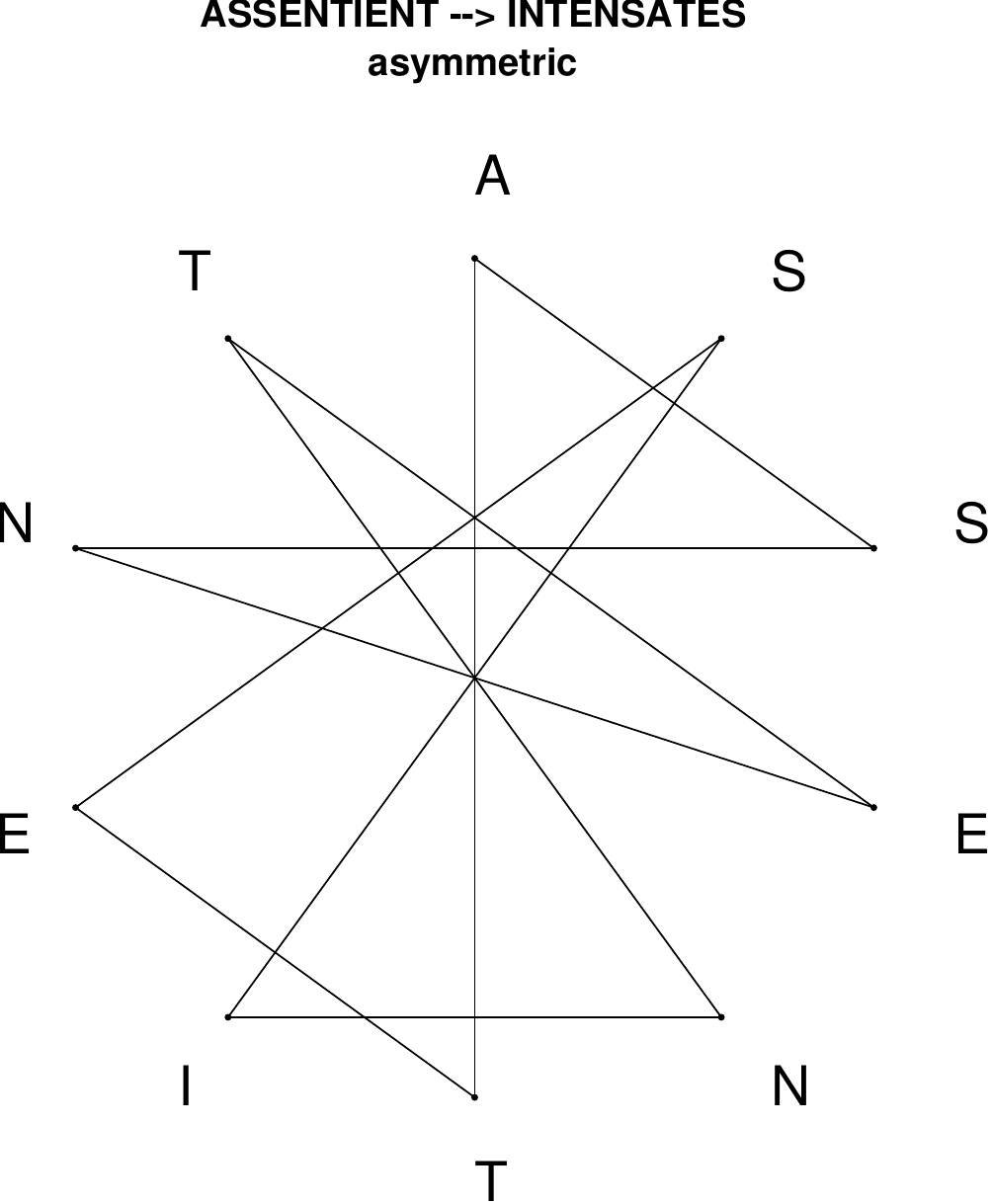}
\end{subfigure}
\hfill
\begin{subfigure}[T]{0.19\textwidth}
\centering
\includegraphics[width=\textwidth]{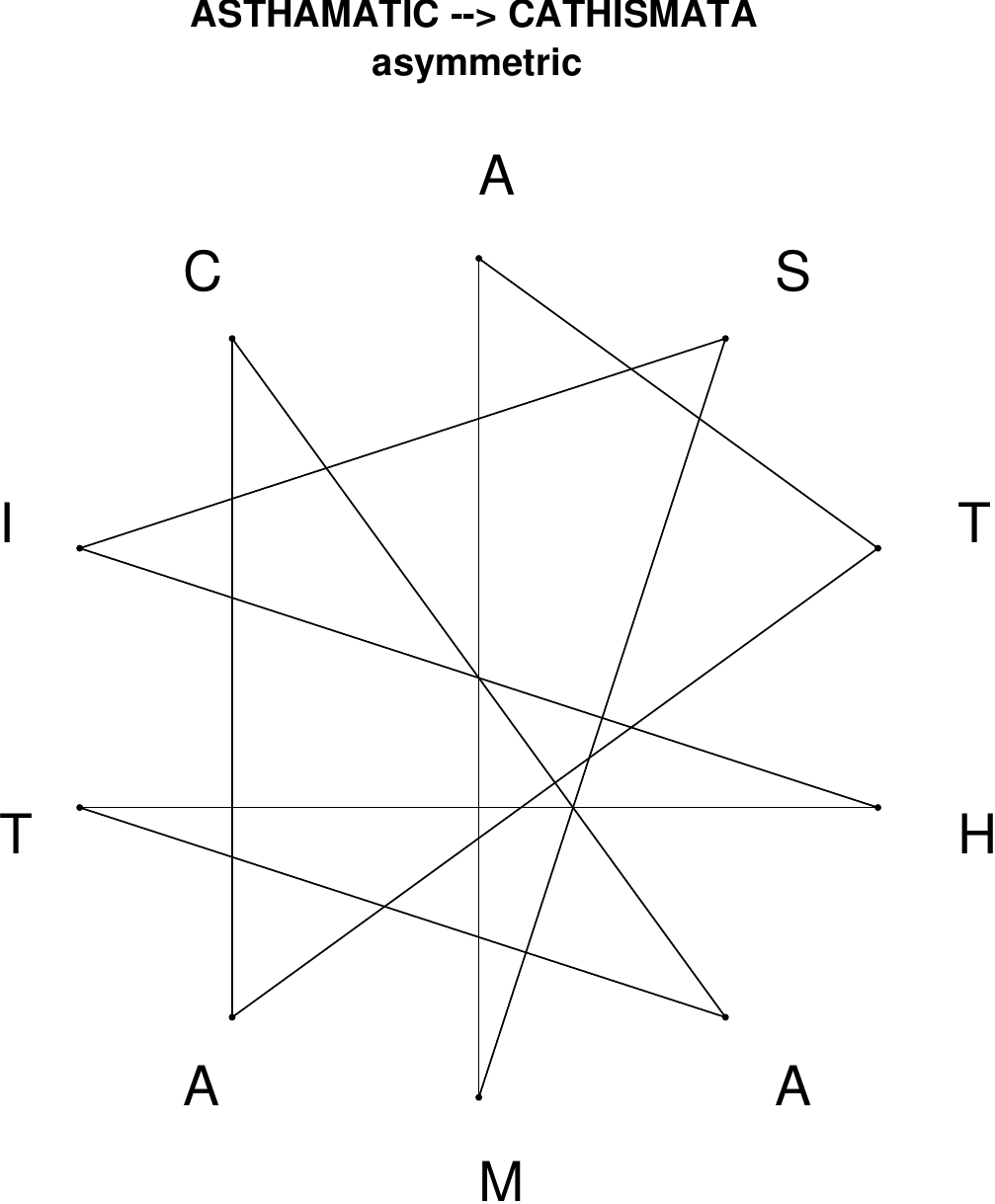}
\end{subfigure}
\hfill
\begin{subfigure}[T]{0.19\textwidth}
\centering
\includegraphics[width=\textwidth]{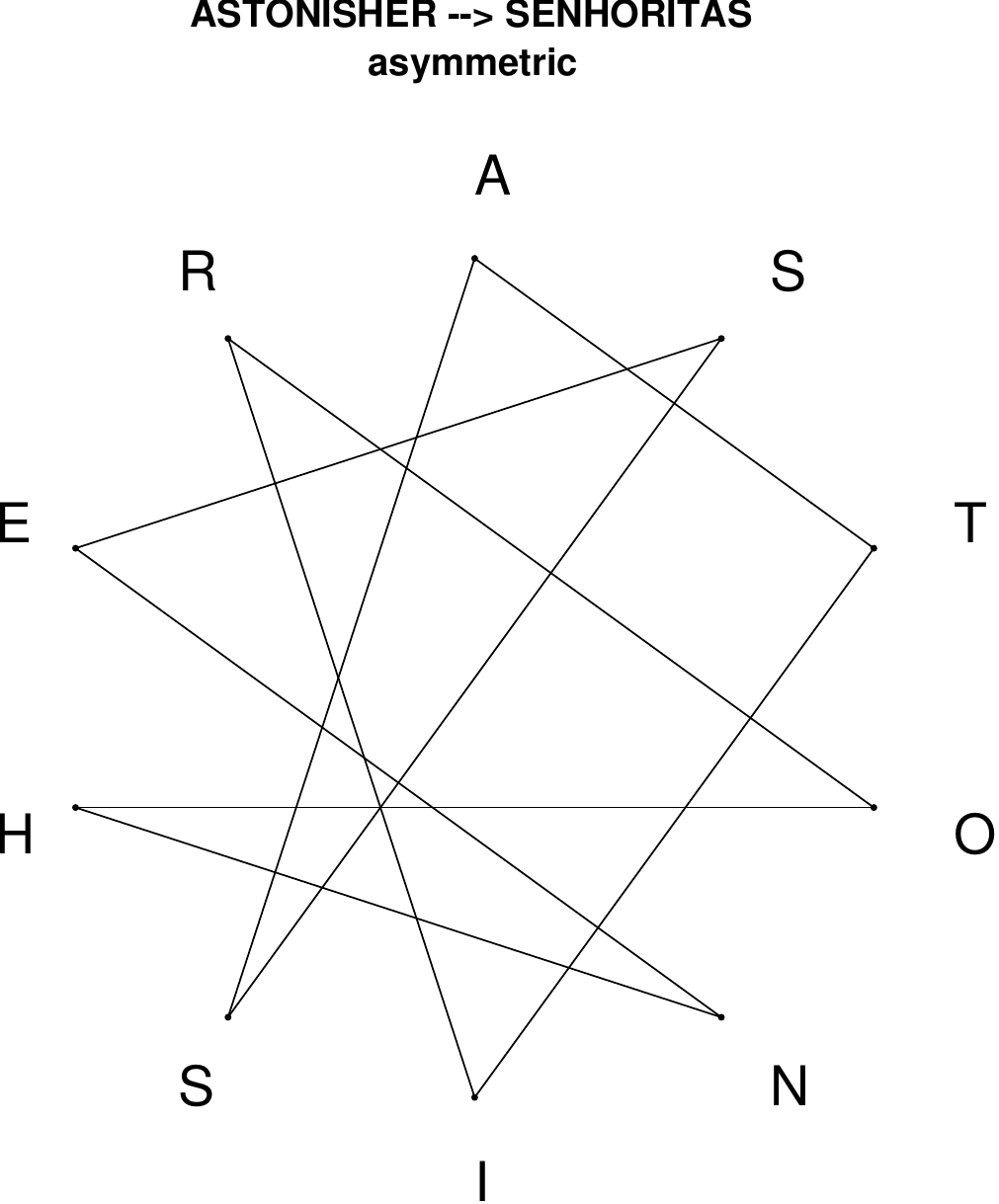}
\end{subfigure}
\hfill
\begin{subfigure}[T]{0.19\textwidth}
\centering
\includegraphics[width=\textwidth]{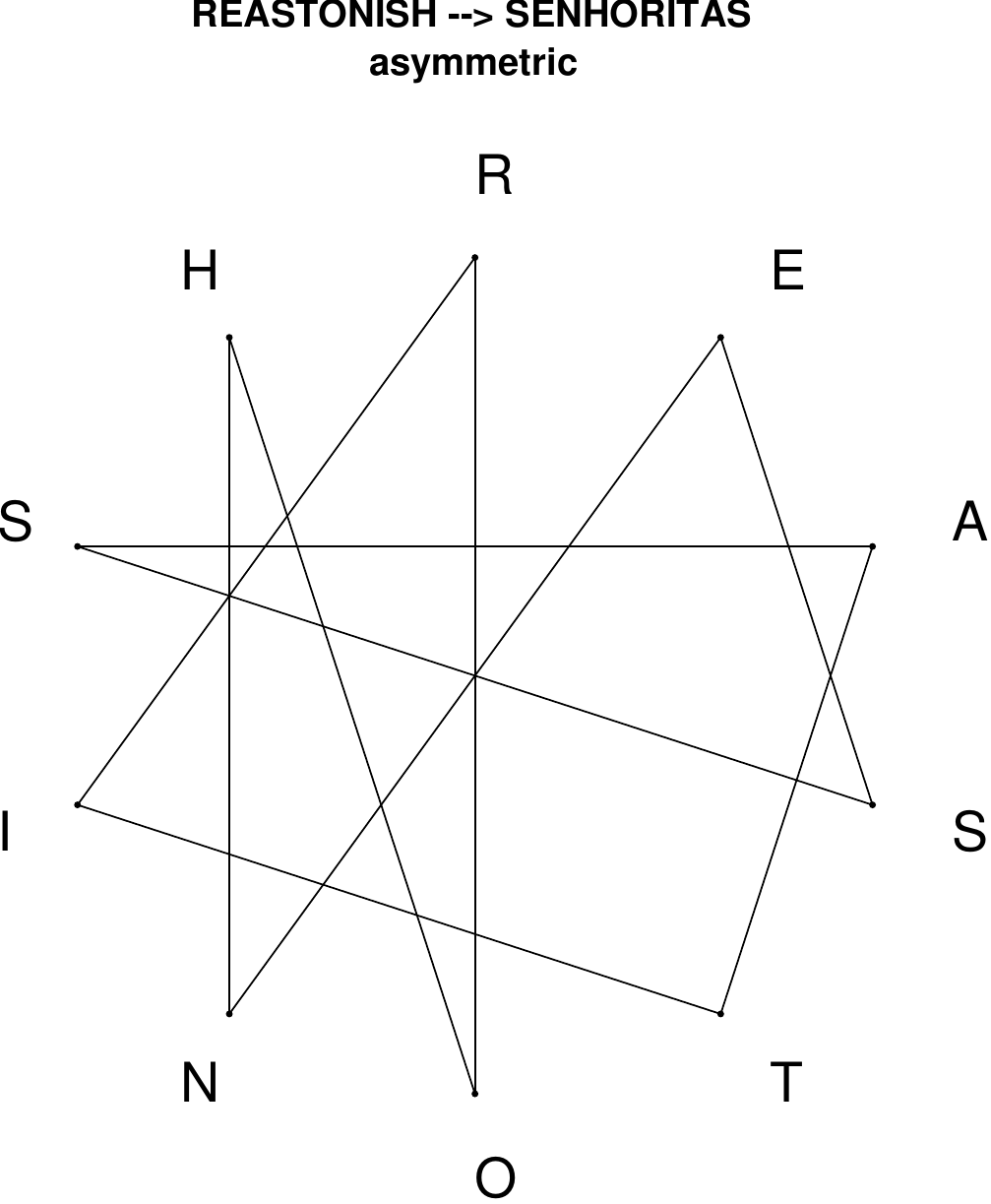}
\end{subfigure}
\end{figure}

\begin{figure}[H]
\centering
\begin{subfigure}[T]{0.19\textwidth}
\centering
\includegraphics[width=\textwidth]{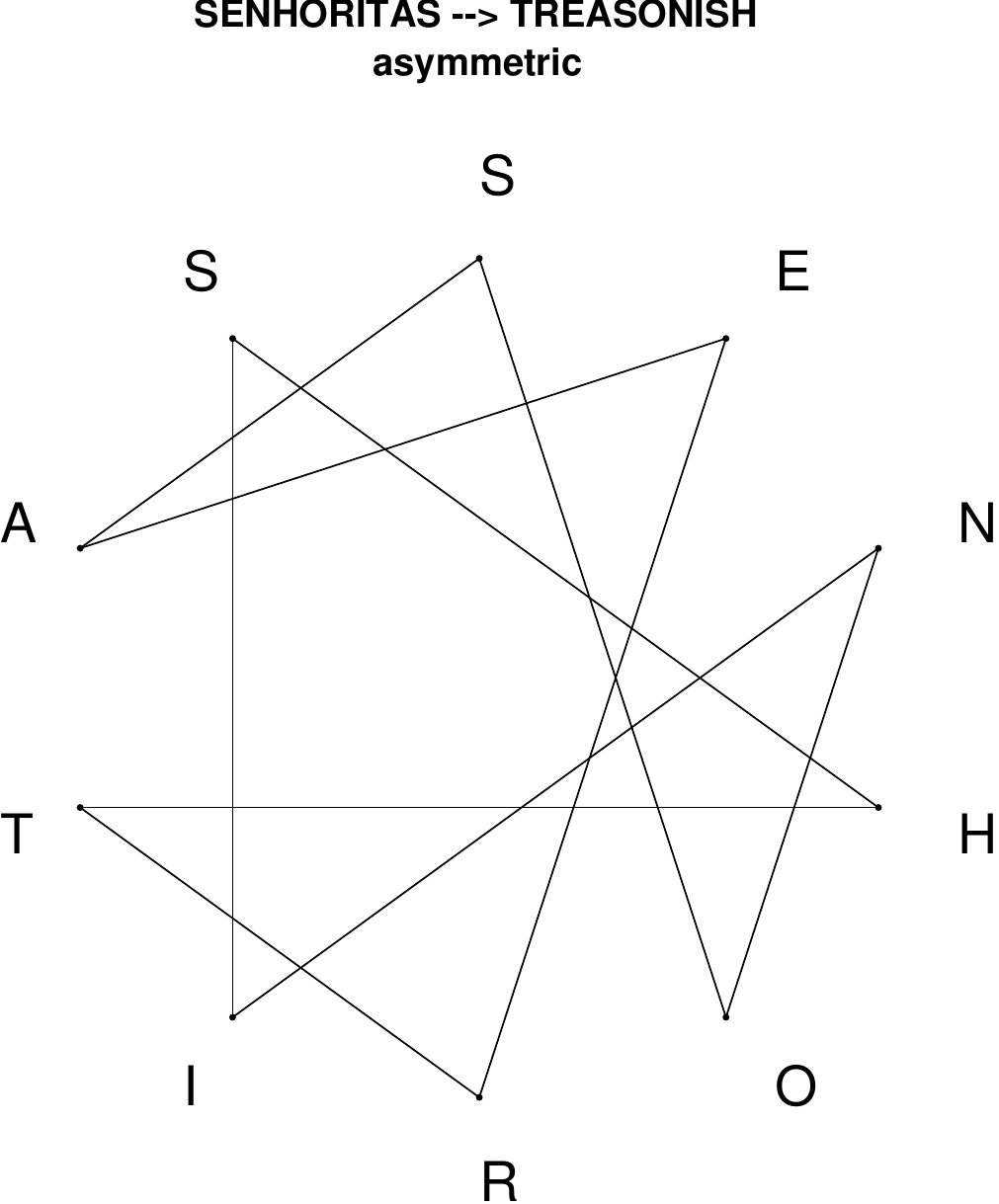}
\end{subfigure}
\hfill
\begin{subfigure}[T]{0.19\textwidth}
\centering
\includegraphics[width=\textwidth]{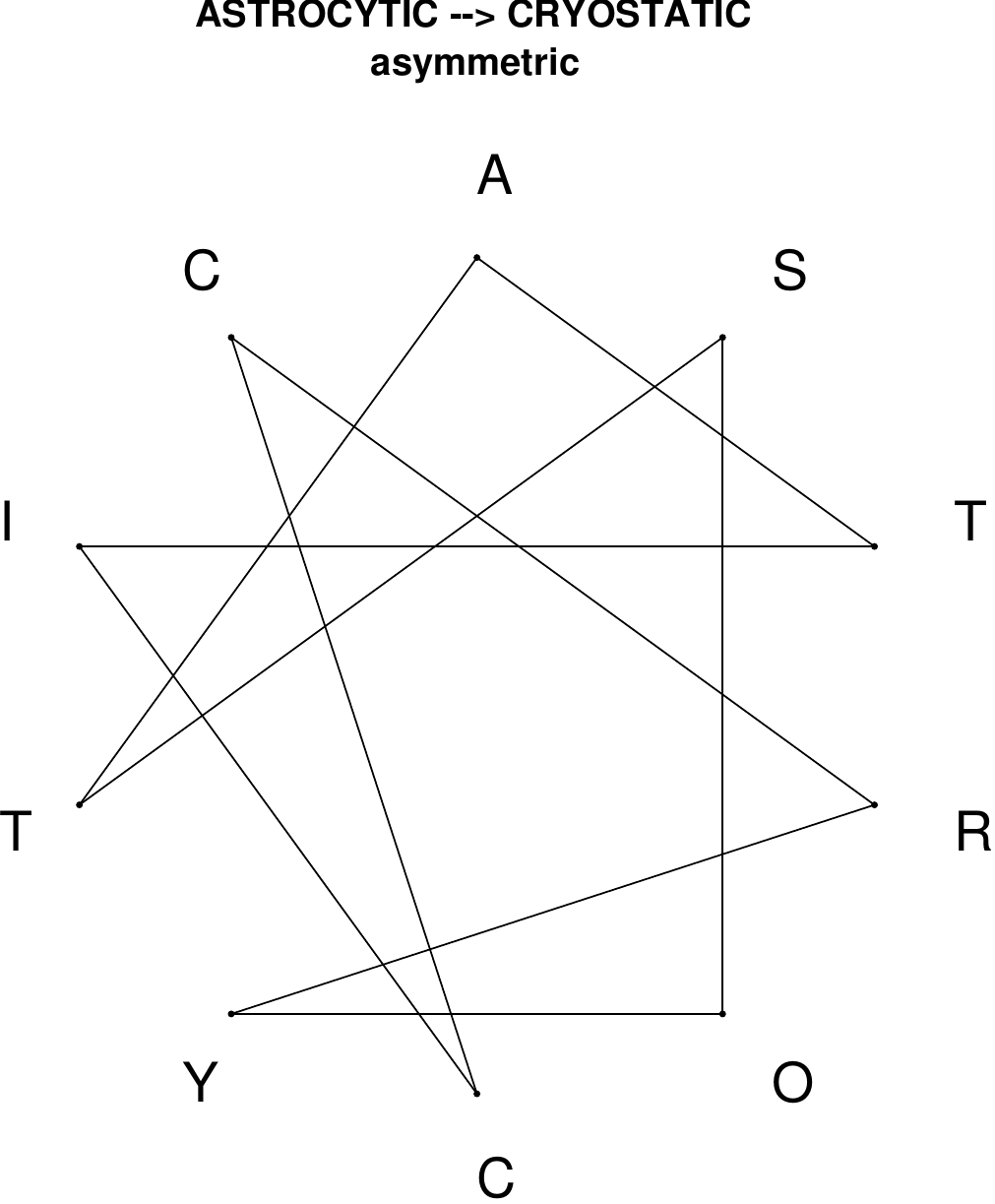}
\end{subfigure}
\hfill
\begin{subfigure}[T]{0.19\textwidth}
\centering
\includegraphics[width=\textwidth]{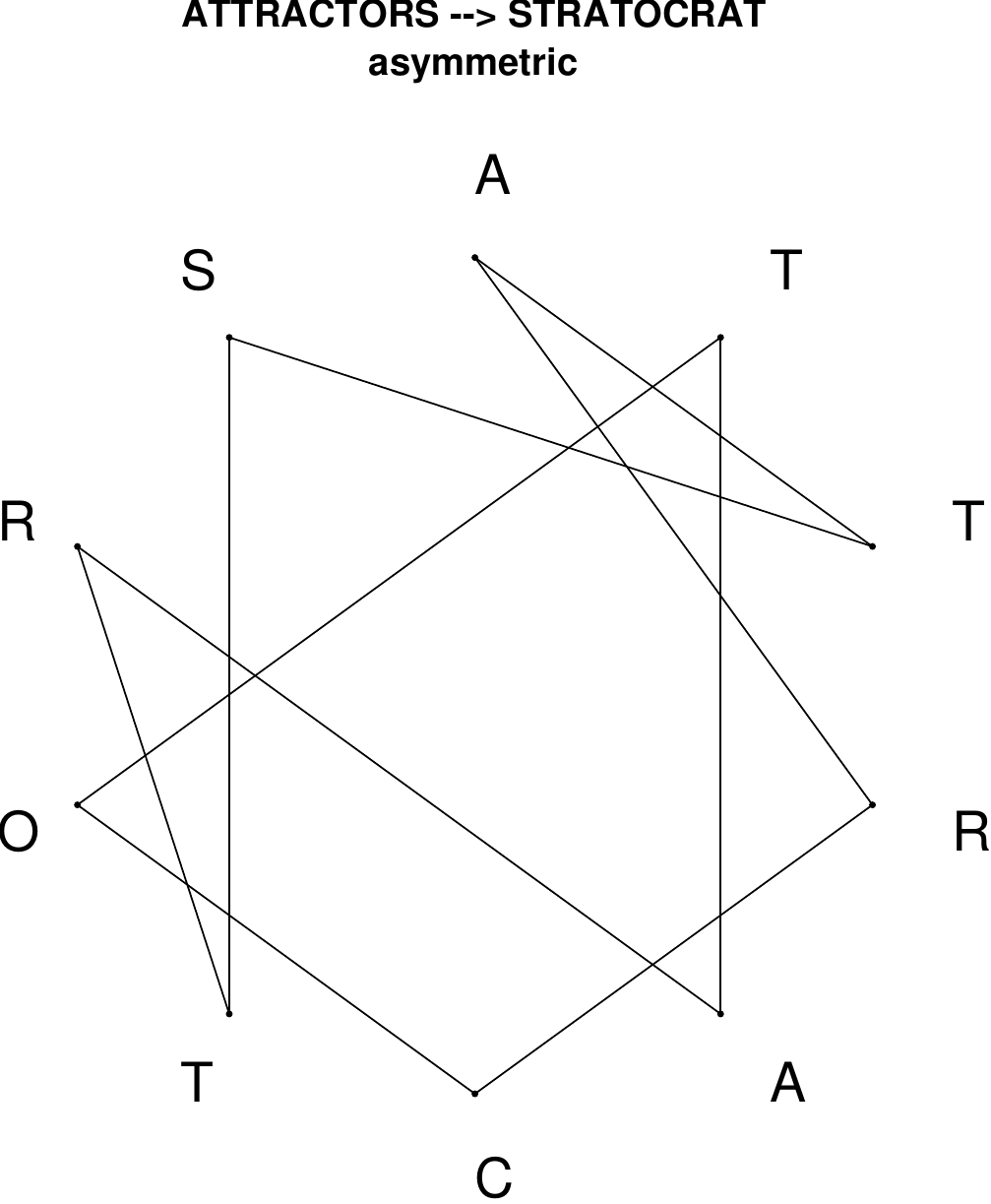}
\end{subfigure}
\hfill
\begin{subfigure}[T]{0.19\textwidth}
\centering
\includegraphics[width=\textwidth]{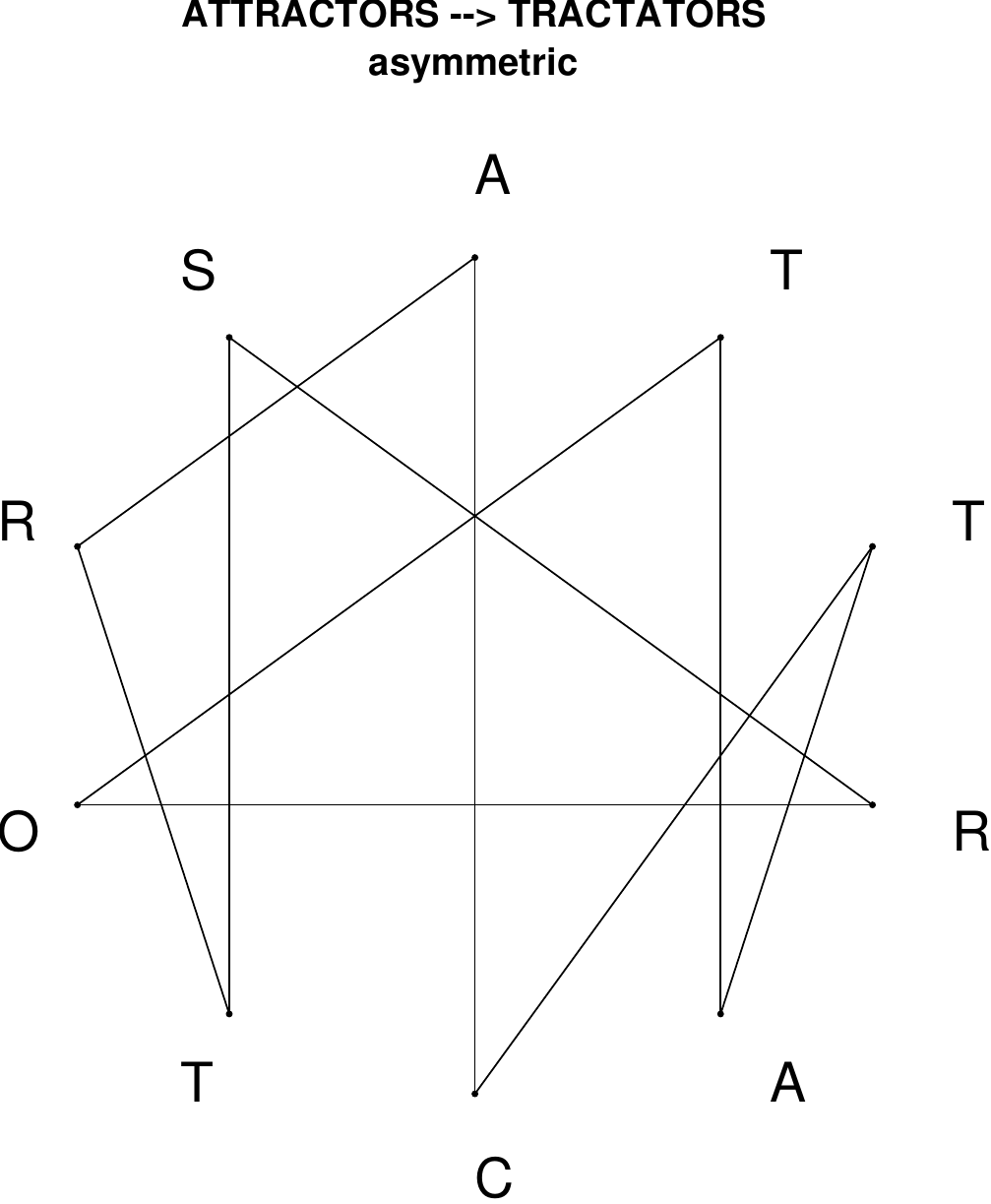}
\end{subfigure}
\hfill
\begin{subfigure}[T]{0.19\textwidth}
\centering
\includegraphics[width=\textwidth]{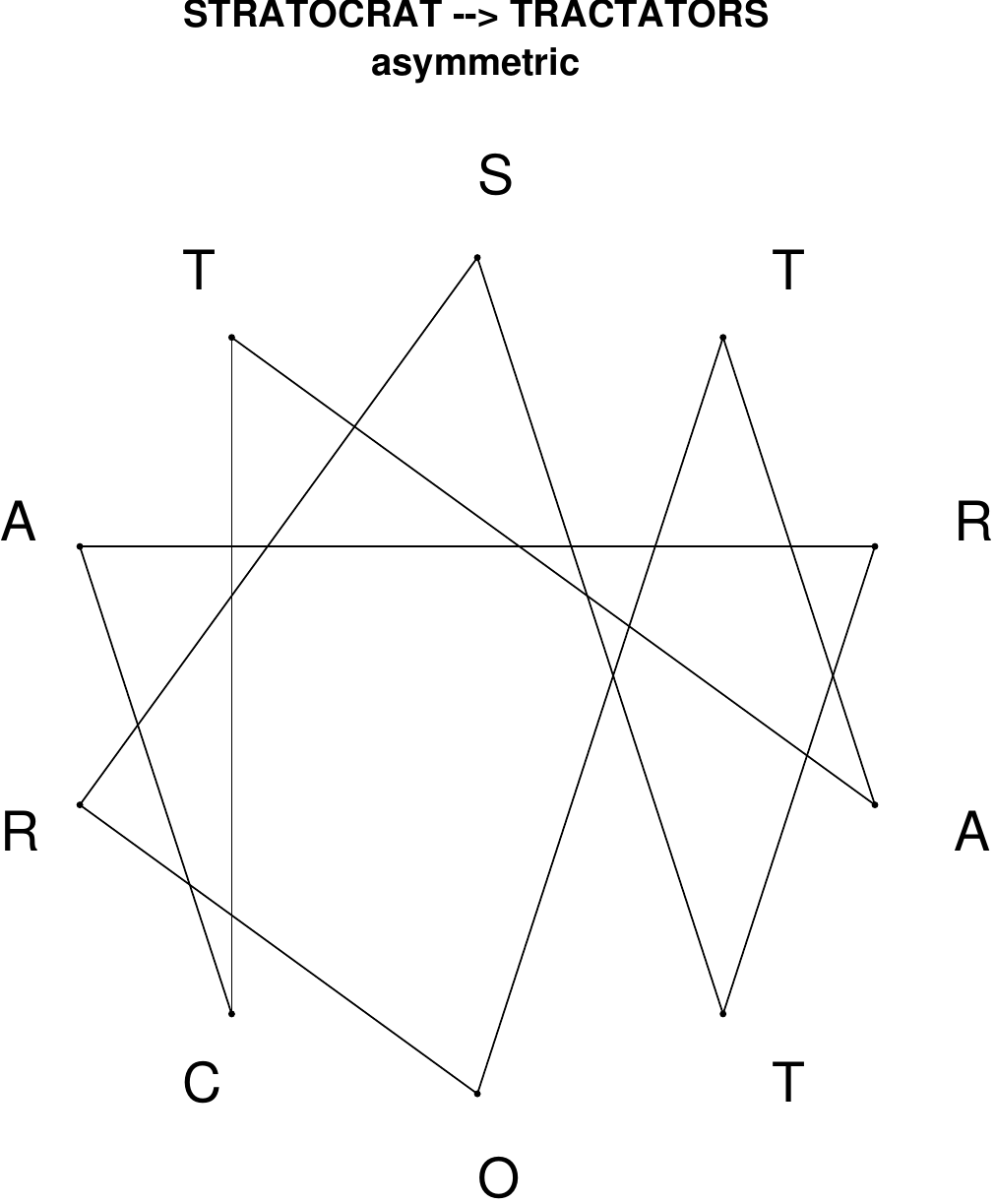}
\end{subfigure}
\end{figure}

\begin{figure}[H]
\centering
\begin{subfigure}[T]{0.19\textwidth}
\centering
\includegraphics[width=\textwidth]{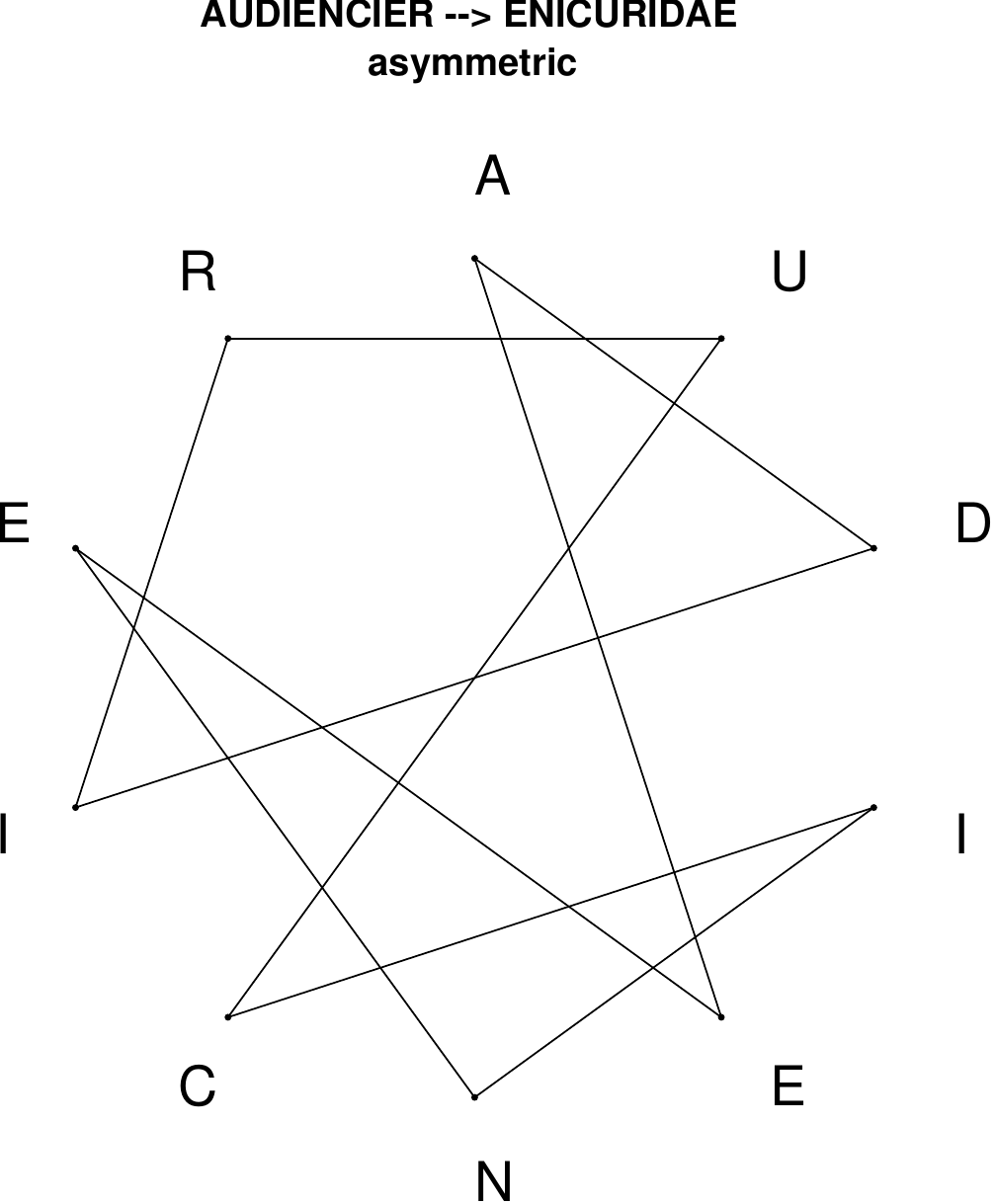}
\end{subfigure}
\hfill
\begin{subfigure}[T]{0.19\textwidth}
\centering
\includegraphics[width=\textwidth]{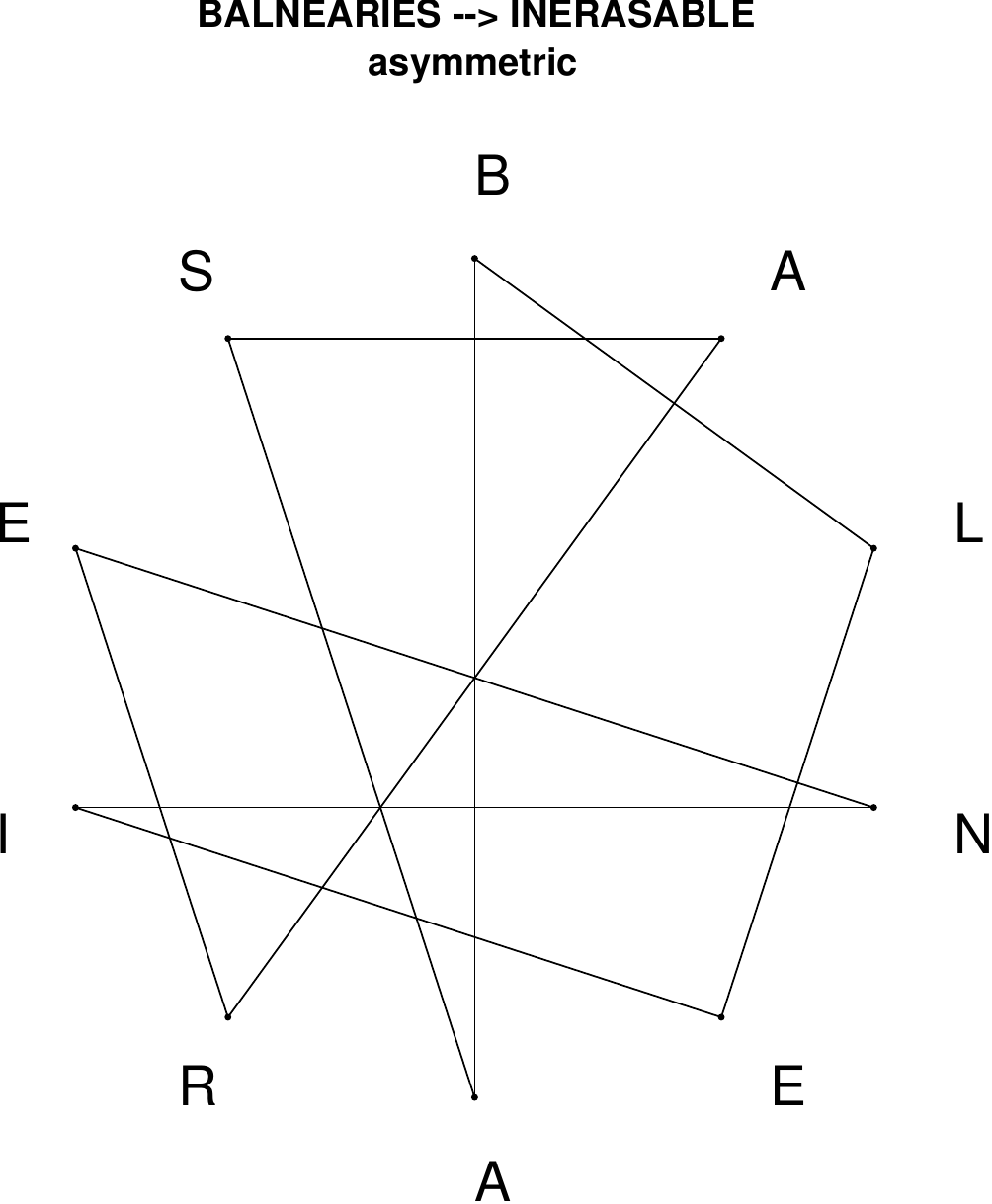}
\end{subfigure}
\hfill
\begin{subfigure}[T]{0.19\textwidth}
\centering
\includegraphics[width=\textwidth]{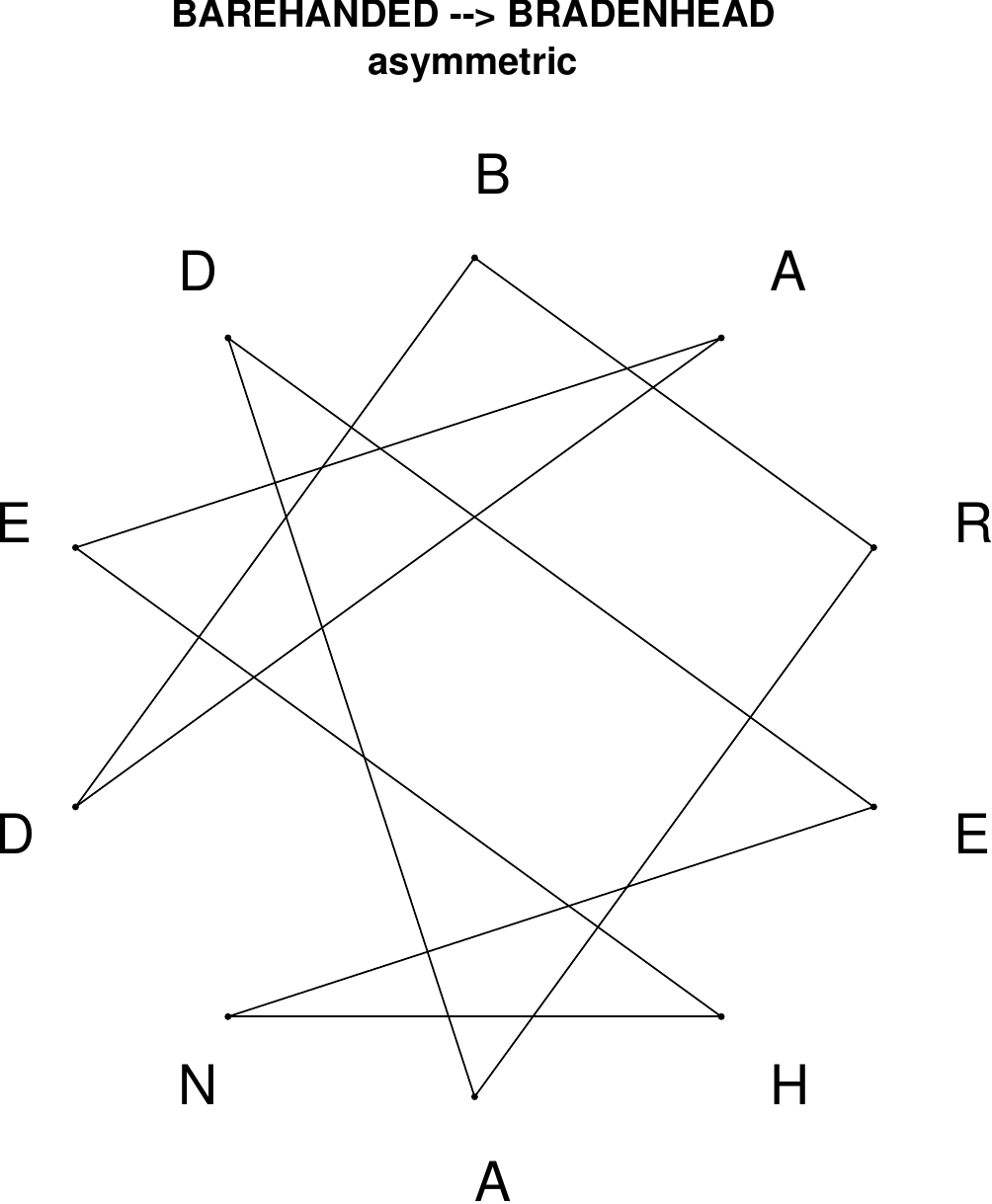}
\end{subfigure}
\hfill
\begin{subfigure}[T]{0.19\textwidth}
\centering
\includegraphics[width=\textwidth]{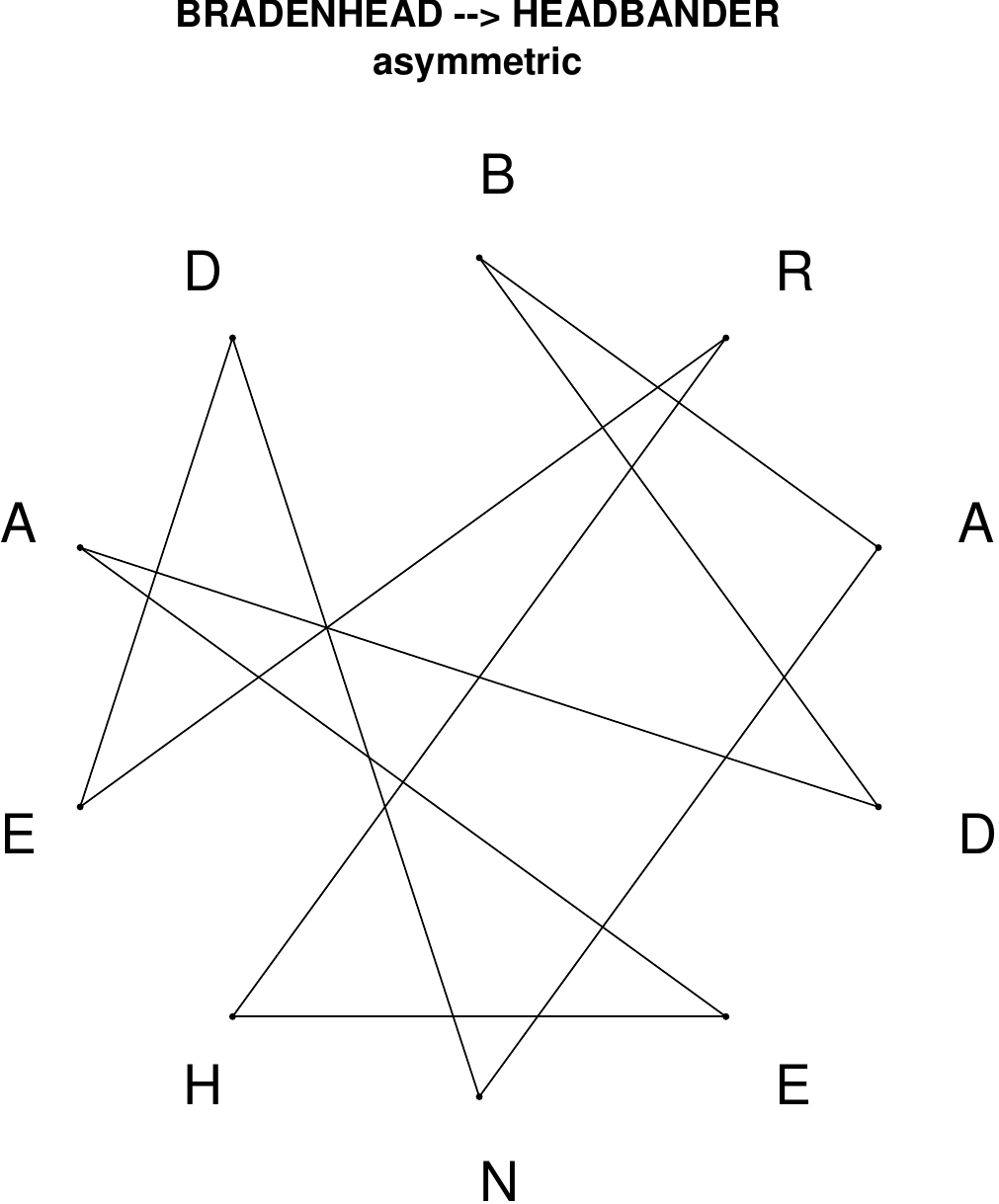}
\end{subfigure}
\hfill
\begin{subfigure}[T]{0.19\textwidth}
\centering
\includegraphics[width=\textwidth]{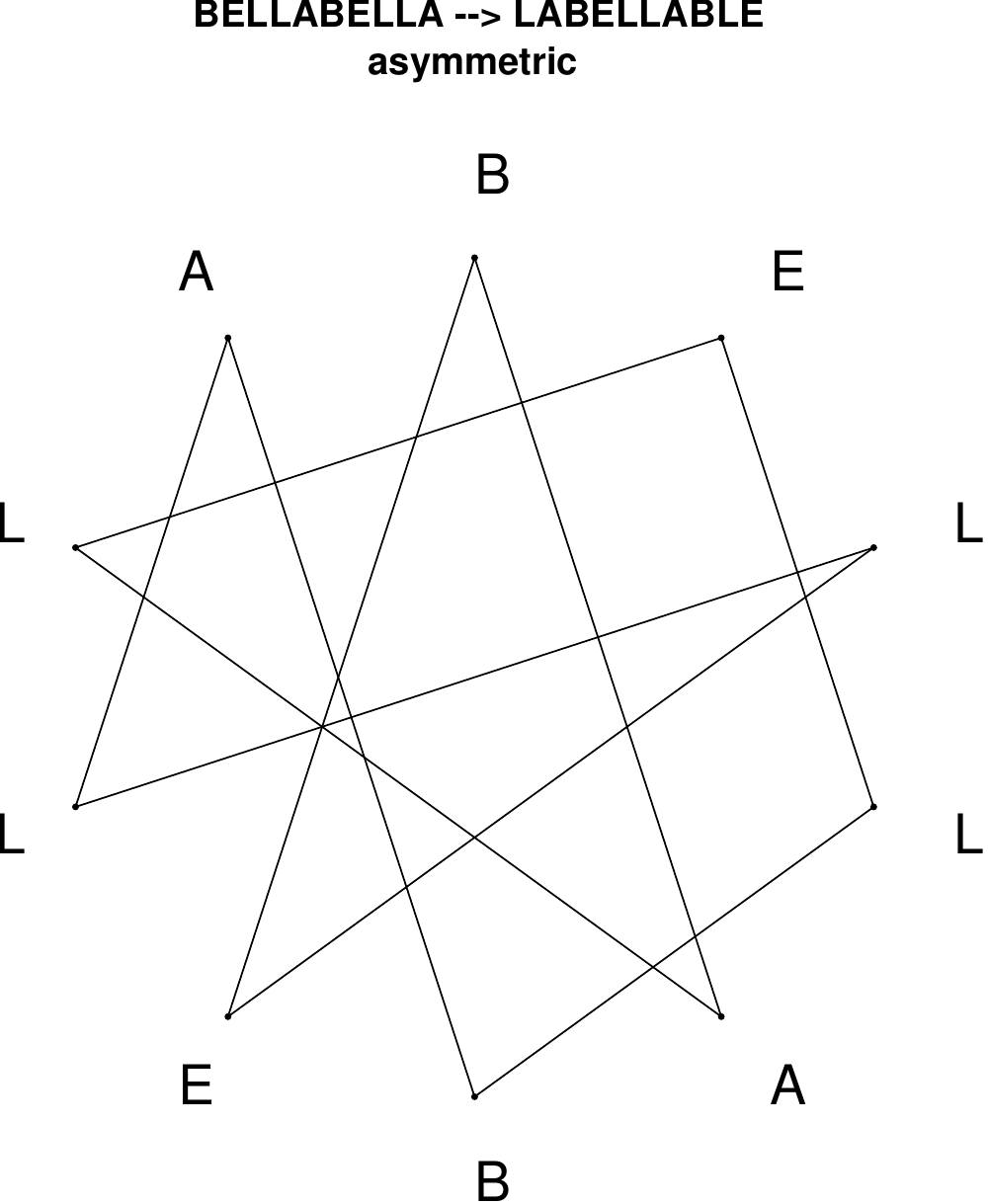}
\end{subfigure}
\end{figure}

\begin{figure}[H]
\centering
\begin{subfigure}[T]{0.19\textwidth}
\centering
\includegraphics[width=\textwidth]{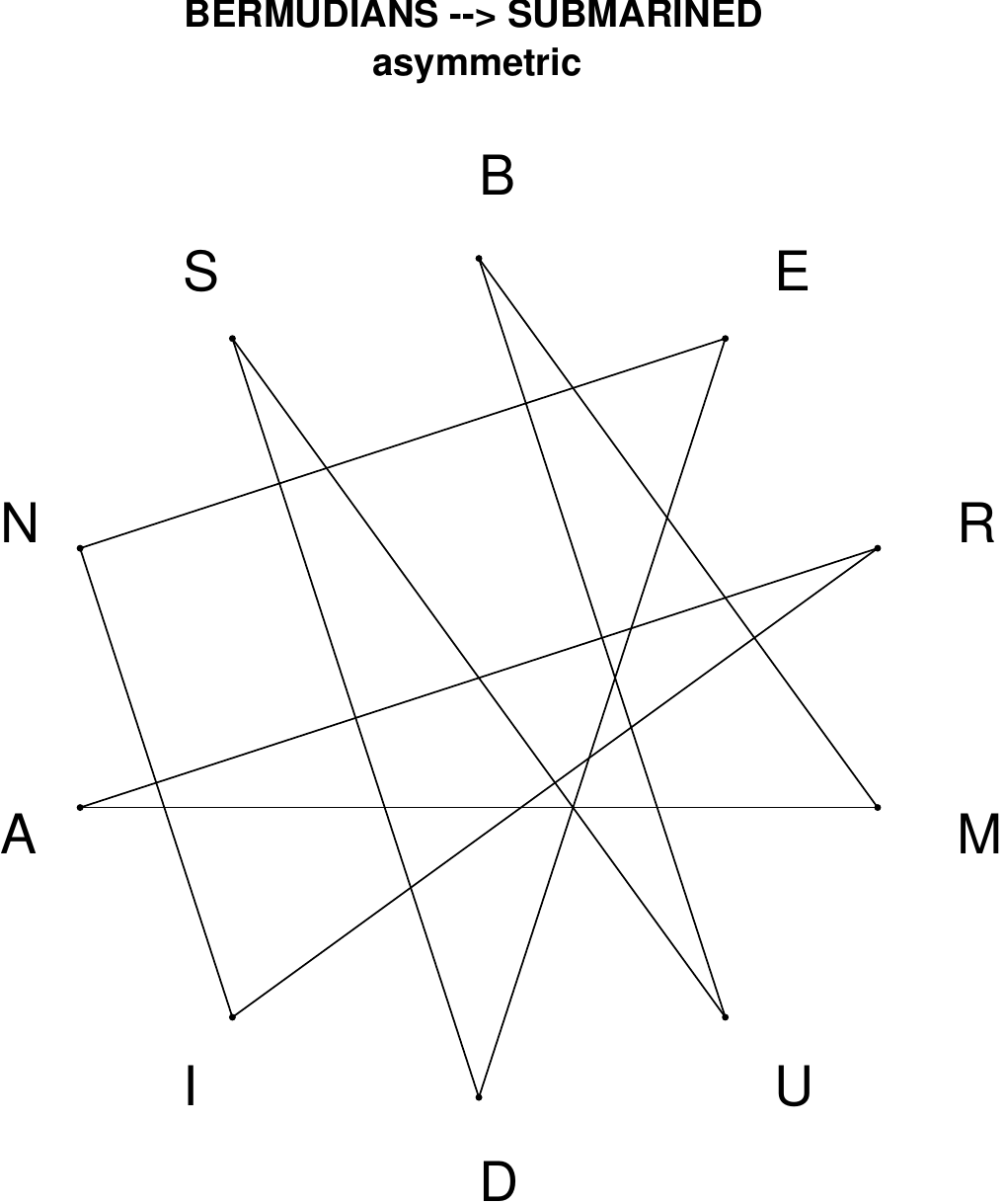}
\end{subfigure}
\hfill
\begin{subfigure}[T]{0.19\textwidth}
\centering
\includegraphics[width=\textwidth]{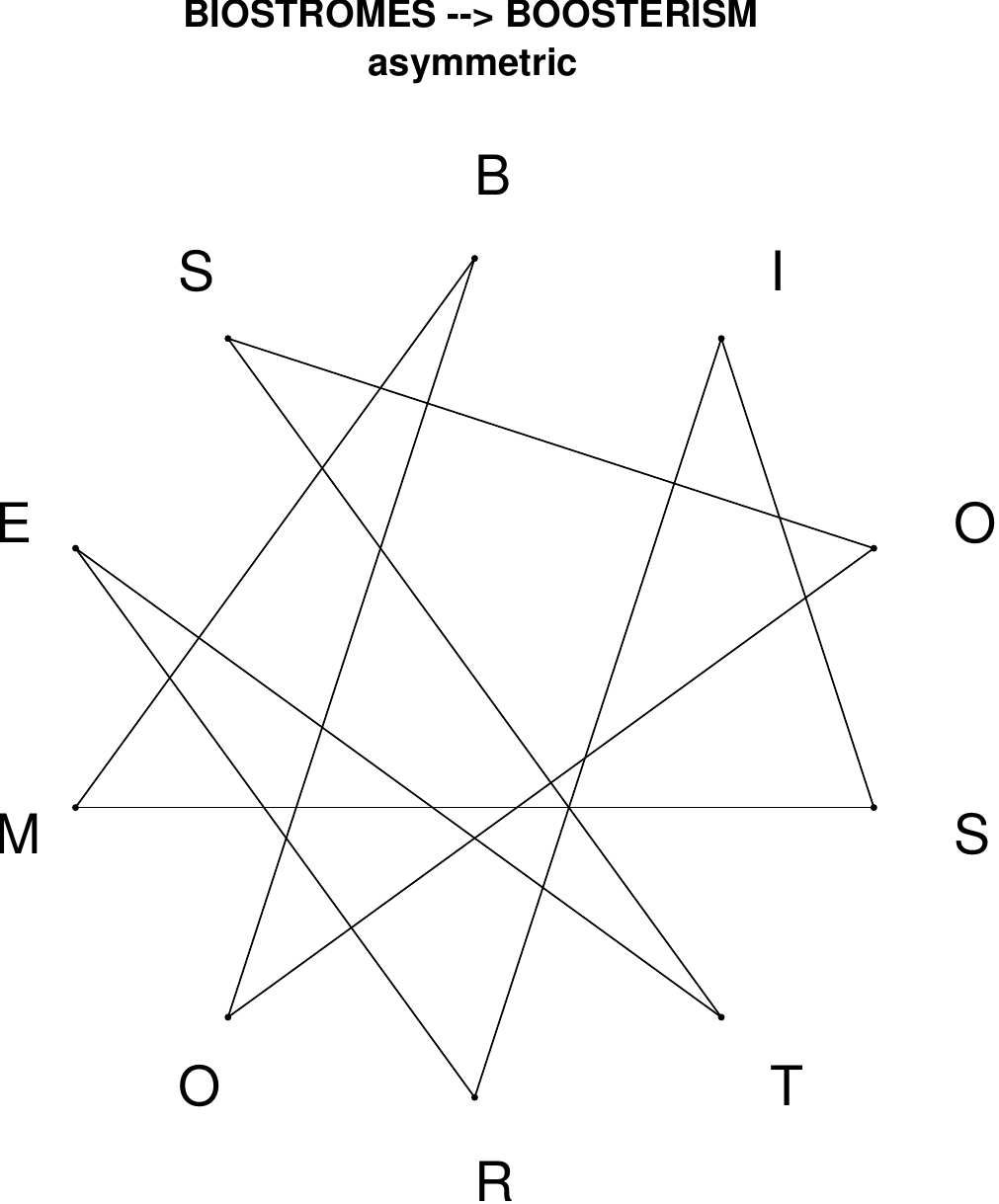}
\end{subfigure}
\hfill
\begin{subfigure}[T]{0.19\textwidth}
\centering
\includegraphics[width=\textwidth]{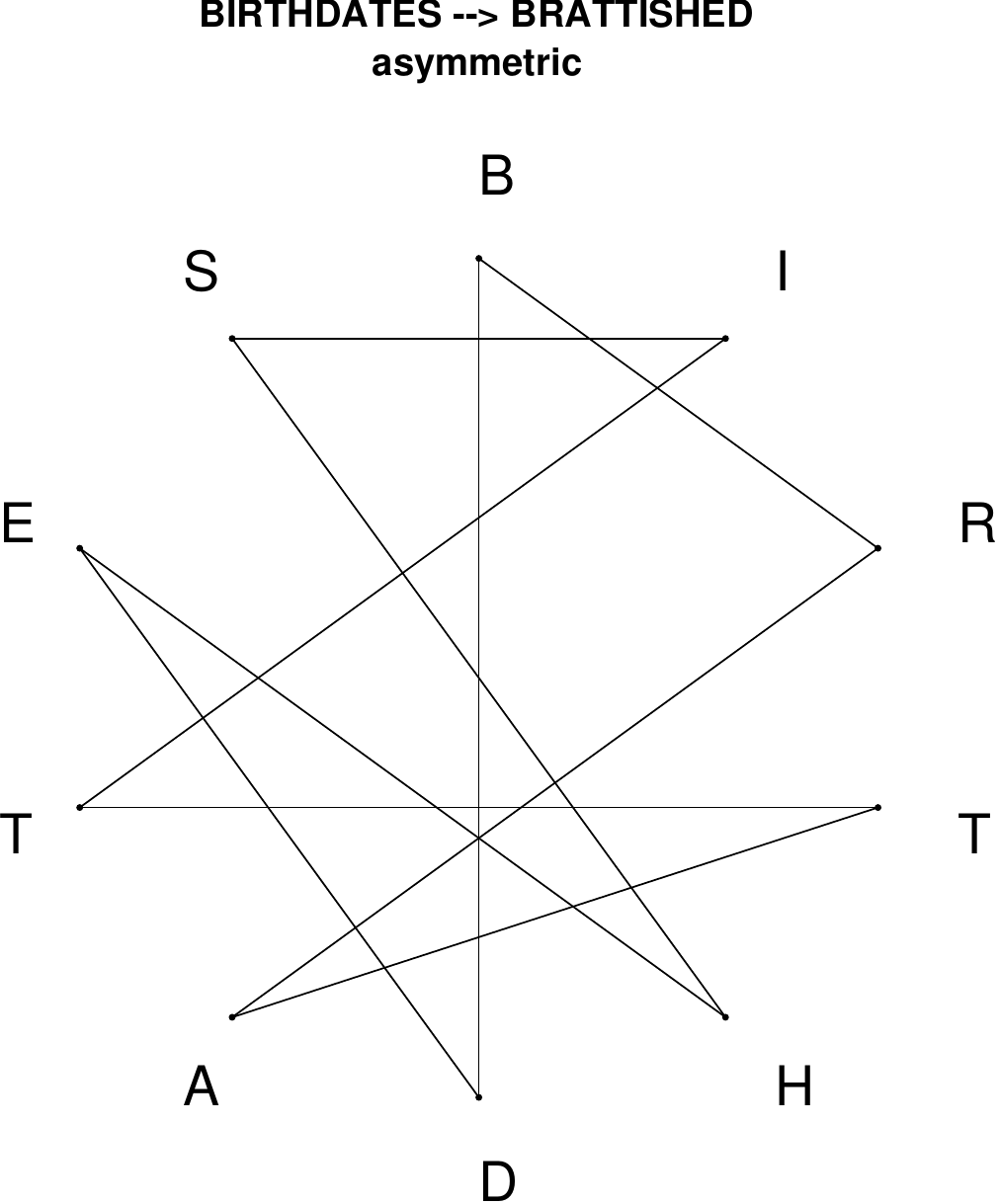}
\end{subfigure}
\hfill
\begin{subfigure}[T]{0.19\textwidth}
\centering
\includegraphics[width=\textwidth]{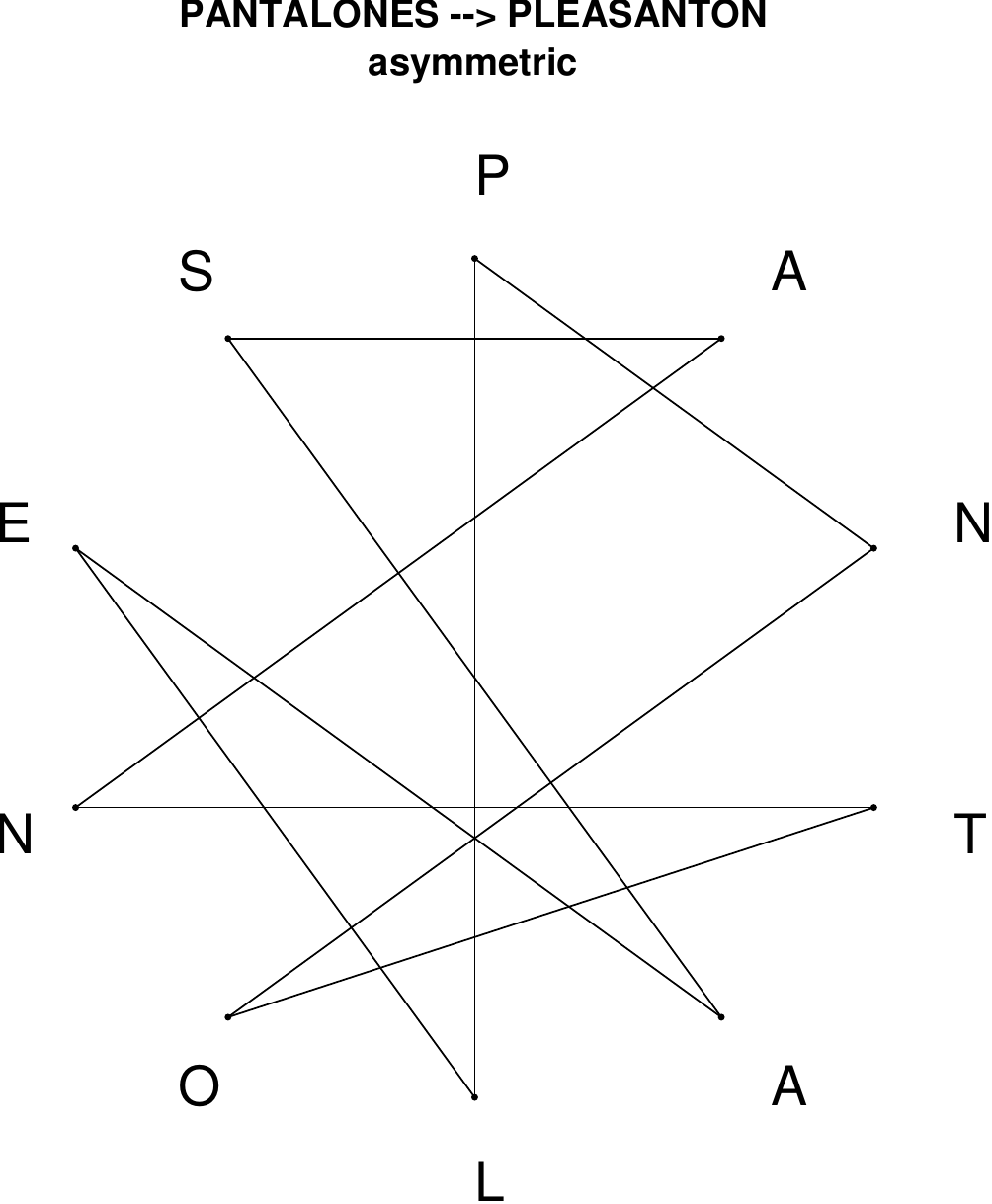}
\end{subfigure}
\hfill
\begin{subfigure}[T]{0.19\textwidth}
\centering
\includegraphics[width=\textwidth]{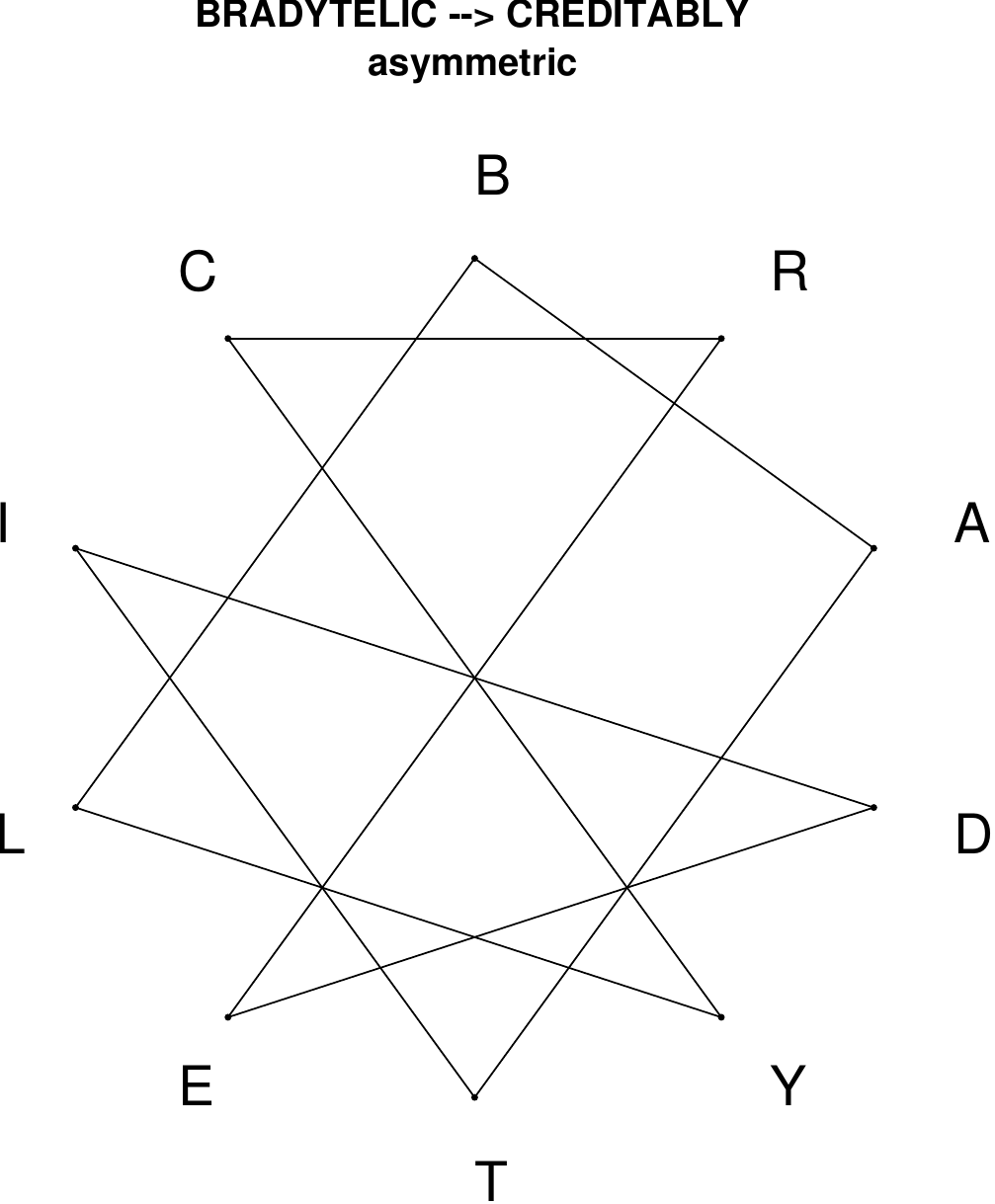}
\end{subfigure}
\end{figure}

\begin{figure}[H]
\centering
\begin{subfigure}[T]{0.19\textwidth}
\centering
\includegraphics[width=\textwidth]{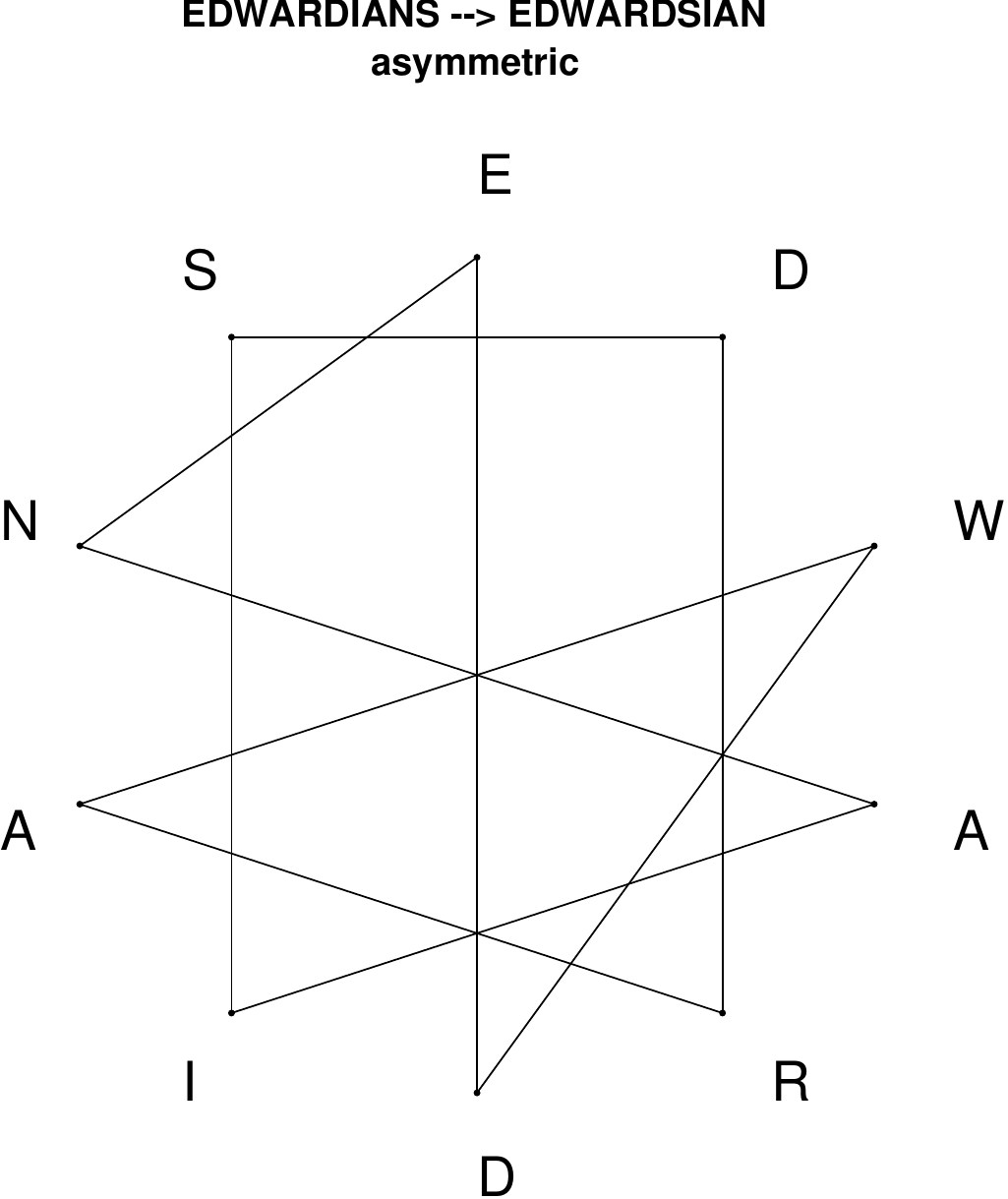}
\end{subfigure}
\hfill
\begin{subfigure}[T]{0.19\textwidth}
\centering
\includegraphics[width=\textwidth]{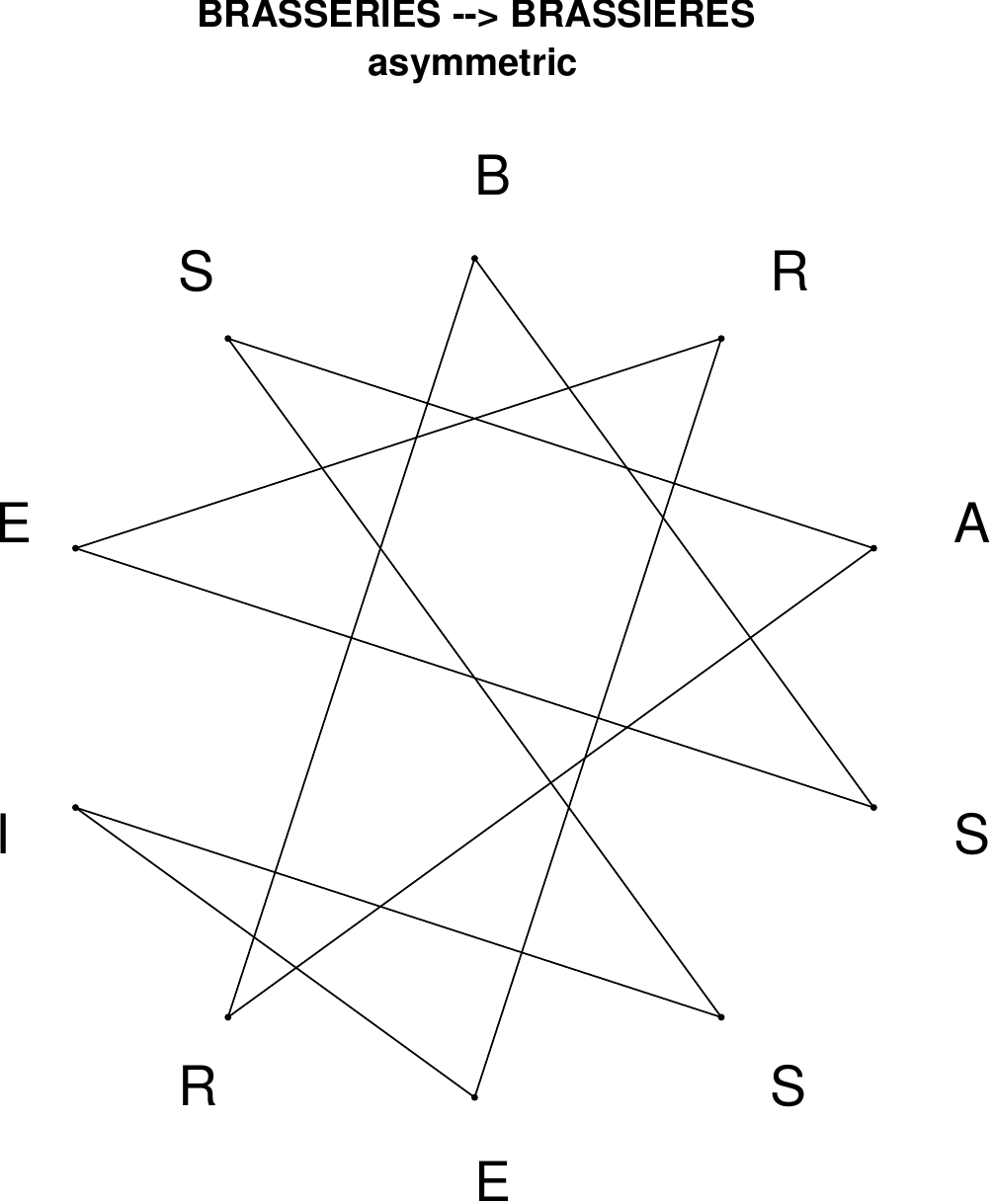}
\end{subfigure}
\hfill
\begin{subfigure}[T]{0.19\textwidth}
\centering
\includegraphics[width=\textwidth]{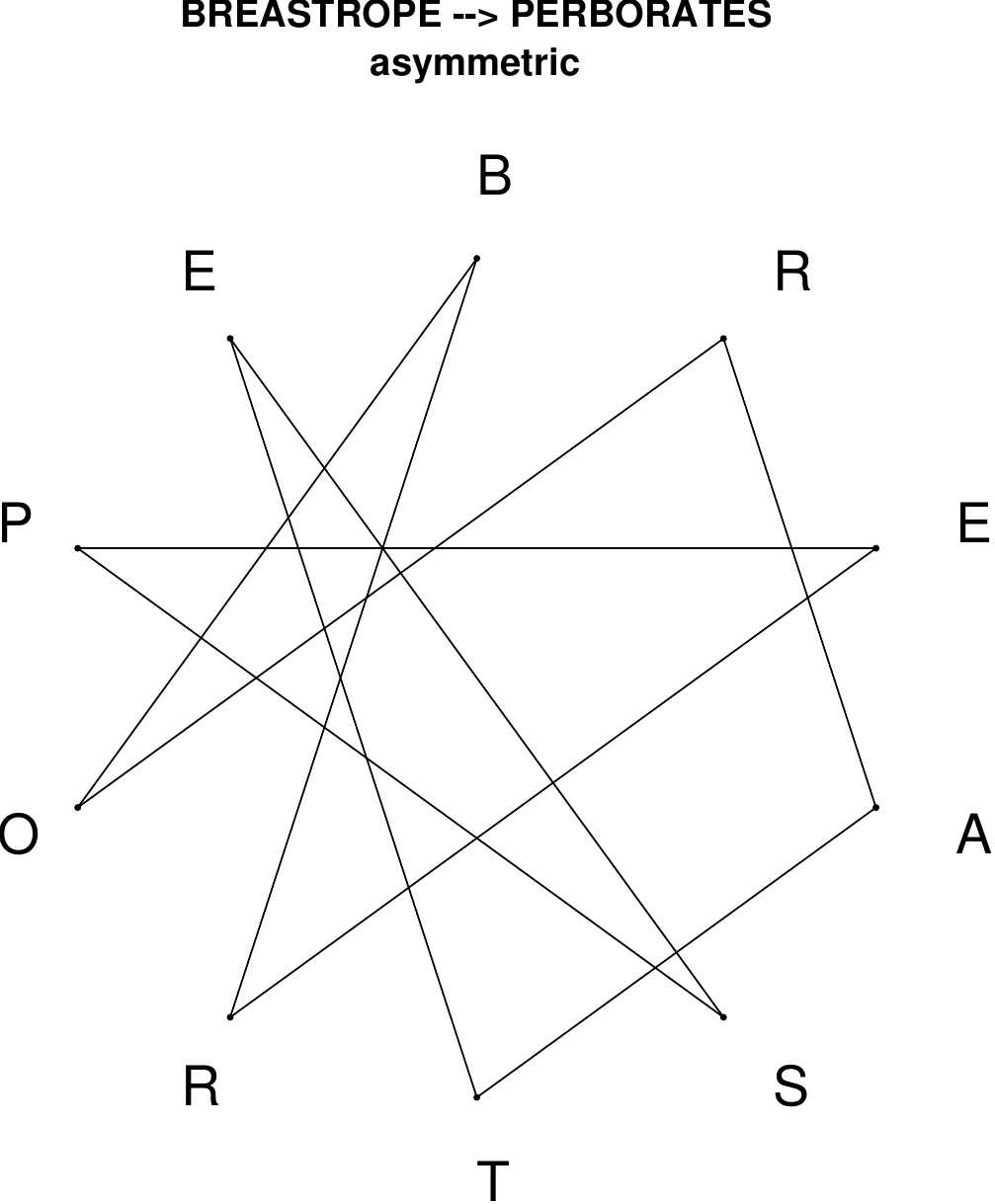}
\end{subfigure}
\hfill
\begin{subfigure}[T]{0.19\textwidth}
\centering
\includegraphics[width=\textwidth]{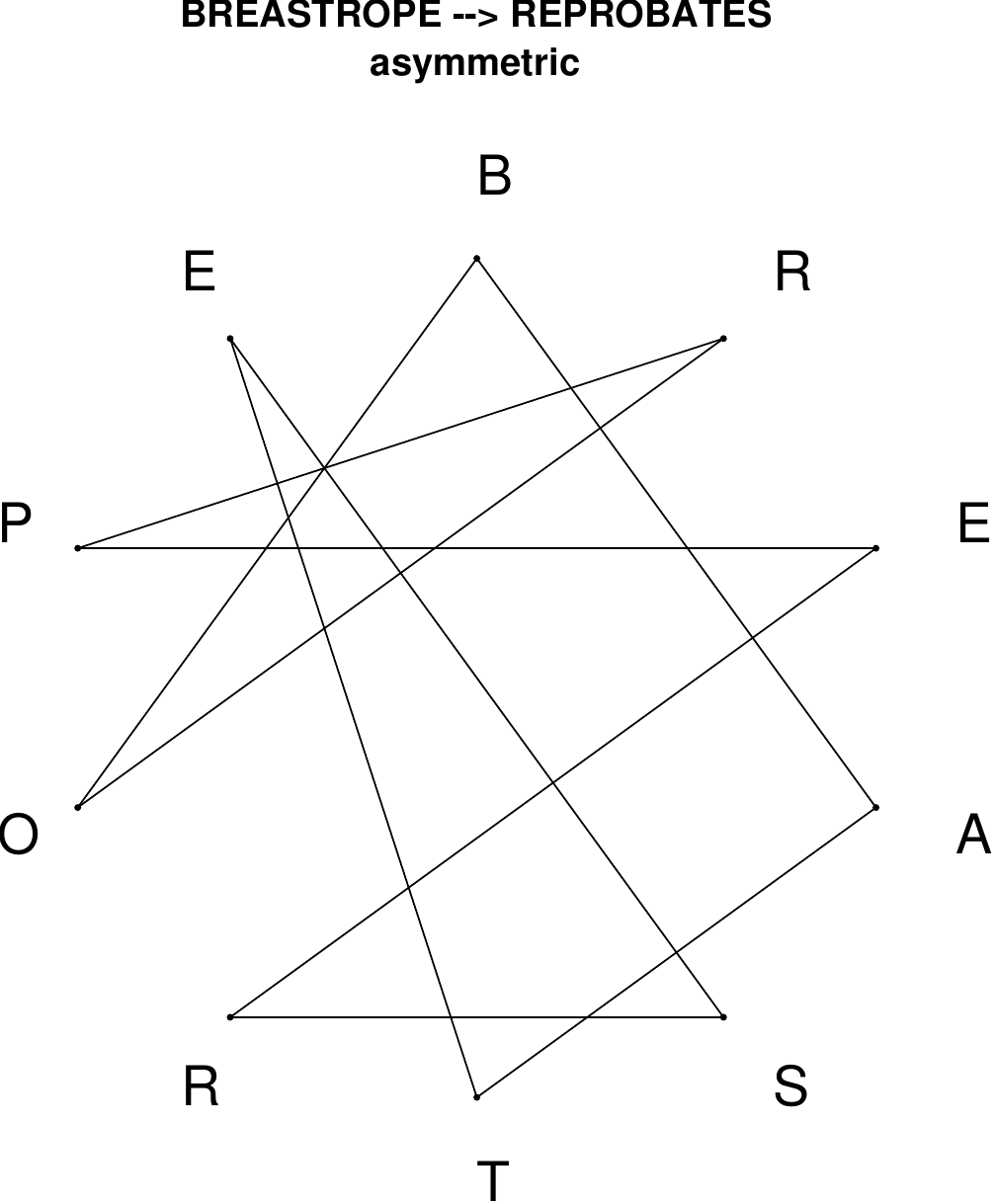}
\end{subfigure}
\hfill
\begin{subfigure}[T]{0.19\textwidth}
\centering
\includegraphics[width=\textwidth]{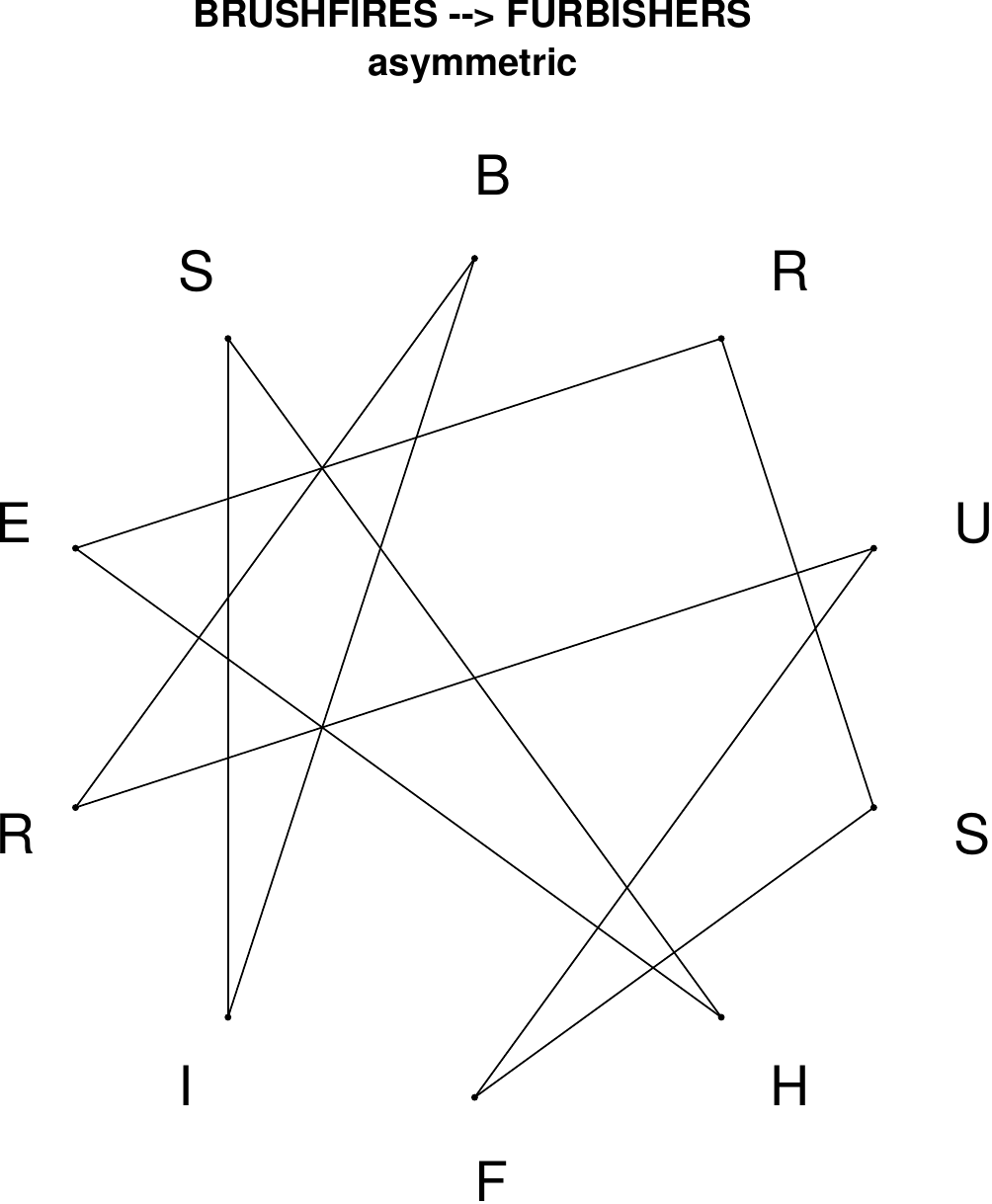}
\end{subfigure}
\end{figure}

\begin{figure}[H]
\centering
\begin{subfigure}[T]{0.19\textwidth}
\centering
\includegraphics[width=\textwidth]{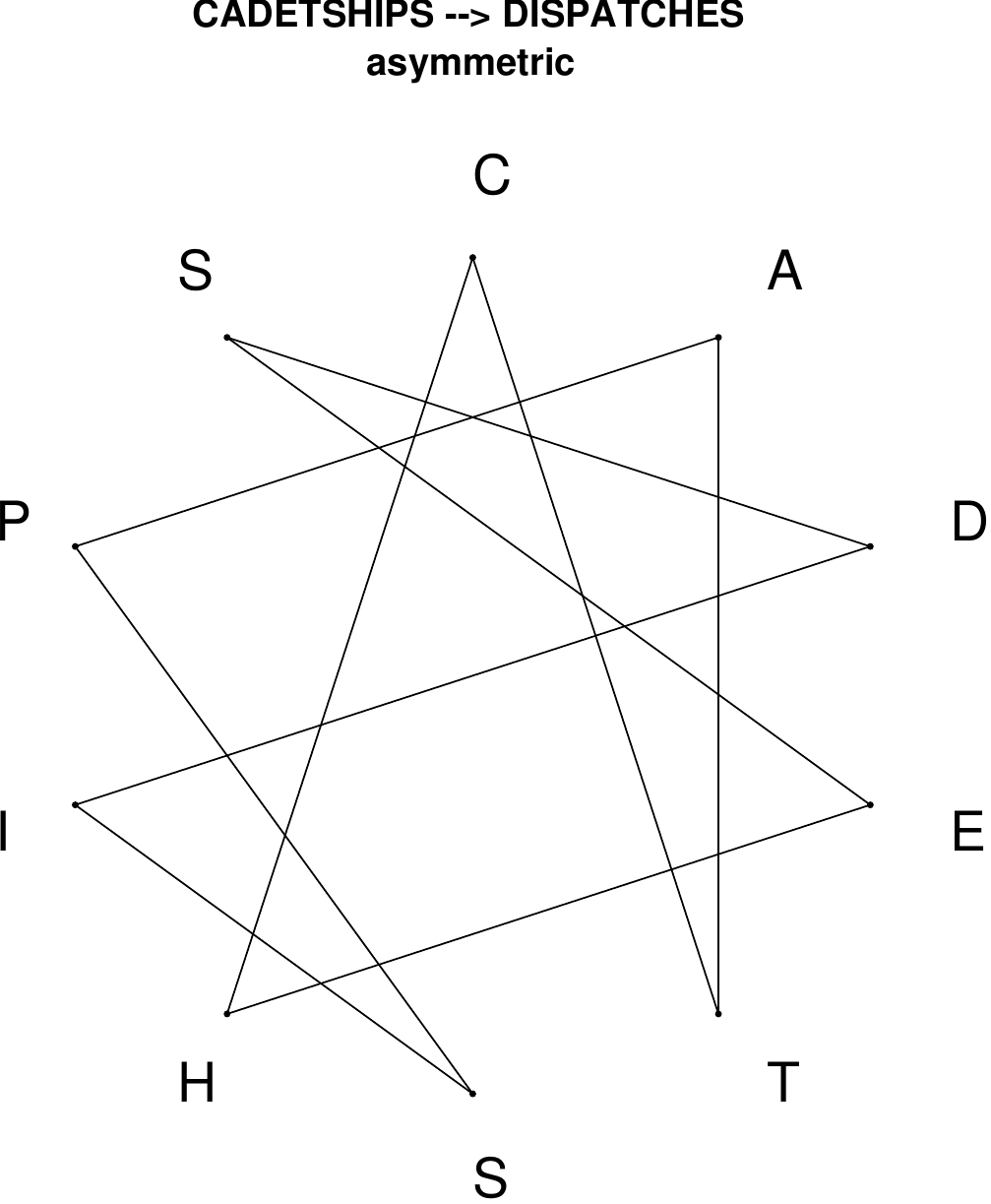}
\end{subfigure}
\hfill
\begin{subfigure}[T]{0.19\textwidth}
\centering
\includegraphics[width=\textwidth]{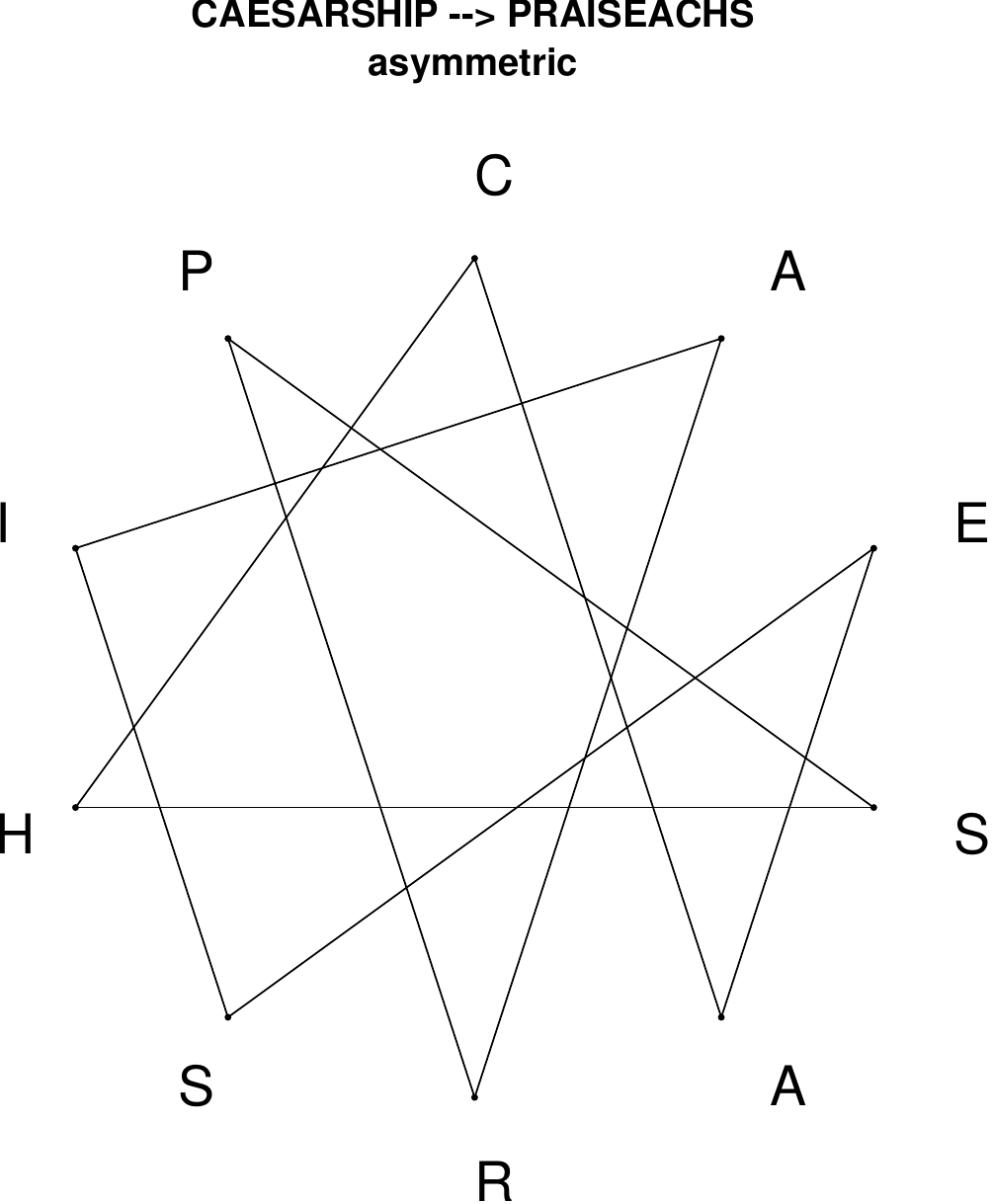}
\end{subfigure}
\hfill
\begin{subfigure}[T]{0.19\textwidth}
\centering
\includegraphics[width=\textwidth]{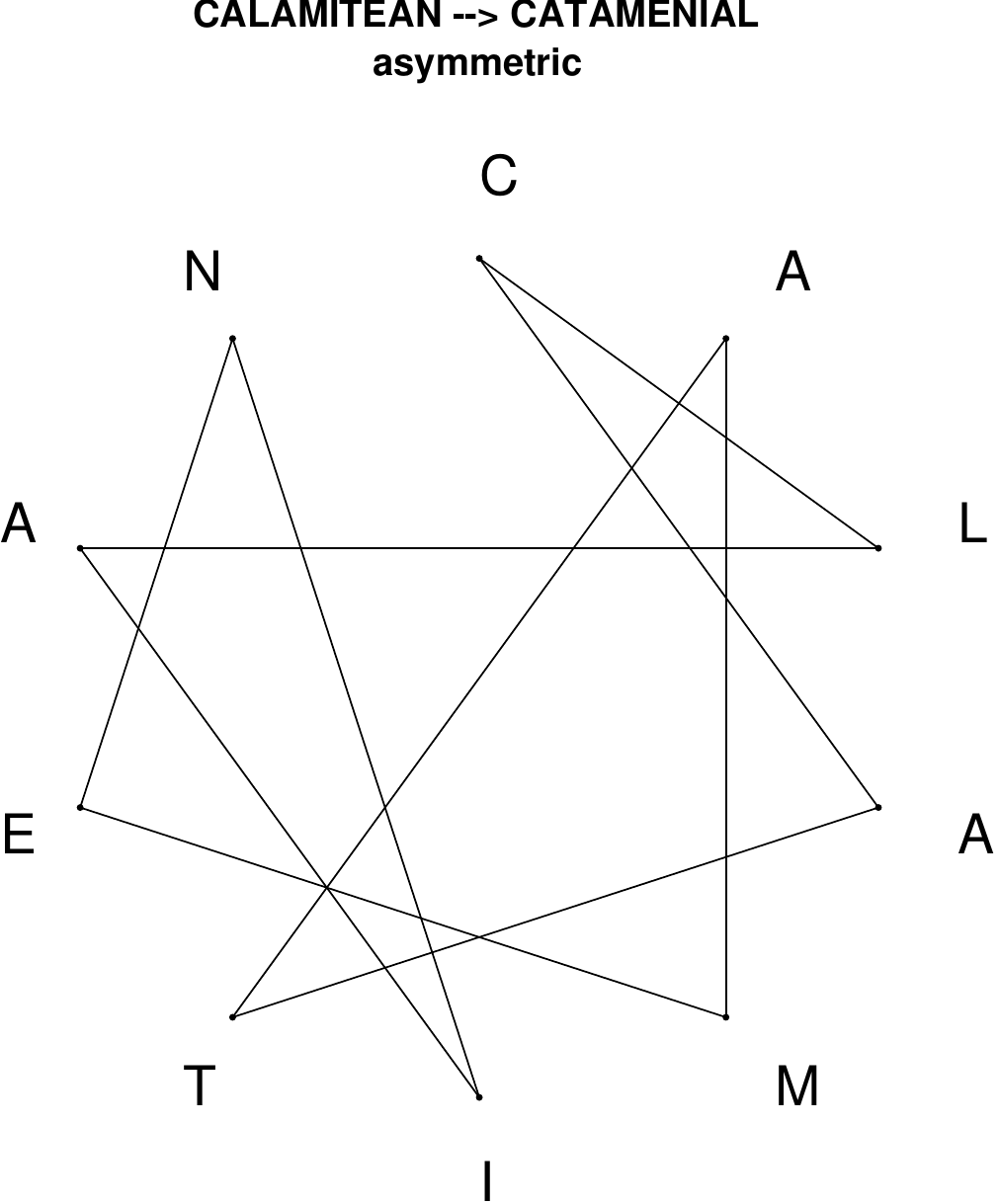}
\end{subfigure}
\hfill
\begin{subfigure}[T]{0.19\textwidth}
\centering
\includegraphics[width=\textwidth]{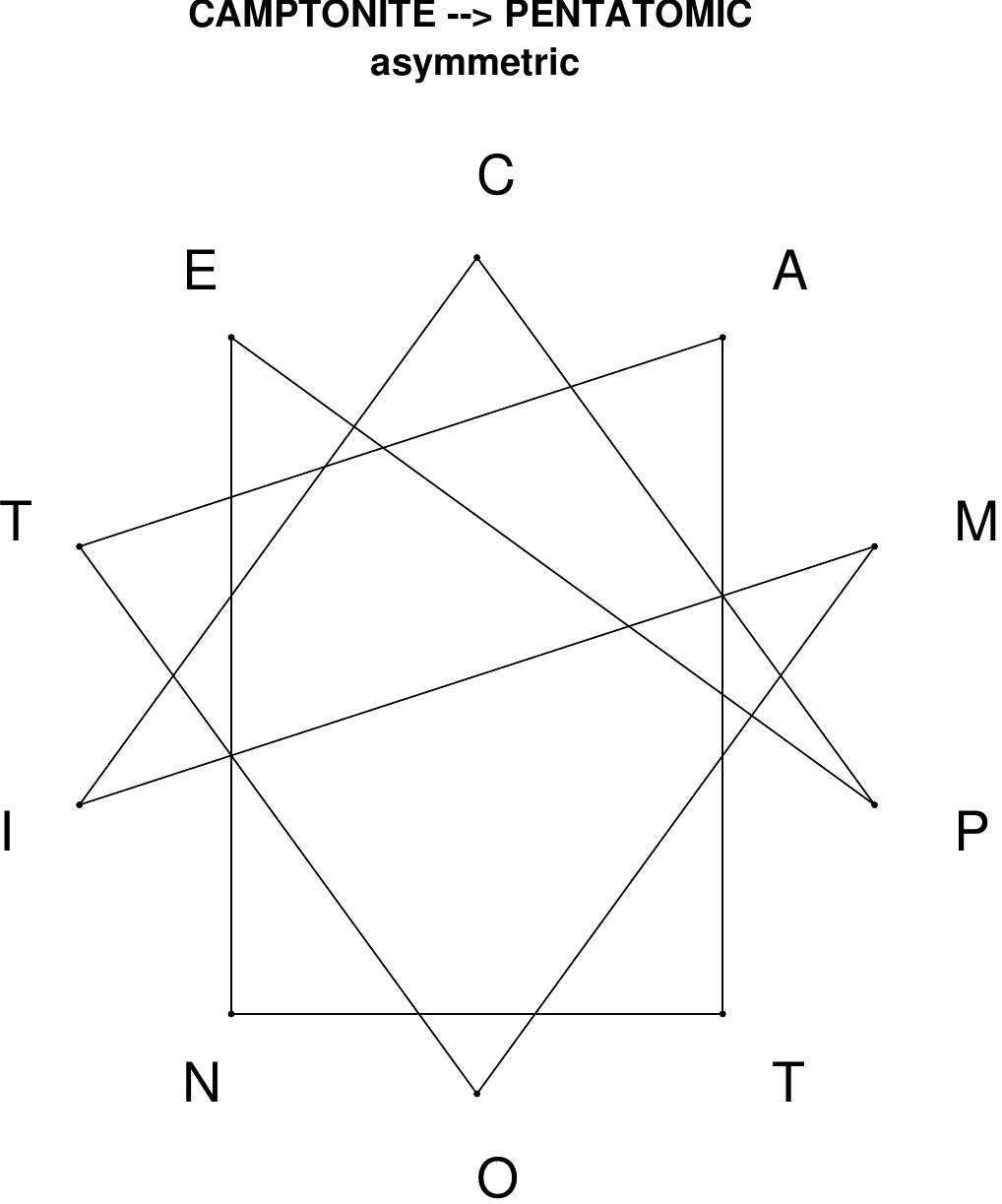}
\end{subfigure}
\hfill
\begin{subfigure}[T]{0.19\textwidth}
\centering
\includegraphics[width=\textwidth]{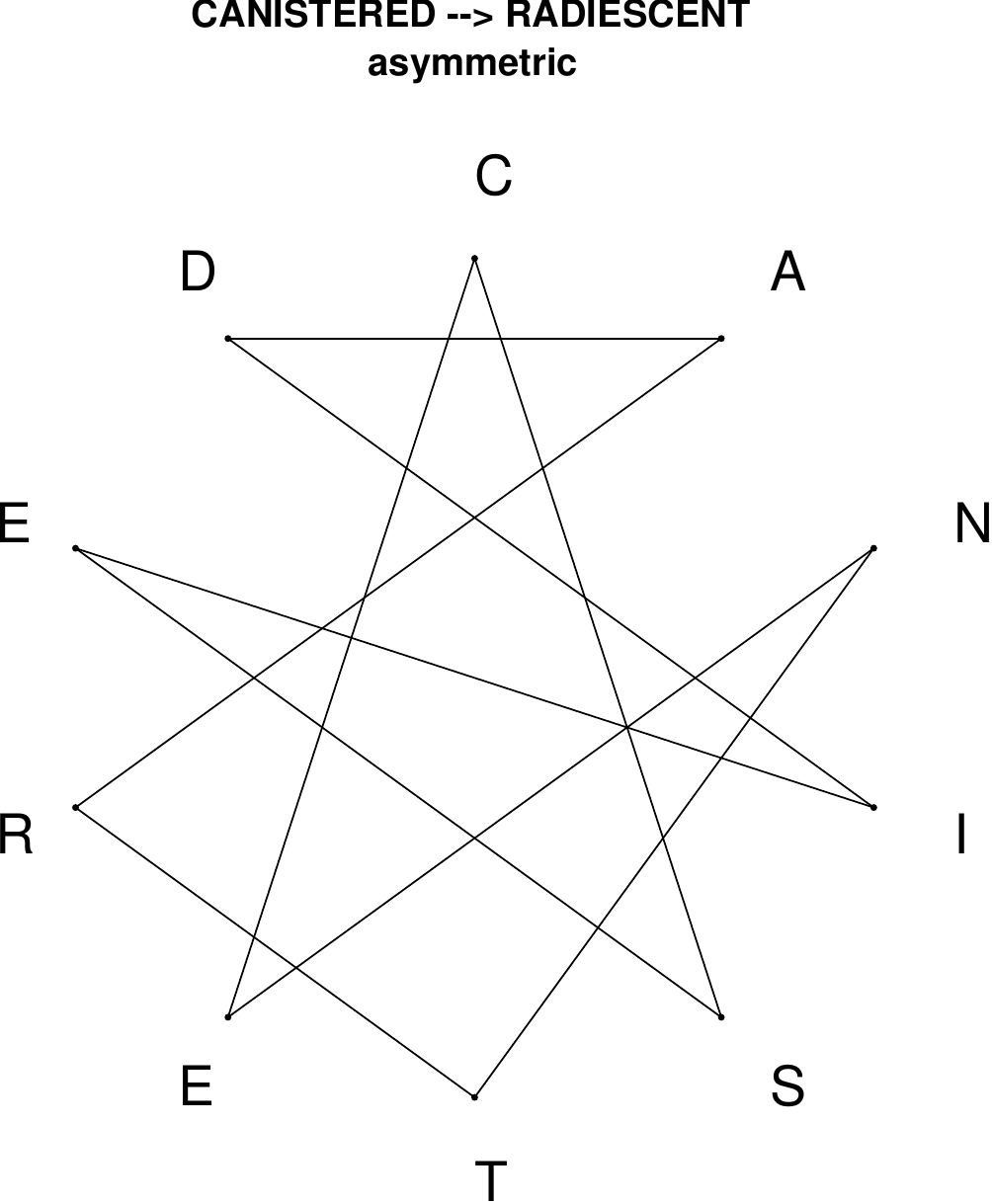}
\end{subfigure}
\end{figure}

\begin{figure}[H]
\centering
\begin{subfigure}[T]{0.19\textwidth}
\centering
\includegraphics[width=\textwidth]{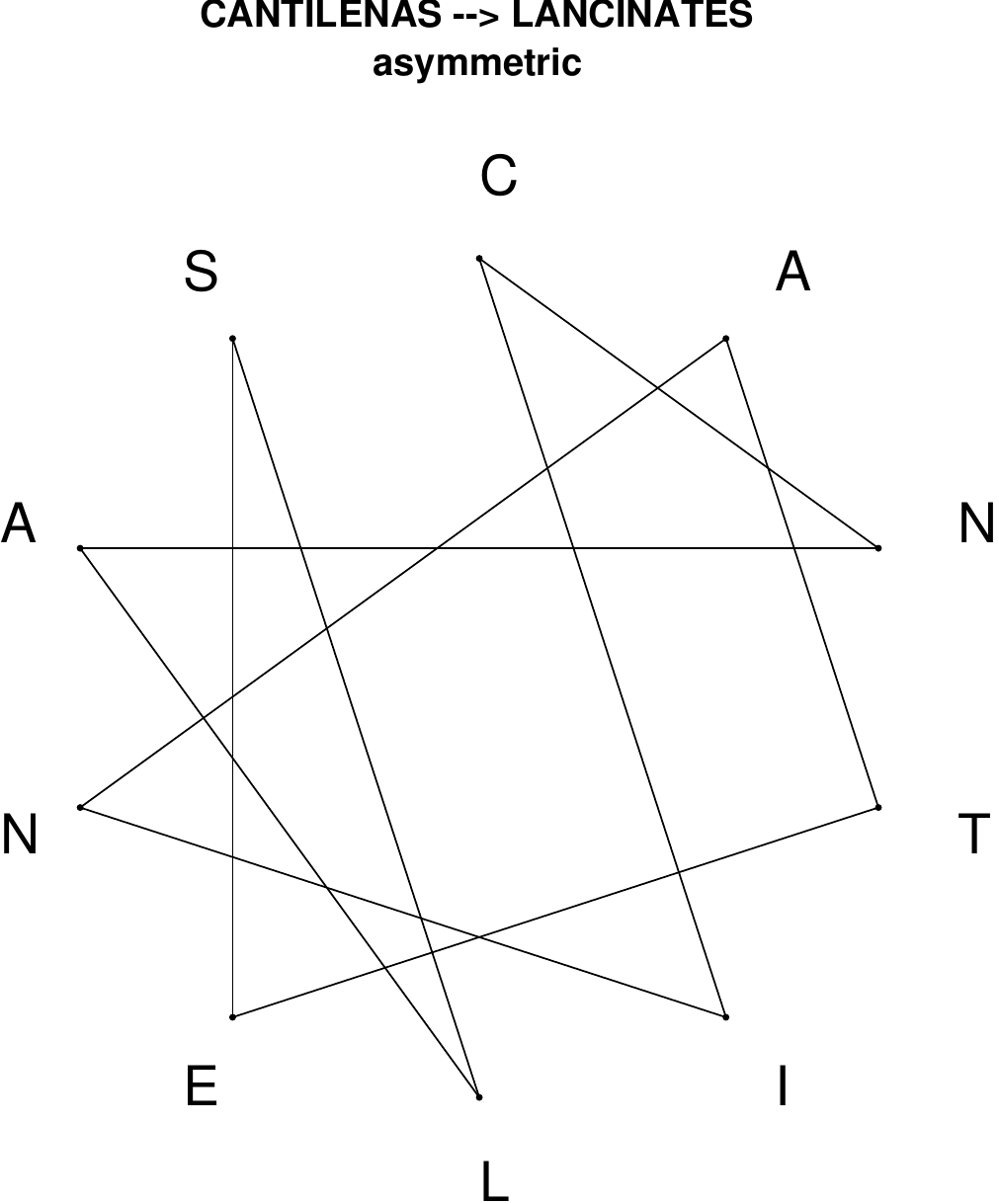}
\end{subfigure}
\hfill
\begin{subfigure}[T]{0.19\textwidth}
\centering
\includegraphics[width=\textwidth]{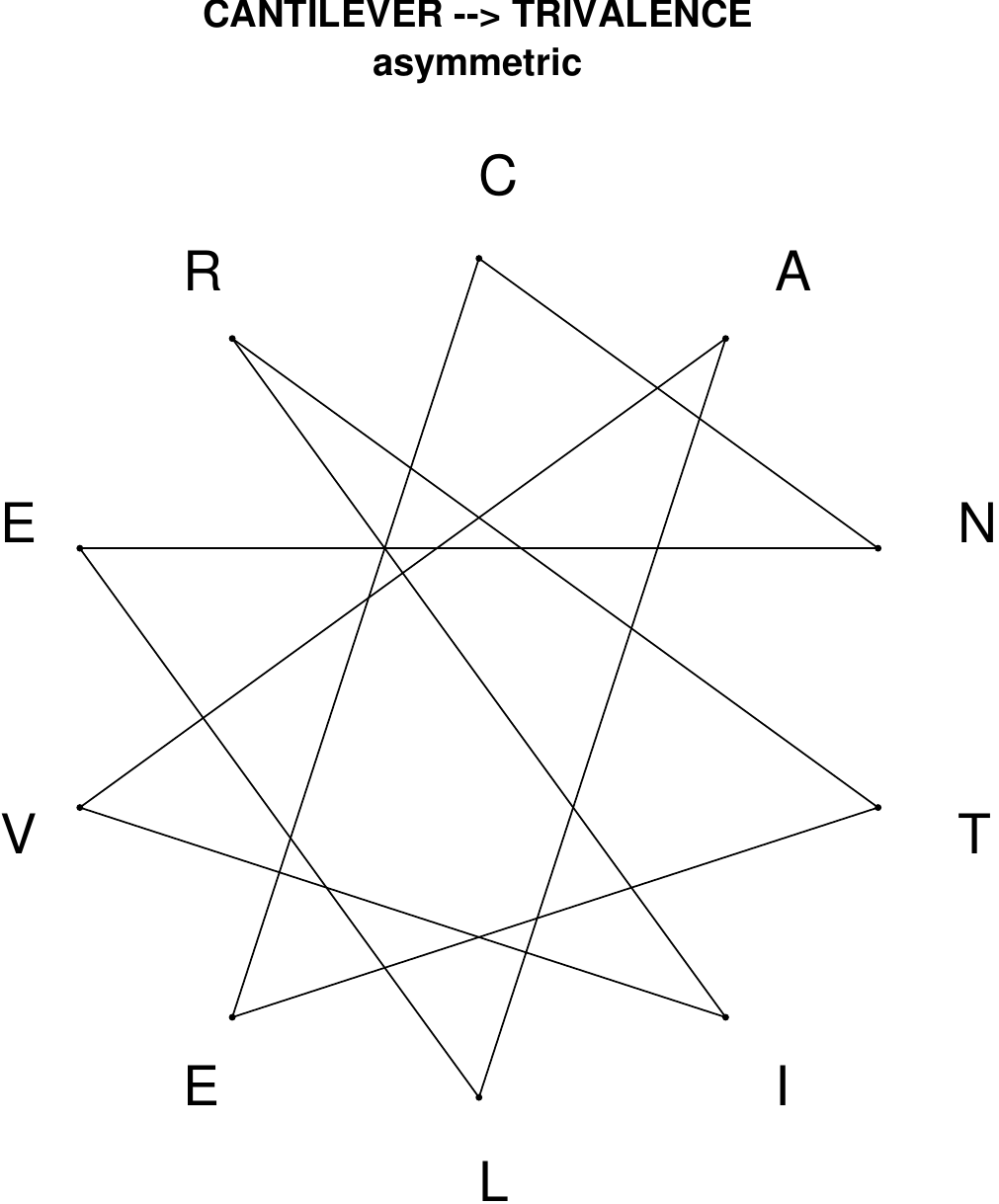}
\end{subfigure}
\hfill
\begin{subfigure}[T]{0.19\textwidth}
\centering
\includegraphics[width=\textwidth]{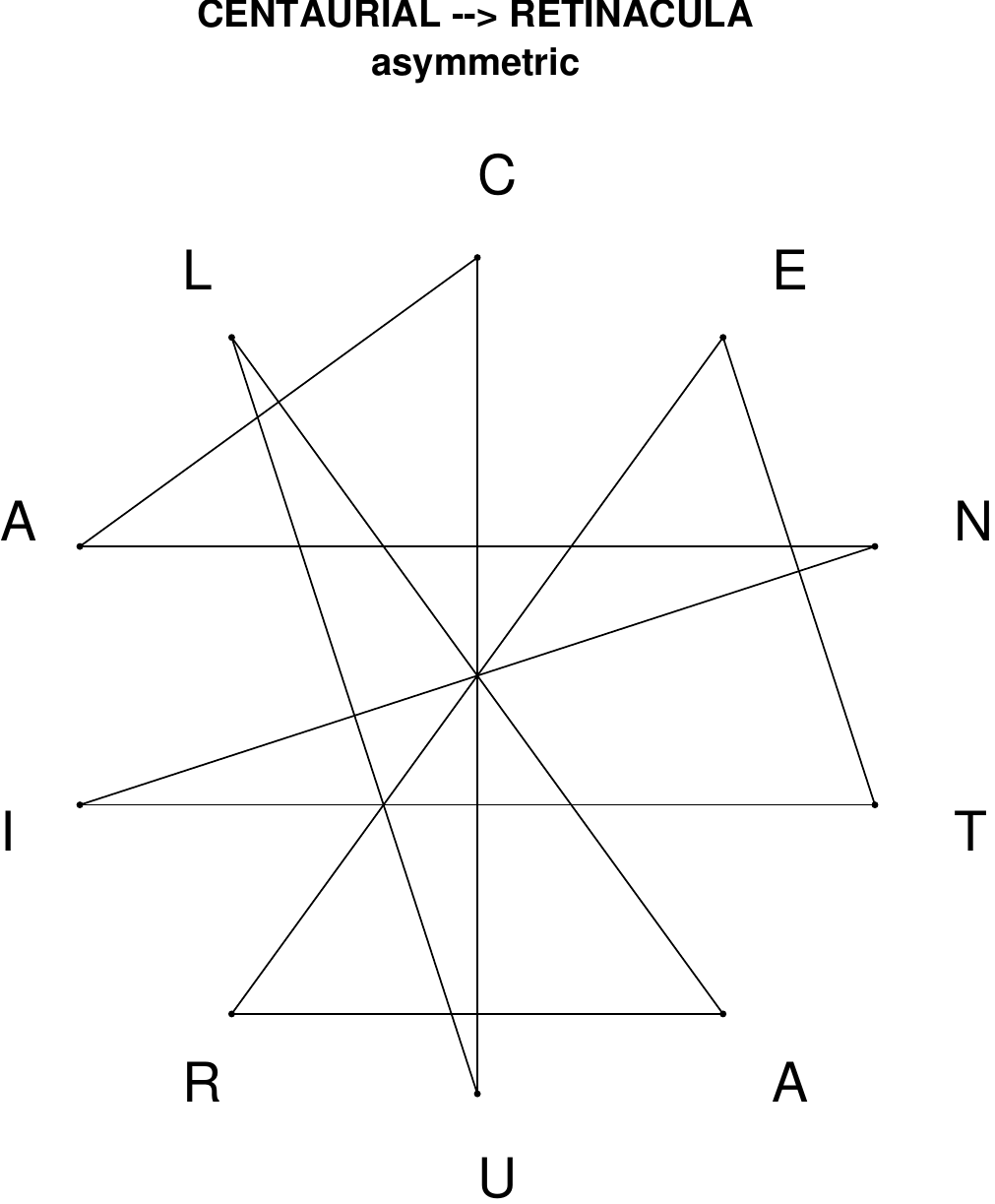}
\end{subfigure}
\hfill
\begin{subfigure}[T]{0.19\textwidth}
\centering
\includegraphics[width=\textwidth]{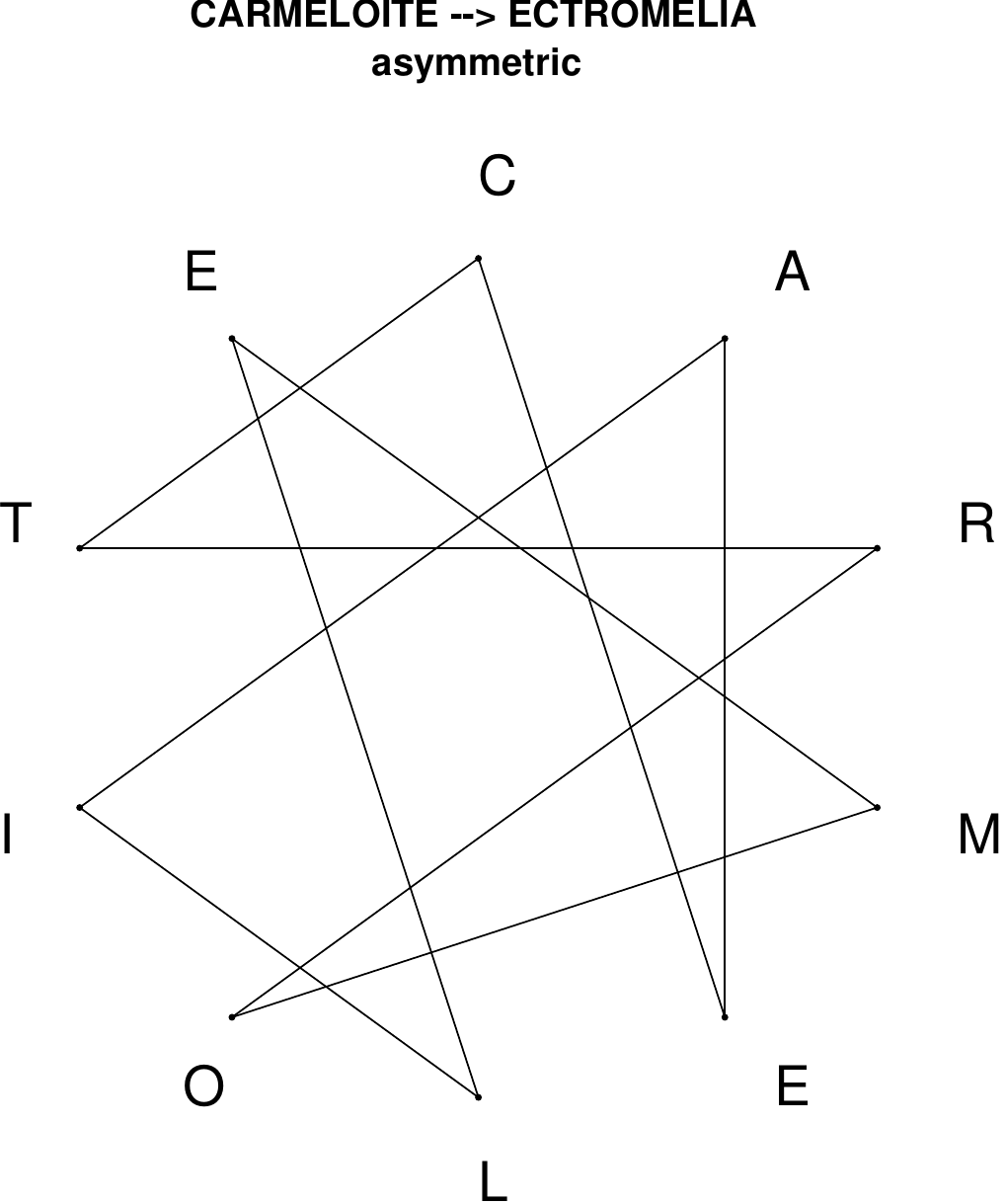}
\end{subfigure}
\hfill
\begin{subfigure}[T]{0.19\textwidth}
\centering
\includegraphics[width=\textwidth]{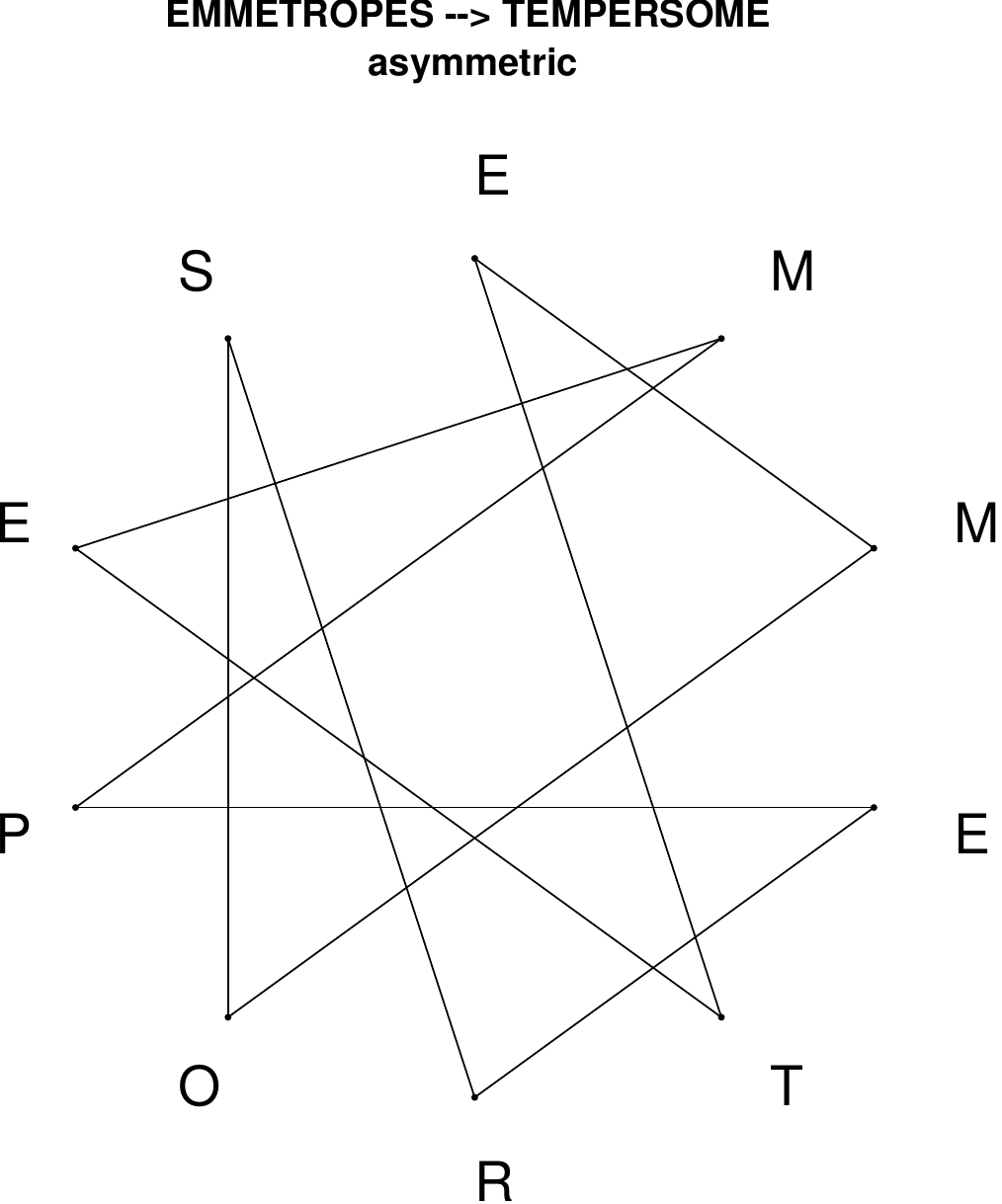}
\end{subfigure}
\end{figure}

\begin{figure}[H]
\centering
\begin{subfigure}[T]{0.19\textwidth}
\centering
\includegraphics[width=\textwidth]{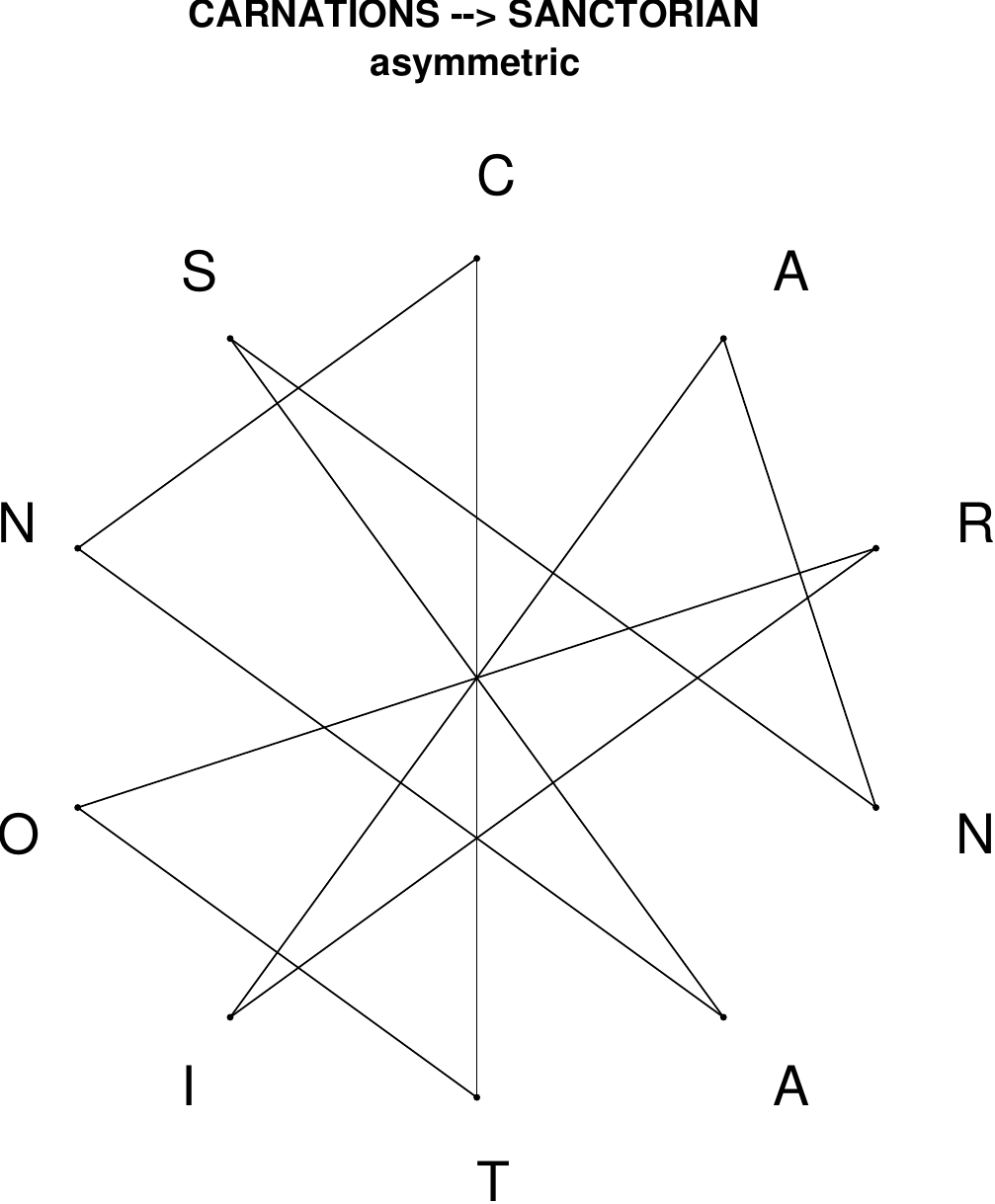}
\end{subfigure}
\hfill
\begin{subfigure}[T]{0.19\textwidth}
\centering
\includegraphics[width=\textwidth]{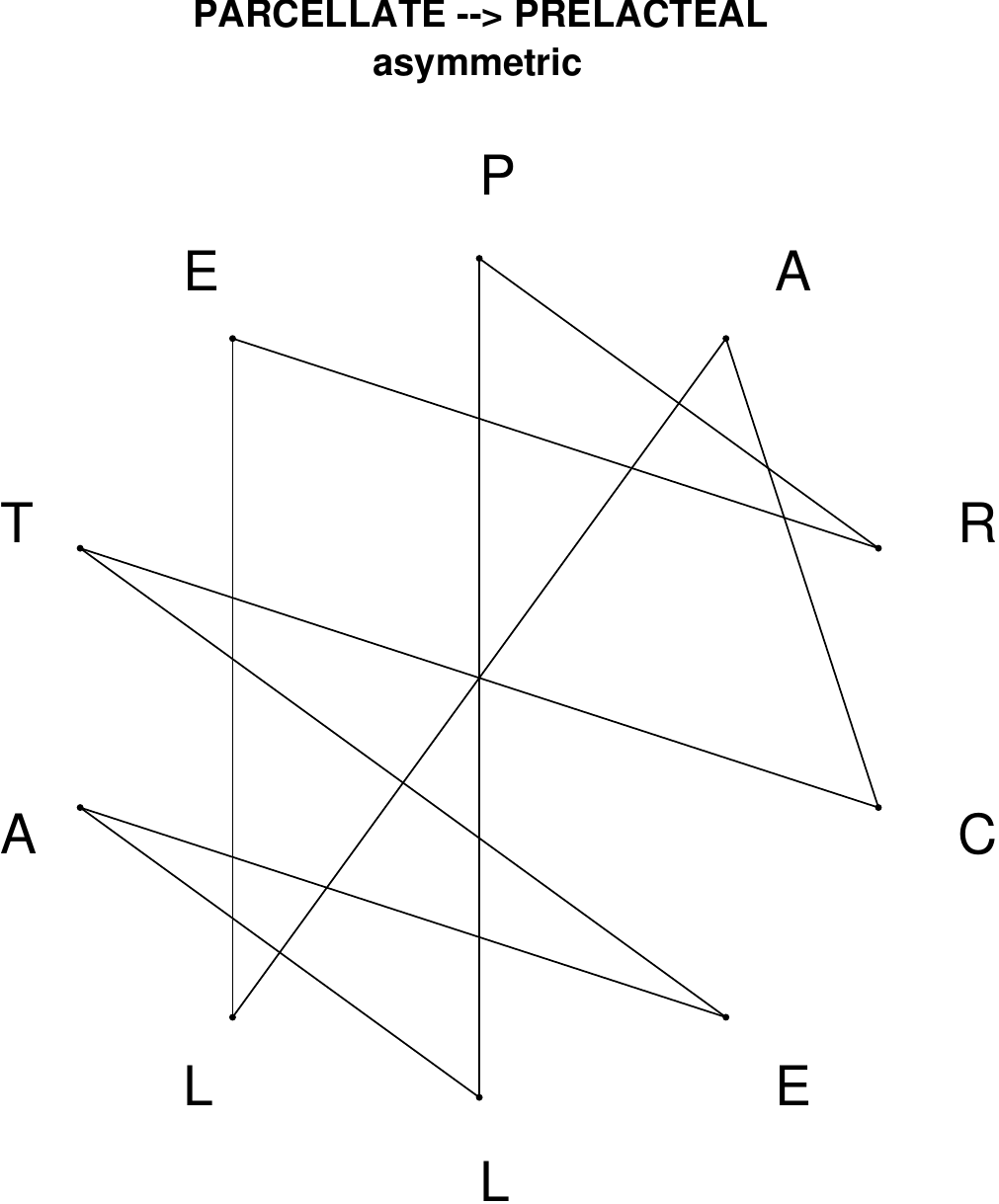}
\end{subfigure}
\hfill
\begin{subfigure}[T]{0.19\textwidth}
\centering
\includegraphics[width=\textwidth]{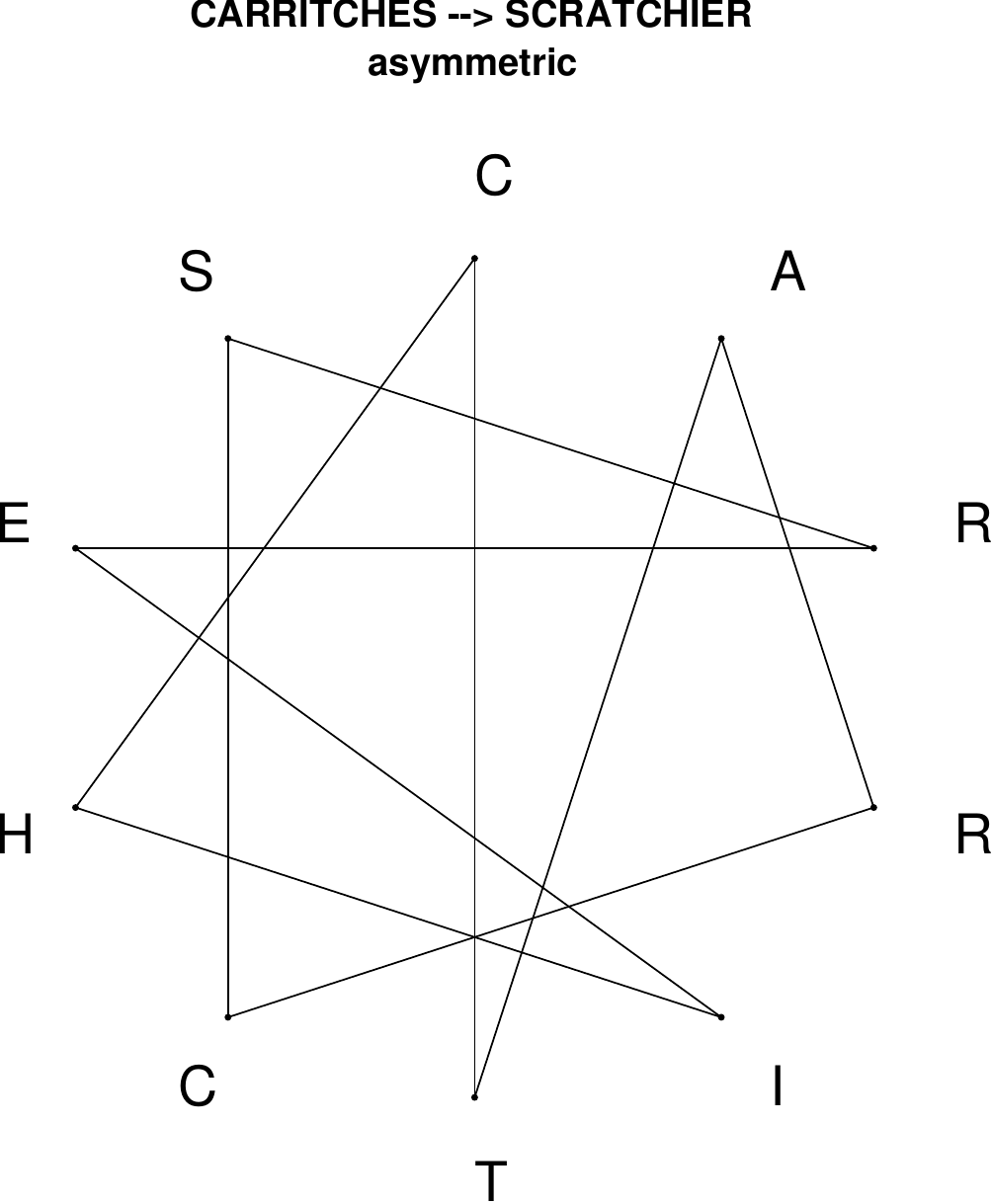}
\end{subfigure}
\hfill
\begin{subfigure}[T]{0.19\textwidth}
\centering
\includegraphics[width=\textwidth]{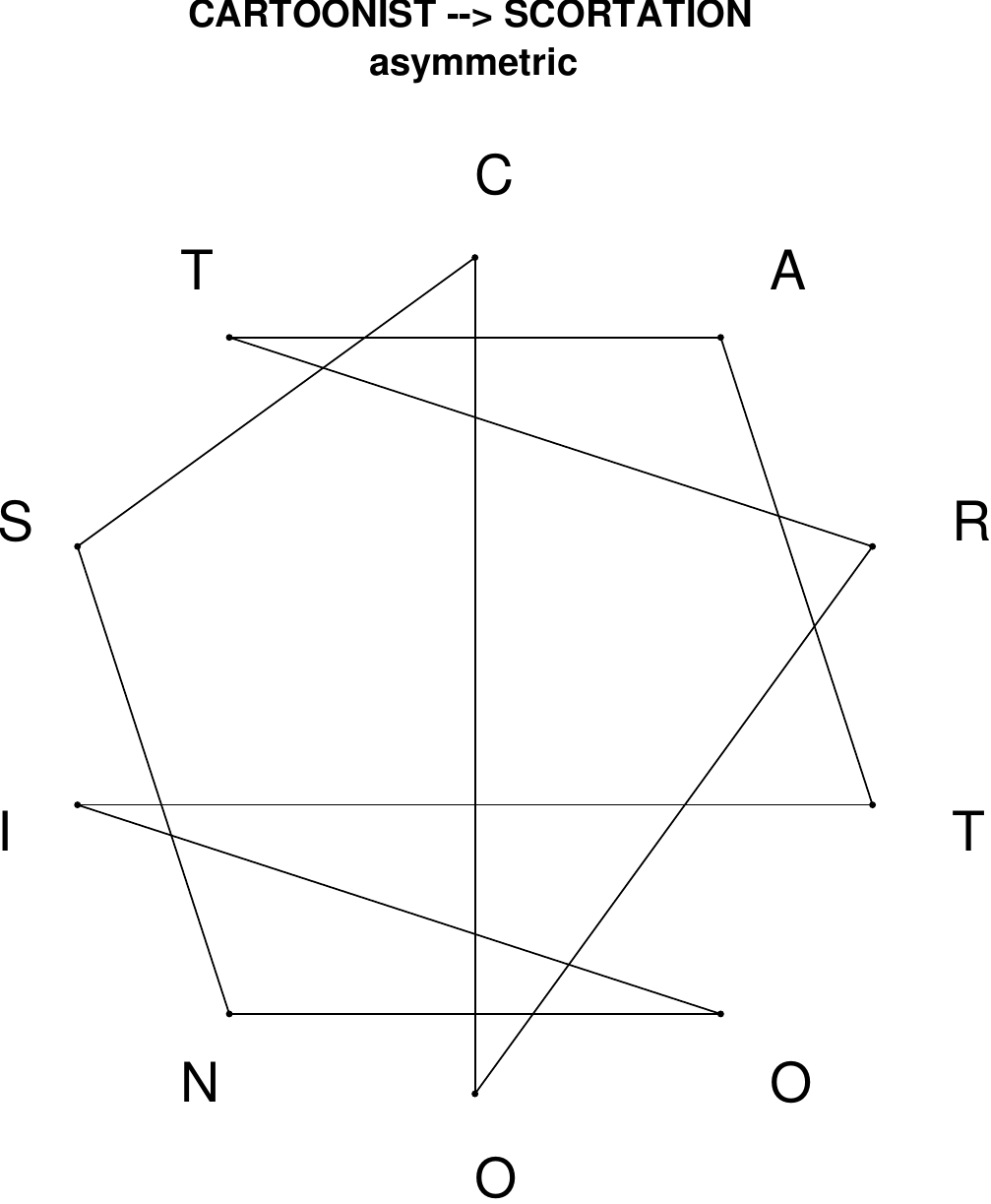}
\end{subfigure}
\hfill
\begin{subfigure}[T]{0.19\textwidth}
\centering
\includegraphics[width=\textwidth]{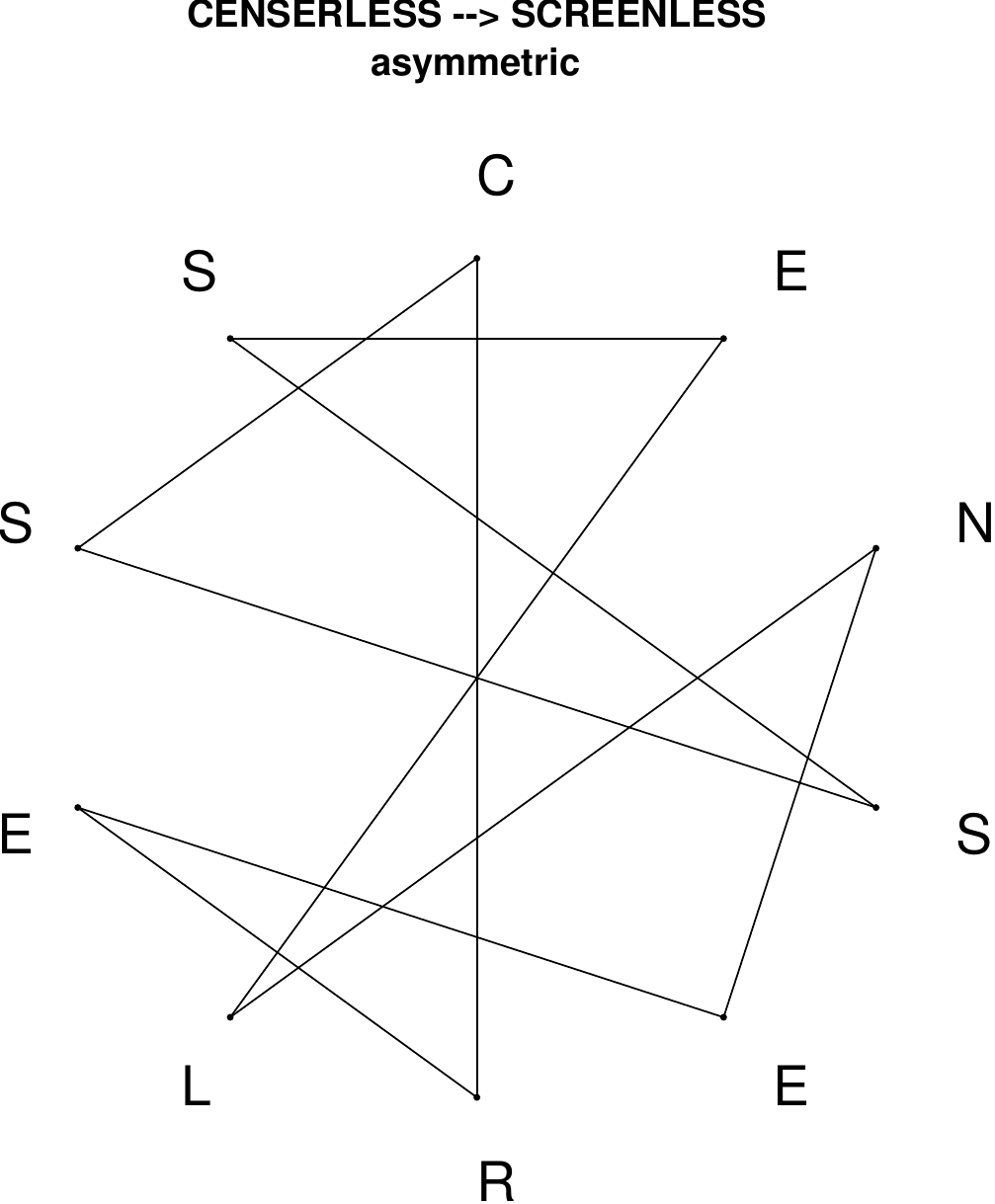}
\end{subfigure}
\end{figure}

\begin{figure}[H]
\centering
\begin{subfigure}[T]{0.19\textwidth}
\centering
\includegraphics[width=\textwidth]{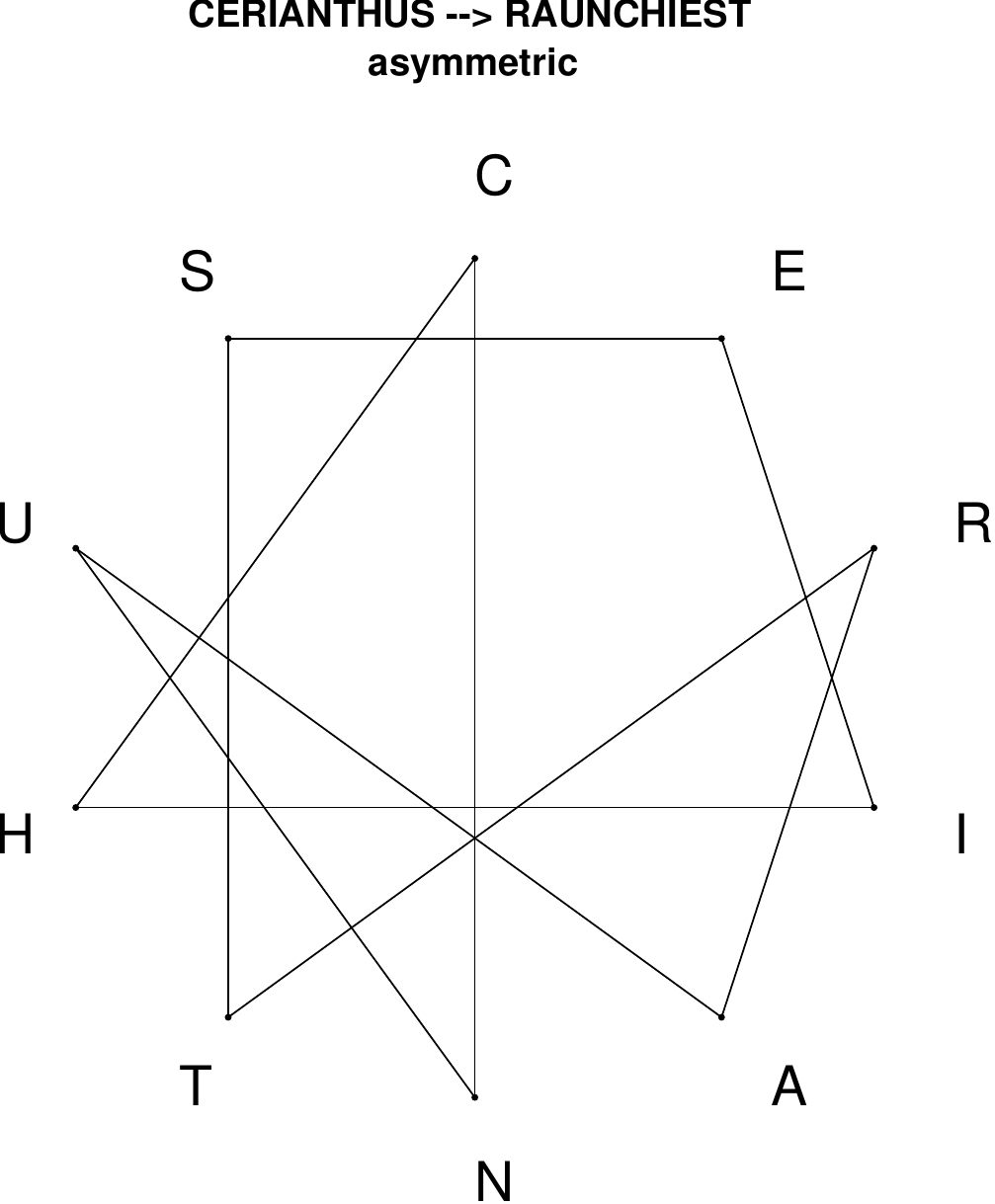}
\end{subfigure}
\hfill
\begin{subfigure}[T]{0.19\textwidth}
\centering
\includegraphics[width=\textwidth]{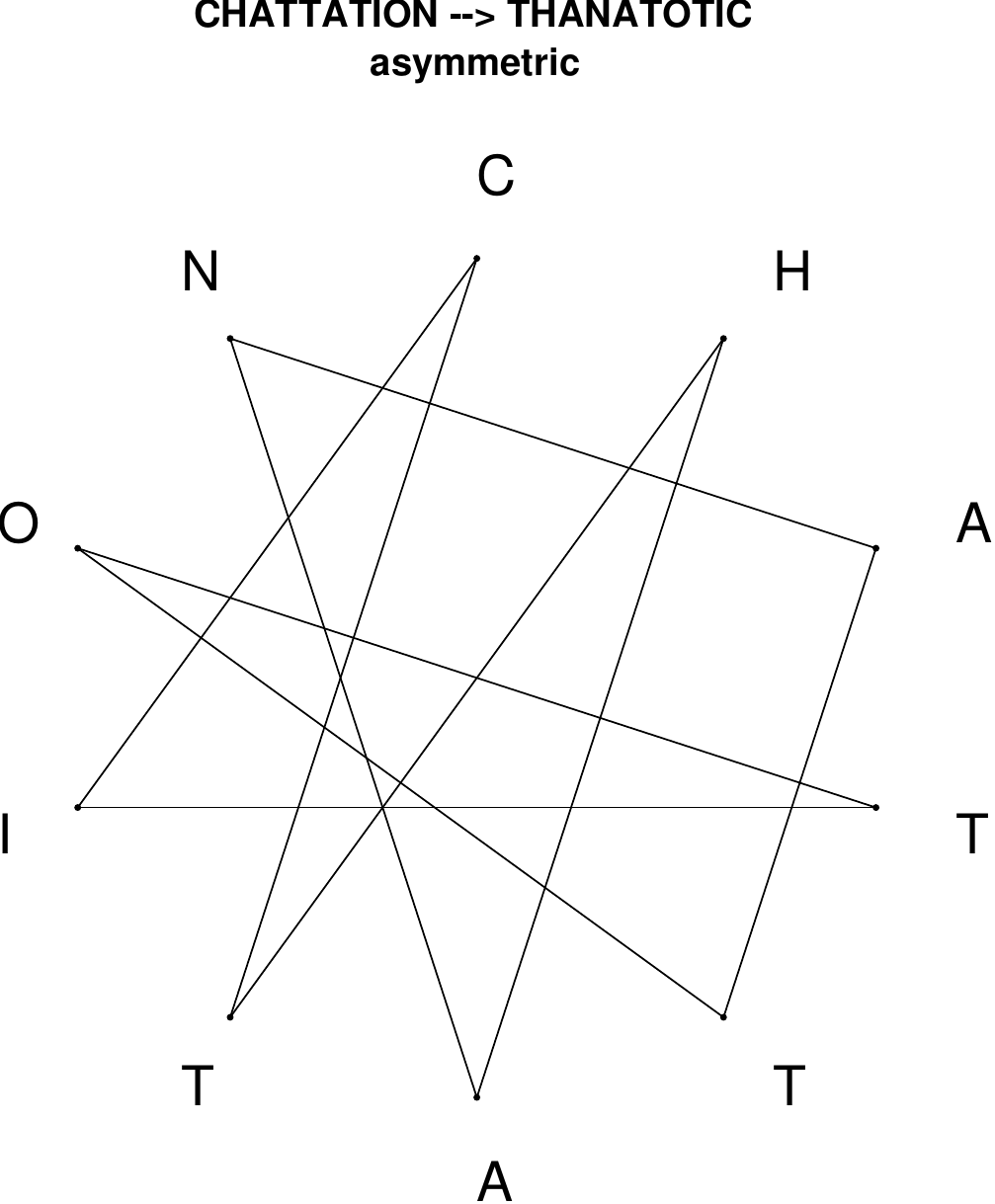}
\end{subfigure}
\hfill
\begin{subfigure}[T]{0.19\textwidth}
\centering
\includegraphics[width=\textwidth]{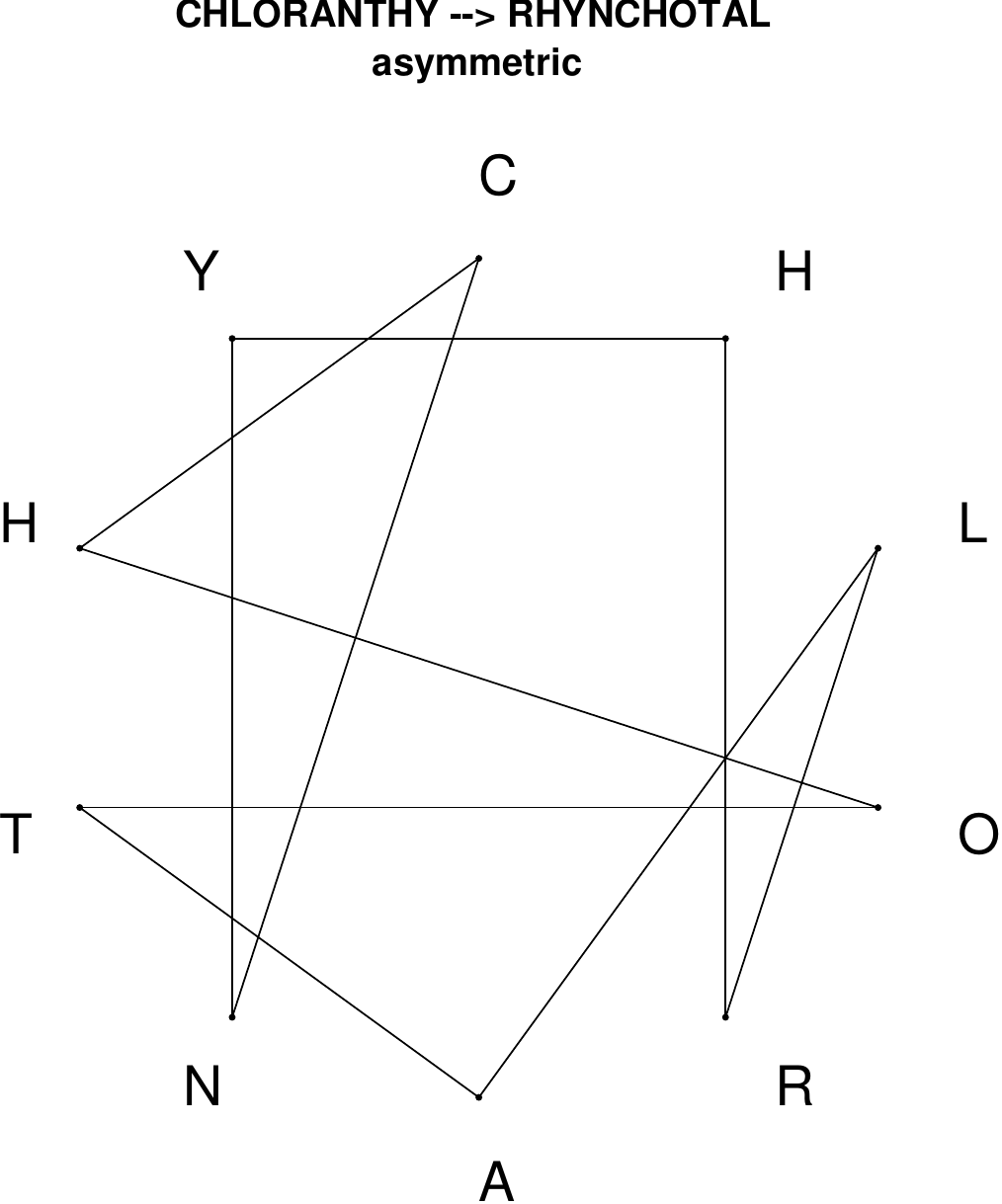}
\end{subfigure}
\hfill
\begin{subfigure}[T]{0.19\textwidth}
\centering
\includegraphics[width=\textwidth]{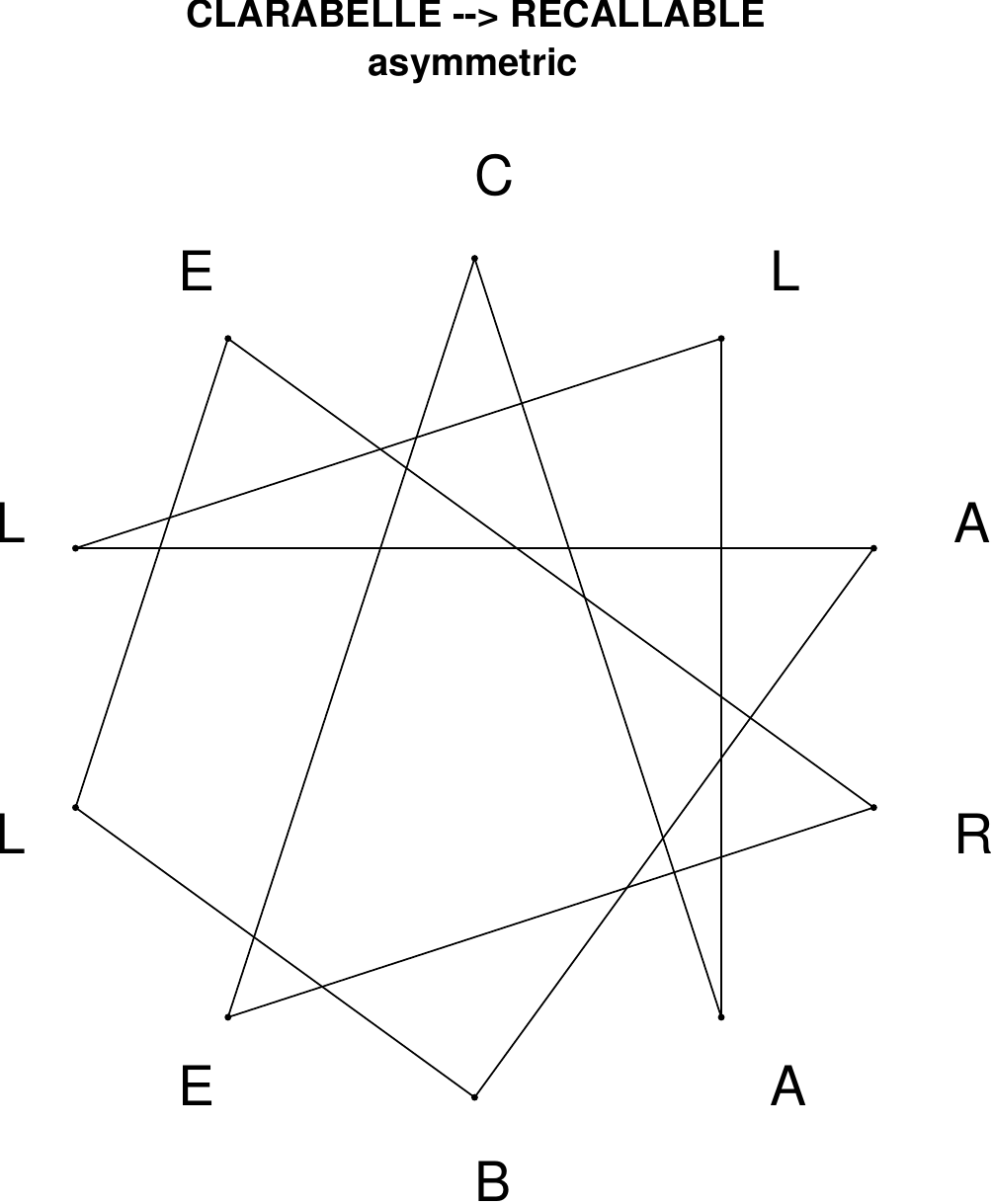}
\end{subfigure}
\hfill
\begin{subfigure}[T]{0.19\textwidth}
\centering
\includegraphics[width=\textwidth]{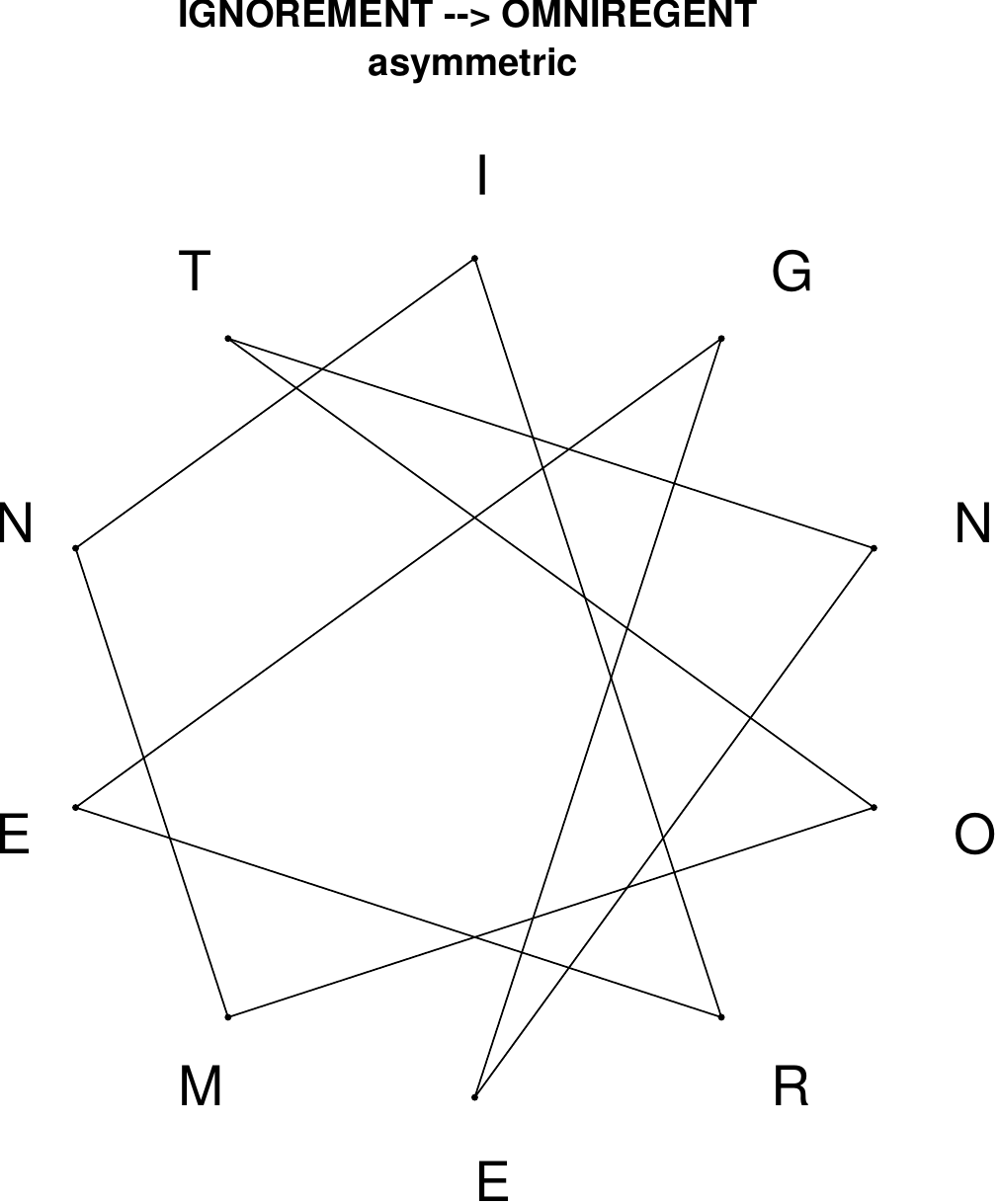}
\end{subfigure}
\end{figure}

\begin{figure}[H]
\centering
\begin{subfigure}[T]{0.19\textwidth}
\centering
\includegraphics[width=\textwidth]{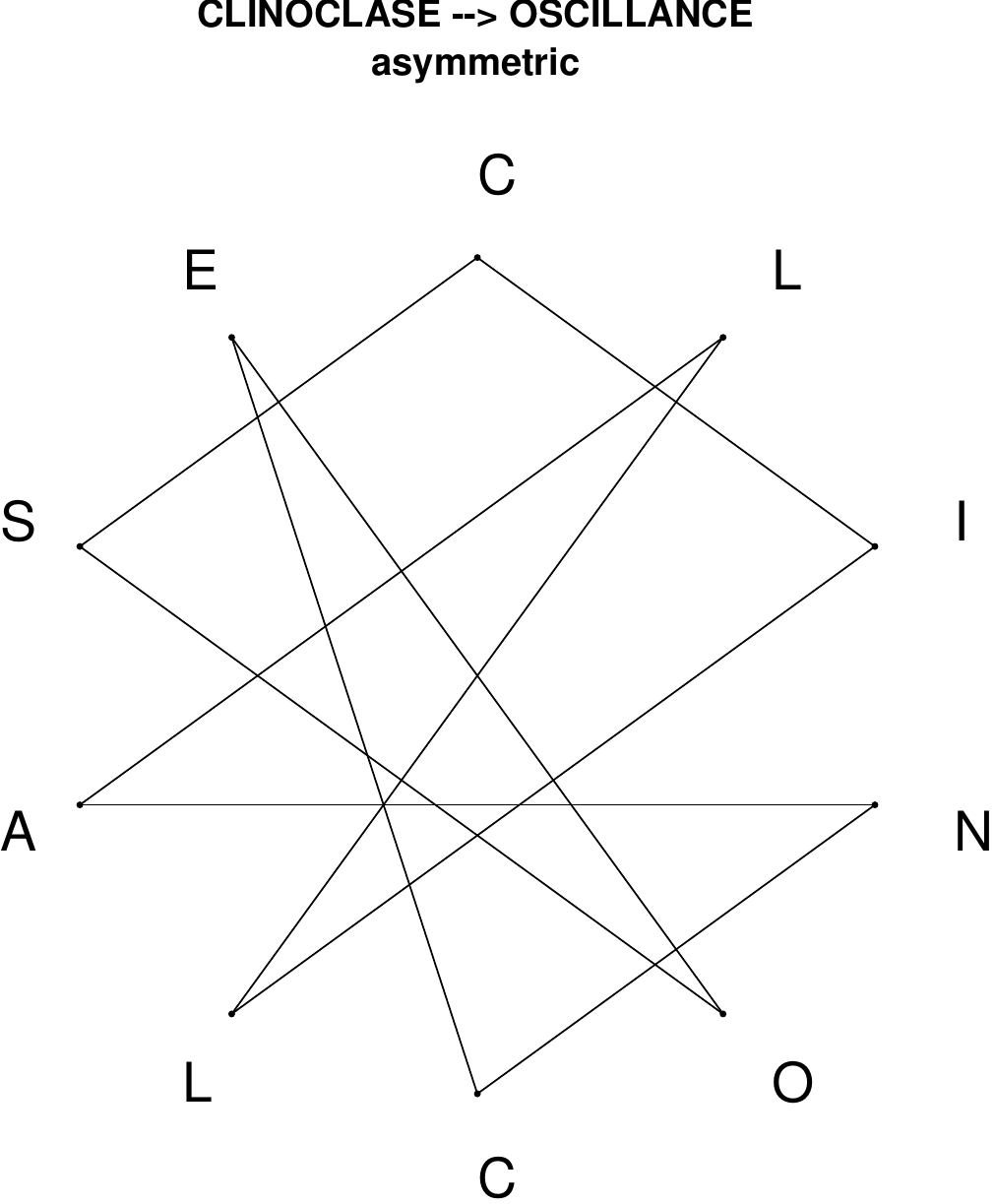}
\end{subfigure}
\hfill
\begin{subfigure}[T]{0.19\textwidth}
\centering
\includegraphics[width=\textwidth]{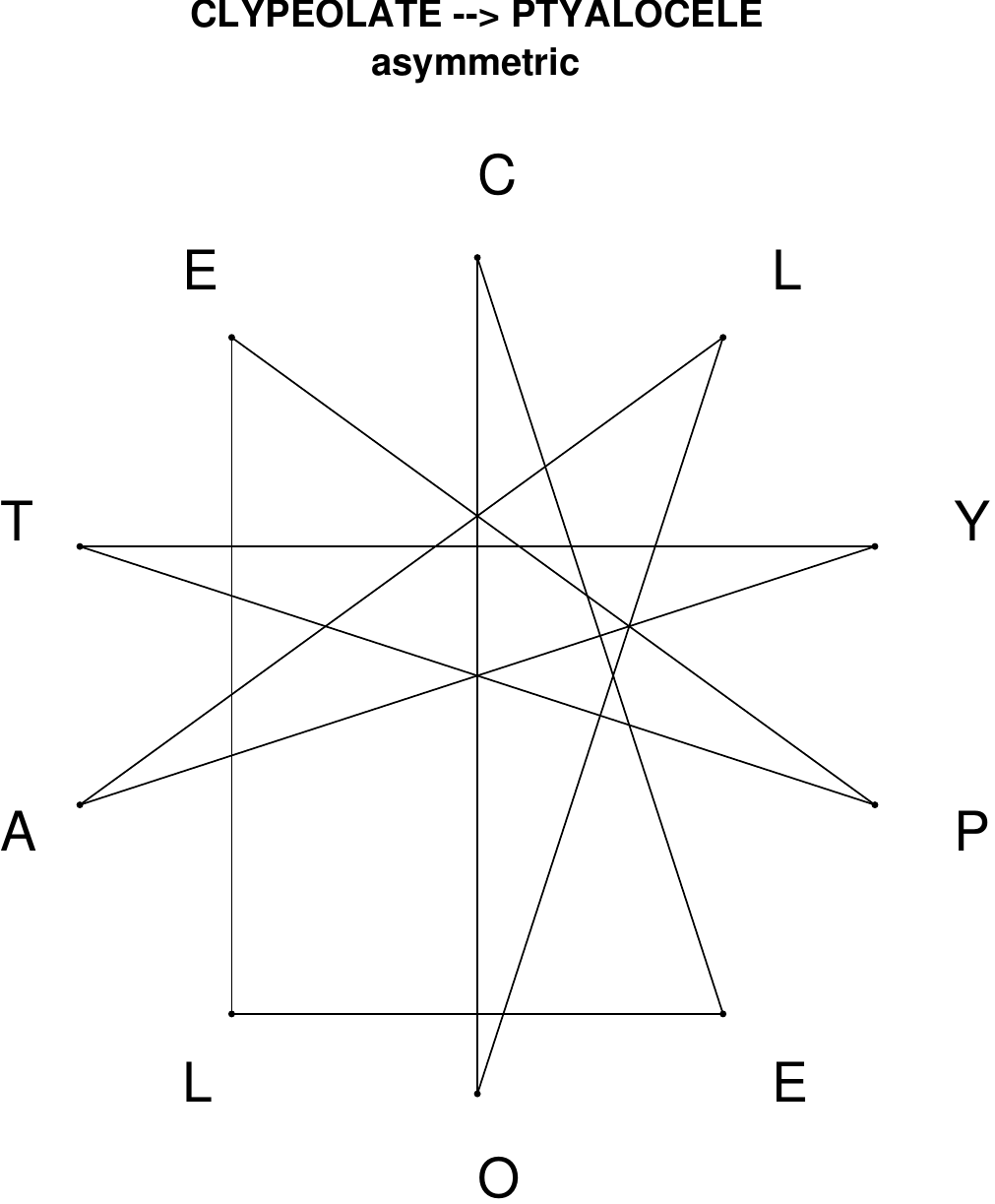}
\end{subfigure}
\hfill
\begin{subfigure}[T]{0.19\textwidth}
\centering
\includegraphics[width=\textwidth]{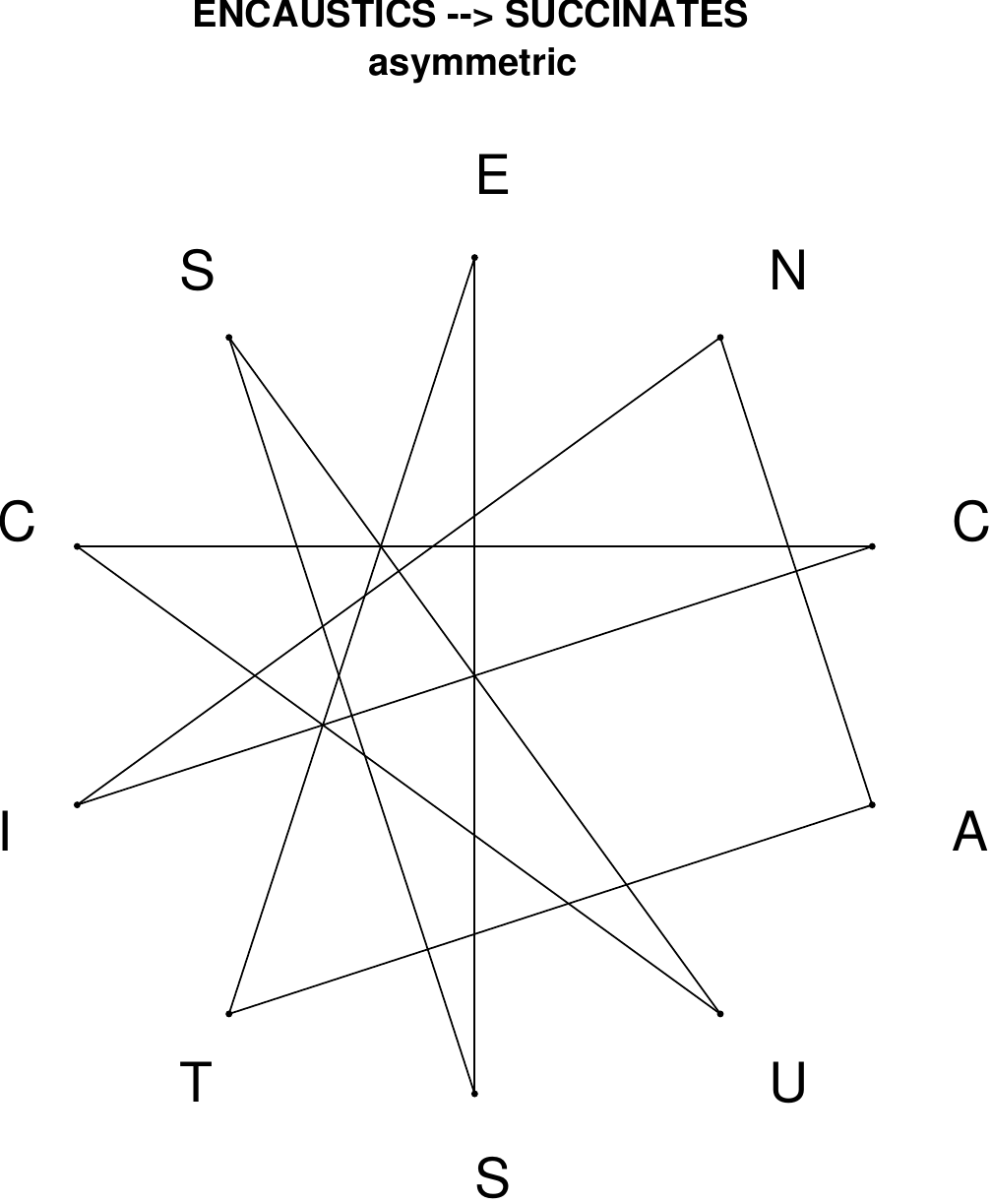}
\end{subfigure}
\hfill
\begin{subfigure}[T]{0.19\textwidth}
\centering
\includegraphics[width=\textwidth]{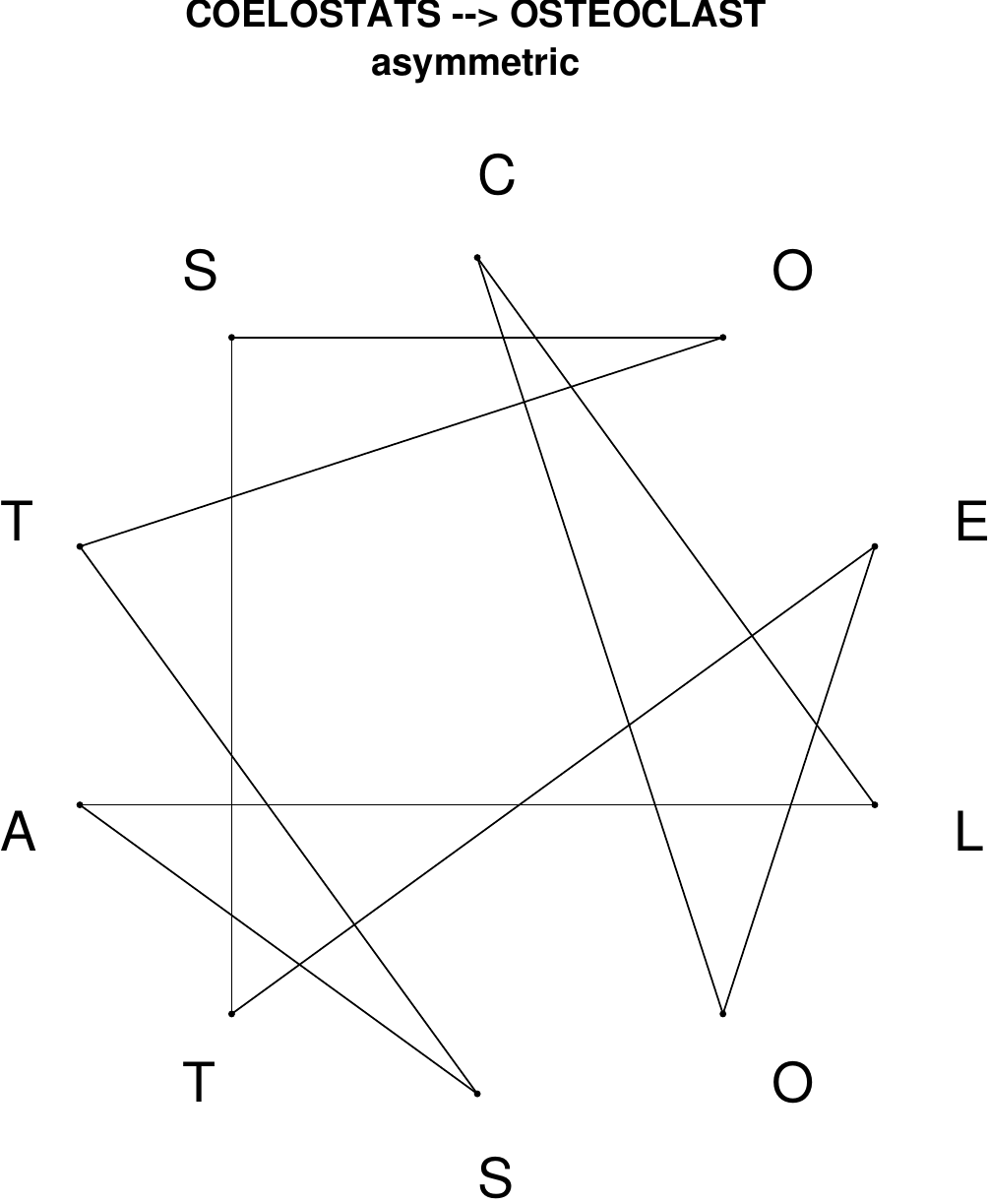}
\end{subfigure}
\hfill
\begin{subfigure}[T]{0.19\textwidth}
\centering
\includegraphics[width=\textwidth]{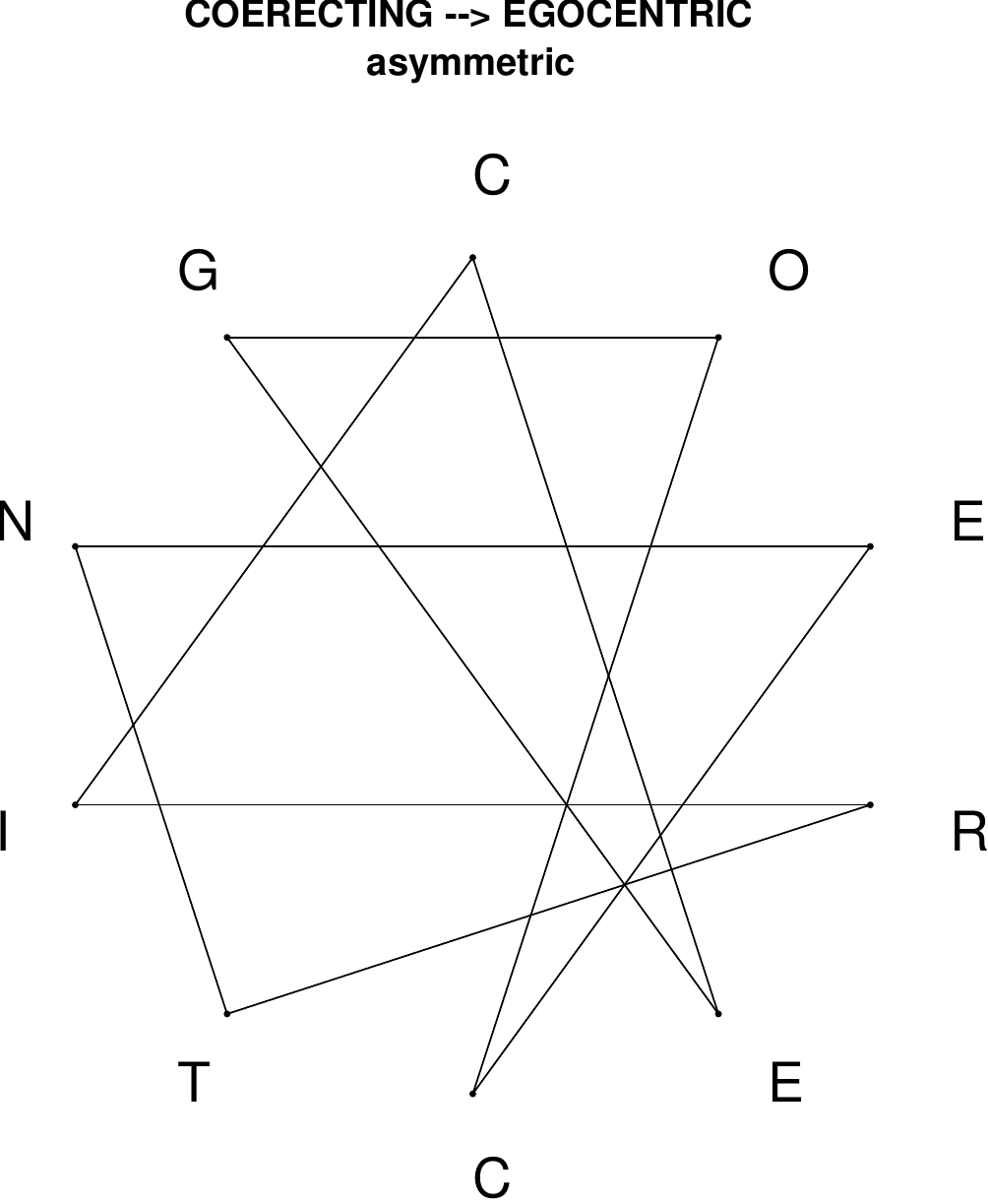}
\end{subfigure}
\end{figure}

\begin{figure}[H]
\centering
\begin{subfigure}[T]{0.19\textwidth}
\centering
\includegraphics[width=\textwidth]{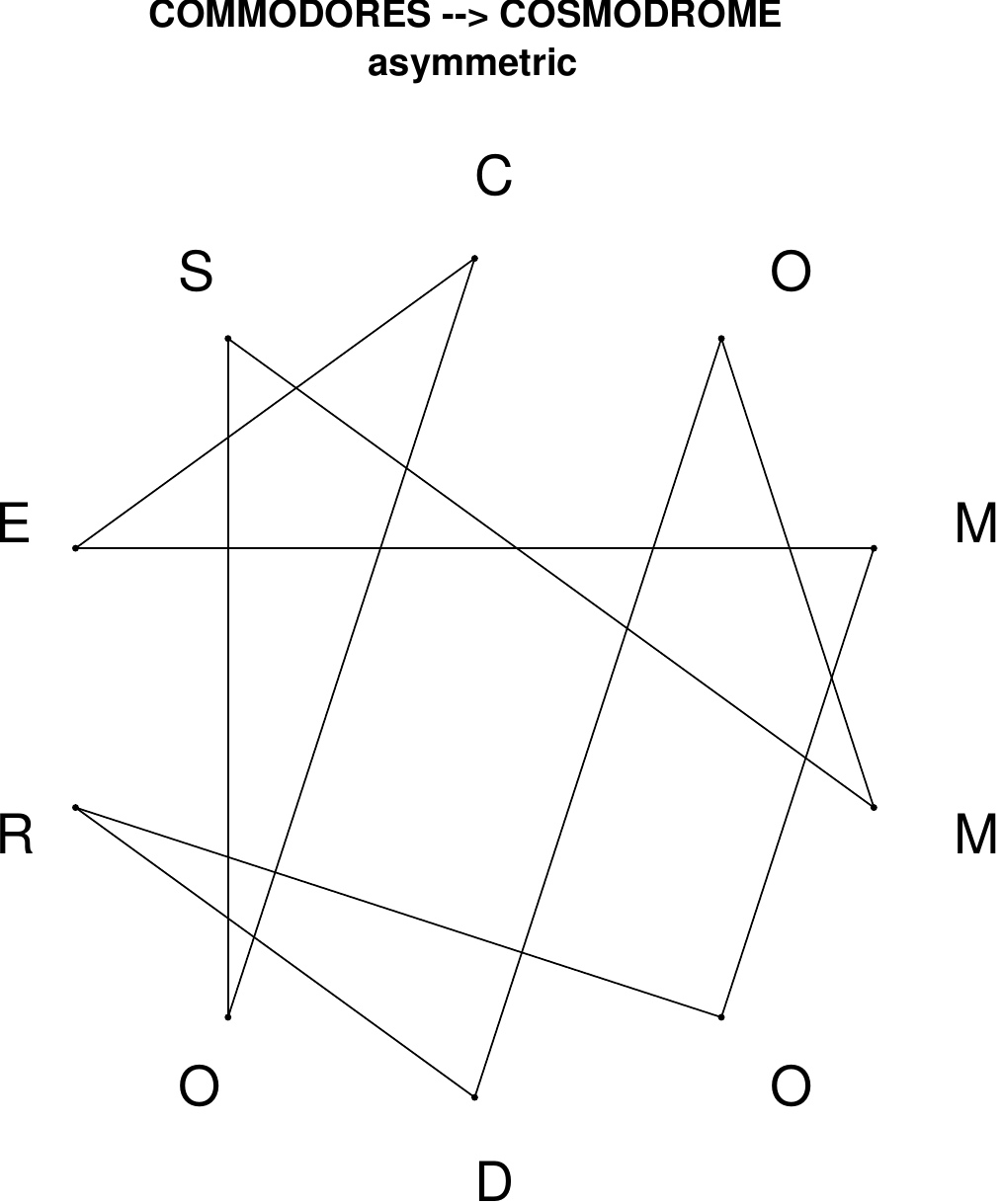}
\end{subfigure}
\hfill
\begin{subfigure}[T]{0.19\textwidth}
\centering
\includegraphics[width=\textwidth]{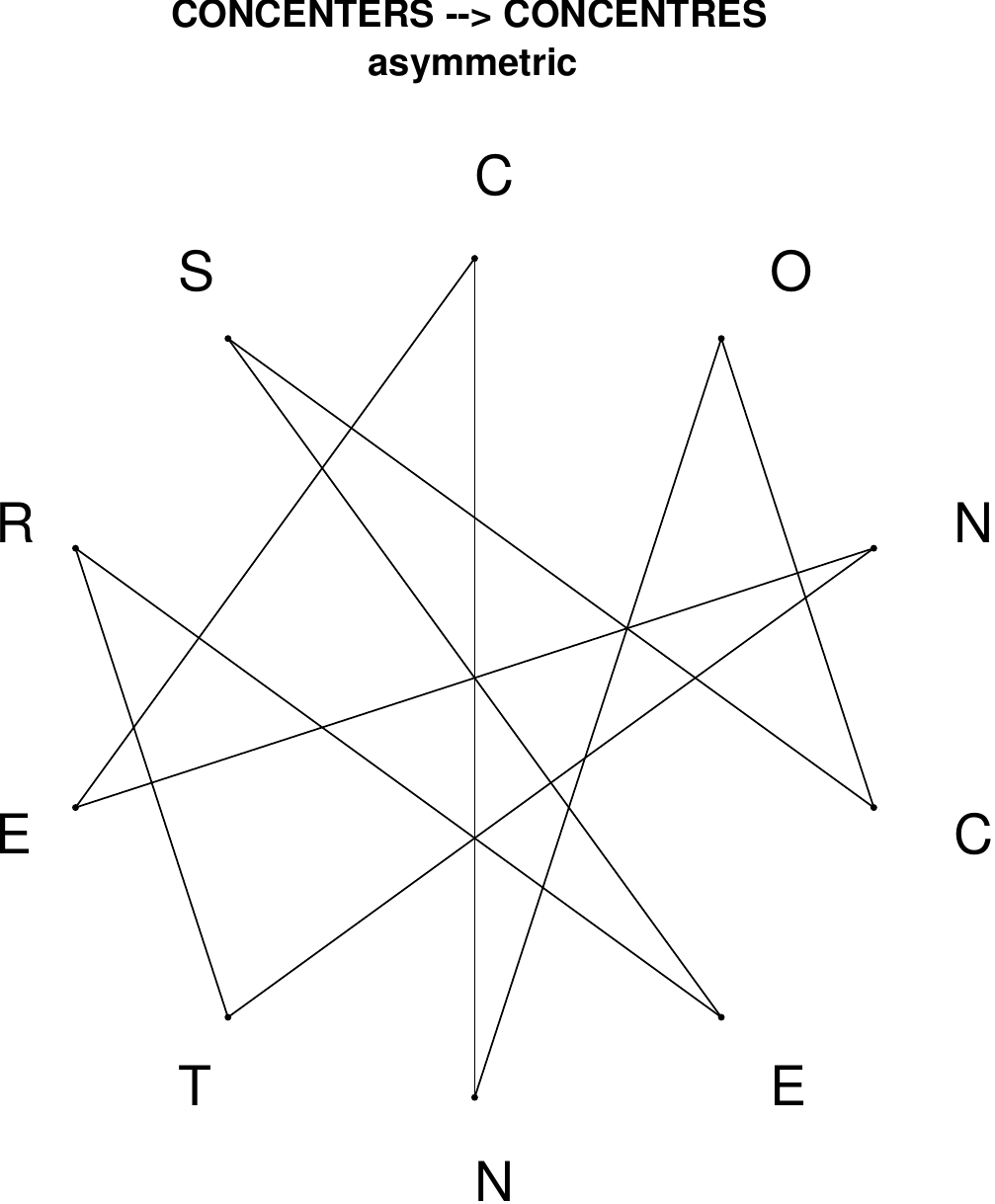}
\end{subfigure}
\hfill
\begin{subfigure}[T]{0.19\textwidth}
\centering
\includegraphics[width=\textwidth]{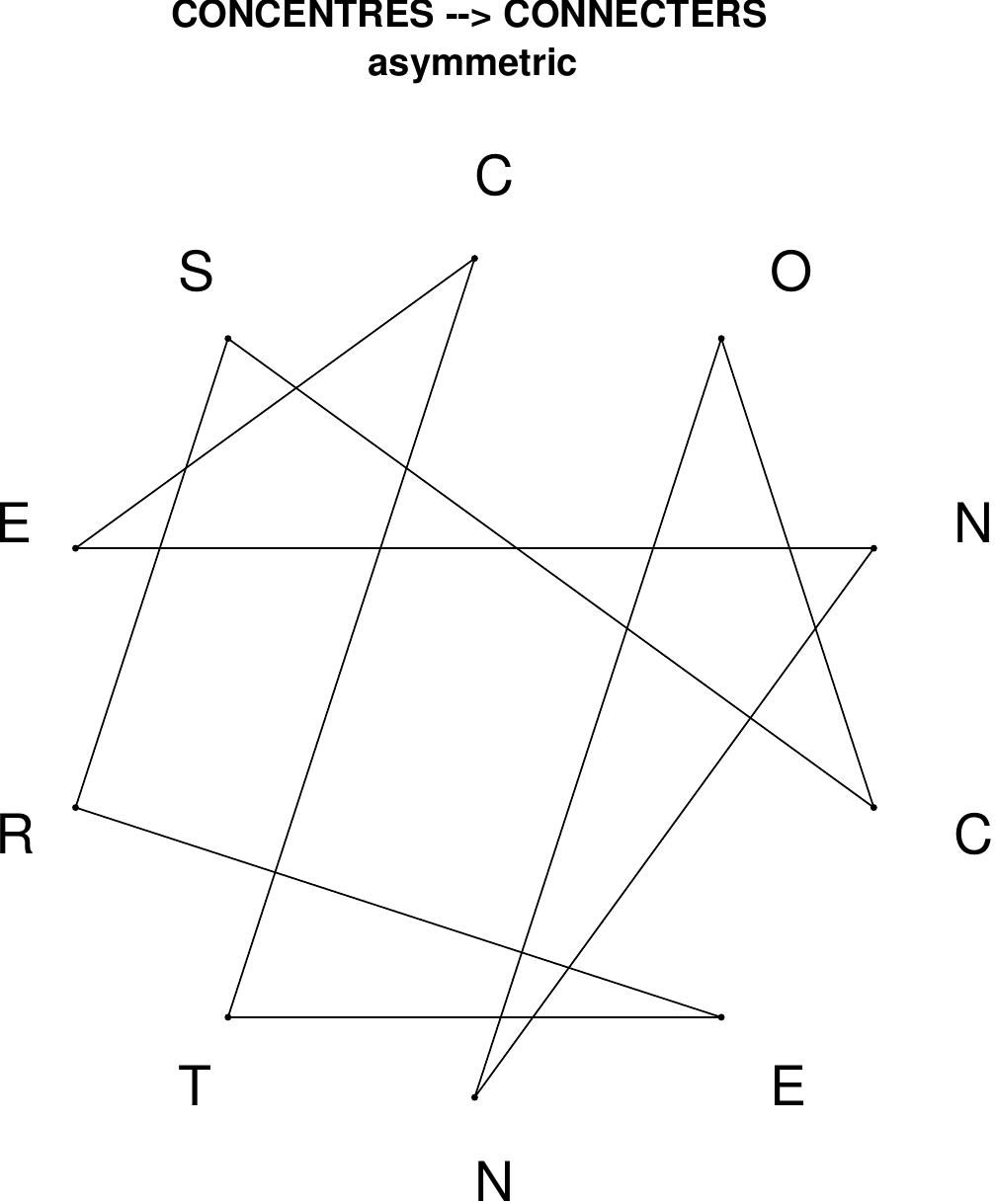}
\end{subfigure}
\hfill
\begin{subfigure}[T]{0.19\textwidth}
\centering
\includegraphics[width=\textwidth]{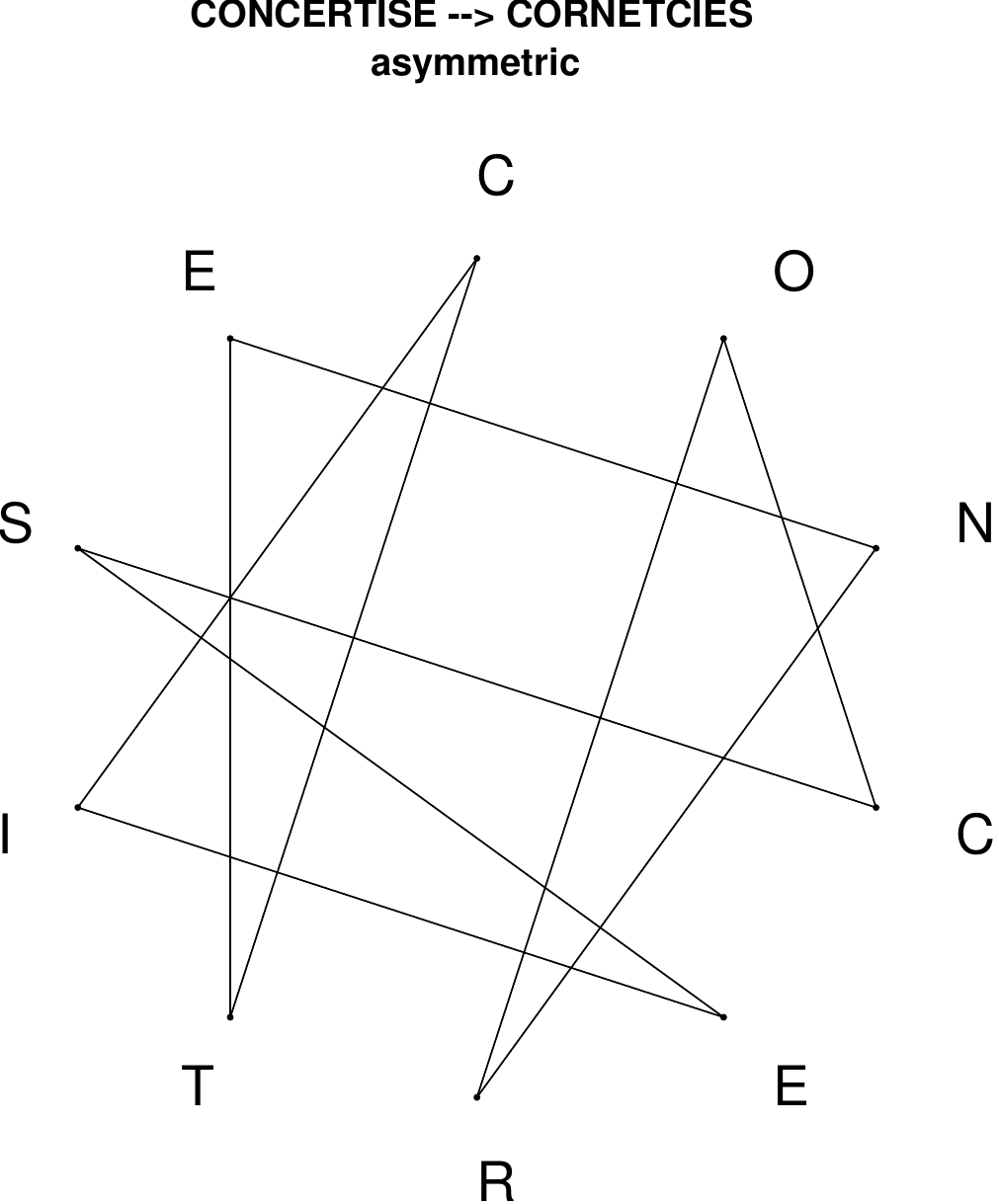}
\end{subfigure}
\hfill
\begin{subfigure}[T]{0.19\textwidth}
\centering
\includegraphics[width=\textwidth]{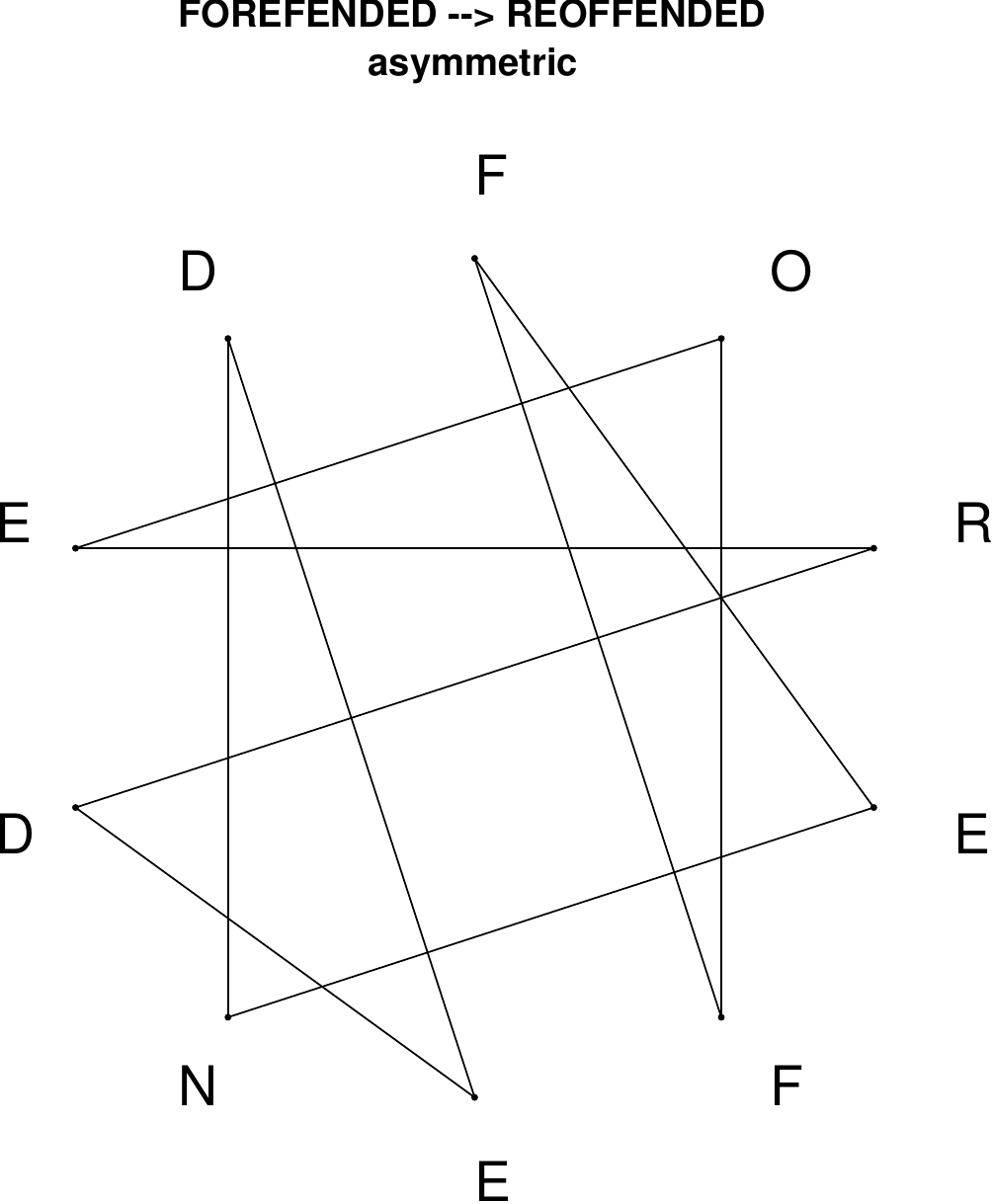}
\end{subfigure}
\end{figure}

\begin{figure}[H]
\centering
\begin{subfigure}[T]{0.19\textwidth}
\centering
\includegraphics[width=\textwidth]{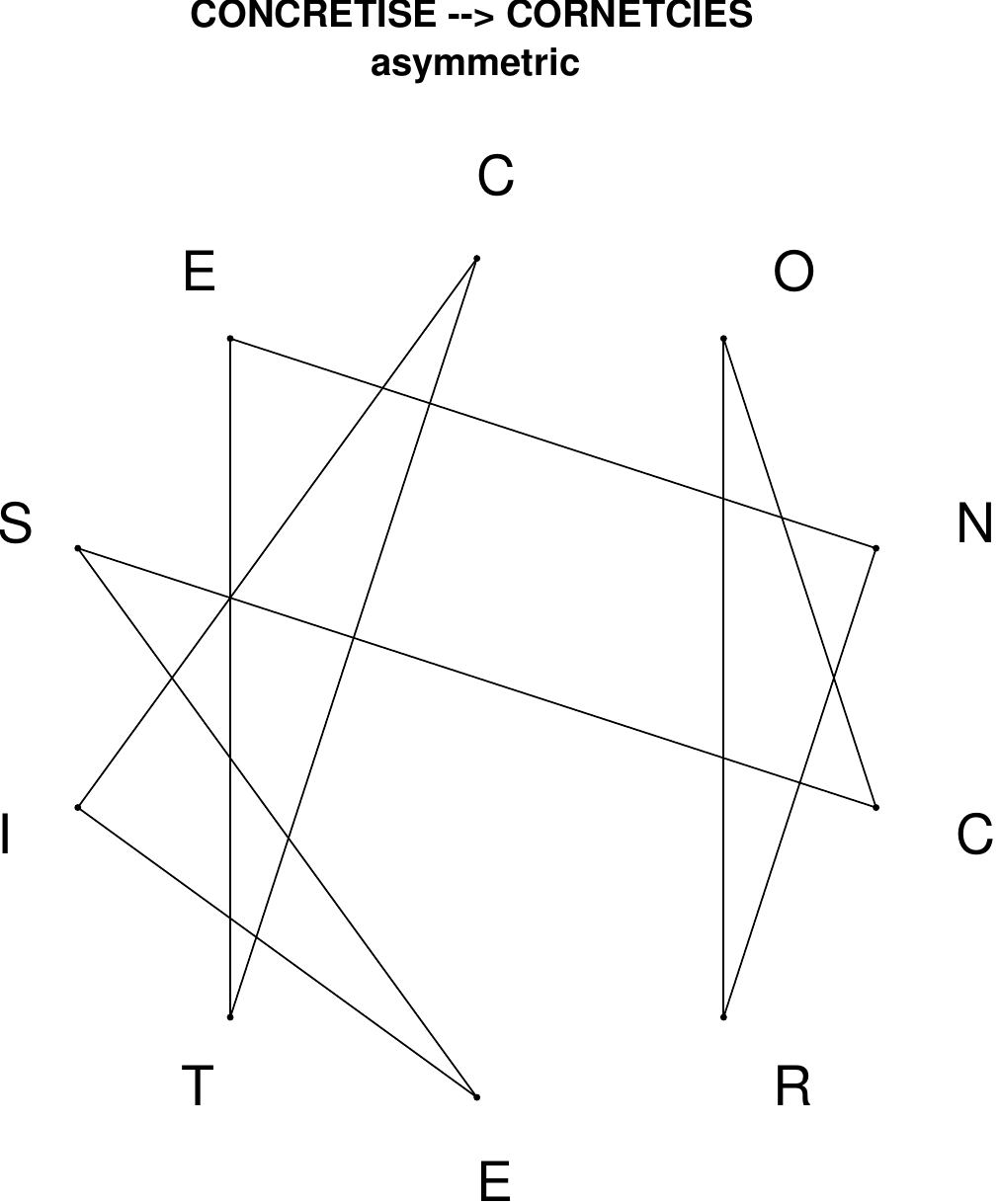}
\end{subfigure}
\hfill
\begin{subfigure}[T]{0.19\textwidth}
\centering
\includegraphics[width=\textwidth]{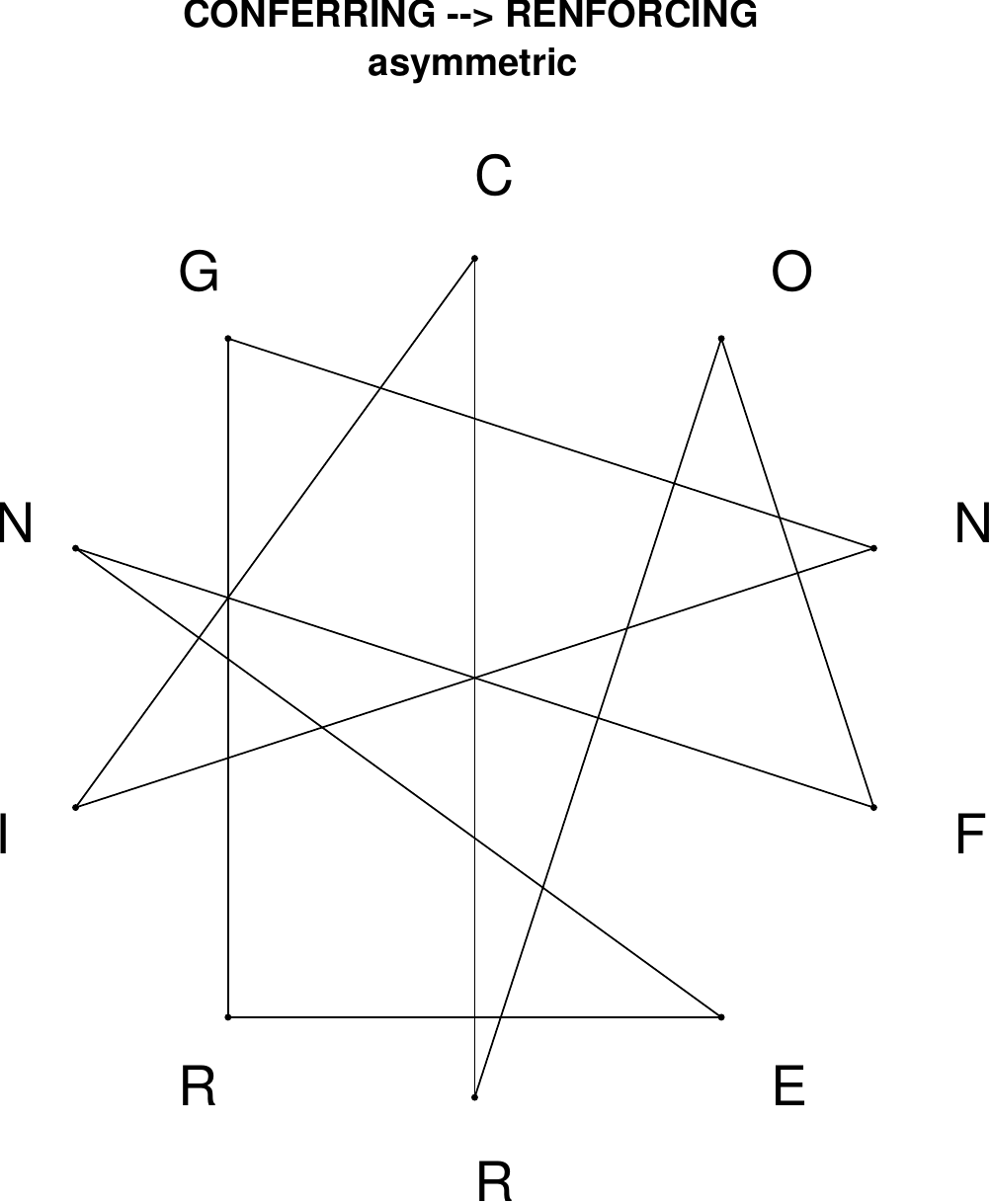}
\end{subfigure}
\hfill
\begin{subfigure}[T]{0.19\textwidth}
\centering
\includegraphics[width=\textwidth]{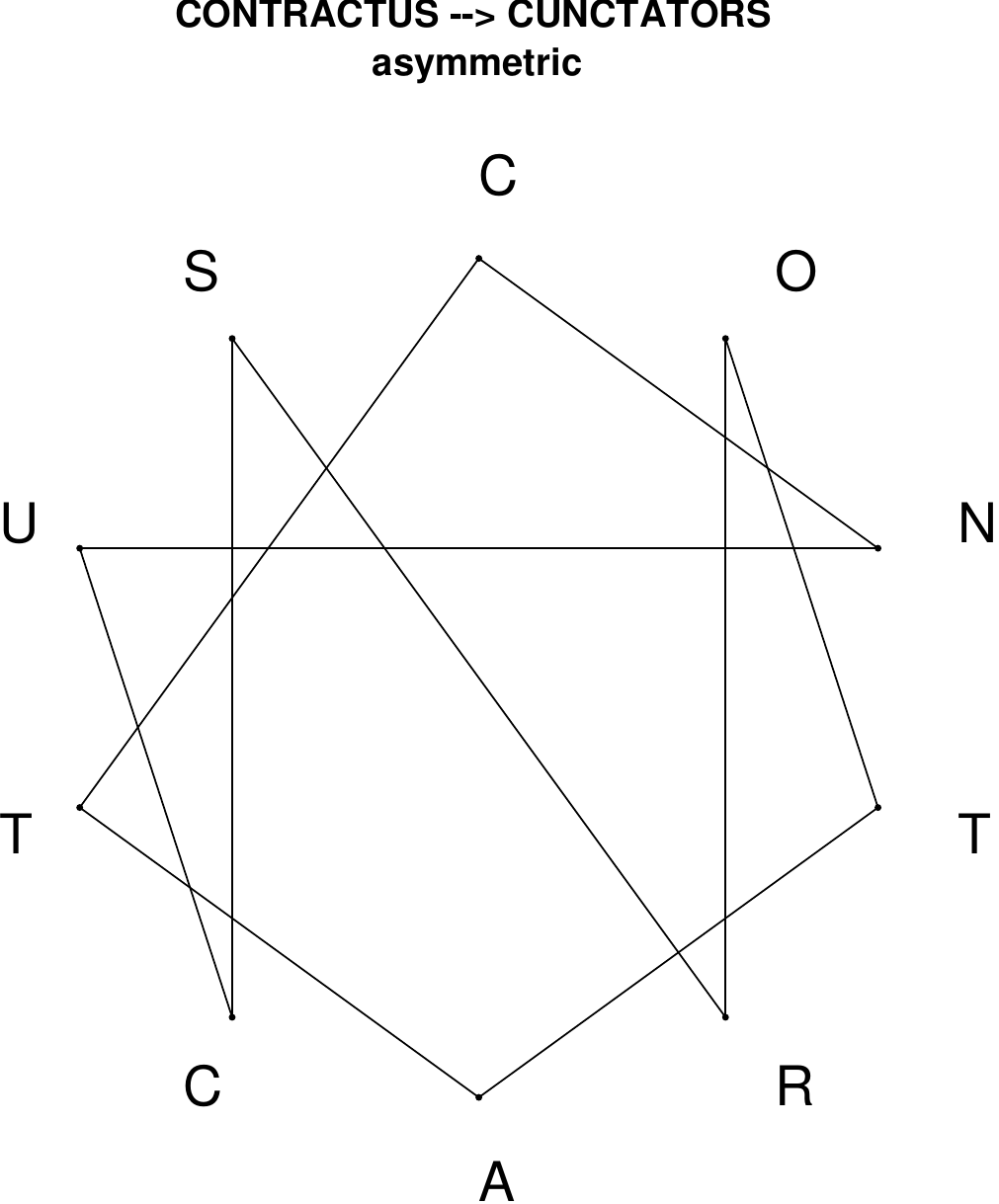}
\end{subfigure}
\hfill
\begin{subfigure}[T]{0.19\textwidth}
\centering
\includegraphics[width=\textwidth]{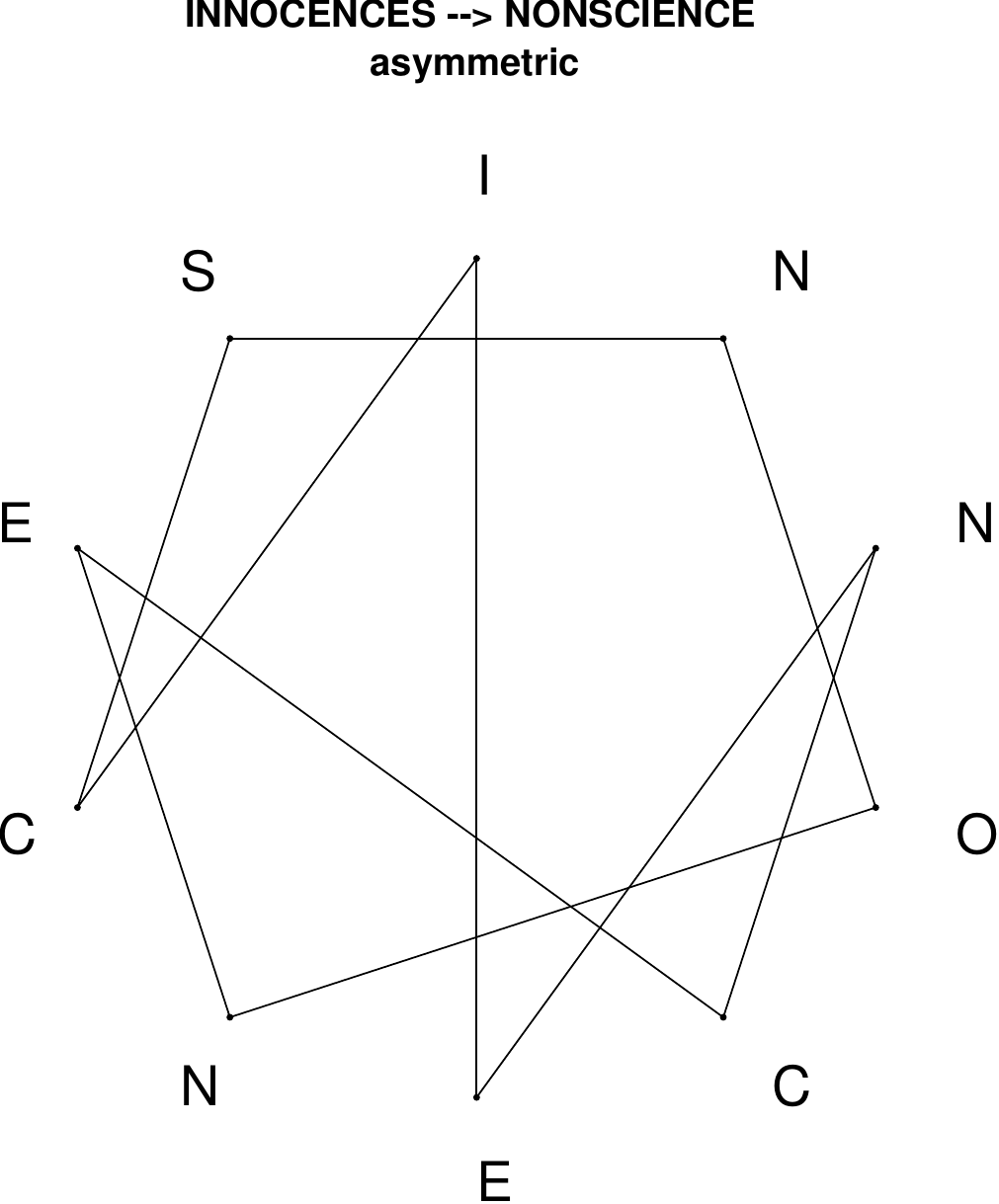}
\end{subfigure}
\hfill
\begin{subfigure}[T]{0.19\textwidth}
\centering
\includegraphics[width=\textwidth]{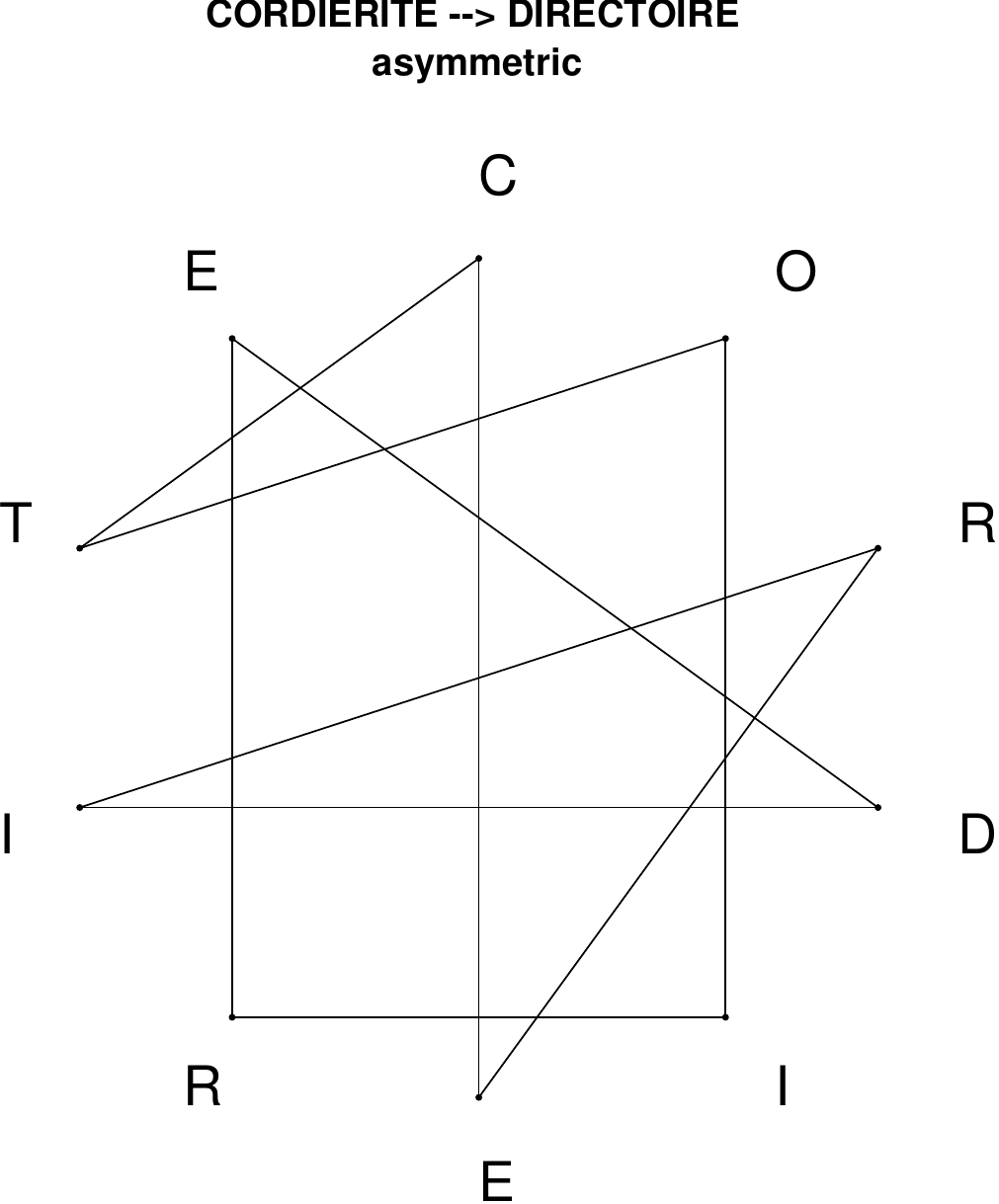}
\end{subfigure}
\end{figure}

\begin{figure}[H]
\centering
\begin{subfigure}[T]{0.19\textwidth}
\centering
\includegraphics[width=\textwidth]{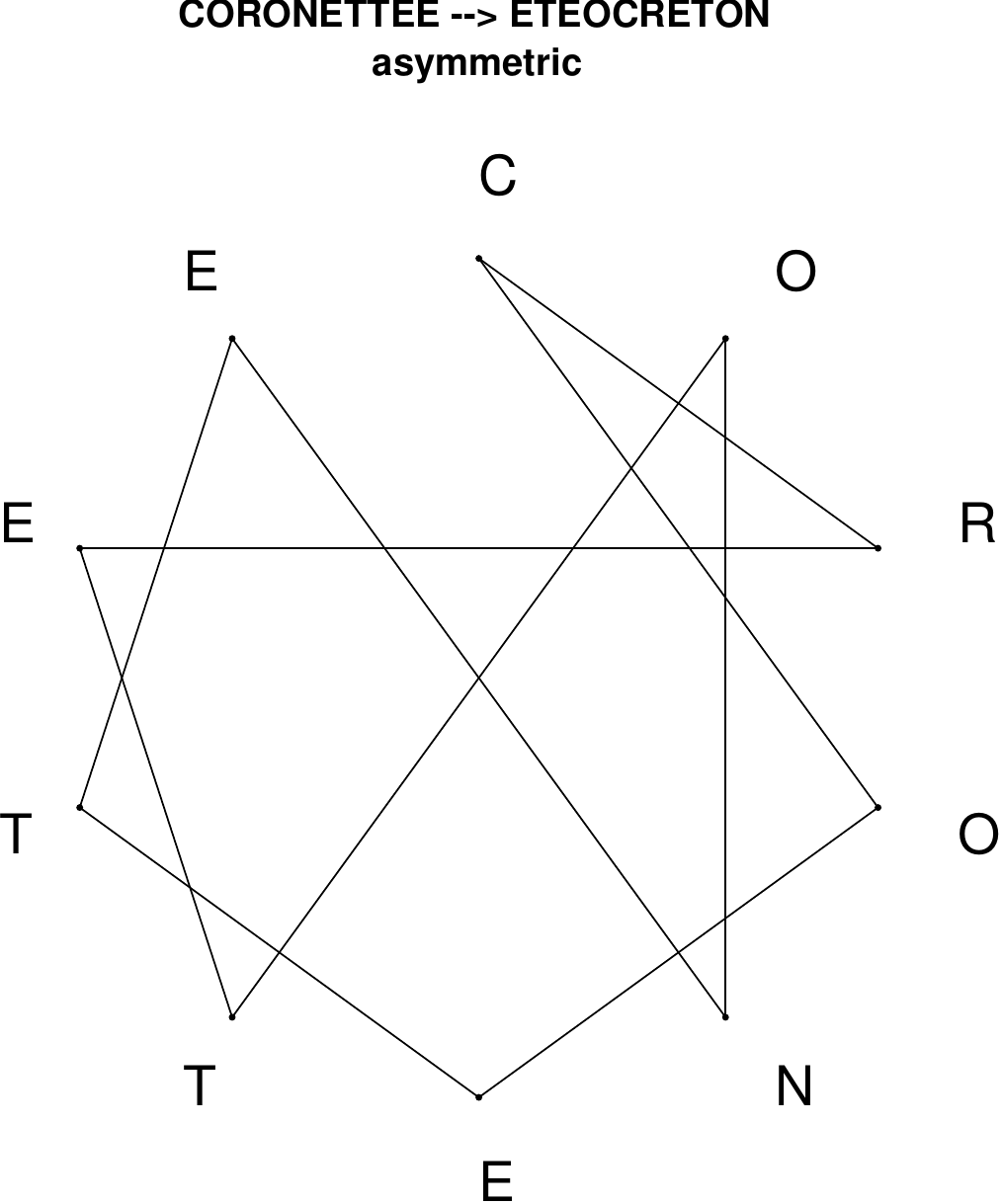}
\end{subfigure}
\hfill
\begin{subfigure}[T]{0.19\textwidth}
\centering
\includegraphics[width=\textwidth]{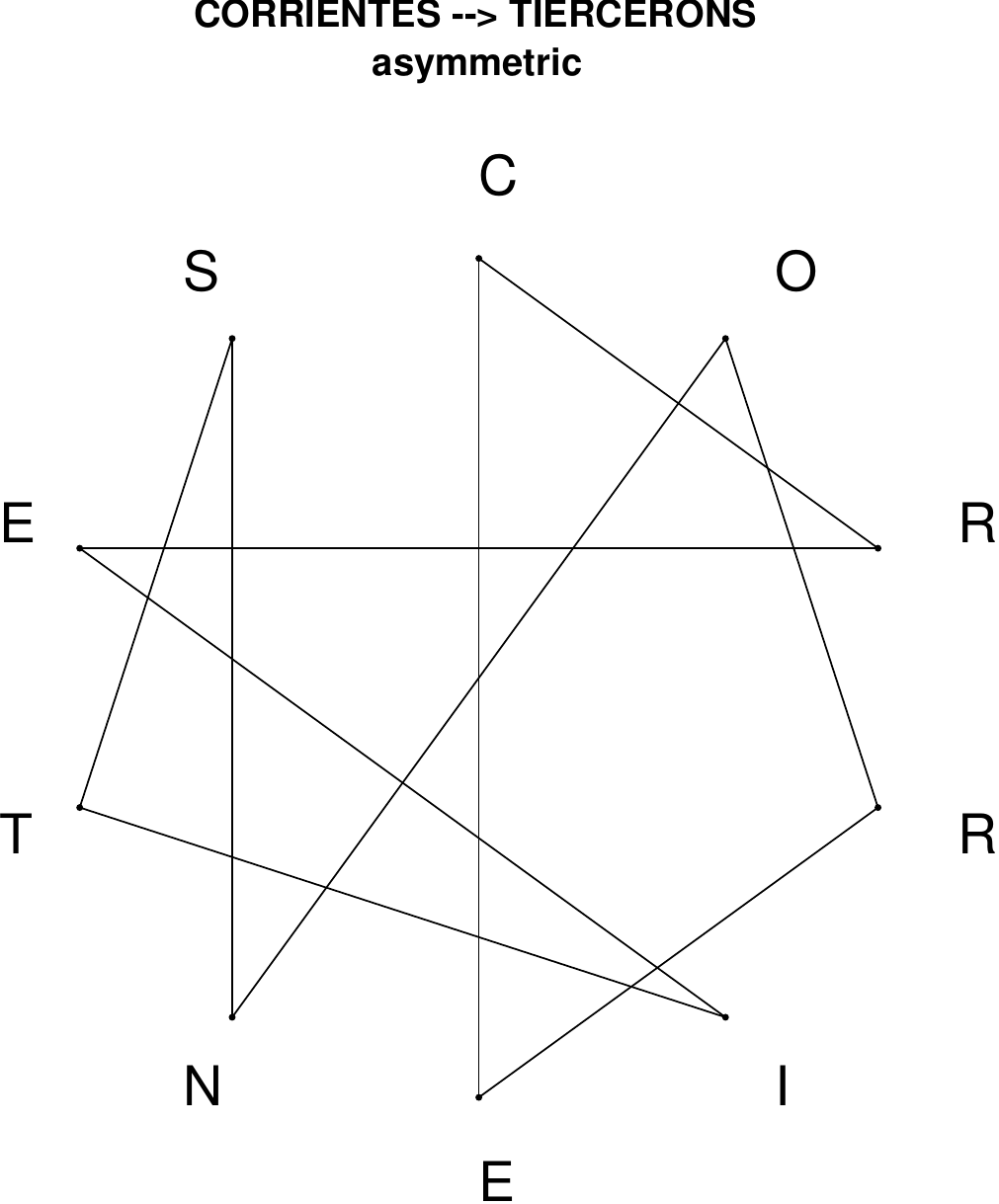}
\end{subfigure}
\hfill
\begin{subfigure}[T]{0.19\textwidth}
\centering
\includegraphics[width=\textwidth]{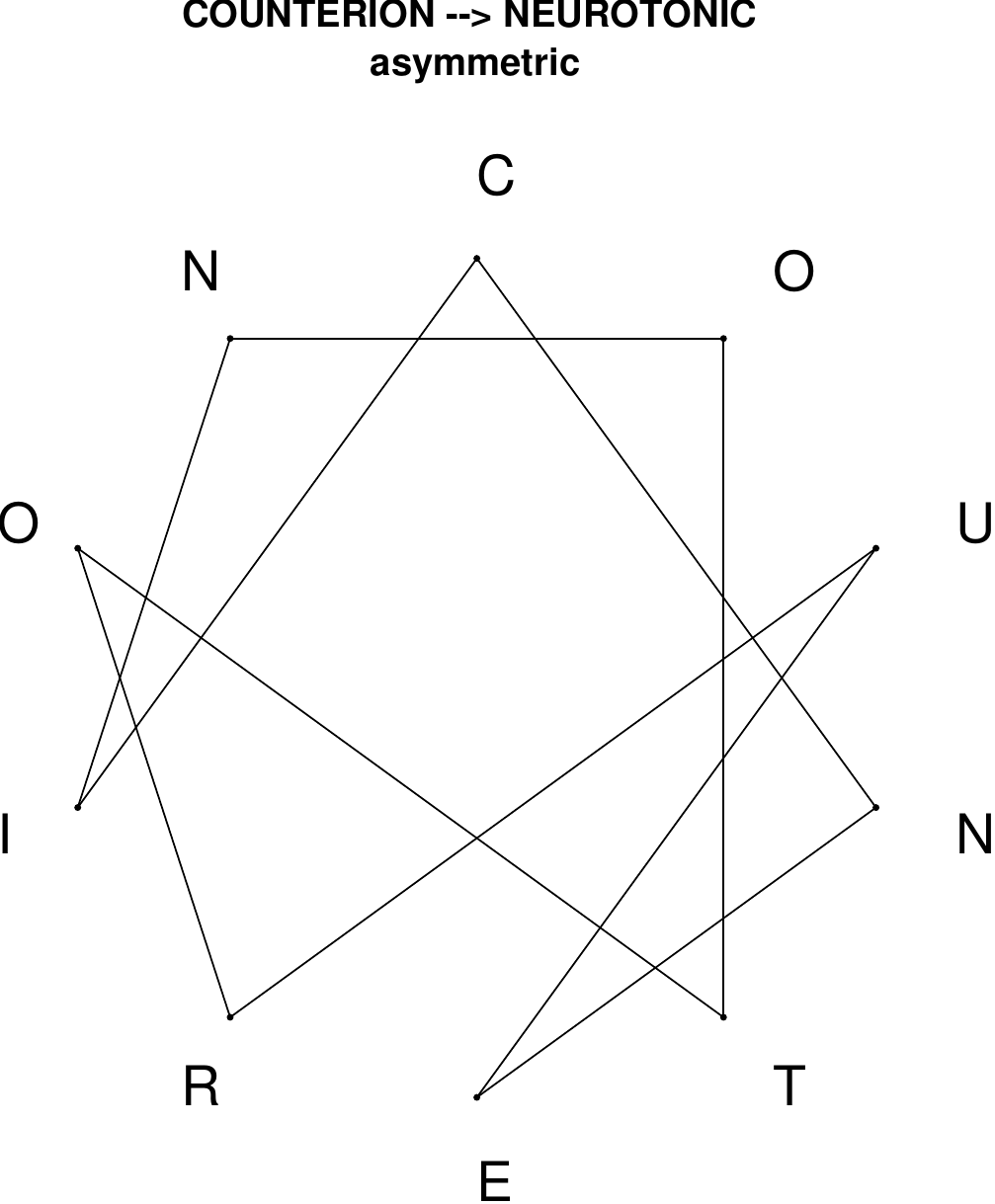}
\end{subfigure}
\hfill
\begin{subfigure}[T]{0.19\textwidth}
\centering
\includegraphics[width=\textwidth]{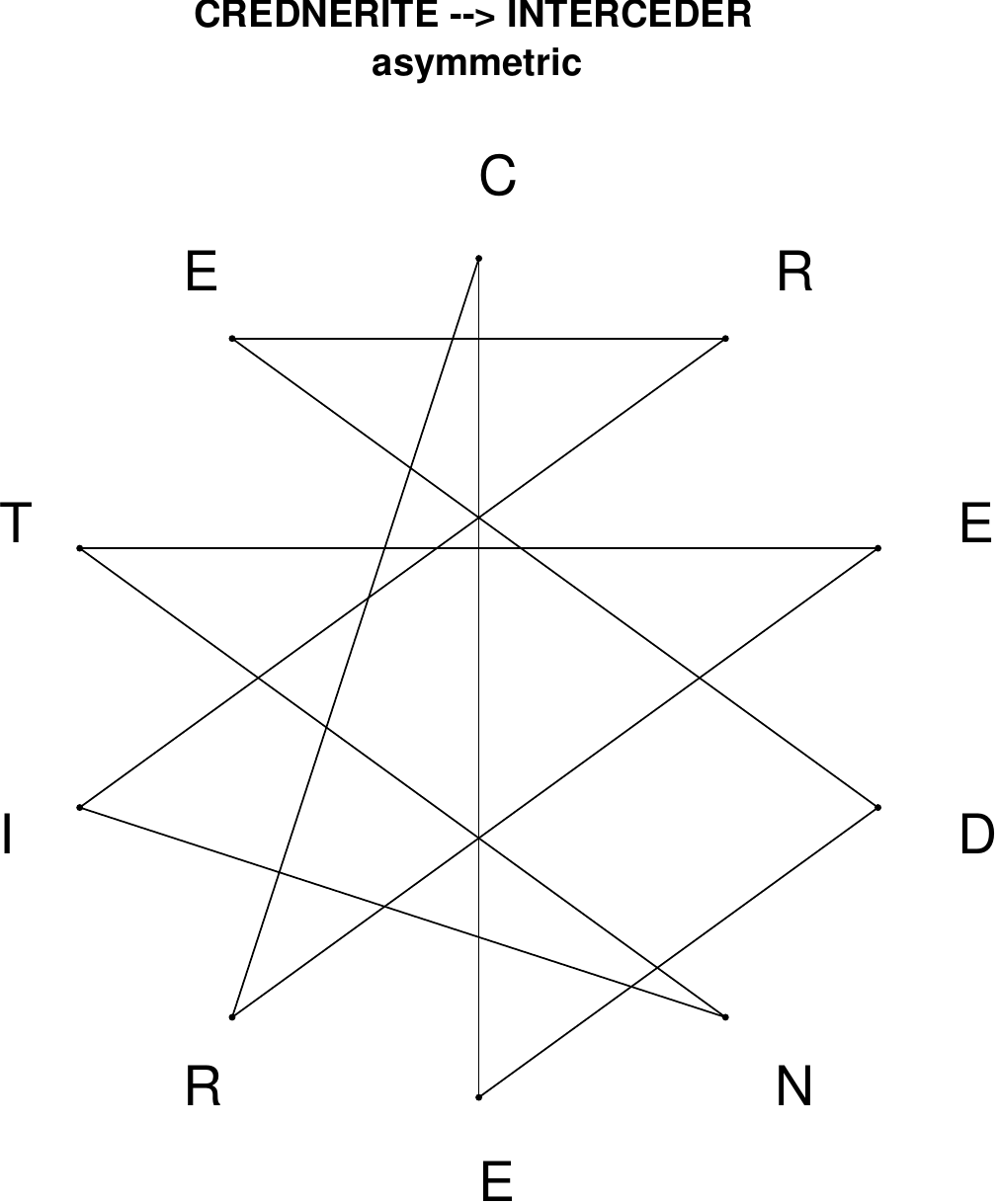}
\end{subfigure}
\hfill
\begin{subfigure}[T]{0.19\textwidth}
\centering
\includegraphics[width=\textwidth]{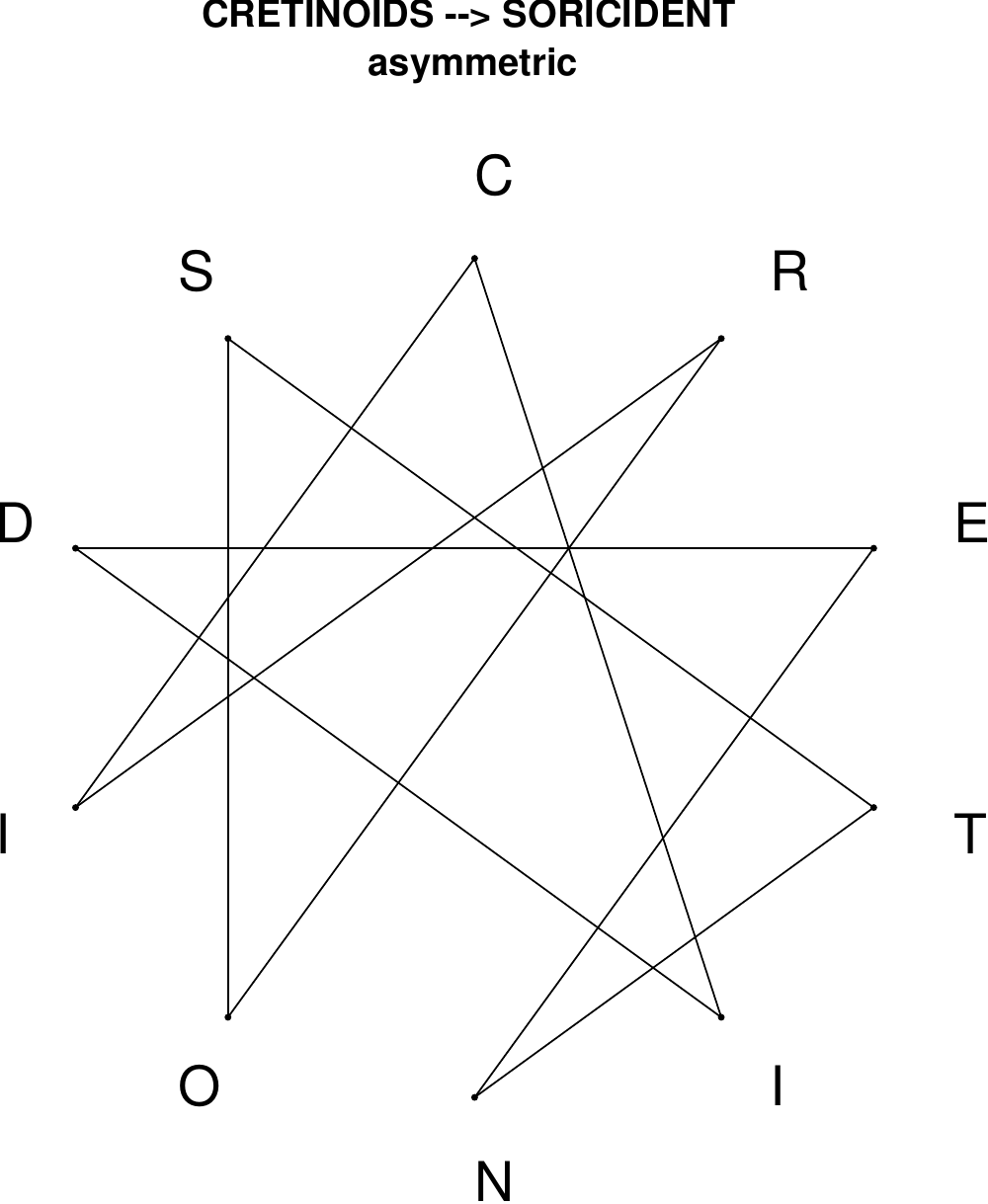}
\end{subfigure}
\end{figure}

\begin{figure}[H]
\centering
\begin{subfigure}[T]{0.19\textwidth}
\centering
\includegraphics[width=\textwidth]{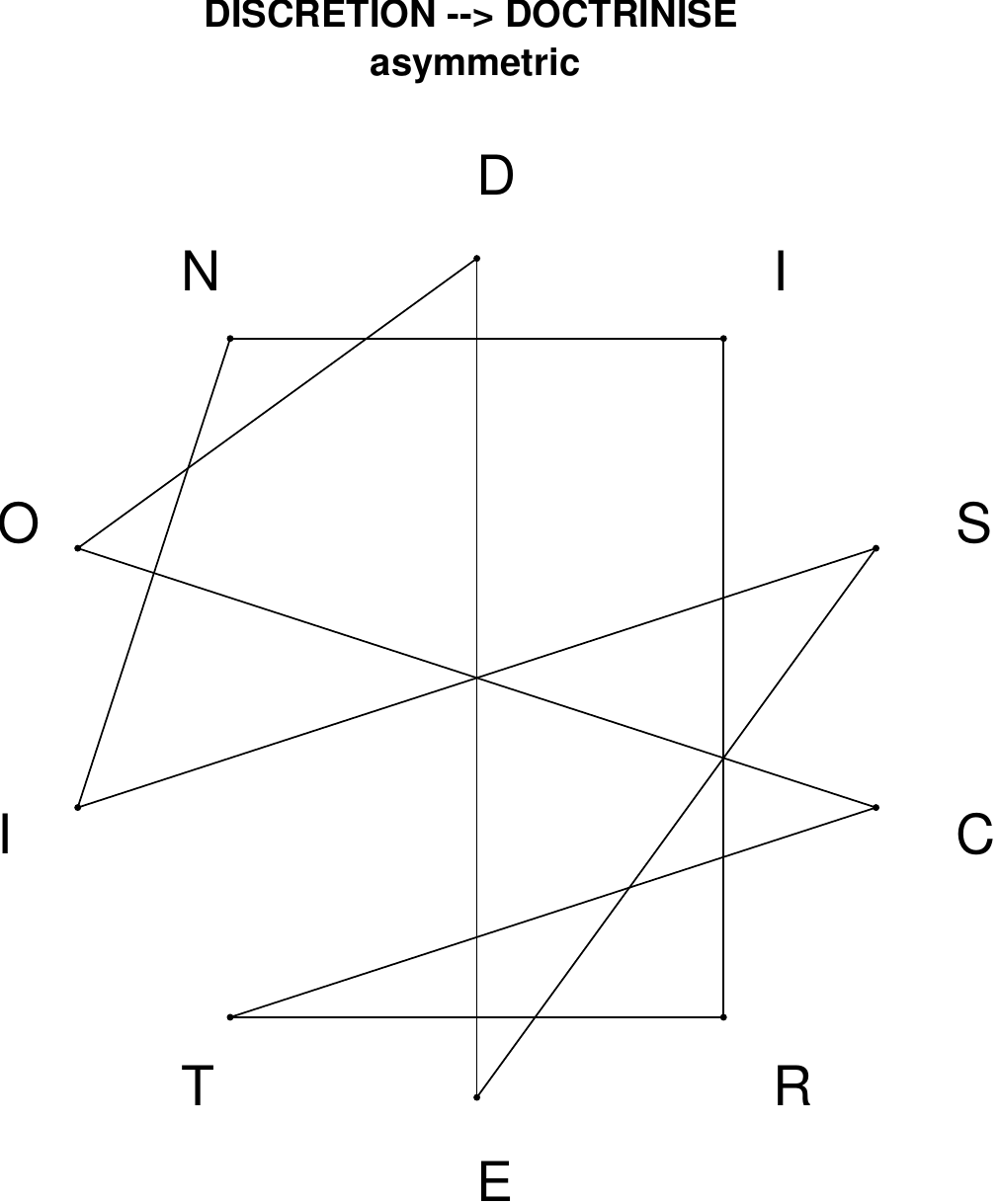}
\end{subfigure}
\hfill
\begin{subfigure}[T]{0.19\textwidth}
\centering
\includegraphics[width=\textwidth]{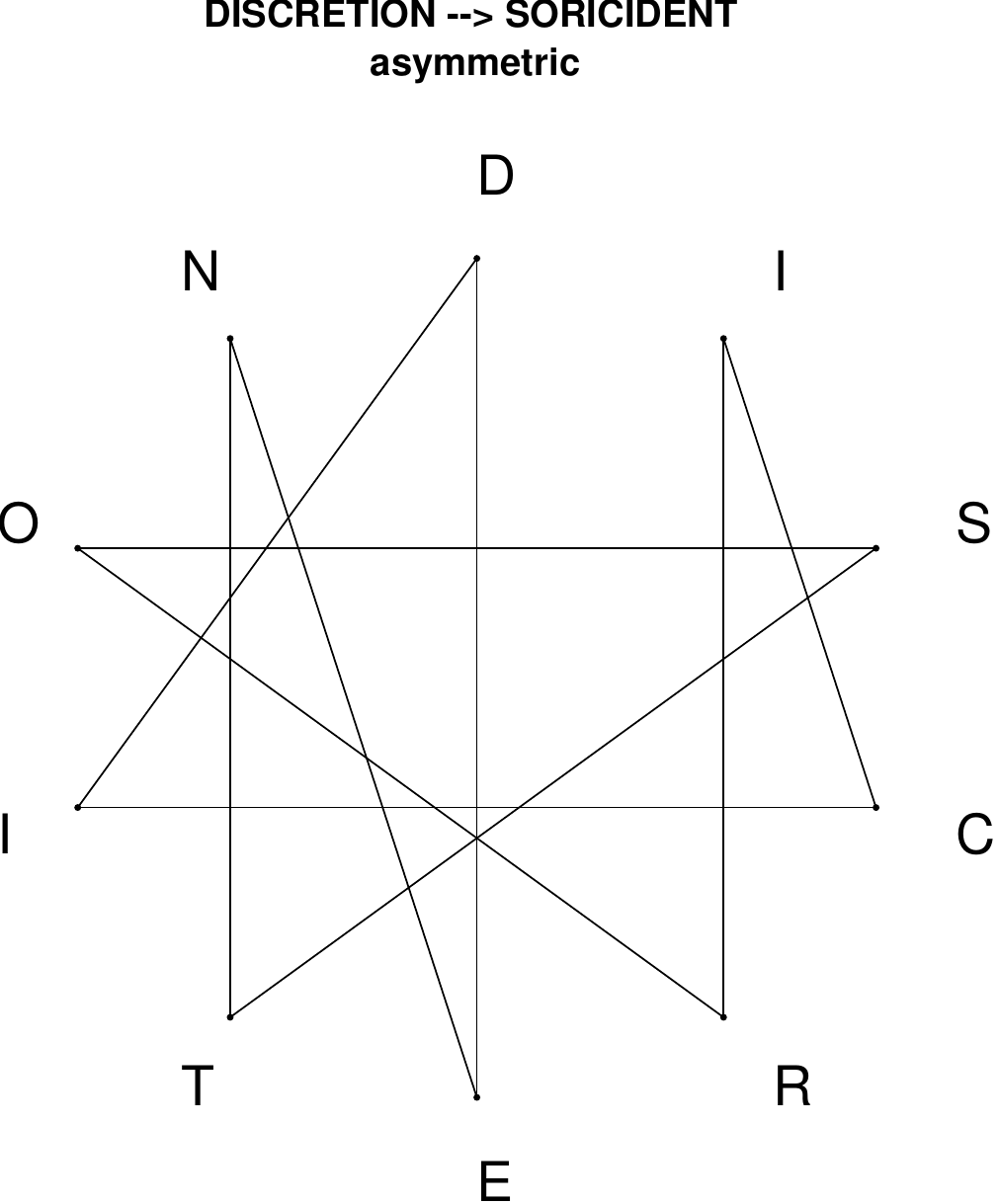}
\end{subfigure}
\hfill
\begin{subfigure}[T]{0.19\textwidth}
\centering
\includegraphics[width=\textwidth]{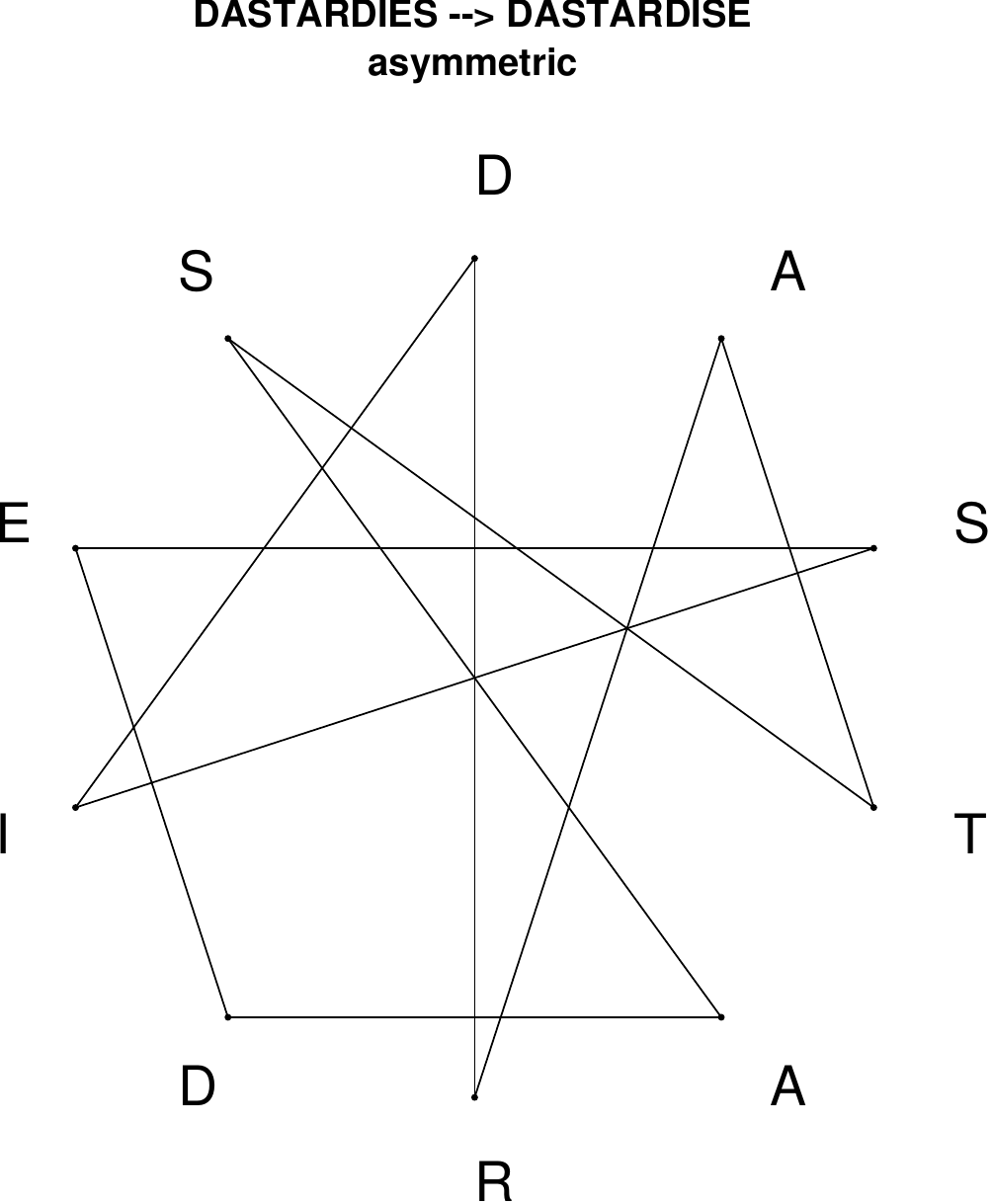}
\end{subfigure}
\hfill
\begin{subfigure}[T]{0.19\textwidth}
\centering
\includegraphics[width=\textwidth]{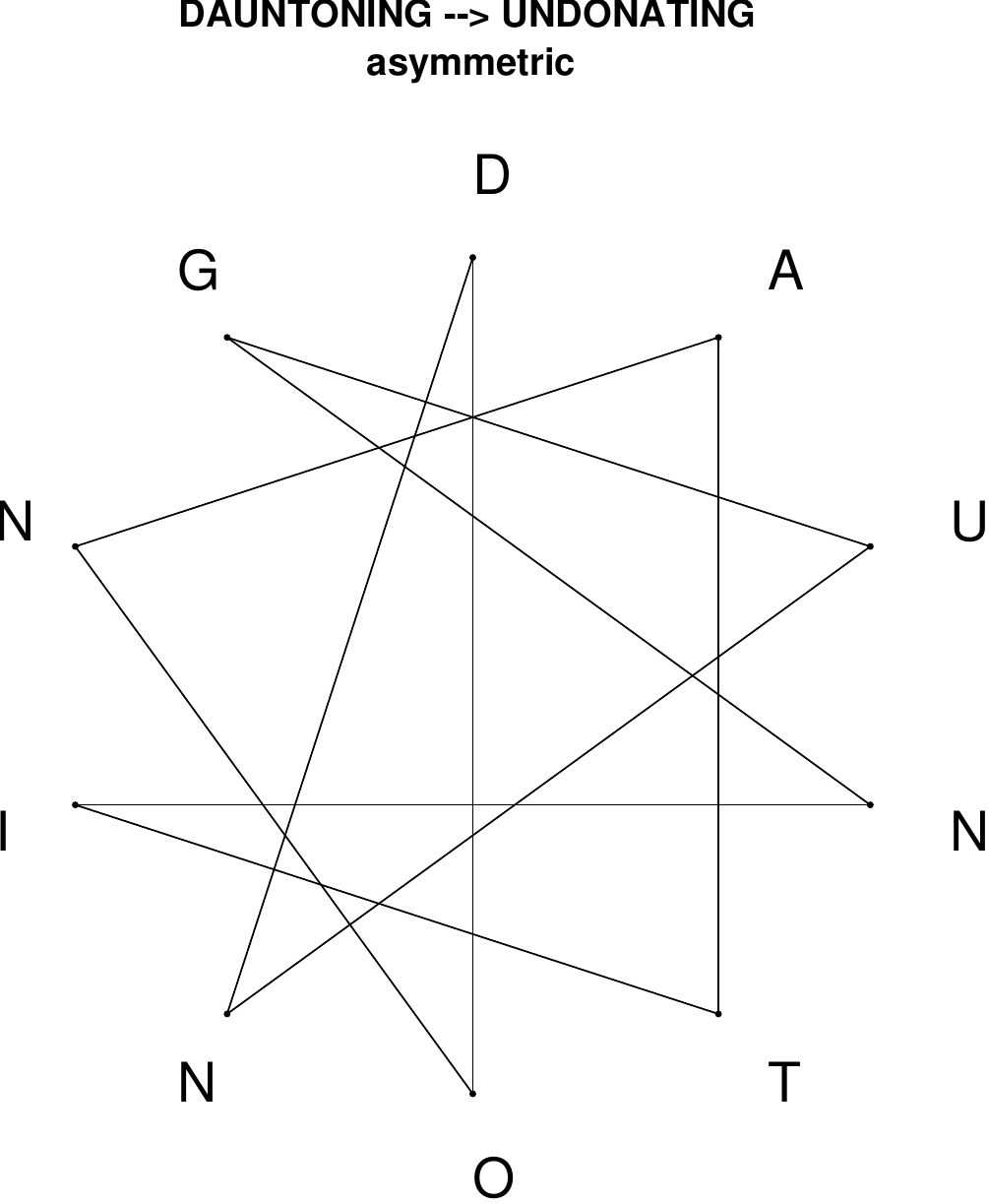}
\end{subfigure}
\hfill
\begin{subfigure}[T]{0.19\textwidth}
\centering
\includegraphics[width=\textwidth]{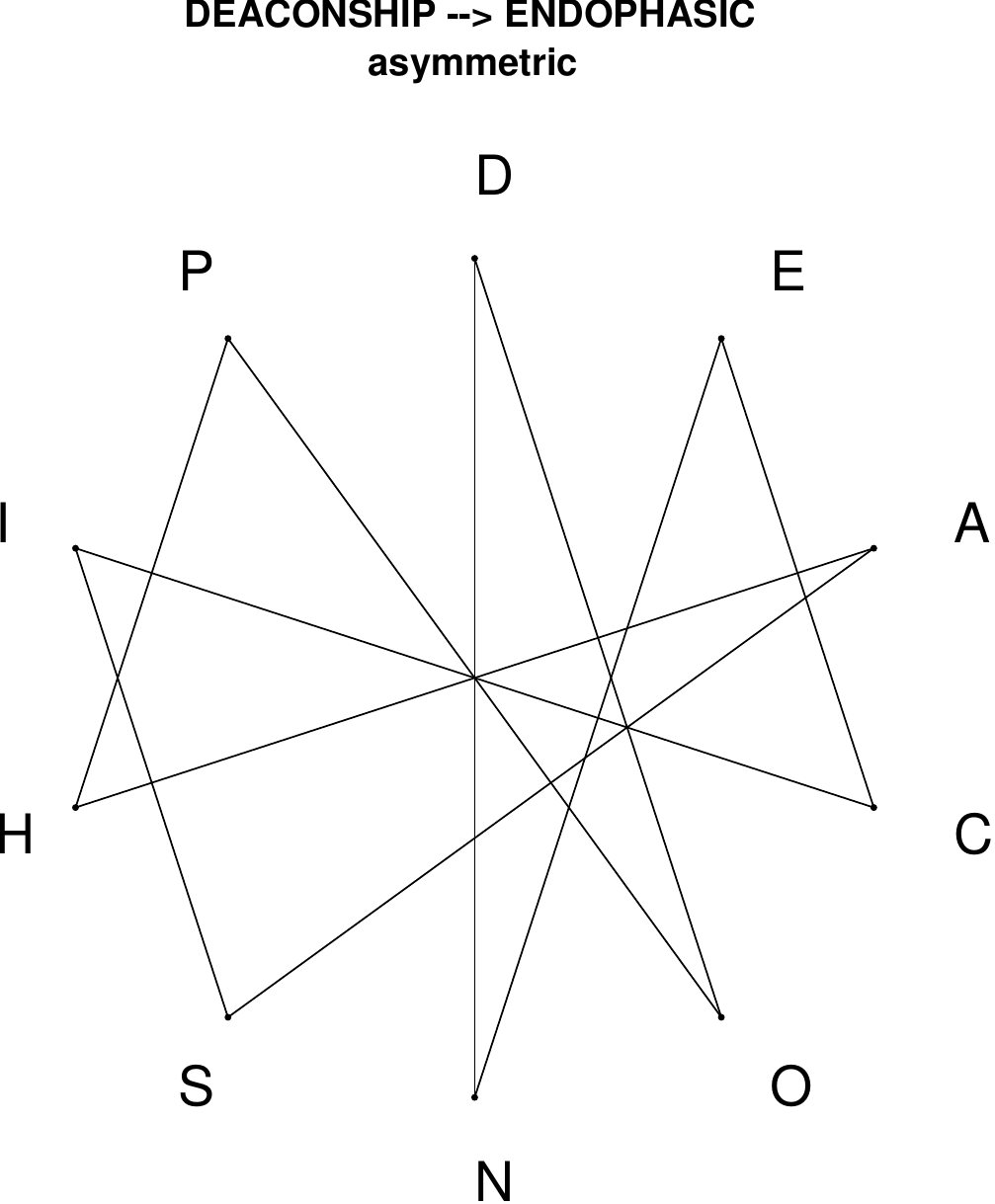}
\end{subfigure}
\end{figure}

\begin{figure}[H]
\centering
\begin{subfigure}[T]{0.19\textwidth}
\centering
\includegraphics[width=\textwidth]{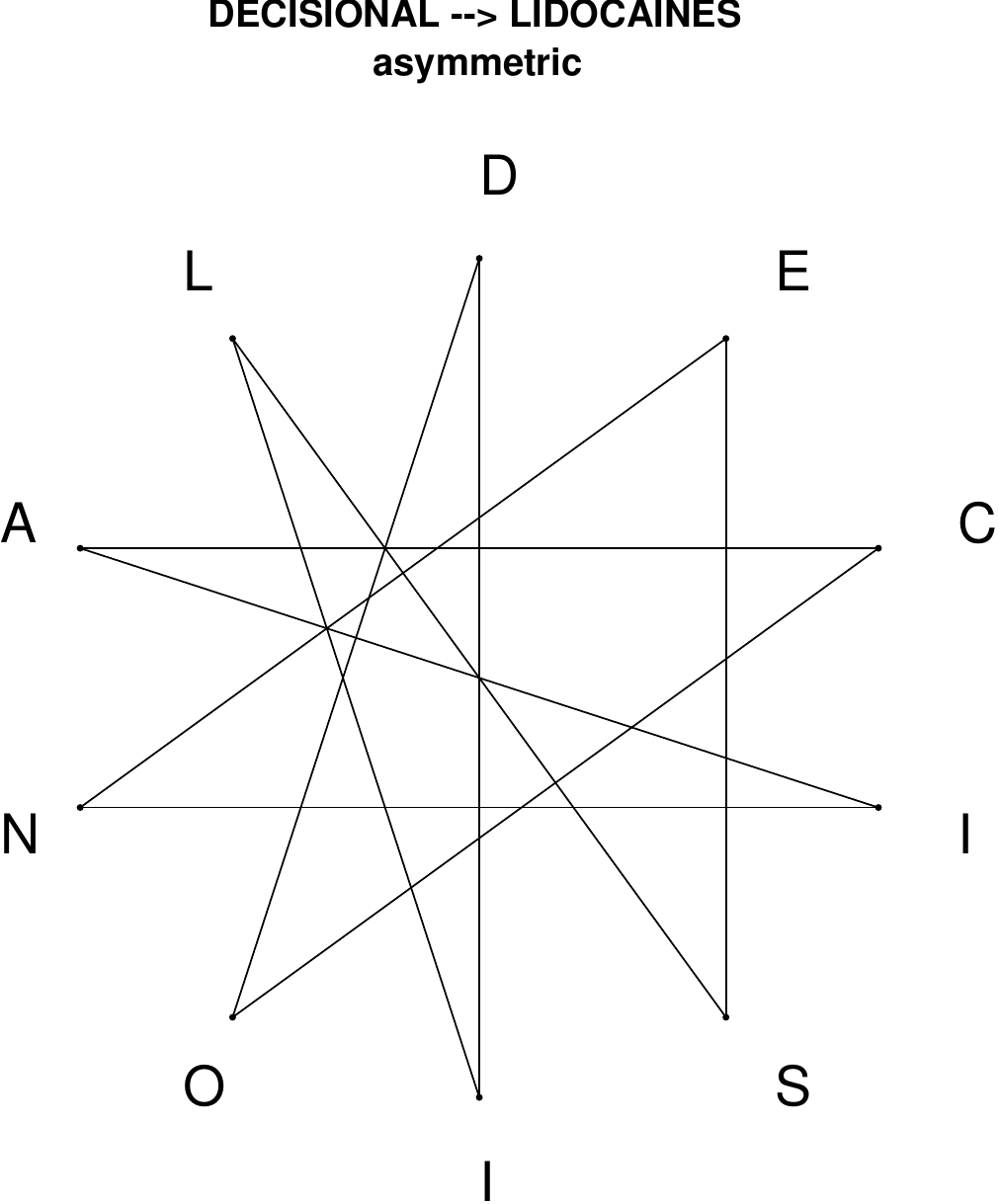}
\end{subfigure}
\hfill
\begin{subfigure}[T]{0.19\textwidth}
\centering
\includegraphics[width=\textwidth]{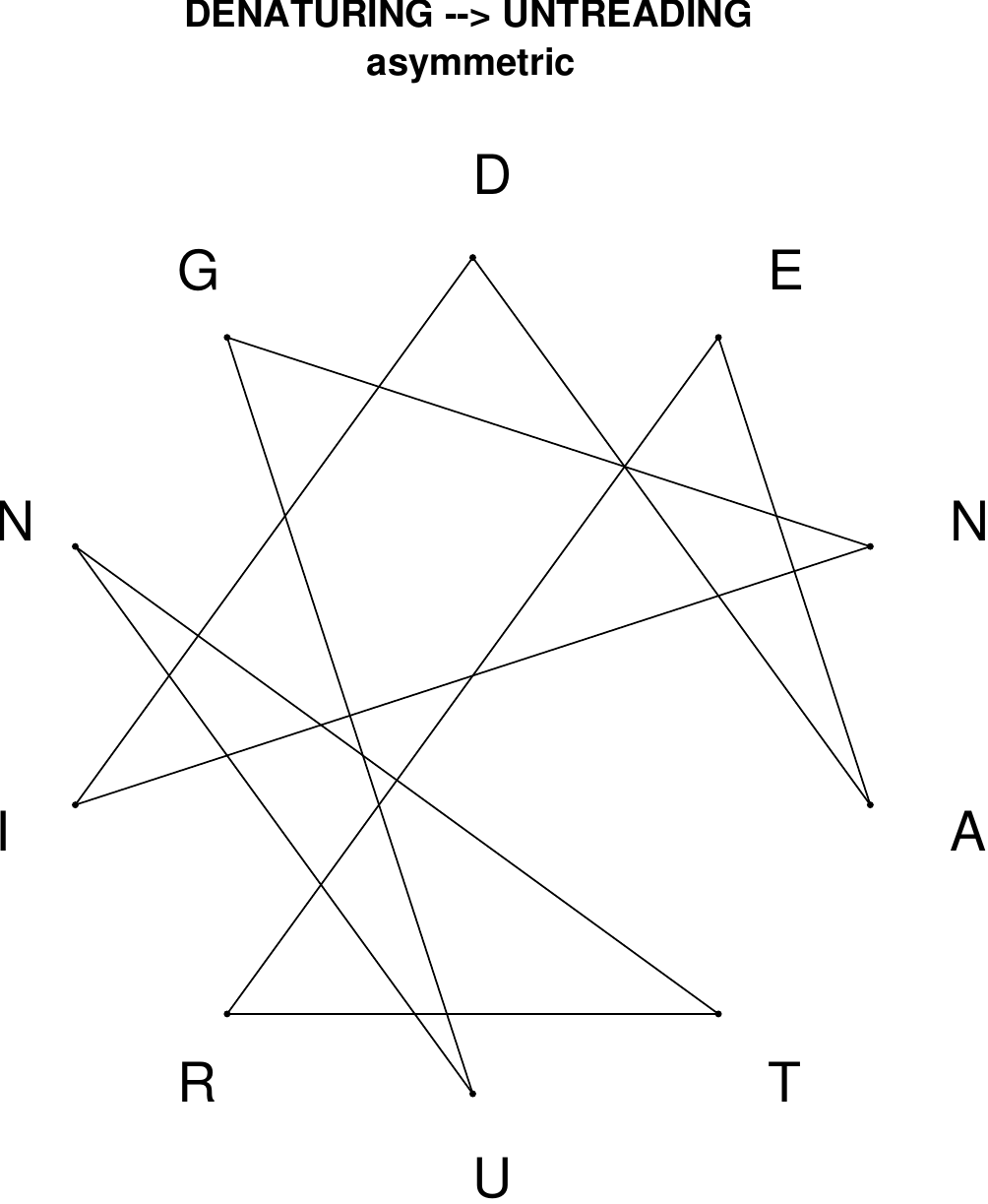}
\end{subfigure}
\hfill
\begin{subfigure}[T]{0.19\textwidth}
\centering
\includegraphics[width=\textwidth]{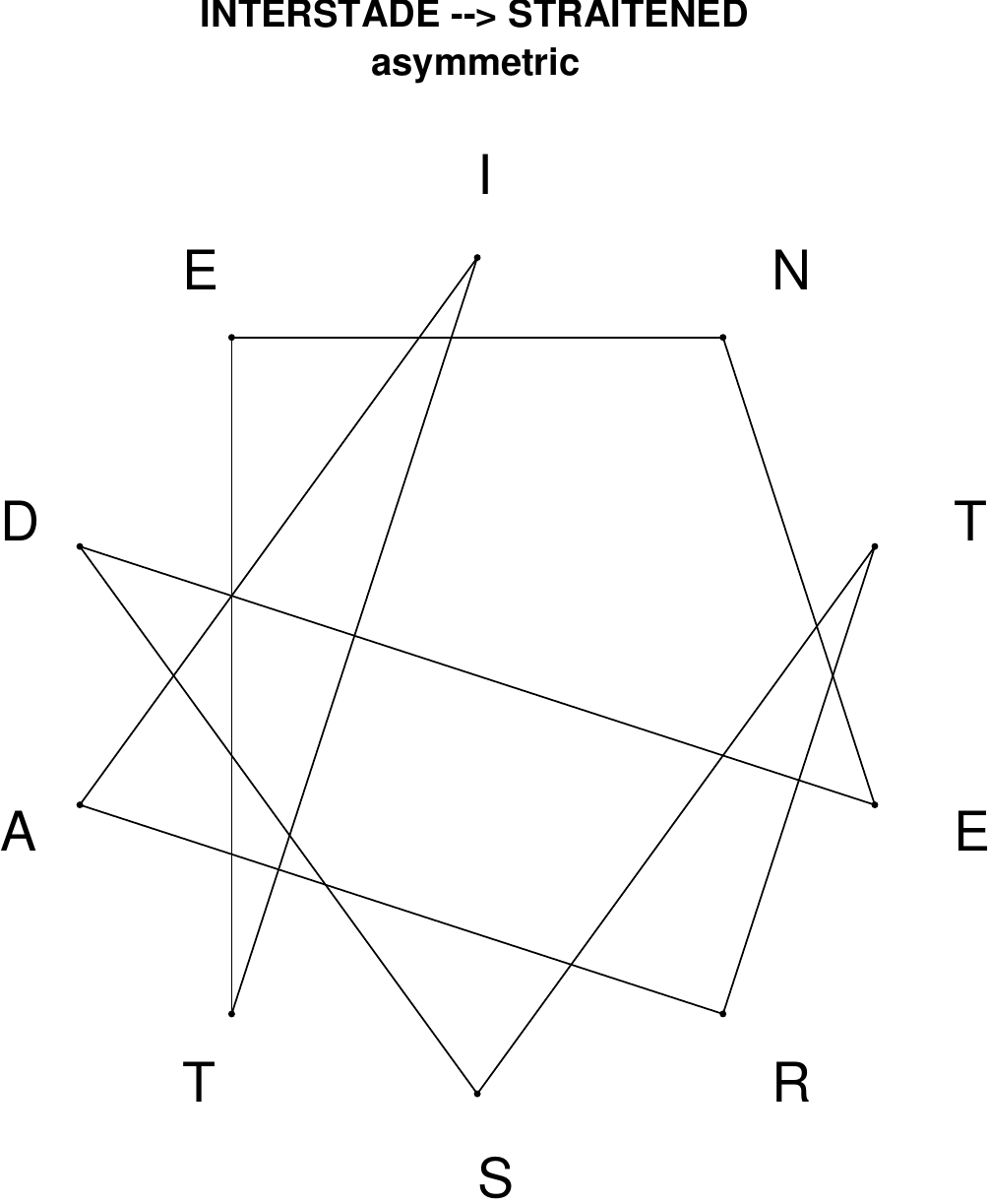}
\end{subfigure}
\hfill
\begin{subfigure}[T]{0.19\textwidth}
\centering
\includegraphics[width=\textwidth]{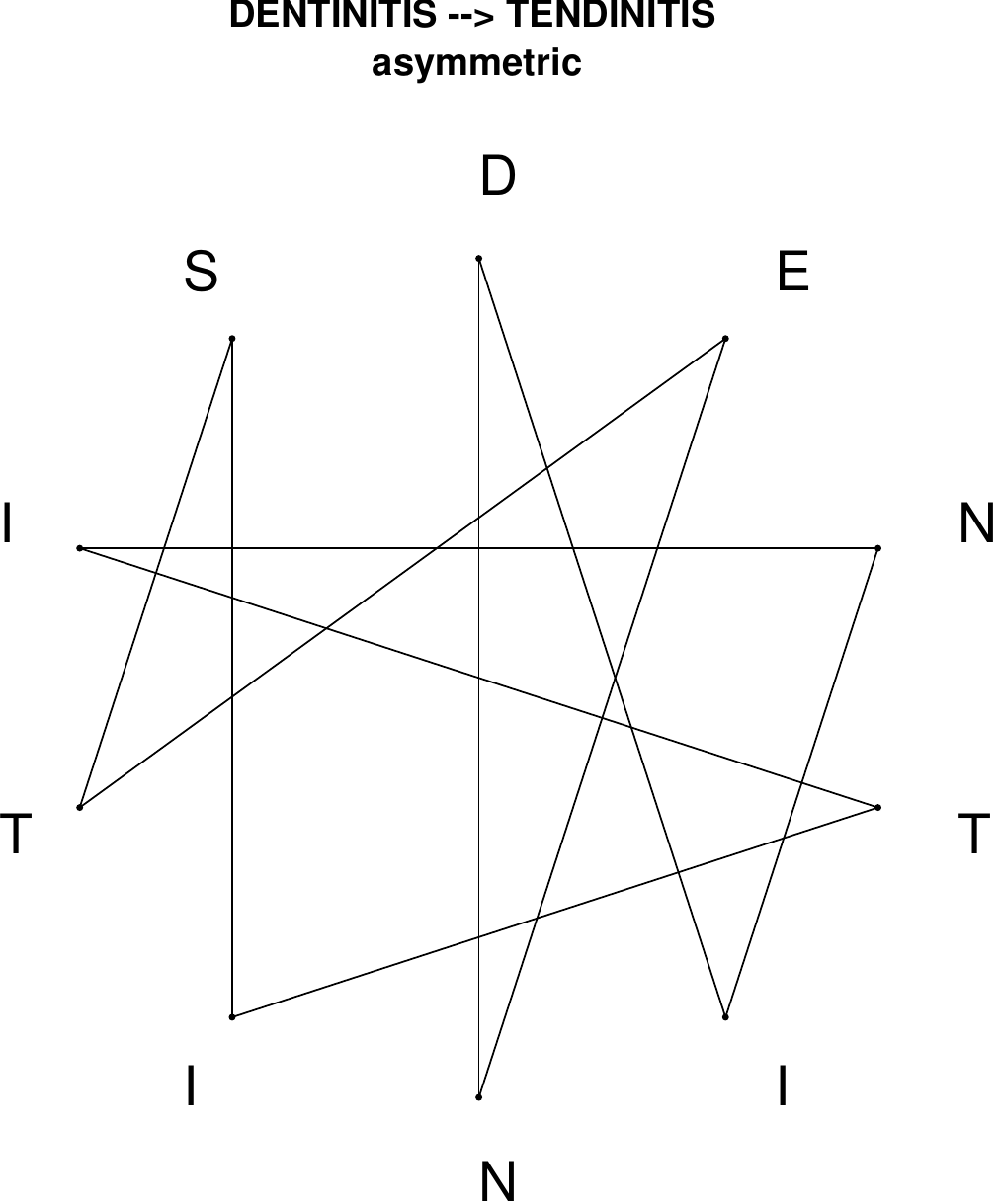}
\end{subfigure}
\hfill
\begin{subfigure}[T]{0.19\textwidth}
\centering
\includegraphics[width=\textwidth]{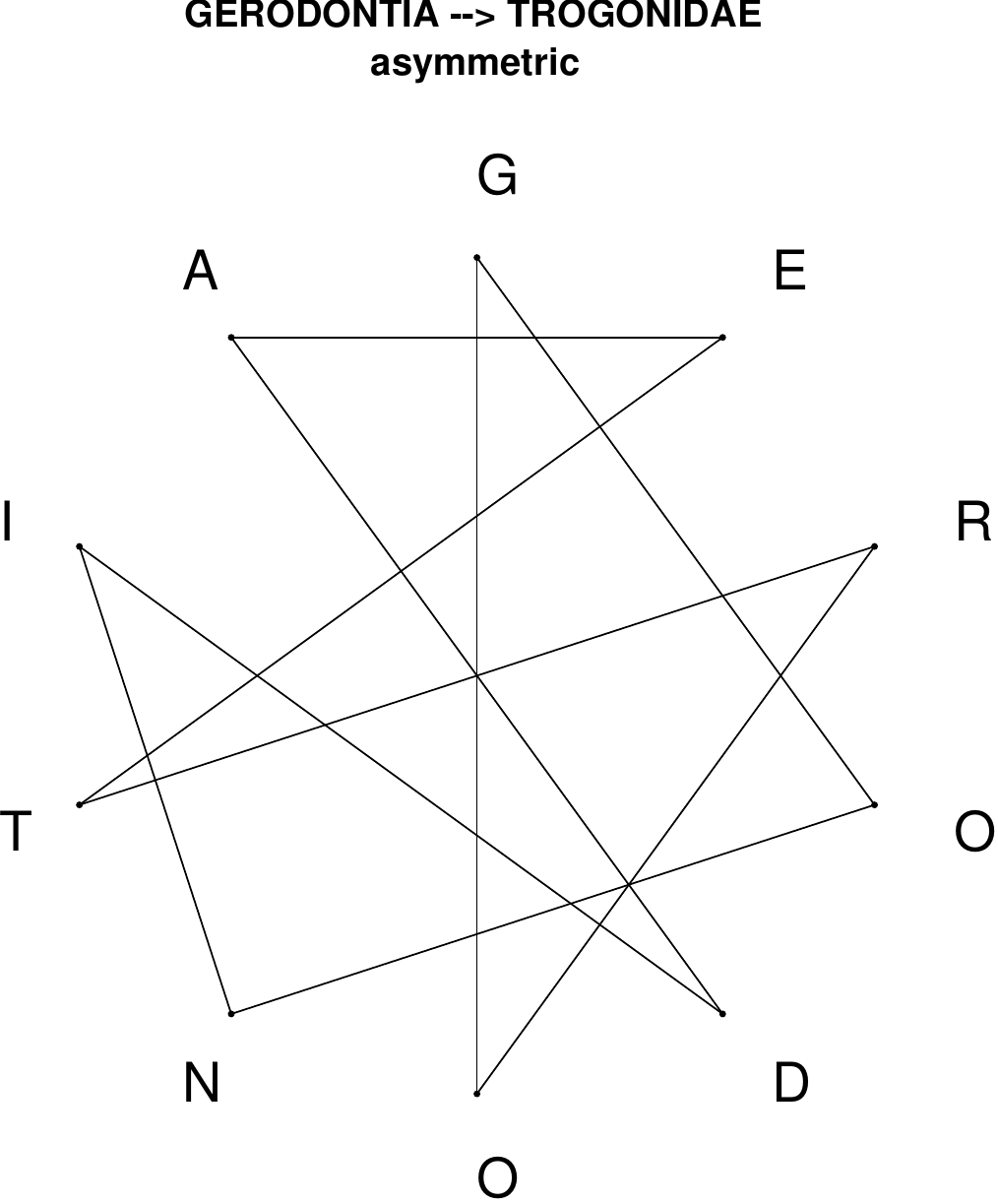}
\end{subfigure}
\end{figure}

\begin{figure}[H]
\centering
\begin{subfigure}[T]{0.19\textwidth}
\centering
\includegraphics[width=\textwidth]{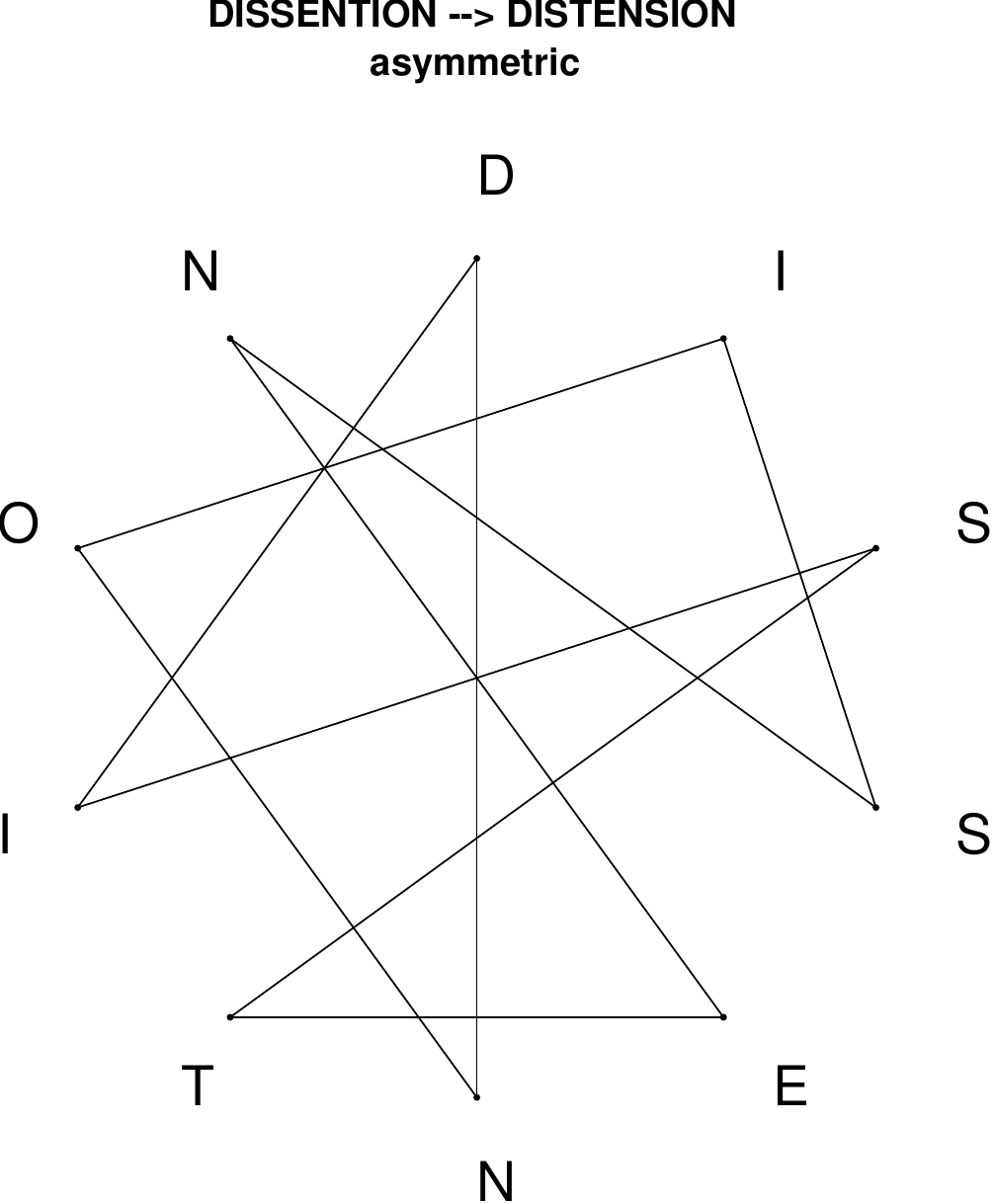}
\end{subfigure}
\hfill
\begin{subfigure}[T]{0.19\textwidth}
\centering
\includegraphics[width=\textwidth]{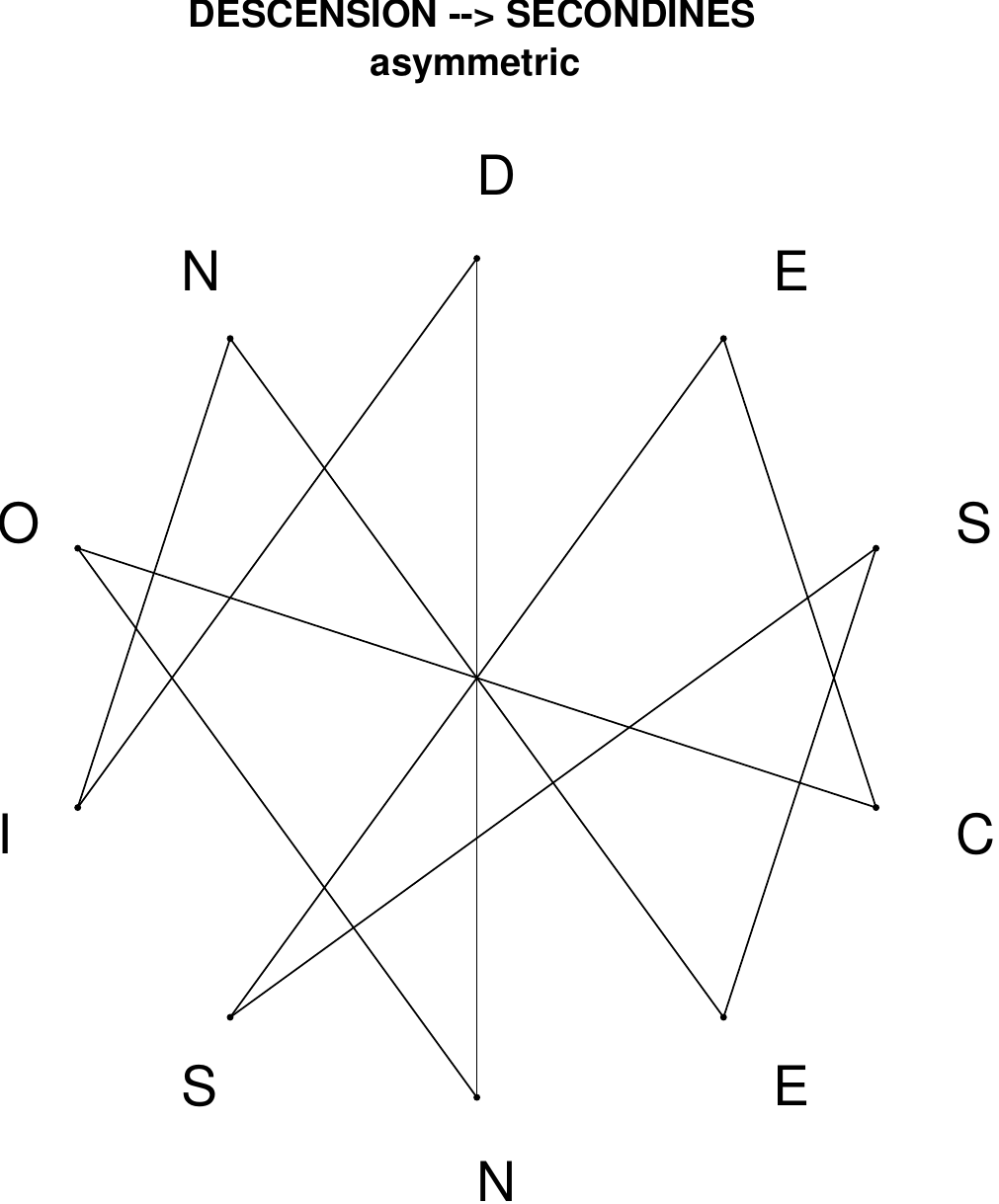}
\end{subfigure}
\hfill
\begin{subfigure}[T]{0.19\textwidth}
\centering
\includegraphics[width=\textwidth]{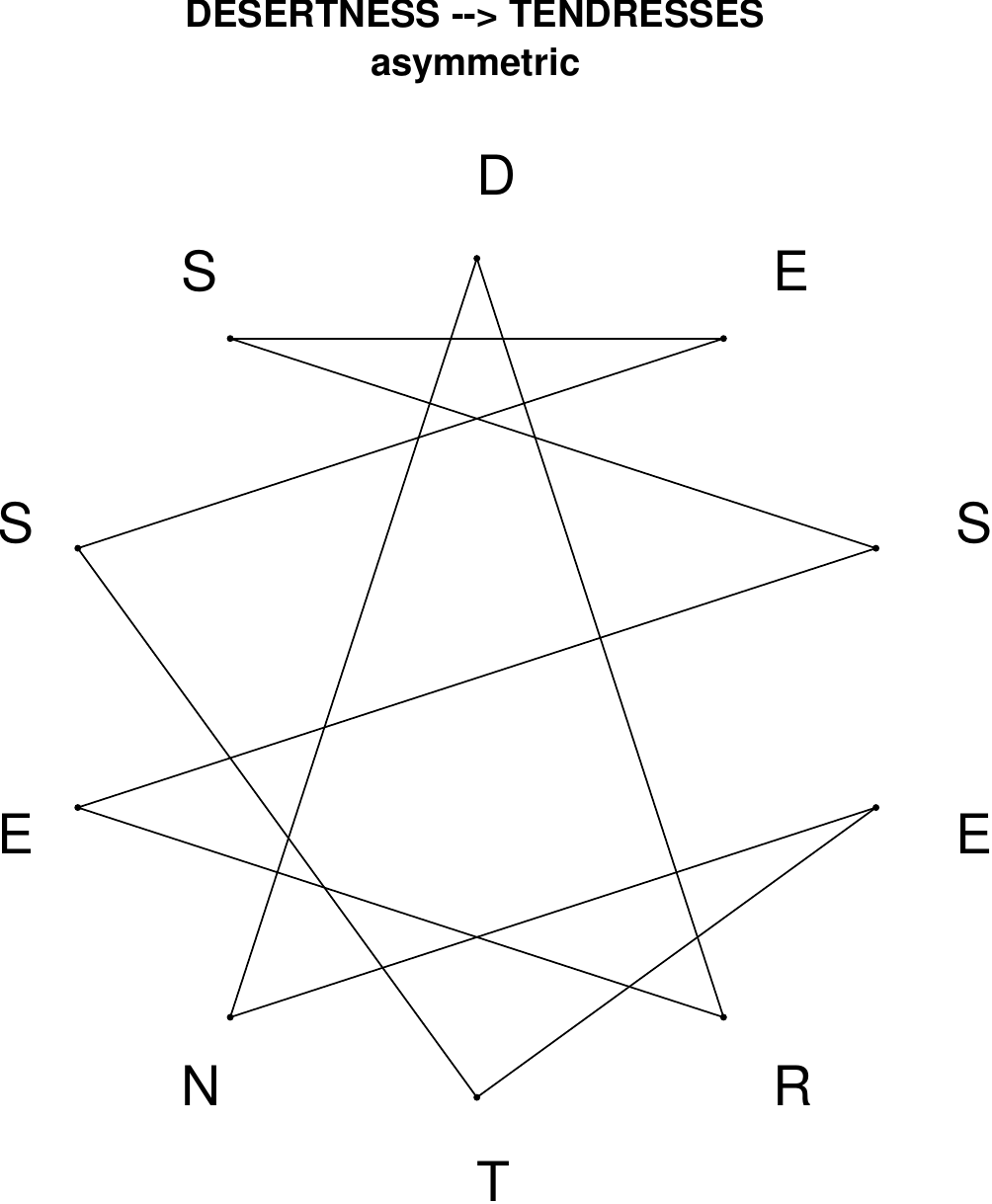}
\end{subfigure}
\hfill
\begin{subfigure}[T]{0.19\textwidth}
\centering
\includegraphics[width=\textwidth]{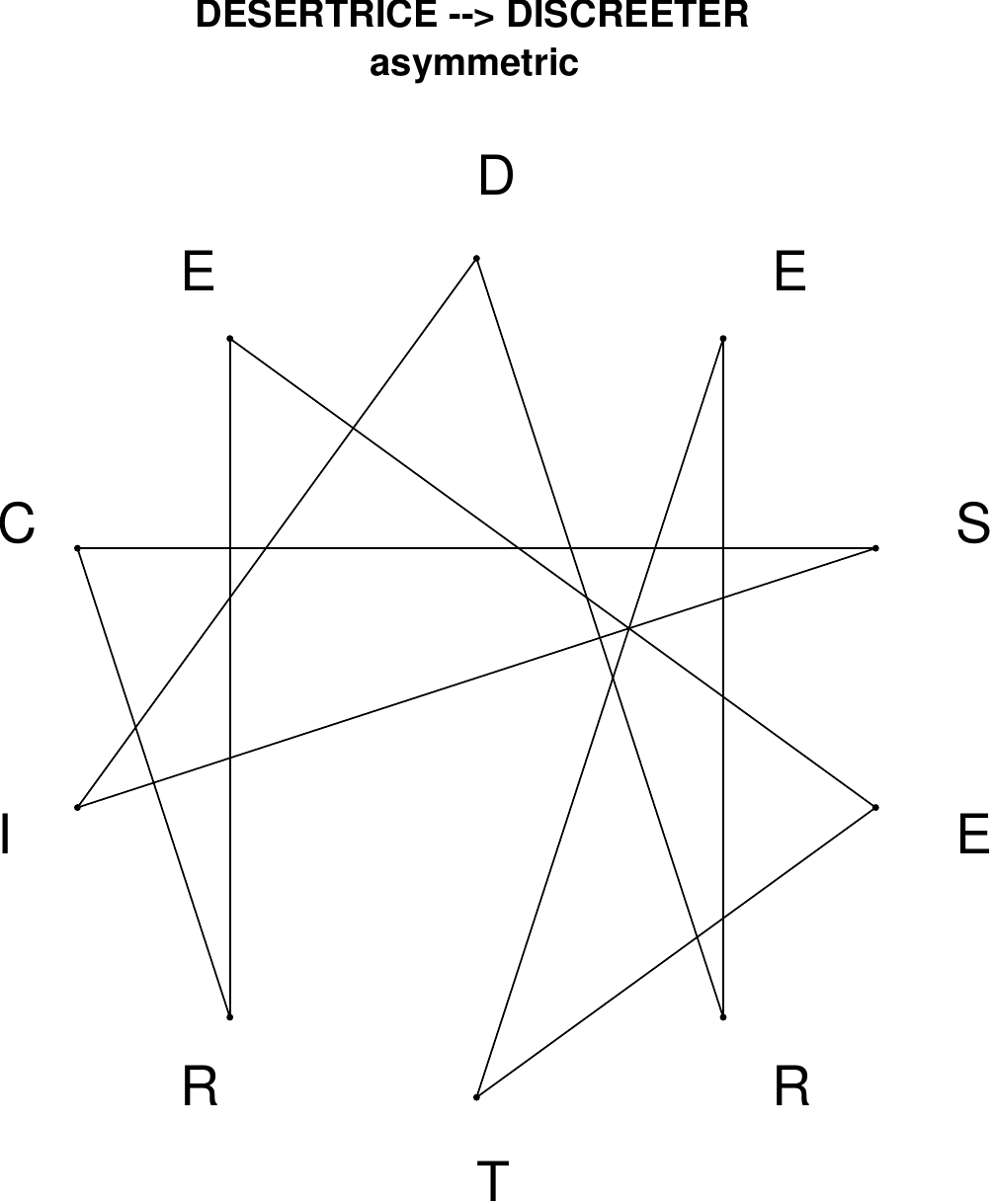}
\end{subfigure}
\hfill
\begin{subfigure}[T]{0.19\textwidth}
\centering
\includegraphics[width=\textwidth]{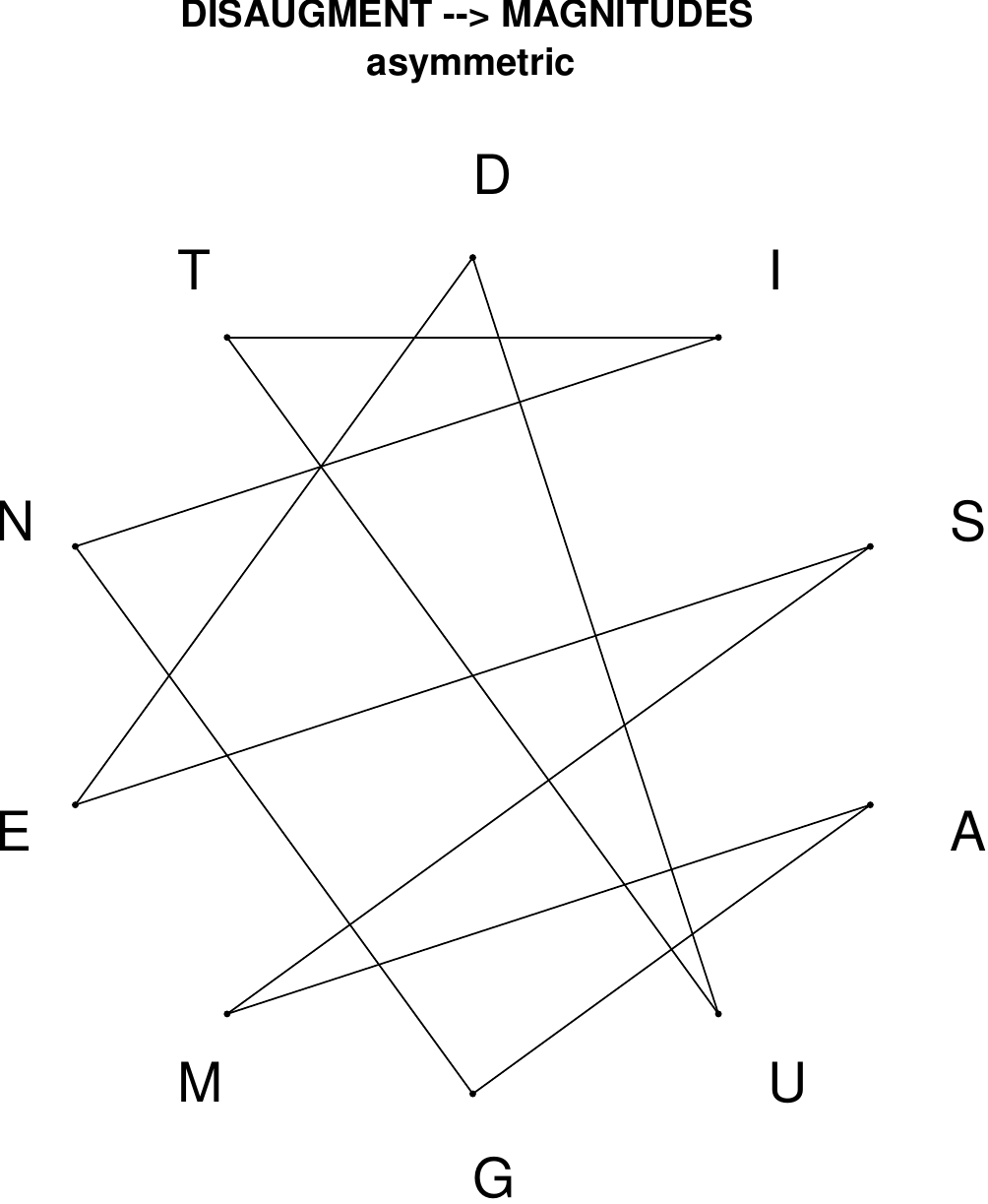}
\end{subfigure}
\end{figure}

\begin{figure}[H]
\centering
\begin{subfigure}[T]{0.19\textwidth}
\centering
\includegraphics[width=\textwidth]{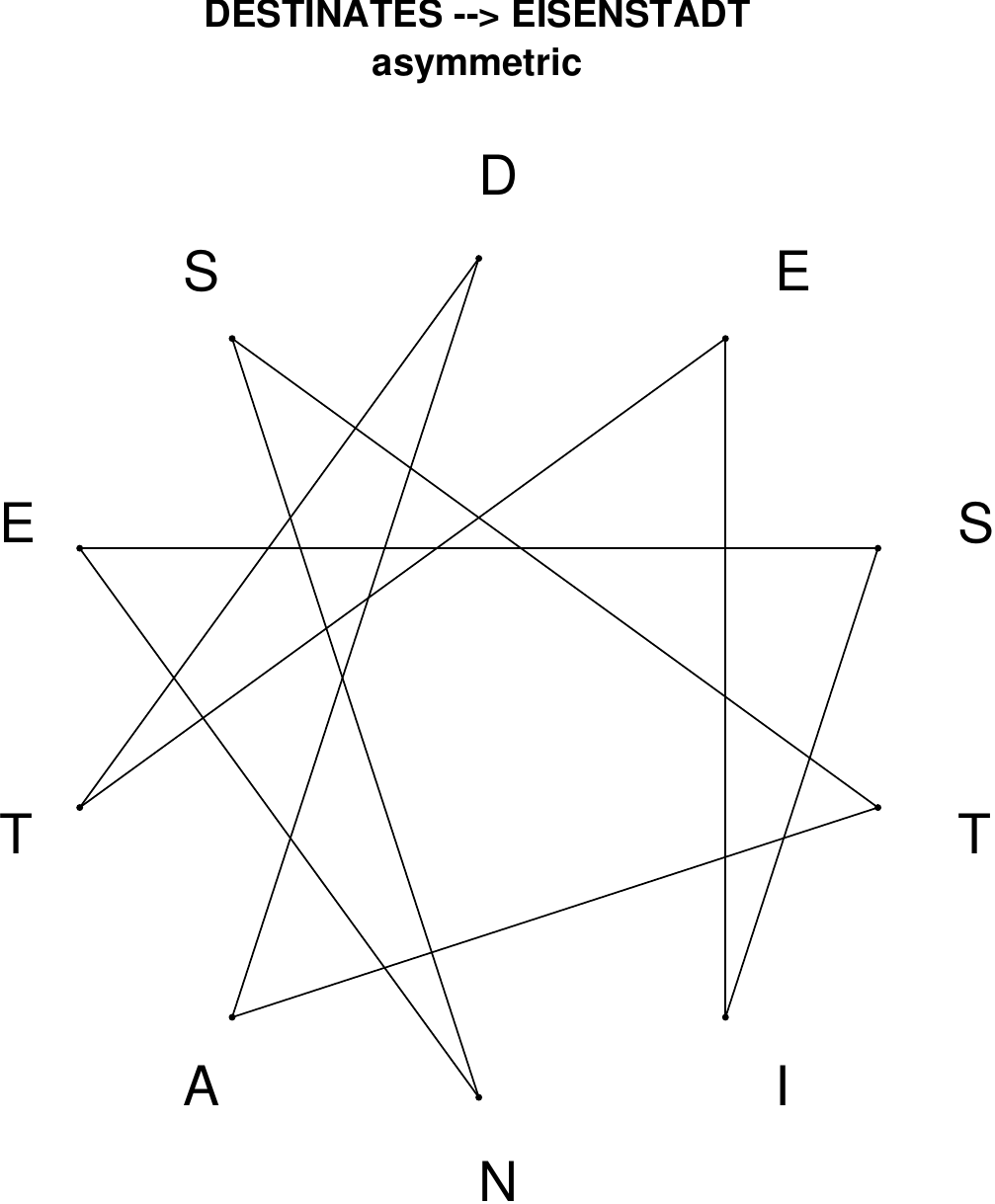}
\end{subfigure}
\hfill
\begin{subfigure}[T]{0.19\textwidth}
\centering
\includegraphics[width=\textwidth]{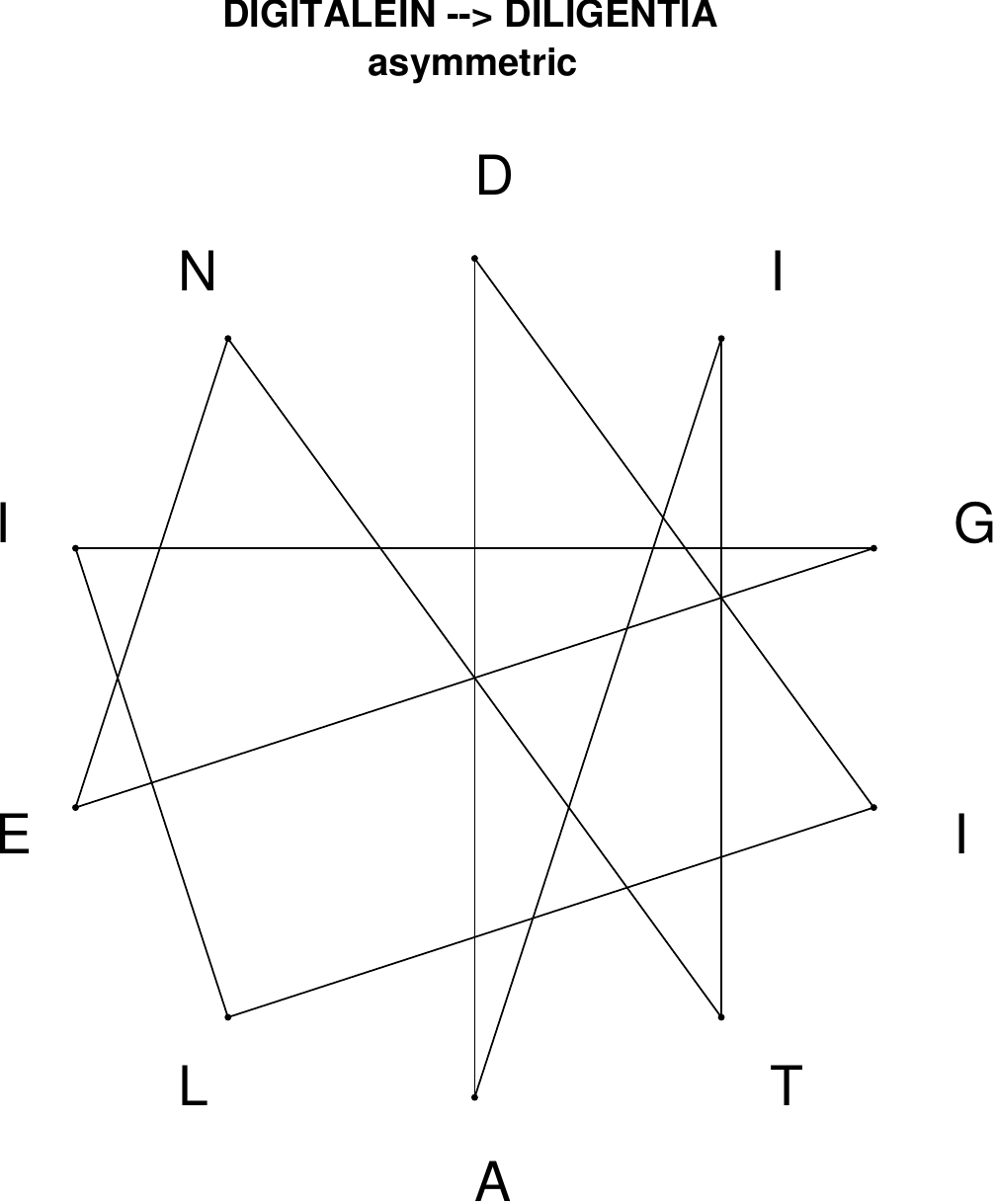}
\end{subfigure}
\hfill
\begin{subfigure}[T]{0.19\textwidth}
\centering
\includegraphics[width=\textwidth]{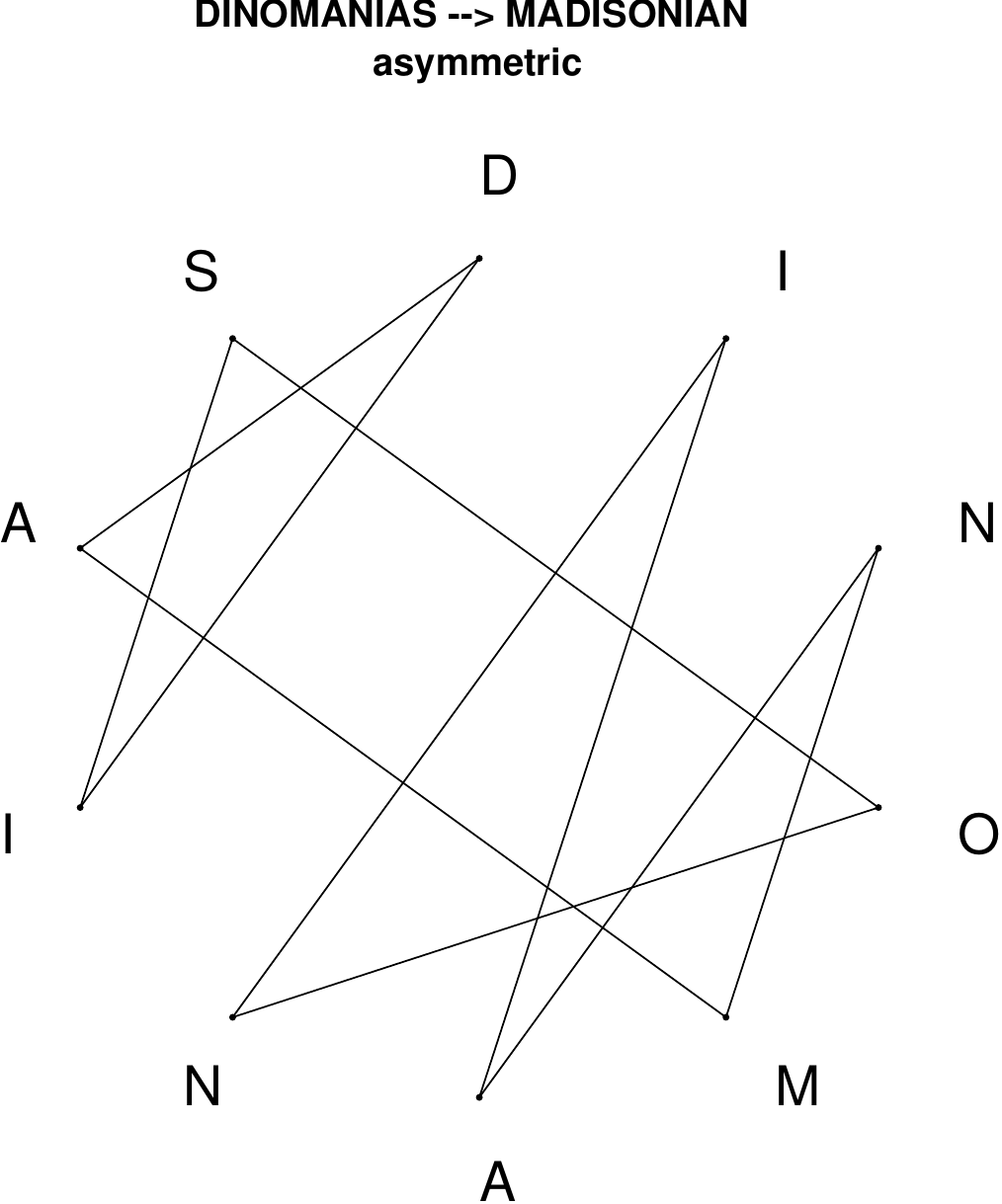}
\end{subfigure}
\hfill
\begin{subfigure}[T]{0.19\textwidth}
\centering
\includegraphics[width=\textwidth]{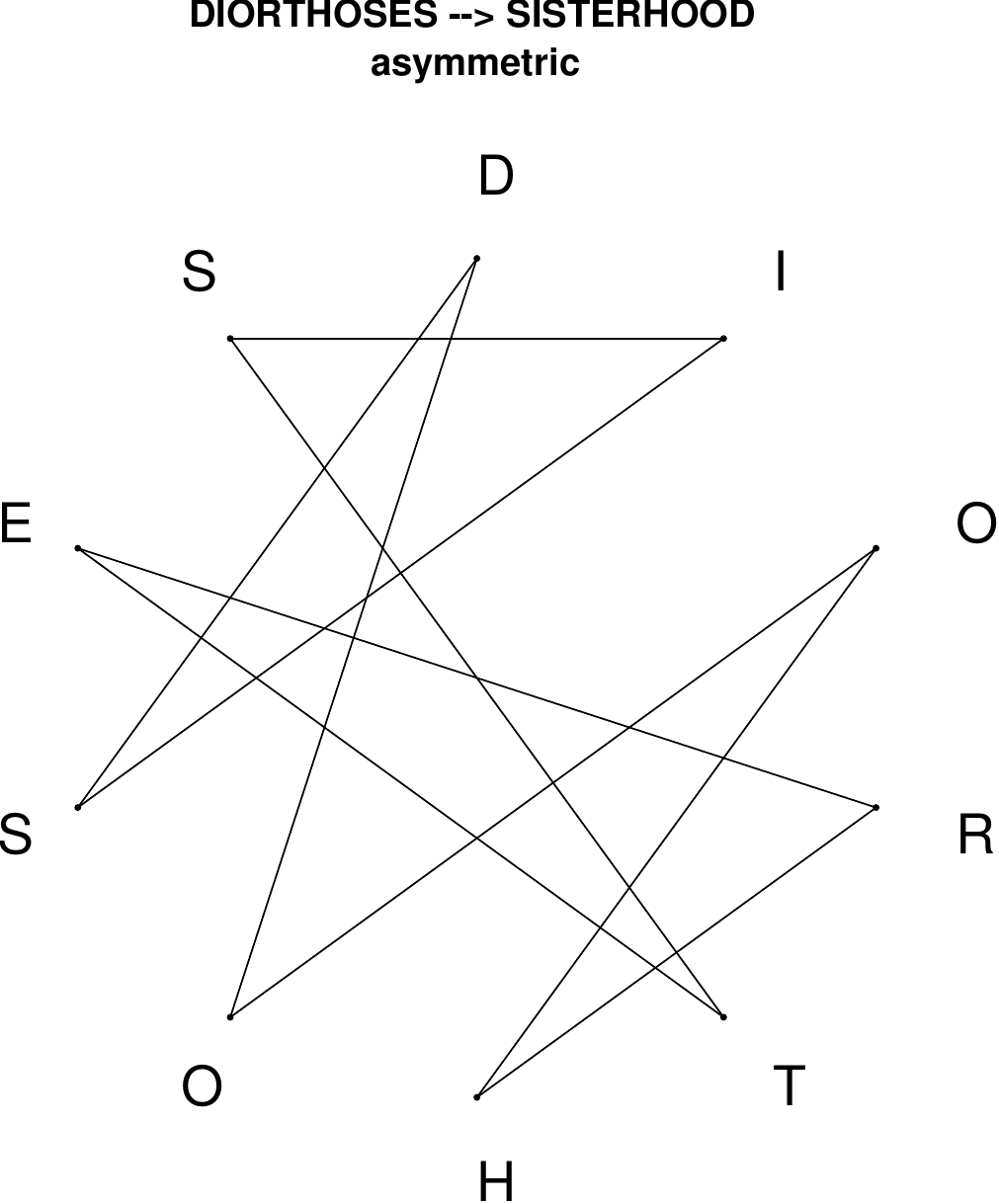}
\end{subfigure}
\hfill
\begin{subfigure}[T]{0.19\textwidth}
\centering
\includegraphics[width=\textwidth]{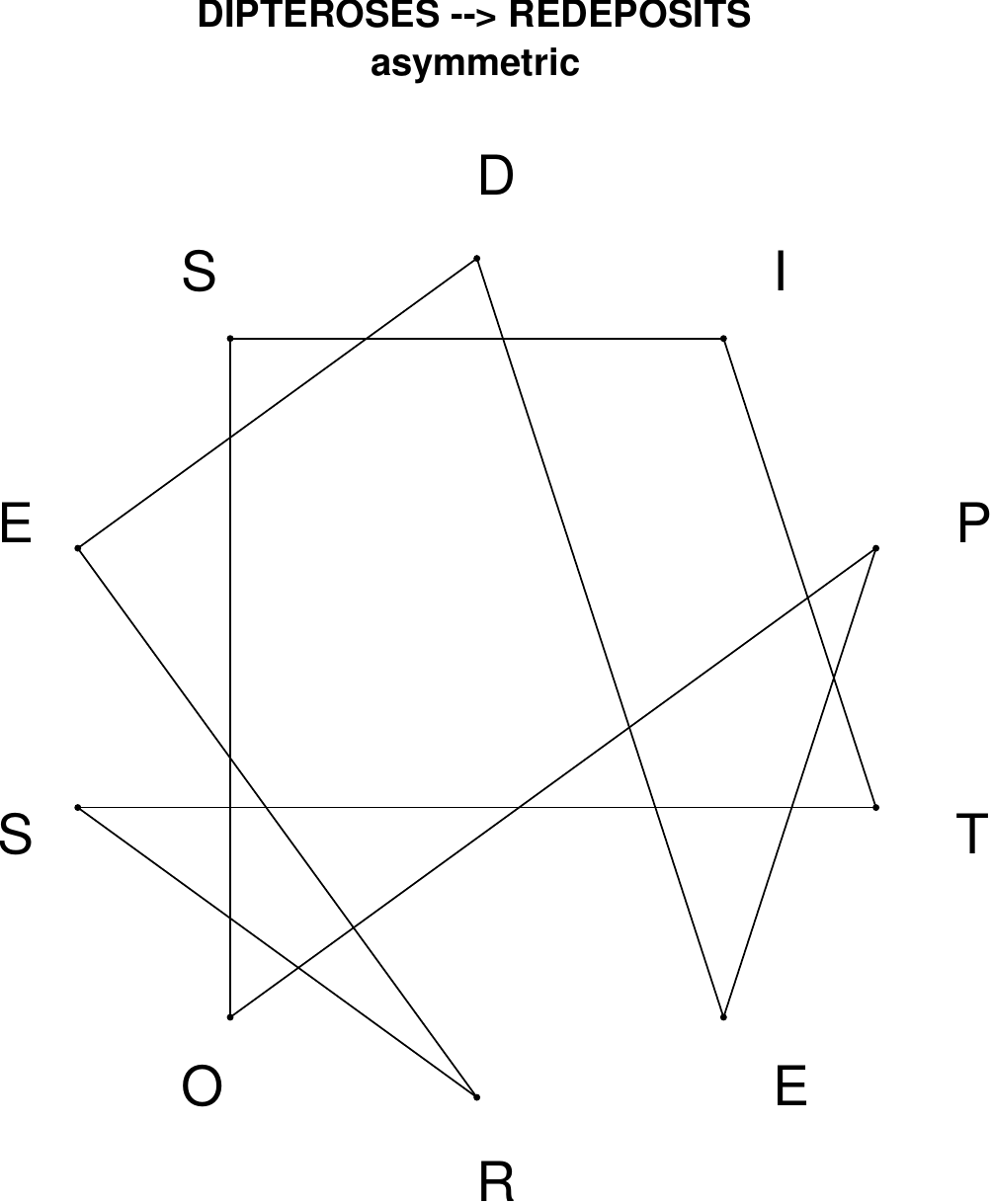}
\end{subfigure}
\end{figure}

\begin{figure}[H]
\centering
\begin{subfigure}[T]{0.19\textwidth}
\centering
\includegraphics[width=\textwidth]{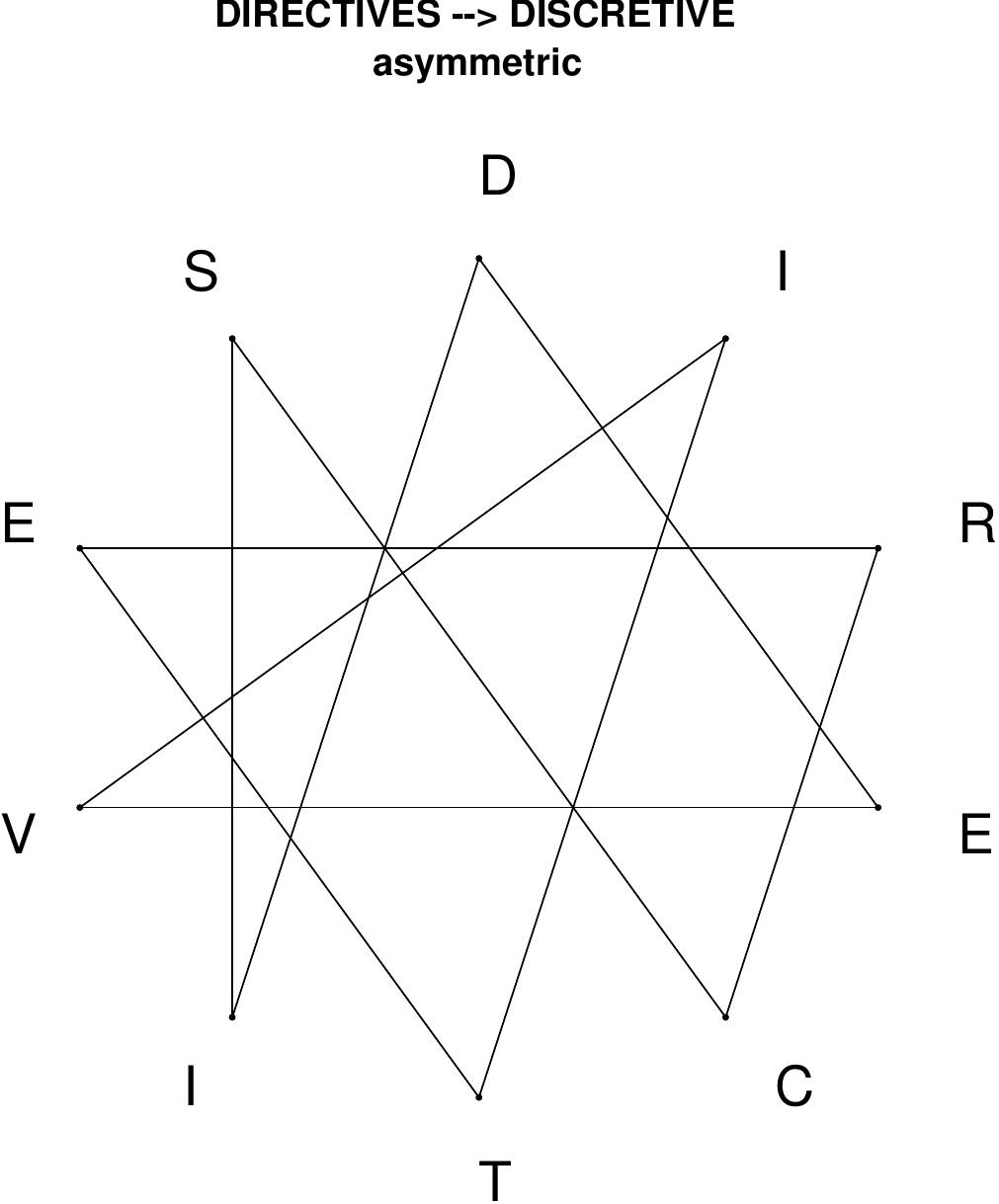}
\end{subfigure}
\hfill
\begin{subfigure}[T]{0.19\textwidth}
\centering
\includegraphics[width=\textwidth]{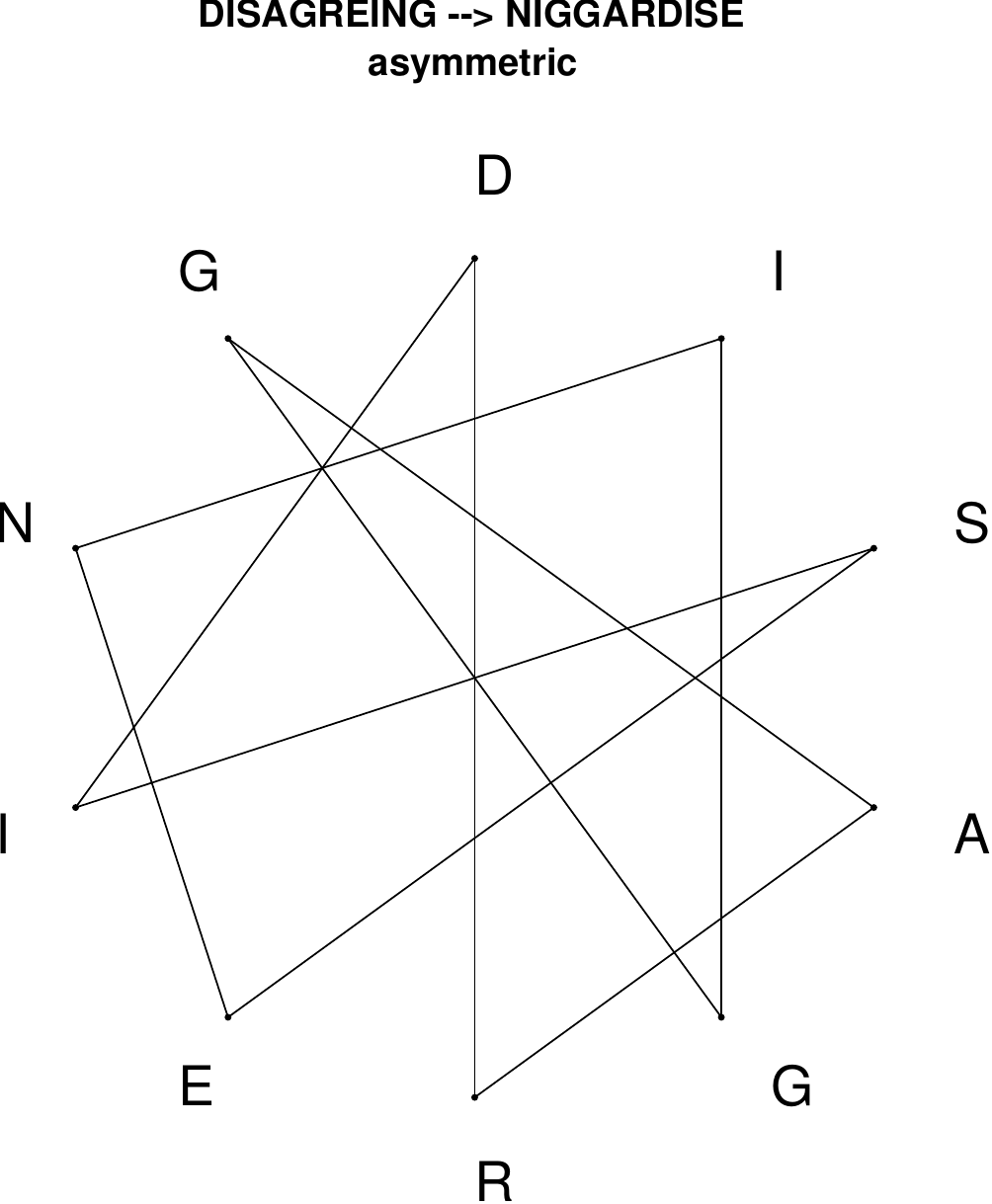}
\end{subfigure}
\hfill
\begin{subfigure}[T]{0.19\textwidth}
\centering
\includegraphics[width=\textwidth]{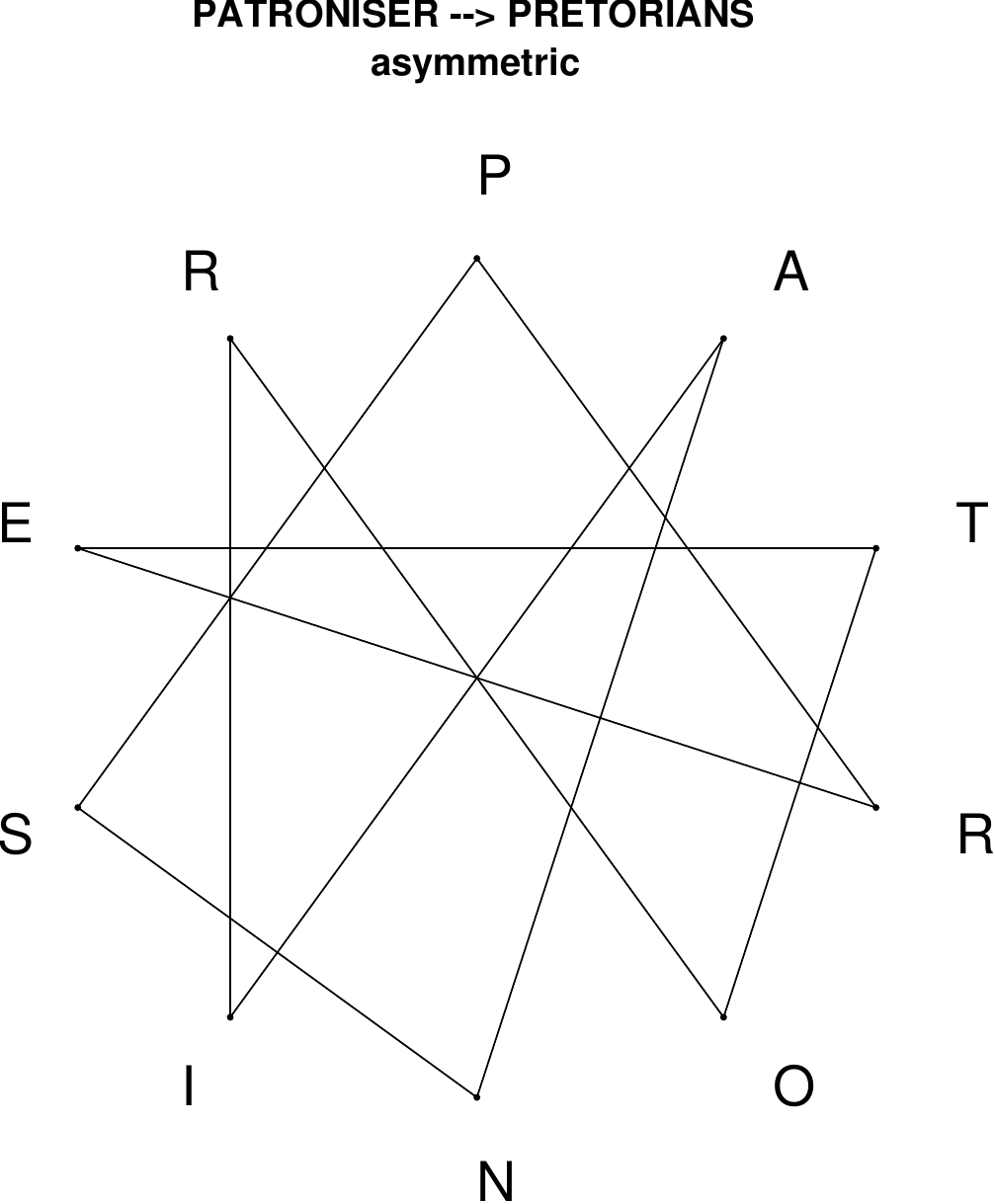}
\end{subfigure}
\hfill
\begin{subfigure}[T]{0.19\textwidth}
\centering
\includegraphics[width=\textwidth]{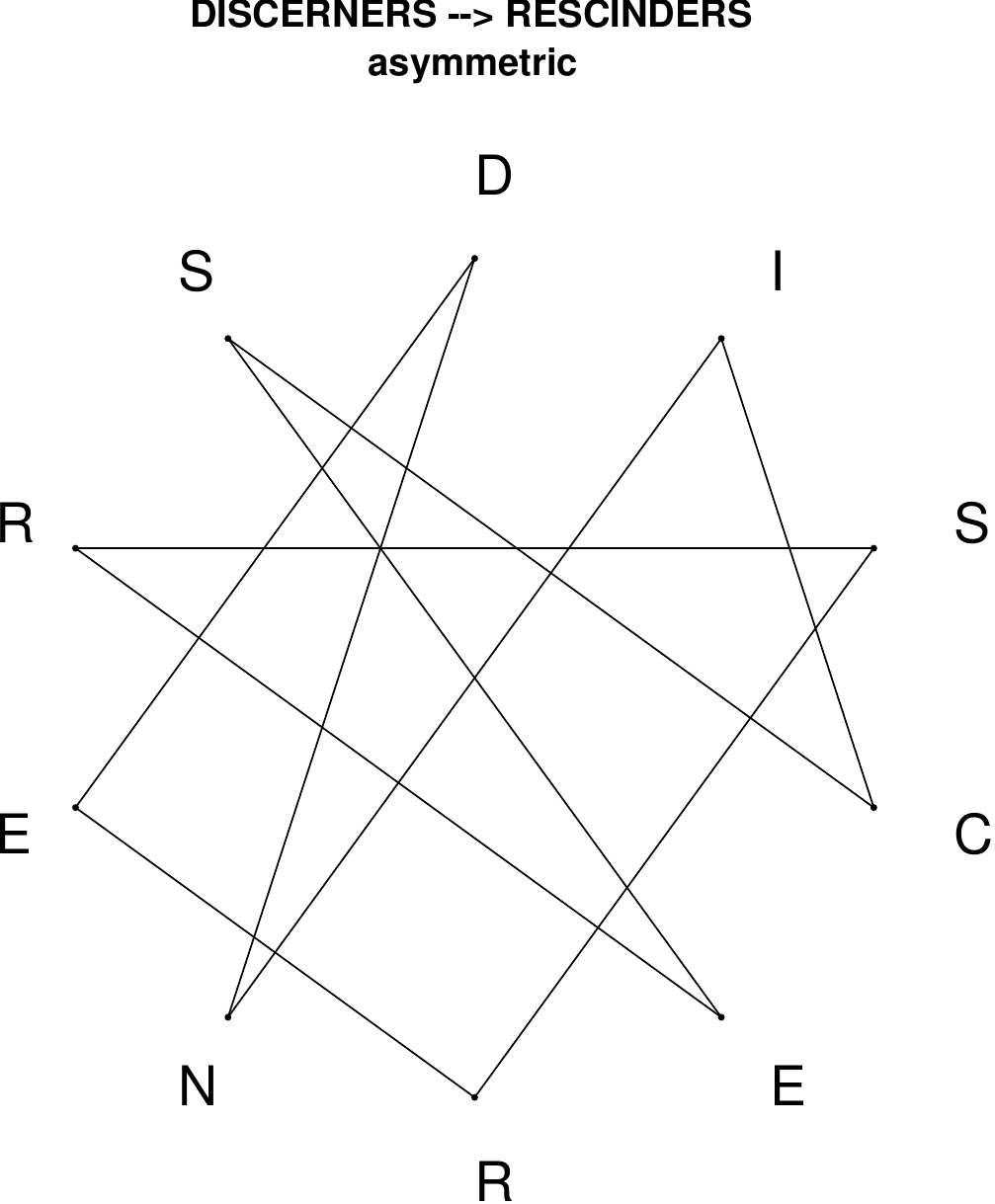}
\end{subfigure}
\hfill
\begin{subfigure}[T]{0.19\textwidth}
\centering
\includegraphics[width=\textwidth]{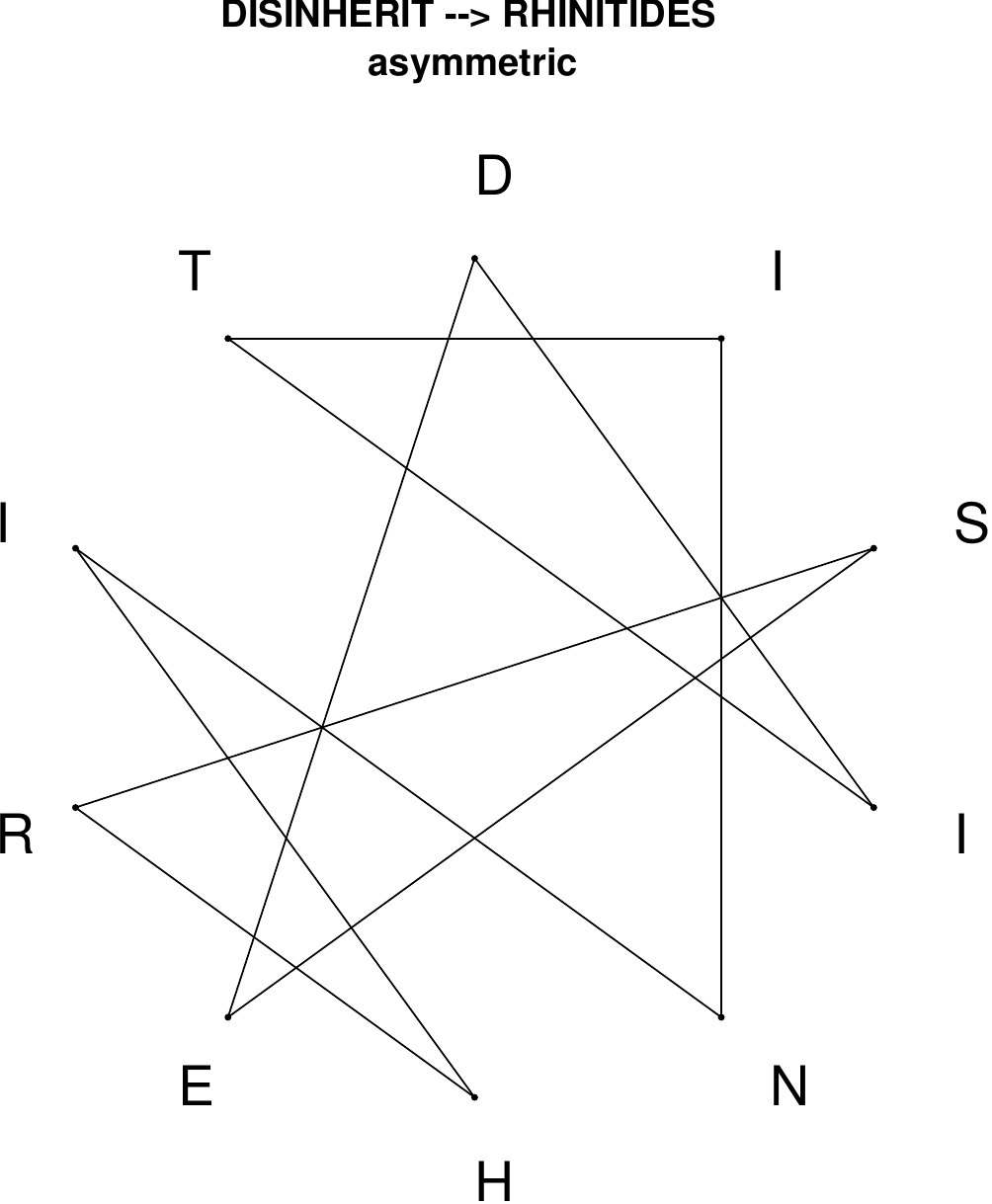}
\end{subfigure}
\end{figure}

\begin{figure}[H]
\centering
\begin{subfigure}[T]{0.19\textwidth}
\centering
\includegraphics[width=\textwidth]{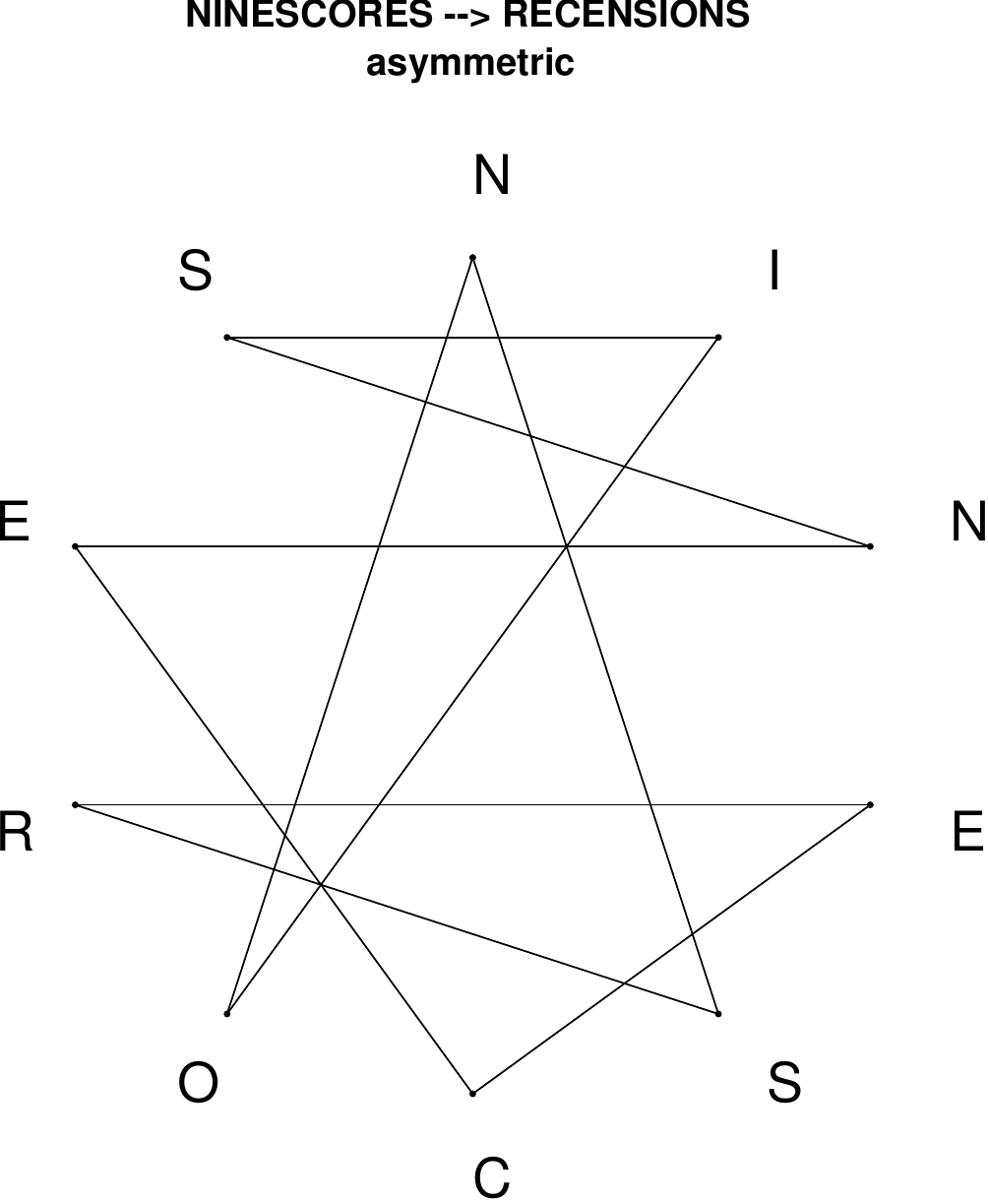}
\end{subfigure}
\hfill
\begin{subfigure}[T]{0.19\textwidth}
\centering
\includegraphics[width=\textwidth]{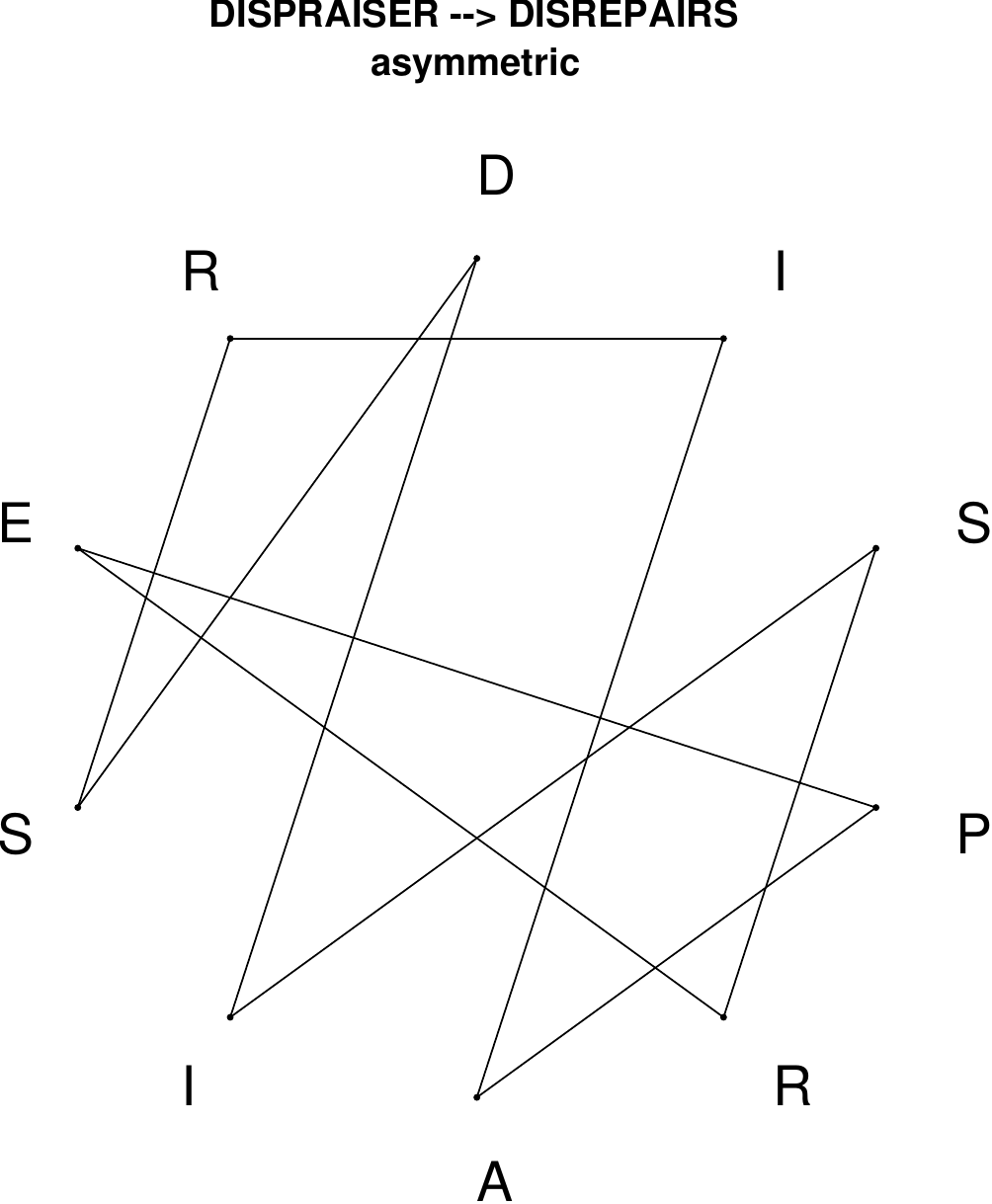}
\end{subfigure}
\hfill
\begin{subfigure}[T]{0.19\textwidth}
\centering
\includegraphics[width=\textwidth]{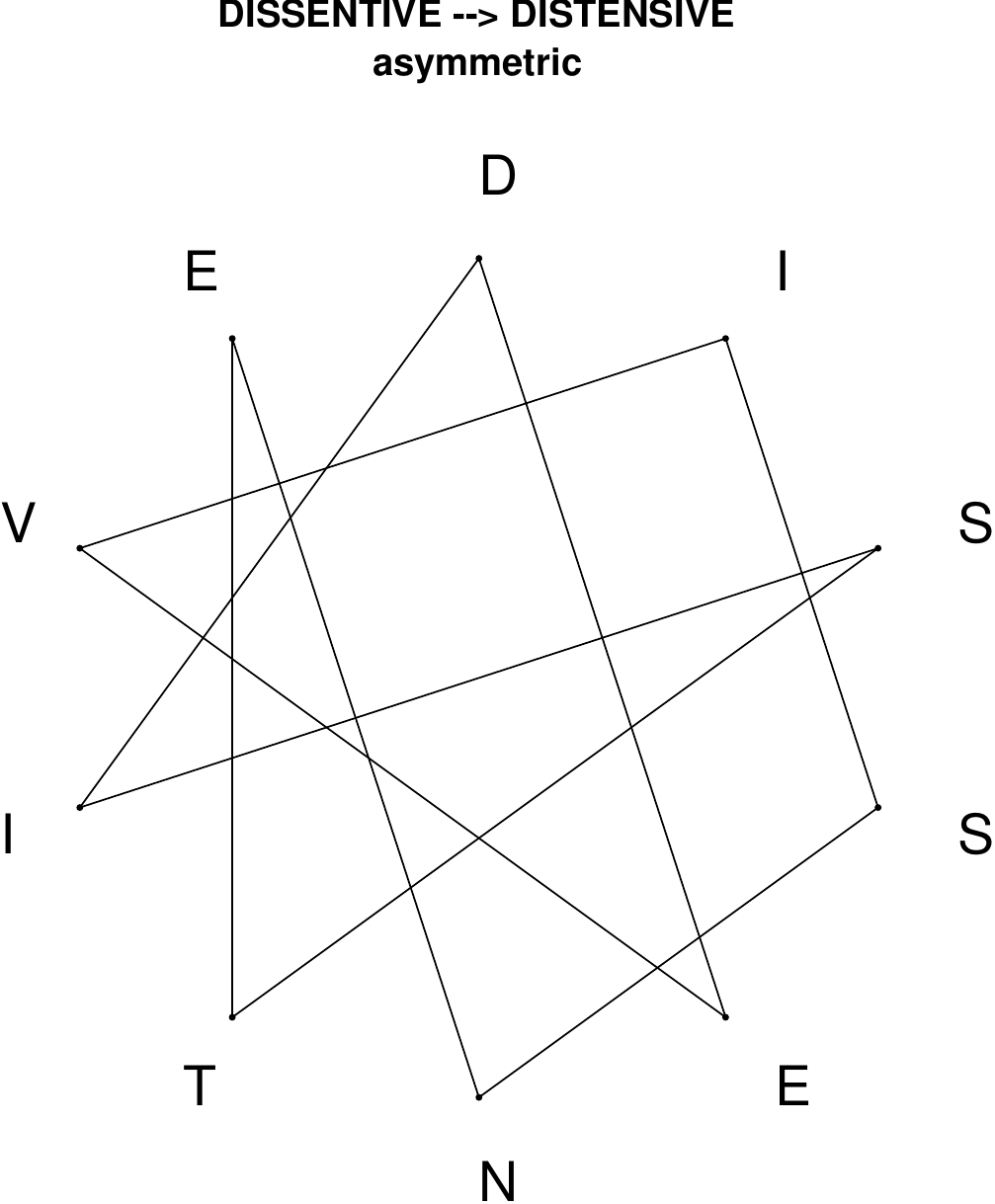}
\end{subfigure}
\hfill
\begin{subfigure}[T]{0.19\textwidth}
\centering
\includegraphics[width=\textwidth]{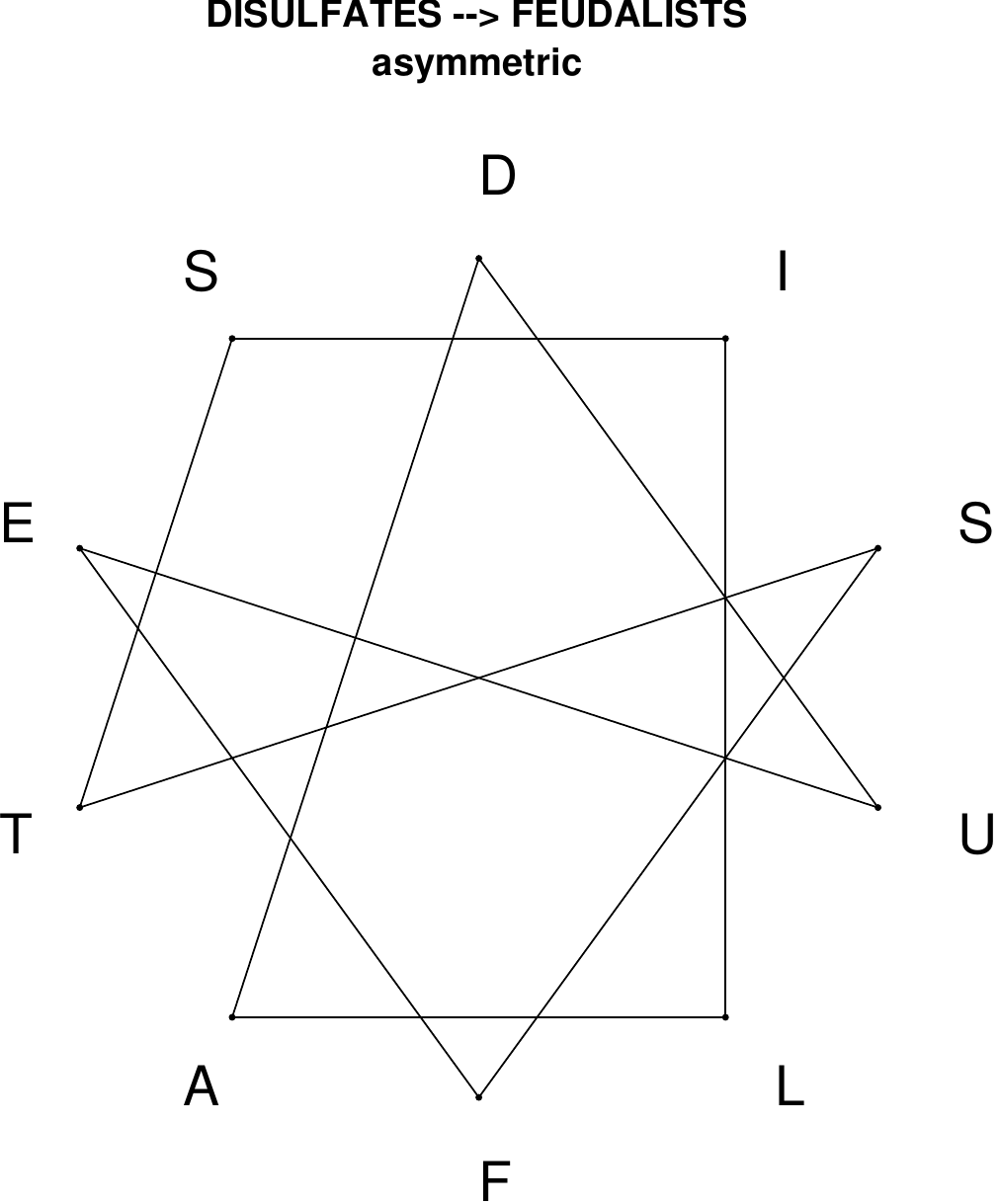}
\end{subfigure}
\hfill
\begin{subfigure}[T]{0.19\textwidth}
\centering
\includegraphics[width=\textwidth]{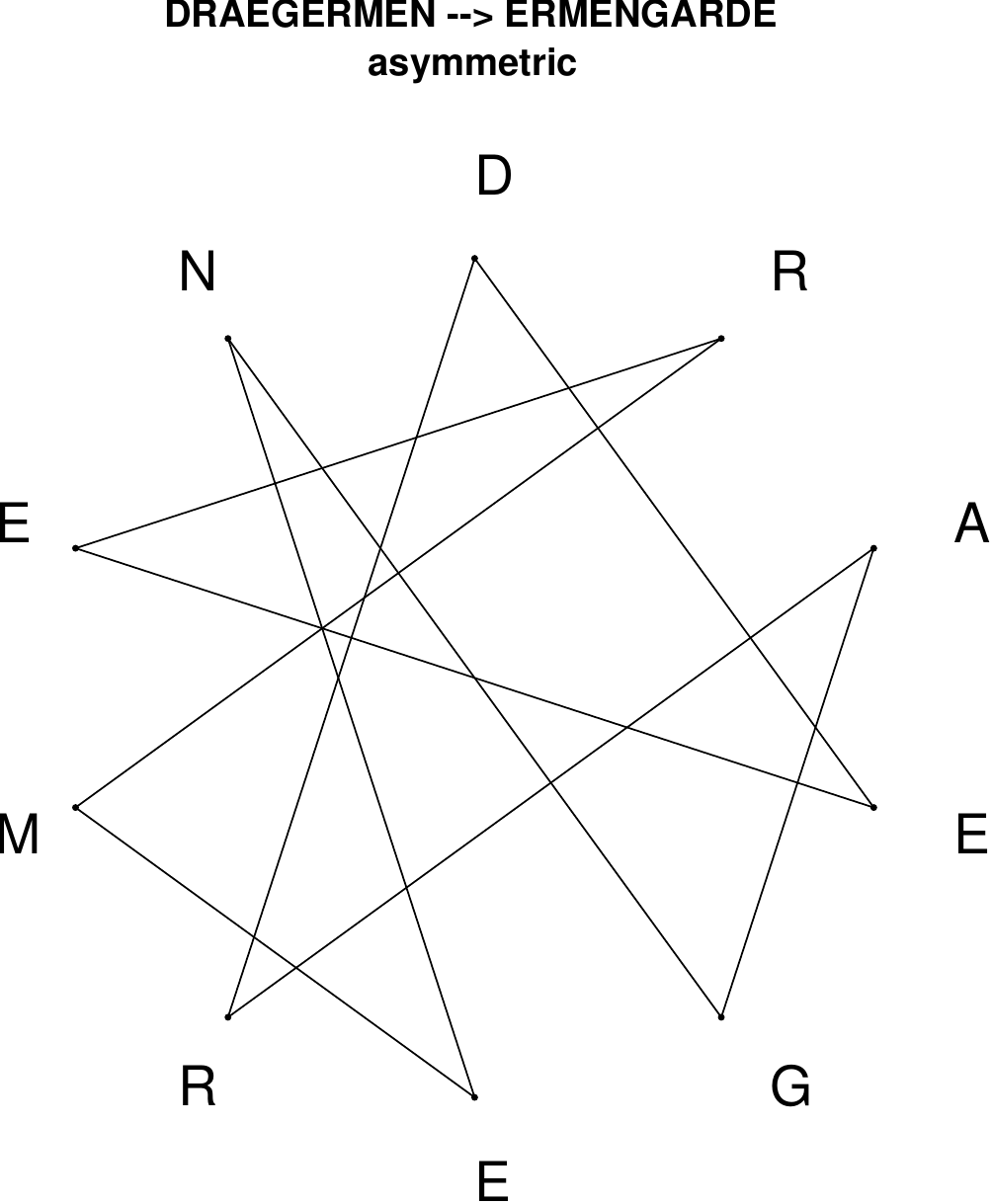}
\end{subfigure}
\end{figure}

\begin{figure}[H]
\centering
\begin{subfigure}[T]{0.19\textwidth}
\centering
\includegraphics[width=\textwidth]{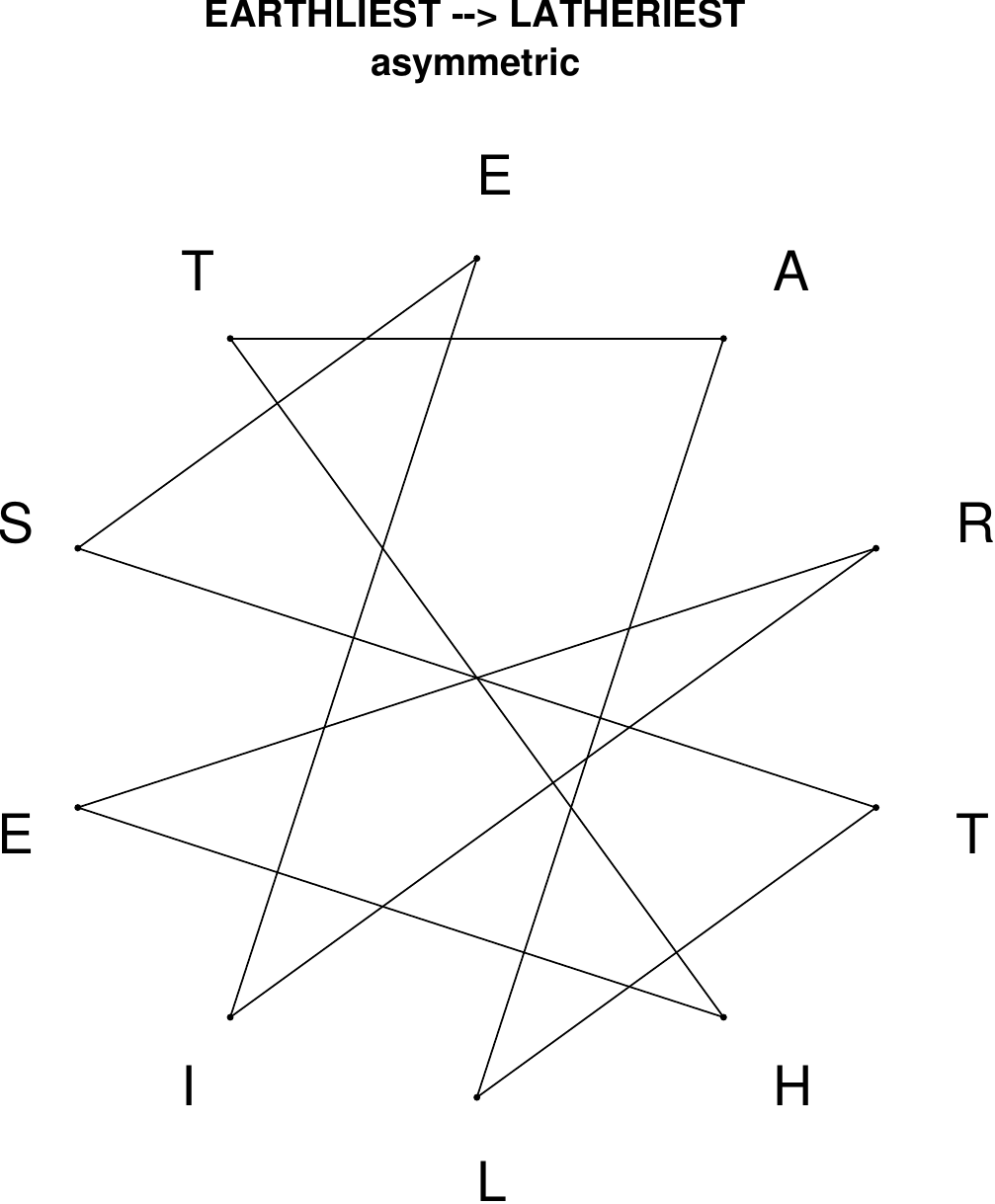}
\end{subfigure}
\hfill
\begin{subfigure}[T]{0.19\textwidth}
\centering
\includegraphics[width=\textwidth]{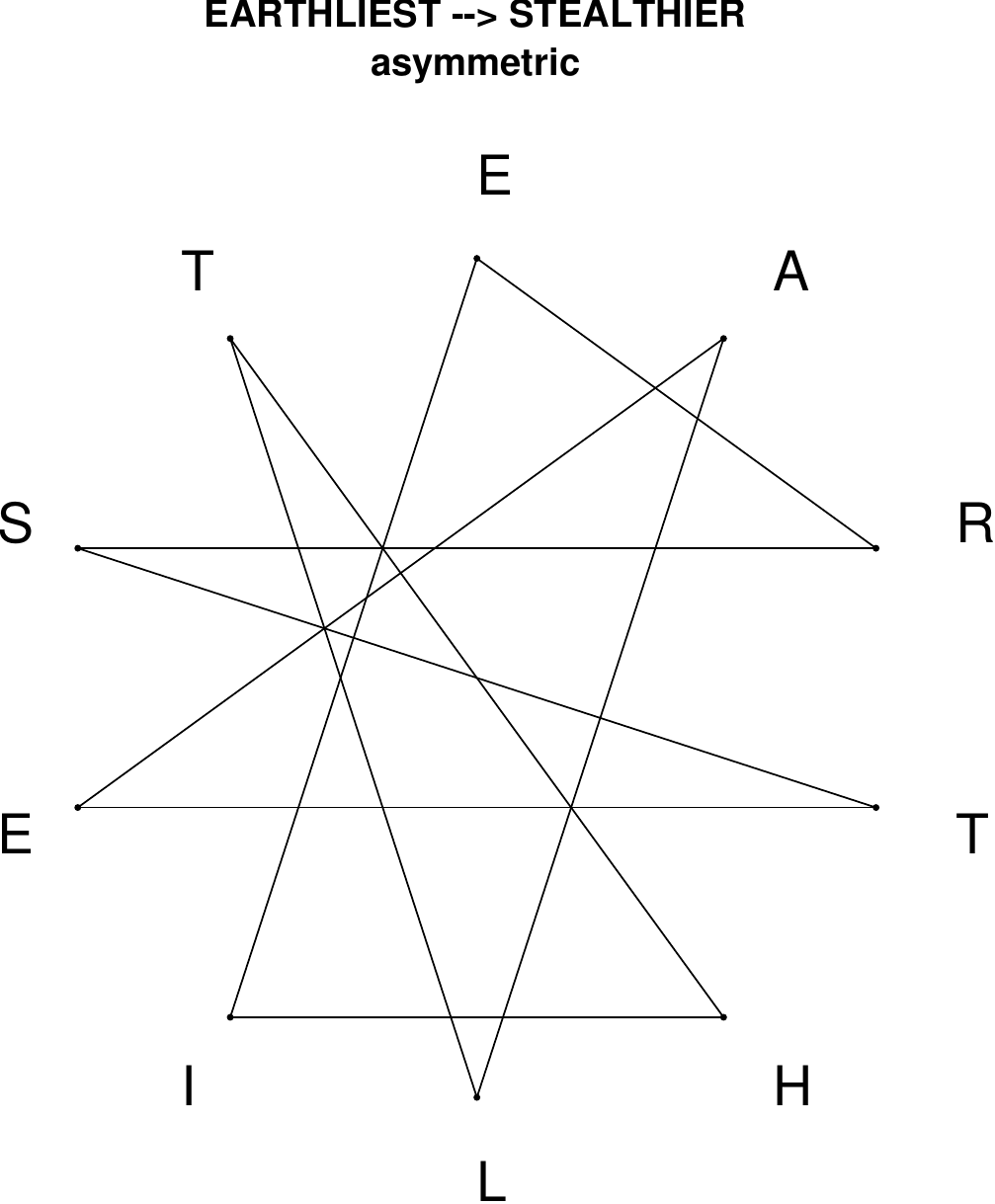}
\end{subfigure}
\hfill
\begin{subfigure}[T]{0.19\textwidth}
\centering
\includegraphics[width=\textwidth]{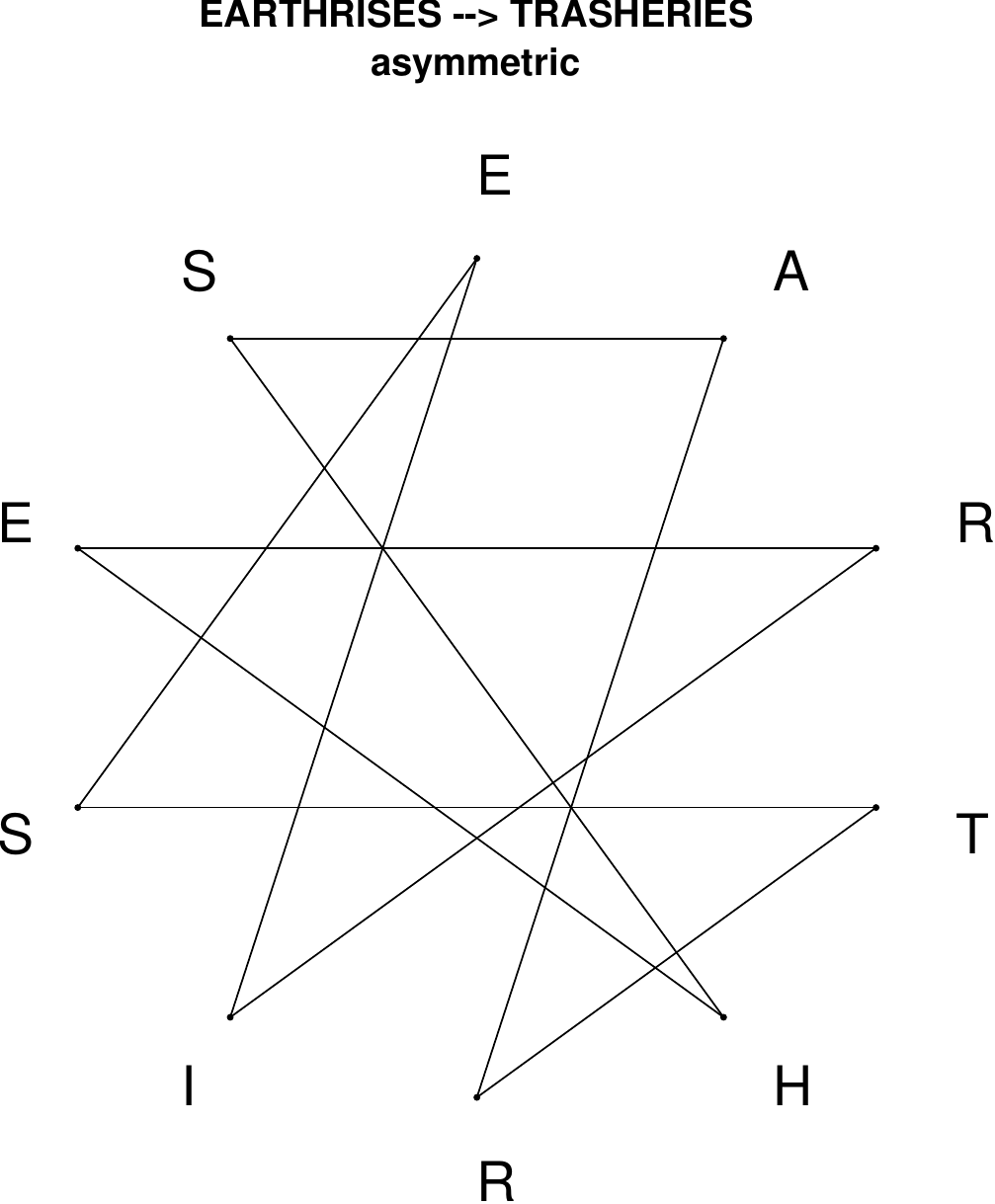}
\end{subfigure}
\hfill
\begin{subfigure}[T]{0.19\textwidth}
\centering
\includegraphics[width=\textwidth]{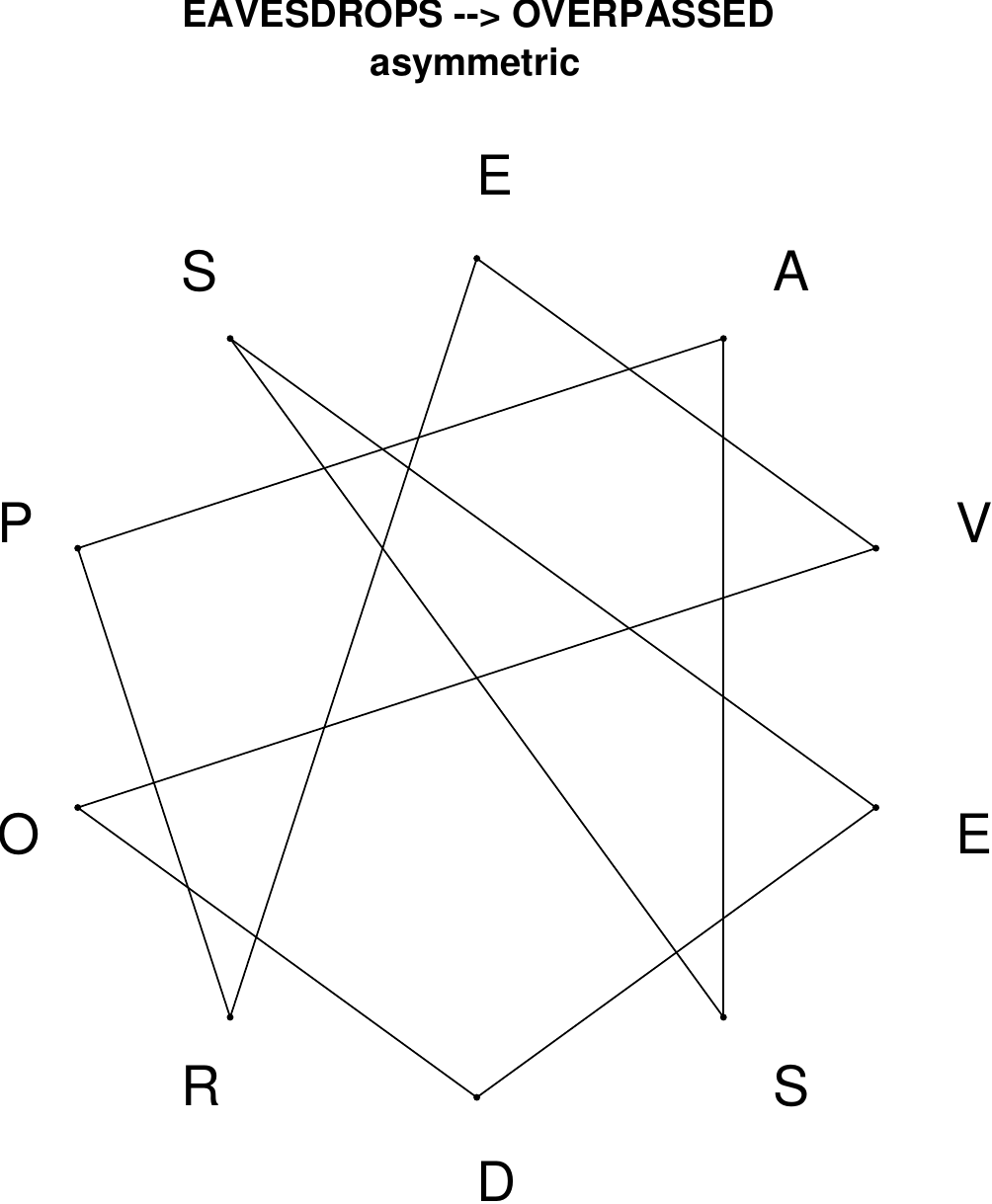}
\end{subfigure}
\hfill
\begin{subfigure}[T]{0.19\textwidth}
\centering
\includegraphics[width=\textwidth]{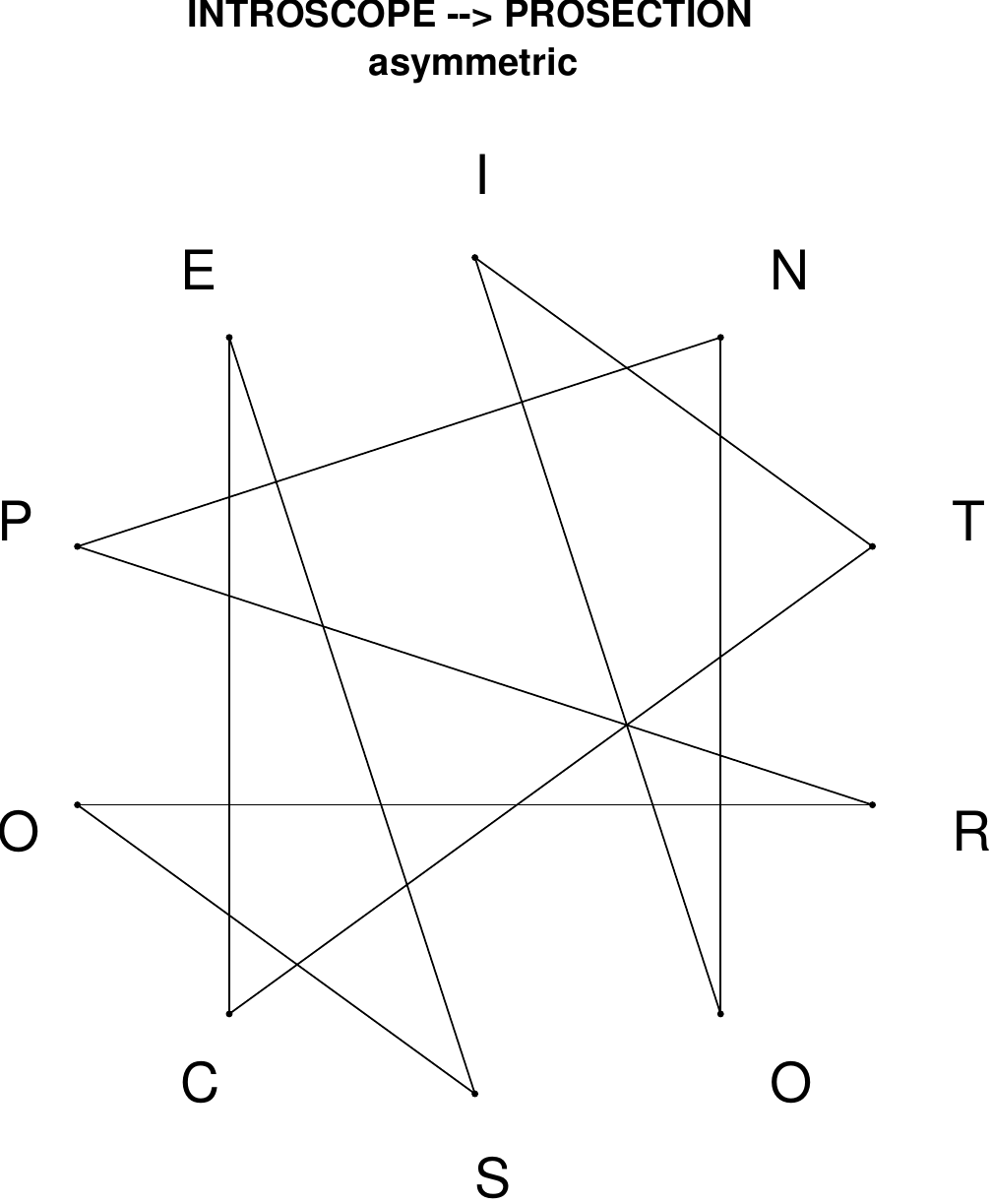}
\end{subfigure}
\end{figure}

\begin{figure}[H]
\centering
\begin{subfigure}[T]{0.19\textwidth}
\centering
\includegraphics[width=\textwidth]{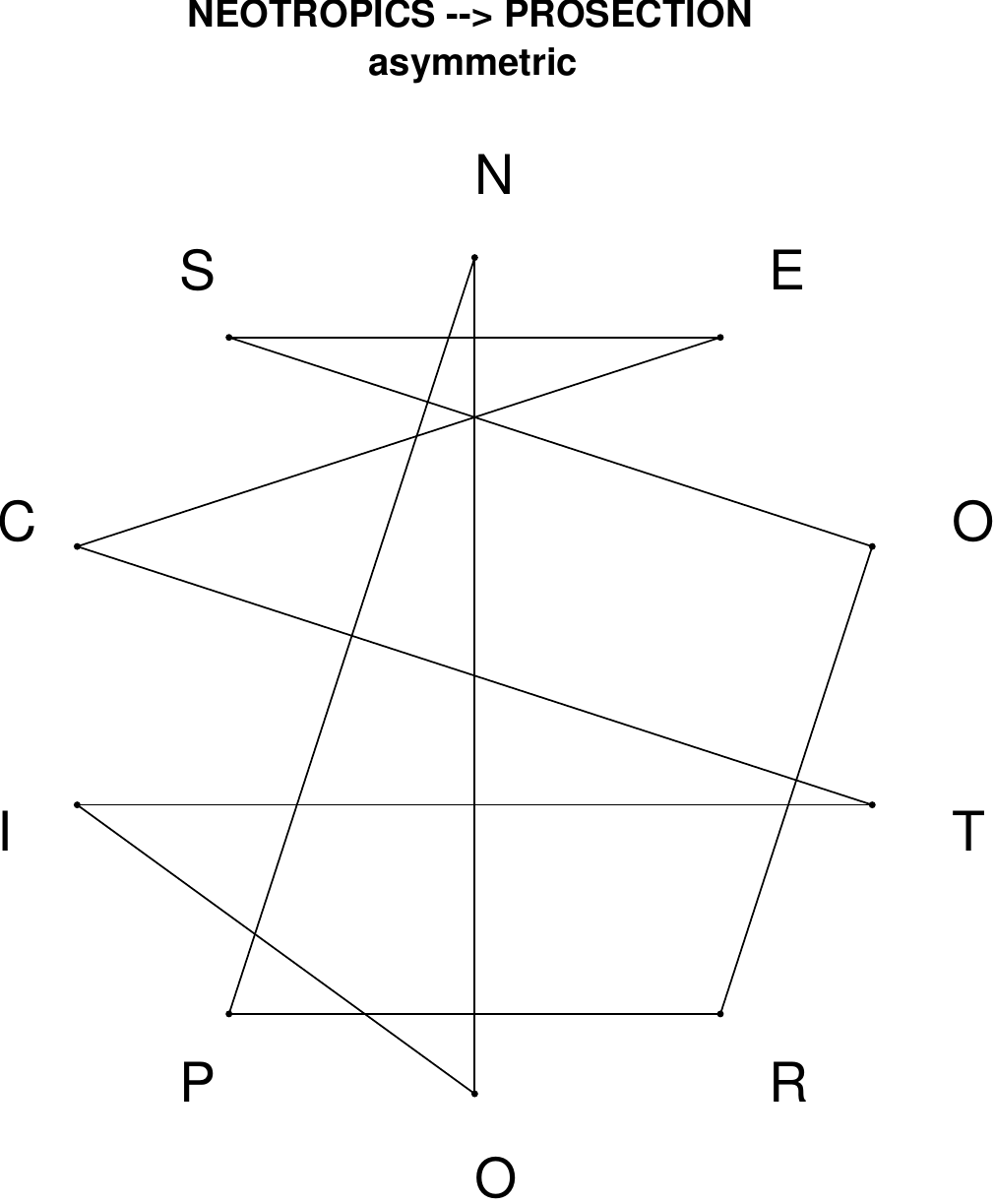}
\end{subfigure}
\hfill
\begin{subfigure}[T]{0.19\textwidth}
\centering
\includegraphics[width=\textwidth]{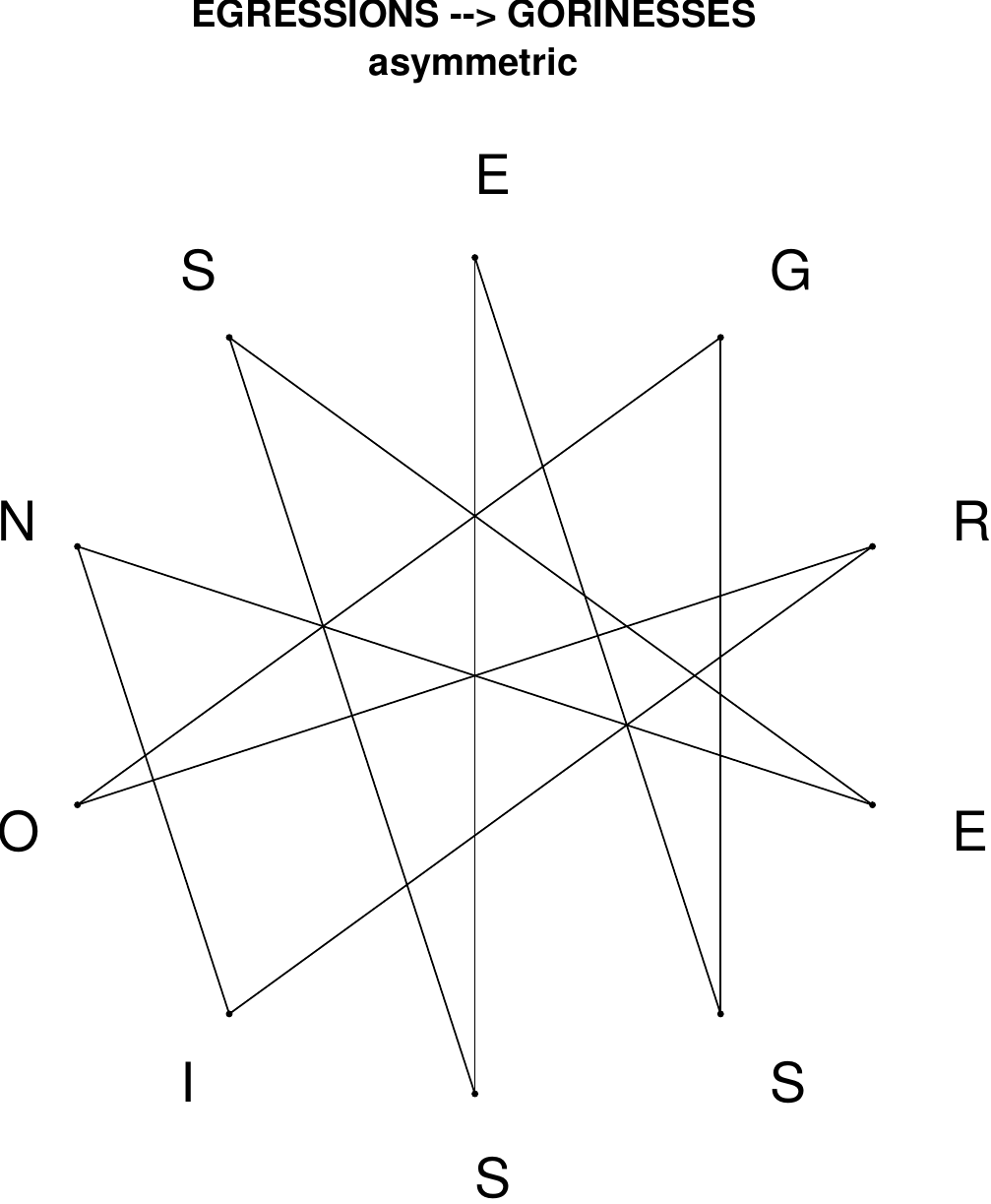}
\end{subfigure}
\hfill
\begin{subfigure}[T]{0.19\textwidth}
\centering
\includegraphics[width=\textwidth]{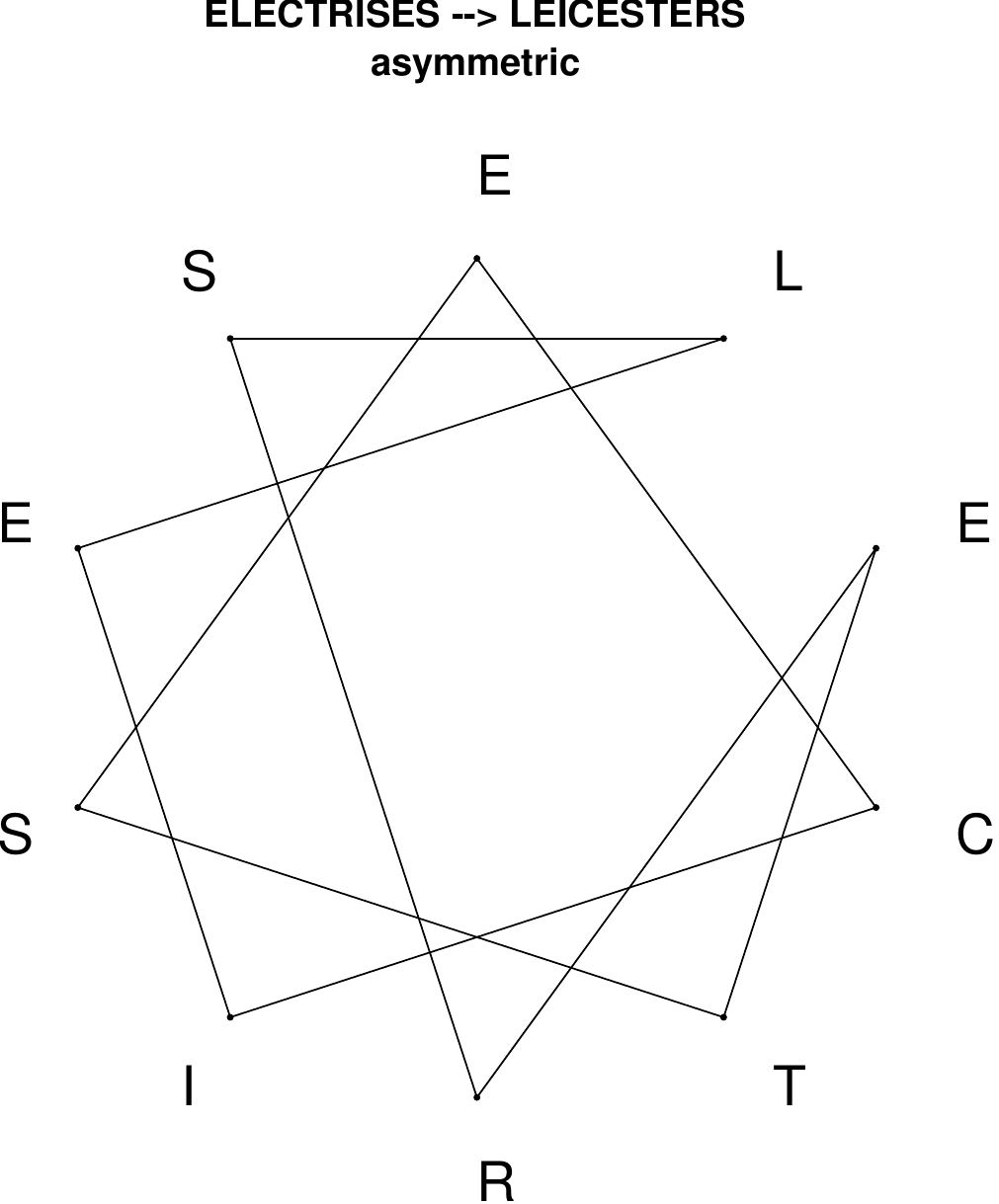}
\end{subfigure}
\hfill
\begin{subfigure}[T]{0.19\textwidth}
\centering
\includegraphics[width=\textwidth]{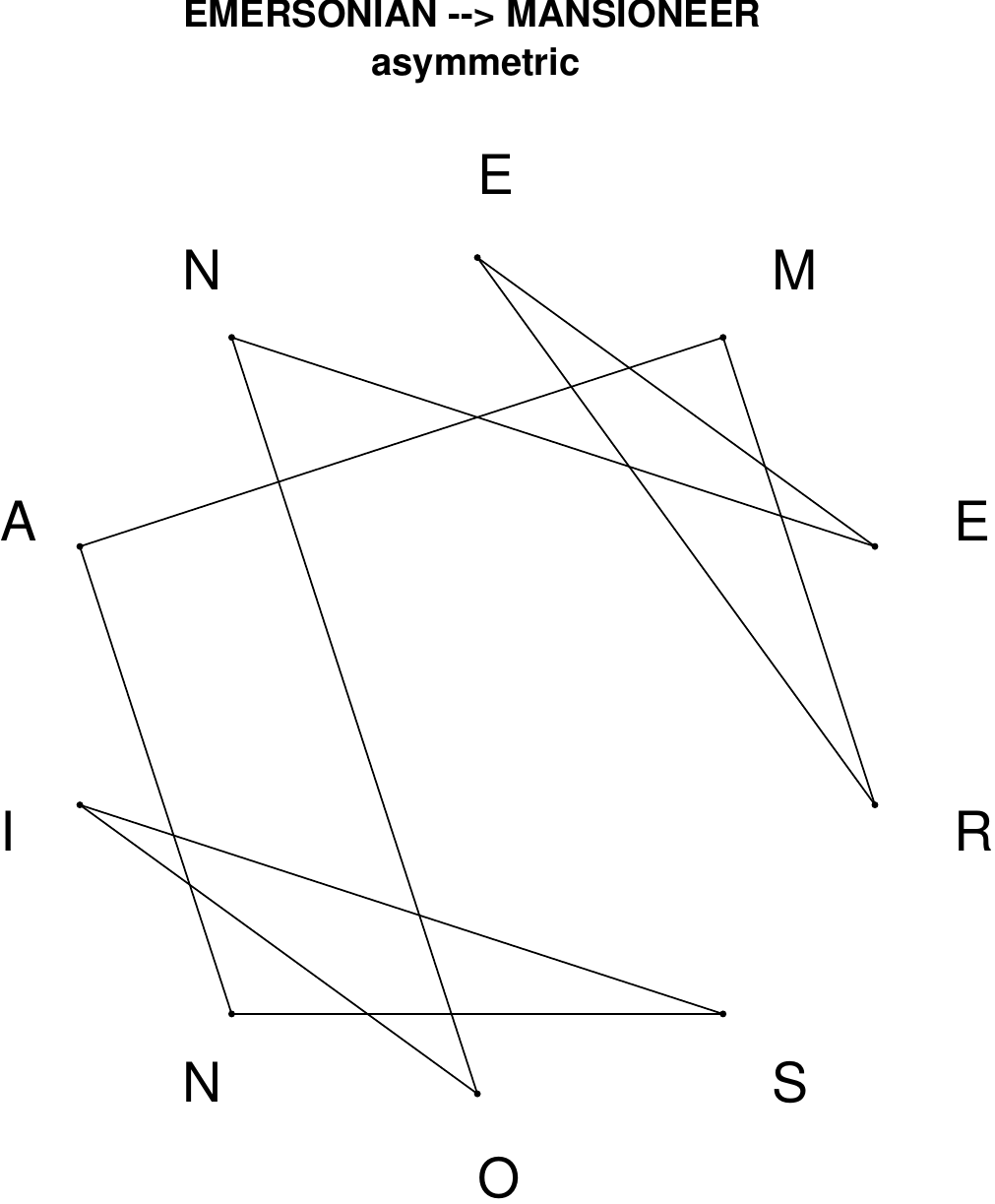}
\end{subfigure}
\hfill
\begin{subfigure}[T]{0.19\textwidth}
\centering
\includegraphics[width=\textwidth]{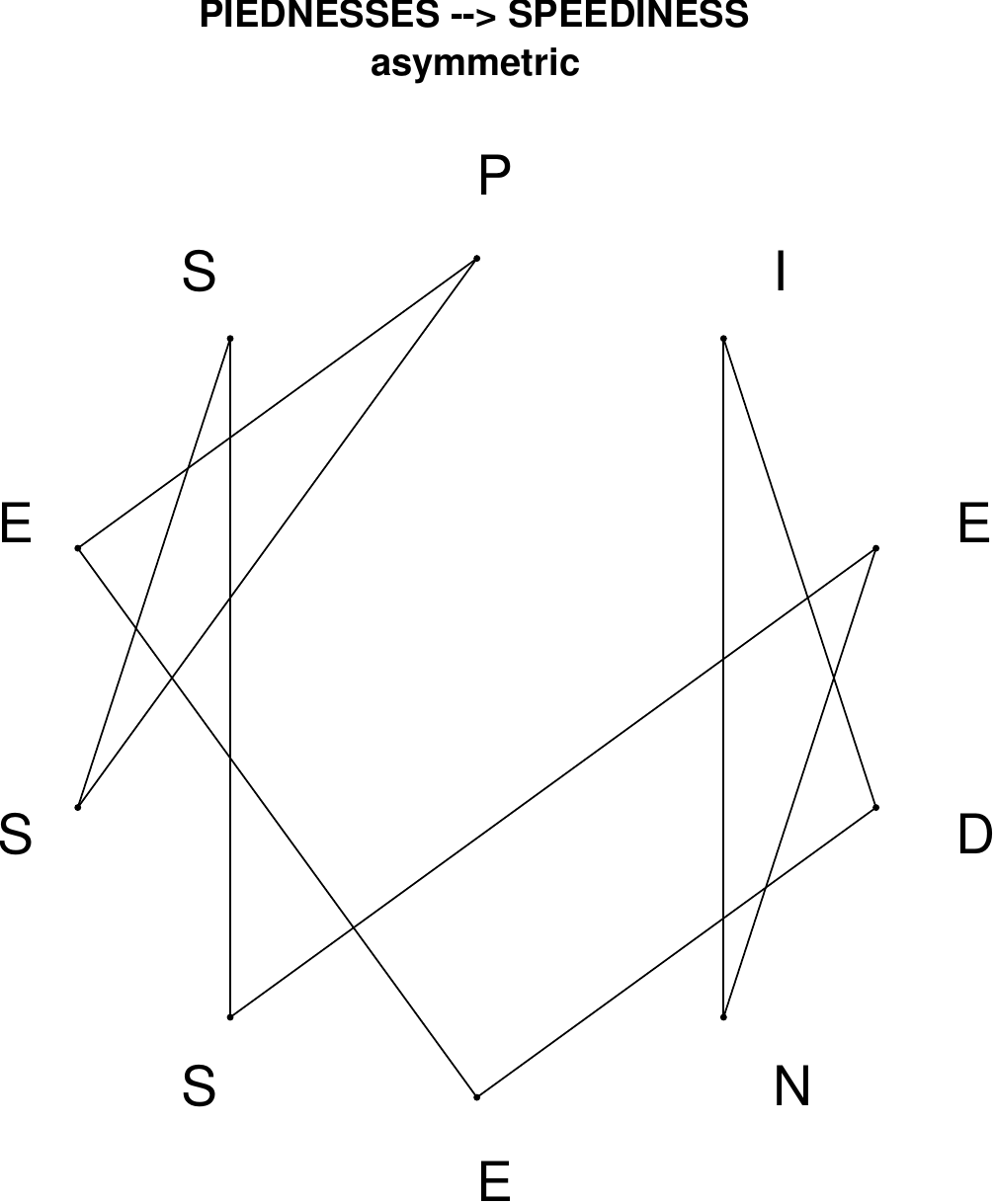}
\end{subfigure}
\end{figure}

\begin{figure}[H]
\centering
\begin{subfigure}[T]{0.19\textwidth}
\centering
\includegraphics[width=\textwidth]{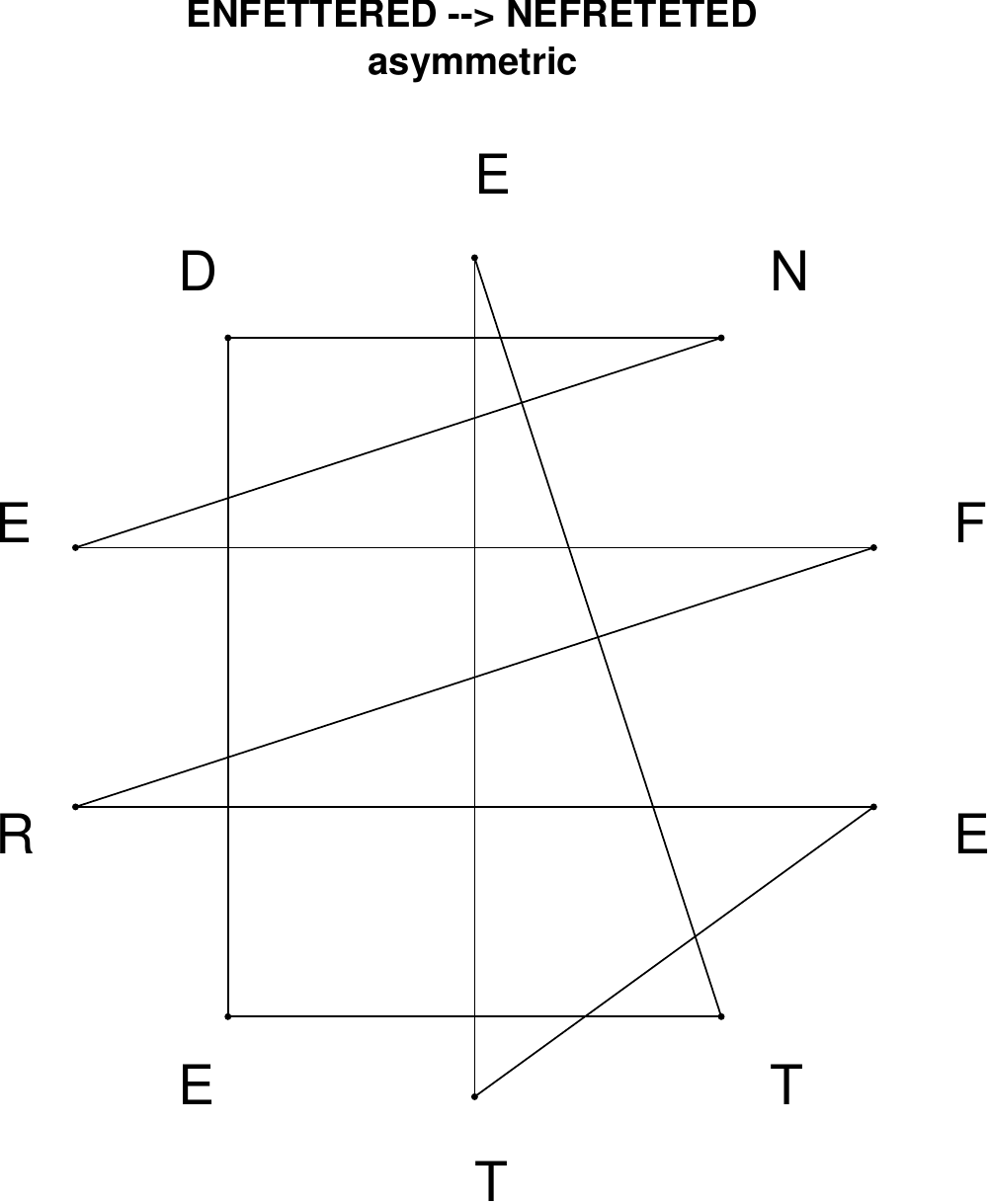}
\end{subfigure}
\hfill
\begin{subfigure}[T]{0.19\textwidth}
\centering
\includegraphics[width=\textwidth]{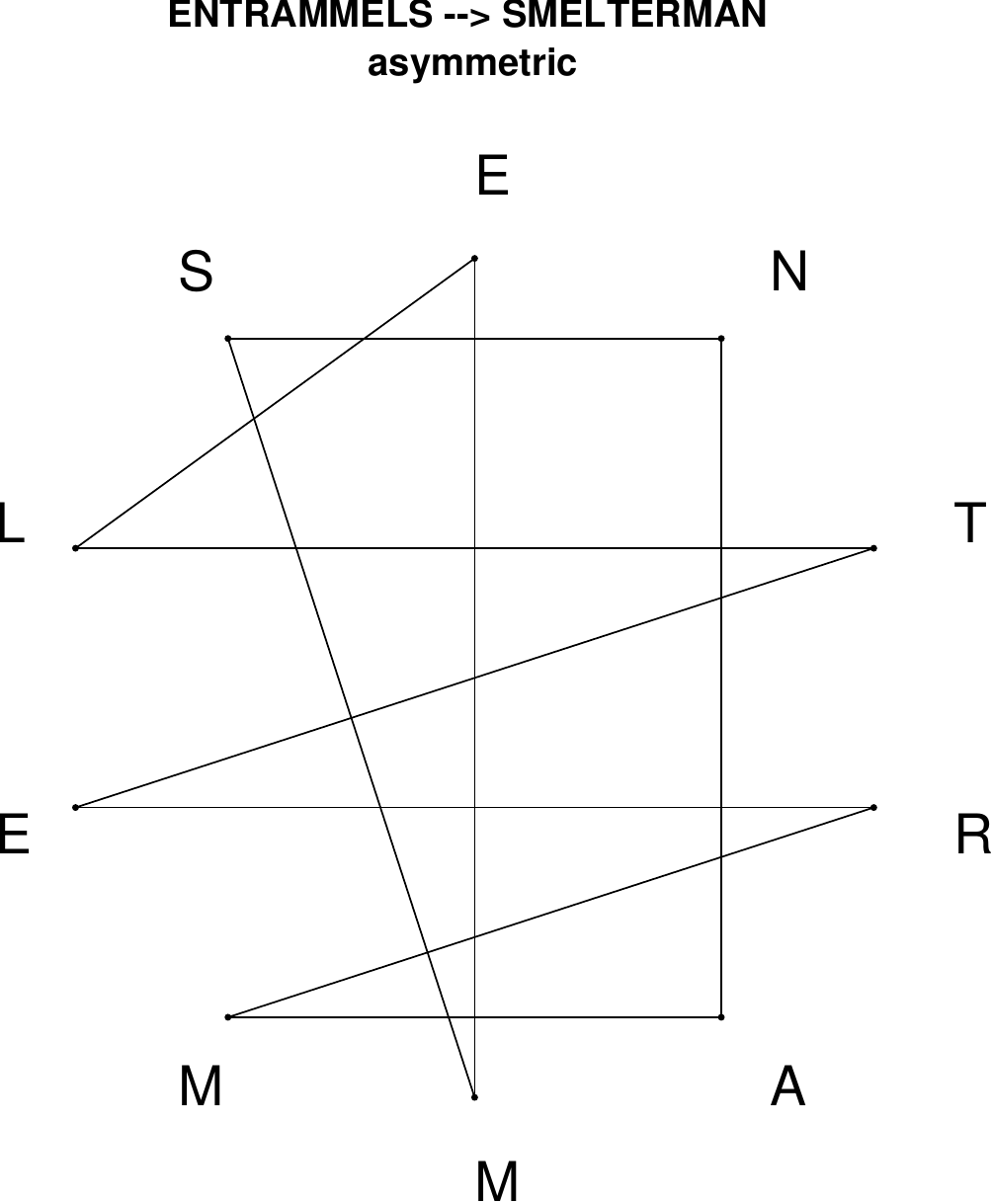}
\end{subfigure}
\hfill
\begin{subfigure}[T]{0.19\textwidth}
\centering
\includegraphics[width=\textwidth]{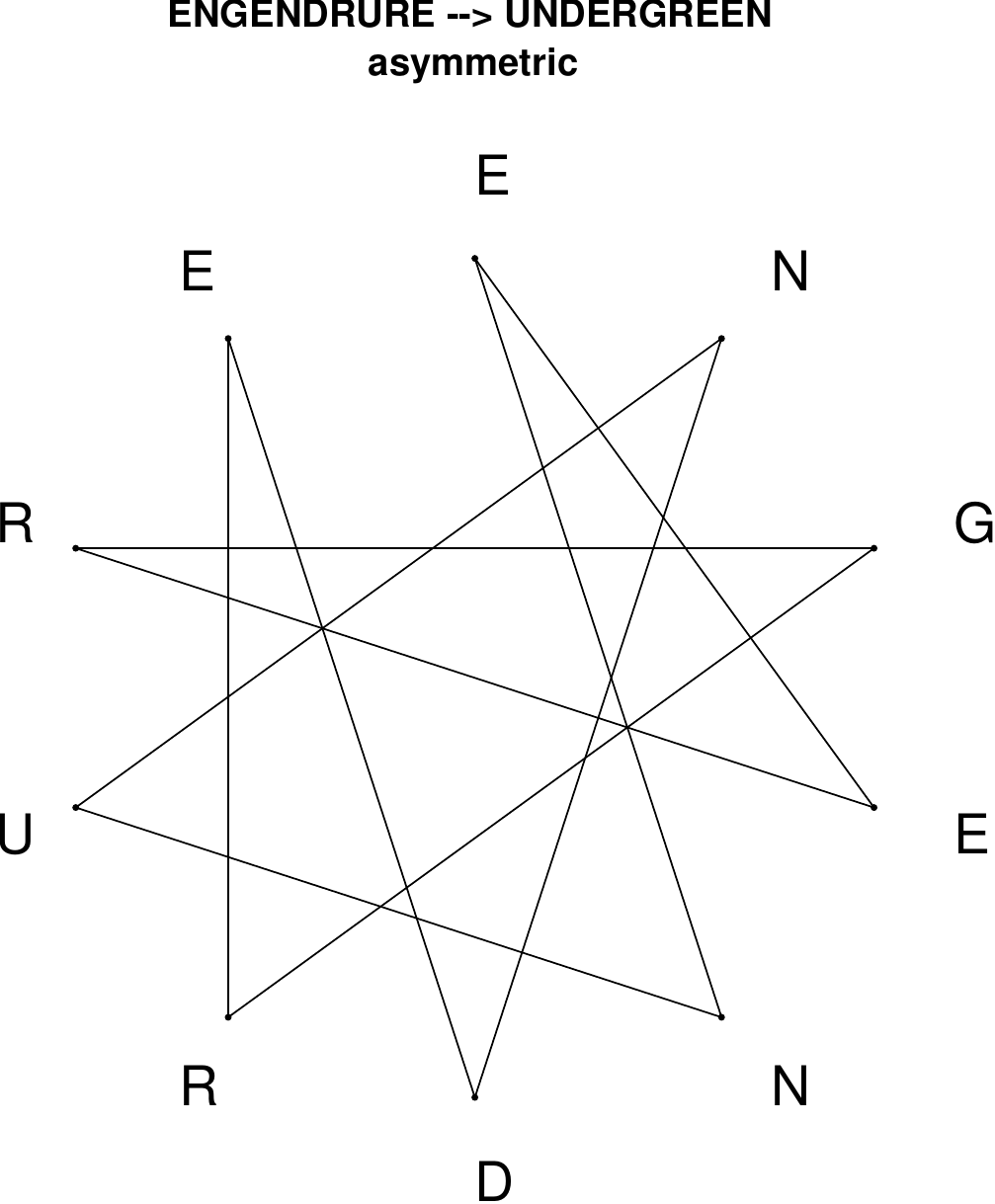}
\end{subfigure}
\hfill
\begin{subfigure}[T]{0.19\textwidth}
\centering
\includegraphics[width=\textwidth]{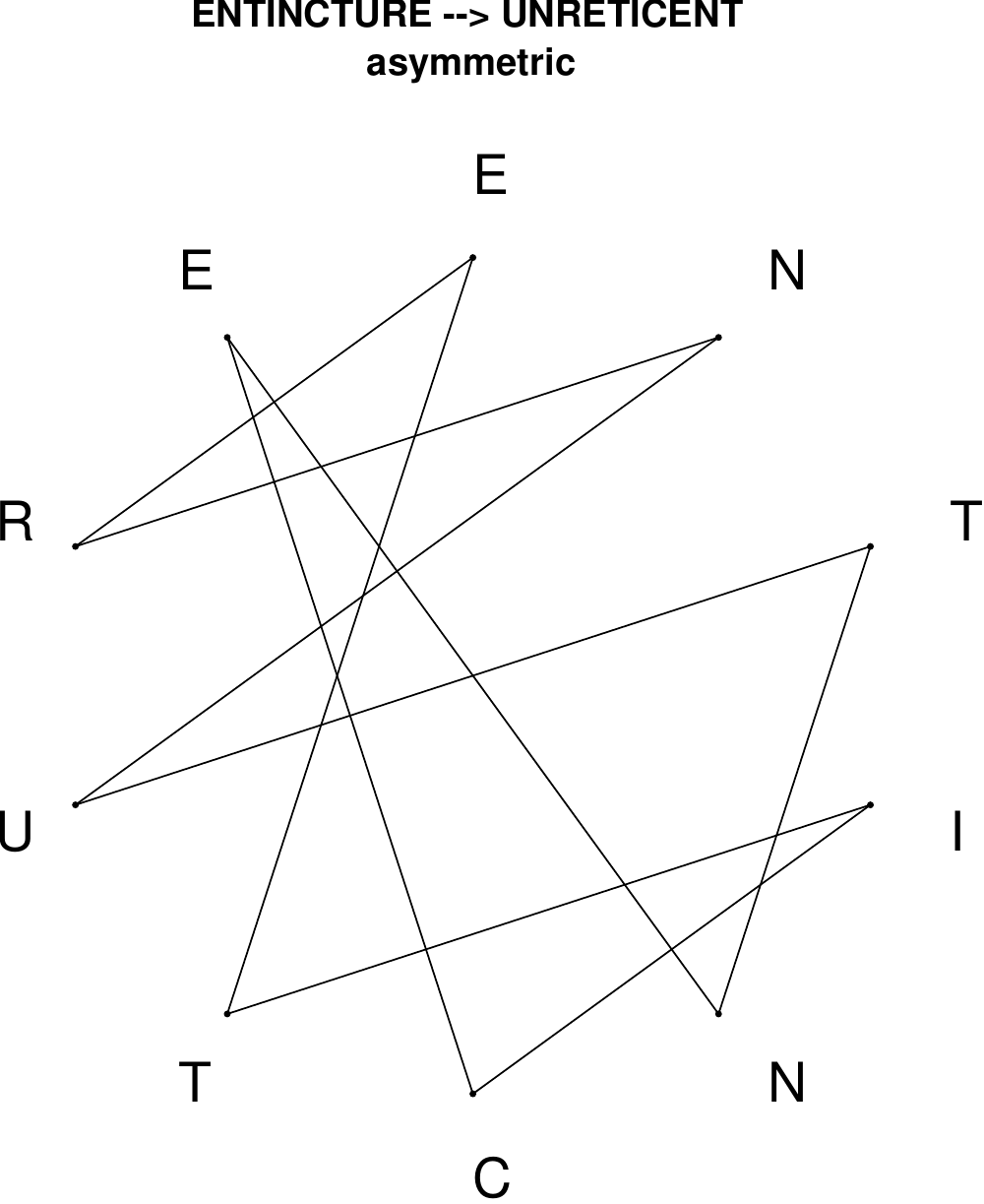}
\end{subfigure}
\hfill
\begin{subfigure}[T]{0.19\textwidth}
\centering
\includegraphics[width=\textwidth]{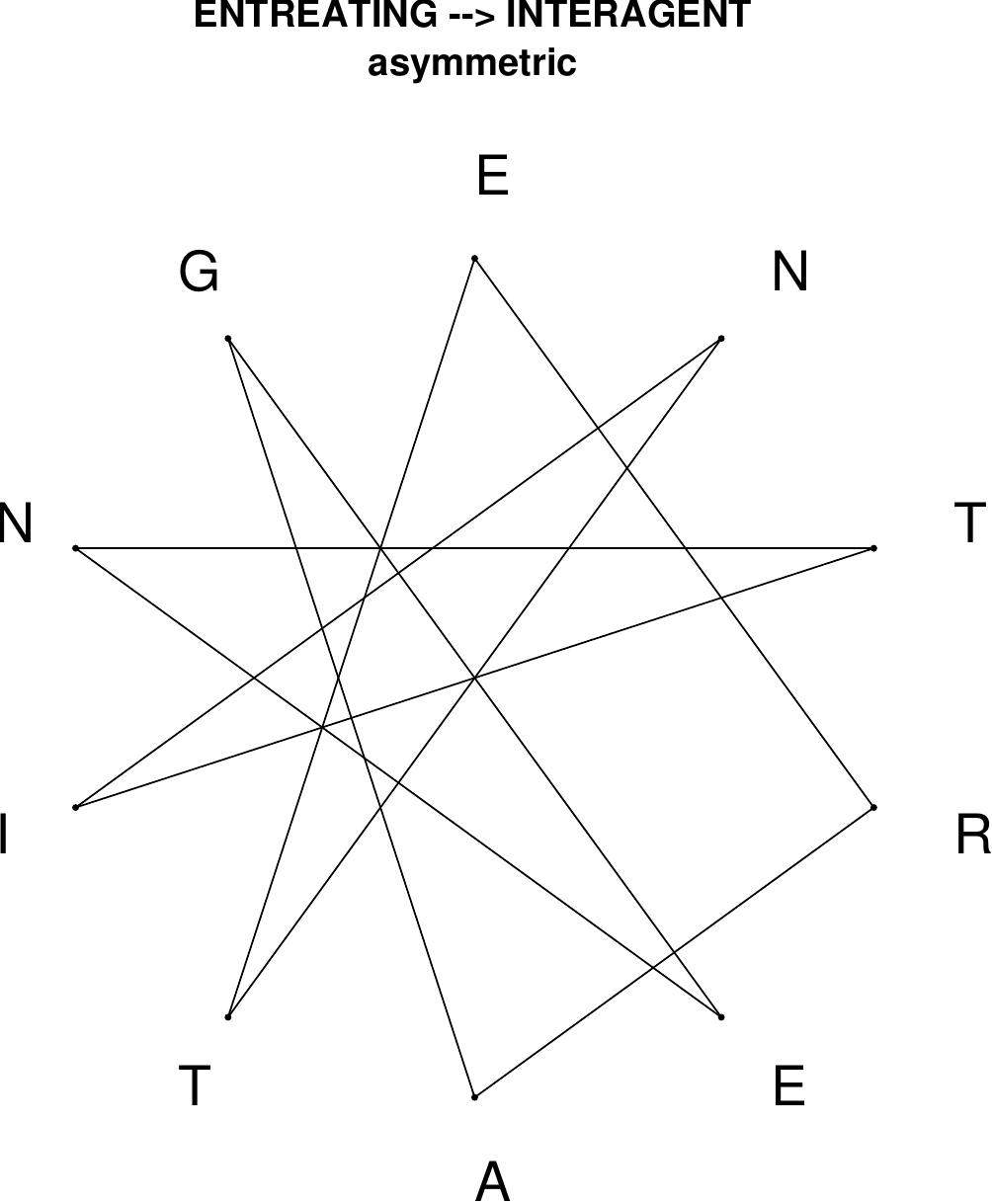}
\end{subfigure}
\end{figure}

\begin{figure}[H]
\centering
\begin{subfigure}[T]{0.19\textwidth}
\centering
\includegraphics[width=\textwidth]{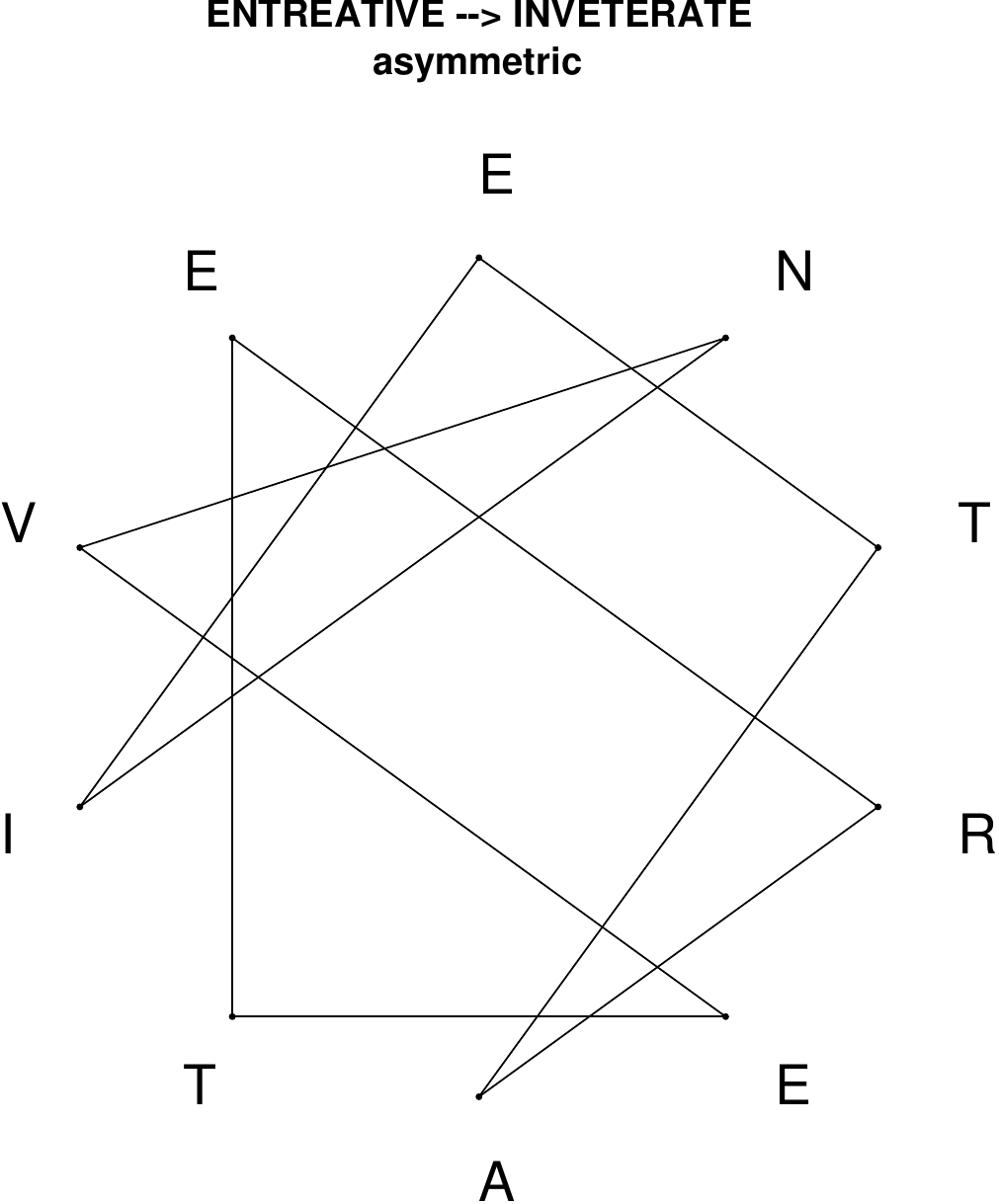}
\end{subfigure}
\hfill
\begin{subfigure}[T]{0.19\textwidth}
\centering
\includegraphics[width=\textwidth]{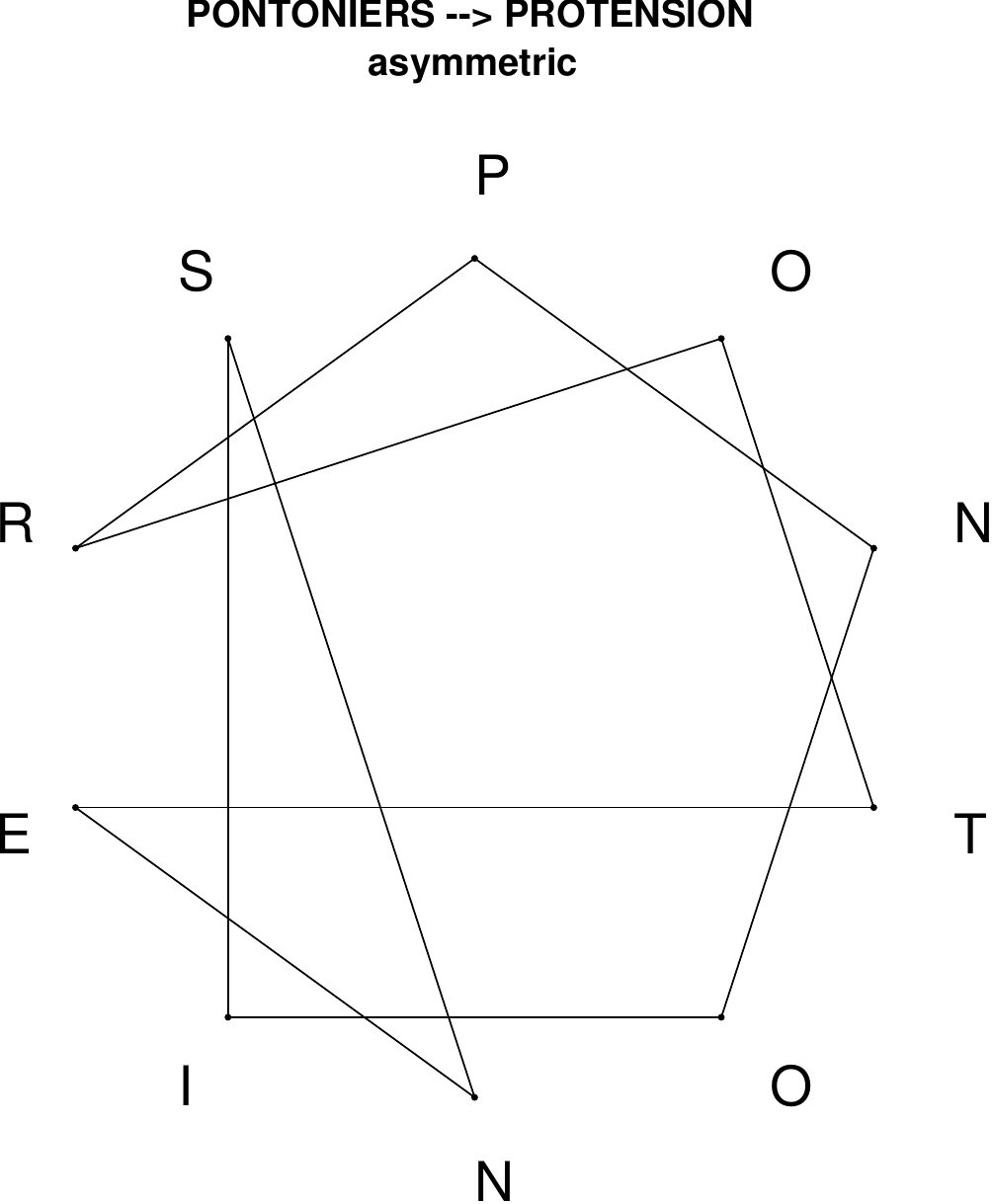}
\end{subfigure}
\hfill
\begin{subfigure}[T]{0.19\textwidth}
\centering
\includegraphics[width=\textwidth]{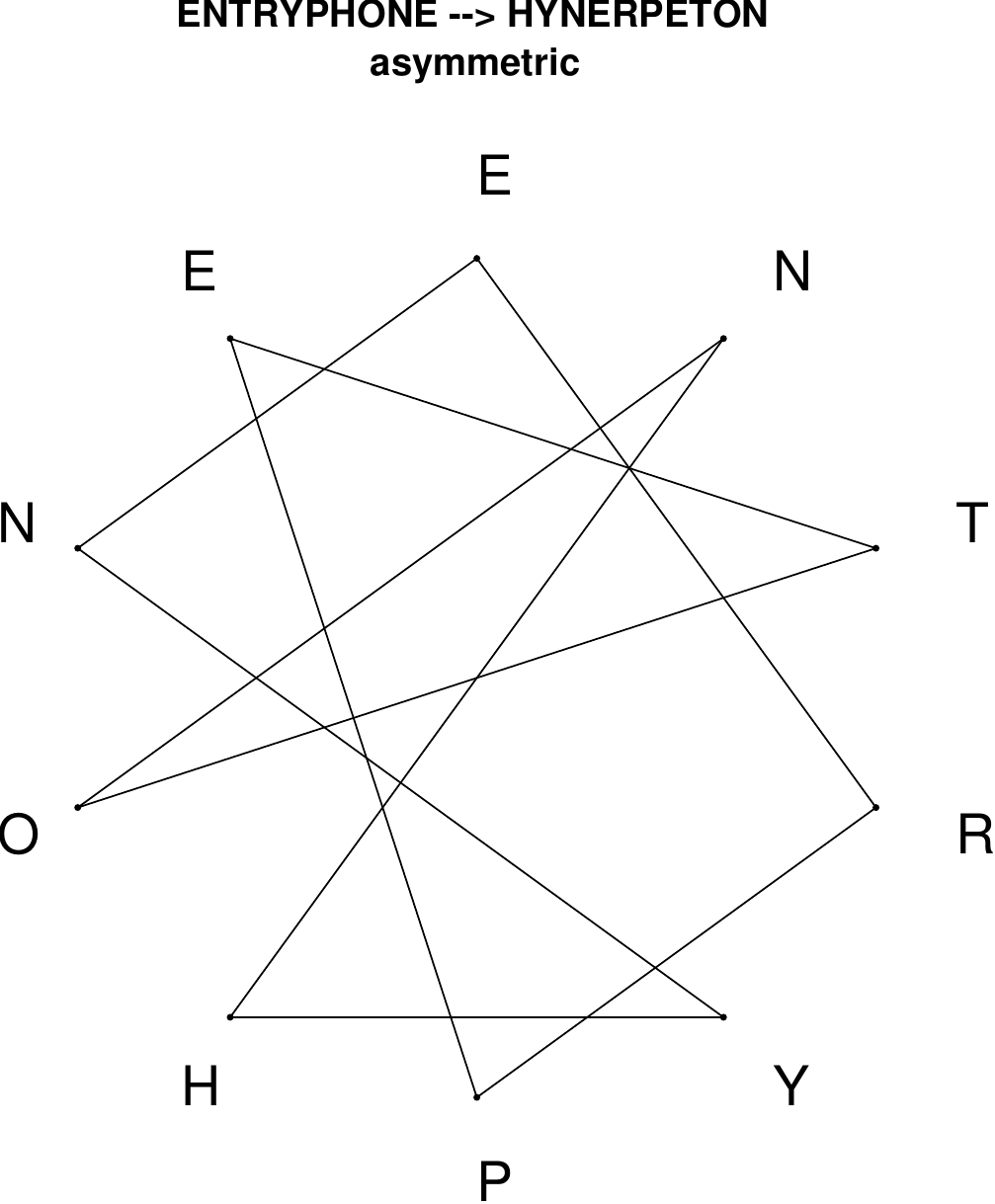}
\end{subfigure}
\hfill
\begin{subfigure}[T]{0.19\textwidth}
\centering
\includegraphics[width=\textwidth]{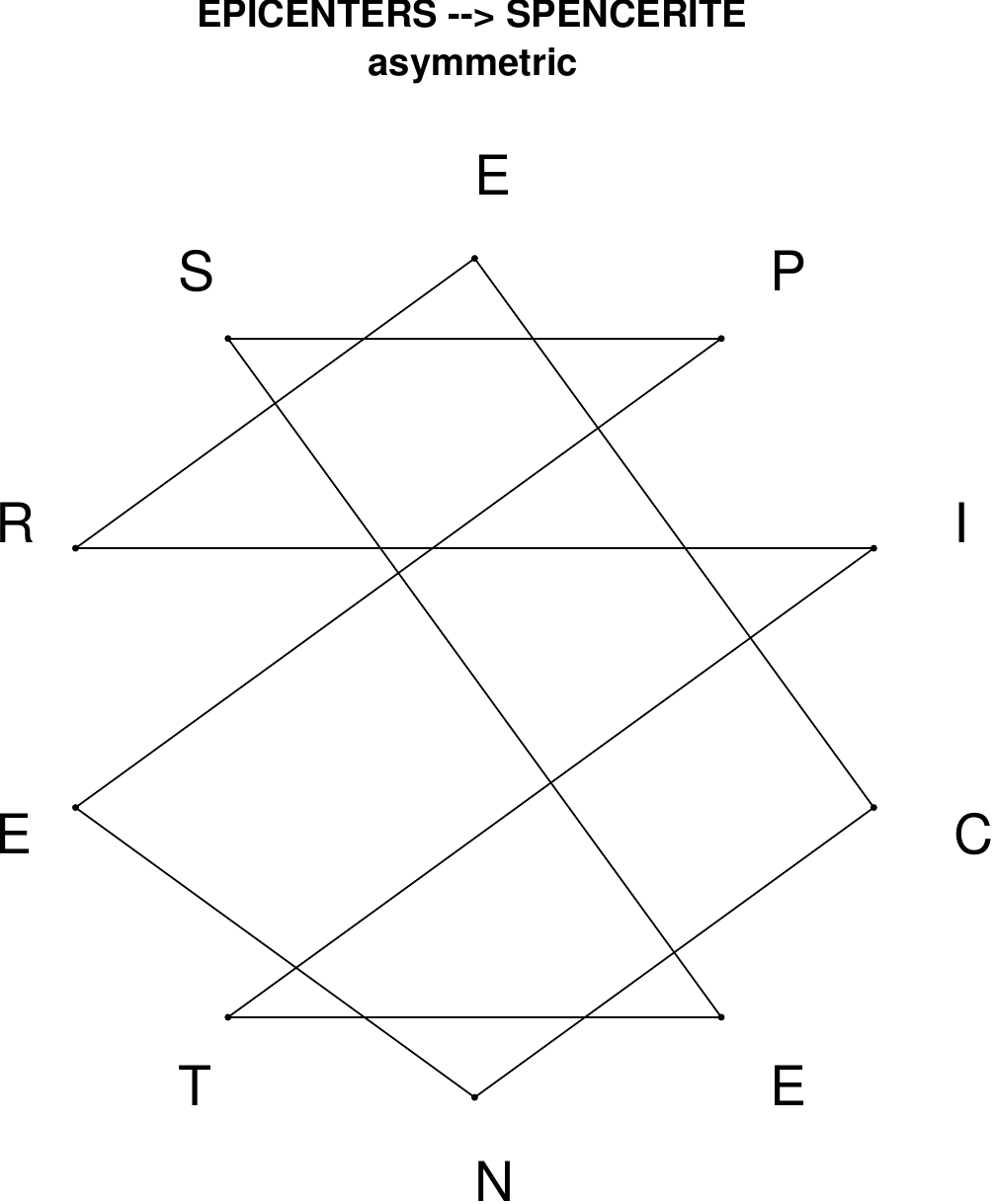}
\end{subfigure}
\hfill
\begin{subfigure}[T]{0.19\textwidth}
\centering
\includegraphics[width=\textwidth]{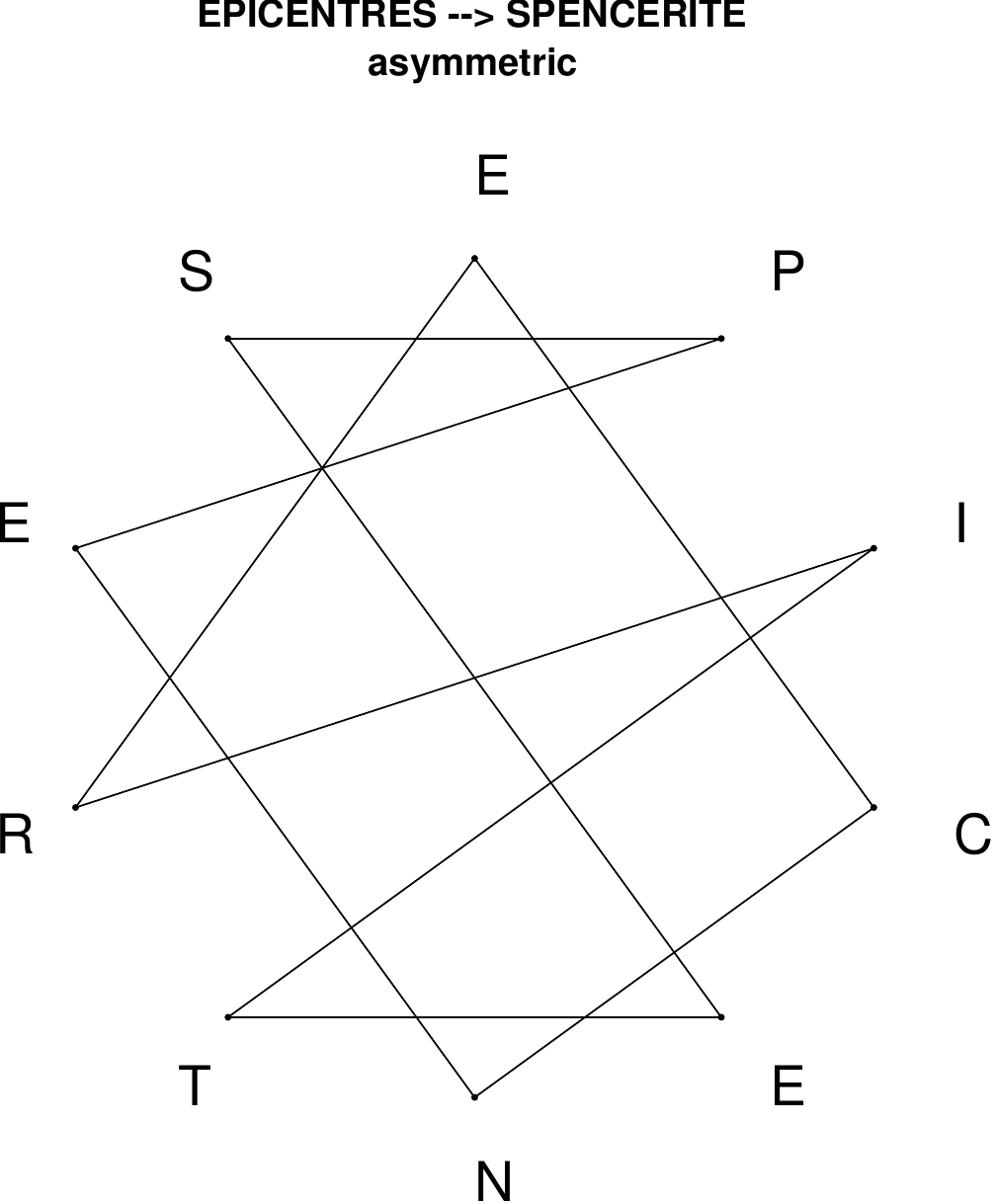}
\end{subfigure}
\end{figure}

\begin{figure}[H]
\centering
\begin{subfigure}[T]{0.19\textwidth}
\centering
\includegraphics[width=\textwidth]{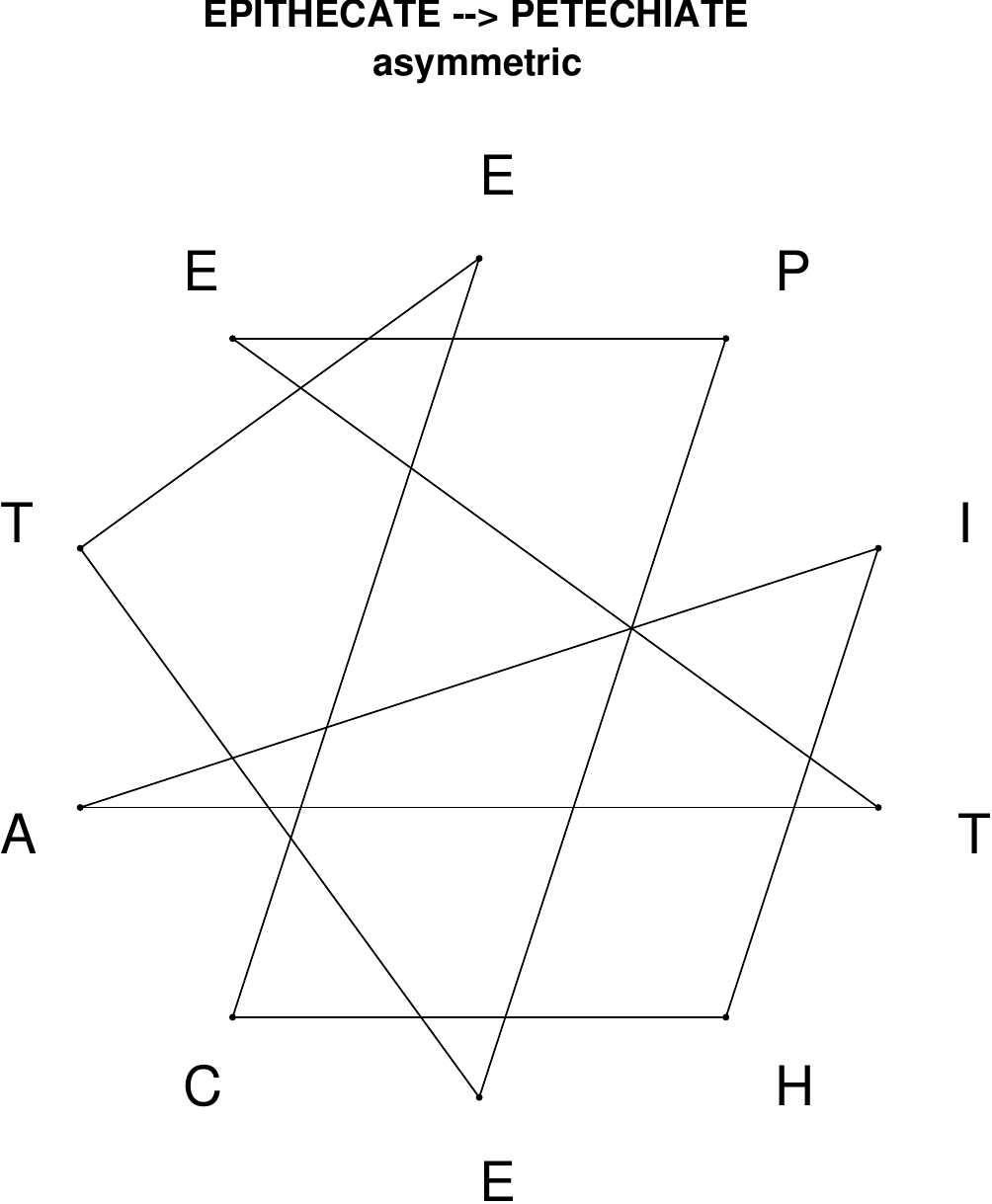}
\end{subfigure}
\hfill
\begin{subfigure}[T]{0.19\textwidth}
\centering
\includegraphics[width=\textwidth]{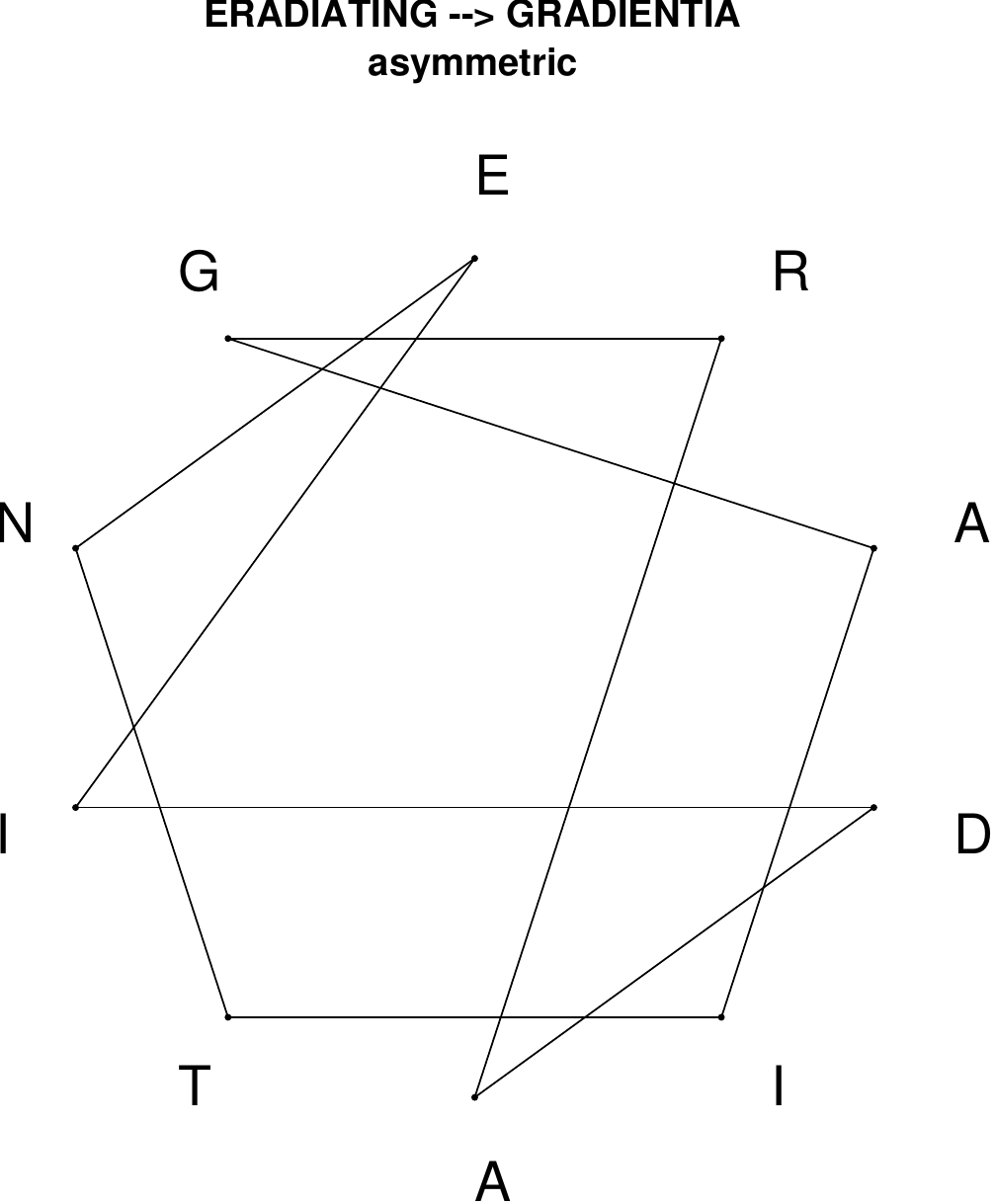}
\end{subfigure}
\hfill
\begin{subfigure}[T]{0.19\textwidth}
\centering
\includegraphics[width=\textwidth]{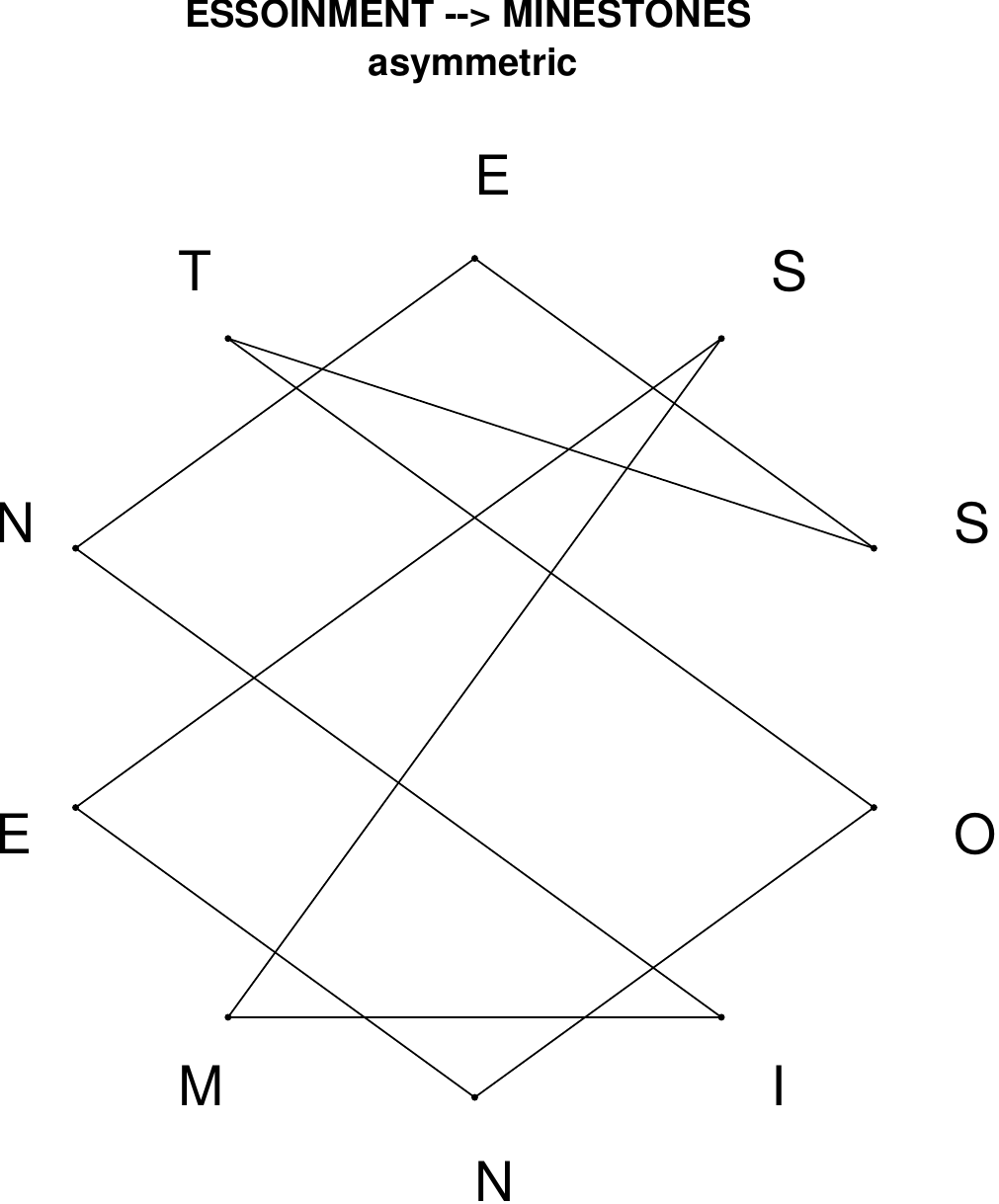}
\end{subfigure}
\hfill
\begin{subfigure}[T]{0.19\textwidth}
\centering
\includegraphics[width=\textwidth]{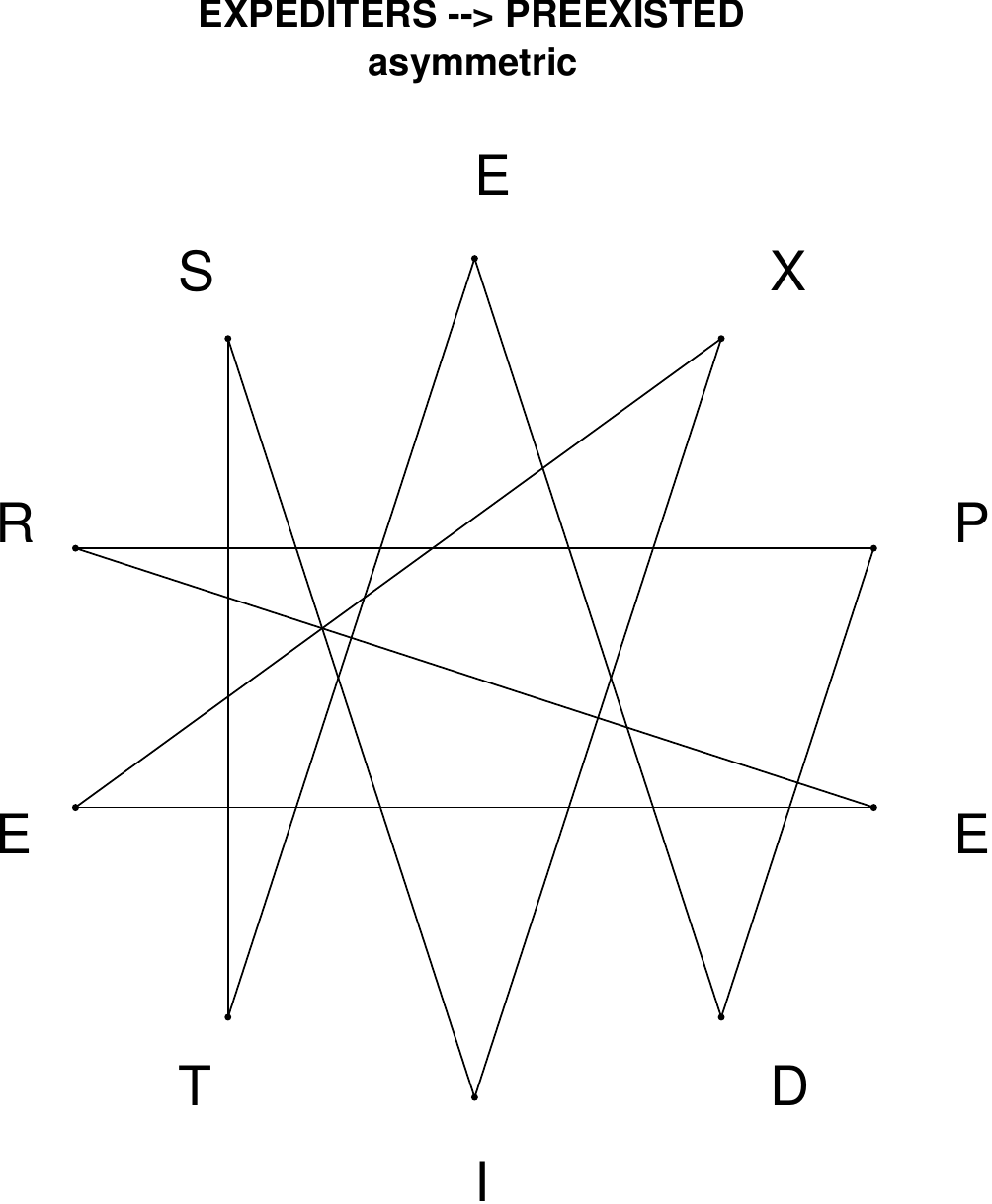}
\end{subfigure}
\hfill
\begin{subfigure}[T]{0.19\textwidth}
\centering
\includegraphics[width=\textwidth]{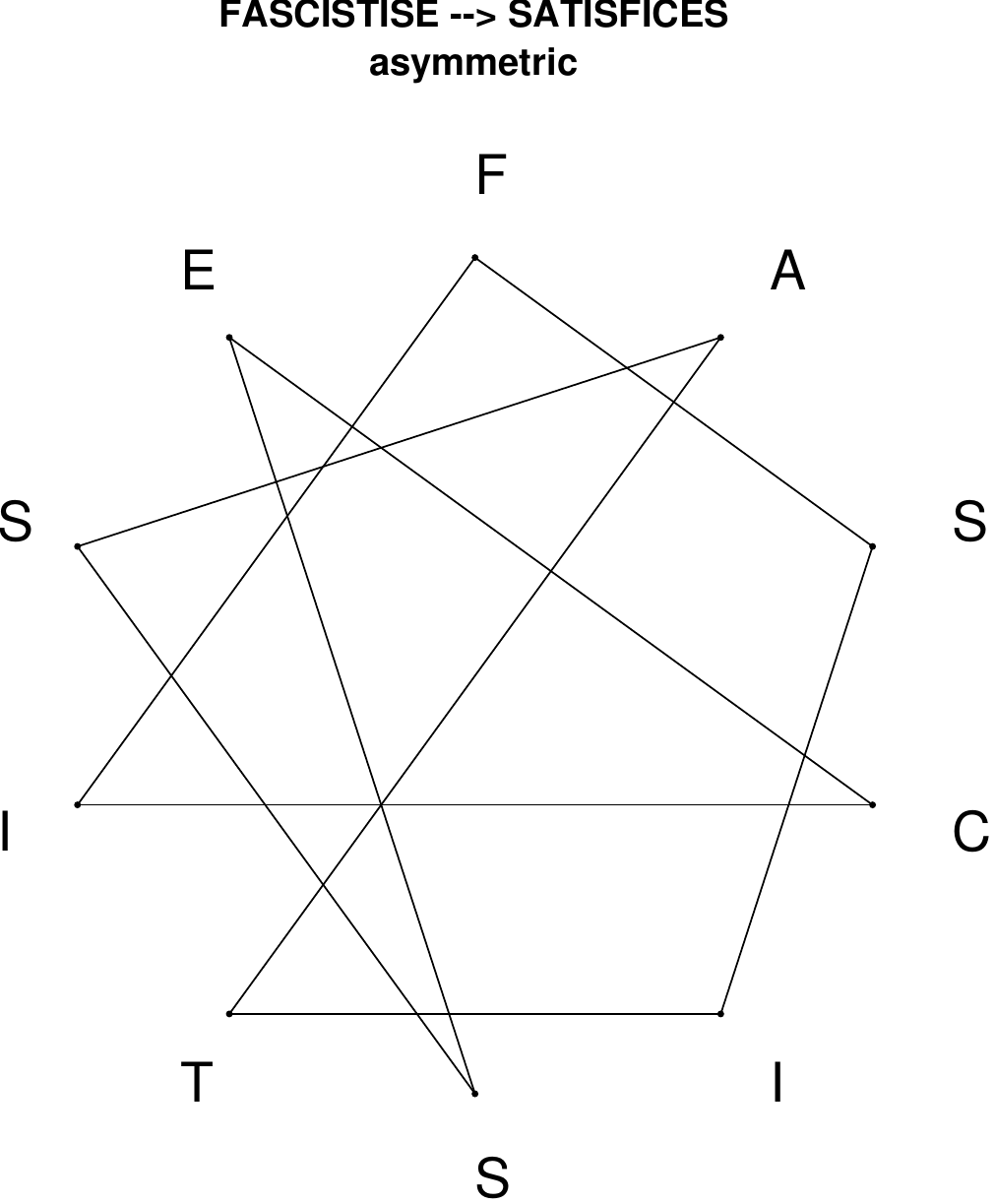}
\end{subfigure}
\end{figure}

\begin{figure}[H]
\centering
\begin{subfigure}[T]{0.19\textwidth}
\centering
\includegraphics[width=\textwidth]{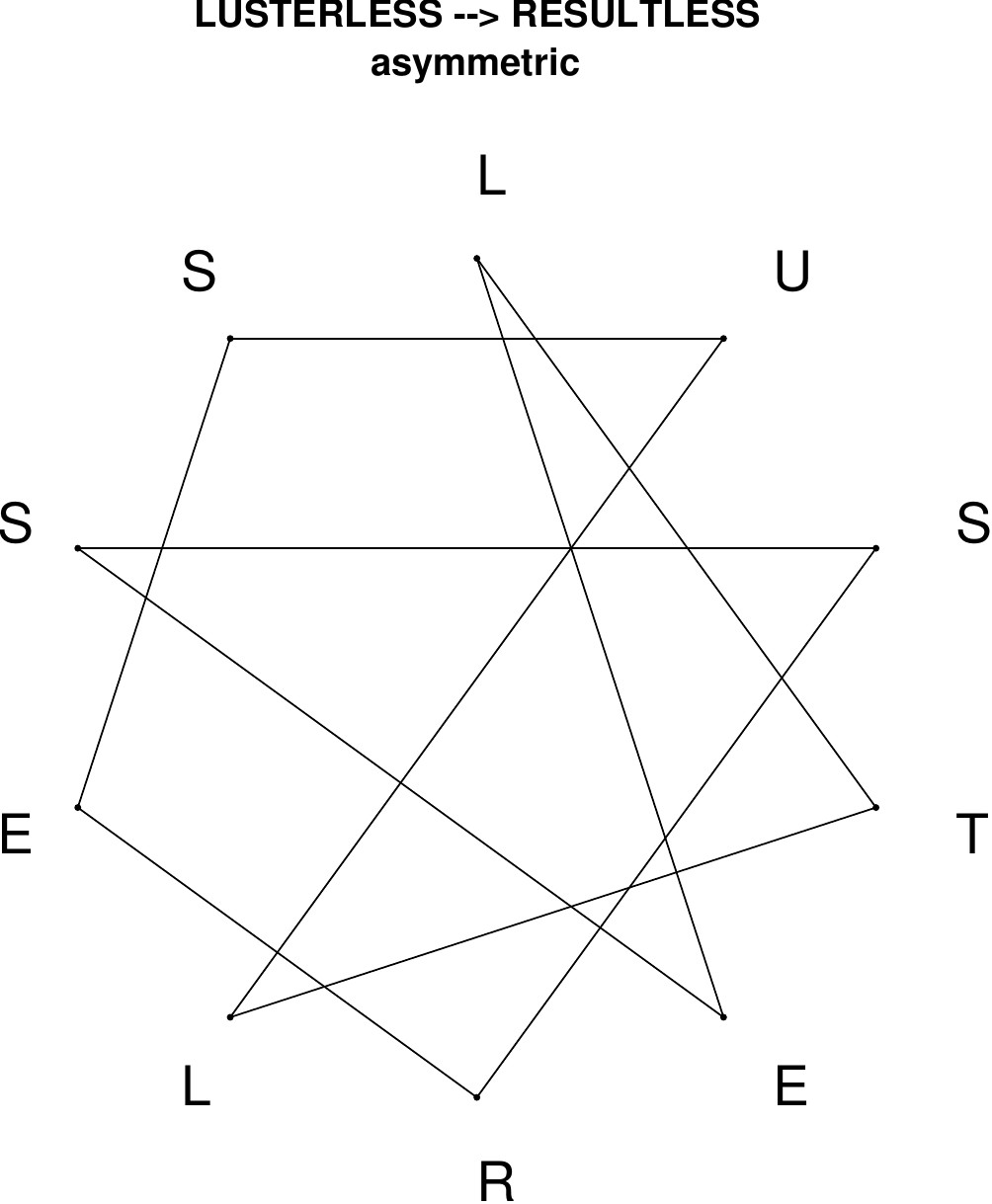}
\end{subfigure}
\hfill
\begin{subfigure}[T]{0.19\textwidth}
\centering
\includegraphics[width=\textwidth]{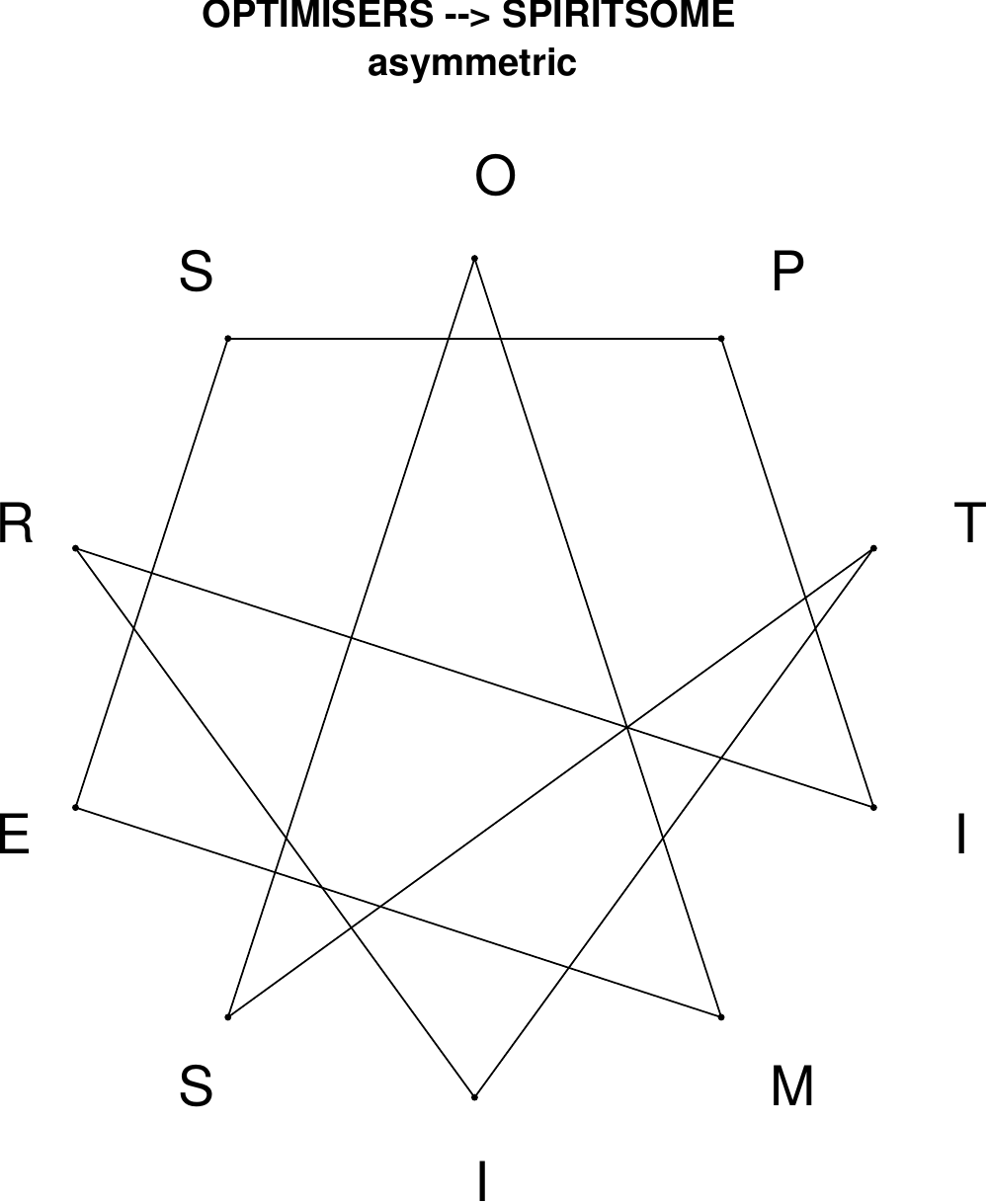}
\end{subfigure}
\hfill
\begin{subfigure}[T]{0.19\textwidth}
\centering
\includegraphics[width=\textwidth]{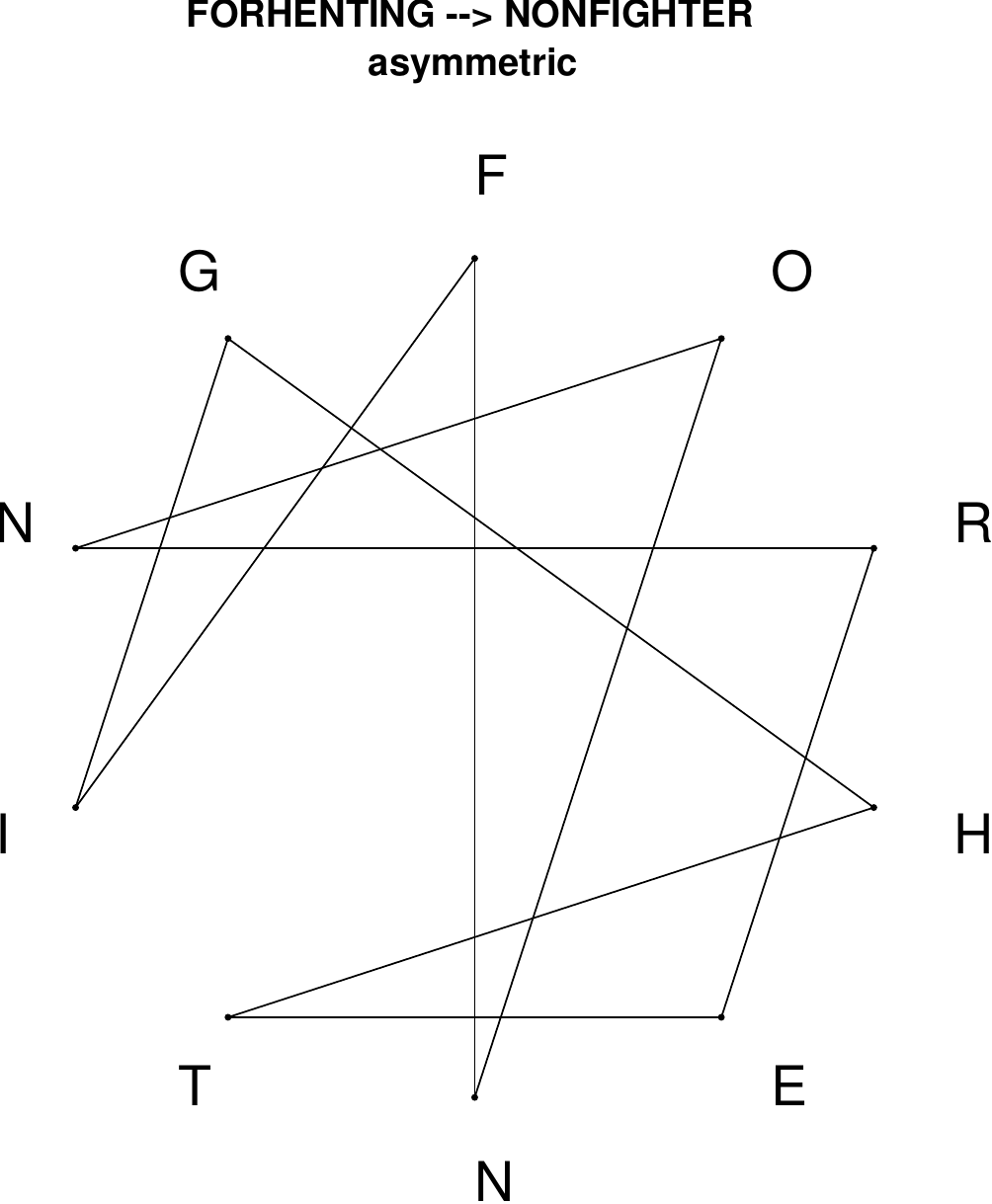}
\end{subfigure}
\hfill
\begin{subfigure}[T]{0.19\textwidth}
\centering
\includegraphics[width=\textwidth]{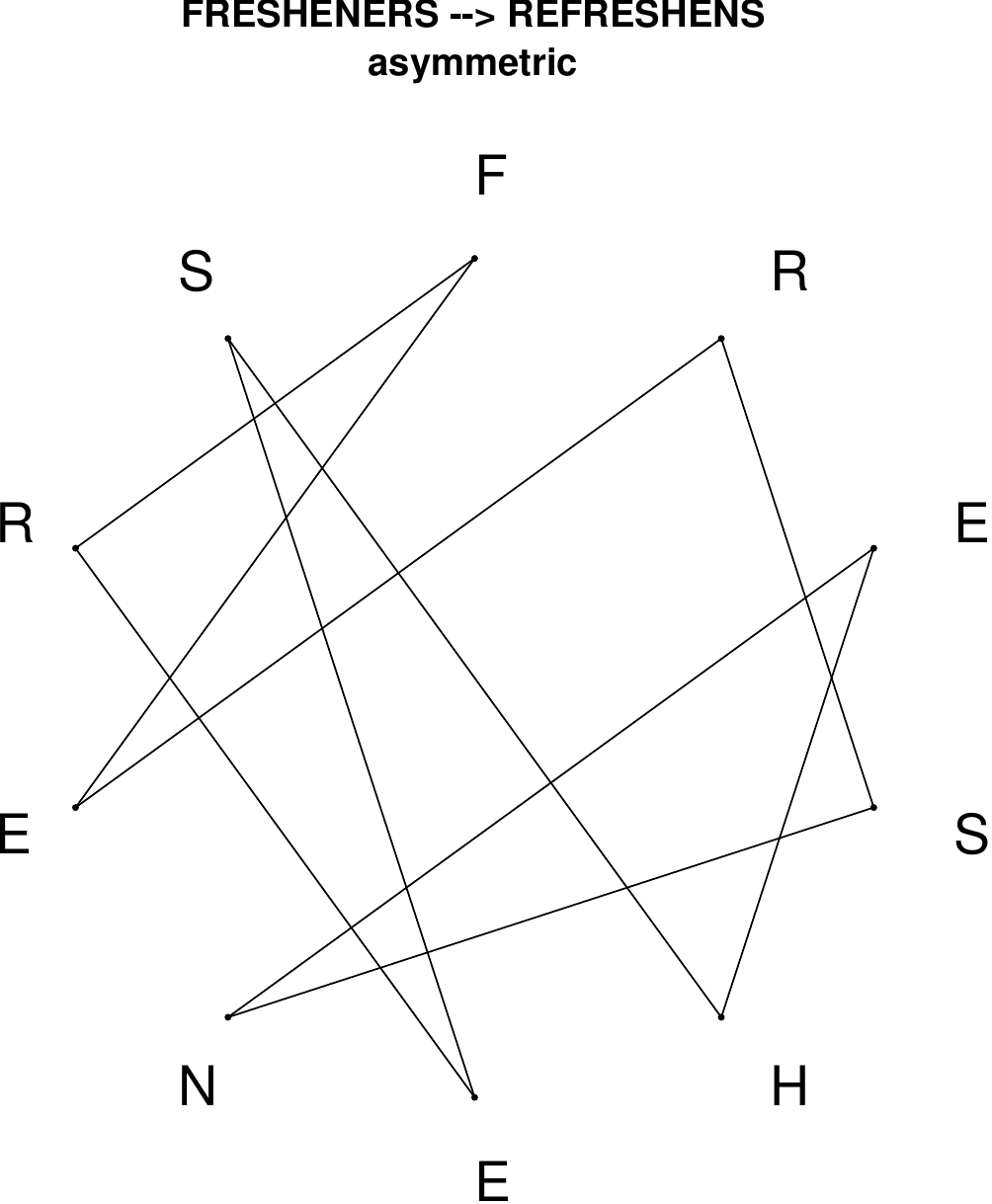}
\end{subfigure}
\hfill
\begin{subfigure}[T]{0.19\textwidth}
\centering
\includegraphics[width=\textwidth]{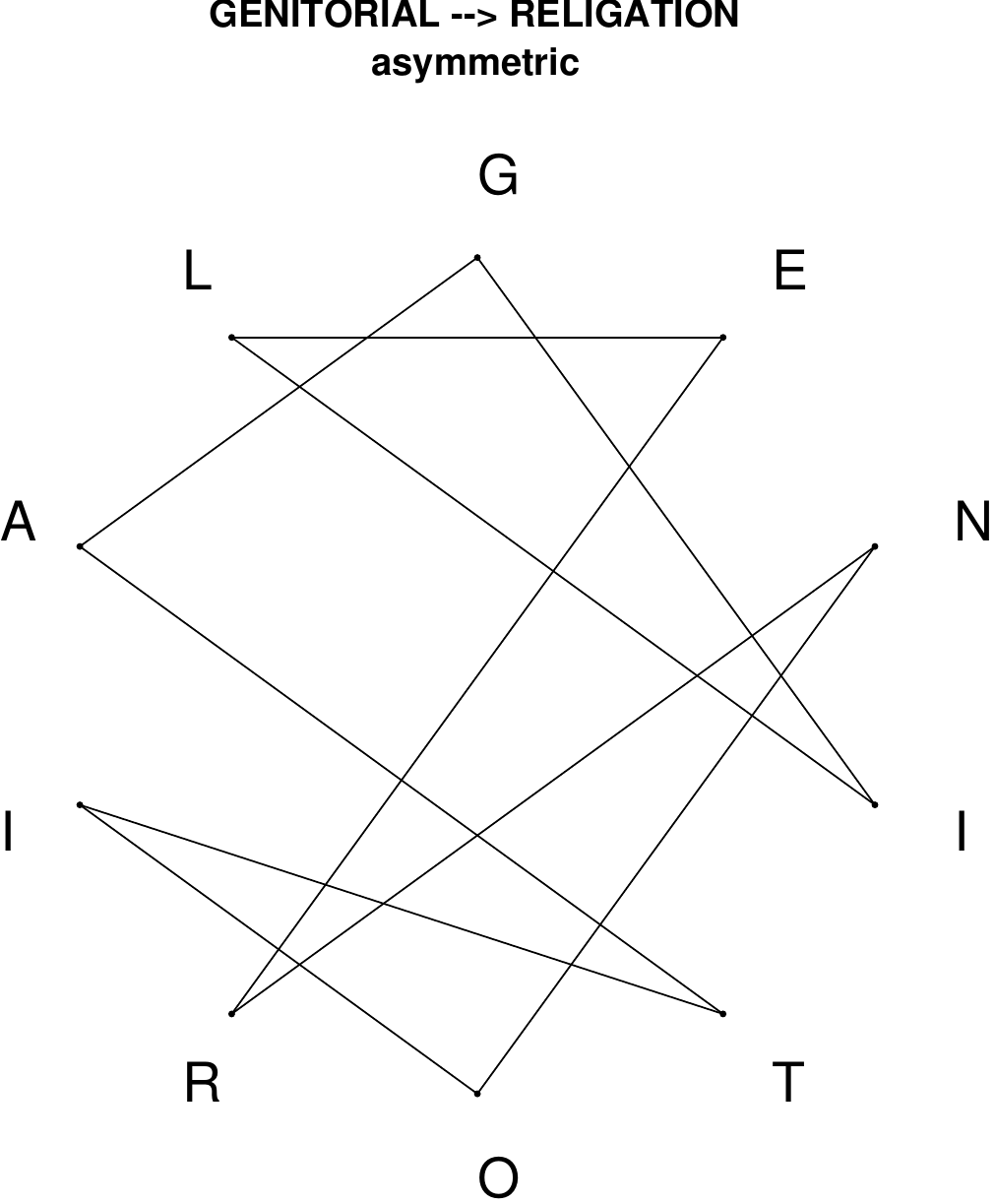}
\end{subfigure}
\end{figure}

\begin{figure}[H]
\centering
\begin{subfigure}[T]{0.19\textwidth}
\centering
\includegraphics[width=\textwidth]{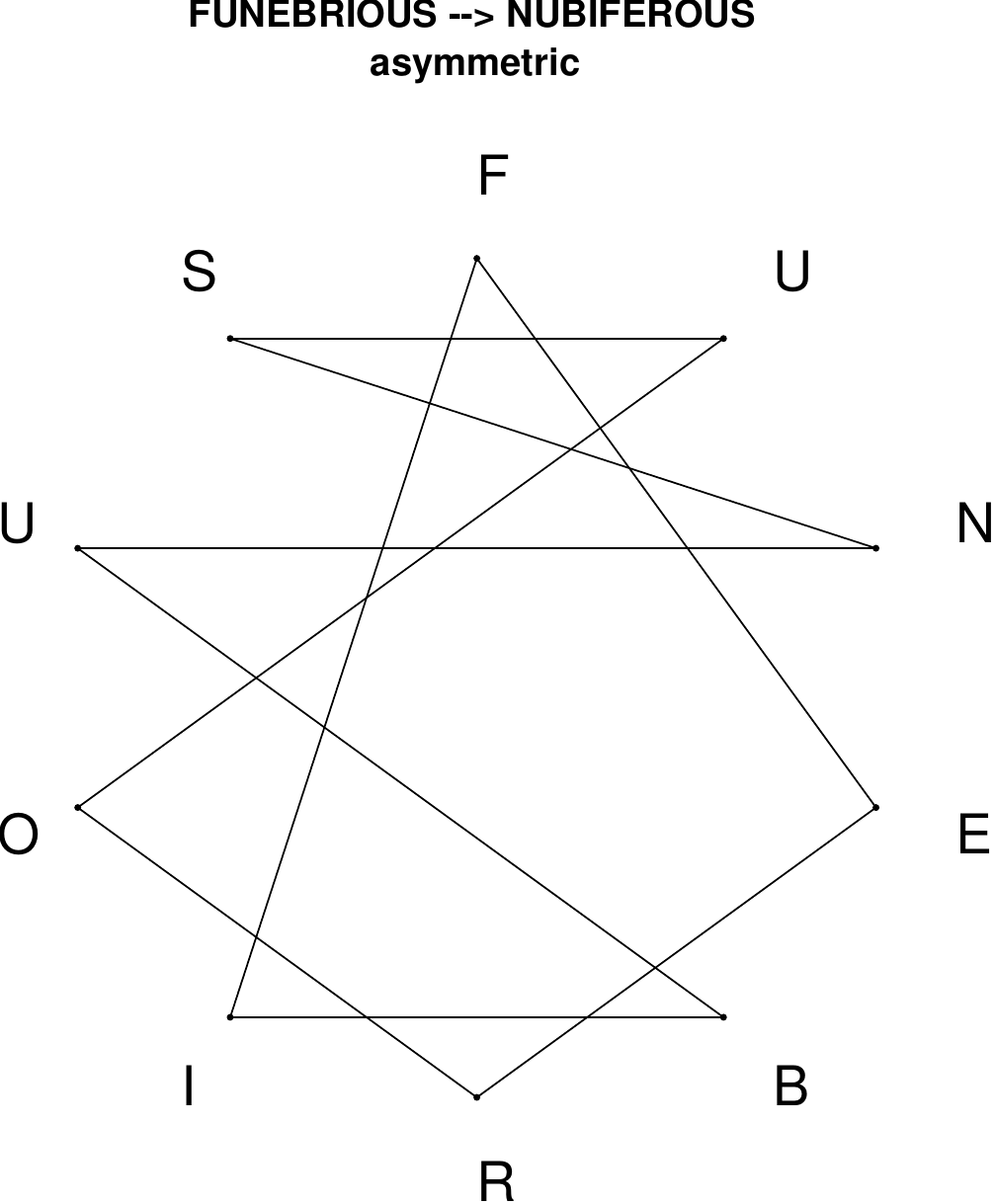}
\end{subfigure}
\hfill
\begin{subfigure}[T]{0.19\textwidth}
\centering
\includegraphics[width=\textwidth]{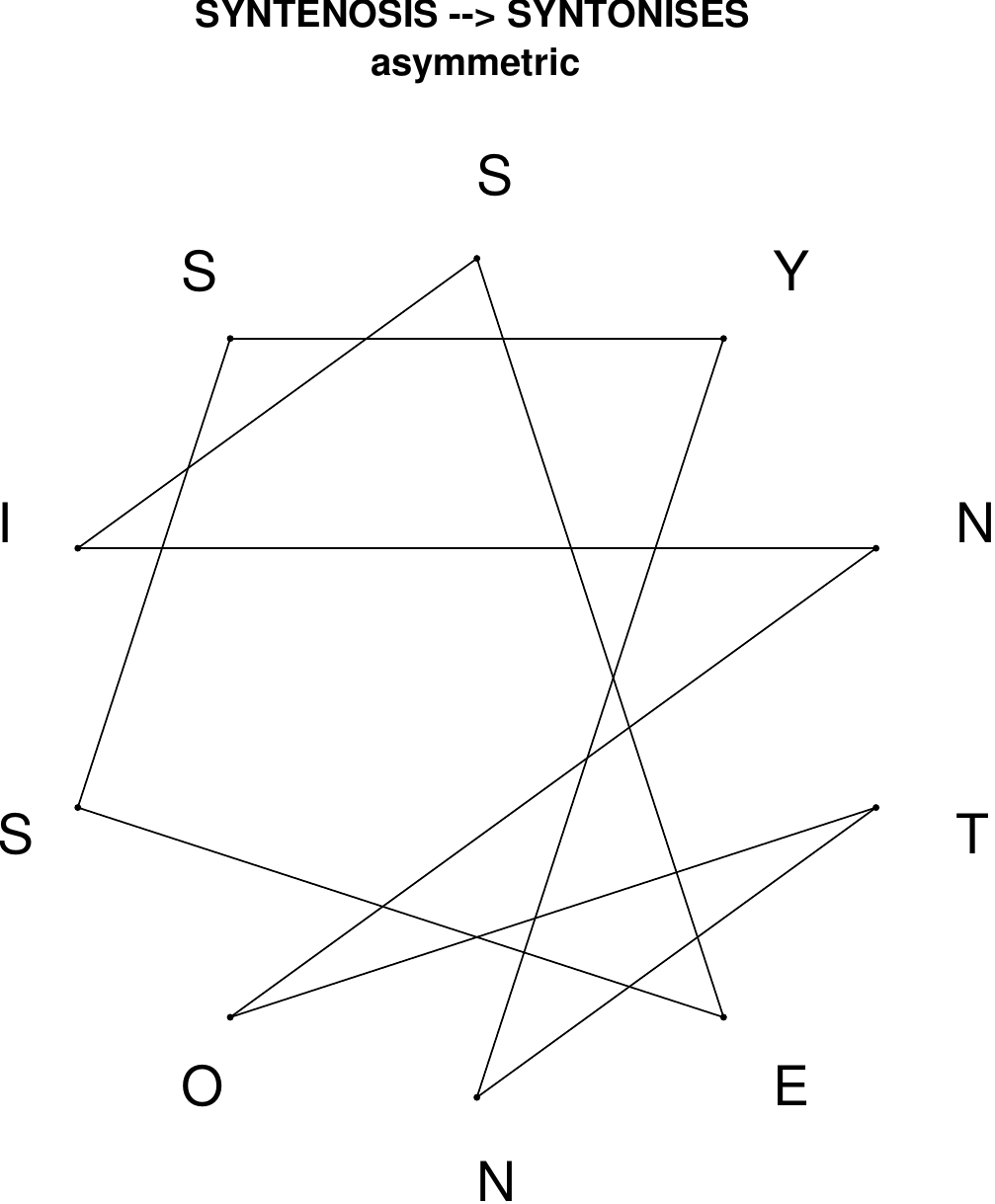}
\end{subfigure}
\hfill
\begin{subfigure}[T]{0.19\textwidth}
\centering
\includegraphics[width=\textwidth]{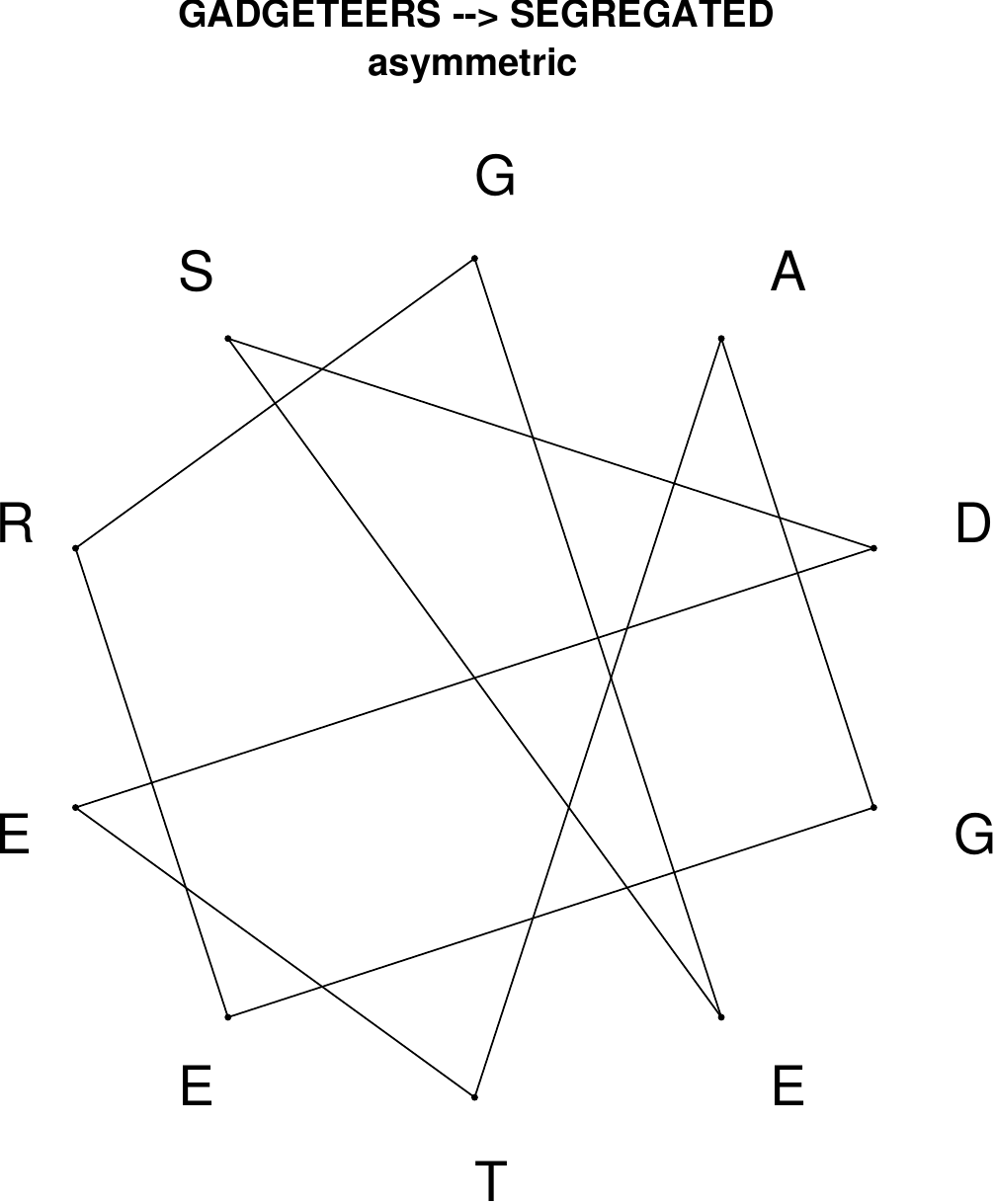}
\end{subfigure}
\hfill
\begin{subfigure}[T]{0.19\textwidth}
\centering
\includegraphics[width=\textwidth]{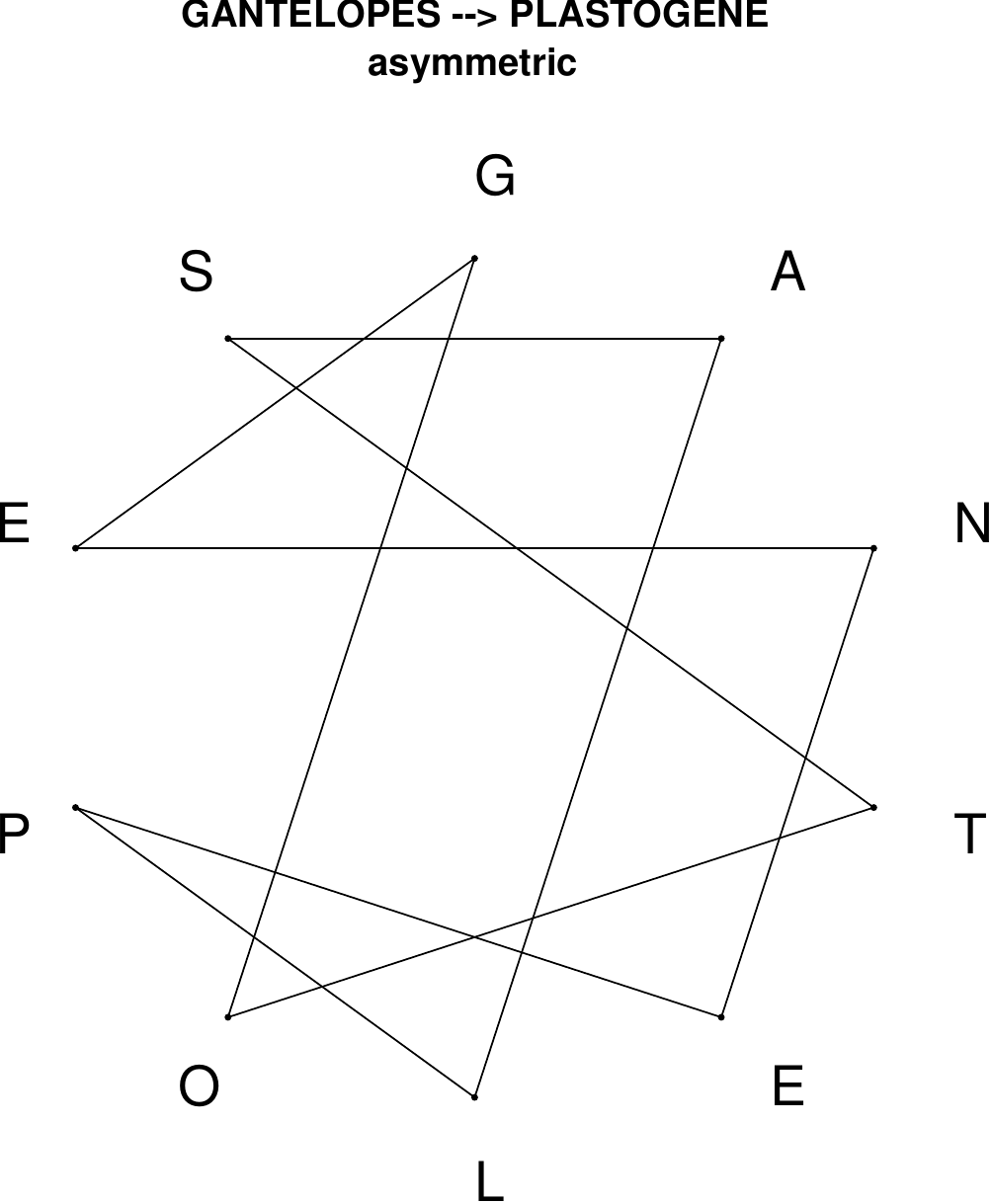}
\end{subfigure}
\hfill
\begin{subfigure}[T]{0.19\textwidth}
\centering
\includegraphics[width=\textwidth]{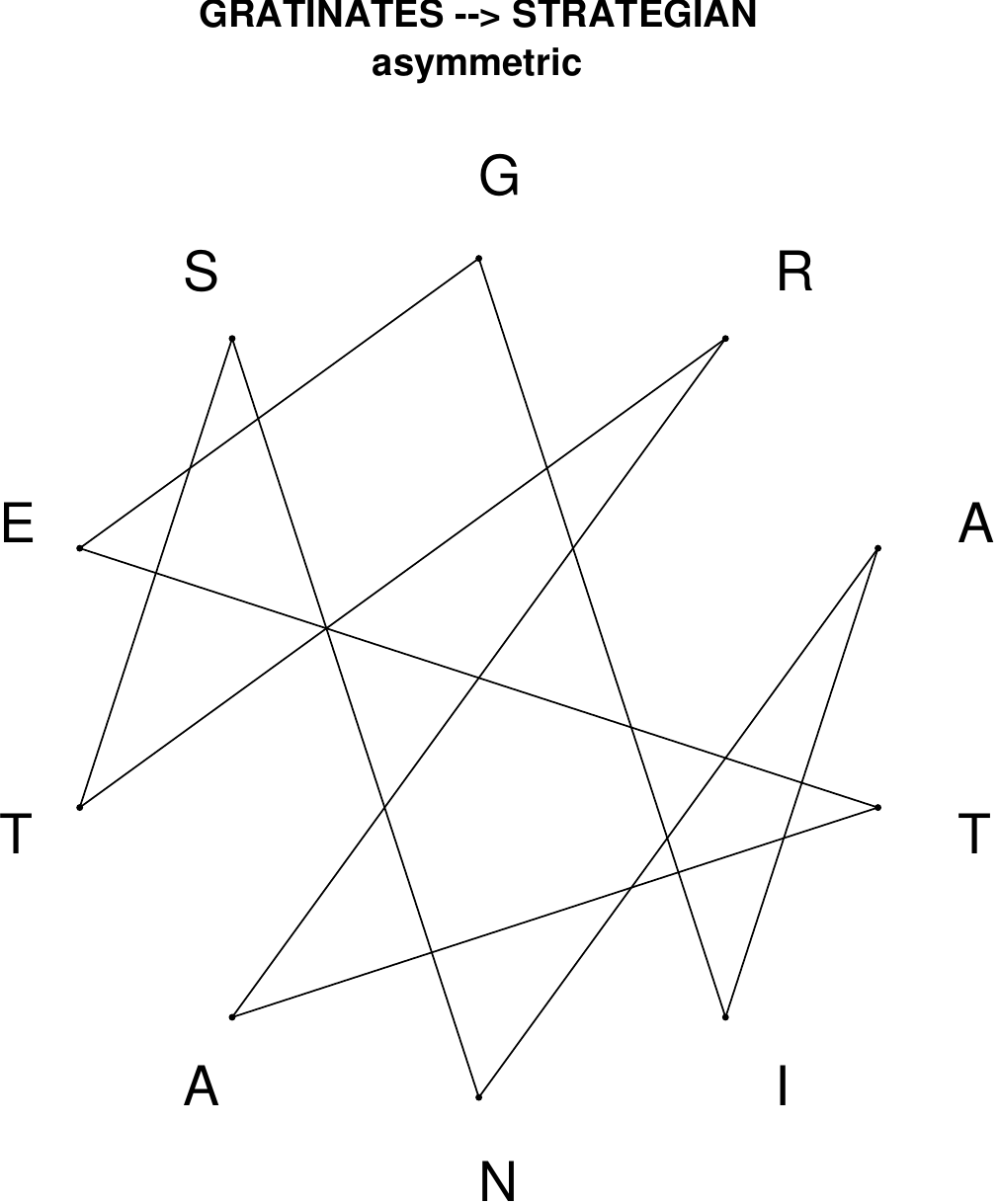}
\end{subfigure}
\end{figure}

\begin{figure}[H]
\centering
\begin{subfigure}[T]{0.19\textwidth}
\centering
\includegraphics[width=\textwidth]{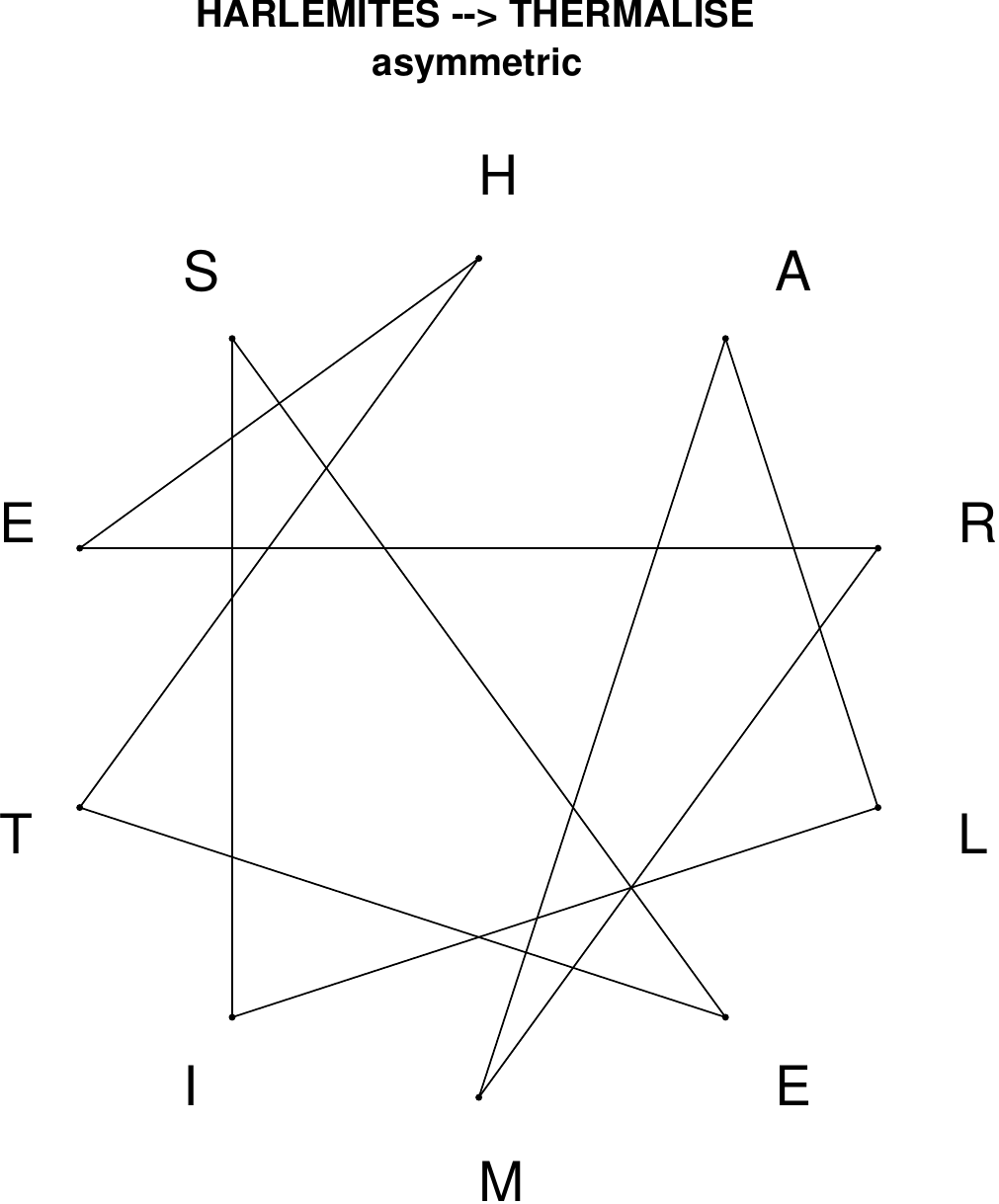}
\end{subfigure}
\hfill
\begin{subfigure}[T]{0.19\textwidth}
\centering
\includegraphics[width=\textwidth]{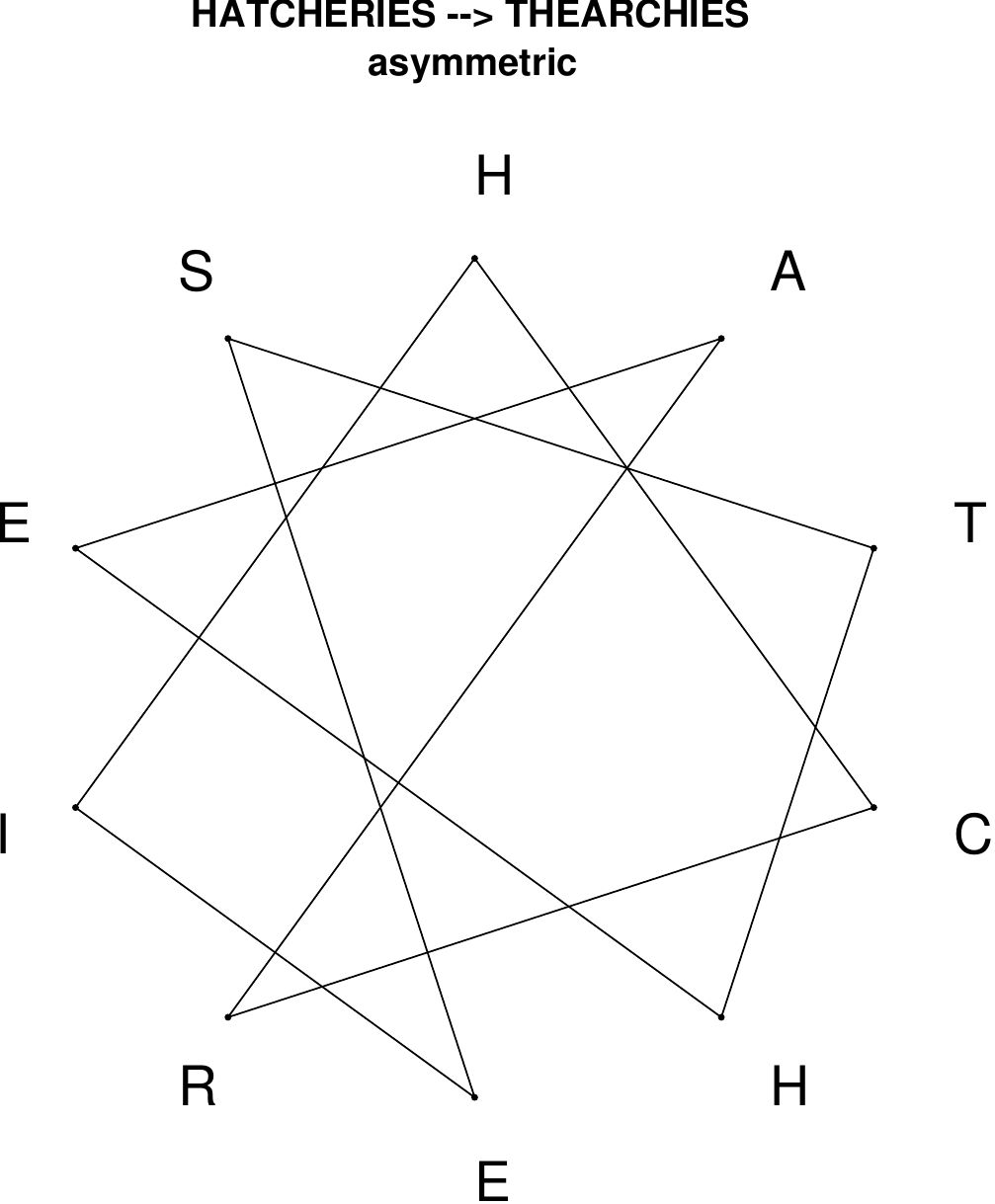}
\end{subfigure}
\hfill
\begin{subfigure}[T]{0.19\textwidth}
\centering
\includegraphics[width=\textwidth]{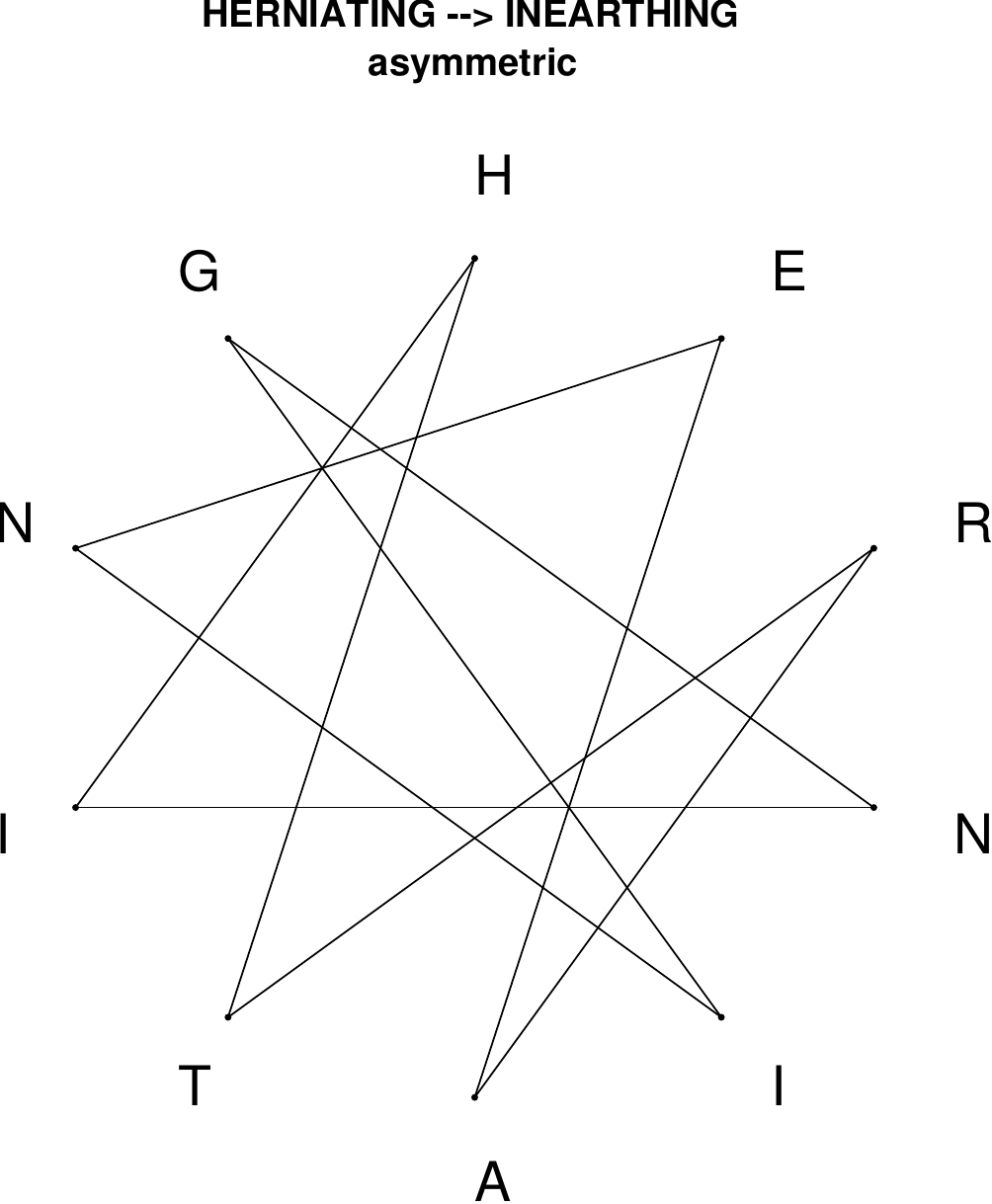}
\end{subfigure}
\hfill
\begin{subfigure}[T]{0.19\textwidth}
\centering
\includegraphics[width=\textwidth]{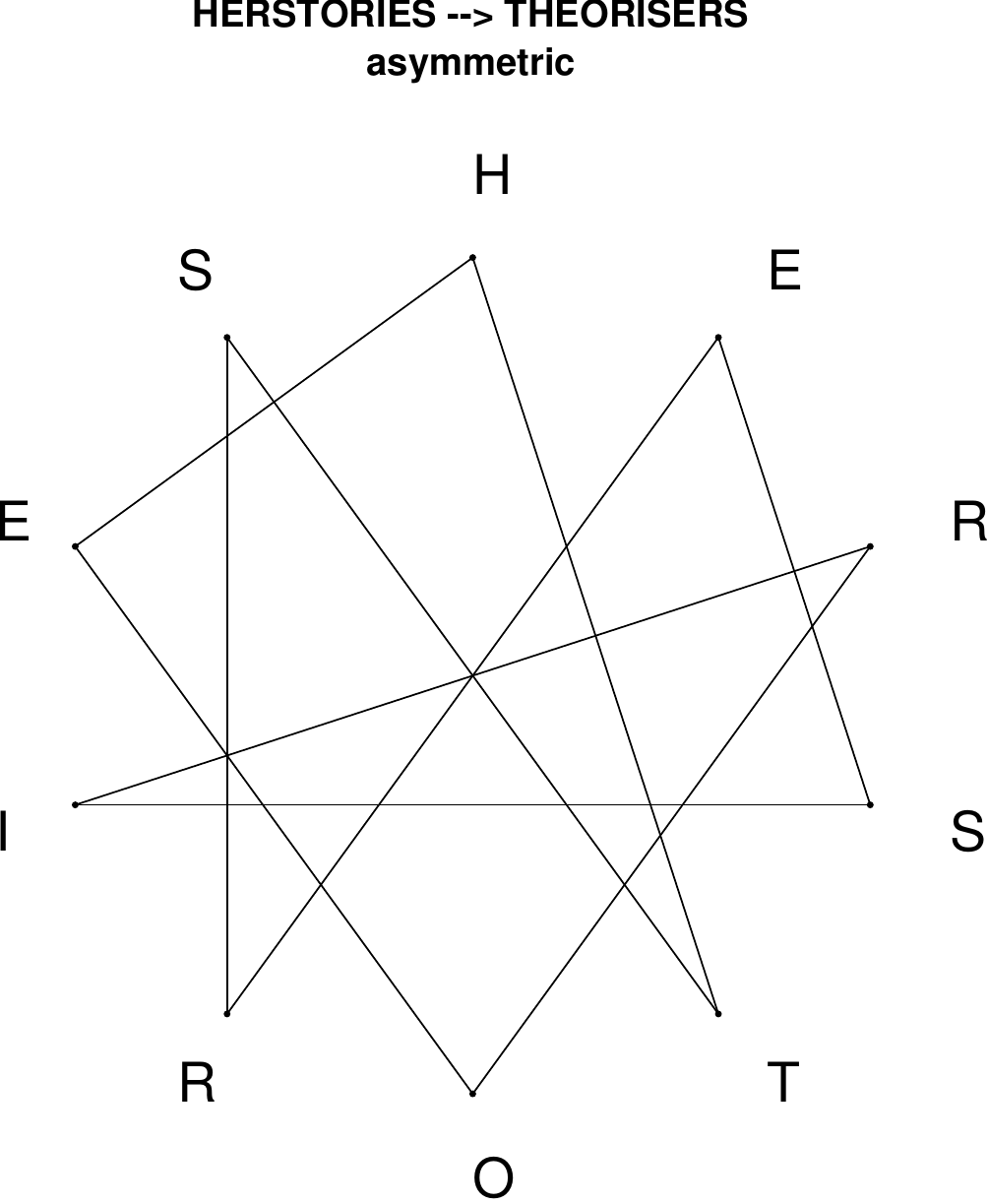}
\end{subfigure}
\hfill
\begin{subfigure}[T]{0.19\textwidth}
\centering
\includegraphics[width=\textwidth]{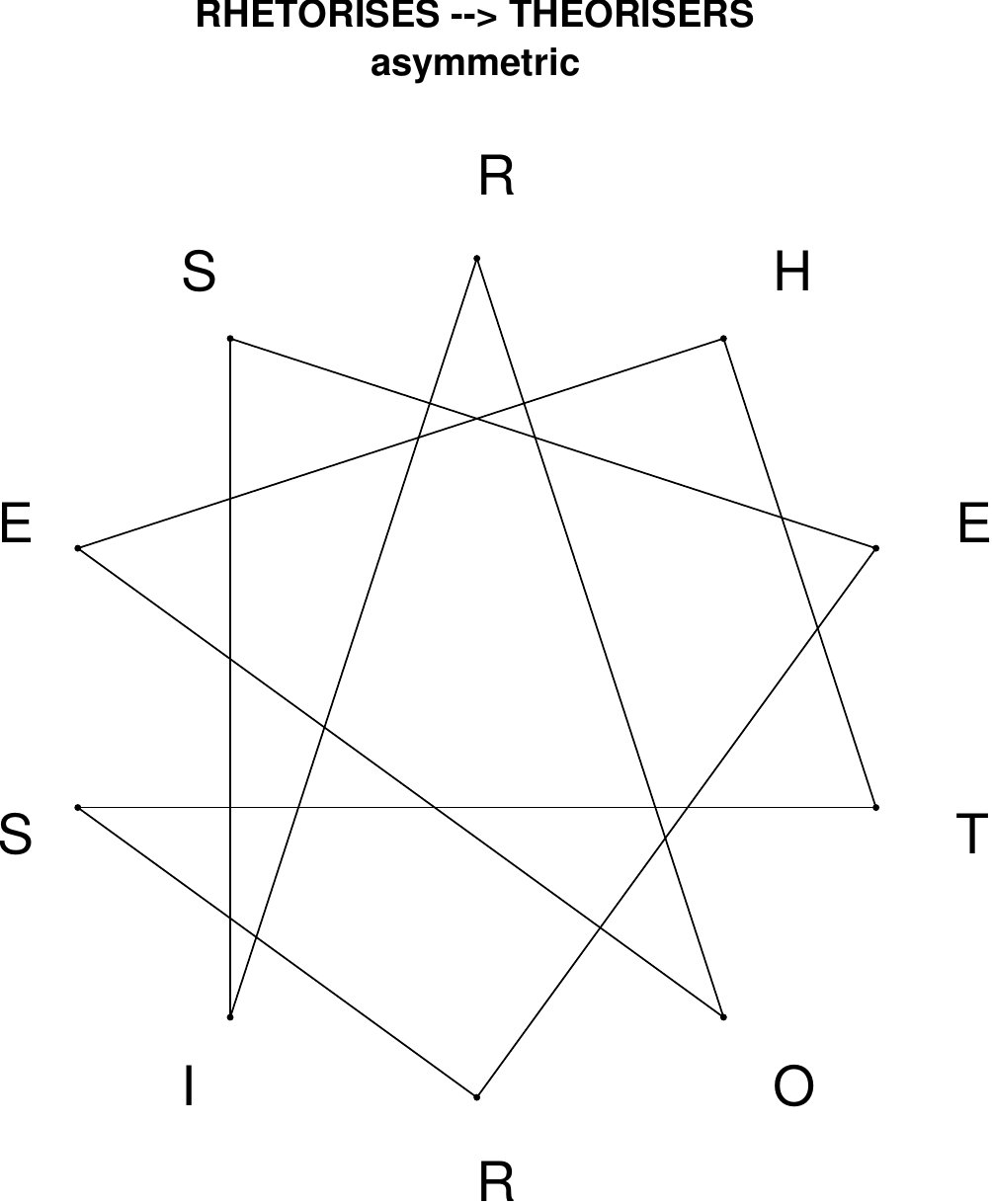}
\end{subfigure}
\end{figure}

\begin{figure}[H]
\centering
\begin{subfigure}[T]{0.19\textwidth}
\centering
\includegraphics[width=\textwidth]{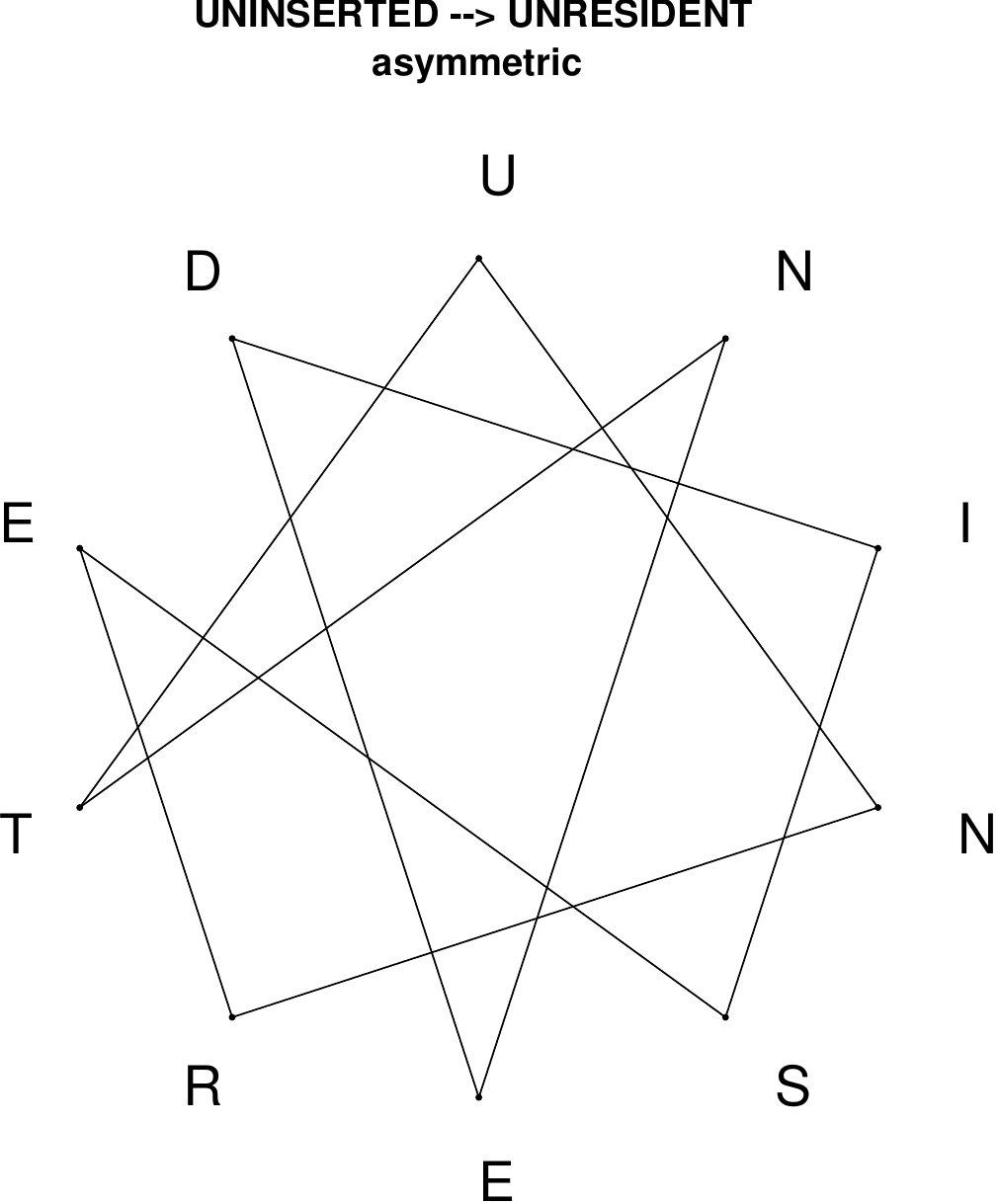}
\end{subfigure}
\hfill
\begin{subfigure}[T]{0.19\textwidth}
\centering
\includegraphics[width=\textwidth]{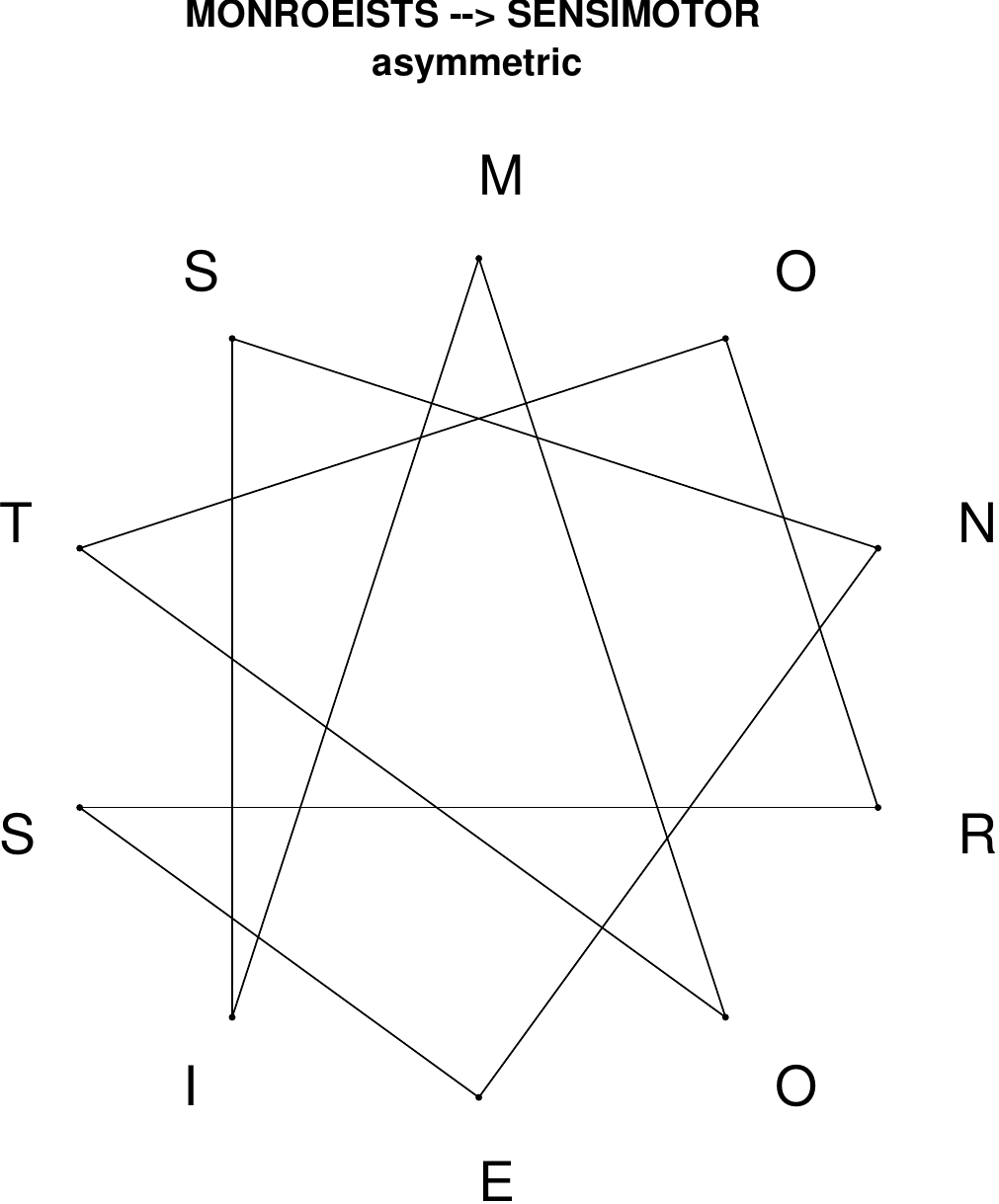}
\end{subfigure}
\hfill
\begin{subfigure}[T]{0.19\textwidth}
\centering
\includegraphics[width=\textwidth]{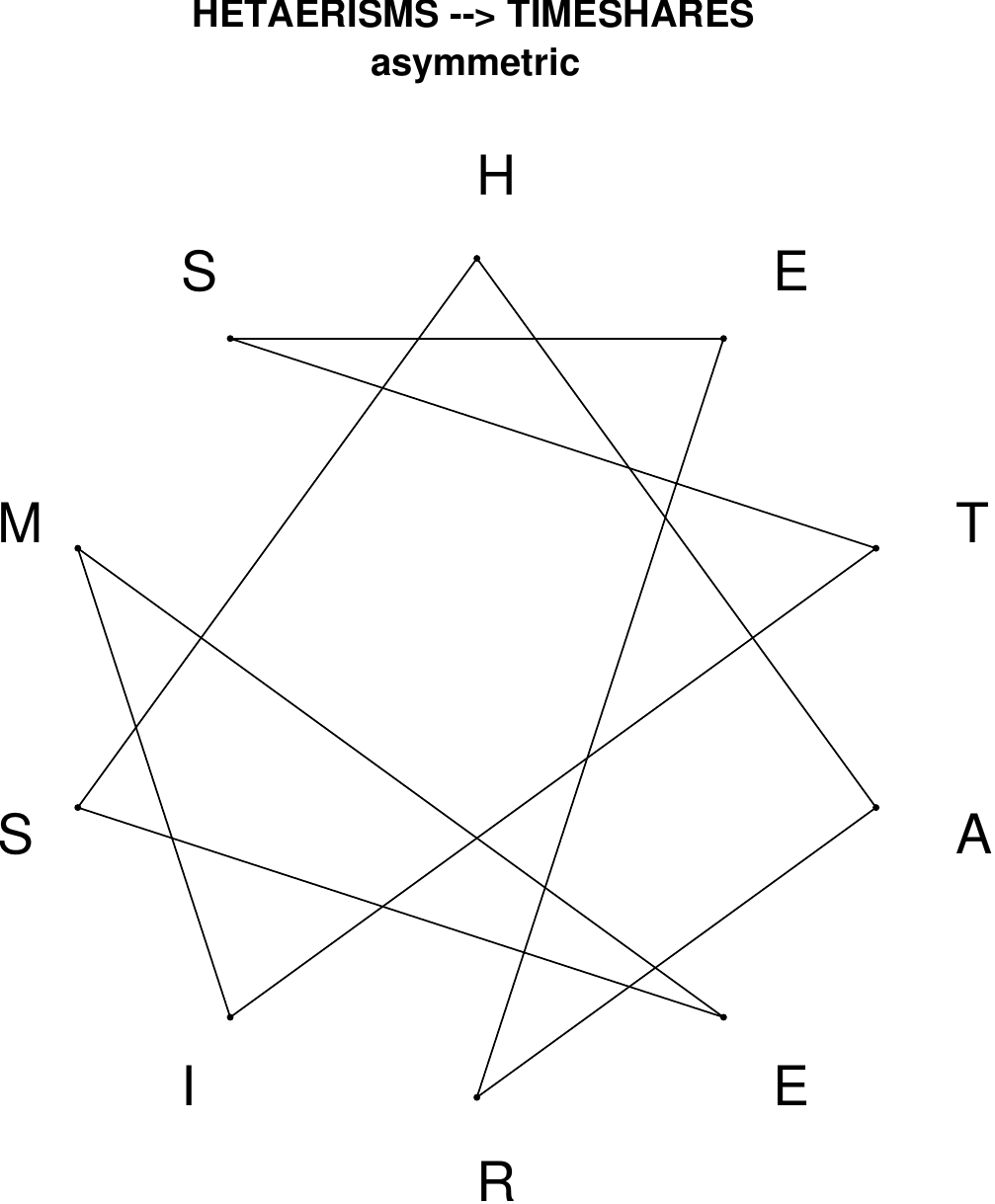}
\end{subfigure}
\hfill
\begin{subfigure}[T]{0.19\textwidth}
\centering
\includegraphics[width=\textwidth]{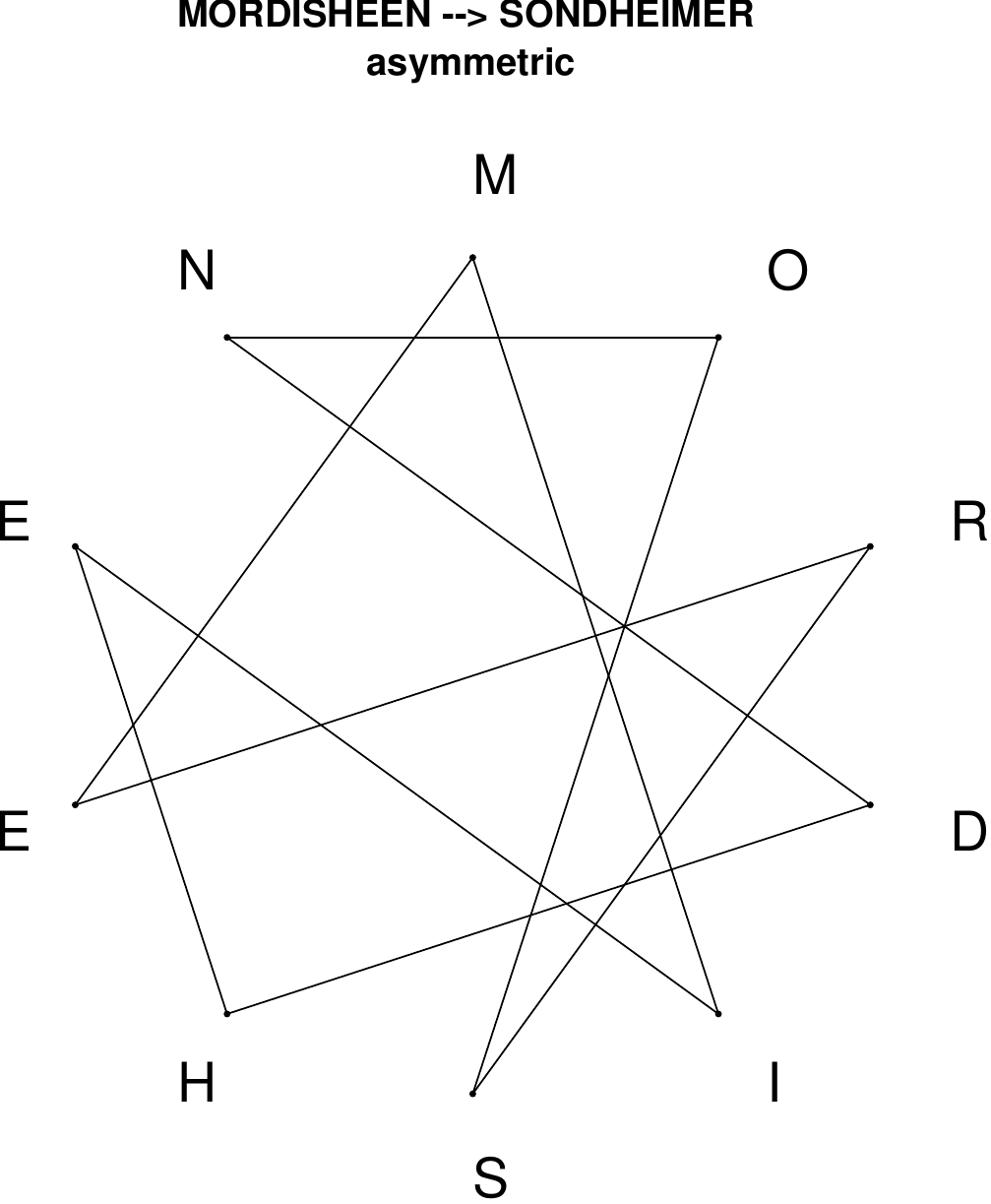}
\end{subfigure}
\hfill
\begin{subfigure}[T]{0.19\textwidth}
\centering
\includegraphics[width=\textwidth]{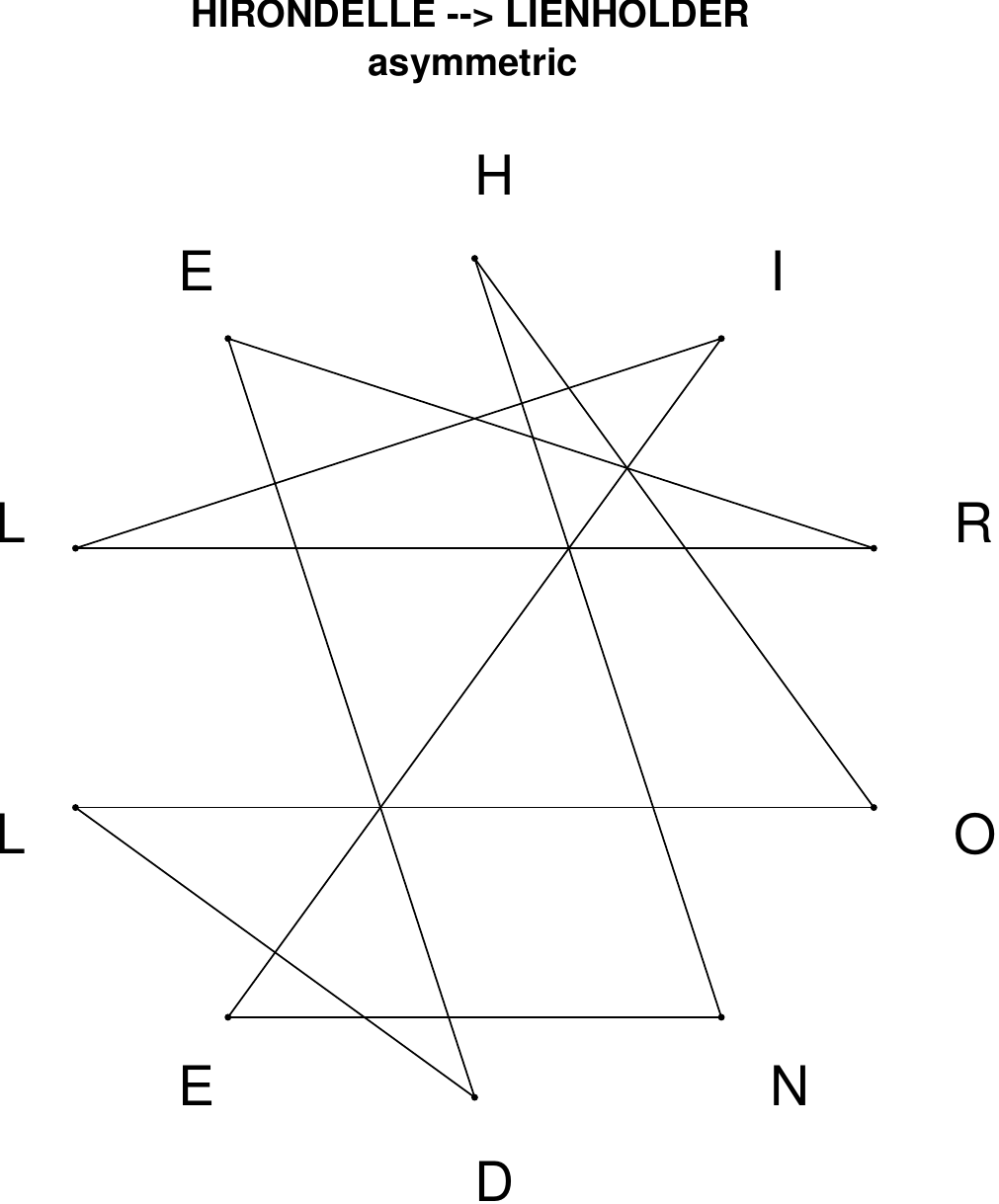}
\end{subfigure}
\end{figure}

\begin{figure}[H]
\centering
\begin{subfigure}[T]{0.19\textwidth}
\centering
\includegraphics[width=\textwidth]{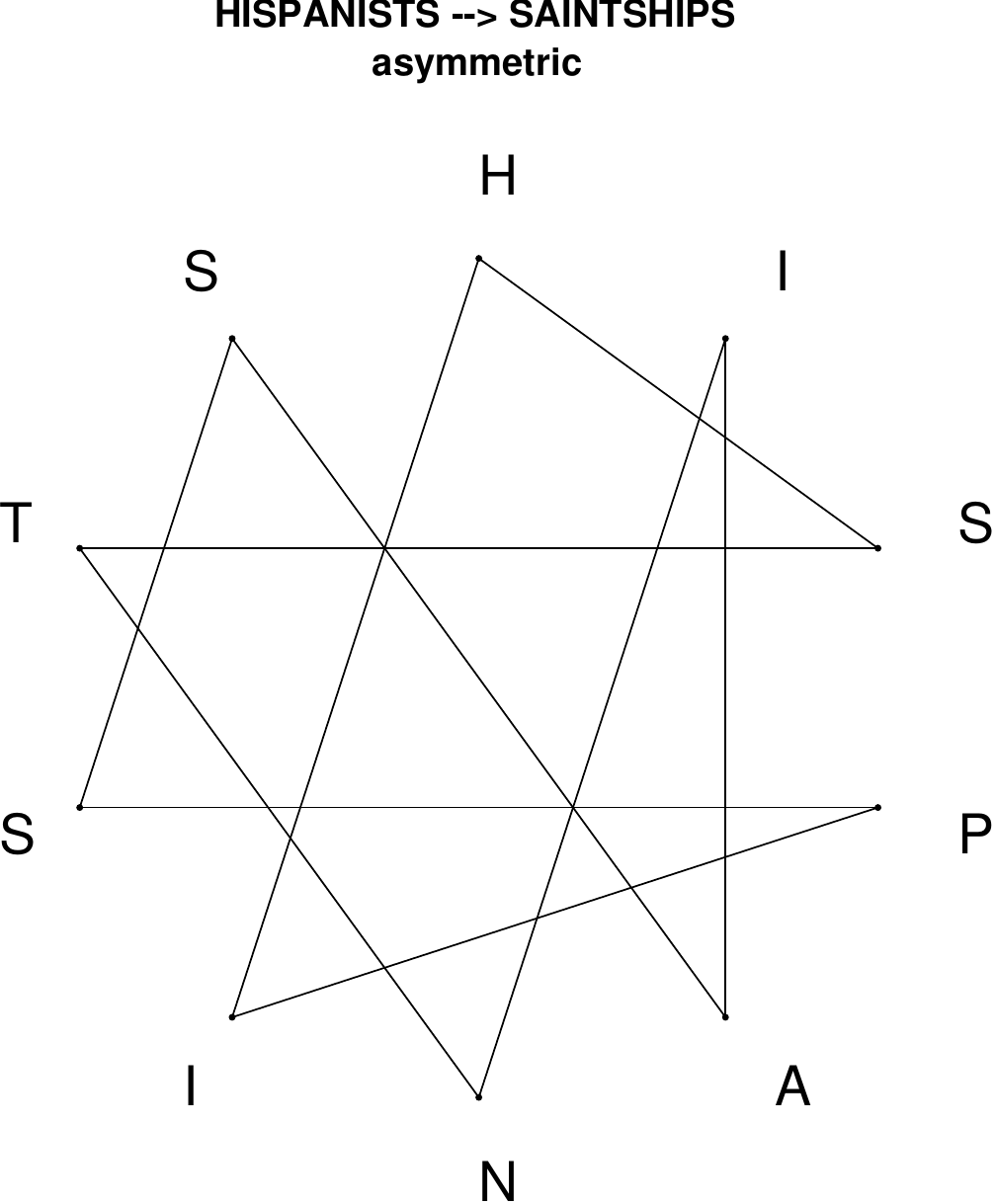}
\end{subfigure}
\hfill
\begin{subfigure}[T]{0.19\textwidth}
\centering
\includegraphics[width=\textwidth]{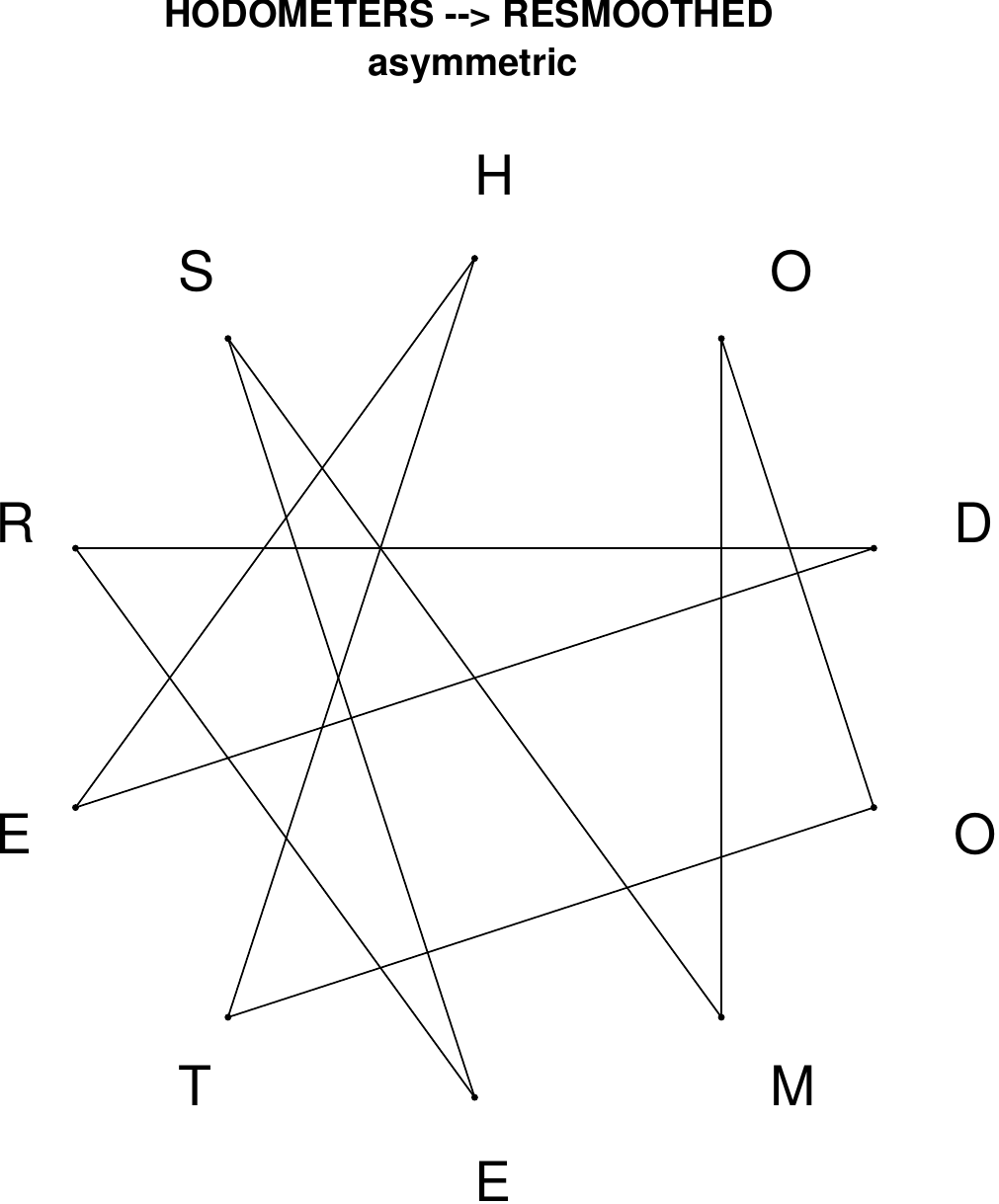}
\end{subfigure}
\hfill
\begin{subfigure}[T]{0.19\textwidth}
\centering
\includegraphics[width=\textwidth]{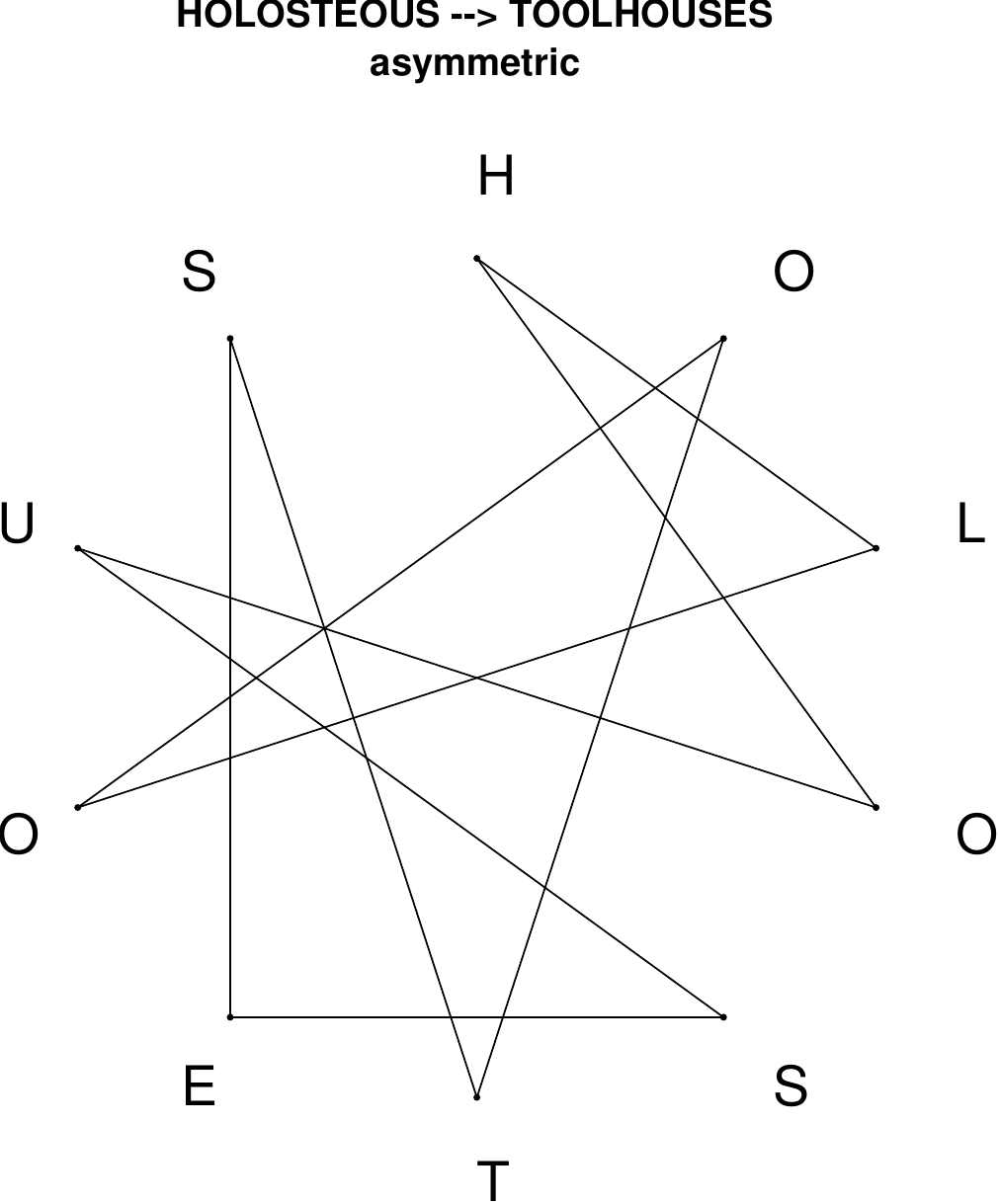}
\end{subfigure}
\hfill
\begin{subfigure}[T]{0.19\textwidth}
\centering
\includegraphics[width=\textwidth]{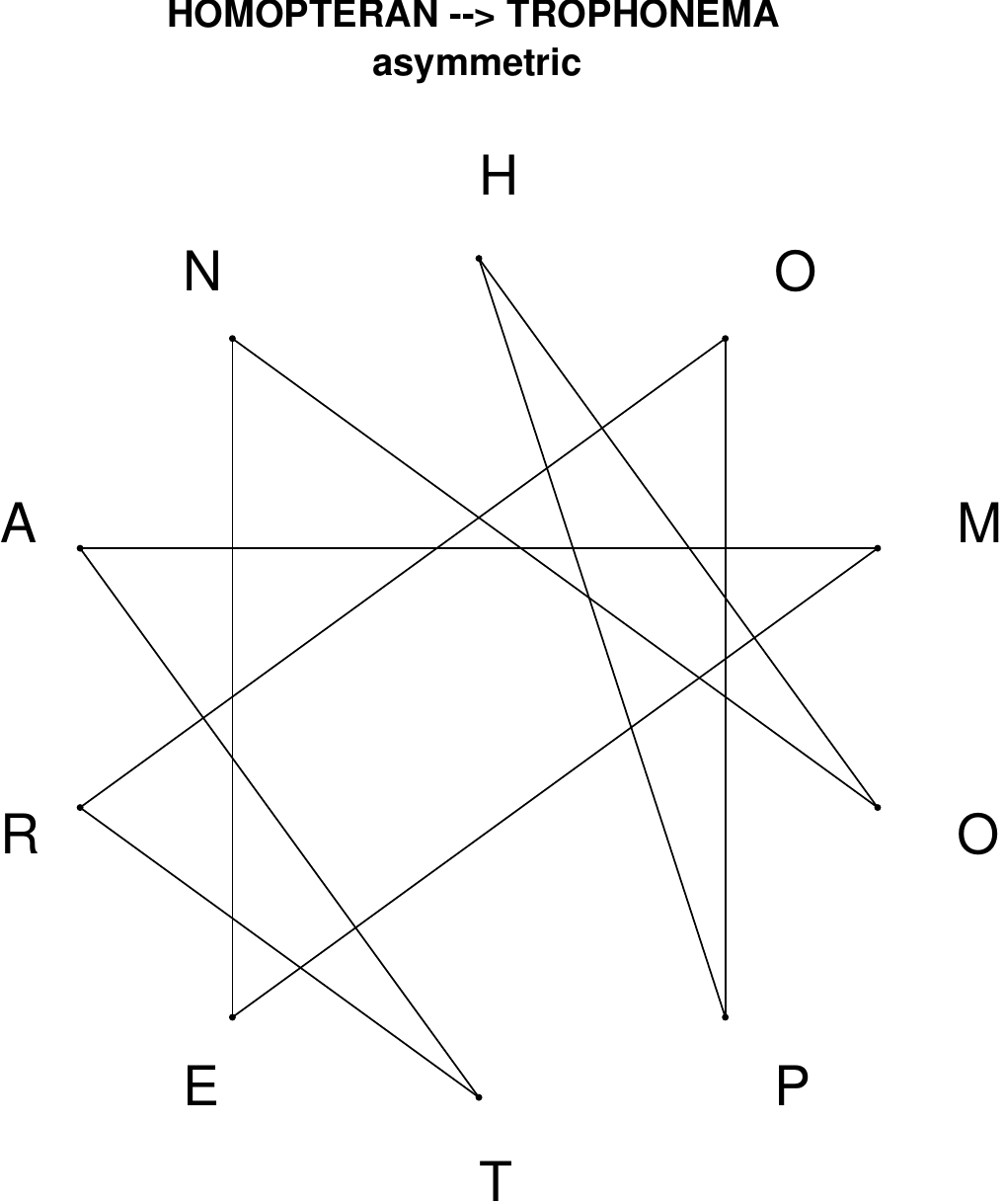}
\end{subfigure}
\hfill
\begin{subfigure}[T]{0.19\textwidth}
\centering
\includegraphics[width=\textwidth]{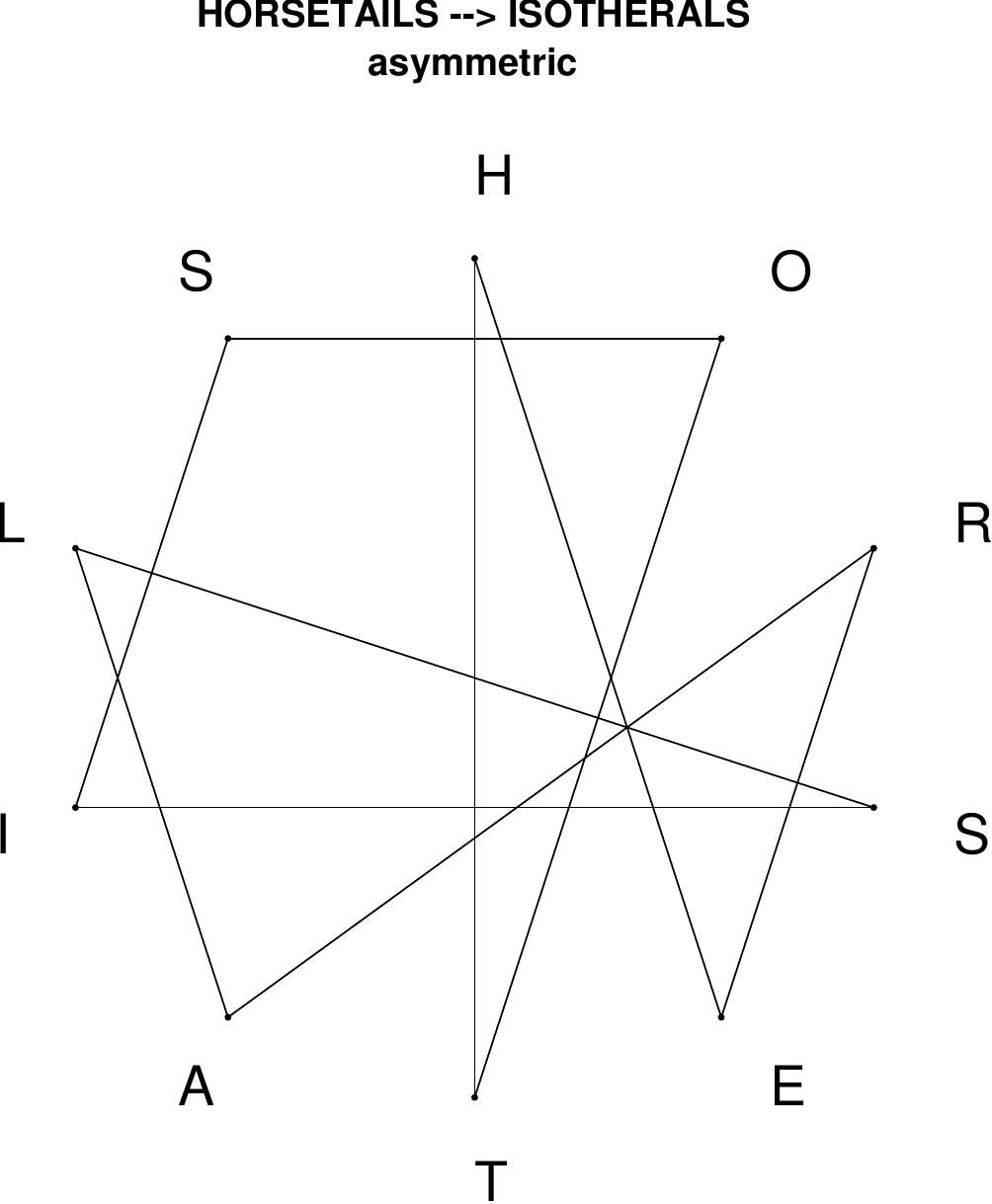}
\end{subfigure}
\end{figure}

\begin{figure}[H]
\centering
\begin{subfigure}[T]{0.19\textwidth}
\centering
\includegraphics[width=\textwidth]{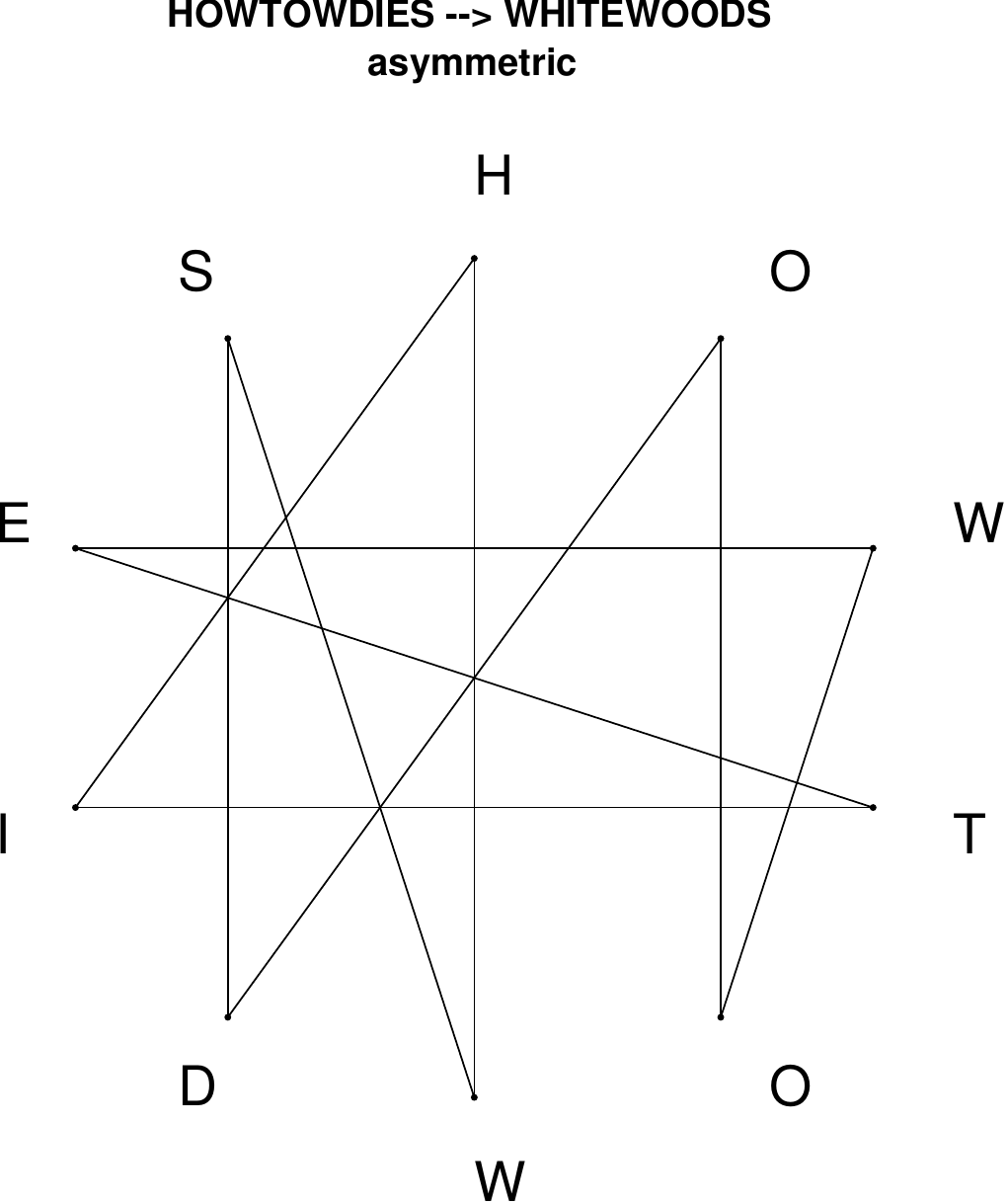}
\end{subfigure}
\hfill
\begin{subfigure}[T]{0.19\textwidth}
\centering
\includegraphics[width=\textwidth]{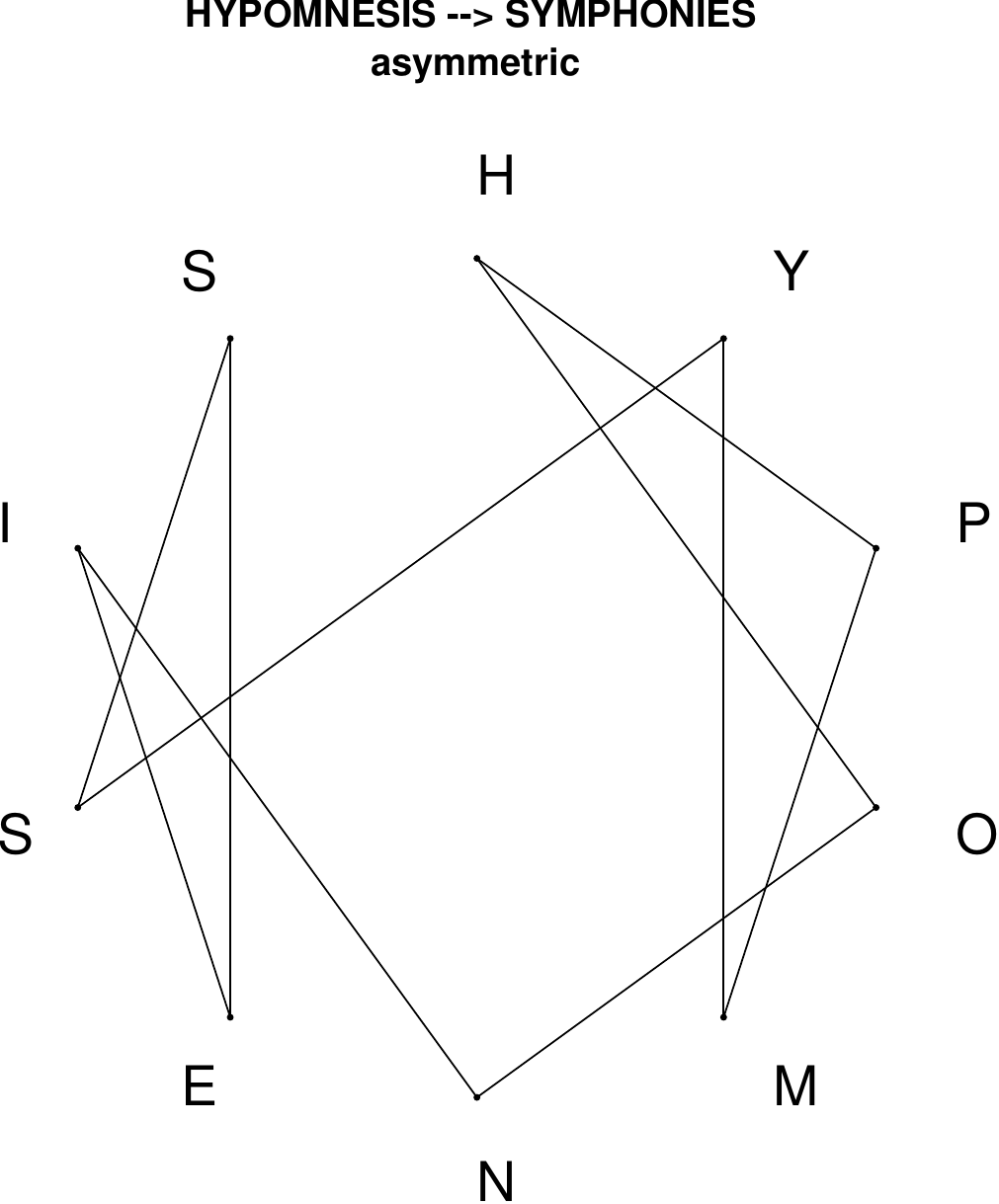}
\end{subfigure}
\hfill
\begin{subfigure}[T]{0.19\textwidth}
\centering
\includegraphics[width=\textwidth]{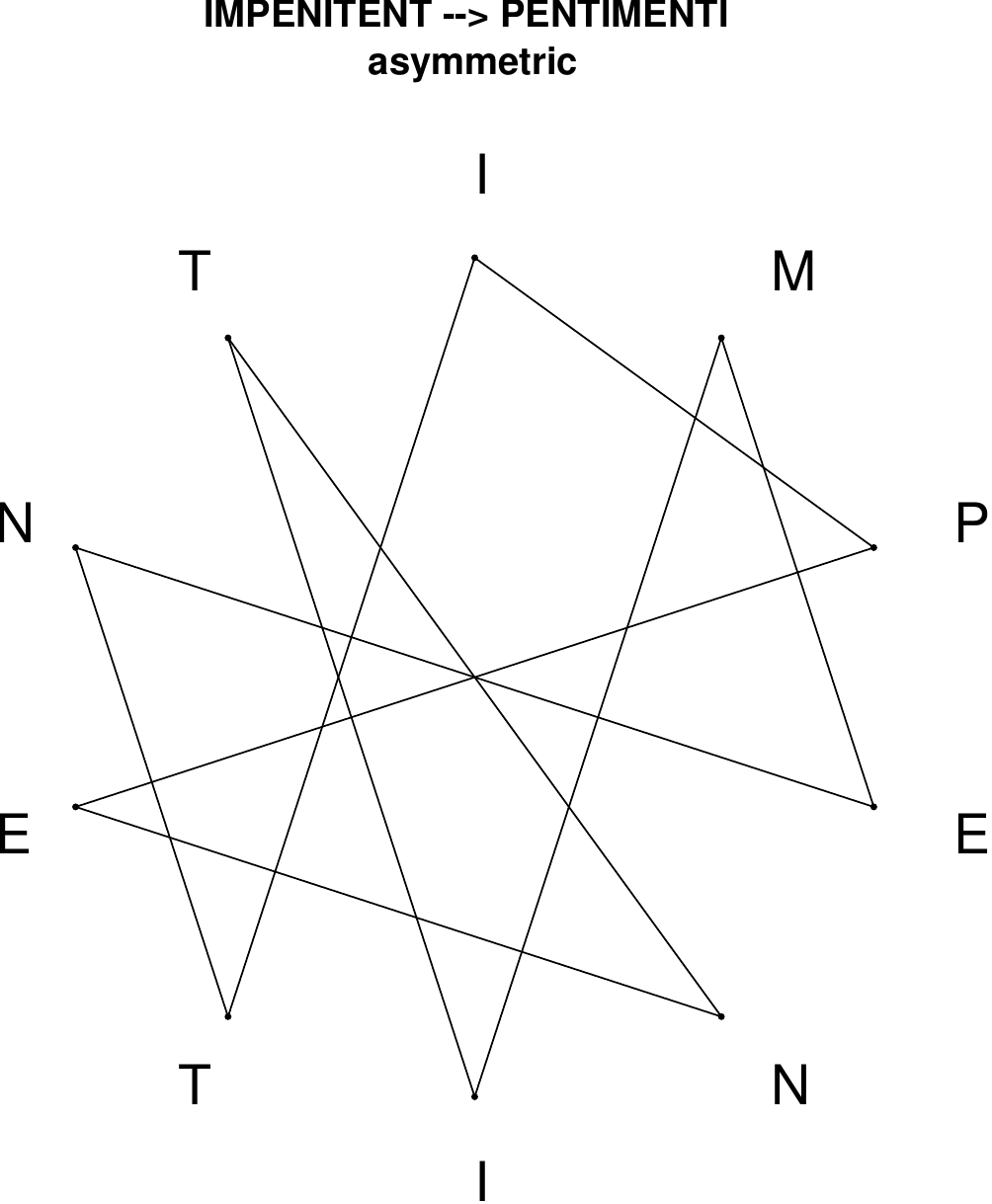}
\end{subfigure}
\hfill
\begin{subfigure}[T]{0.19\textwidth}
\centering
\includegraphics[width=\textwidth]{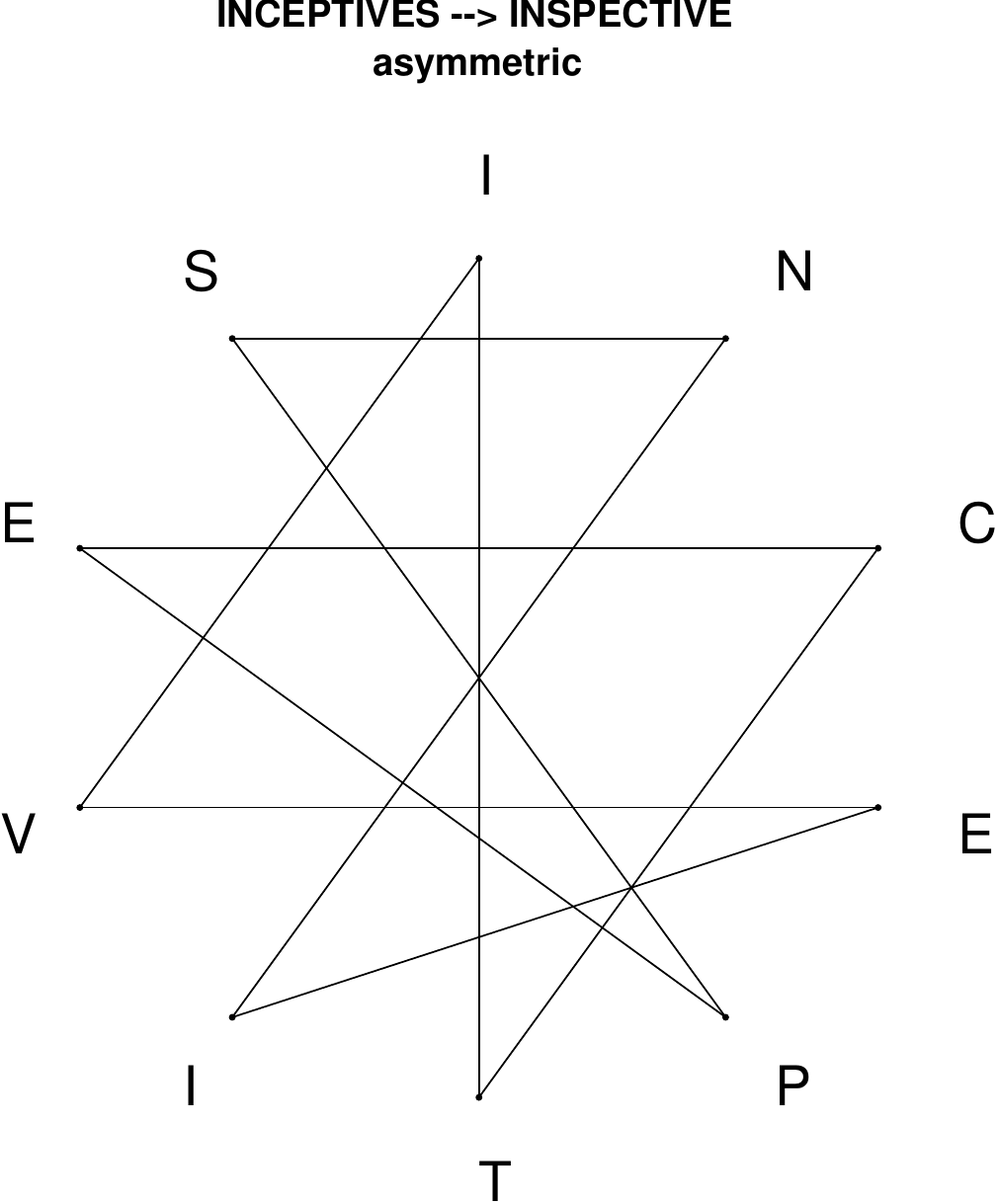}
\end{subfigure}
\hfill
\begin{subfigure}[T]{0.19\textwidth}
\centering
\includegraphics[width=\textwidth]{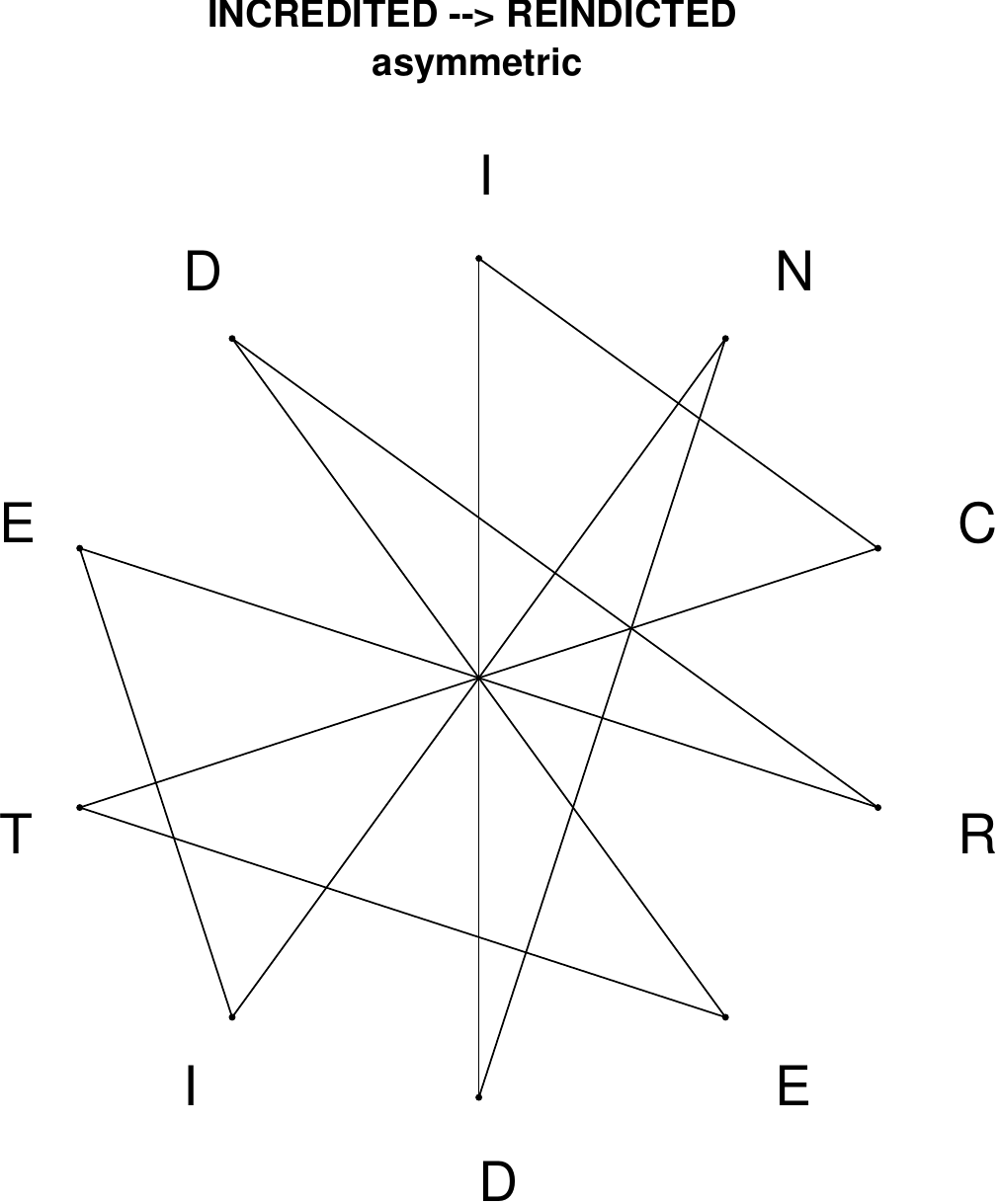}
\end{subfigure}
\end{figure}

\begin{figure}[H]
\centering
\begin{subfigure}[T]{0.19\textwidth}
\centering
\includegraphics[width=\textwidth]{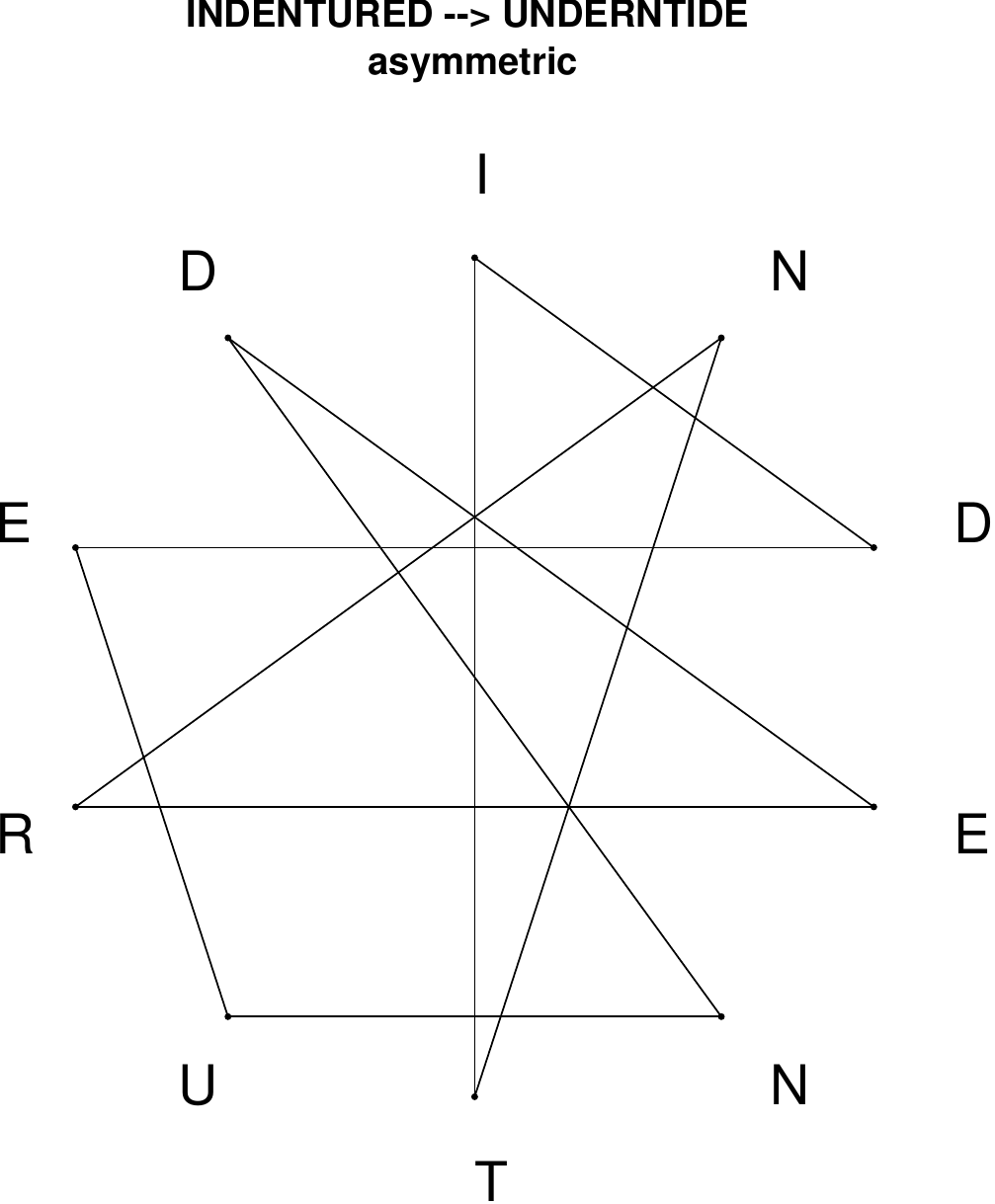}
\end{subfigure}
\hfill
\begin{subfigure}[T]{0.19\textwidth}
\centering
\includegraphics[width=\textwidth]{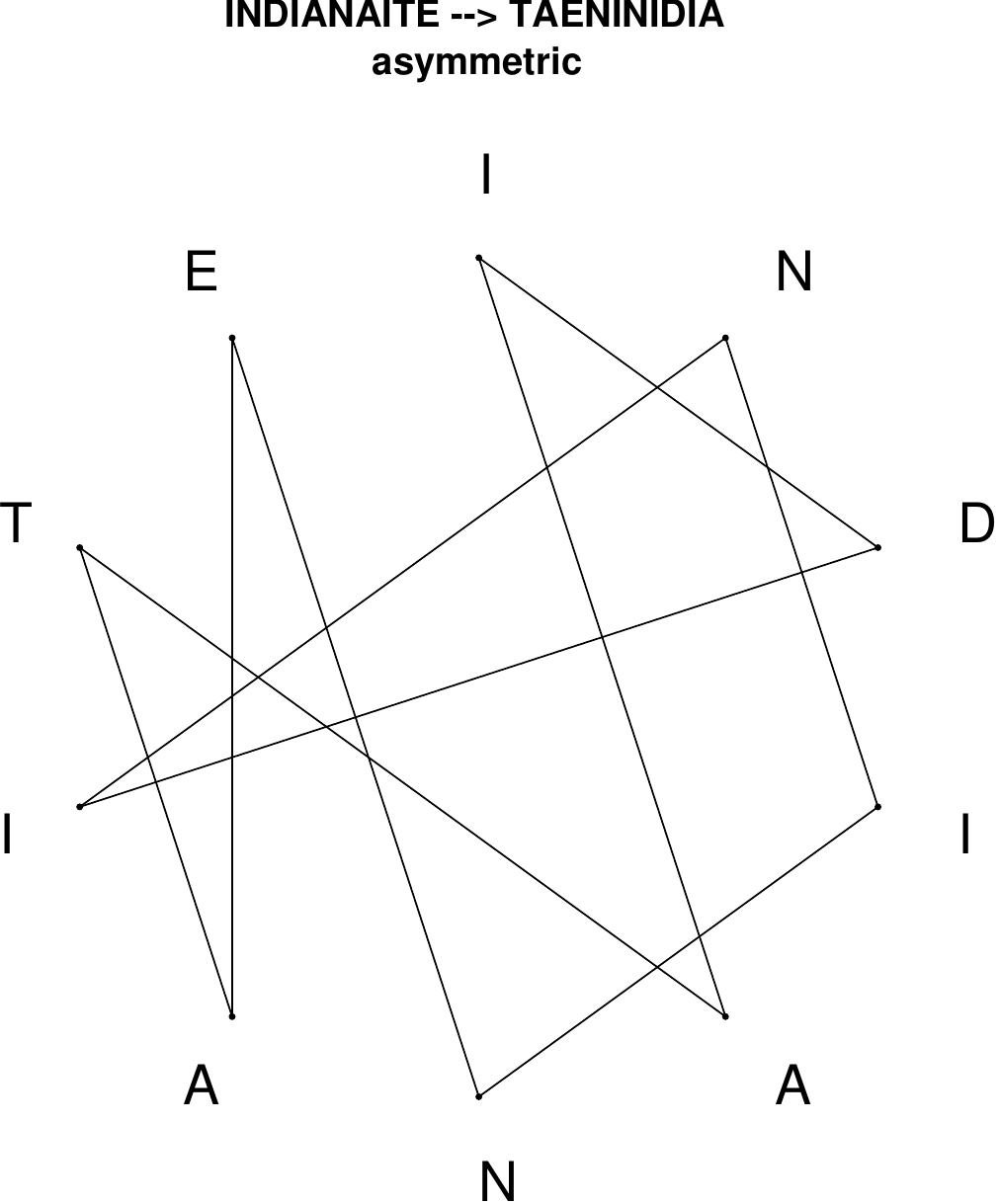}
\end{subfigure}
\hfill
\begin{subfigure}[T]{0.19\textwidth}
\centering
\includegraphics[width=\textwidth]{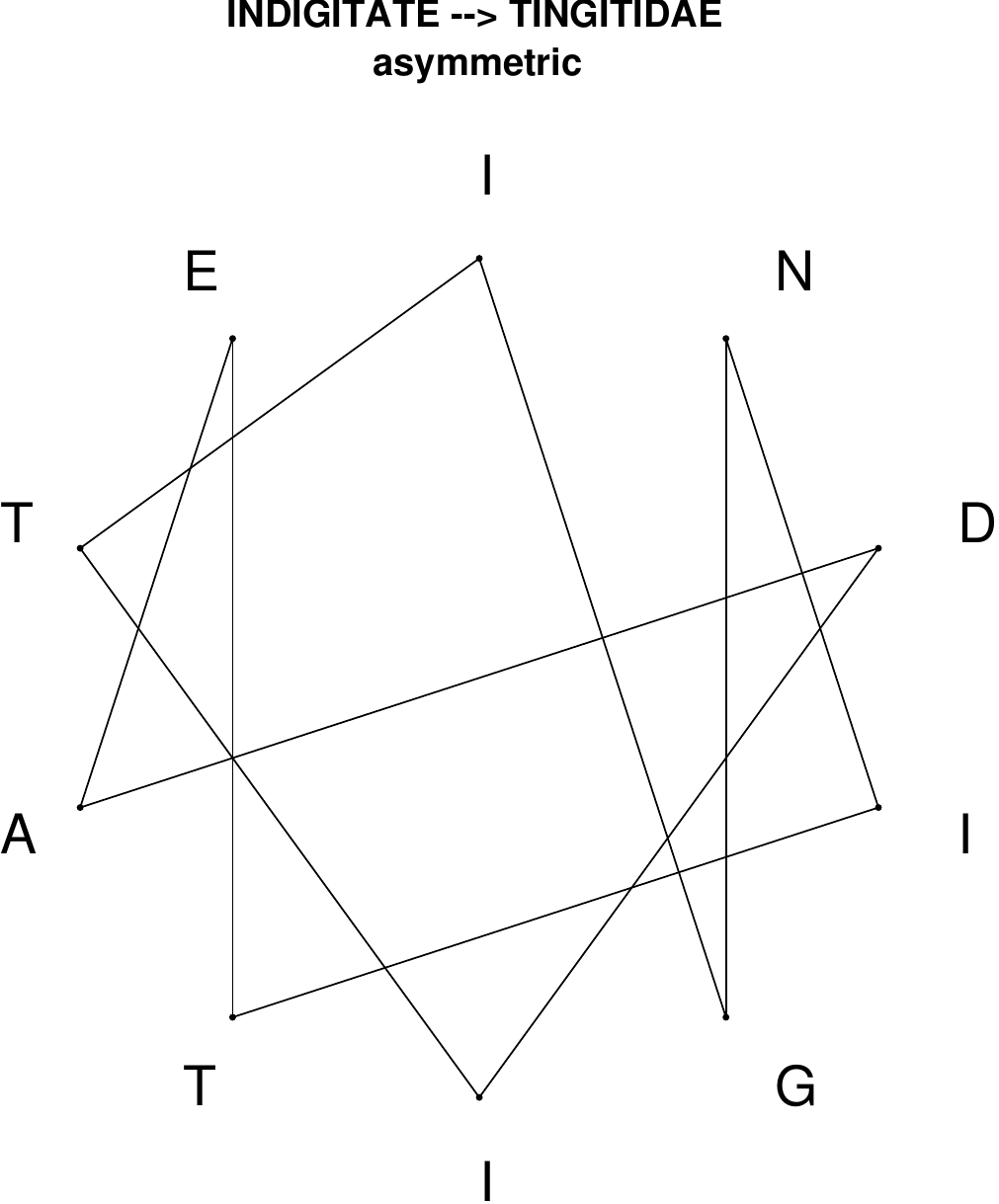}
\end{subfigure}
\hfill
\begin{subfigure}[T]{0.19\textwidth}
\centering
\includegraphics[width=\textwidth]{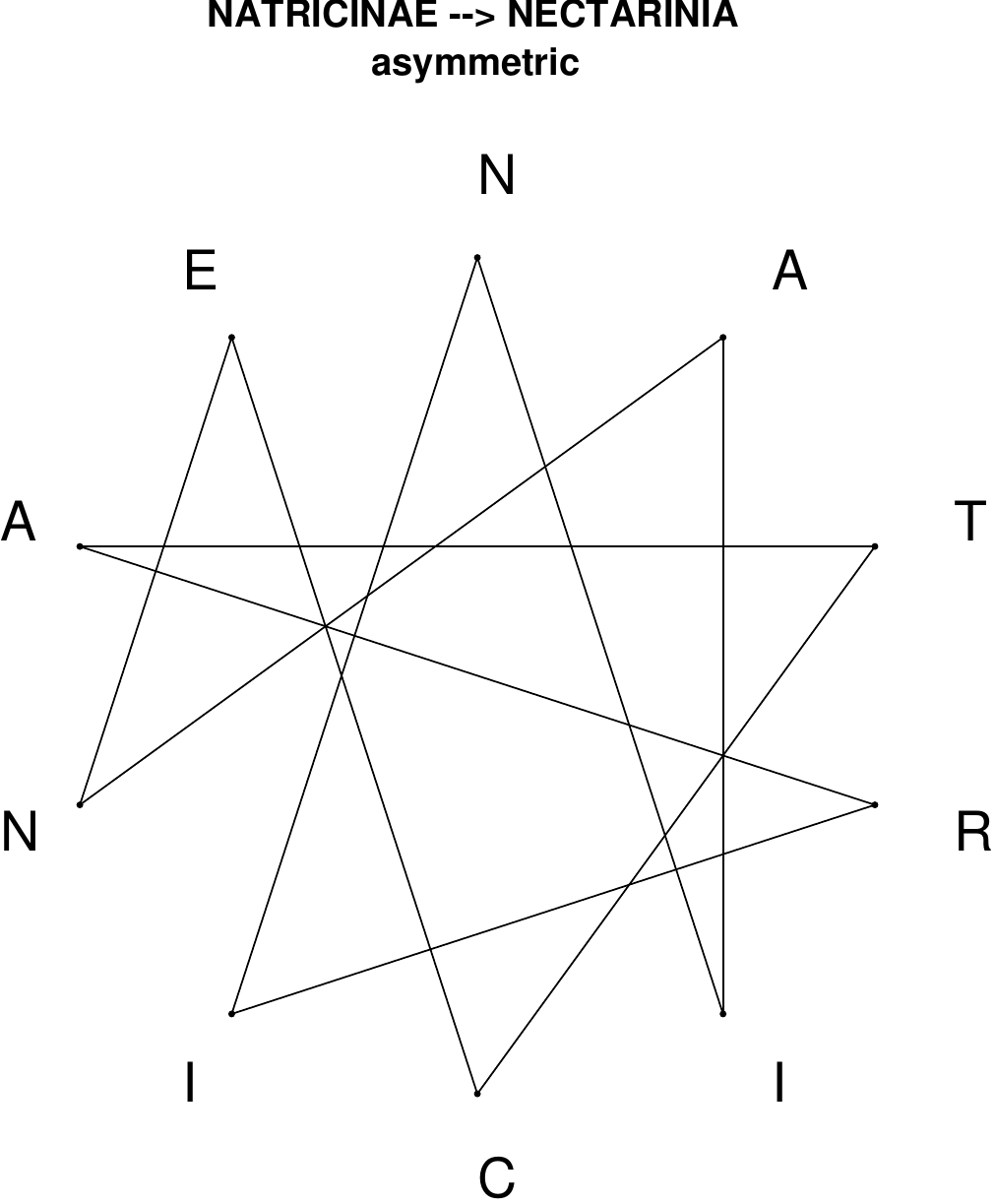}
\end{subfigure}
\hfill
\begin{subfigure}[T]{0.19\textwidth}
\centering
\includegraphics[width=\textwidth]{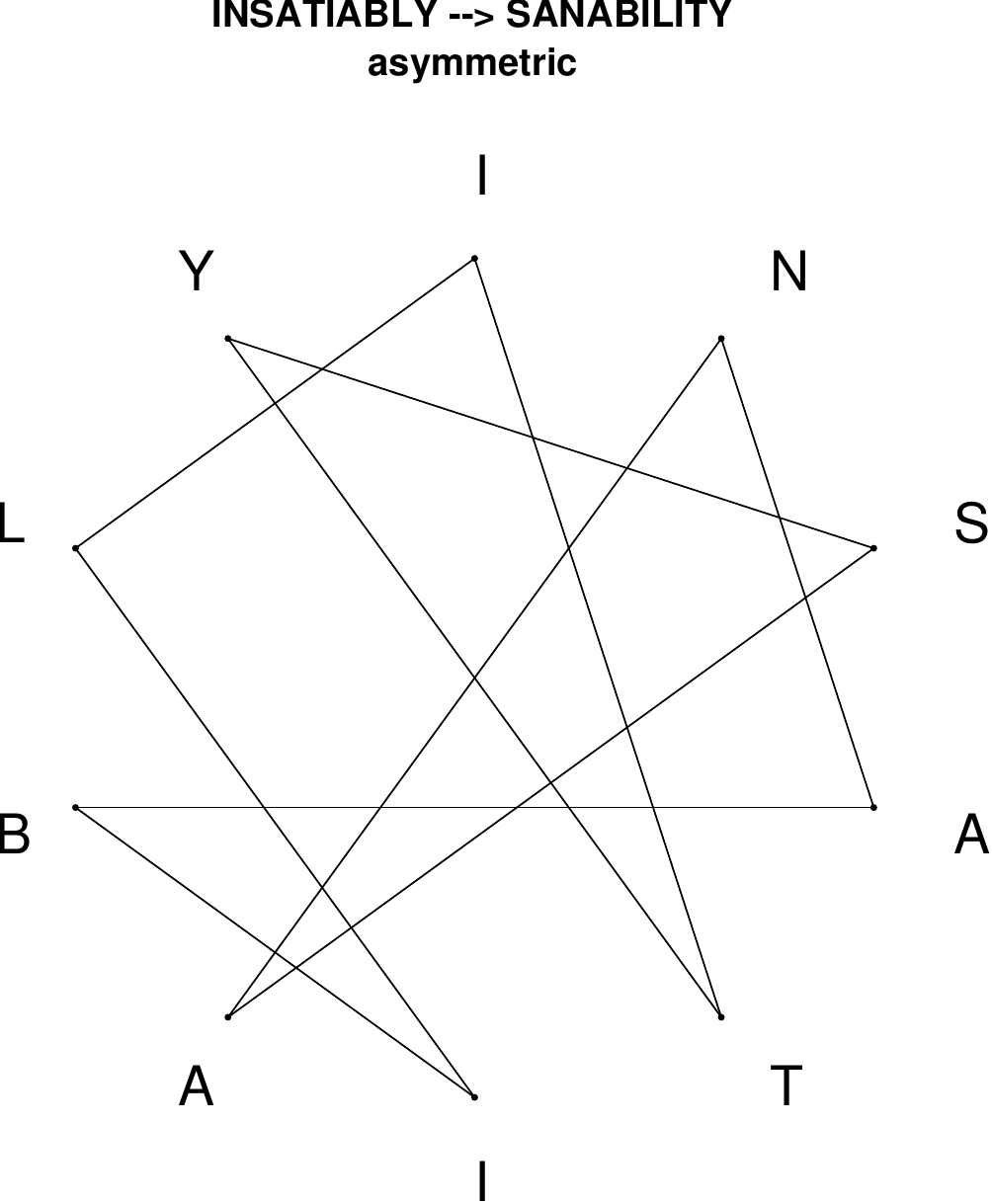}
\end{subfigure}
\end{figure}

\begin{figure}[H]
\centering
\begin{subfigure}[T]{0.19\textwidth}
\centering
\includegraphics[width=\textwidth]{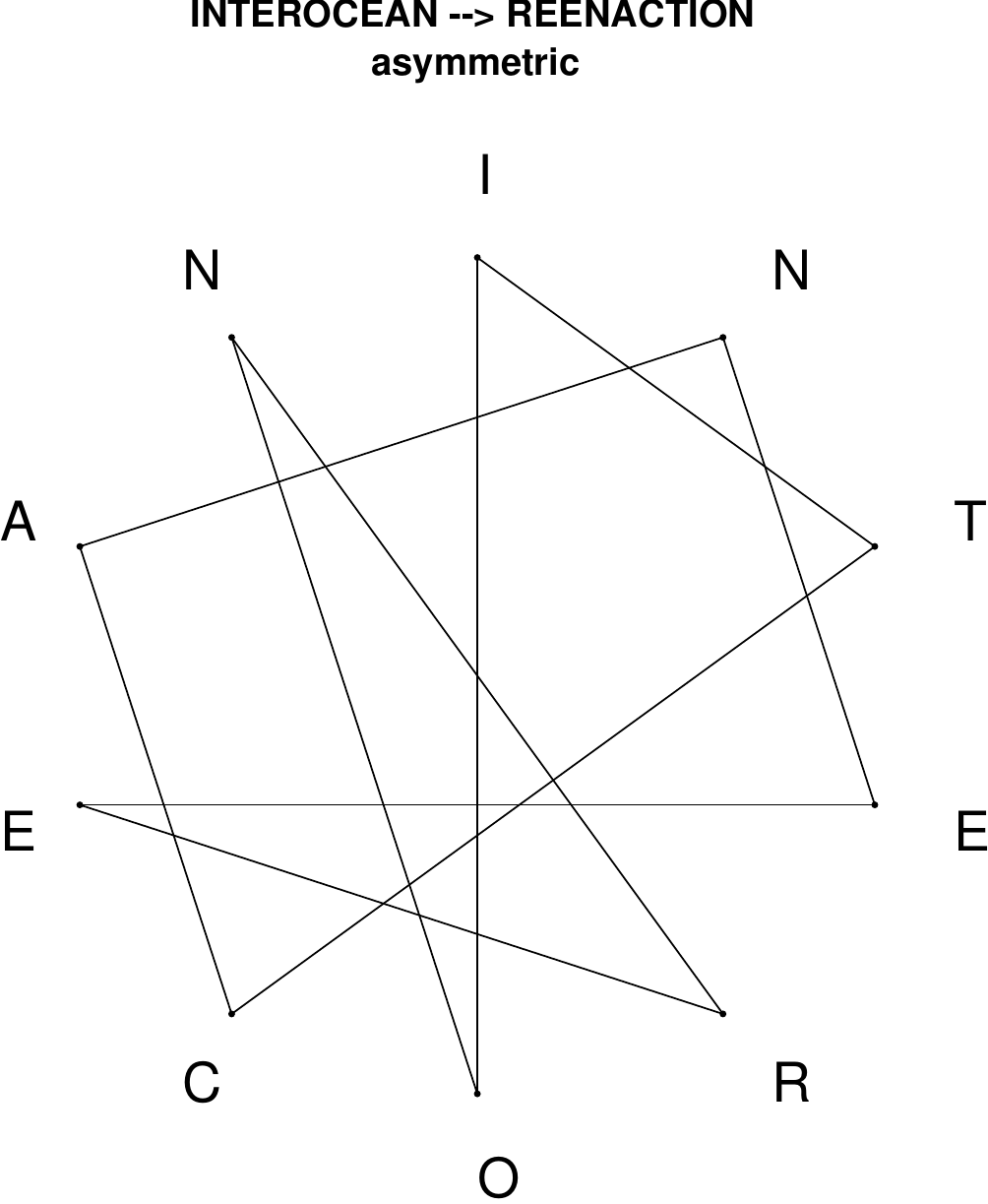}
\end{subfigure}
\hfill
\begin{subfigure}[T]{0.19\textwidth}
\centering
\includegraphics[width=\textwidth]{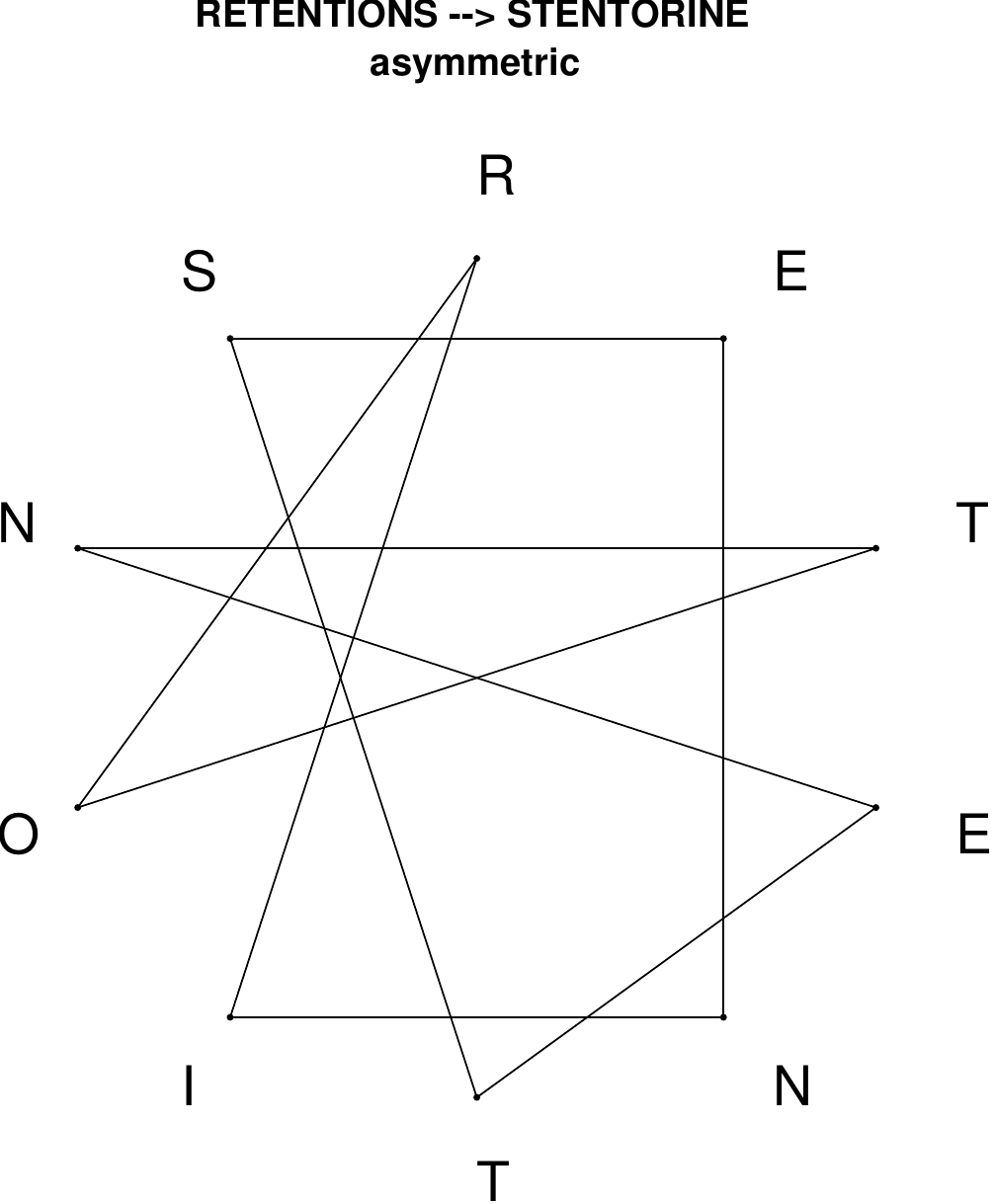}
\end{subfigure}
\hfill
\begin{subfigure}[T]{0.19\textwidth}
\centering
\includegraphics[width=\textwidth]{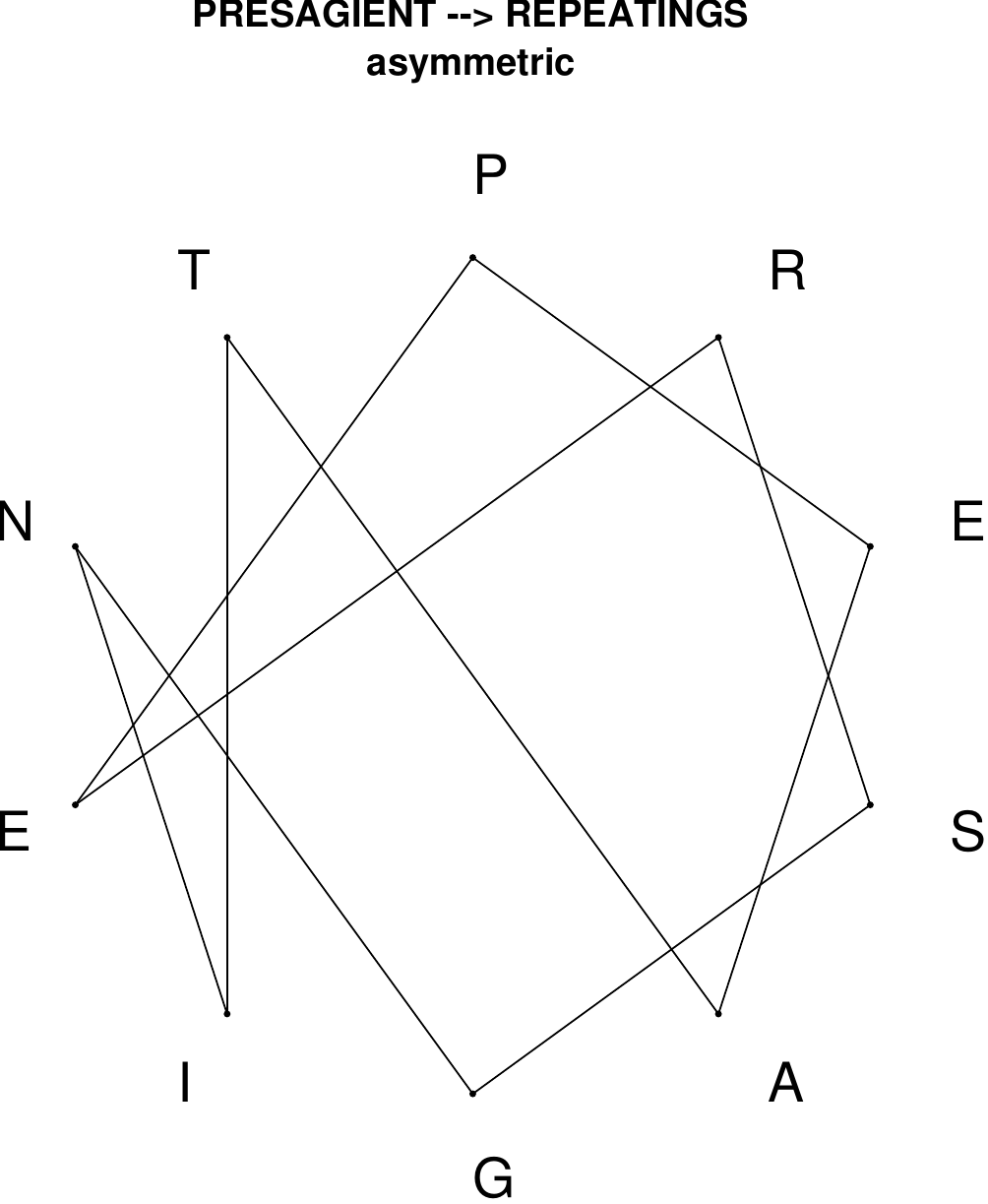}
\end{subfigure}
\hfill
\begin{subfigure}[T]{0.19\textwidth}
\centering
\includegraphics[width=\textwidth]{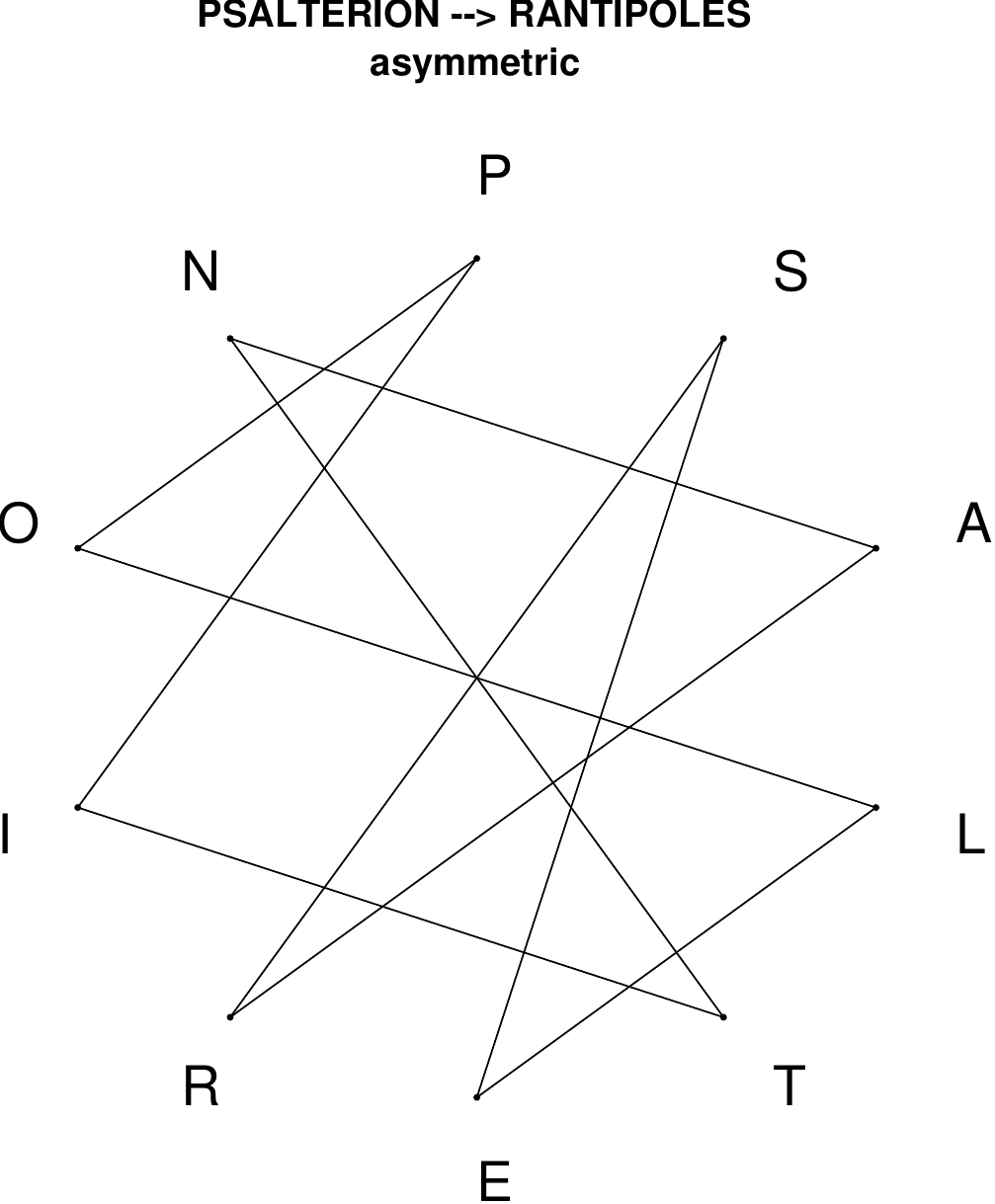}
\end{subfigure}
\hfill
\begin{subfigure}[T]{0.19\textwidth}
\centering
\includegraphics[width=\textwidth]{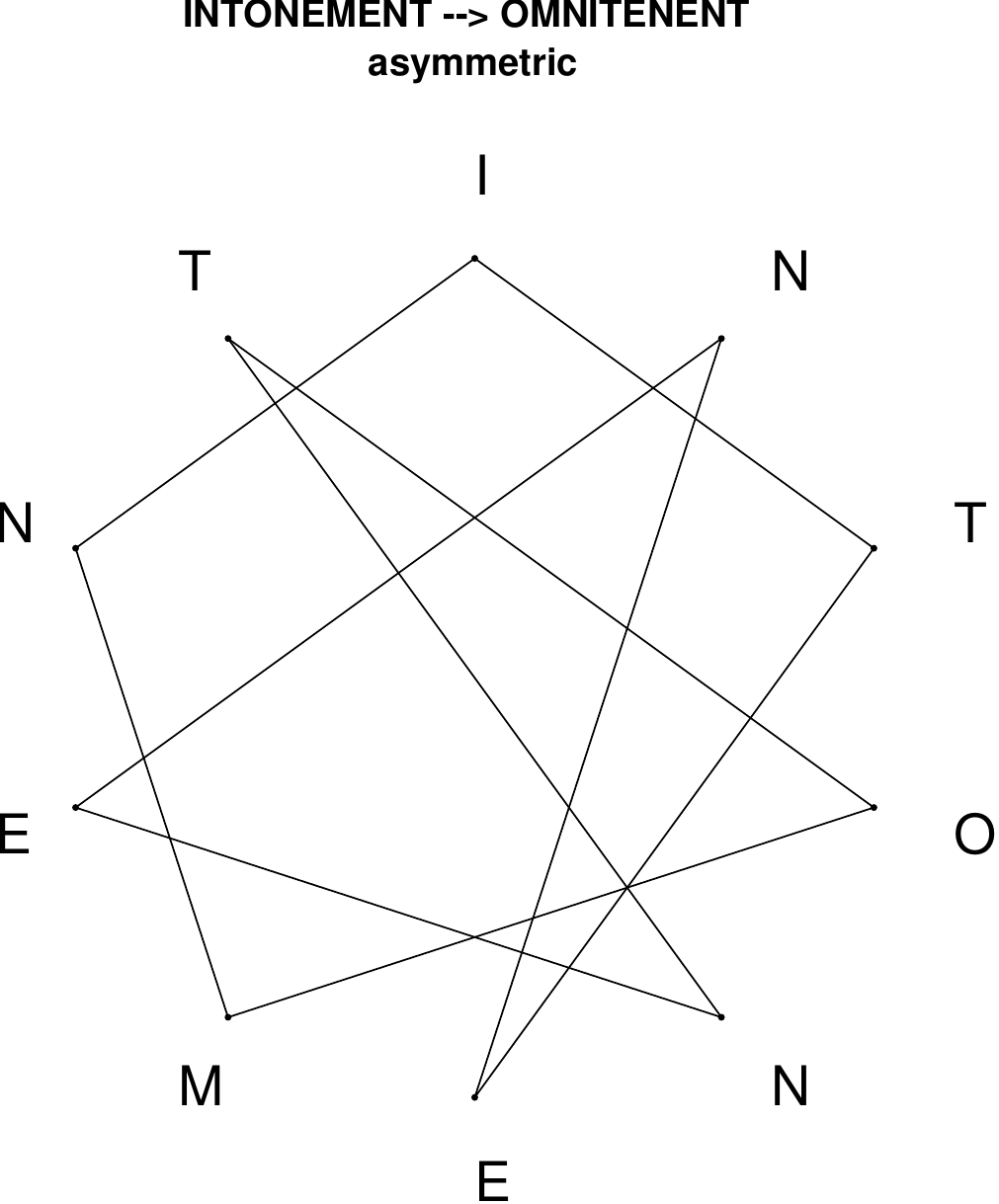}
\end{subfigure}
\end{figure}

\begin{figure}[H]
\centering
\begin{subfigure}[T]{0.19\textwidth}
\centering
\includegraphics[width=\textwidth]{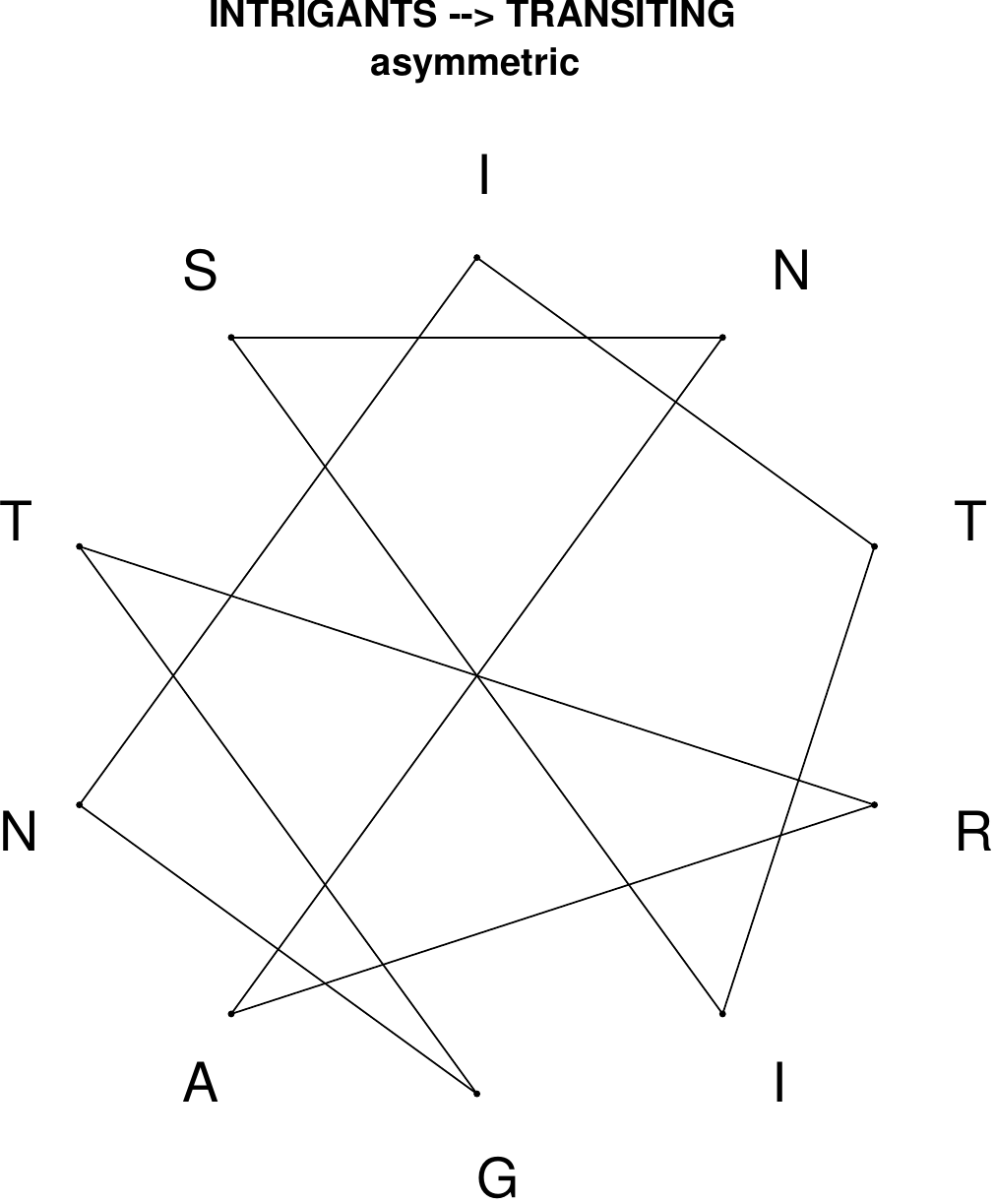}
\end{subfigure}
\hfill
\begin{subfigure}[T]{0.19\textwidth}
\centering
\includegraphics[width=\textwidth]{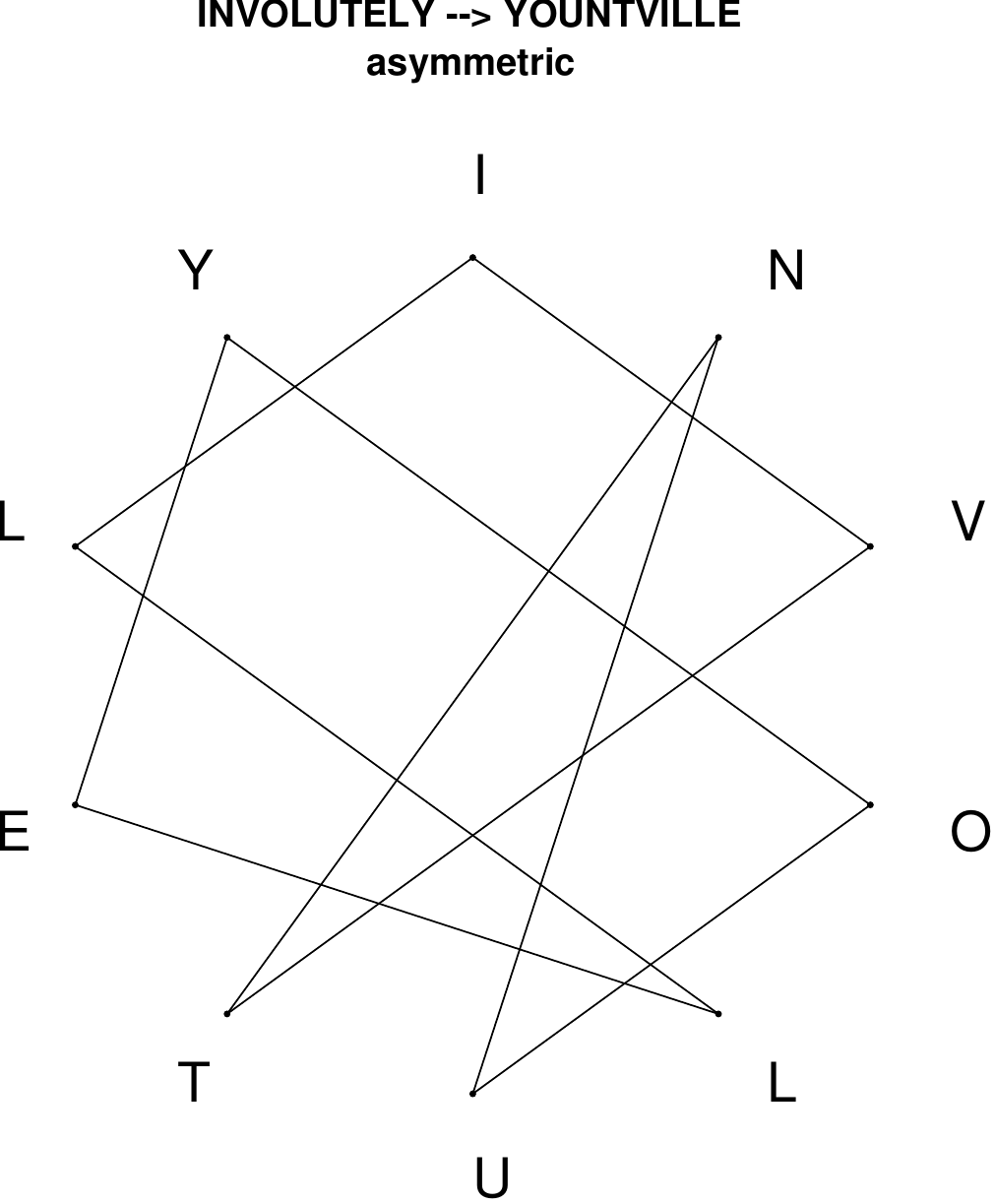}
\end{subfigure}
\hfill
\begin{subfigure}[T]{0.19\textwidth}
\centering
\includegraphics[width=\textwidth]{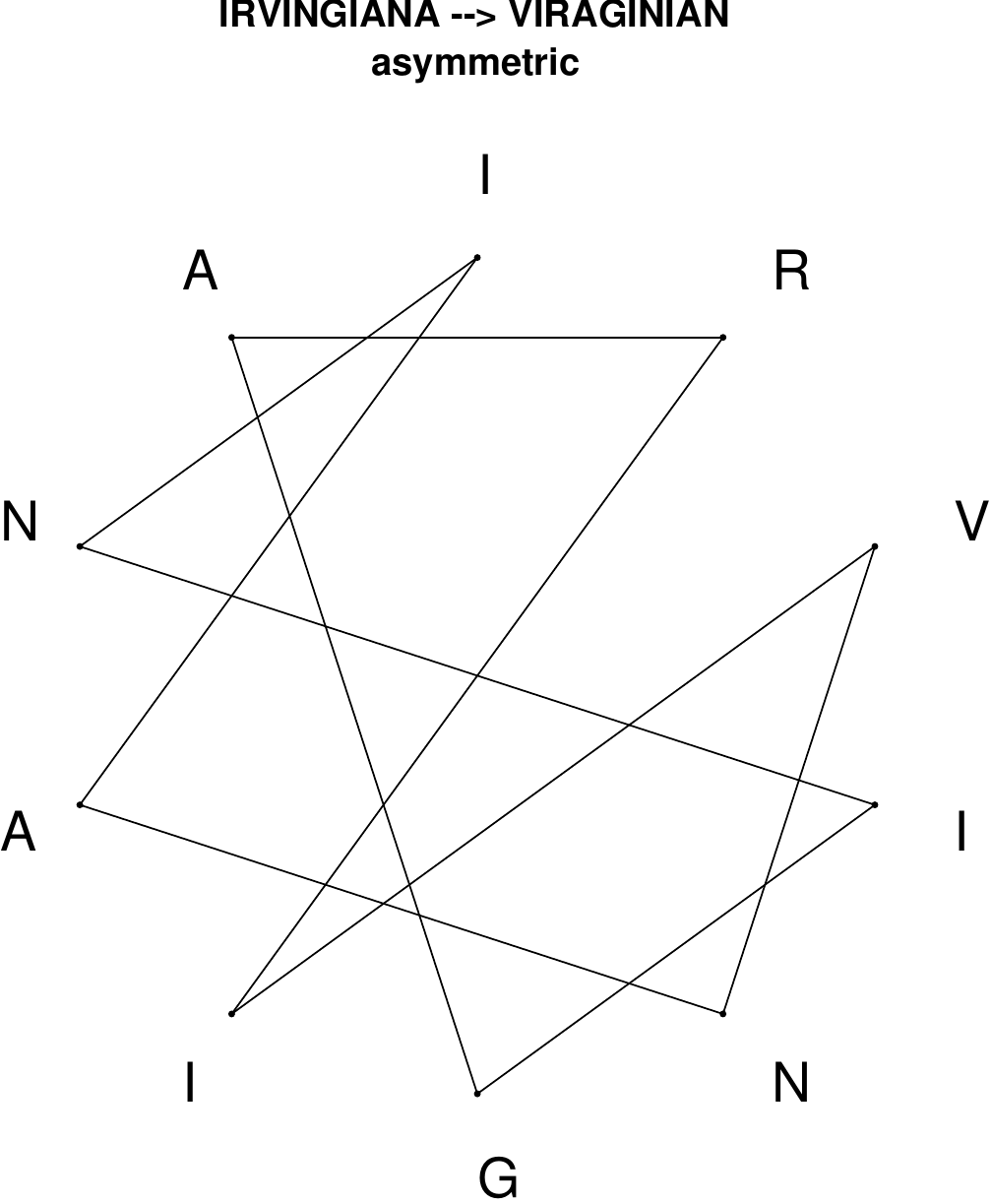}
\end{subfigure}
\hfill
\begin{subfigure}[T]{0.19\textwidth}
\centering
\includegraphics[width=\textwidth]{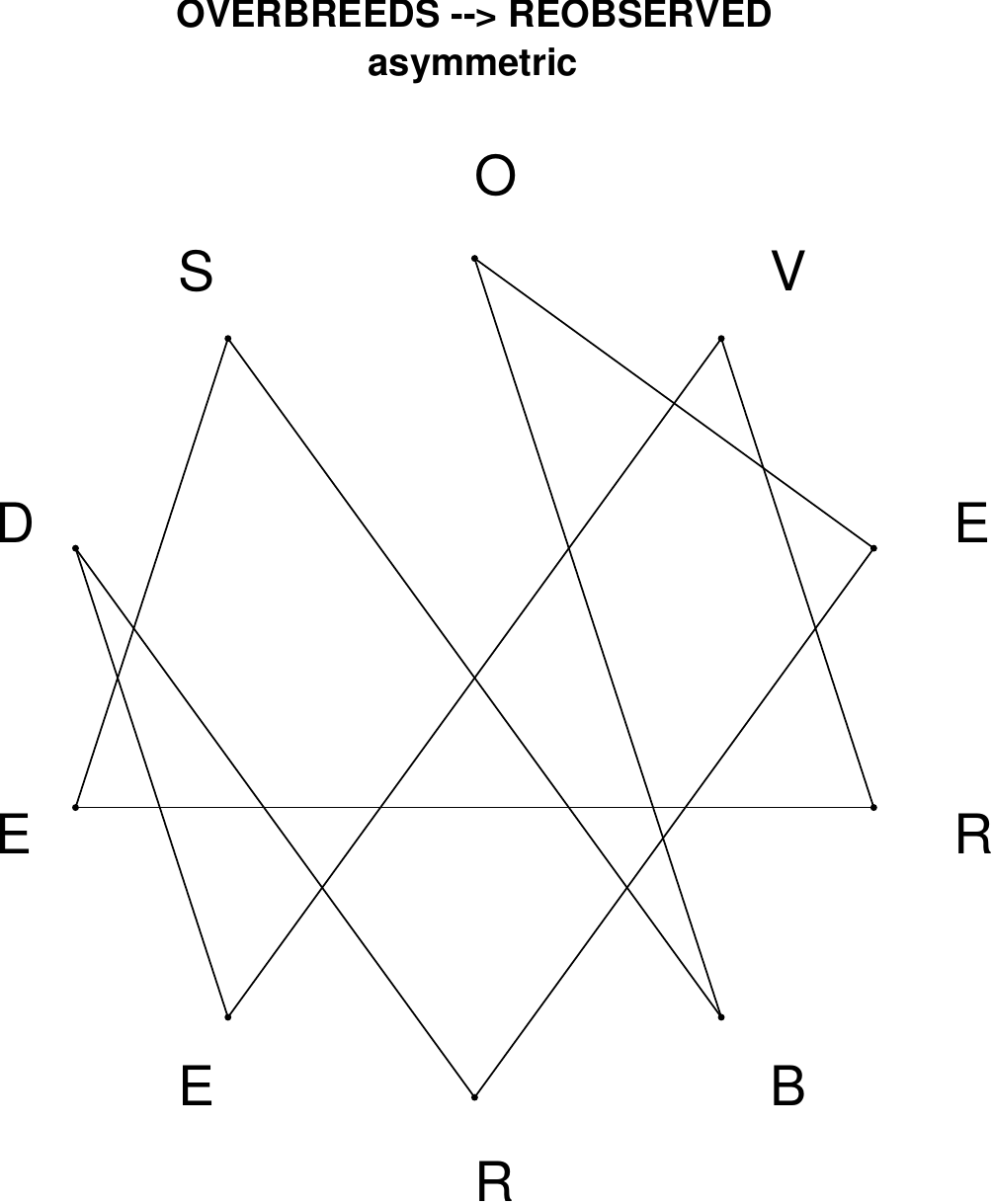}
\end{subfigure}
\hfill
\begin{subfigure}[T]{0.19\textwidth}
\centering
\includegraphics[width=\textwidth]{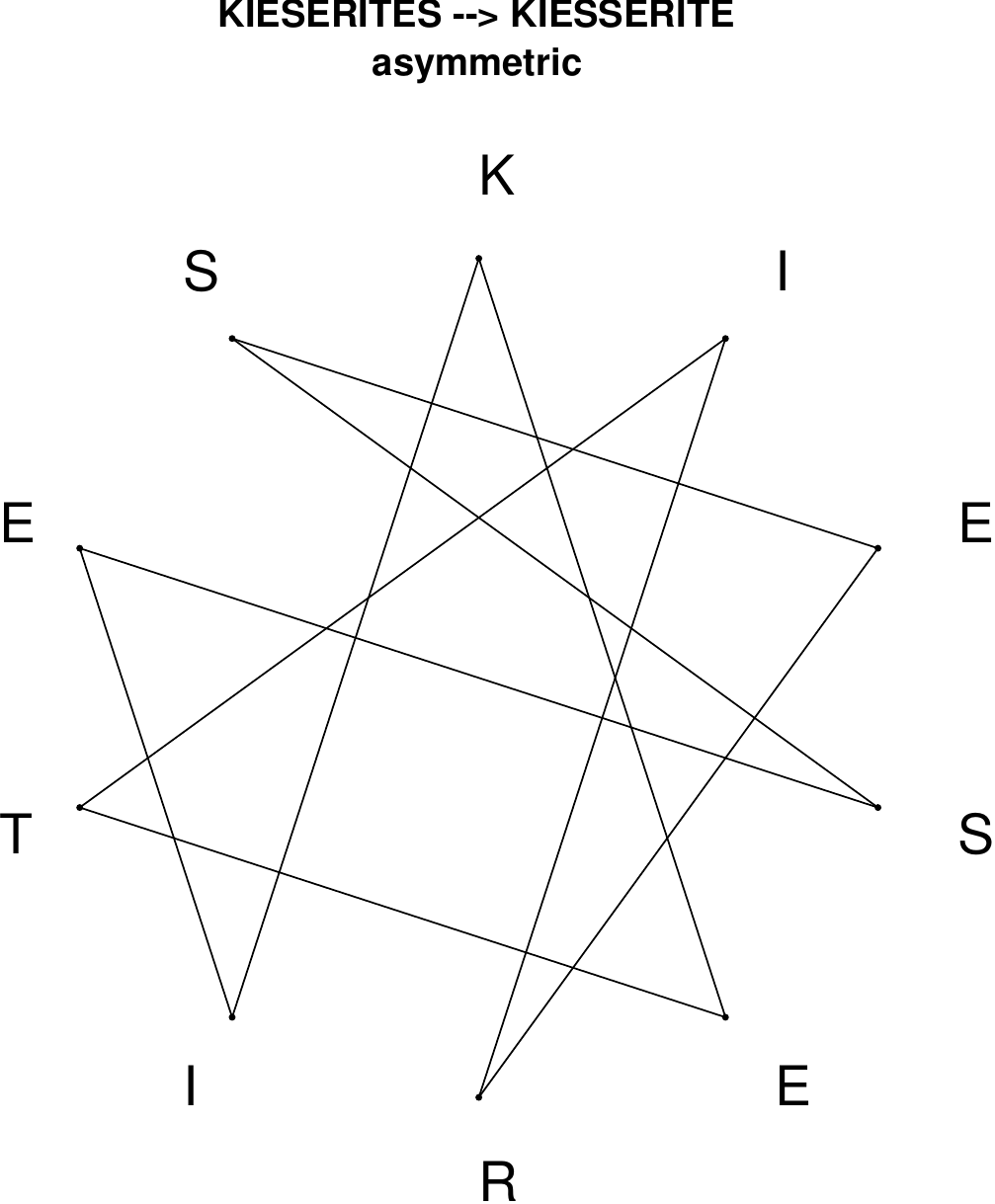}
\end{subfigure}
\end{figure}

\begin{figure}[H]
\centering
\begin{subfigure}[T]{0.19\textwidth}
\centering
\includegraphics[width=\textwidth]{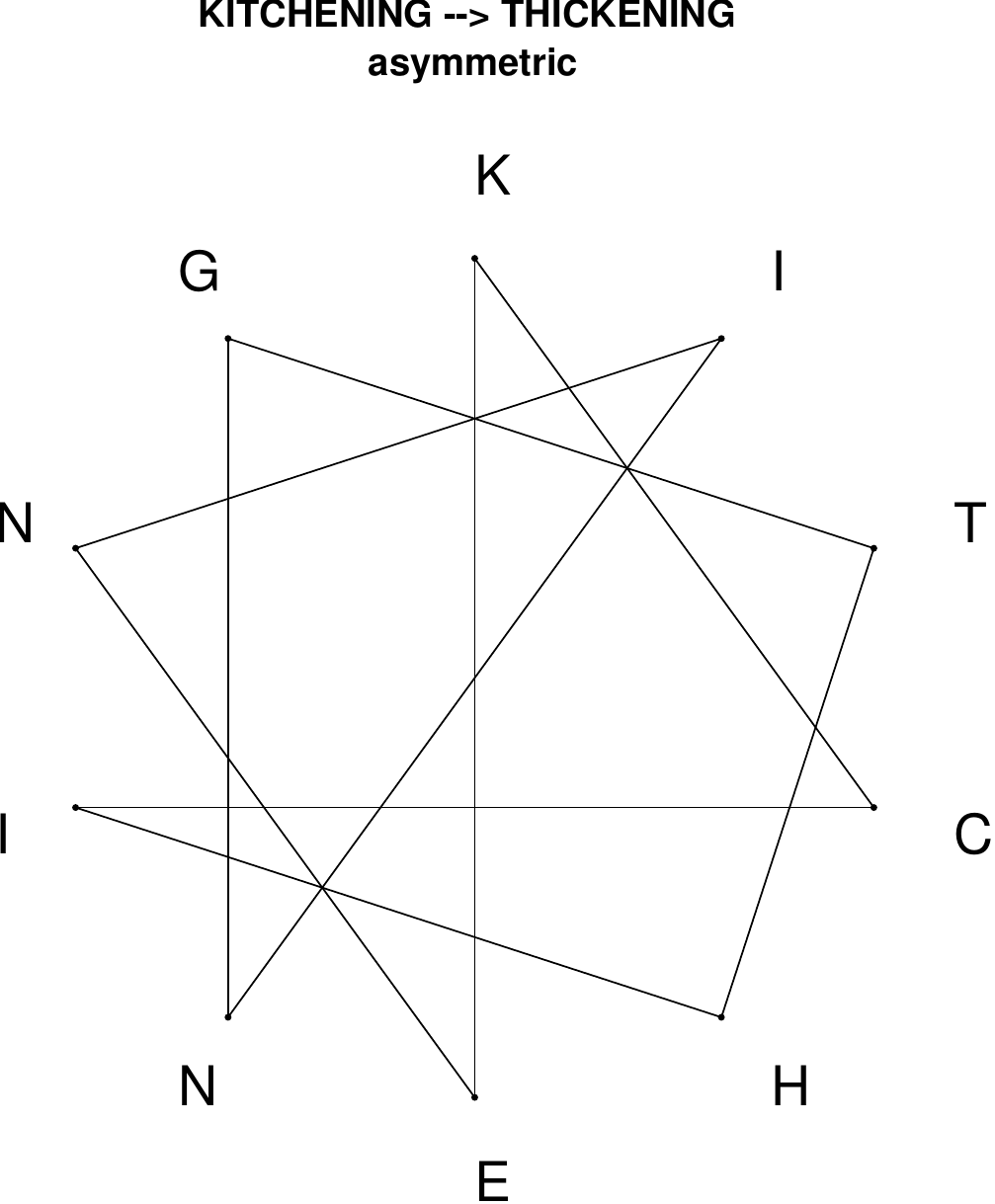}
\end{subfigure}
\hfill
\begin{subfigure}[T]{0.19\textwidth}
\centering
\includegraphics[width=\textwidth]{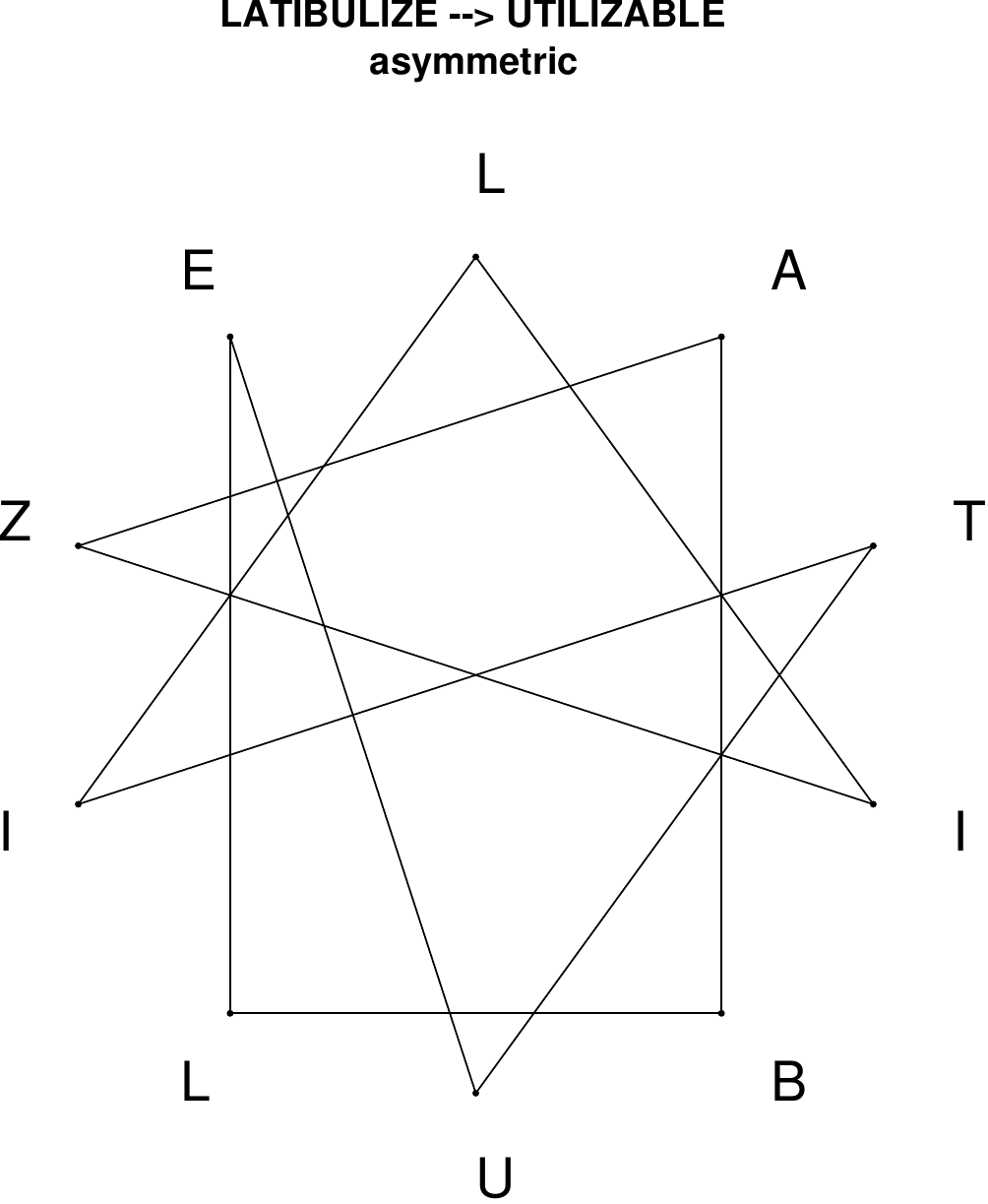}
\end{subfigure}
\hfill
\begin{subfigure}[T]{0.19\textwidth}
\centering
\includegraphics[width=\textwidth]{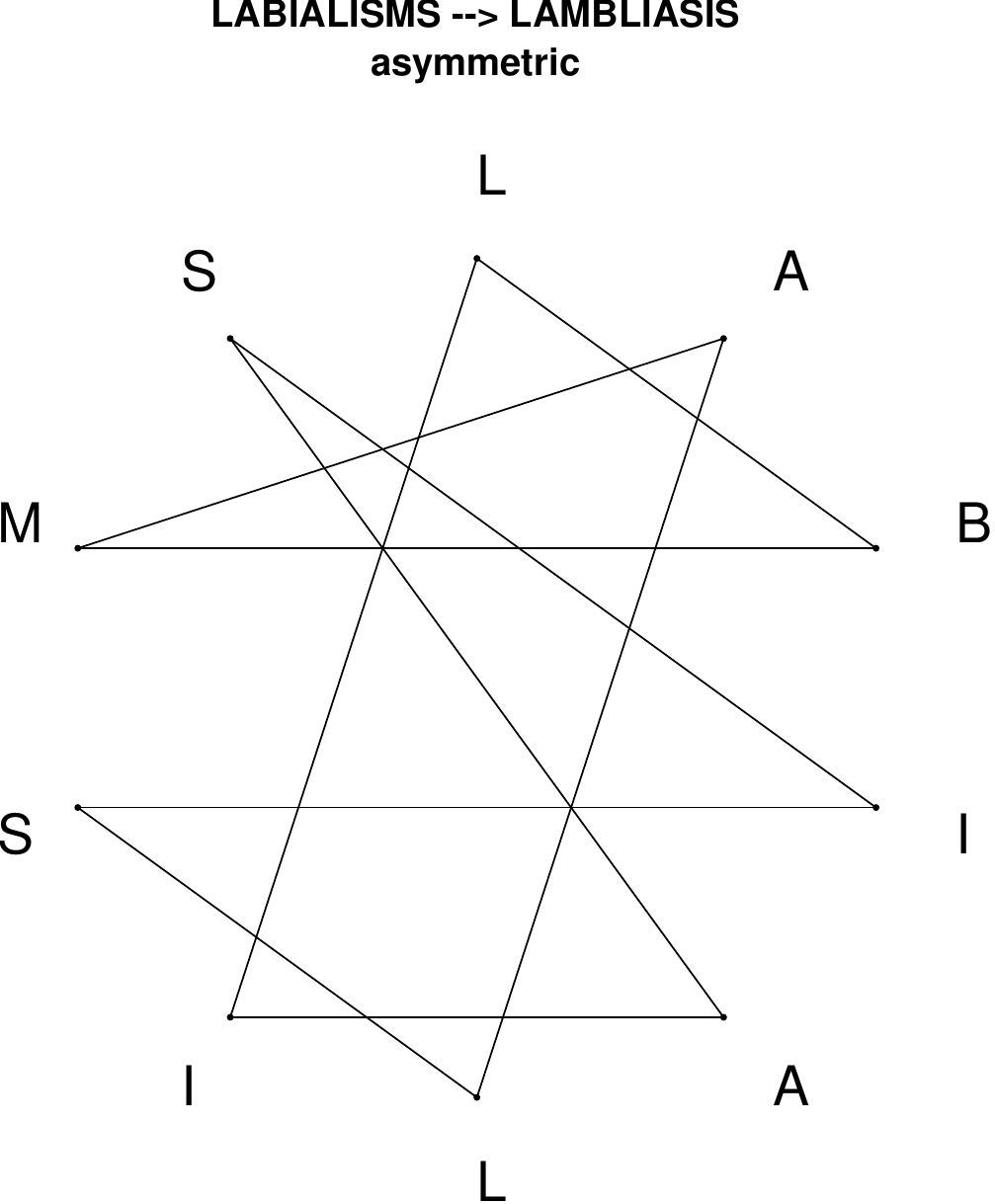}
\end{subfigure}
\hfill
\begin{subfigure}[T]{0.19\textwidth}
\centering
\includegraphics[width=\textwidth]{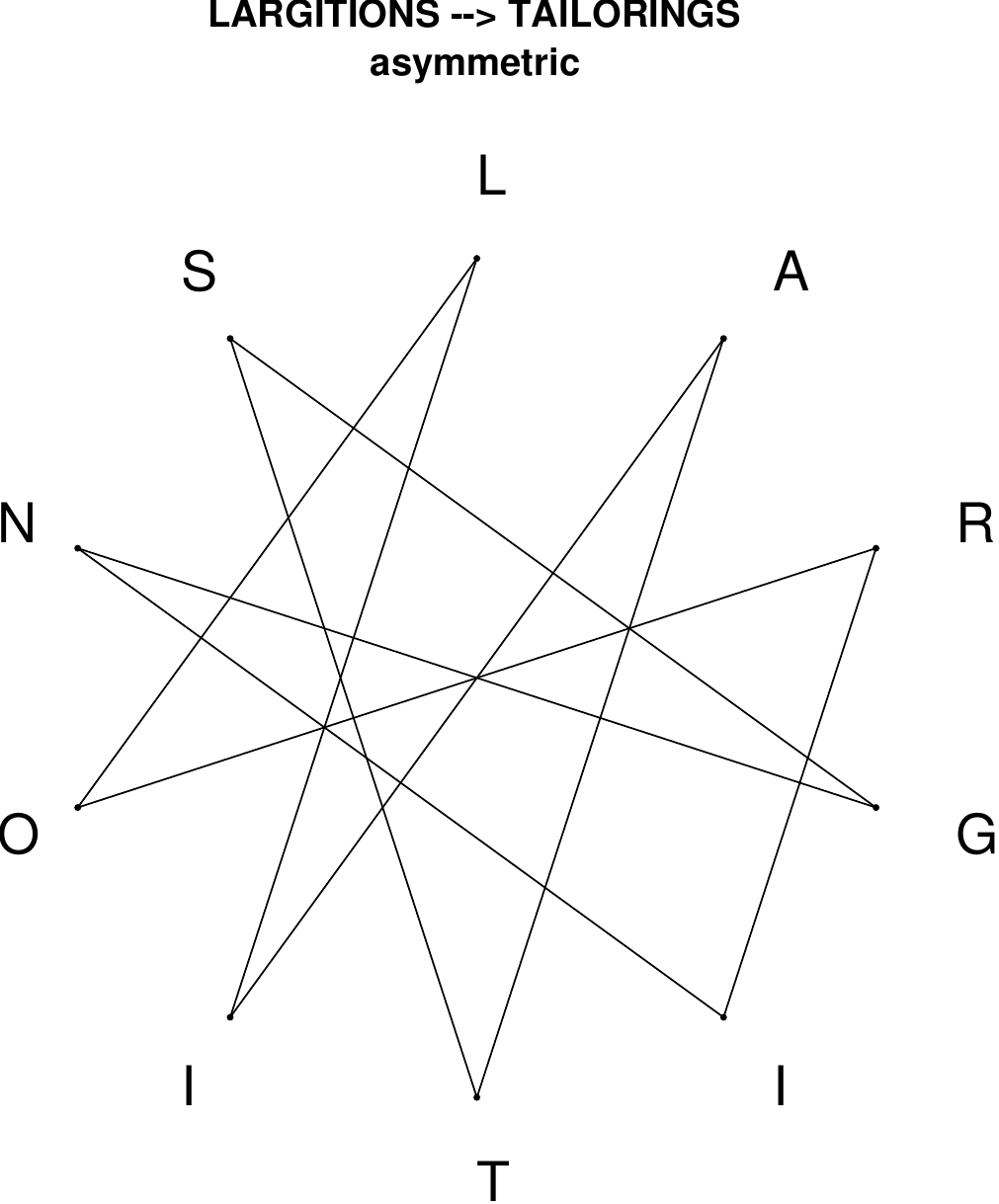}
\end{subfigure}
\hfill
\begin{subfigure}[T]{0.19\textwidth}
\centering
\includegraphics[width=\textwidth]{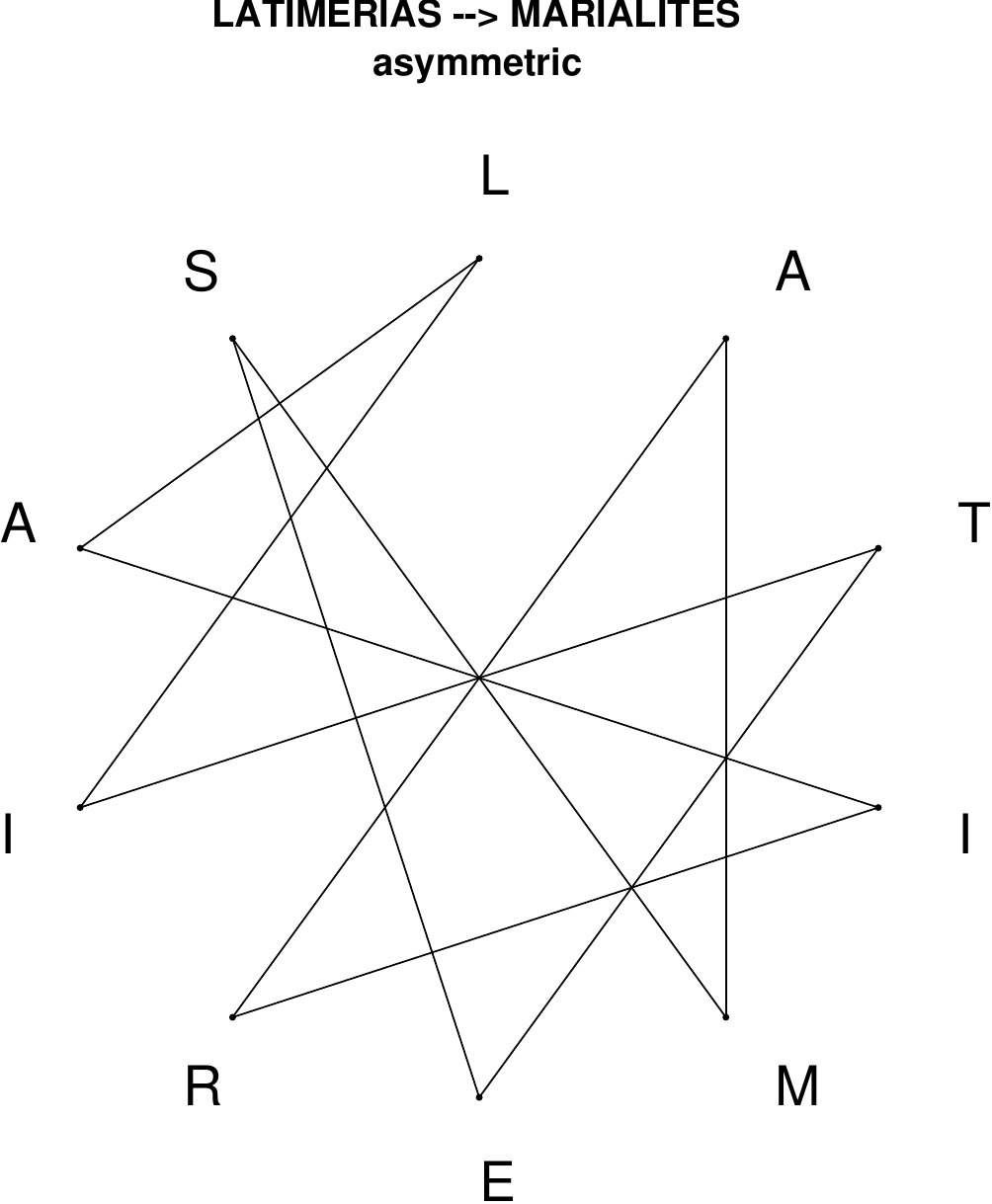}
\end{subfigure}
\end{figure}

\begin{figure}[H]
\centering
\begin{subfigure}[T]{0.19\textwidth}
\centering
\includegraphics[width=\textwidth]{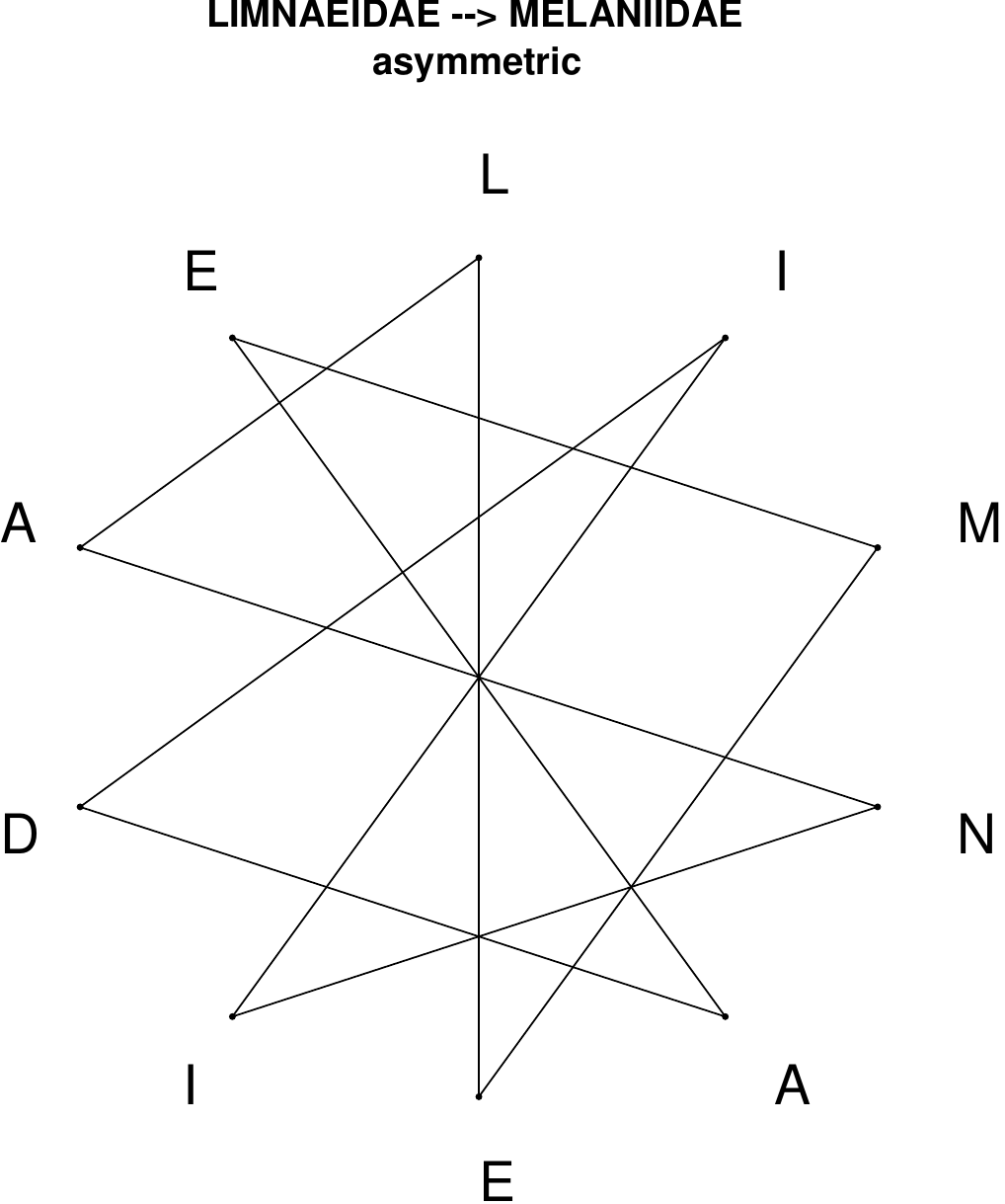}
\end{subfigure}
\hfill
\begin{subfigure}[T]{0.19\textwidth}
\centering
\includegraphics[width=\textwidth]{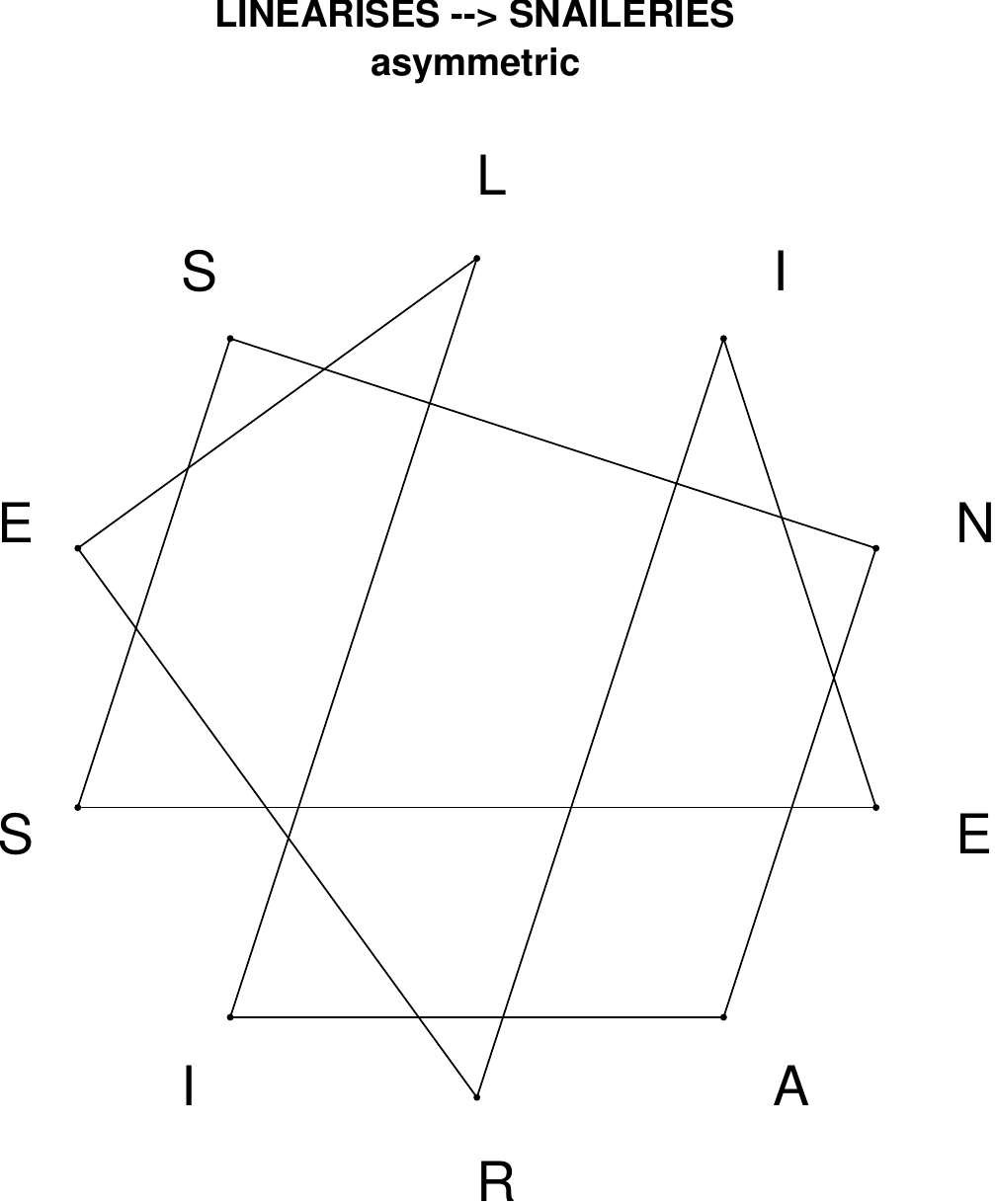}
\end{subfigure}
\hfill
\begin{subfigure}[T]{0.19\textwidth}
\centering
\includegraphics[width=\textwidth]{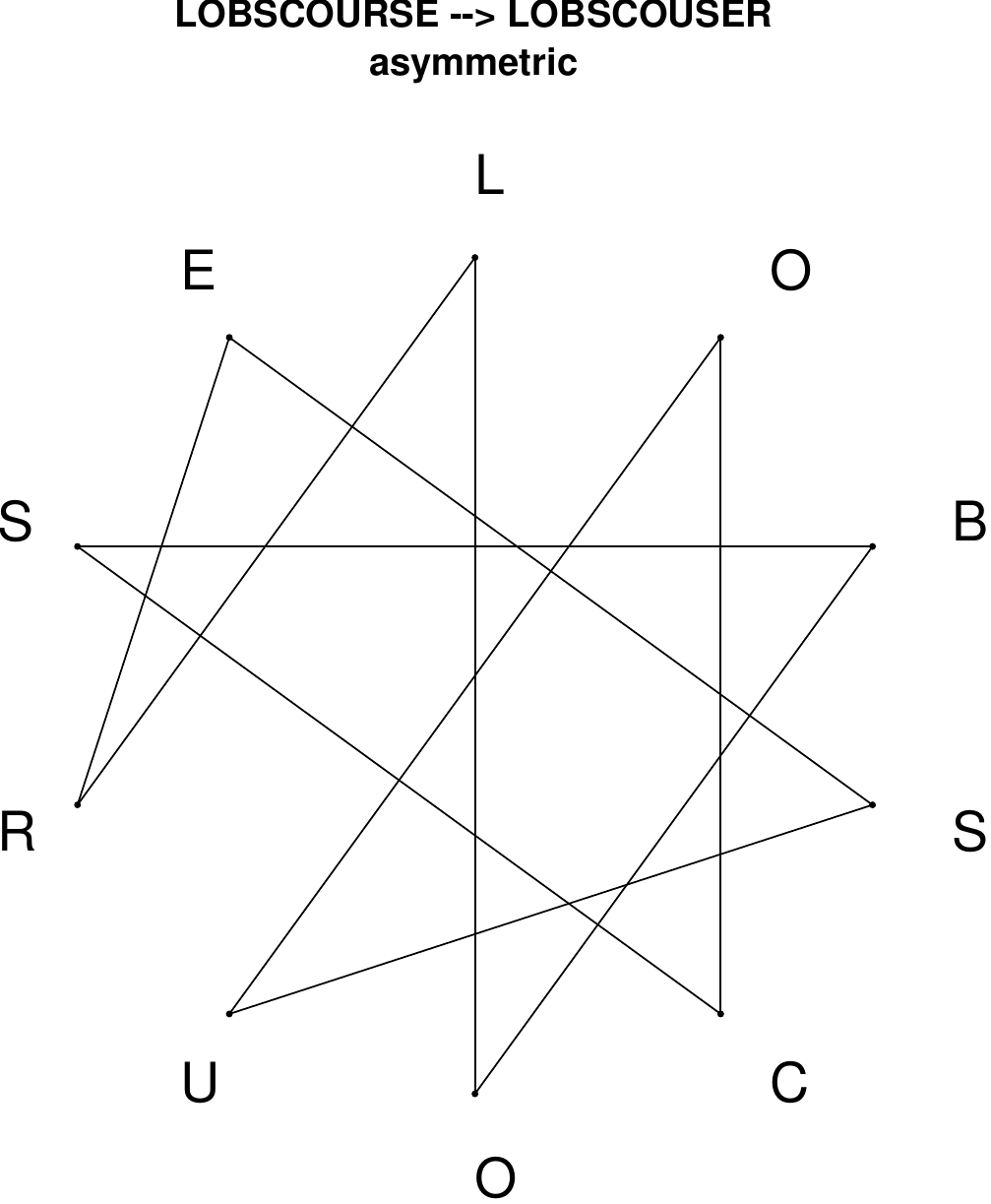}
\end{subfigure}
\hfill
\begin{subfigure}[T]{0.19\textwidth}
\centering
\includegraphics[width=\textwidth]{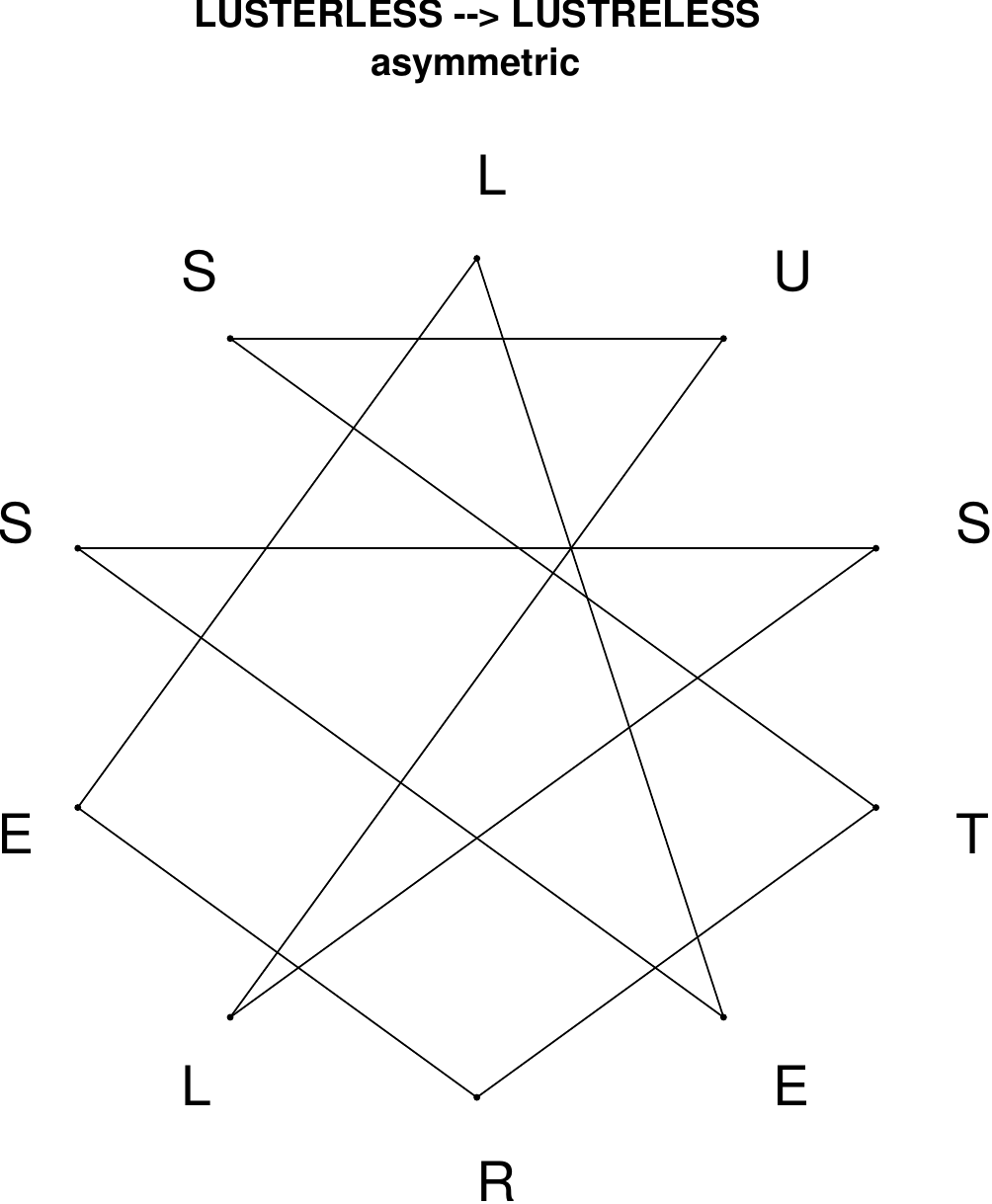}
\end{subfigure}
\hfill
\begin{subfigure}[T]{0.19\textwidth}
\centering
\includegraphics[width=\textwidth]{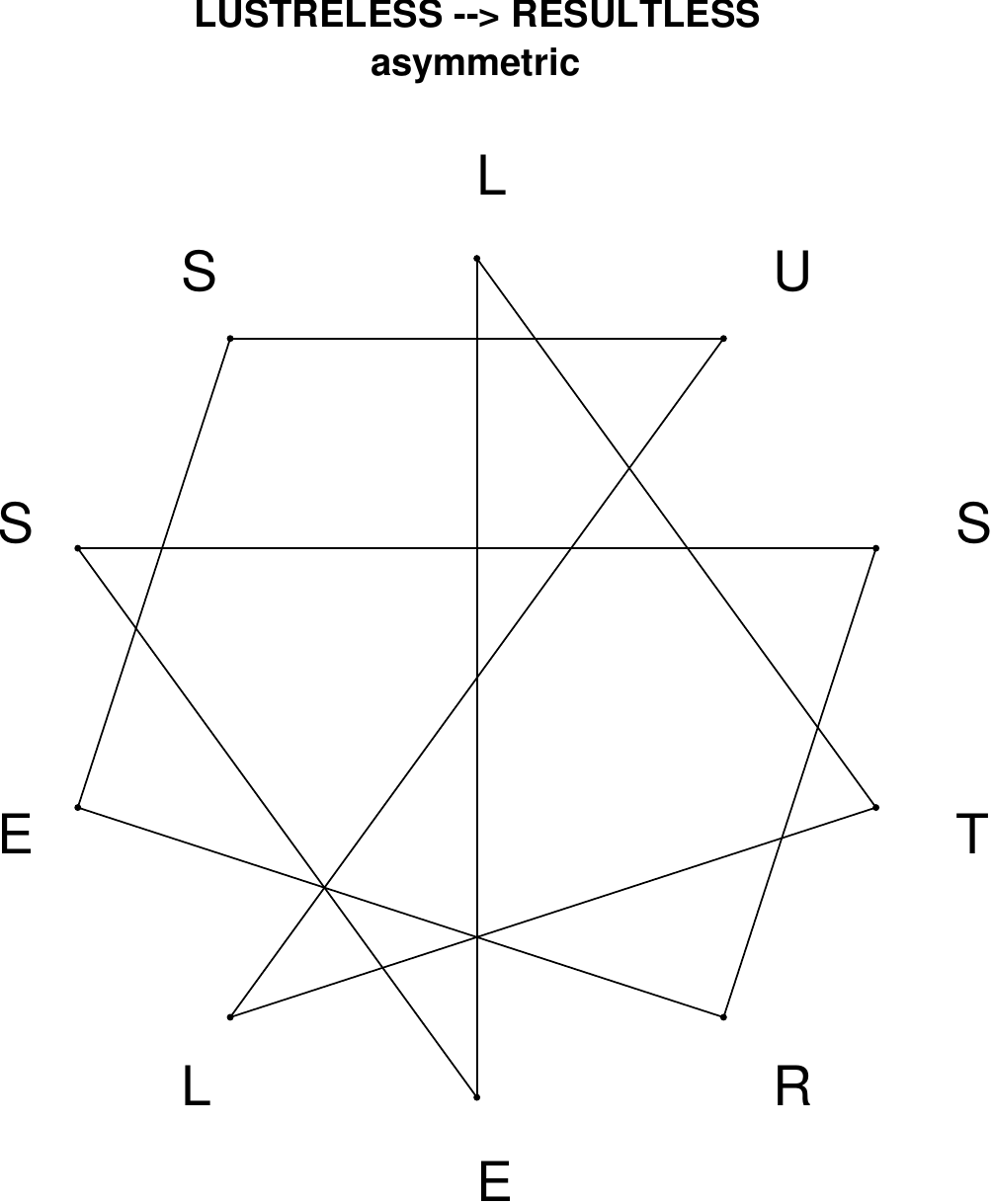}
\end{subfigure}
\end{figure}

\begin{figure}[H]
\centering
\begin{subfigure}[T]{0.19\textwidth}
\centering
\includegraphics[width=\textwidth]{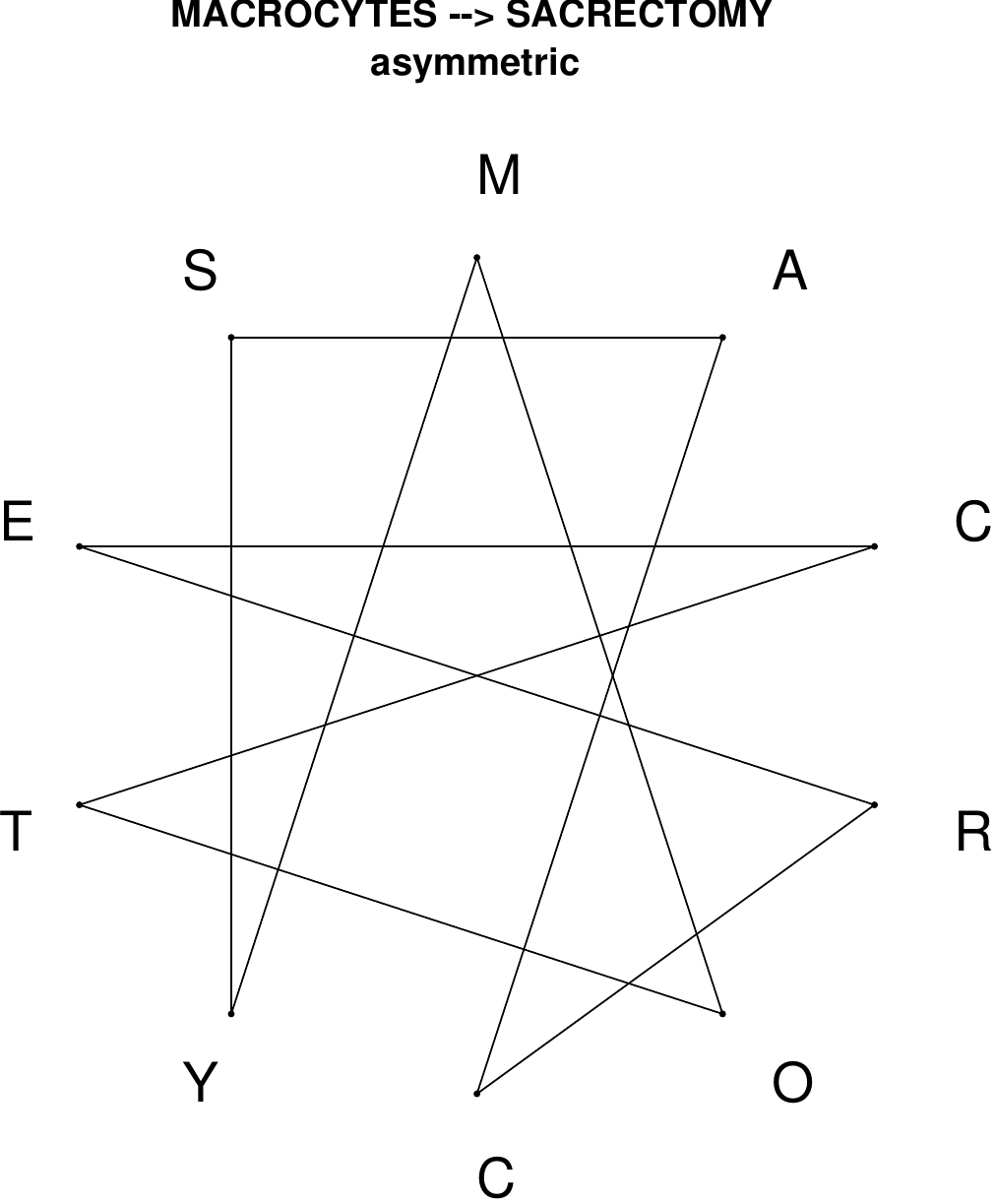}
\end{subfigure}
\hfill
\begin{subfigure}[T]{0.19\textwidth}
\centering
\includegraphics[width=\textwidth]{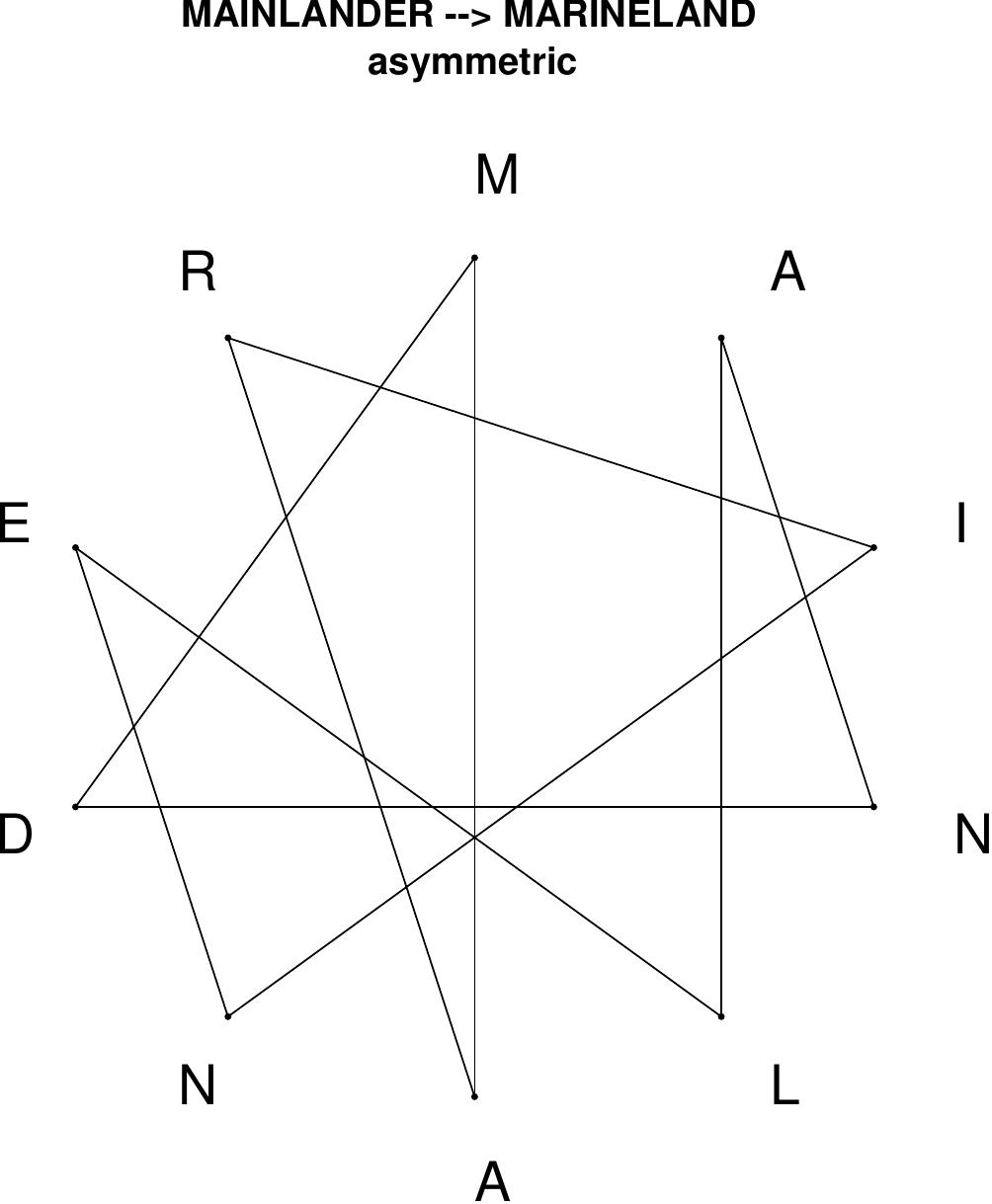}
\end{subfigure}
\hfill
\begin{subfigure}[T]{0.19\textwidth}
\centering
\includegraphics[width=\textwidth]{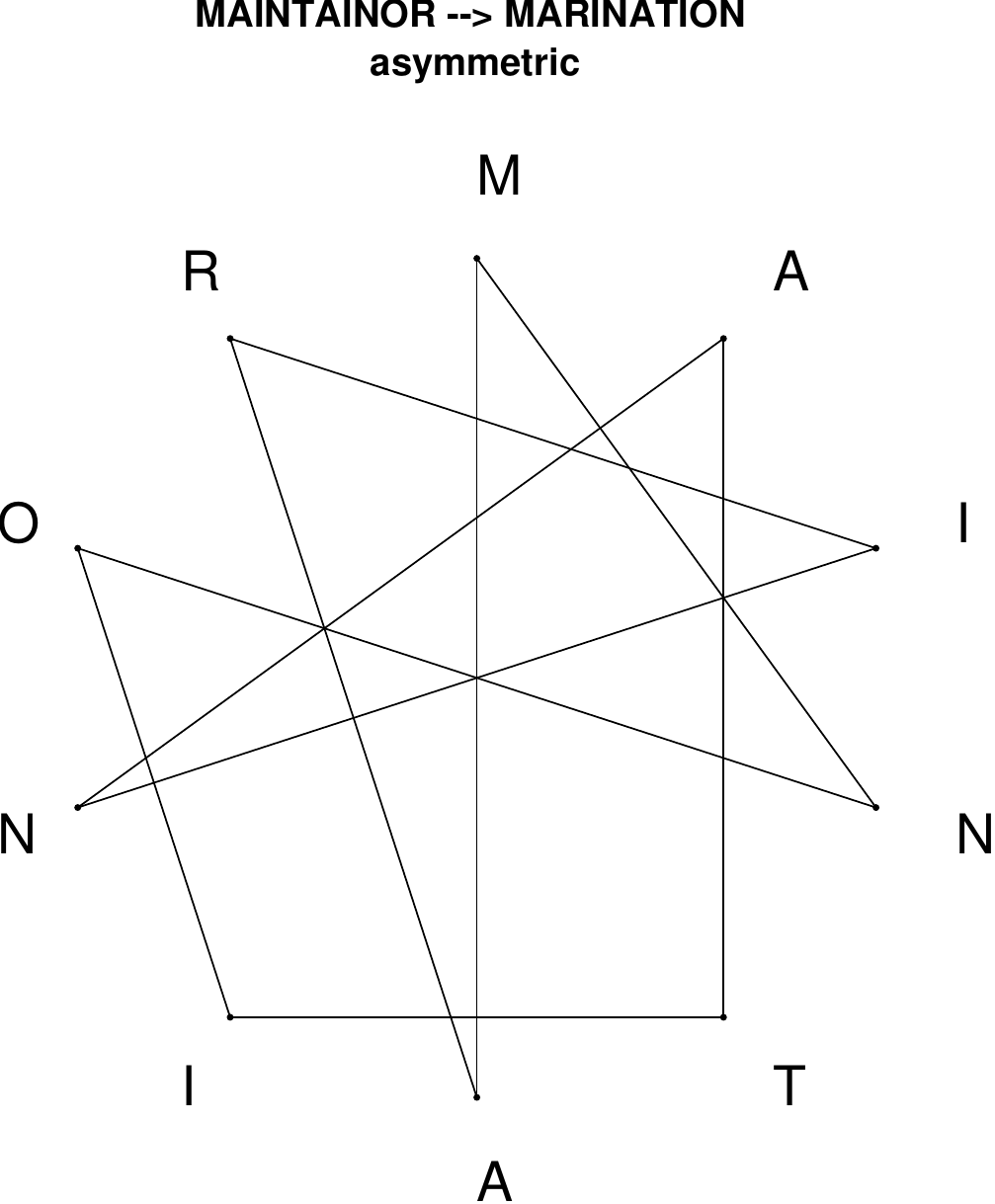}
\end{subfigure}
\hfill
\begin{subfigure}[T]{0.19\textwidth}
\centering
\includegraphics[width=\textwidth]{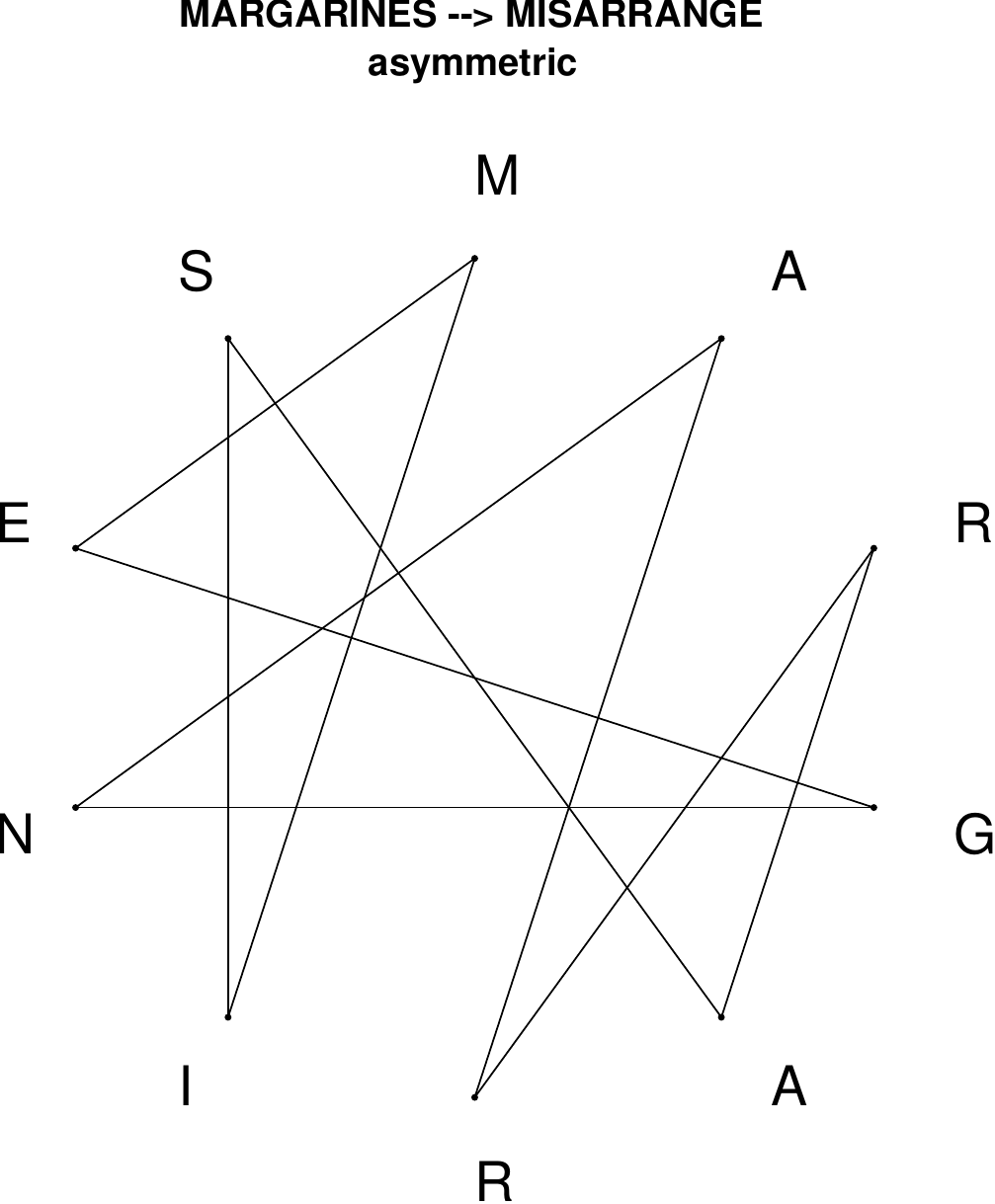}
\end{subfigure}
\hfill
\begin{subfigure}[T]{0.19\textwidth}
\centering
\includegraphics[width=\textwidth]{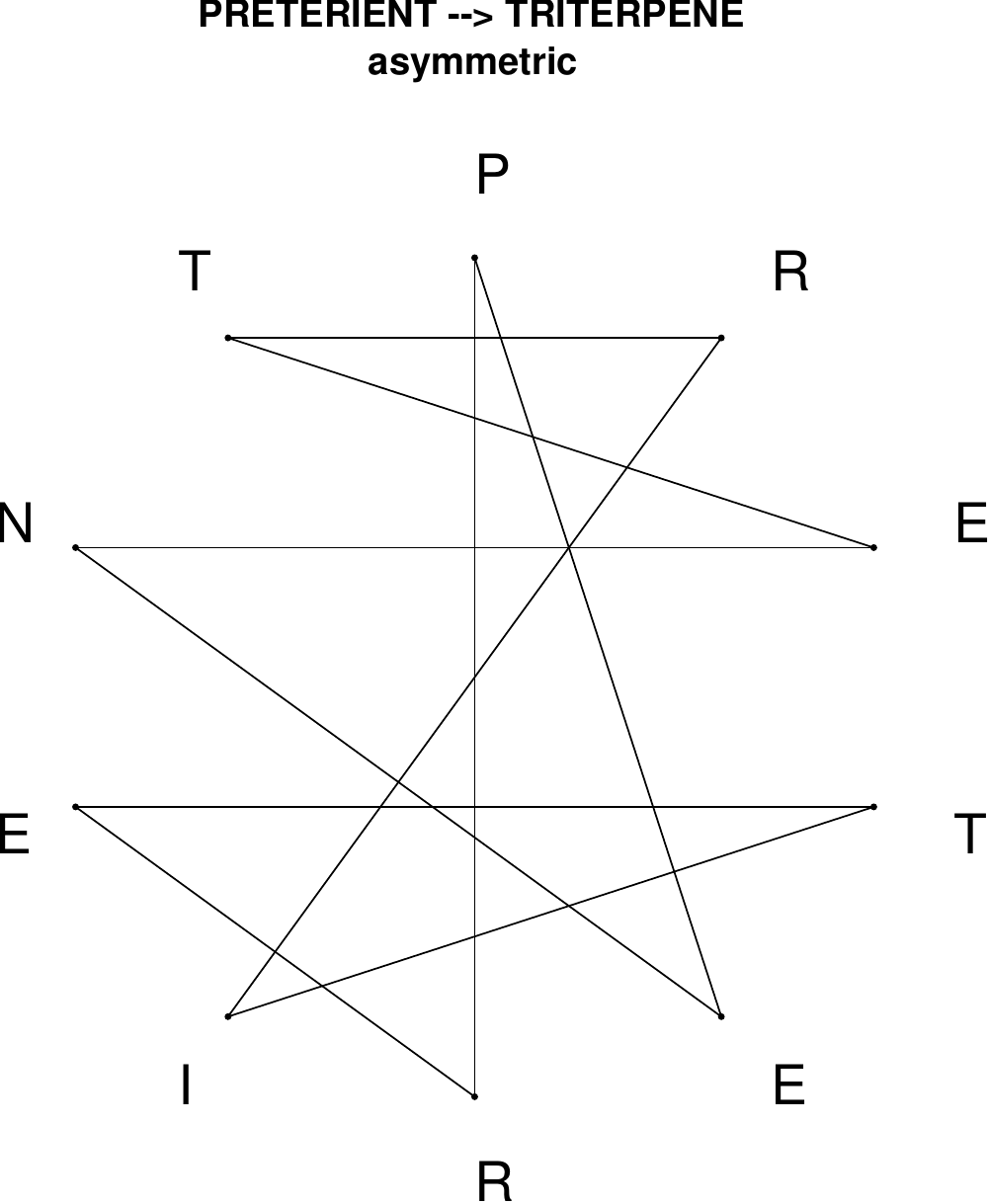}
\end{subfigure}
\end{figure}

\begin{figure}[H]
\centering
\begin{subfigure}[T]{0.19\textwidth}
\centering
\includegraphics[width=\textwidth]{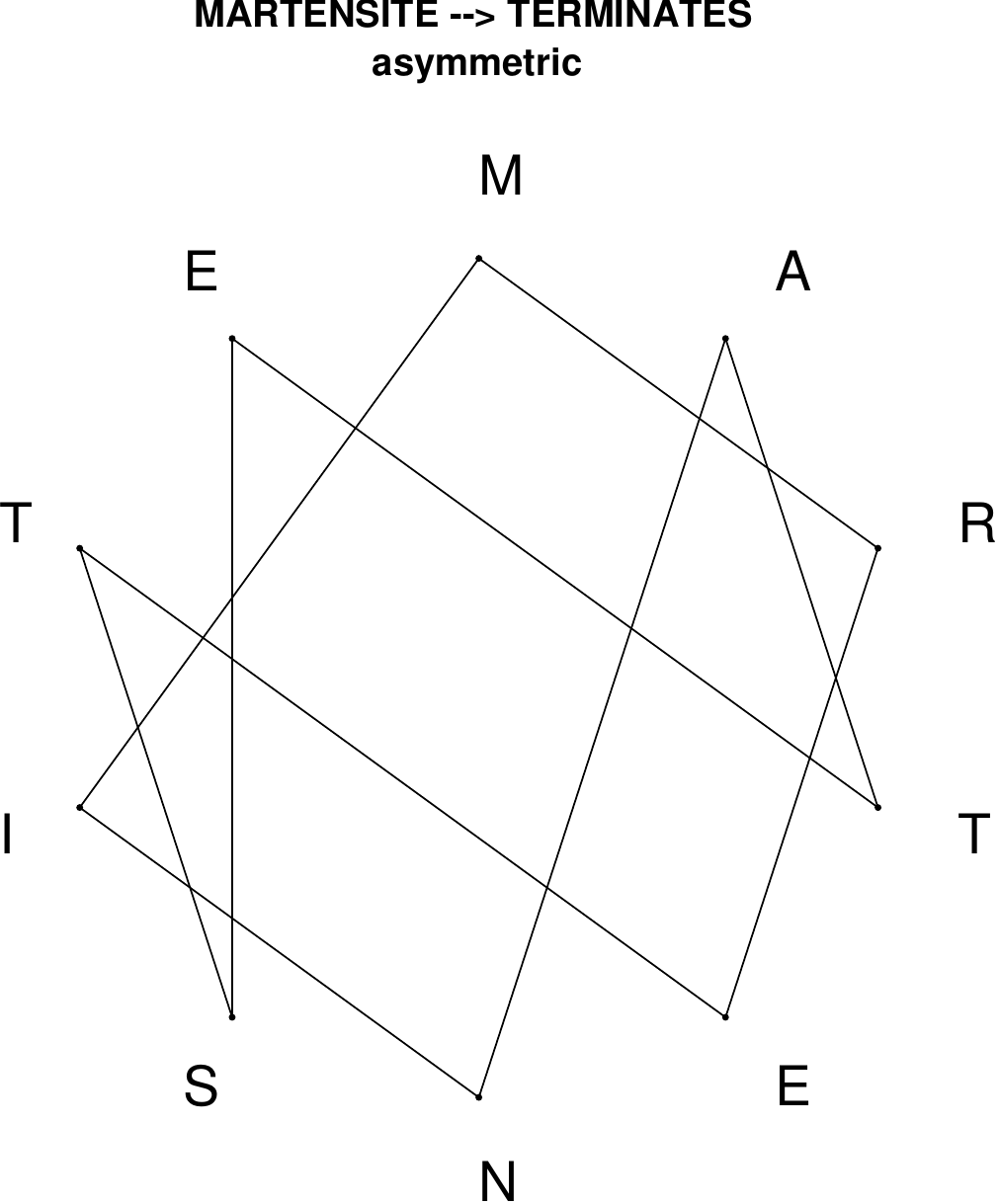}
\end{subfigure}
\hfill
\begin{subfigure}[T]{0.19\textwidth}
\centering
\includegraphics[width=\textwidth]{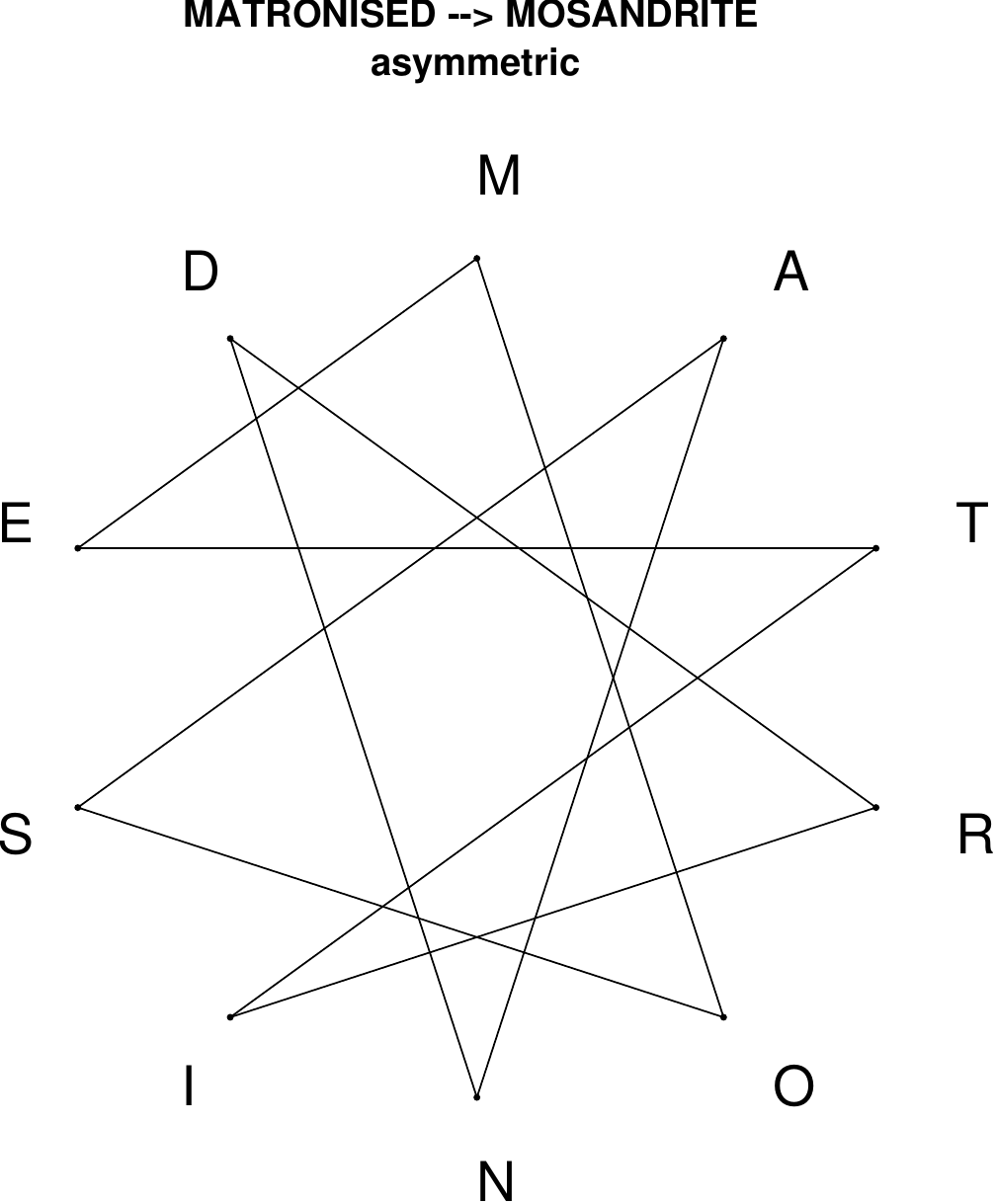}
\end{subfigure}
\hfill
\begin{subfigure}[T]{0.19\textwidth}
\centering
\includegraphics[width=\textwidth]{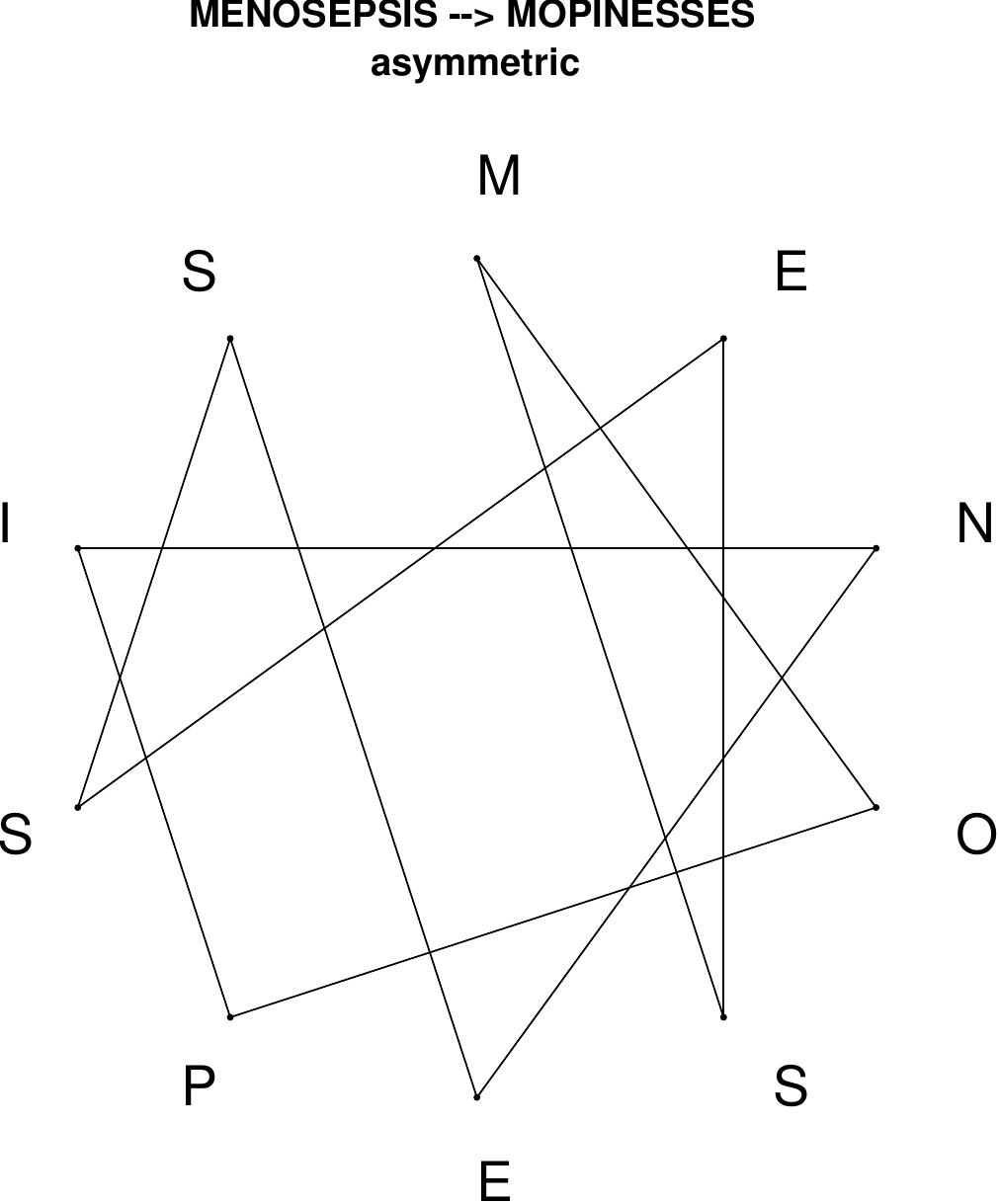}
\end{subfigure}
\hfill
\begin{subfigure}[T]{0.19\textwidth}
\centering
\includegraphics[width=\textwidth]{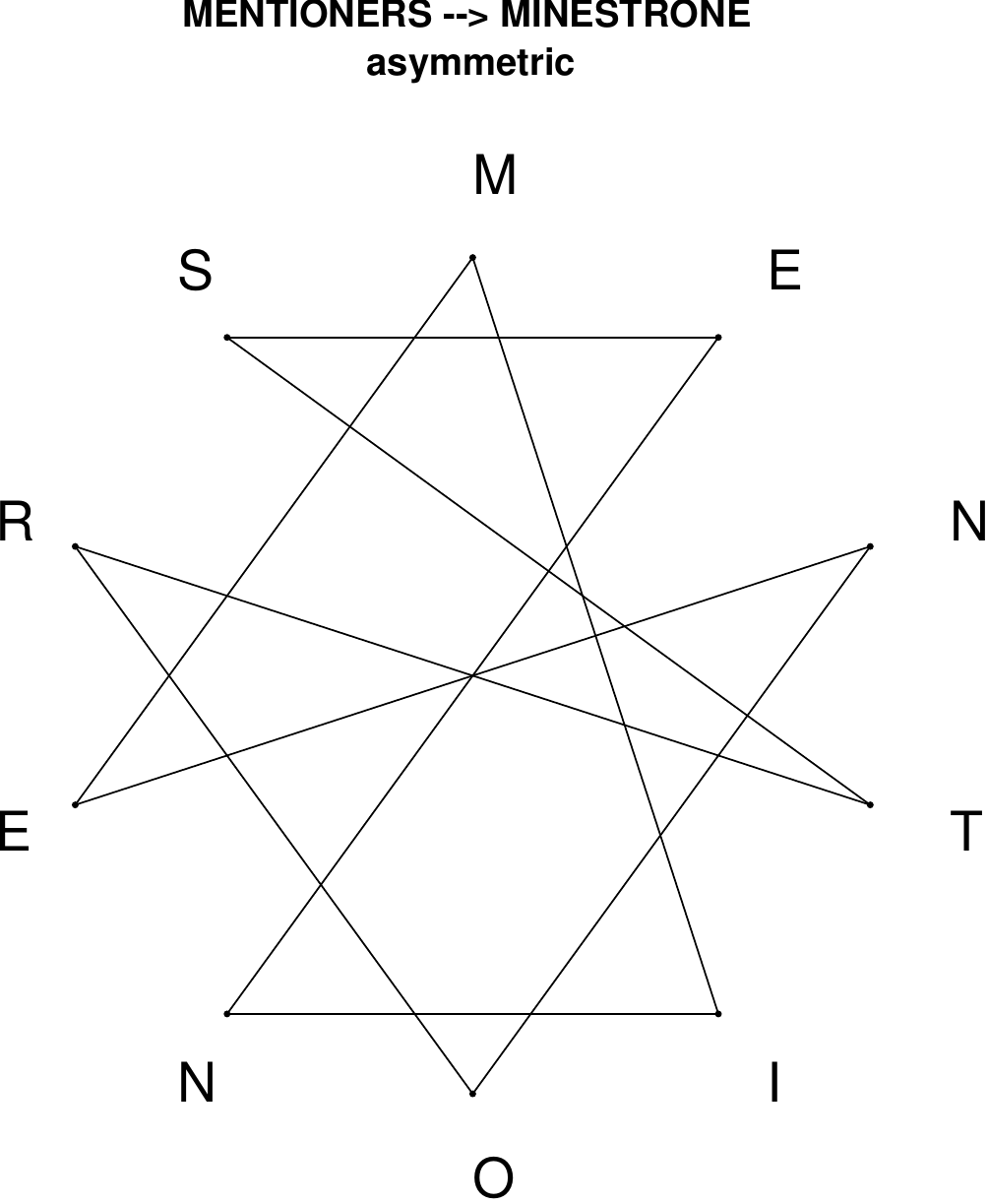}
\end{subfigure}
\hfill
\begin{subfigure}[T]{0.19\textwidth}
\centering
\includegraphics[width=\textwidth]{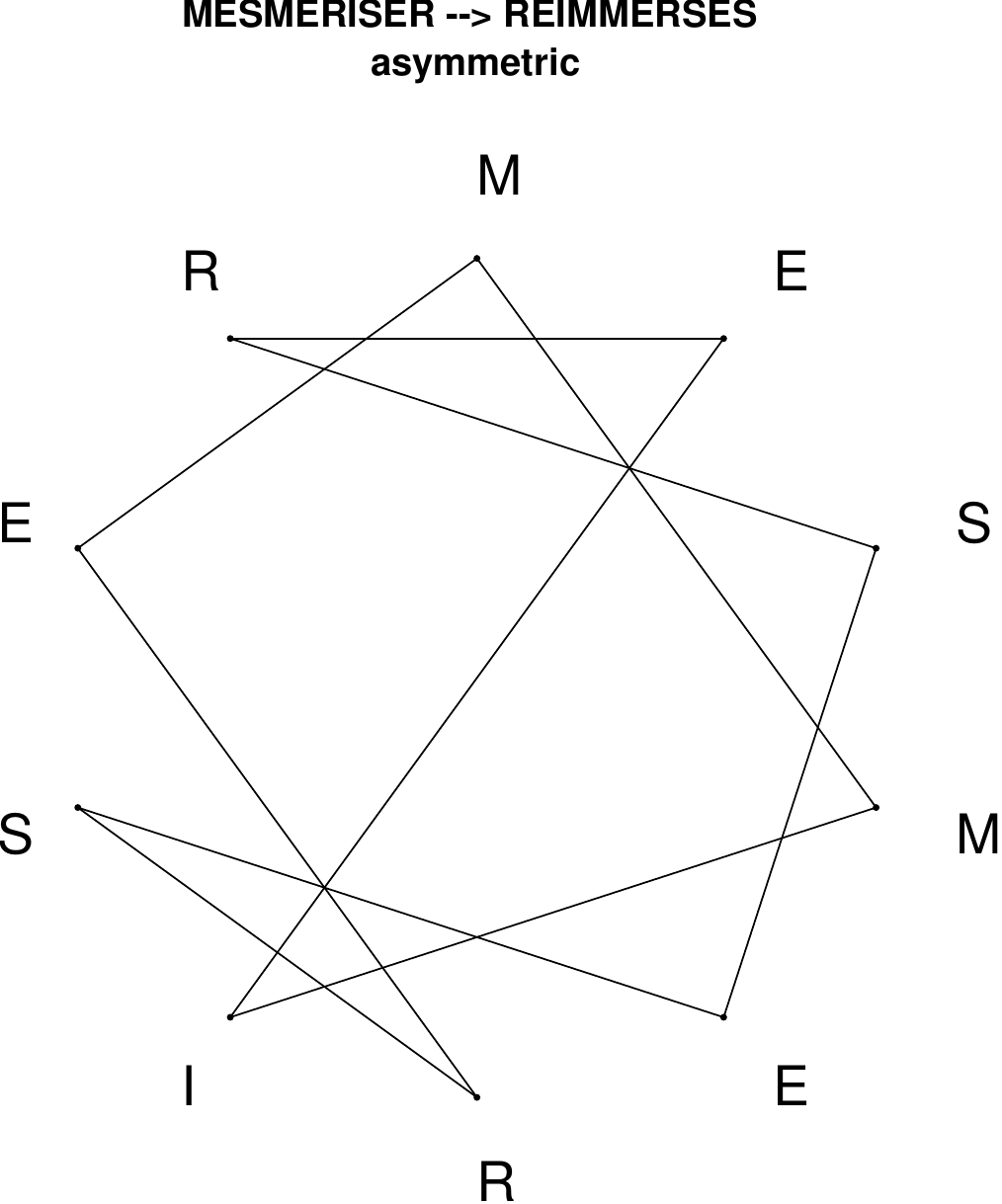}
\end{subfigure}
\end{figure}

\begin{figure}[H]
\centering
\begin{subfigure}[T]{0.19\textwidth}
\centering
\includegraphics[width=\textwidth]{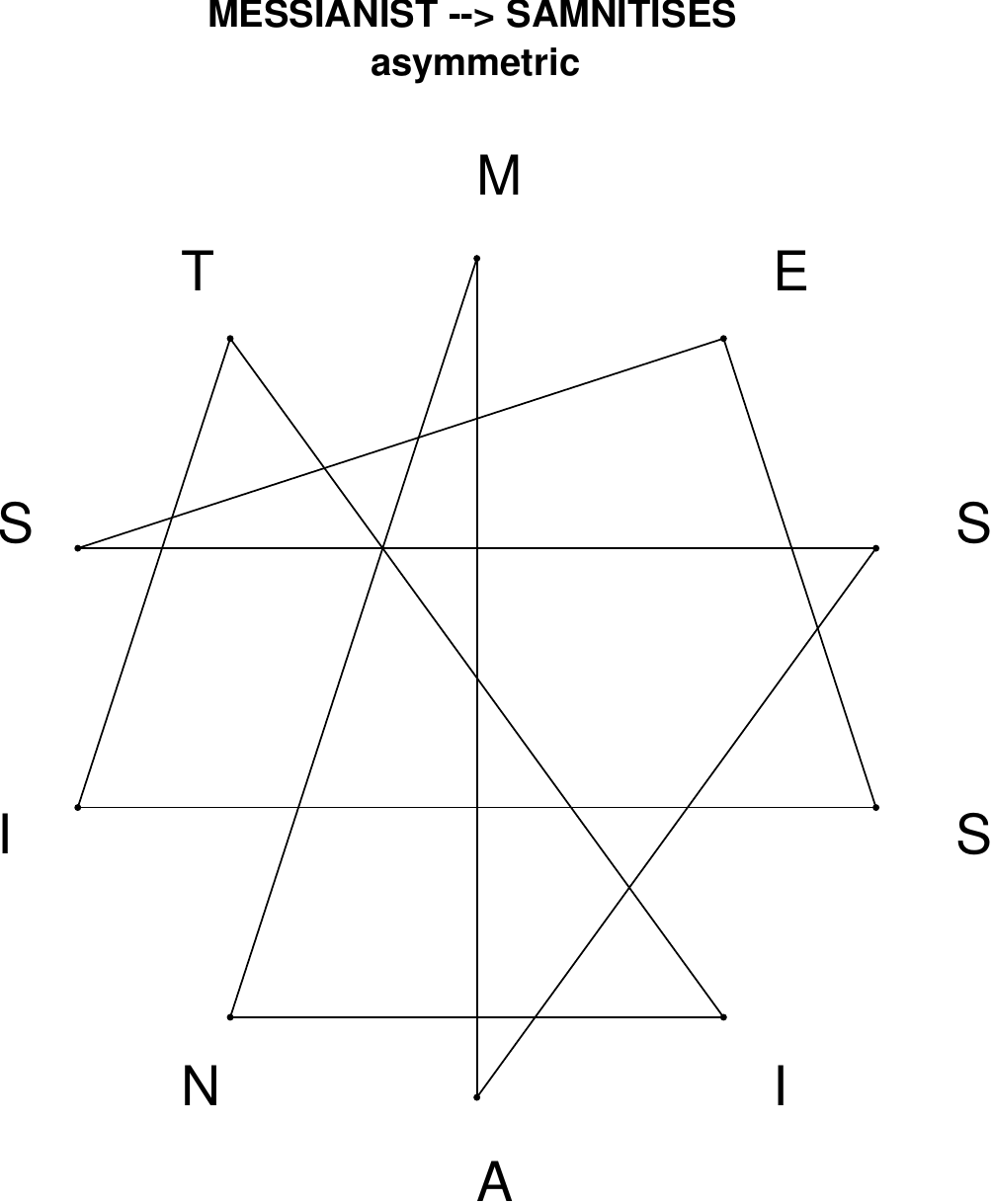}
\end{subfigure}
\hfill
\begin{subfigure}[T]{0.19\textwidth}
\centering
\includegraphics[width=\textwidth]{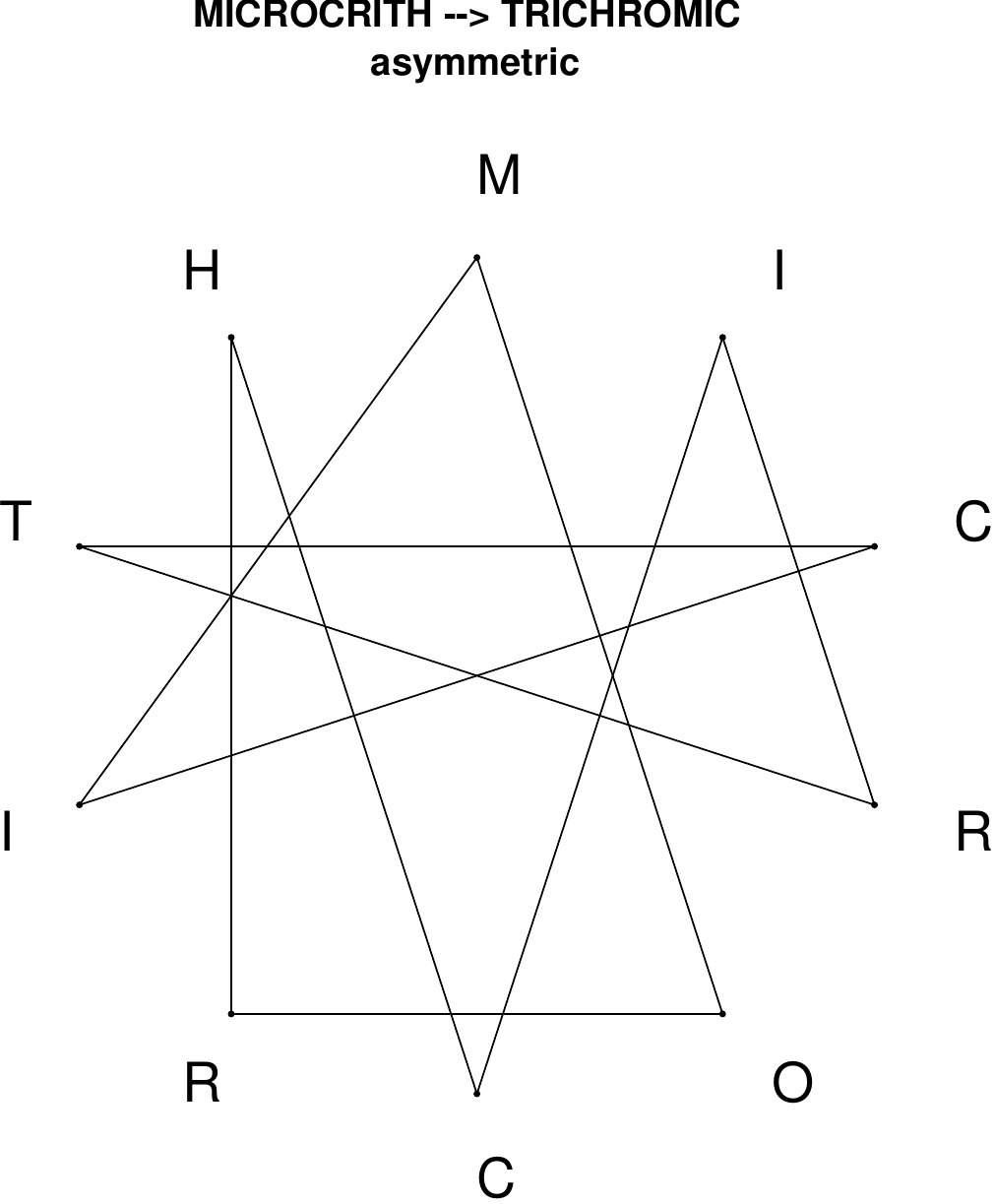}
\end{subfigure}
\hfill
\begin{subfigure}[T]{0.19\textwidth}
\centering
\includegraphics[width=\textwidth]{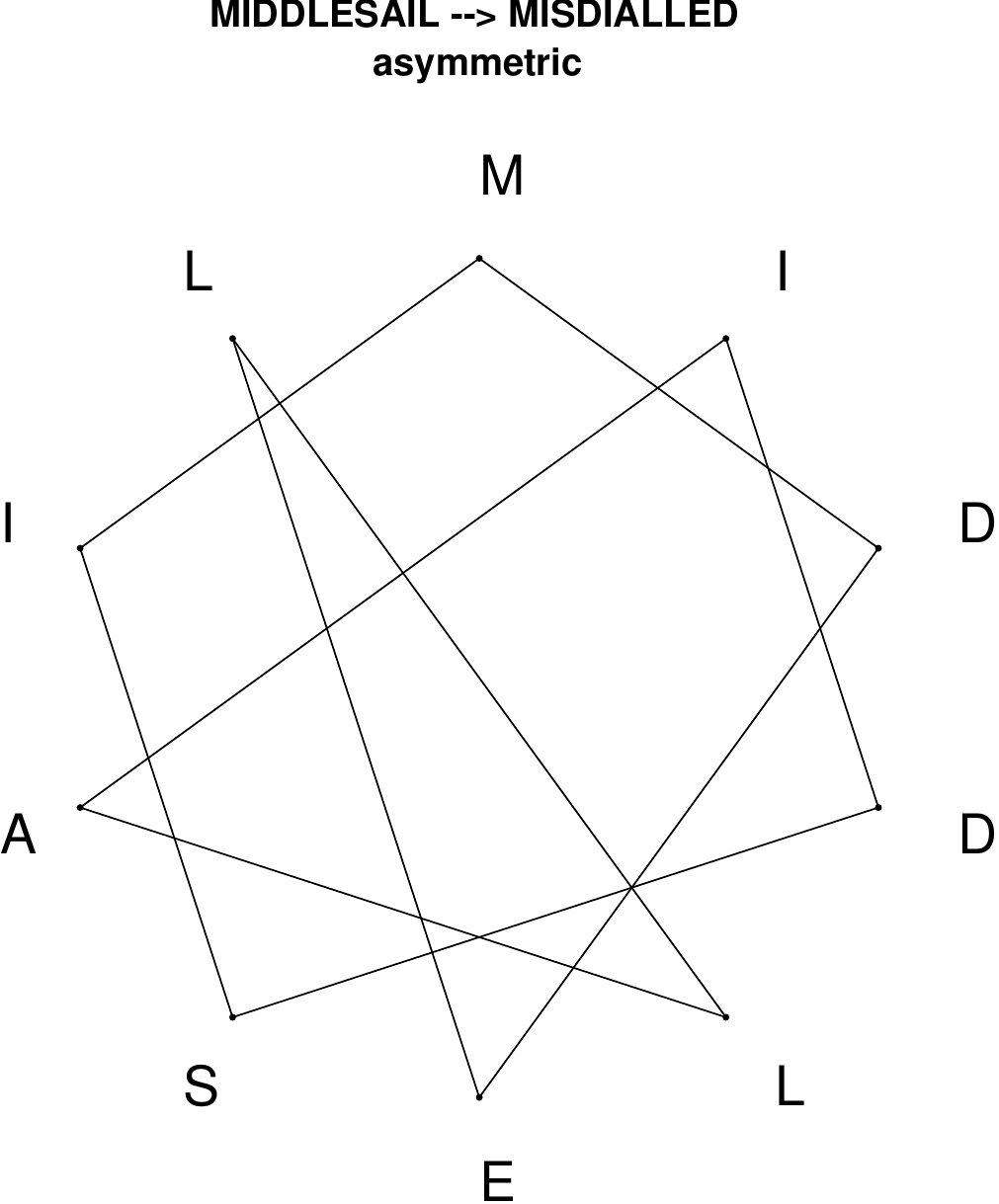}
\end{subfigure}
\hfill
\begin{subfigure}[T]{0.19\textwidth}
\centering
\includegraphics[width=\textwidth]{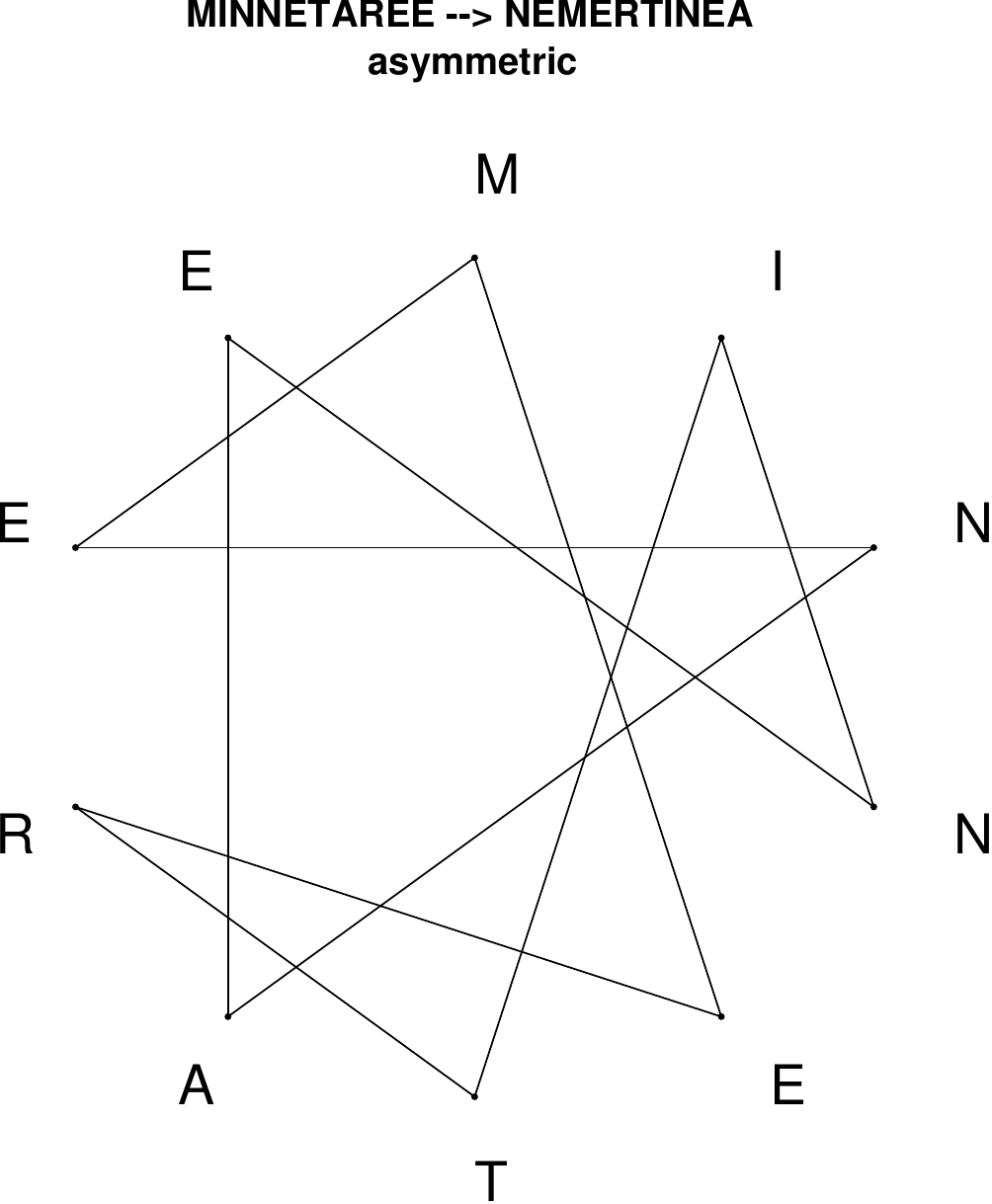}
\end{subfigure}
\hfill
\begin{subfigure}[T]{0.19\textwidth}
\centering
\includegraphics[width=\textwidth]{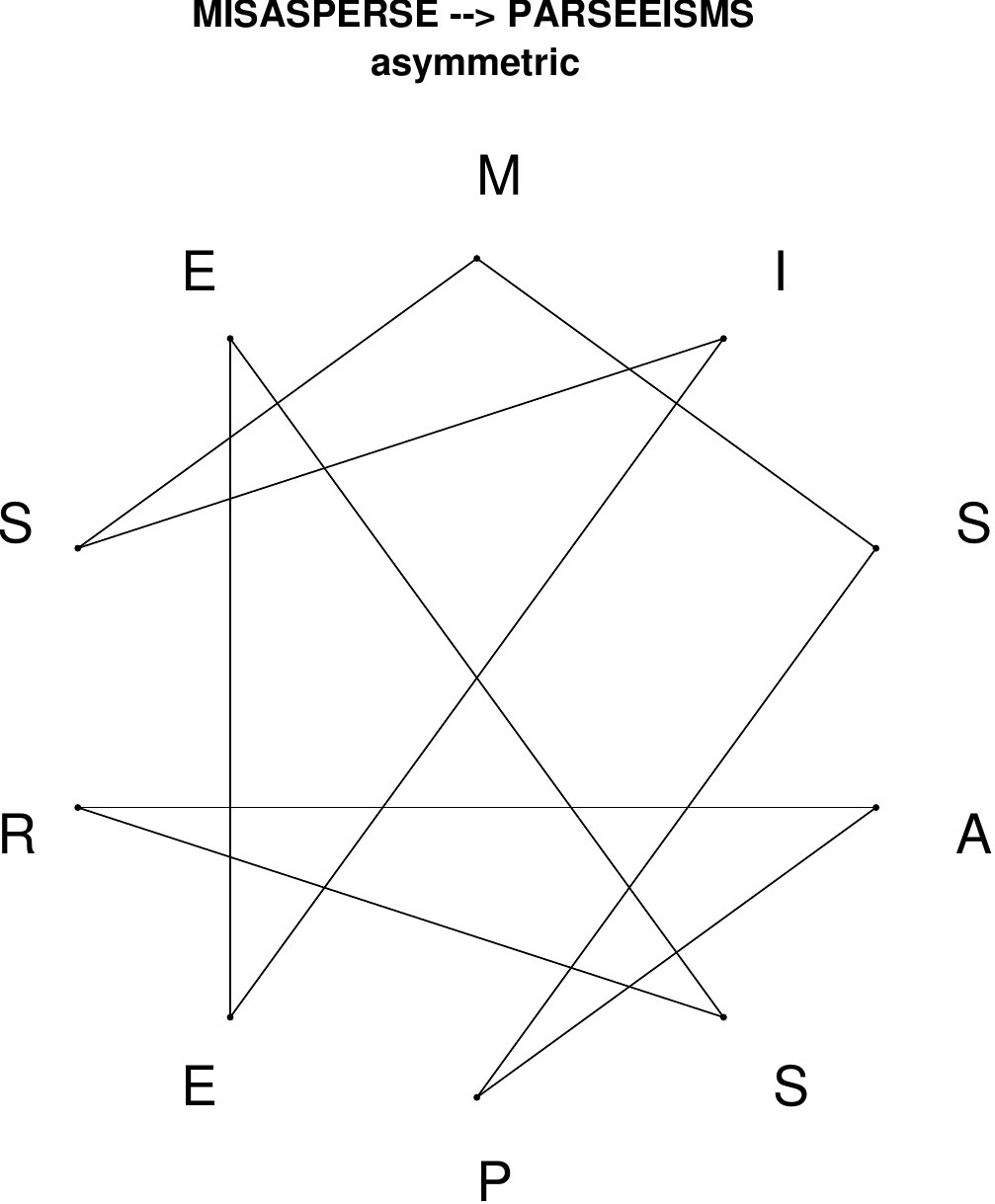}
\end{subfigure}
\end{figure}

\begin{figure}[H]
\centering
\begin{subfigure}[T]{0.19\textwidth}
\centering
\includegraphics[width=\textwidth]{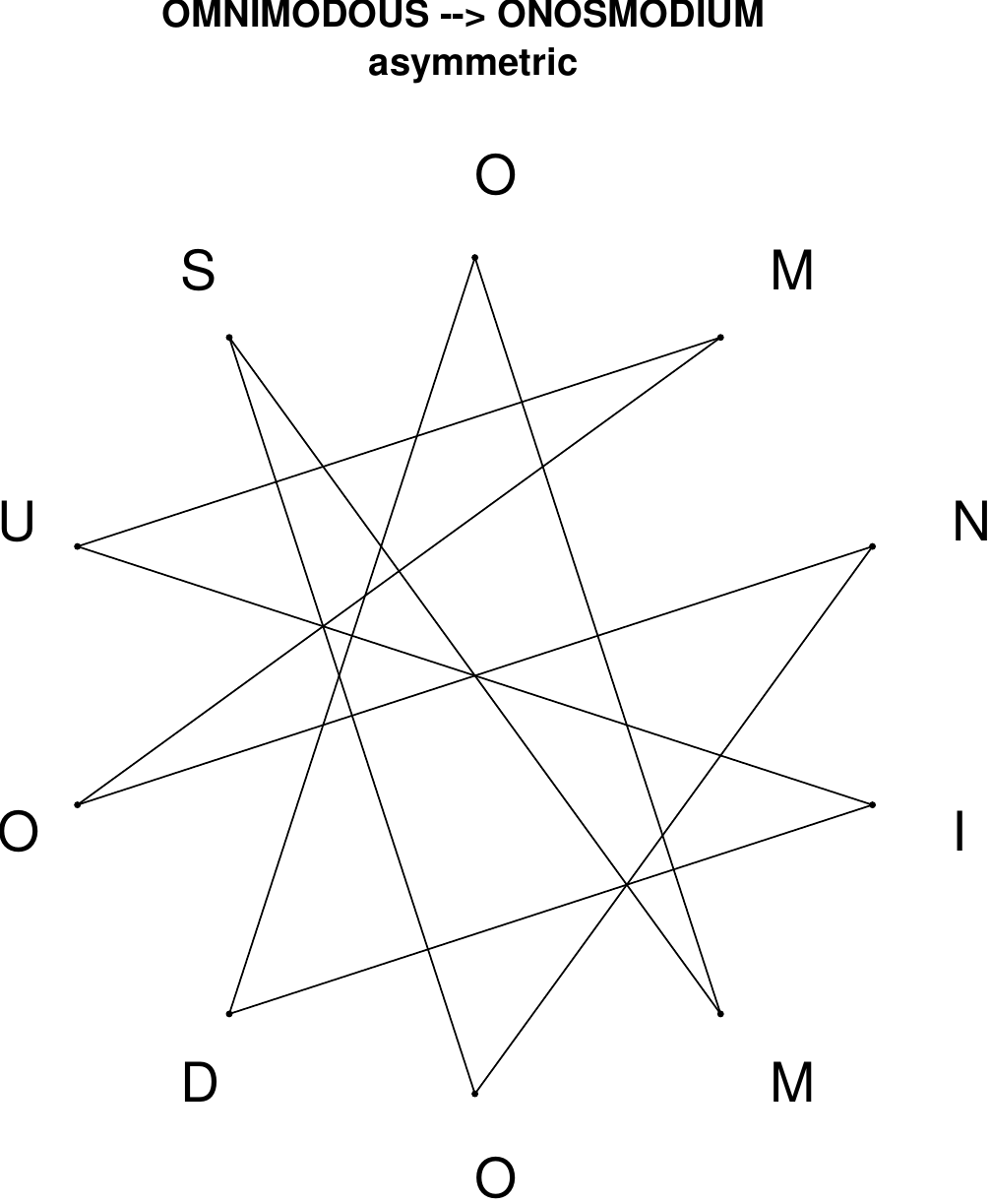}
\end{subfigure}
\hfill
\begin{subfigure}[T]{0.19\textwidth}
\centering
\includegraphics[width=\textwidth]{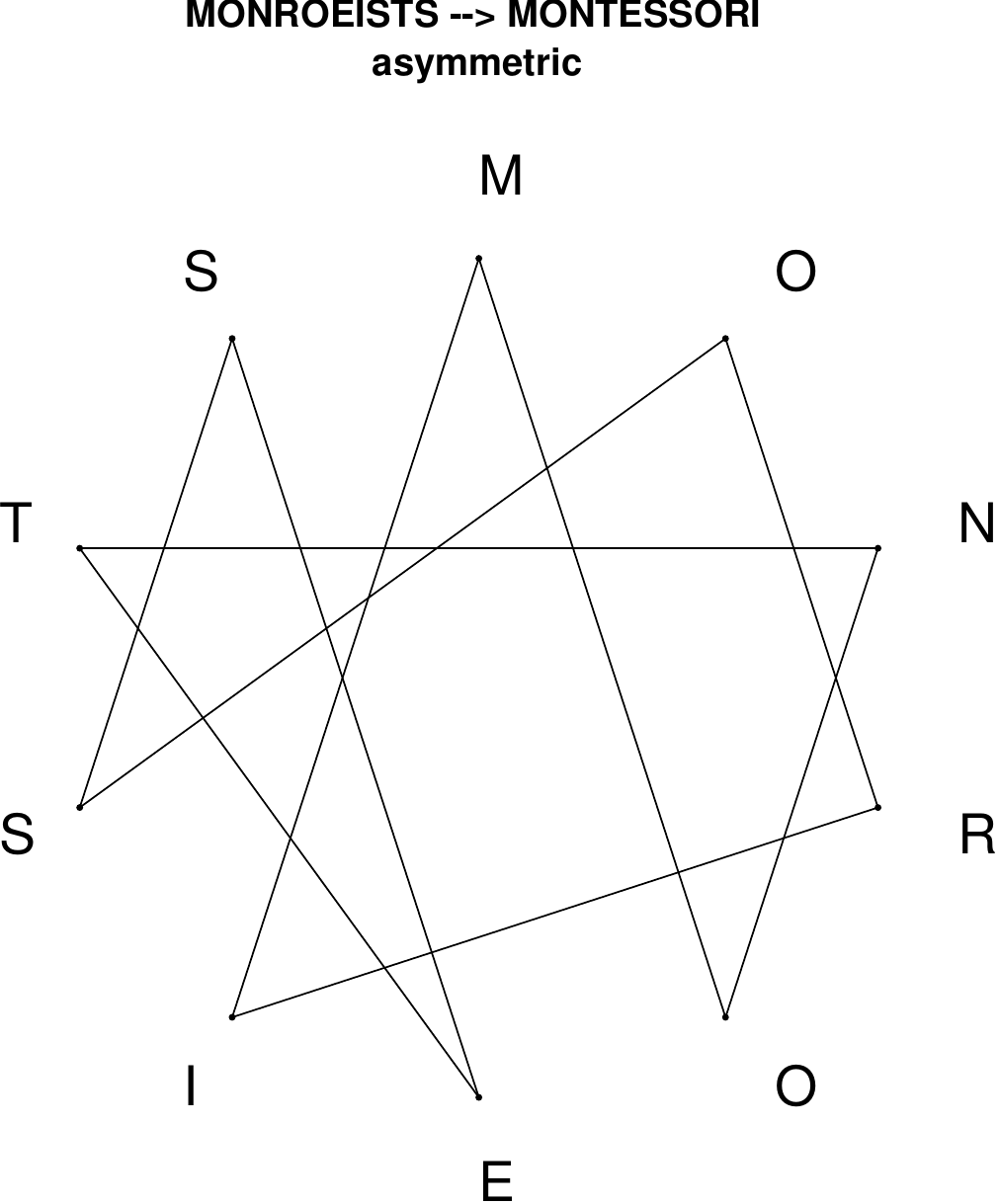}
\end{subfigure}
\hfill
\begin{subfigure}[T]{0.19\textwidth}
\centering
\includegraphics[width=\textwidth]{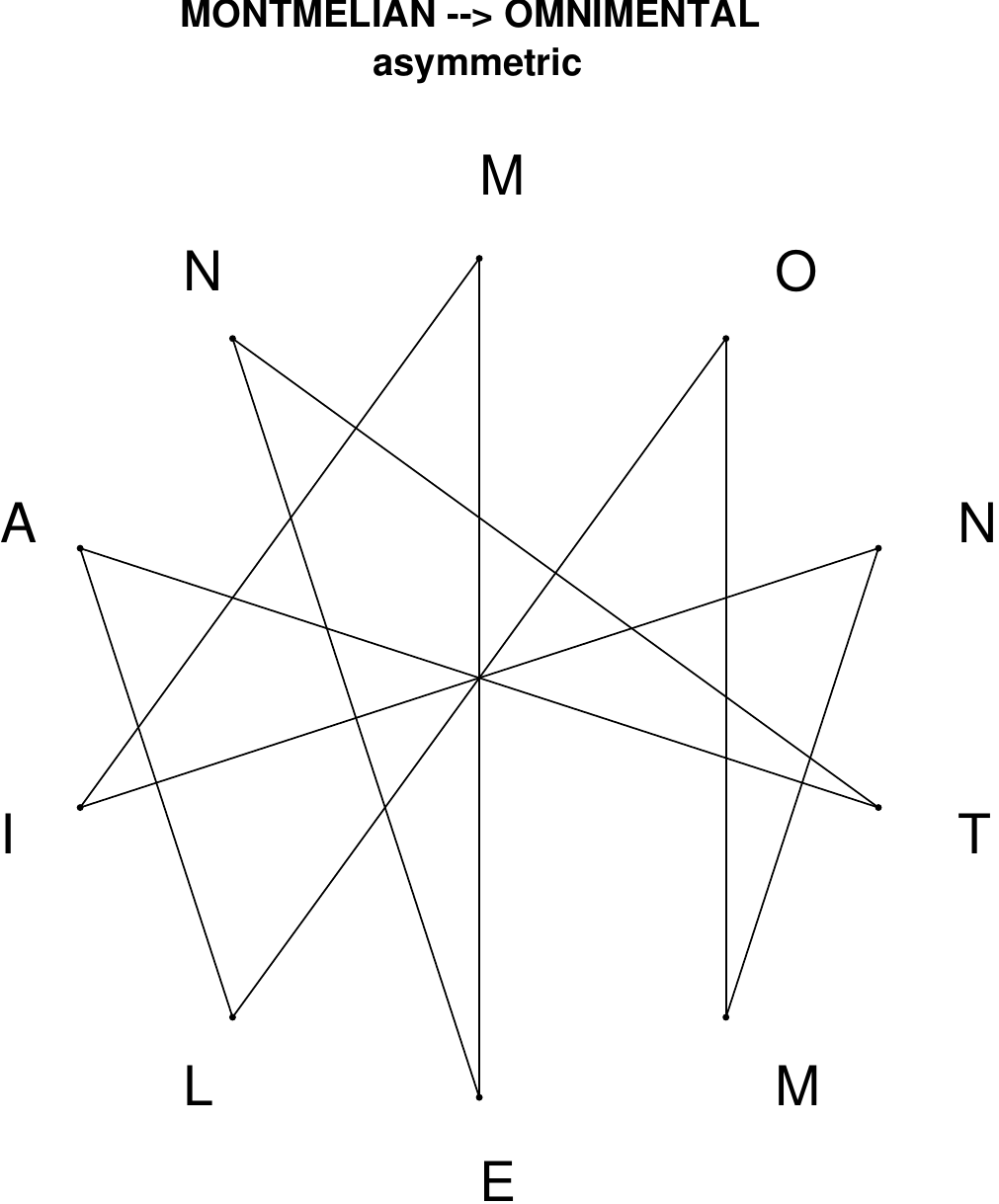}
\end{subfigure}
\hfill
\begin{subfigure}[T]{0.19\textwidth}
\centering
\includegraphics[width=\textwidth]{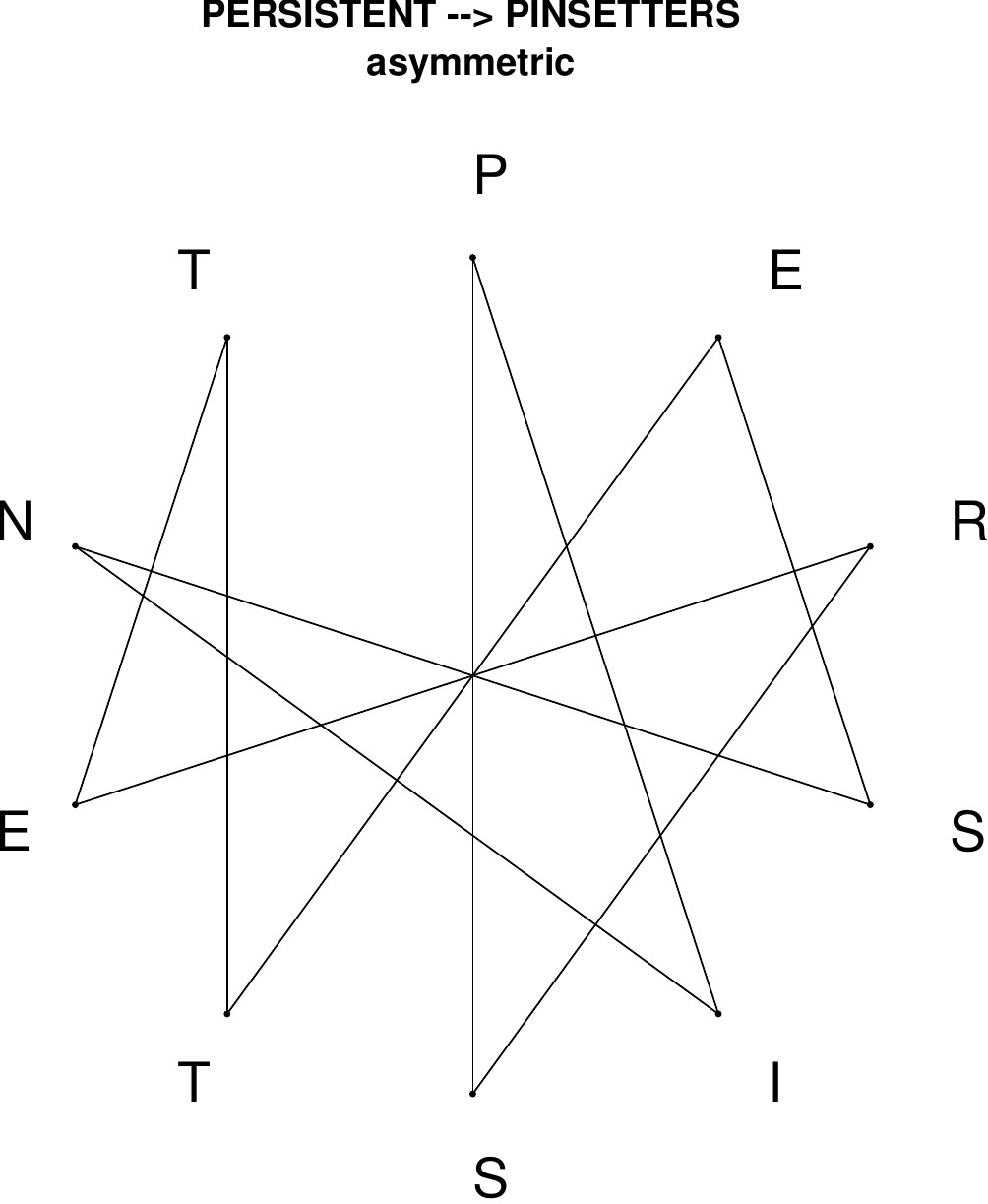}
\end{subfigure}
\hfill
\begin{subfigure}[T]{0.19\textwidth}
\centering
\includegraphics[width=\textwidth]{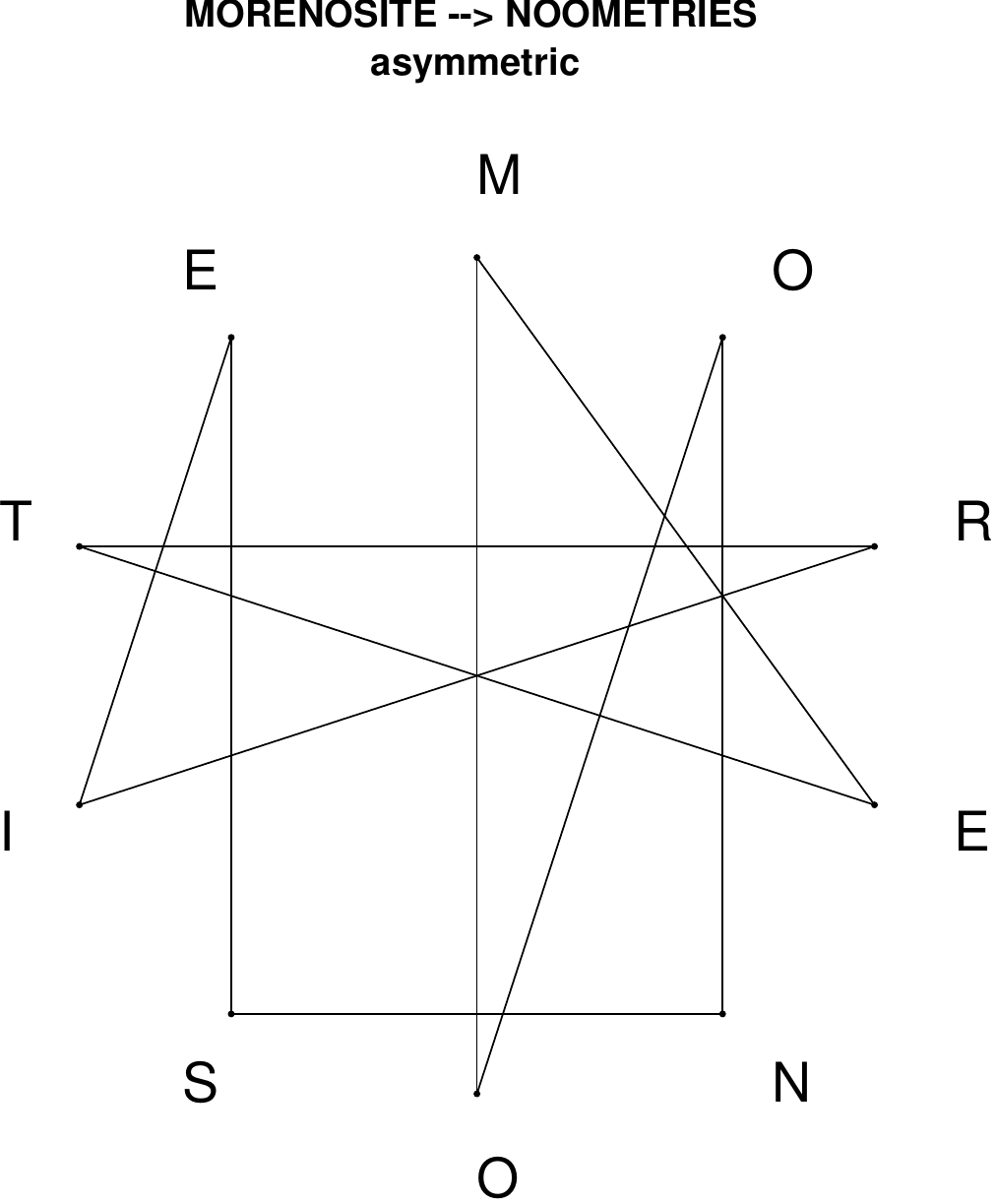}
\end{subfigure}
\end{figure}

\begin{figure}[H]
\centering
\begin{subfigure}[T]{0.19\textwidth}
\centering
\includegraphics[width=\textwidth]{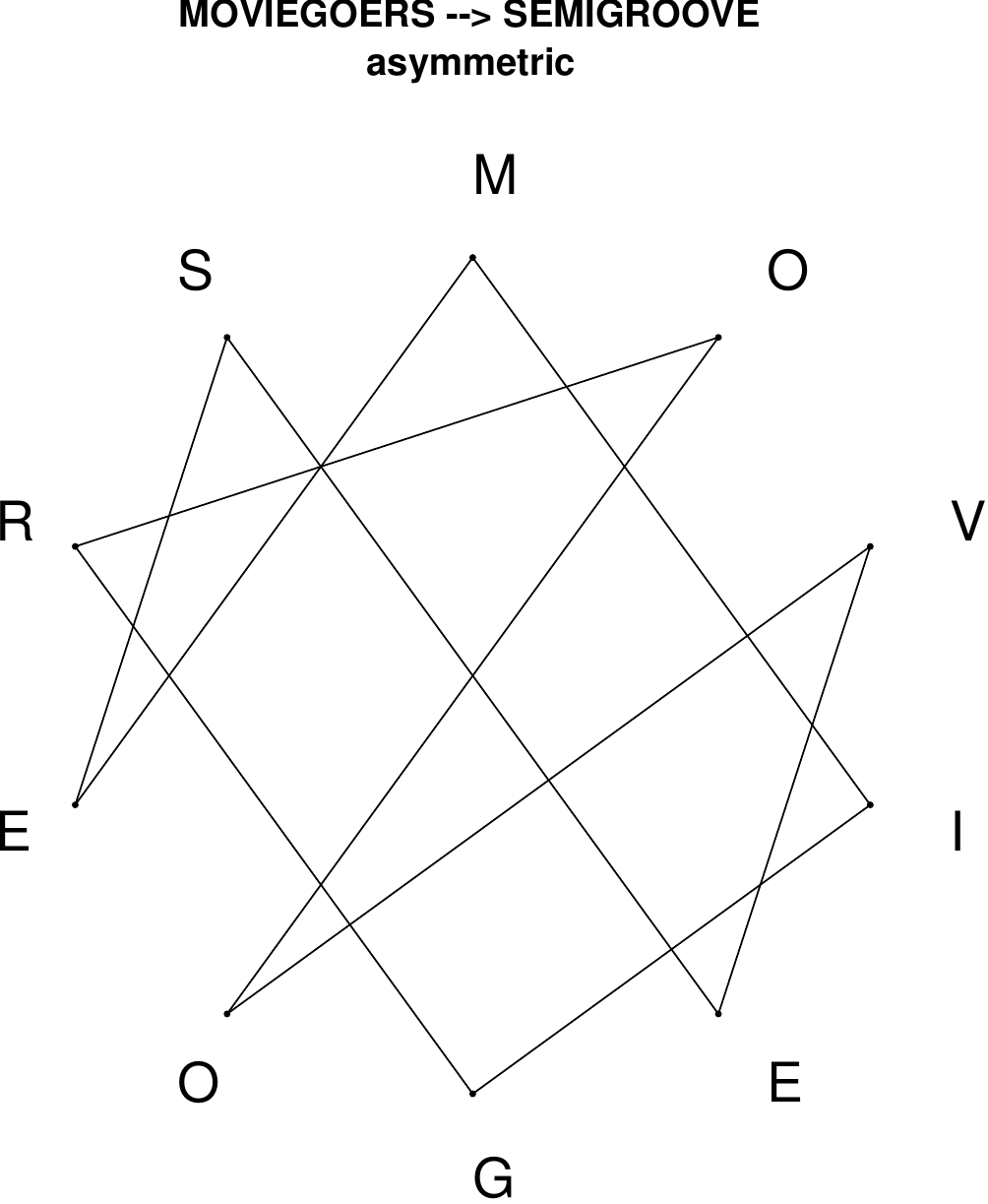}
\end{subfigure}
\hfill
\begin{subfigure}[T]{0.19\textwidth}
\centering
\includegraphics[width=\textwidth]{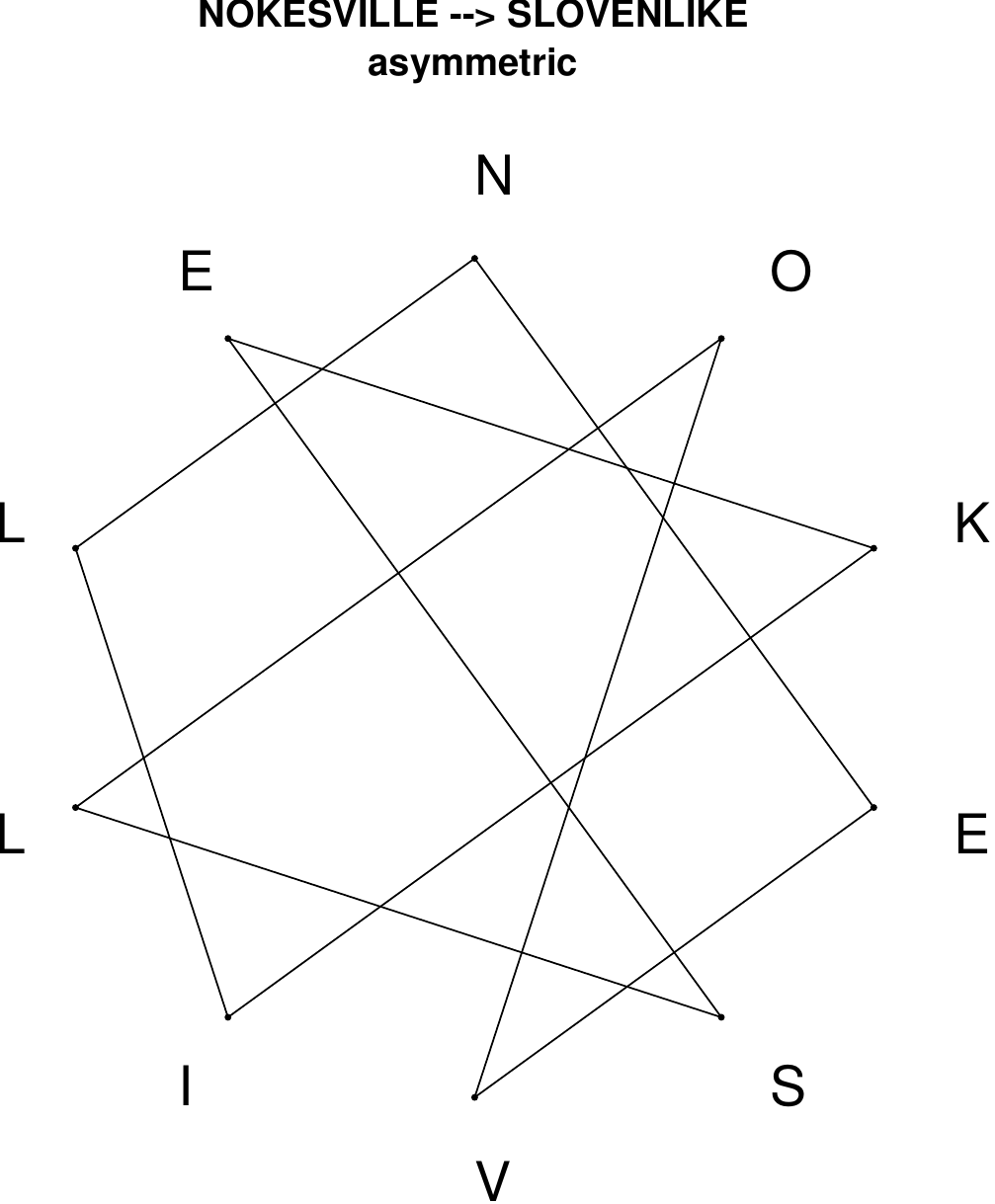}
\end{subfigure}
\hfill
\begin{subfigure}[T]{0.19\textwidth}
\centering
\includegraphics[width=\textwidth]{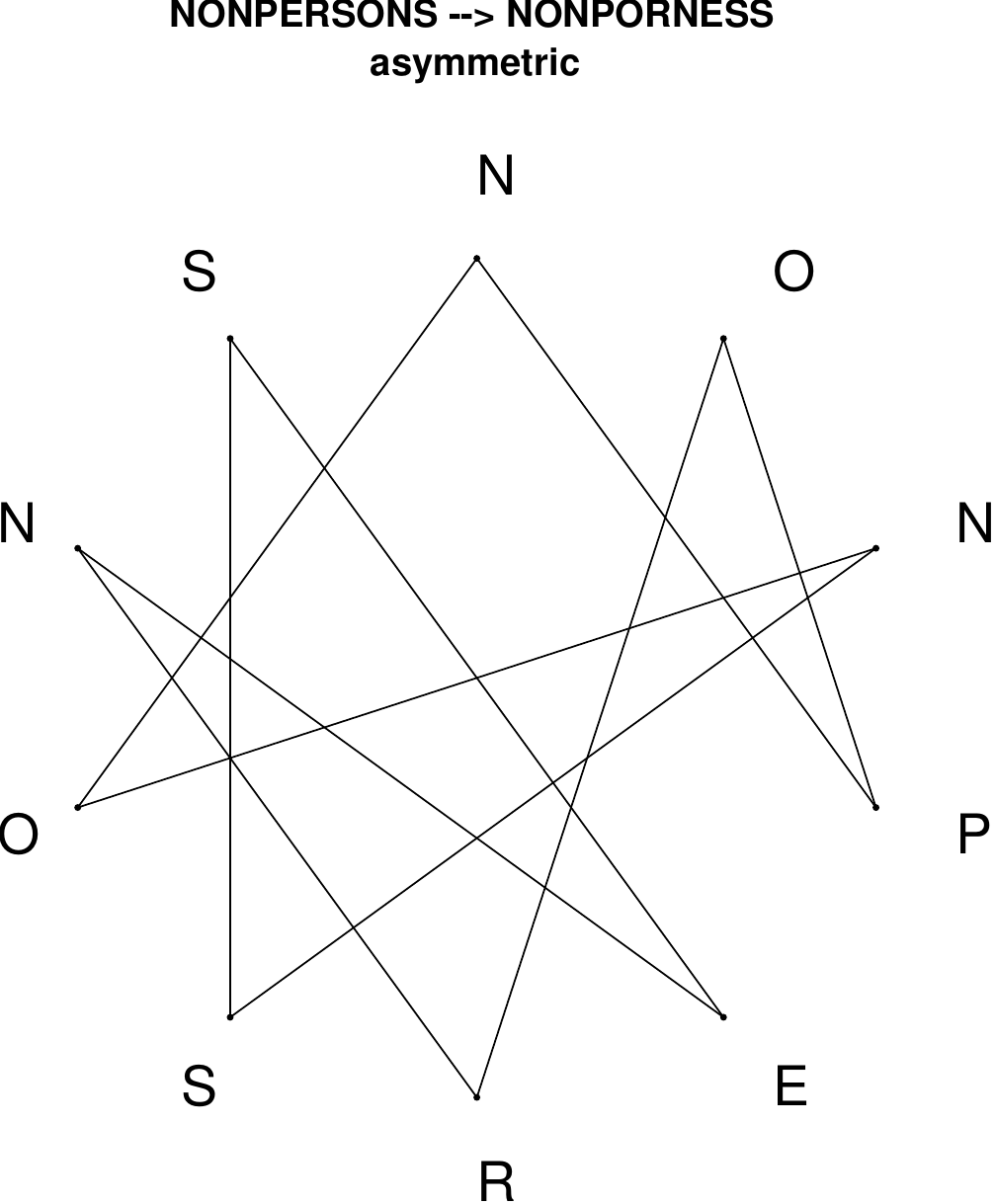}
\end{subfigure}
\hfill
\begin{subfigure}[T]{0.19\textwidth}
\centering
\includegraphics[width=\textwidth]{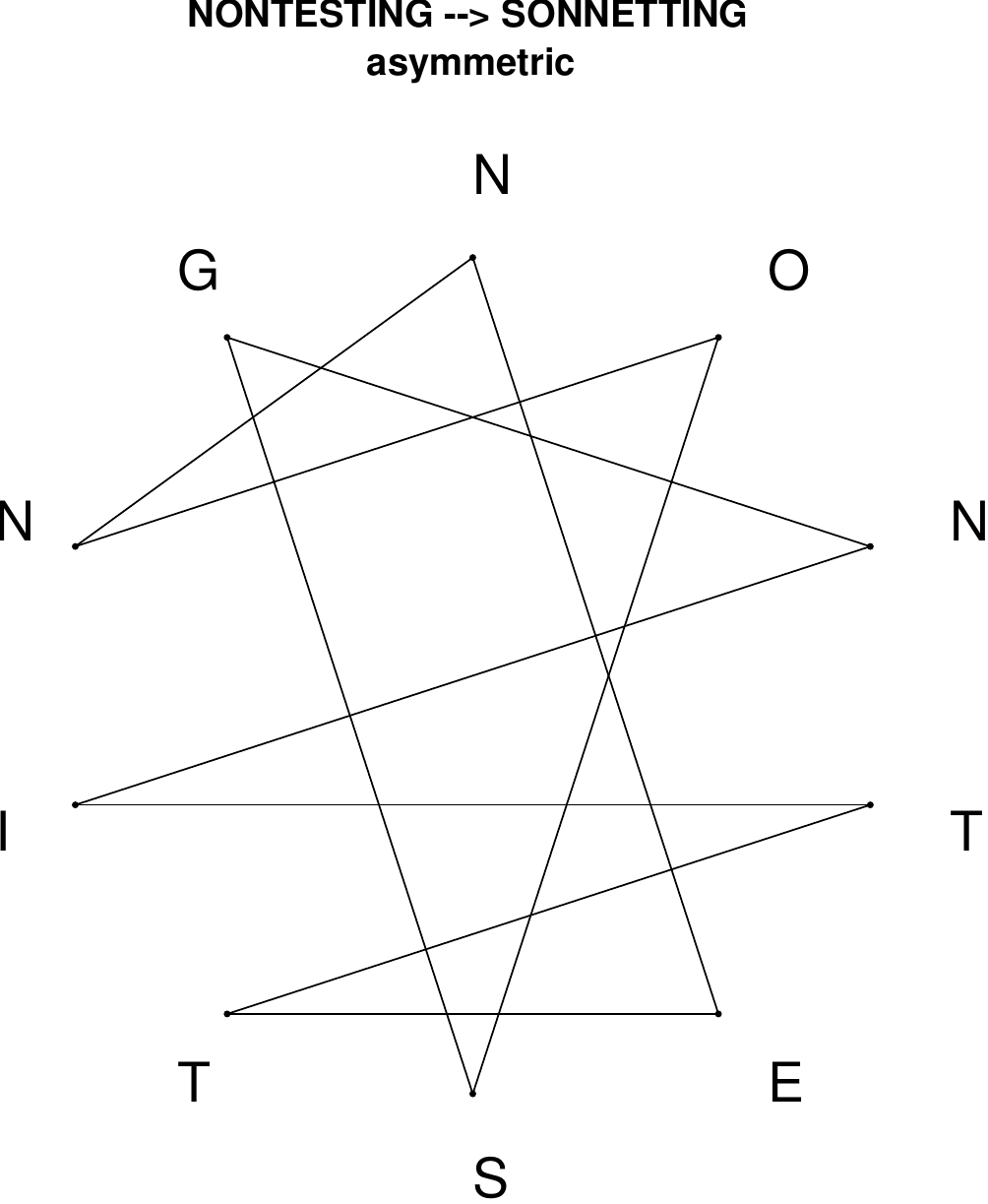}
\end{subfigure}
\hfill
\begin{subfigure}[T]{0.19\textwidth}
\centering
\includegraphics[width=\textwidth]{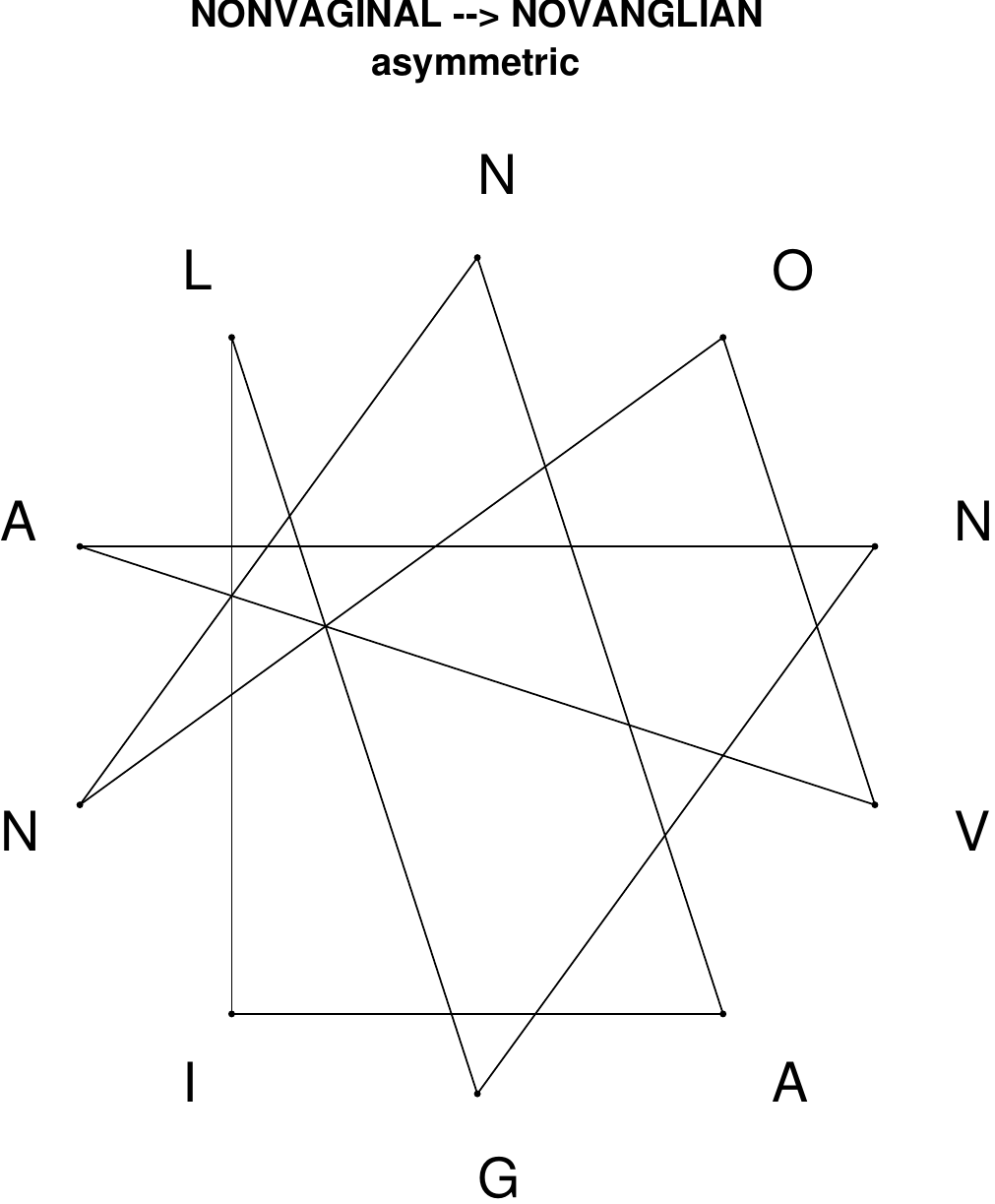}
\end{subfigure}
\end{figure}

\begin{figure}[H]
\centering
\begin{subfigure}[T]{0.19\textwidth}
\centering
\includegraphics[width=\textwidth]{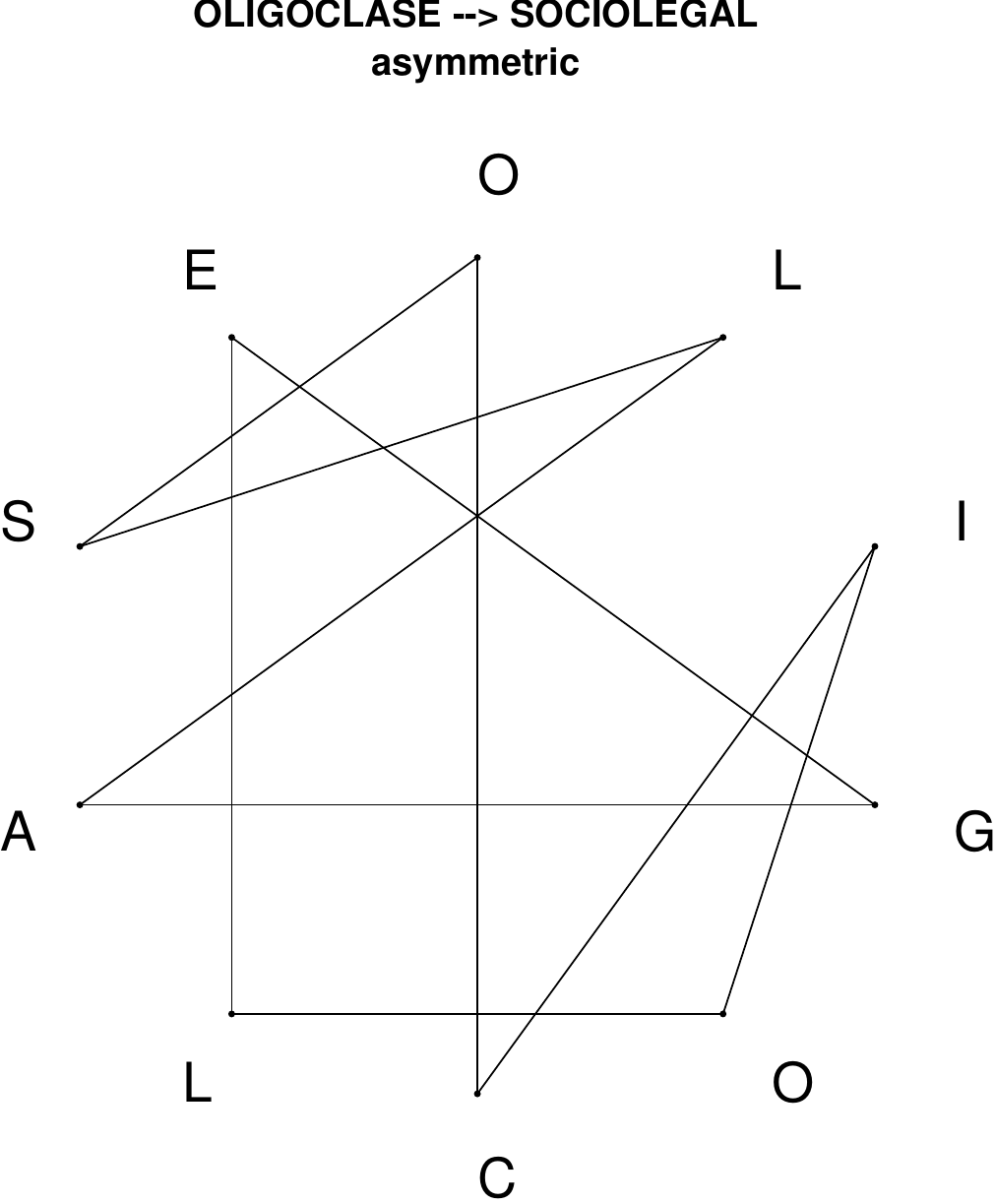}
\end{subfigure}
\hfill
\begin{subfigure}[T]{0.19\textwidth}
\centering
\includegraphics[width=\textwidth]{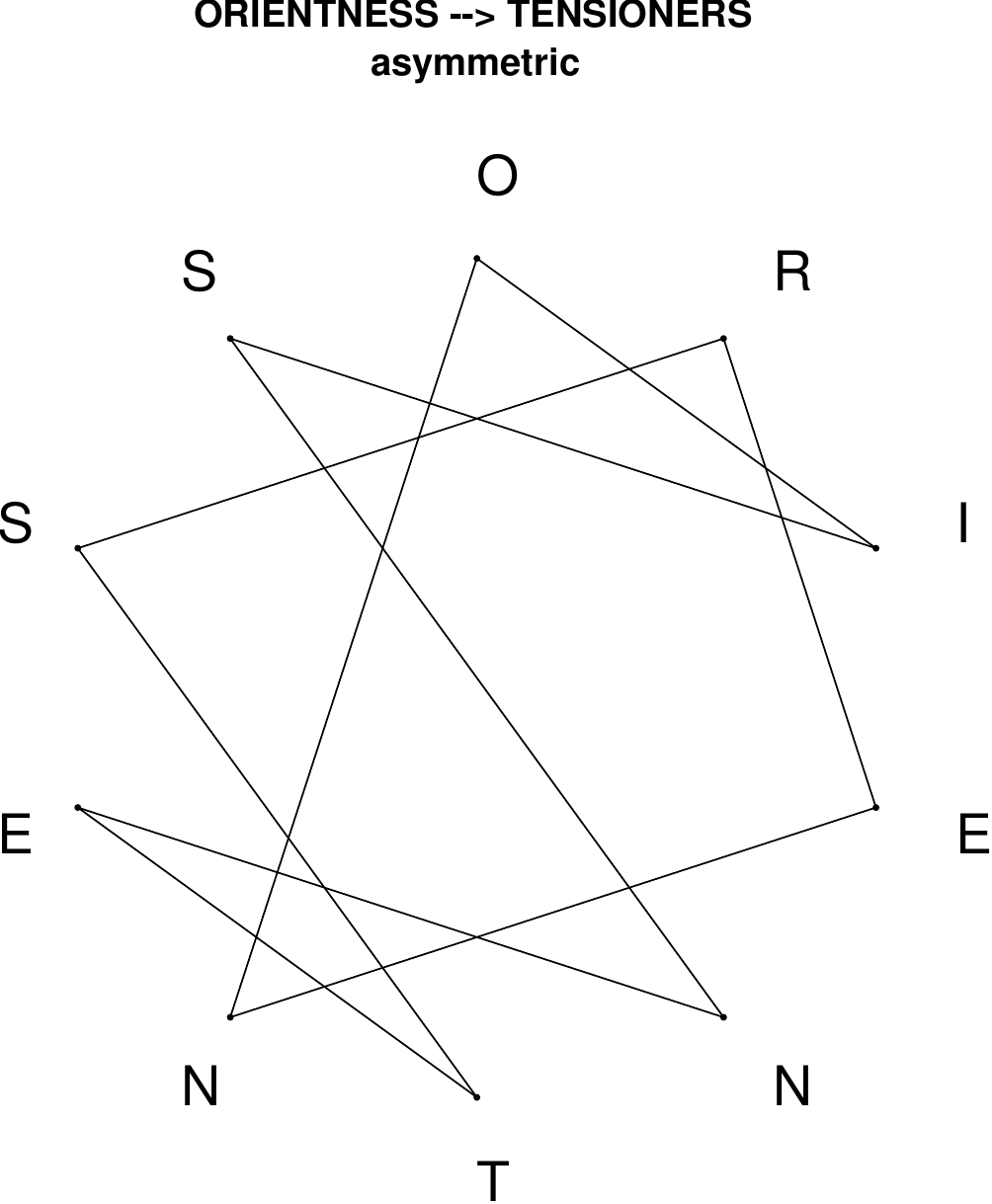}
\end{subfigure}
\hfill
\begin{subfigure}[T]{0.19\textwidth}
\centering
\includegraphics[width=\textwidth]{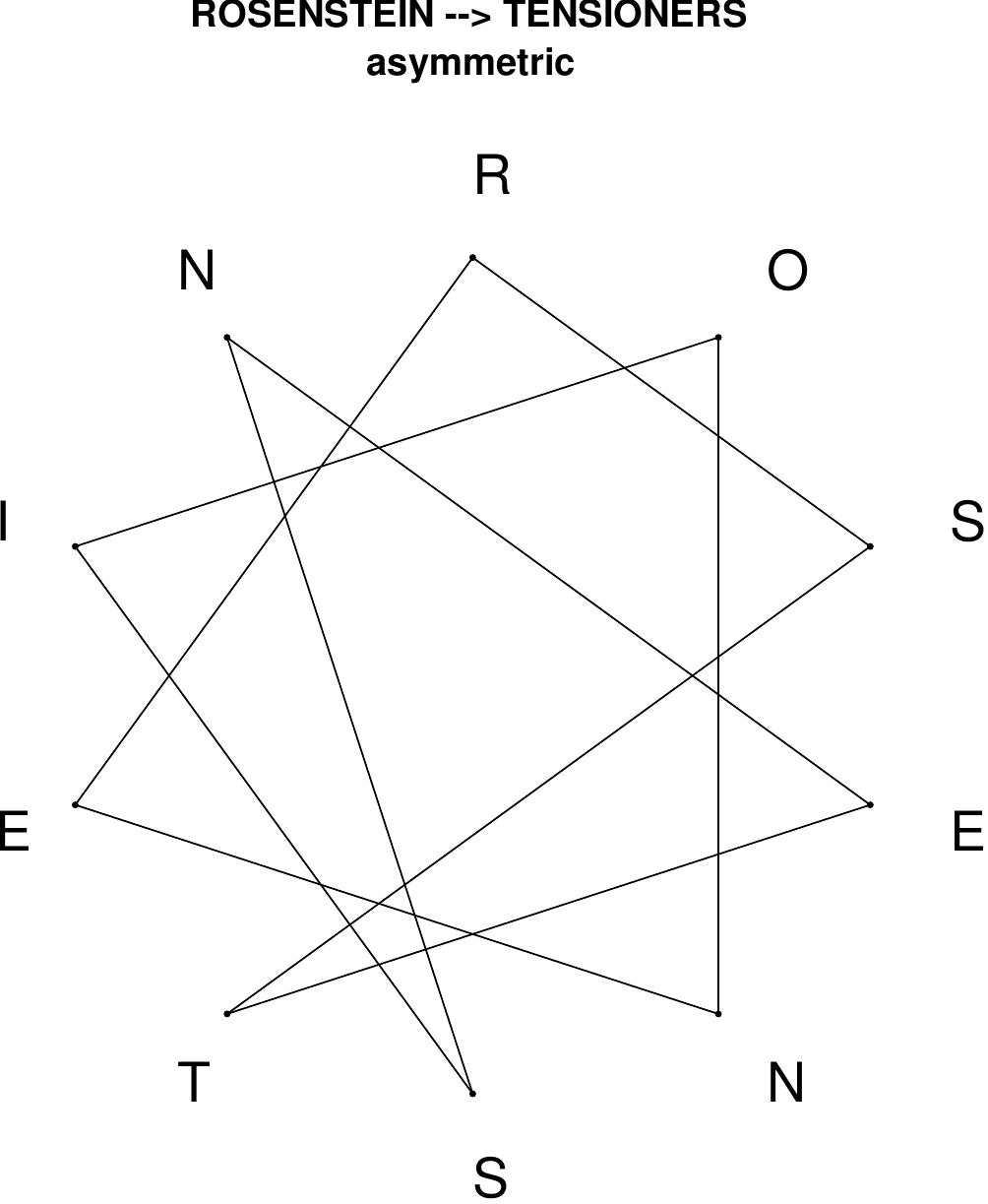}
\end{subfigure}
\hfill
\begin{subfigure}[T]{0.19\textwidth}
\centering
\includegraphics[width=\textwidth]{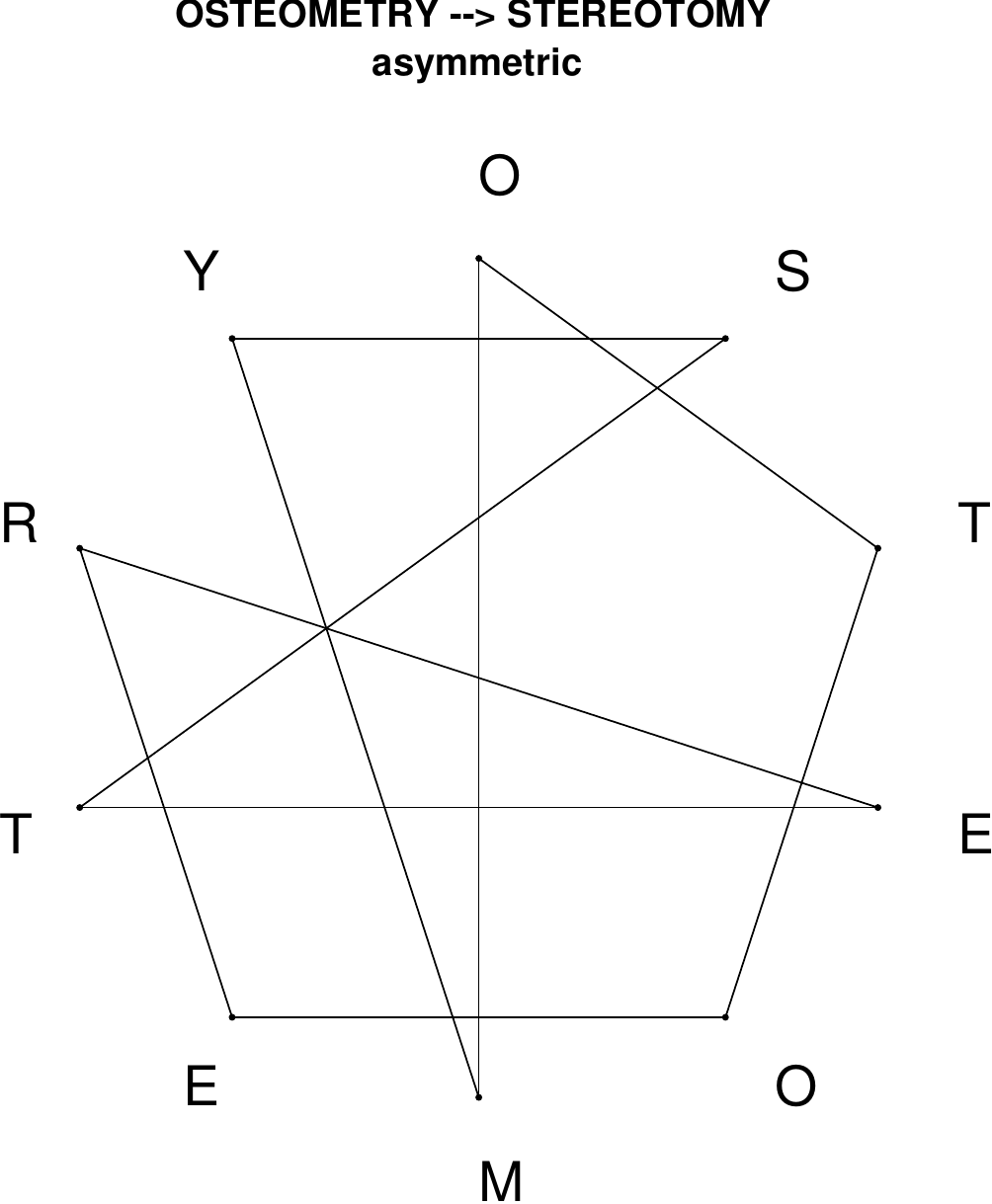}
\end{subfigure}
\hfill
\begin{subfigure}[T]{0.19\textwidth}
\centering
\includegraphics[width=\textwidth]{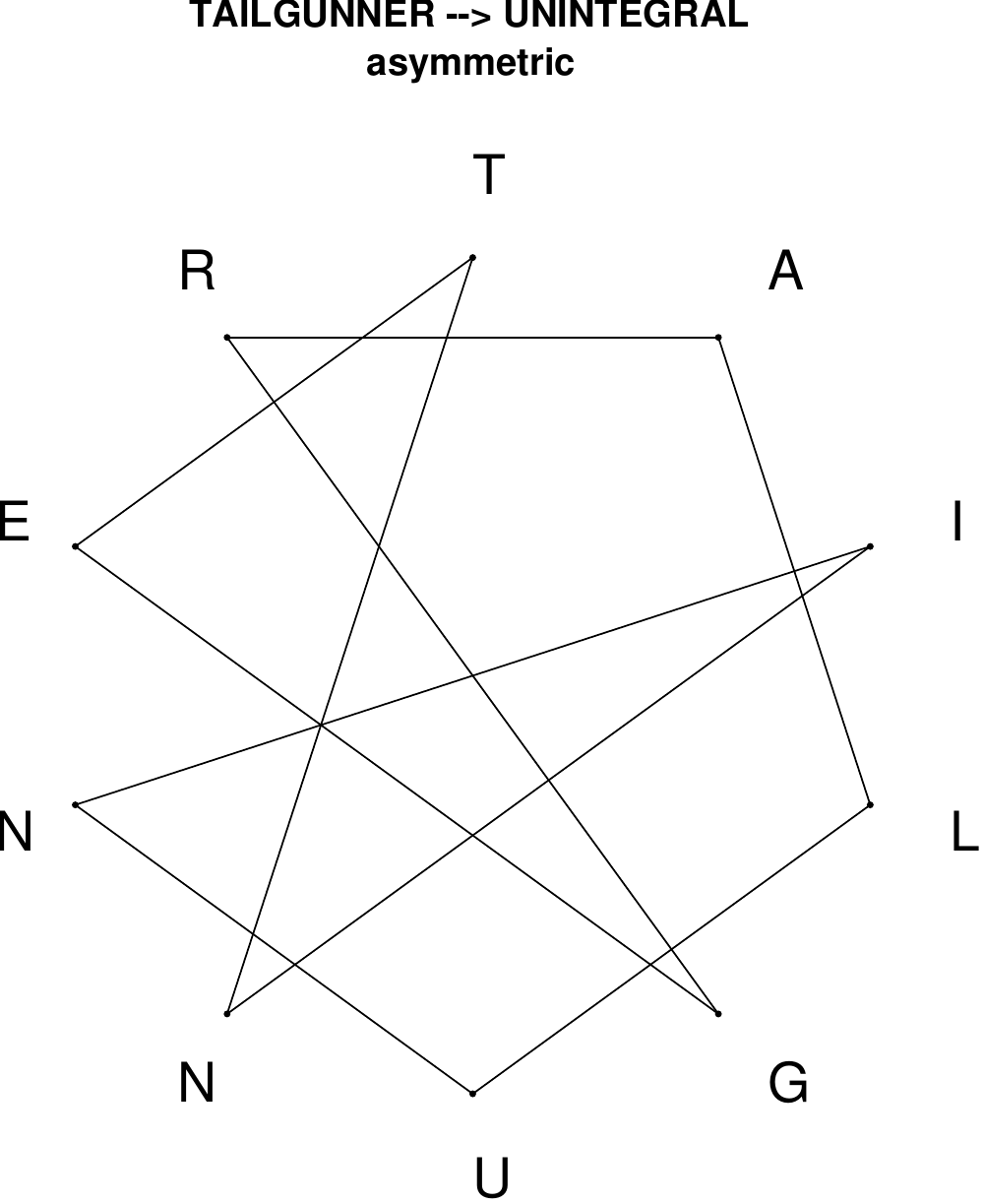}
\end{subfigure}
\end{figure}

\begin{figure}[H]
\centering
\begin{subfigure}[T]{0.19\textwidth}
\centering
\includegraphics[width=\textwidth]{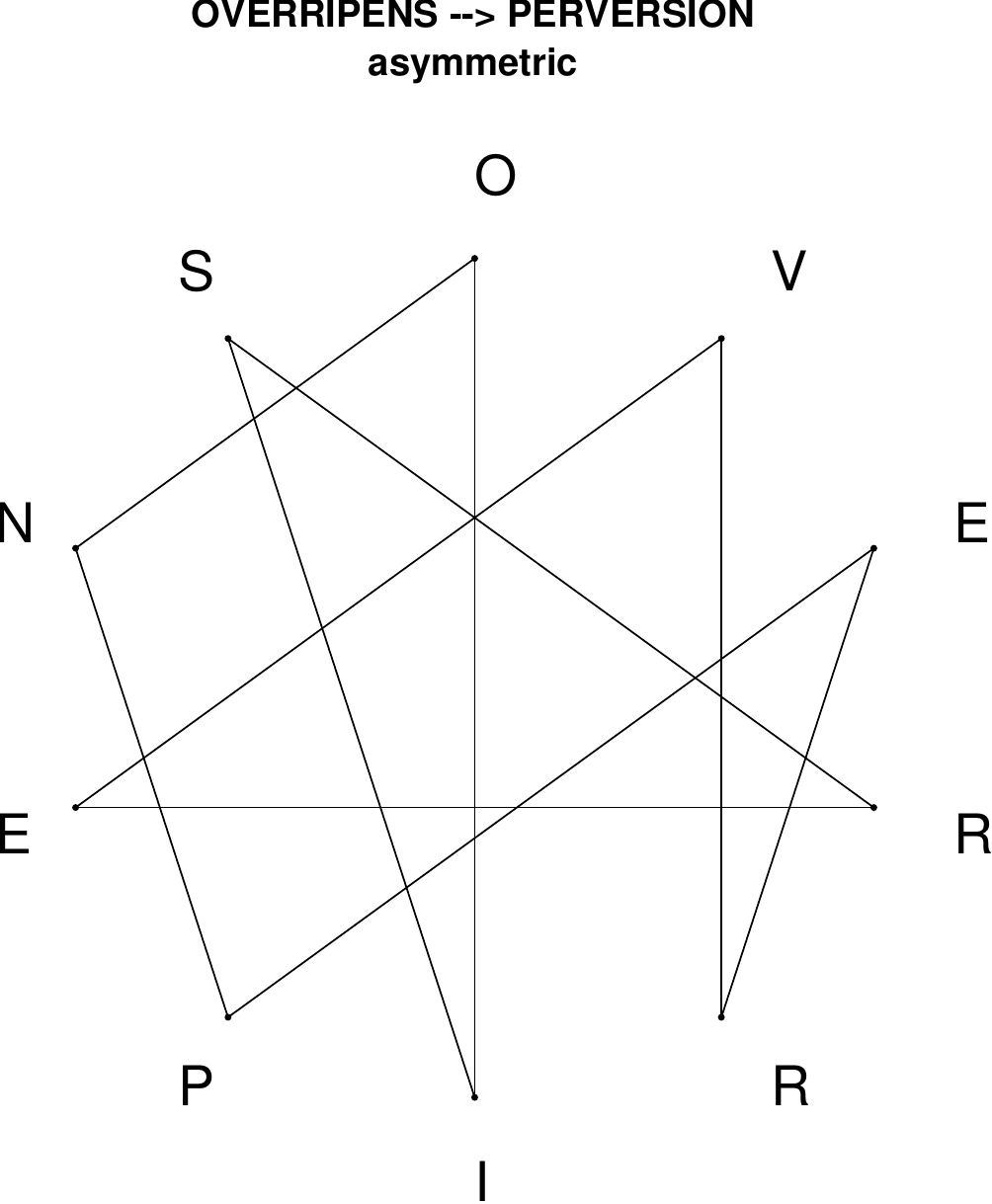}
\end{subfigure}
\hfill
\begin{subfigure}[T]{0.19\textwidth}
\centering
\includegraphics[width=\textwidth]{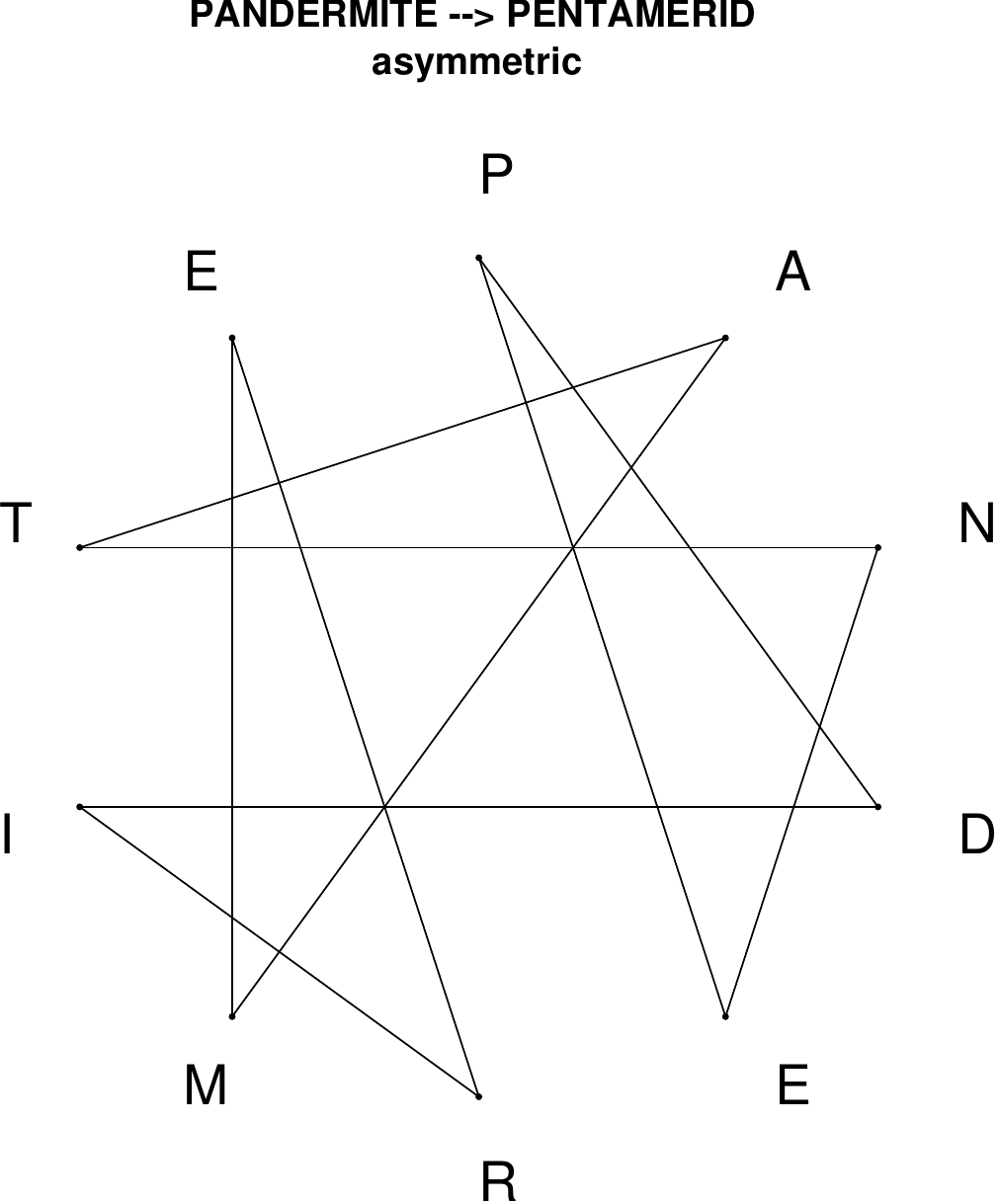}
\end{subfigure}
\hfill
\begin{subfigure}[T]{0.19\textwidth}
\centering
\includegraphics[width=\textwidth]{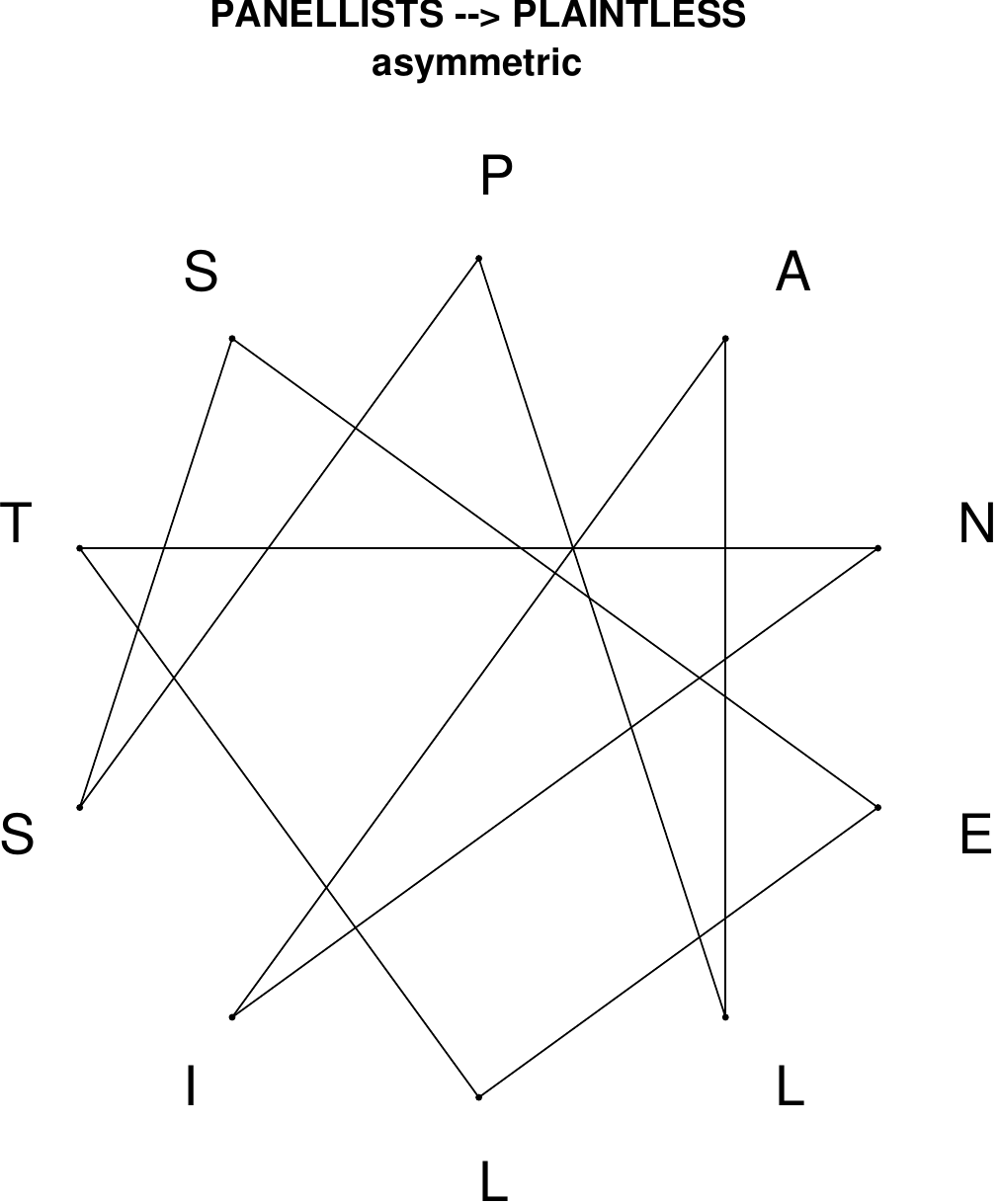}
\end{subfigure}
\hfill
\begin{subfigure}[T]{0.19\textwidth}
\centering
\includegraphics[width=\textwidth]{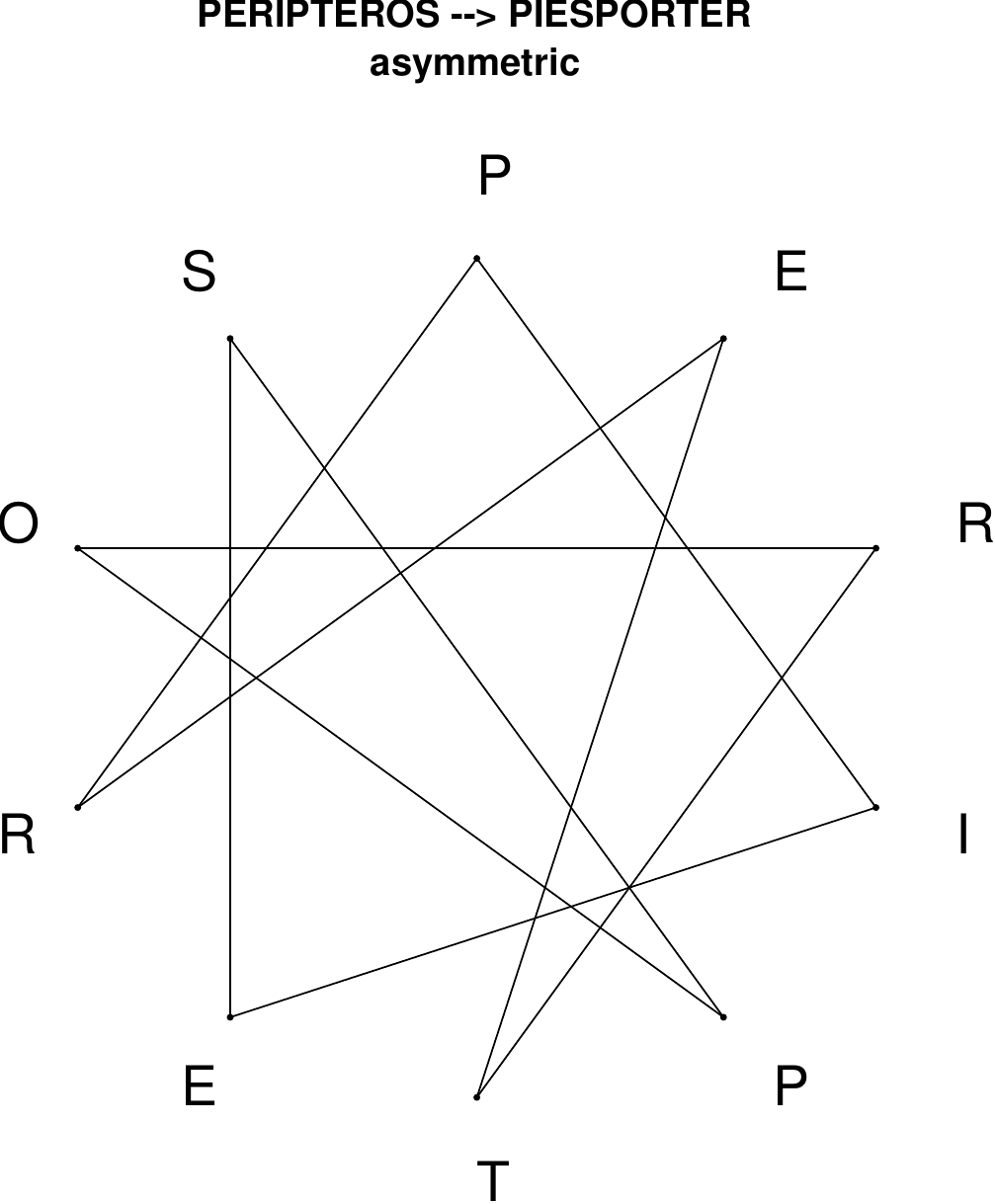}
\end{subfigure}
\hfill
\begin{subfigure}[T]{0.19\textwidth}
\centering
\includegraphics[width=\textwidth]{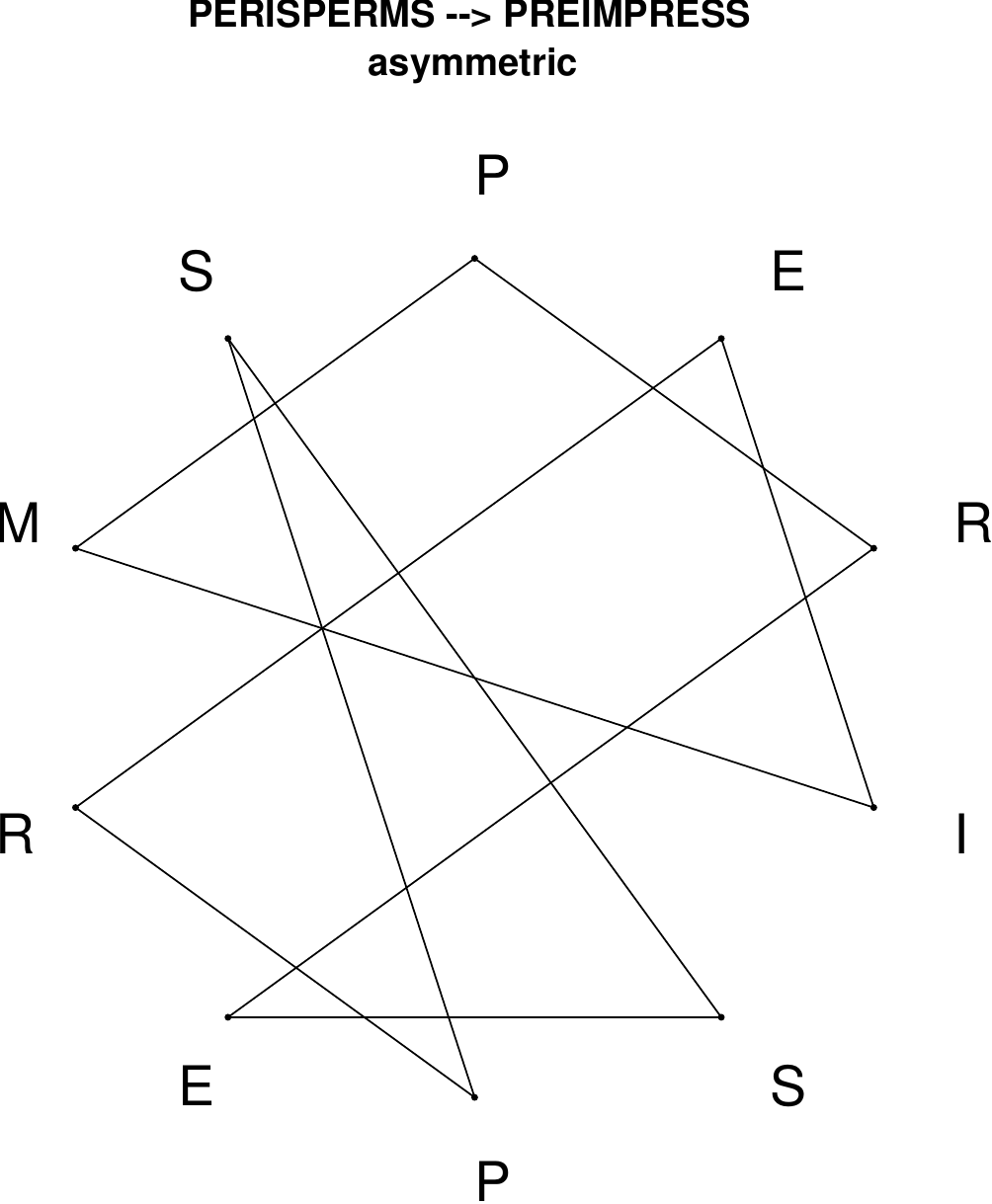}
\end{subfigure}
\end{figure}

\begin{figure}[H]
\centering
\begin{subfigure}[T]{0.19\textwidth}
\centering
\includegraphics[width=\textwidth]{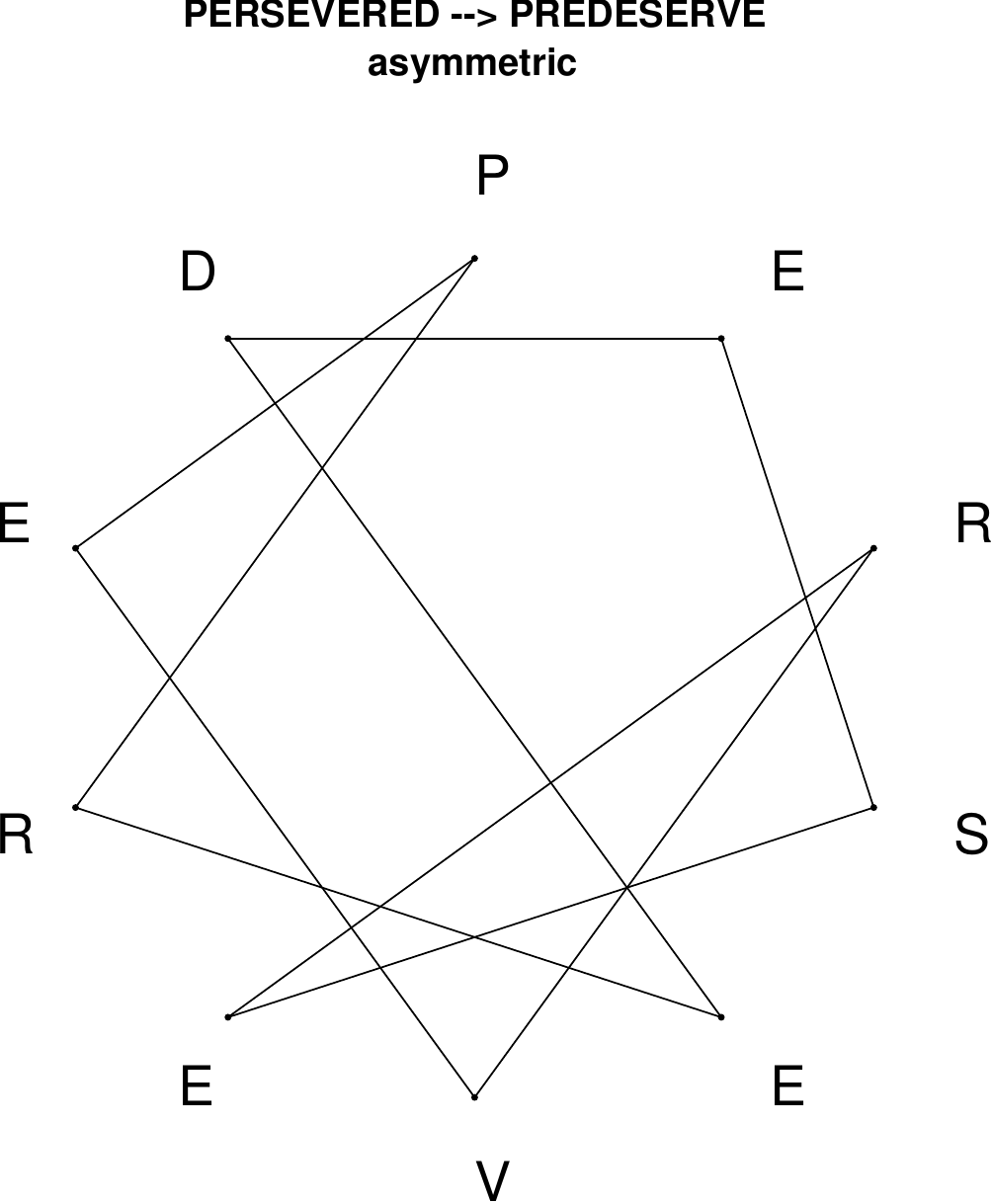}
\end{subfigure}
\hfill
\begin{subfigure}[T]{0.19\textwidth}
\centering
\includegraphics[width=\textwidth]{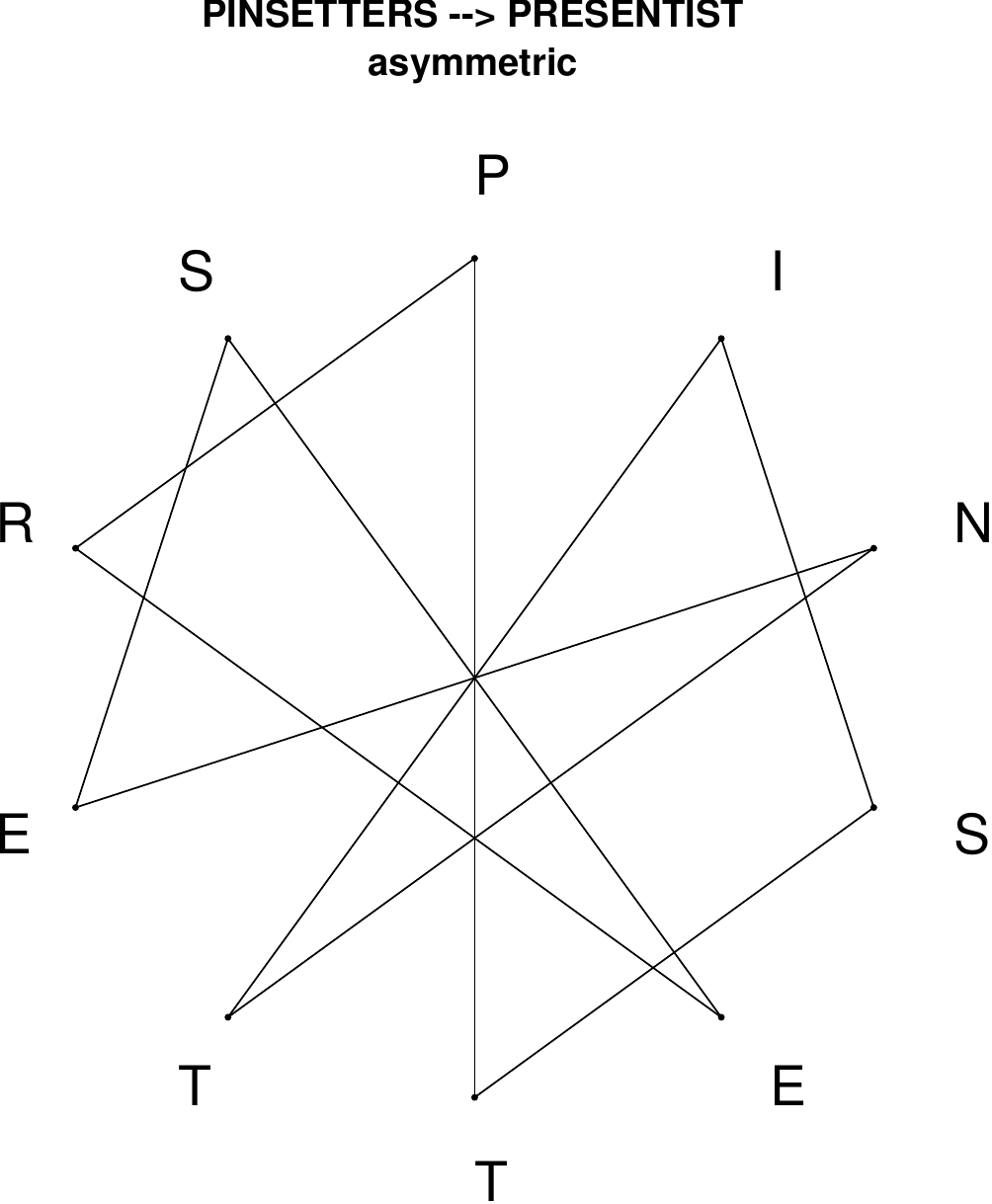}
\end{subfigure}
\hfill
\begin{subfigure}[T]{0.19\textwidth}
\centering
\includegraphics[width=\textwidth]{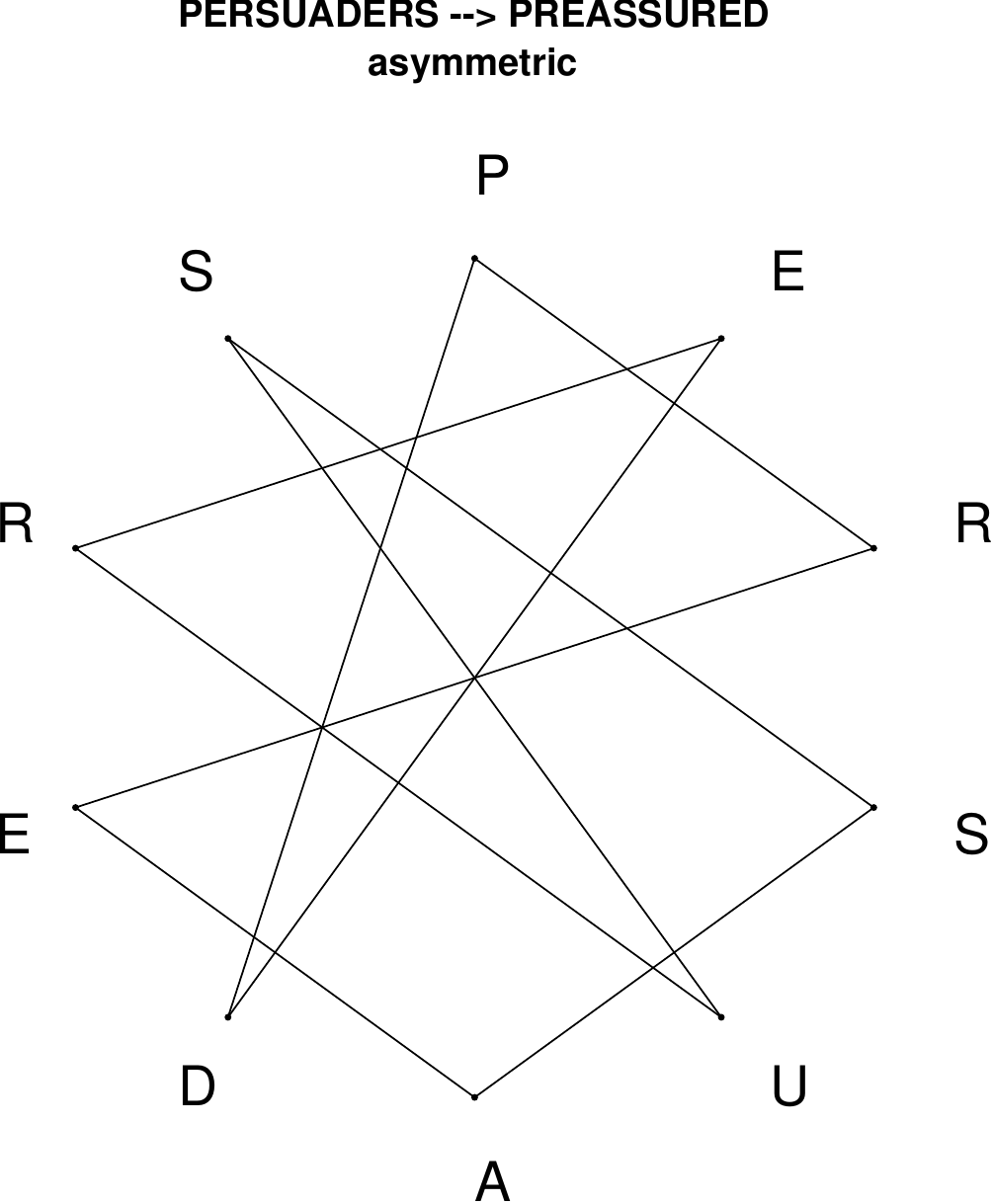}
\end{subfigure}
\hfill
\begin{subfigure}[T]{0.19\textwidth}
\centering
\includegraphics[width=\textwidth]{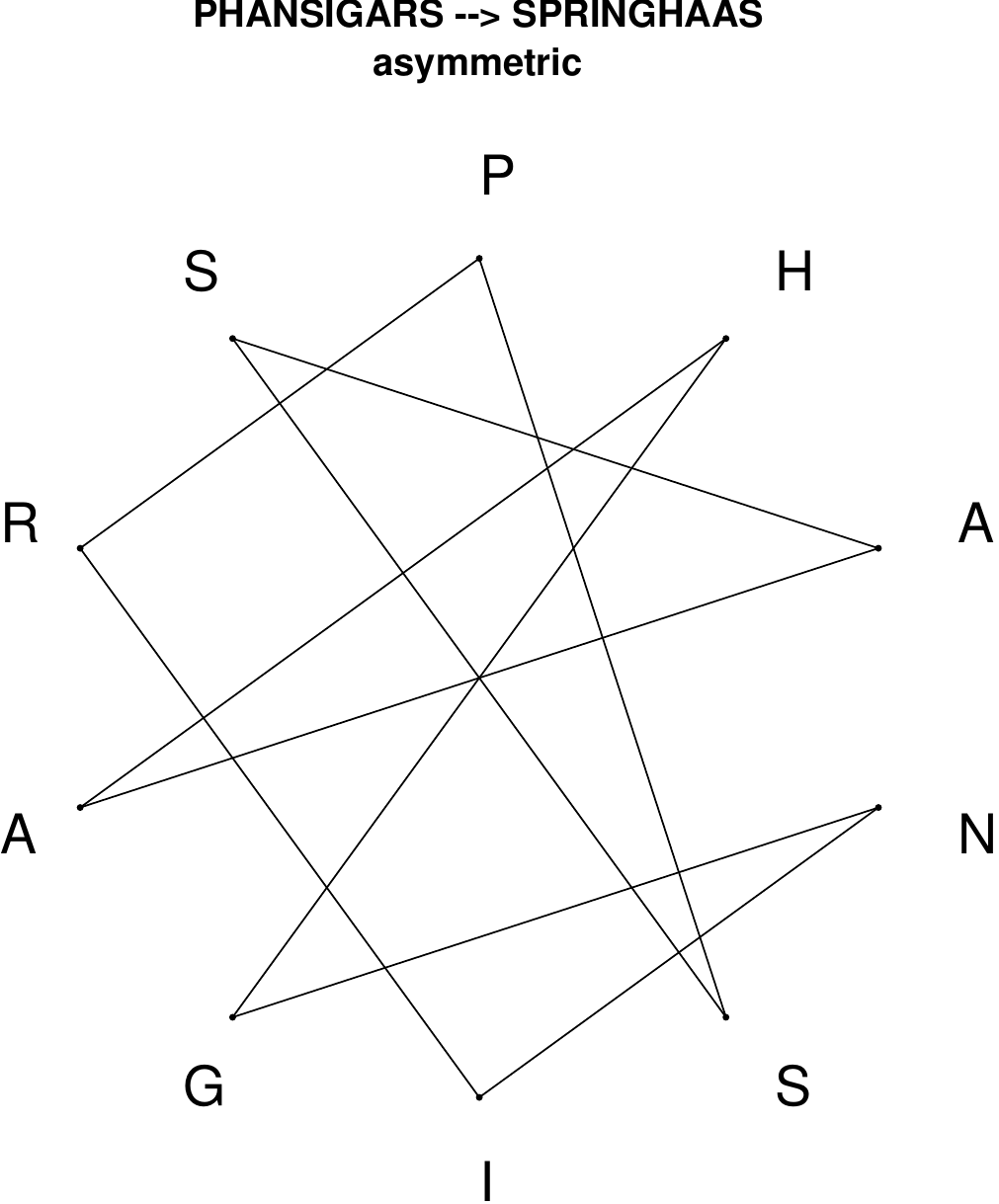}
\end{subfigure}
\hfill
\begin{subfigure}[T]{0.19\textwidth}
\centering
\includegraphics[width=\textwidth]{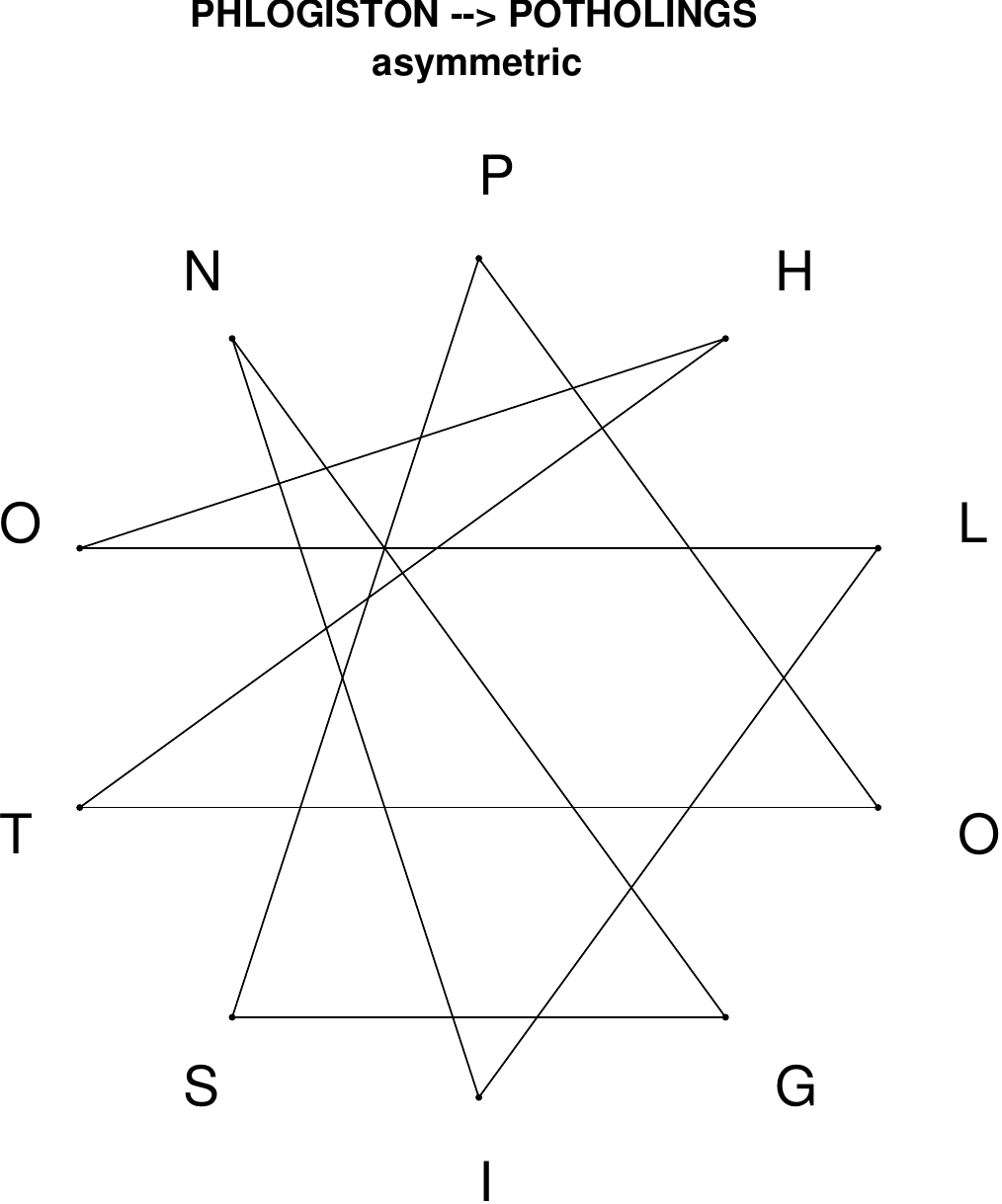}
\end{subfigure}
\end{figure}

\begin{figure}[H]
\centering
\begin{subfigure}[T]{0.19\textwidth}
\centering
\includegraphics[width=\textwidth]{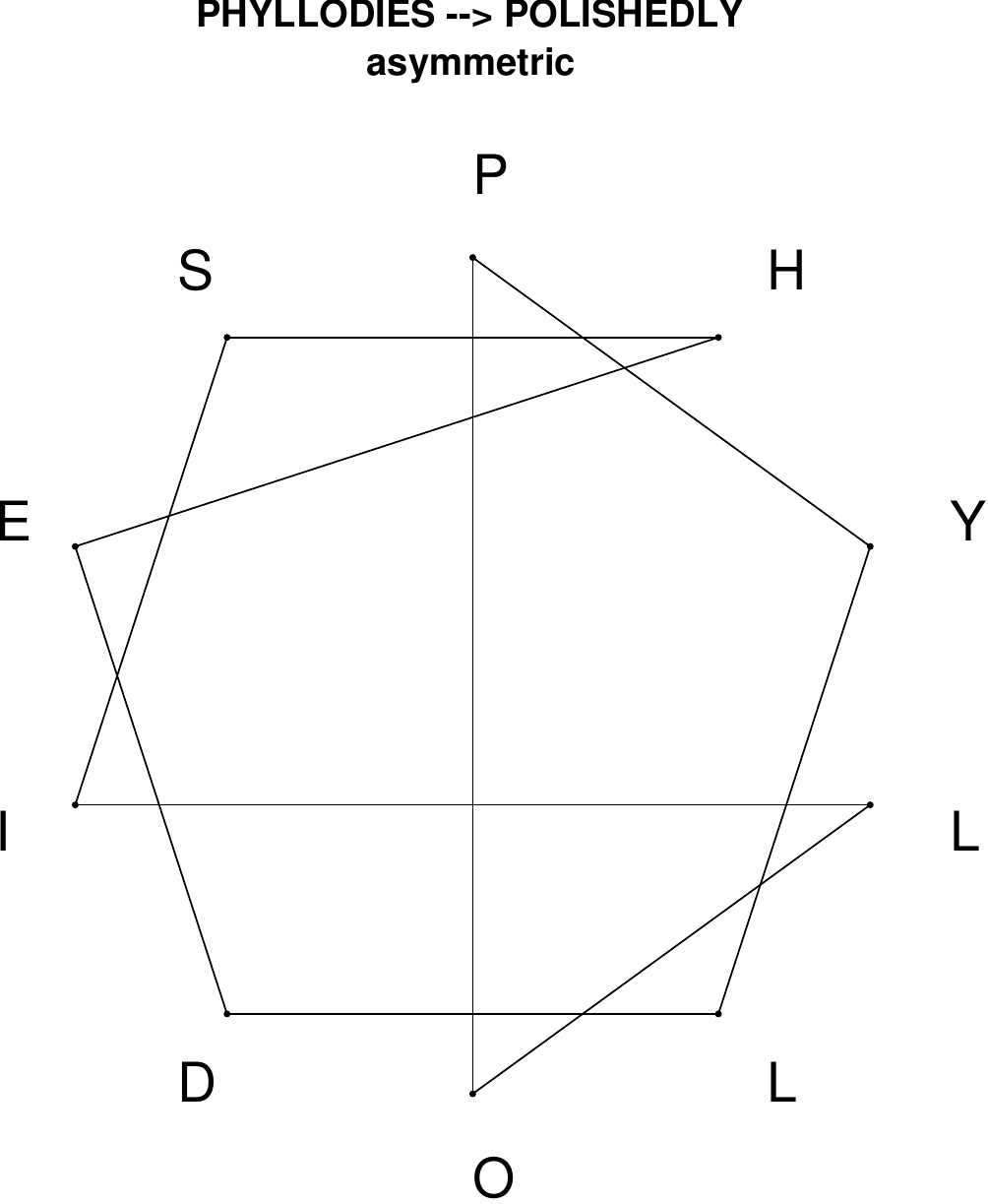}
\end{subfigure}
\hfill
\begin{subfigure}[T]{0.19\textwidth}
\centering
\includegraphics[width=\textwidth]{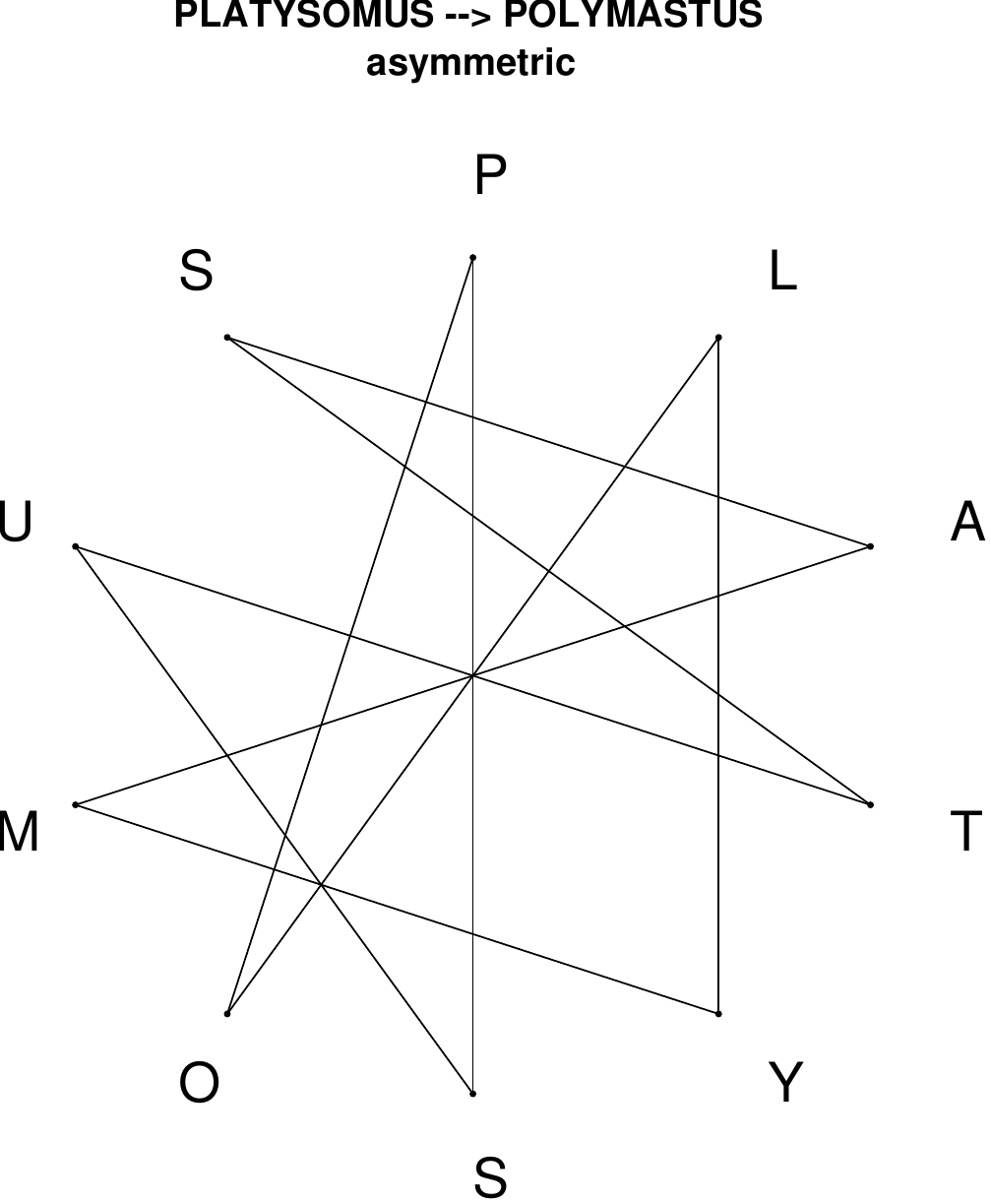}
\end{subfigure}
\hfill
\begin{subfigure}[T]{0.19\textwidth}
\centering
\includegraphics[width=\textwidth]{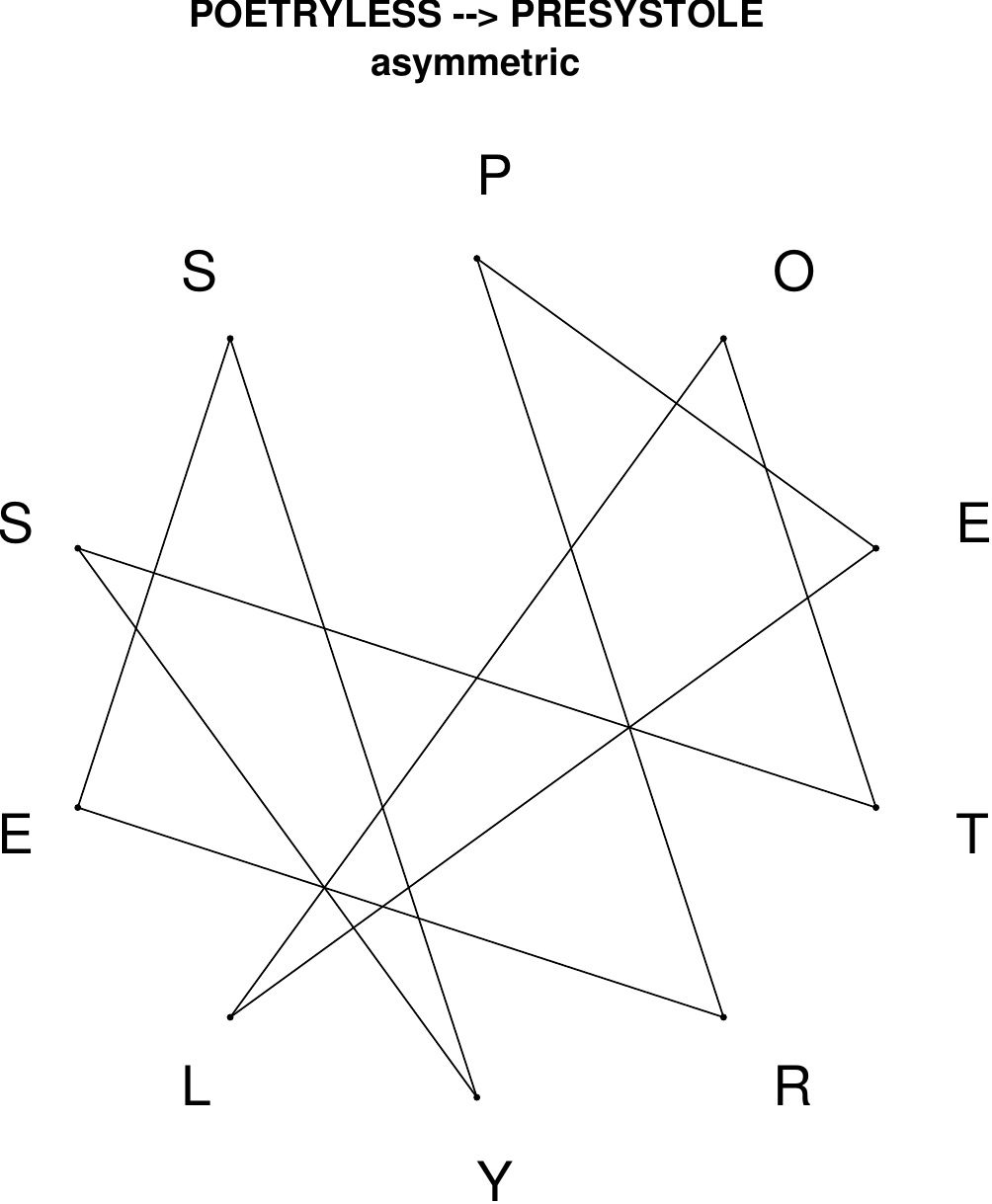}
\end{subfigure}
\hfill
\begin{subfigure}[T]{0.19\textwidth}
\centering
\includegraphics[width=\textwidth]{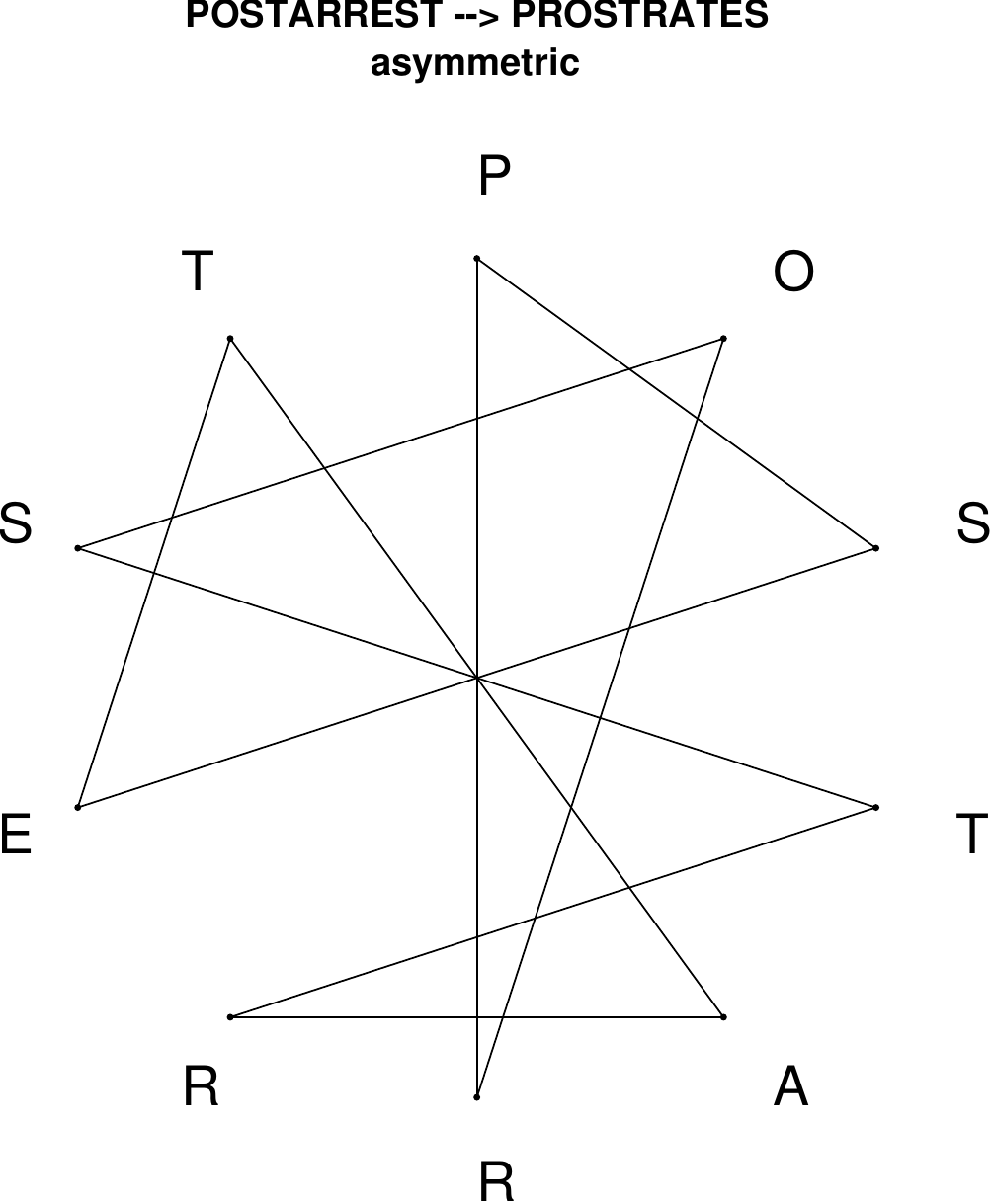}
\end{subfigure}
\hfill
\begin{subfigure}[T]{0.19\textwidth}
\centering
\includegraphics[width=\textwidth]{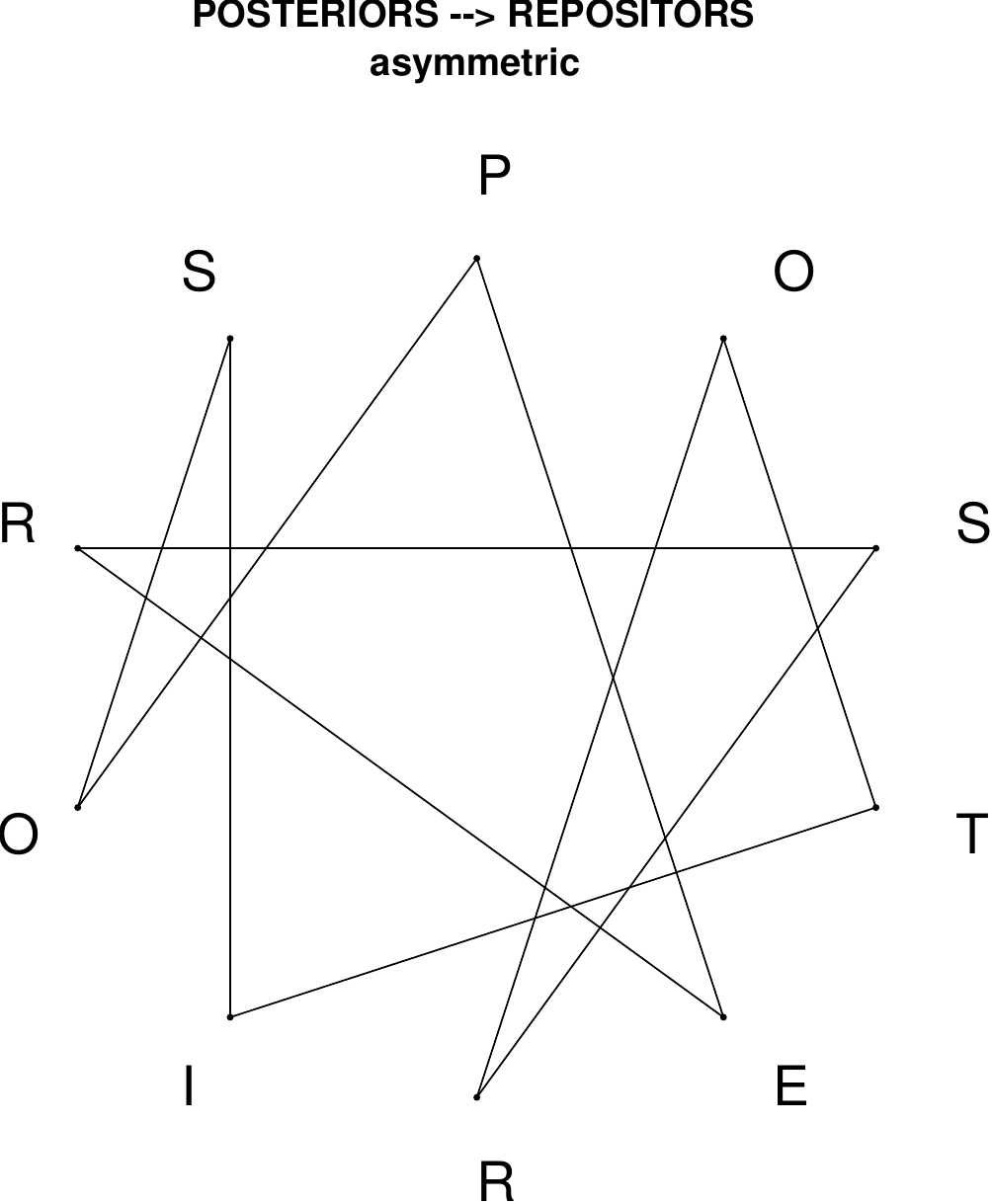}
\end{subfigure}
\end{figure}

\begin{figure}[H]
\centering
\begin{subfigure}[T]{0.19\textwidth}
\centering
\includegraphics[width=\textwidth]{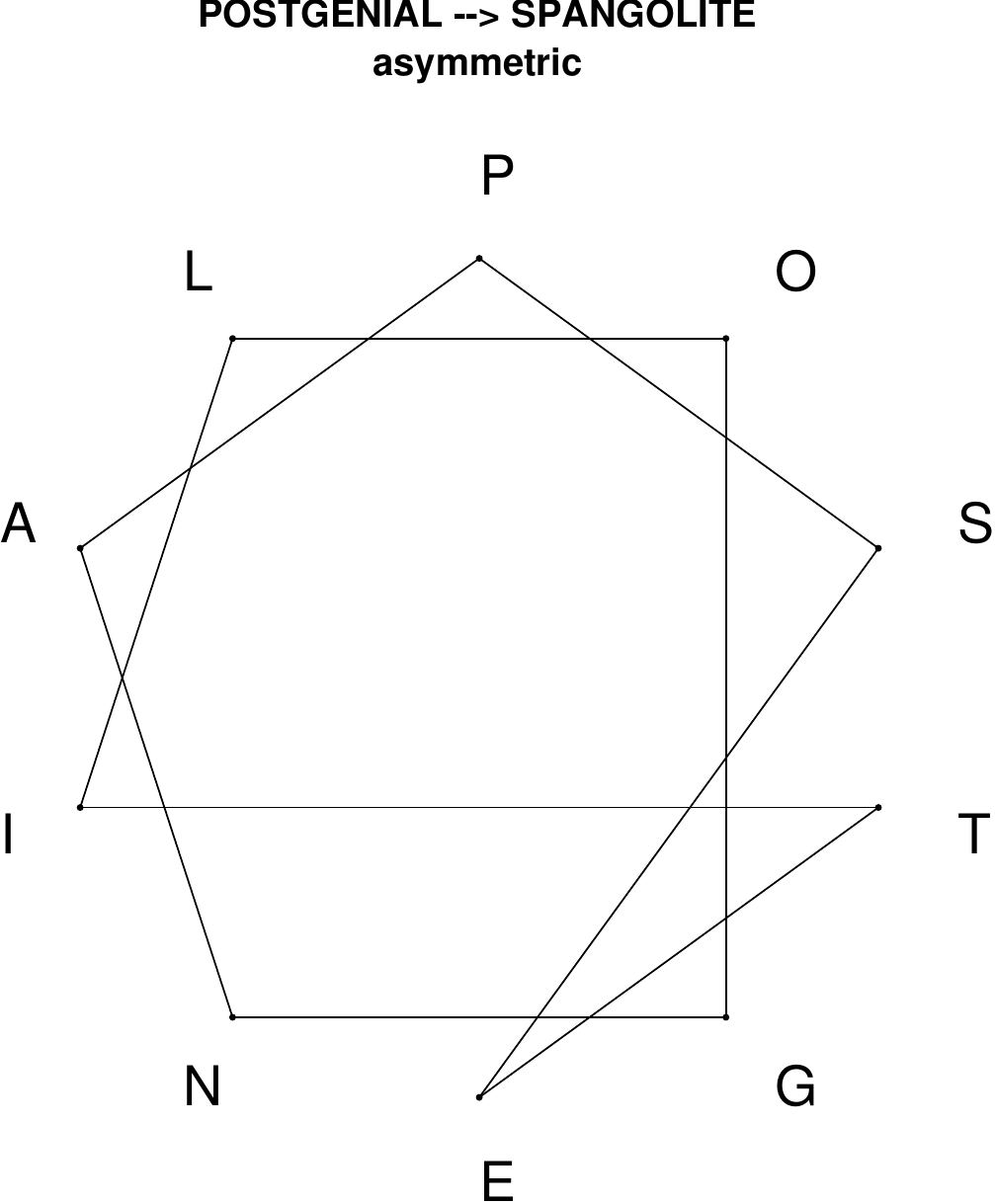}
\end{subfigure}
\hfill
\begin{subfigure}[T]{0.19\textwidth}
\centering
\includegraphics[width=\textwidth]{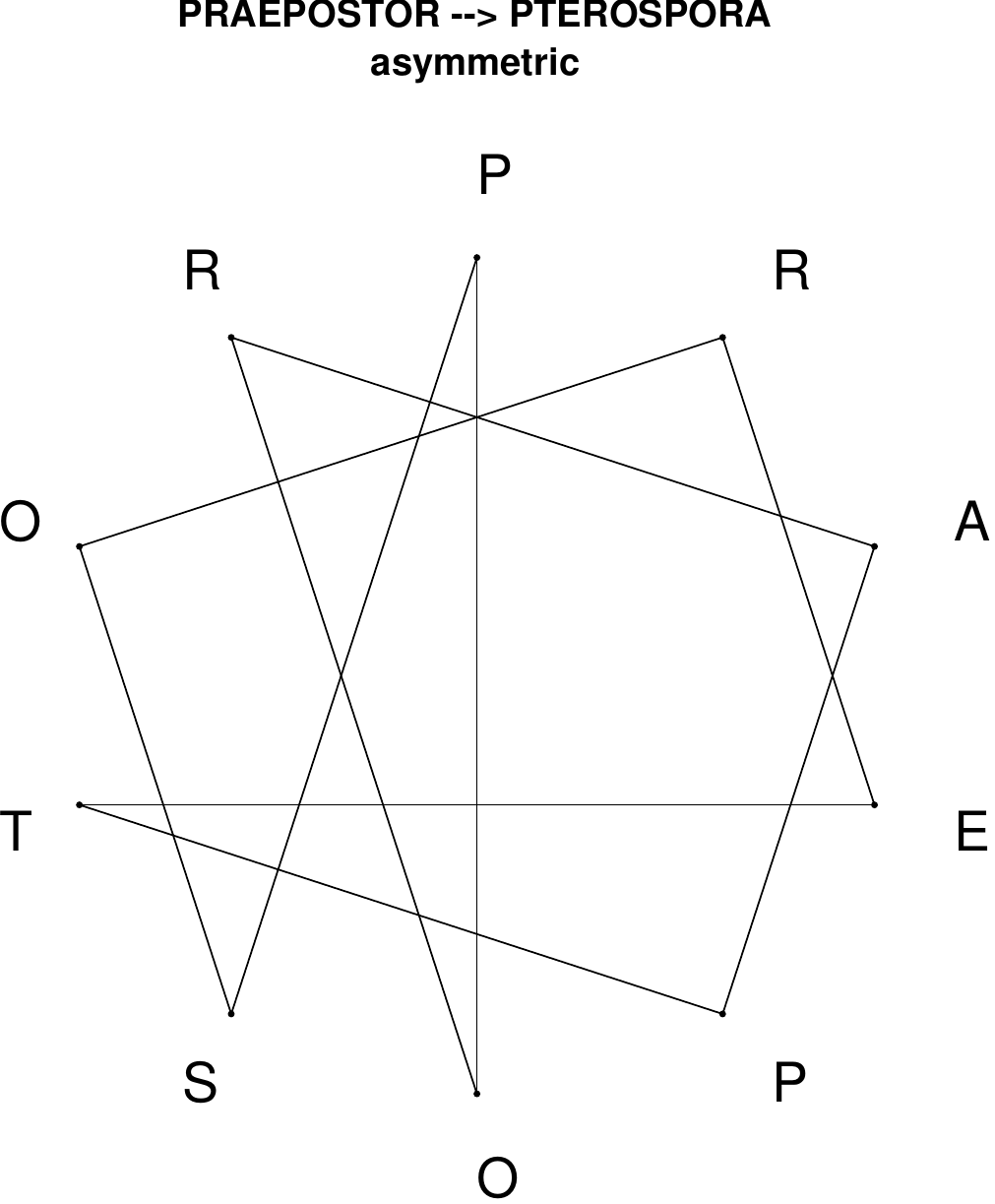}
\end{subfigure}
\hfill
\begin{subfigure}[T]{0.19\textwidth}
\centering
\includegraphics[width=\textwidth]{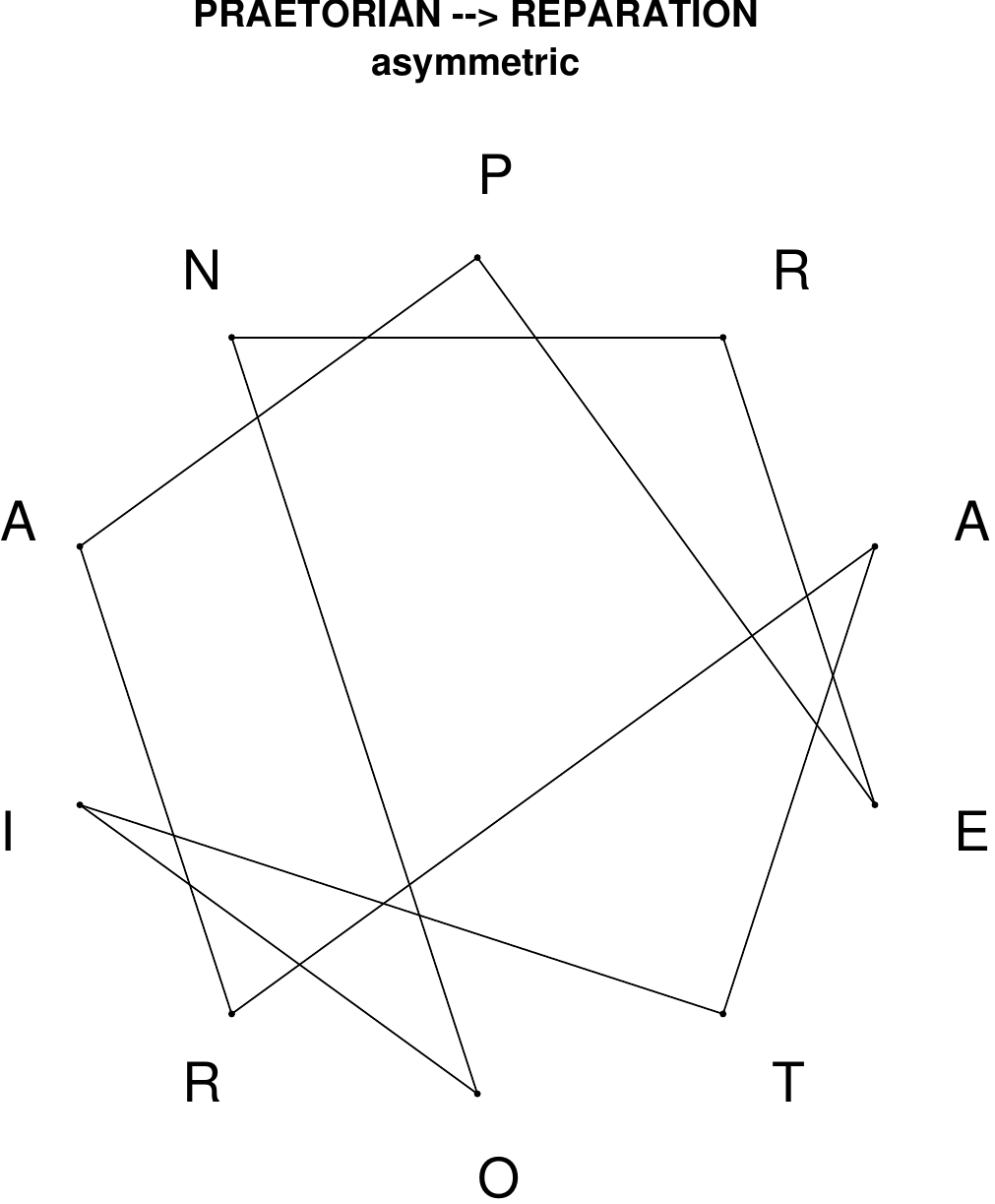}
\end{subfigure}
\hfill
\begin{subfigure}[T]{0.19\textwidth}
\centering
\includegraphics[width=\textwidth]{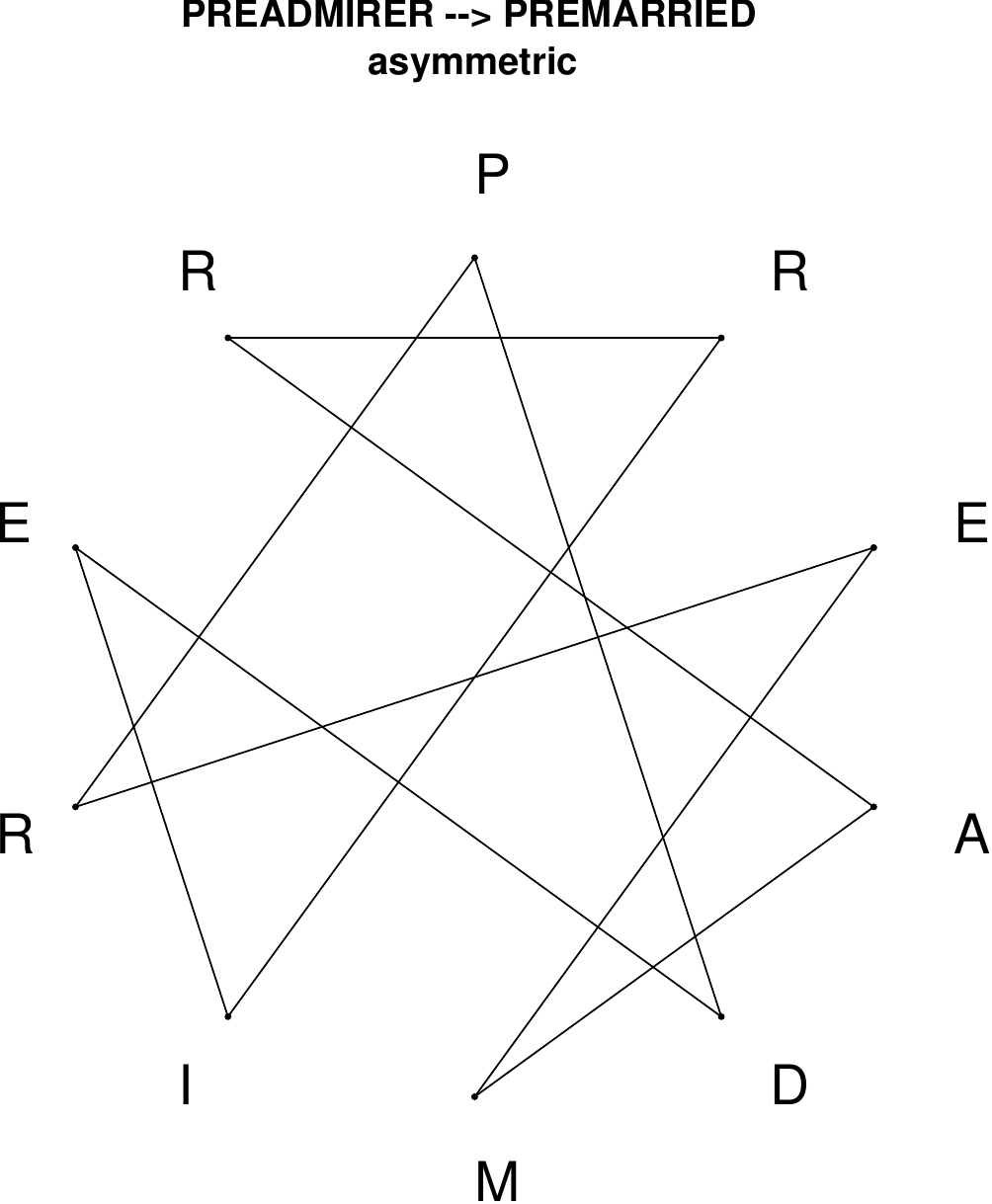}
\end{subfigure}
\hfill
\begin{subfigure}[T]{0.19\textwidth}
\centering
\includegraphics[width=\textwidth]{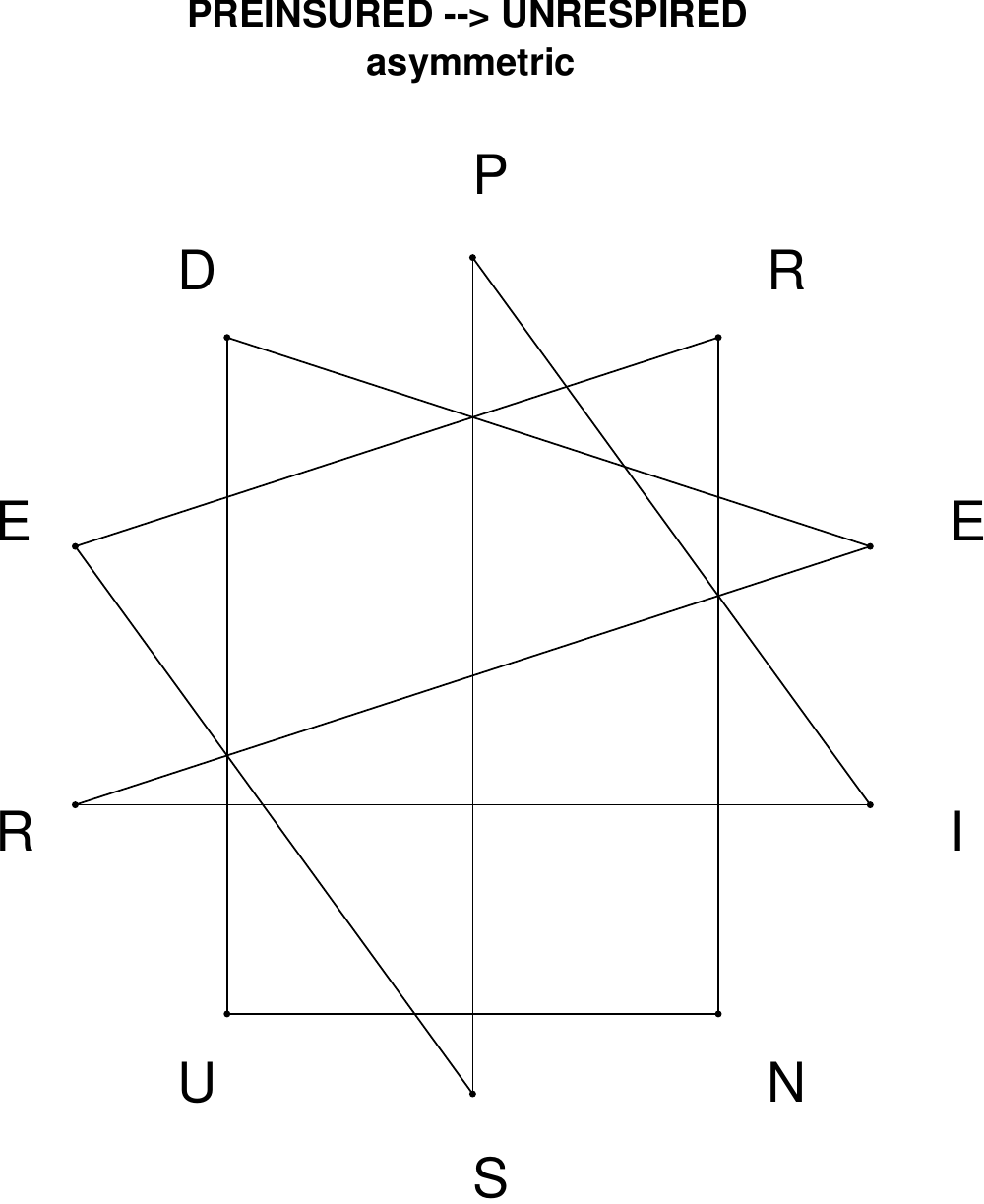}
\end{subfigure}
\end{figure}

\begin{figure}[H]
\centering
\begin{subfigure}[T]{0.19\textwidth}
\centering
\includegraphics[width=\textwidth]{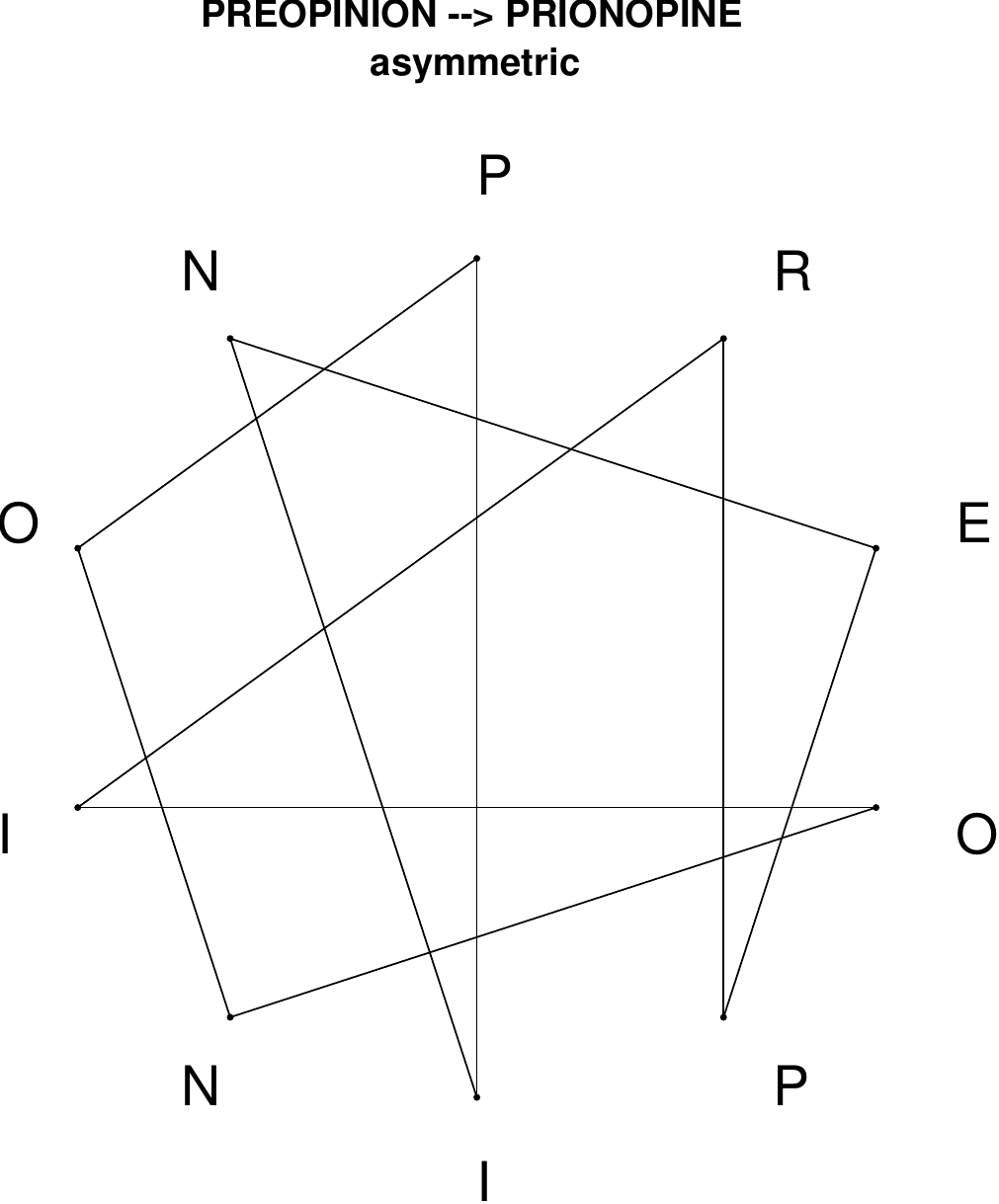}
\end{subfigure}
\hfill
\begin{subfigure}[T]{0.19\textwidth}
\centering
\includegraphics[width=\textwidth]{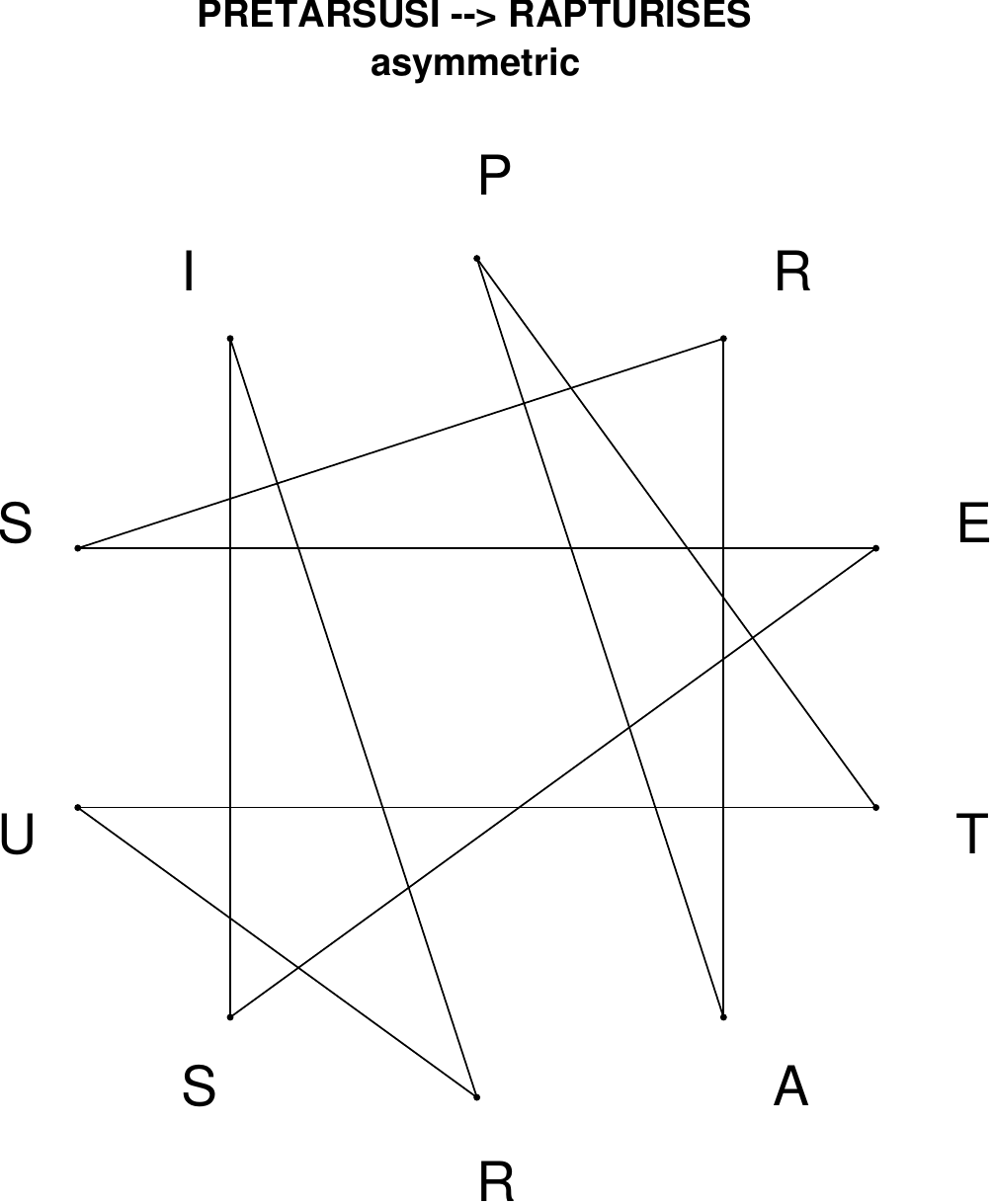}
\end{subfigure}
\hfill
\begin{subfigure}[T]{0.19\textwidth}
\centering
\includegraphics[width=\textwidth]{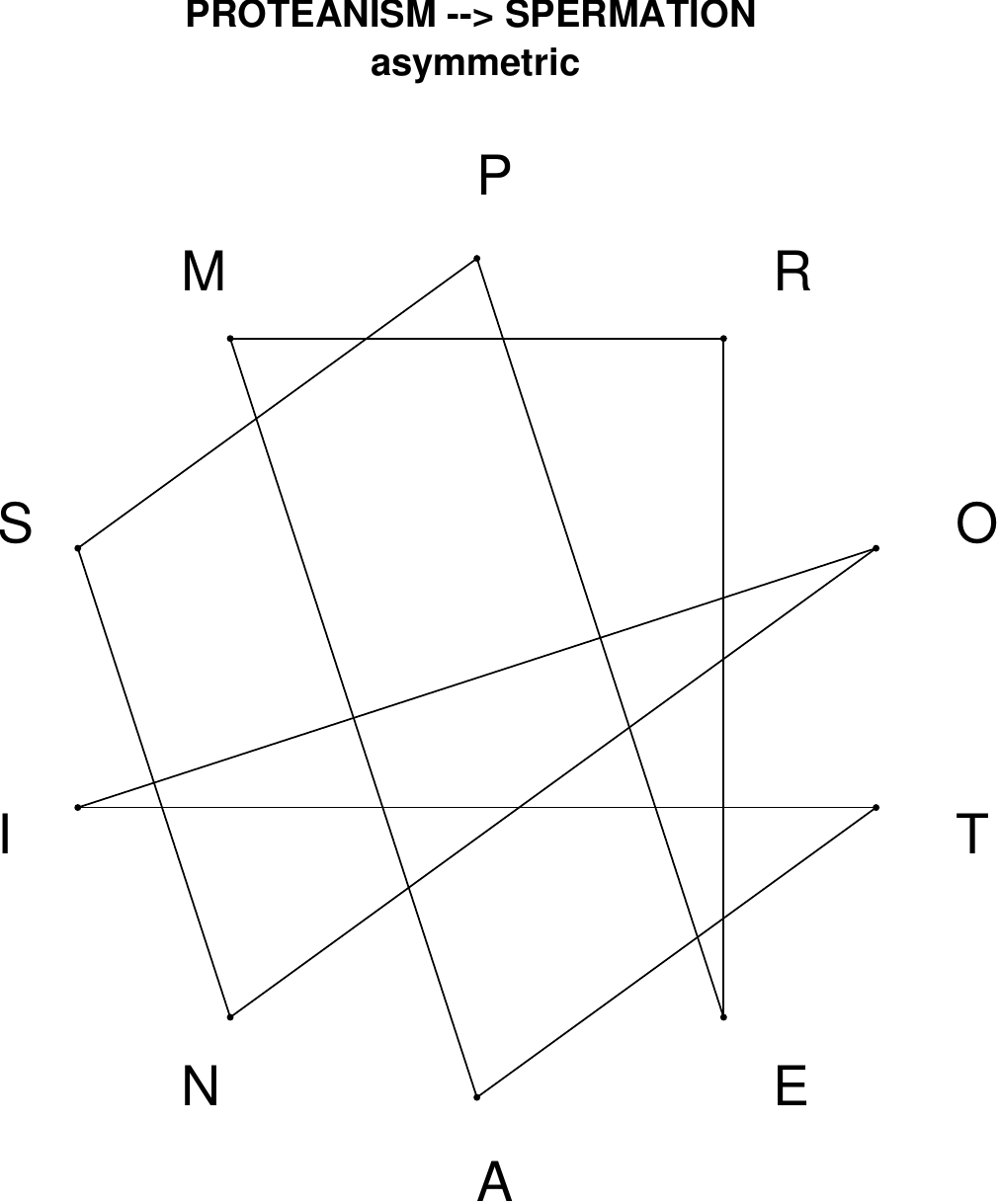}
\end{subfigure}
\hfill
\begin{subfigure}[T]{0.19\textwidth}
\centering
\includegraphics[width=\textwidth]{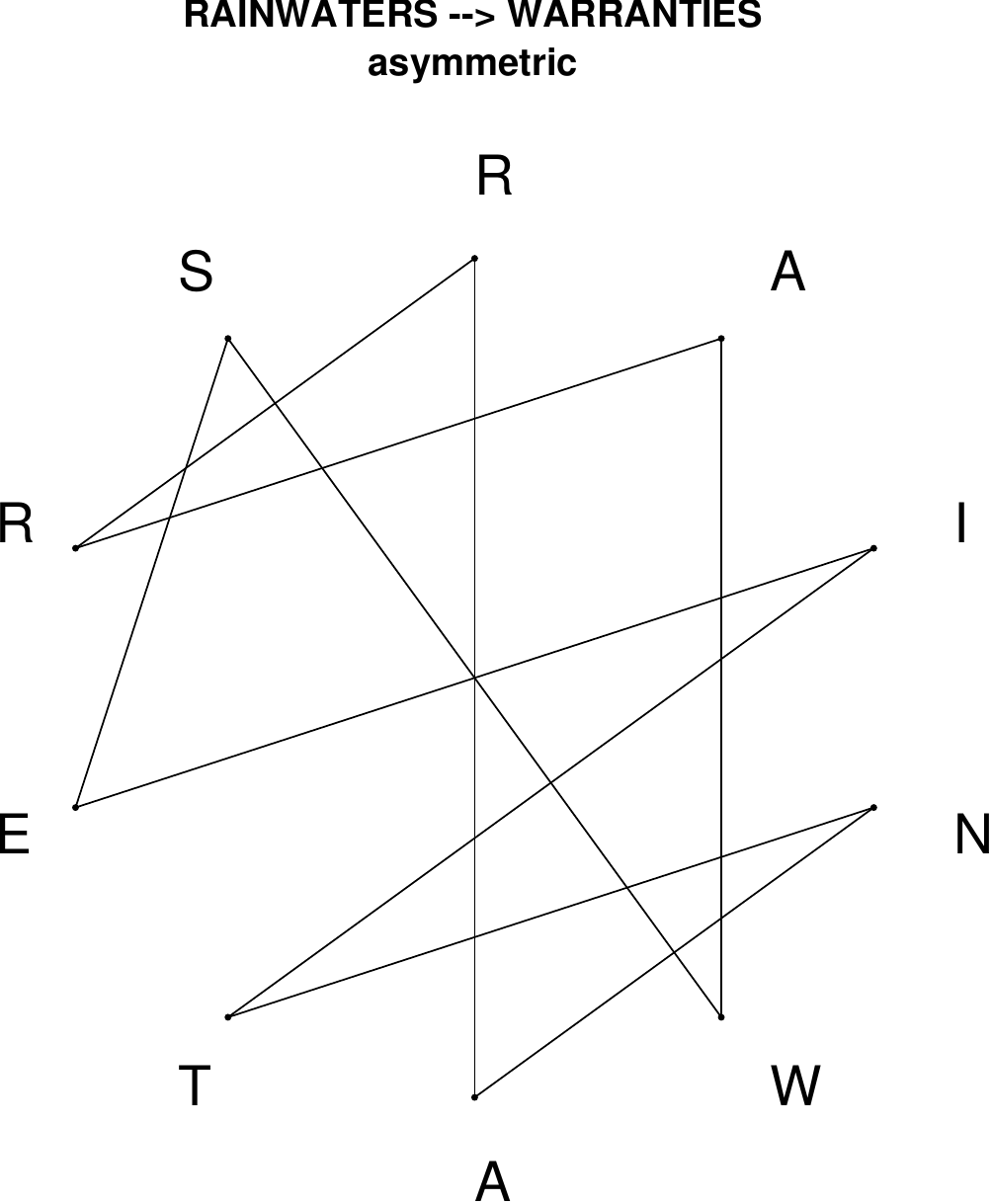}
\end{subfigure}
\hfill
\begin{subfigure}[T]{0.19\textwidth}
\centering
\includegraphics[width=\textwidth]{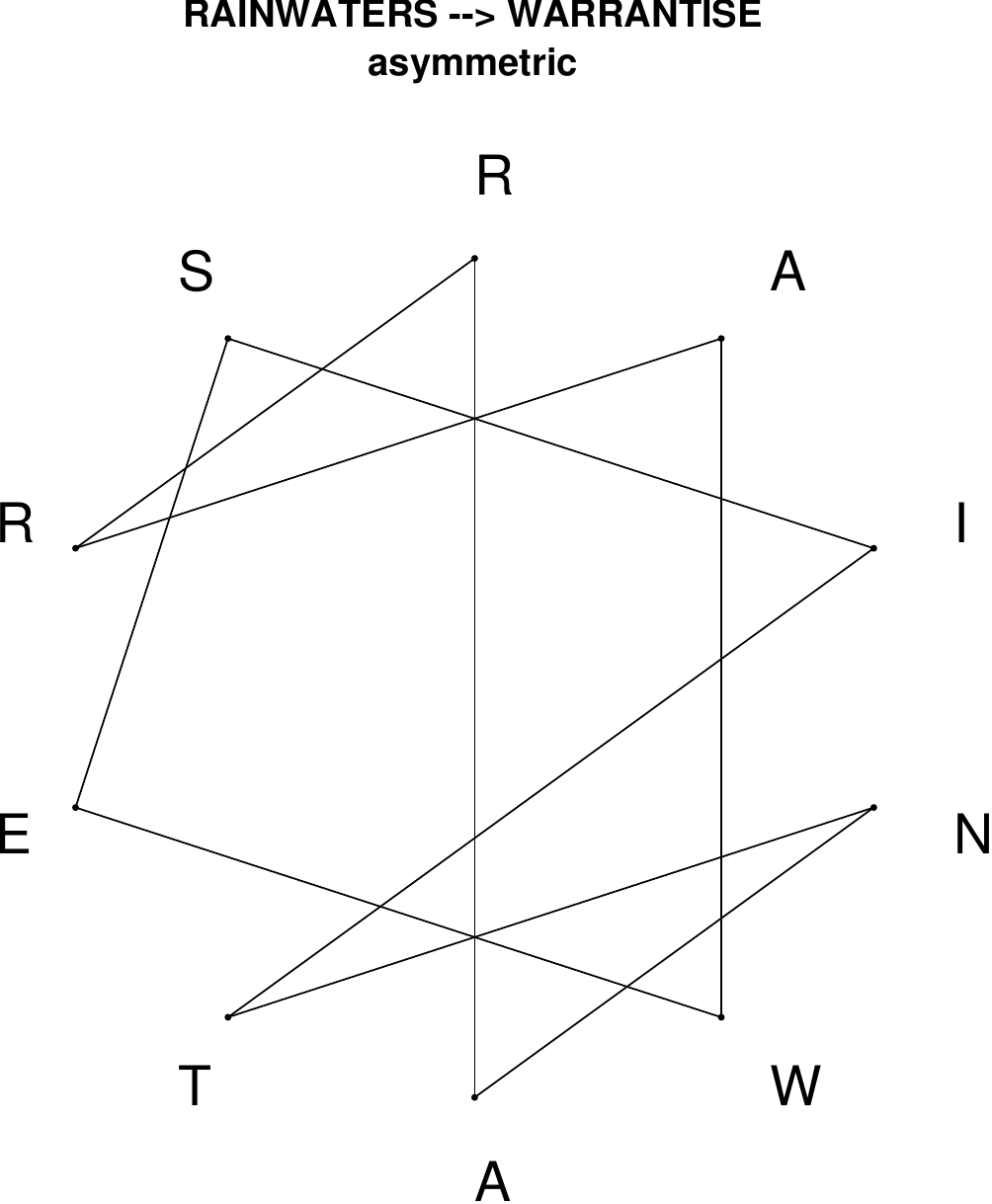}
\end{subfigure}
\end{figure}

\begin{figure}[H]
\centering
\begin{subfigure}[T]{0.19\textwidth}
\centering
\includegraphics[width=\textwidth]{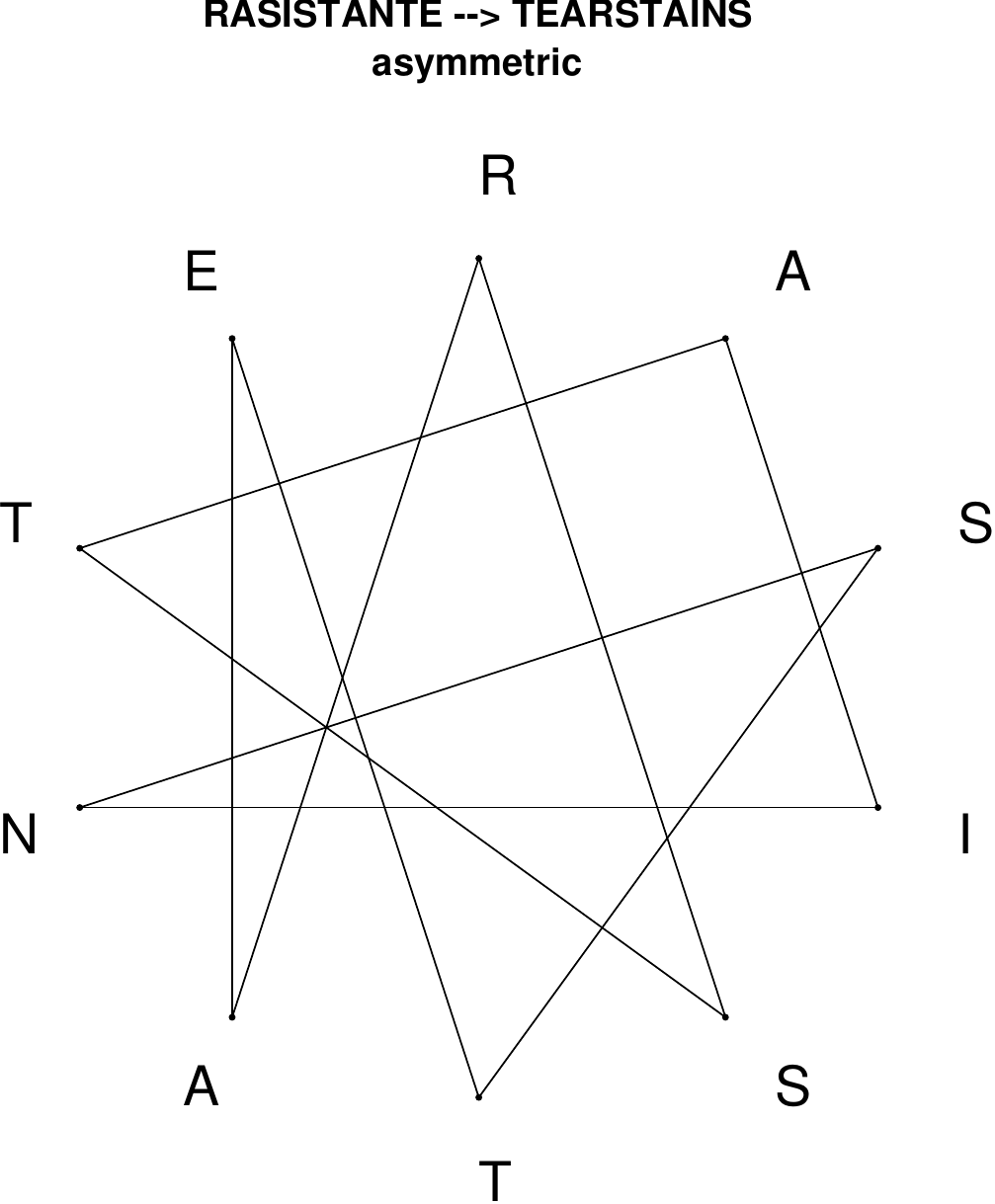}
\end{subfigure}
\hfill
\begin{subfigure}[T]{0.19\textwidth}
\centering
\includegraphics[width=\textwidth]{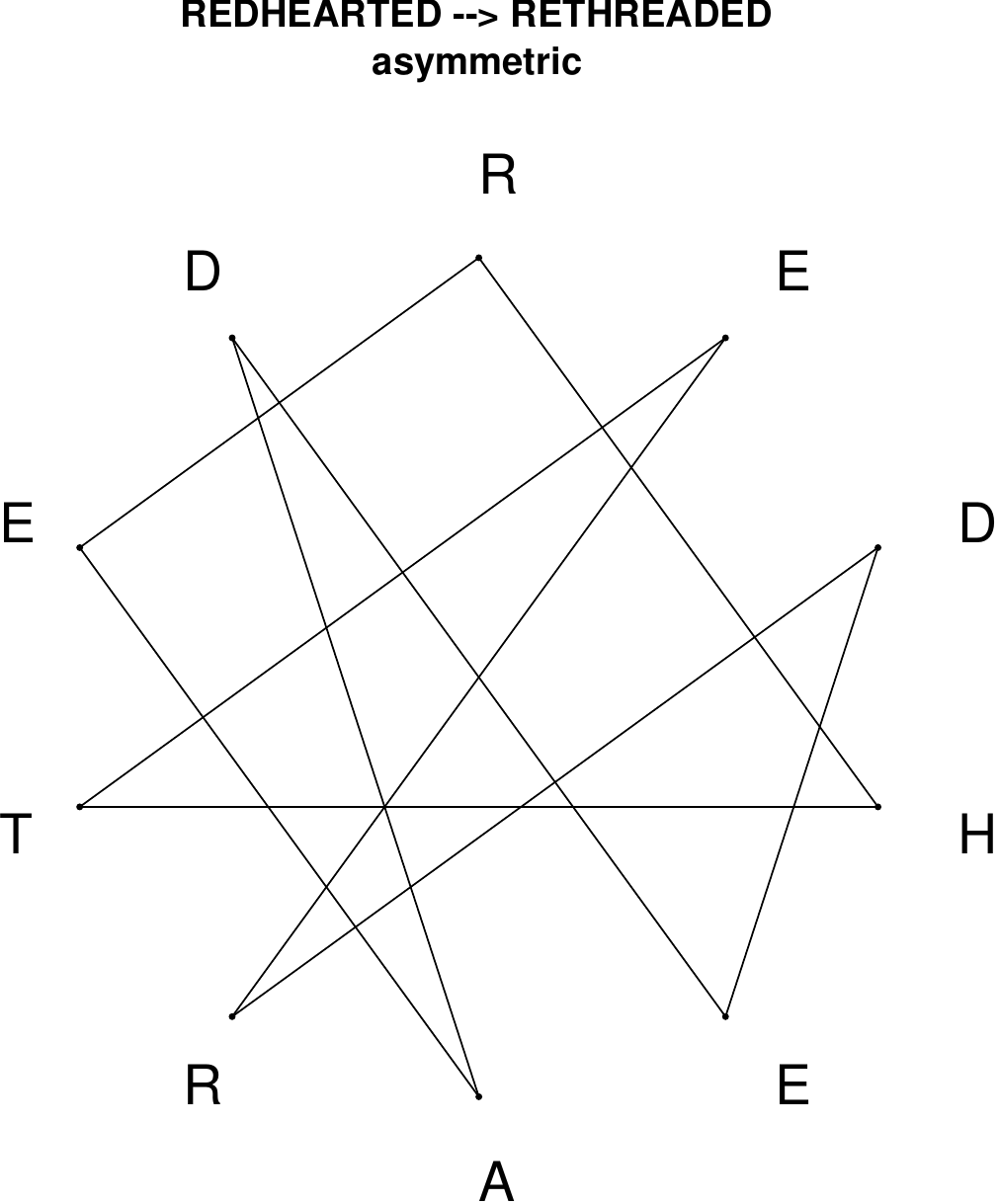}
\end{subfigure}
\hfill
\begin{subfigure}[T]{0.19\textwidth}
\centering
\includegraphics[width=\textwidth]{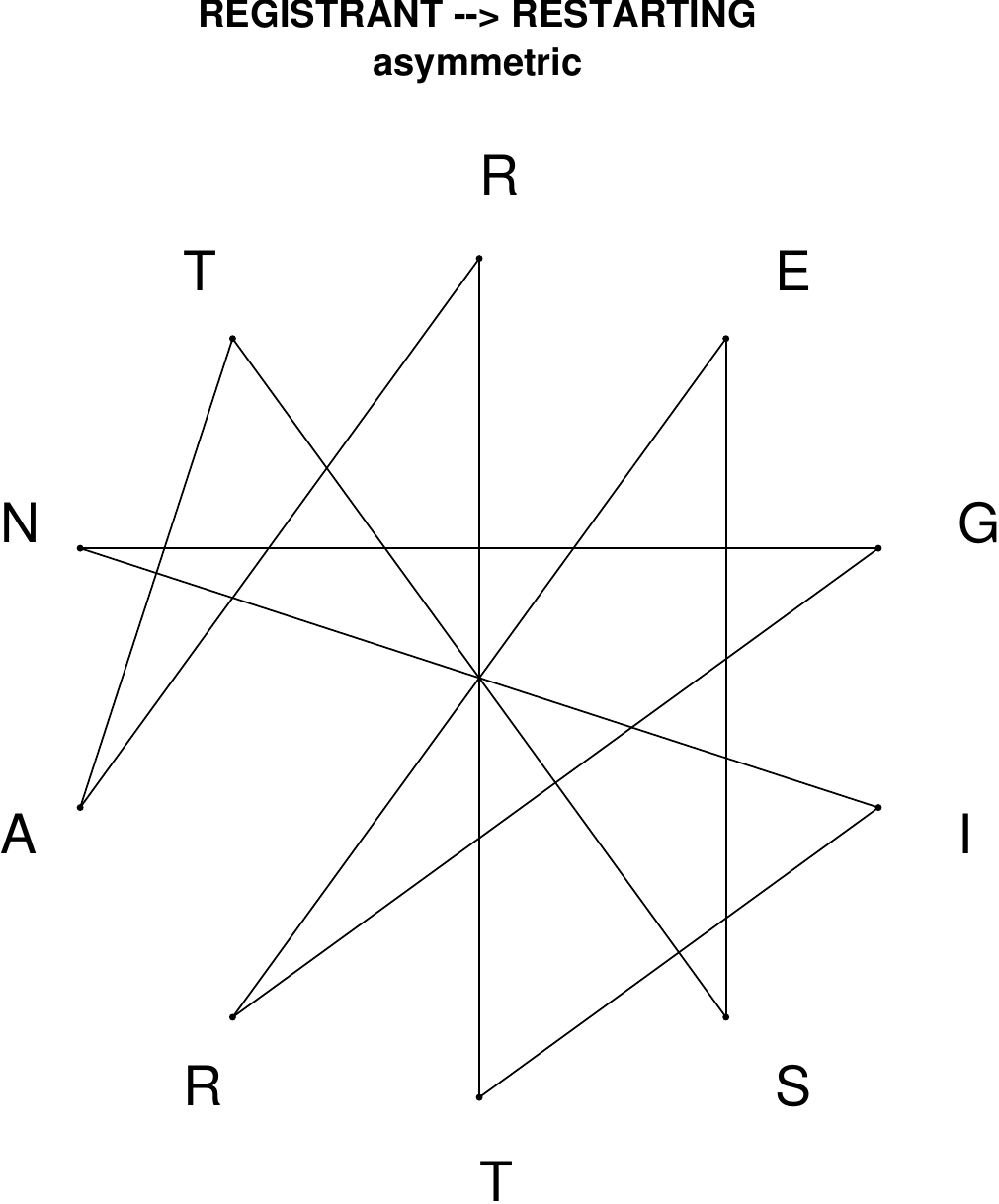}
\end{subfigure}
\hfill
\begin{subfigure}[T]{0.19\textwidth}
\centering
\includegraphics[width=\textwidth]{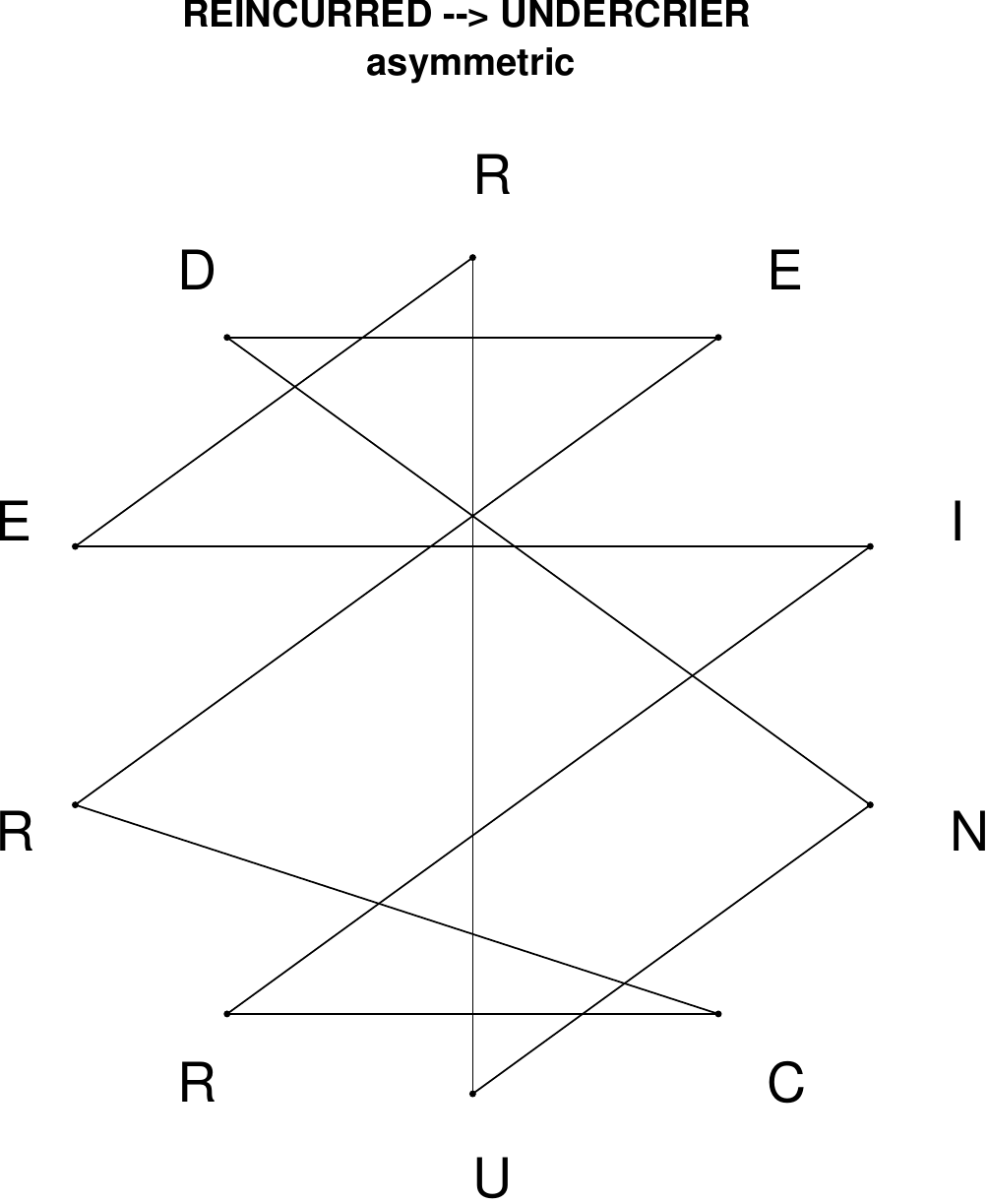}
\end{subfigure}
\hfill
\begin{subfigure}[T]{0.19\textwidth}
\centering
\includegraphics[width=\textwidth]{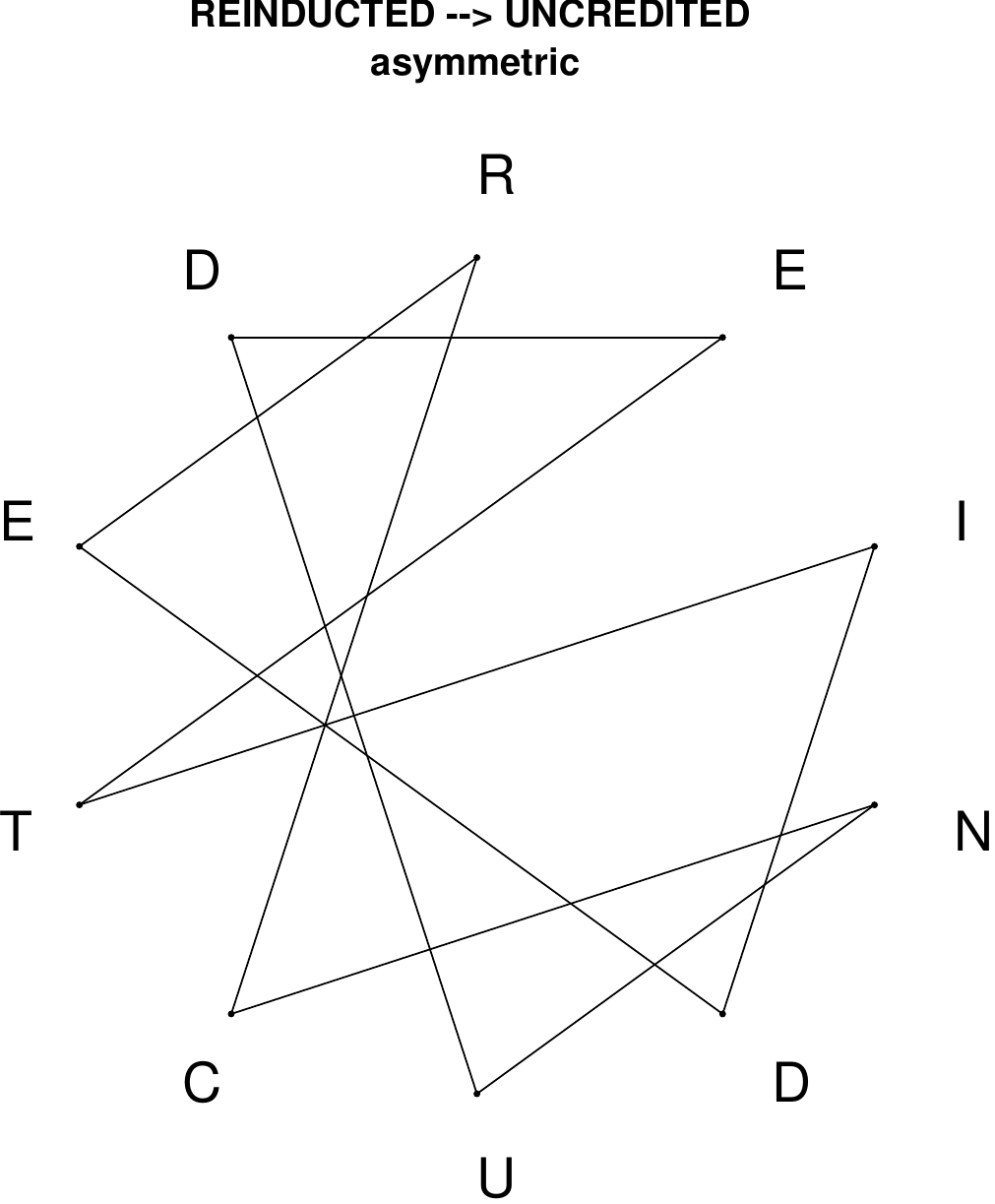}
\end{subfigure}
\end{figure}

\begin{figure}[H]
\centering
\begin{subfigure}[T]{0.19\textwidth}
\centering
\includegraphics[width=\textwidth]{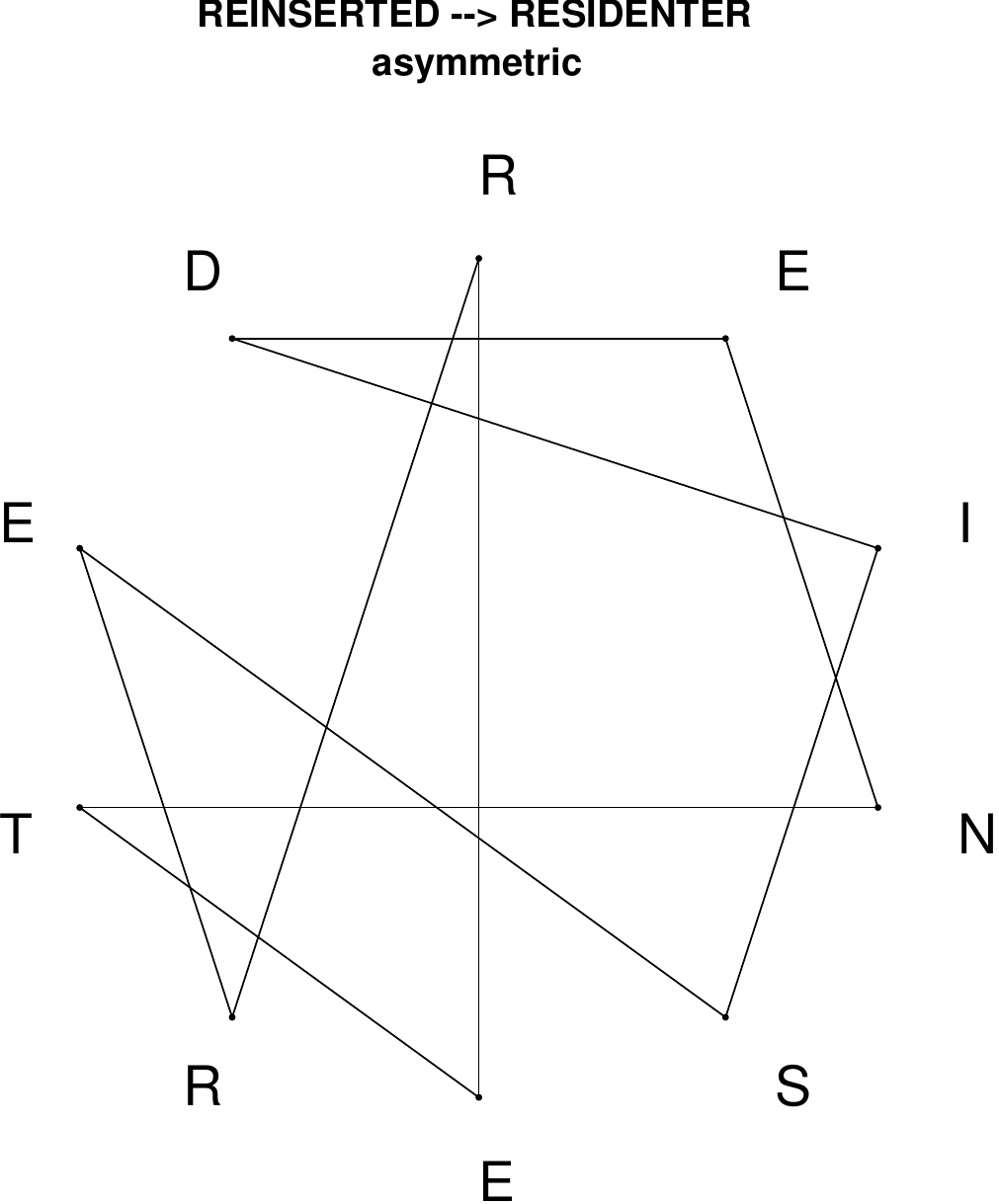}
\end{subfigure}
\hfill
\begin{subfigure}[T]{0.19\textwidth}
\centering
\includegraphics[width=\textwidth]{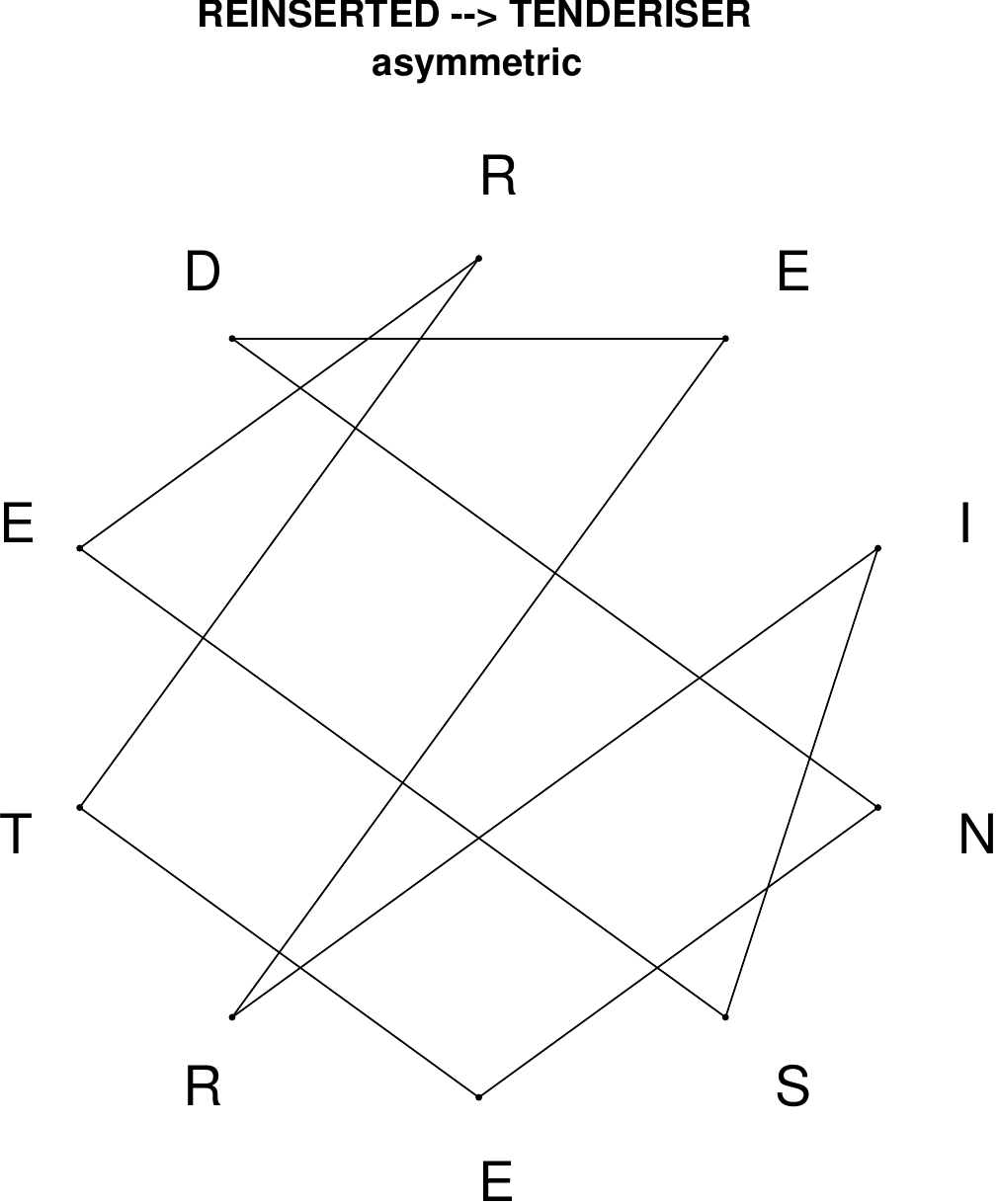}
\end{subfigure}
\hfill
\begin{subfigure}[T]{0.19\textwidth}
\centering
\includegraphics[width=\textwidth]{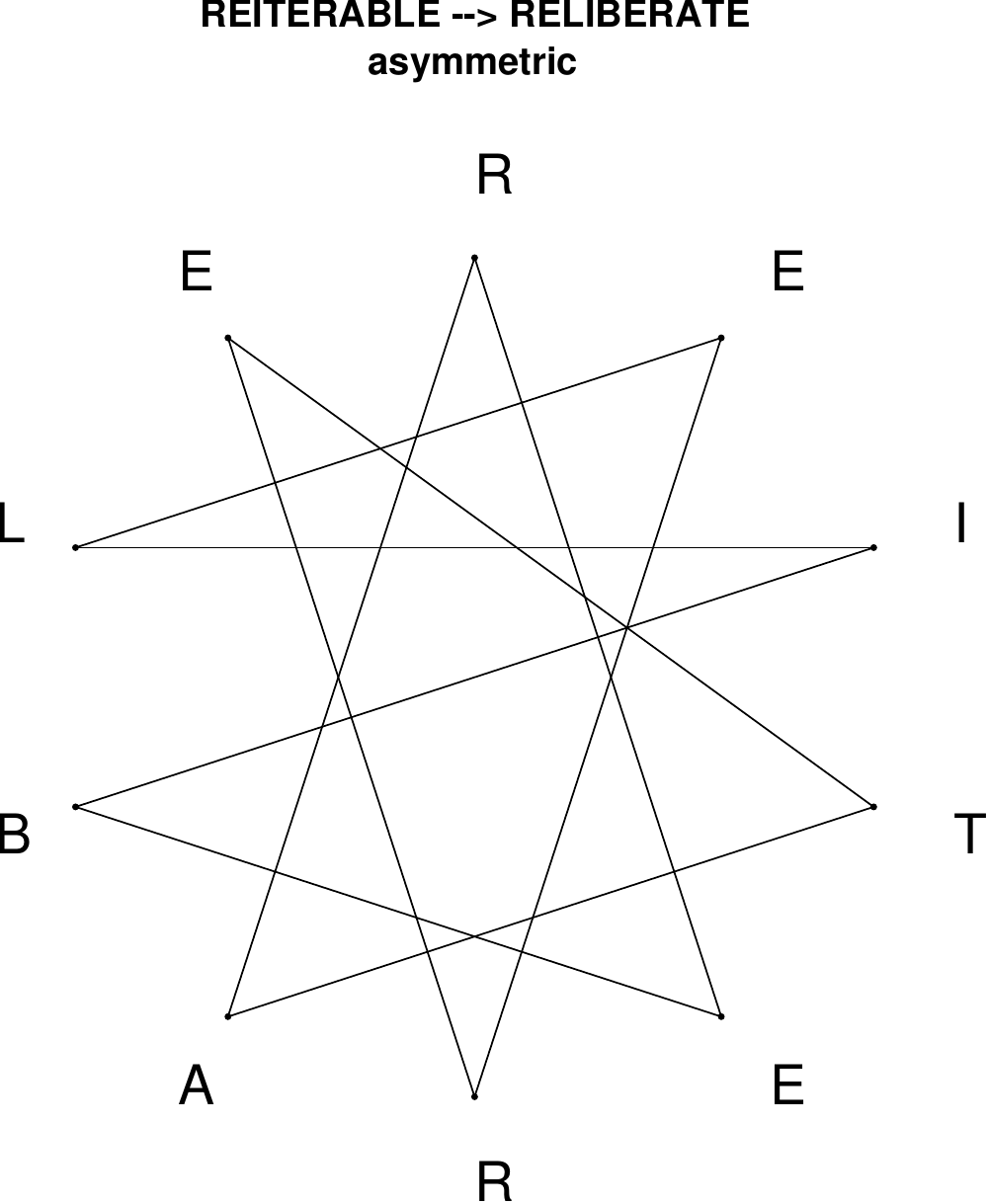}
\end{subfigure}
\hfill
\begin{subfigure}[T]{0.19\textwidth}
\centering
\includegraphics[width=\textwidth]{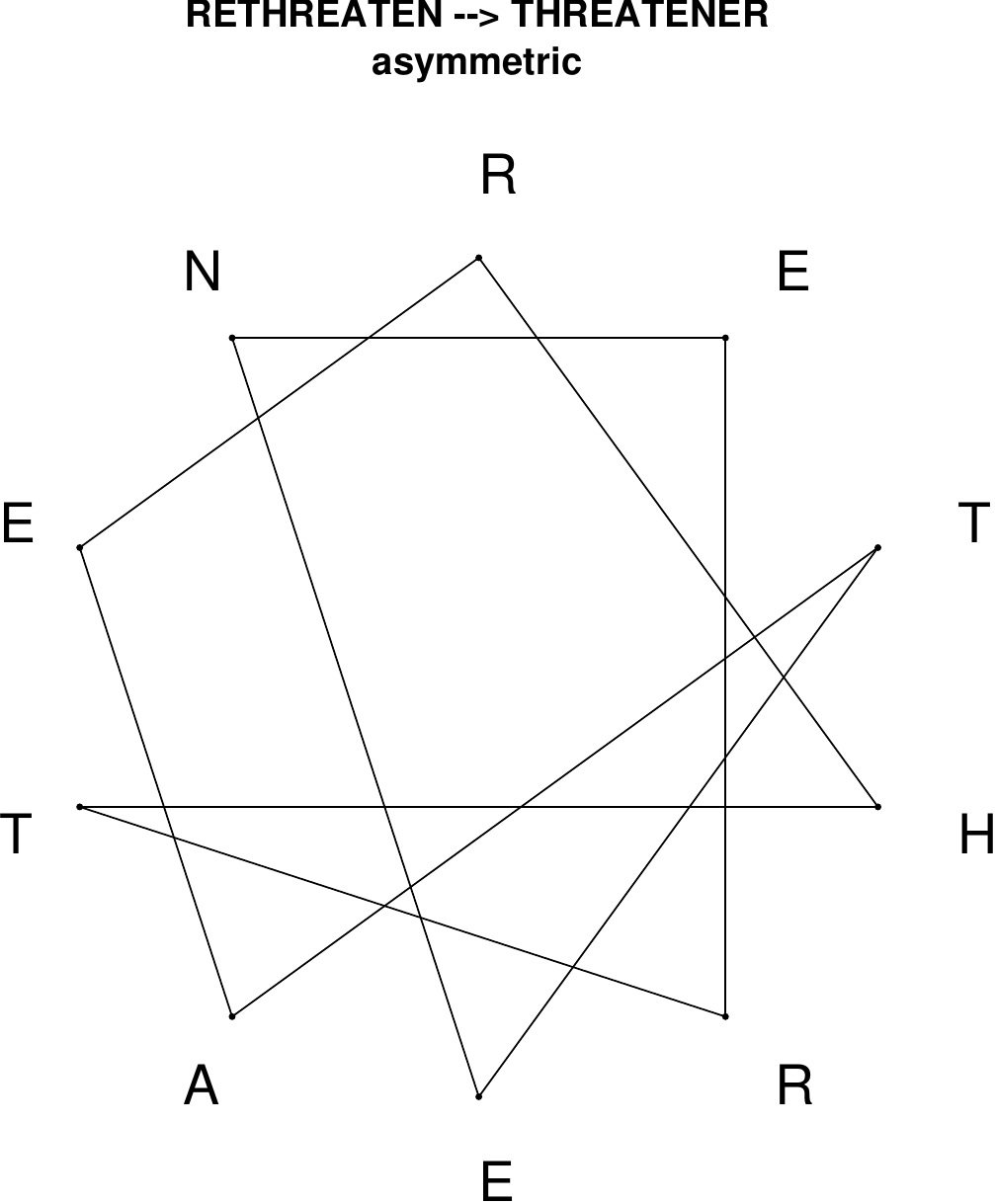}
\end{subfigure}
\hfill
\begin{subfigure}[T]{0.19\textwidth}
\centering
\includegraphics[width=\textwidth]{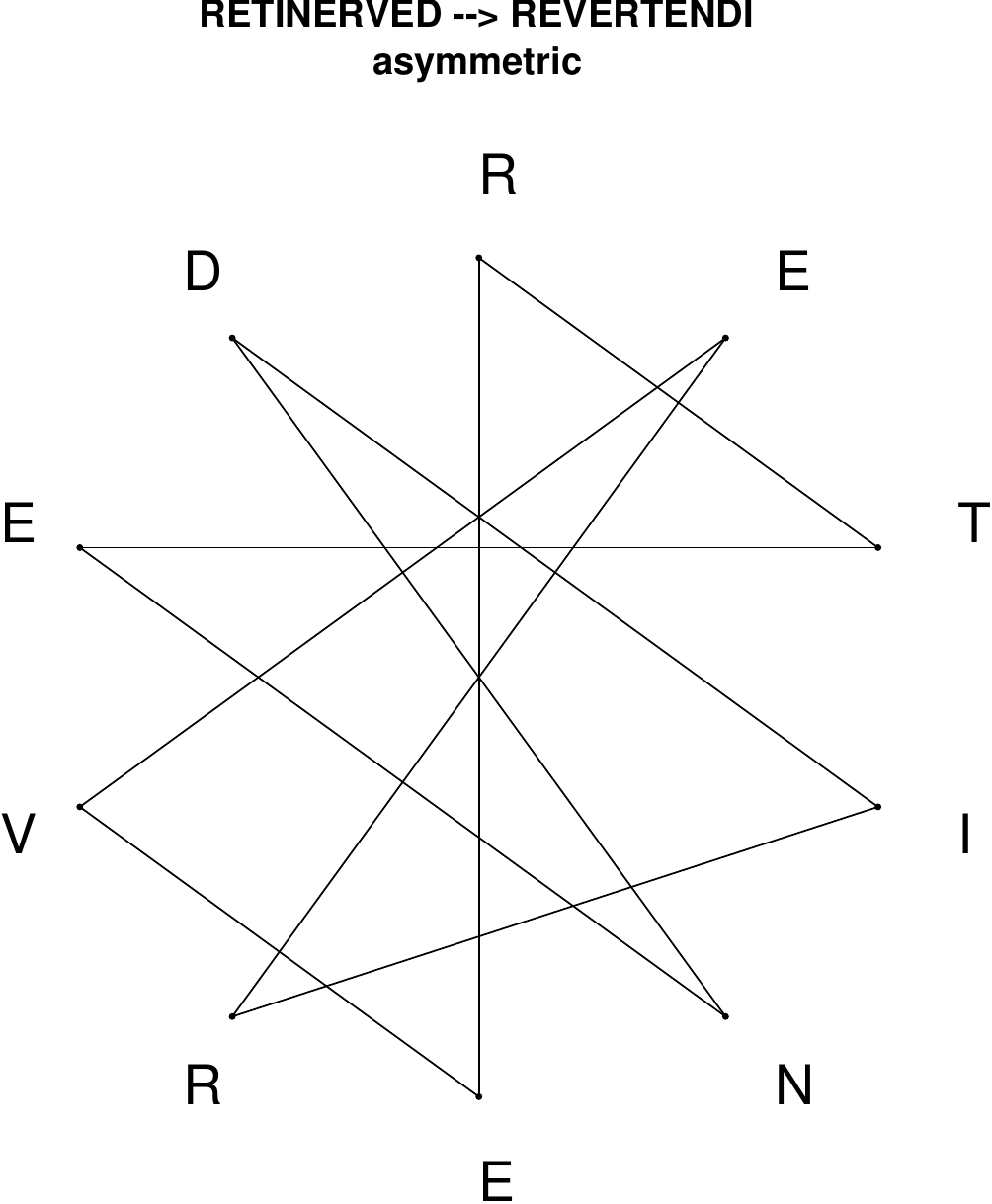}
\end{subfigure}
\end{figure}

\begin{figure}[H]
\centering
\begin{subfigure}[T]{0.19\textwidth}
\centering
\includegraphics[width=\textwidth]{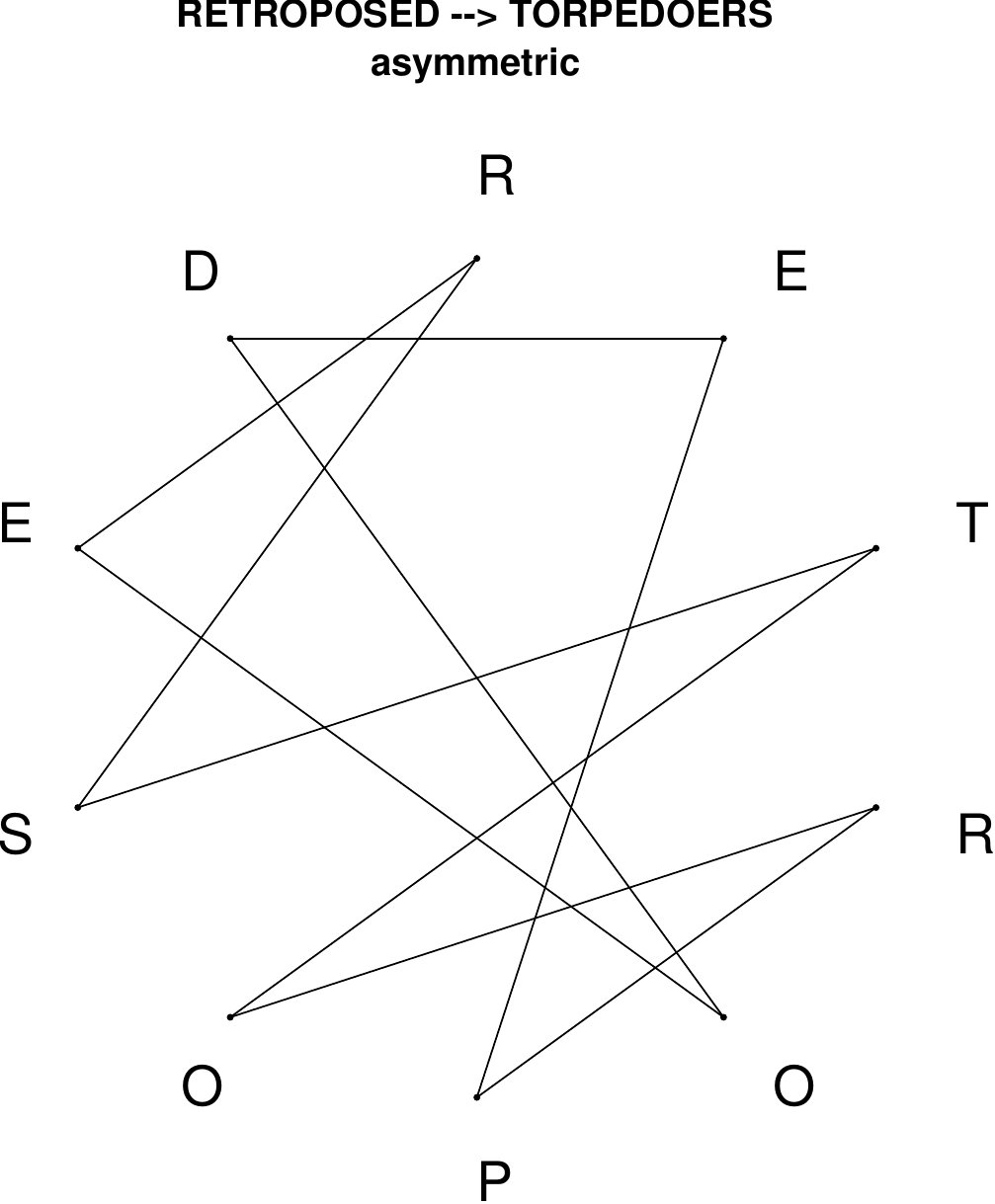}
\end{subfigure}
\hfill
\begin{subfigure}[T]{0.19\textwidth}
\centering
\includegraphics[width=\textwidth]{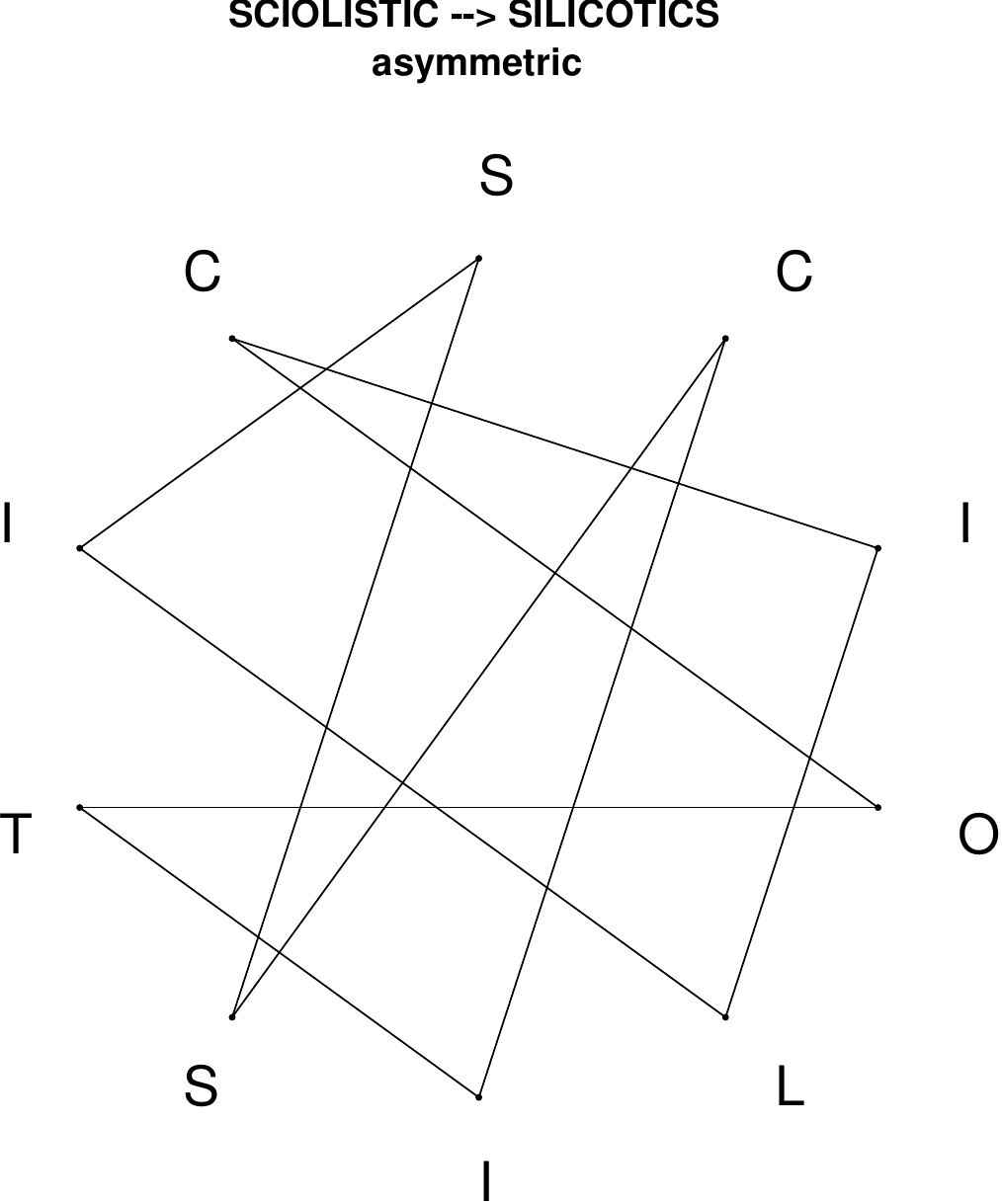}
\end{subfigure}
\hfill
\begin{subfigure}[T]{0.19\textwidth}
\centering
\includegraphics[width=\textwidth]{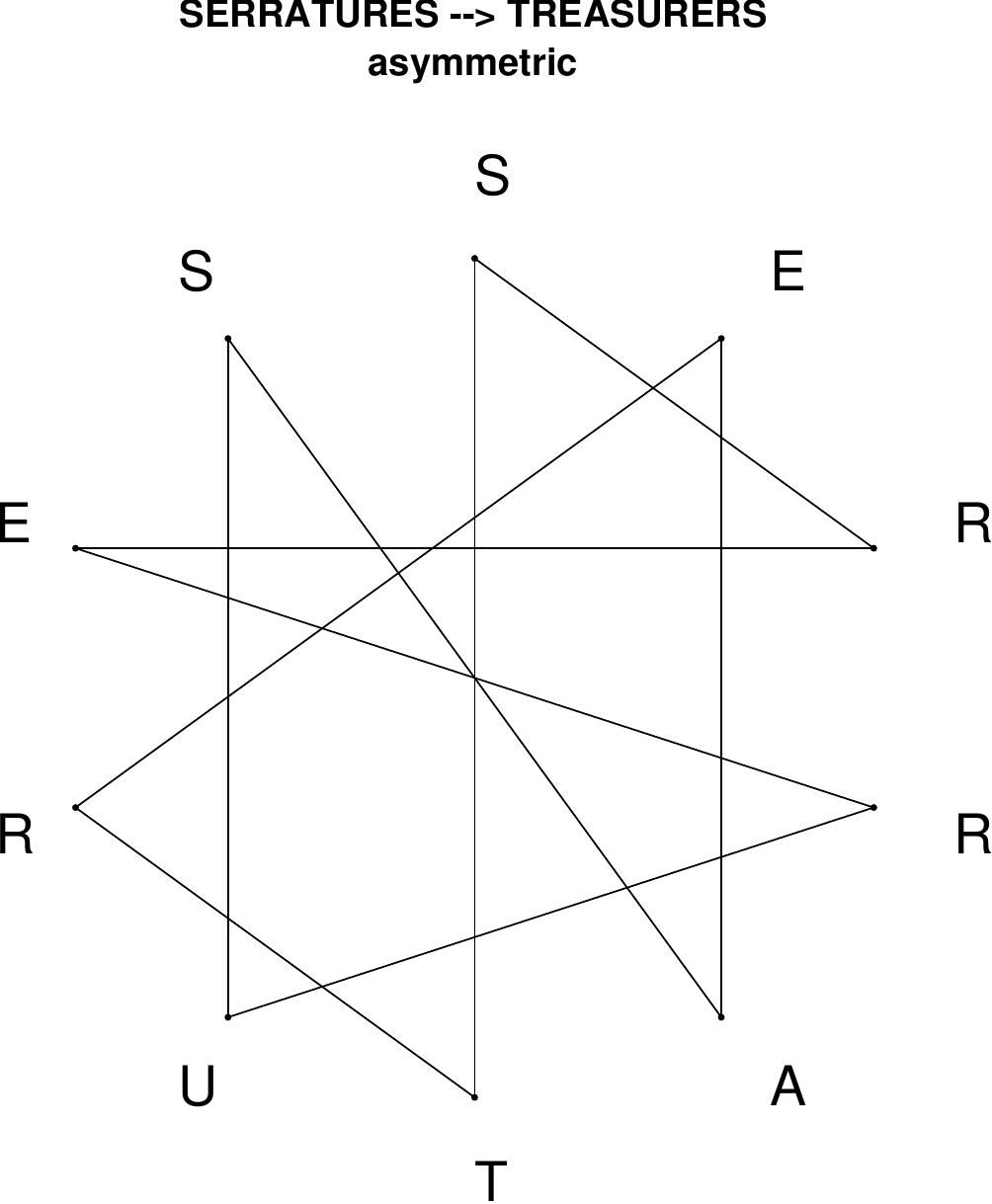}
\end{subfigure}
\hfill
\begin{subfigure}[T]{0.19\textwidth}
\centering
\includegraphics[width=\textwidth]{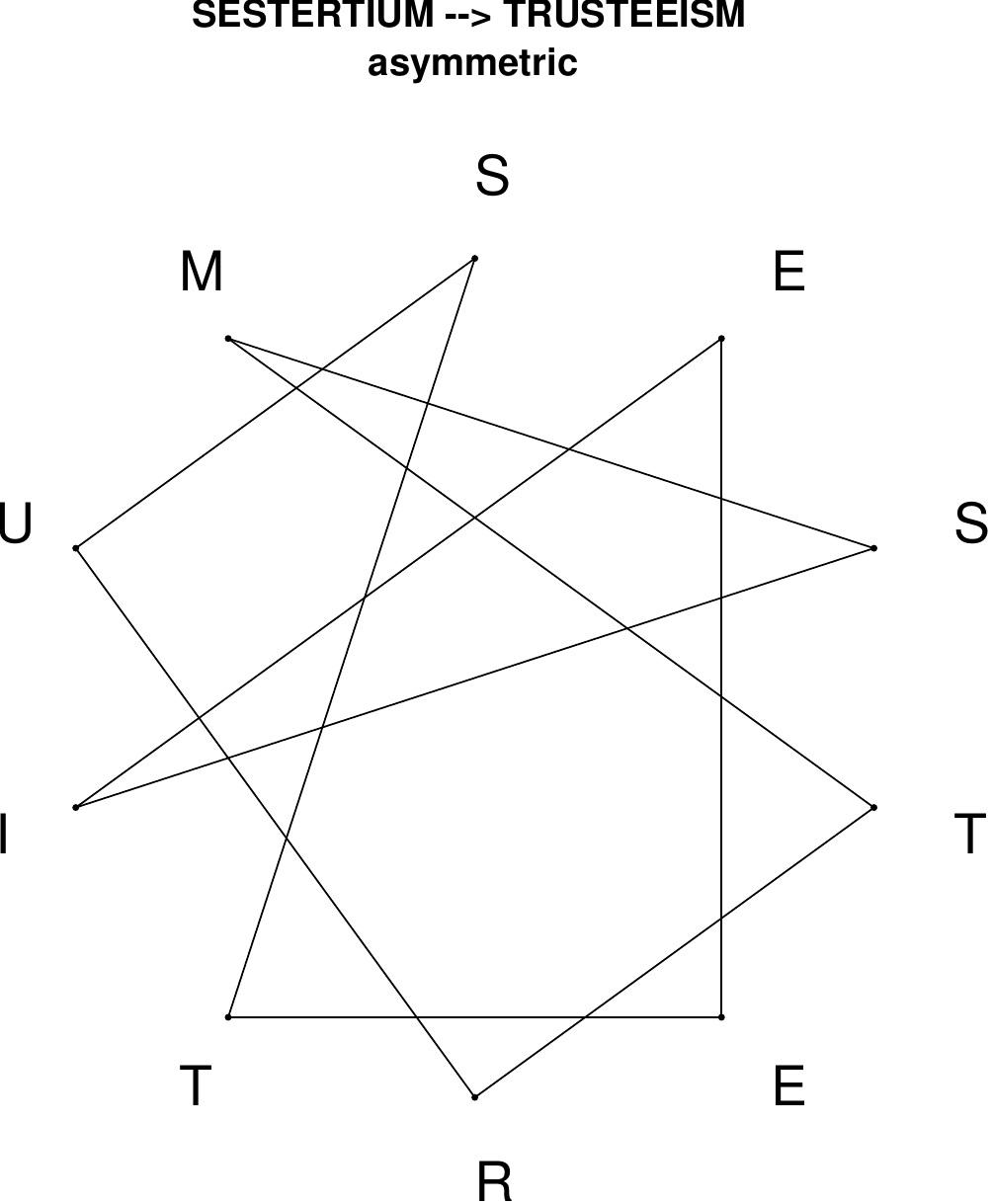}
\end{subfigure}
\hfill
\begin{subfigure}[T]{0.19\textwidth}
\centering
\includegraphics[width=\textwidth]{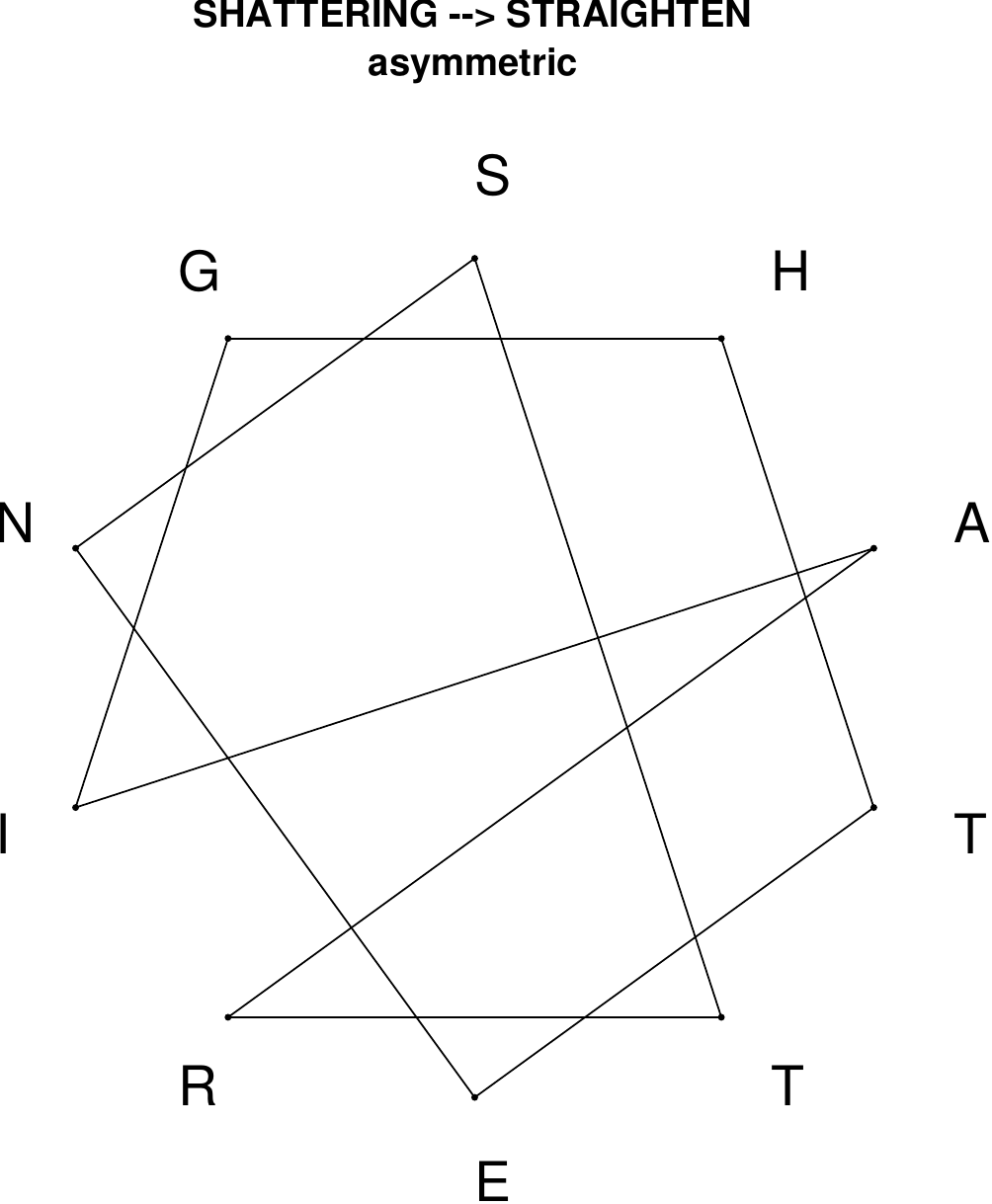}
\end{subfigure}
\end{figure}

\begin{figure}[H]
\centering
\begin{subfigure}[T]{0.19\textwidth}
\centering
\includegraphics[width=\textwidth]{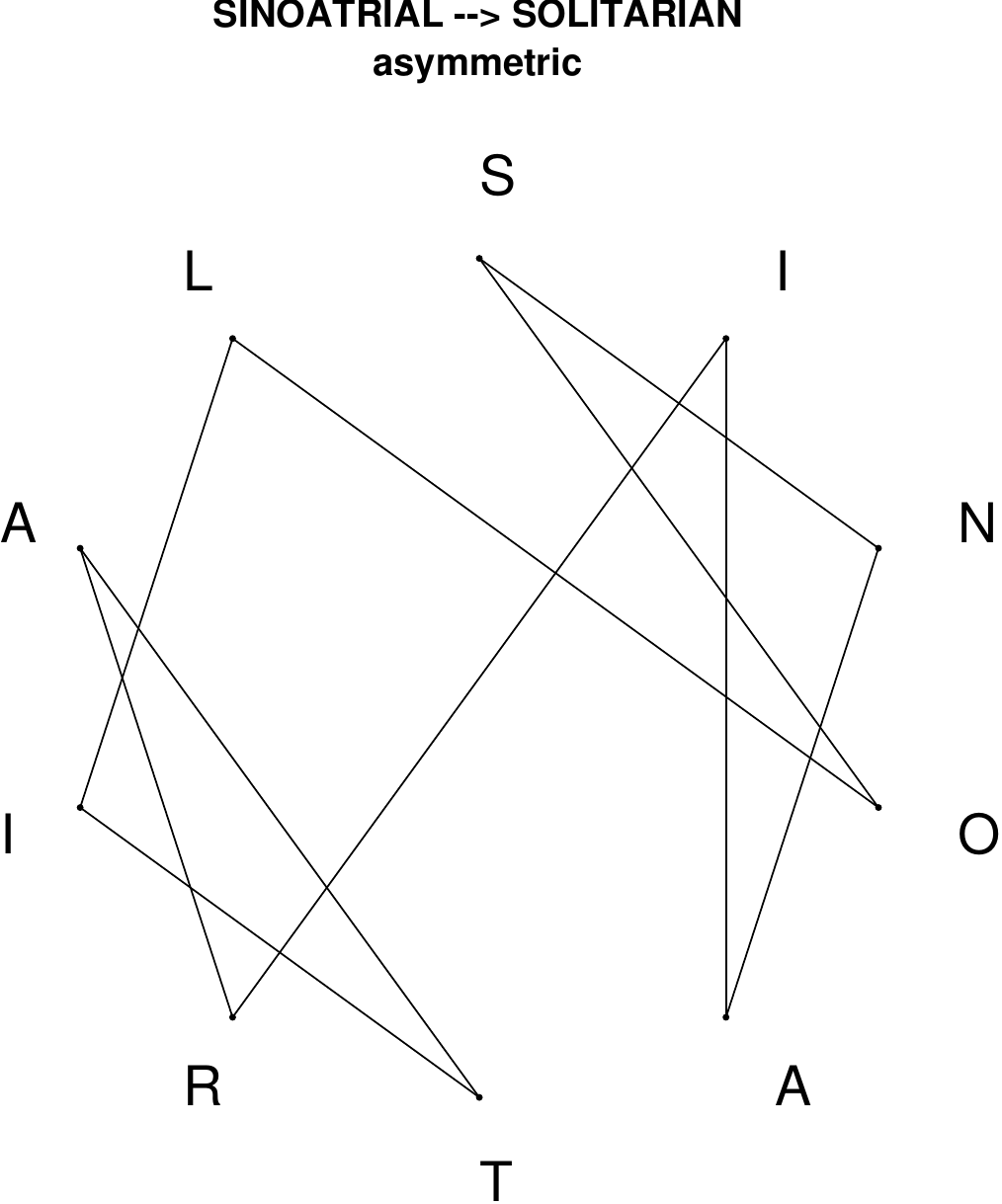}
\end{subfigure}
\hfill
\begin{subfigure}[T]{0.19\textwidth}
\centering
\includegraphics[width=\textwidth]{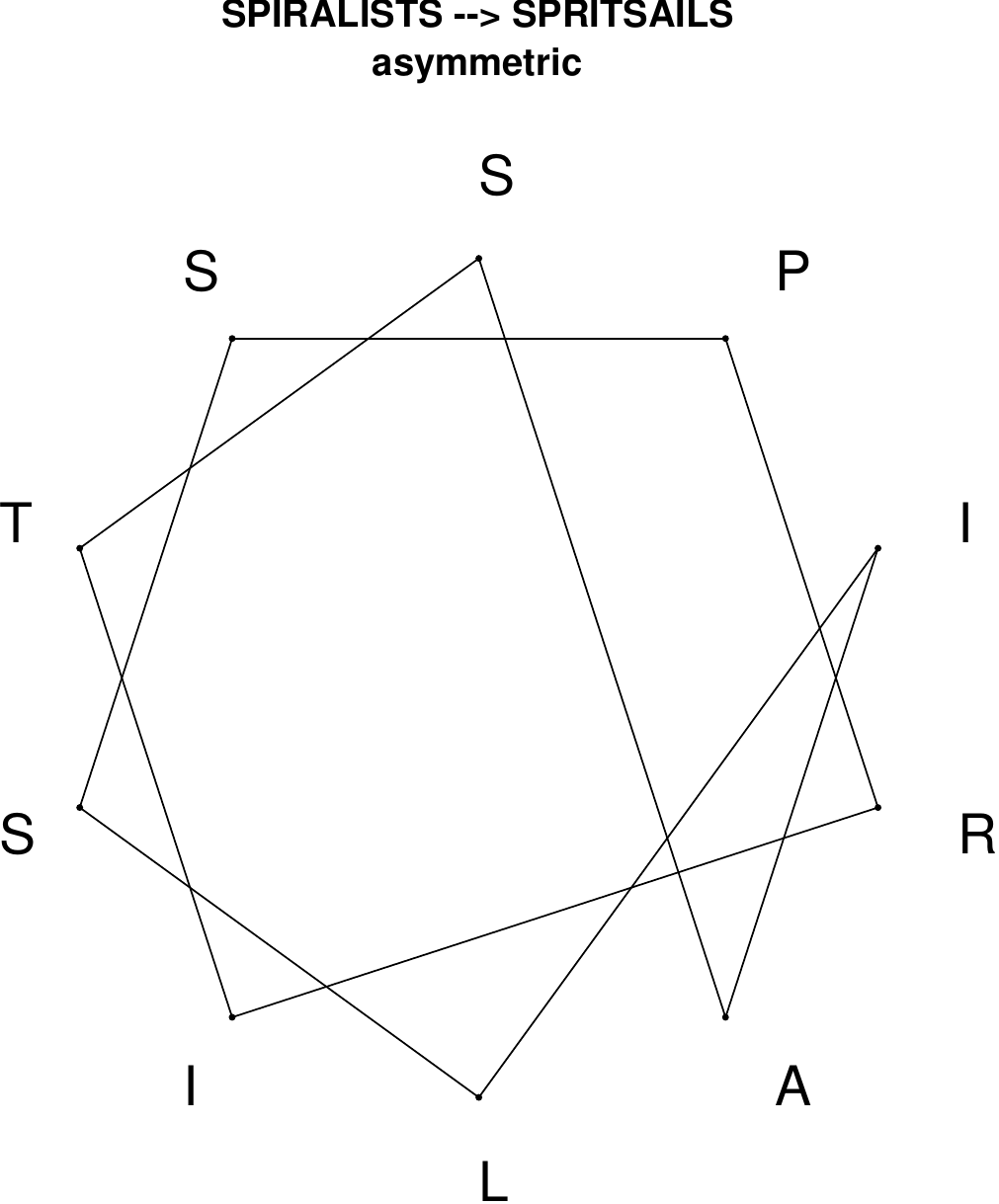}
\end{subfigure}
\hfill
\begin{subfigure}[T]{0.19\textwidth}
\centering
\includegraphics[width=\textwidth]{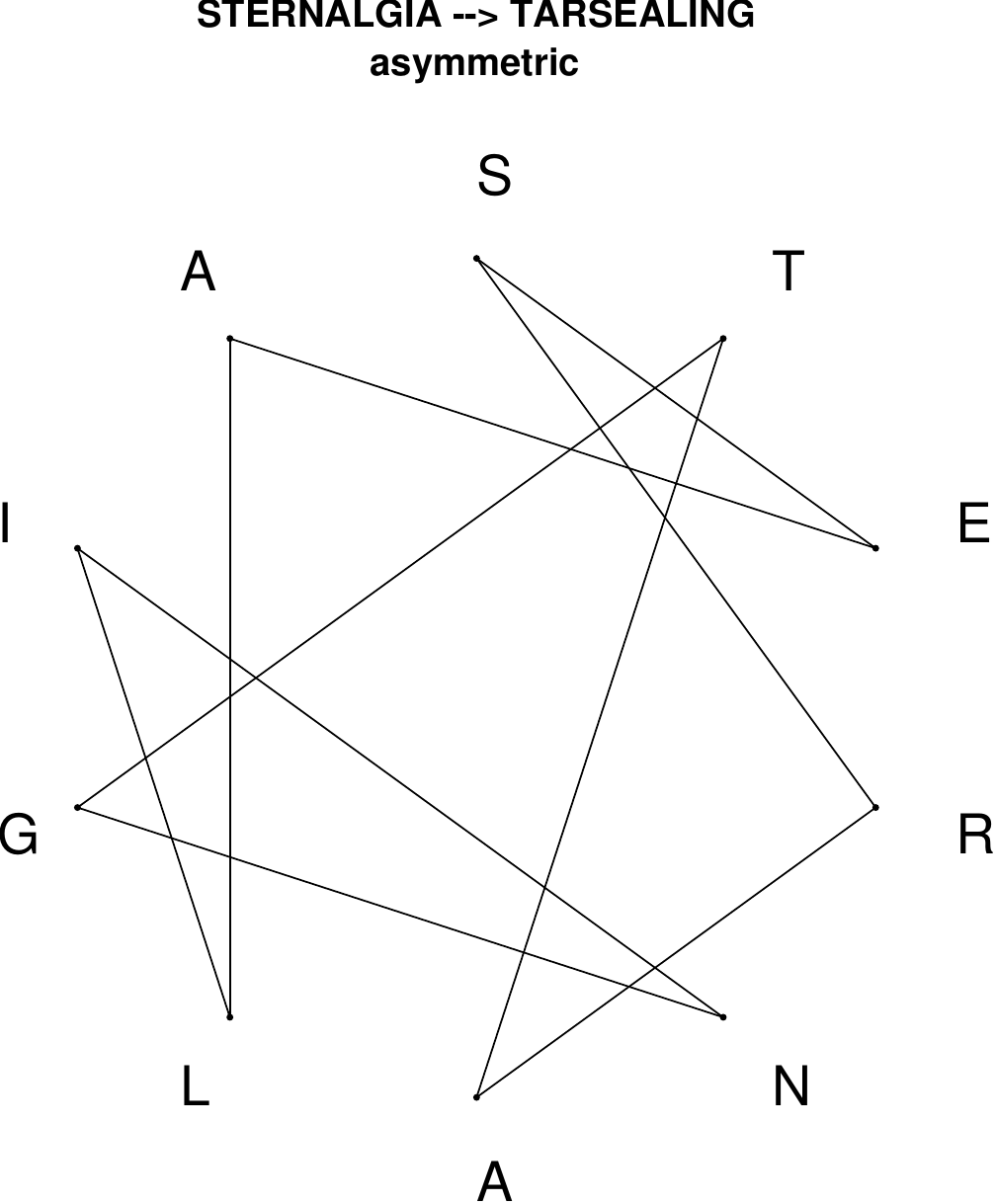}
\end{subfigure}
\hfill
\begin{subfigure}[T]{0.19\textwidth}
\centering
\includegraphics[width=\textwidth]{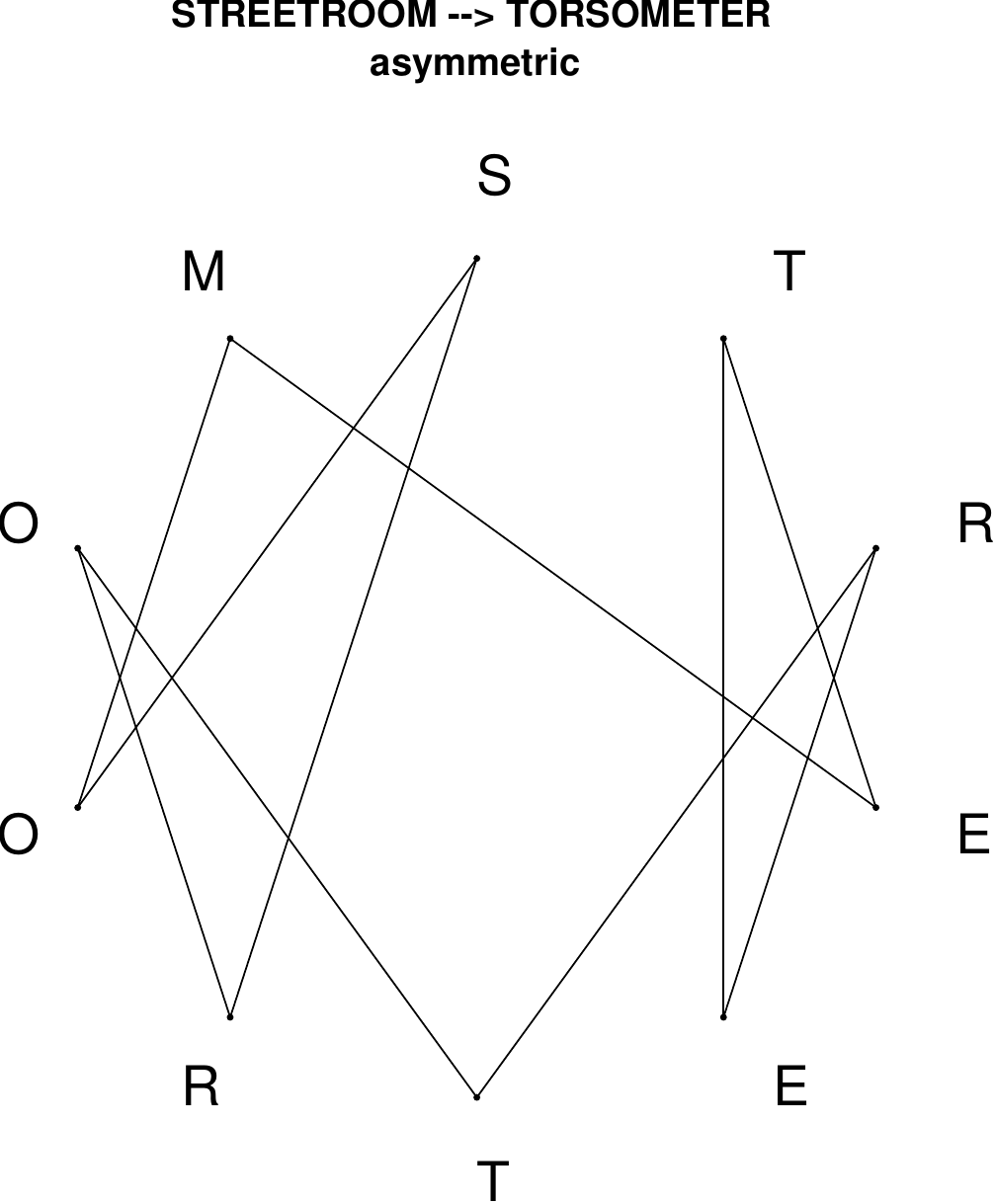}
\end{subfigure}
\hfill
\begin{subfigure}[T]{0.19\textwidth}
\centering
\includegraphics[width=\textwidth]{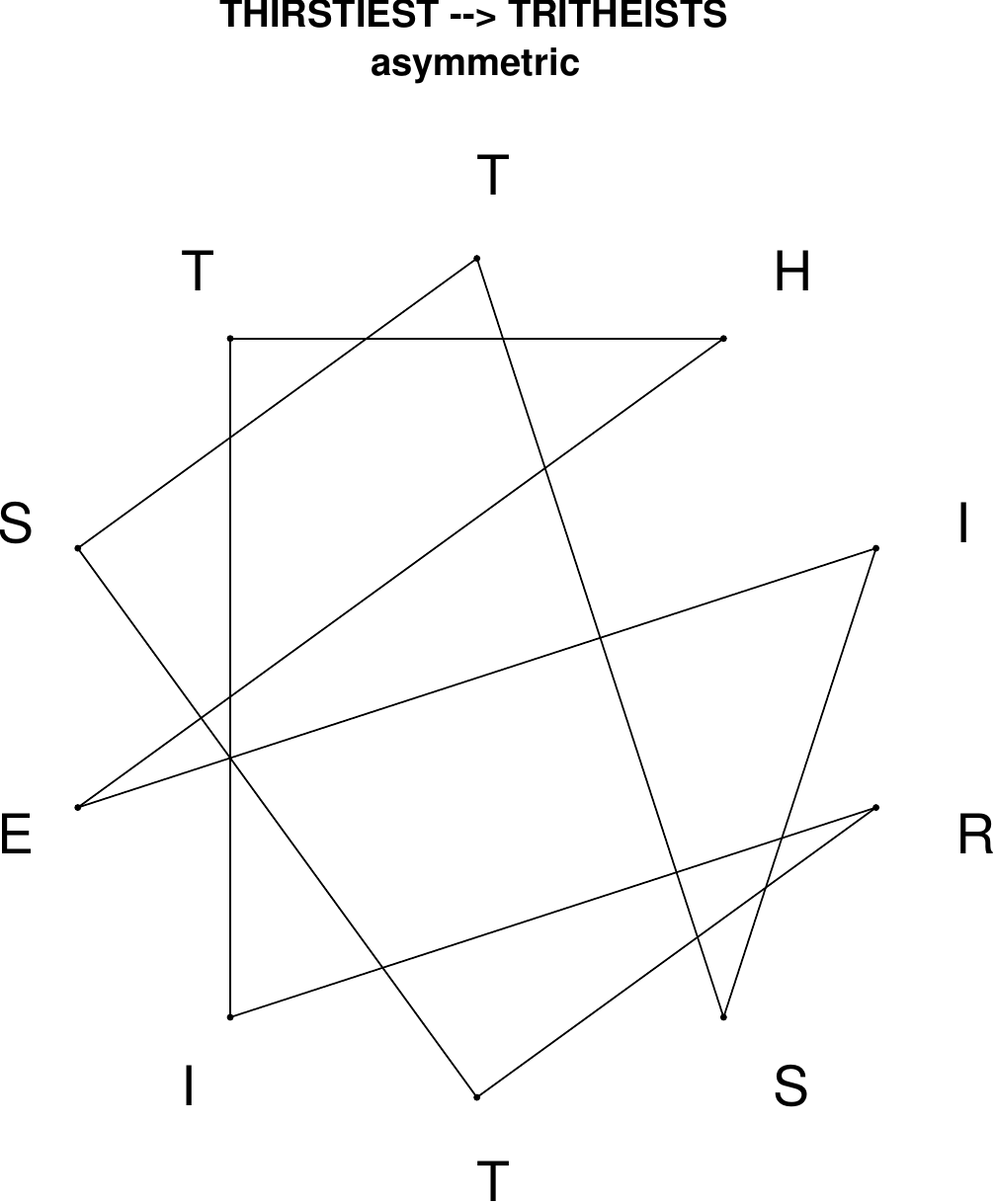}
\end{subfigure}
\end{figure}

\begin{figure}[H]
\centering
\begin{subfigure}[T]{0.19\textwidth}
\centering
\includegraphics[width=\textwidth]{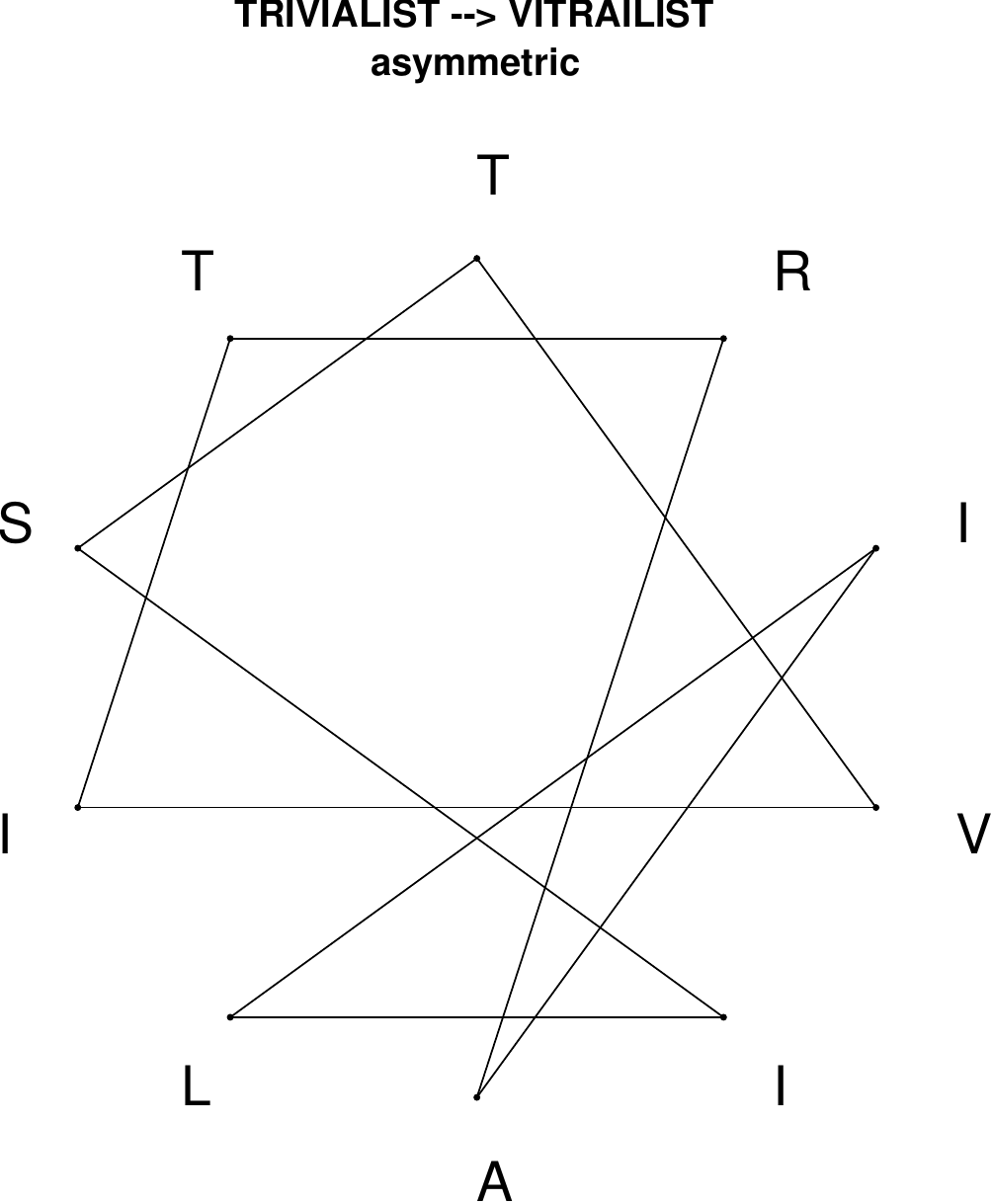}
\end{subfigure}
\hfill
\begin{subfigure}[T]{0.19\textwidth}
\centering
\includegraphics[width=\textwidth]{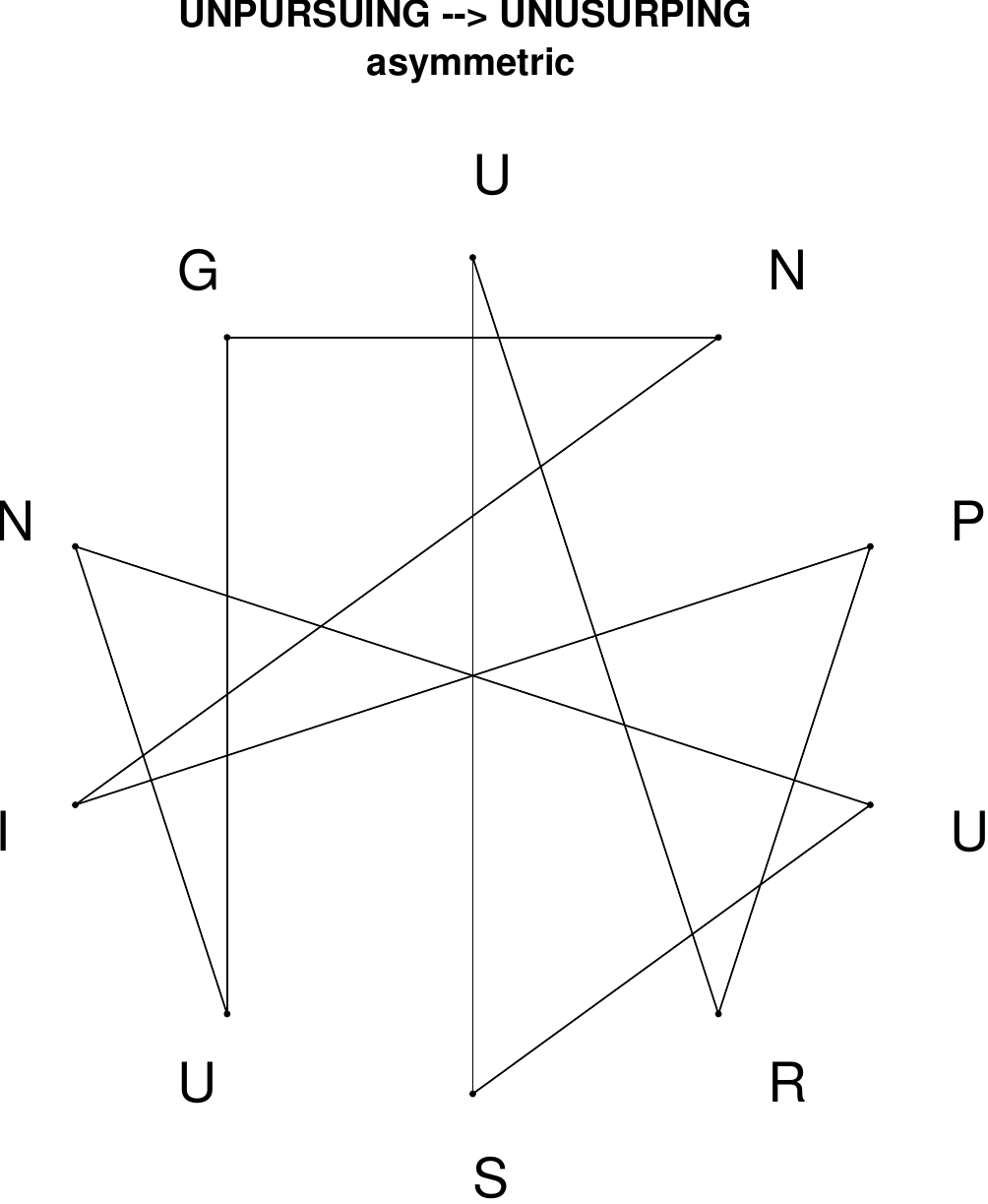}
\end{subfigure}
\hfill
\end{figure}

%%%%%%%%%%%%%%%%%%
\clearpage
\subsection{Star Anagrams $N = 9$}
For $N=9$, we again found stars from all three classes with a single perfect star. 

\subsubsection{Perfect Stars $N=9$}

\begin{figure}[H]
\centering
\begin{subfigure}[T]{0.19\textwidth}
\centering
\includegraphics[width=\textwidth]{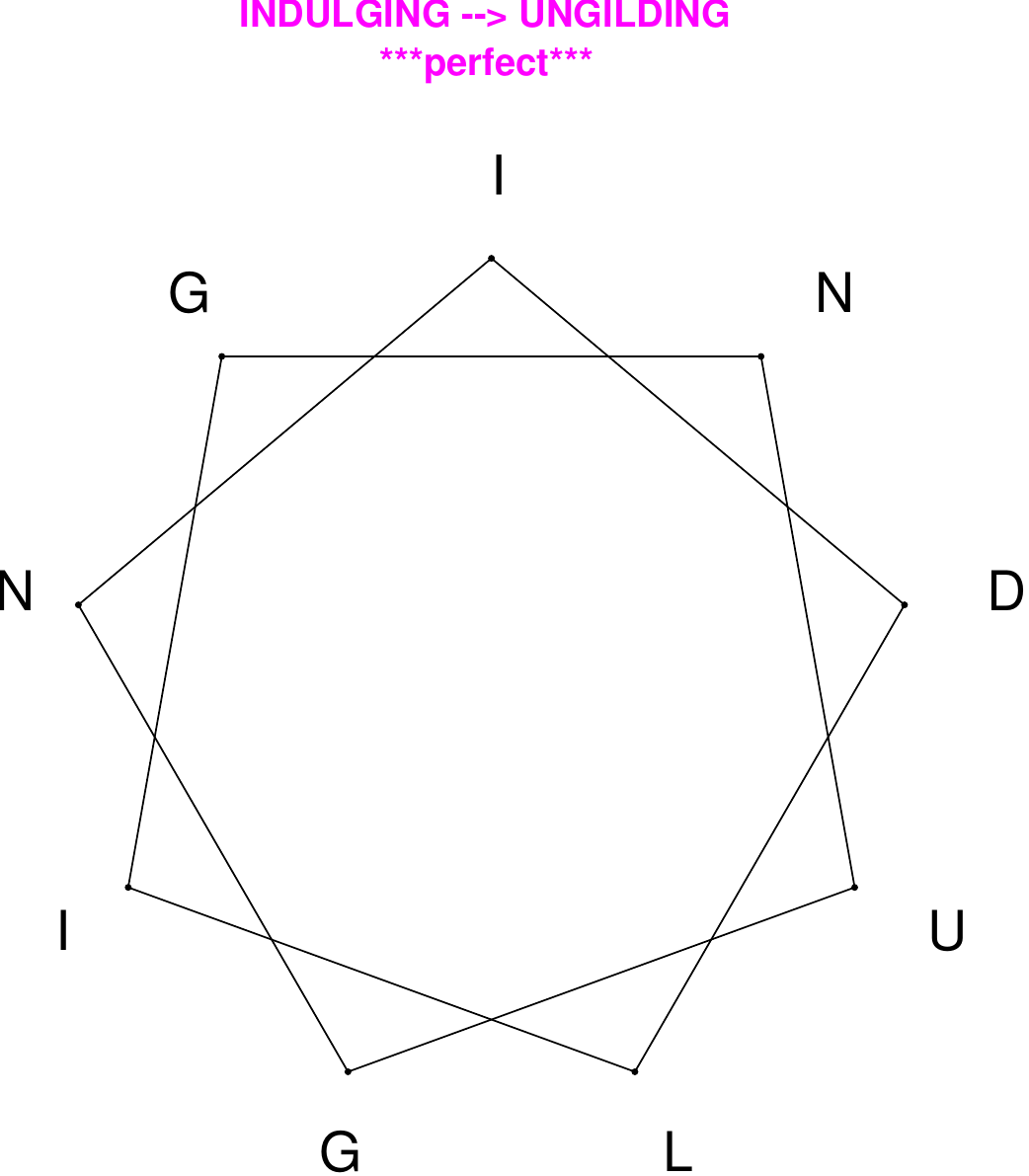}
\end{subfigure}
\hfill
\end{figure}

\subsubsection{Symmetric Stars $N=9$}

\begin{figure}[H]
\centering
\begin{subfigure}[T]{0.19\textwidth}
\centering
\includegraphics[width=\textwidth]{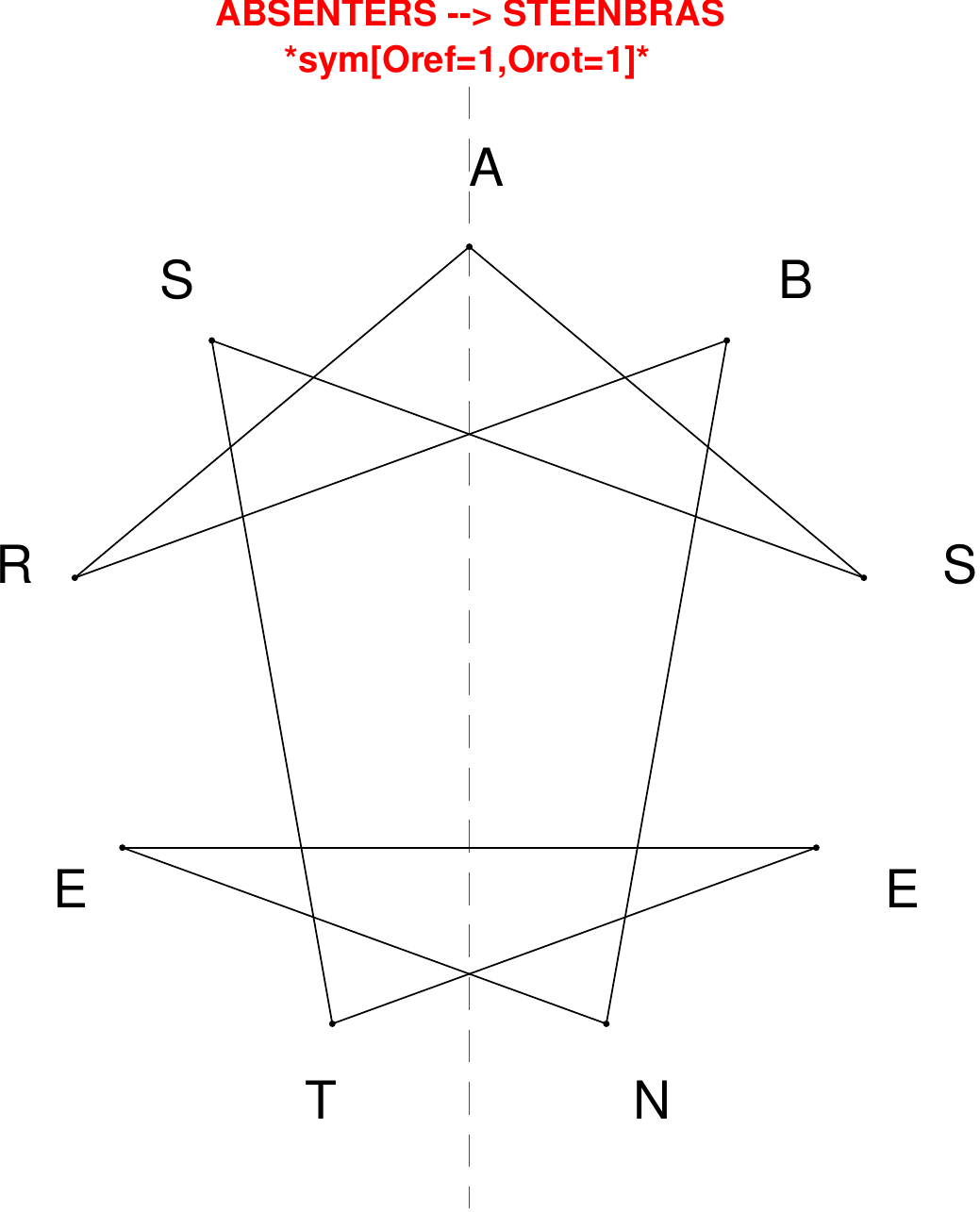}
\end{subfigure}
\hfill
\begin{subfigure}[T]{0.19\textwidth}
\centering
\includegraphics[width=\textwidth]{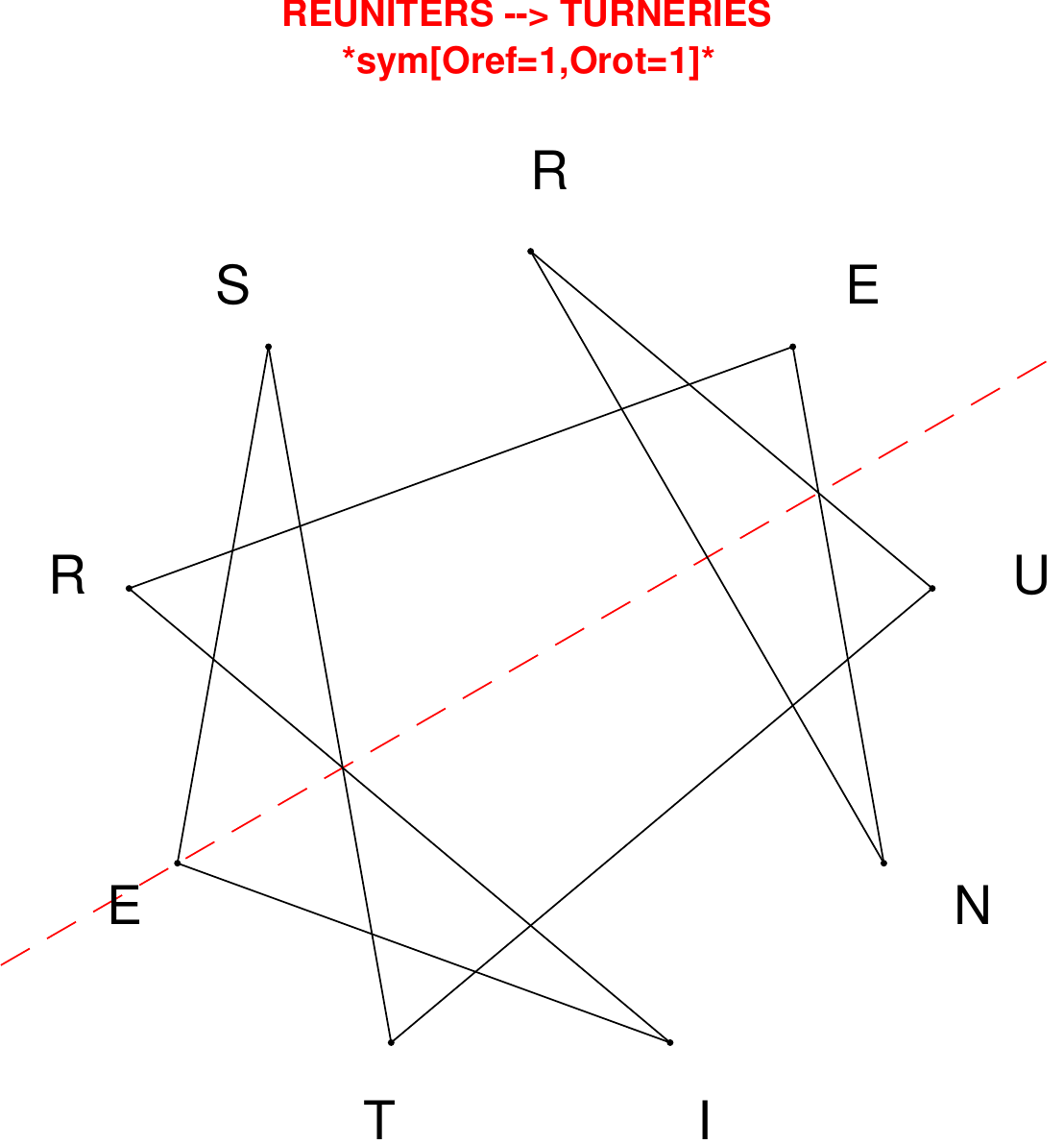}
\end{subfigure}
\hfill
\begin{subfigure}[T]{0.19\textwidth}
\centering
\includegraphics[width=\textwidth]{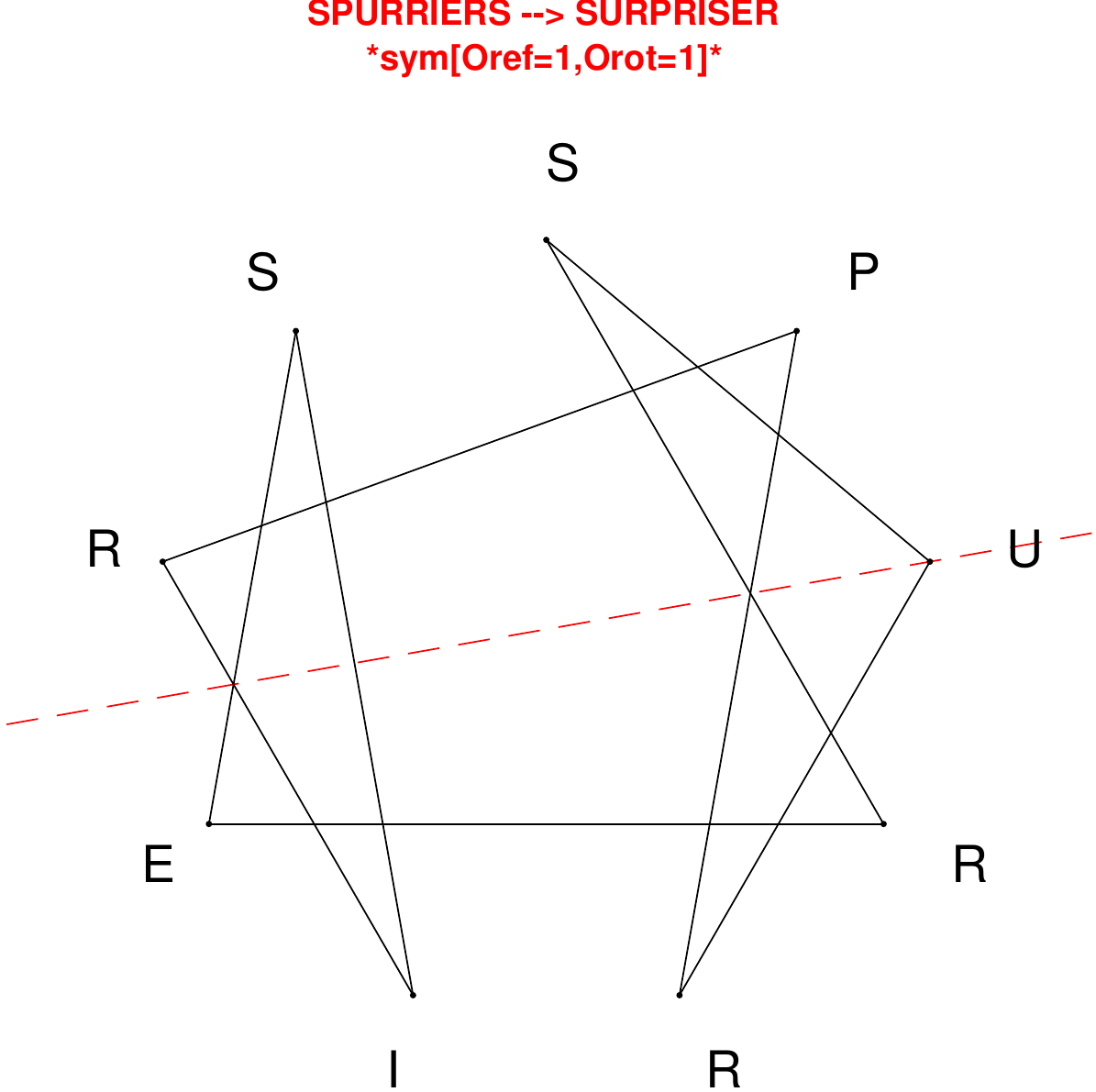}
\end{subfigure}
\hfill
\begin{subfigure}[T]{0.19\textwidth}
\centering
\includegraphics[width=\textwidth]{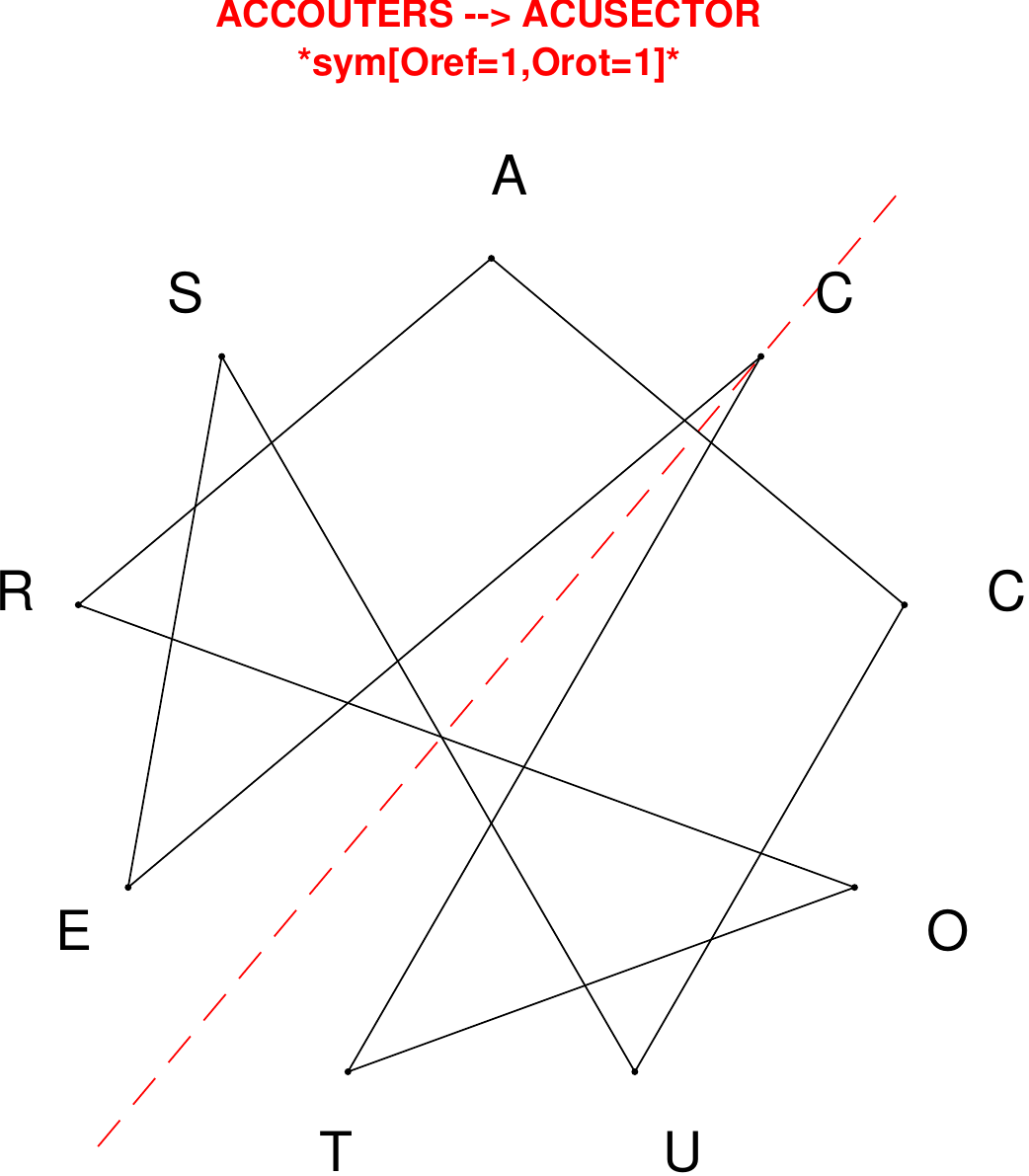}
\end{subfigure}
\hfill
\begin{subfigure}[T]{0.19\textwidth}
\centering
\includegraphics[width=\textwidth]{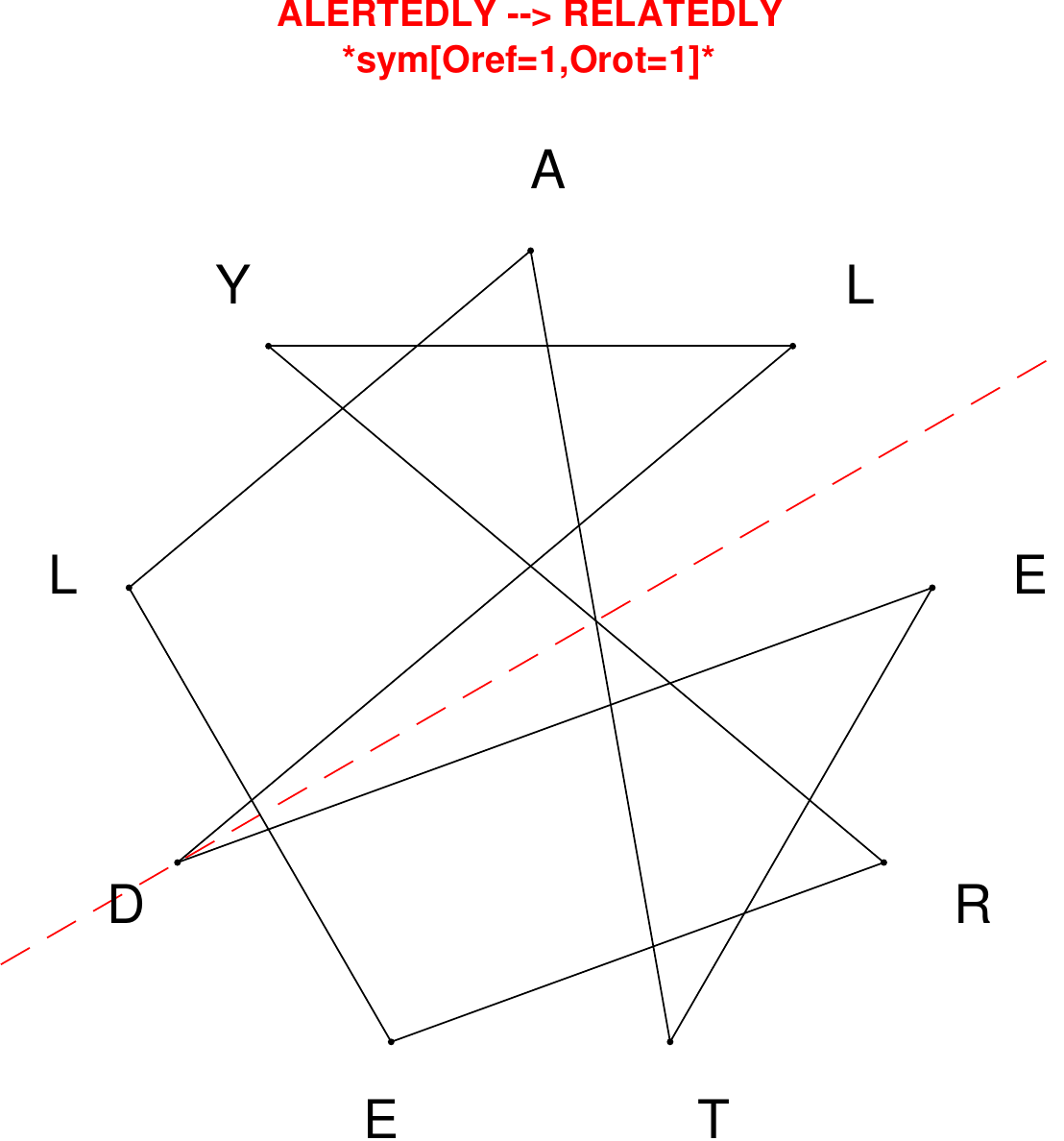}
\end{subfigure}
\end{figure}

\begin{figure}[H]
\centering
\begin{subfigure}[T]{0.19\textwidth}
\centering
\includegraphics[width=\textwidth]{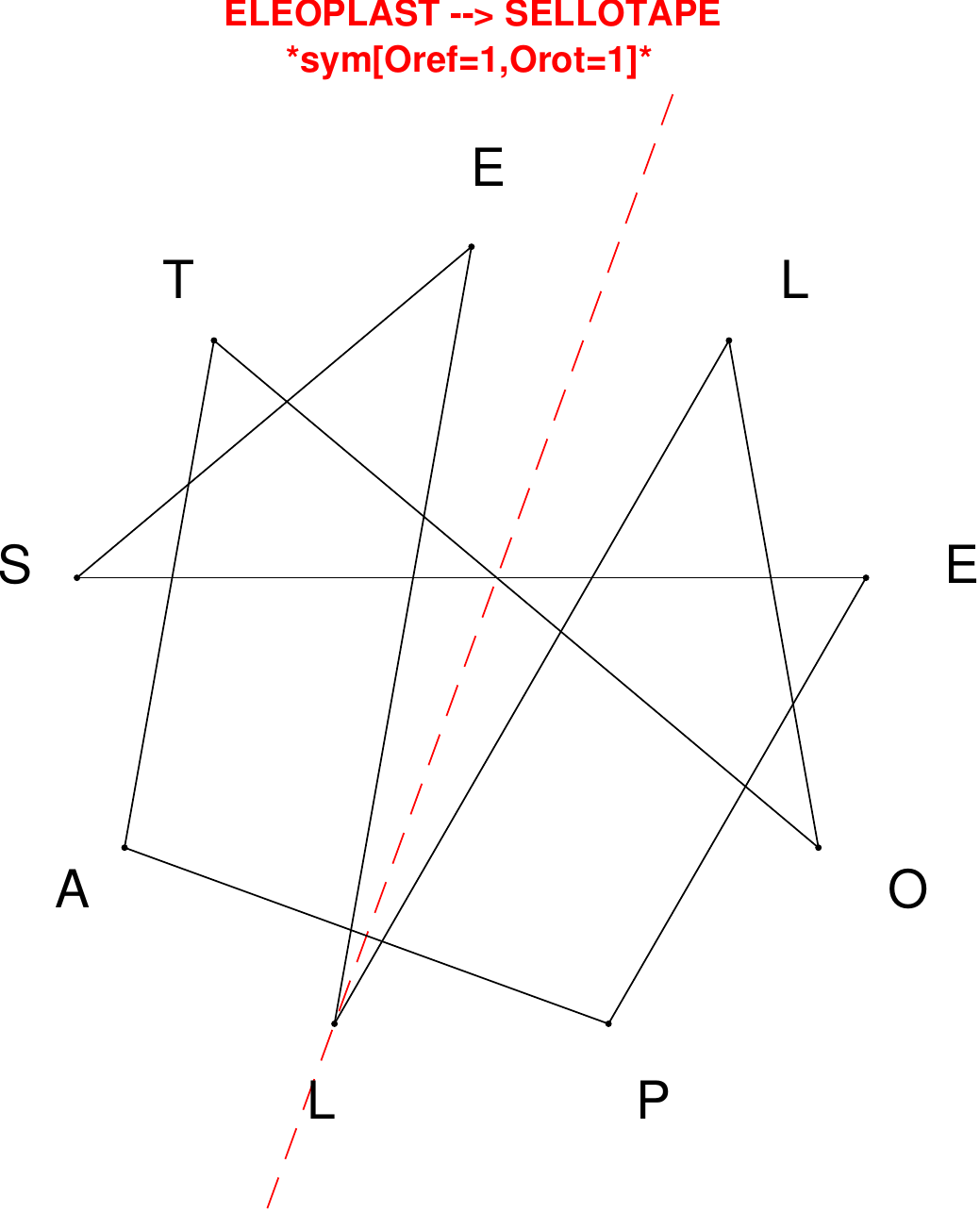}
\end{subfigure}
\hfill
\begin{subfigure}[T]{0.19\textwidth}
\centering
\includegraphics[width=\textwidth]{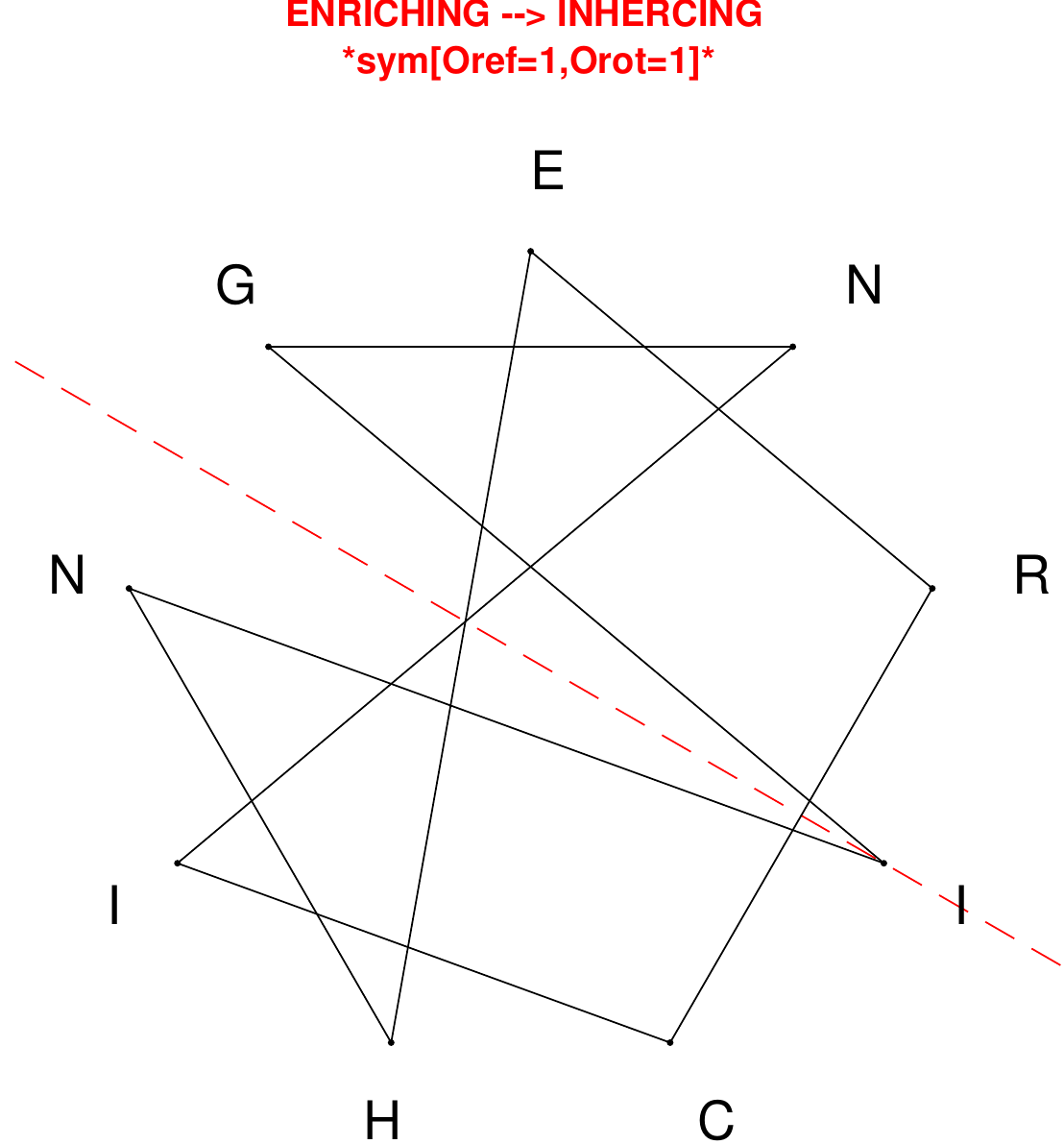}
\end{subfigure}
\hfill
\begin{subfigure}[T]{0.19\textwidth}
\centering
\includegraphics[width=\textwidth]{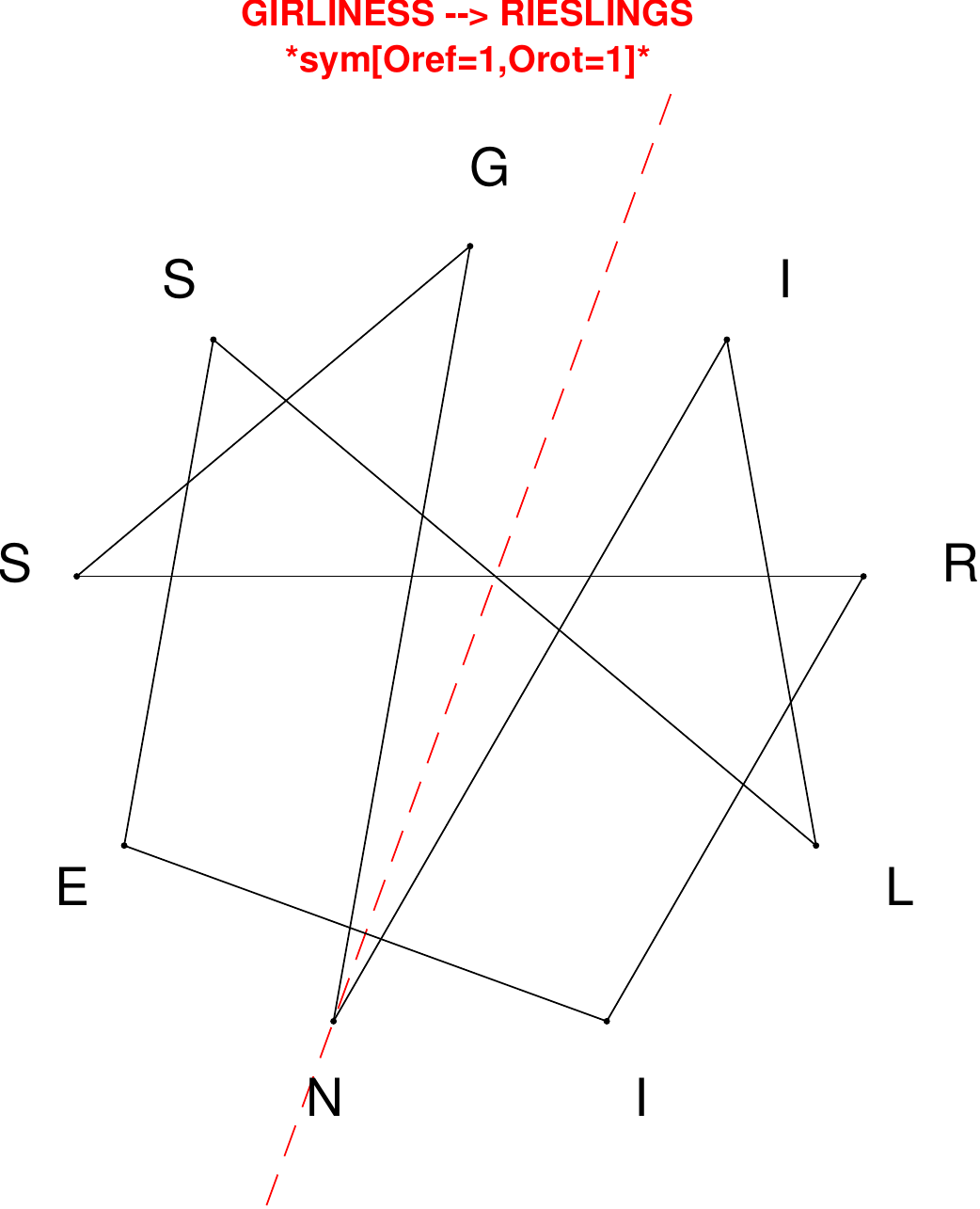}
\end{subfigure}
\hfill
\begin{subfigure}[T]{0.19\textwidth}
\centering
\includegraphics[width=\textwidth]{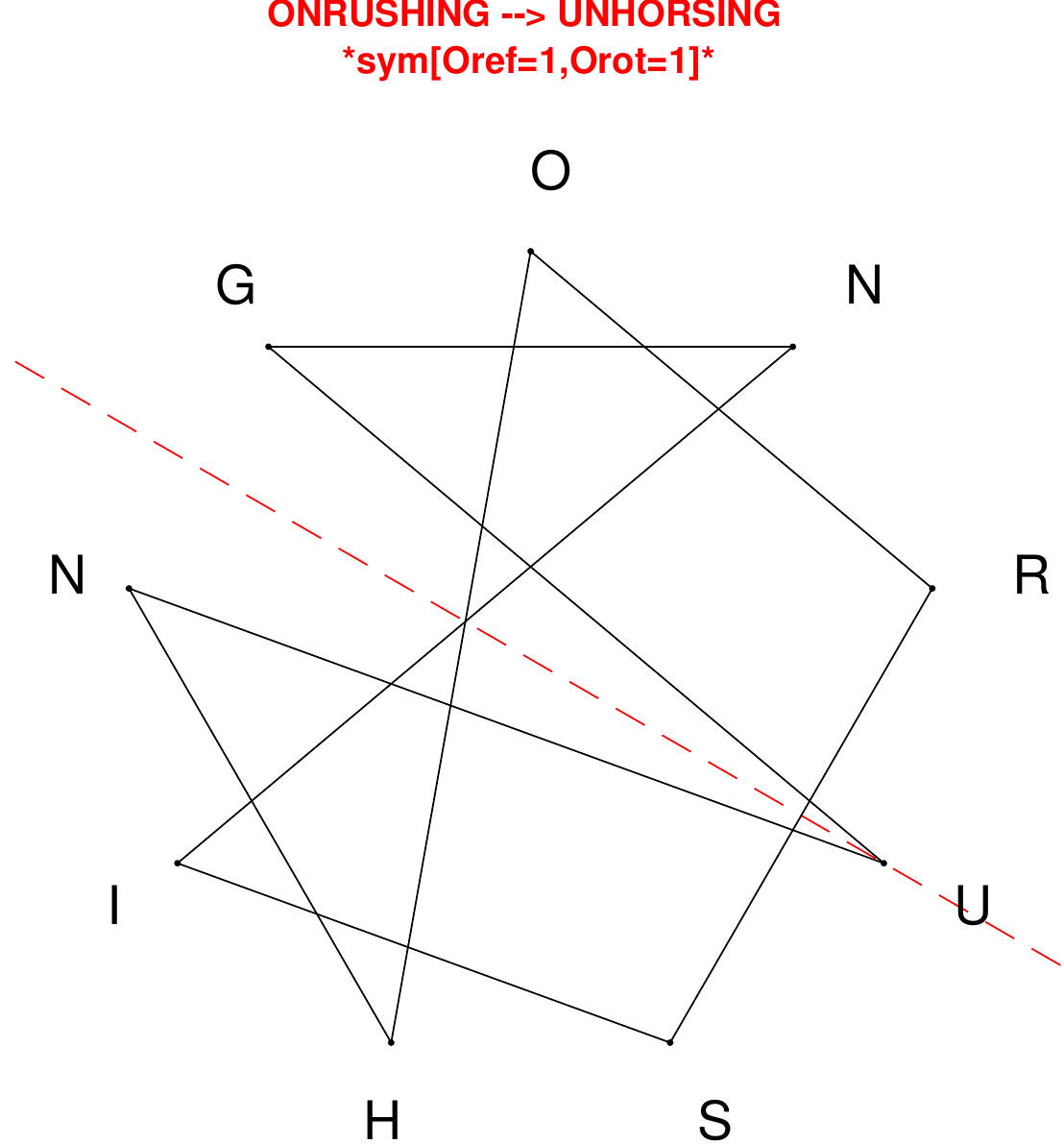}
\end{subfigure}
\hfill
\begin{subfigure}[T]{0.19\textwidth}
\centering
\includegraphics[width=\textwidth]{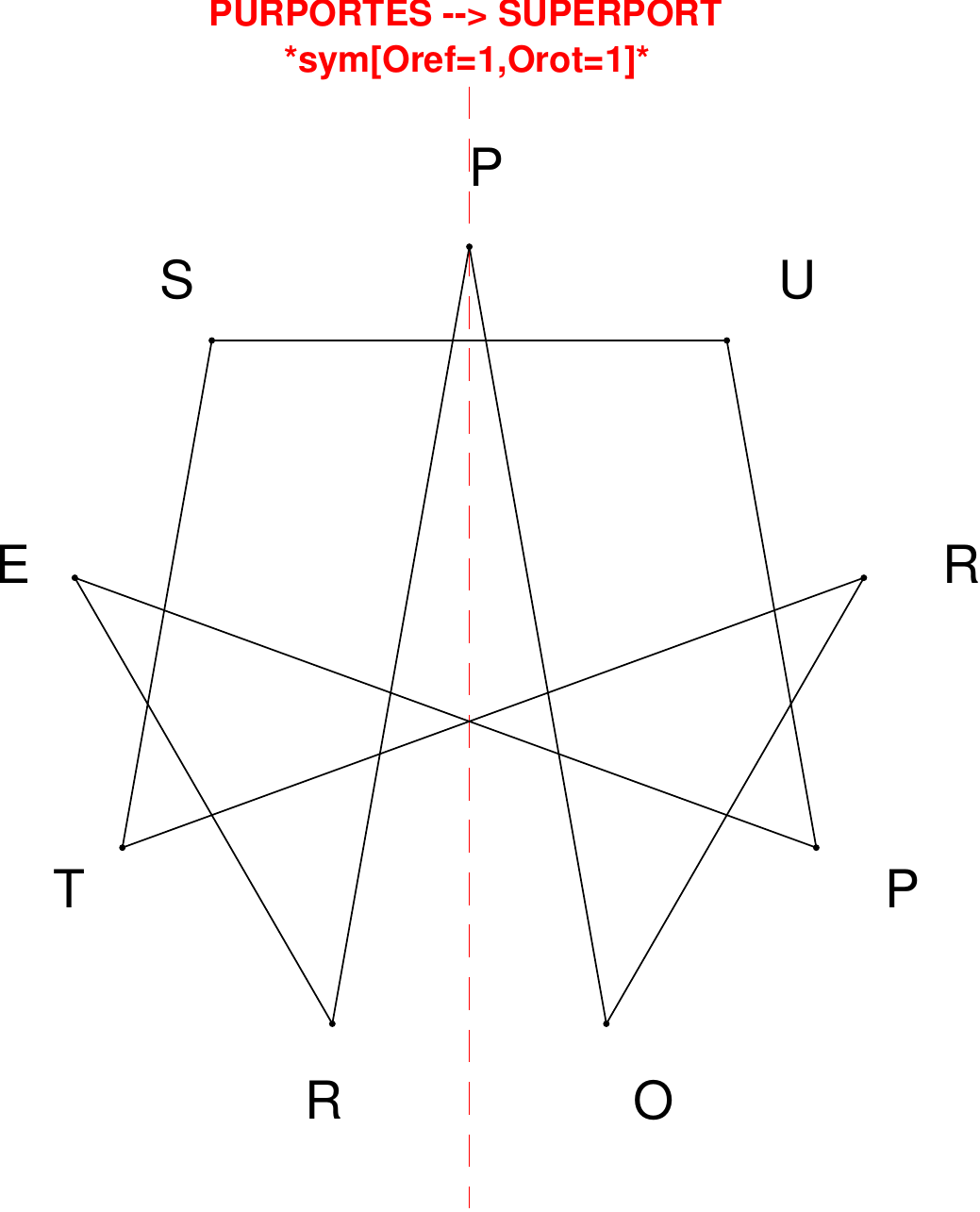}
\end{subfigure}
\end{figure}

\begin{figure}[H]
\centering
\begin{subfigure}[T]{0.19\textwidth}
\centering
\includegraphics[width=\textwidth]{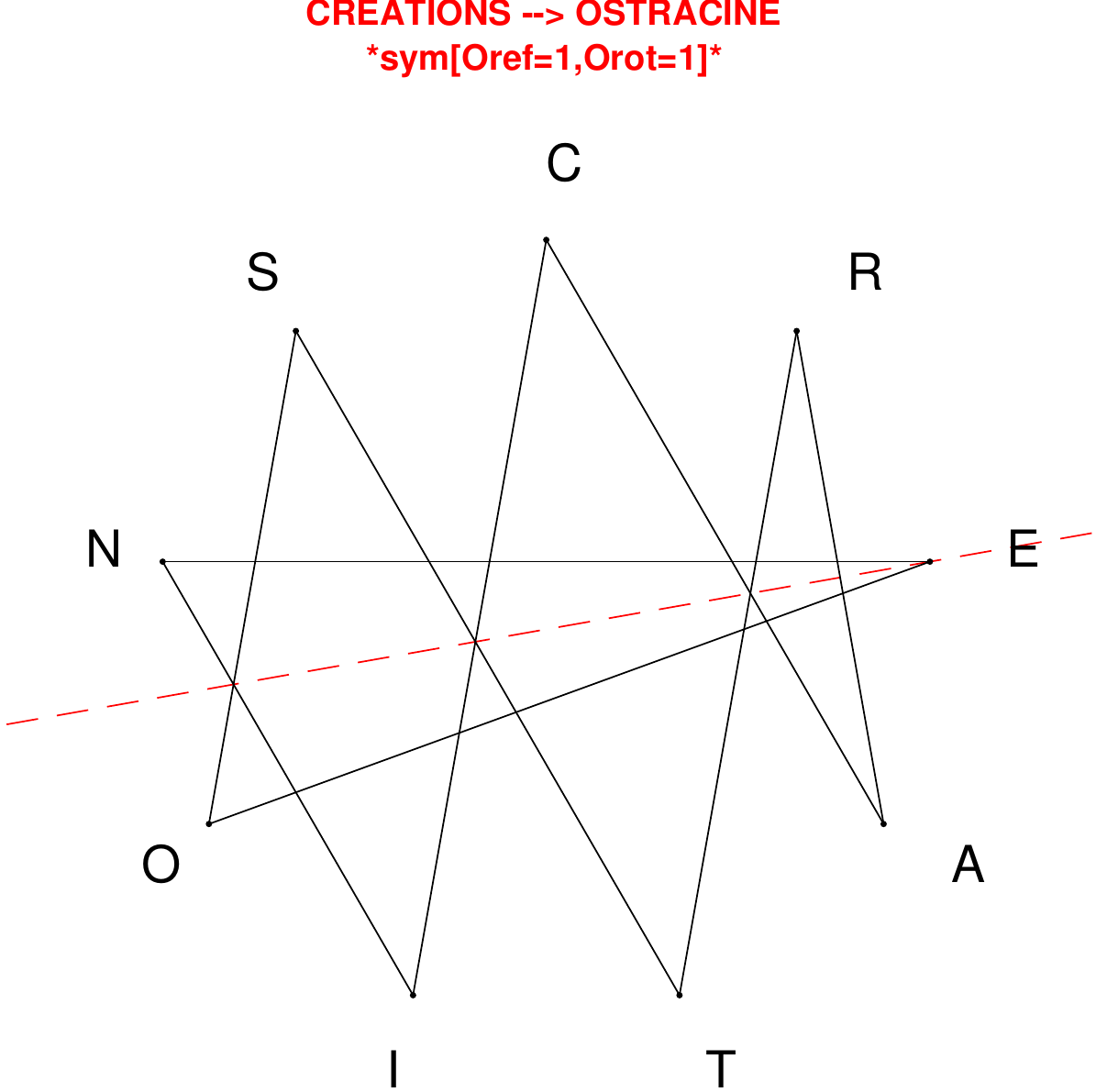}
\end{subfigure}
\hfill
\begin{subfigure}[T]{0.19\textwidth}
\centering
\includegraphics[width=\textwidth]{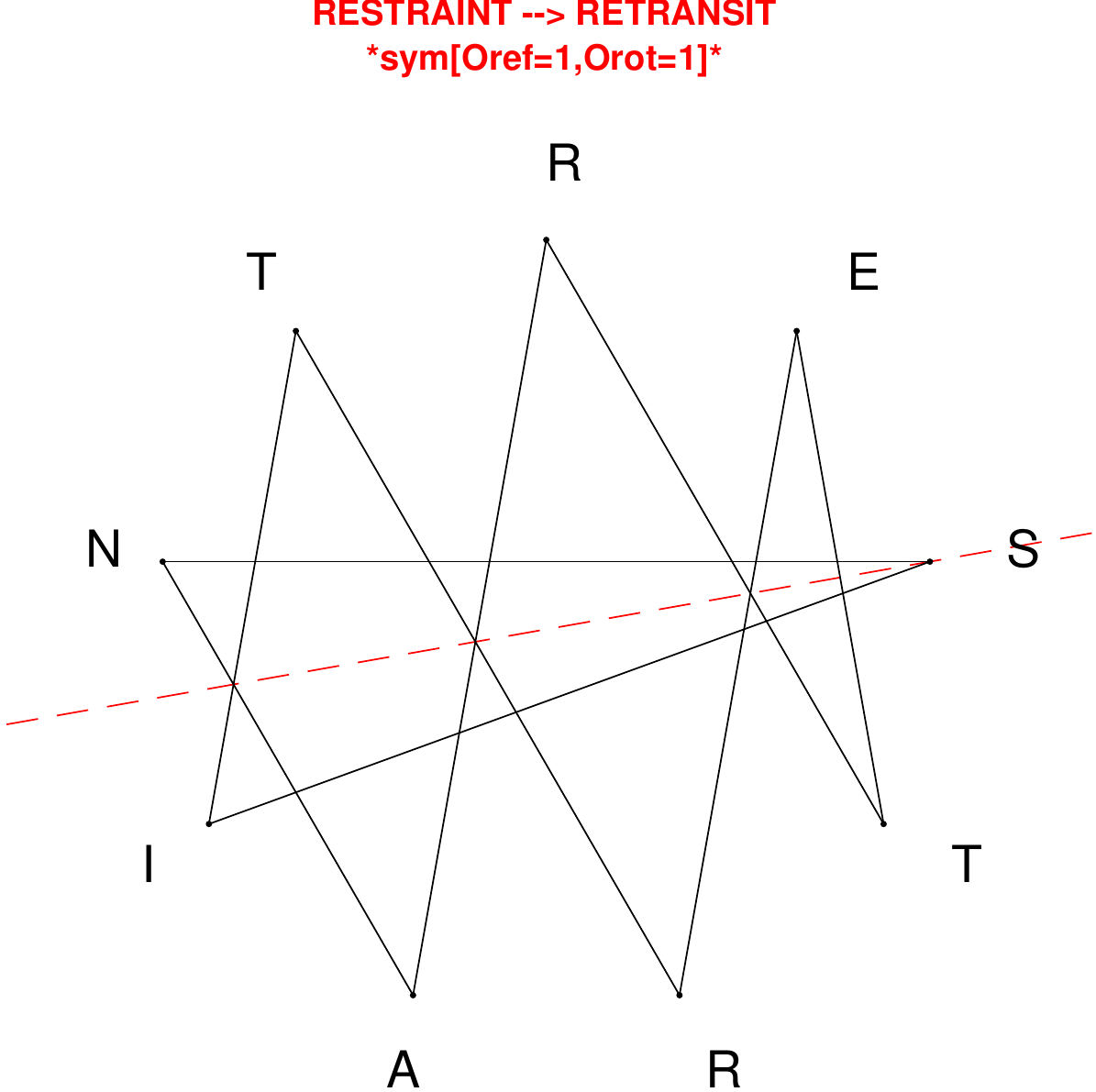}
\end{subfigure}
\hfill
\begin{subfigure}[T]{0.19\textwidth}
\centering
\includegraphics[width=\textwidth]{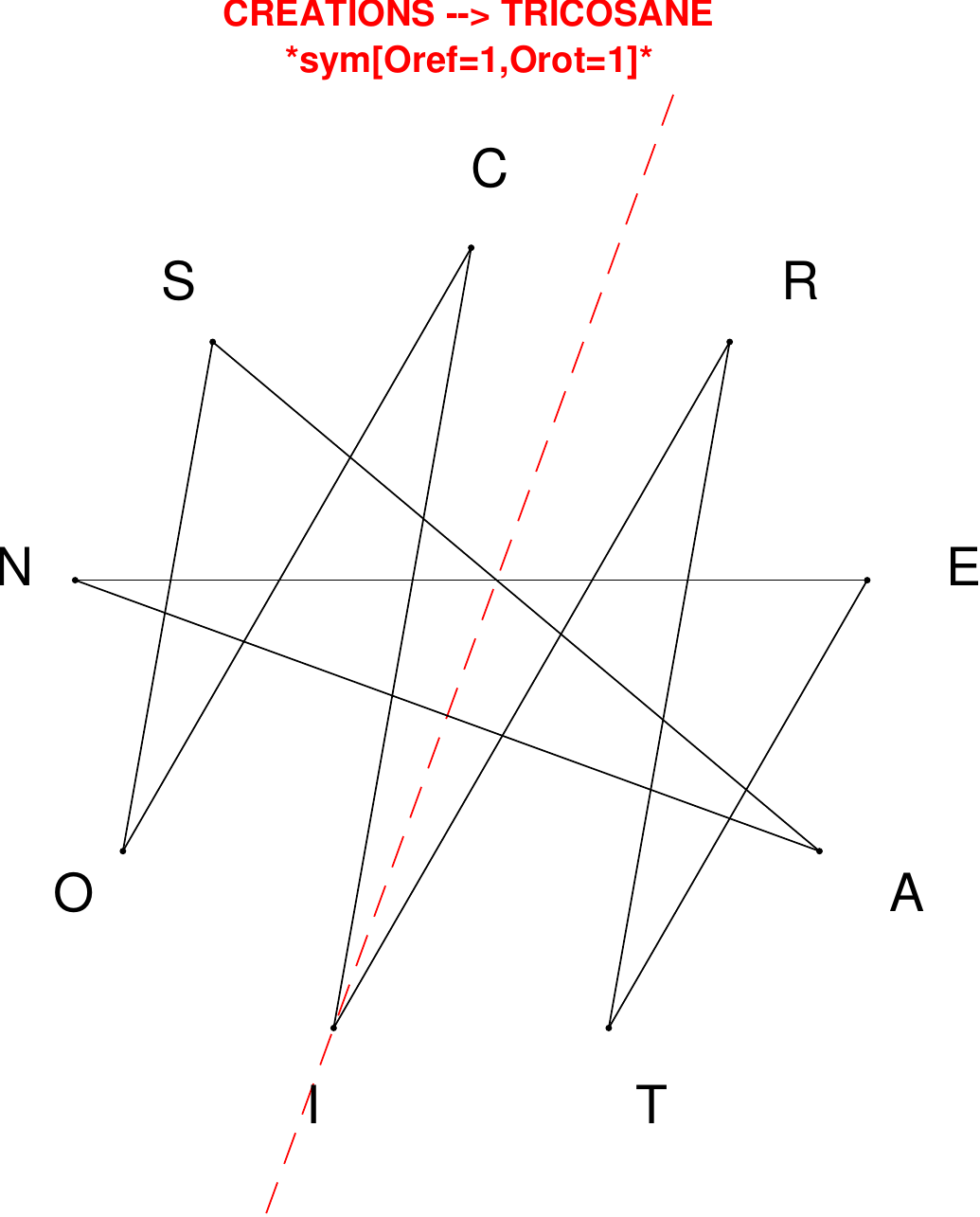}
\end{subfigure}
\hfill
\begin{subfigure}[T]{0.19\textwidth}
\centering
\includegraphics[width=\textwidth]{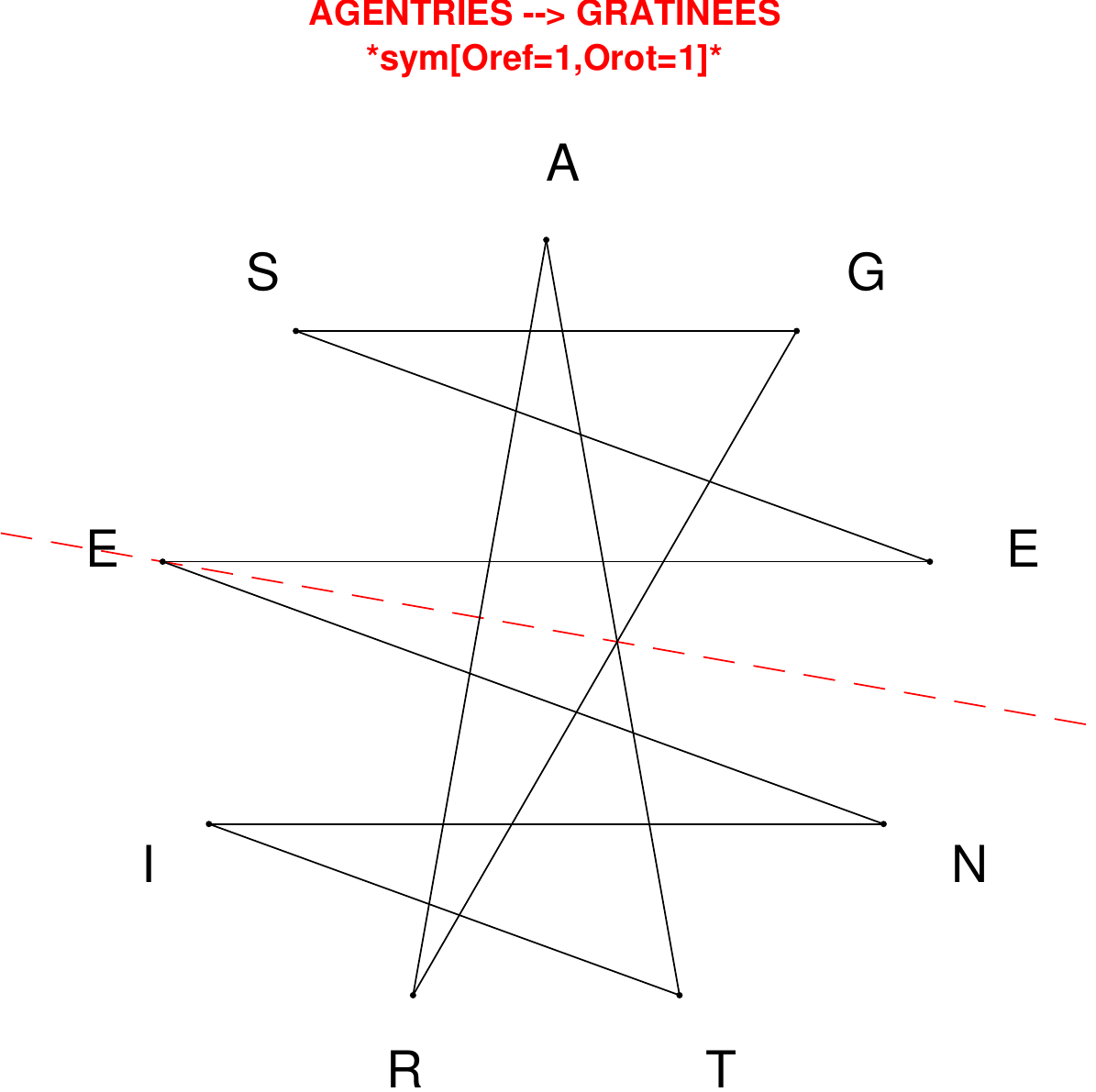}
\end{subfigure}
\hfill
\begin{subfigure}[T]{0.19\textwidth}
\centering
\includegraphics[width=\textwidth]{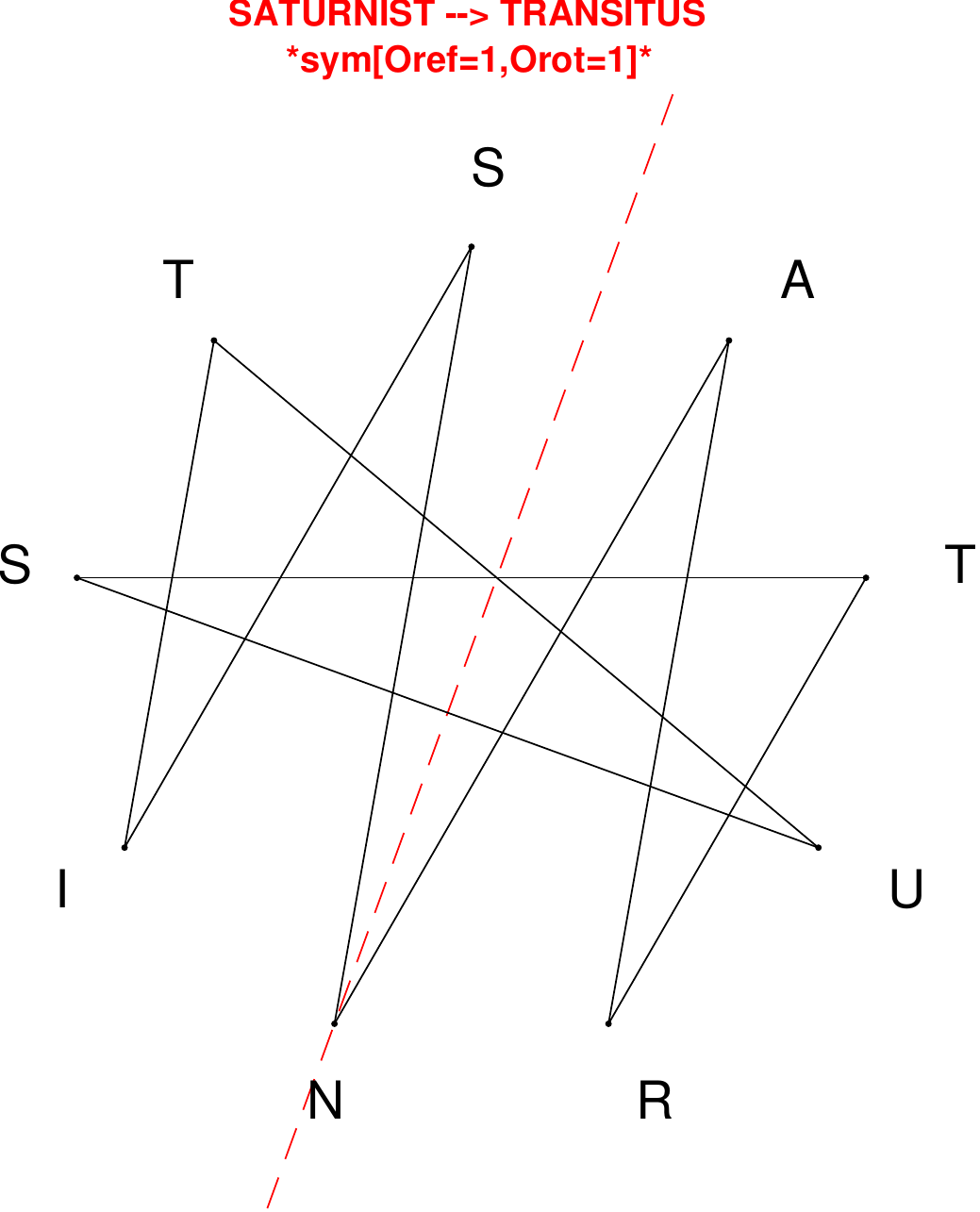}
\end{subfigure}
\end{figure}

\begin{figure}[H]
\centering
\begin{subfigure}[T]{0.19\textwidth}
\centering
\includegraphics[width=\textwidth]{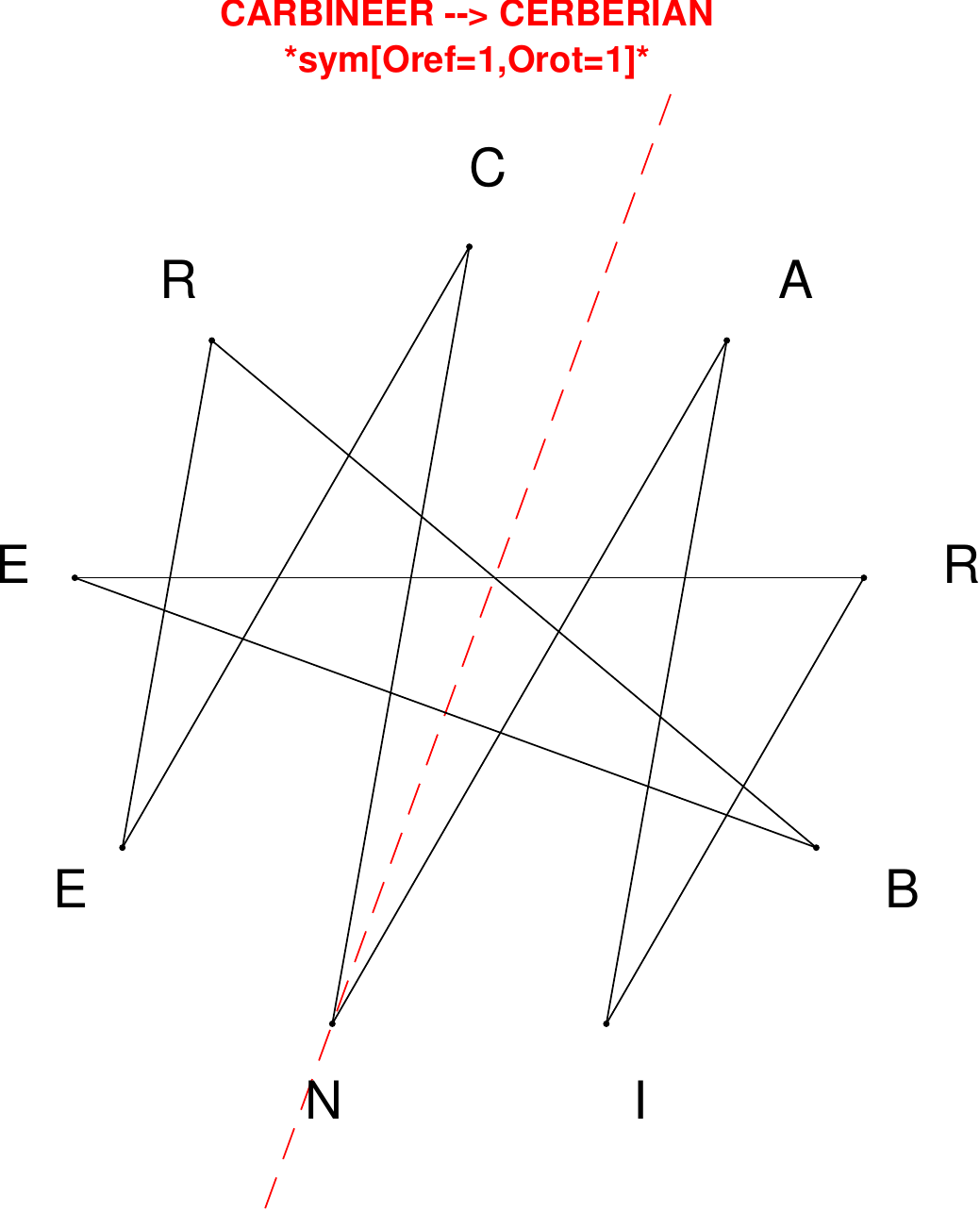}
\end{subfigure}
\hfill
\begin{subfigure}[T]{0.19\textwidth}
\centering
\includegraphics[width=\textwidth]{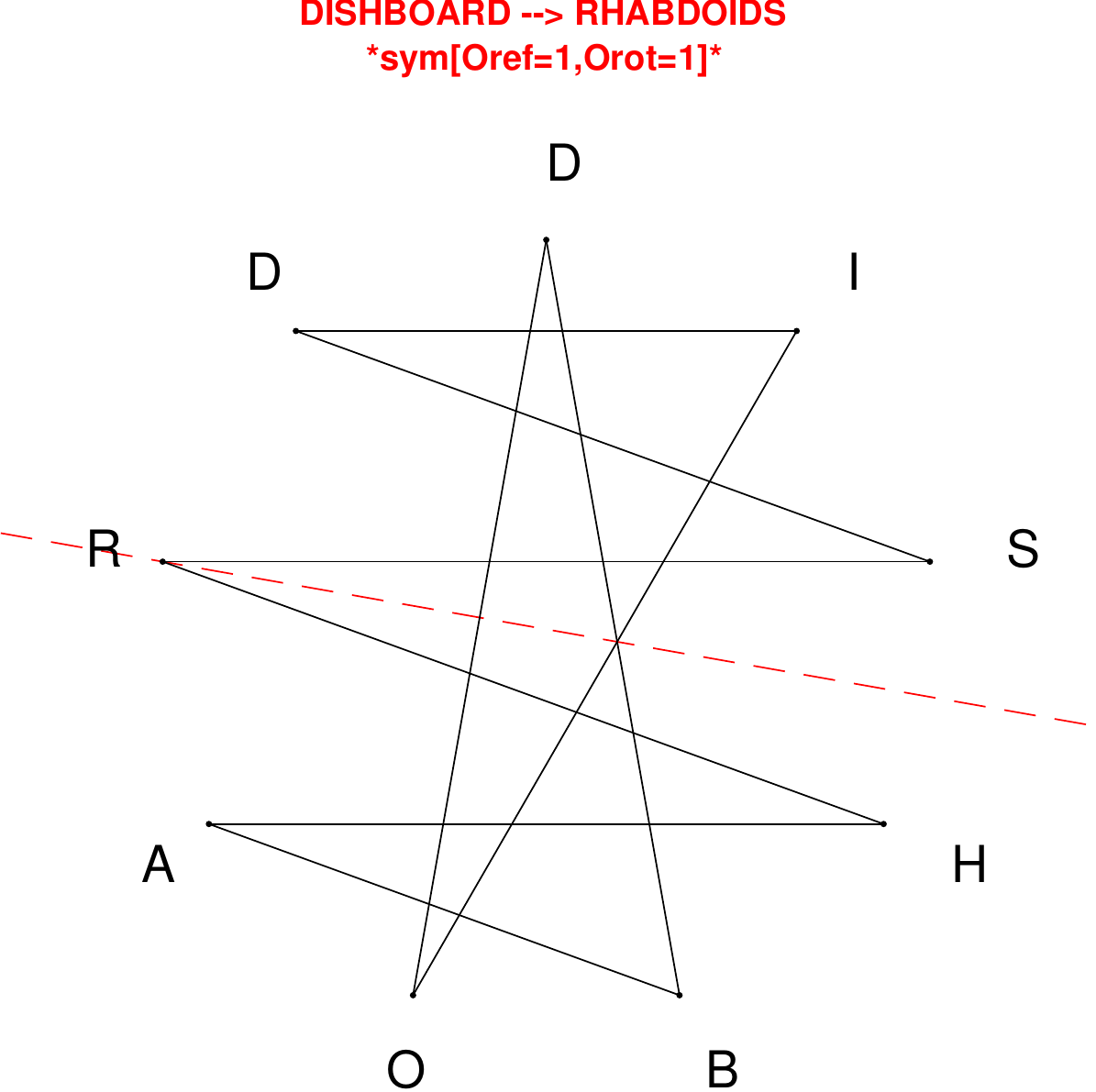}
\end{subfigure}
\hfill
\begin{subfigure}[T]{0.19\textwidth}
\centering
\includegraphics[width=\textwidth]{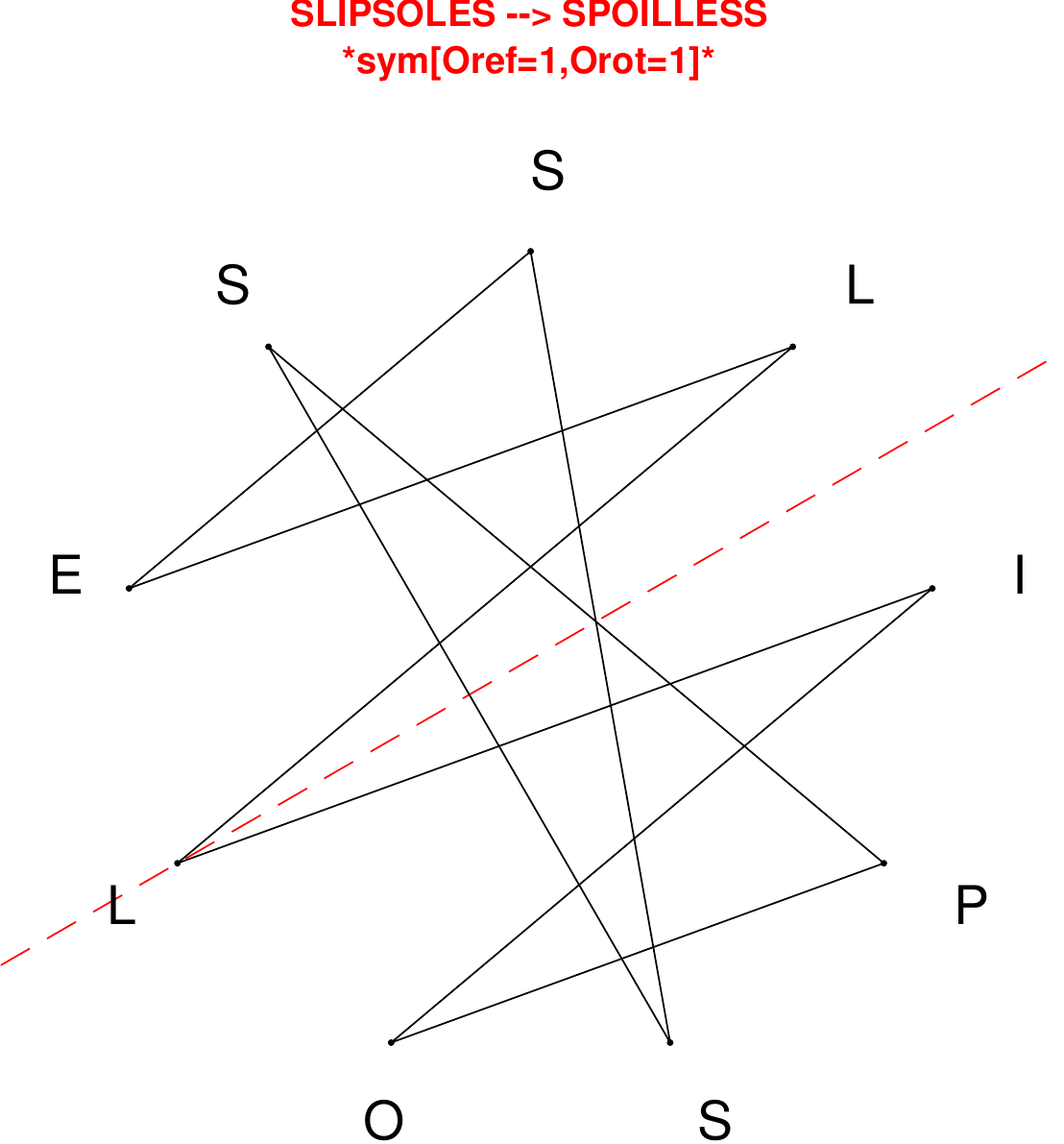}
\end{subfigure}
\hfill
\begin{subfigure}[T]{0.19\textwidth}
\centering
\includegraphics[width=\textwidth]{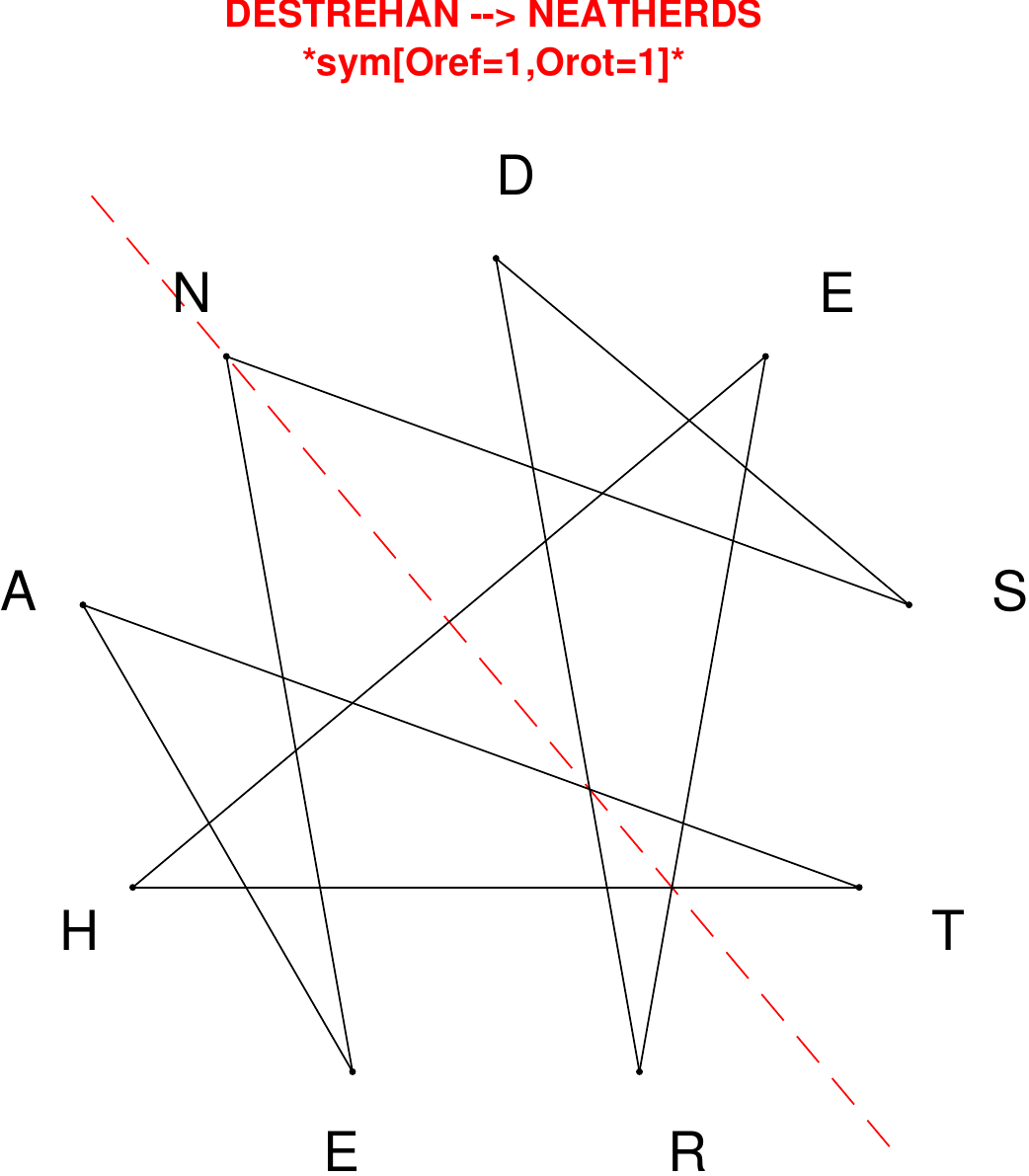}
\end{subfigure}
\hfill
\begin{subfigure}[T]{0.19\textwidth}
\centering
\includegraphics[width=\textwidth]{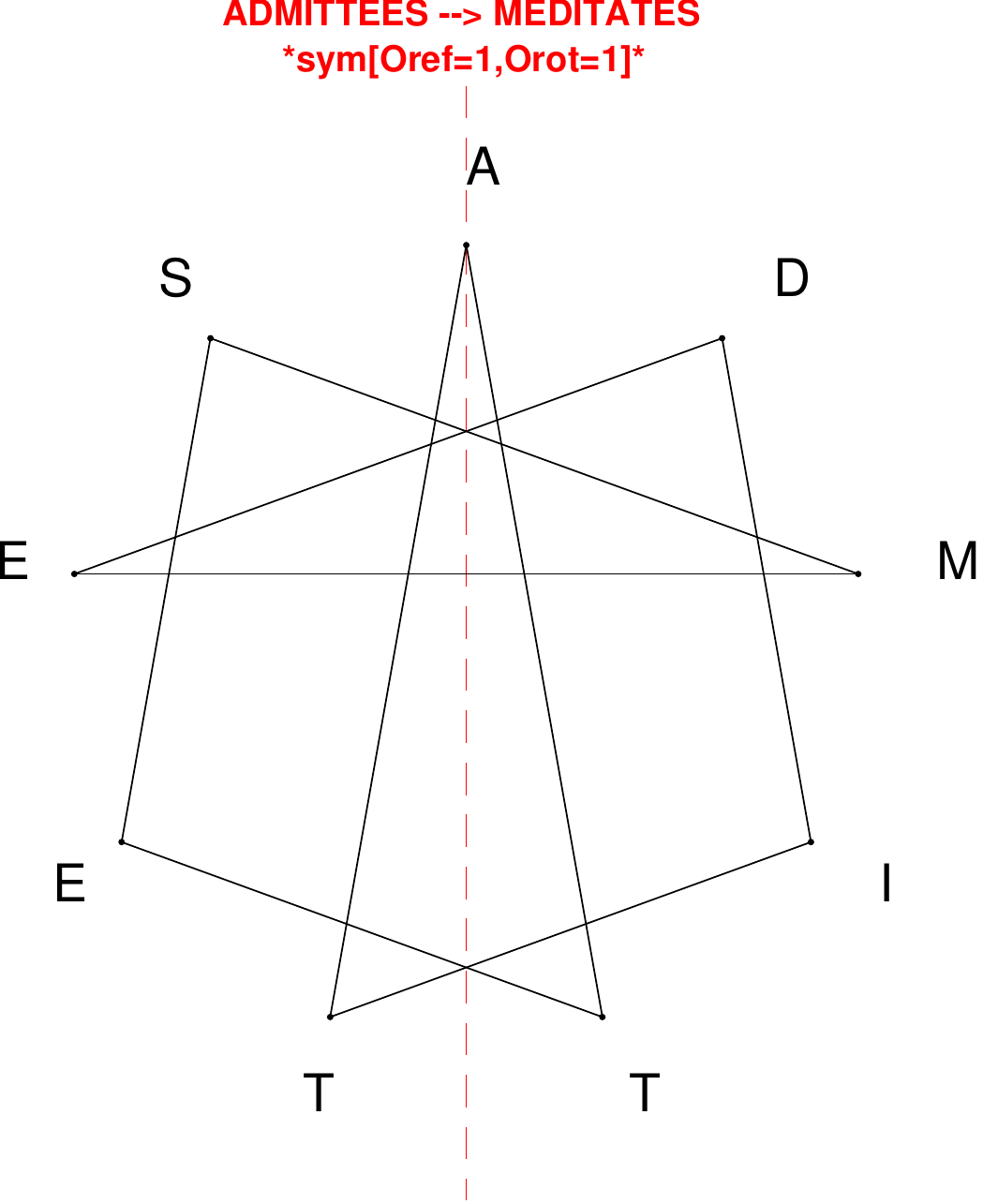}
\end{subfigure}
\end{figure}

\begin{figure}[H]
\centering
\begin{subfigure}[T]{0.19\textwidth}
\centering
\includegraphics[width=\textwidth]{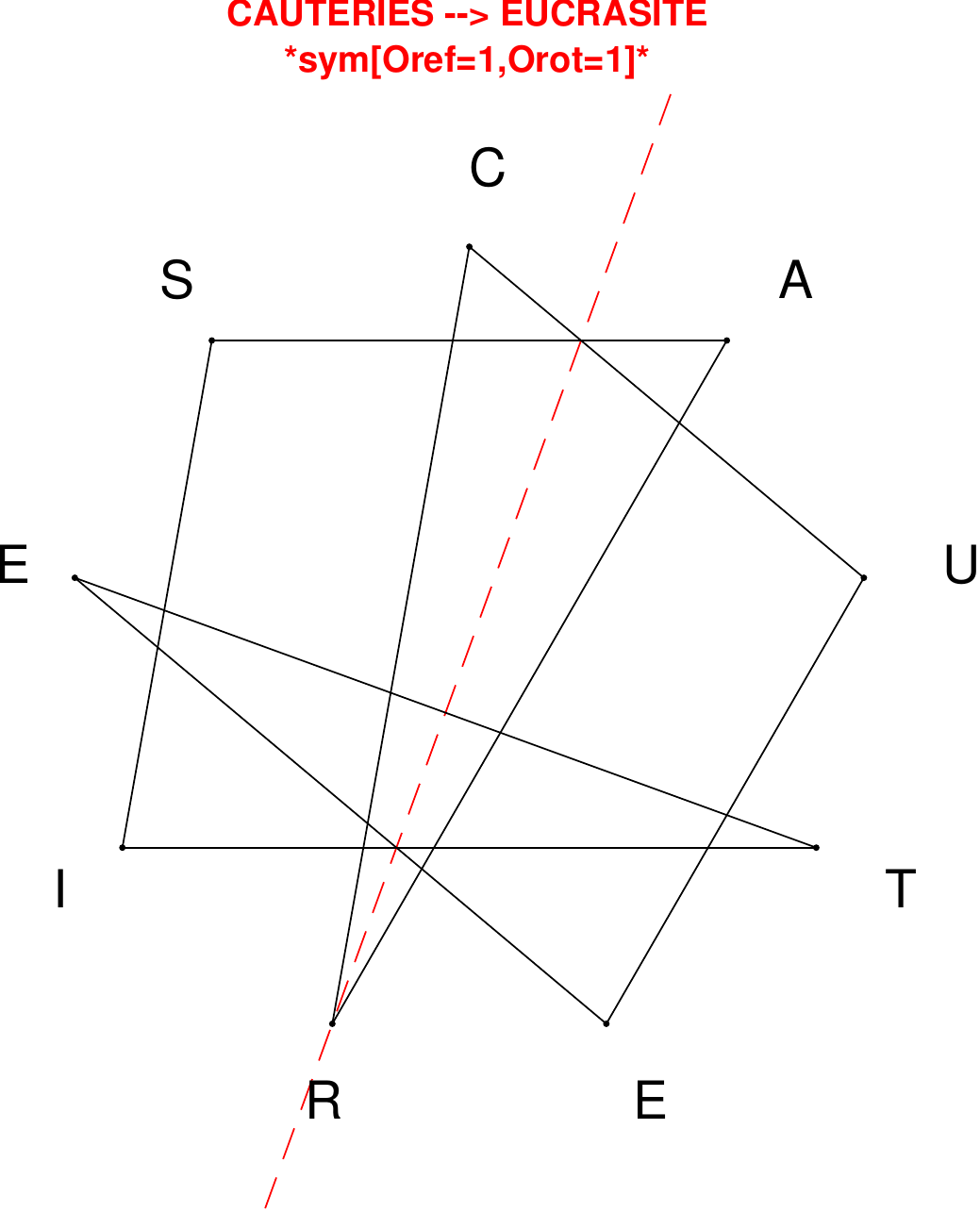}
\end{subfigure}
\hfill
\begin{subfigure}[T]{0.19\textwidth}
\centering
\includegraphics[width=\textwidth]{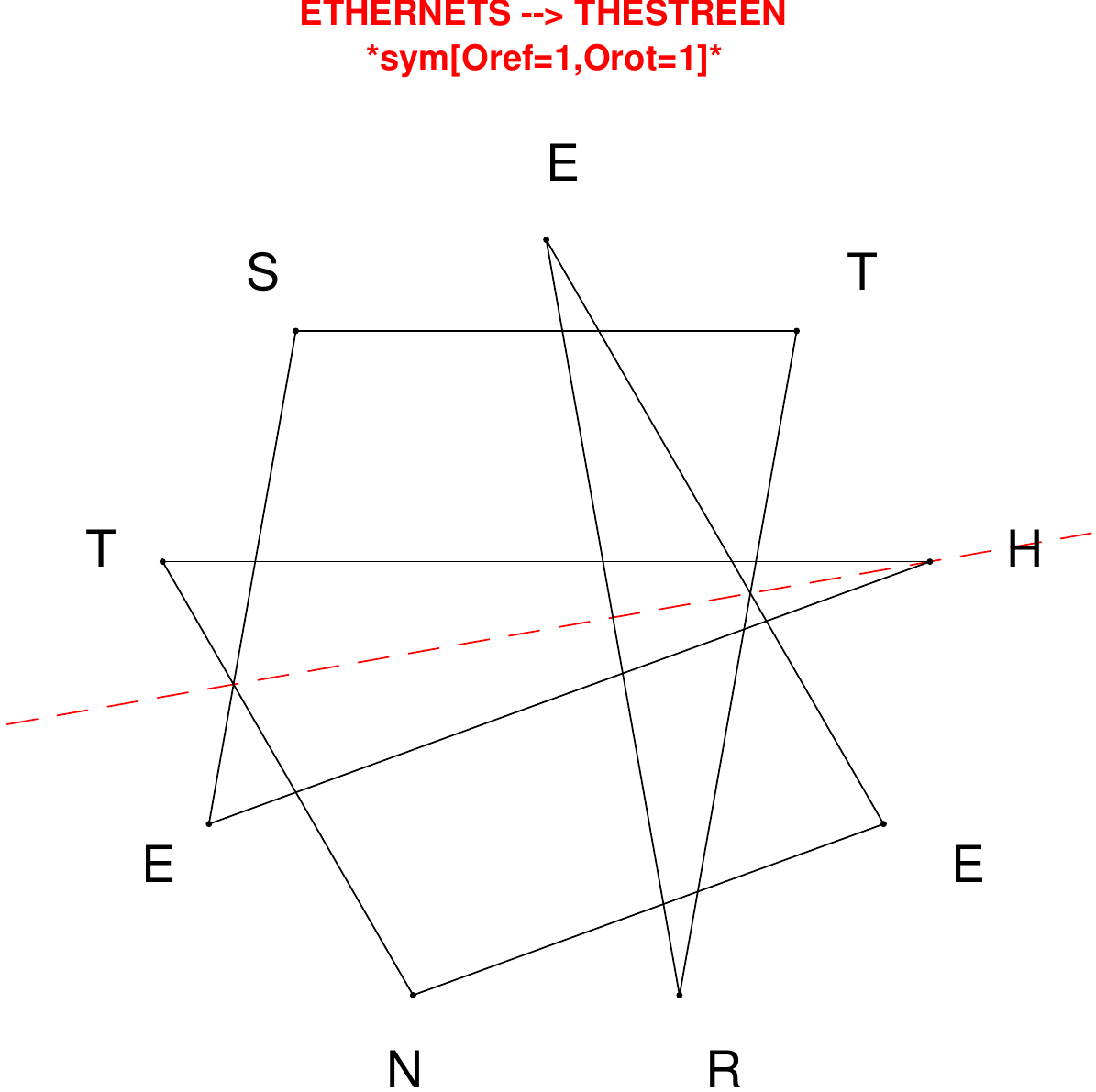}
\end{subfigure}
\hfill
\begin{subfigure}[T]{0.19\textwidth}
\centering
\includegraphics[width=\textwidth]{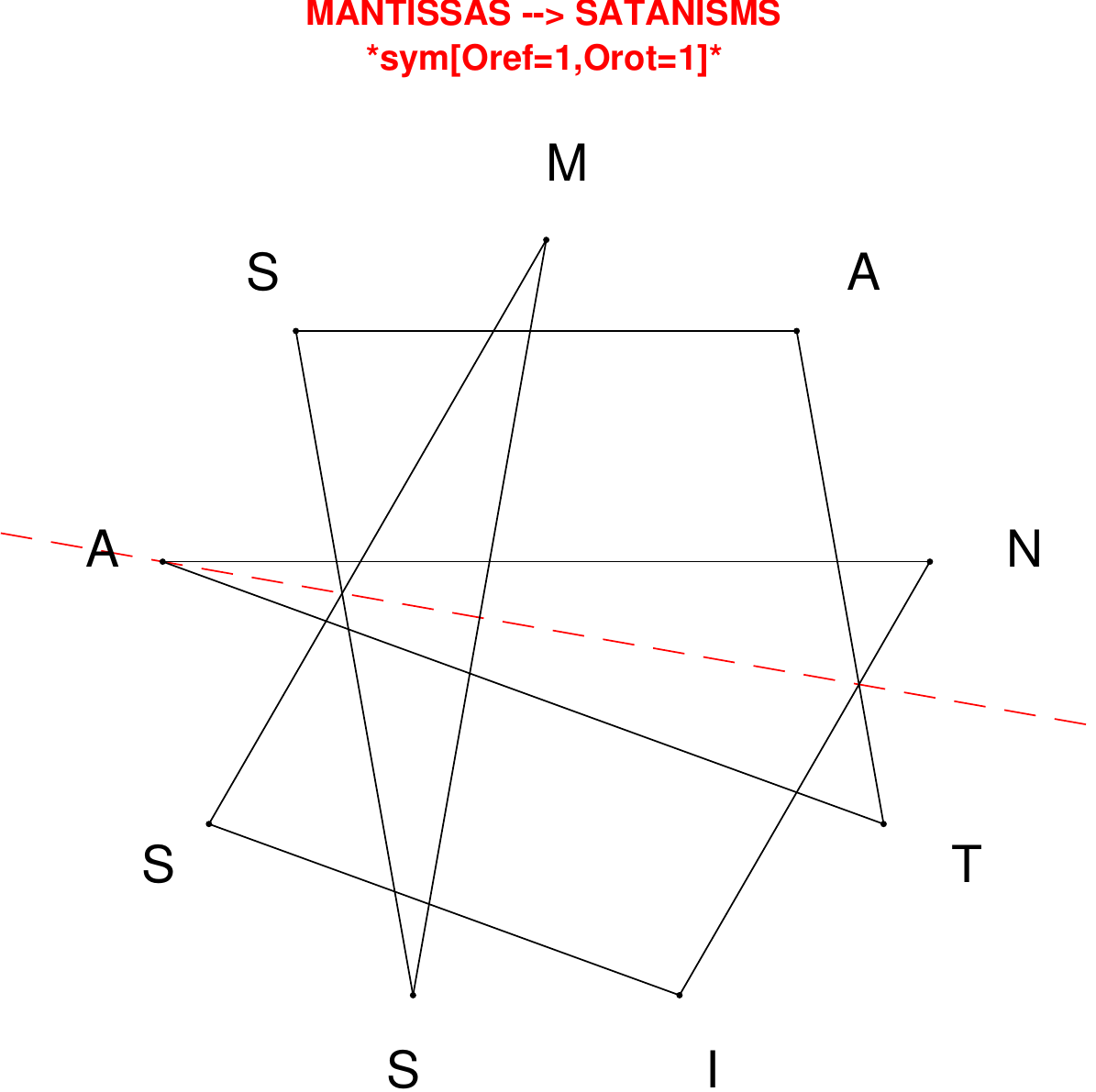}
\end{subfigure}
\hfill
\begin{subfigure}[T]{0.19\textwidth}
\centering
\includegraphics[width=\textwidth]{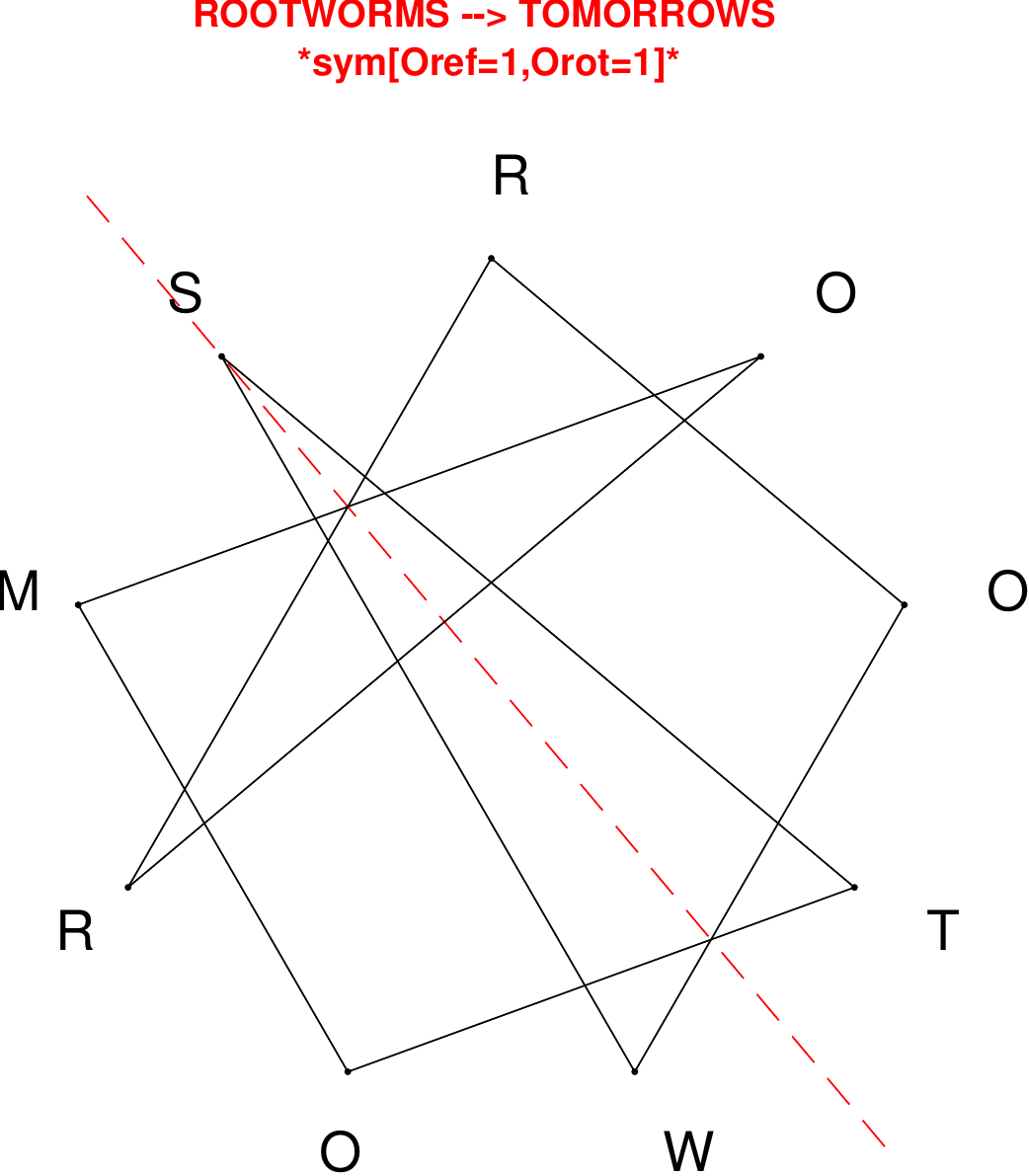}
\end{subfigure}
\hfill
\begin{subfigure}[T]{0.19\textwidth}
\centering
\includegraphics[width=\textwidth]{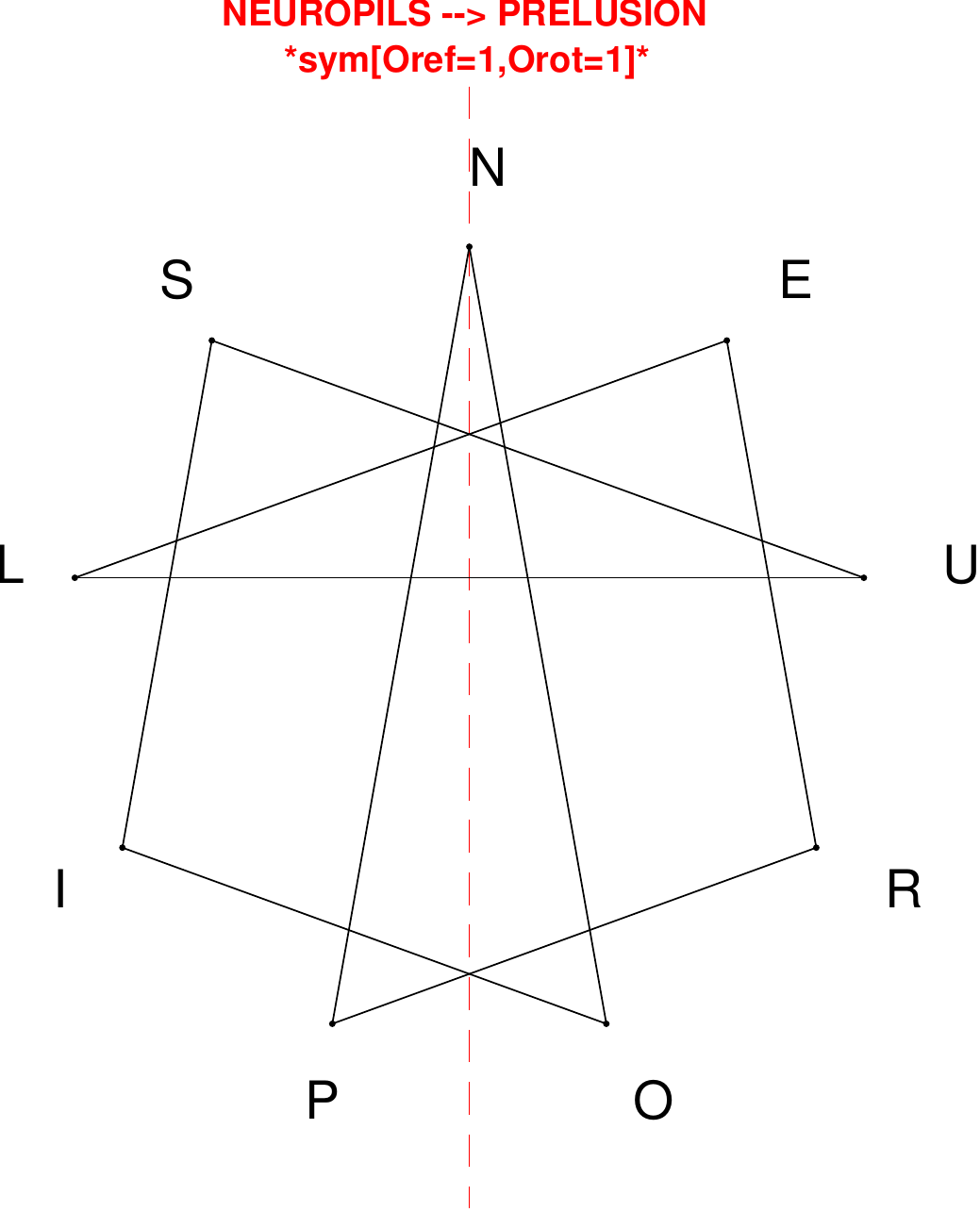}
\end{subfigure}
\end{figure}

\begin{figure}[H]
\centering
\begin{subfigure}[T]{0.19\textwidth}
\centering
\includegraphics[width=\textwidth]{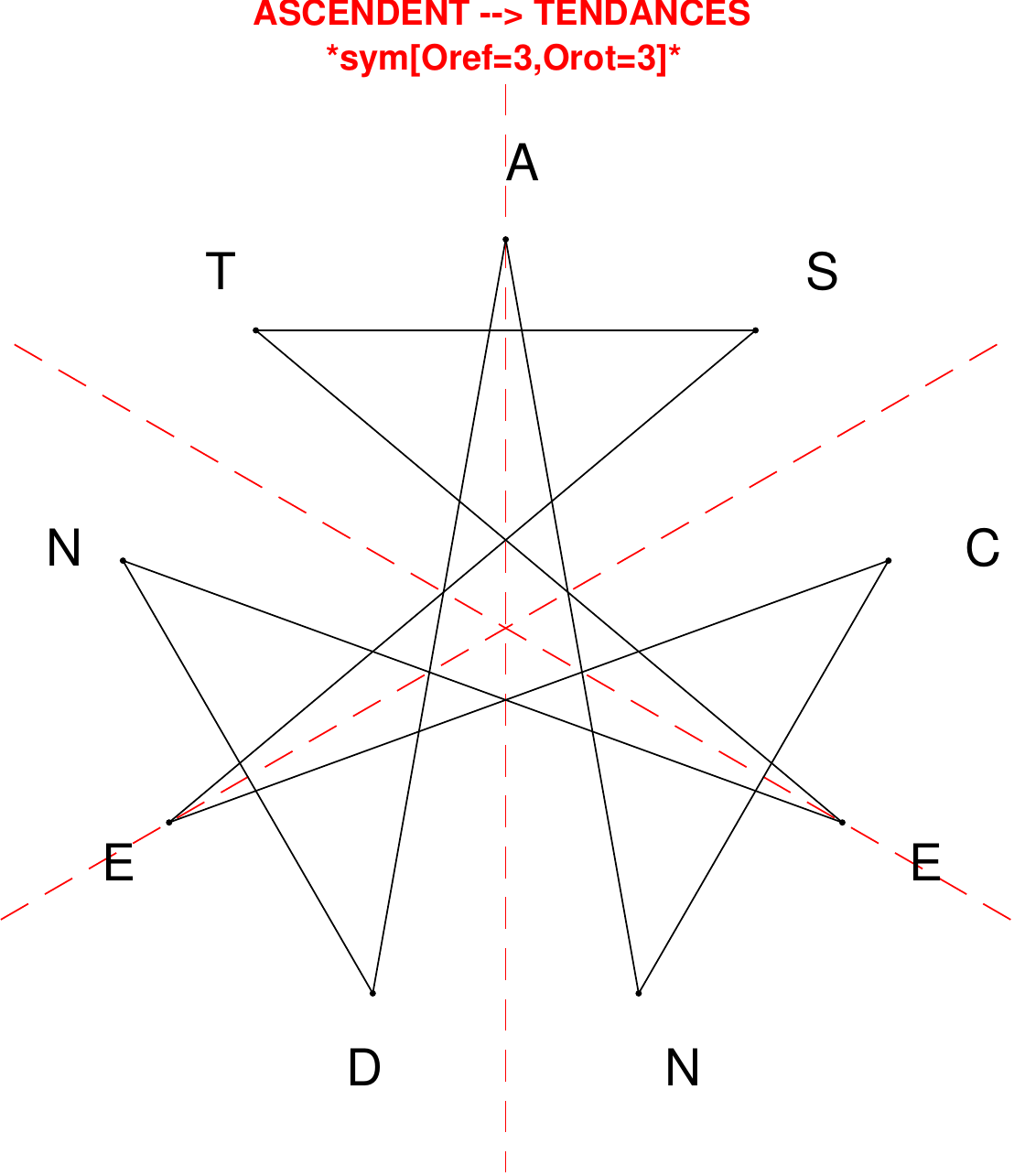}
\end{subfigure}
\hfill
\begin{subfigure}[T]{0.19\textwidth}
\centering
\includegraphics[width=\textwidth]{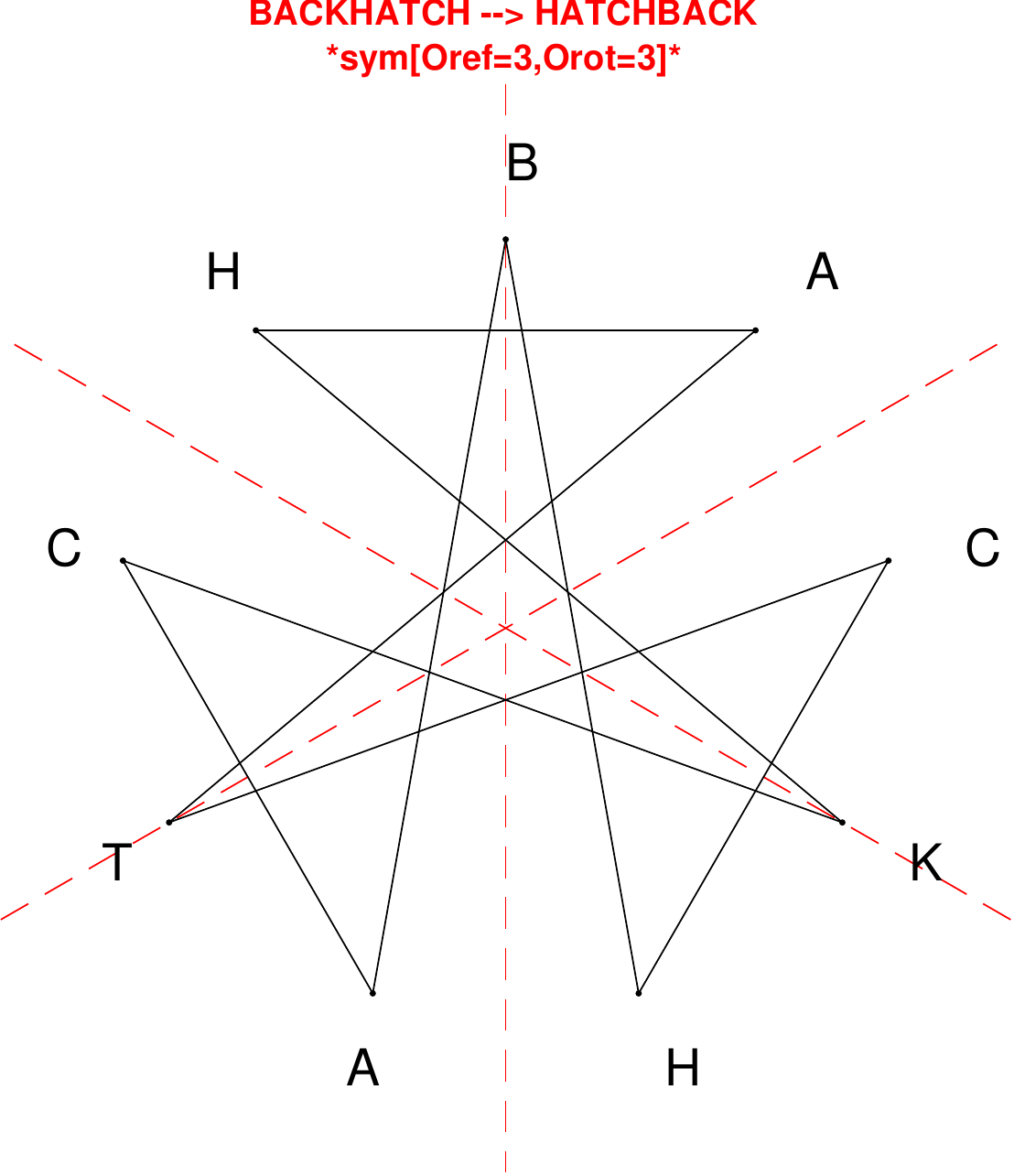}
\end{subfigure}
\hfill
\begin{subfigure}[T]{0.19\textwidth}
\centering
\includegraphics[width=\textwidth]{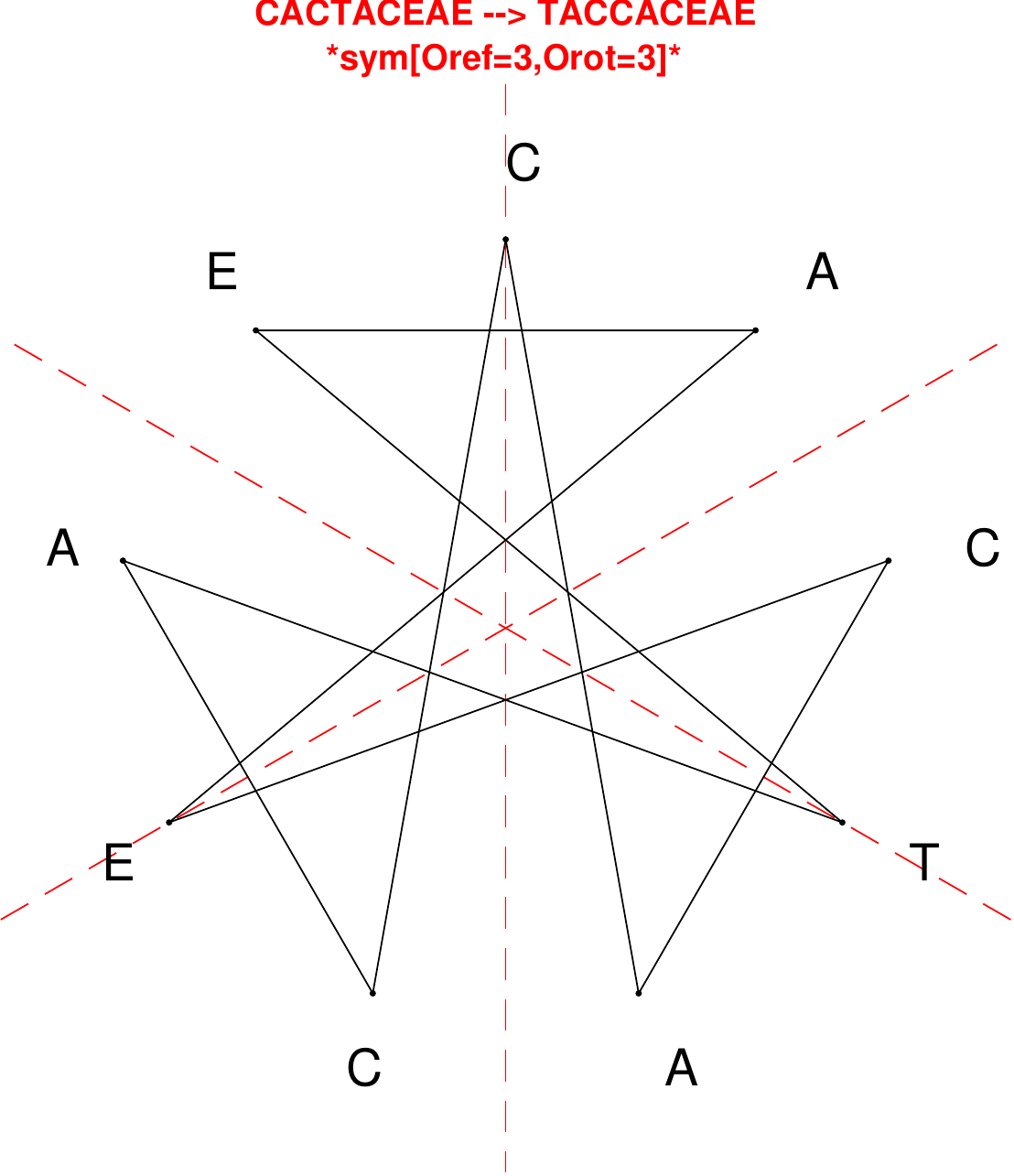}
\end{subfigure}
\hfill
\begin{subfigure}[T]{0.19\textwidth}
\centering
\includegraphics[width=\textwidth]{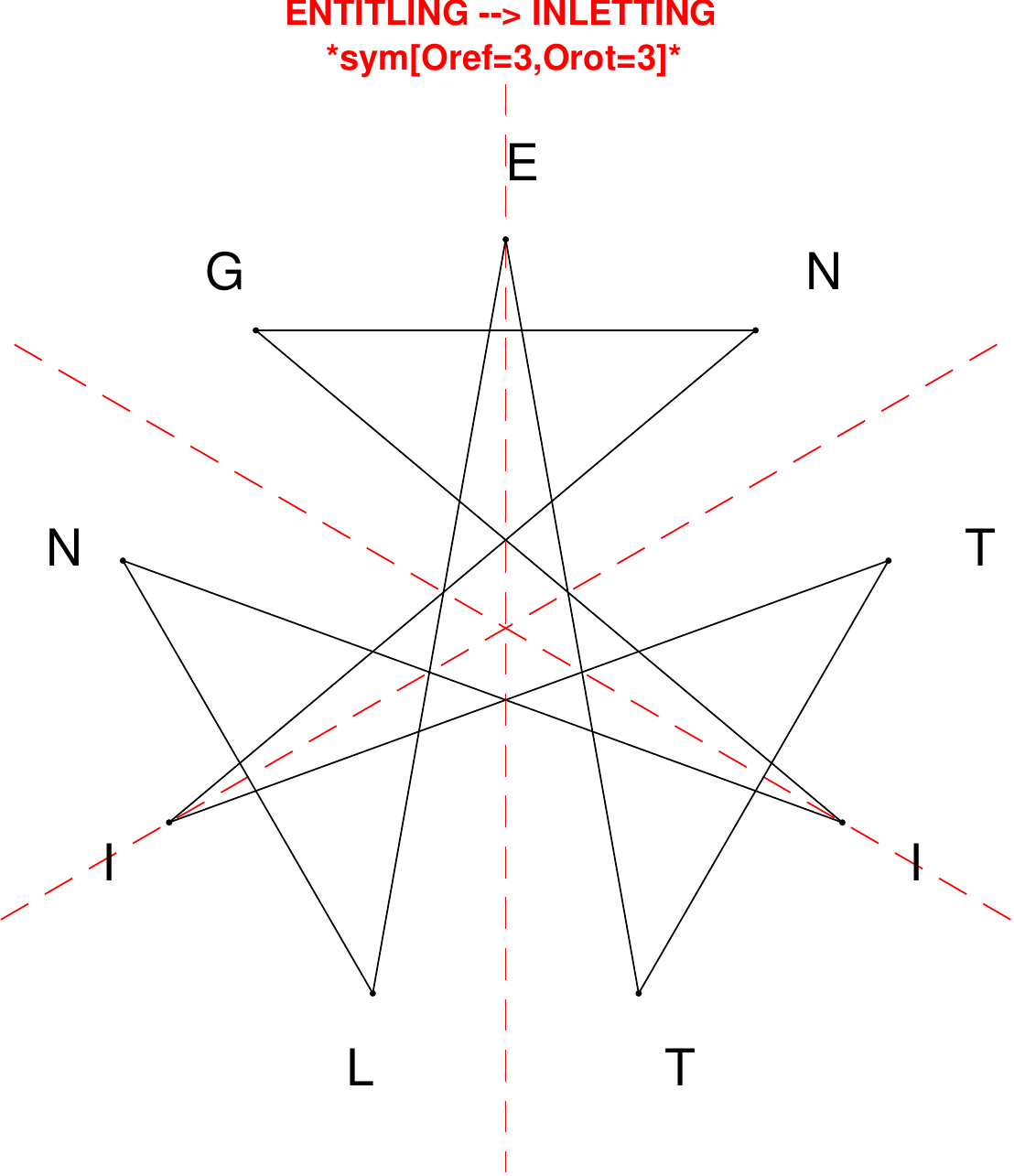}
\end{subfigure}
\hfill
\begin{subfigure}[T]{0.19\textwidth}
\centering
\includegraphics[width=\textwidth]{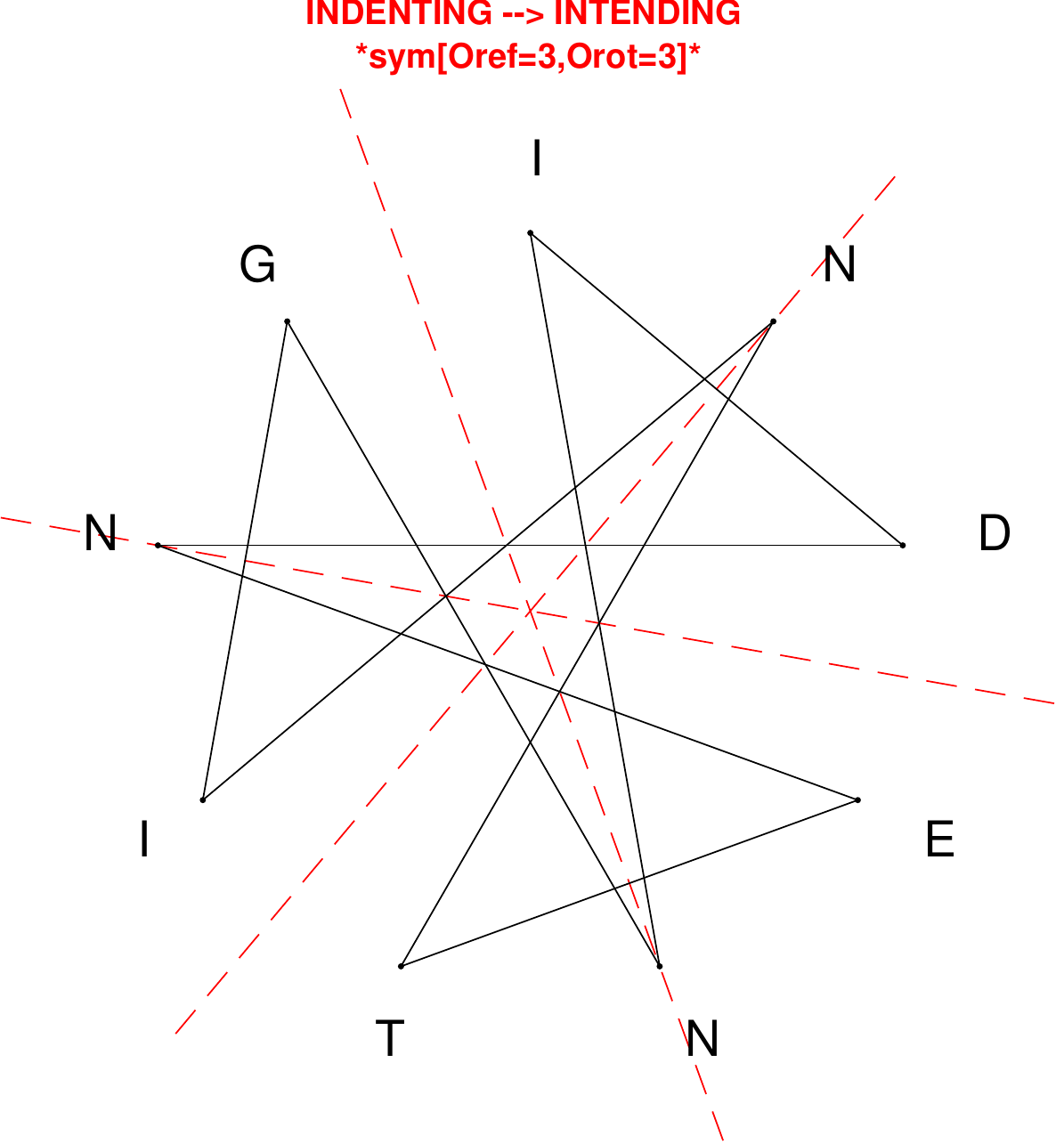}
\end{subfigure}
\end{figure}

\begin{figure}[H]
\centering
\begin{subfigure}[T]{0.19\textwidth}
\centering
\includegraphics[width=\textwidth]{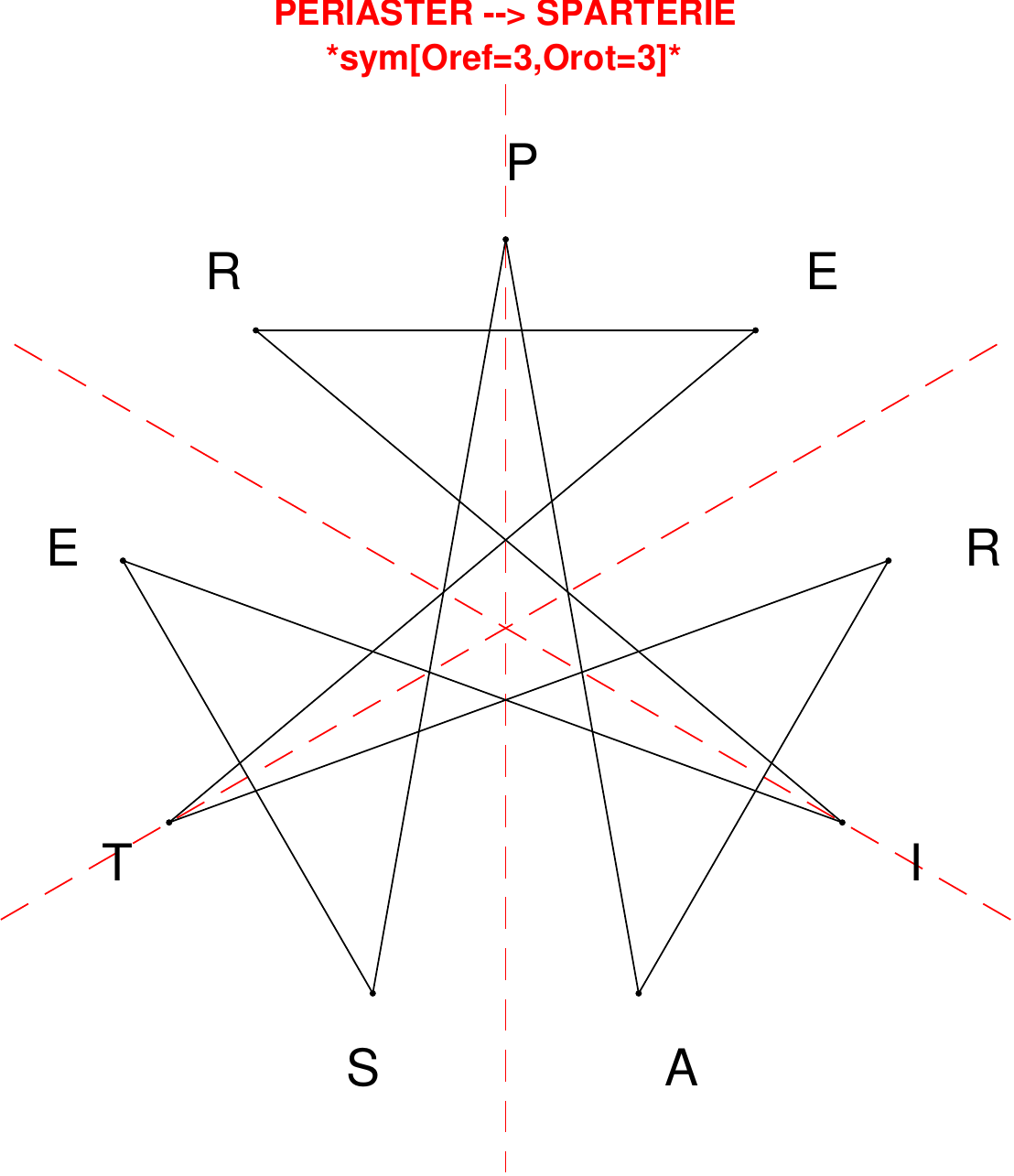}
\end{subfigure}
\hfill
\begin{subfigure}[T]{0.19\textwidth}
\centering
\includegraphics[width=\textwidth]{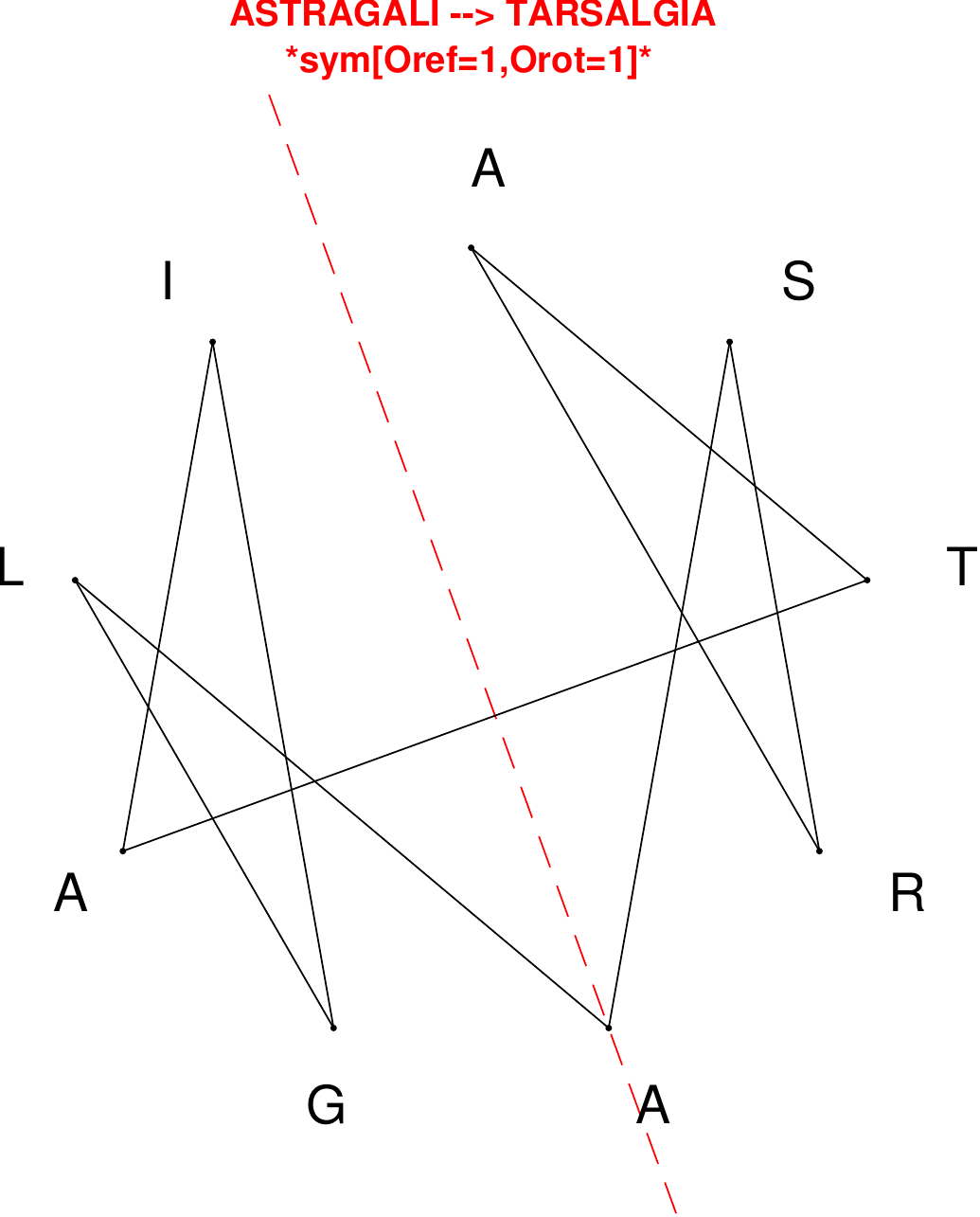}
\end{subfigure}
\hfill
\begin{subfigure}[T]{0.19\textwidth}
\centering
\includegraphics[width=\textwidth]{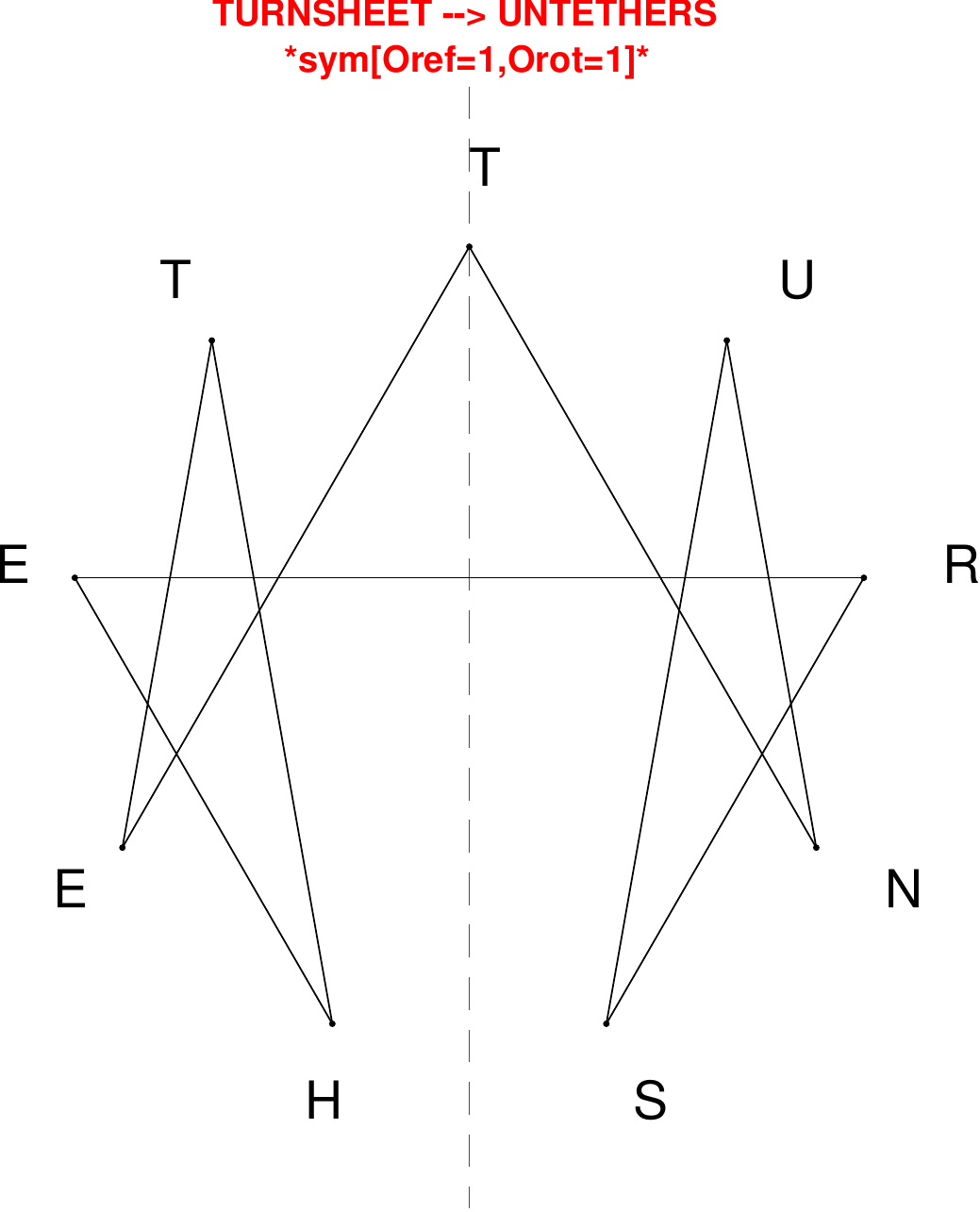}
\end{subfigure}
\hfill
\begin{subfigure}[T]{0.19\textwidth}
\centering
\includegraphics[width=\textwidth]{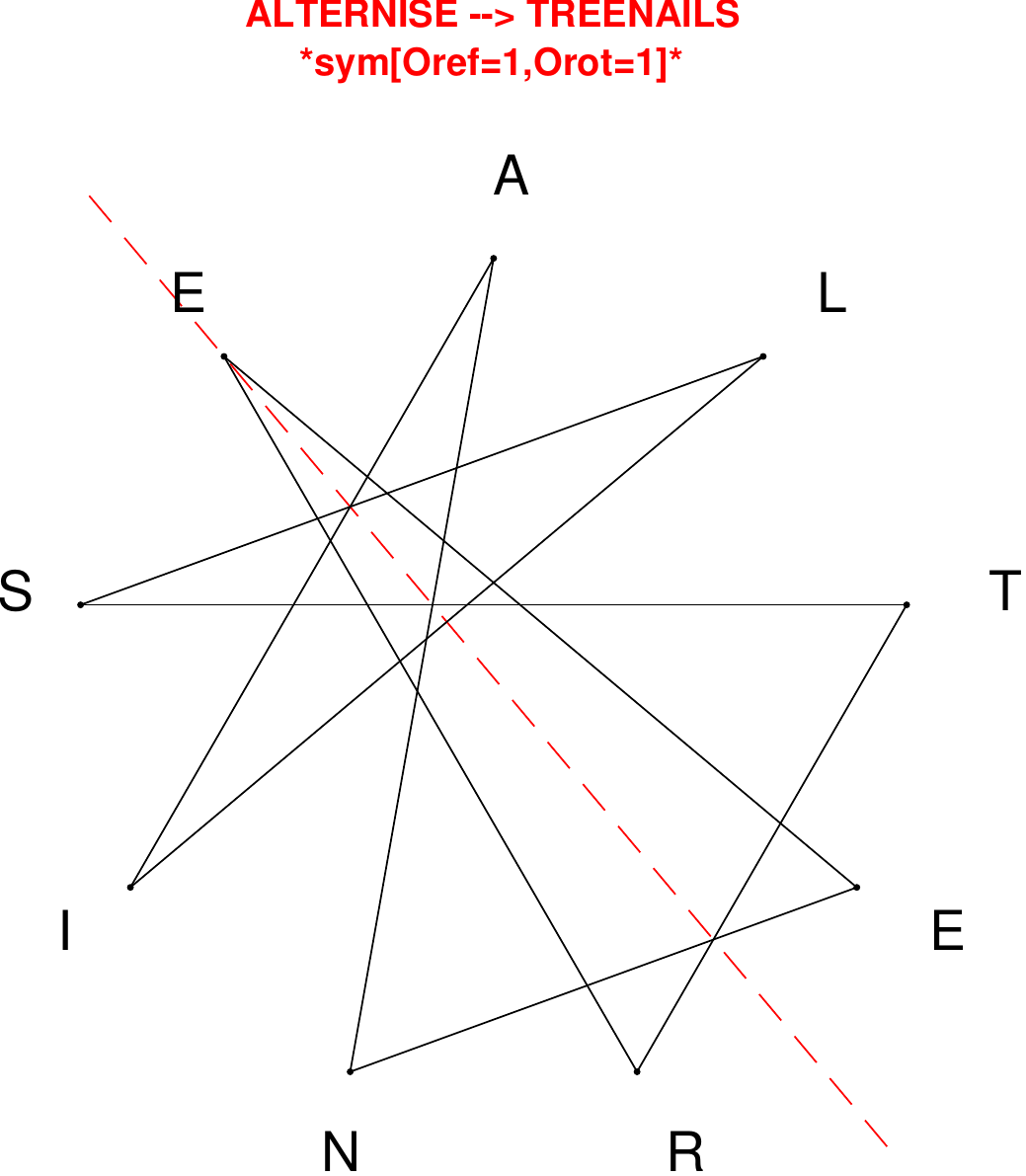}
\end{subfigure}
\hfill
\begin{subfigure}[T]{0.19\textwidth}
\centering
\includegraphics[width=\textwidth]{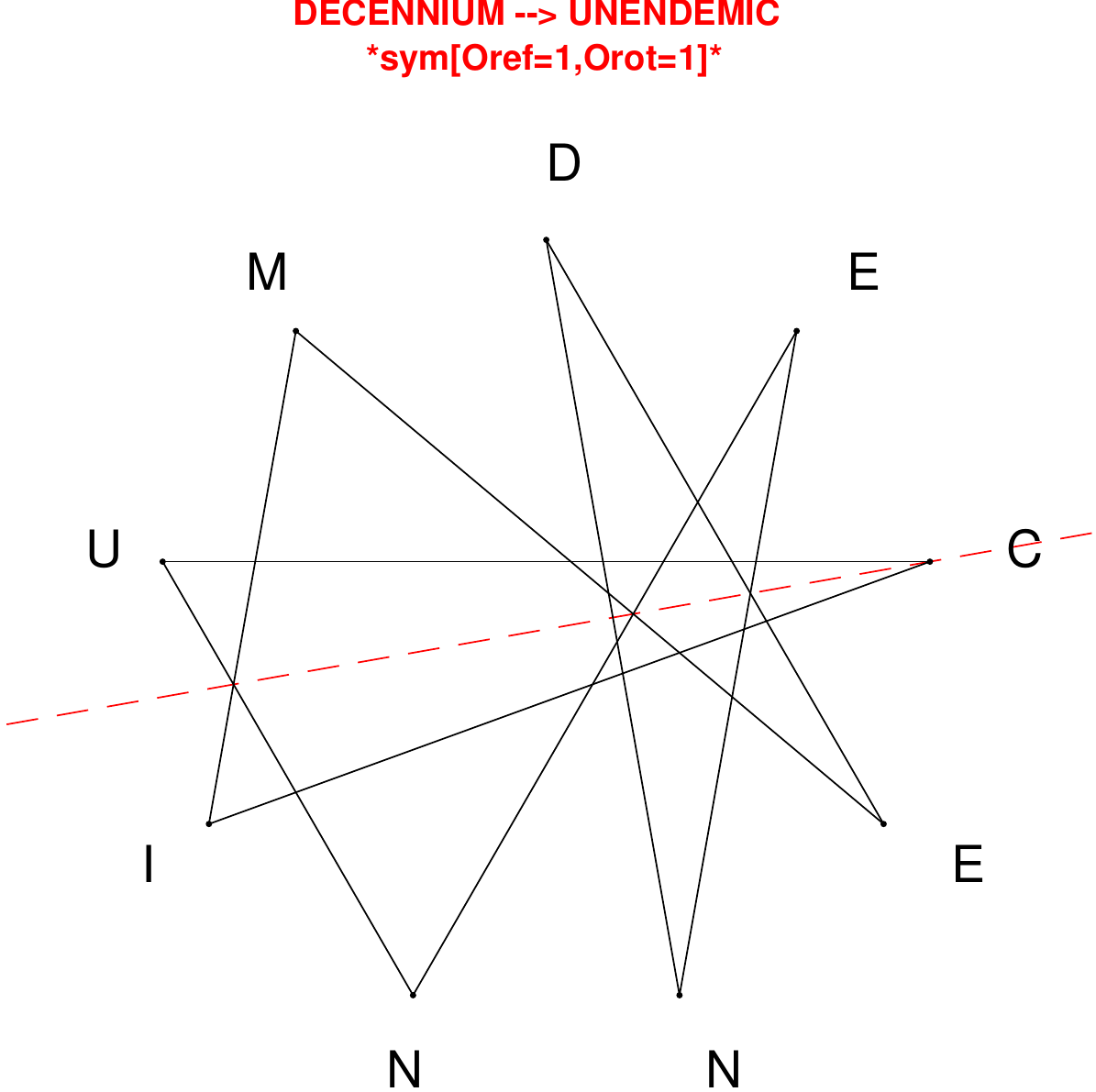}
\end{subfigure}
\end{figure}

\begin{figure}[H]
\centering
\begin{subfigure}[T]{0.19\textwidth}
\centering
\includegraphics[width=\textwidth]{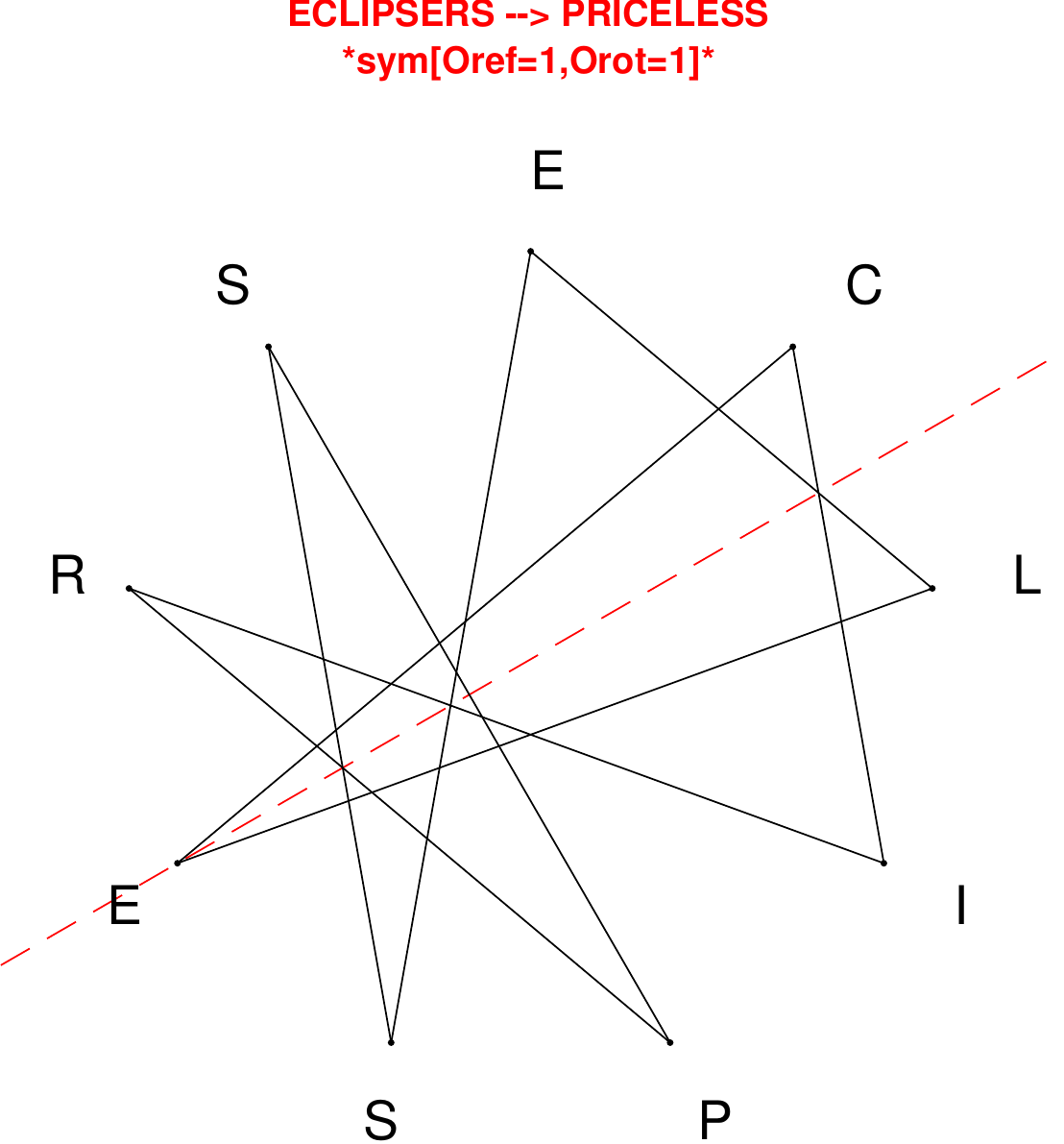}
\end{subfigure}
\hfill
\begin{subfigure}[T]{0.19\textwidth}
\centering
\includegraphics[width=\textwidth]{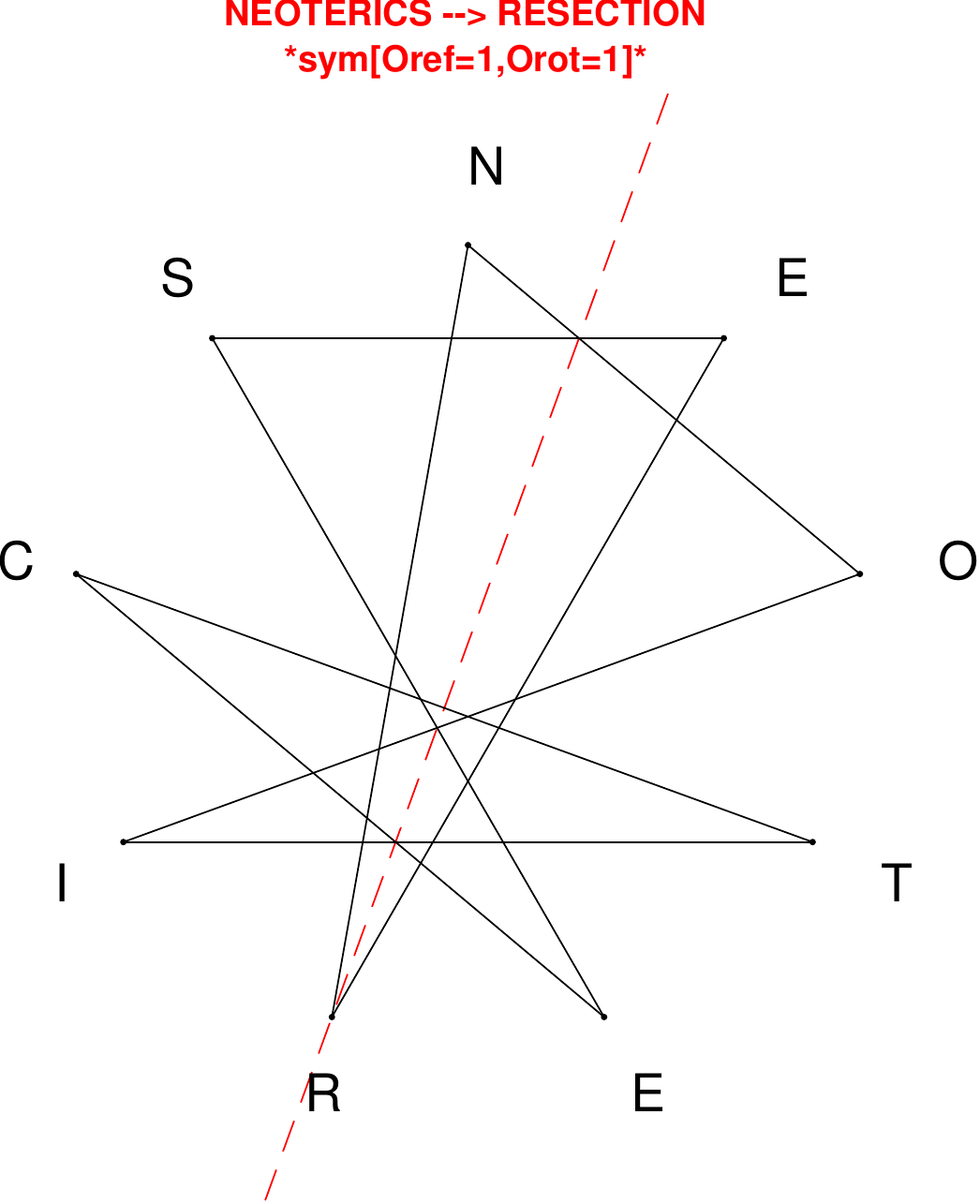}
\end{subfigure}
\hfill
\begin{subfigure}[T]{0.19\textwidth}
\centering
\includegraphics[width=\textwidth]{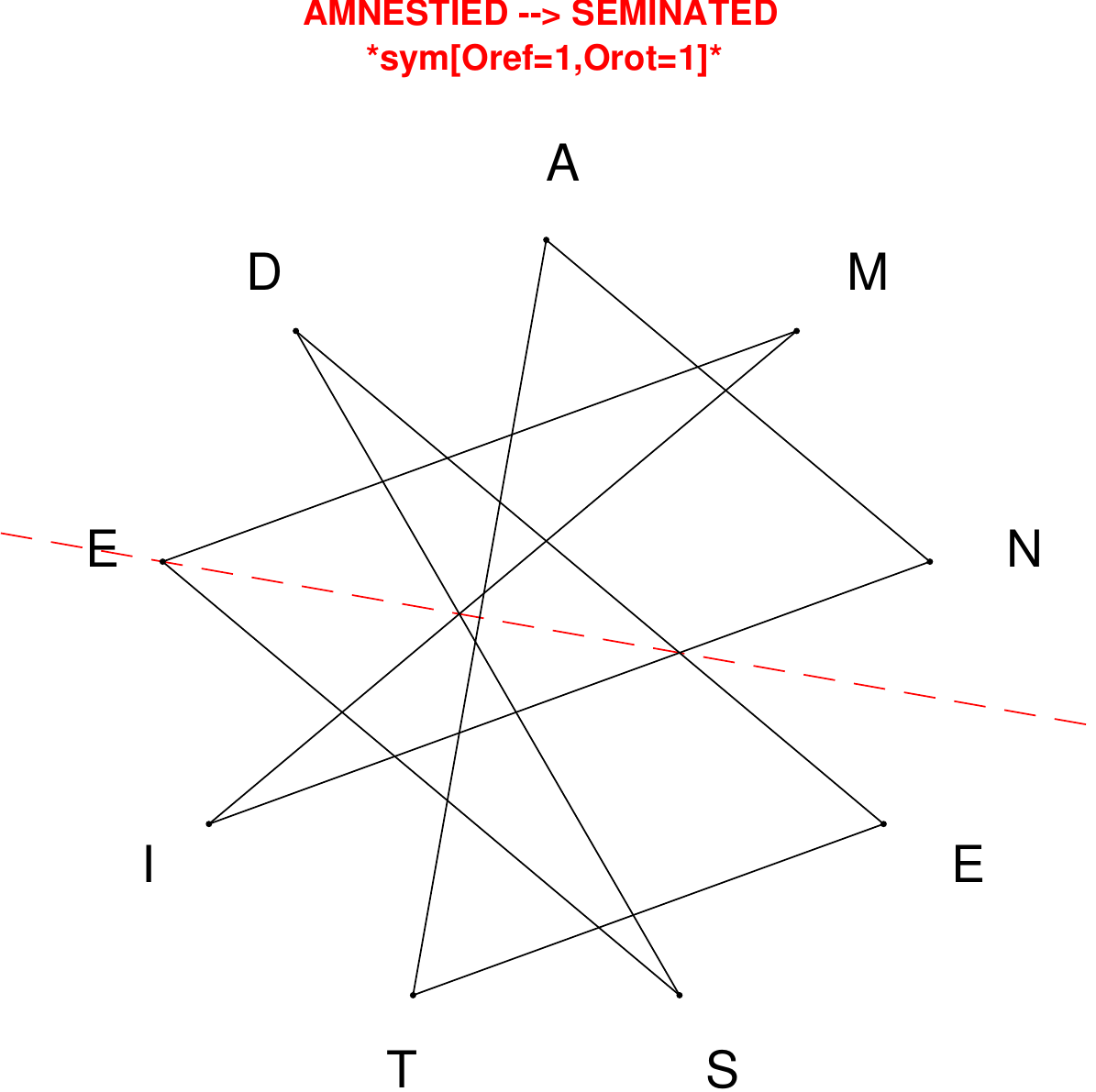}
\end{subfigure}
\hfill
\begin{subfigure}[T]{0.19\textwidth}
\centering
\includegraphics[width=\textwidth]{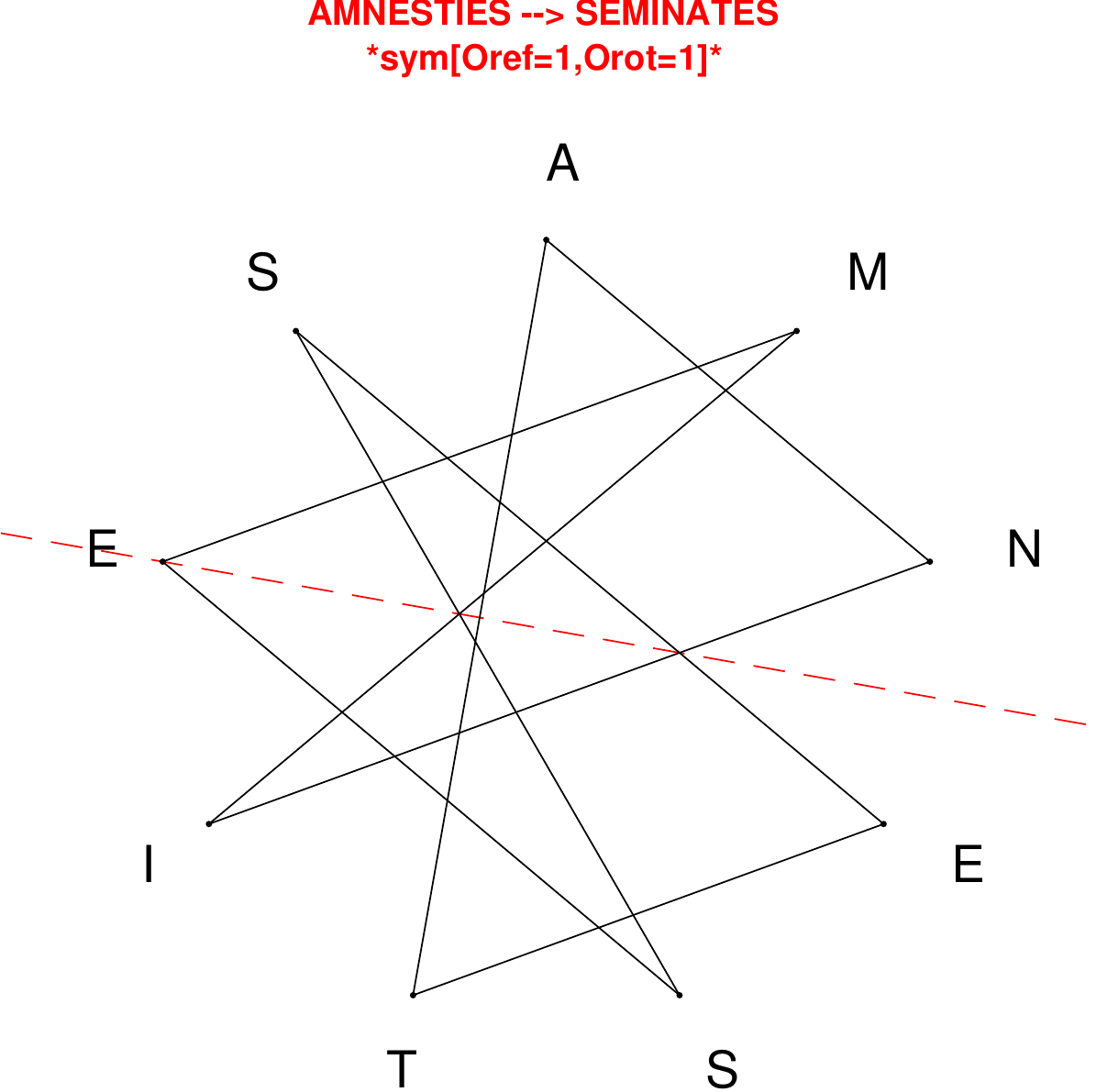}
\end{subfigure}
\hfill
\begin{subfigure}[T]{0.19\textwidth}
\centering
\includegraphics[width=\textwidth]{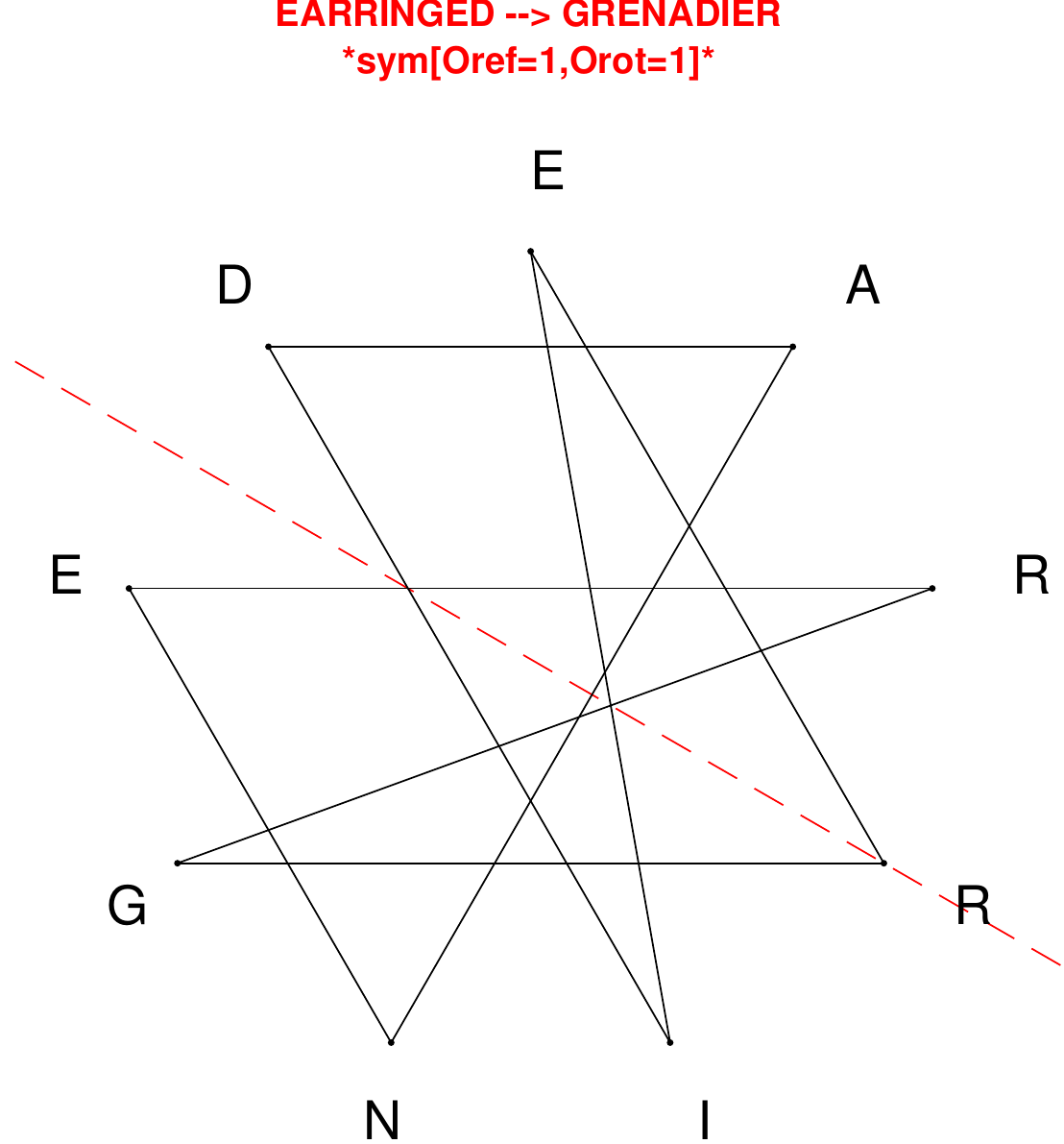}
\end{subfigure}
\end{figure}

\begin{figure}[H]
\centering
\begin{subfigure}[T]{0.19\textwidth}
\centering
\includegraphics[width=\textwidth]{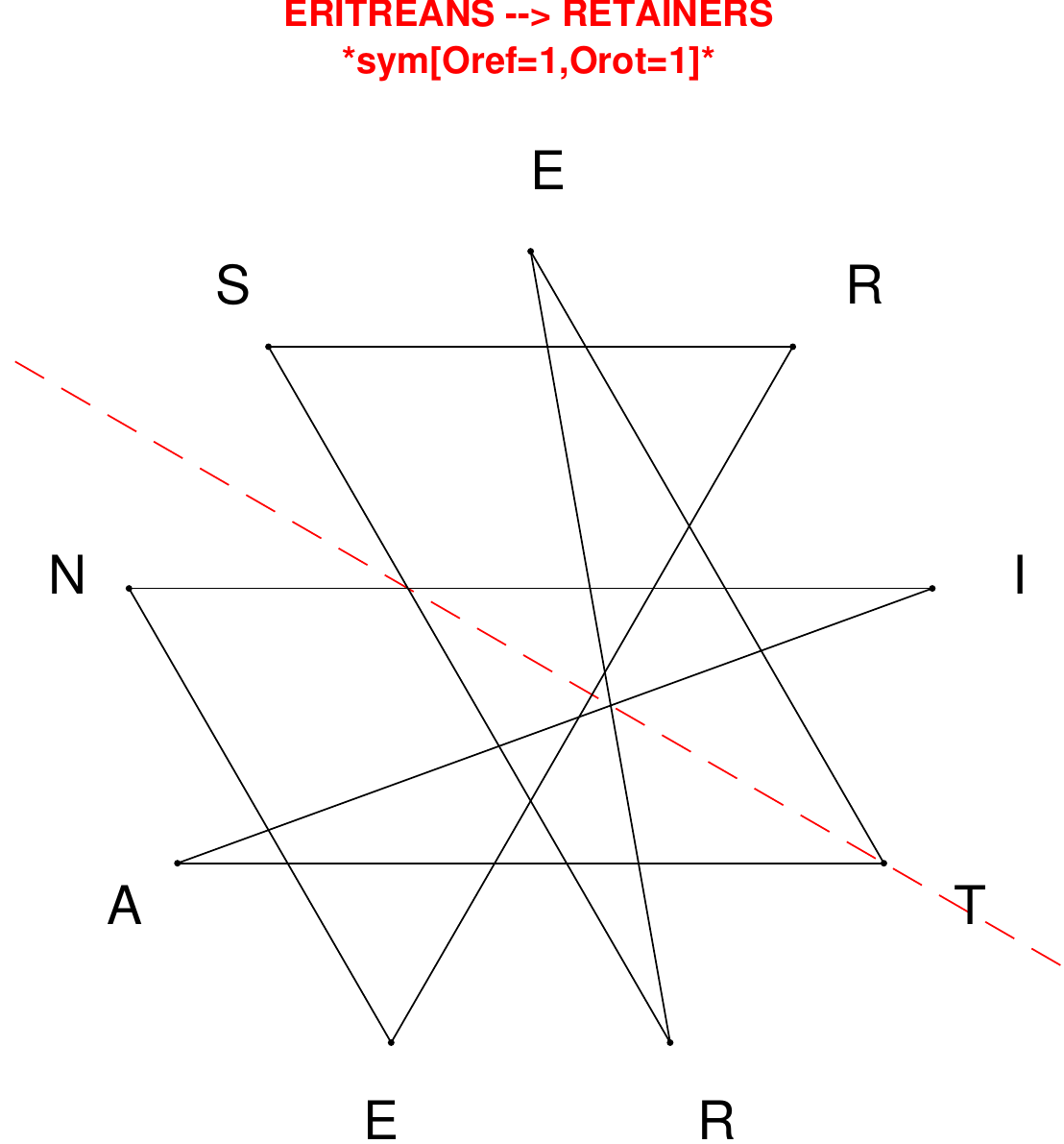}
\end{subfigure}
\hfill
\begin{subfigure}[T]{0.19\textwidth}
\centering
\includegraphics[width=\textwidth]{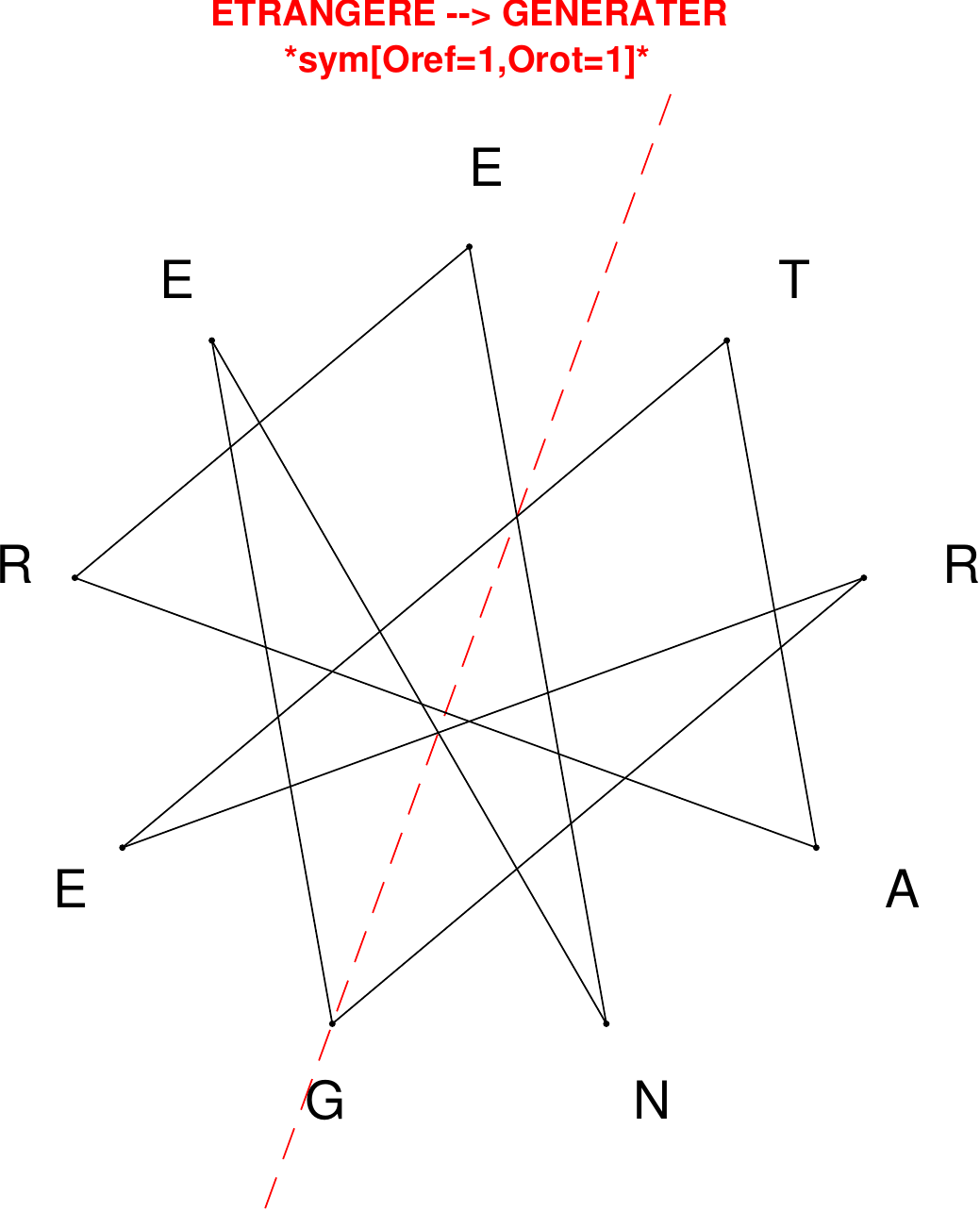}
\end{subfigure}
\hfill
\begin{subfigure}[T]{0.19\textwidth}
\centering
\includegraphics[width=\textwidth]{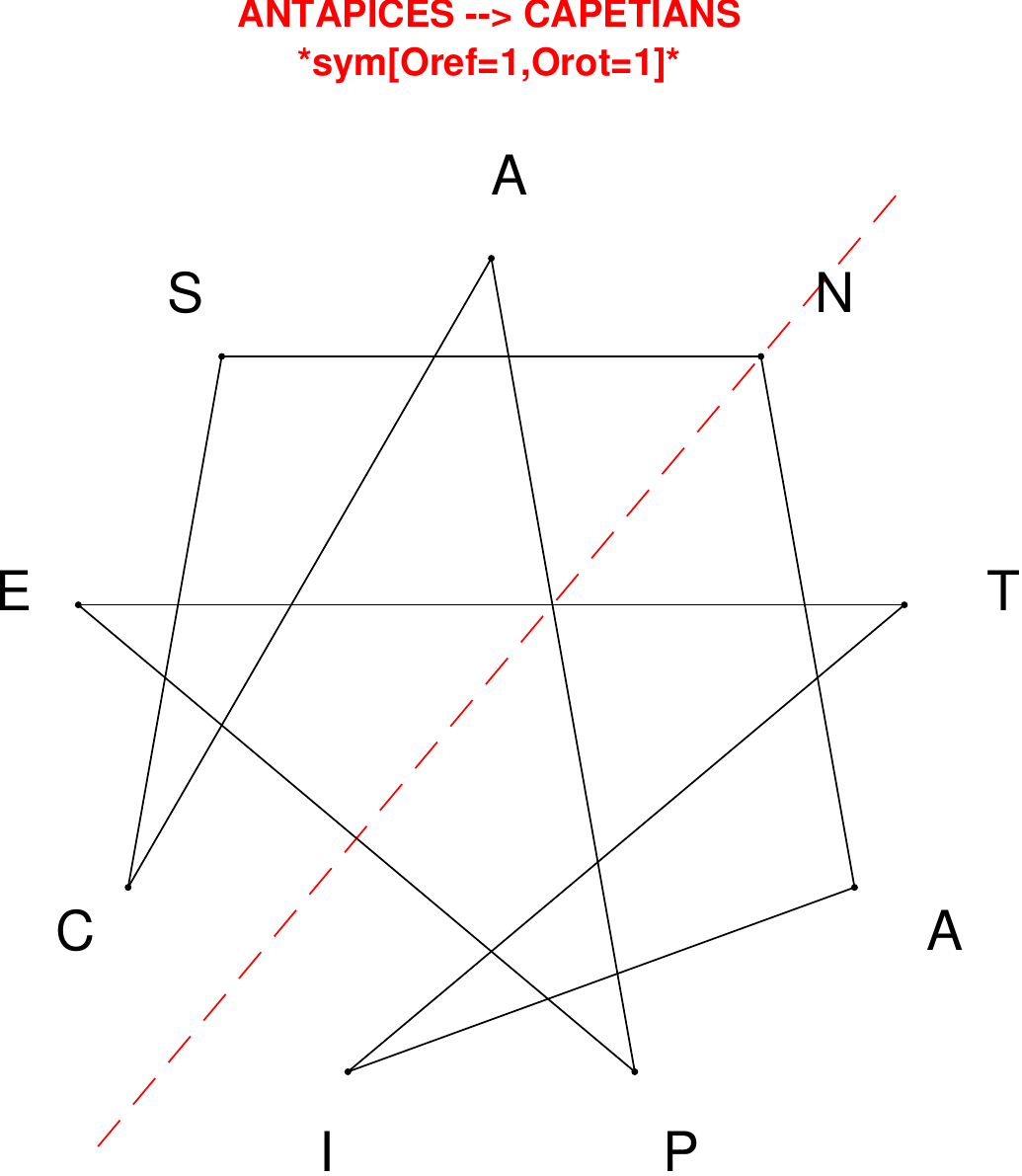}
\end{subfigure}
\hfill
\begin{subfigure}[T]{0.19\textwidth}
\centering
\includegraphics[width=\textwidth]{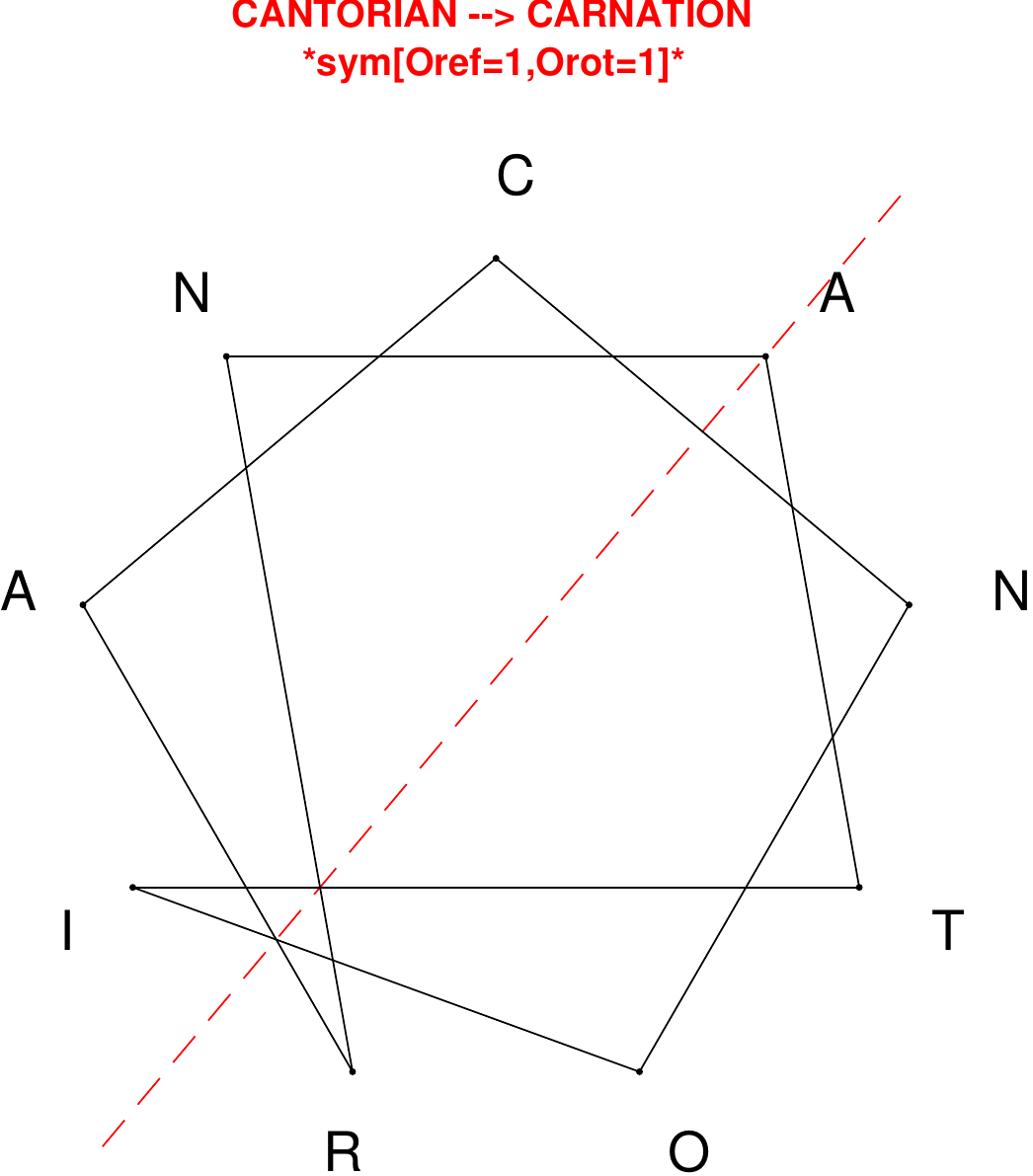}
\end{subfigure}
\hfill
\begin{subfigure}[T]{0.19\textwidth}
\centering
\includegraphics[width=\textwidth]{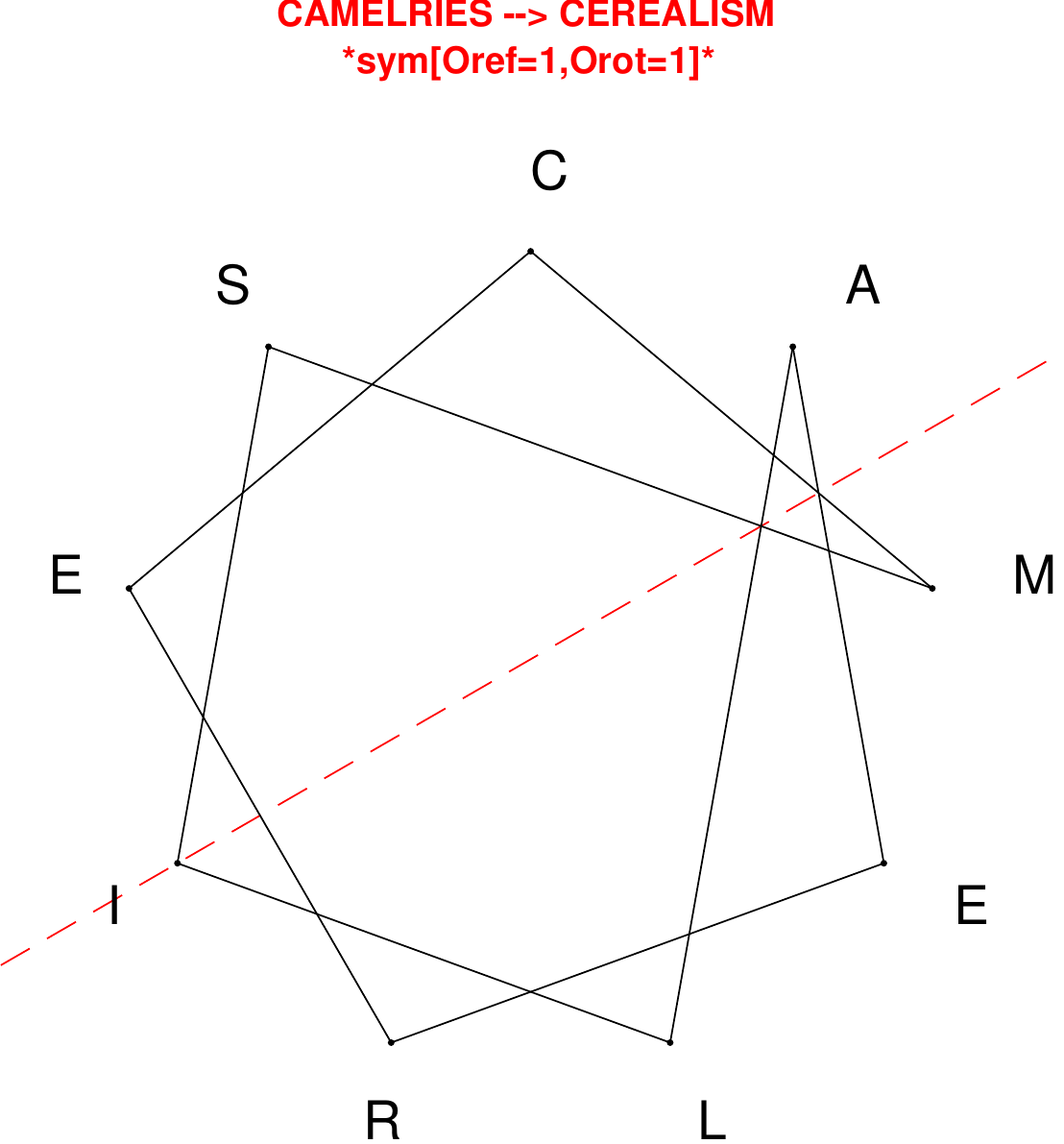}
\end{subfigure}
\end{figure}

\begin{figure}[H]
\centering
\begin{subfigure}[T]{0.19\textwidth}
\centering
\includegraphics[width=\textwidth]{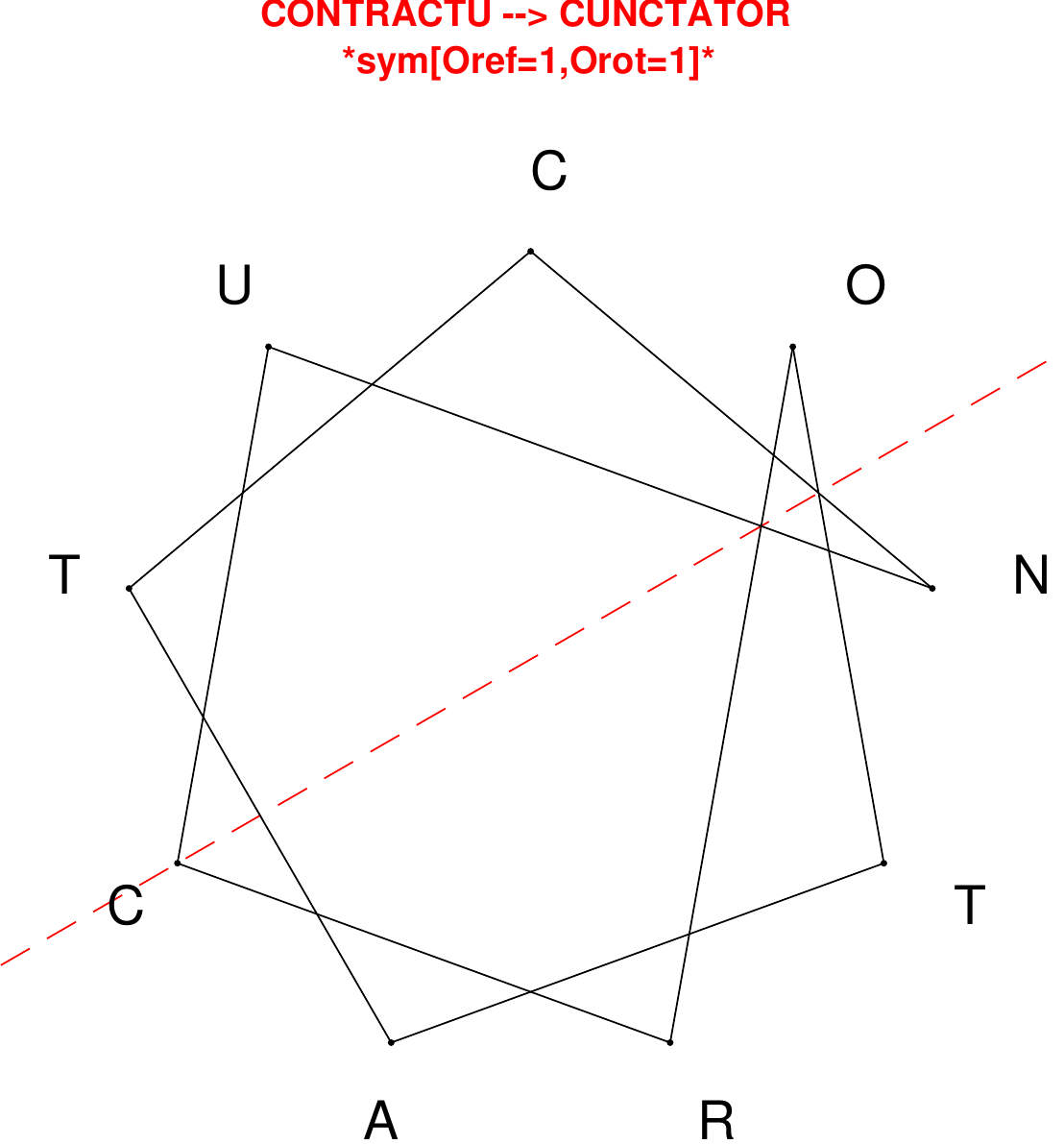}
\end{subfigure}
\hfill
\begin{subfigure}[T]{0.19\textwidth}
\centering
\includegraphics[width=\textwidth]{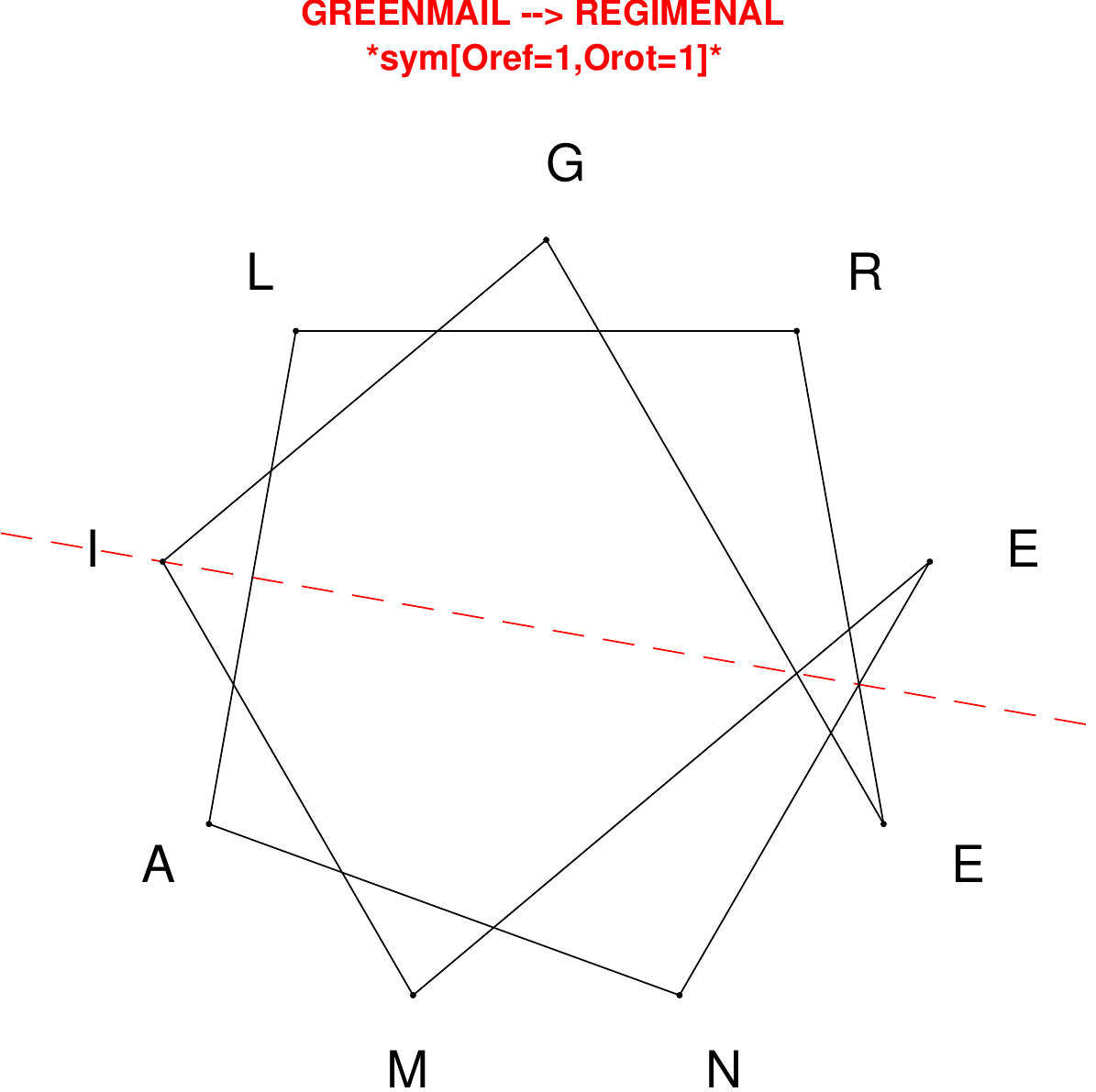}
\end{subfigure}
\hfill
\begin{subfigure}[T]{0.19\textwidth}
\centering
\includegraphics[width=\textwidth]{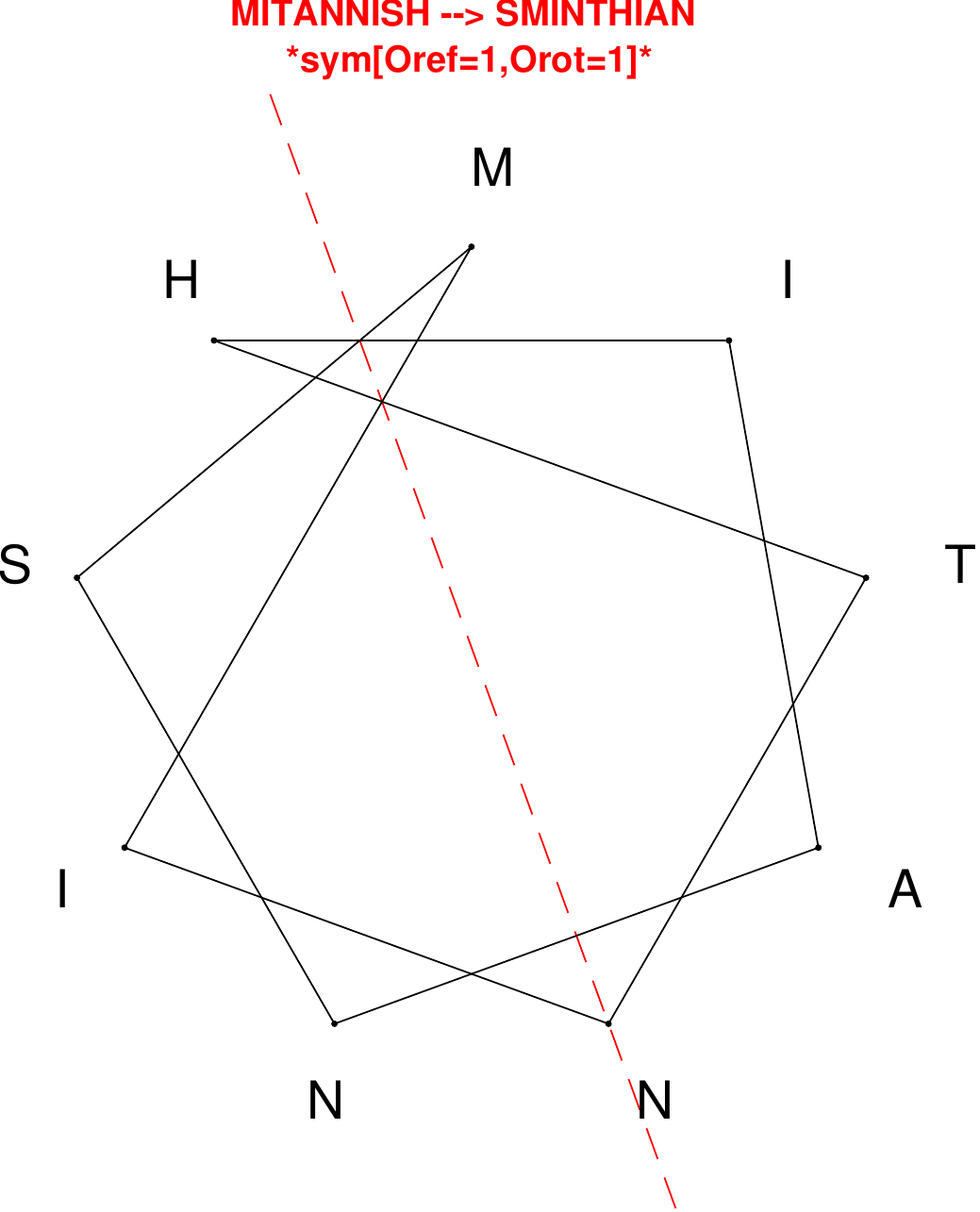}
\end{subfigure}
\hfill
\begin{subfigure}[T]{0.19\textwidth}
\centering
\includegraphics[width=\textwidth]{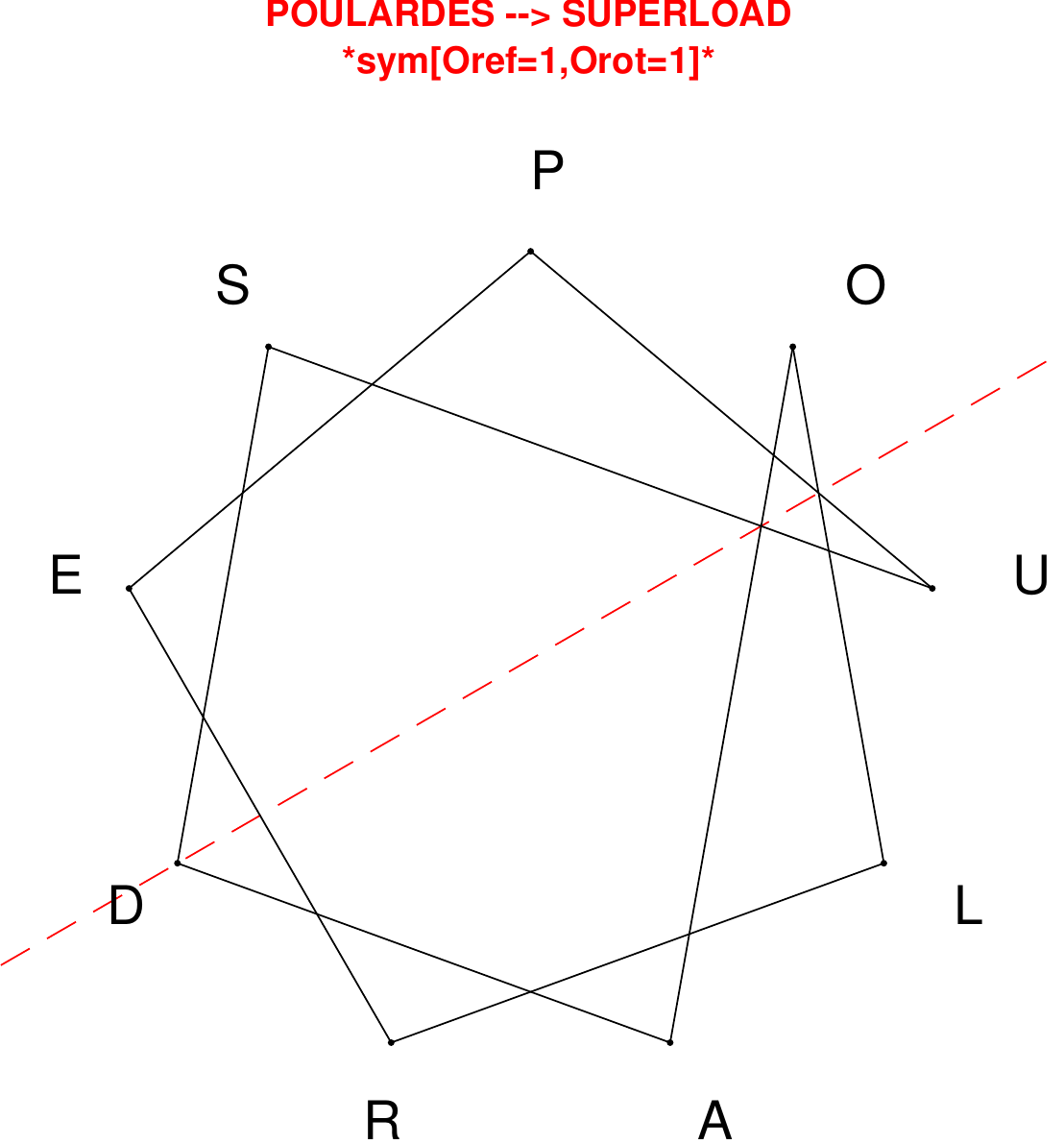}
\end{subfigure}
\hfill
\begin{subfigure}[T]{0.19\textwidth}
\centering
\includegraphics[width=\textwidth]{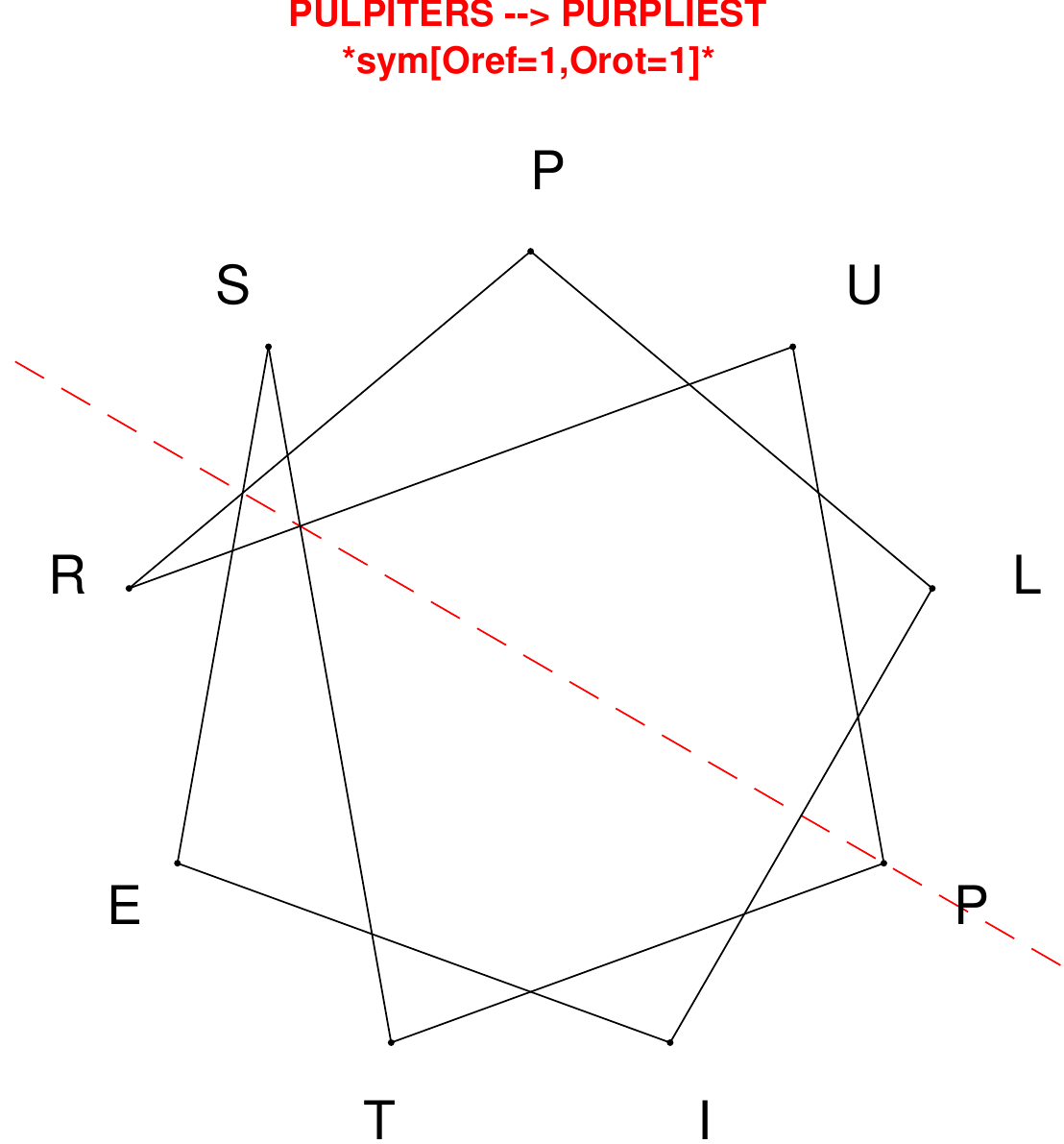}
\end{subfigure}
\end{figure}

\begin{figure}[H]
\centering
\begin{subfigure}[T]{0.19\textwidth}
\centering
\includegraphics[width=\textwidth]{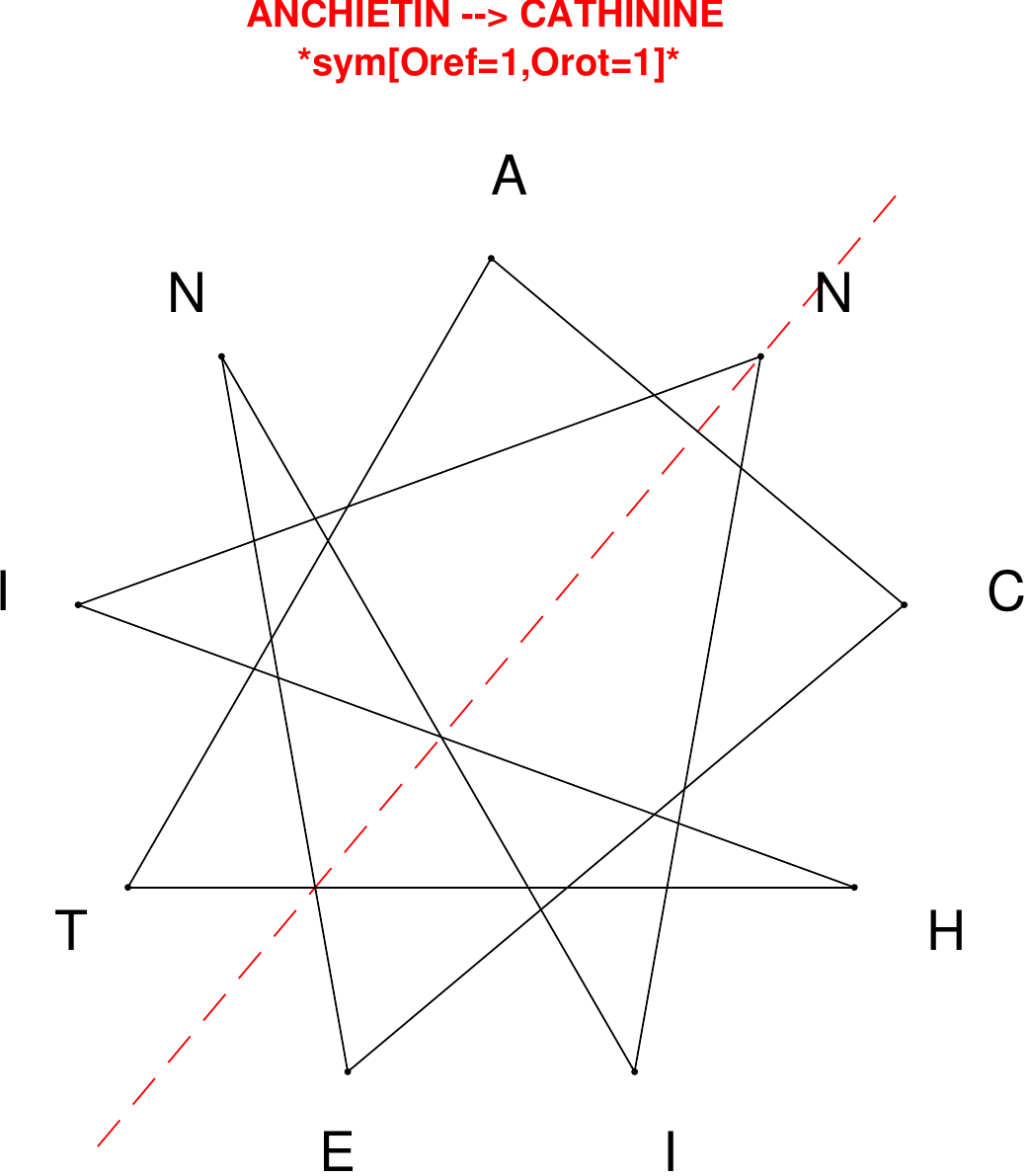}
\end{subfigure}
\hfill
\begin{subfigure}[T]{0.19\textwidth}
\centering
\includegraphics[width=\textwidth]{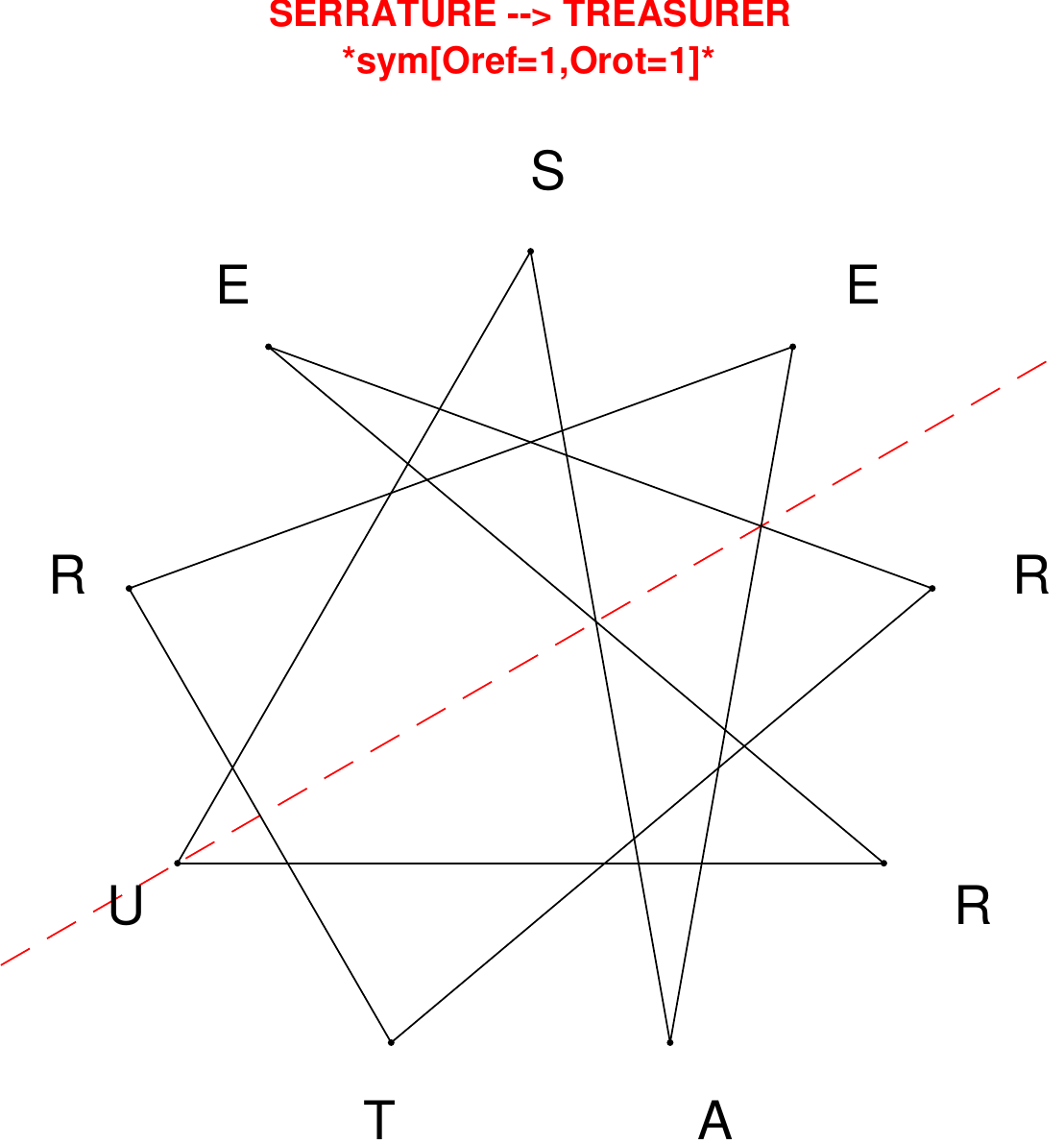}
\end{subfigure}
\hfill
\begin{subfigure}[T]{0.19\textwidth}
\centering
\includegraphics[width=\textwidth]{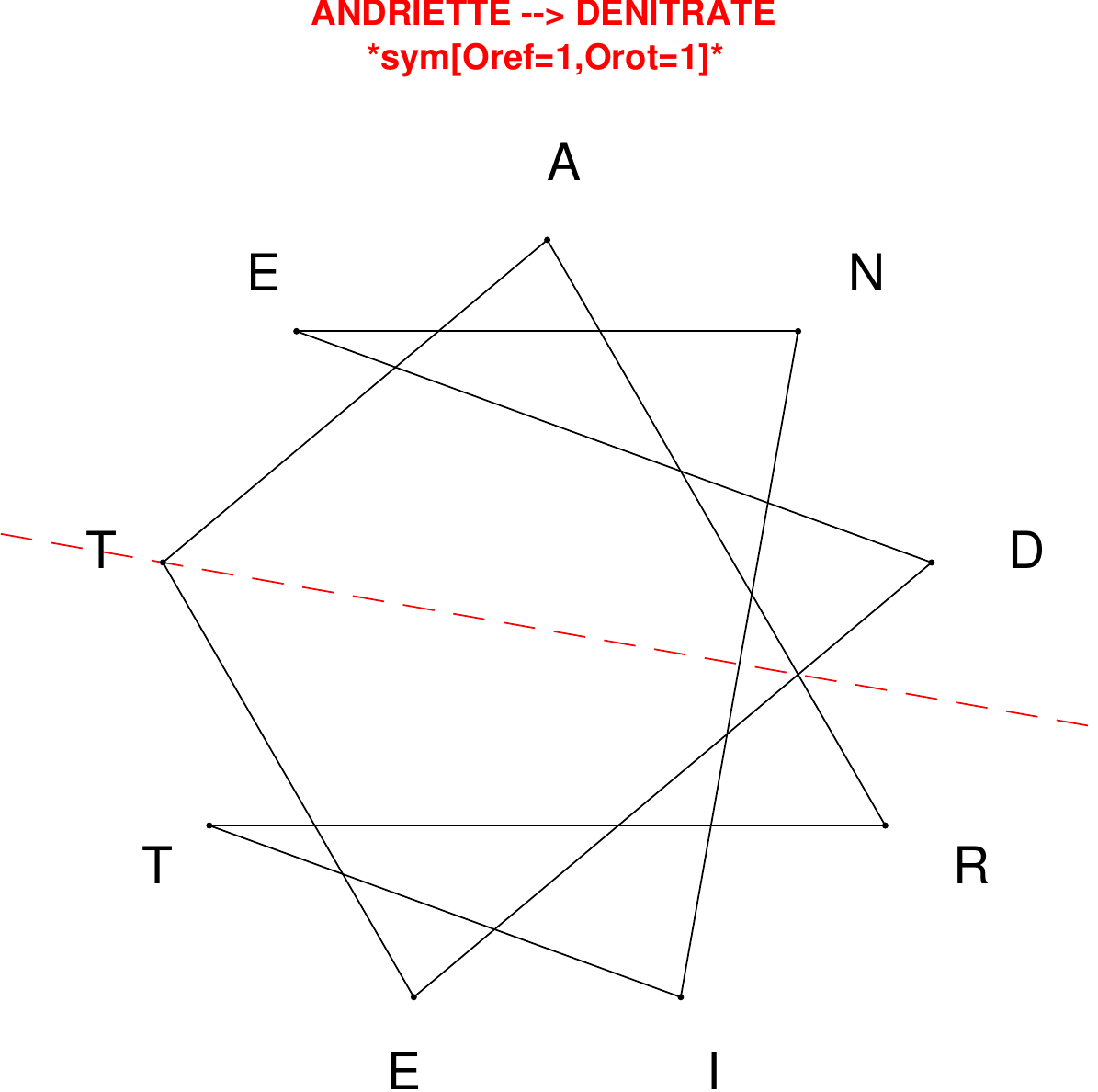}
\end{subfigure}
\hfill
\begin{subfigure}[T]{0.19\textwidth}
\centering
\includegraphics[width=\textwidth]{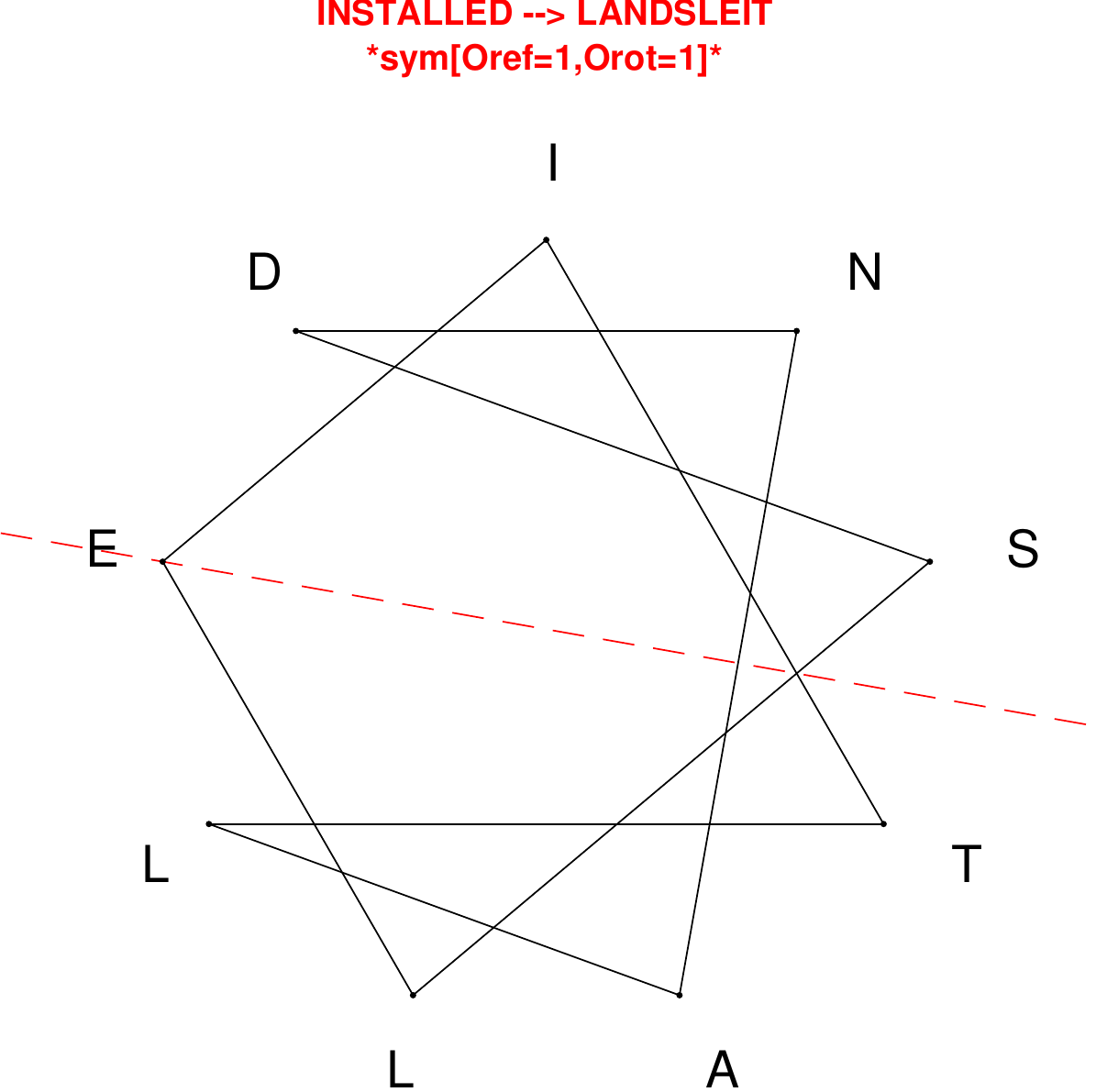}
\end{subfigure}
\hfill
\begin{subfigure}[T]{0.19\textwidth}
\centering
\includegraphics[width=\textwidth]{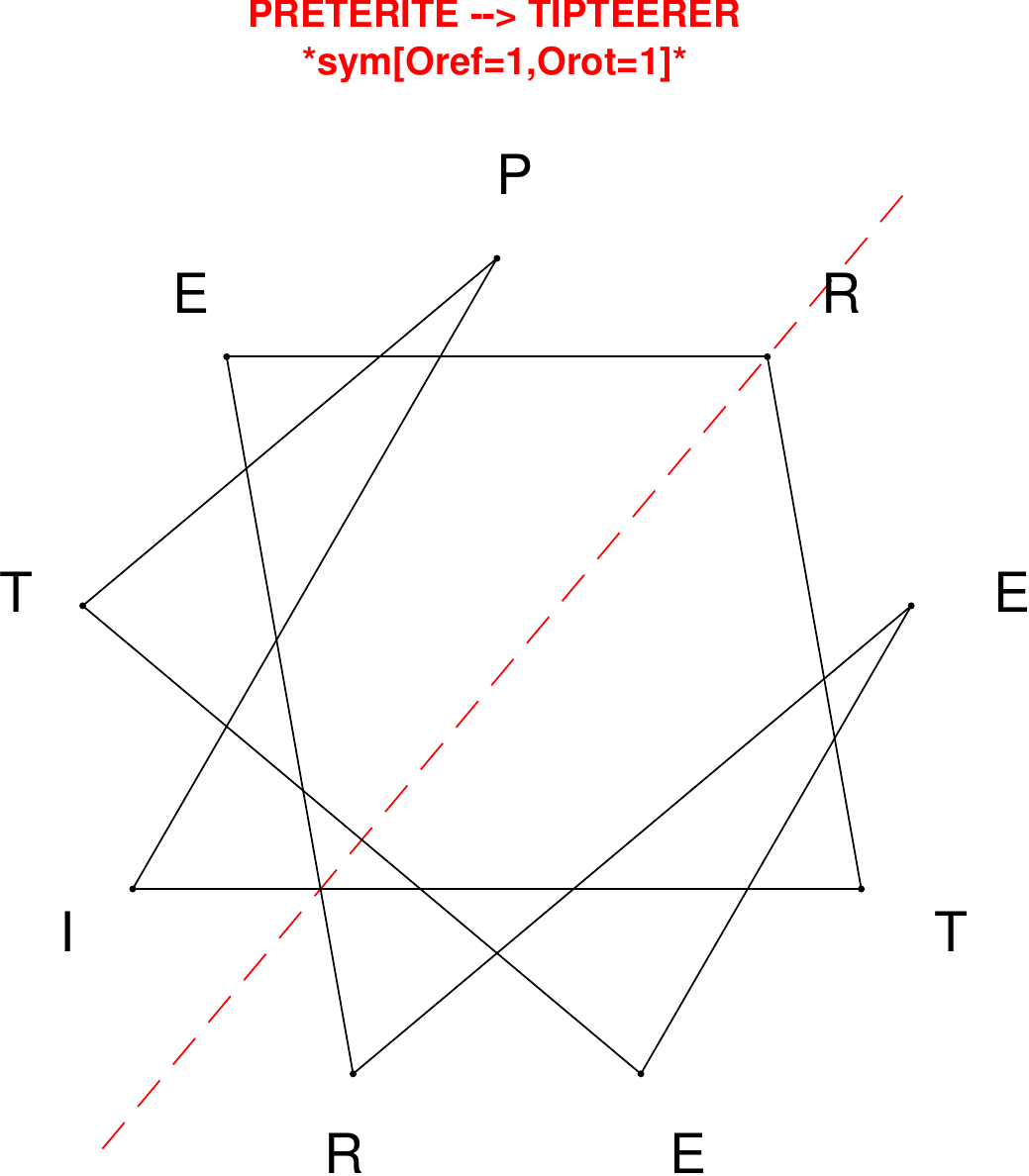}
\end{subfigure}
\end{figure}

\begin{figure}[H]
\centering
\begin{subfigure}[T]{0.19\textwidth}
\centering
\includegraphics[width=\textwidth]{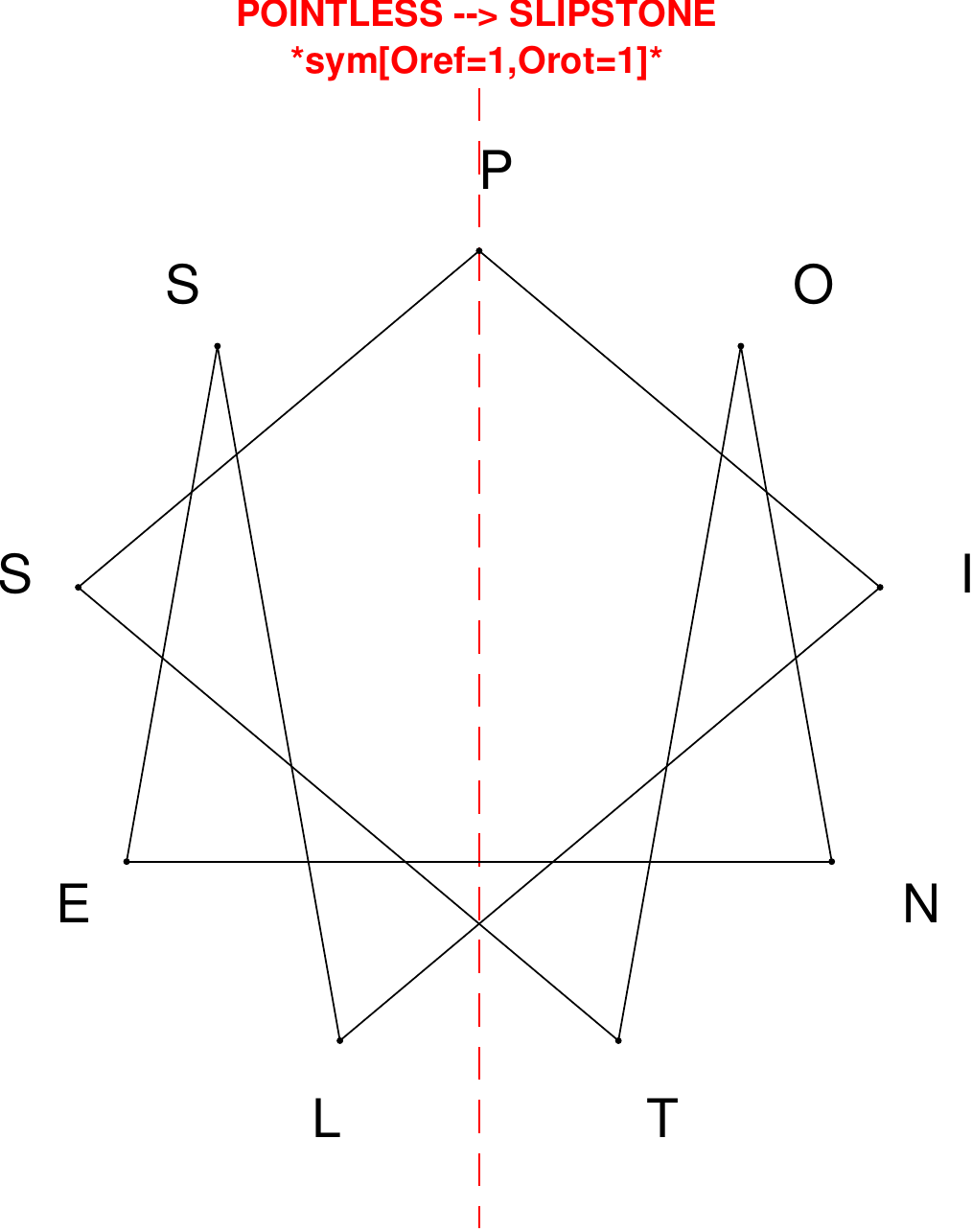}
\end{subfigure}
\hfill
\begin{subfigure}[T]{0.19\textwidth}
\centering
\includegraphics[width=\textwidth]{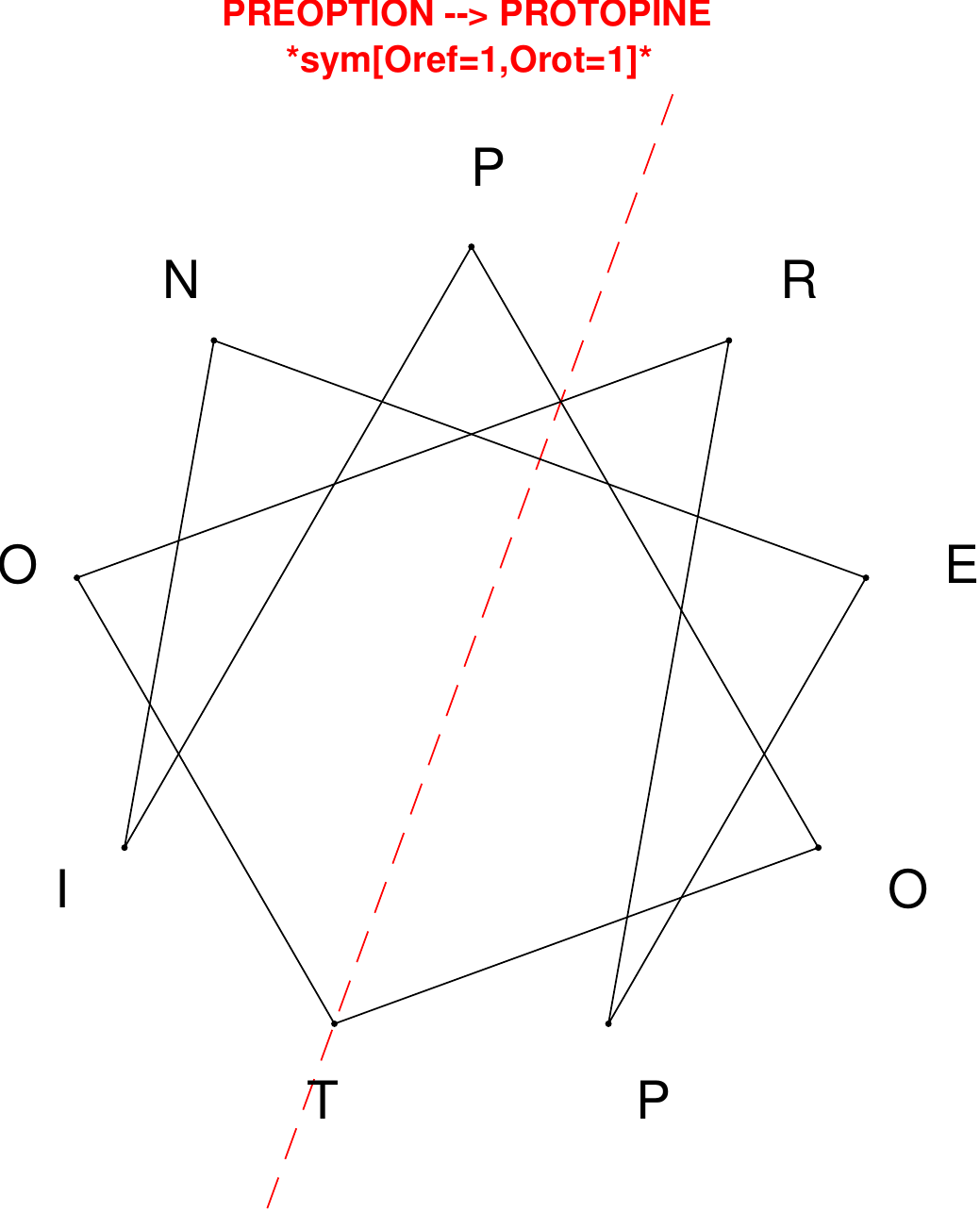}
\end{subfigure}
\hfill
\begin{subfigure}[T]{0.19\textwidth}
\centering
\includegraphics[width=\textwidth]{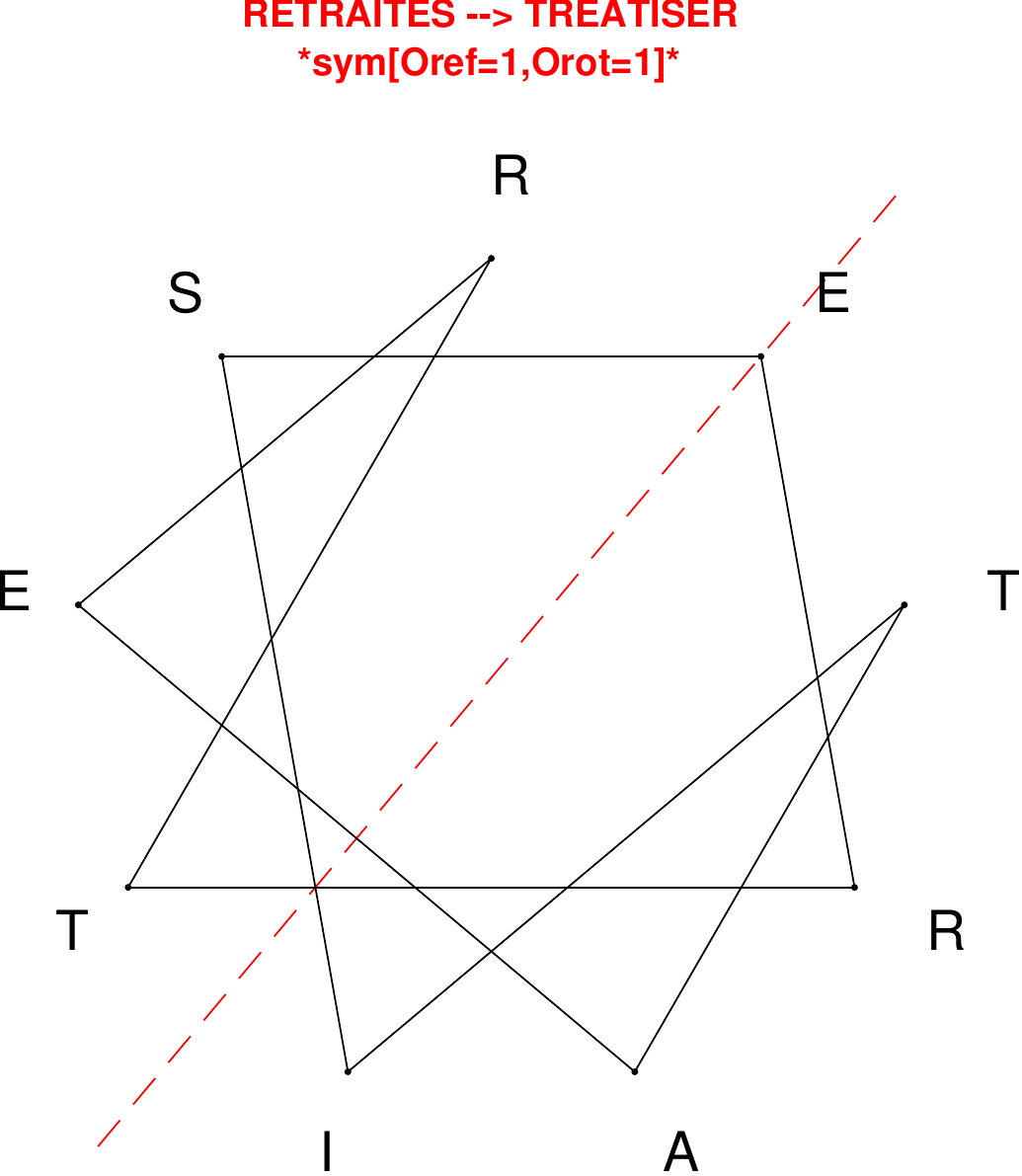}
\end{subfigure}
\hfill
\begin{subfigure}[T]{0.19\textwidth}
\centering
\includegraphics[width=\textwidth]{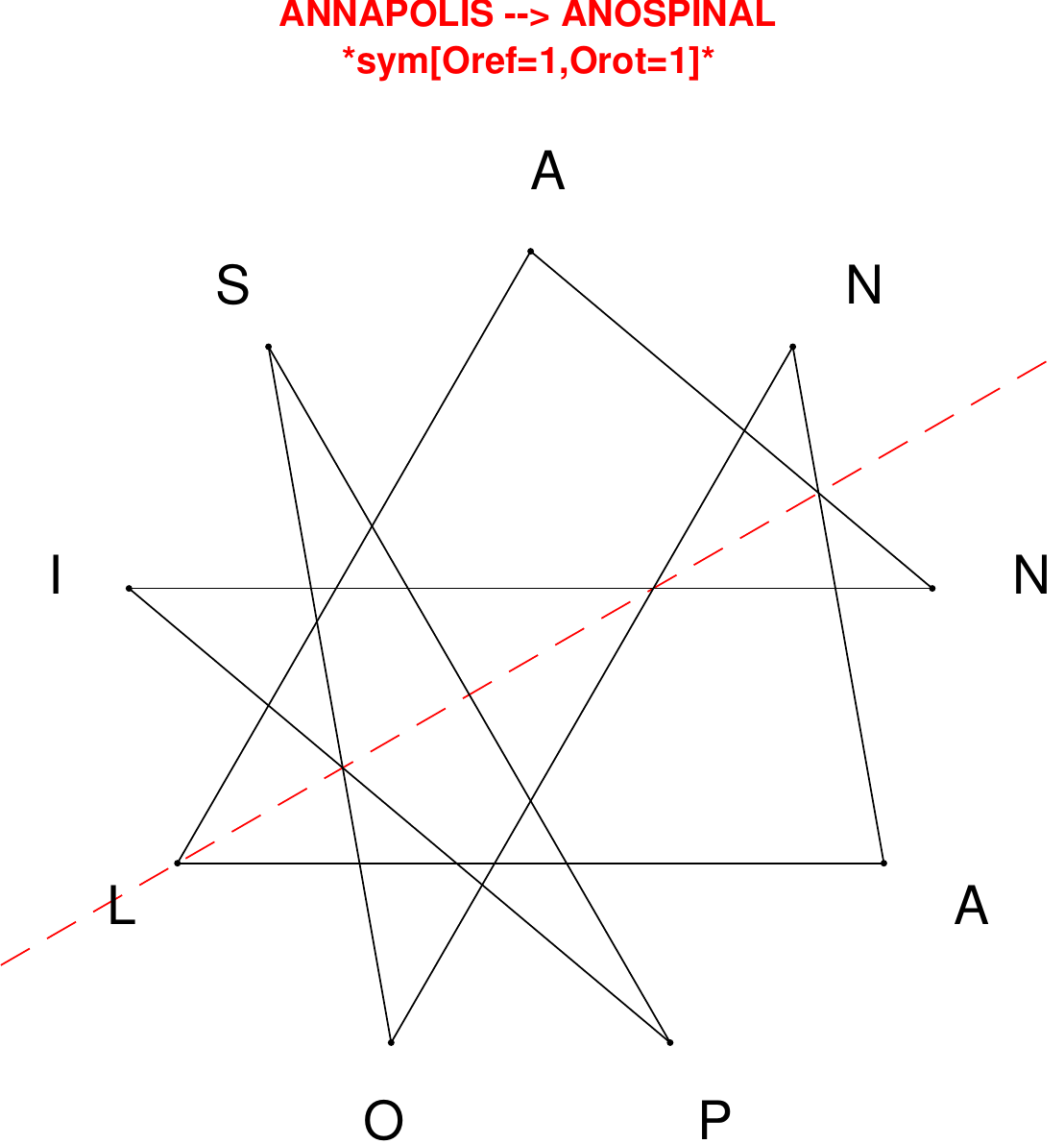}
\end{subfigure}
\hfill
\begin{subfigure}[T]{0.19\textwidth}
\centering
\includegraphics[width=\textwidth]{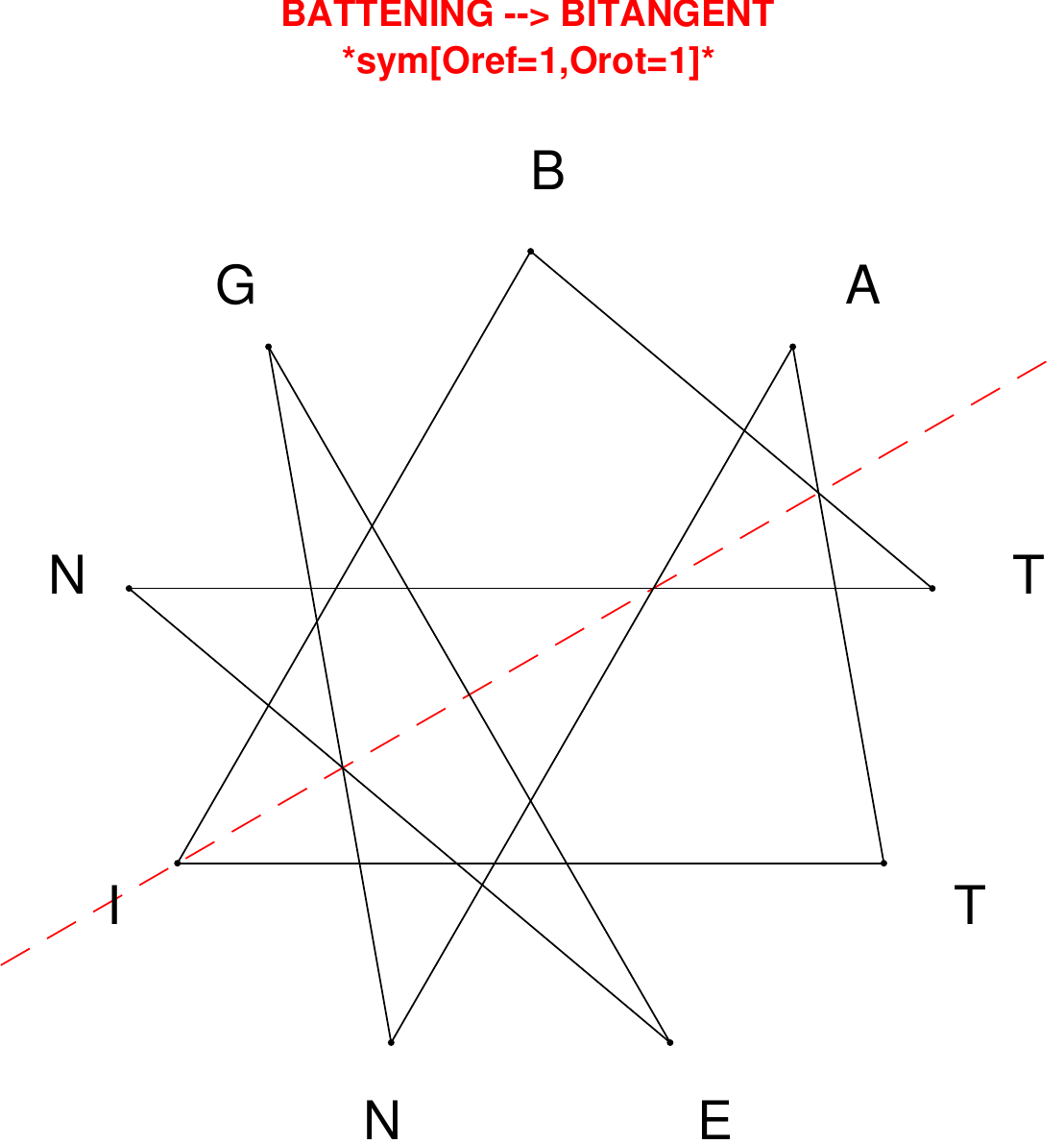}
\end{subfigure}
\end{figure}

\begin{figure}[H]
\centering
\begin{subfigure}[T]{0.19\textwidth}
\centering
\includegraphics[width=\textwidth]{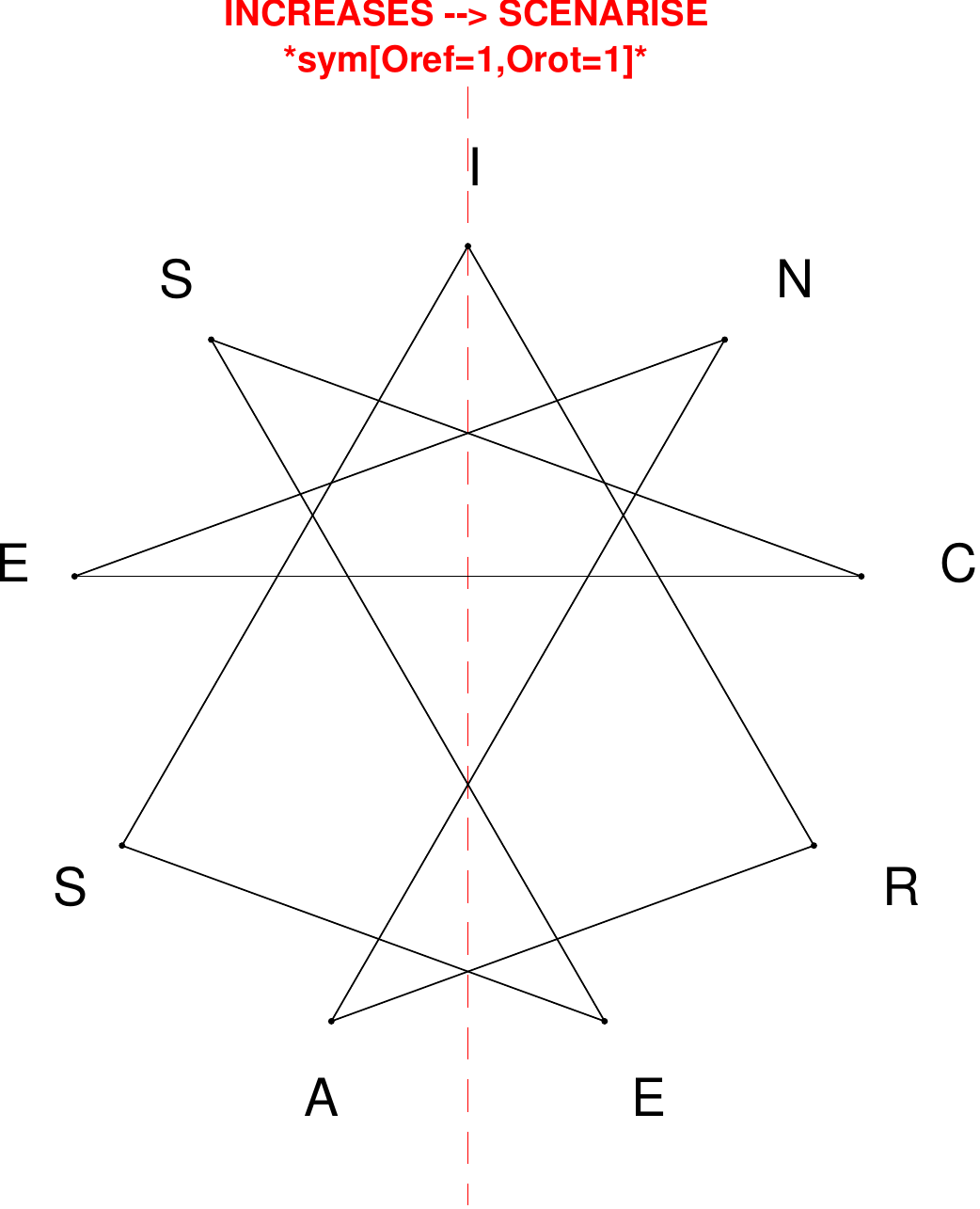}
\end{subfigure}
\hfill
\begin{subfigure}[T]{0.19\textwidth}
\centering
\includegraphics[width=\textwidth]{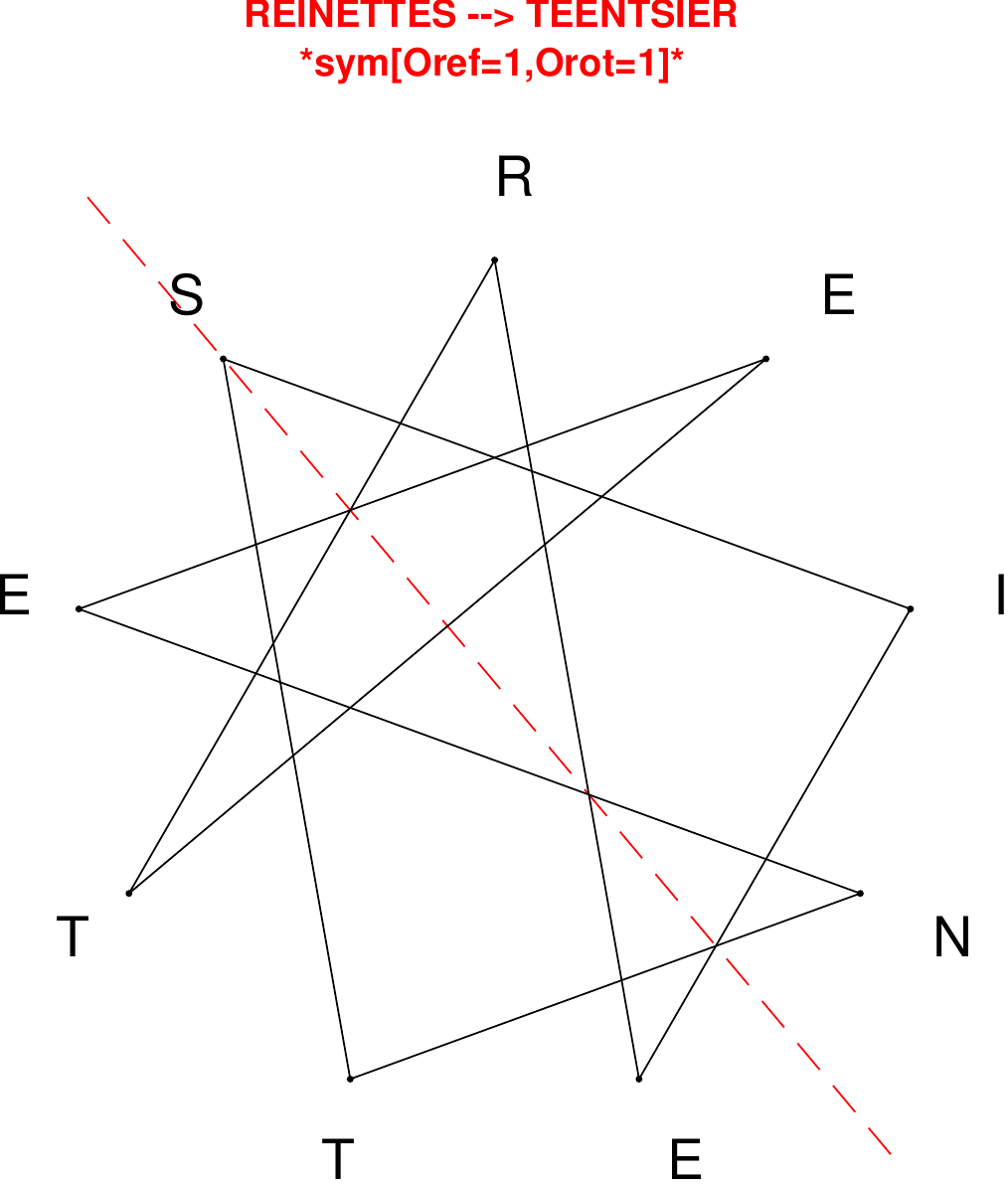}
\end{subfigure}
\hfill
\begin{subfigure}[T]{0.19\textwidth}
\centering
\includegraphics[width=\textwidth]{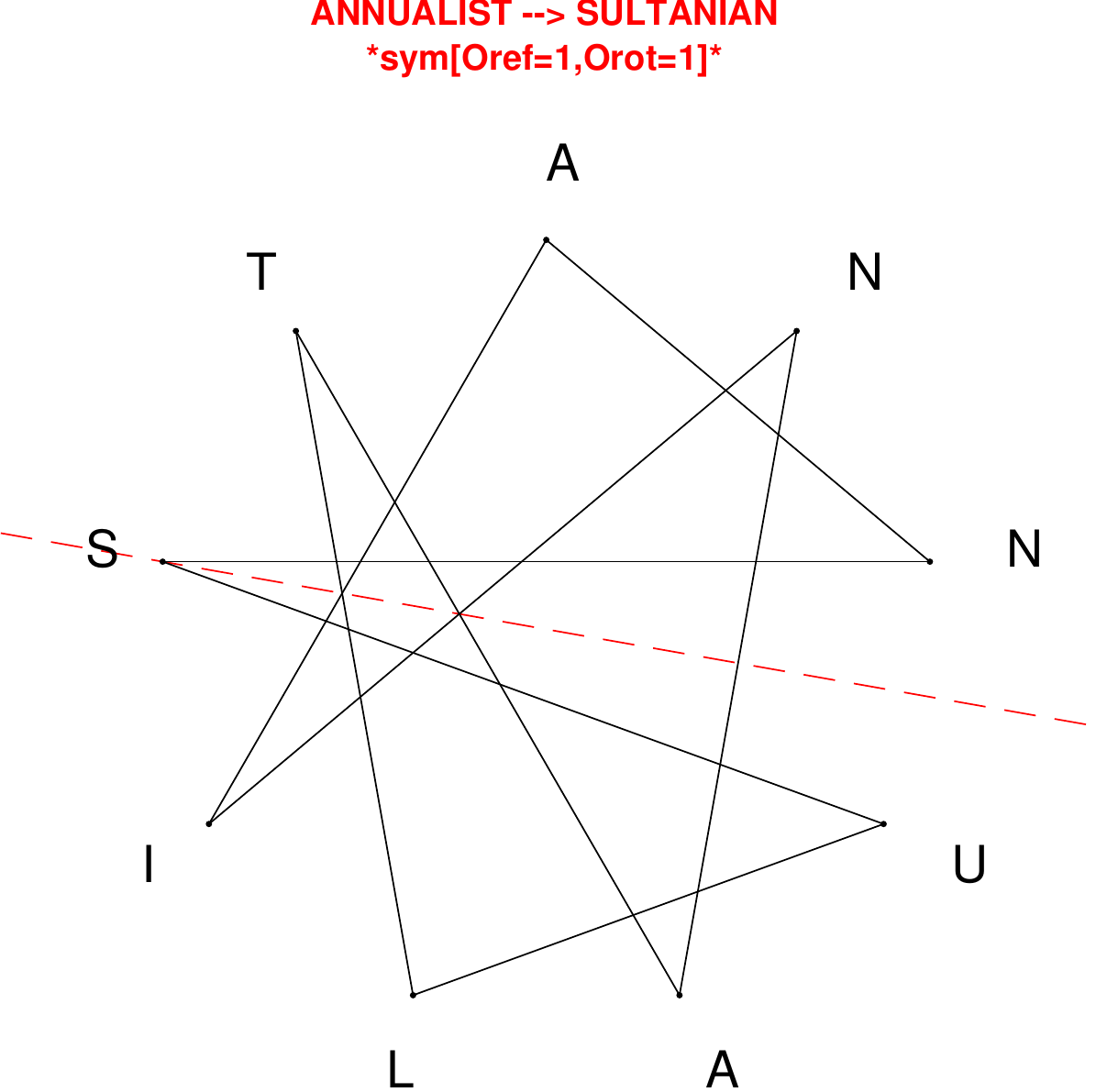}
\end{subfigure}
\hfill
\begin{subfigure}[T]{0.19\textwidth}
\centering
\includegraphics[width=\textwidth]{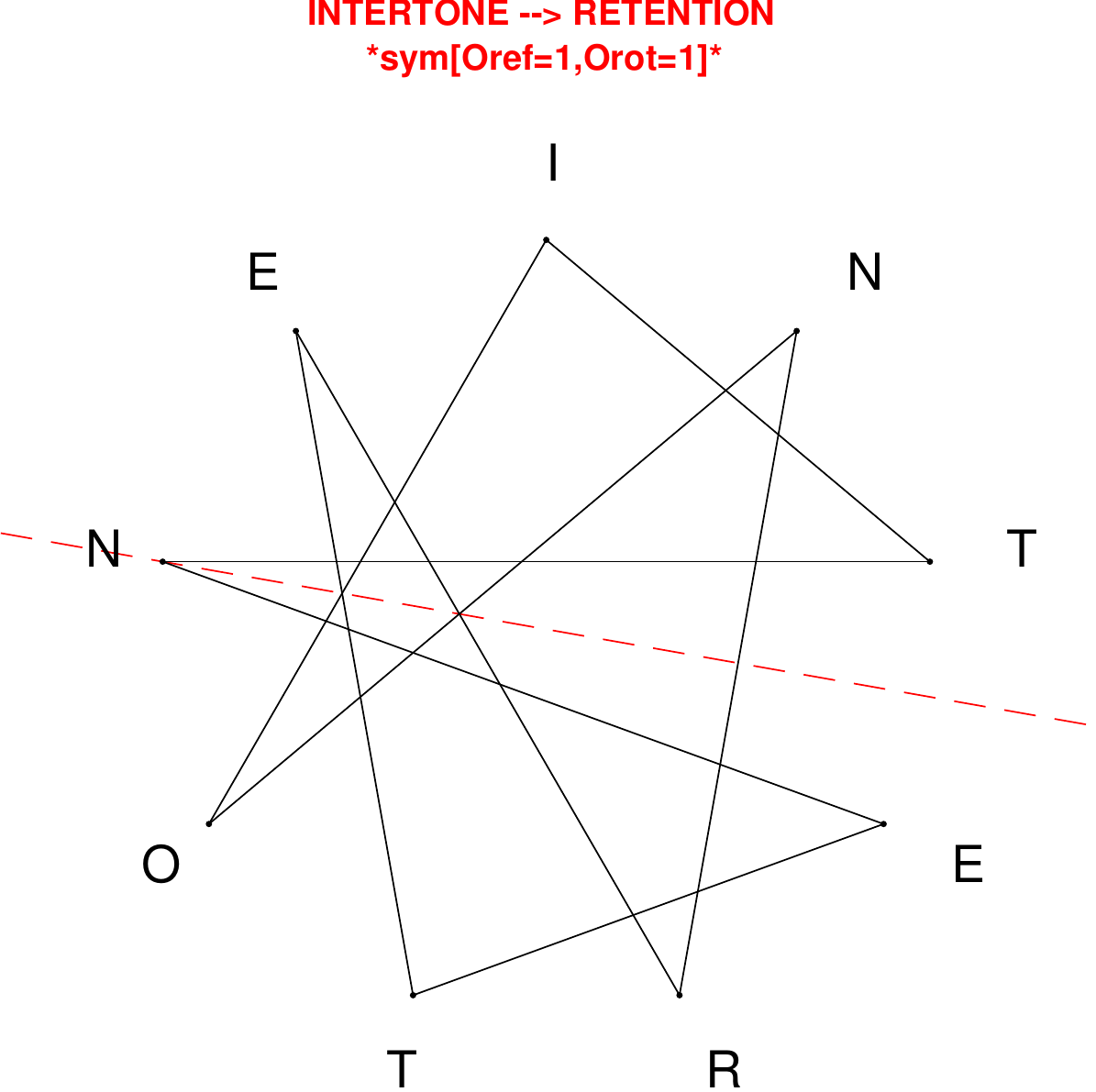}
\end{subfigure}
\hfill
\begin{subfigure}[T]{0.19\textwidth}
\centering
\includegraphics[width=\textwidth]{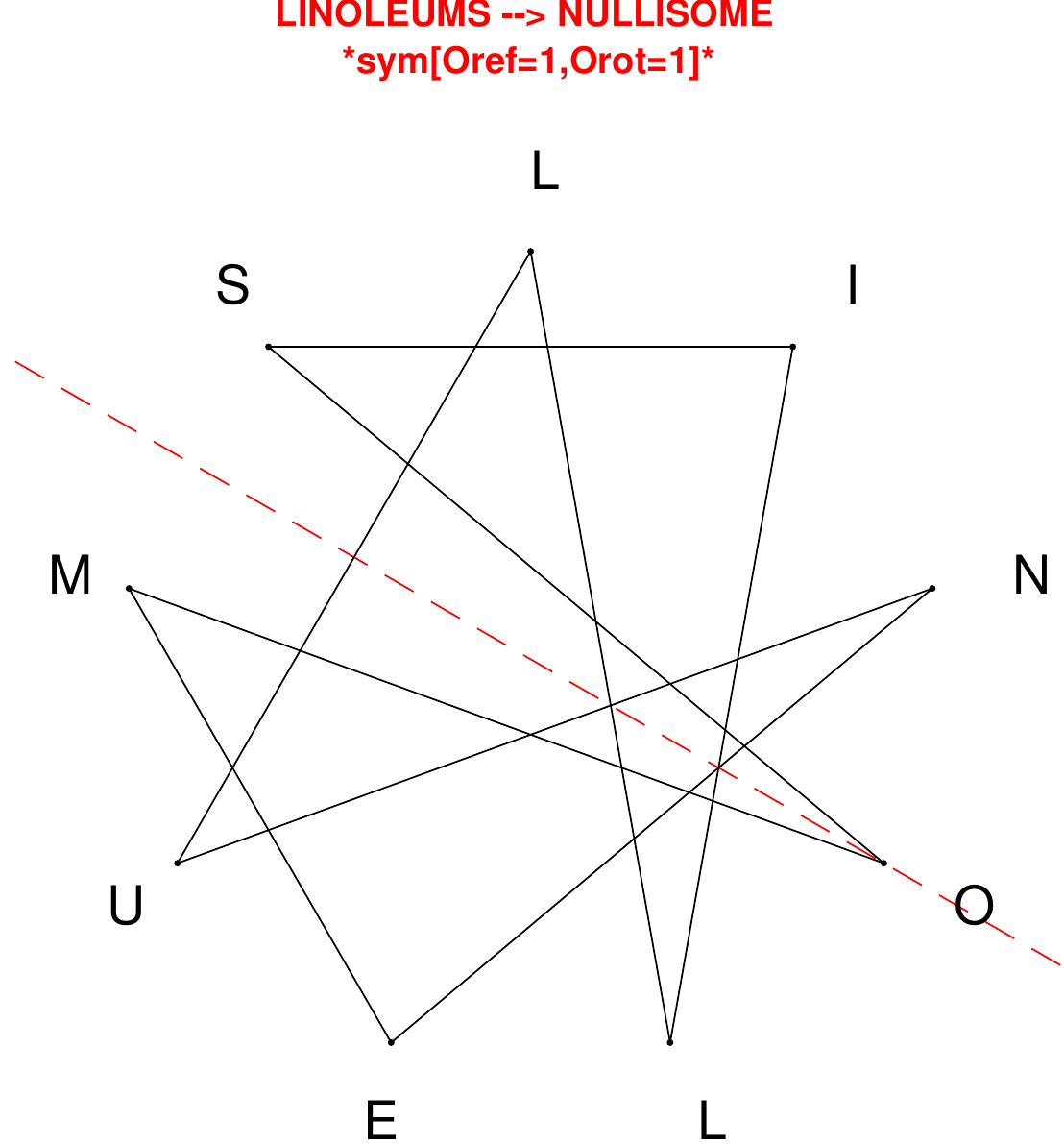}
\end{subfigure}
\end{figure}

\begin{figure}[H]
\centering
\begin{subfigure}[T]{0.19\textwidth}
\centering
\includegraphics[width=\textwidth]{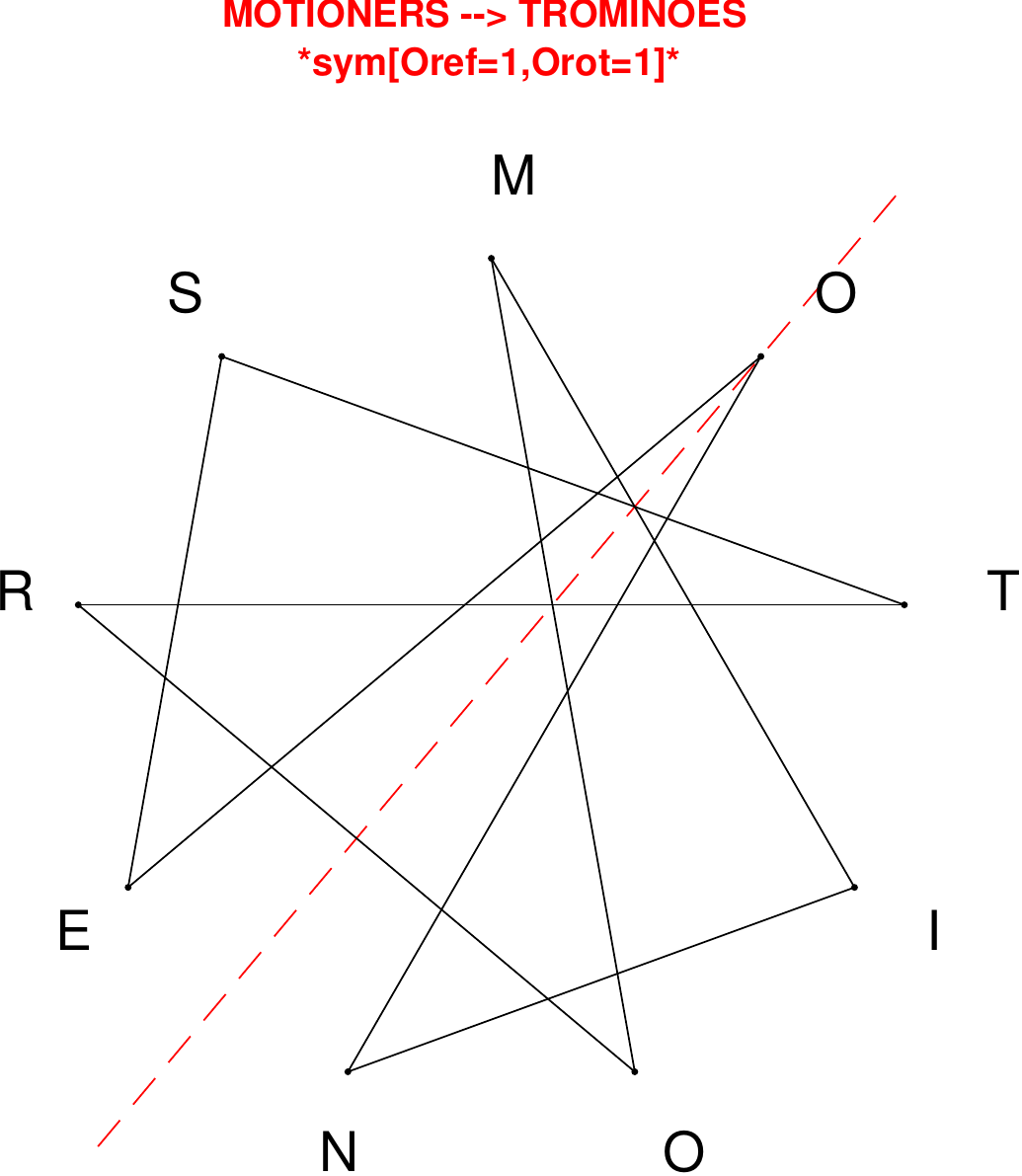}
\end{subfigure}
\hfill
\begin{subfigure}[T]{0.19\textwidth}
\centering
\includegraphics[width=\textwidth]{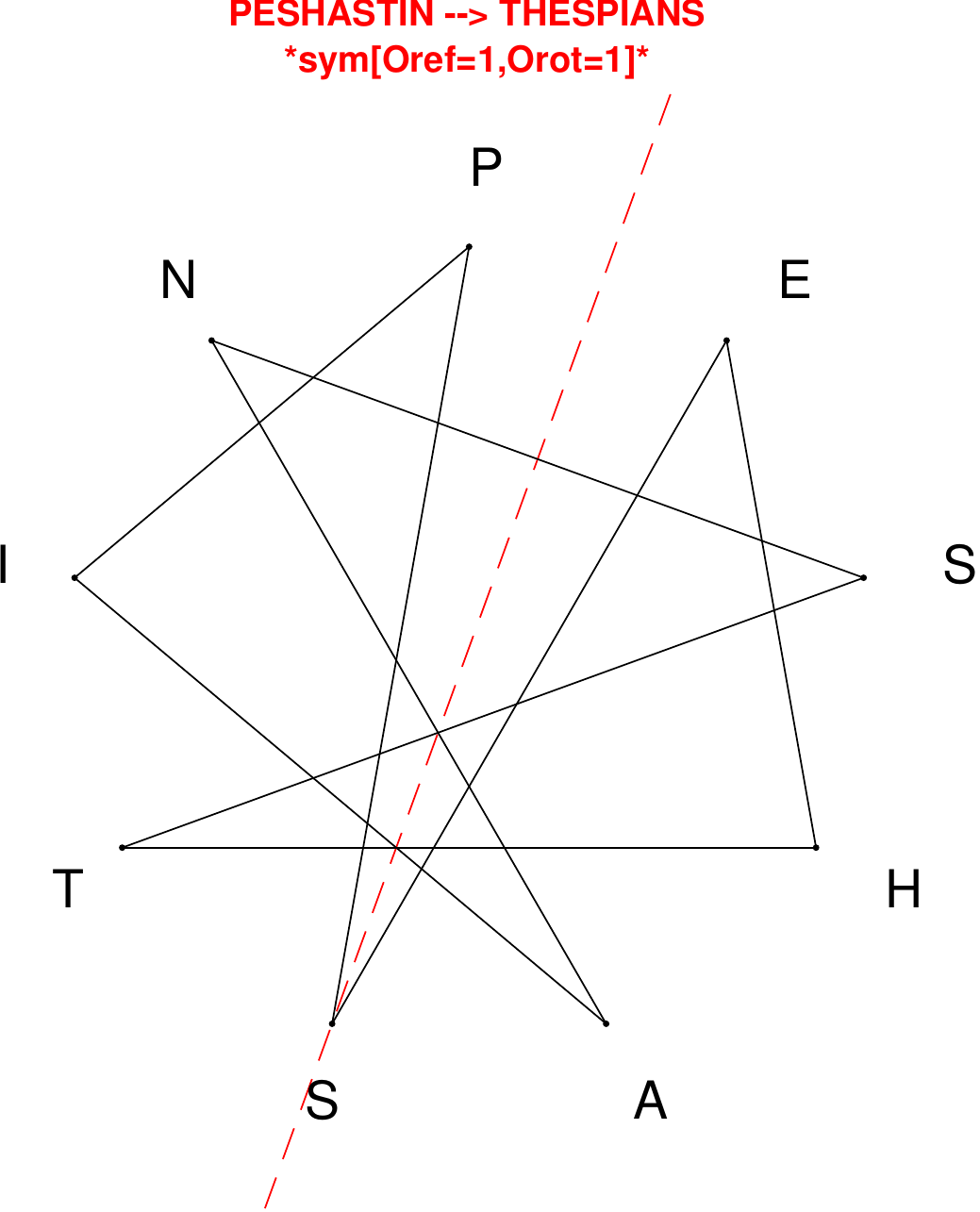}
\end{subfigure}
\hfill
\begin{subfigure}[T]{0.19\textwidth}
\centering
\includegraphics[width=\textwidth]{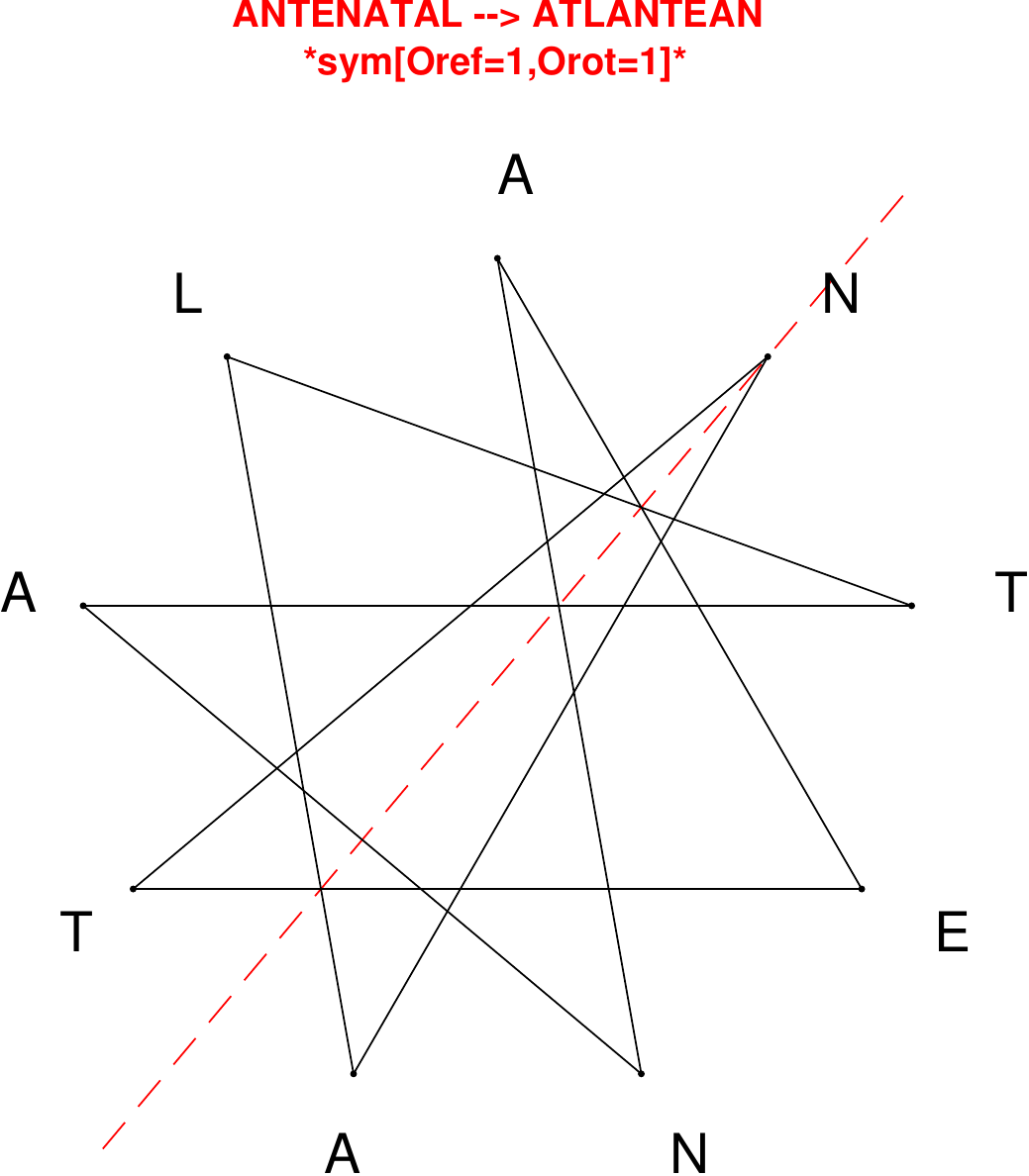}
\end{subfigure}
\hfill
\begin{subfigure}[T]{0.19\textwidth}
\centering
\includegraphics[width=\textwidth]{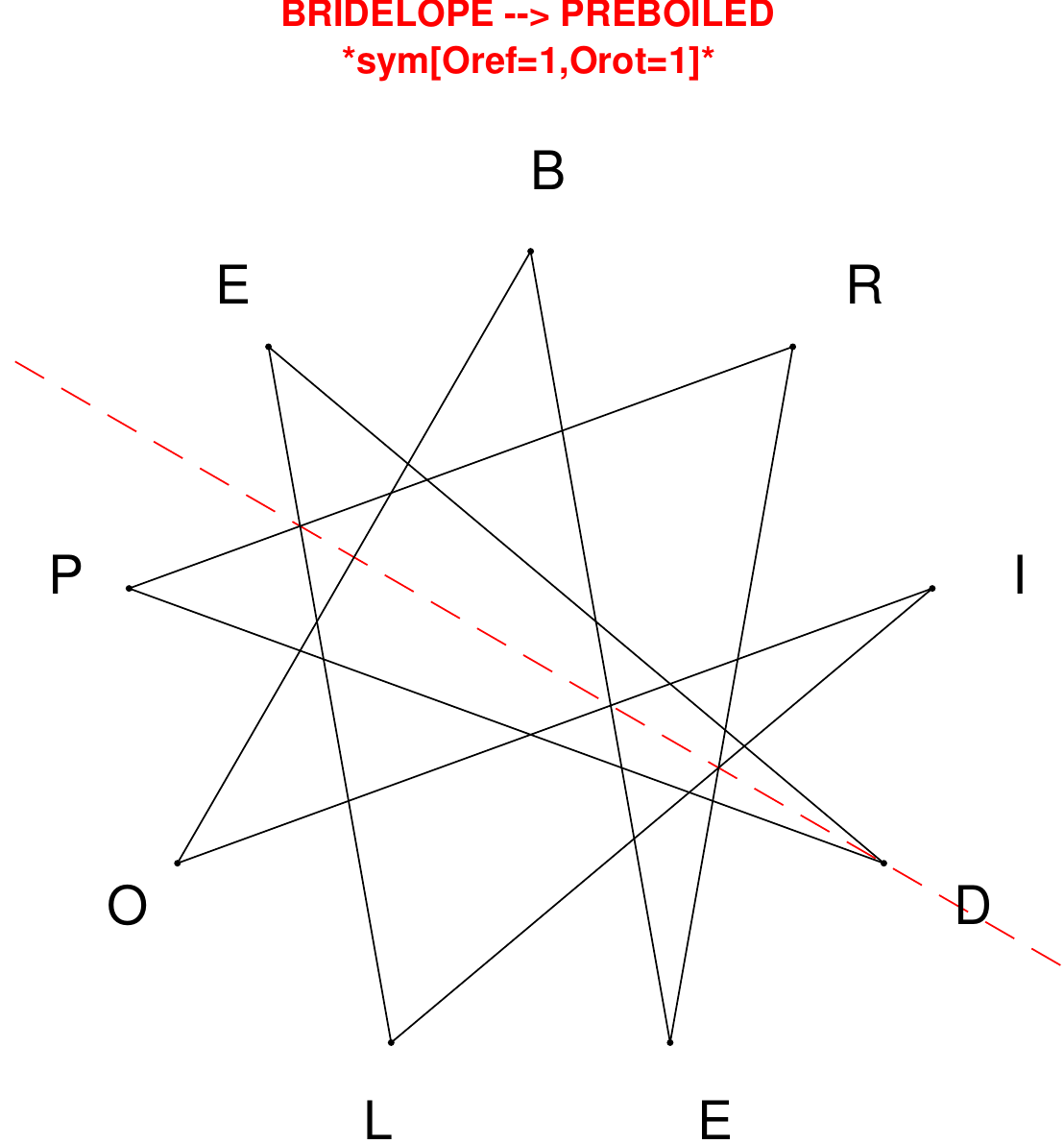}
\end{subfigure}
\hfill
\begin{subfigure}[T]{0.19\textwidth}
\centering
\includegraphics[width=\textwidth]{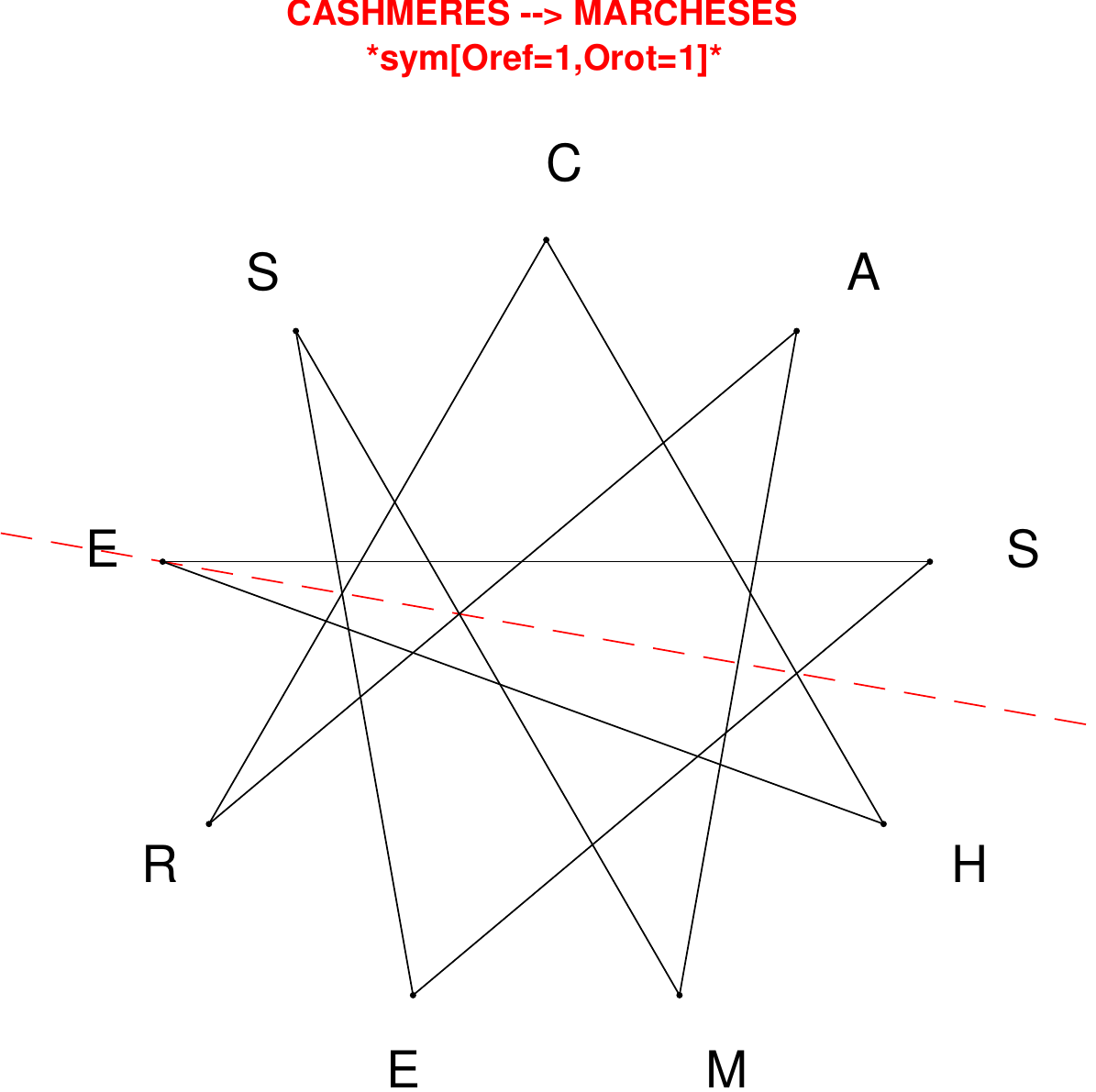}
\end{subfigure}
\end{figure}

\begin{figure}[H]
\centering
\begin{subfigure}[T]{0.19\textwidth}
\centering
\includegraphics[width=\textwidth]{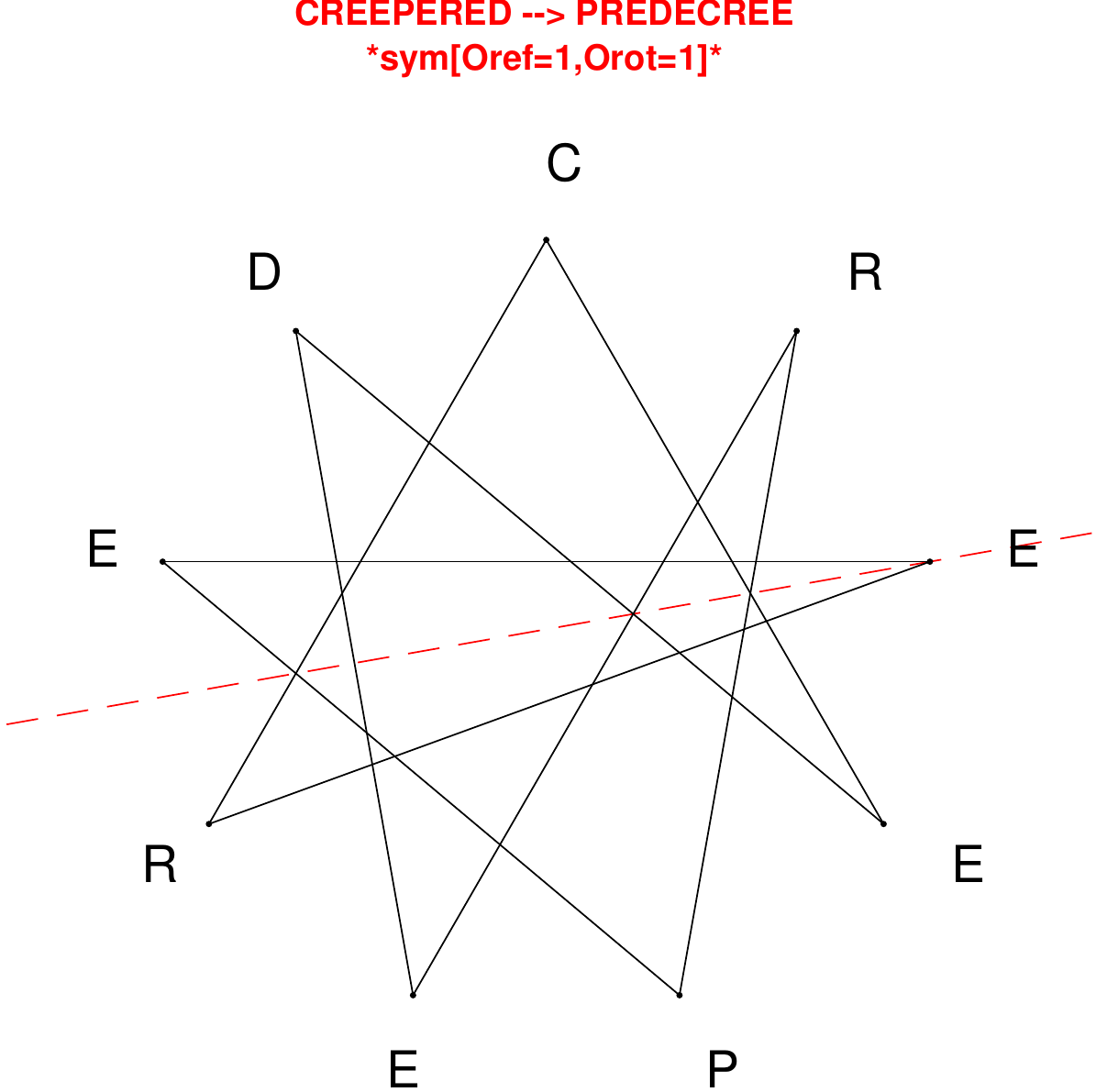}
\end{subfigure}
\hfill
\begin{subfigure}[T]{0.19\textwidth}
\centering
\includegraphics[width=\textwidth]{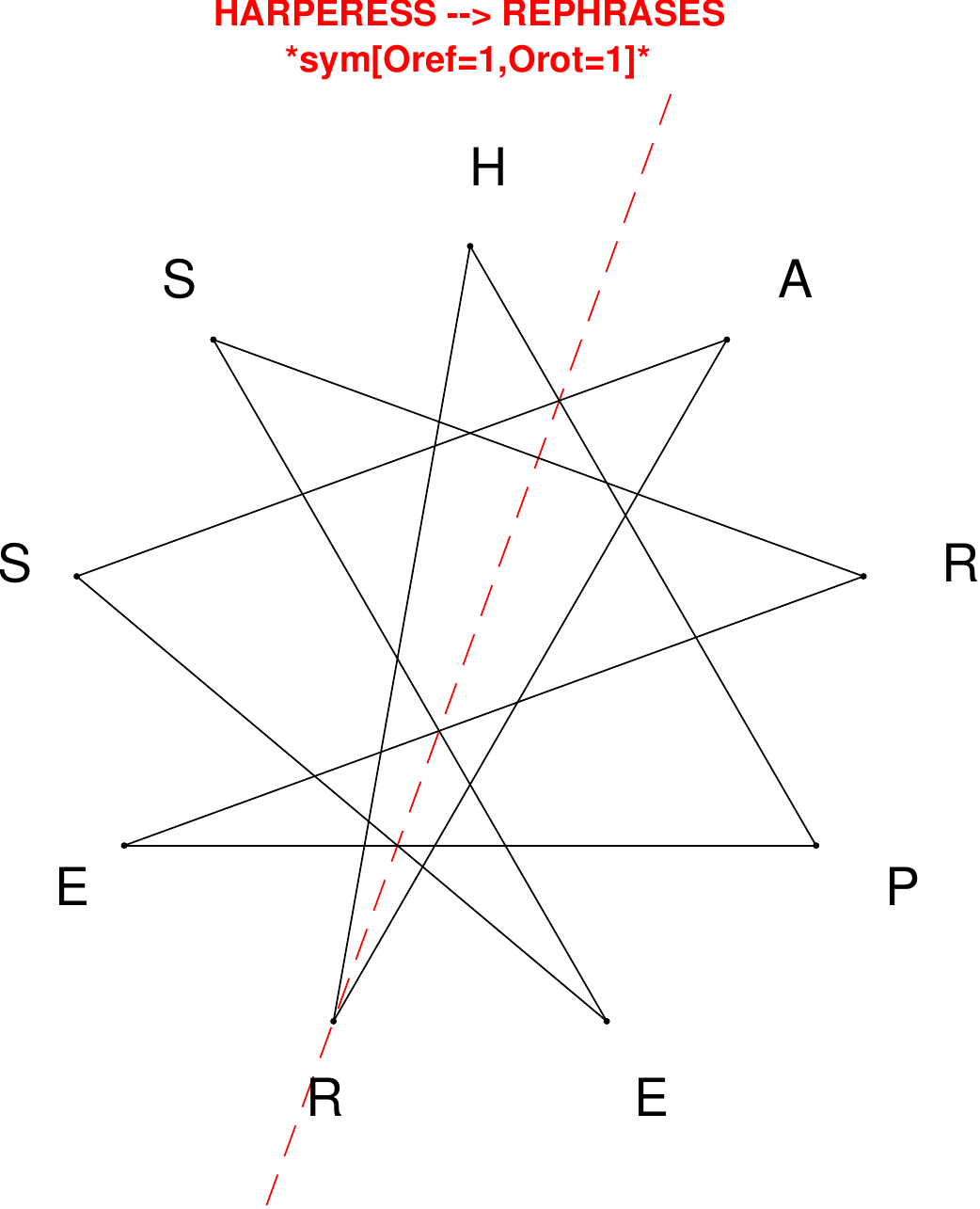}
\end{subfigure}
\hfill
\begin{subfigure}[T]{0.19\textwidth}
\centering
\includegraphics[width=\textwidth]{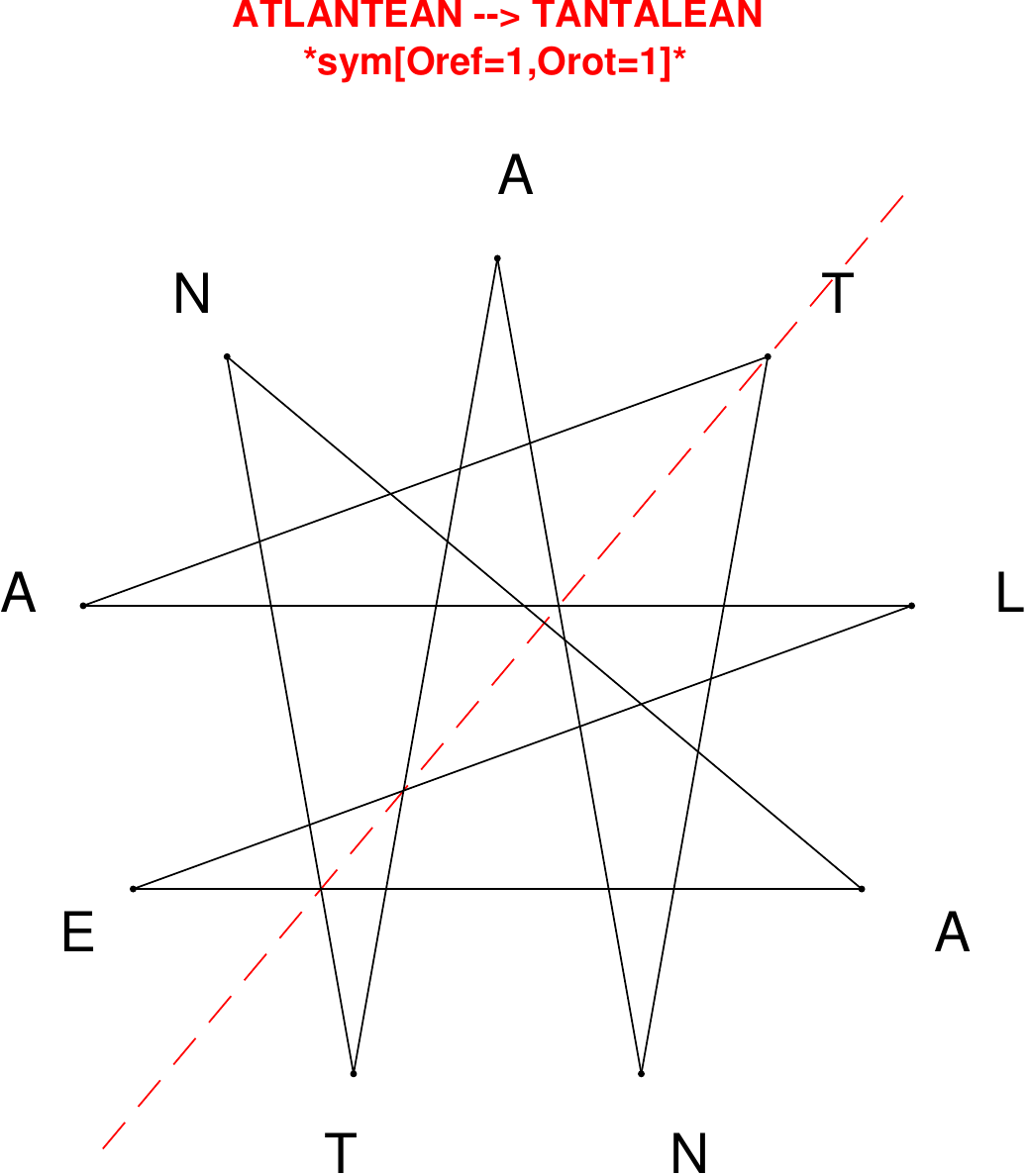}
\end{subfigure}
\hfill
\begin{subfigure}[T]{0.19\textwidth}
\centering
\includegraphics[width=\textwidth]{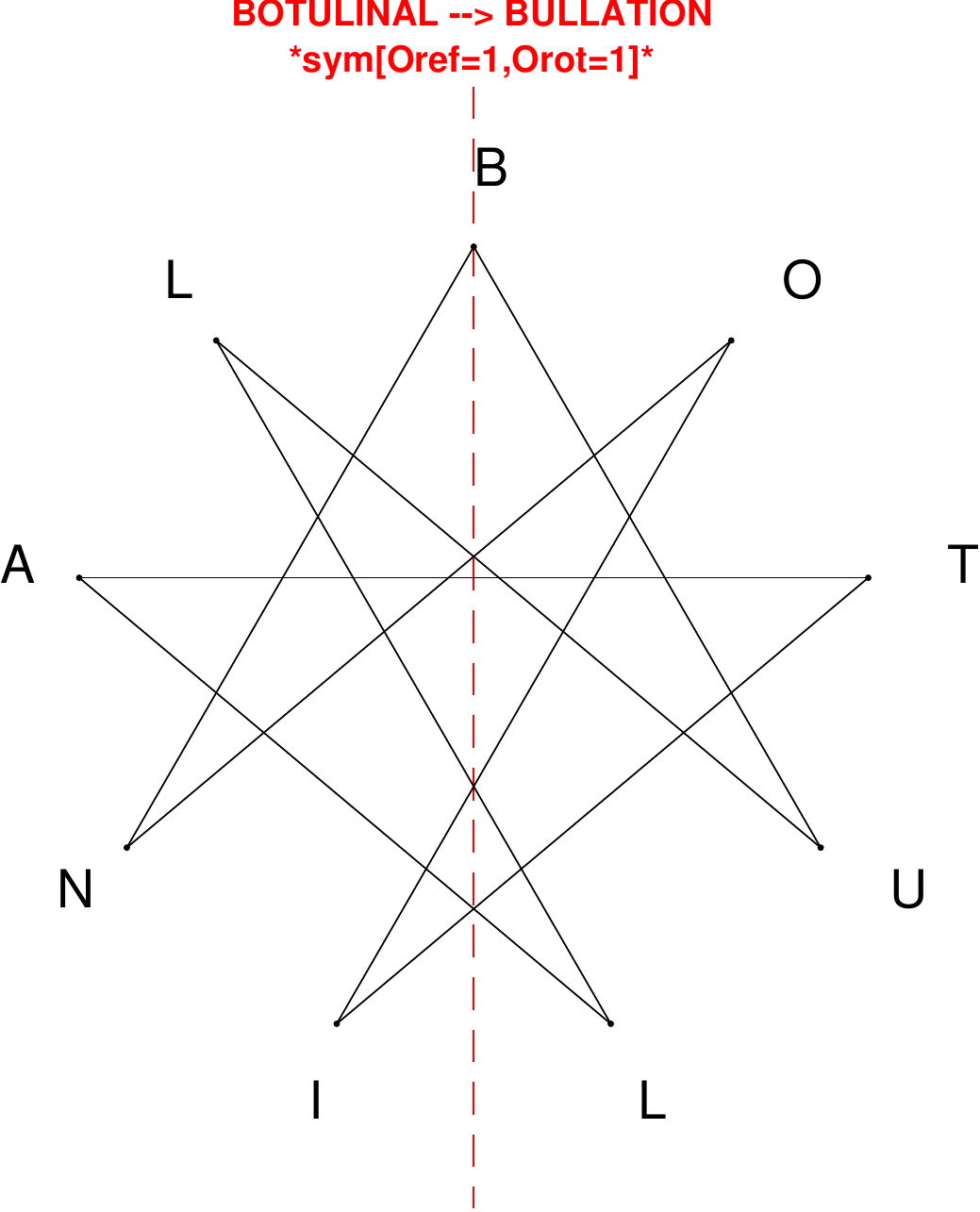}
\end{subfigure}
\hfill
\begin{subfigure}[T]{0.19\textwidth}
\centering
\includegraphics[width=\textwidth]{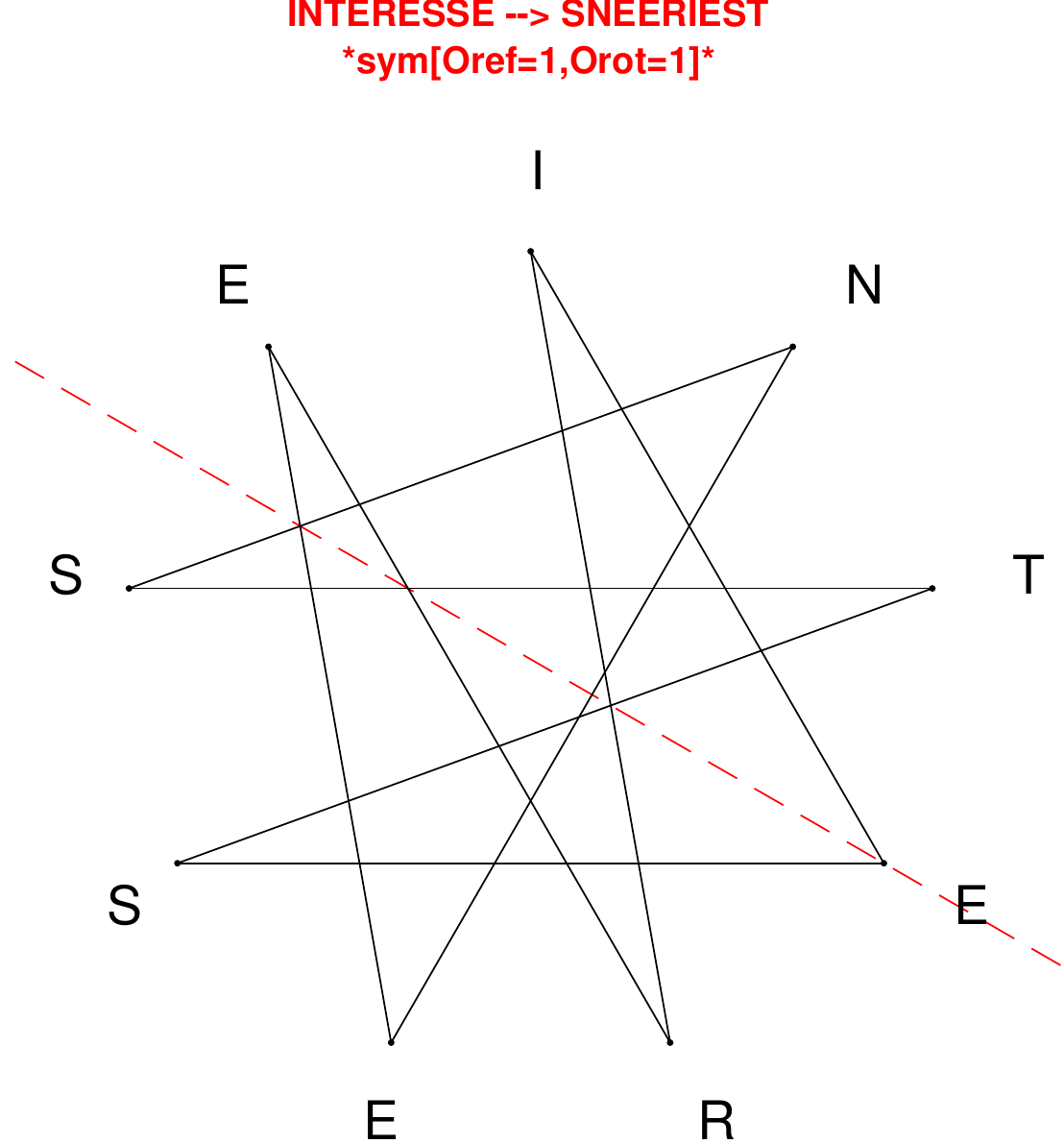}
\end{subfigure}
\end{figure}

\begin{figure}[H]
\centering
\begin{subfigure}[T]{0.19\textwidth}
\centering
\includegraphics[width=\textwidth]{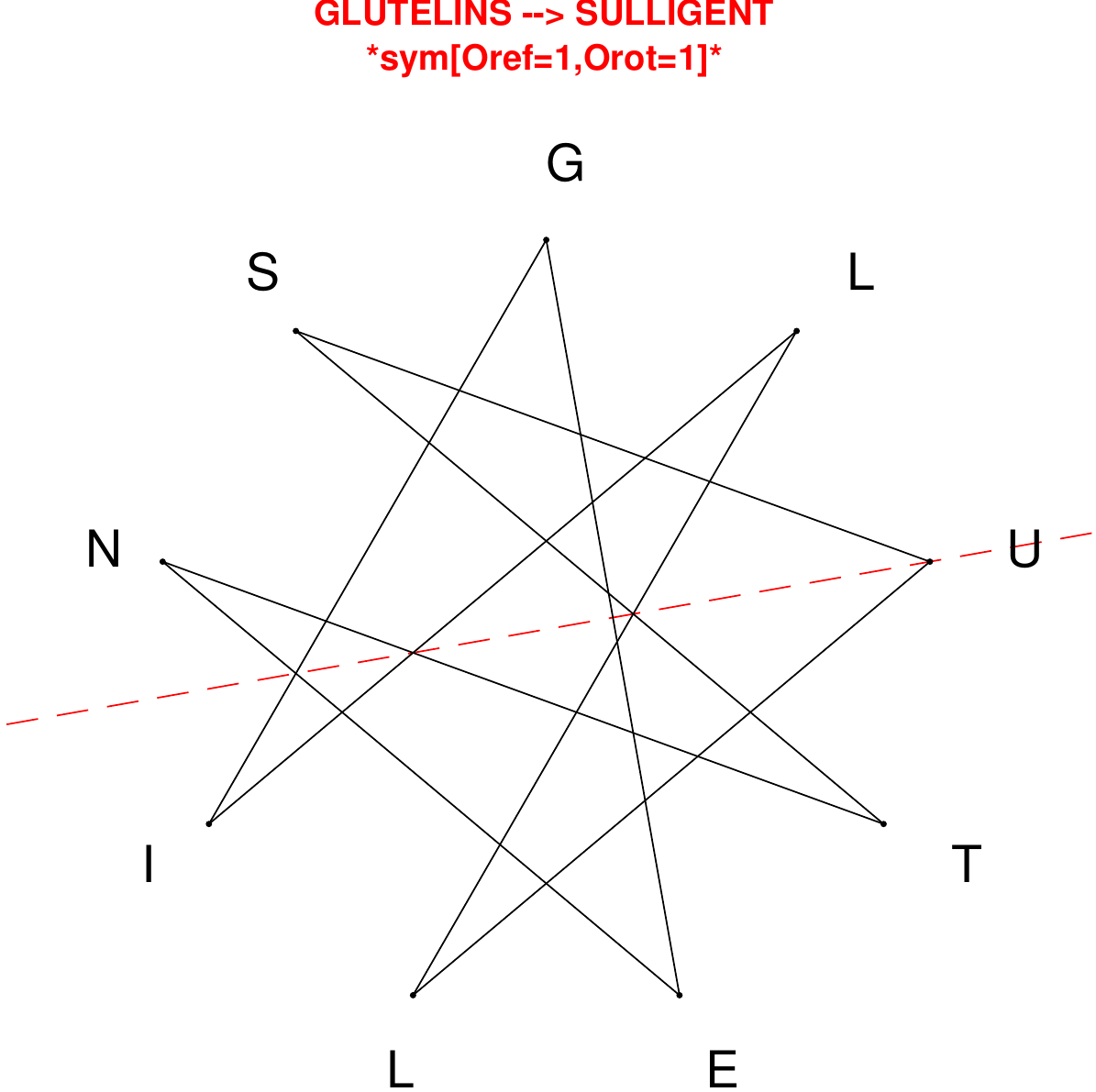}
\end{subfigure}
\hfill
\begin{subfigure}[T]{0.19\textwidth}
\centering
\includegraphics[width=\textwidth]{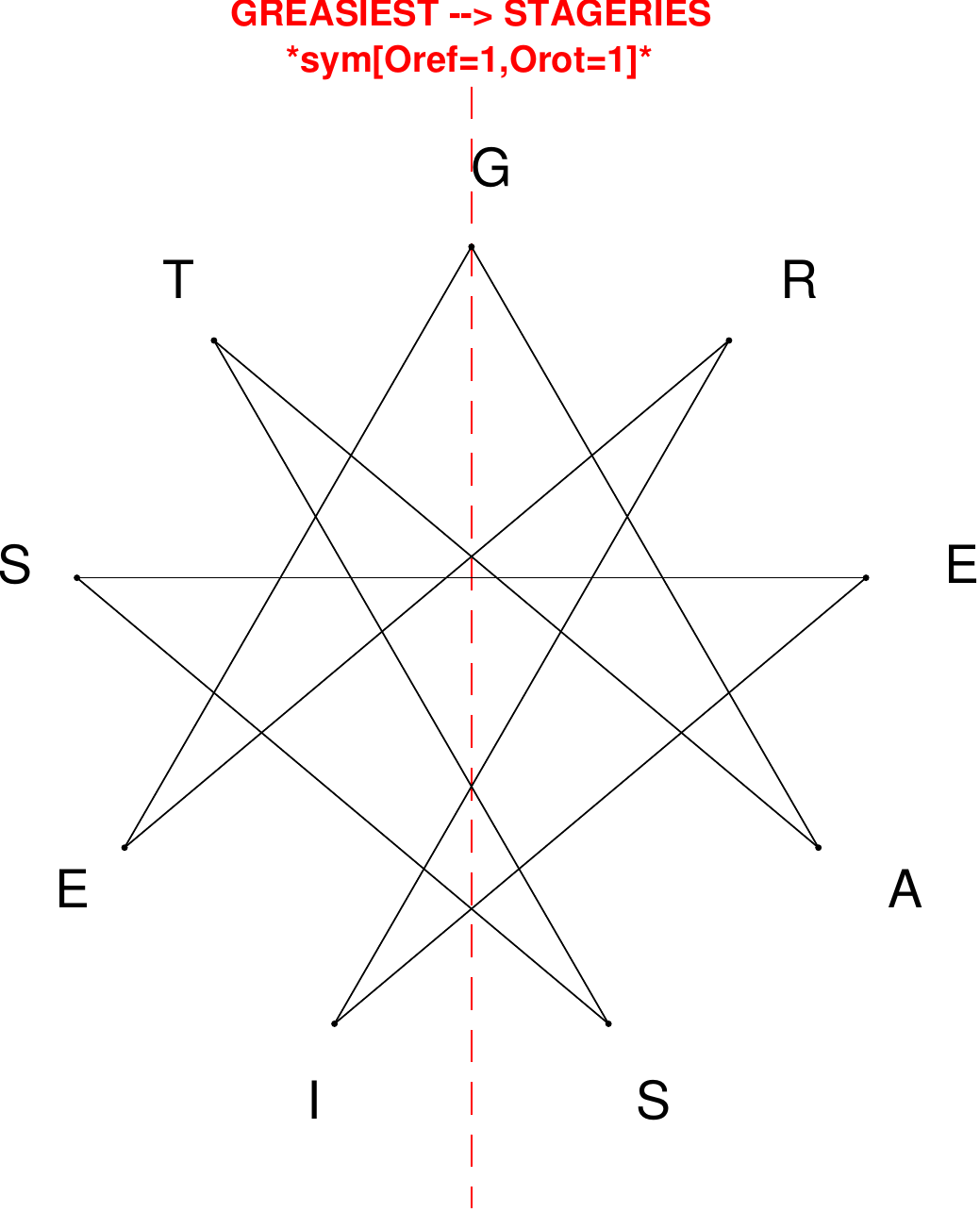}
\end{subfigure}
\hfill
\begin{subfigure}[T]{0.19\textwidth}
\centering
\includegraphics[width=\textwidth]{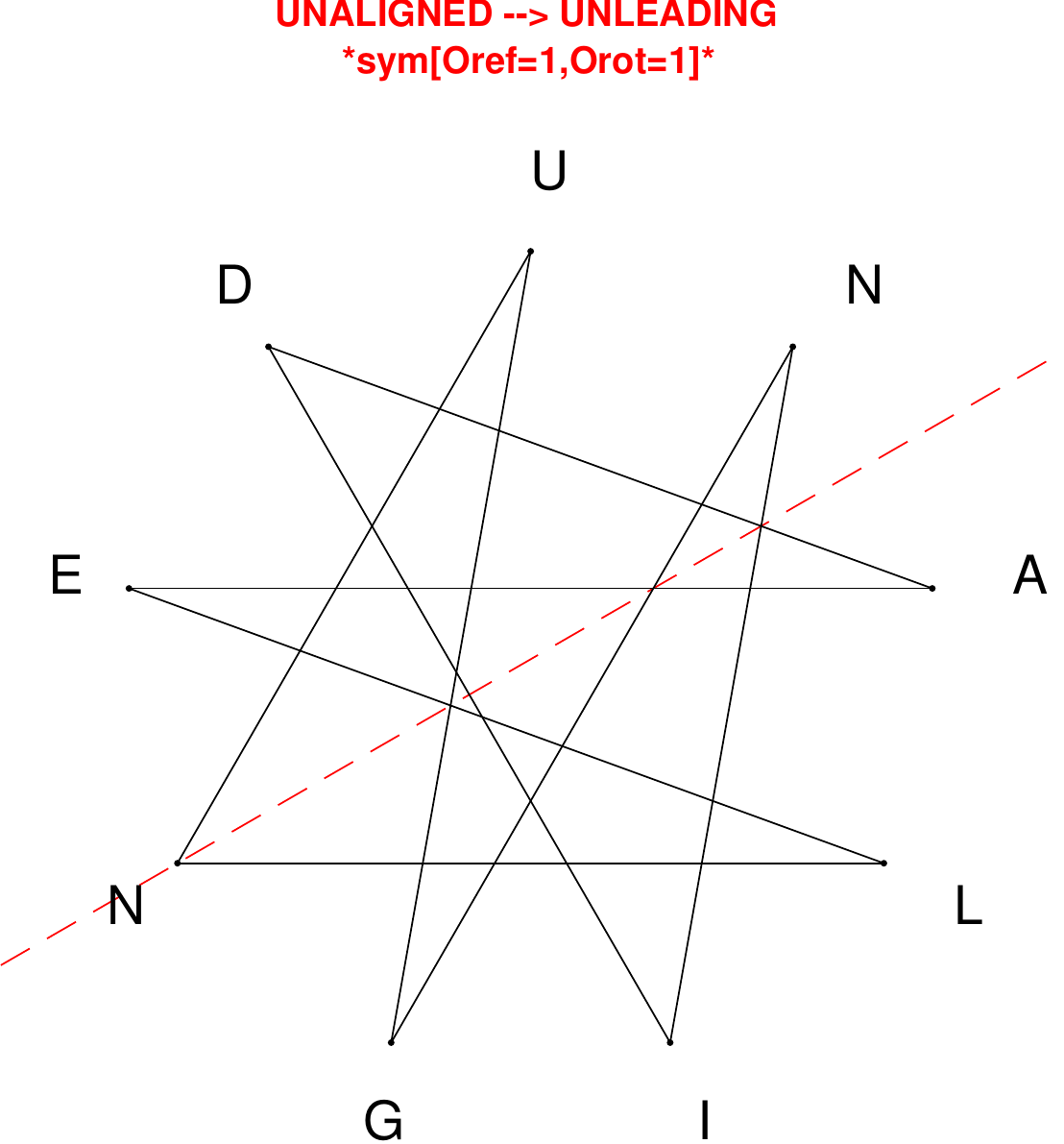}
\end{subfigure}
\hfill
\begin{subfigure}[T]{0.19\textwidth}
\centering
\includegraphics[width=\textwidth]{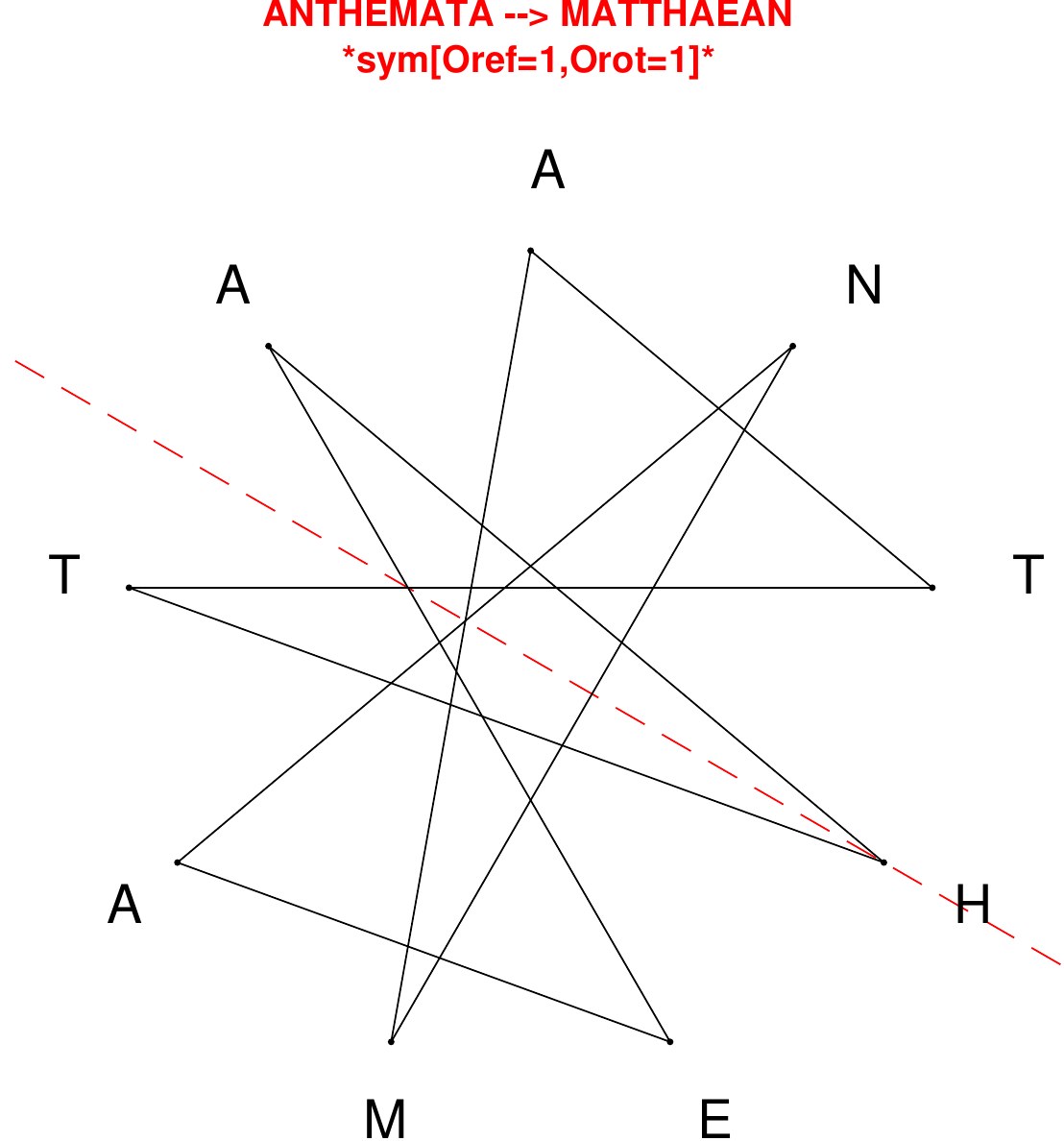}
\end{subfigure}
\hfill
\begin{subfigure}[T]{0.19\textwidth}
\centering
\includegraphics[width=\textwidth]{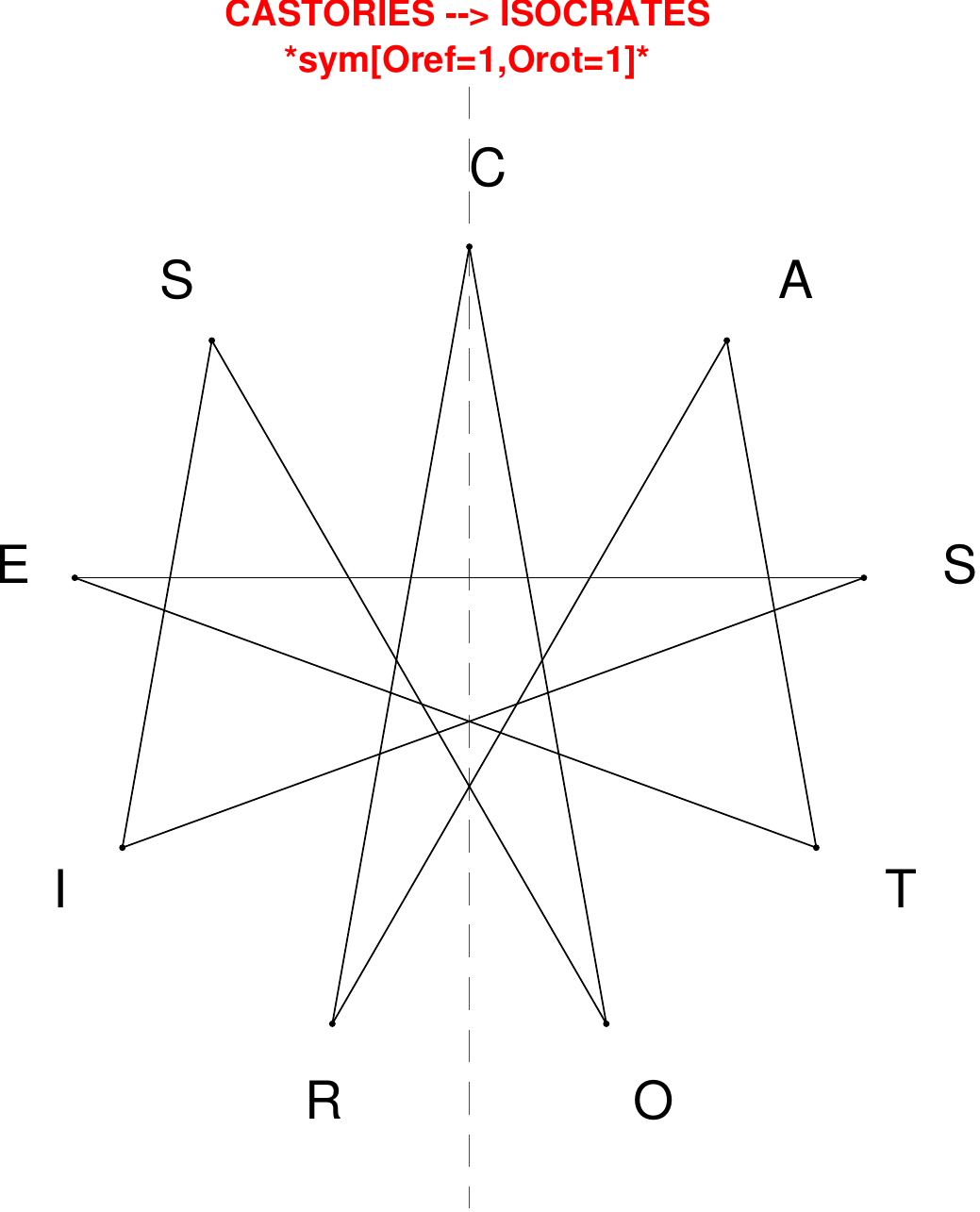}
\end{subfigure}
\end{figure}

\begin{figure}[H]
\centering
\begin{subfigure}[T]{0.19\textwidth}
\centering
\includegraphics[width=\textwidth]{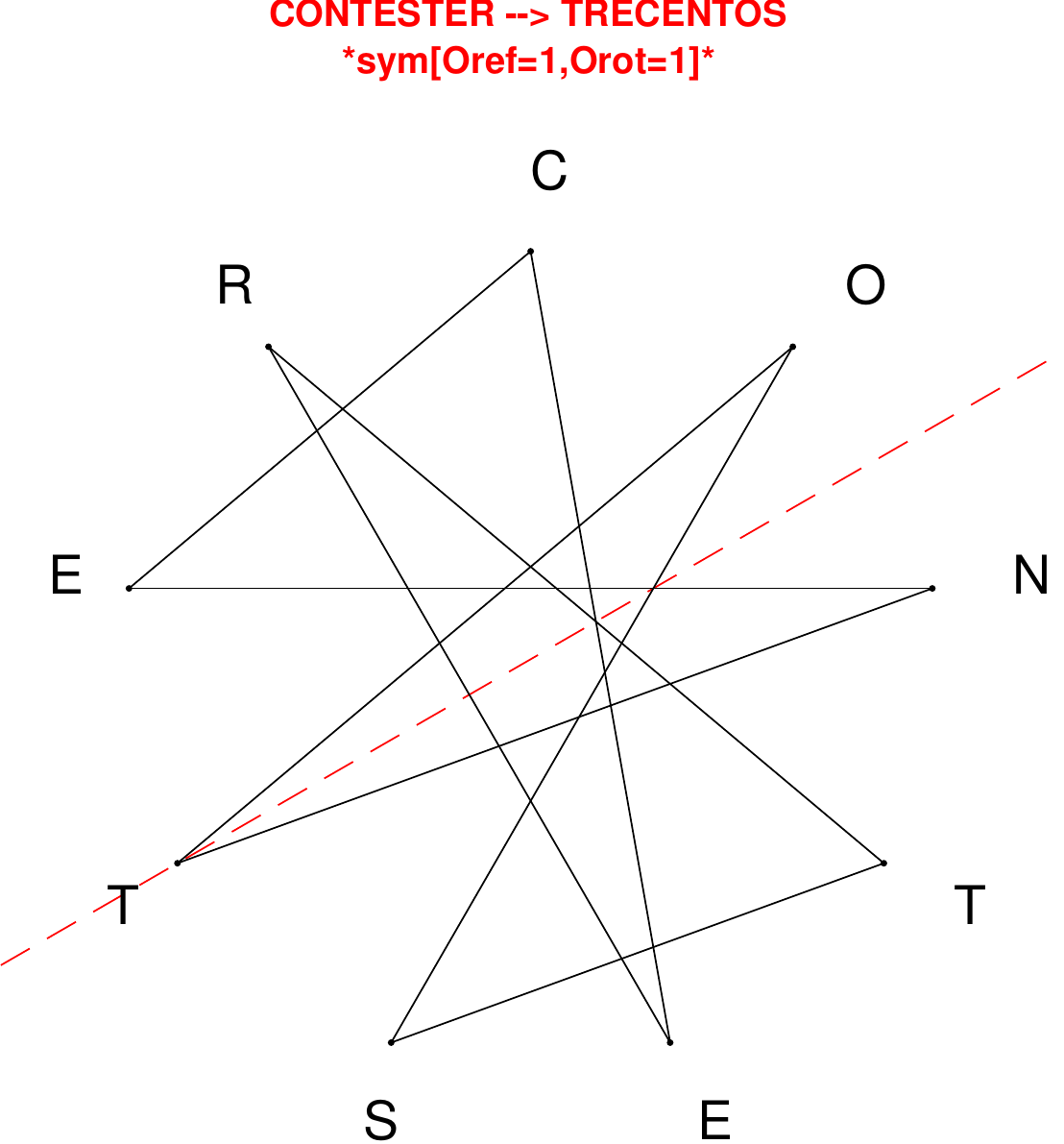}
\end{subfigure}
\hfill
\begin{subfigure}[T]{0.19\textwidth}
\centering
\includegraphics[width=\textwidth]{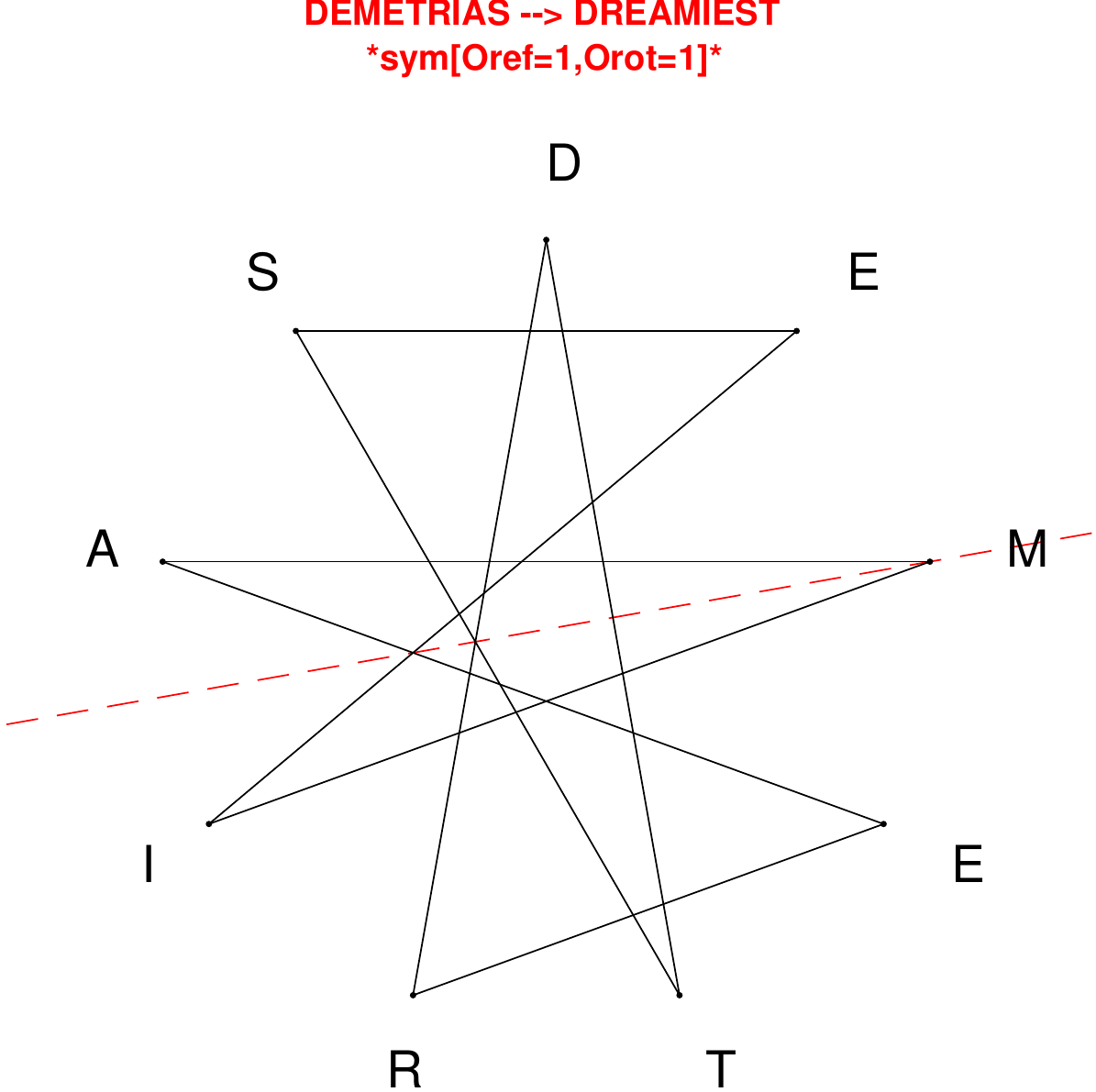}
\end{subfigure}
\hfill
\begin{subfigure}[T]{0.19\textwidth}
\centering
\includegraphics[width=\textwidth]{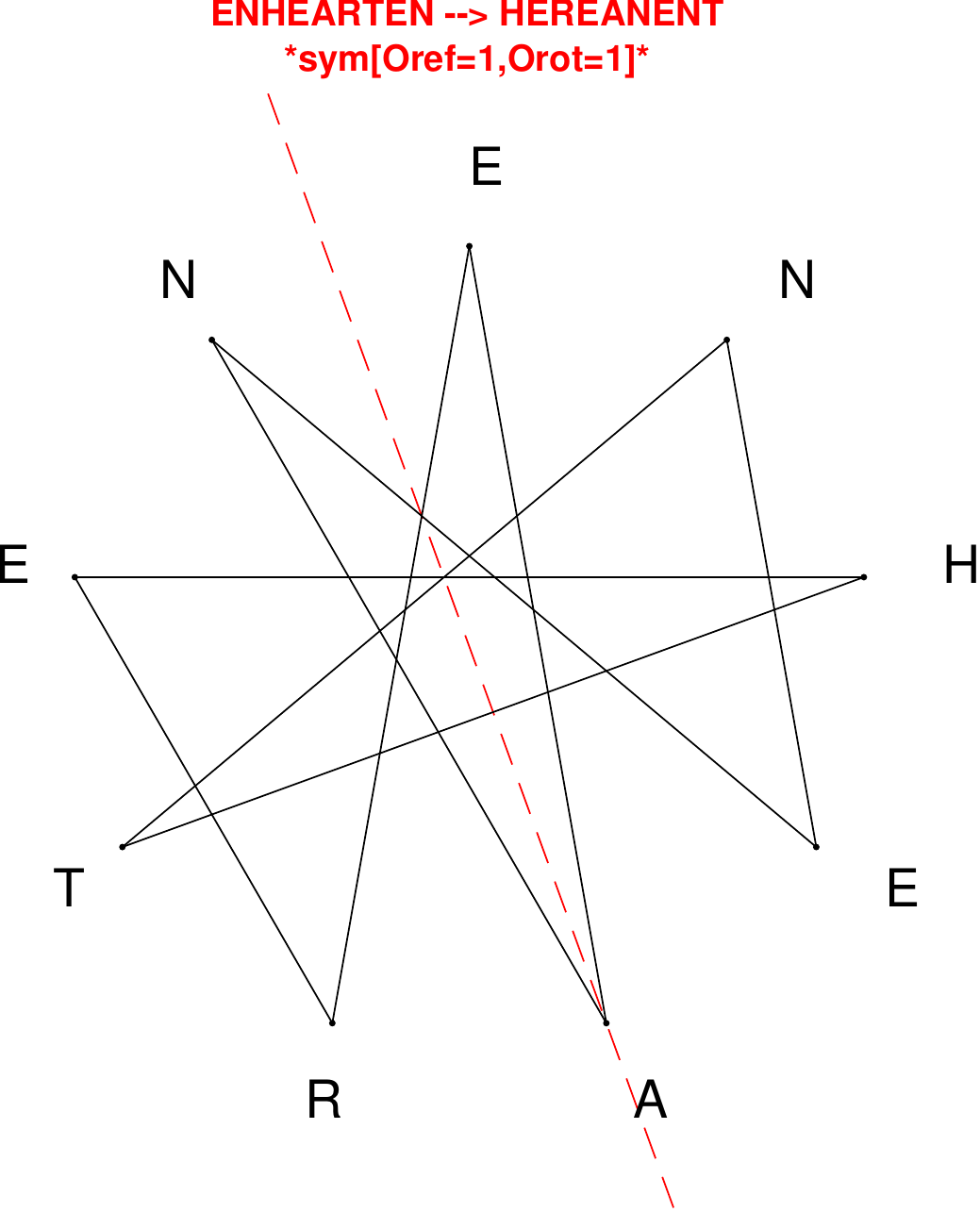}
\end{subfigure}
\hfill
\begin{subfigure}[T]{0.19\textwidth}
\centering
\includegraphics[width=\textwidth]{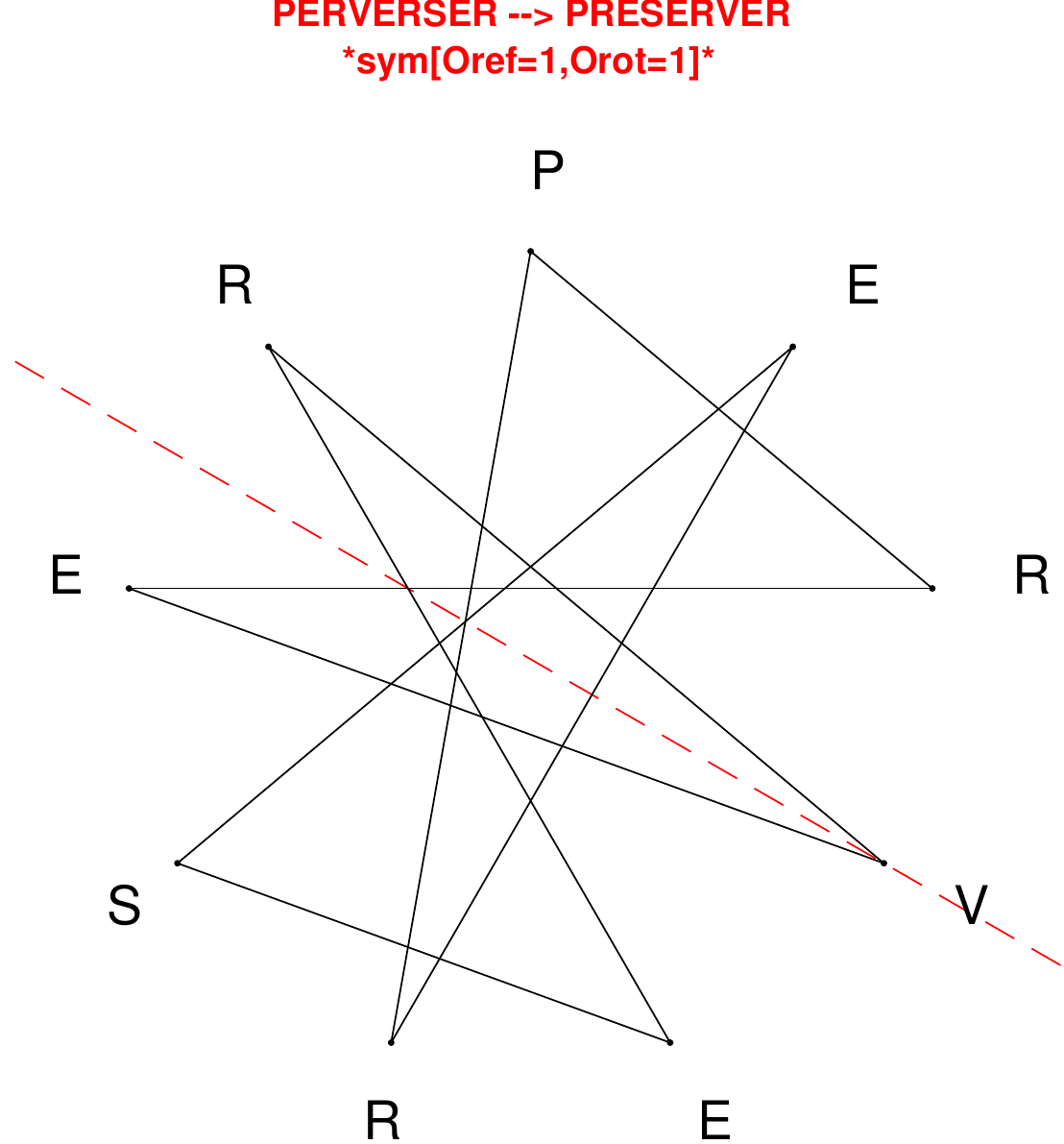}
\end{subfigure}
\hfill
\begin{subfigure}[T]{0.19\textwidth}
\centering
\includegraphics[width=\textwidth]{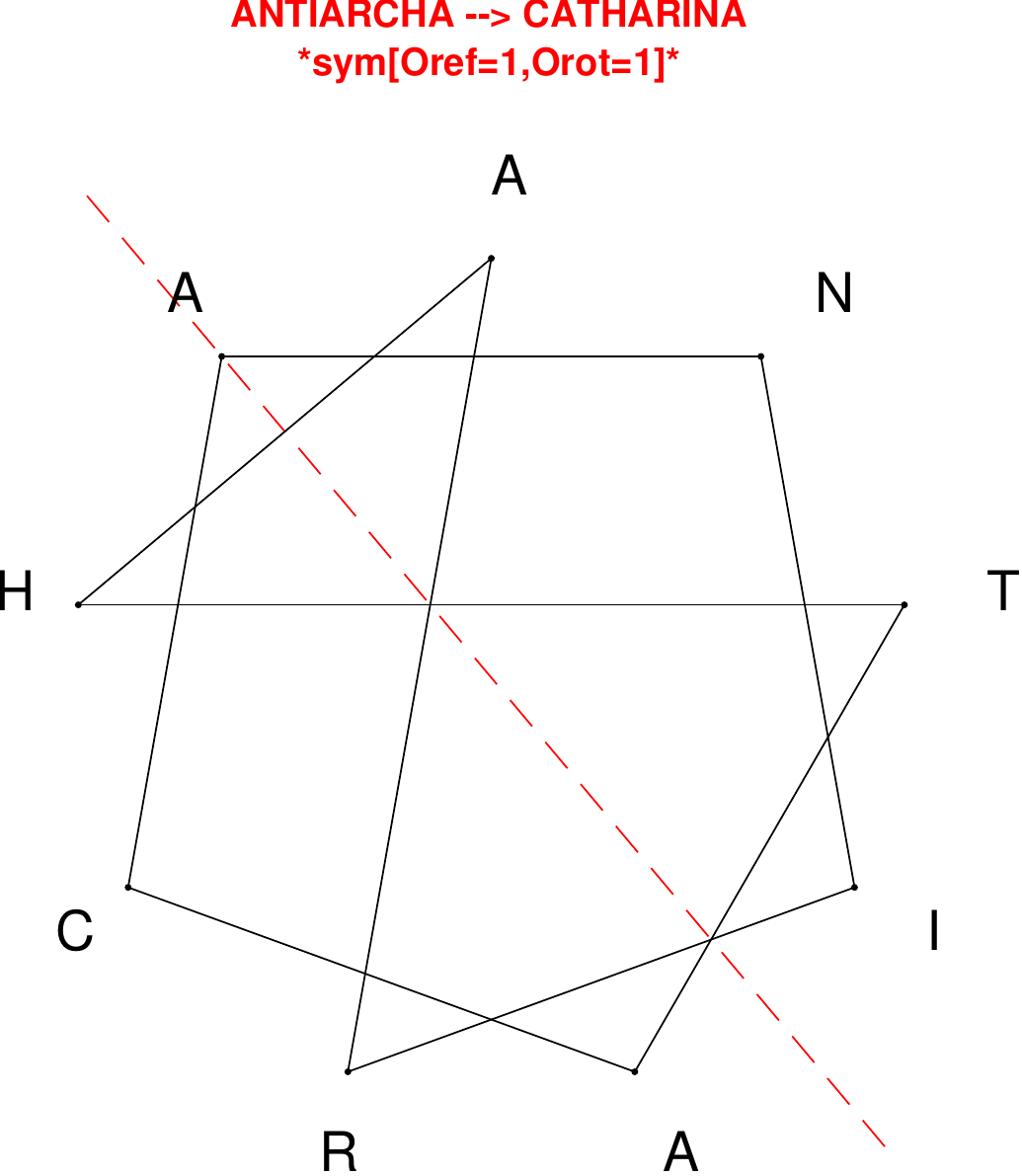}
\end{subfigure}
\end{figure}

\begin{figure}[H]
\centering
\begin{subfigure}[T]{0.19\textwidth}
\centering
\includegraphics[width=\textwidth]{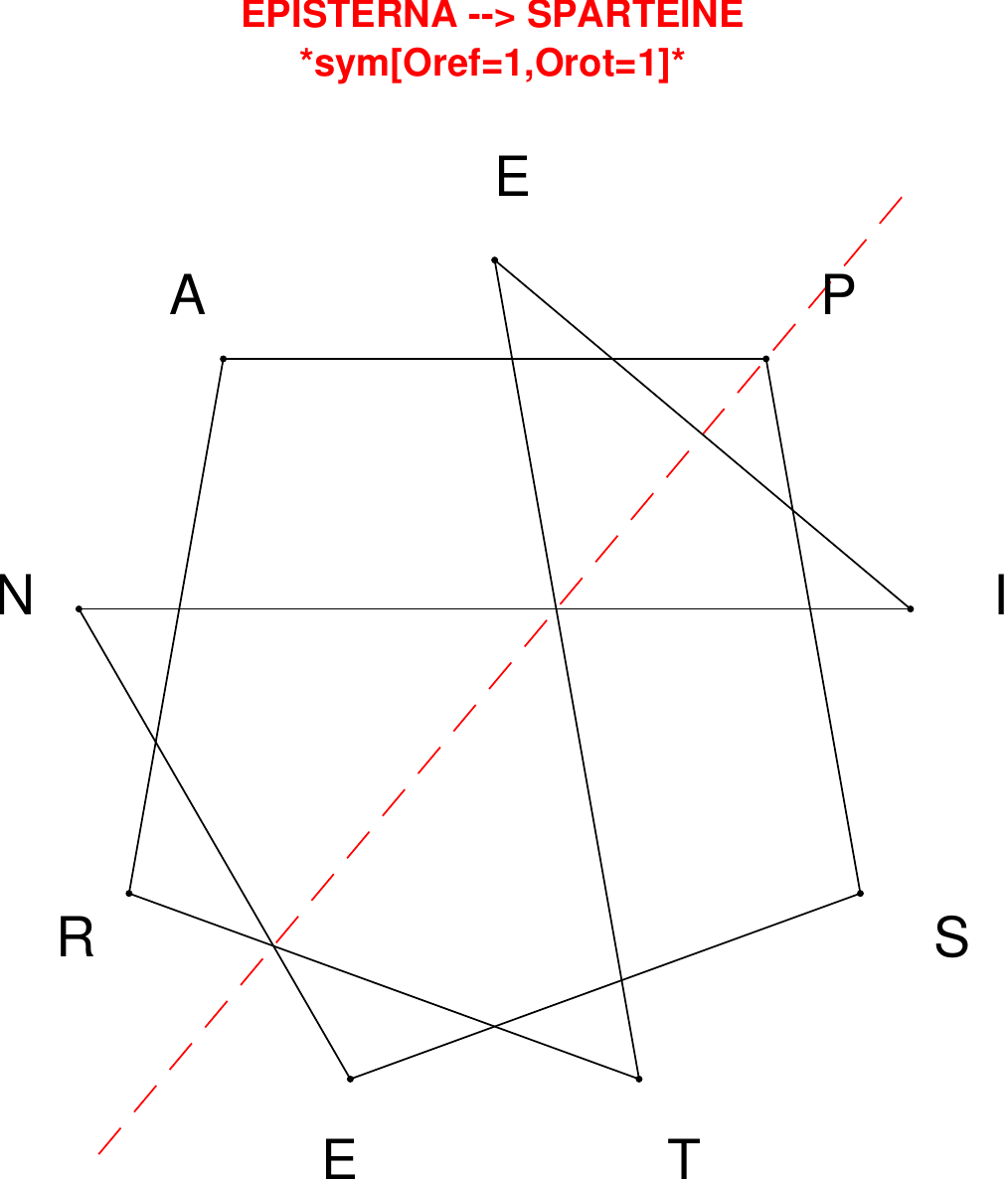}
\end{subfigure}
\hfill
\begin{subfigure}[T]{0.19\textwidth}
\centering
\includegraphics[width=\textwidth]{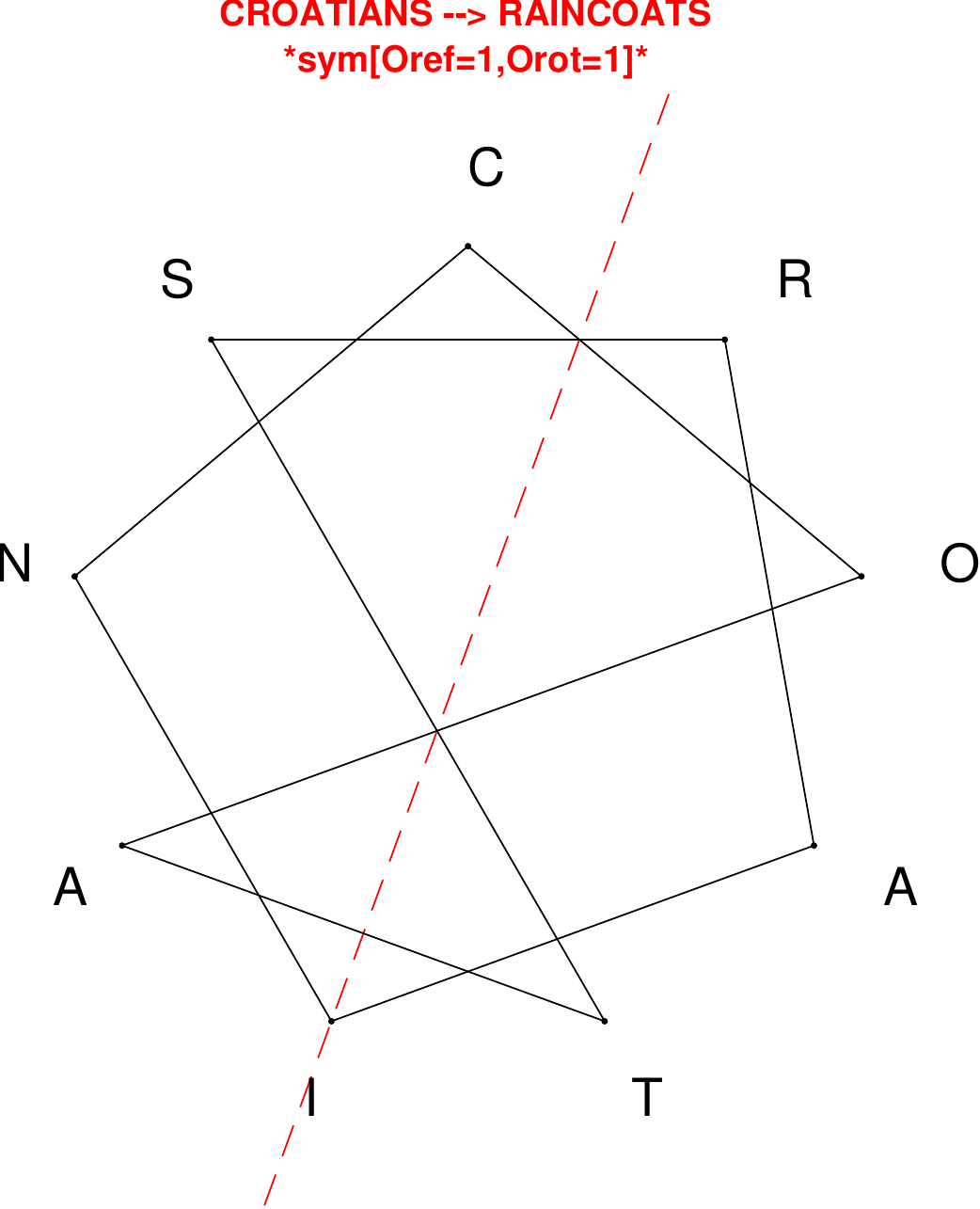}
\end{subfigure}
\hfill
\begin{subfigure}[T]{0.19\textwidth}
\centering
\includegraphics[width=\textwidth]{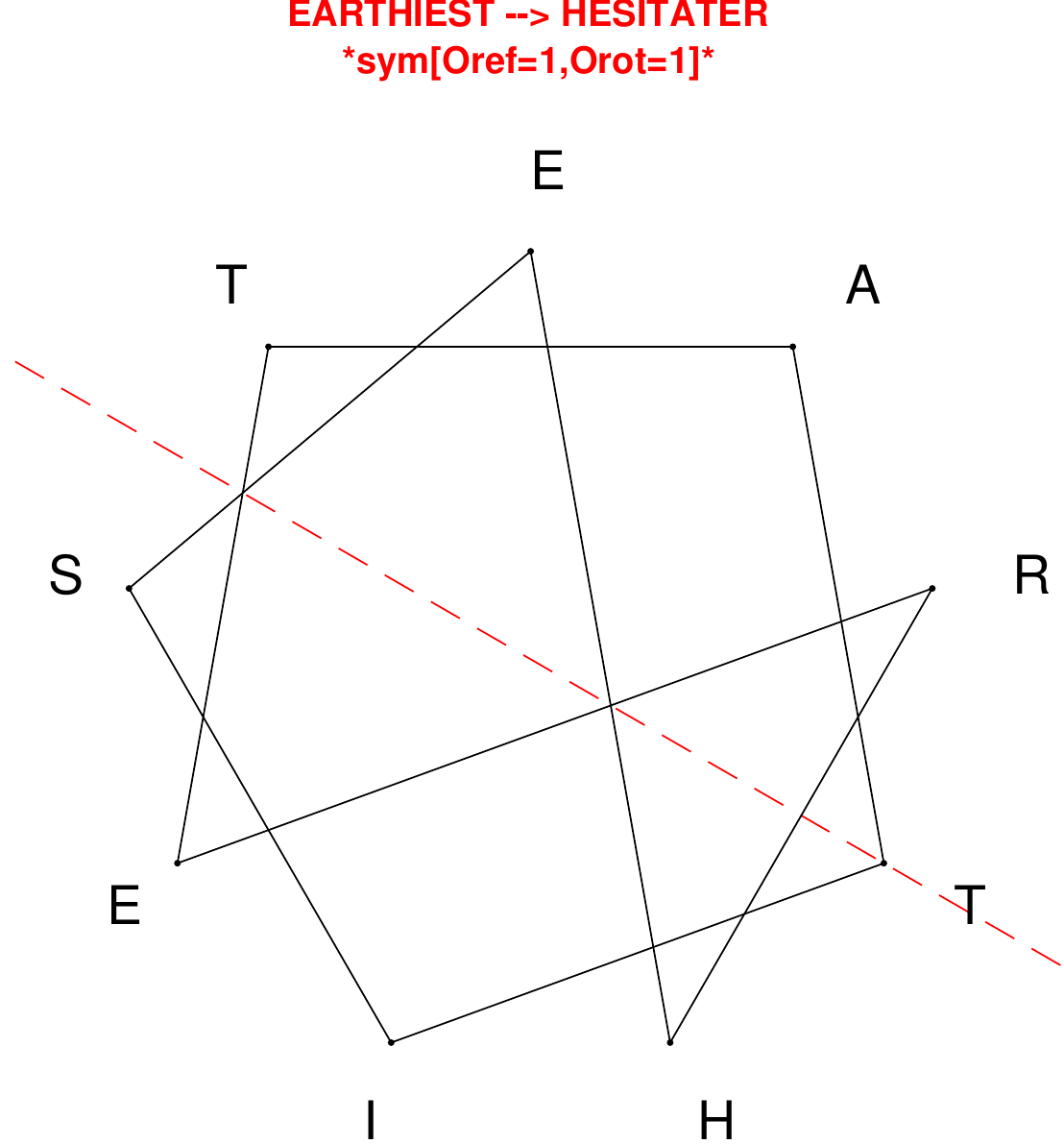}
\end{subfigure}
\hfill
\begin{subfigure}[T]{0.19\textwidth}
\centering
\includegraphics[width=\textwidth]{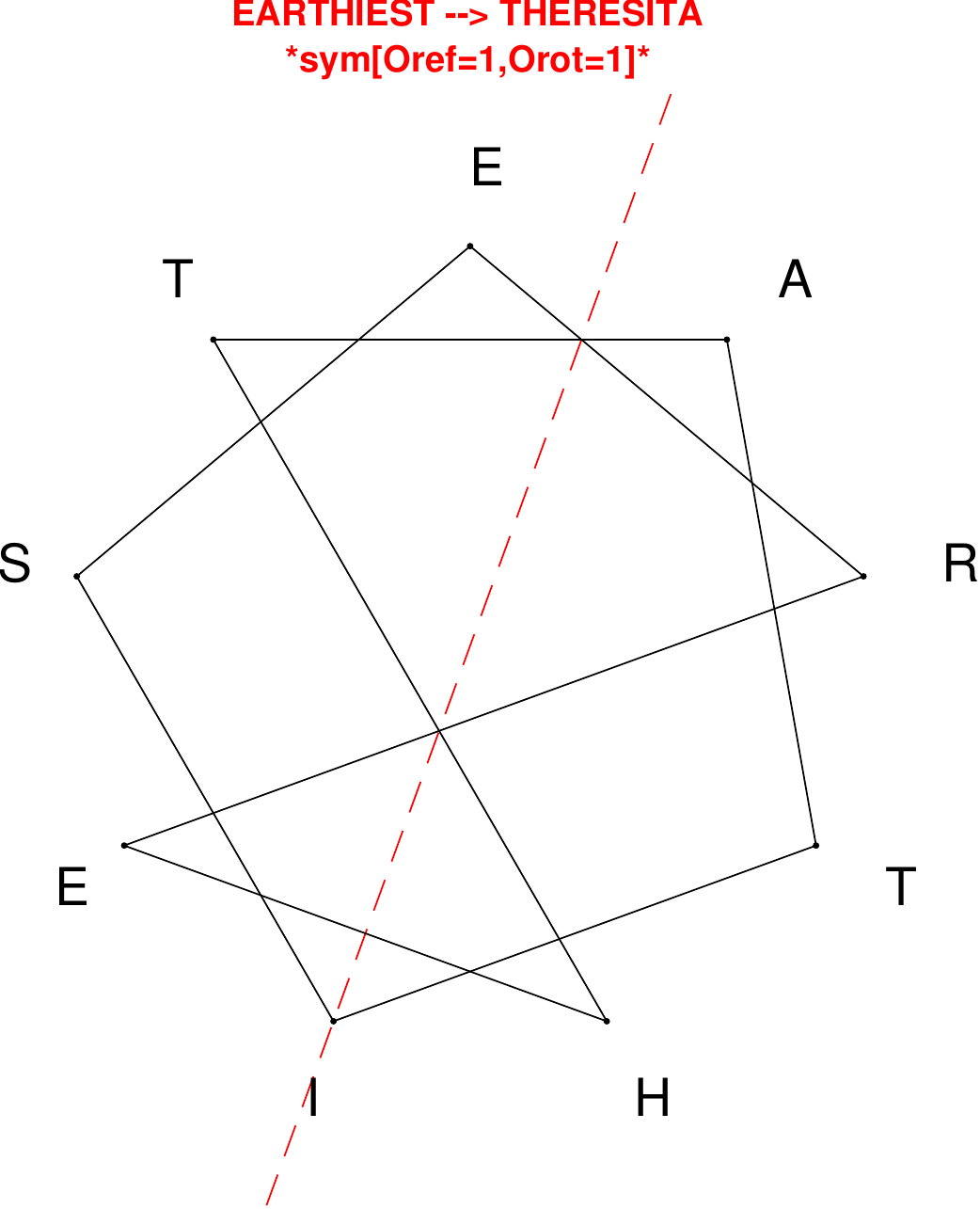}
\end{subfigure}
\hfill
\begin{subfigure}[T]{0.19\textwidth}
\centering
\includegraphics[width=\textwidth]{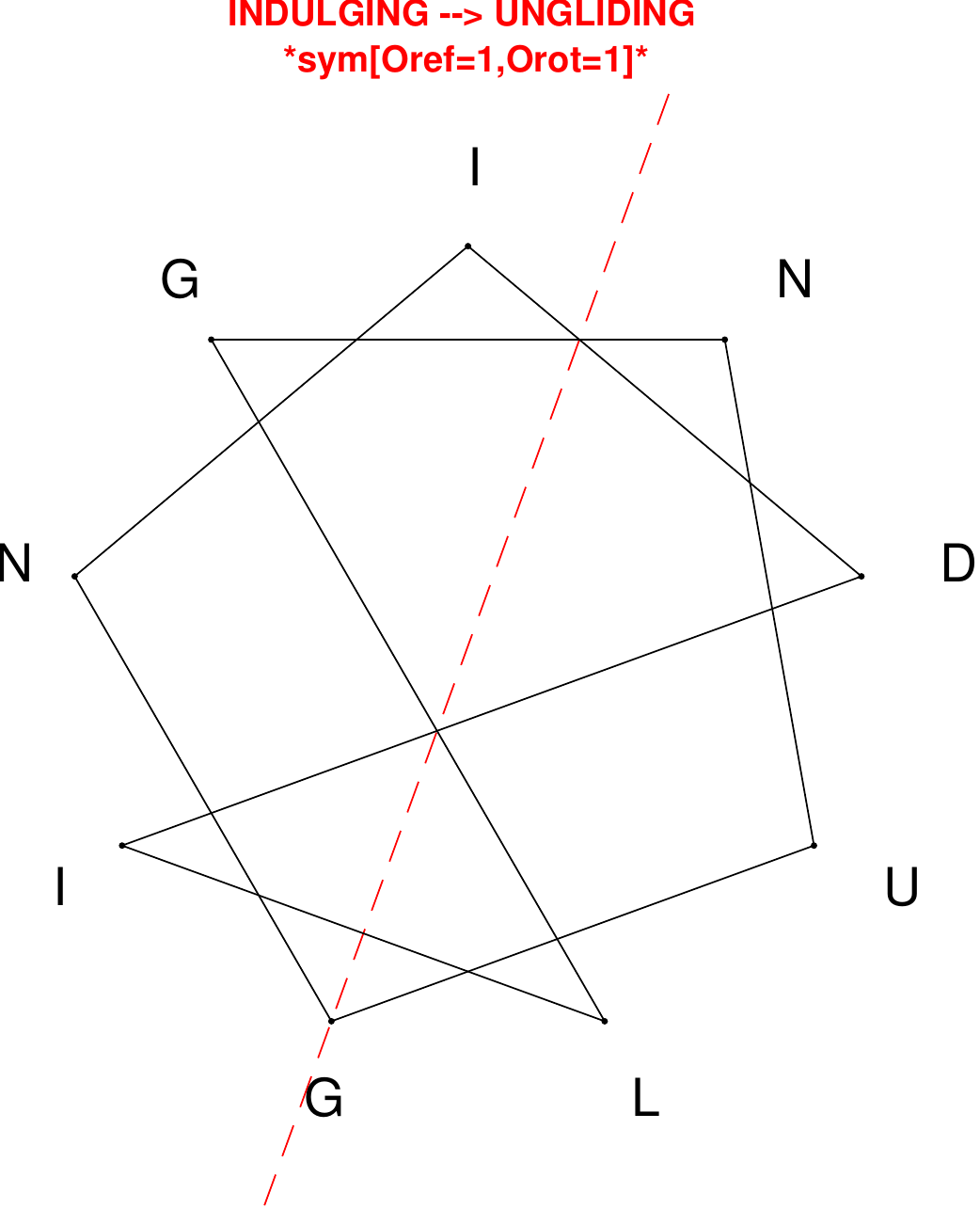}
\end{subfigure}
\end{figure}

\begin{figure}[H]
\centering
\begin{subfigure}[T]{0.19\textwidth}
\centering
\includegraphics[width=\textwidth]{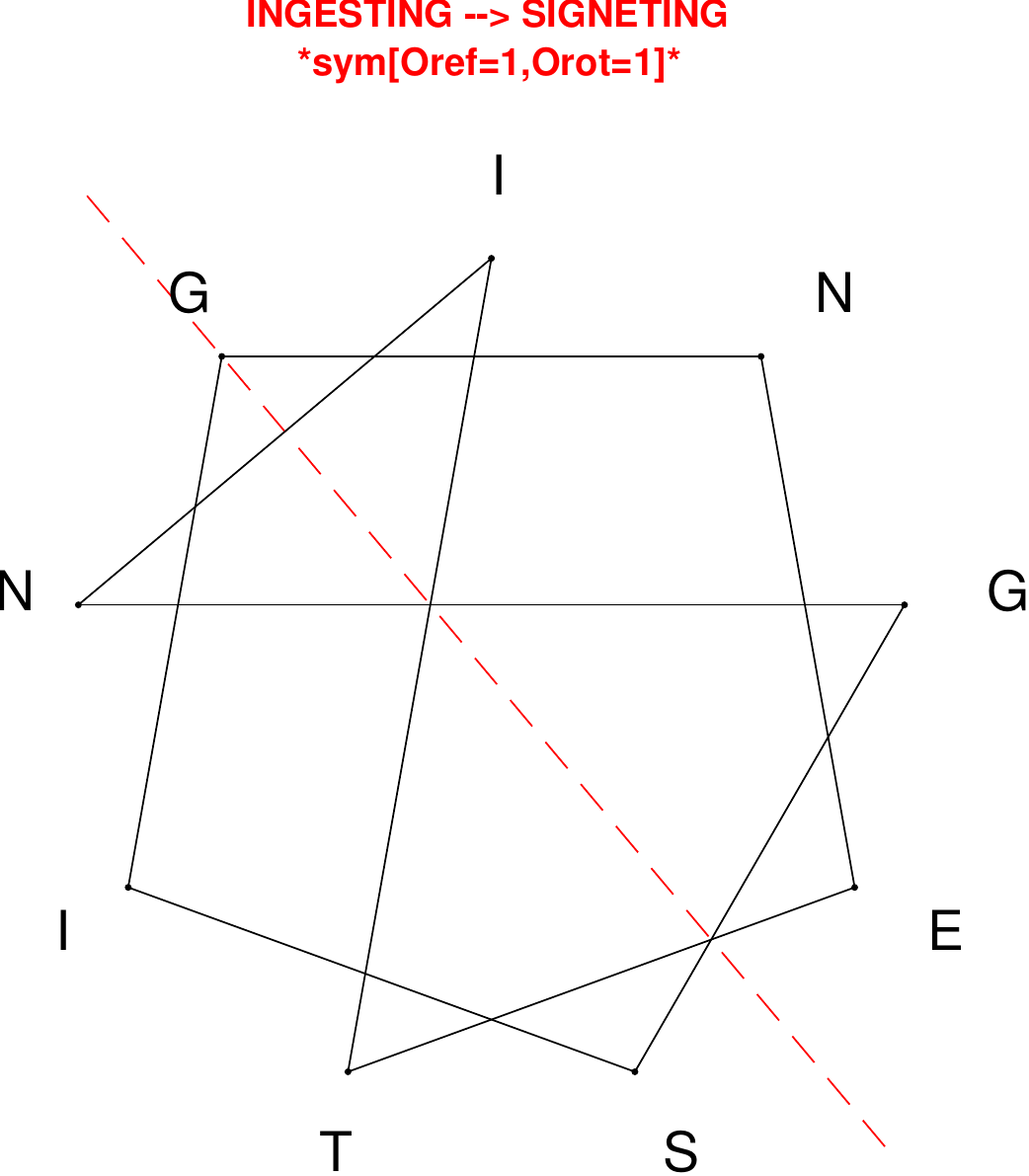}
\end{subfigure}
\hfill
\begin{subfigure}[T]{0.19\textwidth}
\centering
\includegraphics[width=\textwidth]{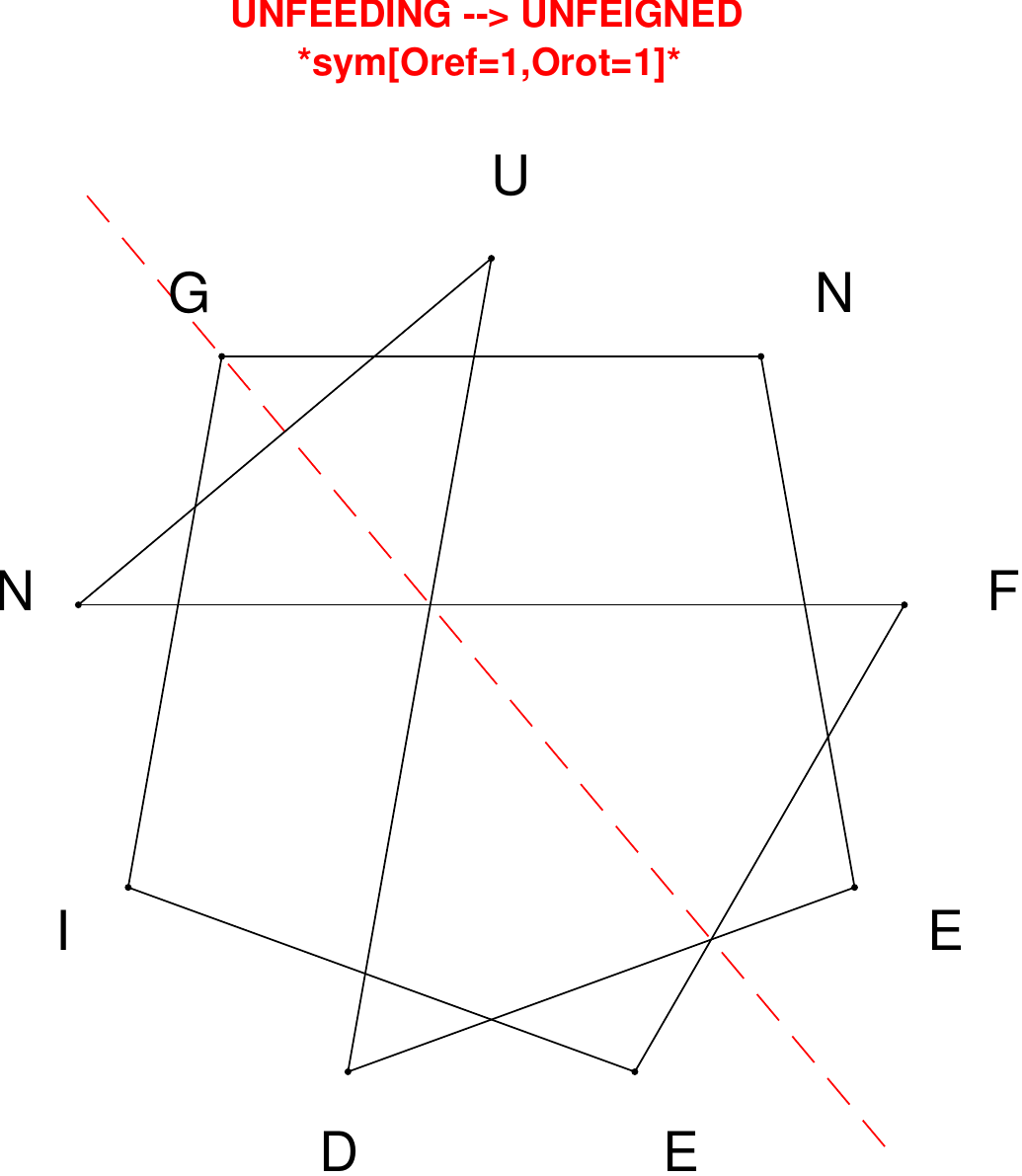}
\end{subfigure}
\hfill
\begin{subfigure}[T]{0.19\textwidth}
\centering
\includegraphics[width=\textwidth]{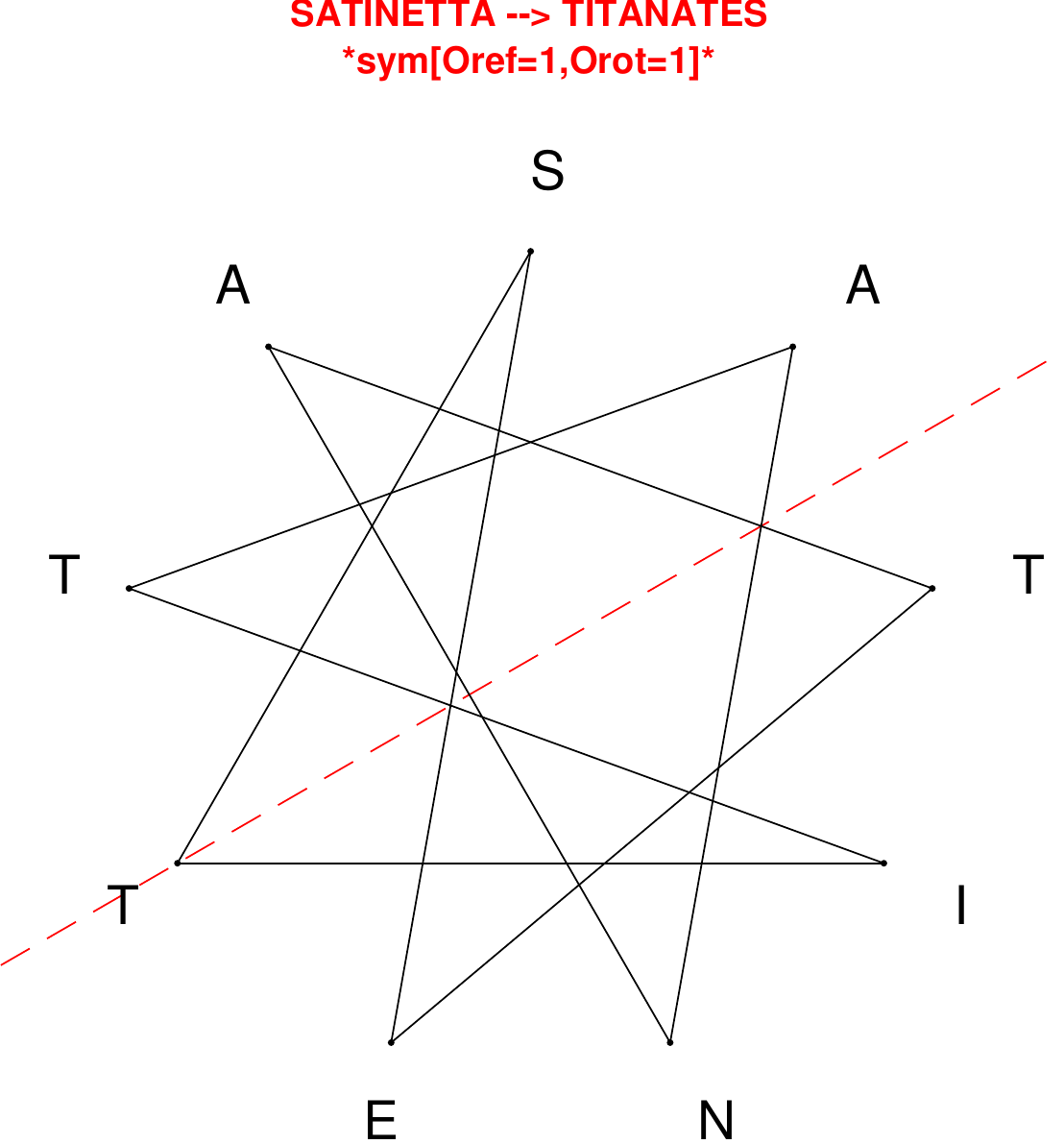}
\end{subfigure}
\hfill
\begin{subfigure}[T]{0.19\textwidth}
\centering
\includegraphics[width=\textwidth]{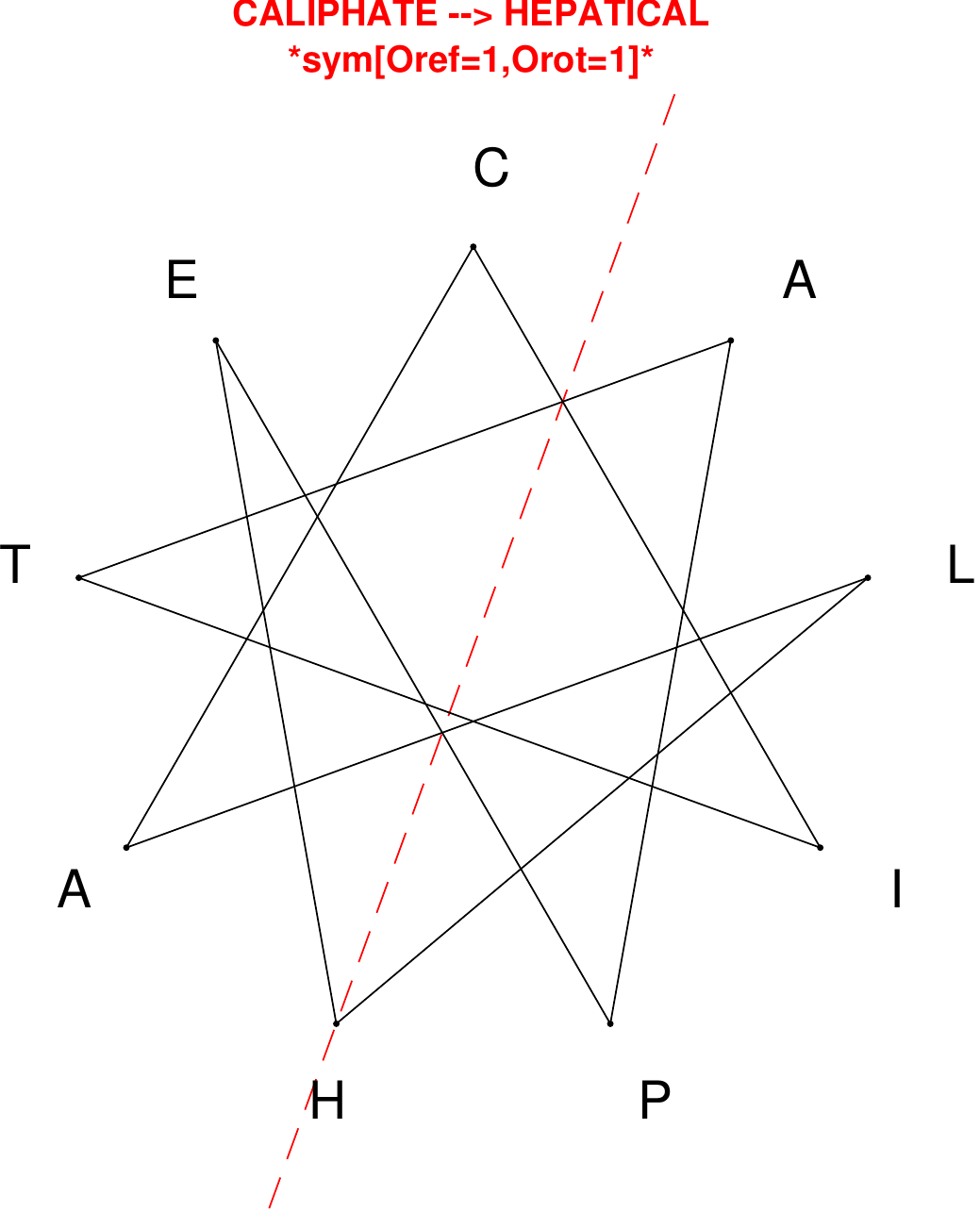}
\end{subfigure}
\hfill
\begin{subfigure}[T]{0.19\textwidth}
\centering
\includegraphics[width=\textwidth]{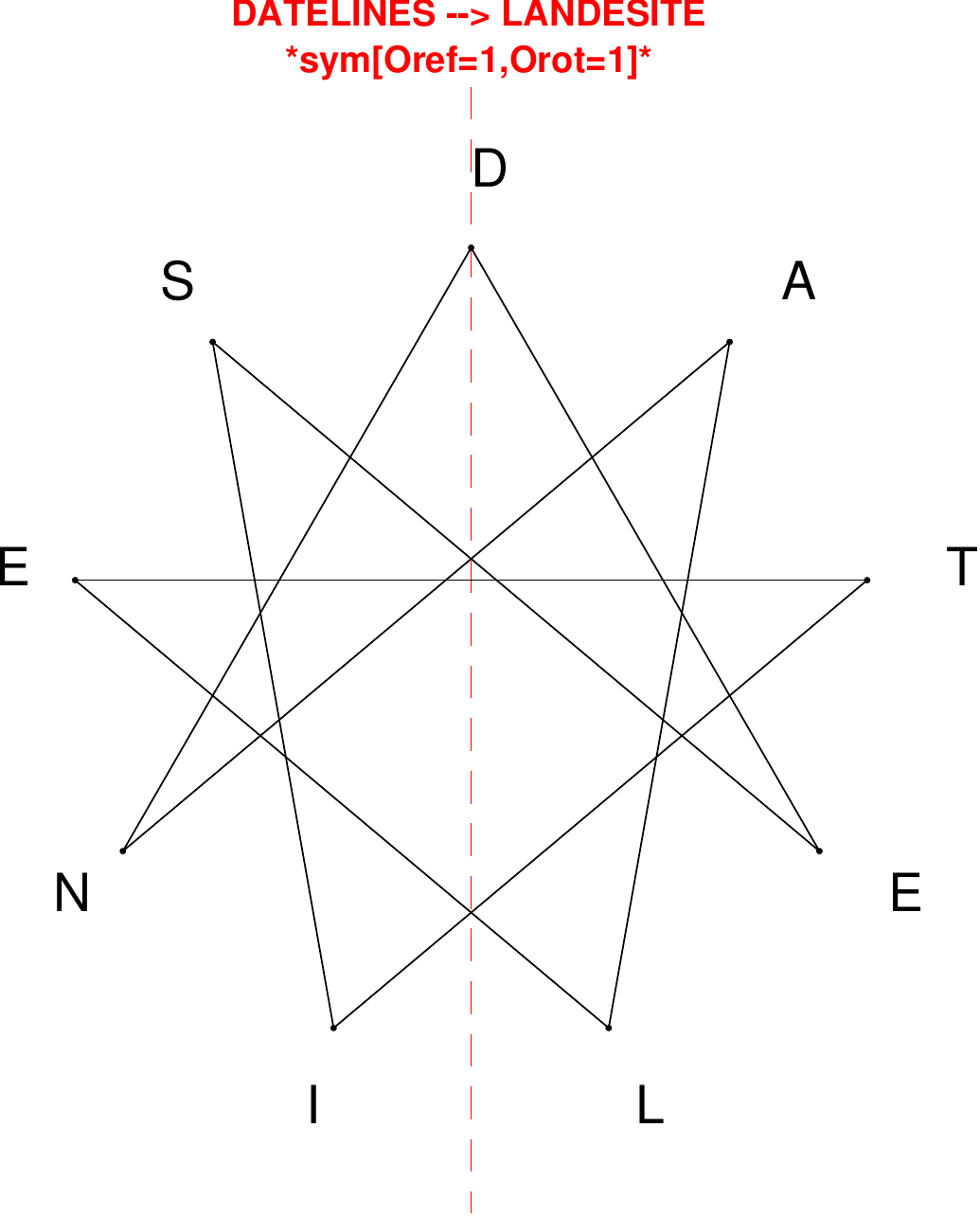}
\end{subfigure}
\end{figure}

\begin{figure}[H]
\centering
\begin{subfigure}[T]{0.19\textwidth}
\centering
\includegraphics[width=\textwidth]{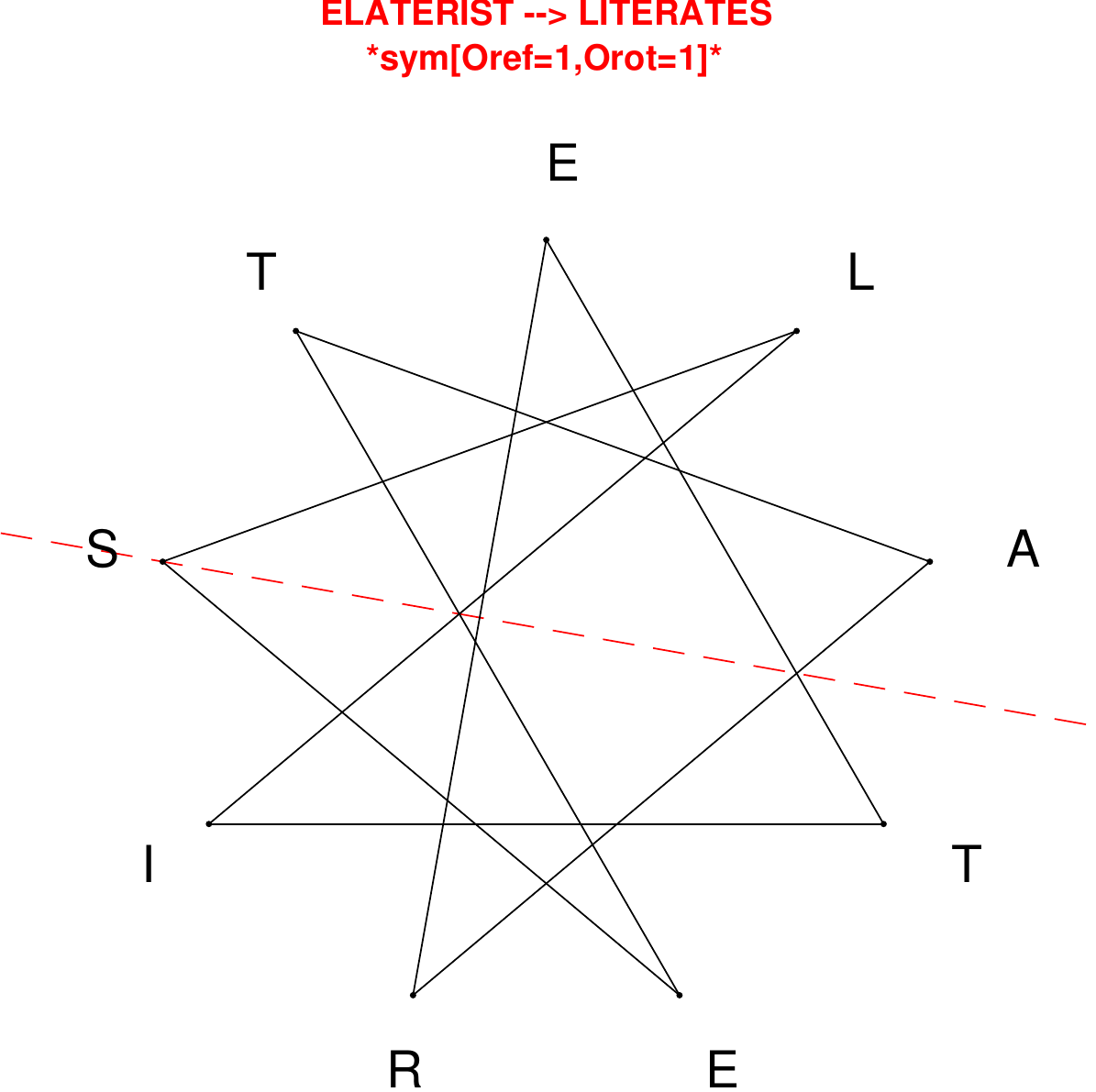}
\end{subfigure}
\hfill
\begin{subfigure}[T]{0.19\textwidth}
\centering
\includegraphics[width=\textwidth]{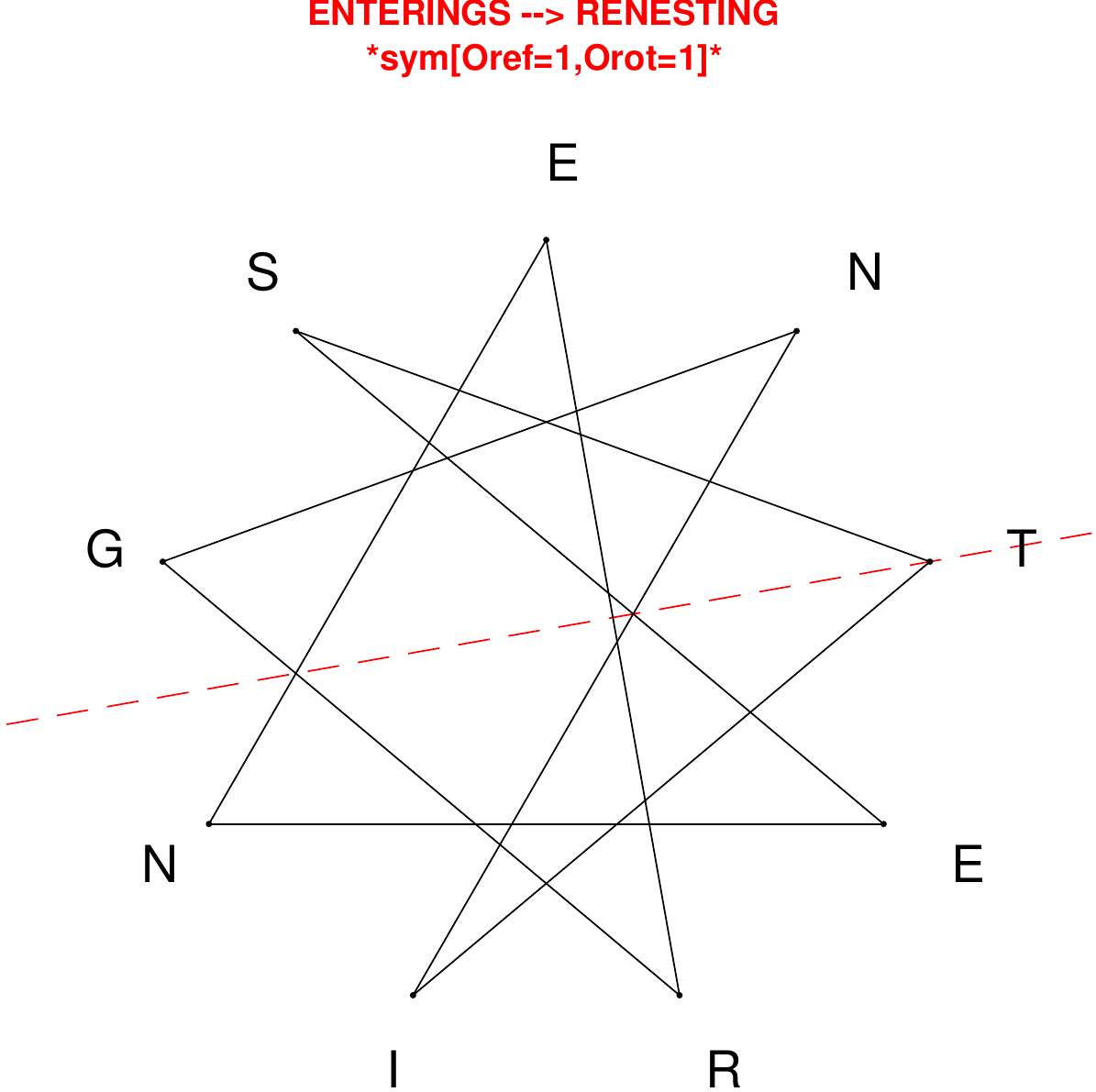}
\end{subfigure}
\hfill
\begin{subfigure}[T]{0.19\textwidth}
\centering
\includegraphics[width=\textwidth]{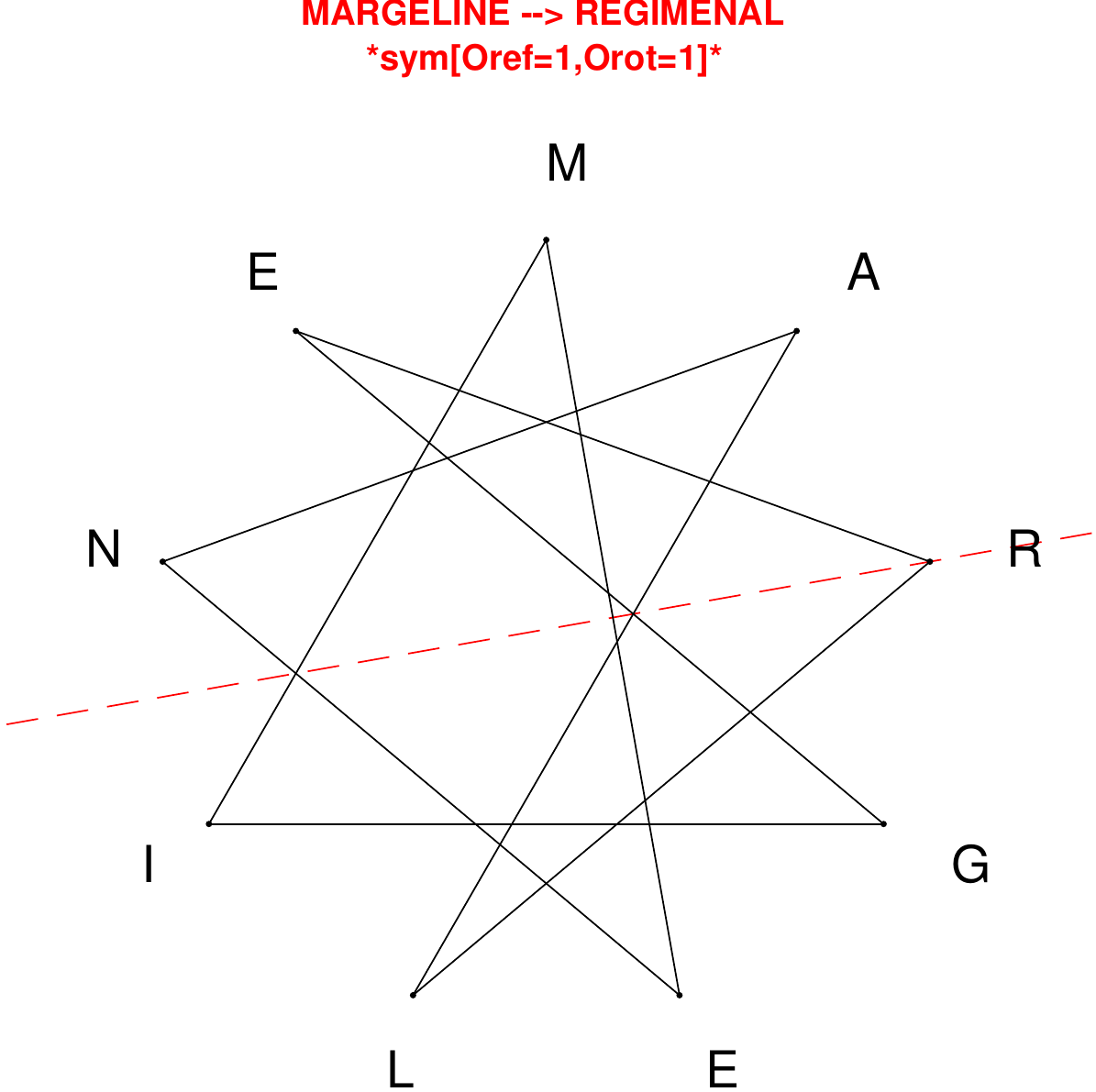}
\end{subfigure}
\hfill
\begin{subfigure}[T]{0.19\textwidth}
\centering
\includegraphics[width=\textwidth]{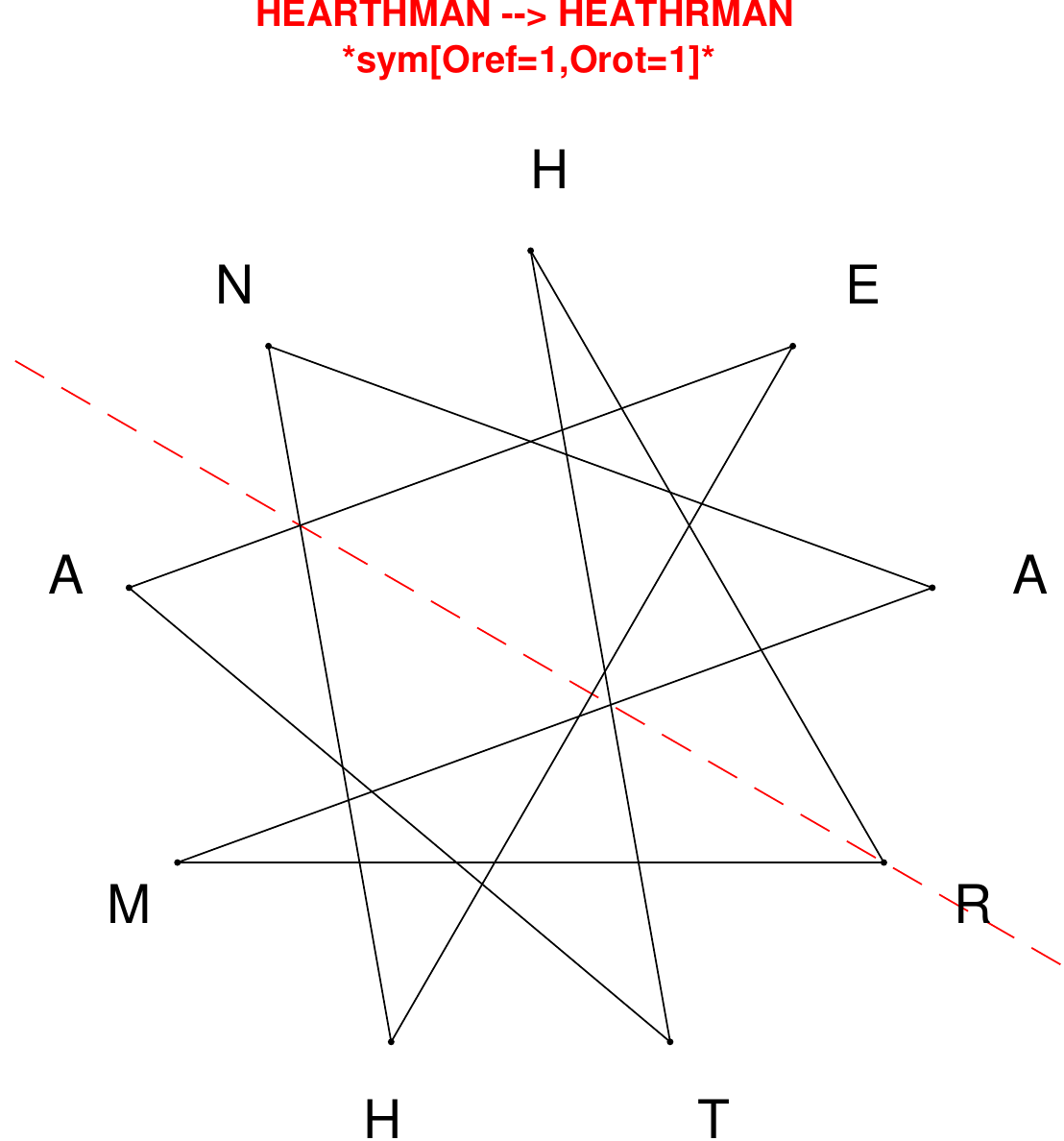}
\end{subfigure}
\hfill
\begin{subfigure}[T]{0.19\textwidth}
\centering
\includegraphics[width=\textwidth]{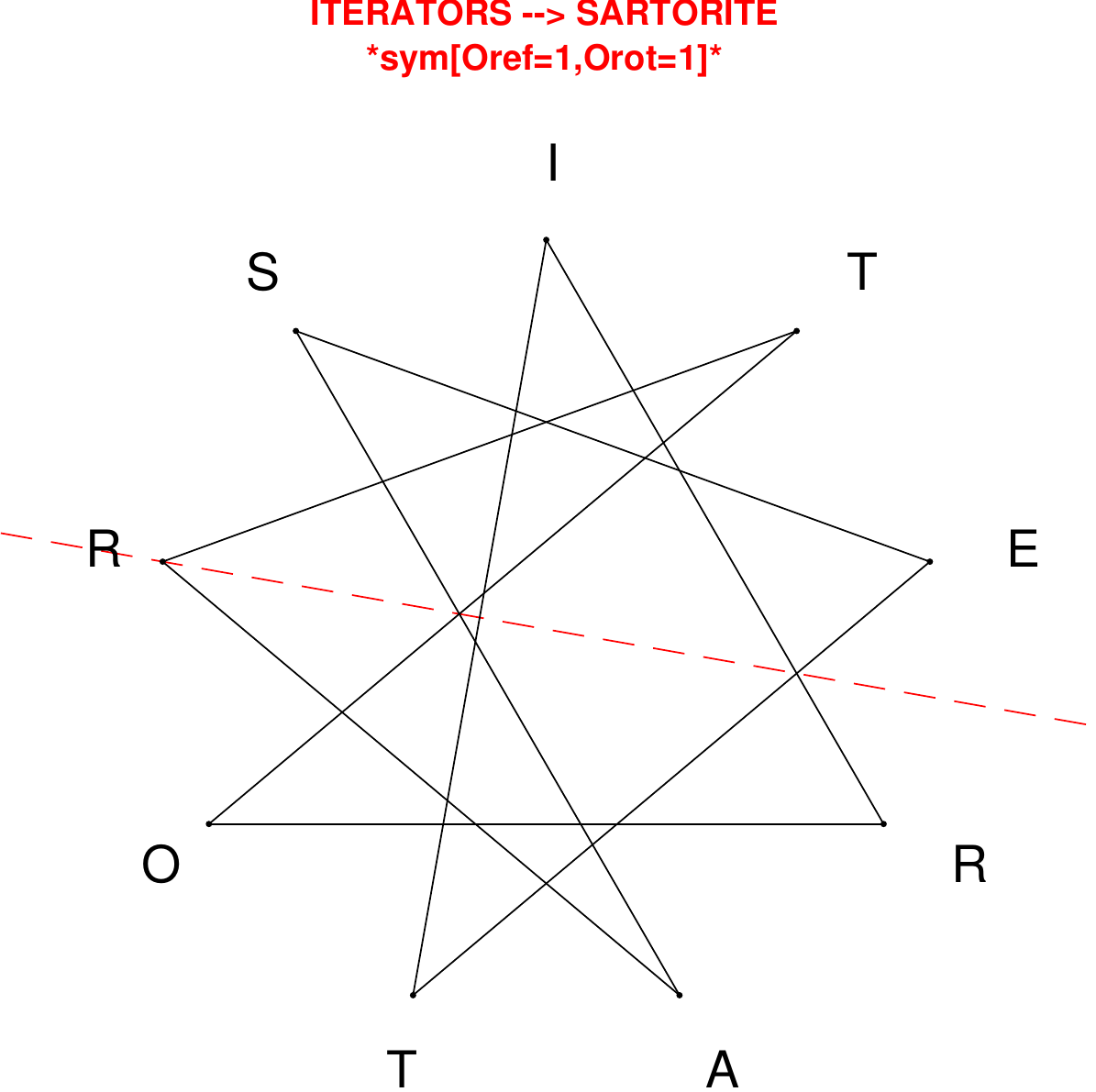}
\end{subfigure}
\end{figure}

\begin{figure}[H]
\centering
\begin{subfigure}[T]{0.19\textwidth}
\centering
\includegraphics[width=\textwidth]{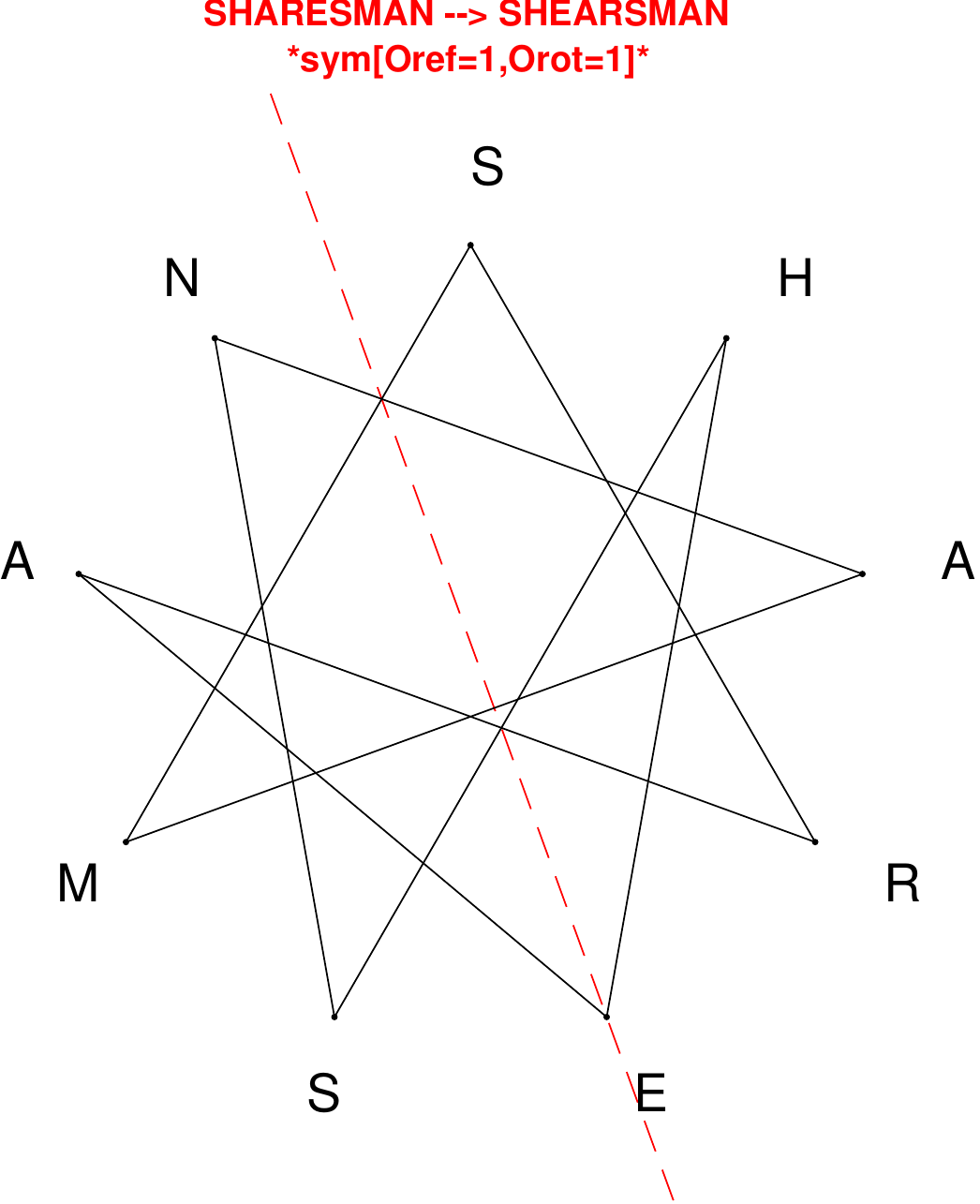}
\end{subfigure}
\hfill
\begin{subfigure}[T]{0.19\textwidth}
\centering
\includegraphics[width=\textwidth]{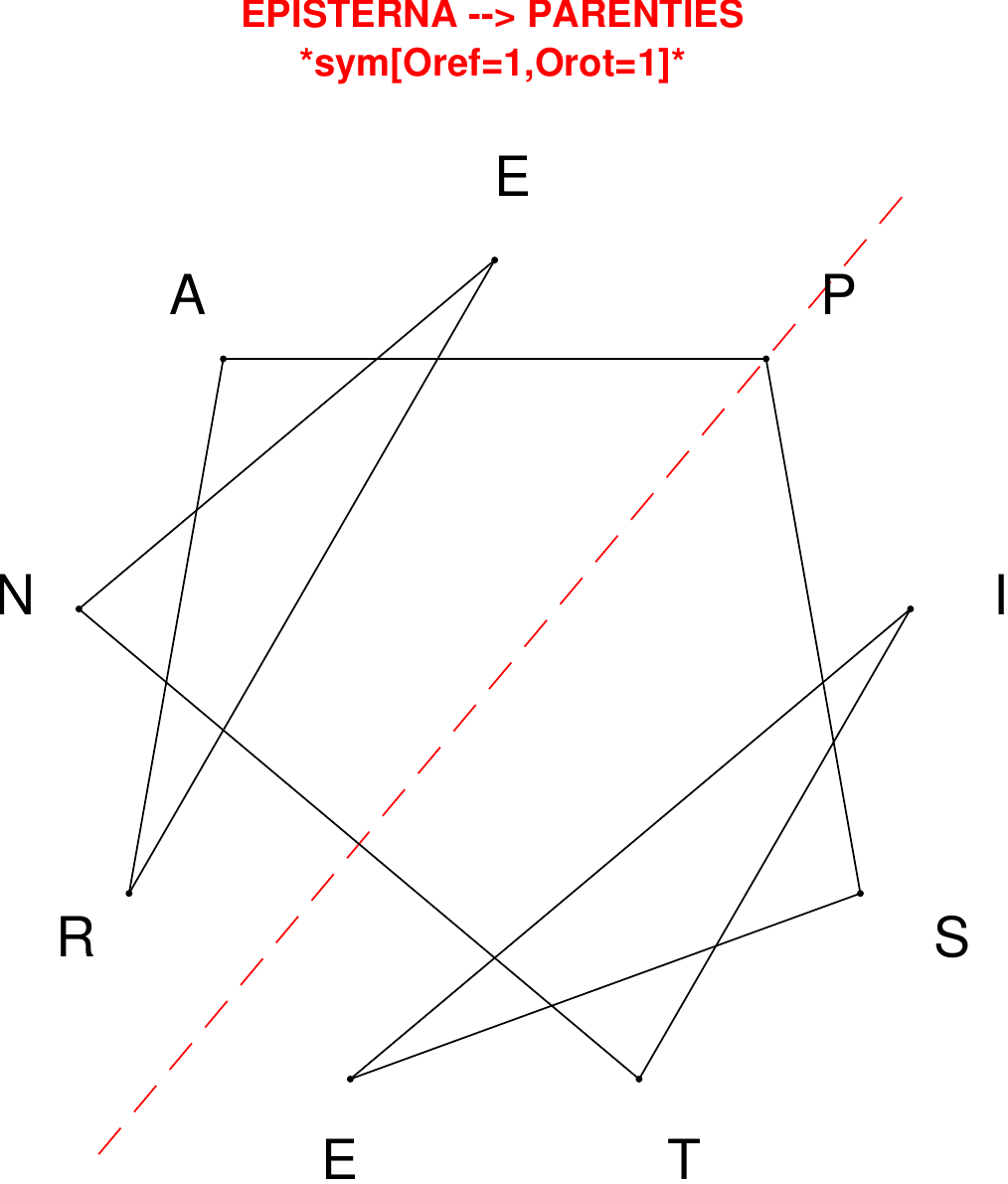}
\end{subfigure}
\hfill
\begin{subfigure}[T]{0.19\textwidth}
\centering
\includegraphics[width=\textwidth]{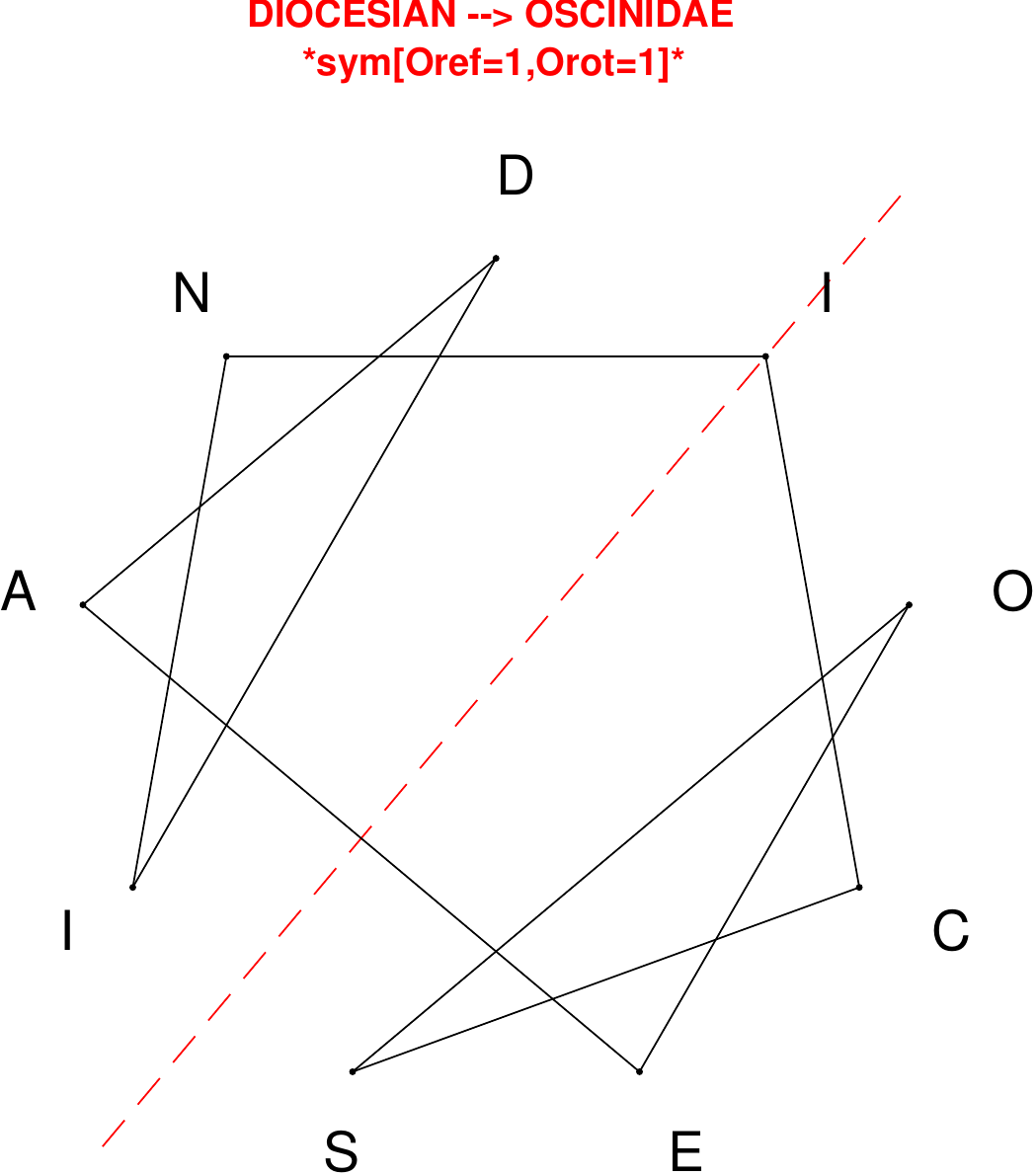}
\end{subfigure}
\hfill
\begin{subfigure}[T]{0.19\textwidth}
\centering
\includegraphics[width=\textwidth]{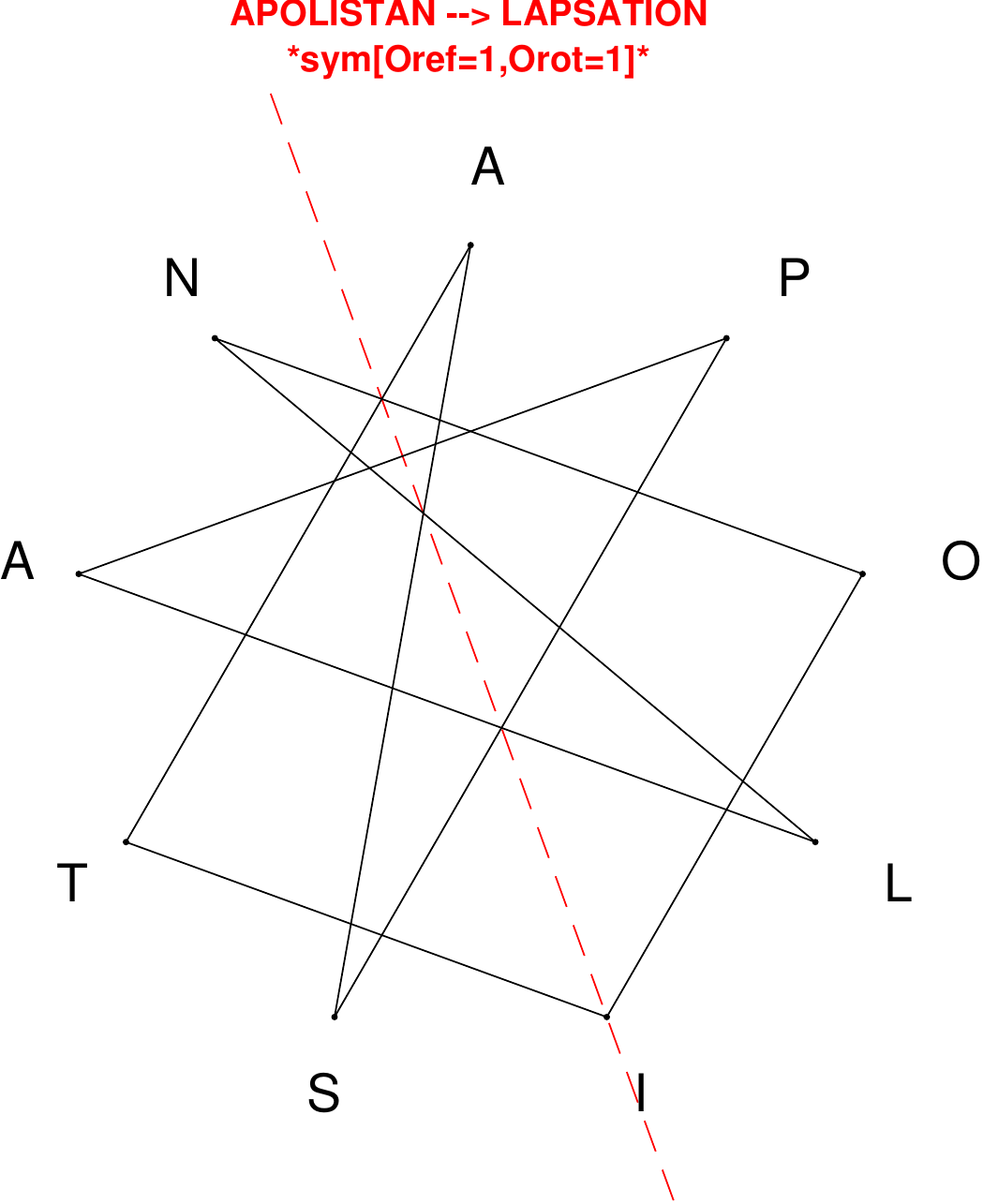}
\end{subfigure}
\hfill
\begin{subfigure}[T]{0.19\textwidth}
\centering
\includegraphics[width=\textwidth]{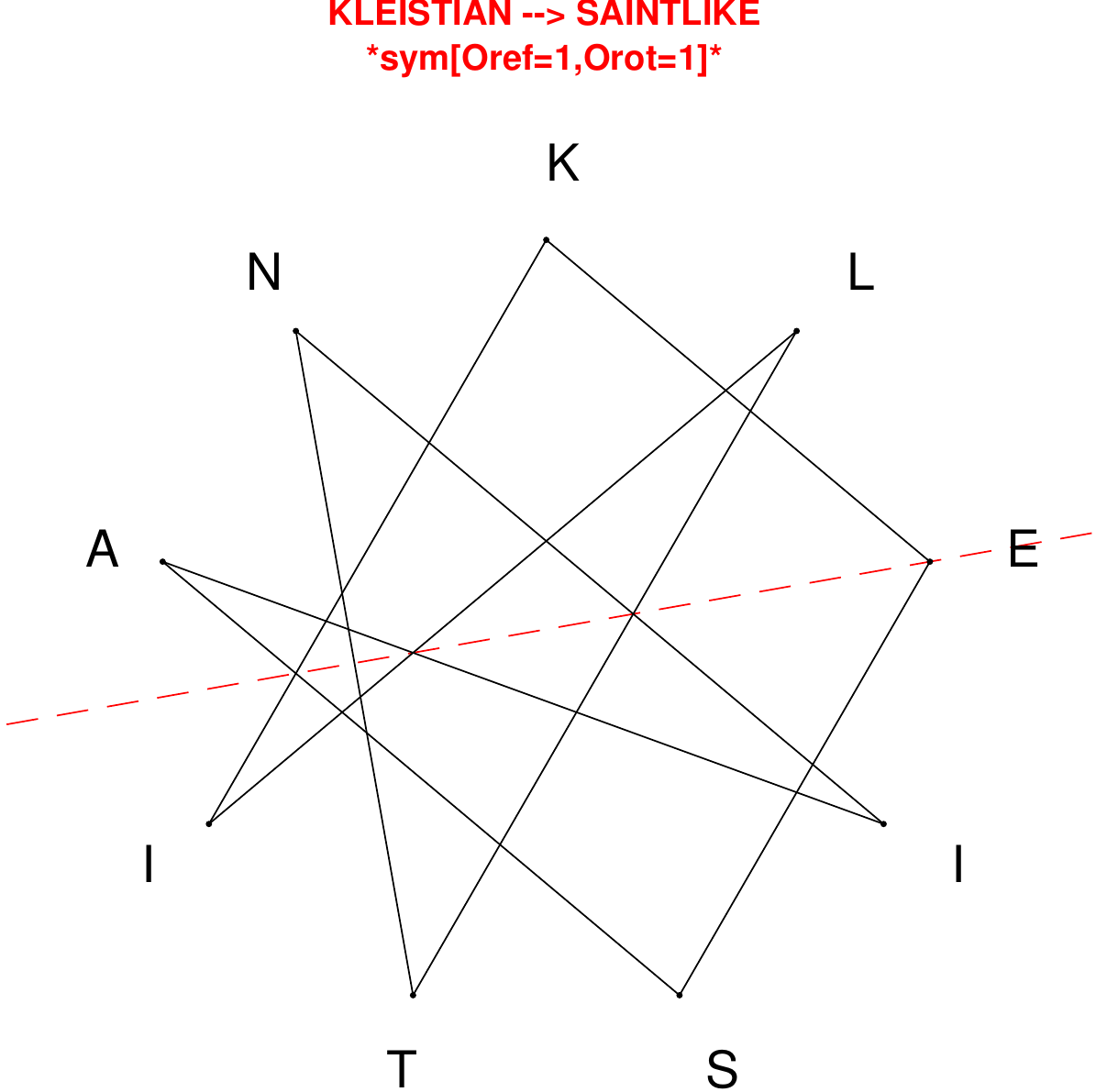}
\end{subfigure}
\end{figure}

\begin{figure}[H]
\centering
\begin{subfigure}[T]{0.19\textwidth}
\centering
\includegraphics[width=\textwidth]{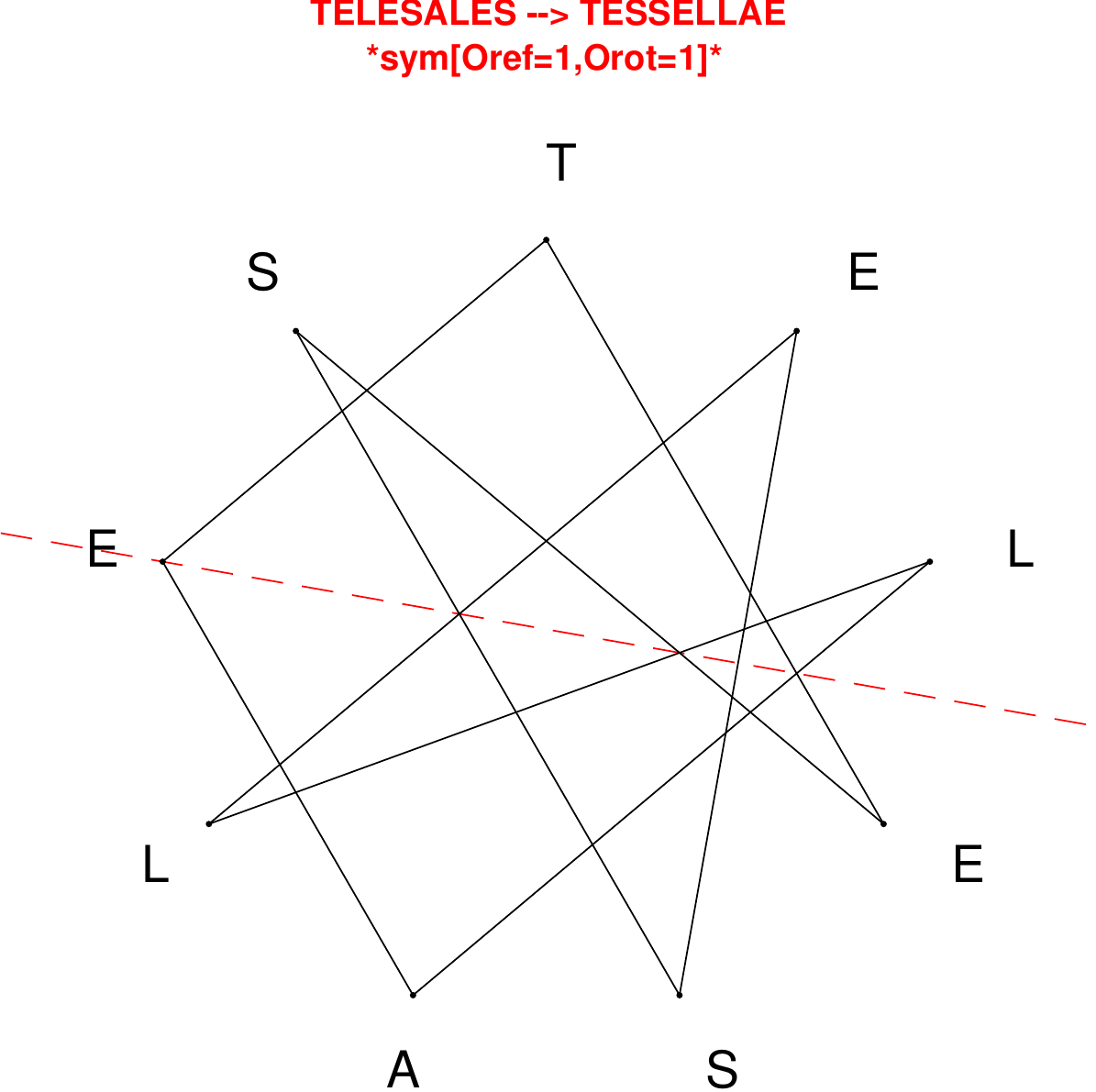}
\end{subfigure}
\hfill
\begin{subfigure}[T]{0.19\textwidth}
\centering
\includegraphics[width=\textwidth]{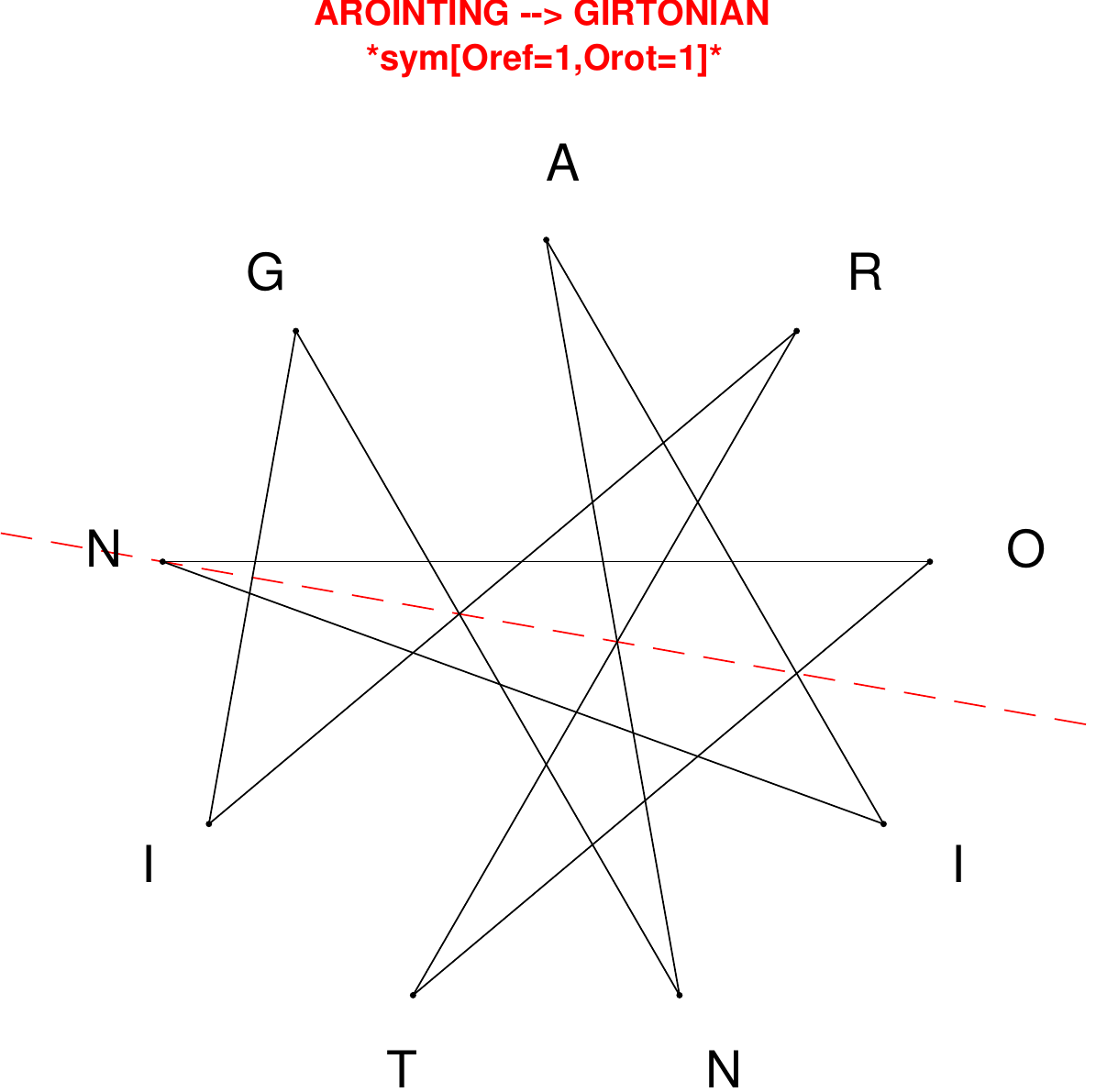}
\end{subfigure}
\hfill
\begin{subfigure}[T]{0.19\textwidth}
\centering
\includegraphics[width=\textwidth]{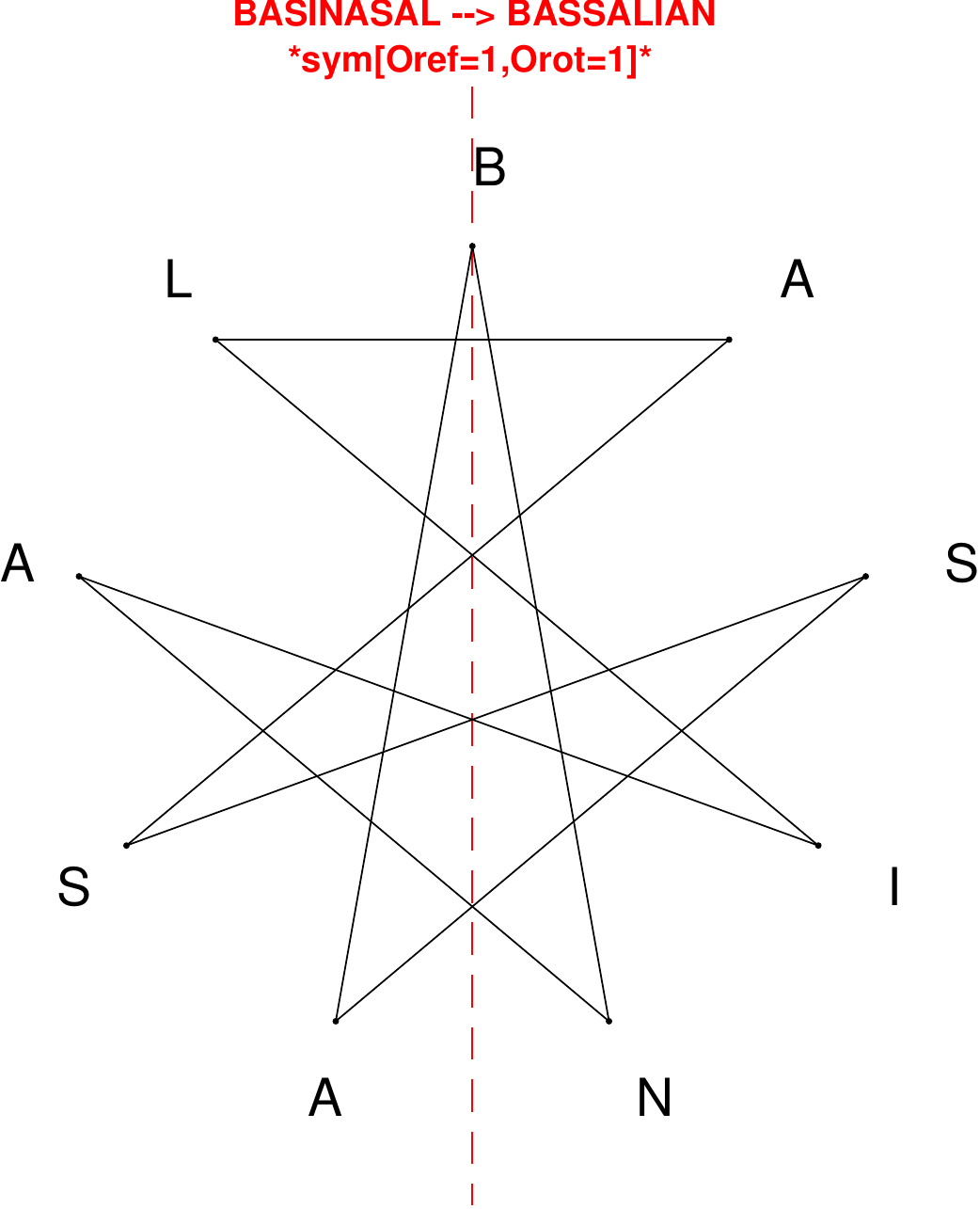}
\end{subfigure}
\hfill
\begin{subfigure}[T]{0.19\textwidth}
\centering
\includegraphics[width=\textwidth]{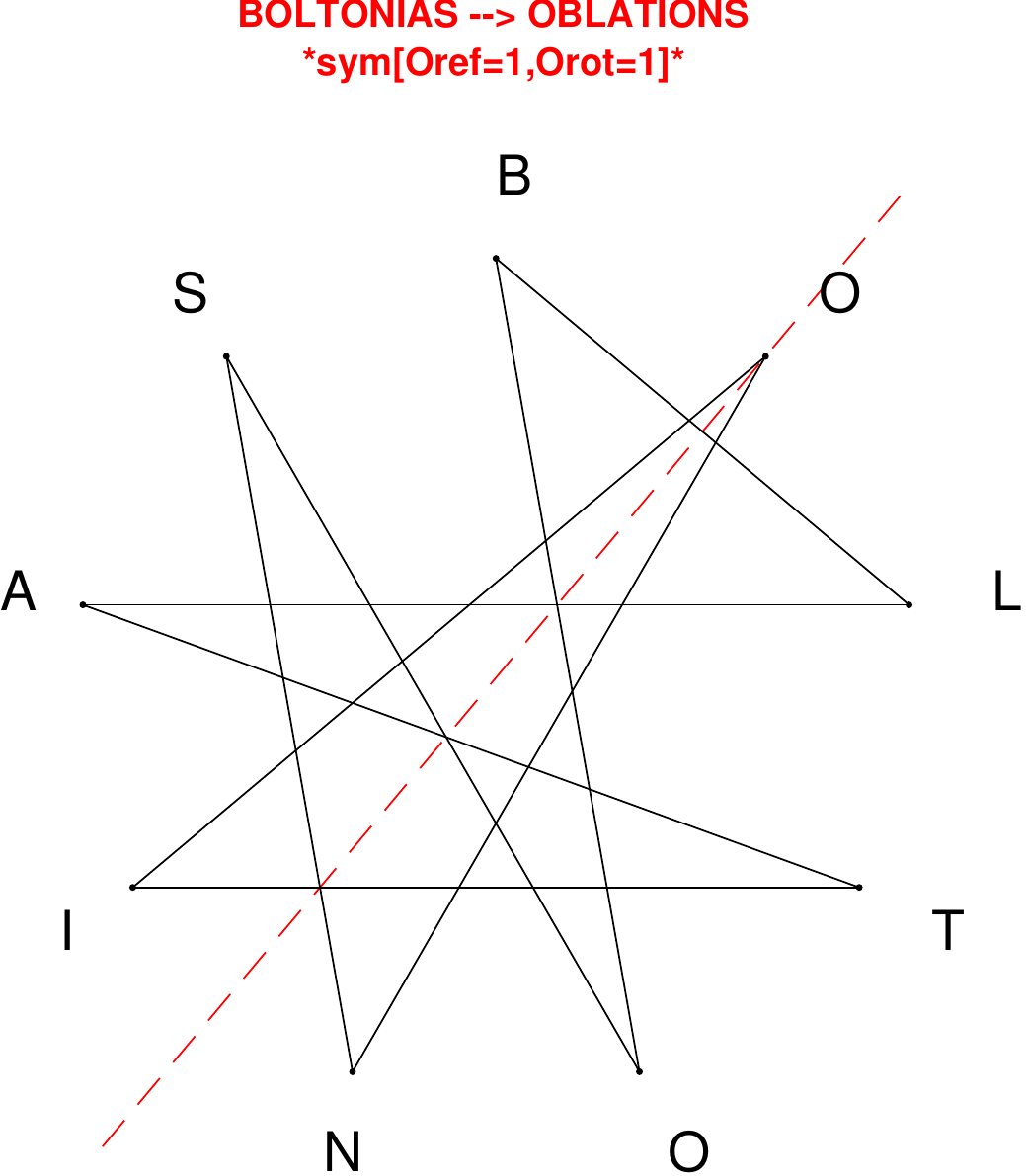}
\end{subfigure}
\hfill
\begin{subfigure}[T]{0.19\textwidth}
\centering
\includegraphics[width=\textwidth]{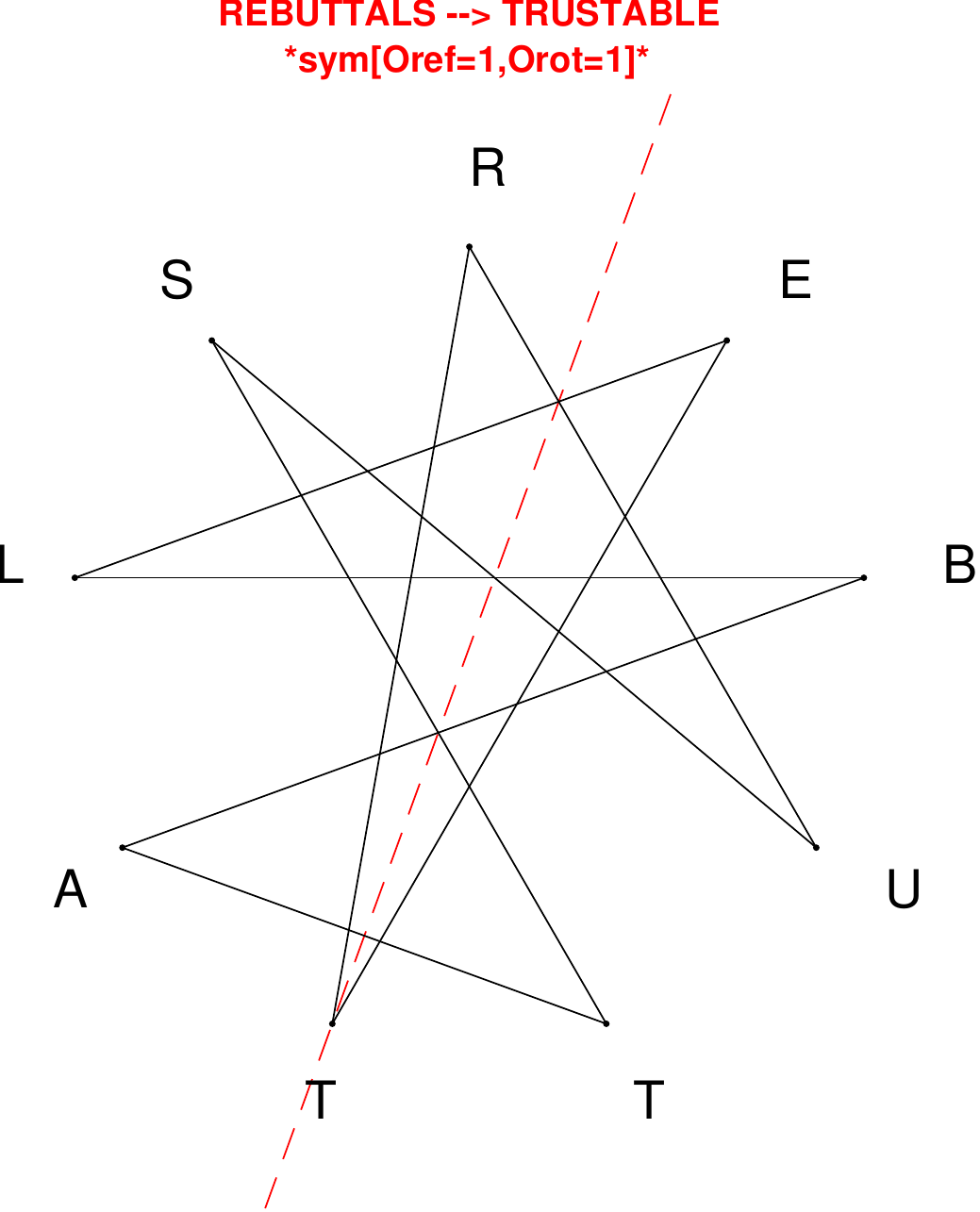}
\end{subfigure}
\end{figure}

\begin{figure}[H]
\centering
\begin{subfigure}[T]{0.19\textwidth}
\centering
\includegraphics[width=\textwidth]{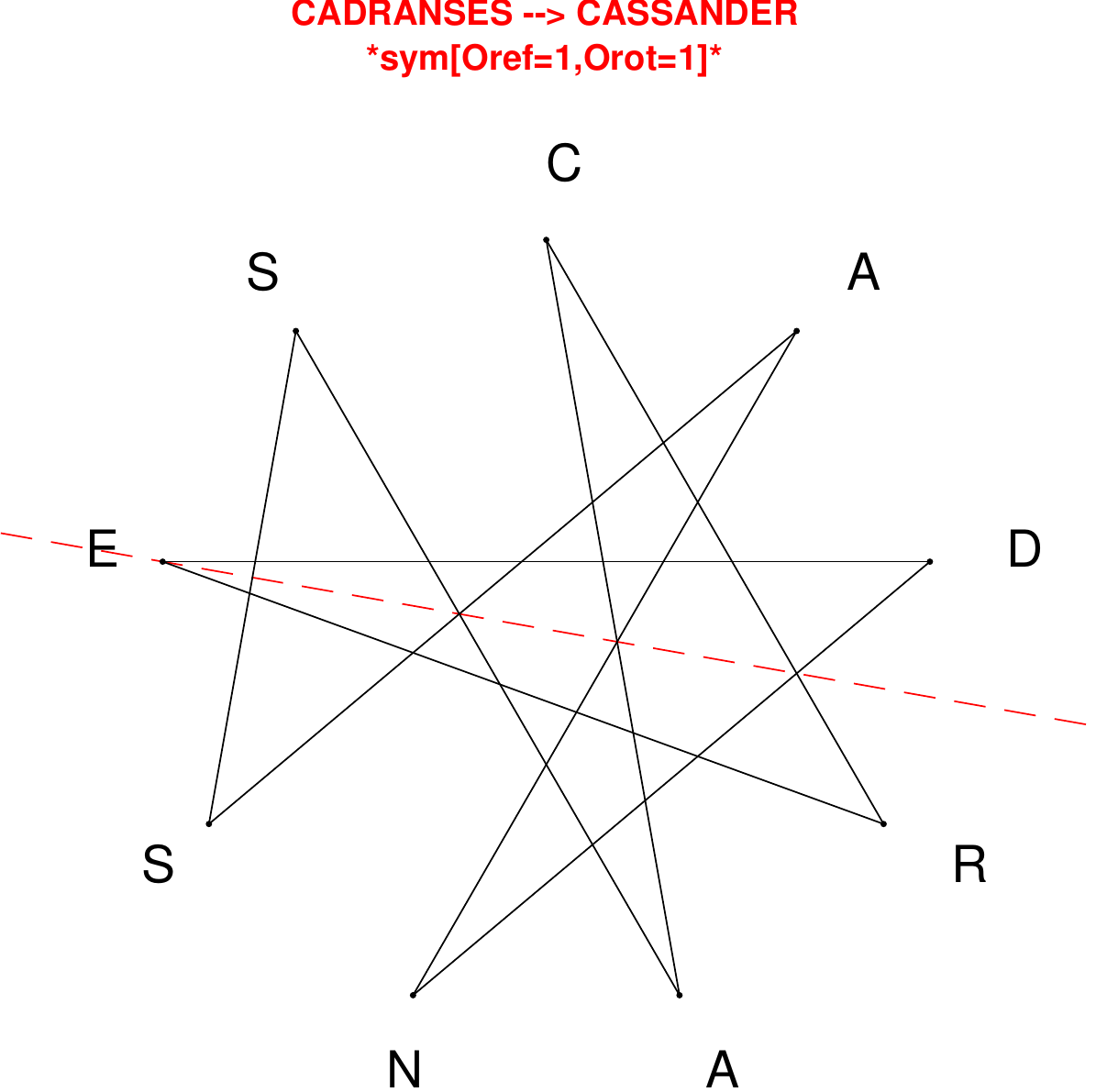}
\end{subfigure}
\hfill
\begin{subfigure}[T]{0.19\textwidth}
\centering
\includegraphics[width=\textwidth]{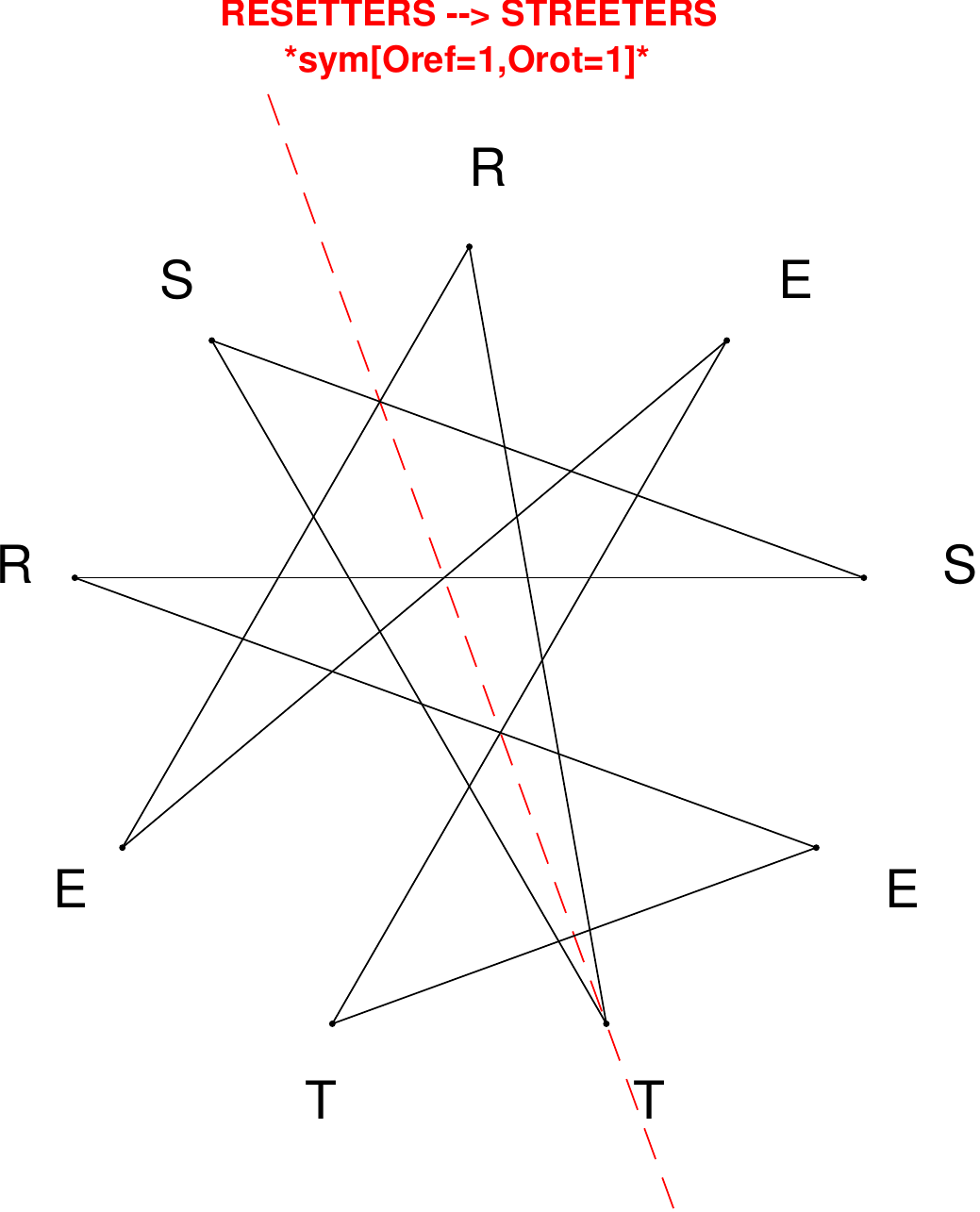}
\end{subfigure}
\hfill
\begin{subfigure}[T]{0.19\textwidth}
\centering
\includegraphics[width=\textwidth]{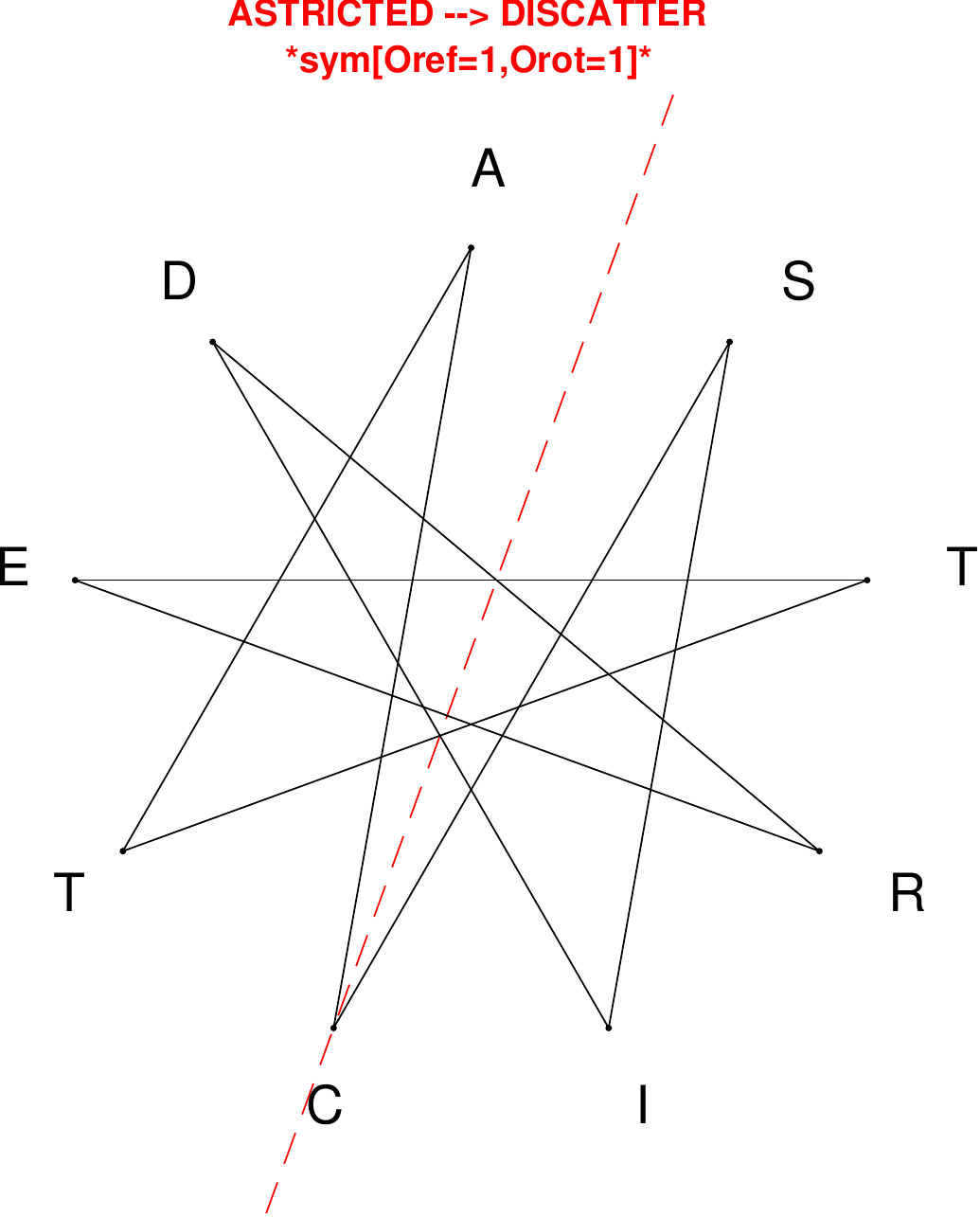}
\end{subfigure}
\hfill
\begin{subfigure}[T]{0.19\textwidth}
\centering
\includegraphics[width=\textwidth]{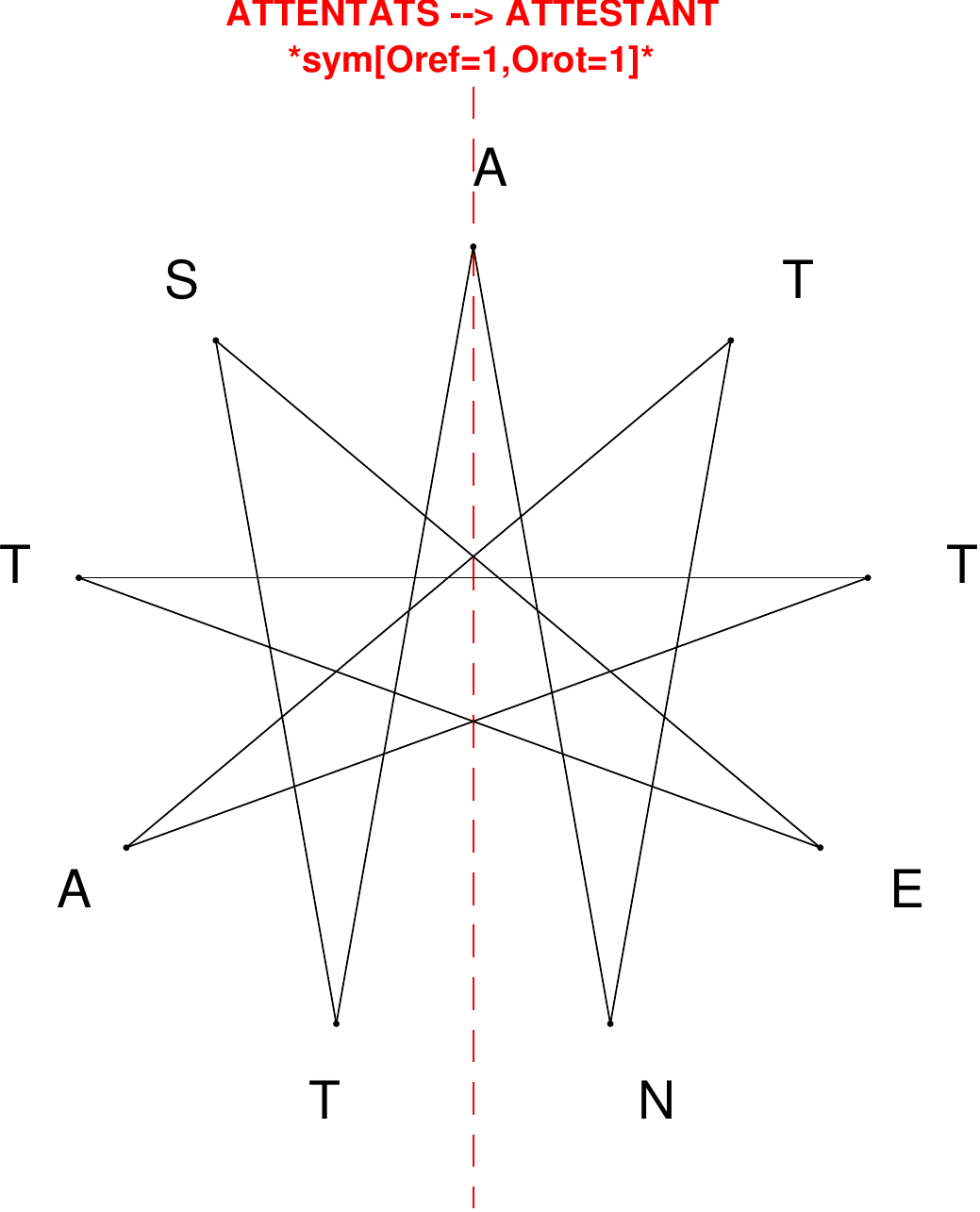}
\end{subfigure}
\hfill
\begin{subfigure}[T]{0.19\textwidth}
\centering
\includegraphics[width=\textwidth]{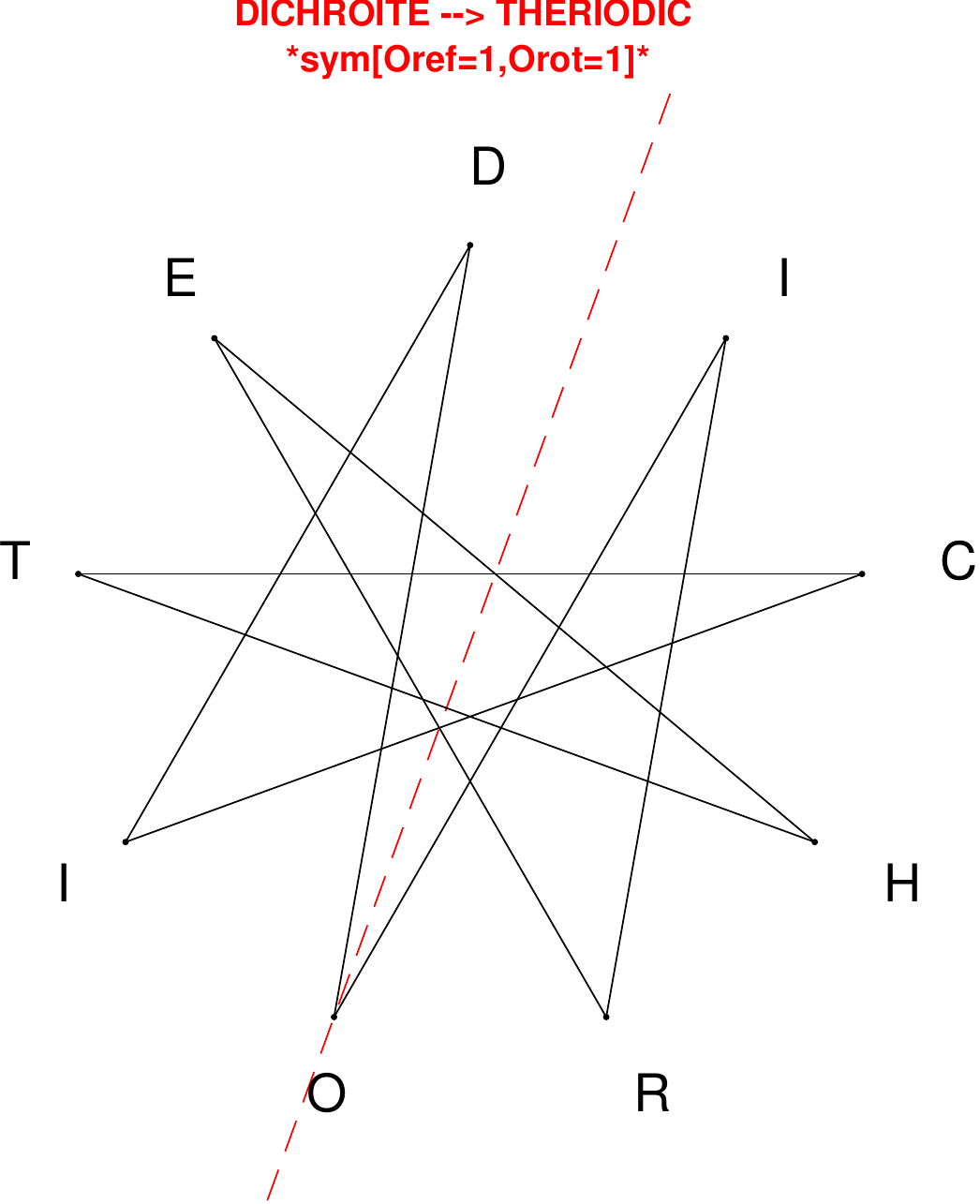}
\end{subfigure}
\end{figure}

\begin{figure}[H]
\centering
\begin{subfigure}[T]{0.19\textwidth}
\centering
\includegraphics[width=\textwidth]{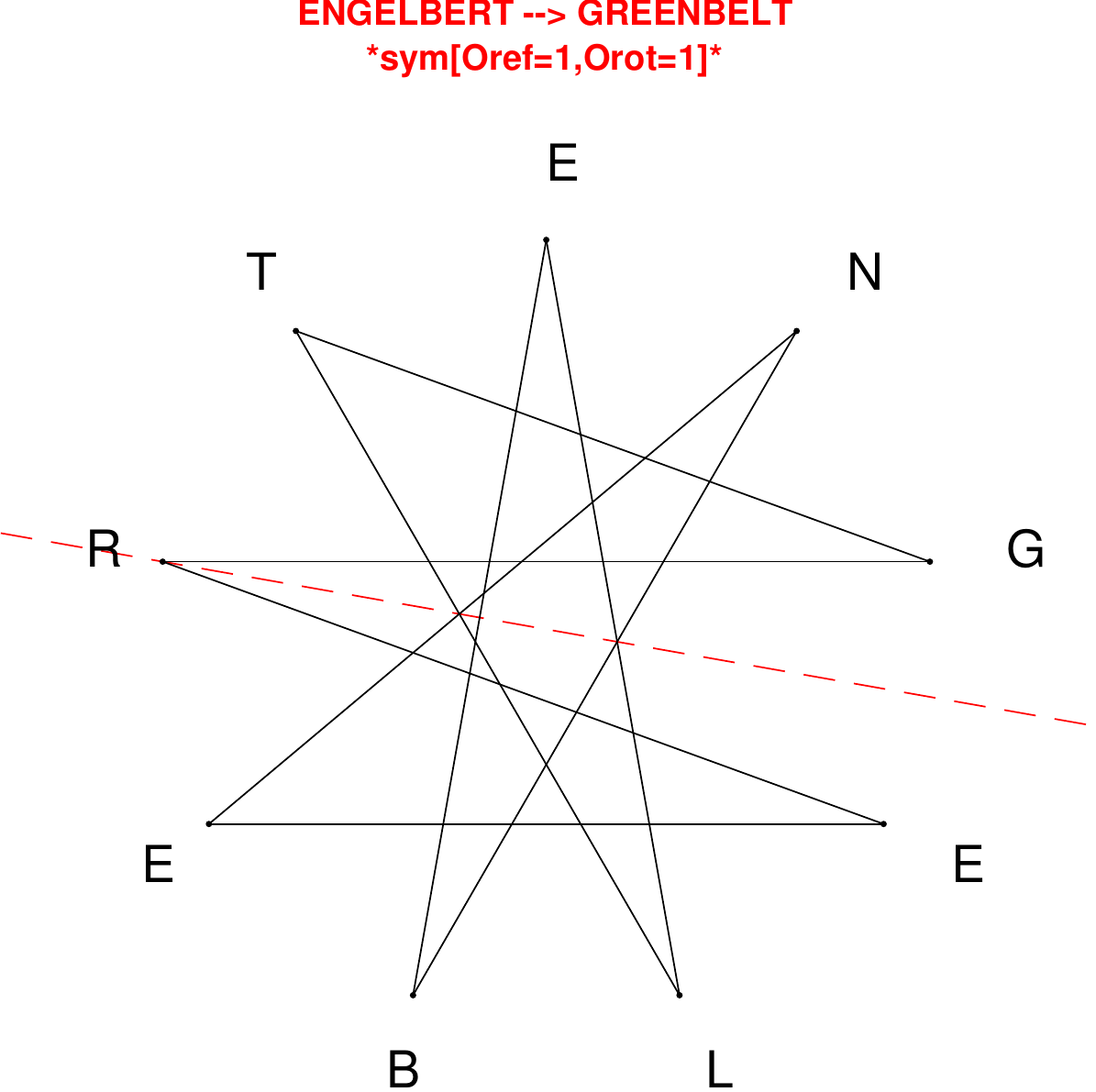}
\end{subfigure}
\hfill
\begin{subfigure}[T]{0.19\textwidth}
\centering
\includegraphics[width=\textwidth]{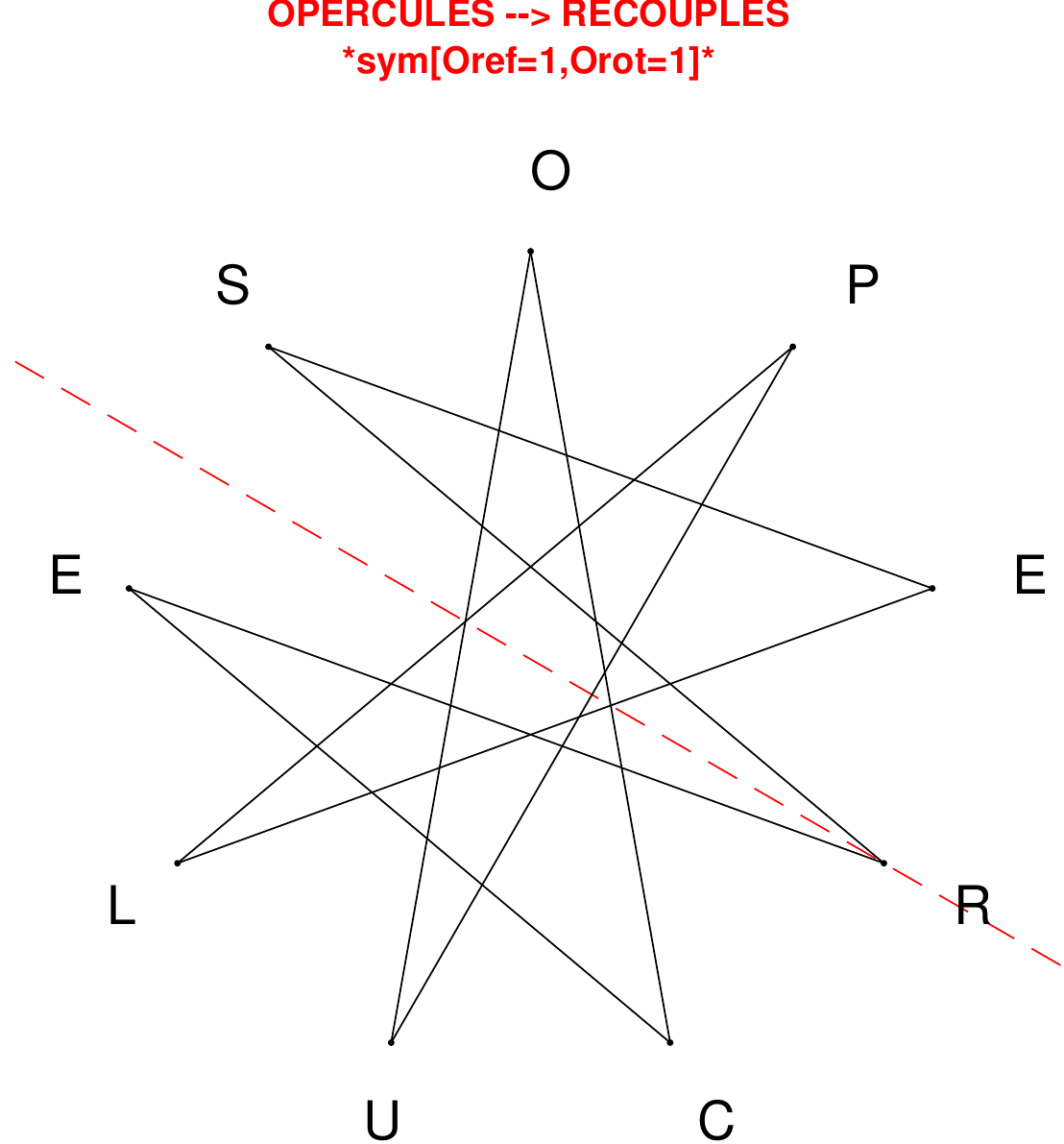}
\end{subfigure}
\hfill
\begin{subfigure}[T]{0.19\textwidth}
\centering
\includegraphics[width=\textwidth]{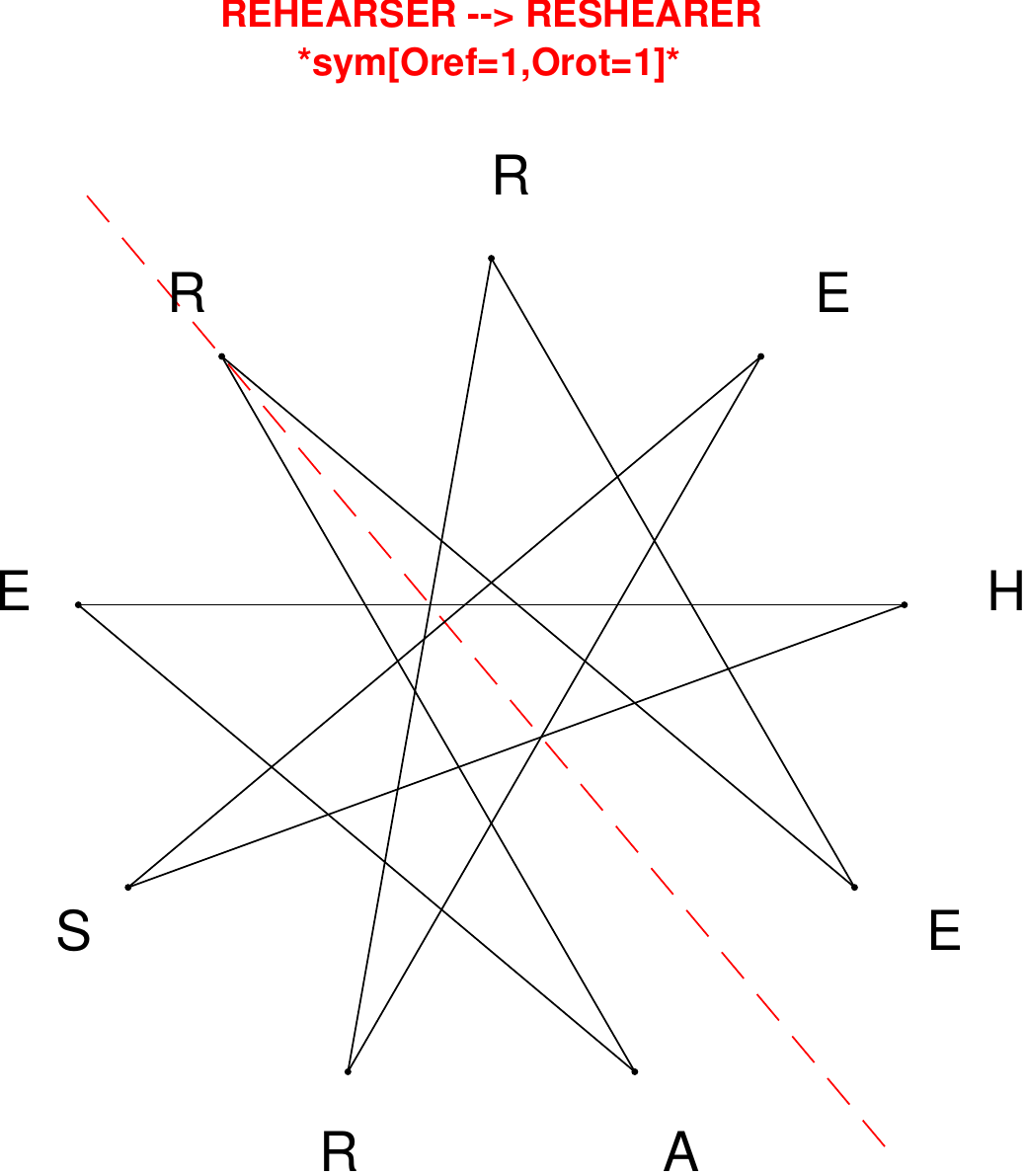}
\end{subfigure}
\hfill
\begin{subfigure}[T]{0.19\textwidth}
\centering
\includegraphics[width=\textwidth]{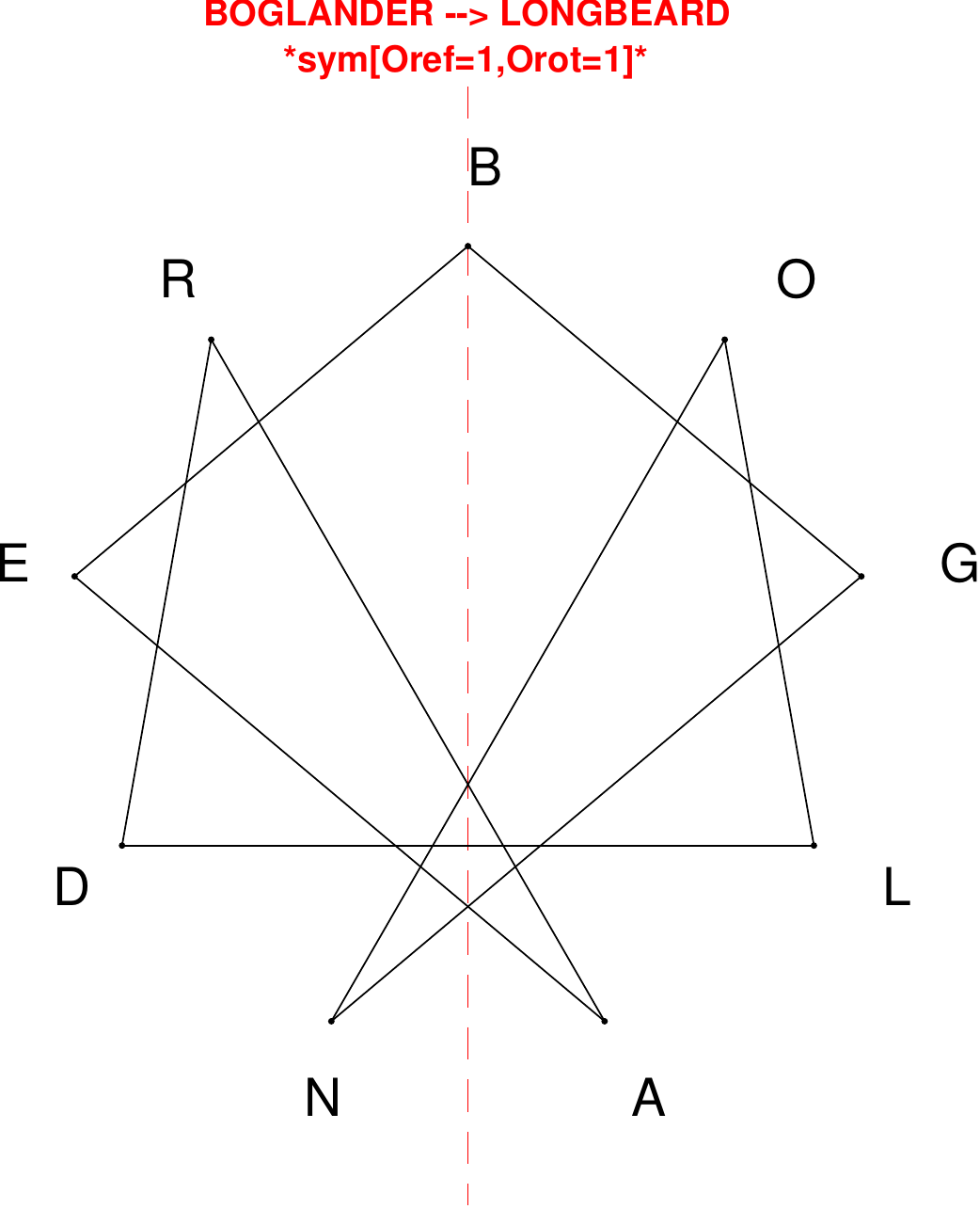}
\end{subfigure}
\hfill
\begin{subfigure}[T]{0.19\textwidth}
\centering
\includegraphics[width=\textwidth]{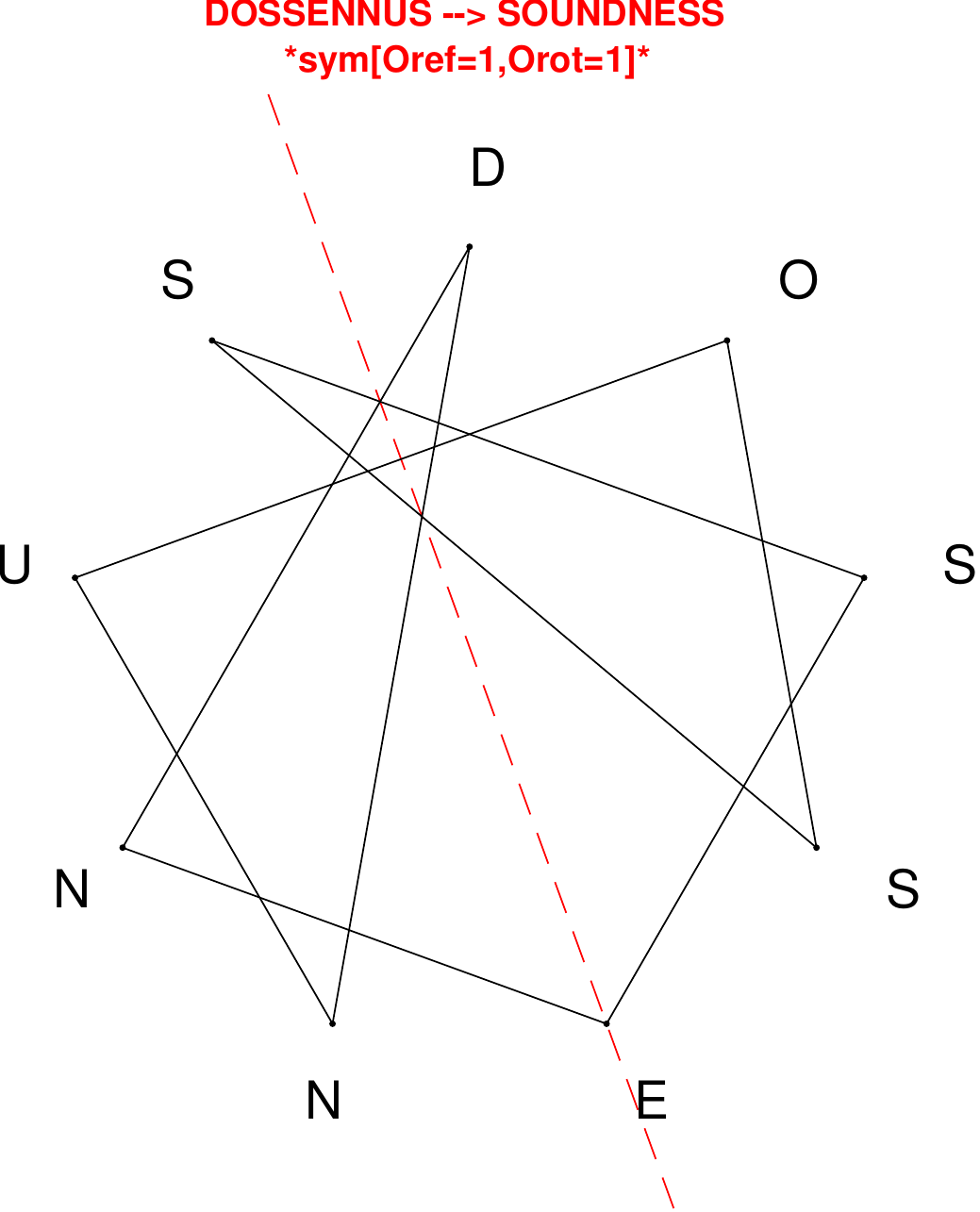}
\end{subfigure}
\end{figure}

\begin{figure}[H]
\centering
\begin{subfigure}[T]{0.19\textwidth}
\centering
\includegraphics[width=\textwidth]{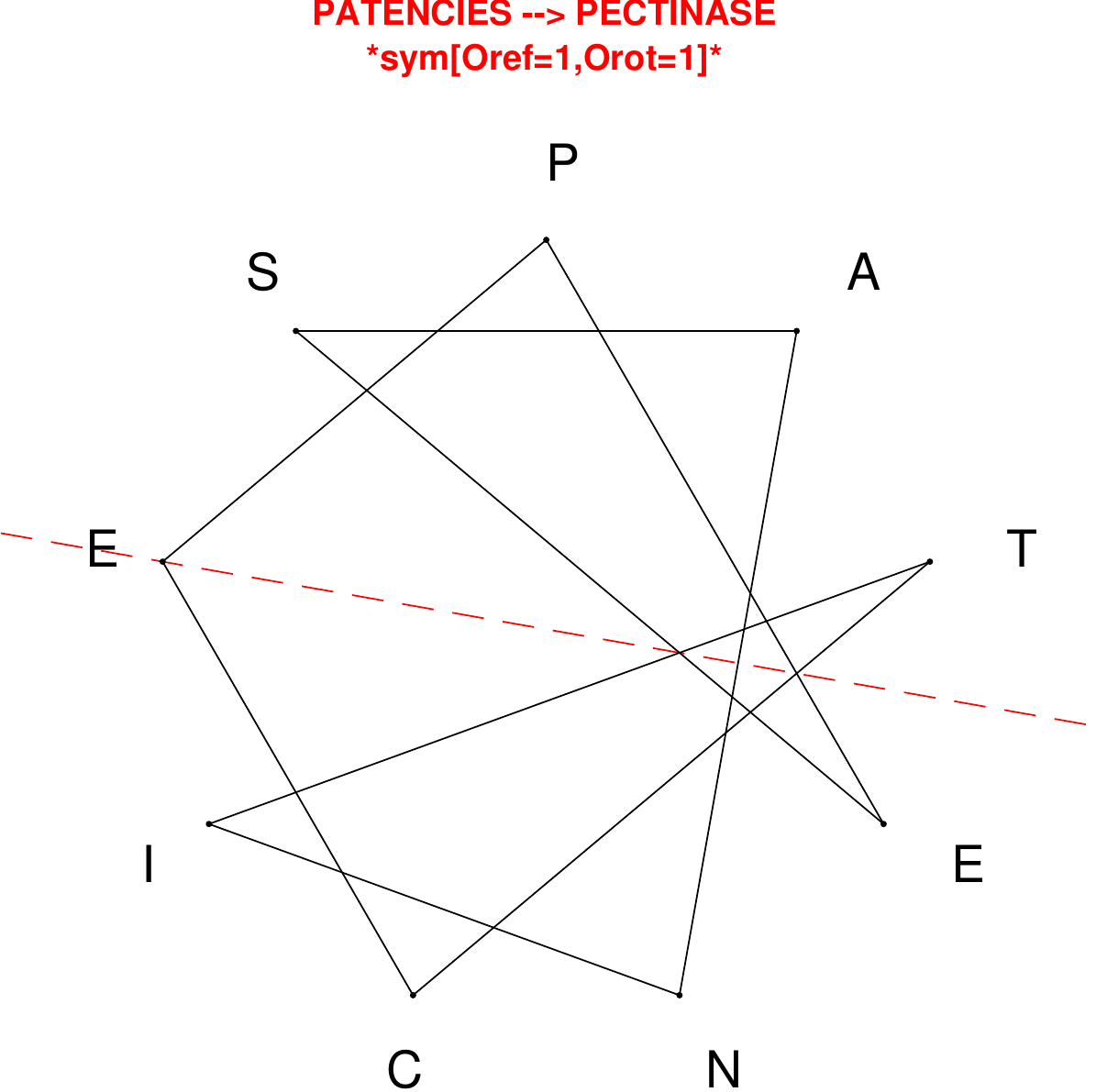}
\end{subfigure}
\hfill
\begin{subfigure}[T]{0.19\textwidth}
\centering
\includegraphics[width=\textwidth]{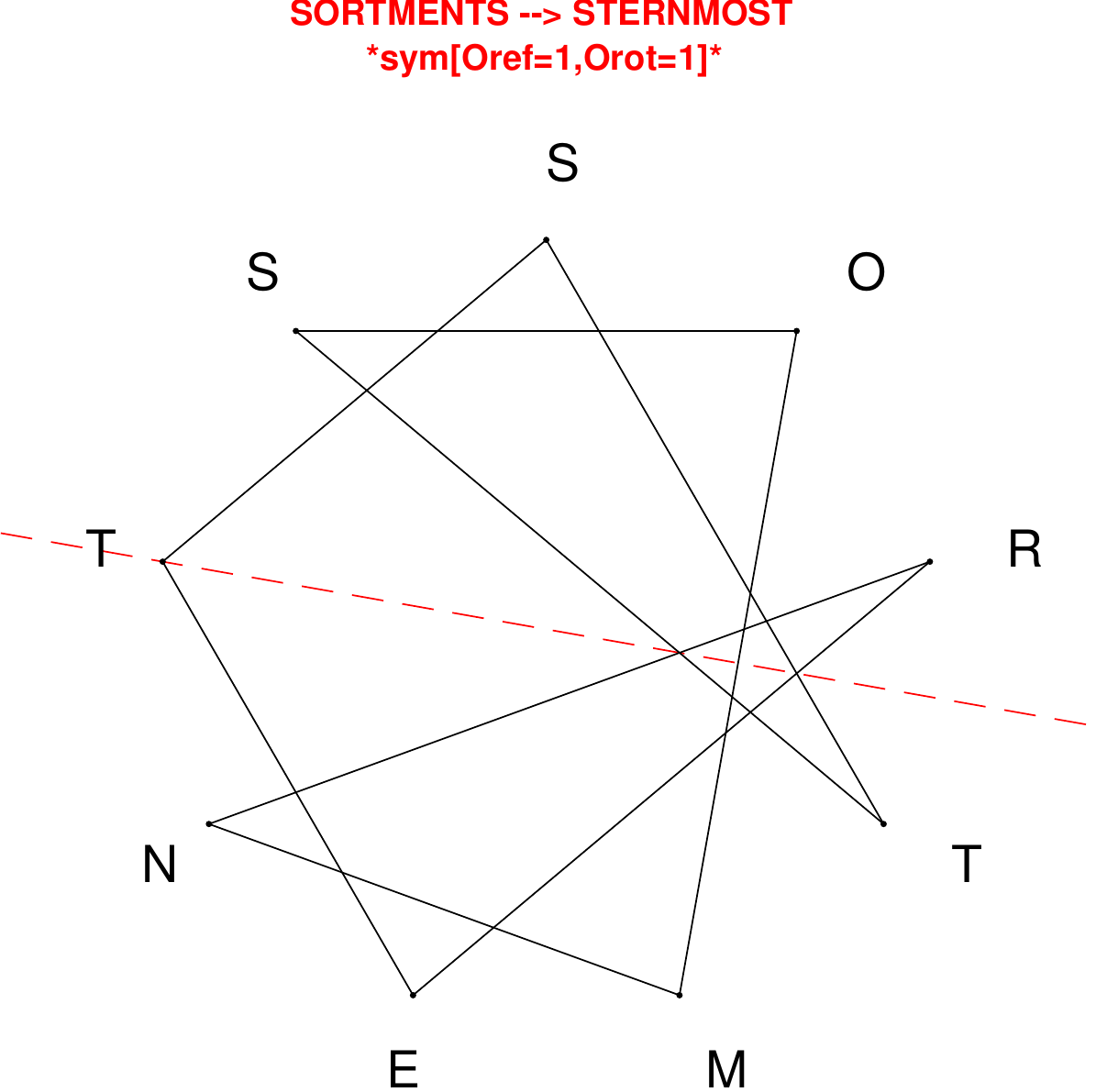}
\end{subfigure}
\hfill
\begin{subfigure}[T]{0.19\textwidth}
\centering
\includegraphics[width=\textwidth]{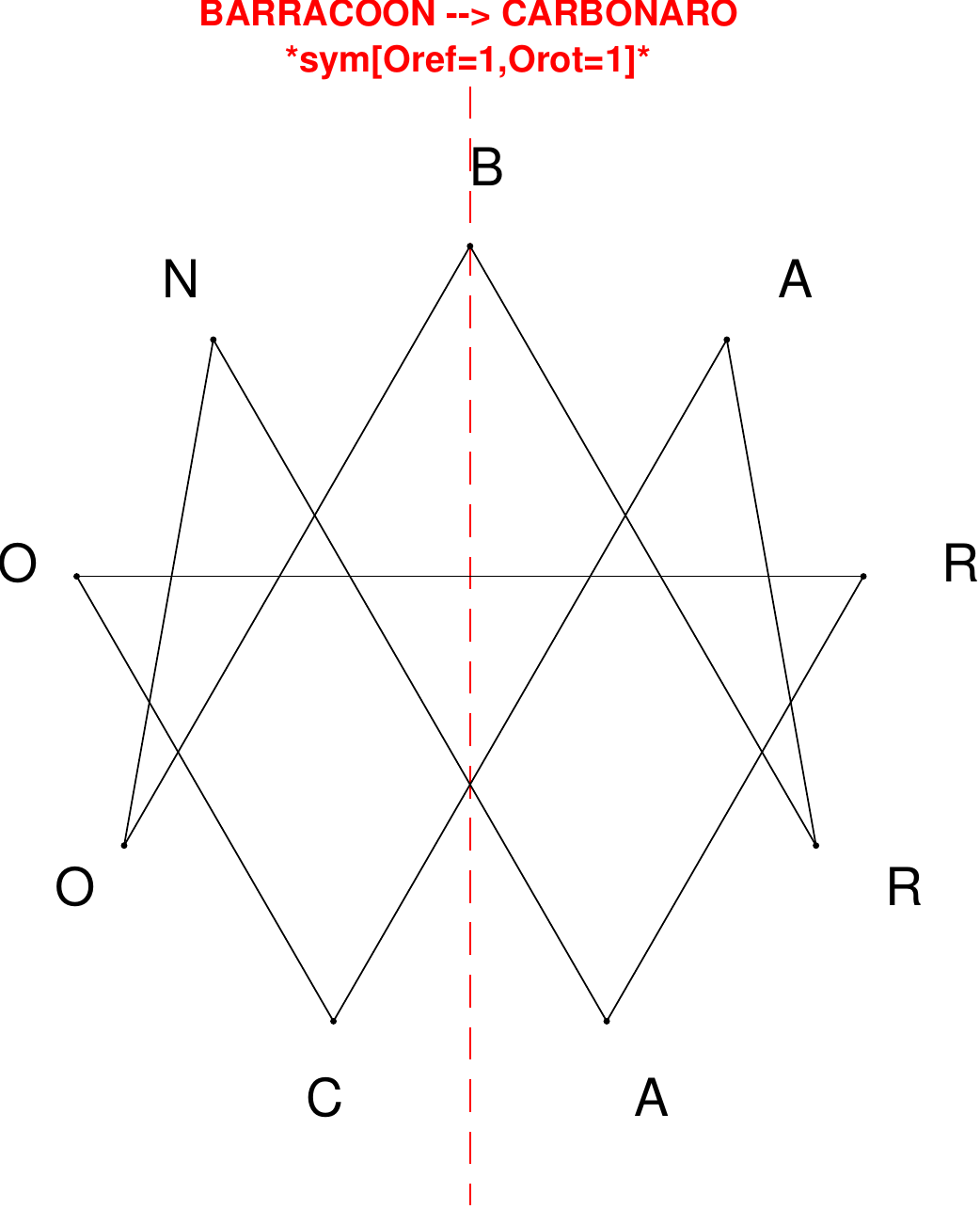}
\end{subfigure}
\hfill
\begin{subfigure}[T]{0.19\textwidth}
\centering
\includegraphics[width=\textwidth]{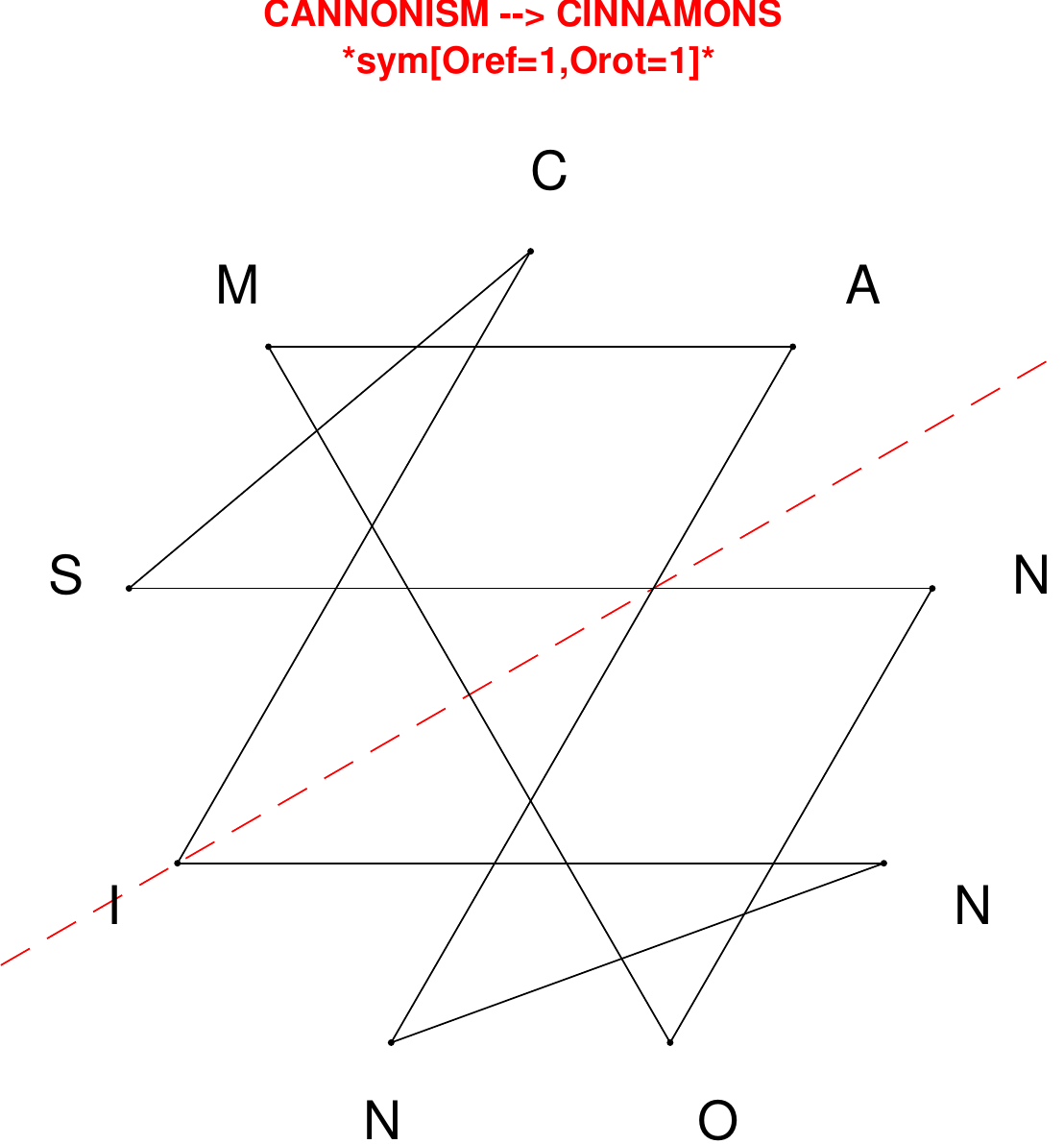}
\end{subfigure}
\hfill
\begin{subfigure}[T]{0.19\textwidth}
\centering
\includegraphics[width=\textwidth]{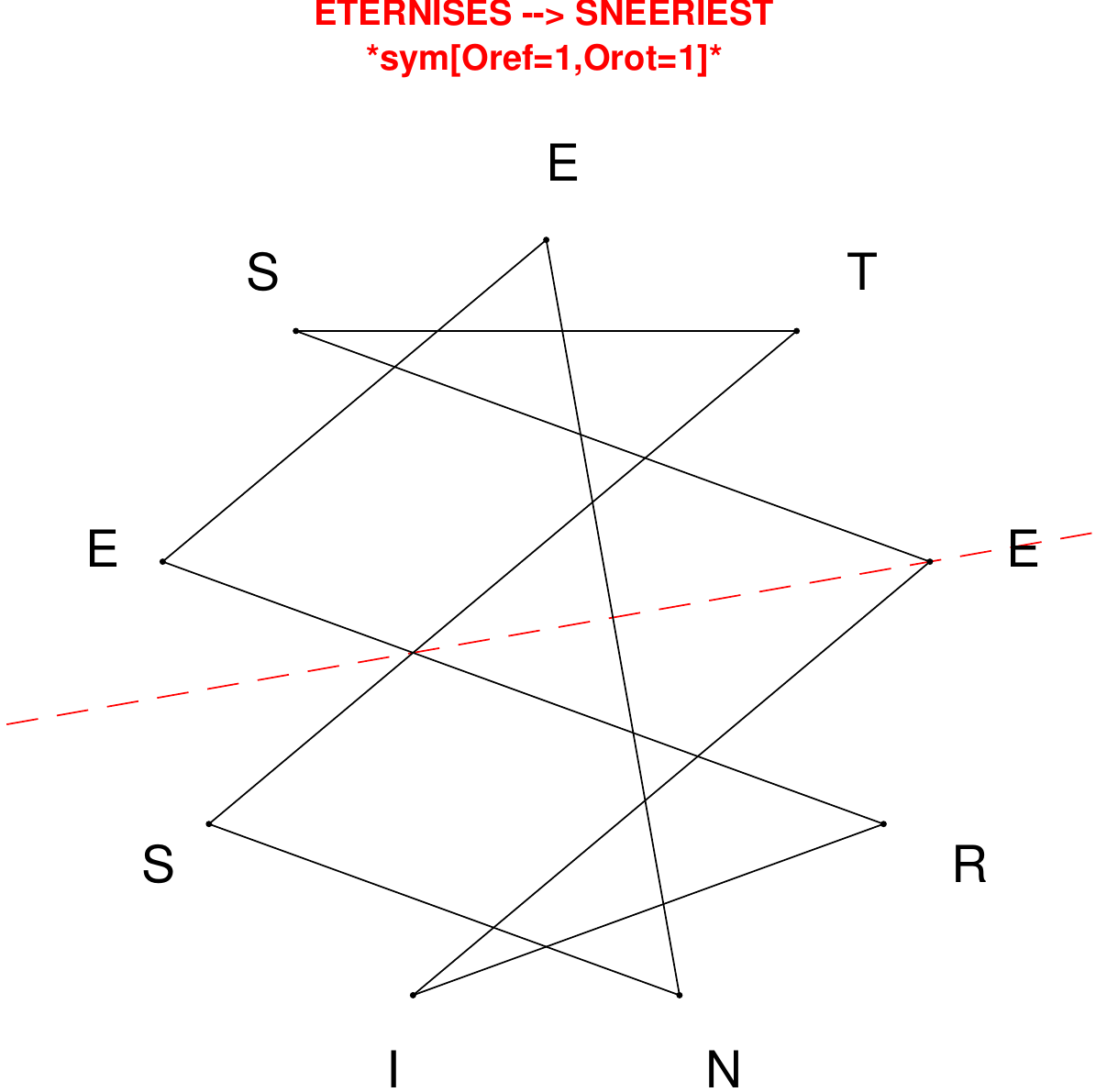}
\end{subfigure}
\end{figure}

\begin{figure}[H]
\centering
\begin{subfigure}[T]{0.19\textwidth}
\centering
\includegraphics[width=\textwidth]{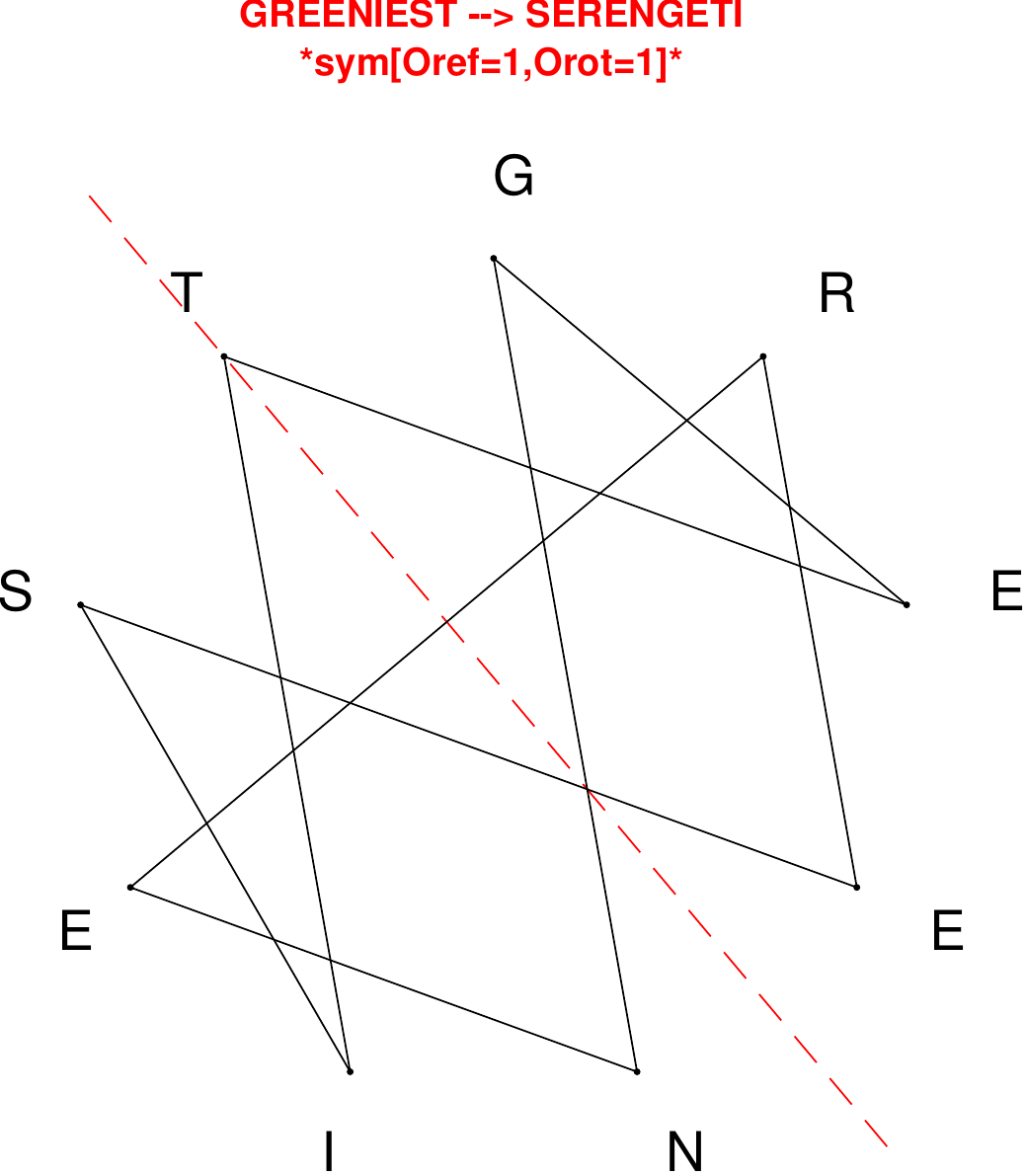}
\end{subfigure}
\hfill
\begin{subfigure}[T]{0.19\textwidth}
\centering
\includegraphics[width=\textwidth]{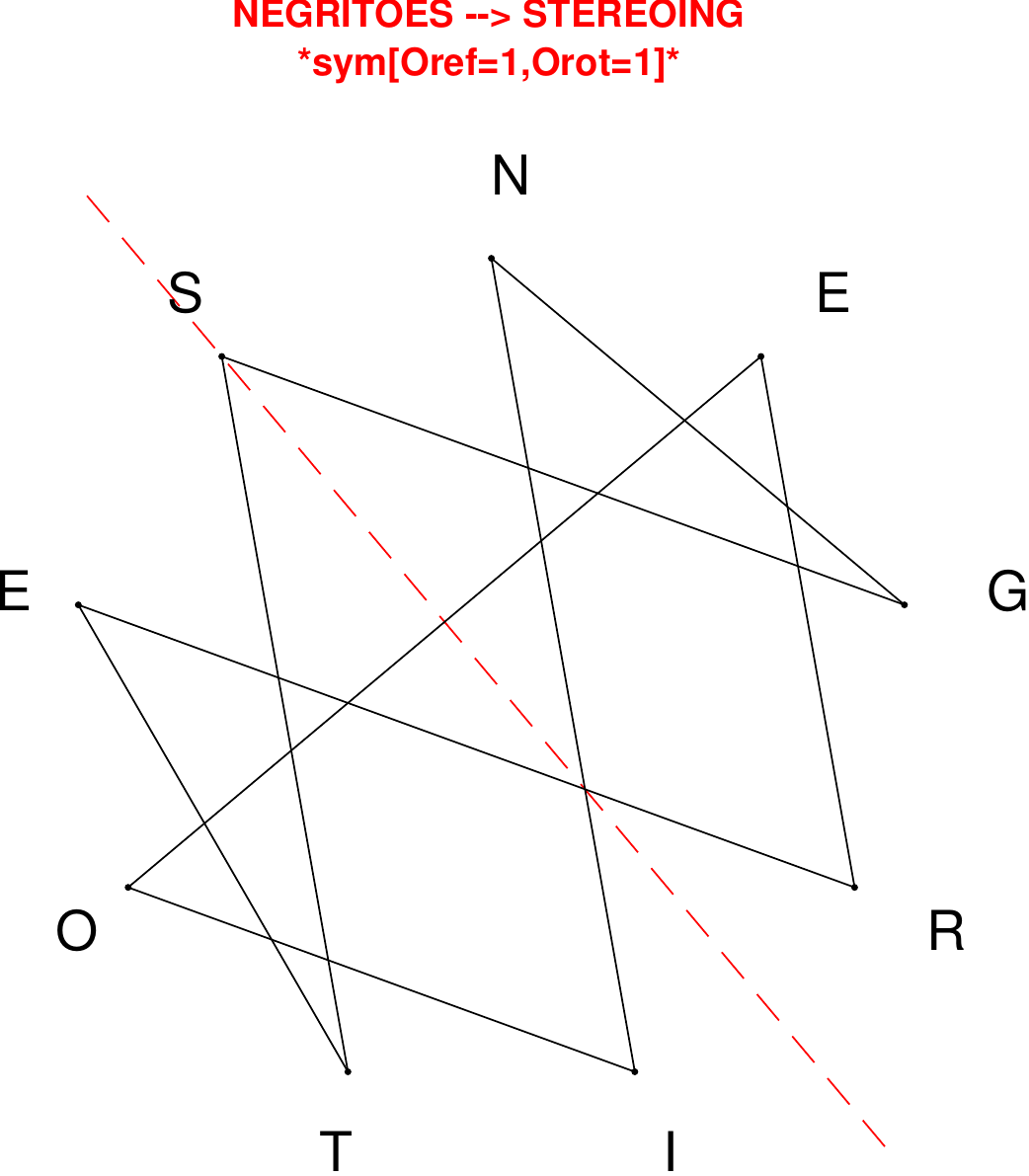}
\end{subfigure}
\hfill
\begin{subfigure}[T]{0.19\textwidth}
\centering
\includegraphics[width=\textwidth]{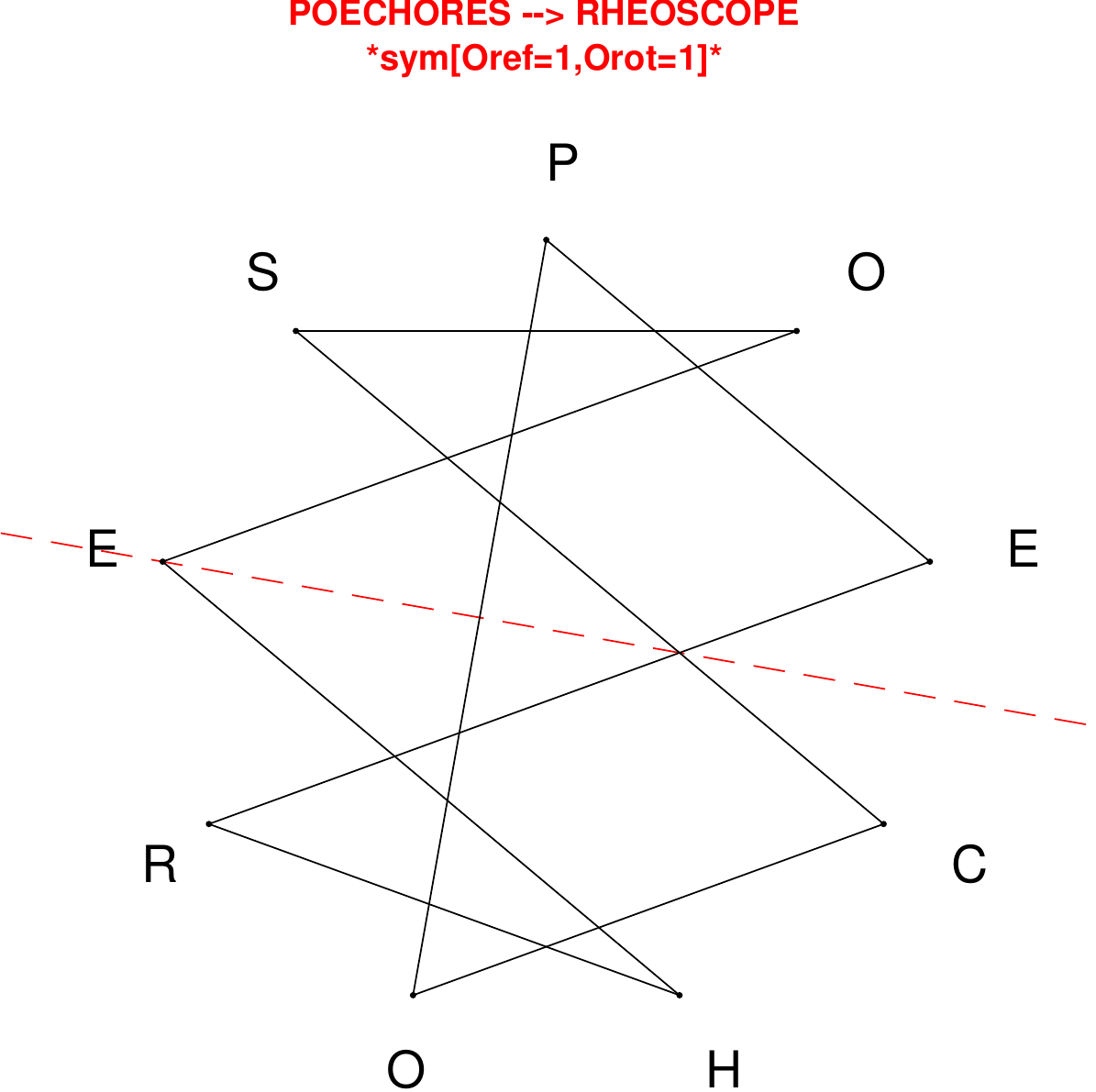}
\end{subfigure}
\hfill
\begin{subfigure}[T]{0.19\textwidth}
\centering
\includegraphics[width=\textwidth]{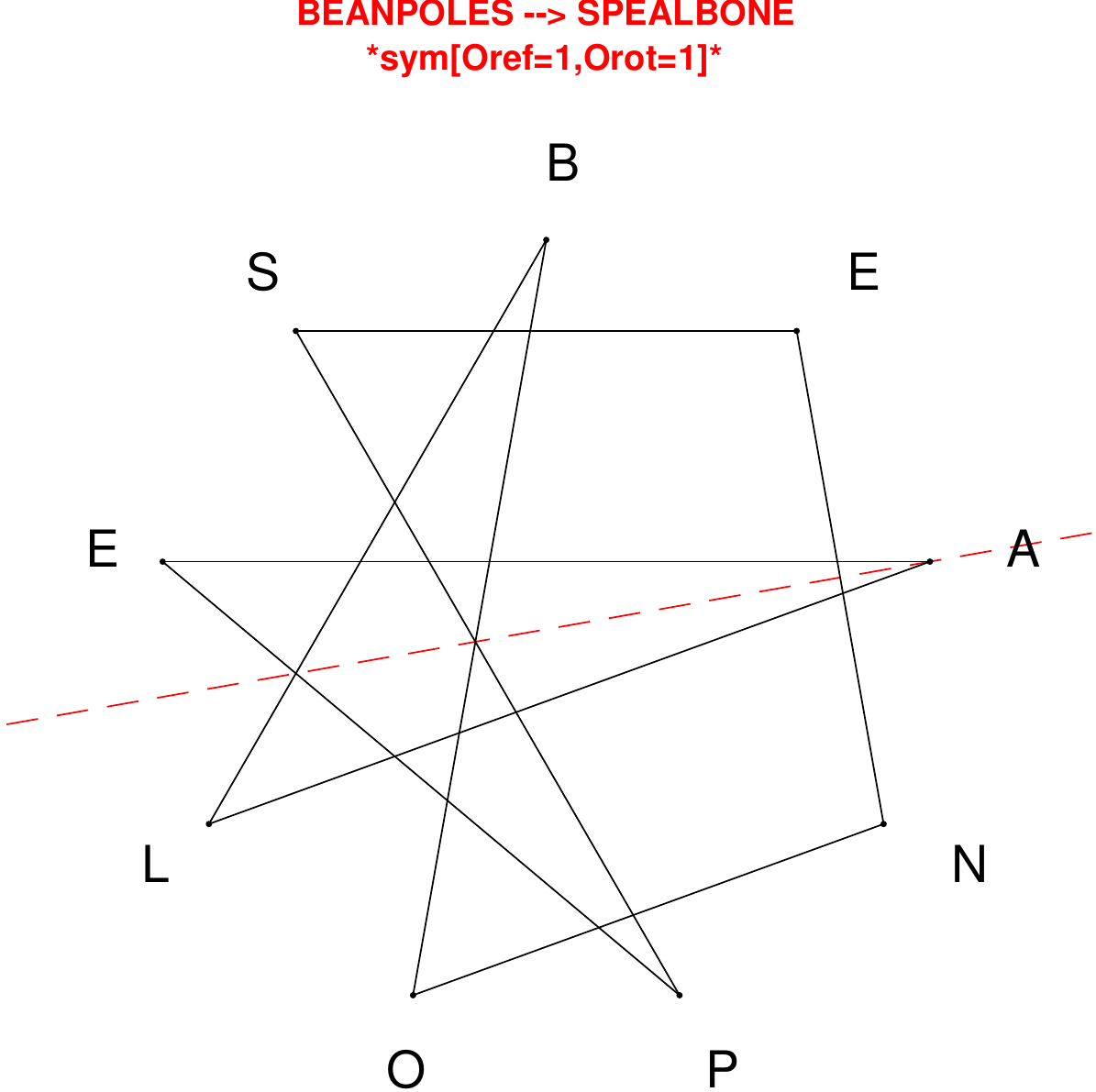}
\end{subfigure}
\hfill
\begin{subfigure}[T]{0.19\textwidth}
\centering
\includegraphics[width=\textwidth]{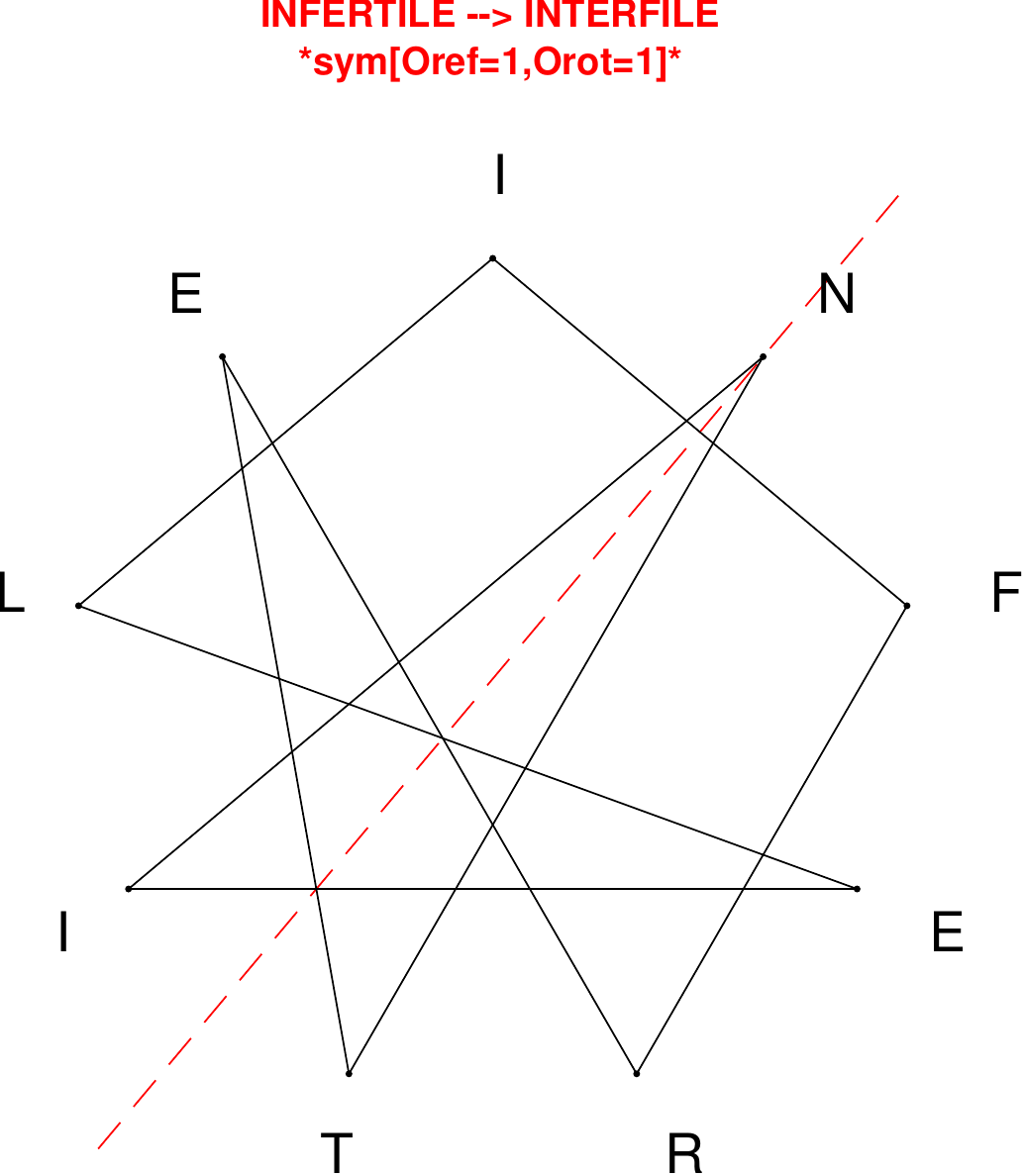}
\end{subfigure}
\end{figure}

\begin{figure}[H]
\centering
\begin{subfigure}[T]{0.19\textwidth}
\centering
\includegraphics[width=\textwidth]{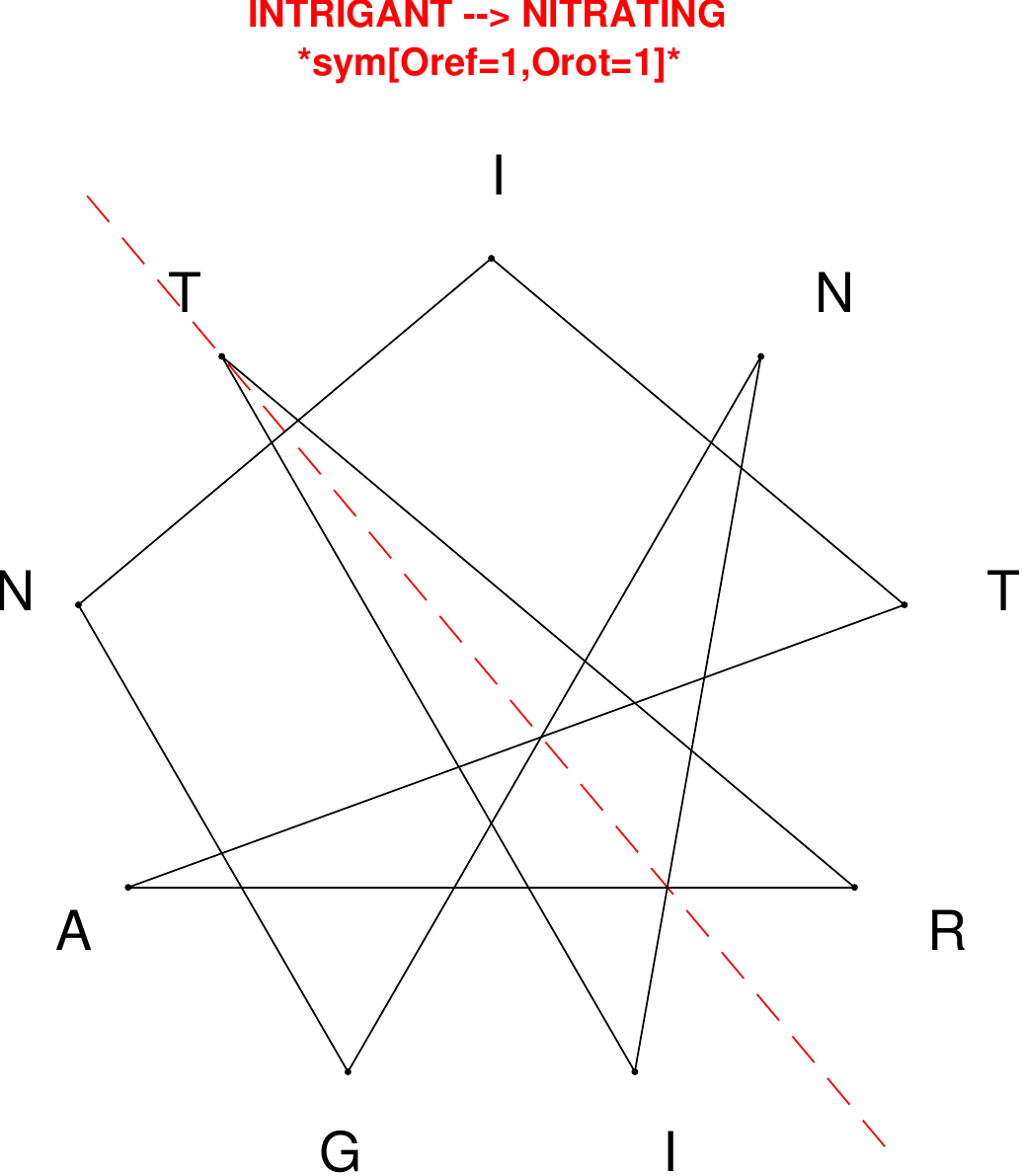}
\end{subfigure}
\hfill
\begin{subfigure}[T]{0.19\textwidth}
\centering
\includegraphics[width=\textwidth]{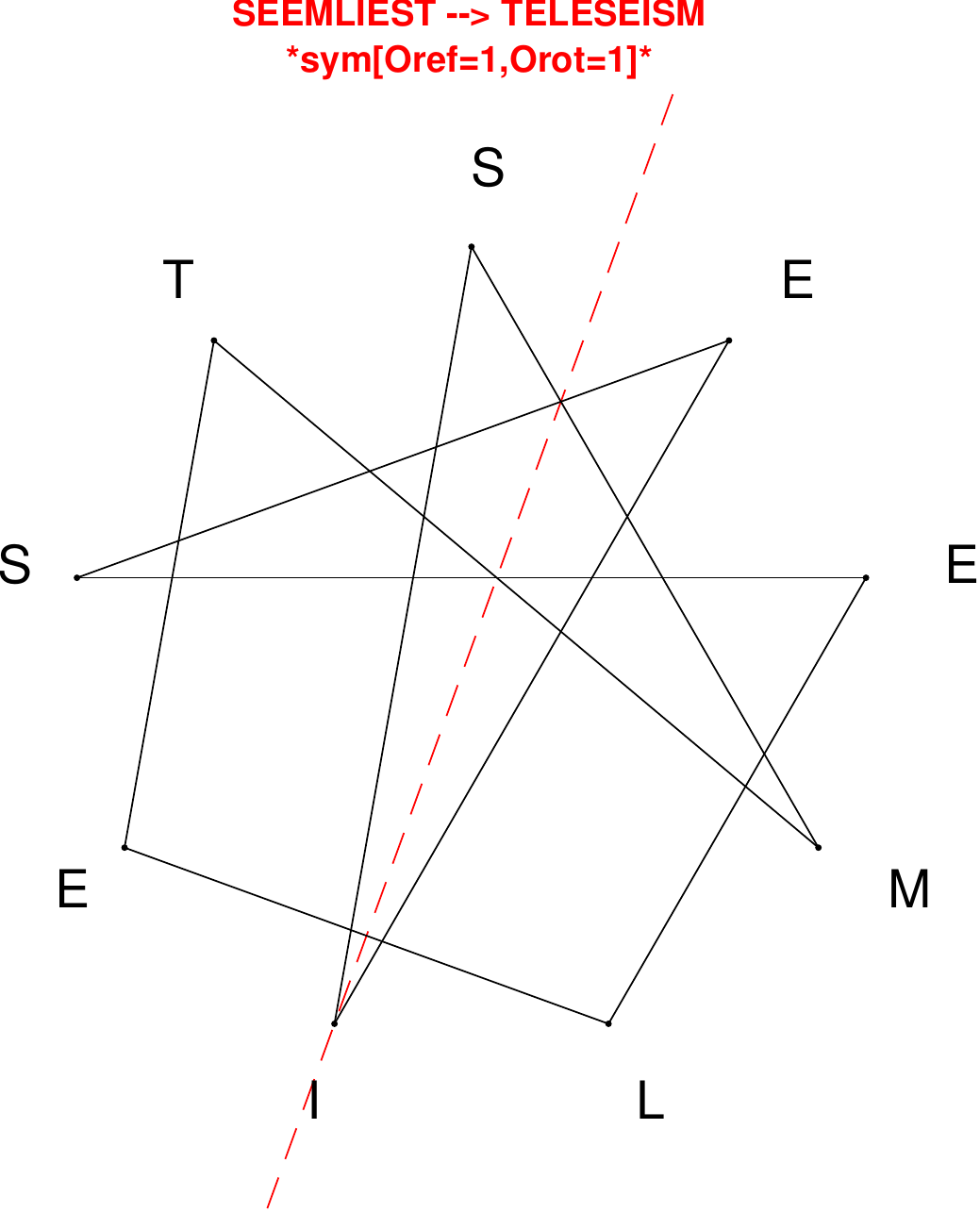}
\end{subfigure}
\hfill
\begin{subfigure}[T]{0.19\textwidth}
\centering
\includegraphics[width=\textwidth]{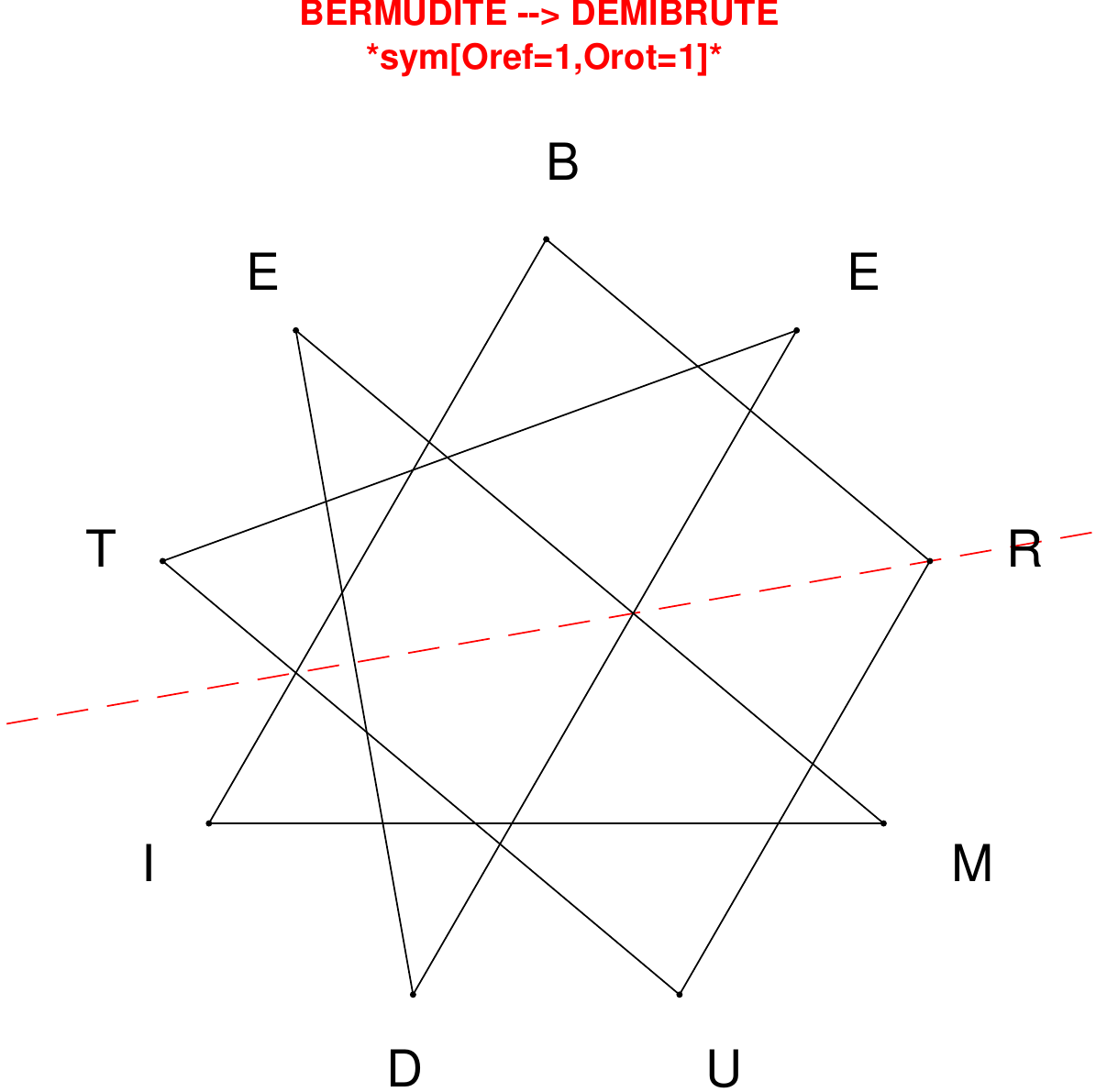}
\end{subfigure}
\hfill
\begin{subfigure}[T]{0.19\textwidth}
\centering
\includegraphics[width=\textwidth]{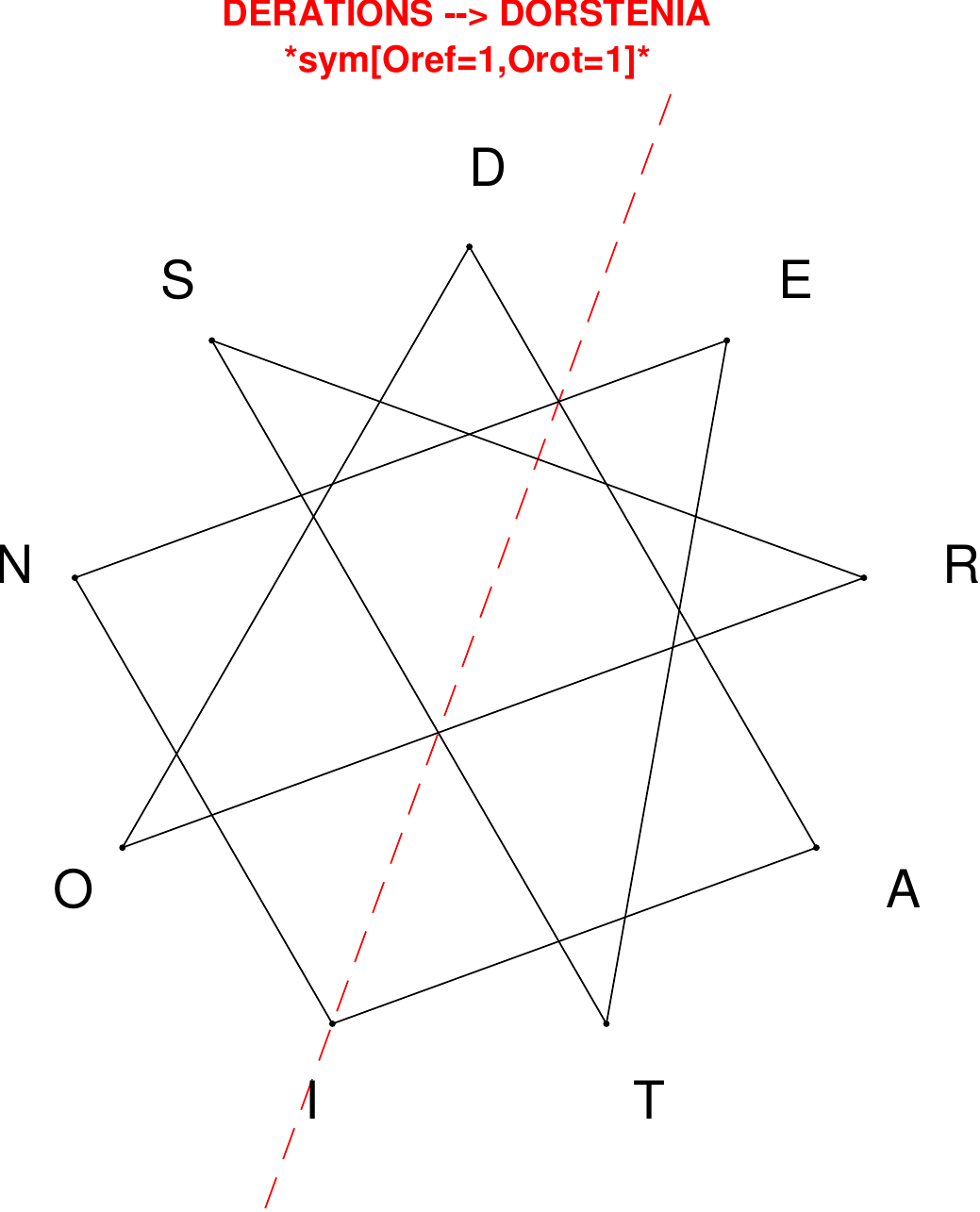}
\end{subfigure}
\hfill
\begin{subfigure}[T]{0.19\textwidth}
\centering
\includegraphics[width=\textwidth]{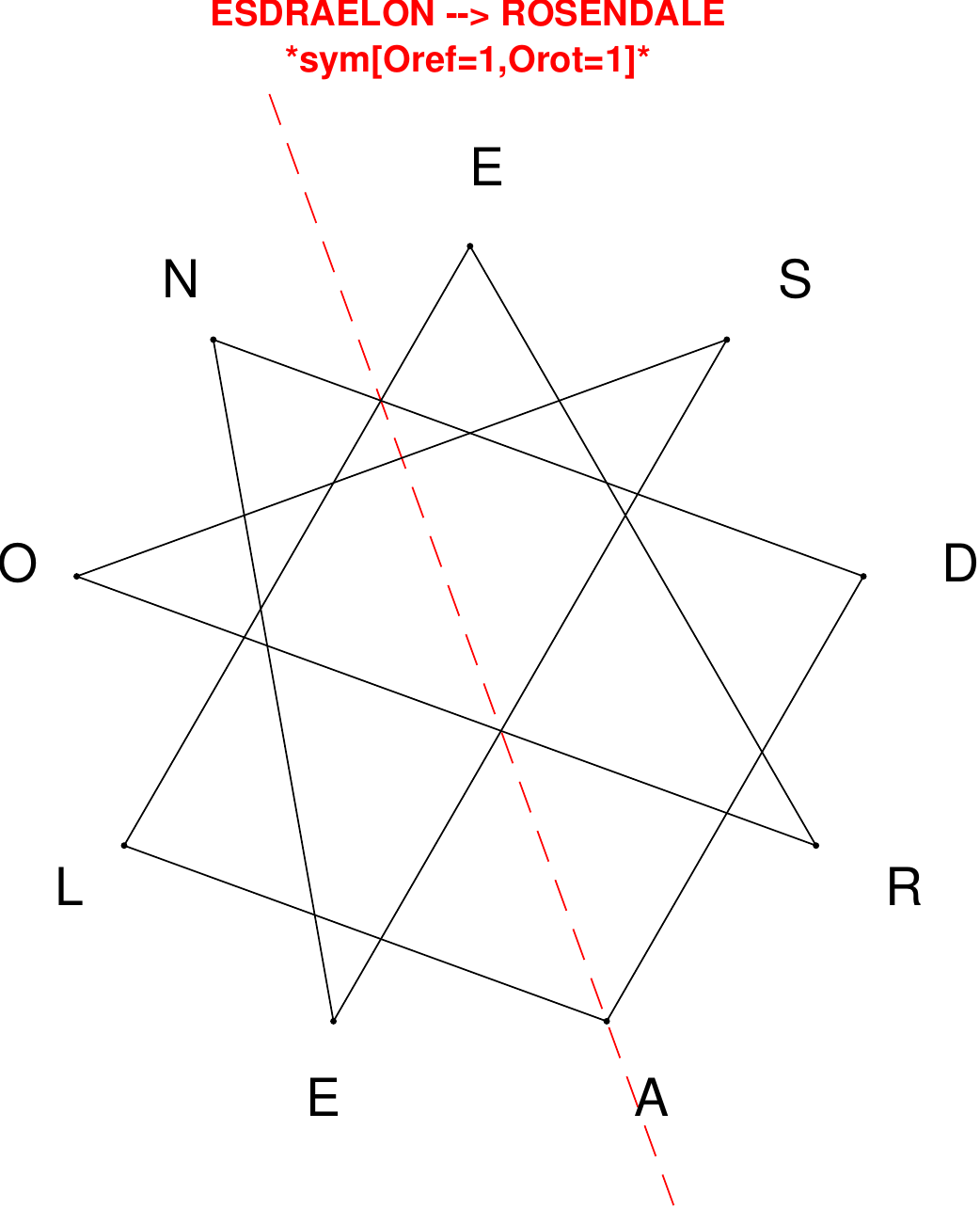}
\end{subfigure}
\end{figure}

\begin{figure}[H]
\centering
\begin{subfigure}[T]{0.19\textwidth}
\centering
\includegraphics[width=\textwidth]{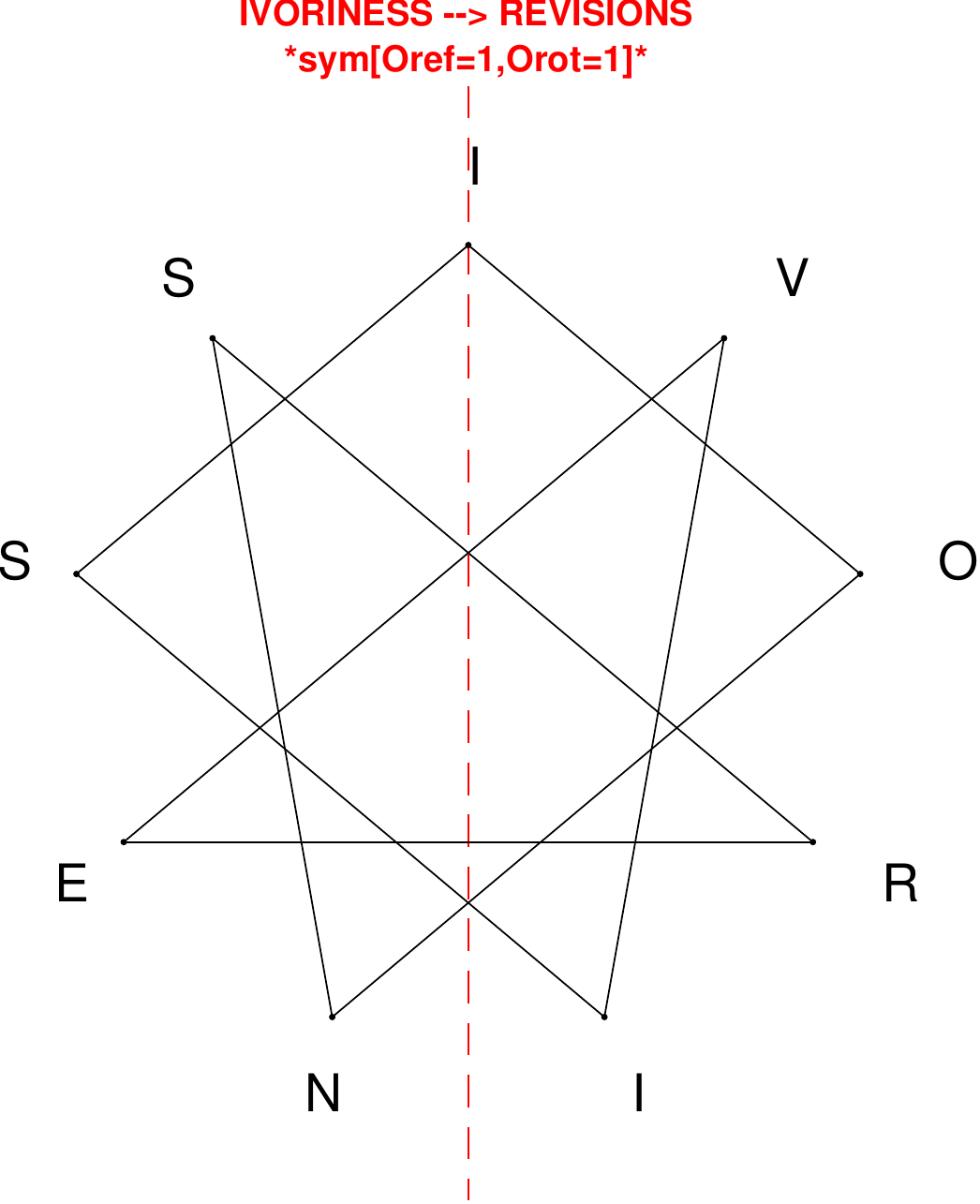}
\end{subfigure}
\hfill
\begin{subfigure}[T]{0.19\textwidth}
\centering
\includegraphics[width=\textwidth]{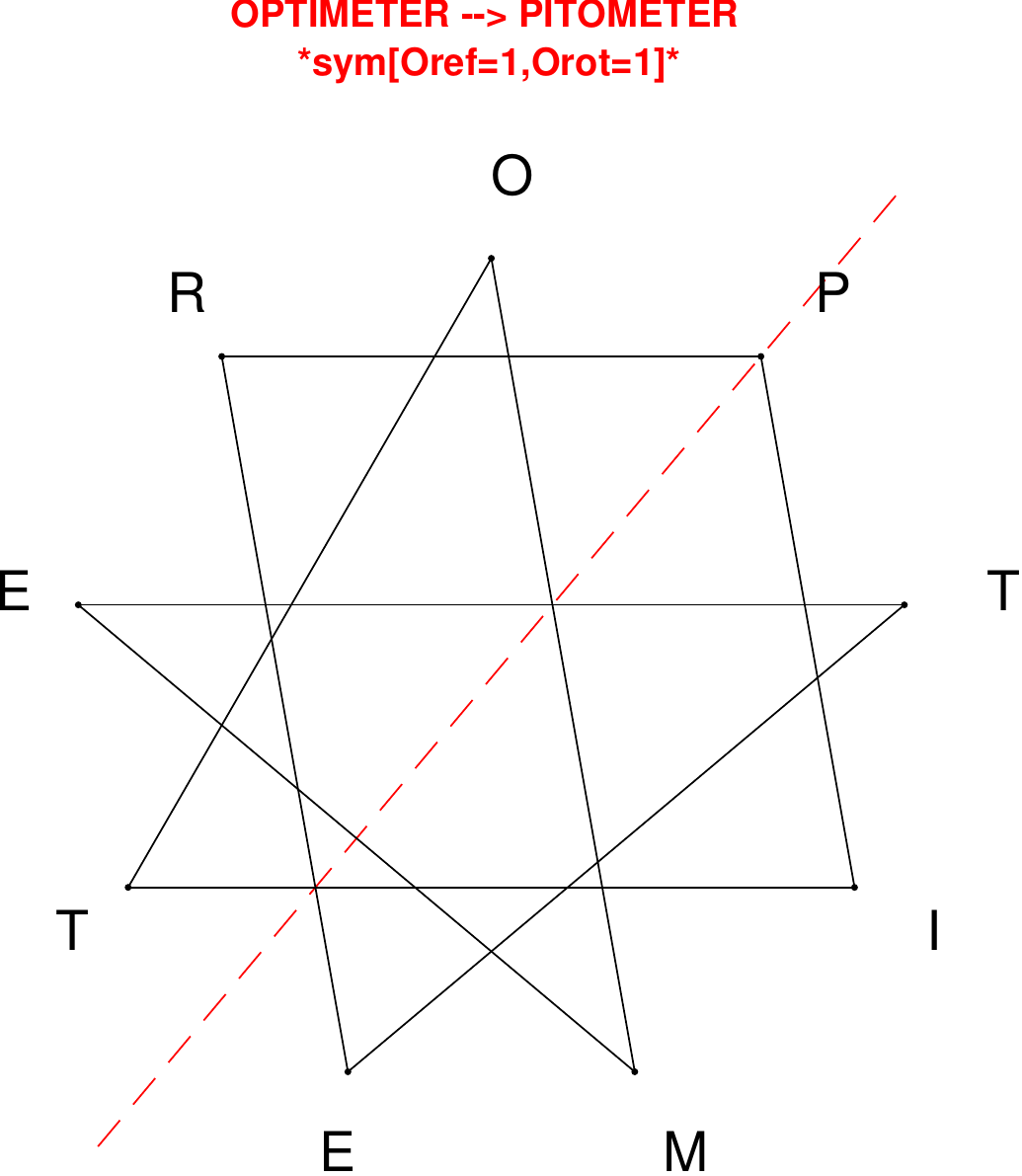}
\end{subfigure}
\hfill
\begin{subfigure}[T]{0.19\textwidth}
\centering
\includegraphics[width=\textwidth]{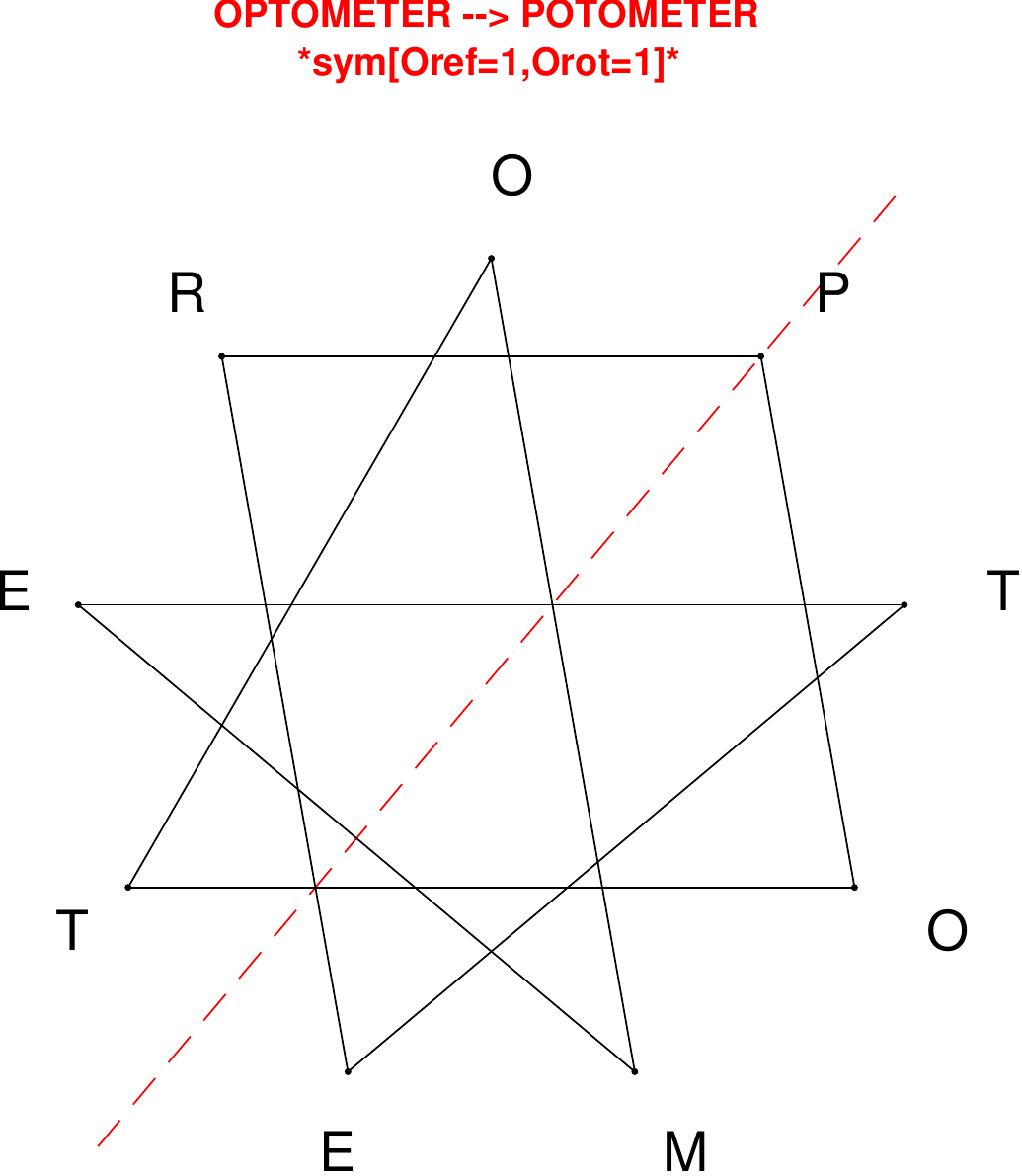}
\end{subfigure}
\hfill
\begin{subfigure}[T]{0.19\textwidth}
\centering
\includegraphics[width=\textwidth]{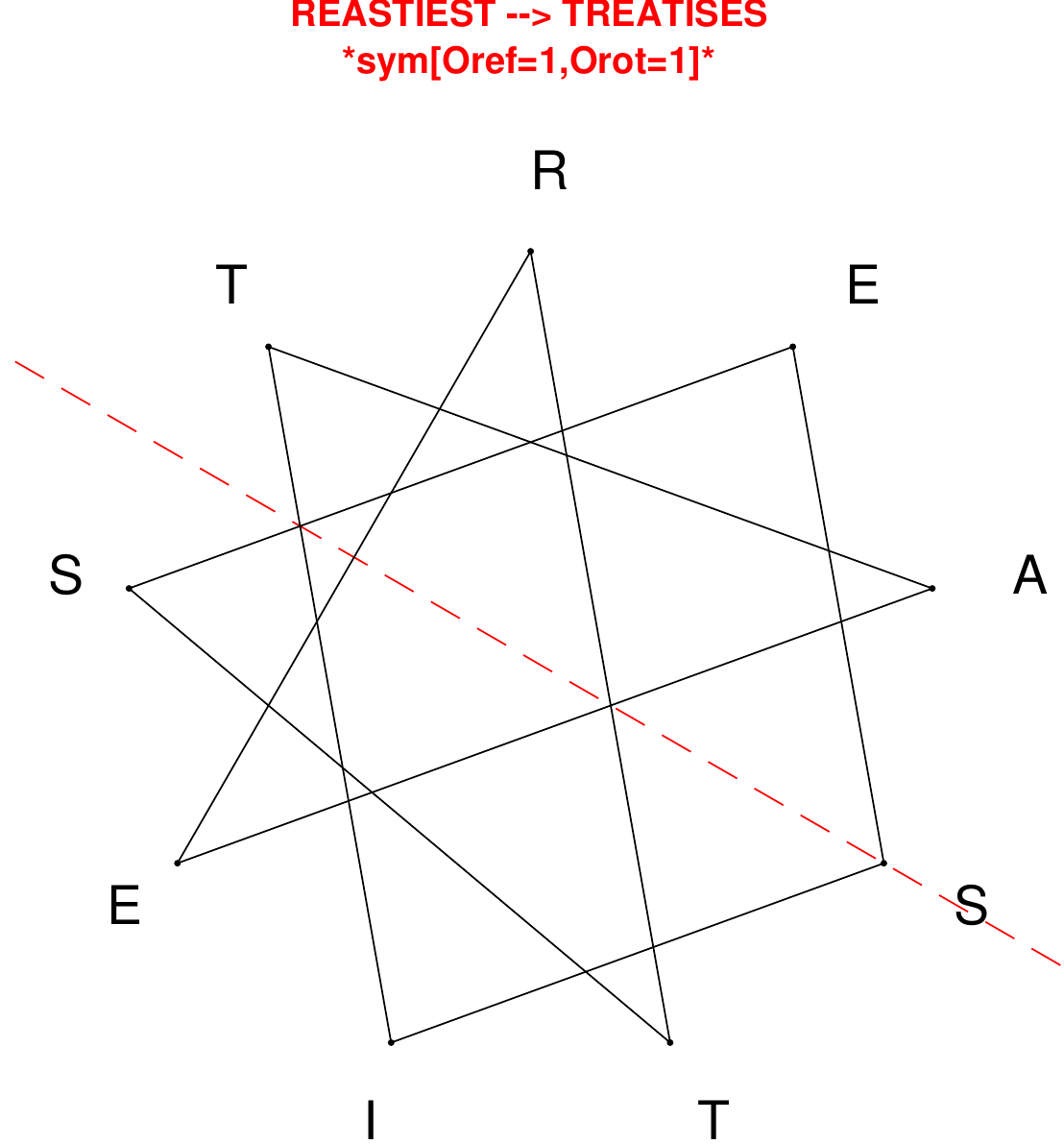}
\end{subfigure}
\hfill
\begin{subfigure}[T]{0.19\textwidth}
\centering
\includegraphics[width=\textwidth]{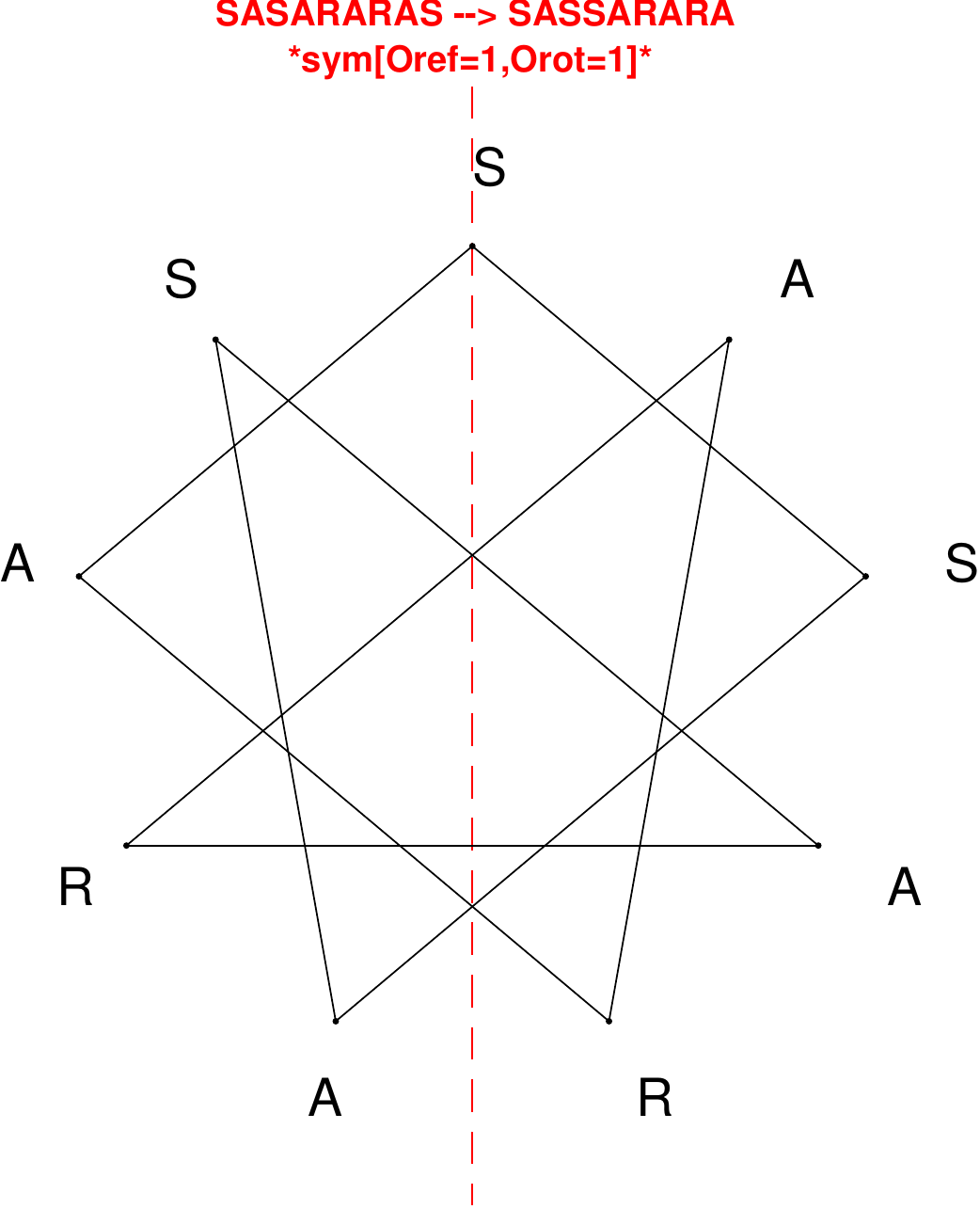}
\end{subfigure}
\end{figure}

\begin{figure}[H]
\centering
\begin{subfigure}[T]{0.19\textwidth}
\centering
\includegraphics[width=\textwidth]{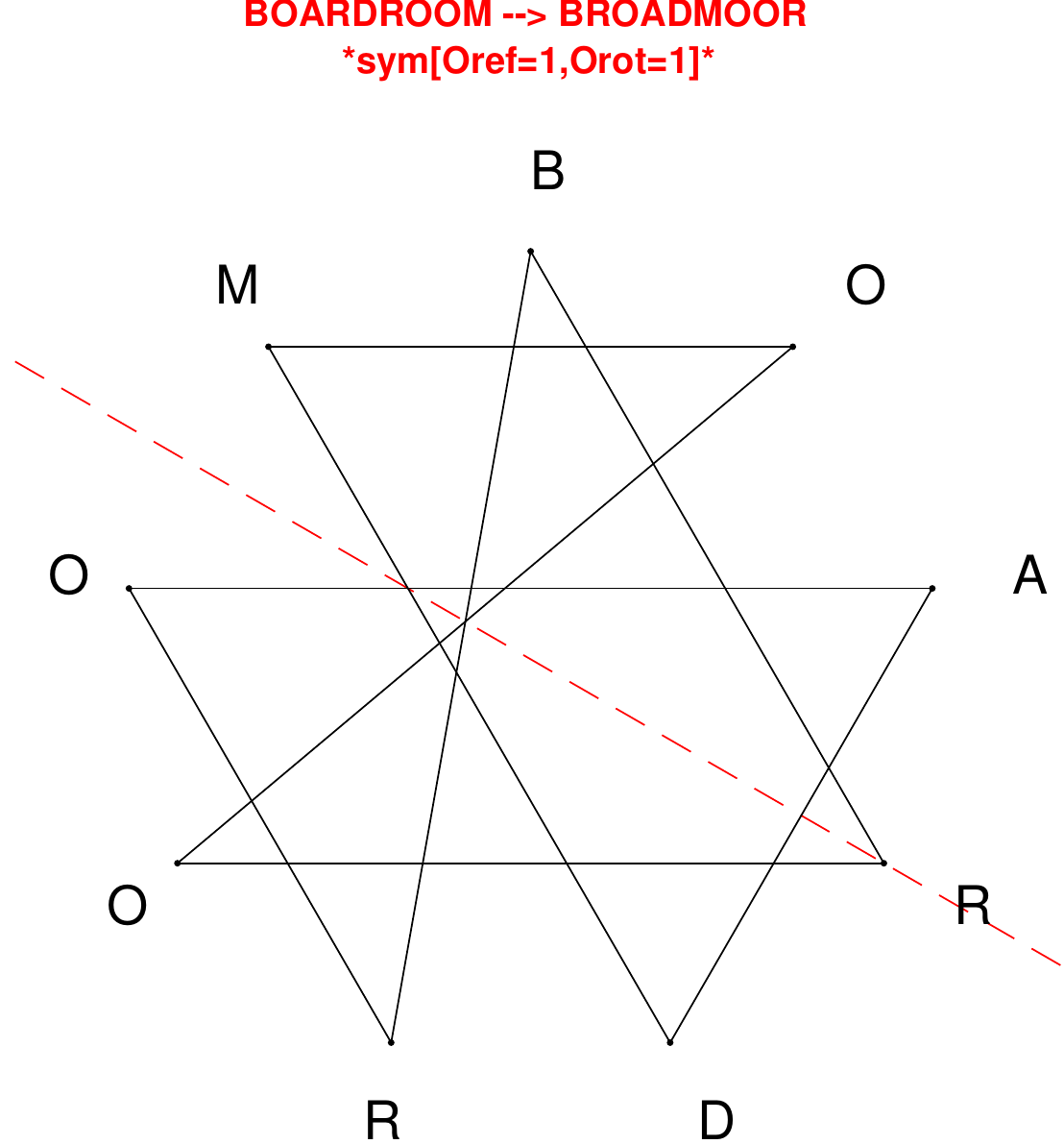}
\end{subfigure}
\hfill
\begin{subfigure}[T]{0.19\textwidth}
\centering
\includegraphics[width=\textwidth]{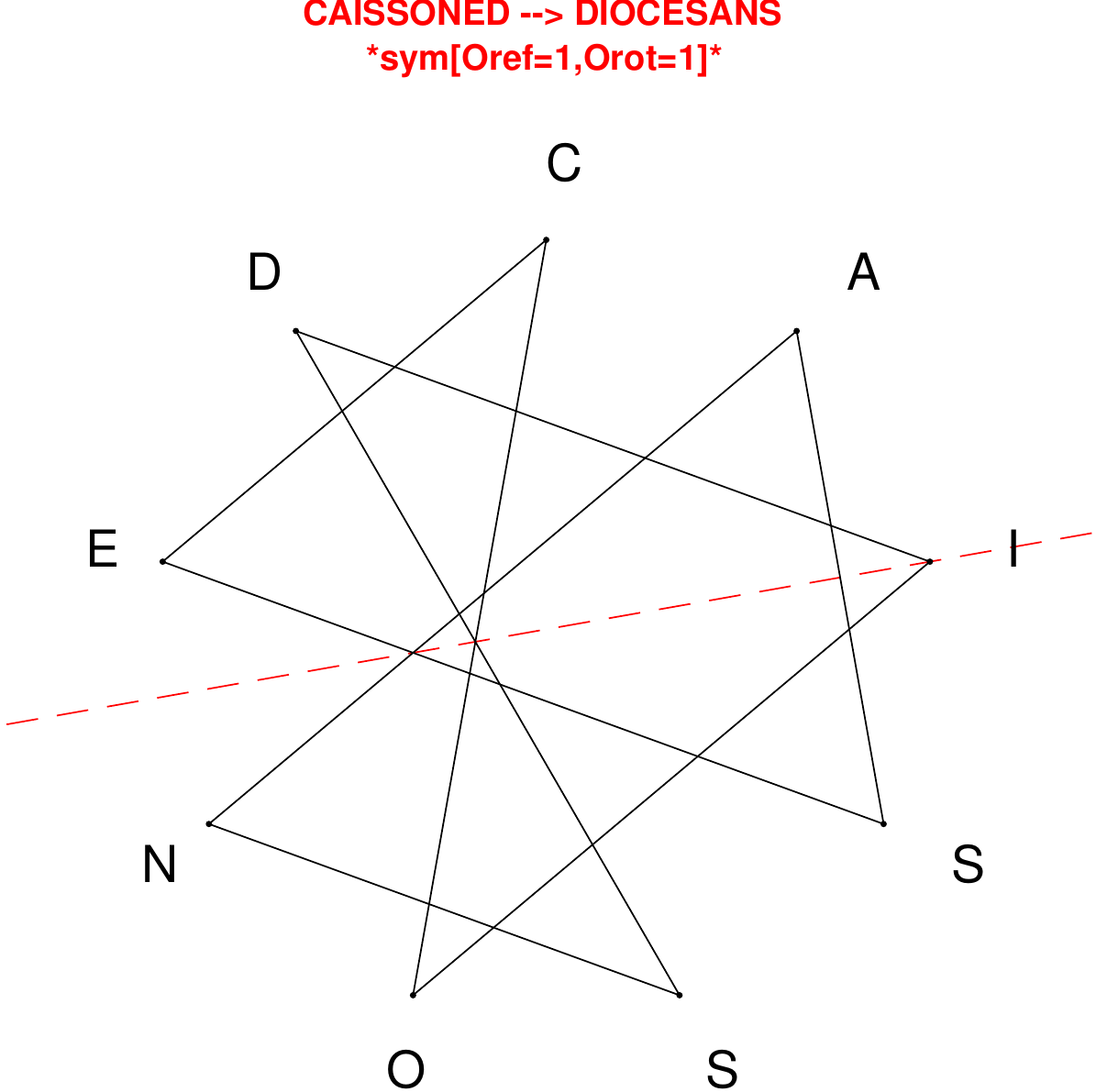}
\end{subfigure}
\hfill
\begin{subfigure}[T]{0.19\textwidth}
\centering
\includegraphics[width=\textwidth]{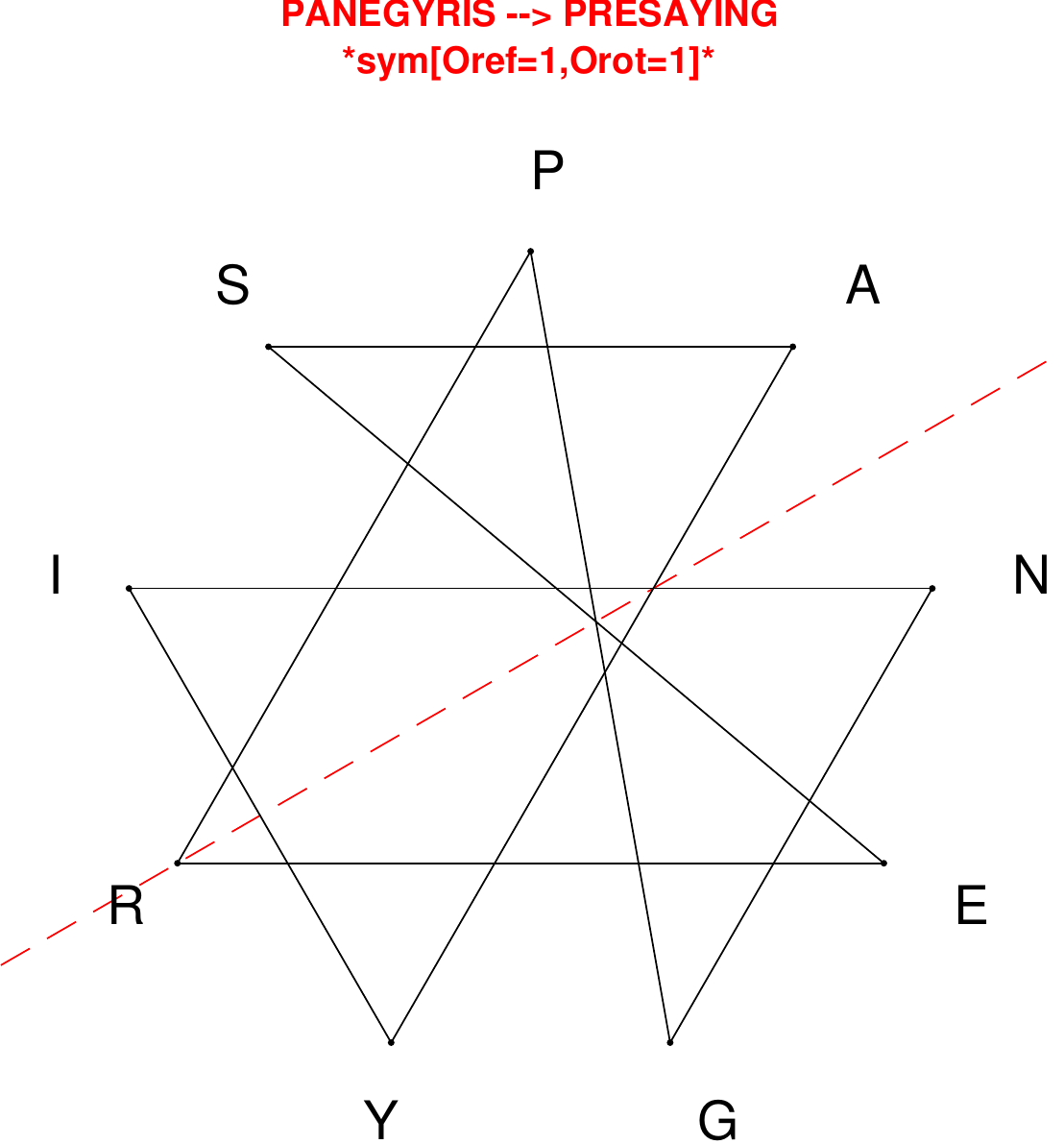}
\end{subfigure}
\hfill
\begin{subfigure}[T]{0.19\textwidth}
\centering
\includegraphics[width=\textwidth]{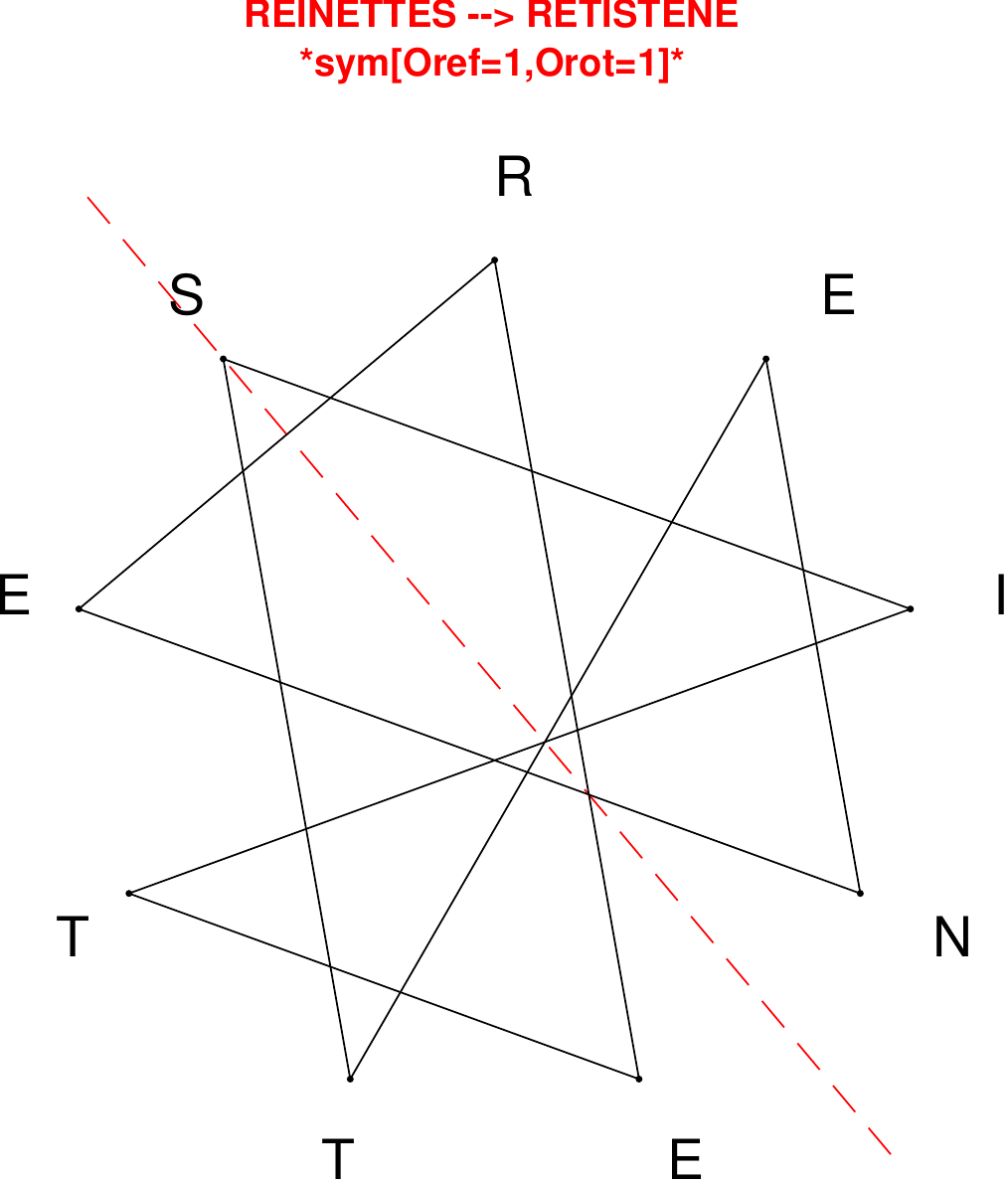}
\end{subfigure}
\hfill
\begin{subfigure}[T]{0.19\textwidth}
\centering
\includegraphics[width=\textwidth]{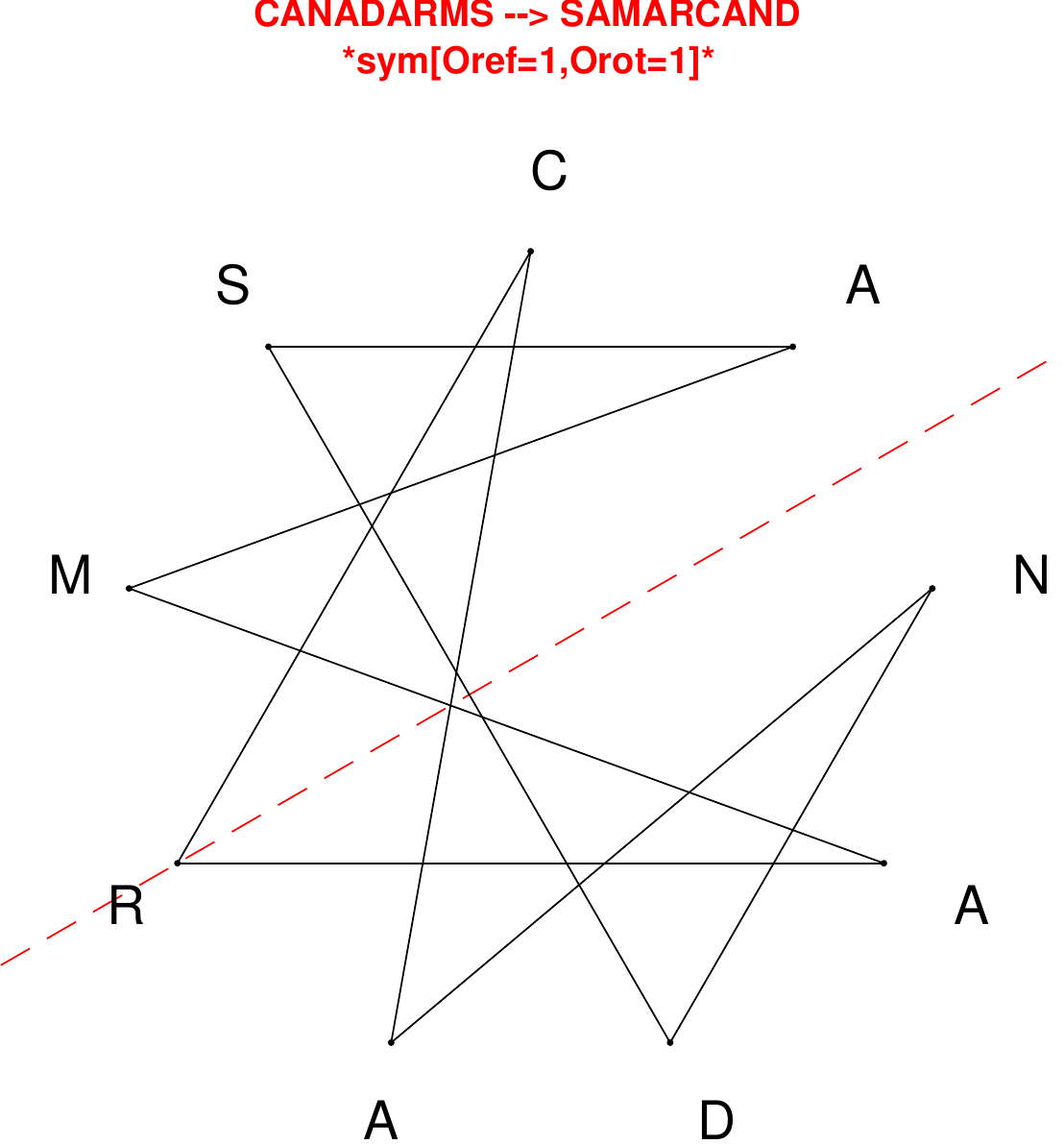}
\end{subfigure}
\end{figure}

\begin{figure}[H]
\centering
\begin{subfigure}[T]{0.19\textwidth}
\centering
\includegraphics[width=\textwidth]{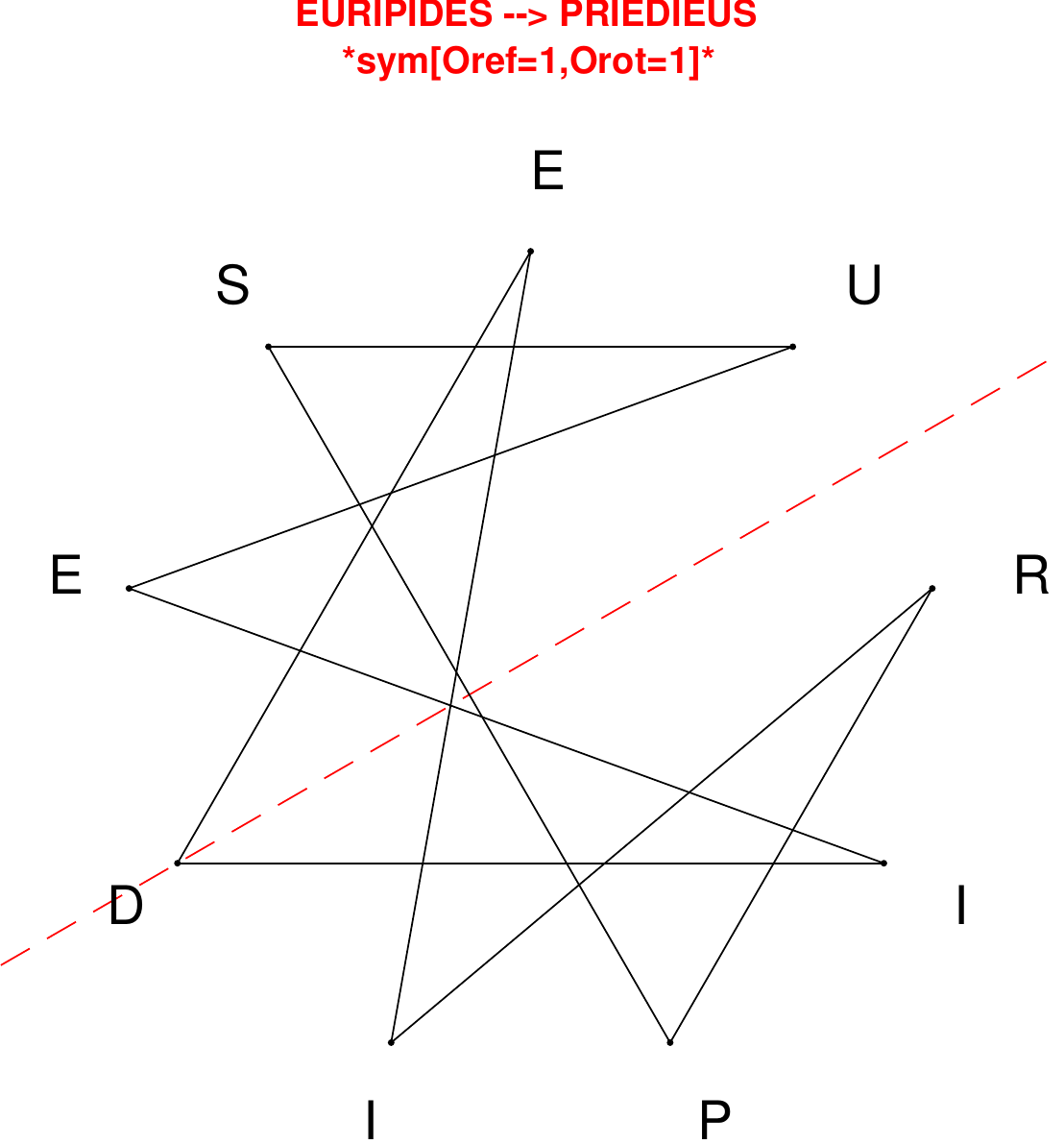}
\end{subfigure}
\hfill
\begin{subfigure}[T]{0.19\textwidth}
\centering
\includegraphics[width=\textwidth]{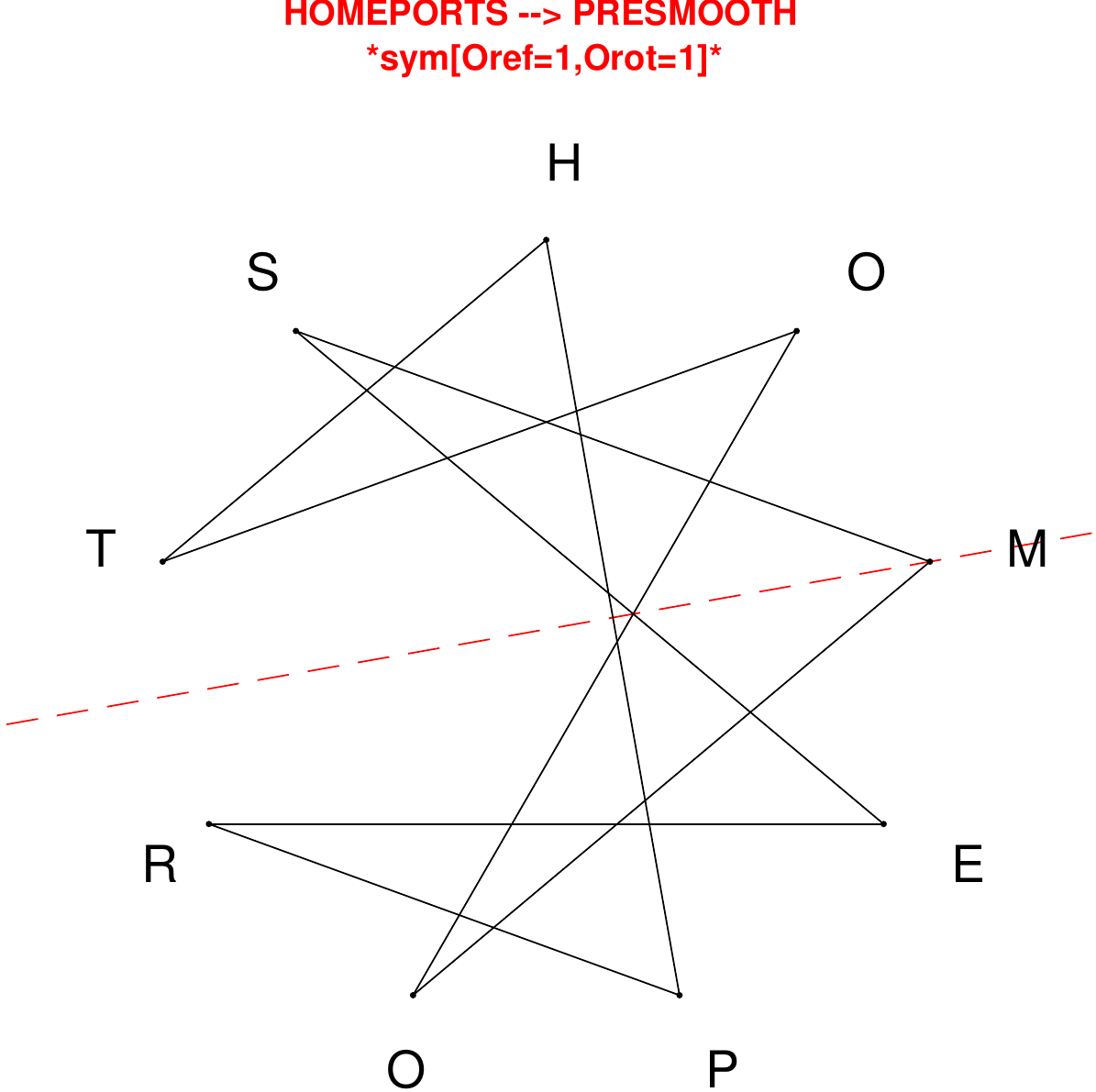}
\end{subfigure}
\hfill
\begin{subfigure}[T]{0.19\textwidth}
\centering
\includegraphics[width=\textwidth]{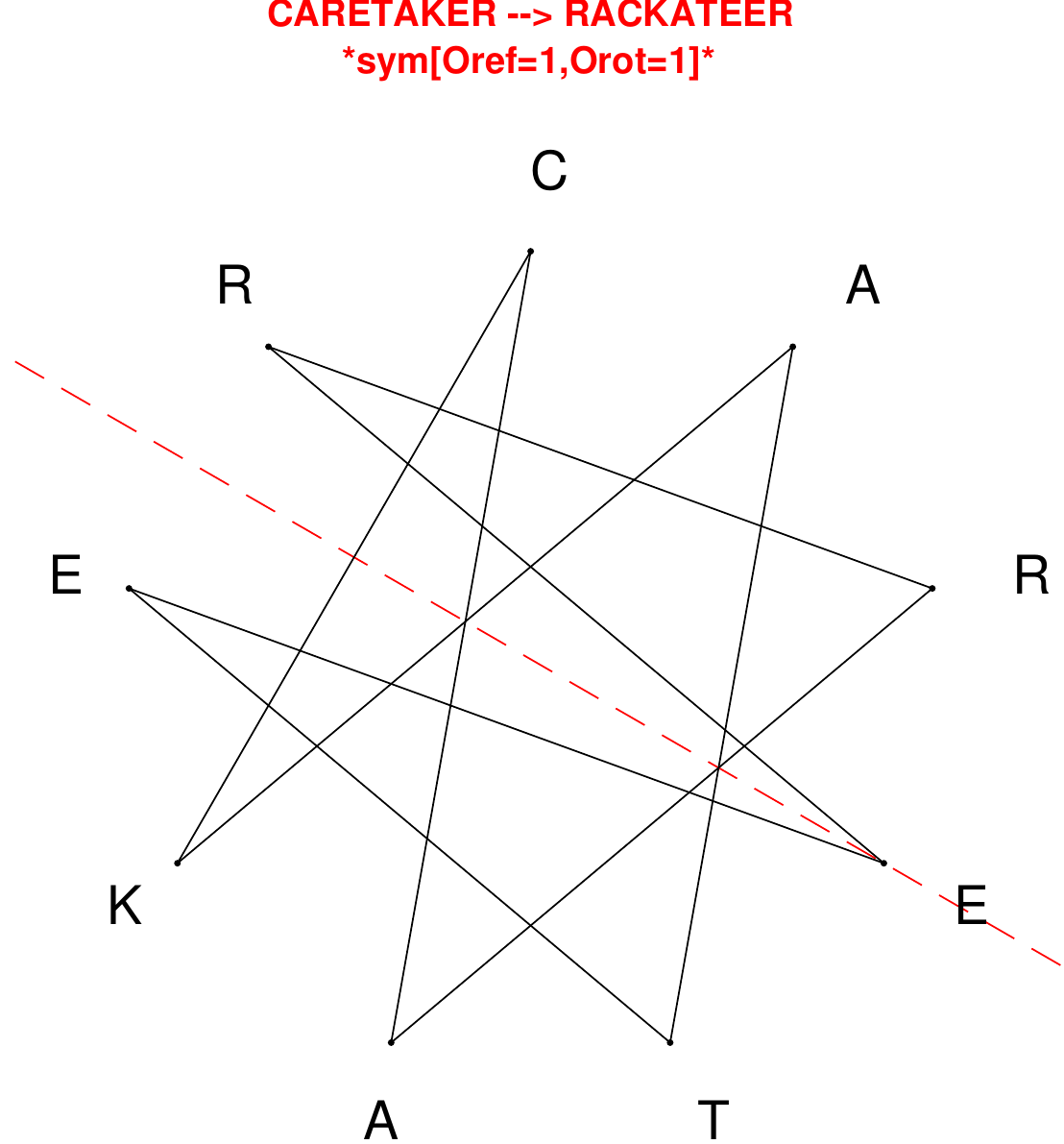}
\end{subfigure}
\hfill
\begin{subfigure}[T]{0.19\textwidth}
\centering
\includegraphics[width=\textwidth]{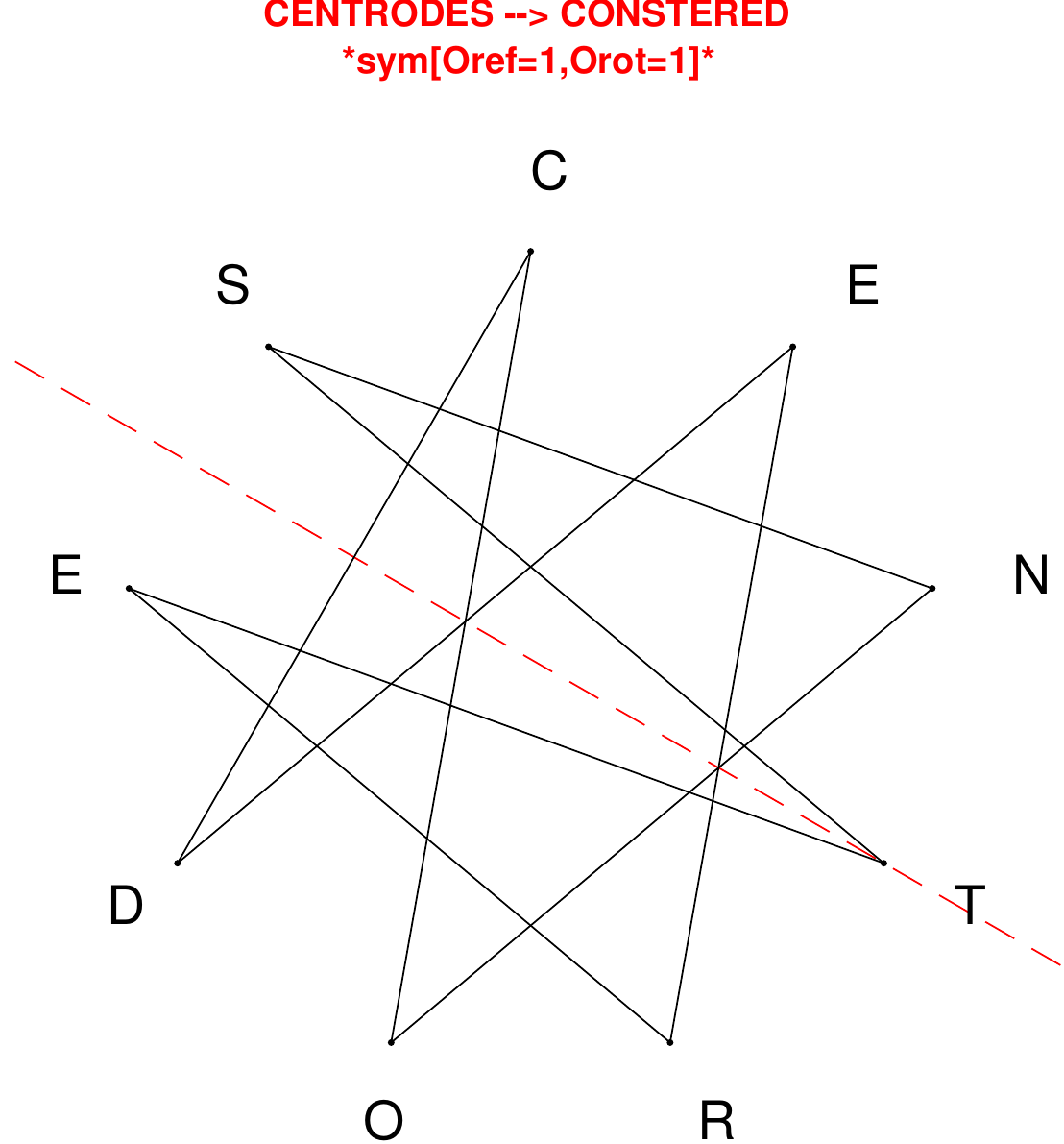}
\end{subfigure}
\hfill
\begin{subfigure}[T]{0.19\textwidth}
\centering
\includegraphics[width=\textwidth]{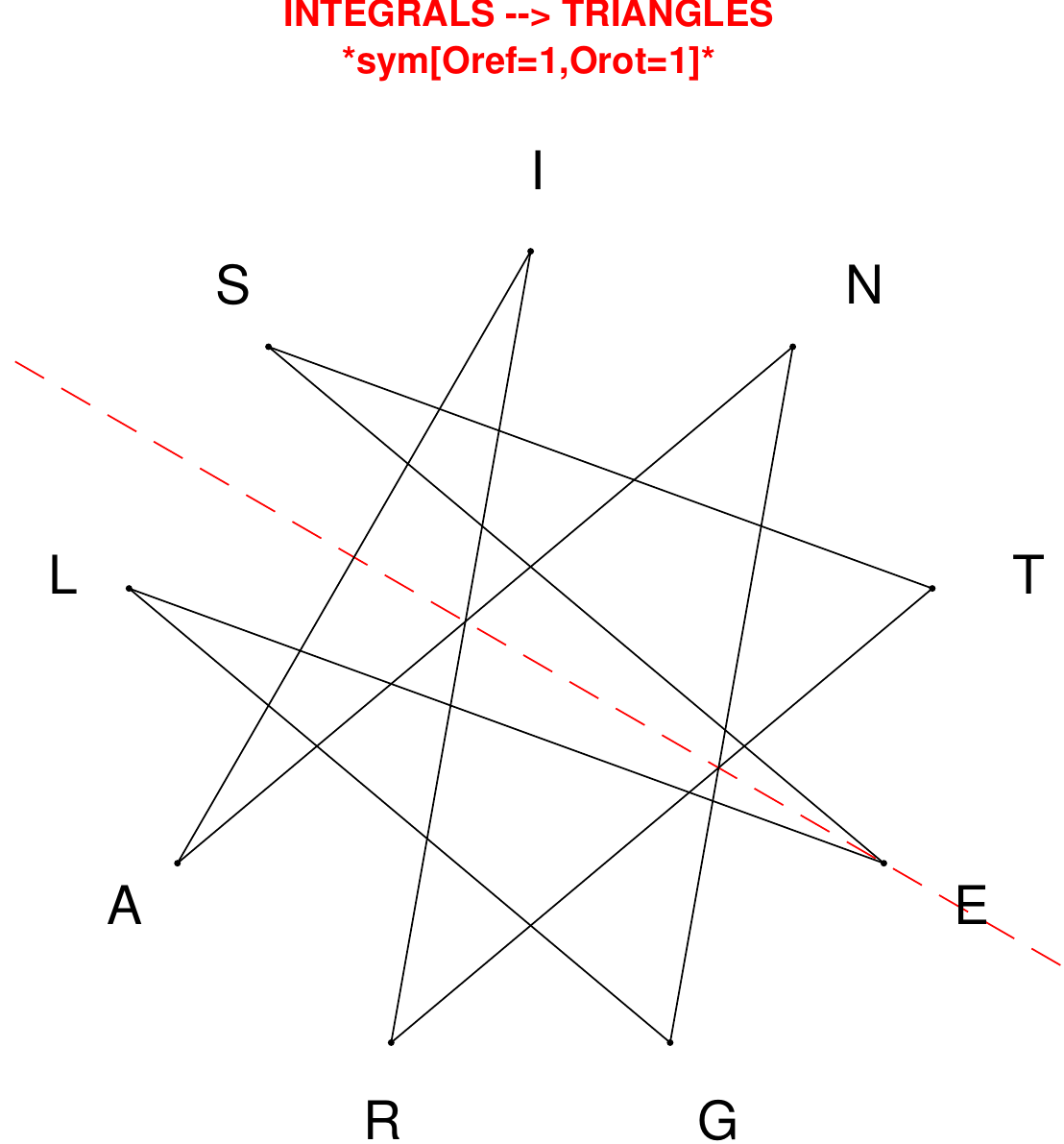}
\end{subfigure}
\end{figure}

\begin{figure}[H]
\centering
\begin{subfigure}[T]{0.19\textwidth}
\centering
\includegraphics[width=\textwidth]{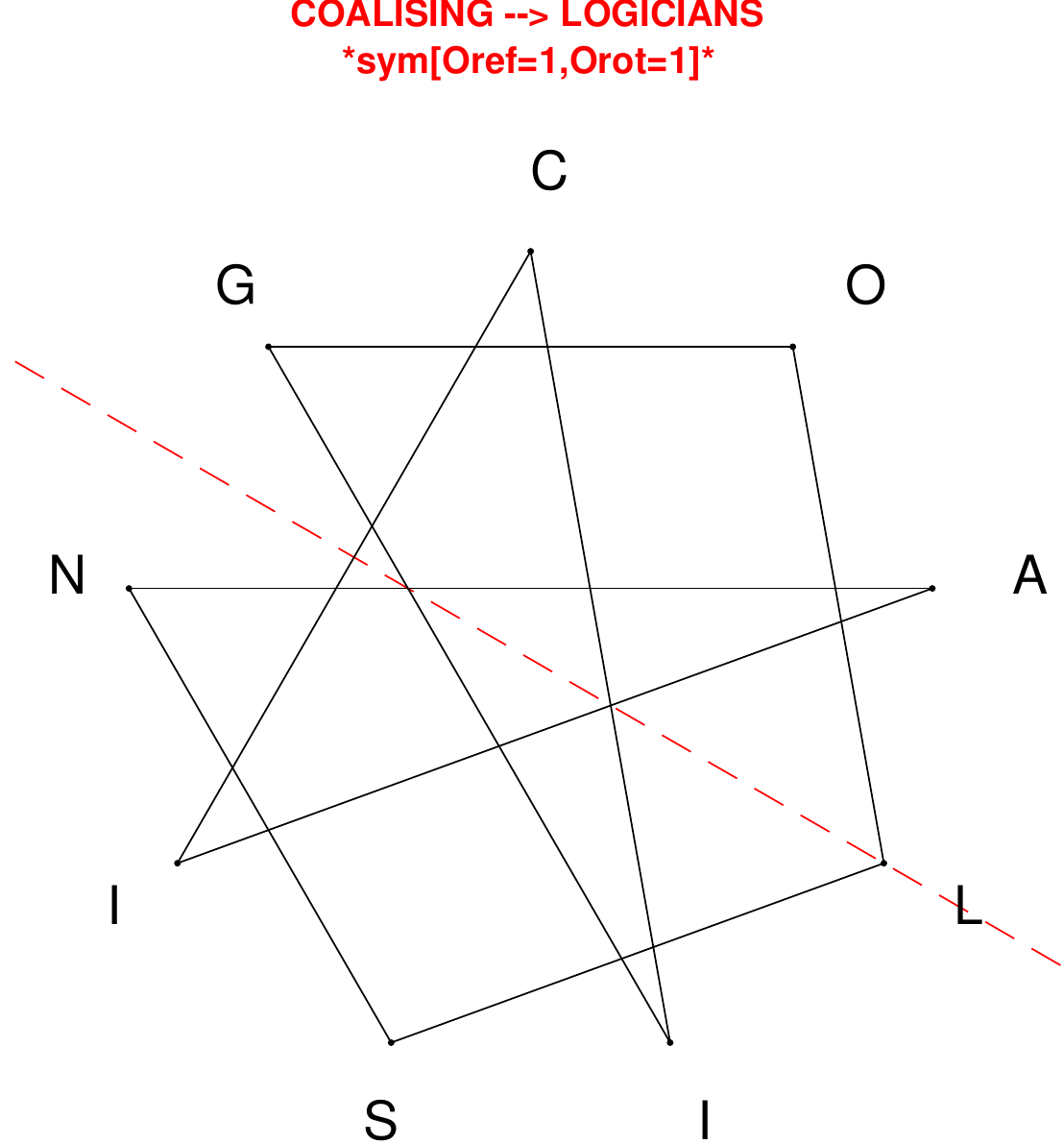}
\end{subfigure}
\hfill
\begin{subfigure}[T]{0.19\textwidth}
\centering
\includegraphics[width=\textwidth]{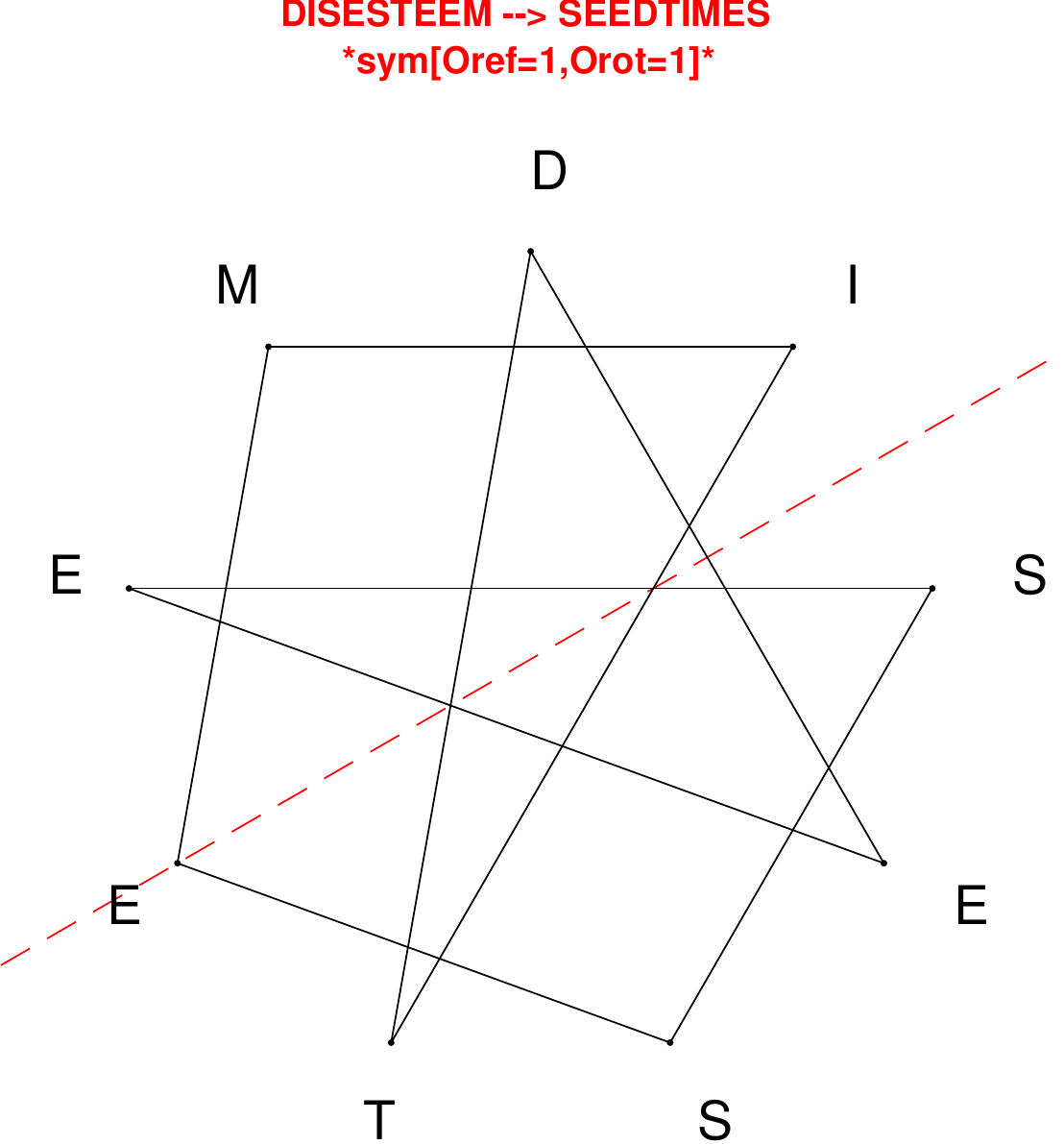}
\end{subfigure}
\hfill
\begin{subfigure}[T]{0.19\textwidth}
\centering
\includegraphics[width=\textwidth]{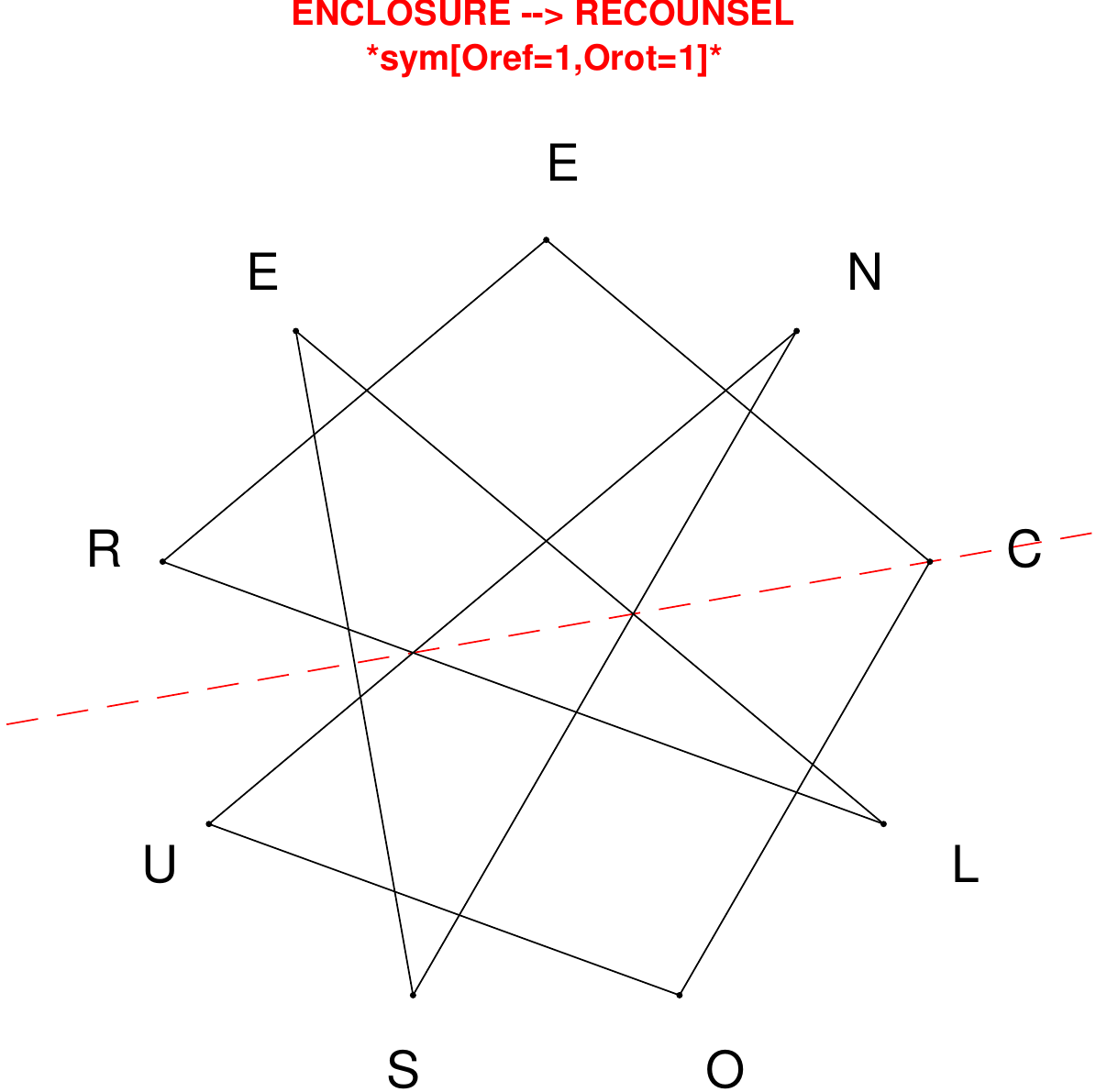}
\end{subfigure}
\hfill
\begin{subfigure}[T]{0.19\textwidth}
\centering
\includegraphics[width=\textwidth]{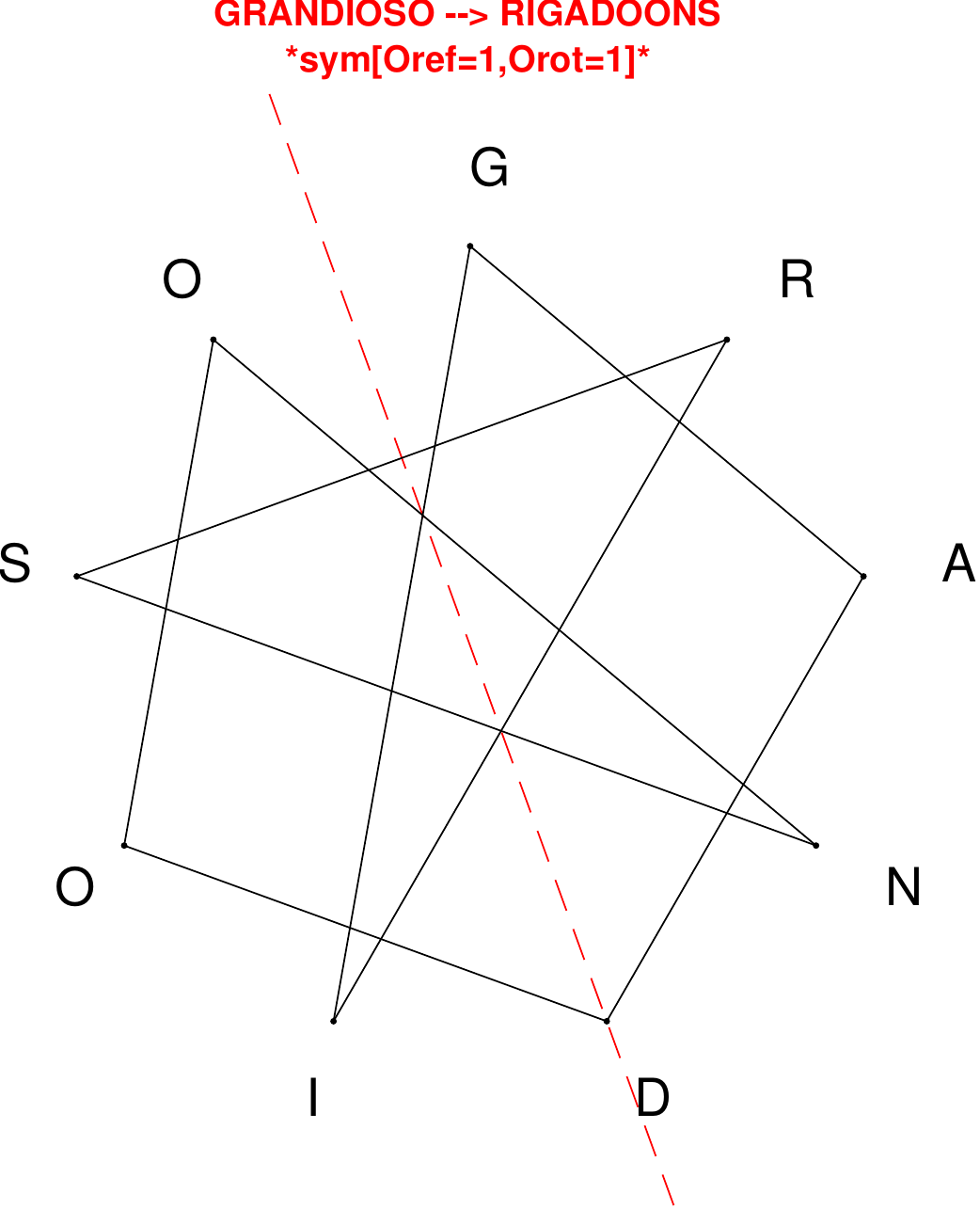}
\end{subfigure}
\hfill
\begin{subfigure}[T]{0.19\textwidth}
\centering
\includegraphics[width=\textwidth]{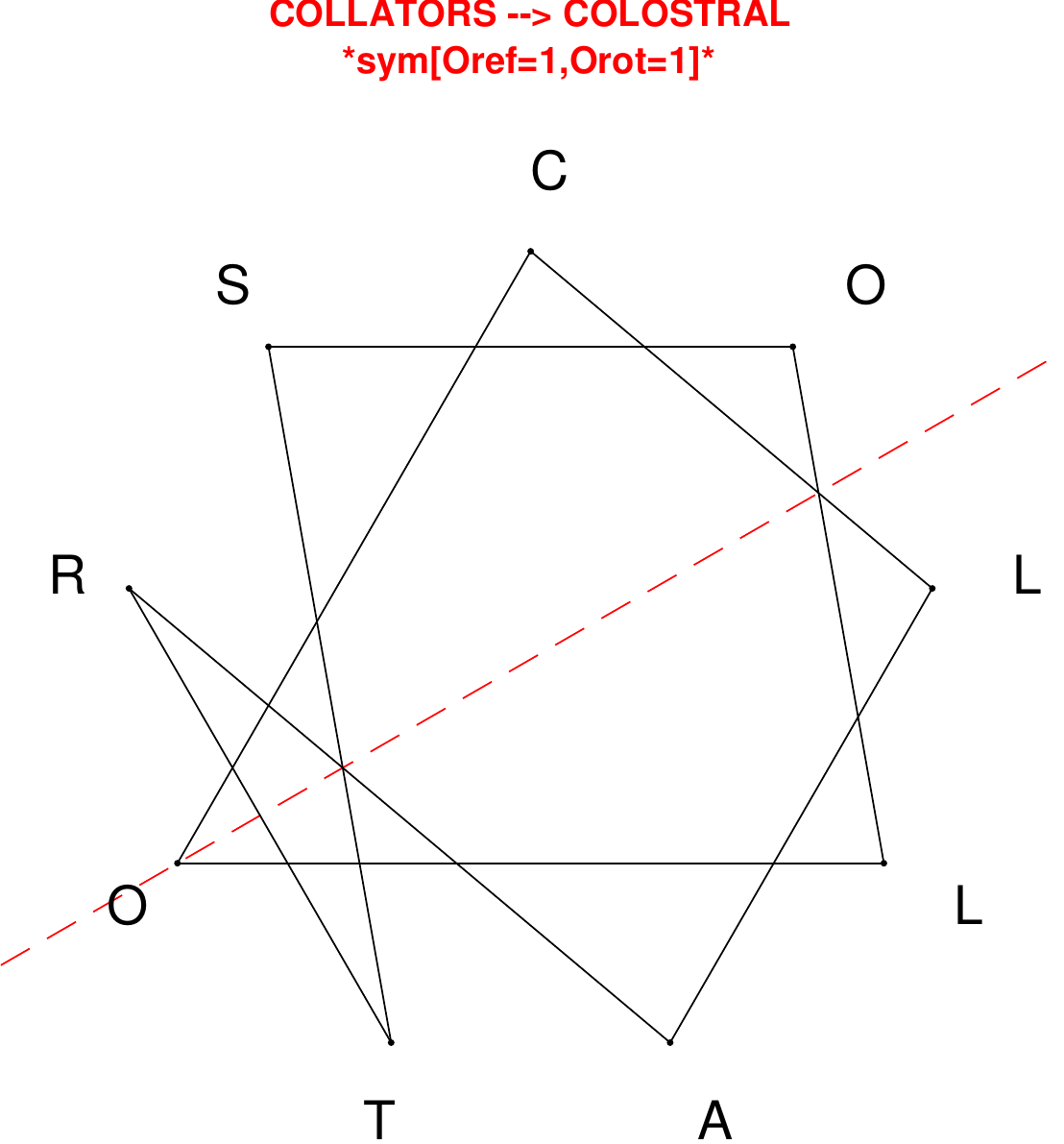}
\end{subfigure}
\end{figure}

\begin{figure}[H]
\centering
\begin{subfigure}[T]{0.19\textwidth}
\centering
\includegraphics[width=\textwidth]{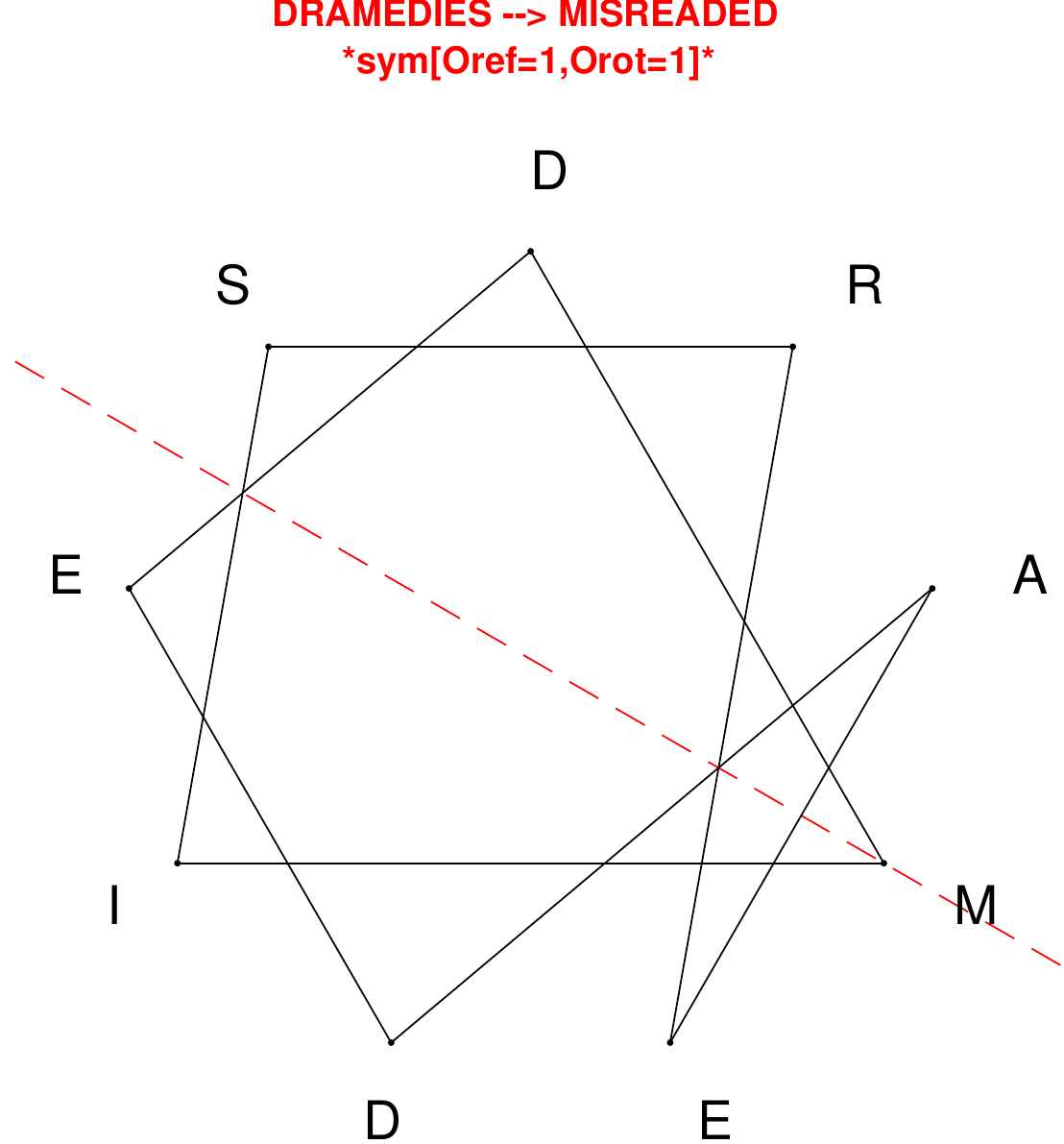}
\end{subfigure}
\hfill
\begin{subfigure}[T]{0.19\textwidth}
\centering
\includegraphics[width=\textwidth]{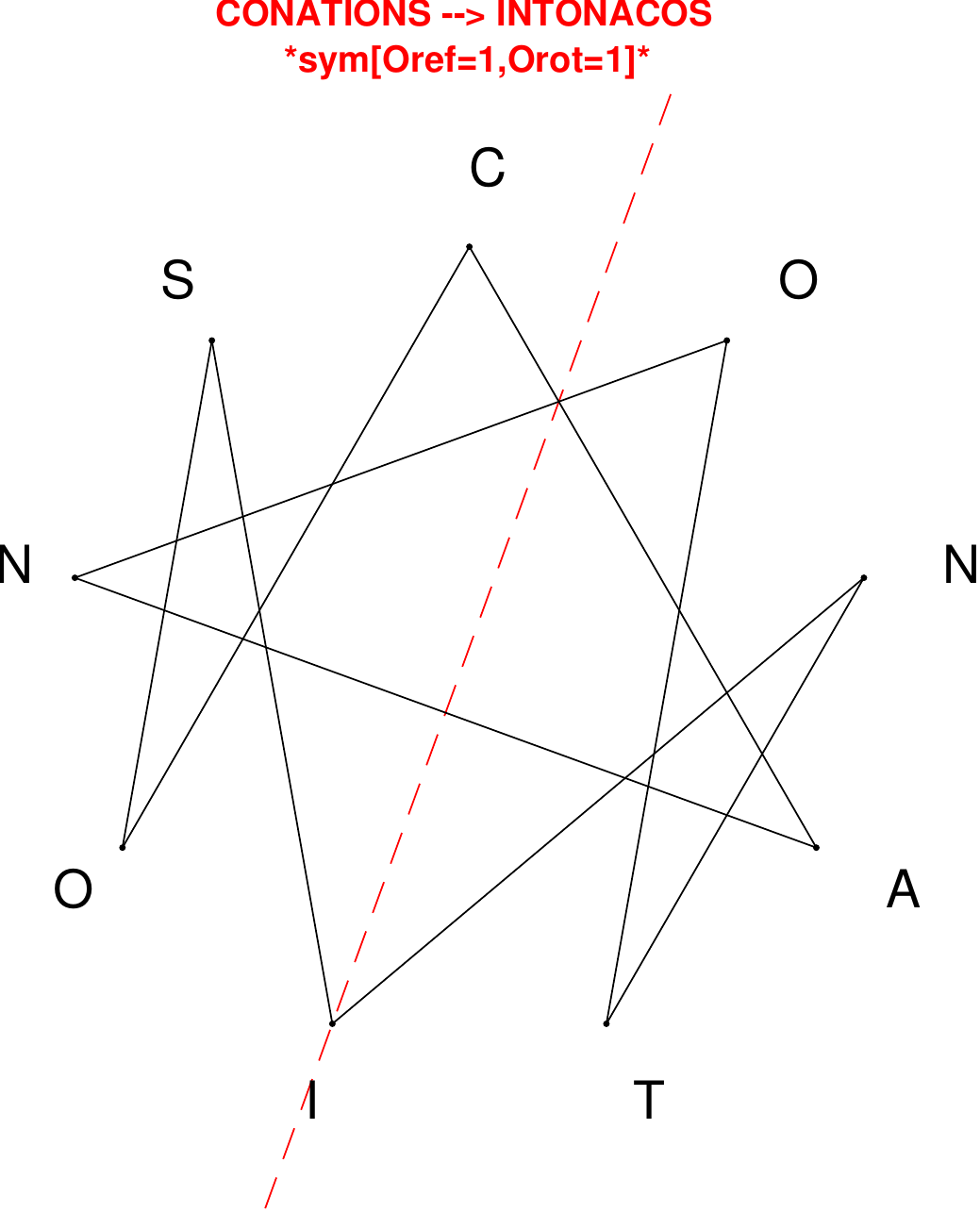}
\end{subfigure}
\hfill
\begin{subfigure}[T]{0.19\textwidth}
\centering
\includegraphics[width=\textwidth]{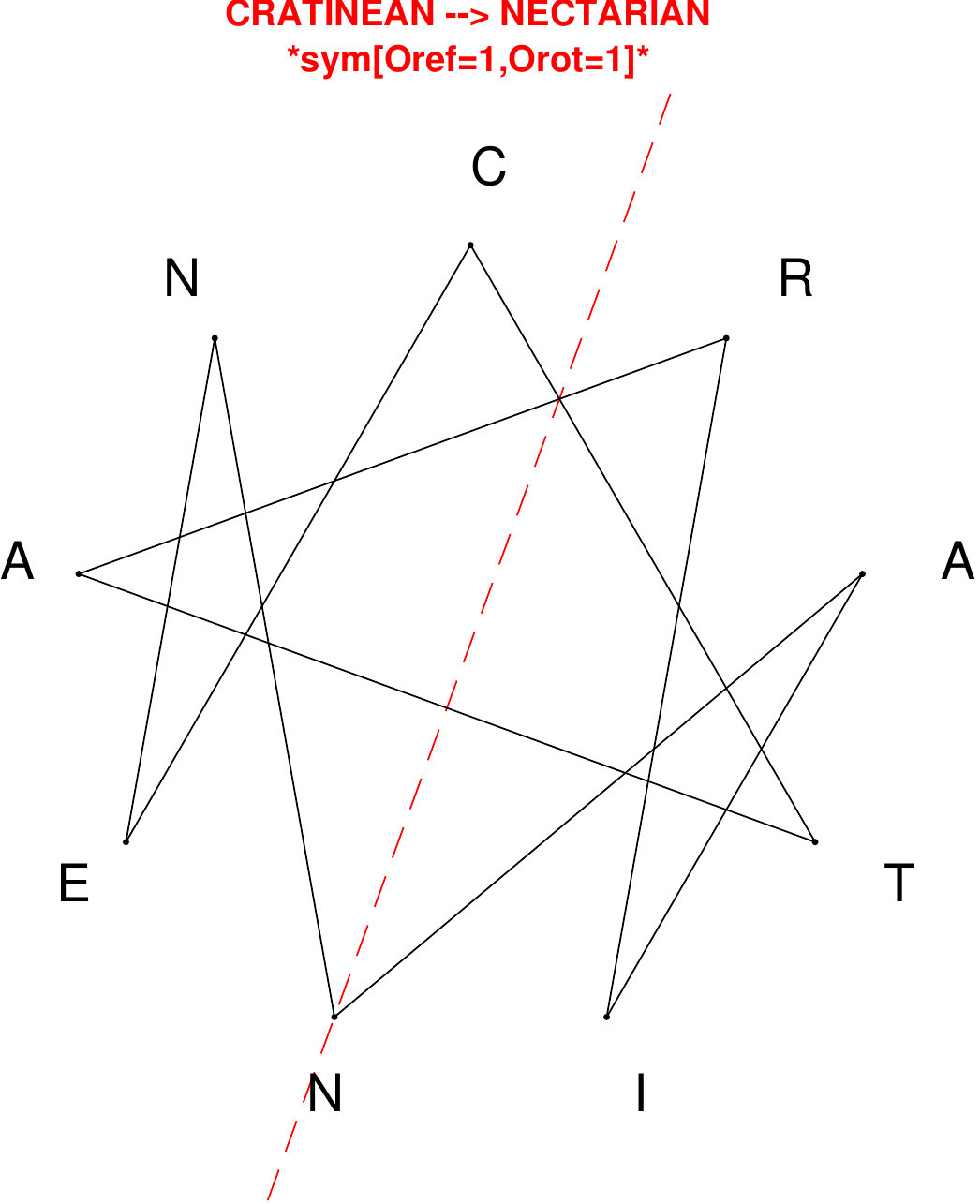}
\end{subfigure}
\hfill
\begin{subfigure}[T]{0.19\textwidth}
\centering
\includegraphics[width=\textwidth]{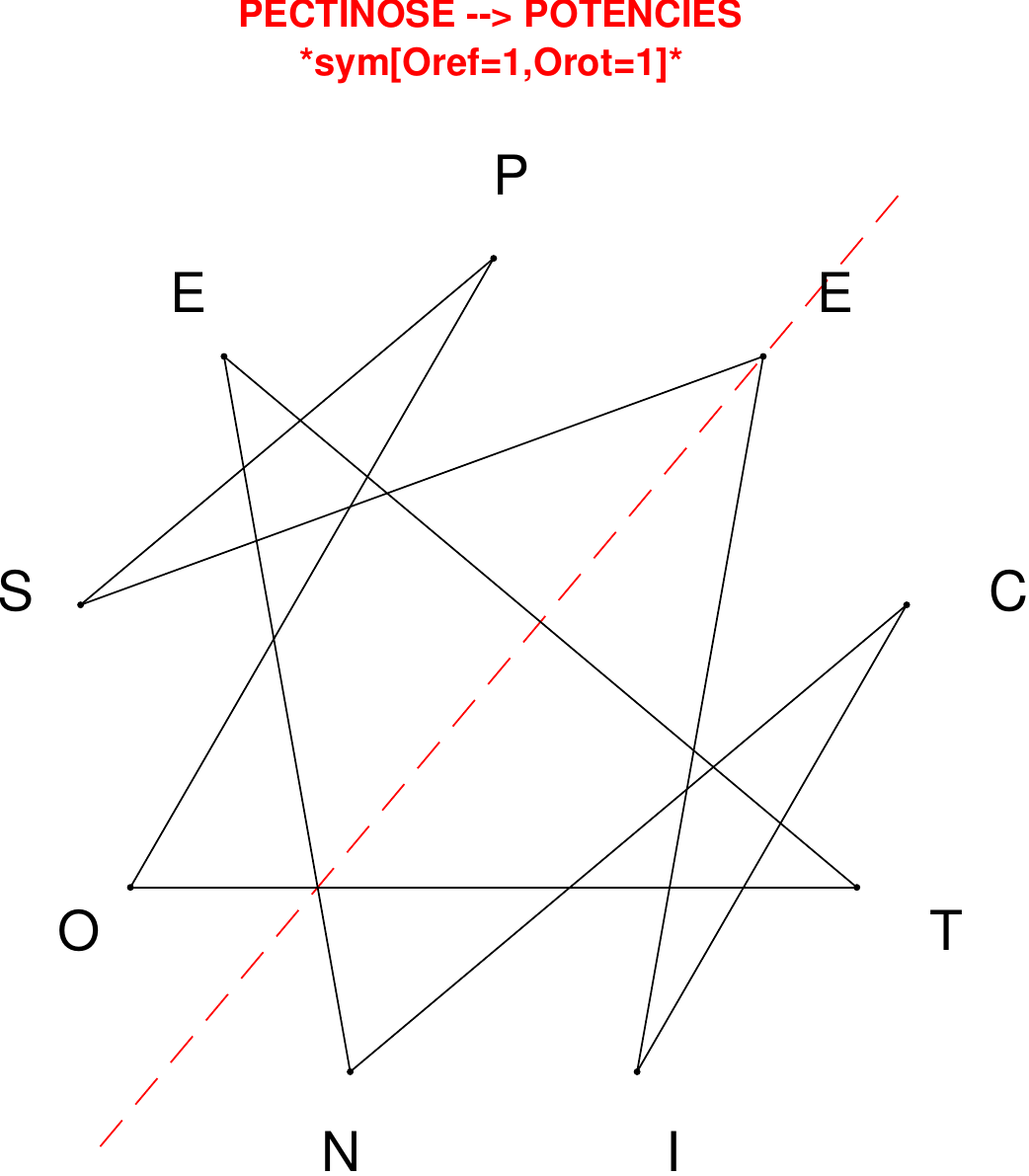}
\end{subfigure}
\hfill
\begin{subfigure}[T]{0.19\textwidth}
\centering
\includegraphics[width=\textwidth]{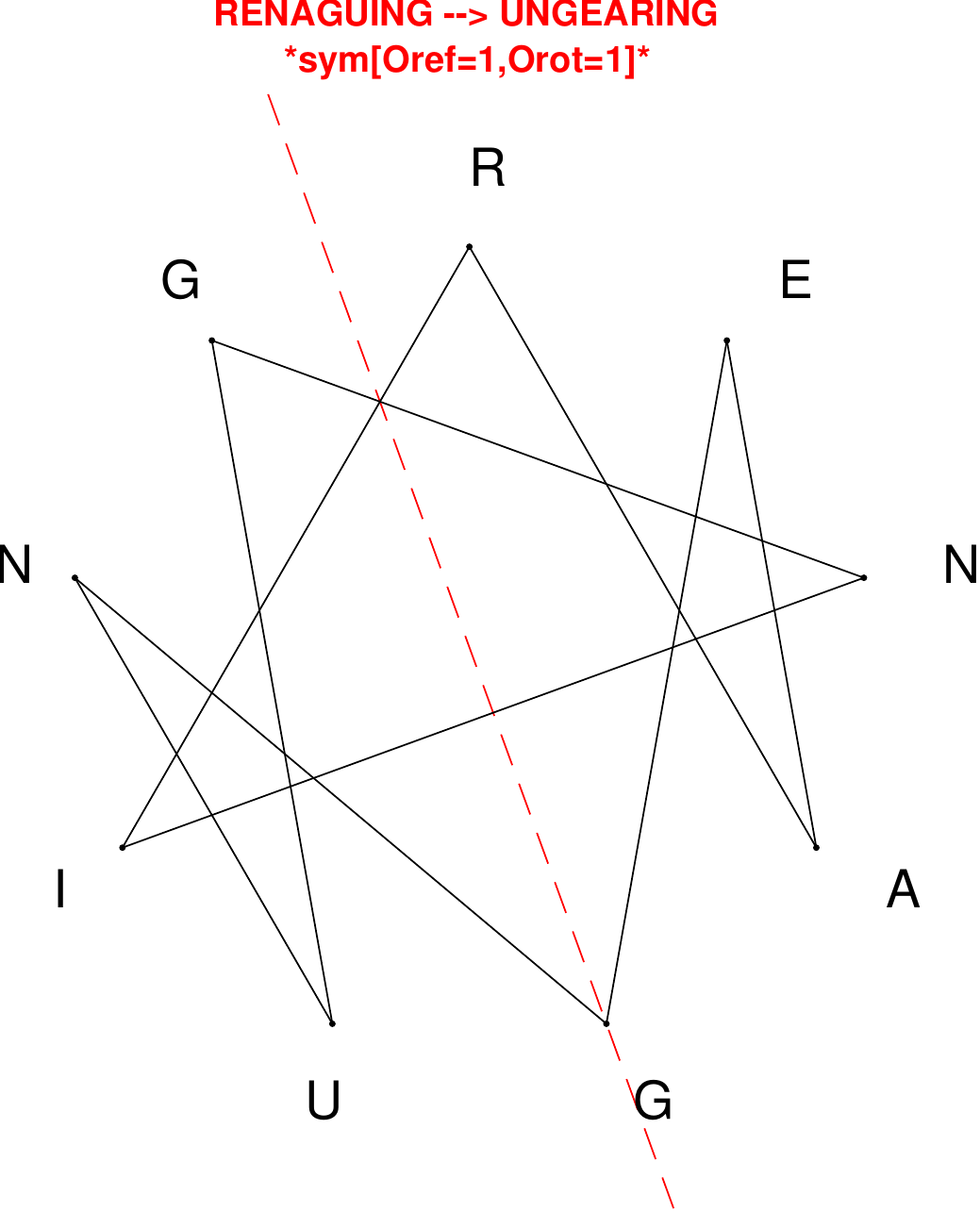}
\end{subfigure}
\end{figure}

\begin{figure}[H]
\centering
\begin{subfigure}[T]{0.19\textwidth}
\centering
\includegraphics[width=\textwidth]{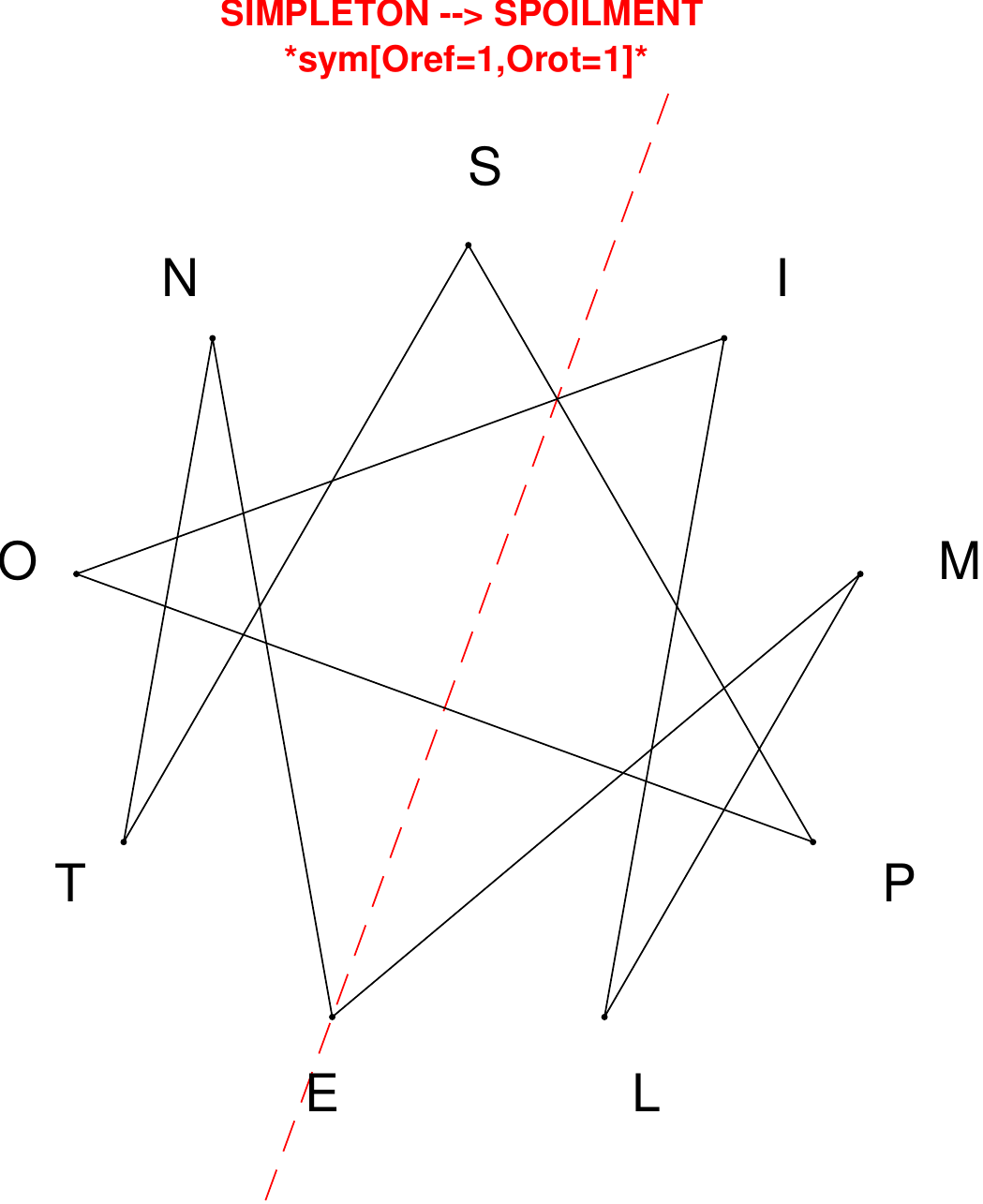}
\end{subfigure}
\hfill
\begin{subfigure}[T]{0.19\textwidth}
\centering
\includegraphics[width=\textwidth]{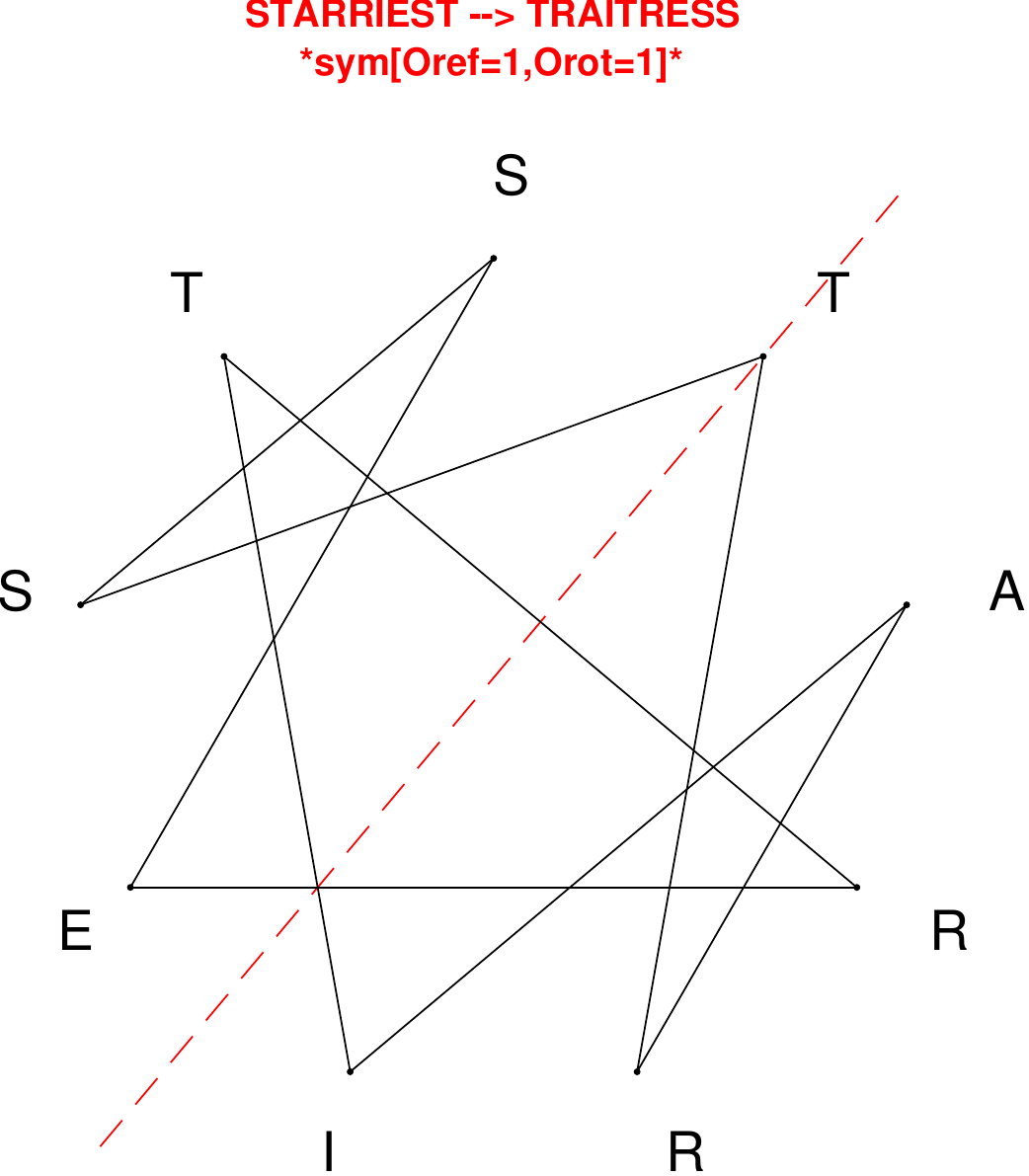}
\end{subfigure}
\hfill
\begin{subfigure}[T]{0.19\textwidth}
\centering
\includegraphics[width=\textwidth]{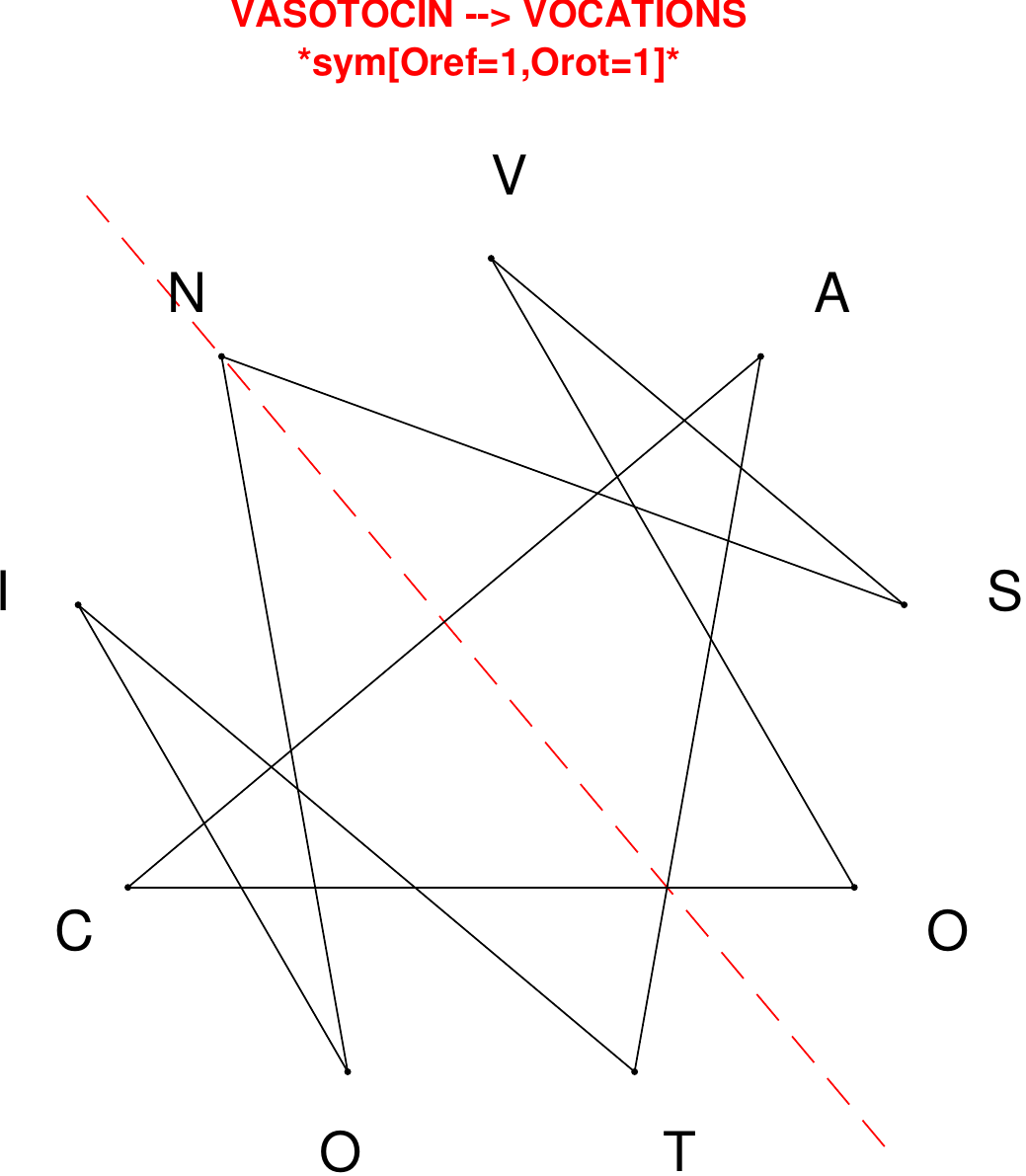}
\end{subfigure}
\hfill
\begin{subfigure}[T]{0.19\textwidth}
\centering
\includegraphics[width=\textwidth]{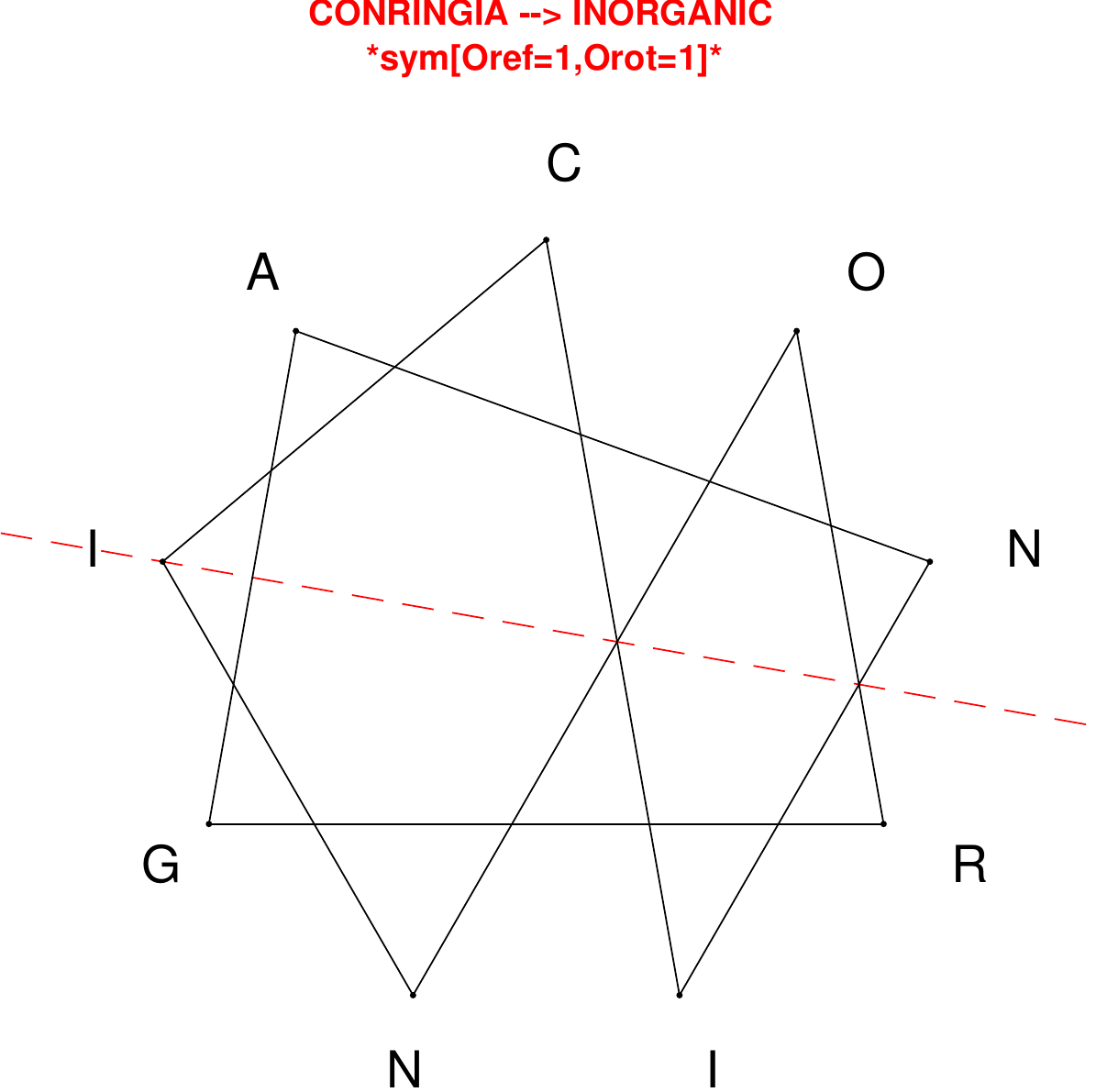}
\end{subfigure}
\hfill
\begin{subfigure}[T]{0.19\textwidth}
\centering
\includegraphics[width=\textwidth]{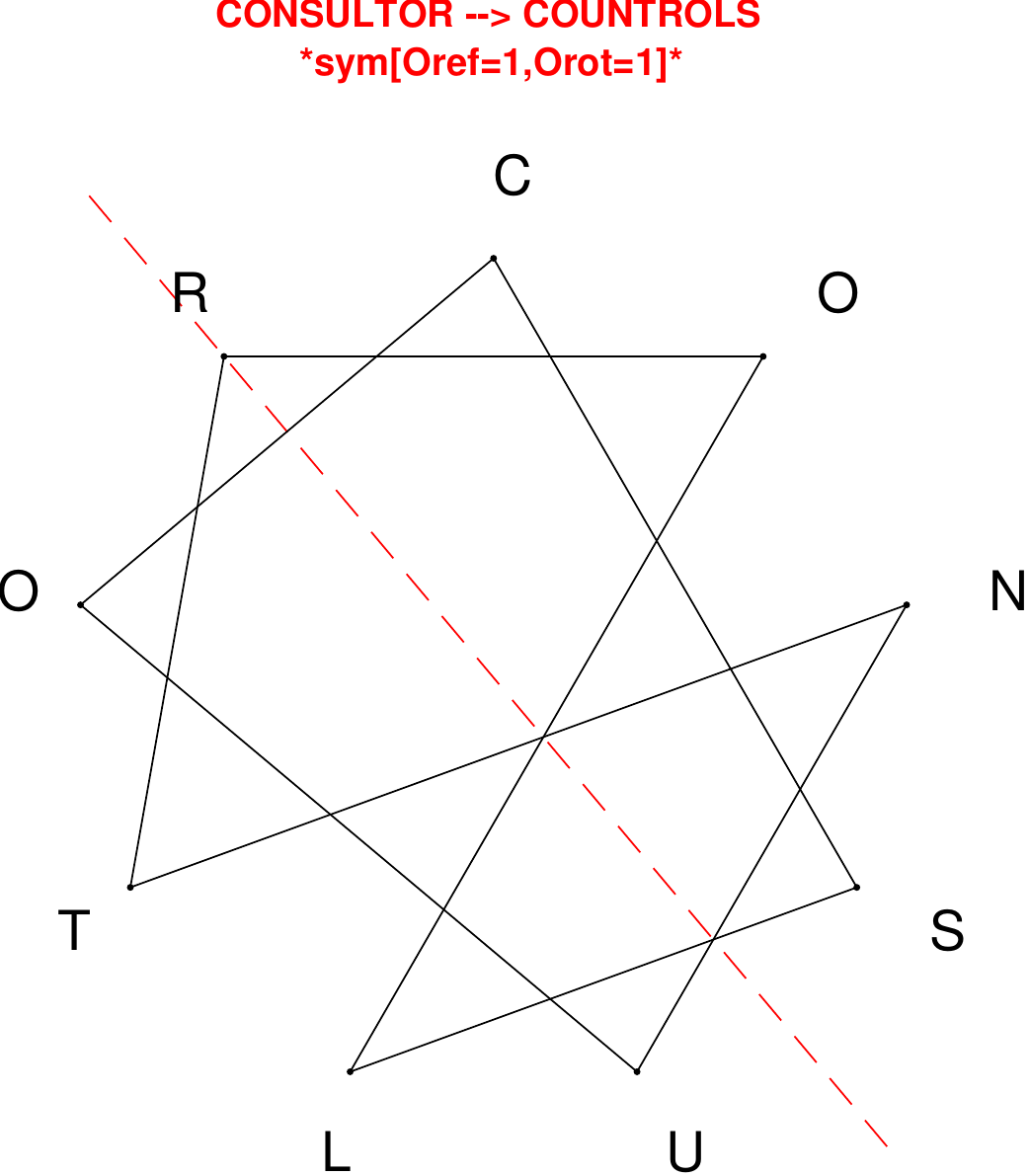}
\end{subfigure}
\end{figure}

\begin{figure}[H]
\centering
\begin{subfigure}[T]{0.19\textwidth}
\centering
\includegraphics[width=\textwidth]{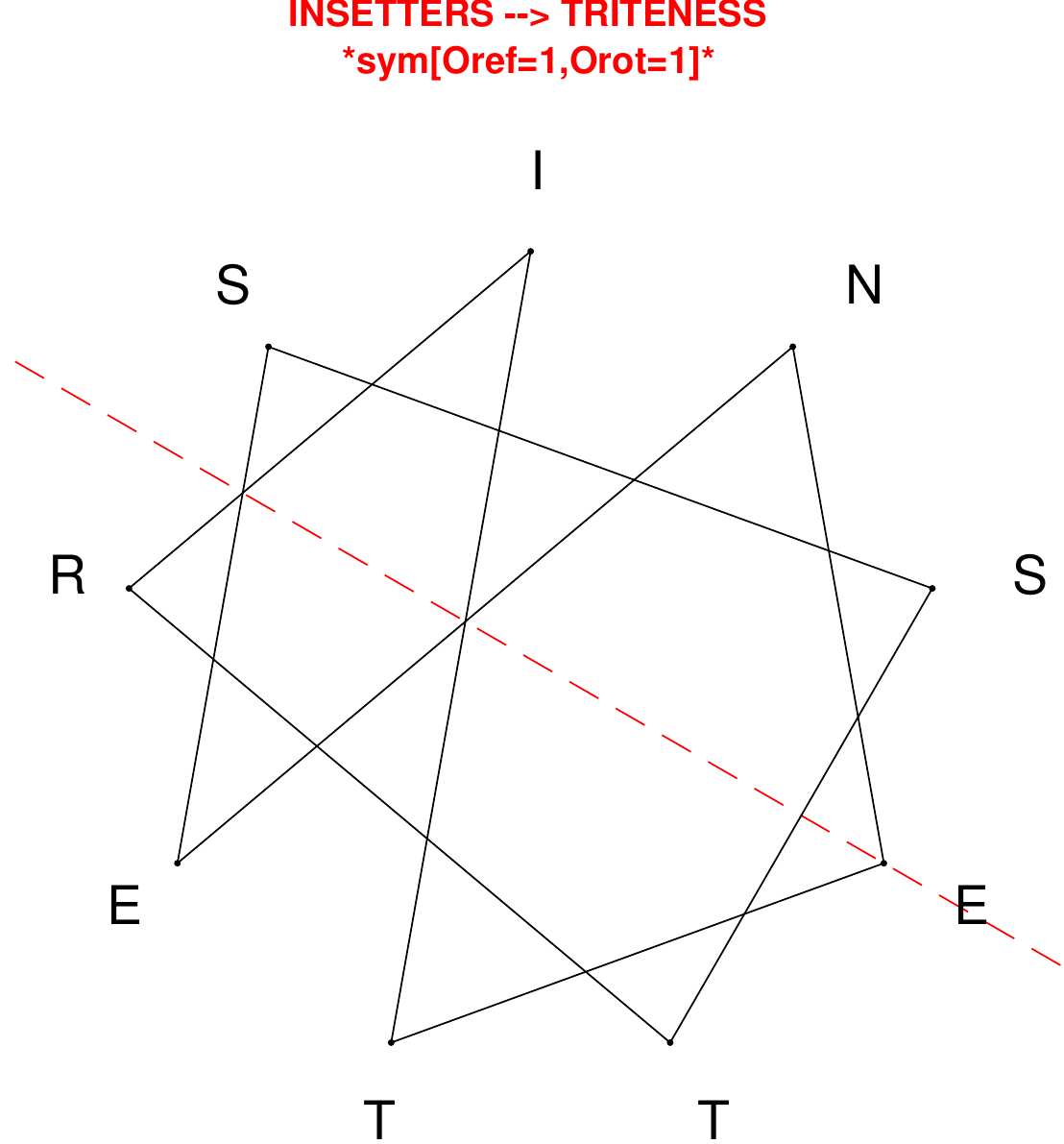}
\end{subfigure}
\hfill
\begin{subfigure}[T]{0.19\textwidth}
\centering
\includegraphics[width=\textwidth]{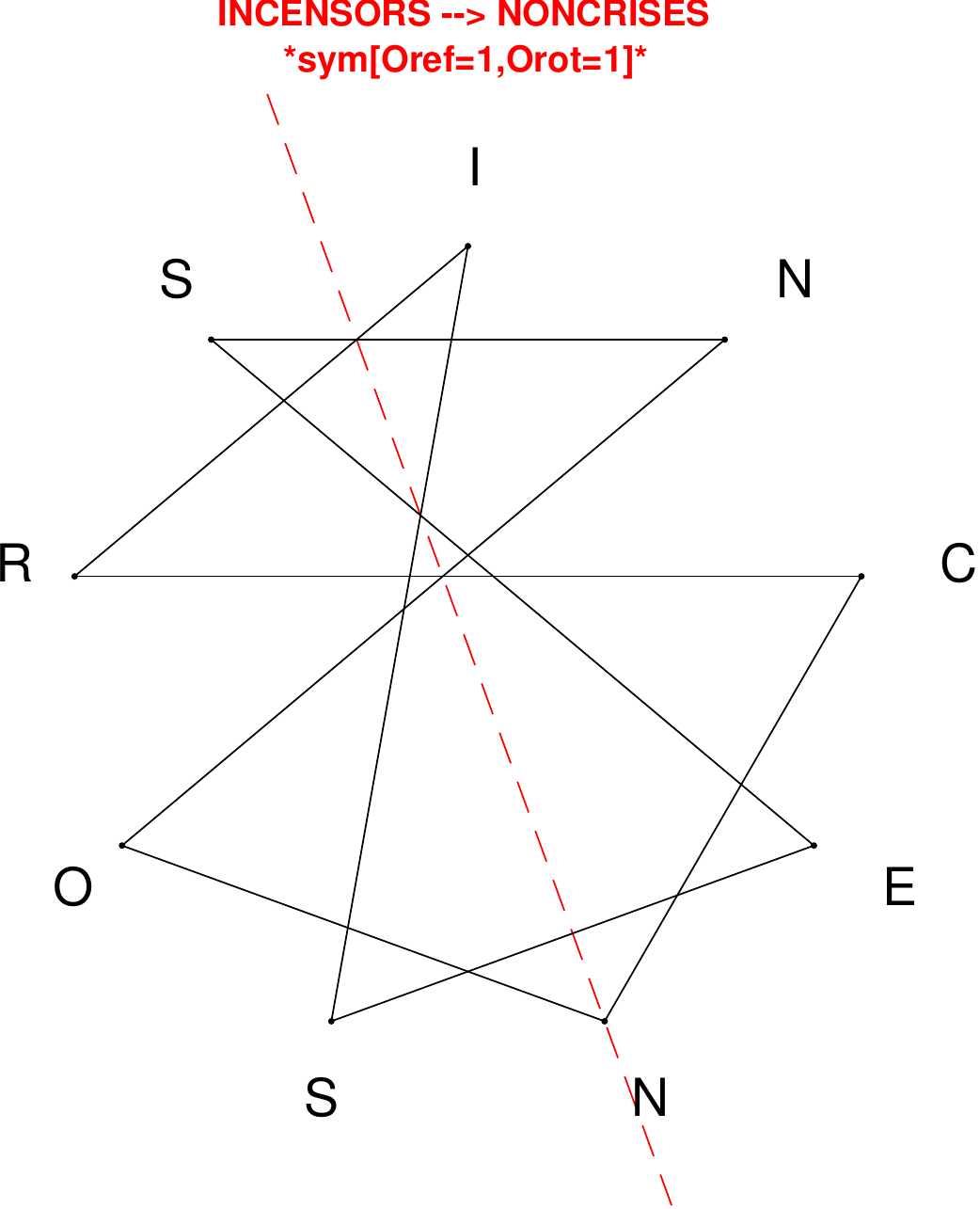}
\end{subfigure}
\hfill
\begin{subfigure}[T]{0.19\textwidth}
\centering
\includegraphics[width=\textwidth]{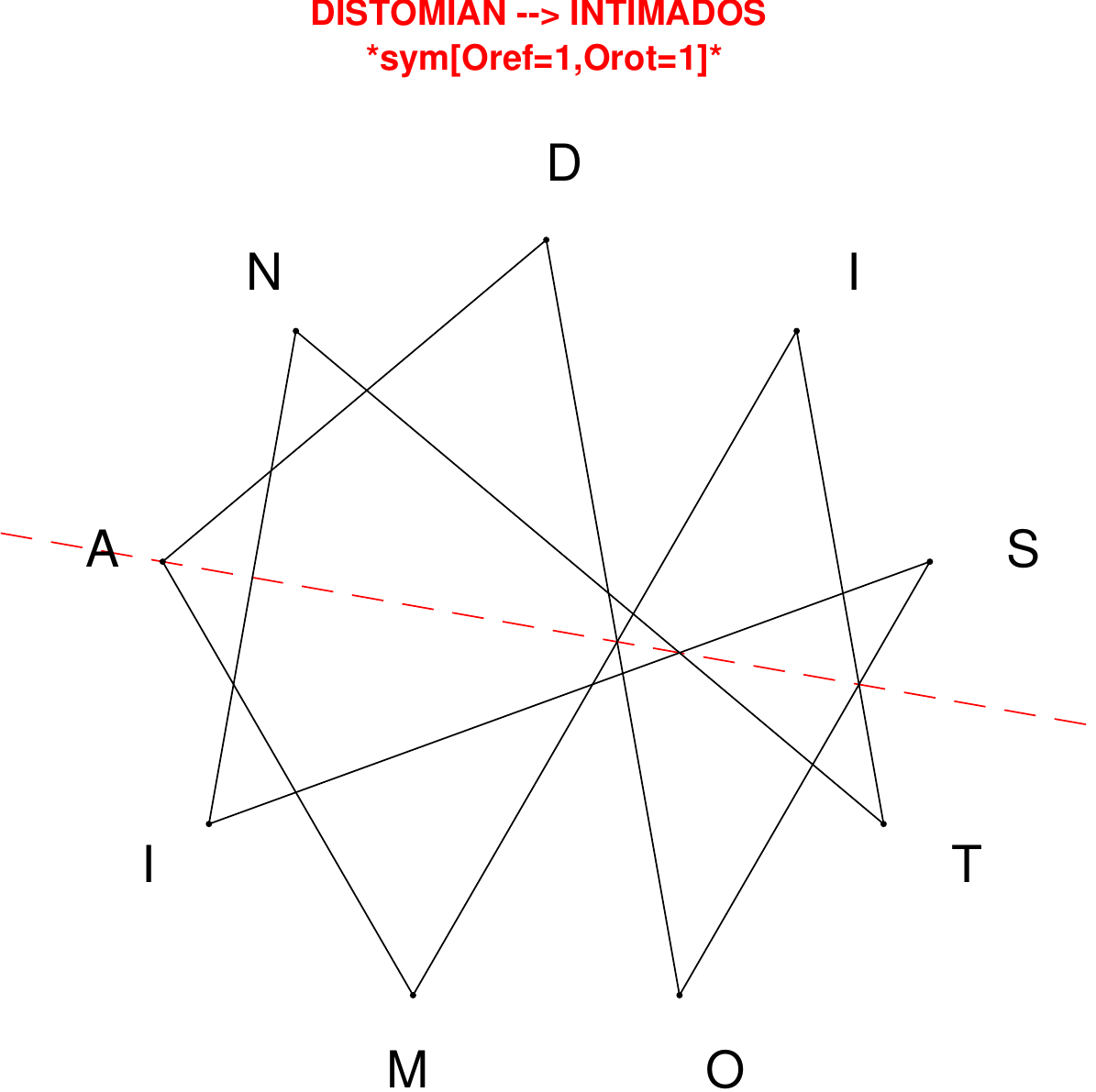}
\end{subfigure}
\hfill
\begin{subfigure}[T]{0.19\textwidth}
\centering
\includegraphics[width=\textwidth]{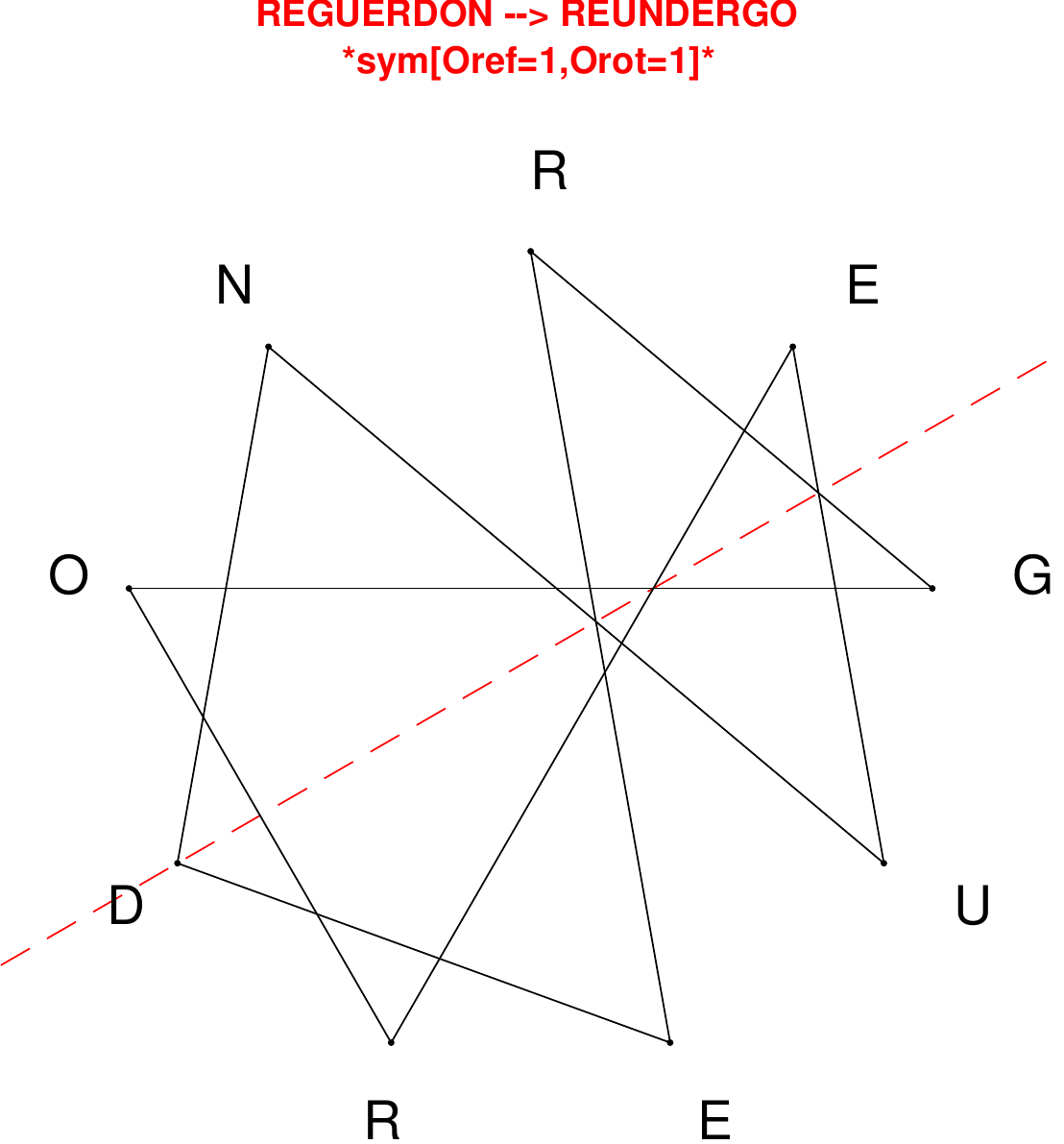}
\end{subfigure}
\hfill
\begin{subfigure}[T]{0.19\textwidth}
\centering
\includegraphics[width=\textwidth]{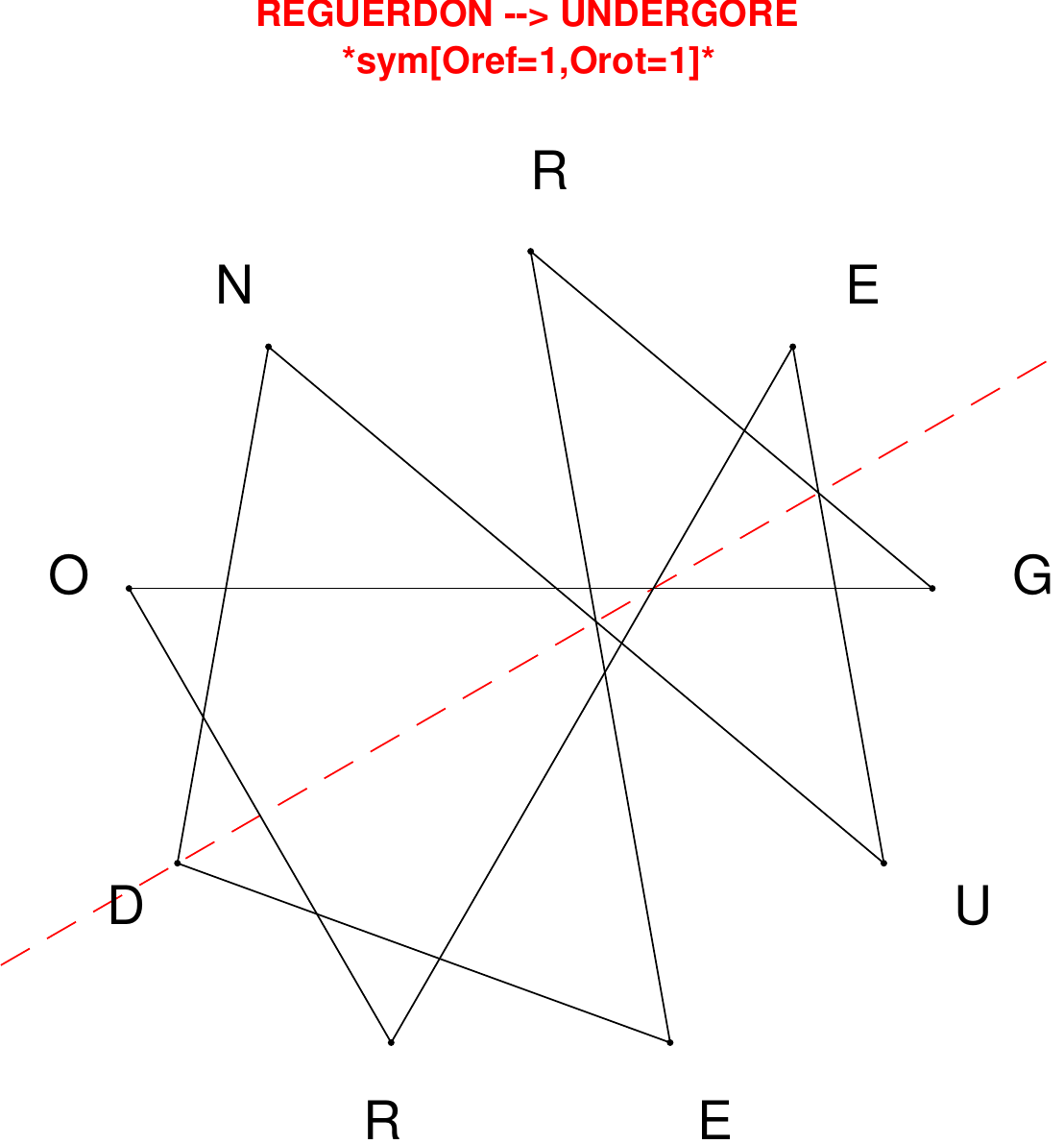}
\end{subfigure}
\end{figure}

\begin{figure}[H]
\centering
\begin{subfigure}[T]{0.19\textwidth}
\centering
\includegraphics[width=\textwidth]{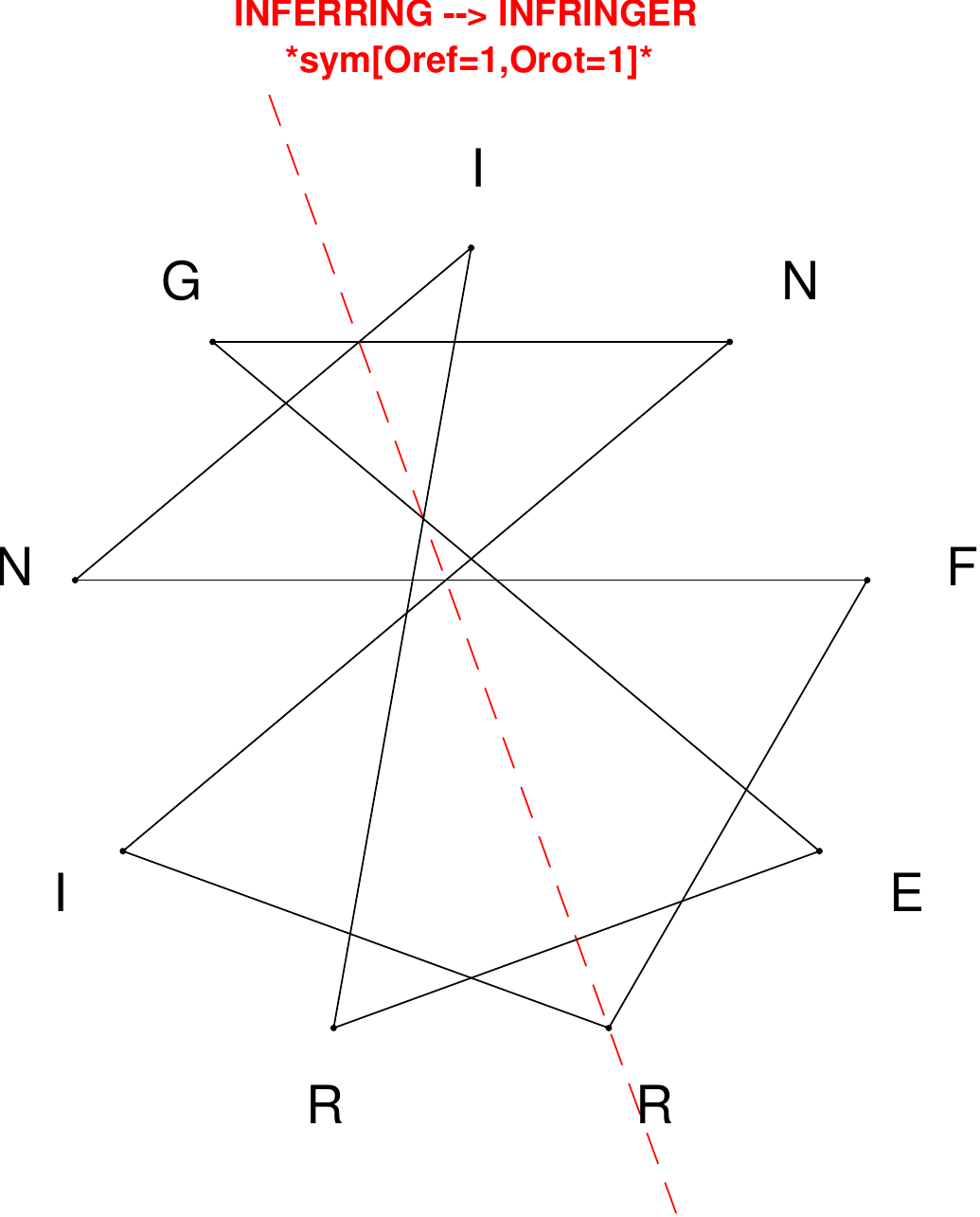}
\end{subfigure}
\hfill
\begin{subfigure}[T]{0.19\textwidth}
\centering
\includegraphics[width=\textwidth]{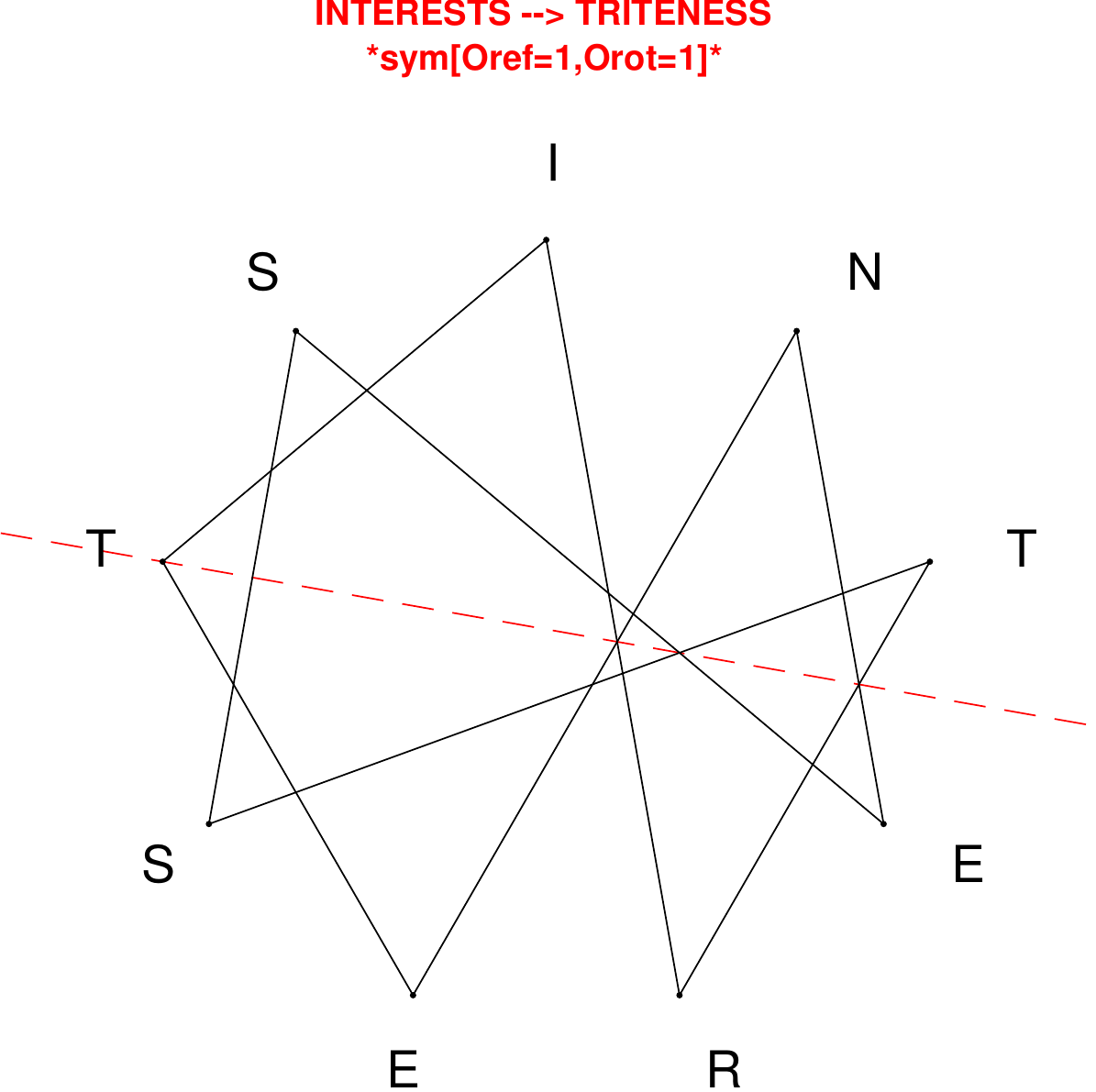}
\end{subfigure}
\hfill
\begin{subfigure}[T]{0.19\textwidth}
\centering
\includegraphics[width=\textwidth]{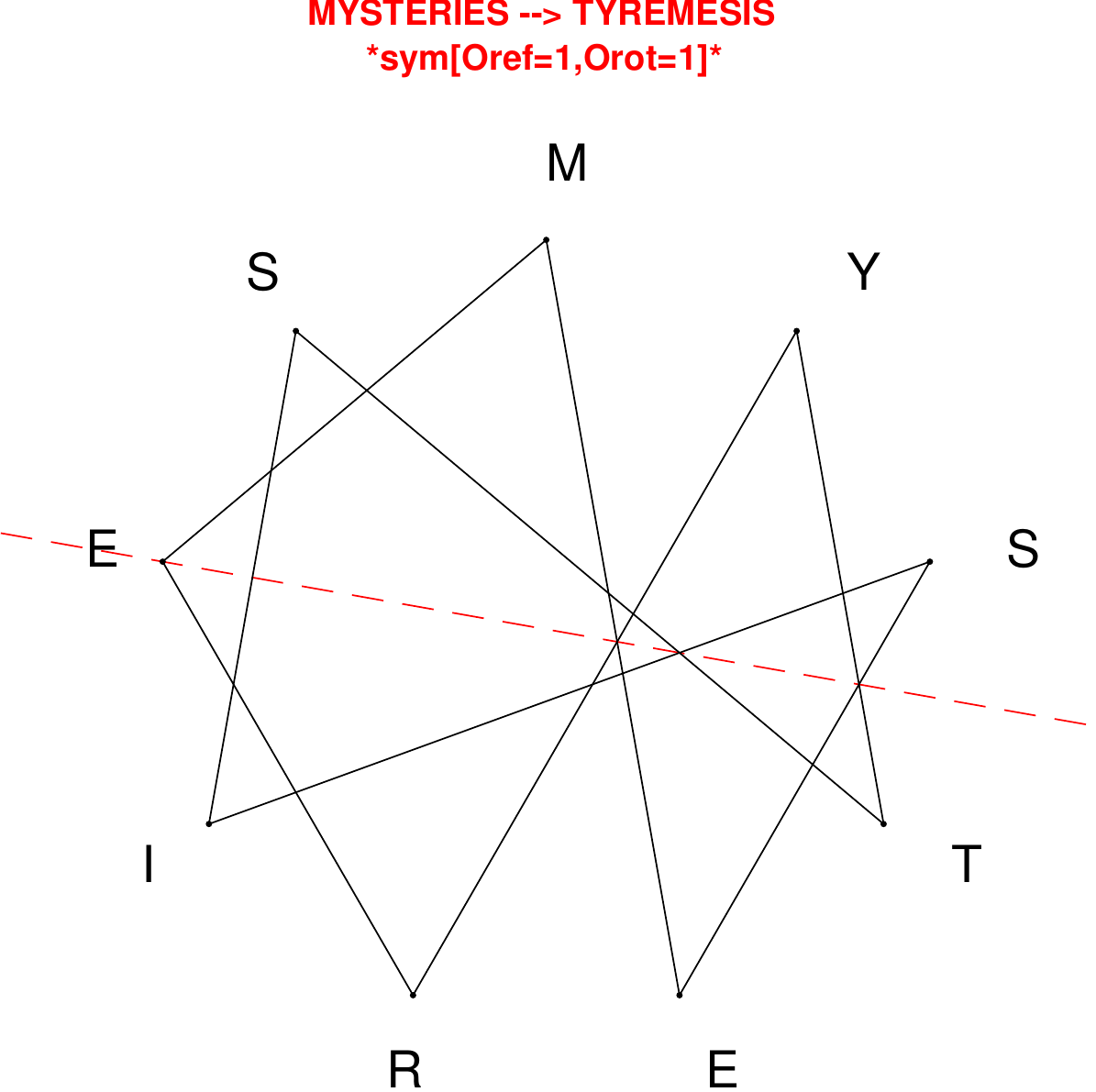}
\end{subfigure}
\hfill
\begin{subfigure}[T]{0.19\textwidth}
\centering
\includegraphics[width=\textwidth]{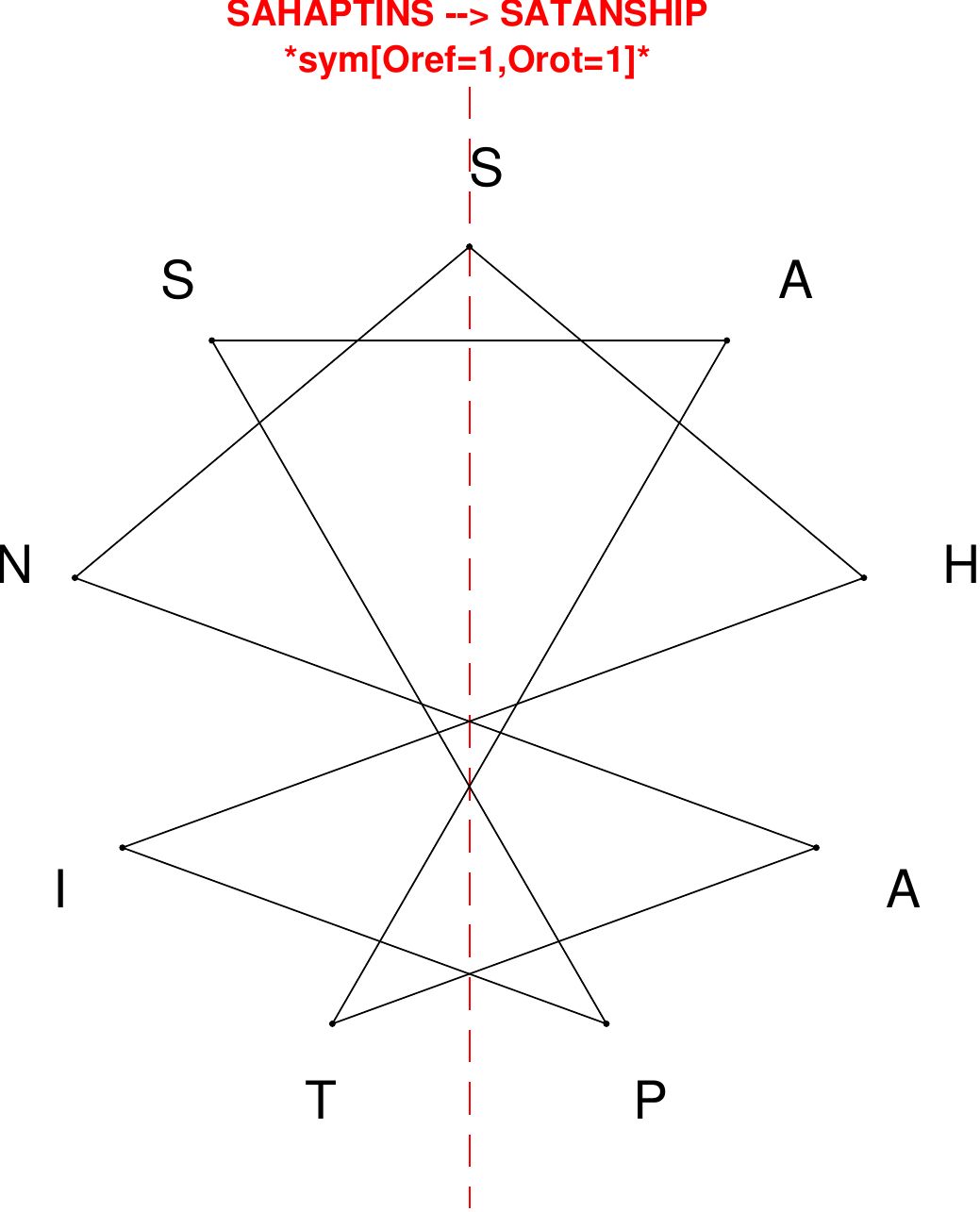}
\end{subfigure}
\hfill
\begin{subfigure}[T]{0.19\textwidth}
\centering
\includegraphics[width=\textwidth]{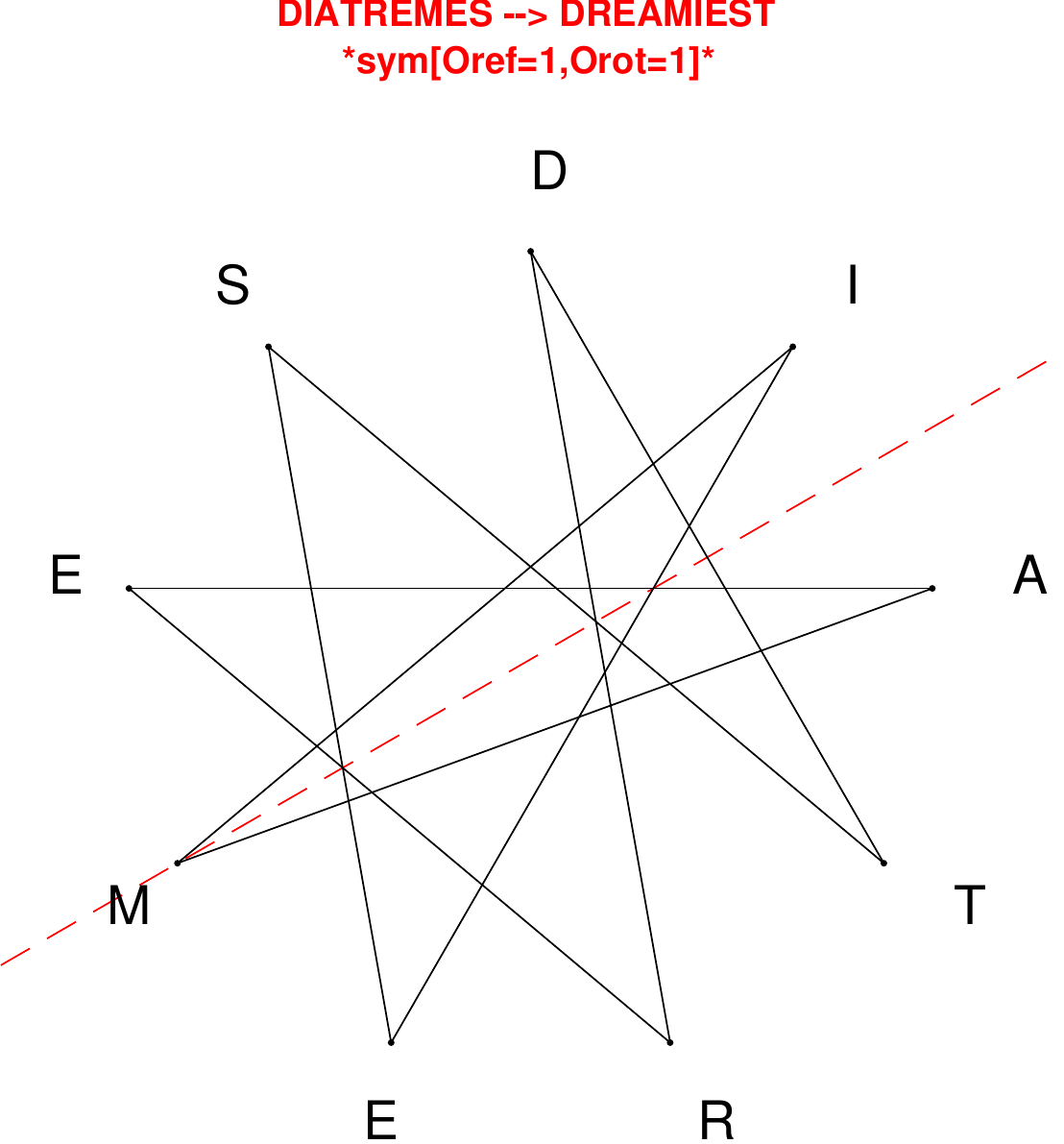}
\end{subfigure}
\end{figure}

\begin{figure}[H]
\centering
\begin{subfigure}[T]{0.19\textwidth}
\centering
\includegraphics[width=\textwidth]{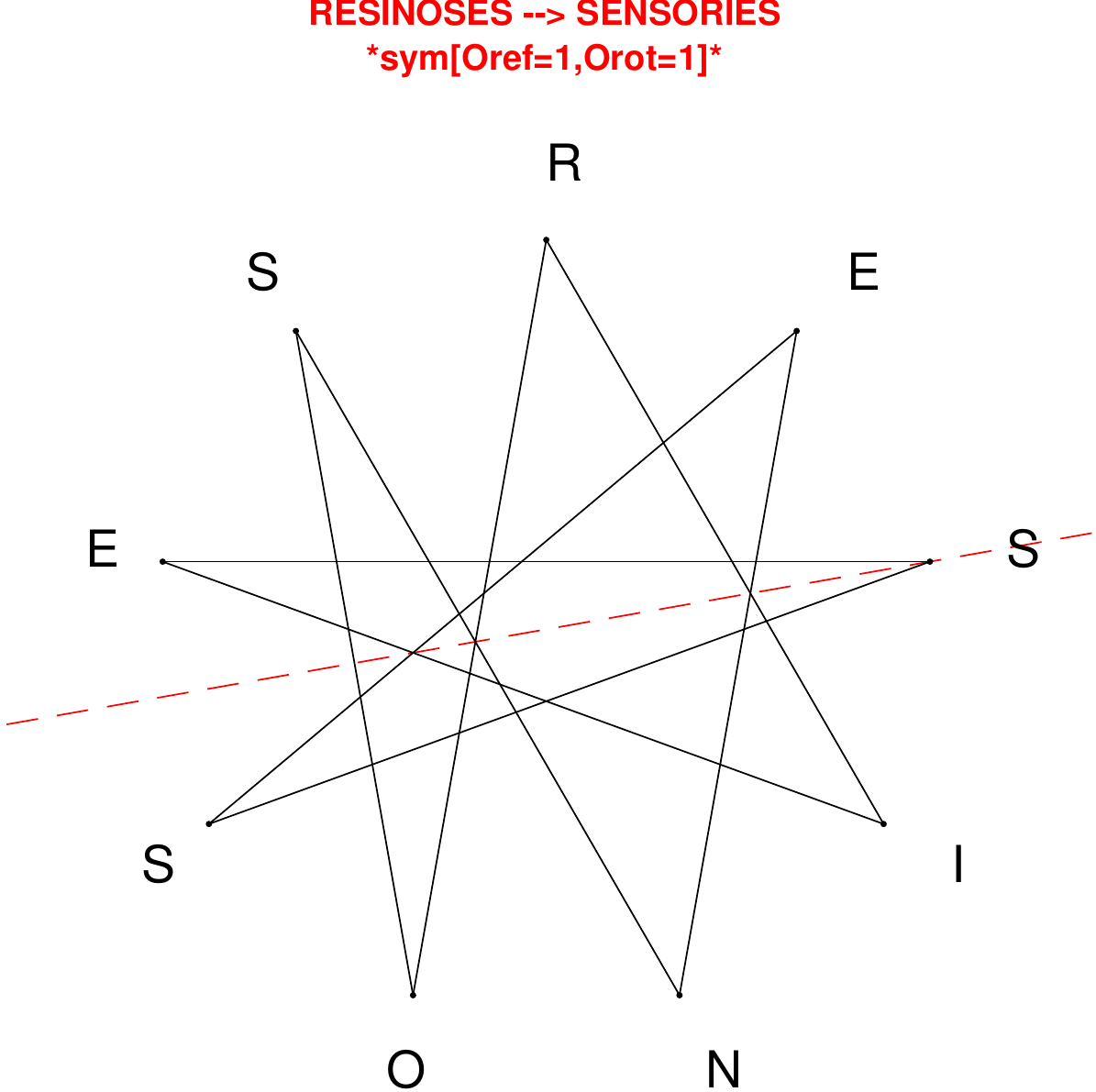}
\end{subfigure}
\hfill
\begin{subfigure}[T]{0.19\textwidth}
\centering
\includegraphics[width=\textwidth]{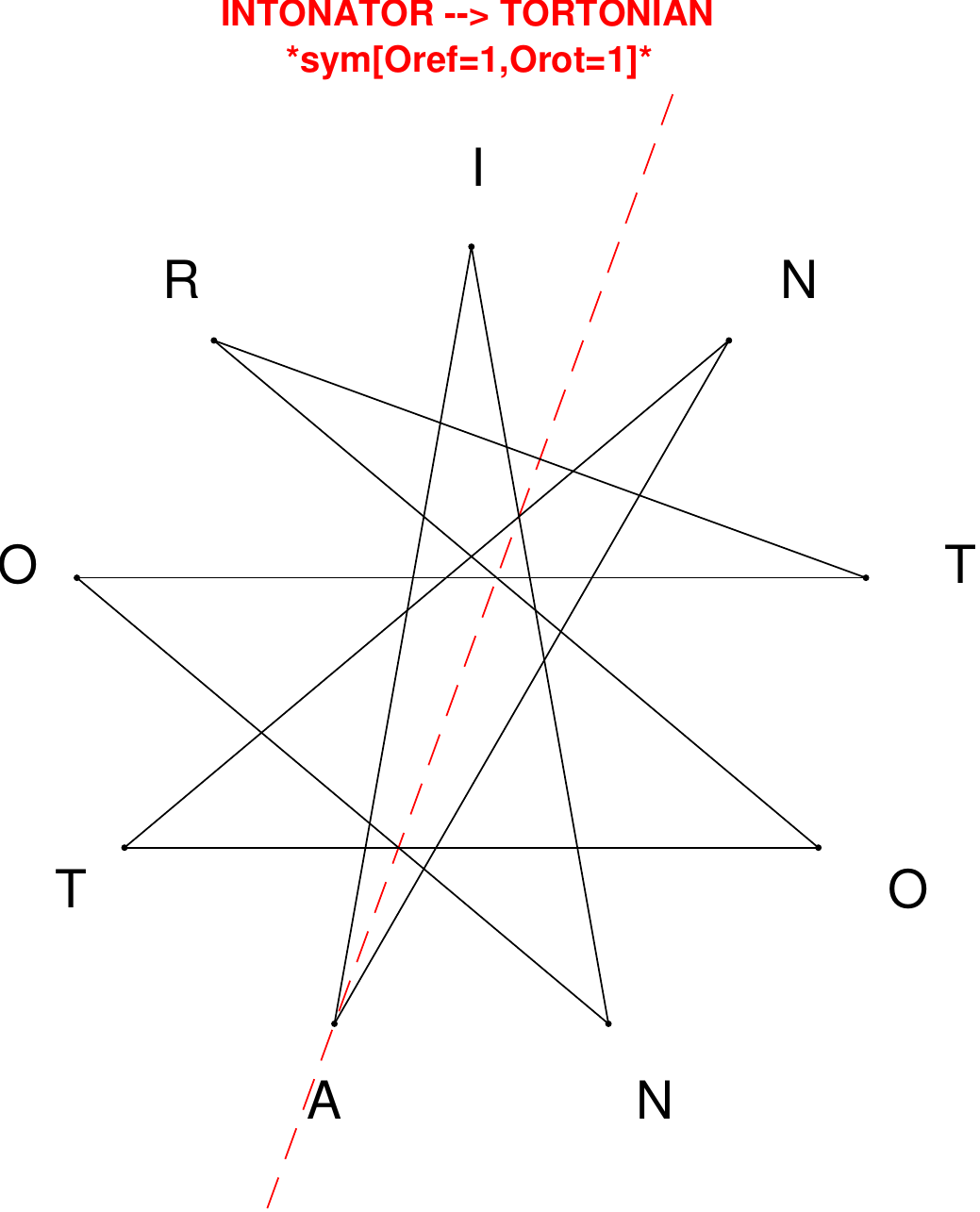}
\end{subfigure}
\hfill
\begin{subfigure}[T]{0.19\textwidth}
\centering
\includegraphics[width=\textwidth]{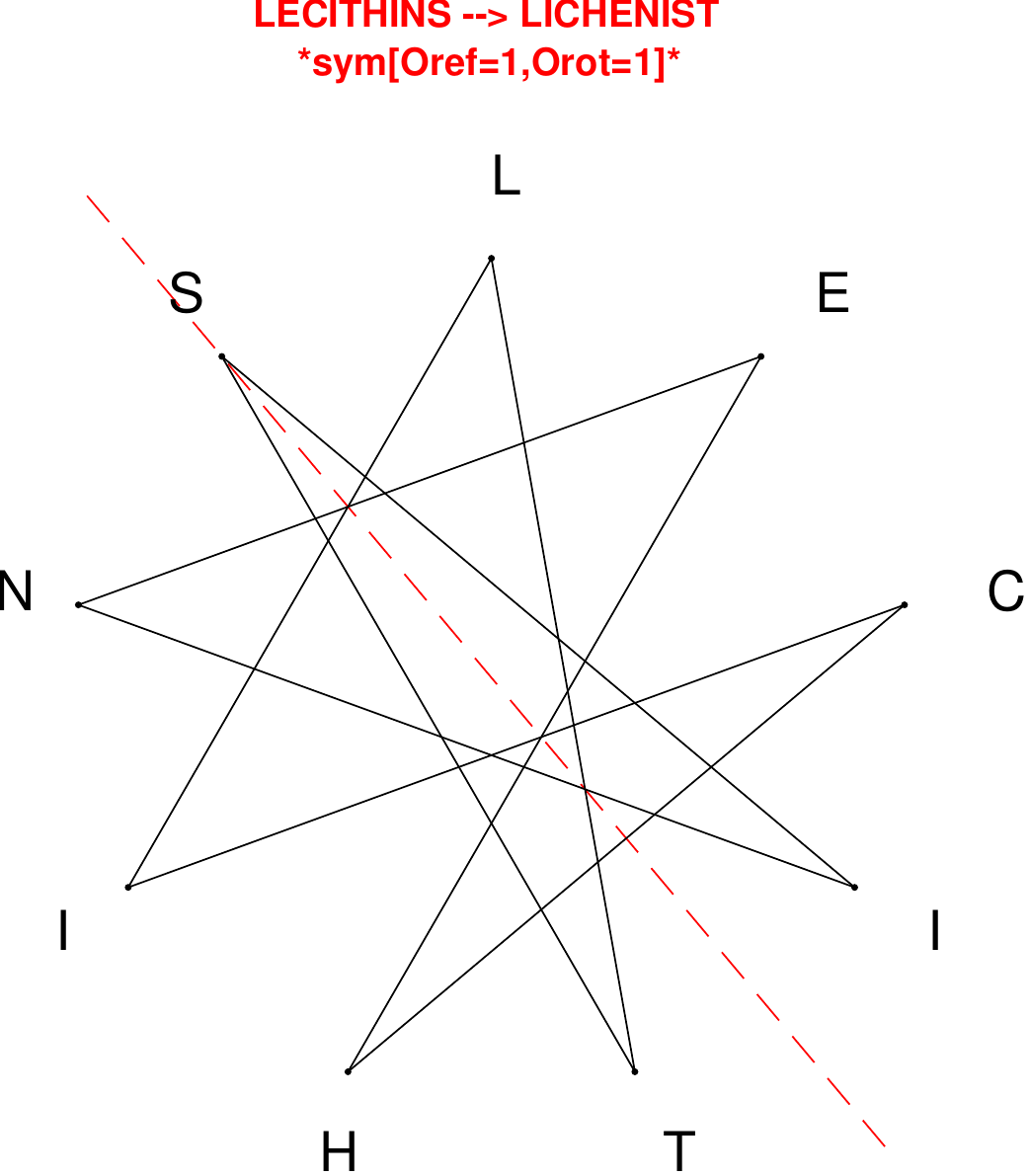}
\end{subfigure}
\hfill
\begin{subfigure}[T]{0.19\textwidth}
\centering
\includegraphics[width=\textwidth]{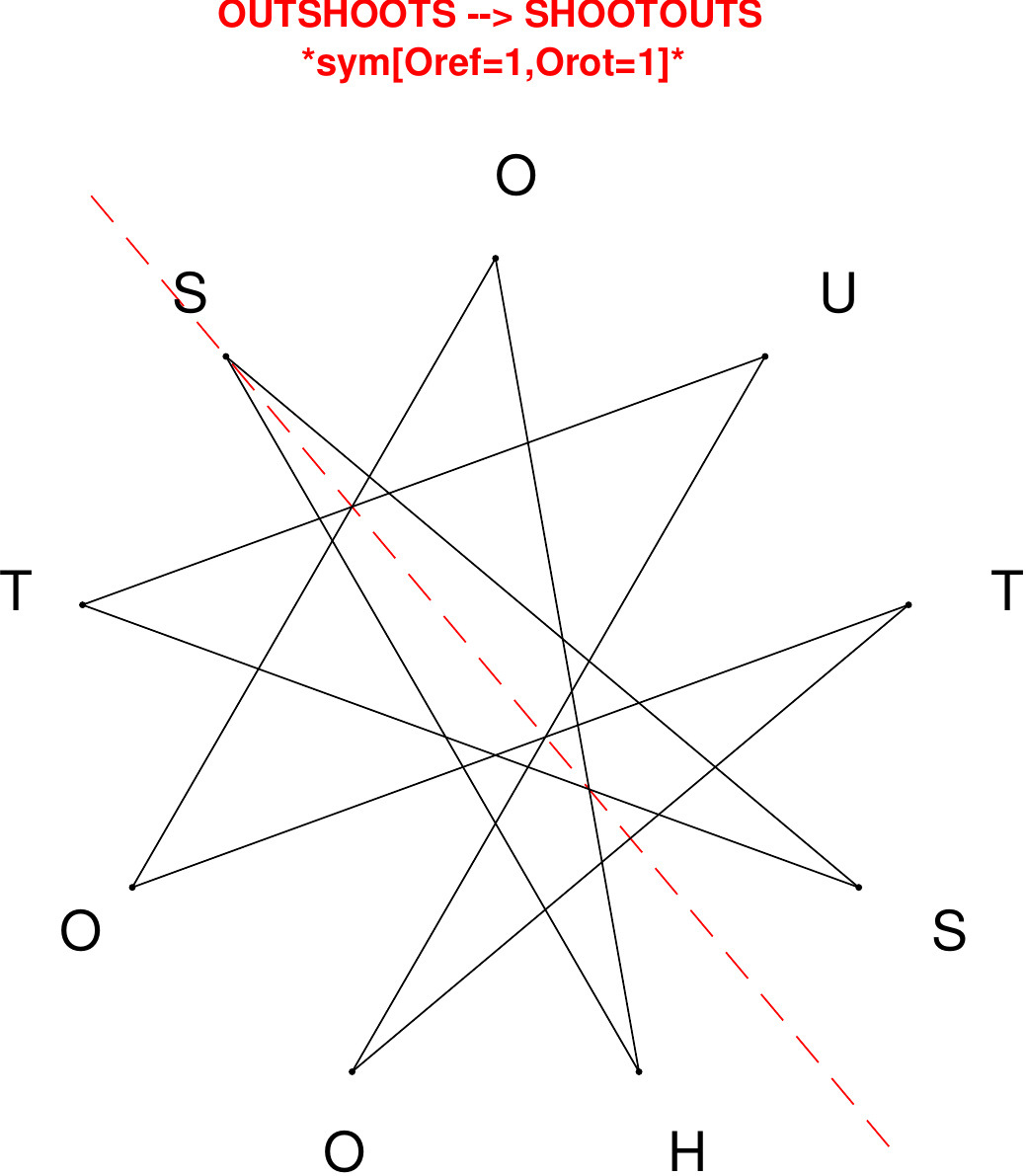}
\end{subfigure}
\hfill
\begin{subfigure}[T]{0.19\textwidth}
\centering
\includegraphics[width=\textwidth]{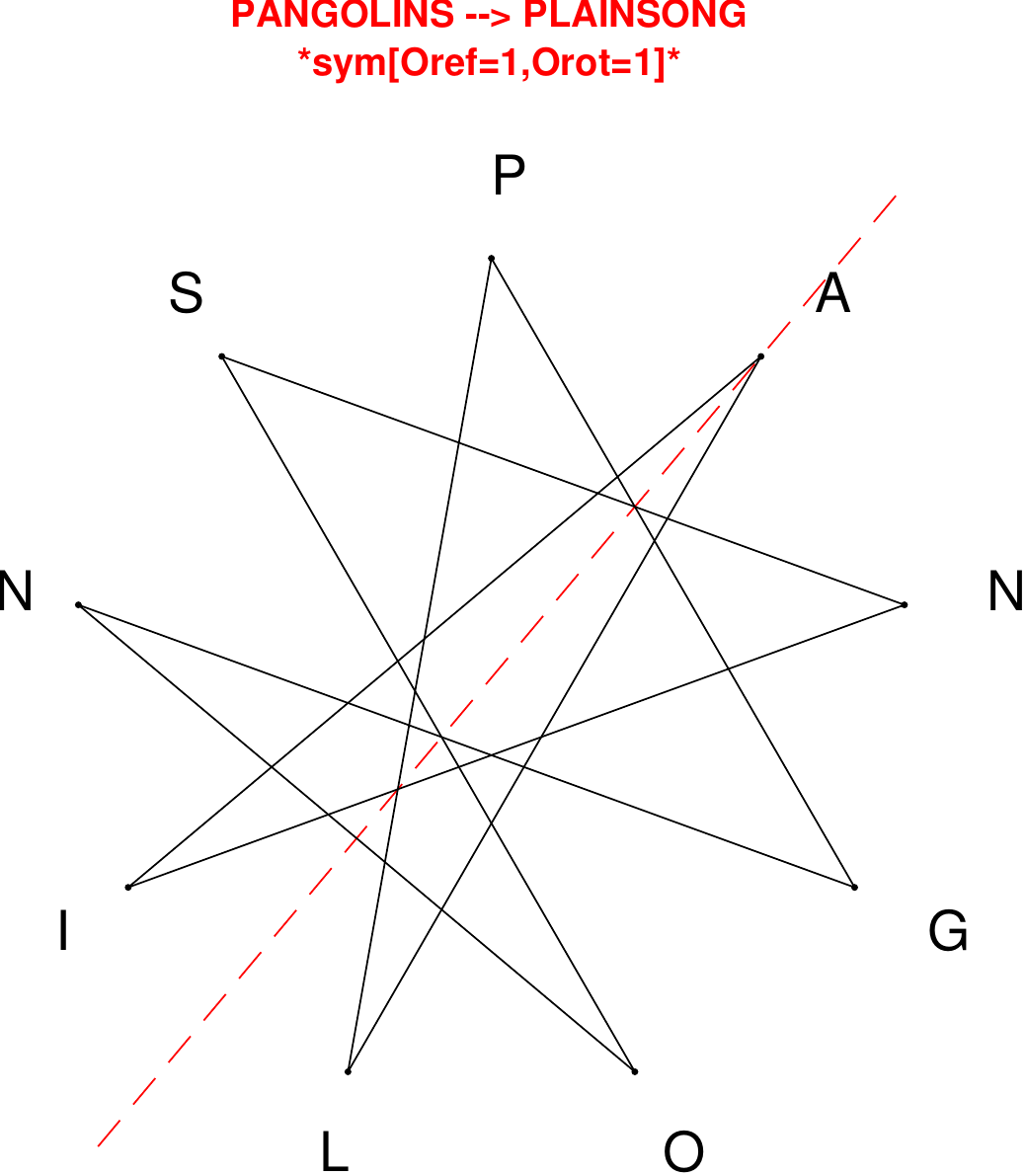}
\end{subfigure}
\end{figure}

\begin{figure}[H]
\centering
\begin{subfigure}[T]{0.19\textwidth}
\centering
\includegraphics[width=\textwidth]{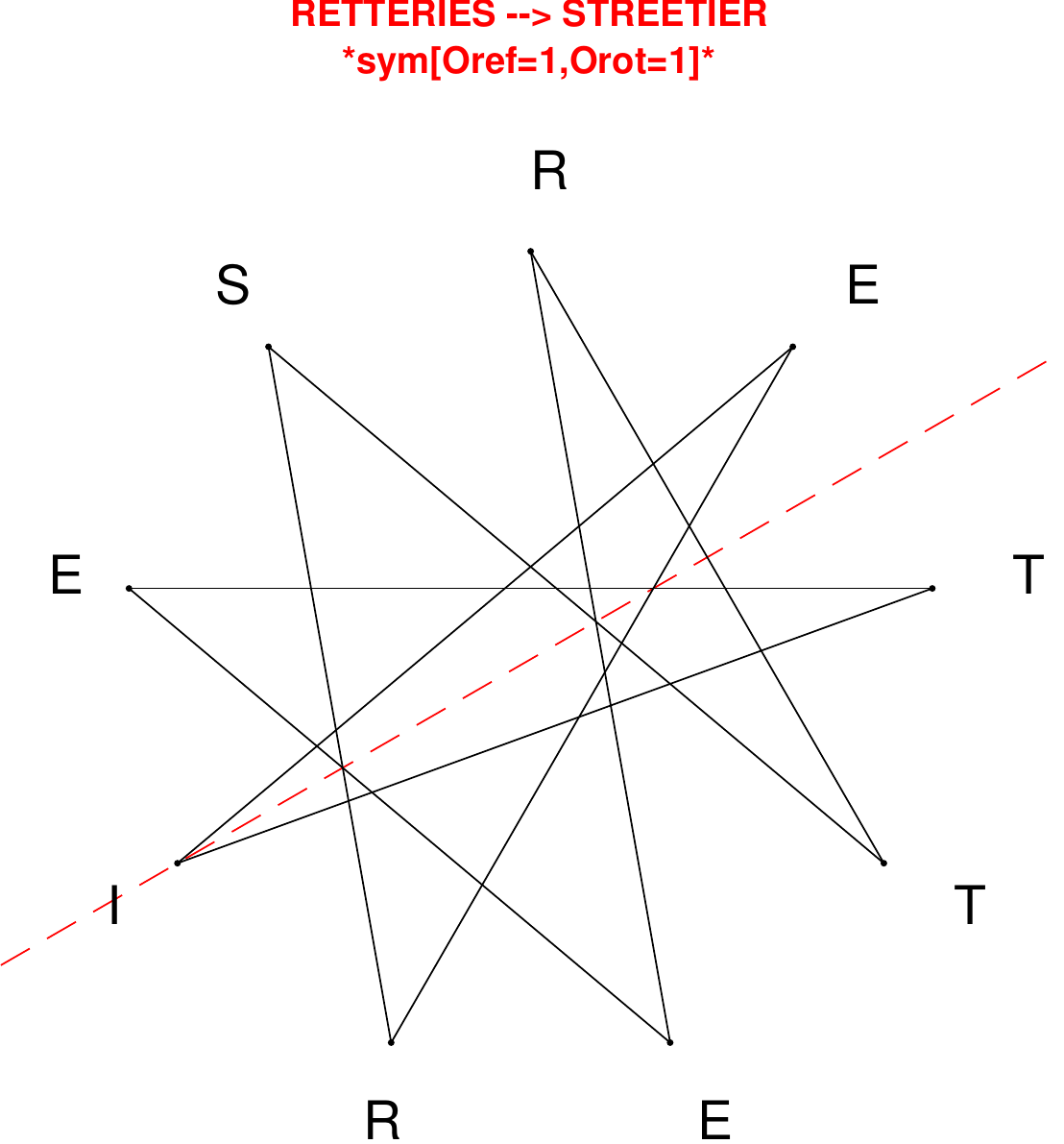}
\end{subfigure}
\hfill
\begin{subfigure}[T]{0.19\textwidth}
\centering
\includegraphics[width=\textwidth]{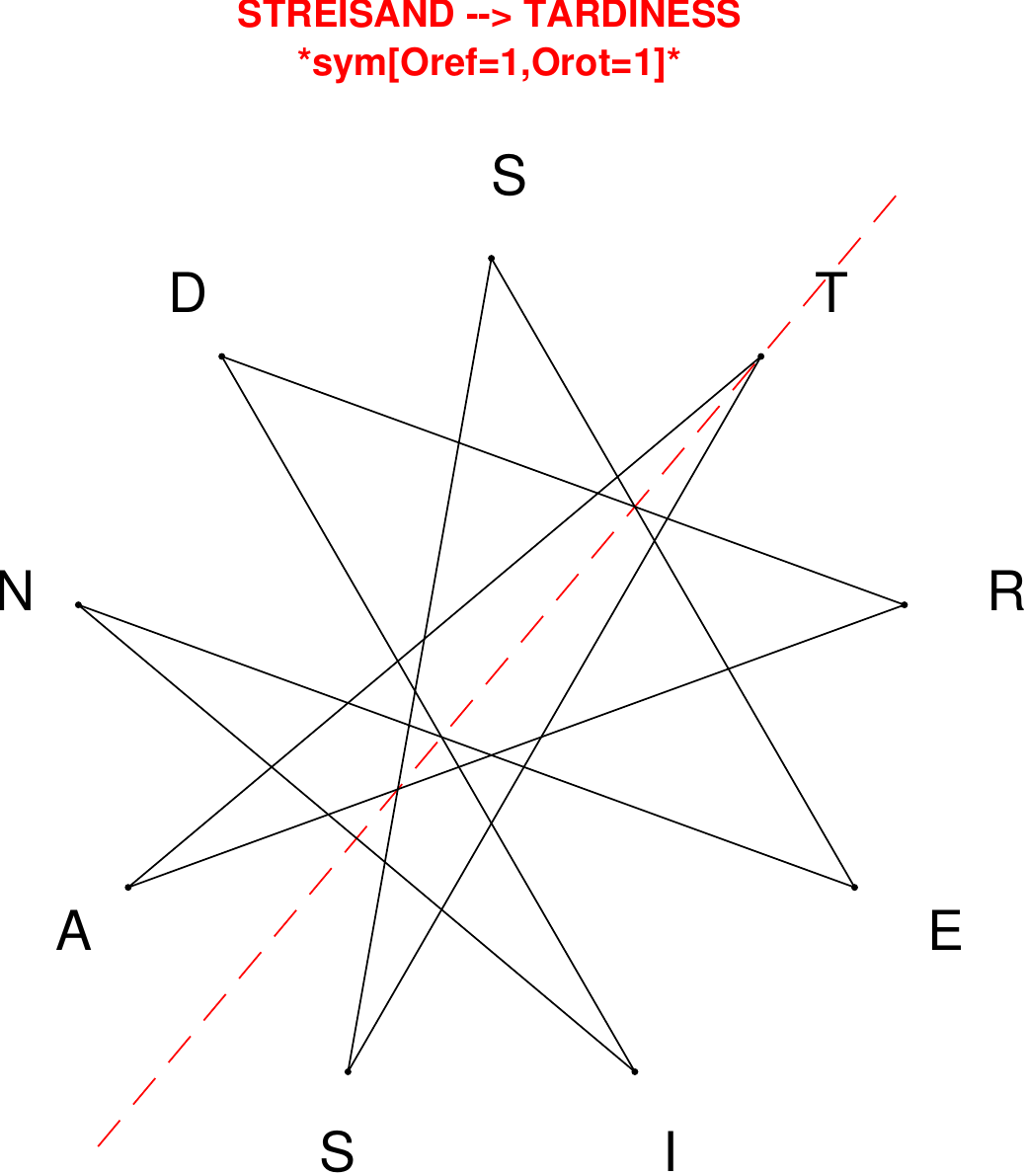}
\end{subfigure}
\hfill
\begin{subfigure}[T]{0.19\textwidth}
\centering
\includegraphics[width=\textwidth]{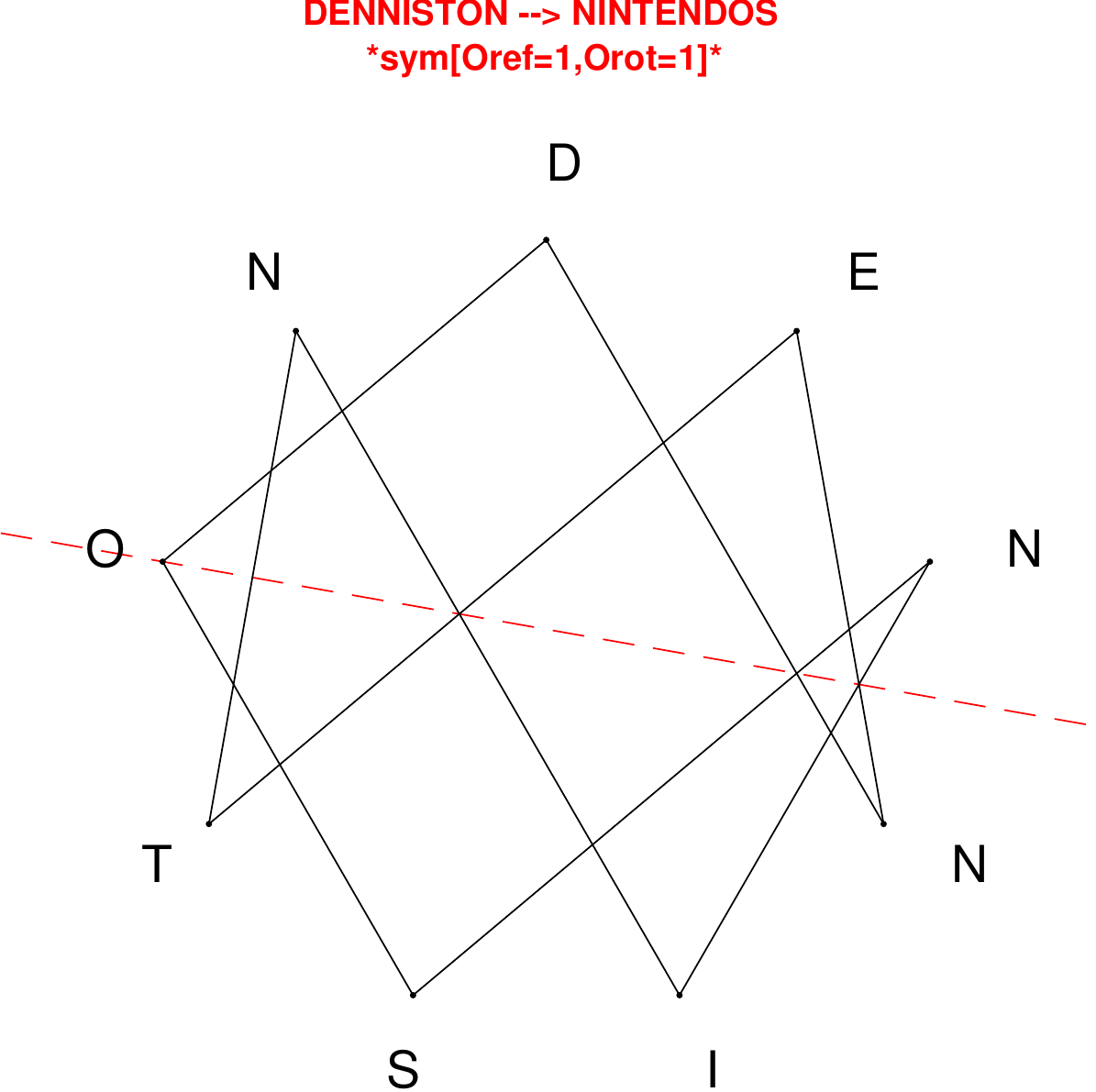}
\end{subfigure}
\hfill
\begin{subfigure}[T]{0.19\textwidth}
\centering
\includegraphics[width=\textwidth]{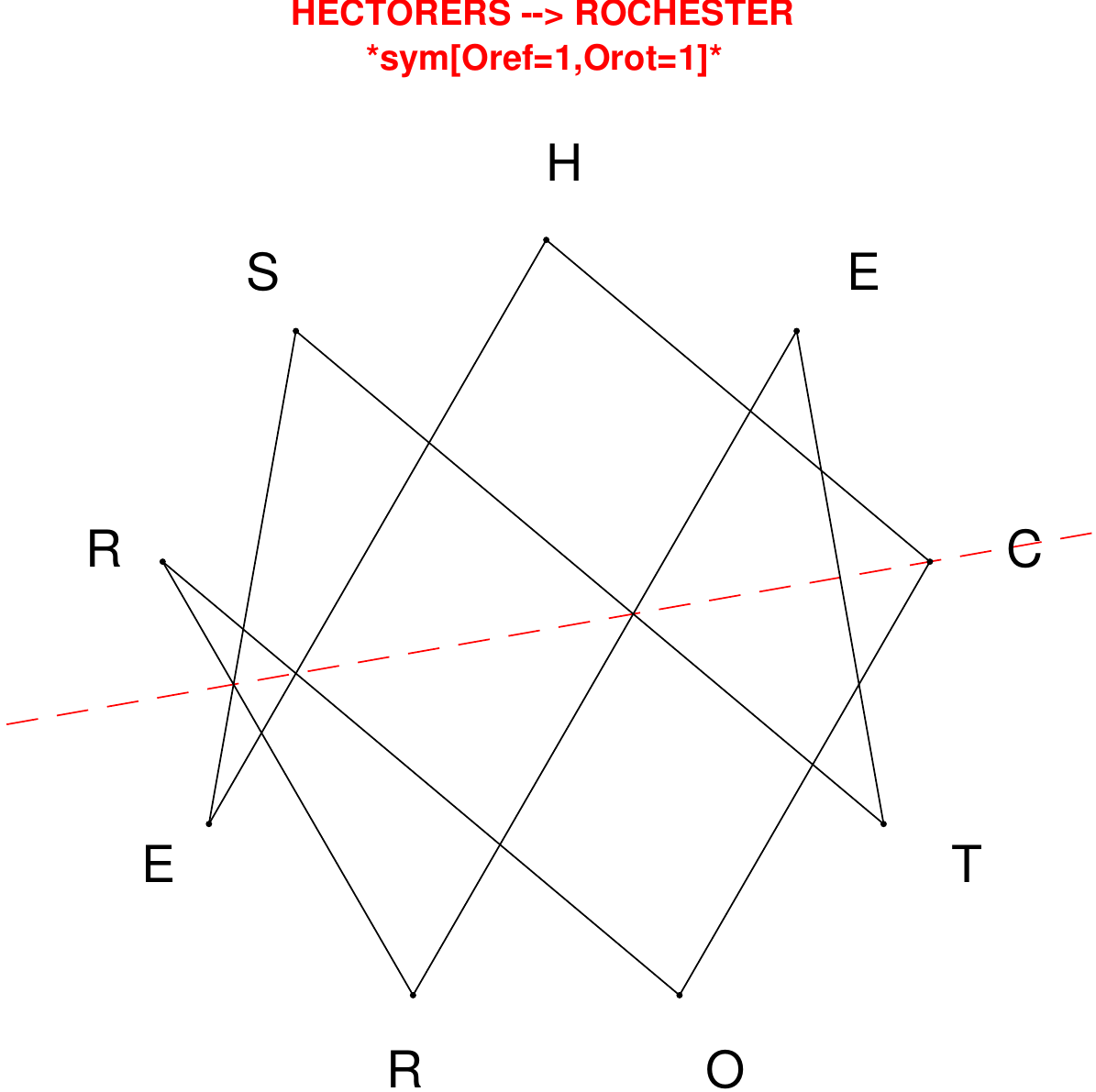}
\end{subfigure}
\hfill
\begin{subfigure}[T]{0.19\textwidth}
\centering
\includegraphics[width=\textwidth]{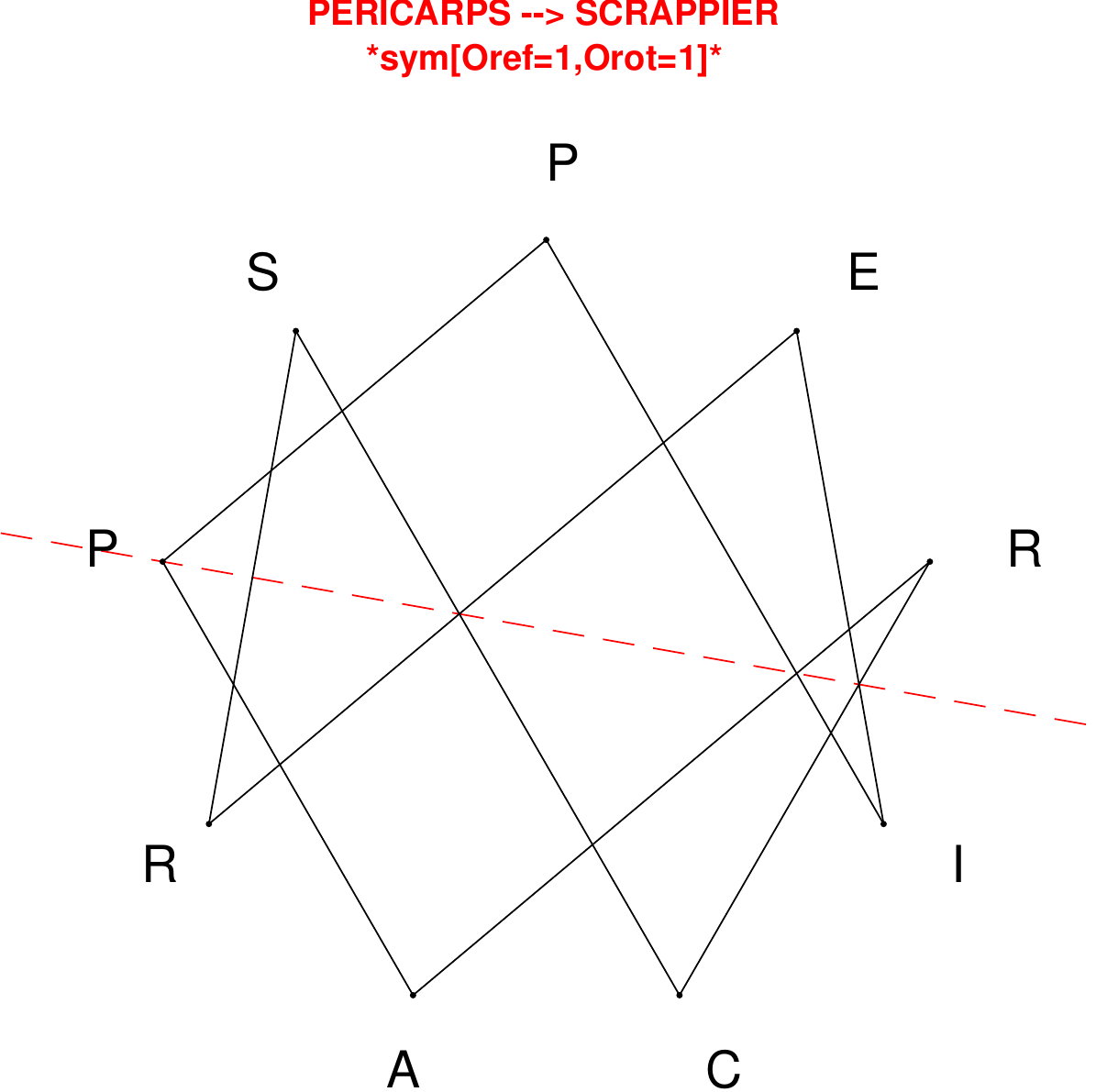}
\end{subfigure}
\end{figure}

\begin{figure}[H]
\centering
\begin{subfigure}[T]{0.19\textwidth}
\centering
\includegraphics[width=\textwidth]{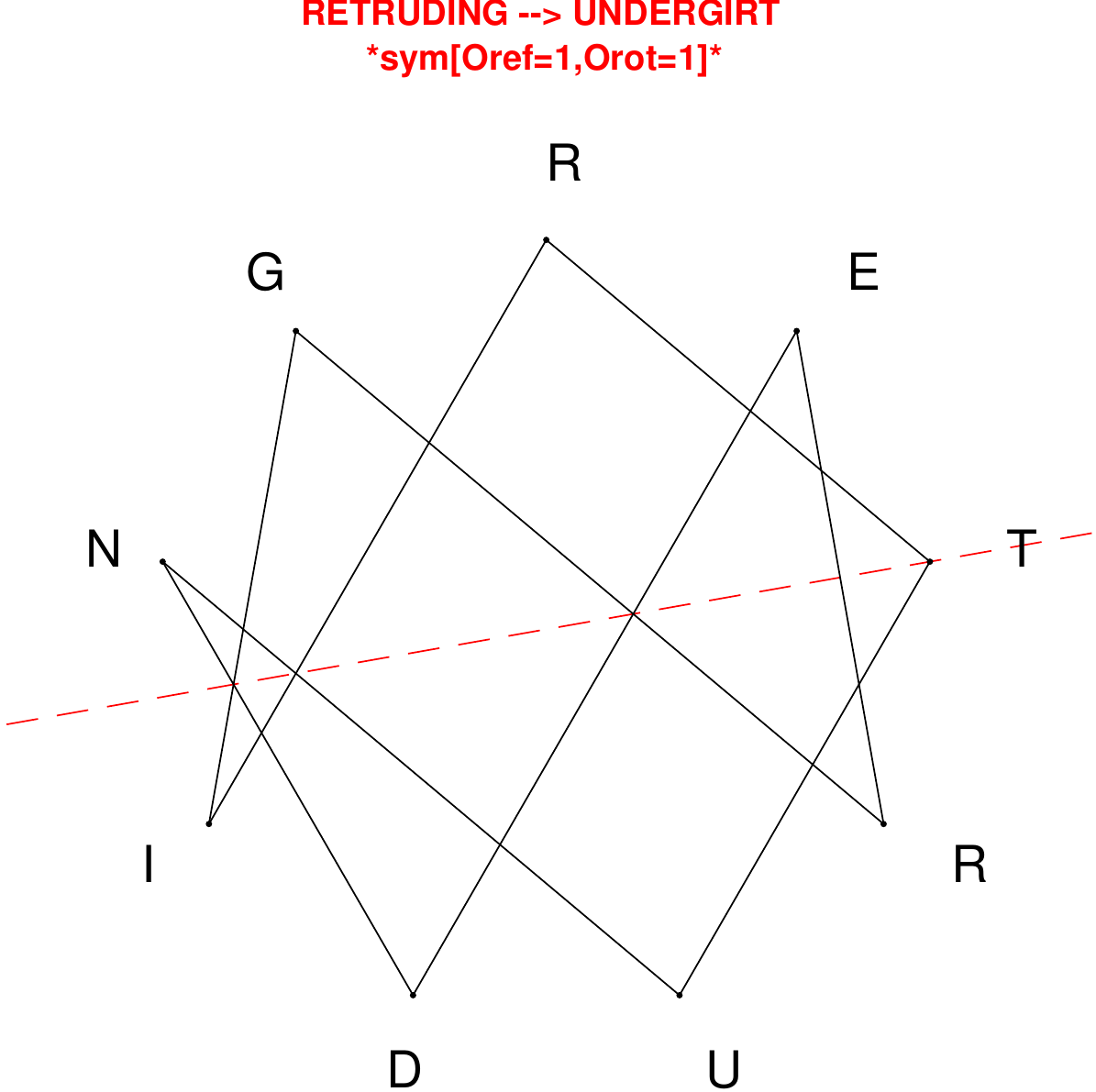}
\end{subfigure}
\hfill
\begin{subfigure}[T]{0.19\textwidth}
\centering
\includegraphics[width=\textwidth]{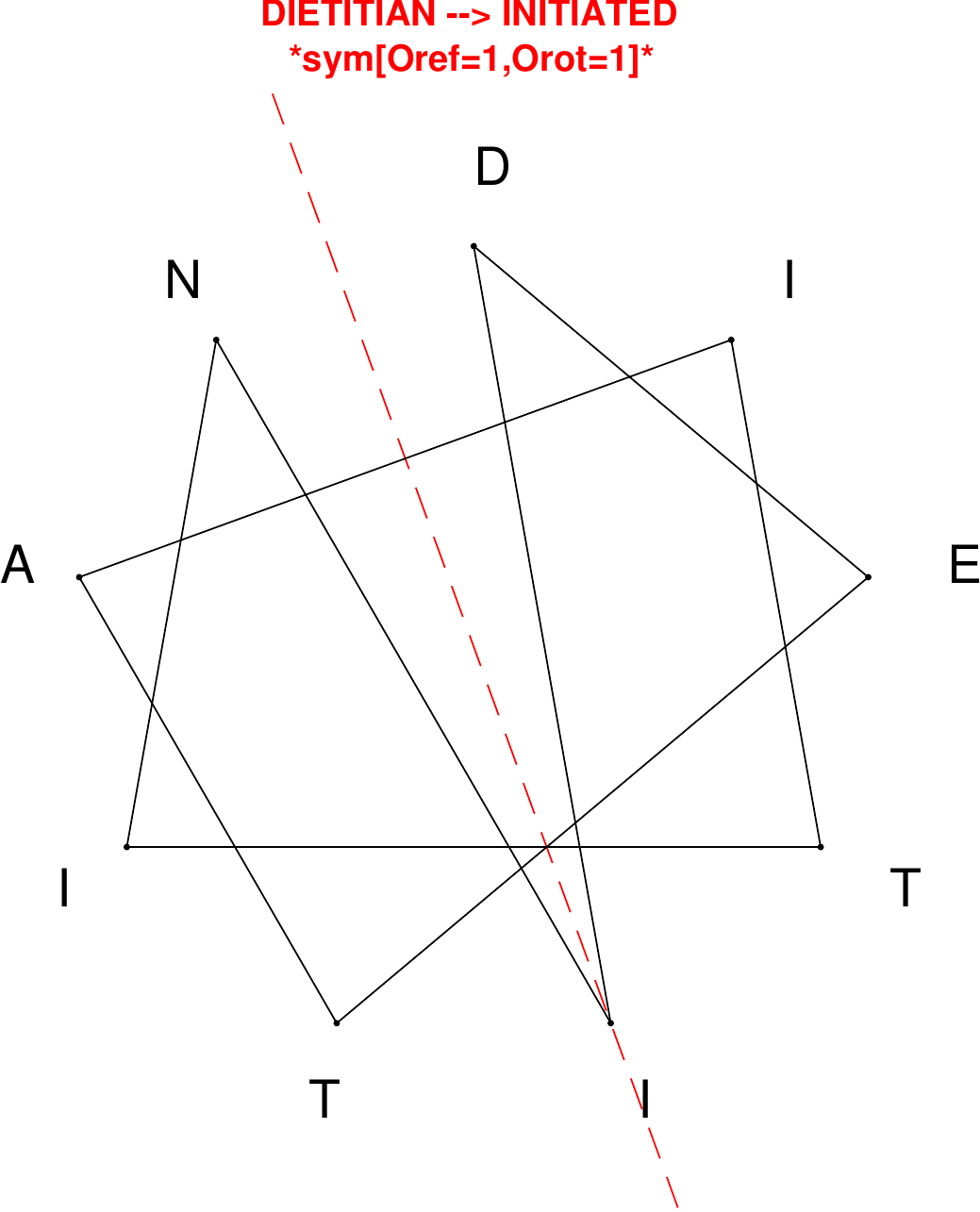}
\end{subfigure}
\hfill
\begin{subfigure}[T]{0.19\textwidth}
\centering
\includegraphics[width=\textwidth]{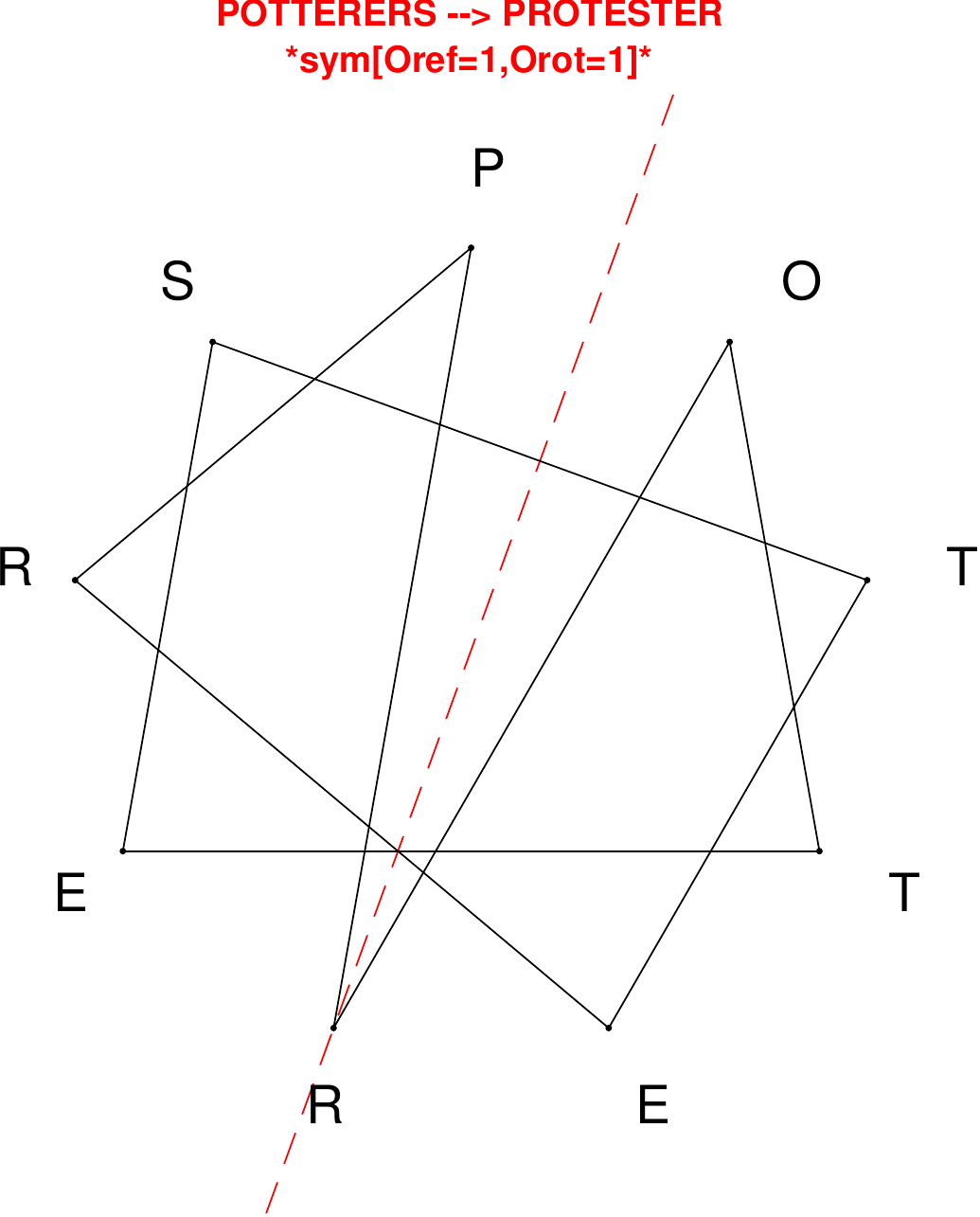}
\end{subfigure}
\hfill
\begin{subfigure}[T]{0.19\textwidth}
\centering
\includegraphics[width=\textwidth]{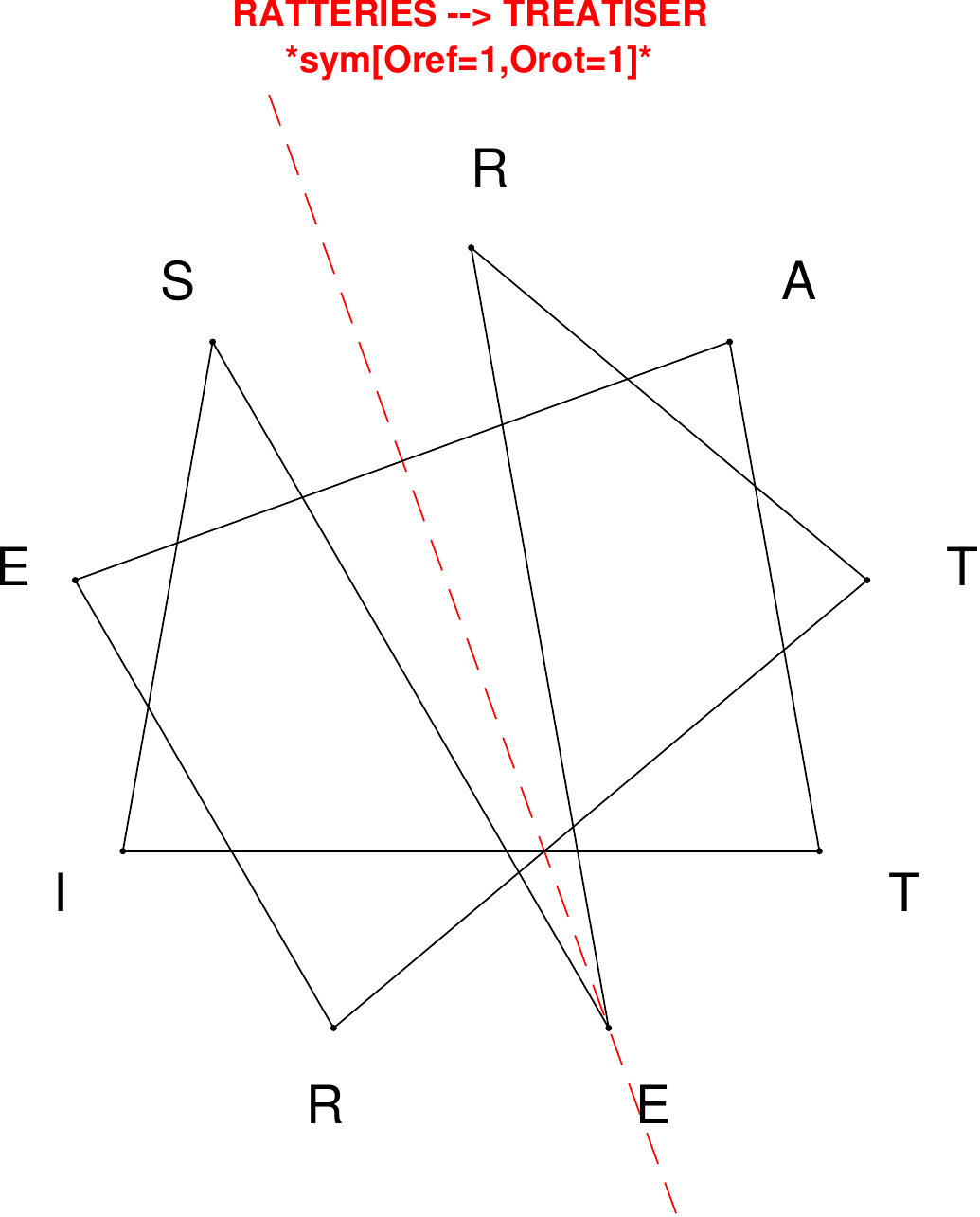}
\end{subfigure}
\hfill
\begin{subfigure}[T]{0.19\textwidth}
\centering
\includegraphics[width=\textwidth]{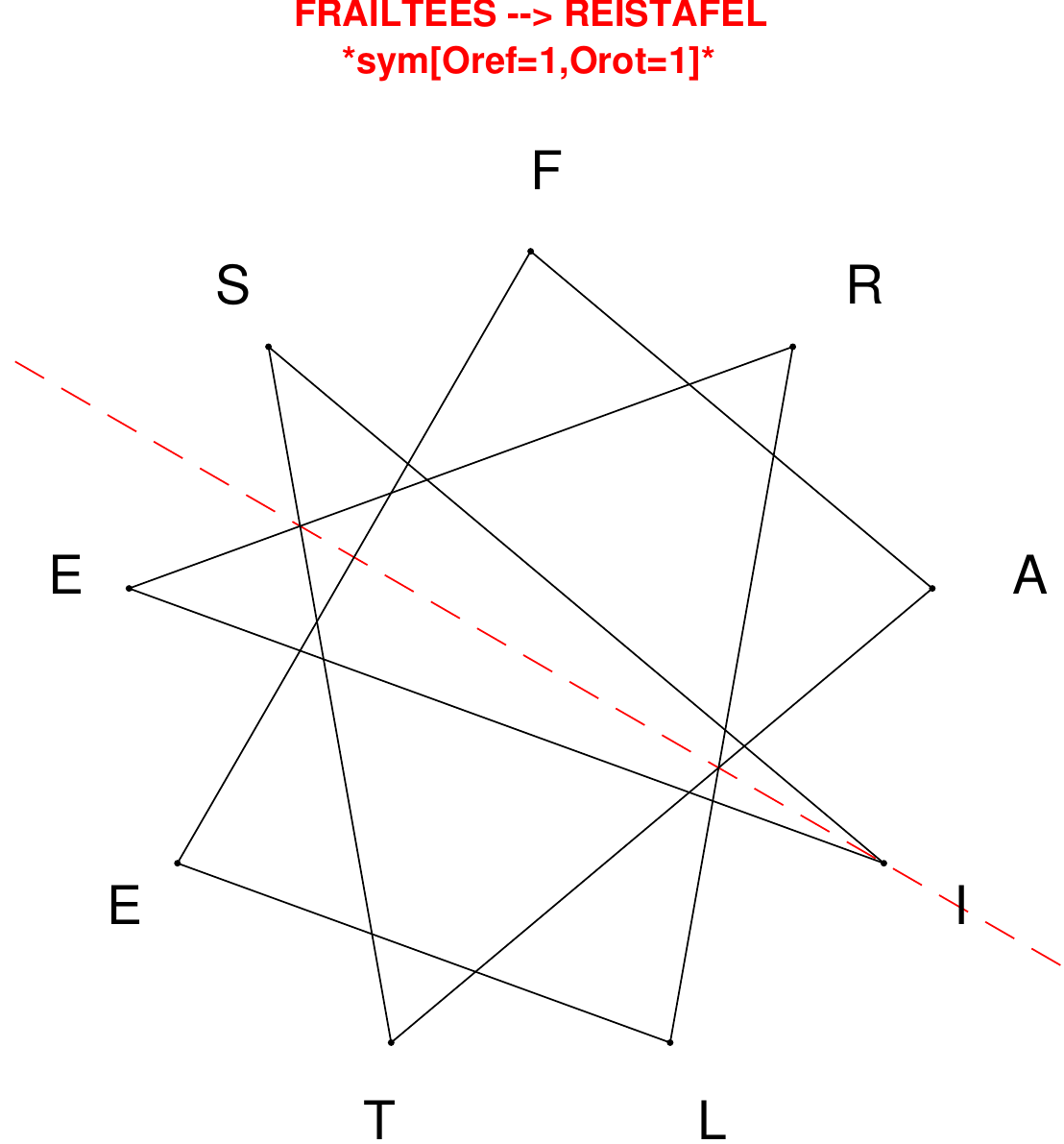}
\end{subfigure}
\end{figure}

\begin{figure}[H]
\centering
\begin{subfigure}[T]{0.19\textwidth}
\centering
\includegraphics[width=\textwidth]{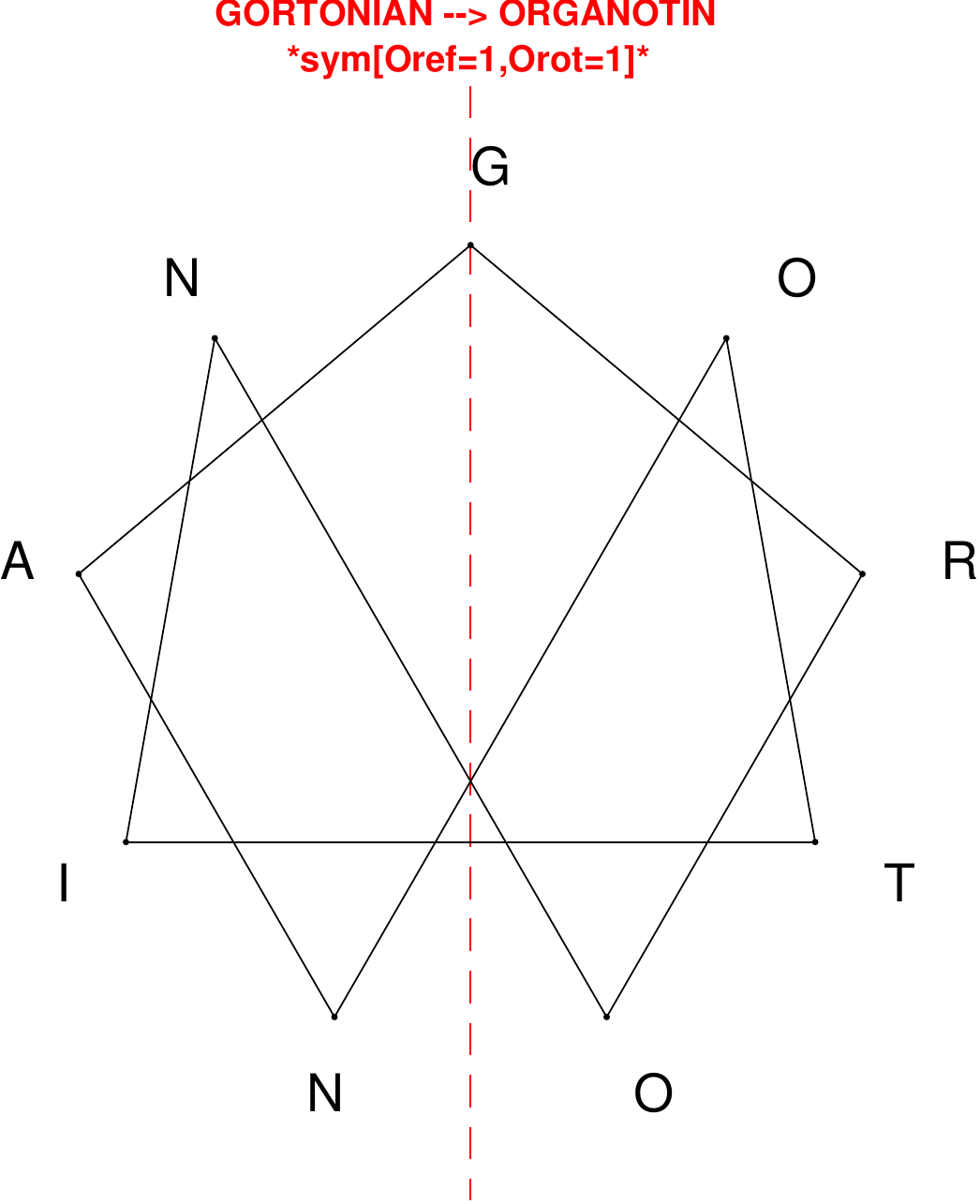}
\end{subfigure}
\hfill
\begin{subfigure}[T]{0.19\textwidth}
\centering
\includegraphics[width=\textwidth]{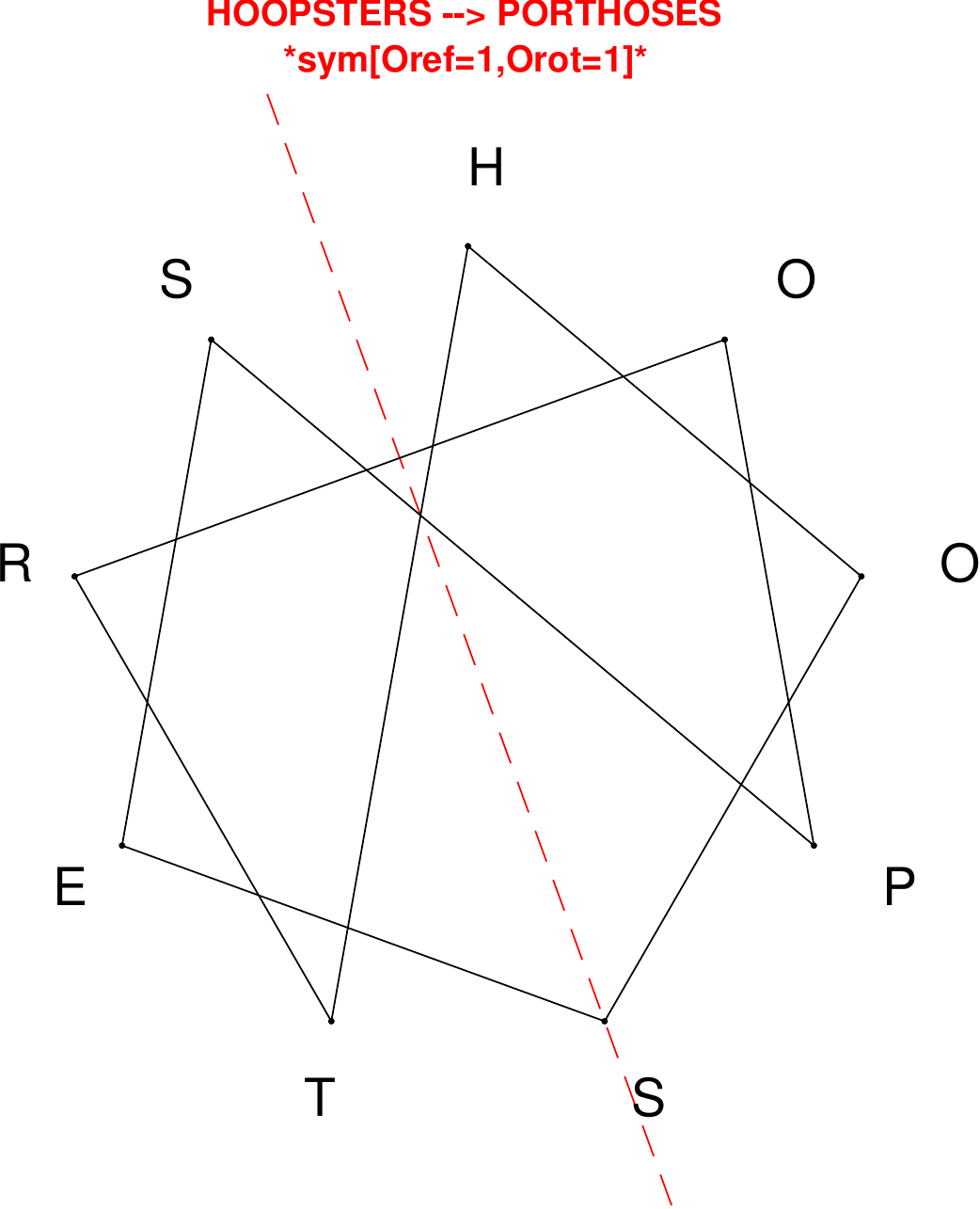}
\end{subfigure}
\hfill
\begin{subfigure}[T]{0.19\textwidth}
\centering
\includegraphics[width=\textwidth]{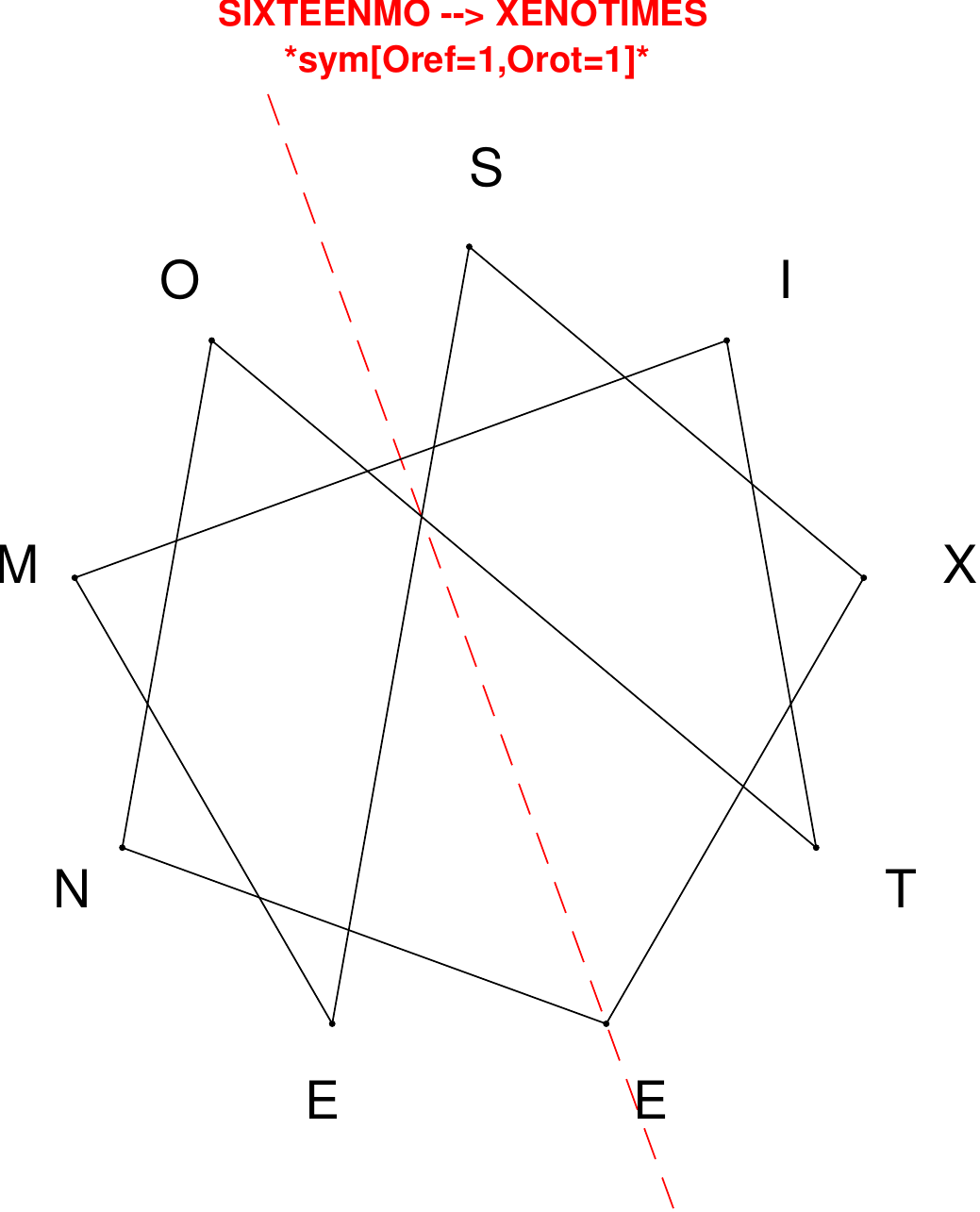}
\end{subfigure}
\hfill
\begin{subfigure}[T]{0.19\textwidth}
\centering
\includegraphics[width=\textwidth]{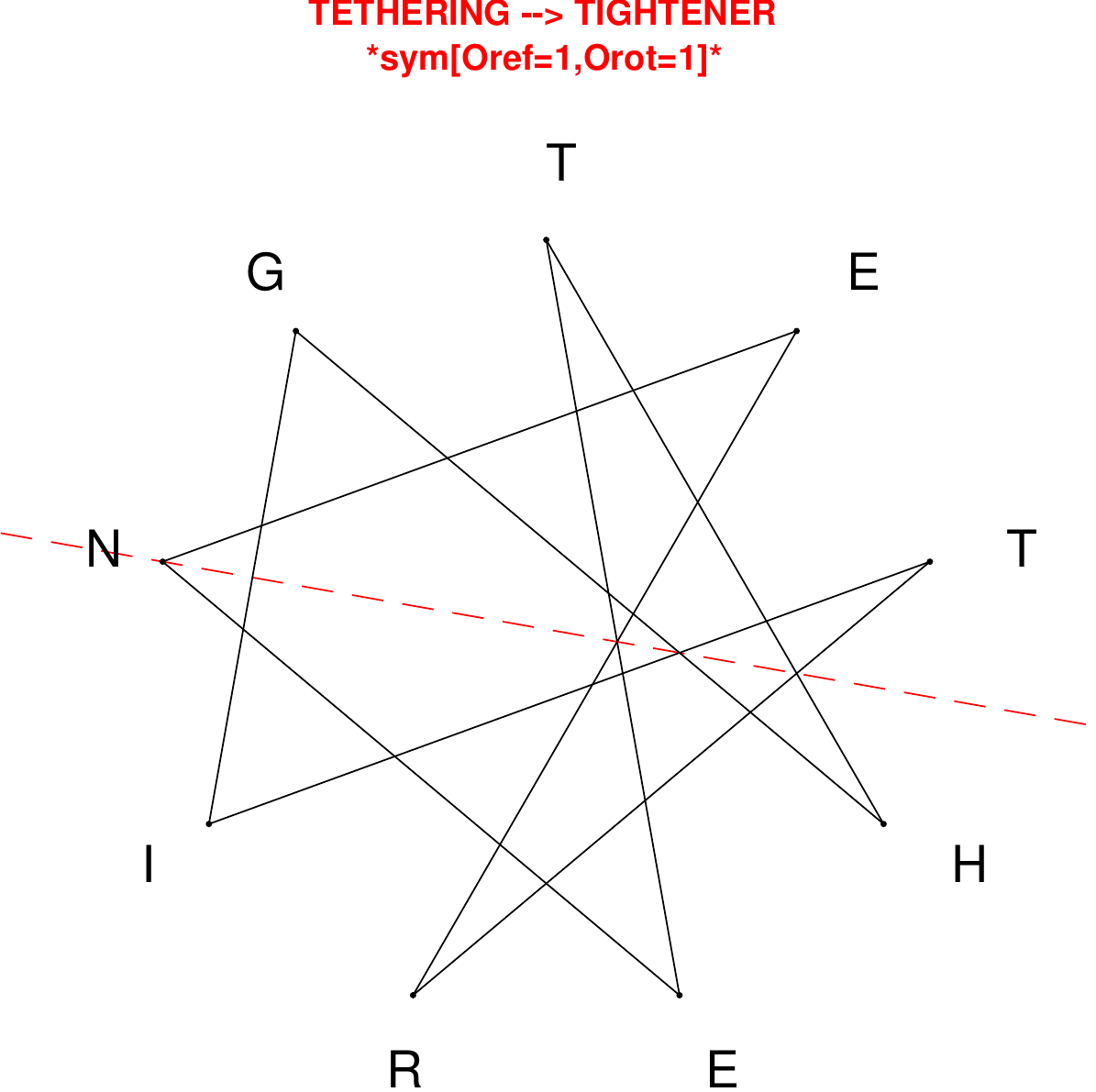}
\end{subfigure}
\hfill
\begin{subfigure}[T]{0.19\textwidth}
\centering
\includegraphics[width=\textwidth]{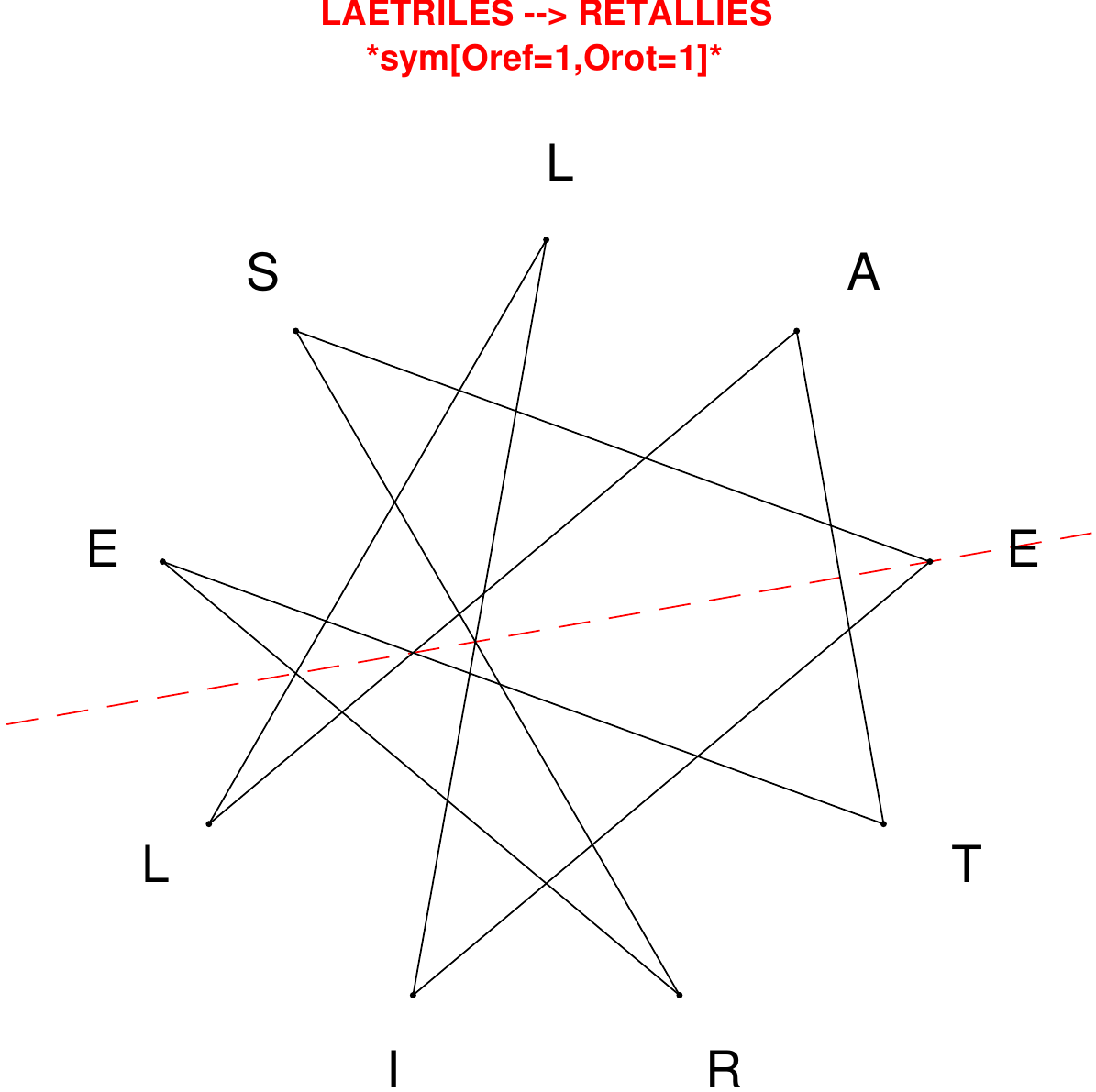}
\end{subfigure}
\end{figure}

\begin{figure}[H]
\centering
\begin{subfigure}[T]{0.19\textwidth}
\centering
\includegraphics[width=\textwidth]{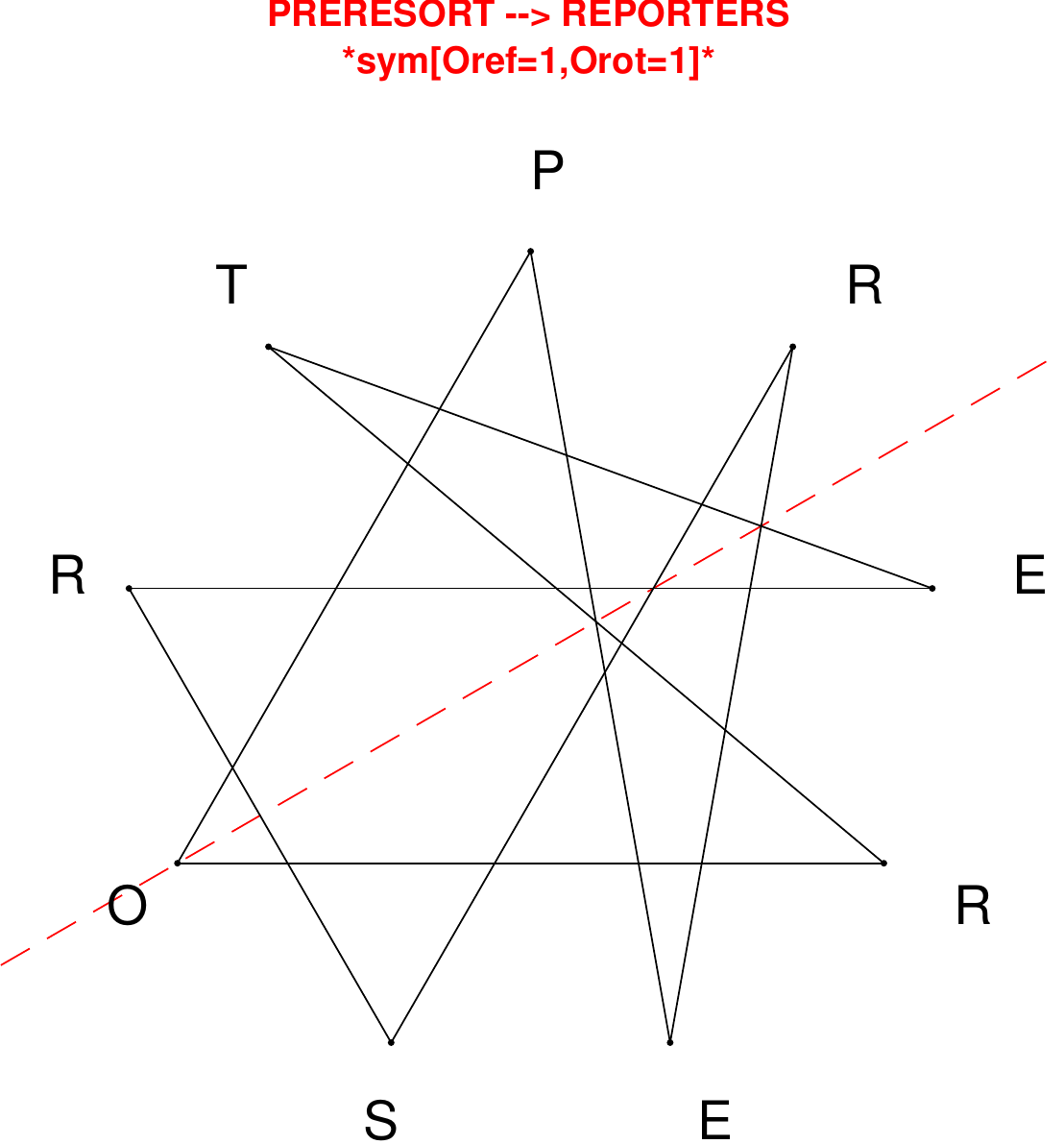}
\end{subfigure}
\hfill
\begin{subfigure}[T]{0.19\textwidth}
\centering
\includegraphics[width=\textwidth]{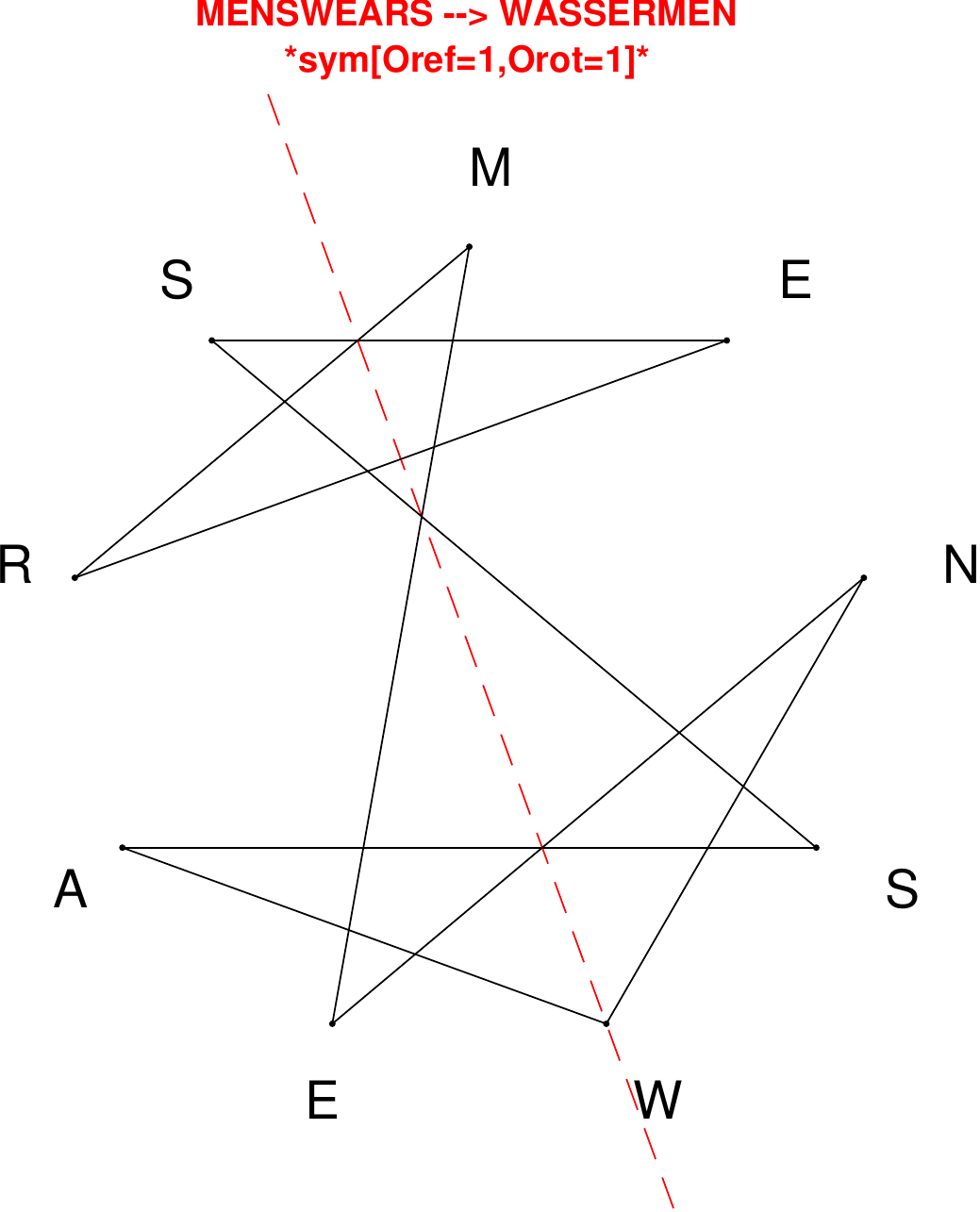}
\end{subfigure}
\hfill
\begin{subfigure}[T]{0.19\textwidth}
\centering
\includegraphics[width=\textwidth]{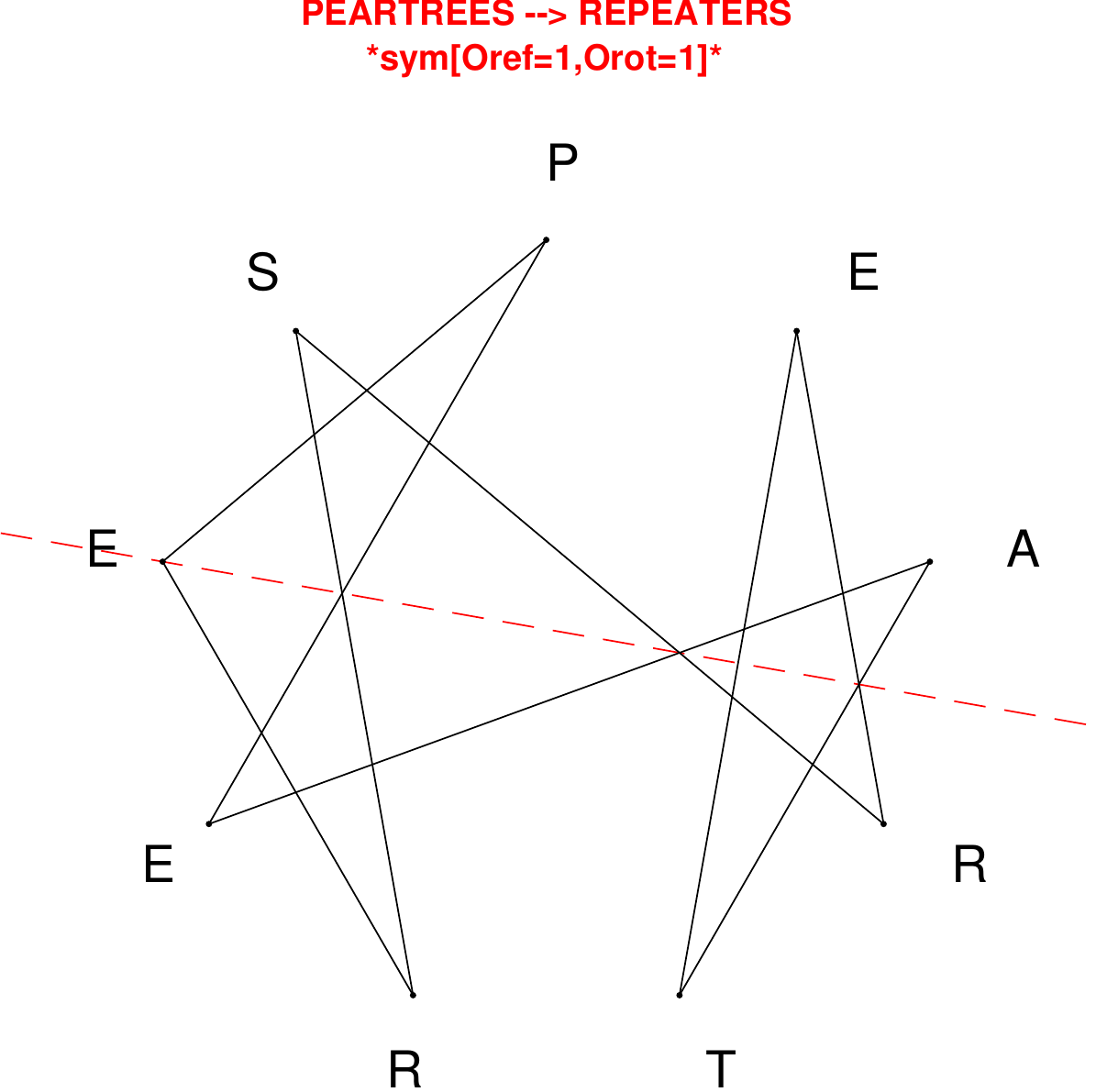}
\end{subfigure}
\hfill
\begin{subfigure}[T]{0.19\textwidth}
\centering
\includegraphics[width=\textwidth]{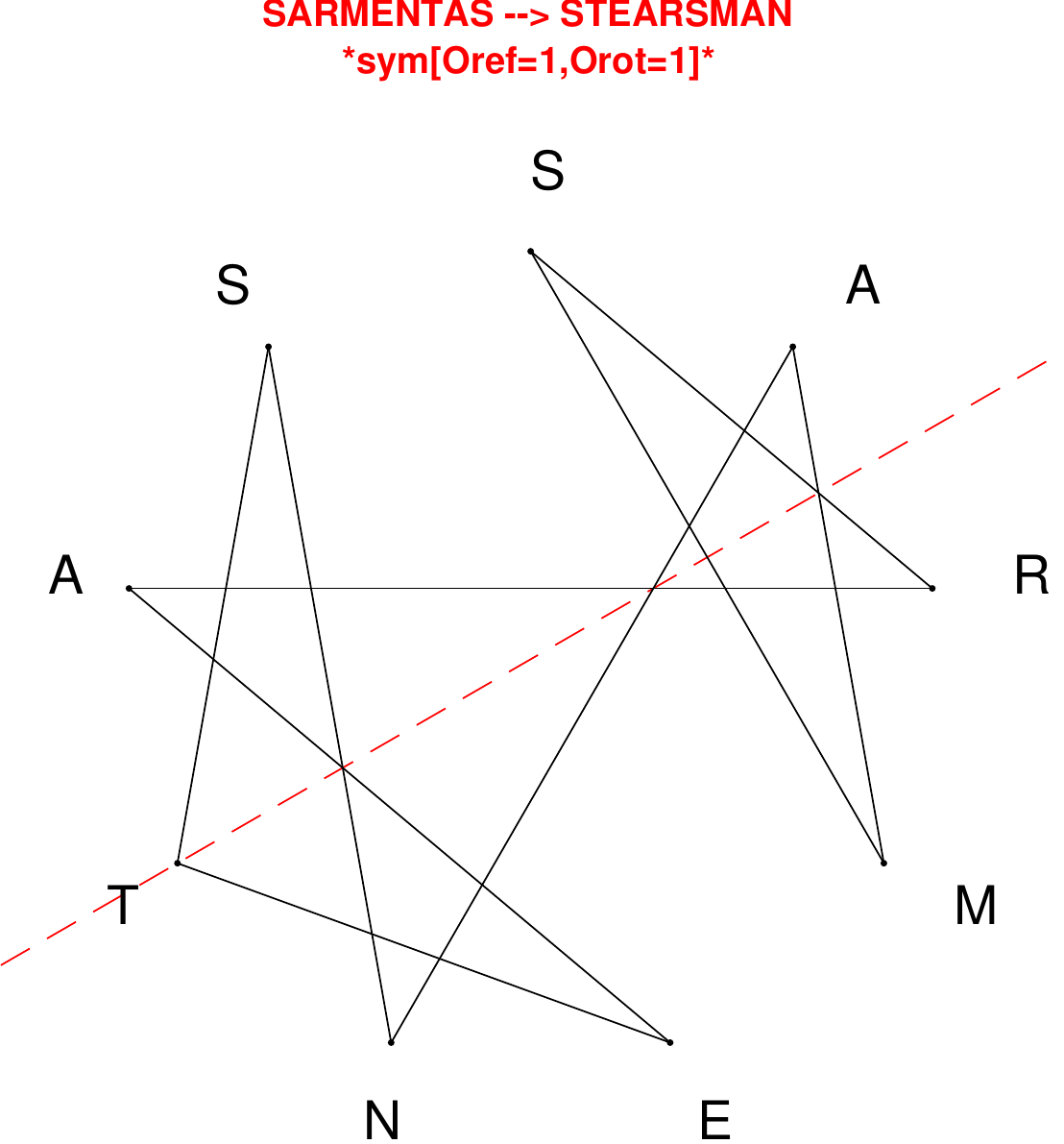}
\end{subfigure}
\hfill
\begin{subfigure}[T]{0.19\textwidth}
\centering
\includegraphics[width=\textwidth]{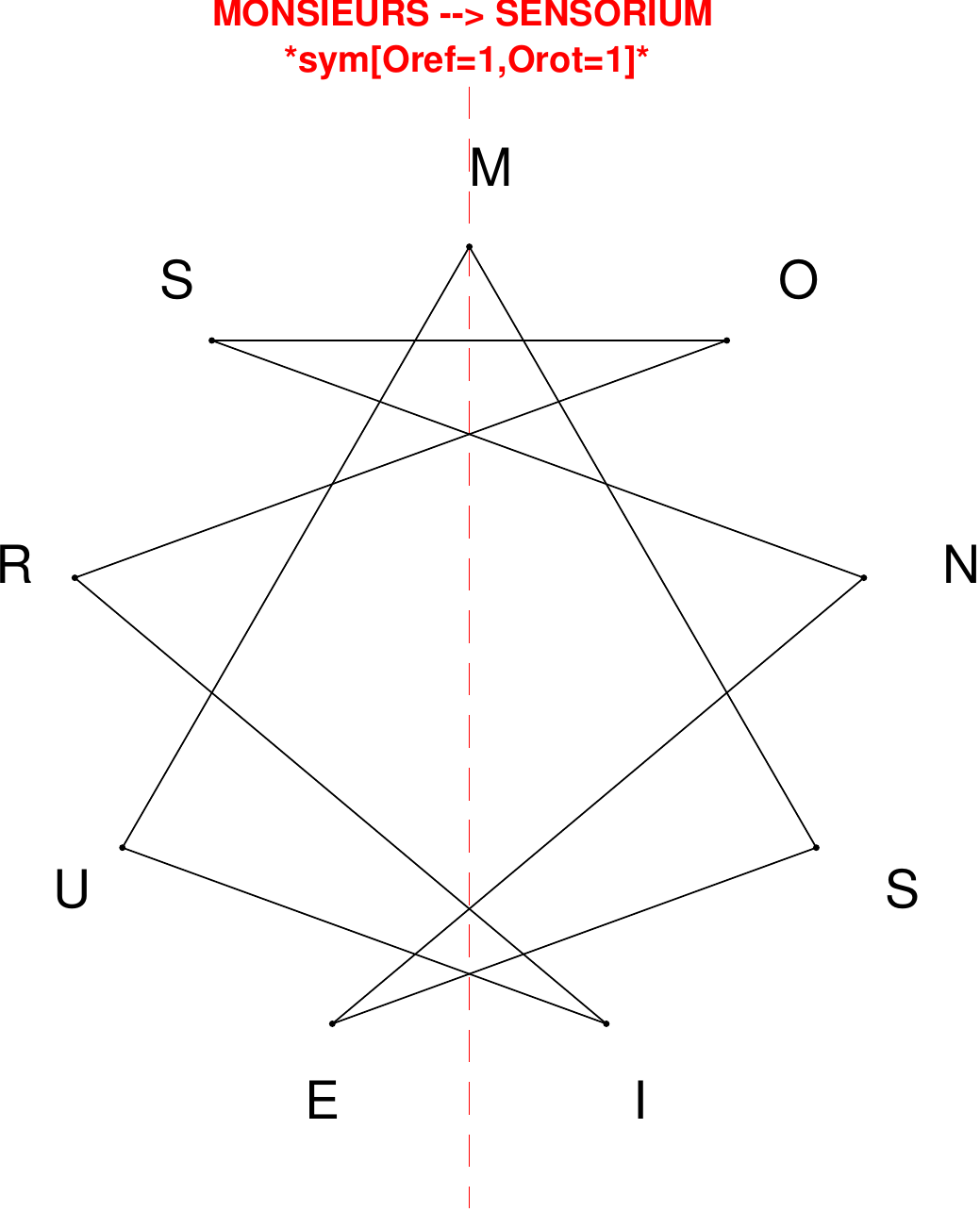}
\end{subfigure}
\end{figure}

\begin{figure}[H]
\centering
\begin{subfigure}[T]{0.19\textwidth}
\centering
\includegraphics[width=\textwidth]{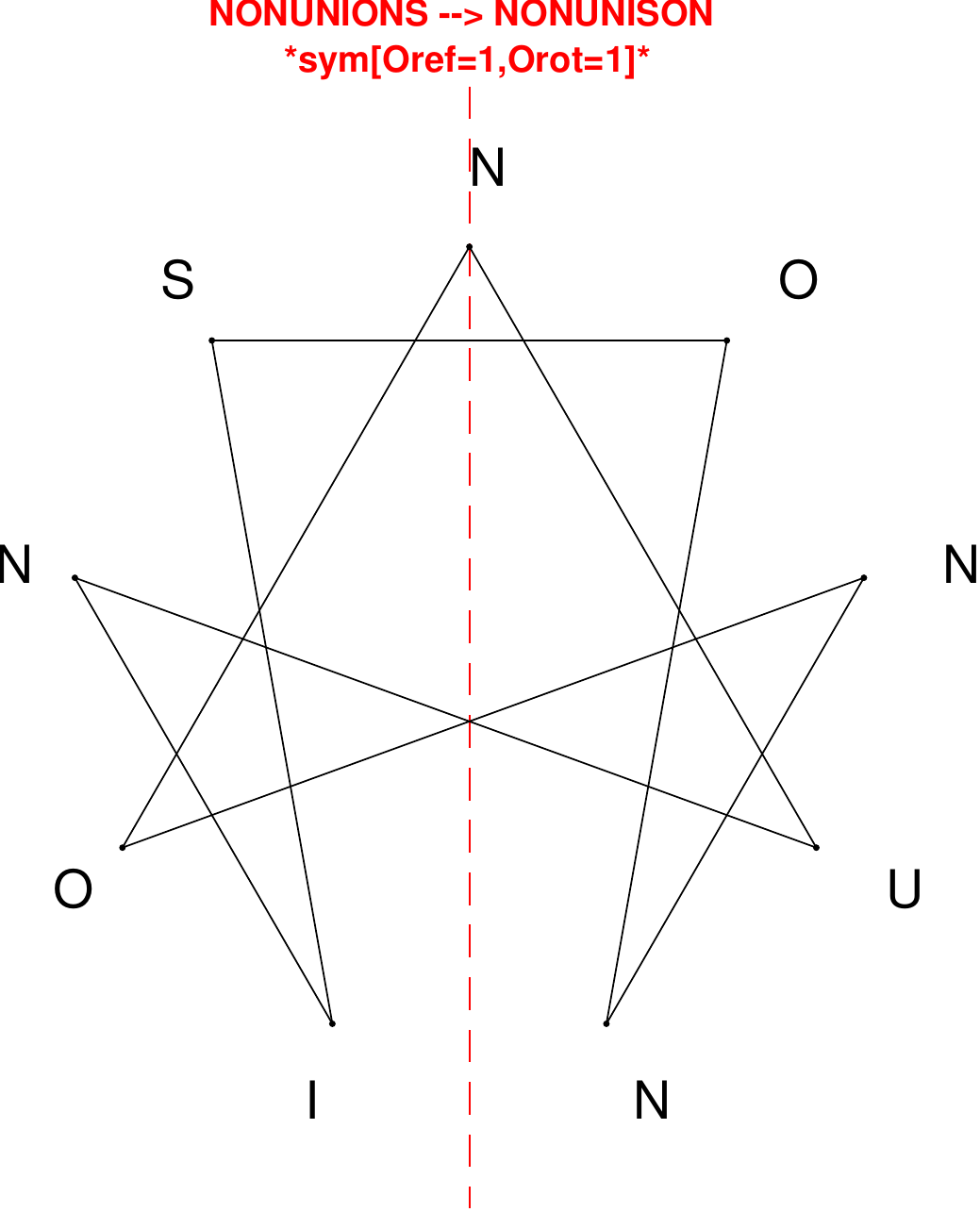}
\end{subfigure}
\hfill
\begin{subfigure}[T]{0.19\textwidth}
\centering
\includegraphics[width=\textwidth]{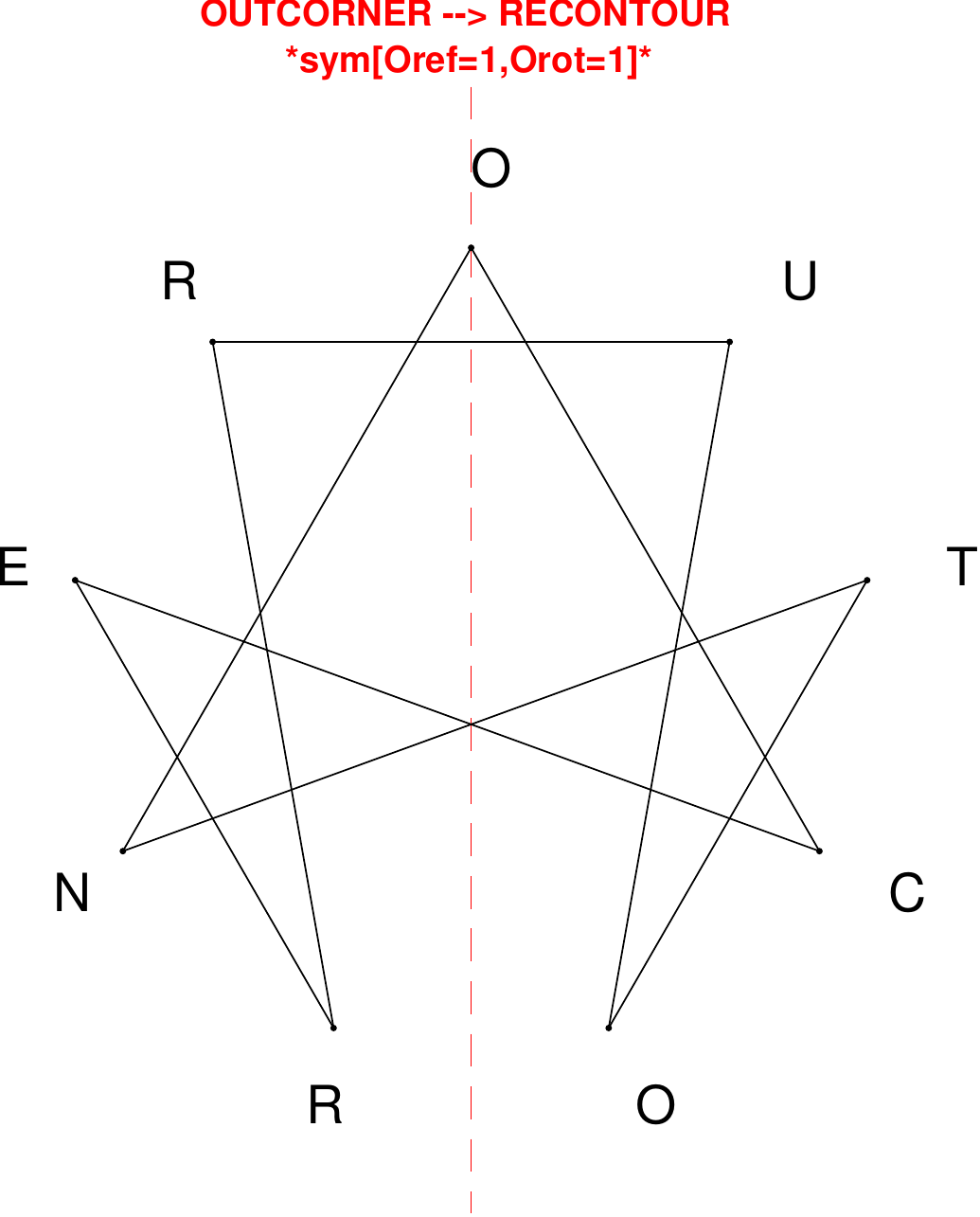}
\end{subfigure}
\hfill
\begin{subfigure}[T]{0.19\textwidth}
\centering
\includegraphics[width=\textwidth]{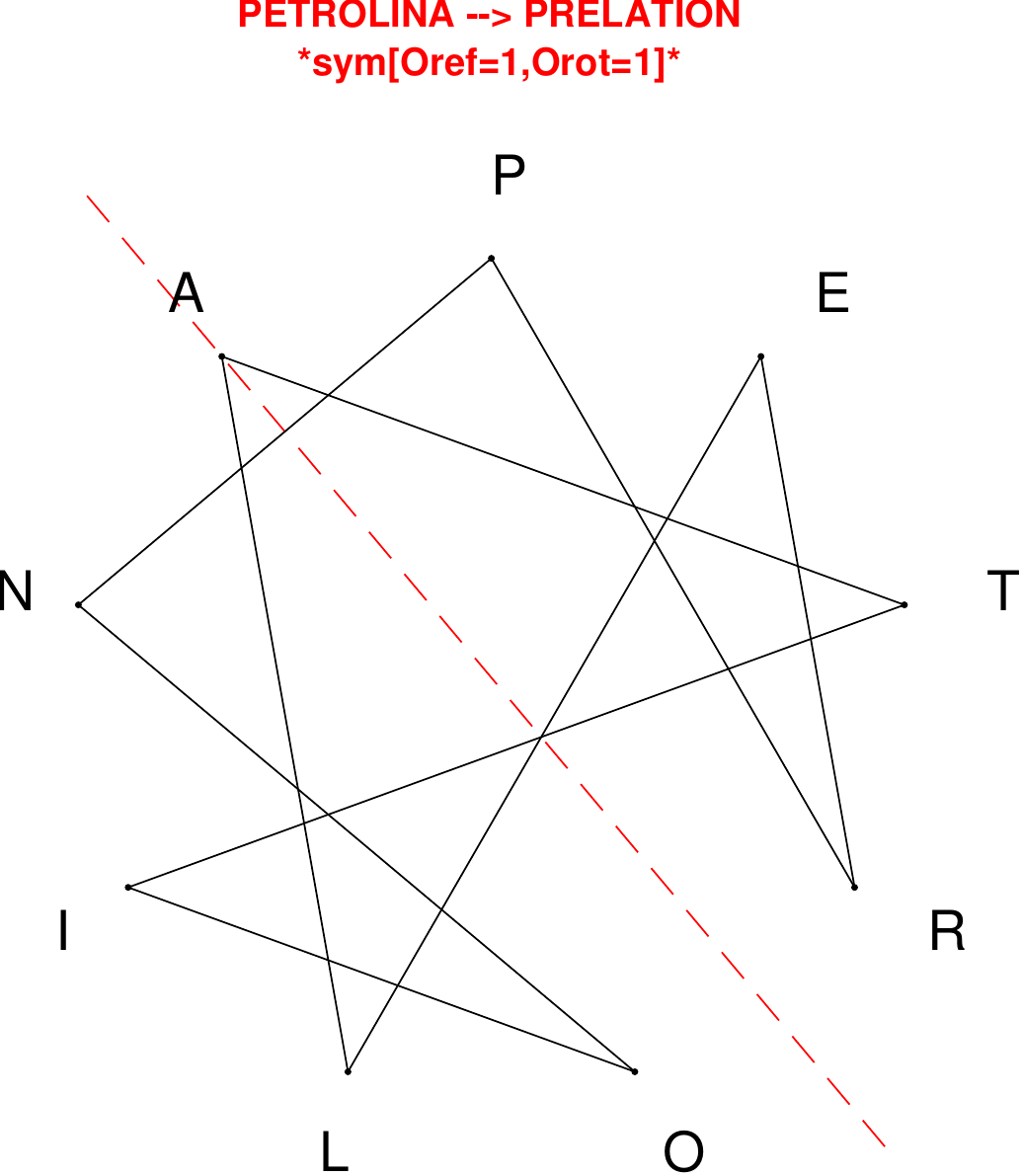}
\end{subfigure}
\hfill
\begin{subfigure}[T]{0.19\textwidth}
\centering
\includegraphics[width=\textwidth]{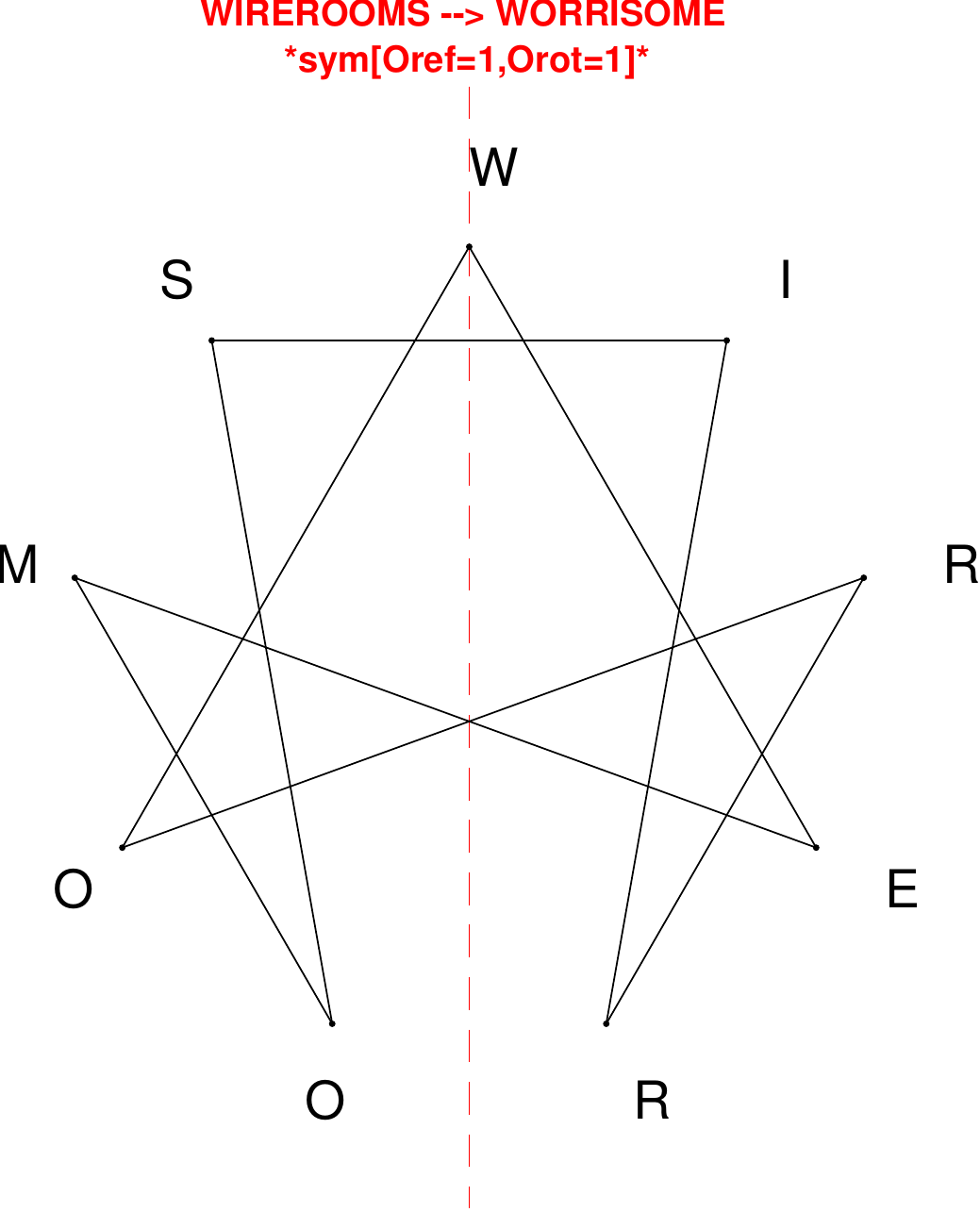}
\end{subfigure}
\hfill
\end{figure}

\subsubsection{Asymmetric Stars $N=9$}

\begin{figure}[H]
\centering
\begin{subfigure}[T]{0.19\textwidth}
\centering
\includegraphics[width=\textwidth]{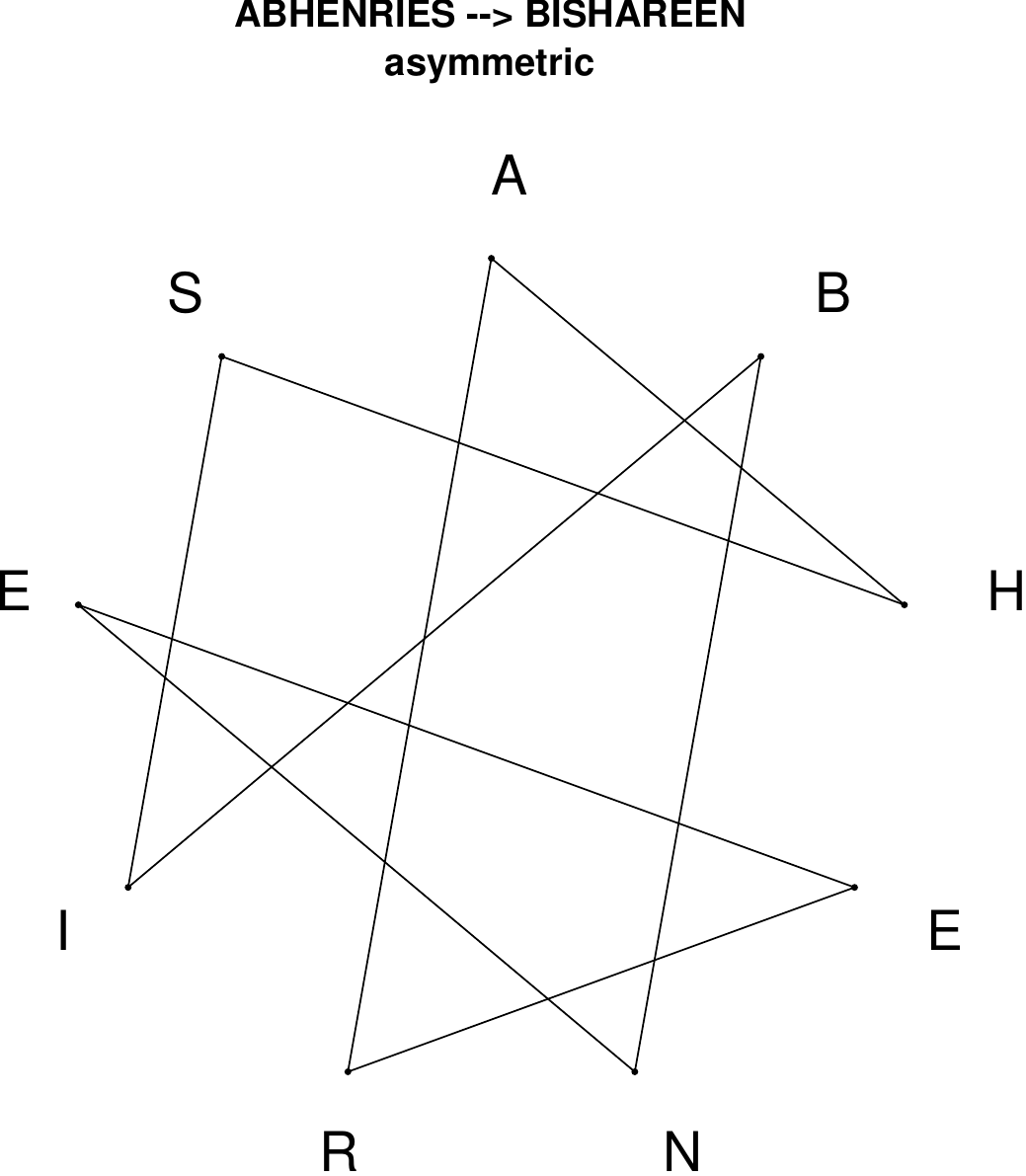}
\end{subfigure}
\hfill
\begin{subfigure}[T]{0.19\textwidth}
\centering
\includegraphics[width=\textwidth]{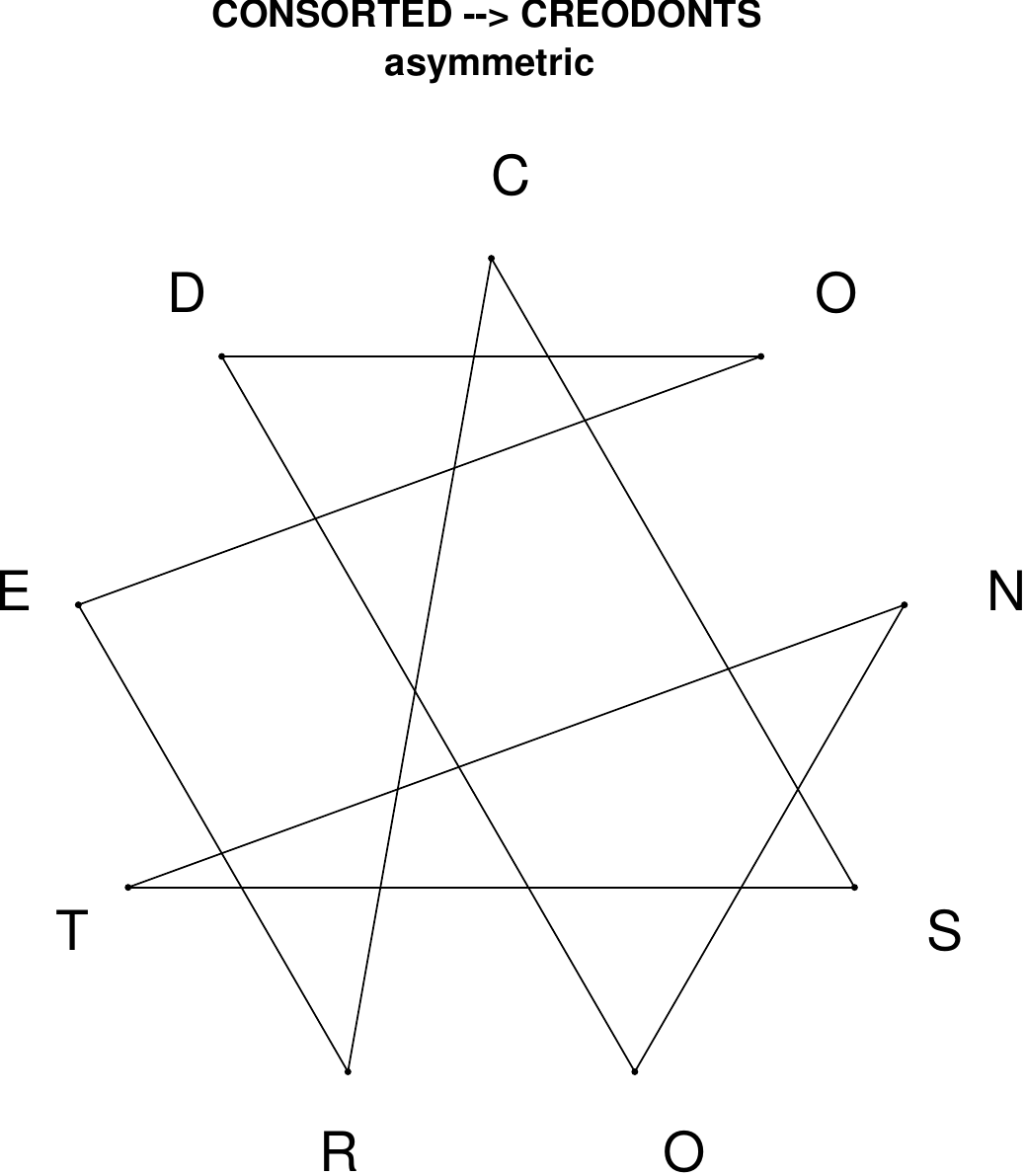}
\end{subfigure}
\hfill
\begin{subfigure}[T]{0.19\textwidth}
\centering
\includegraphics[width=\textwidth]{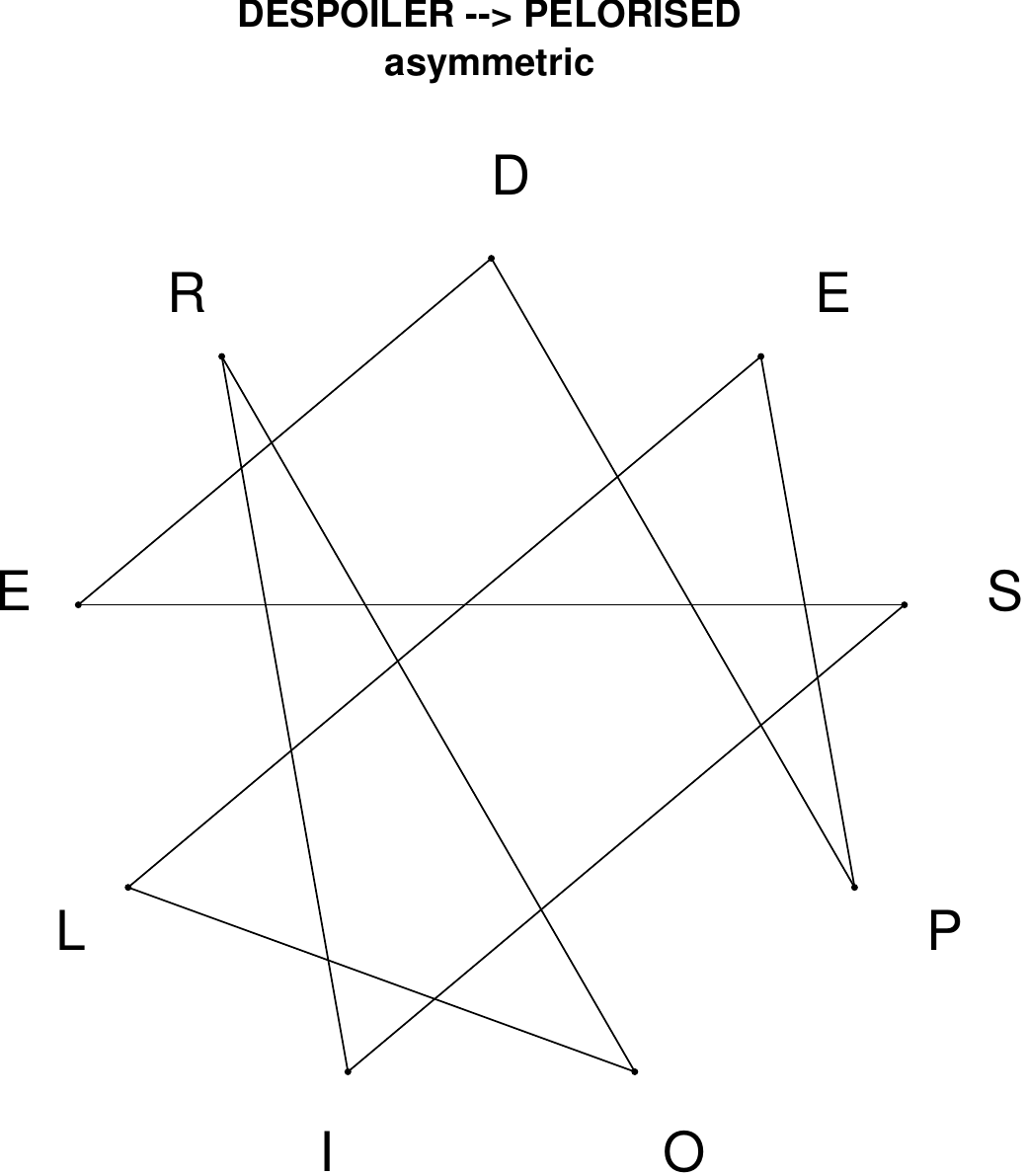}
\end{subfigure}
\hfill
\begin{subfigure}[T]{0.19\textwidth}
\centering
\includegraphics[width=\textwidth]{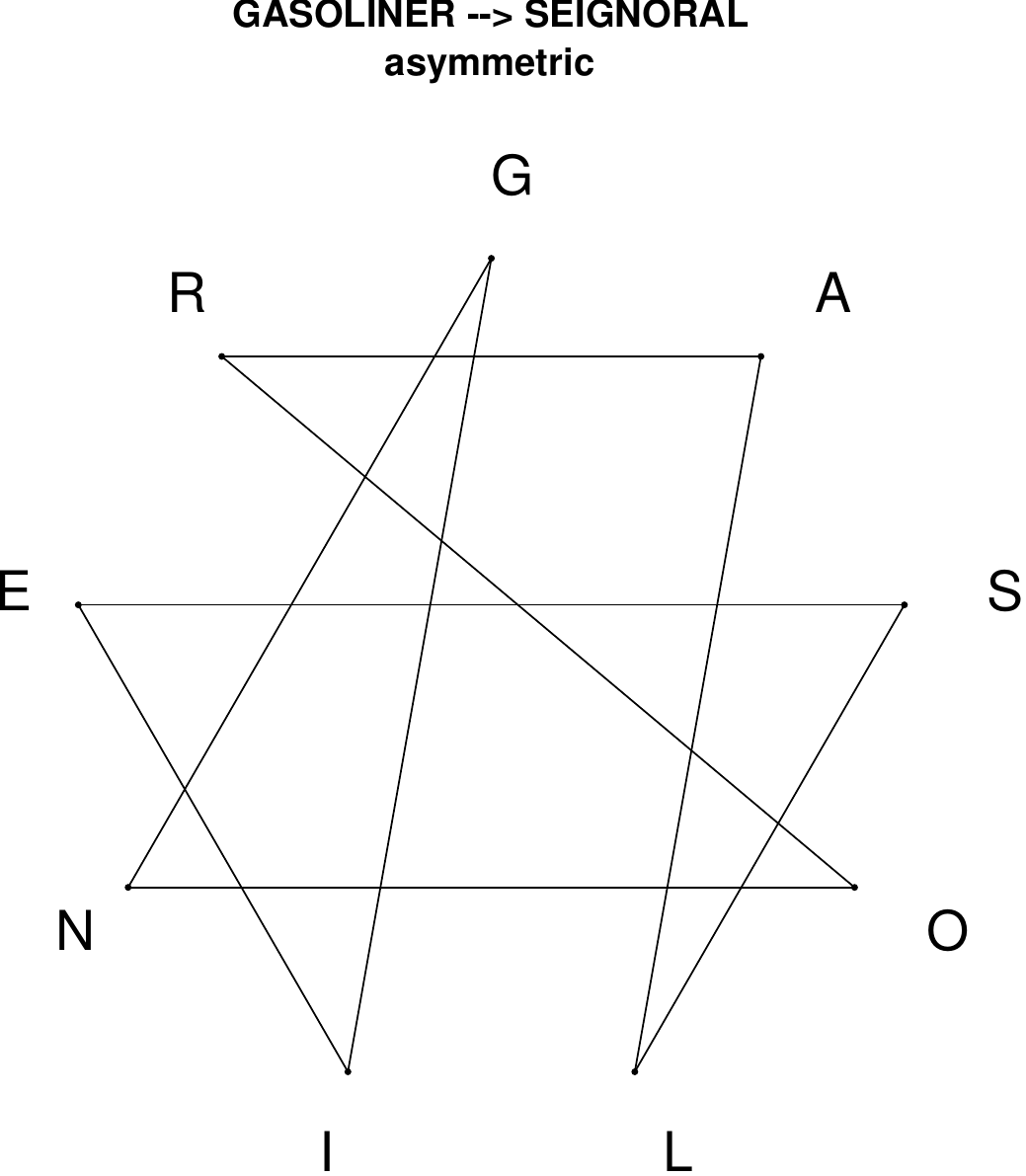}
\end{subfigure}
\hfill
\begin{subfigure}[T]{0.19\textwidth}
\centering
\includegraphics[width=\textwidth]{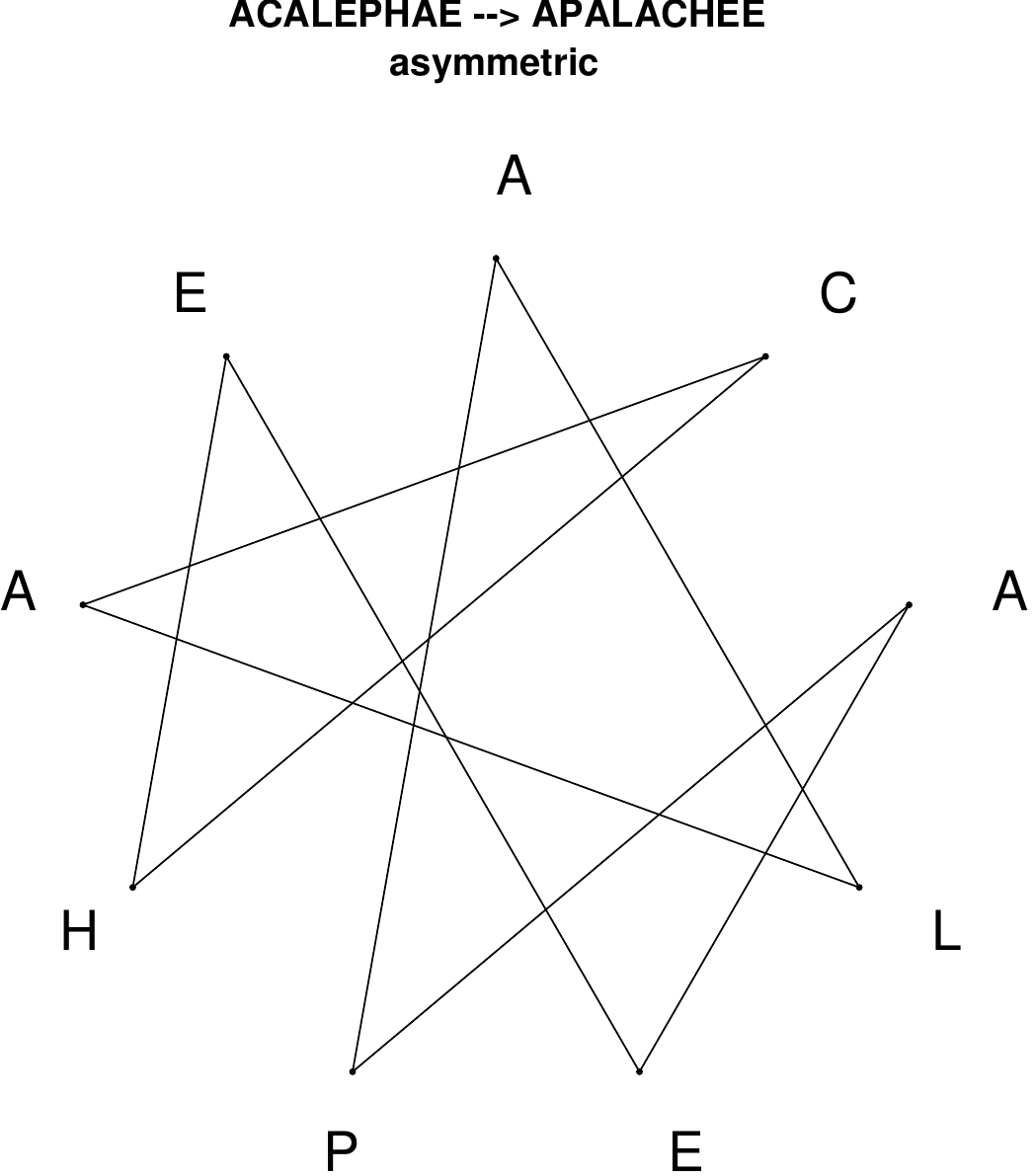}
\end{subfigure}
\end{figure}

\begin{figure}[H]
\centering
\begin{subfigure}[T]{0.19\textwidth}
\centering
\includegraphics[width=\textwidth]{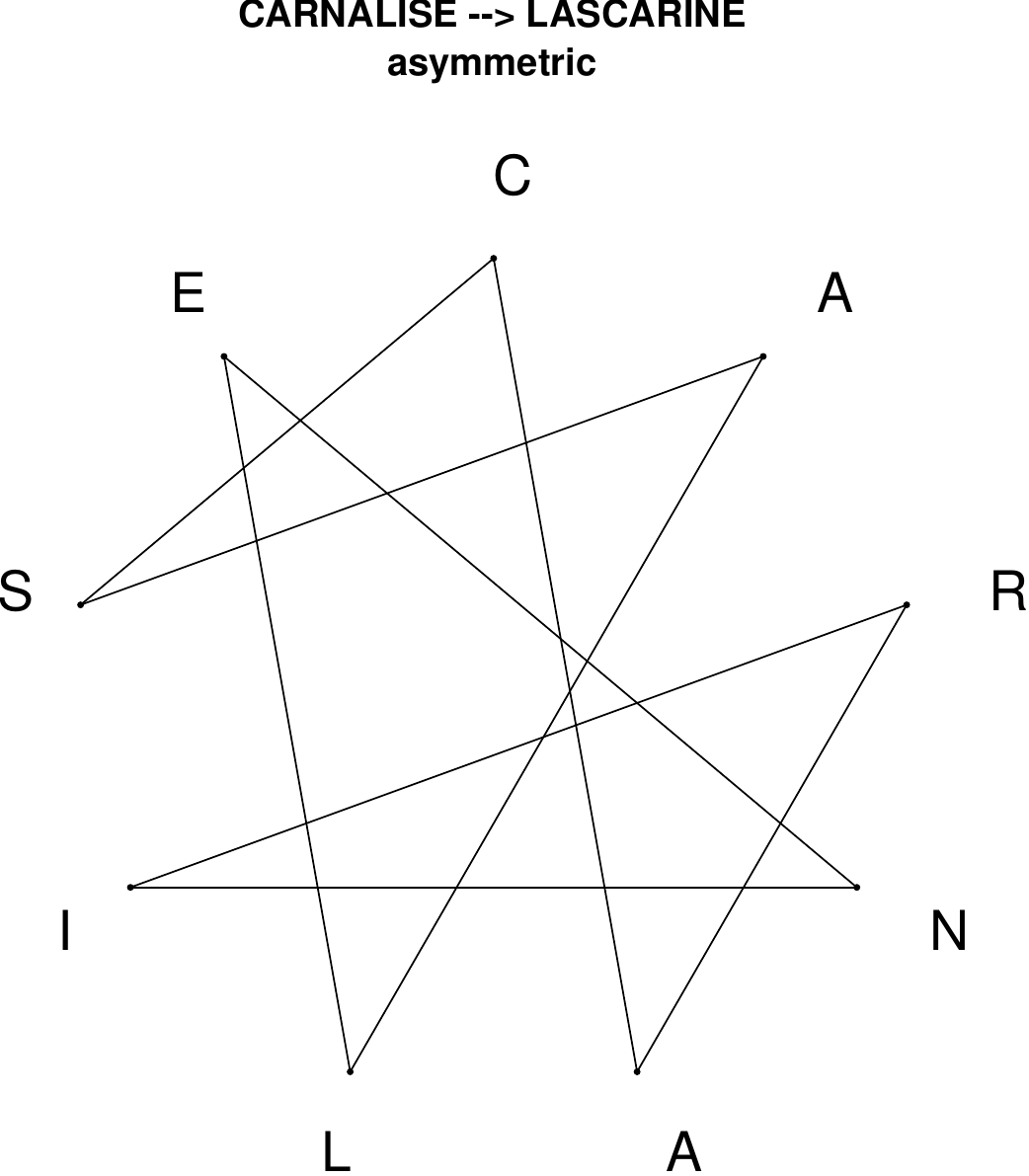}
\end{subfigure}
\hfill
\begin{subfigure}[T]{0.19\textwidth}
\centering
\includegraphics[width=\textwidth]{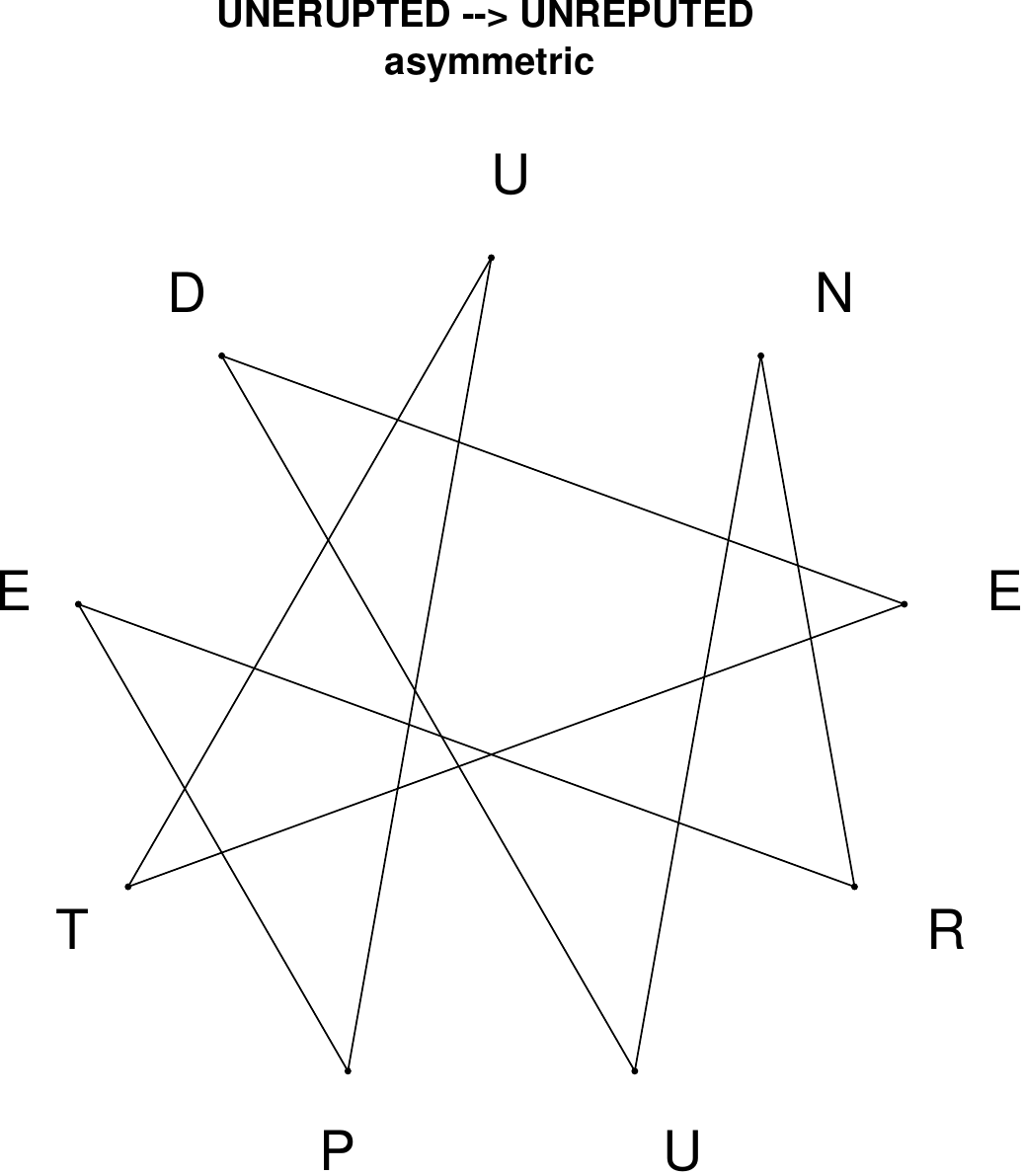}
\end{subfigure}
\hfill
\begin{subfigure}[T]{0.19\textwidth}
\centering
\includegraphics[width=\textwidth]{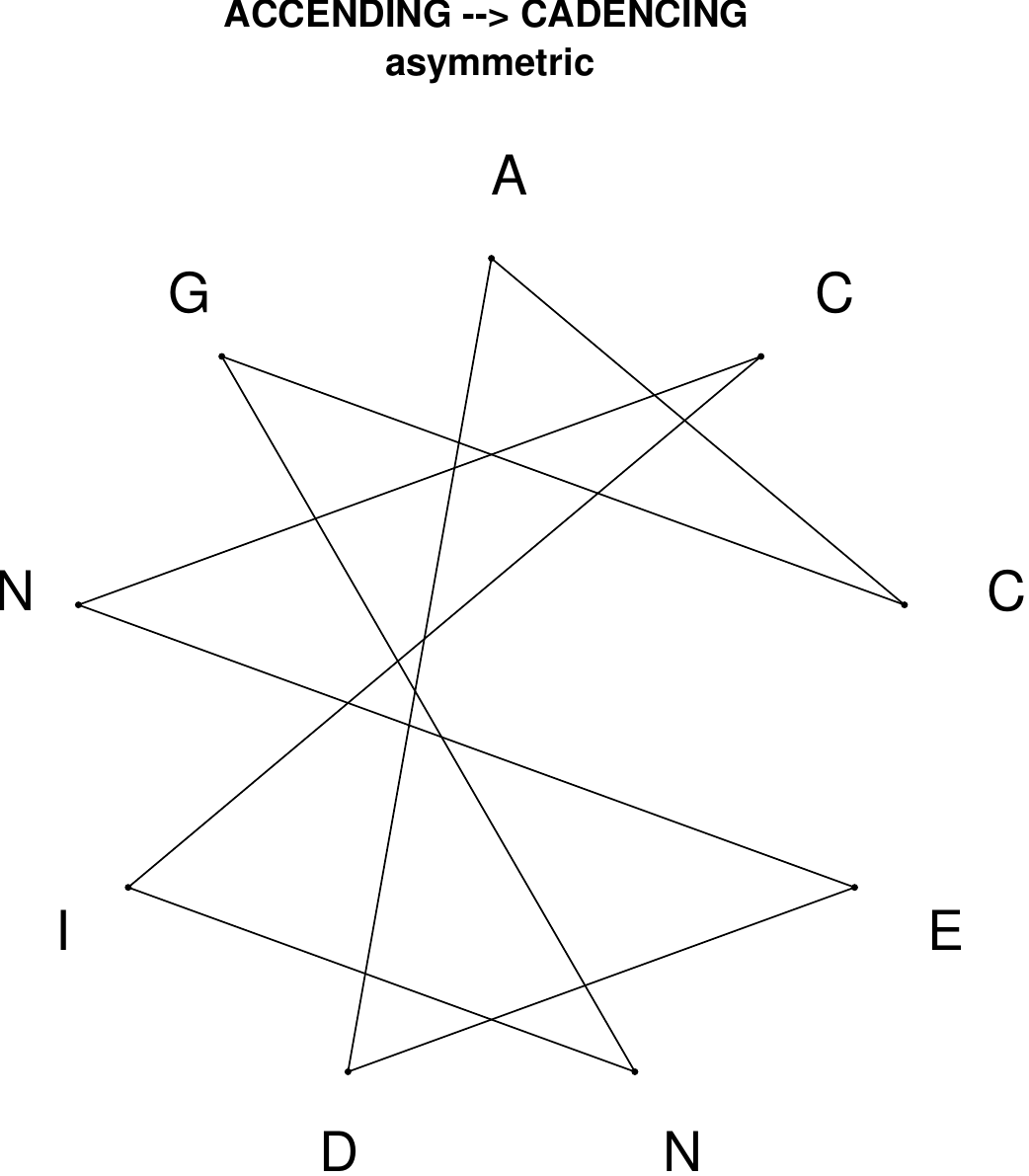}
\end{subfigure}
\hfill
\begin{subfigure}[T]{0.19\textwidth}
\centering
\includegraphics[width=\textwidth]{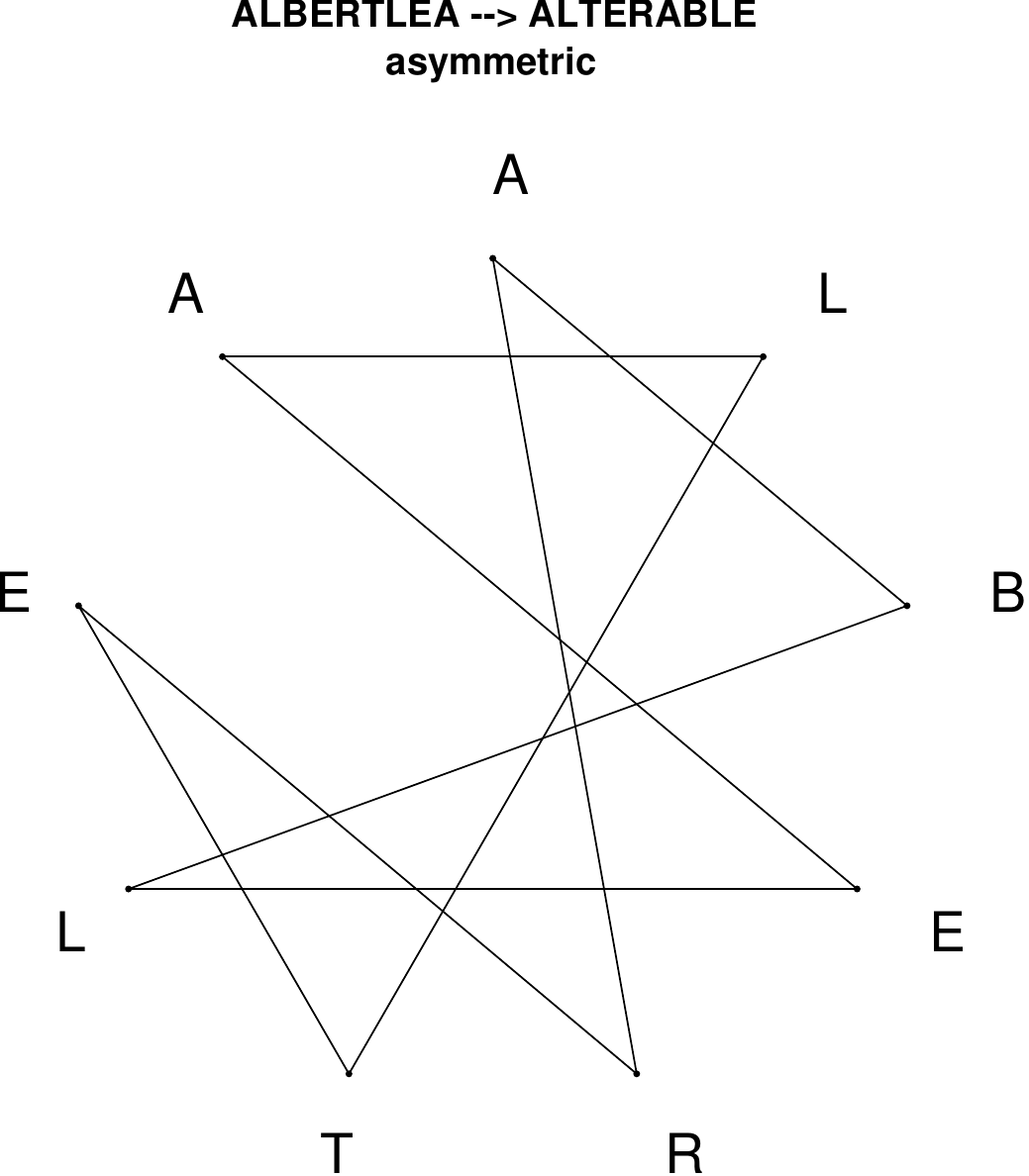}
\end{subfigure}
\hfill
\begin{subfigure}[T]{0.19\textwidth}
\centering
\includegraphics[width=\textwidth]{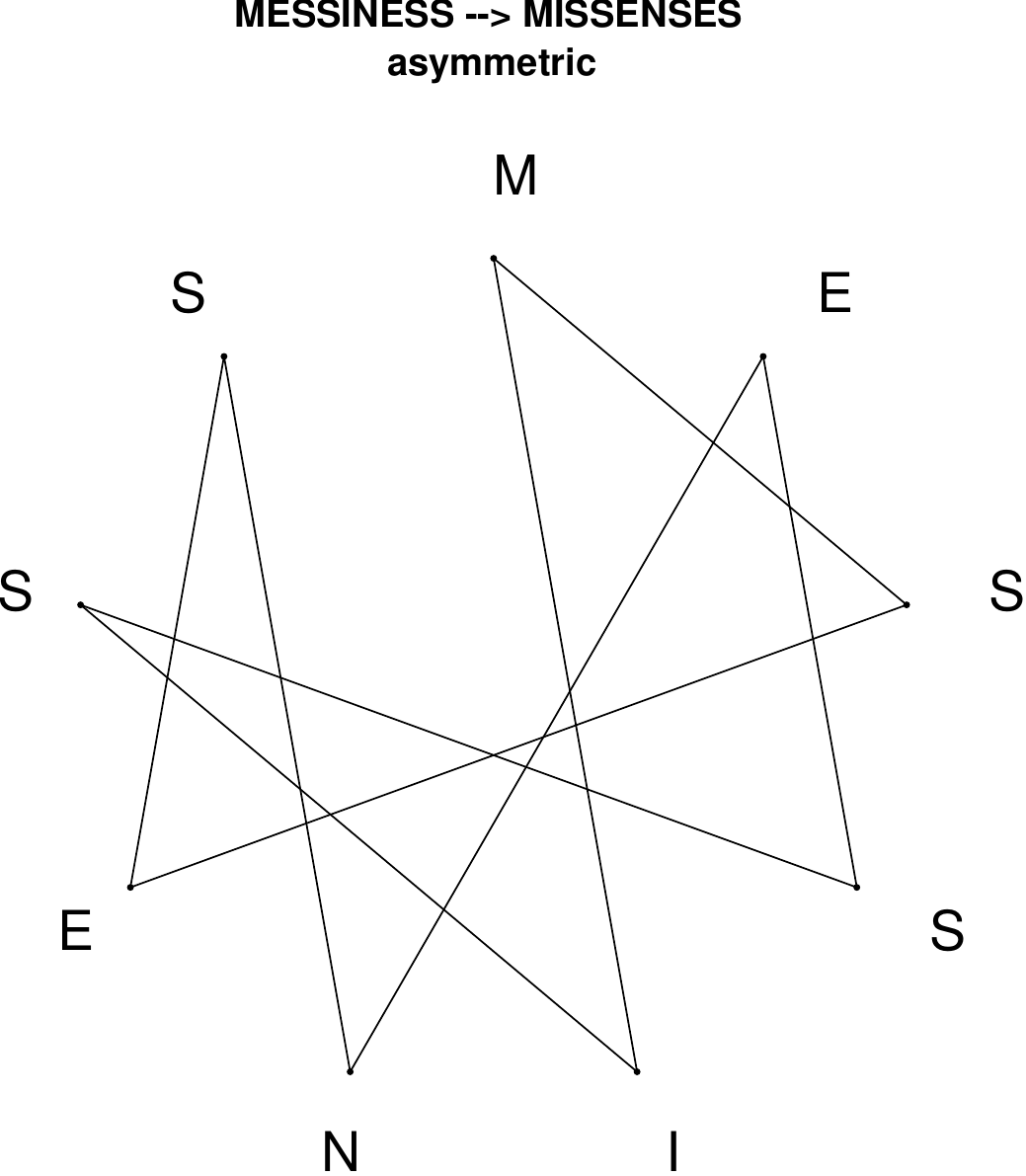}
\end{subfigure}
\end{figure}

\begin{figure}[H]
\centering
\begin{subfigure}[T]{0.19\textwidth}
\centering
\includegraphics[width=\textwidth]{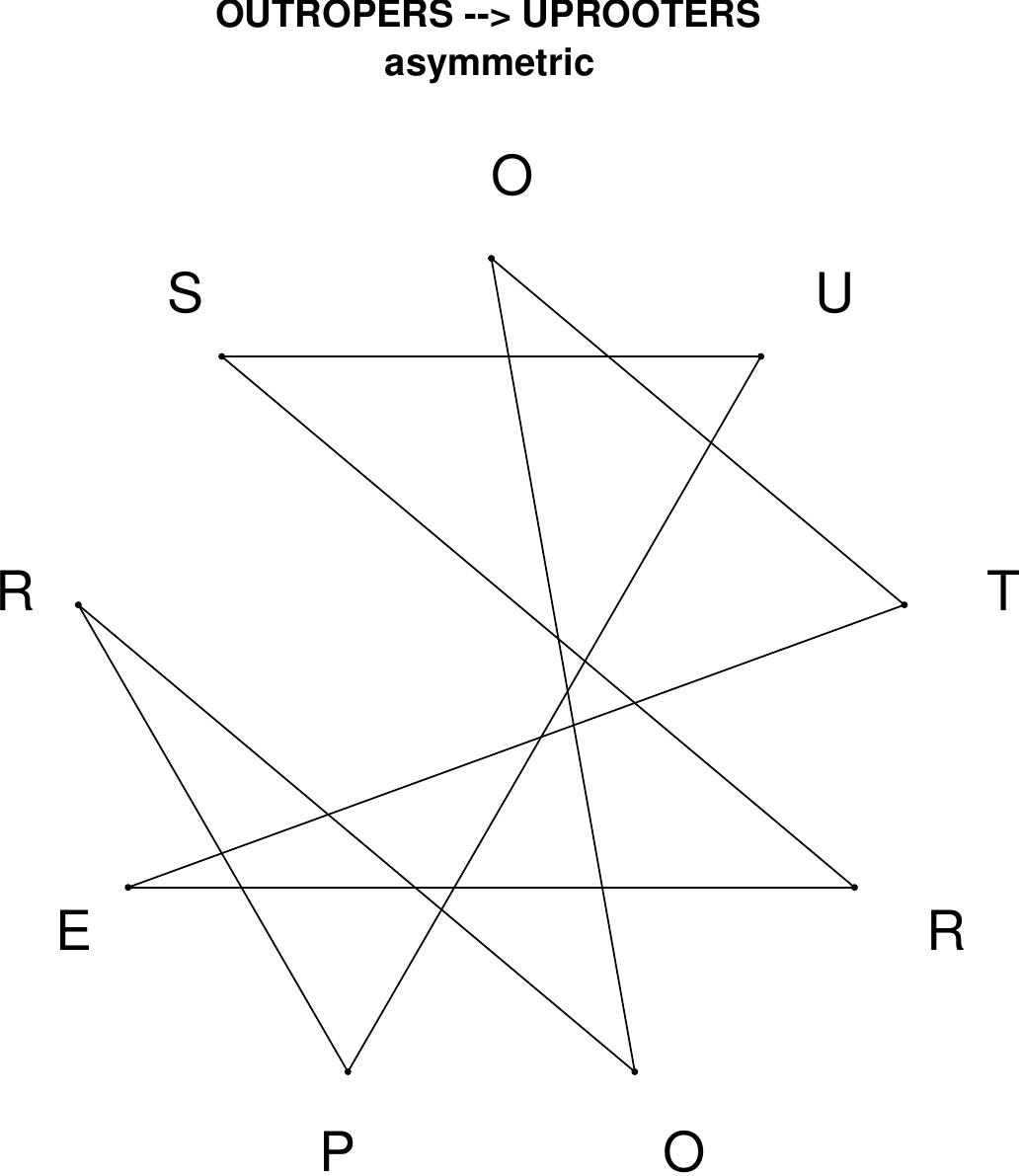}
\end{subfigure}
\hfill
\begin{subfigure}[T]{0.19\textwidth}
\centering
\includegraphics[width=\textwidth]{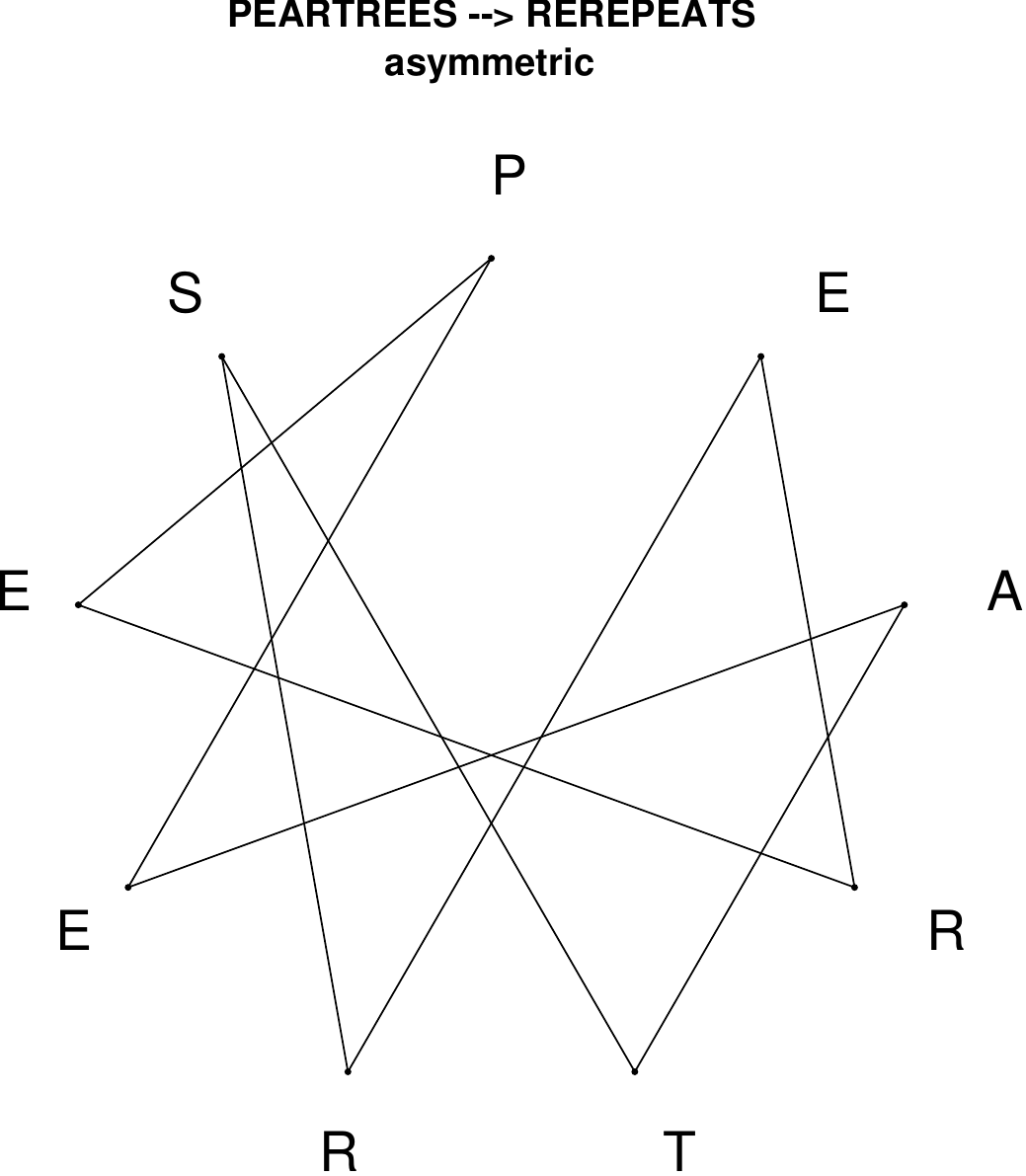}
\end{subfigure}
\hfill
\begin{subfigure}[T]{0.19\textwidth}
\centering
\includegraphics[width=\textwidth]{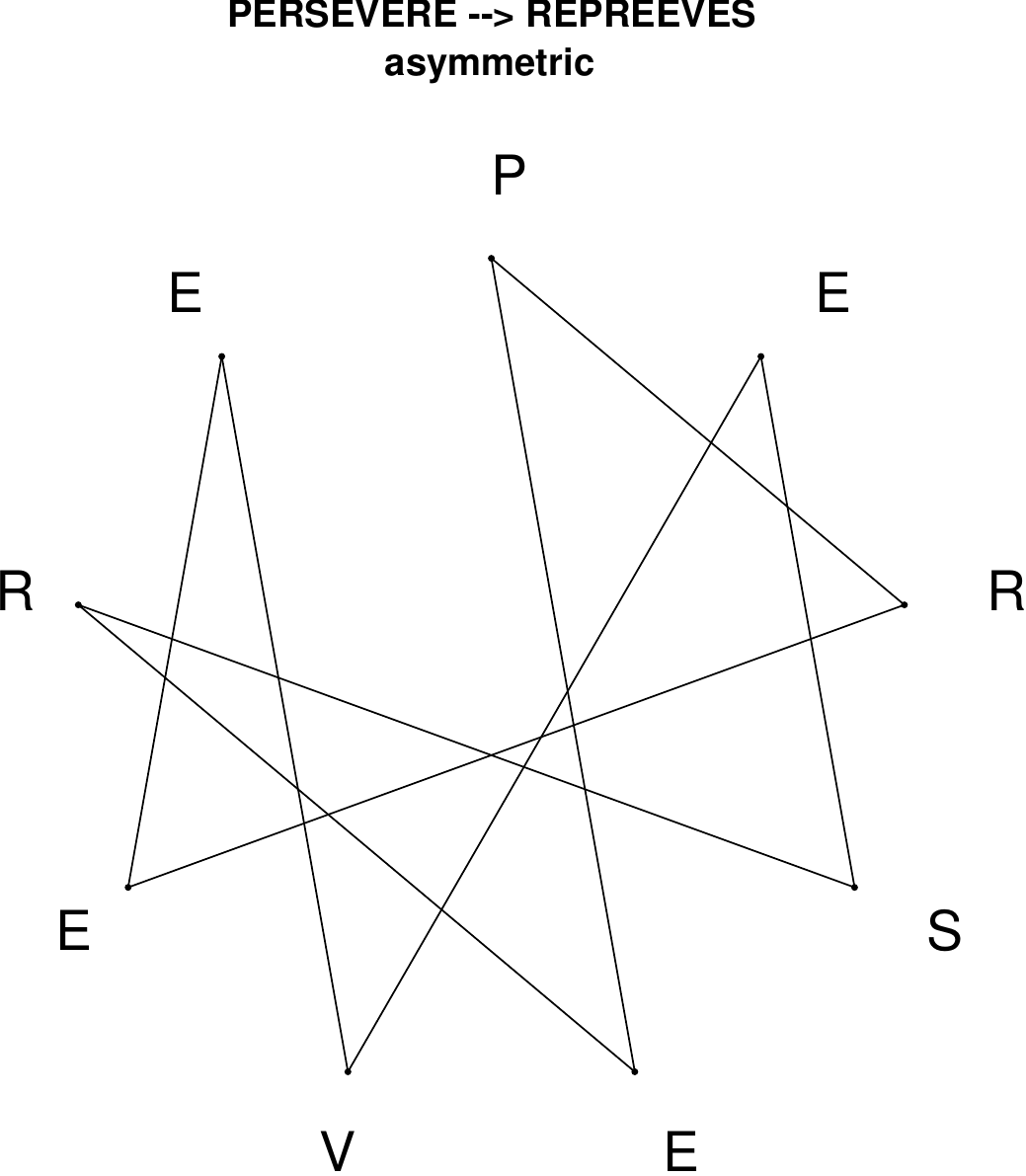}
\end{subfigure}
\hfill
\begin{subfigure}[T]{0.19\textwidth}
\centering
\includegraphics[width=\textwidth]{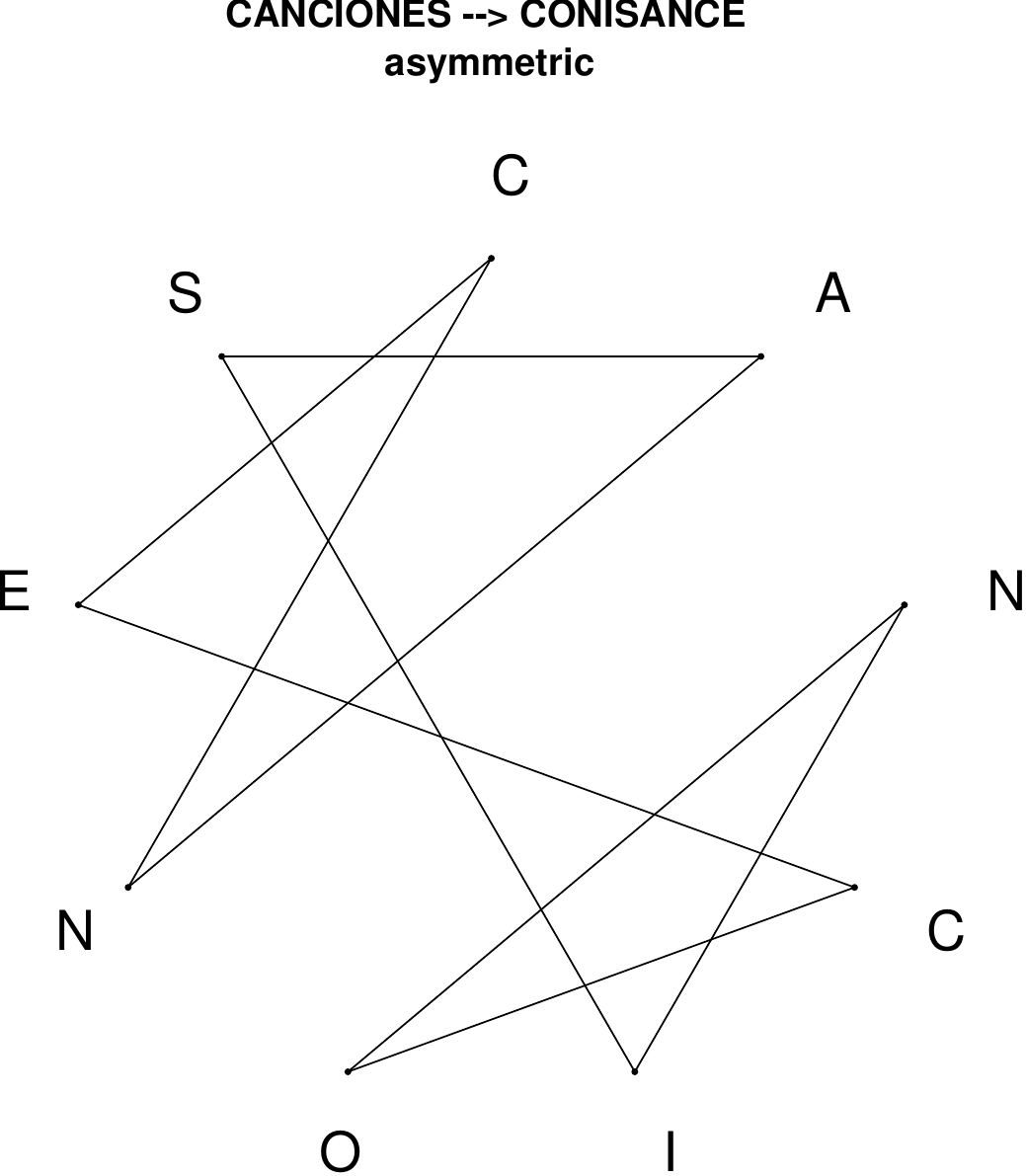}
\end{subfigure}
\hfill
\begin{subfigure}[T]{0.19\textwidth}
\centering
\includegraphics[width=\textwidth]{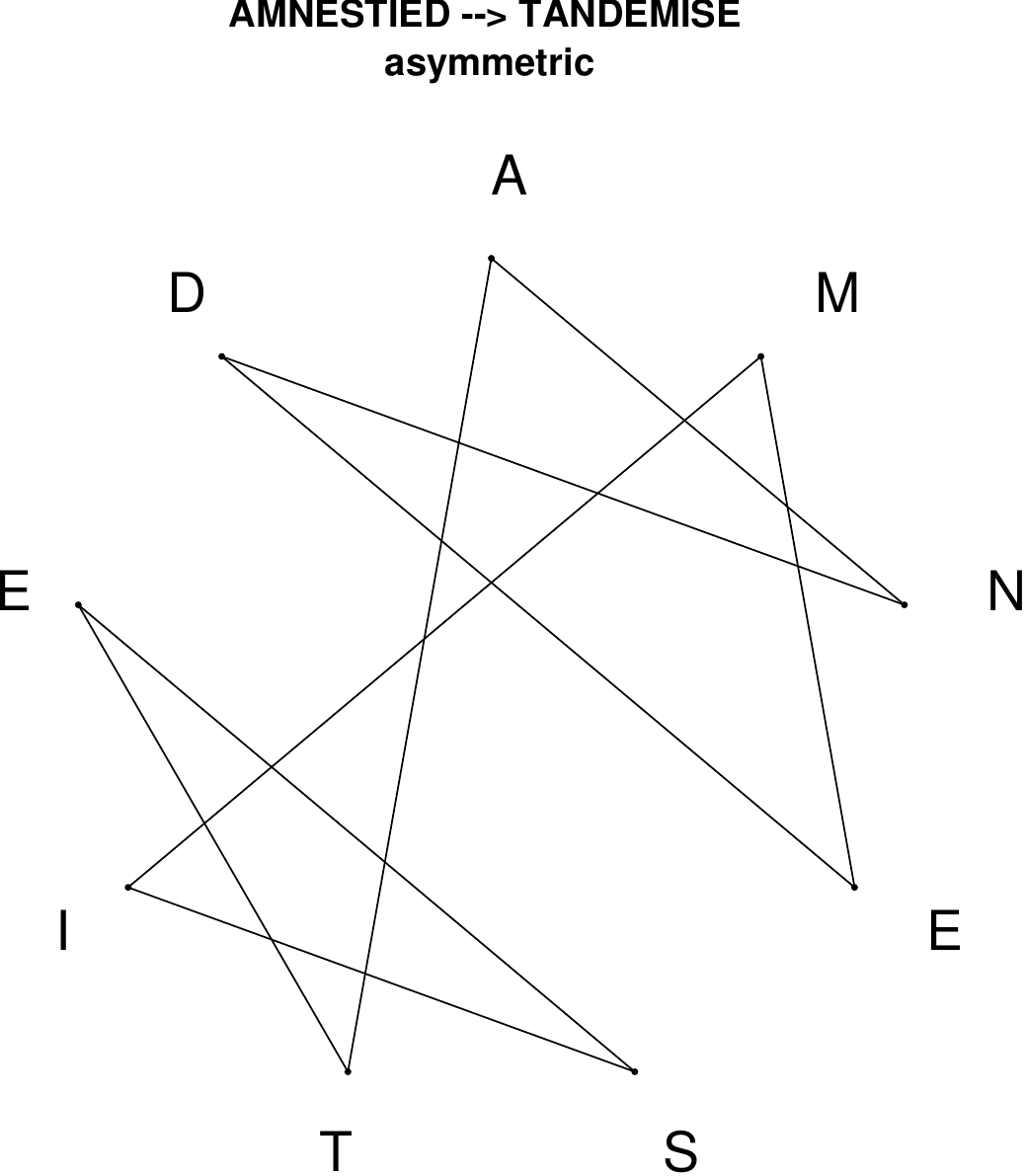}
\end{subfigure}
\end{figure}

\begin{figure}[H]
\centering
\begin{subfigure}[T]{0.19\textwidth}
\centering
\includegraphics[width=\textwidth]{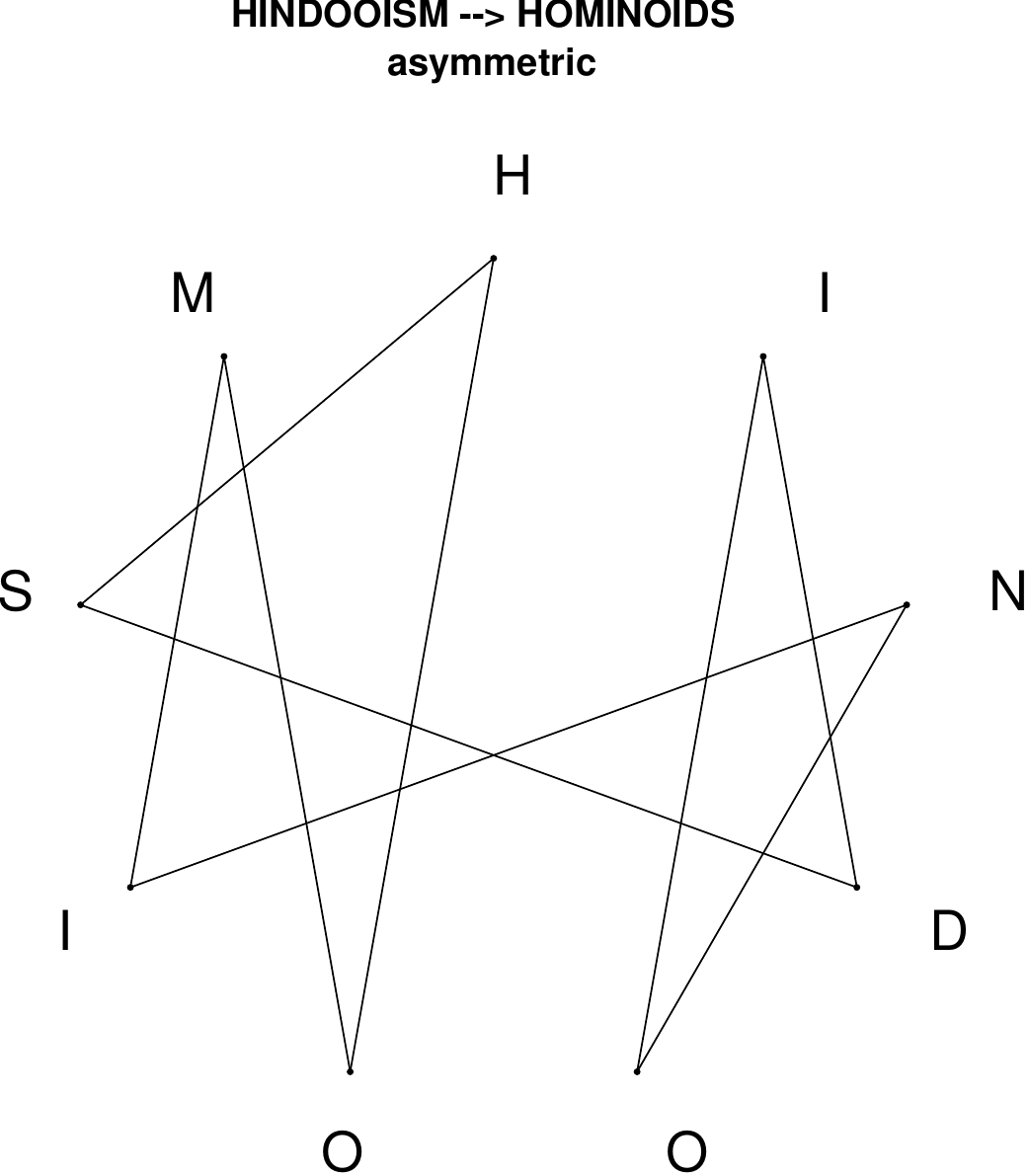}
\end{subfigure}
\hfill
\begin{subfigure}[T]{0.19\textwidth}
\centering
\includegraphics[width=\textwidth]{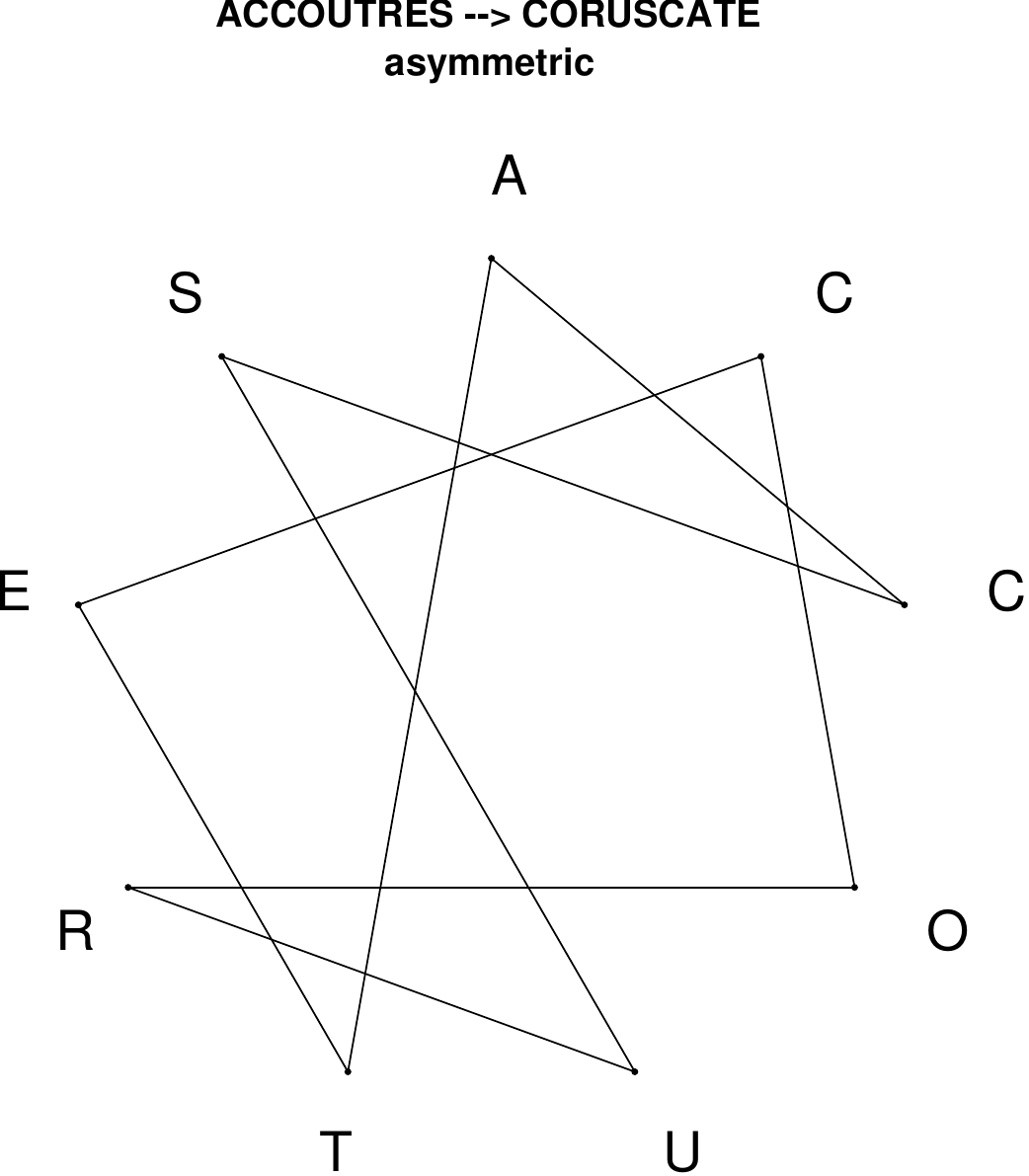}
\end{subfigure}
\hfill
\begin{subfigure}[T]{0.19\textwidth}
\centering
\includegraphics[width=\textwidth]{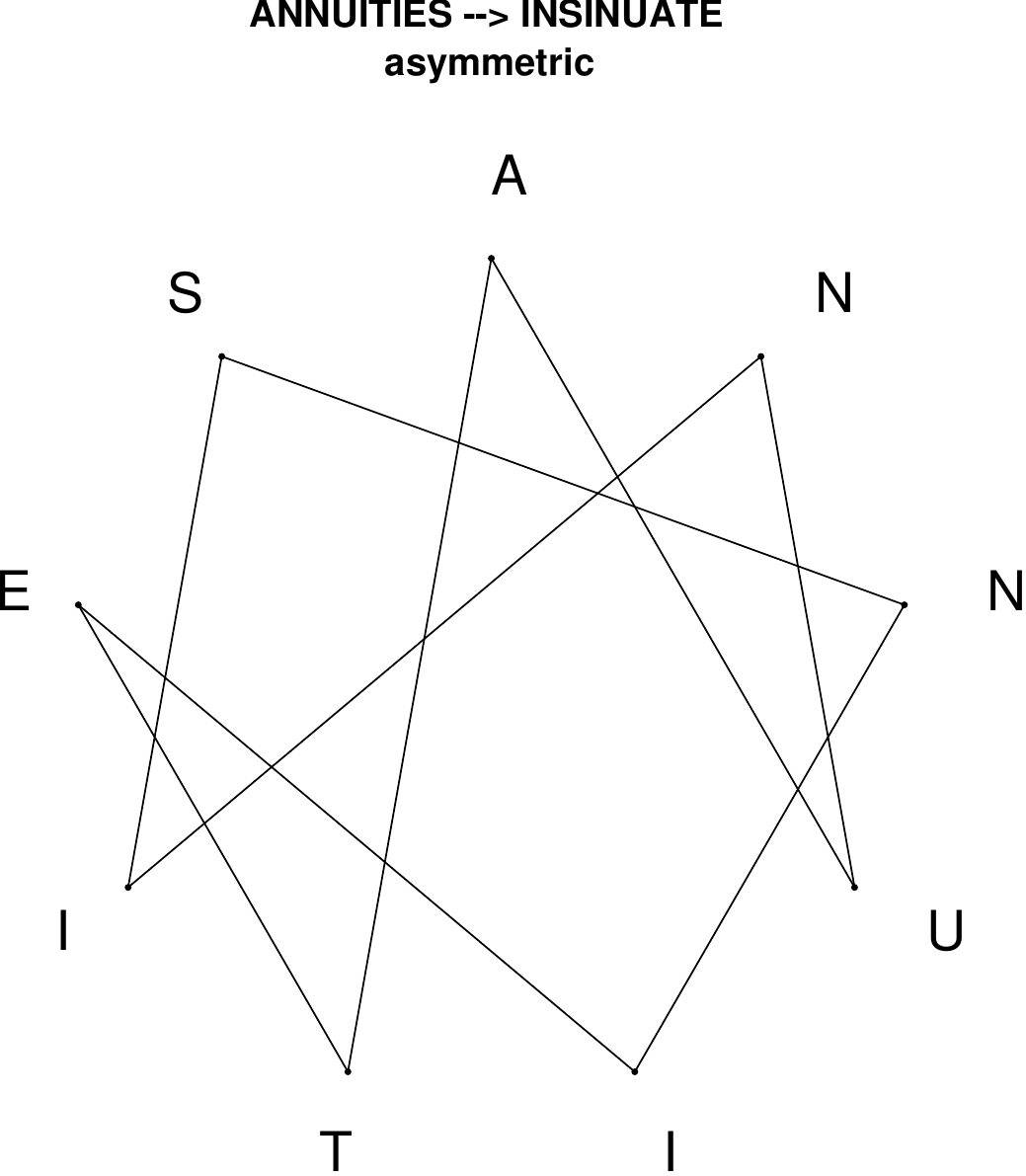}
\end{subfigure}
\hfill
\begin{subfigure}[T]{0.19\textwidth}
\centering
\includegraphics[width=\textwidth]{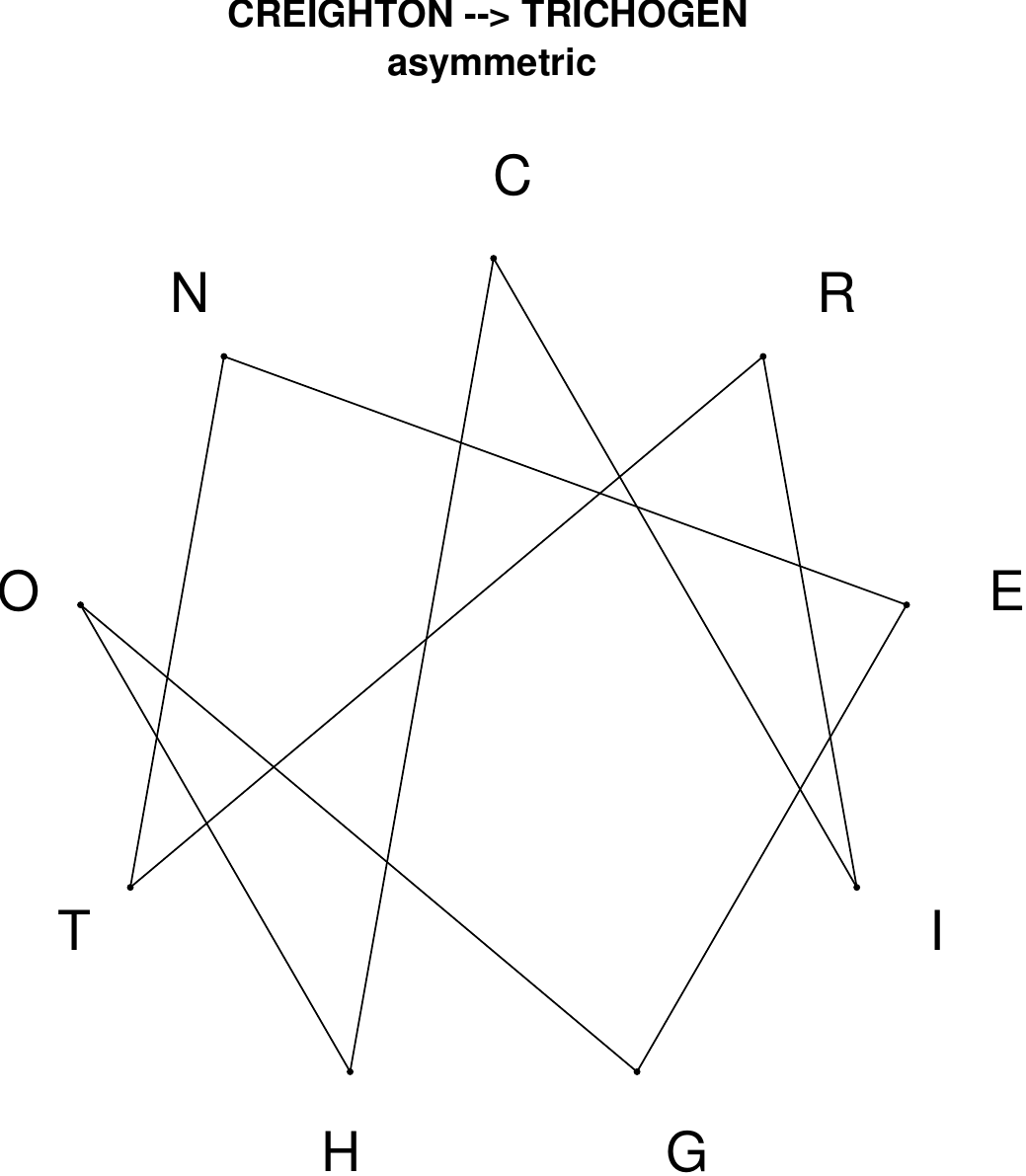}
\end{subfigure}
\hfill
\begin{subfigure}[T]{0.19\textwidth}
\centering
\includegraphics[width=\textwidth]{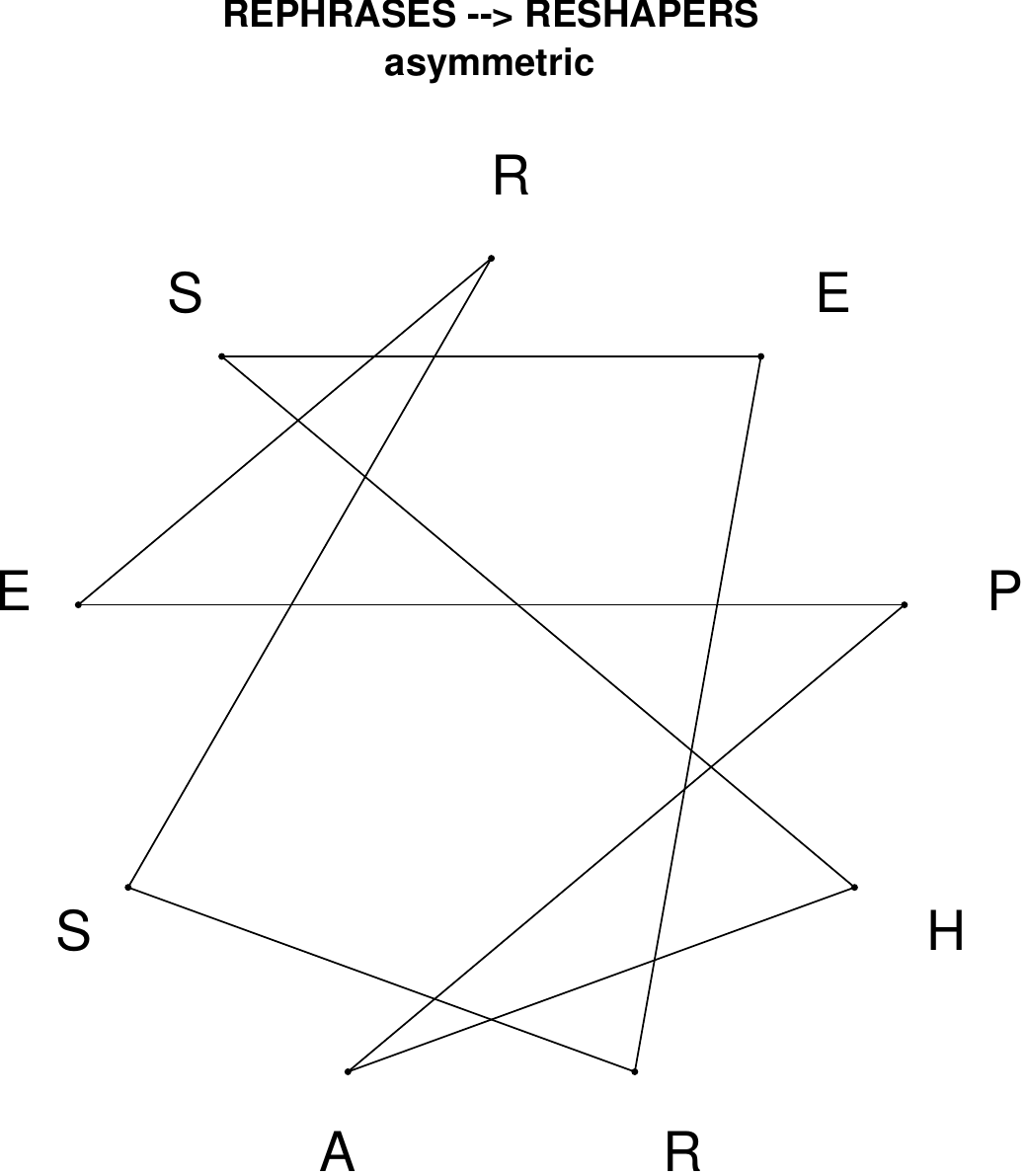}
\end{subfigure}
\end{figure}

\begin{figure}[H]
\centering
\begin{subfigure}[T]{0.19\textwidth}
\centering
\includegraphics[width=\textwidth]{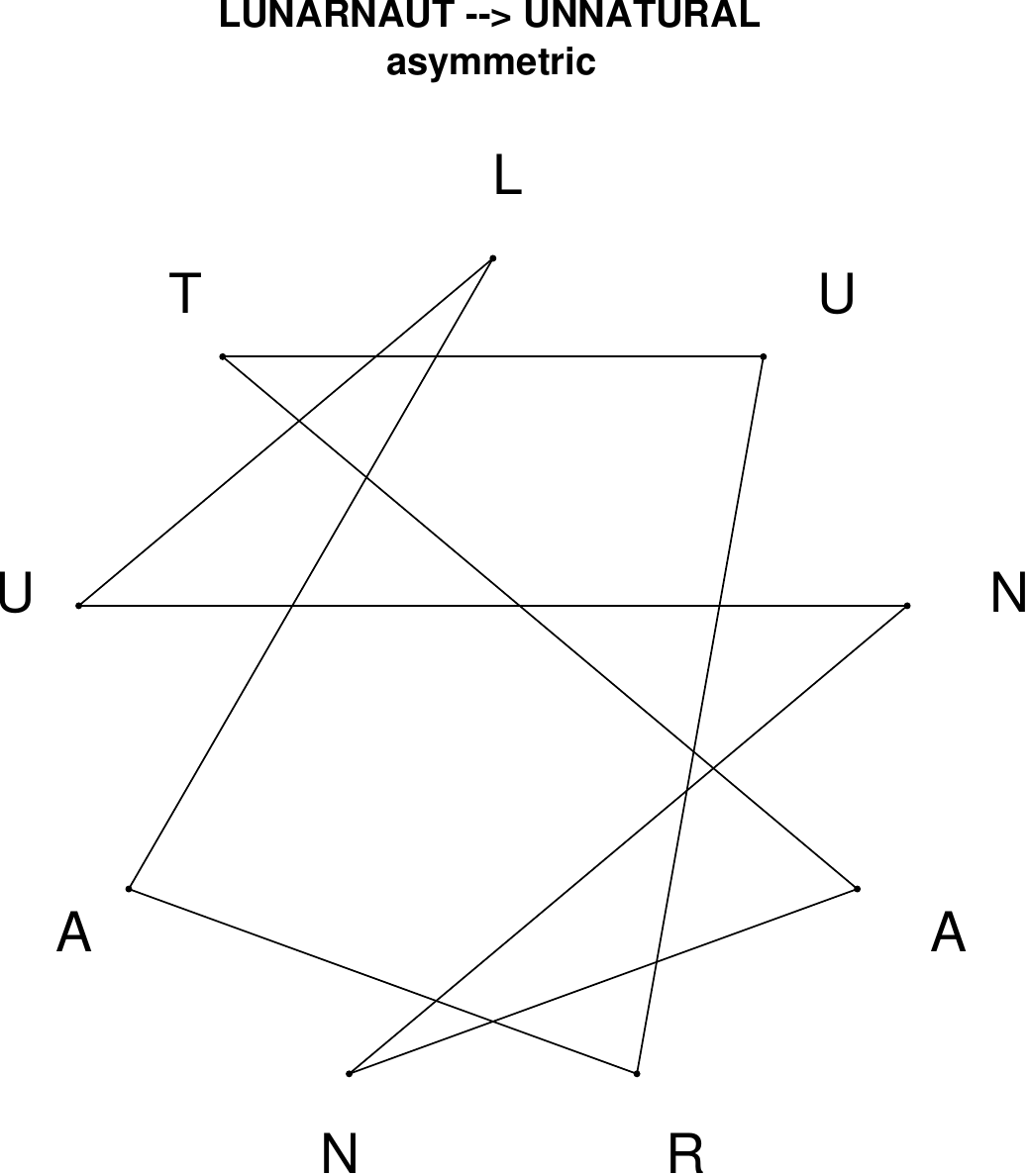}
\end{subfigure}
\hfill
\begin{subfigure}[T]{0.19\textwidth}
\centering
\includegraphics[width=\textwidth]{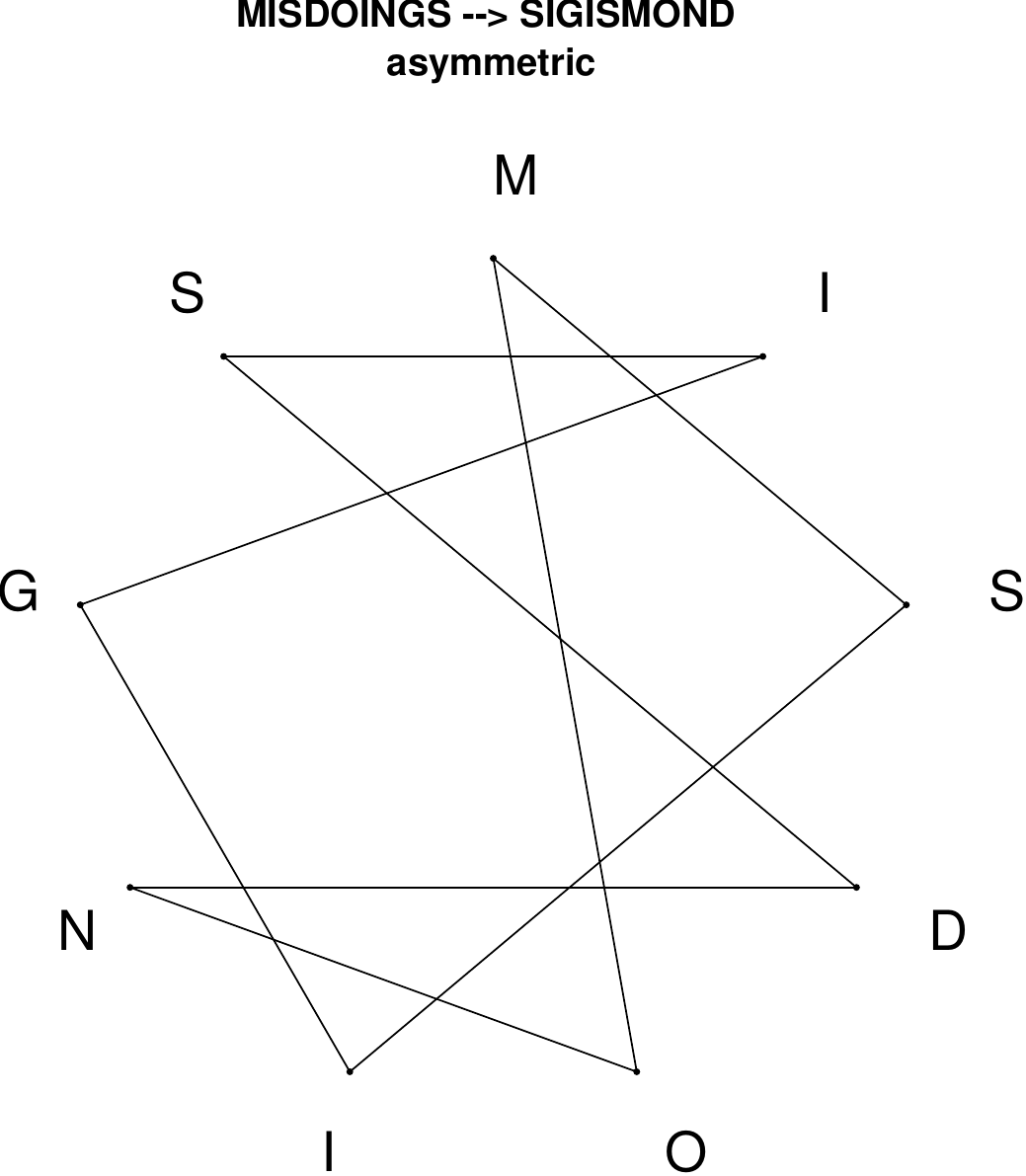}
\end{subfigure}
\hfill
\begin{subfigure}[T]{0.19\textwidth}
\centering
\includegraphics[width=\textwidth]{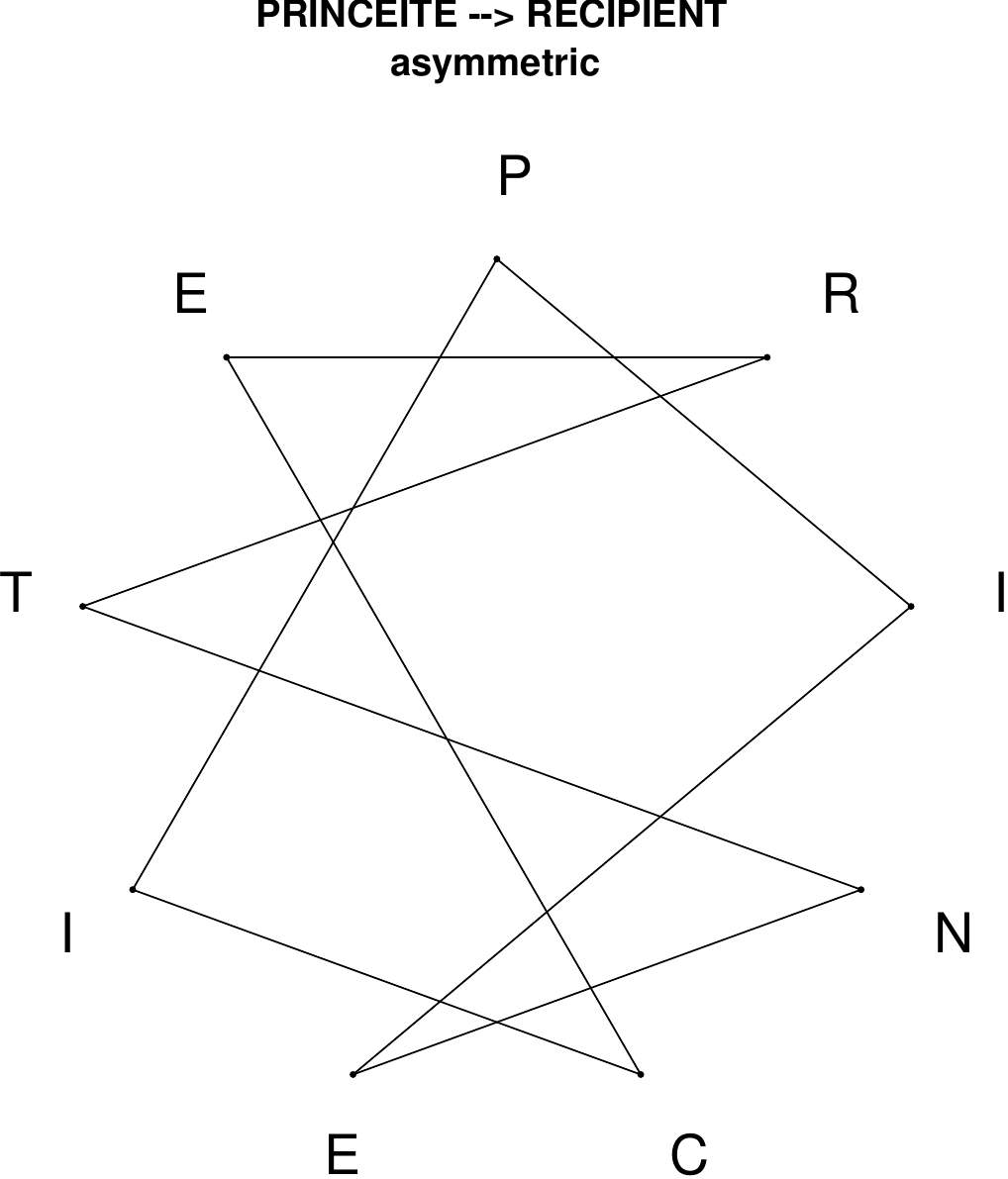}
\end{subfigure}
\hfill
\begin{subfigure}[T]{0.19\textwidth}
\centering
\includegraphics[width=\textwidth]{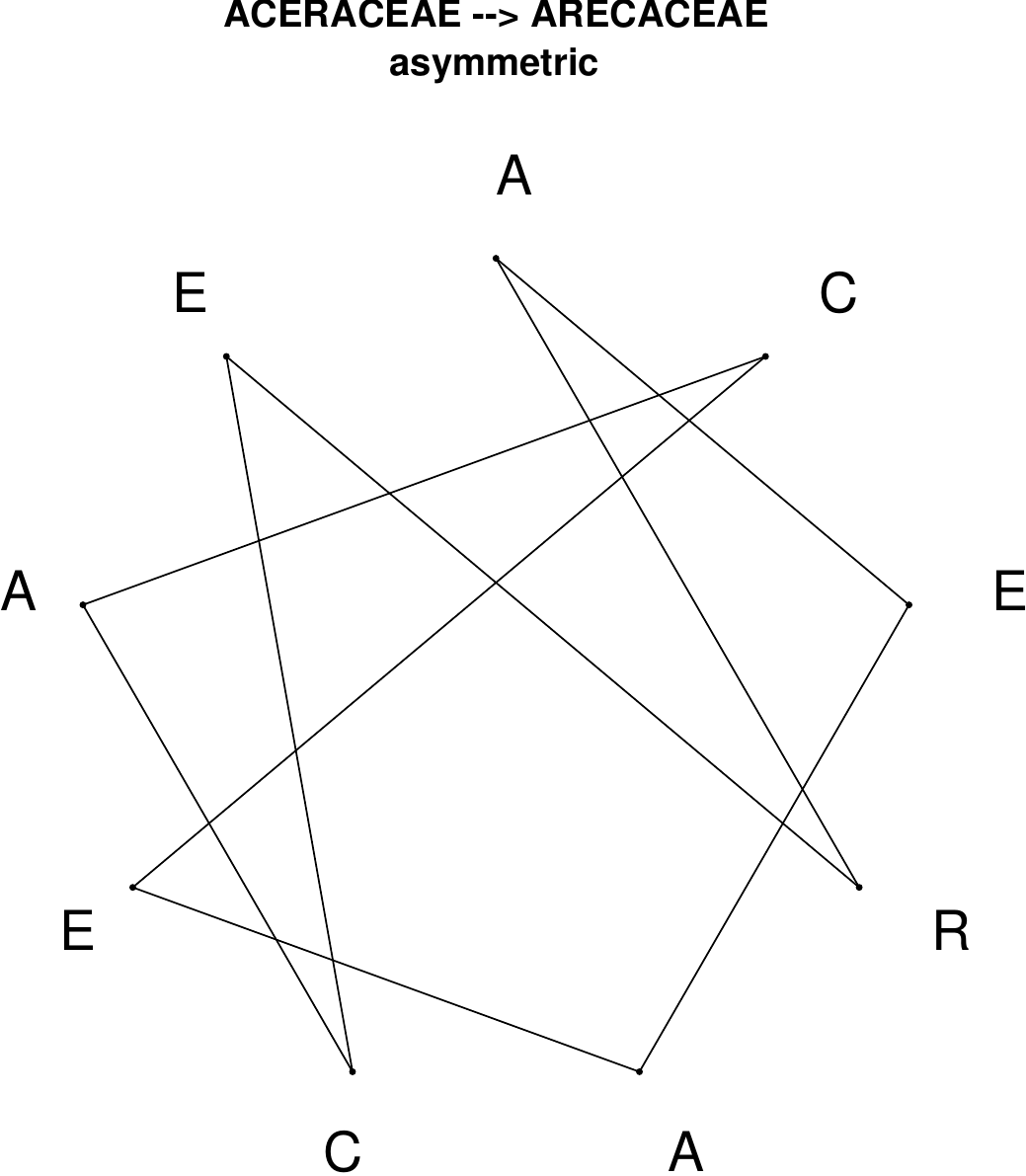}
\end{subfigure}
\hfill
\begin{subfigure}[T]{0.19\textwidth}
\centering
\includegraphics[width=\textwidth]{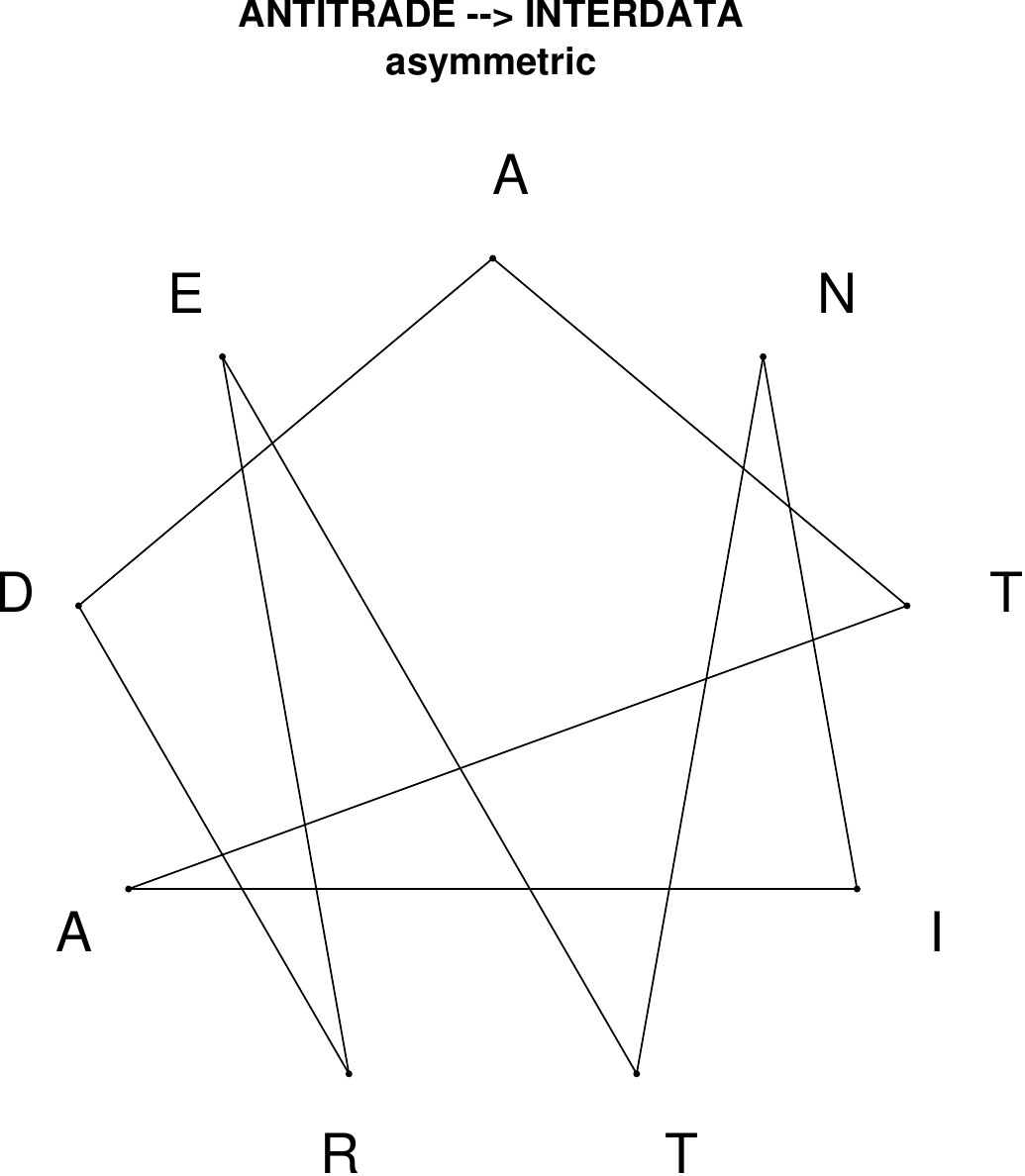}
\end{subfigure}
\end{figure}

\begin{figure}[H]
\centering
\begin{subfigure}[T]{0.19\textwidth}
\centering
\includegraphics[width=\textwidth]{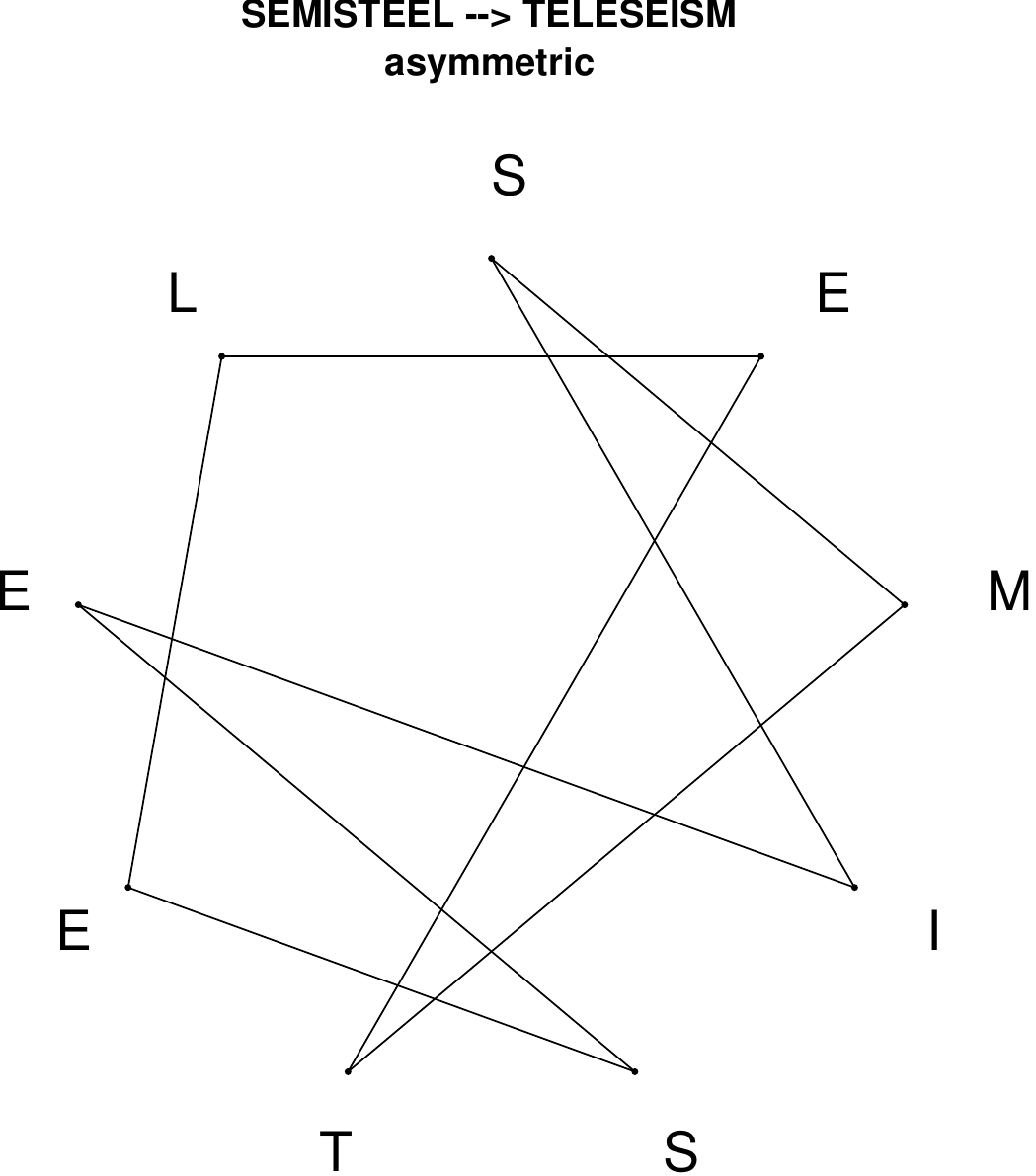}
\end{subfigure}
\hfill
\begin{subfigure}[T]{0.19\textwidth}
\centering
\includegraphics[width=\textwidth]{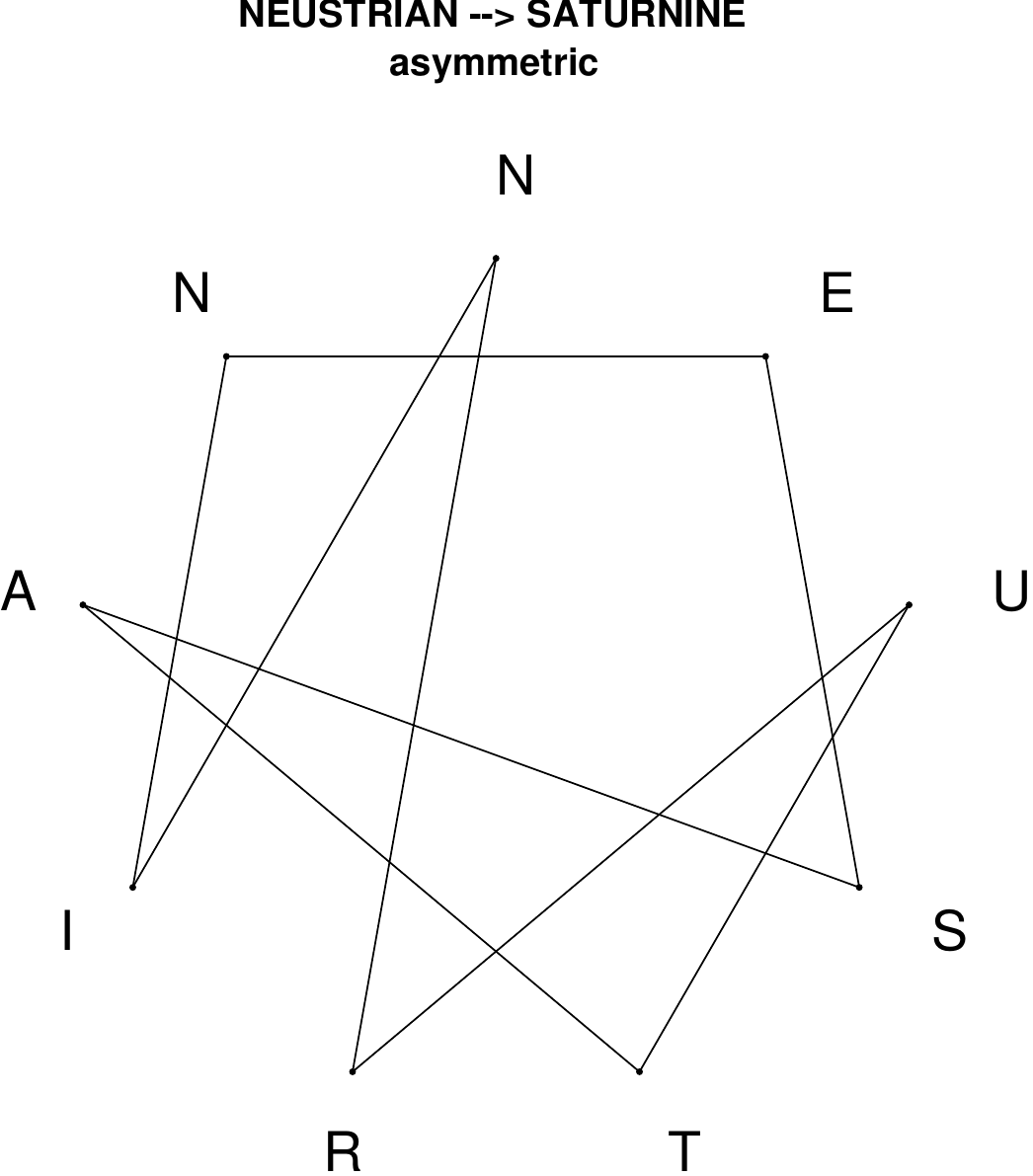}
\end{subfigure}
\hfill
\begin{subfigure}[T]{0.19\textwidth}
\centering
\includegraphics[width=\textwidth]{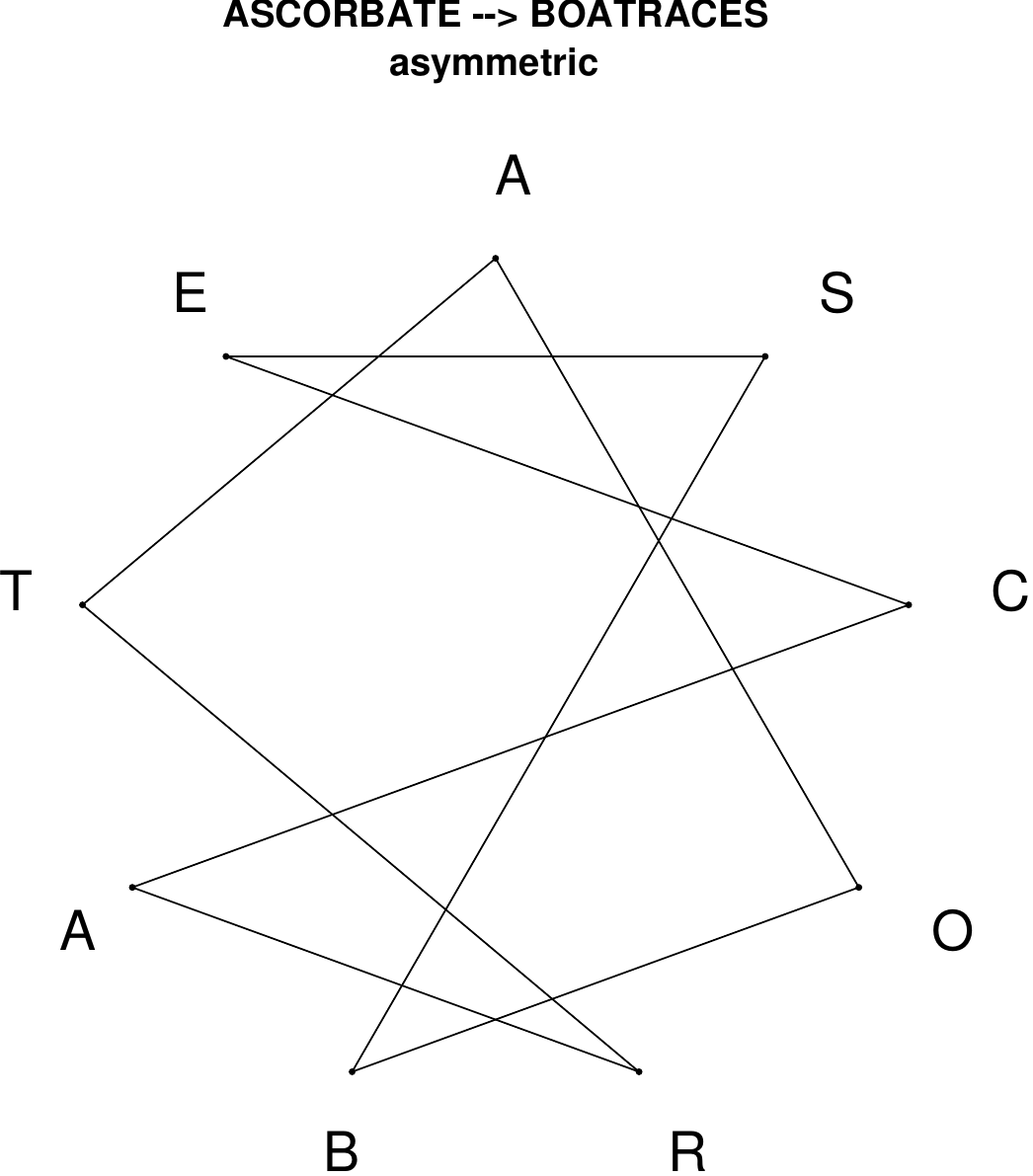}
\end{subfigure}
\hfill
\begin{subfigure}[T]{0.19\textwidth}
\centering
\includegraphics[width=\textwidth]{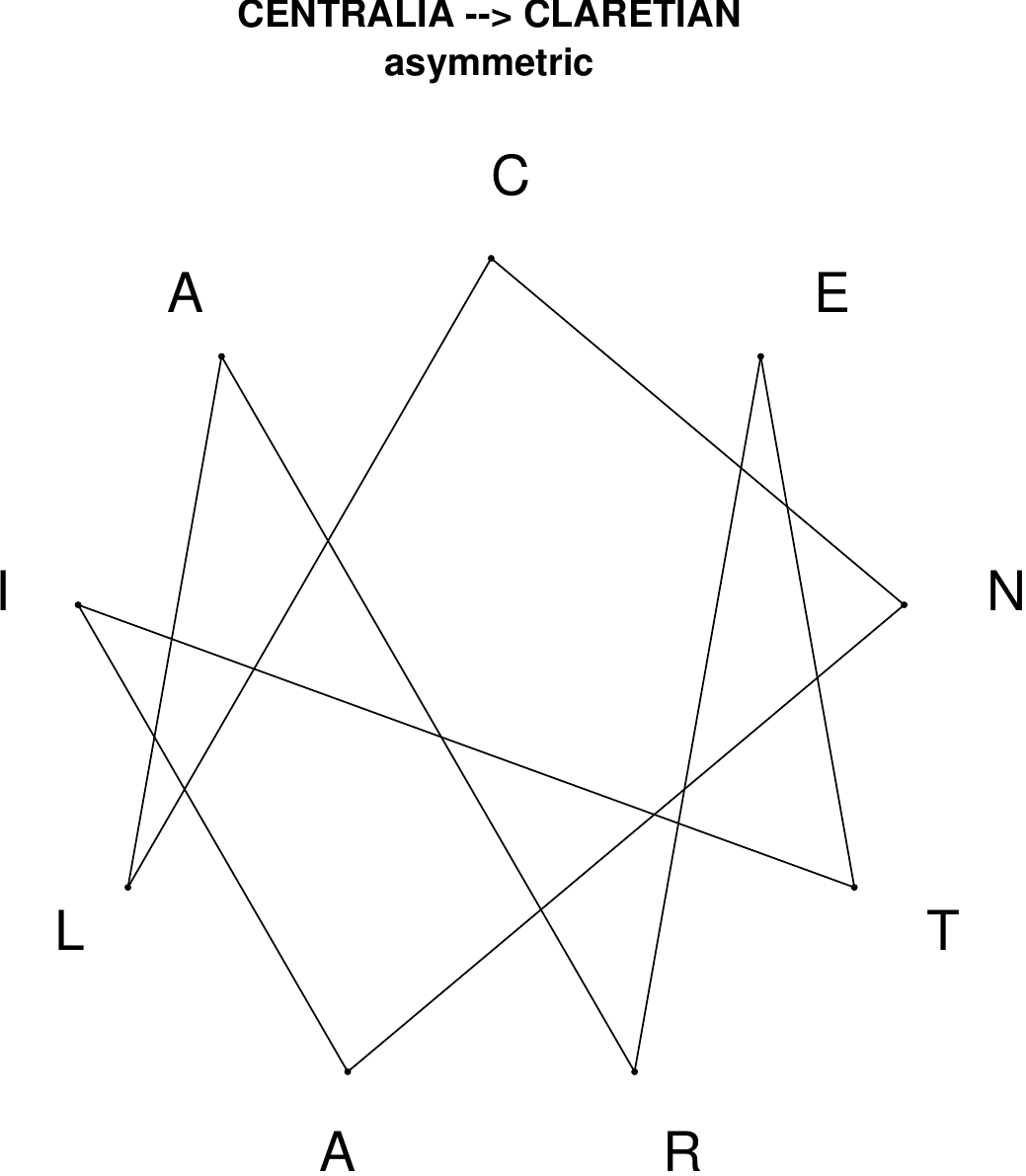}
\end{subfigure}
\hfill
\begin{subfigure}[T]{0.19\textwidth}
\centering
\includegraphics[width=\textwidth]{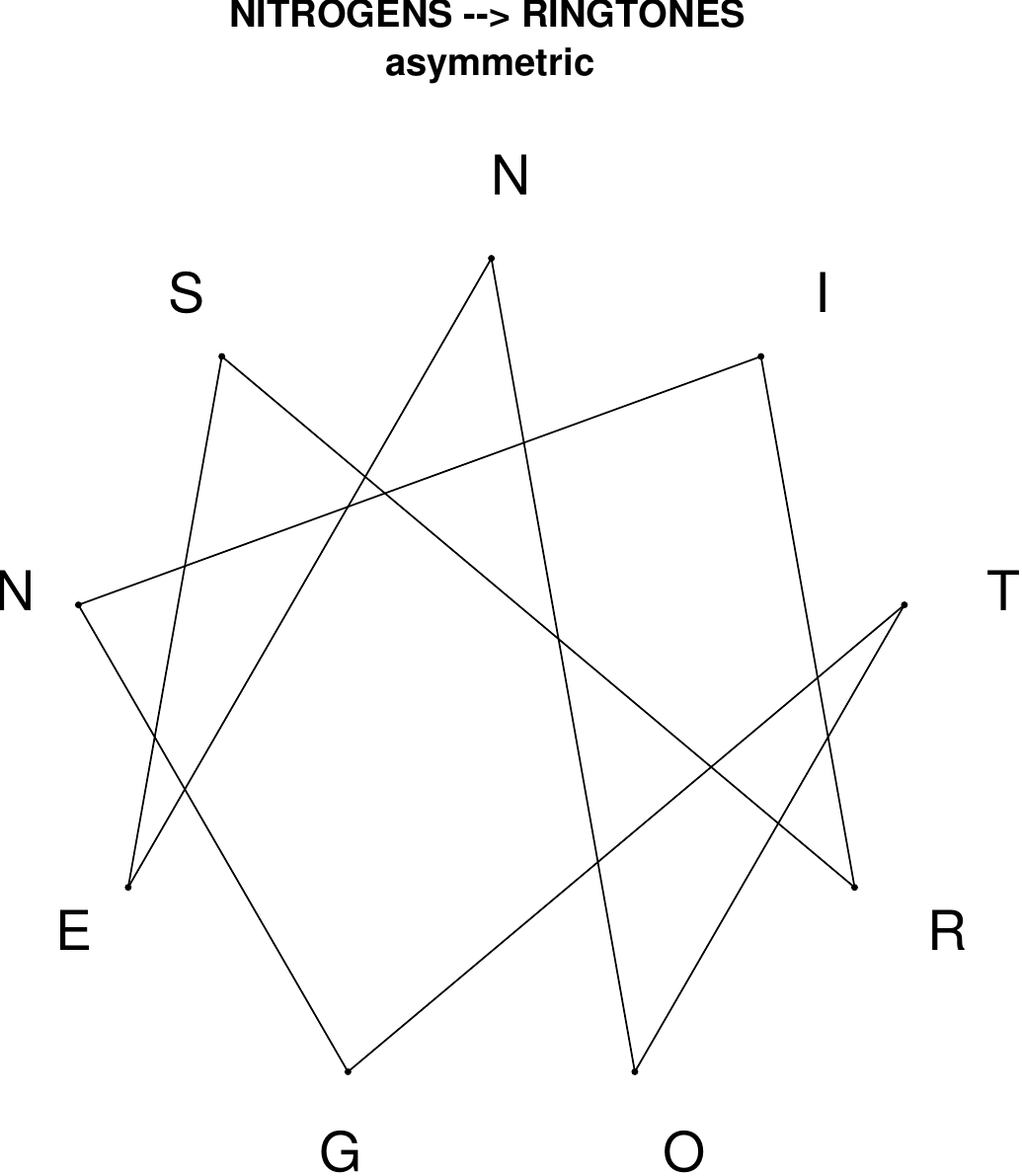}
\end{subfigure}
\end{figure}

\begin{figure}[H]
\centering
\begin{subfigure}[T]{0.19\textwidth}
\centering
\includegraphics[width=\textwidth]{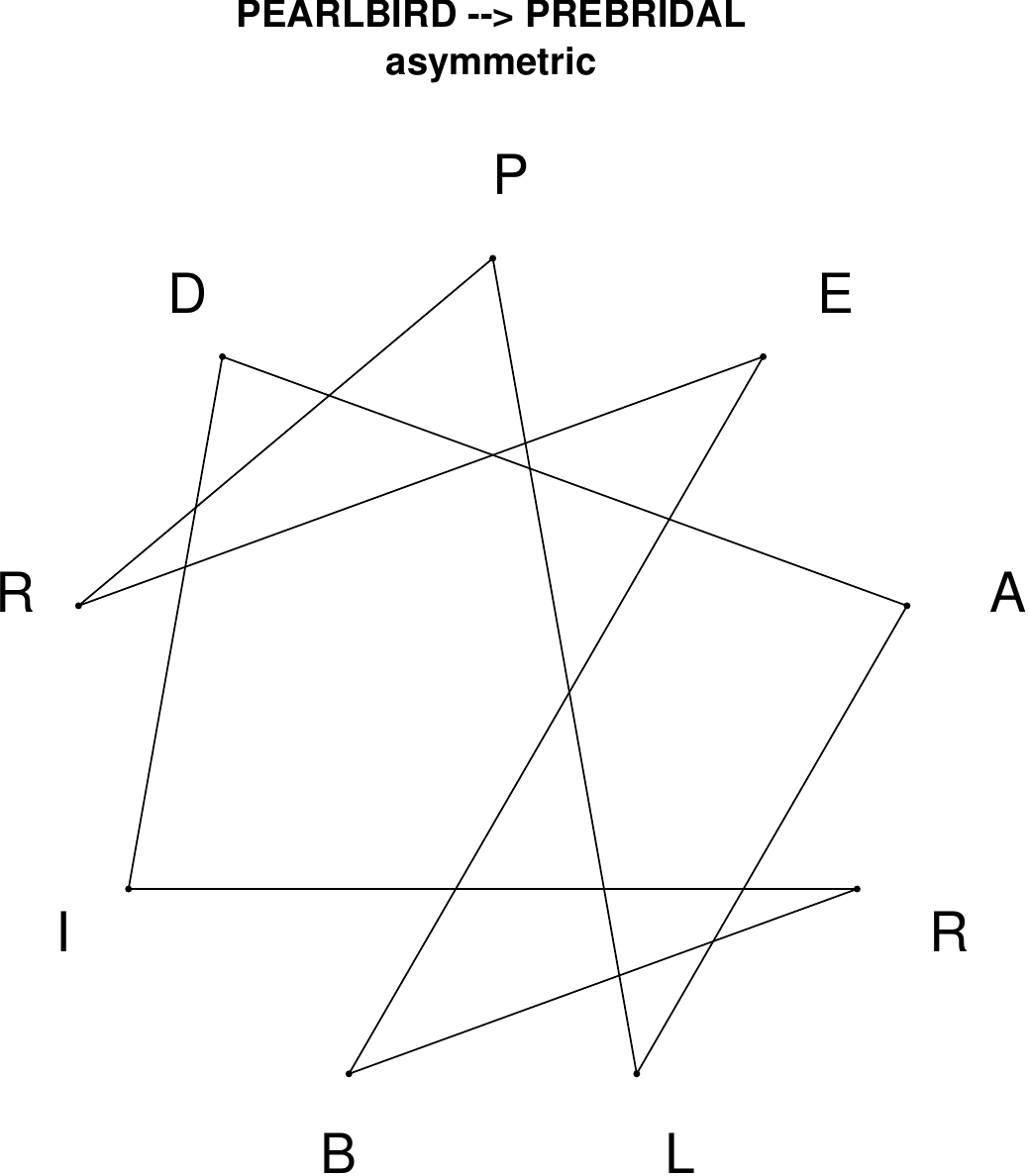}
\end{subfigure}
\hfill
\begin{subfigure}[T]{0.19\textwidth}
\centering
\includegraphics[width=\textwidth]{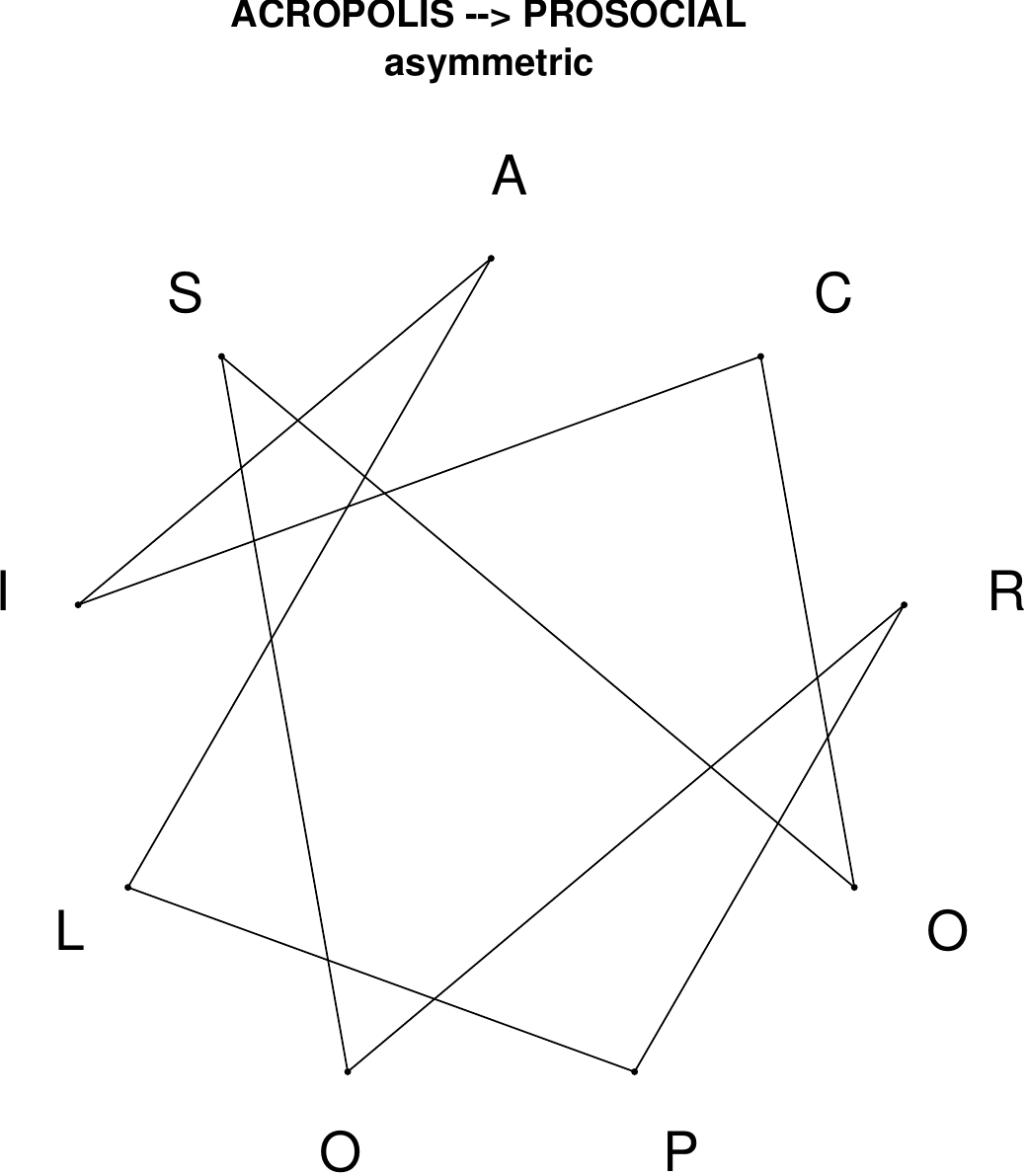}
\end{subfigure}
\hfill
\begin{subfigure}[T]{0.19\textwidth}
\centering
\includegraphics[width=\textwidth]{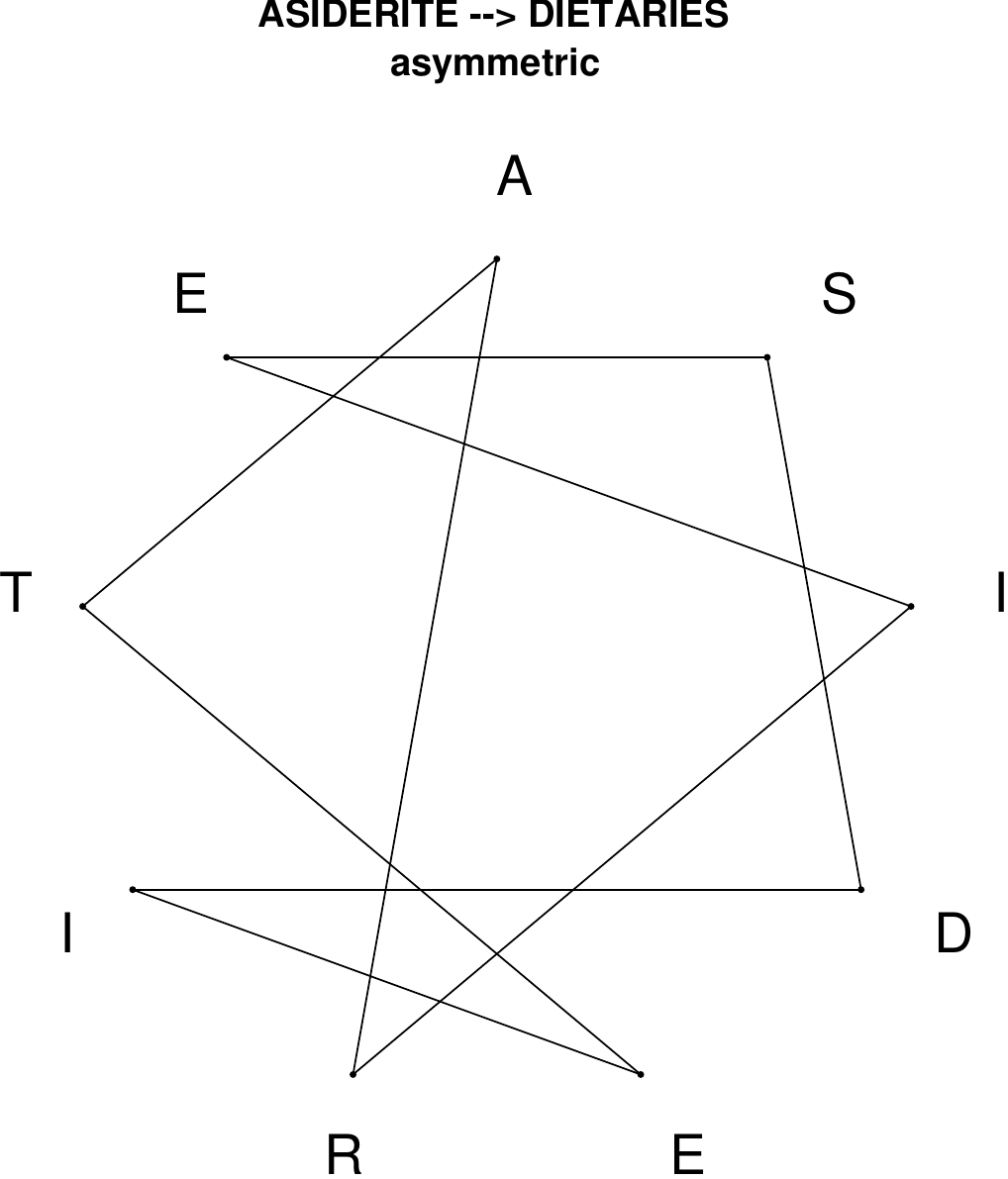}
\end{subfigure}
\hfill
\begin{subfigure}[T]{0.19\textwidth}
\centering
\includegraphics[width=\textwidth]{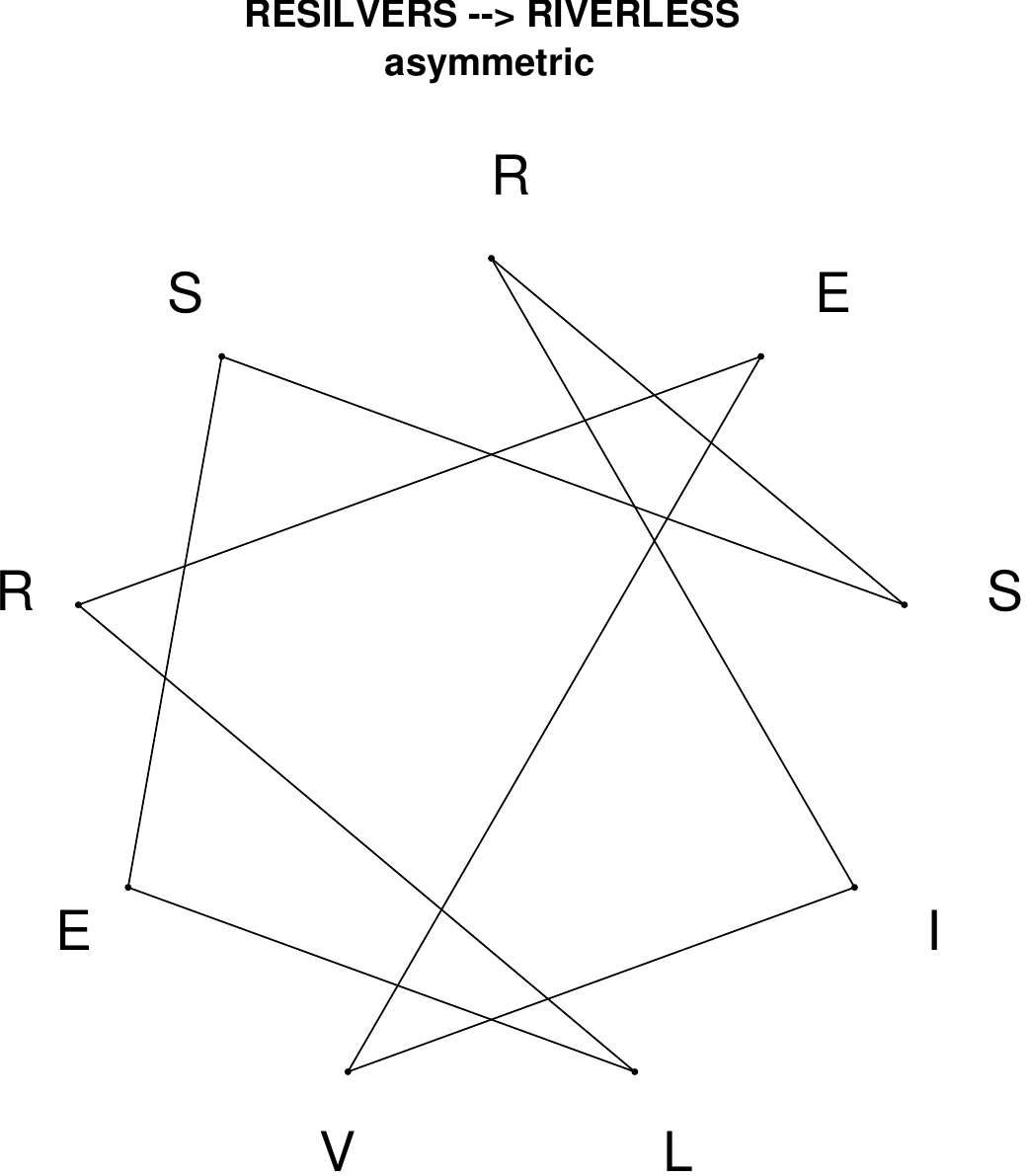}
\end{subfigure}
\hfill
\begin{subfigure}[T]{0.19\textwidth}
\centering
\includegraphics[width=\textwidth]{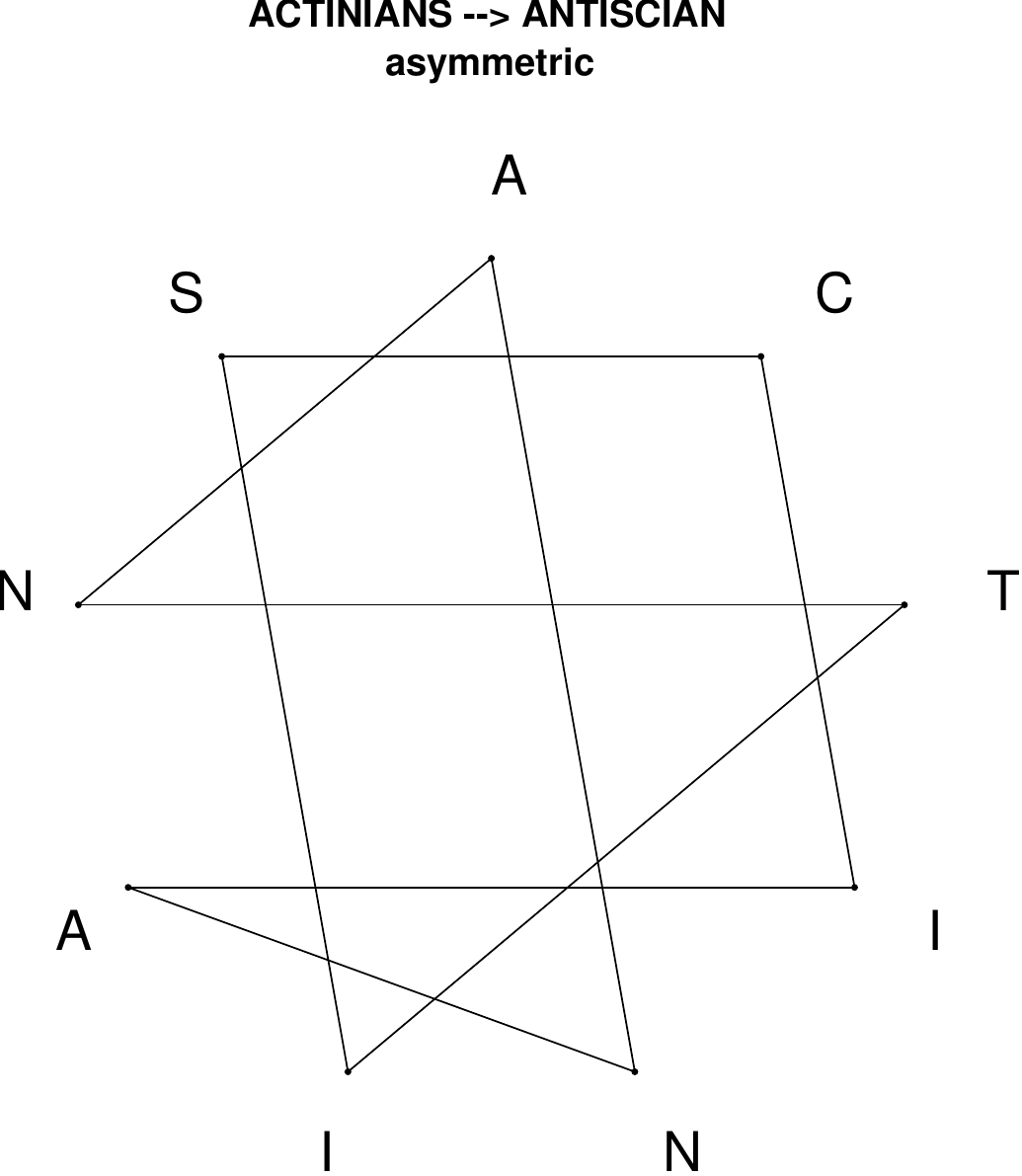}
\end{subfigure}
\end{figure}

\begin{figure}[H]
\centering
\begin{subfigure}[T]{0.19\textwidth}
\centering
\includegraphics[width=\textwidth]{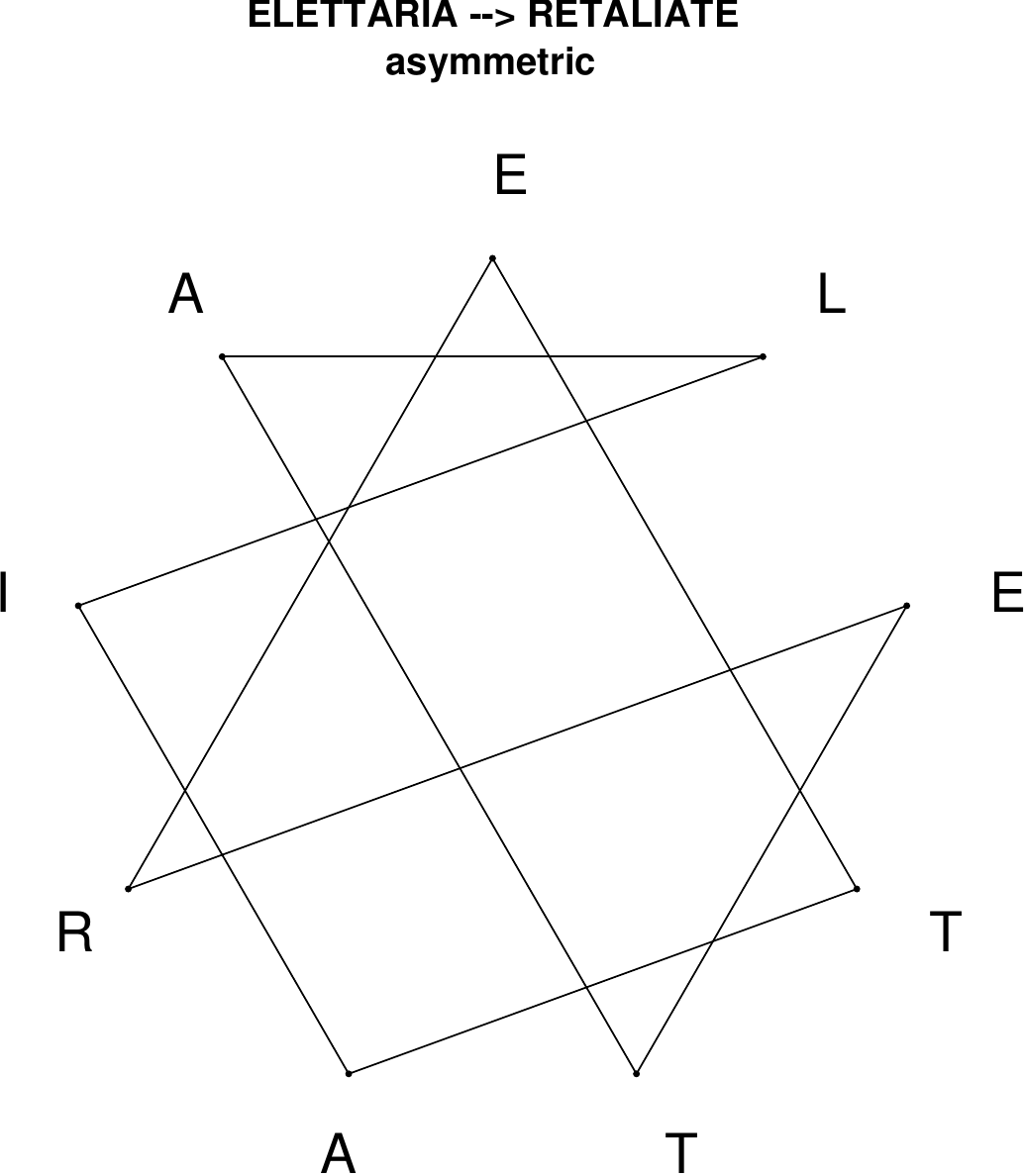}
\end{subfigure}
\hfill
\begin{subfigure}[T]{0.19\textwidth}
\centering
\includegraphics[width=\textwidth]{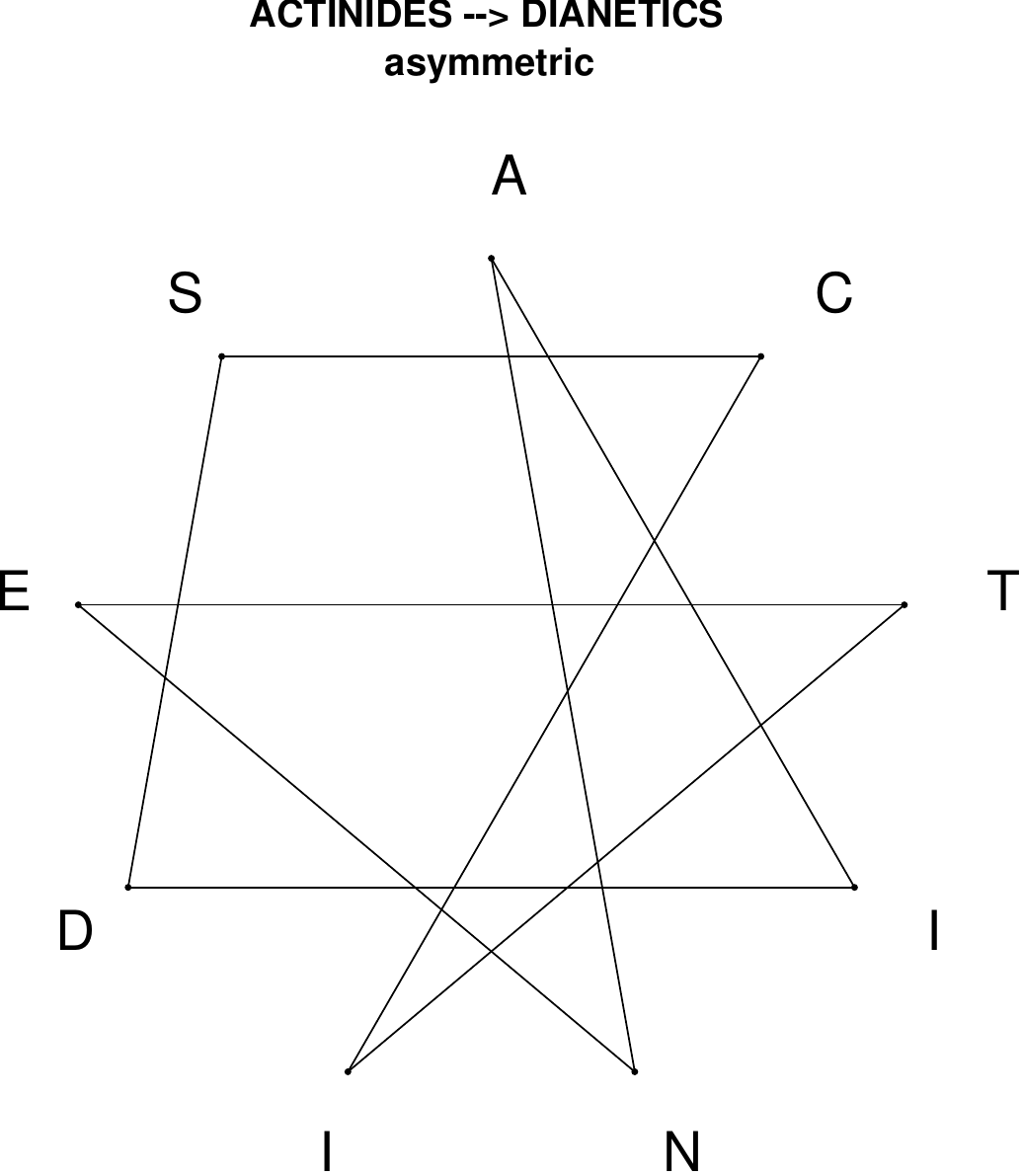}
\end{subfigure}
\hfill
\begin{subfigure}[T]{0.19\textwidth}
\centering
\includegraphics[width=\textwidth]{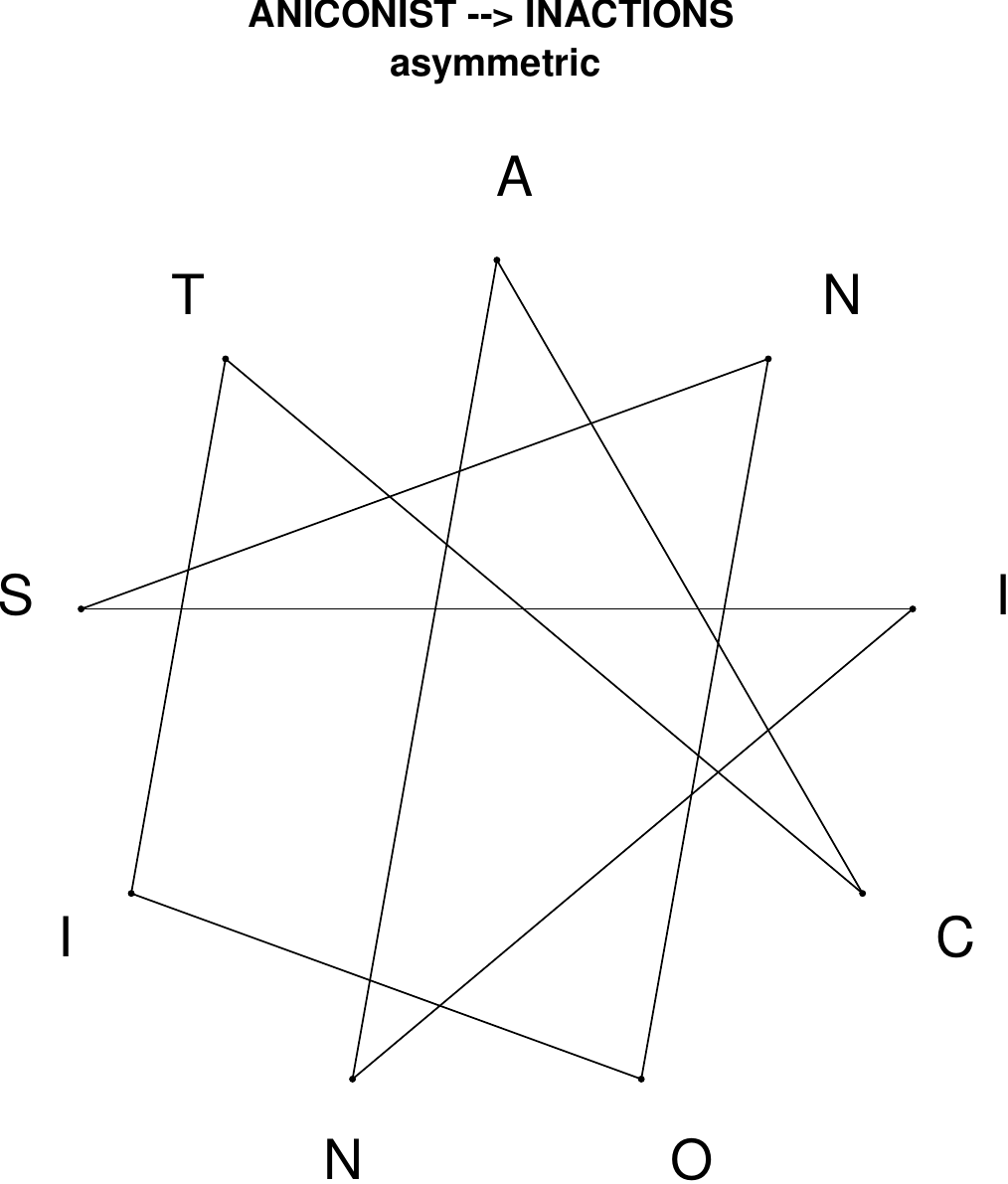}
\end{subfigure}
\hfill
\begin{subfigure}[T]{0.19\textwidth}
\centering
\includegraphics[width=\textwidth]{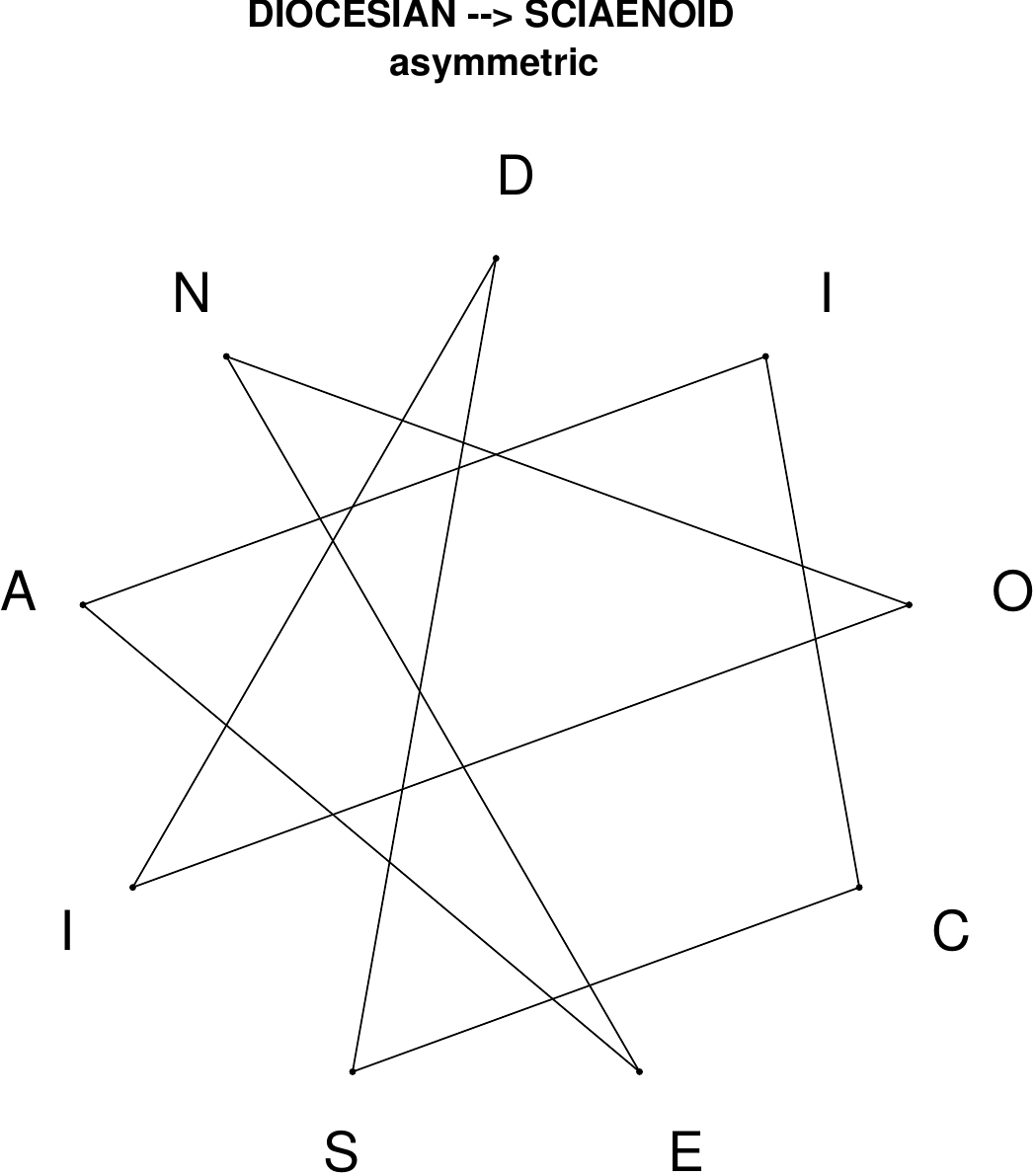}
\end{subfigure}
\hfill
\begin{subfigure}[T]{0.19\textwidth}
\centering
\includegraphics[width=\textwidth]{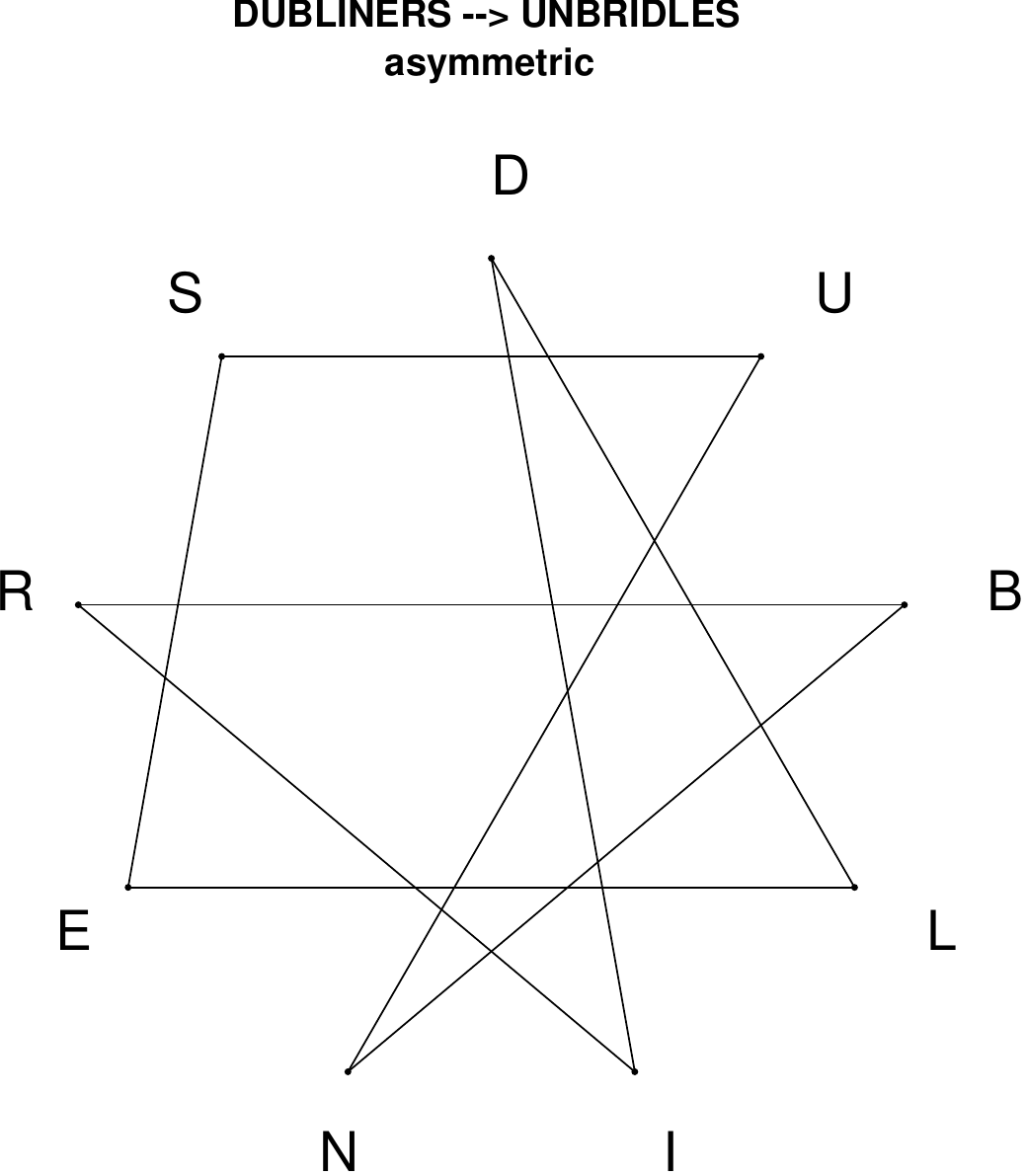}
\end{subfigure}
\end{figure}

\begin{figure}[H]
\centering
\begin{subfigure}[T]{0.19\textwidth}
\centering
\includegraphics[width=\textwidth]{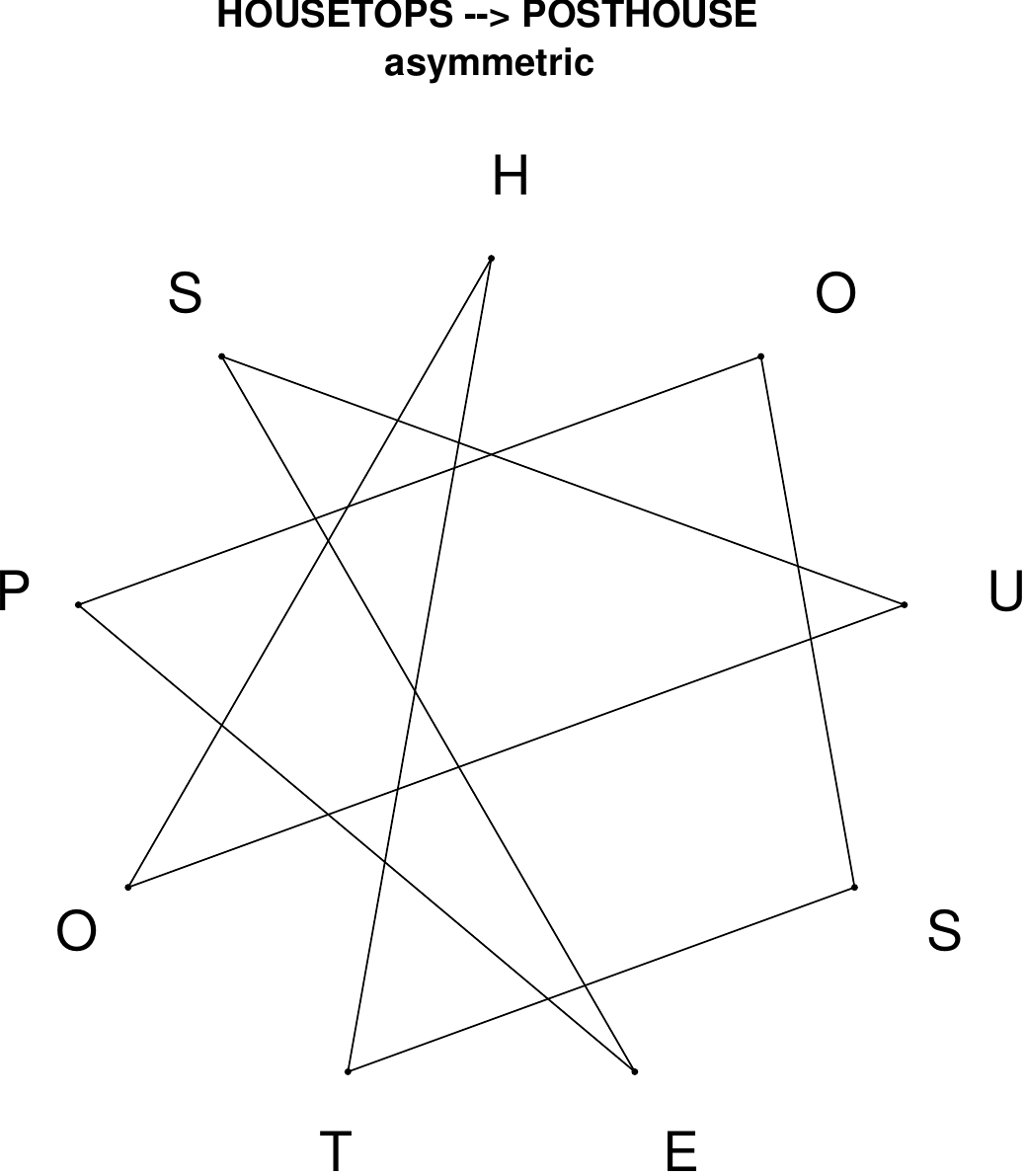}
\end{subfigure}
\hfill
\begin{subfigure}[T]{0.19\textwidth}
\centering
\includegraphics[width=\textwidth]{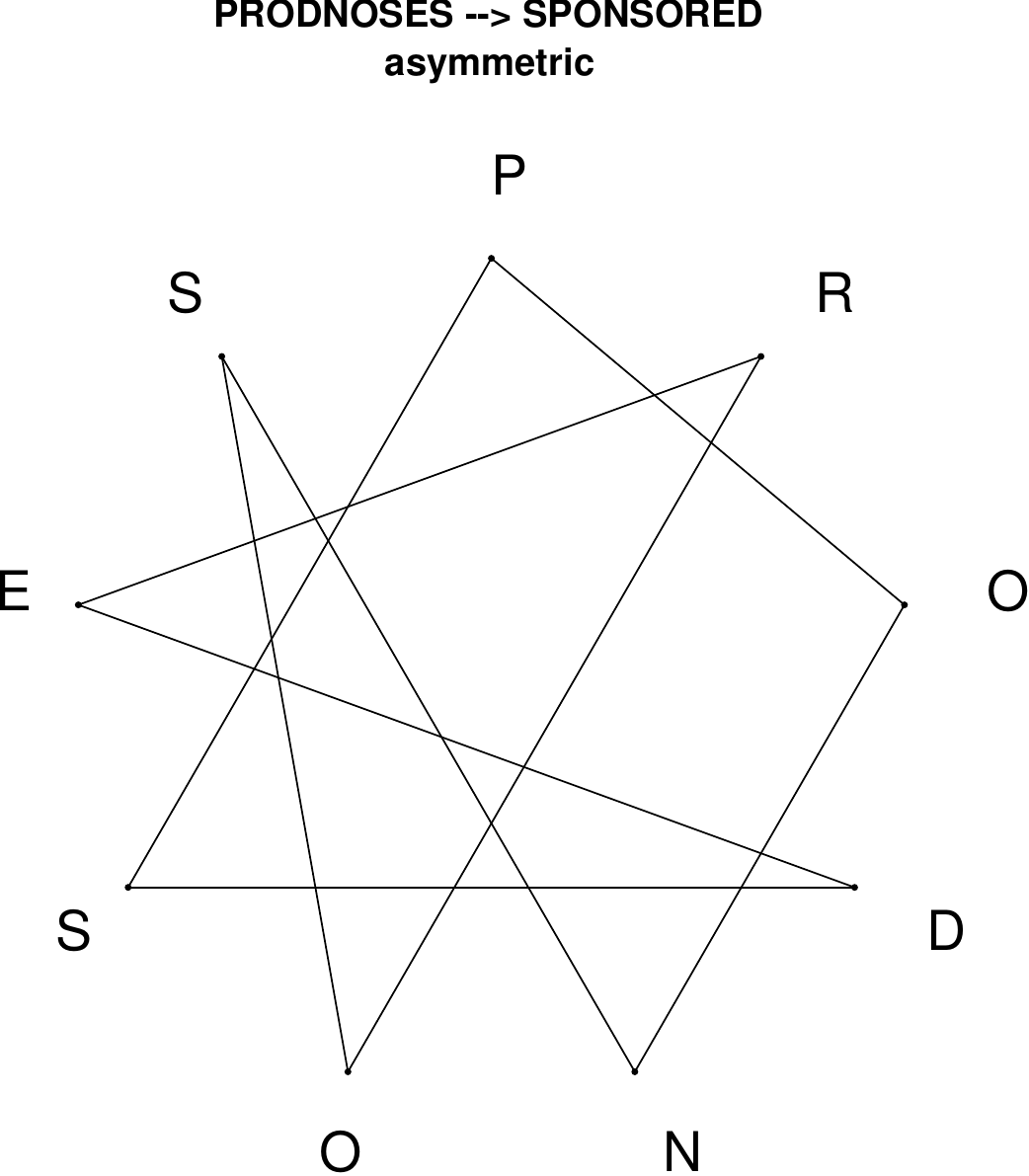}
\end{subfigure}
\hfill
\begin{subfigure}[T]{0.19\textwidth}
\centering
\includegraphics[width=\textwidth]{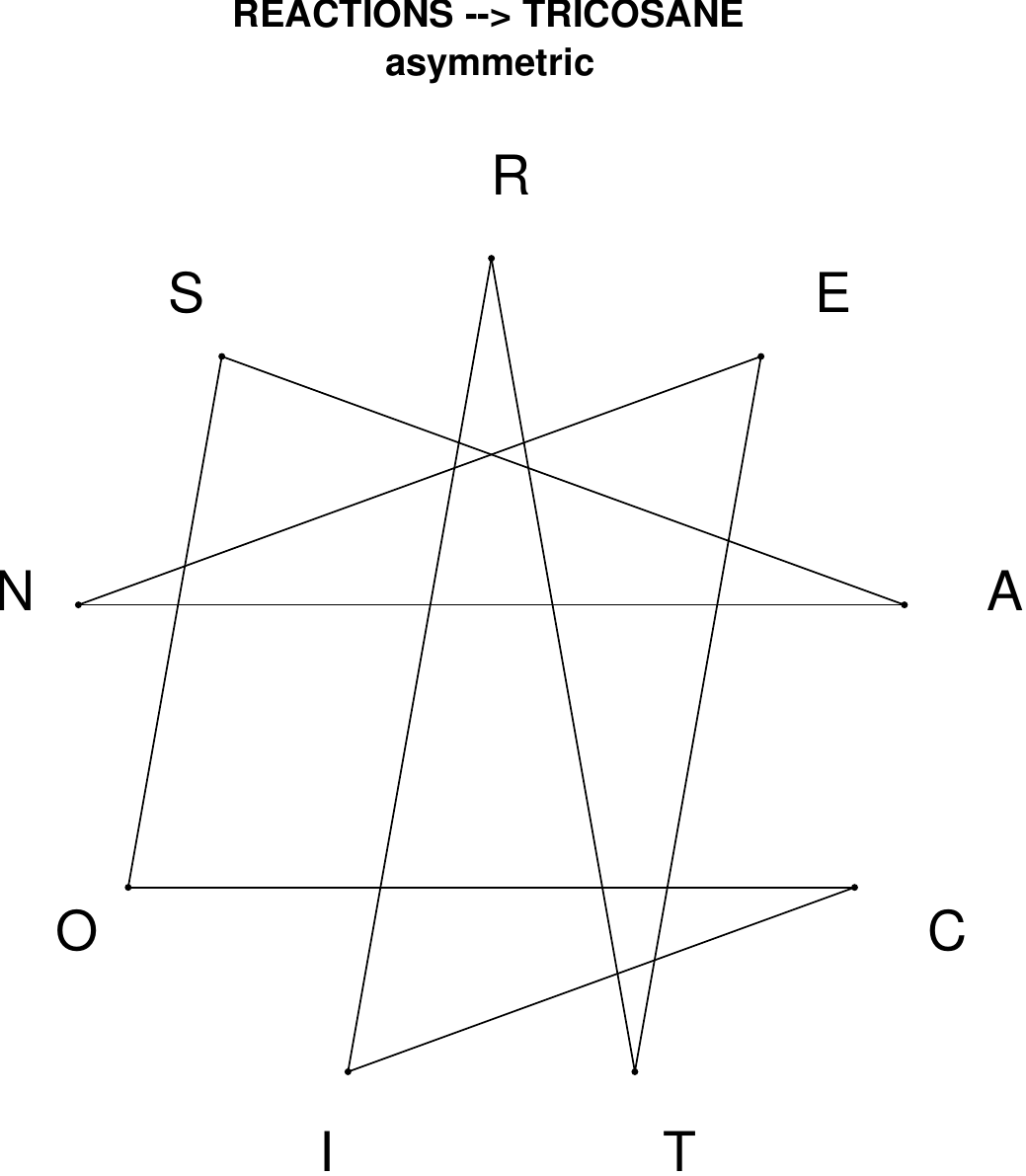}
\end{subfigure}
\hfill
\begin{subfigure}[T]{0.19\textwidth}
\centering
\includegraphics[width=\textwidth]{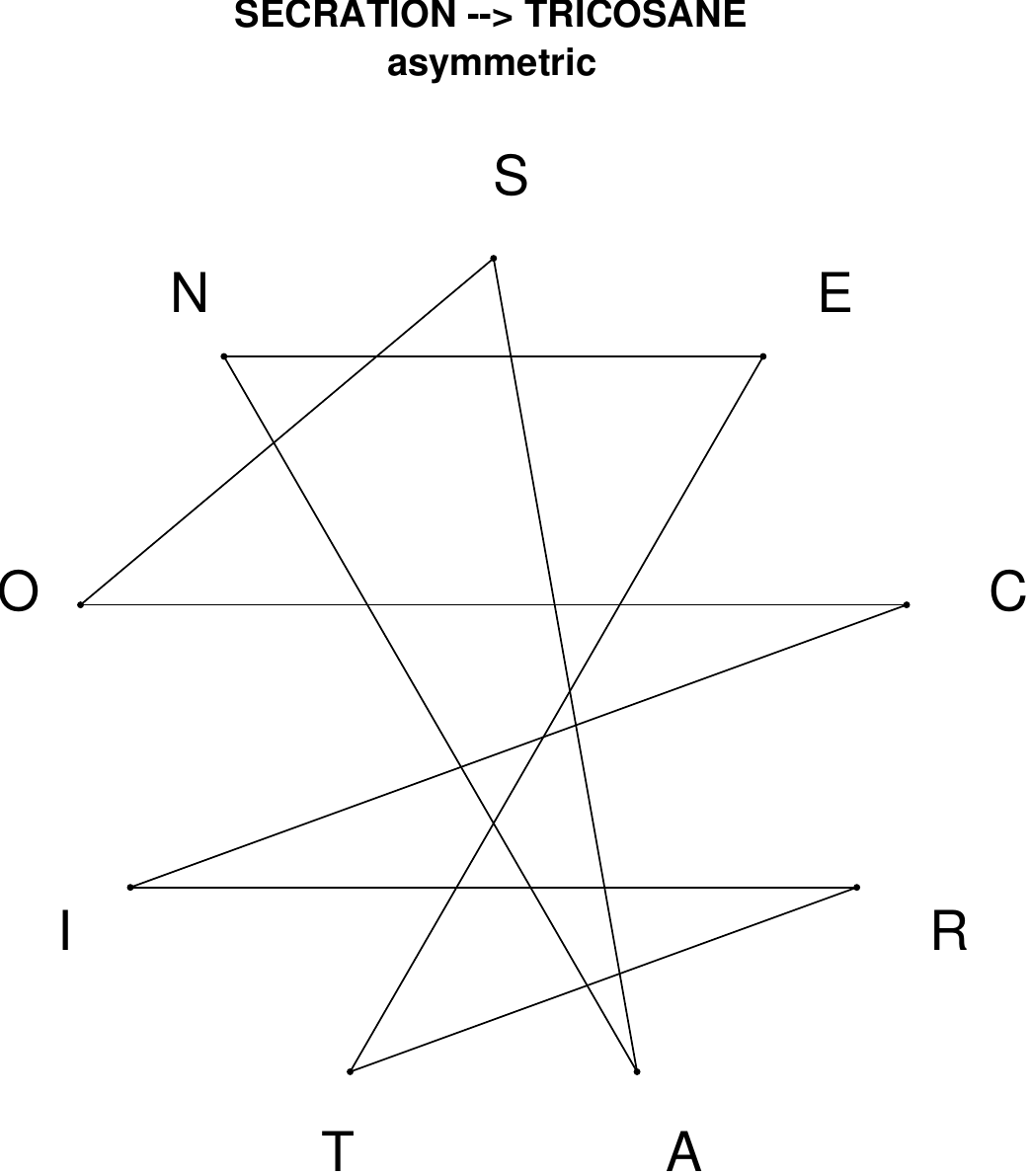}
\end{subfigure}
\hfill
\begin{subfigure}[T]{0.19\textwidth}
\centering
\includegraphics[width=\textwidth]{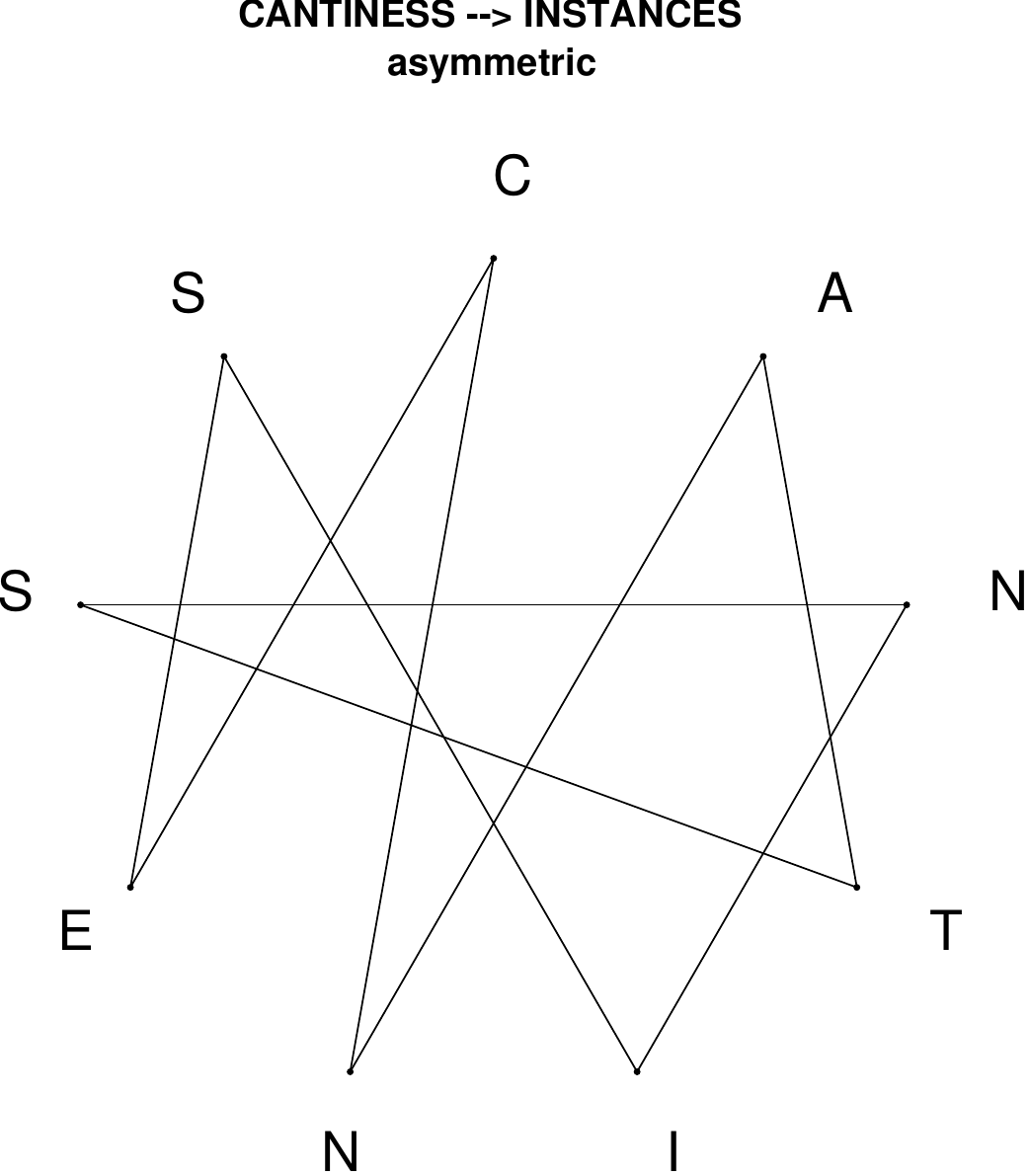}
\end{subfigure}
\end{figure}

\begin{figure}[H]
\centering
\begin{subfigure}[T]{0.19\textwidth}
\centering
\includegraphics[width=\textwidth]{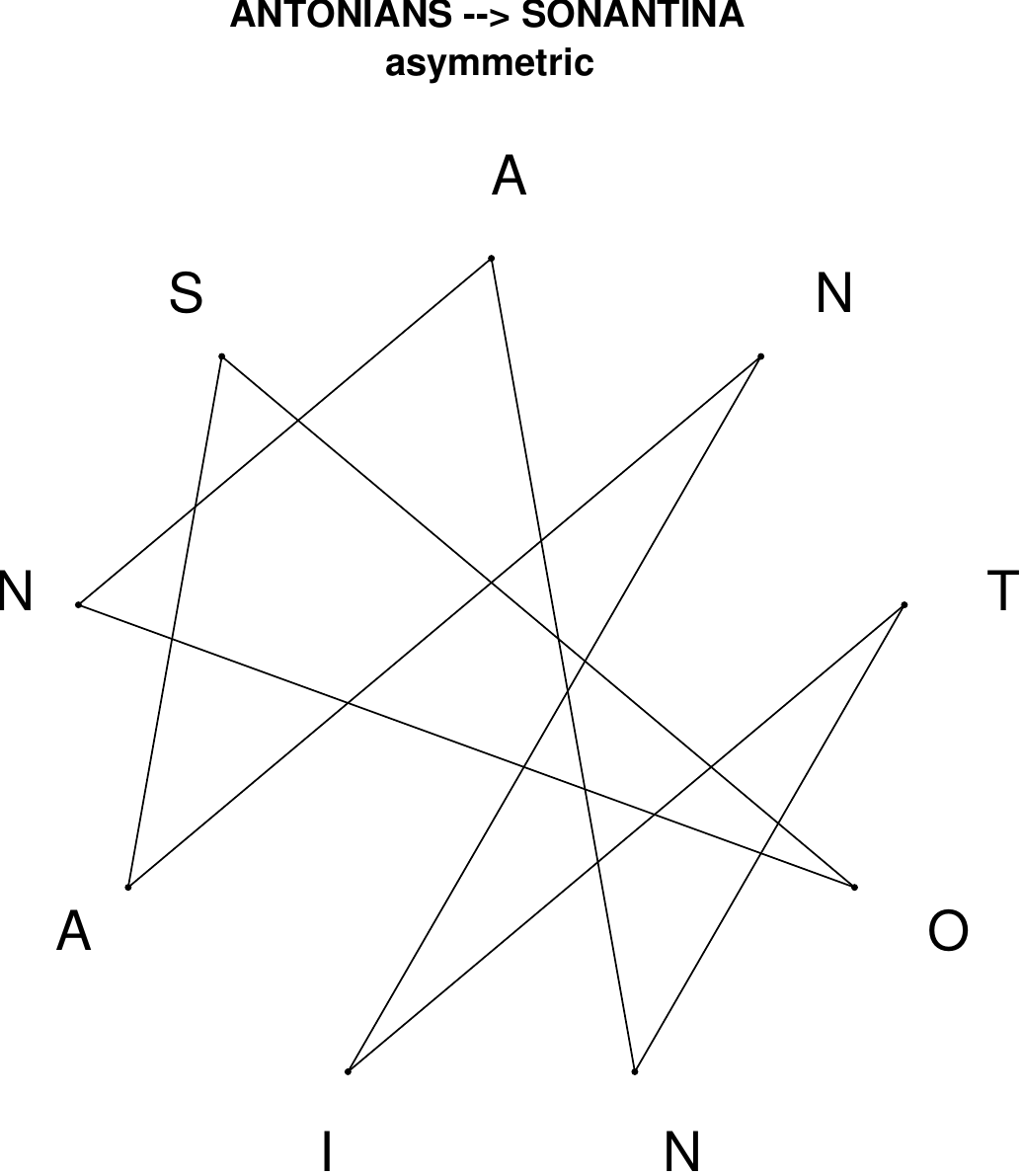}
\end{subfigure}
\hfill
\begin{subfigure}[T]{0.19\textwidth}
\centering
\includegraphics[width=\textwidth]{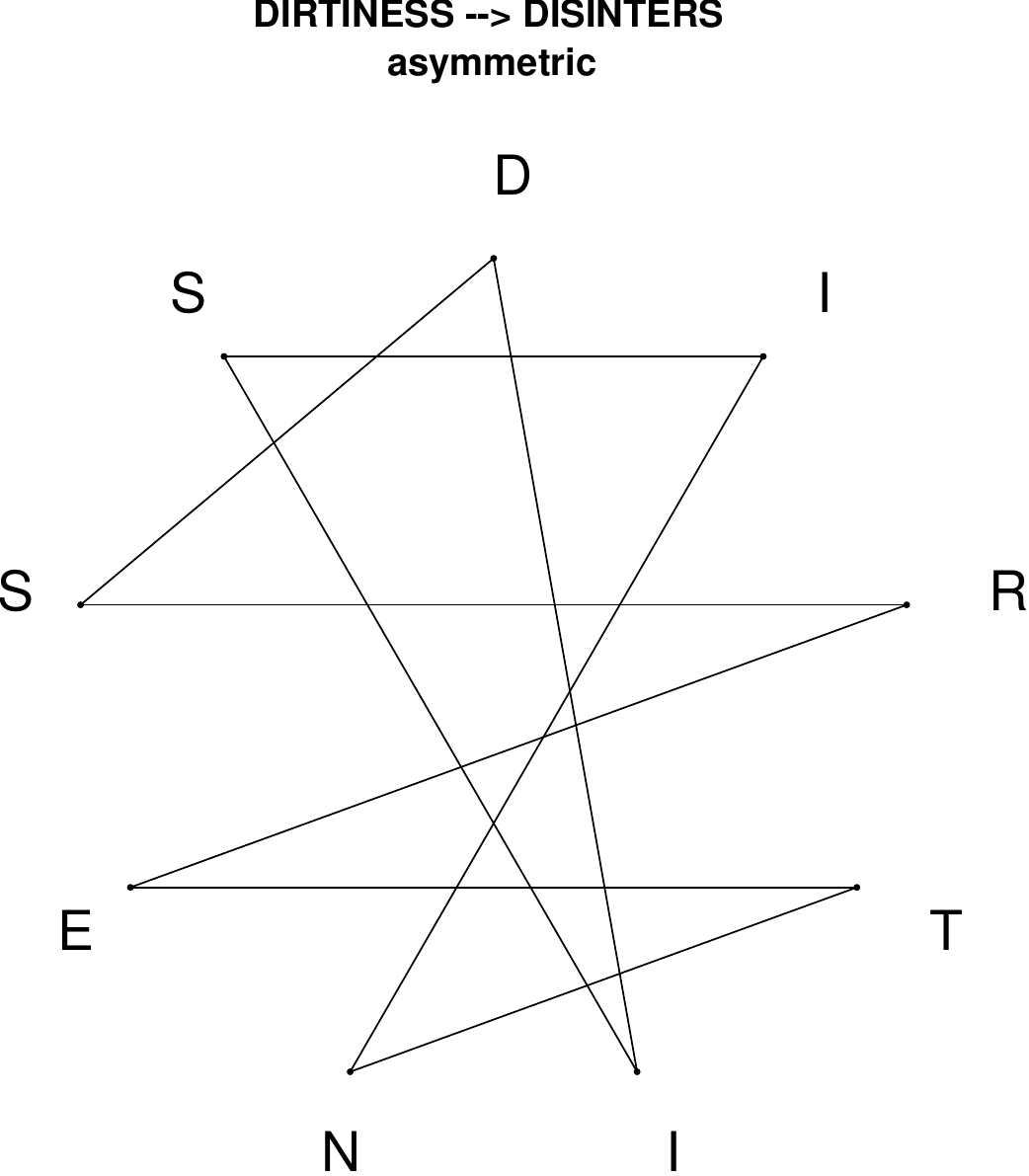}
\end{subfigure}
\hfill
\begin{subfigure}[T]{0.19\textwidth}
\centering
\includegraphics[width=\textwidth]{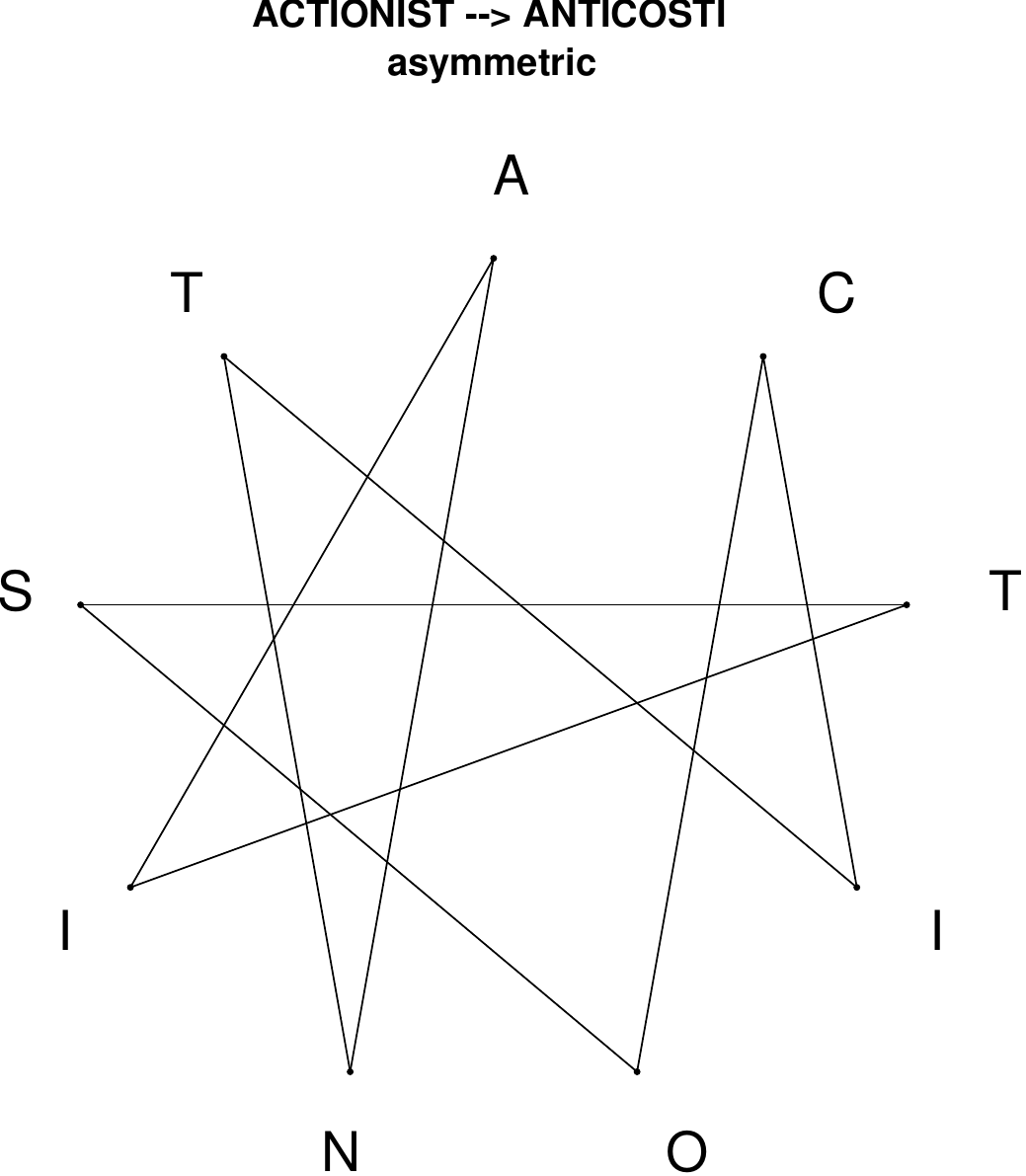}
\end{subfigure}
\hfill
\begin{subfigure}[T]{0.19\textwidth}
\centering
\includegraphics[width=\textwidth]{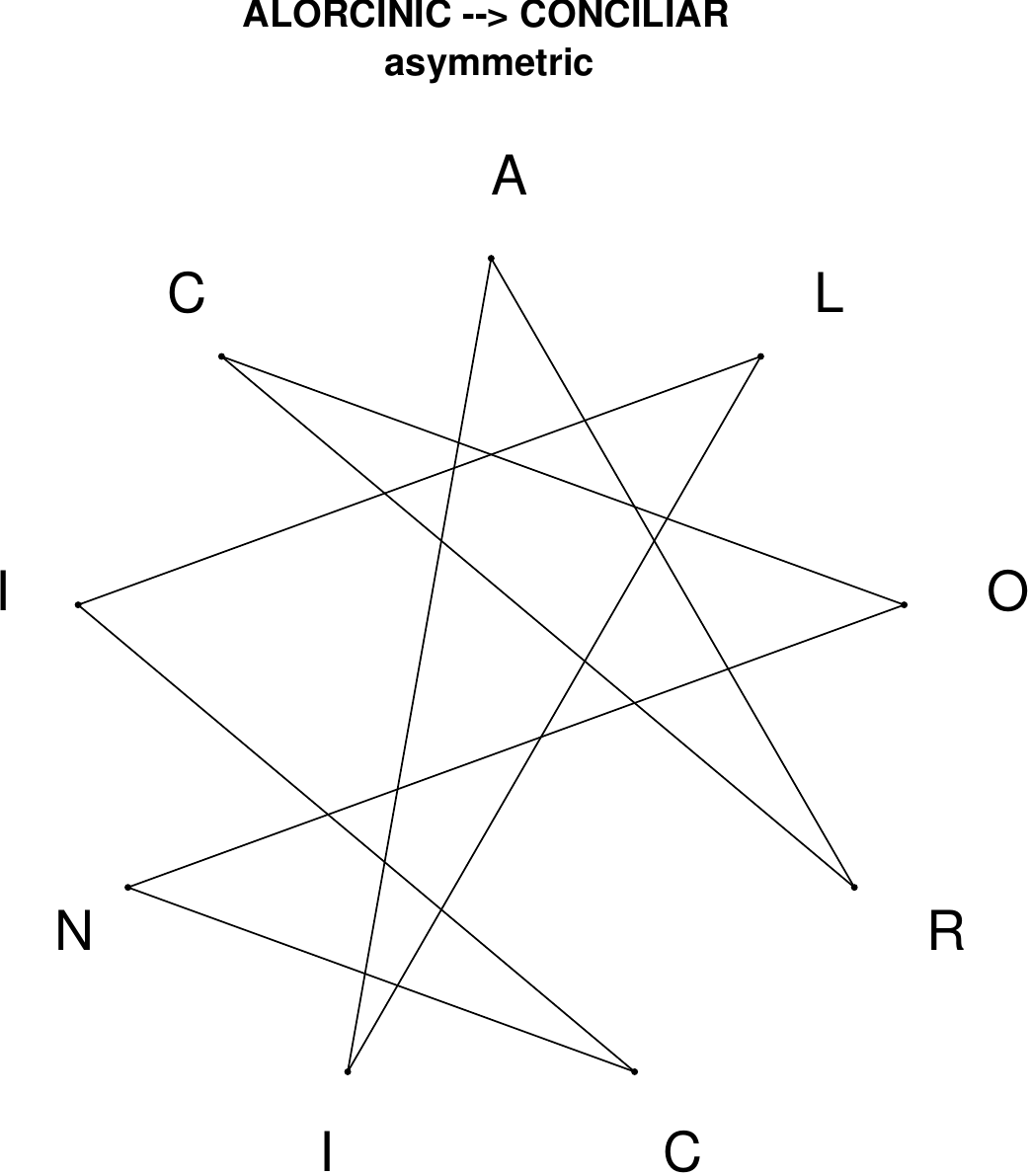}
\end{subfigure}
\hfill
\begin{subfigure}[T]{0.19\textwidth}
\centering
\includegraphics[width=\textwidth]{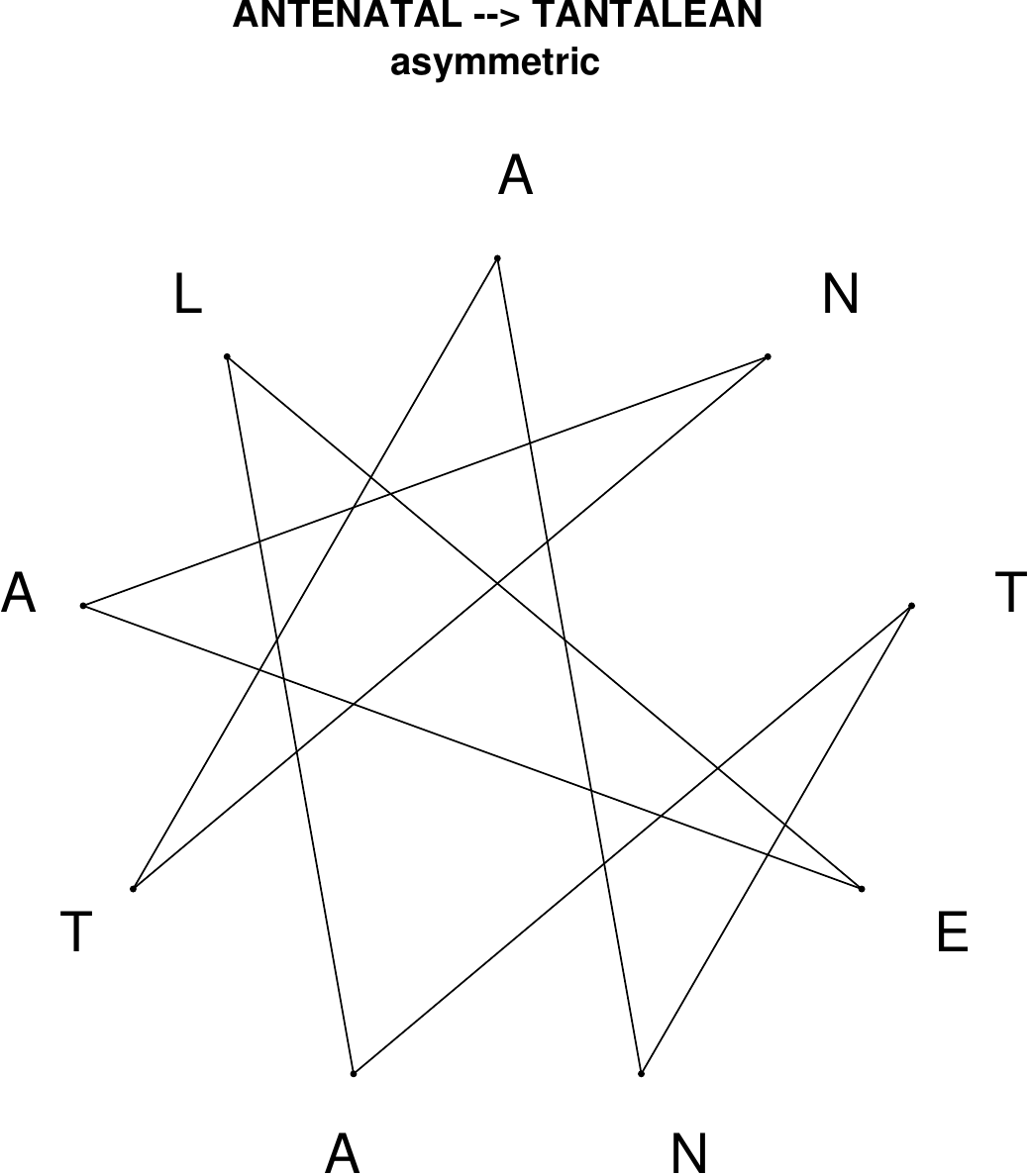}
\end{subfigure}
\end{figure}

\begin{figure}[H]
\centering
\begin{subfigure}[T]{0.19\textwidth}
\centering
\includegraphics[width=\textwidth]{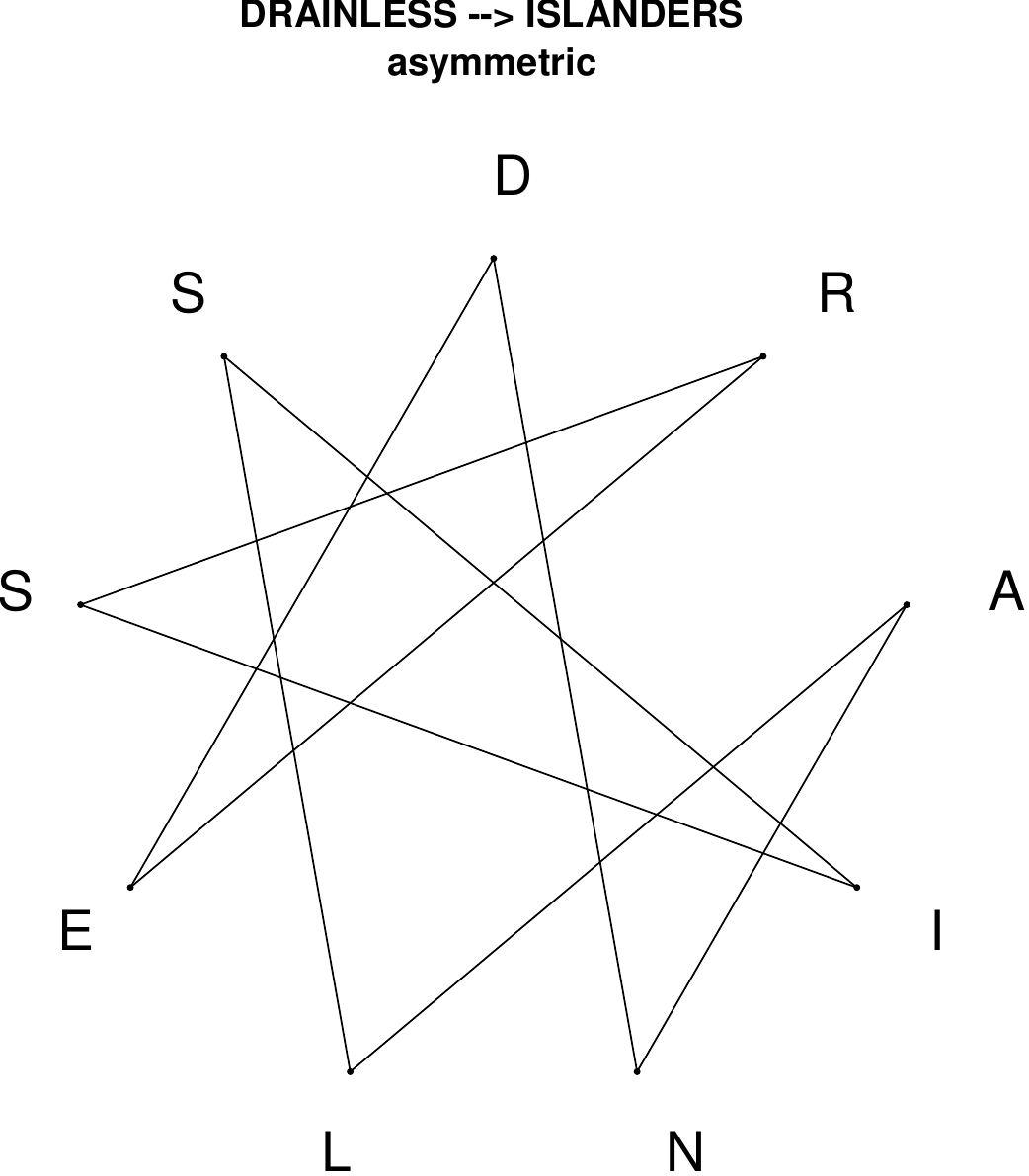}
\end{subfigure}
\hfill
\begin{subfigure}[T]{0.19\textwidth}
\centering
\includegraphics[width=\textwidth]{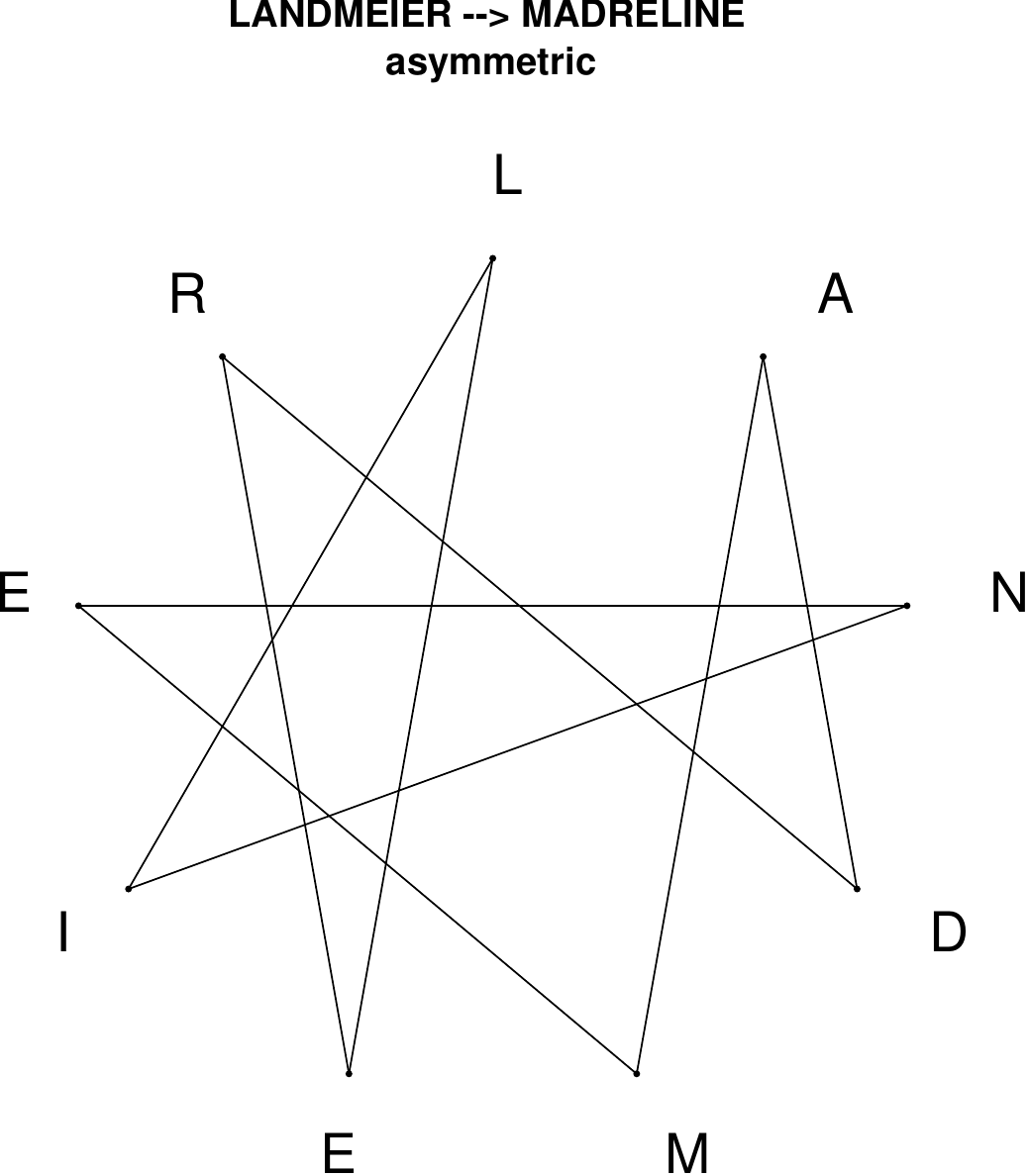}
\end{subfigure}
\hfill
\begin{subfigure}[T]{0.19\textwidth}
\centering
\includegraphics[width=\textwidth]{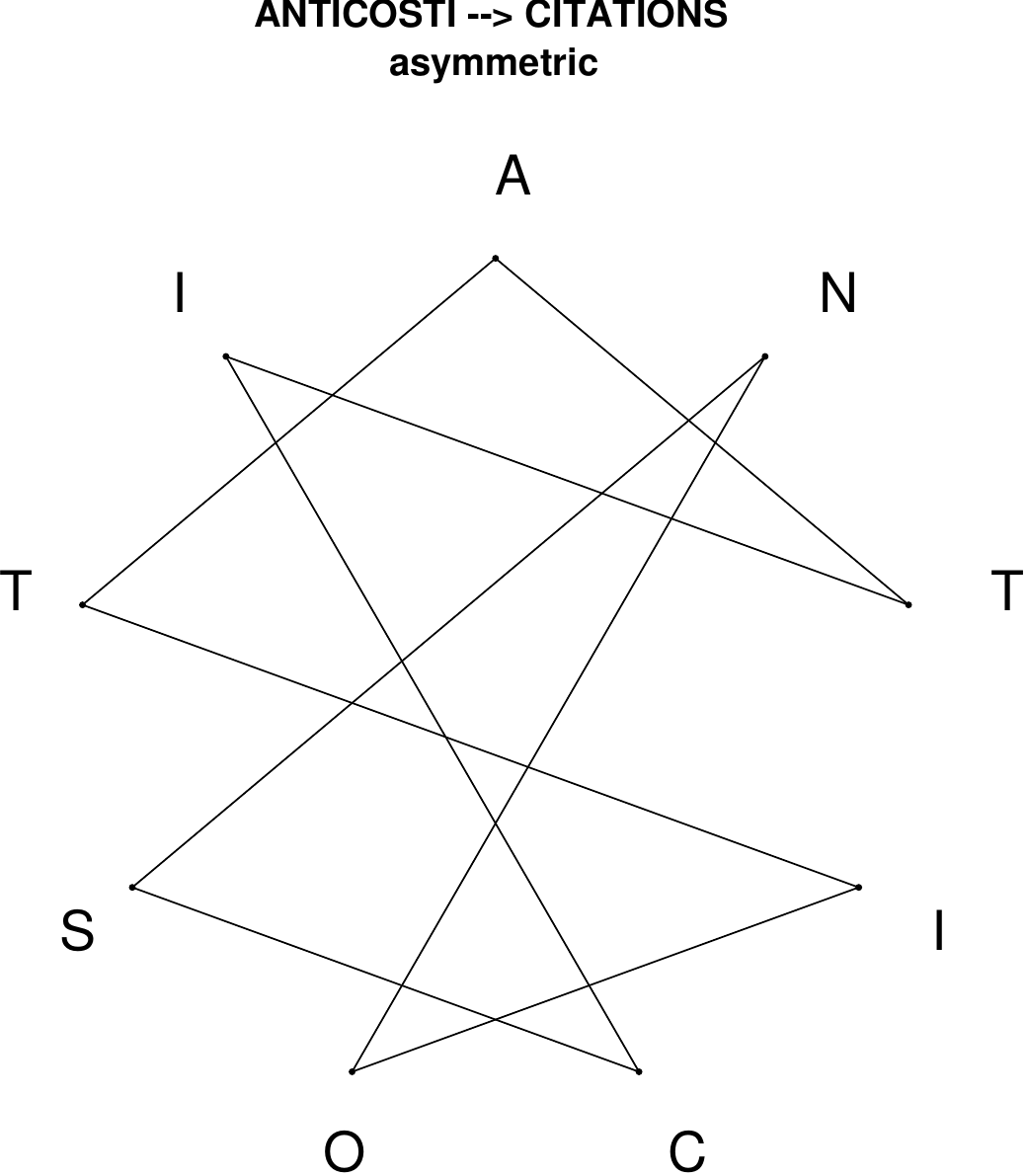}
\end{subfigure}
\hfill
\begin{subfigure}[T]{0.19\textwidth}
\centering
\includegraphics[width=\textwidth]{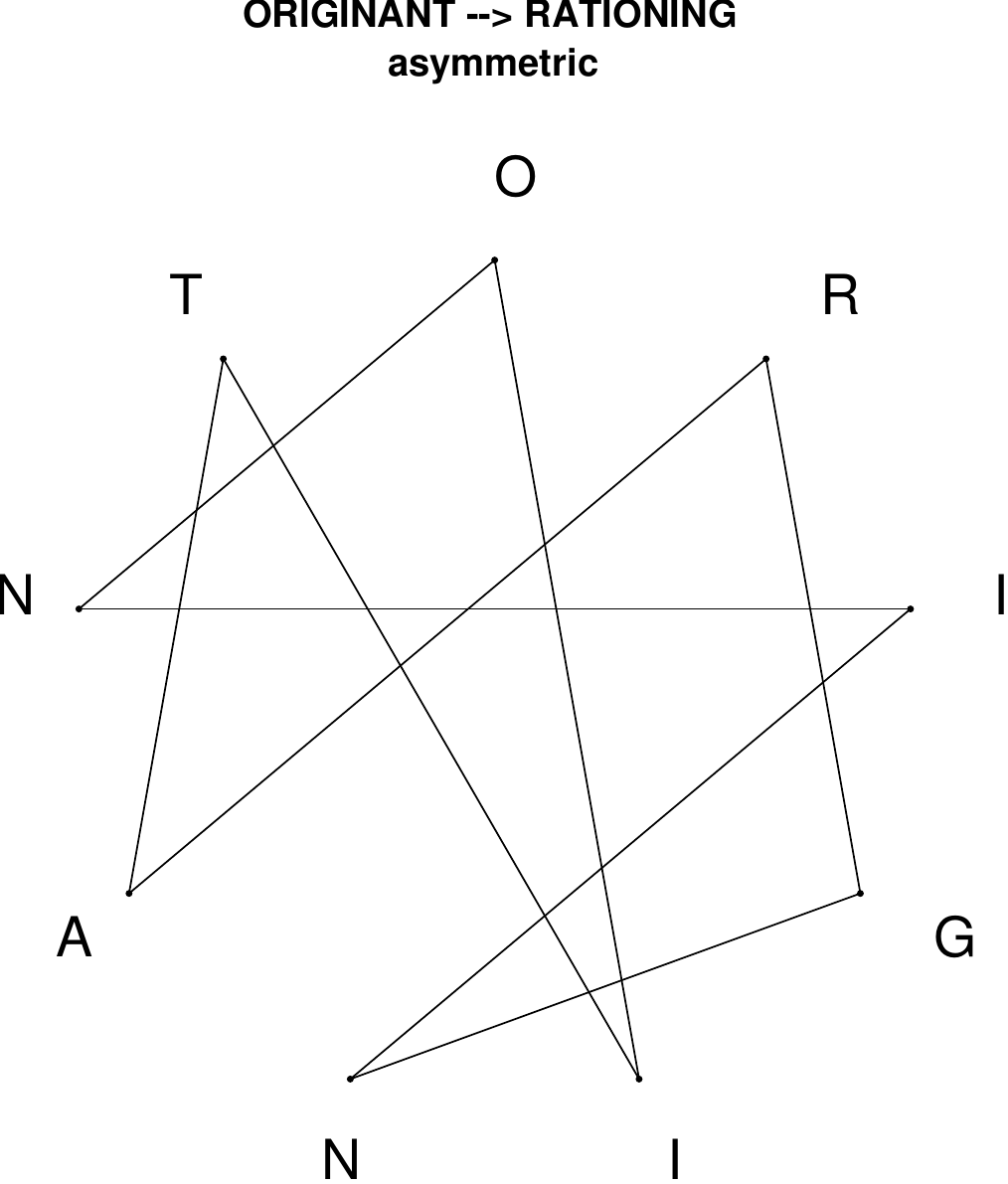}
\end{subfigure}
\hfill
\begin{subfigure}[T]{0.19\textwidth}
\centering
\includegraphics[width=\textwidth]{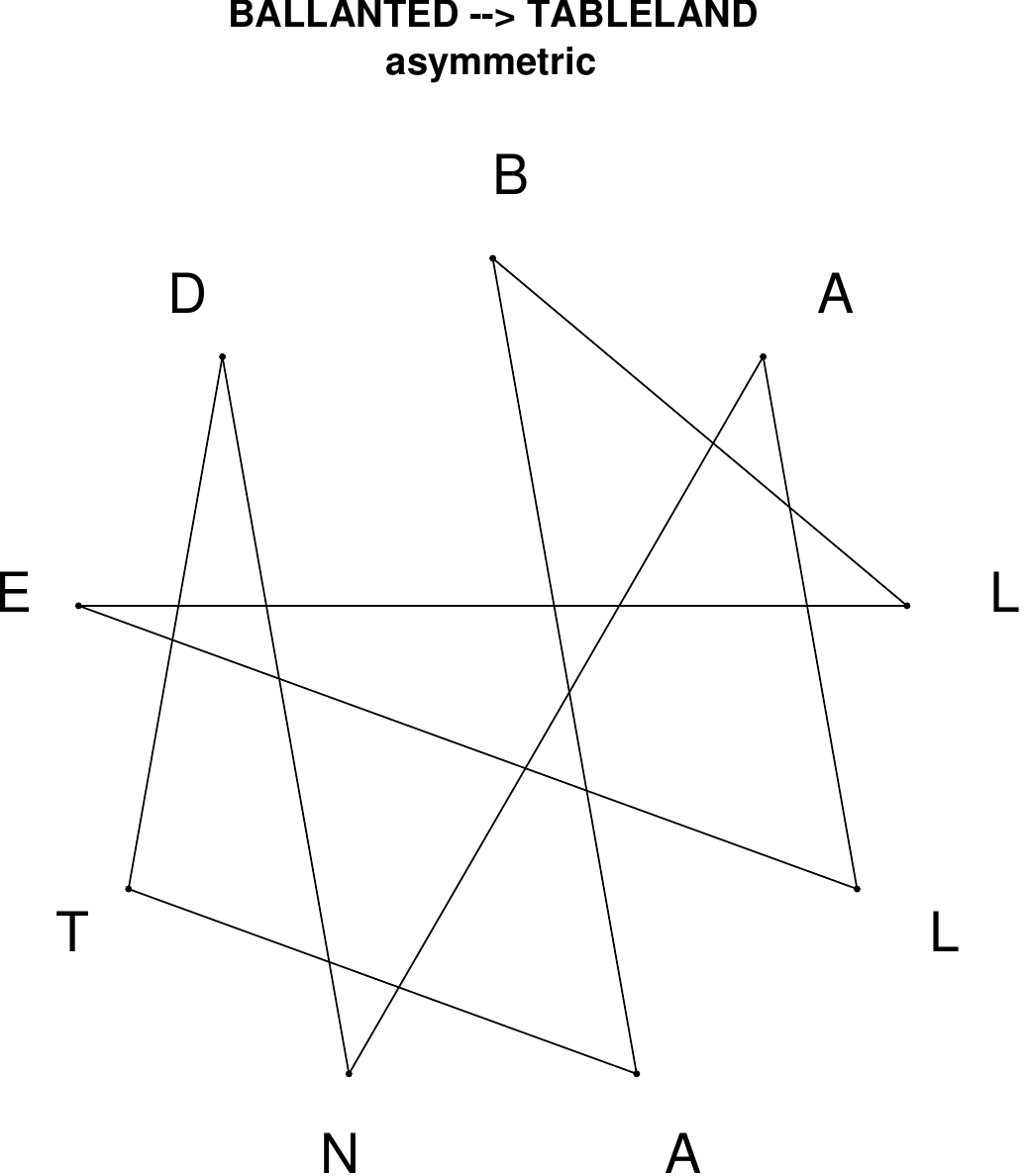}
\end{subfigure}
\end{figure}

\begin{figure}[H]
\centering
\begin{subfigure}[T]{0.19\textwidth}
\centering
\includegraphics[width=\textwidth]{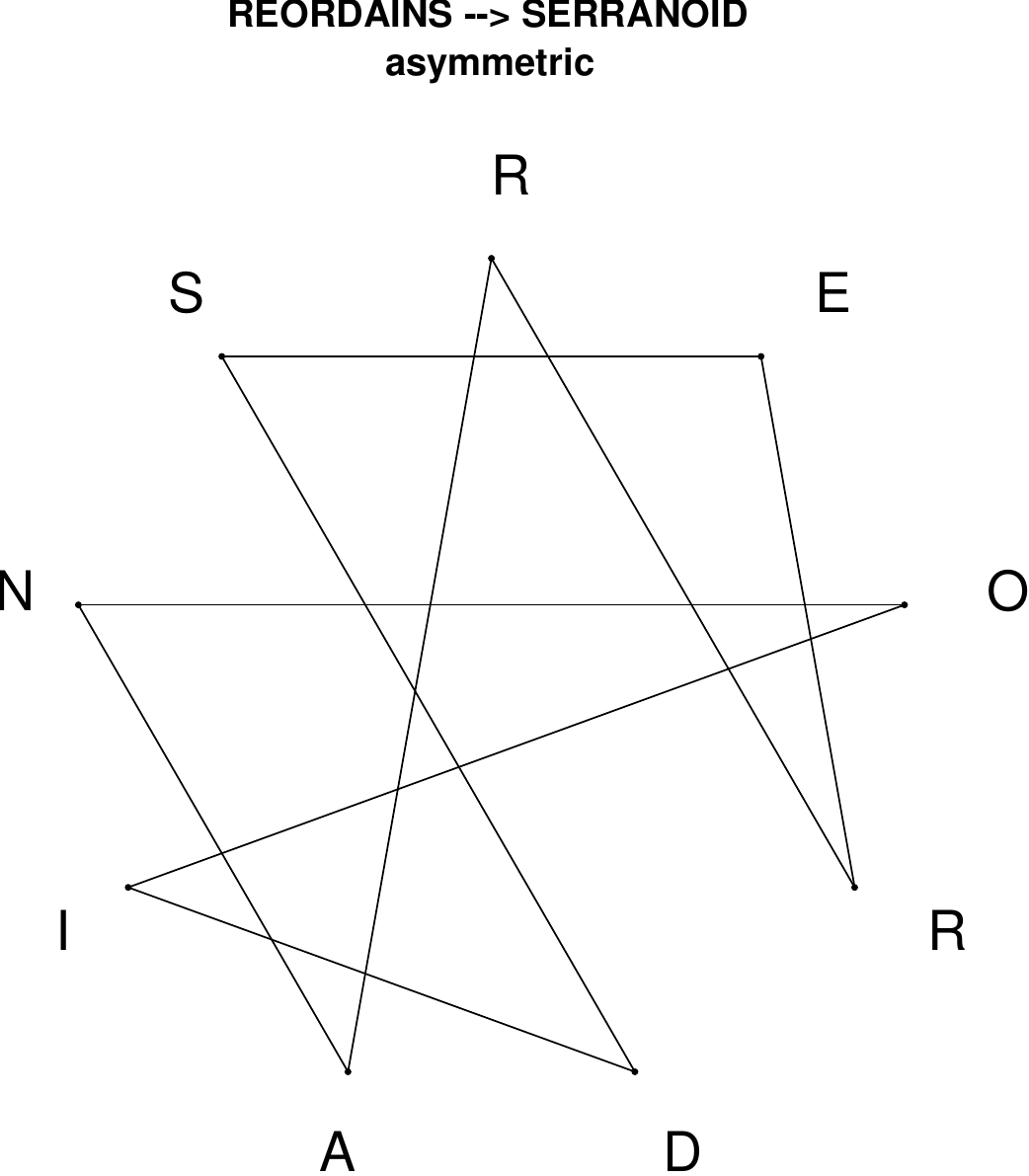}
\end{subfigure}
\hfill
\begin{subfigure}[T]{0.19\textwidth}
\centering
\includegraphics[width=\textwidth]{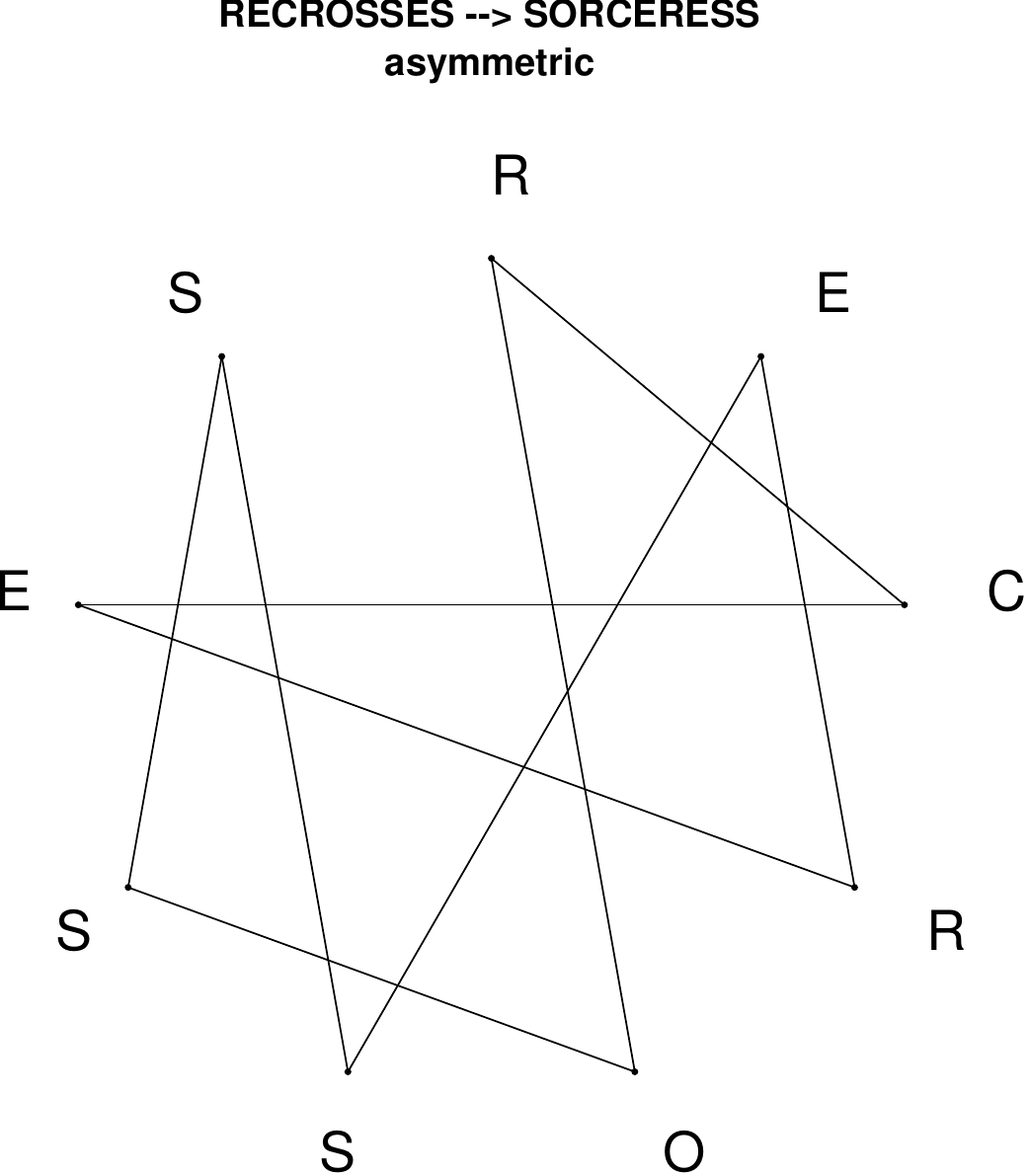}
\end{subfigure}
\hfill
\begin{subfigure}[T]{0.19\textwidth}
\centering
\includegraphics[width=\textwidth]{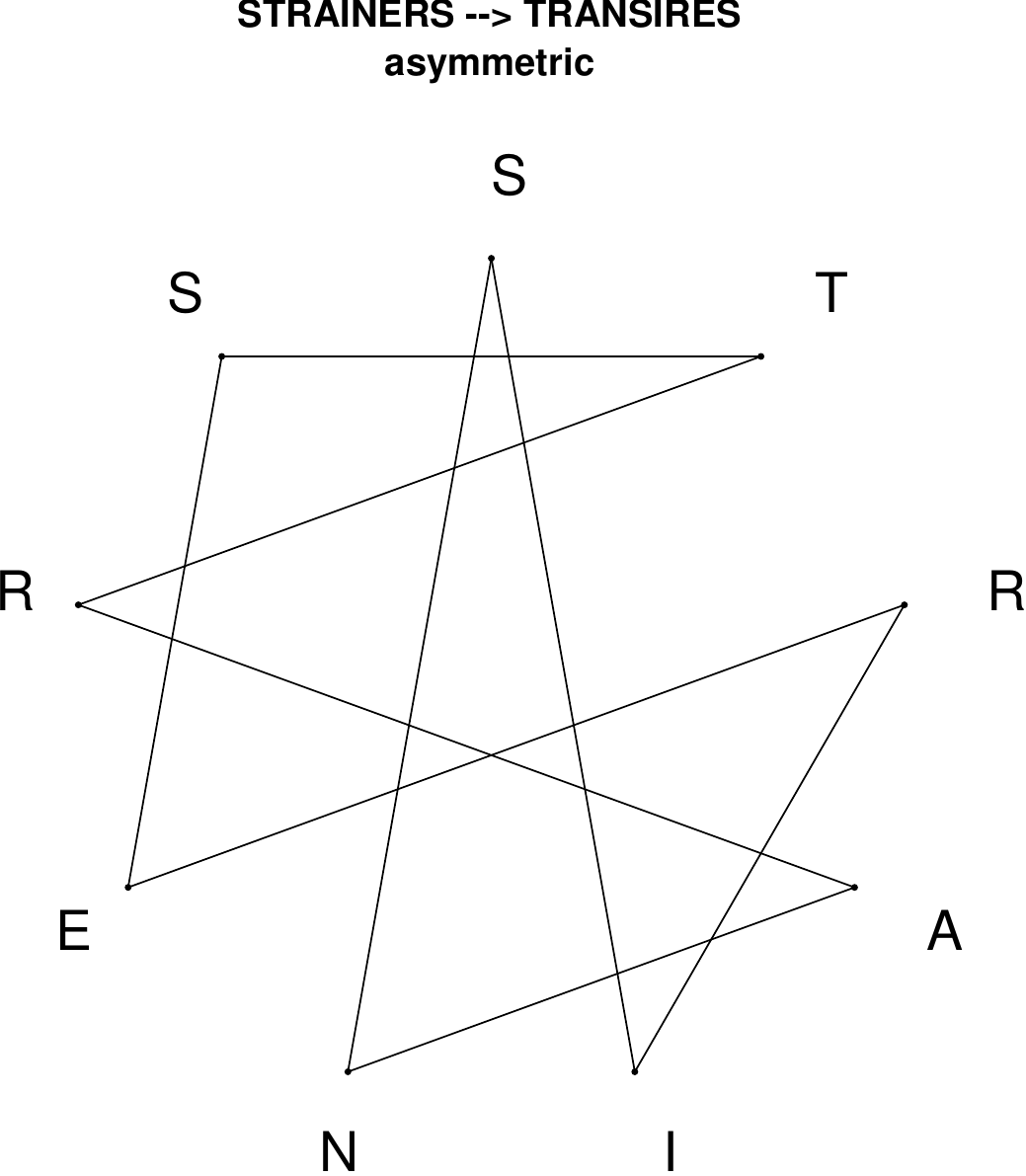}
\end{subfigure}
\hfill
\begin{subfigure}[T]{0.19\textwidth}
\centering
\includegraphics[width=\textwidth]{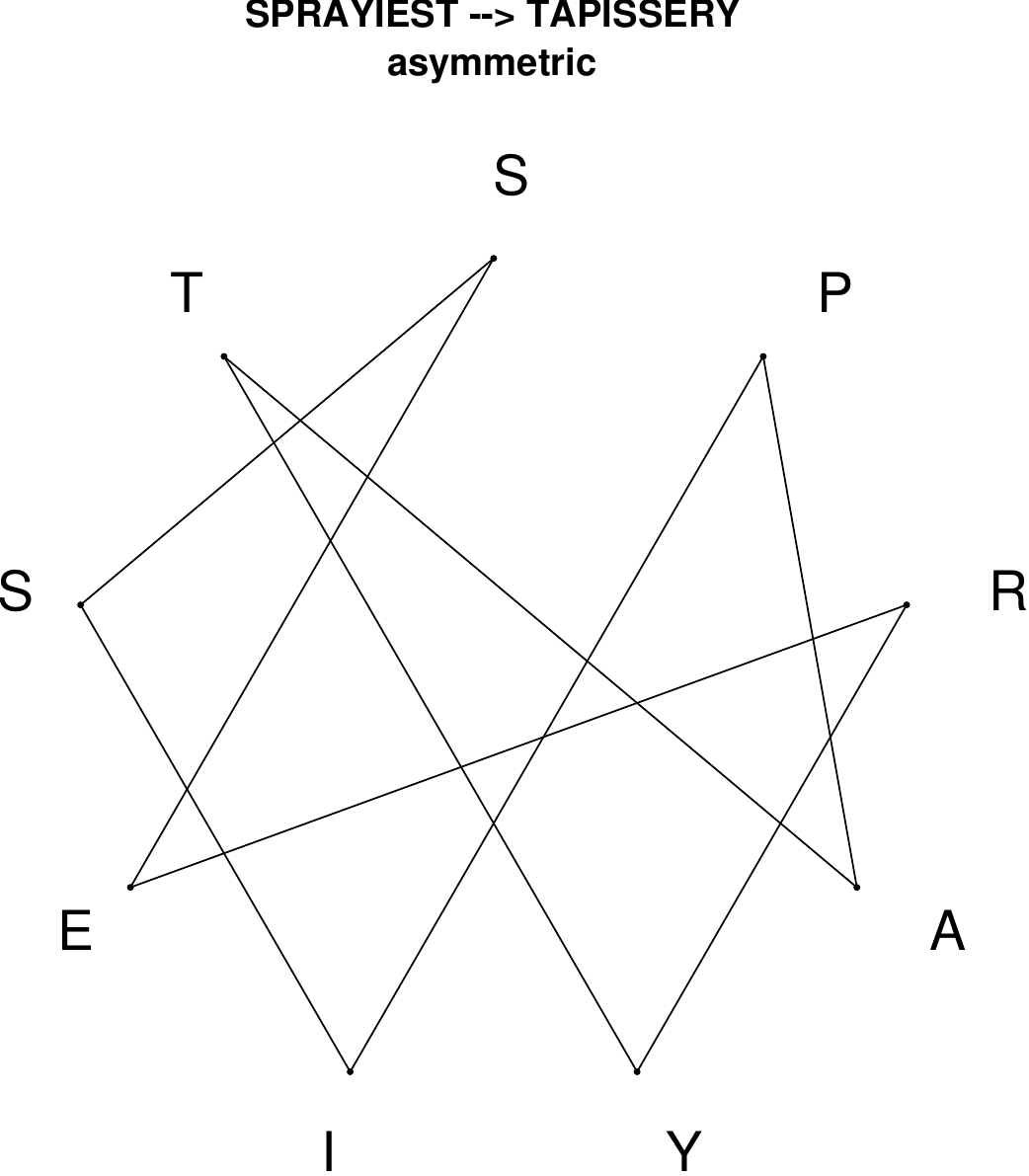}
\end{subfigure}
\hfill
\begin{subfigure}[T]{0.19\textwidth}
\centering
\includegraphics[width=\textwidth]{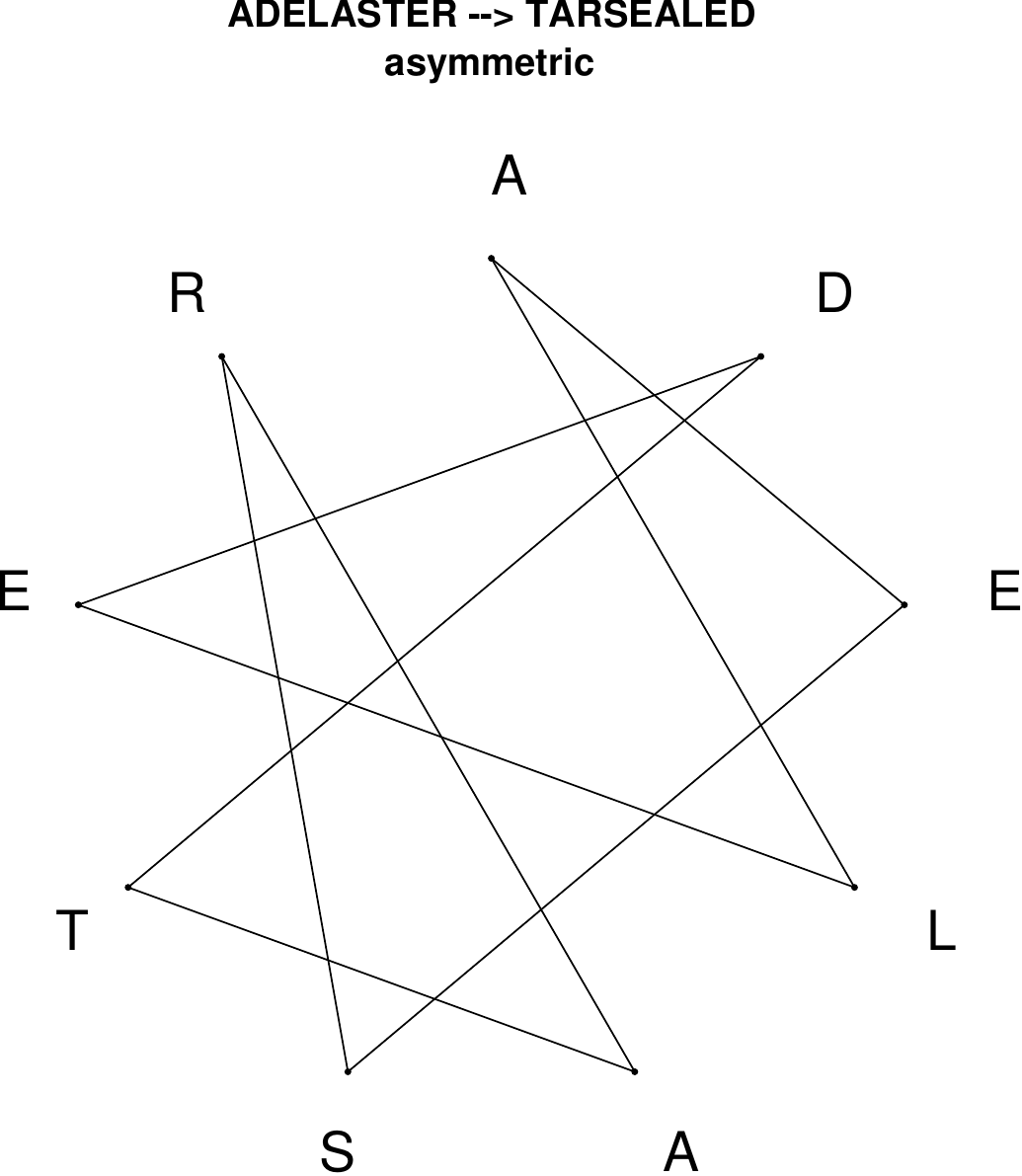}
\end{subfigure}
\end{figure}

\begin{figure}[H]
\centering
\begin{subfigure}[T]{0.19\textwidth}
\centering
\includegraphics[width=\textwidth]{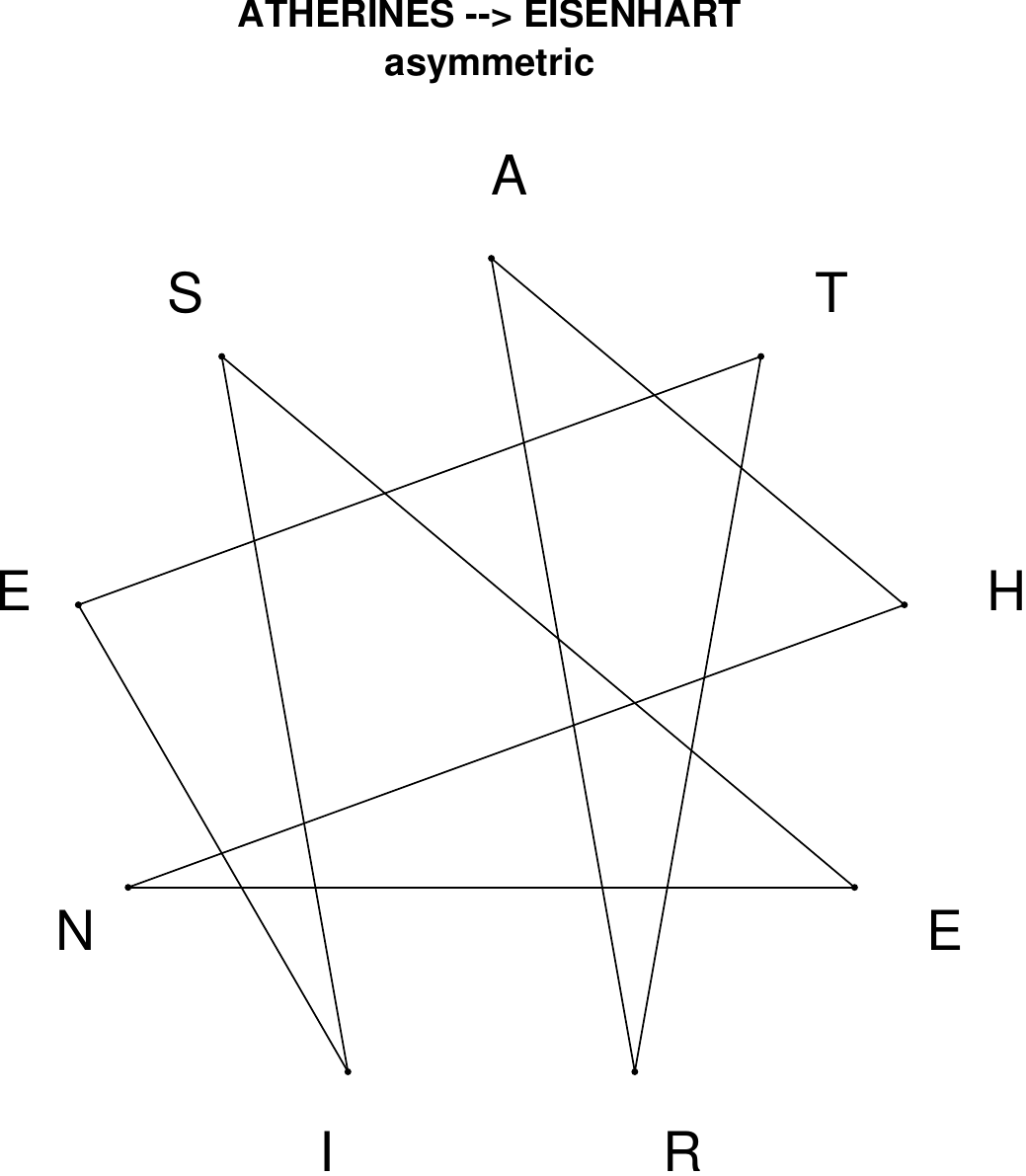}
\end{subfigure}
\hfill
\begin{subfigure}[T]{0.19\textwidth}
\centering
\includegraphics[width=\textwidth]{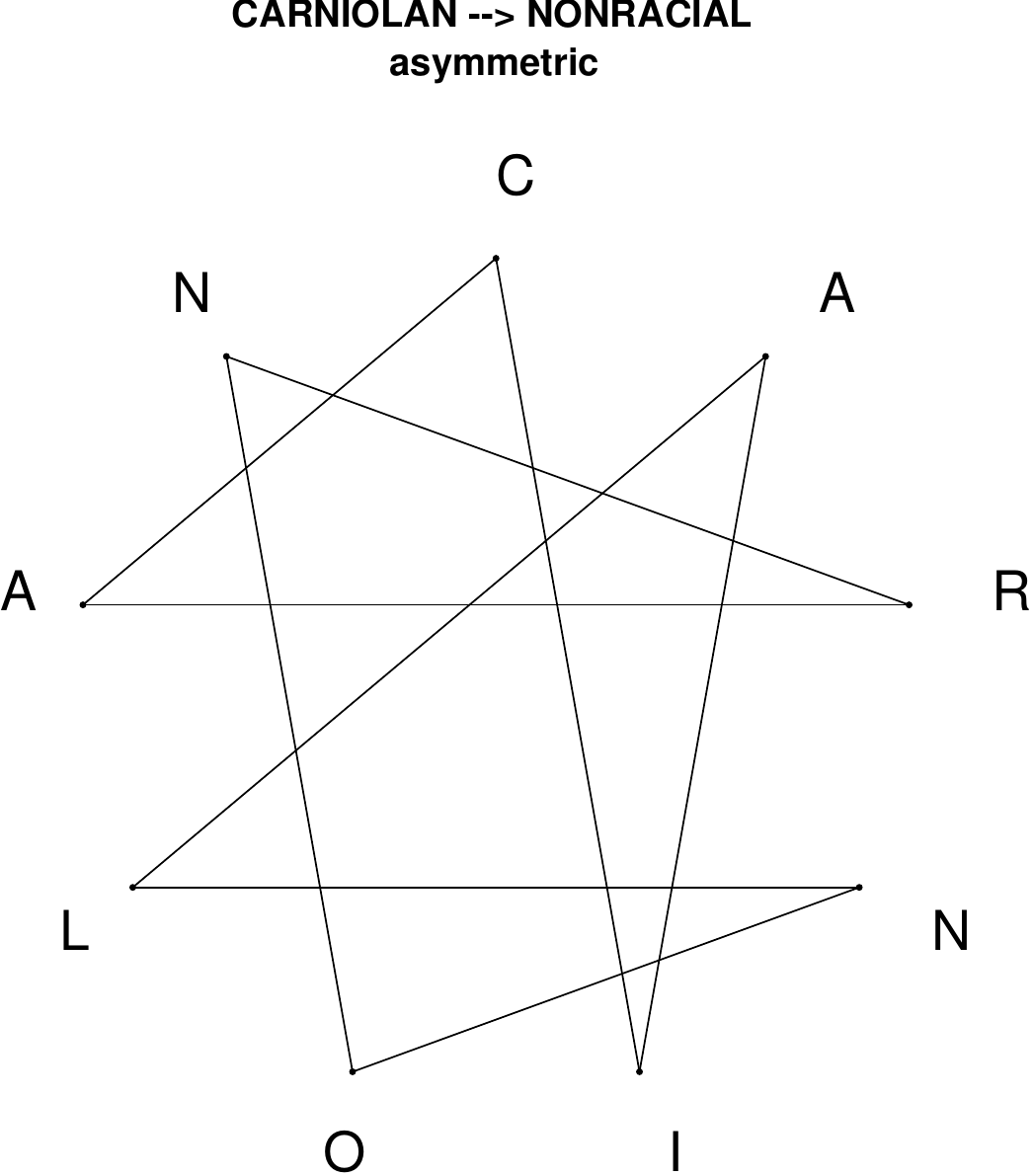}
\end{subfigure}
\hfill
\begin{subfigure}[T]{0.19\textwidth}
\centering
\includegraphics[width=\textwidth]{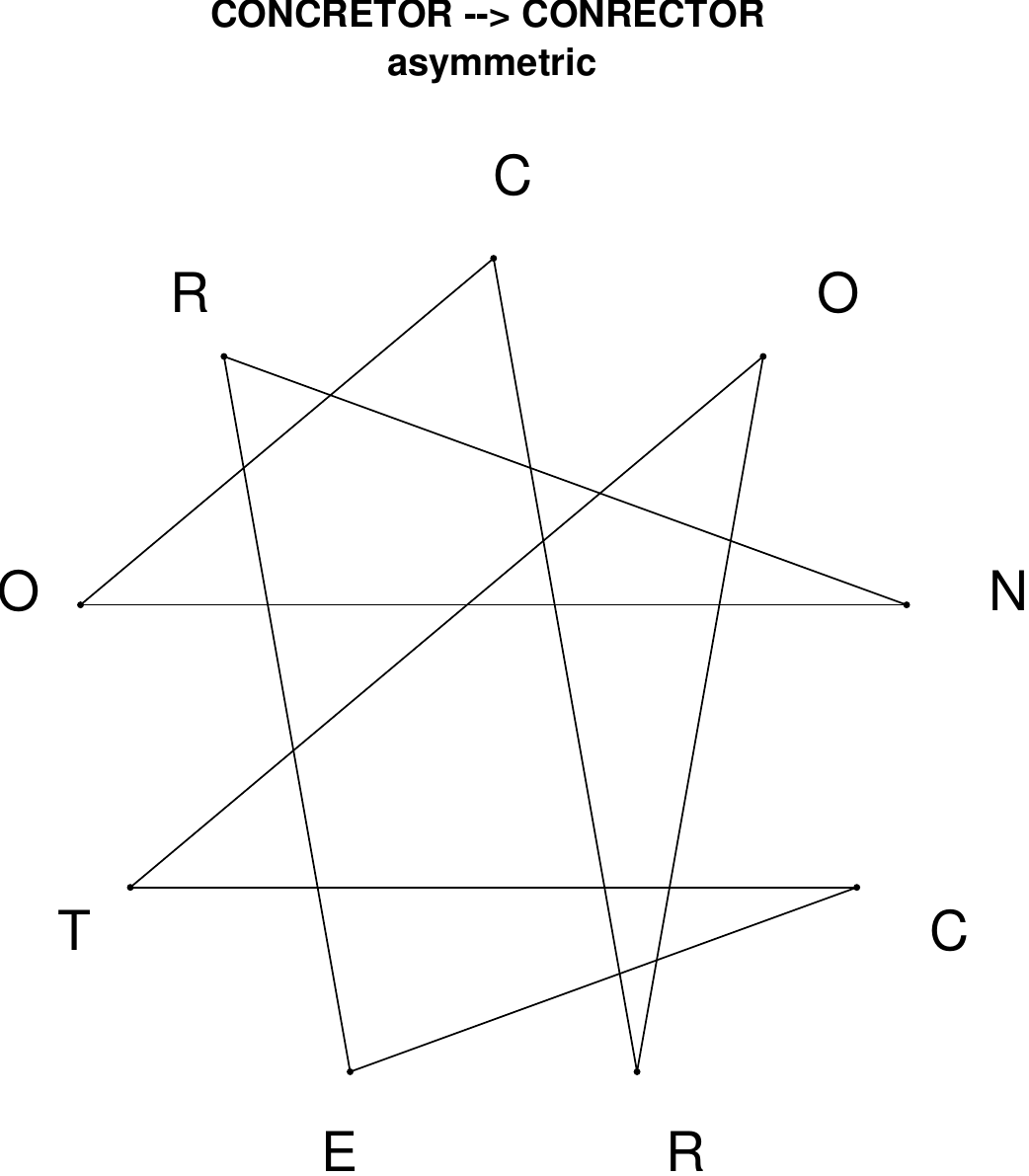}
\end{subfigure}
\hfill
\begin{subfigure}[T]{0.19\textwidth}
\centering
\includegraphics[width=\textwidth]{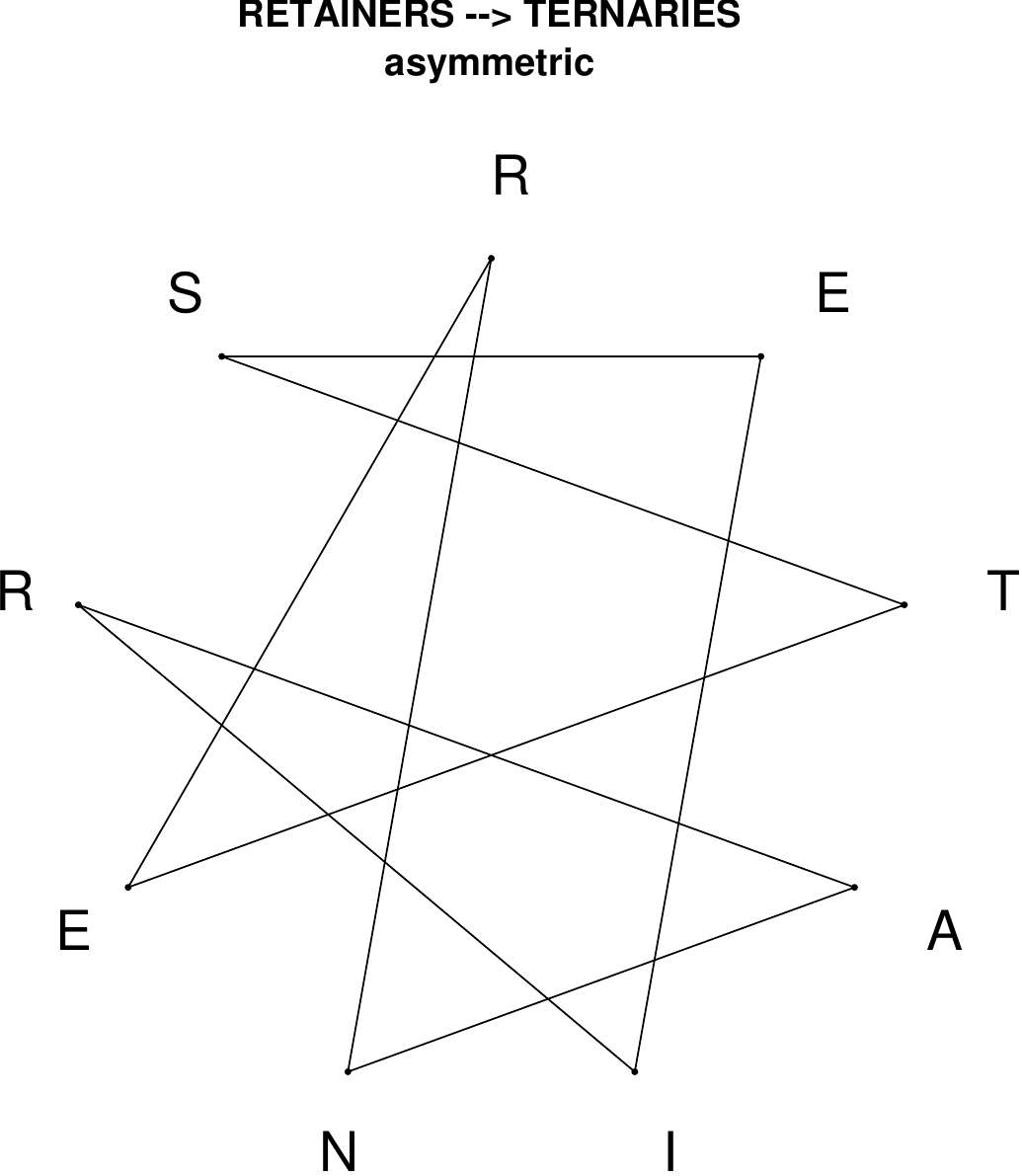}
\end{subfigure}
\hfill
\begin{subfigure}[T]{0.19\textwidth}
\centering
\includegraphics[width=\textwidth]{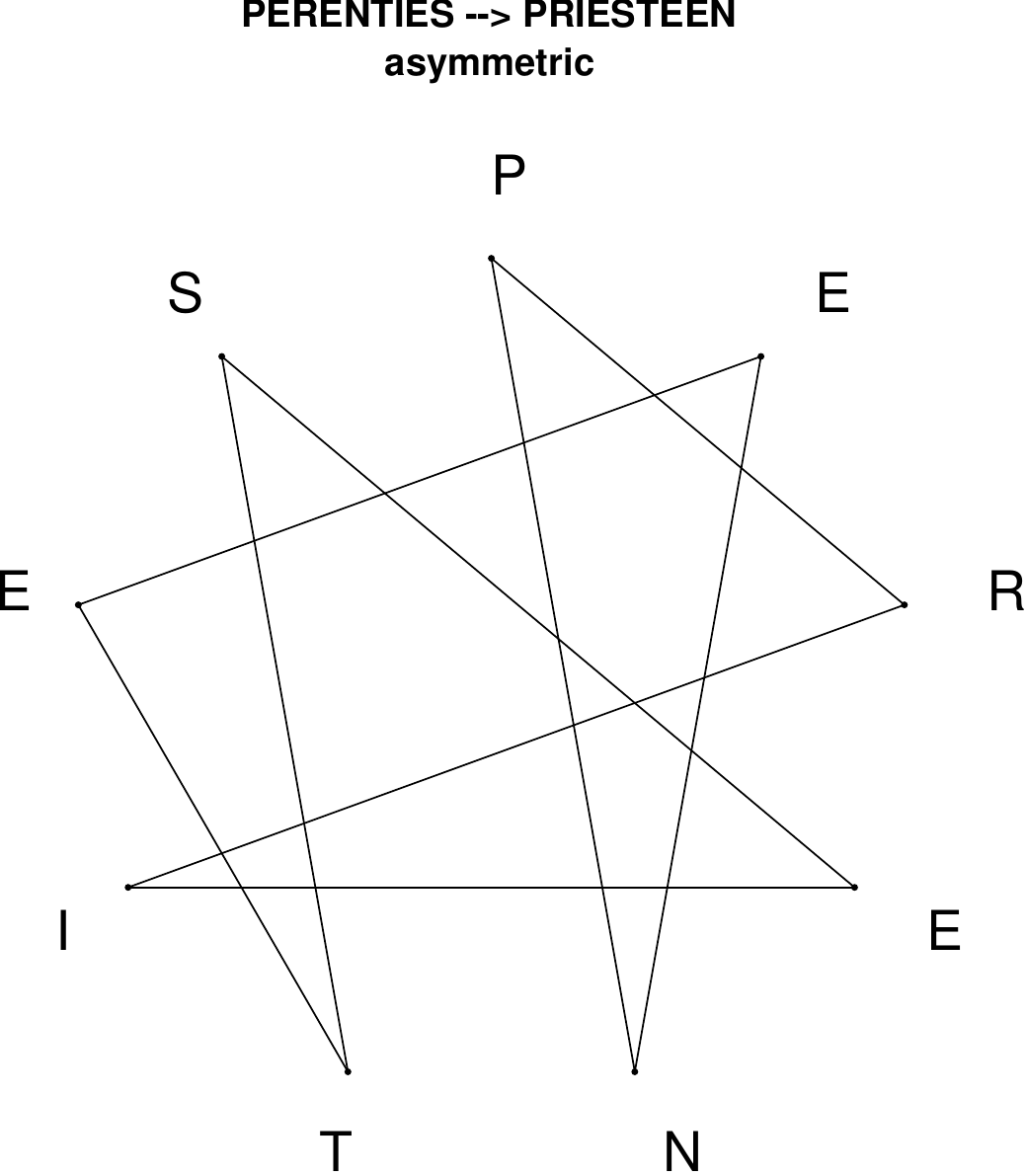}
\end{subfigure}
\end{figure}

\begin{figure}[H]
\centering
\begin{subfigure}[T]{0.19\textwidth}
\centering
\includegraphics[width=\textwidth]{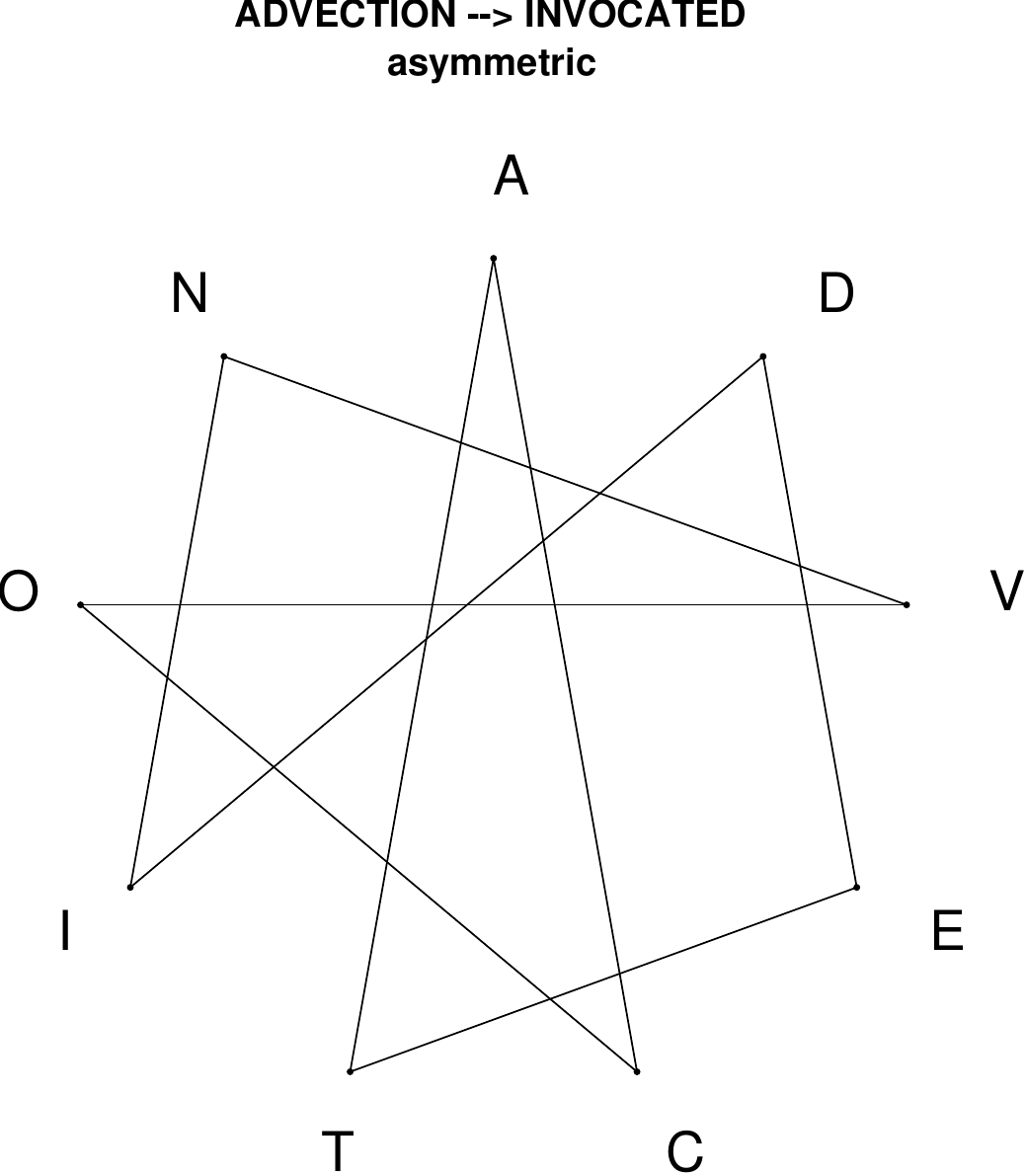}
\end{subfigure}
\hfill
\begin{subfigure}[T]{0.19\textwidth}
\centering
\includegraphics[width=\textwidth]{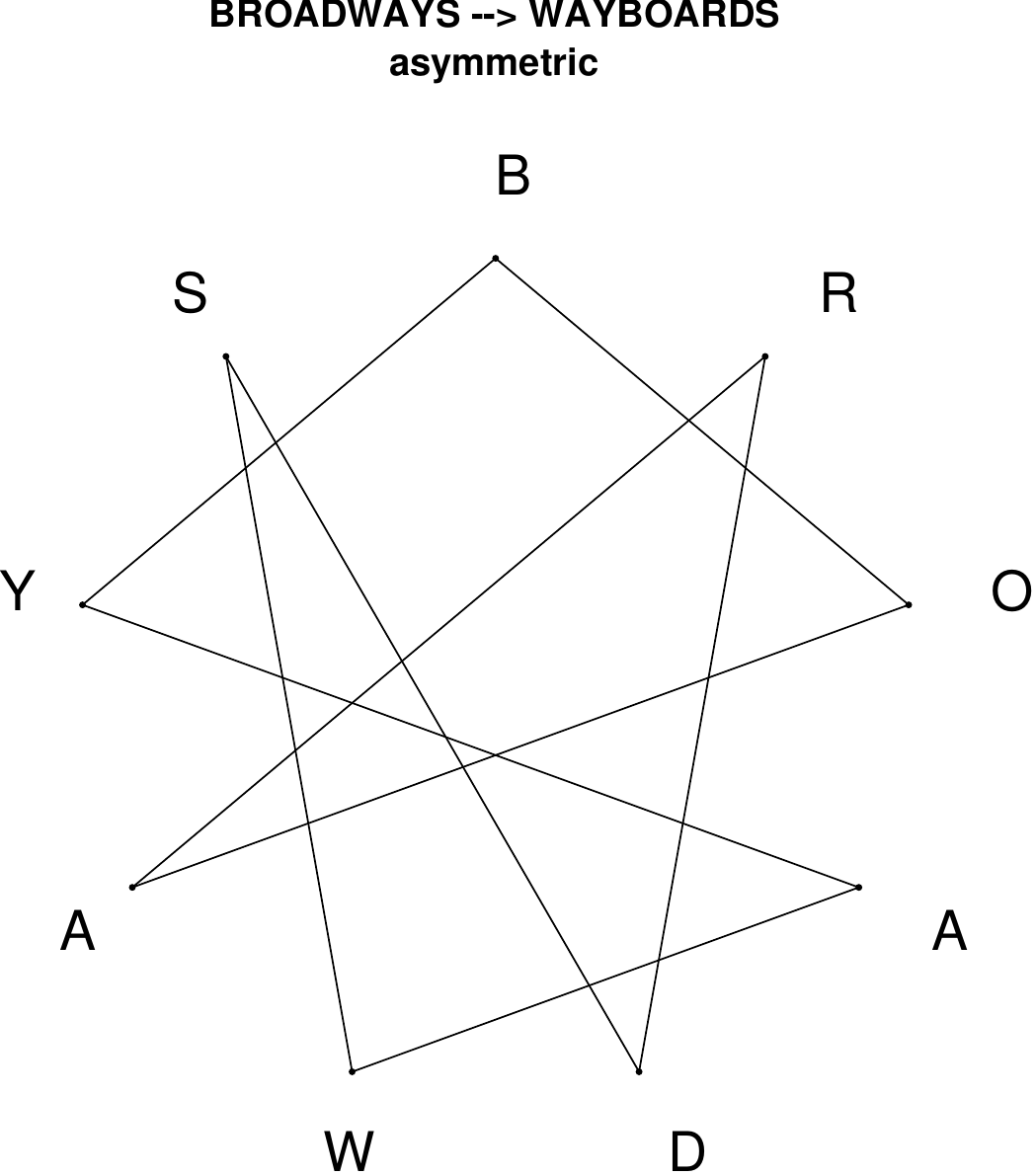}
\end{subfigure}
\hfill
\begin{subfigure}[T]{0.19\textwidth}
\centering
\includegraphics[width=\textwidth]{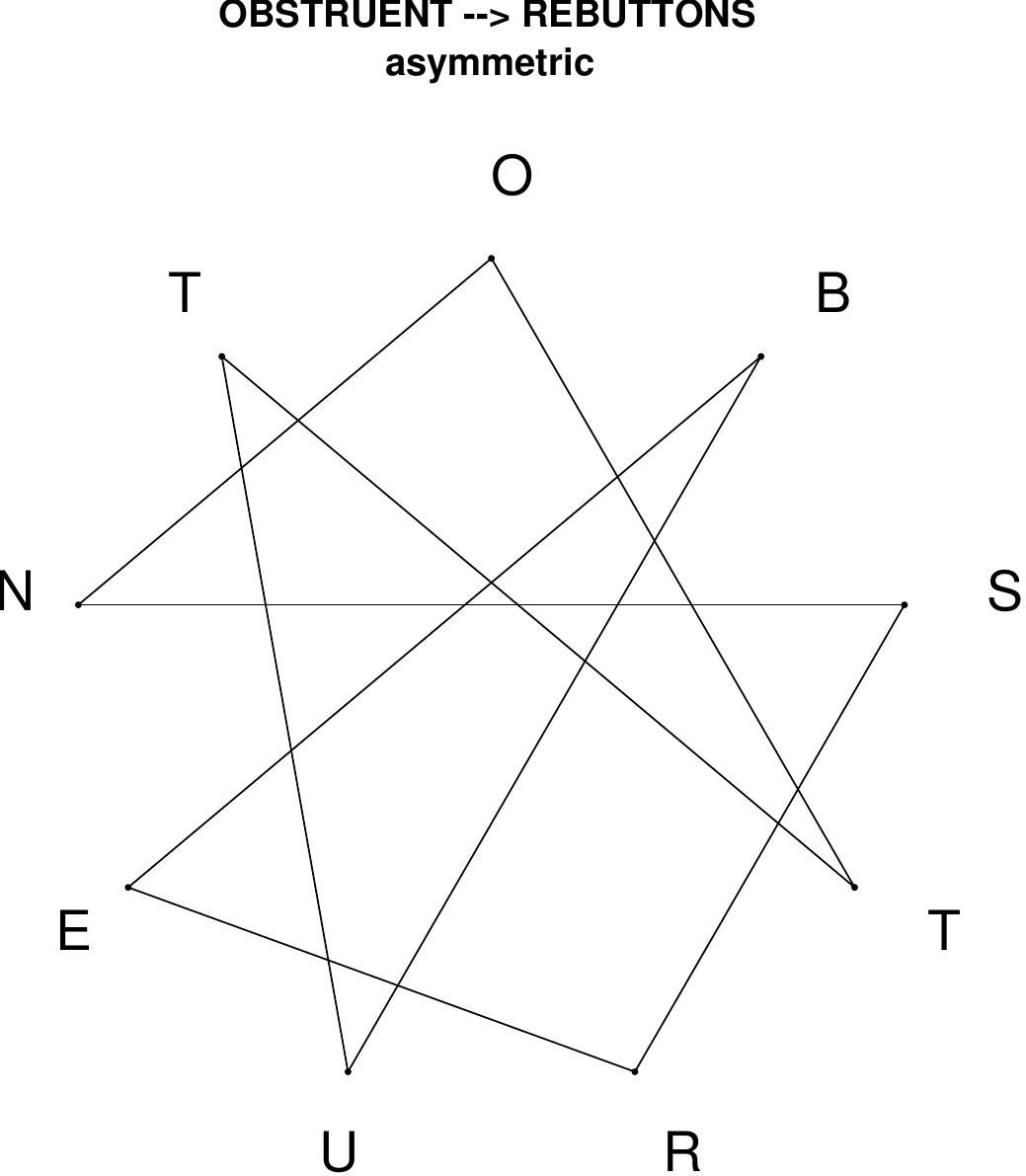}
\end{subfigure}
\hfill
\begin{subfigure}[T]{0.19\textwidth}
\centering
\includegraphics[width=\textwidth]{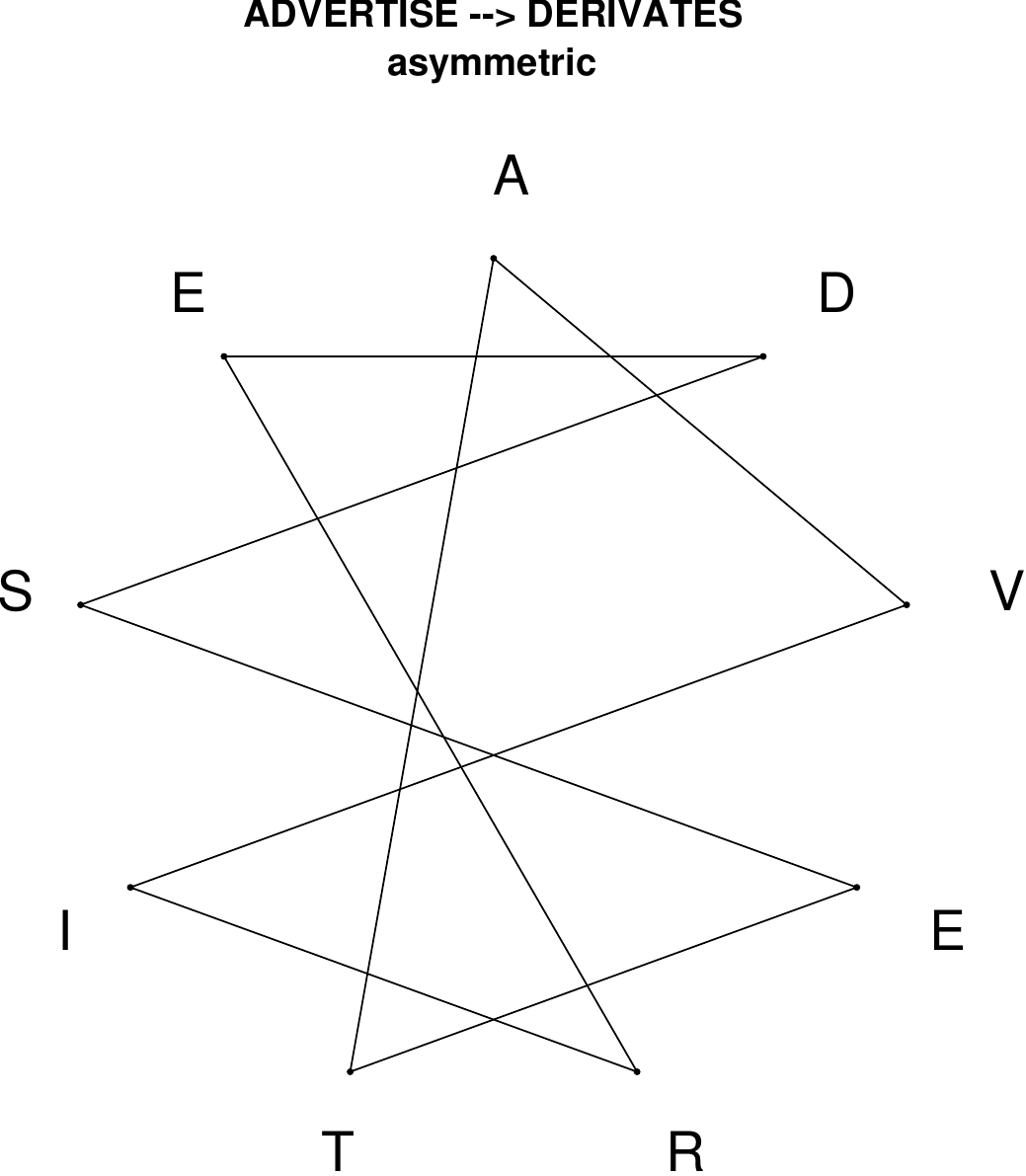}
\end{subfigure}
\hfill
\begin{subfigure}[T]{0.19\textwidth}
\centering
\includegraphics[width=\textwidth]{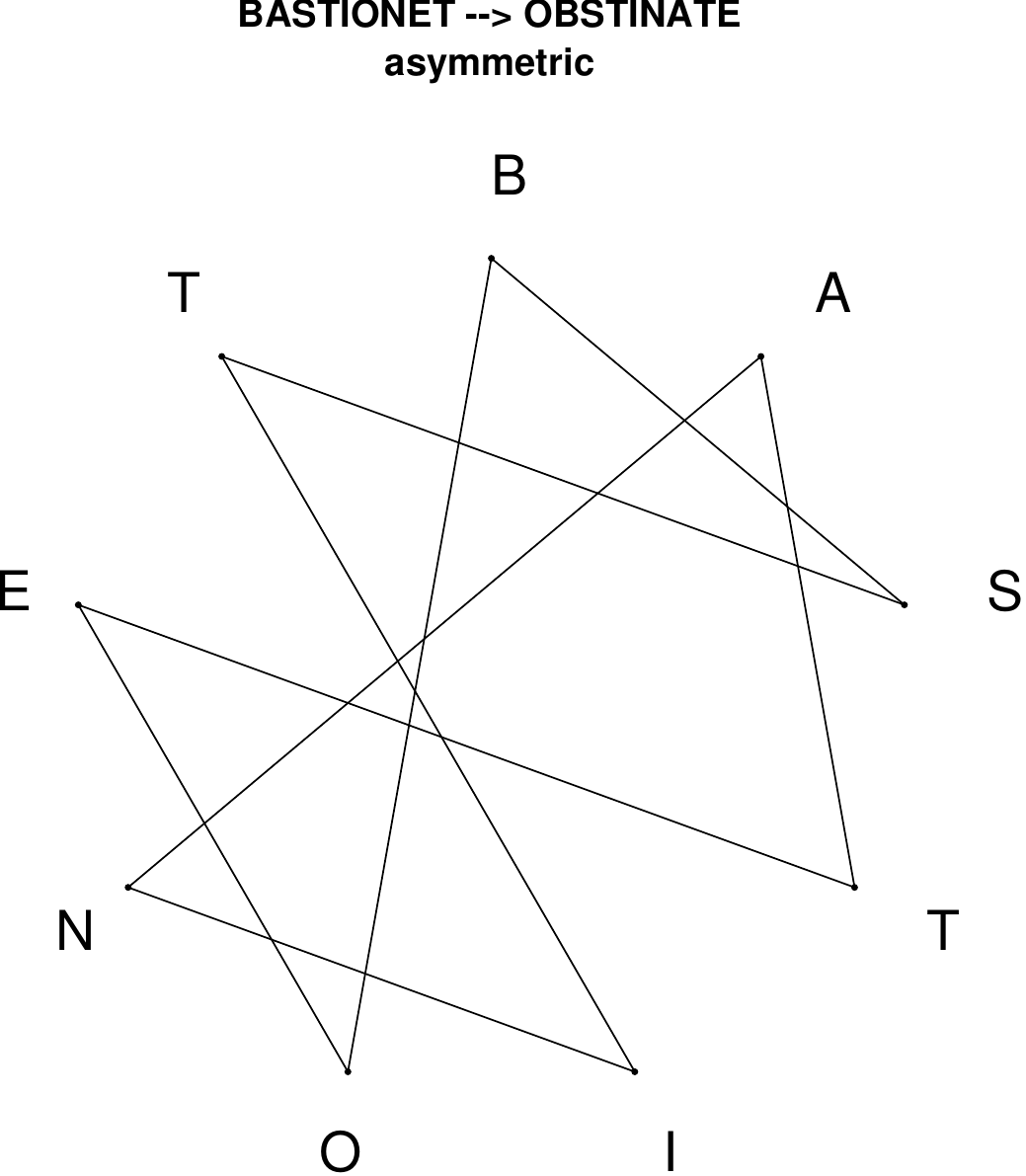}
\end{subfigure}
\end{figure}

\begin{figure}[H]
\centering
\begin{subfigure}[T]{0.19\textwidth}
\centering
\includegraphics[width=\textwidth]{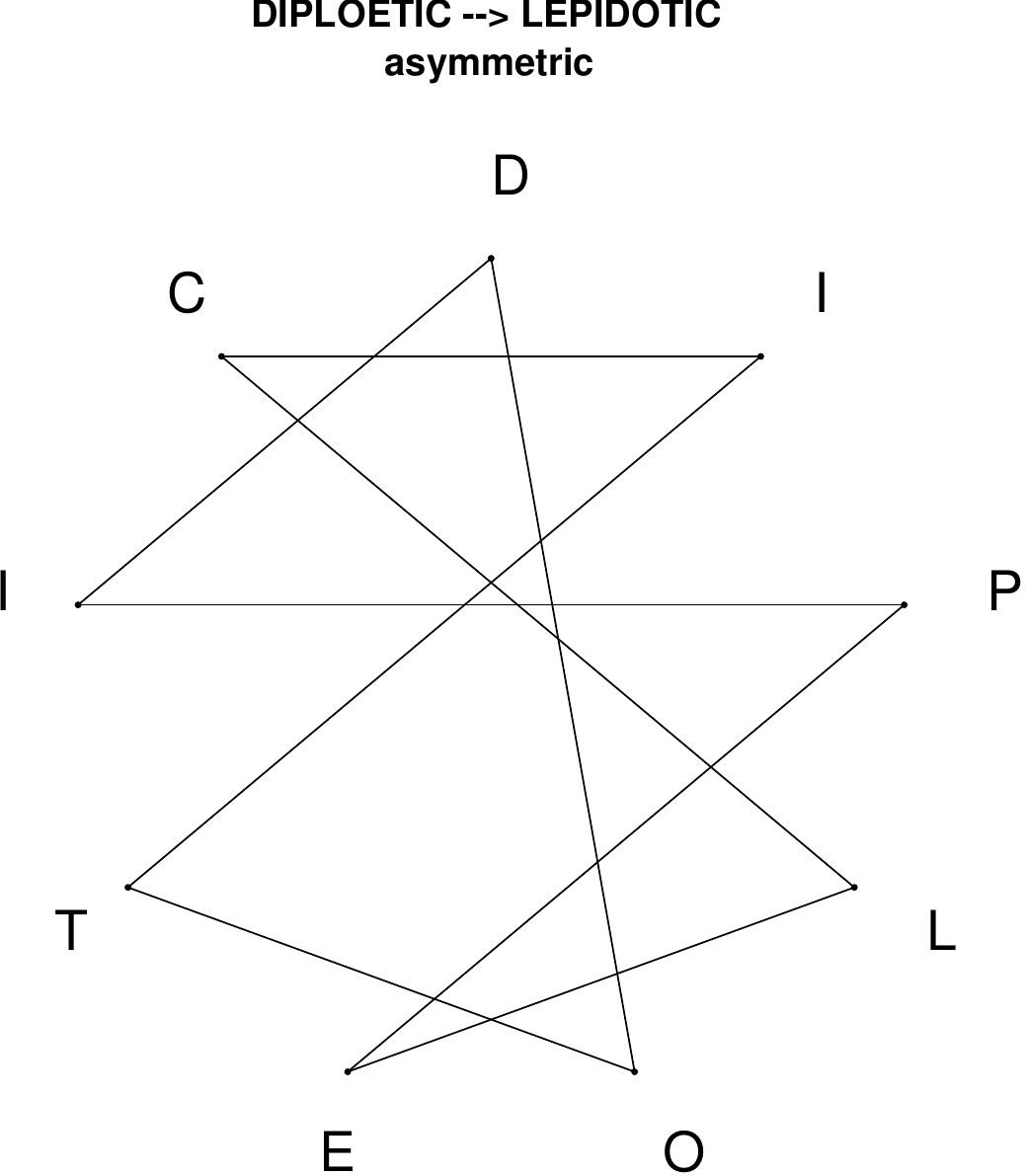}
\end{subfigure}
\hfill
\begin{subfigure}[T]{0.19\textwidth}
\centering
\includegraphics[width=\textwidth]{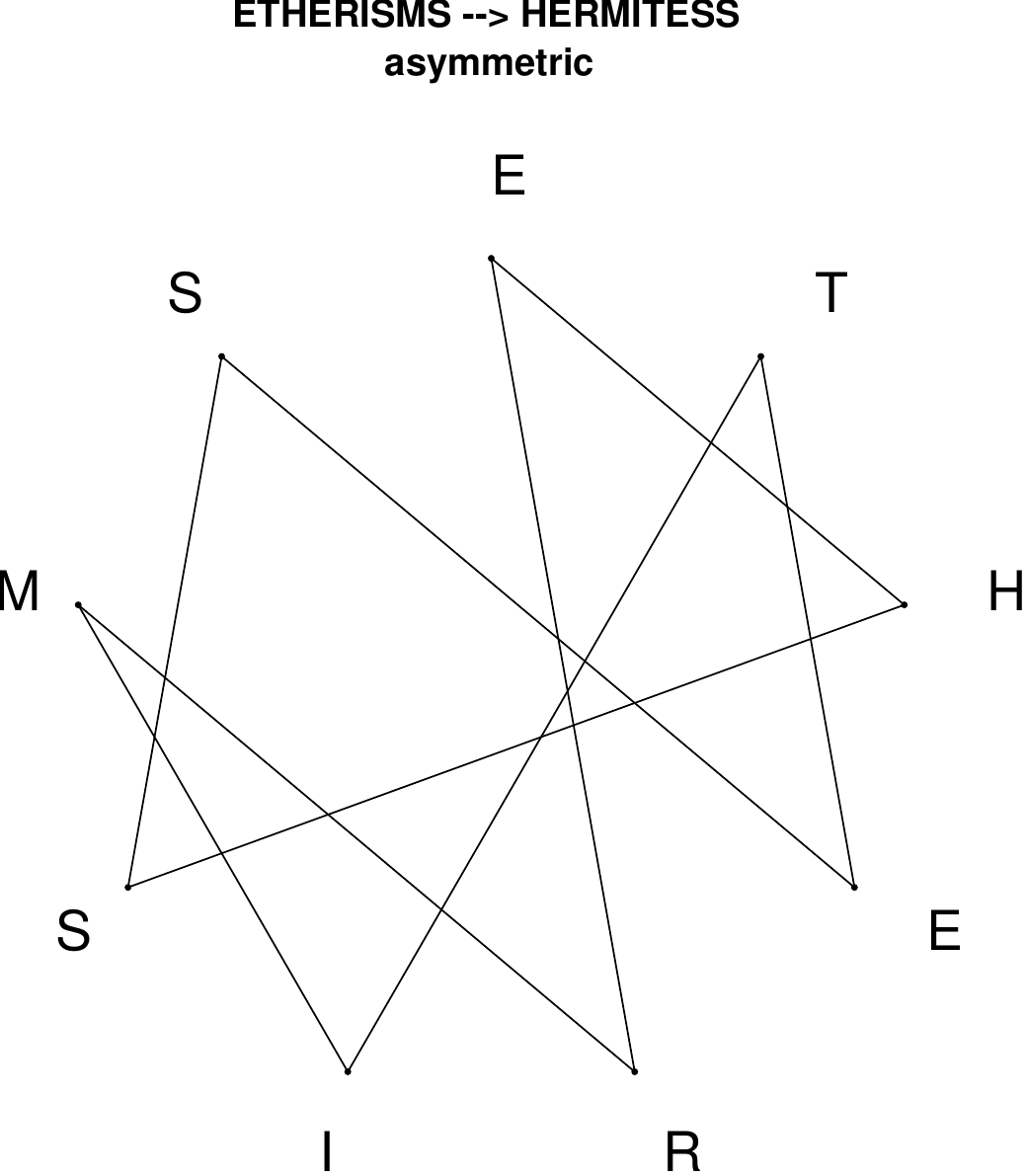}
\end{subfigure}
\hfill
\begin{subfigure}[T]{0.19\textwidth}
\centering
\includegraphics[width=\textwidth]{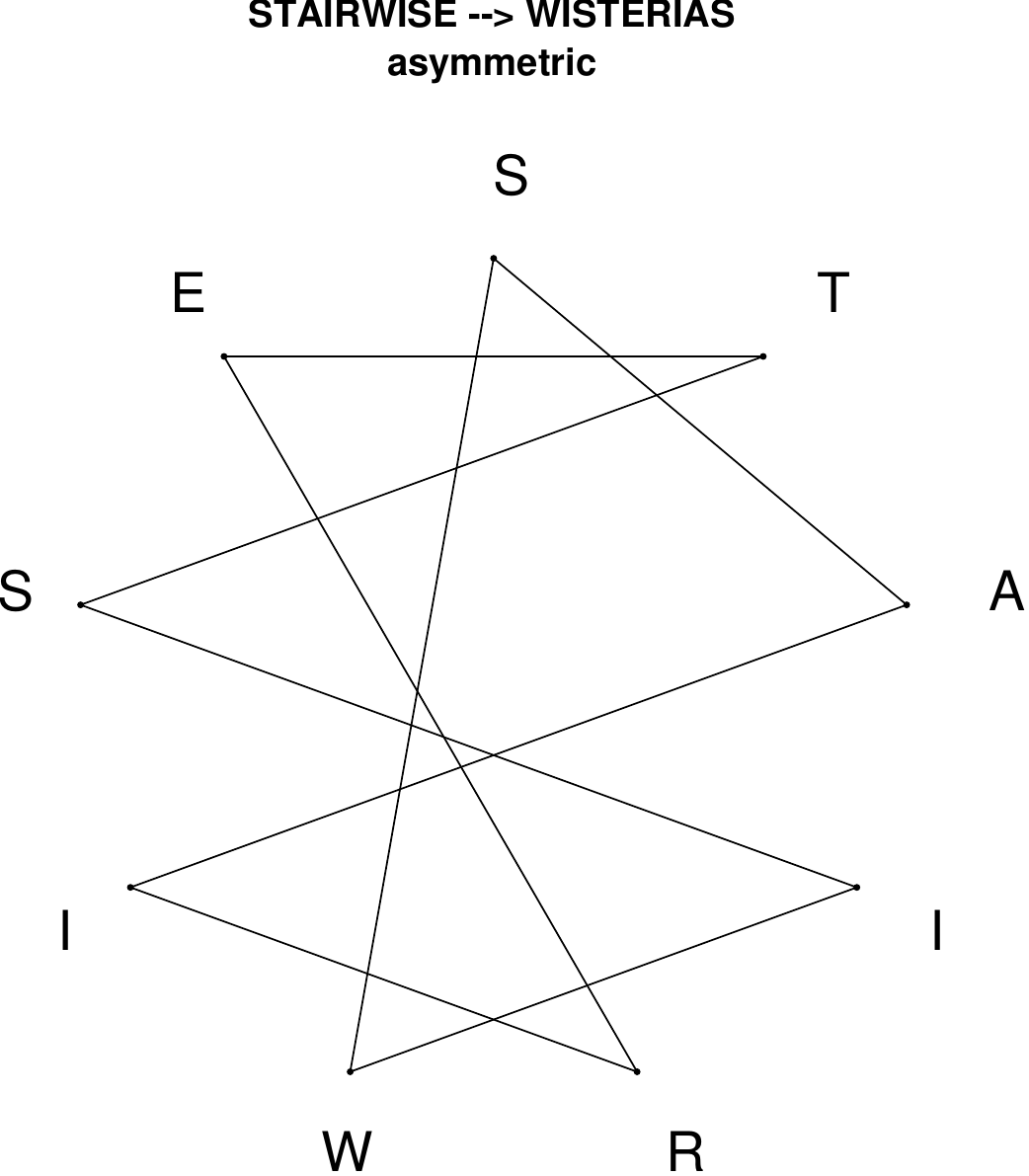}
\end{subfigure}
\hfill
\begin{subfigure}[T]{0.19\textwidth}
\centering
\includegraphics[width=\textwidth]{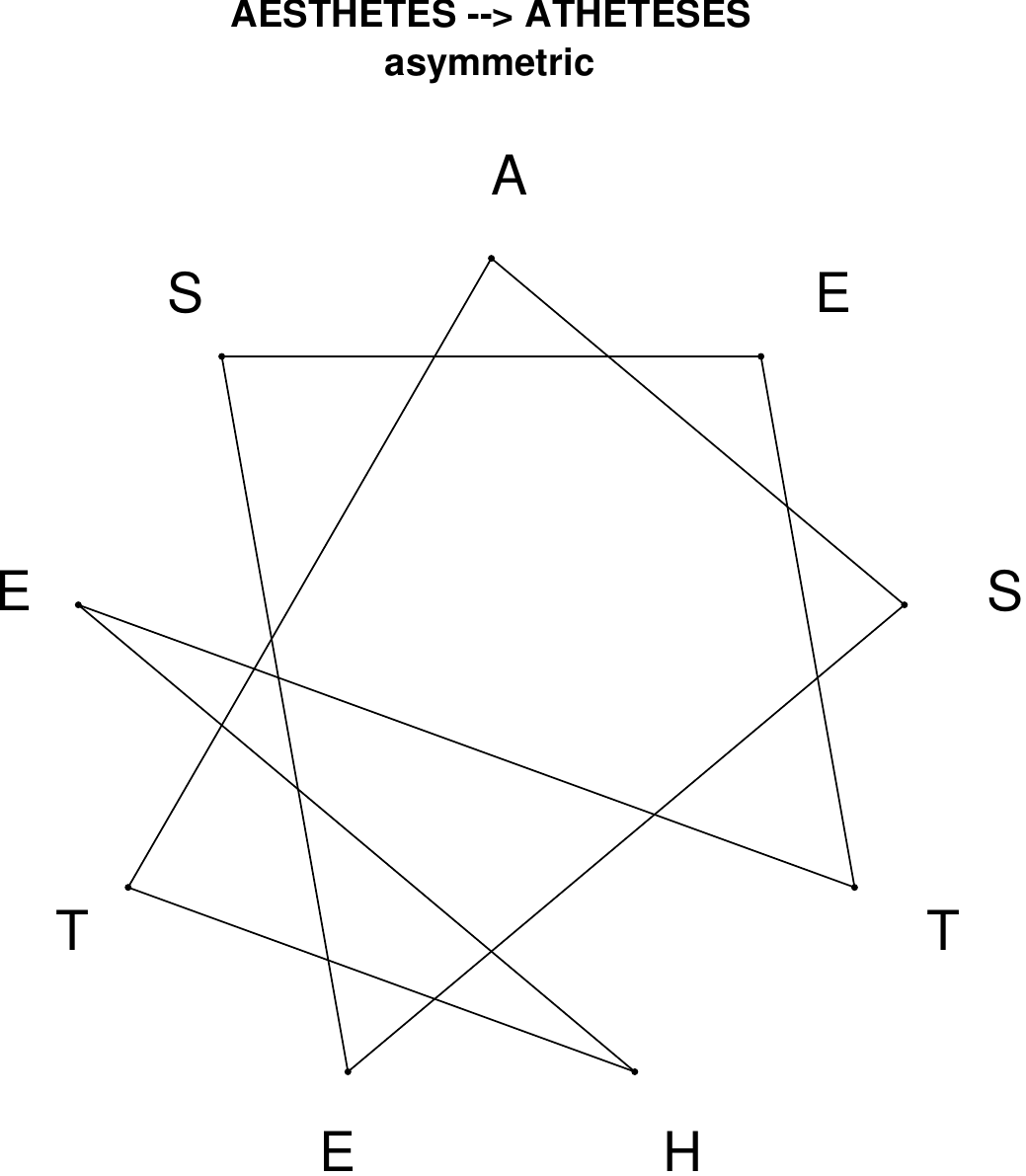}
\end{subfigure}
\hfill
\begin{subfigure}[T]{0.19\textwidth}
\centering
\includegraphics[width=\textwidth]{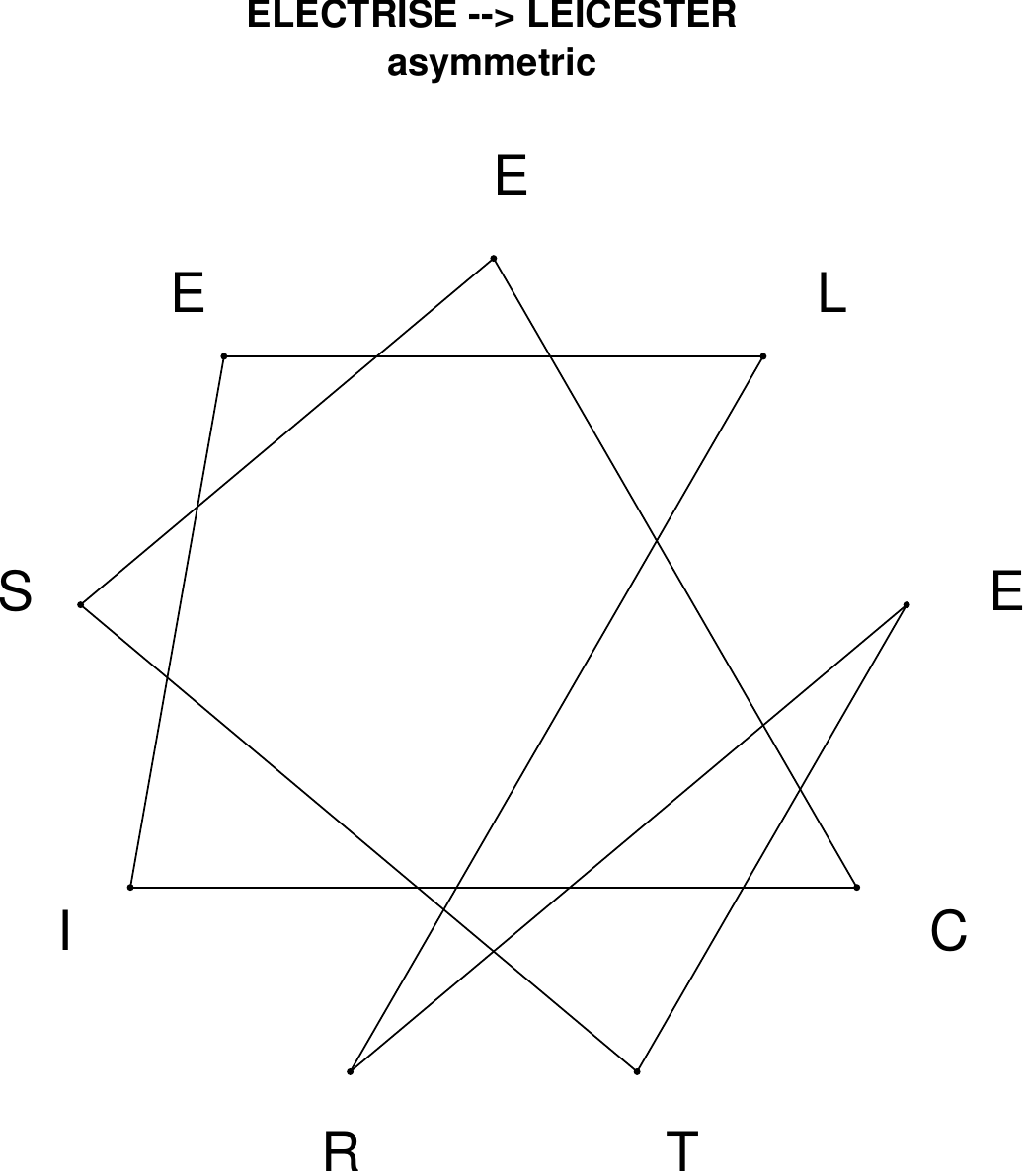}
\end{subfigure}
\end{figure}

\begin{figure}[H]
\centering
\begin{subfigure}[T]{0.19\textwidth}
\centering
\includegraphics[width=\textwidth]{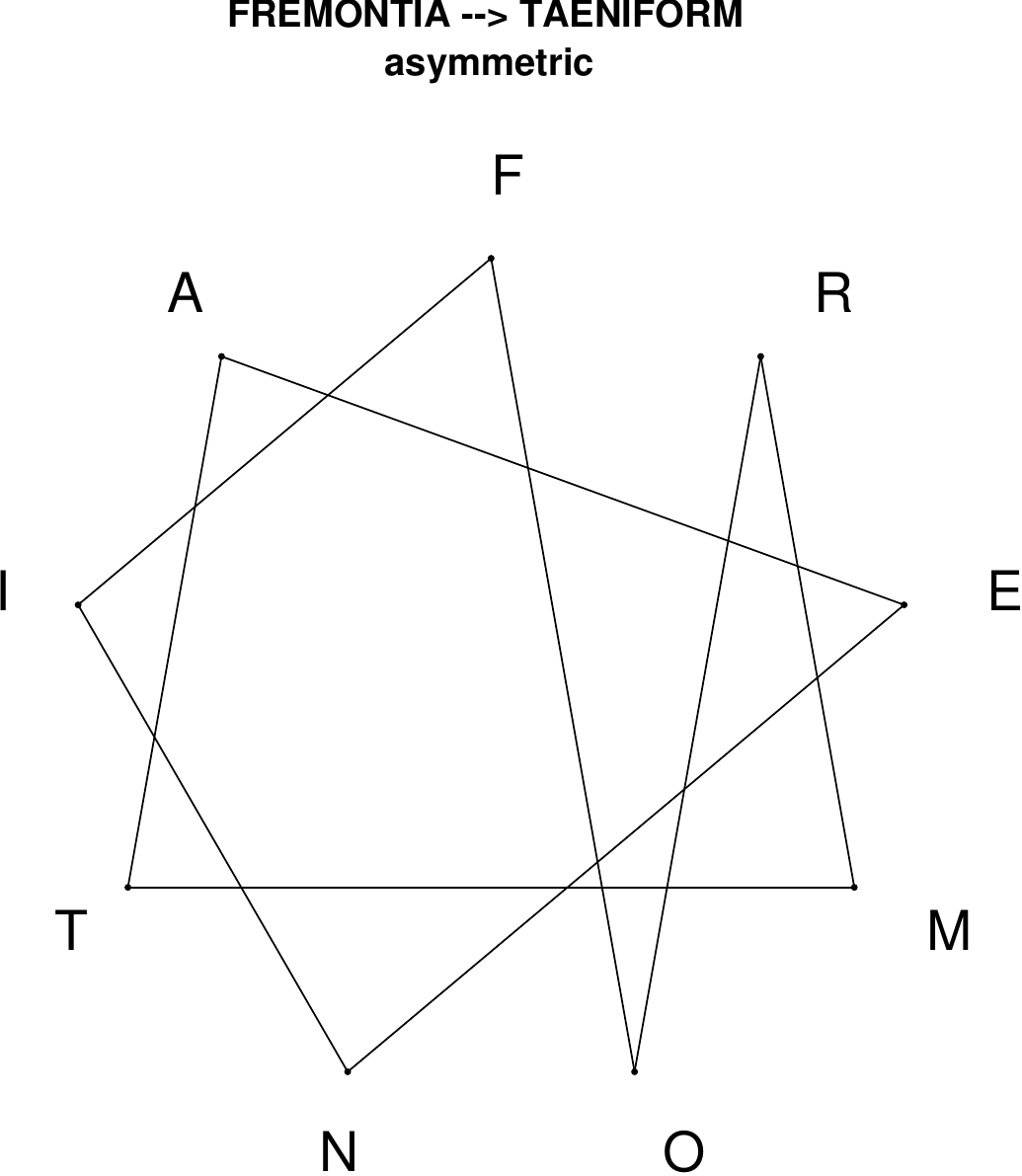}
\end{subfigure}
\hfill
\begin{subfigure}[T]{0.19\textwidth}
\centering
\includegraphics[width=\textwidth]{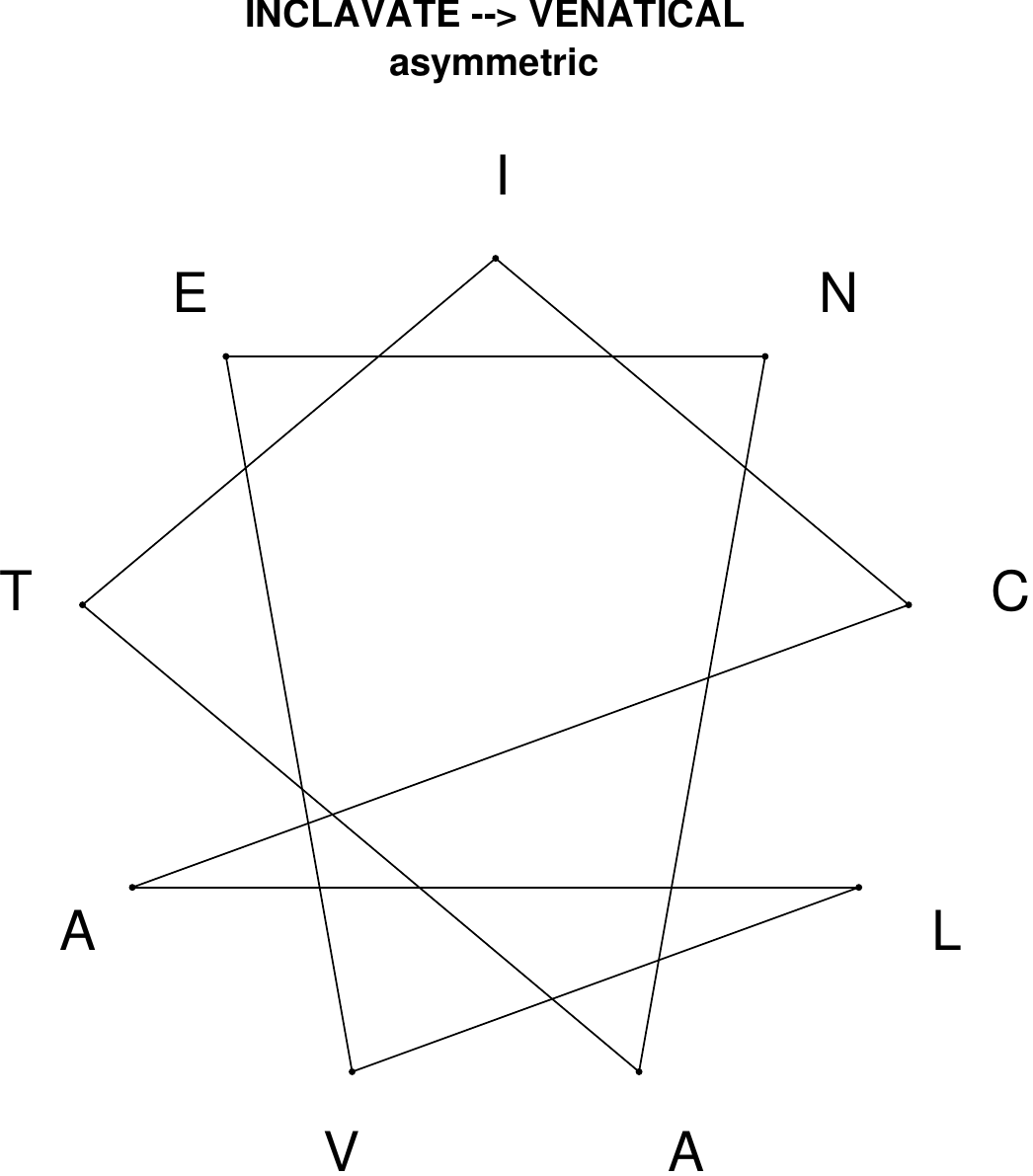}
\end{subfigure}
\hfill
\begin{subfigure}[T]{0.19\textwidth}
\centering
\includegraphics[width=\textwidth]{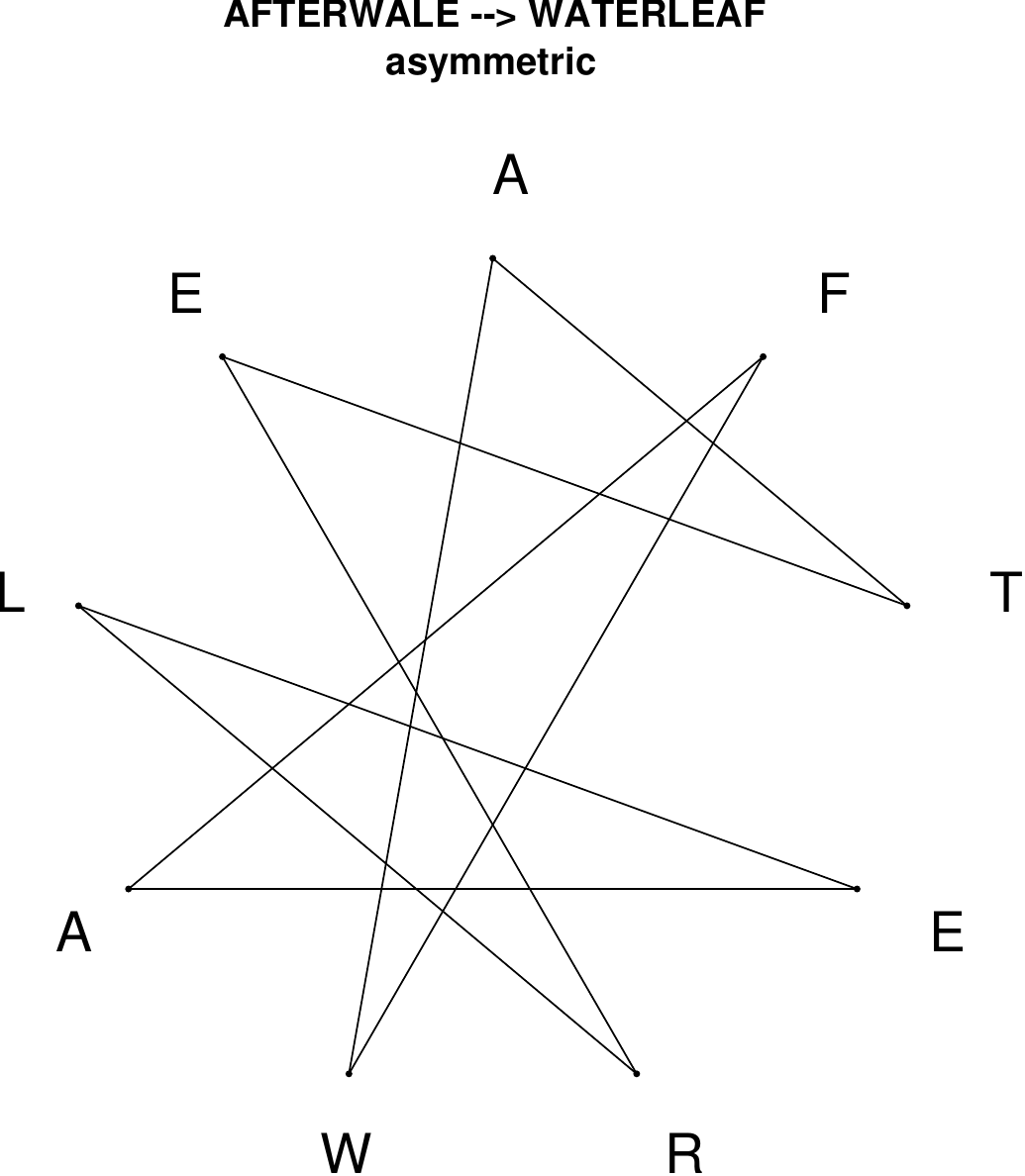}
\end{subfigure}
\hfill
\begin{subfigure}[T]{0.19\textwidth}
\centering
\includegraphics[width=\textwidth]{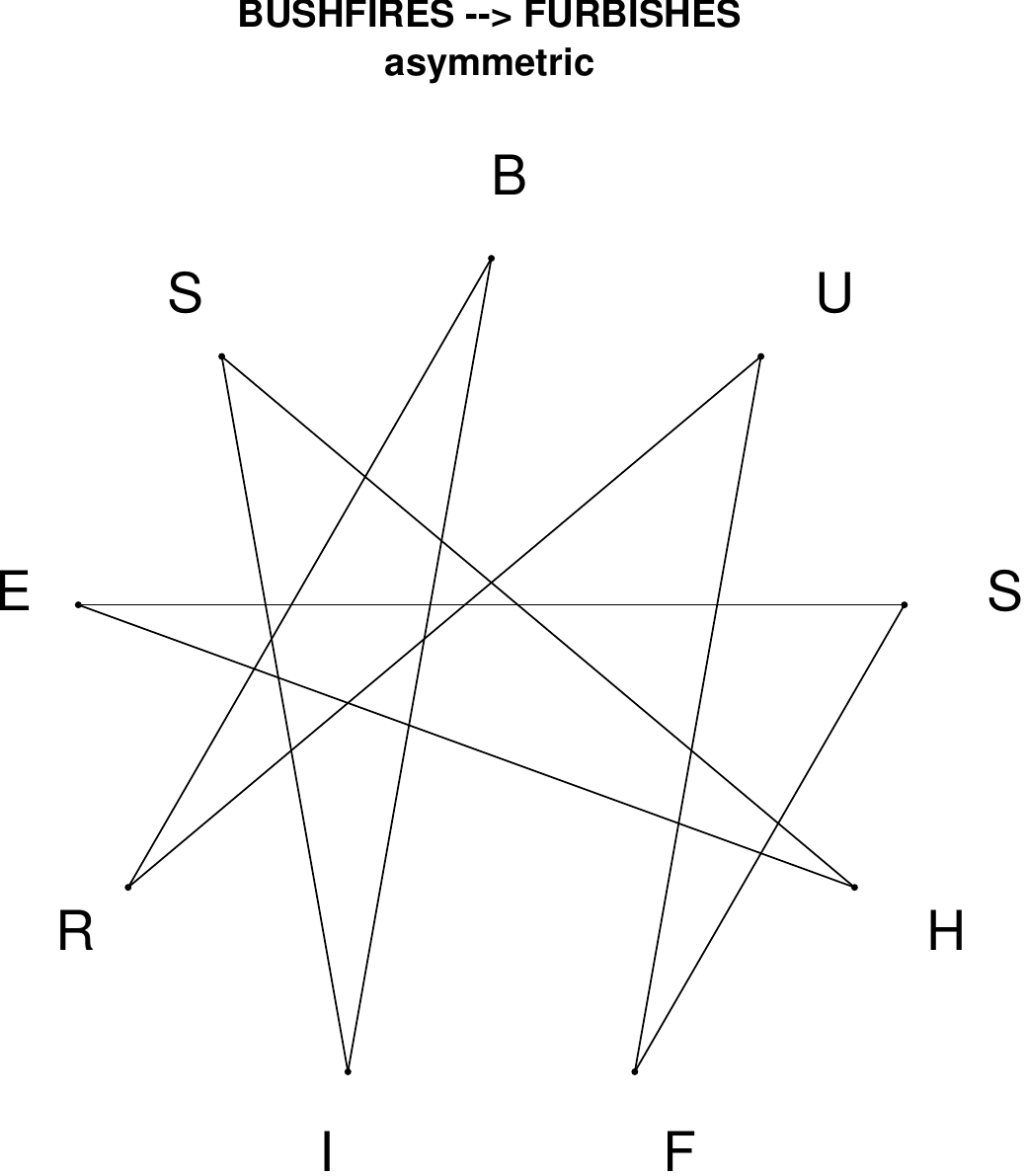}
\end{subfigure}
\hfill
\begin{subfigure}[T]{0.19\textwidth}
\centering
\includegraphics[width=\textwidth]{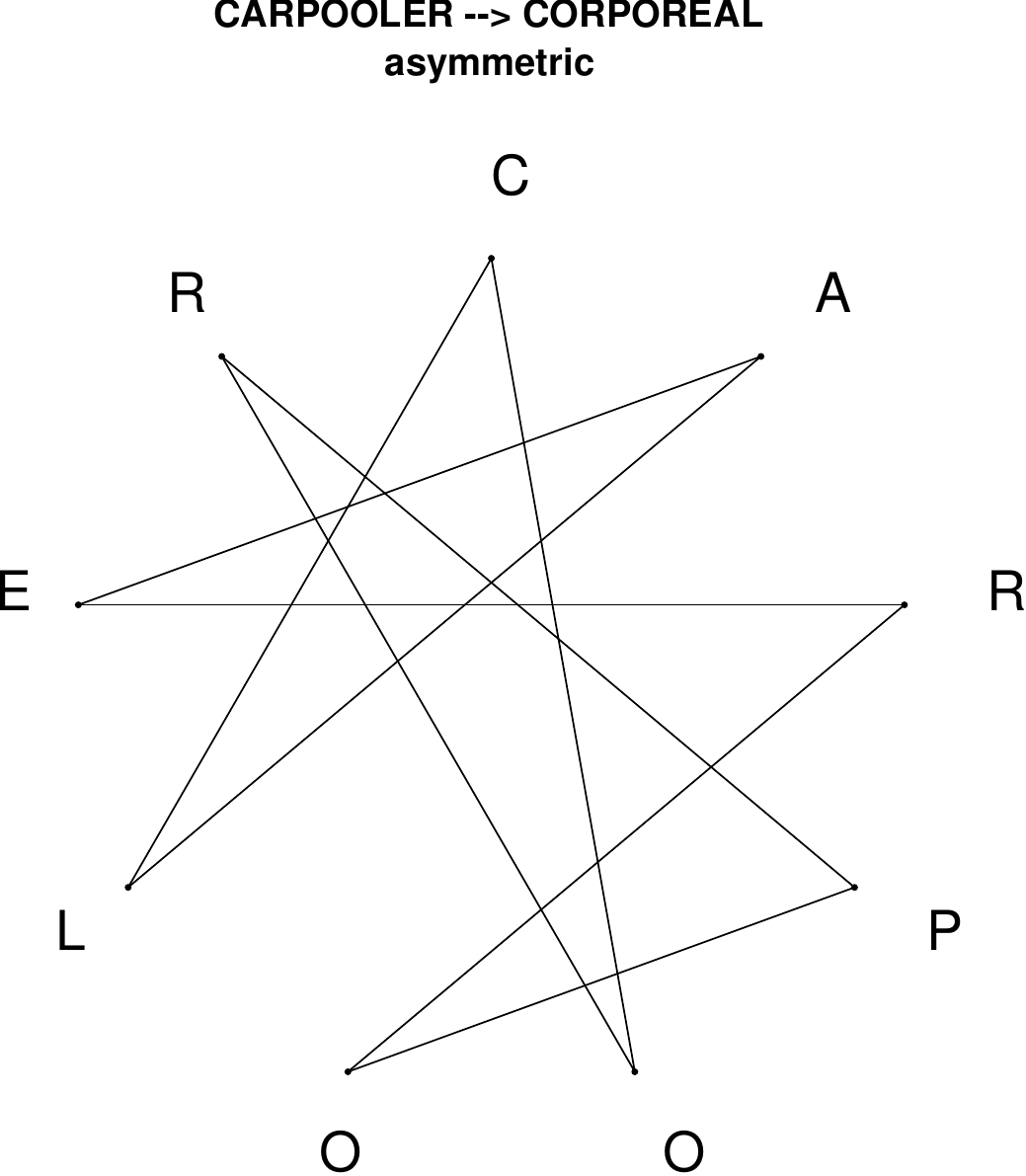}
\end{subfigure}
\end{figure}

\begin{figure}[H]
\centering
\begin{subfigure}[T]{0.19\textwidth}
\centering
\includegraphics[width=\textwidth]{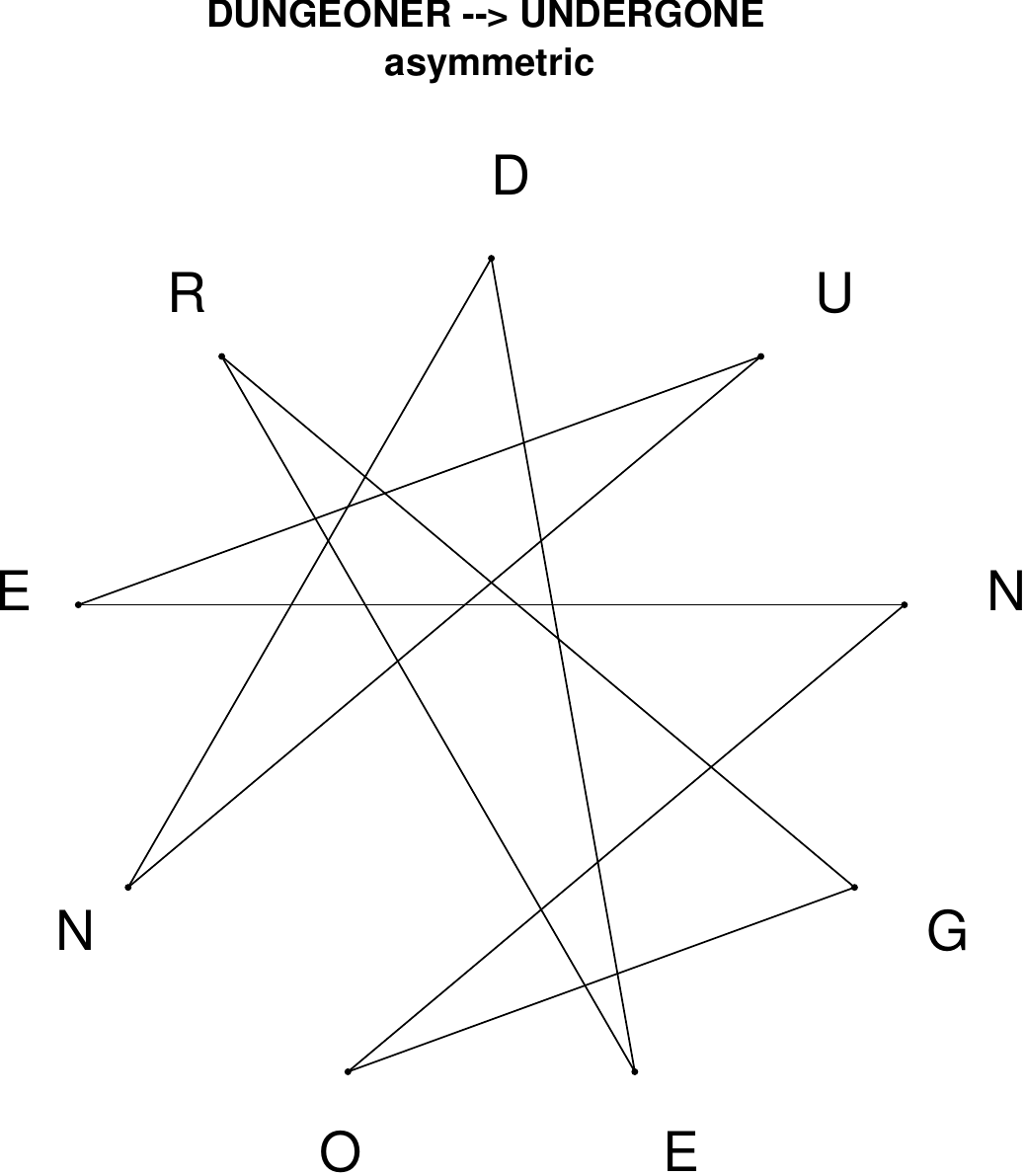}
\end{subfigure}
\hfill
\begin{subfigure}[T]{0.19\textwidth}
\centering
\includegraphics[width=\textwidth]{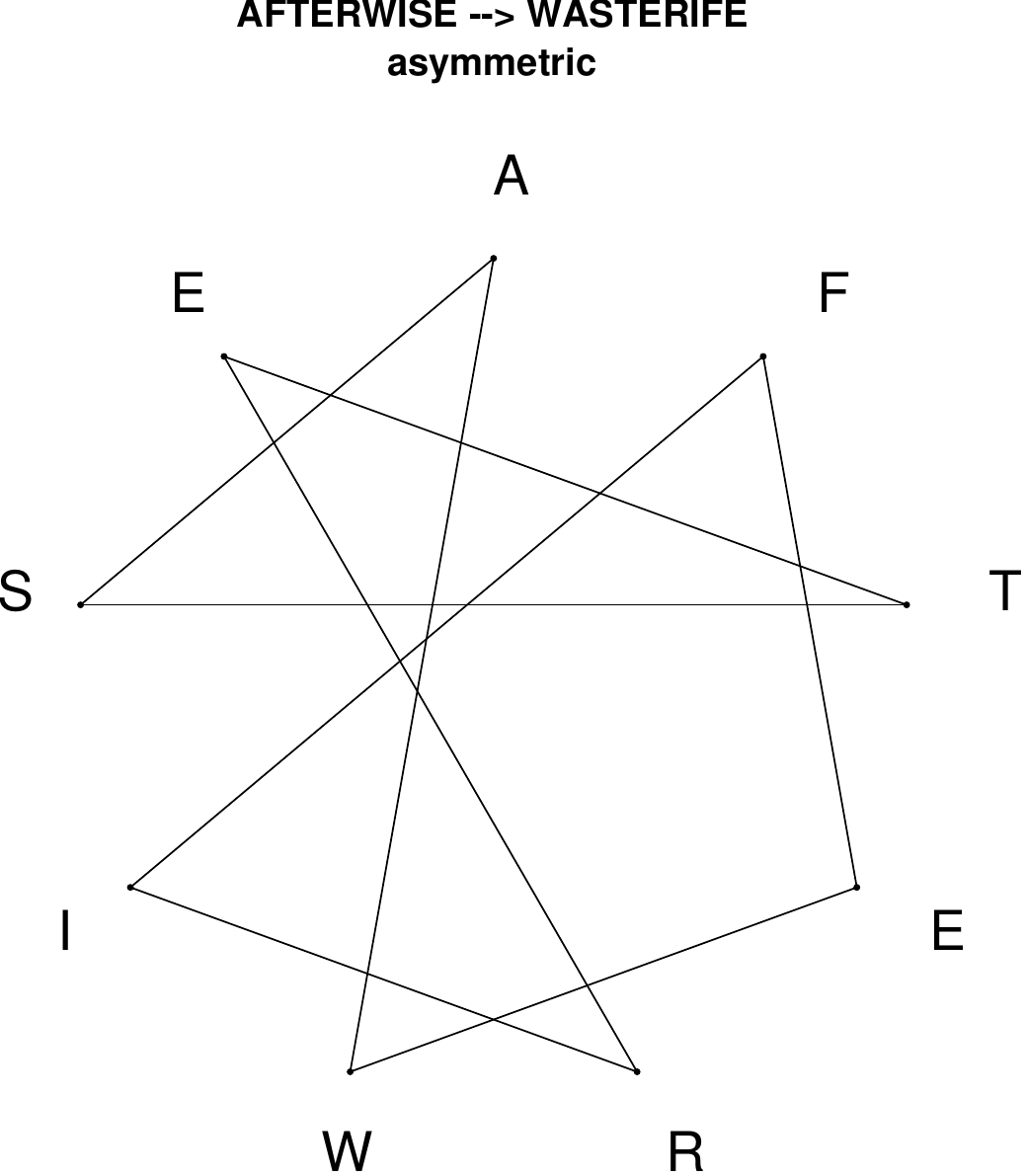}
\end{subfigure}
\hfill
\begin{subfigure}[T]{0.19\textwidth}
\centering
\includegraphics[width=\textwidth]{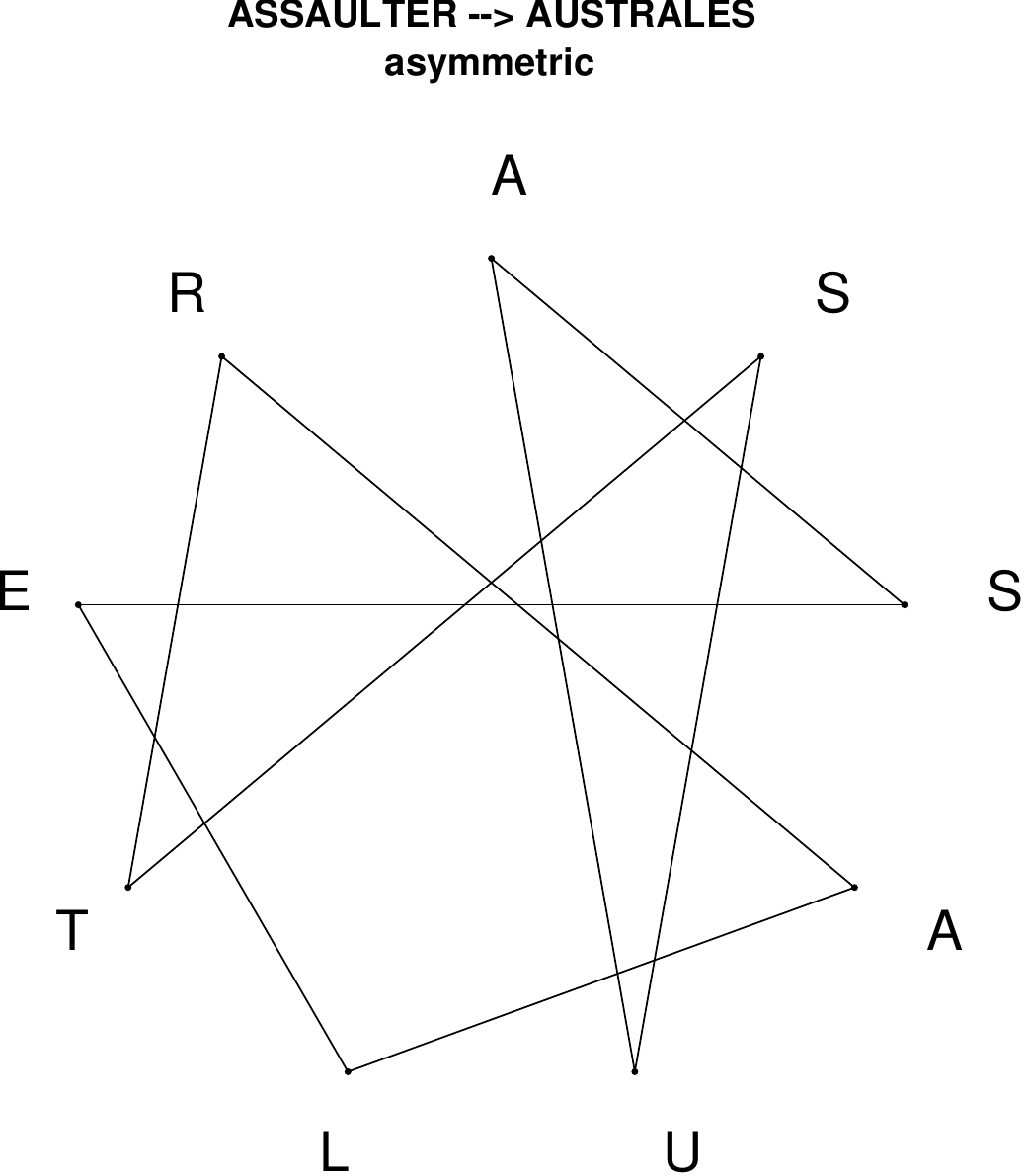}
\end{subfigure}
\hfill
\begin{subfigure}[T]{0.19\textwidth}
\centering
\includegraphics[width=\textwidth]{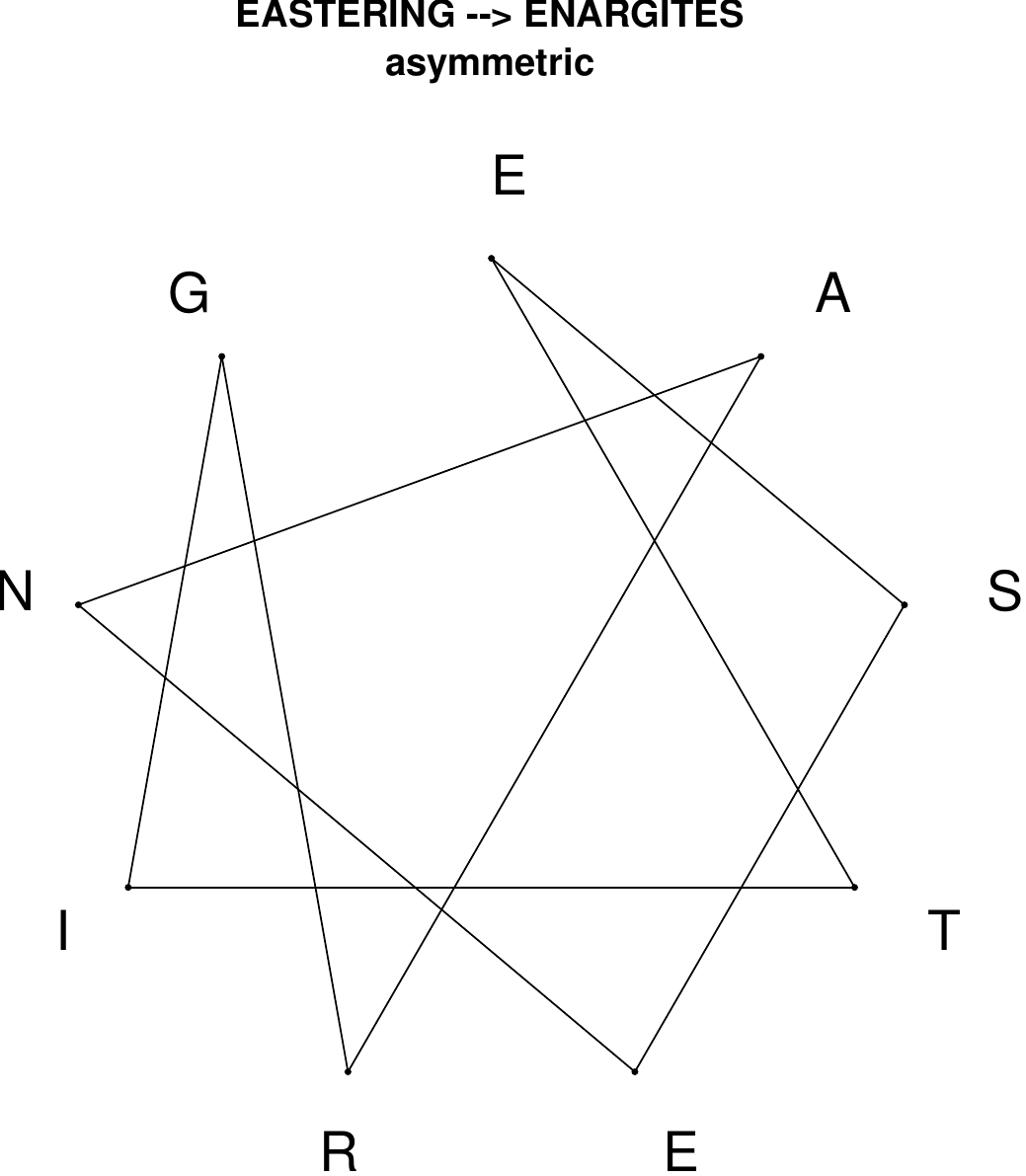}
\end{subfigure}
\hfill
\begin{subfigure}[T]{0.19\textwidth}
\centering
\includegraphics[width=\textwidth]{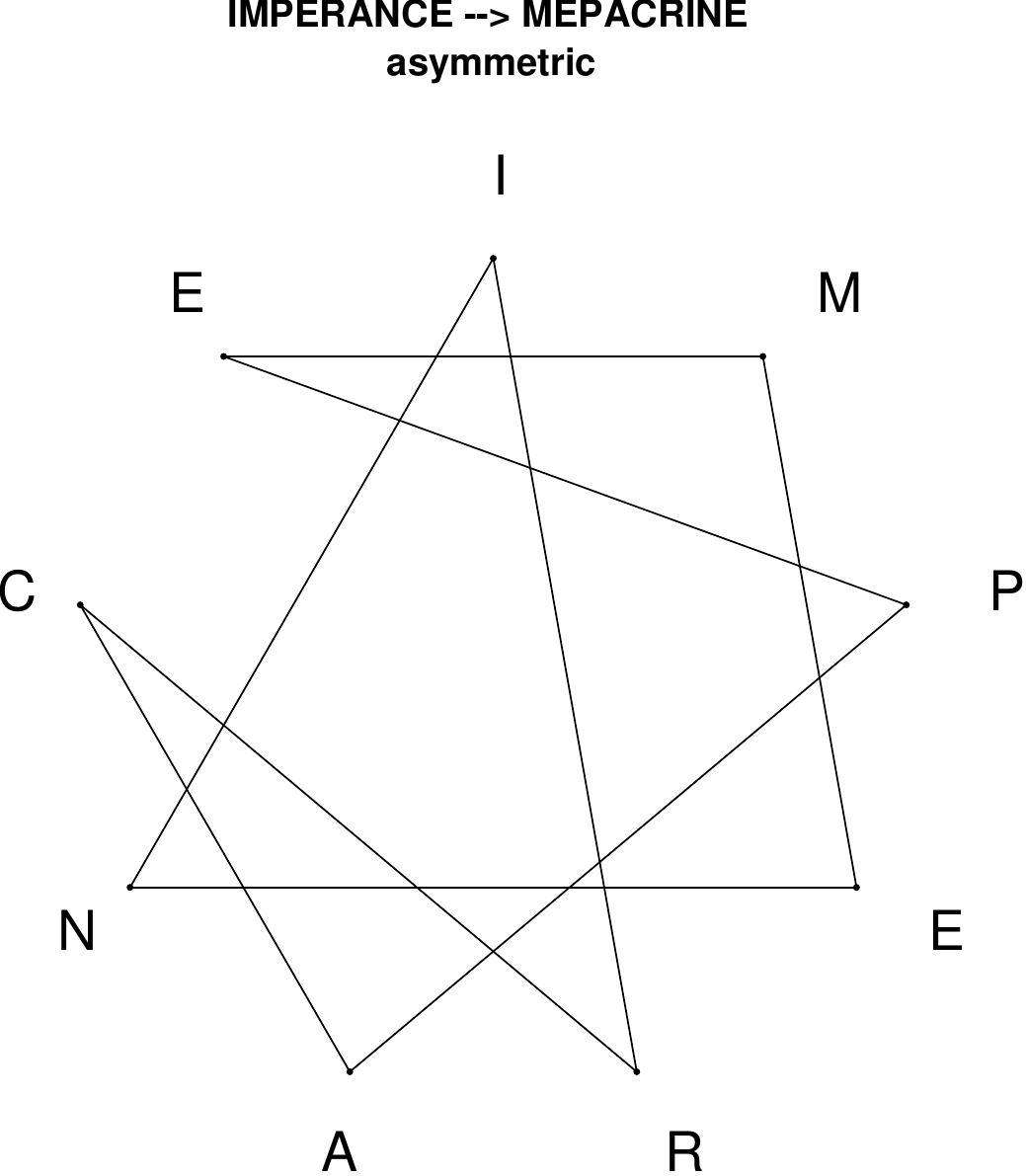}
\end{subfigure}
\end{figure}

\begin{figure}[H]
\centering
\begin{subfigure}[T]{0.19\textwidth}
\centering
\includegraphics[width=\textwidth]{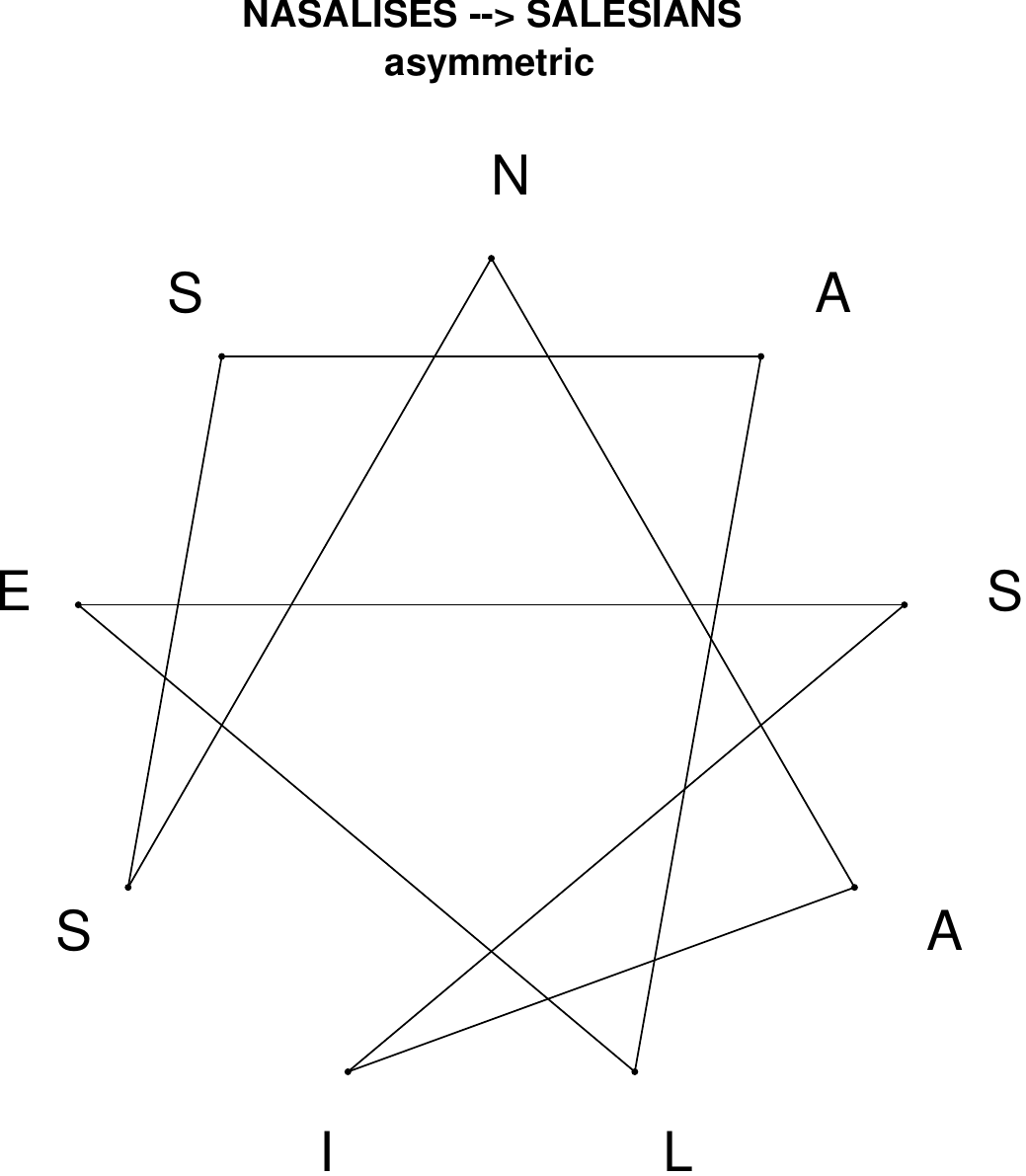}
\end{subfigure}
\hfill
\begin{subfigure}[T]{0.19\textwidth}
\centering
\includegraphics[width=\textwidth]{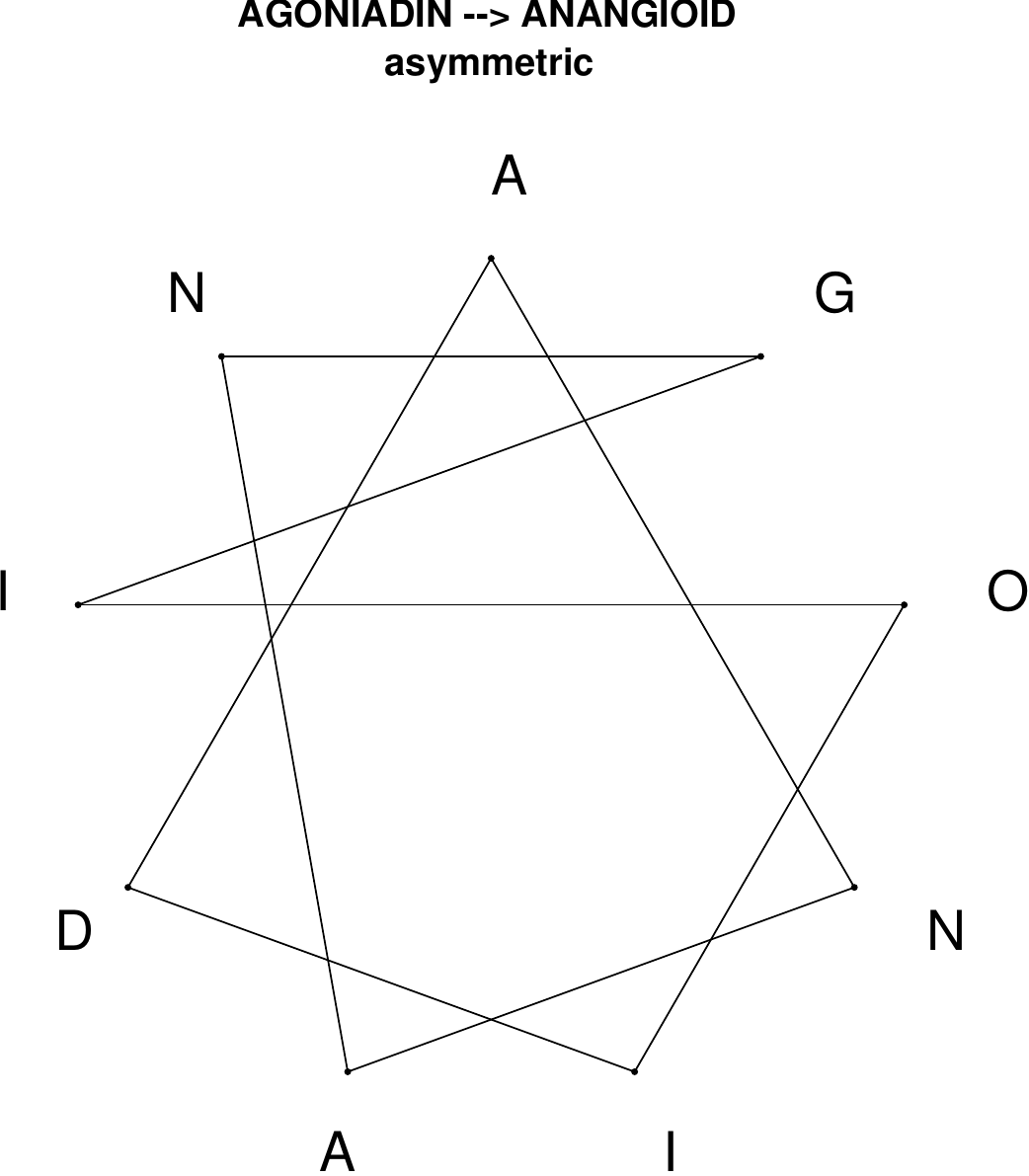}
\end{subfigure}
\hfill
\begin{subfigure}[T]{0.19\textwidth}
\centering
\includegraphics[width=\textwidth]{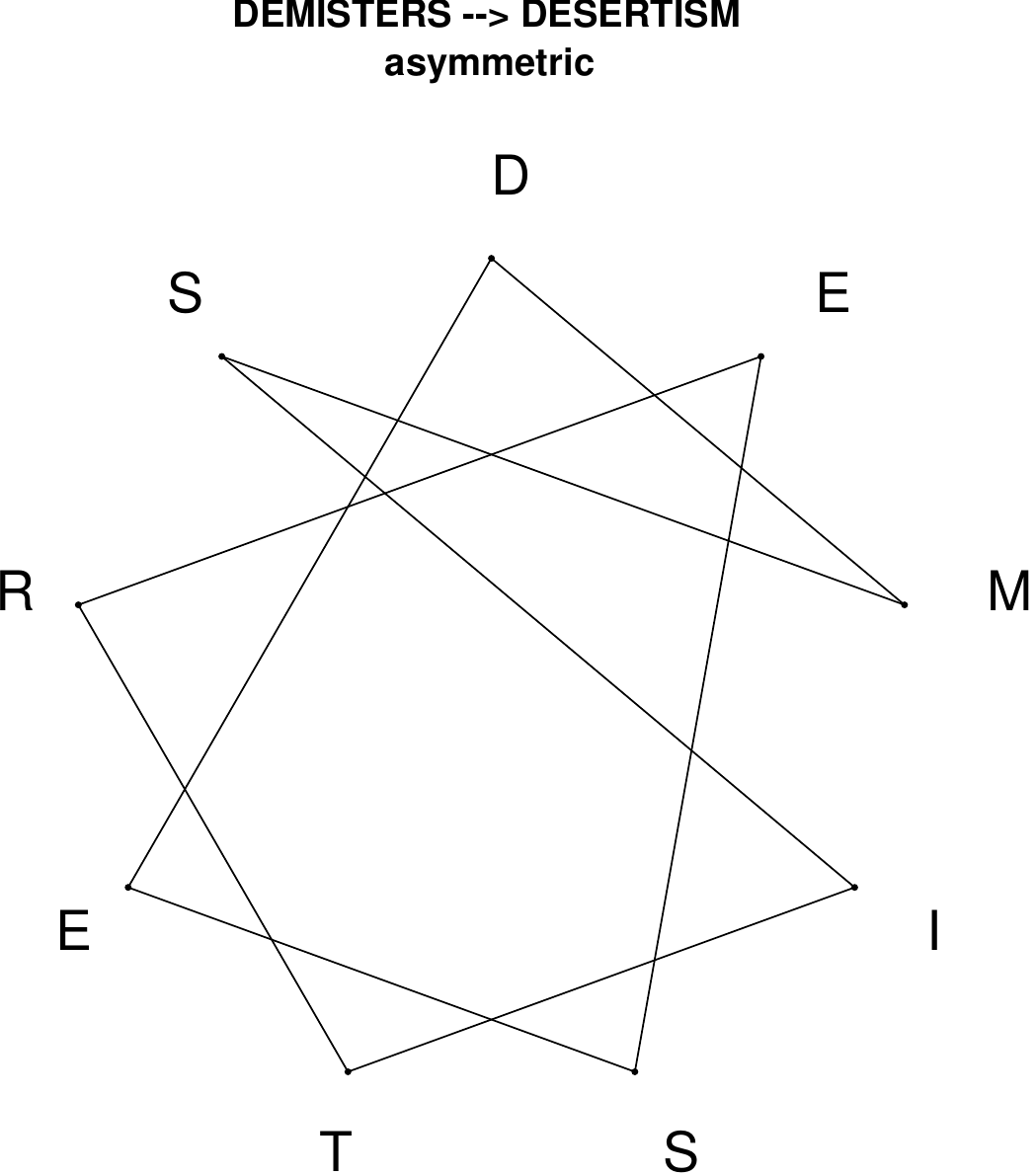}
\end{subfigure}
\hfill
\begin{subfigure}[T]{0.19\textwidth}
\centering
\includegraphics[width=\textwidth]{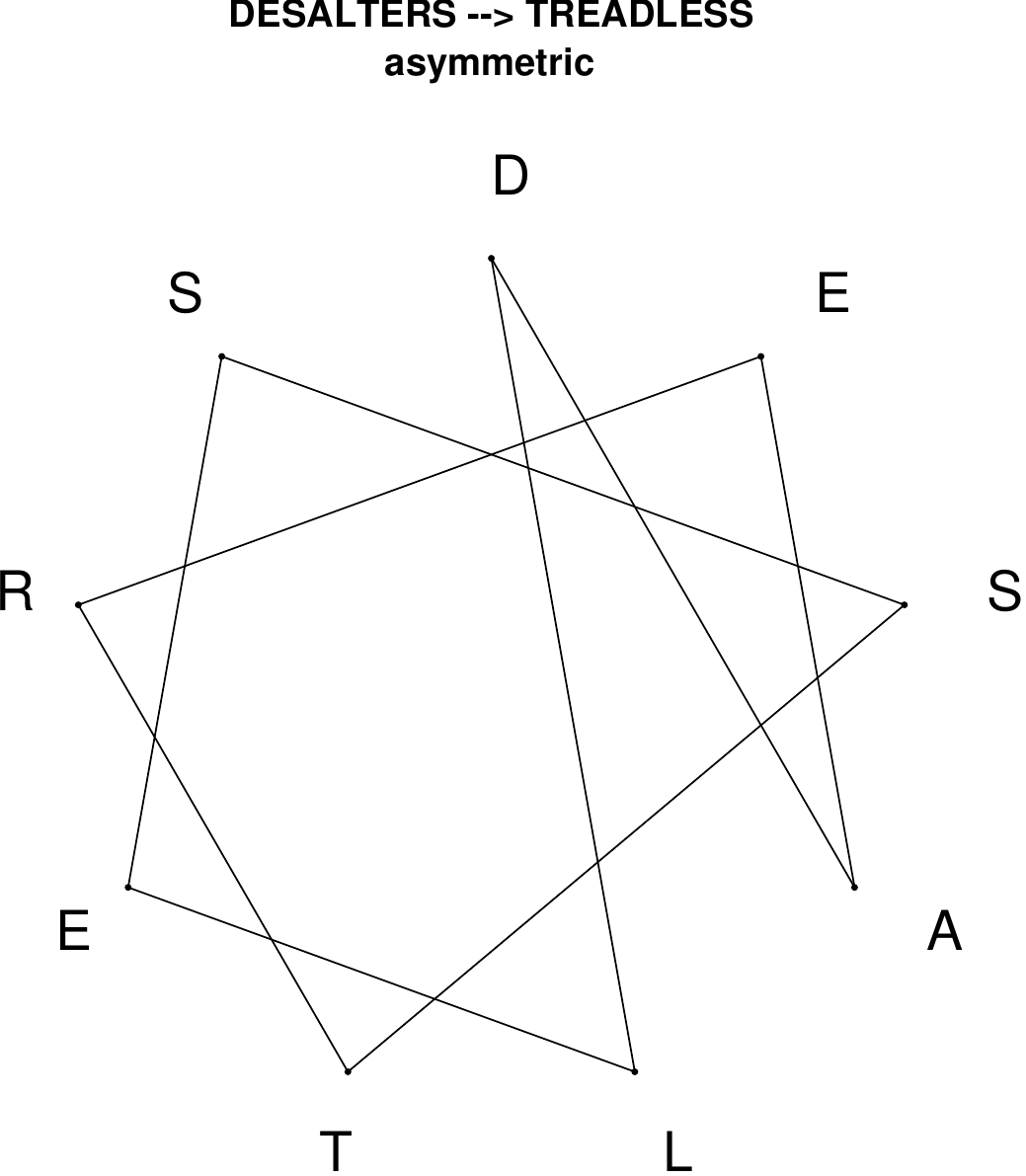}
\end{subfigure}
\hfill
\begin{subfigure}[T]{0.19\textwidth}
\centering
\includegraphics[width=\textwidth]{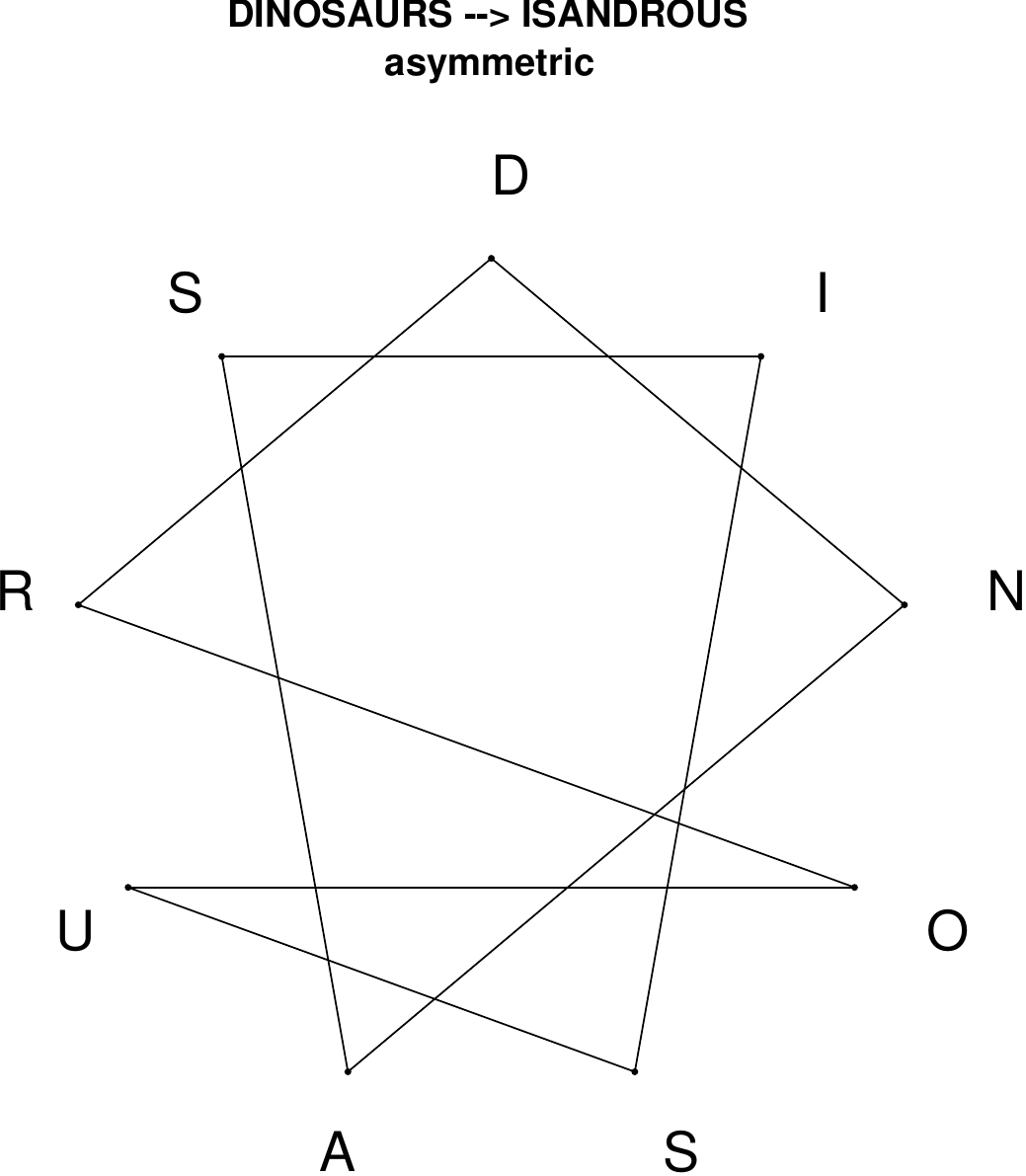}
\end{subfigure}
\end{figure}

\begin{figure}[H]
\centering
\begin{subfigure}[T]{0.19\textwidth}
\centering
\includegraphics[width=\textwidth]{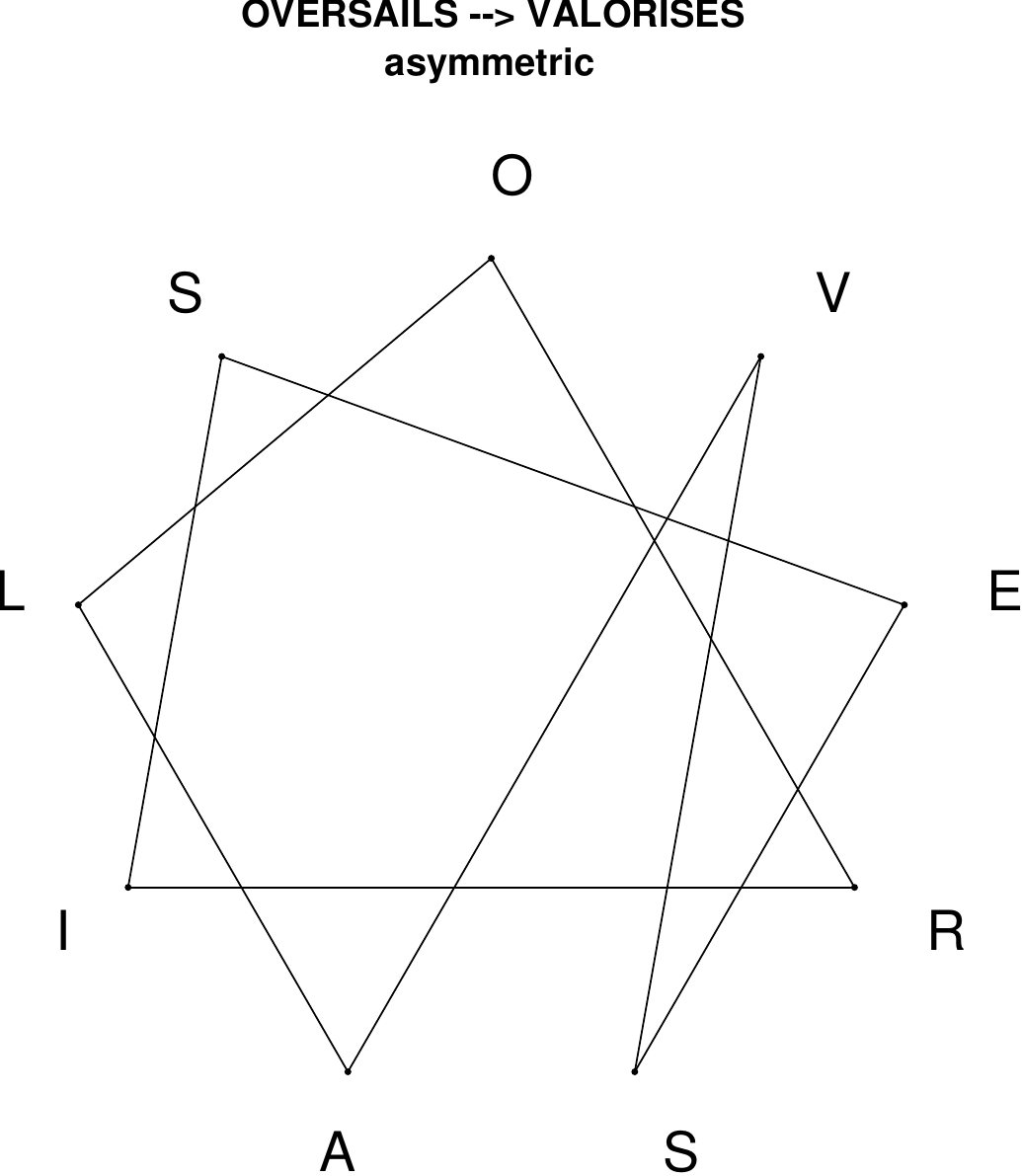}
\end{subfigure}
\hfill
\begin{subfigure}[T]{0.19\textwidth}
\centering
\includegraphics[width=\textwidth]{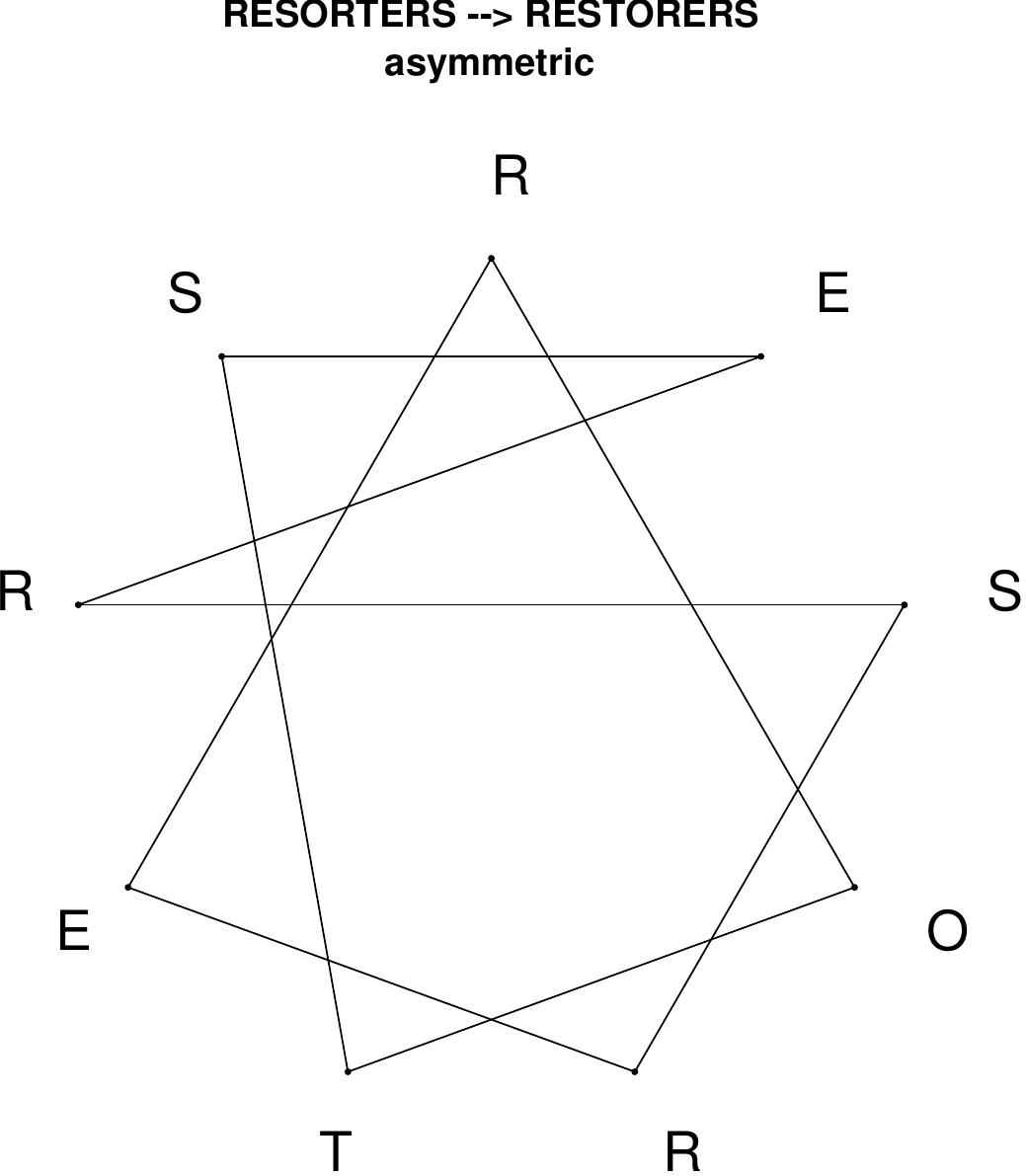}
\end{subfigure}
\hfill
\begin{subfigure}[T]{0.19\textwidth}
\centering
\includegraphics[width=\textwidth]{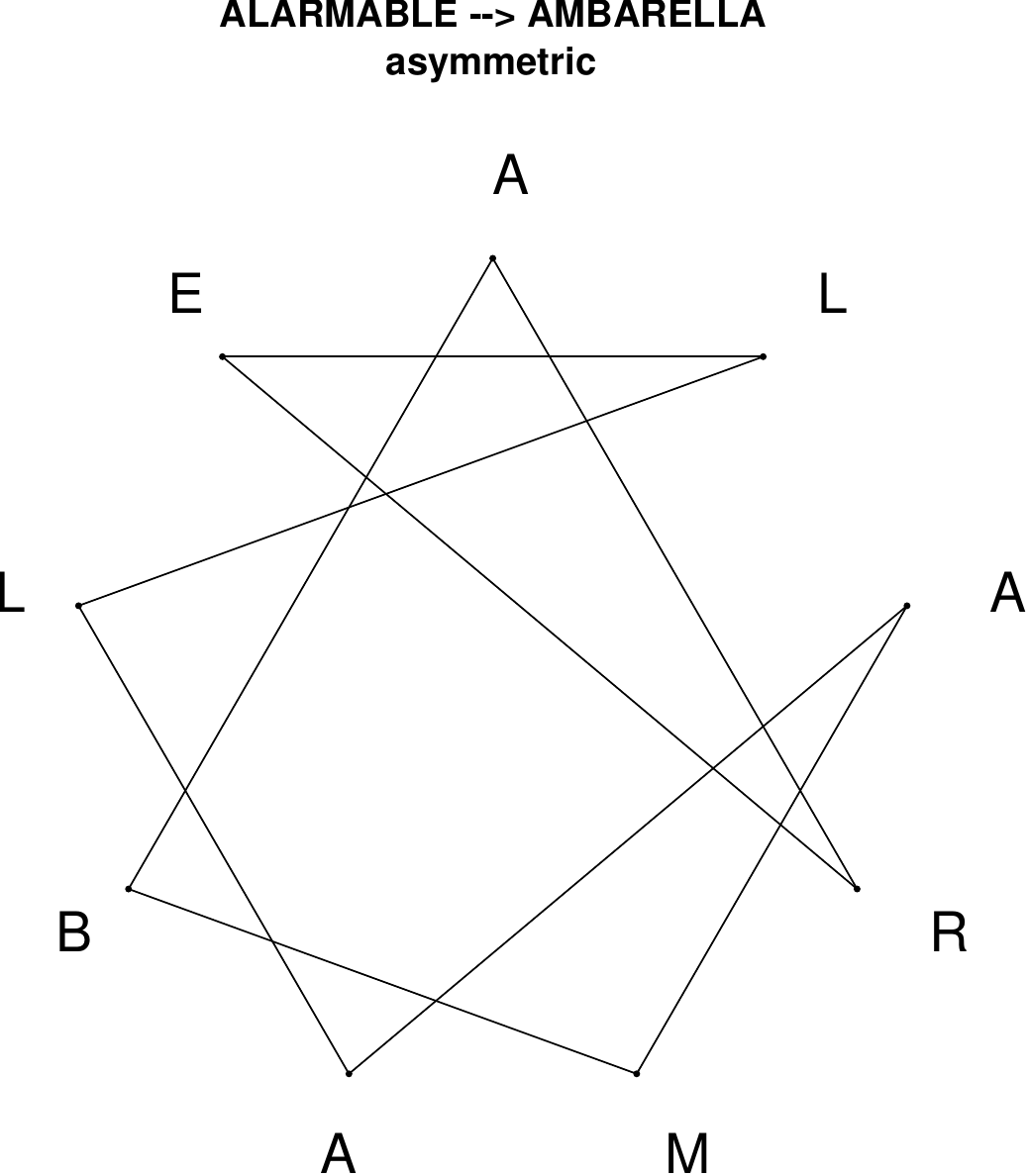}
\end{subfigure}
\hfill
\begin{subfigure}[T]{0.19\textwidth}
\centering
\includegraphics[width=\textwidth]{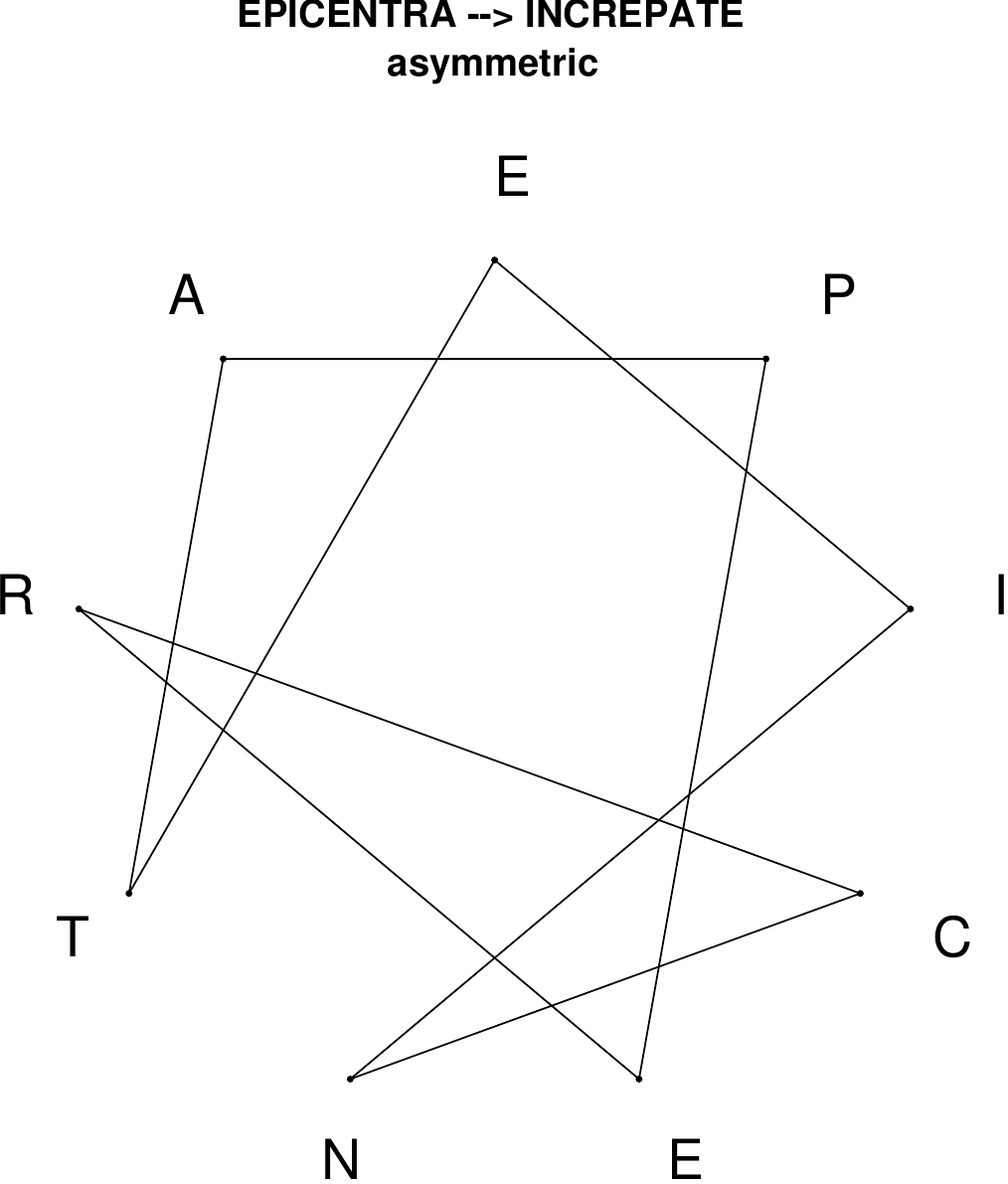}
\end{subfigure}
\hfill
\begin{subfigure}[T]{0.19\textwidth}
\centering
\includegraphics[width=\textwidth]{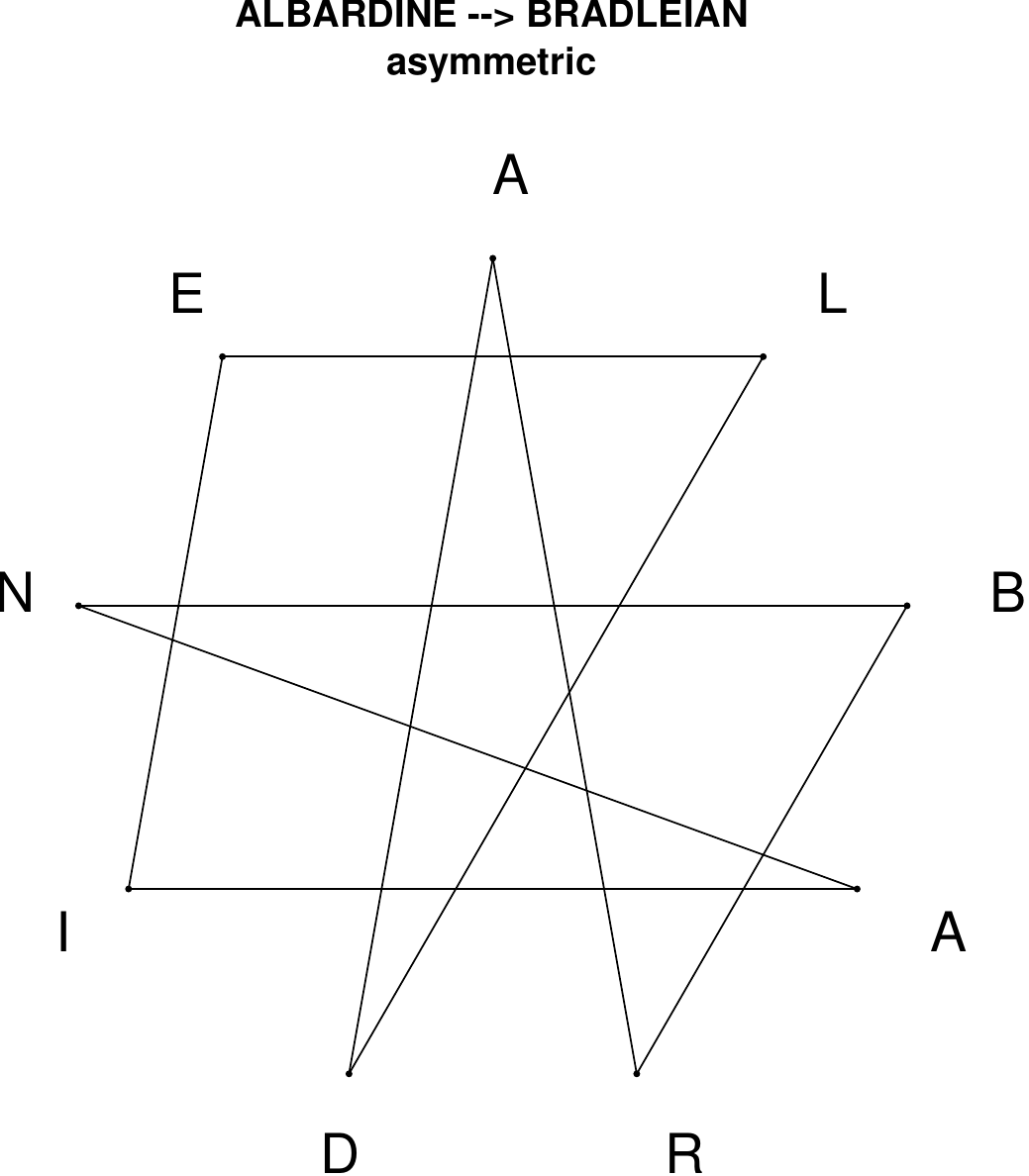}
\end{subfigure}
\end{figure}

\begin{figure}[H]
\centering
\begin{subfigure}[T]{0.19\textwidth}
\centering
\includegraphics[width=\textwidth]{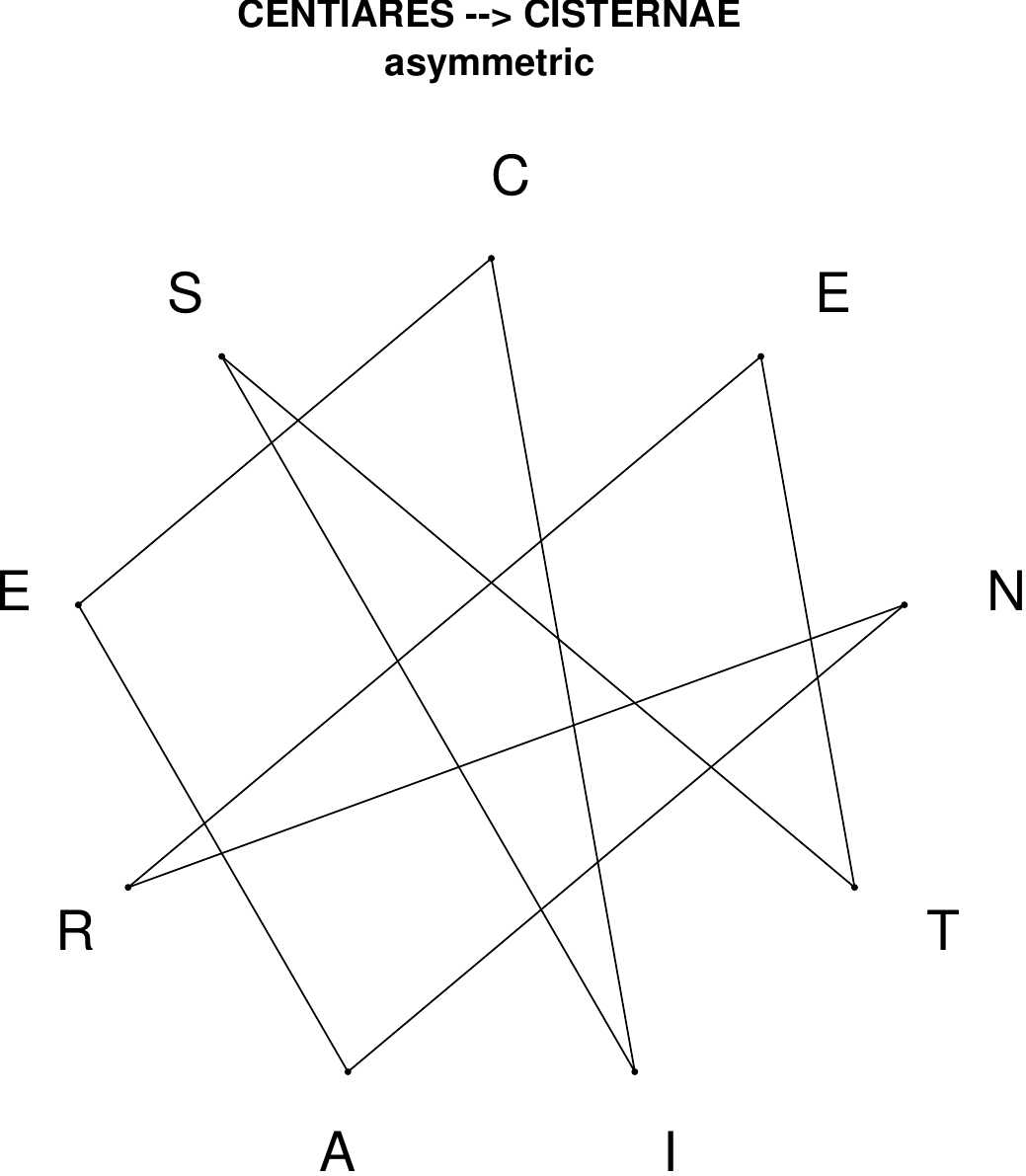}
\end{subfigure}
\hfill
\begin{subfigure}[T]{0.19\textwidth}
\centering
\includegraphics[width=\textwidth]{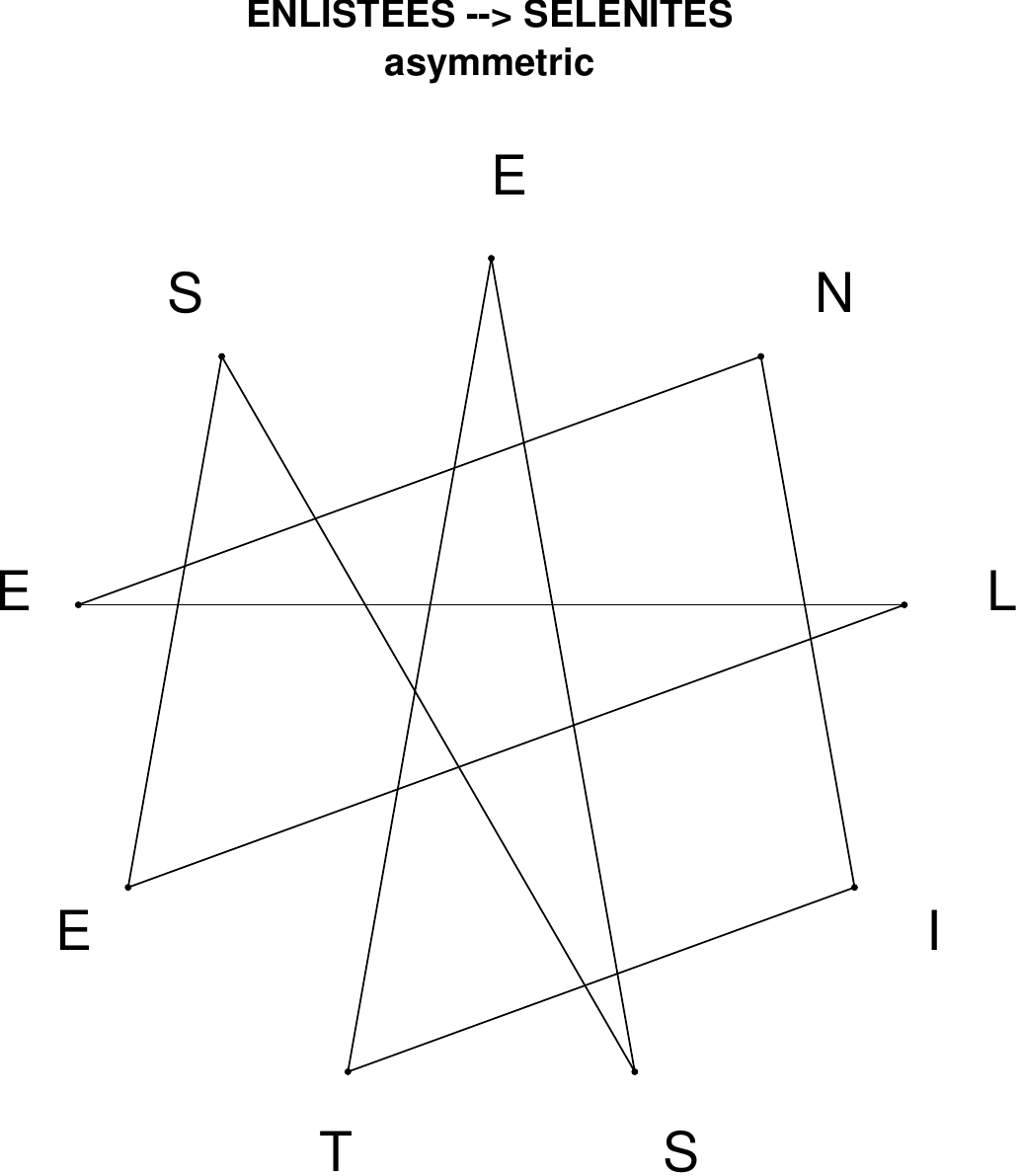}
\end{subfigure}
\hfill
\begin{subfigure}[T]{0.19\textwidth}
\centering
\includegraphics[width=\textwidth]{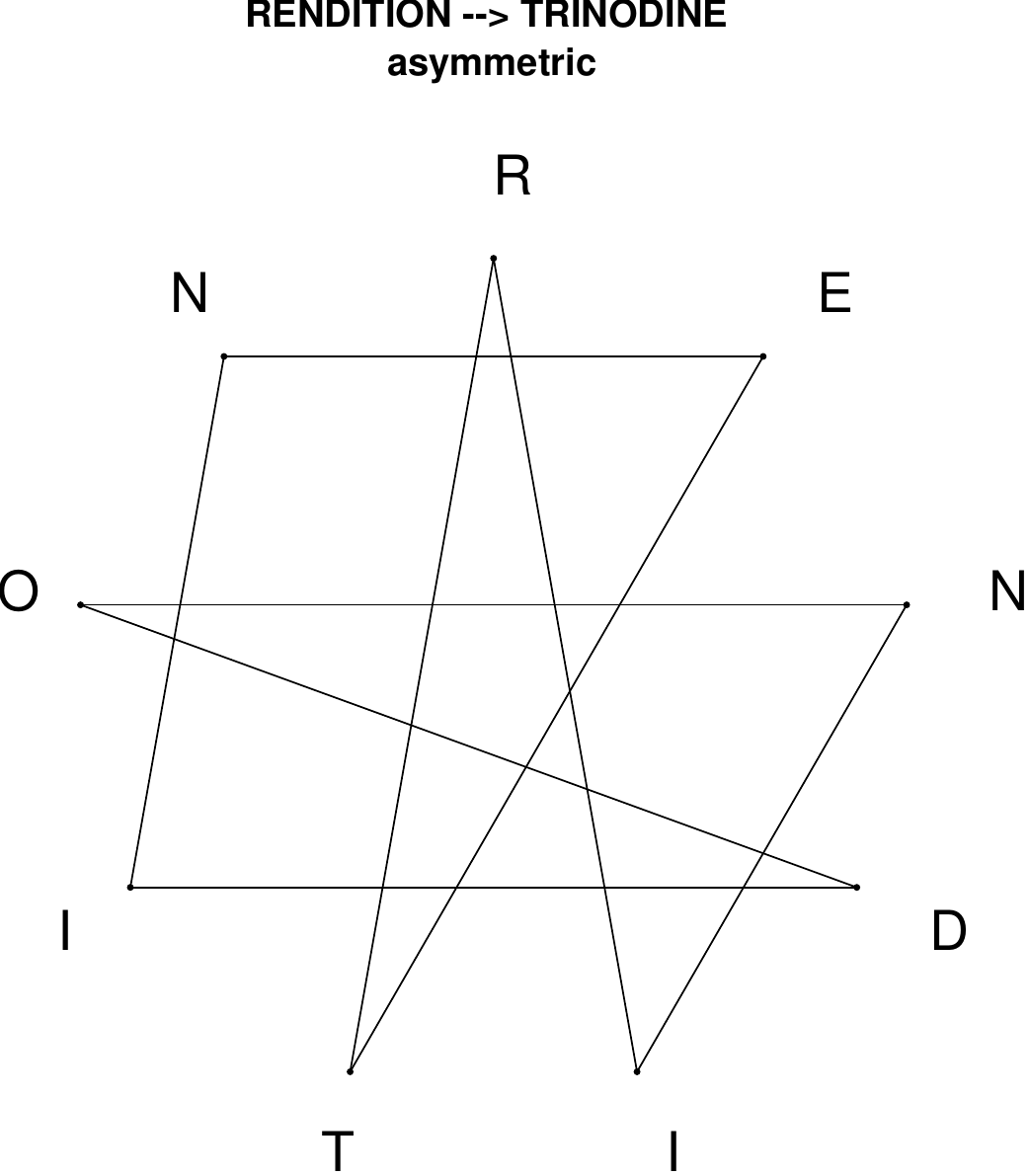}
\end{subfigure}
\hfill
\begin{subfigure}[T]{0.19\textwidth}
\centering
\includegraphics[width=\textwidth]{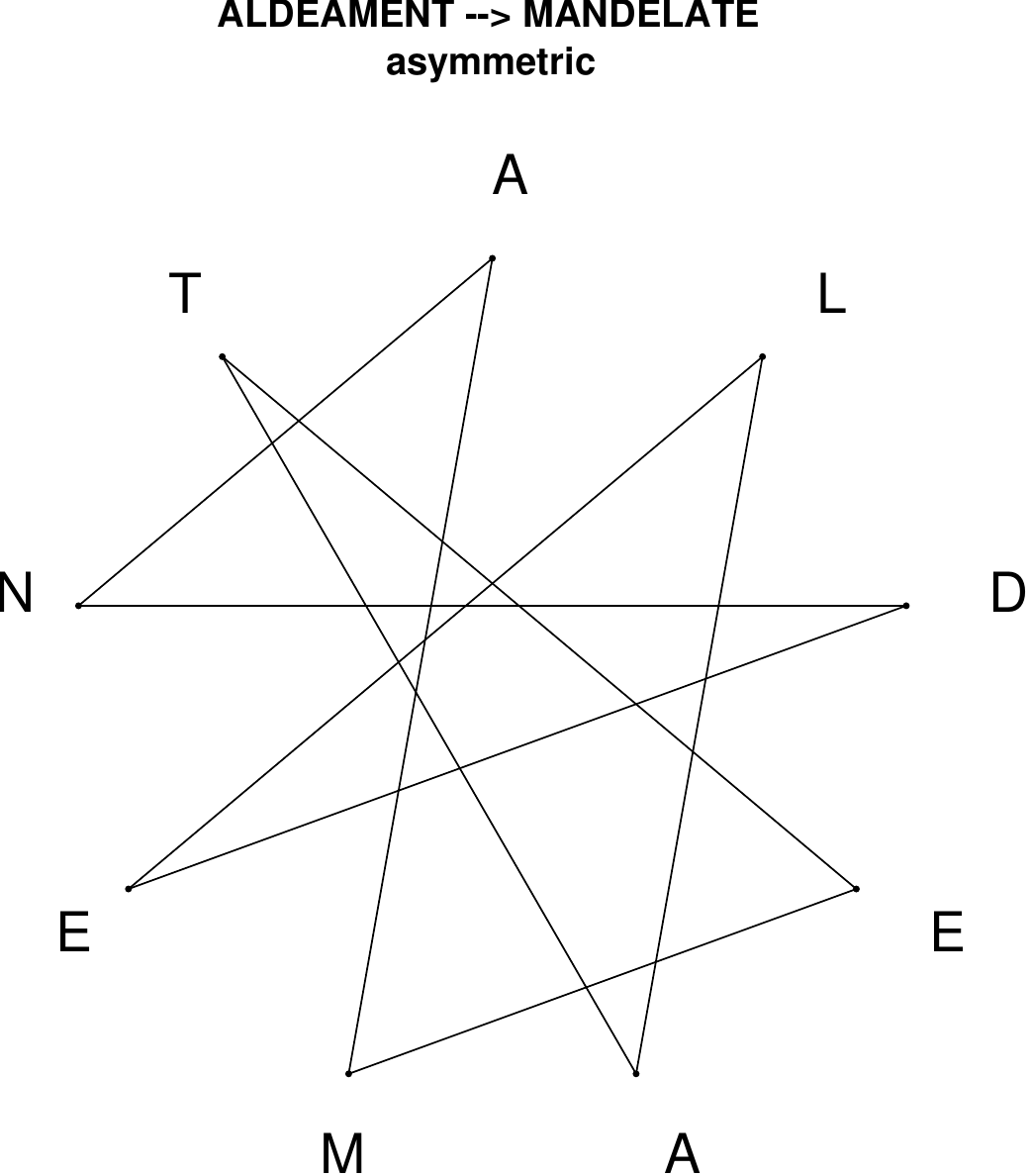}
\end{subfigure}
\hfill
\begin{subfigure}[T]{0.19\textwidth}
\centering
\includegraphics[width=\textwidth]{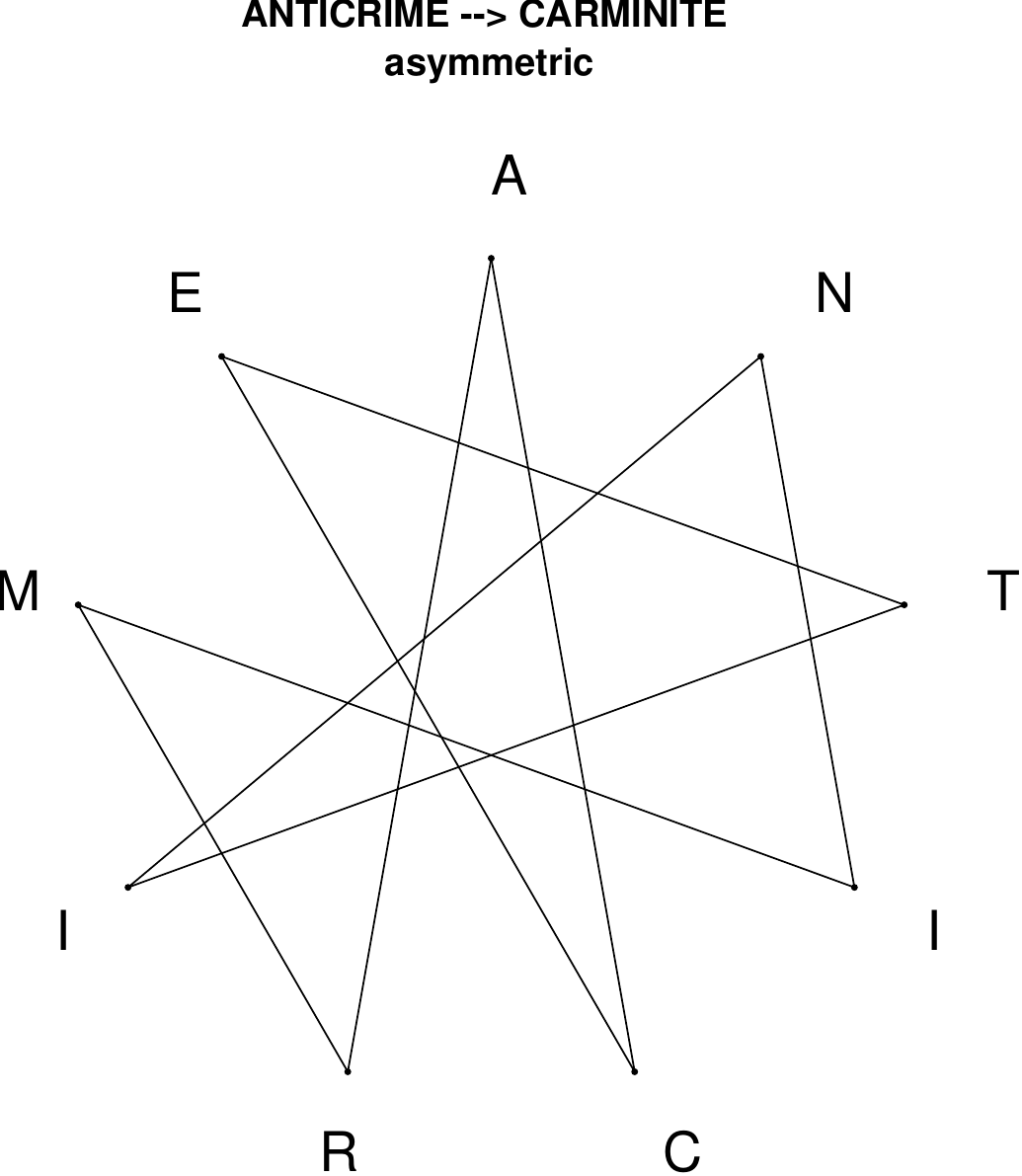}
\end{subfigure}
\end{figure}

\begin{figure}[H]
\centering
\begin{subfigure}[T]{0.19\textwidth}
\centering
\includegraphics[width=\textwidth]{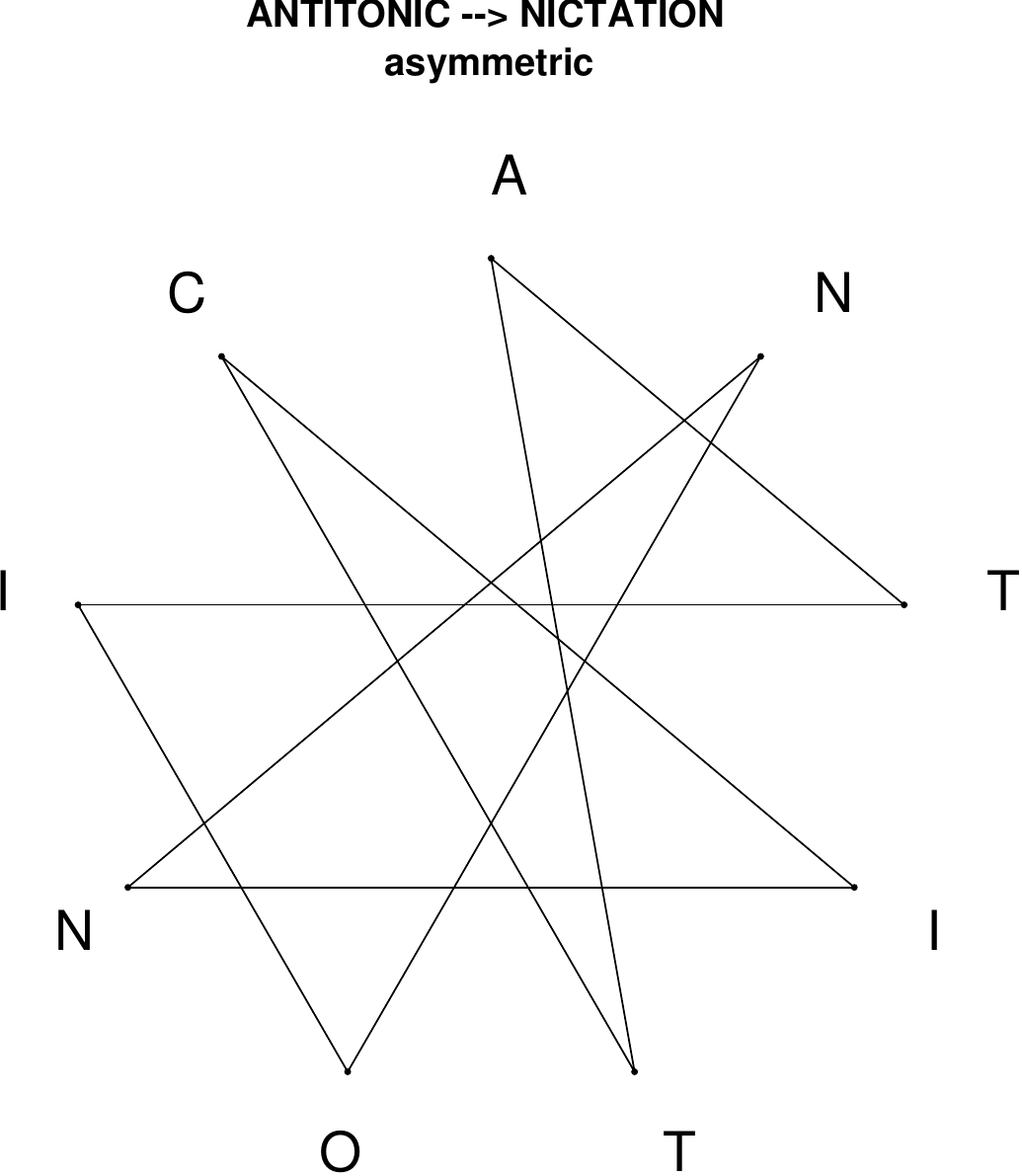}
\end{subfigure}
\hfill
\begin{subfigure}[T]{0.19\textwidth}
\centering
\includegraphics[width=\textwidth]{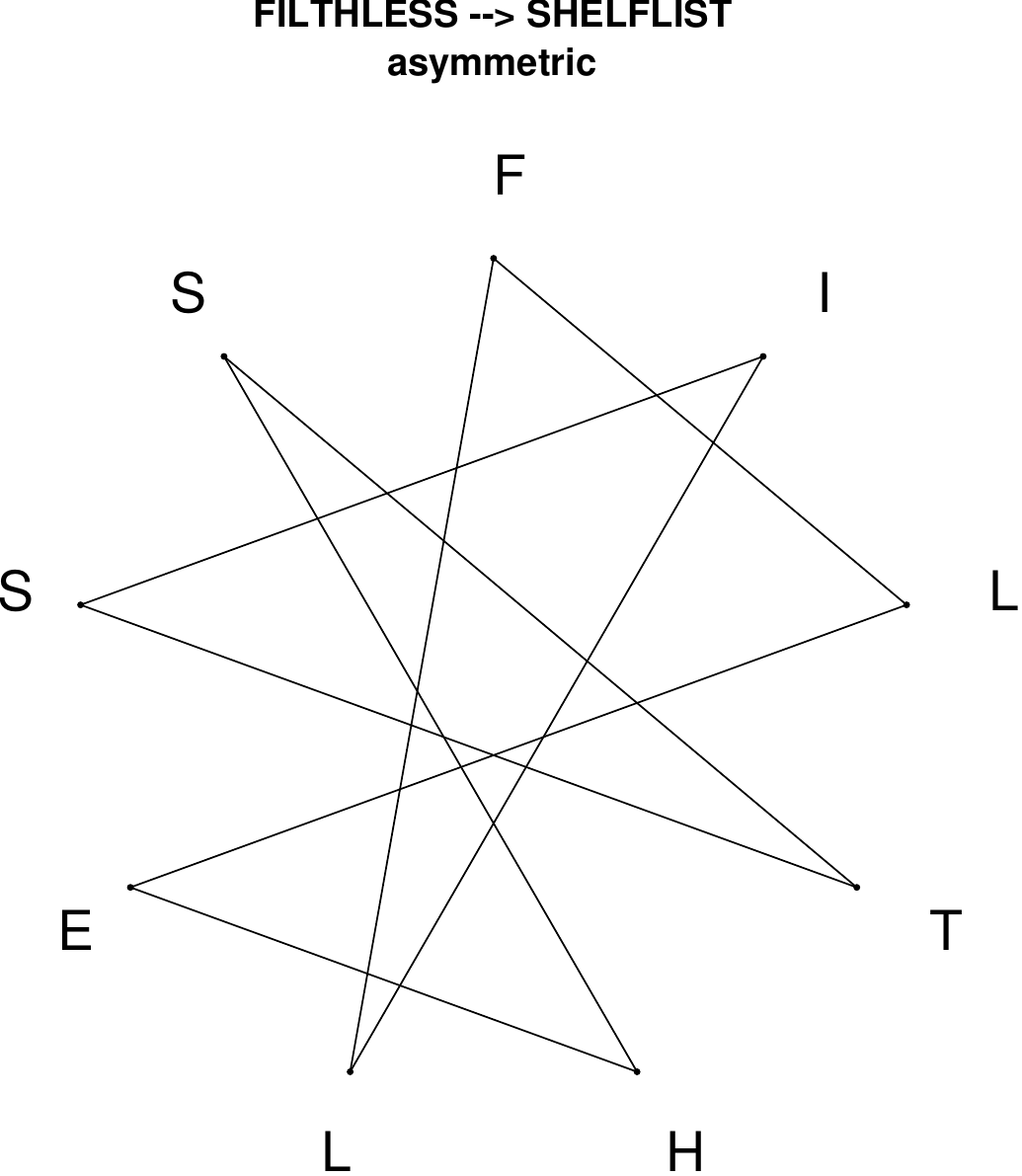}
\end{subfigure}
\hfill
\begin{subfigure}[T]{0.19\textwidth}
\centering
\includegraphics[width=\textwidth]{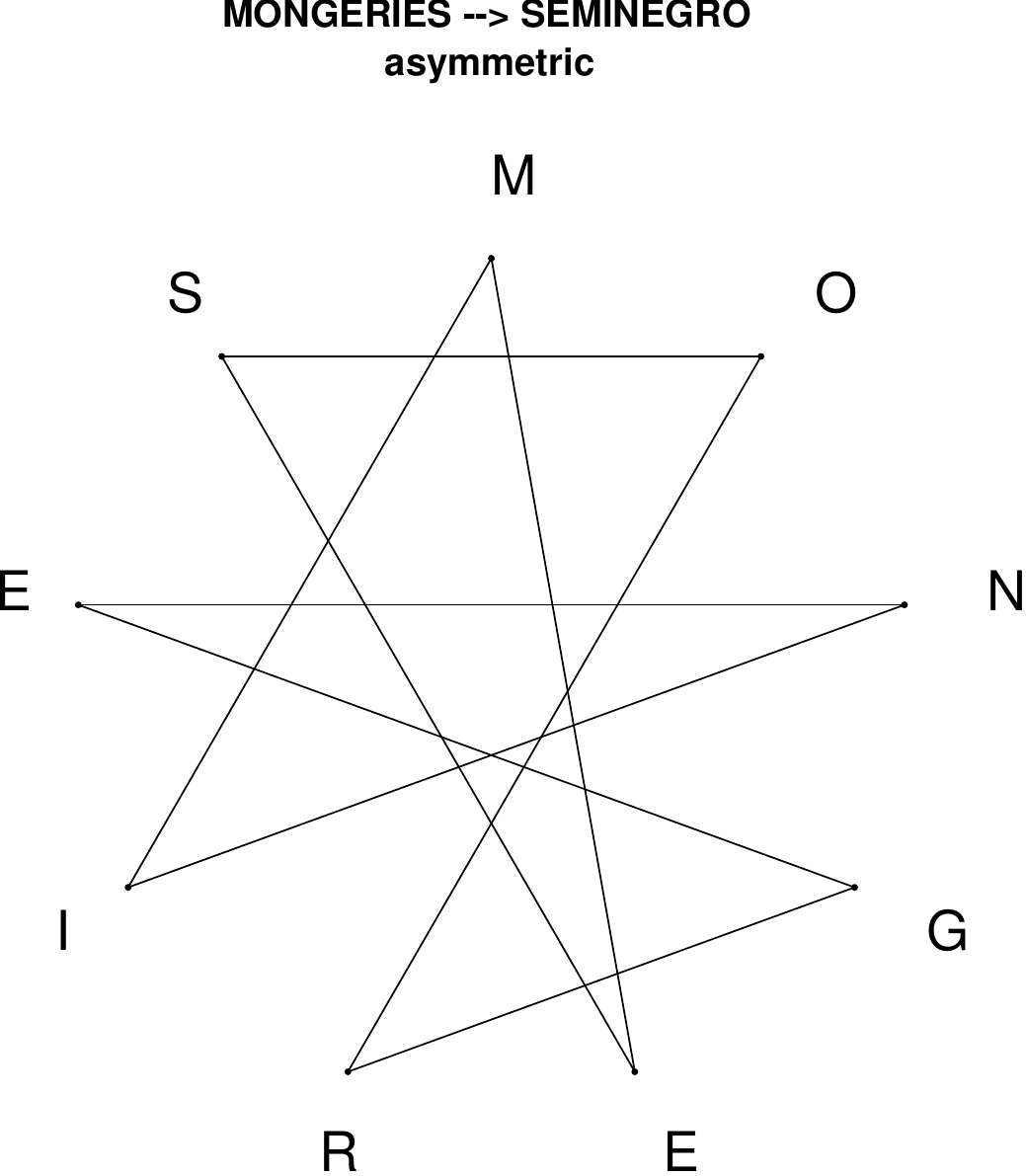}
\end{subfigure}
\hfill
\begin{subfigure}[T]{0.19\textwidth}
\centering
\includegraphics[width=\textwidth]{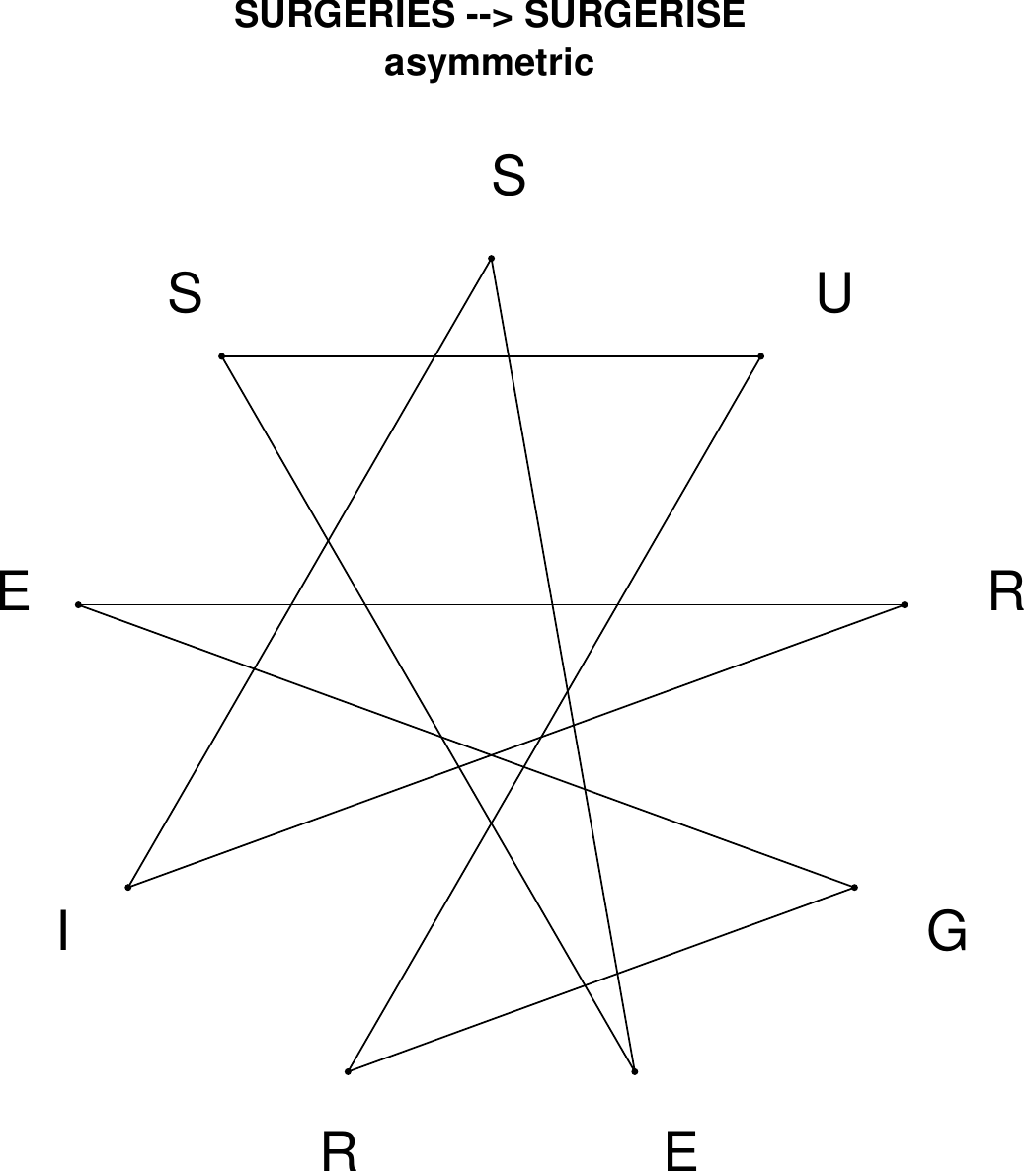}
\end{subfigure}
\hfill
\begin{subfigure}[T]{0.19\textwidth}
\centering
\includegraphics[width=\textwidth]{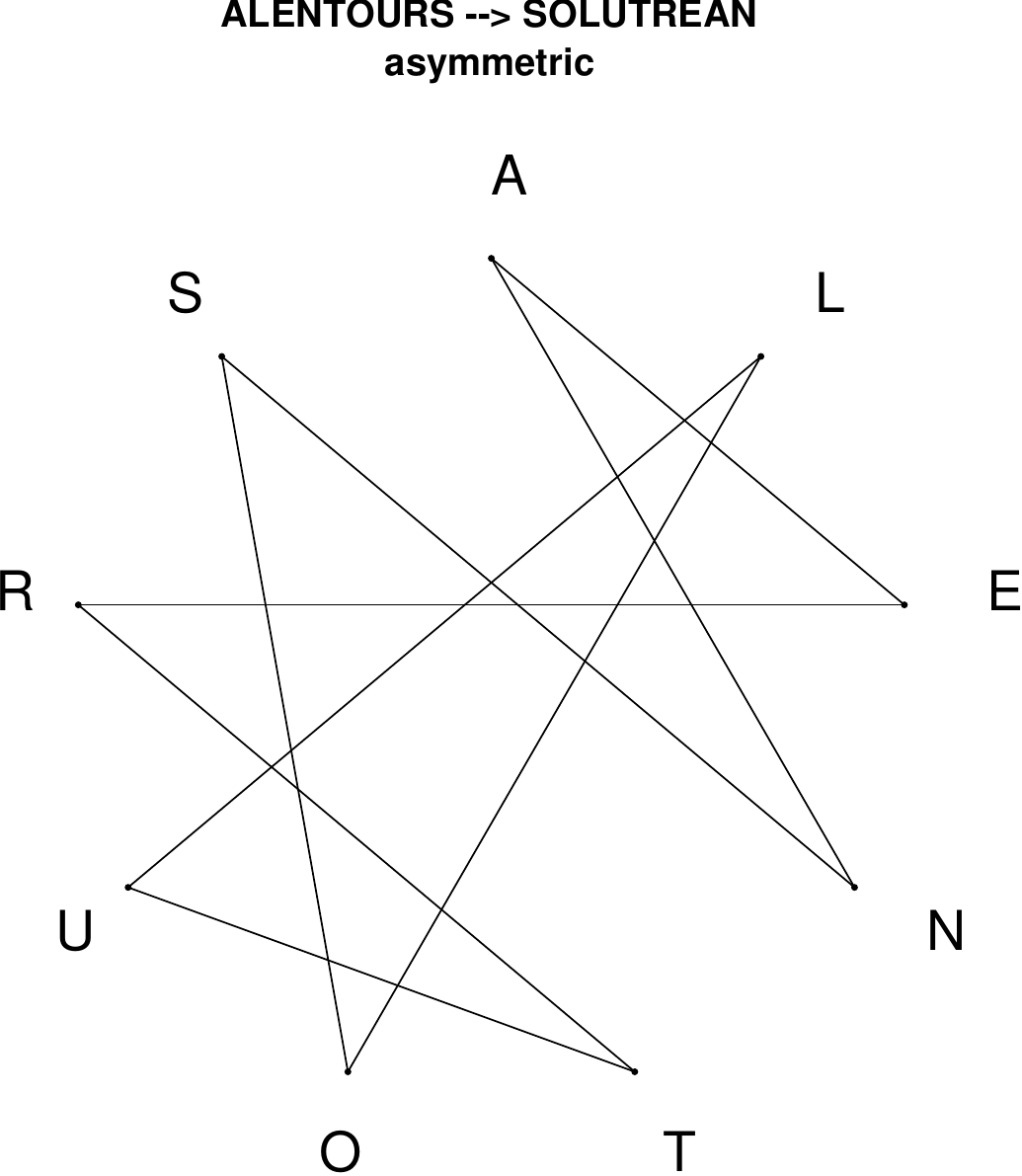}
\end{subfigure}
\end{figure}

\begin{figure}[H]
\centering
\begin{subfigure}[T]{0.19\textwidth}
\centering
\includegraphics[width=\textwidth]{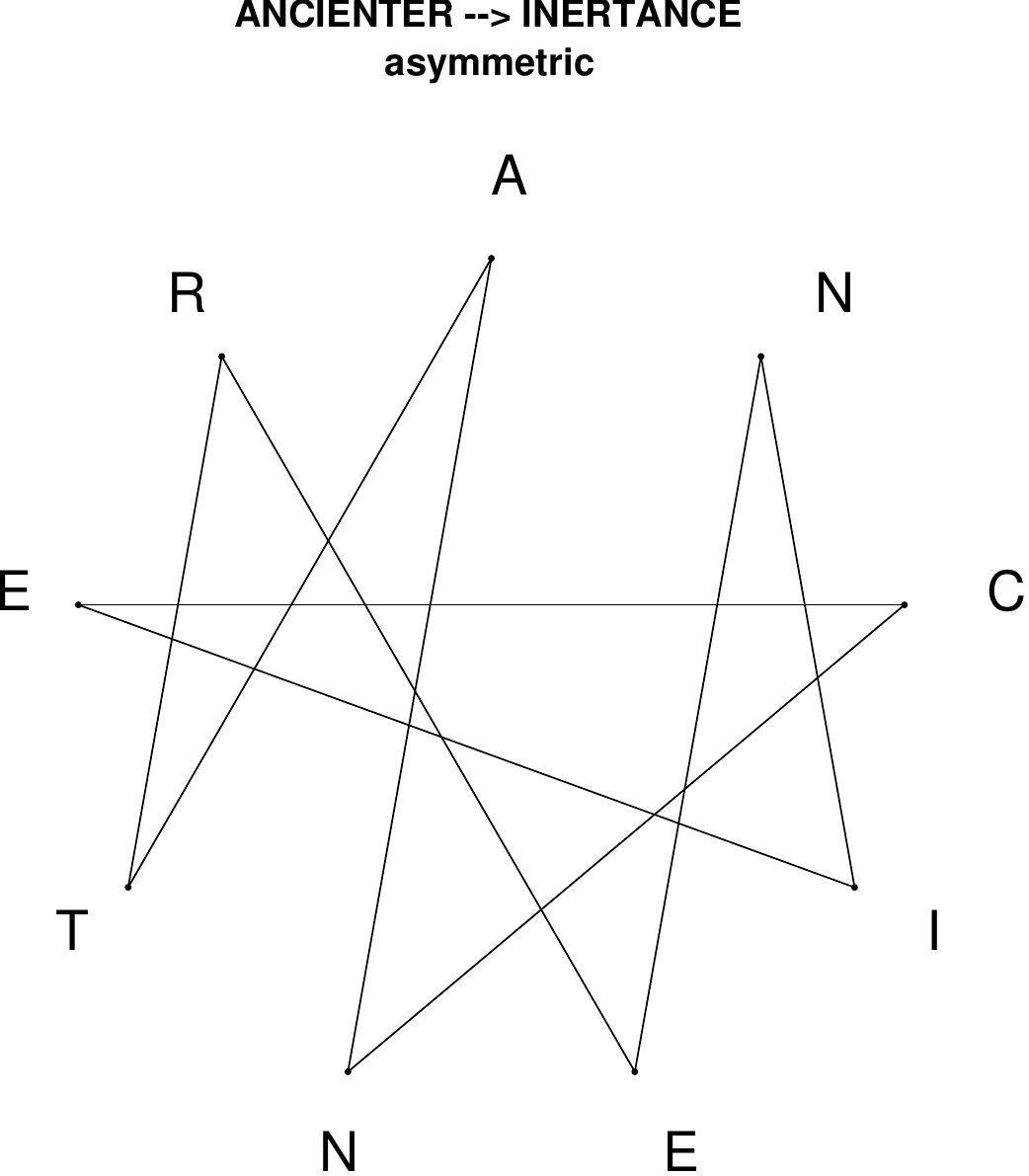}
\end{subfigure}
\hfill
\begin{subfigure}[T]{0.19\textwidth}
\centering
\includegraphics[width=\textwidth]{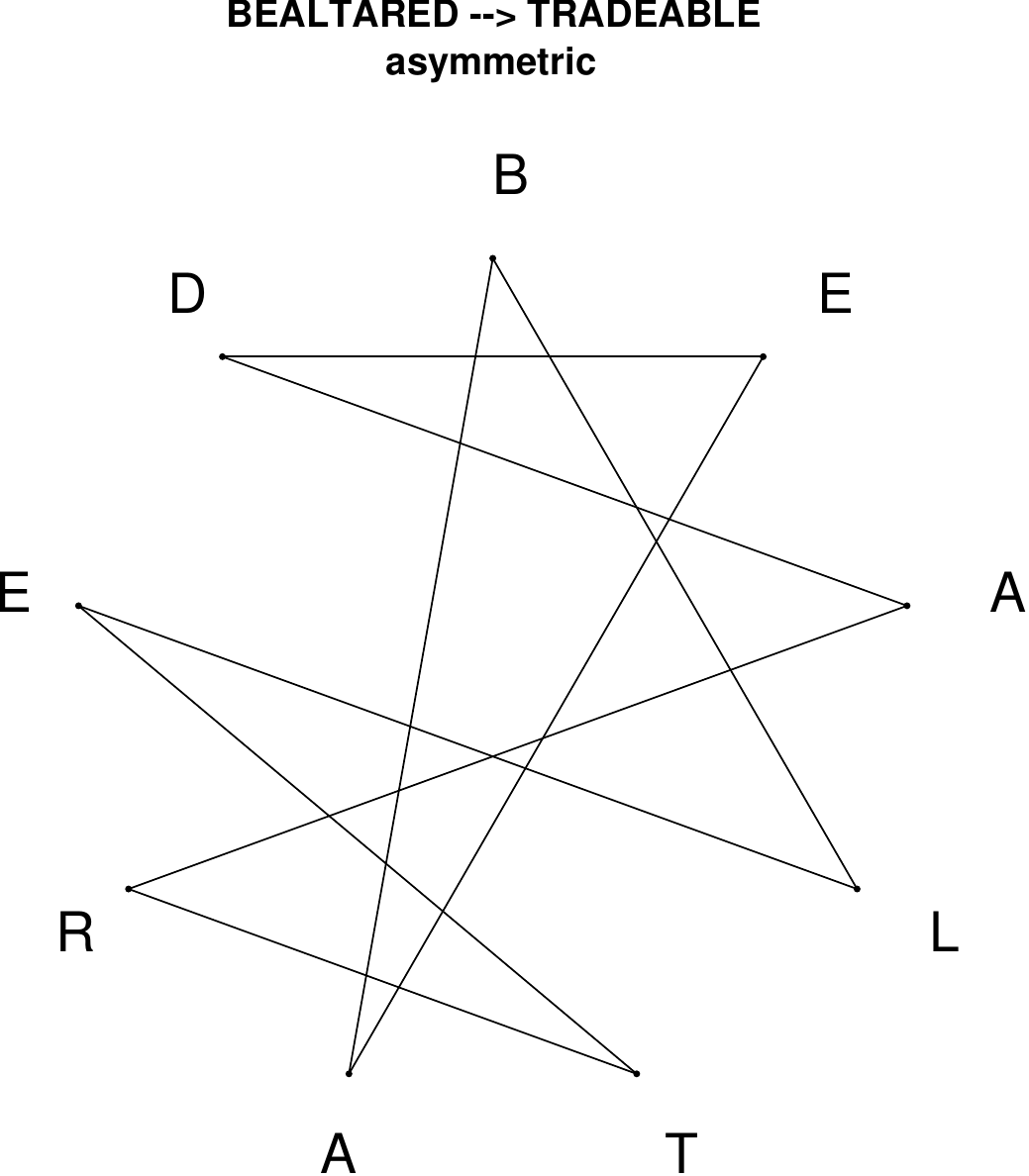}
\end{subfigure}
\hfill
\begin{subfigure}[T]{0.19\textwidth}
\centering
\includegraphics[width=\textwidth]{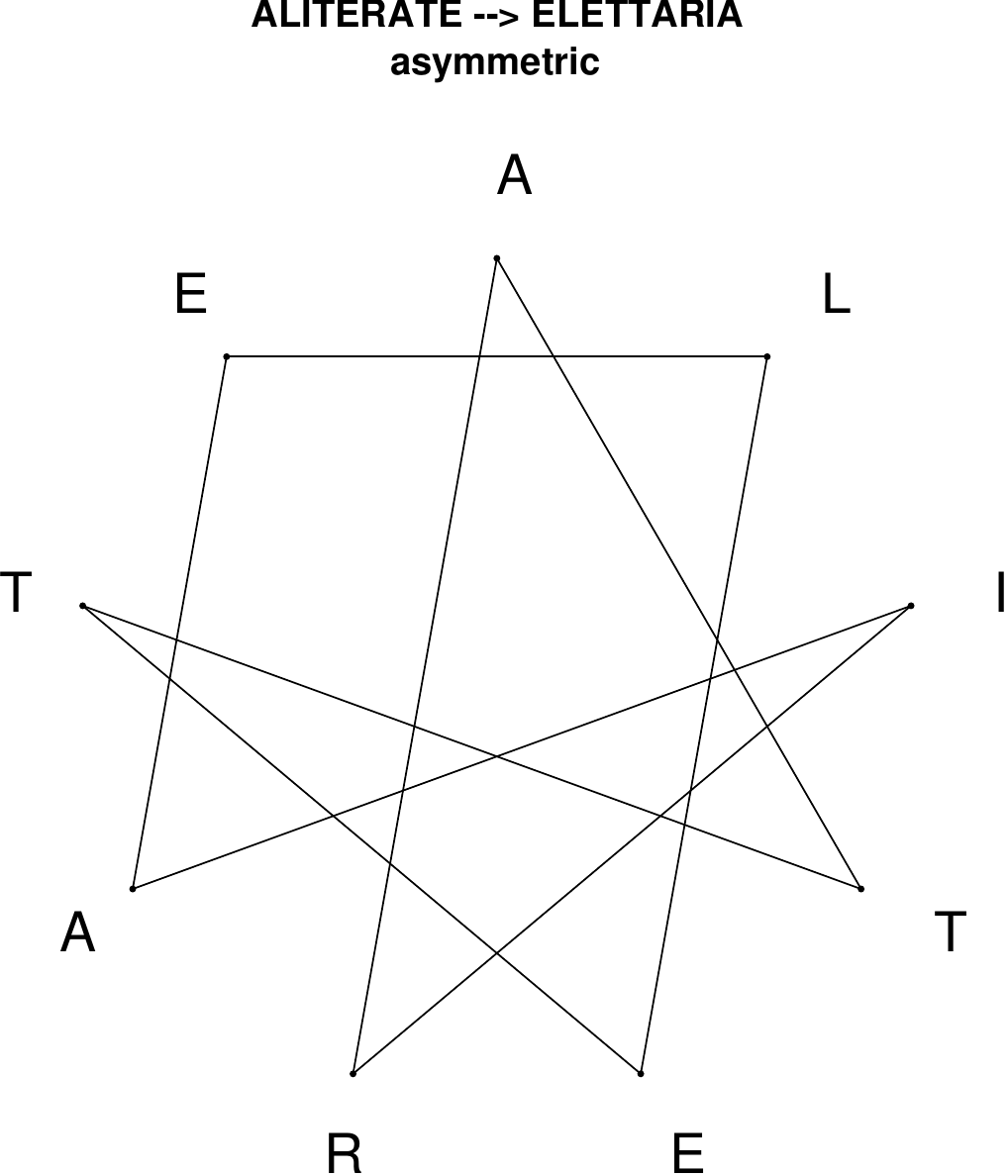}
\end{subfigure}
\hfill
\begin{subfigure}[T]{0.19\textwidth}
\centering
\includegraphics[width=\textwidth]{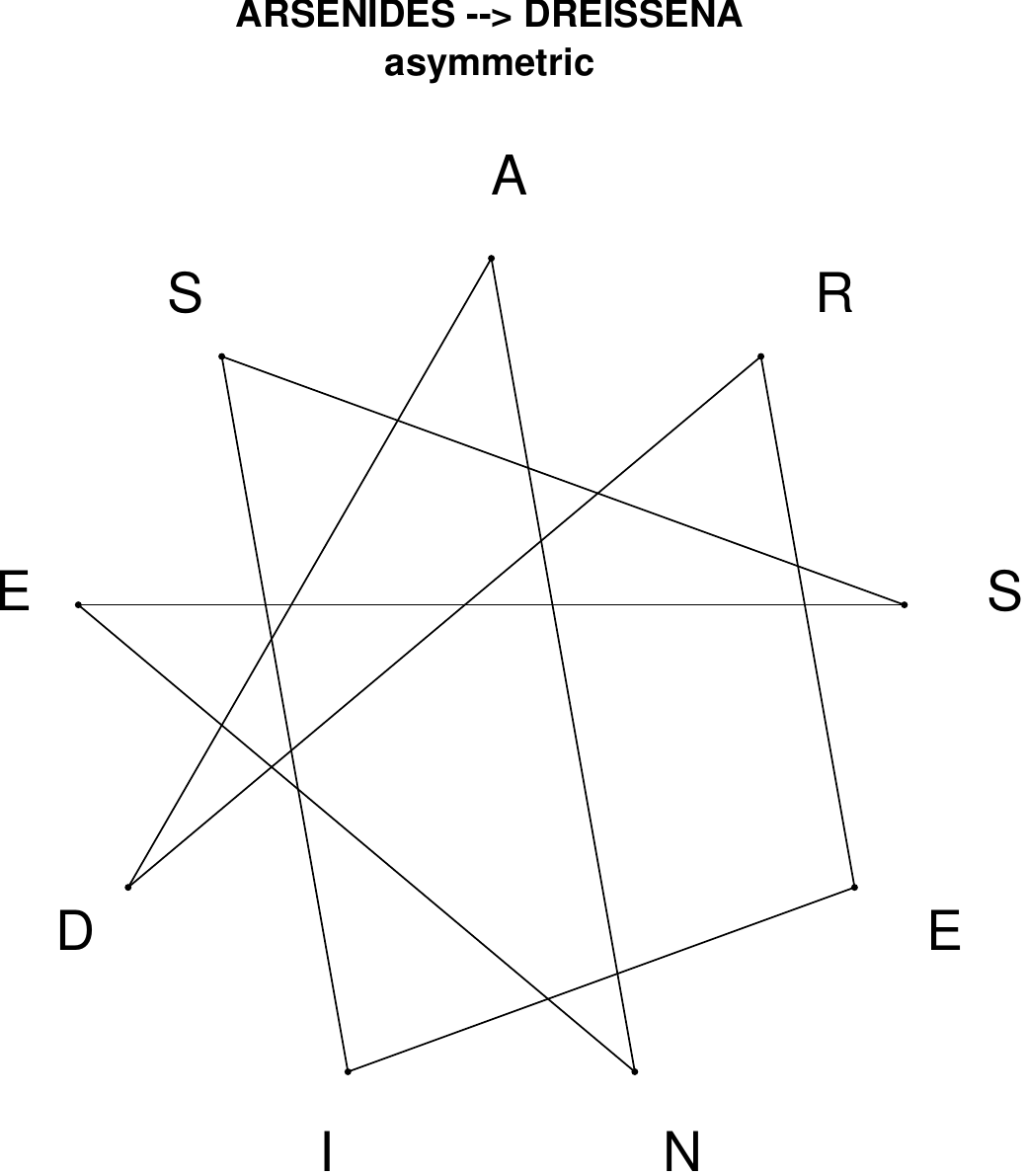}
\end{subfigure}
\hfill
\begin{subfigure}[T]{0.19\textwidth}
\centering
\includegraphics[width=\textwidth]{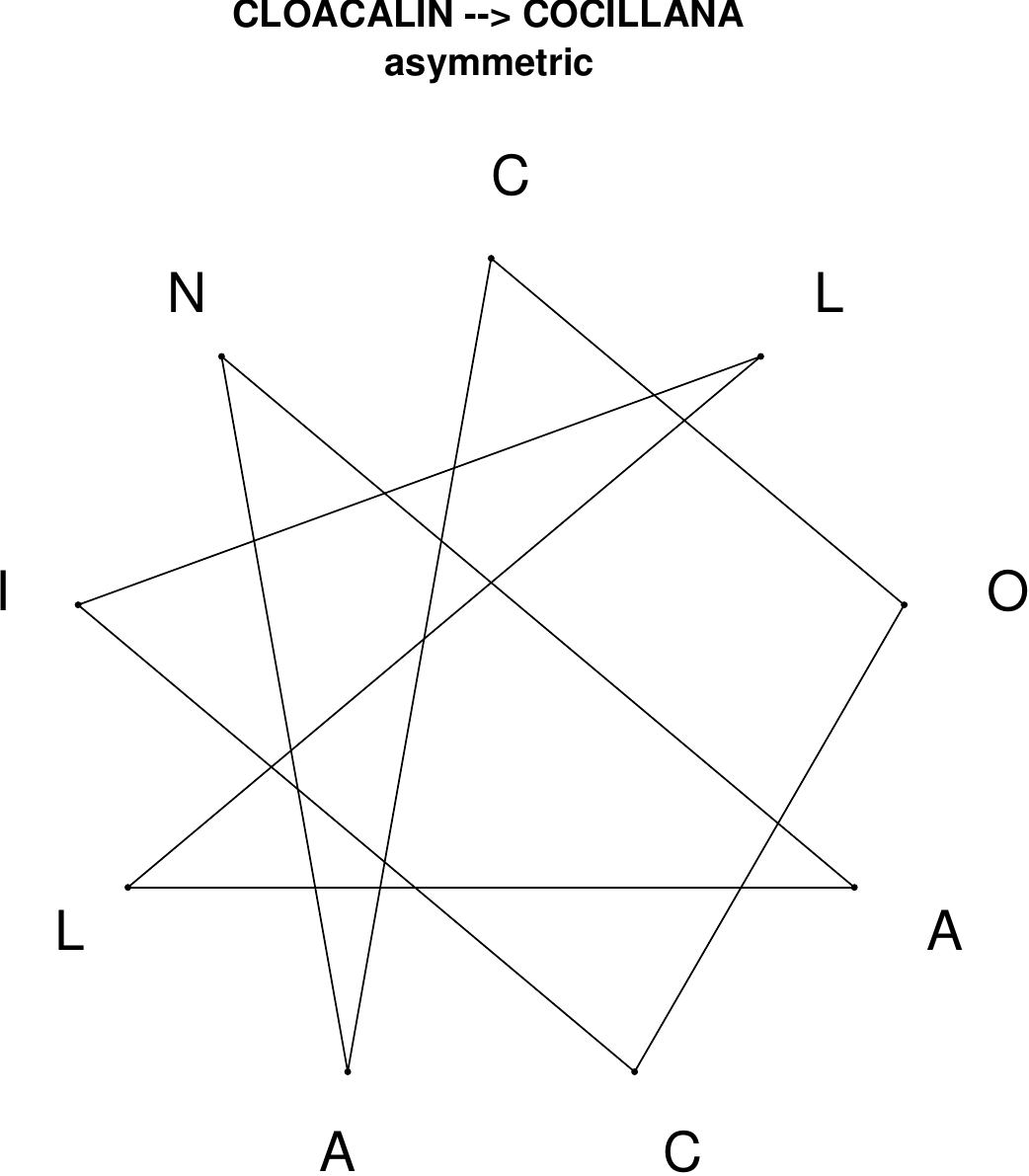}
\end{subfigure}
\end{figure}

\begin{figure}[H]
\centering
\begin{subfigure}[T]{0.19\textwidth}
\centering
\includegraphics[width=\textwidth]{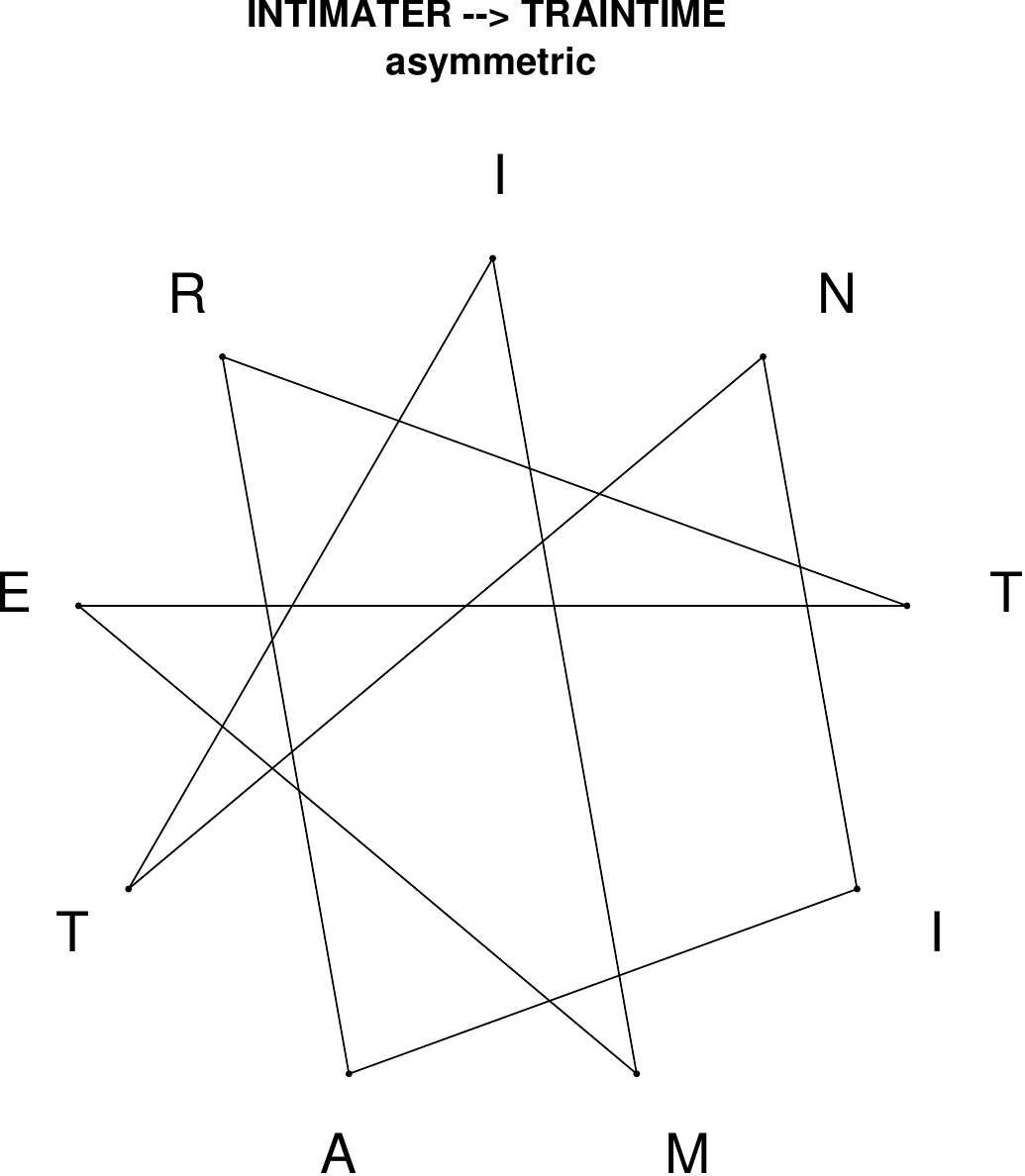}
\end{subfigure}
\hfill
\begin{subfigure}[T]{0.19\textwidth}
\centering
\includegraphics[width=\textwidth]{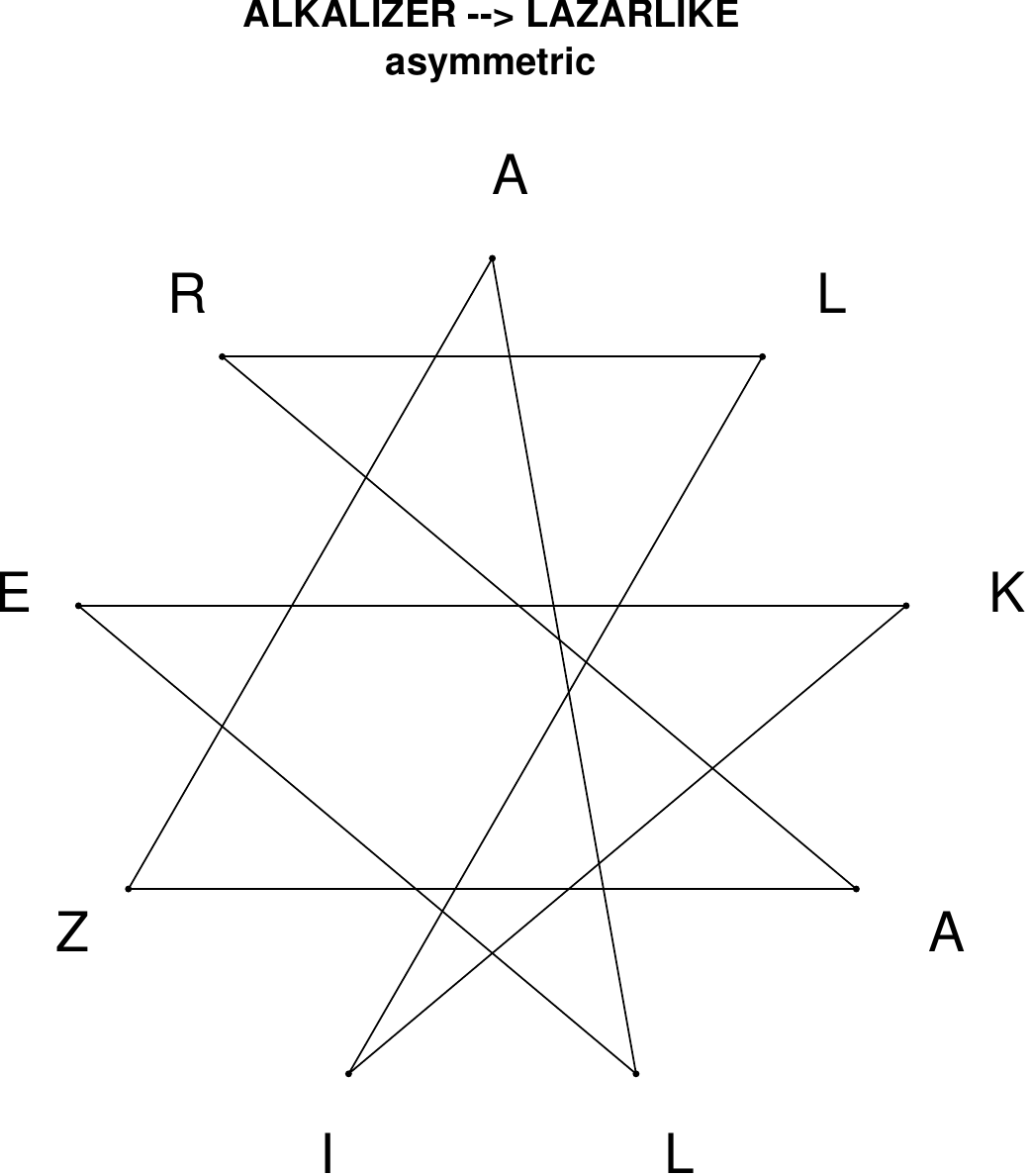}
\end{subfigure}
\hfill
\begin{subfigure}[T]{0.19\textwidth}
\centering
\includegraphics[width=\textwidth]{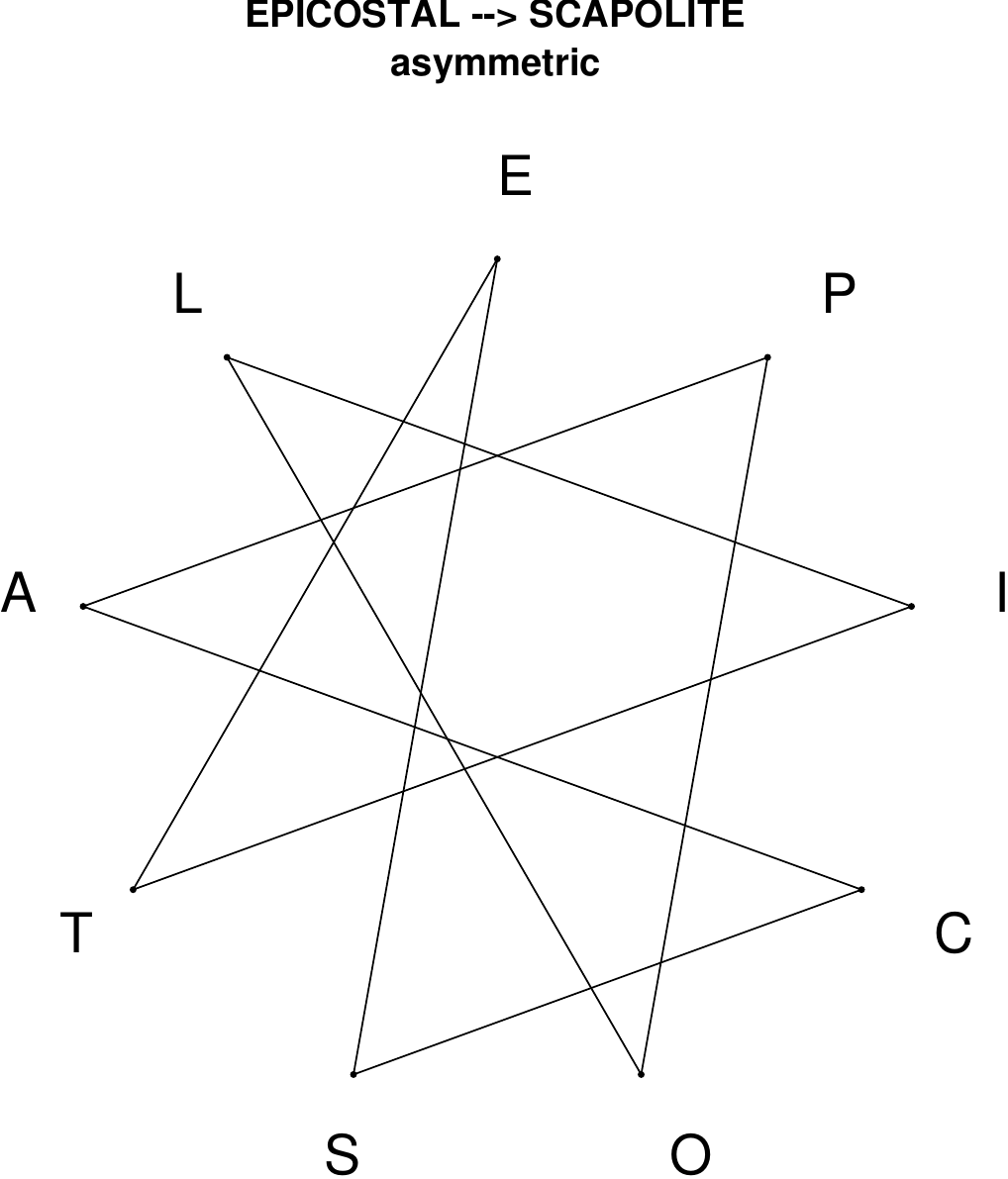}
\end{subfigure}
\hfill
\begin{subfigure}[T]{0.19\textwidth}
\centering
\includegraphics[width=\textwidth]{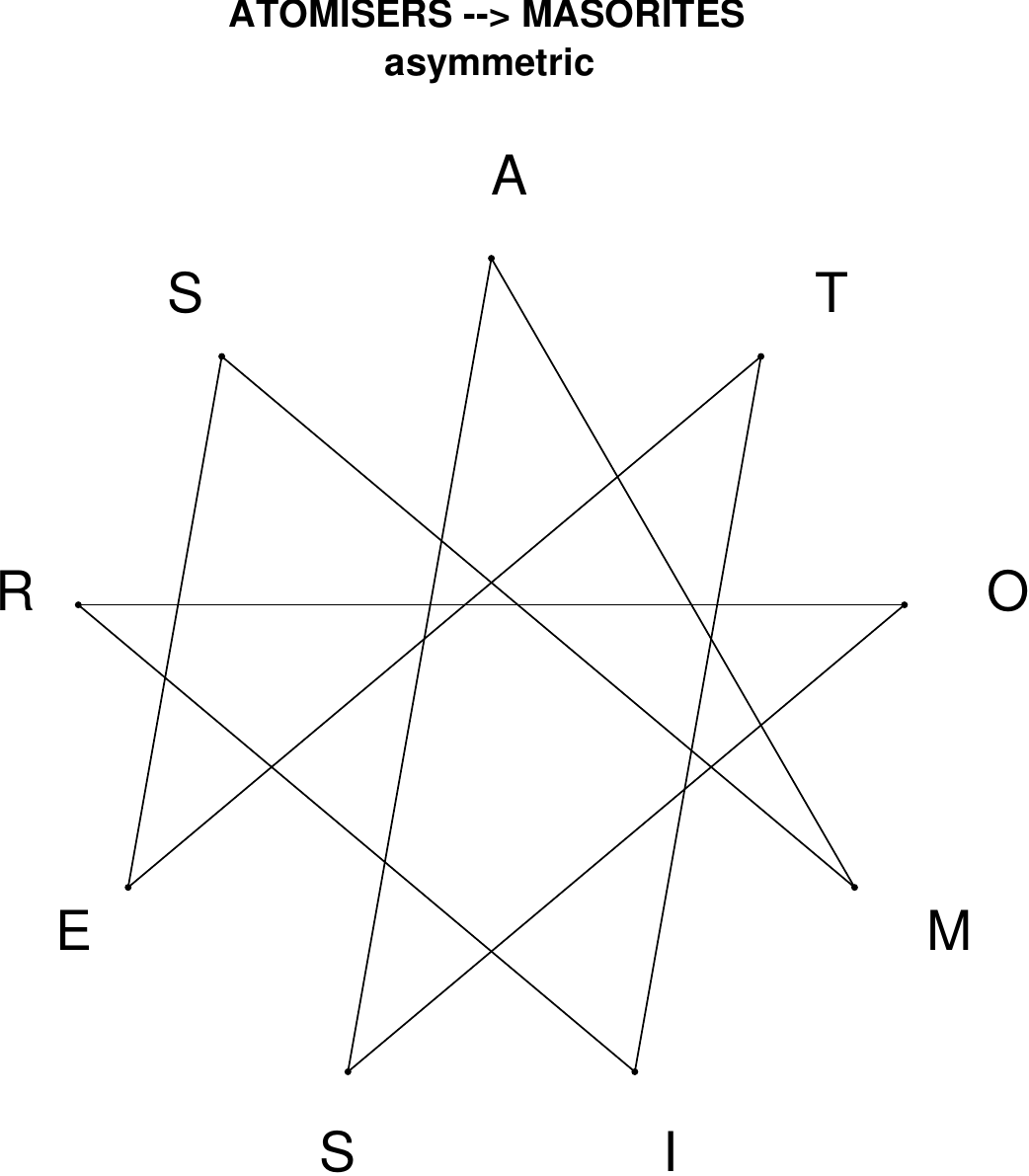}
\end{subfigure}
\hfill
\begin{subfigure}[T]{0.19\textwidth}
\centering
\includegraphics[width=\textwidth]{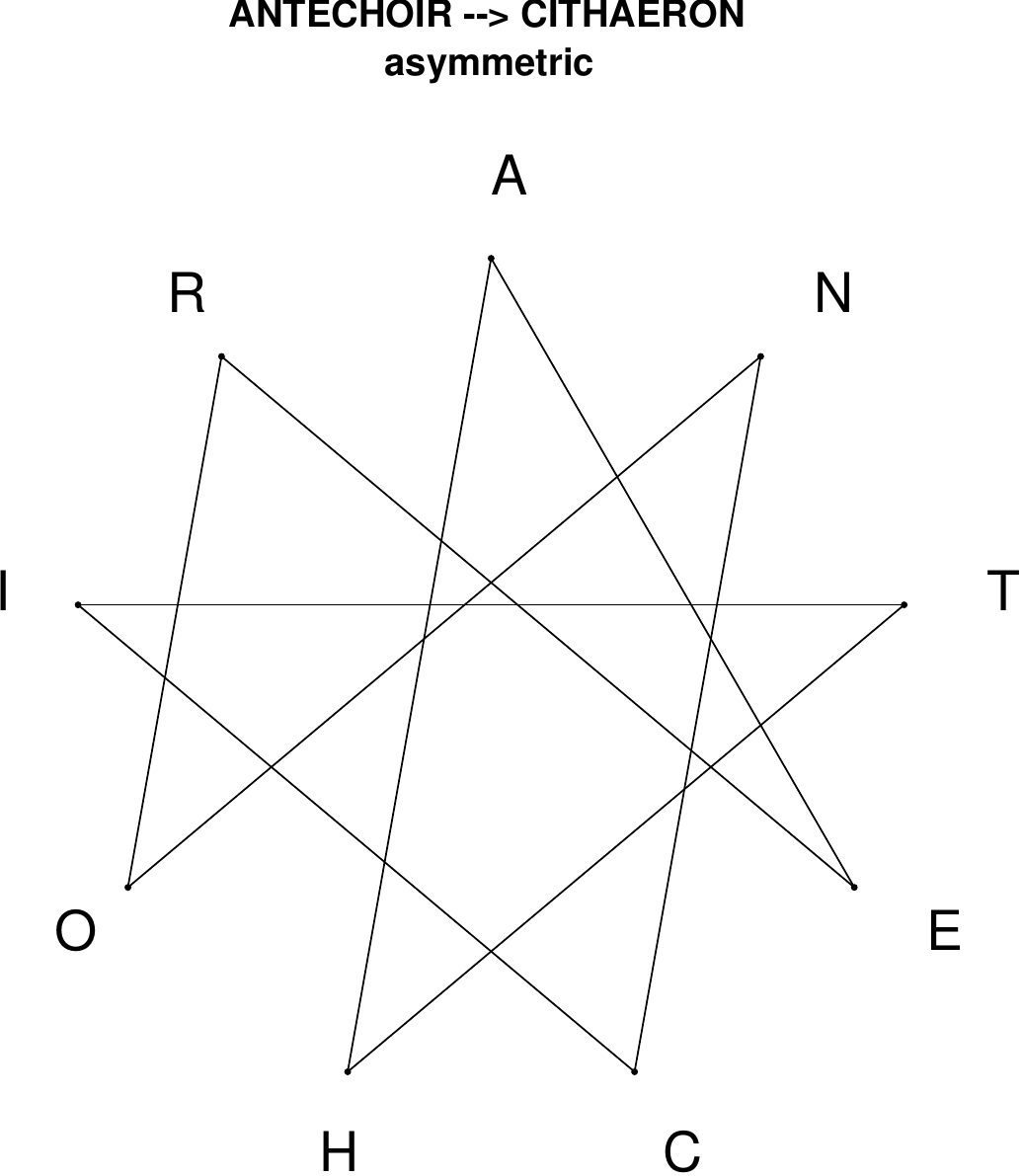}
\end{subfigure}
\end{figure}

\begin{figure}[H]
\centering
\begin{subfigure}[T]{0.19\textwidth}
\centering
\includegraphics[width=\textwidth]{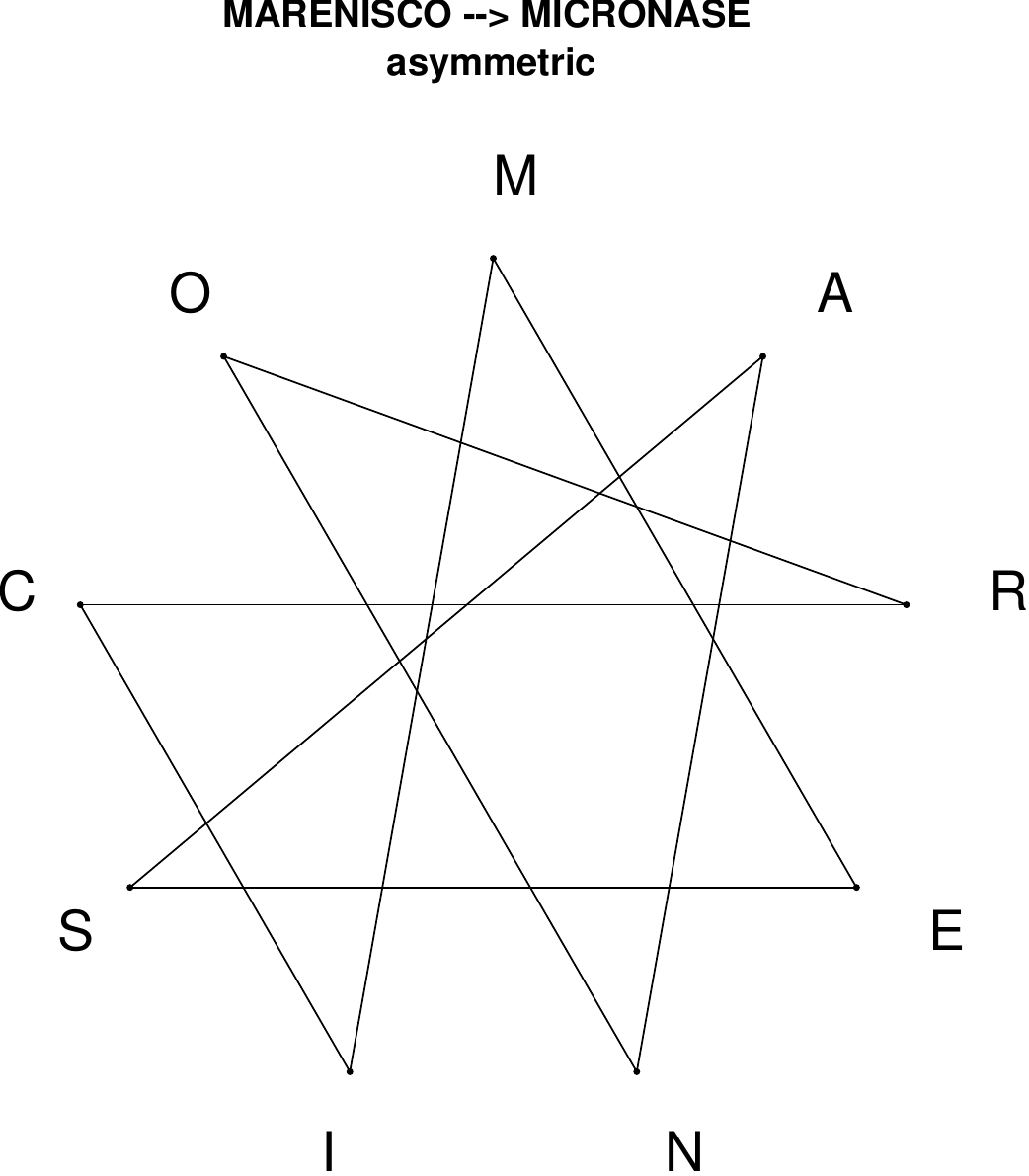}
\end{subfigure}
\hfill
\begin{subfigure}[T]{0.19\textwidth}
\centering
\includegraphics[width=\textwidth]{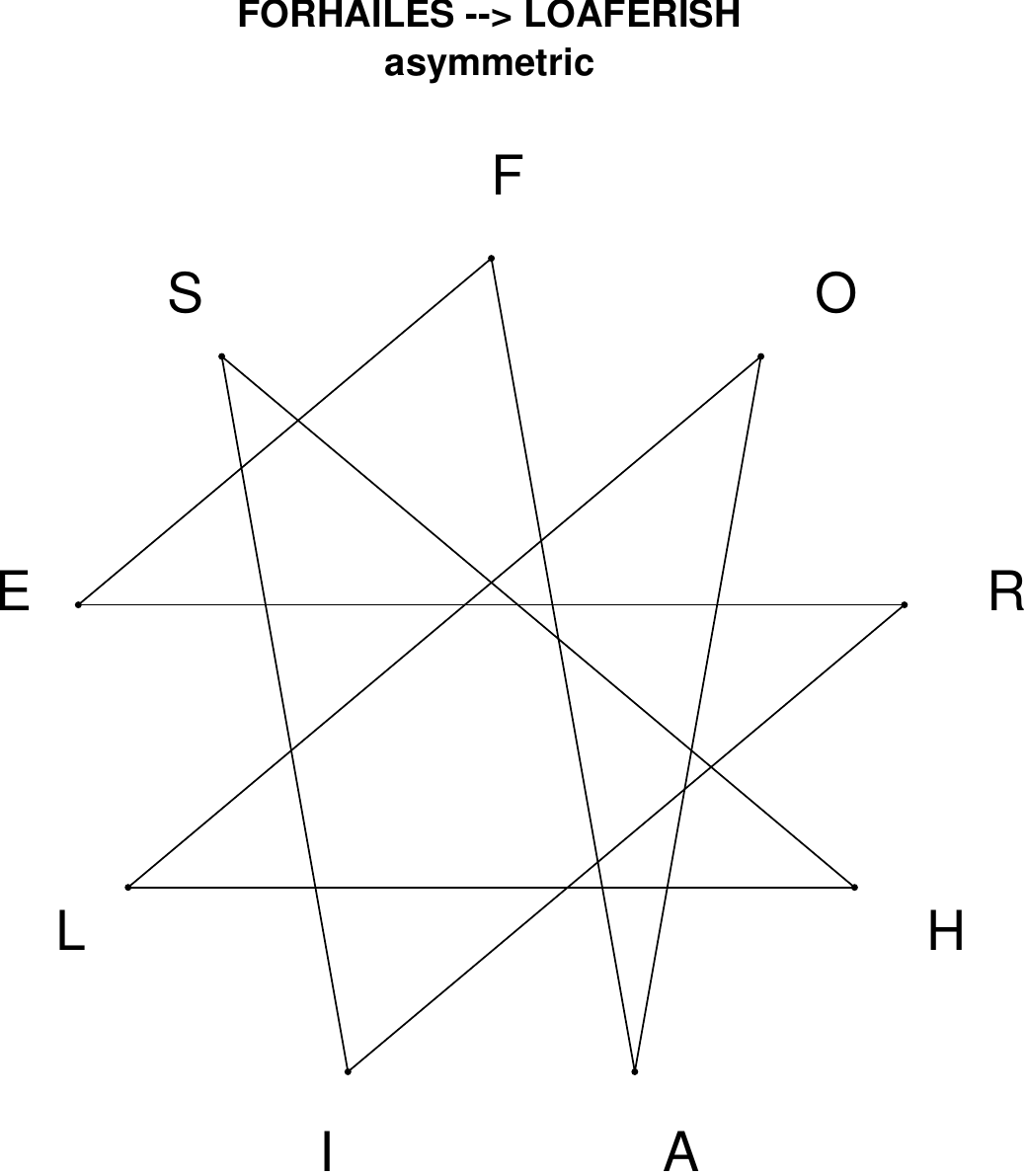}
\end{subfigure}
\hfill
\begin{subfigure}[T]{0.19\textwidth}
\centering
\includegraphics[width=\textwidth]{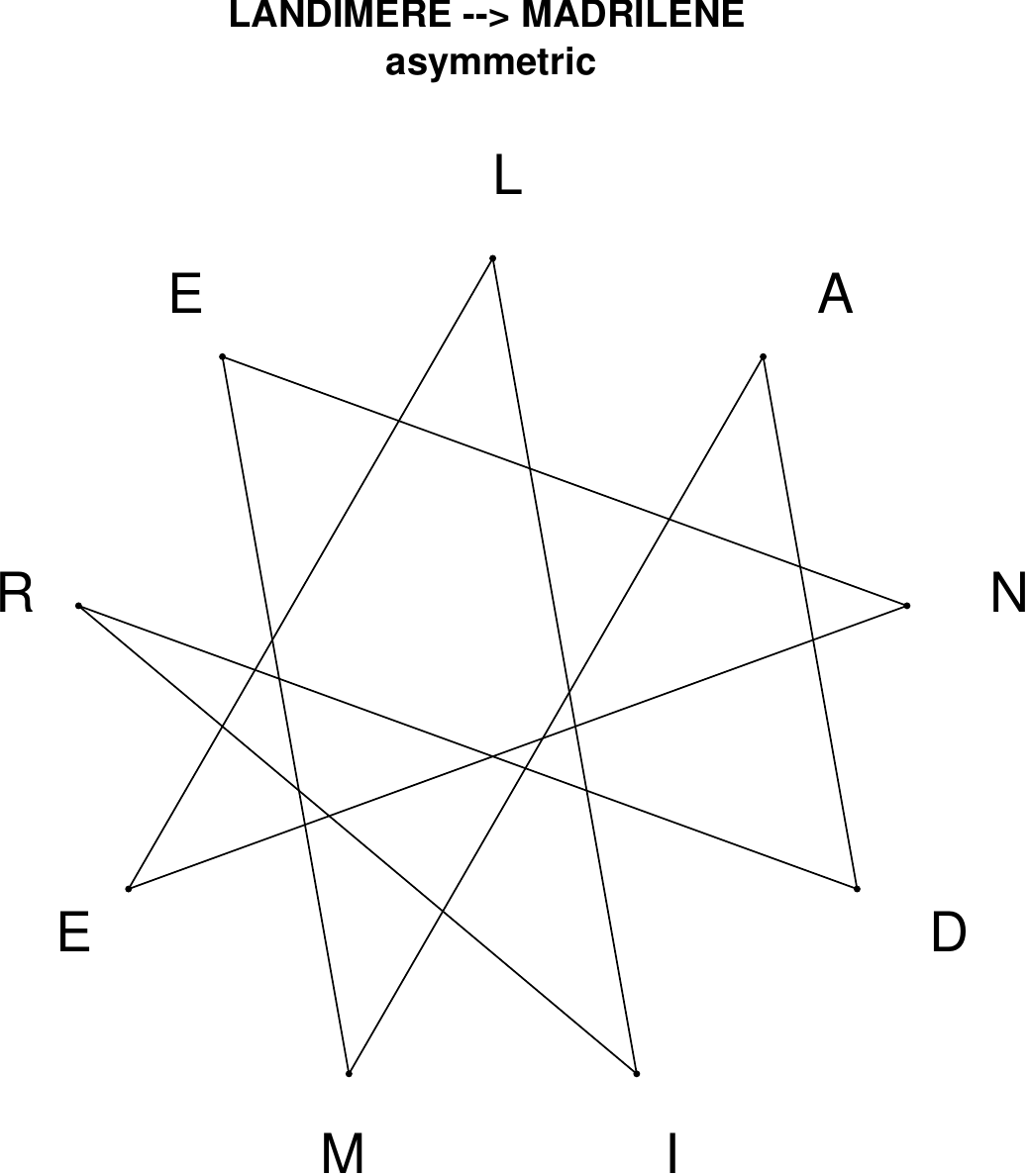}
\end{subfigure}
\hfill
\begin{subfigure}[T]{0.19\textwidth}
\centering
\includegraphics[width=\textwidth]{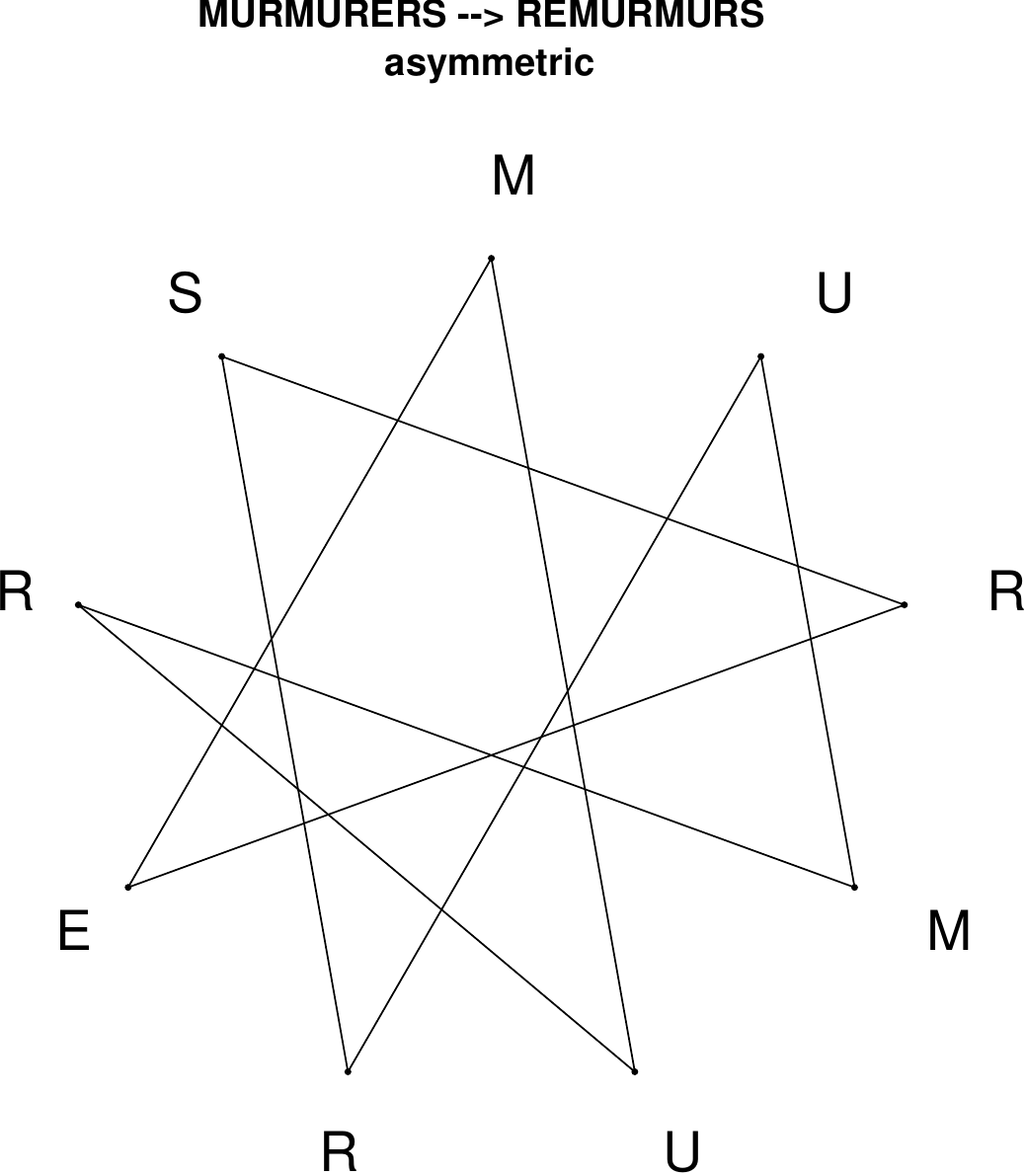}
\end{subfigure}
\hfill
\begin{subfigure}[T]{0.19\textwidth}
\centering
\includegraphics[width=\textwidth]{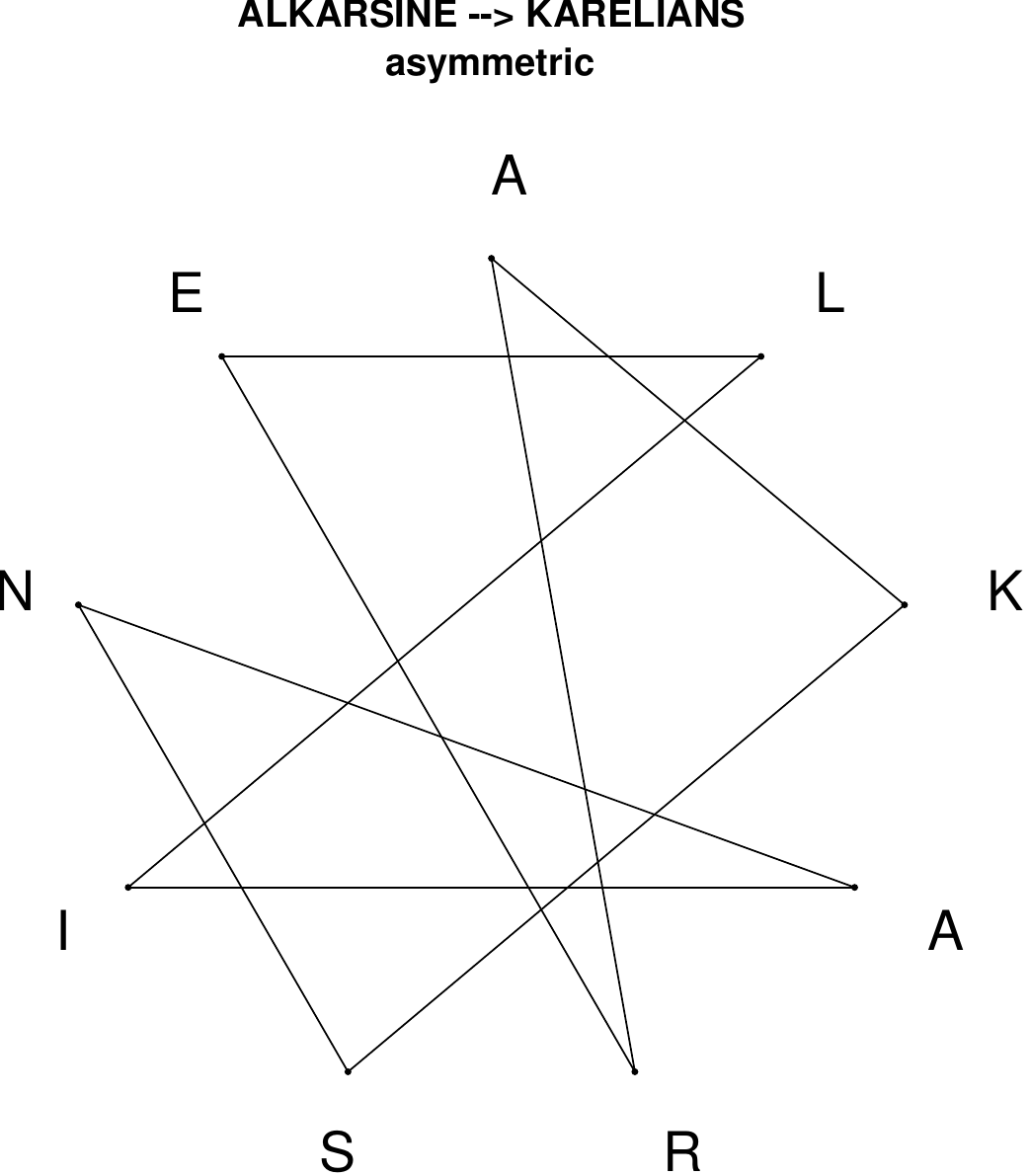}
\end{subfigure}
\end{figure}

\begin{figure}[H]
\centering
\begin{subfigure}[T]{0.19\textwidth}
\centering
\includegraphics[width=\textwidth]{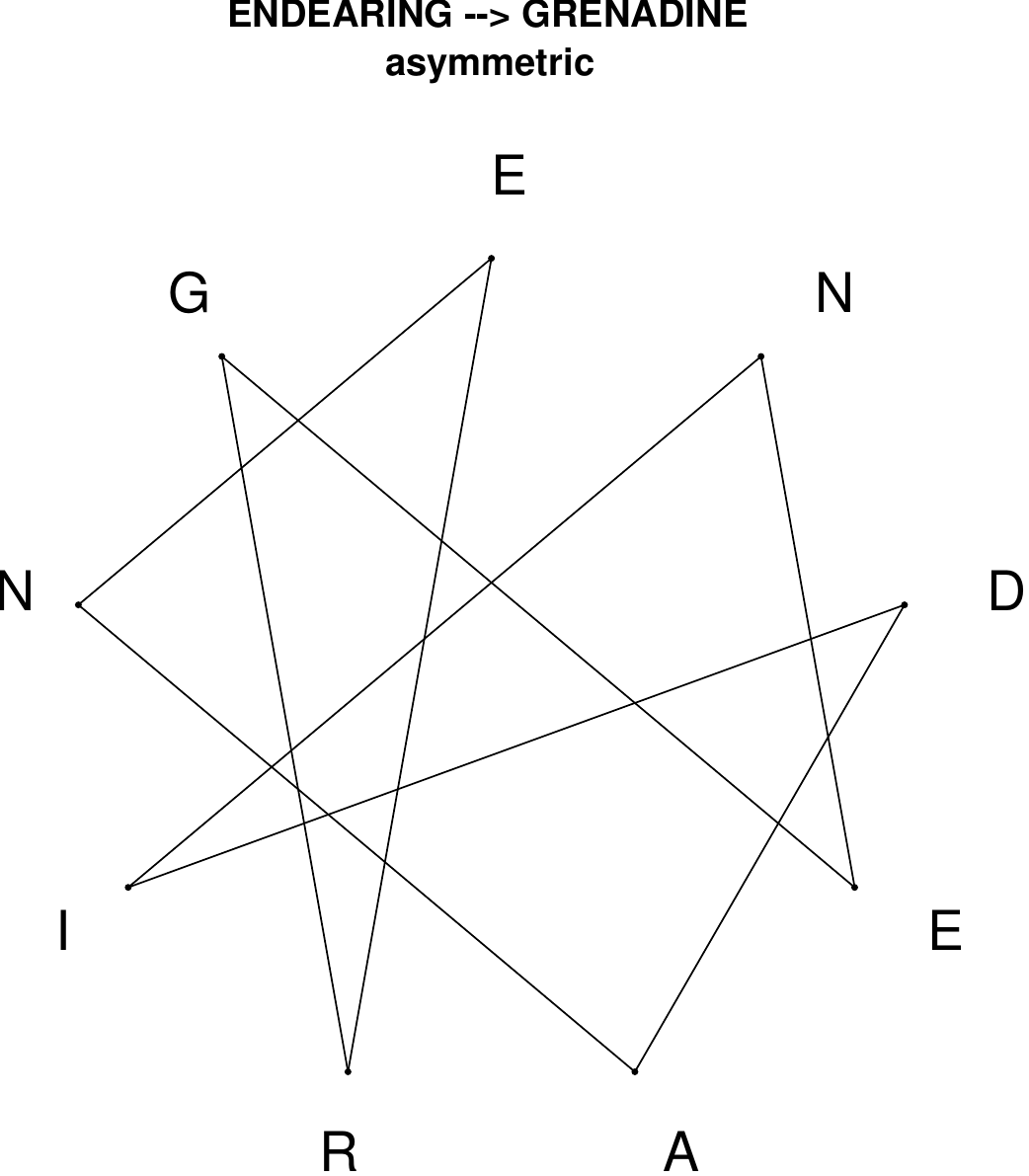}
\end{subfigure}
\hfill
\begin{subfigure}[T]{0.19\textwidth}
\centering
\includegraphics[width=\textwidth]{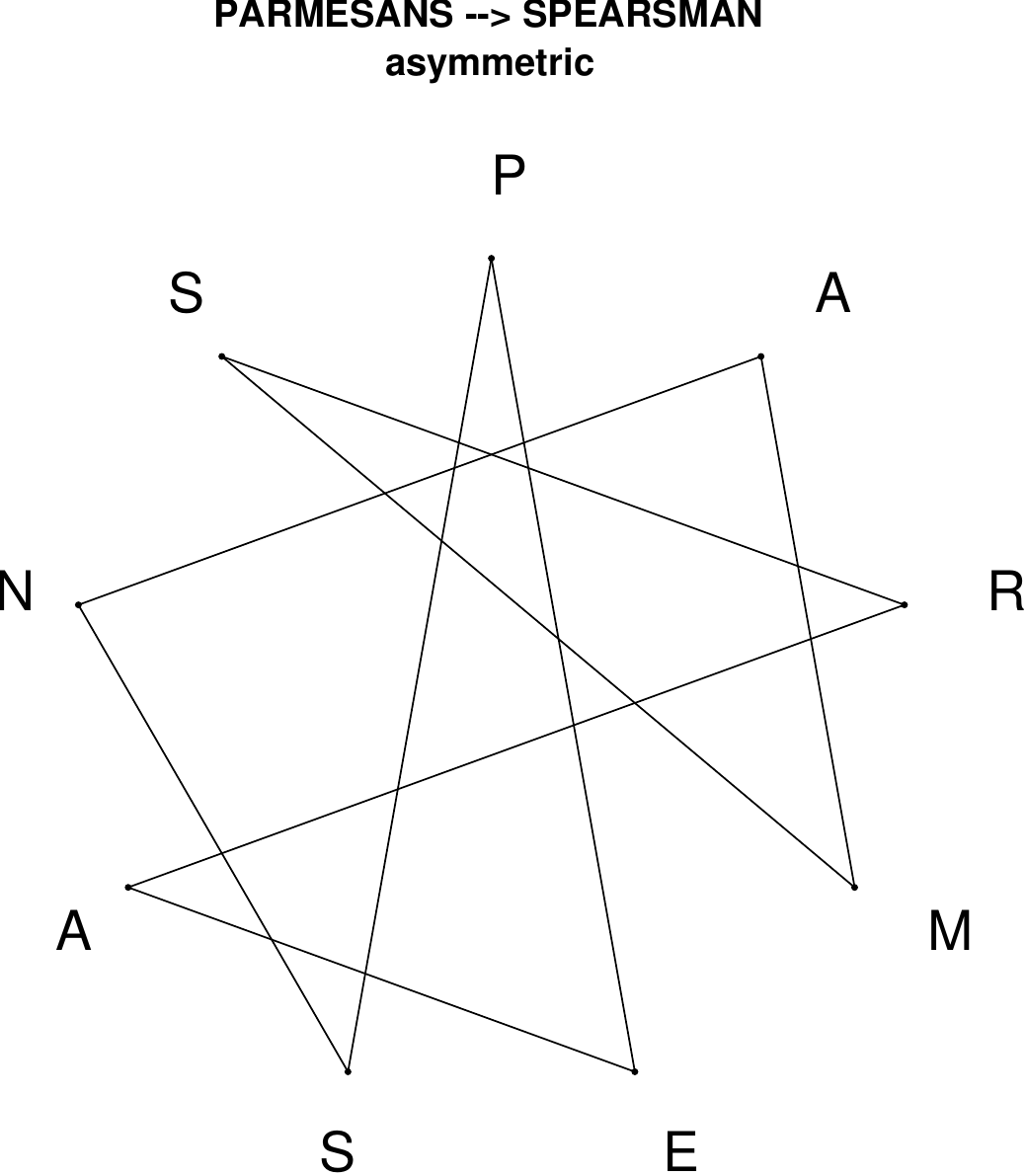}
\end{subfigure}
\hfill
\begin{subfigure}[T]{0.19\textwidth}
\centering
\includegraphics[width=\textwidth]{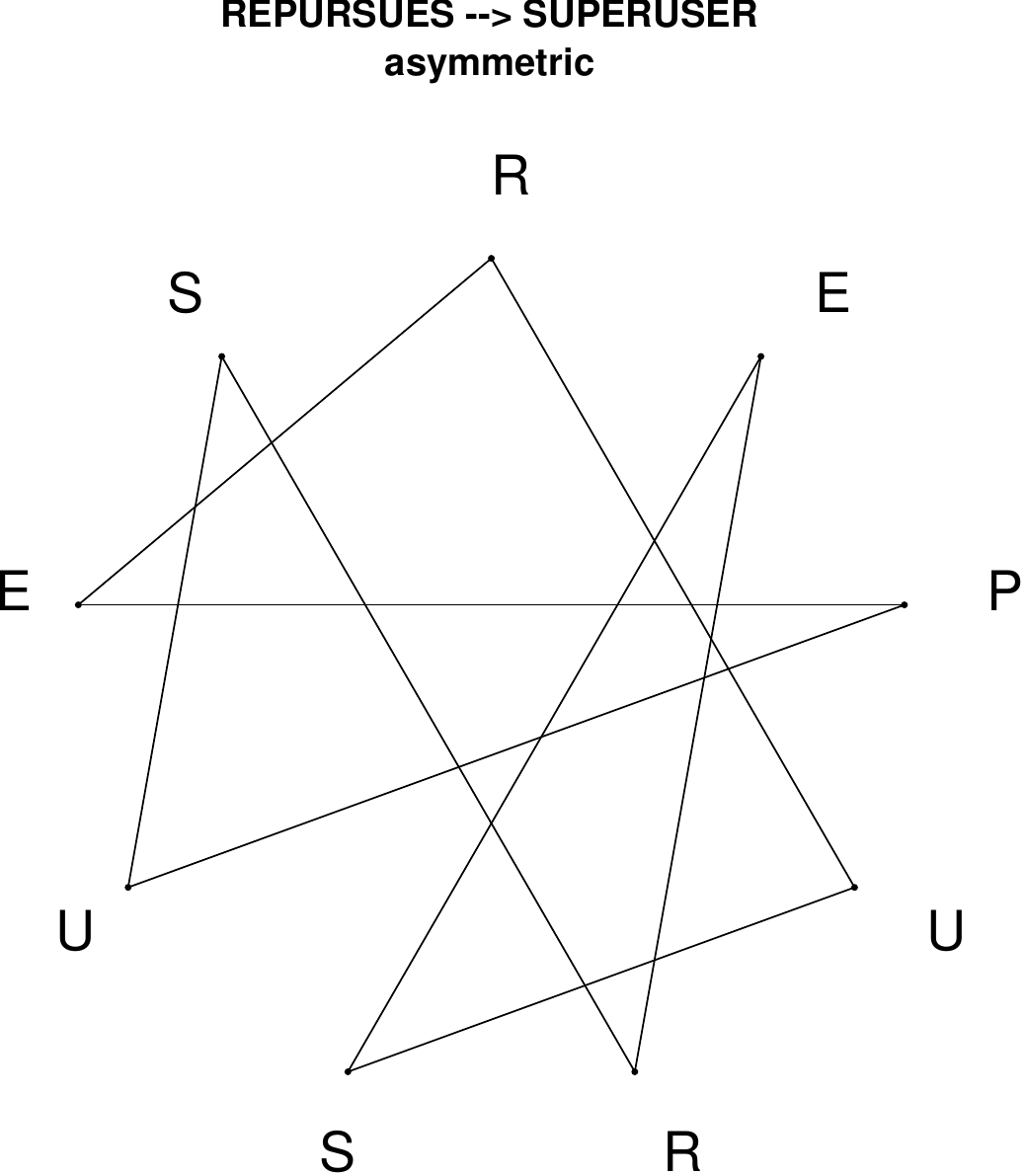}
\end{subfigure}
\hfill
\begin{subfigure}[T]{0.19\textwidth}
\centering
\includegraphics[width=\textwidth]{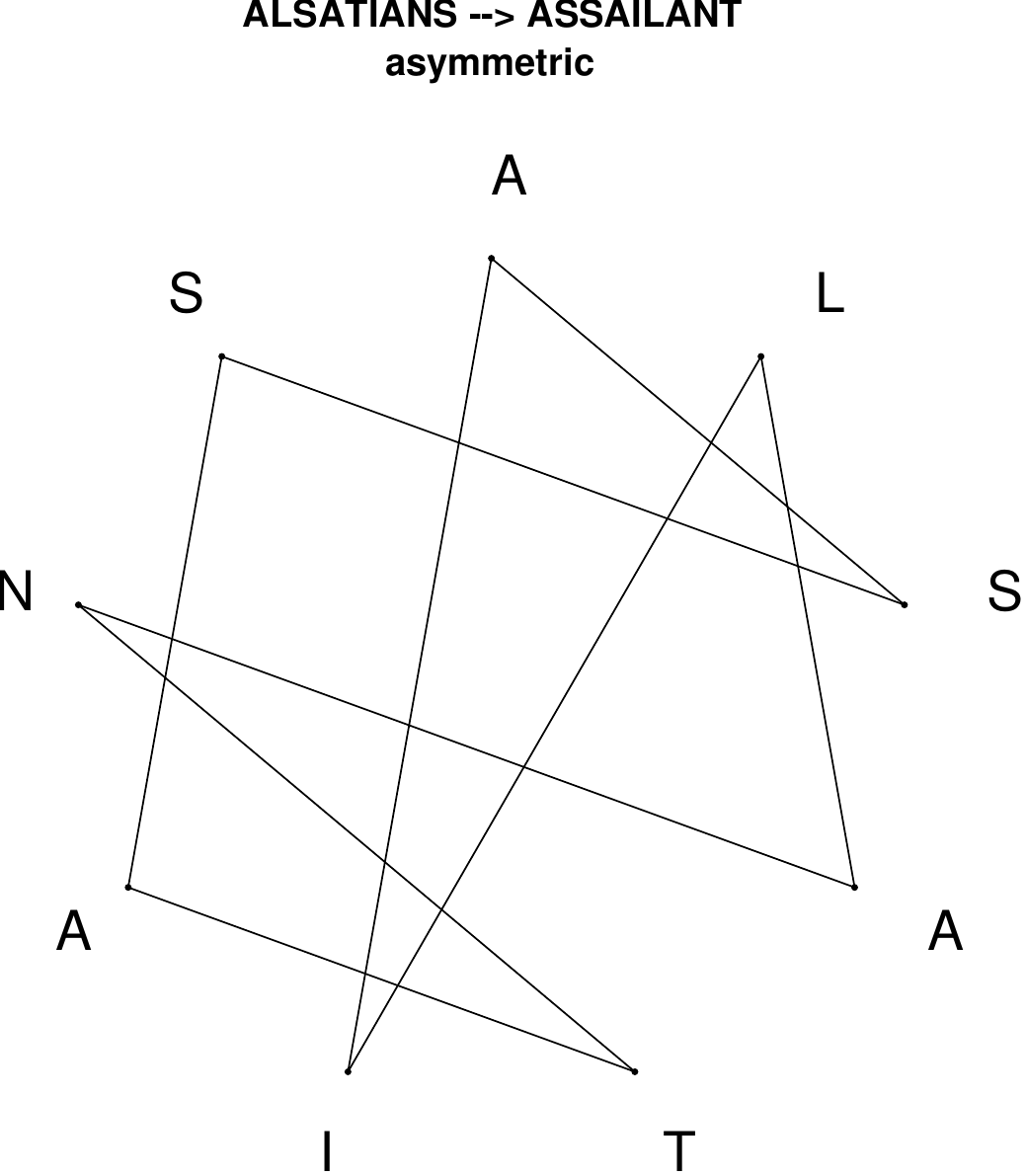}
\end{subfigure}
\hfill
\begin{subfigure}[T]{0.19\textwidth}
\centering
\includegraphics[width=\textwidth]{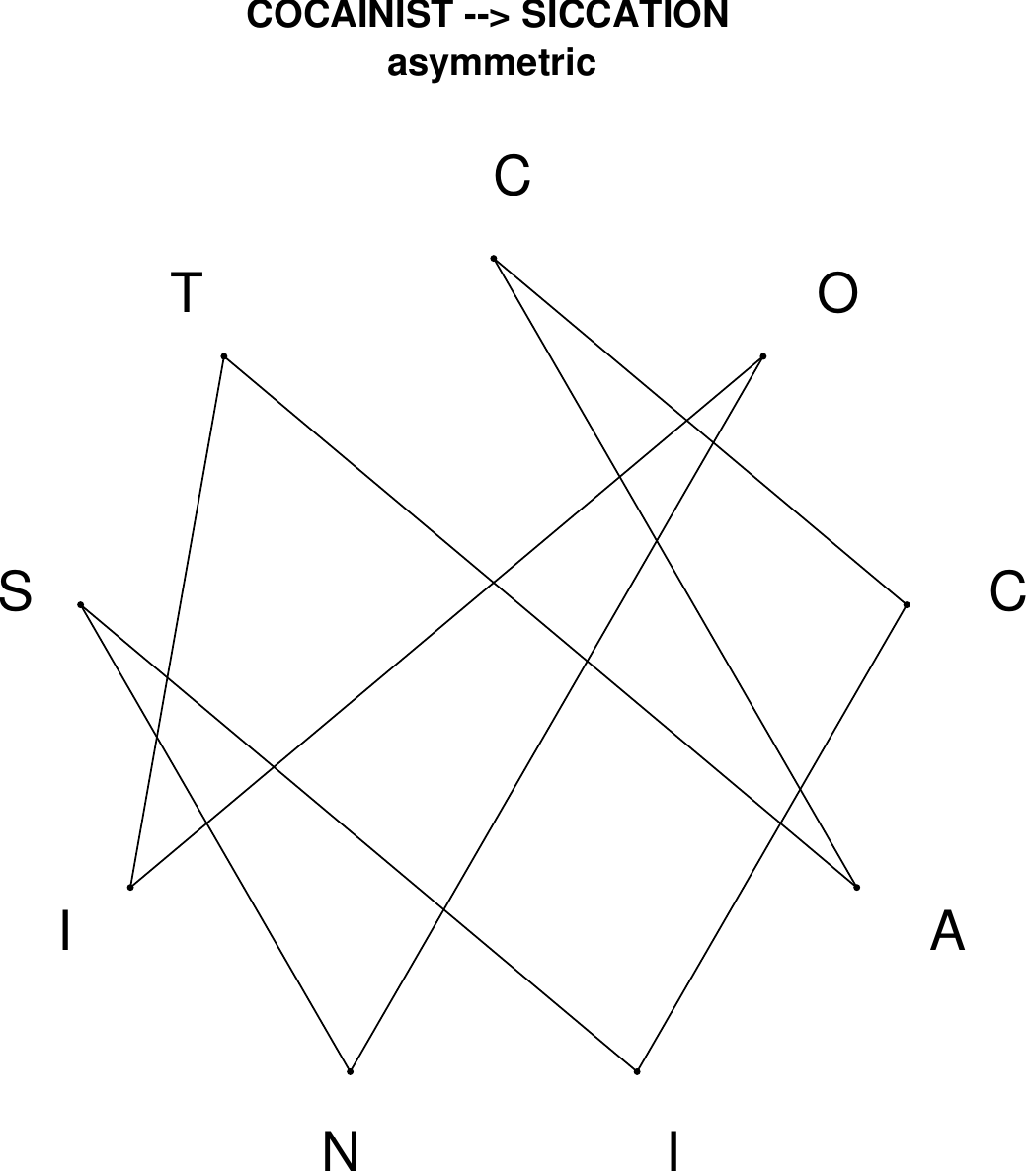}
\end{subfigure}
\end{figure}

\begin{figure}[H]
\centering
\begin{subfigure}[T]{0.19\textwidth}
\centering
\includegraphics[width=\textwidth]{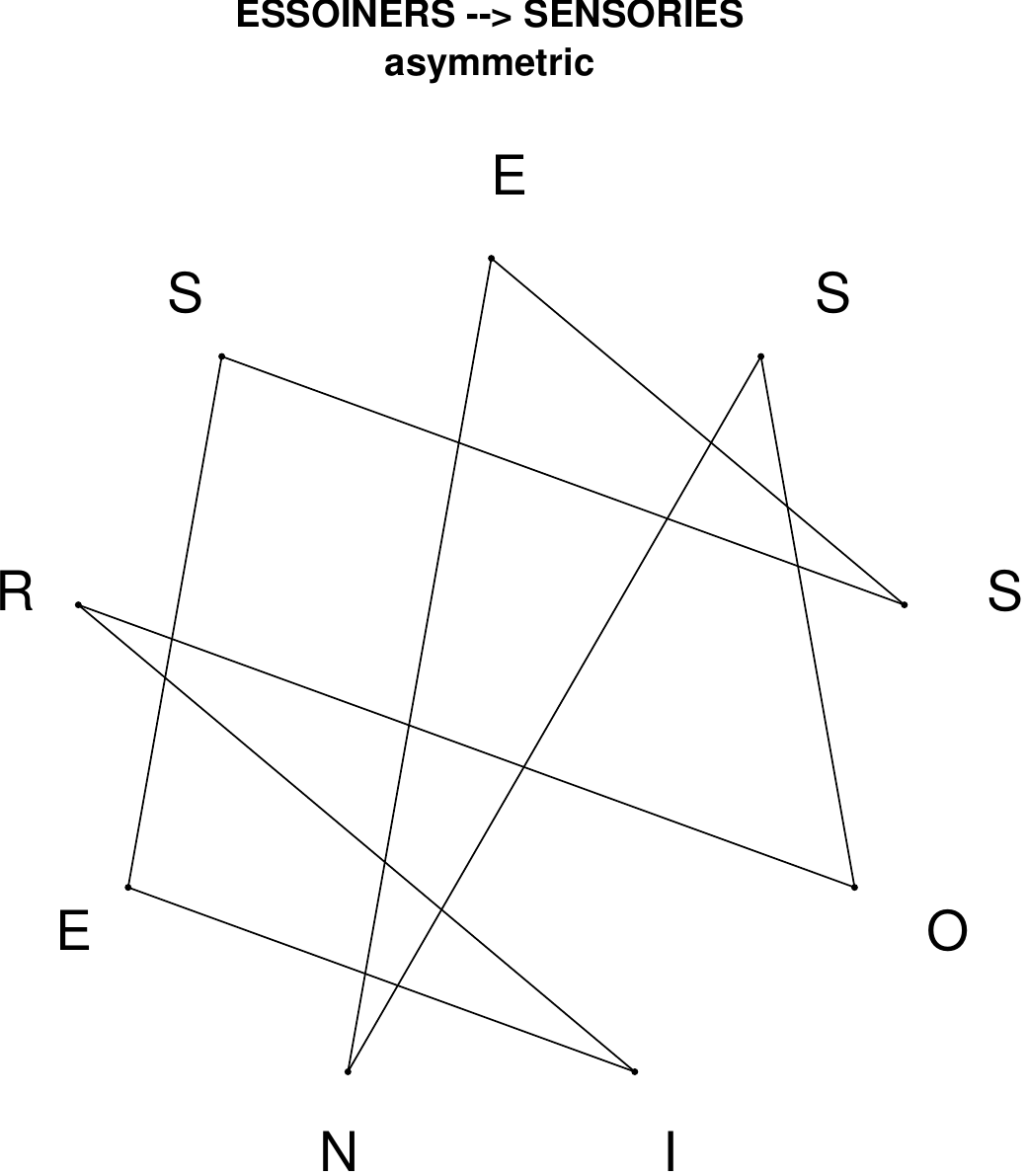}
\end{subfigure}
\hfill
\begin{subfigure}[T]{0.19\textwidth}
\centering
\includegraphics[width=\textwidth]{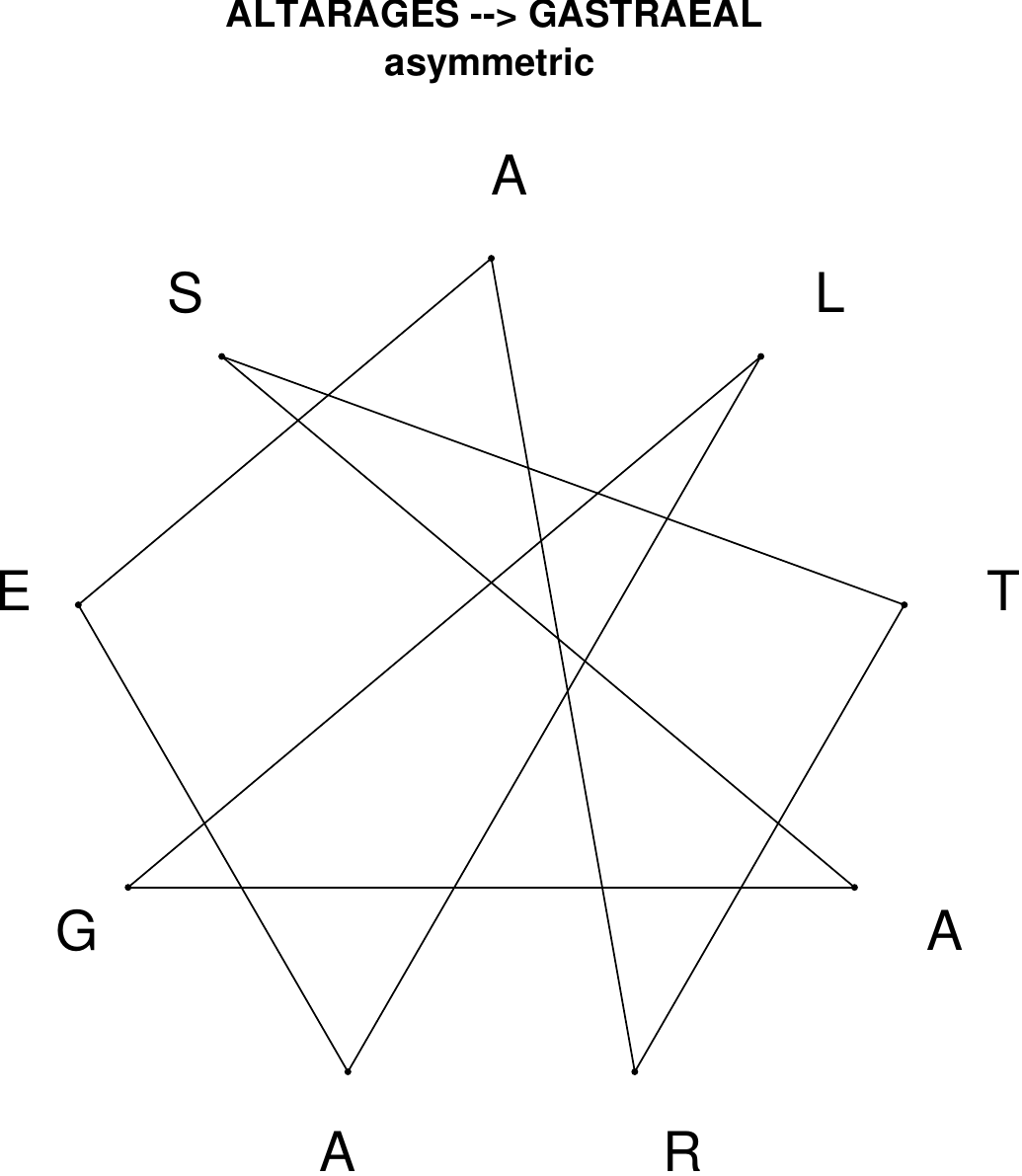}
\end{subfigure}
\hfill
\begin{subfigure}[T]{0.19\textwidth}
\centering
\includegraphics[width=\textwidth]{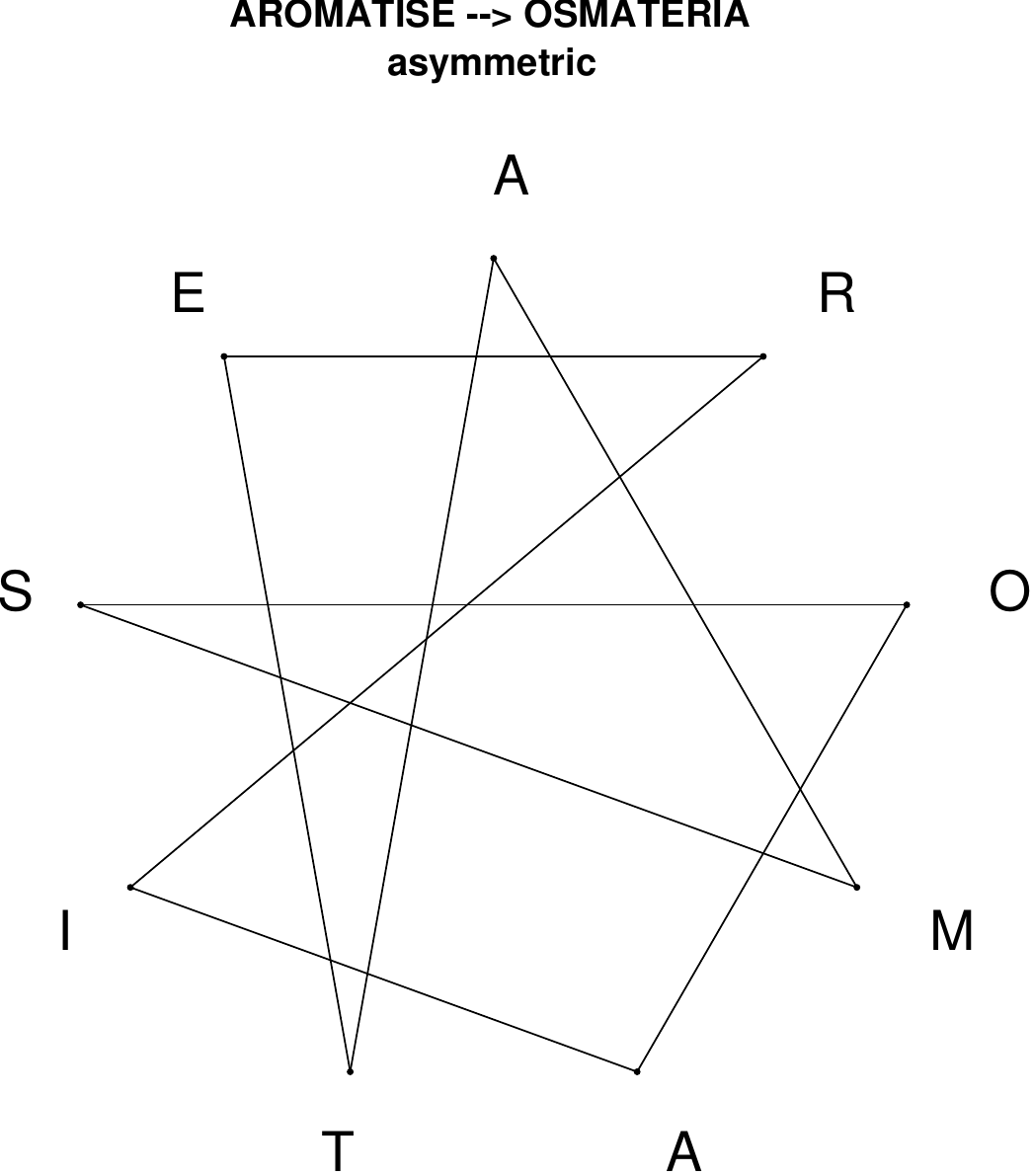}
\end{subfigure}
\hfill
\begin{subfigure}[T]{0.19\textwidth}
\centering
\includegraphics[width=\textwidth]{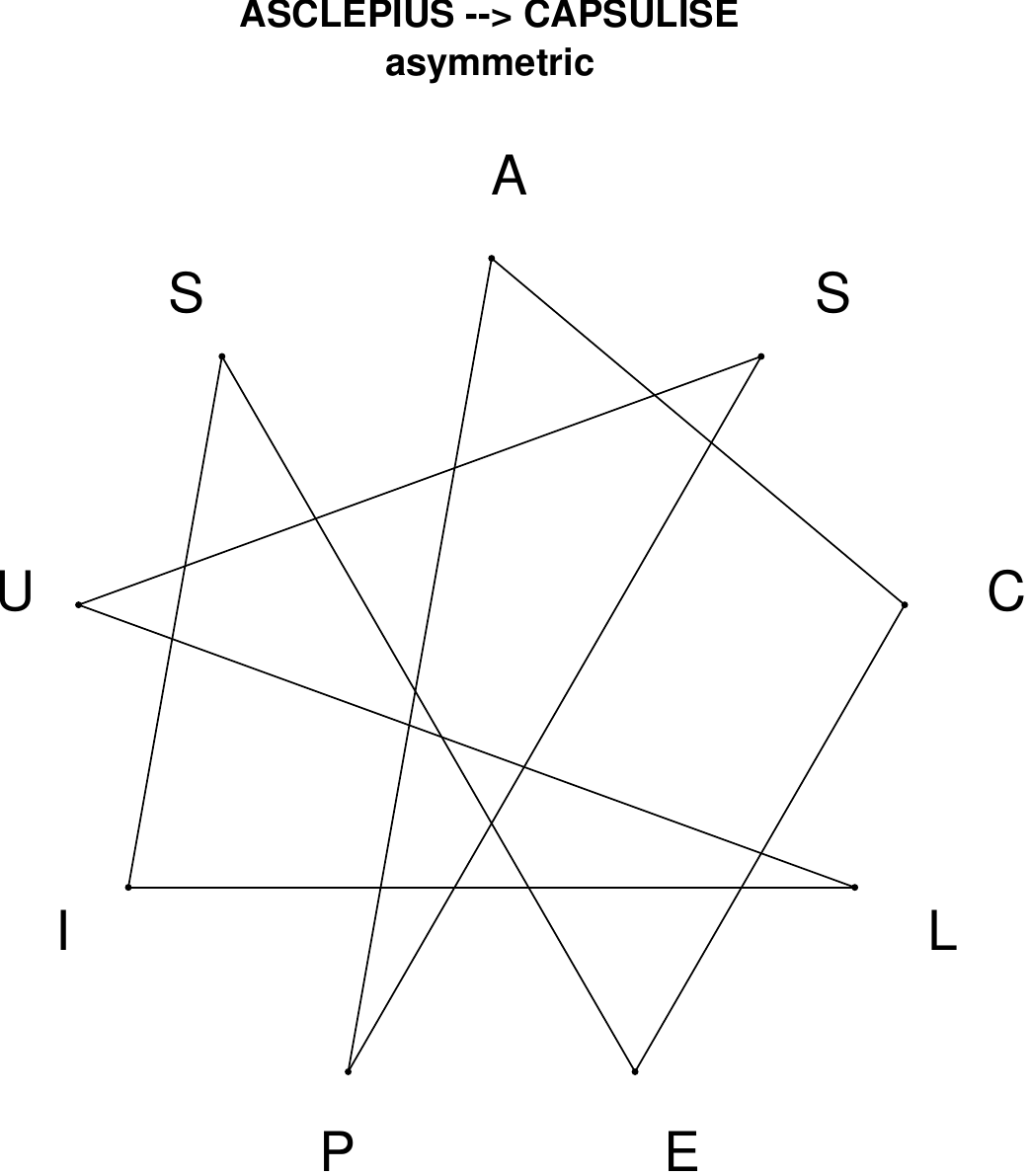}
\end{subfigure}
\hfill
\begin{subfigure}[T]{0.19\textwidth}
\centering
\includegraphics[width=\textwidth]{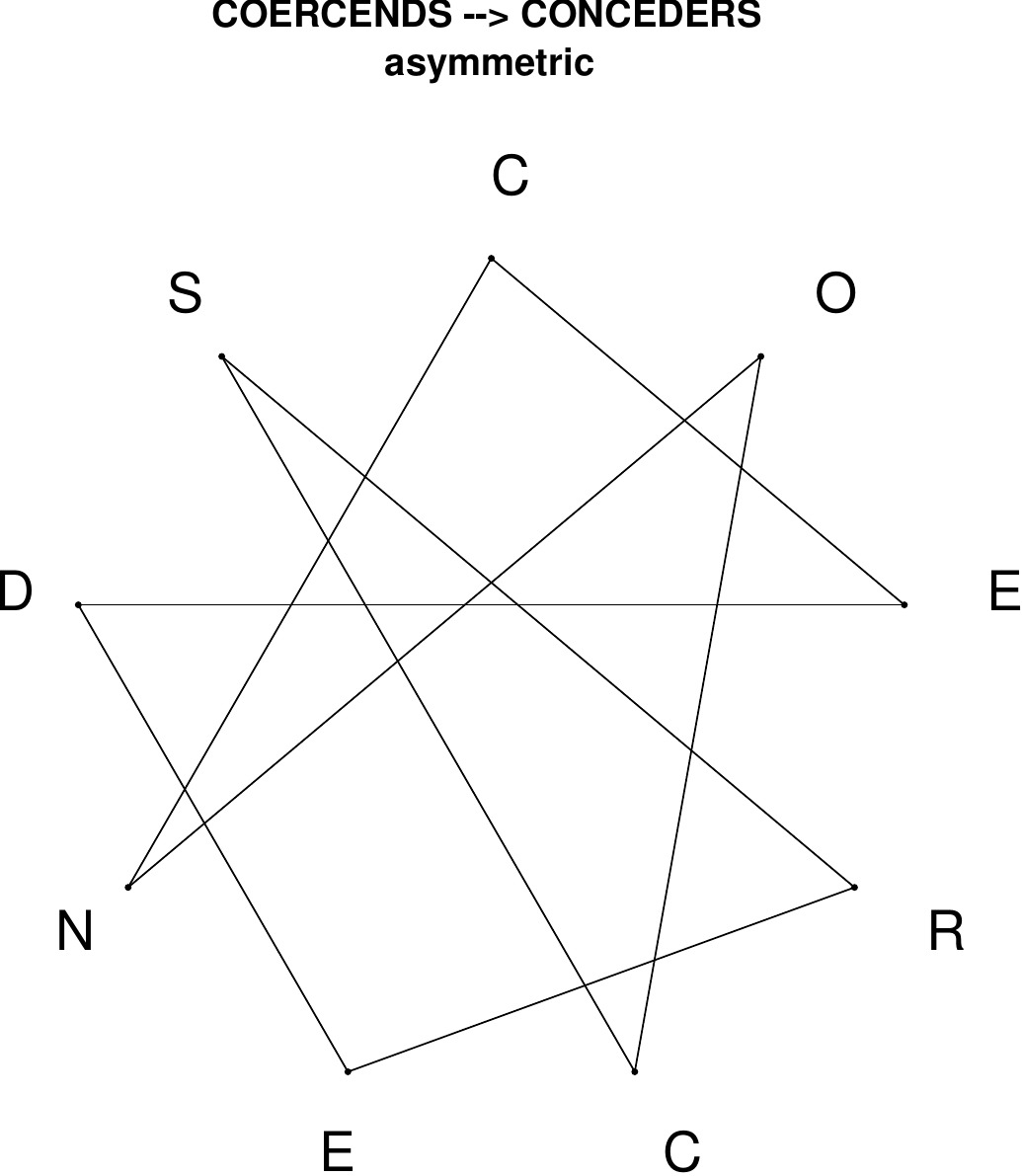}
\end{subfigure}
\end{figure}

\begin{figure}[H]
\centering
\begin{subfigure}[T]{0.19\textwidth}
\centering
\includegraphics[width=\textwidth]{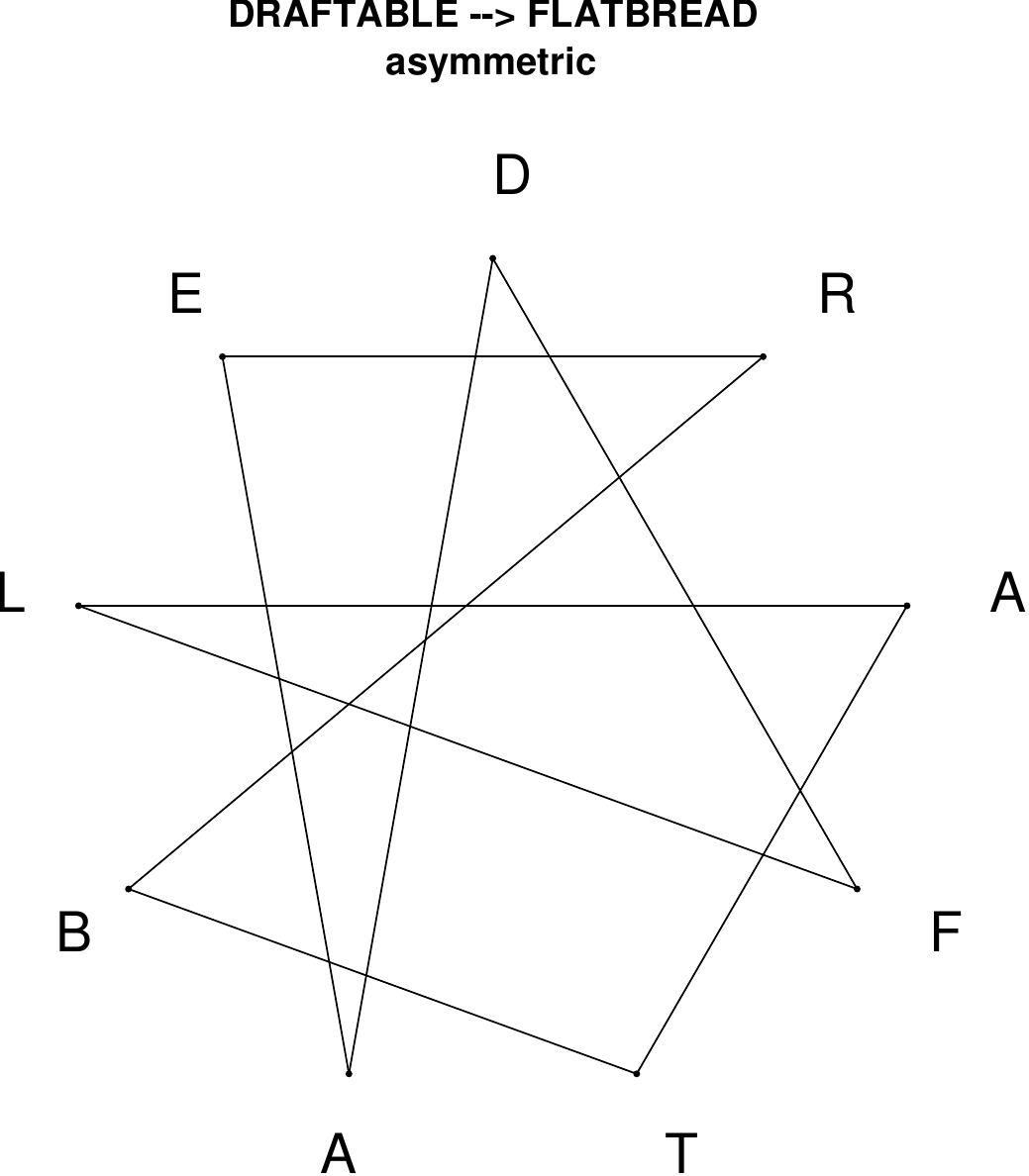}
\end{subfigure}
\hfill
\begin{subfigure}[T]{0.19\textwidth}
\centering
\includegraphics[width=\textwidth]{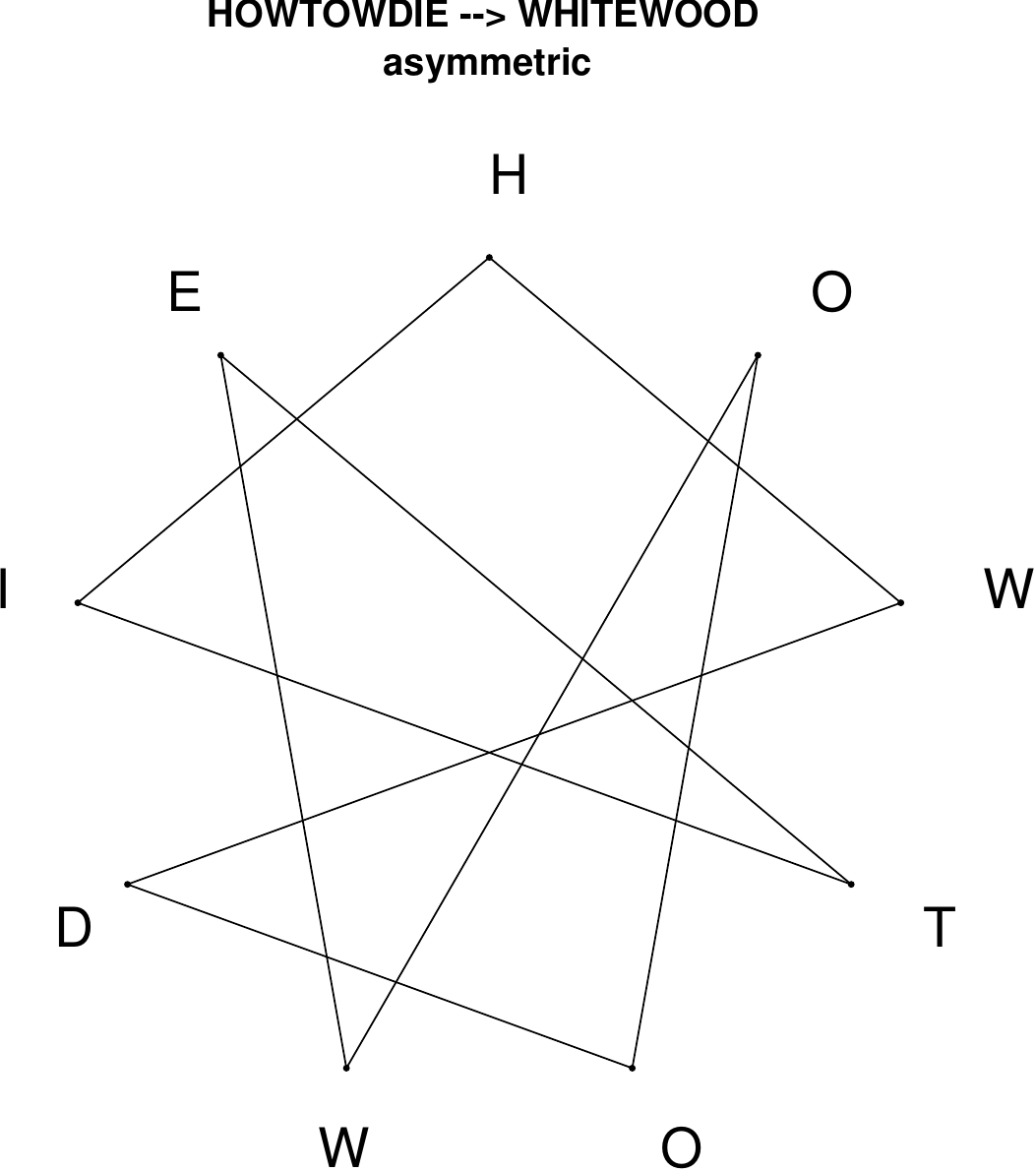}
\end{subfigure}
\hfill
\begin{subfigure}[T]{0.19\textwidth}
\centering
\includegraphics[width=\textwidth]{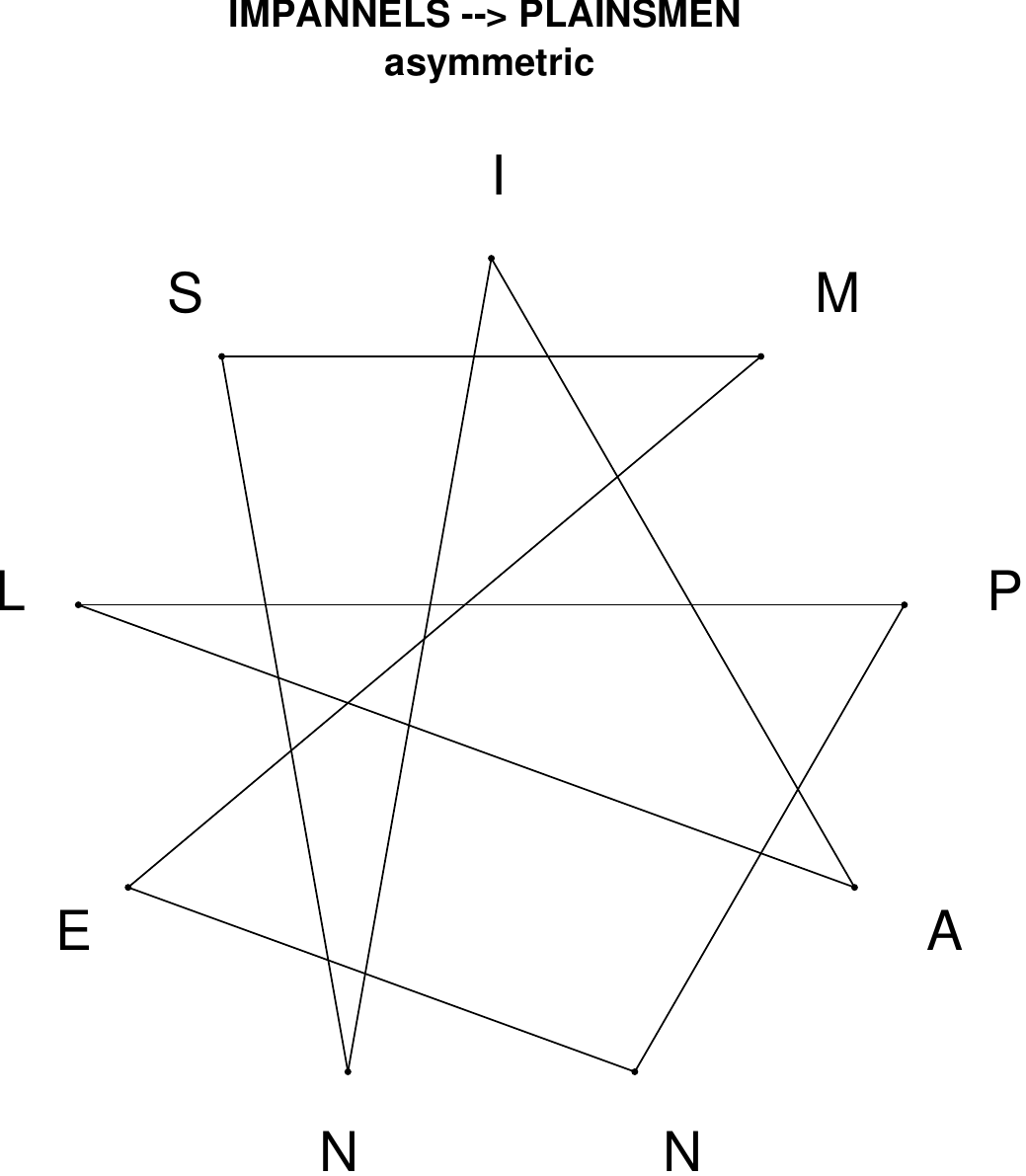}
\end{subfigure}
\hfill
\begin{subfigure}[T]{0.19\textwidth}
\centering
\includegraphics[width=\textwidth]{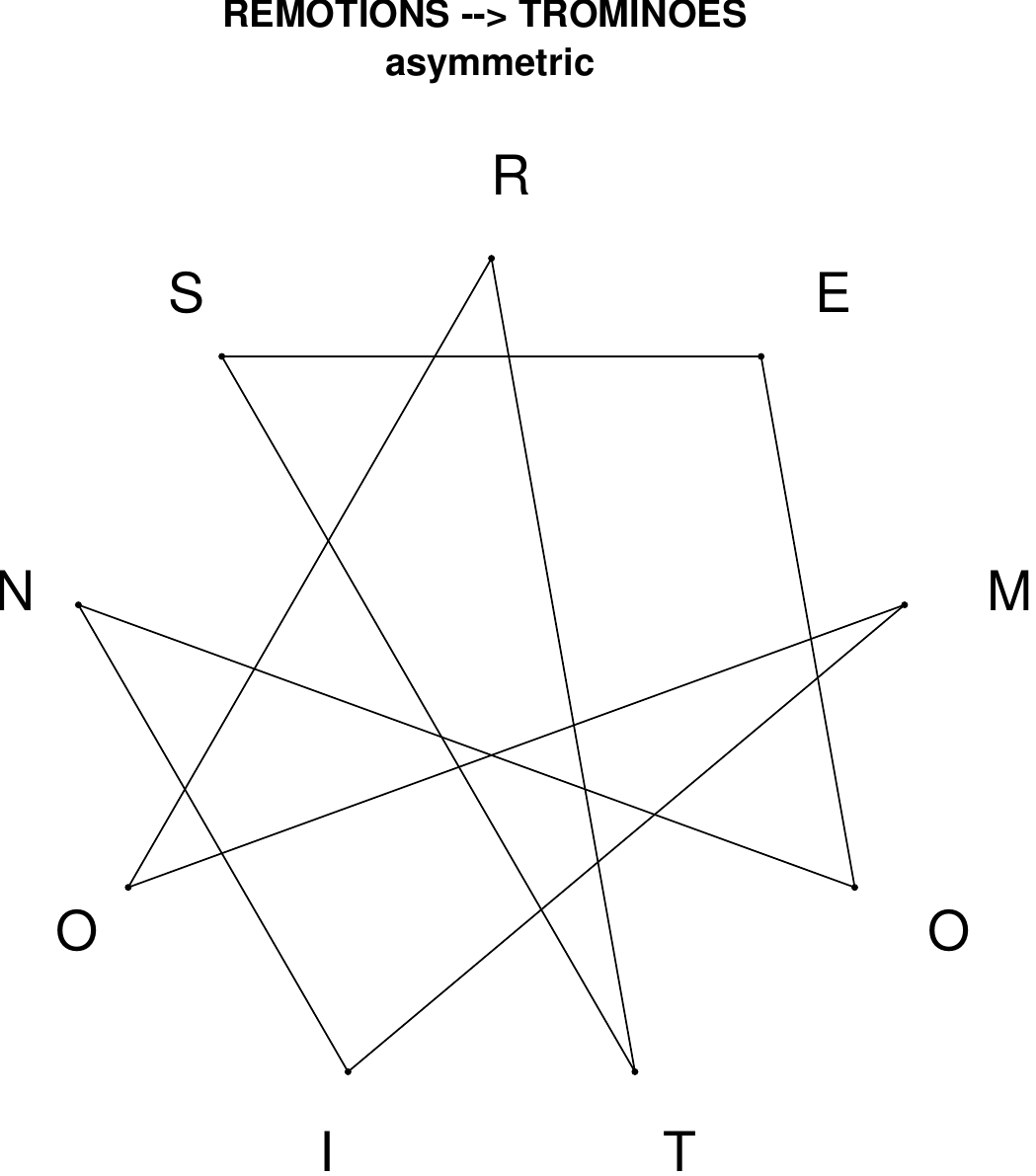}
\end{subfigure}
\hfill
\begin{subfigure}[T]{0.19\textwidth}
\centering
\includegraphics[width=\textwidth]{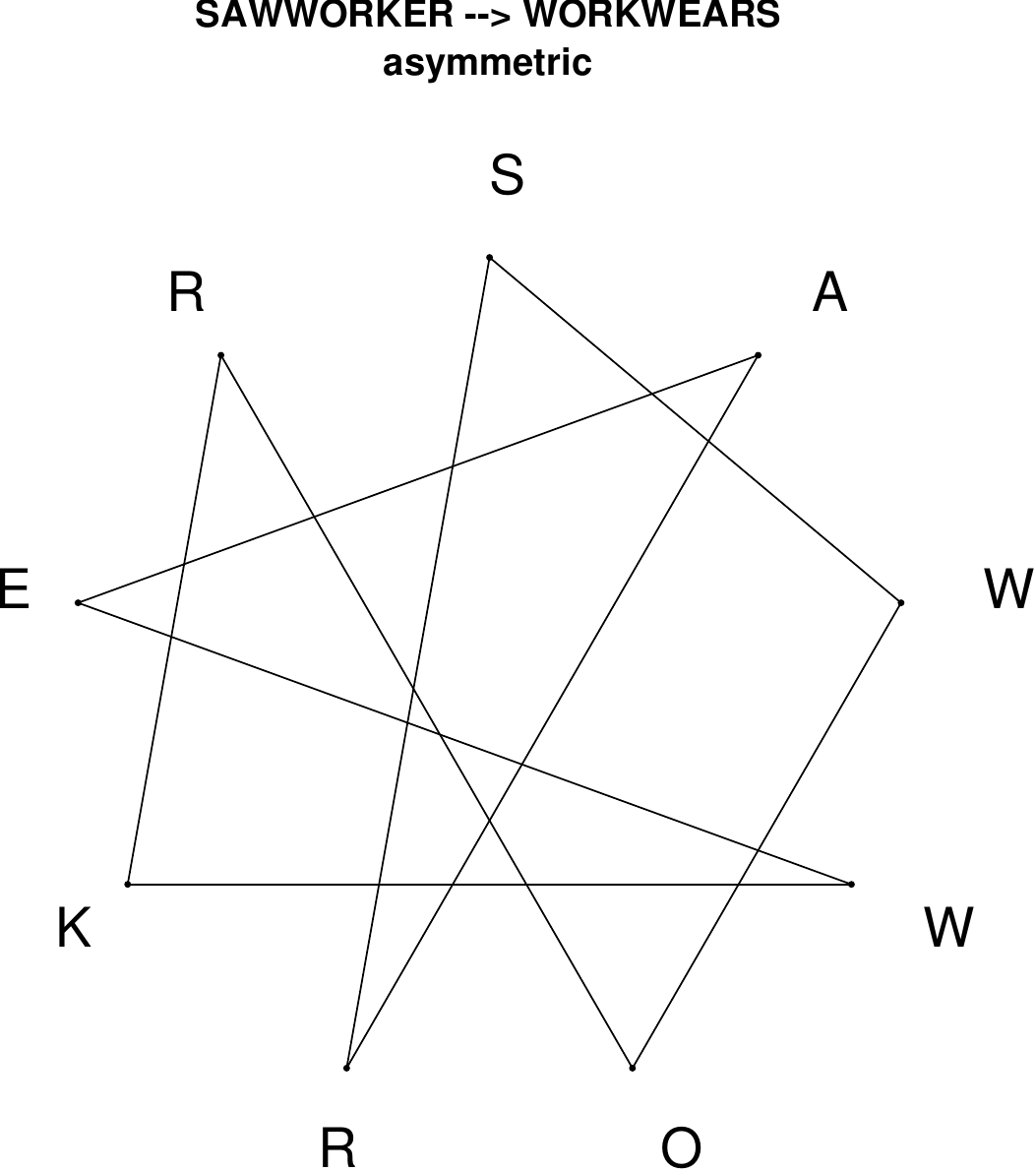}
\end{subfigure}
\end{figure}

\begin{figure}[H]
\centering
\begin{subfigure}[T]{0.19\textwidth}
\centering
\includegraphics[width=\textwidth]{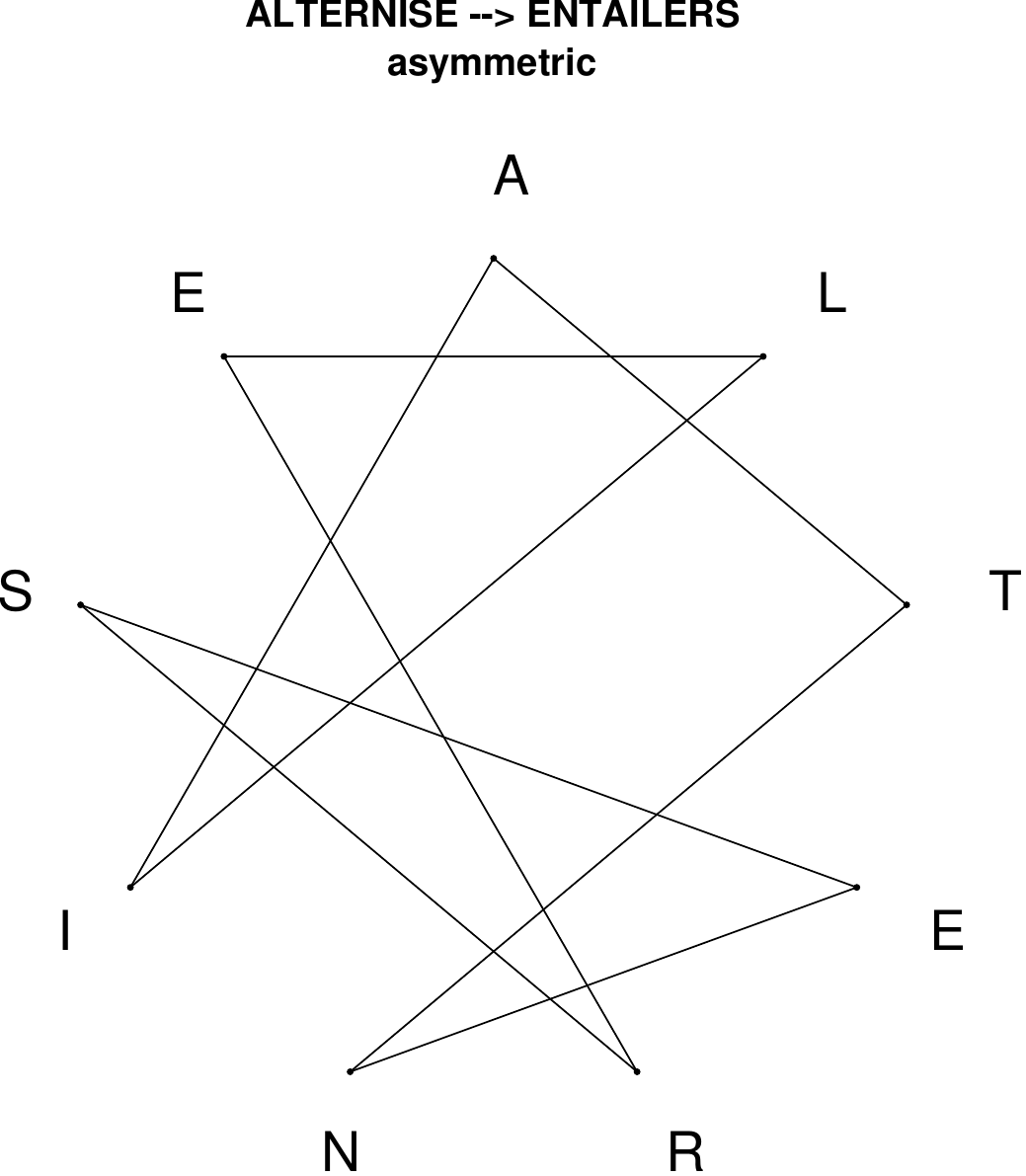}
\end{subfigure}
\hfill
\begin{subfigure}[T]{0.19\textwidth}
\centering
\includegraphics[width=\textwidth]{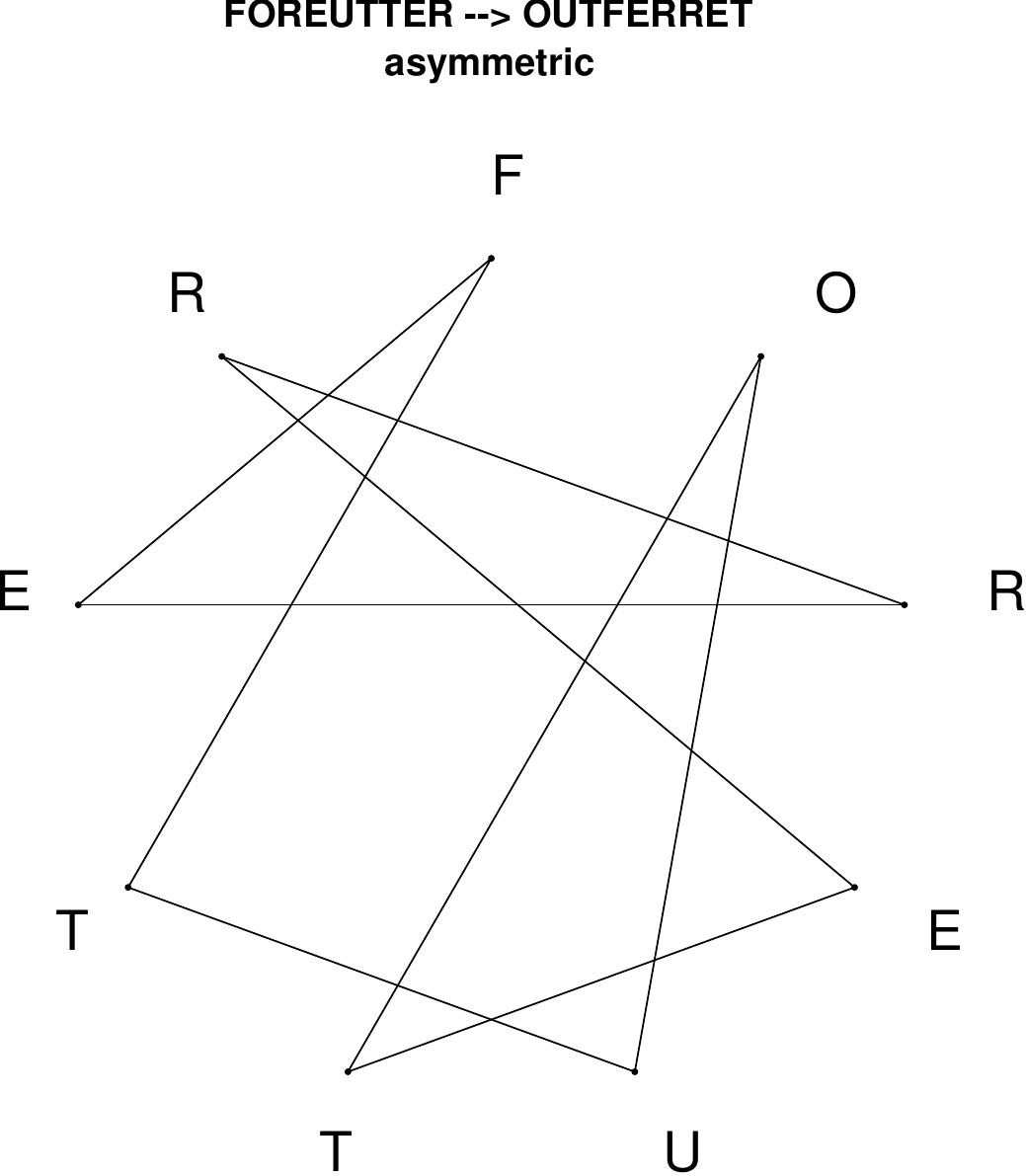}
\end{subfigure}
\hfill
\begin{subfigure}[T]{0.19\textwidth}
\centering
\includegraphics[width=\textwidth]{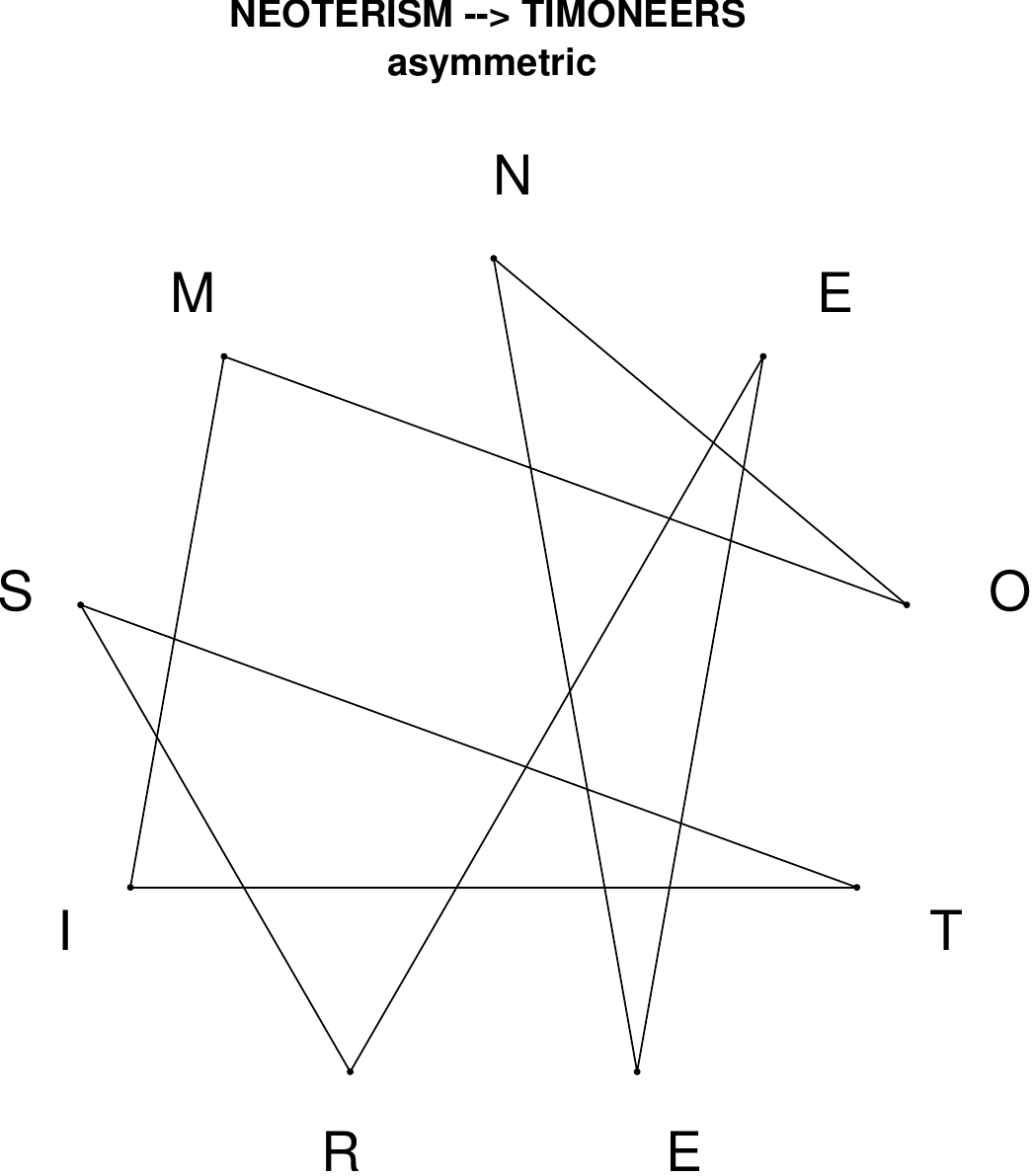}
\end{subfigure}
\hfill
\begin{subfigure}[T]{0.19\textwidth}
\centering
\includegraphics[width=\textwidth]{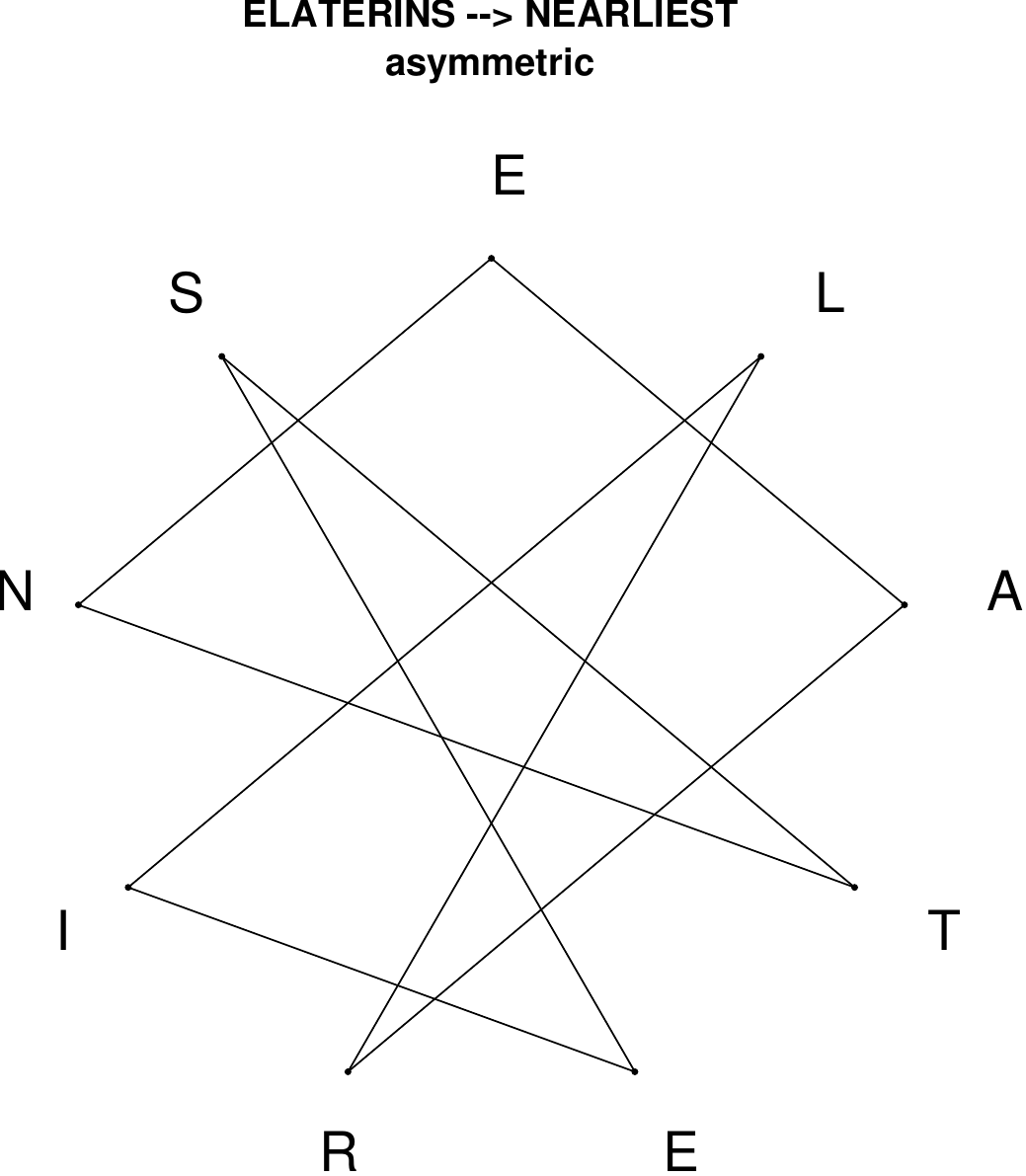}
\end{subfigure}
\hfill
\begin{subfigure}[T]{0.19\textwidth}
\centering
\includegraphics[width=\textwidth]{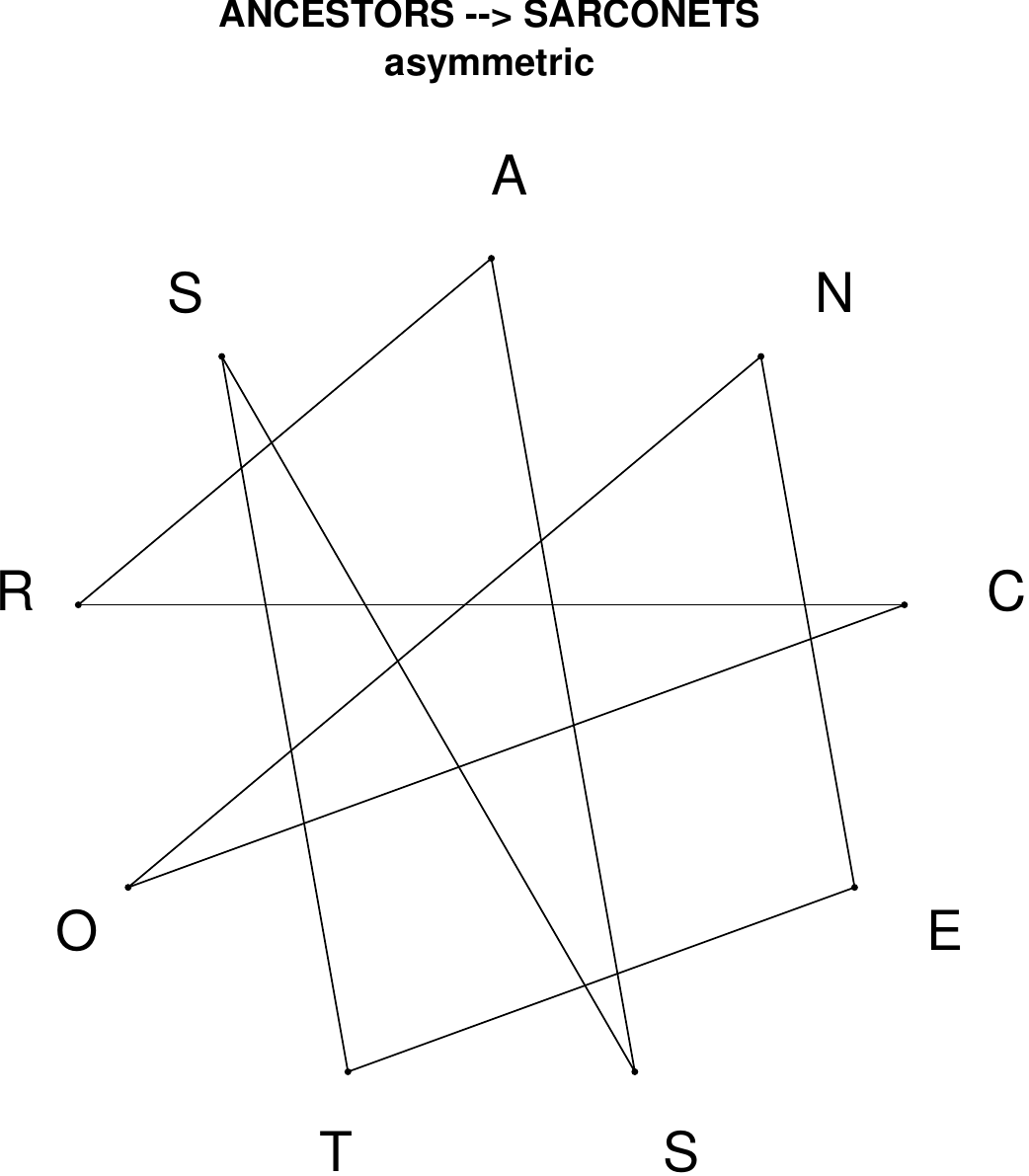}
\end{subfigure}
\end{figure}

\begin{figure}[H]
\centering
\begin{subfigure}[T]{0.19\textwidth}
\centering
\includegraphics[width=\textwidth]{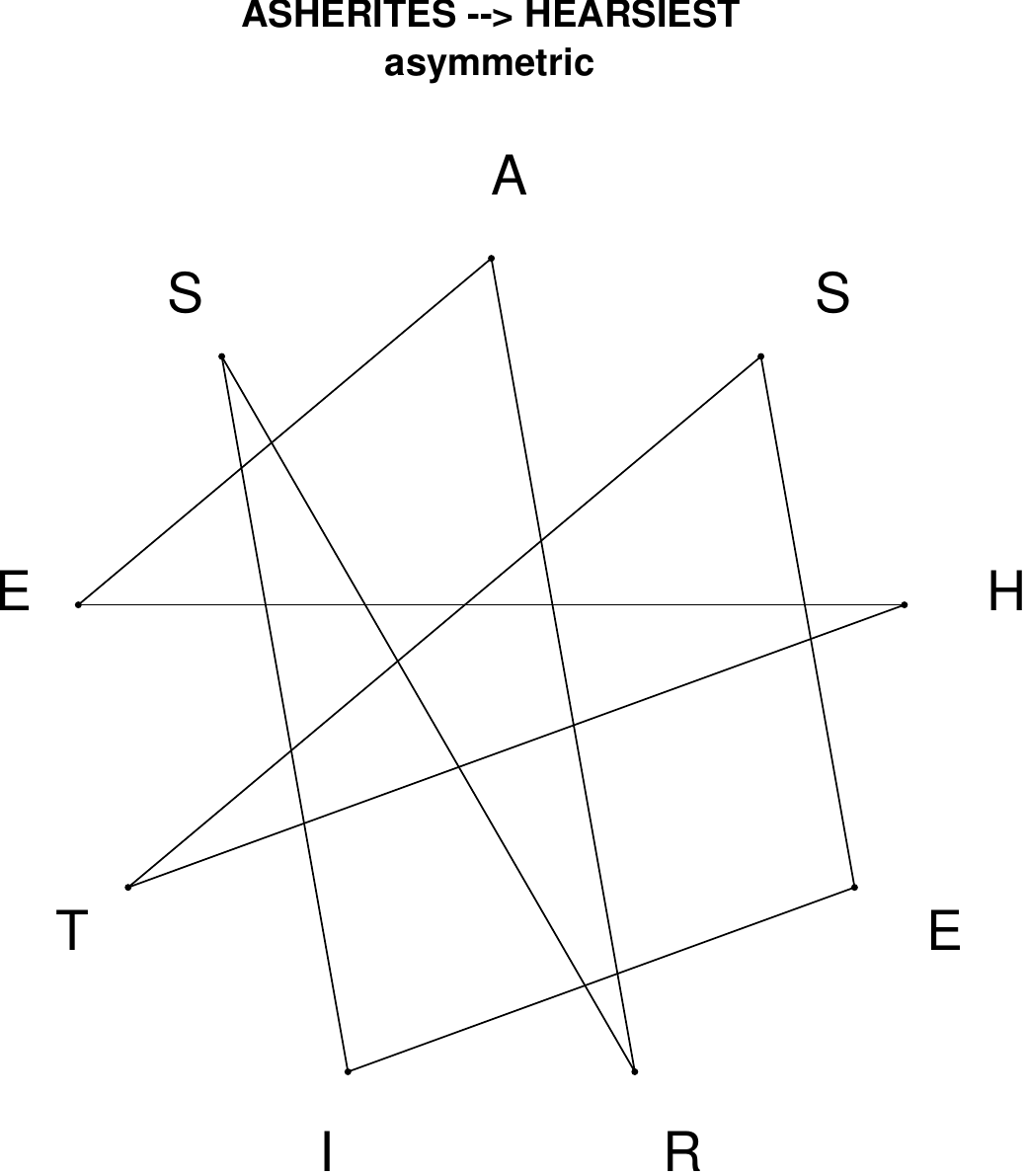}
\end{subfigure}
\hfill
\begin{subfigure}[T]{0.19\textwidth}
\centering
\includegraphics[width=\textwidth]{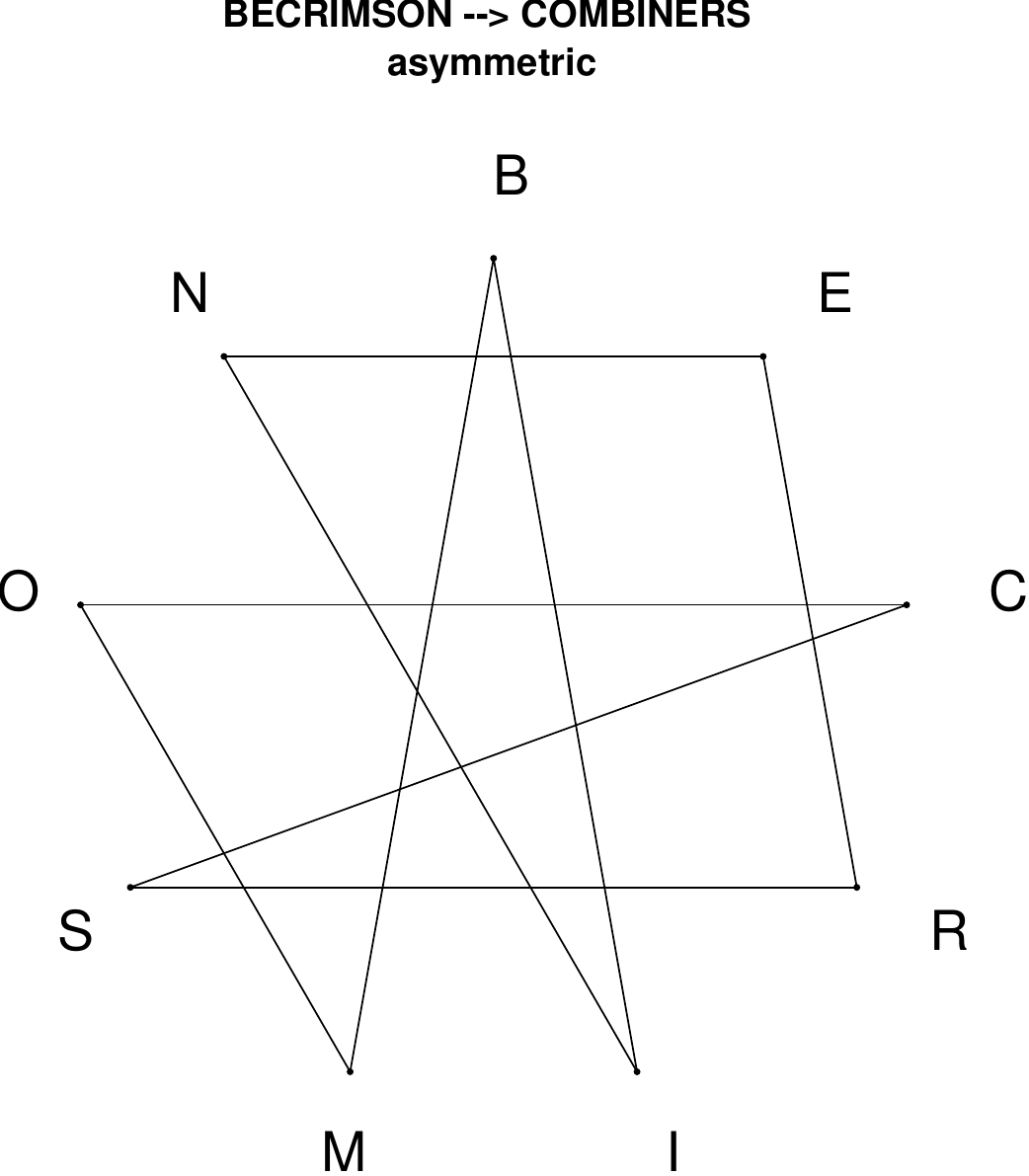}
\end{subfigure}
\hfill
\begin{subfigure}[T]{0.19\textwidth}
\centering
\includegraphics[width=\textwidth]{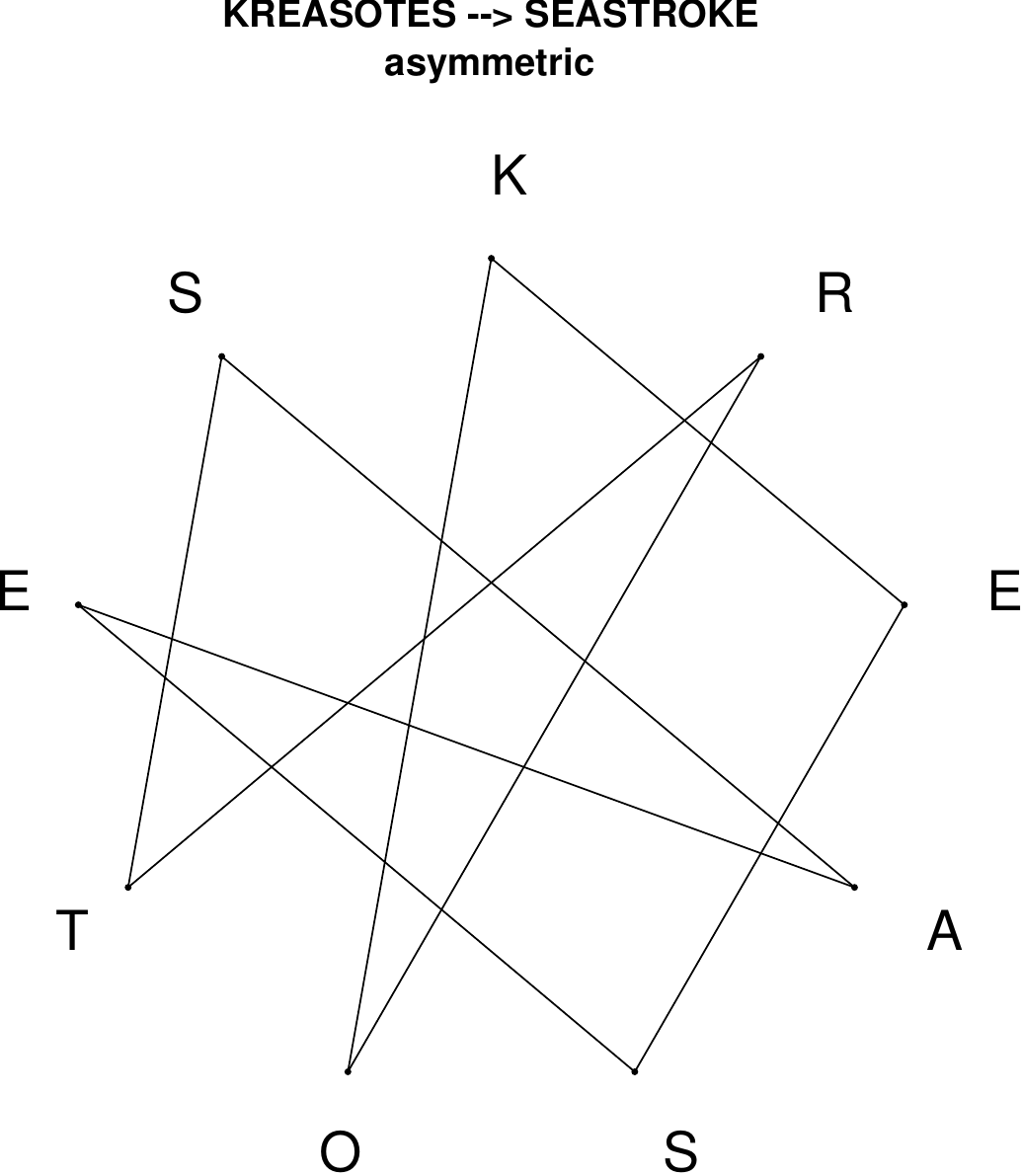}
\end{subfigure}
\hfill
\begin{subfigure}[T]{0.19\textwidth}
\centering
\includegraphics[width=\textwidth]{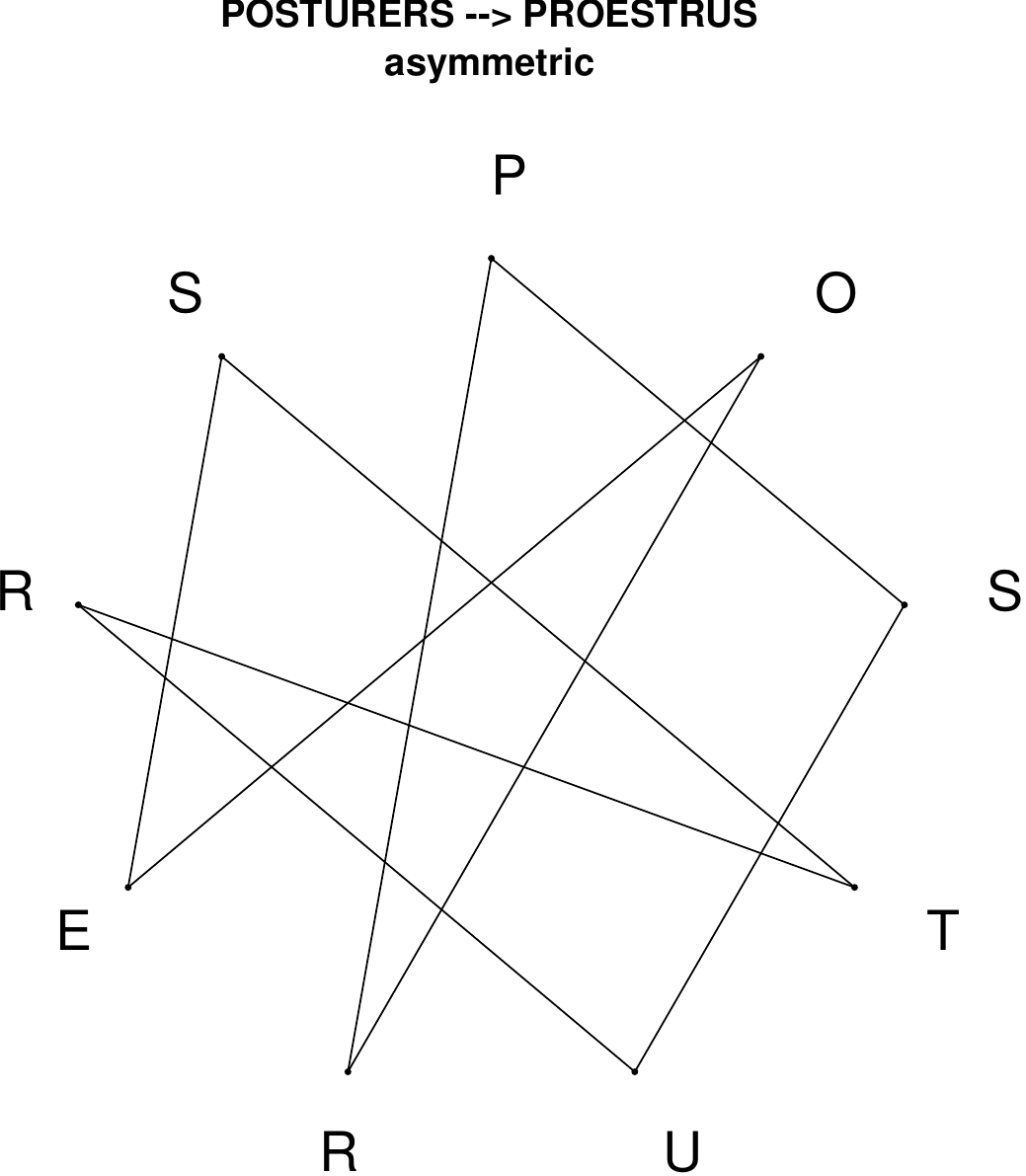}
\end{subfigure}
\hfill
\begin{subfigure}[T]{0.19\textwidth}
\centering
\includegraphics[width=\textwidth]{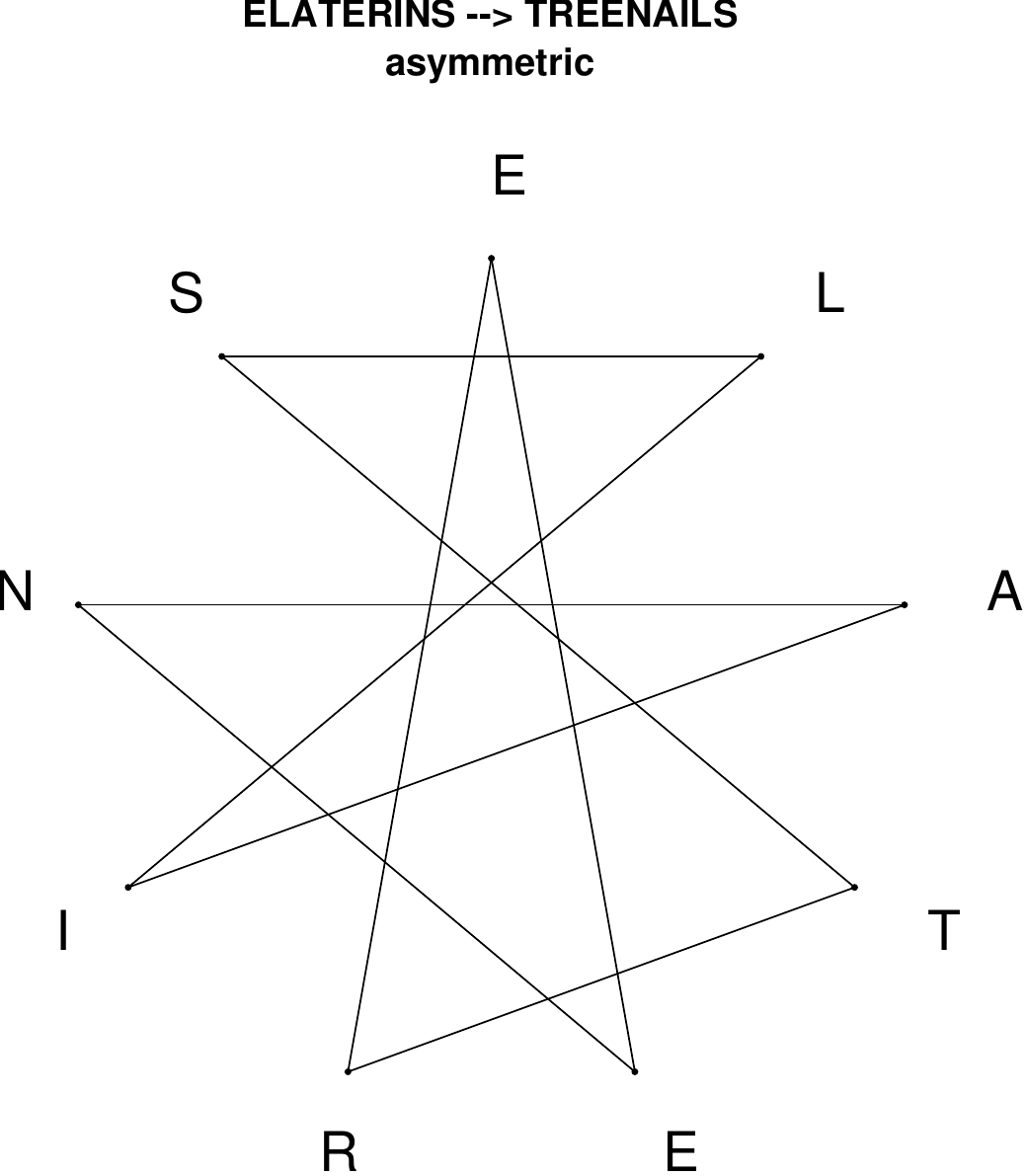}
\end{subfigure}
\end{figure}

\begin{figure}[H]
\centering
\begin{subfigure}[T]{0.19\textwidth}
\centering
\includegraphics[width=\textwidth]{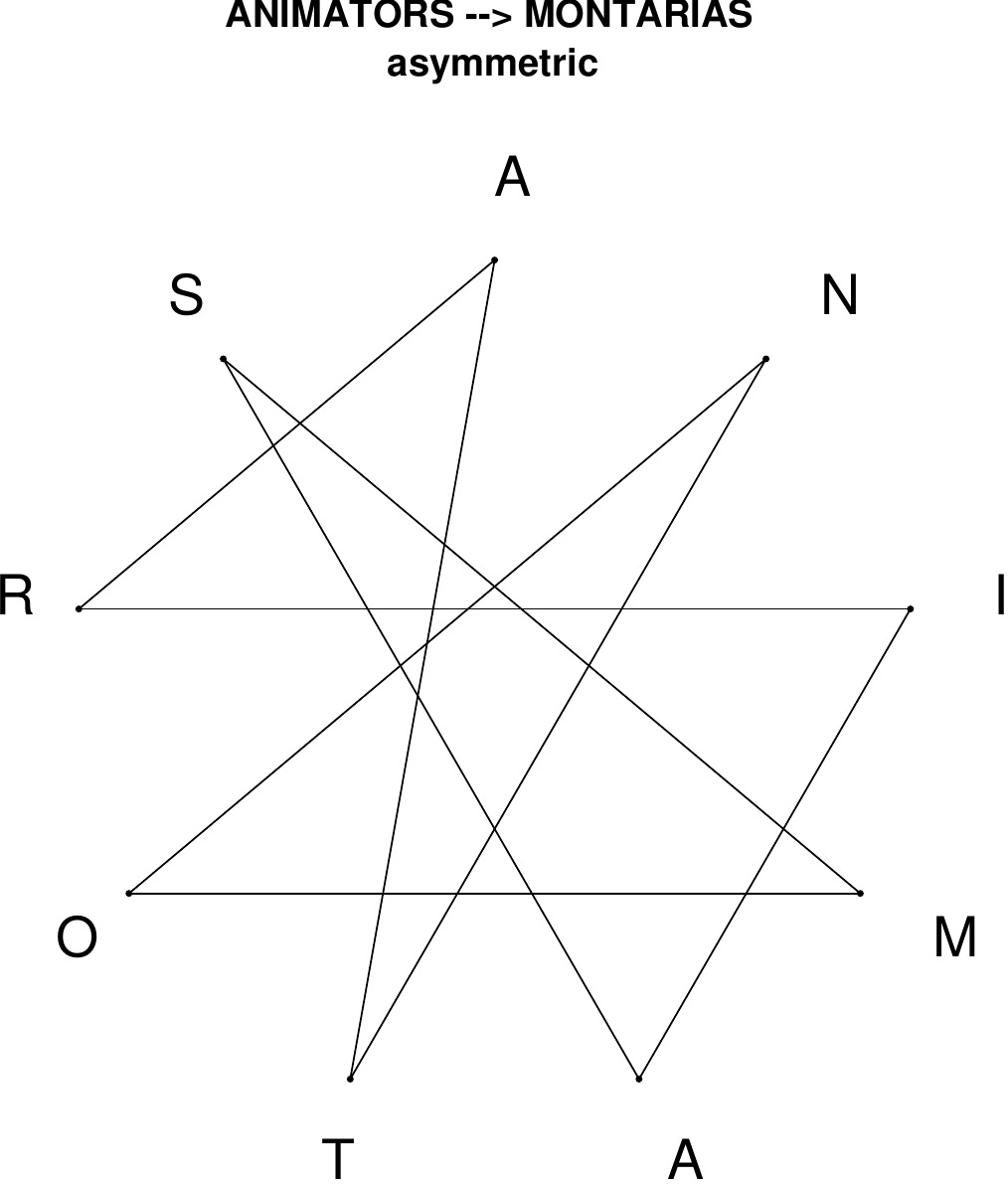}
\end{subfigure}
\hfill
\begin{subfigure}[T]{0.19\textwidth}
\centering
\includegraphics[width=\textwidth]{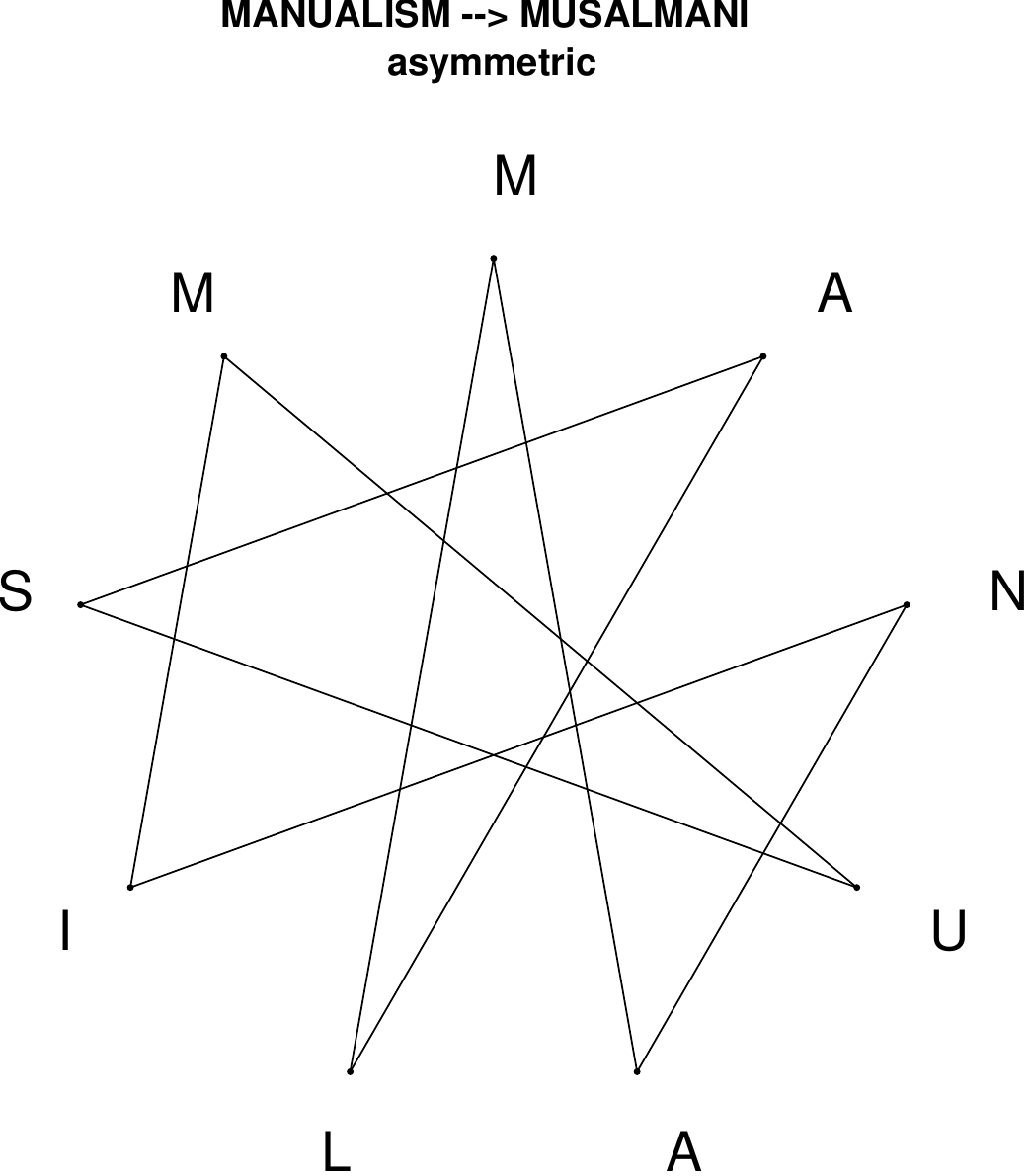}
\end{subfigure}
\hfill
\begin{subfigure}[T]{0.19\textwidth}
\centering
\includegraphics[width=\textwidth]{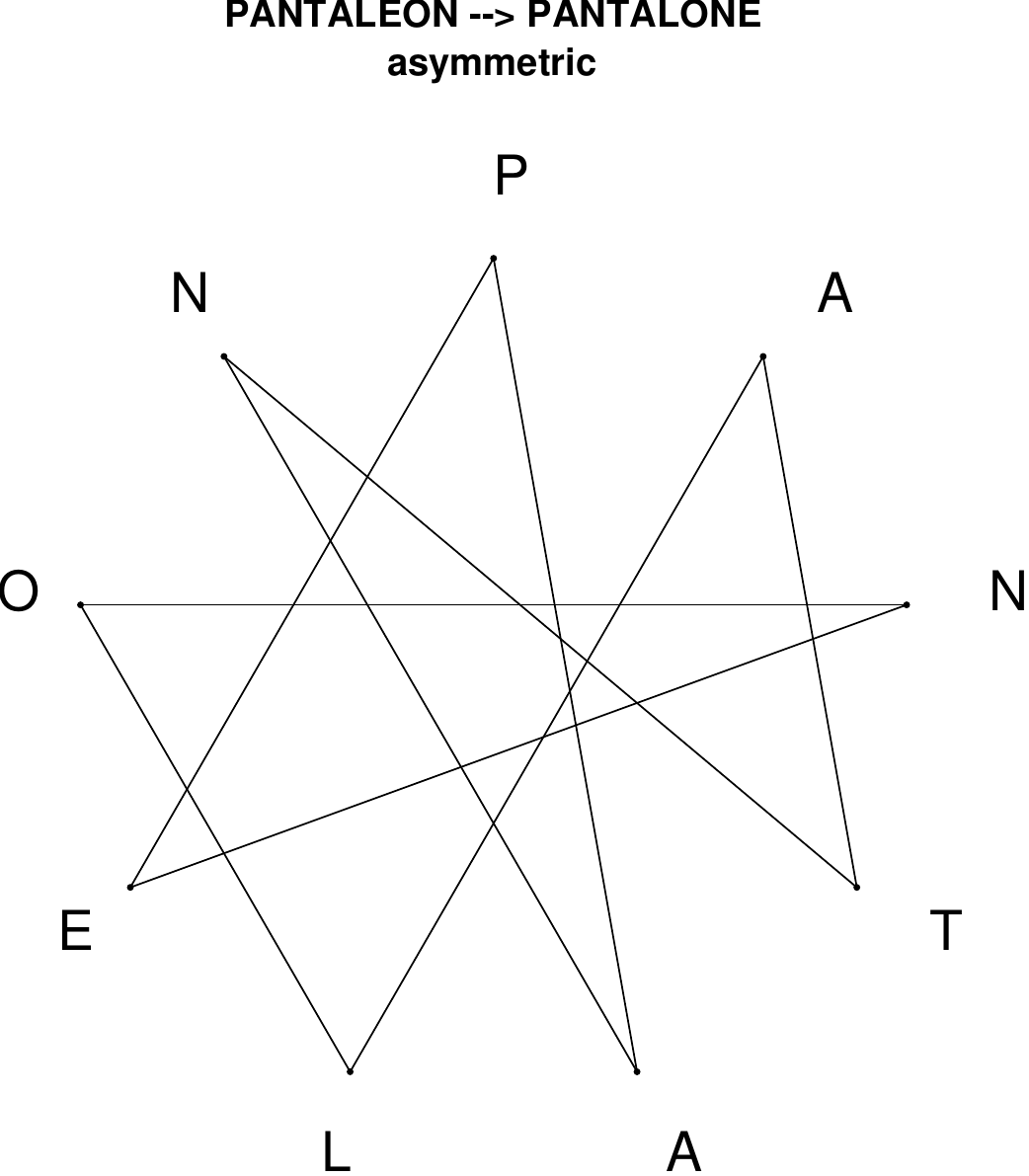}
\end{subfigure}
\hfill
\begin{subfigure}[T]{0.19\textwidth}
\centering
\includegraphics[width=\textwidth]{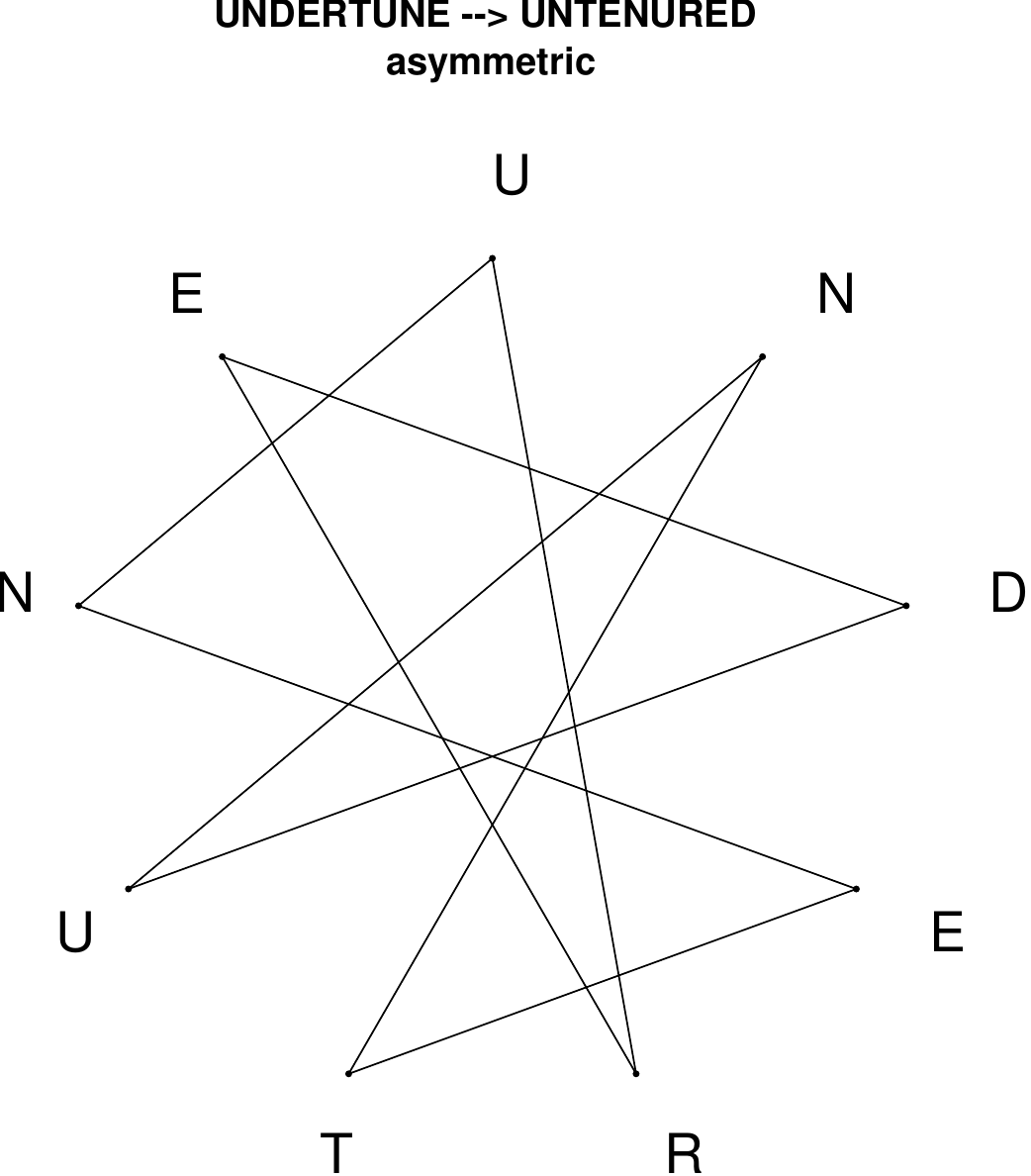}
\end{subfigure}
\hfill
\begin{subfigure}[T]{0.19\textwidth}
\centering
\includegraphics[width=\textwidth]{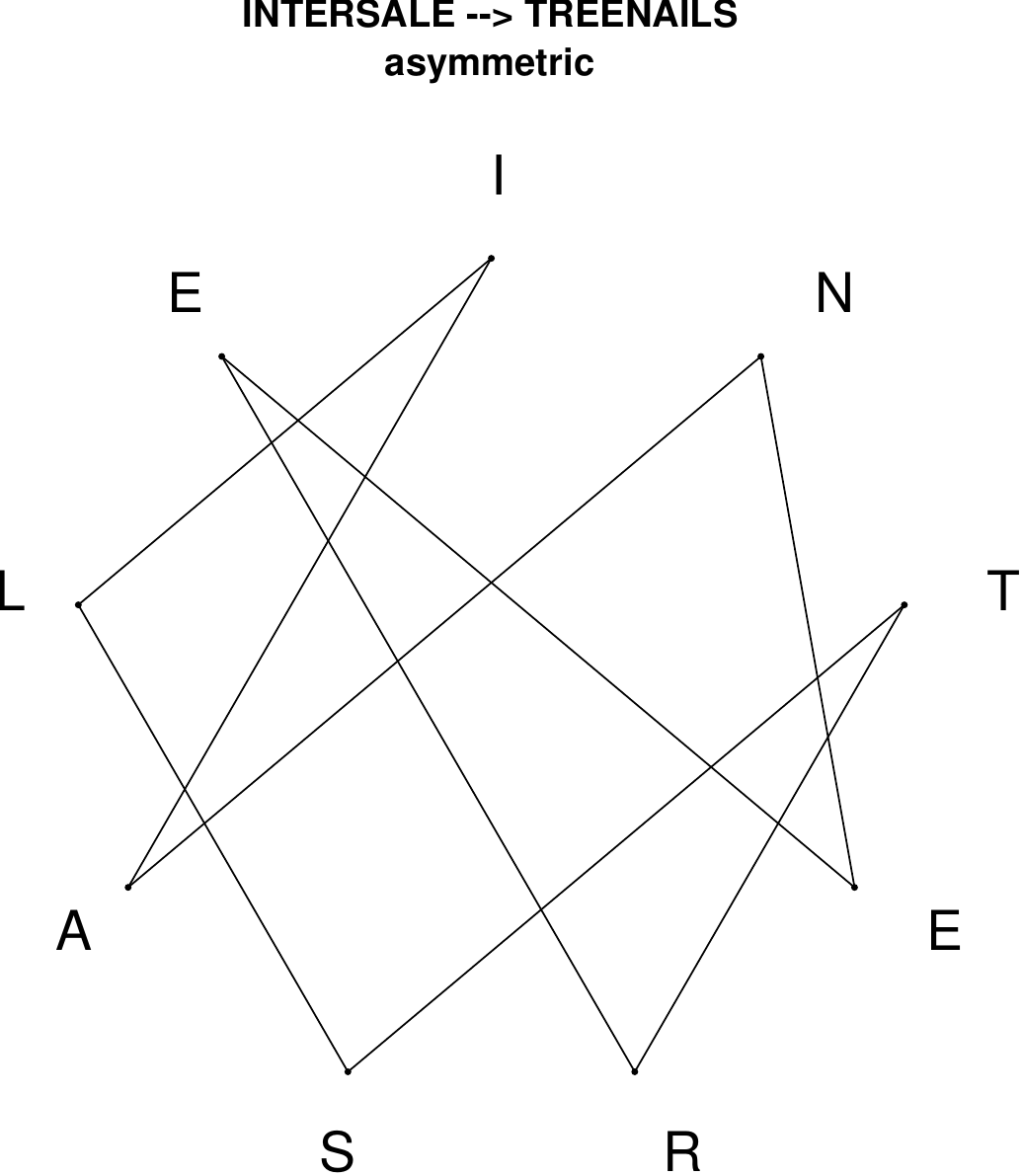}
\end{subfigure}
\end{figure}

\begin{figure}[H]
\centering
\begin{subfigure}[T]{0.19\textwidth}
\centering
\includegraphics[width=\textwidth]{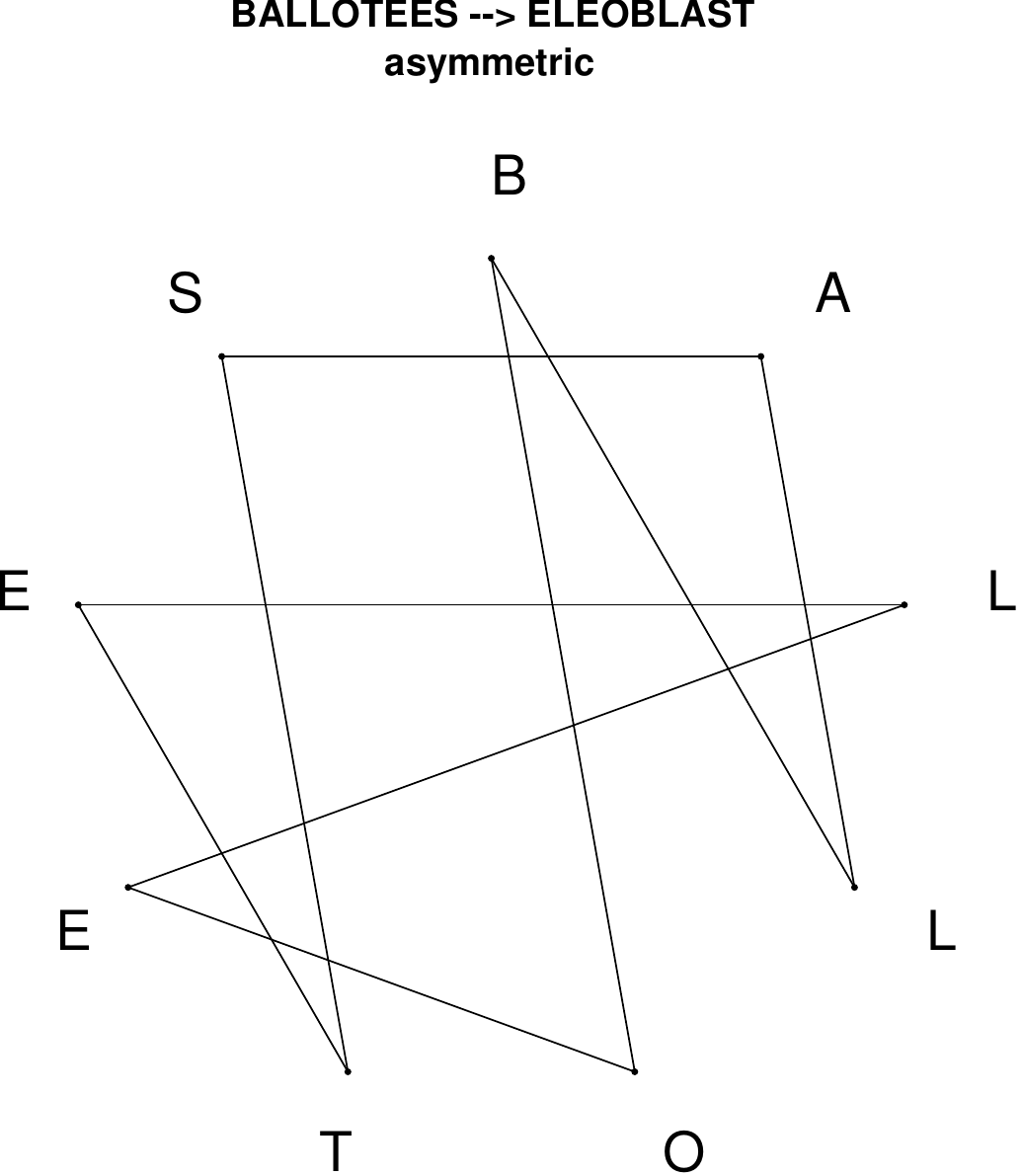}
\end{subfigure}
\hfill
\begin{subfigure}[T]{0.19\textwidth}
\centering
\includegraphics[width=\textwidth]{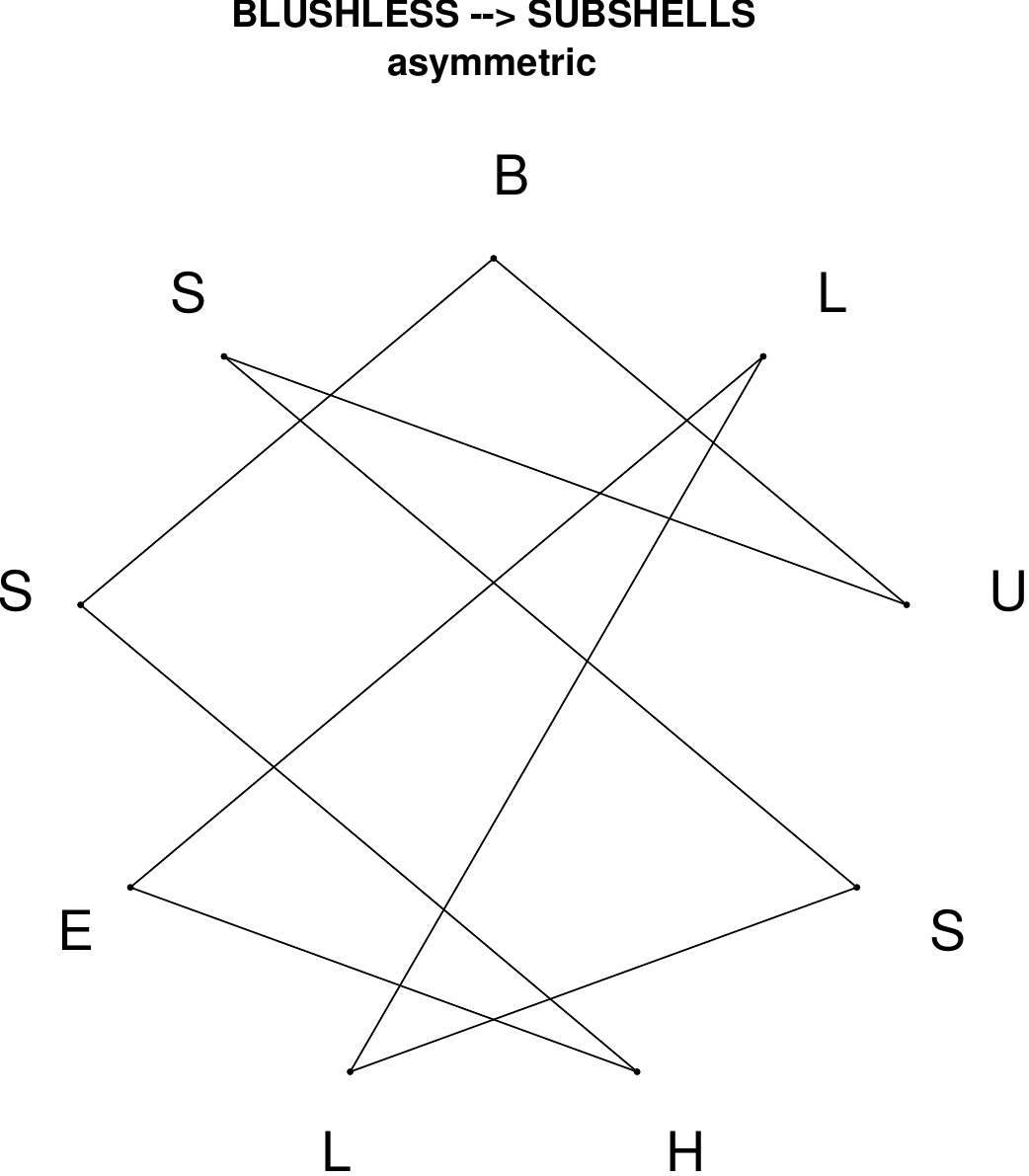}
\end{subfigure}
\hfill
\begin{subfigure}[T]{0.19\textwidth}
\centering
\includegraphics[width=\textwidth]{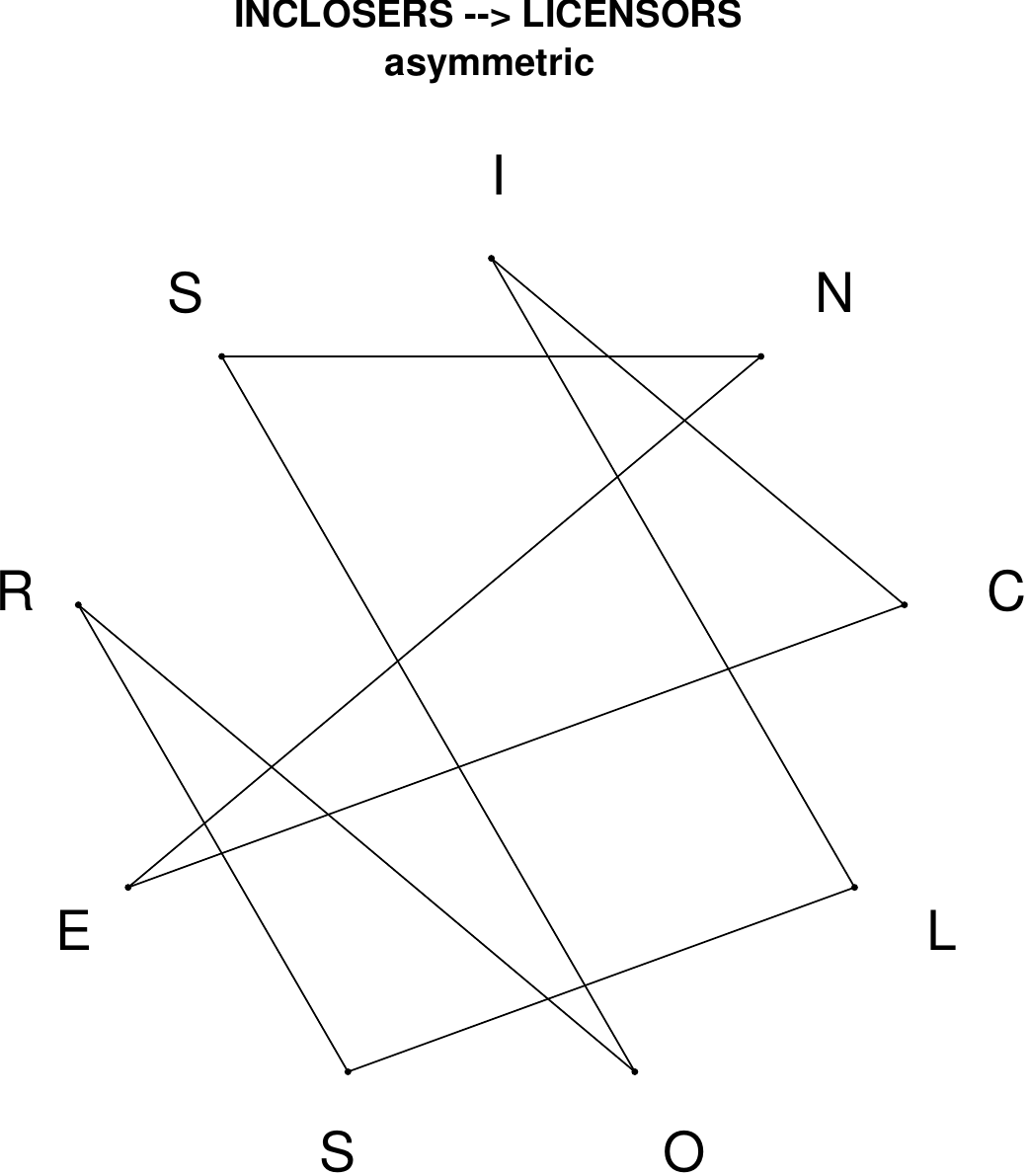}
\end{subfigure}
\hfill
\begin{subfigure}[T]{0.19\textwidth}
\centering
\includegraphics[width=\textwidth]{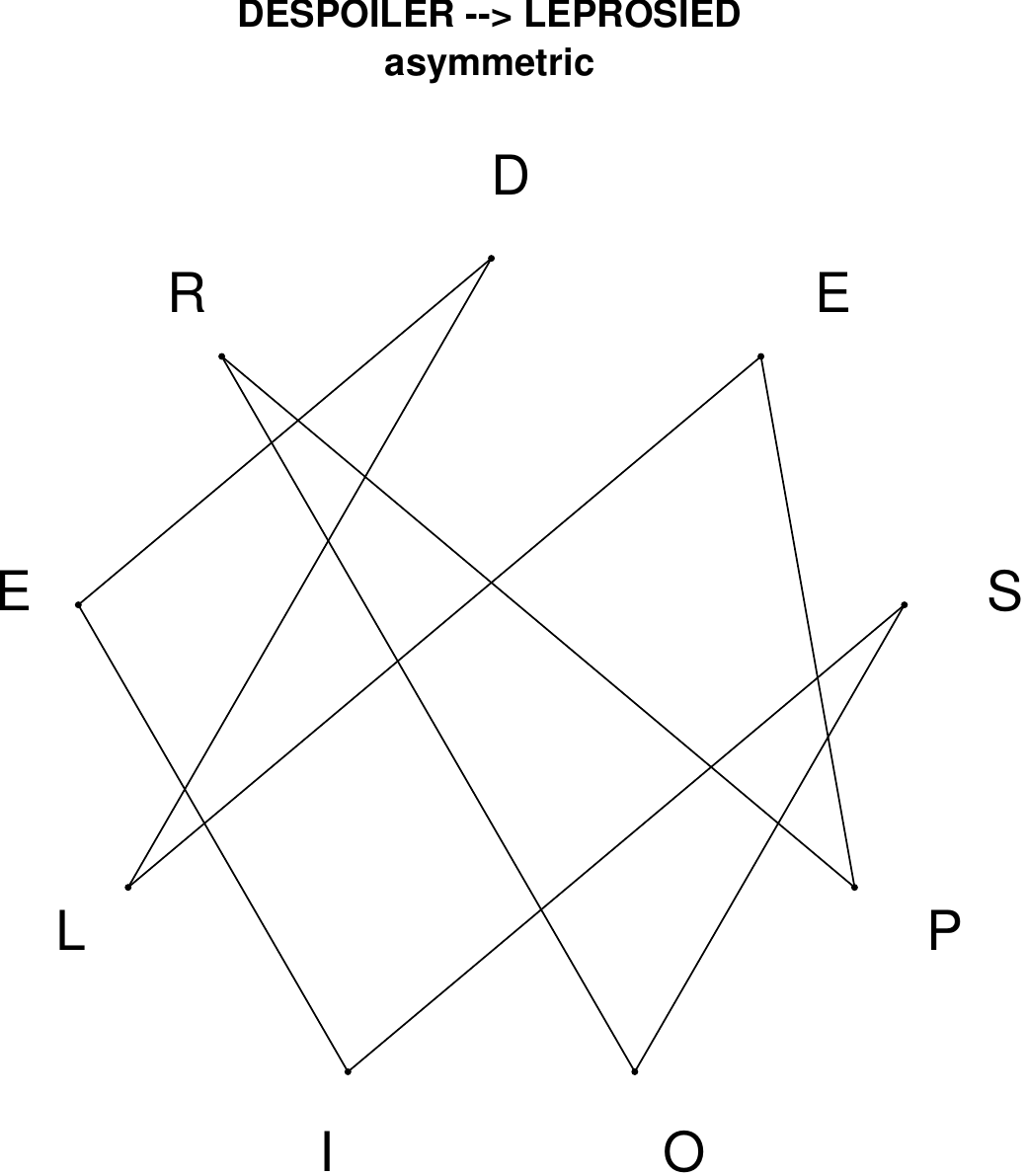}
\end{subfigure}
\hfill
\begin{subfigure}[T]{0.19\textwidth}
\centering
\includegraphics[width=\textwidth]{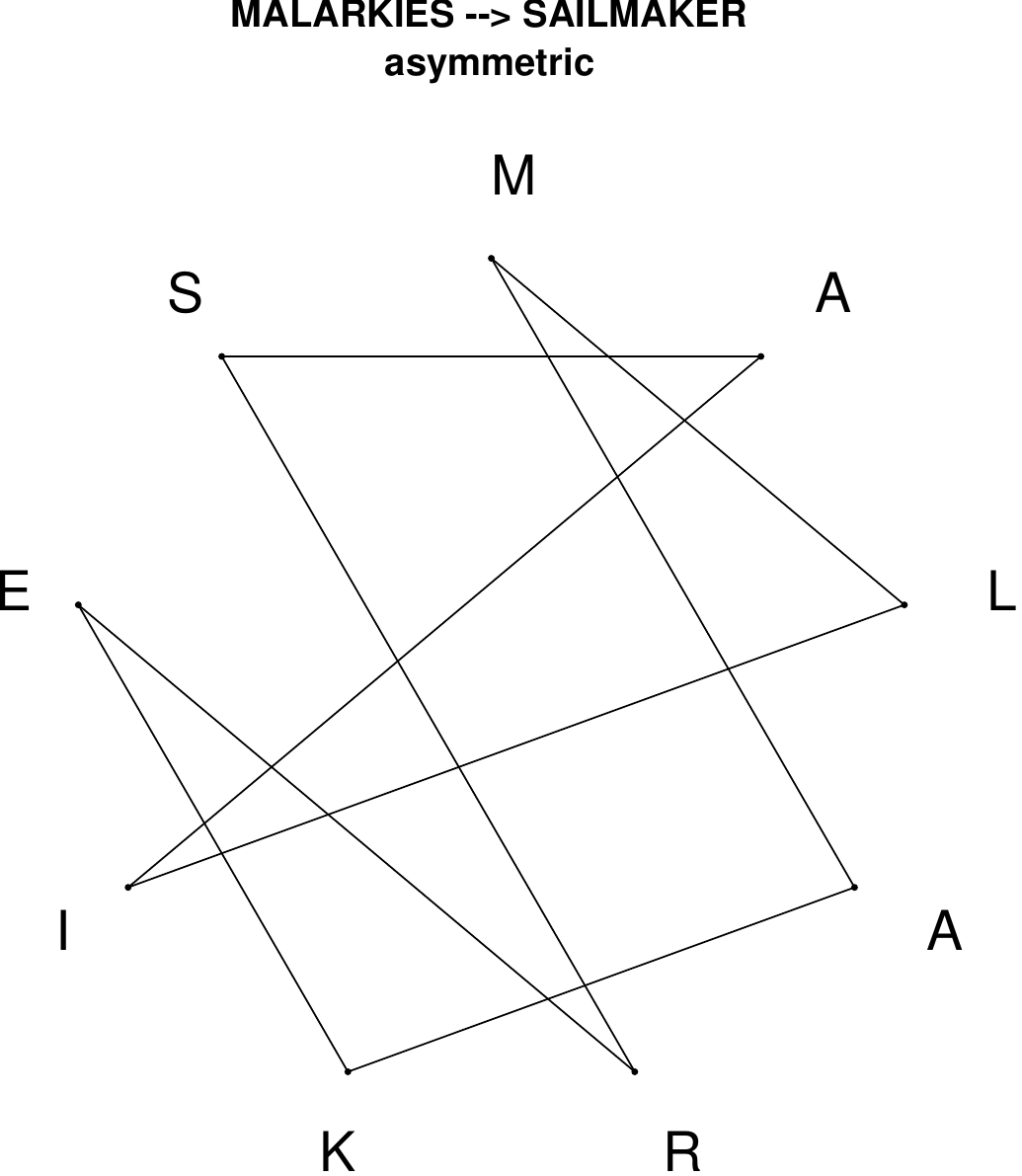}
\end{subfigure}
\end{figure}

\begin{figure}[H]
\centering
\begin{subfigure}[T]{0.19\textwidth}
\centering
\includegraphics[width=\textwidth]{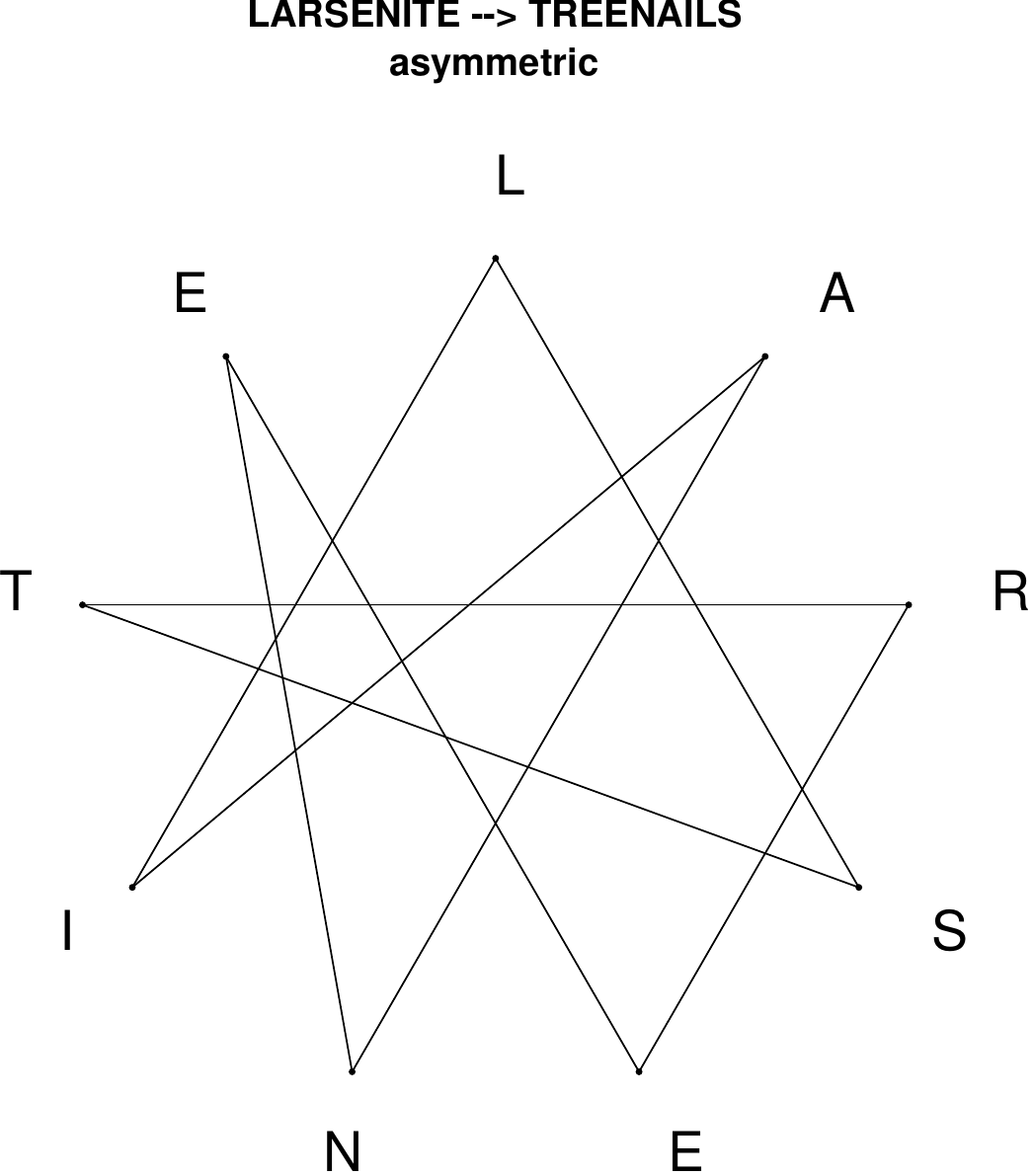}
\end{subfigure}
\hfill
\begin{subfigure}[T]{0.19\textwidth}
\centering
\includegraphics[width=\textwidth]{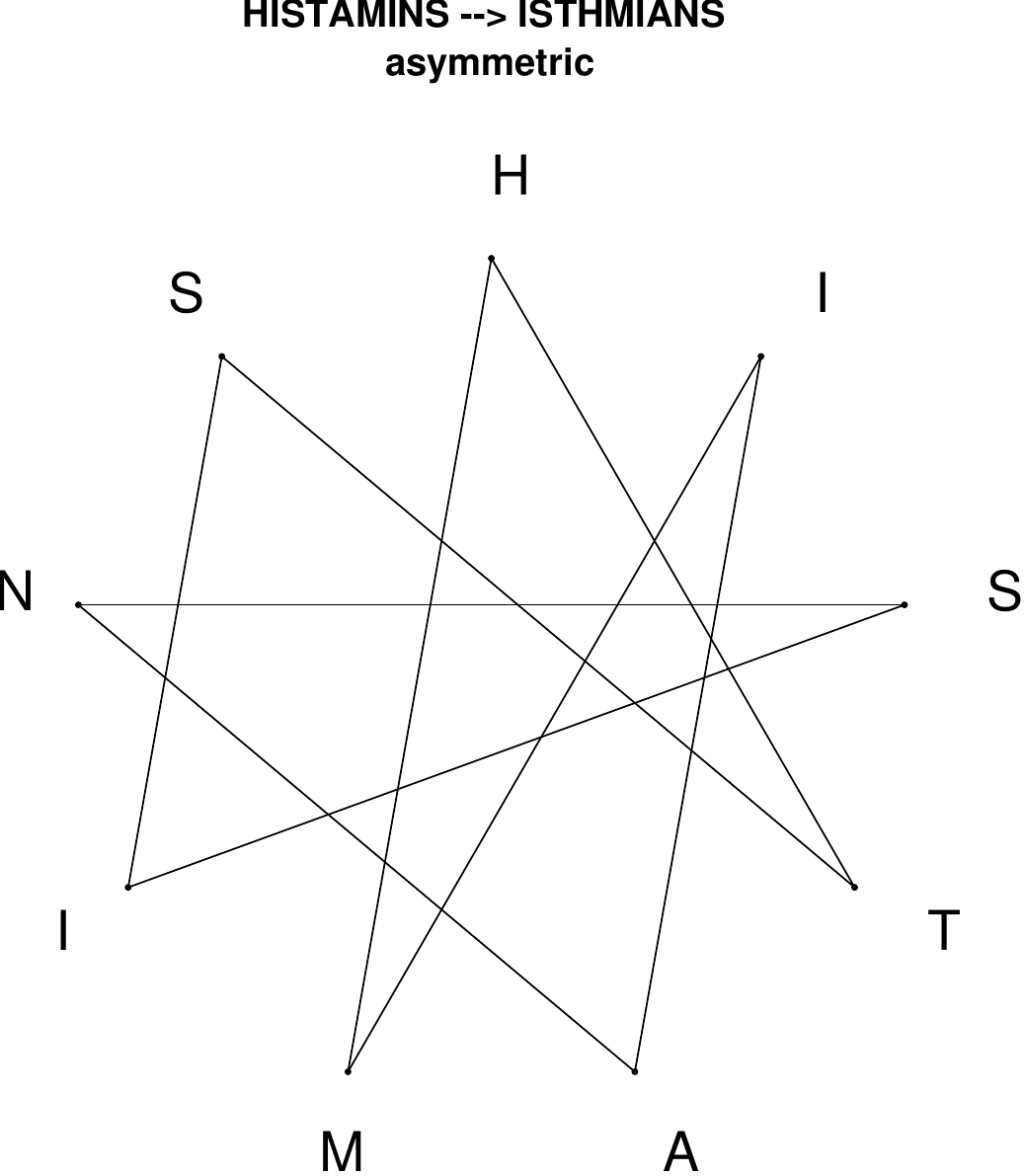}
\end{subfigure}
\hfill
\begin{subfigure}[T]{0.19\textwidth}
\centering
\includegraphics[width=\textwidth]{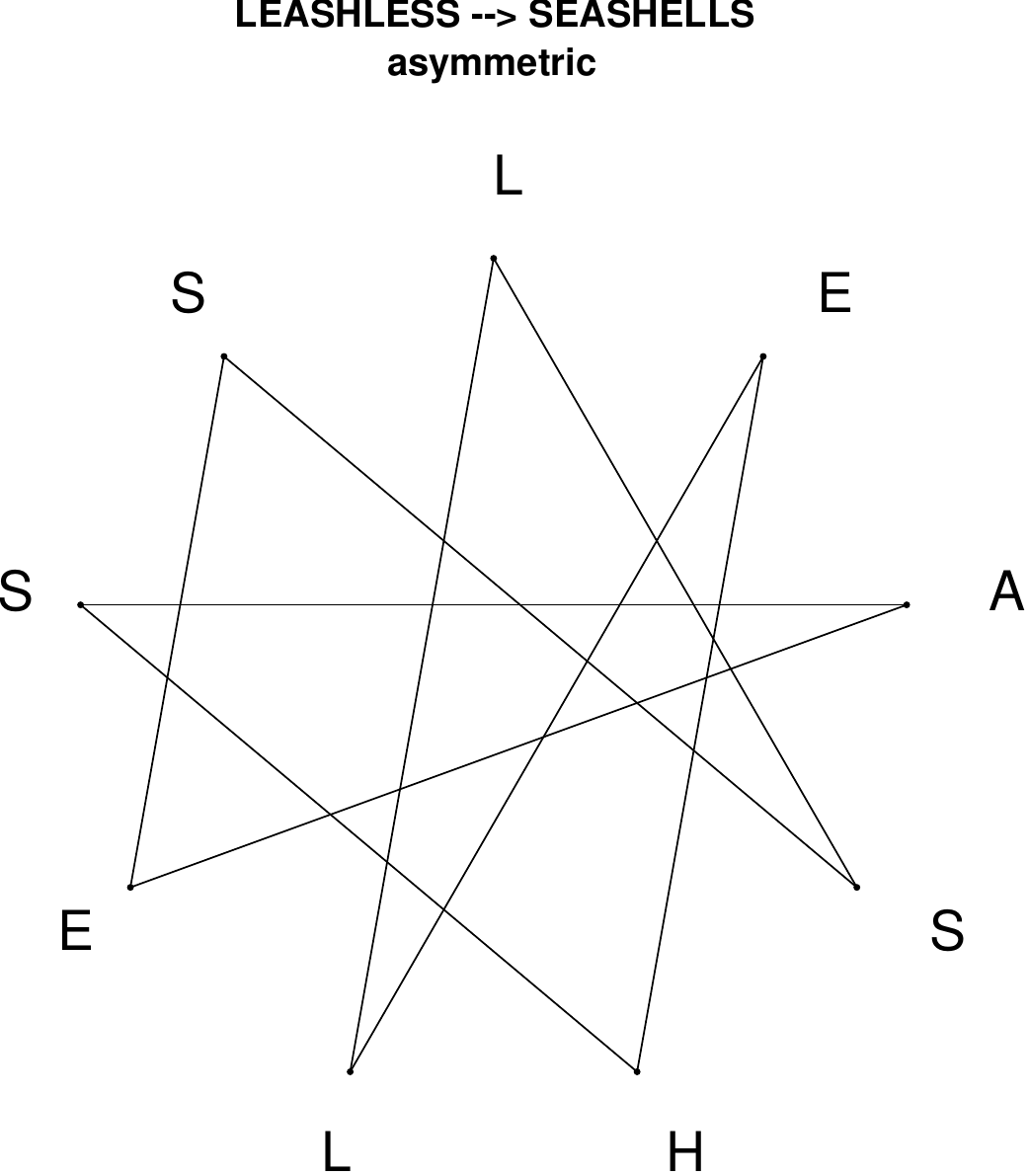}
\end{subfigure}
\hfill
\begin{subfigure}[T]{0.19\textwidth}
\centering
\includegraphics[width=\textwidth]{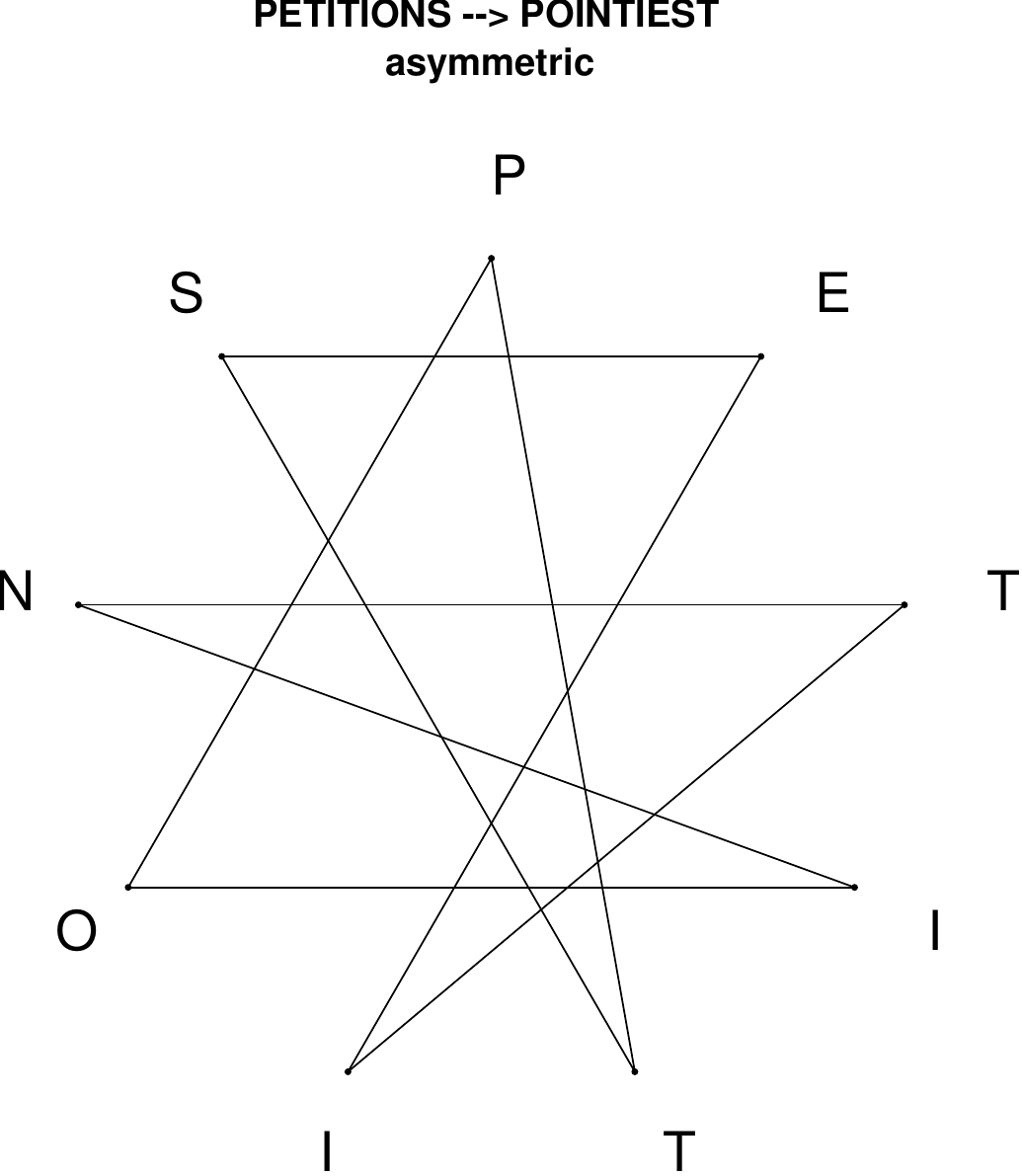}
\end{subfigure}
\hfill
\begin{subfigure}[T]{0.19\textwidth}
\centering
\includegraphics[width=\textwidth]{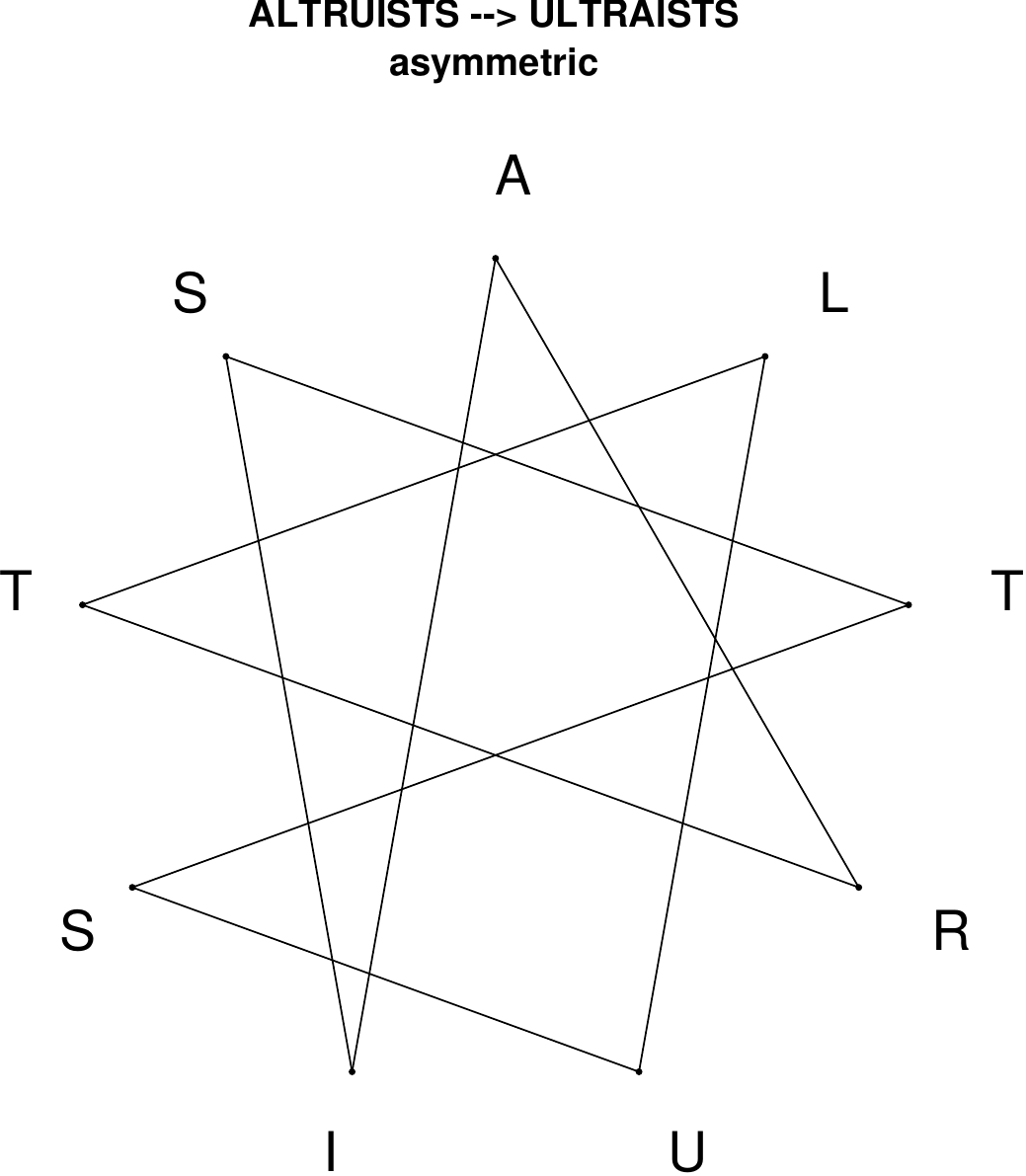}
\end{subfigure}
\end{figure}

\begin{figure}[H]
\centering
\begin{subfigure}[T]{0.19\textwidth}
\centering
\includegraphics[width=\textwidth]{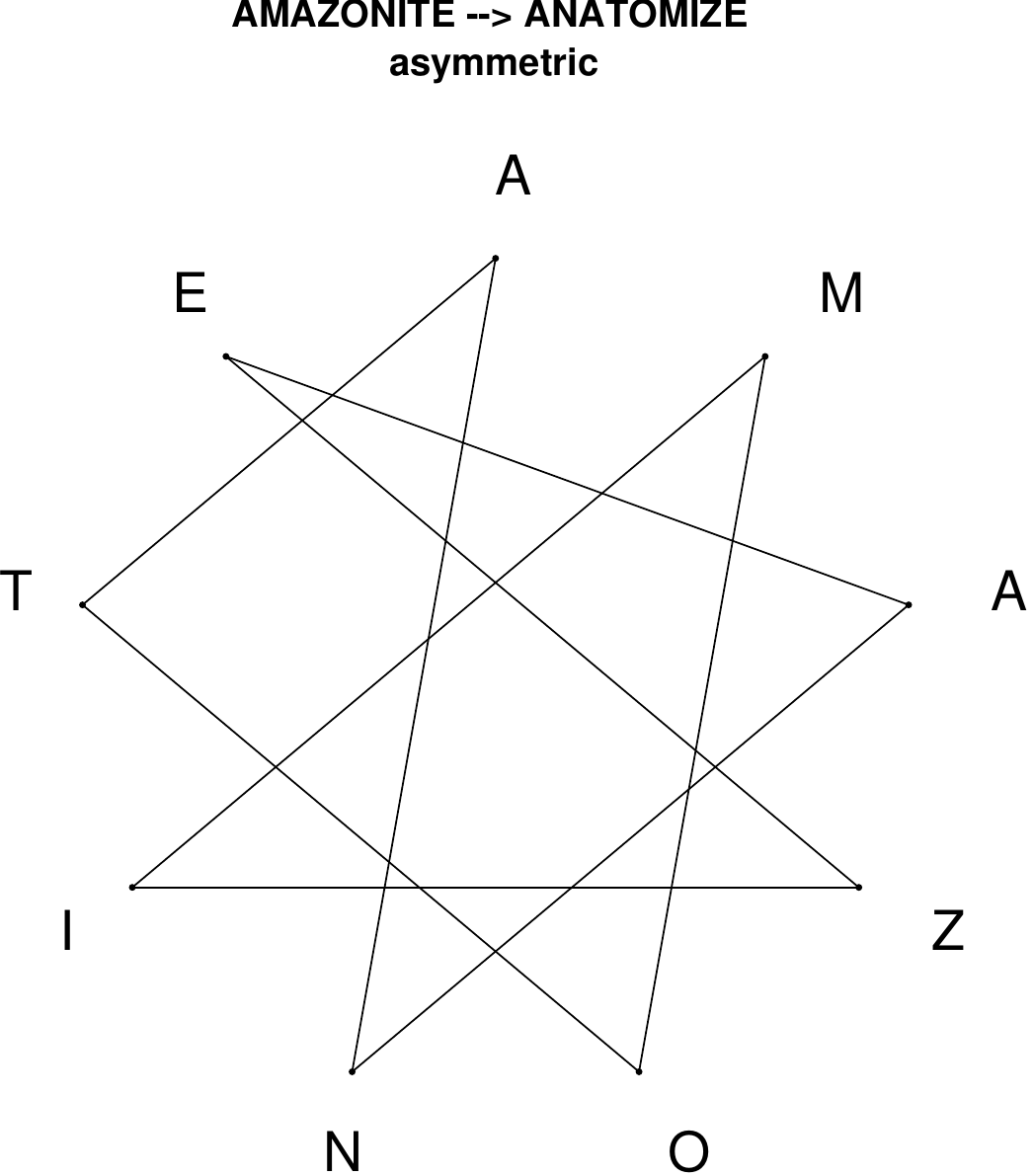}
\end{subfigure}
\hfill
\begin{subfigure}[T]{0.19\textwidth}
\centering
\includegraphics[width=\textwidth]{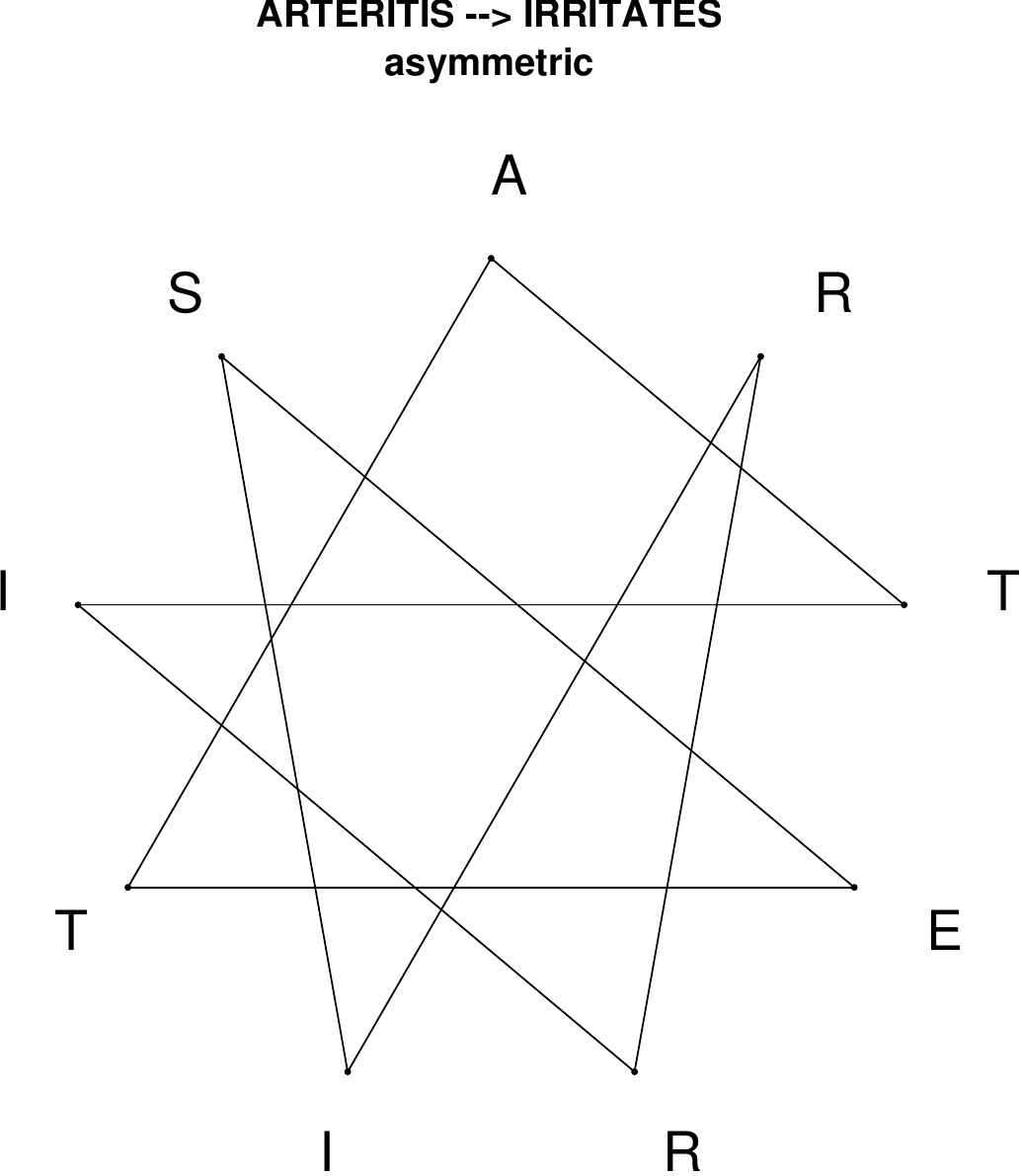}
\end{subfigure}
\hfill
\begin{subfigure}[T]{0.19\textwidth}
\centering
\includegraphics[width=\textwidth]{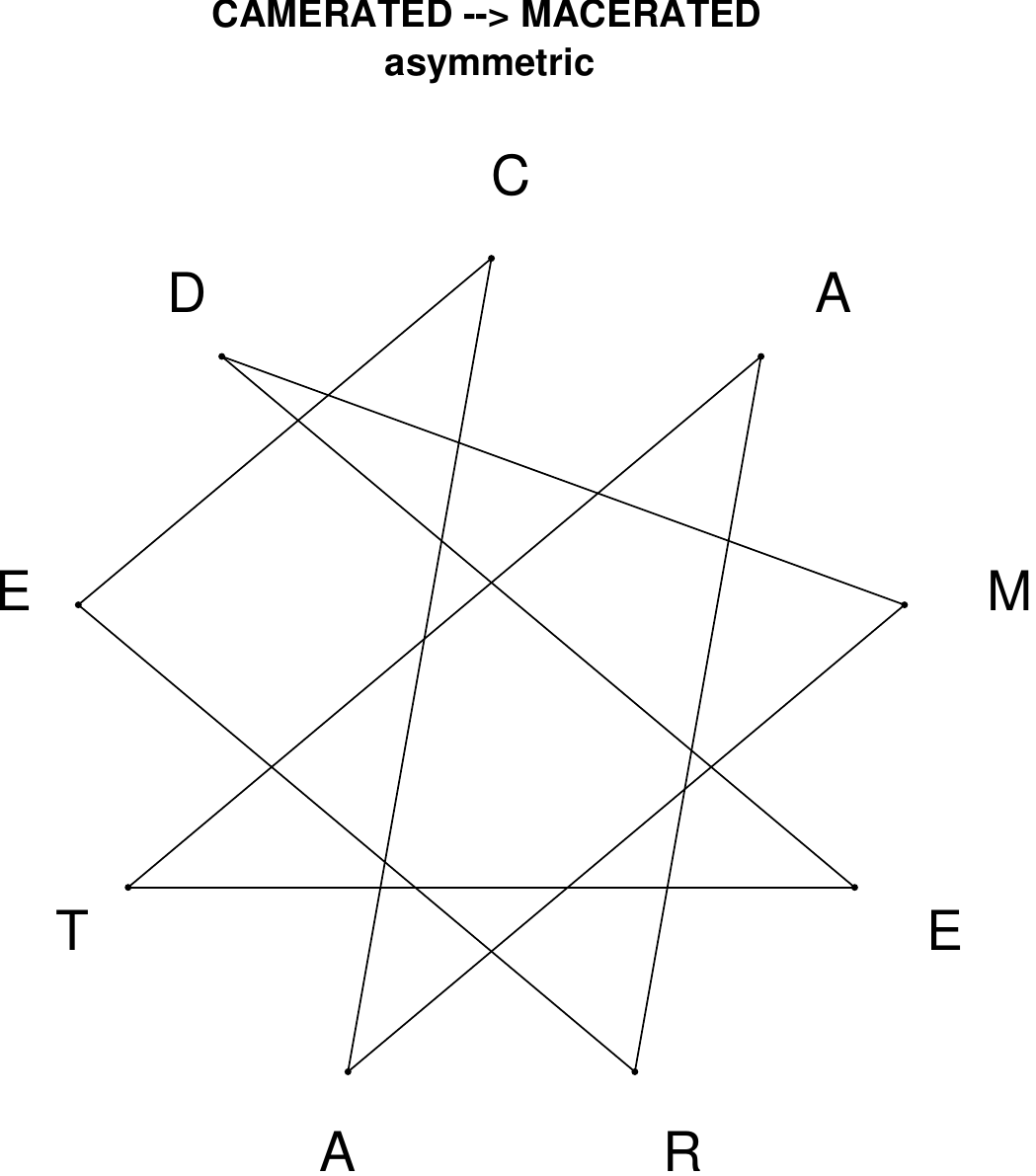}
\end{subfigure}
\hfill
\begin{subfigure}[T]{0.19\textwidth}
\centering
\includegraphics[width=\textwidth]{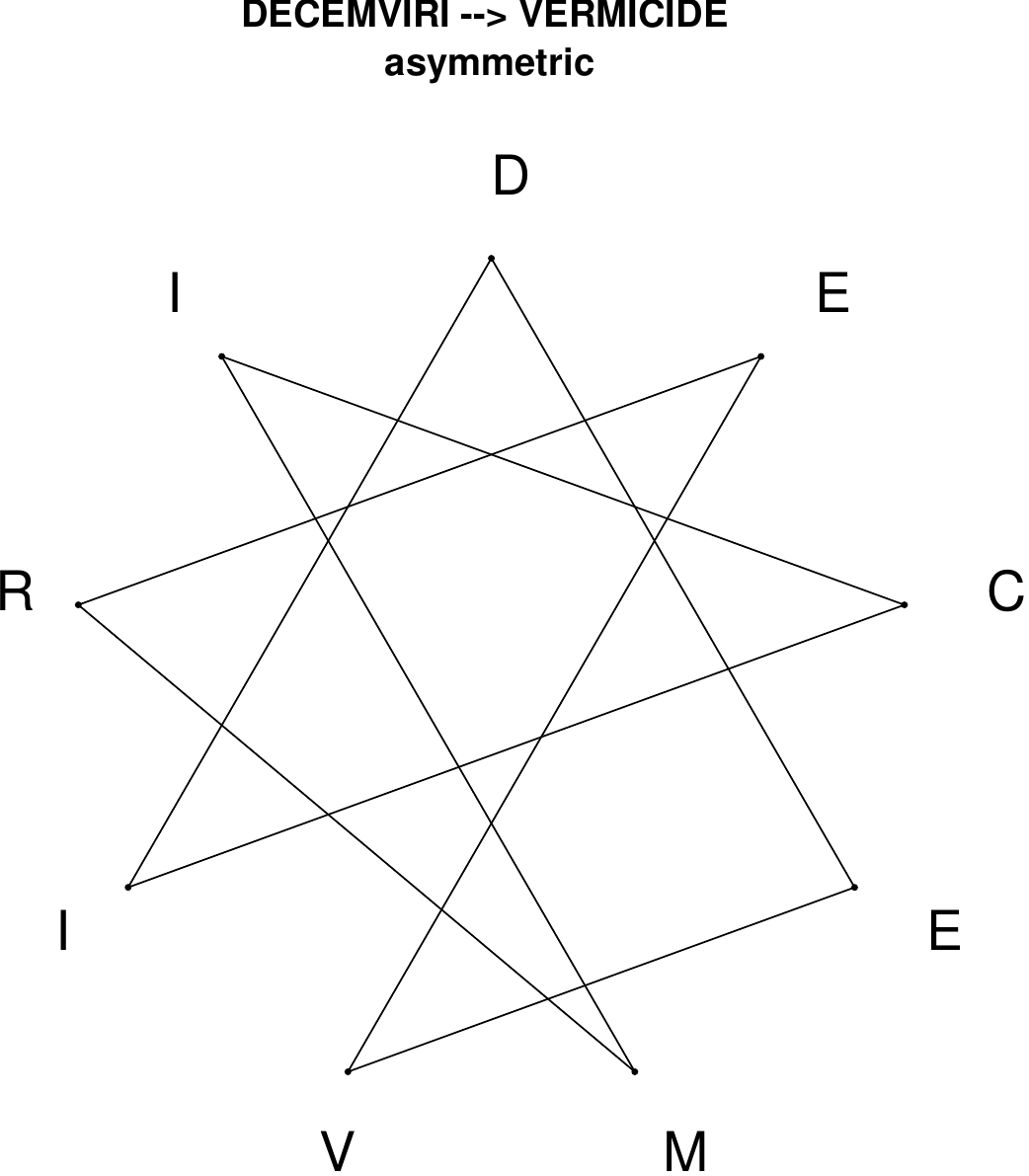}
\end{subfigure}
\hfill
\begin{subfigure}[T]{0.19\textwidth}
\centering
\includegraphics[width=\textwidth]{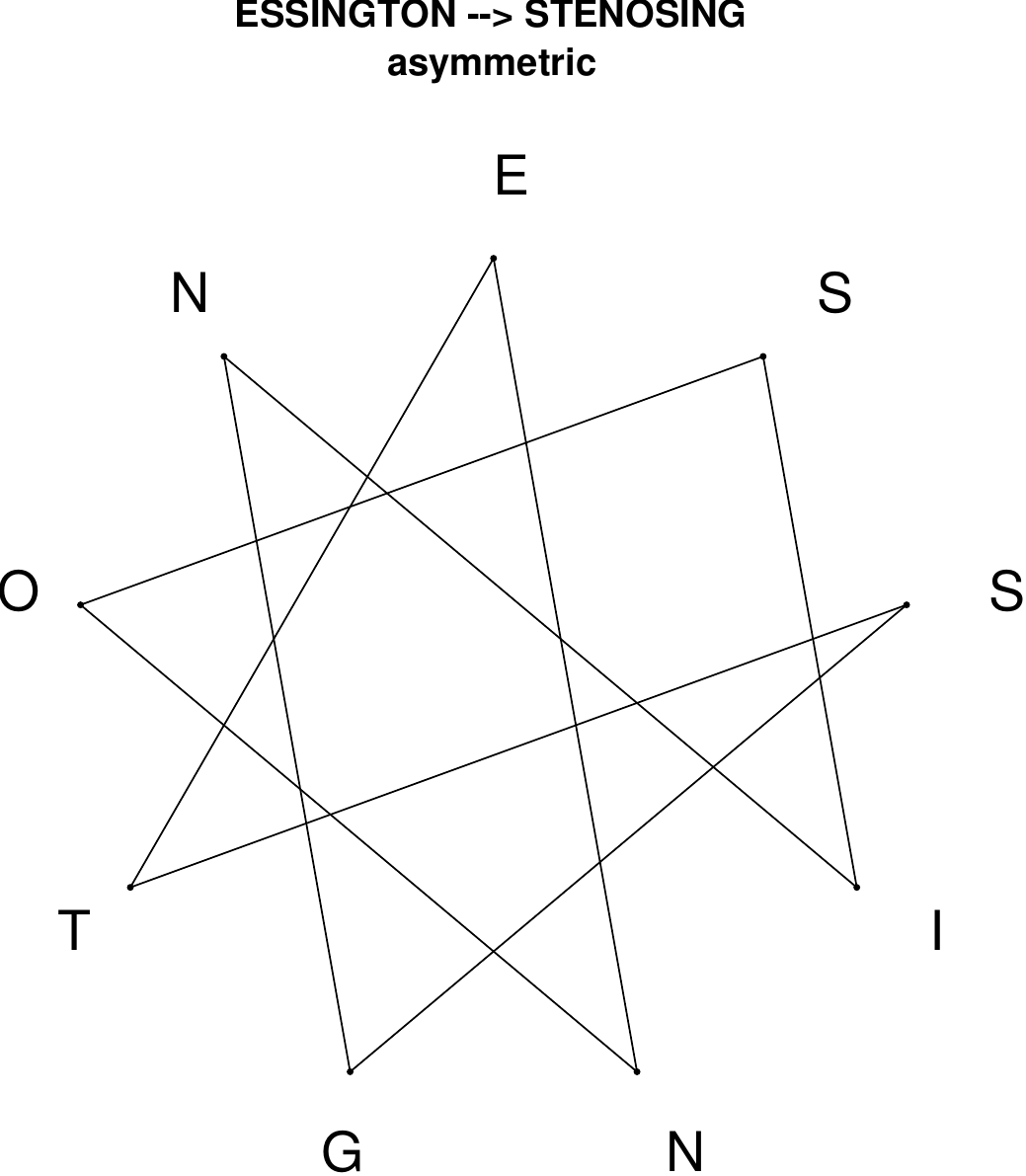}
\end{subfigure}
\end{figure}

\begin{figure}[H]
\centering
\begin{subfigure}[T]{0.19\textwidth}
\centering
\includegraphics[width=\textwidth]{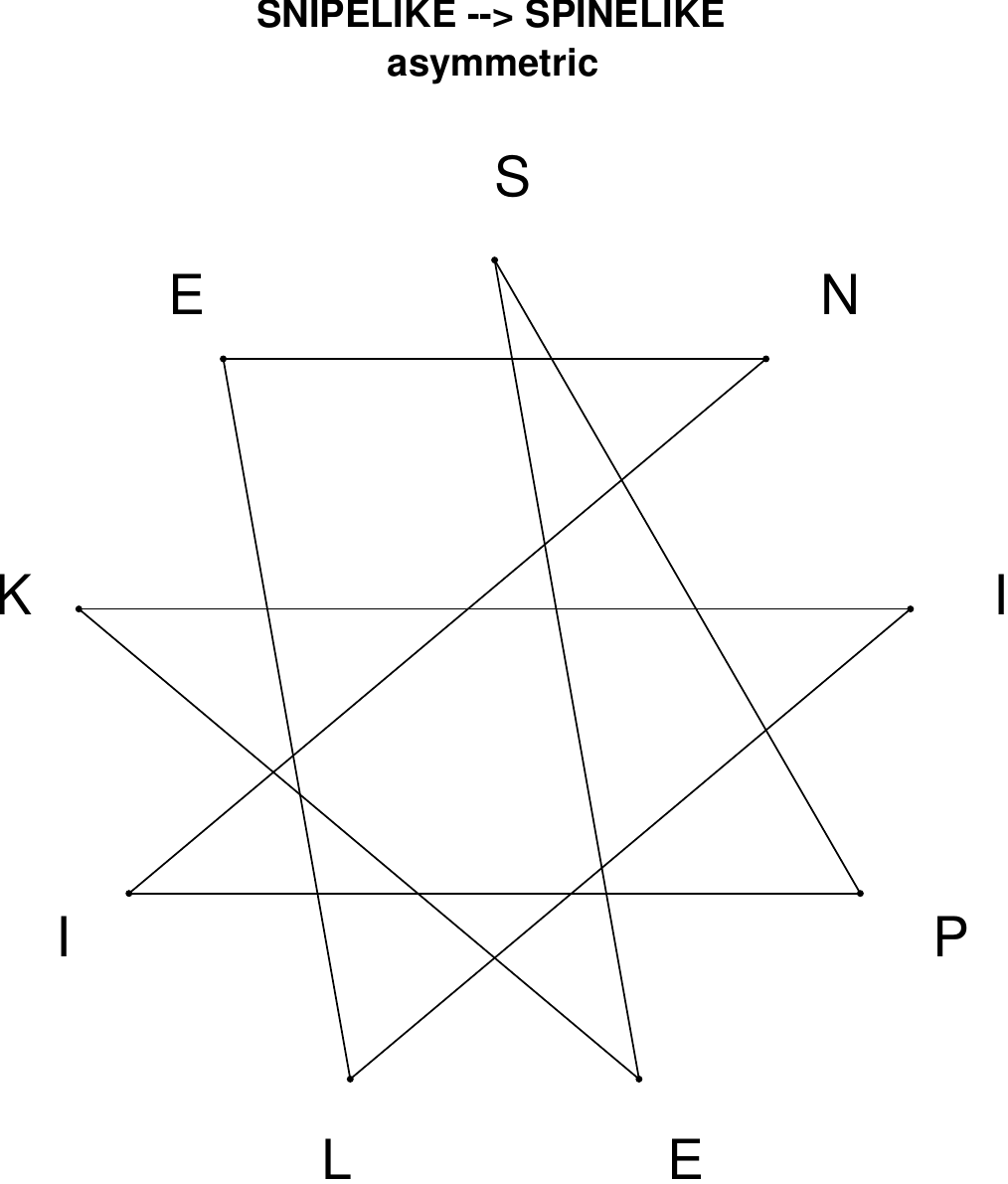}
\end{subfigure}
\hfill
\begin{subfigure}[T]{0.19\textwidth}
\centering
\includegraphics[width=\textwidth]{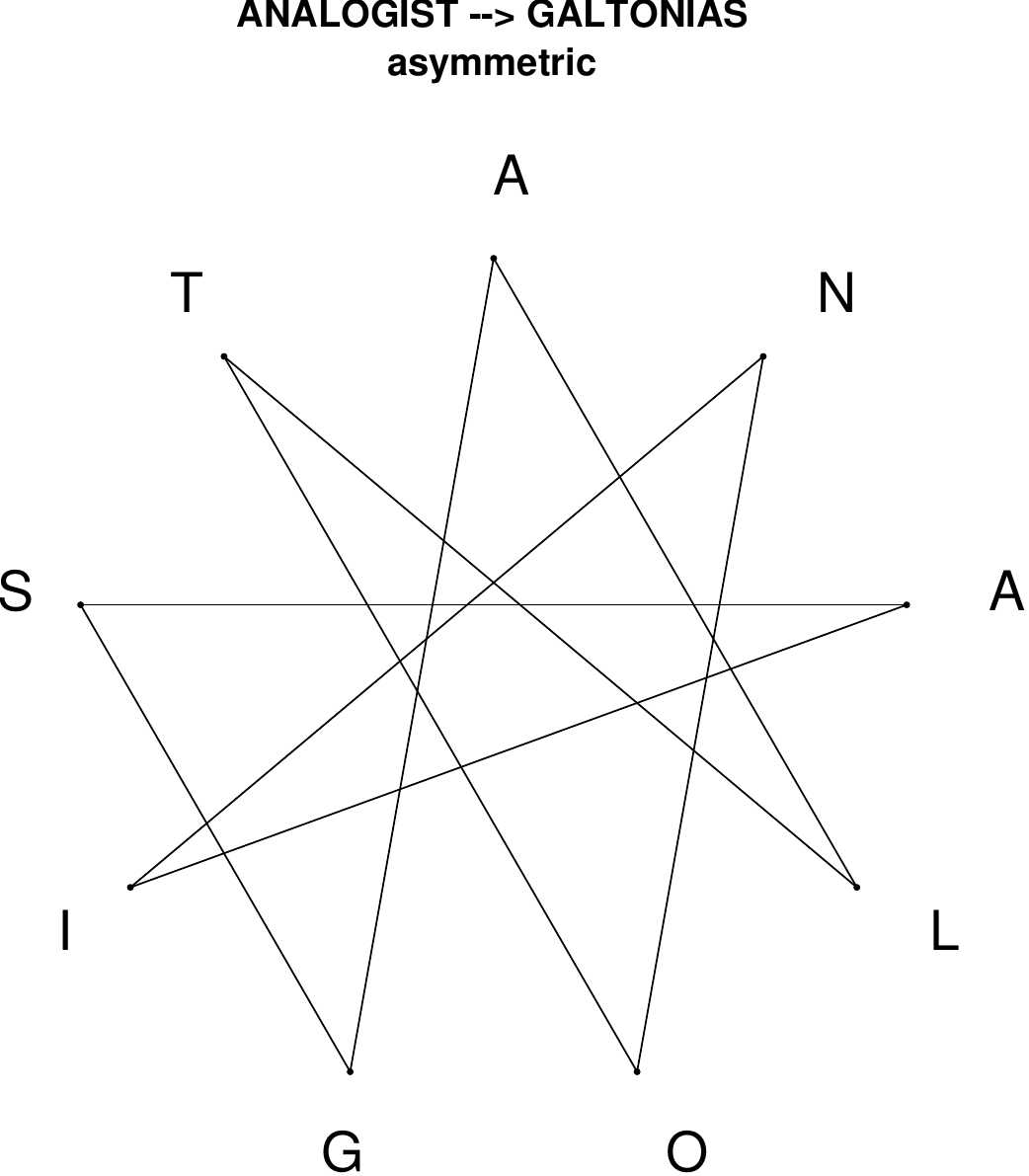}
\end{subfigure}
\hfill
\begin{subfigure}[T]{0.19\textwidth}
\centering
\includegraphics[width=\textwidth]{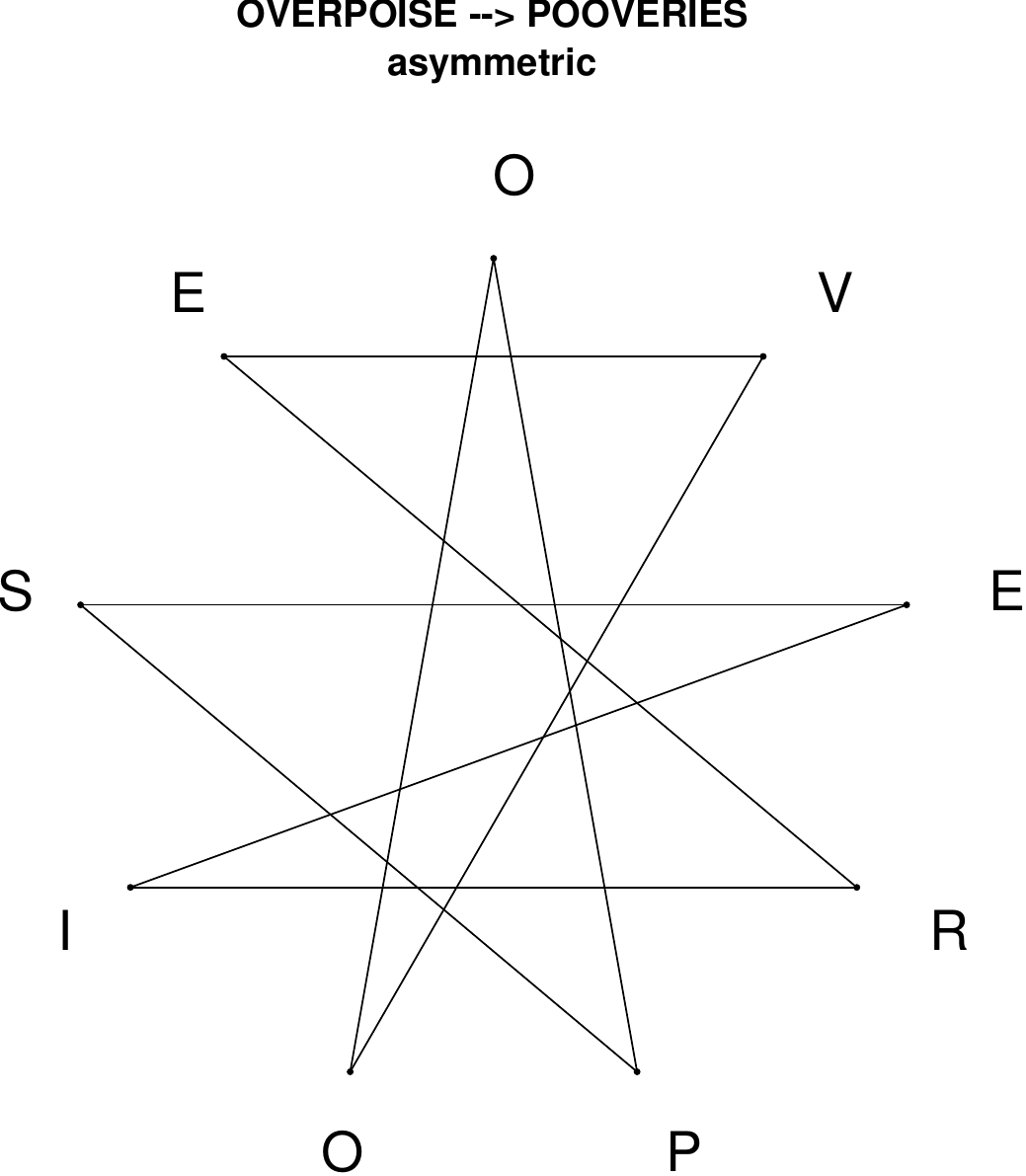}
\end{subfigure}
\hfill
\begin{subfigure}[T]{0.19\textwidth}
\centering
\includegraphics[width=\textwidth]{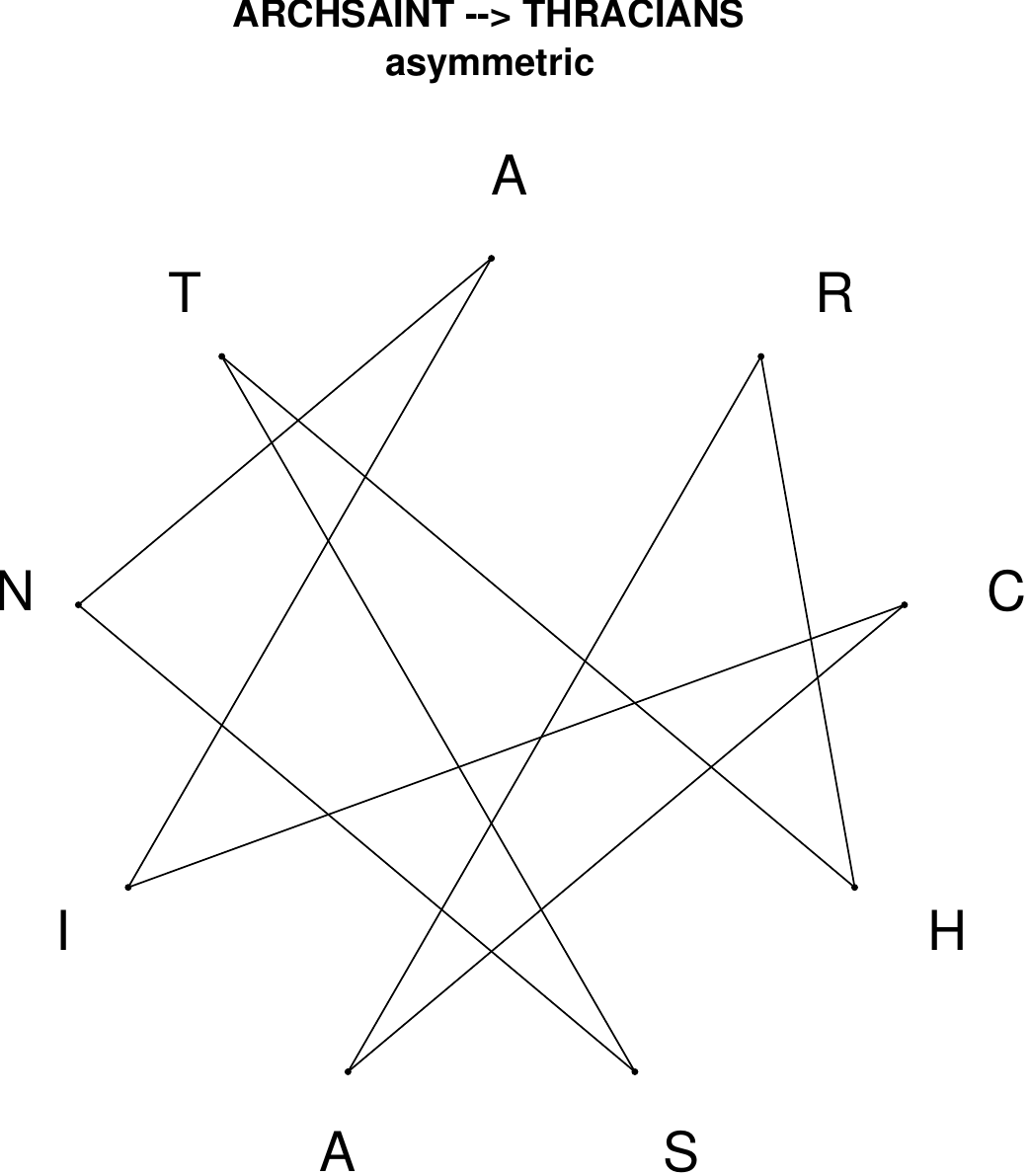}
\end{subfigure}
\hfill
\begin{subfigure}[T]{0.19\textwidth}
\centering
\includegraphics[width=\textwidth]{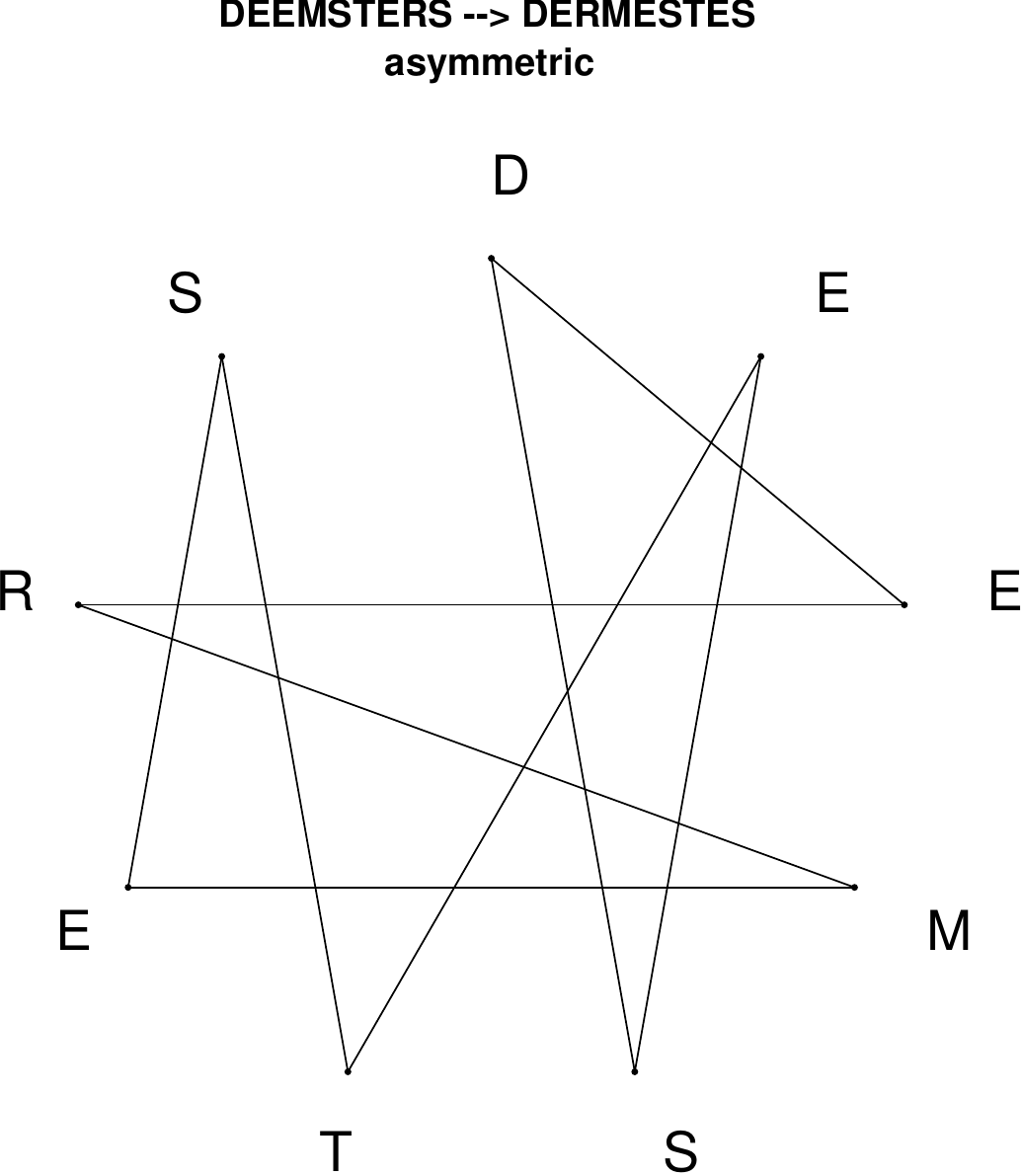}
\end{subfigure}
\end{figure}

\begin{figure}[H]
\centering
\begin{subfigure}[T]{0.19\textwidth}
\centering
\includegraphics[width=\textwidth]{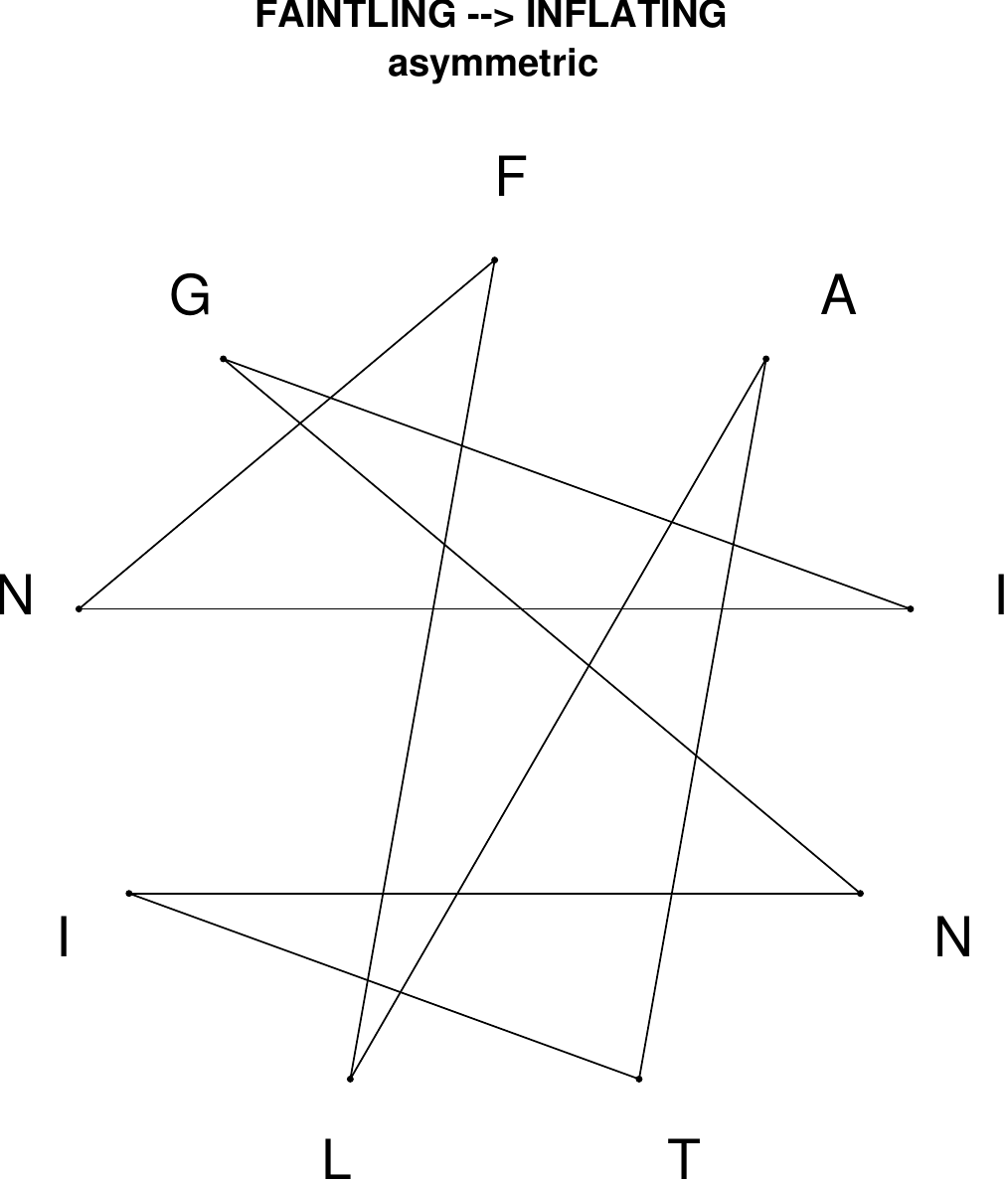}
\end{subfigure}
\hfill
\begin{subfigure}[T]{0.19\textwidth}
\centering
\includegraphics[width=\textwidth]{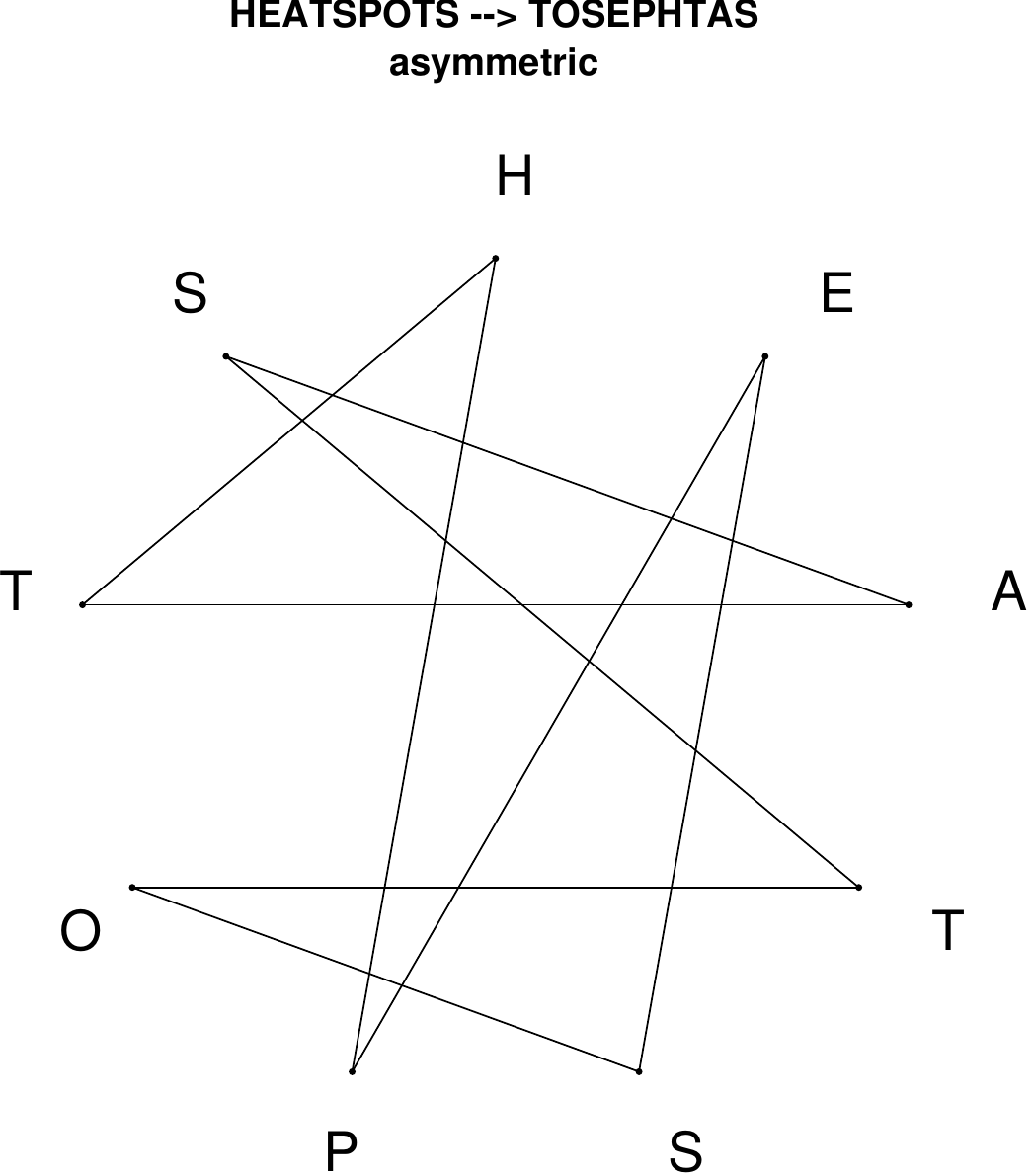}
\end{subfigure}
\hfill
\begin{subfigure}[T]{0.19\textwidth}
\centering
\includegraphics[width=\textwidth]{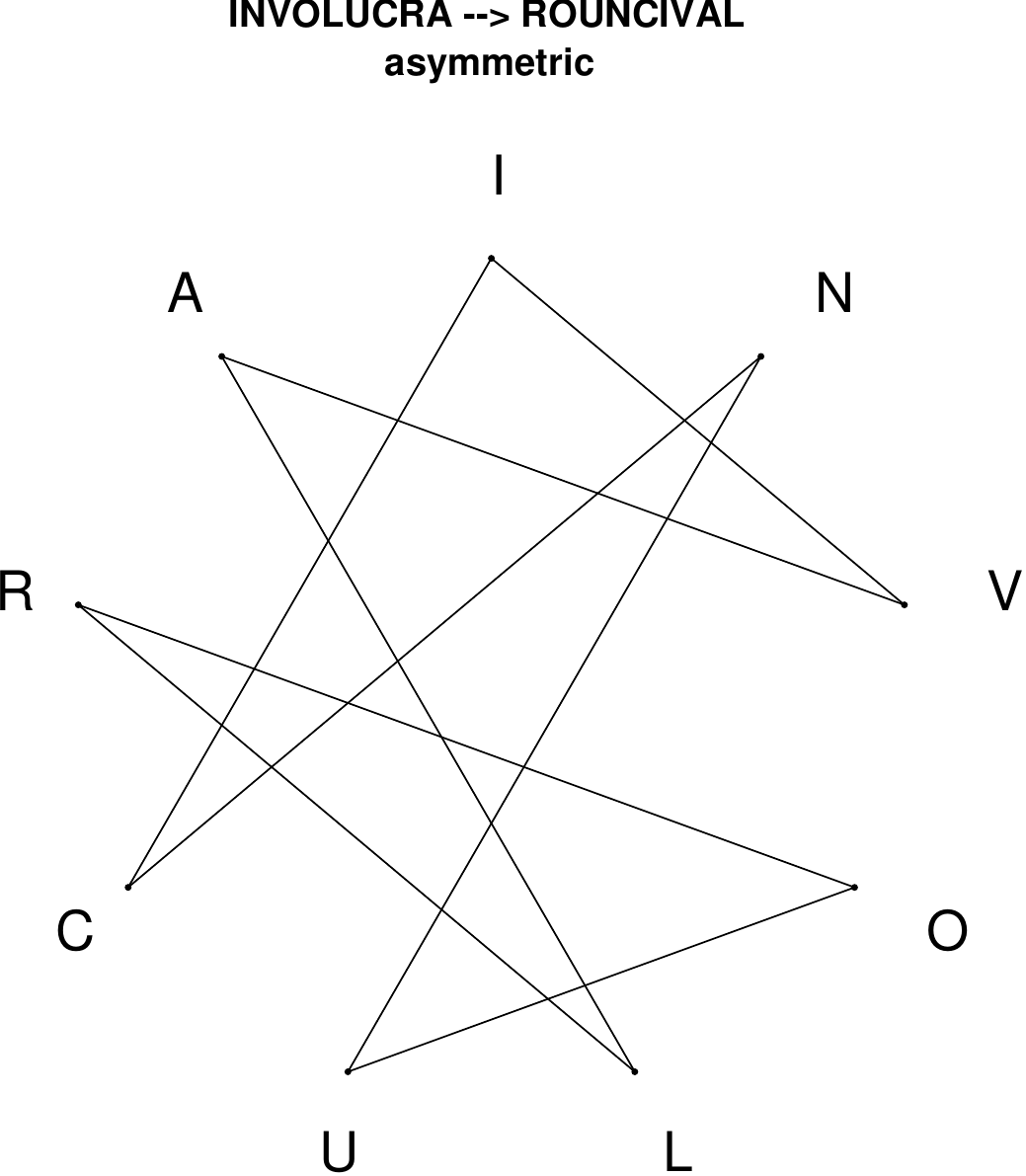}
\end{subfigure}
\hfill
\begin{subfigure}[T]{0.19\textwidth}
\centering
\includegraphics[width=\textwidth]{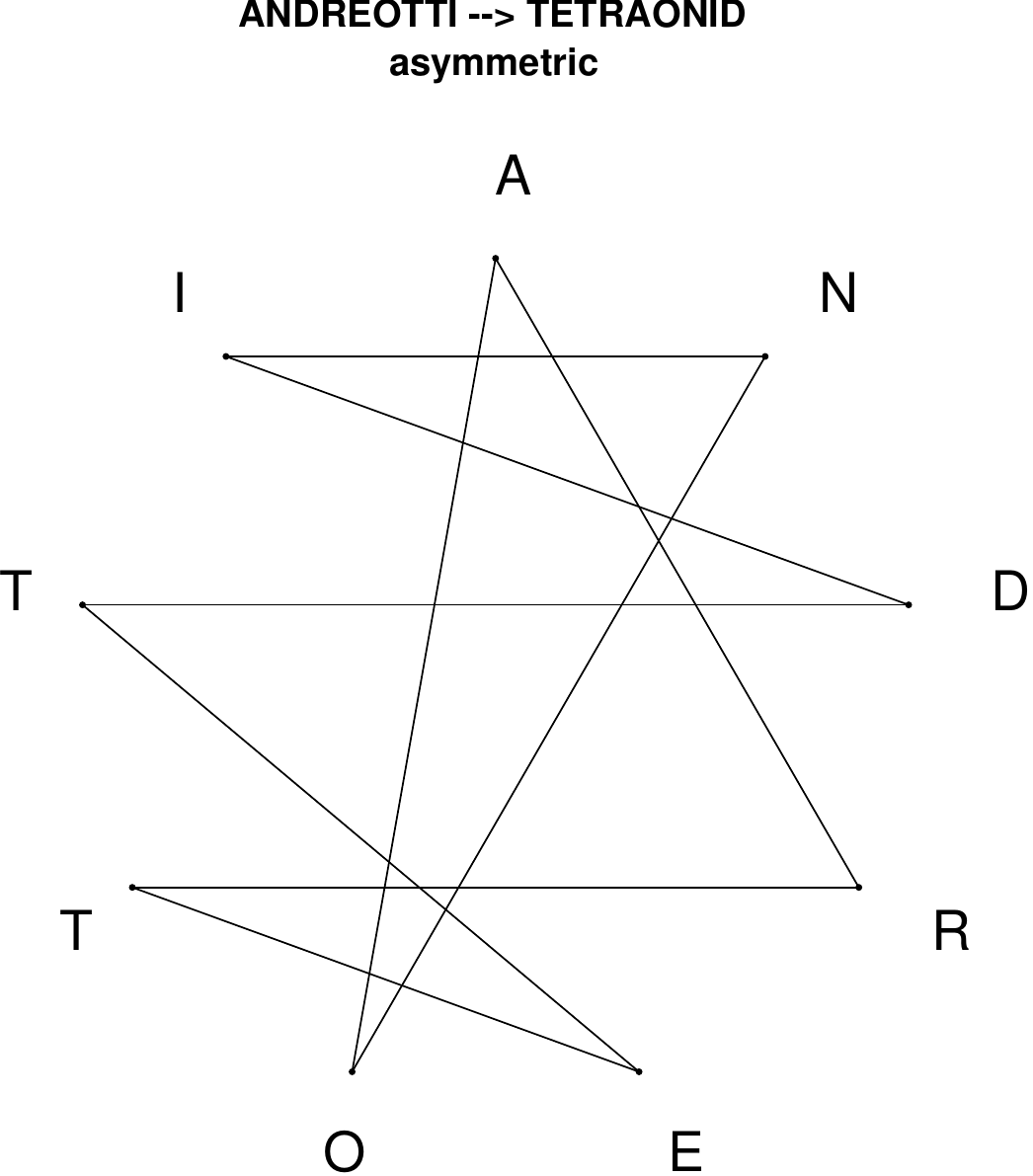}
\end{subfigure}
\hfill
\begin{subfigure}[T]{0.19\textwidth}
\centering
\includegraphics[width=\textwidth]{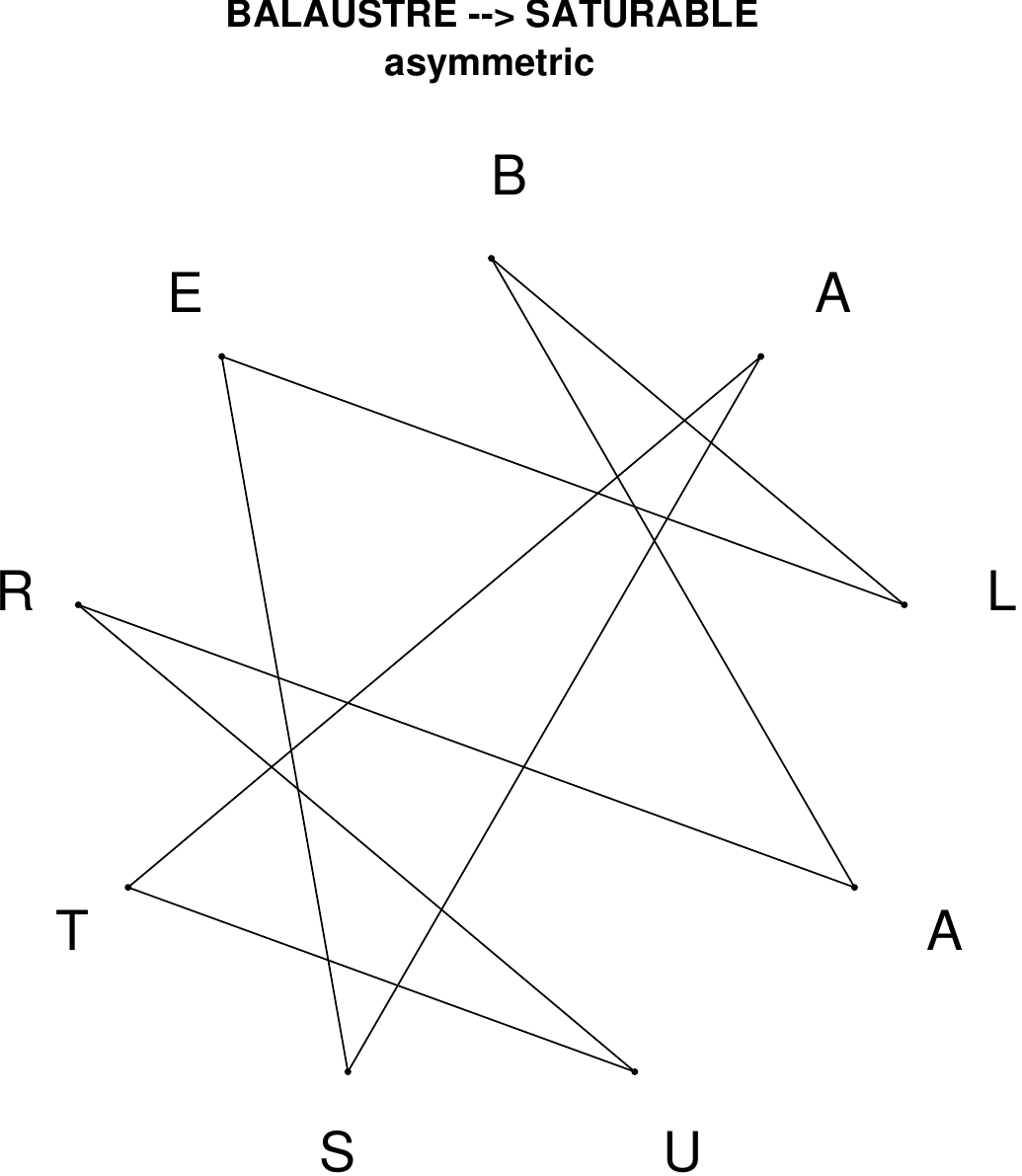}
\end{subfigure}
\end{figure}

\begin{figure}[H]
\centering
\begin{subfigure}[T]{0.19\textwidth}
\centering
\includegraphics[width=\textwidth]{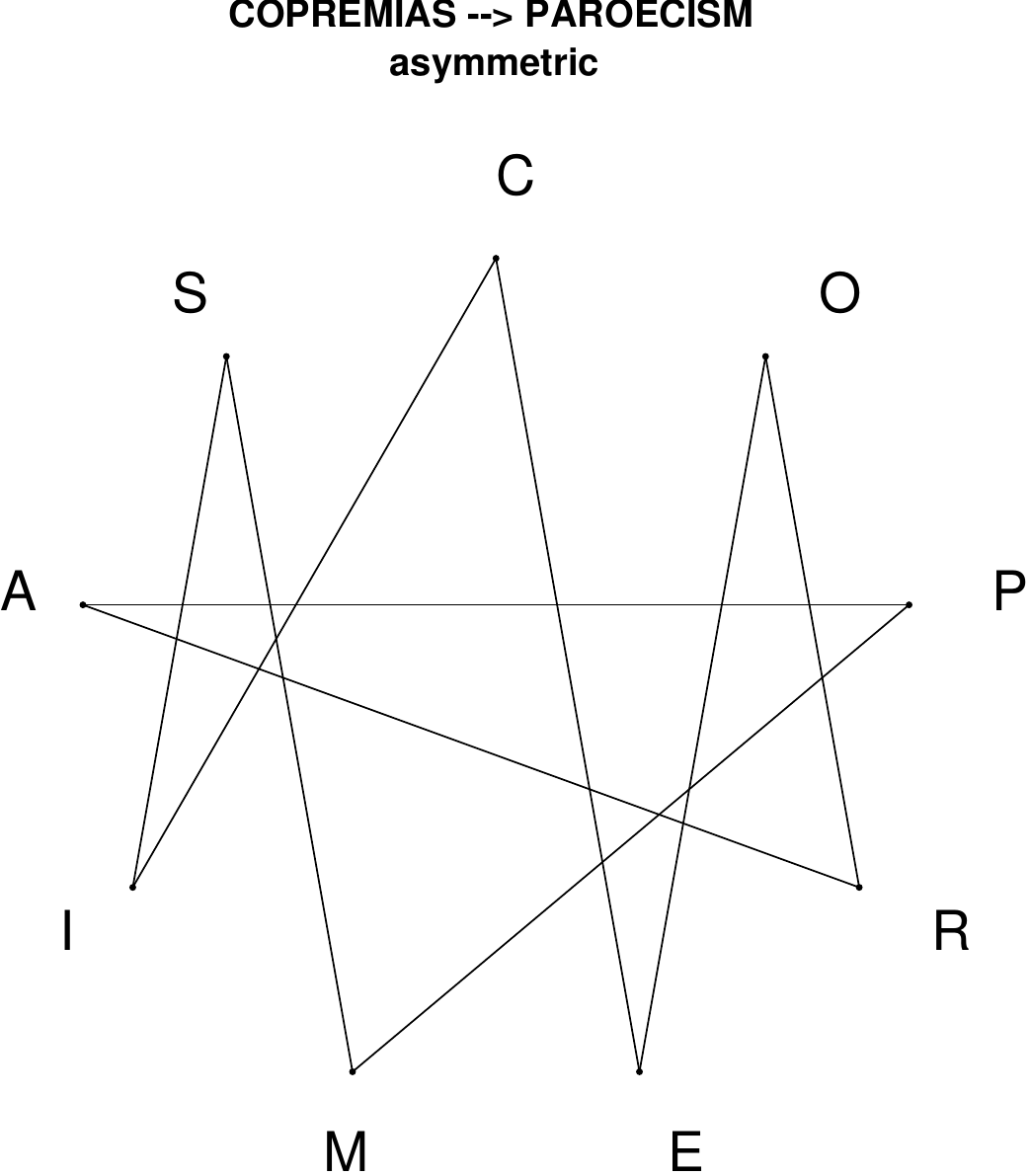}
\end{subfigure}
\hfill
\begin{subfigure}[T]{0.19\textwidth}
\centering
\includegraphics[width=\textwidth]{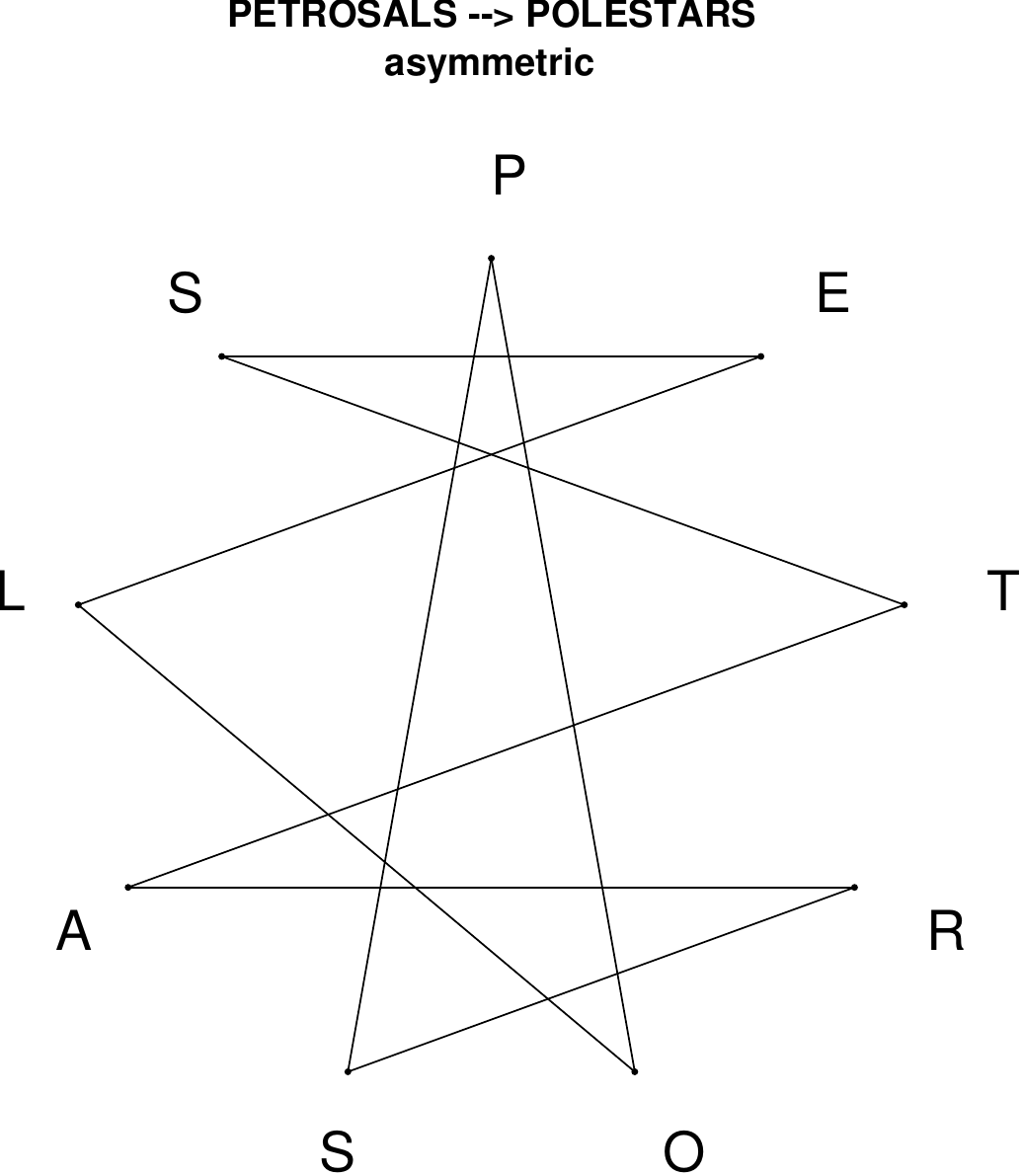}
\end{subfigure}
\hfill
\begin{subfigure}[T]{0.19\textwidth}
\centering
\includegraphics[width=\textwidth]{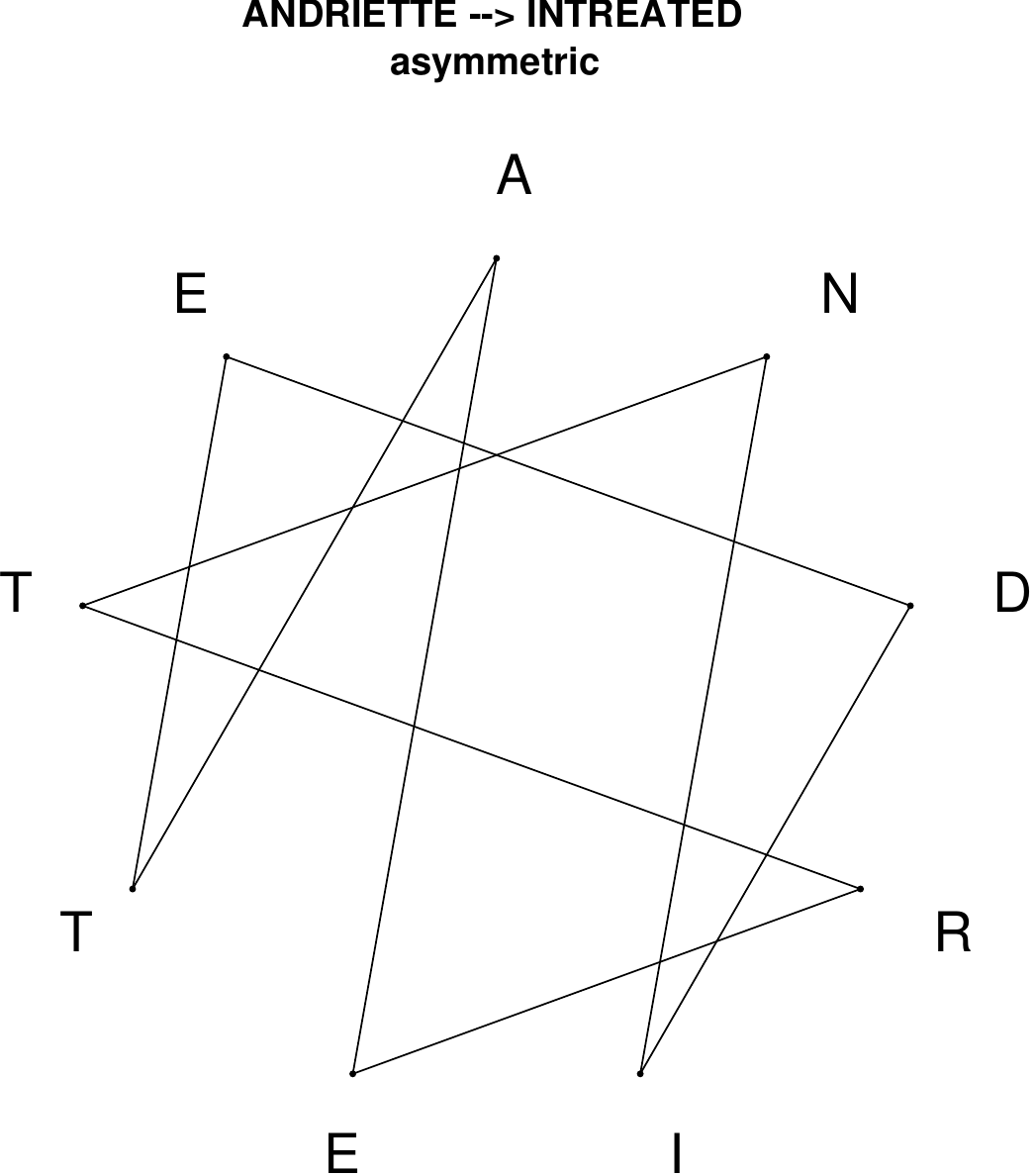}
\end{subfigure}
\hfill
\begin{subfigure}[T]{0.19\textwidth}
\centering
\includegraphics[width=\textwidth]{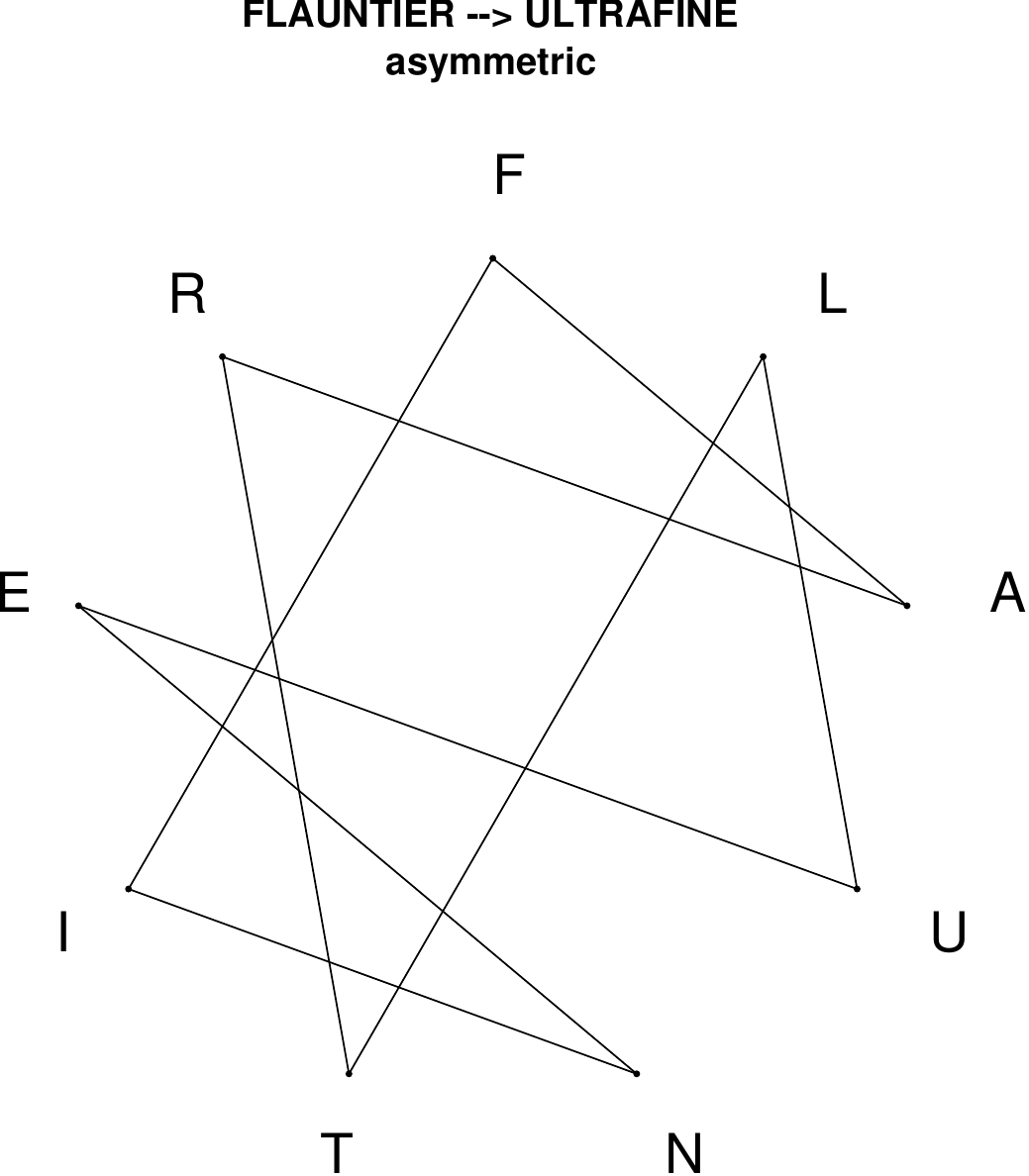}
\end{subfigure}
\hfill
\begin{subfigure}[T]{0.19\textwidth}
\centering
\includegraphics[width=\textwidth]{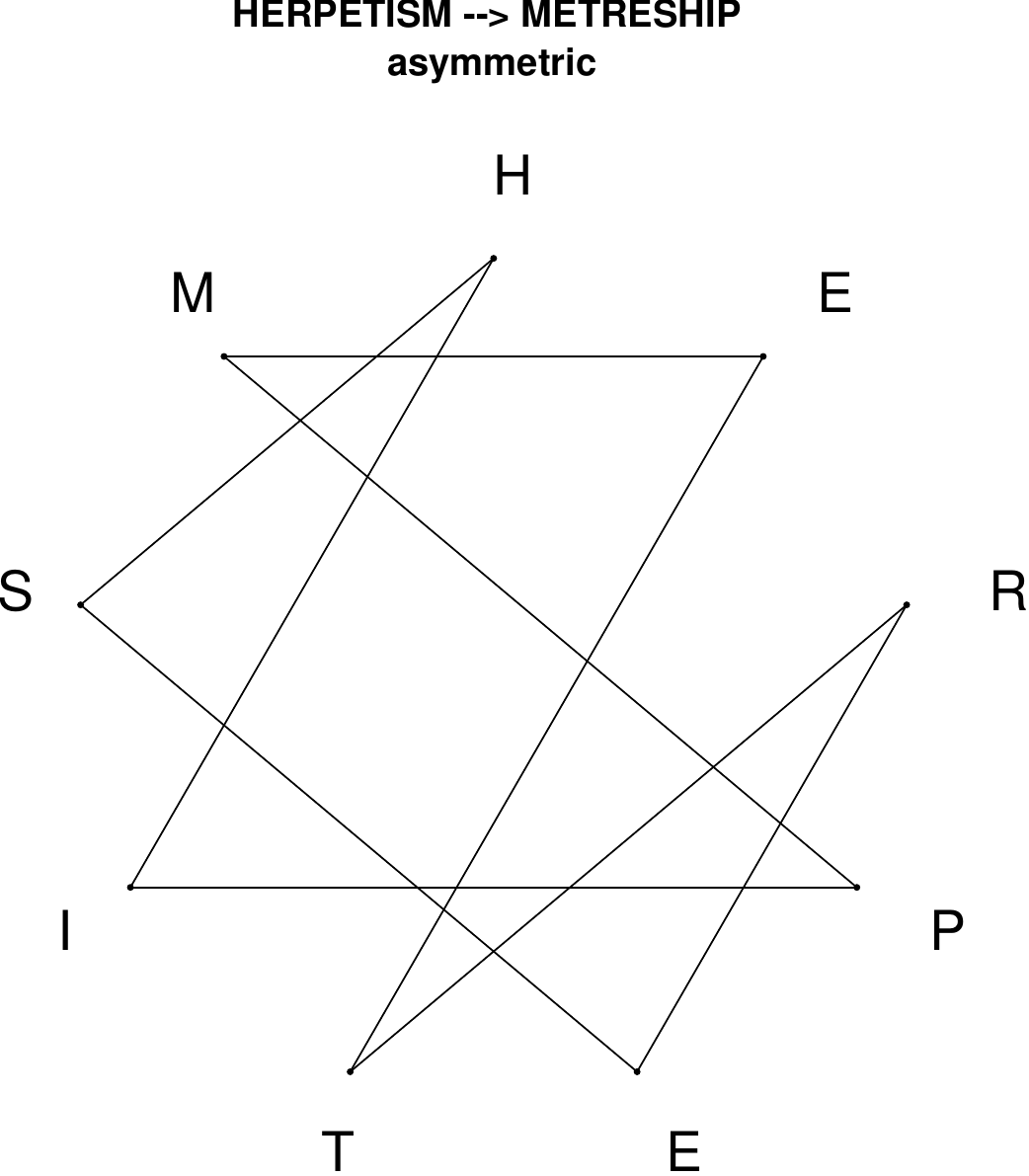}
\end{subfigure}
\end{figure}

\begin{figure}[H]
\centering
\begin{subfigure}[T]{0.19\textwidth}
\centering
\includegraphics[width=\textwidth]{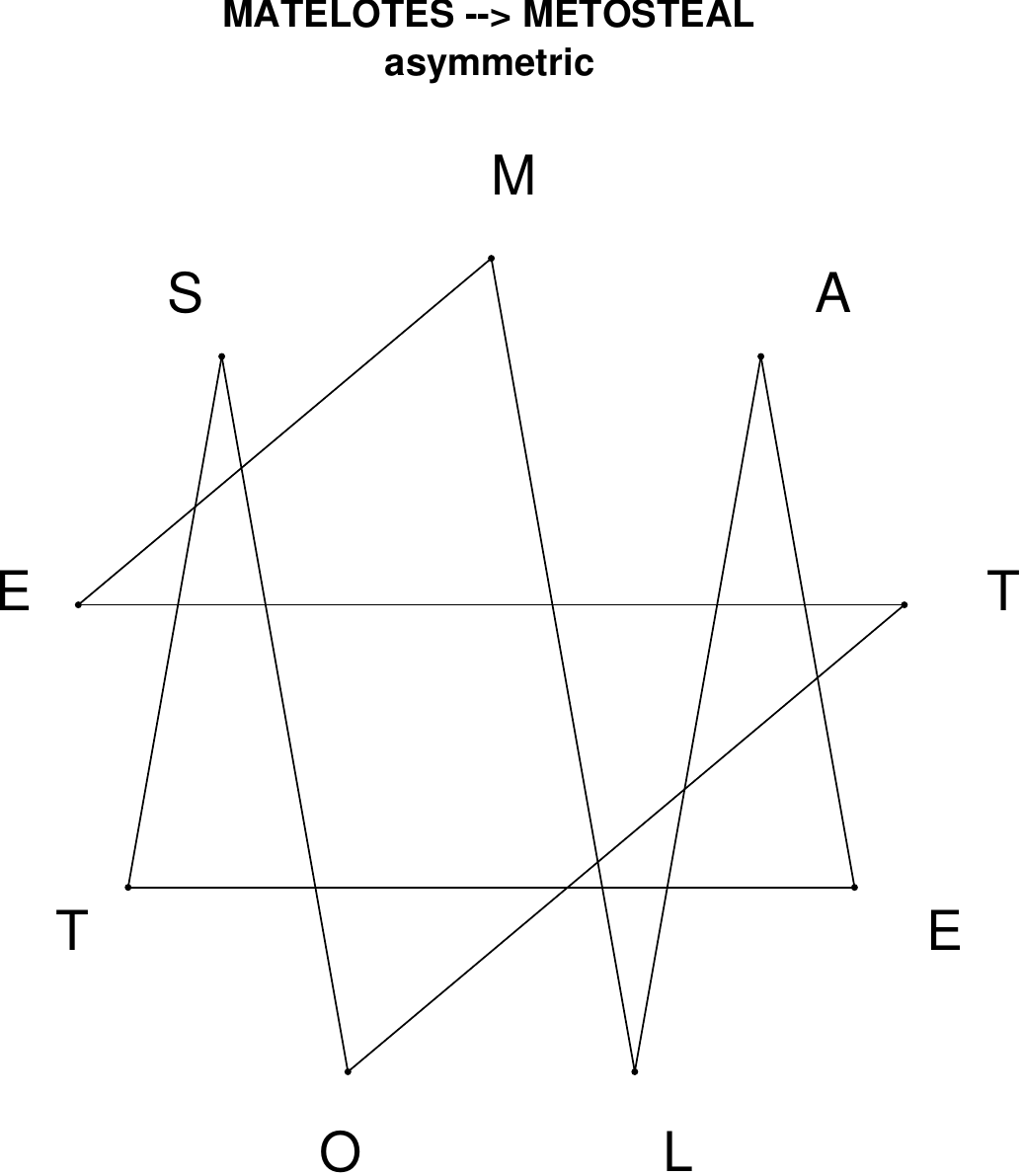}
\end{subfigure}
\hfill
\begin{subfigure}[T]{0.19\textwidth}
\centering
\includegraphics[width=\textwidth]{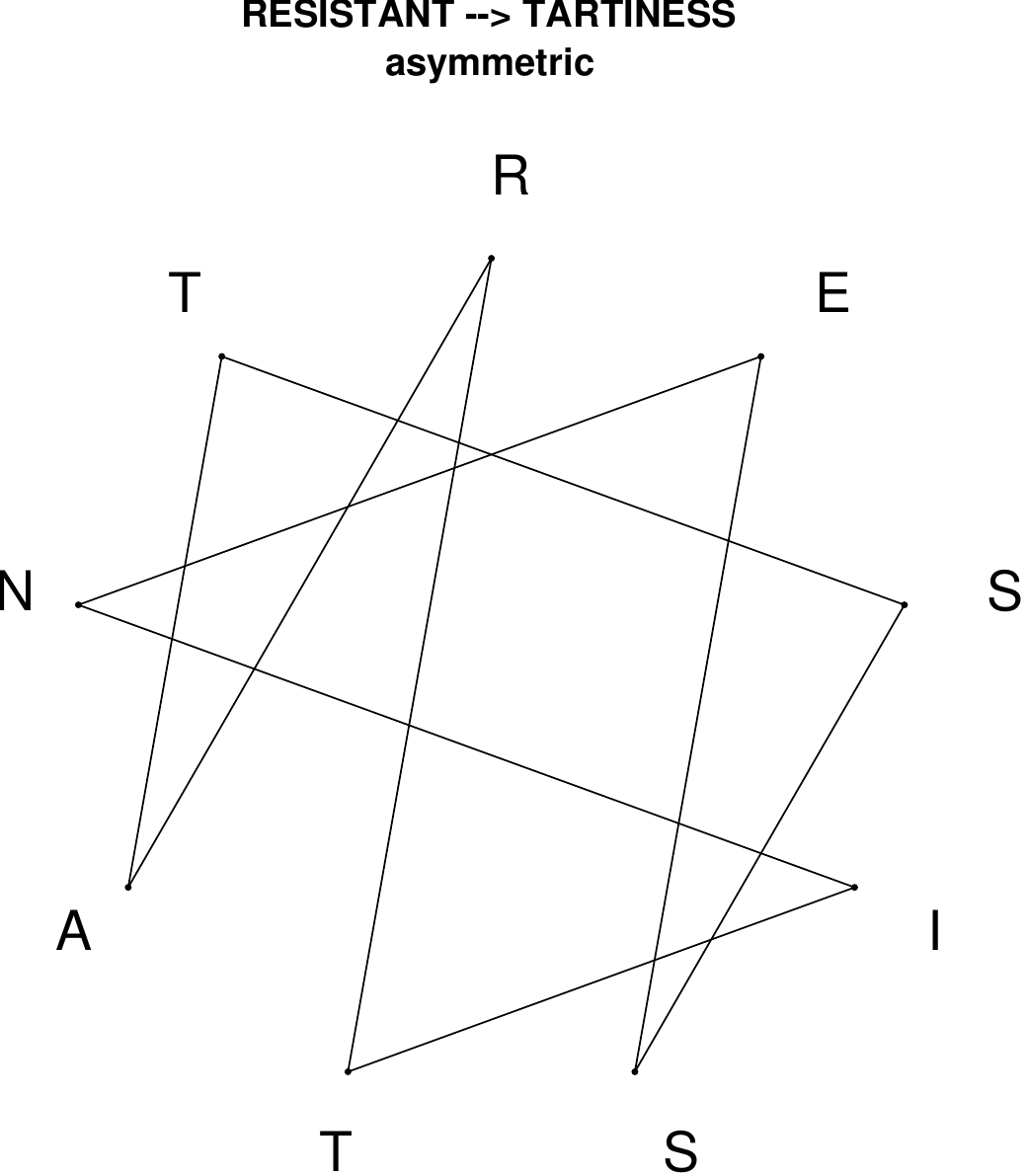}
\end{subfigure}
\hfill
\begin{subfigure}[T]{0.19\textwidth}
\centering
\includegraphics[width=\textwidth]{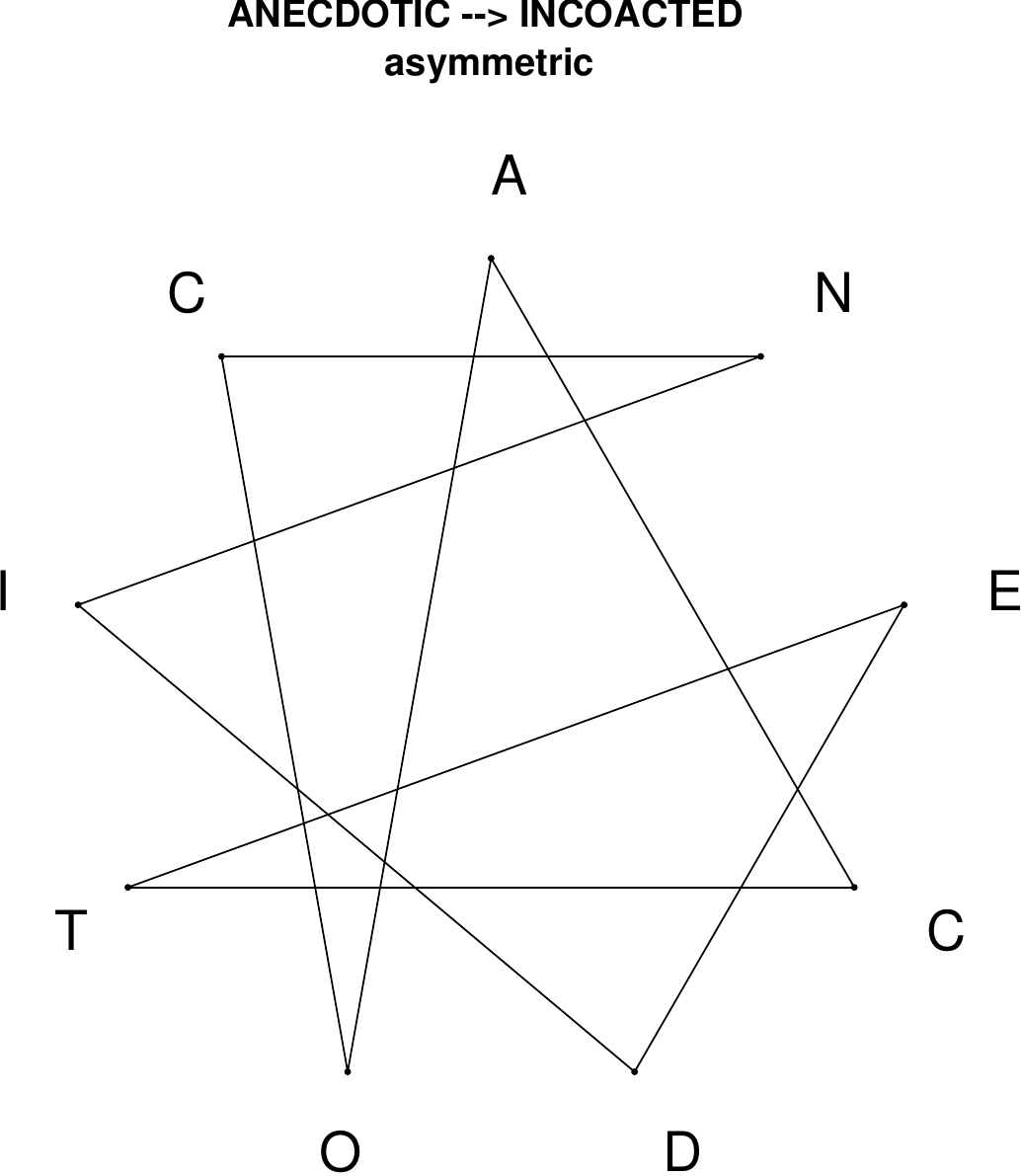}
\end{subfigure}
\hfill
\begin{subfigure}[T]{0.19\textwidth}
\centering
\includegraphics[width=\textwidth]{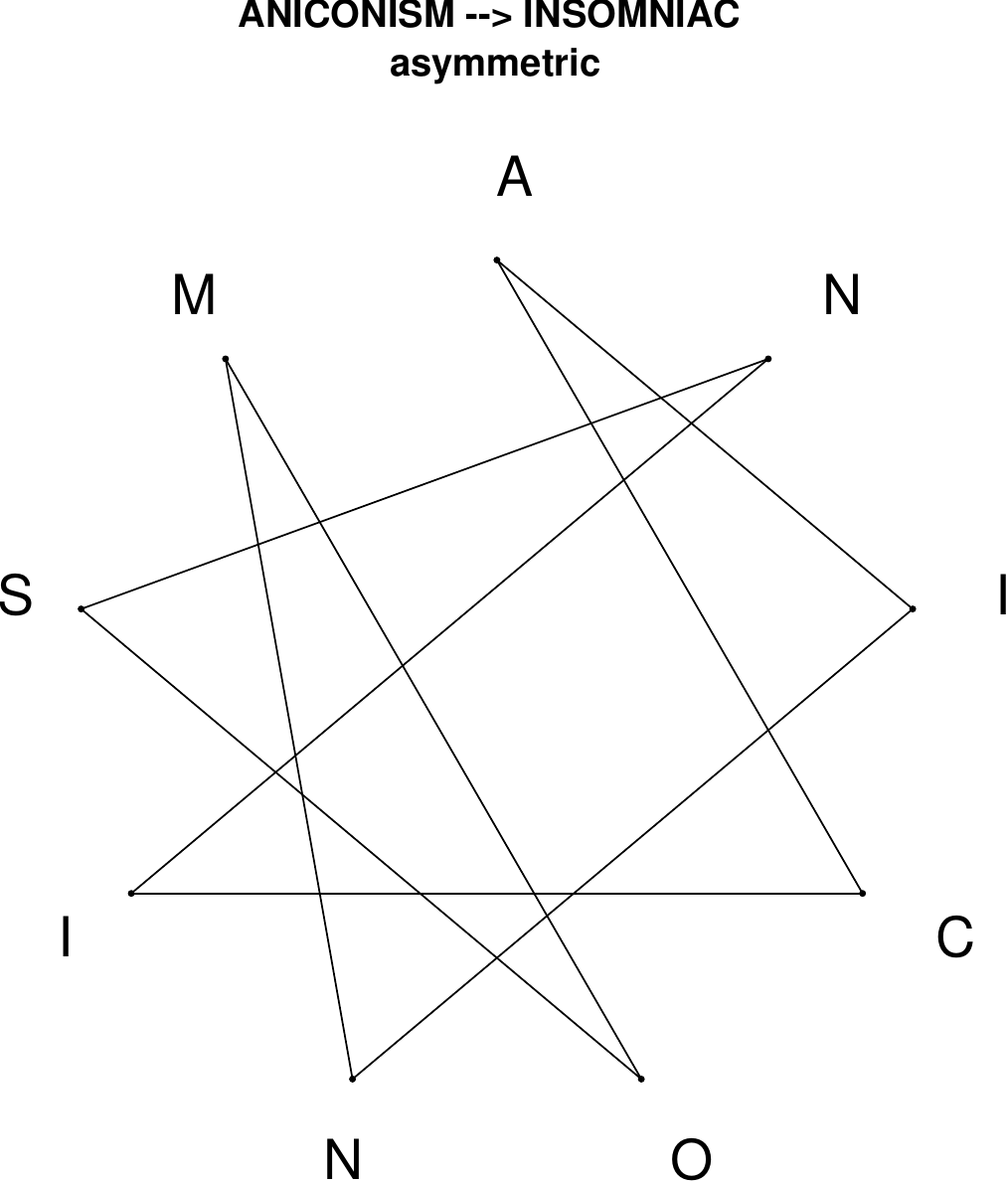}
\end{subfigure}
\hfill
\begin{subfigure}[T]{0.19\textwidth}
\centering
\includegraphics[width=\textwidth]{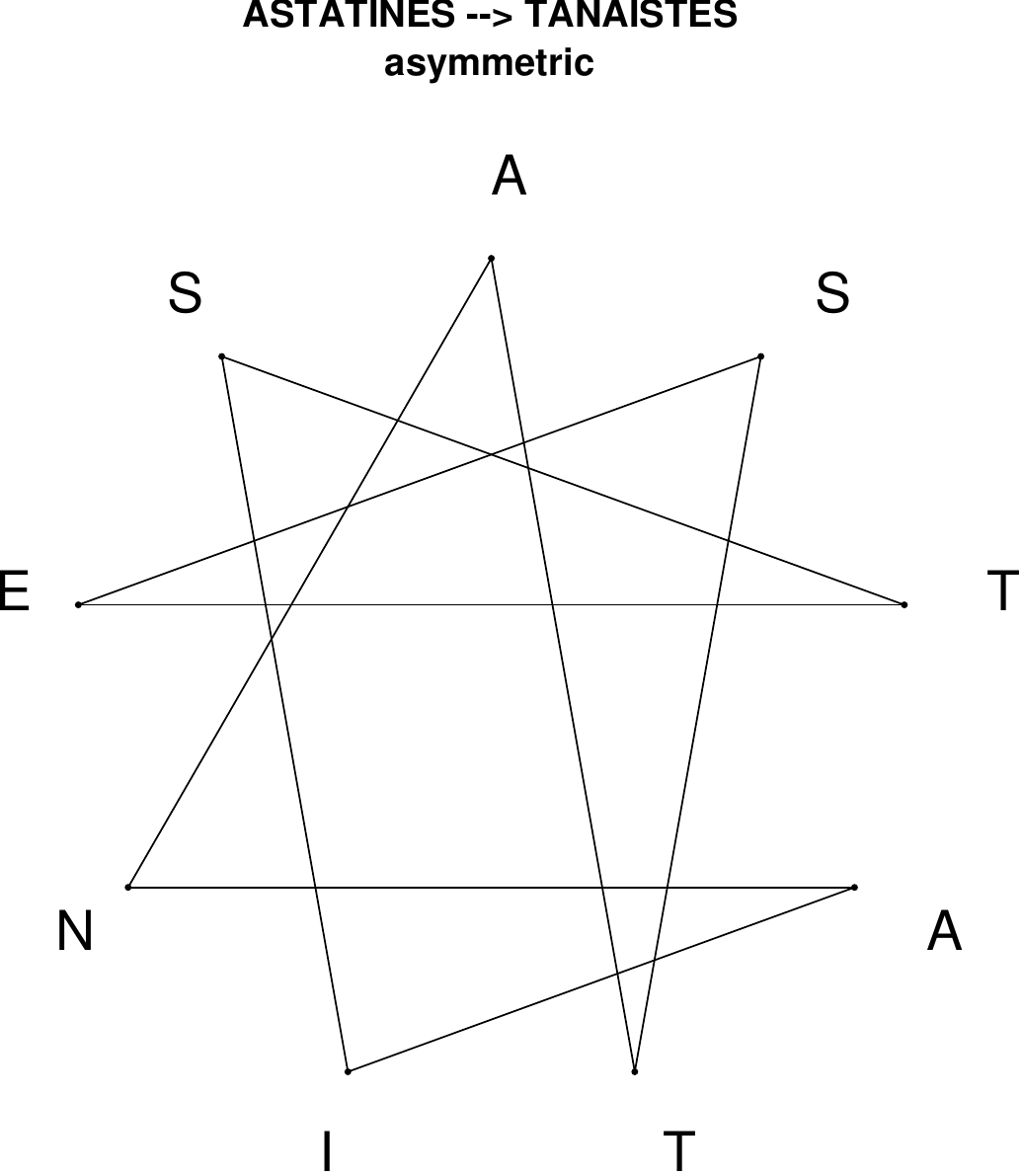}
\end{subfigure}
\end{figure}

\begin{figure}[H]
\centering
\begin{subfigure}[T]{0.19\textwidth}
\centering
\includegraphics[width=\textwidth]{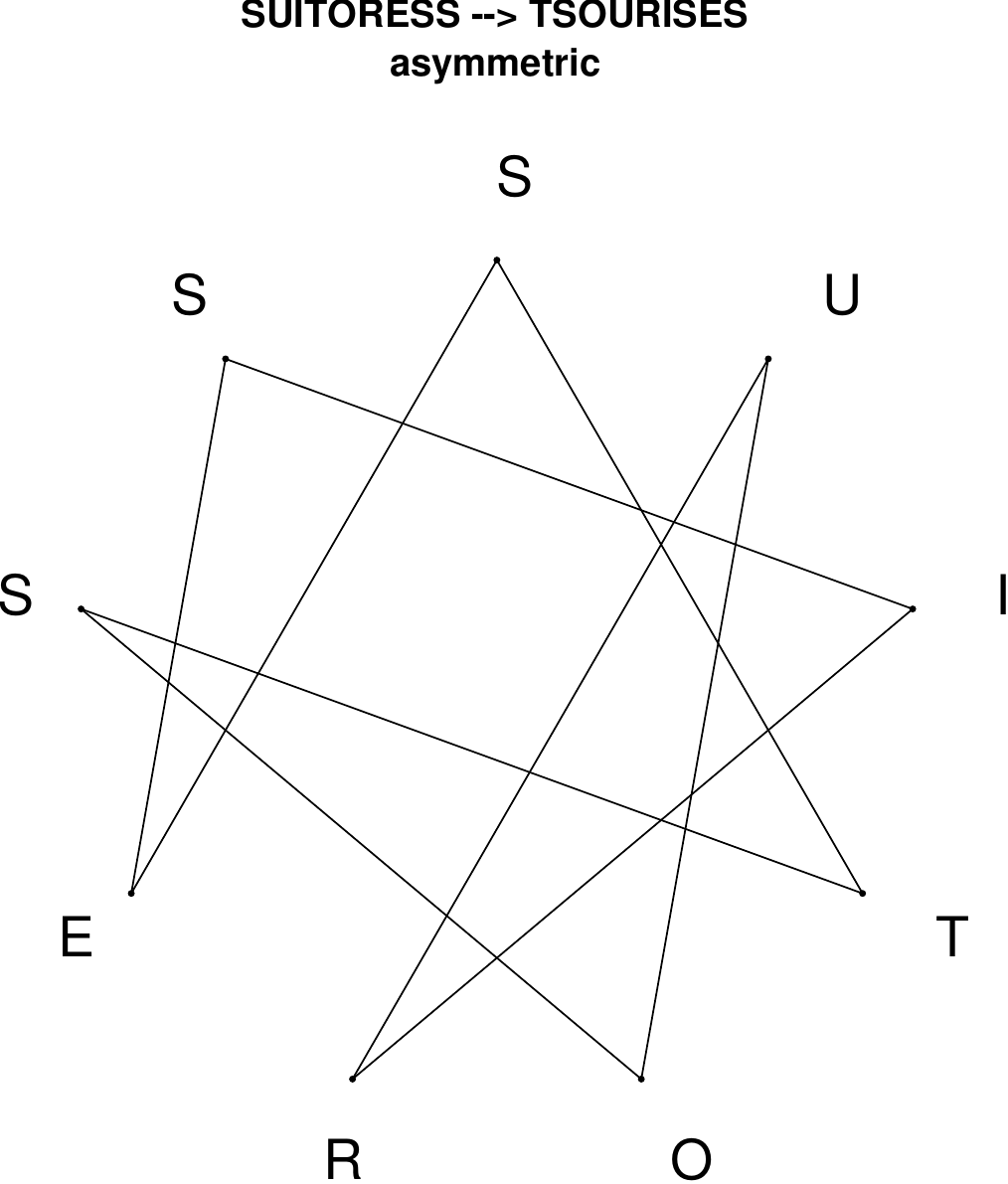}
\end{subfigure}
\hfill
\begin{subfigure}[T]{0.19\textwidth}
\centering
\includegraphics[width=\textwidth]{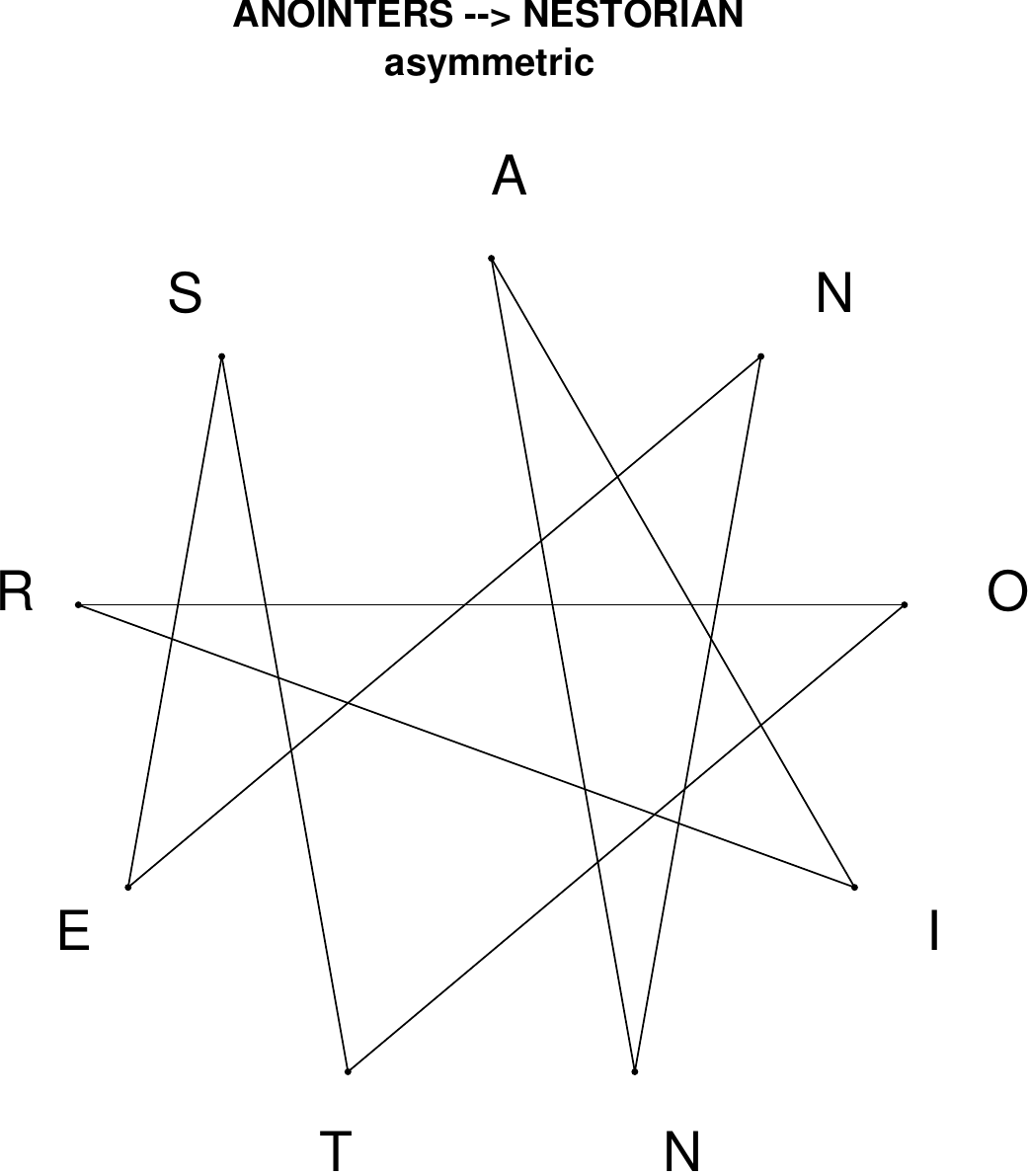}
\end{subfigure}
\hfill
\begin{subfigure}[T]{0.19\textwidth}
\centering
\includegraphics[width=\textwidth]{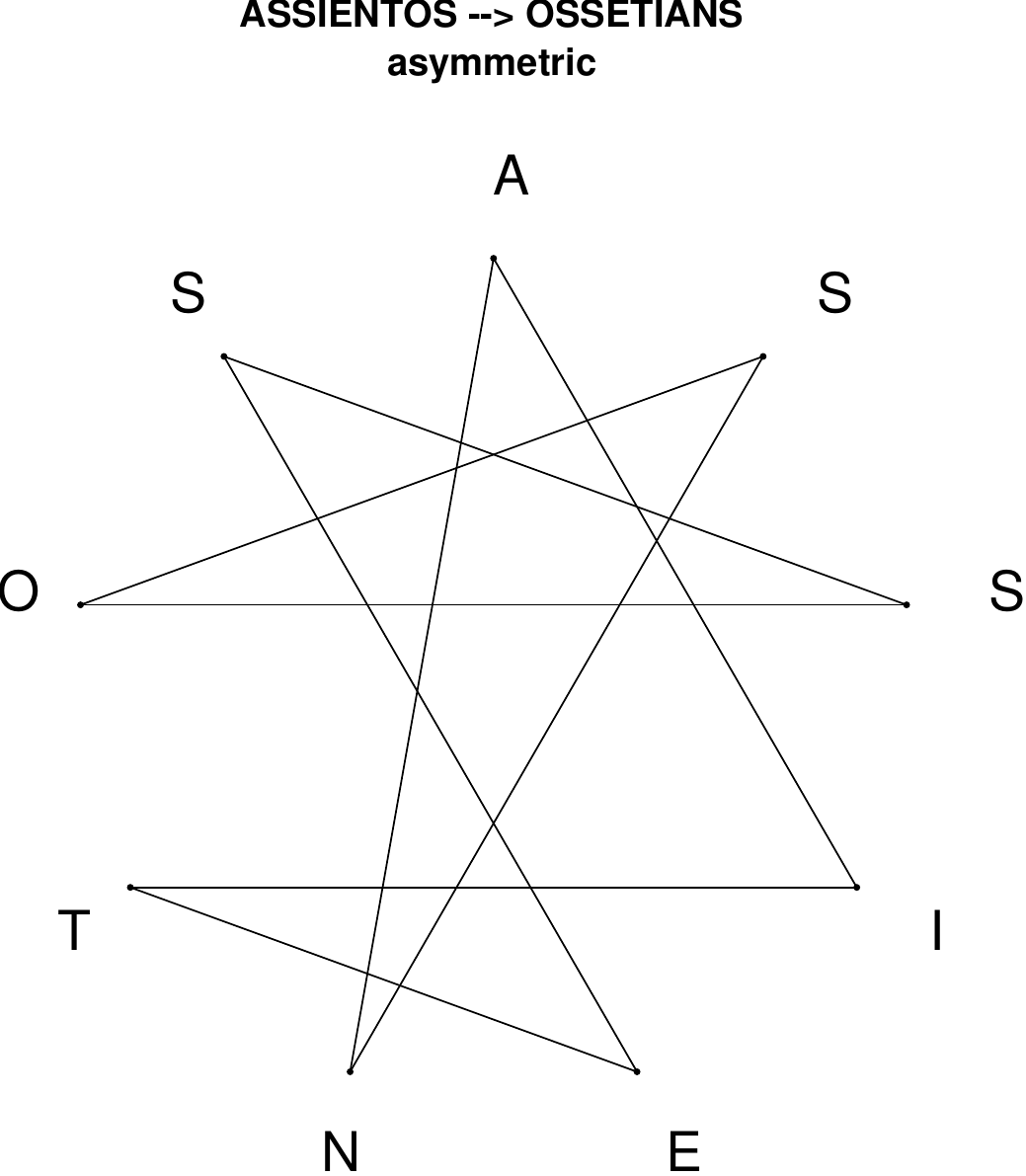}
\end{subfigure}
\hfill
\begin{subfigure}[T]{0.19\textwidth}
\centering
\includegraphics[width=\textwidth]{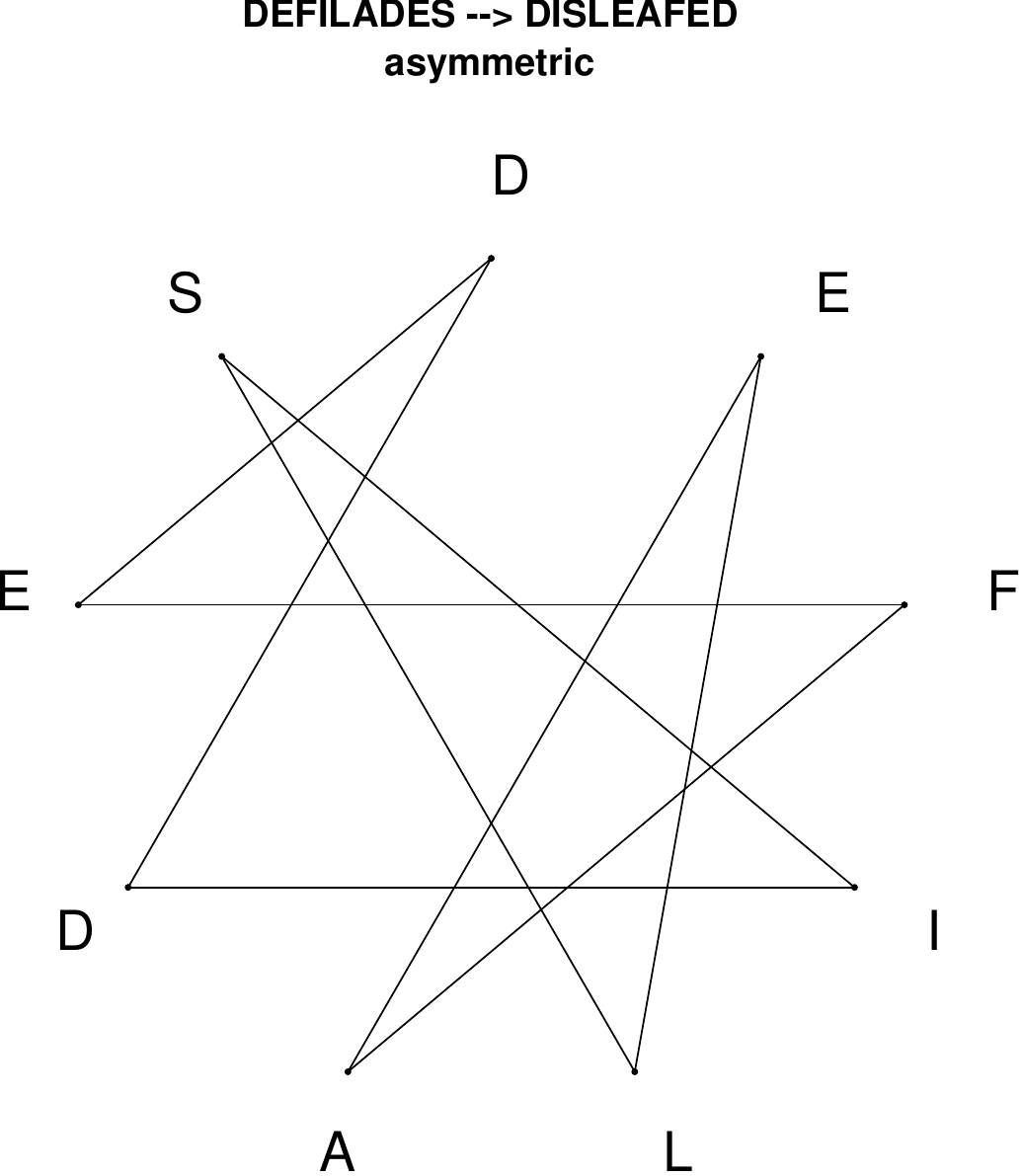}
\end{subfigure}
\hfill
\begin{subfigure}[T]{0.19\textwidth}
\centering
\includegraphics[width=\textwidth]{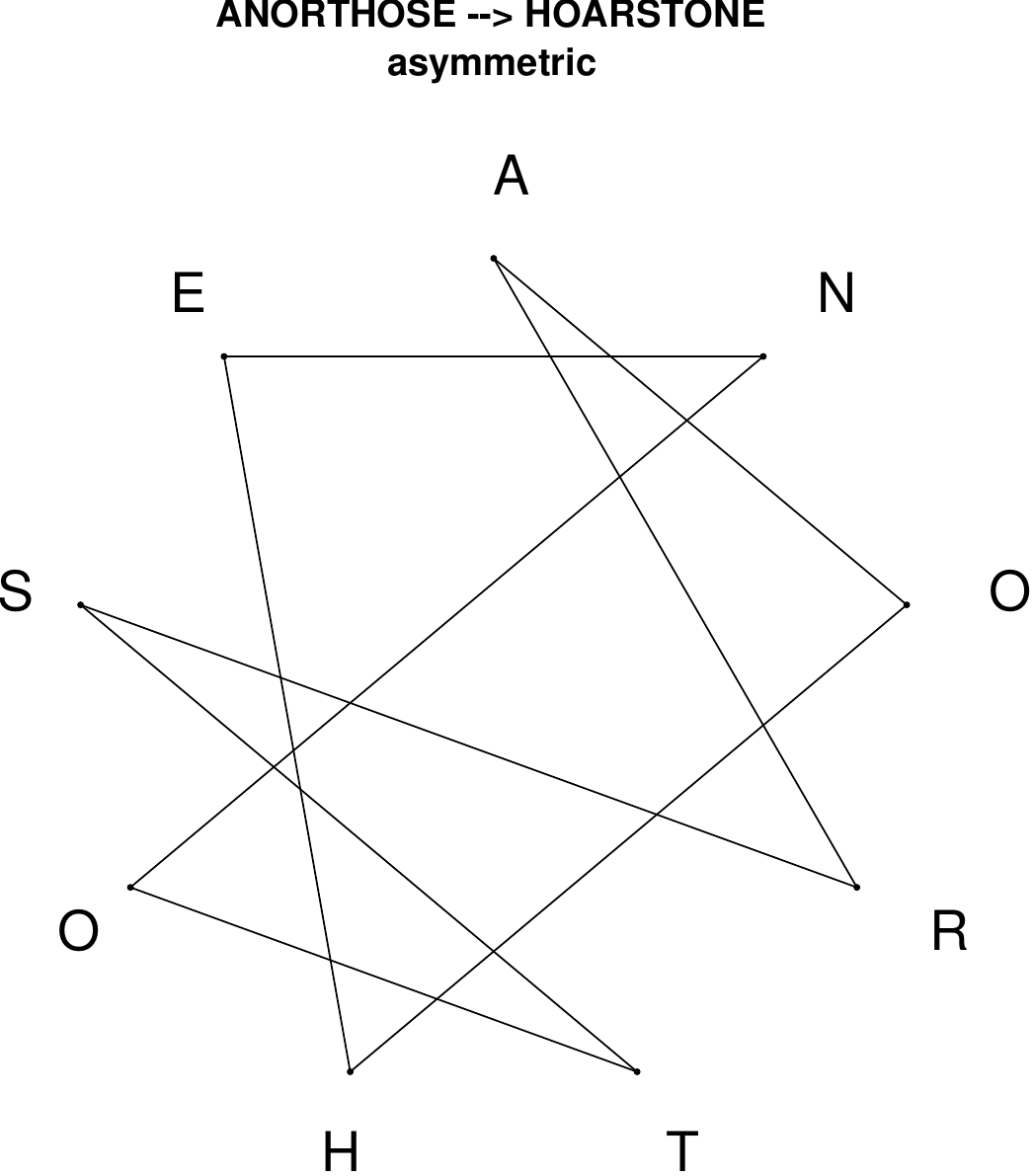}
\end{subfigure}
\end{figure}

\begin{figure}[H]
\centering
\begin{subfigure}[T]{0.19\textwidth}
\centering
\includegraphics[width=\textwidth]{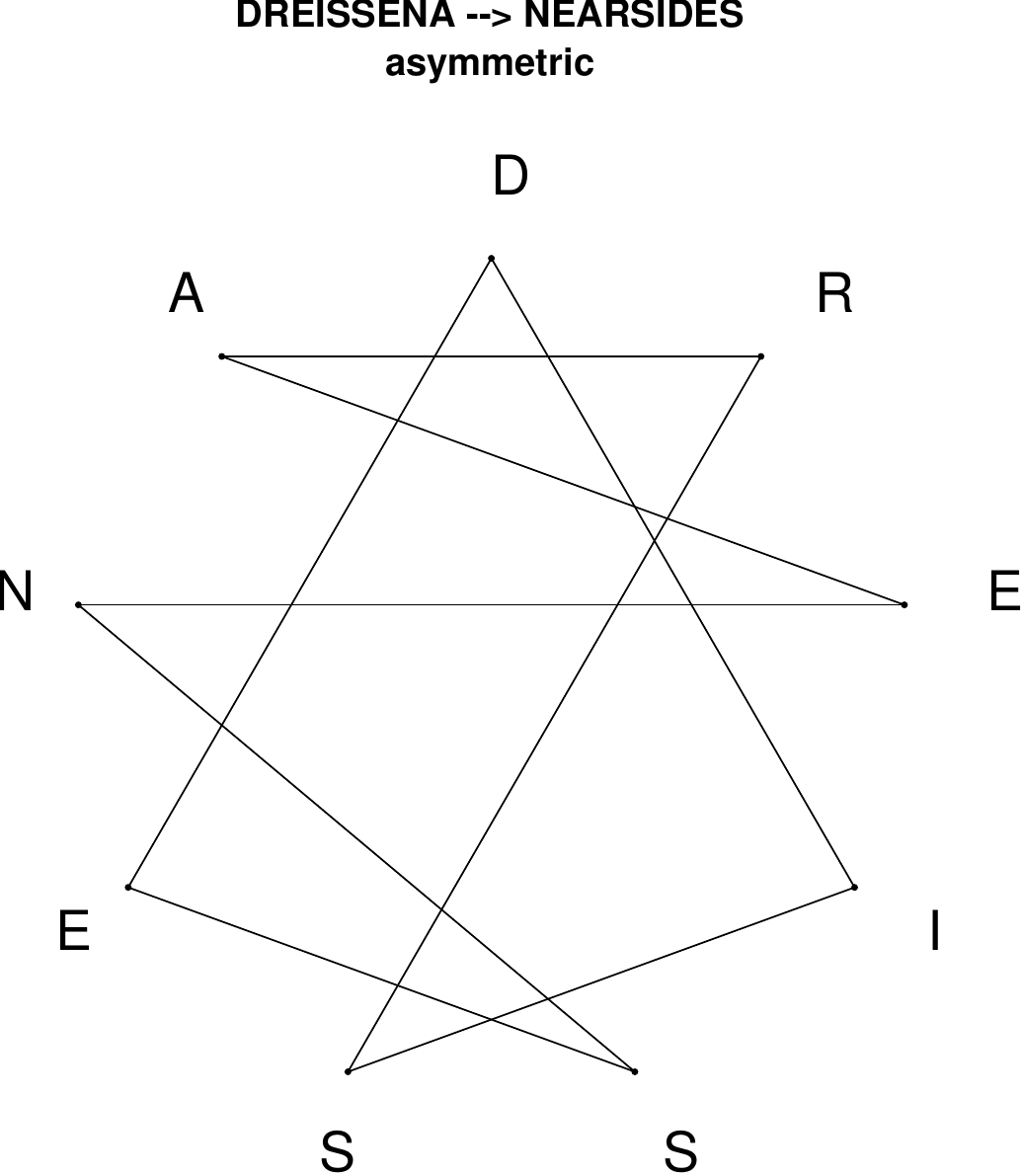}
\end{subfigure}
\hfill
\begin{subfigure}[T]{0.19\textwidth}
\centering
\includegraphics[width=\textwidth]{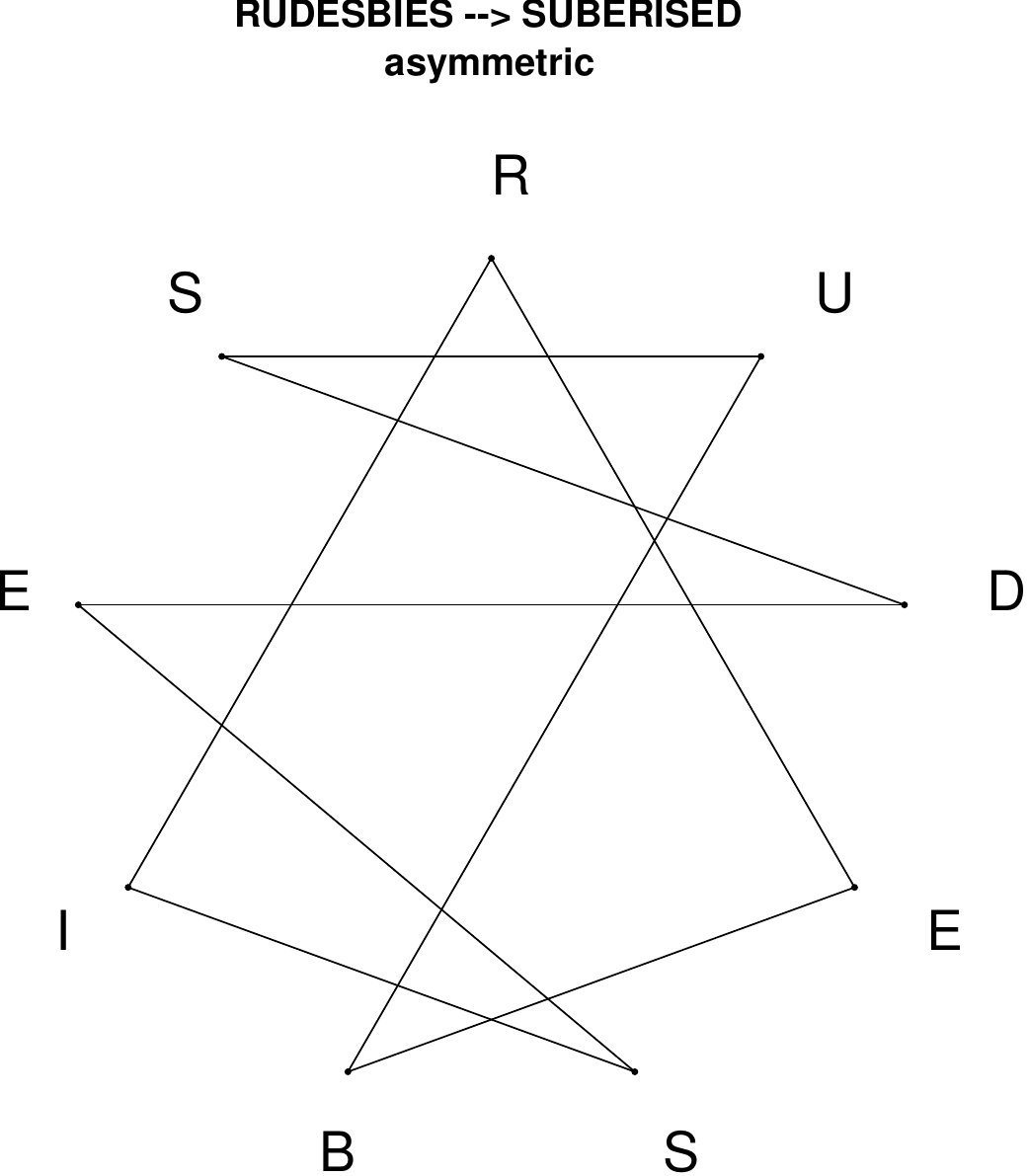}
\end{subfigure}
\hfill
\begin{subfigure}[T]{0.19\textwidth}
\centering
\includegraphics[width=\textwidth]{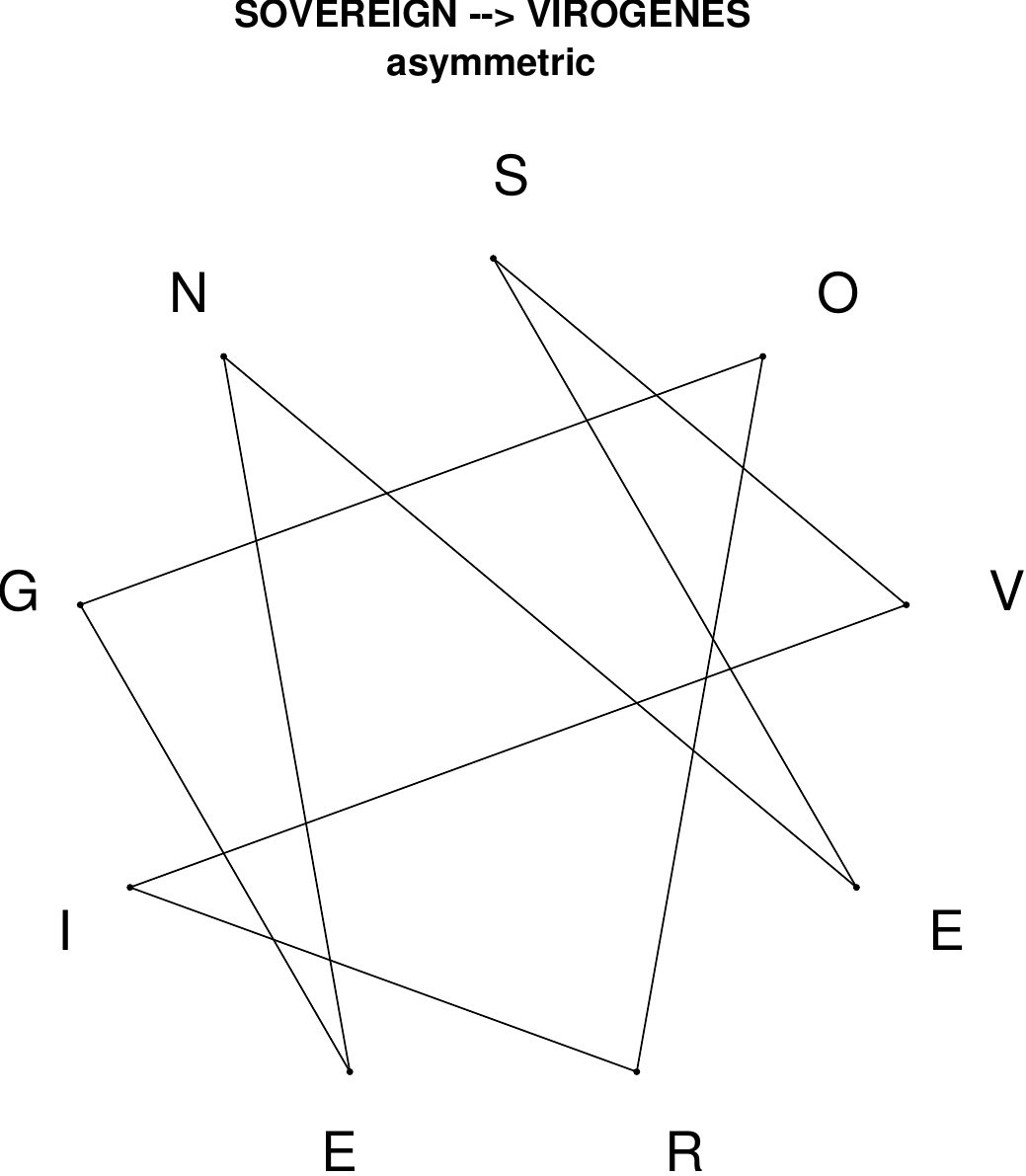}
\end{subfigure}
\hfill
\begin{subfigure}[T]{0.19\textwidth}
\centering
\includegraphics[width=\textwidth]{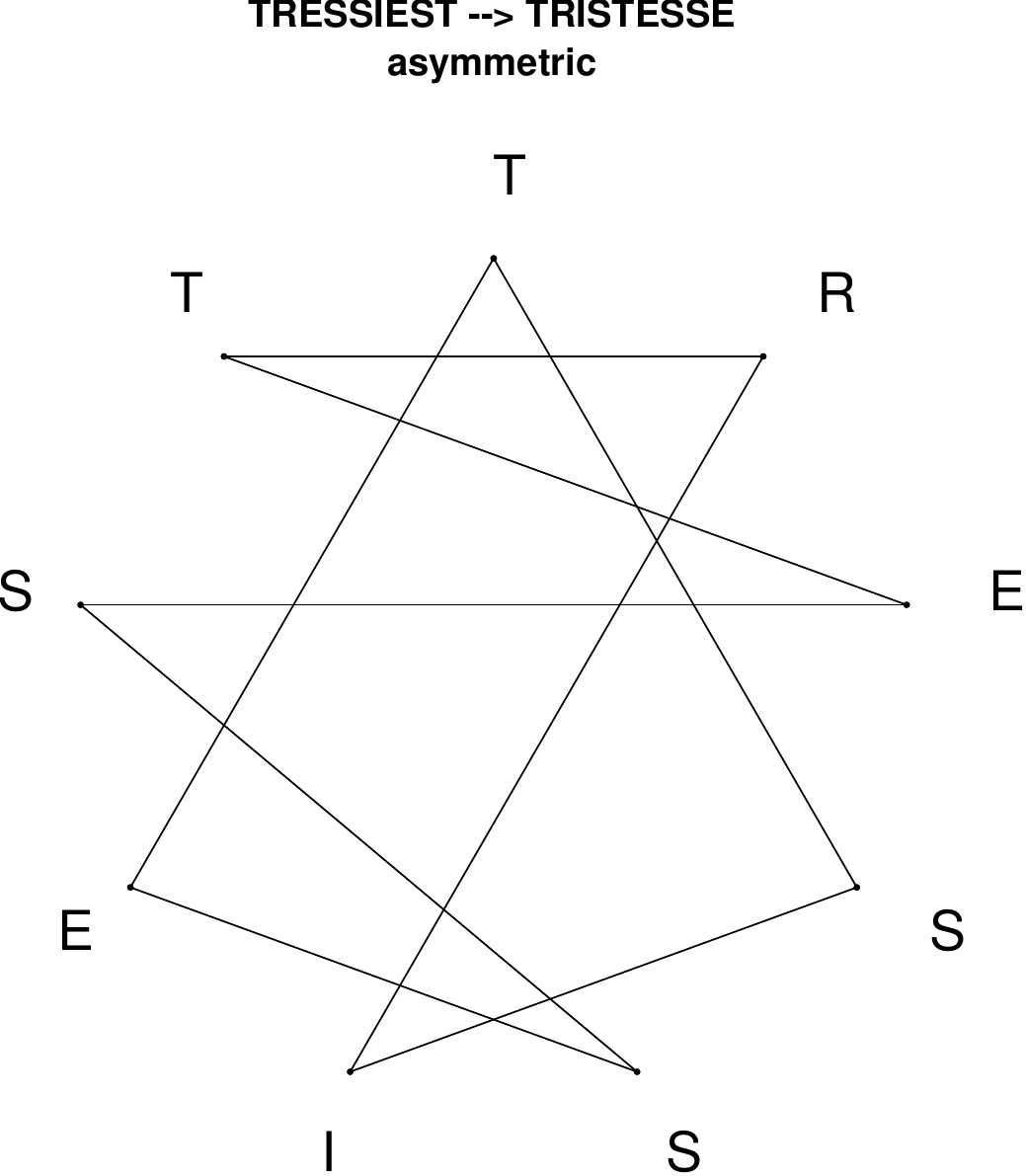}
\end{subfigure}
\hfill
\begin{subfigure}[T]{0.19\textwidth}
\centering
\includegraphics[width=\textwidth]{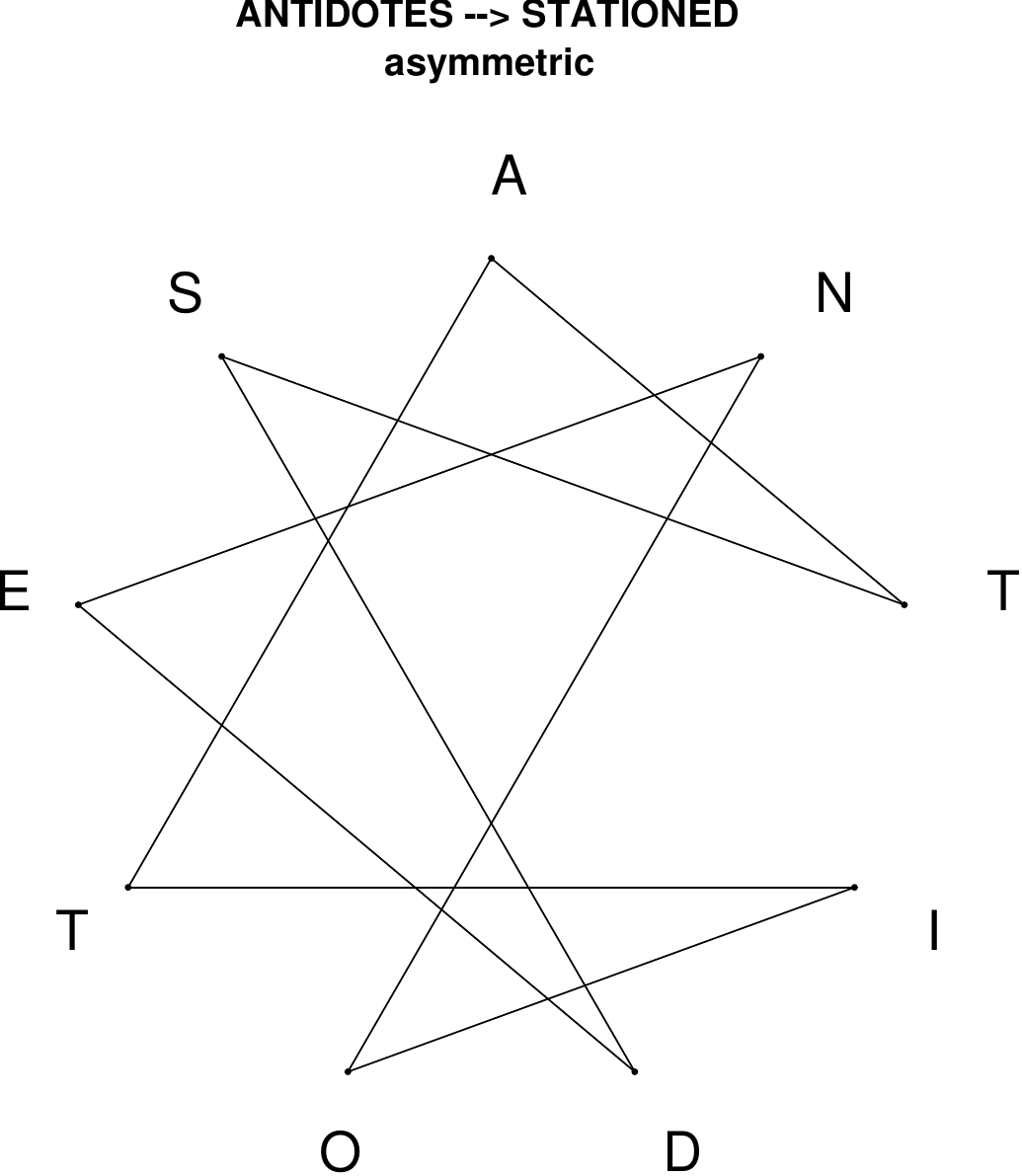}
\end{subfigure}
\end{figure}

\begin{figure}[H]
\centering
\begin{subfigure}[T]{0.19\textwidth}
\centering
\includegraphics[width=\textwidth]{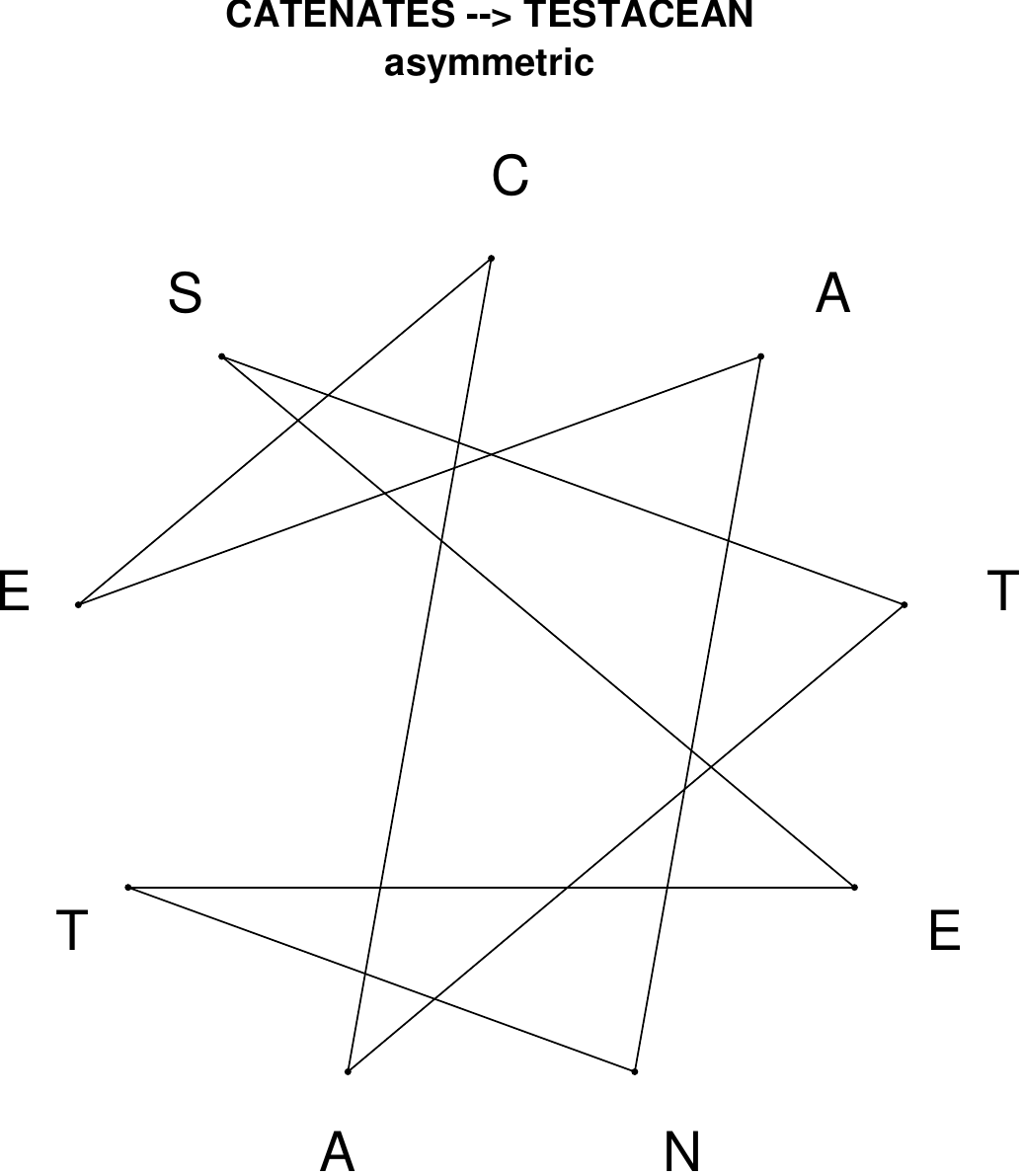}
\end{subfigure}
\hfill
\begin{subfigure}[T]{0.19\textwidth}
\centering
\includegraphics[width=\textwidth]{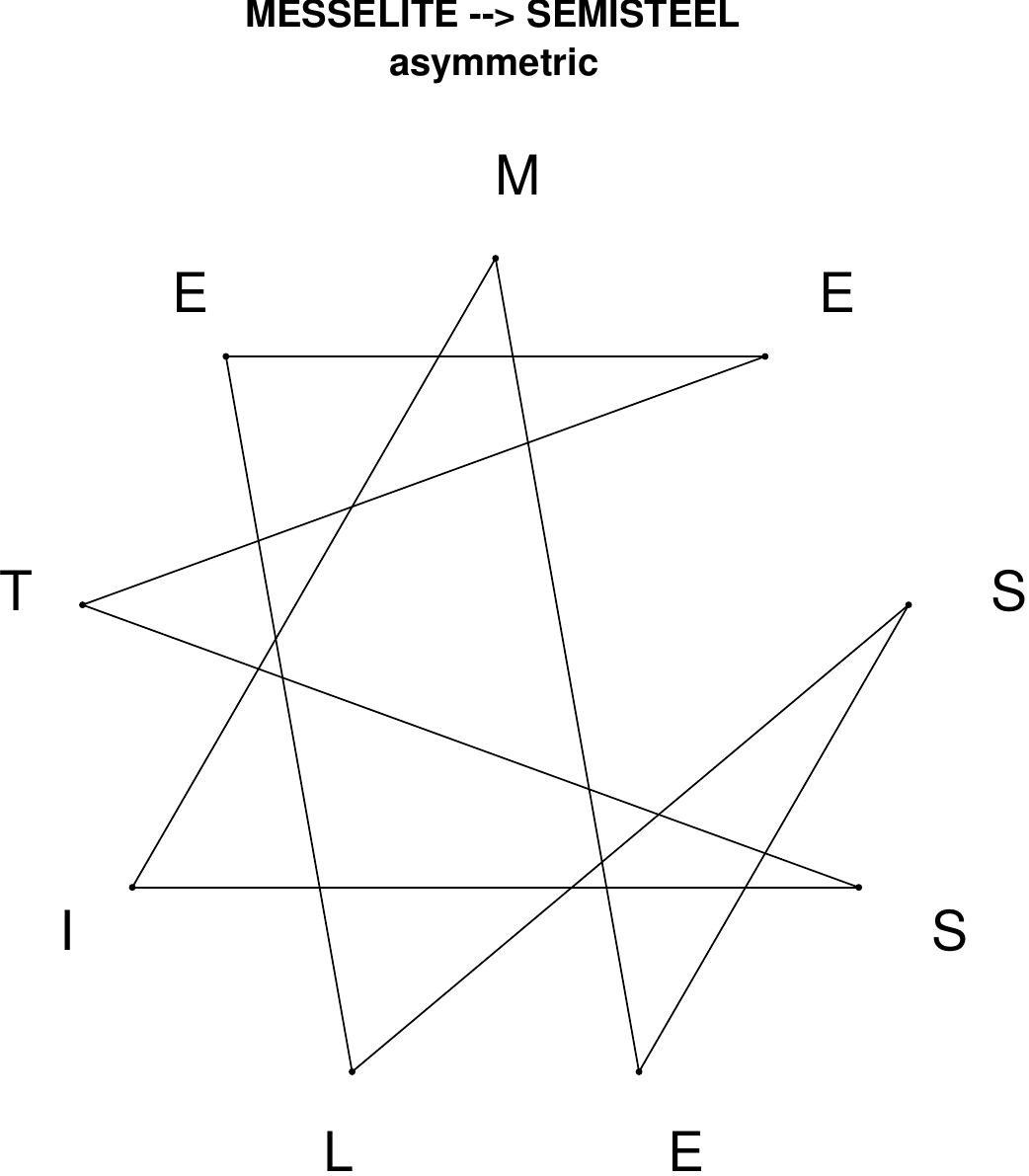}
\end{subfigure}
\hfill
\begin{subfigure}[T]{0.19\textwidth}
\centering
\includegraphics[width=\textwidth]{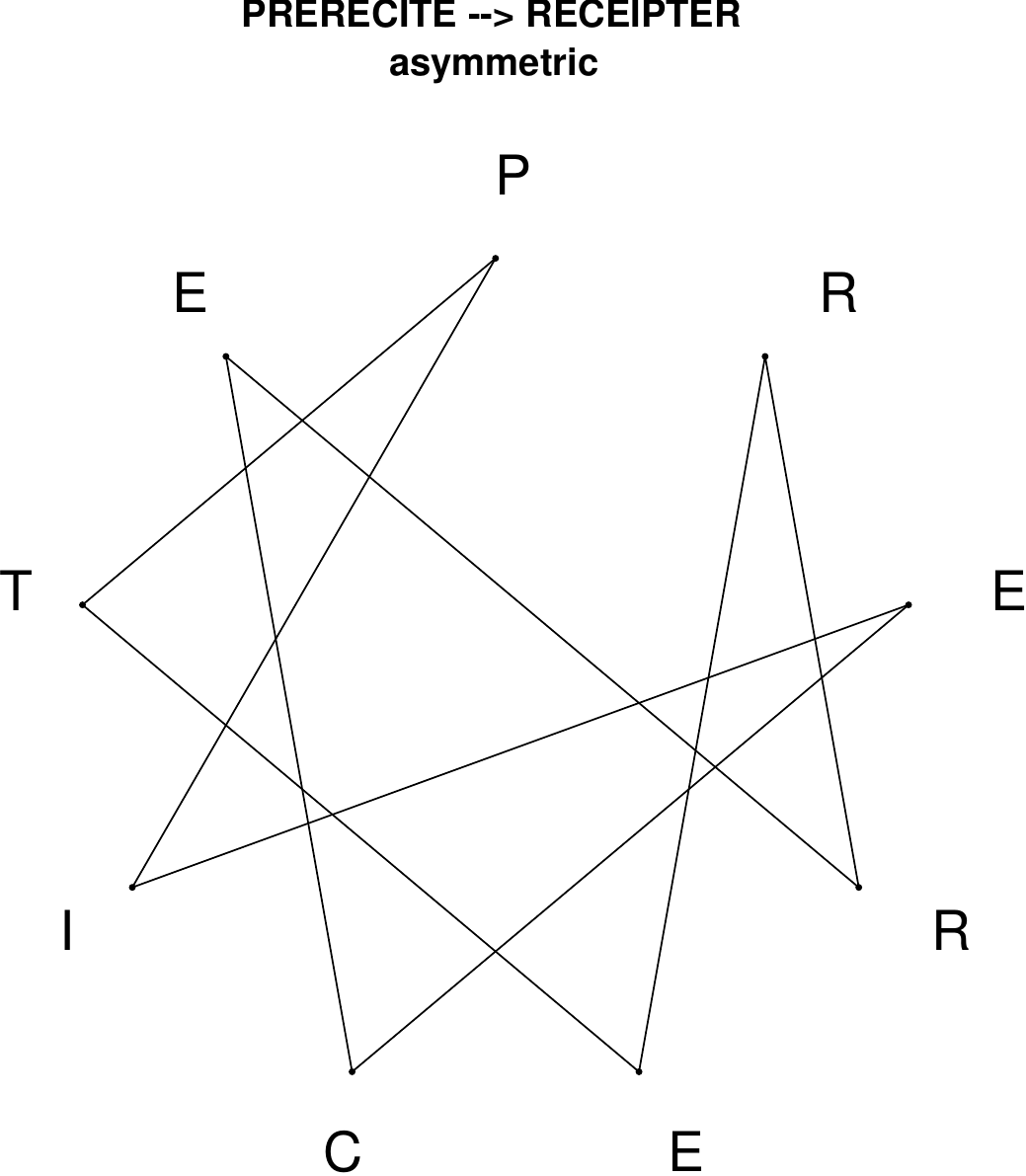}
\end{subfigure}
\hfill
\begin{subfigure}[T]{0.19\textwidth}
\centering
\includegraphics[width=\textwidth]{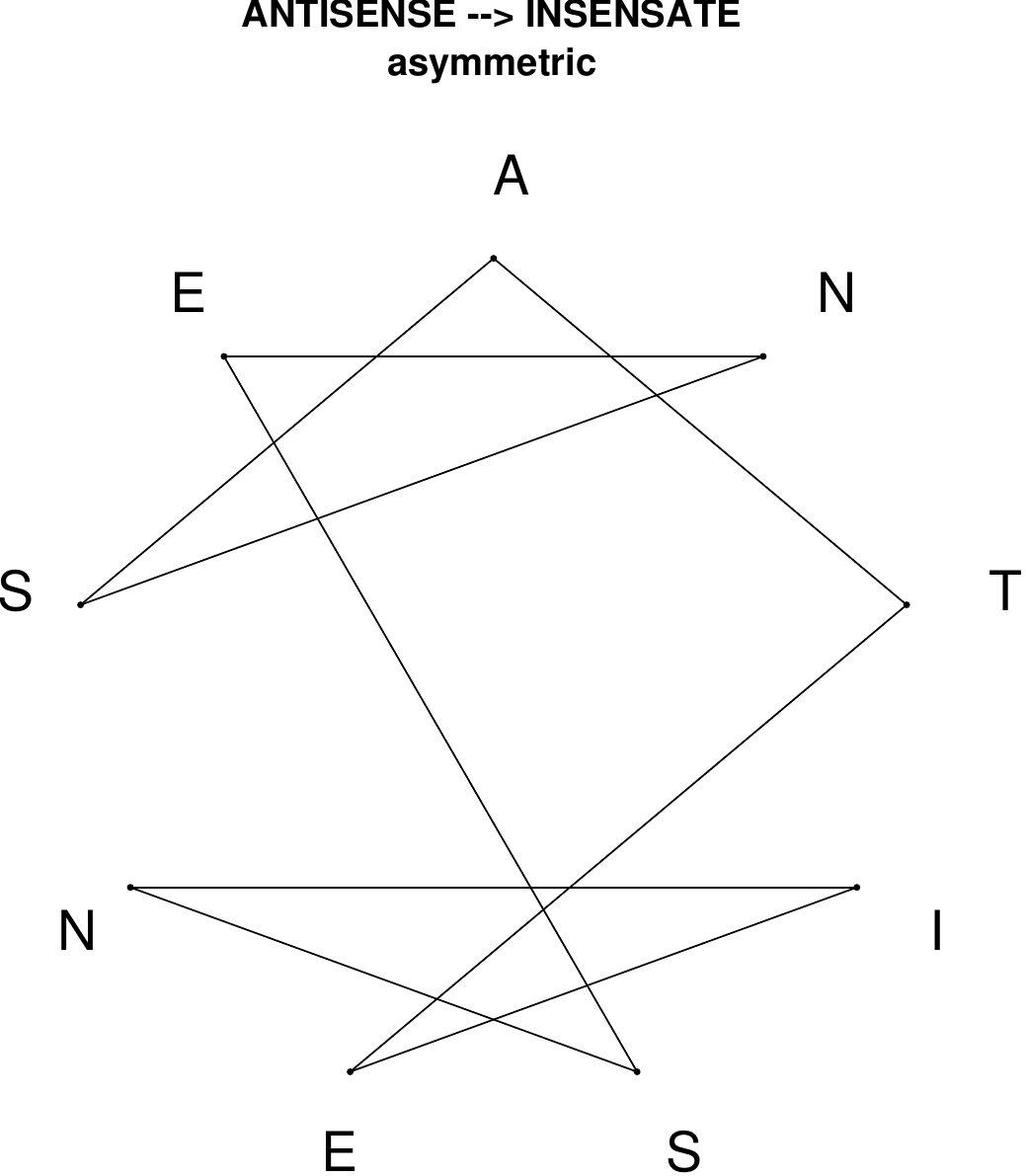}
\end{subfigure}
\hfill
\begin{subfigure}[T]{0.19\textwidth}
\centering
\includegraphics[width=\textwidth]{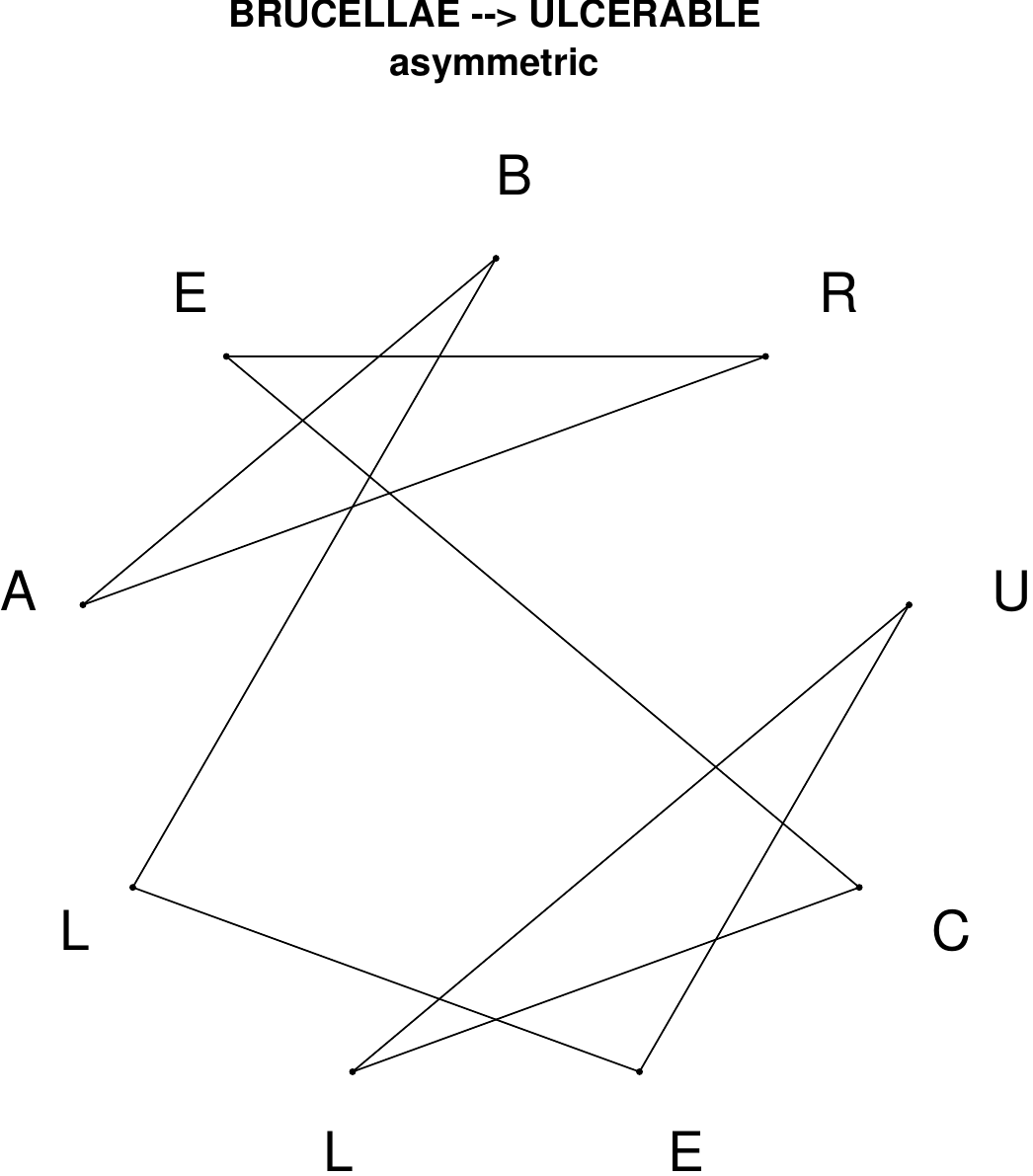}
\end{subfigure}
\end{figure}

\begin{figure}[H]
\centering
\begin{subfigure}[T]{0.19\textwidth}
\centering
\includegraphics[width=\textwidth]{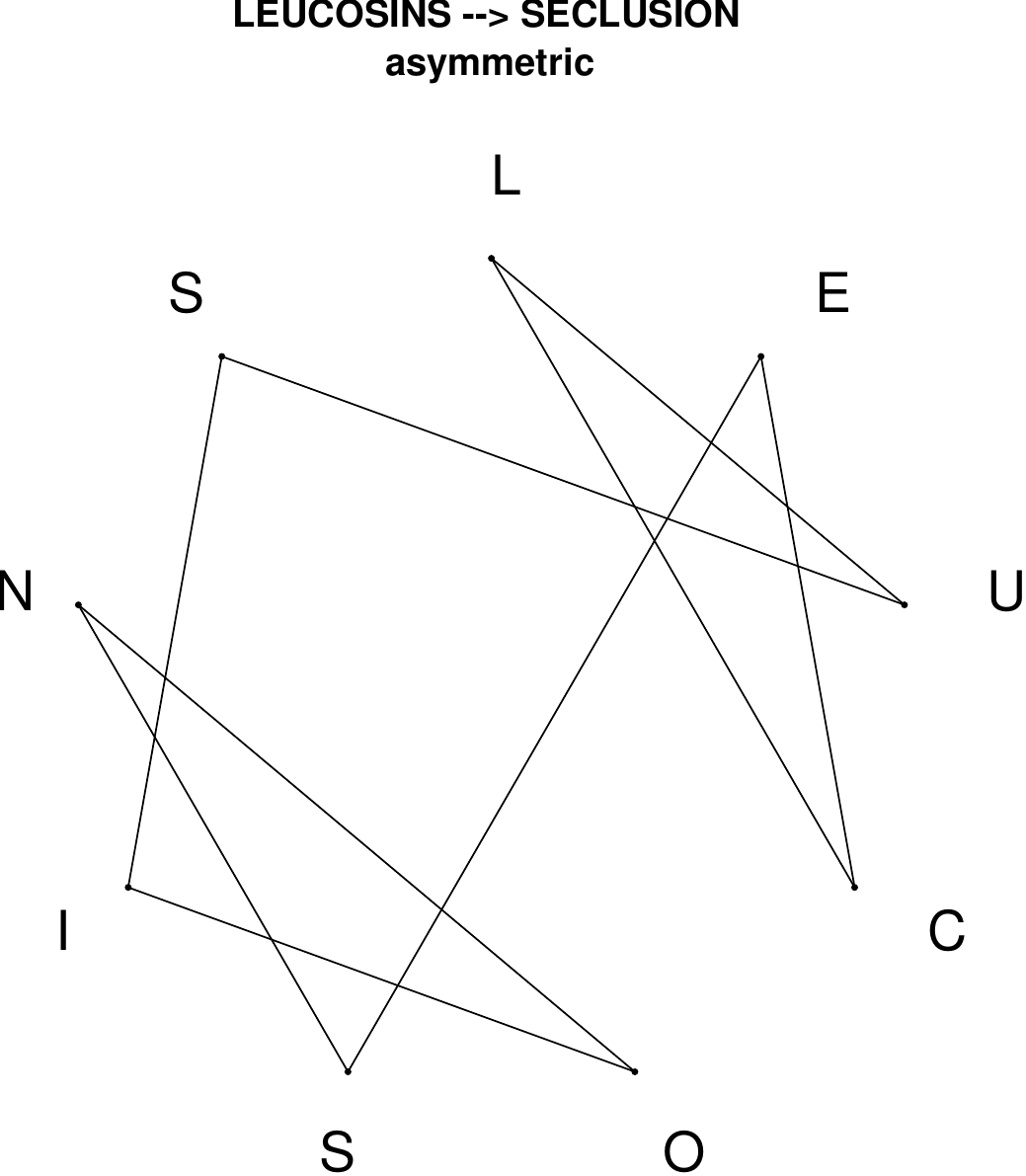}
\end{subfigure}
\hfill
\begin{subfigure}[T]{0.19\textwidth}
\centering
\includegraphics[width=\textwidth]{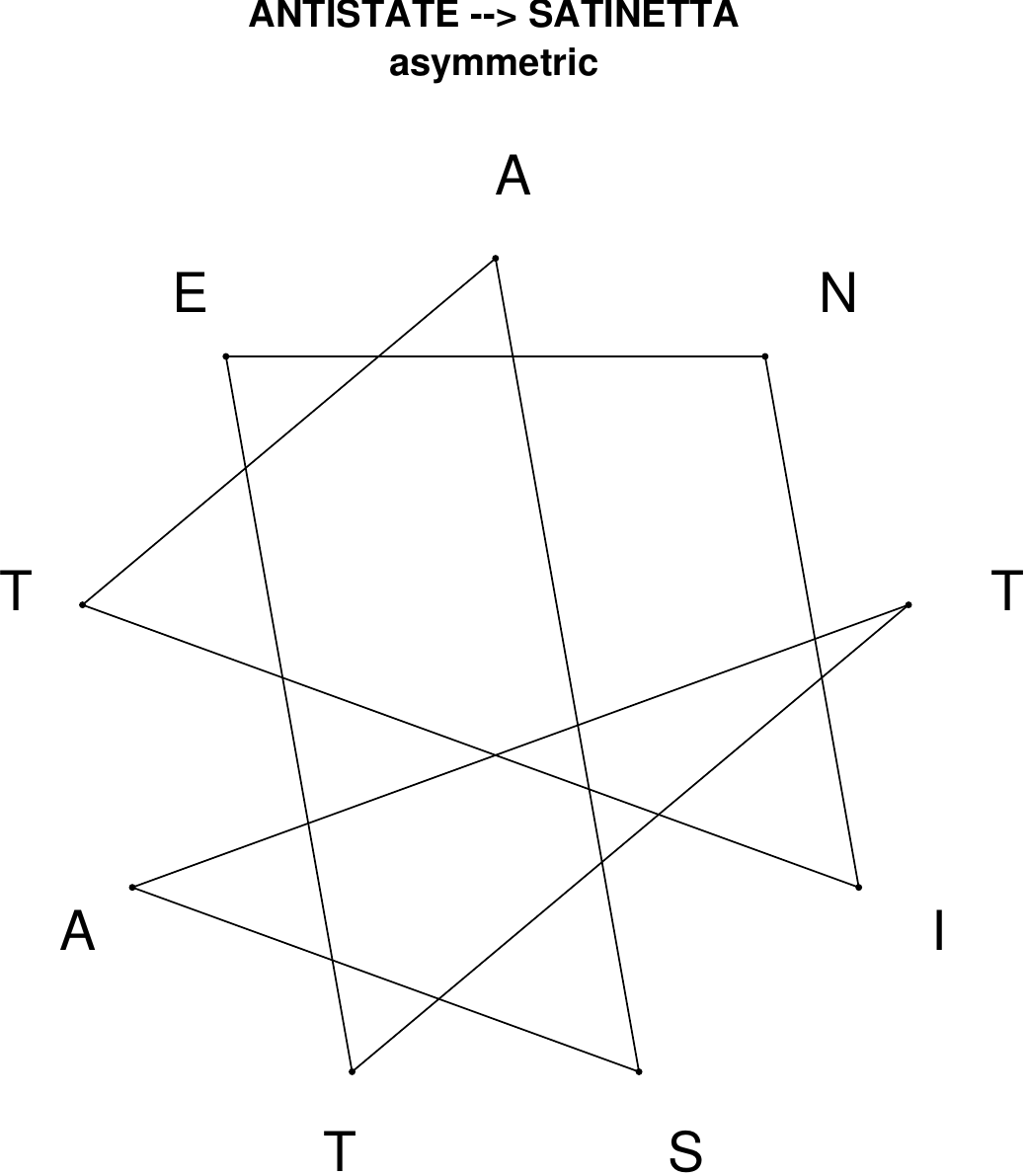}
\end{subfigure}
\hfill
\begin{subfigure}[T]{0.19\textwidth}
\centering
\includegraphics[width=\textwidth]{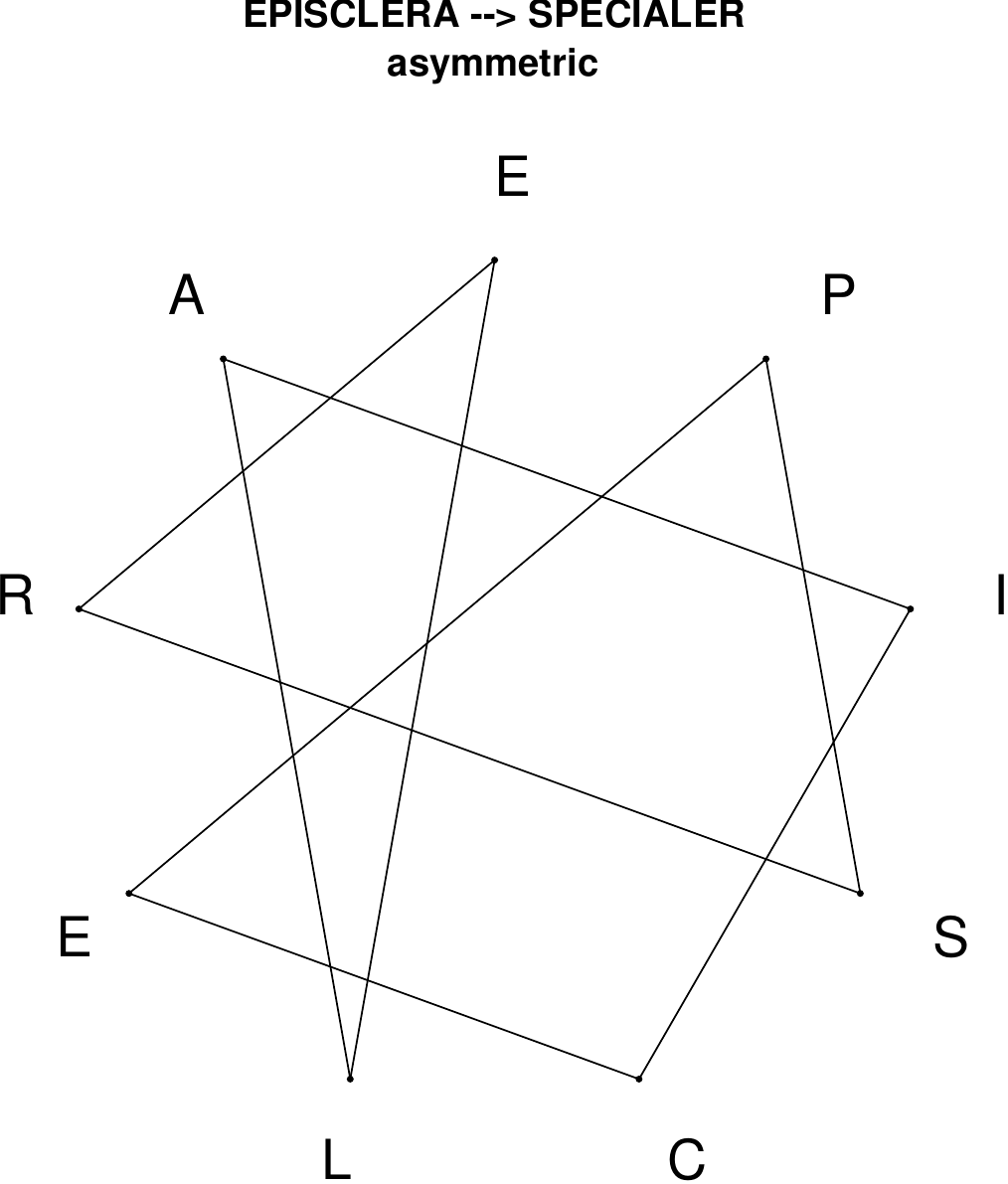}
\end{subfigure}
\hfill
\begin{subfigure}[T]{0.19\textwidth}
\centering
\includegraphics[width=\textwidth]{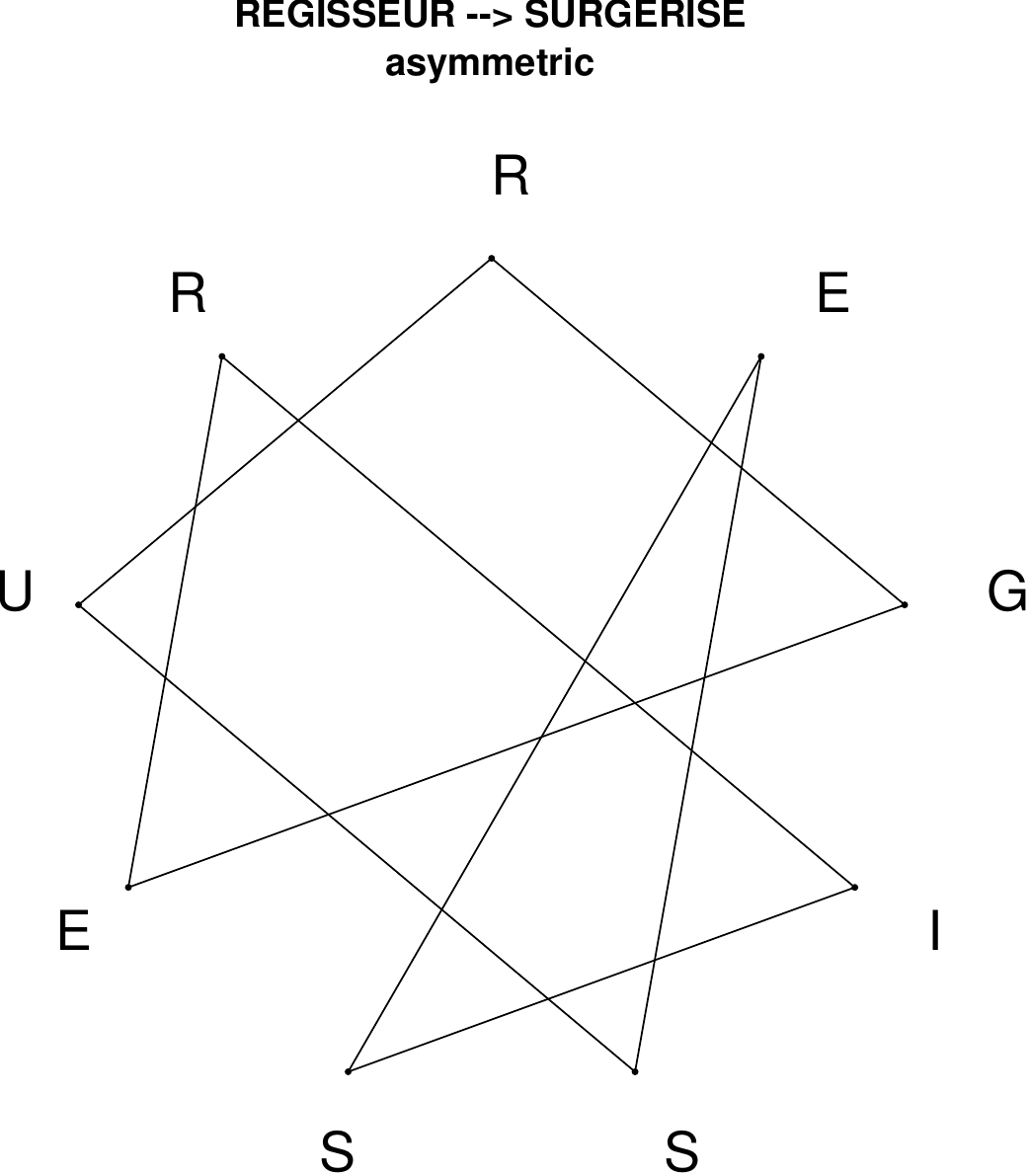}
\end{subfigure}
\hfill
\begin{subfigure}[T]{0.19\textwidth}
\centering
\includegraphics[width=\textwidth]{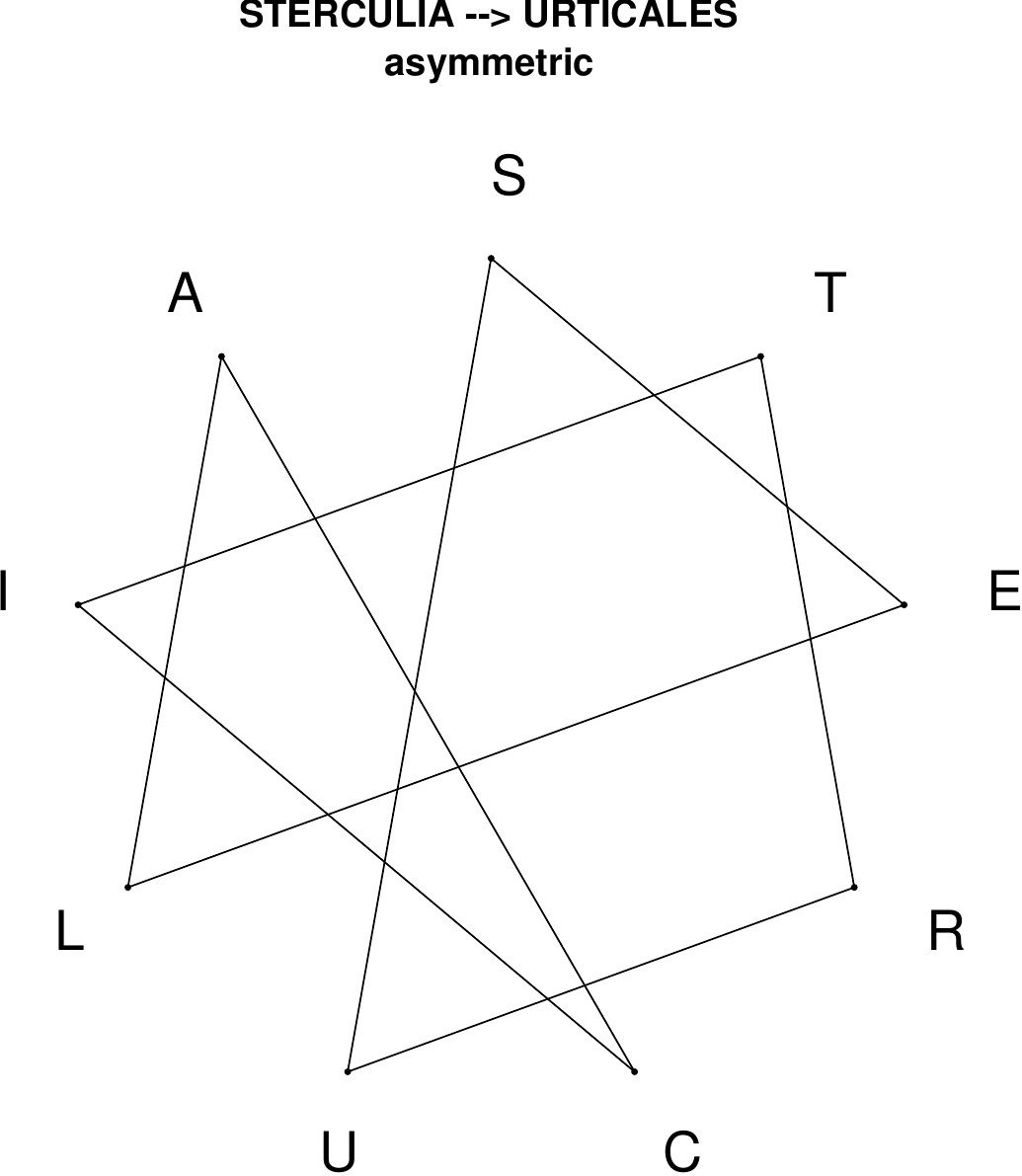}
\end{subfigure}
\end{figure}

\begin{figure}[H]
\centering
\begin{subfigure}[T]{0.19\textwidth}
\centering
\includegraphics[width=\textwidth]{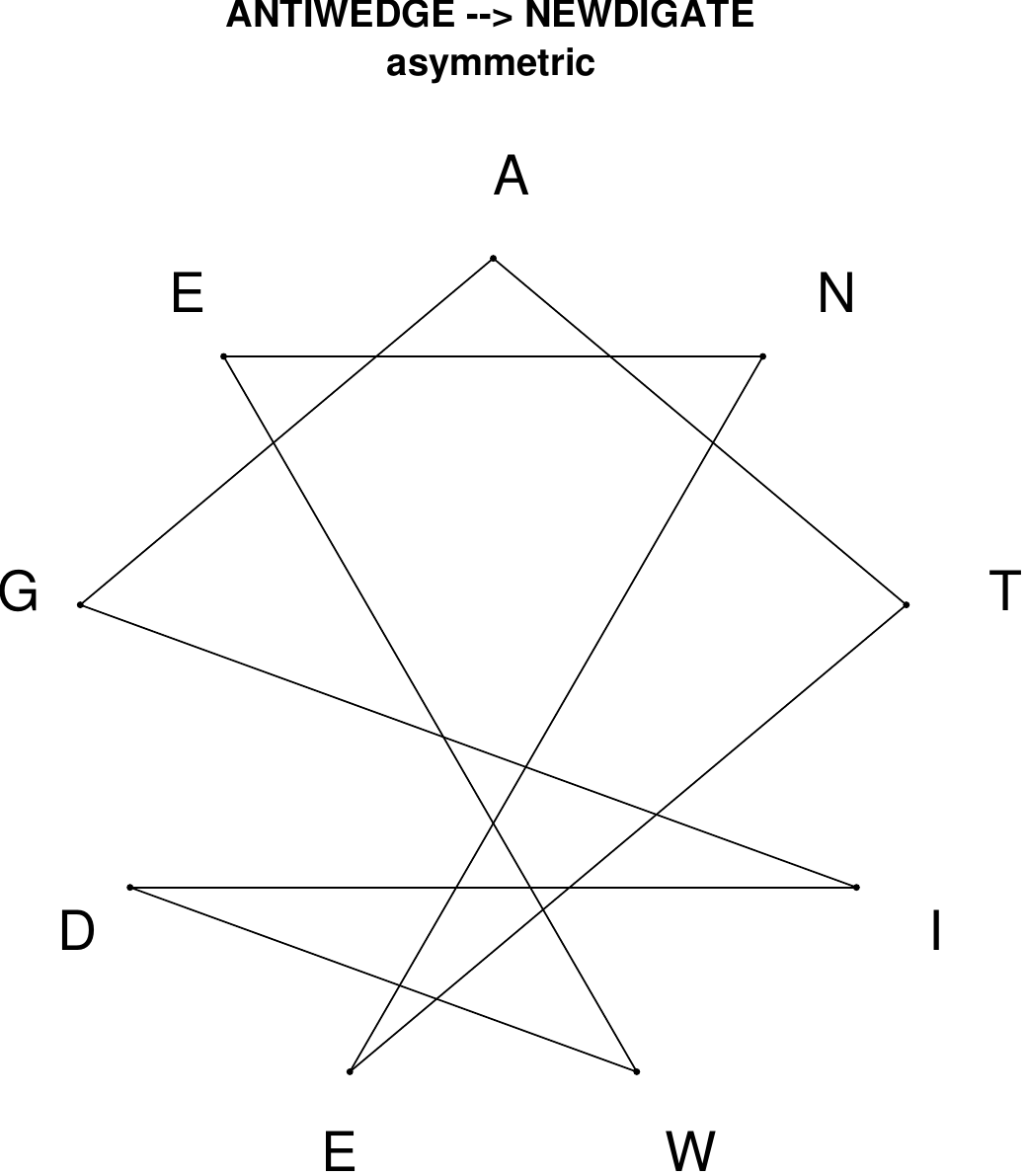}
\end{subfigure}
\hfill
\begin{subfigure}[T]{0.19\textwidth}
\centering
\includegraphics[width=\textwidth]{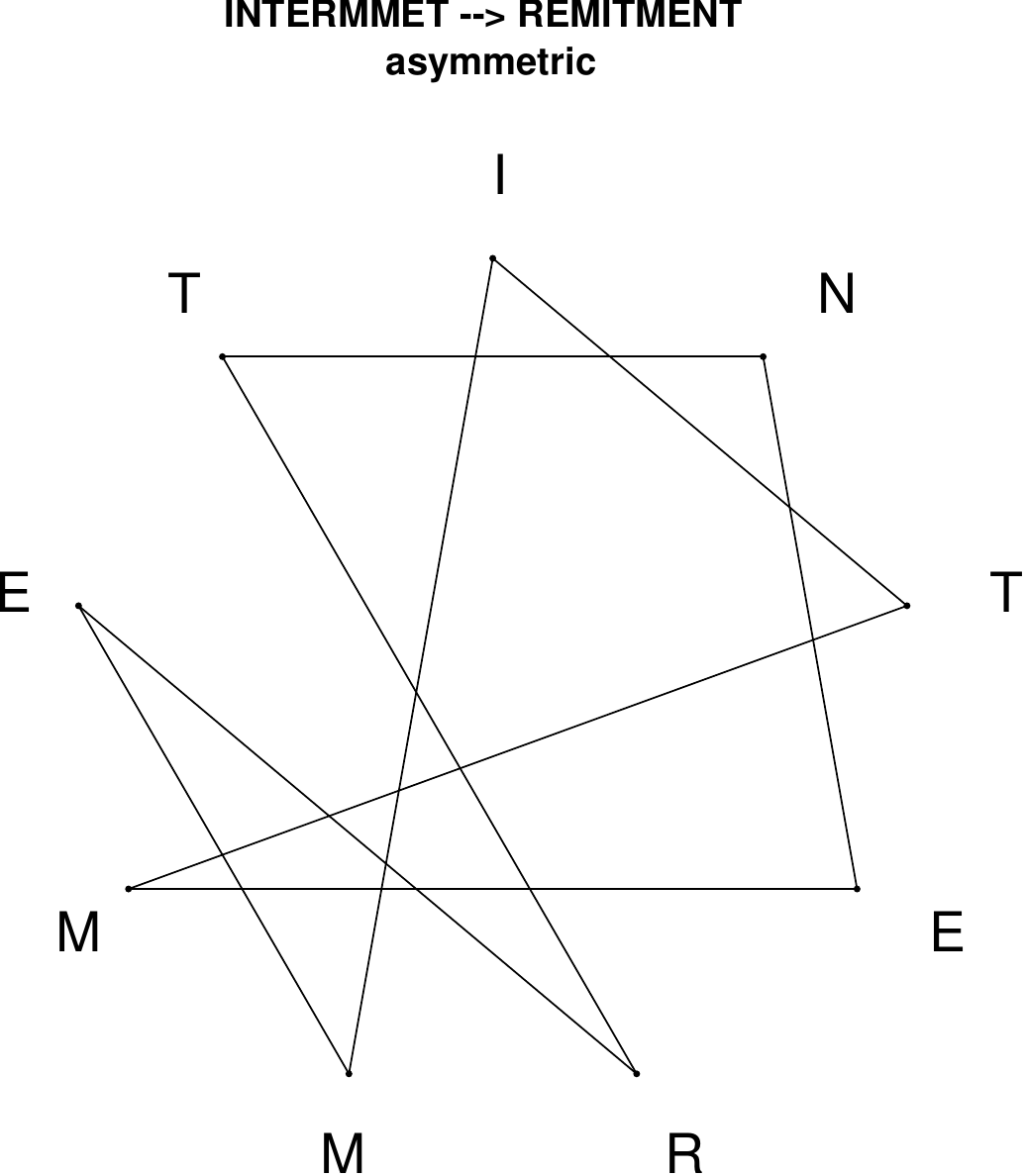}
\end{subfigure}
\hfill
\begin{subfigure}[T]{0.19\textwidth}
\centering
\includegraphics[width=\textwidth]{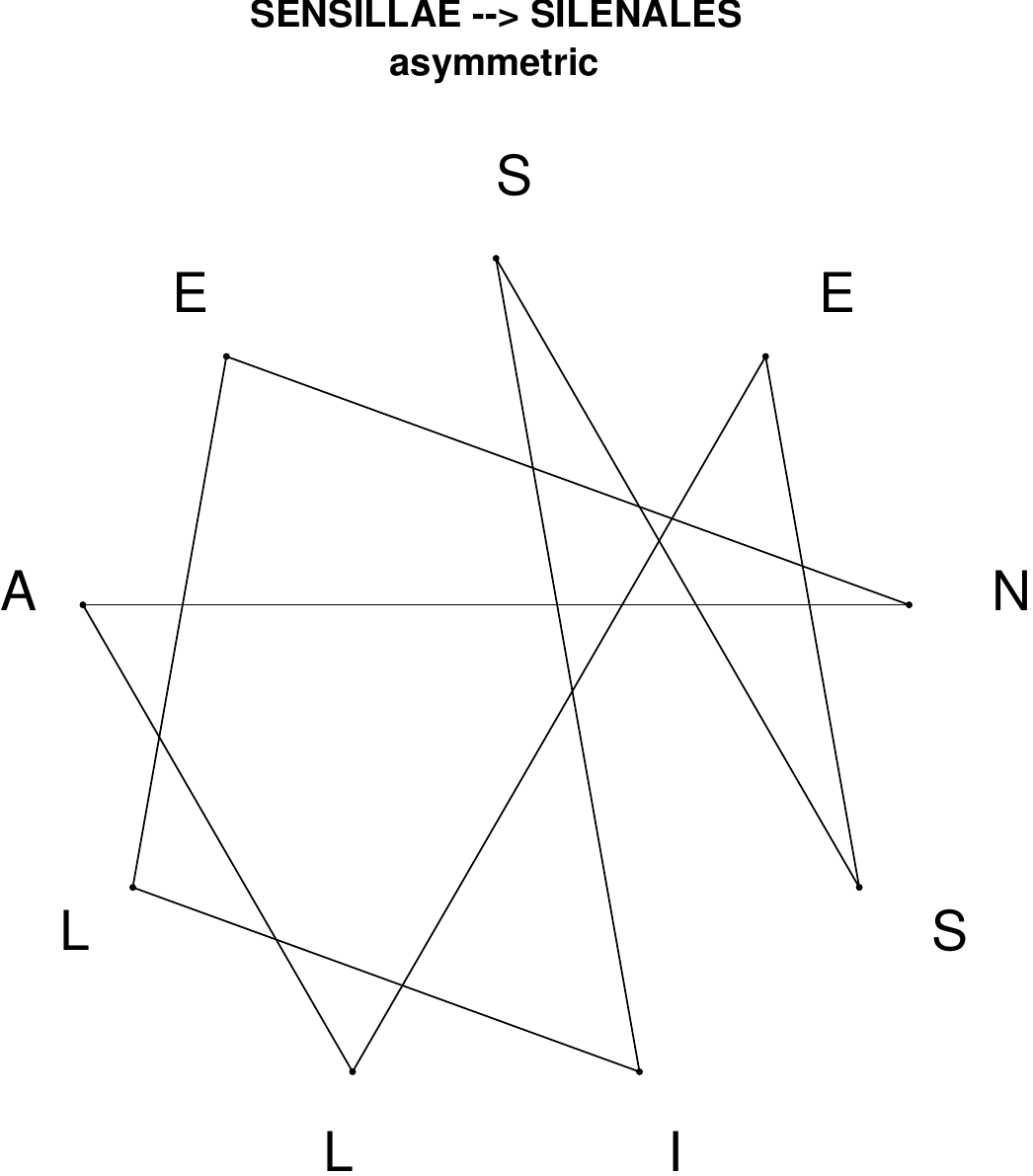}
\end{subfigure}
\hfill
\begin{subfigure}[T]{0.19\textwidth}
\centering
\includegraphics[width=\textwidth]{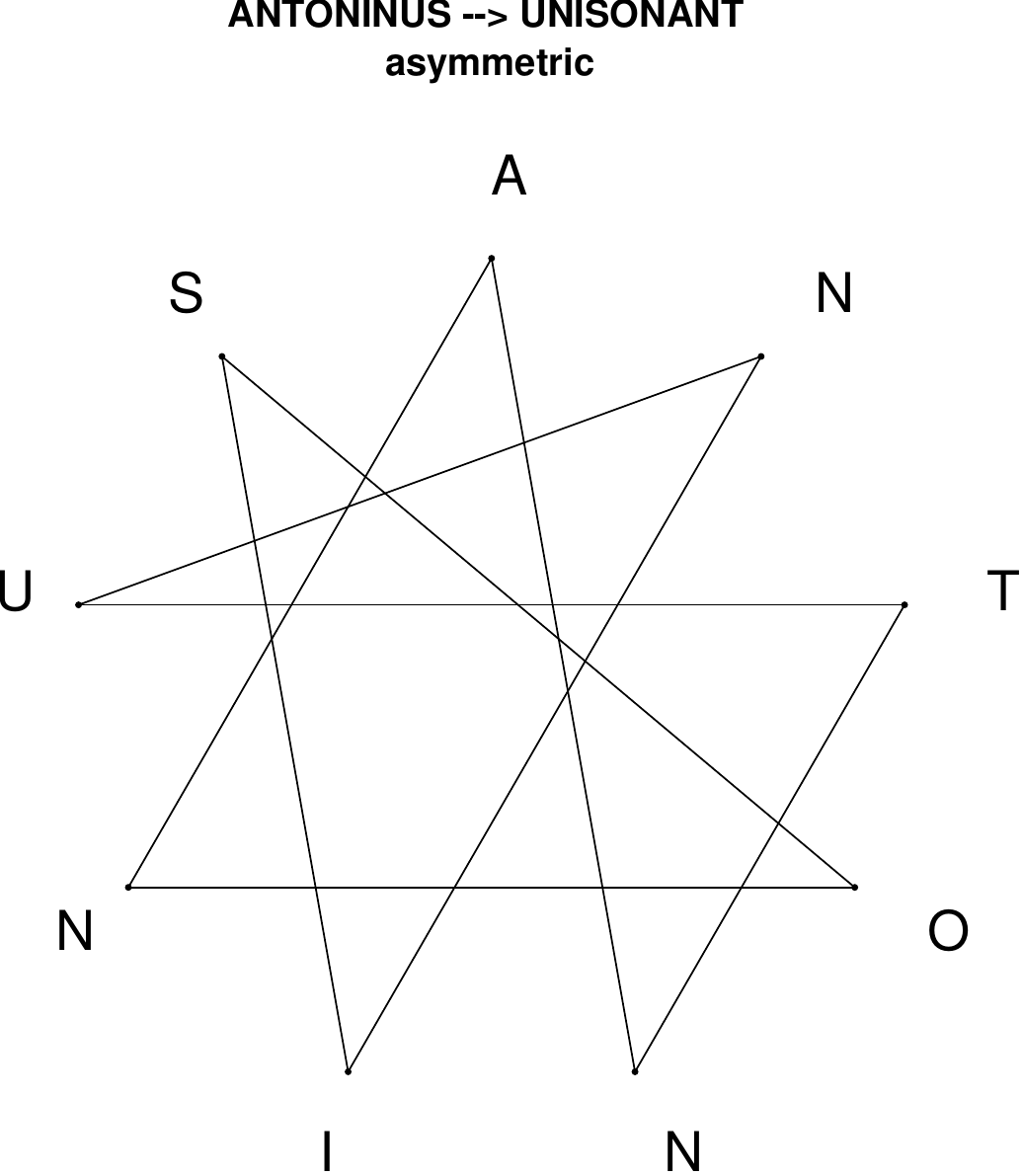}
\end{subfigure}
\hfill
\begin{subfigure}[T]{0.19\textwidth}
\centering
\includegraphics[width=\textwidth]{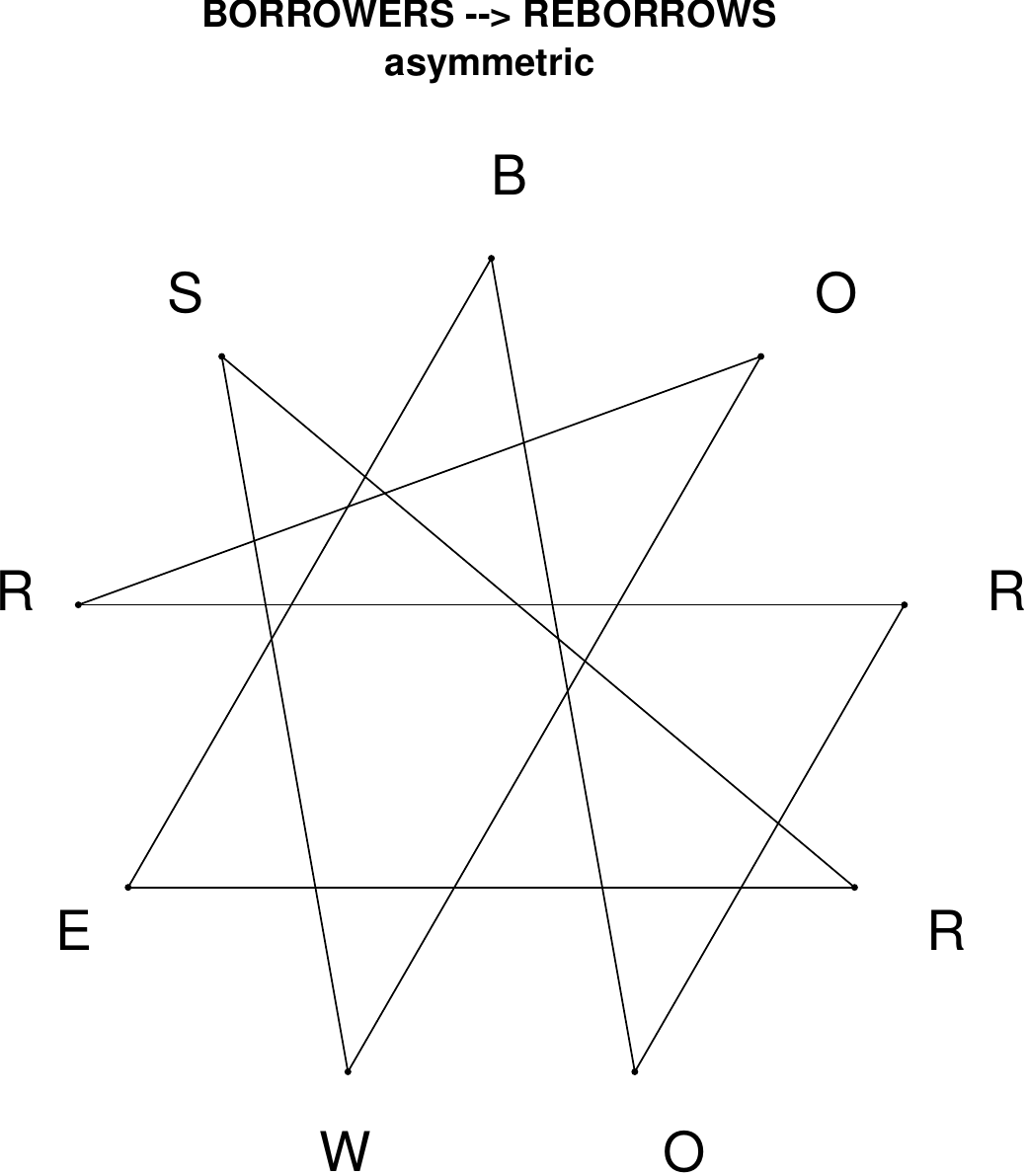}
\end{subfigure}
\end{figure}

\begin{figure}[H]
\centering
\begin{subfigure}[T]{0.19\textwidth}
\centering
\includegraphics[width=\textwidth]{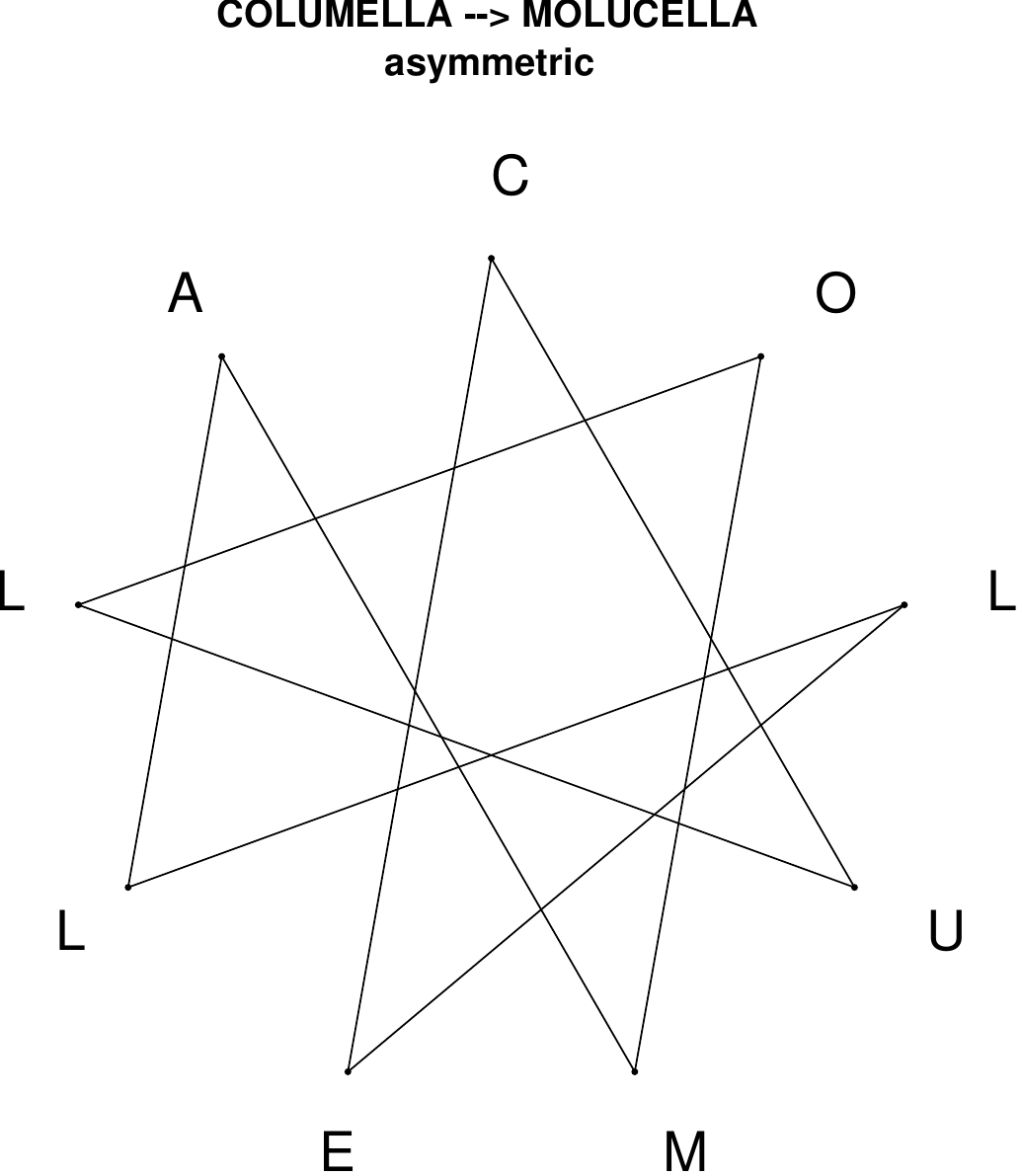}
\end{subfigure}
\hfill
\begin{subfigure}[T]{0.19\textwidth}
\centering
\includegraphics[width=\textwidth]{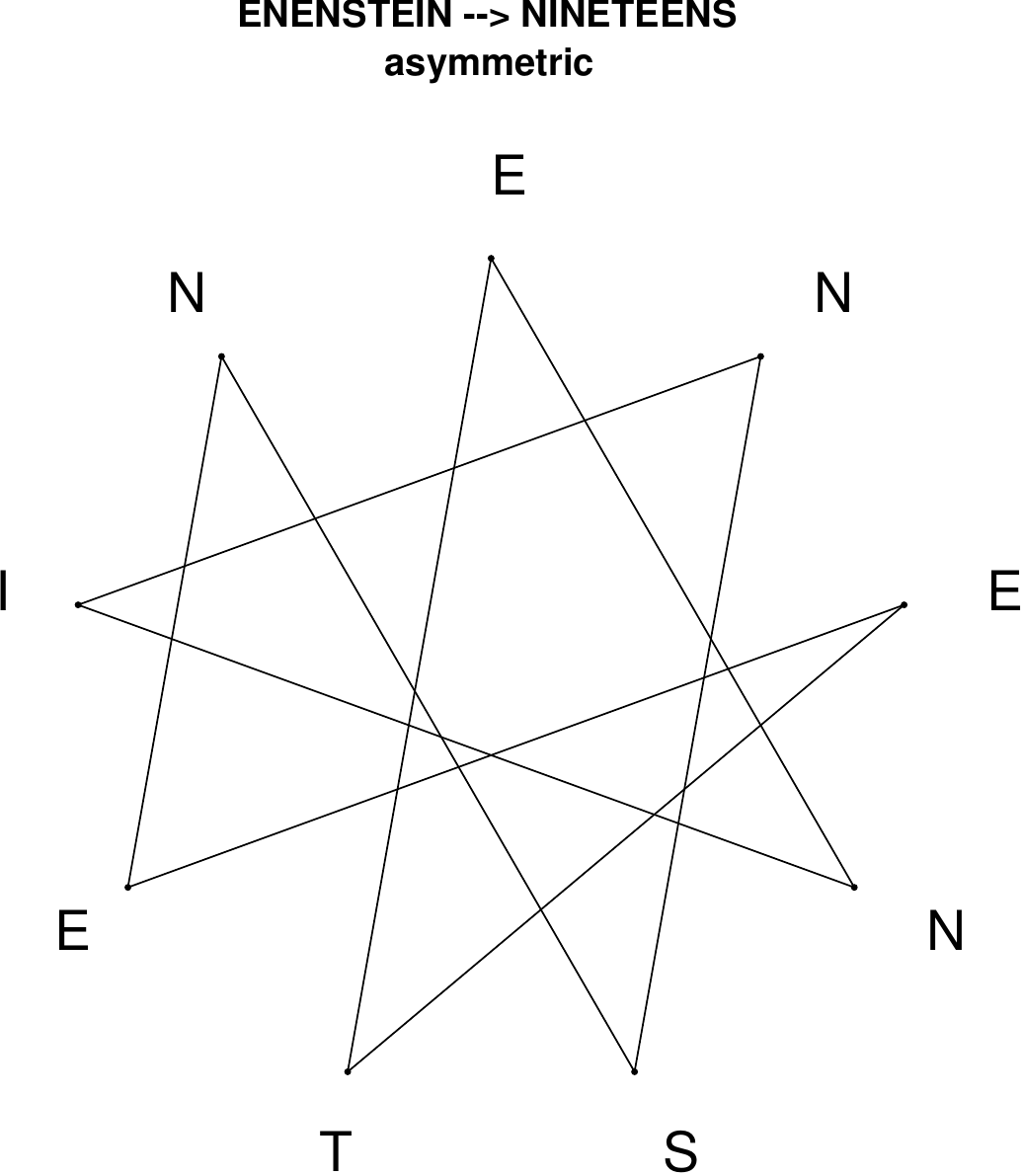}
\end{subfigure}
\hfill
\begin{subfigure}[T]{0.19\textwidth}
\centering
\includegraphics[width=\textwidth]{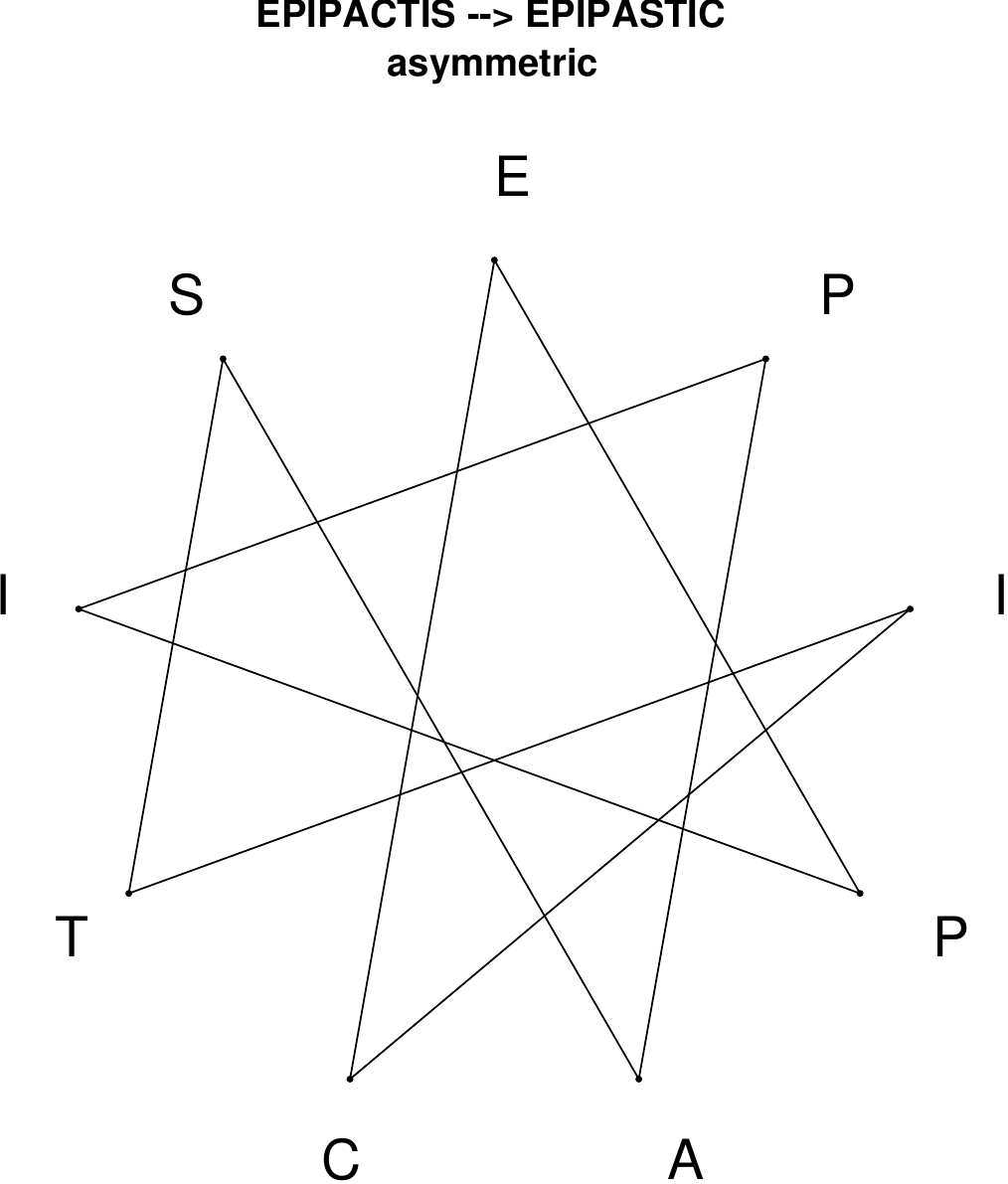}
\end{subfigure}
\hfill
\begin{subfigure}[T]{0.19\textwidth}
\centering
\includegraphics[width=\textwidth]{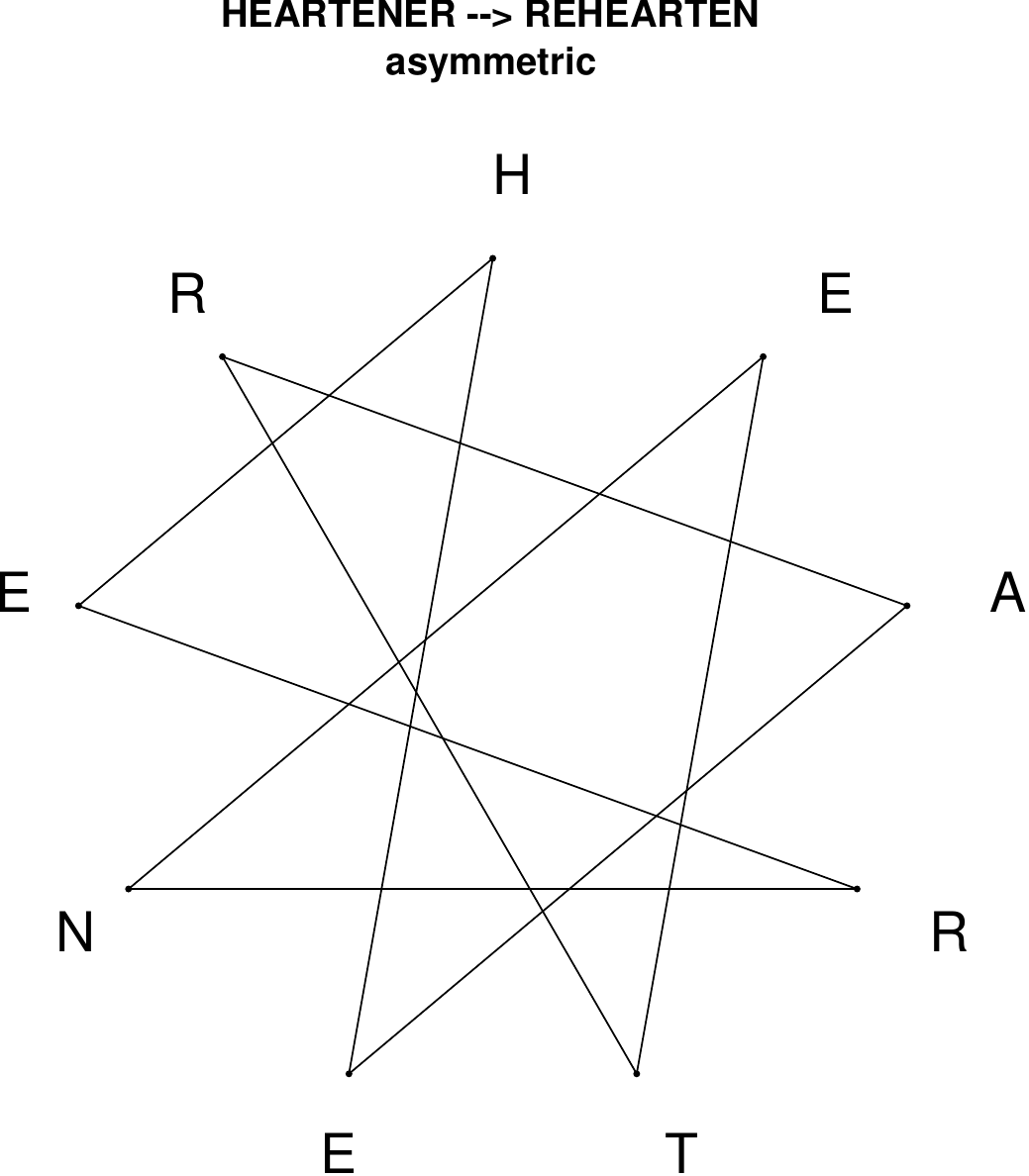}
\end{subfigure}
\hfill
\begin{subfigure}[T]{0.19\textwidth}
\centering
\includegraphics[width=\textwidth]{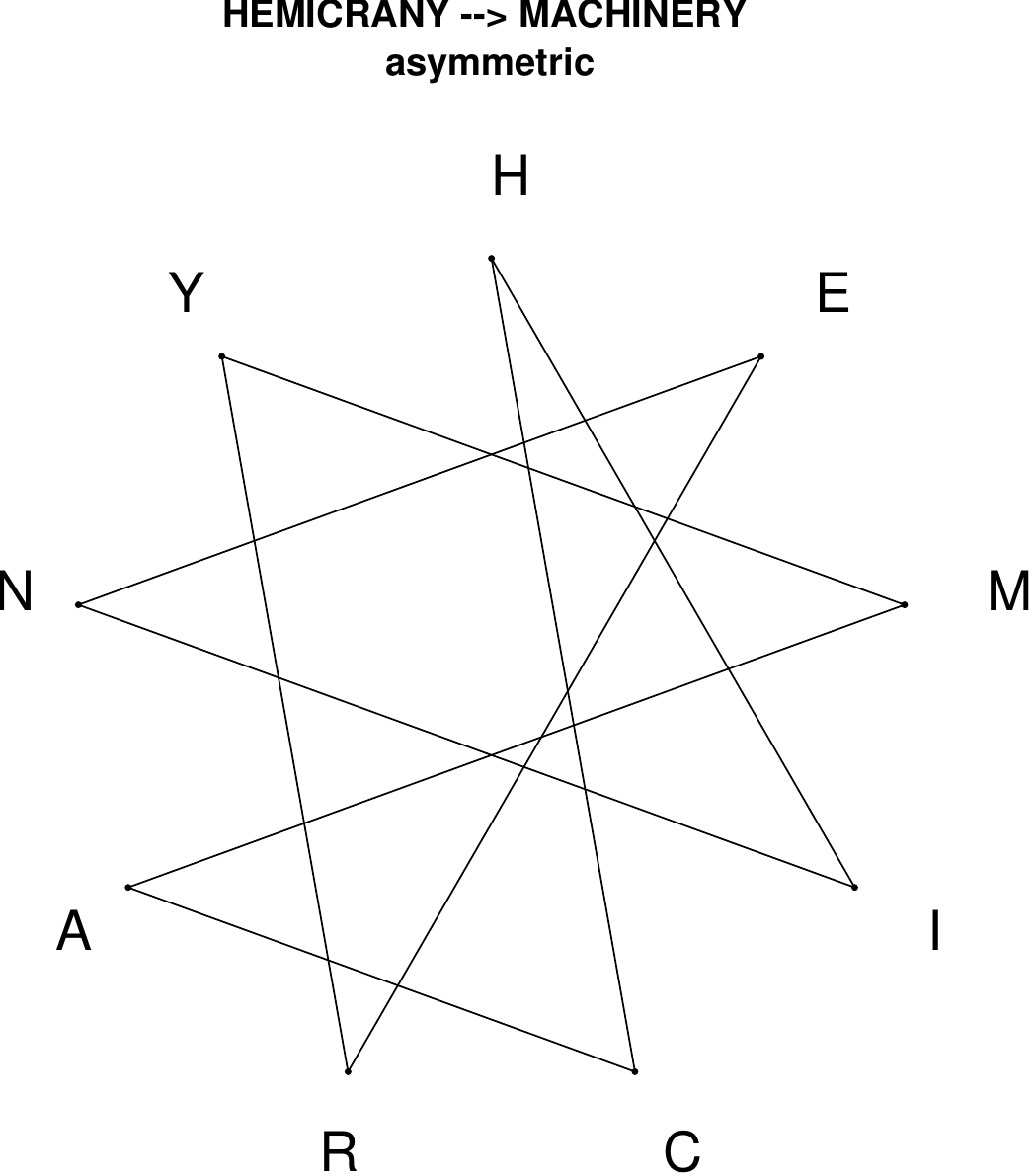}
\end{subfigure}
\end{figure}

\begin{figure}[H]
\centering
\begin{subfigure}[T]{0.19\textwidth}
\centering
\includegraphics[width=\textwidth]{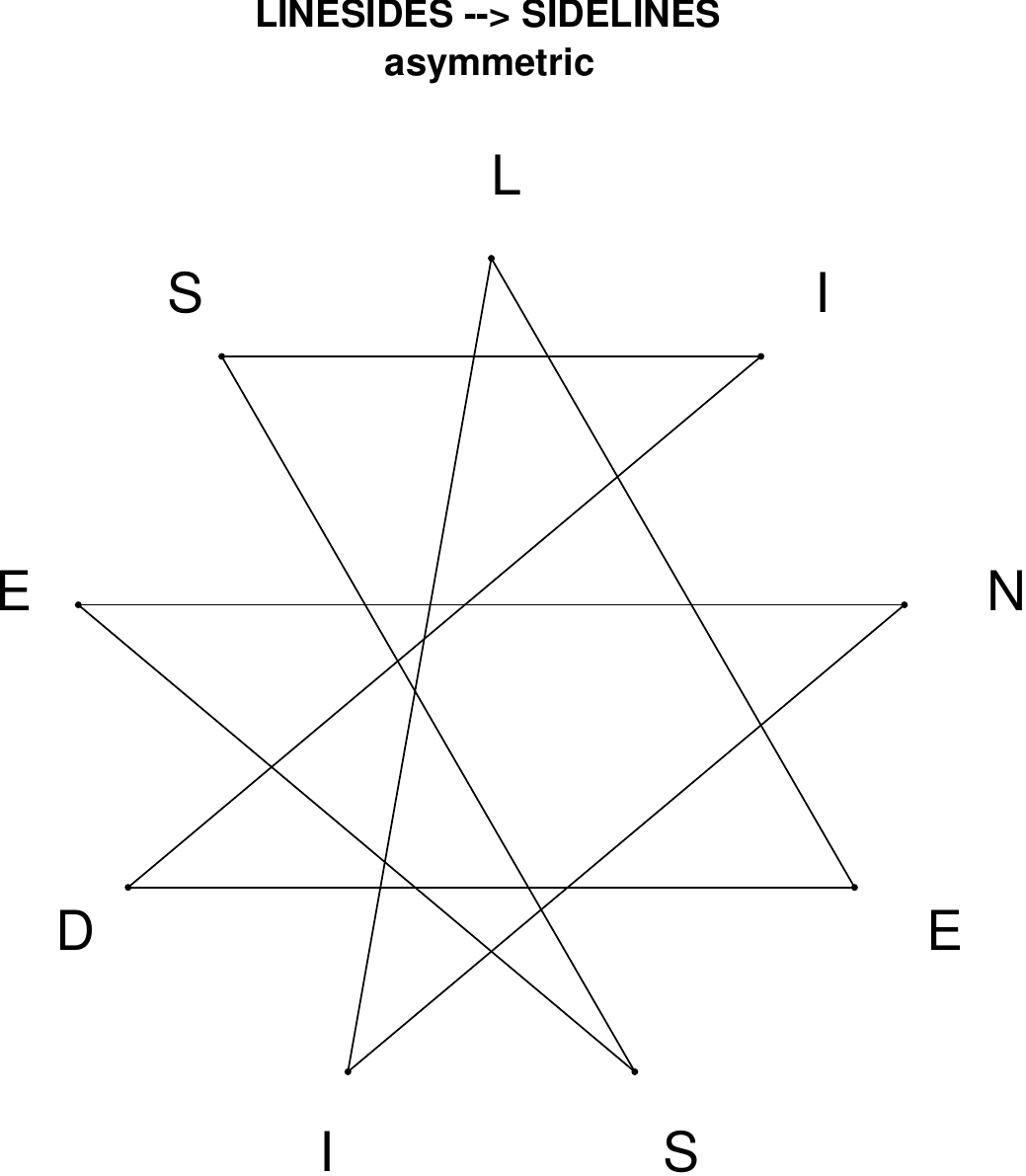}
\end{subfigure}
\hfill
\begin{subfigure}[T]{0.19\textwidth}
\centering
\includegraphics[width=\textwidth]{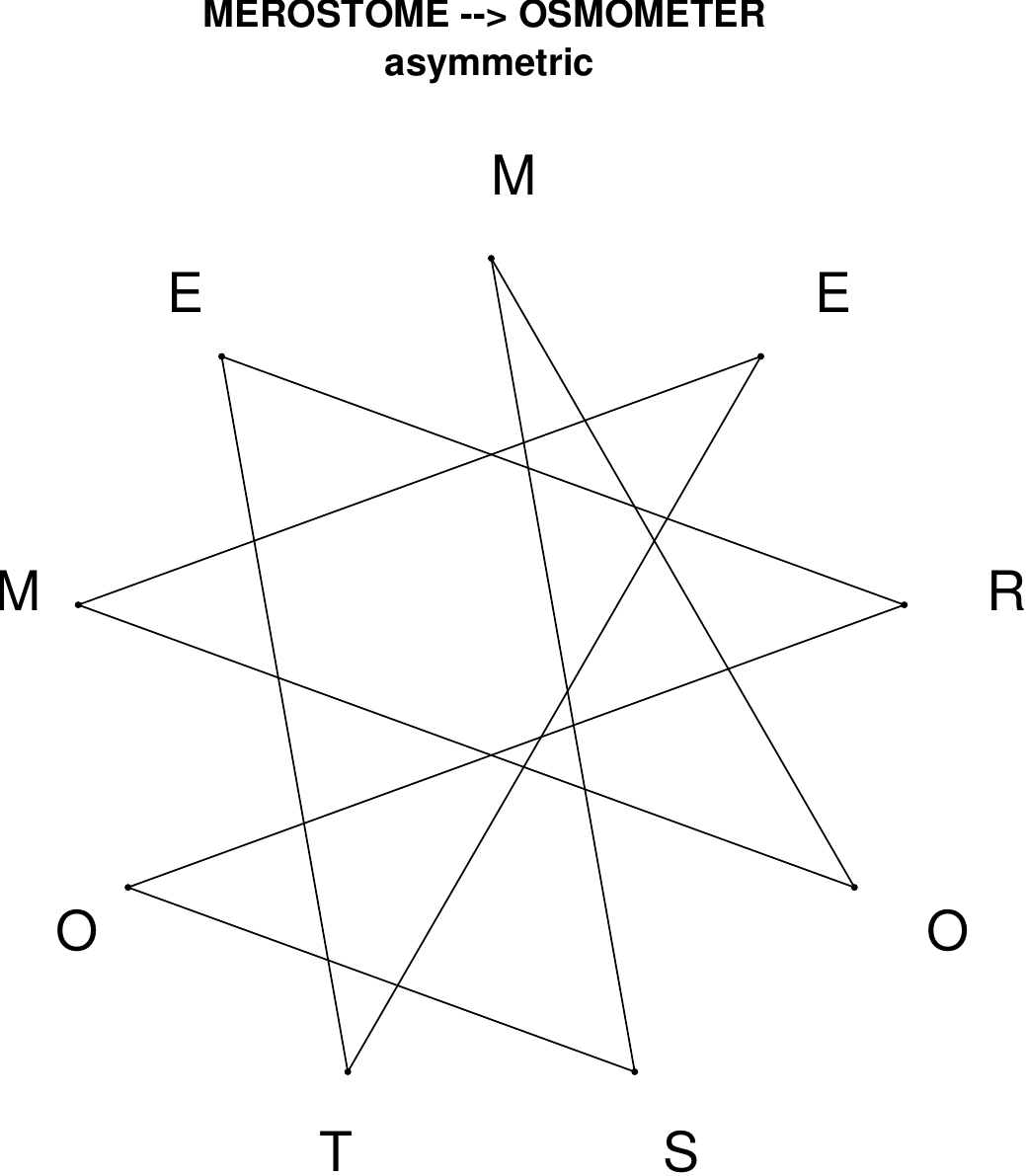}
\end{subfigure}
\hfill
\begin{subfigure}[T]{0.19\textwidth}
\centering
\includegraphics[width=\textwidth]{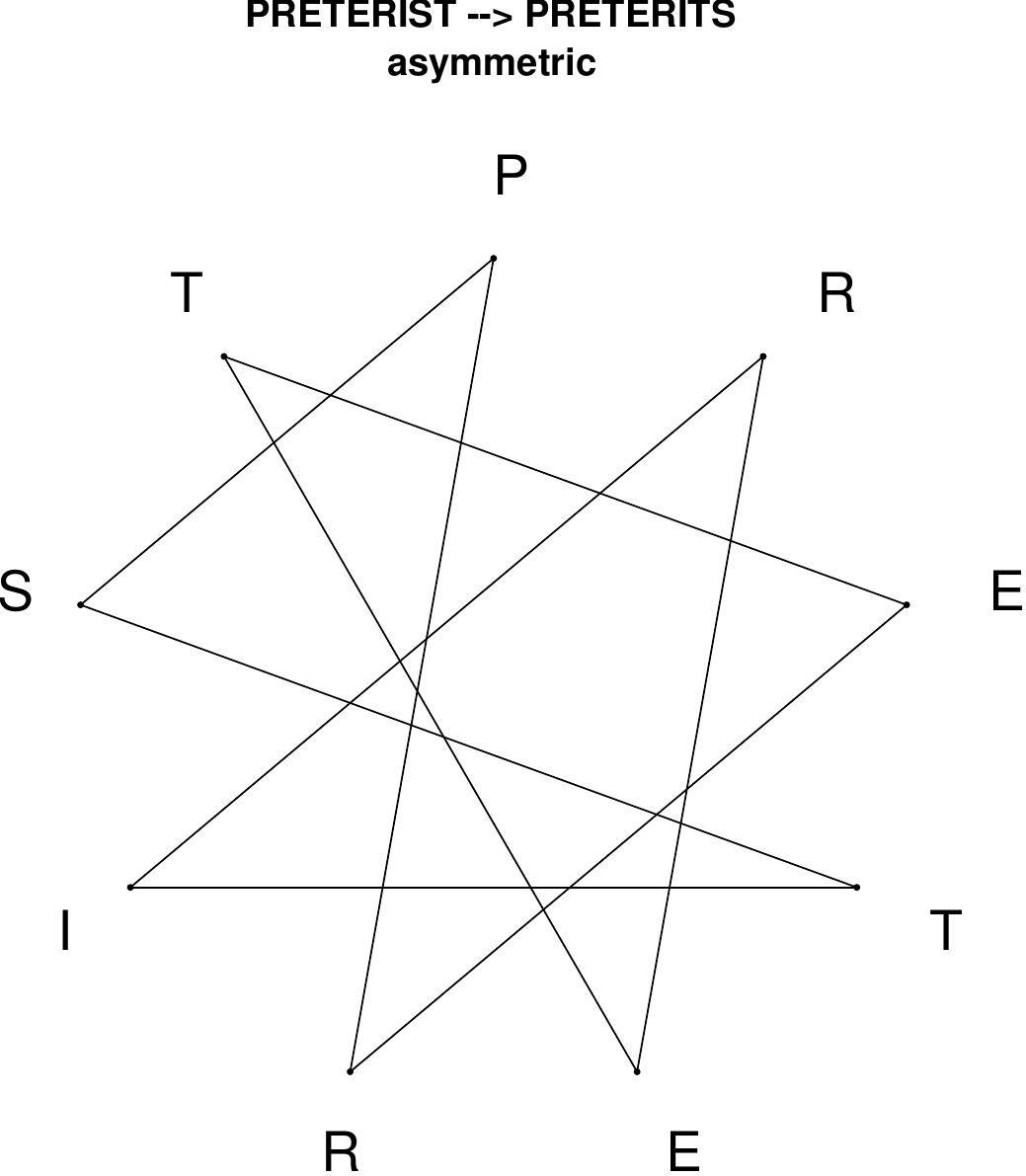}
\end{subfigure}
\hfill
\begin{subfigure}[T]{0.19\textwidth}
\centering
\includegraphics[width=\textwidth]{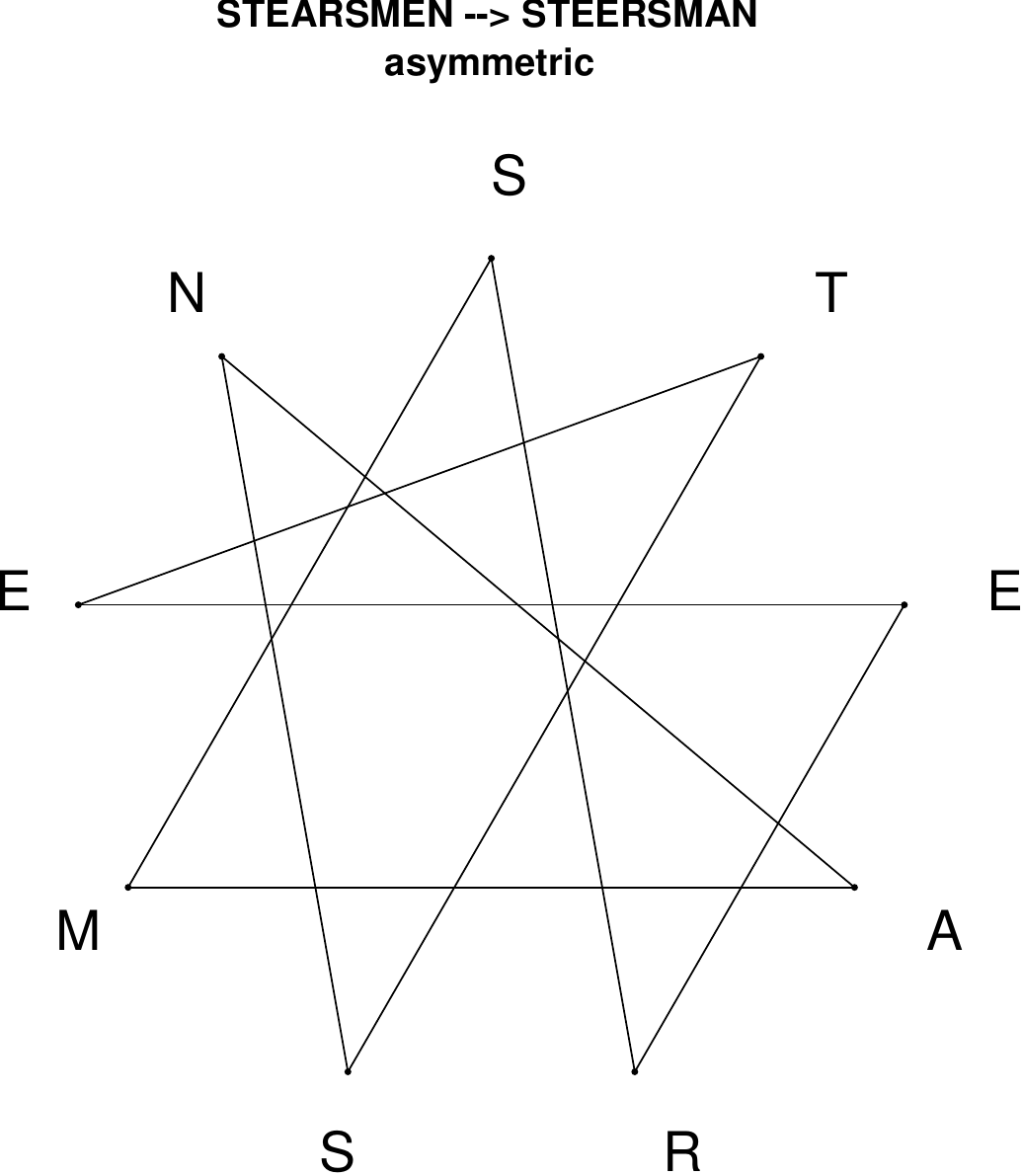}
\end{subfigure}
\hfill
\begin{subfigure}[T]{0.19\textwidth}
\centering
\includegraphics[width=\textwidth]{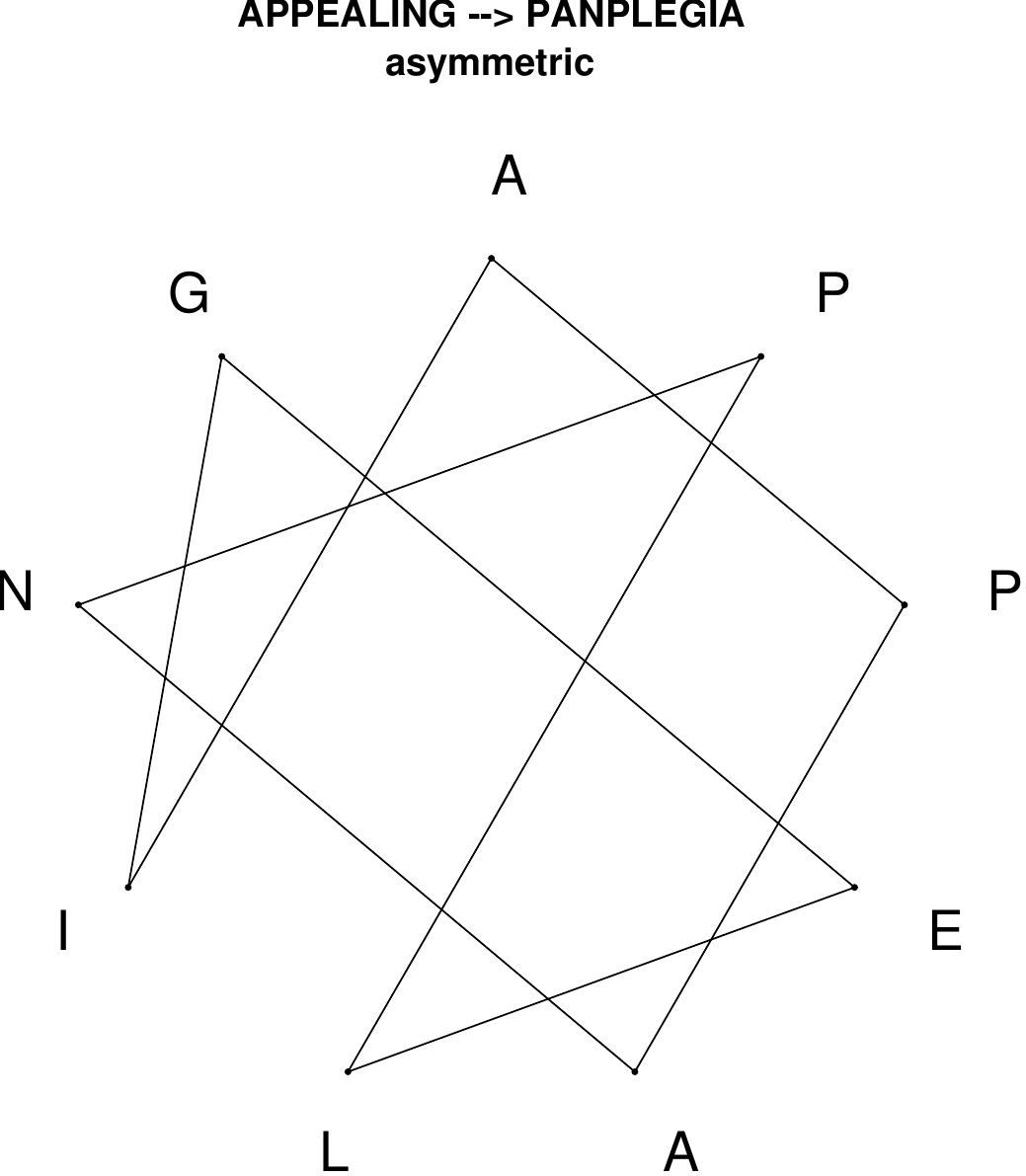}
\end{subfigure}
\end{figure}

\begin{figure}[H]
\centering
\begin{subfigure}[T]{0.19\textwidth}
\centering
\includegraphics[width=\textwidth]{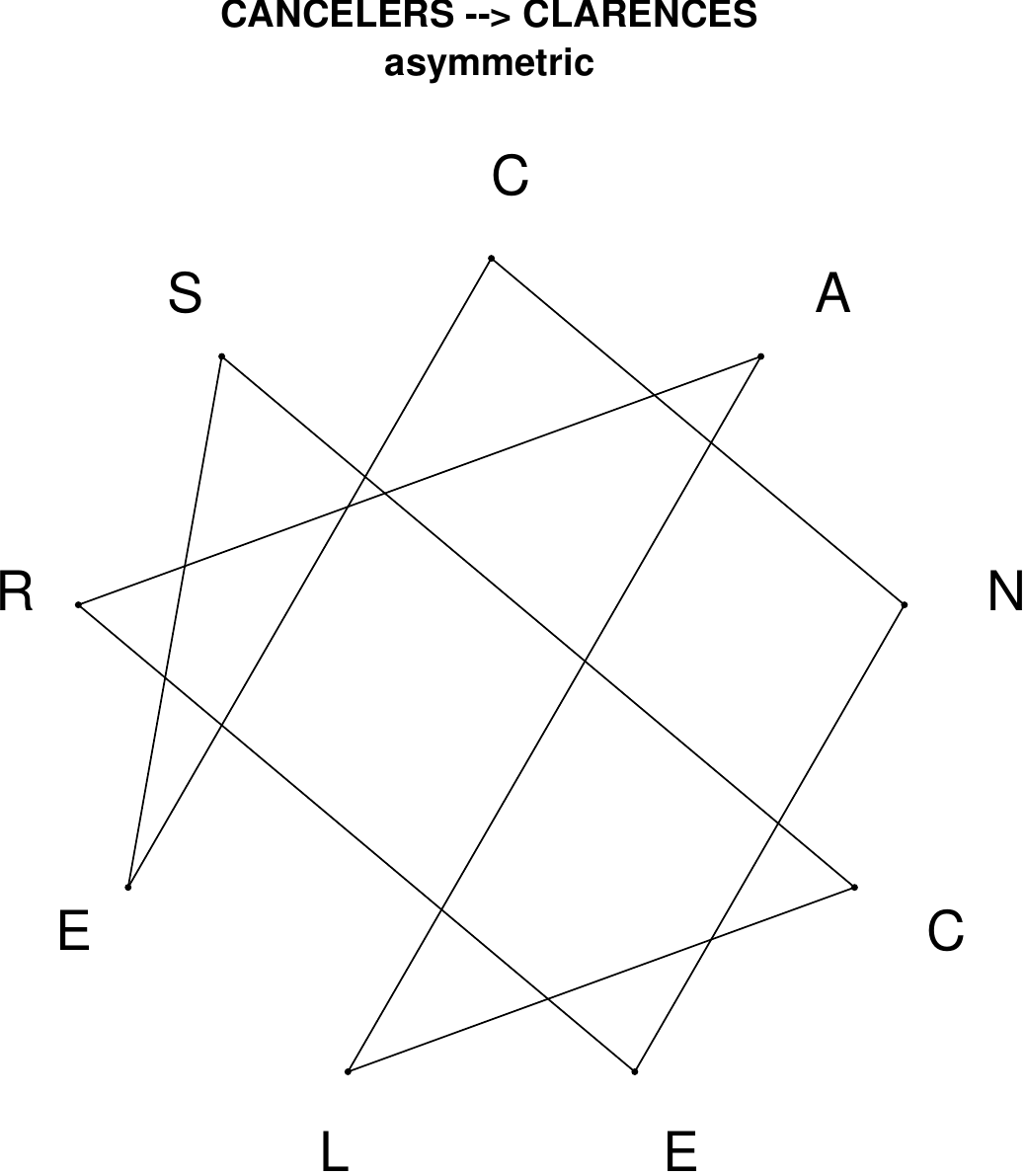}
\end{subfigure}
\hfill
\begin{subfigure}[T]{0.19\textwidth}
\centering
\includegraphics[width=\textwidth]{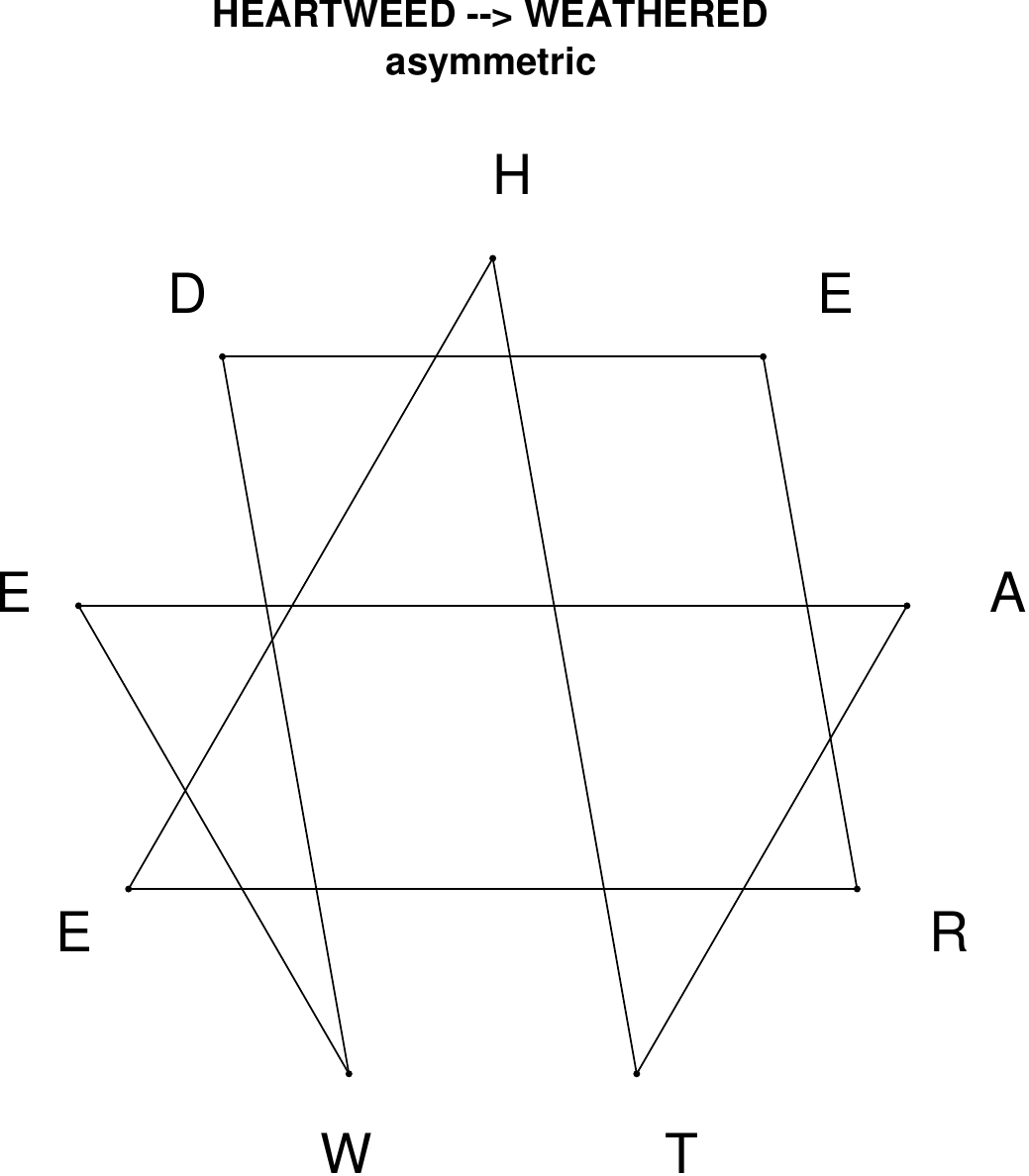}
\end{subfigure}
\hfill
\begin{subfigure}[T]{0.19\textwidth}
\centering
\includegraphics[width=\textwidth]{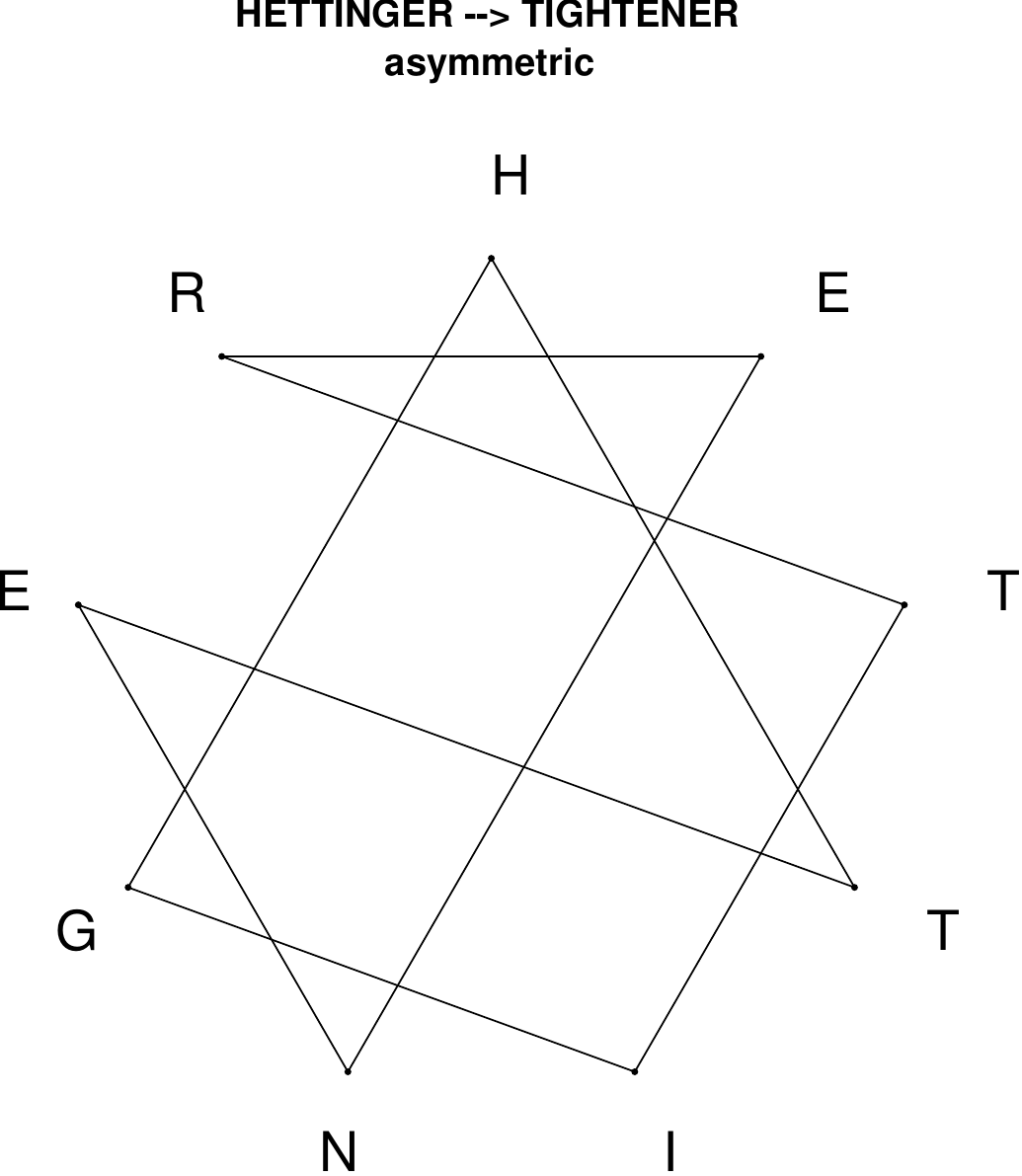}
\end{subfigure}
\hfill
\begin{subfigure}[T]{0.19\textwidth}
\centering
\includegraphics[width=\textwidth]{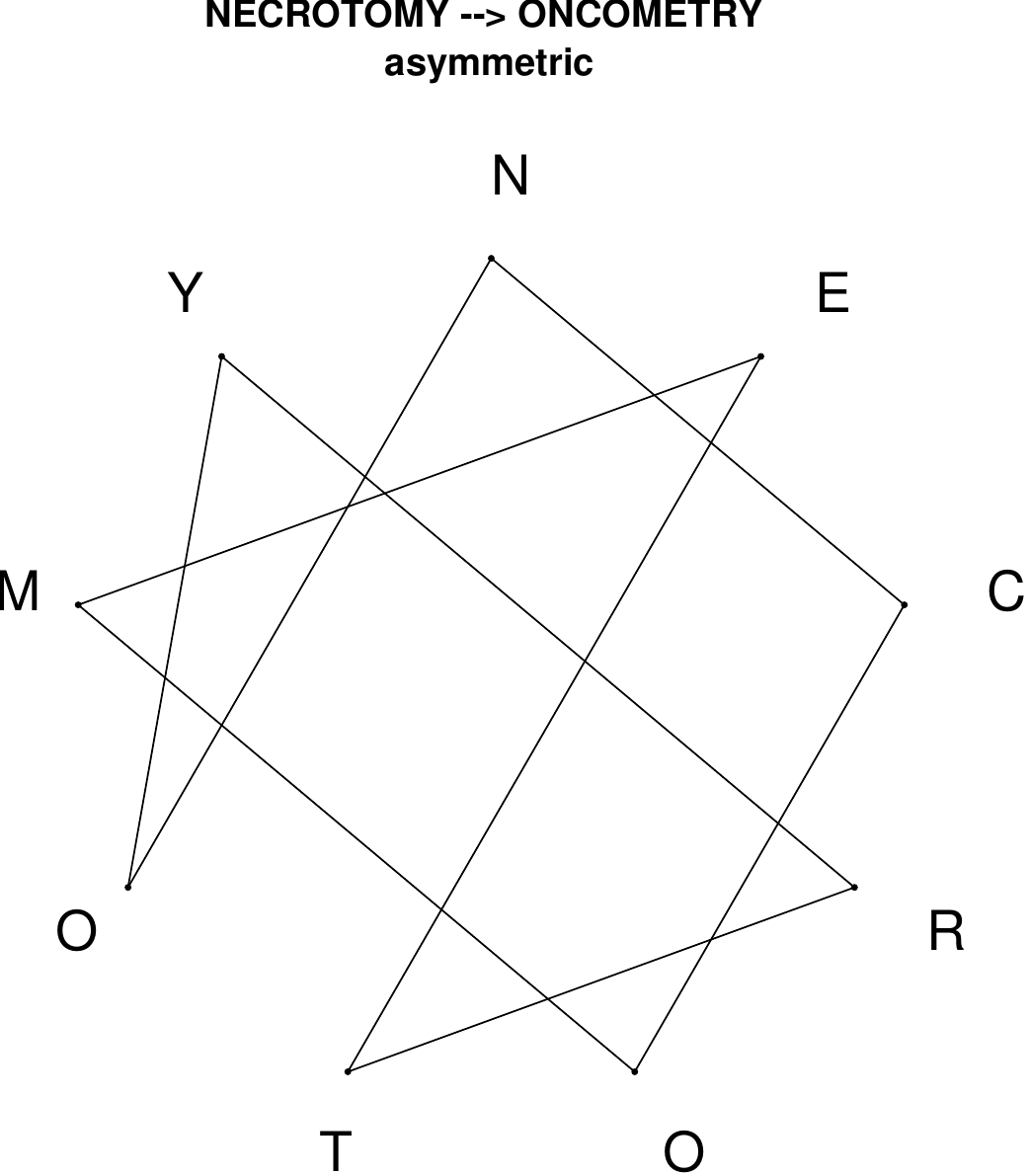}
\end{subfigure}
\hfill
\begin{subfigure}[T]{0.19\textwidth}
\centering
\includegraphics[width=\textwidth]{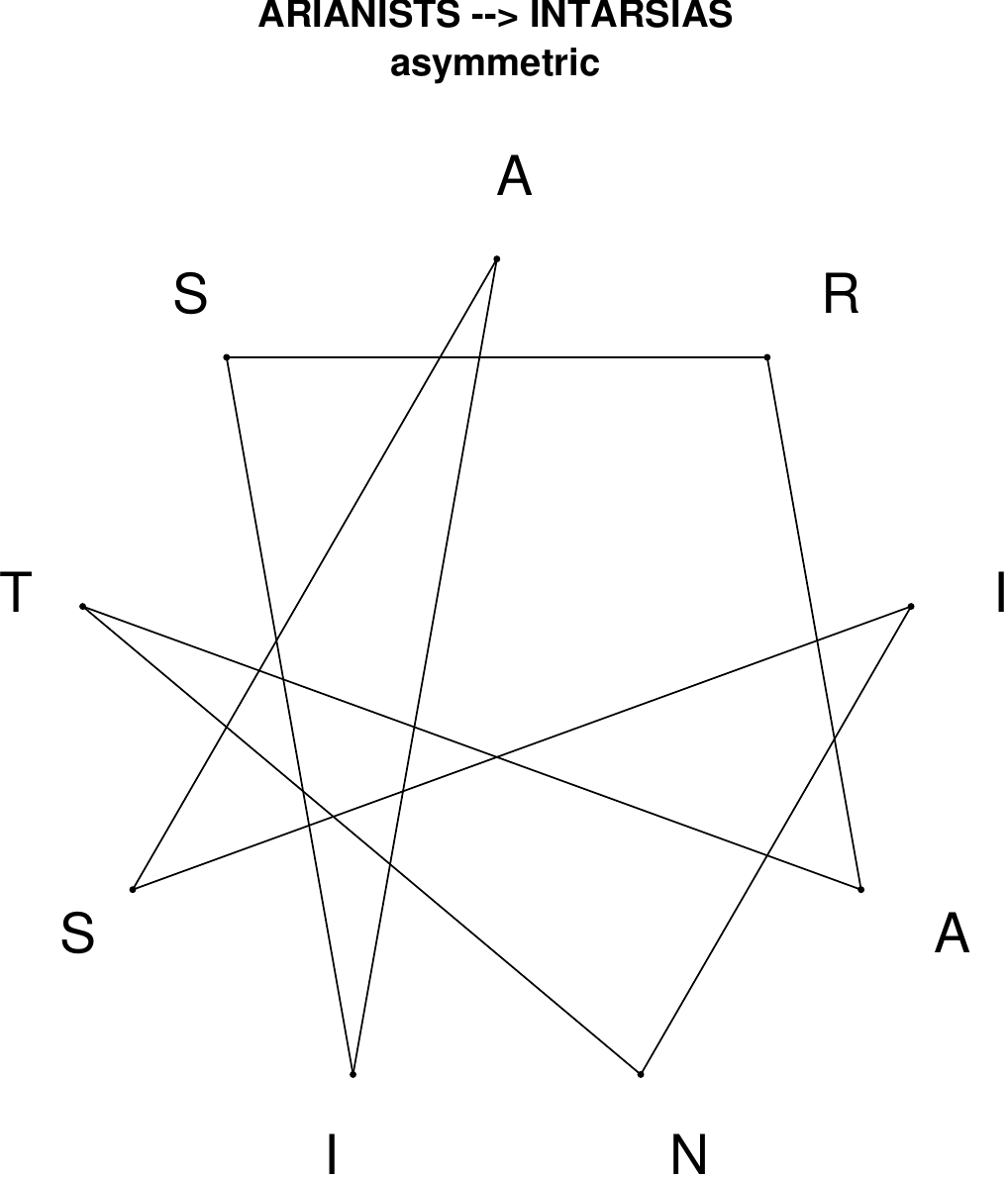}
\end{subfigure}
\end{figure}

\begin{figure}[H]
\centering
\begin{subfigure}[T]{0.19\textwidth}
\centering
\includegraphics[width=\textwidth]{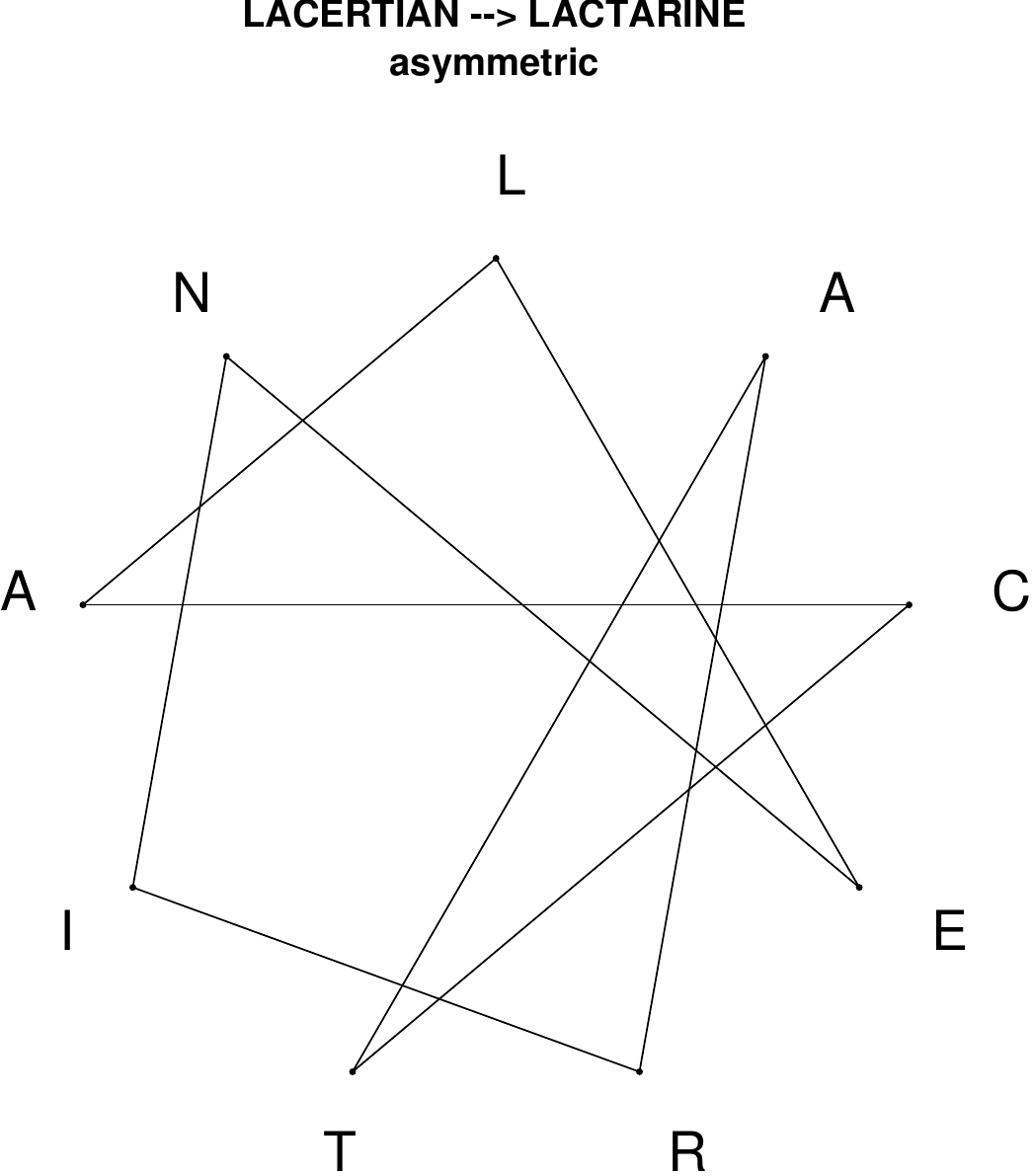}
\end{subfigure}
\hfill
\begin{subfigure}[T]{0.19\textwidth}
\centering
\includegraphics[width=\textwidth]{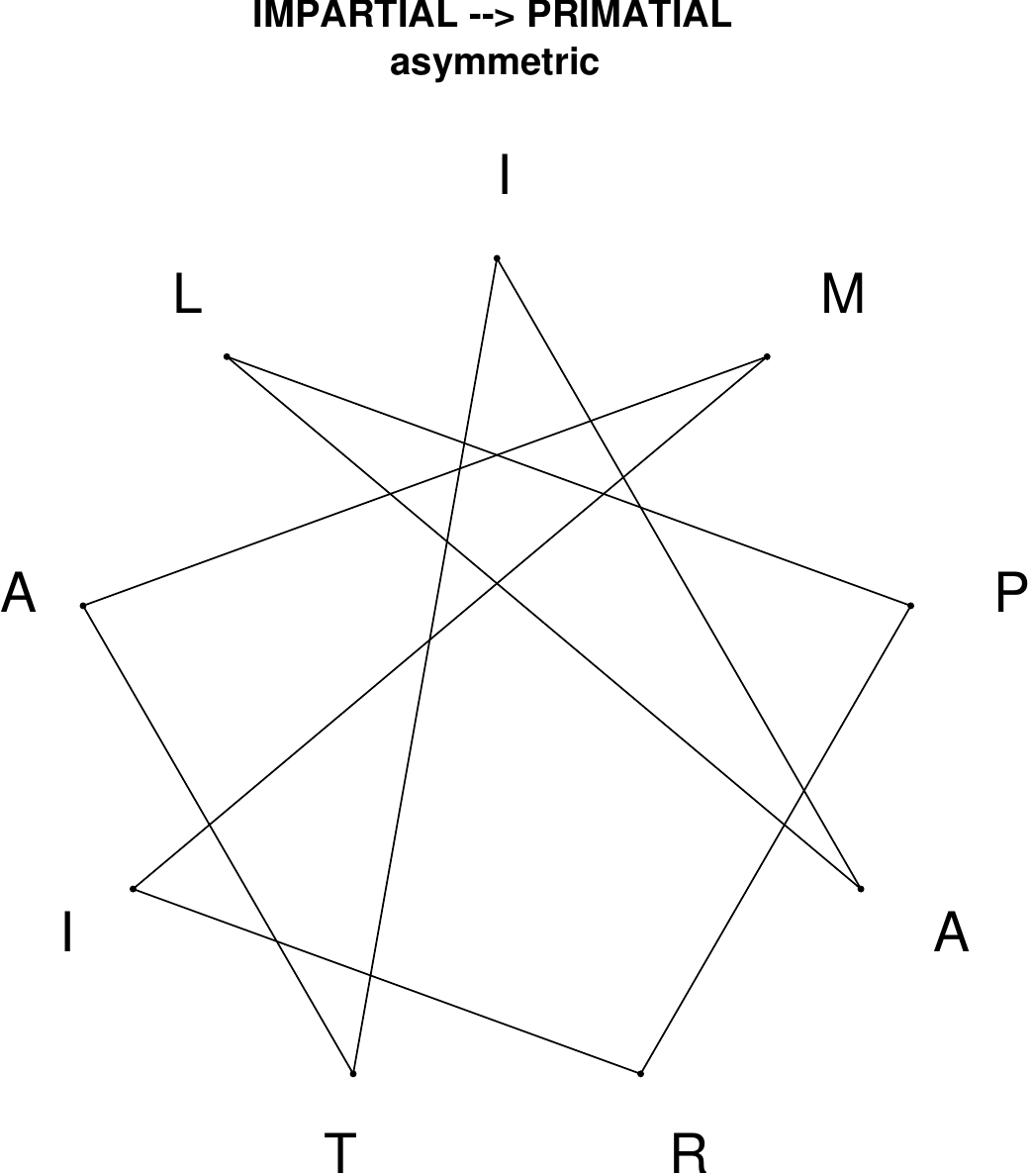}
\end{subfigure}
\hfill
\begin{subfigure}[T]{0.19\textwidth}
\centering
\includegraphics[width=\textwidth]{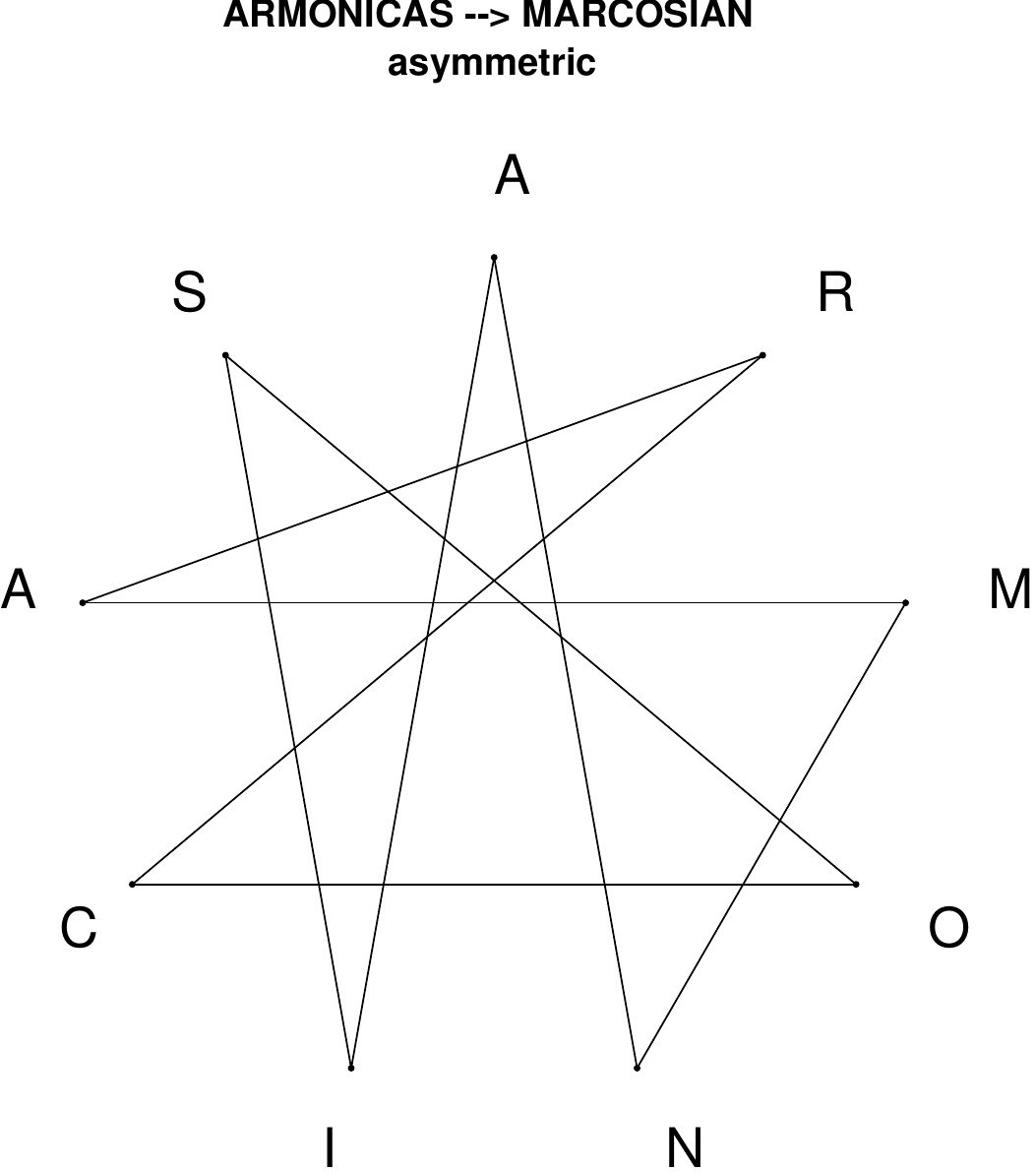}
\end{subfigure}
\hfill
\begin{subfigure}[T]{0.19\textwidth}
\centering
\includegraphics[width=\textwidth]{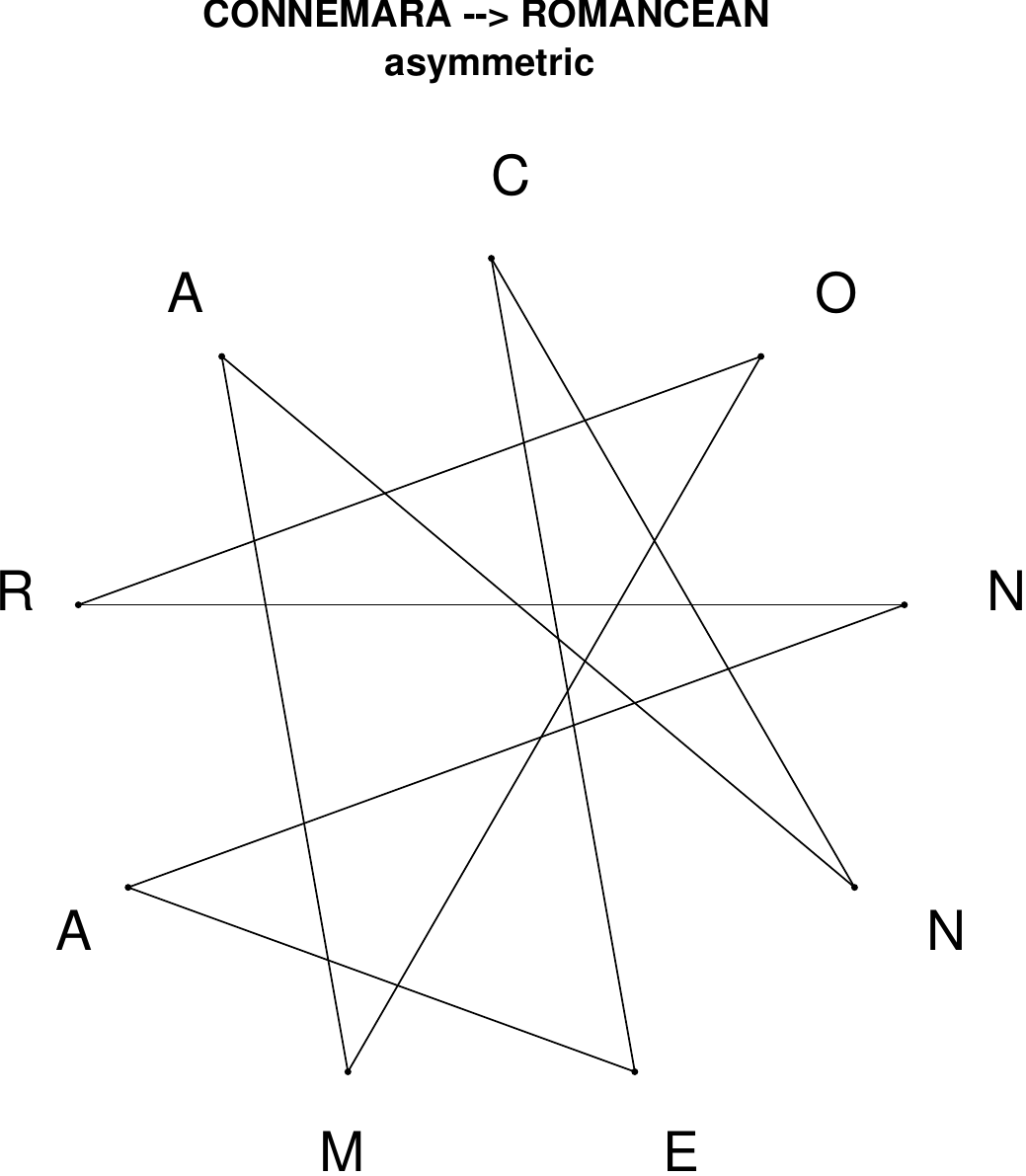}
\end{subfigure}
\hfill
\begin{subfigure}[T]{0.19\textwidth}
\centering
\includegraphics[width=\textwidth]{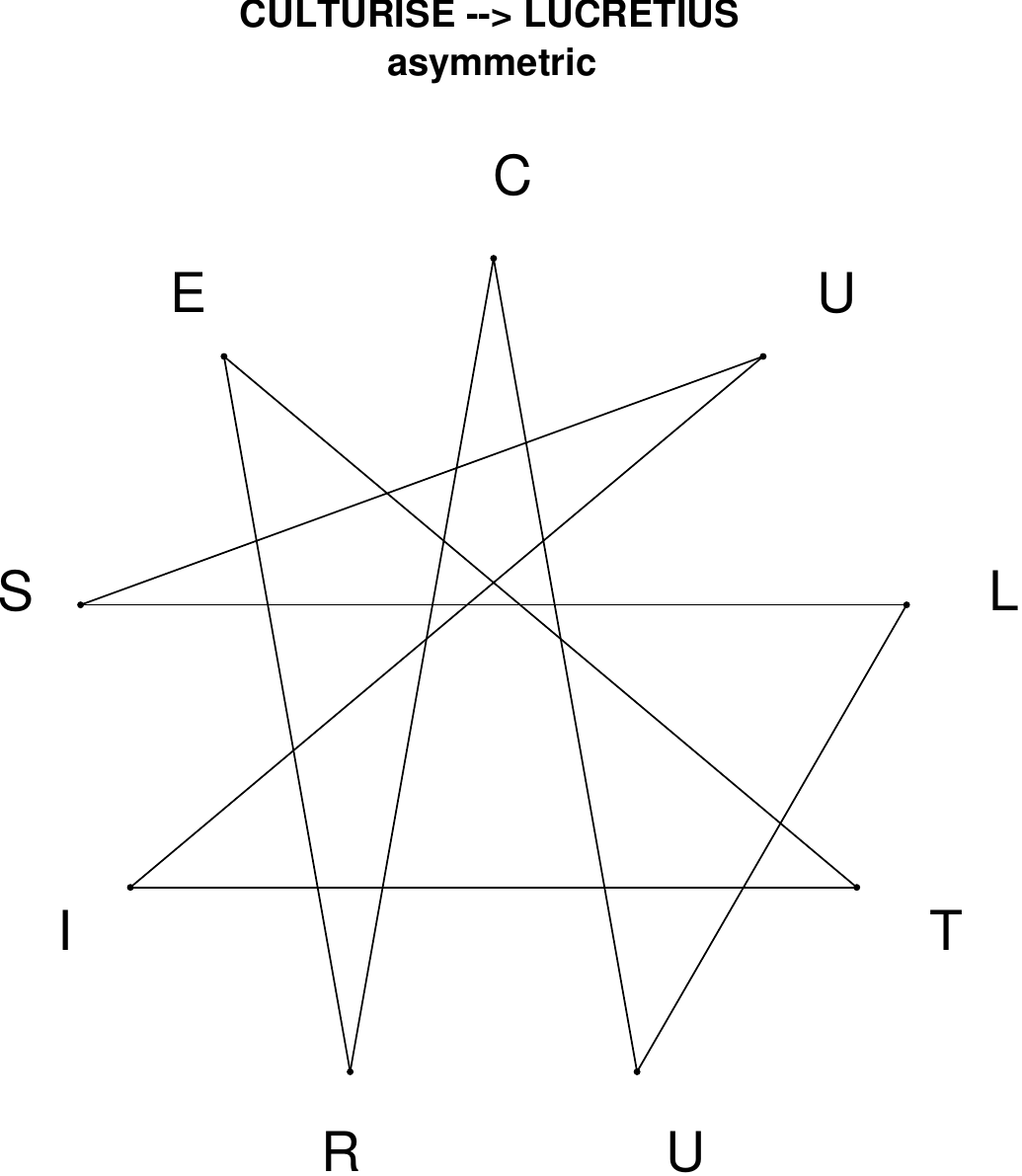}
\end{subfigure}
\end{figure}

\begin{figure}[H]
\centering
\begin{subfigure}[T]{0.19\textwidth}
\centering
\includegraphics[width=\textwidth]{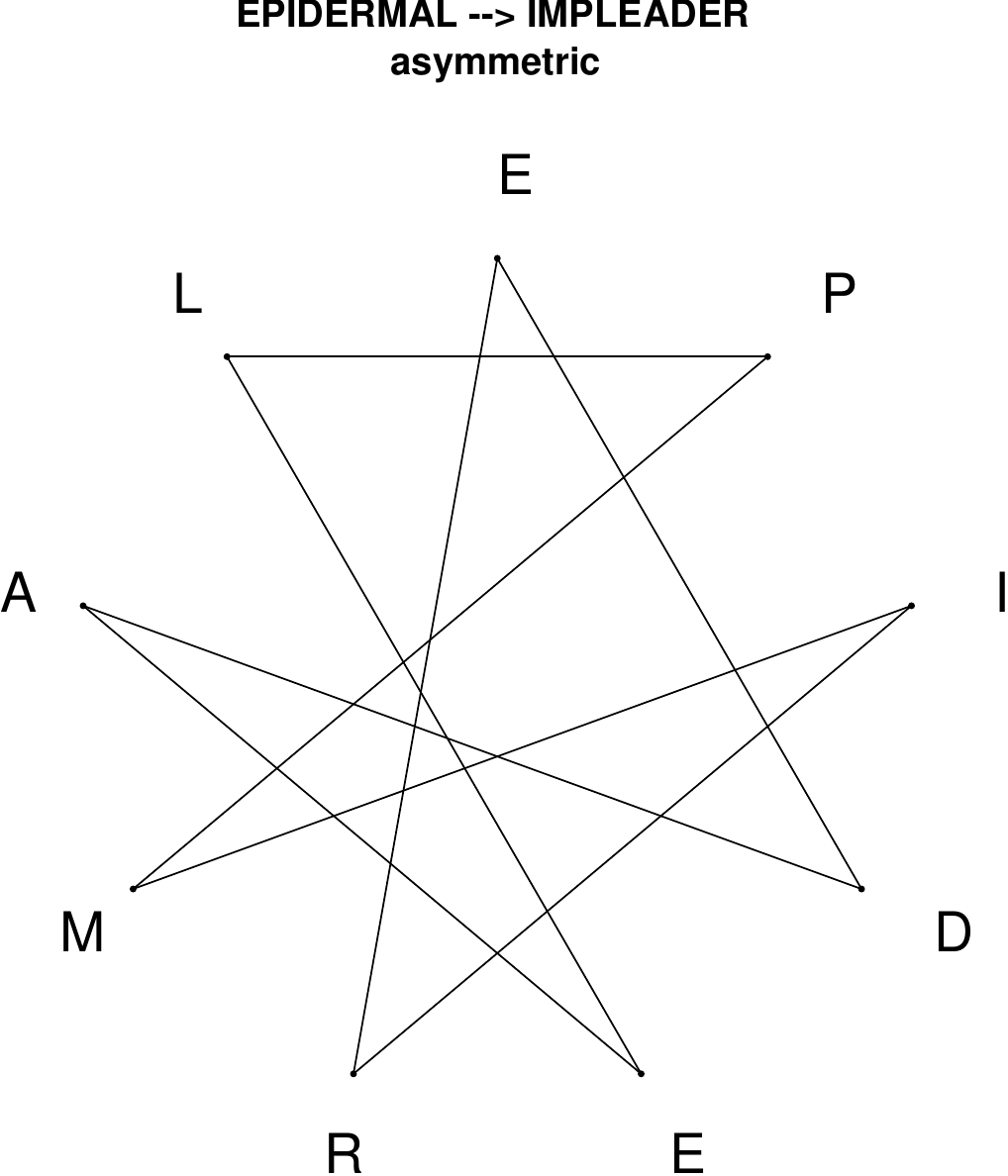}
\end{subfigure}
\hfill
\begin{subfigure}[T]{0.19\textwidth}
\centering
\includegraphics[width=\textwidth]{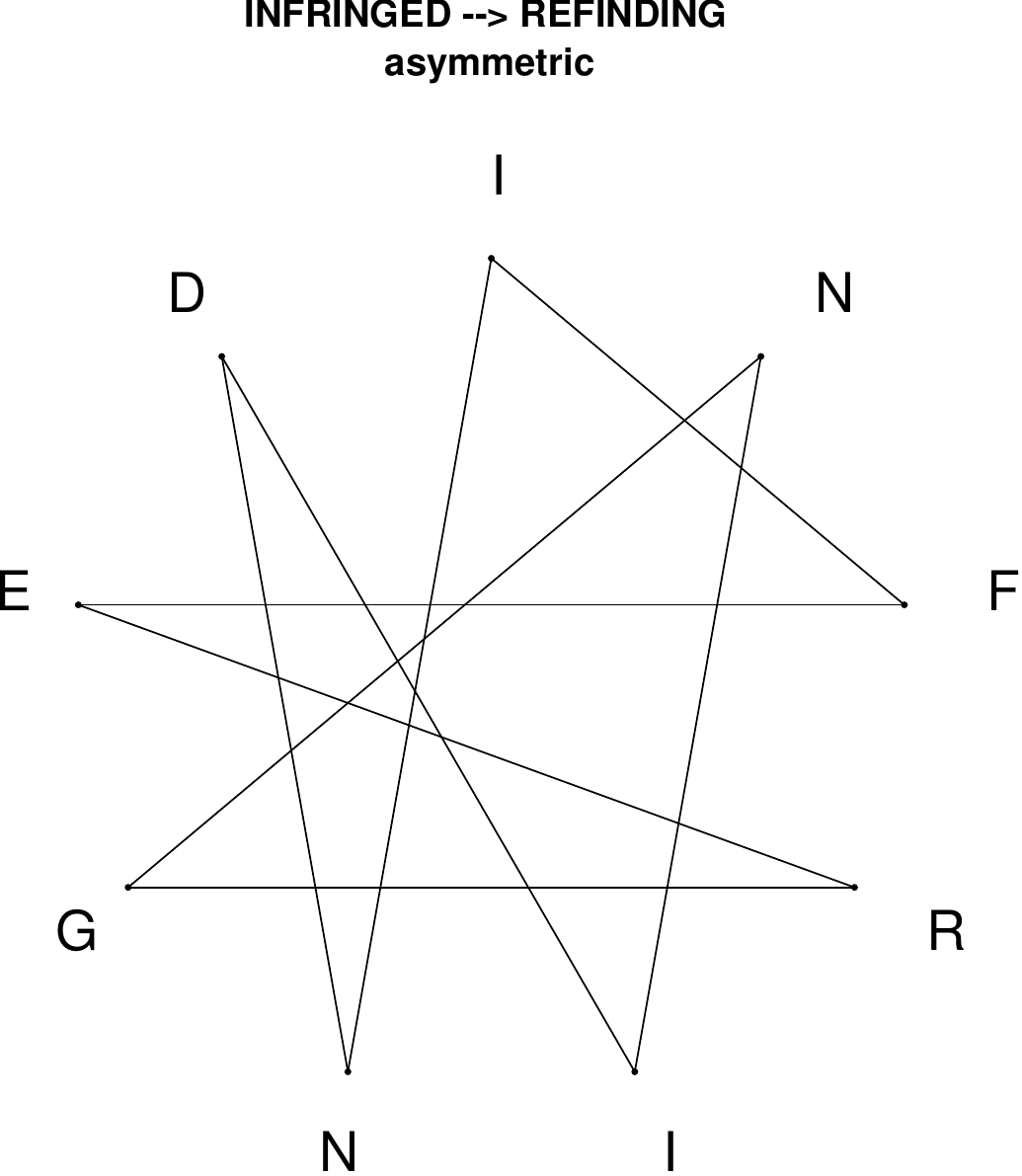}
\end{subfigure}
\hfill
\begin{subfigure}[T]{0.19\textwidth}
\centering
\includegraphics[width=\textwidth]{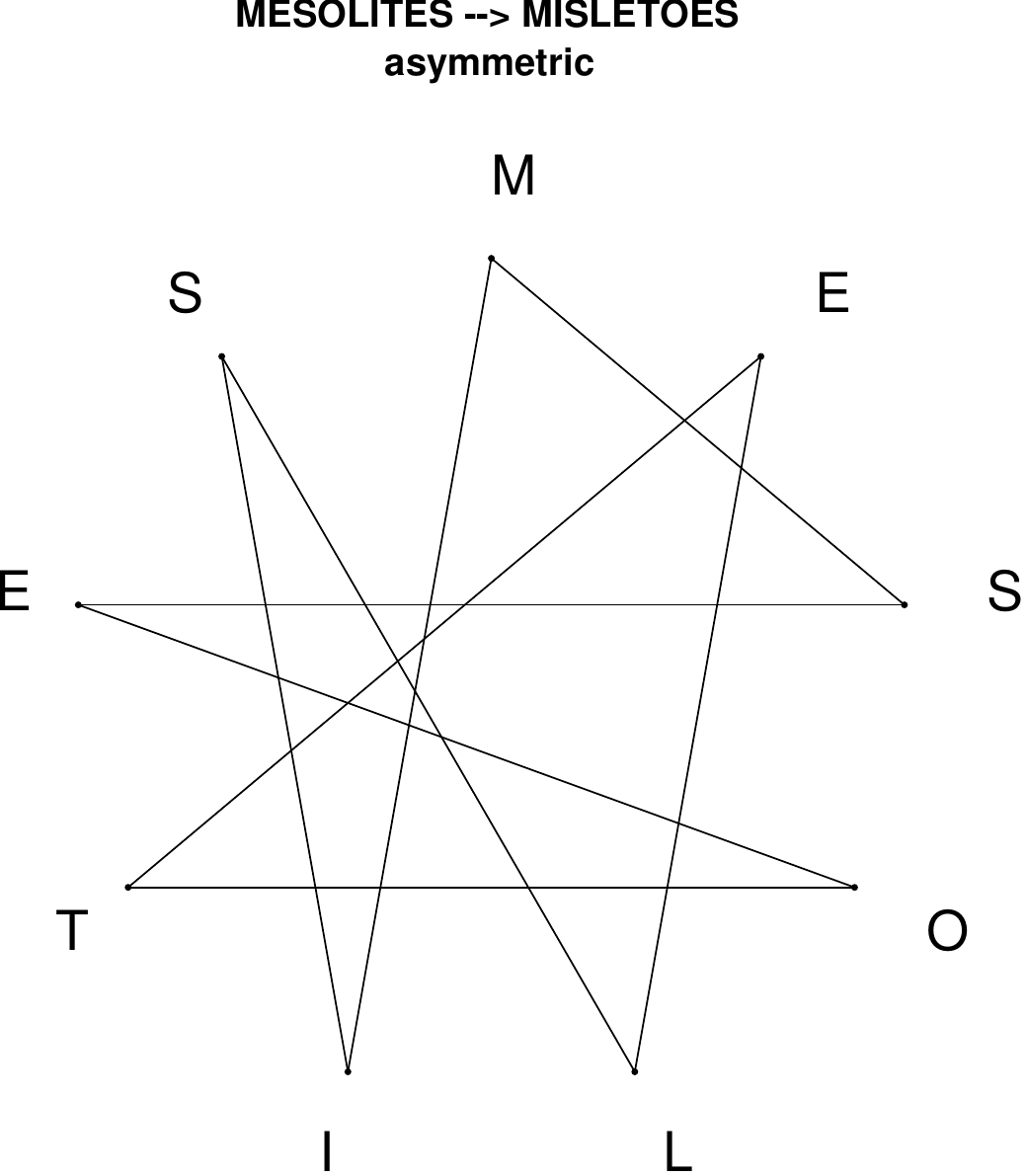}
\end{subfigure}
\hfill
\begin{subfigure}[T]{0.19\textwidth}
\centering
\includegraphics[width=\textwidth]{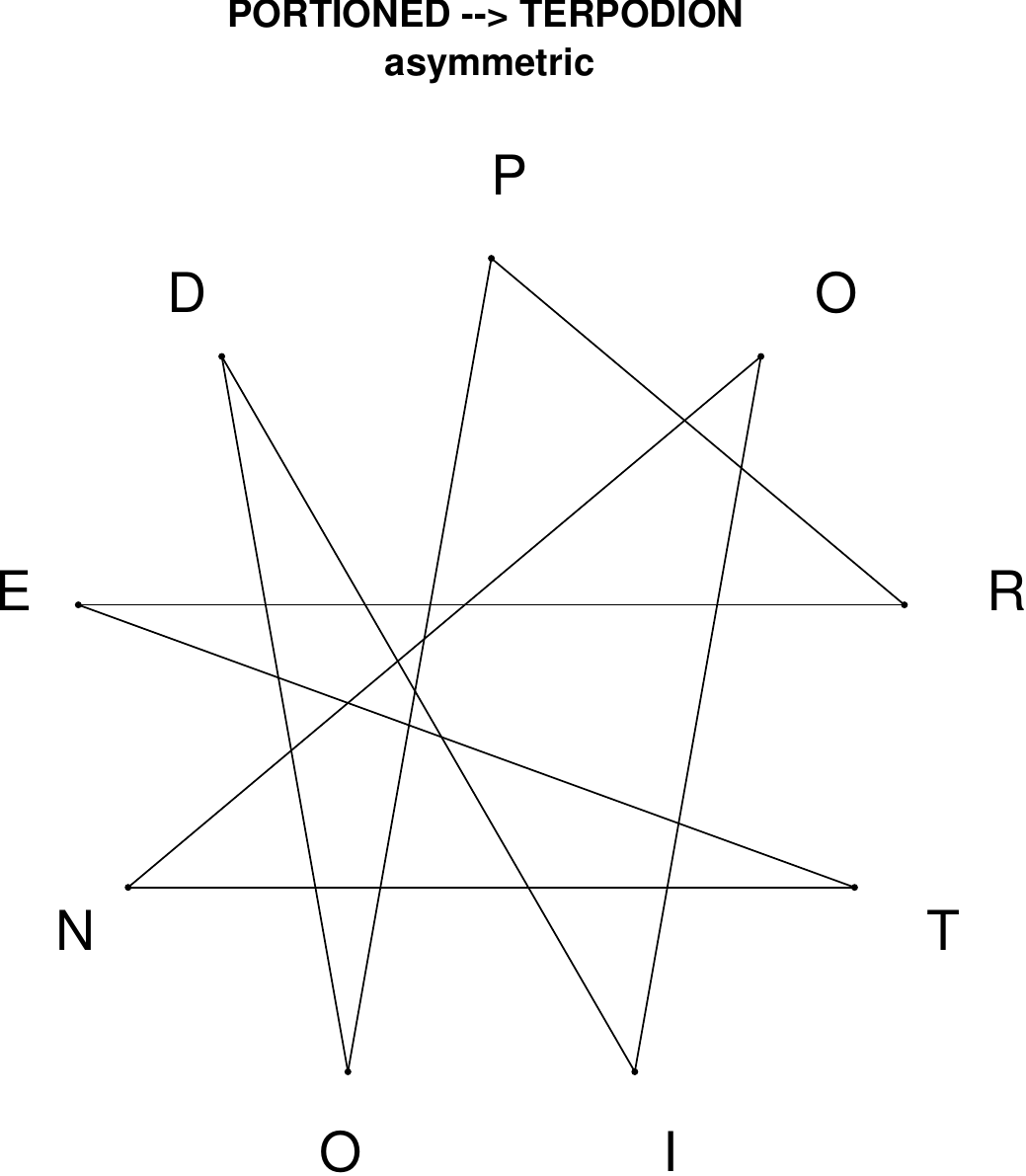}
\end{subfigure}
\hfill
\begin{subfigure}[T]{0.19\textwidth}
\centering
\includegraphics[width=\textwidth]{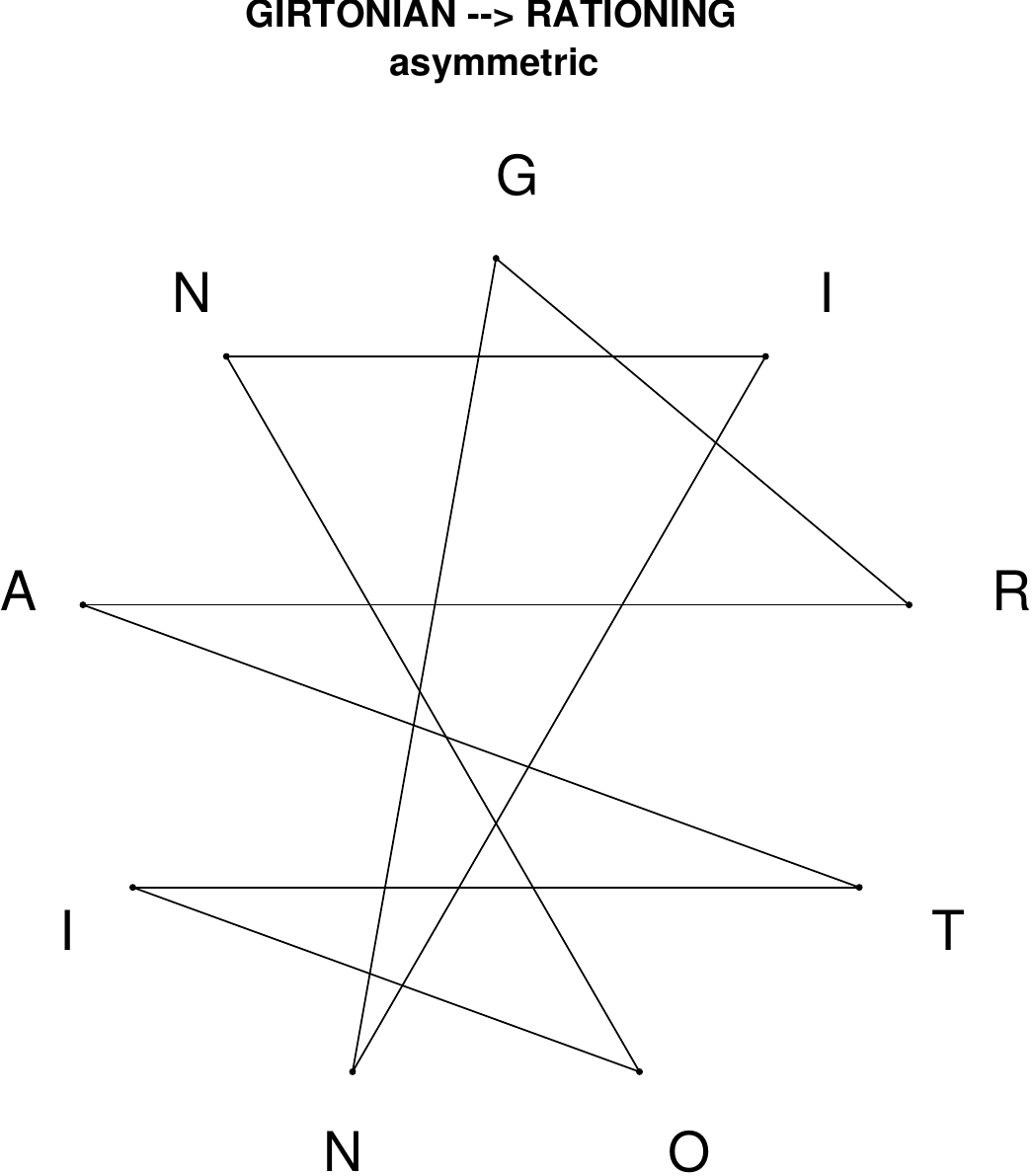}
\end{subfigure}
\end{figure}

\begin{figure}[H]
\centering
\begin{subfigure}[T]{0.19\textwidth}
\centering
\includegraphics[width=\textwidth]{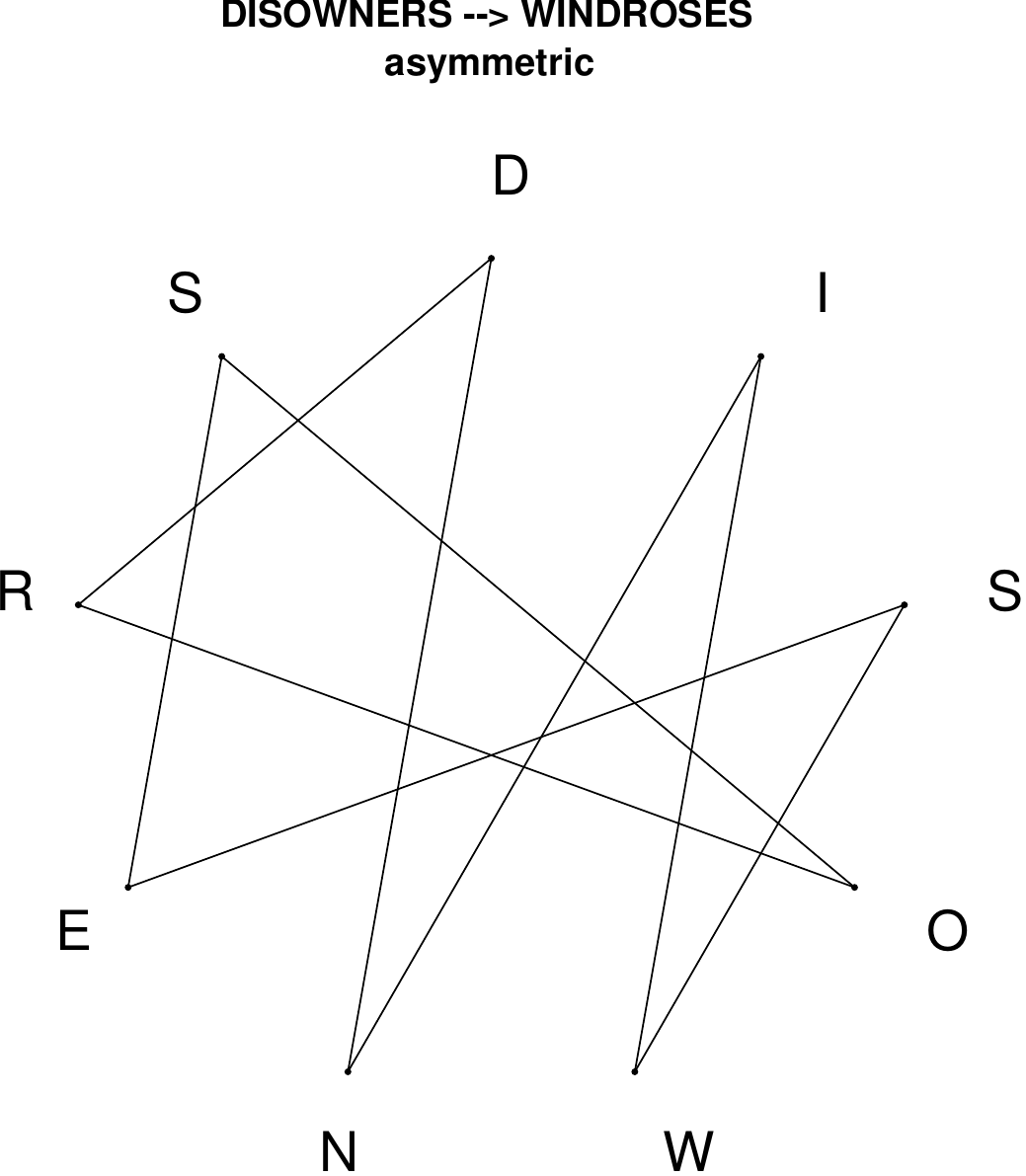}
\end{subfigure}
\hfill
\begin{subfigure}[T]{0.19\textwidth}
\centering
\includegraphics[width=\textwidth]{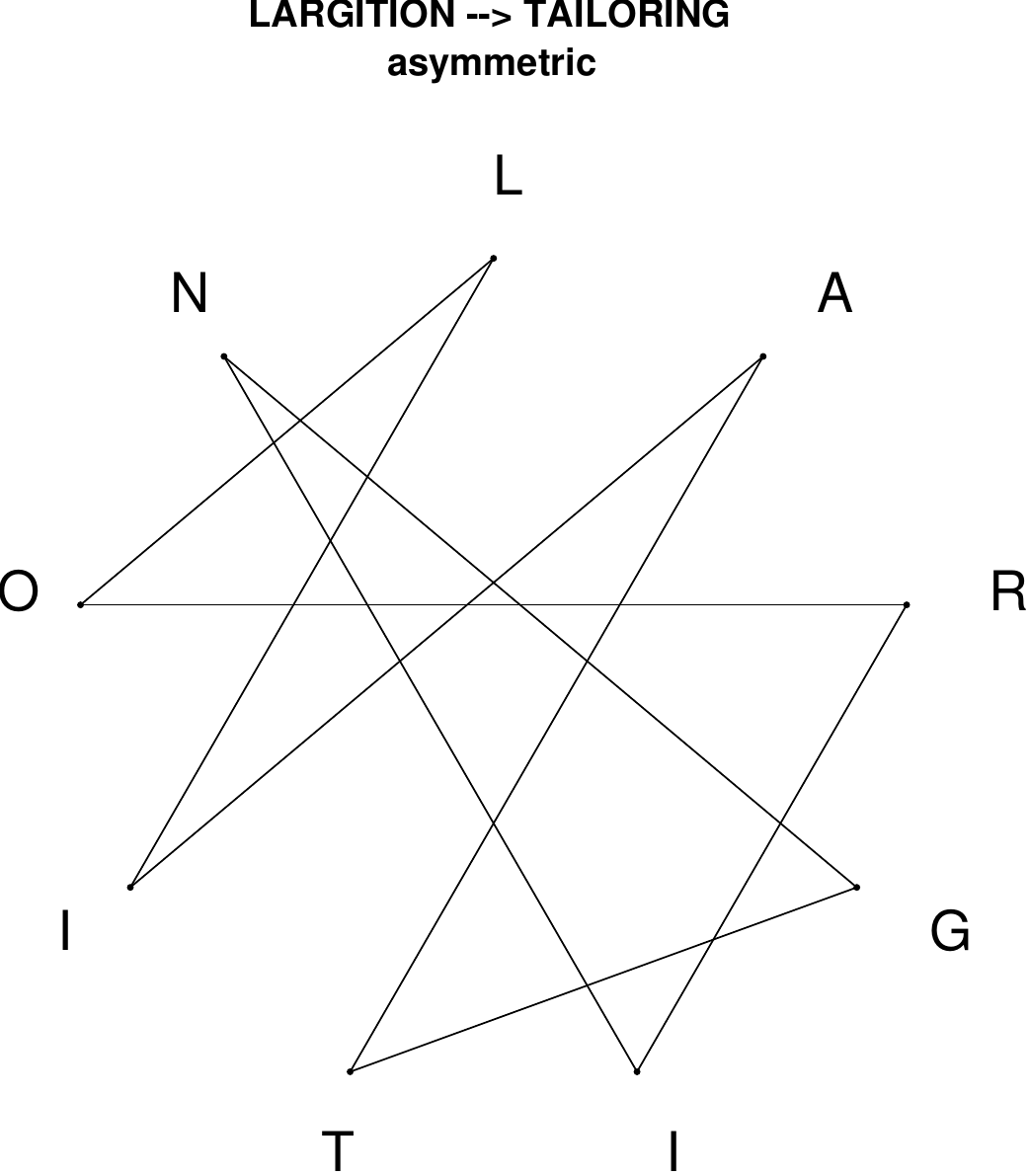}
\end{subfigure}
\hfill
\begin{subfigure}[T]{0.19\textwidth}
\centering
\includegraphics[width=\textwidth]{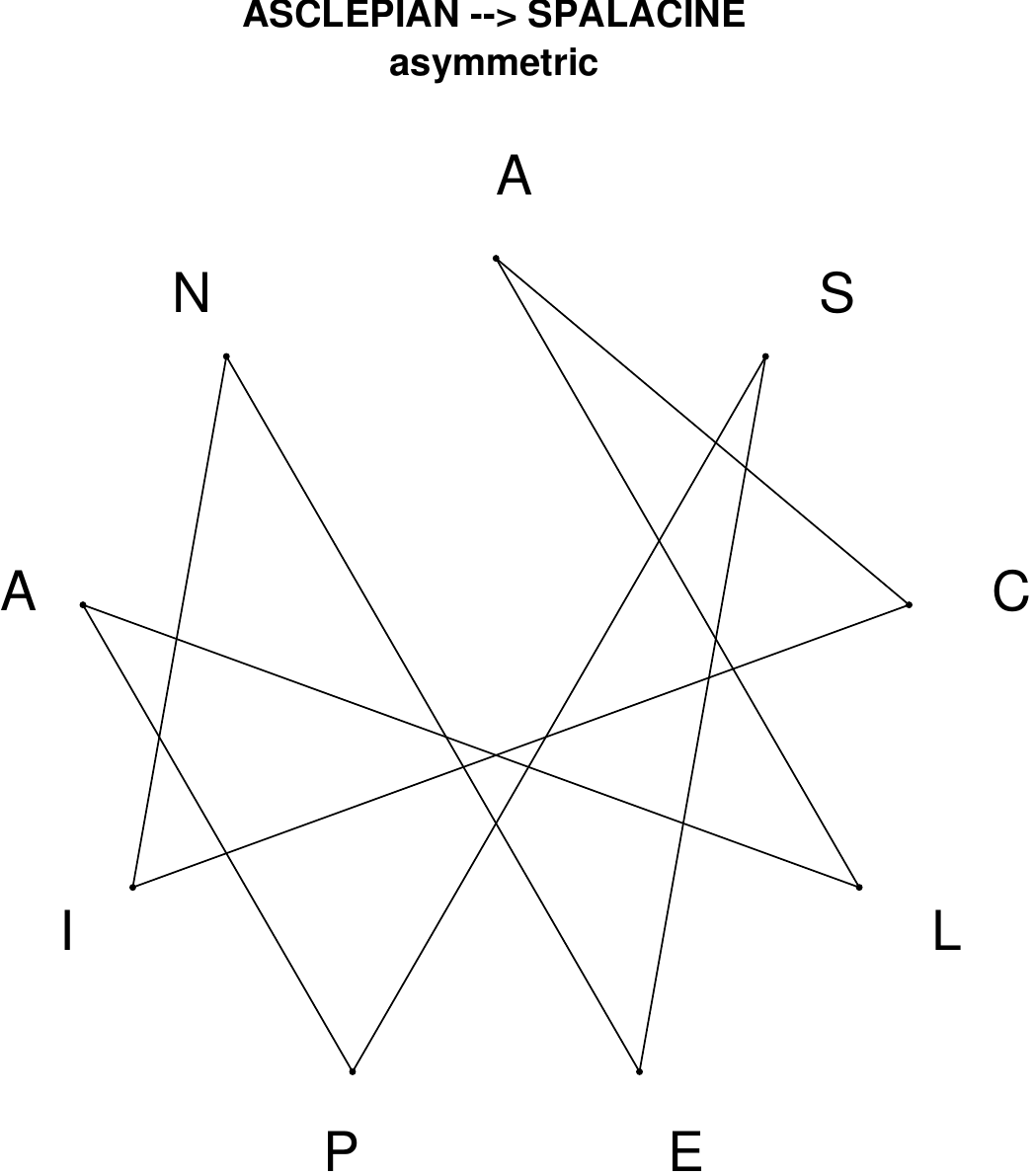}
\end{subfigure}
\hfill
\begin{subfigure}[T]{0.19\textwidth}
\centering
\includegraphics[width=\textwidth]{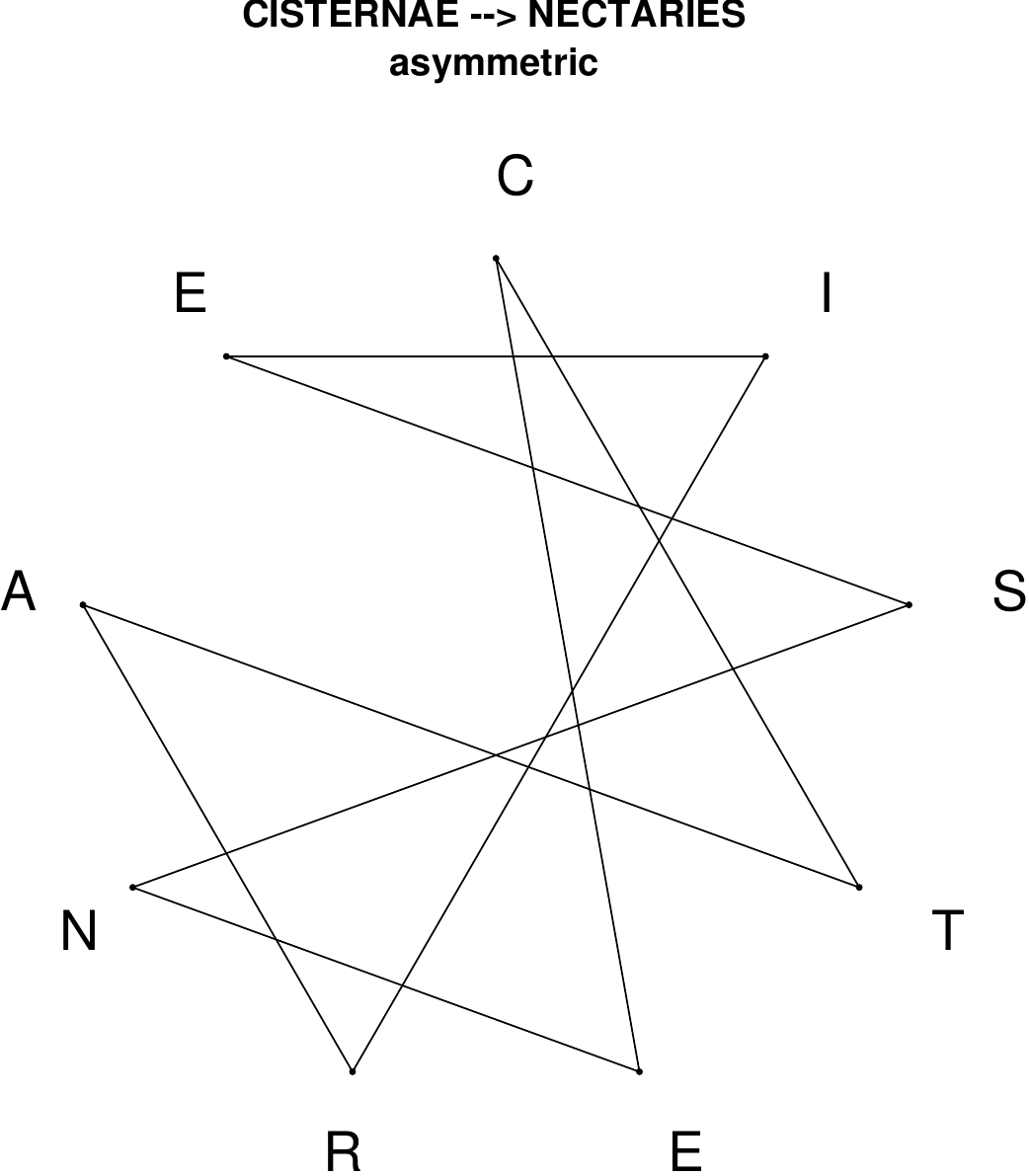}
\end{subfigure}
\hfill
\begin{subfigure}[T]{0.19\textwidth}
\centering
\includegraphics[width=\textwidth]{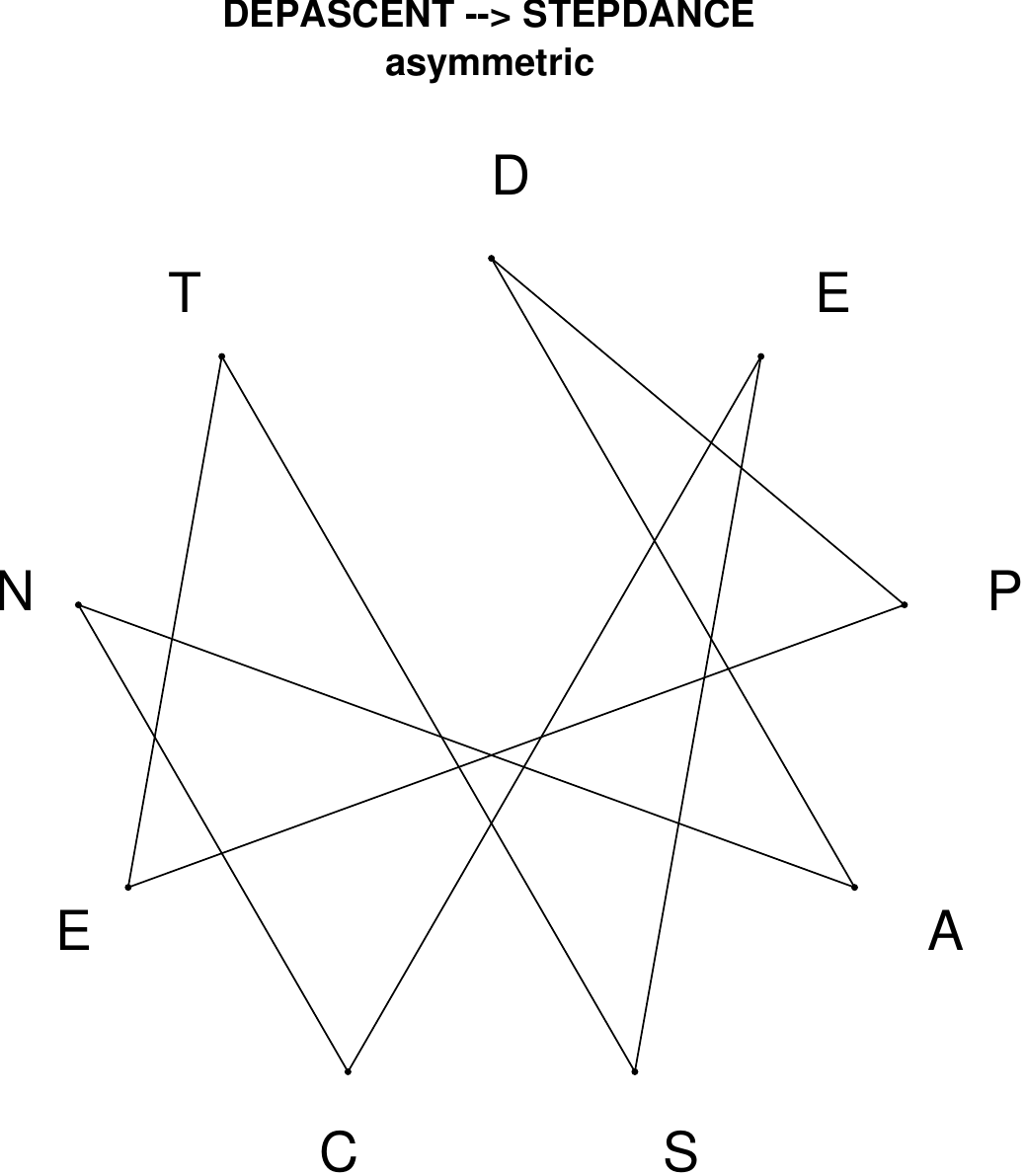}
\end{subfigure}
\end{figure}

\begin{figure}[H]
\centering
\begin{subfigure}[T]{0.19\textwidth}
\centering
\includegraphics[width=\textwidth]{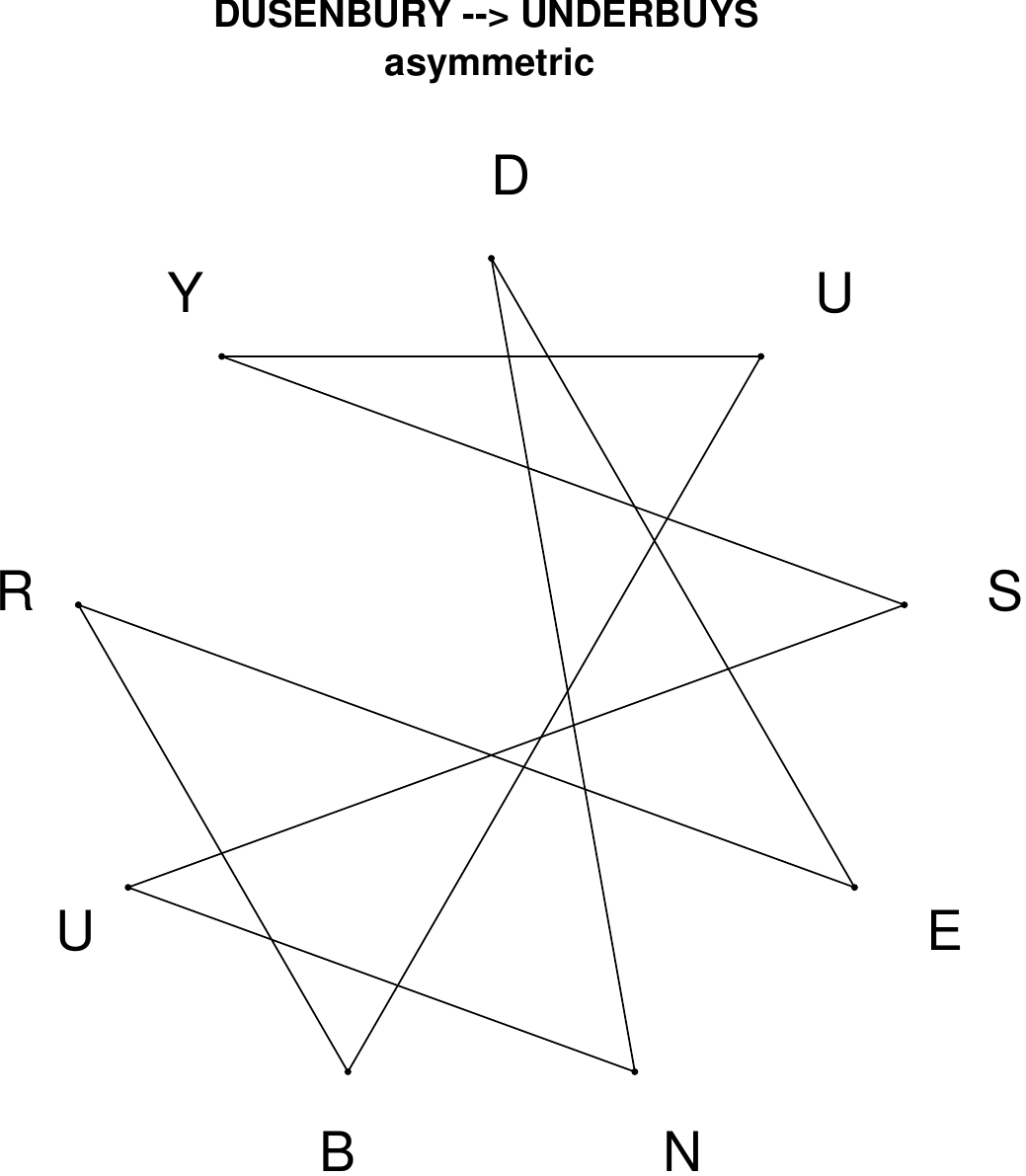}
\end{subfigure}
\hfill
\begin{subfigure}[T]{0.19\textwidth}
\centering
\includegraphics[width=\textwidth]{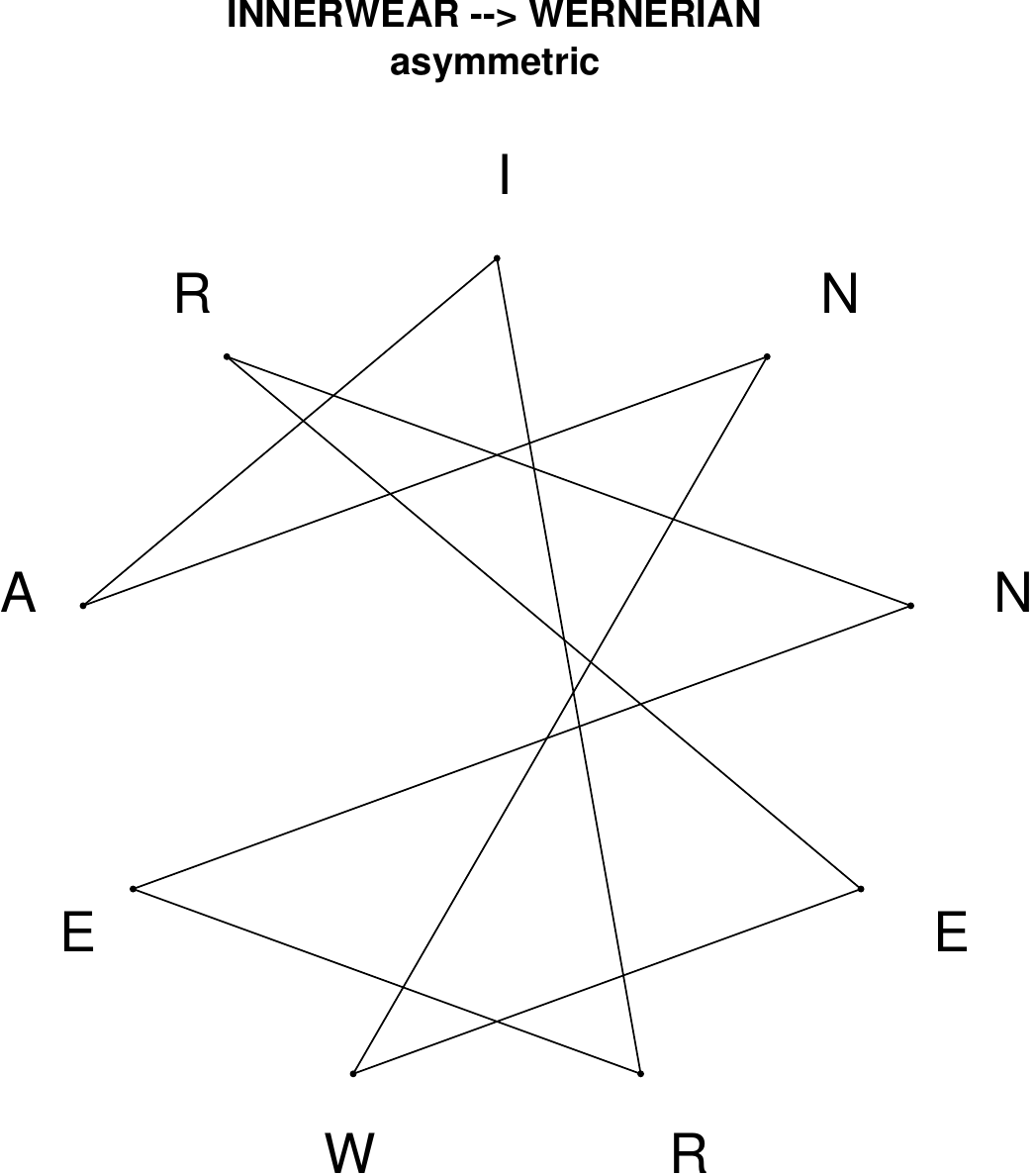}
\end{subfigure}
\hfill
\begin{subfigure}[T]{0.19\textwidth}
\centering
\includegraphics[width=\textwidth]{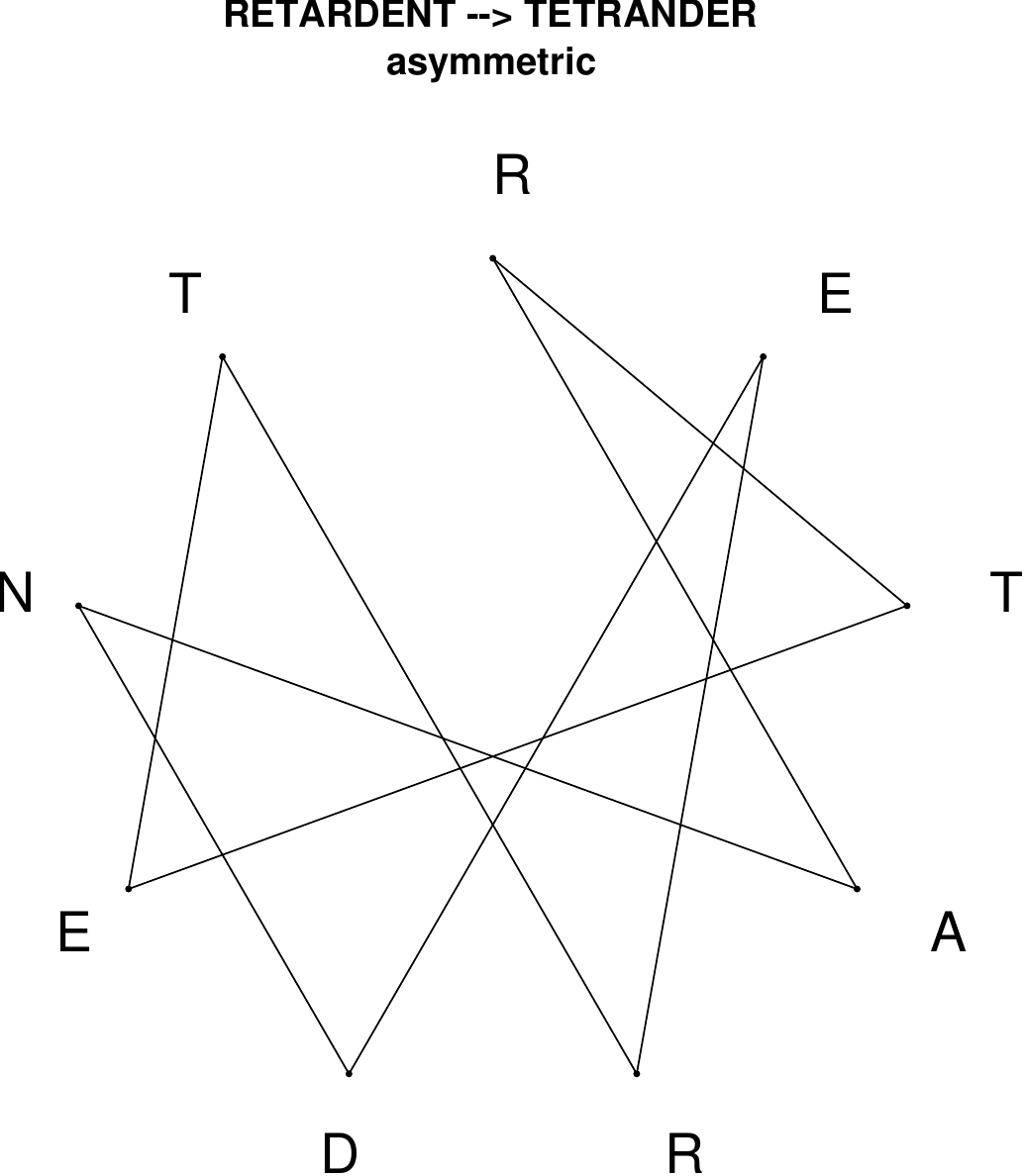}
\end{subfigure}
\hfill
\begin{subfigure}[T]{0.19\textwidth}
\centering
\includegraphics[width=\textwidth]{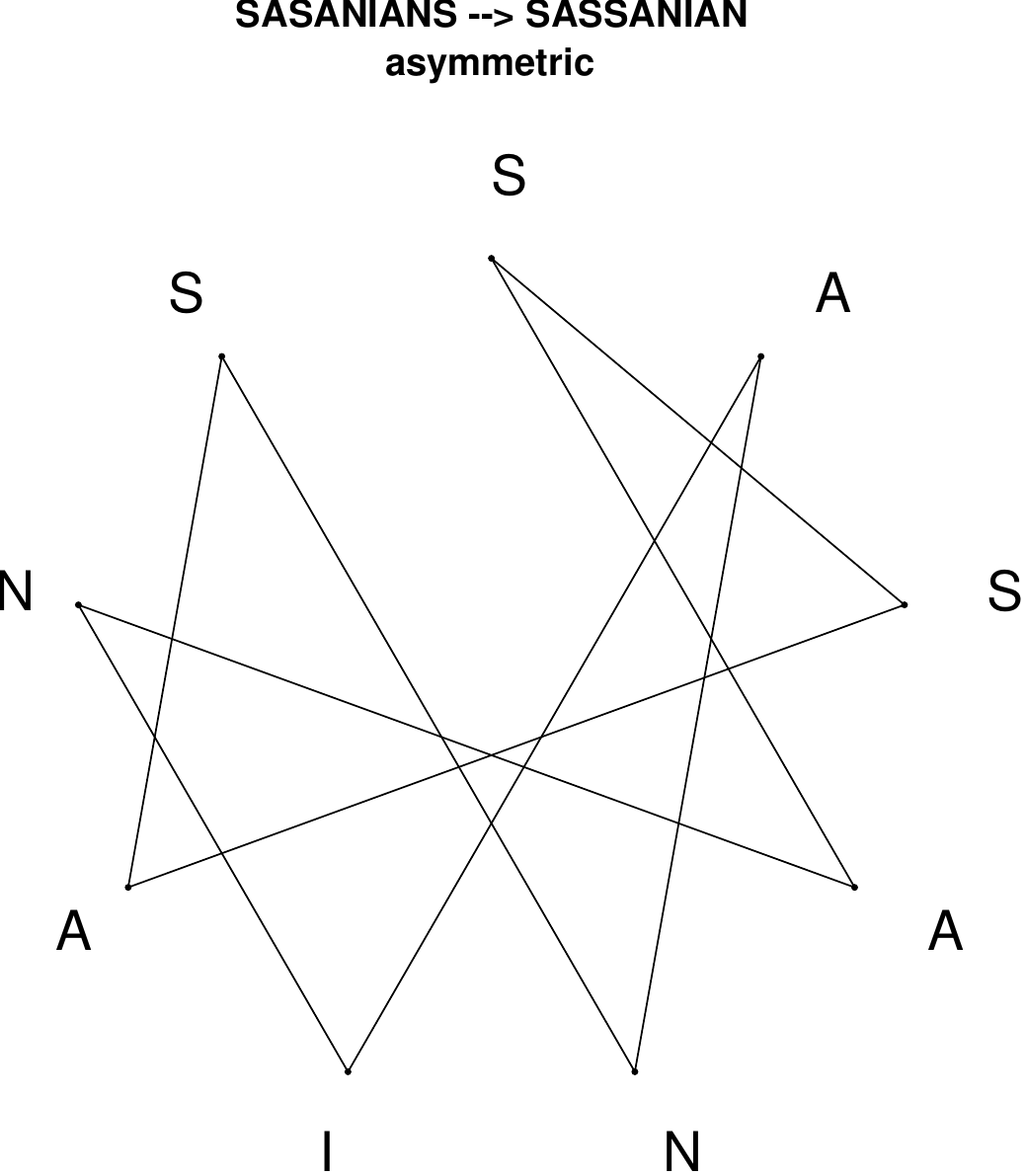}
\end{subfigure}
\hfill
\begin{subfigure}[T]{0.19\textwidth}
\centering
\includegraphics[width=\textwidth]{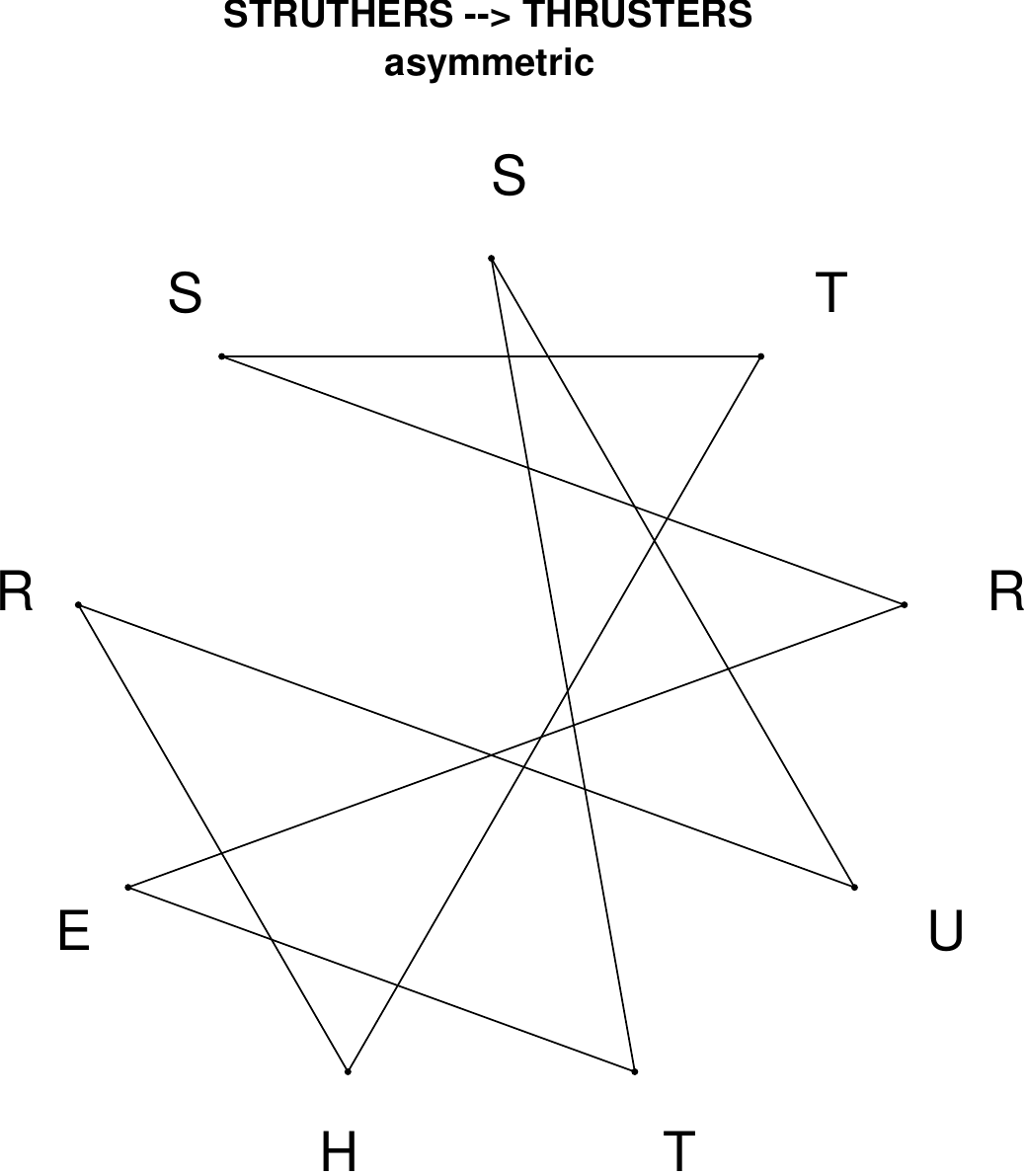}
\end{subfigure}
\end{figure}

\begin{figure}[H]
\centering
\begin{subfigure}[T]{0.19\textwidth}
\centering
\includegraphics[width=\textwidth]{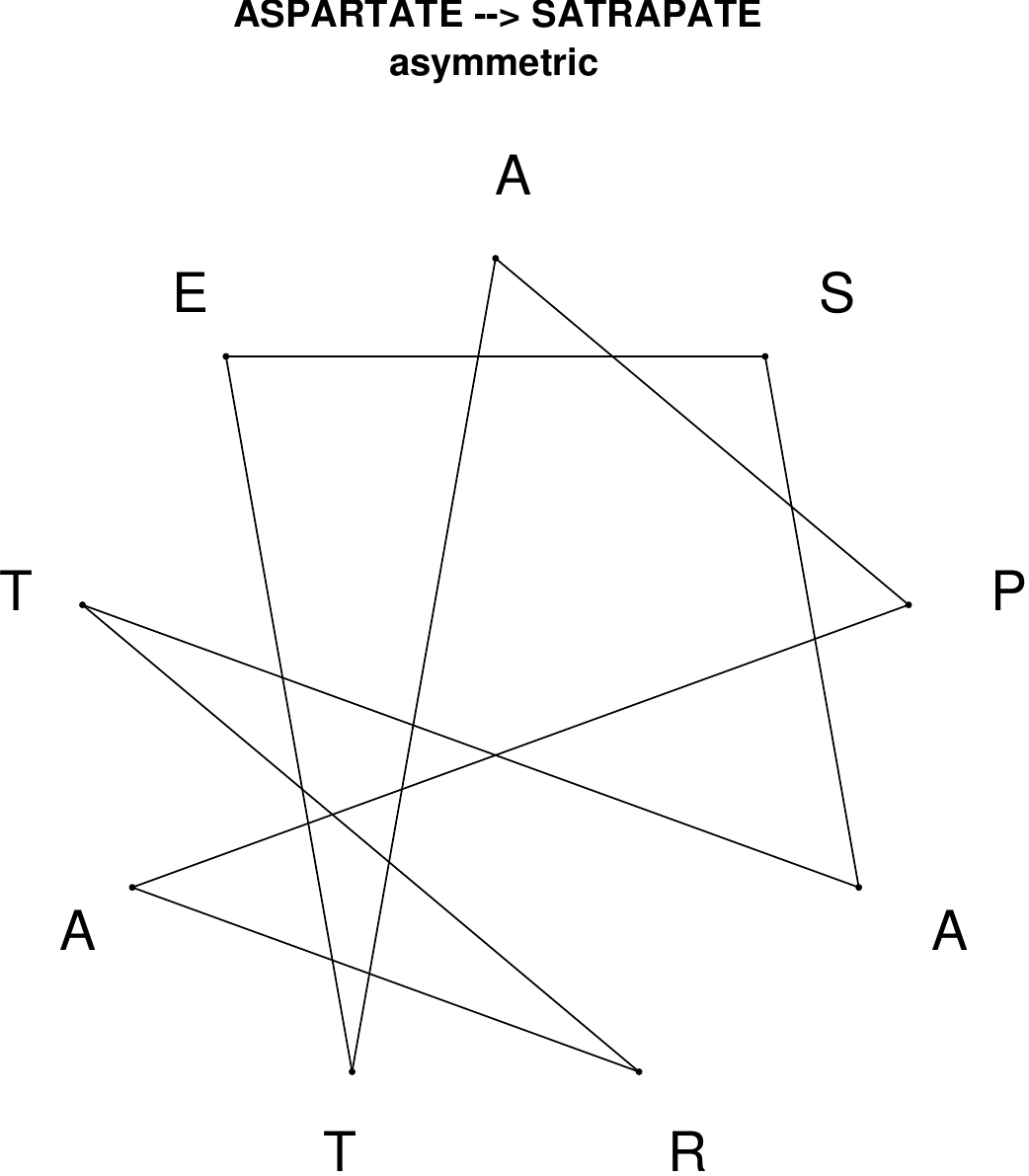}
\end{subfigure}
\hfill
\begin{subfigure}[T]{0.19\textwidth}
\centering
\includegraphics[width=\textwidth]{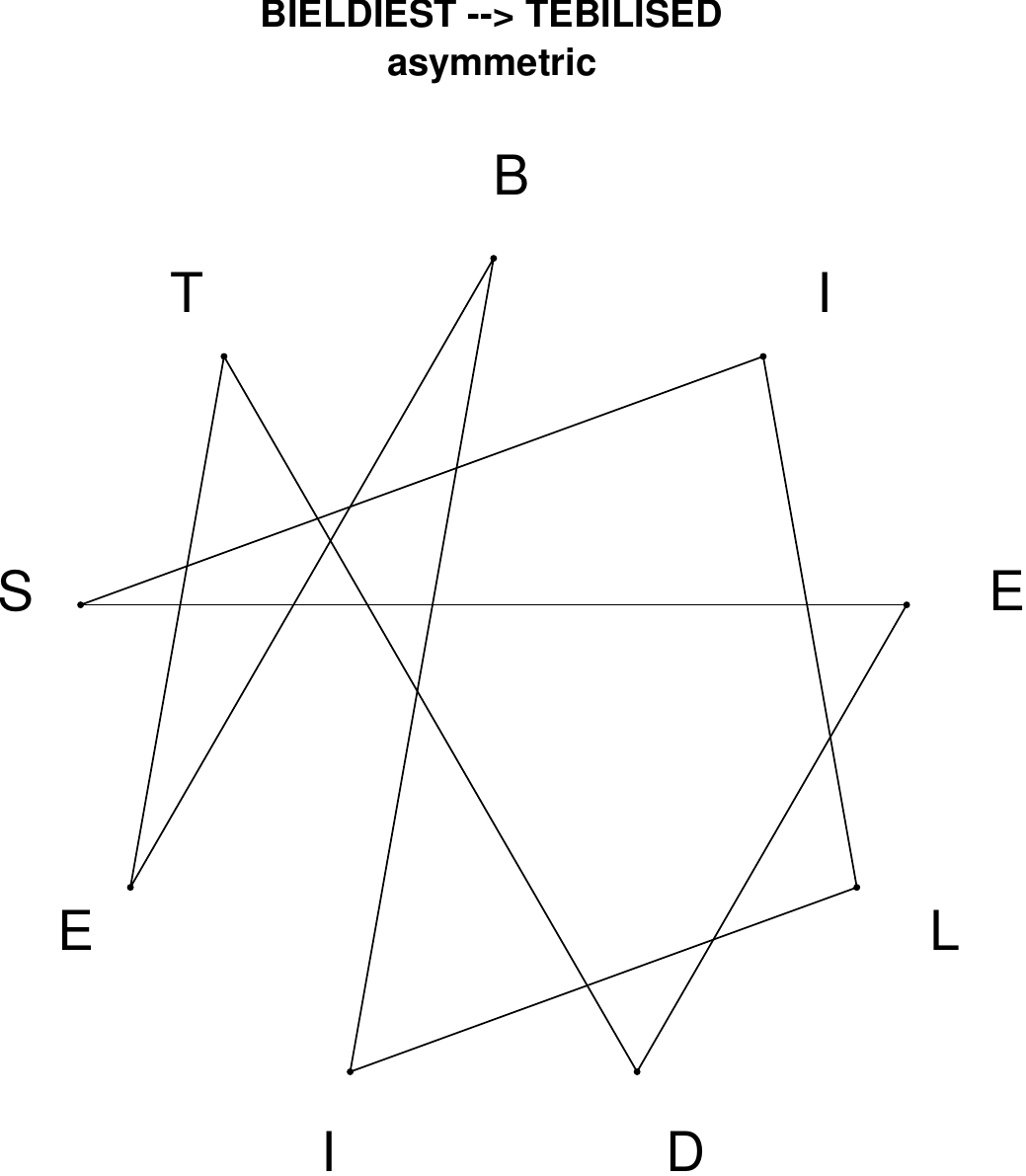}
\end{subfigure}
\hfill
\begin{subfigure}[T]{0.19\textwidth}
\centering
\includegraphics[width=\textwidth]{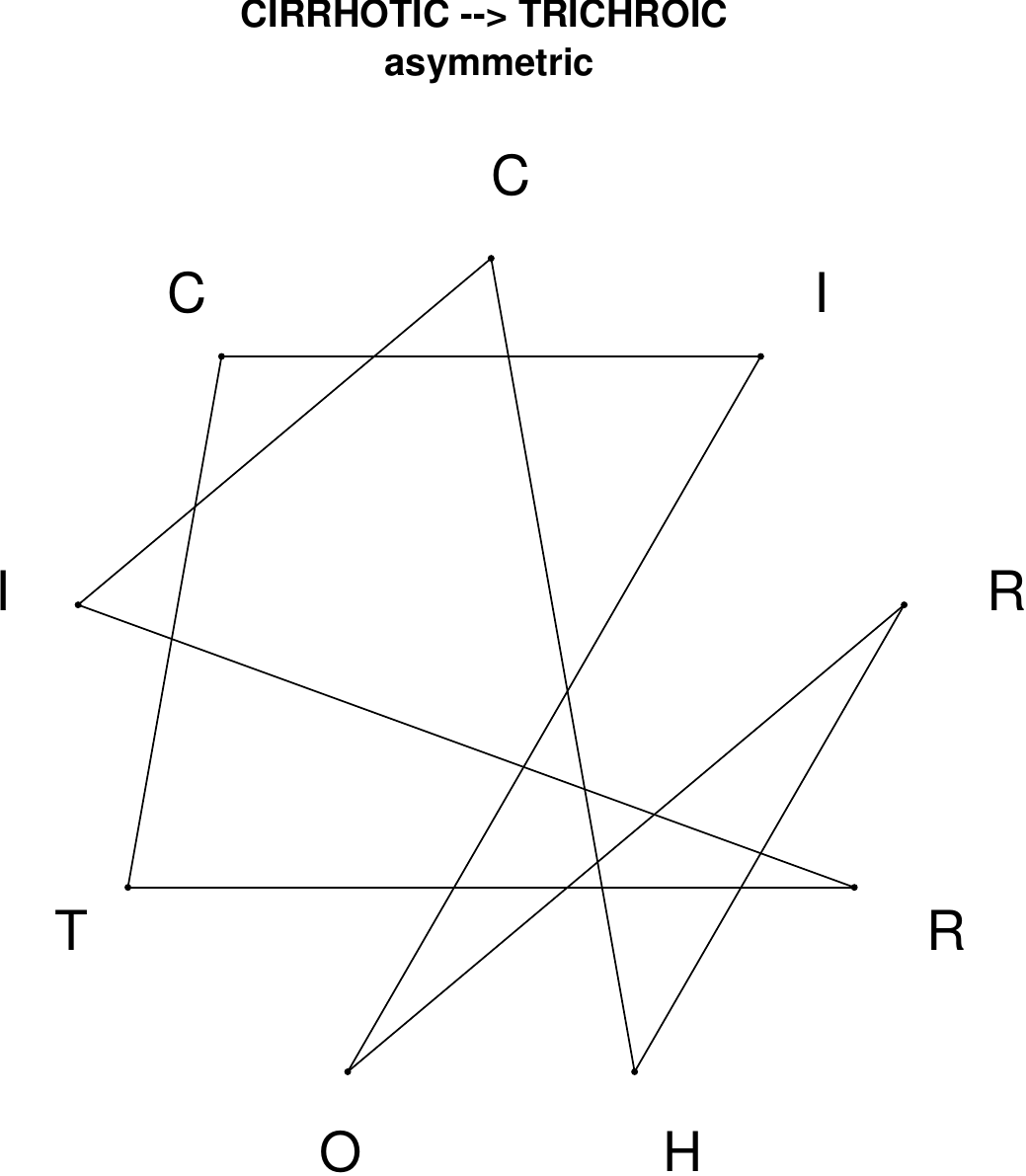}
\end{subfigure}
\hfill
\begin{subfigure}[T]{0.19\textwidth}
\centering
\includegraphics[width=\textwidth]{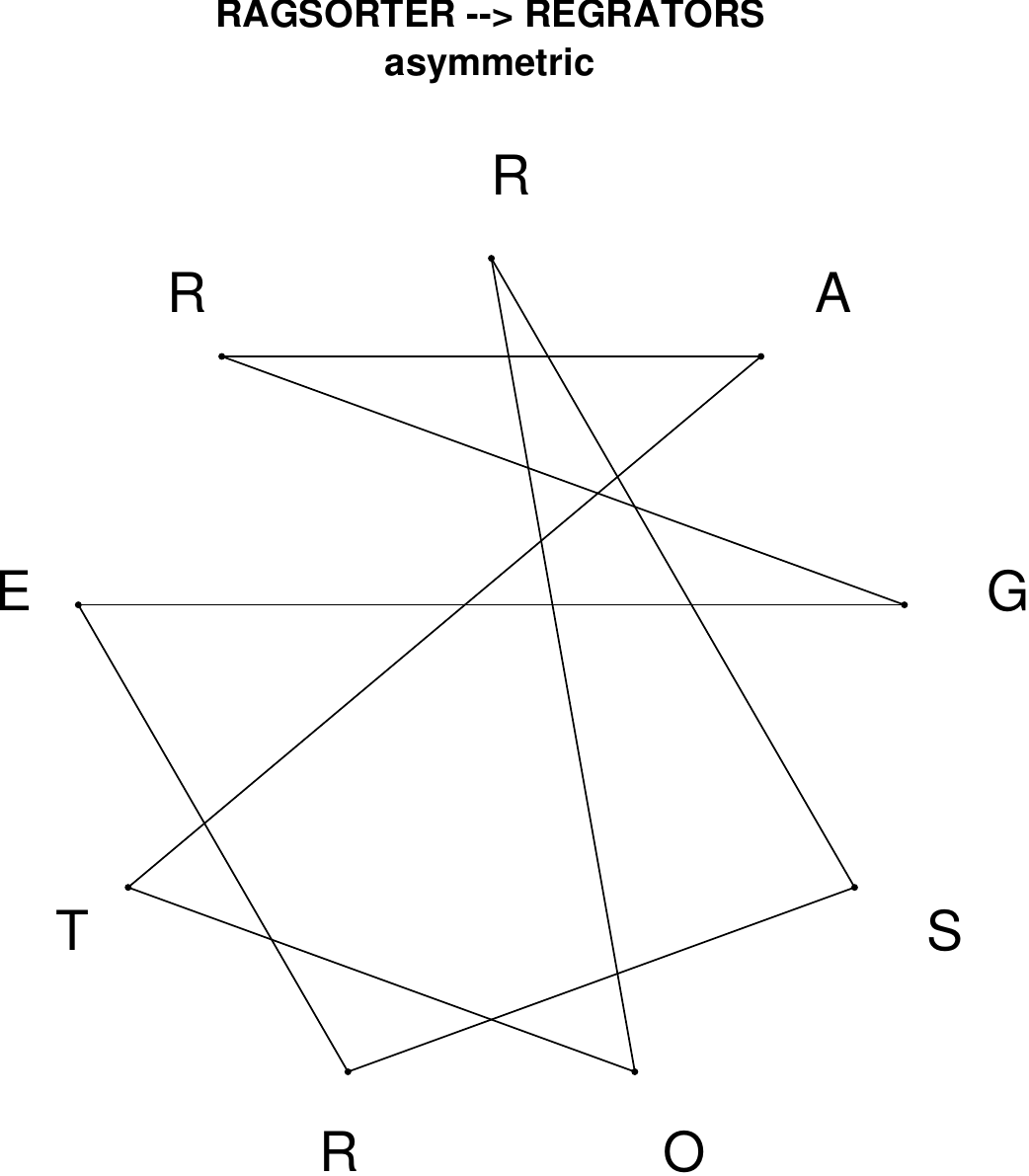}
\end{subfigure}
\hfill
\begin{subfigure}[T]{0.19\textwidth}
\centering
\includegraphics[width=\textwidth]{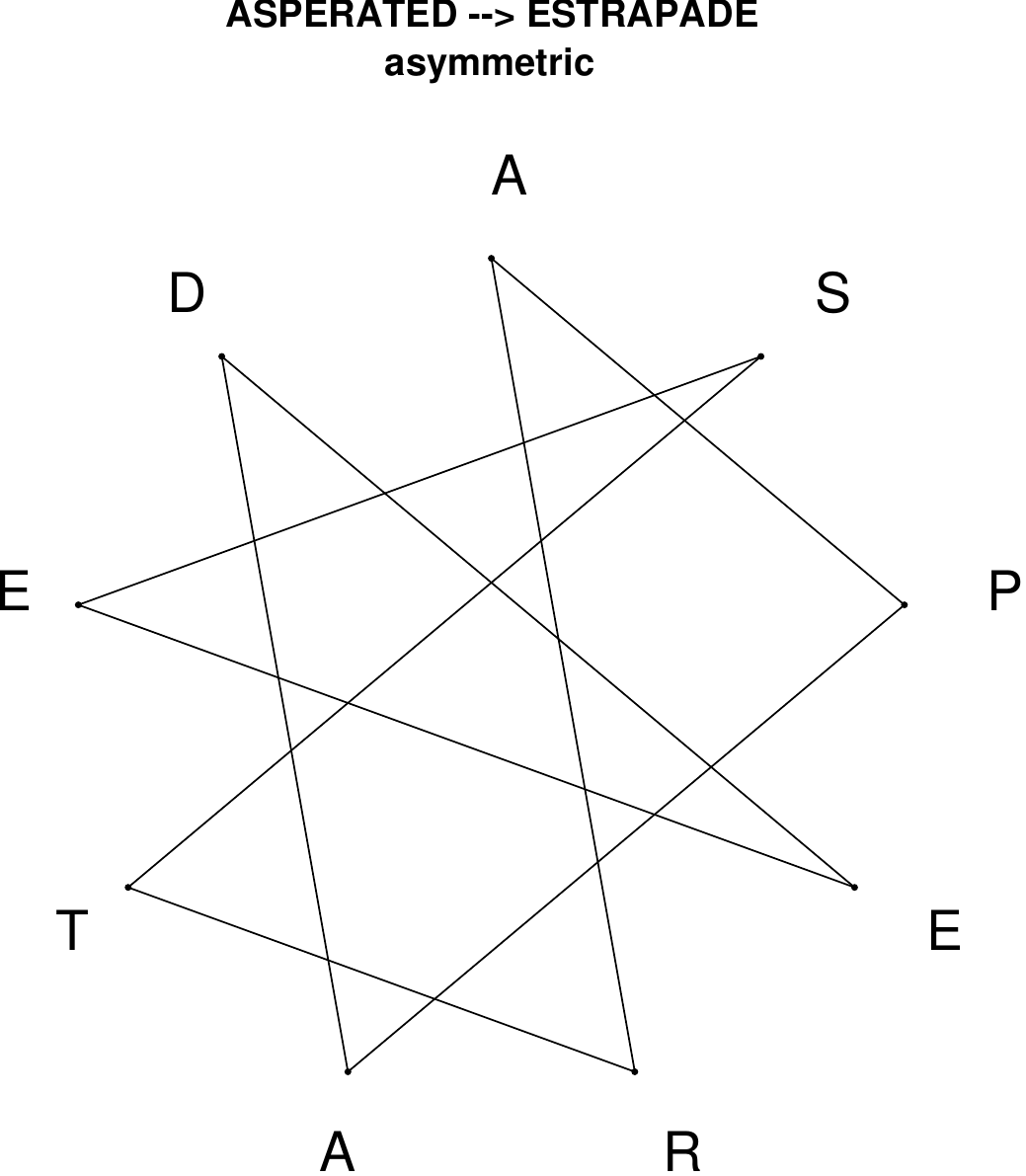}
\end{subfigure}
\end{figure}

\begin{figure}[H]
\centering
\begin{subfigure}[T]{0.19\textwidth}
\centering
\includegraphics[width=\textwidth]{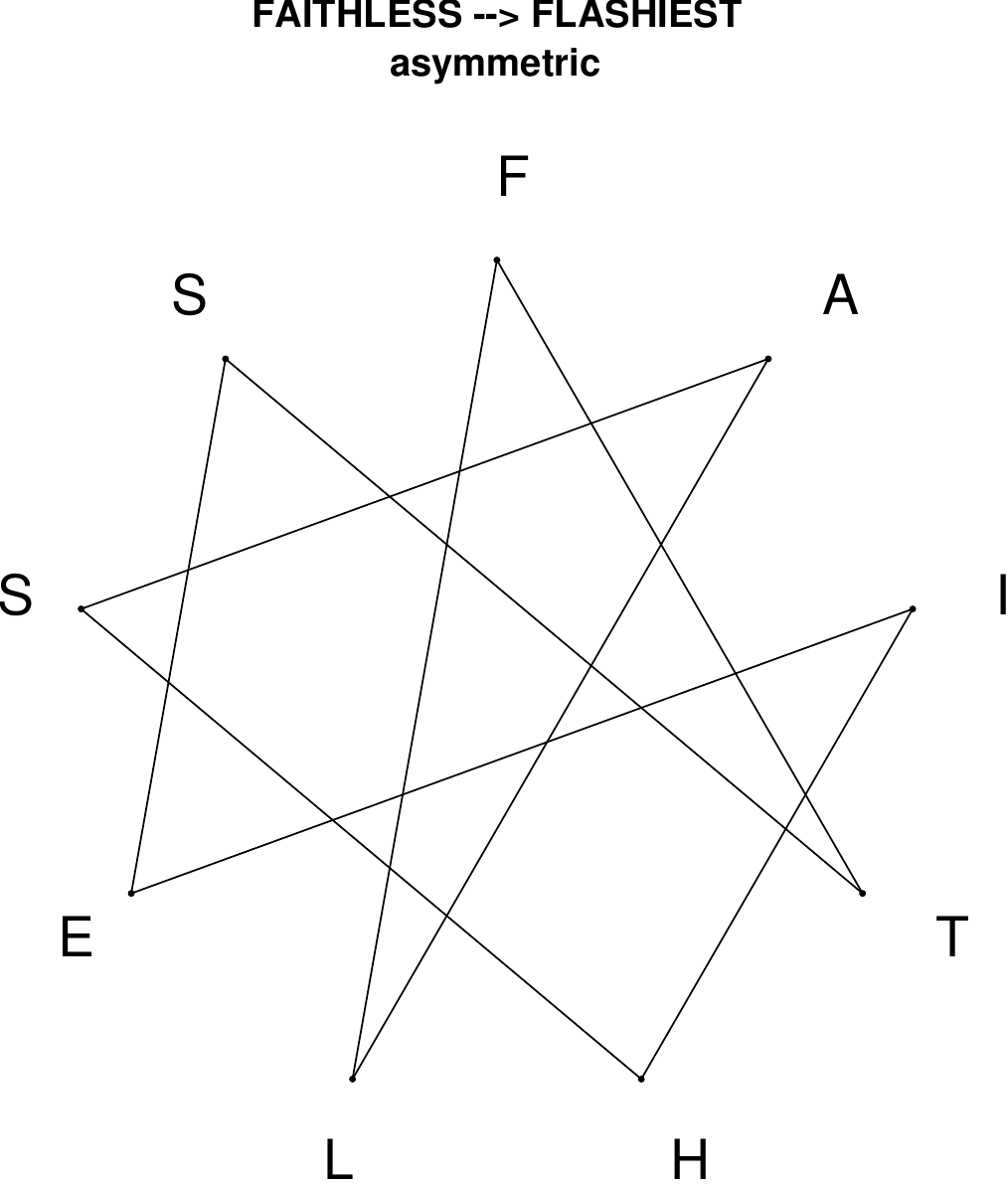}
\end{subfigure}
\hfill
\begin{subfigure}[T]{0.19\textwidth}
\centering
\includegraphics[width=\textwidth]{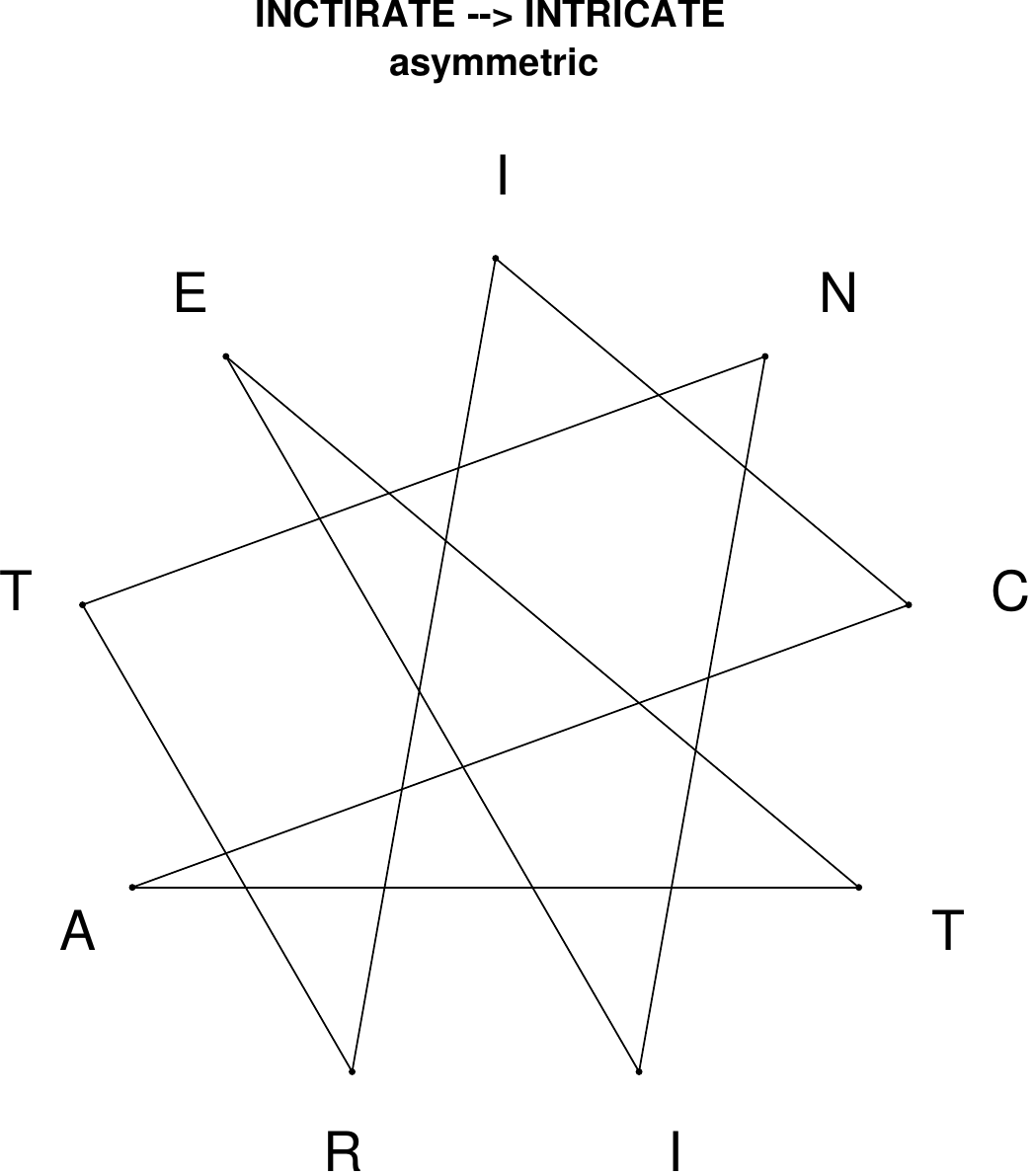}
\end{subfigure}
\hfill
\begin{subfigure}[T]{0.19\textwidth}
\centering
\includegraphics[width=\textwidth]{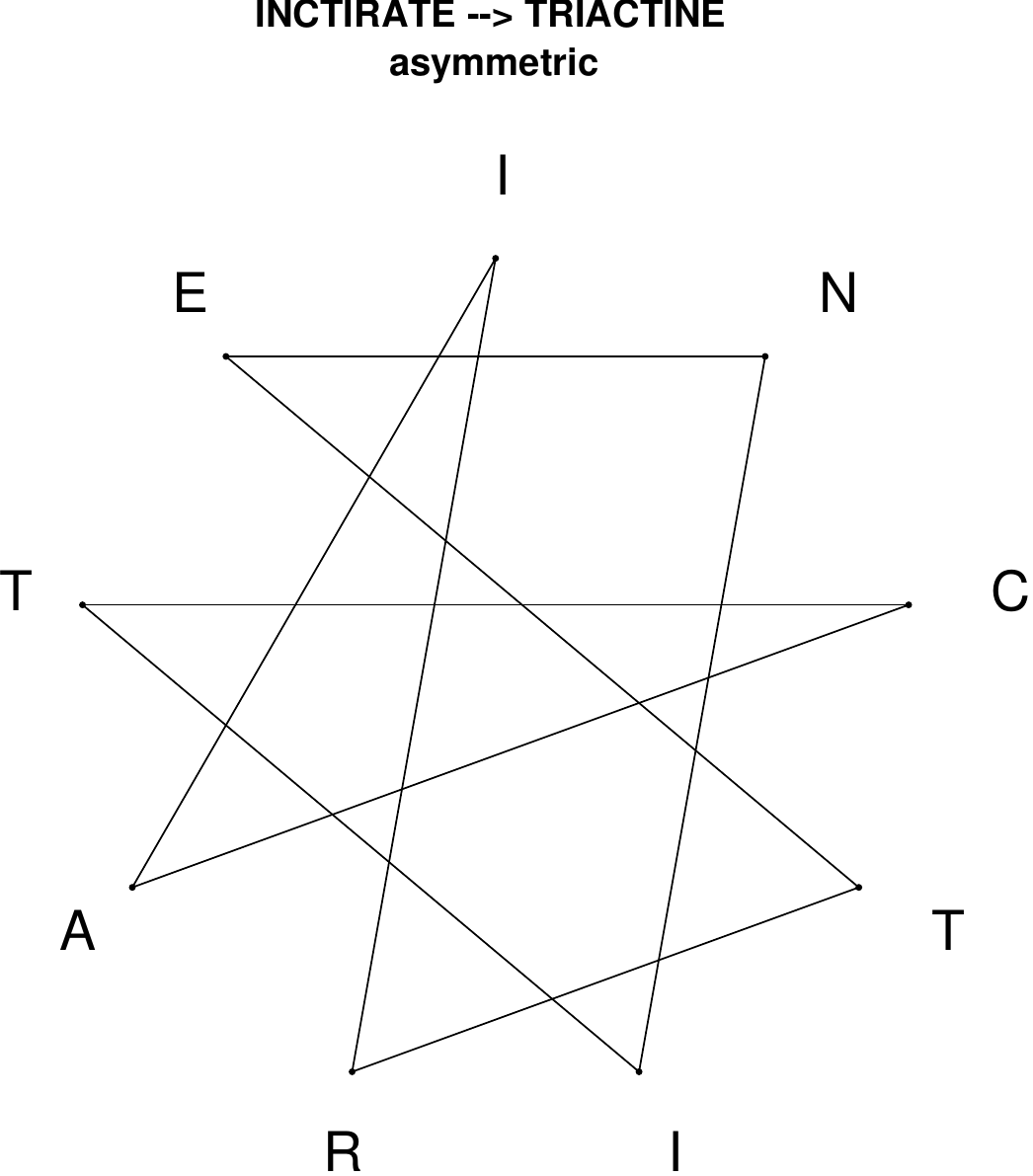}
\end{subfigure}
\hfill
\begin{subfigure}[T]{0.19\textwidth}
\centering
\includegraphics[width=\textwidth]{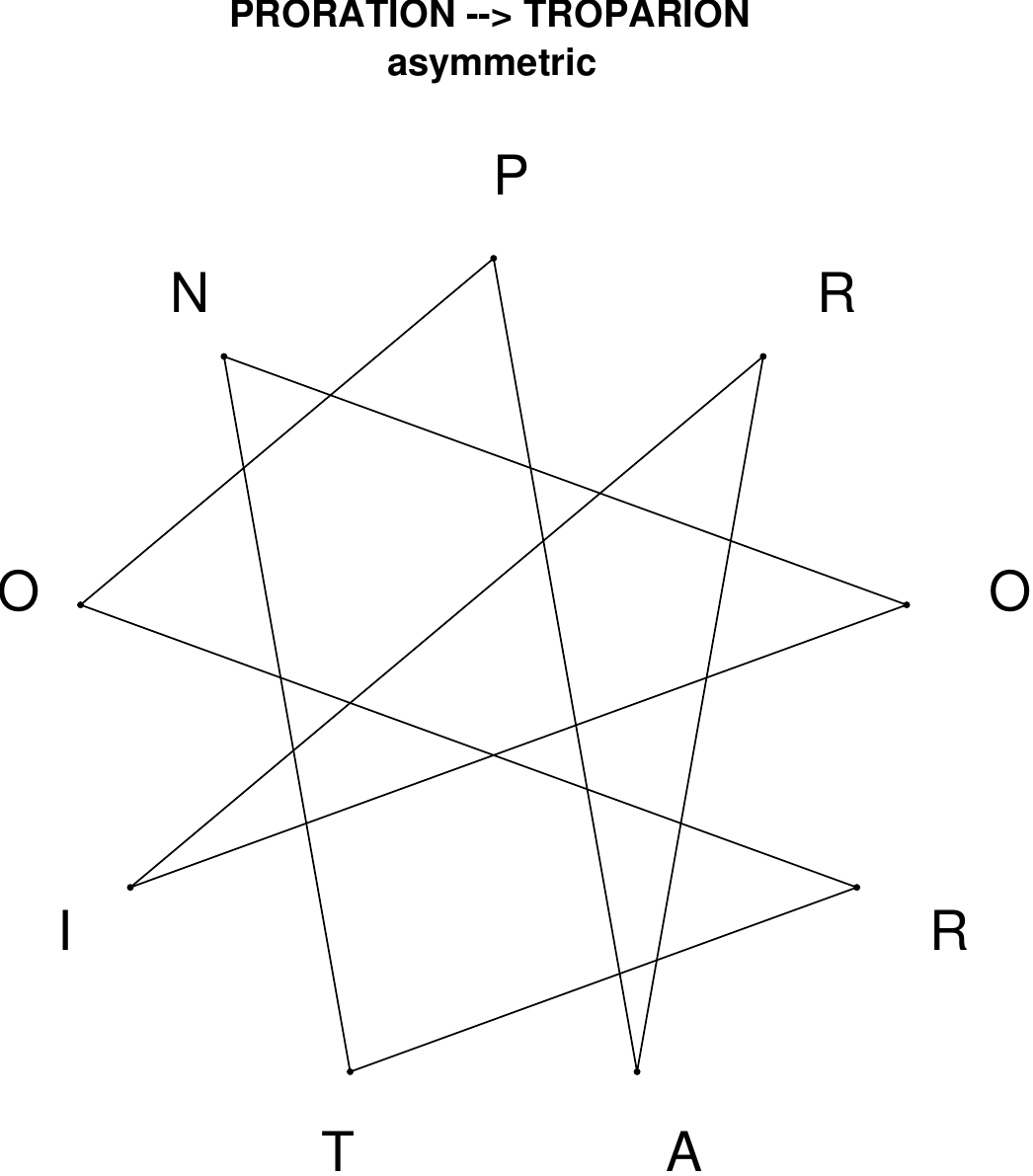}
\end{subfigure}
\hfill
\begin{subfigure}[T]{0.19\textwidth}
\centering
\includegraphics[width=\textwidth]{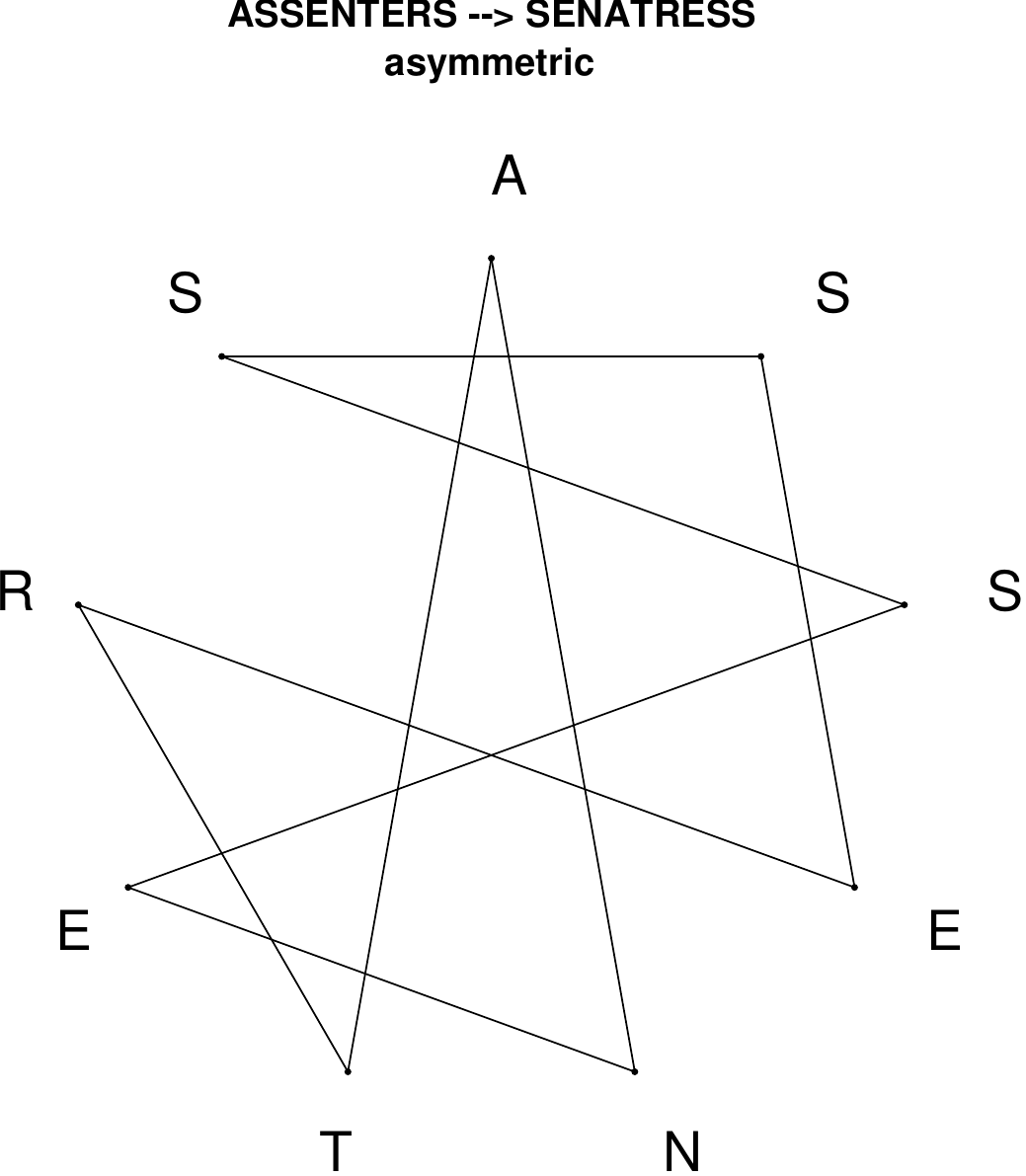}
\end{subfigure}
\end{figure}

\begin{figure}[H]
\centering
\begin{subfigure}[T]{0.19\textwidth}
\centering
\includegraphics[width=\textwidth]{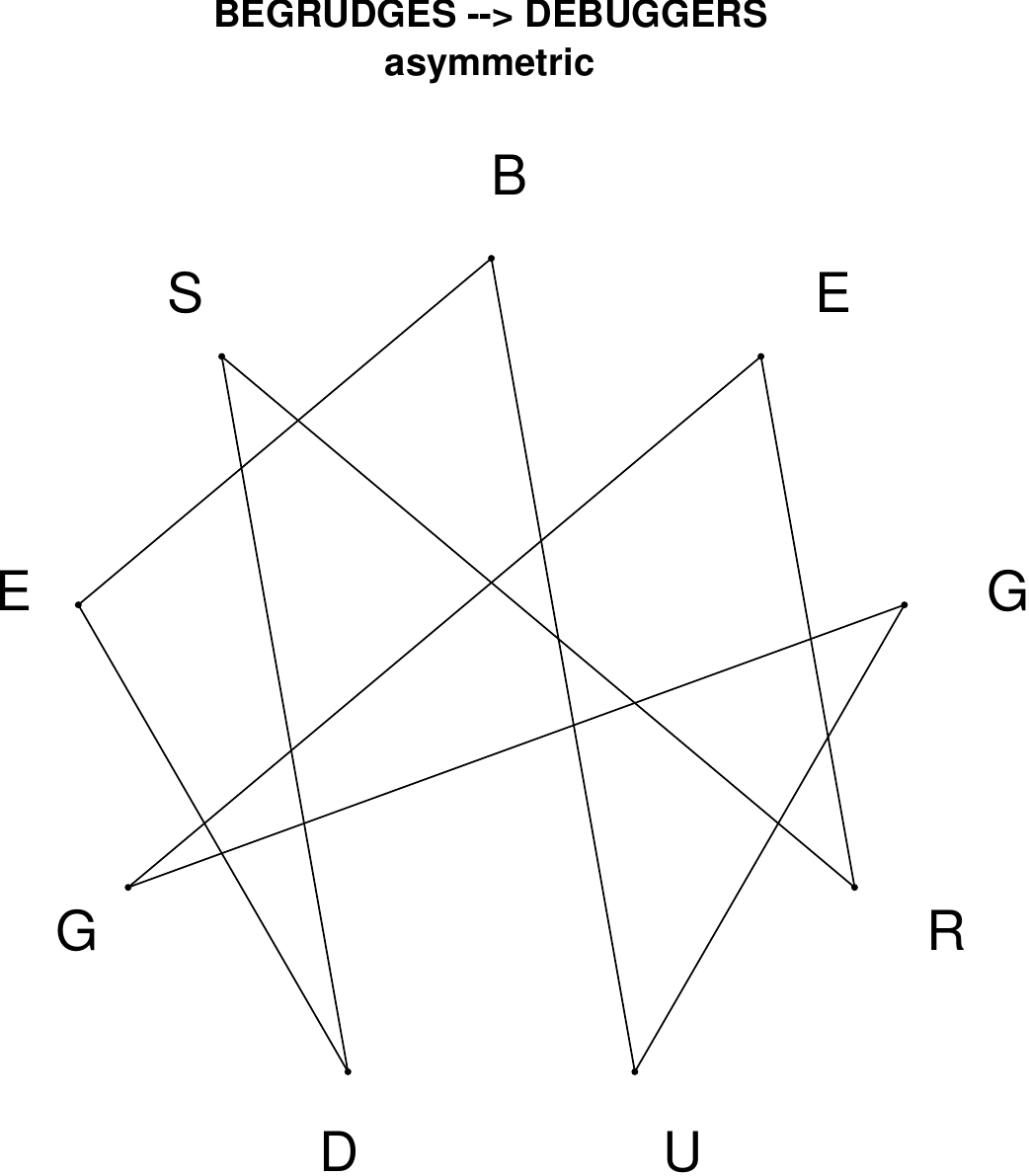}
\end{subfigure}
\hfill
\begin{subfigure}[T]{0.19\textwidth}
\centering
\includegraphics[width=\textwidth]{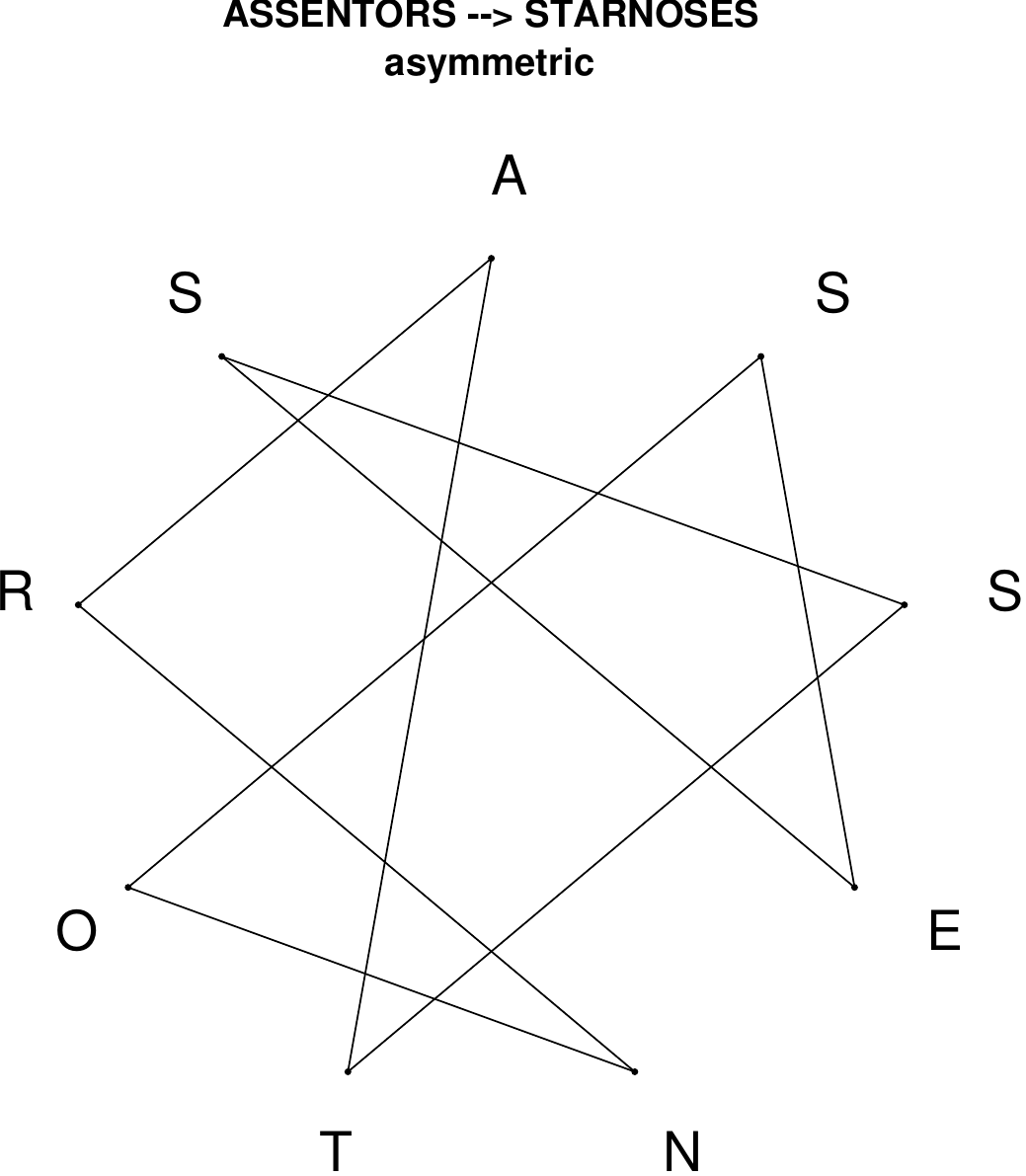}
\end{subfigure}
\hfill
\begin{subfigure}[T]{0.19\textwidth}
\centering
\includegraphics[width=\textwidth]{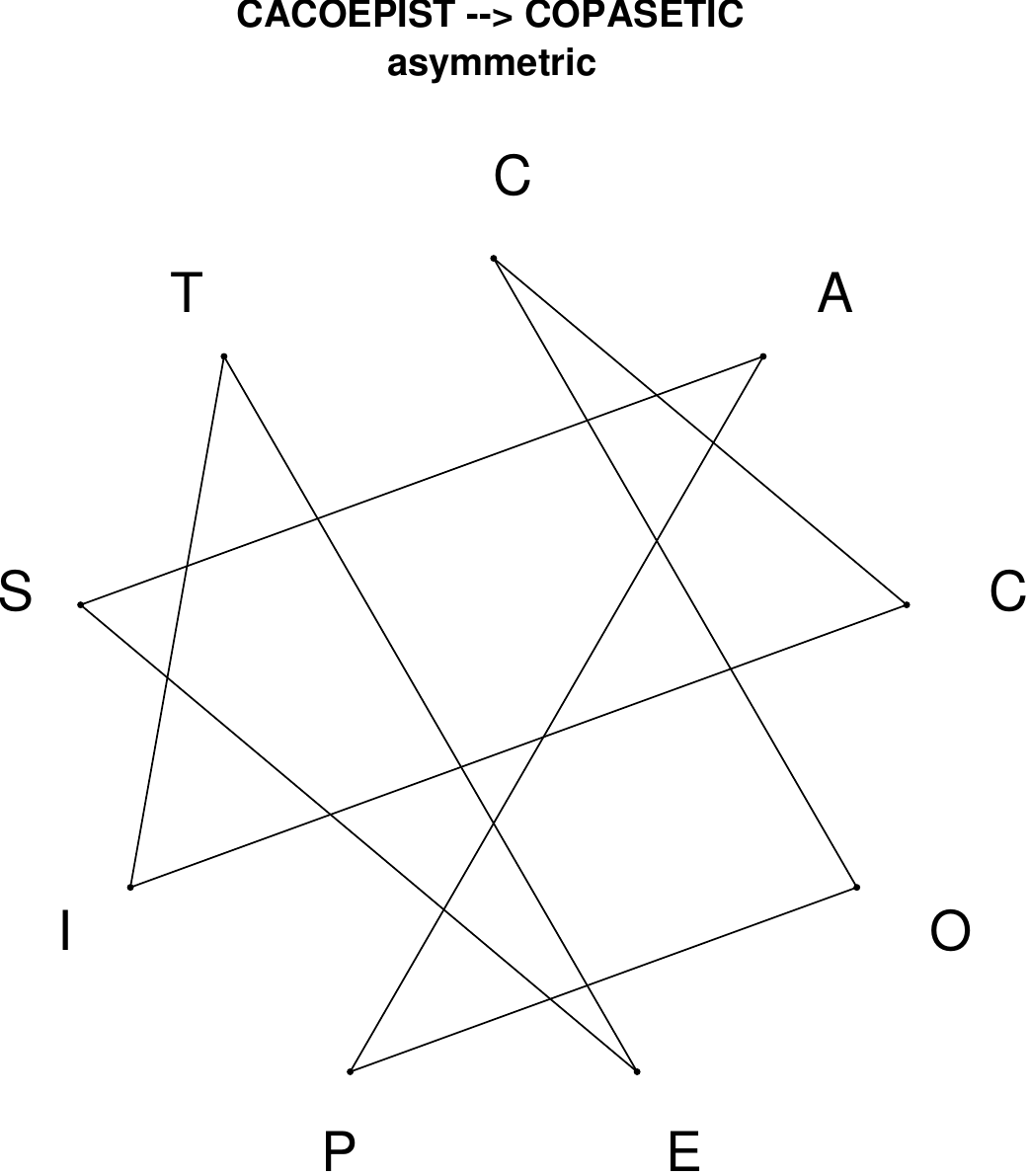}
\end{subfigure}
\hfill
\begin{subfigure}[T]{0.19\textwidth}
\centering
\includegraphics[width=\textwidth]{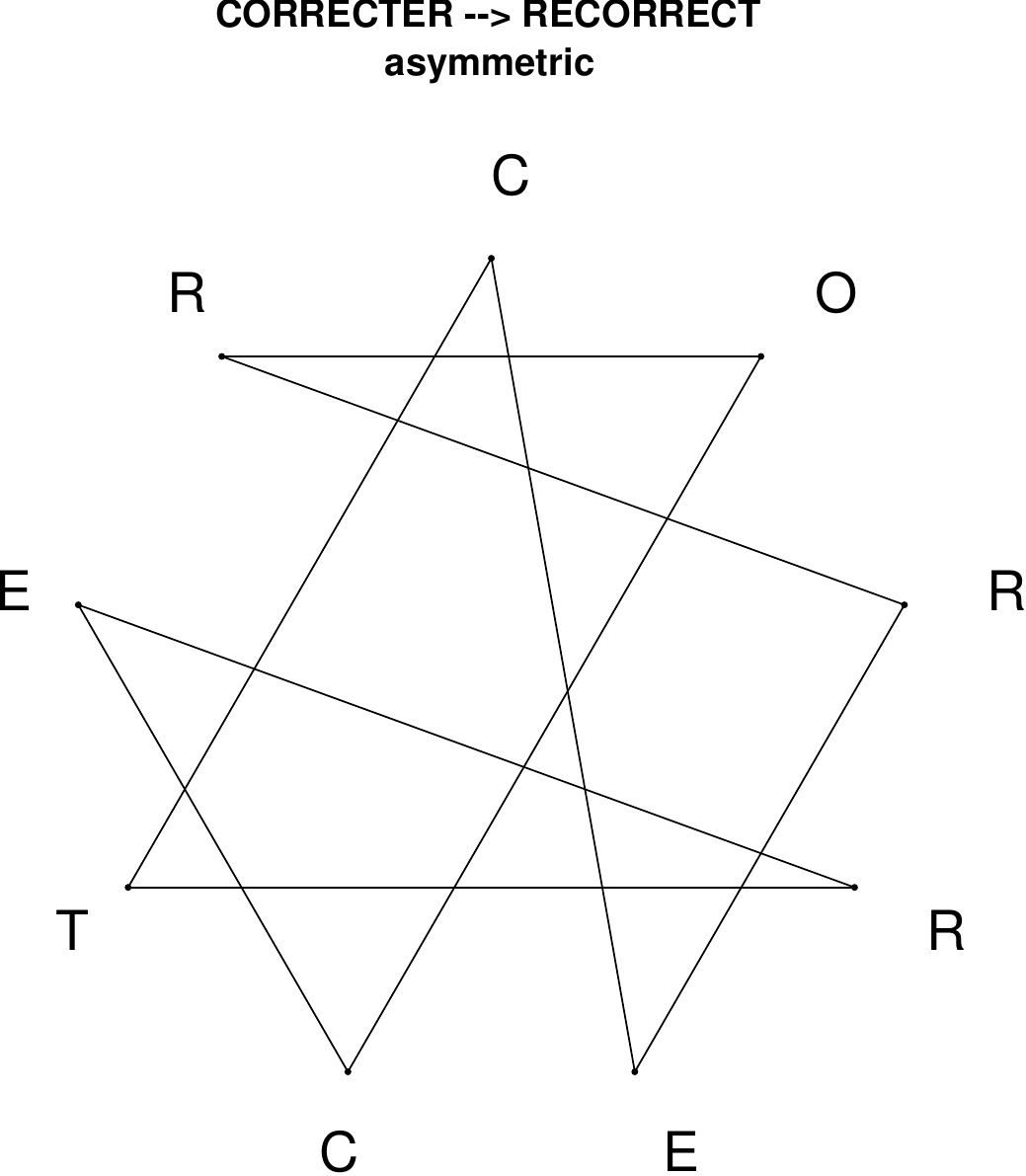}
\end{subfigure}
\hfill
\begin{subfigure}[T]{0.19\textwidth}
\centering
\includegraphics[width=\textwidth]{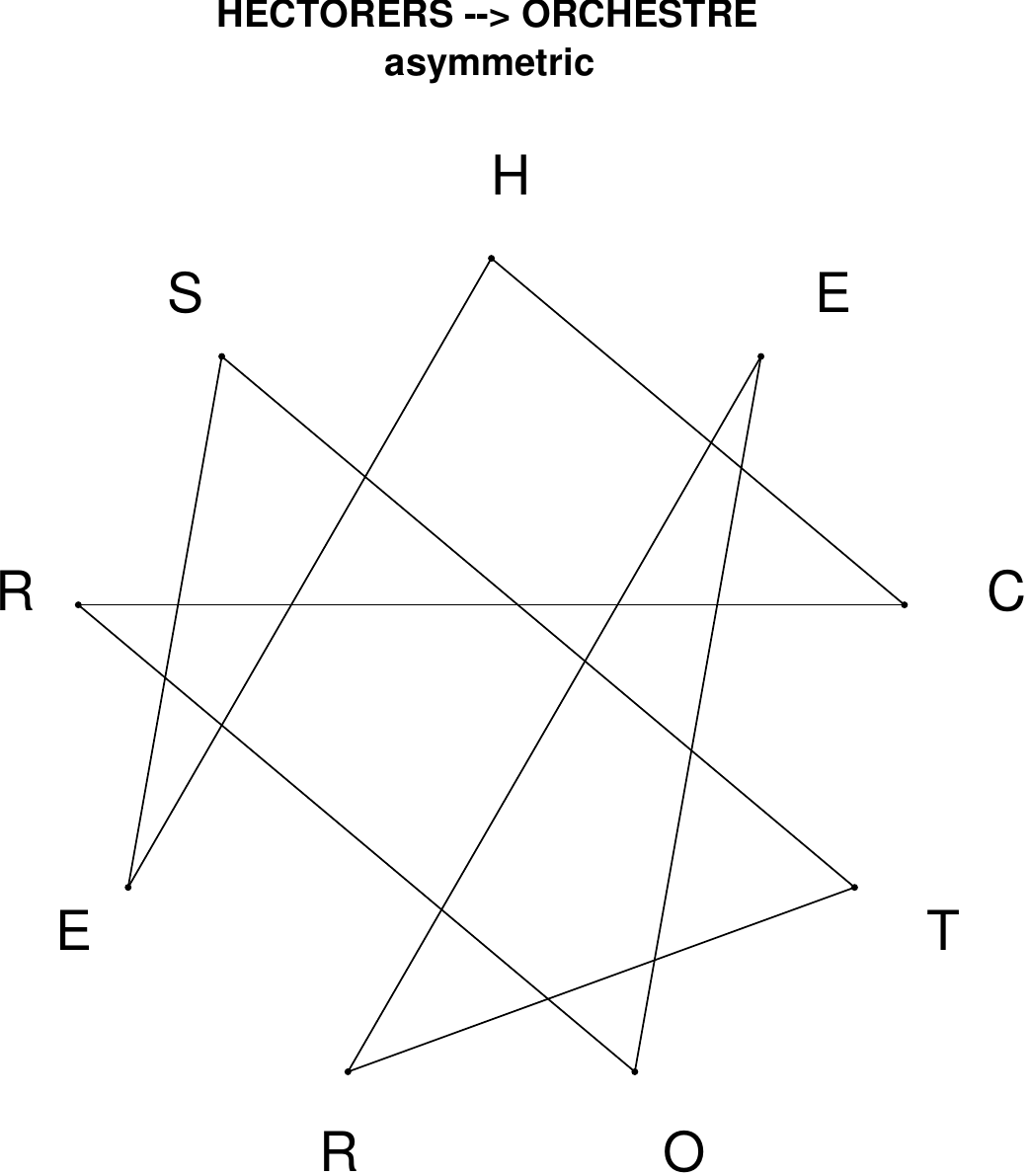}
\end{subfigure}
\end{figure}

\begin{figure}[H]
\centering
\begin{subfigure}[T]{0.19\textwidth}
\centering
\includegraphics[width=\textwidth]{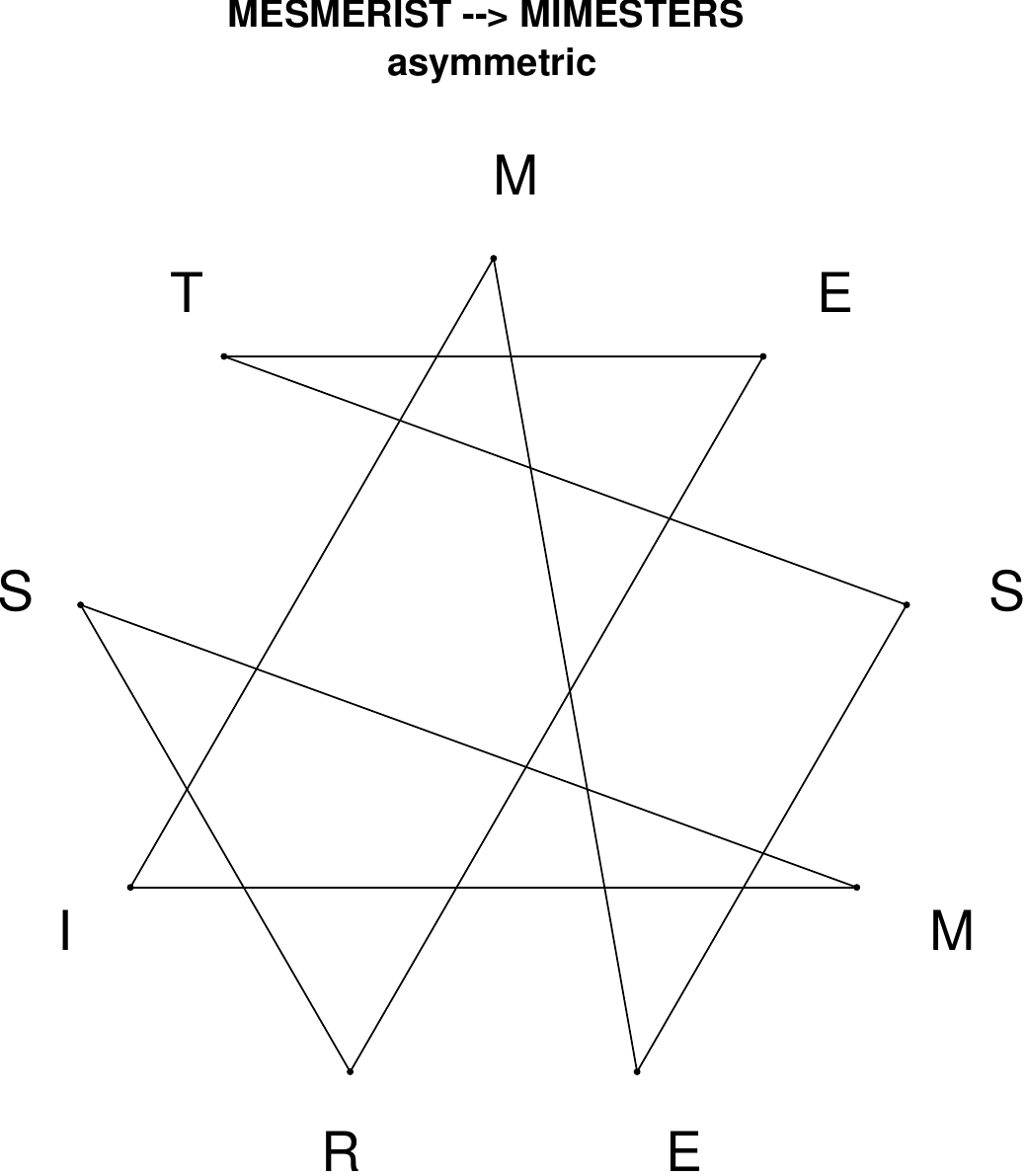}
\end{subfigure}
\hfill
\begin{subfigure}[T]{0.19\textwidth}
\centering
\includegraphics[width=\textwidth]{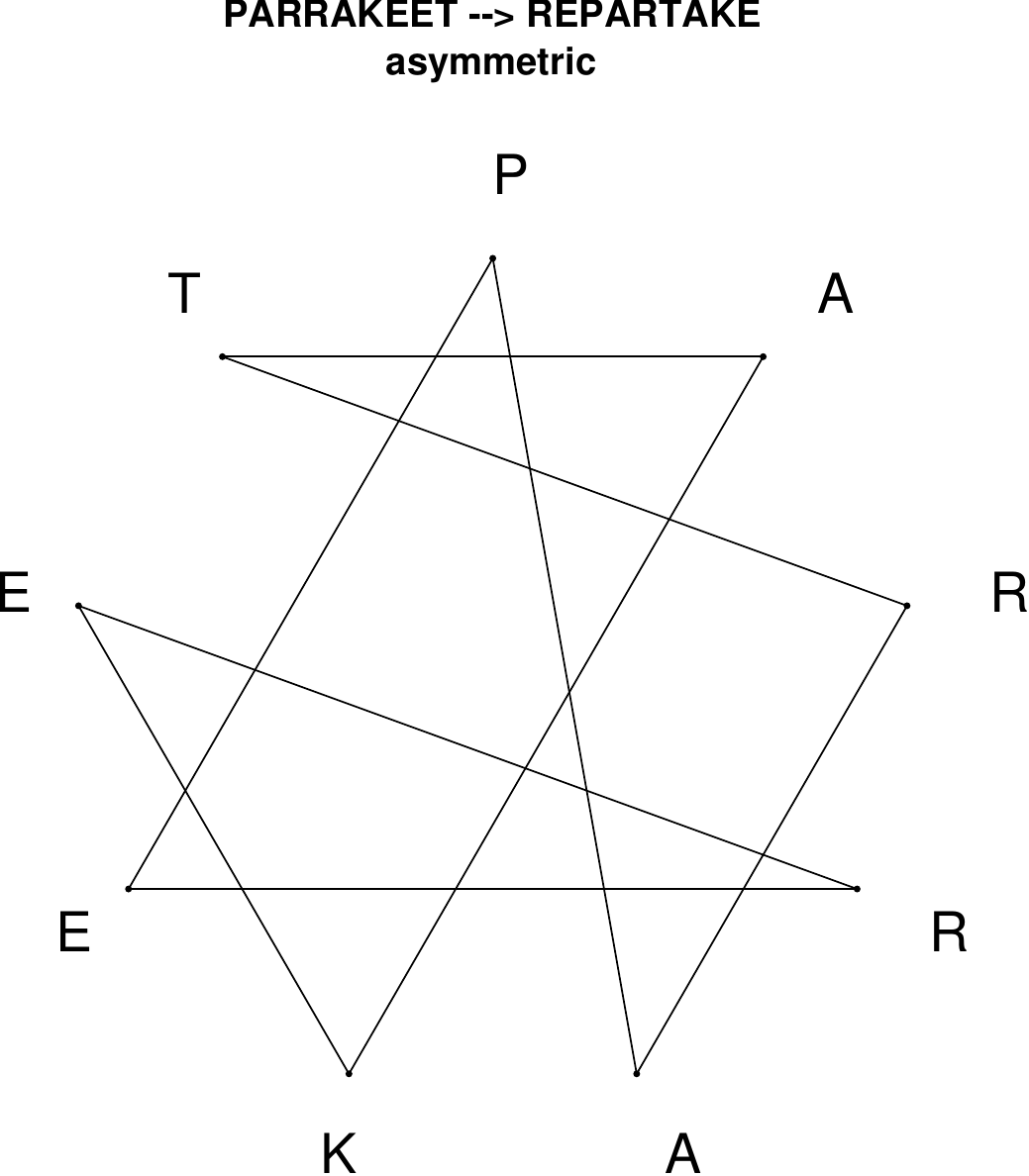}
\end{subfigure}
\hfill
\begin{subfigure}[T]{0.19\textwidth}
\centering
\includegraphics[width=\textwidth]{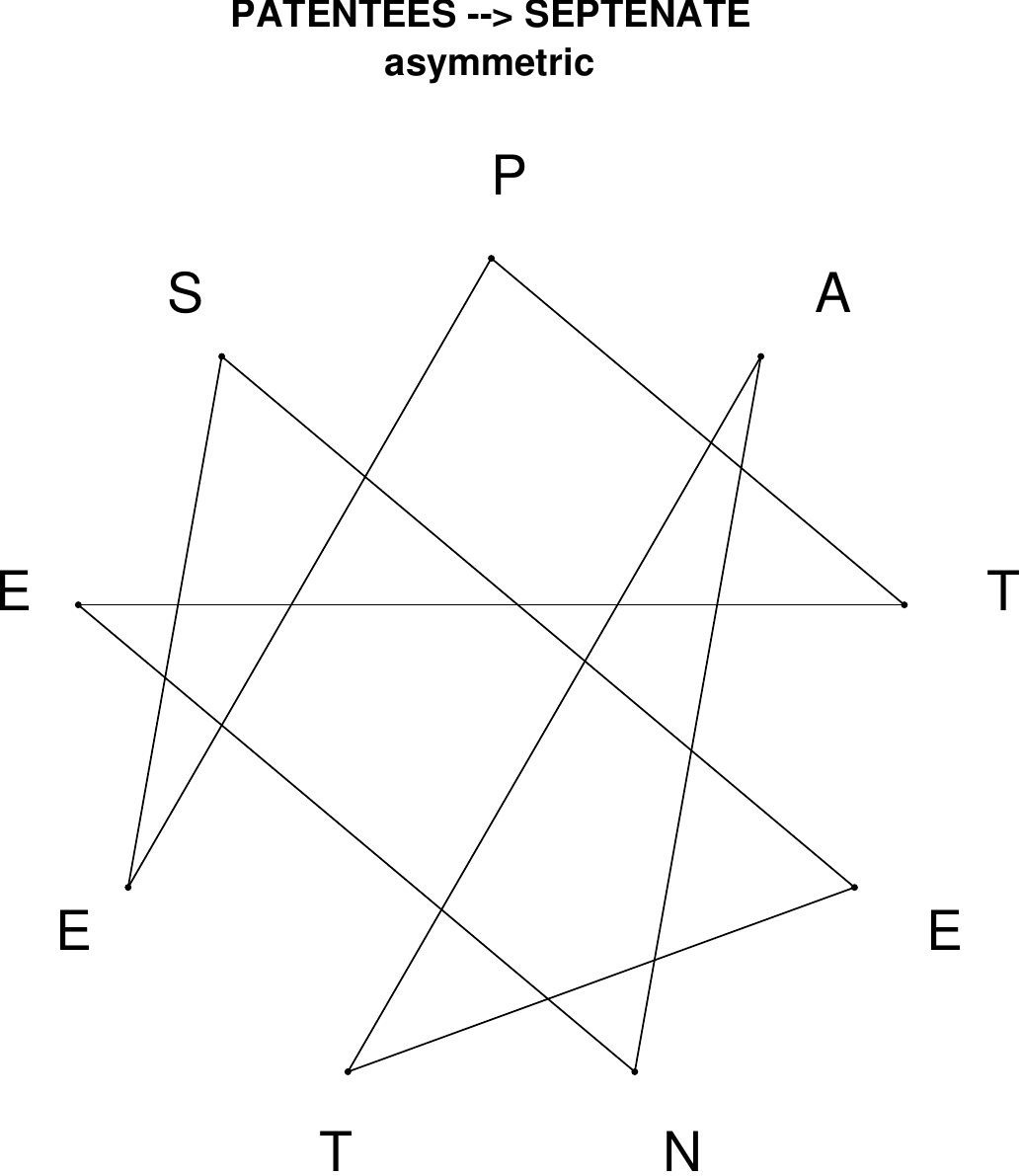}
\end{subfigure}
\hfill
\begin{subfigure}[T]{0.19\textwidth}
\centering
\includegraphics[width=\textwidth]{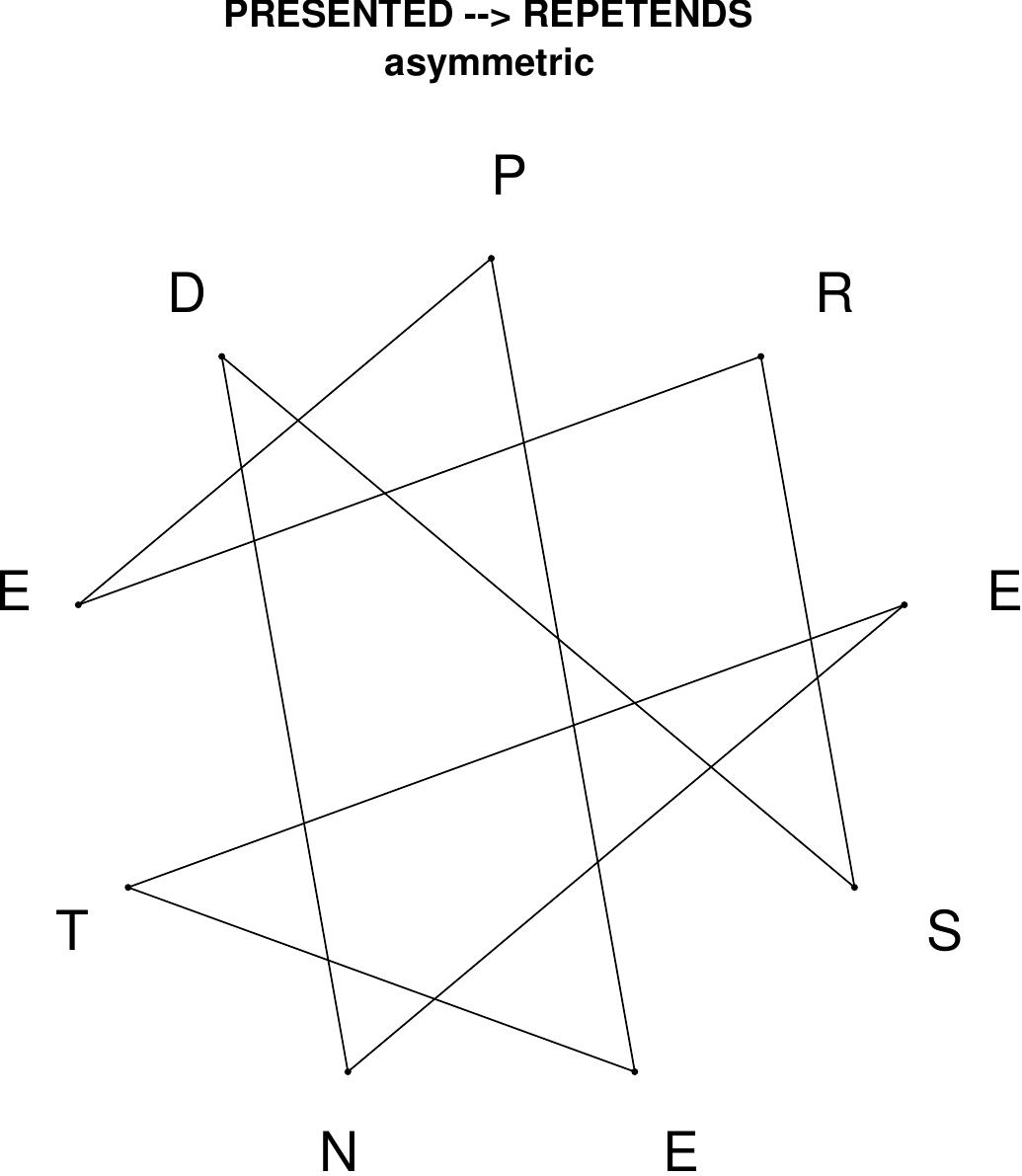}
\end{subfigure}
\hfill
\begin{subfigure}[T]{0.19\textwidth}
\centering
\includegraphics[width=\textwidth]{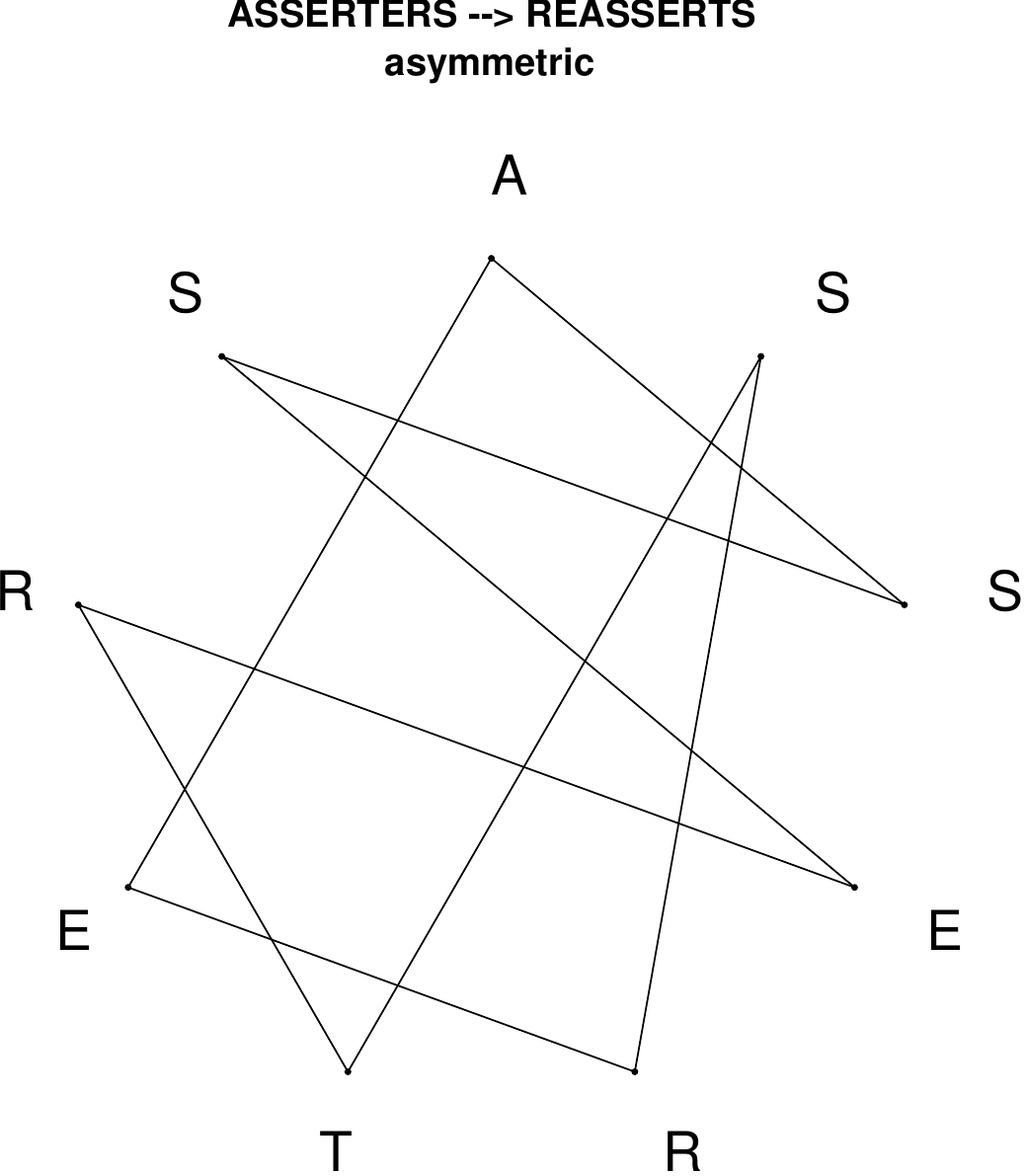}
\end{subfigure}
\end{figure}

\begin{figure}[H]
\centering
\begin{subfigure}[T]{0.19\textwidth}
\centering
\includegraphics[width=\textwidth]{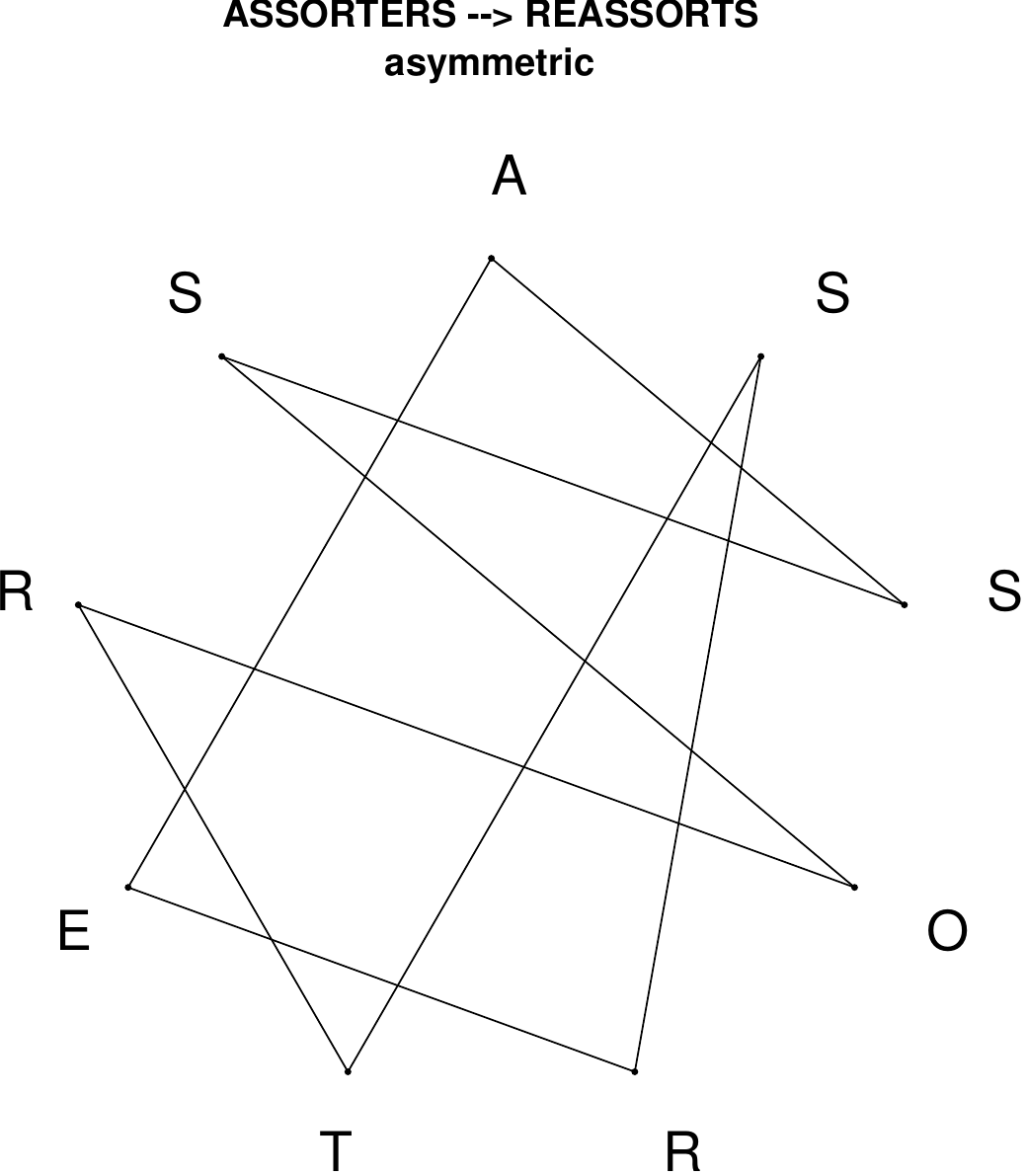}
\end{subfigure}
\hfill
\begin{subfigure}[T]{0.19\textwidth}
\centering
\includegraphics[width=\textwidth]{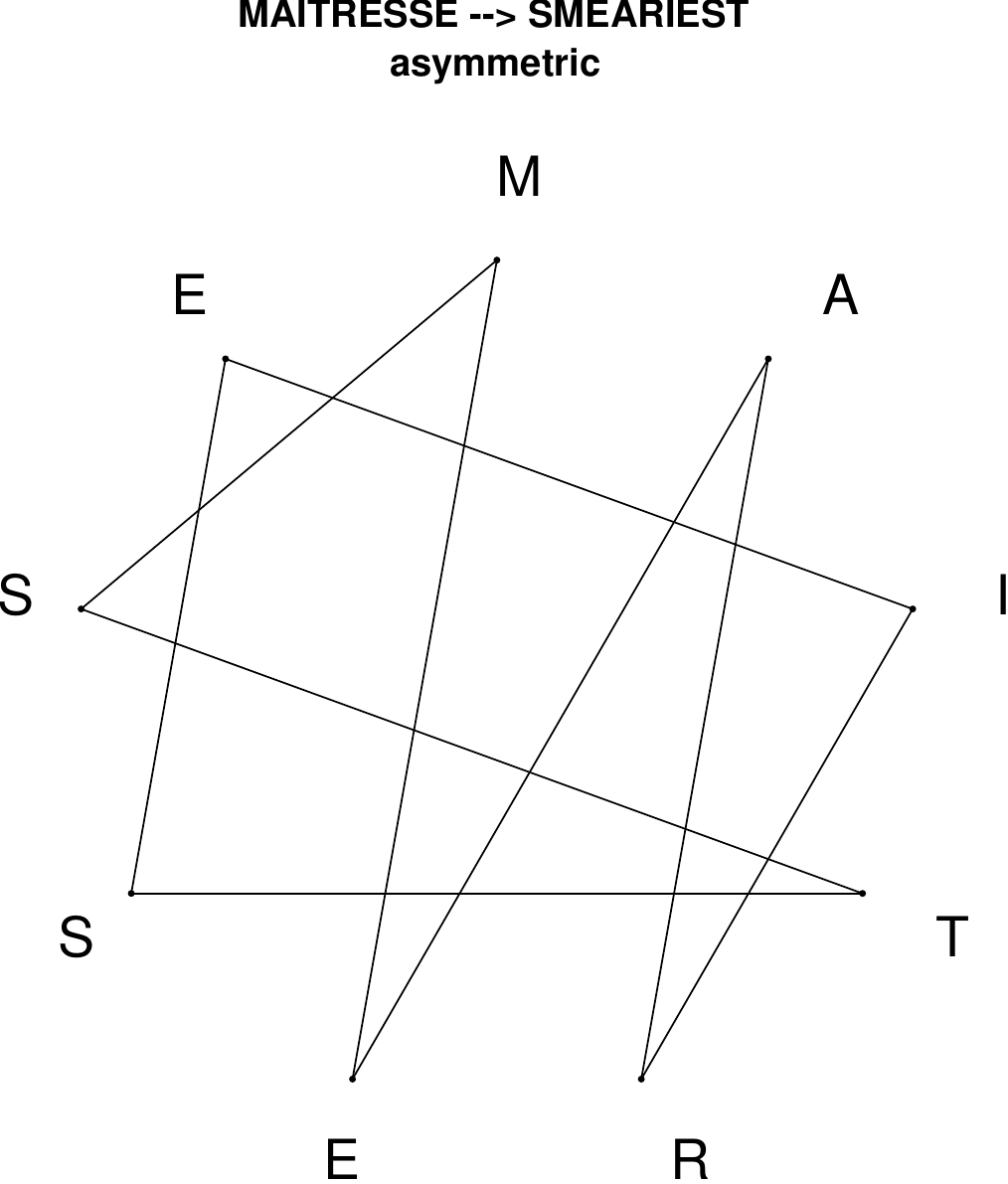}
\end{subfigure}
\hfill
\begin{subfigure}[T]{0.19\textwidth}
\centering
\includegraphics[width=\textwidth]{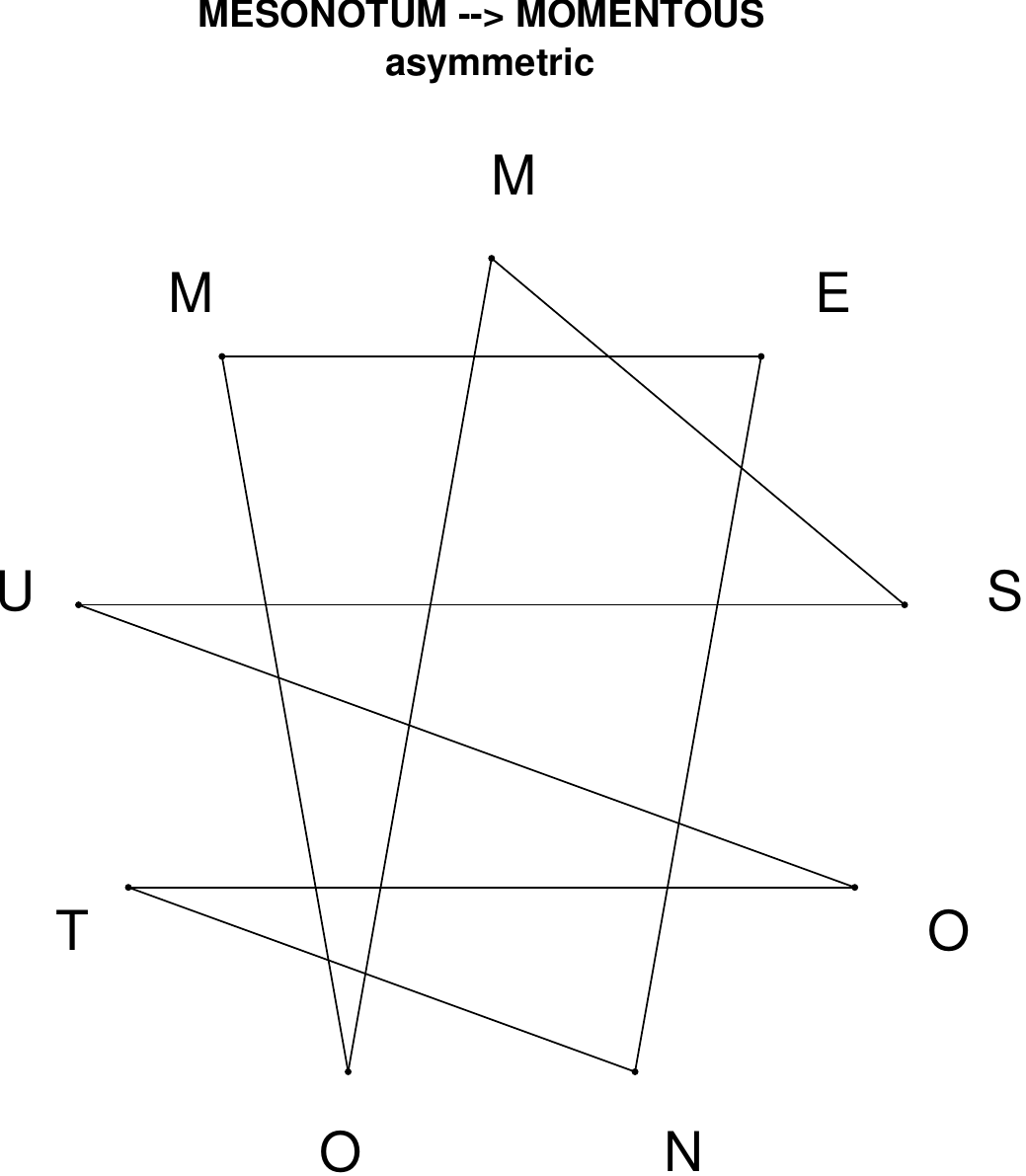}
\end{subfigure}
\hfill
\begin{subfigure}[T]{0.19\textwidth}
\centering
\includegraphics[width=\textwidth]{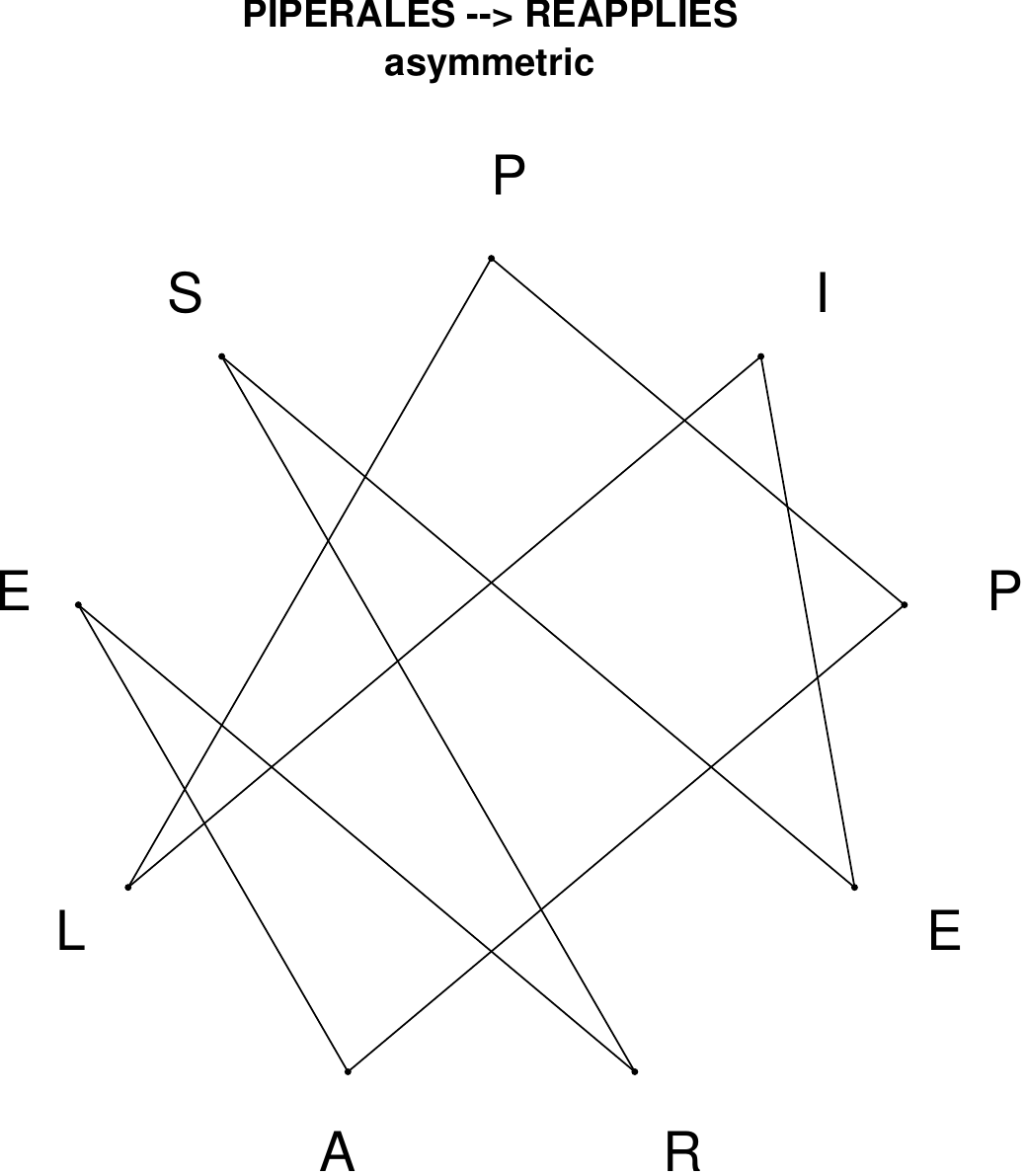}
\end{subfigure}
\hfill
\begin{subfigure}[T]{0.19\textwidth}
\centering
\includegraphics[width=\textwidth]{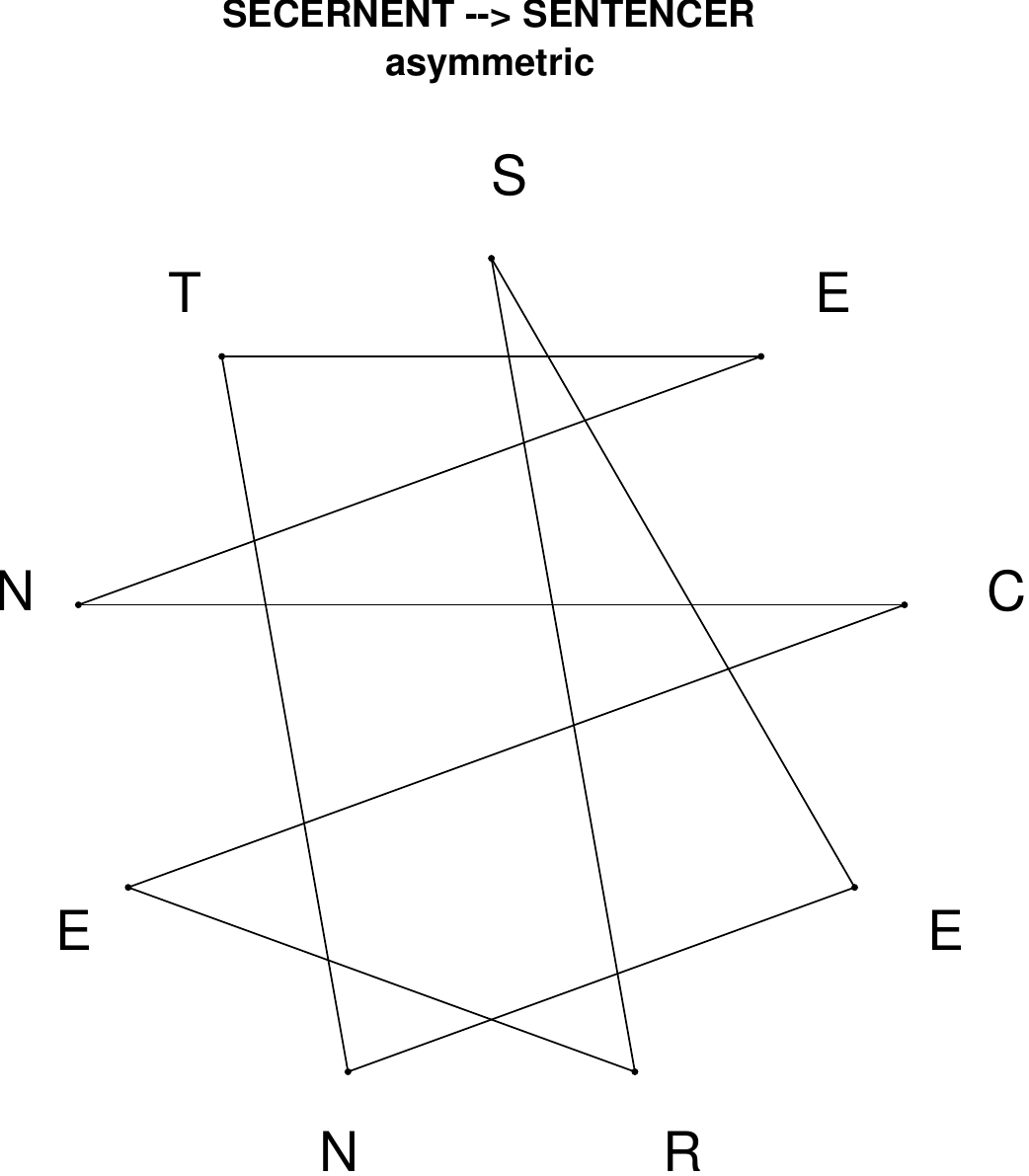}
\end{subfigure}
\end{figure}

\begin{figure}[H]
\centering
\begin{subfigure}[T]{0.19\textwidth}
\centering
\includegraphics[width=\textwidth]{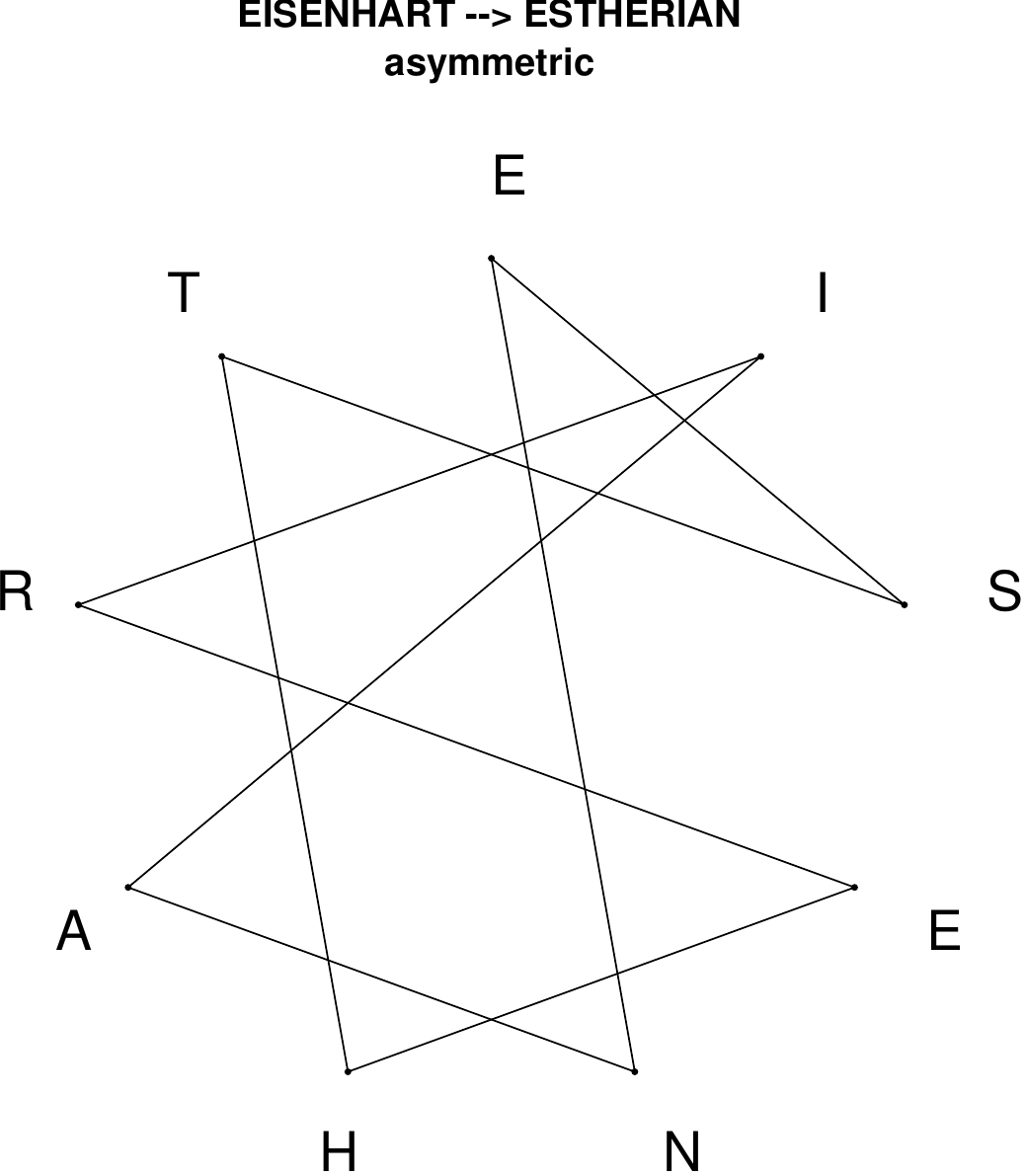}
\end{subfigure}
\hfill
\begin{subfigure}[T]{0.19\textwidth}
\centering
\includegraphics[width=\textwidth]{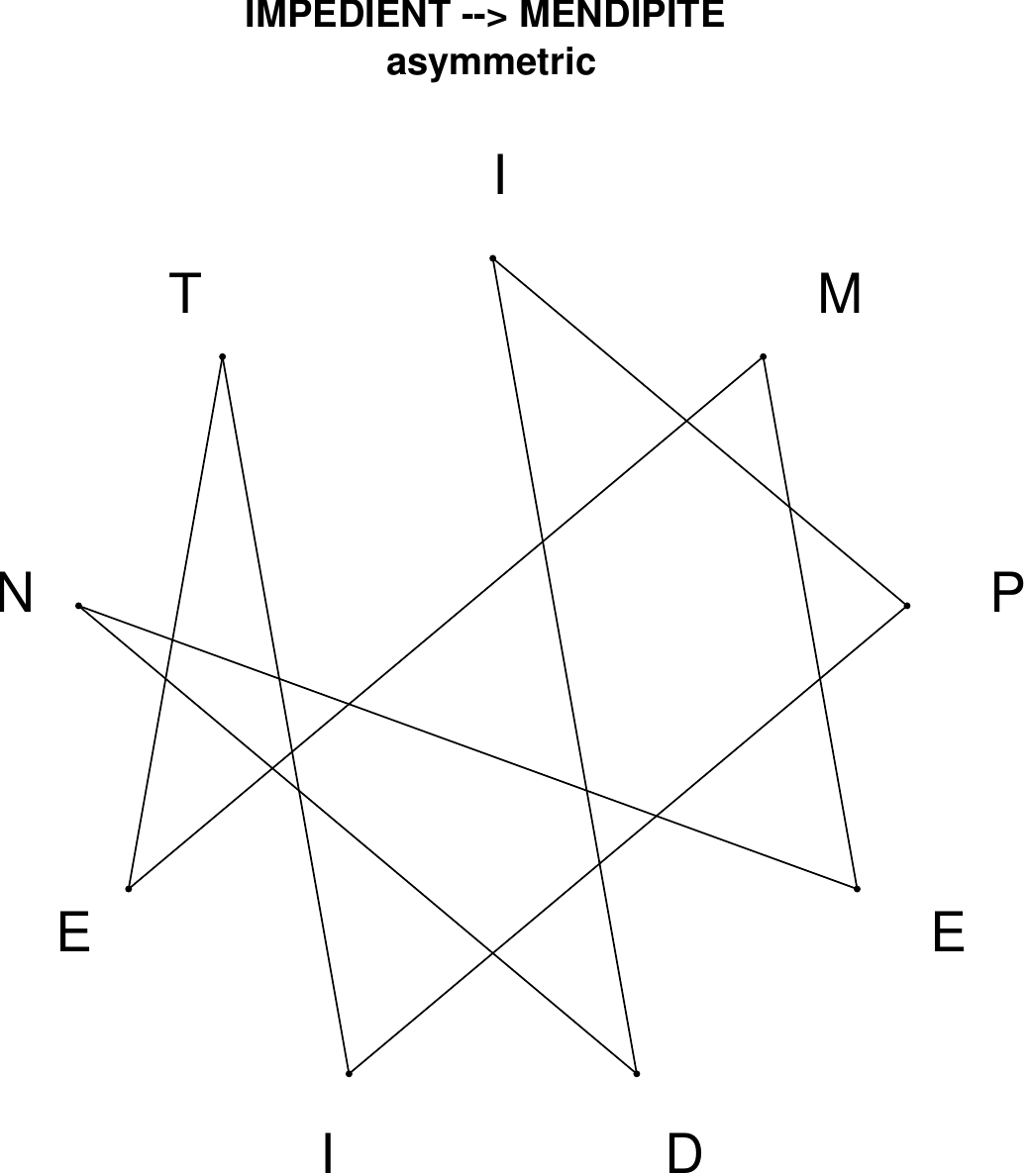}
\end{subfigure}
\hfill
\begin{subfigure}[T]{0.19\textwidth}
\centering
\includegraphics[width=\textwidth]{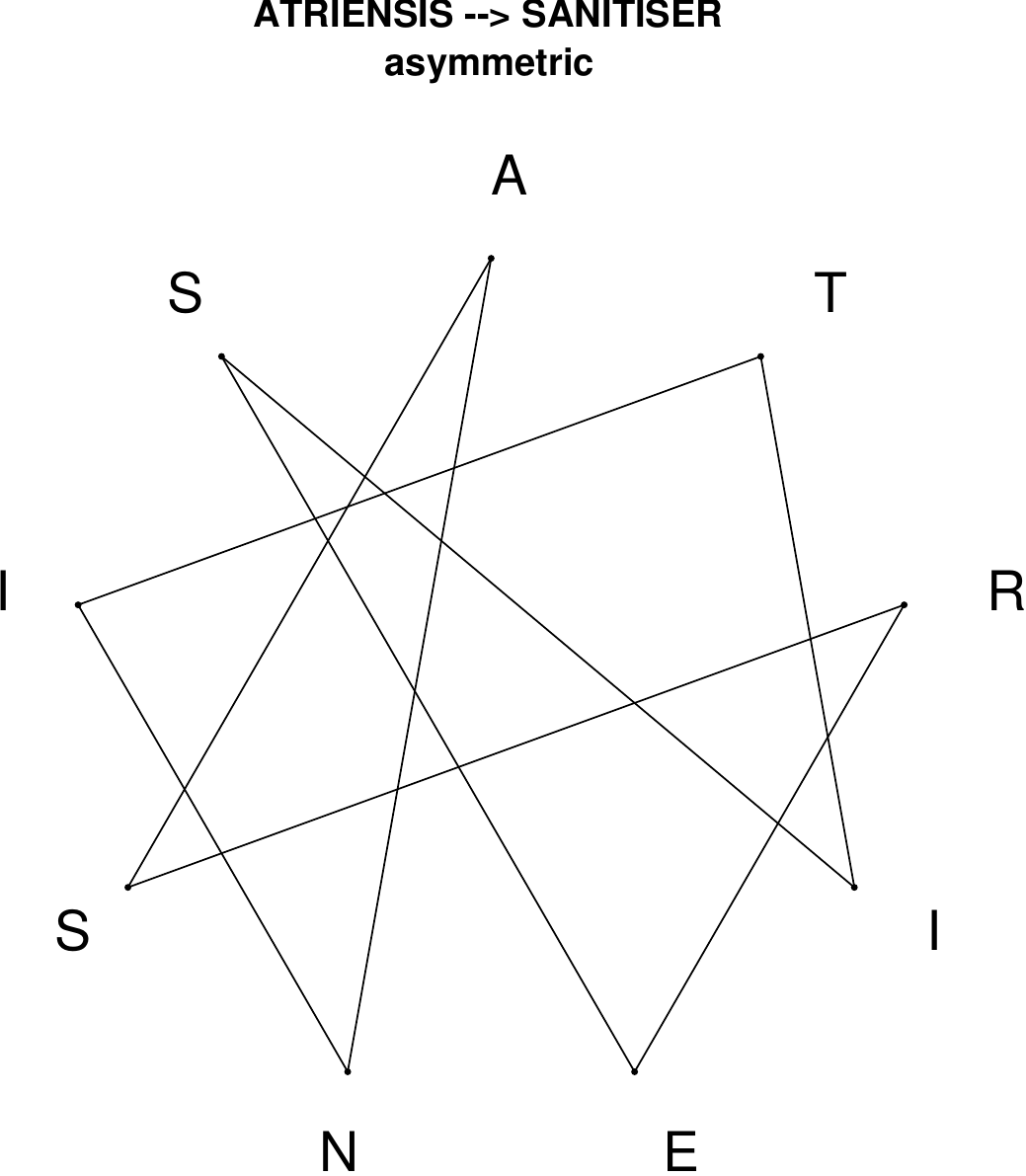}
\end{subfigure}
\hfill
\begin{subfigure}[T]{0.19\textwidth}
\centering
\includegraphics[width=\textwidth]{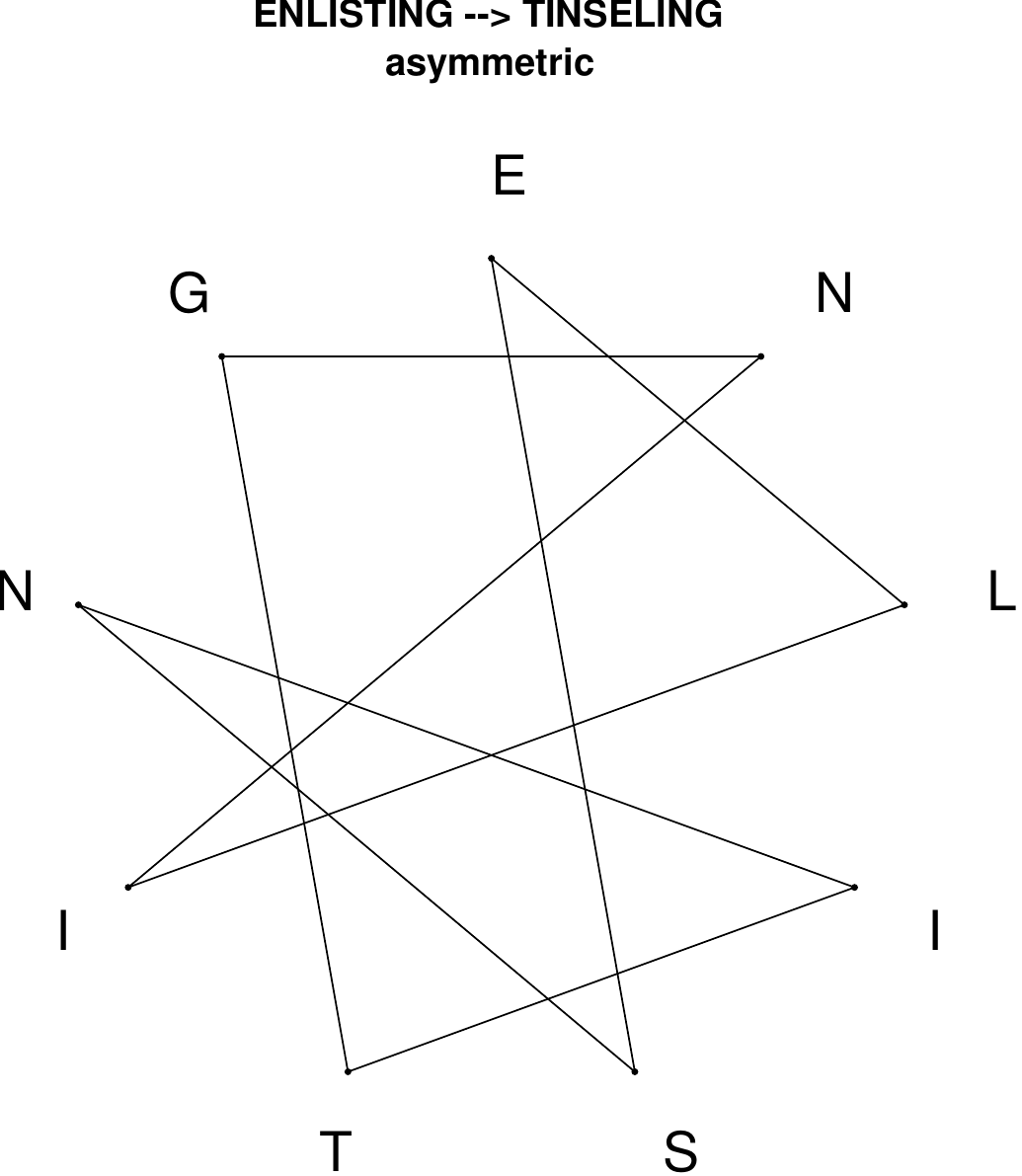}
\end{subfigure}
\hfill
\begin{subfigure}[T]{0.19\textwidth}
\centering
\includegraphics[width=\textwidth]{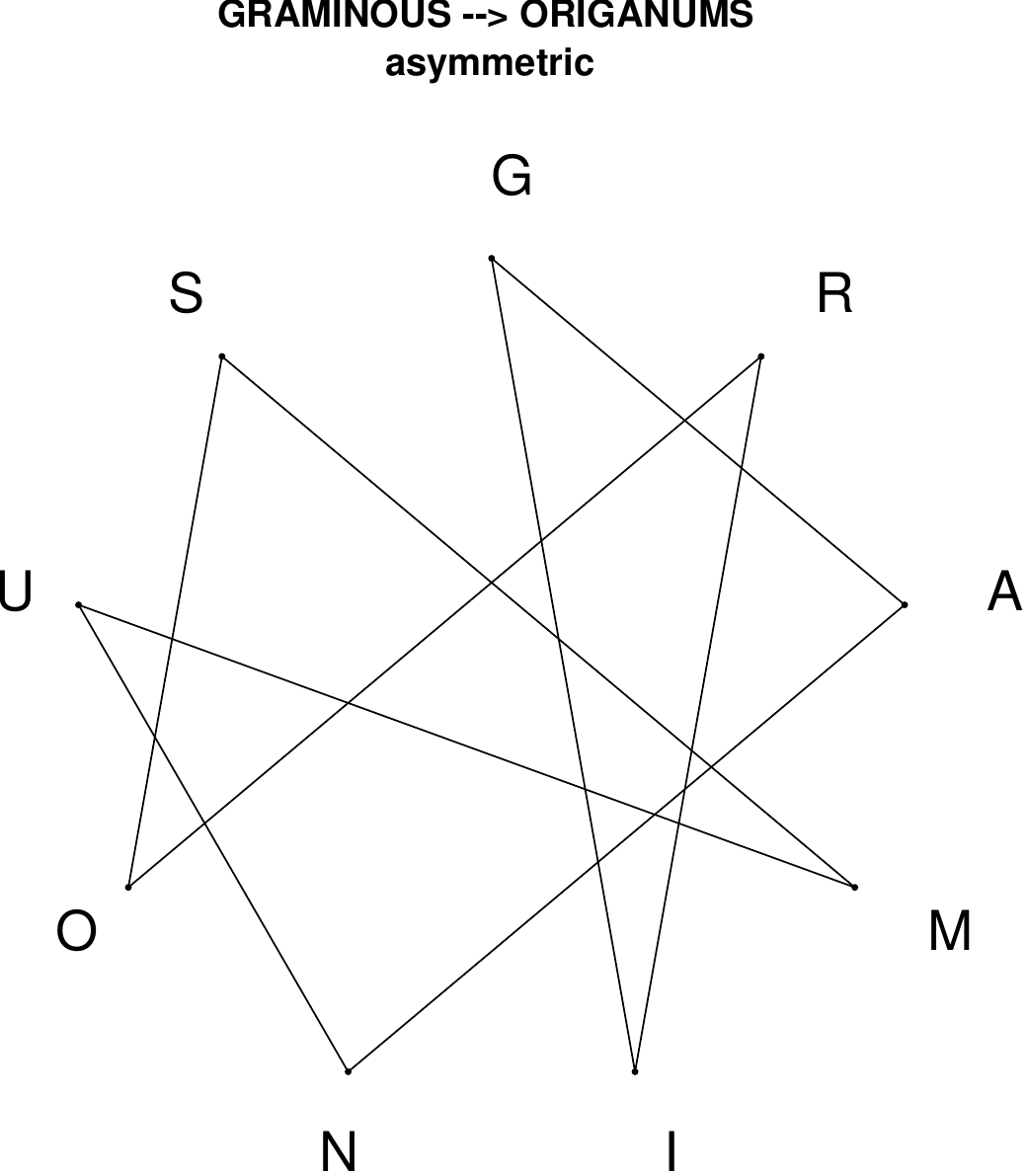}
\end{subfigure}
\end{figure}

\begin{figure}[H]
\centering
\begin{subfigure}[T]{0.19\textwidth}
\centering
\includegraphics[width=\textwidth]{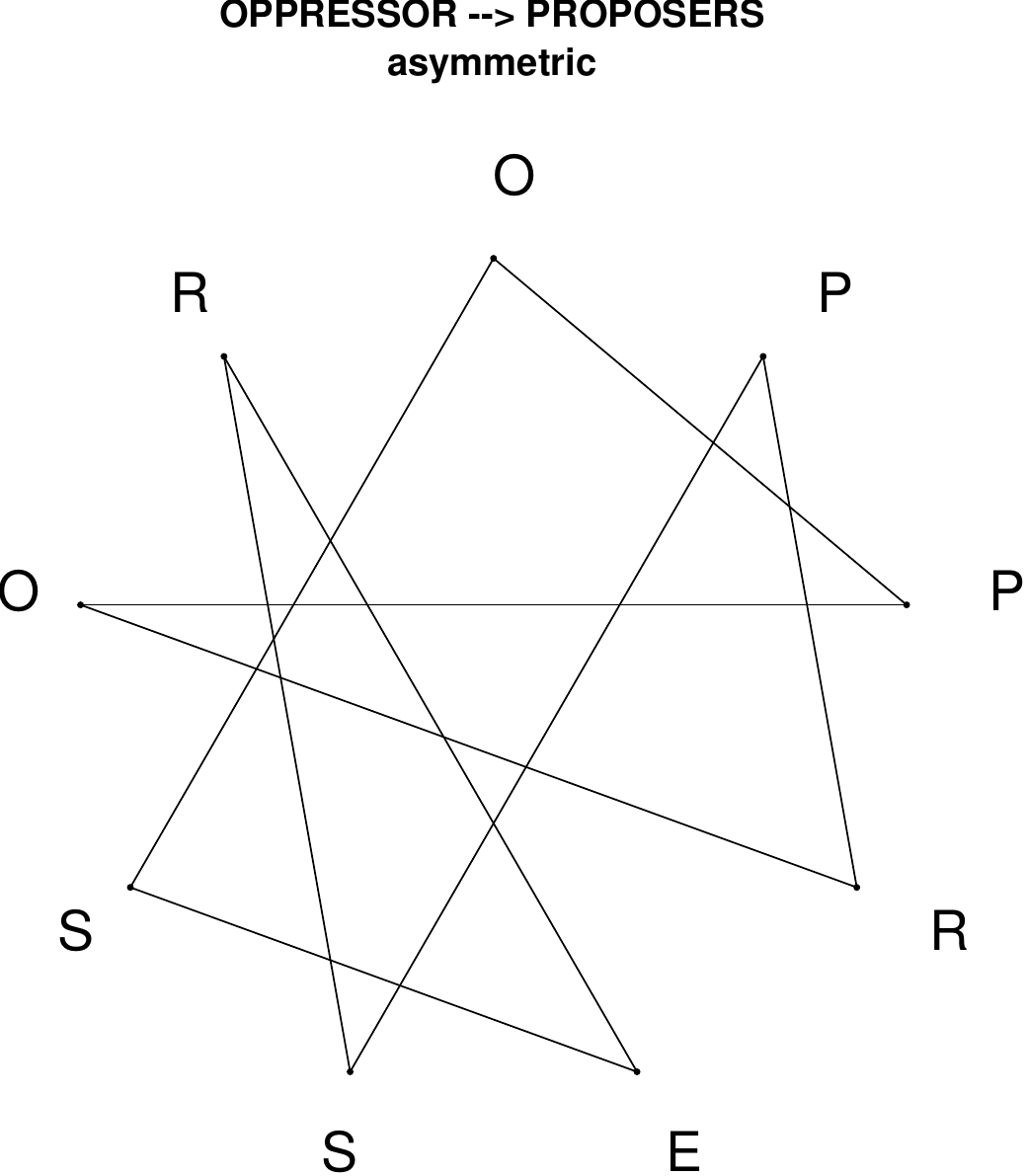}
\end{subfigure}
\hfill
\begin{subfigure}[T]{0.19\textwidth}
\centering
\includegraphics[width=\textwidth]{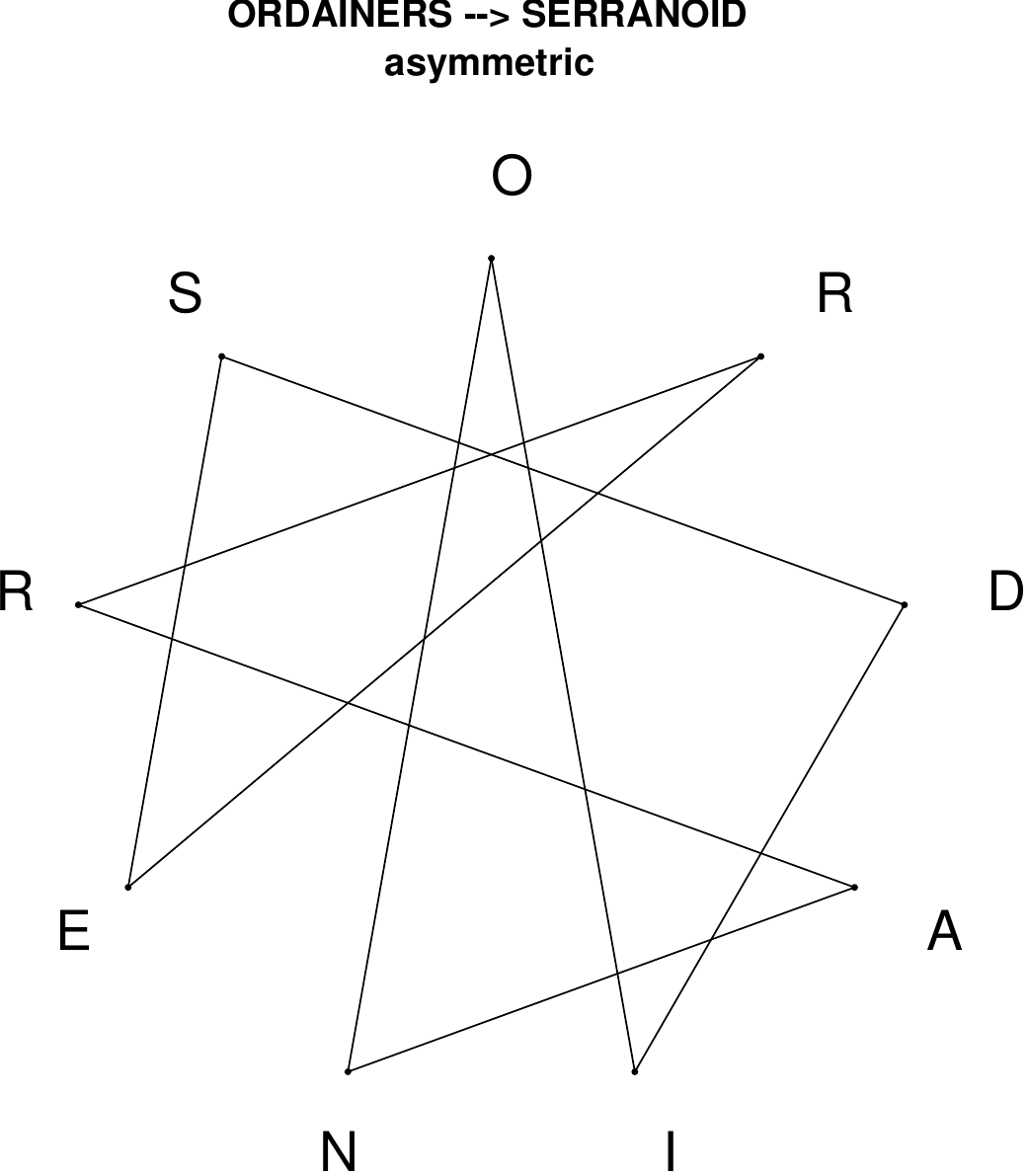}
\end{subfigure}
\hfill
\begin{subfigure}[T]{0.19\textwidth}
\centering
\includegraphics[width=\textwidth]{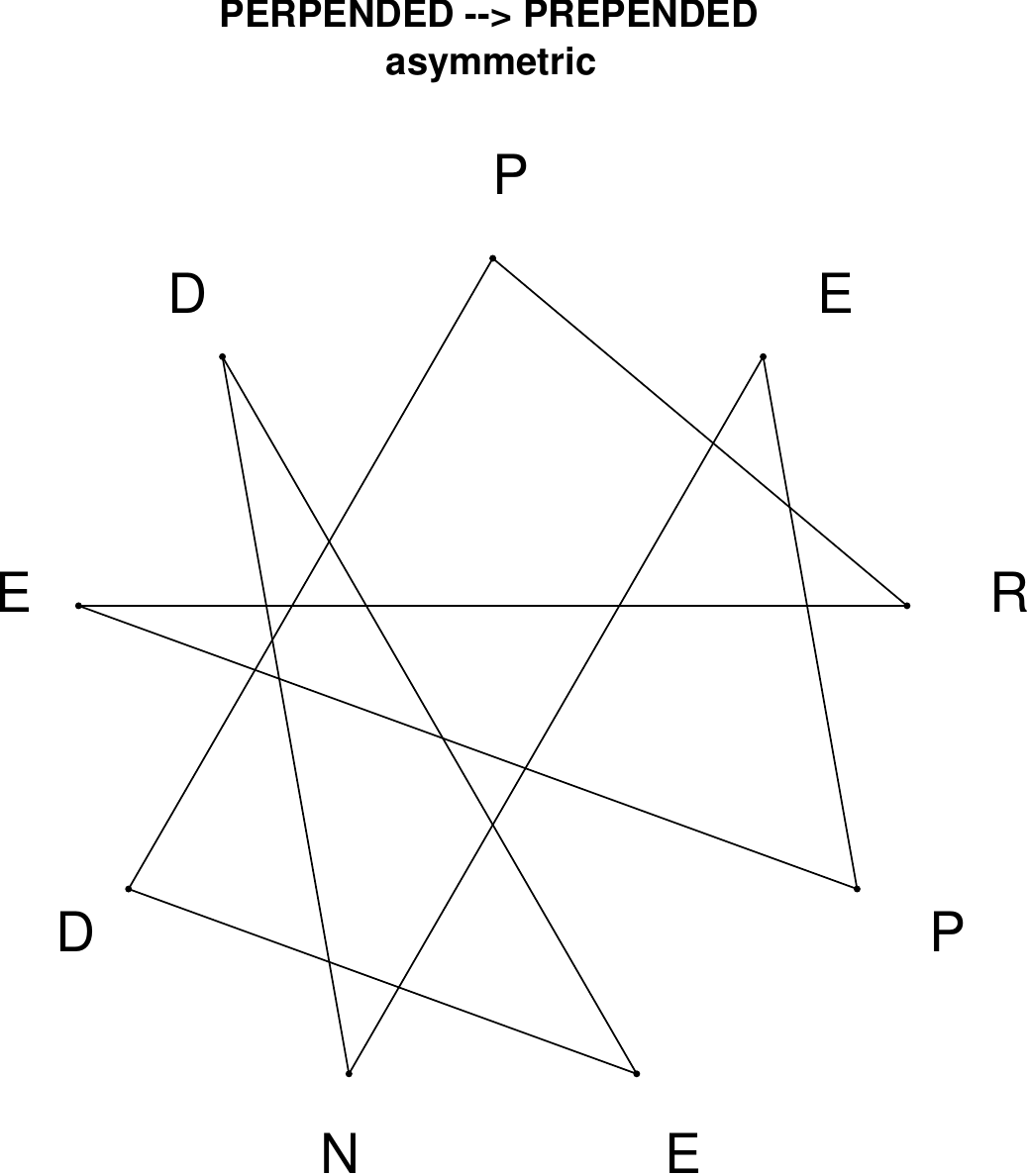}
\end{subfigure}
\hfill
\begin{subfigure}[T]{0.19\textwidth}
\centering
\includegraphics[width=\textwidth]{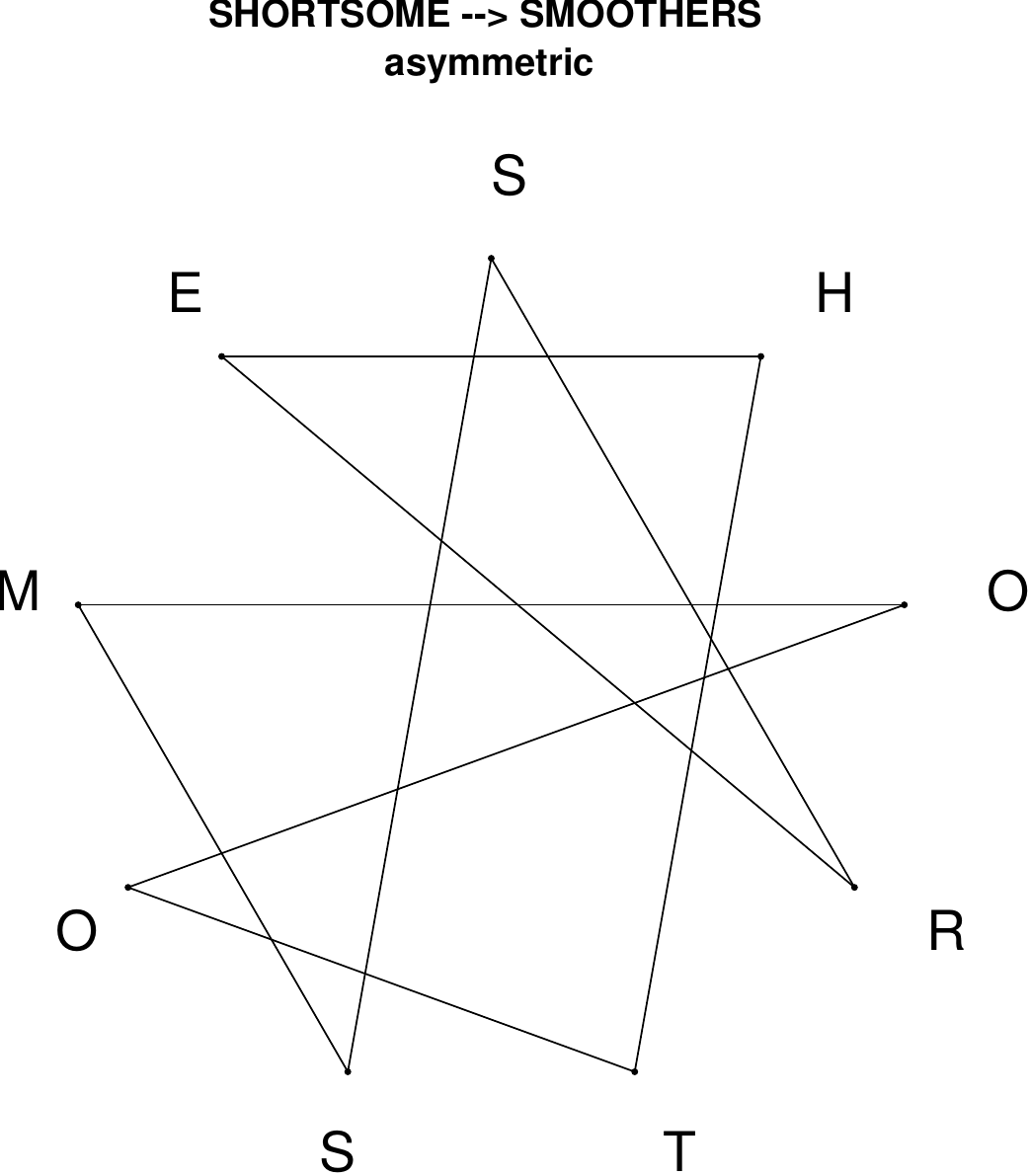}
\end{subfigure}
\hfill
\begin{subfigure}[T]{0.19\textwidth}
\centering
\includegraphics[width=\textwidth]{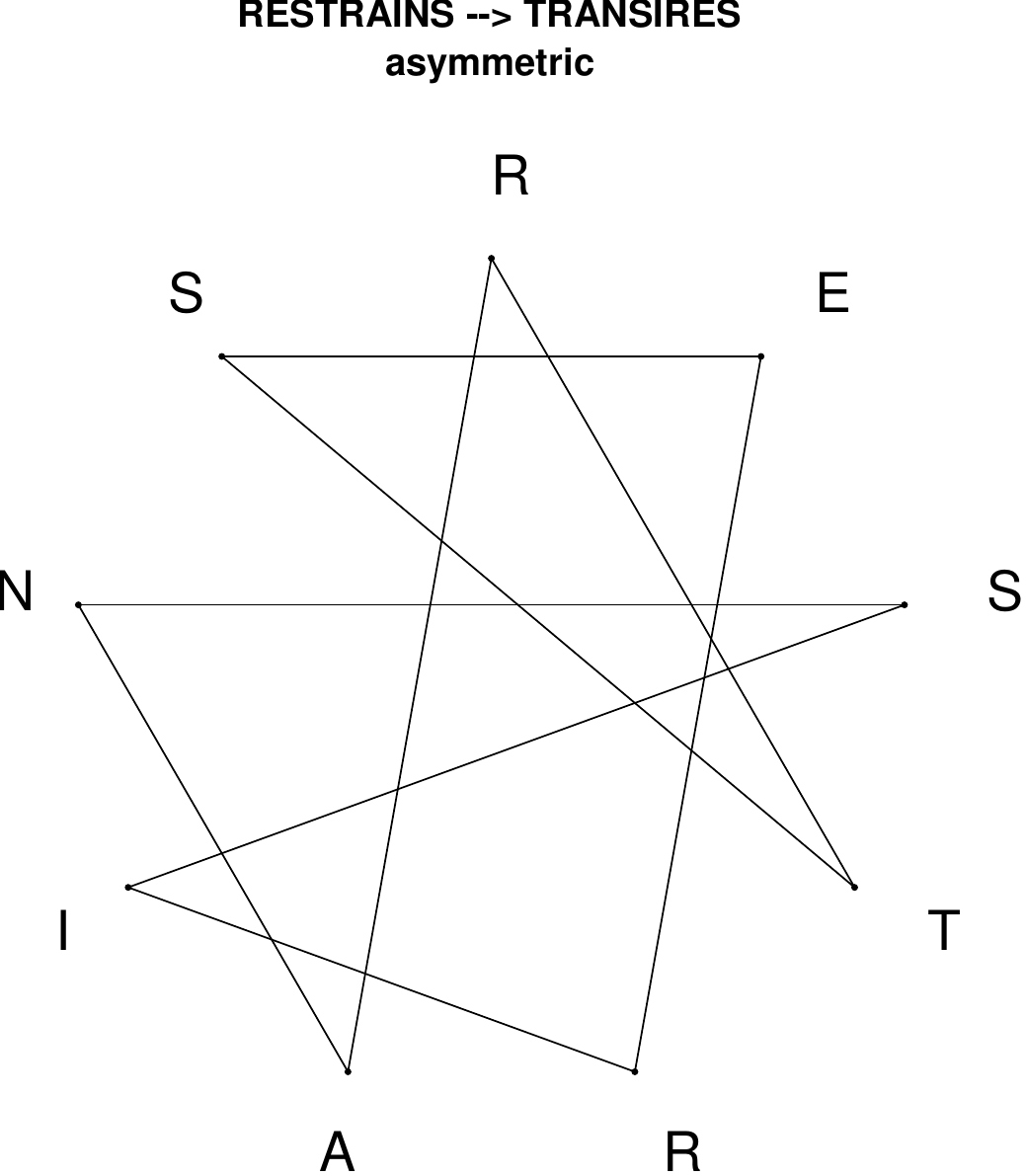}
\end{subfigure}
\end{figure}

\begin{figure}[H]
\centering
\begin{subfigure}[T]{0.19\textwidth}
\centering
\includegraphics[width=\textwidth]{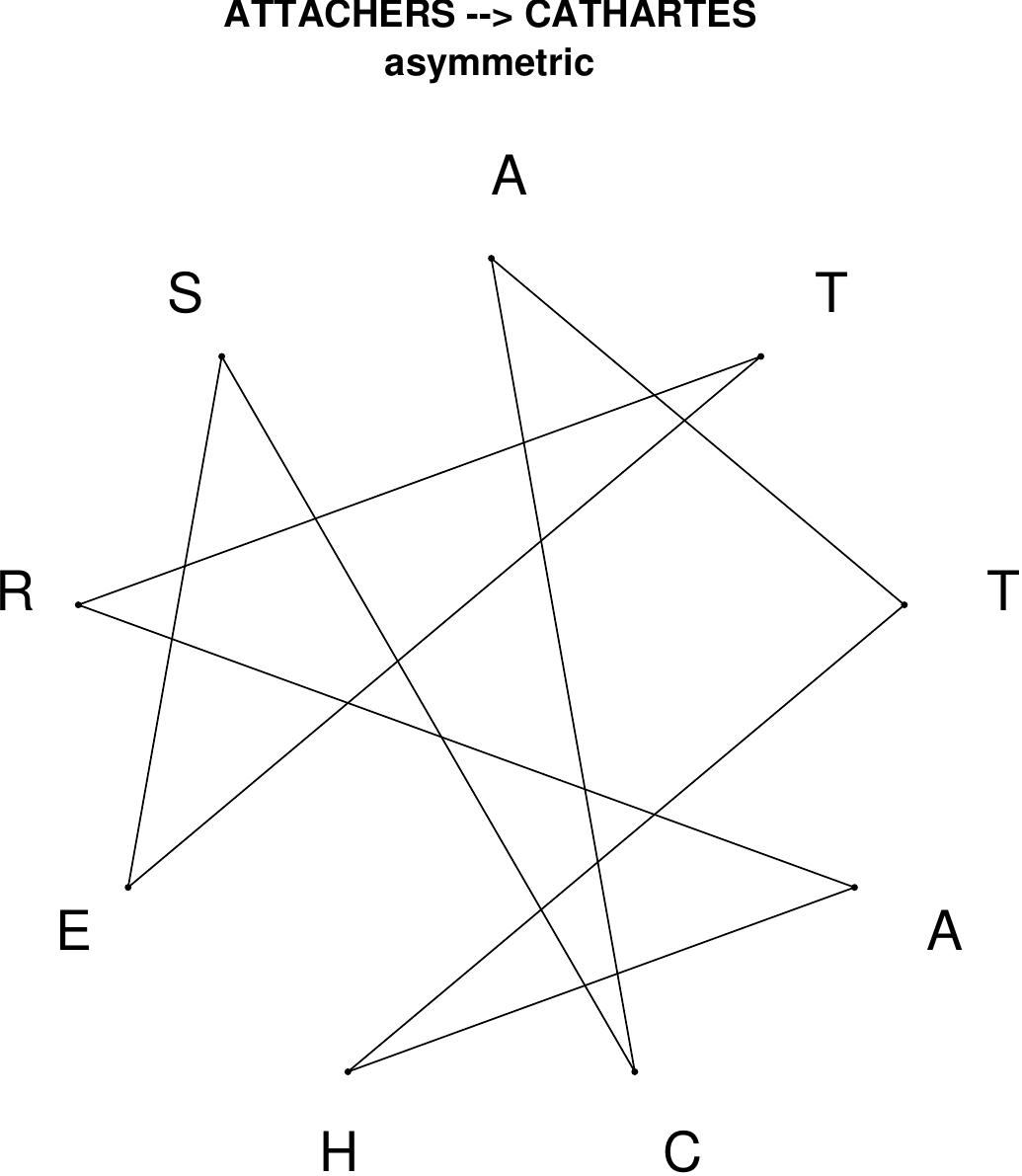}
\end{subfigure}
\hfill
\begin{subfigure}[T]{0.19\textwidth}
\centering
\includegraphics[width=\textwidth]{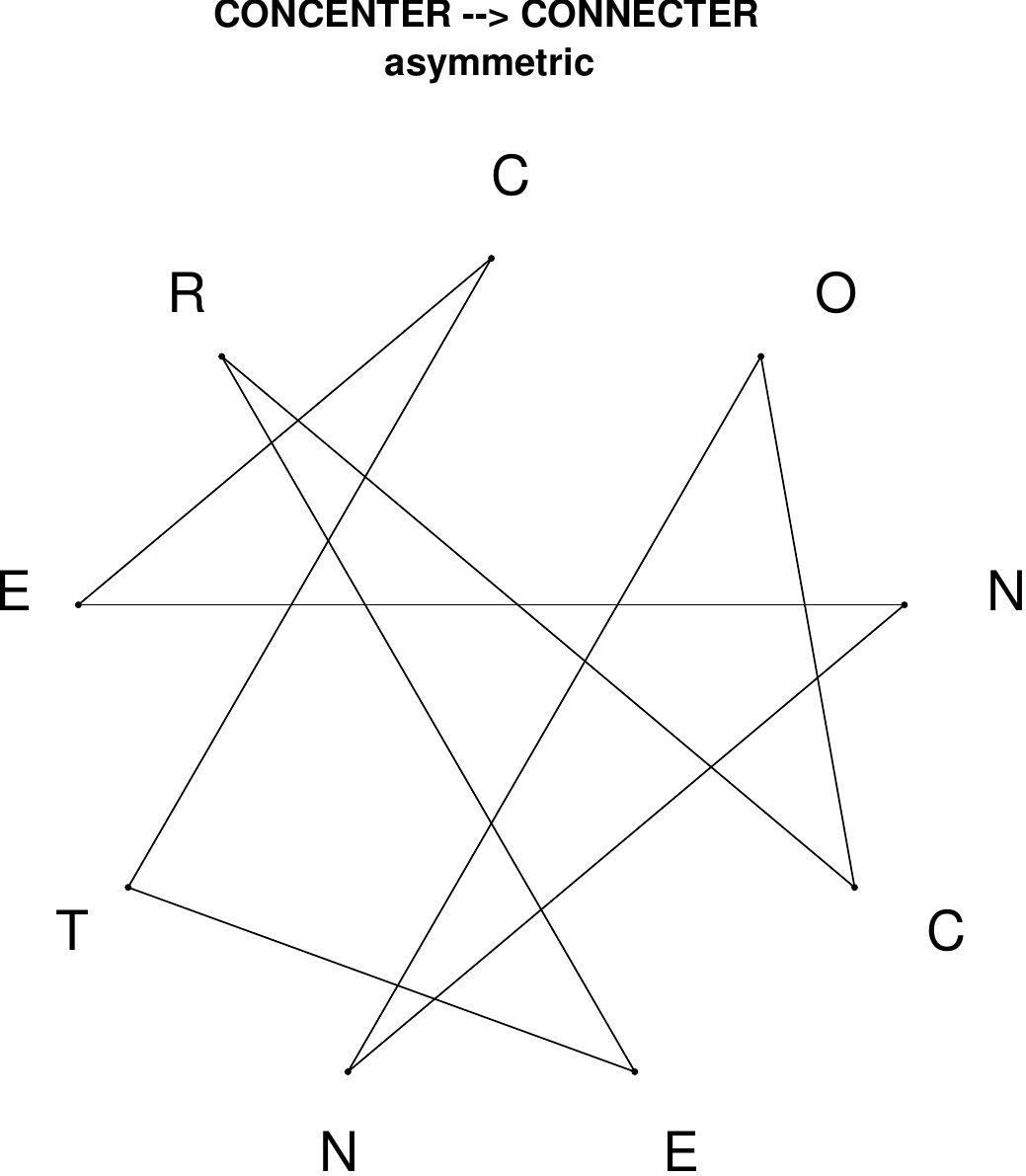}
\end{subfigure}
\hfill
\begin{subfigure}[T]{0.19\textwidth}
\centering
\includegraphics[width=\textwidth]{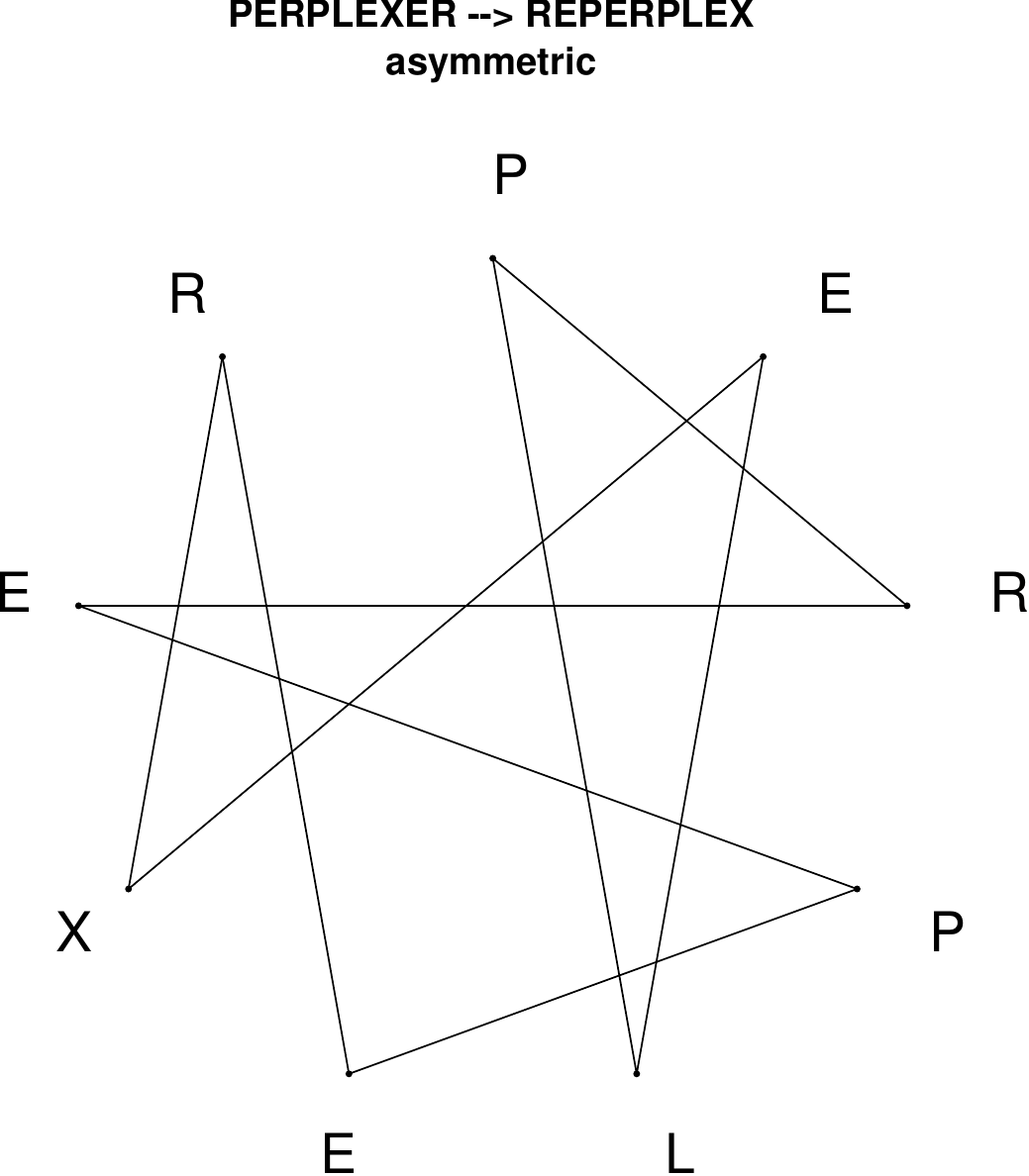}
\end{subfigure}
\hfill
\begin{subfigure}[T]{0.19\textwidth}
\centering
\includegraphics[width=\textwidth]{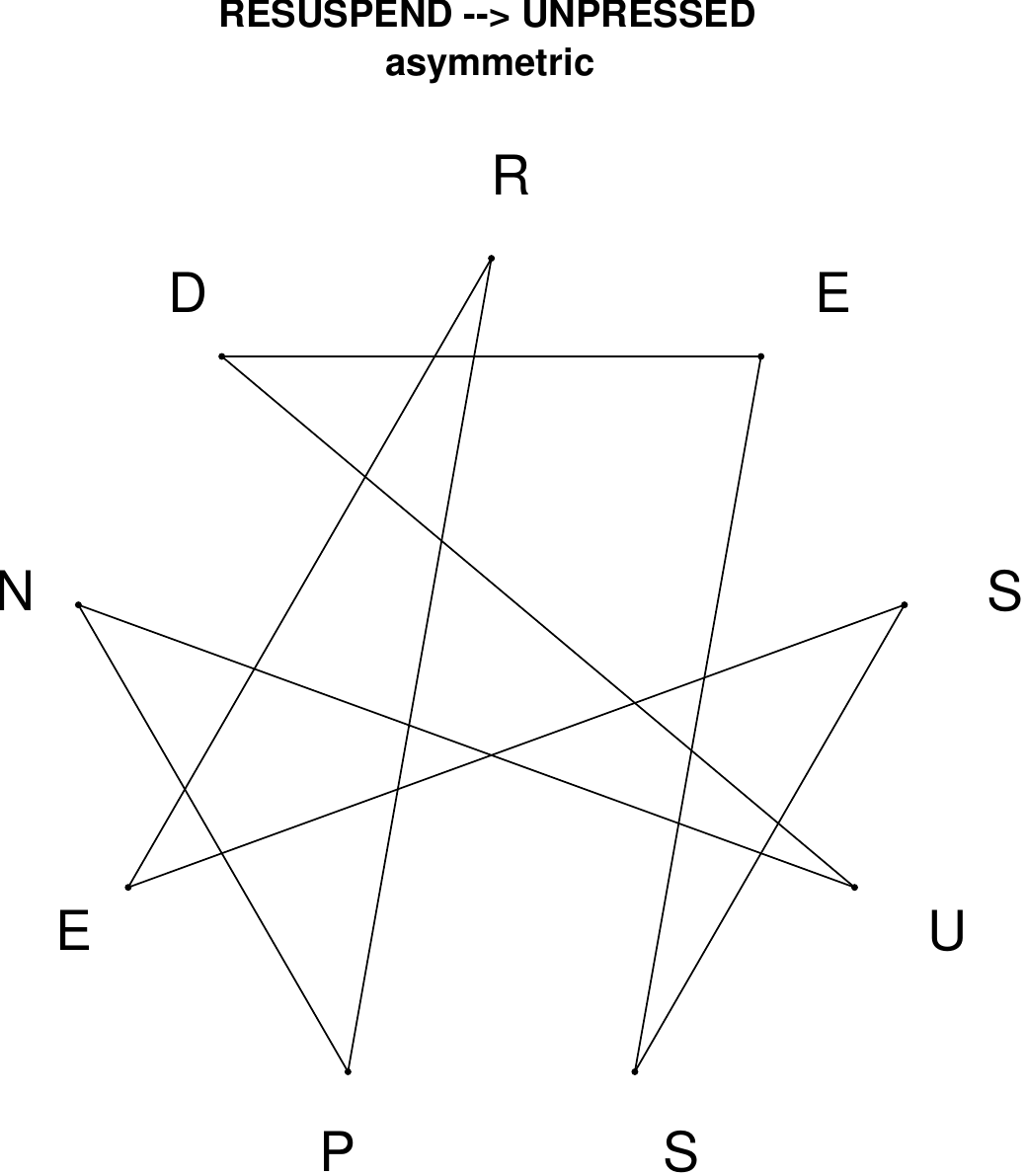}
\end{subfigure}
\hfill
\begin{subfigure}[T]{0.19\textwidth}
\centering
\includegraphics[width=\textwidth]{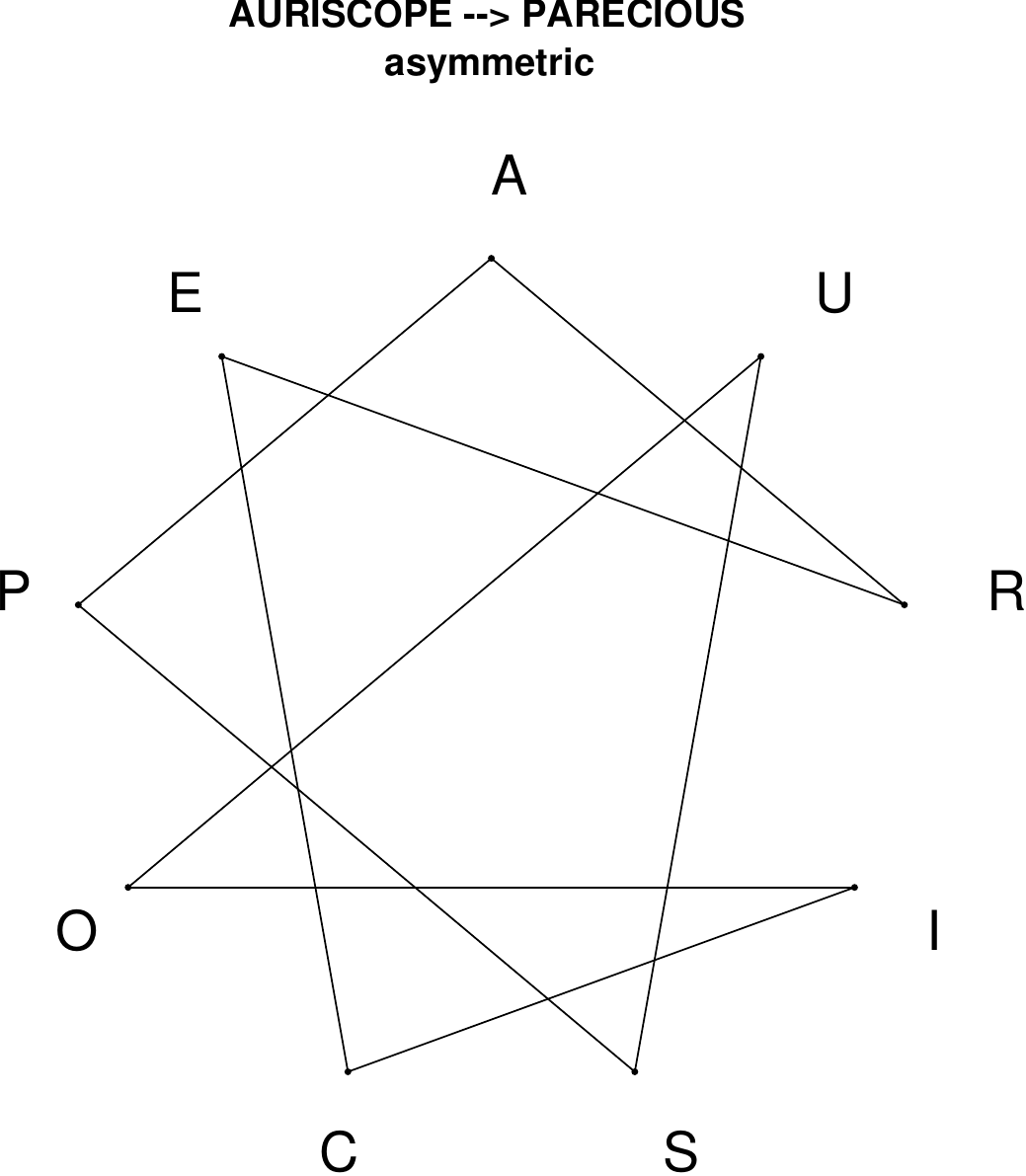}
\end{subfigure}
\end{figure}

\begin{figure}[H]
\centering
\begin{subfigure}[T]{0.19\textwidth}
\centering
\includegraphics[width=\textwidth]{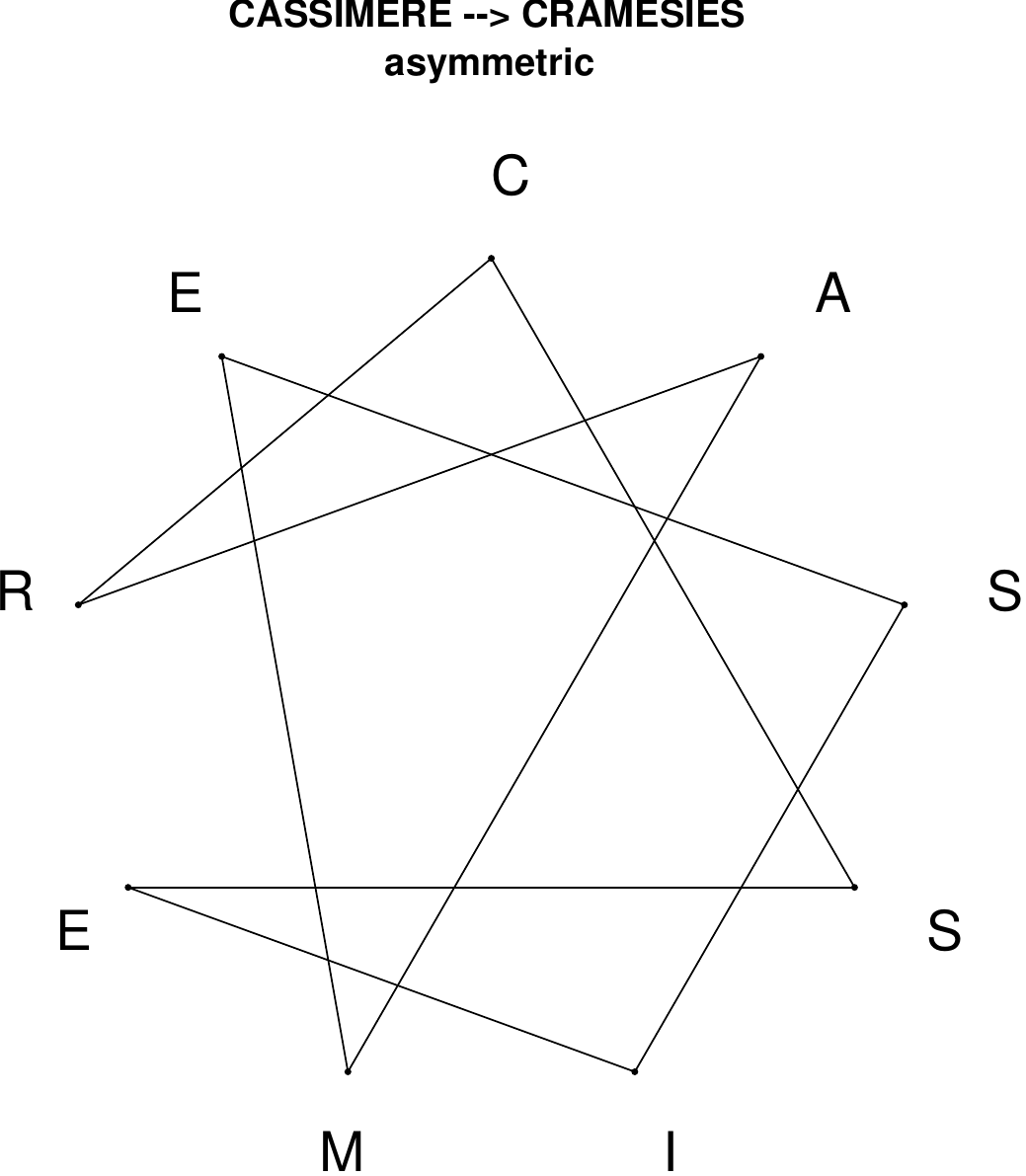}
\end{subfigure}
\hfill
\begin{subfigure}[T]{0.19\textwidth}
\centering
\includegraphics[width=\textwidth]{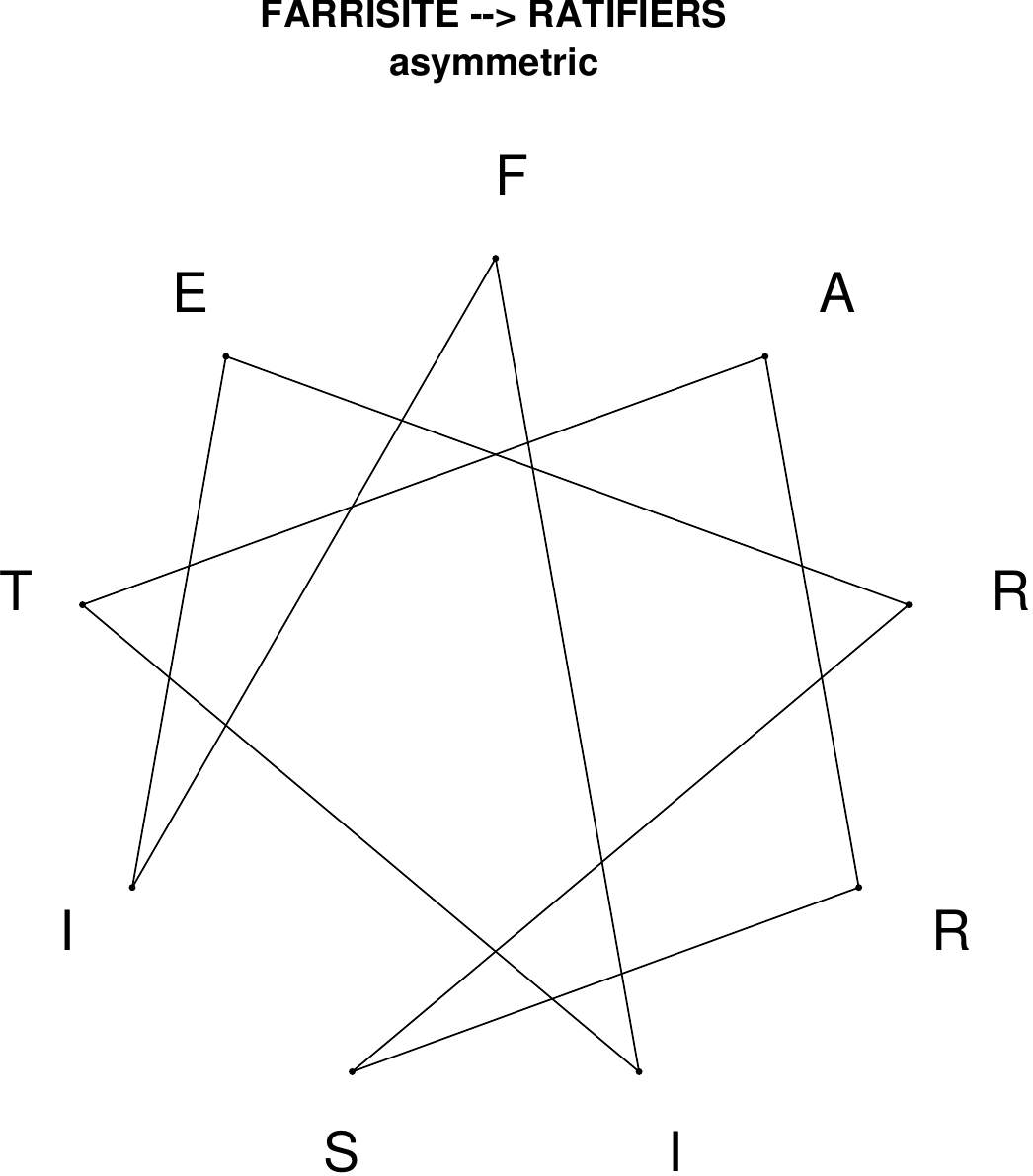}
\end{subfigure}
\hfill
\begin{subfigure}[T]{0.19\textwidth}
\centering
\includegraphics[width=\textwidth]{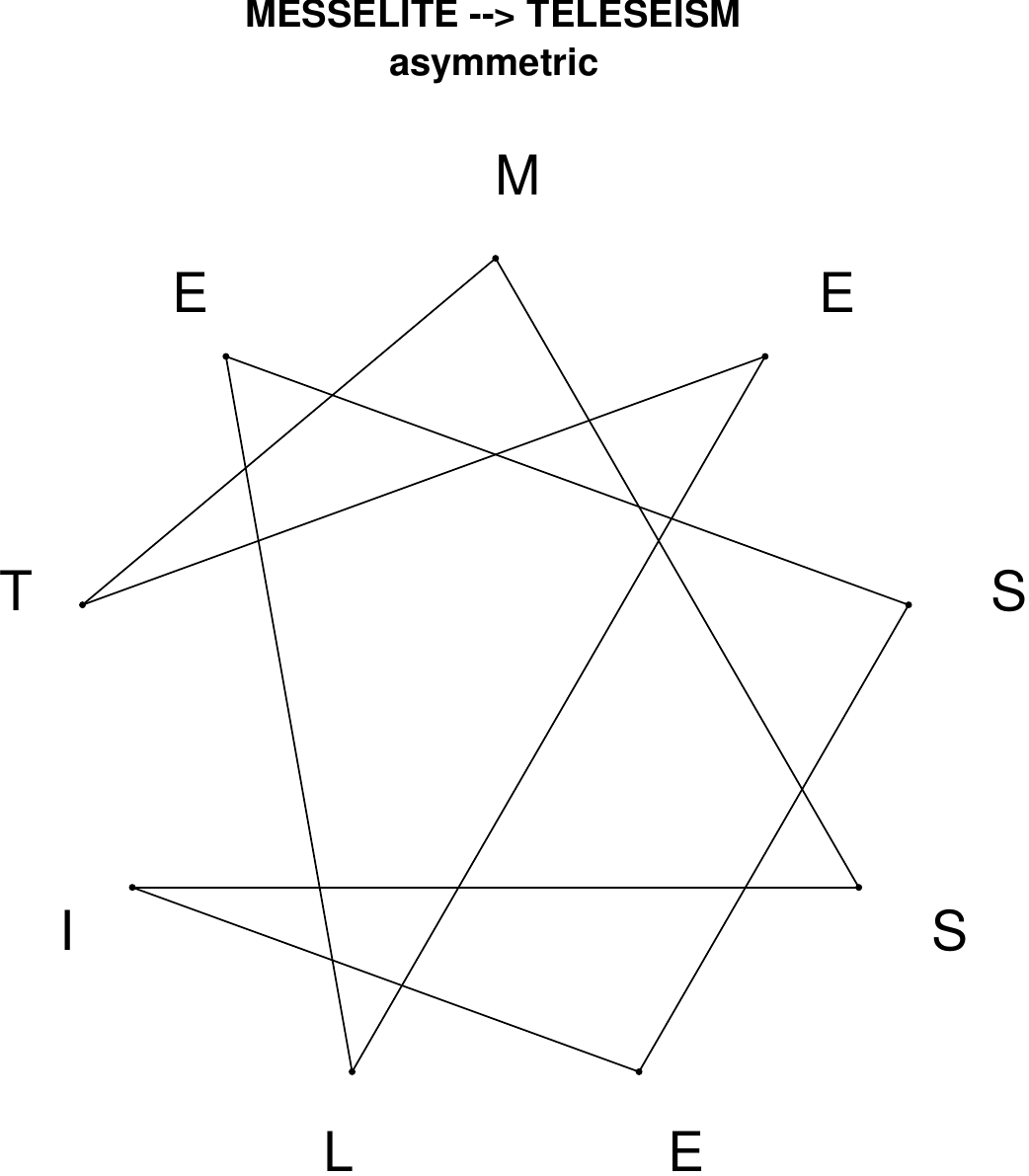}
\end{subfigure}
\hfill
\begin{subfigure}[T]{0.19\textwidth}
\centering
\includegraphics[width=\textwidth]{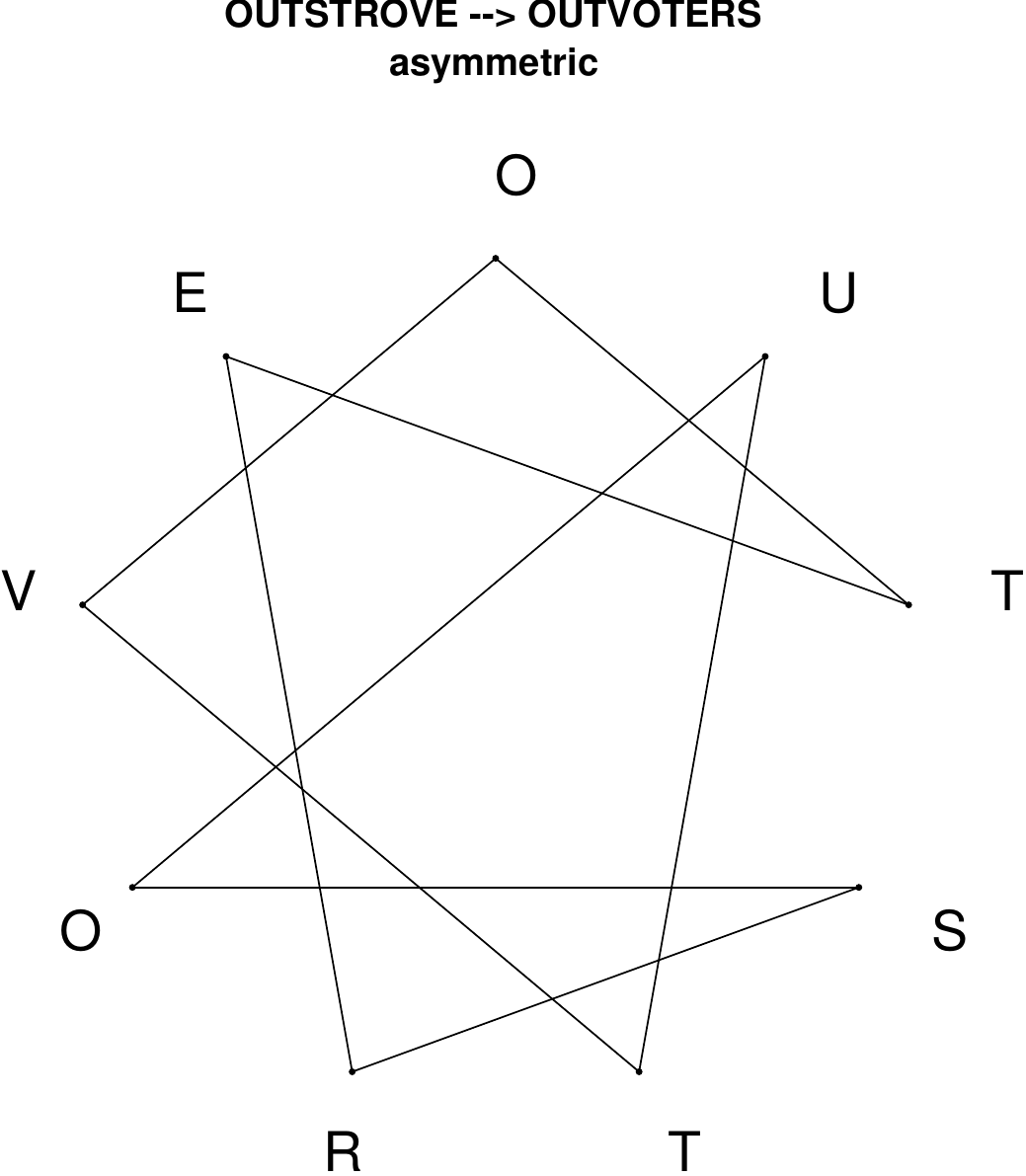}
\end{subfigure}
\hfill
\begin{subfigure}[T]{0.19\textwidth}
\centering
\includegraphics[width=\textwidth]{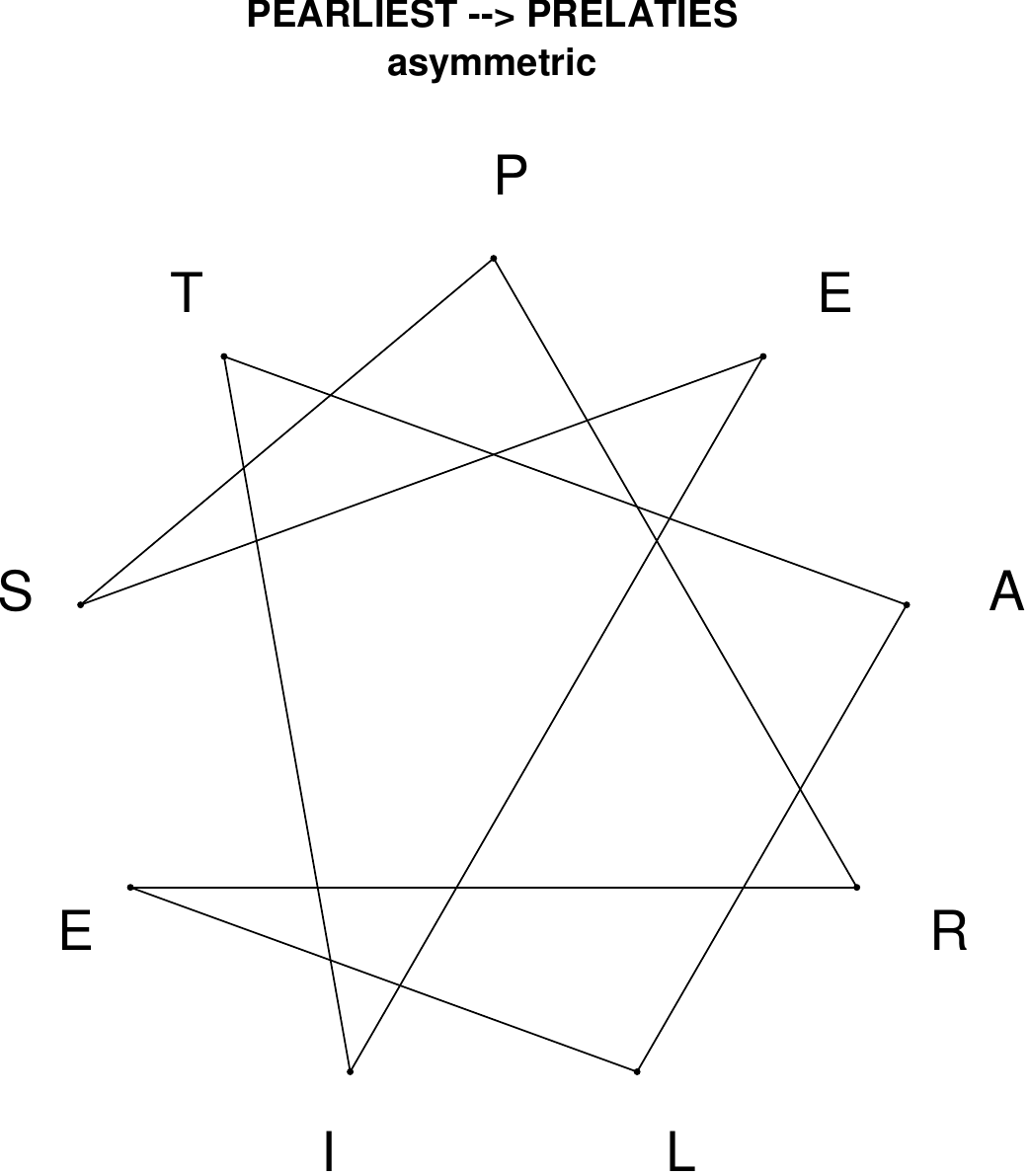}
\end{subfigure}
\end{figure}

\begin{figure}[H]
\centering
\begin{subfigure}[T]{0.19\textwidth}
\centering
\includegraphics[width=\textwidth]{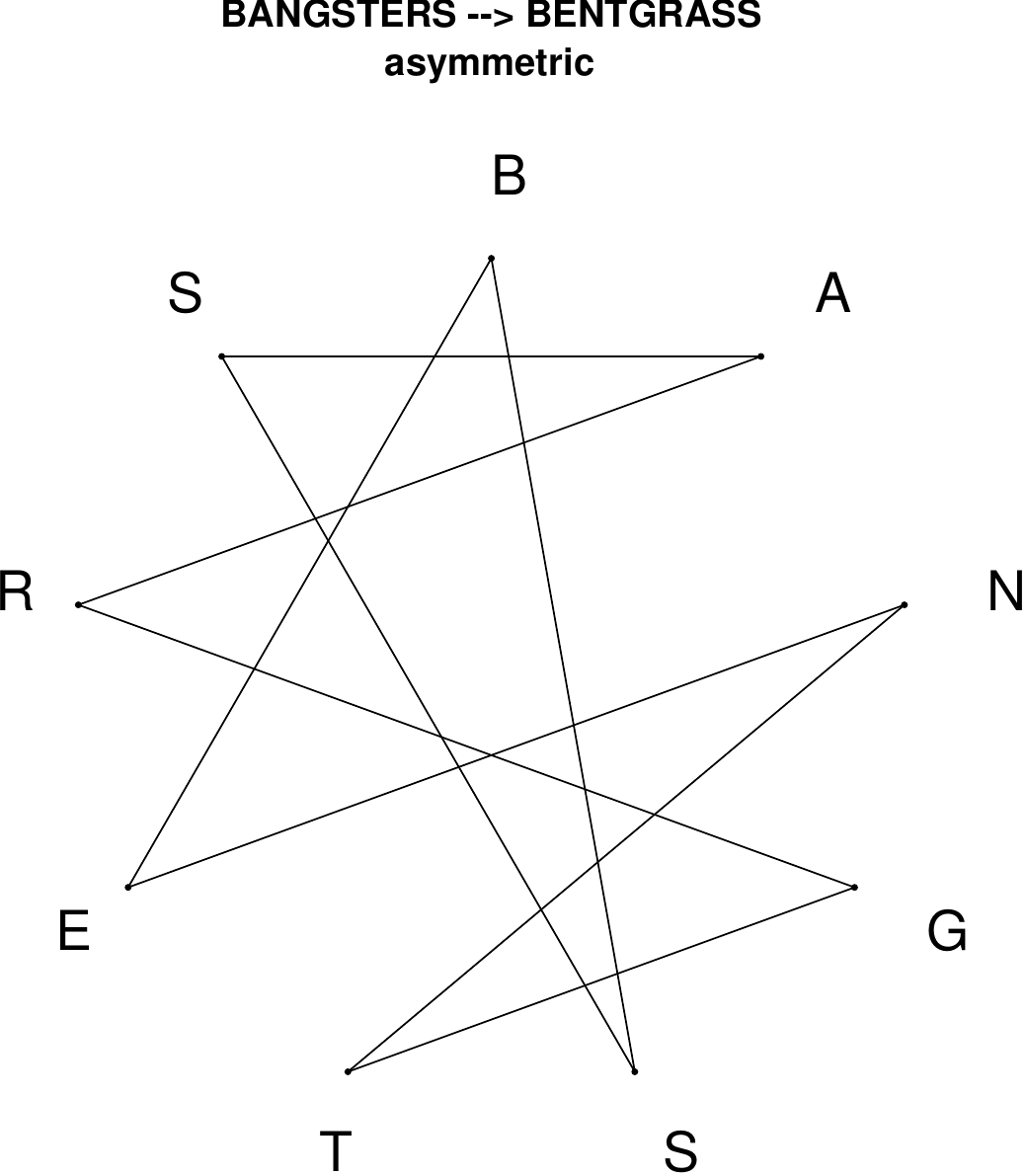}
\end{subfigure}
\hfill
\begin{subfigure}[T]{0.19\textwidth}
\centering
\includegraphics[width=\textwidth]{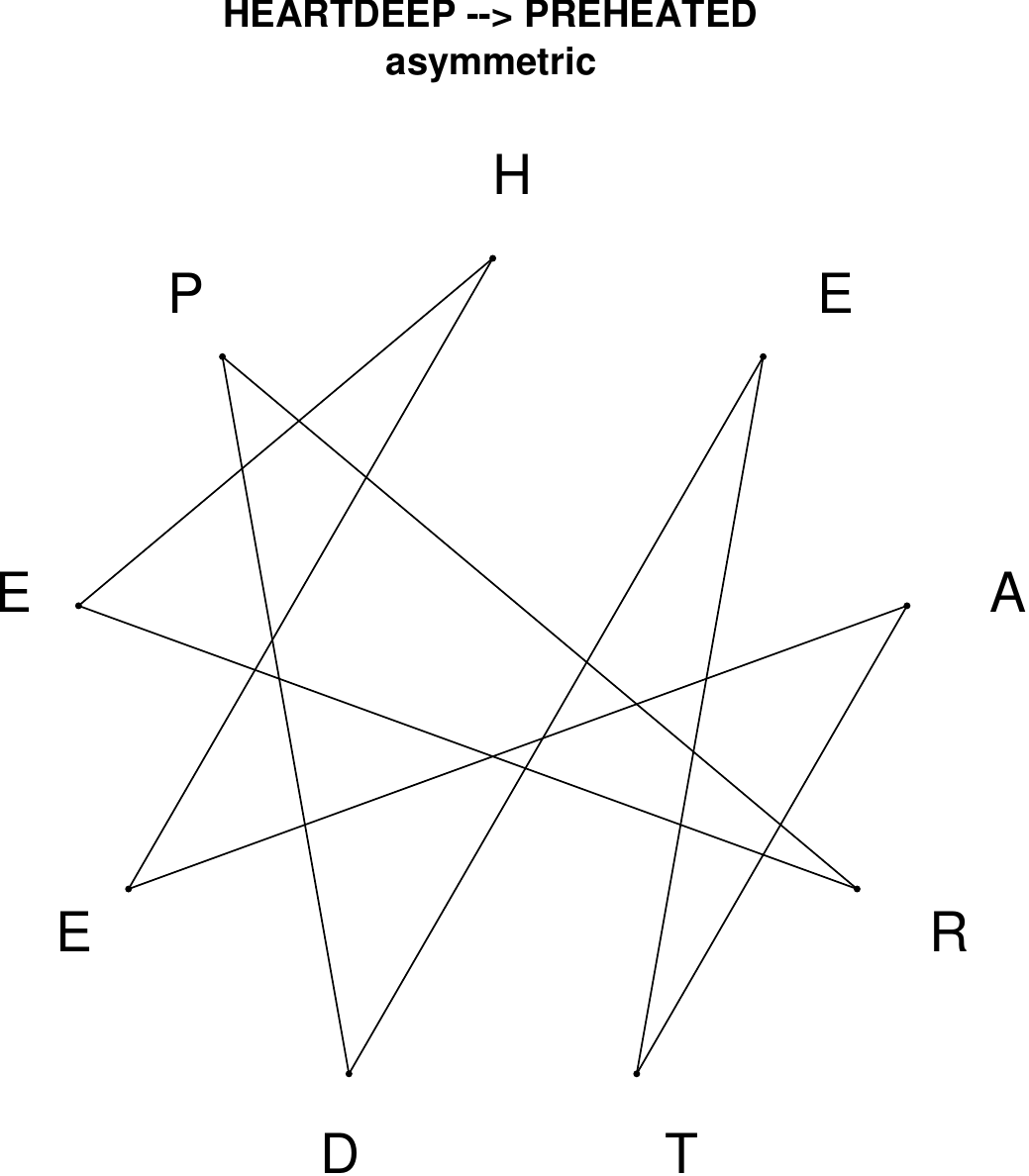}
\end{subfigure}
\hfill
\begin{subfigure}[T]{0.19\textwidth}
\centering
\includegraphics[width=\textwidth]{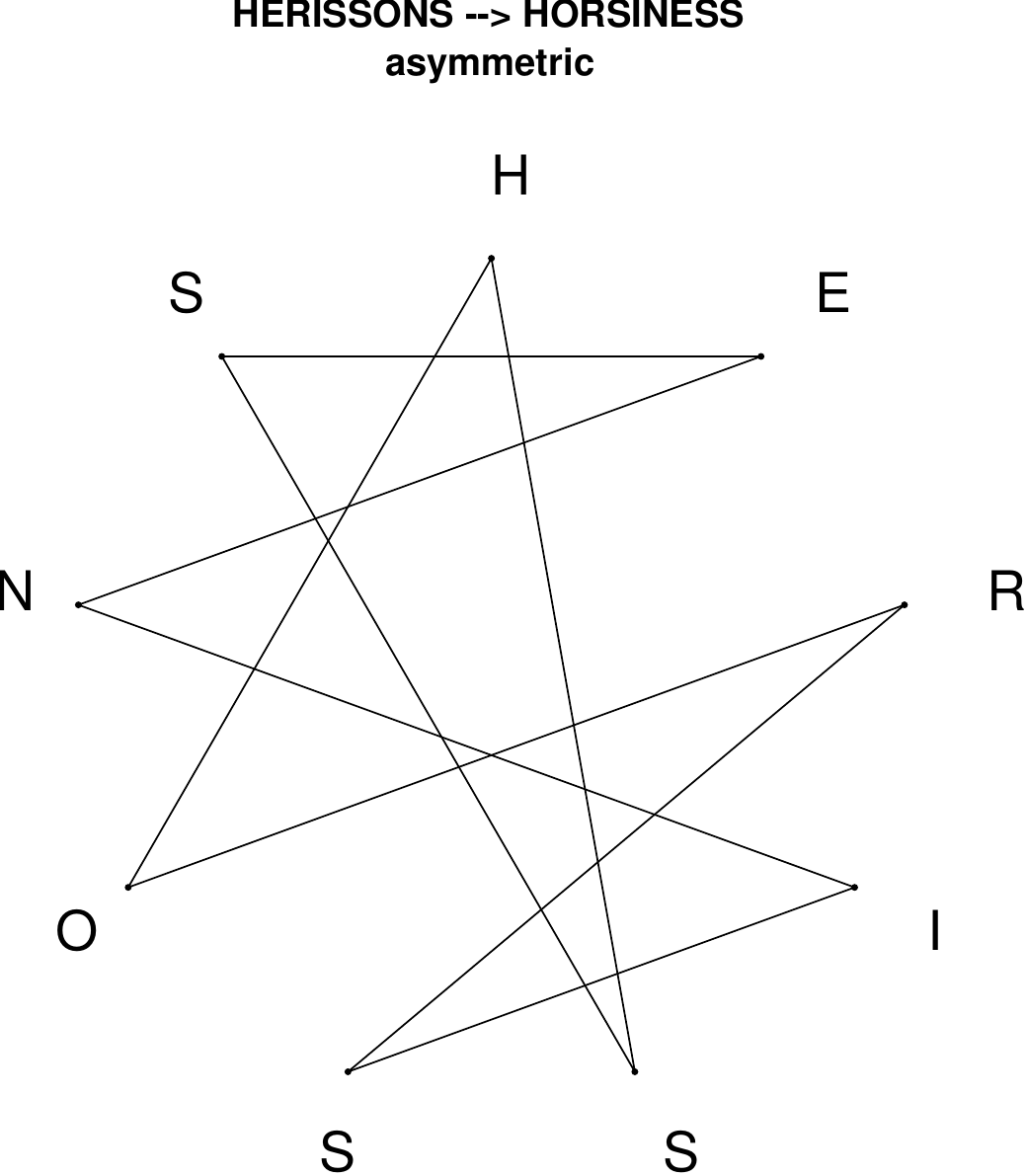}
\end{subfigure}
\hfill
\begin{subfigure}[T]{0.19\textwidth}
\centering
\includegraphics[width=\textwidth]{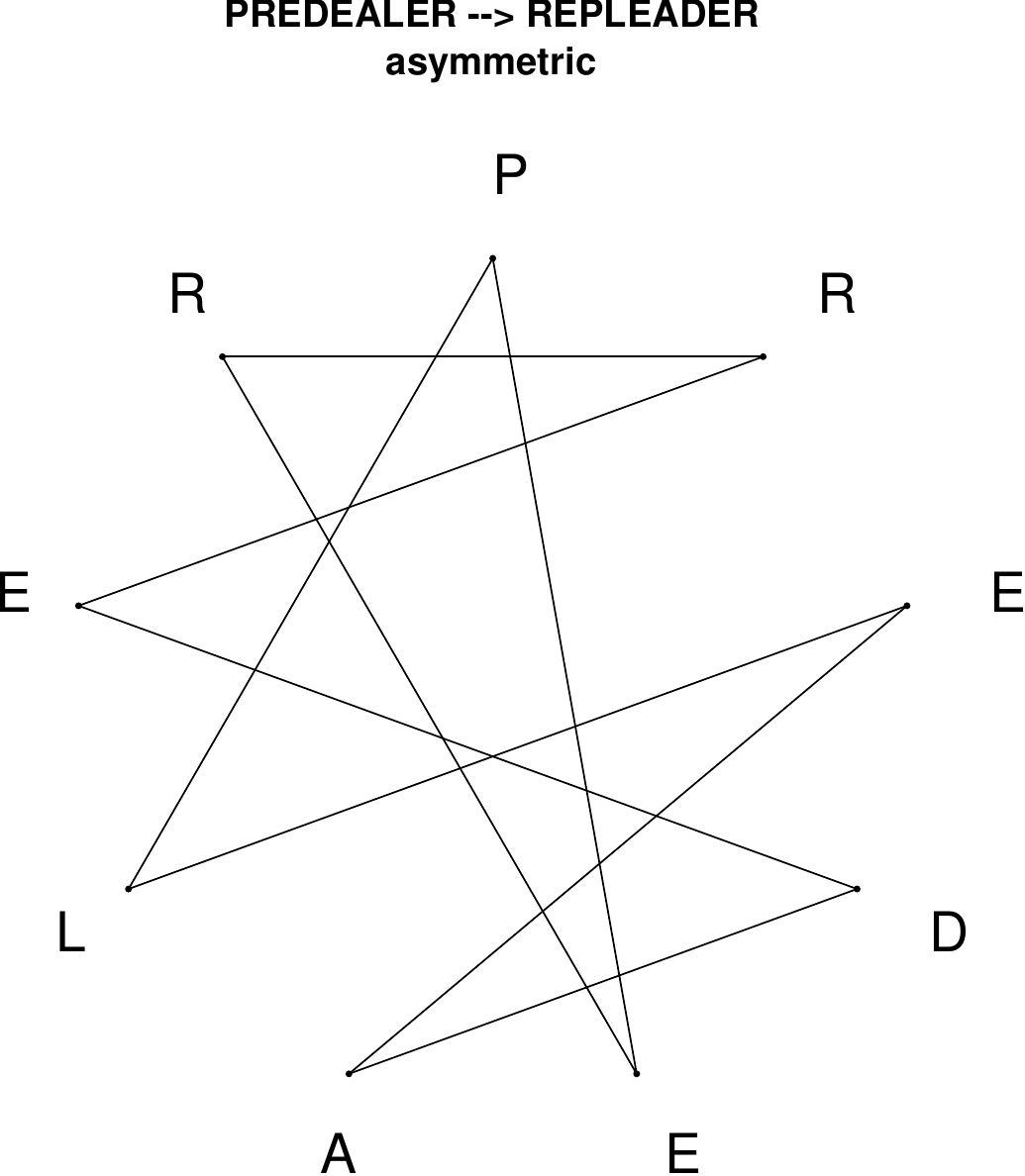}
\end{subfigure}
\hfill
\begin{subfigure}[T]{0.19\textwidth}
\centering
\includegraphics[width=\textwidth]{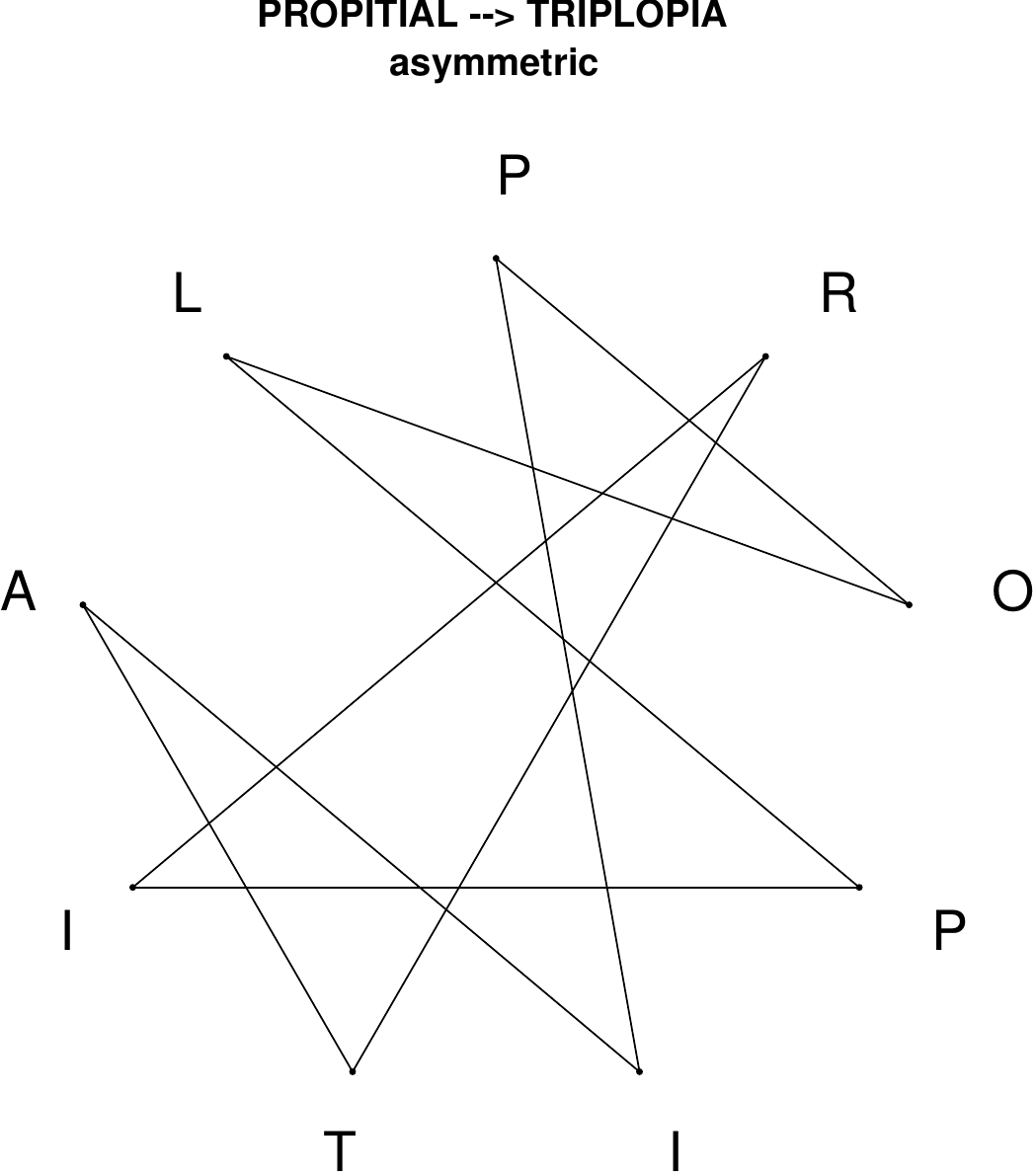}
\end{subfigure}
\end{figure}

\begin{figure}[H]
\centering
\begin{subfigure}[T]{0.19\textwidth}
\centering
\includegraphics[width=\textwidth]{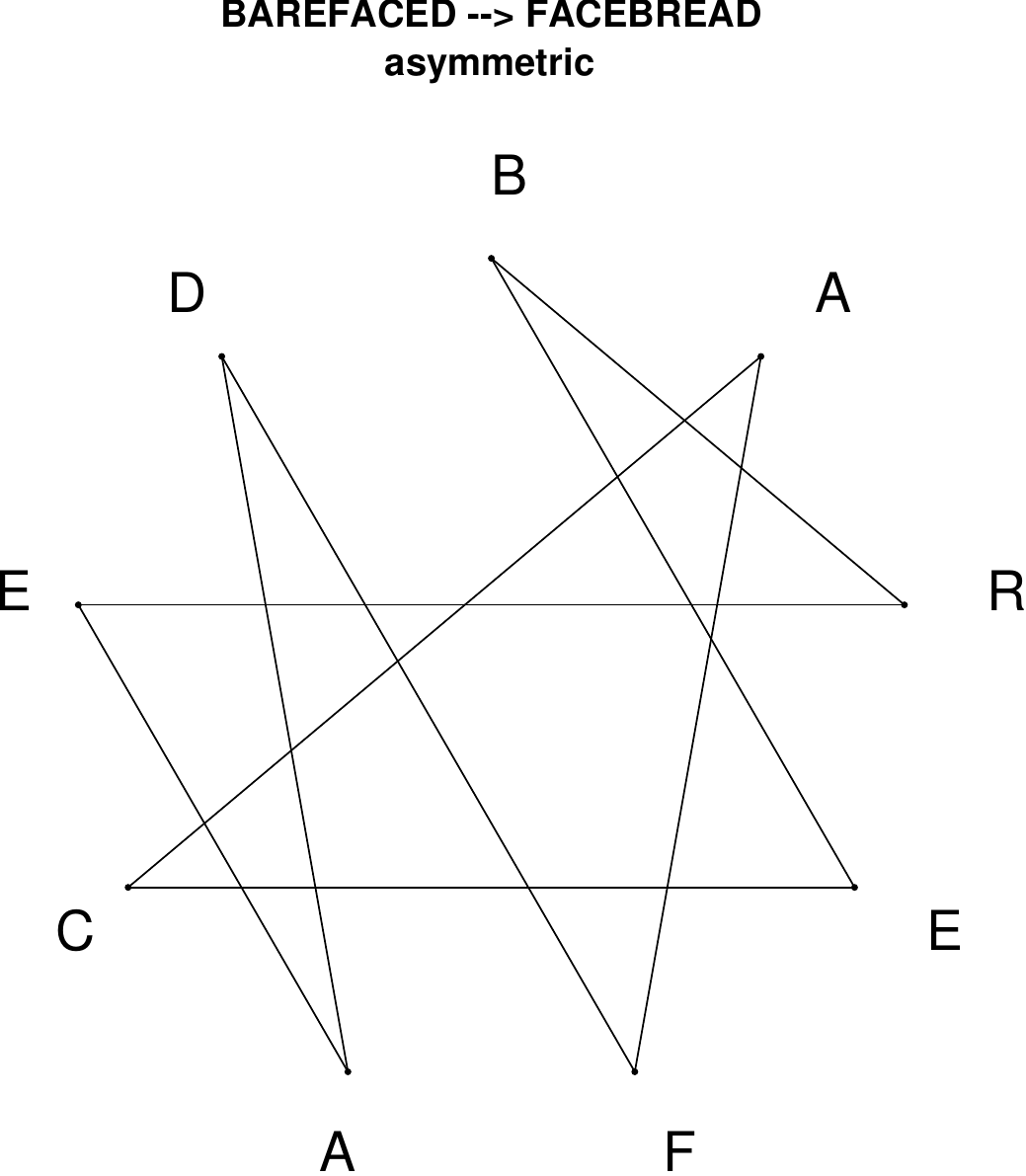}
\end{subfigure}
\hfill
\begin{subfigure}[T]{0.19\textwidth}
\centering
\includegraphics[width=\textwidth]{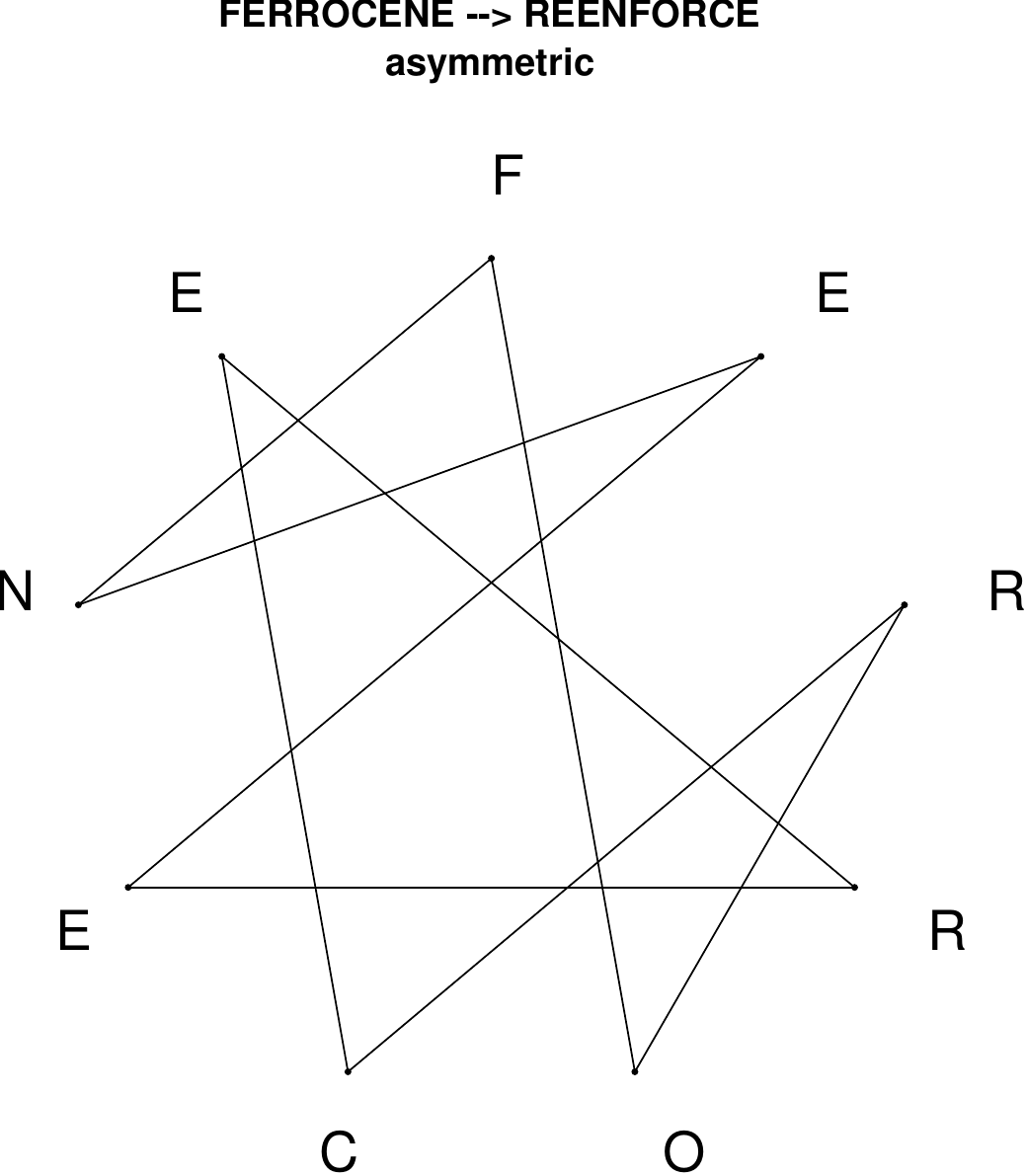}
\end{subfigure}
\hfill
\begin{subfigure}[T]{0.19\textwidth}
\centering
\includegraphics[width=\textwidth]{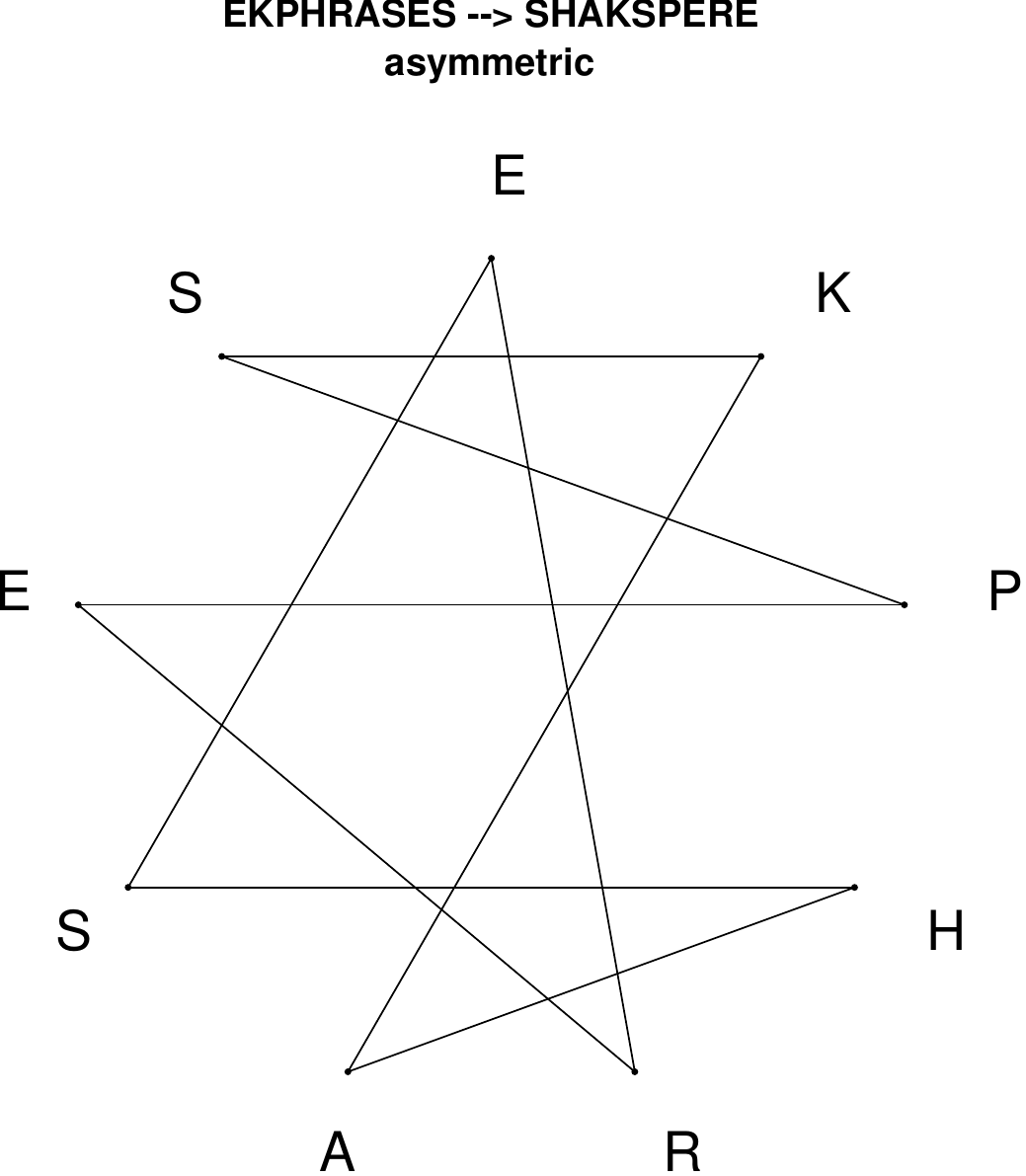}
\end{subfigure}
\hfill
\begin{subfigure}[T]{0.19\textwidth}
\centering
\includegraphics[width=\textwidth]{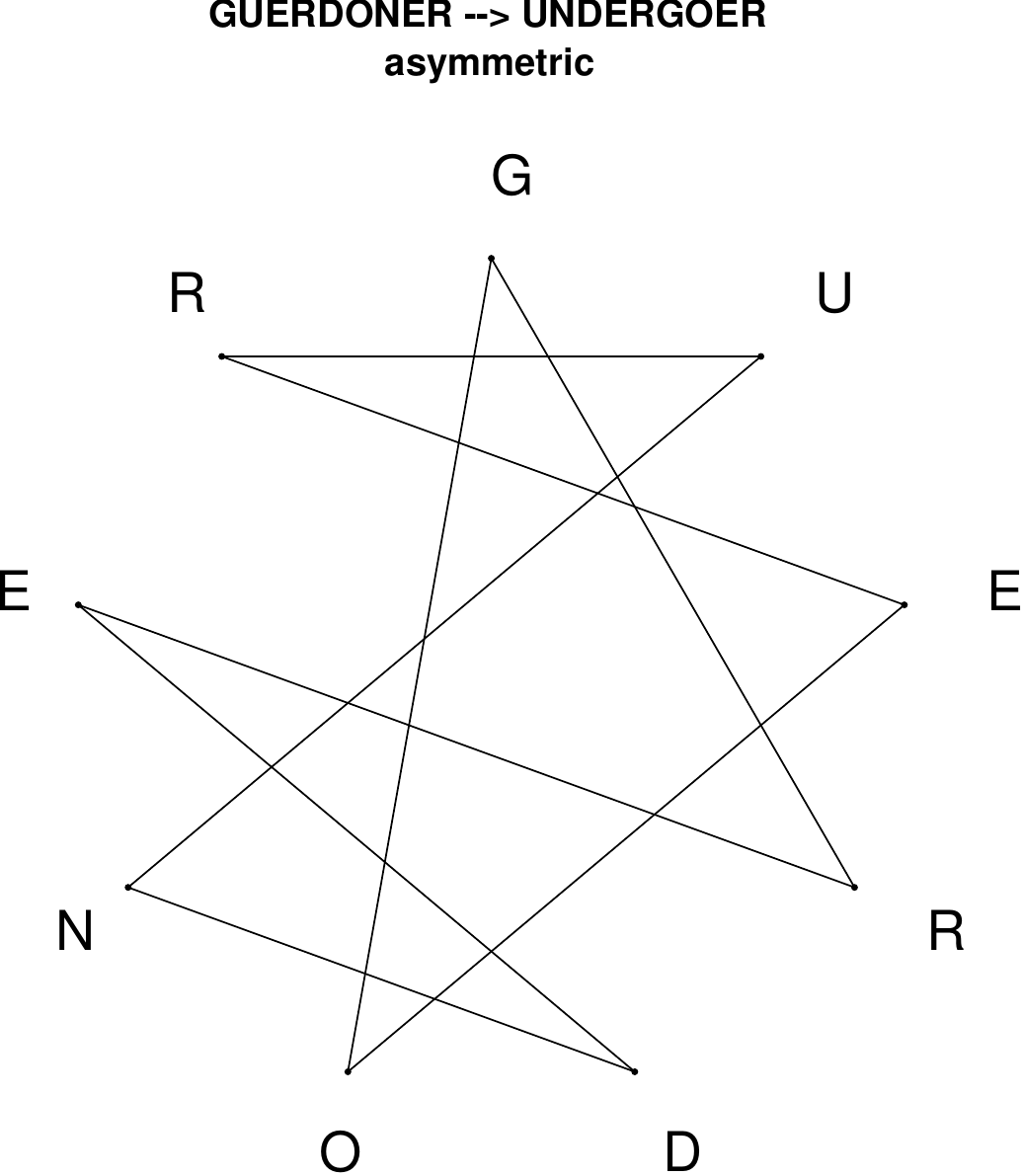}
\end{subfigure}
\hfill
\begin{subfigure}[T]{0.19\textwidth}
\centering
\includegraphics[width=\textwidth]{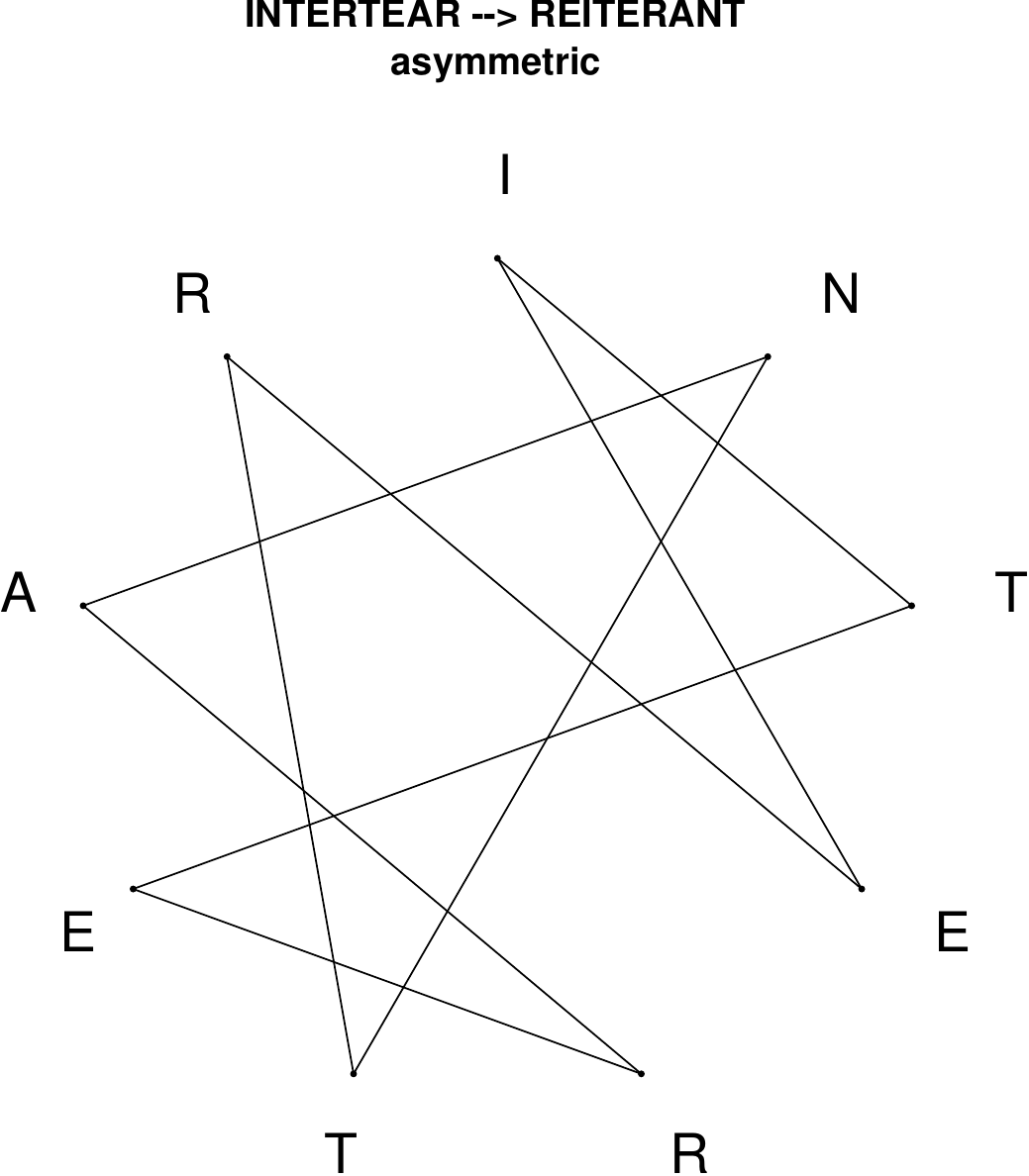}
\end{subfigure}
\end{figure}

\begin{figure}[H]
\centering
\begin{subfigure}[T]{0.19\textwidth}
\centering
\includegraphics[width=\textwidth]{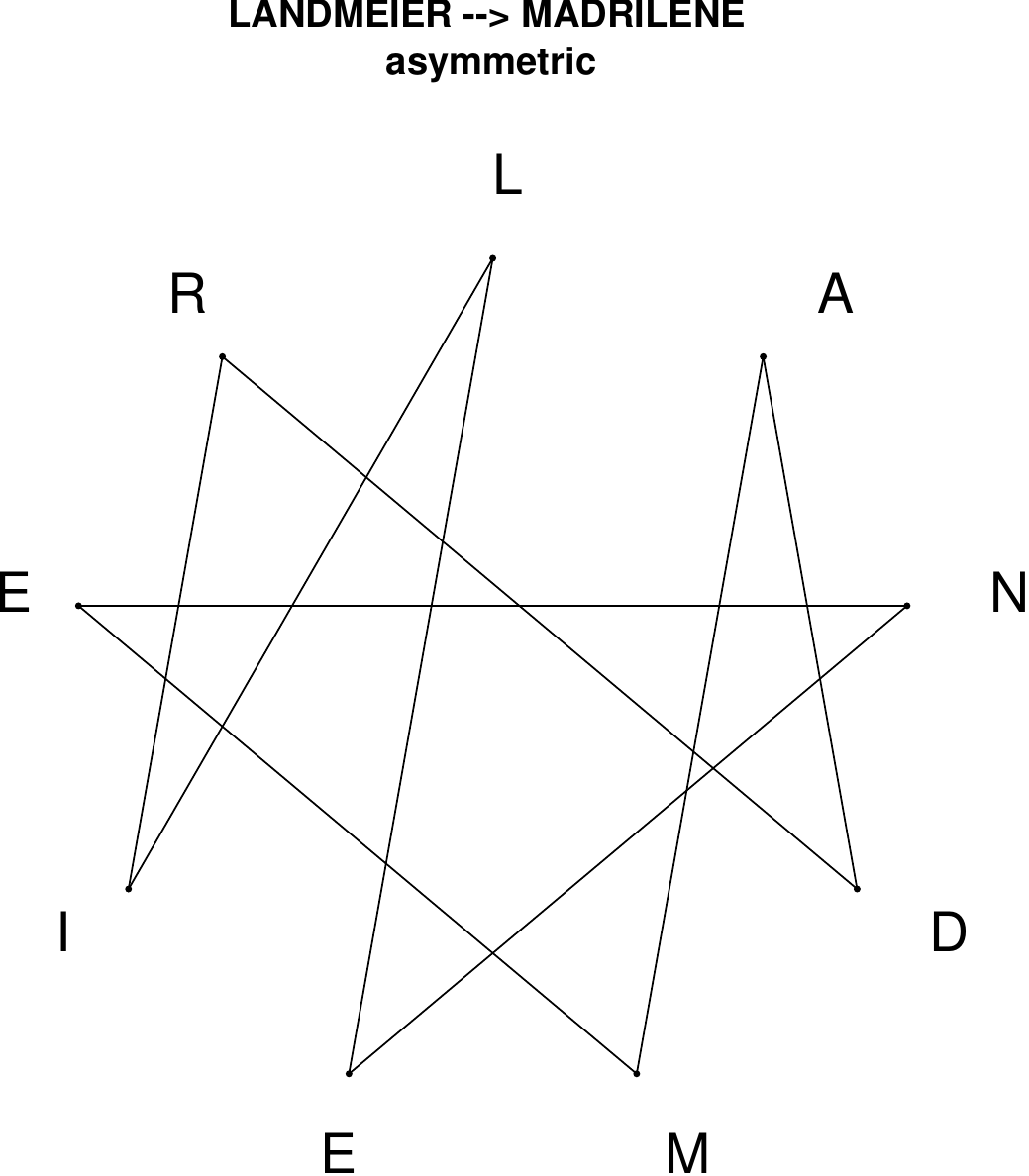}
\end{subfigure}
\hfill
\begin{subfigure}[T]{0.19\textwidth}
\centering
\includegraphics[width=\textwidth]{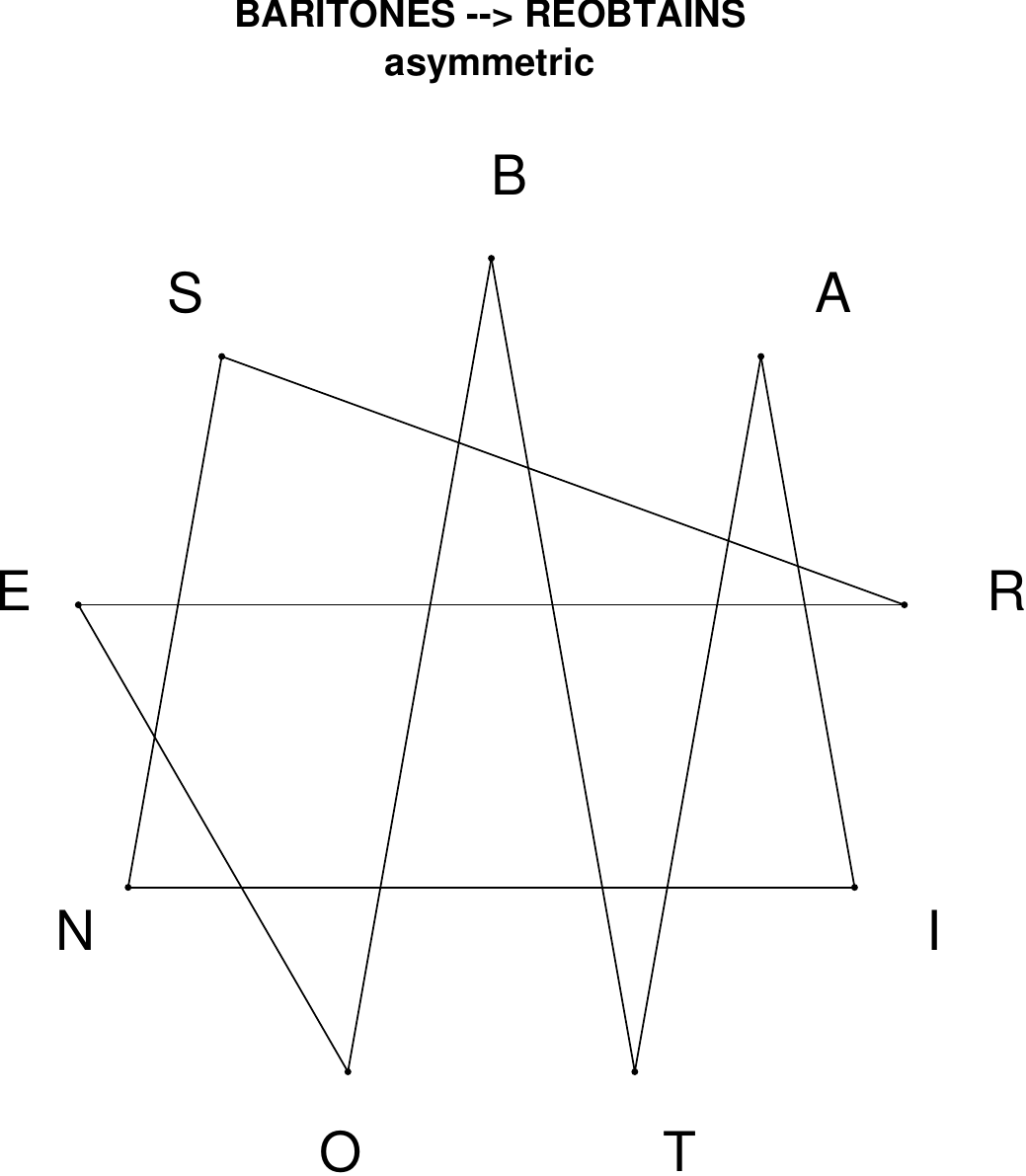}
\end{subfigure}
\hfill
\begin{subfigure}[T]{0.19\textwidth}
\centering
\includegraphics[width=\textwidth]{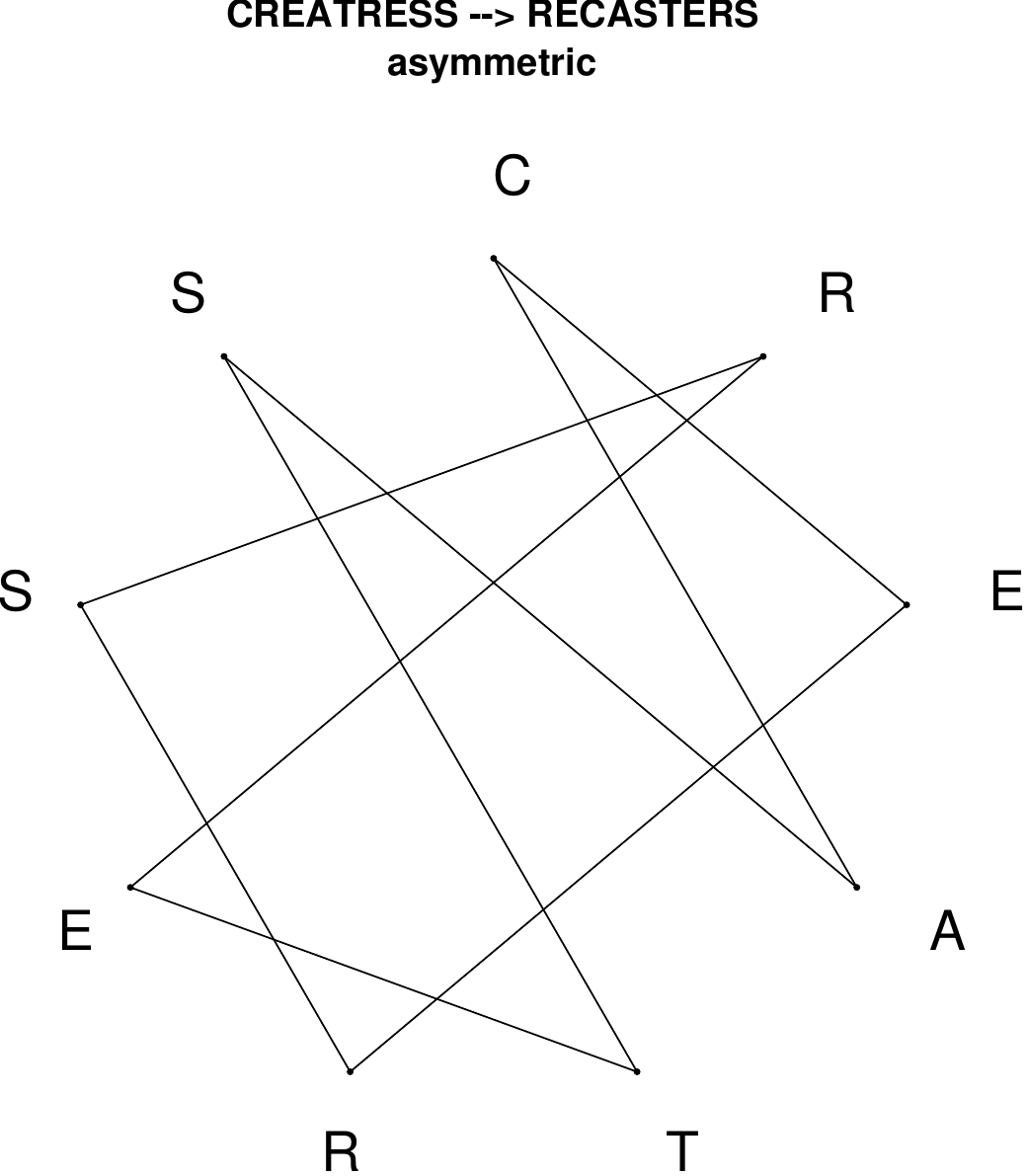}
\end{subfigure}
\hfill
\begin{subfigure}[T]{0.19\textwidth}
\centering
\includegraphics[width=\textwidth]{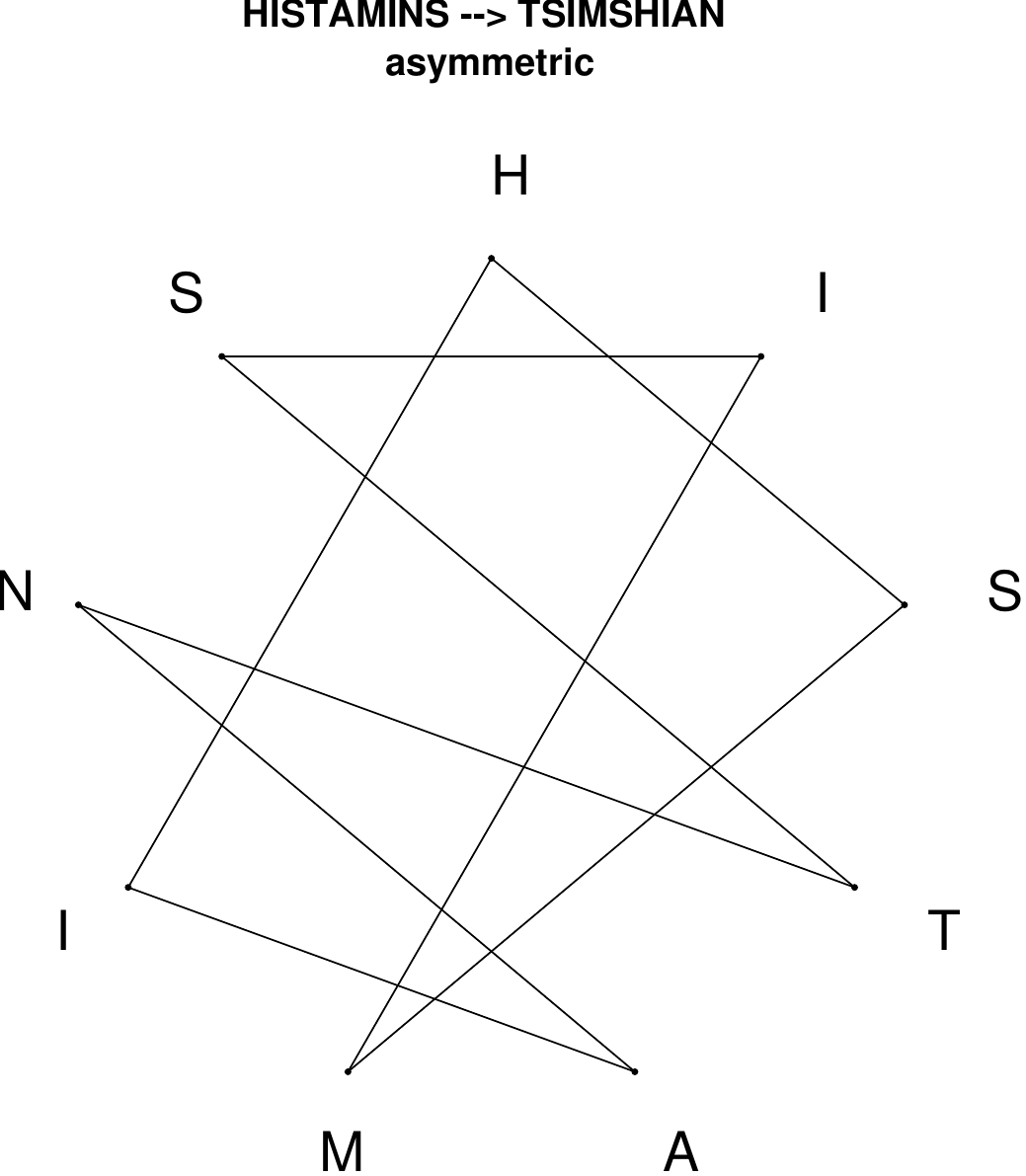}
\end{subfigure}
\hfill
\begin{subfigure}[T]{0.19\textwidth}
\centering
\includegraphics[width=\textwidth]{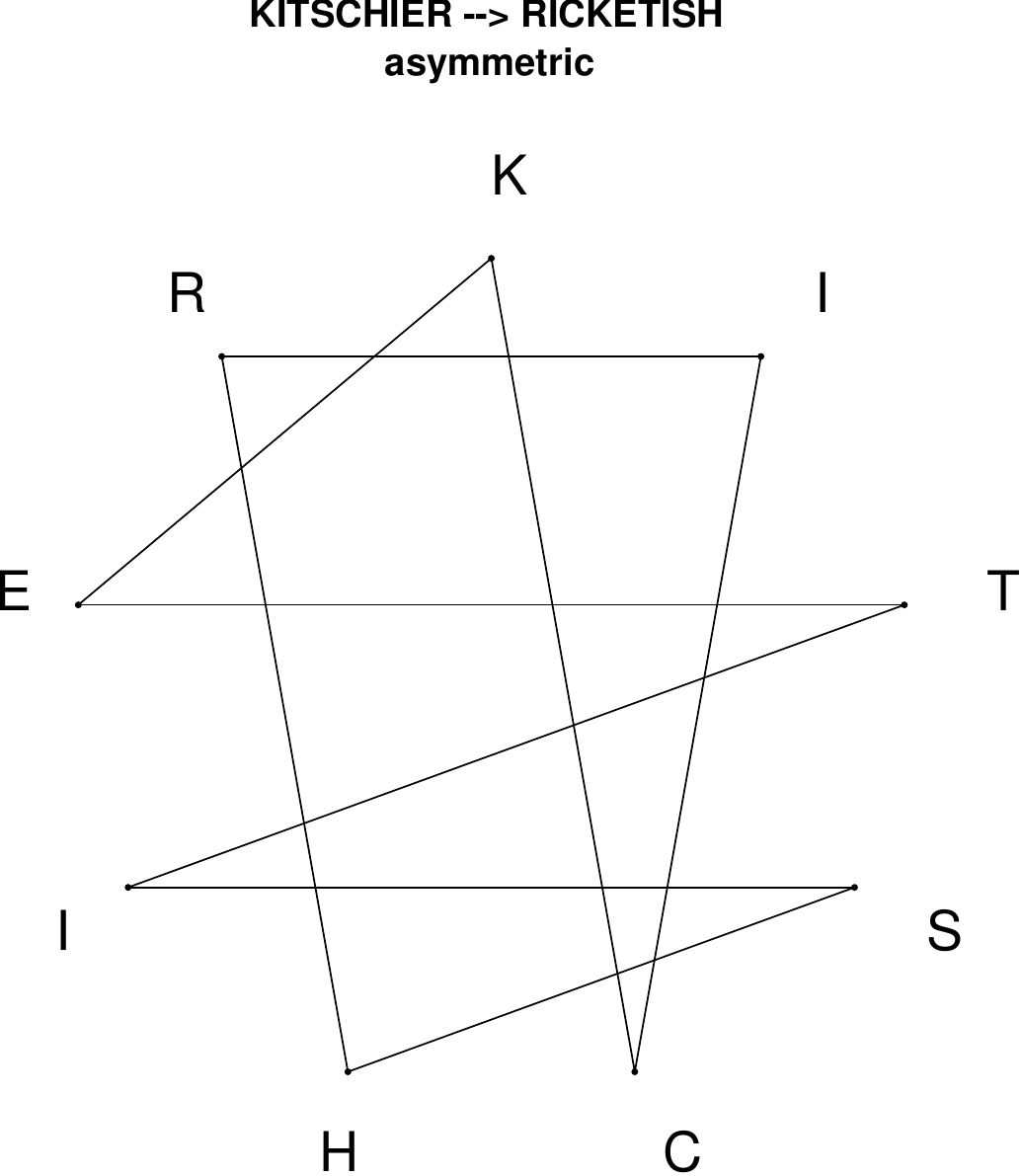}
\end{subfigure}
\end{figure}

\begin{figure}[H]
\centering
\begin{subfigure}[T]{0.19\textwidth}
\centering
\includegraphics[width=\textwidth]{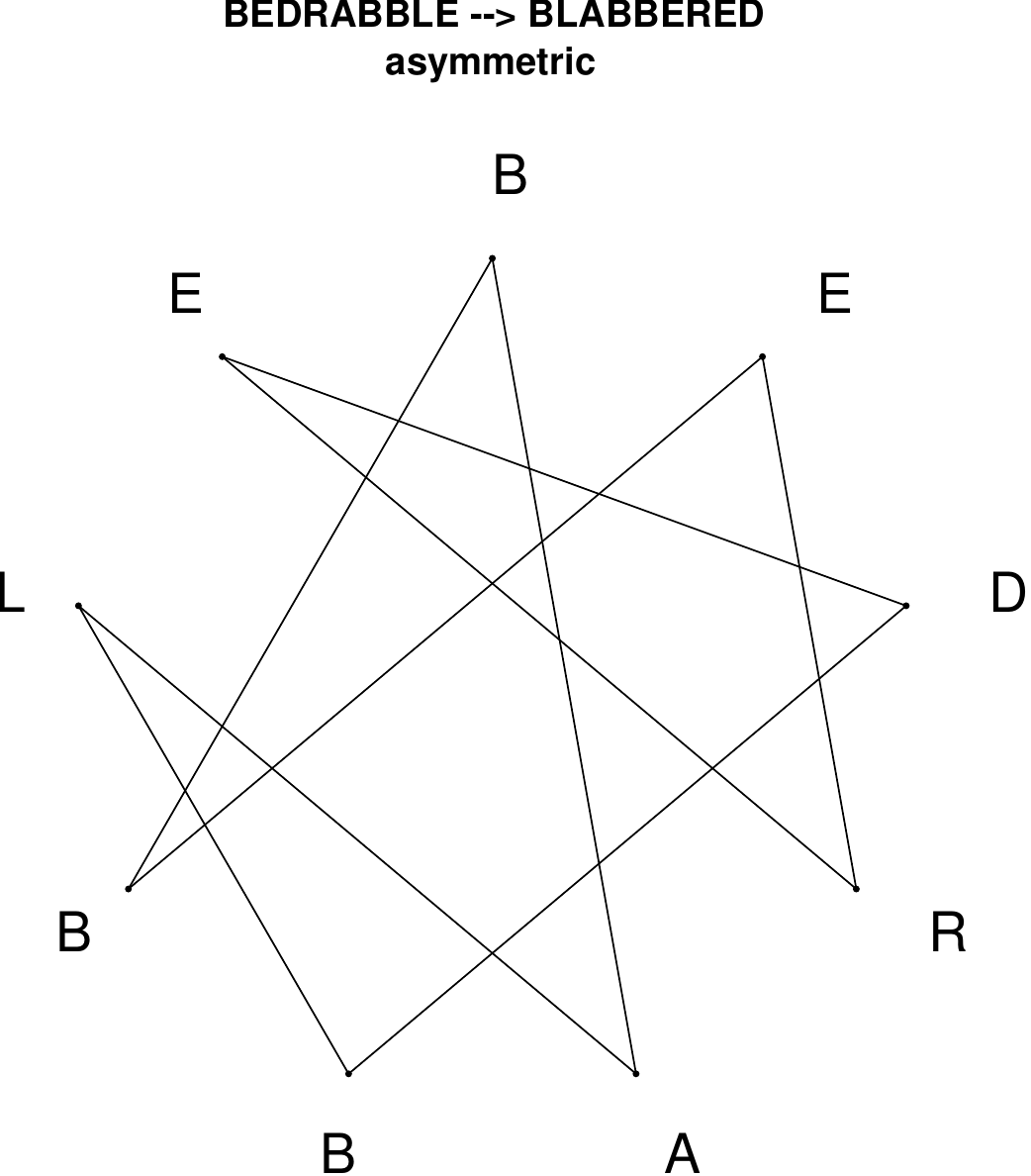}
\end{subfigure}
\hfill
\begin{subfigure}[T]{0.19\textwidth}
\centering
\includegraphics[width=\textwidth]{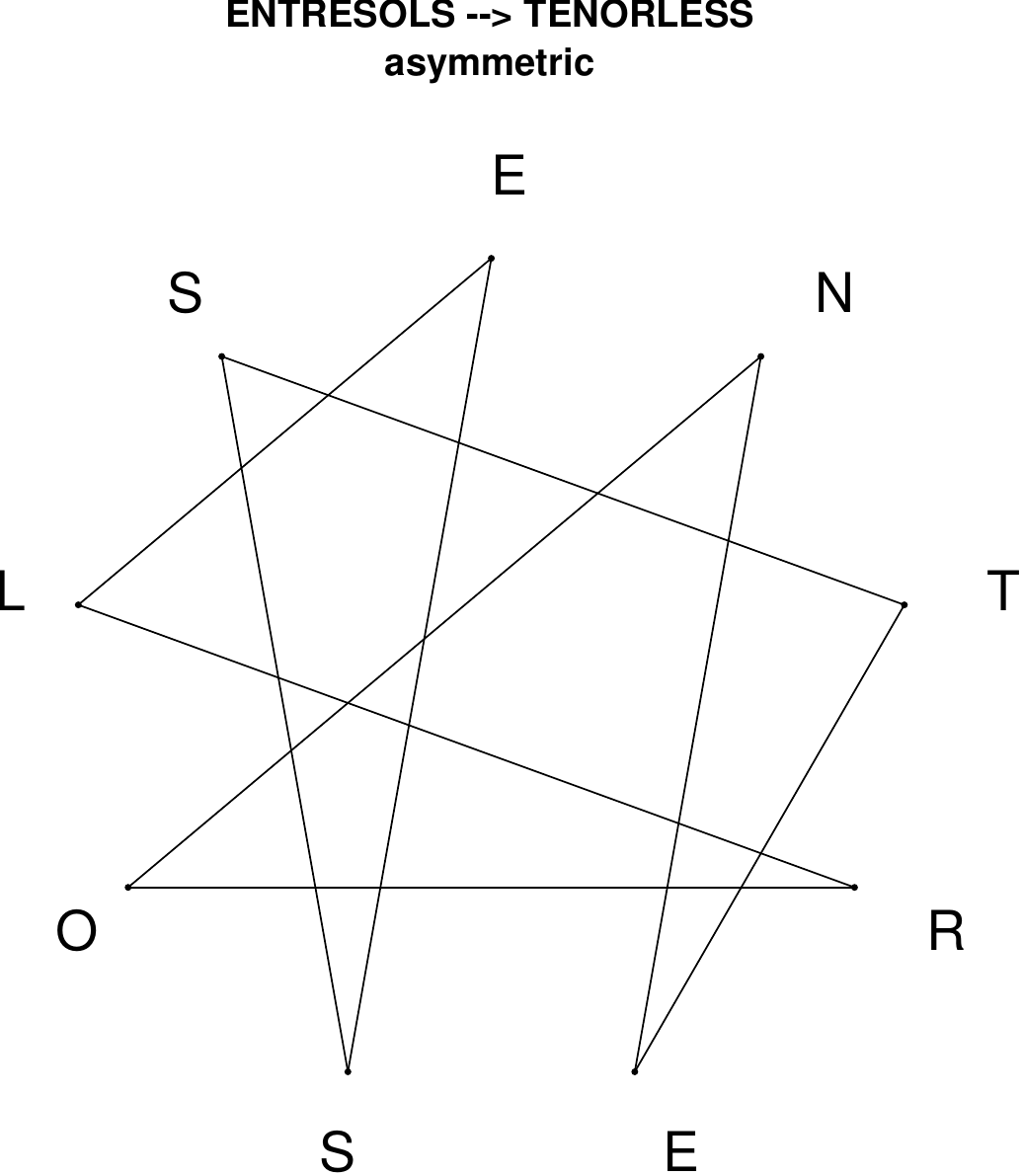}
\end{subfigure}
\hfill
\begin{subfigure}[T]{0.19\textwidth}
\centering
\includegraphics[width=\textwidth]{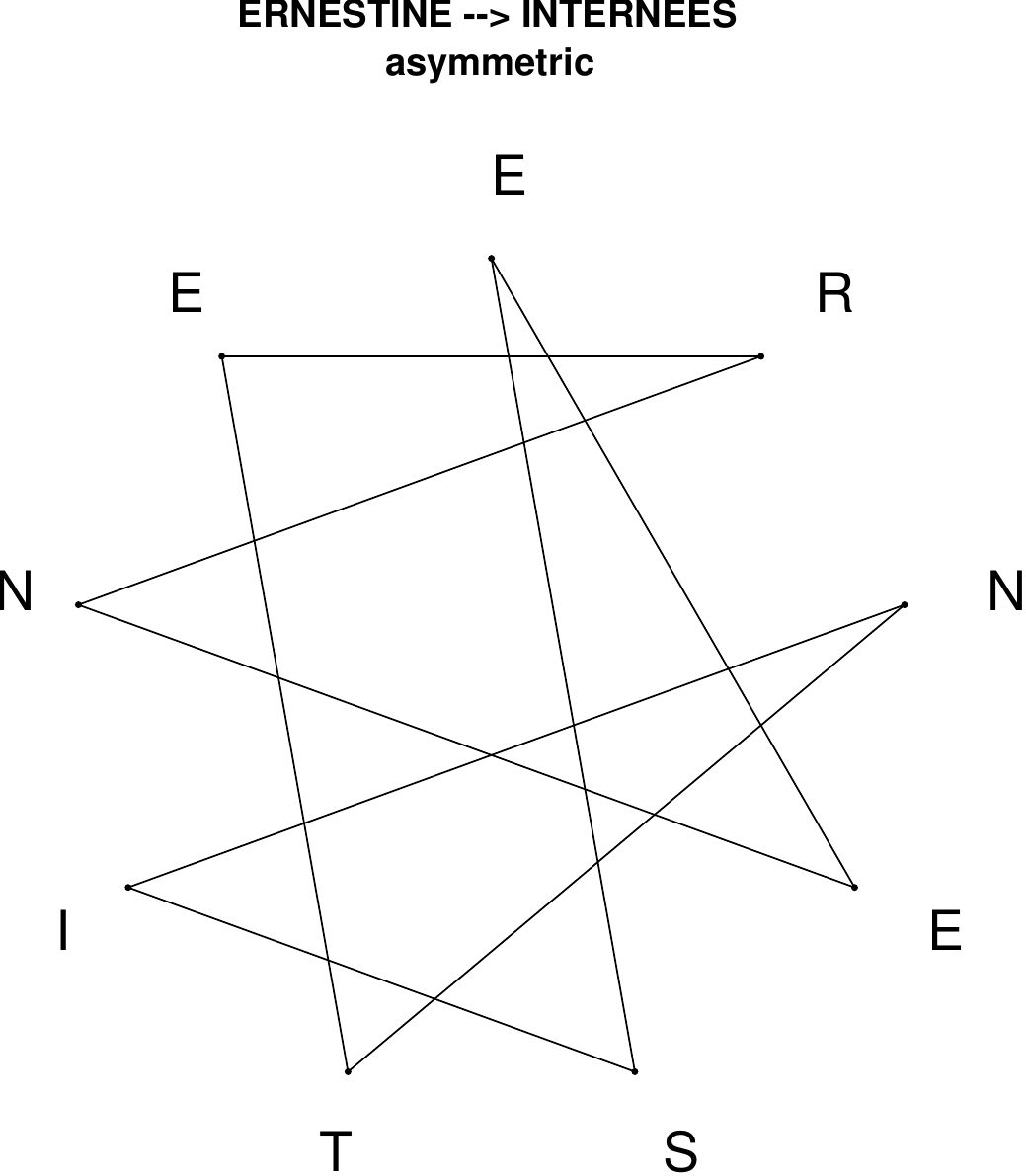}
\end{subfigure}
\hfill
\begin{subfigure}[T]{0.19\textwidth}
\centering
\includegraphics[width=\textwidth]{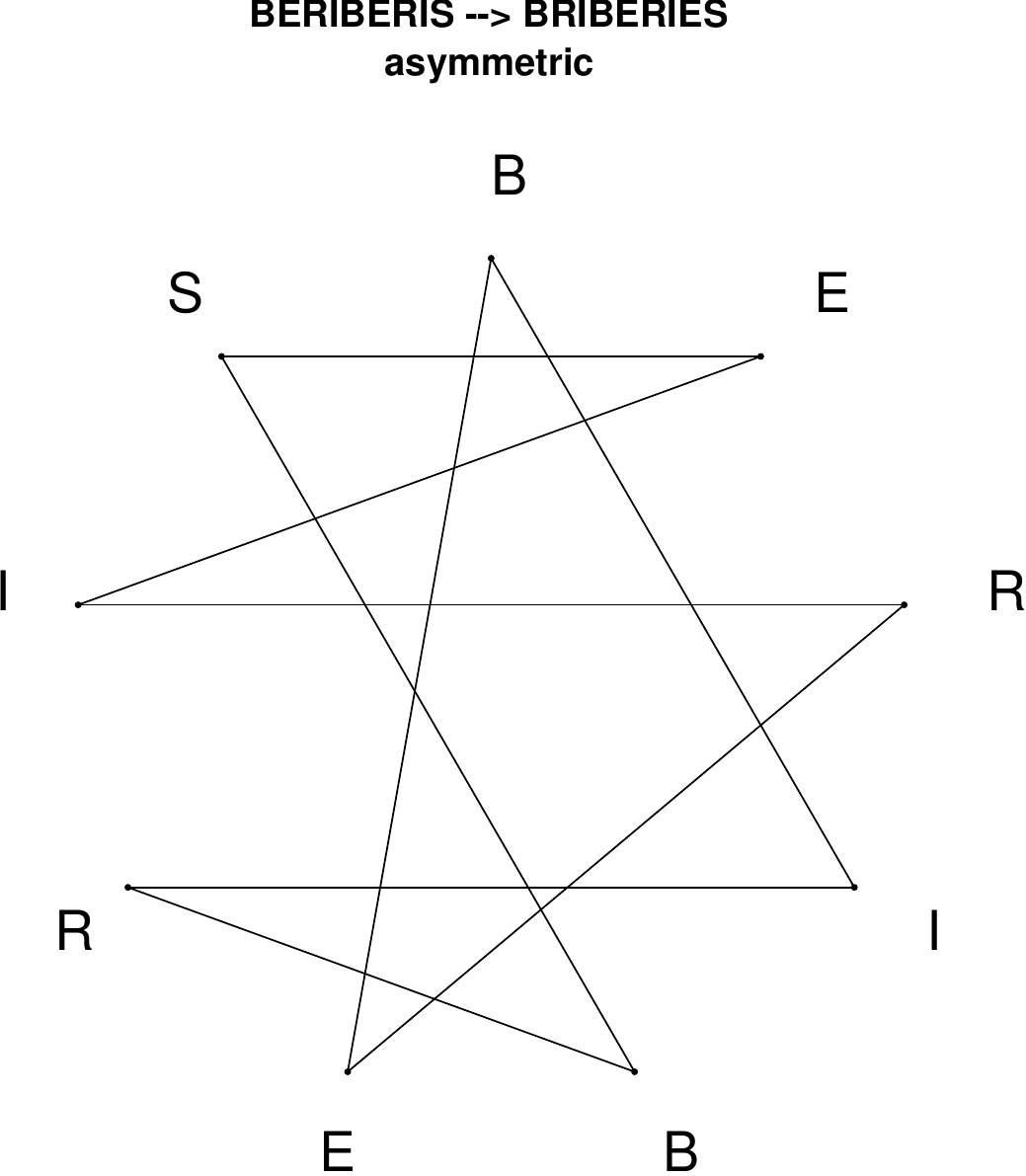}
\end{subfigure}
\hfill
\begin{subfigure}[T]{0.19\textwidth}
\centering
\includegraphics[width=\textwidth]{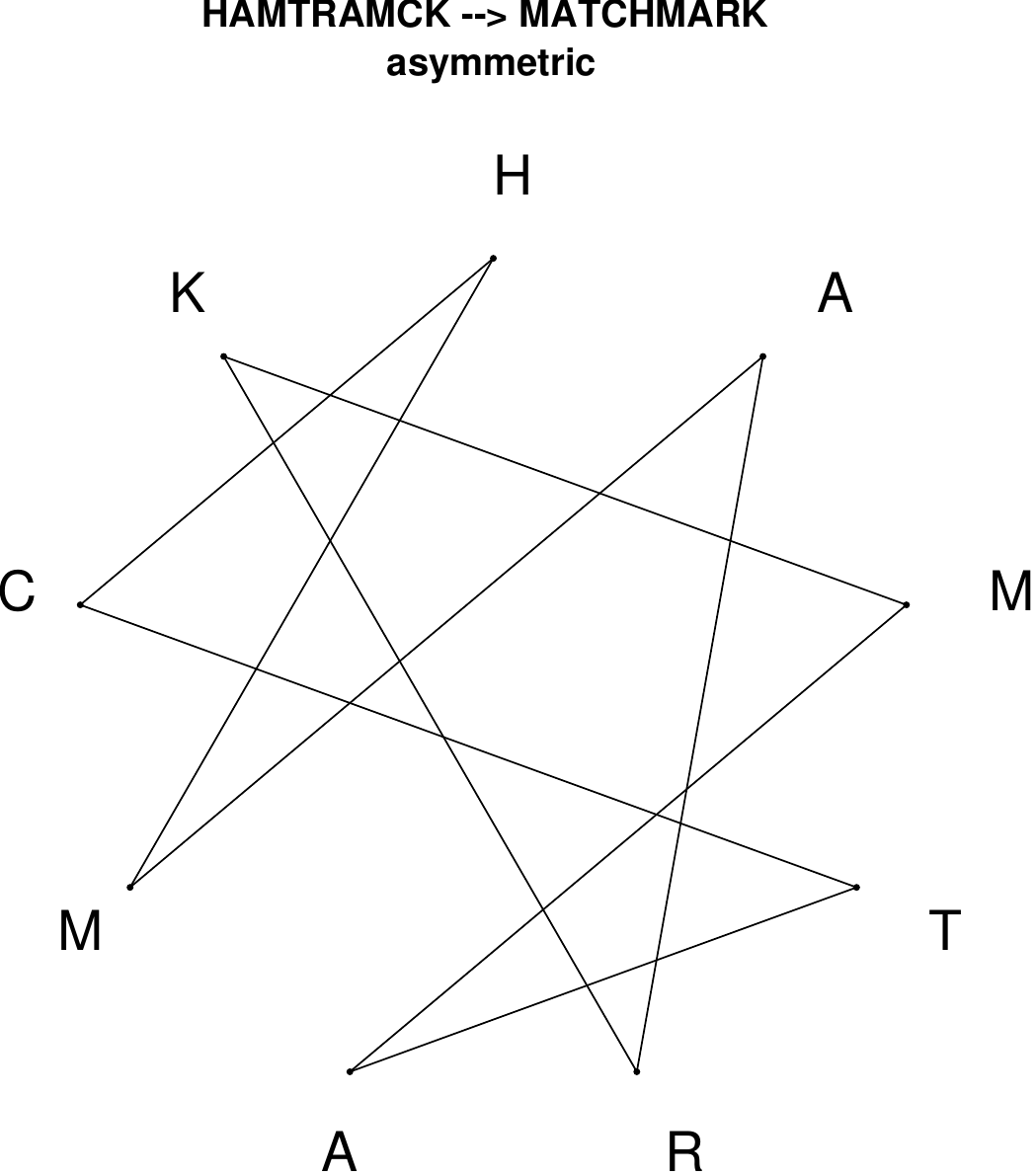}
\end{subfigure}
\end{figure}

\begin{figure}[H]
\centering
\begin{subfigure}[T]{0.19\textwidth}
\centering
\includegraphics[width=\textwidth]{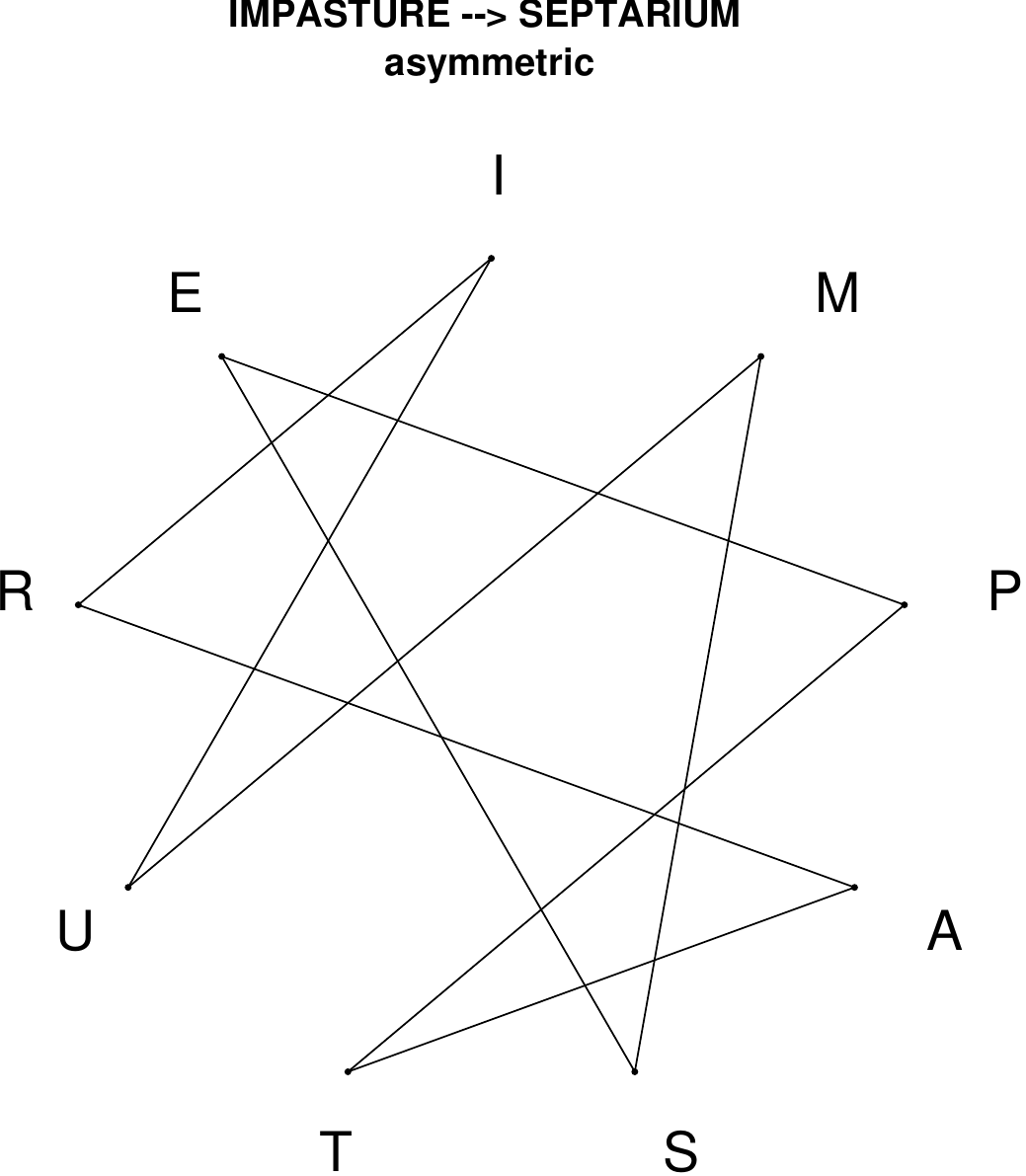}
\end{subfigure}
\hfill
\begin{subfigure}[T]{0.19\textwidth}
\centering
\includegraphics[width=\textwidth]{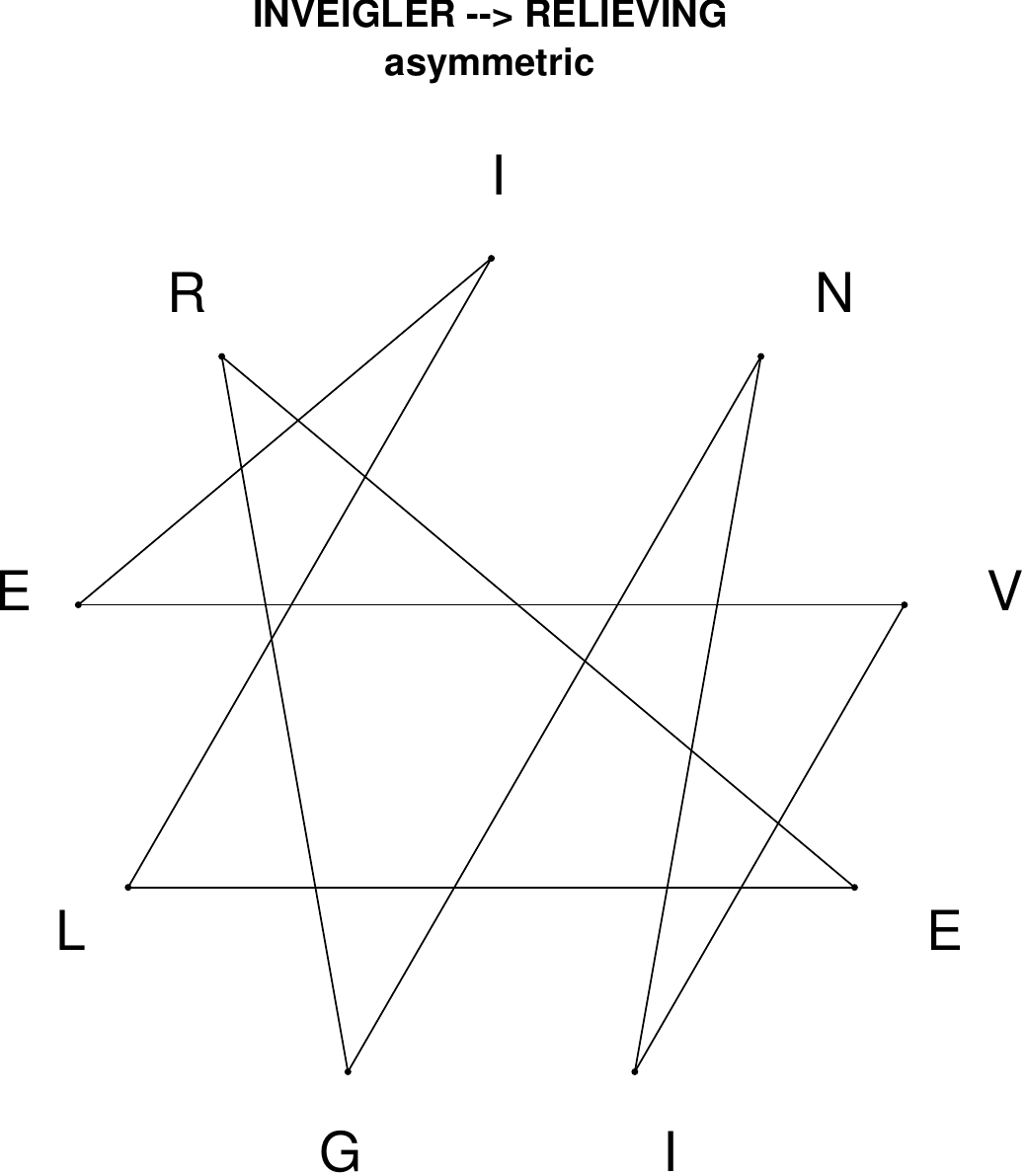}
\end{subfigure}
\hfill
\begin{subfigure}[T]{0.19\textwidth}
\centering
\includegraphics[width=\textwidth]{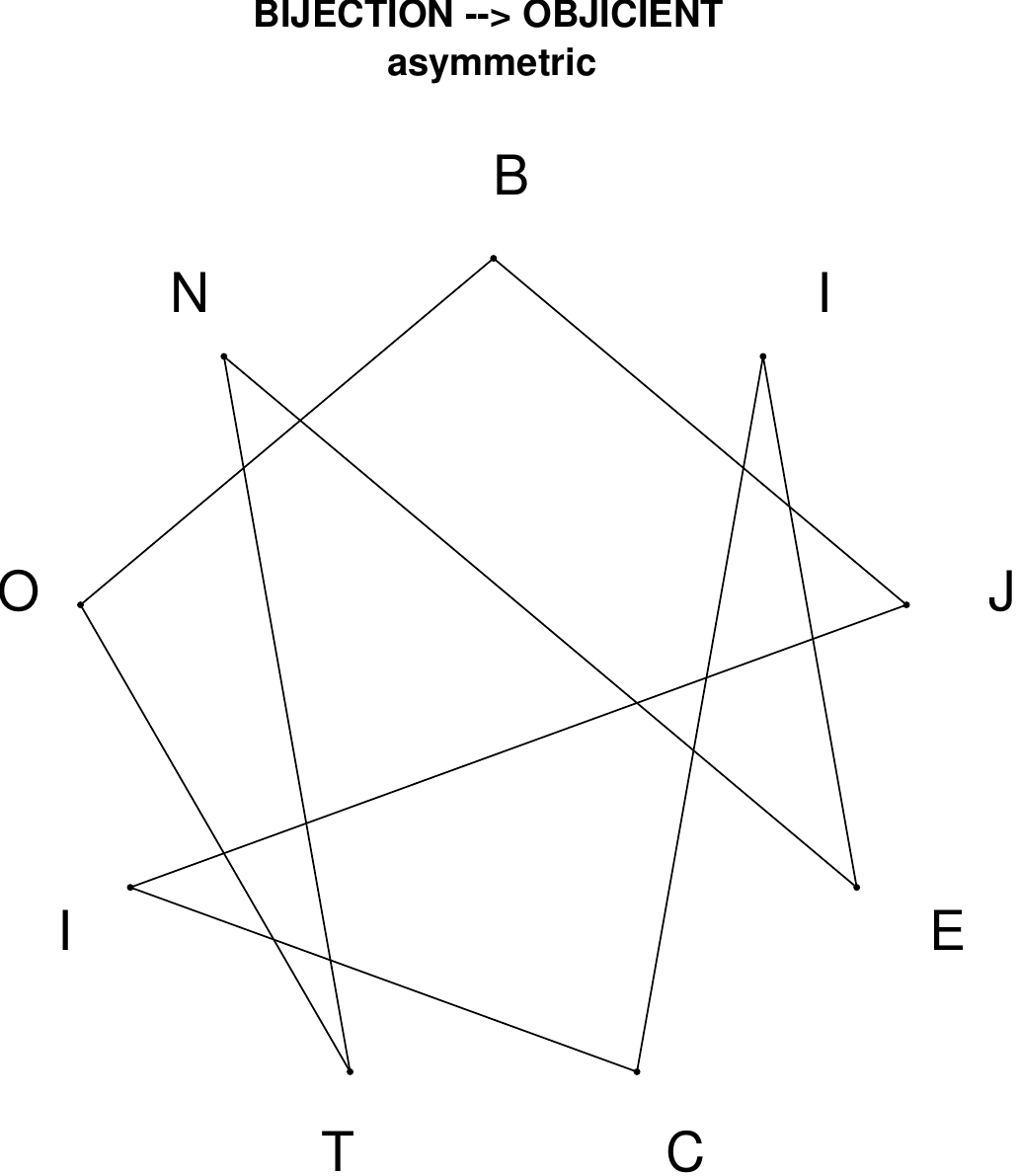}
\end{subfigure}
\hfill
\begin{subfigure}[T]{0.19\textwidth}
\centering
\includegraphics[width=\textwidth]{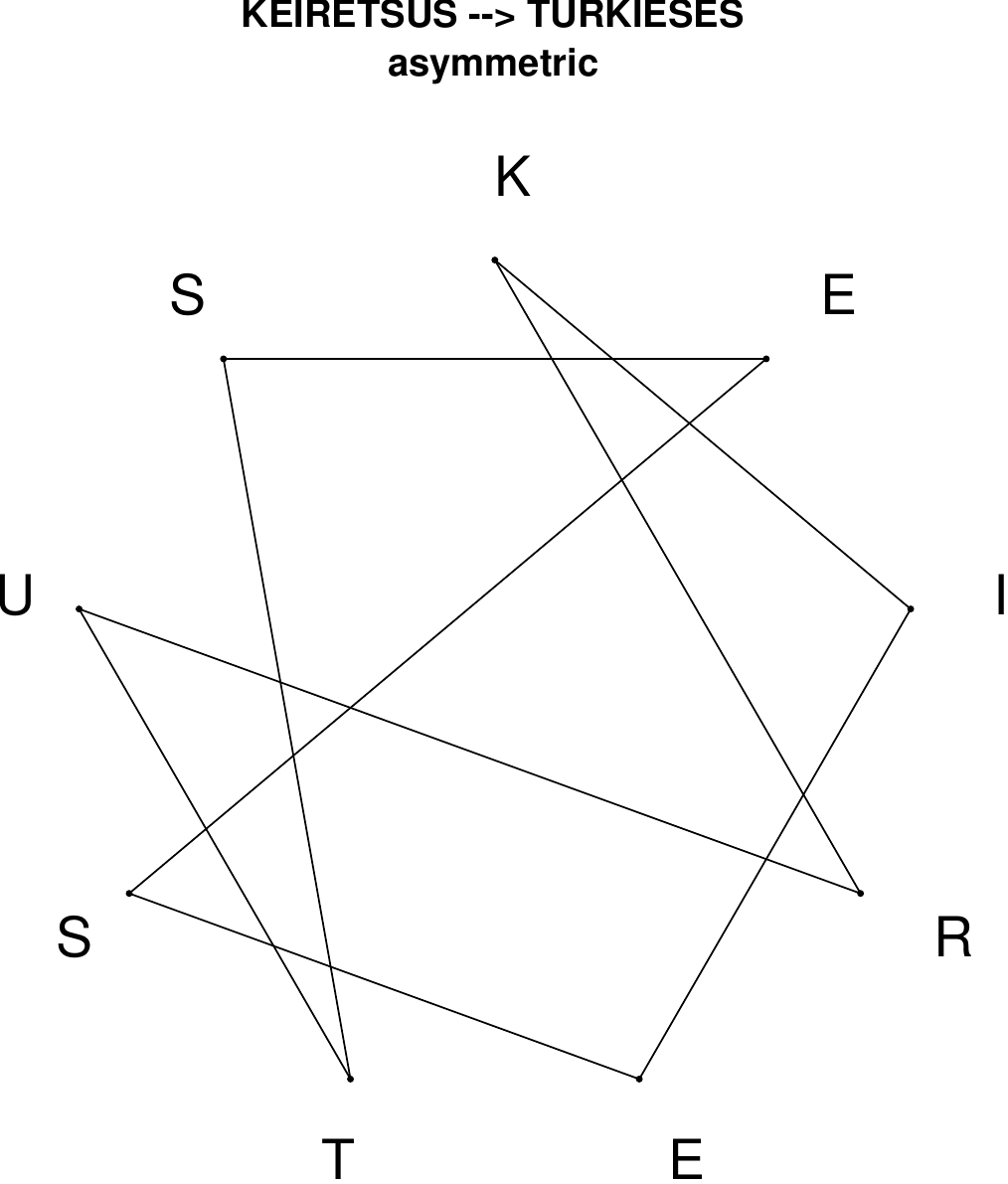}
\end{subfigure}
\hfill
\begin{subfigure}[T]{0.19\textwidth}
\centering
\includegraphics[width=\textwidth]{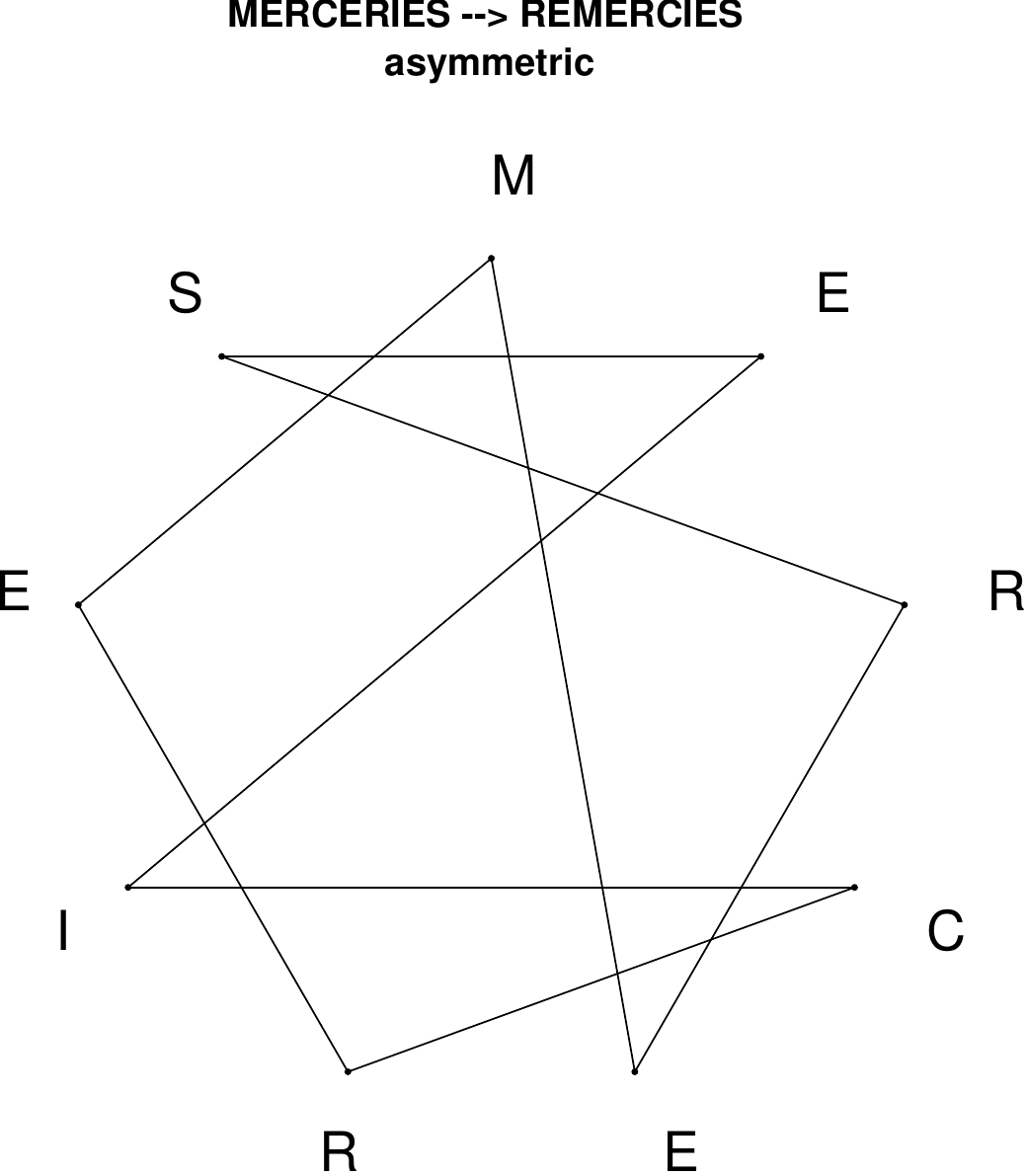}
\end{subfigure}
\end{figure}

\begin{figure}[H]
\centering
\begin{subfigure}[T]{0.19\textwidth}
\centering
\includegraphics[width=\textwidth]{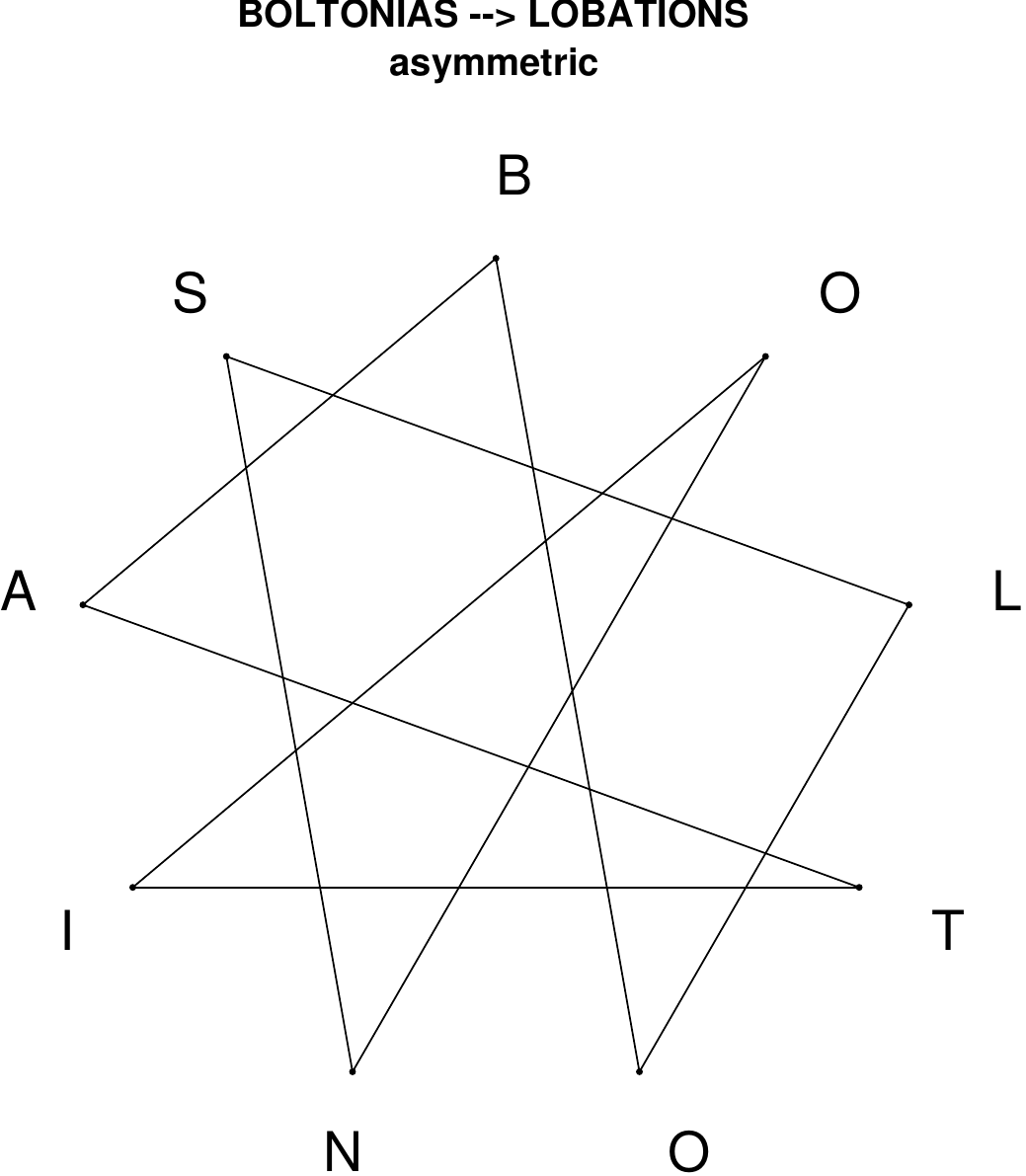}
\end{subfigure}
\hfill
\begin{subfigure}[T]{0.19\textwidth}
\centering
\includegraphics[width=\textwidth]{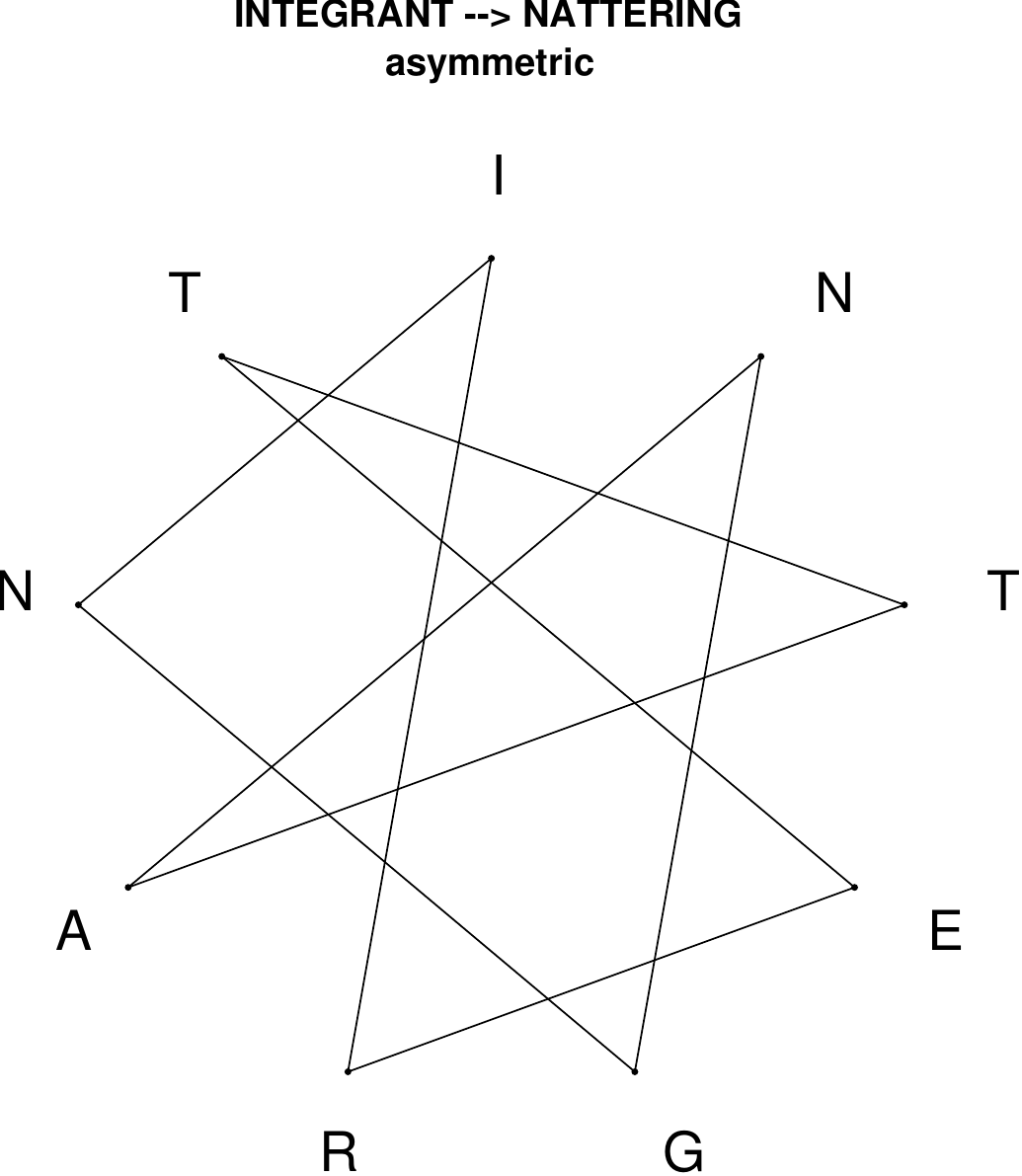}
\end{subfigure}
\hfill
\begin{subfigure}[T]{0.19\textwidth}
\centering
\includegraphics[width=\textwidth]{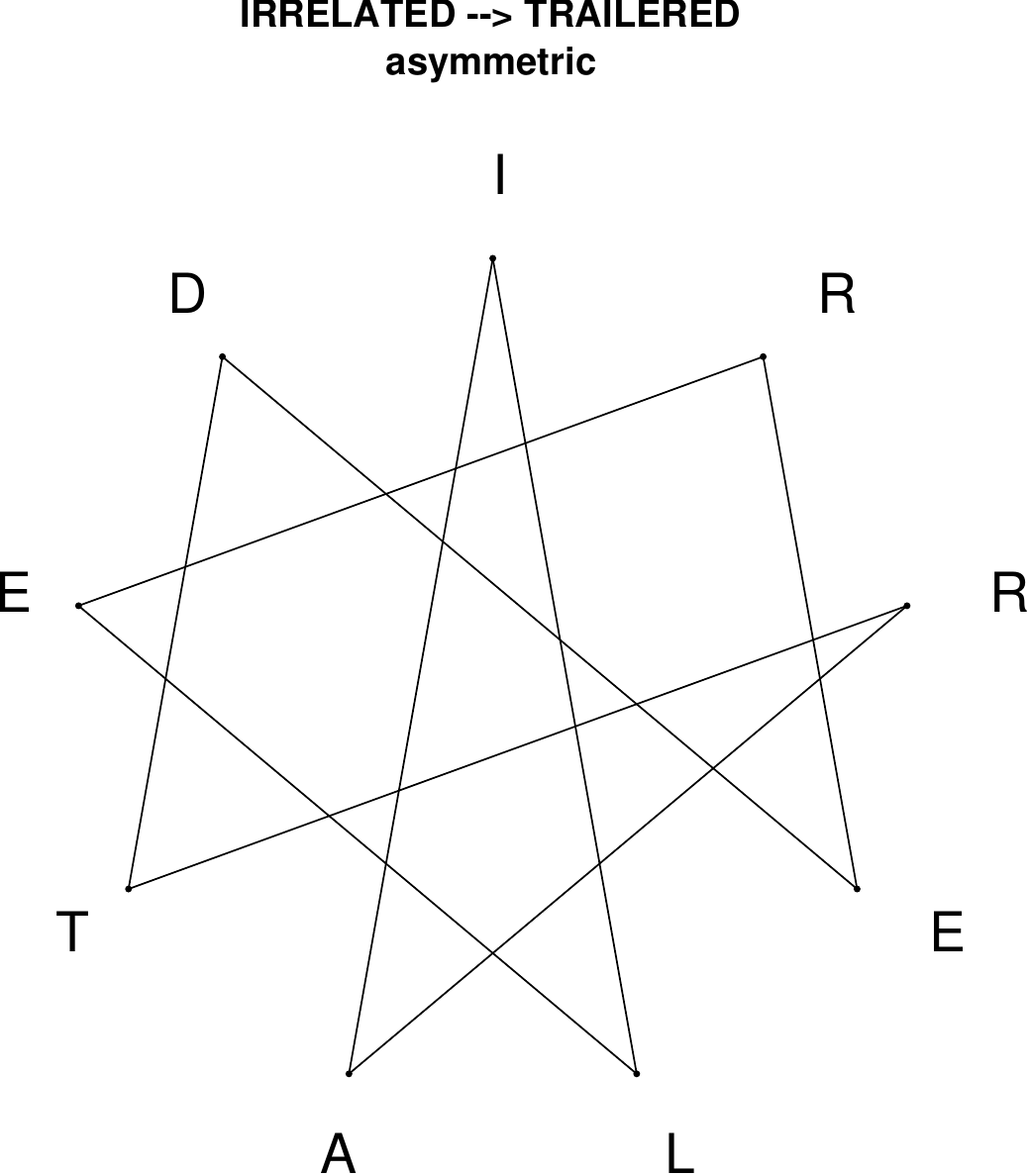}
\end{subfigure}
\hfill
\begin{subfigure}[T]{0.19\textwidth}
\centering
\includegraphics[width=\textwidth]{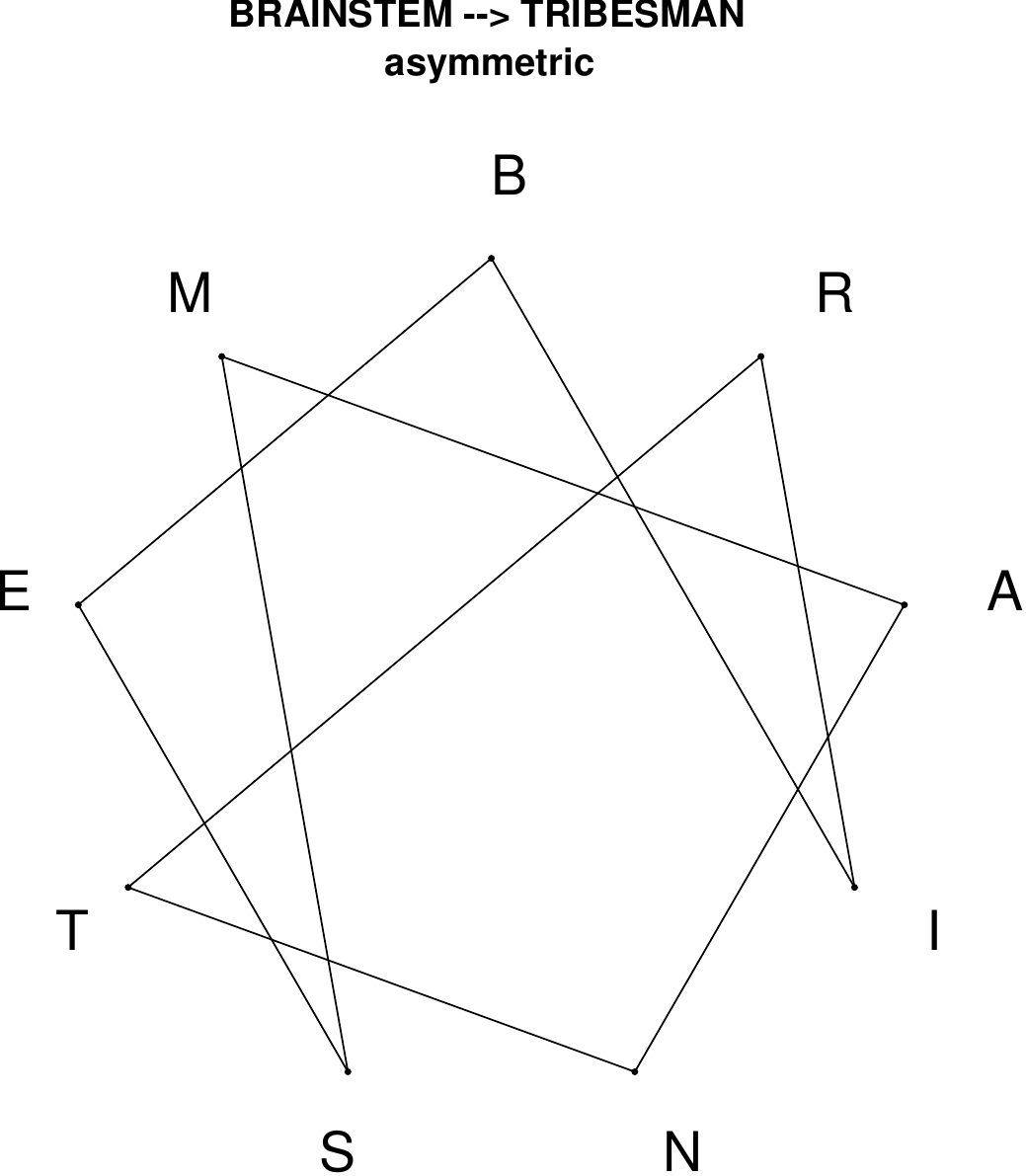}
\end{subfigure}
\hfill
\begin{subfigure}[T]{0.19\textwidth}
\centering
\includegraphics[width=\textwidth]{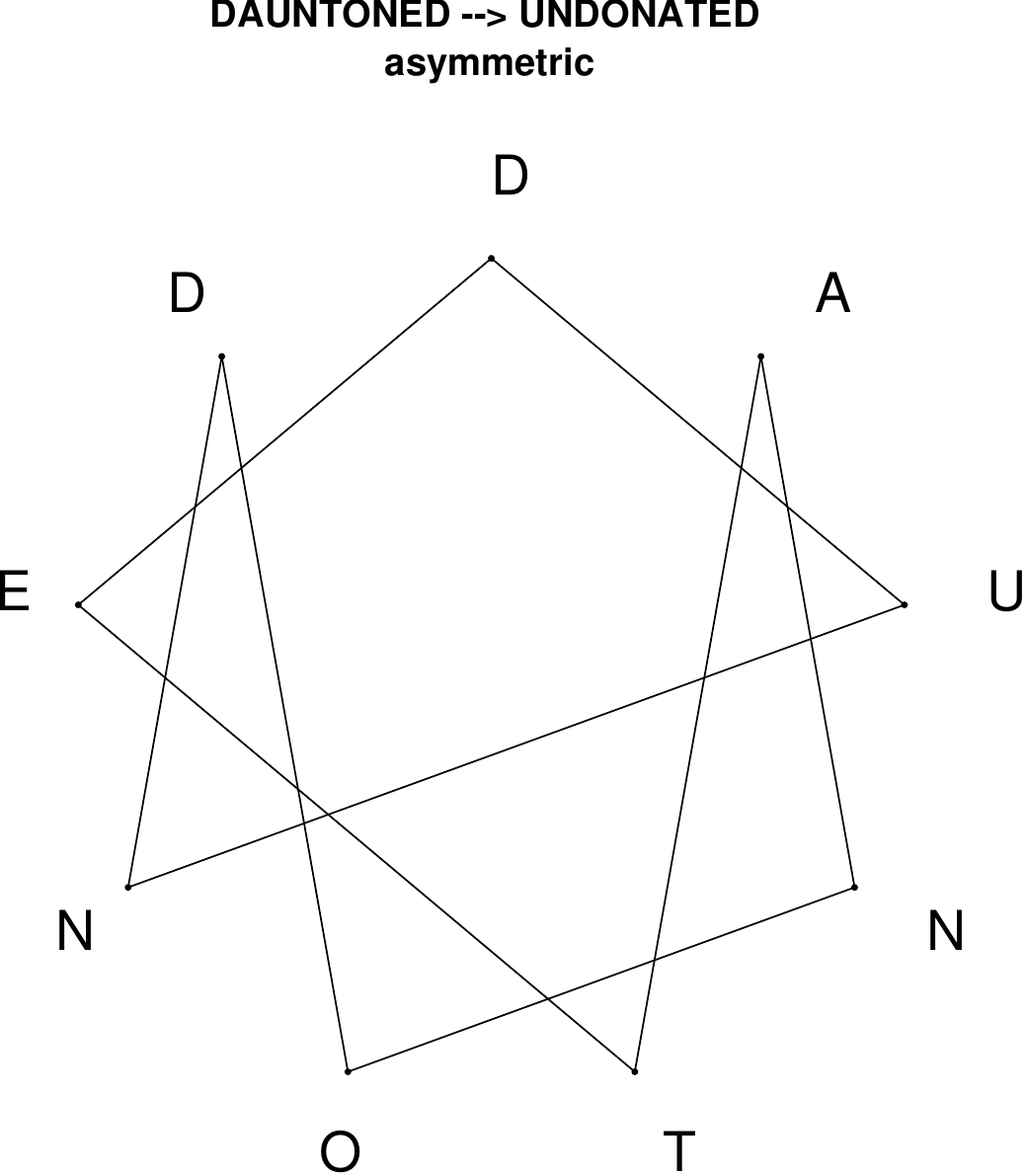}
\end{subfigure}
\end{figure}

\begin{figure}[H]
\centering
\begin{subfigure}[T]{0.19\textwidth}
\centering
\includegraphics[width=\textwidth]{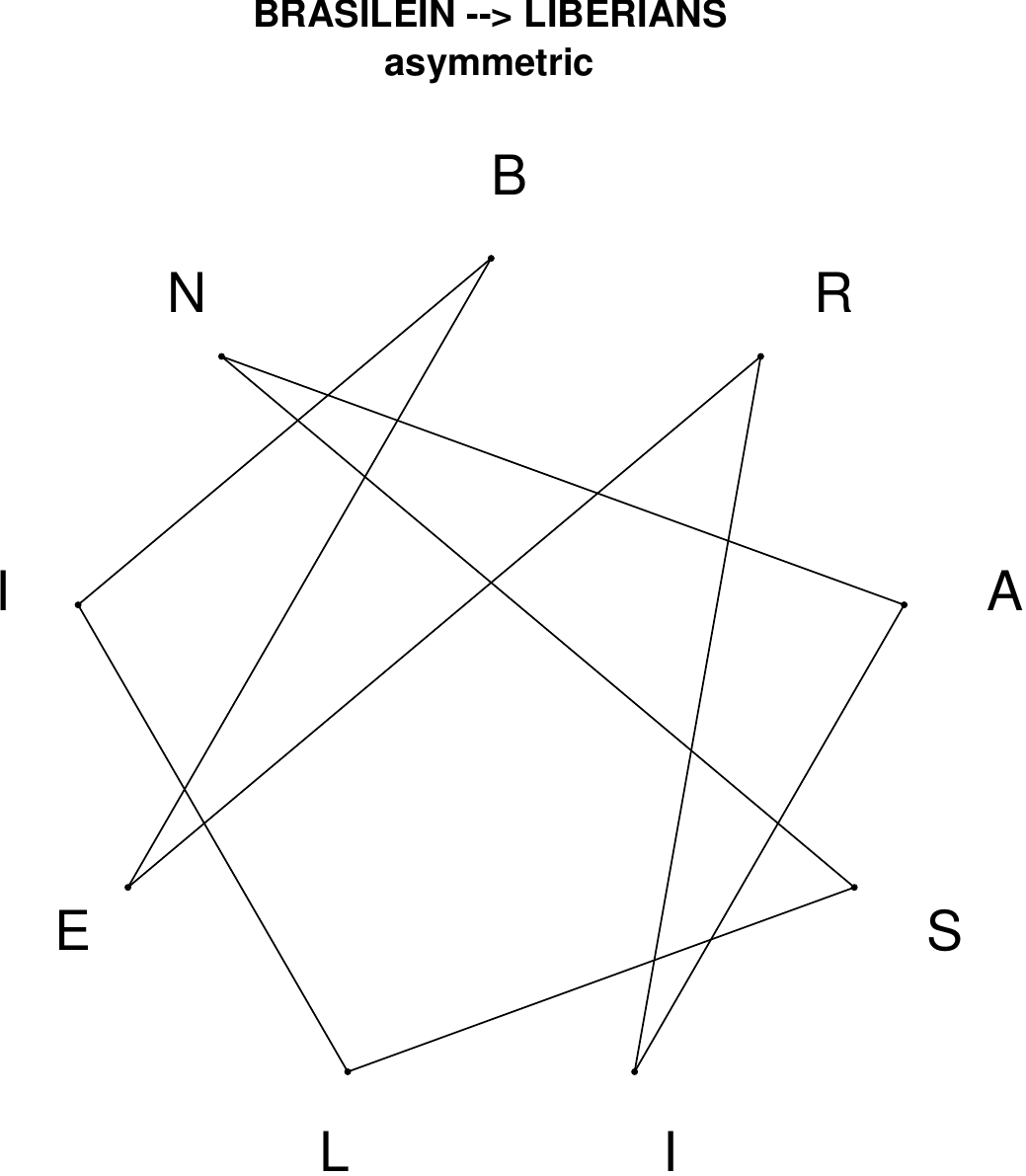}
\end{subfigure}
\hfill
\begin{subfigure}[T]{0.19\textwidth}
\centering
\includegraphics[width=\textwidth]{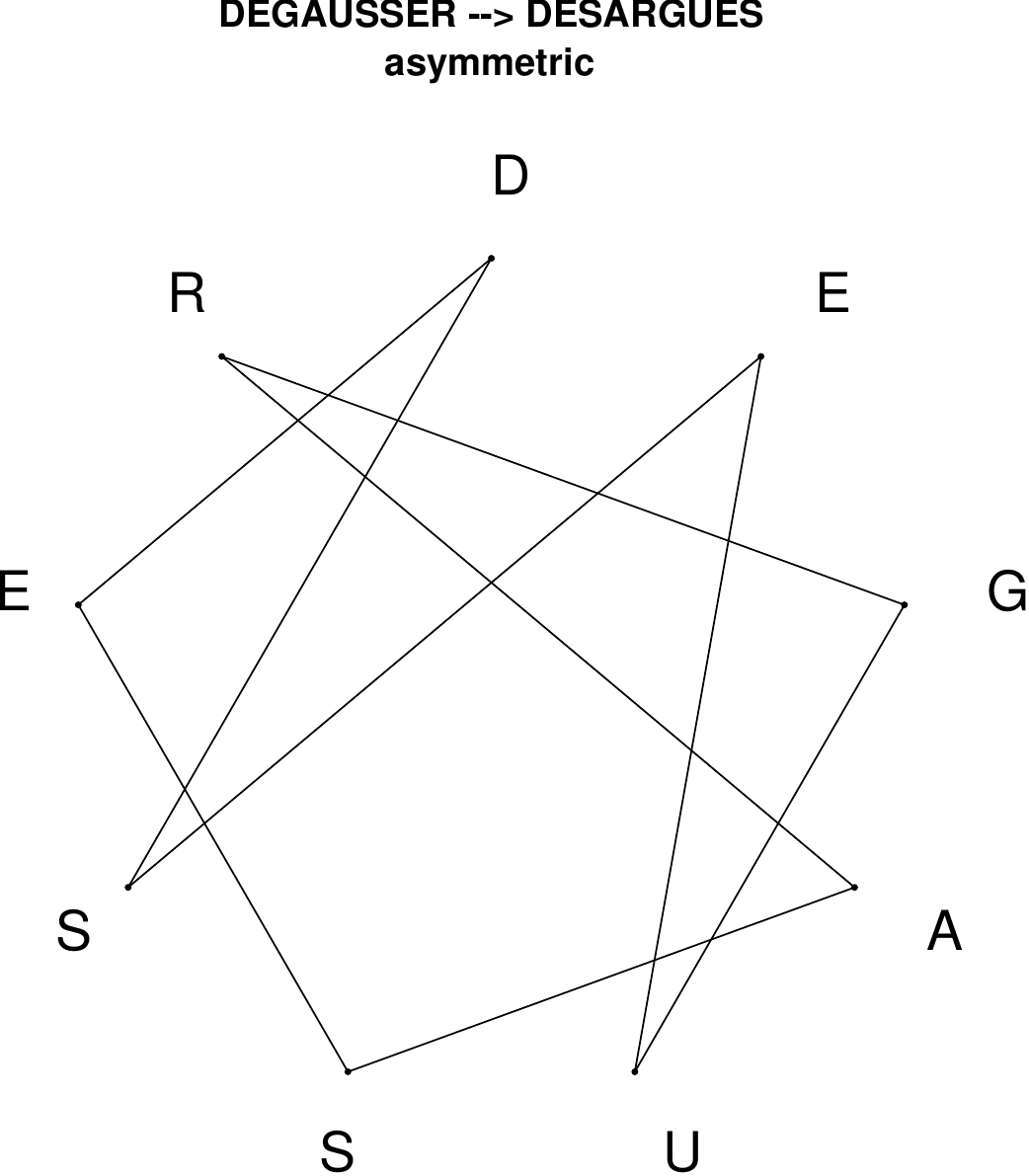}
\end{subfigure}
\hfill
\begin{subfigure}[T]{0.19\textwidth}
\centering
\includegraphics[width=\textwidth]{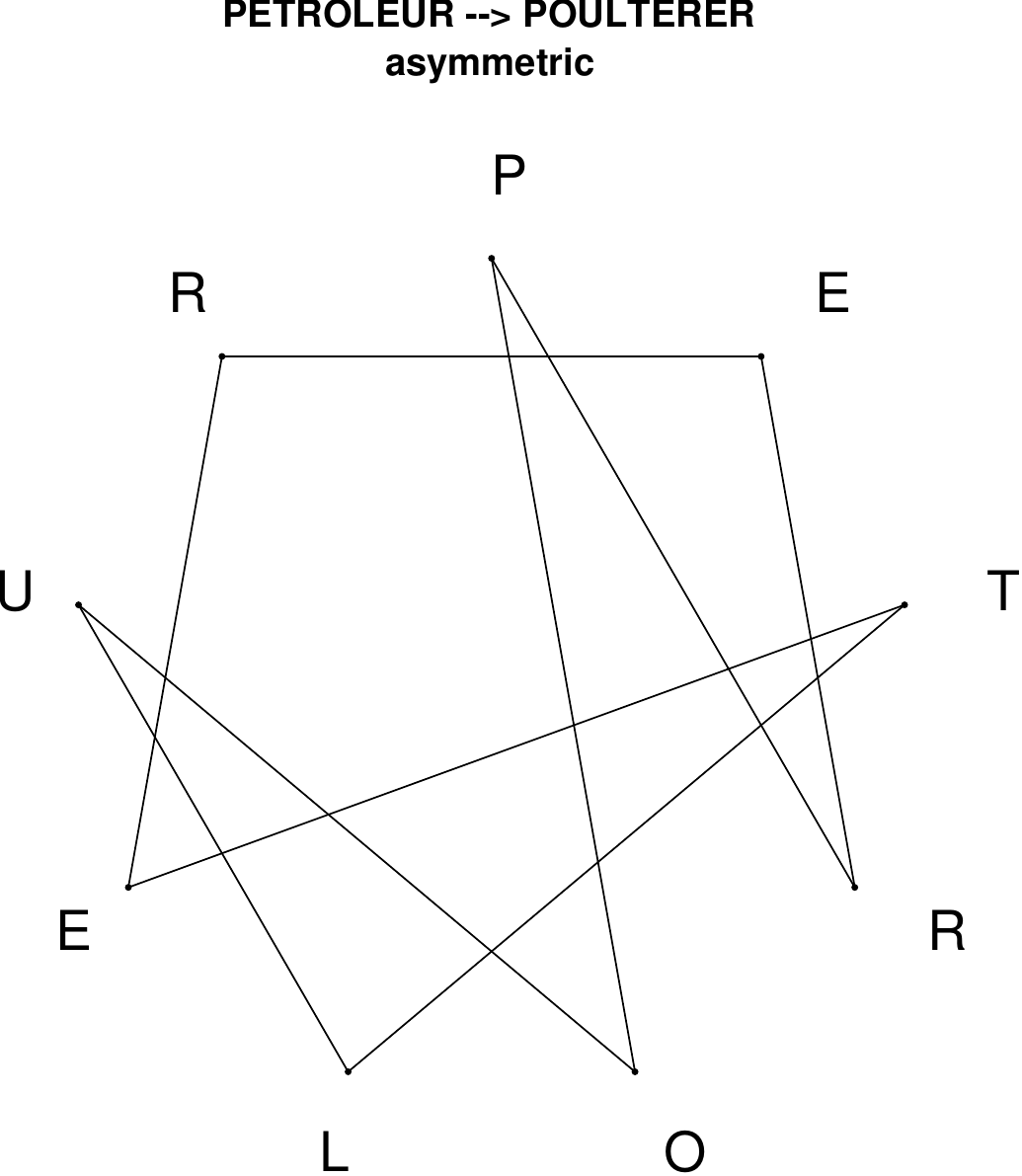}
\end{subfigure}
\hfill
\begin{subfigure}[T]{0.19\textwidth}
\centering
\includegraphics[width=\textwidth]{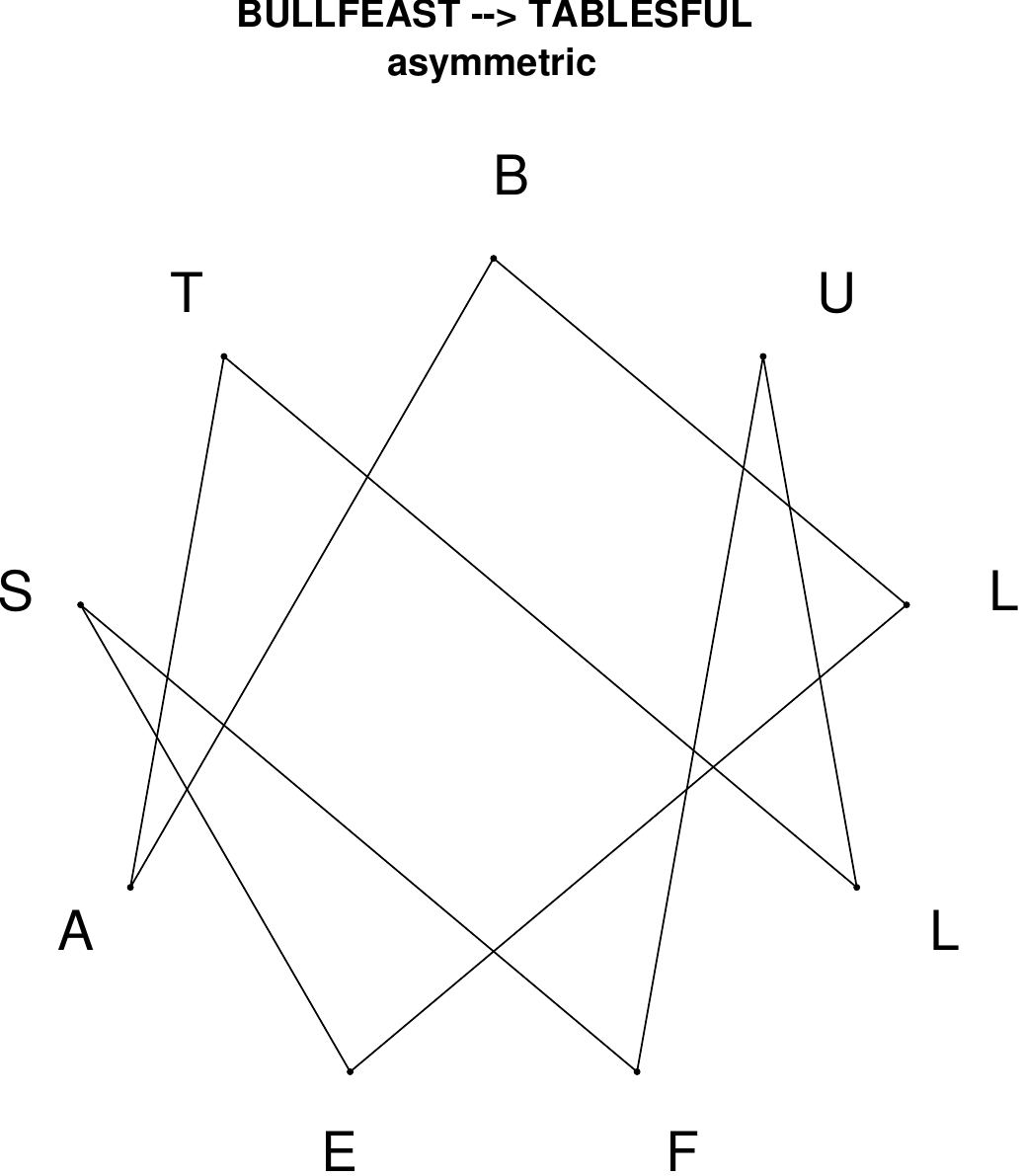}
\end{subfigure}
\hfill
\begin{subfigure}[T]{0.19\textwidth}
\centering
\includegraphics[width=\textwidth]{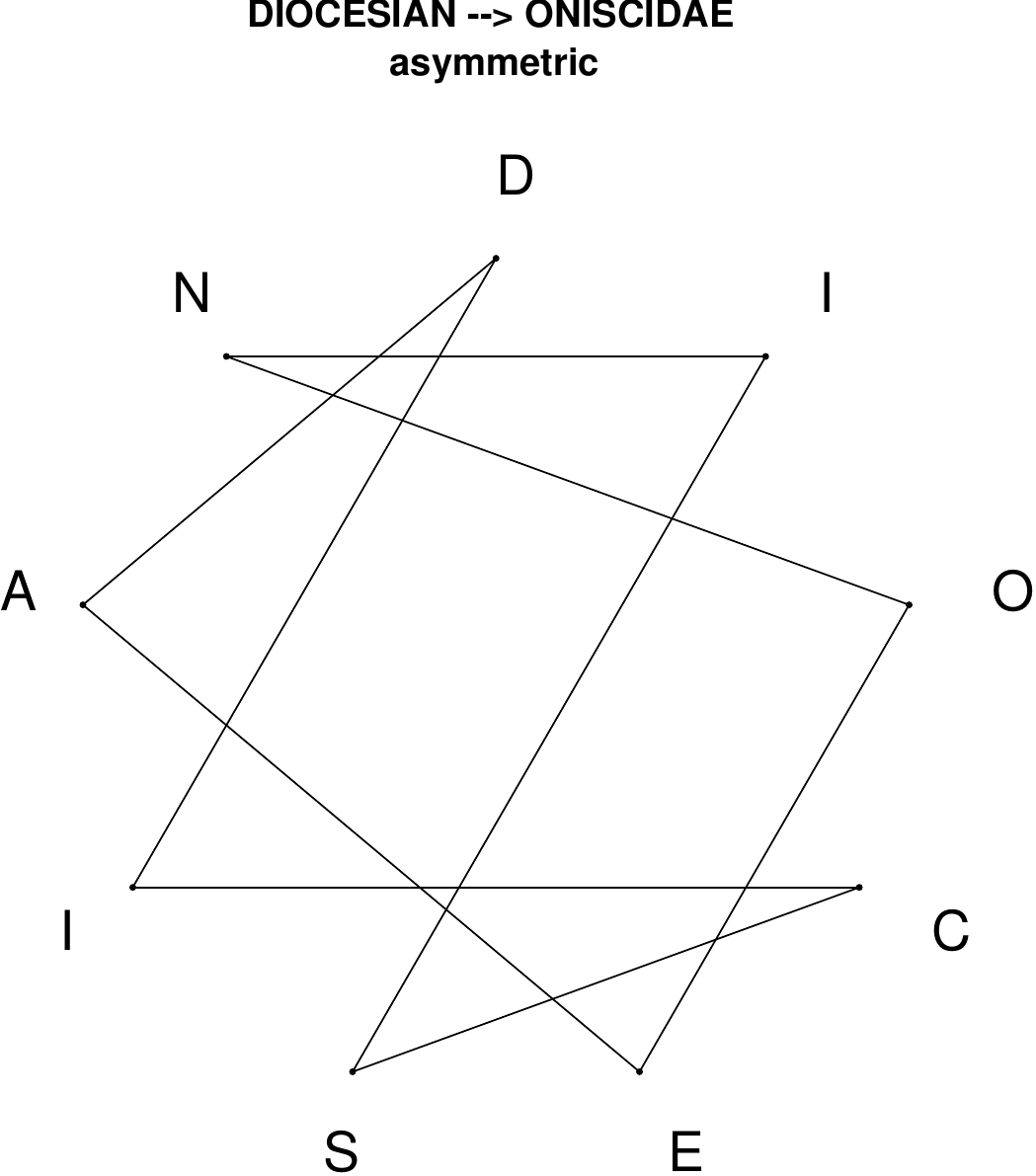}
\end{subfigure}
\end{figure}

\begin{figure}[H]
\centering
\begin{subfigure}[T]{0.19\textwidth}
\centering
\includegraphics[width=\textwidth]{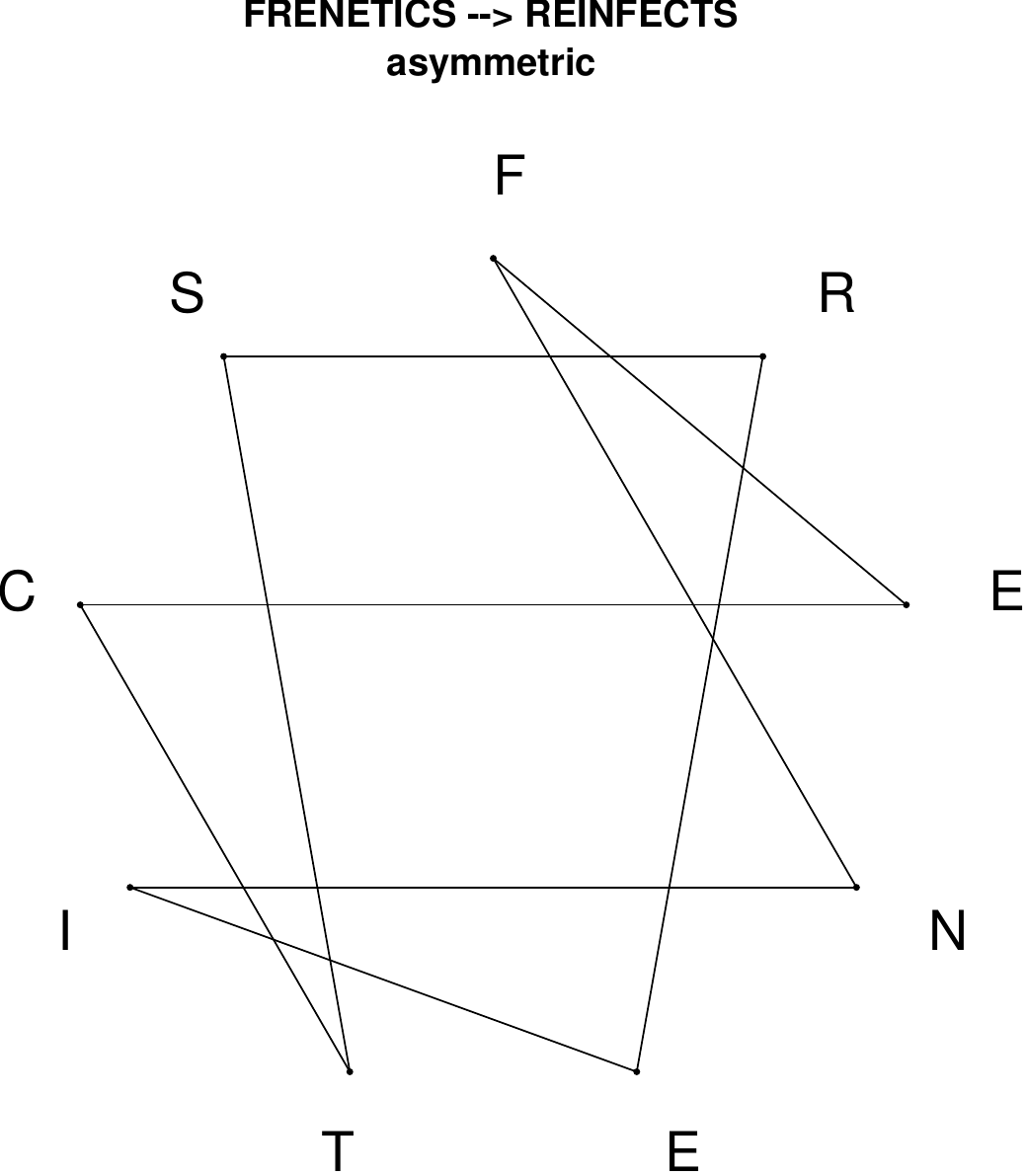}
\end{subfigure}
\hfill
\begin{subfigure}[T]{0.19\textwidth}
\centering
\includegraphics[width=\textwidth]{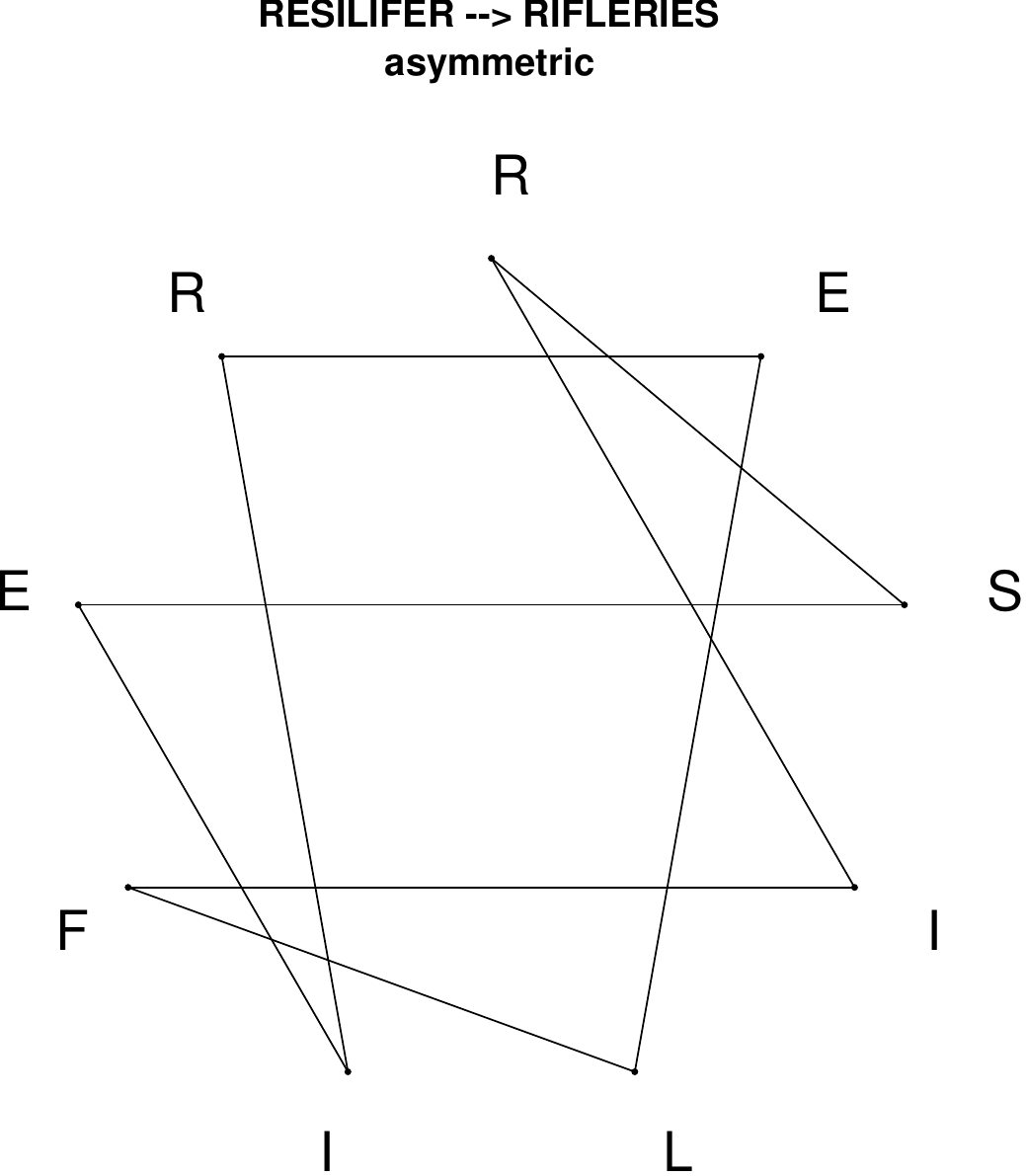}
\end{subfigure}
\hfill
\begin{subfigure}[T]{0.19\textwidth}
\centering
\includegraphics[width=\textwidth]{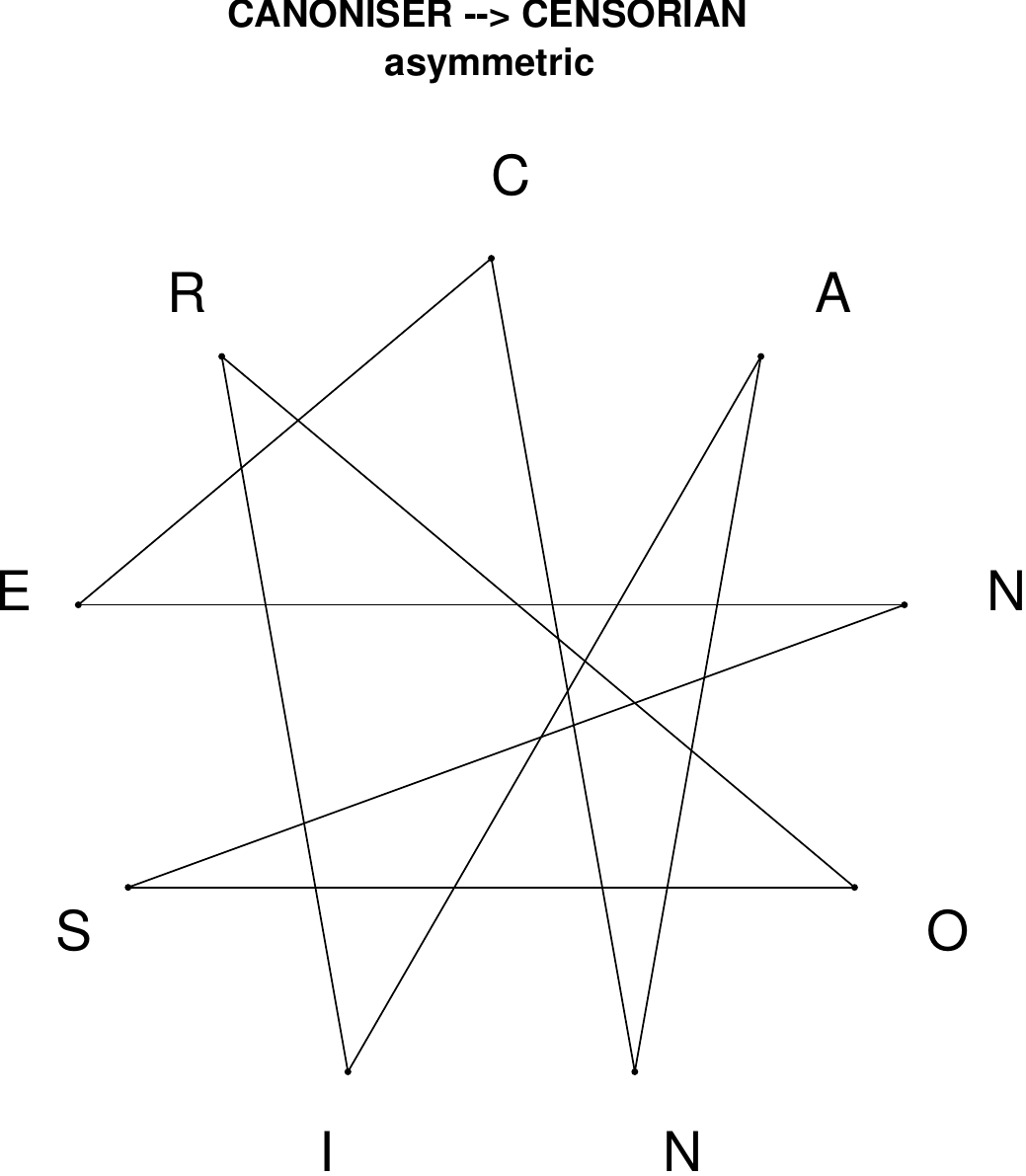}
\end{subfigure}
\hfill
\begin{subfigure}[T]{0.19\textwidth}
\centering
\includegraphics[width=\textwidth]{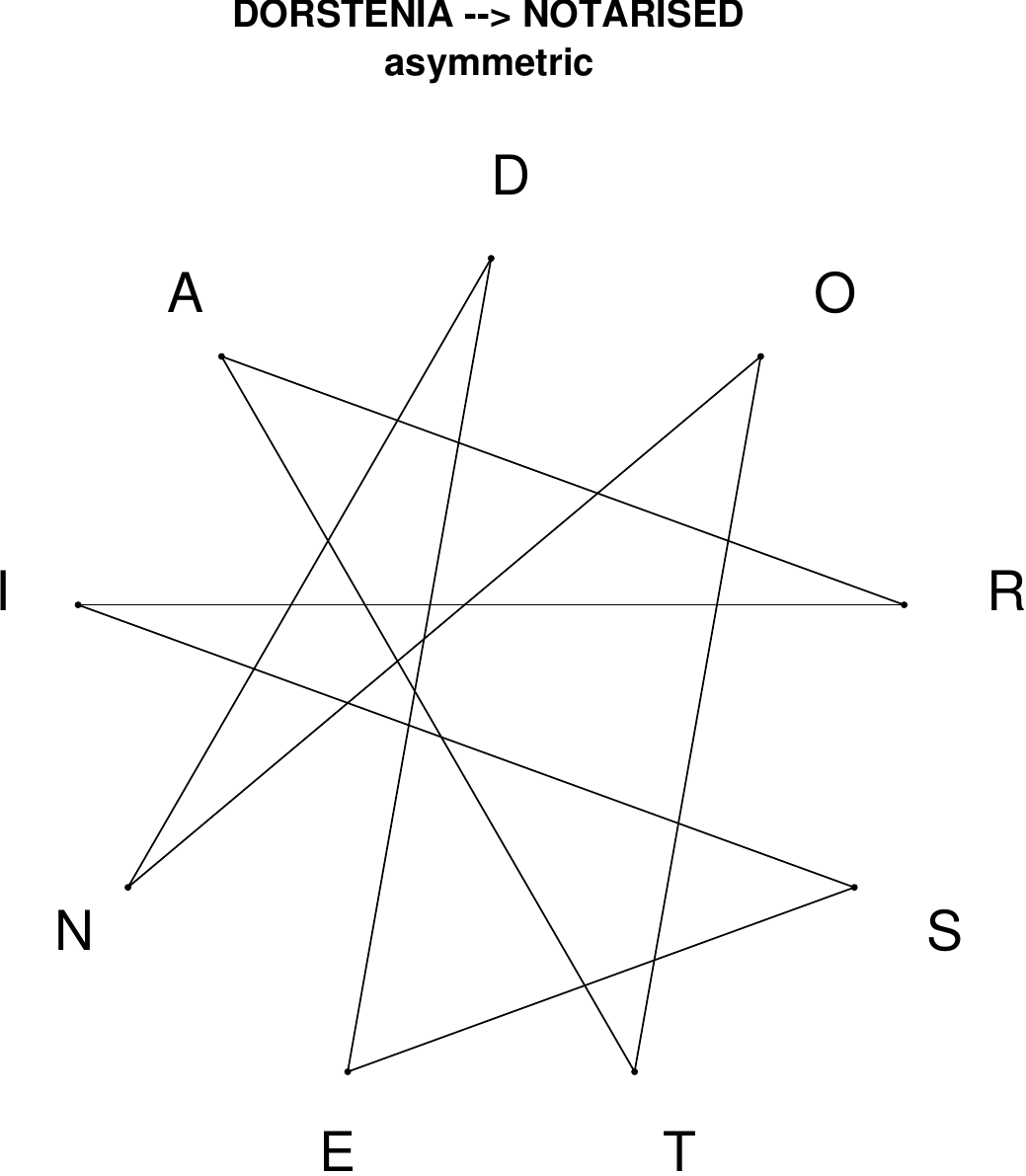}
\end{subfigure}
\hfill
\begin{subfigure}[T]{0.19\textwidth}
\centering
\includegraphics[width=\textwidth]{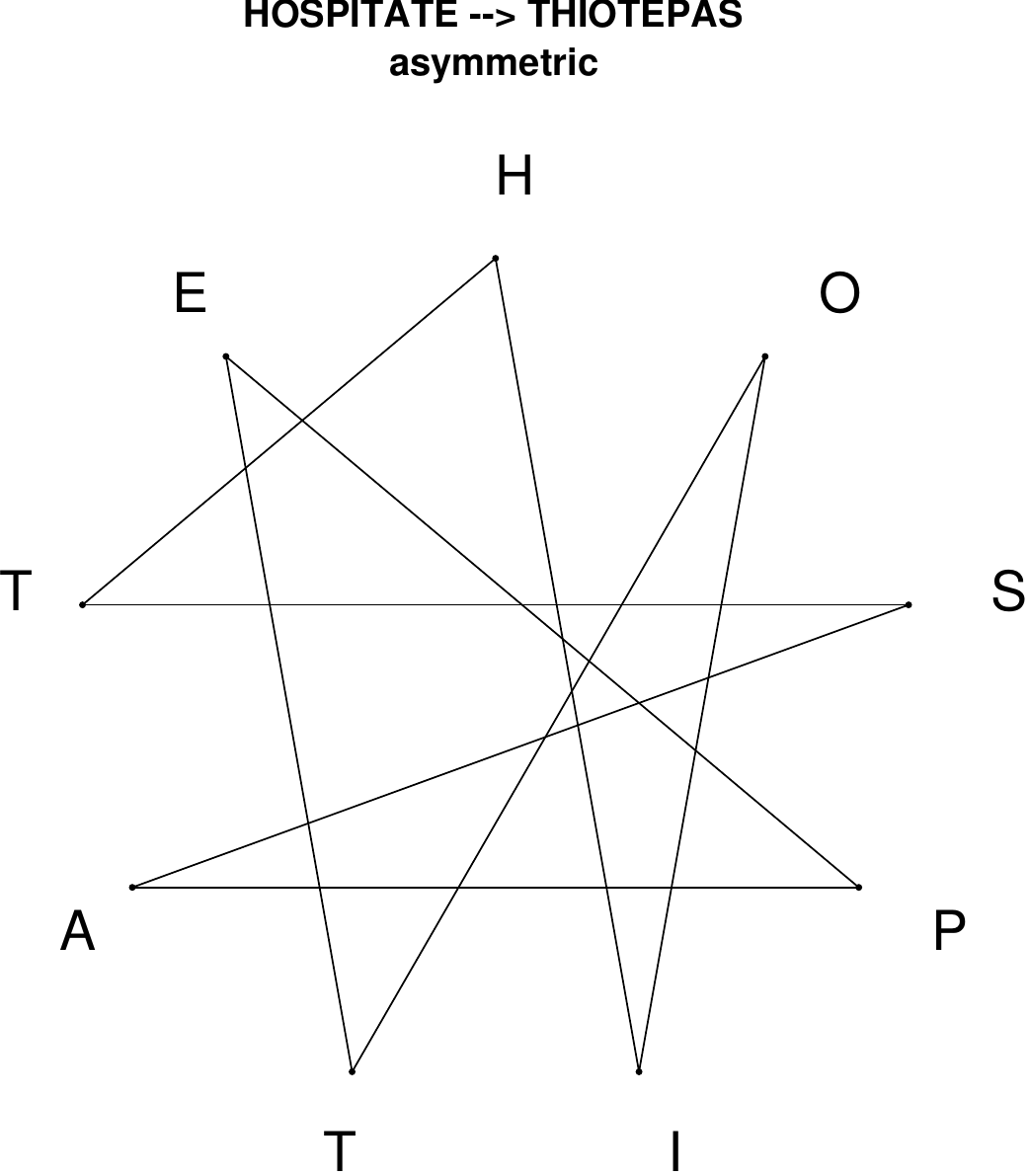}
\end{subfigure}
\end{figure}

\begin{figure}[H]
\centering
\begin{subfigure}[T]{0.19\textwidth}
\centering
\includegraphics[width=\textwidth]{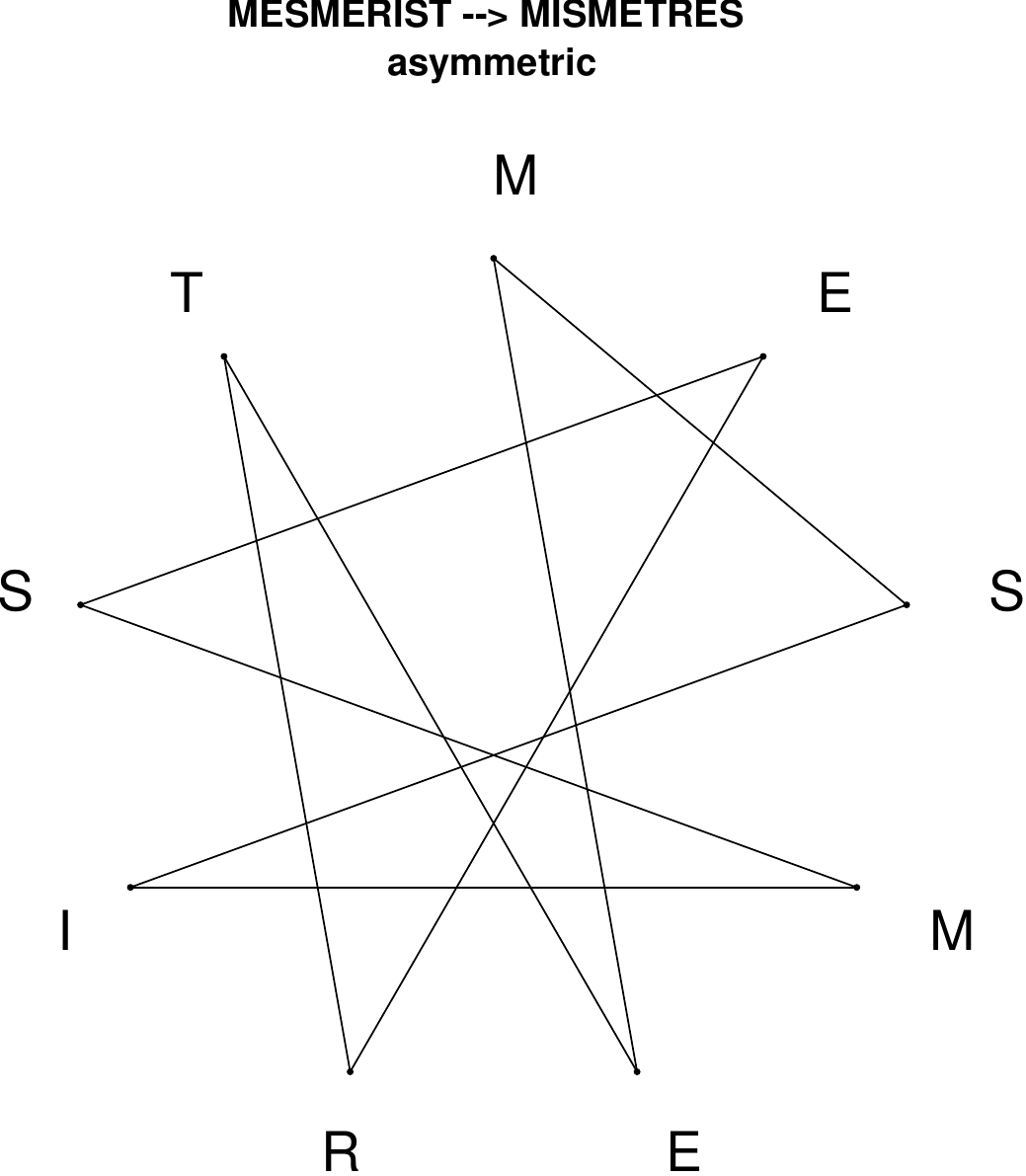}
\end{subfigure}
\hfill
\begin{subfigure}[T]{0.19\textwidth}
\centering
\includegraphics[width=\textwidth]{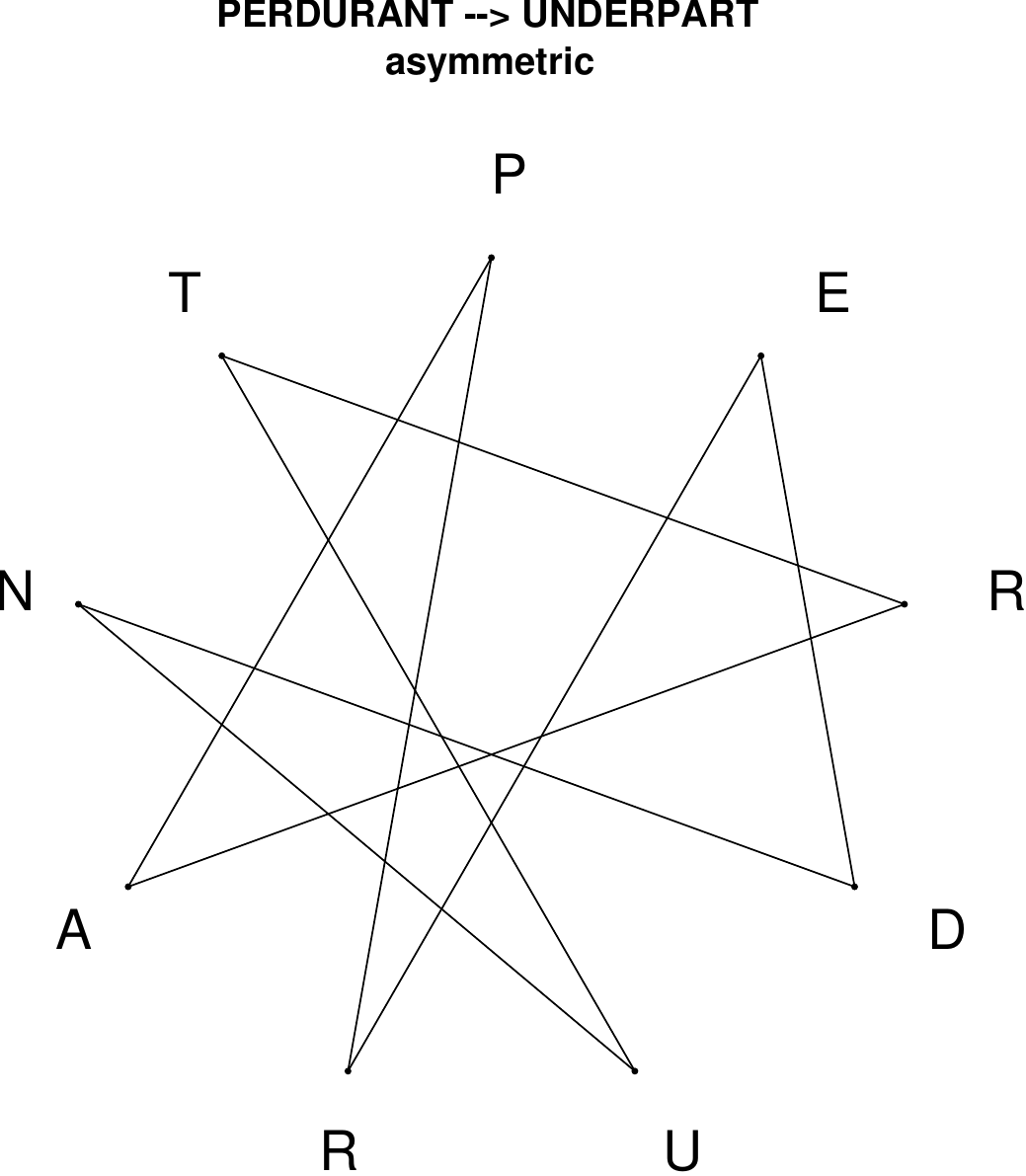}
\end{subfigure}
\hfill
\begin{subfigure}[T]{0.19\textwidth}
\centering
\includegraphics[width=\textwidth]{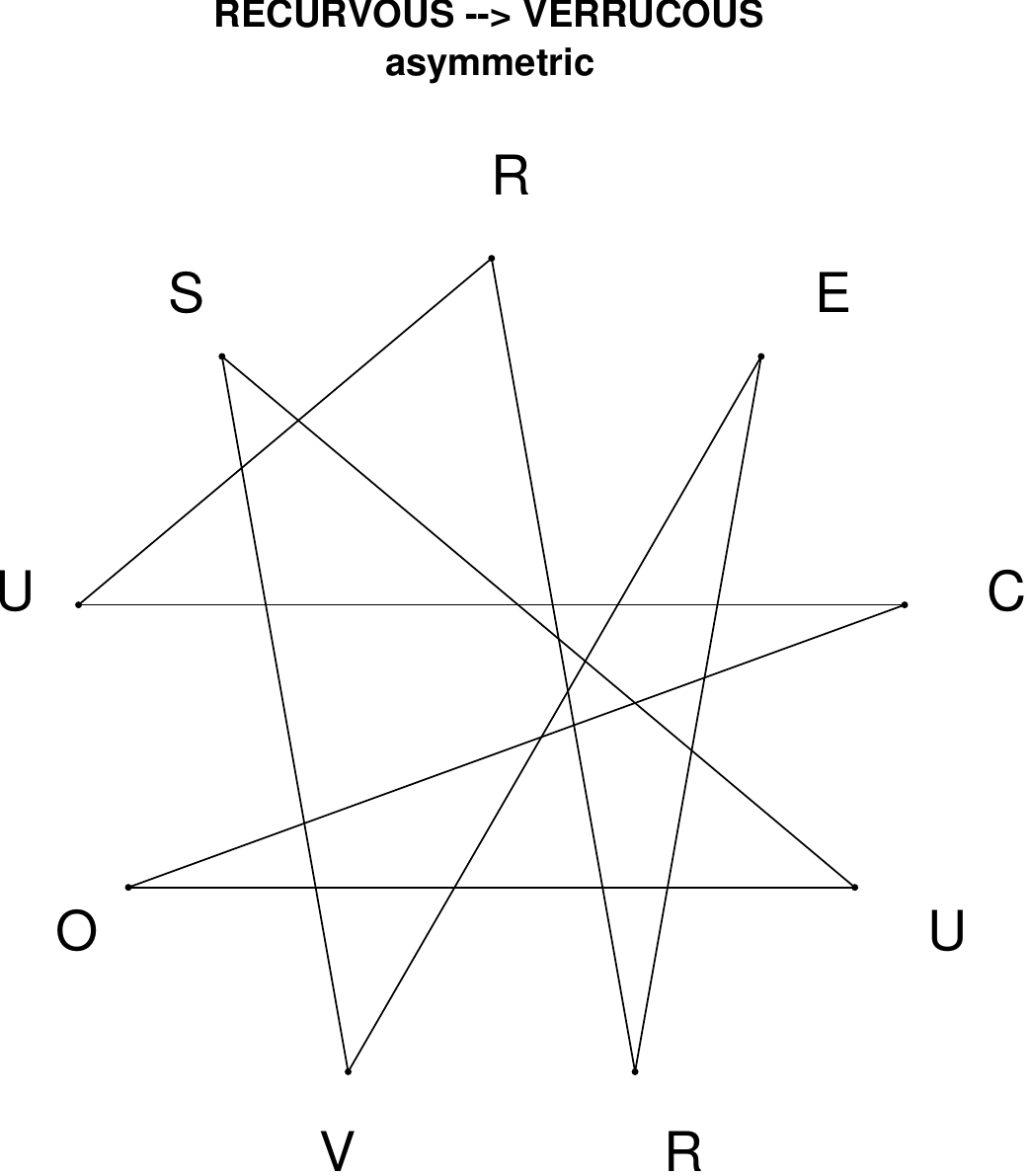}
\end{subfigure}
\hfill
\begin{subfigure}[T]{0.19\textwidth}
\centering
\includegraphics[width=\textwidth]{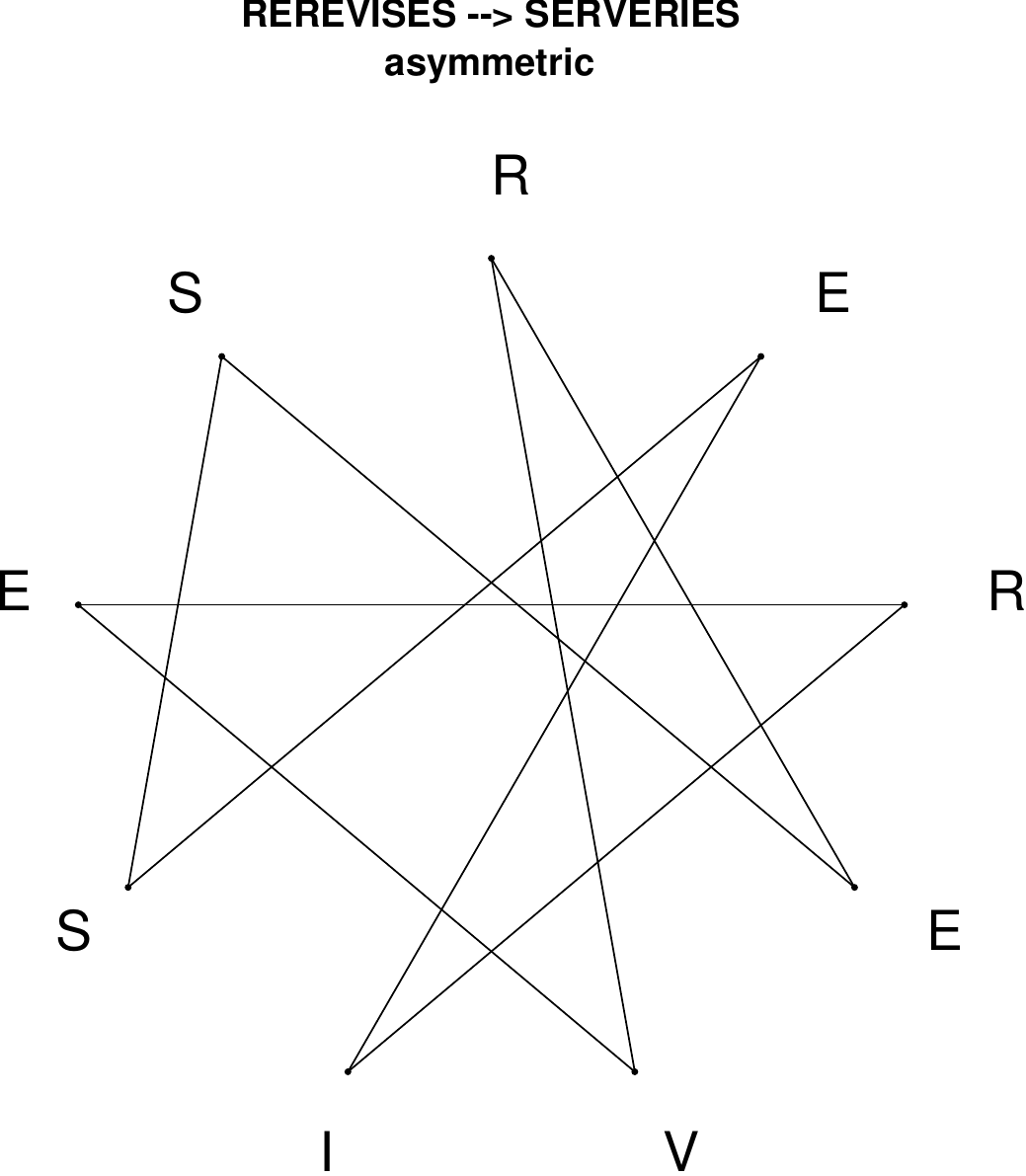}
\end{subfigure}
\hfill
\begin{subfigure}[T]{0.19\textwidth}
\centering
\includegraphics[width=\textwidth]{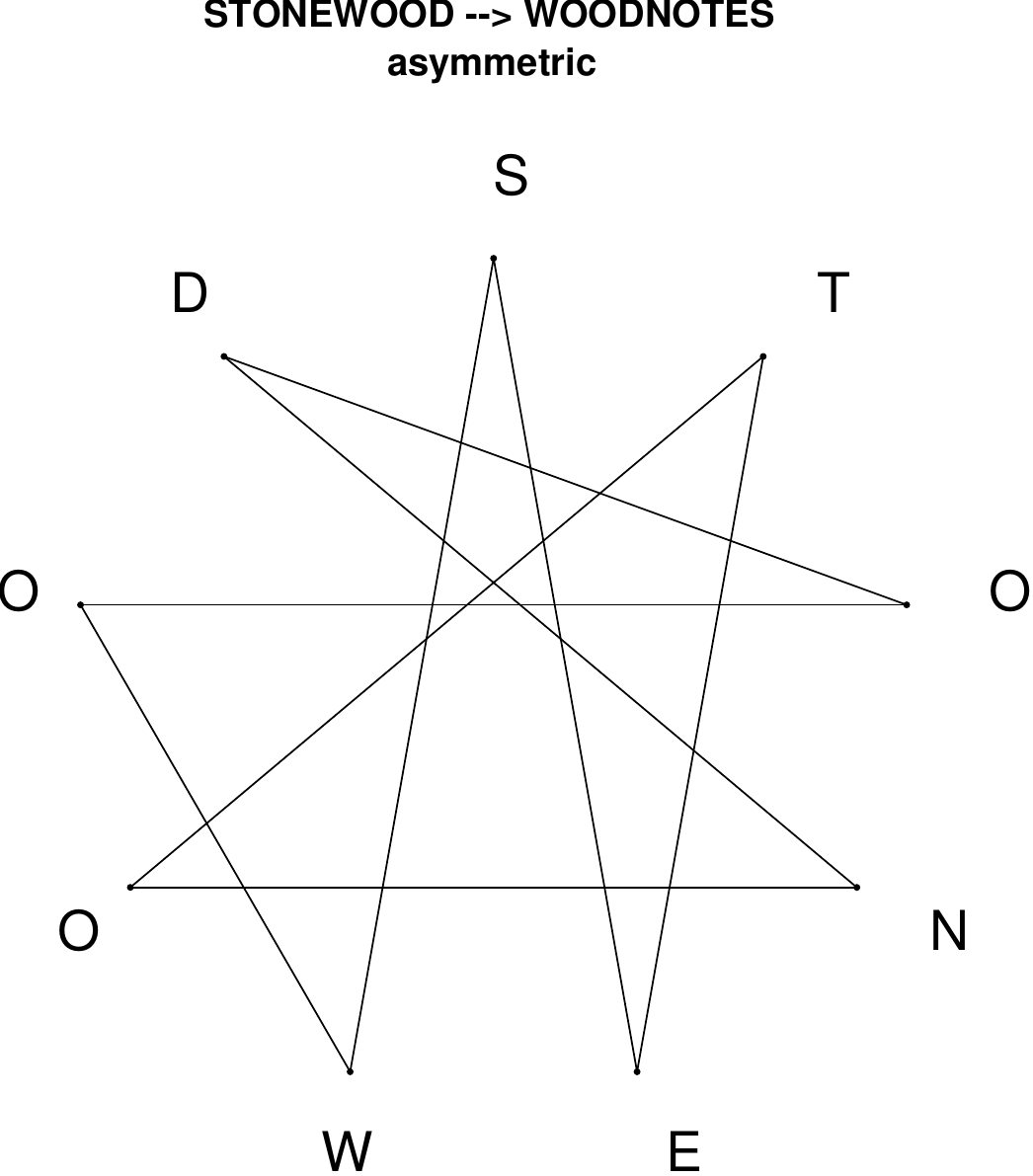}
\end{subfigure}
\end{figure}

\begin{figure}[H]
\centering
\begin{subfigure}[T]{0.19\textwidth}
\centering
\includegraphics[width=\textwidth]{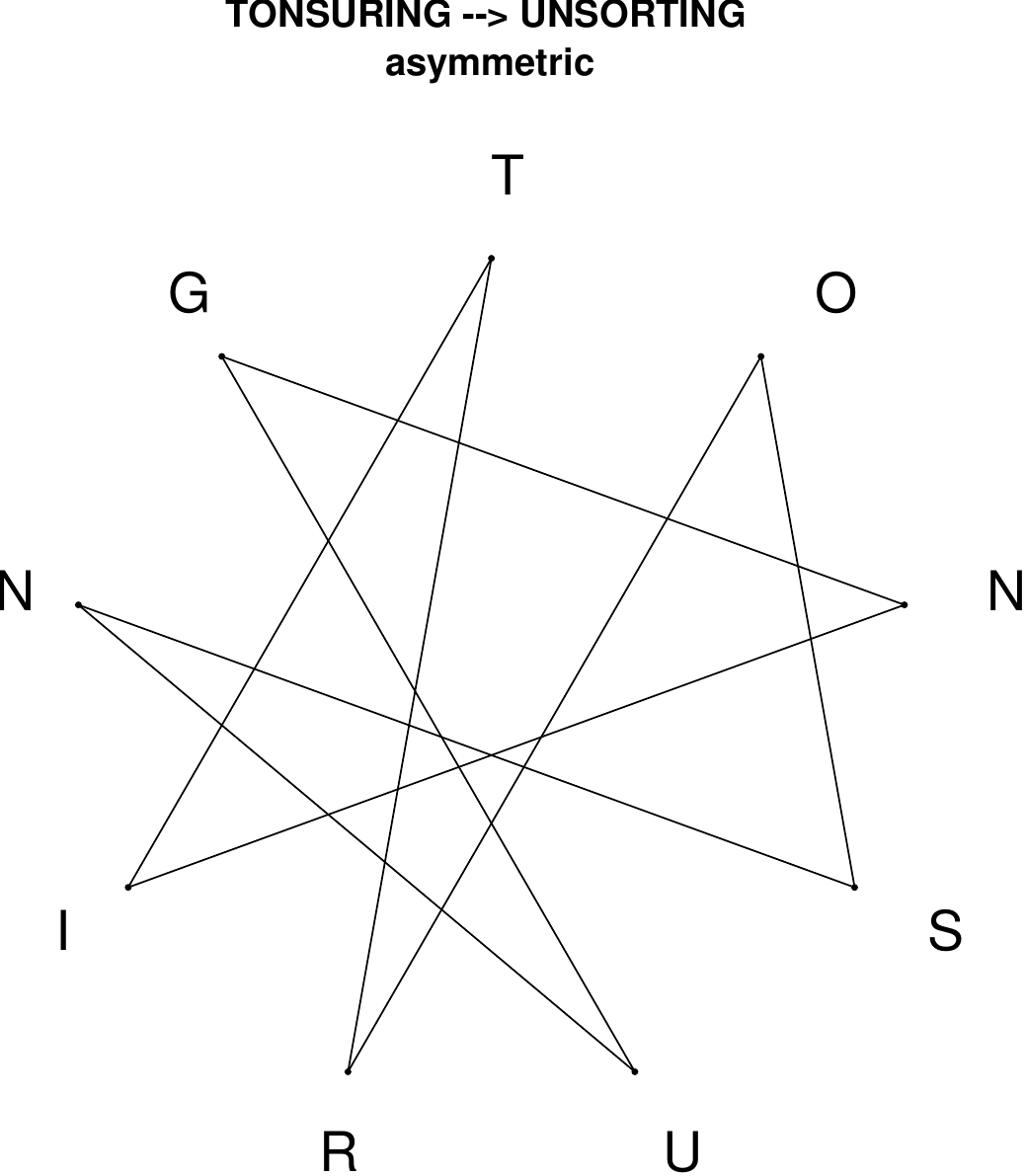}
\end{subfigure}
\hfill
\begin{subfigure}[T]{0.19\textwidth}
\centering
\includegraphics[width=\textwidth]{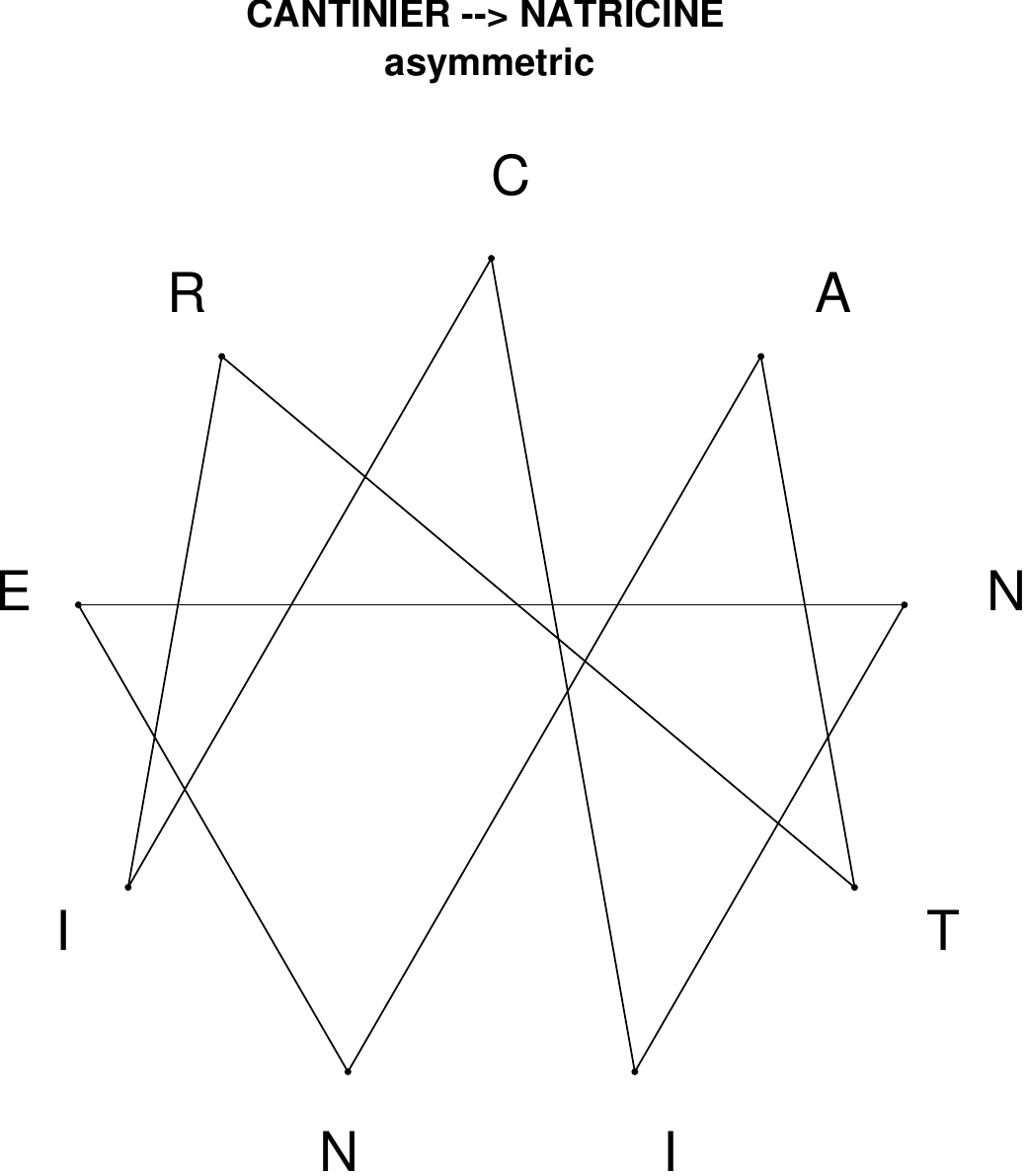}
\end{subfigure}
\hfill
\begin{subfigure}[T]{0.19\textwidth}
\centering
\includegraphics[width=\textwidth]{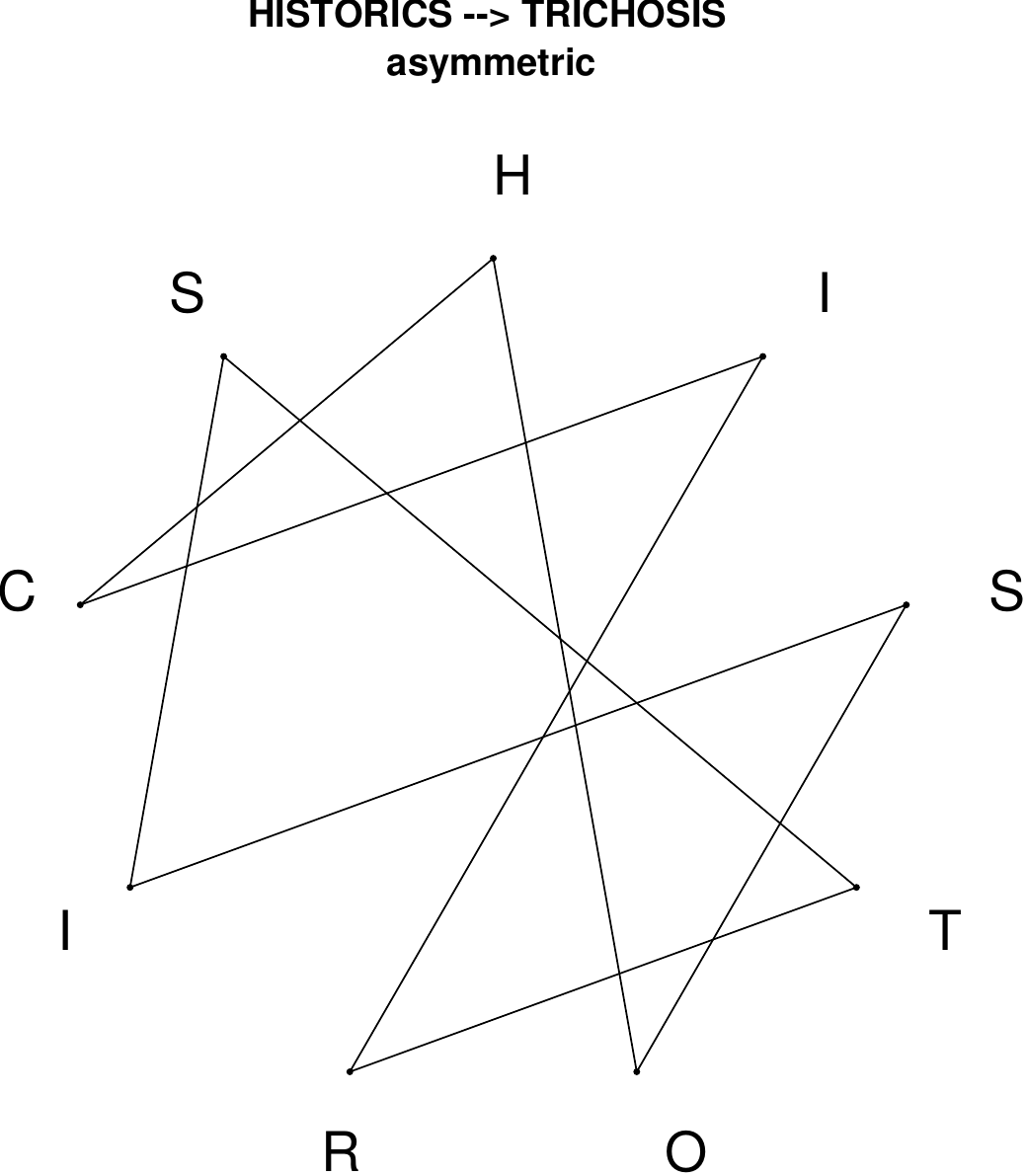}
\end{subfigure}
\hfill
\begin{subfigure}[T]{0.19\textwidth}
\centering
\includegraphics[width=\textwidth]{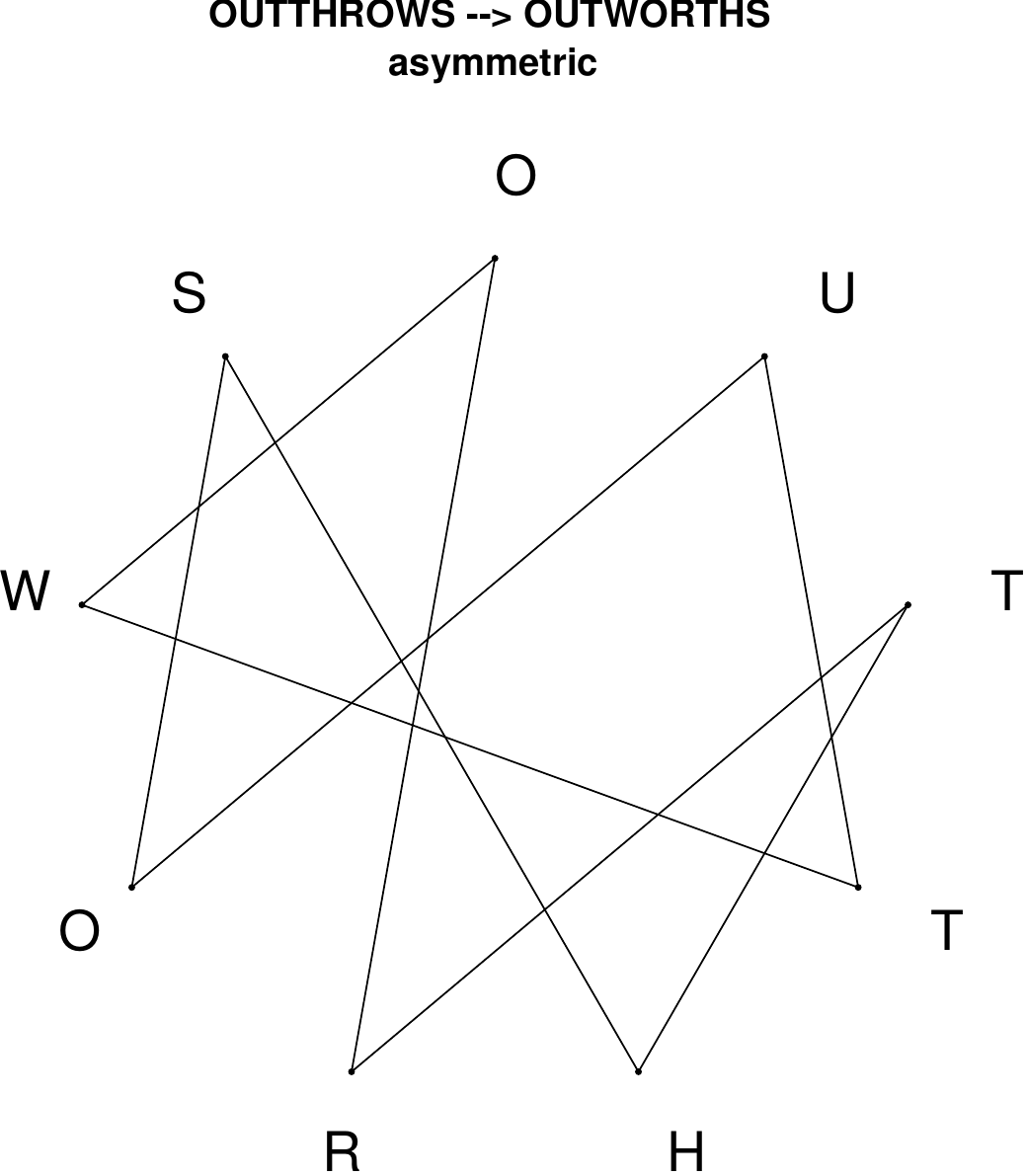}
\end{subfigure}
\hfill
\begin{subfigure}[T]{0.19\textwidth}
\centering
\includegraphics[width=\textwidth]{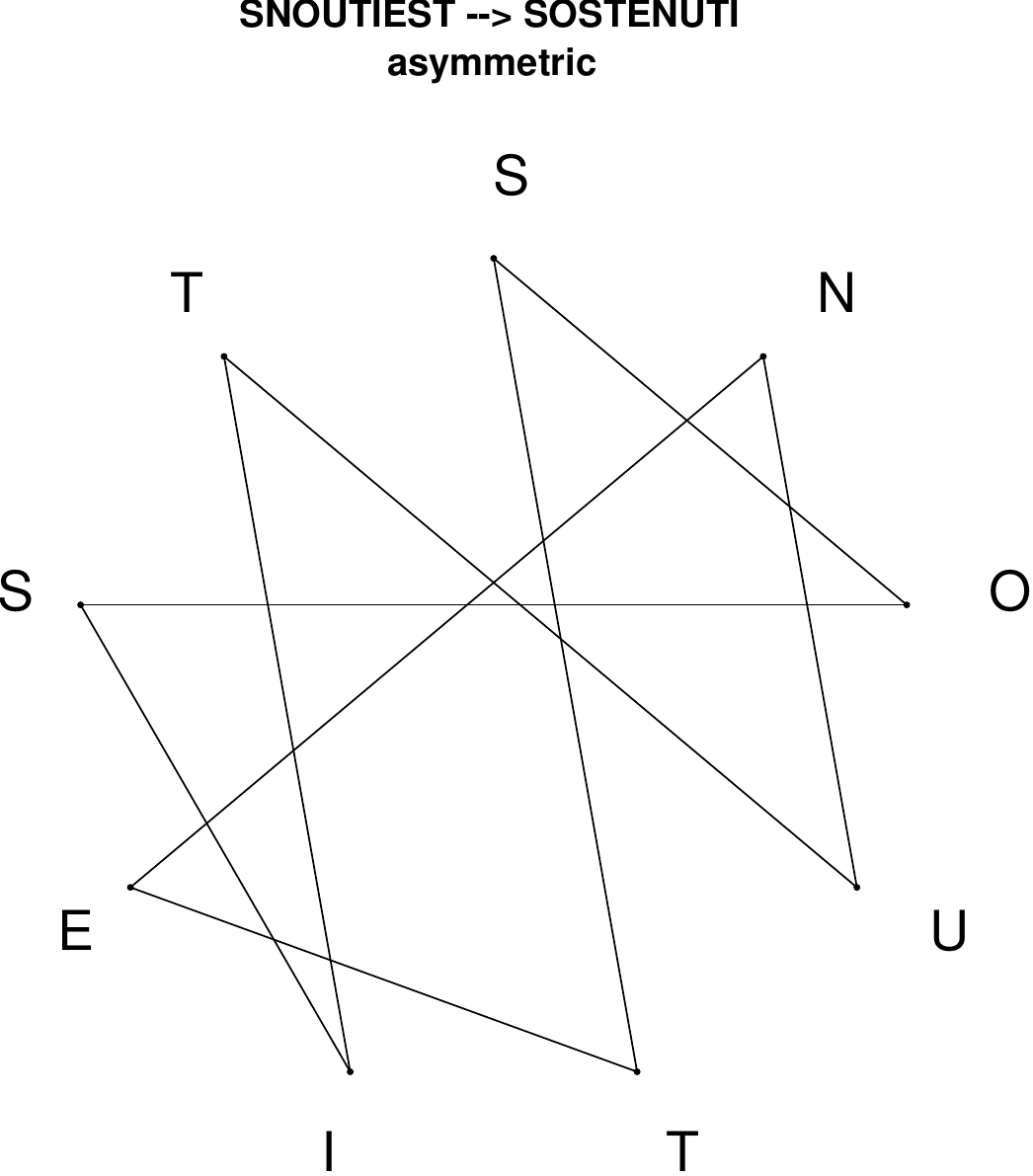}
\end{subfigure}
\end{figure}

\begin{figure}[H]
\centering
\begin{subfigure}[T]{0.19\textwidth}
\centering
\includegraphics[width=\textwidth]{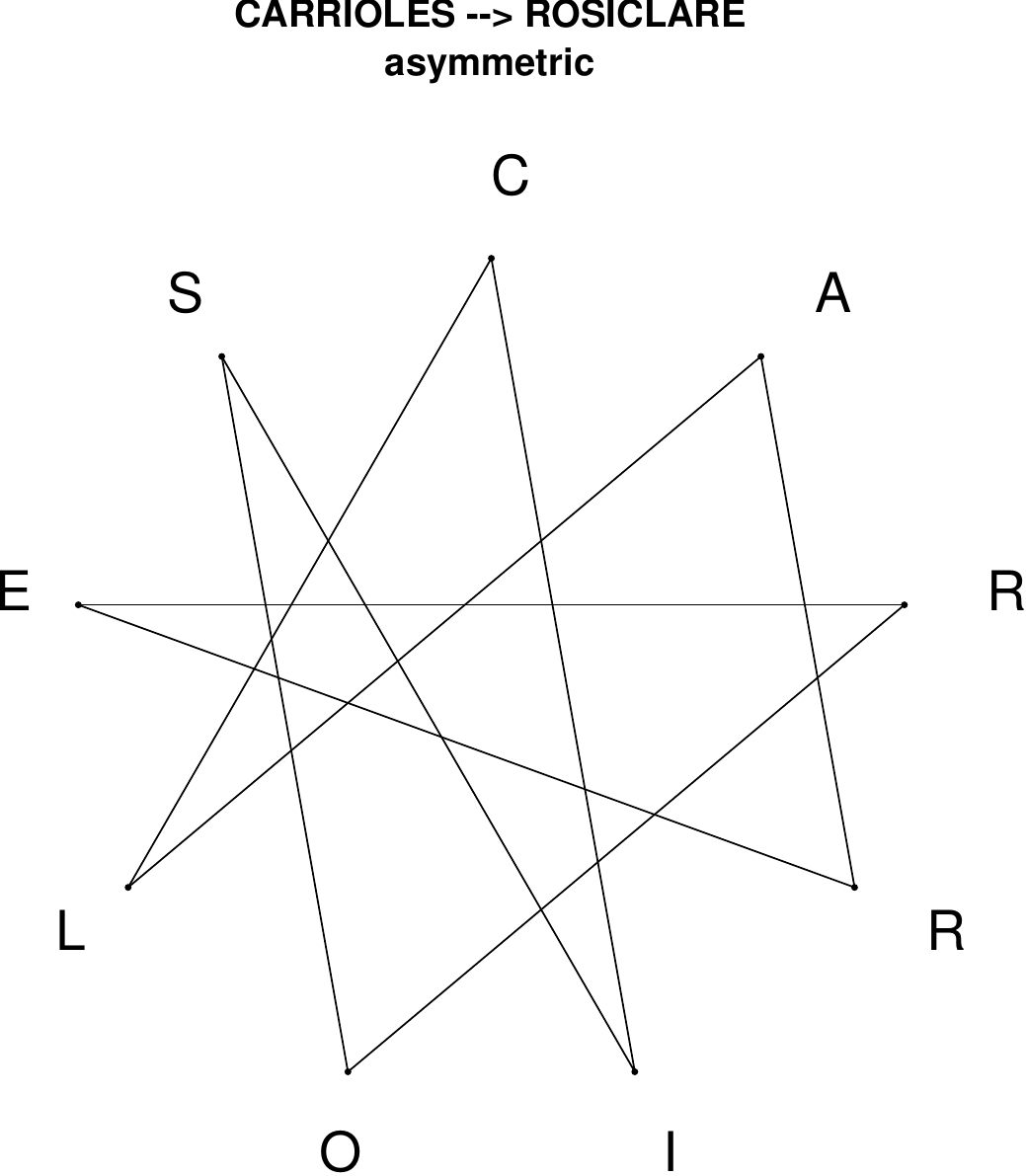}
\end{subfigure}
\hfill
\begin{subfigure}[T]{0.19\textwidth}
\centering
\includegraphics[width=\textwidth]{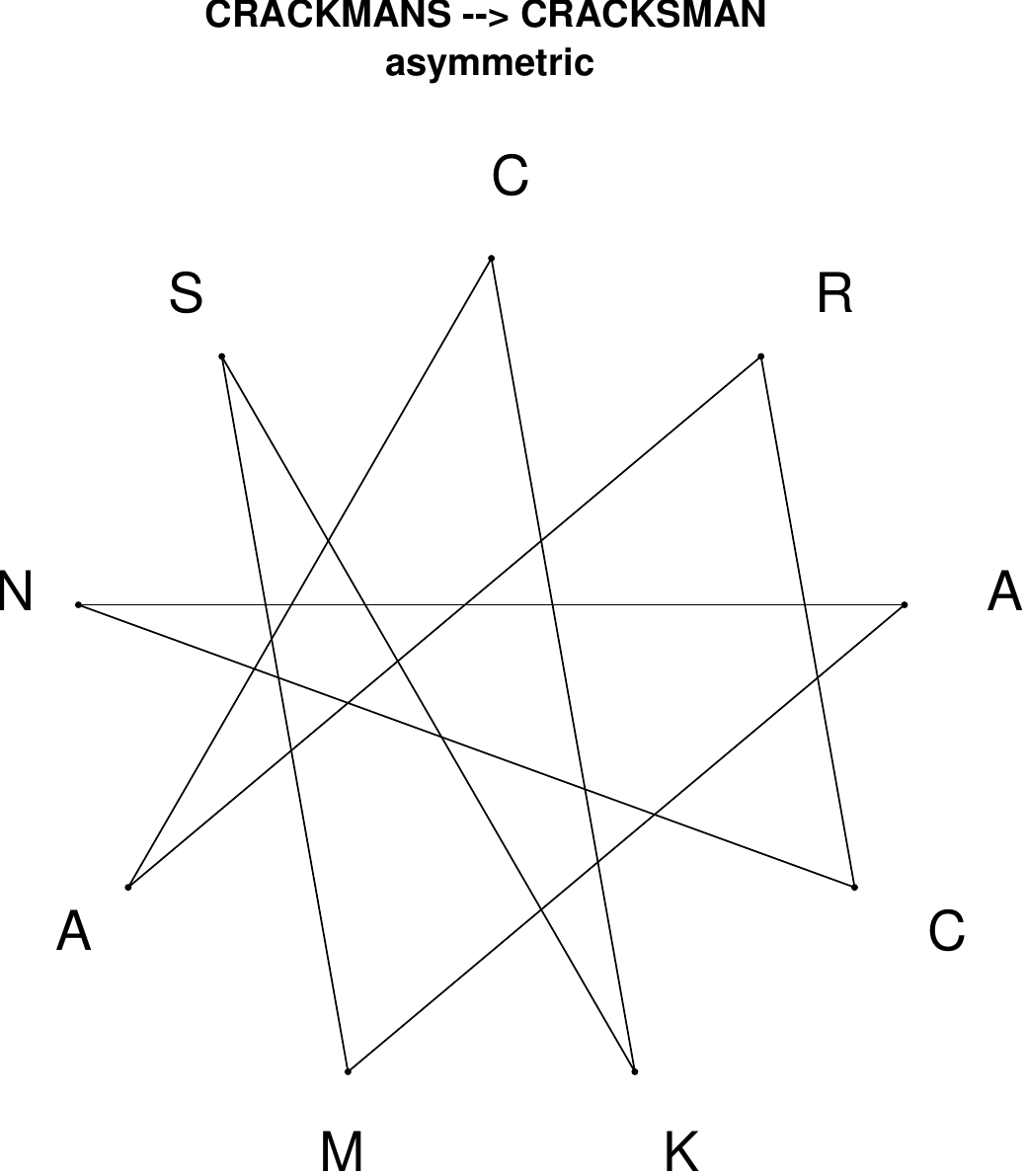}
\end{subfigure}
\hfill
\begin{subfigure}[T]{0.19\textwidth}
\centering
\includegraphics[width=\textwidth]{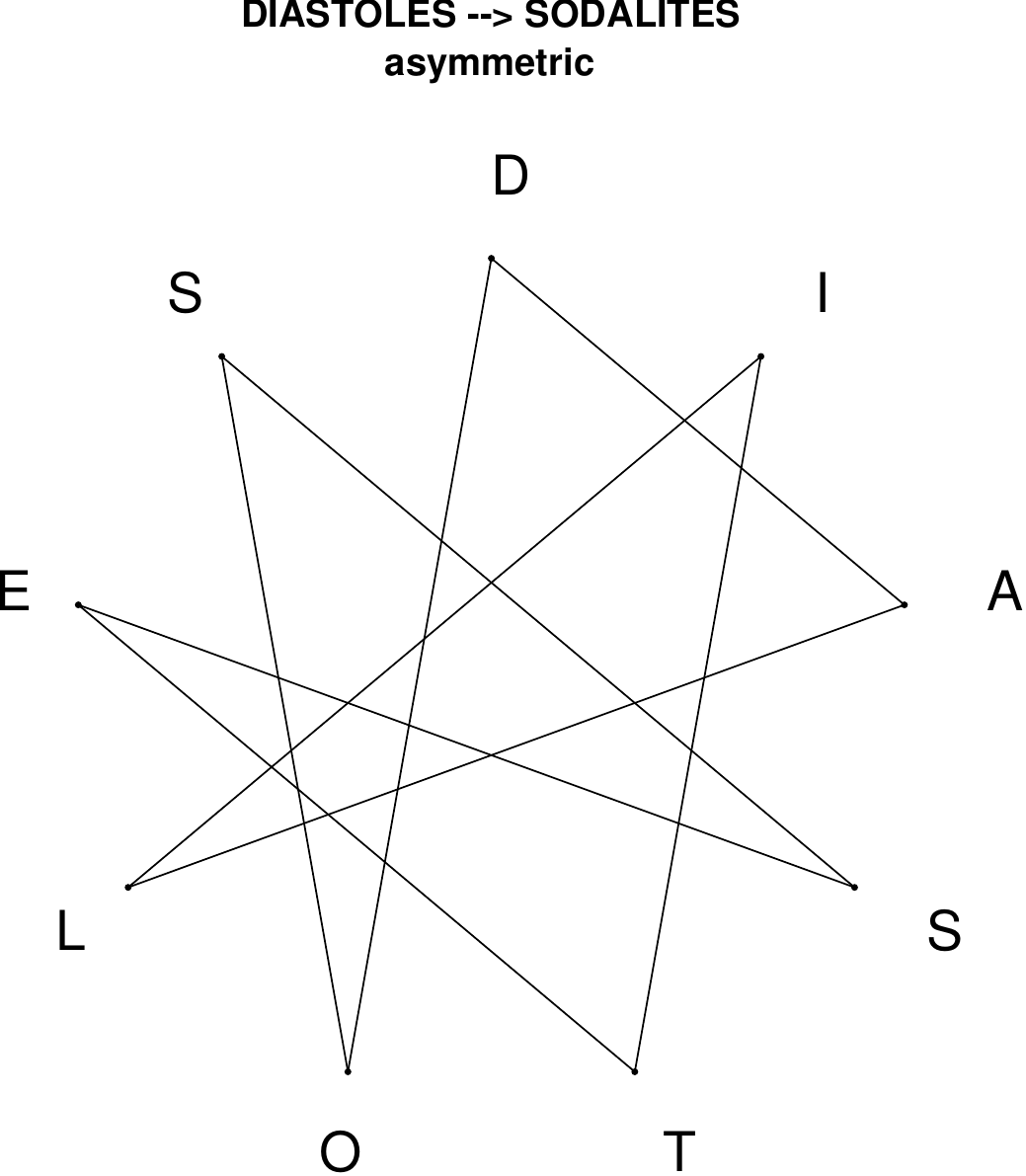}
\end{subfigure}
\hfill
\begin{subfigure}[T]{0.19\textwidth}
\centering
\includegraphics[width=\textwidth]{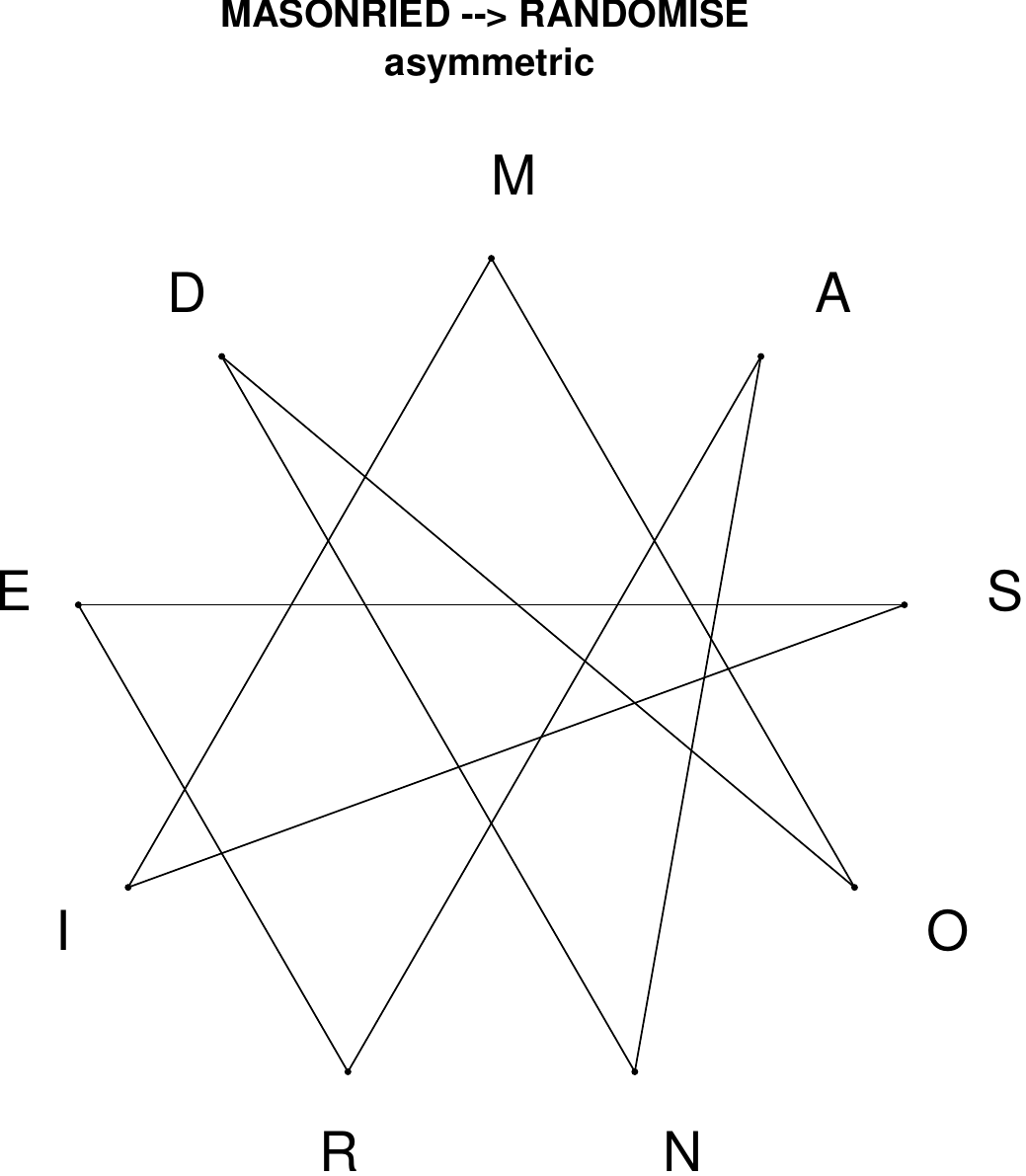}
\end{subfigure}
\hfill
\begin{subfigure}[T]{0.19\textwidth}
\centering
\includegraphics[width=\textwidth]{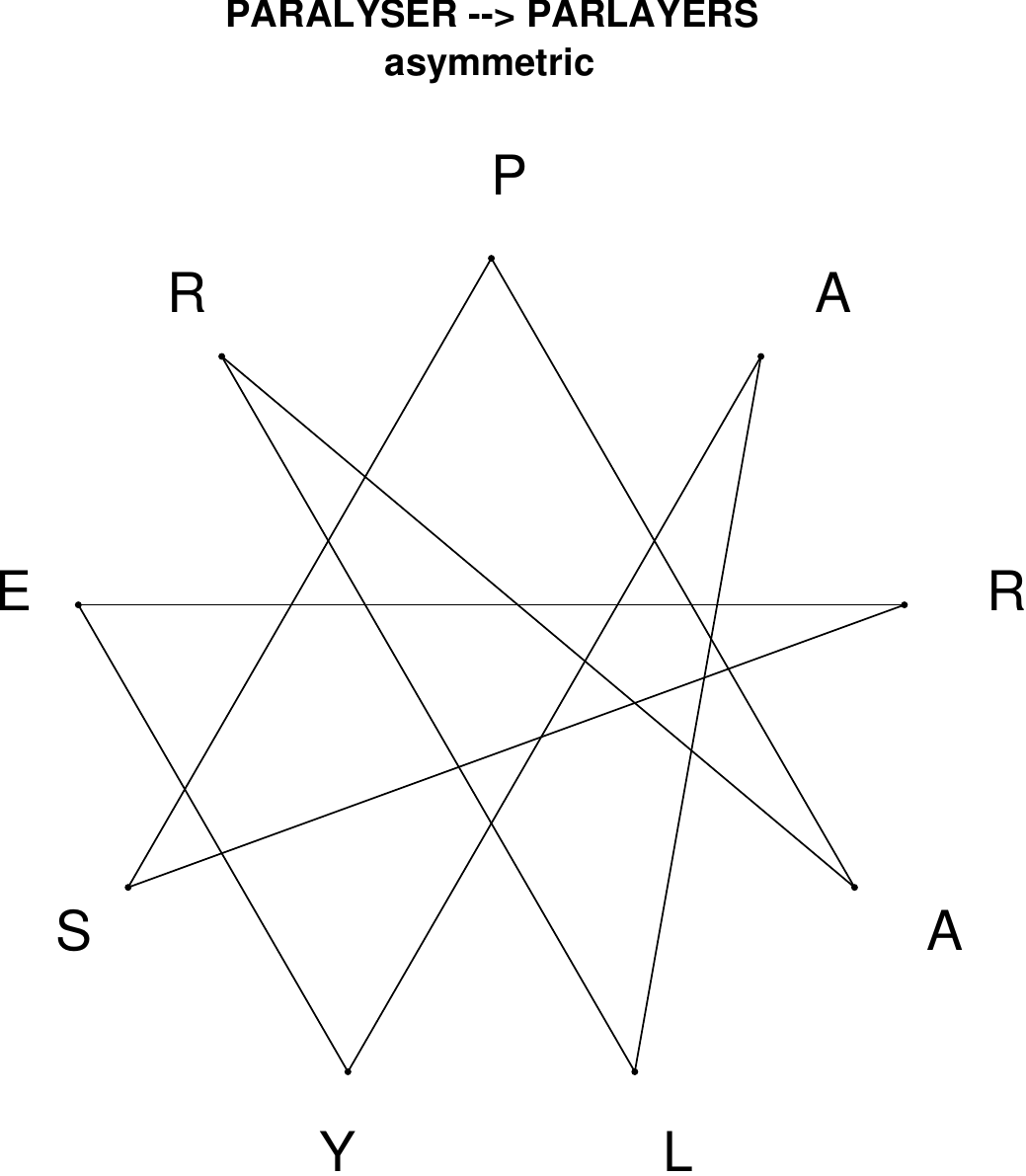}
\end{subfigure}
\end{figure}

\begin{figure}[H]
\centering
\begin{subfigure}[T]{0.19\textwidth}
\centering
\includegraphics[width=\textwidth]{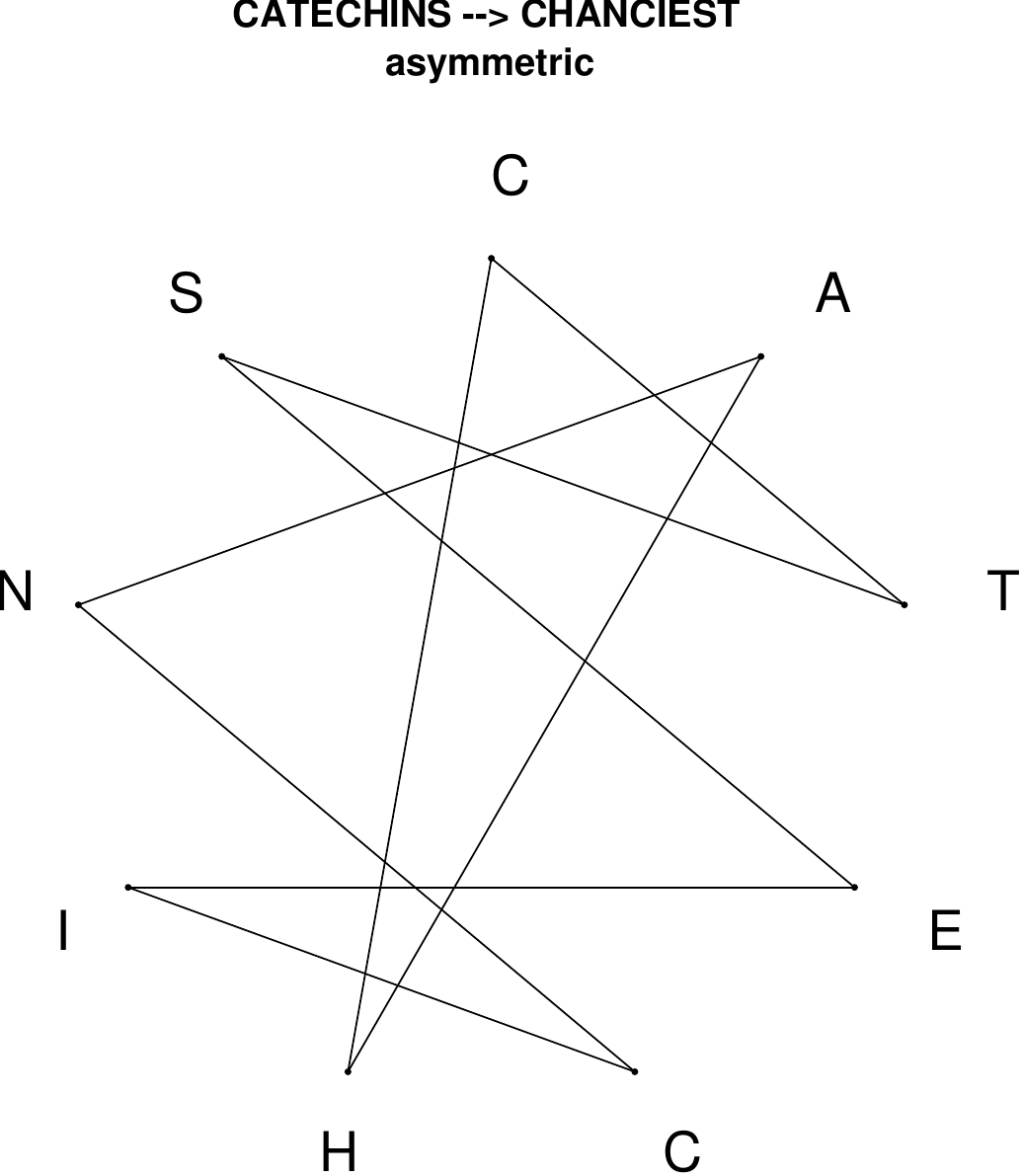}
\end{subfigure}
\hfill
\begin{subfigure}[T]{0.19\textwidth}
\centering
\includegraphics[width=\textwidth]{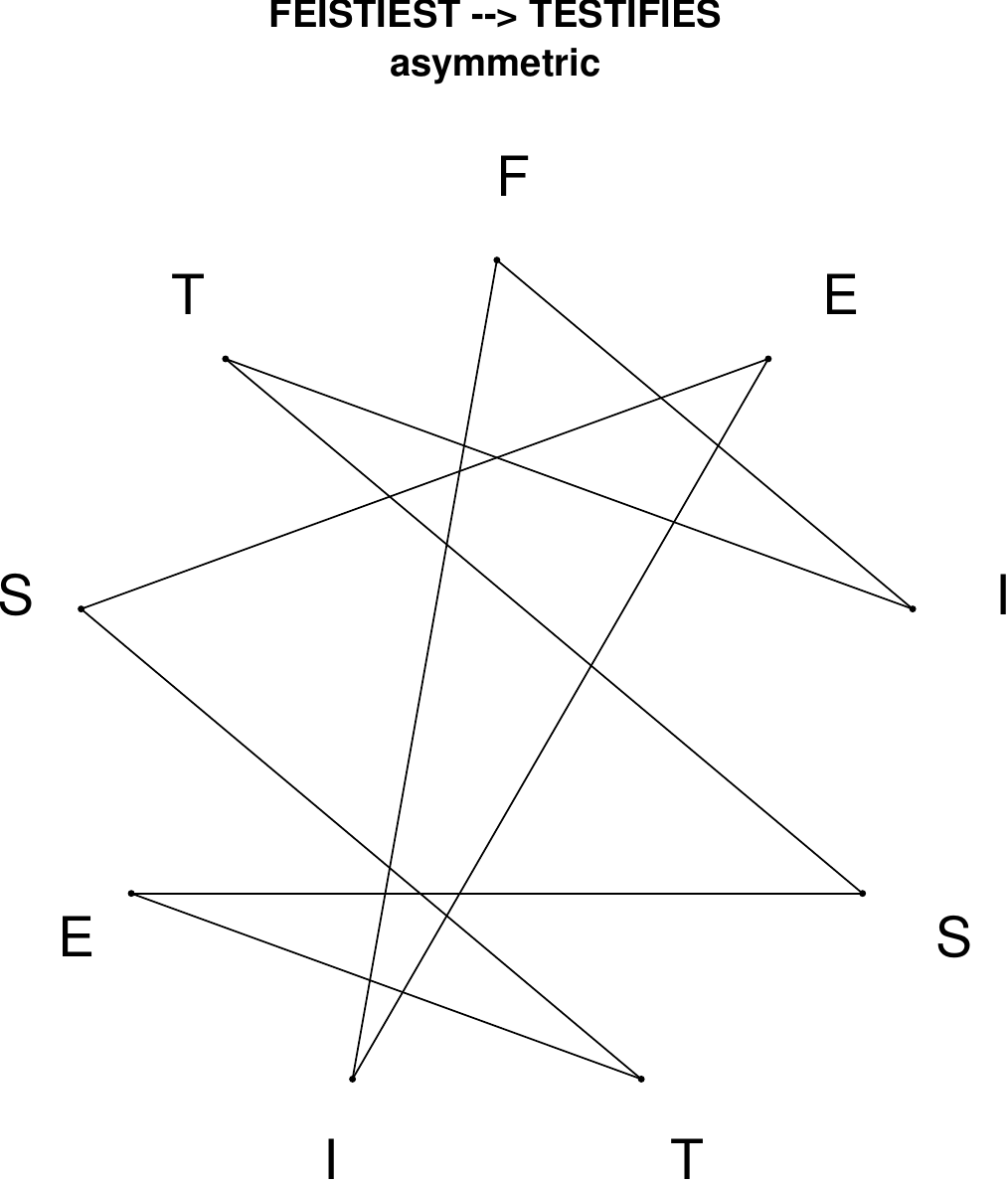}
\end{subfigure}
\hfill
\begin{subfigure}[T]{0.19\textwidth}
\centering
\includegraphics[width=\textwidth]{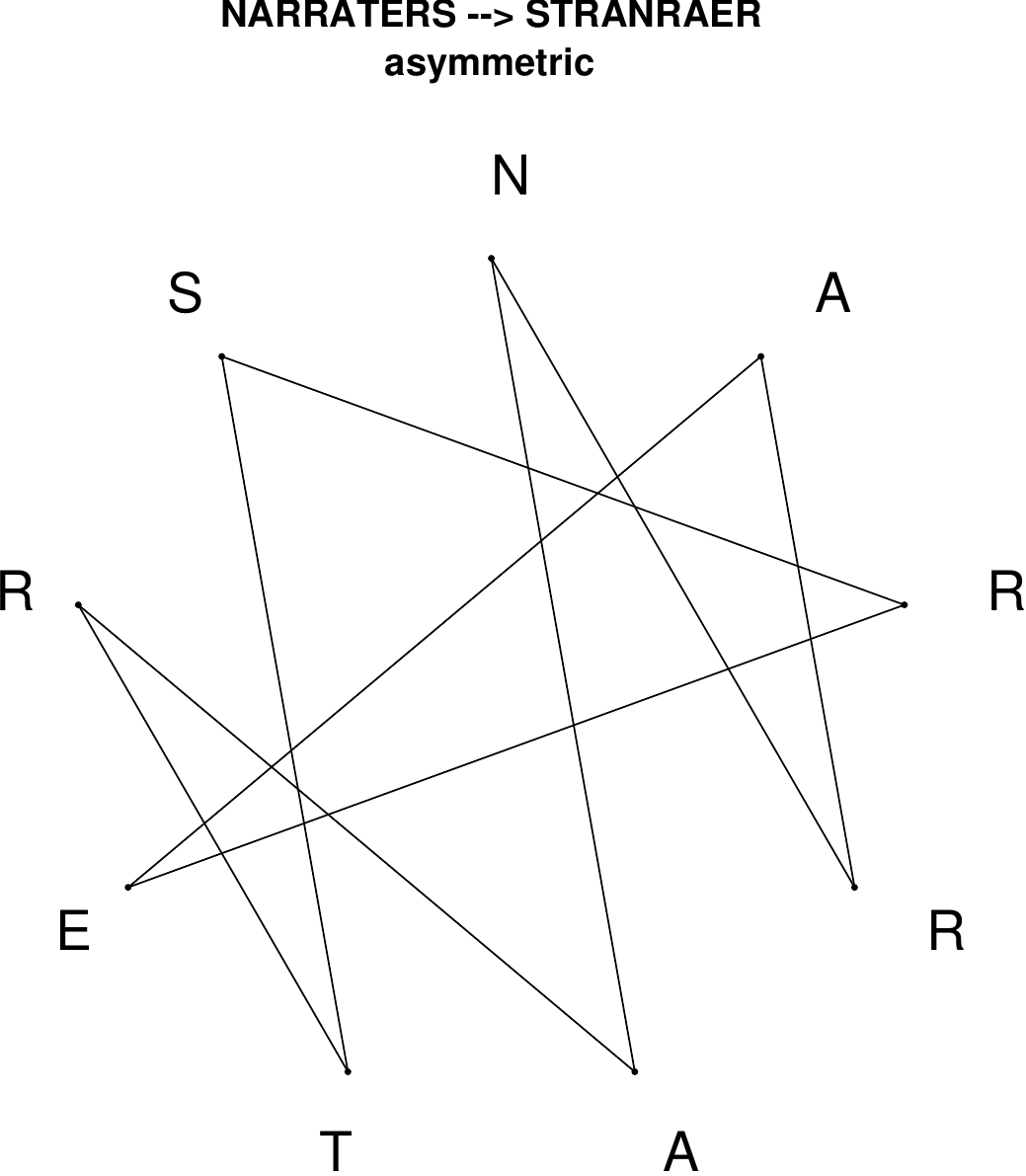}
\end{subfigure}
\hfill
\begin{subfigure}[T]{0.19\textwidth}
\centering
\includegraphics[width=\textwidth]{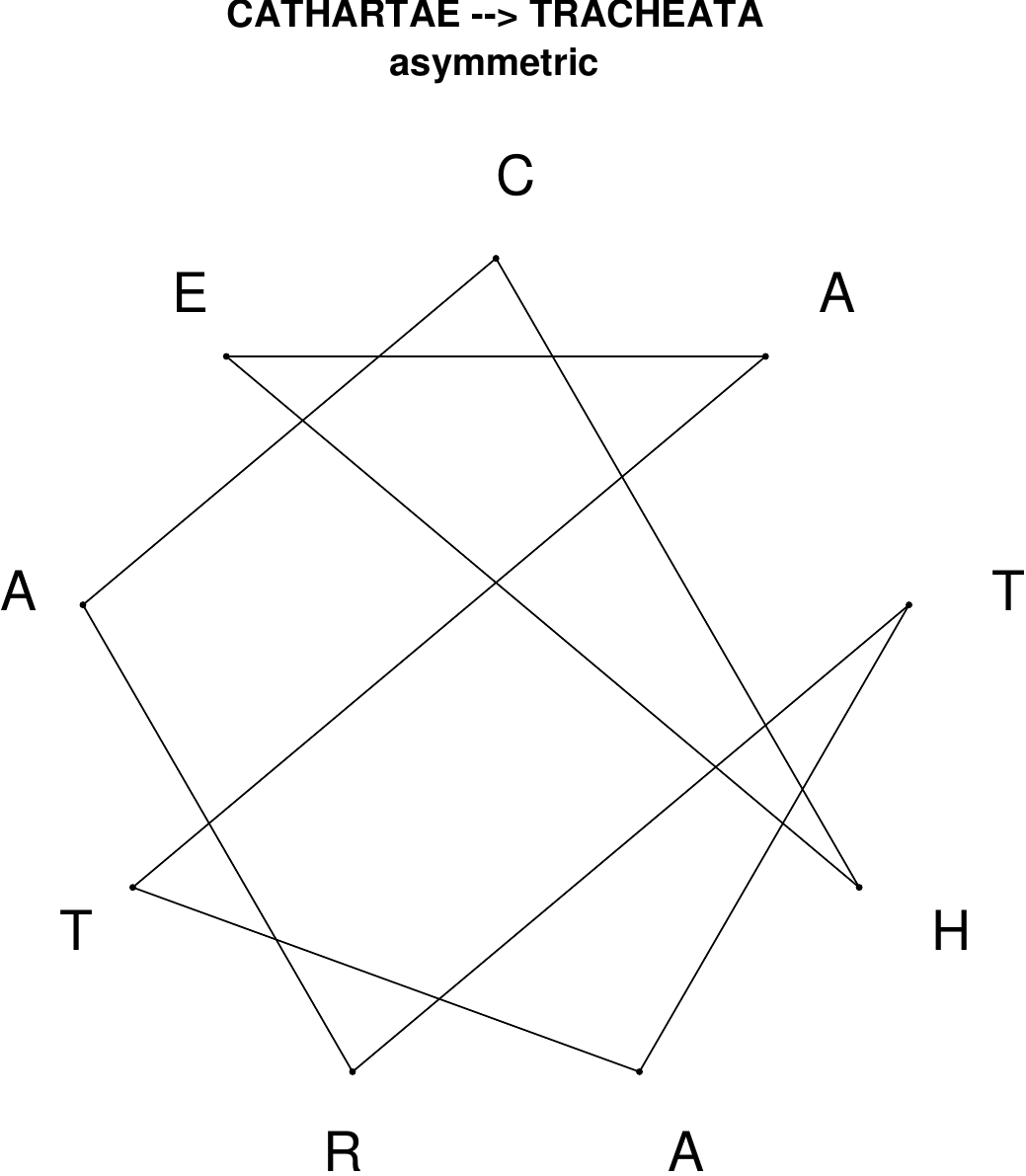}
\end{subfigure}
\hfill
\begin{subfigure}[T]{0.19\textwidth}
\centering
\includegraphics[width=\textwidth]{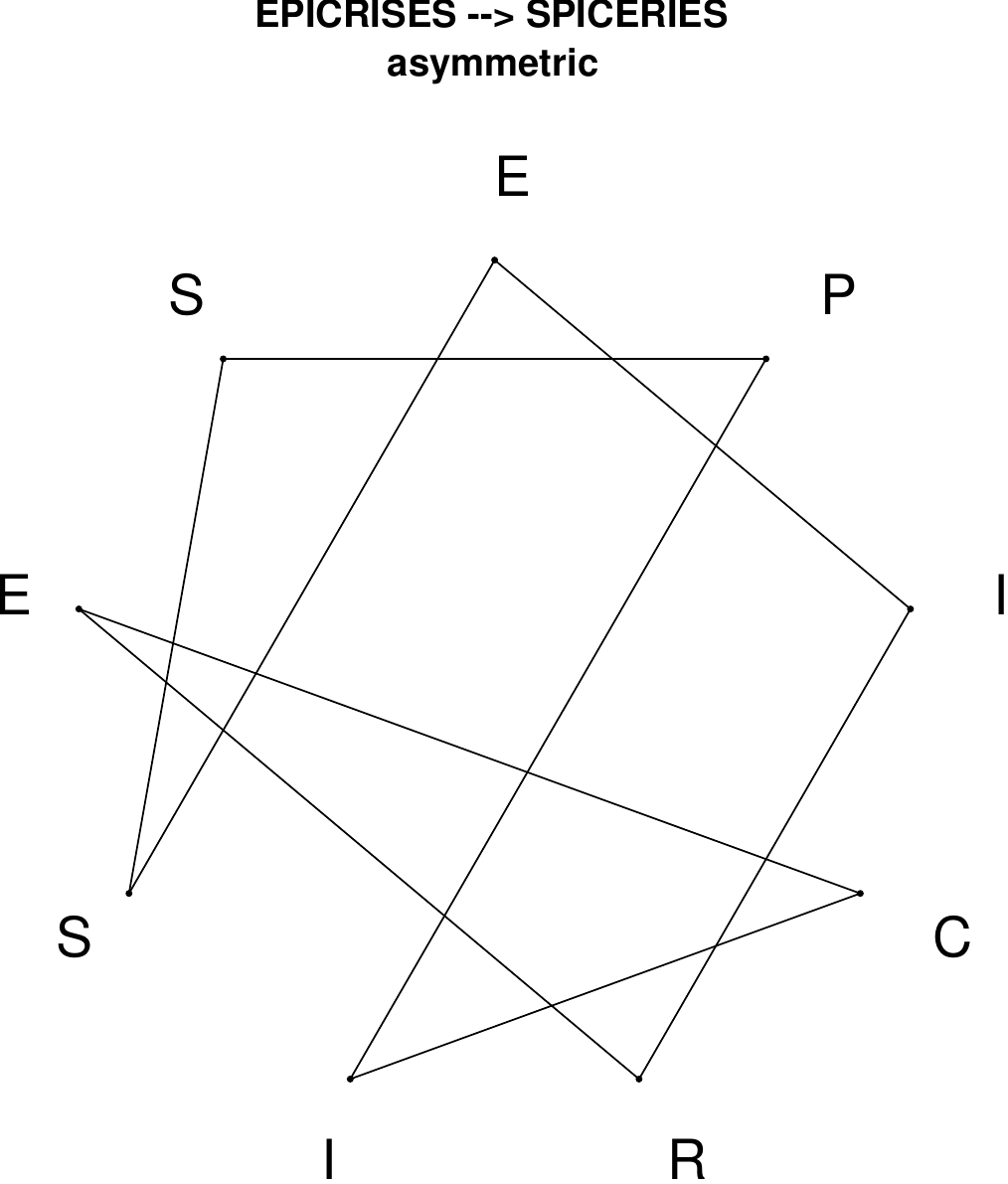}
\end{subfigure}
\end{figure}

\begin{figure}[H]
\centering
\begin{subfigure}[T]{0.19\textwidth}
\centering
\includegraphics[width=\textwidth]{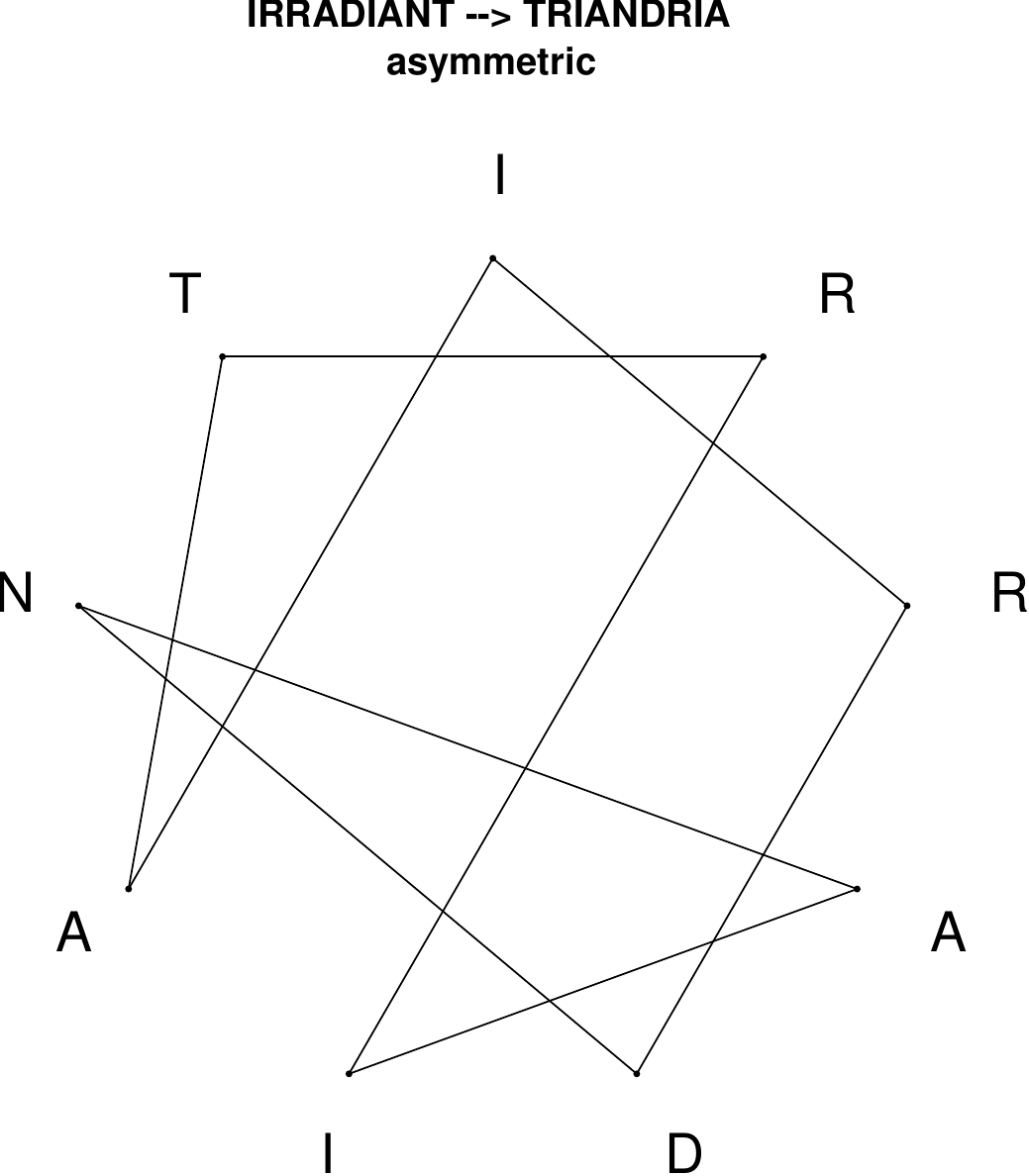}
\end{subfigure}
\hfill
\begin{subfigure}[T]{0.19\textwidth}
\centering
\includegraphics[width=\textwidth]{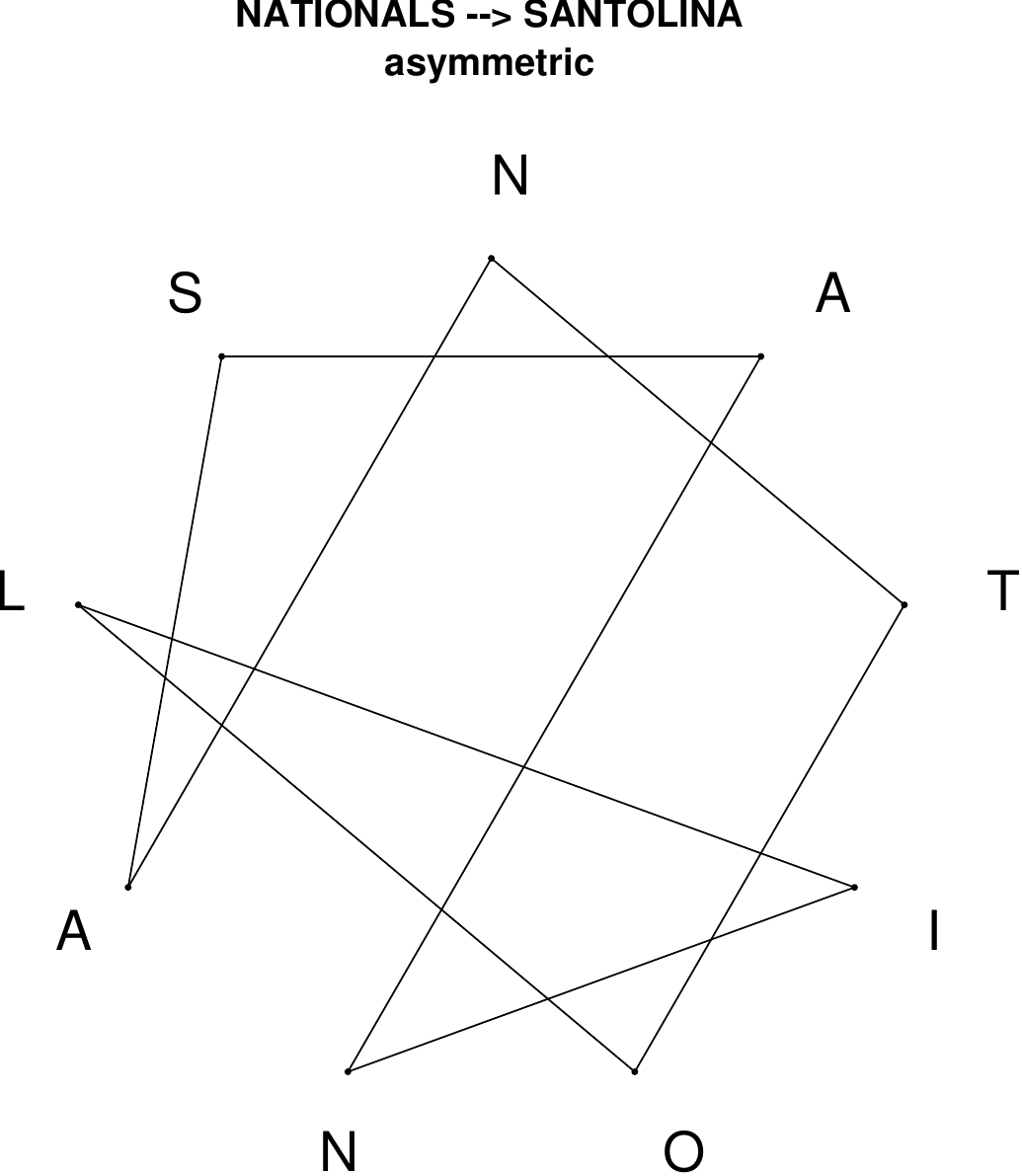}
\end{subfigure}
\hfill
\begin{subfigure}[T]{0.19\textwidth}
\centering
\includegraphics[width=\textwidth]{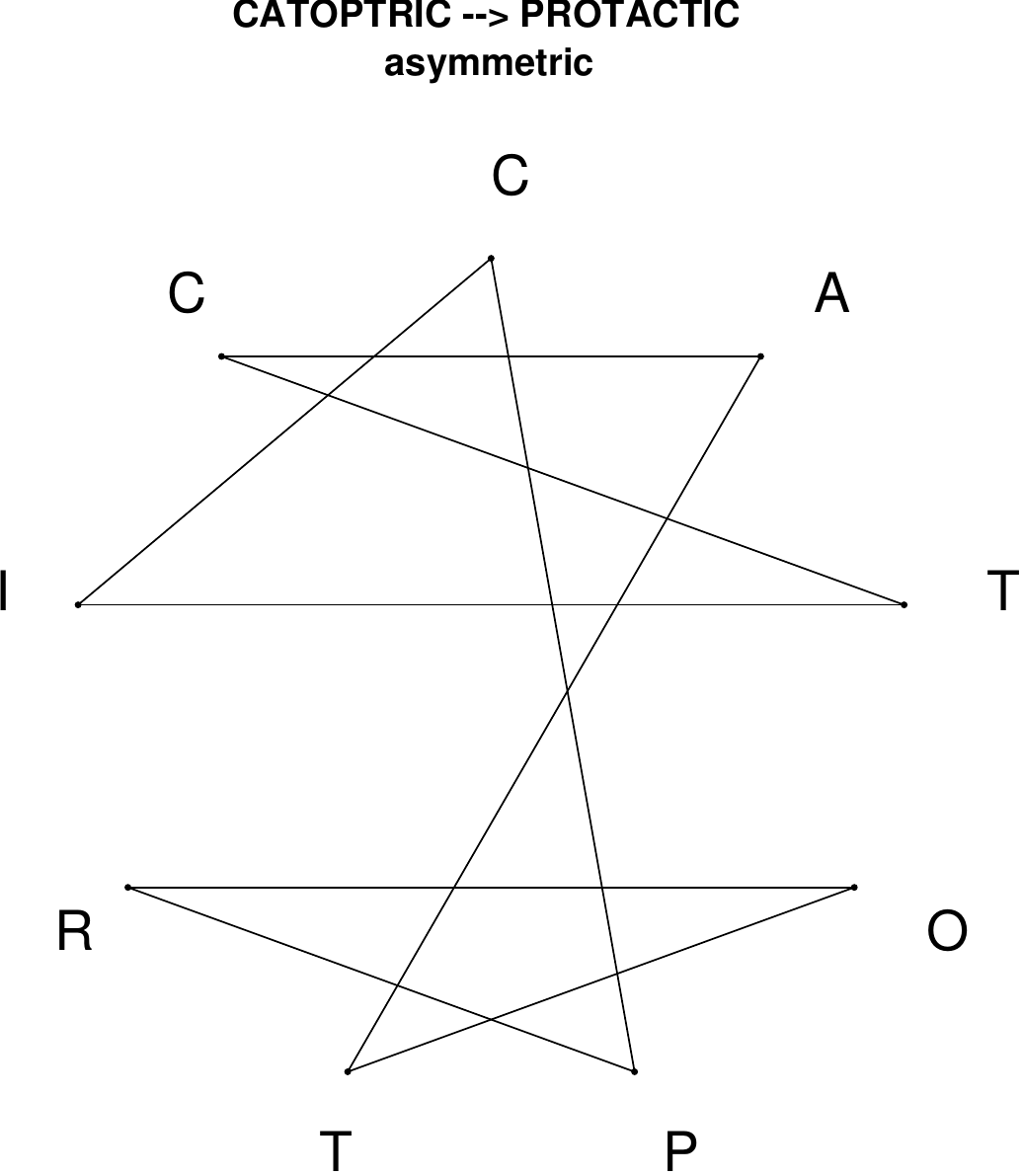}
\end{subfigure}
\hfill
\begin{subfigure}[T]{0.19\textwidth}
\centering
\includegraphics[width=\textwidth]{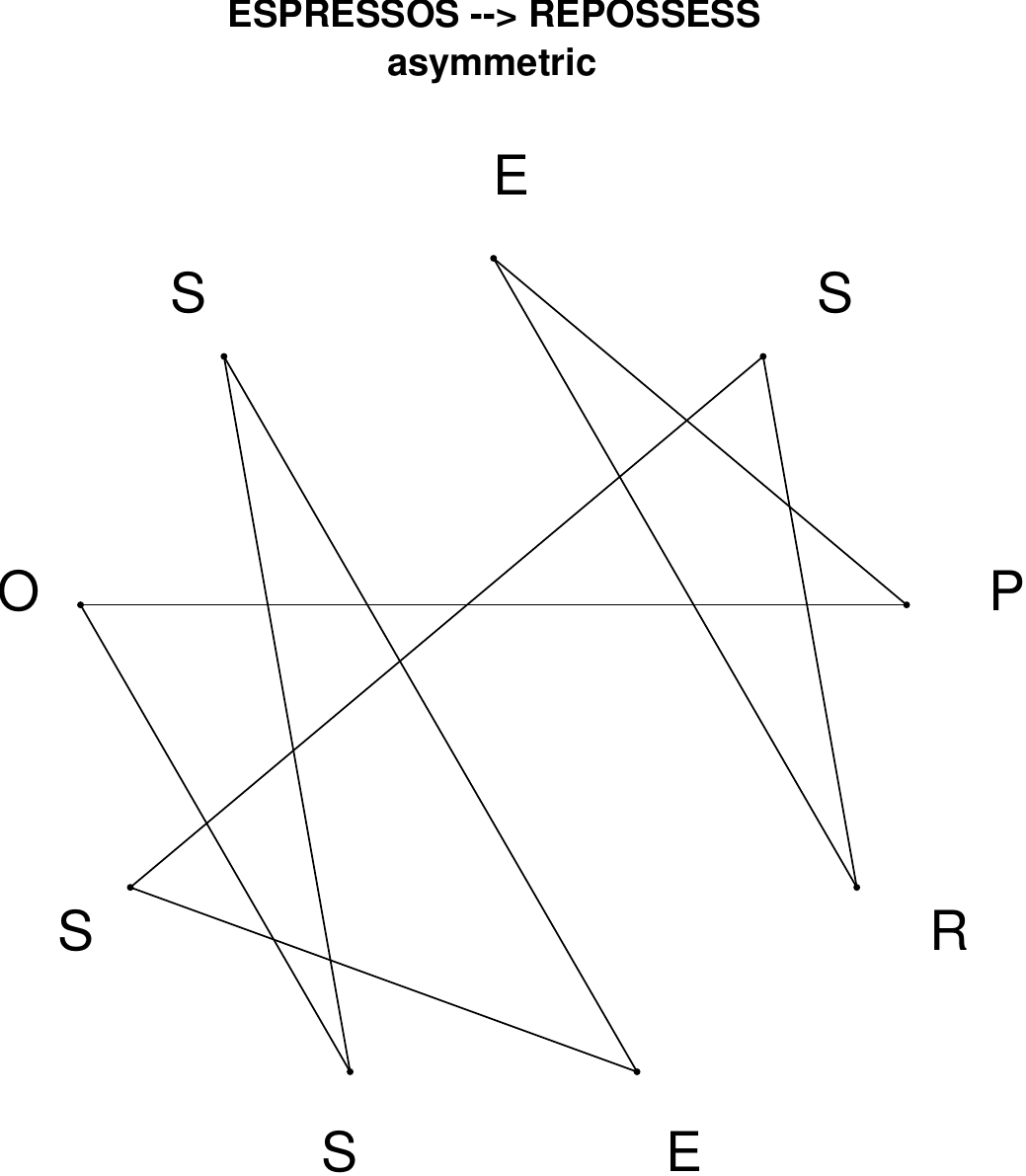}
\end{subfigure}
\hfill
\begin{subfigure}[T]{0.19\textwidth}
\centering
\includegraphics[width=\textwidth]{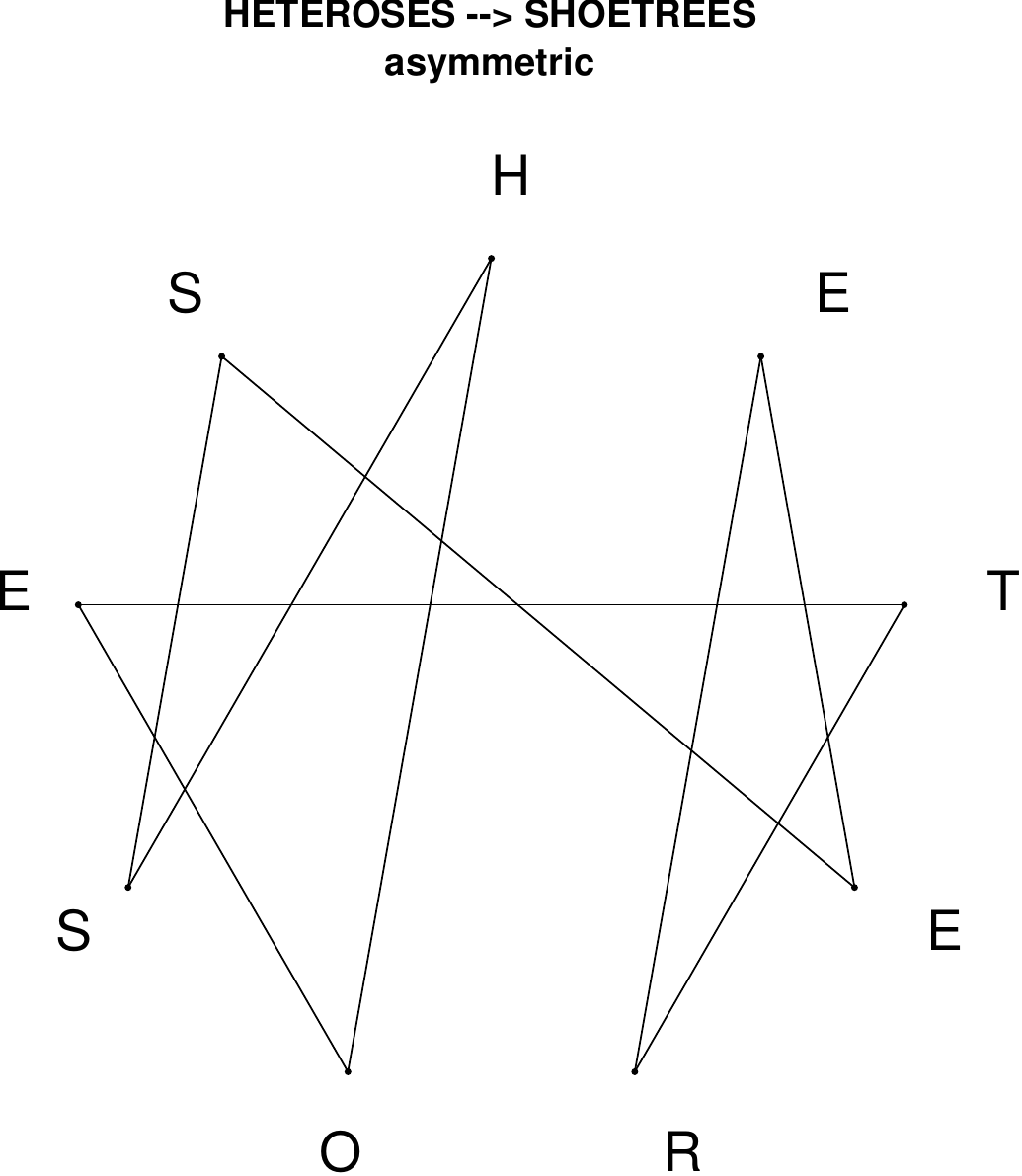}
\end{subfigure}
\end{figure}

\begin{figure}[H]
\centering
\begin{subfigure}[T]{0.19\textwidth}
\centering
\includegraphics[width=\textwidth]{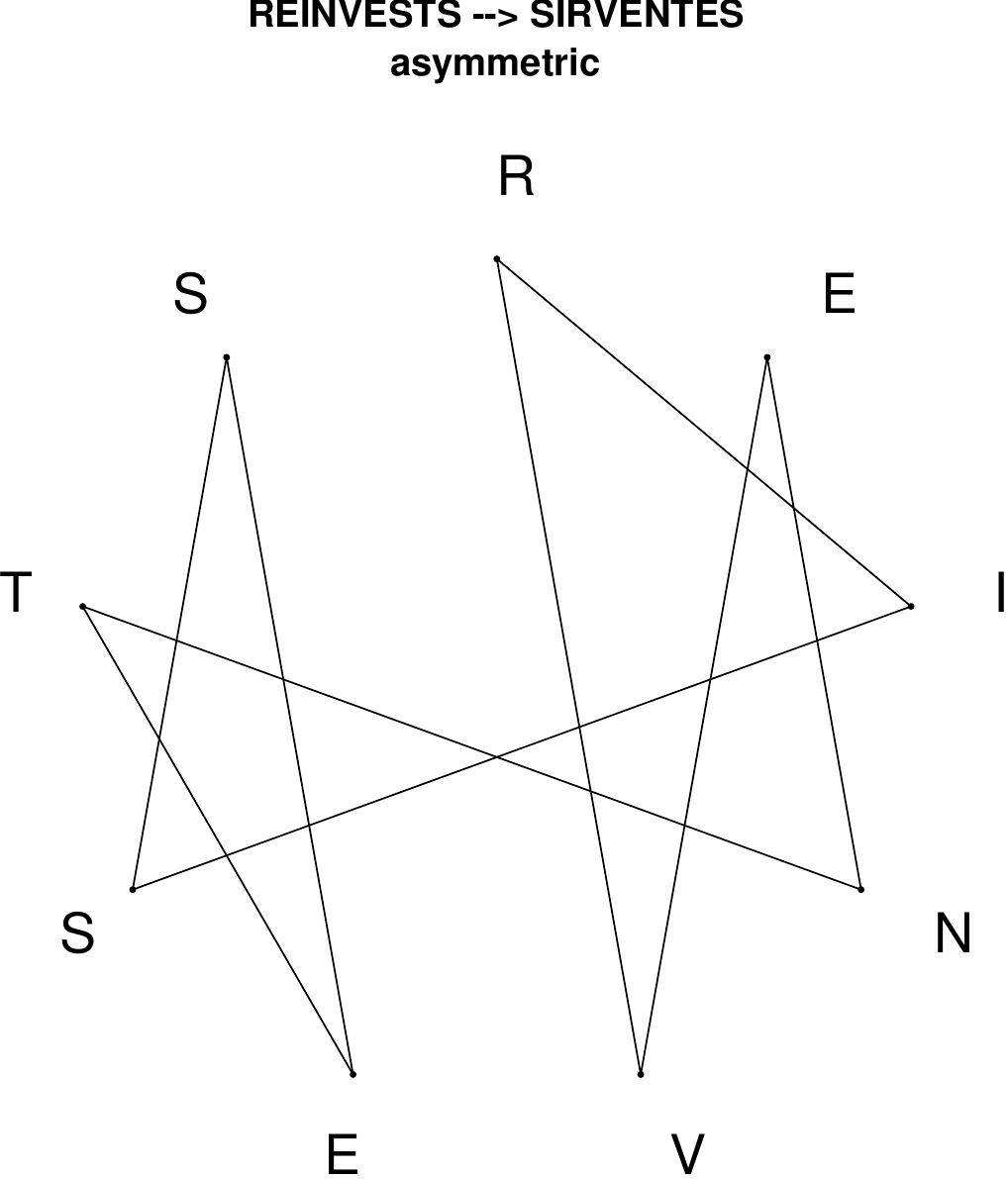}
\end{subfigure}
\hfill
\begin{subfigure}[T]{0.19\textwidth}
\centering
\includegraphics[width=\textwidth]{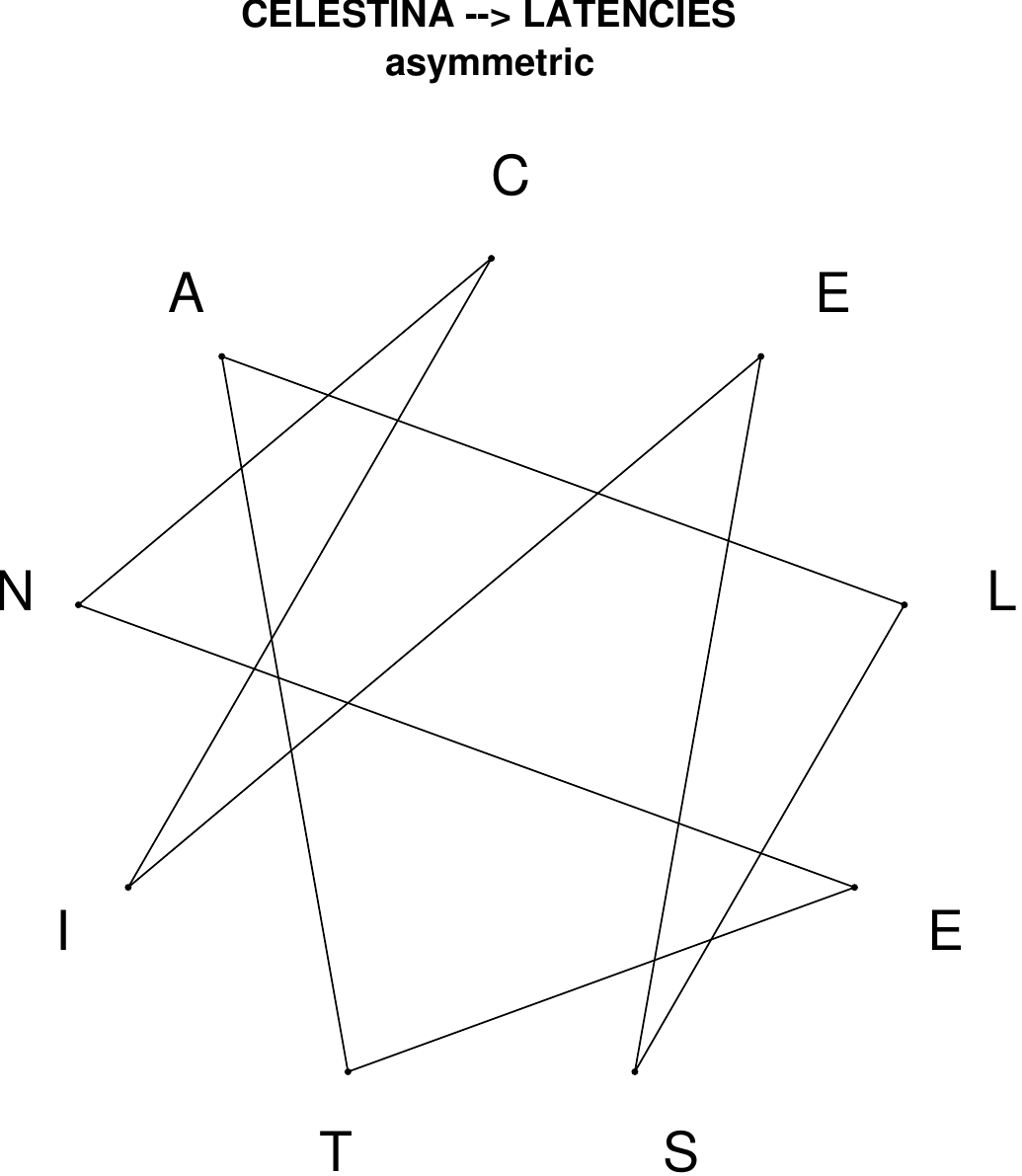}
\end{subfigure}
\hfill
\begin{subfigure}[T]{0.19\textwidth}
\centering
\includegraphics[width=\textwidth]{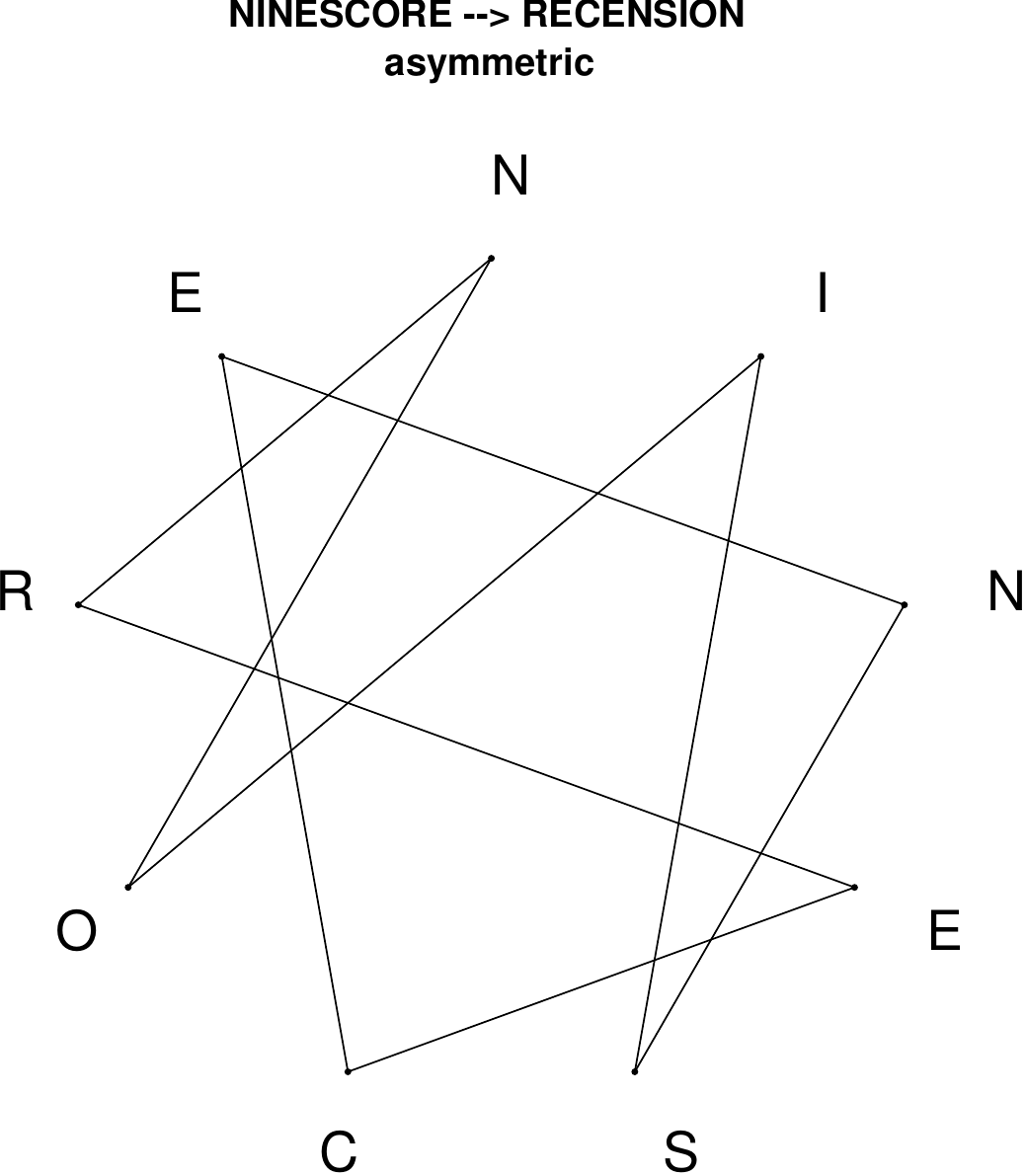}
\end{subfigure}
\hfill
\begin{subfigure}[T]{0.19\textwidth}
\centering
\includegraphics[width=\textwidth]{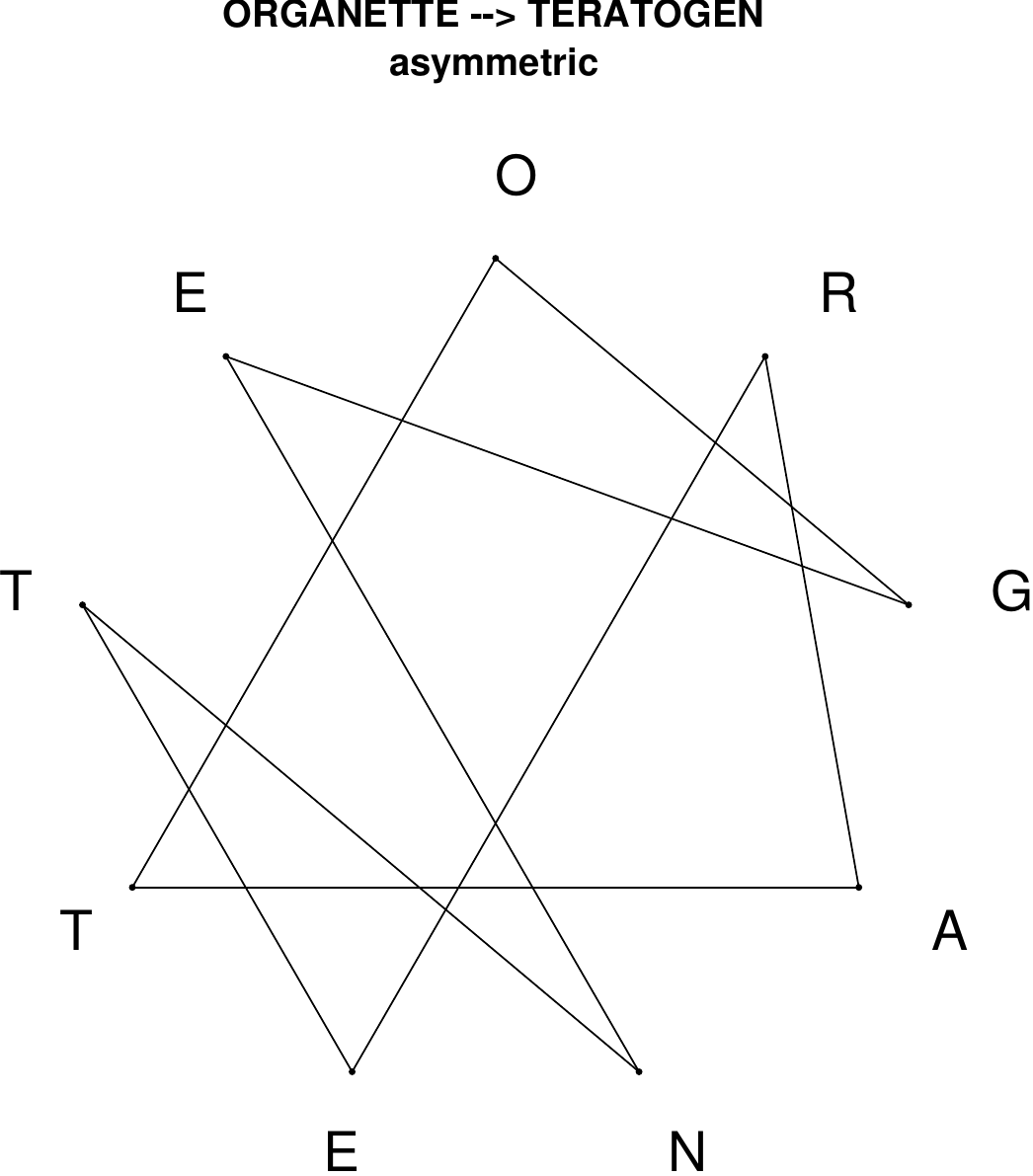}
\end{subfigure}
\hfill
\begin{subfigure}[T]{0.19\textwidth}
\centering
\includegraphics[width=\textwidth]{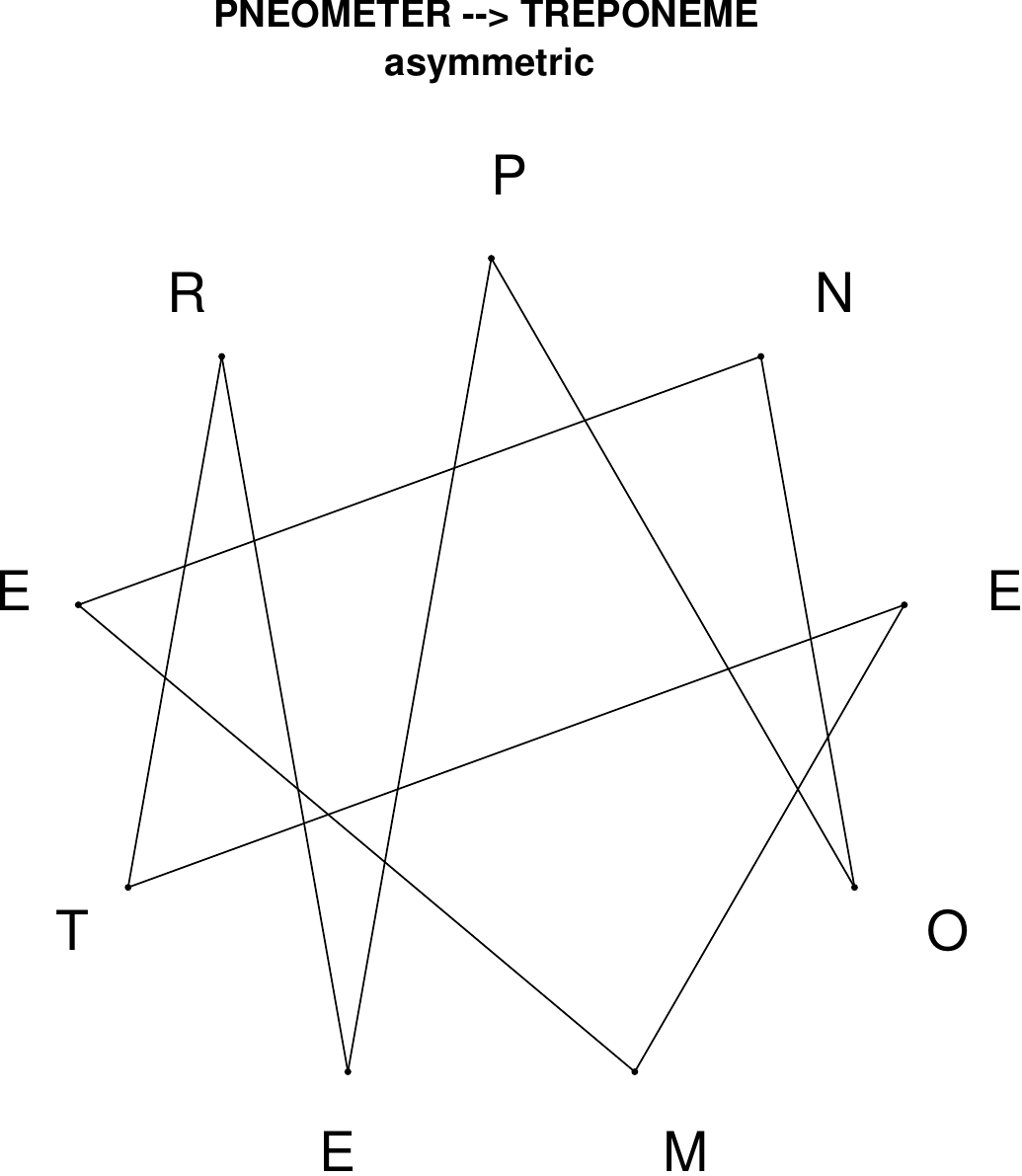}
\end{subfigure}
\end{figure}

\begin{figure}[H]
\centering
\begin{subfigure}[T]{0.19\textwidth}
\centering
\includegraphics[width=\textwidth]{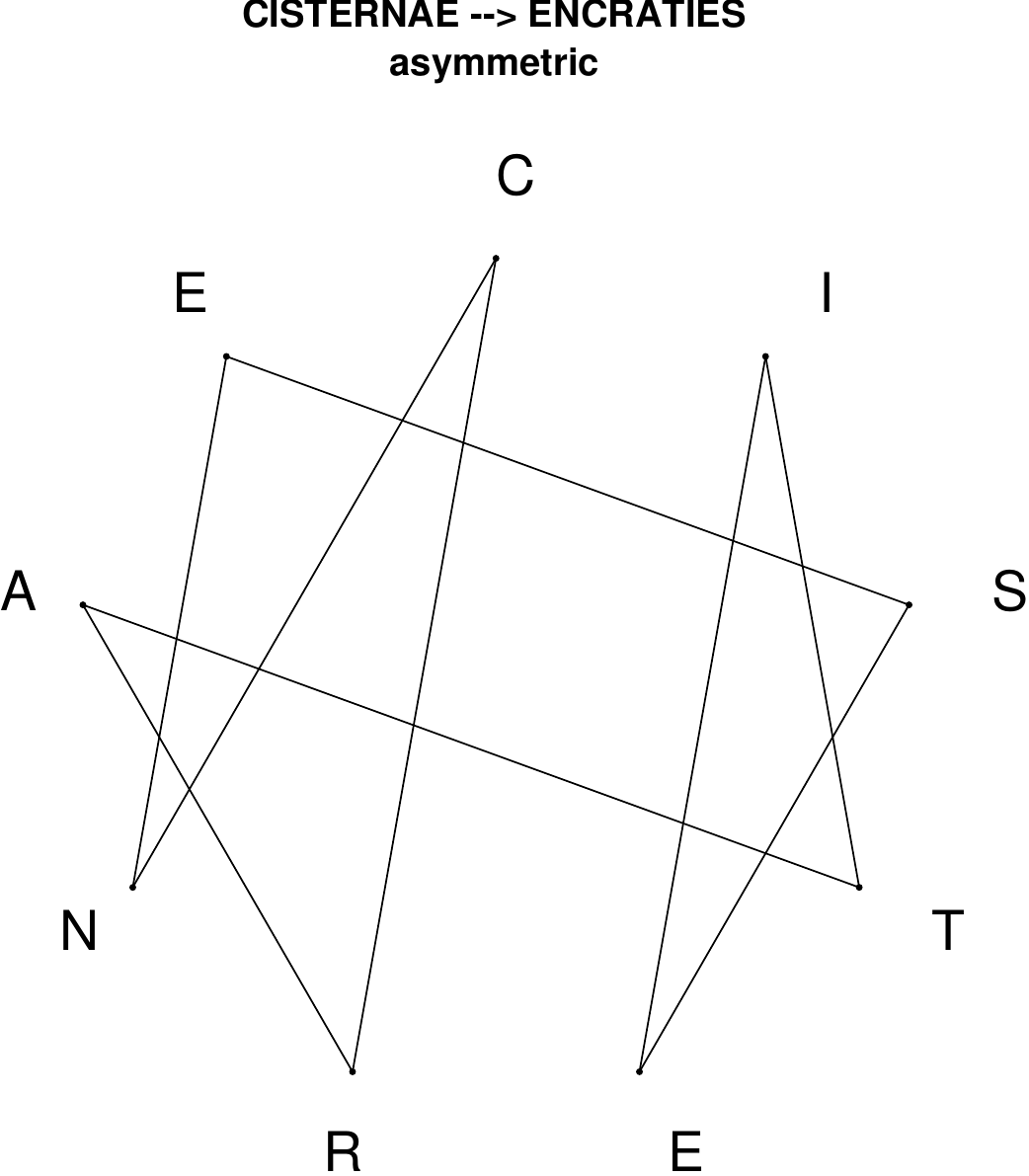}
\end{subfigure}
\hfill
\begin{subfigure}[T]{0.19\textwidth}
\centering
\includegraphics[width=\textwidth]{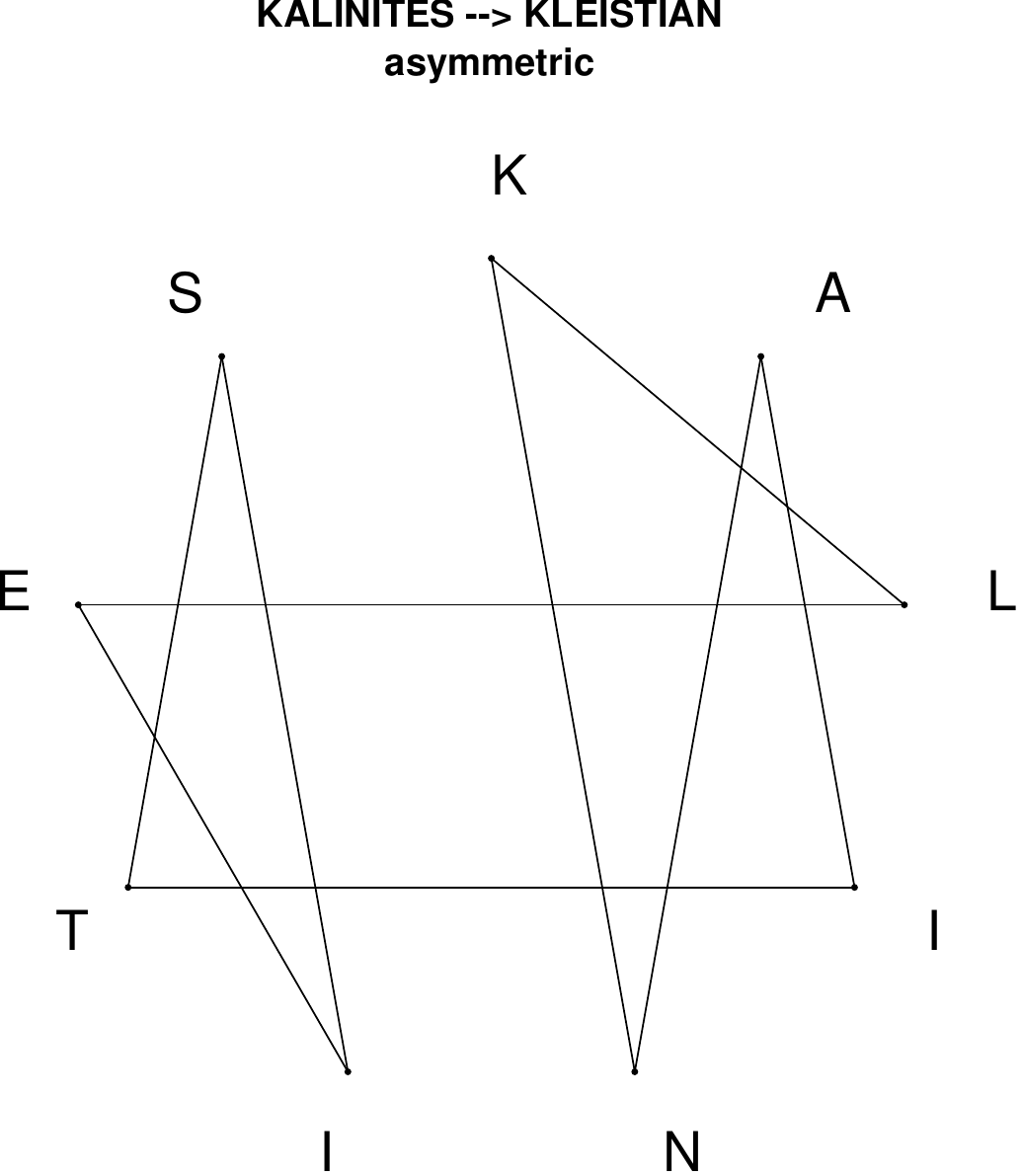}
\end{subfigure}
\hfill
\begin{subfigure}[T]{0.19\textwidth}
\centering
\includegraphics[width=\textwidth]{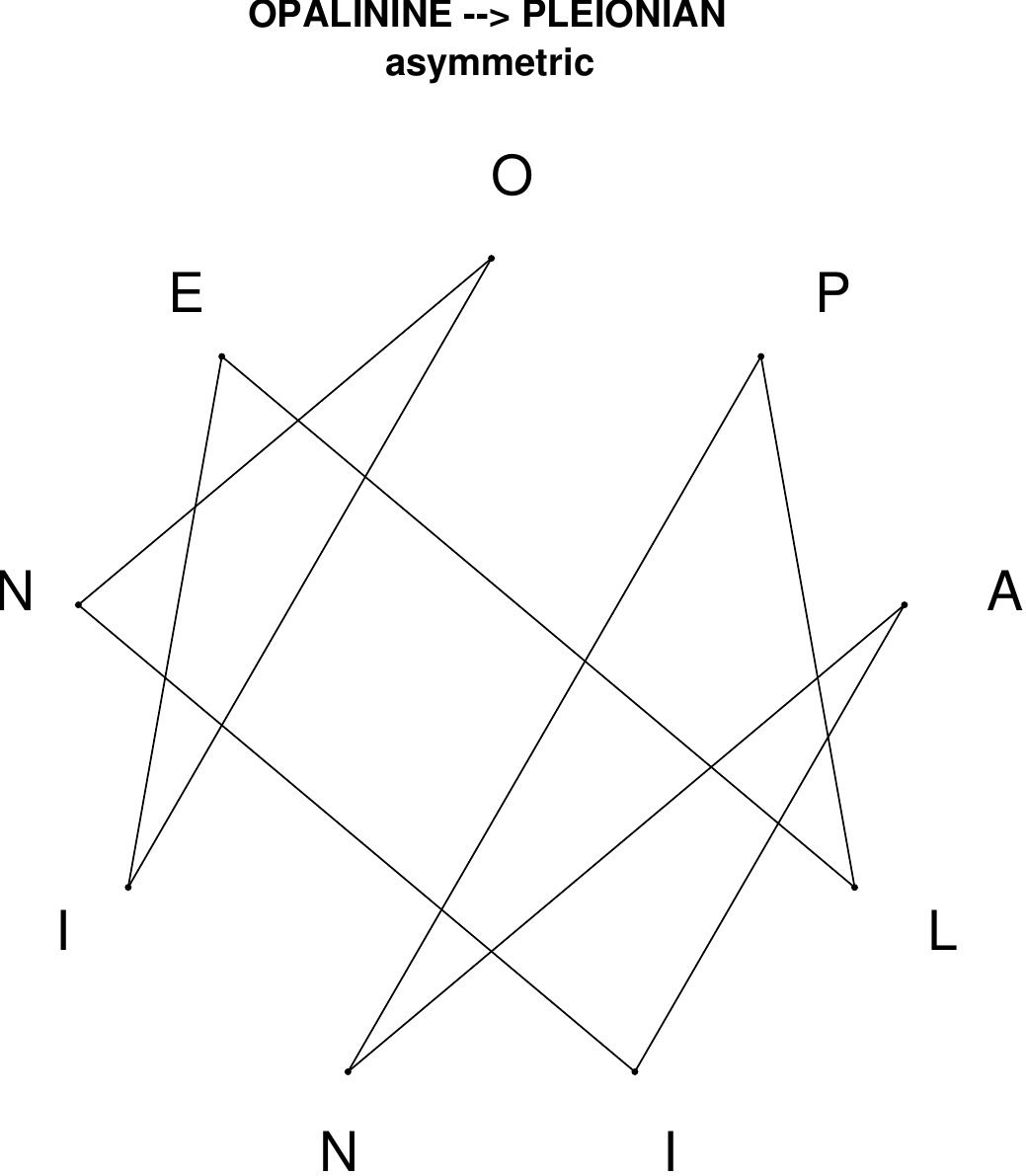}
\end{subfigure}
\hfill
\begin{subfigure}[T]{0.19\textwidth}
\centering
\includegraphics[width=\textwidth]{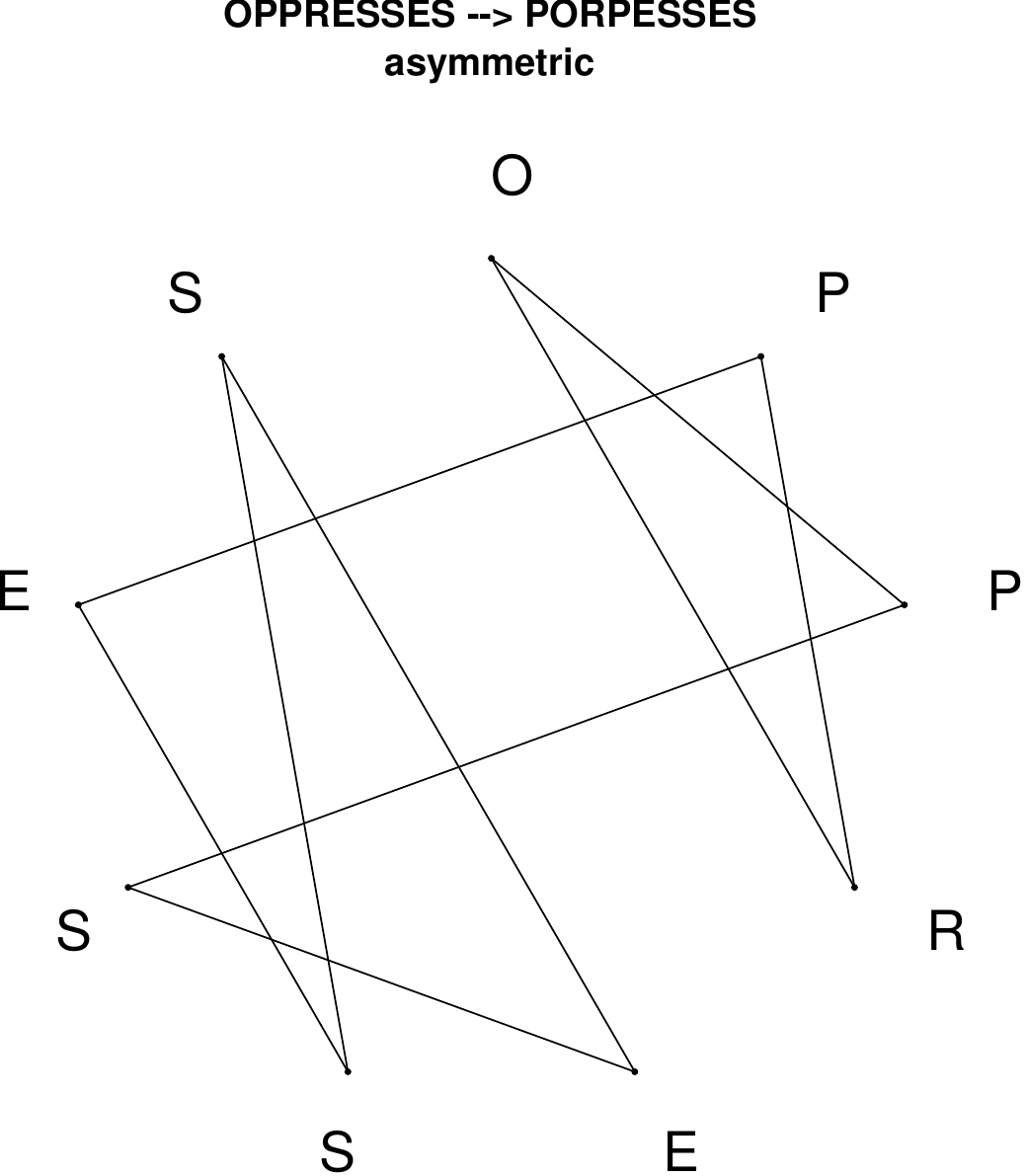}
\end{subfigure}
\hfill
\begin{subfigure}[T]{0.19\textwidth}
\centering
\includegraphics[width=\textwidth]{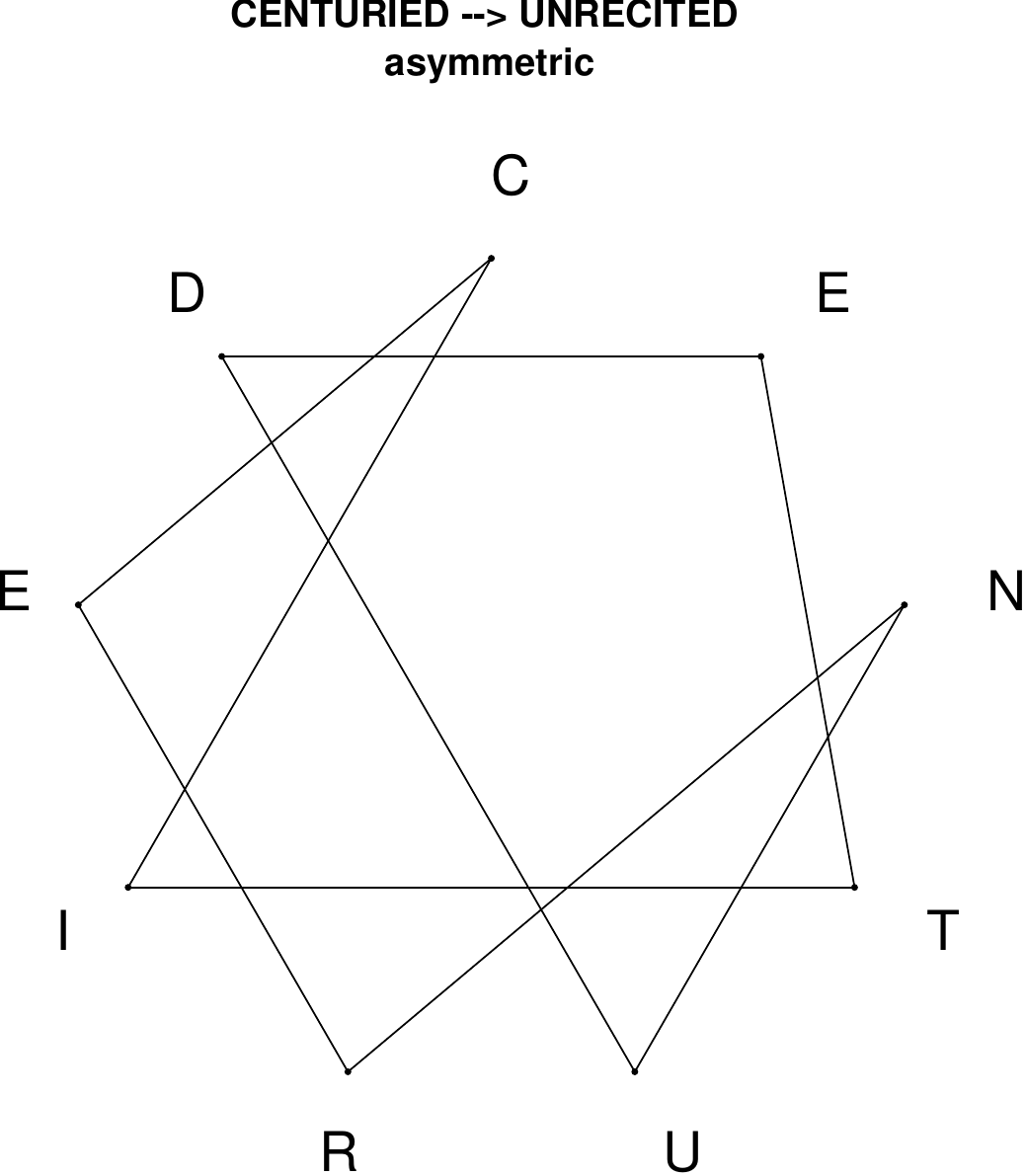}
\end{subfigure}
\end{figure}

\begin{figure}[H]
\centering
\begin{subfigure}[T]{0.19\textwidth}
\centering
\includegraphics[width=\textwidth]{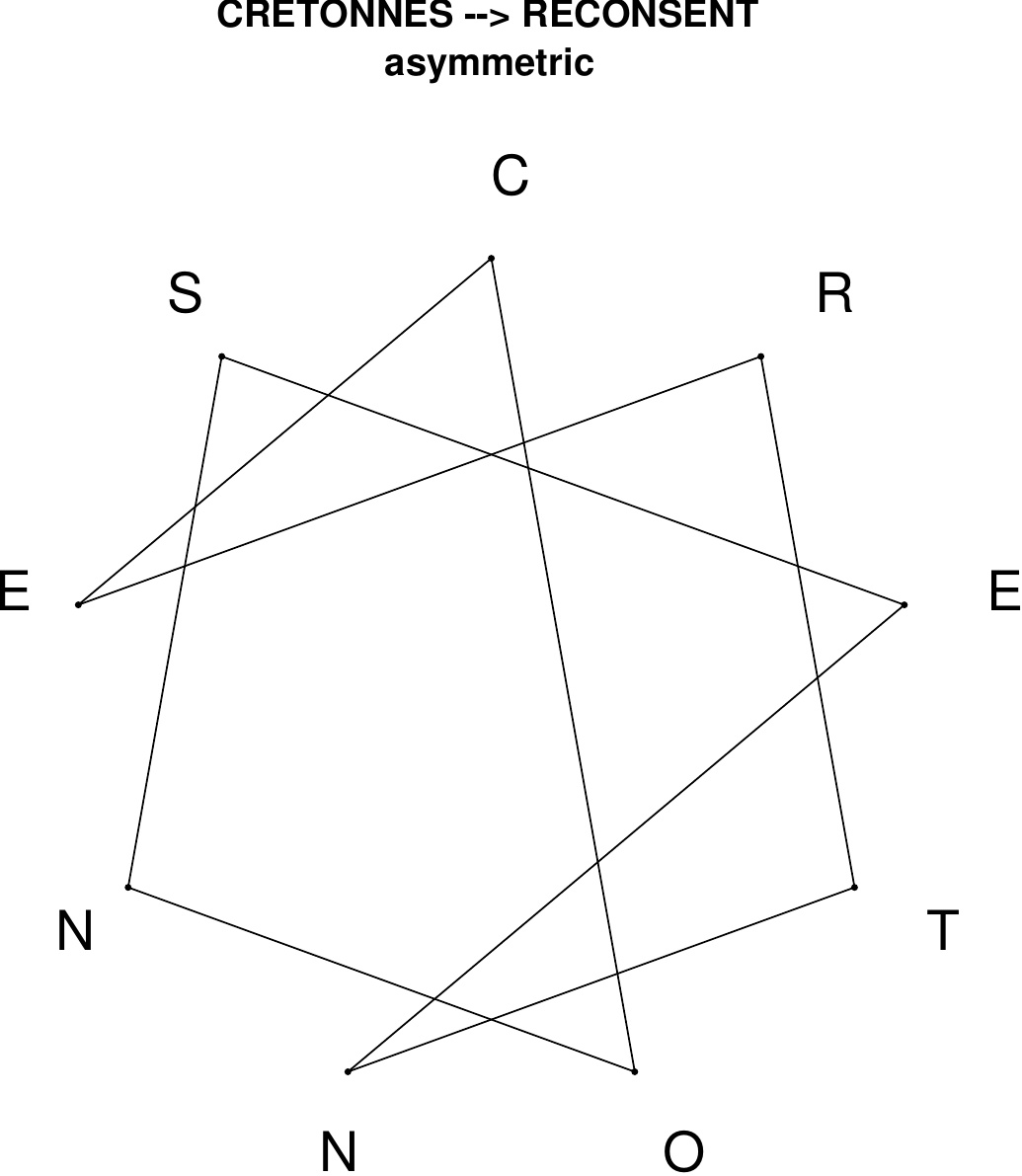}
\end{subfigure}
\hfill
\begin{subfigure}[T]{0.19\textwidth}
\centering
\includegraphics[width=\textwidth]{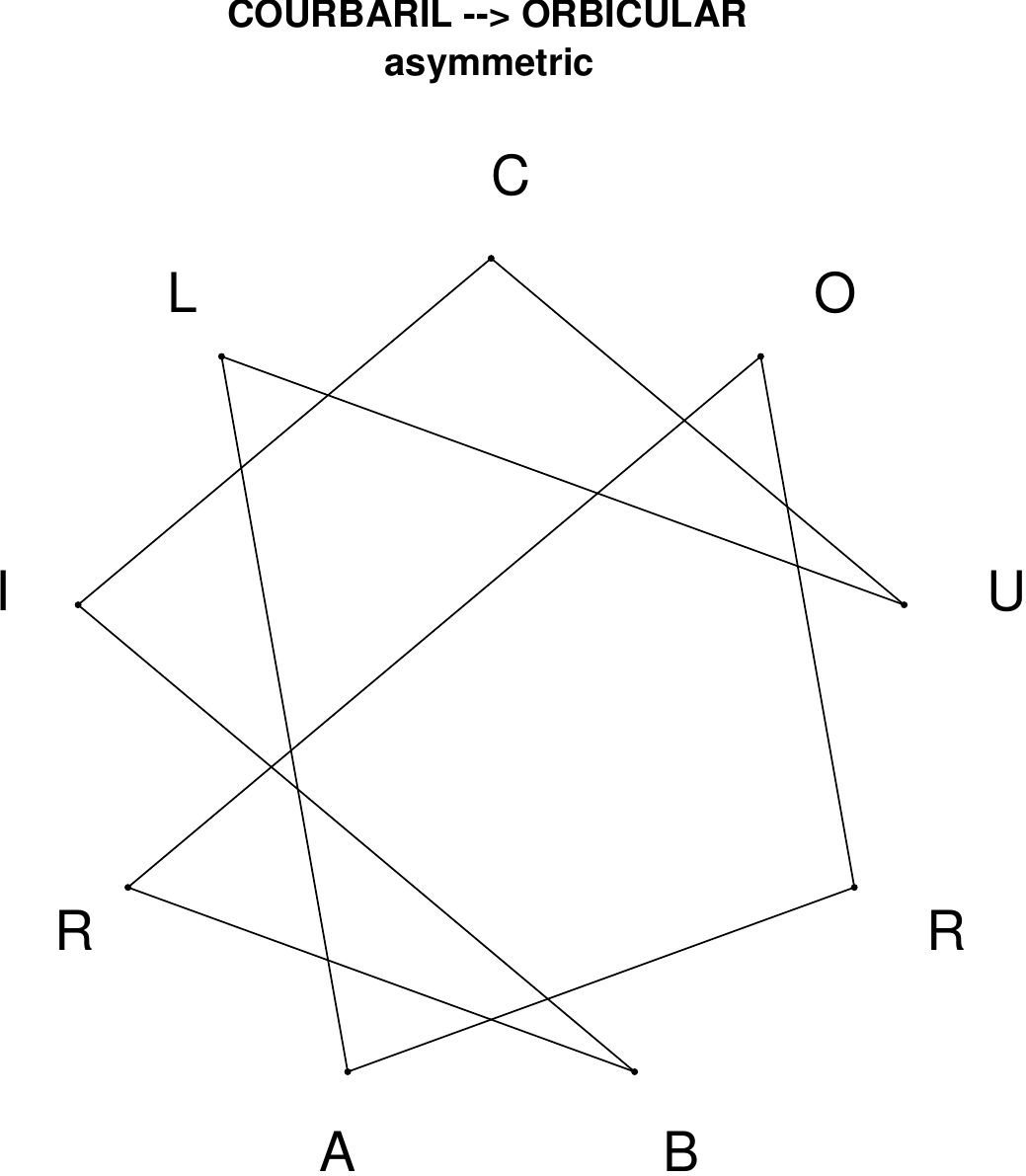}
\end{subfigure}
\hfill
\begin{subfigure}[T]{0.19\textwidth}
\centering
\includegraphics[width=\textwidth]{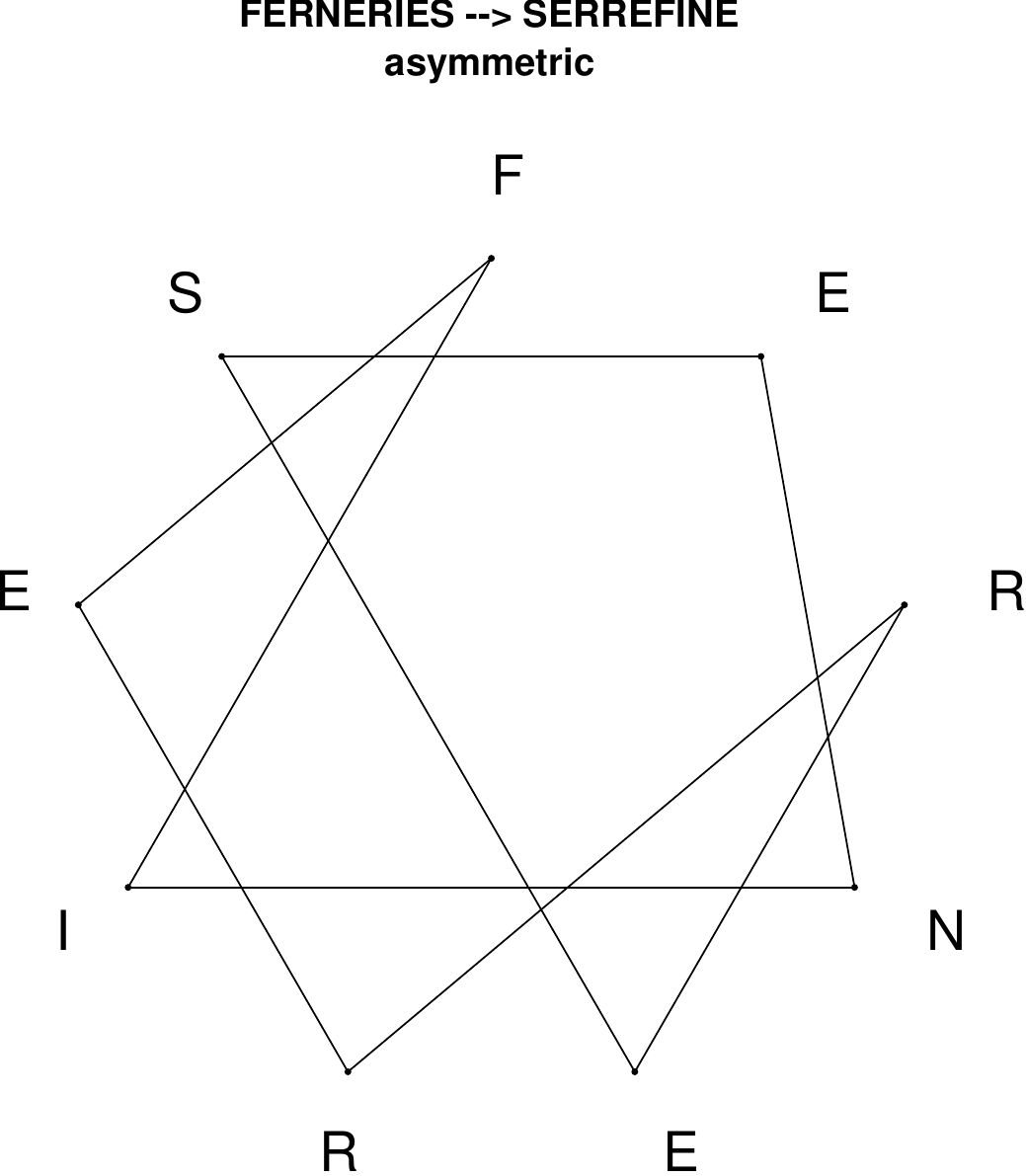}
\end{subfigure}
\hfill
\begin{subfigure}[T]{0.19\textwidth}
\centering
\includegraphics[width=\textwidth]{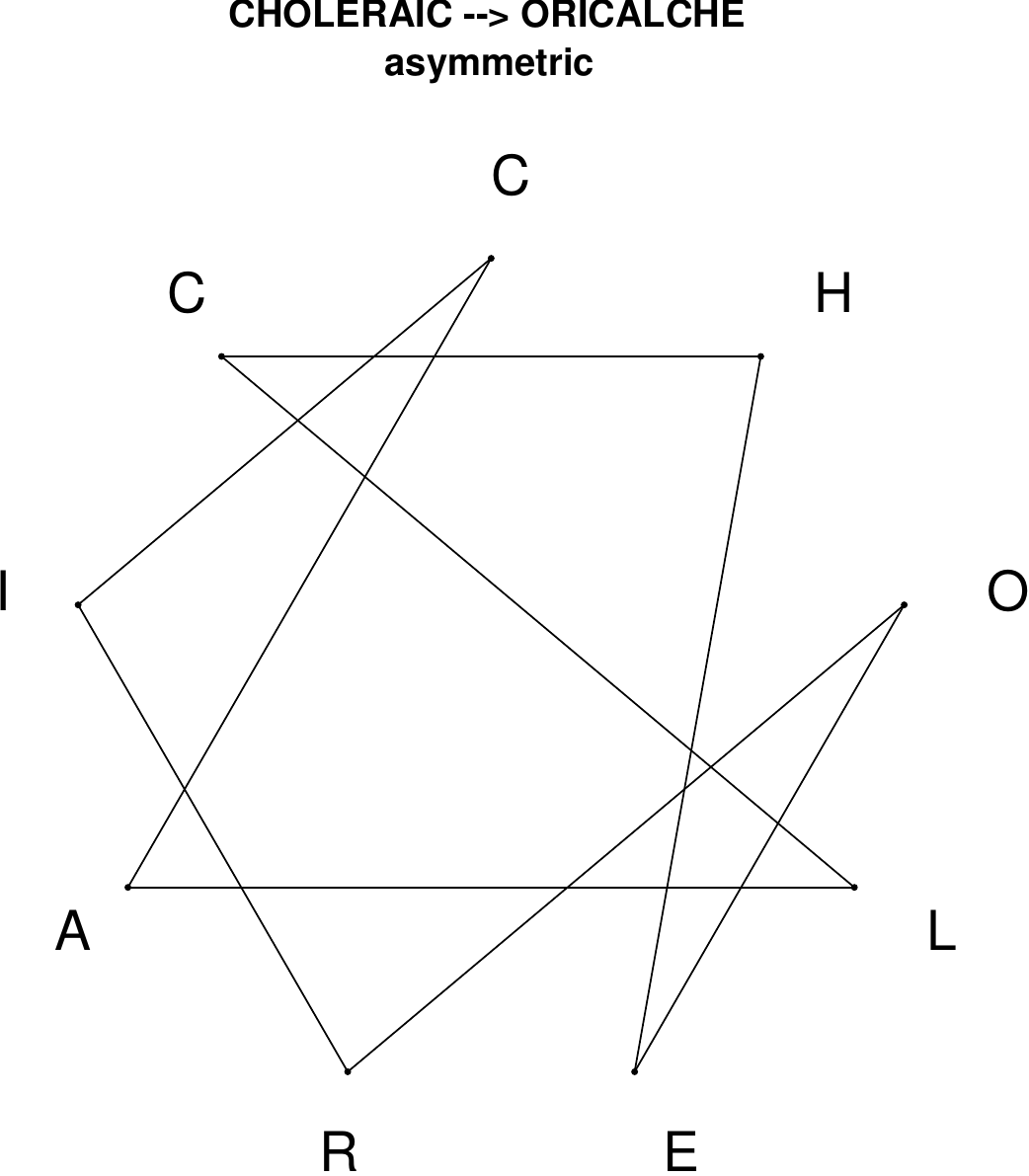}
\end{subfigure}
\hfill
\begin{subfigure}[T]{0.19\textwidth}
\centering
\includegraphics[width=\textwidth]{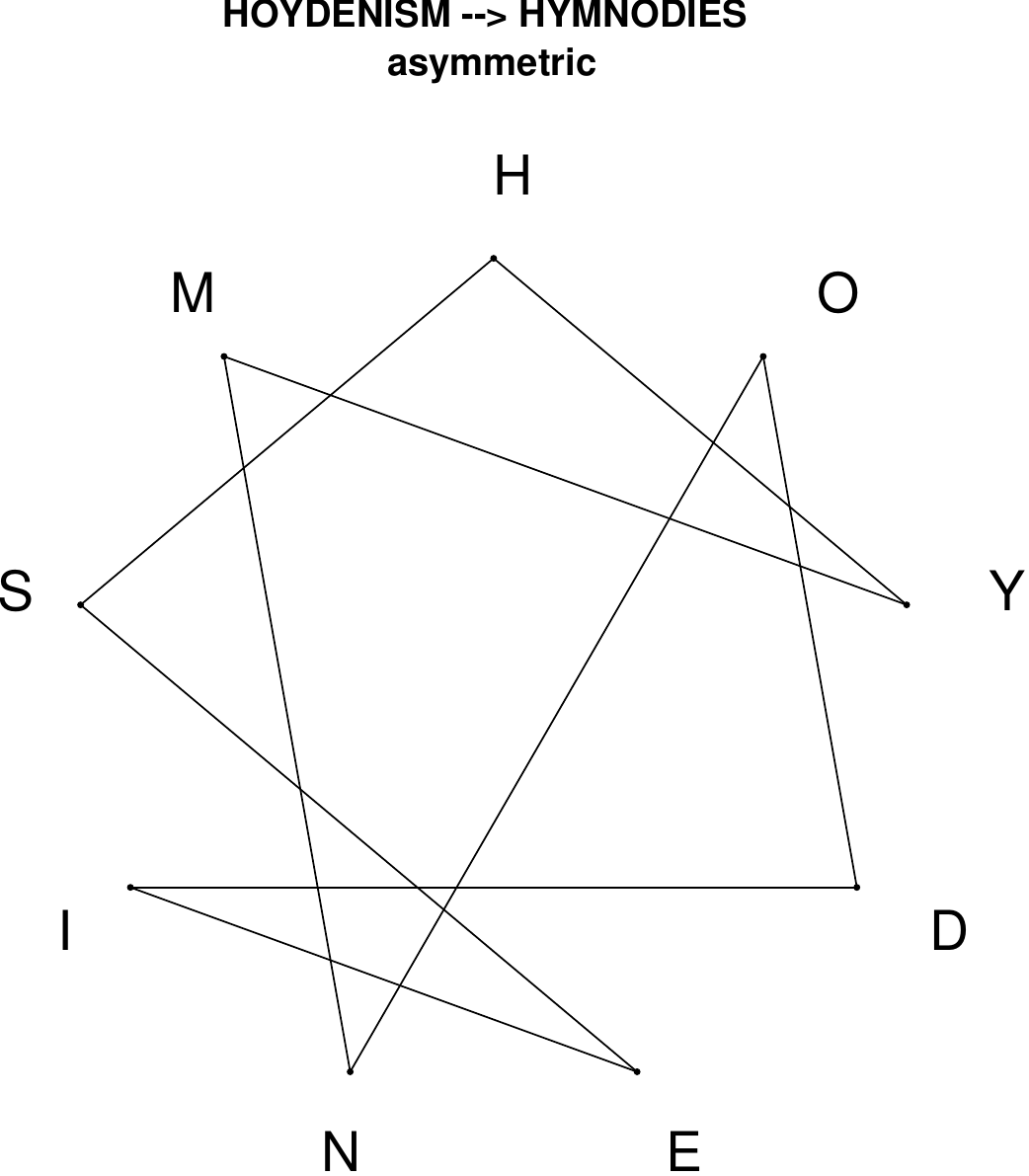}
\end{subfigure}
\end{figure}

\begin{figure}[H]
\centering
\begin{subfigure}[T]{0.19\textwidth}
\centering
\includegraphics[width=\textwidth]{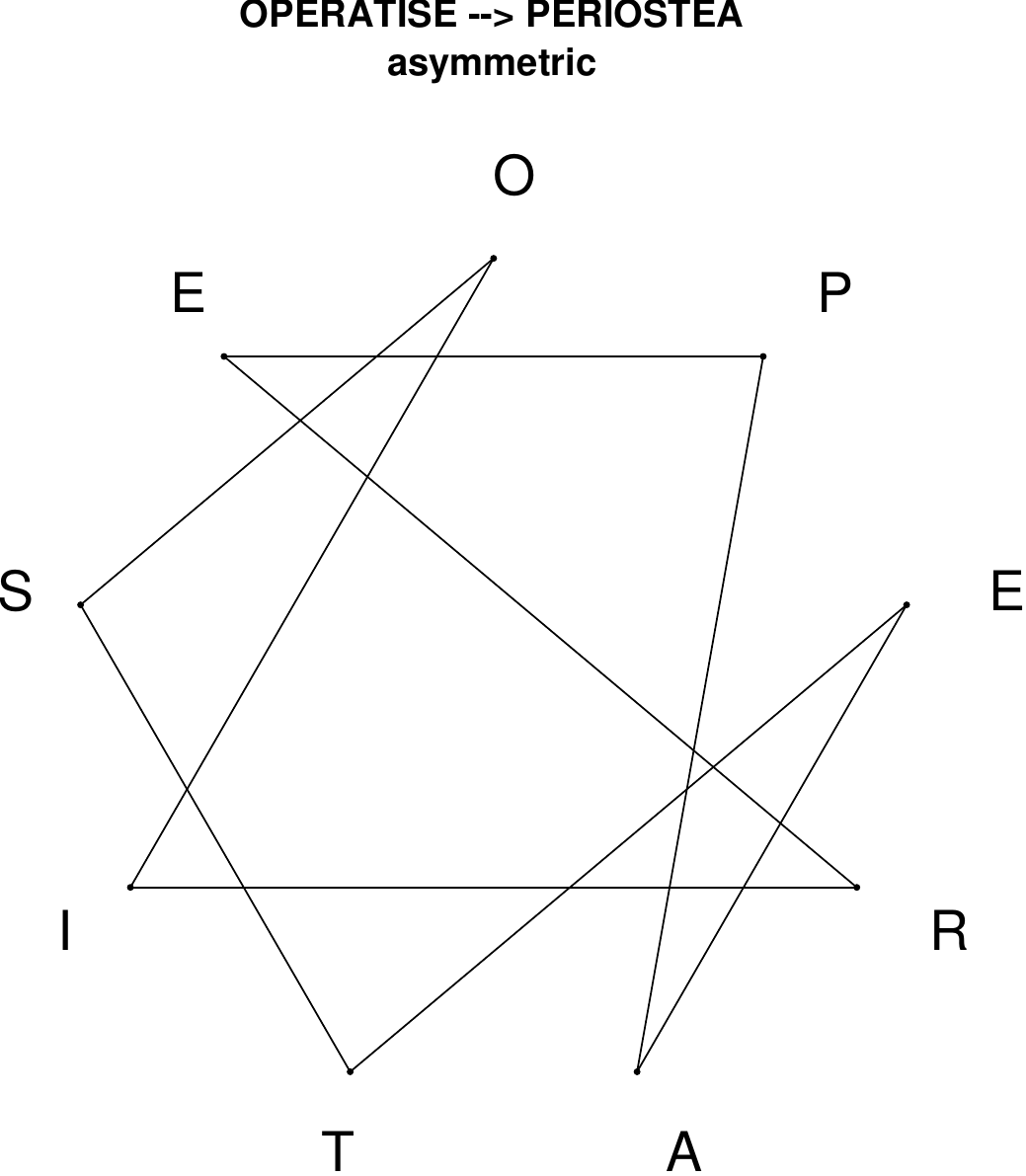}
\end{subfigure}
\hfill
\begin{subfigure}[T]{0.19\textwidth}
\centering
\includegraphics[width=\textwidth]{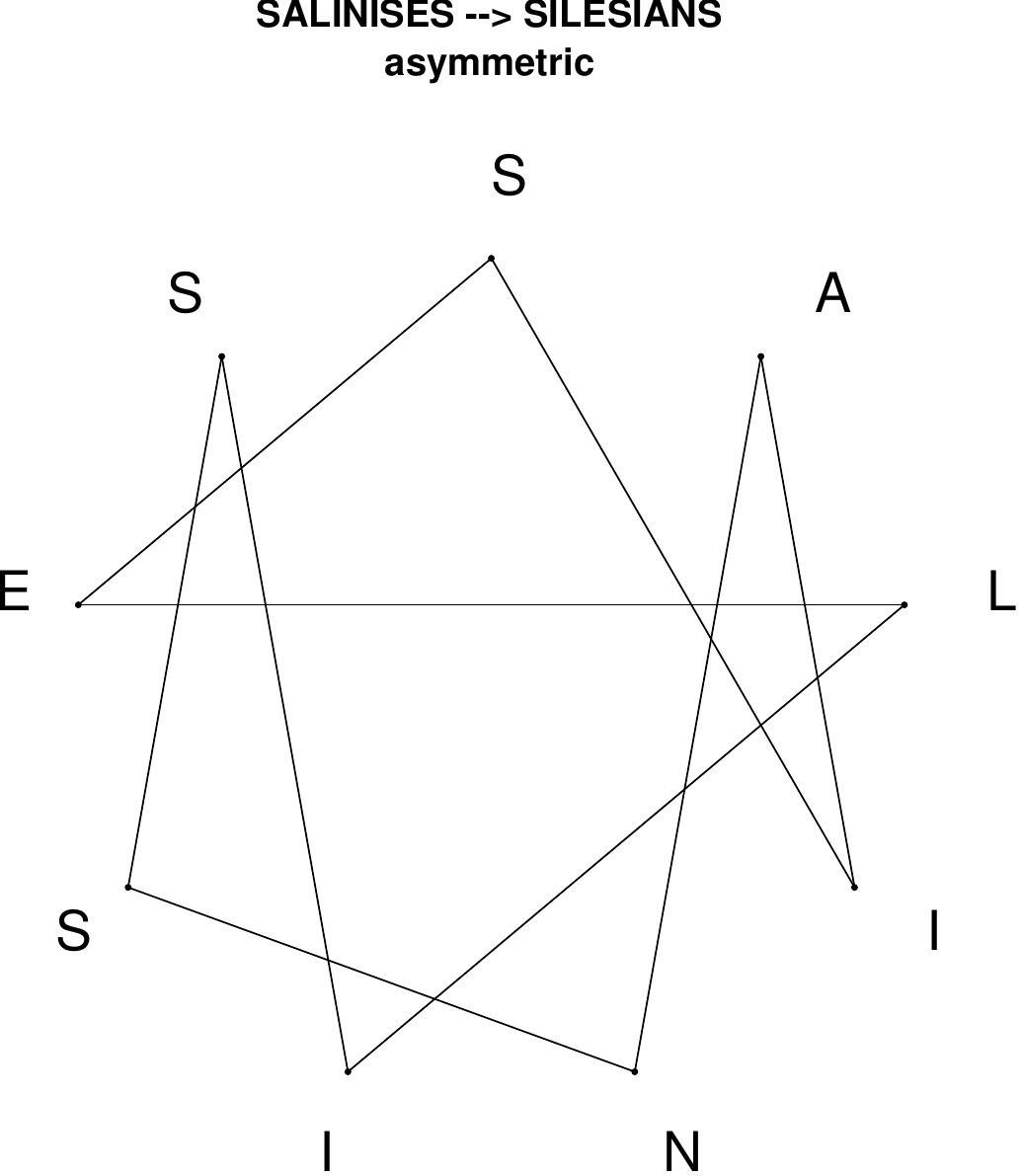}
\end{subfigure}
\hfill
\begin{subfigure}[T]{0.19\textwidth}
\centering
\includegraphics[width=\textwidth]{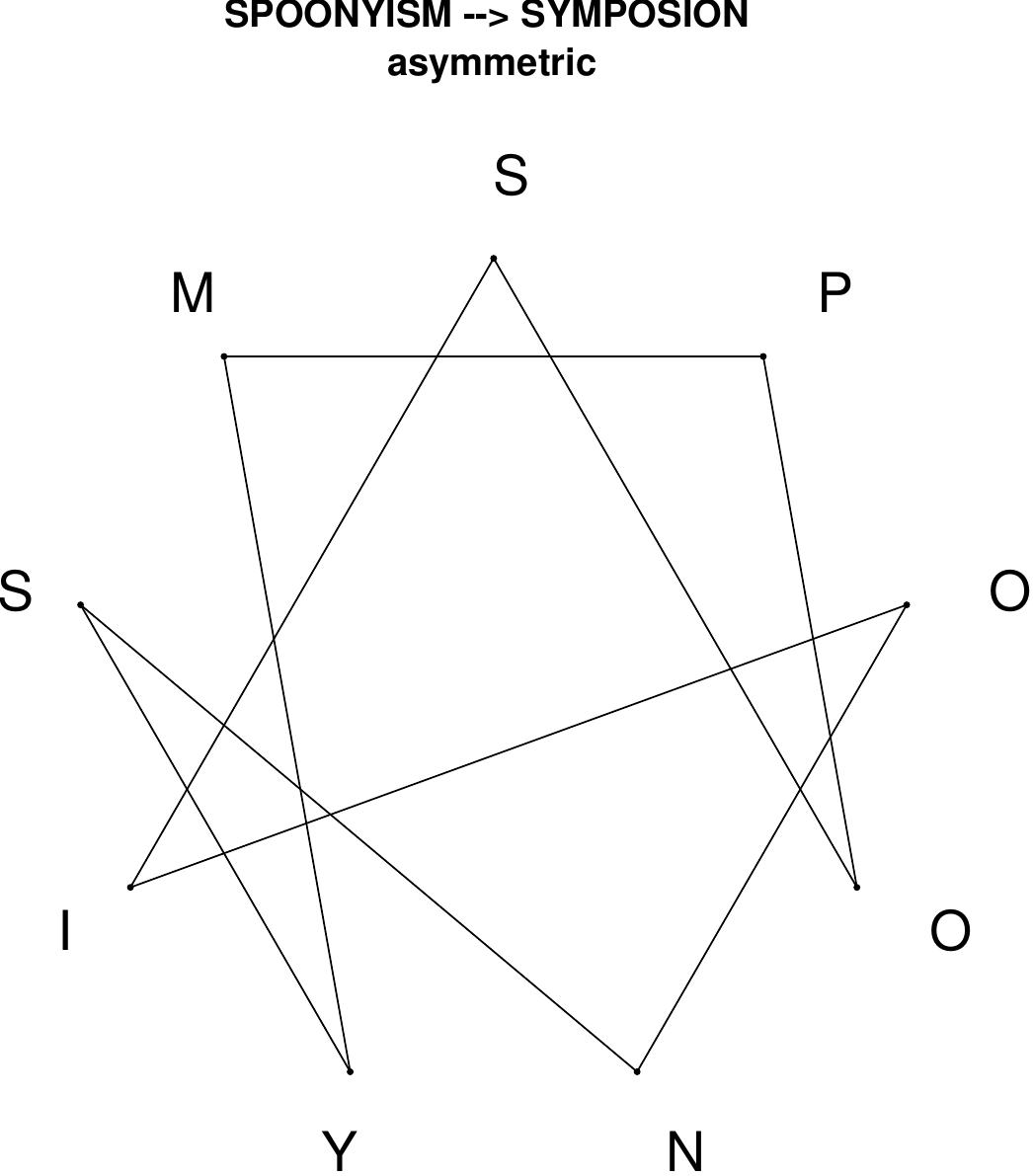}
\end{subfigure}
\hfill
\begin{subfigure}[T]{0.19\textwidth}
\centering
\includegraphics[width=\textwidth]{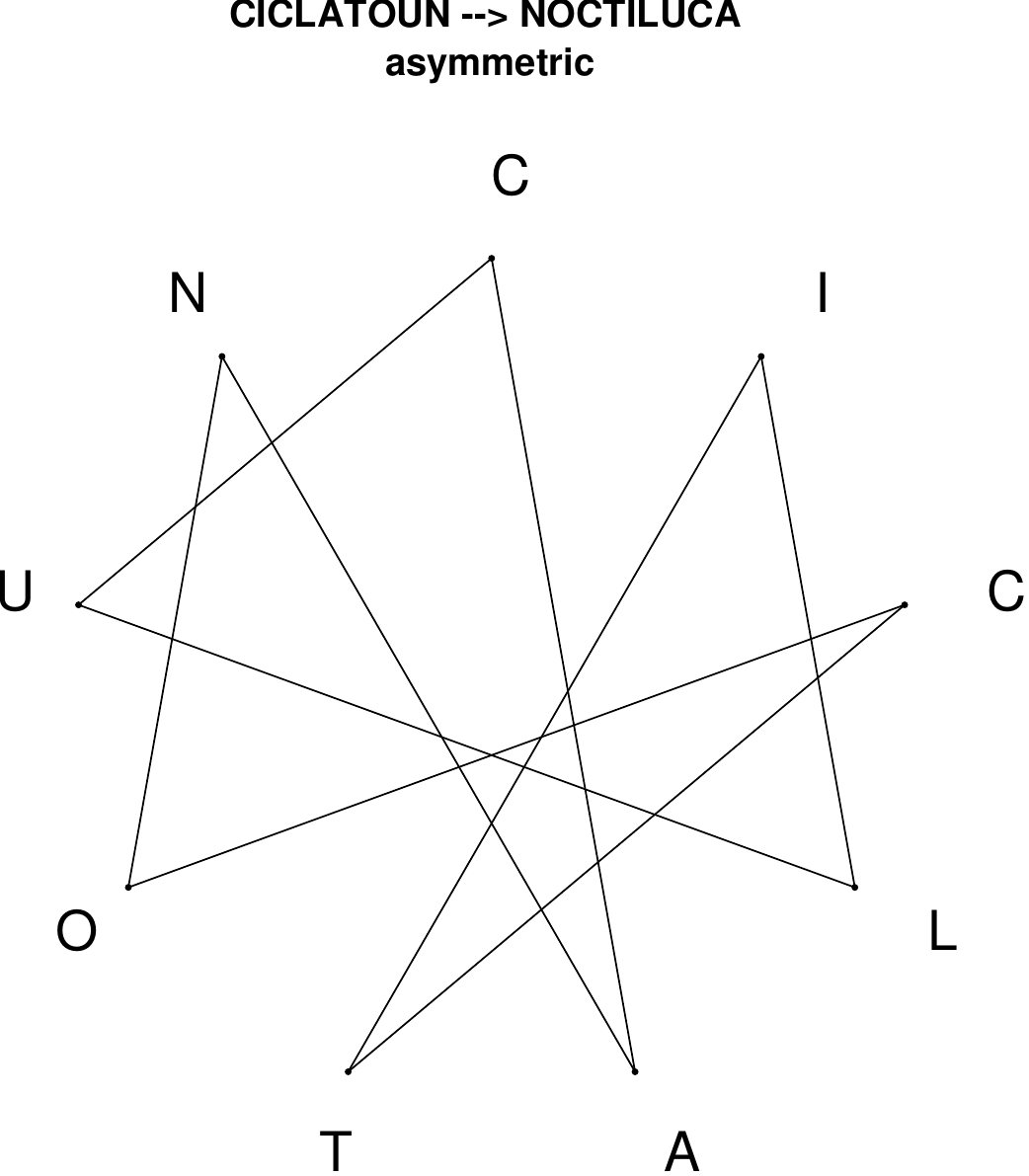}
\end{subfigure}
\hfill
\begin{subfigure}[T]{0.19\textwidth}
\centering
\includegraphics[width=\textwidth]{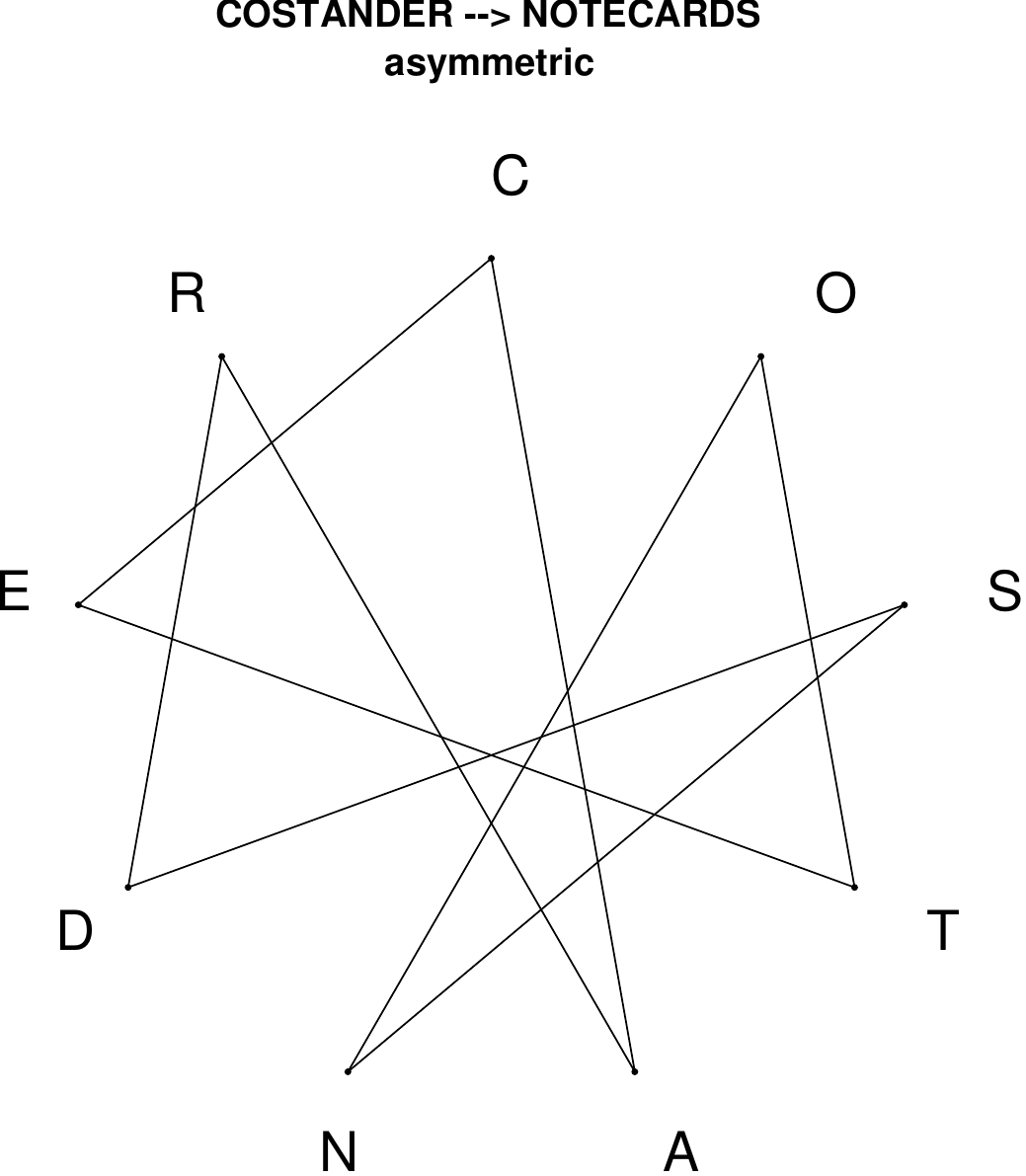}
\end{subfigure}
\end{figure}

\begin{figure}[H]
\centering
\begin{subfigure}[T]{0.19\textwidth}
\centering
\includegraphics[width=\textwidth]{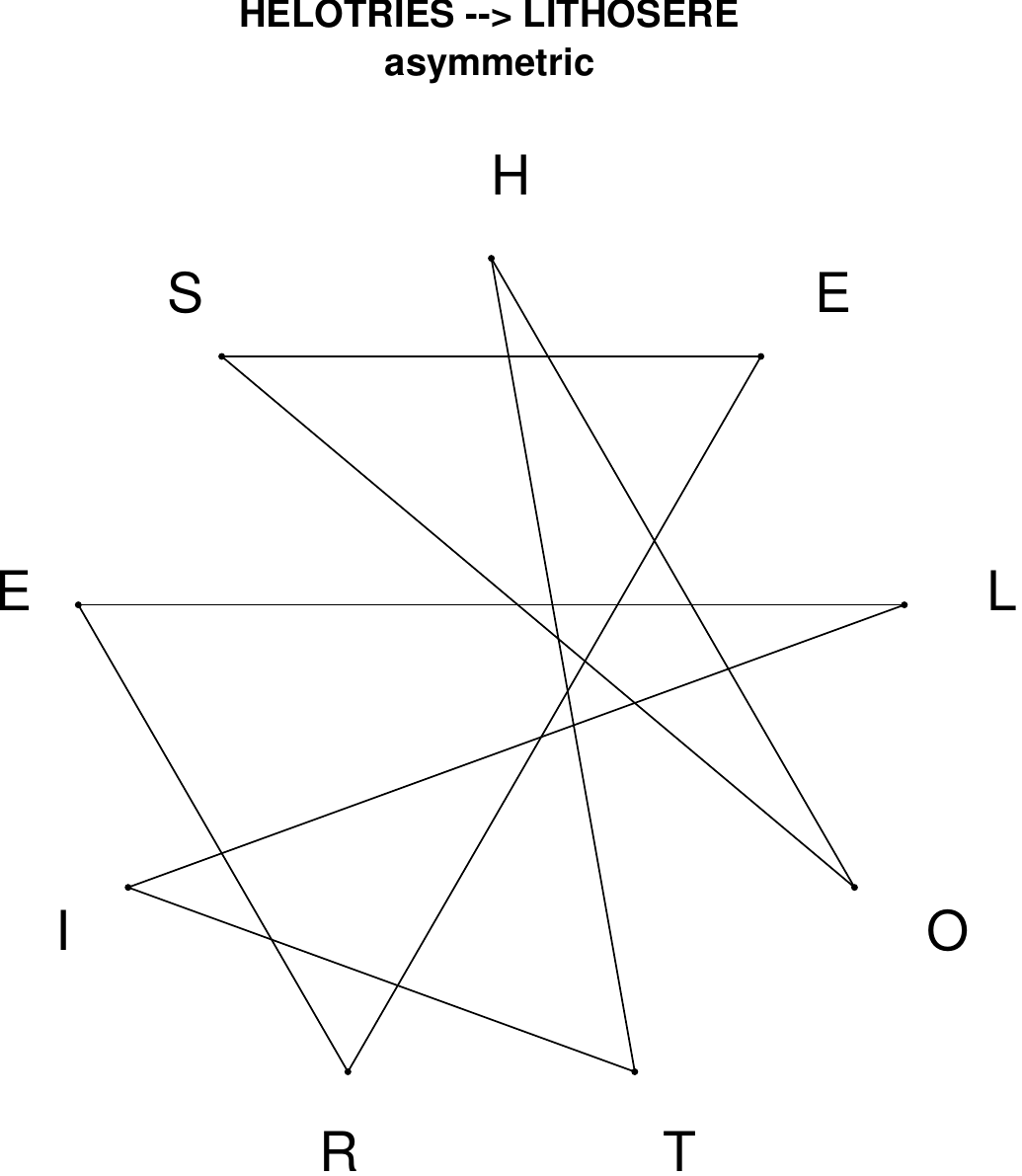}
\end{subfigure}
\hfill
\begin{subfigure}[T]{0.19\textwidth}
\centering
\includegraphics[width=\textwidth]{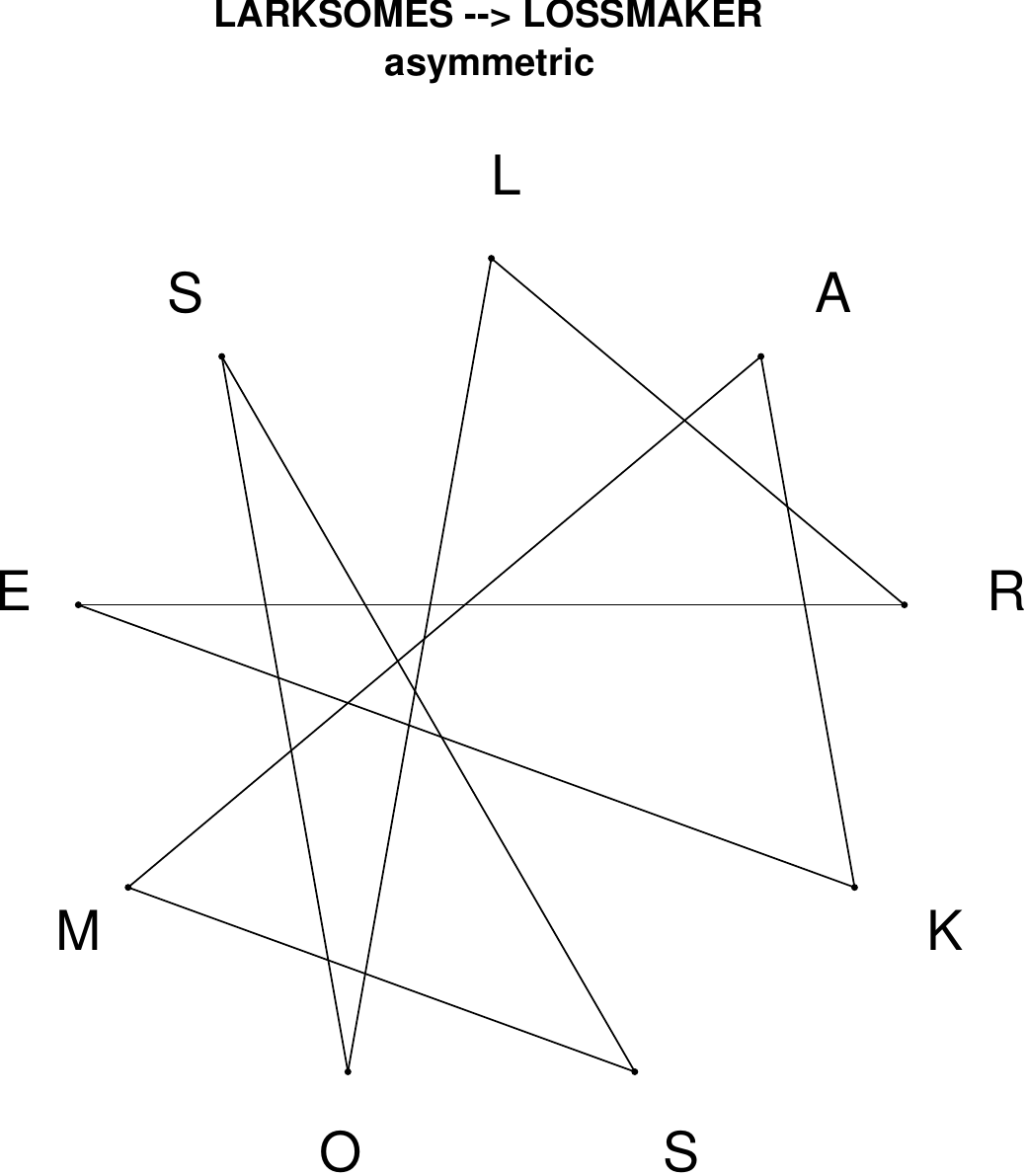}
\end{subfigure}
\hfill
\begin{subfigure}[T]{0.19\textwidth}
\centering
\includegraphics[width=\textwidth]{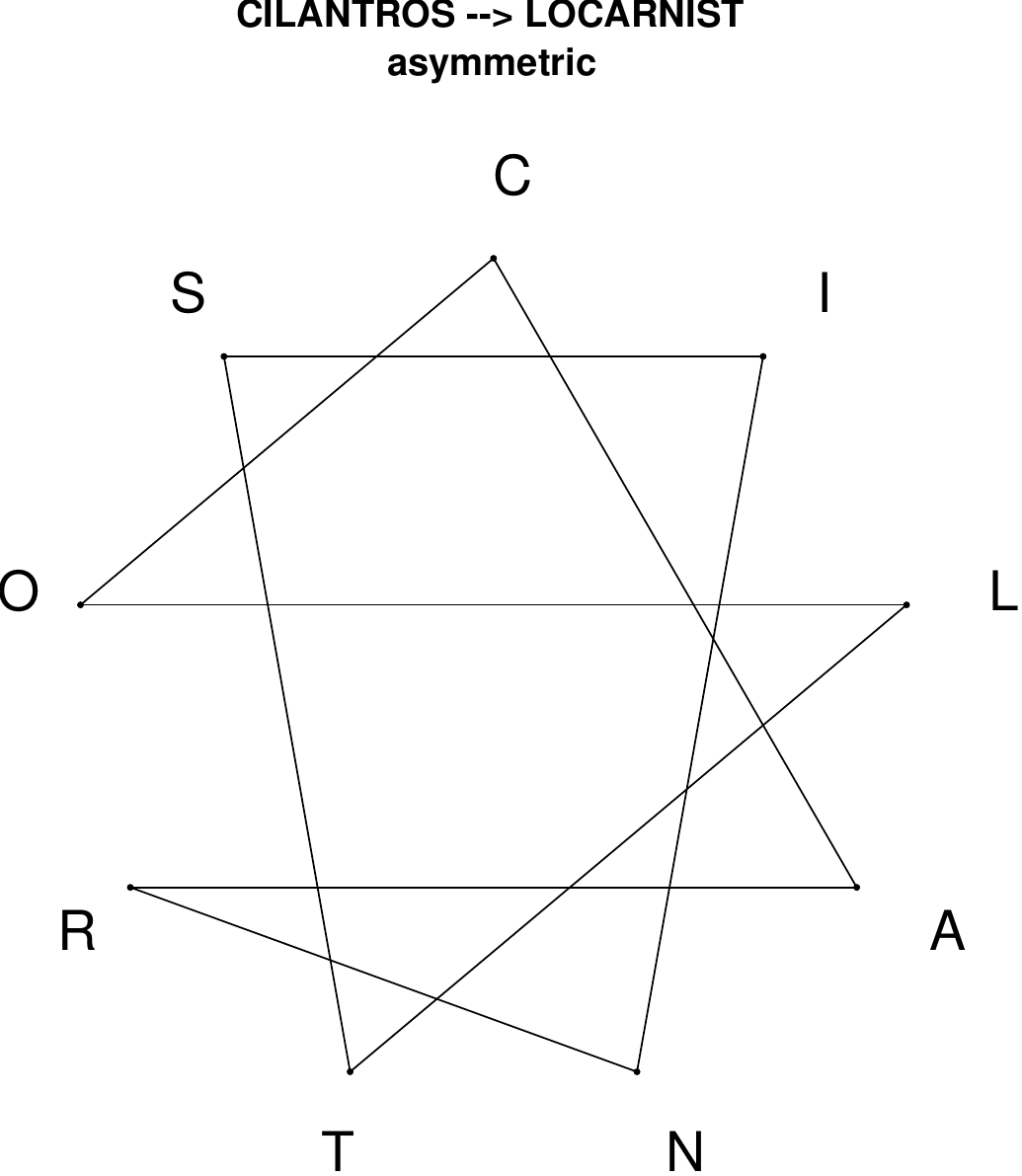}
\end{subfigure}
\hfill
\begin{subfigure}[T]{0.19\textwidth}
\centering
\includegraphics[width=\textwidth]{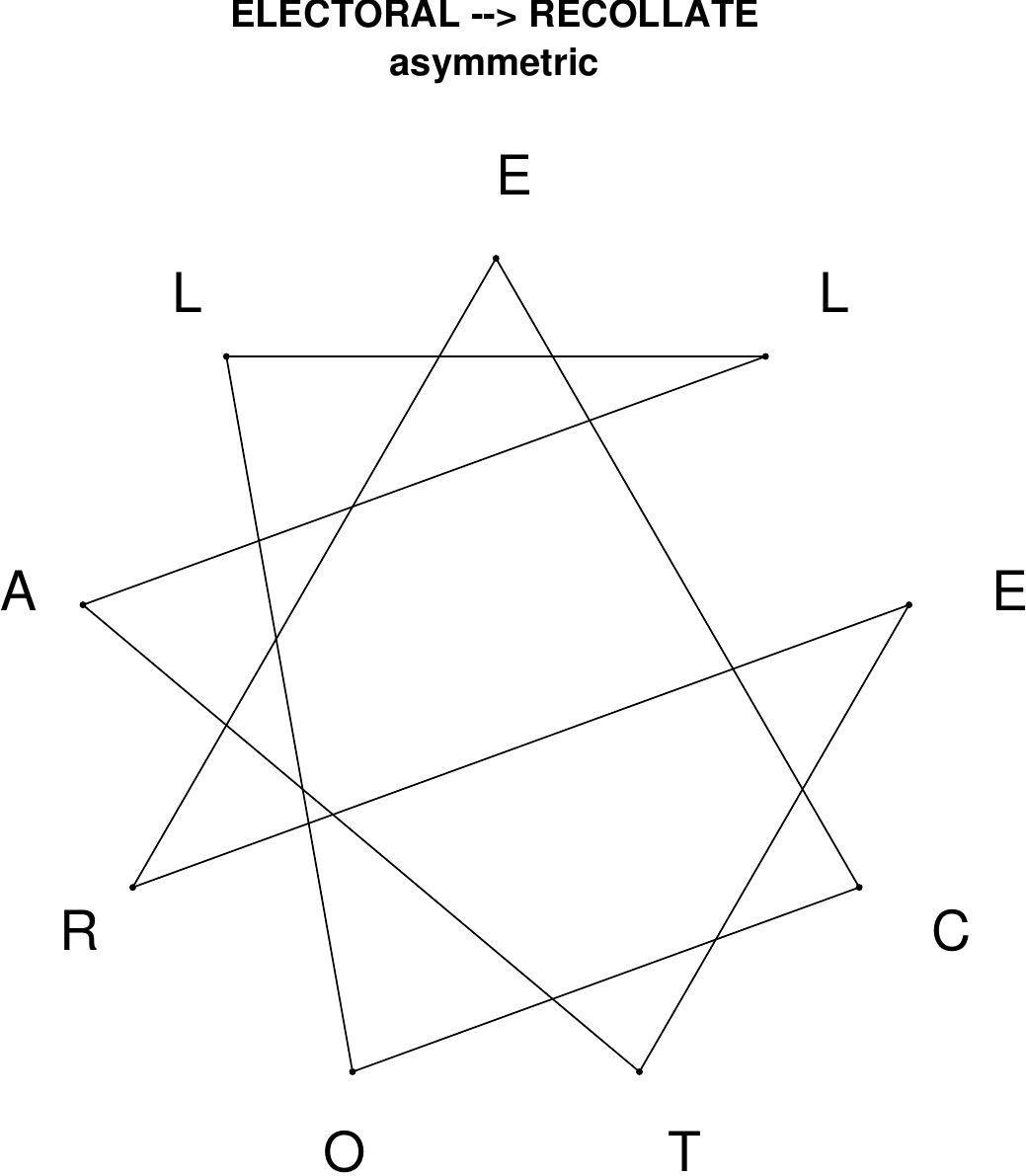}
\end{subfigure}
\hfill
\begin{subfigure}[T]{0.19\textwidth}
\centering
\includegraphics[width=\textwidth]{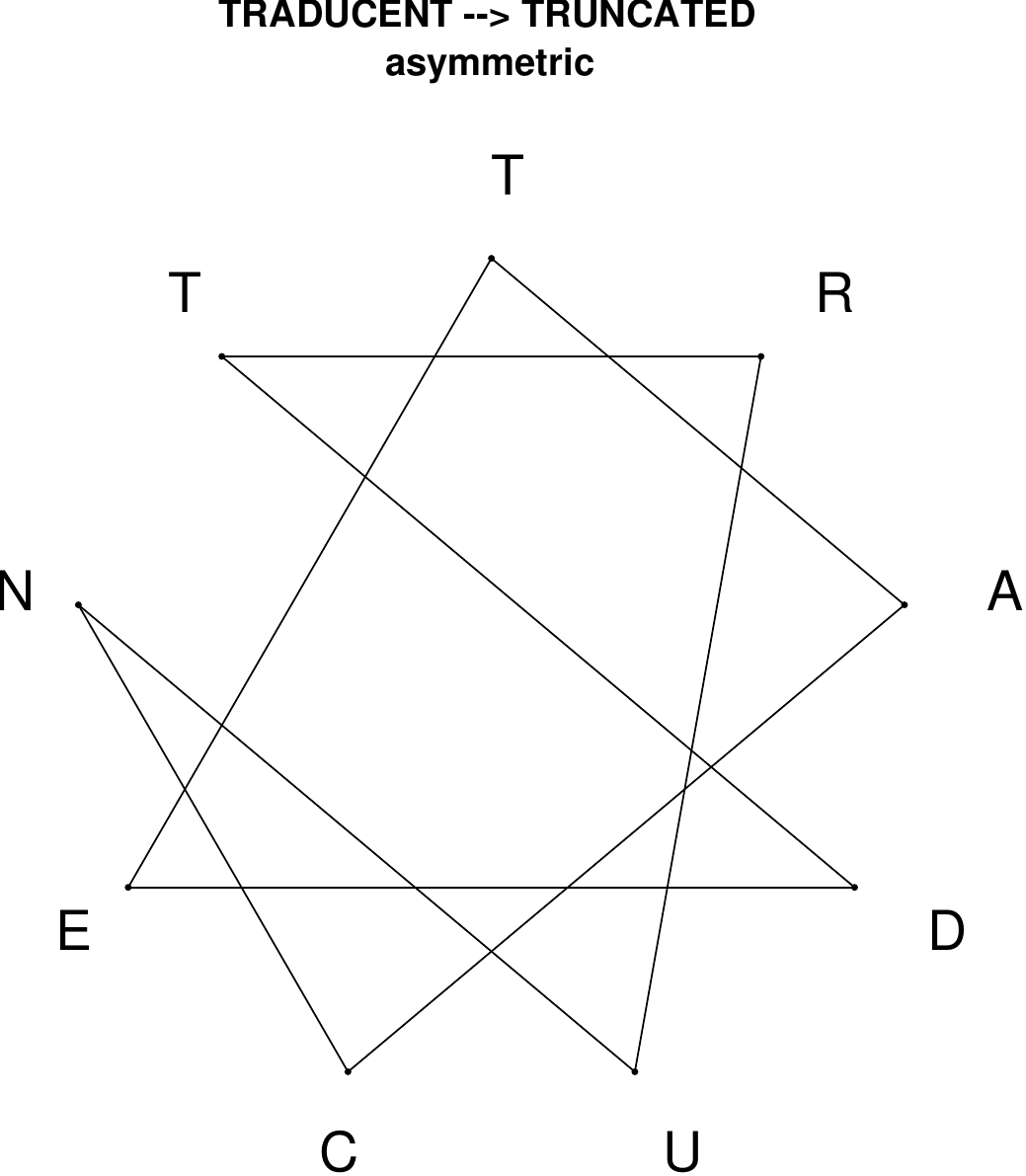}
\end{subfigure}
\end{figure}

\begin{figure}[H]
\centering
\begin{subfigure}[T]{0.19\textwidth}
\centering
\includegraphics[width=\textwidth]{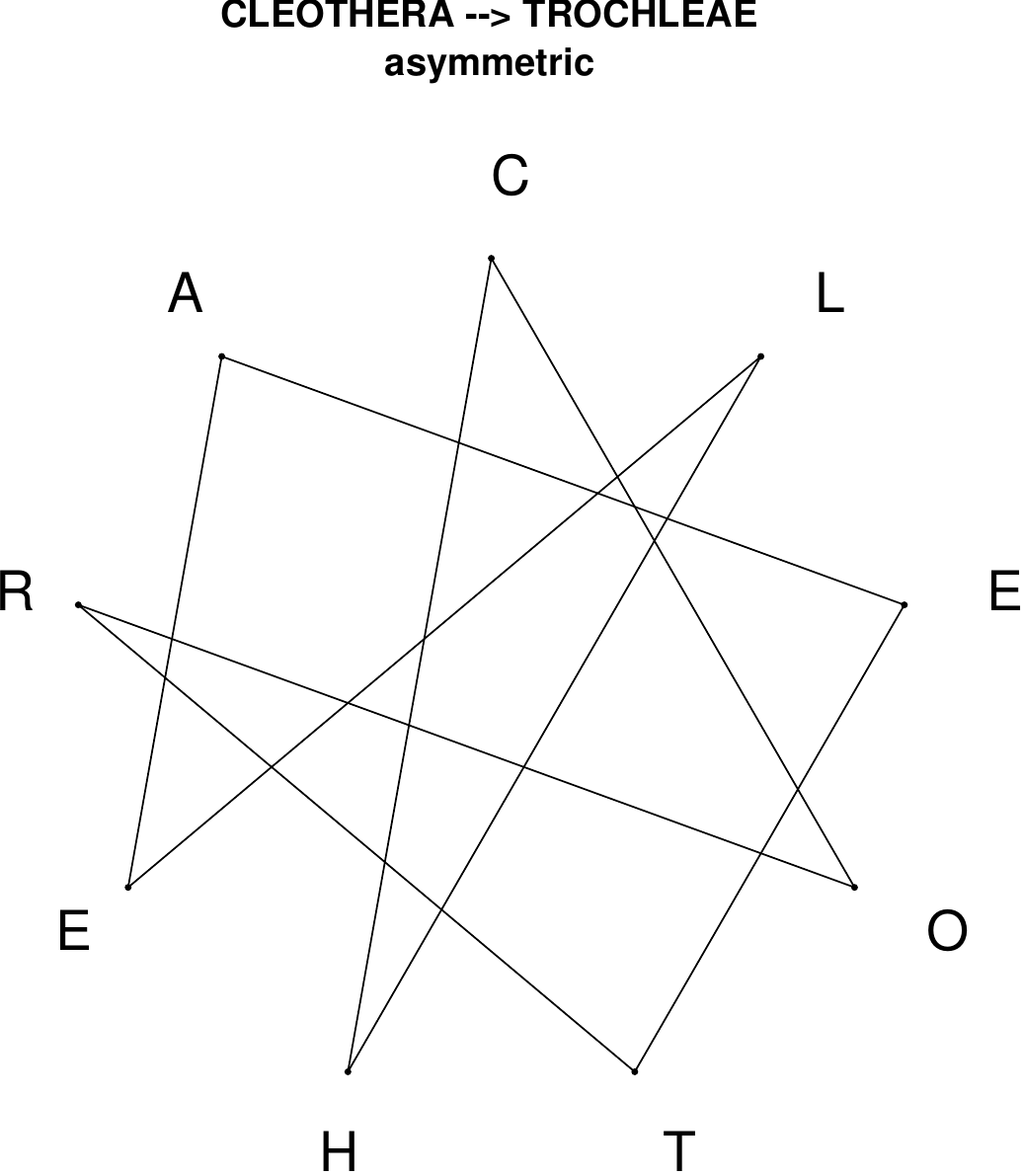}
\end{subfigure}
\hfill
\begin{subfigure}[T]{0.19\textwidth}
\centering
\includegraphics[width=\textwidth]{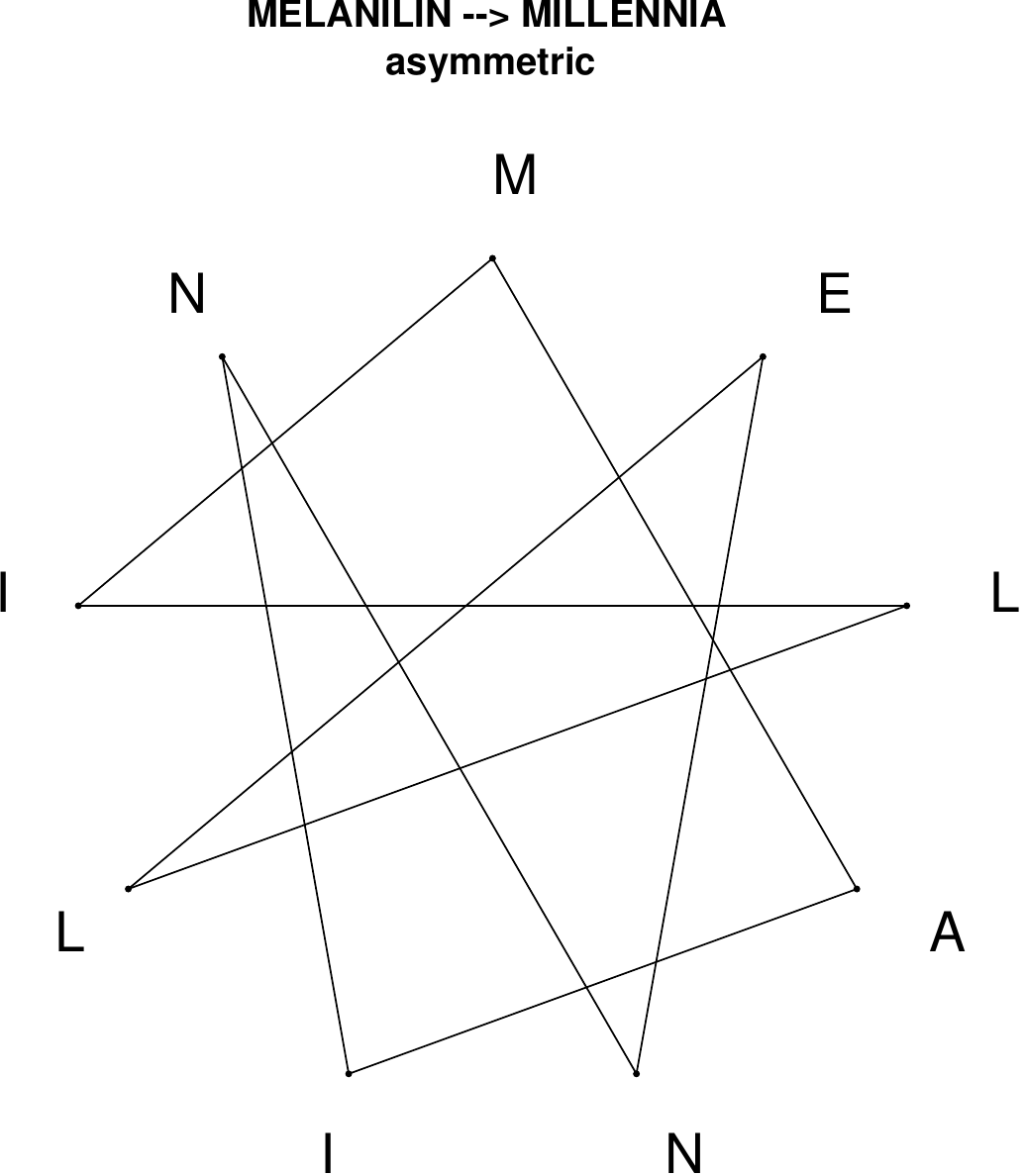}
\end{subfigure}
\hfill
\begin{subfigure}[T]{0.19\textwidth}
\centering
\includegraphics[width=\textwidth]{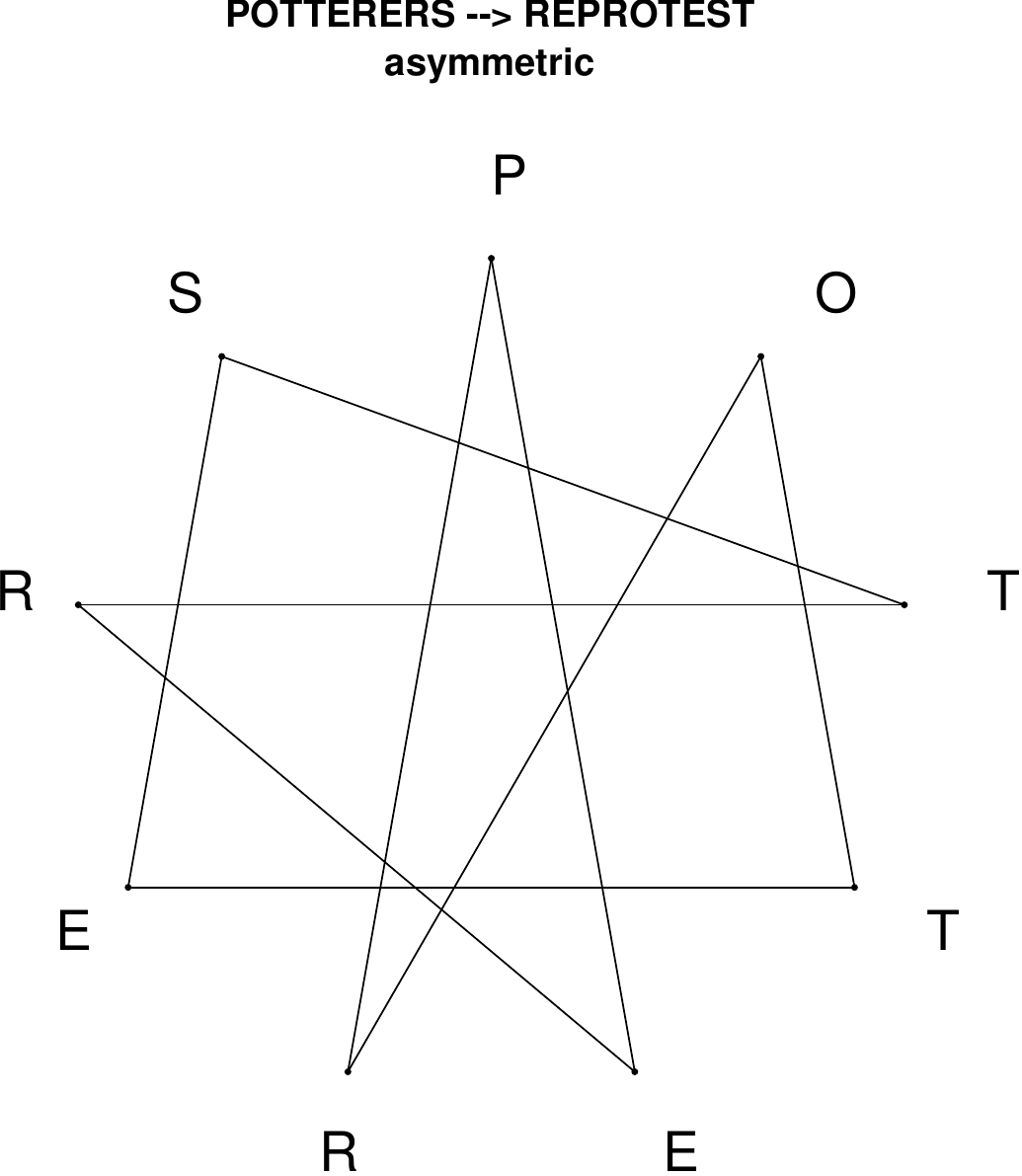}
\end{subfigure}
\hfill
\begin{subfigure}[T]{0.19\textwidth}
\centering
\includegraphics[width=\textwidth]{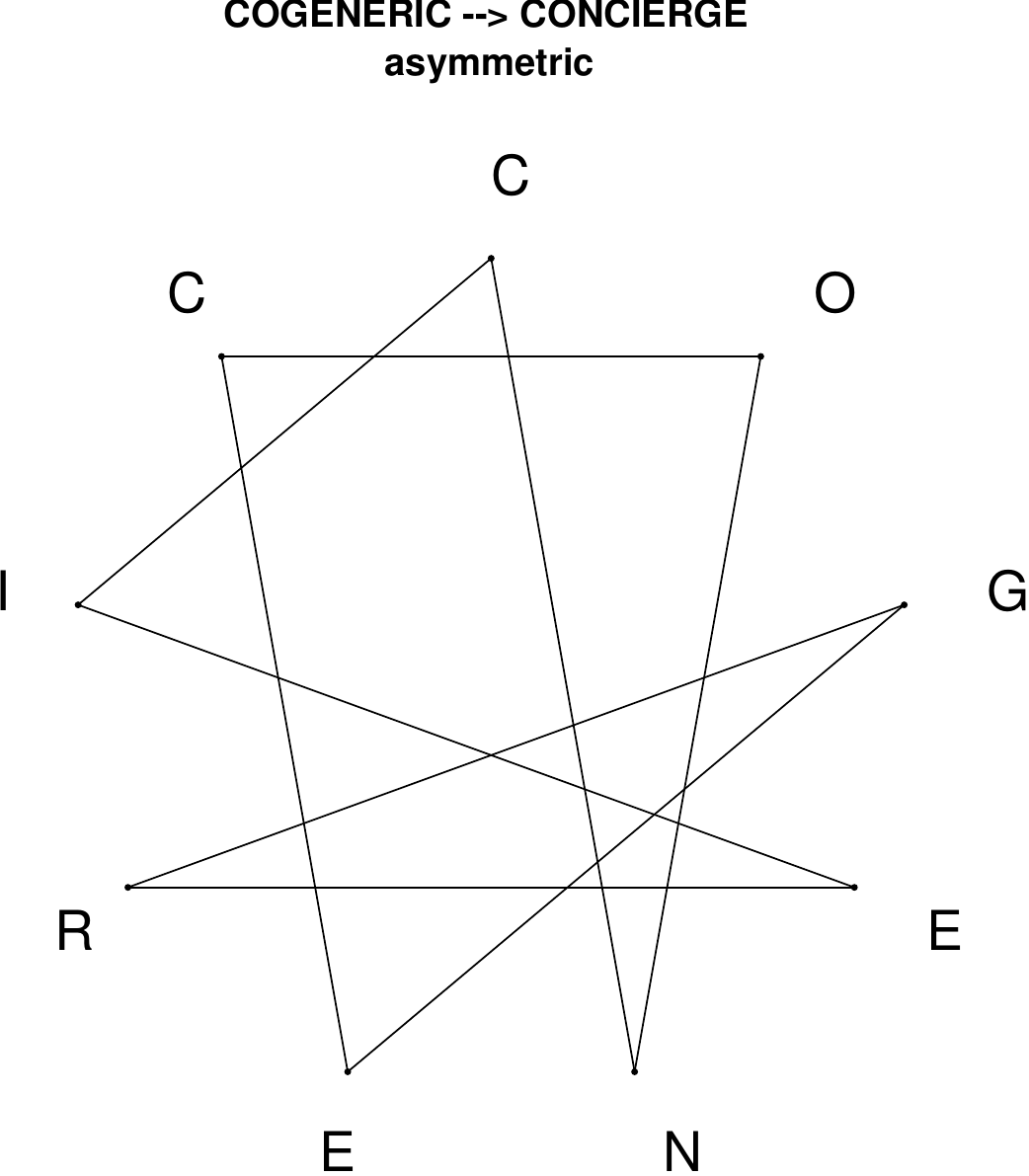}
\end{subfigure}
\hfill
\begin{subfigure}[T]{0.19\textwidth}
\centering
\includegraphics[width=\textwidth]{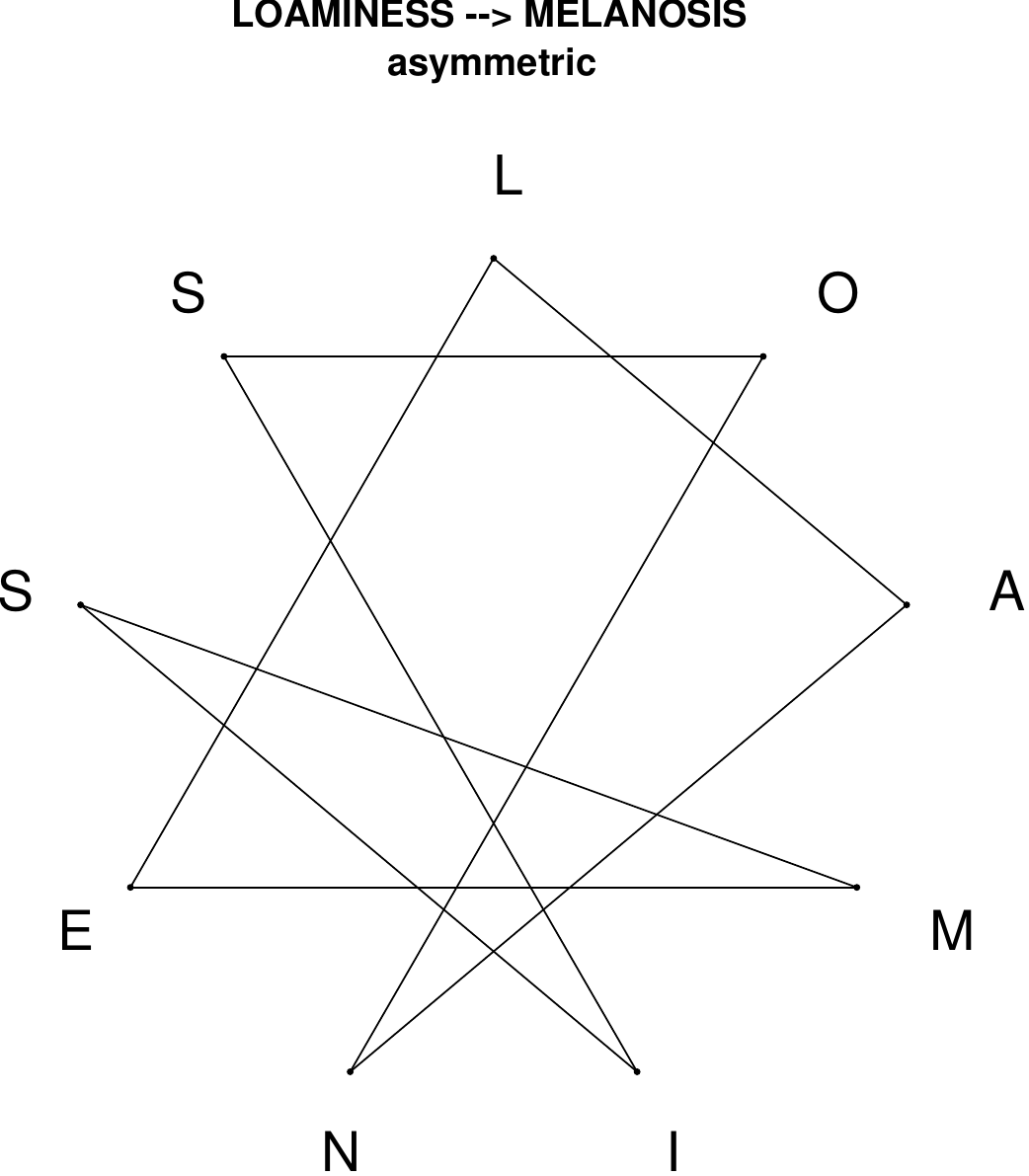}
\end{subfigure}
\end{figure}

\begin{figure}[H]
\centering
\begin{subfigure}[T]{0.19\textwidth}
\centering
\includegraphics[width=\textwidth]{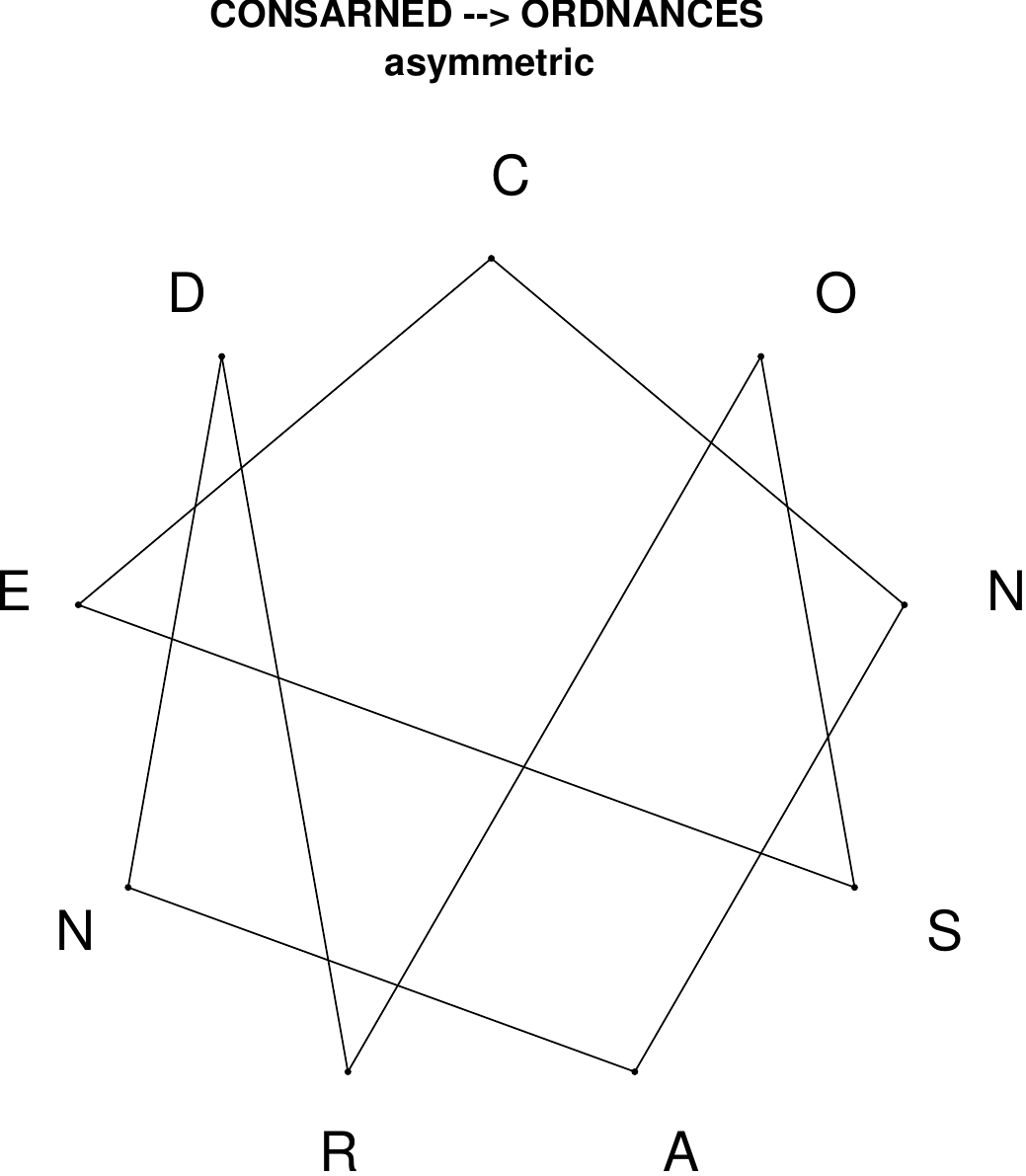}
\end{subfigure}
\hfill
\begin{subfigure}[T]{0.19\textwidth}
\centering
\includegraphics[width=\textwidth]{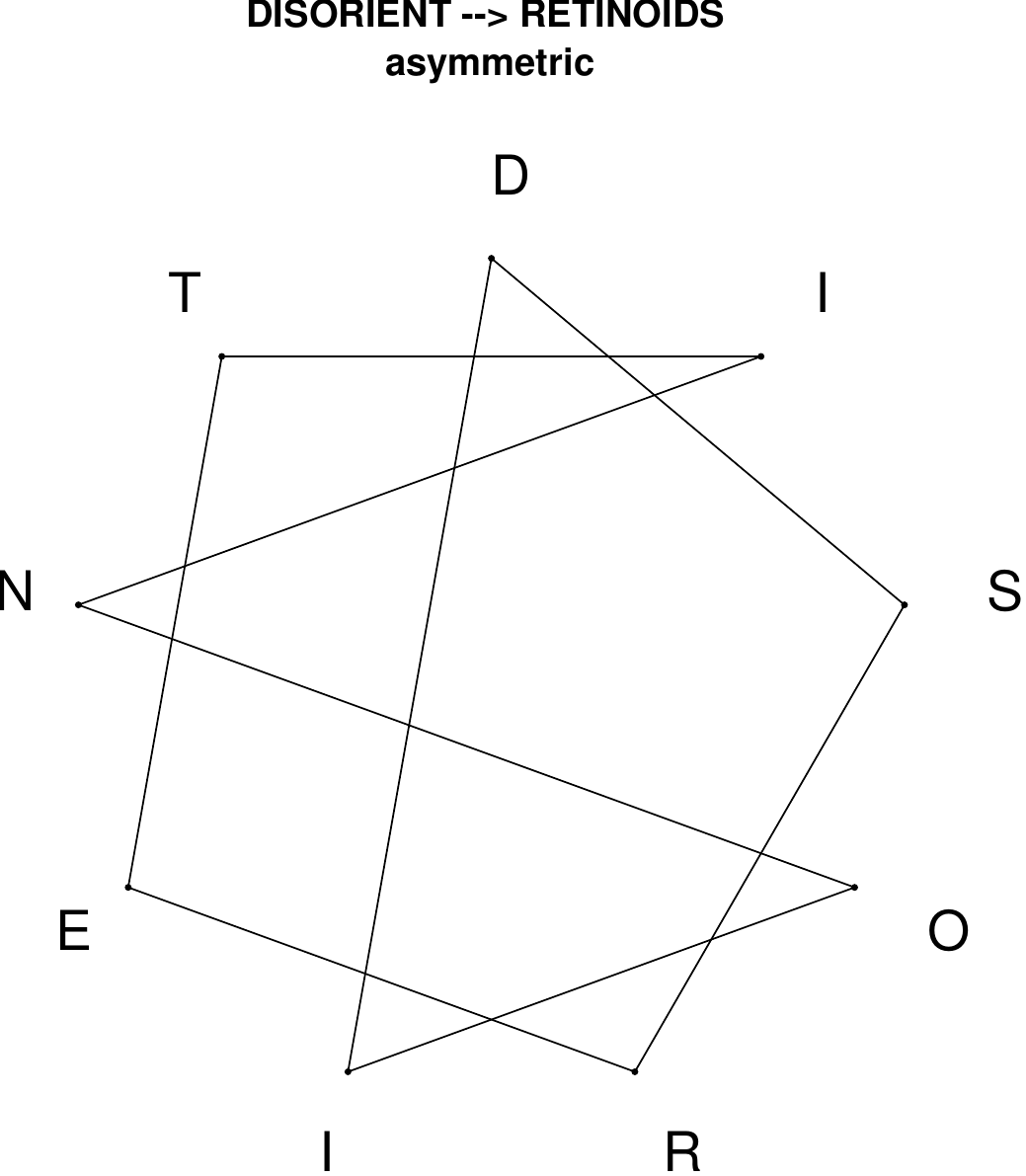}
\end{subfigure}
\hfill
\begin{subfigure}[T]{0.19\textwidth}
\centering
\includegraphics[width=\textwidth]{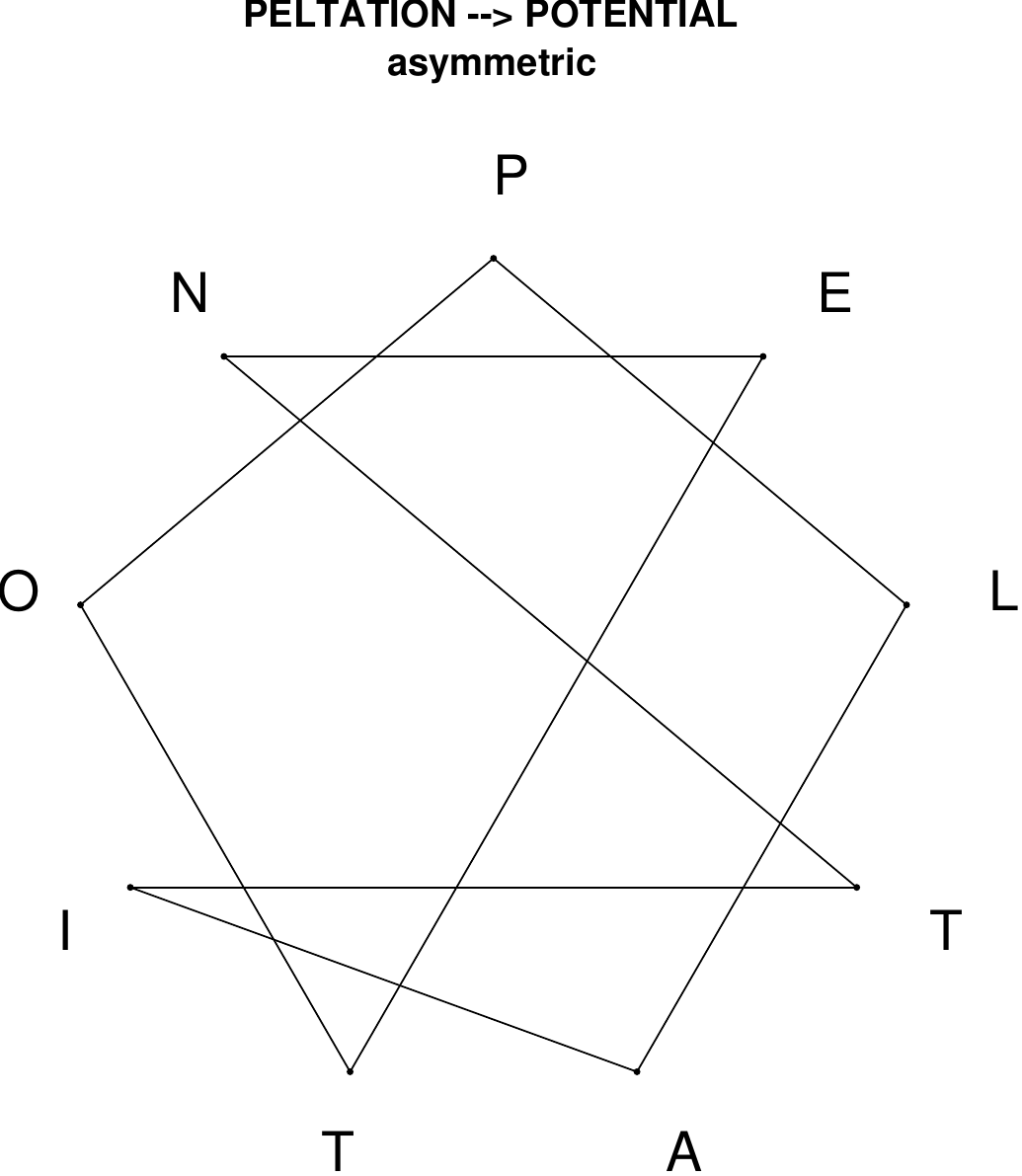}
\end{subfigure}
\hfill
\begin{subfigure}[T]{0.19\textwidth}
\centering
\includegraphics[width=\textwidth]{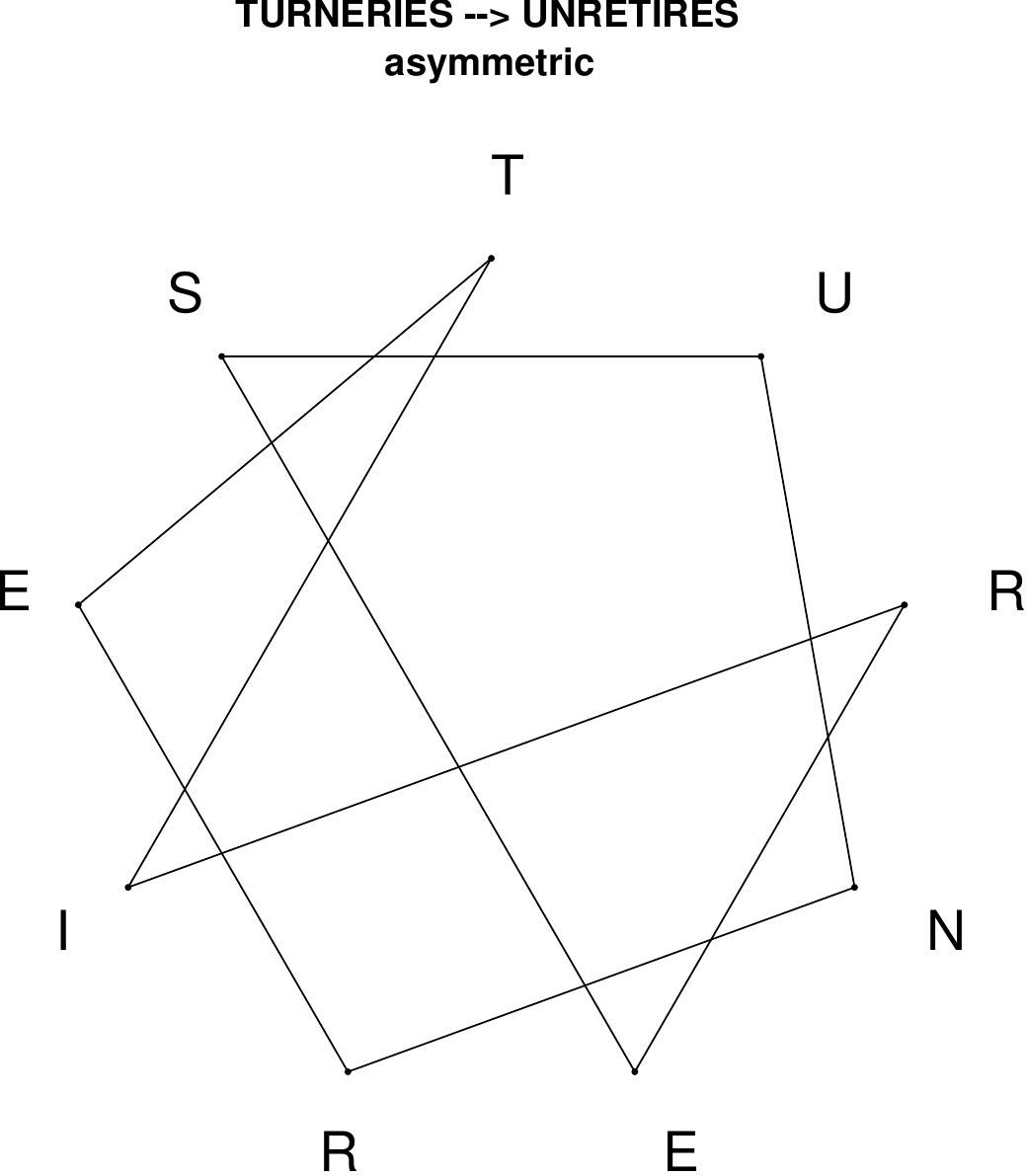}
\end{subfigure}
\hfill
\begin{subfigure}[T]{0.19\textwidth}
\centering
\includegraphics[width=\textwidth]{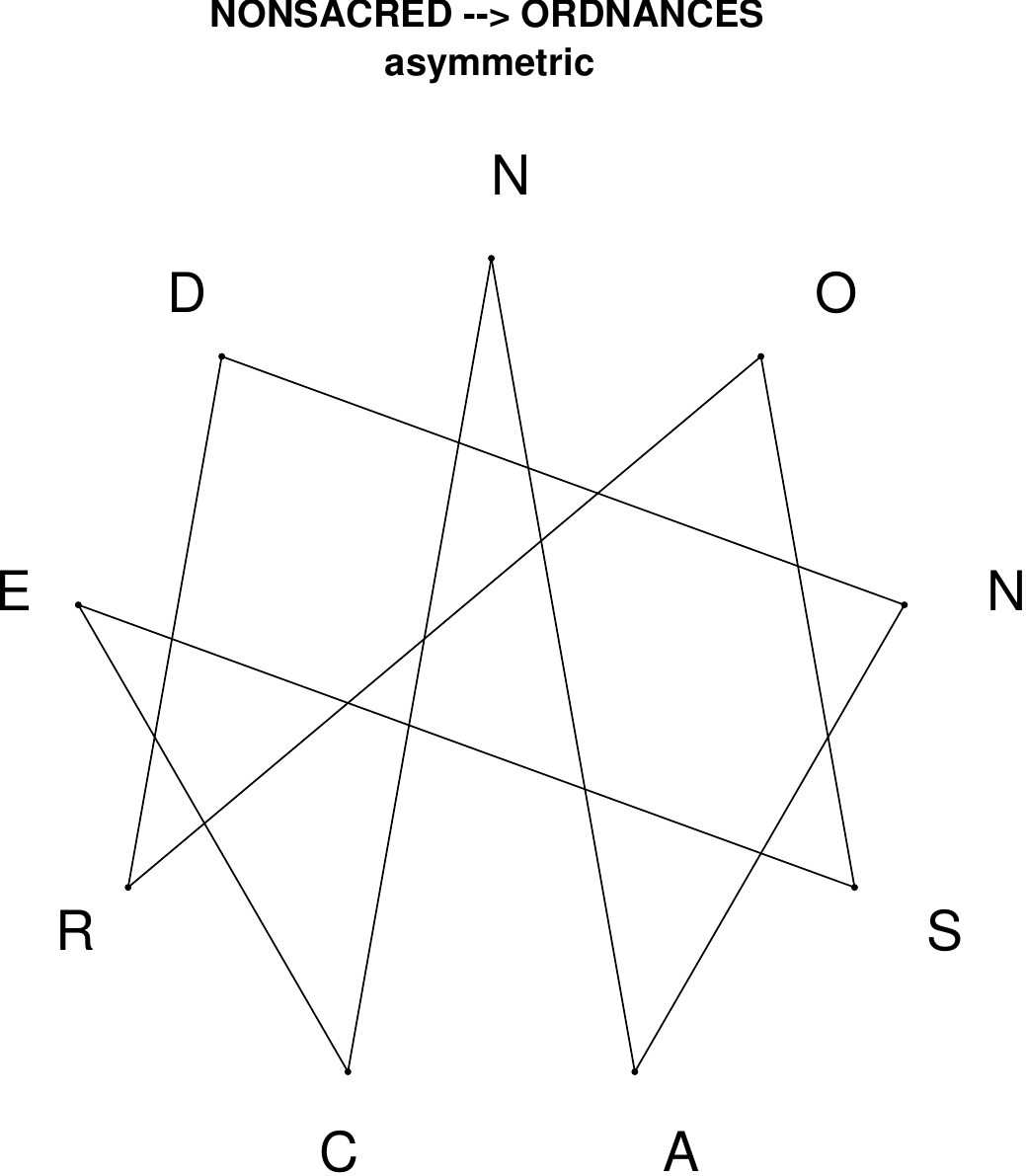}
\end{subfigure}
\end{figure}

\begin{figure}[H]
\centering
\begin{subfigure}[T]{0.19\textwidth}
\centering
\includegraphics[width=\textwidth]{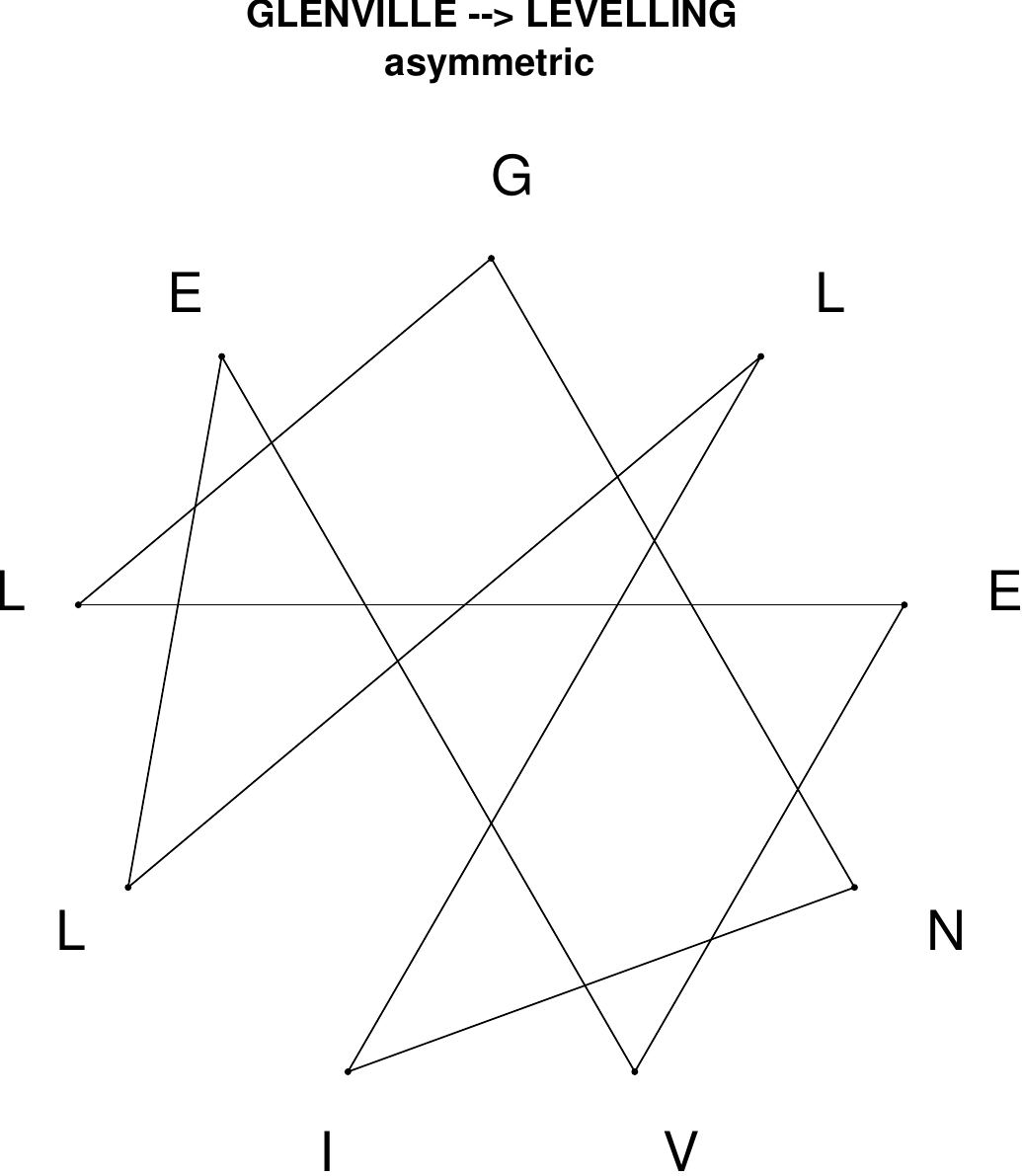}
\end{subfigure}
\hfill
\begin{subfigure}[T]{0.19\textwidth}
\centering
\includegraphics[width=\textwidth]{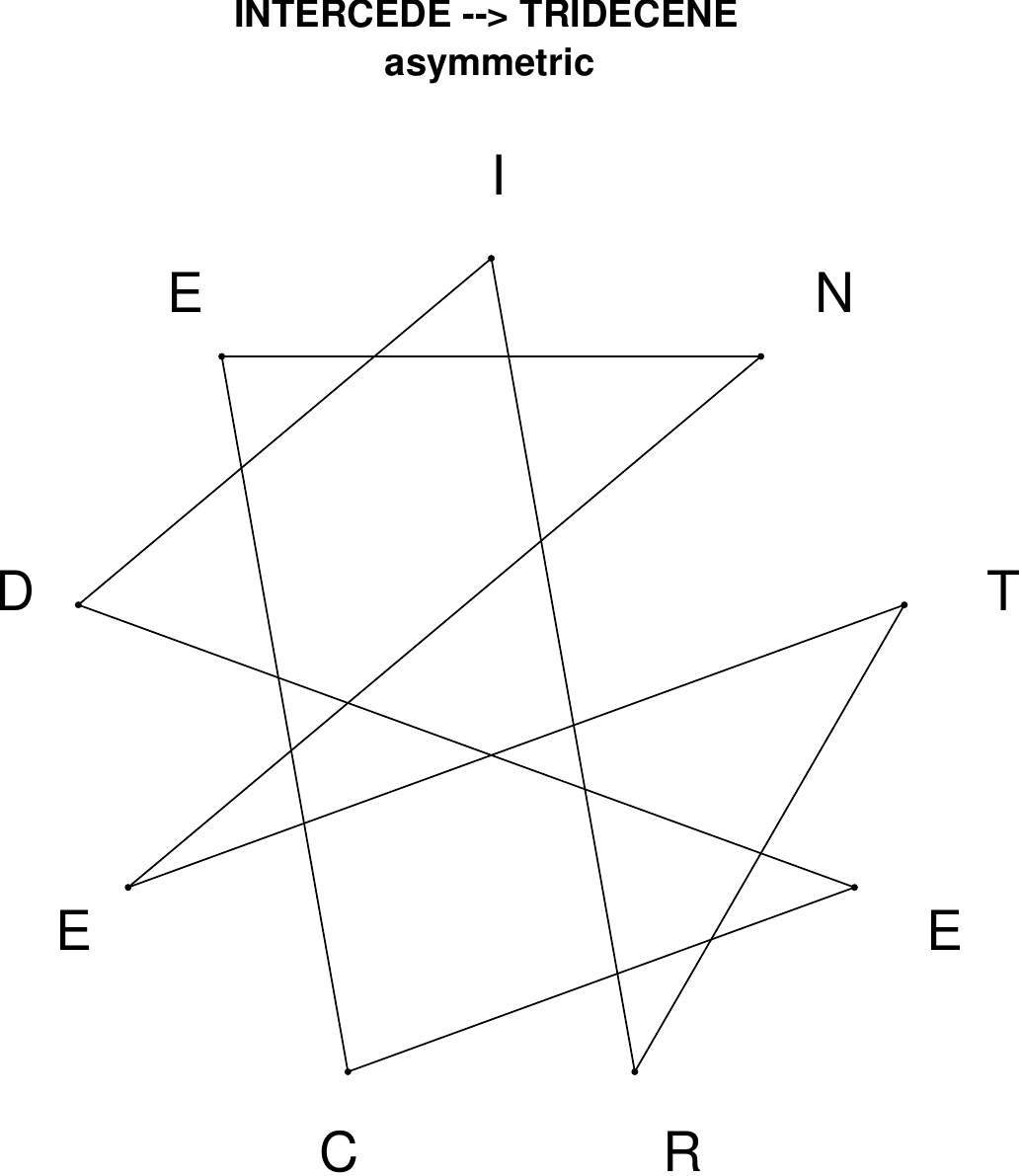}
\end{subfigure}
\hfill
\begin{subfigure}[T]{0.19\textwidth}
\centering
\includegraphics[width=\textwidth]{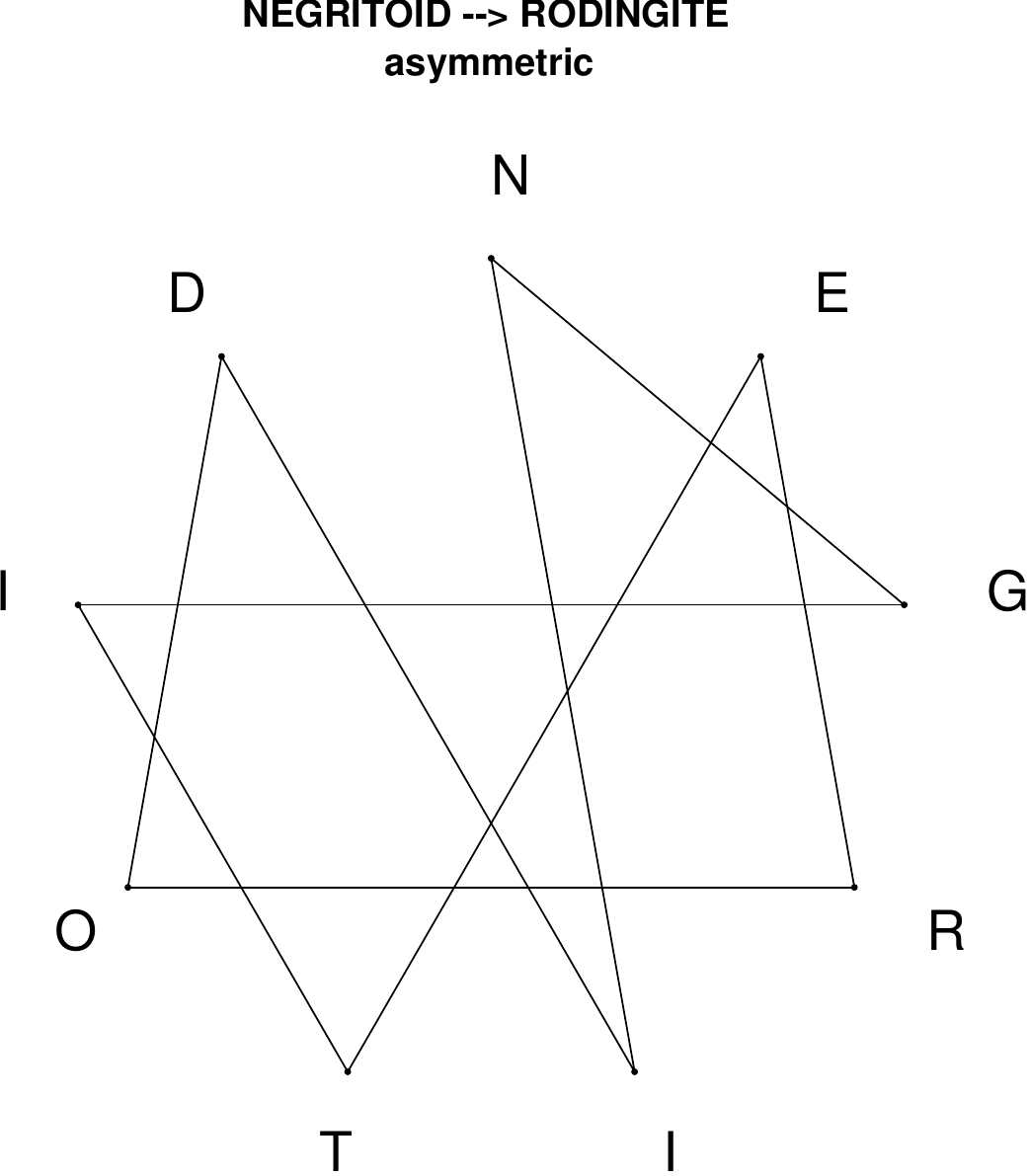}
\end{subfigure}
\hfill
\begin{subfigure}[T]{0.19\textwidth}
\centering
\includegraphics[width=\textwidth]{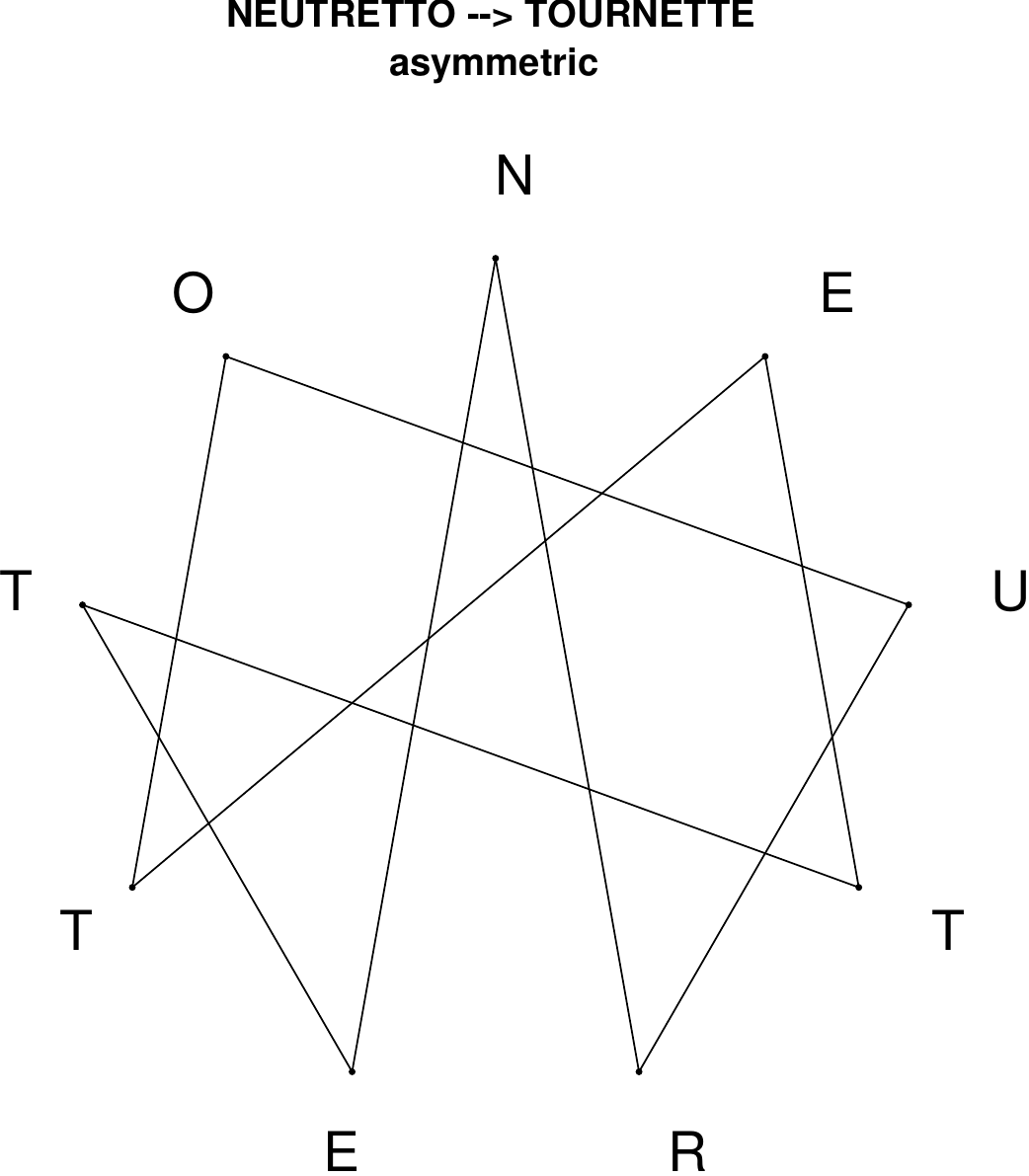}
\end{subfigure}
\hfill
\begin{subfigure}[T]{0.19\textwidth}
\centering
\includegraphics[width=\textwidth]{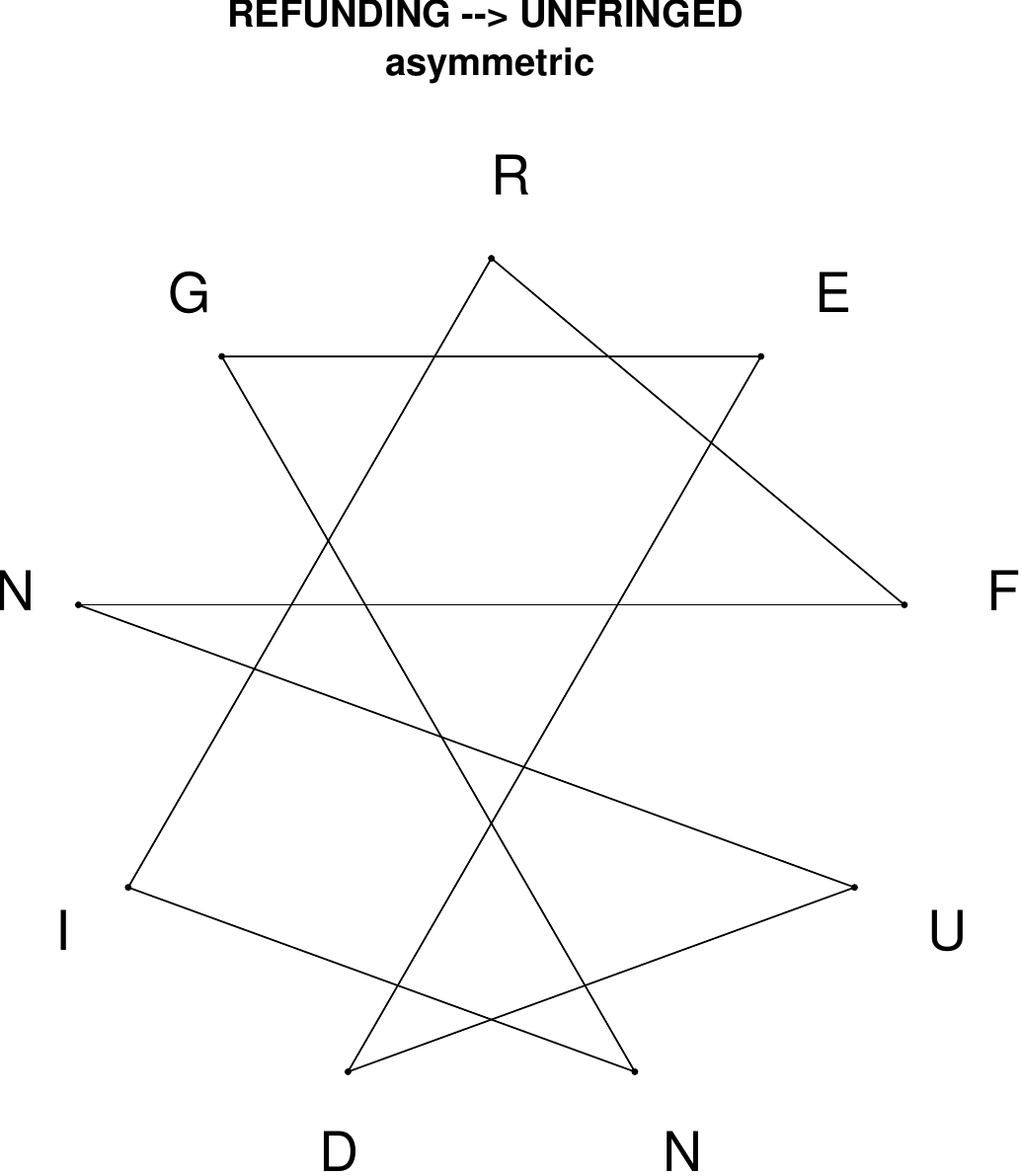}
\end{subfigure}
\end{figure}

\begin{figure}[H]
\centering
\begin{subfigure}[T]{0.19\textwidth}
\centering
\includegraphics[width=\textwidth]{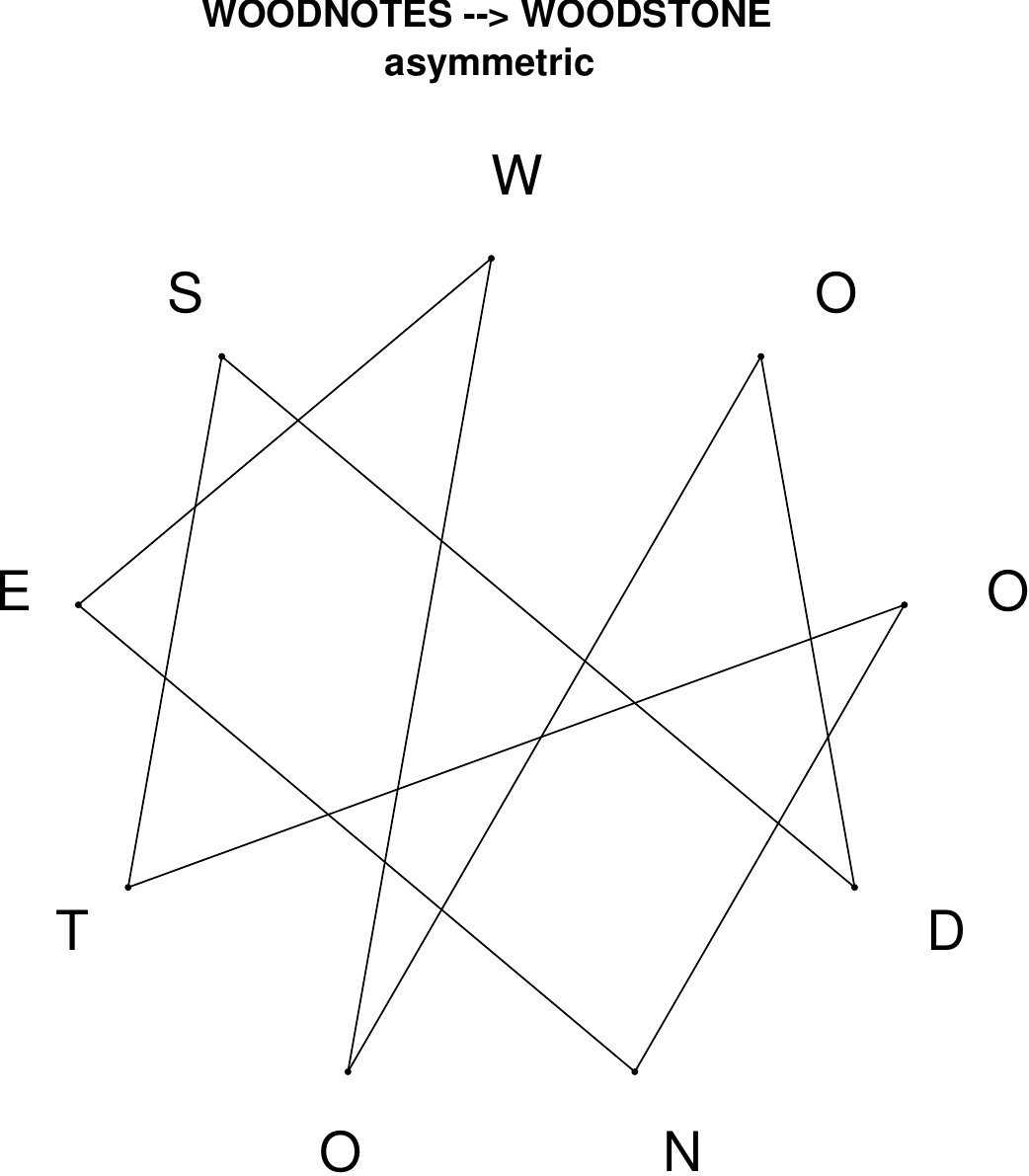}
\end{subfigure}
\hfill
\begin{subfigure}[T]{0.19\textwidth}
\centering
\includegraphics[width=\textwidth]{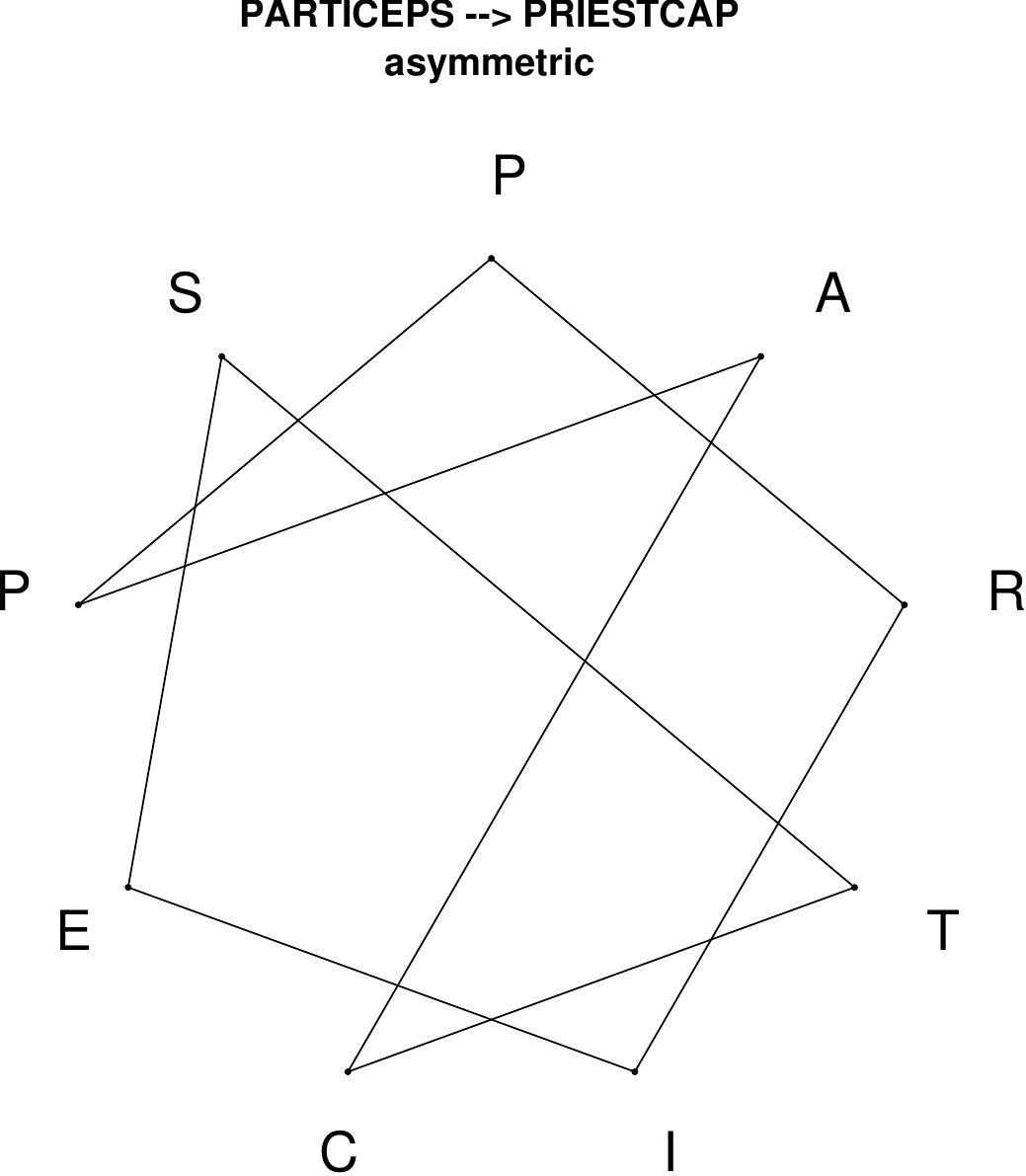}
\end{subfigure}
\hfill
\begin{subfigure}[T]{0.19\textwidth}
\centering
\includegraphics[width=\textwidth]{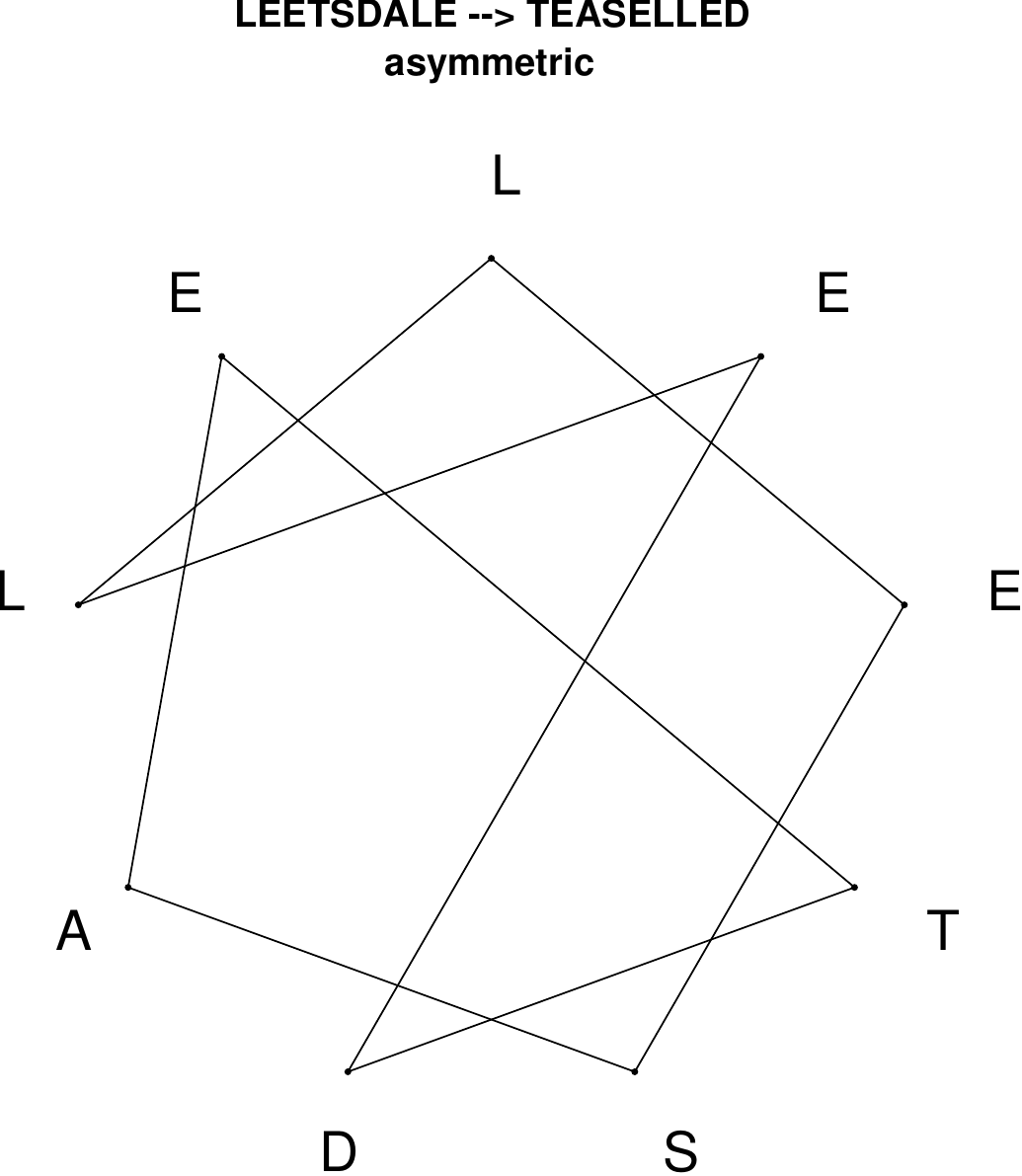}
\end{subfigure}
\hfill
\begin{subfigure}[T]{0.19\textwidth}
\centering
\includegraphics[width=\textwidth]{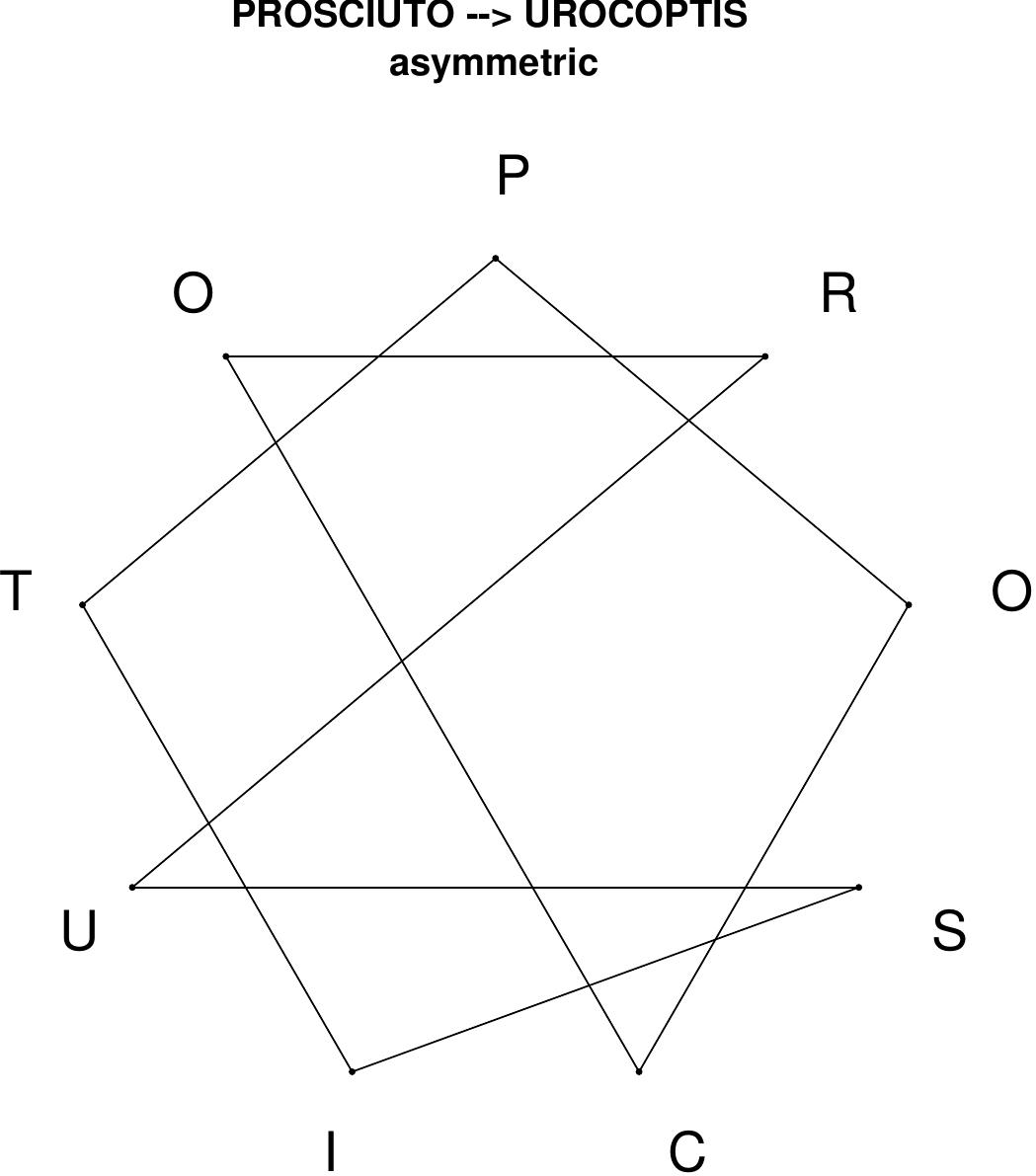}
\end{subfigure}
\hfill
\begin{subfigure}[T]{0.19\textwidth}
\centering
\includegraphics[width=\textwidth]{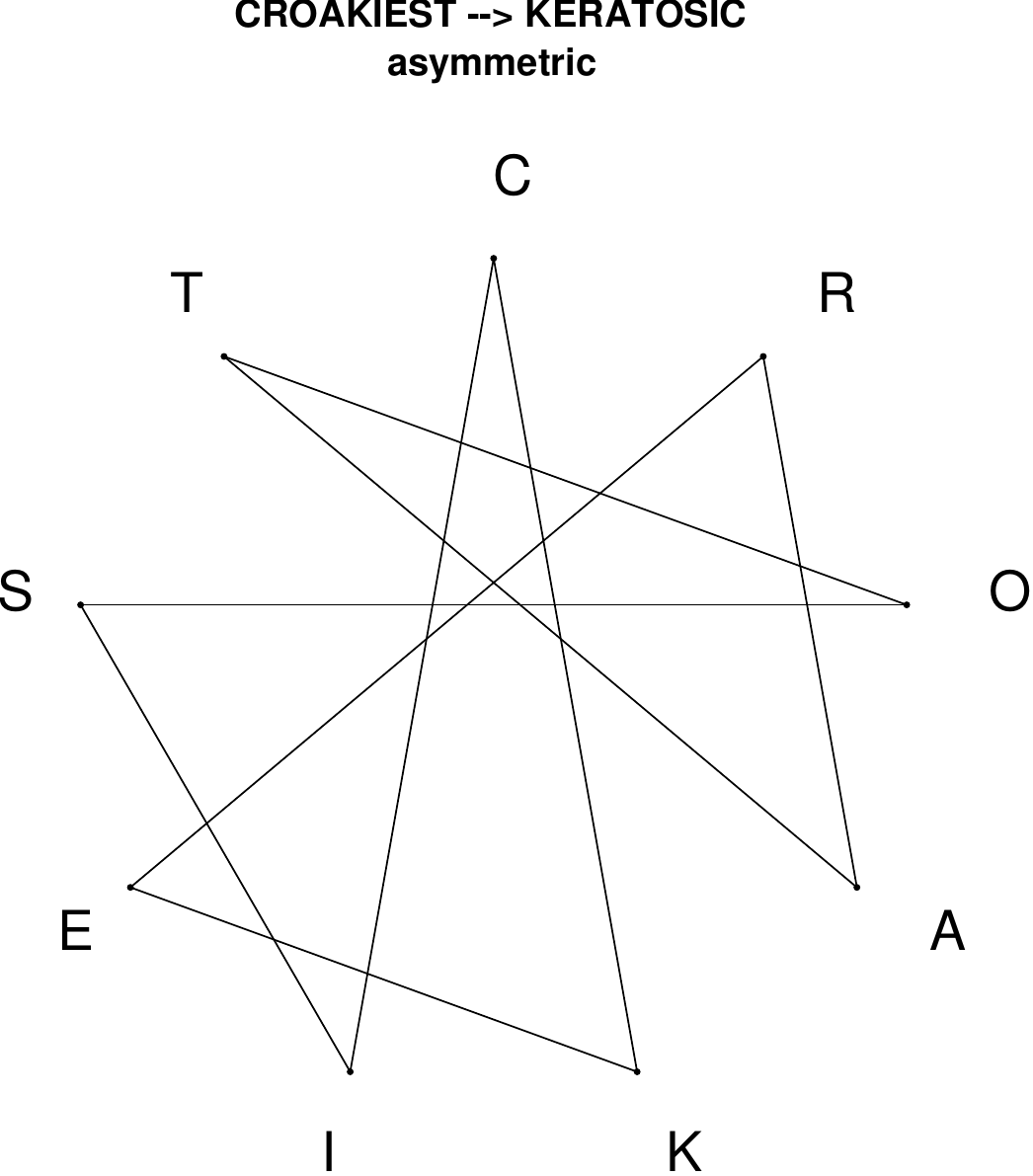}
\end{subfigure}
\end{figure}

\begin{figure}[H]
\centering
\begin{subfigure}[T]{0.19\textwidth}
\centering
\includegraphics[width=\textwidth]{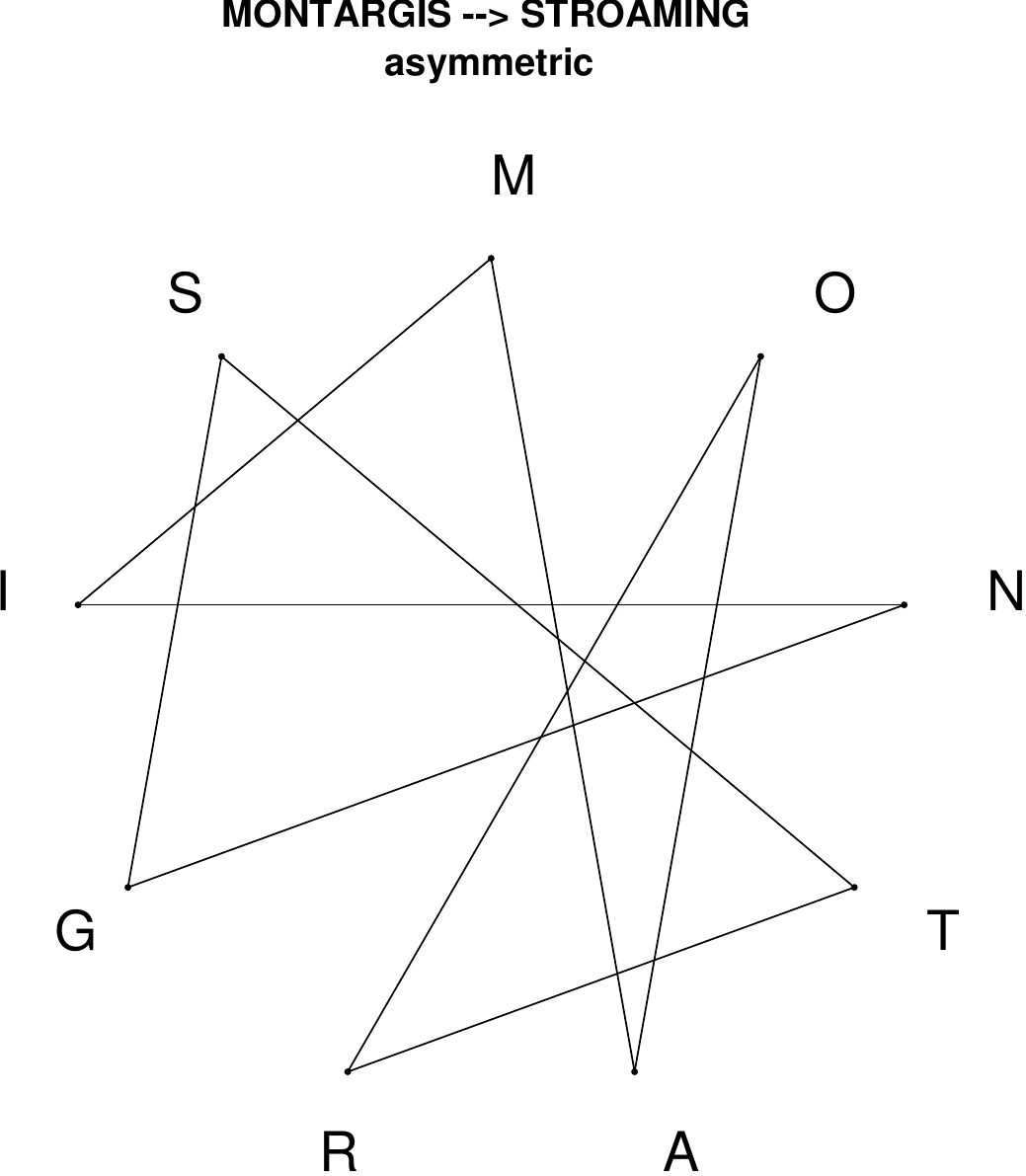}
\end{subfigure}
\hfill
\begin{subfigure}[T]{0.19\textwidth}
\centering
\includegraphics[width=\textwidth]{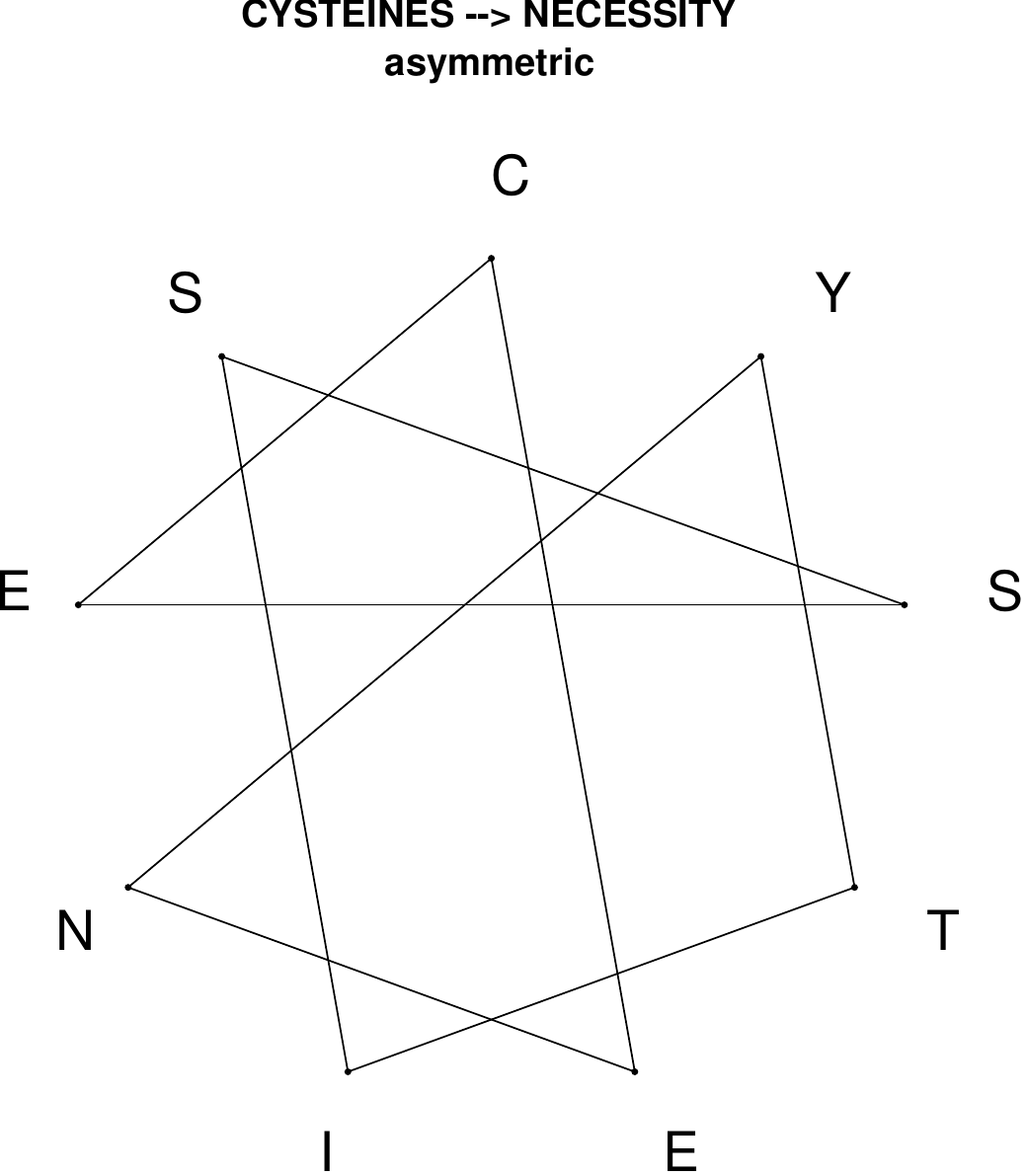}
\end{subfigure}
\hfill
\begin{subfigure}[T]{0.19\textwidth}
\centering
\includegraphics[width=\textwidth]{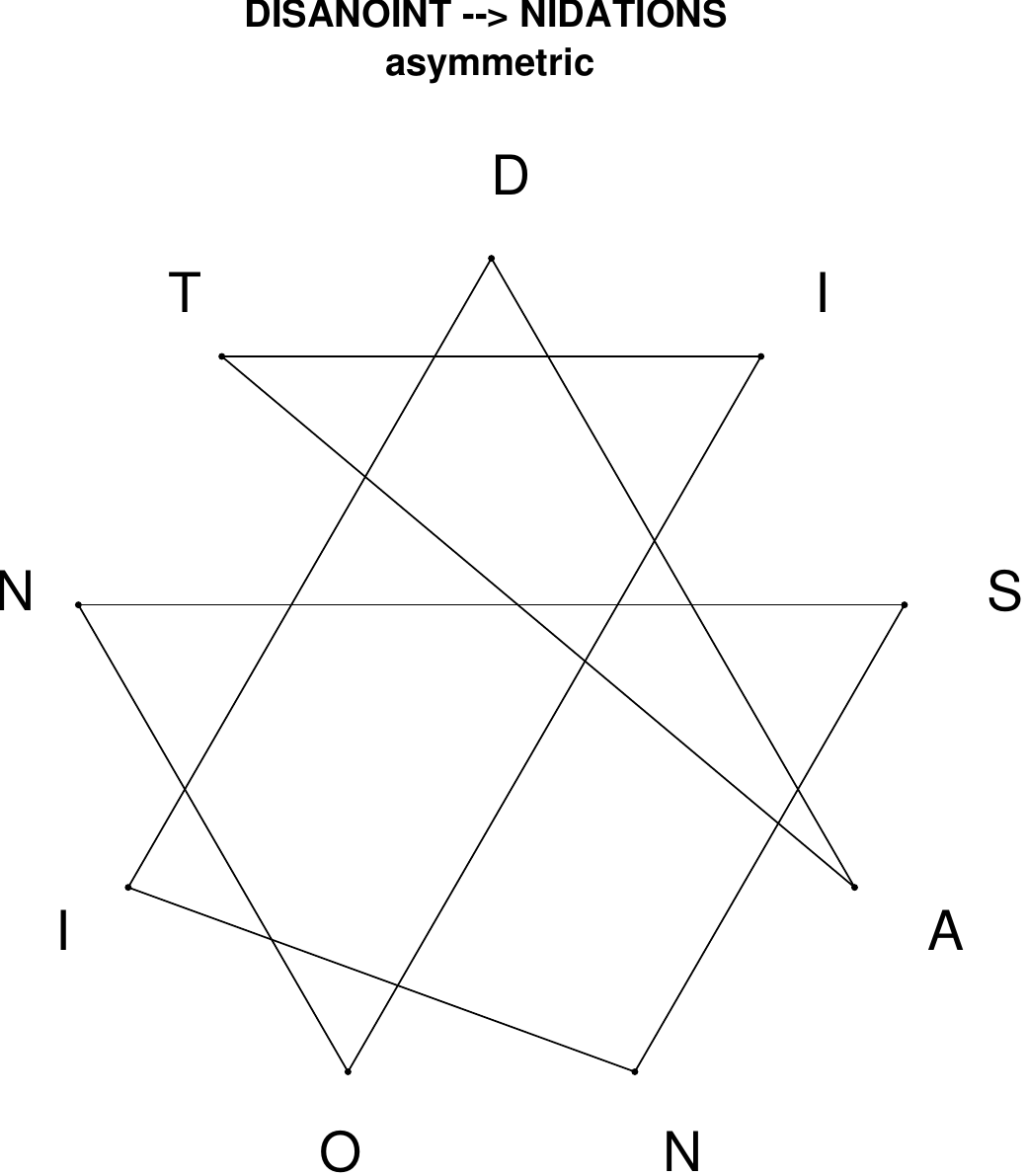}
\end{subfigure}
\hfill
\begin{subfigure}[T]{0.19\textwidth}
\centering
\includegraphics[width=\textwidth]{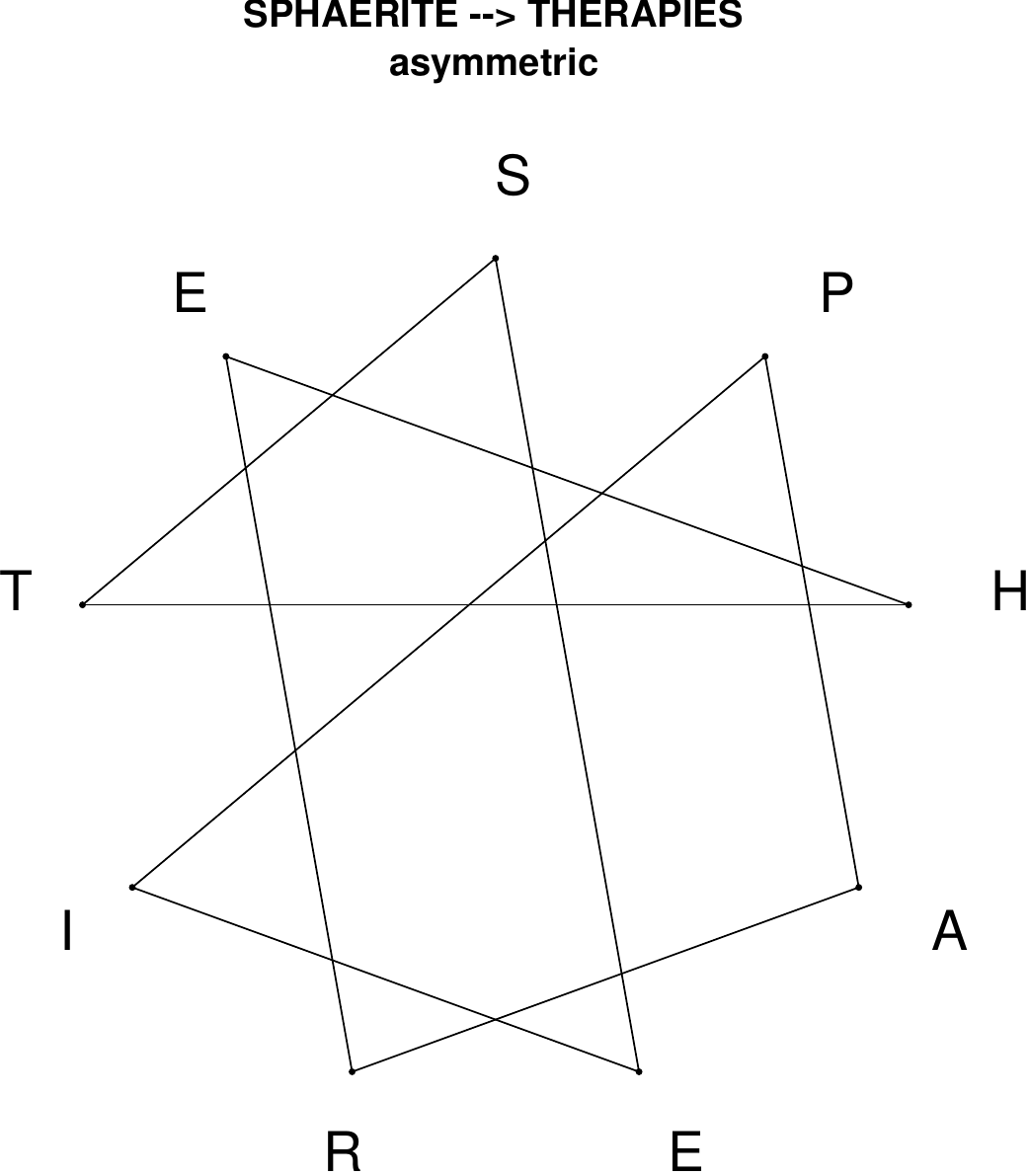}
\end{subfigure}
\hfill
\begin{subfigure}[T]{0.19\textwidth}
\centering
\includegraphics[width=\textwidth]{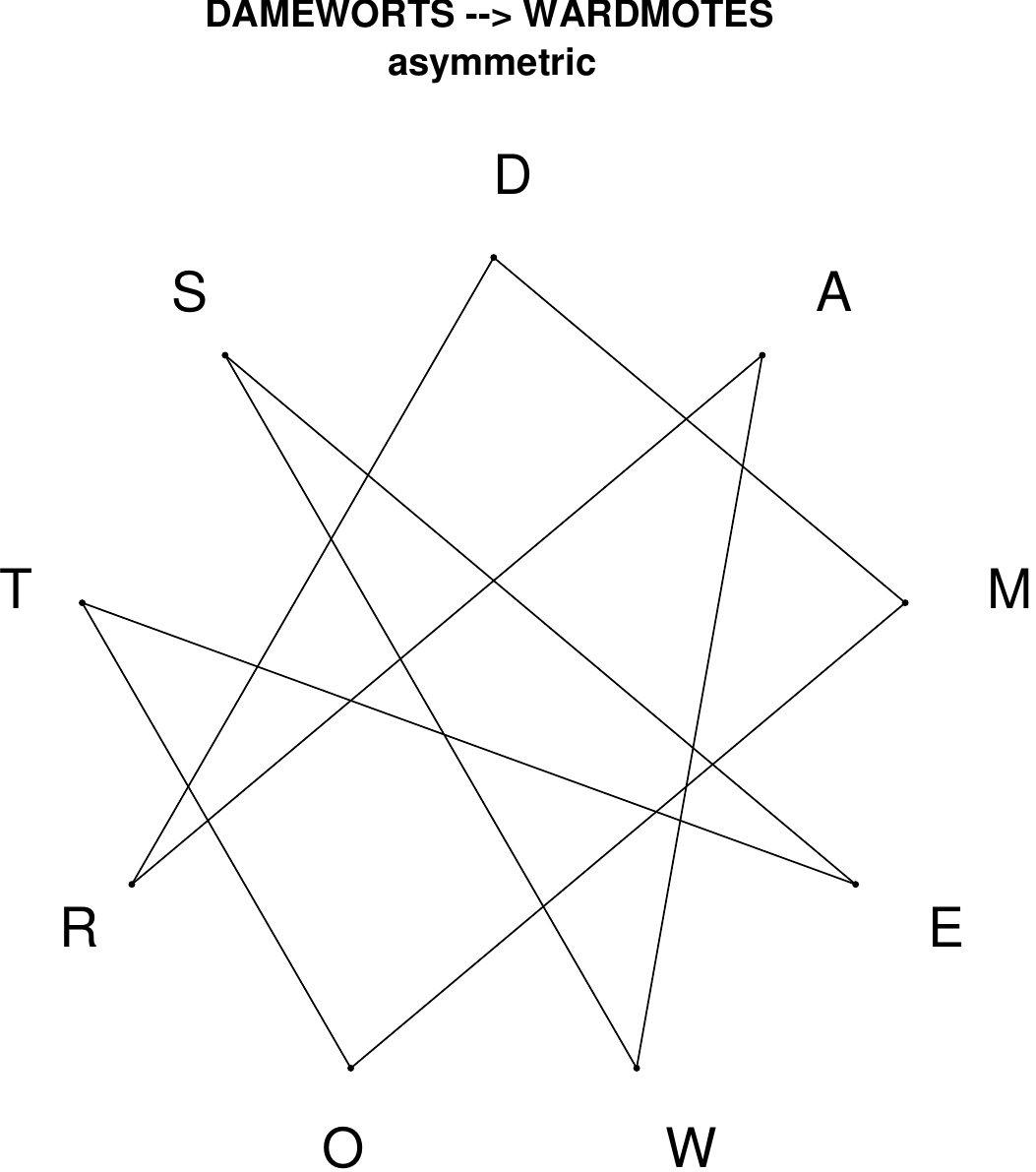}
\end{subfigure}
\end{figure}

\begin{figure}[H]
\centering
\begin{subfigure}[T]{0.19\textwidth}
\centering
\includegraphics[width=\textwidth]{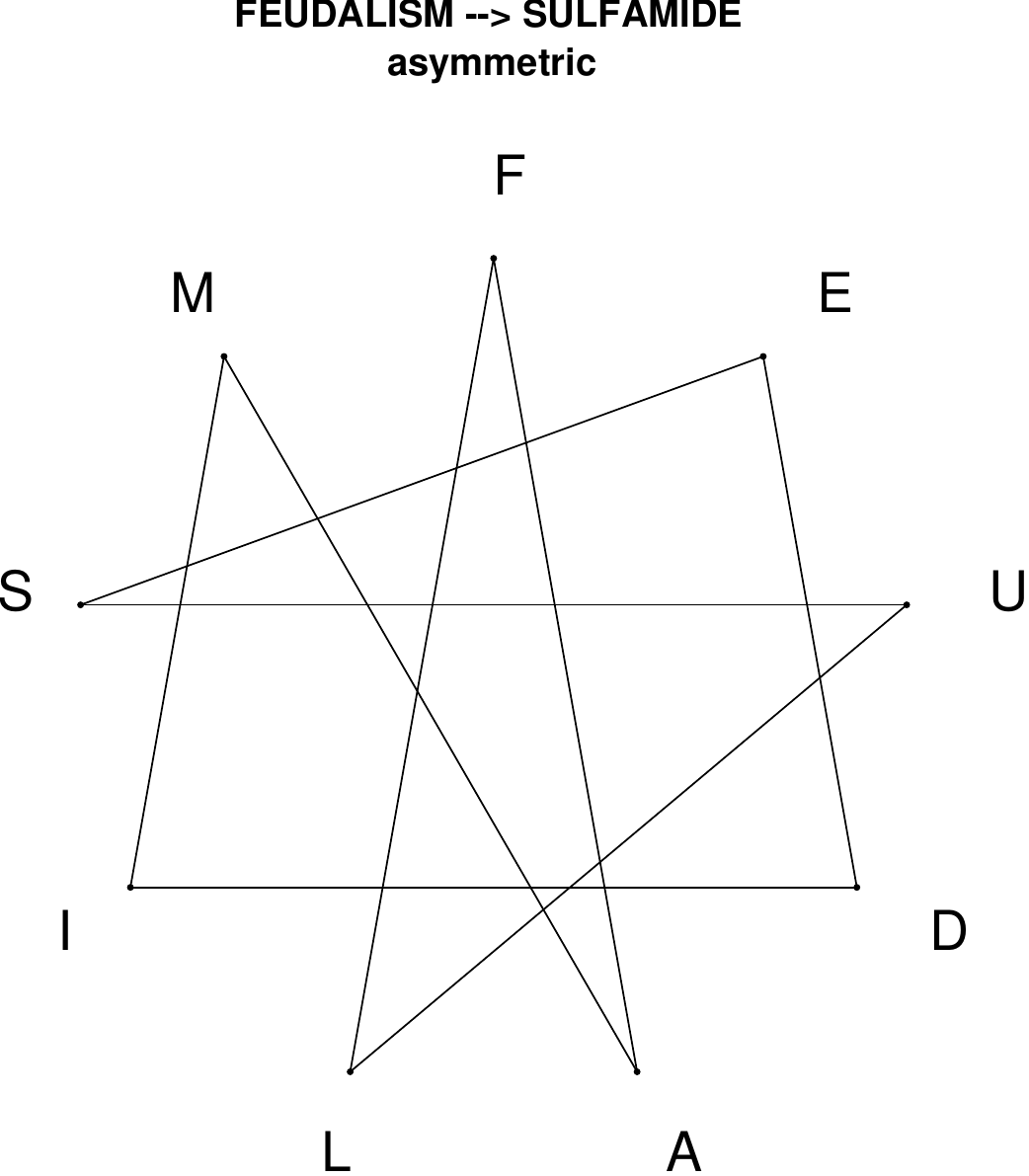}
\end{subfigure}
\hfill
\begin{subfigure}[T]{0.19\textwidth}
\centering
\includegraphics[width=\textwidth]{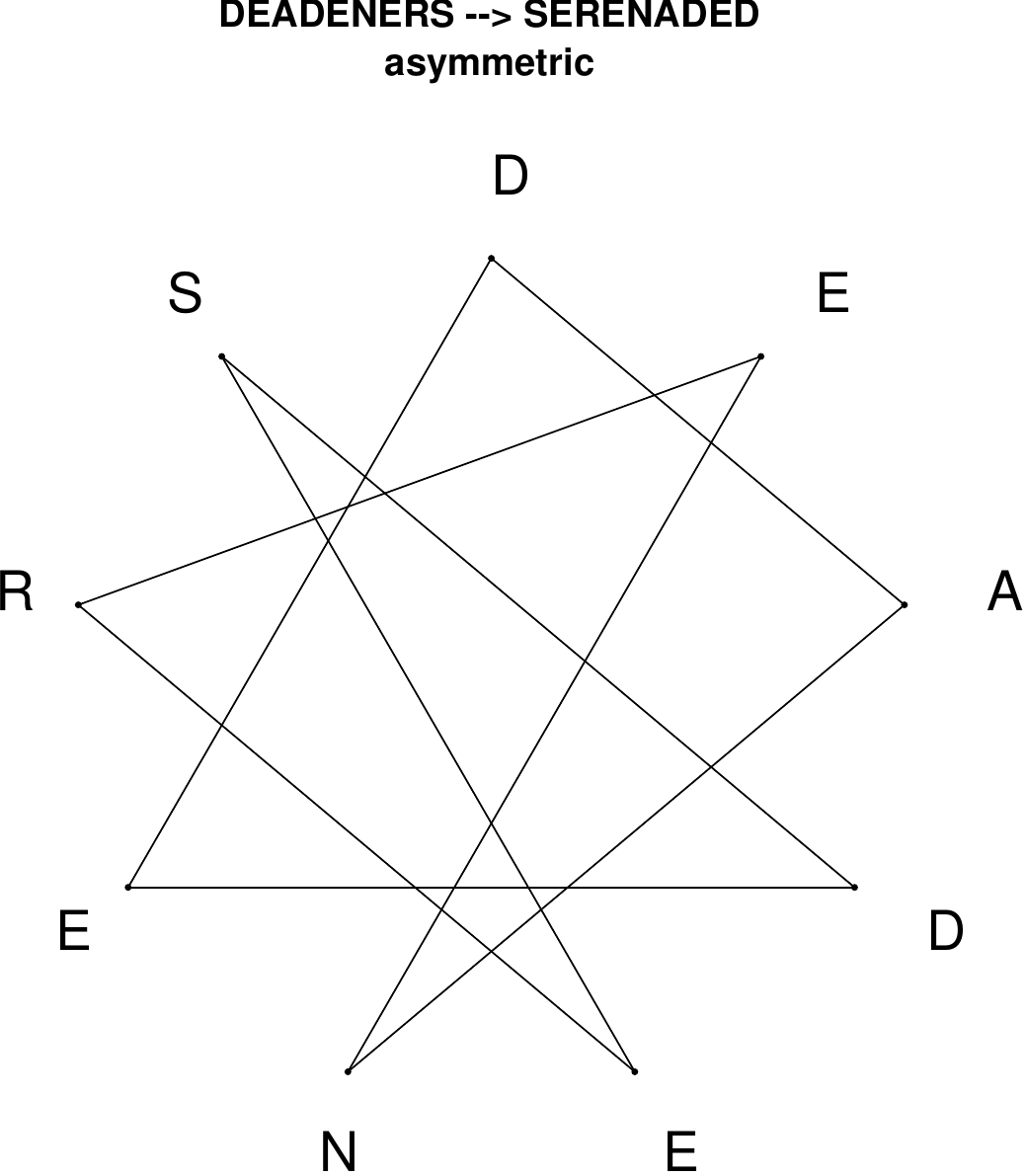}
\end{subfigure}
\hfill
\begin{subfigure}[T]{0.19\textwidth}
\centering
\includegraphics[width=\textwidth]{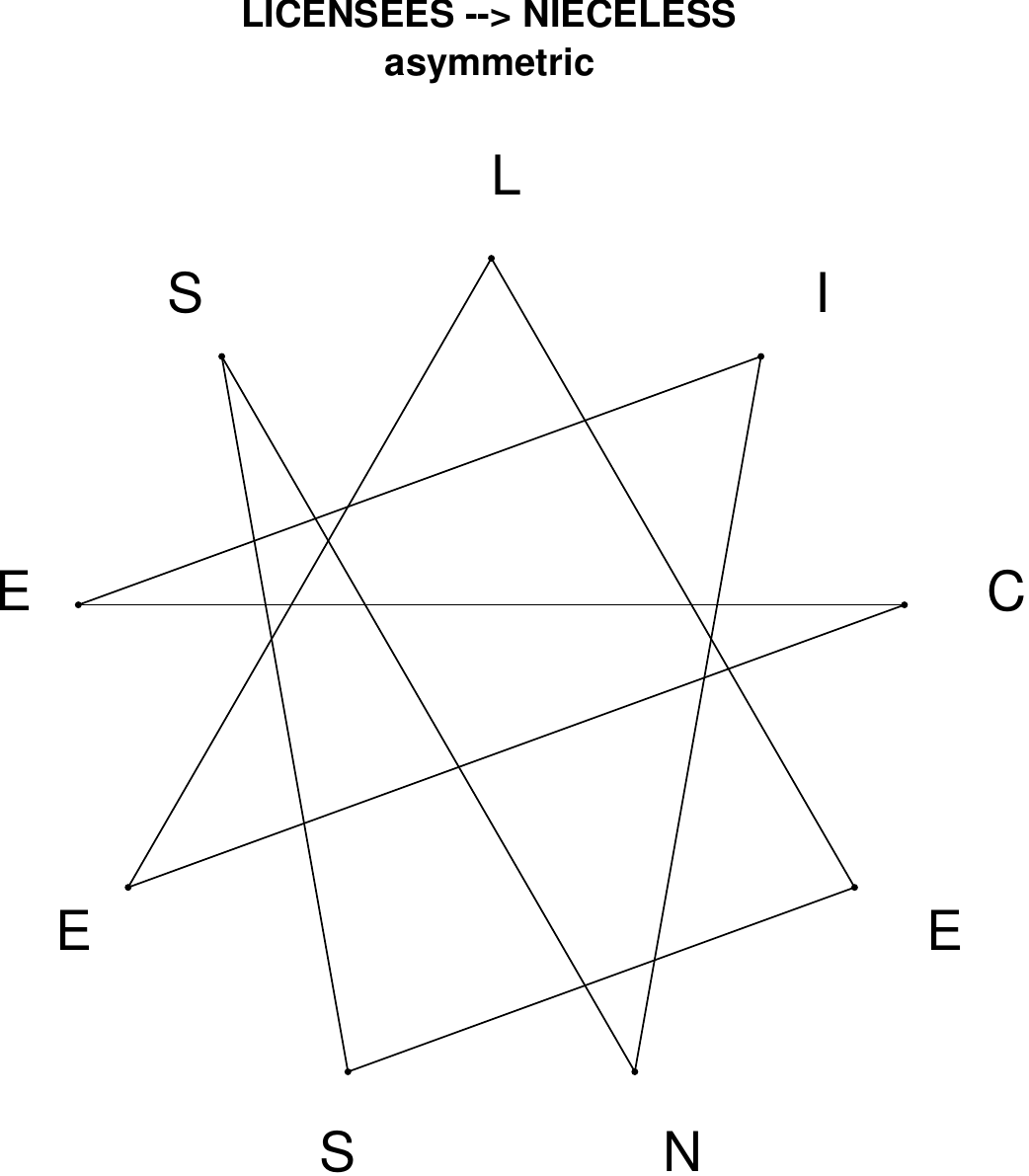}
\end{subfigure}
\hfill
\begin{subfigure}[T]{0.19\textwidth}
\centering
\includegraphics[width=\textwidth]{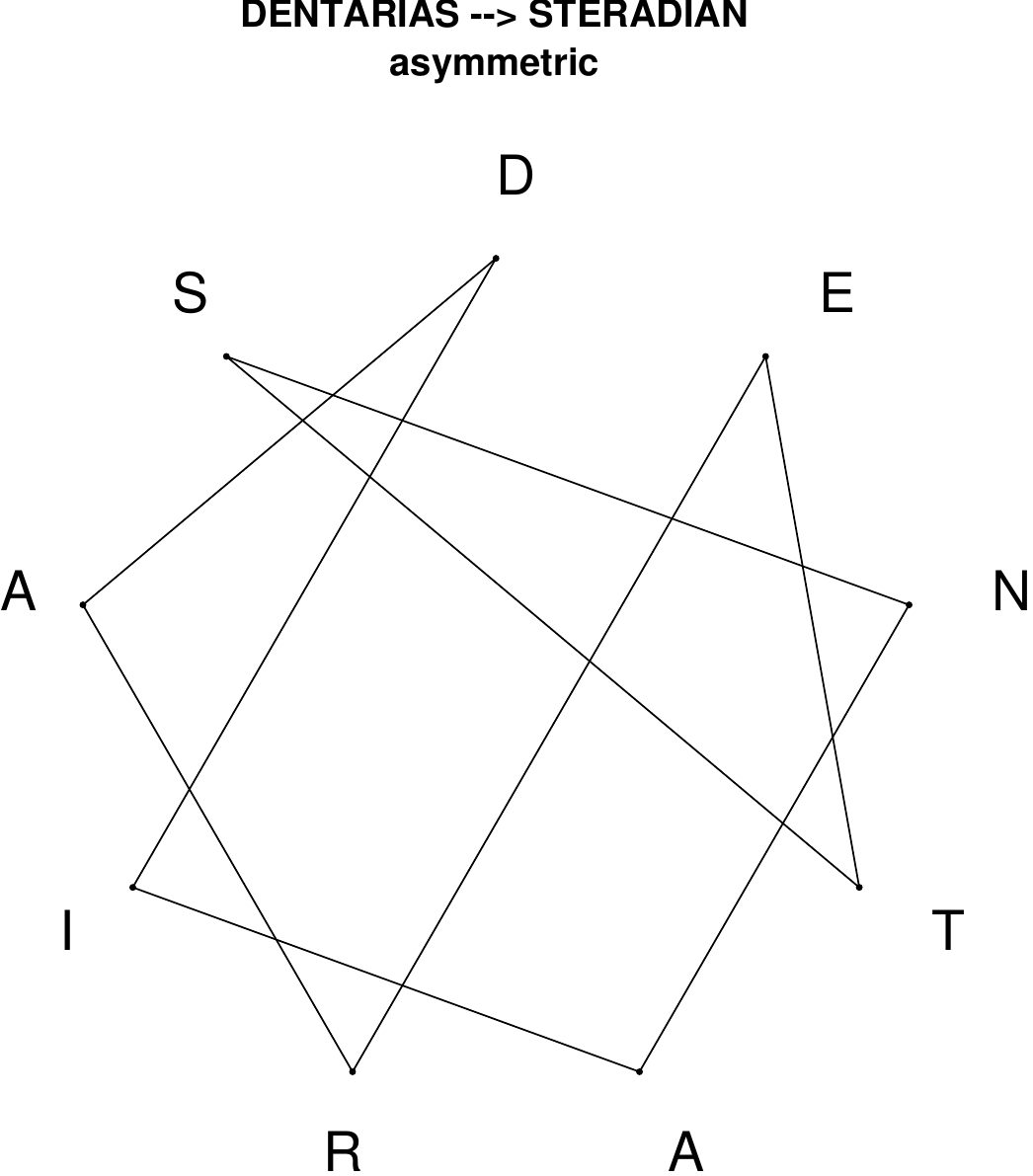}
\end{subfigure}
\hfill
\begin{subfigure}[T]{0.19\textwidth}
\centering
\includegraphics[width=\textwidth]{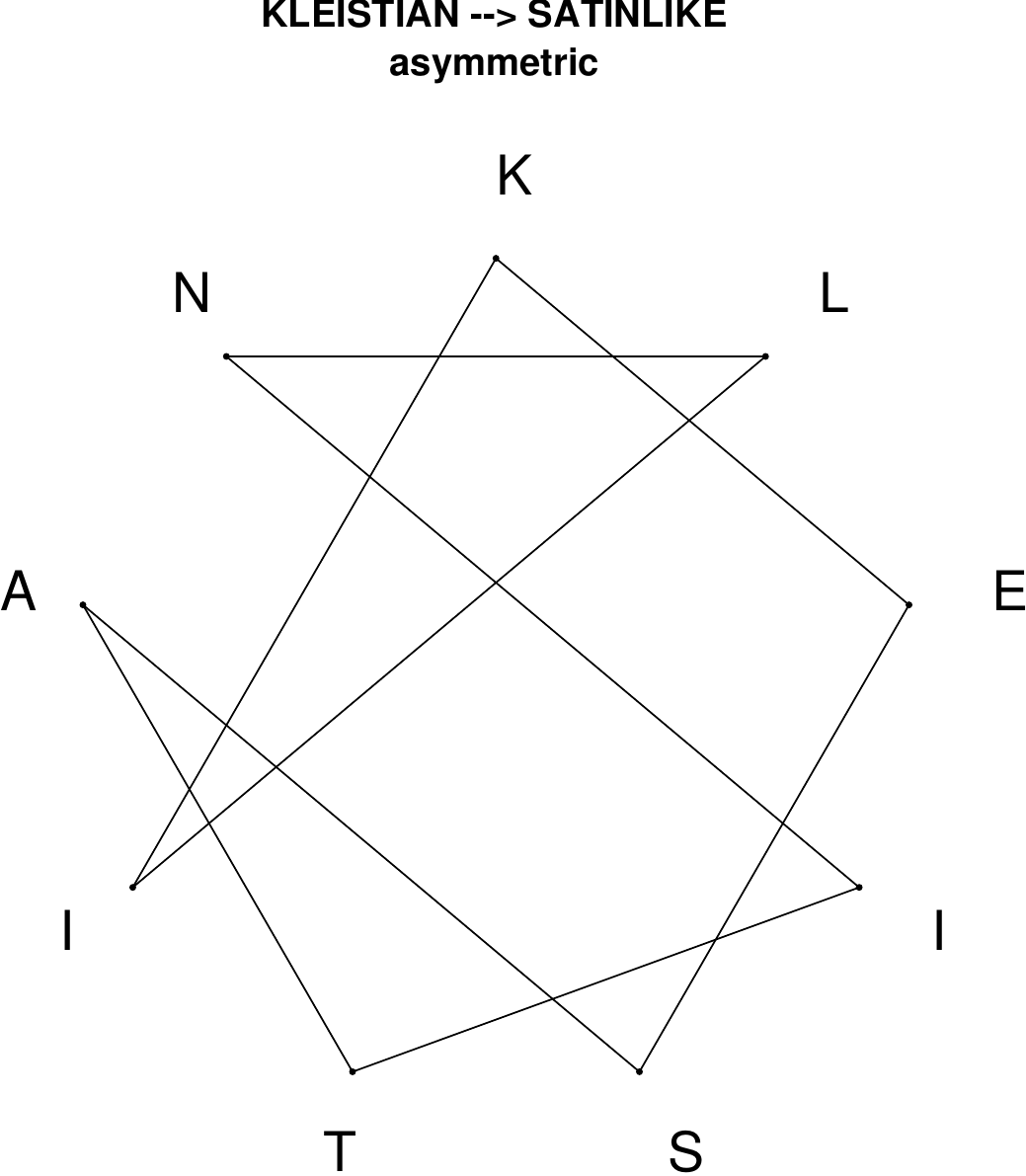}
\end{subfigure}
\end{figure}

\begin{figure}[H]
\centering
\begin{subfigure}[T]{0.19\textwidth}
\centering
\includegraphics[width=\textwidth]{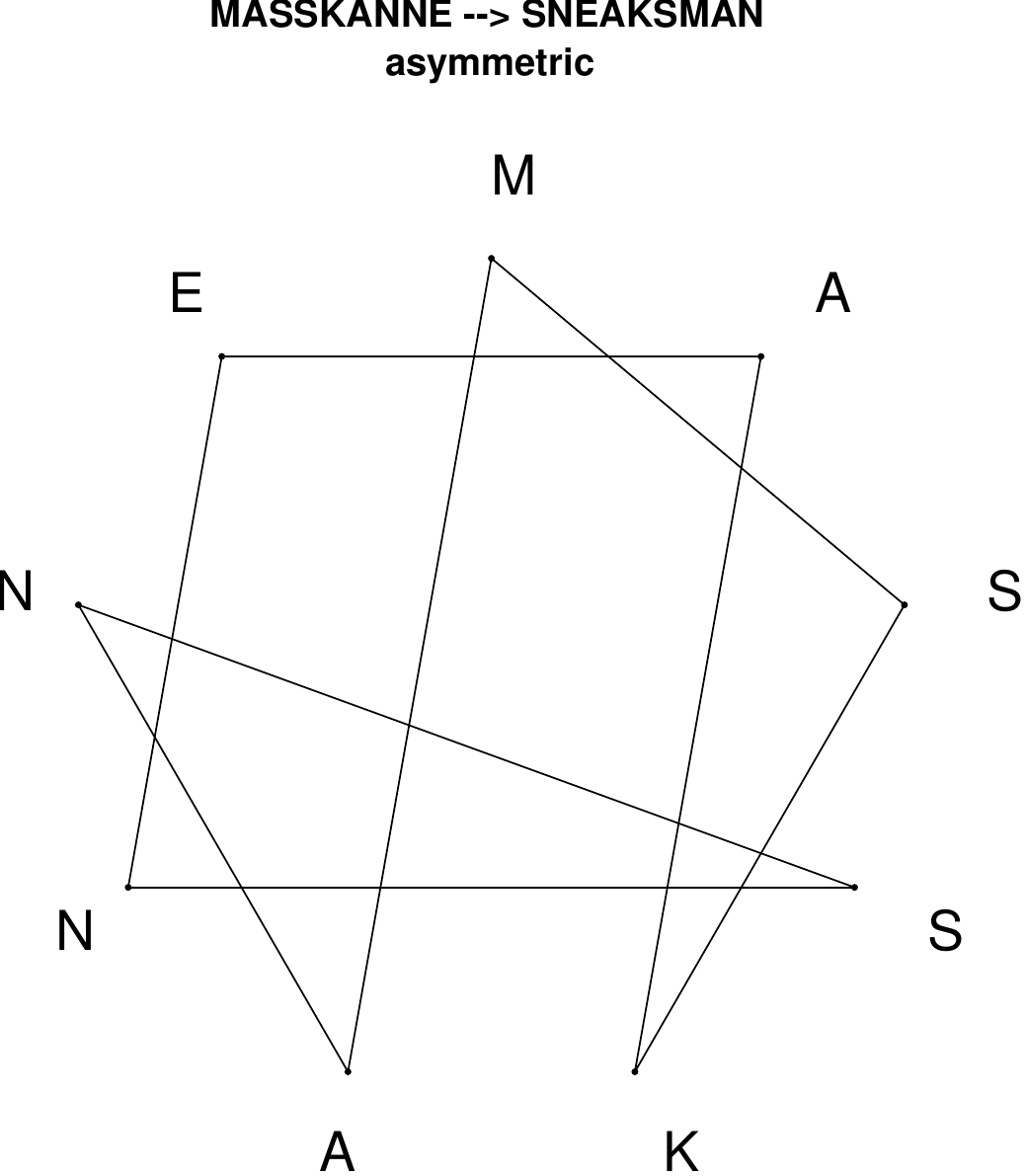}
\end{subfigure}
\hfill
\begin{subfigure}[T]{0.19\textwidth}
\centering
\includegraphics[width=\textwidth]{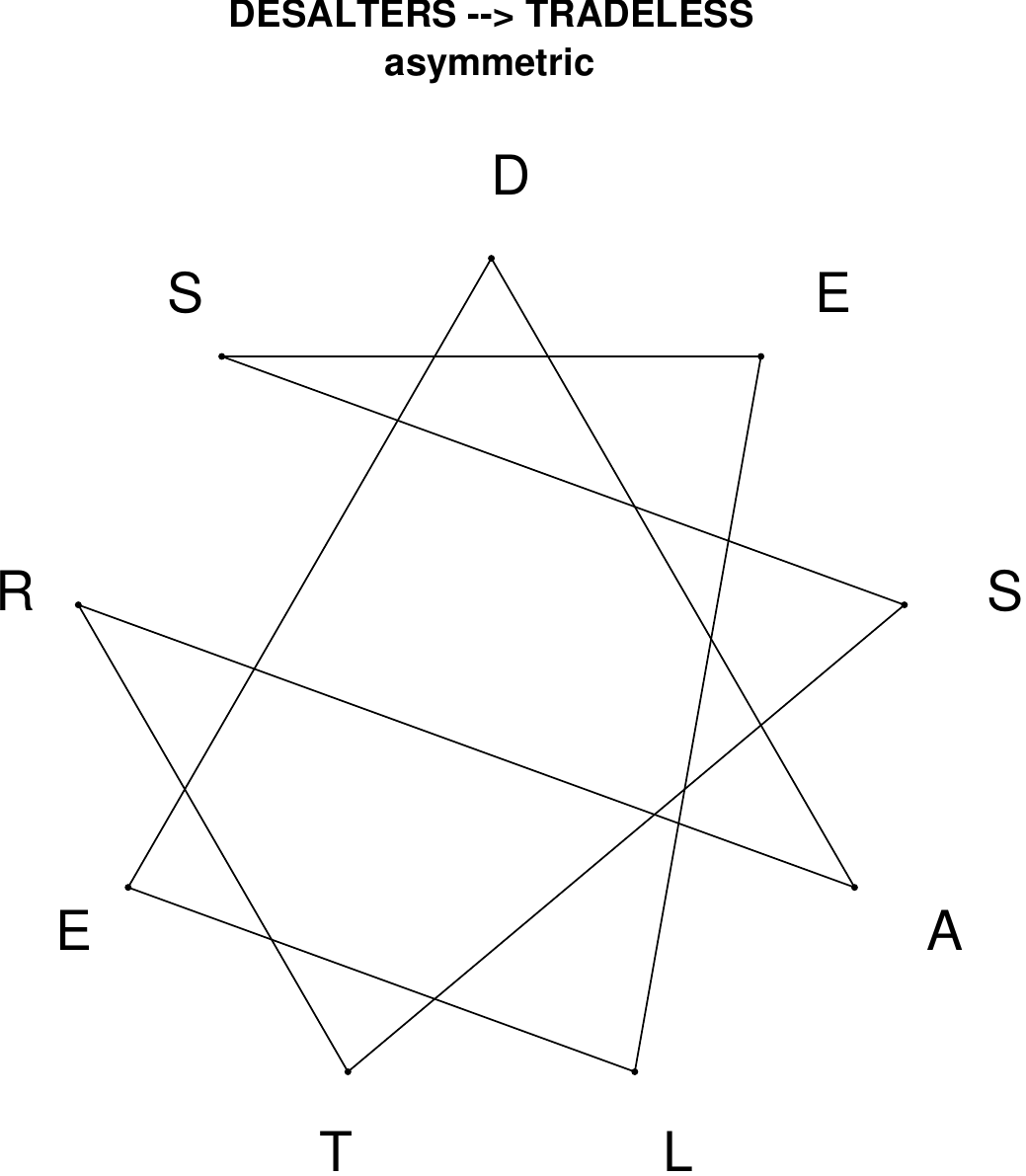}
\end{subfigure}
\hfill
\begin{subfigure}[T]{0.19\textwidth}
\centering
\includegraphics[width=\textwidth]{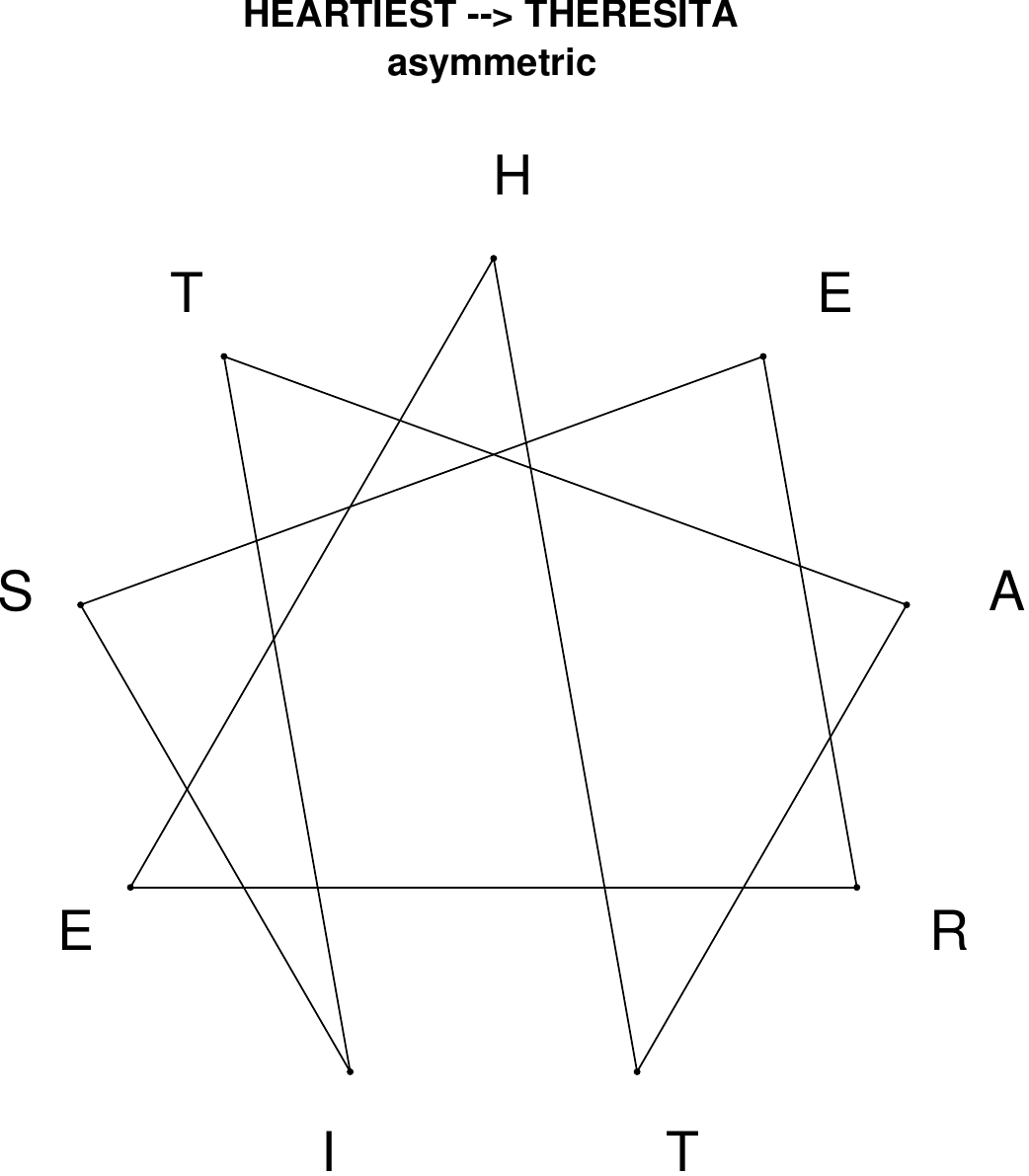}
\end{subfigure}
\hfill
\begin{subfigure}[T]{0.19\textwidth}
\centering
\includegraphics[width=\textwidth]{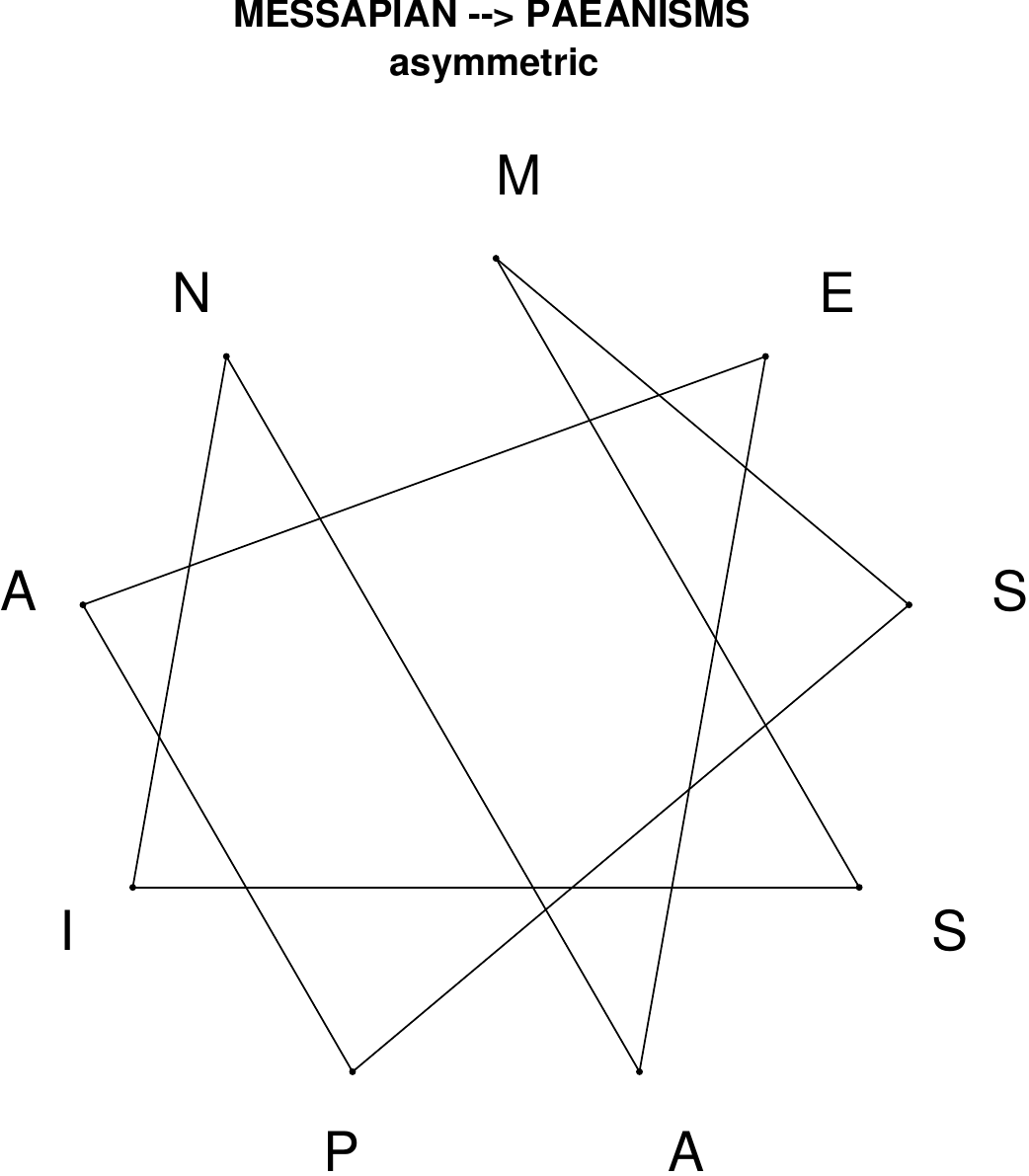}
\end{subfigure}
\hfill
\begin{subfigure}[T]{0.19\textwidth}
\centering
\includegraphics[width=\textwidth]{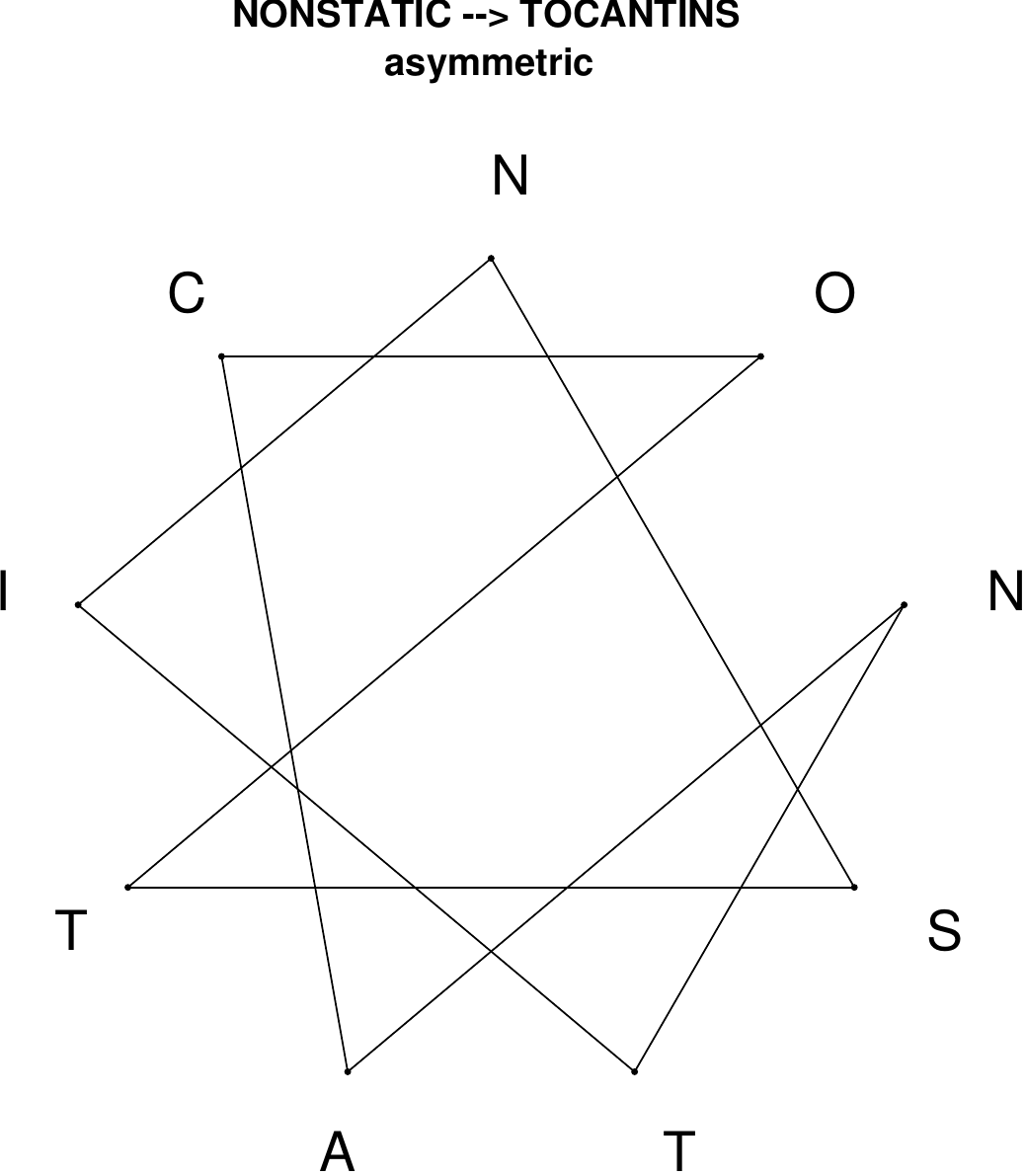}
\end{subfigure}
\end{figure}

\begin{figure}[H]
\centering
\begin{subfigure}[T]{0.19\textwidth}
\centering
\includegraphics[width=\textwidth]{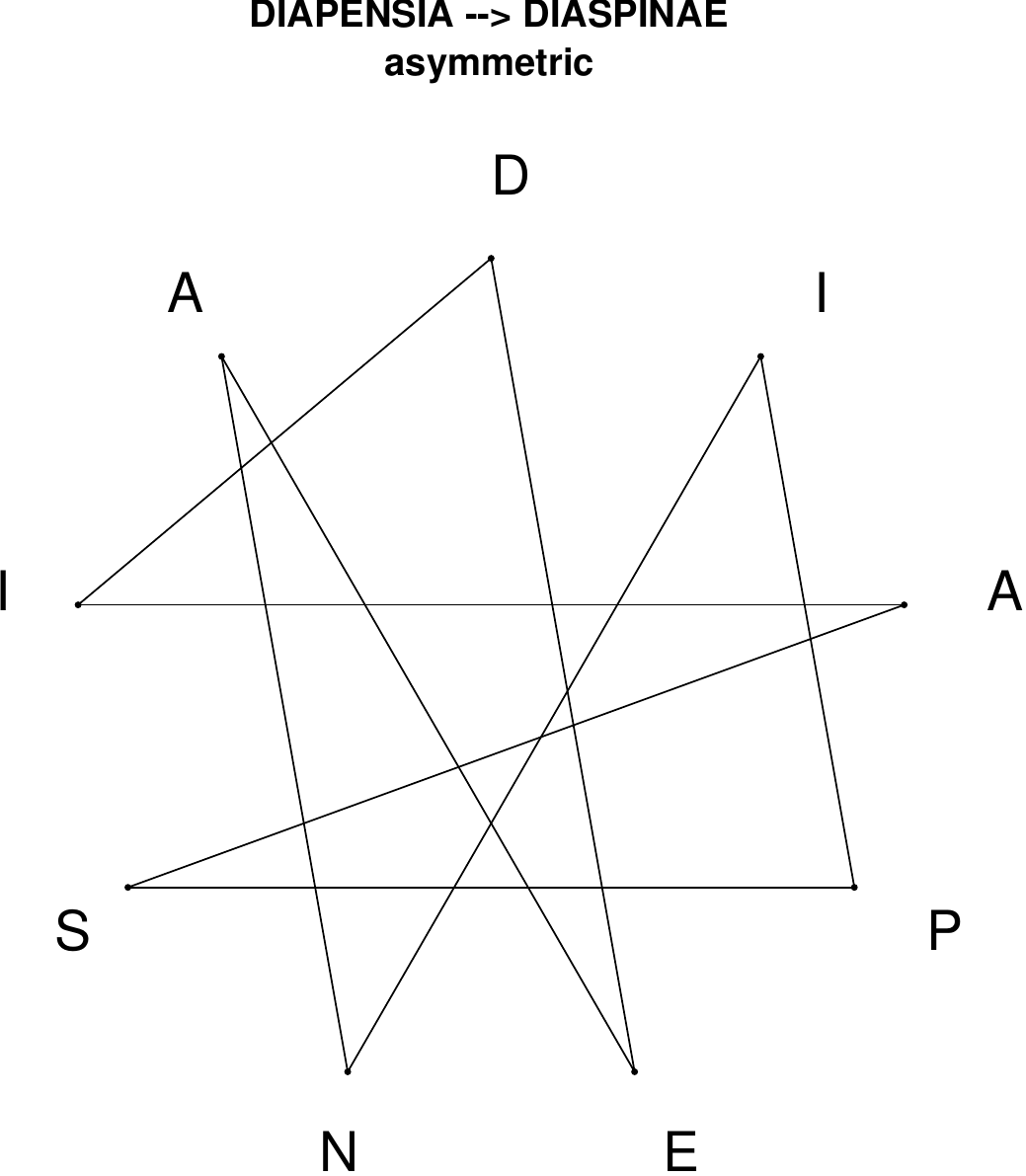}
\end{subfigure}
\hfill
\begin{subfigure}[T]{0.19\textwidth}
\centering
\includegraphics[width=\textwidth]{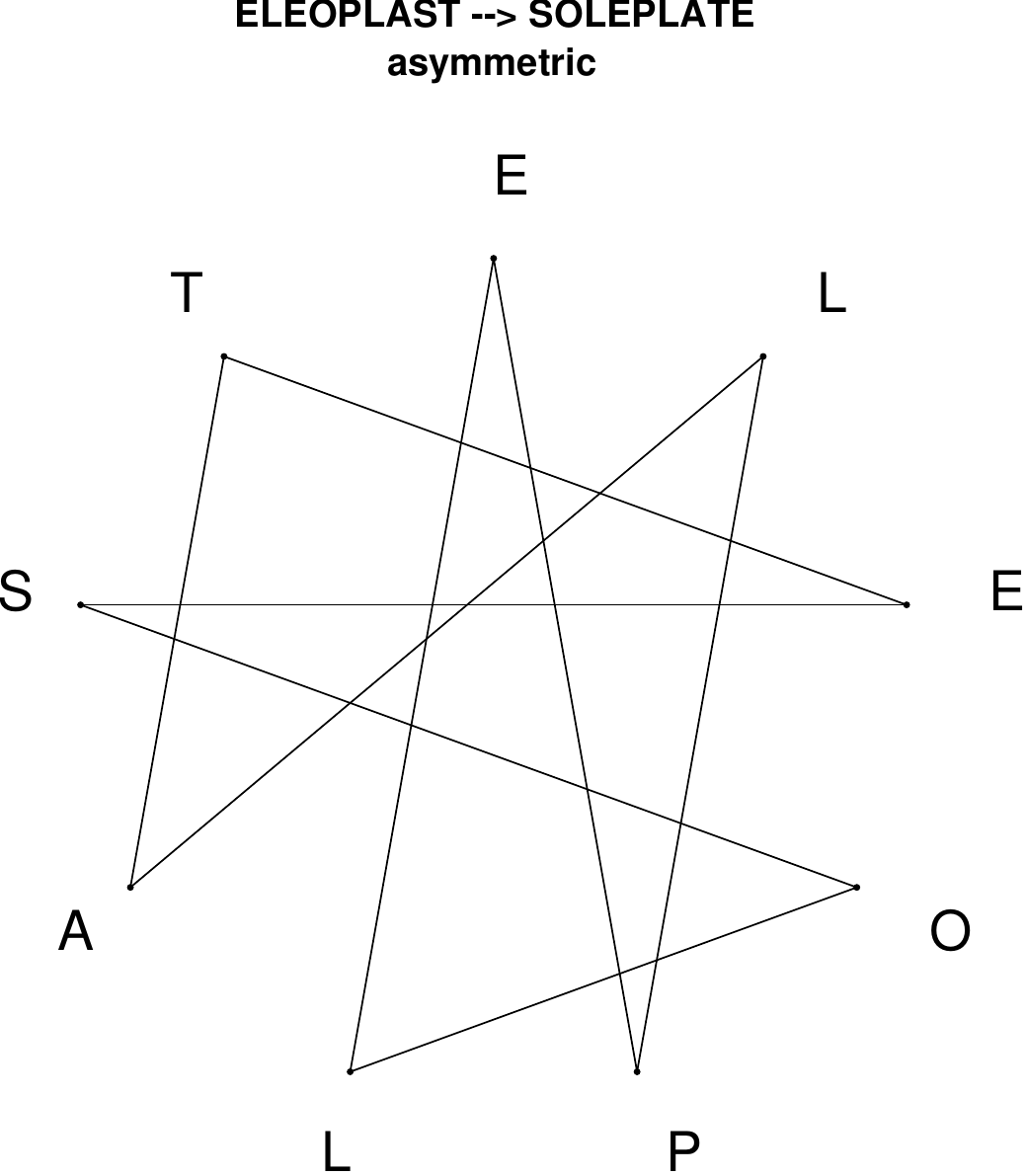}
\end{subfigure}
\hfill
\begin{subfigure}[T]{0.19\textwidth}
\centering
\includegraphics[width=\textwidth]{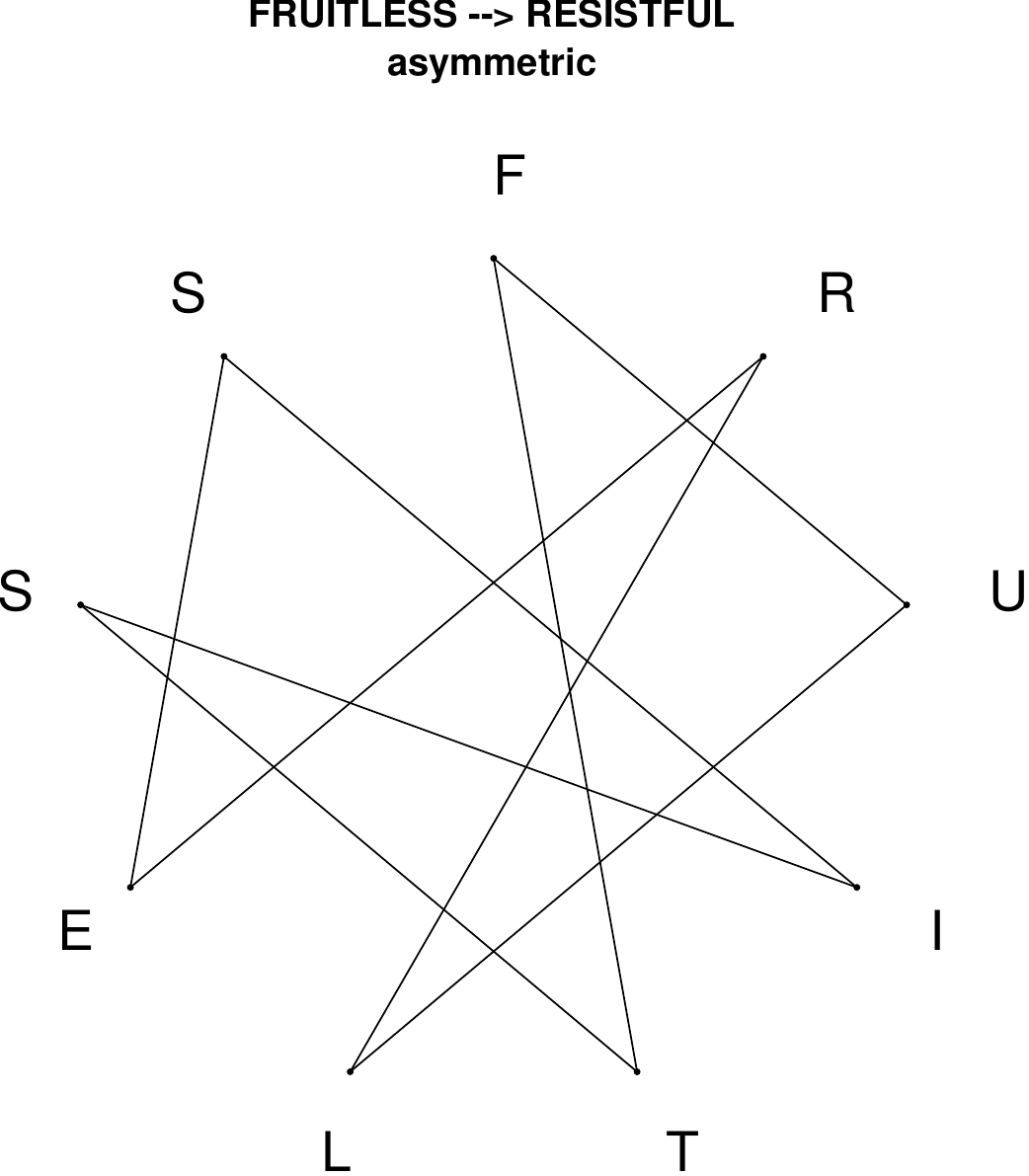}
\end{subfigure}
\hfill
\begin{subfigure}[T]{0.19\textwidth}
\centering
\includegraphics[width=\textwidth]{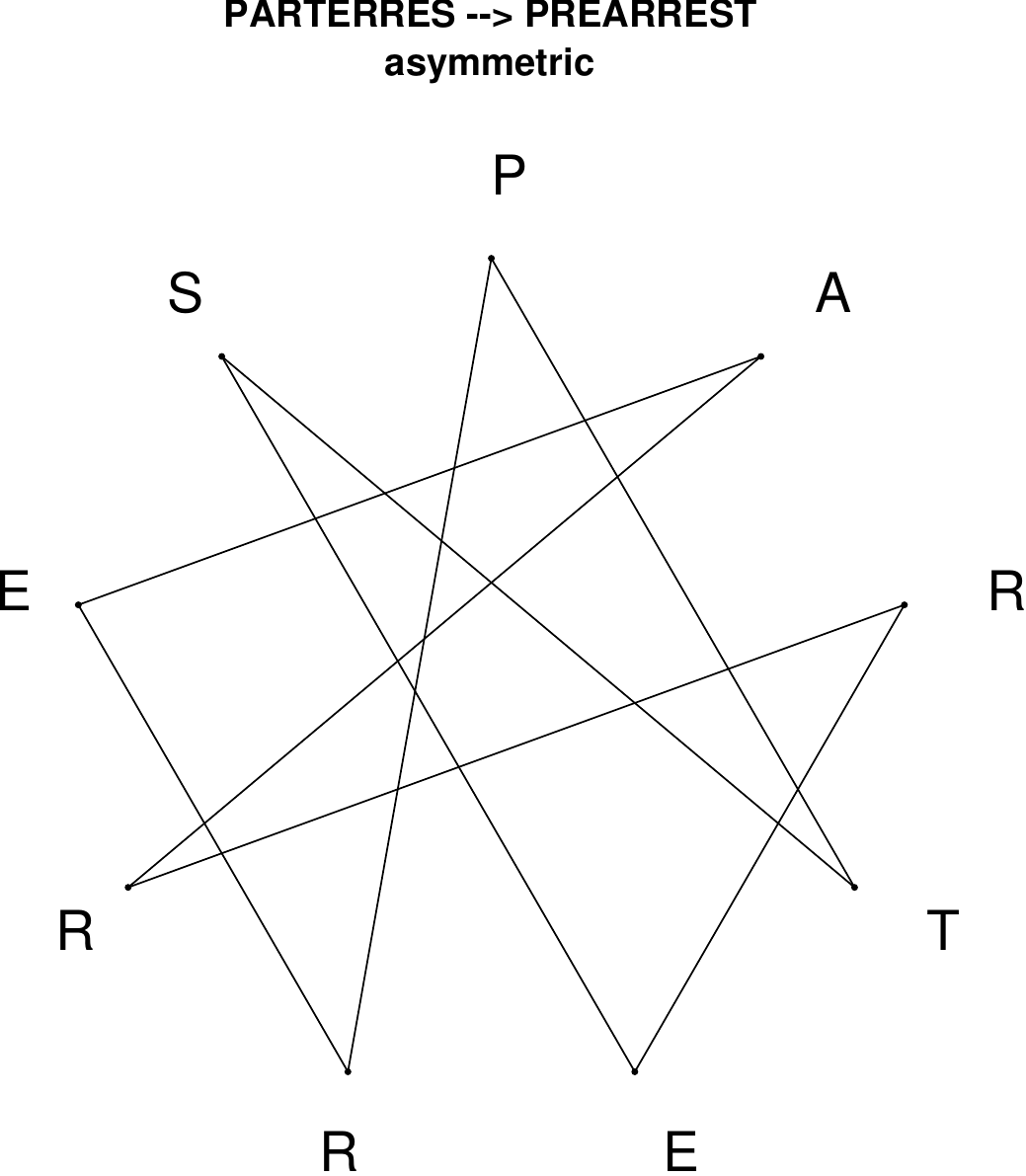}
\end{subfigure}
\hfill
\begin{subfigure}[T]{0.19\textwidth}
\centering
\includegraphics[width=\textwidth]{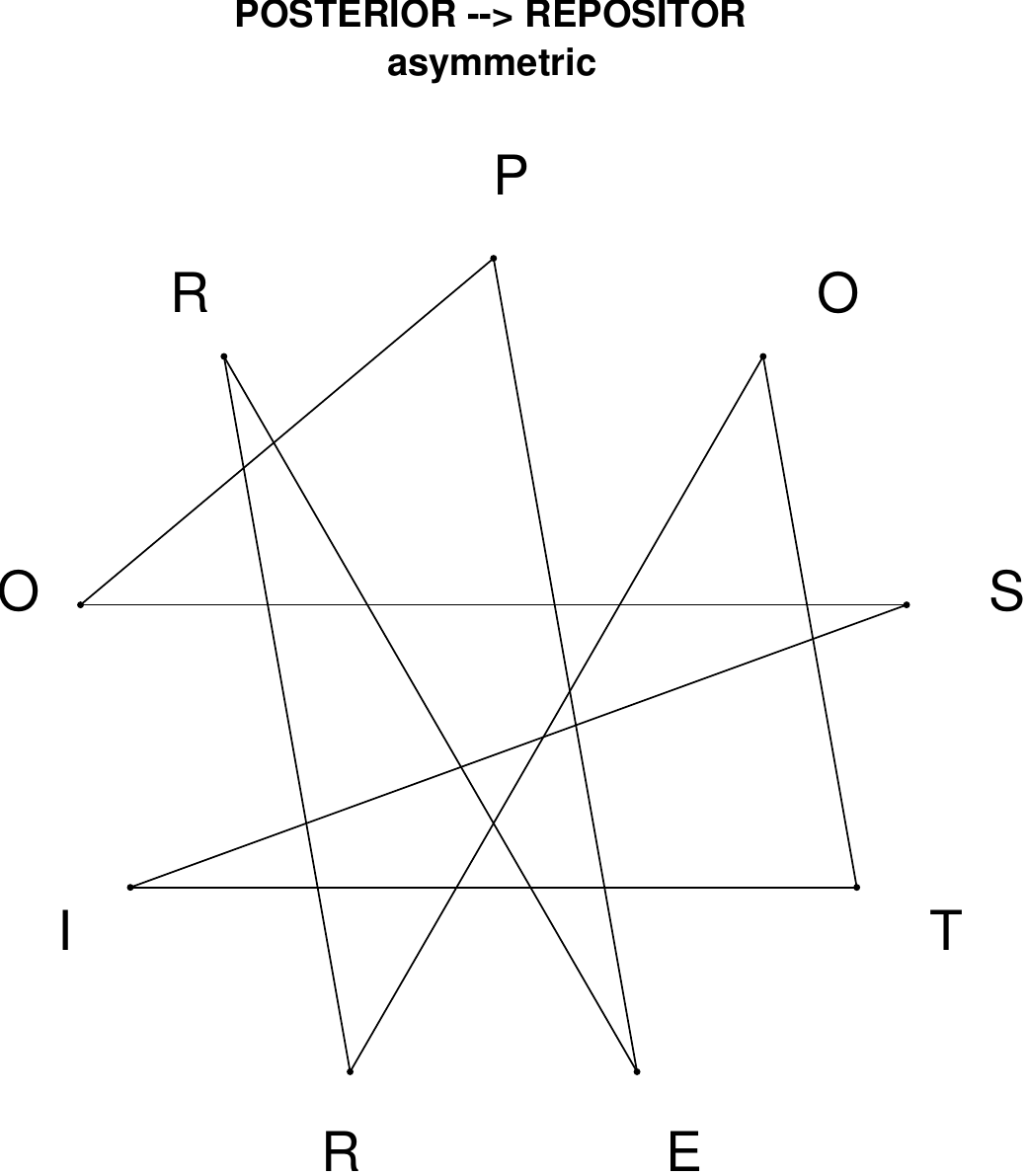}
\end{subfigure}
\end{figure}

\begin{figure}[H]
\centering
\begin{subfigure}[T]{0.19\textwidth}
\centering
\includegraphics[width=\textwidth]{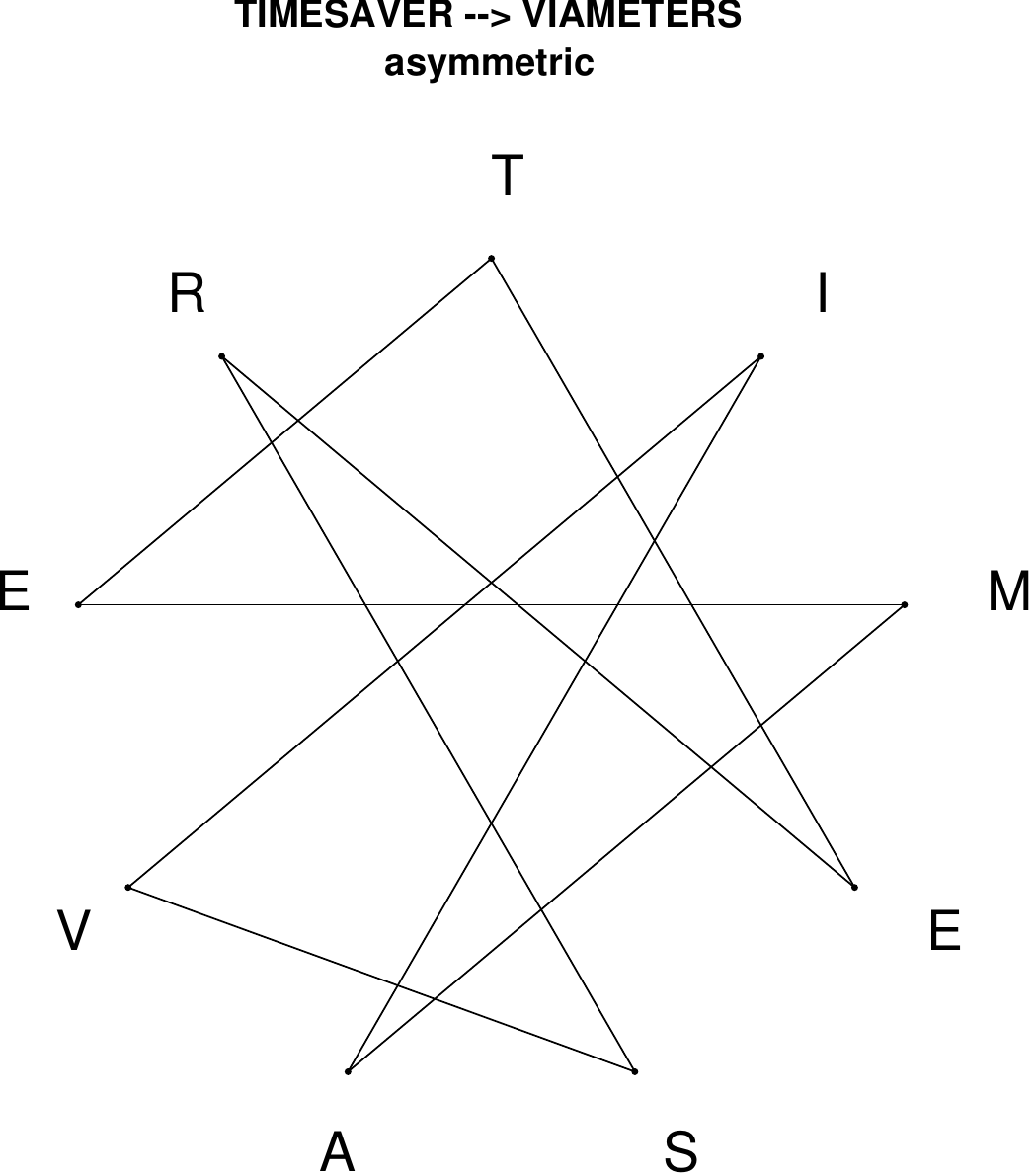}
\end{subfigure}
\hfill
\begin{subfigure}[T]{0.19\textwidth}
\centering
\includegraphics[width=\textwidth]{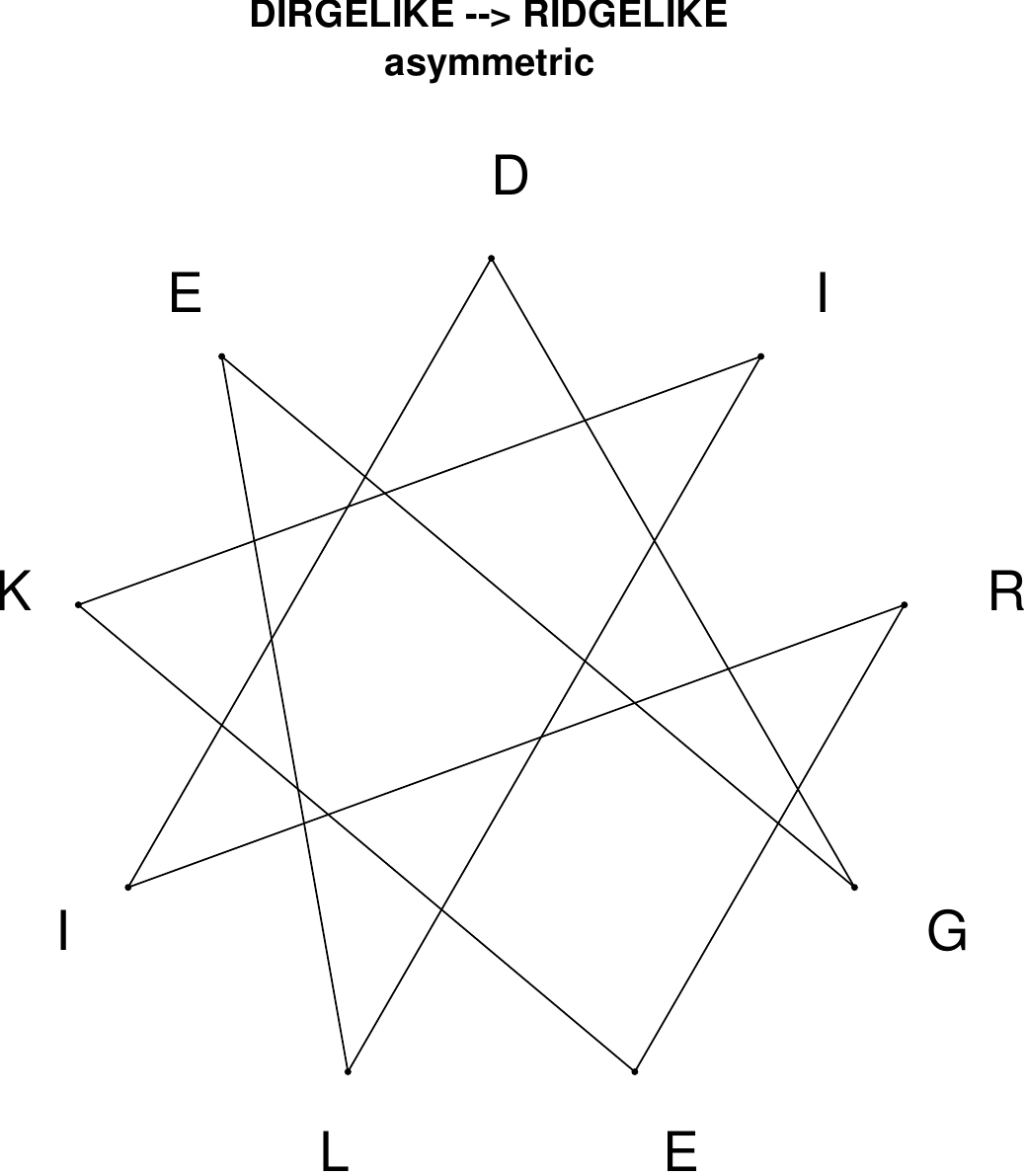}
\end{subfigure}
\hfill
\begin{subfigure}[T]{0.19\textwidth}
\centering
\includegraphics[width=\textwidth]{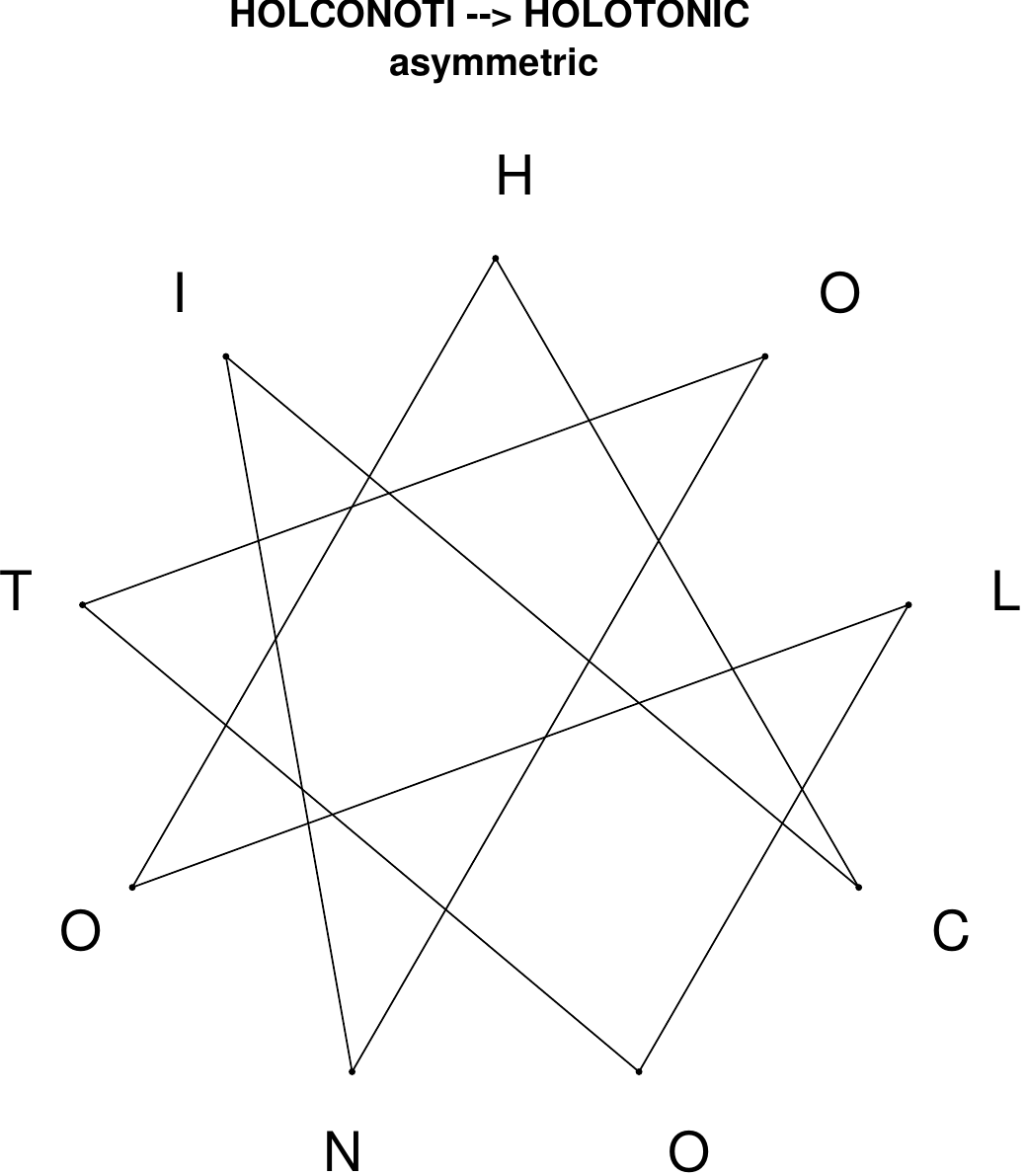}
\end{subfigure}
\hfill
\begin{subfigure}[T]{0.19\textwidth}
\centering
\includegraphics[width=\textwidth]{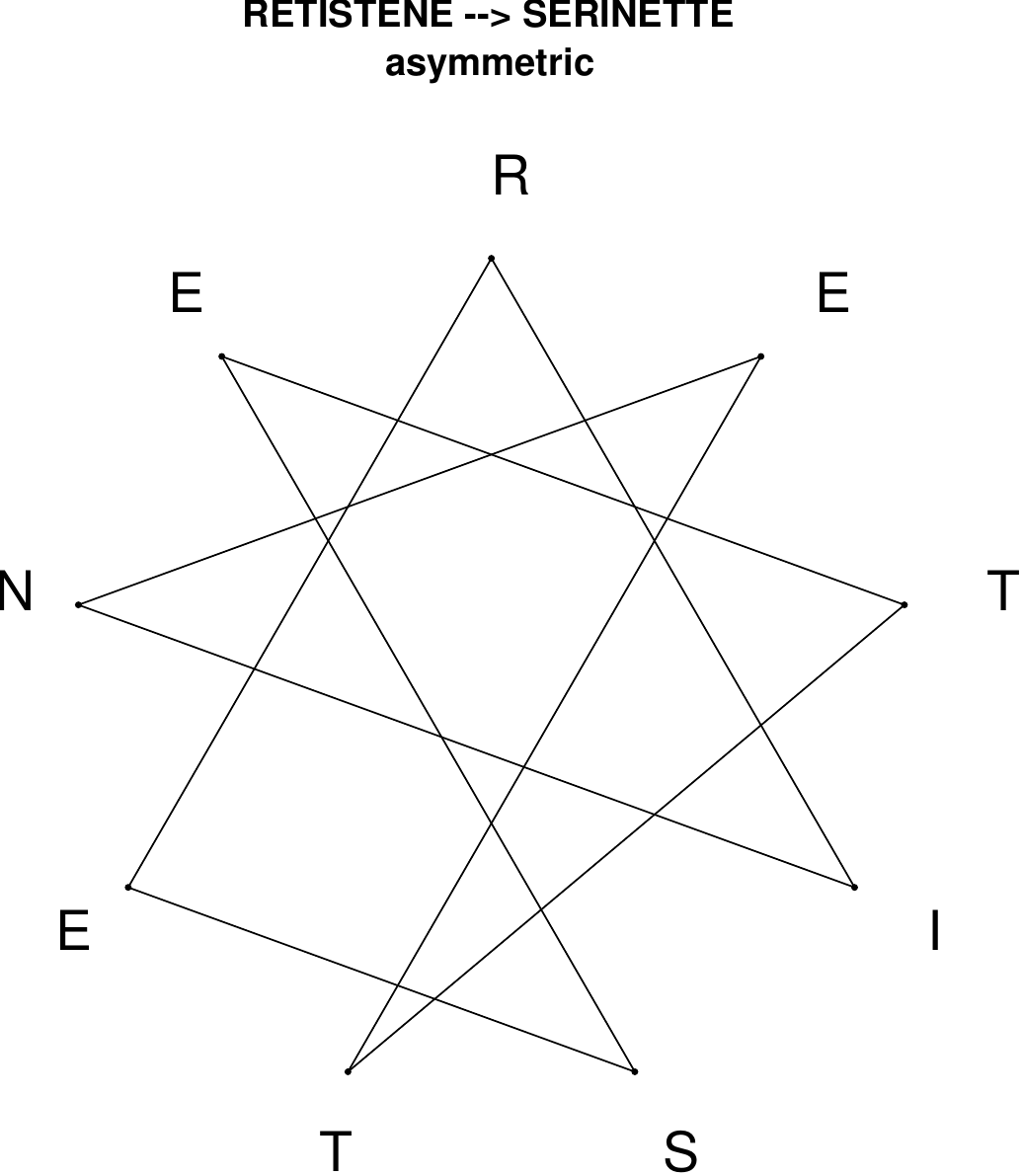}
\end{subfigure}
\hfill
\begin{subfigure}[T]{0.19\textwidth}
\centering
\includegraphics[width=\textwidth]{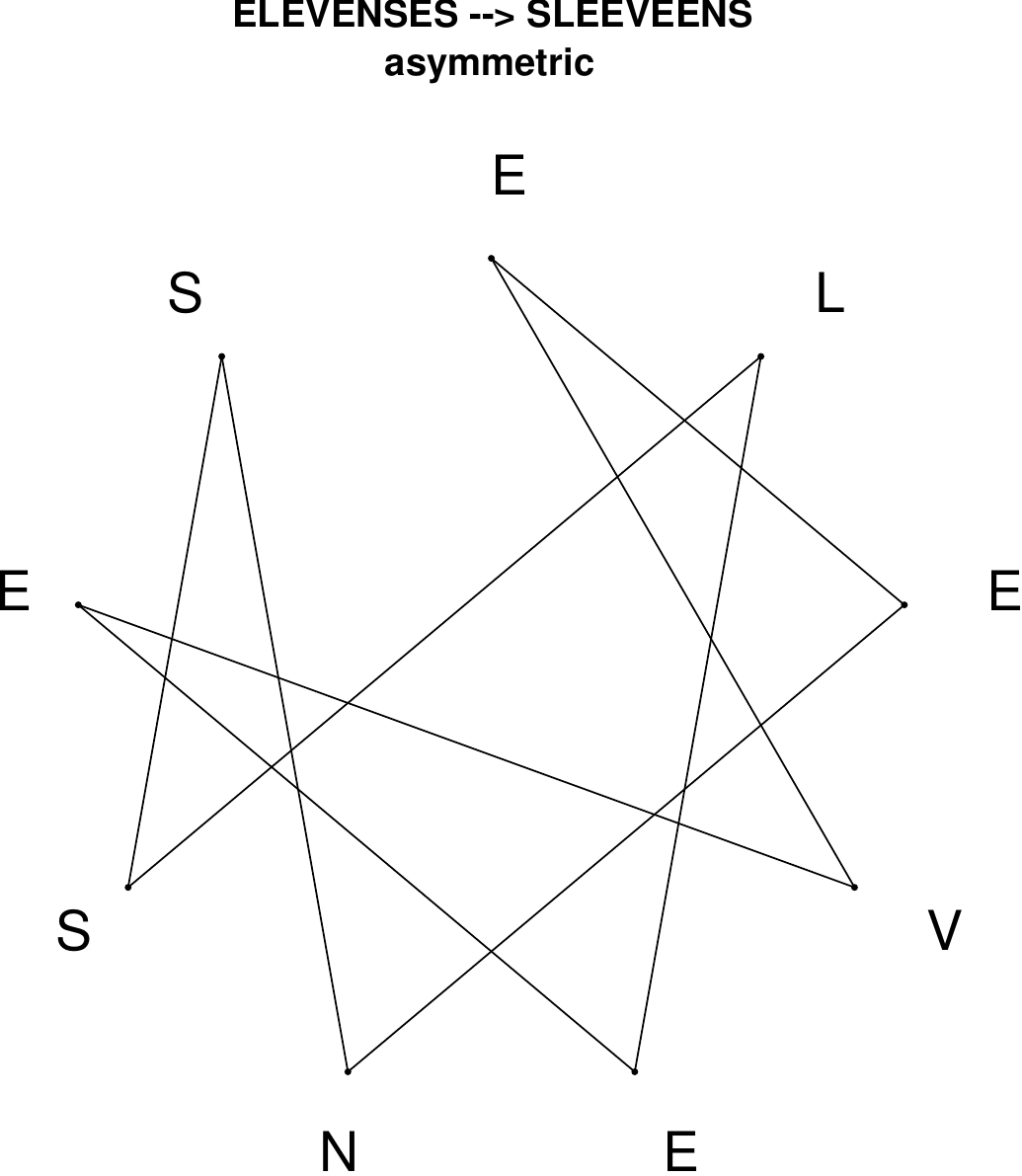}
\end{subfigure}
\end{figure}

\begin{figure}[H]
\centering
\begin{subfigure}[T]{0.19\textwidth}
\centering
\includegraphics[width=\textwidth]{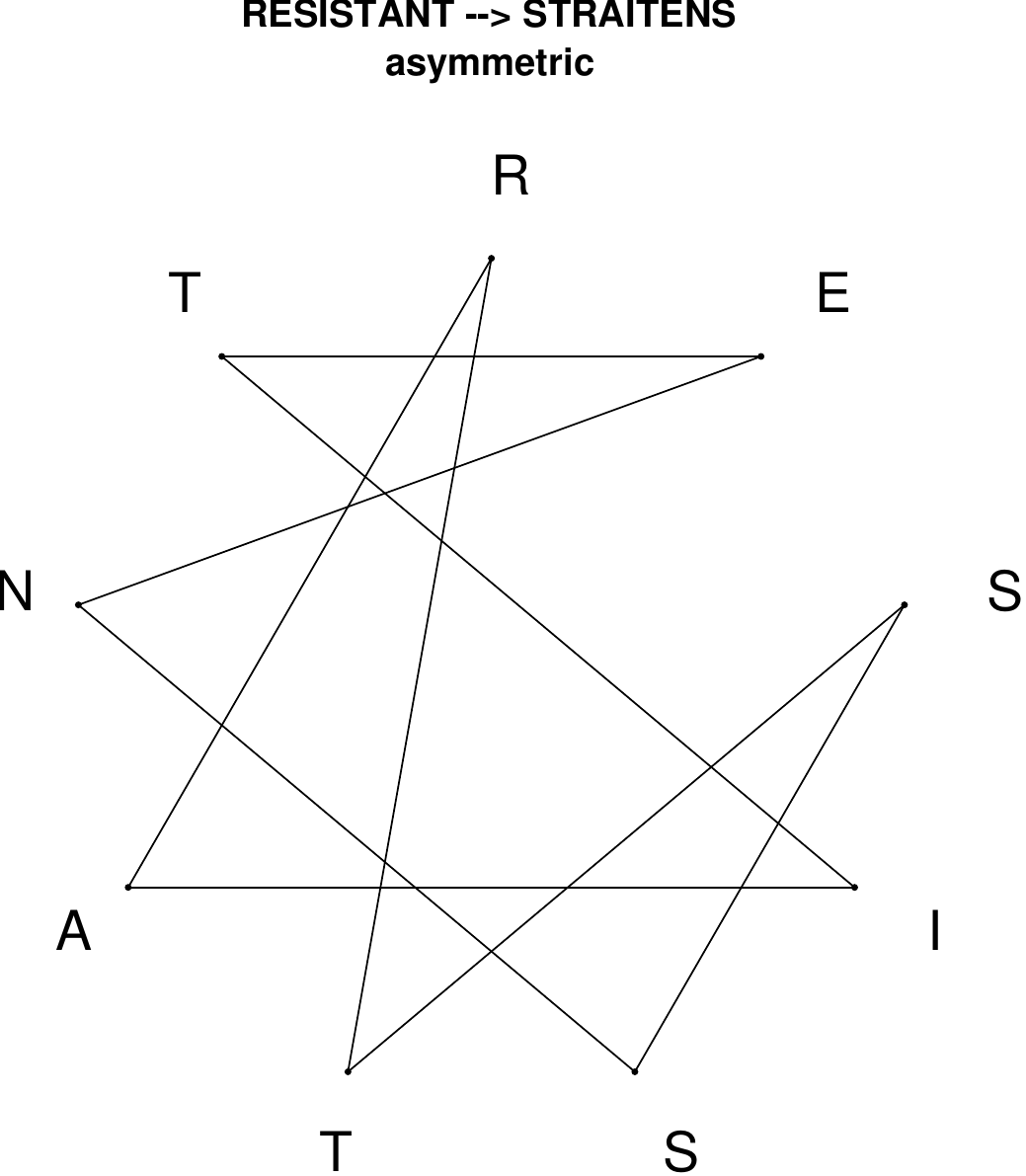}
\end{subfigure}
\hfill
\begin{subfigure}[T]{0.19\textwidth}
\centering
\includegraphics[width=\textwidth]{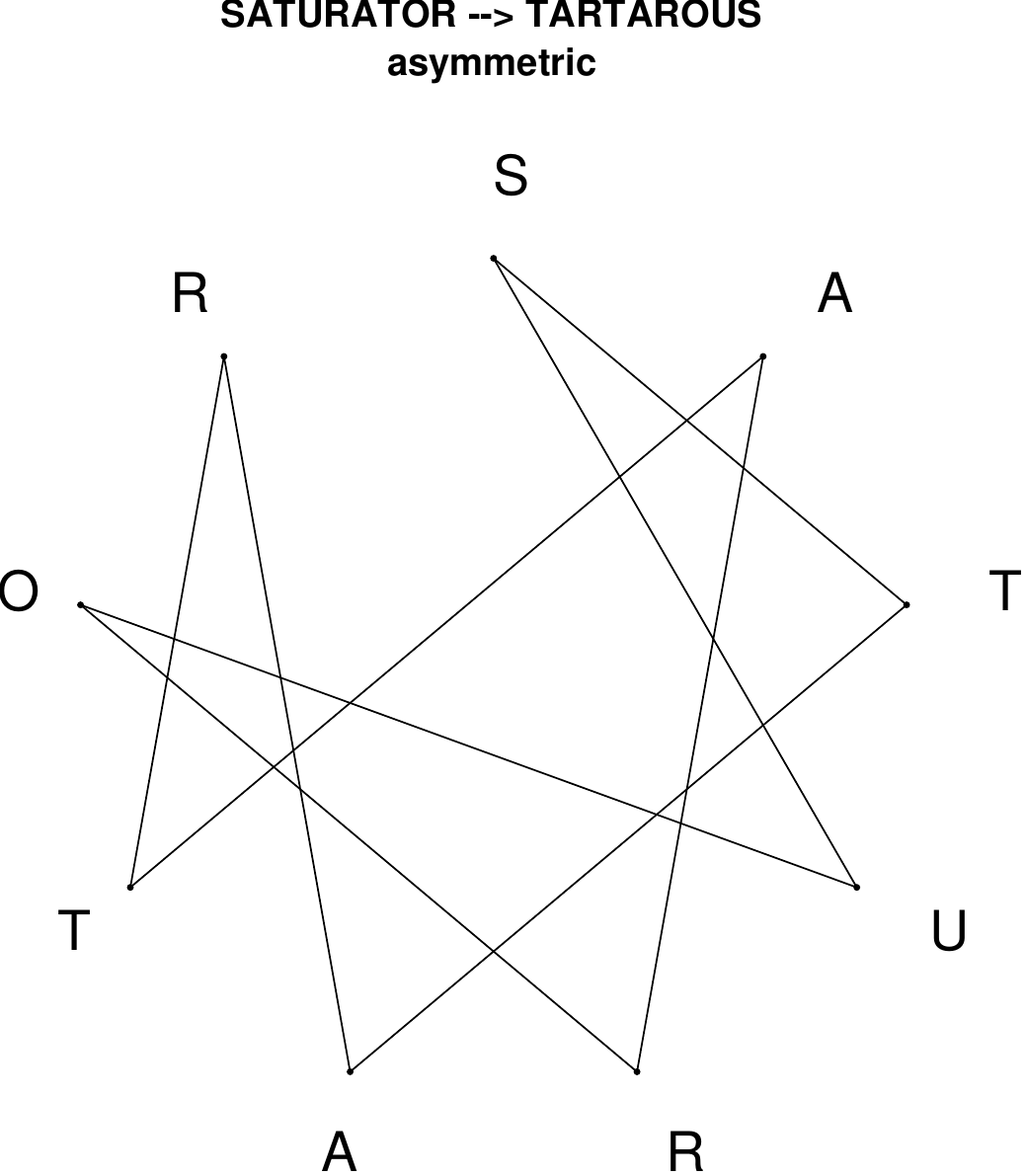}
\end{subfigure}
\hfill
\begin{subfigure}[T]{0.19\textwidth}
\centering
\includegraphics[width=\textwidth]{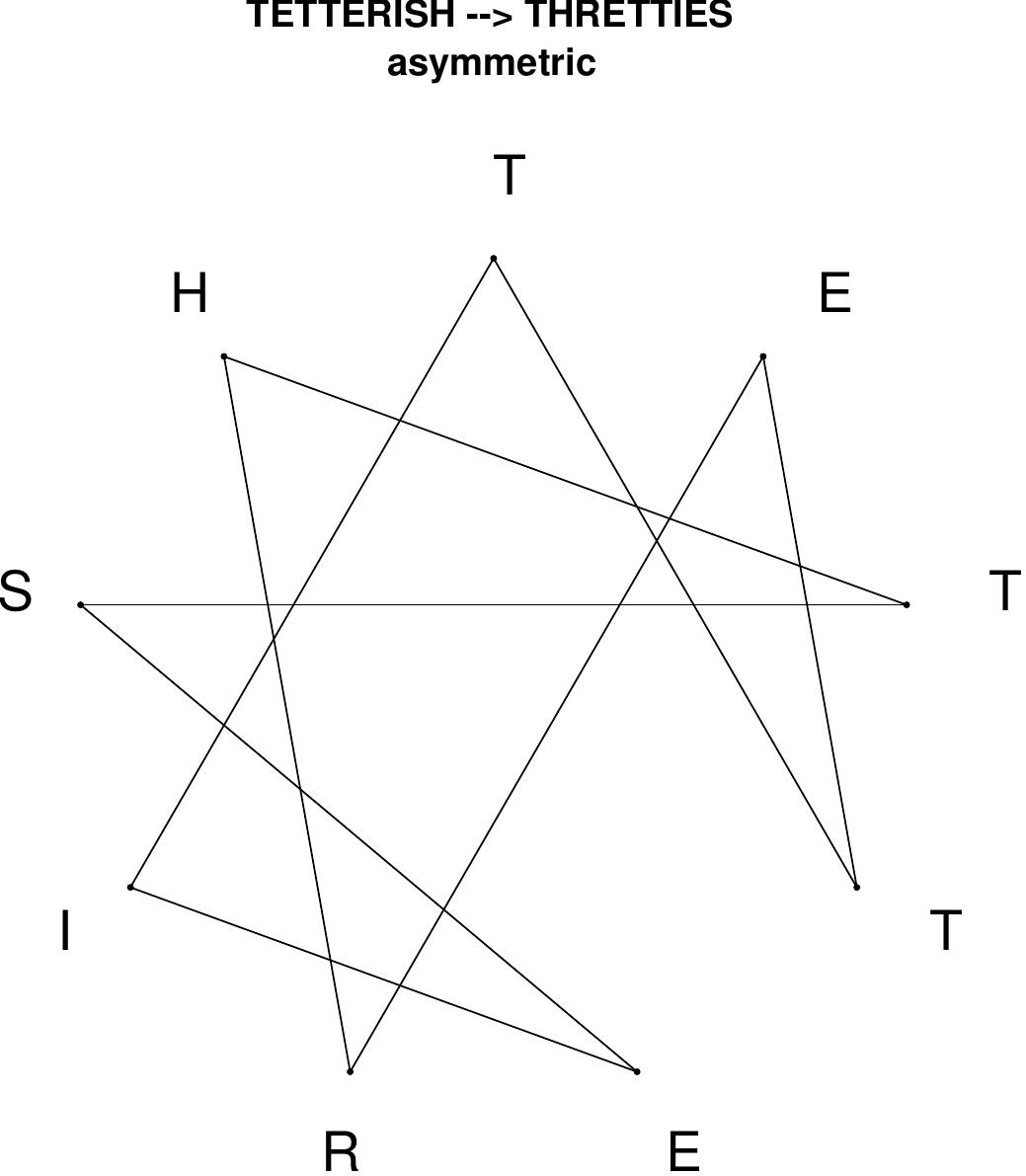}
\end{subfigure}
\hfill
\begin{subfigure}[T]{0.19\textwidth}
\centering
\includegraphics[width=\textwidth]{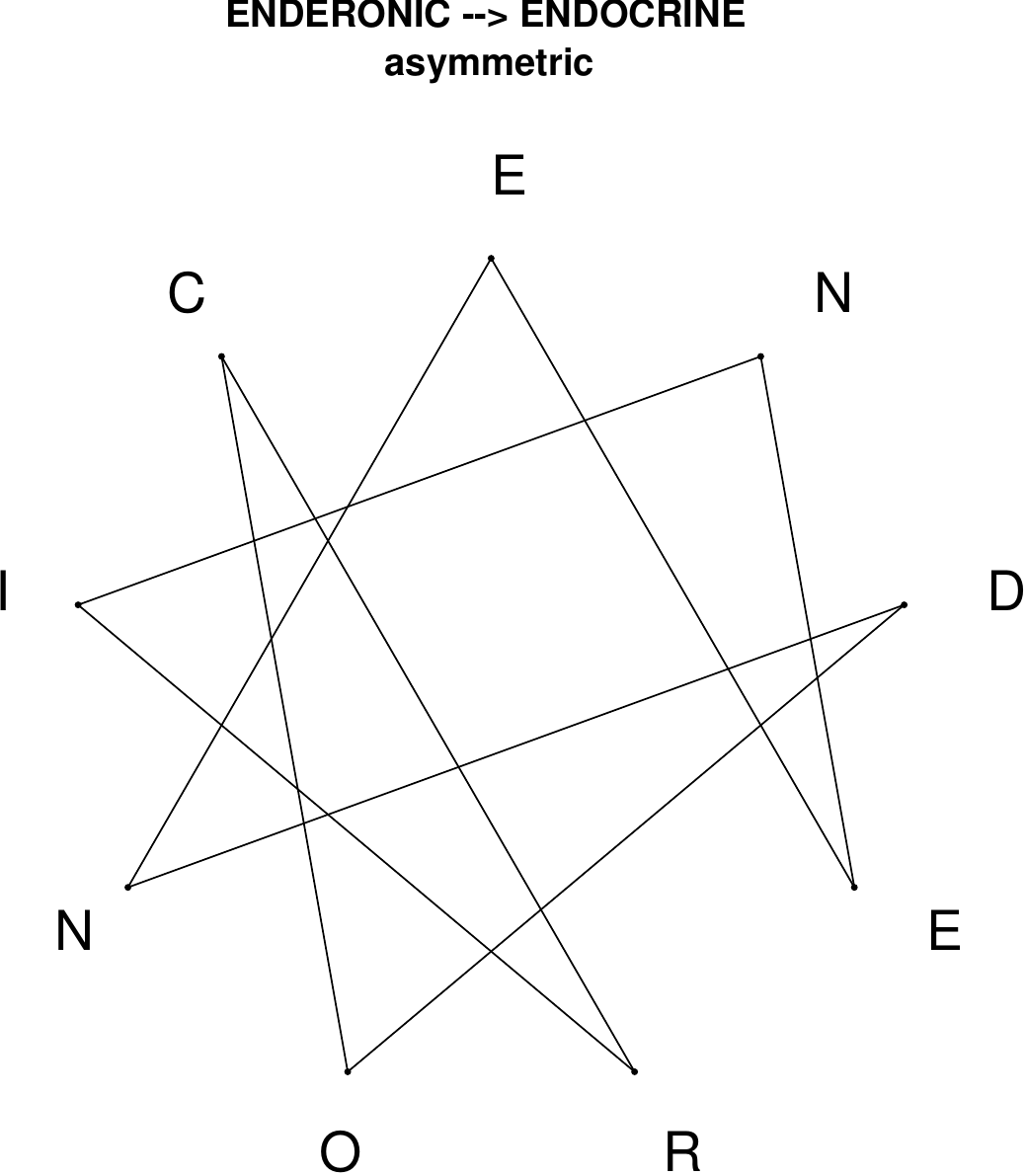}
\end{subfigure}
\hfill
\begin{subfigure}[T]{0.19\textwidth}
\centering
\includegraphics[width=\textwidth]{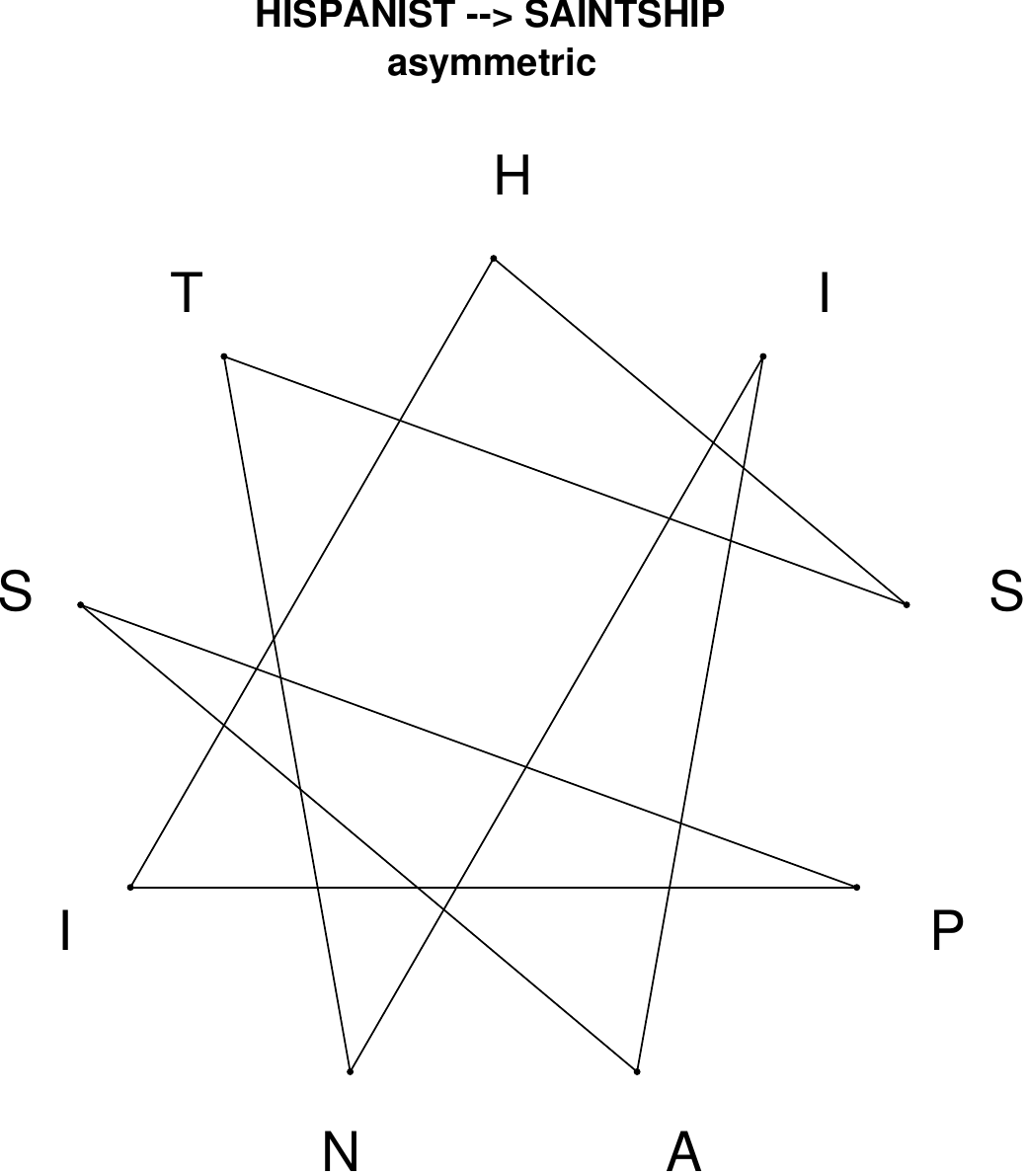}
\end{subfigure}
\end{figure}

\begin{figure}[H]
\centering
\begin{subfigure}[T]{0.19\textwidth}
\centering
\includegraphics[width=\textwidth]{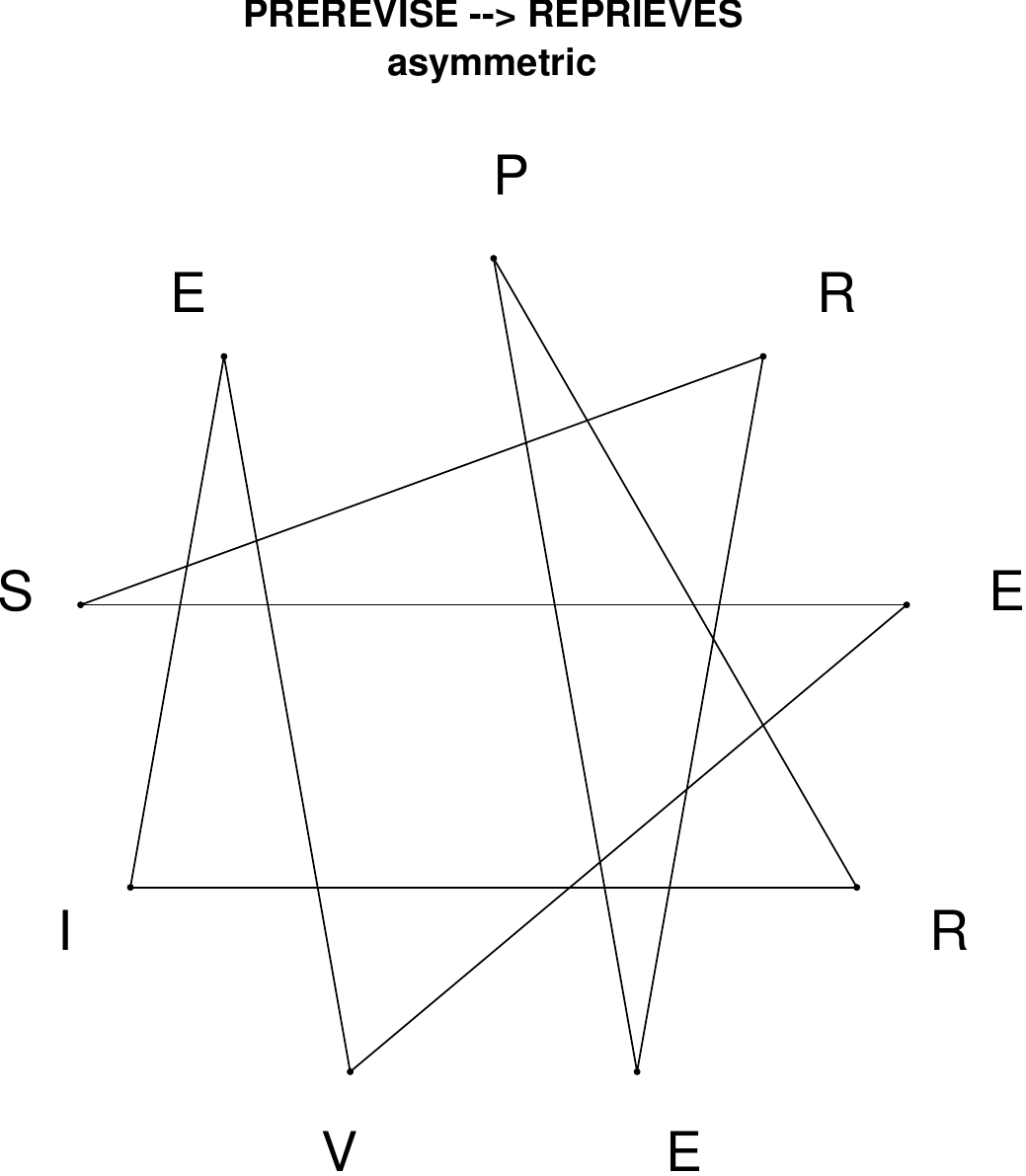}
\end{subfigure}
\hfill
\begin{subfigure}[T]{0.19\textwidth}
\centering
\includegraphics[width=\textwidth]{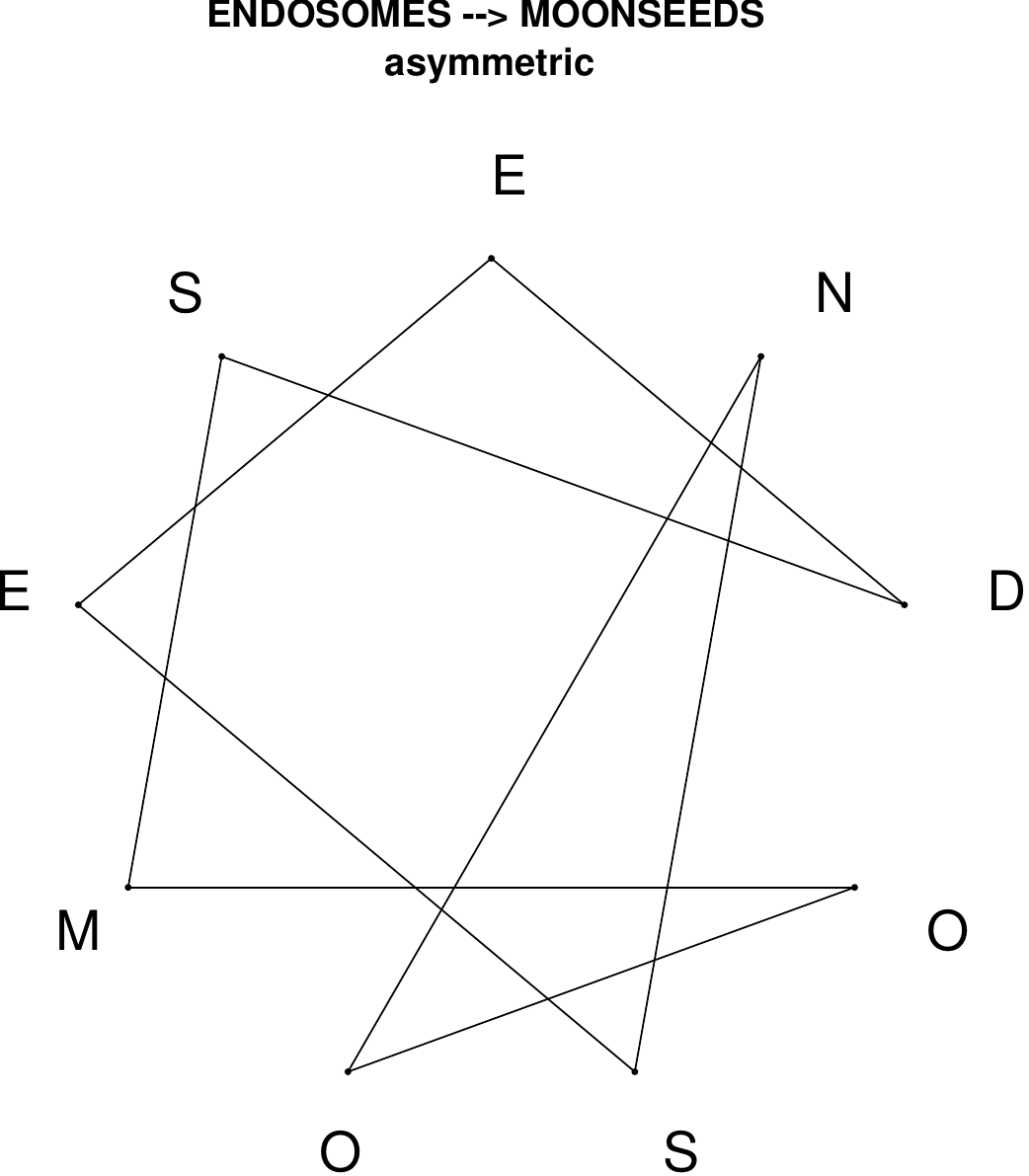}
\end{subfigure}
\hfill
\begin{subfigure}[T]{0.19\textwidth}
\centering
\includegraphics[width=\textwidth]{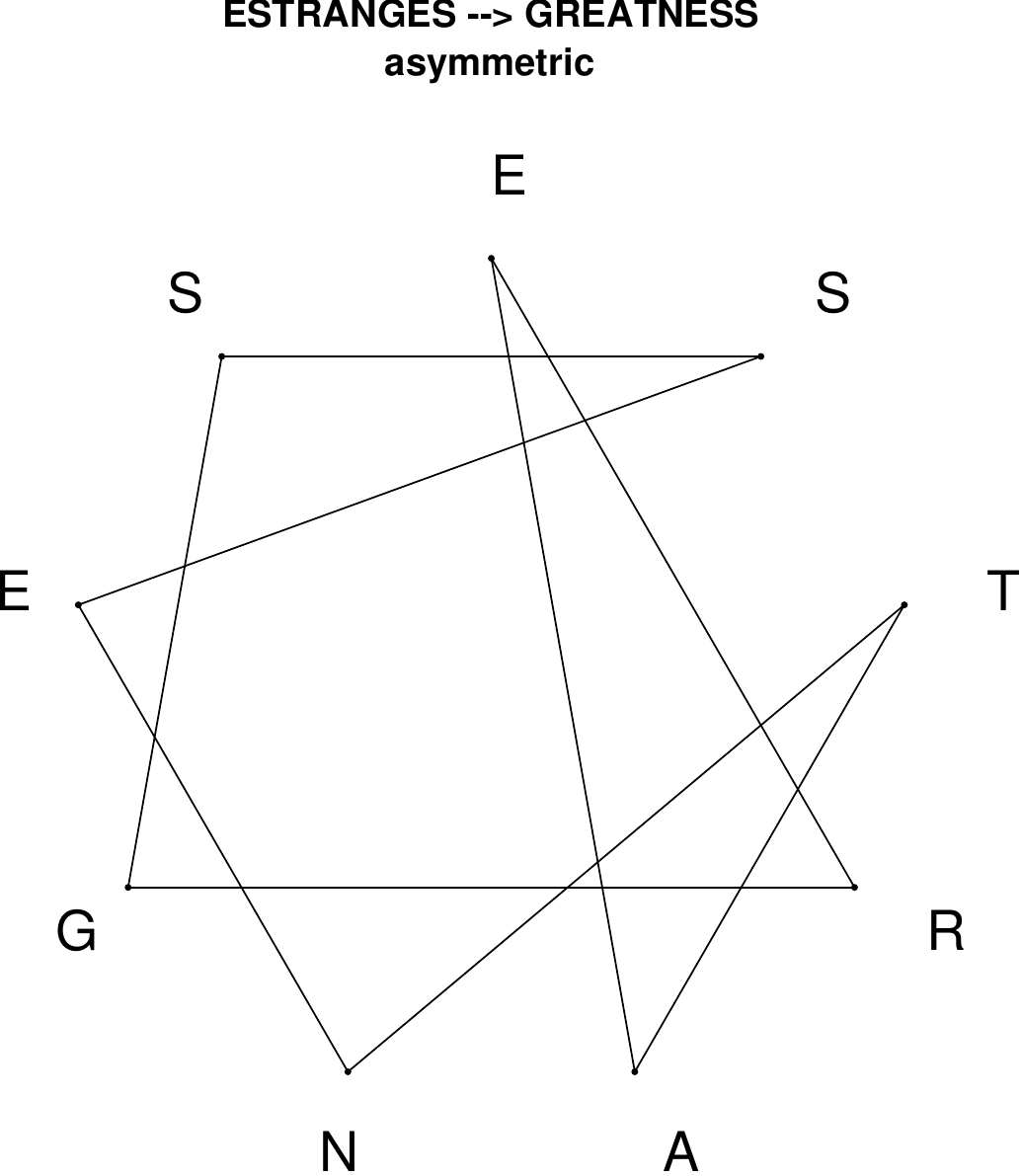}
\end{subfigure}
\hfill
\begin{subfigure}[T]{0.19\textwidth}
\centering
\includegraphics[width=\textwidth]{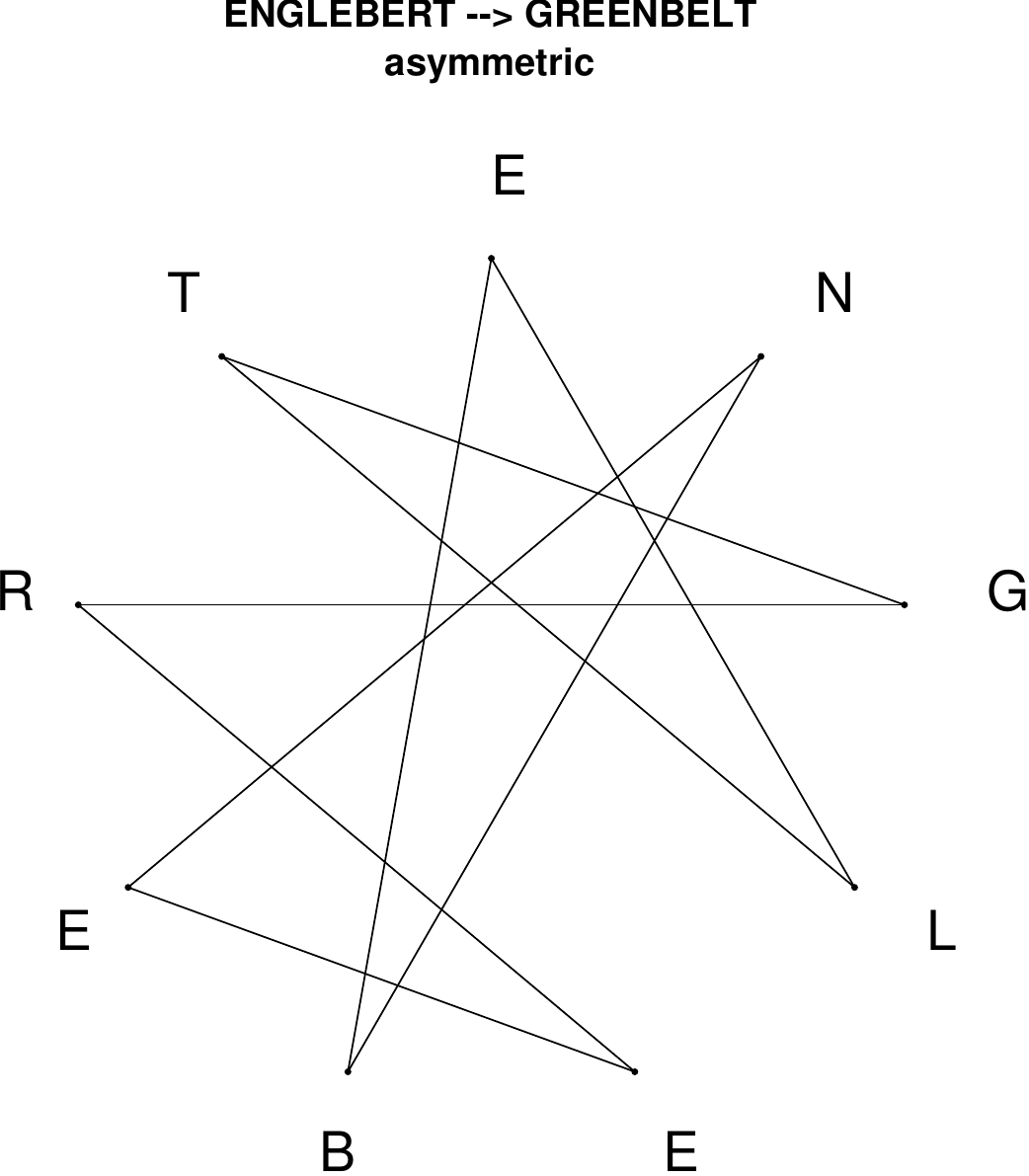}
\end{subfigure}
\hfill
\begin{subfigure}[T]{0.19\textwidth}
\centering
\includegraphics[width=\textwidth]{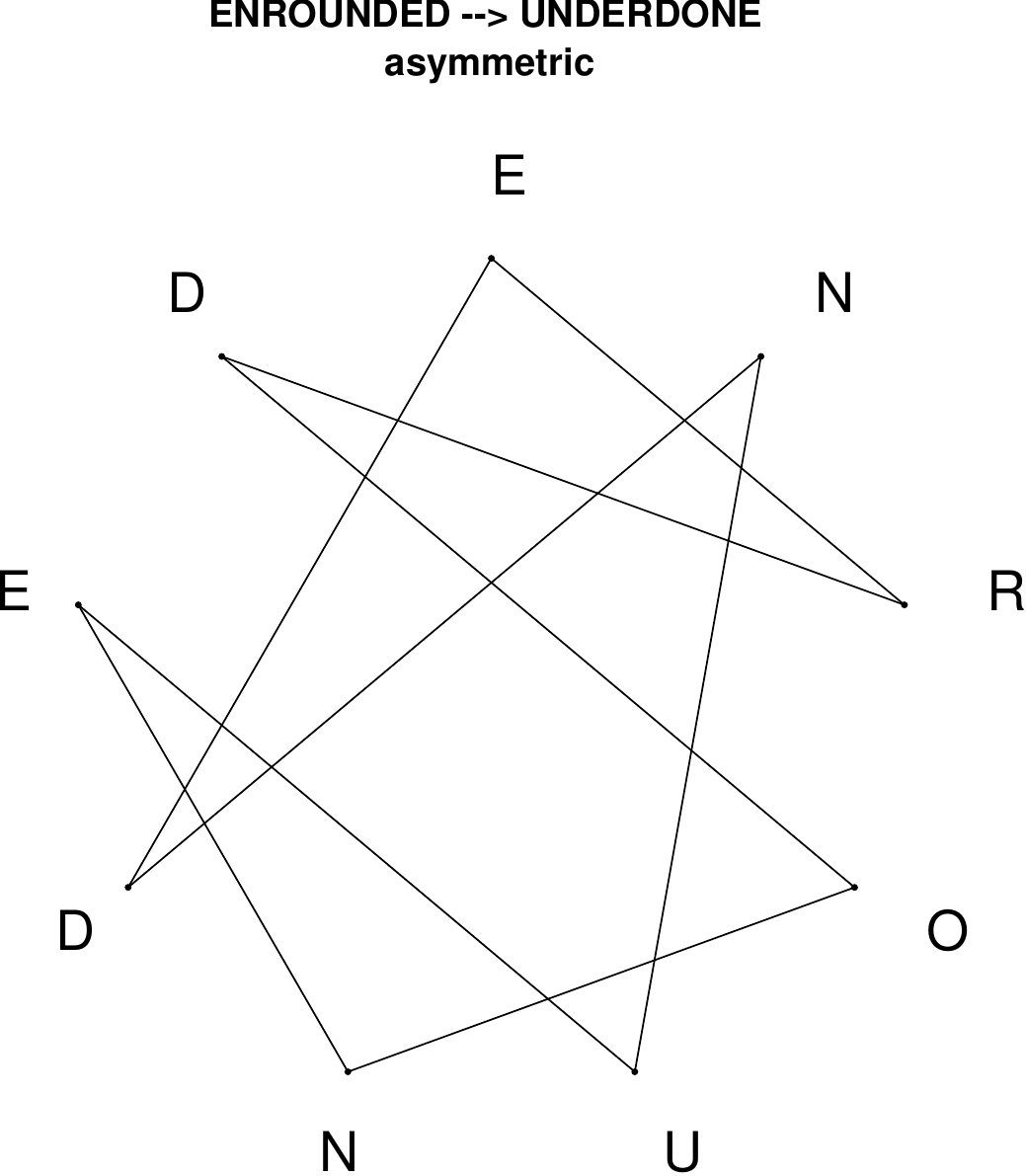}
\end{subfigure}
\end{figure}

\begin{figure}[H]
\centering
\begin{subfigure}[T]{0.19\textwidth}
\centering
\includegraphics[width=\textwidth]{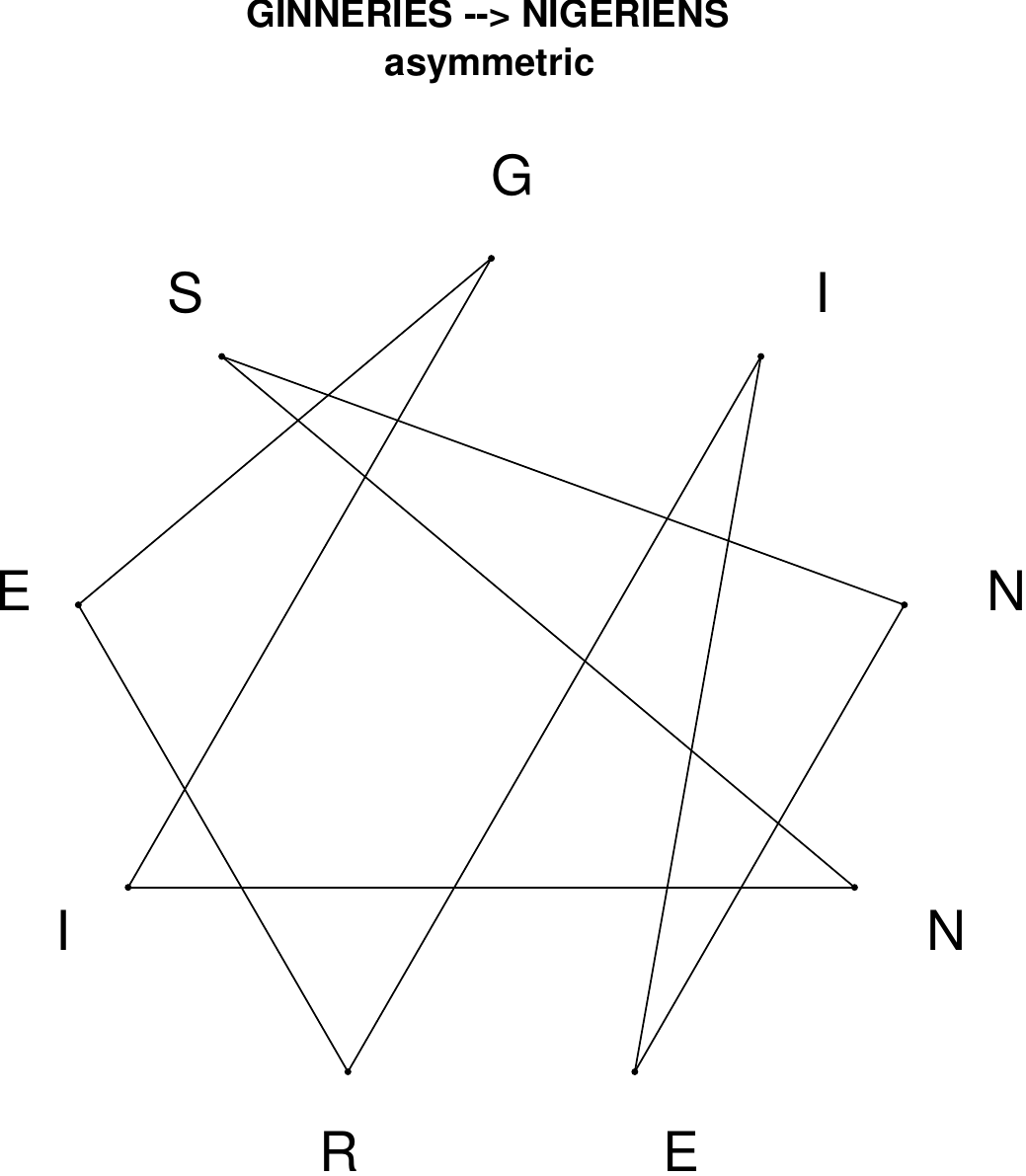}
\end{subfigure}
\hfill
\begin{subfigure}[T]{0.19\textwidth}
\centering
\includegraphics[width=\textwidth]{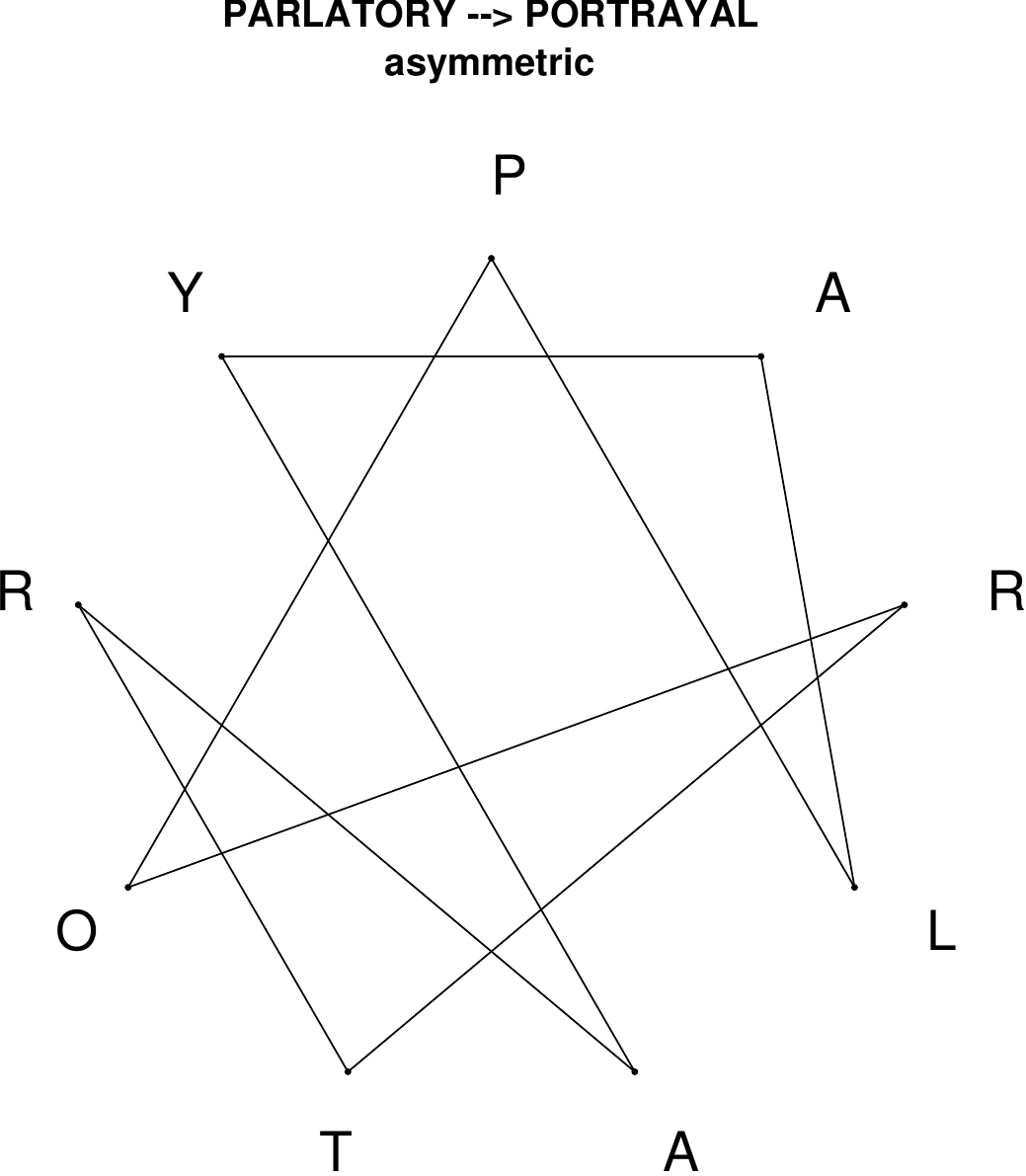}
\end{subfigure}
\hfill
\begin{subfigure}[T]{0.19\textwidth}
\centering
\includegraphics[width=\textwidth]{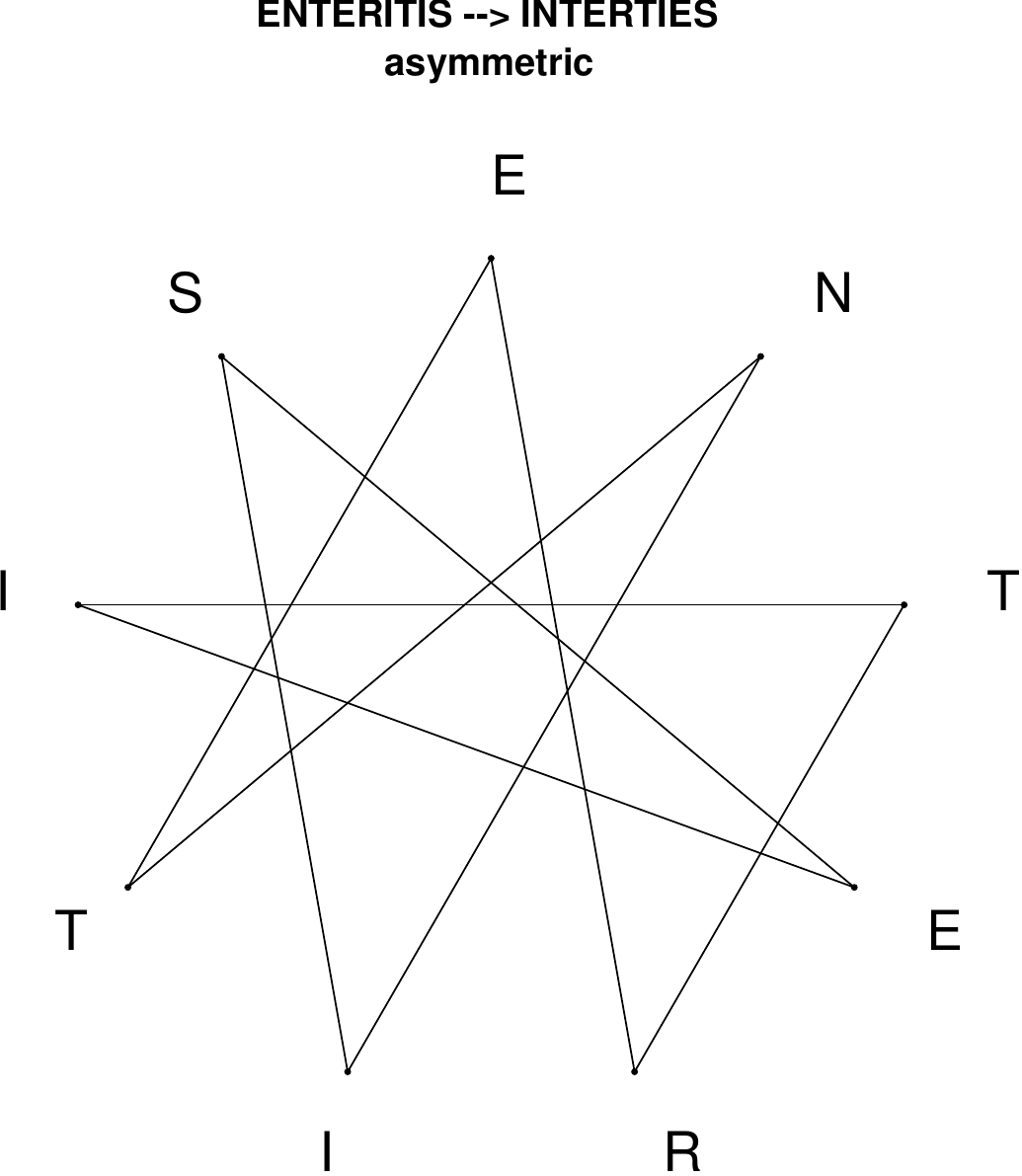}
\end{subfigure}
\hfill
\begin{subfigure}[T]{0.19\textwidth}
\centering
\includegraphics[width=\textwidth]{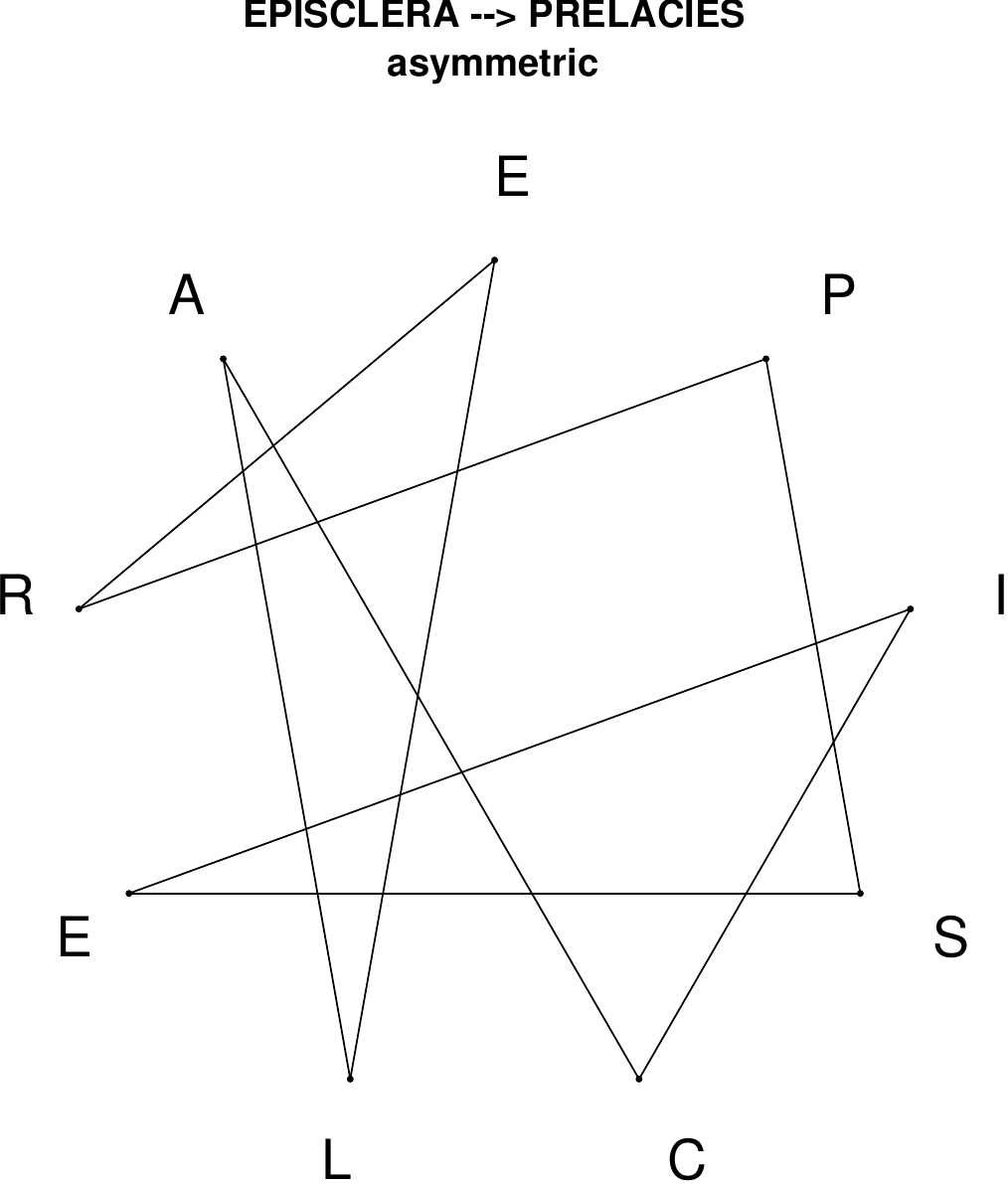}
\end{subfigure}
\hfill
\begin{subfigure}[T]{0.19\textwidth}
\centering
\includegraphics[width=\textwidth]{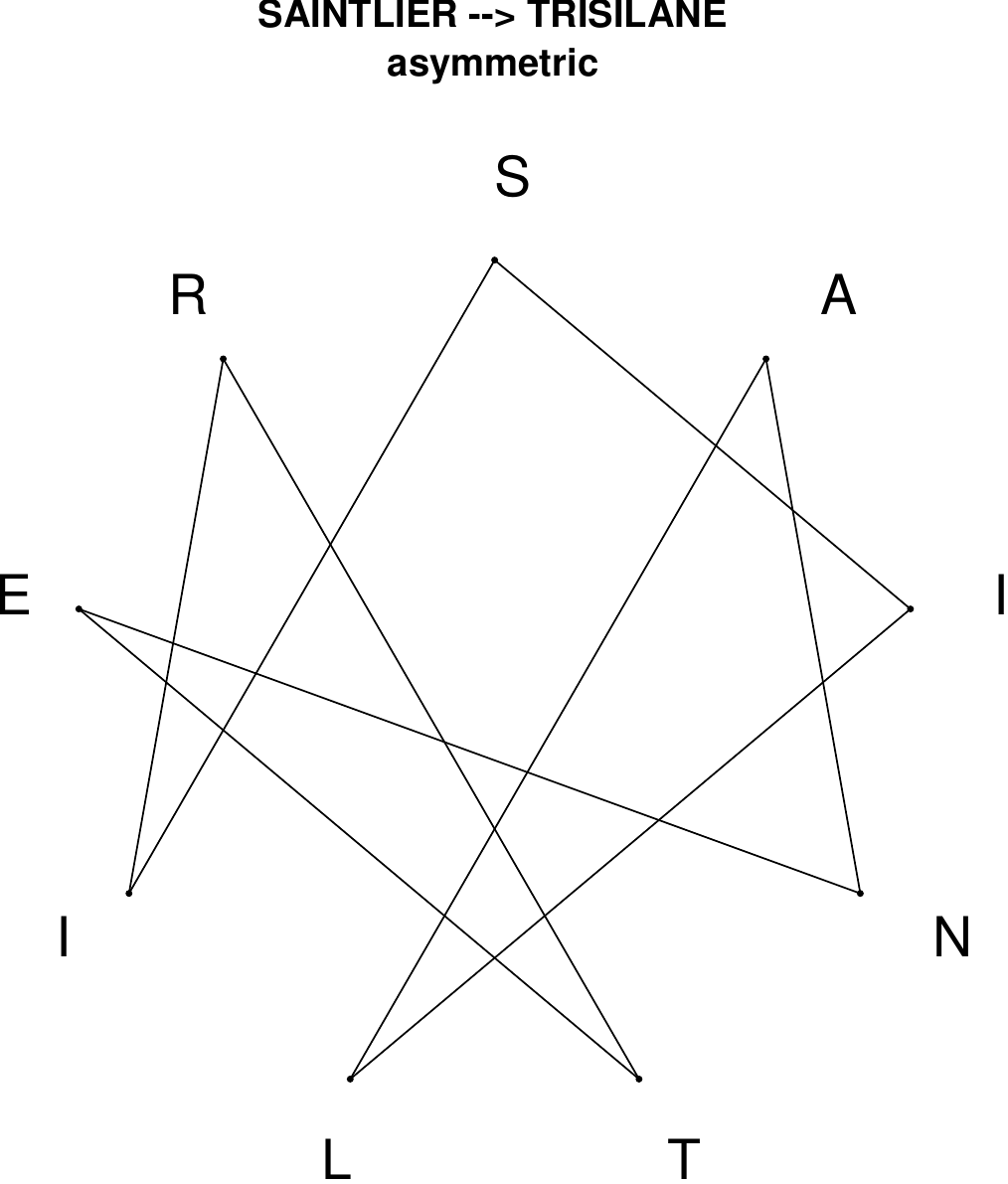}
\end{subfigure}
\end{figure}

\begin{figure}[H]
\centering
\begin{subfigure}[T]{0.19\textwidth}
\centering
\includegraphics[width=\textwidth]{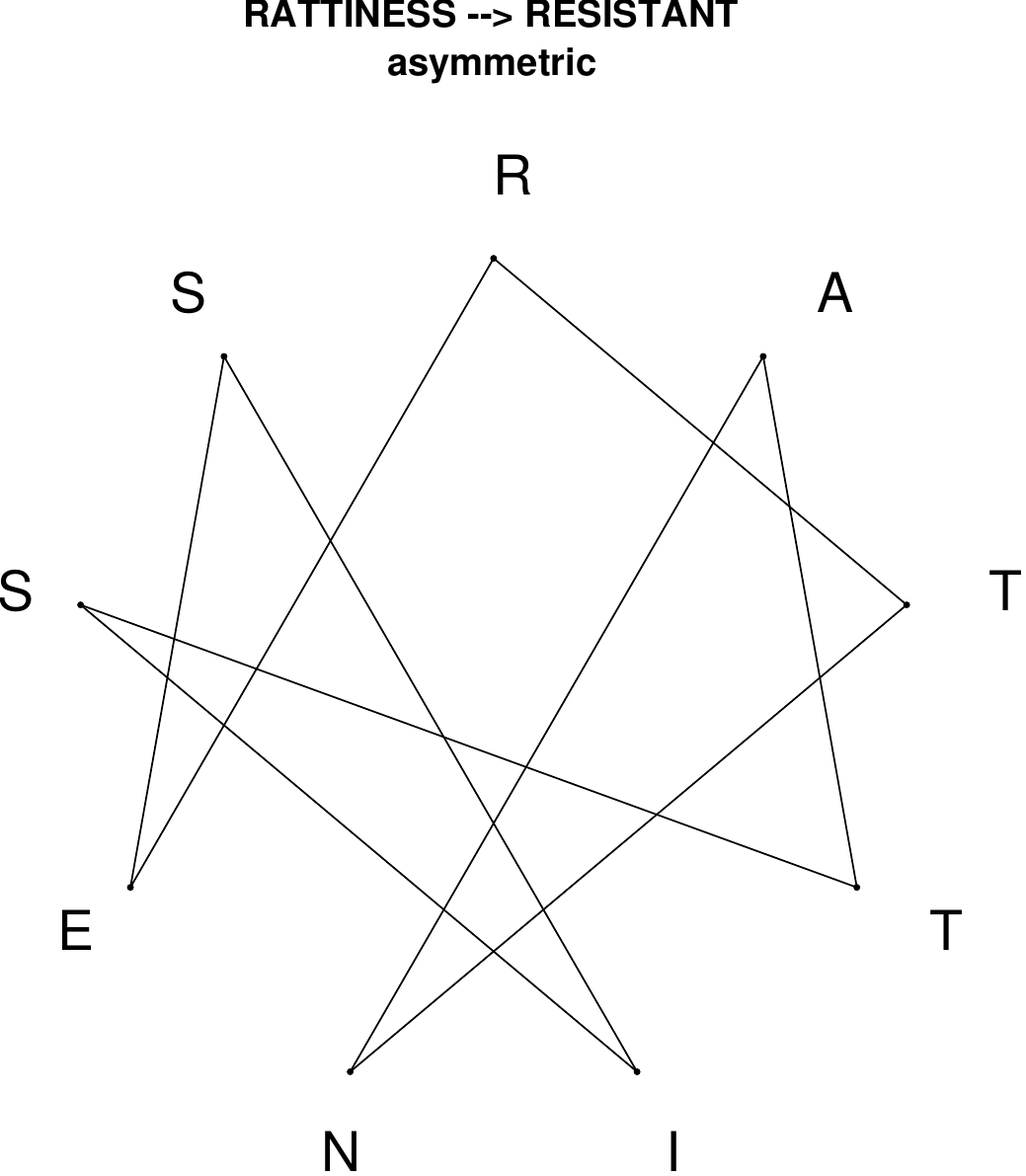}
\end{subfigure}
\hfill
\begin{subfigure}[T]{0.19\textwidth}
\centering
\includegraphics[width=\textwidth]{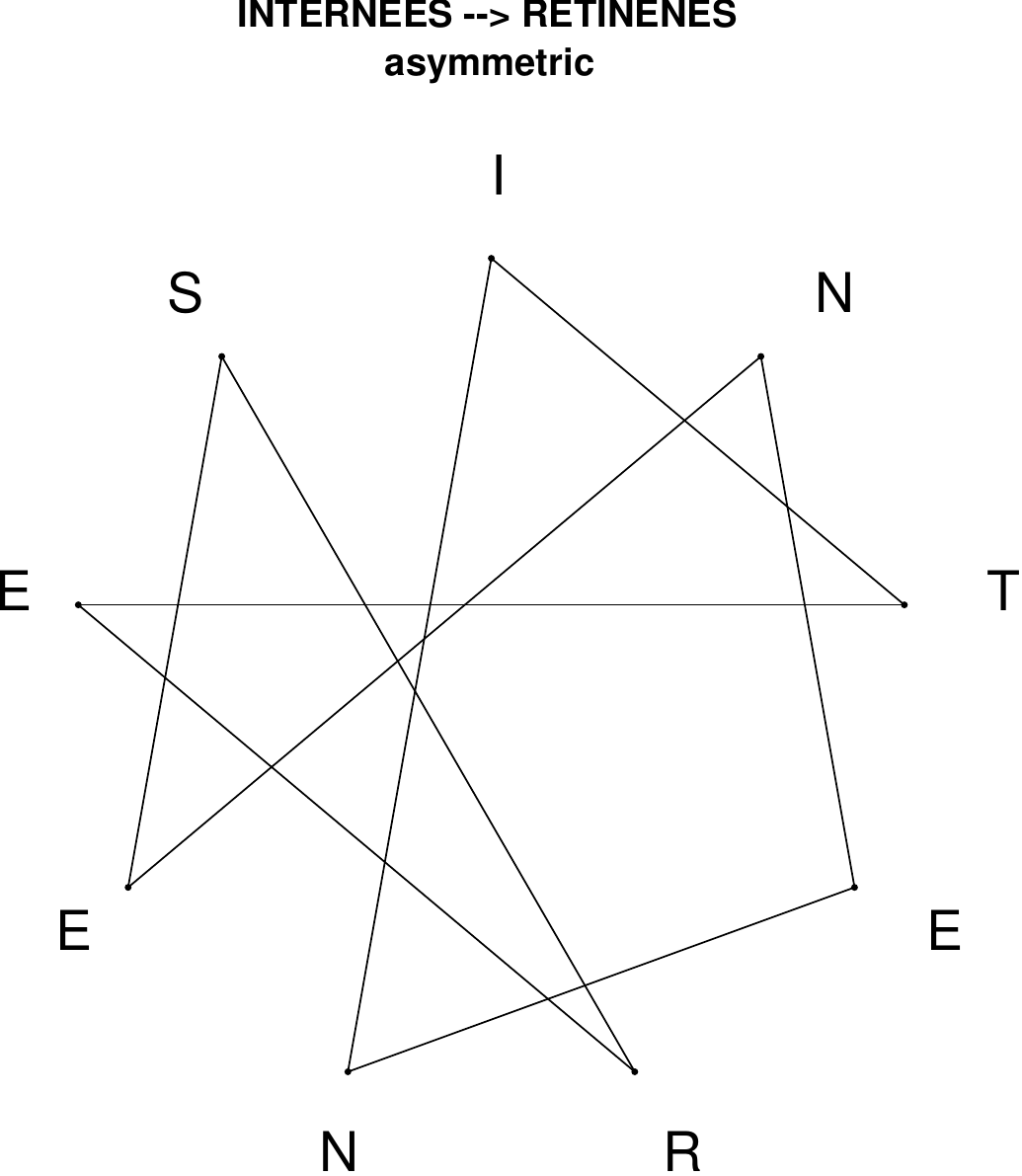}
\end{subfigure}
\hfill
\begin{subfigure}[T]{0.19\textwidth}
\centering
\includegraphics[width=\textwidth]{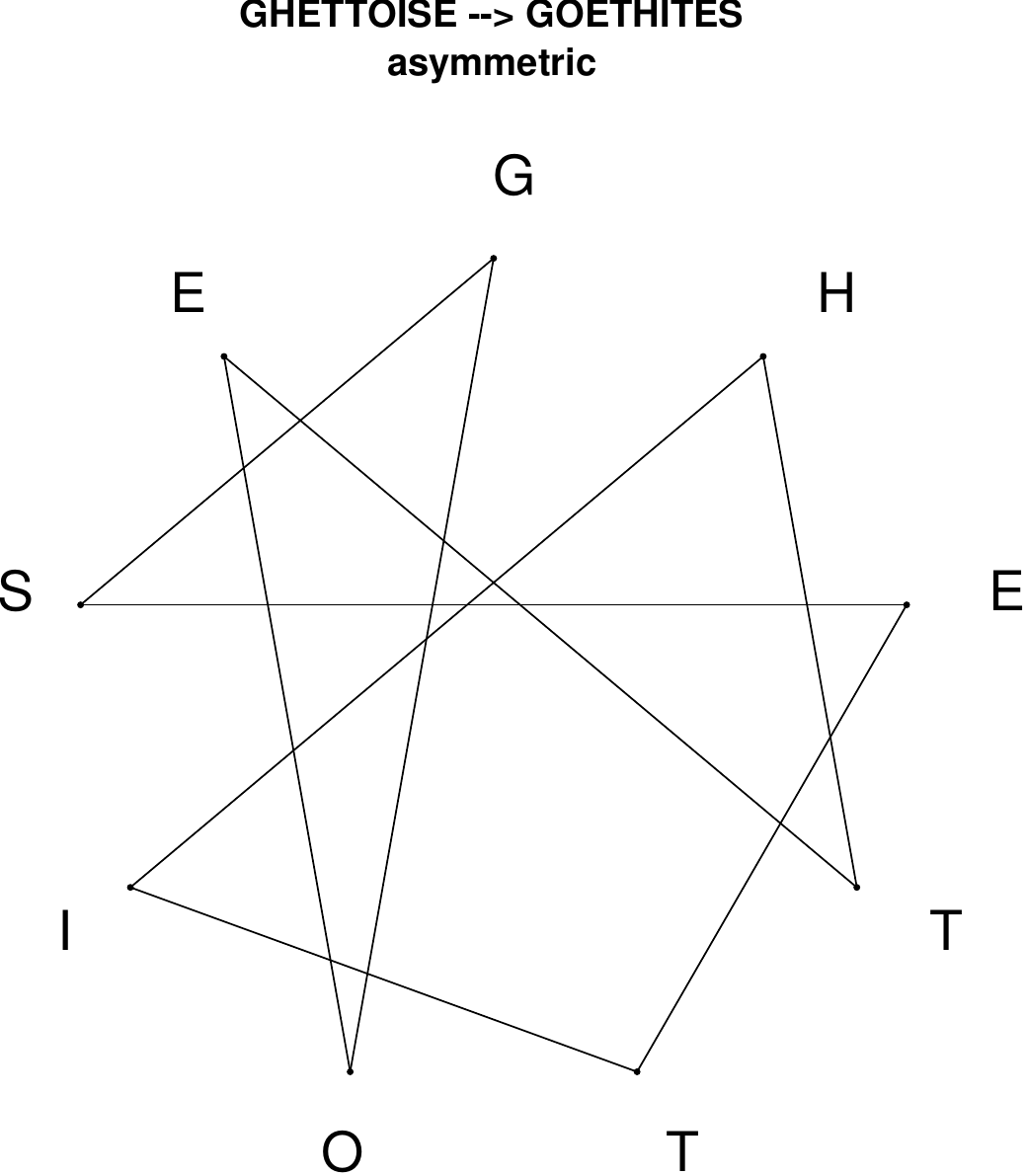}
\end{subfigure}
\hfill
\begin{subfigure}[T]{0.19\textwidth}
\centering
\includegraphics[width=\textwidth]{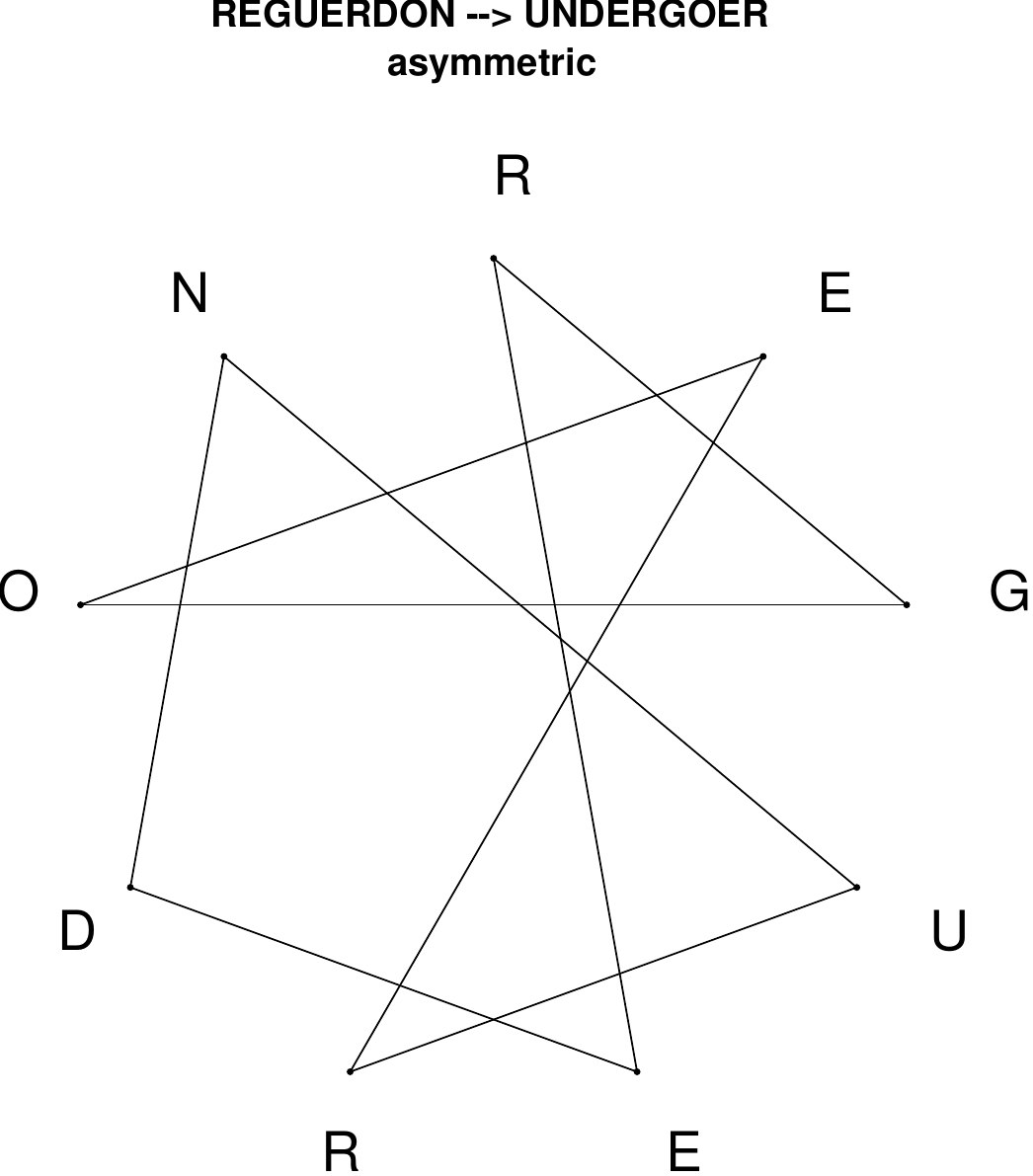}
\end{subfigure}
\hfill
\begin{subfigure}[T]{0.19\textwidth}
\centering
\includegraphics[width=\textwidth]{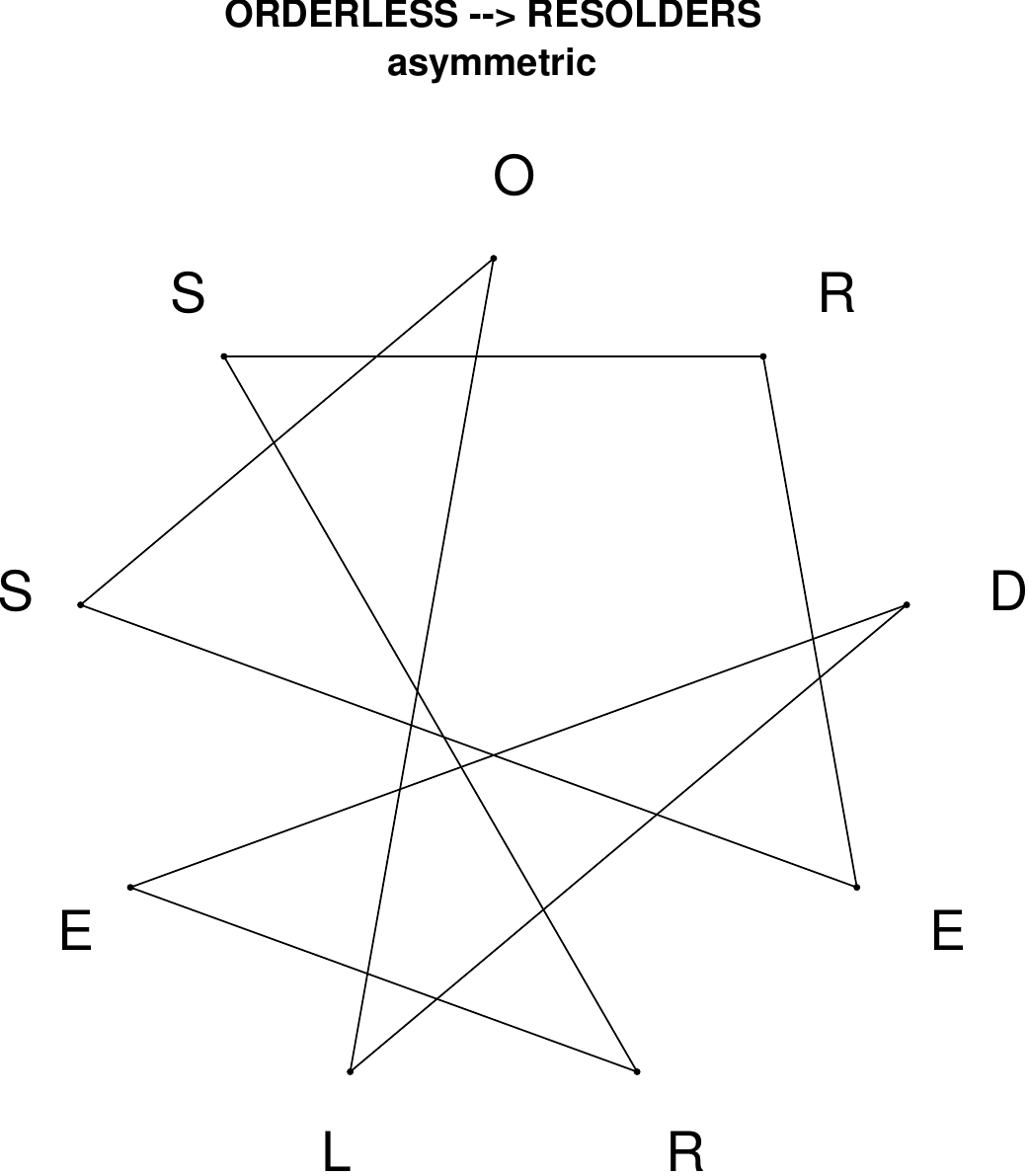}
\end{subfigure}
\end{figure}

\begin{figure}[H]
\centering
\begin{subfigure}[T]{0.19\textwidth}
\centering
\includegraphics[width=\textwidth]{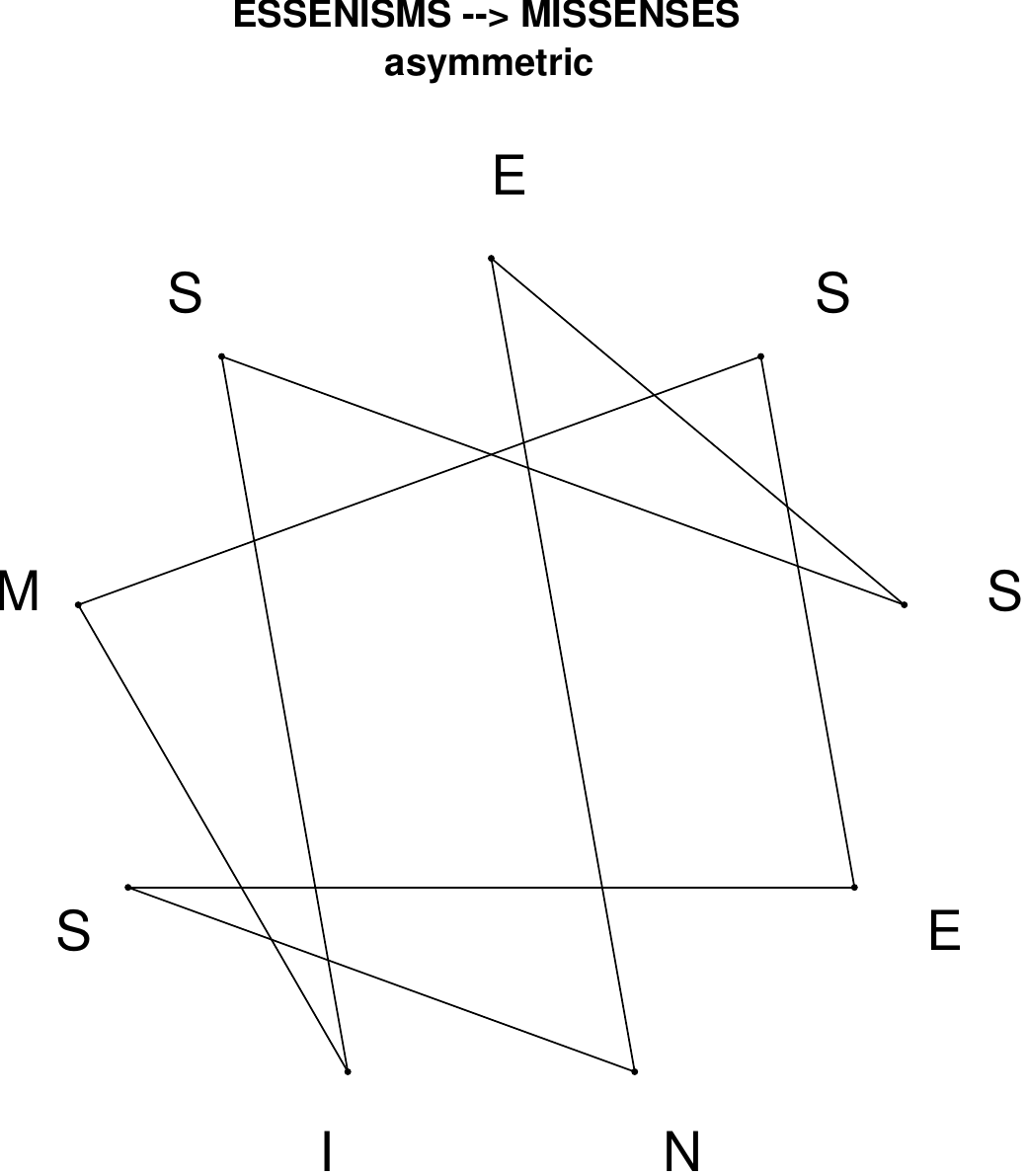}
\end{subfigure}
\hfill
\begin{subfigure}[T]{0.19\textwidth}
\centering
\includegraphics[width=\textwidth]{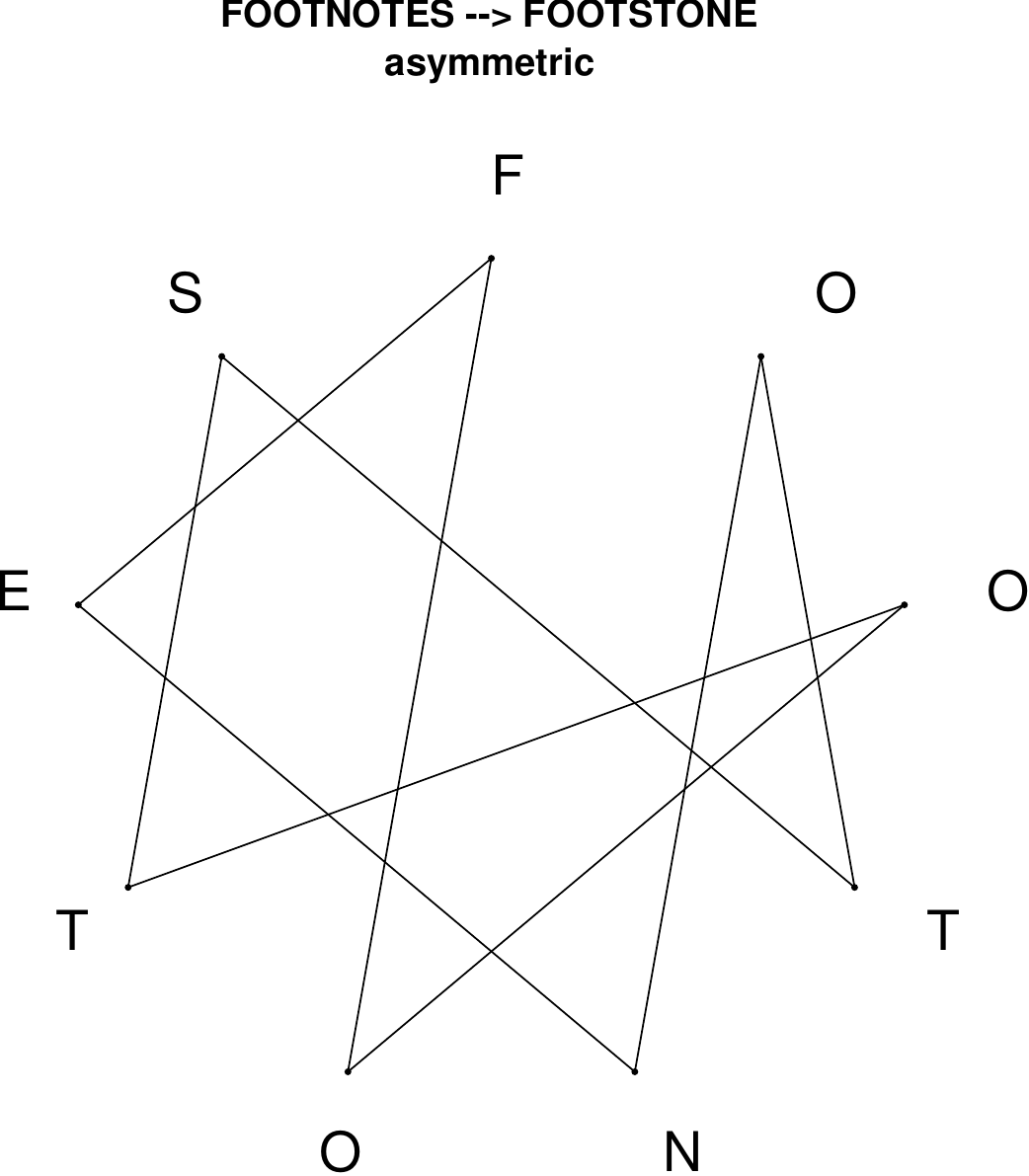}
\end{subfigure}
\hfill
\begin{subfigure}[T]{0.19\textwidth}
\centering
\includegraphics[width=\textwidth]{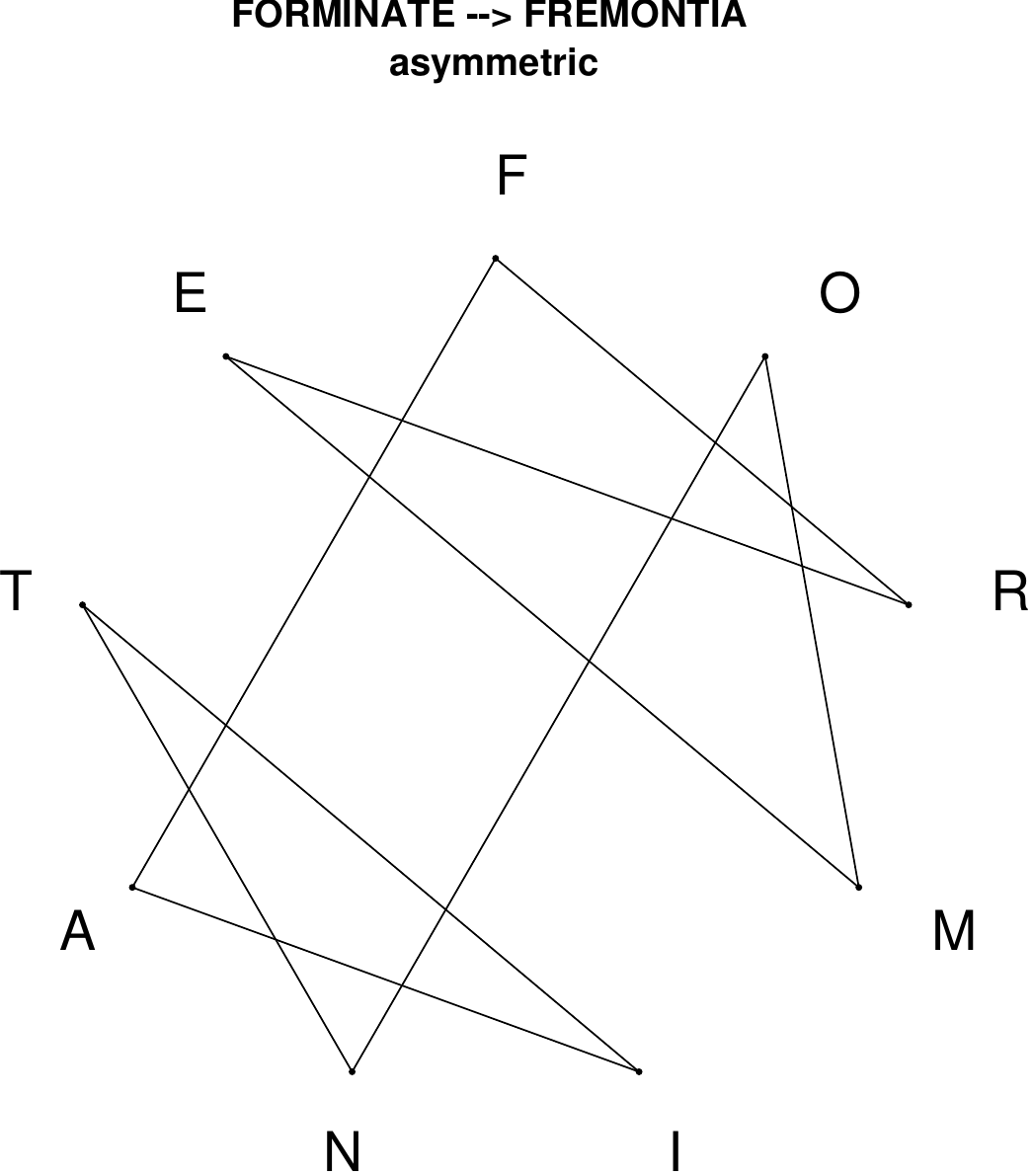}
\end{subfigure}
\hfill
\begin{subfigure}[T]{0.19\textwidth}
\centering
\includegraphics[width=\textwidth]{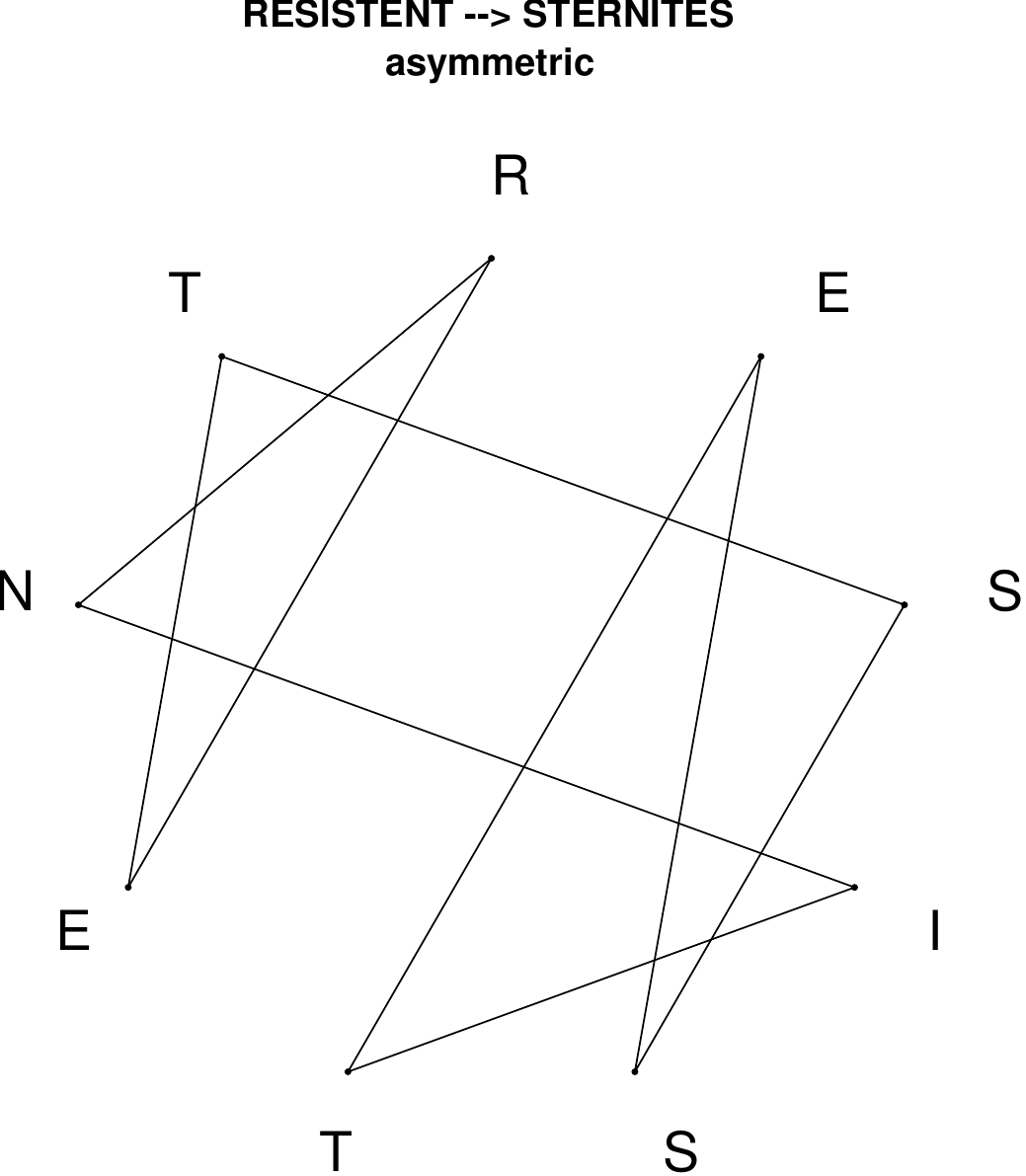}
\end{subfigure}
\hfill
\begin{subfigure}[T]{0.19\textwidth}
\centering
\includegraphics[width=\textwidth]{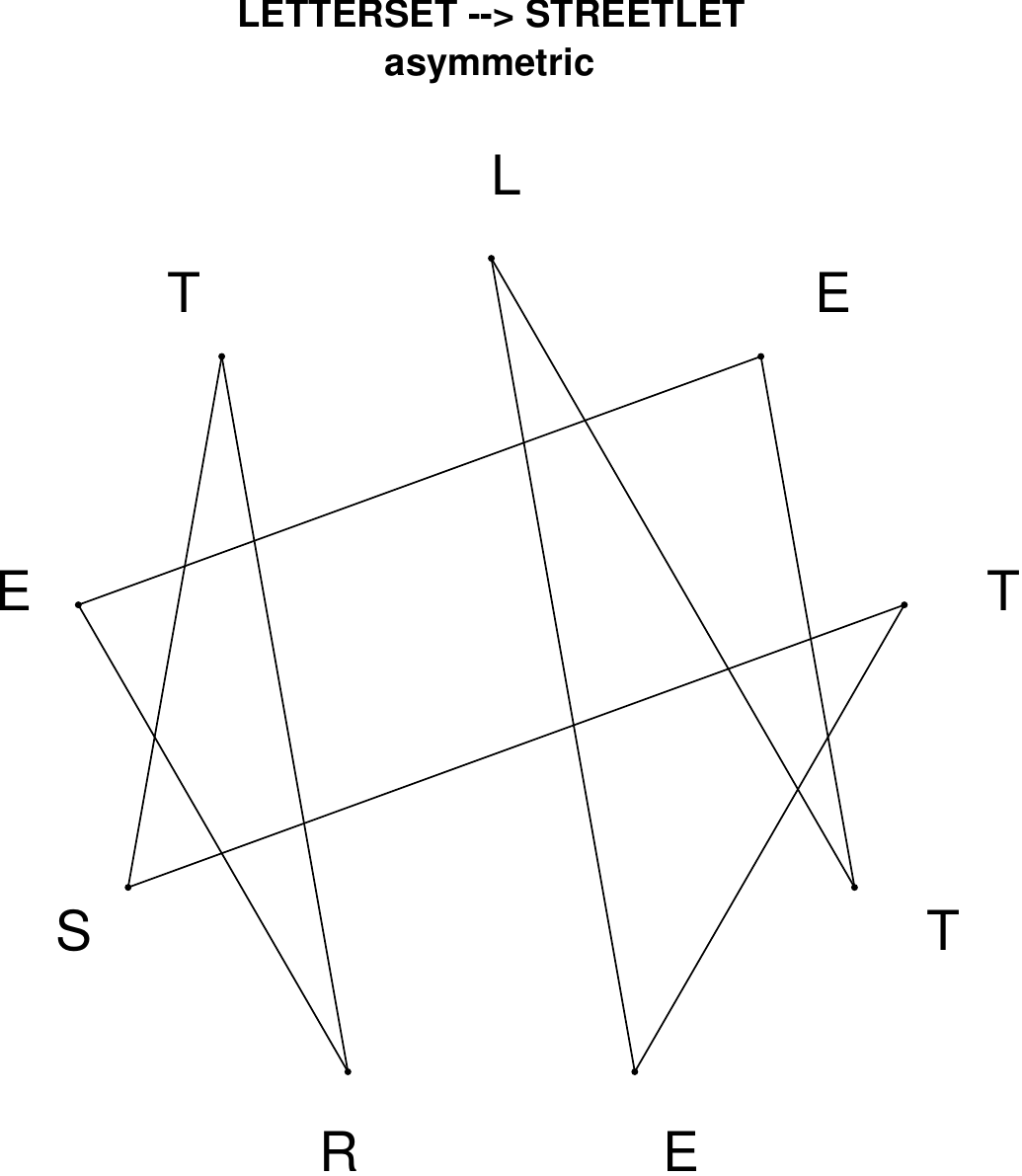}
\end{subfigure}
\end{figure}

\begin{figure}[H]
\centering
\begin{subfigure}[T]{0.19\textwidth}
\centering
\includegraphics[width=\textwidth]{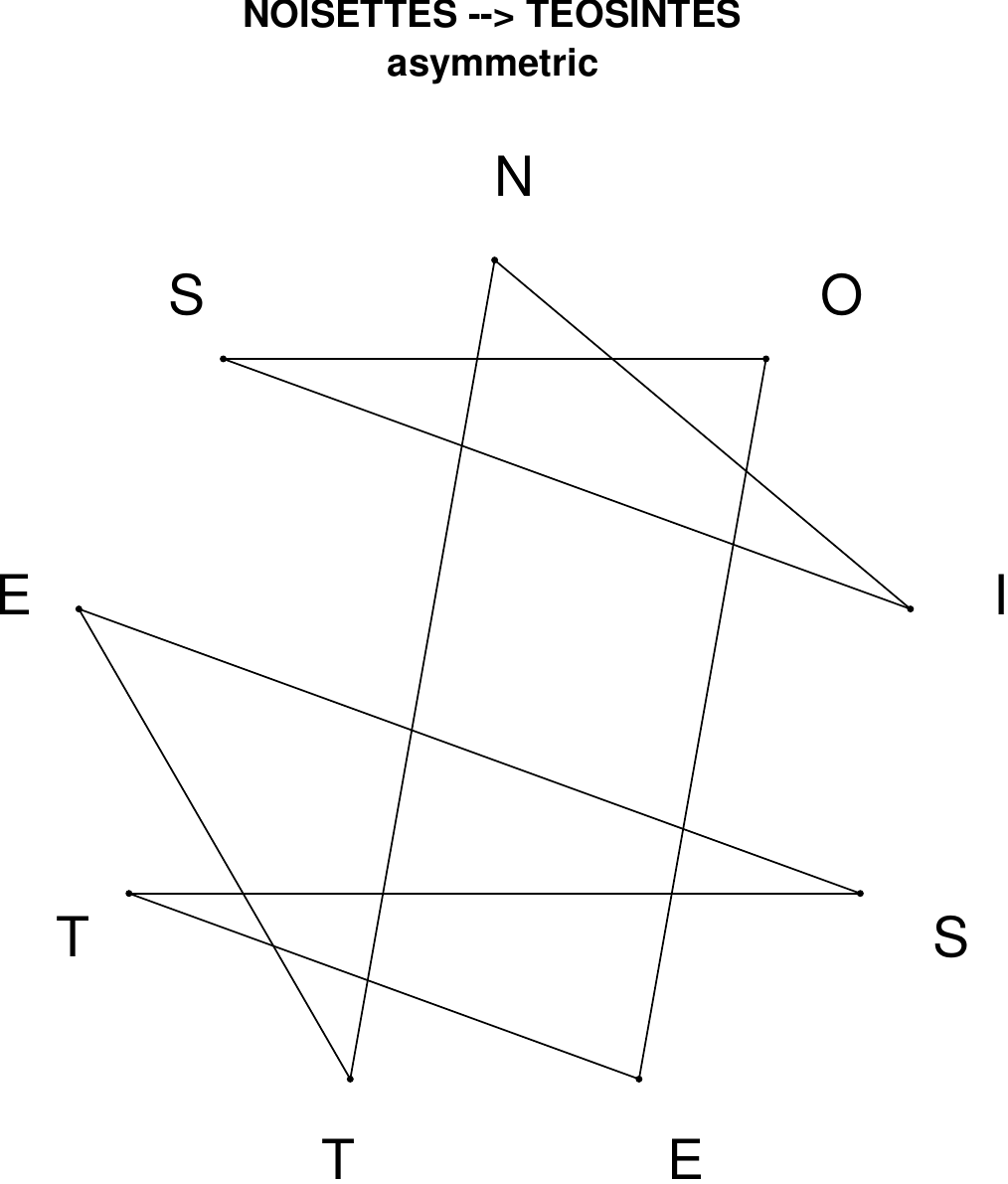}
\end{subfigure}
\hfill
\begin{subfigure}[T]{0.19\textwidth}
\centering
\includegraphics[width=\textwidth]{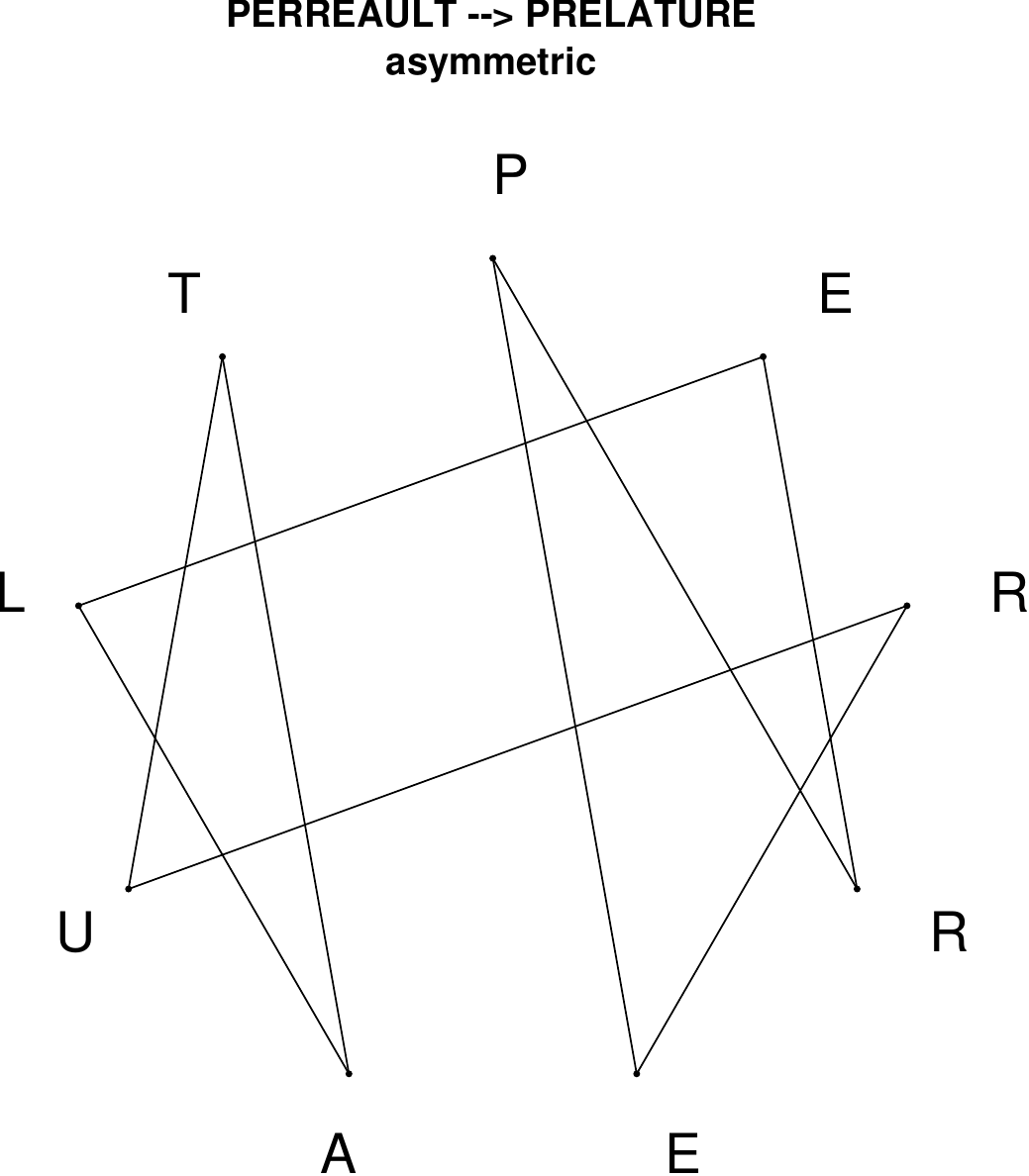}
\end{subfigure}
\hfill
\begin{subfigure}[T]{0.19\textwidth}
\centering
\includegraphics[width=\textwidth]{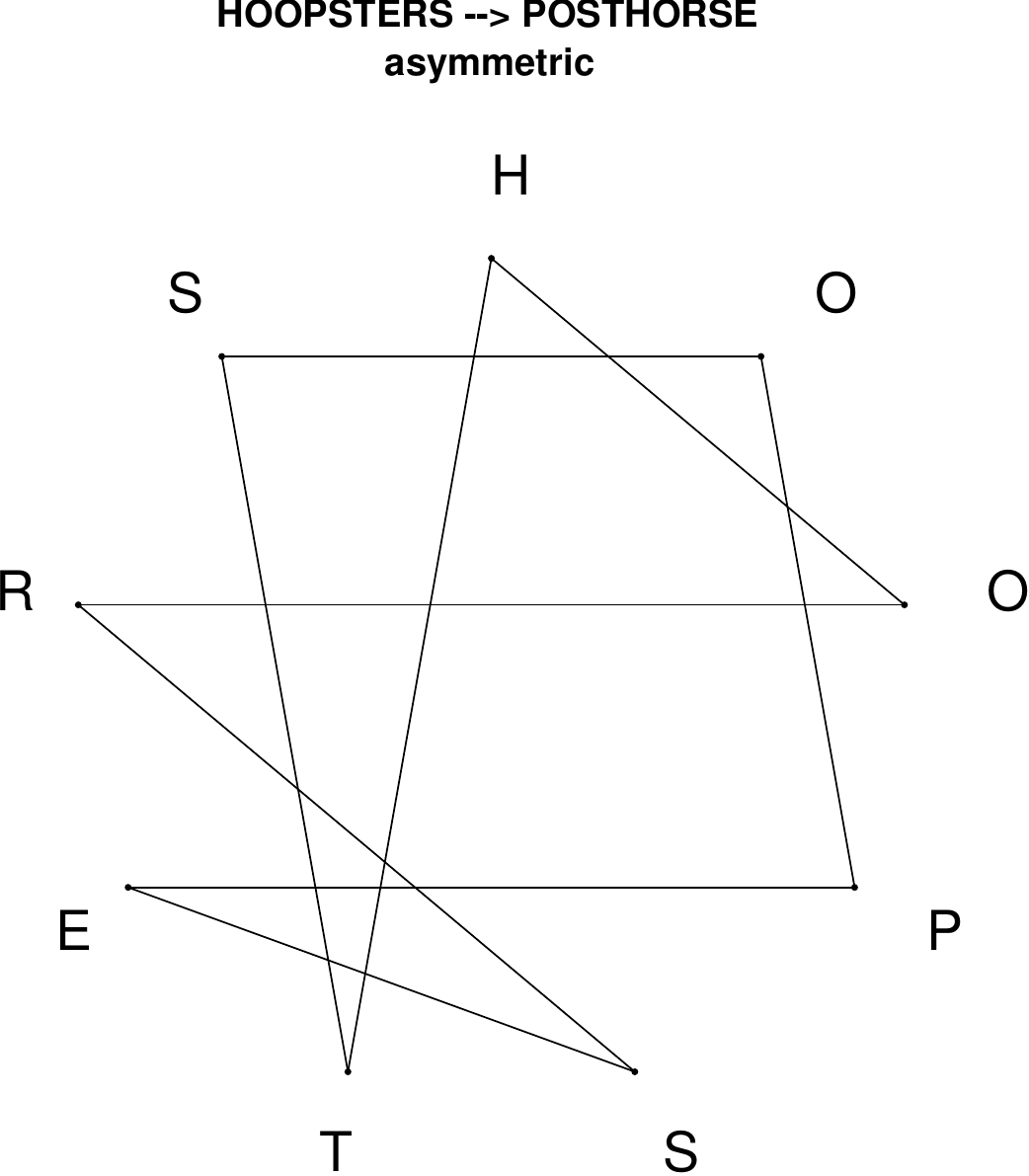}
\end{subfigure}
\hfill
\begin{subfigure}[T]{0.19\textwidth}
\centering
\includegraphics[width=\textwidth]{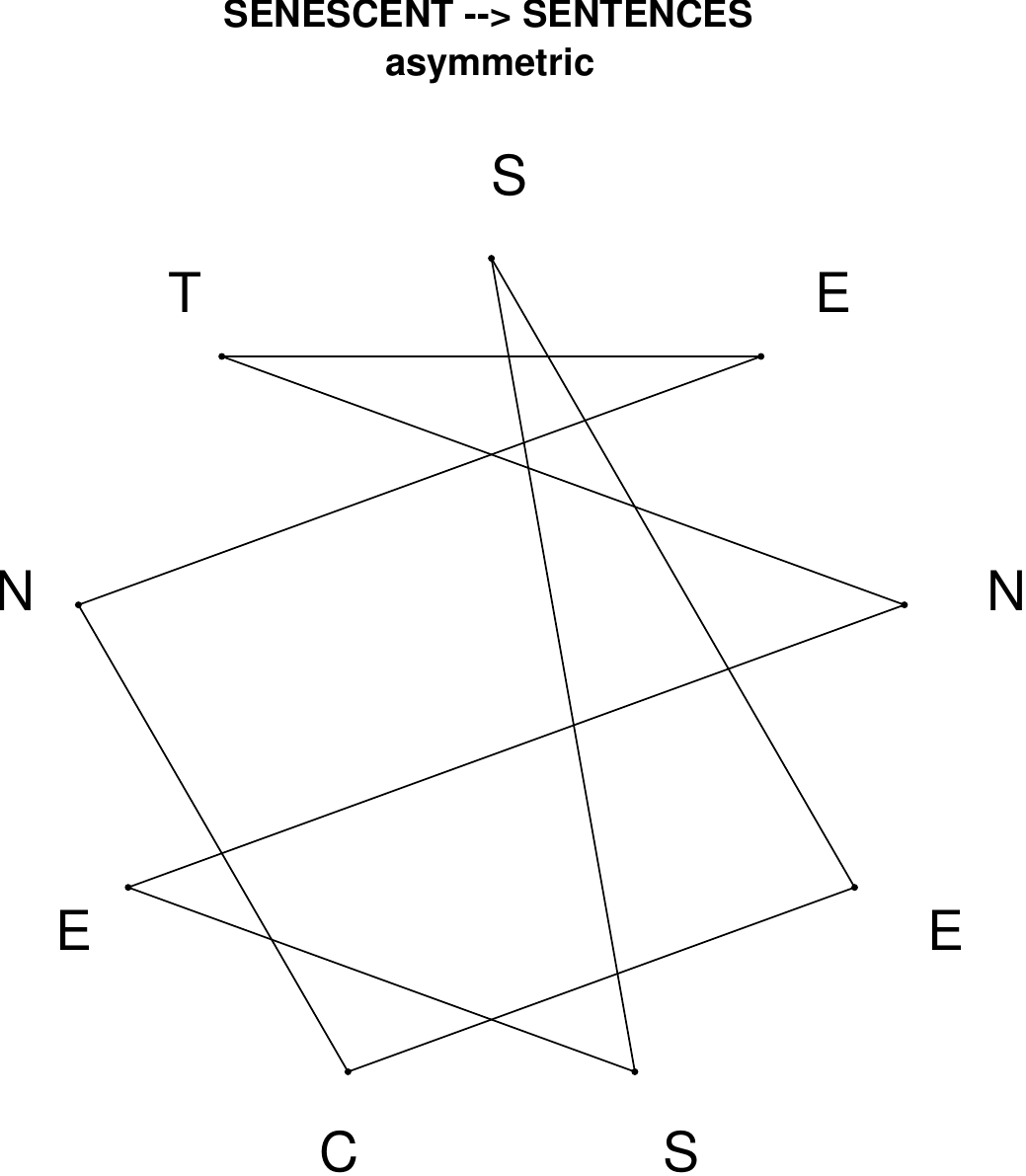}
\end{subfigure}
\hfill
\begin{subfigure}[T]{0.19\textwidth}
\centering
\includegraphics[width=\textwidth]{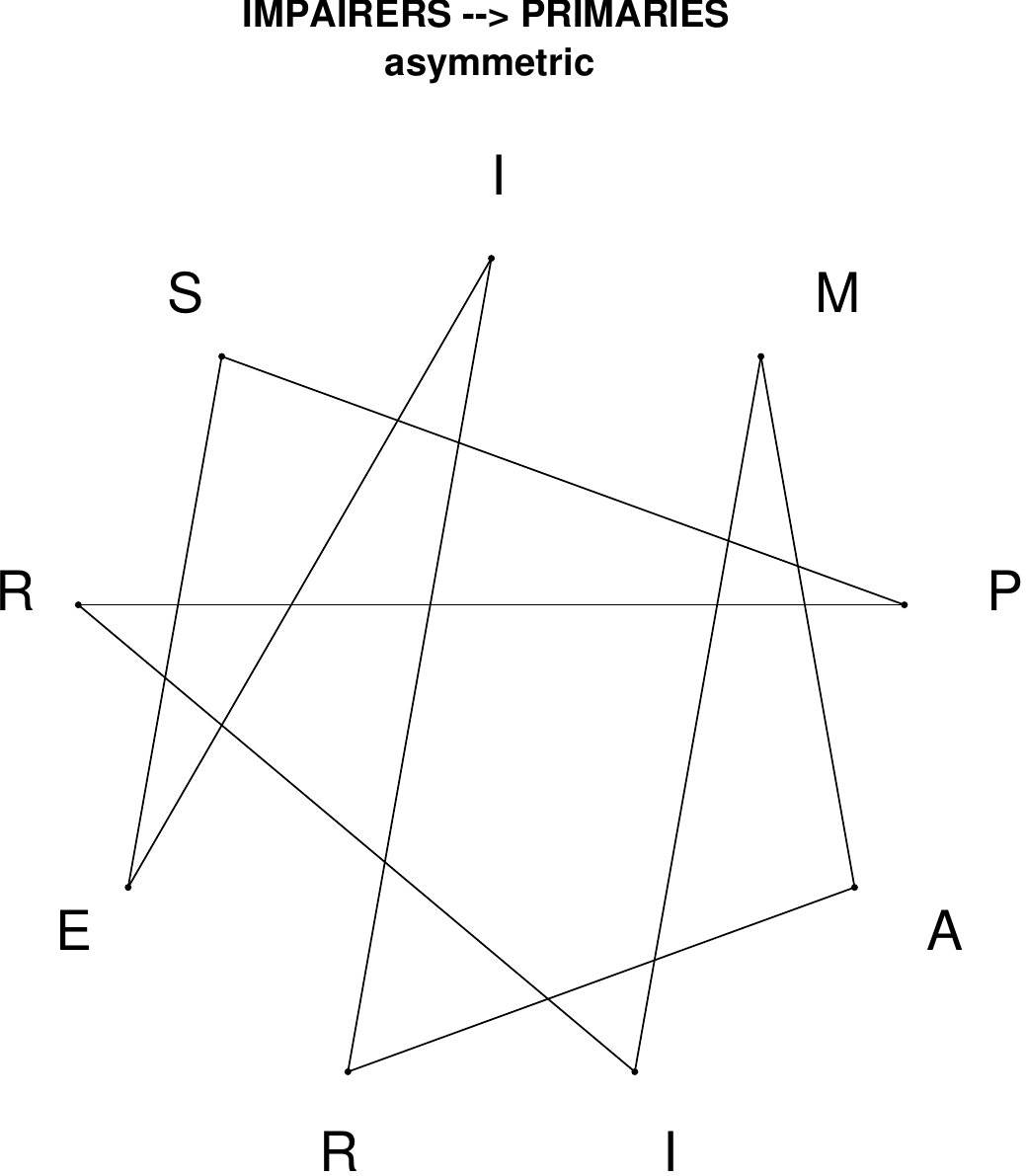}
\end{subfigure}
\end{figure}

\begin{figure}[H]
\centering
\begin{subfigure}[T]{0.19\textwidth}
\centering
\includegraphics[width=\textwidth]{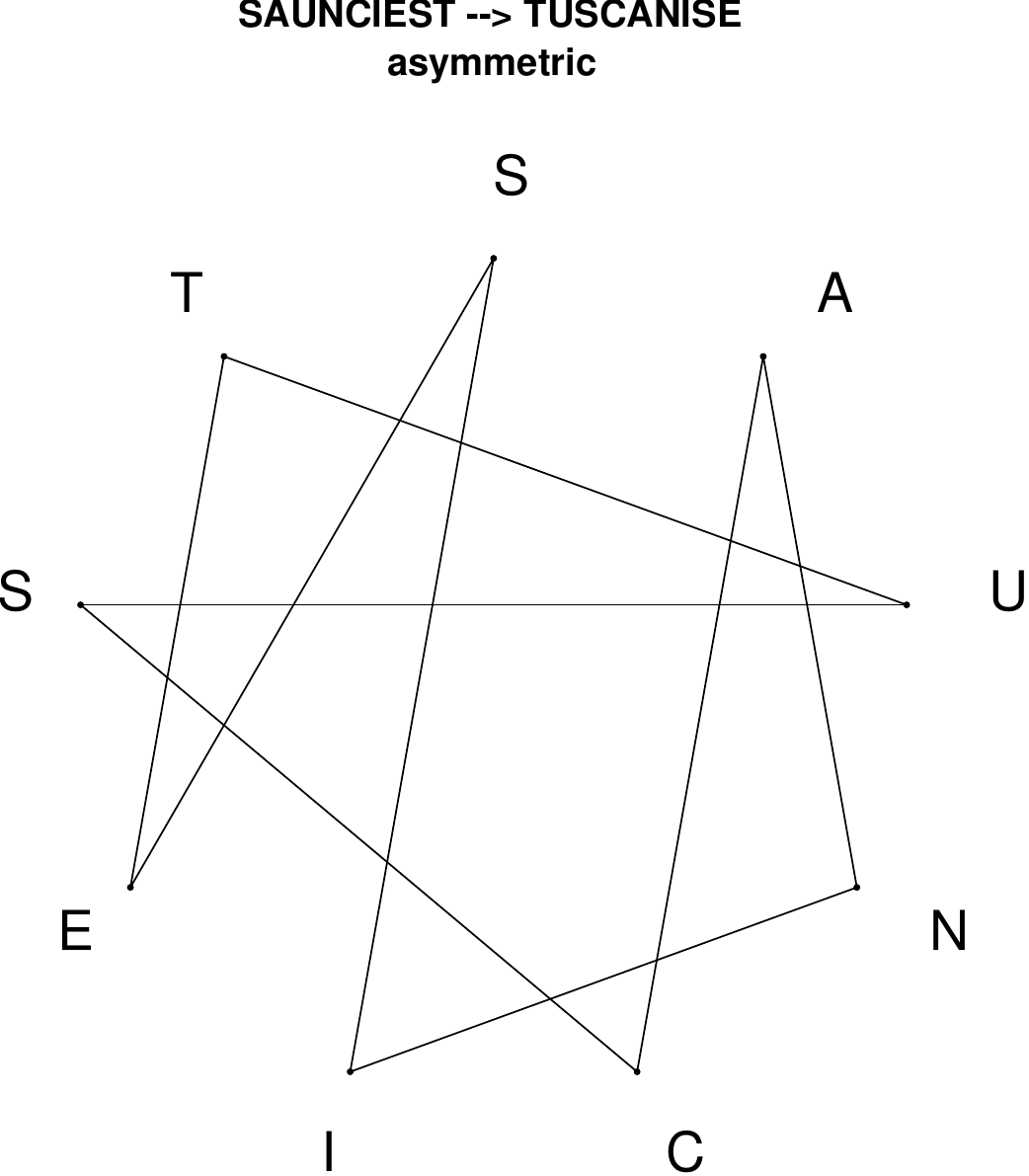}
\end{subfigure}
\hfill
\begin{subfigure}[T]{0.19\textwidth}
\centering
\includegraphics[width=\textwidth]{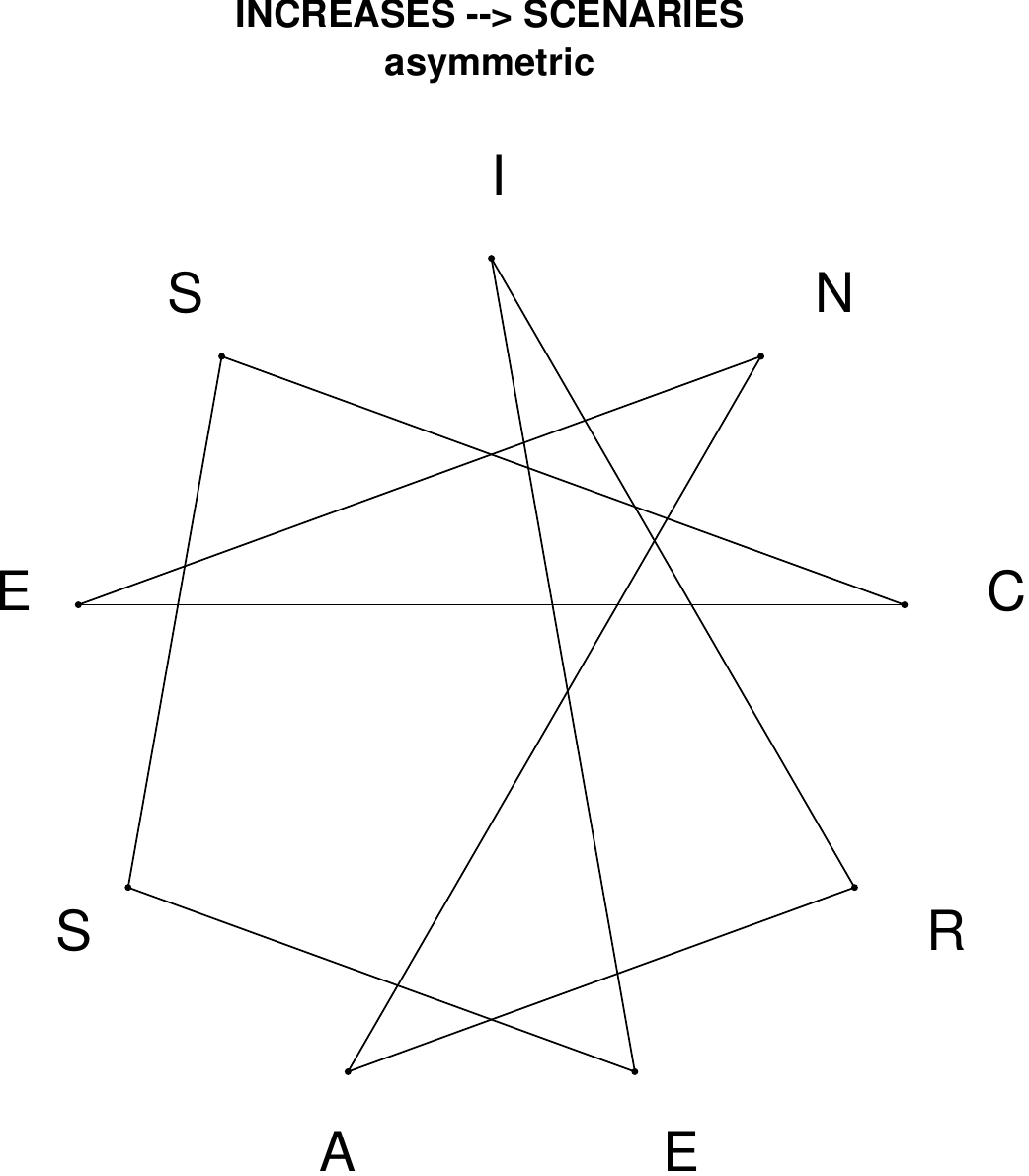}
\end{subfigure}
\hfill
\begin{subfigure}[T]{0.19\textwidth}
\centering
\includegraphics[width=\textwidth]{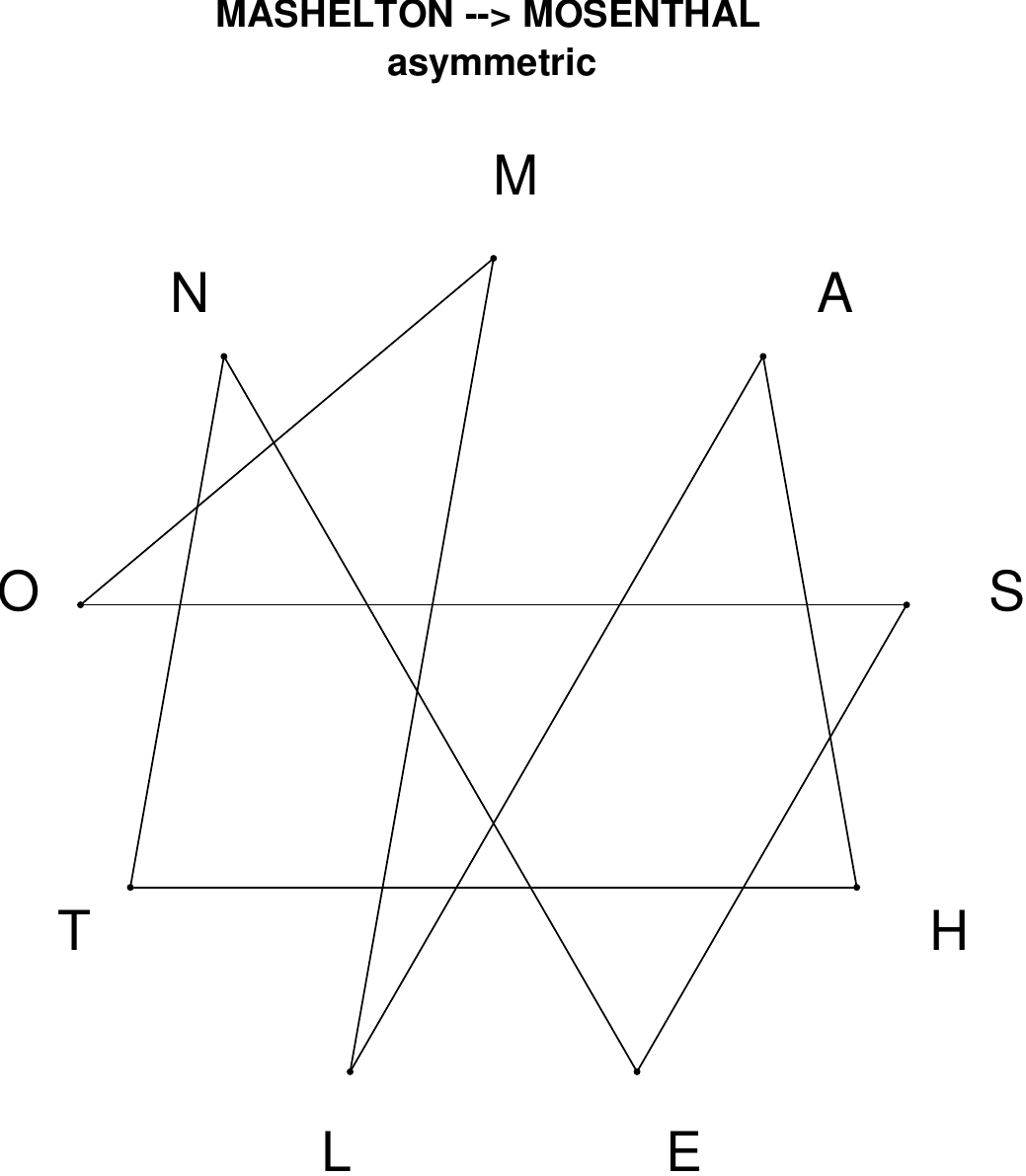}
\end{subfigure}
\hfill
\begin{subfigure}[T]{0.19\textwidth}
\centering
\includegraphics[width=\textwidth]{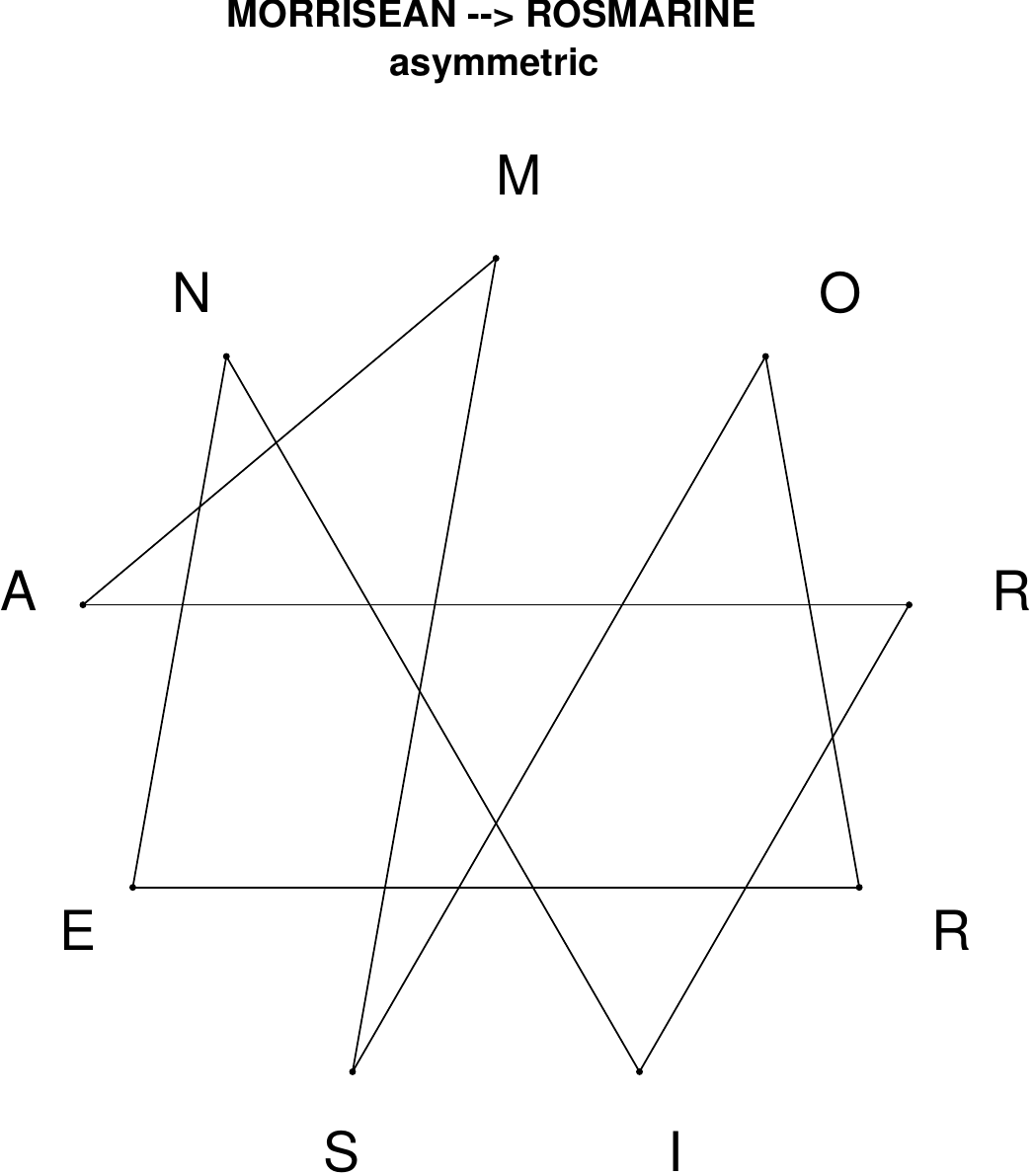}
\end{subfigure}
\hfill
\begin{subfigure}[T]{0.19\textwidth}
\centering
\includegraphics[width=\textwidth]{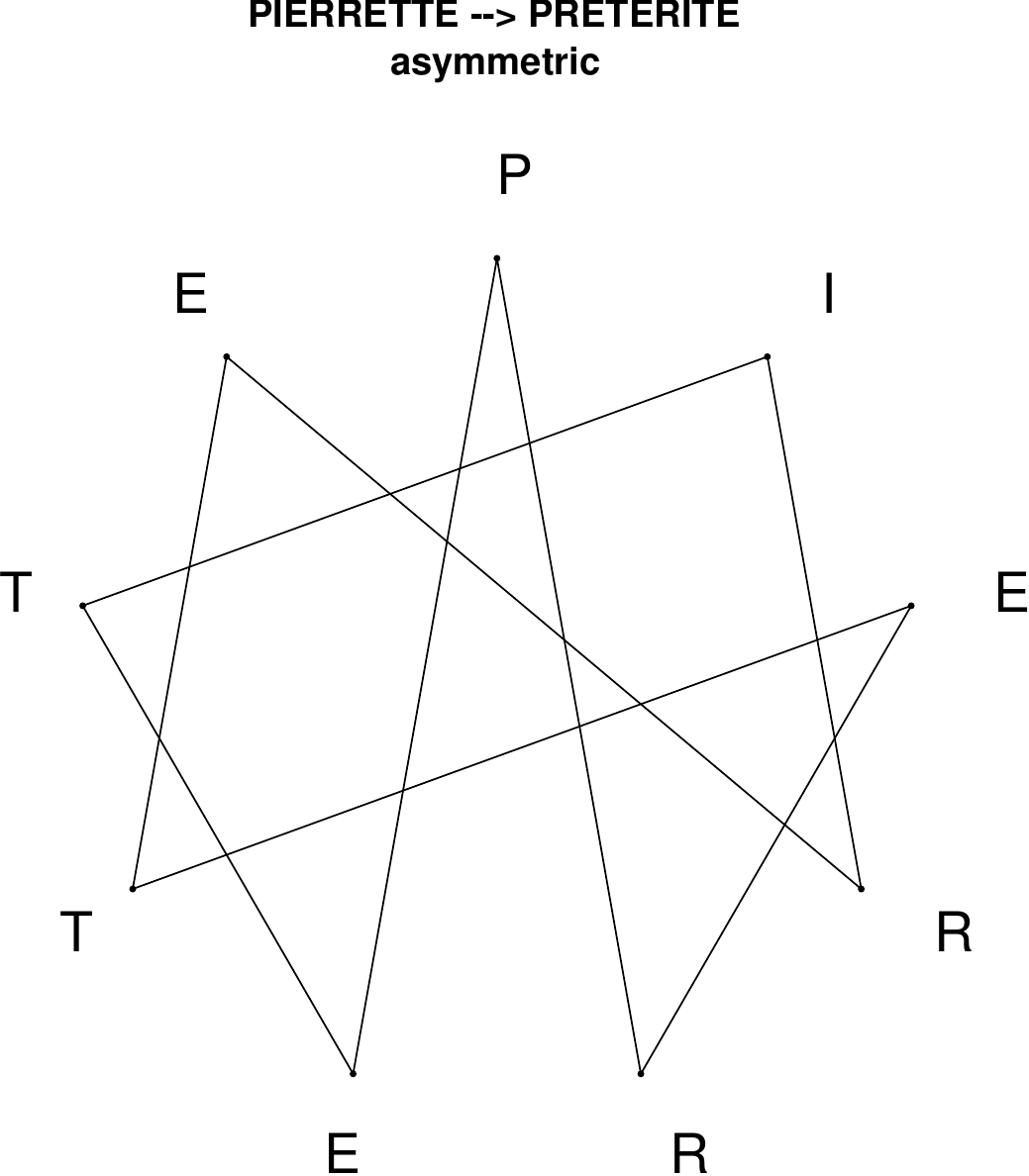}
\end{subfigure}
\end{figure}

\begin{figure}[H]
\centering
\begin{subfigure}[T]{0.19\textwidth}
\centering
\includegraphics[width=\textwidth]{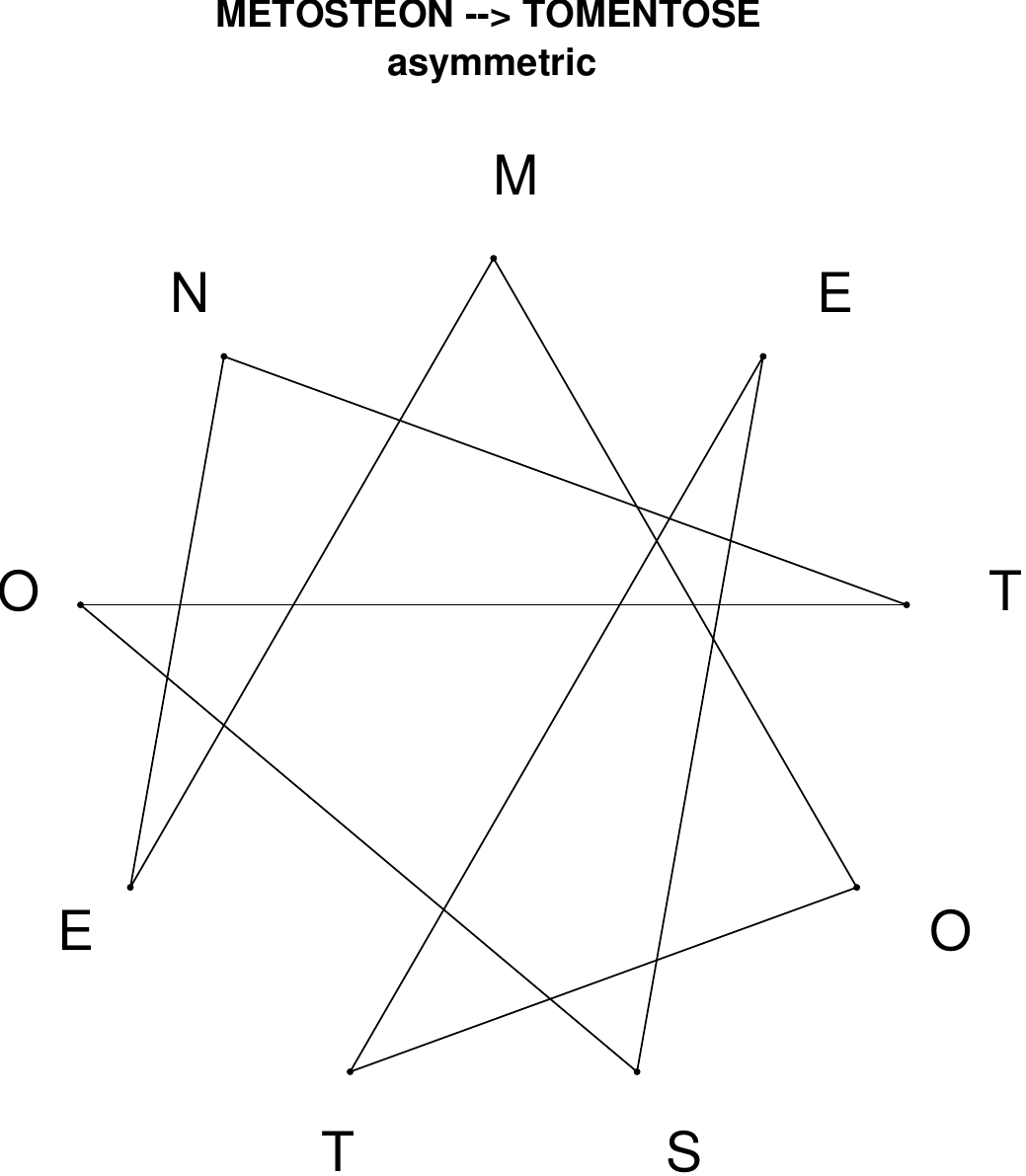}
\end{subfigure}
\hfill
\begin{subfigure}[T]{0.19\textwidth}
\centering
\includegraphics[width=\textwidth]{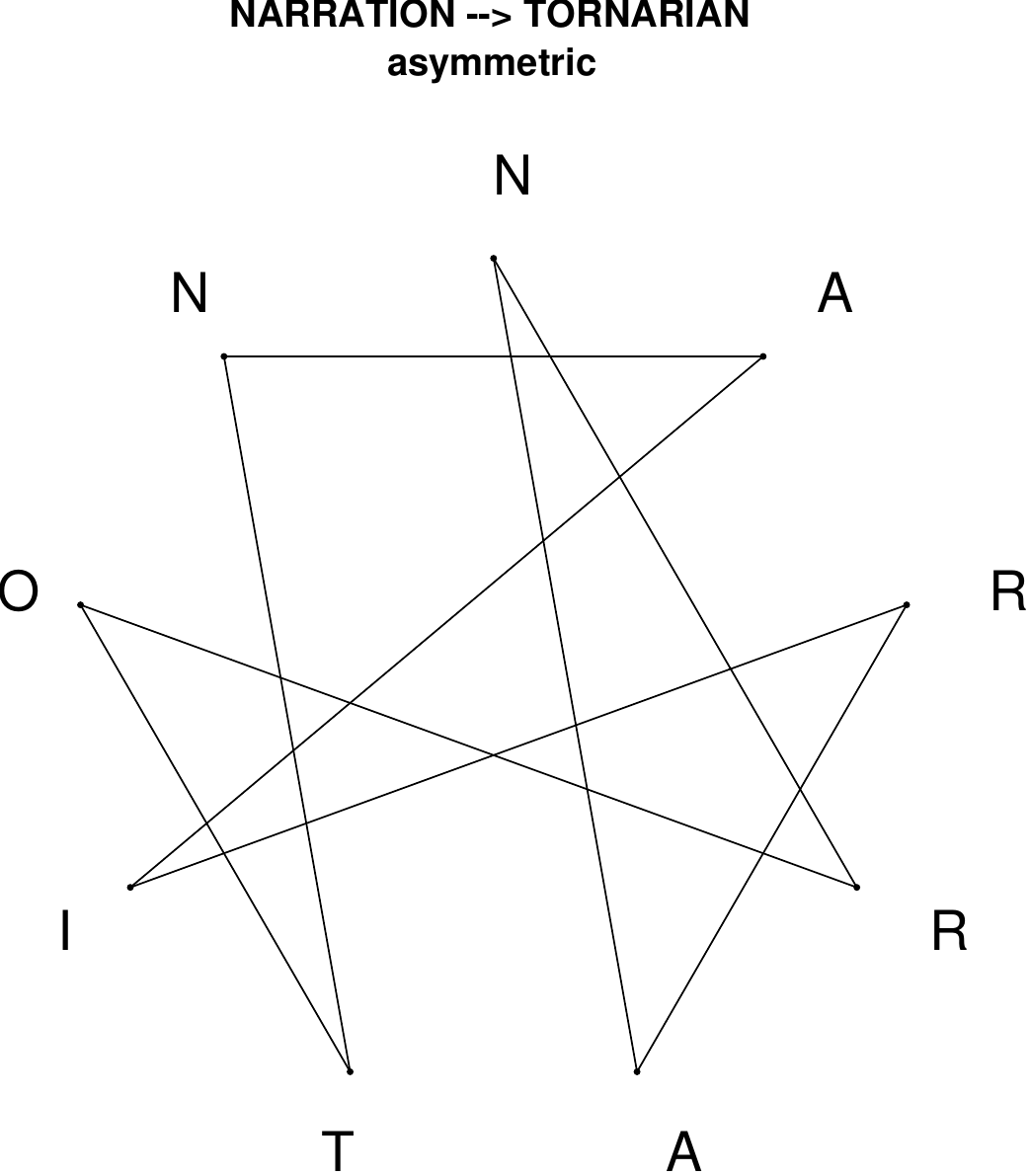}
\end{subfigure}
\hfill
\begin{subfigure}[T]{0.19\textwidth}
\centering
\includegraphics[width=\textwidth]{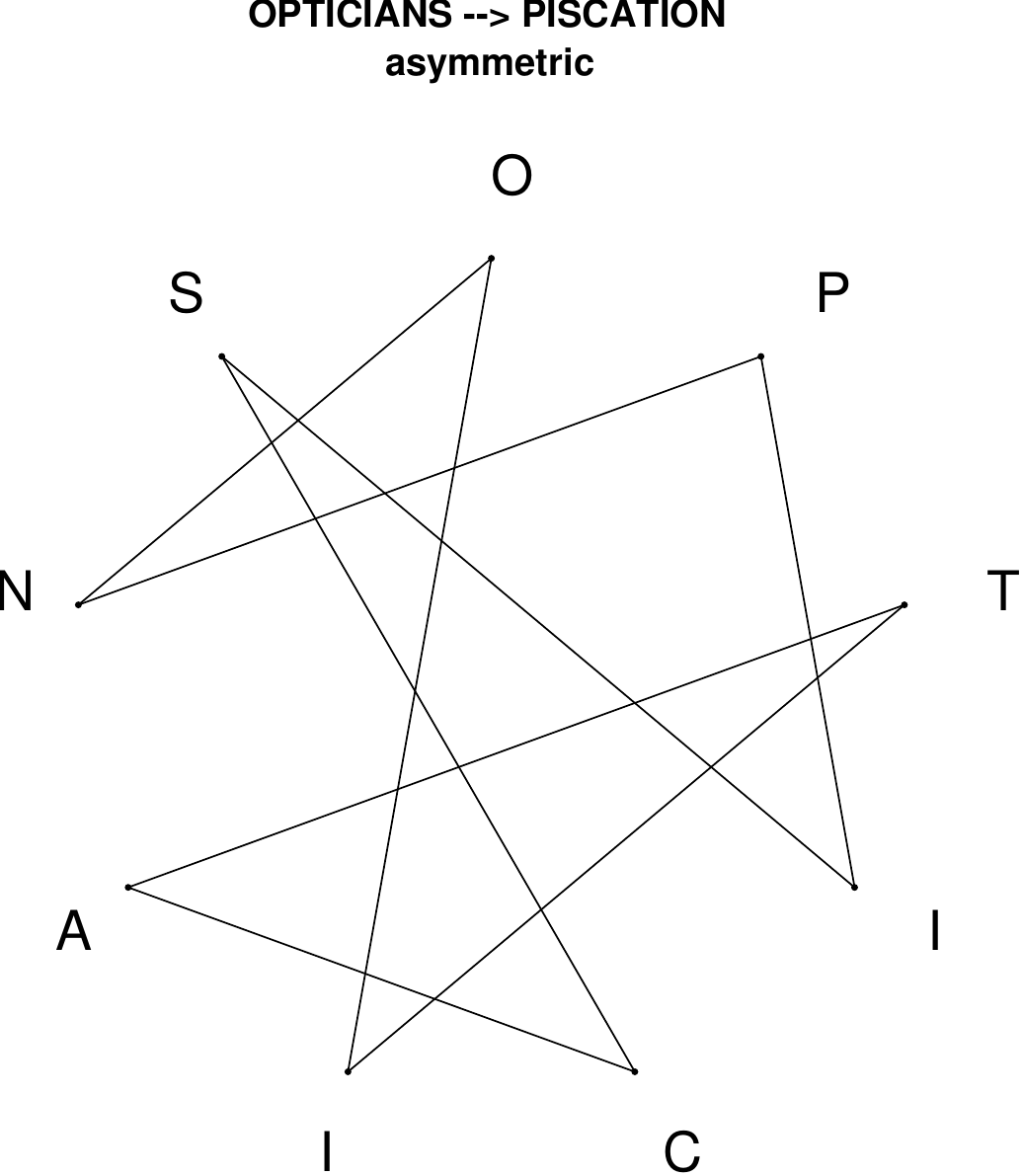}
\end{subfigure}
\hfill
\begin{subfigure}[T]{0.19\textwidth}
\centering
\includegraphics[width=\textwidth]{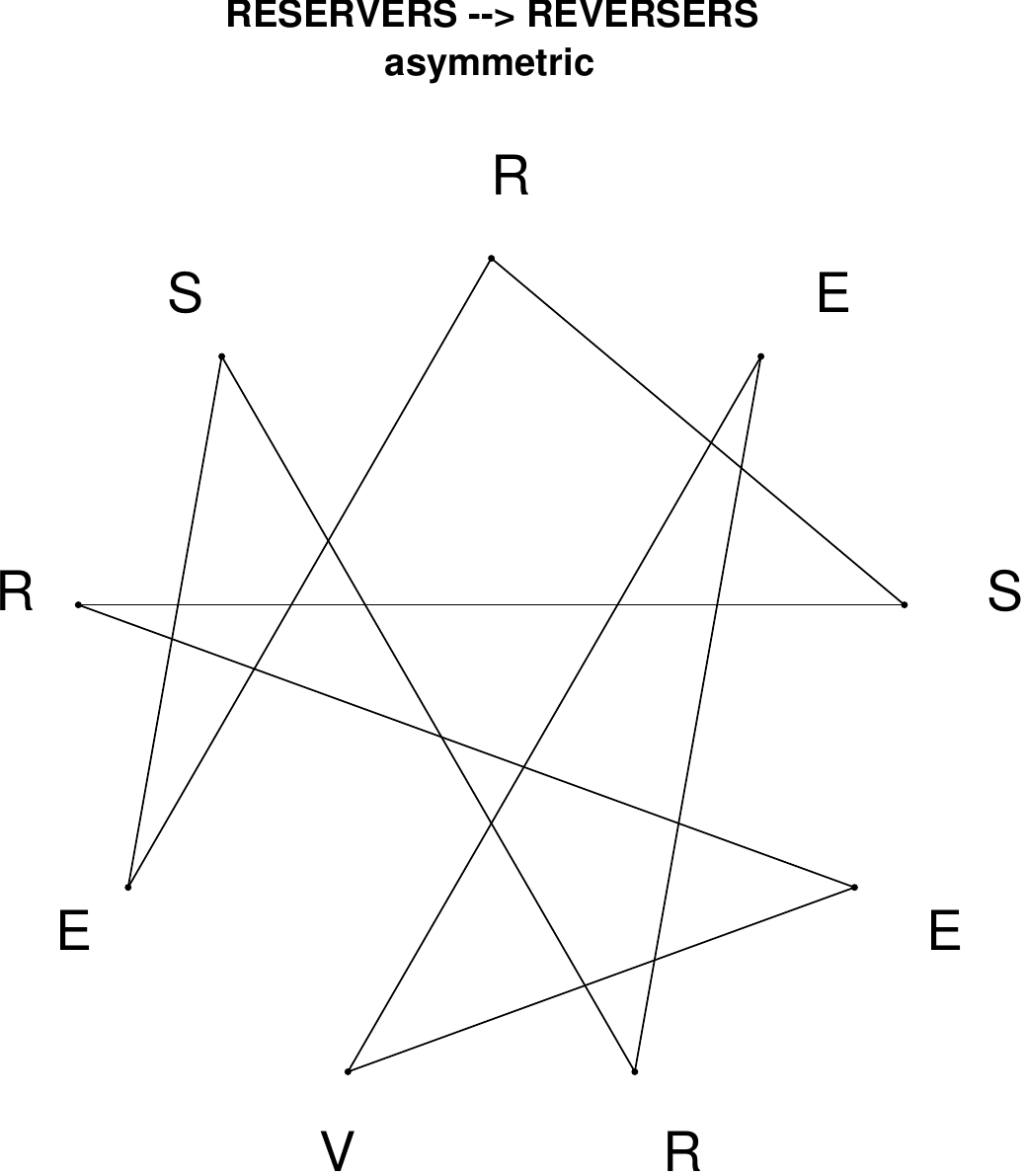}
\end{subfigure}
\hfill
\begin{subfigure}[T]{0.19\textwidth}
\centering
\includegraphics[width=\textwidth]{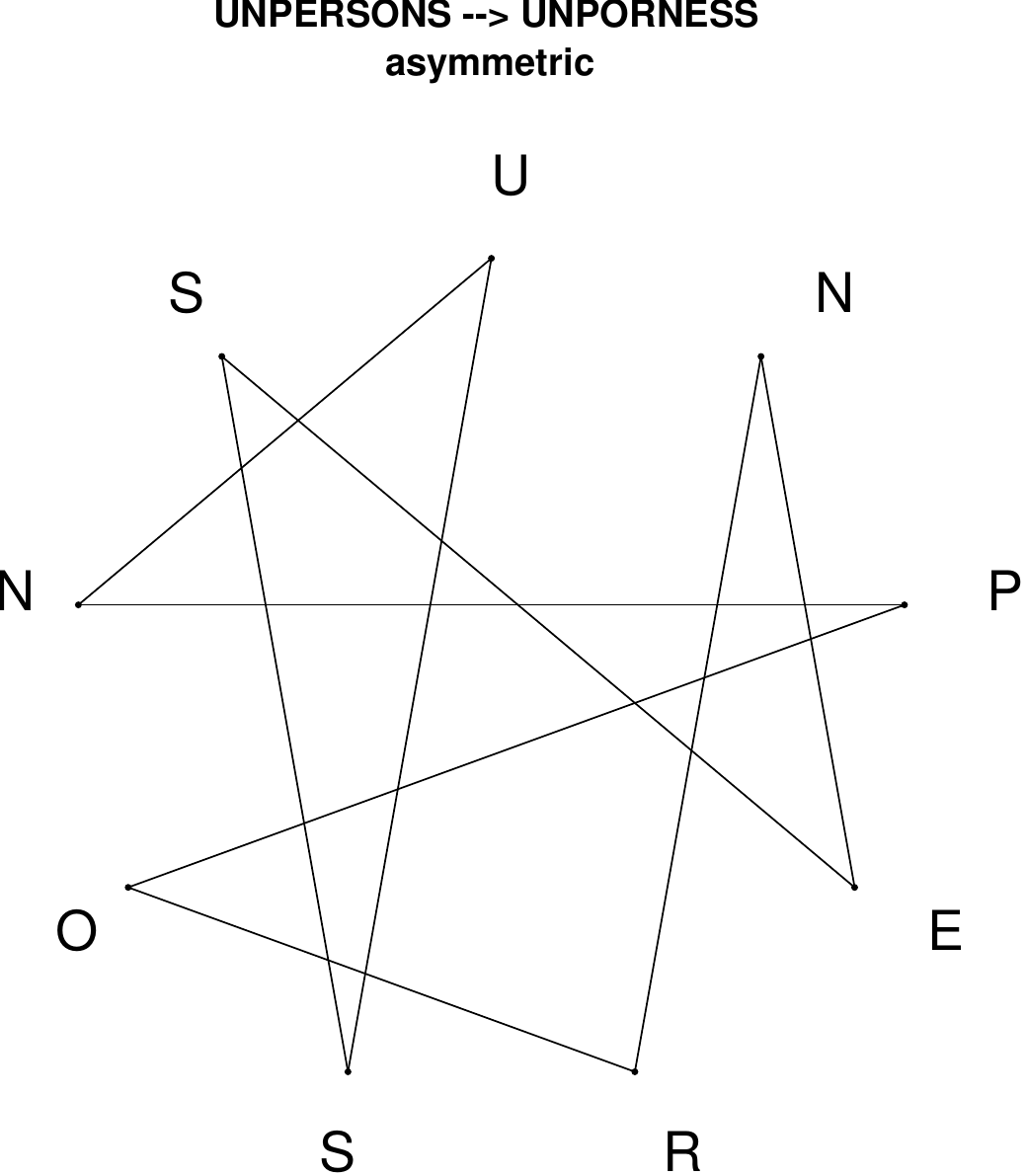}
\end{subfigure}
\end{figure}

\begin{figure}[H]
\centering
\begin{subfigure}[T]{0.19\textwidth}
\centering
\includegraphics[width=\textwidth]{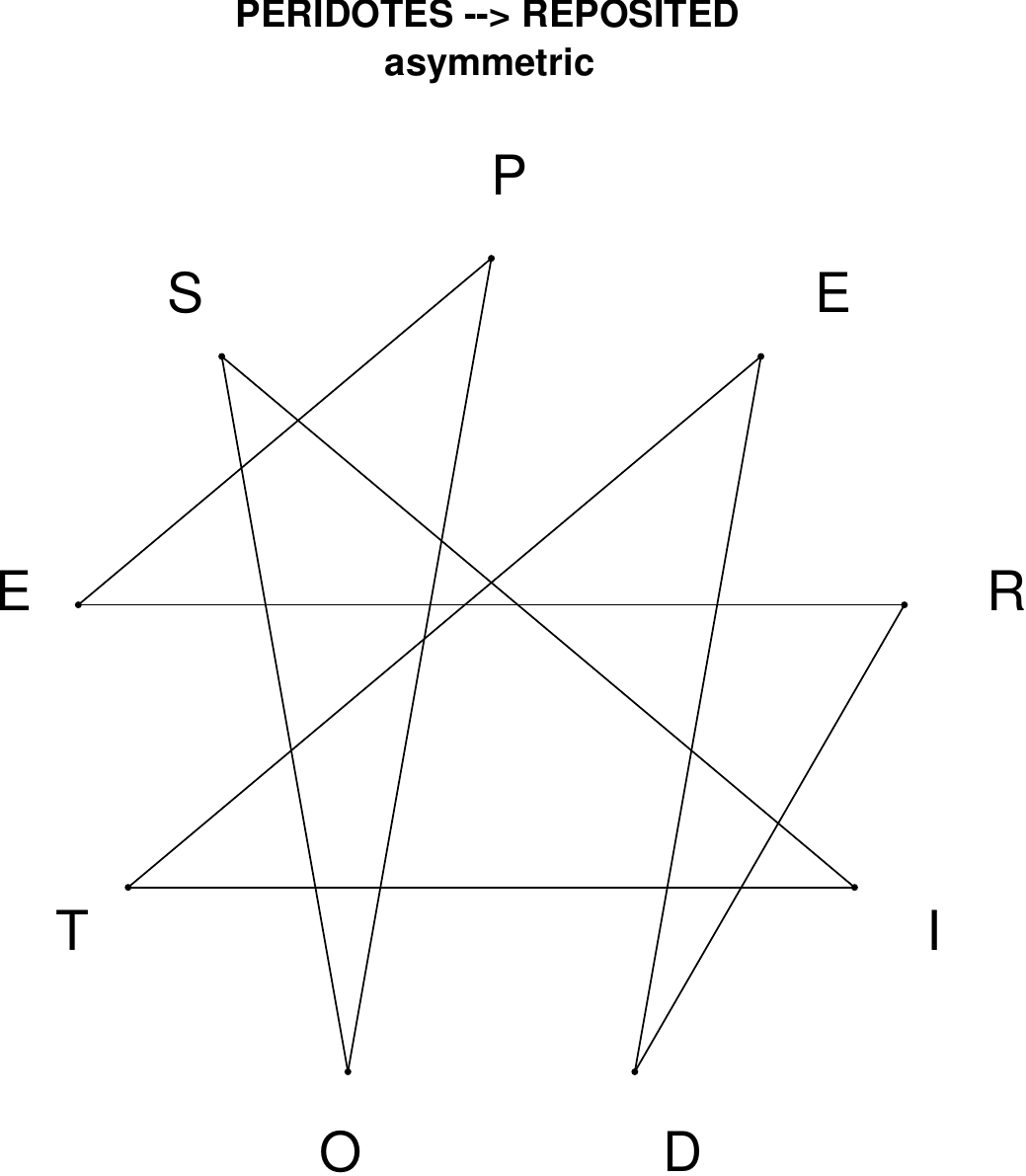}
\end{subfigure}
\hfill
\begin{subfigure}[T]{0.19\textwidth}
\centering
\includegraphics[width=\textwidth]{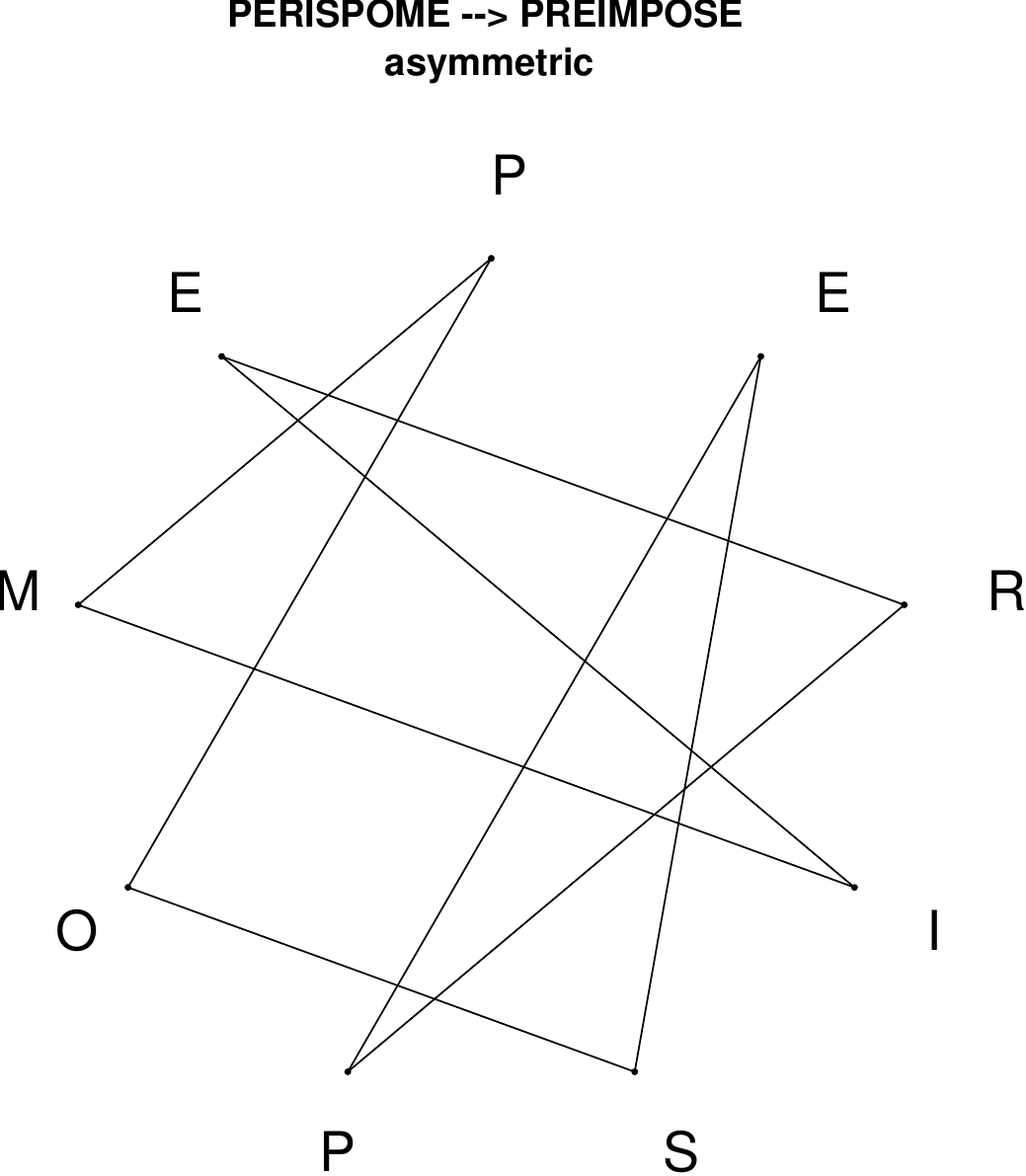}
\end{subfigure}
\hfill
\begin{subfigure}[T]{0.19\textwidth}
\centering
\includegraphics[width=\textwidth]{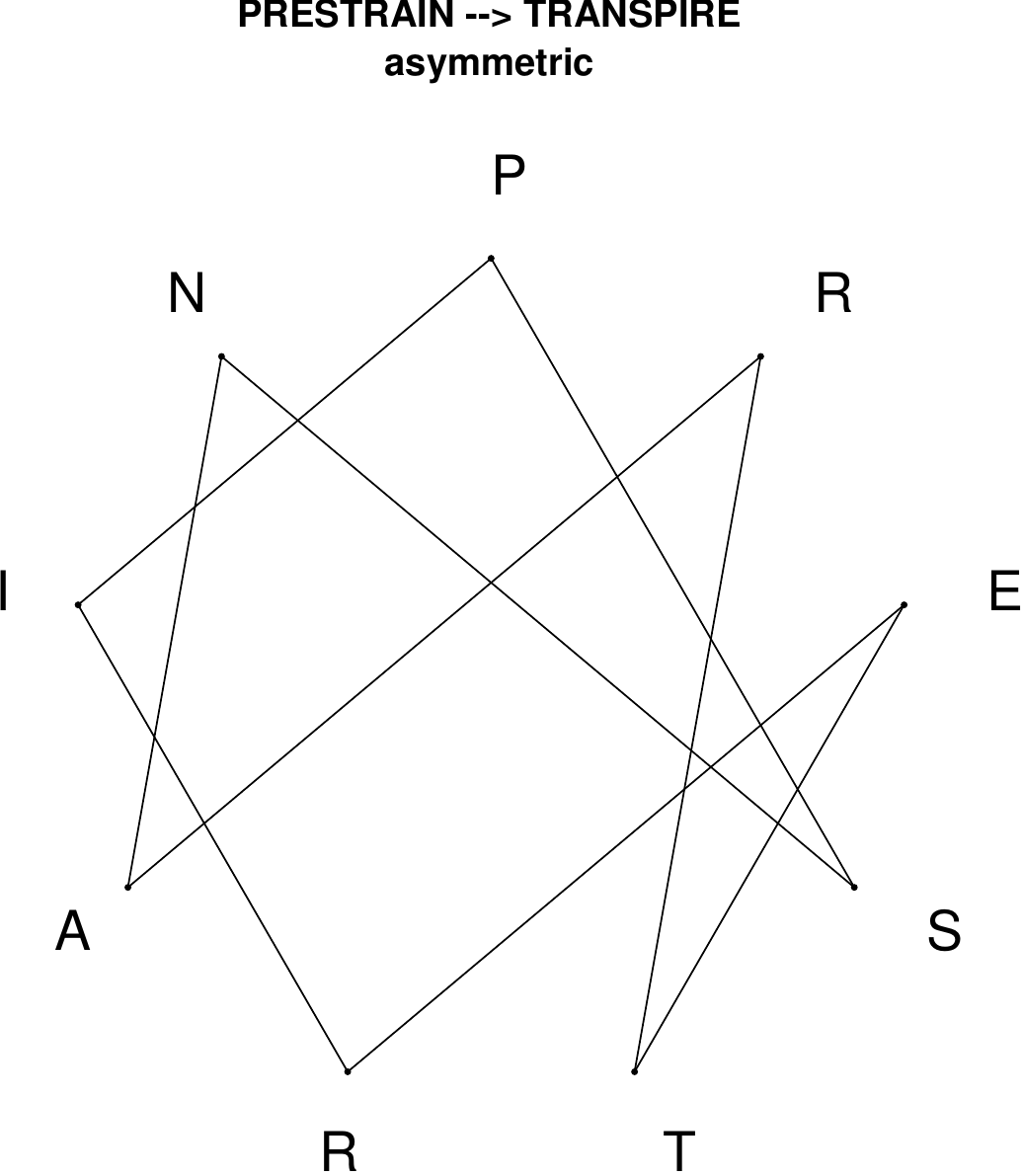}
\end{subfigure}
\hfill
\begin{subfigure}[T]{0.19\textwidth}
\centering
\includegraphics[width=\textwidth]{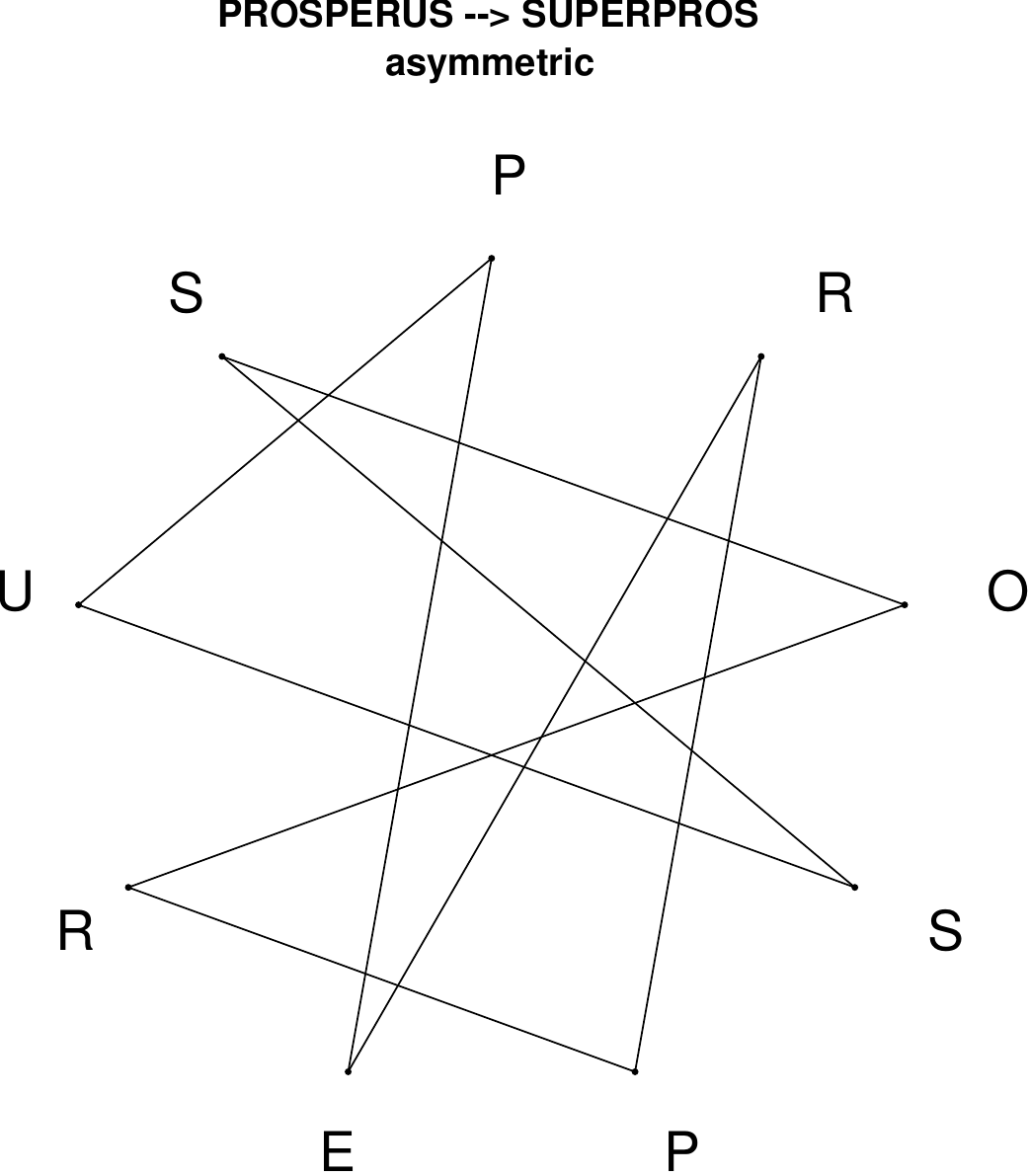}
\end{subfigure}
\hfill
\begin{subfigure}[T]{0.19\textwidth}
\centering
\includegraphics[width=\textwidth]{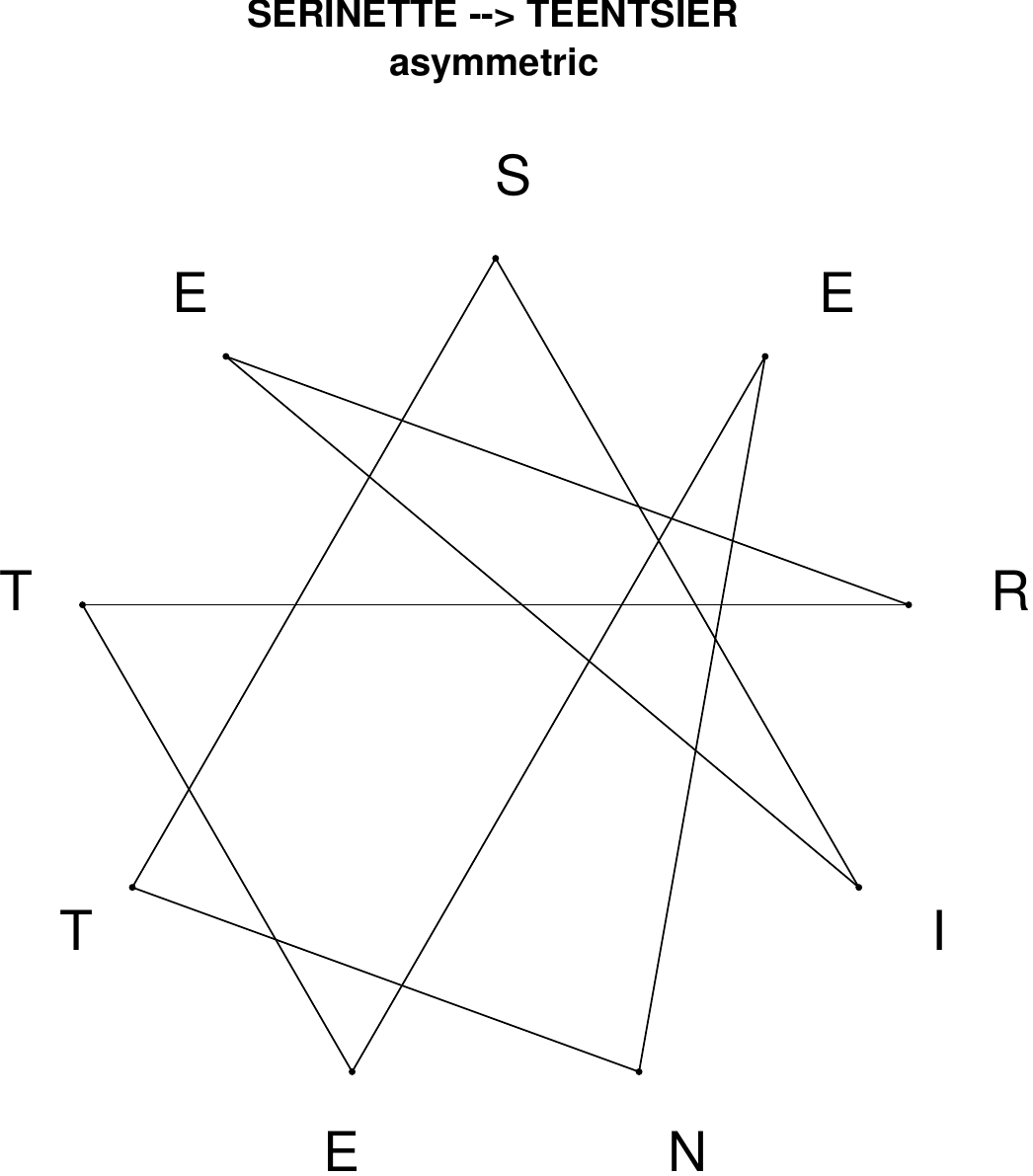}
\end{subfigure}
\end{figure}

\begin{figure}[H]
\centering
\begin{subfigure}[T]{0.19\textwidth}
\centering
\includegraphics[width=\textwidth]{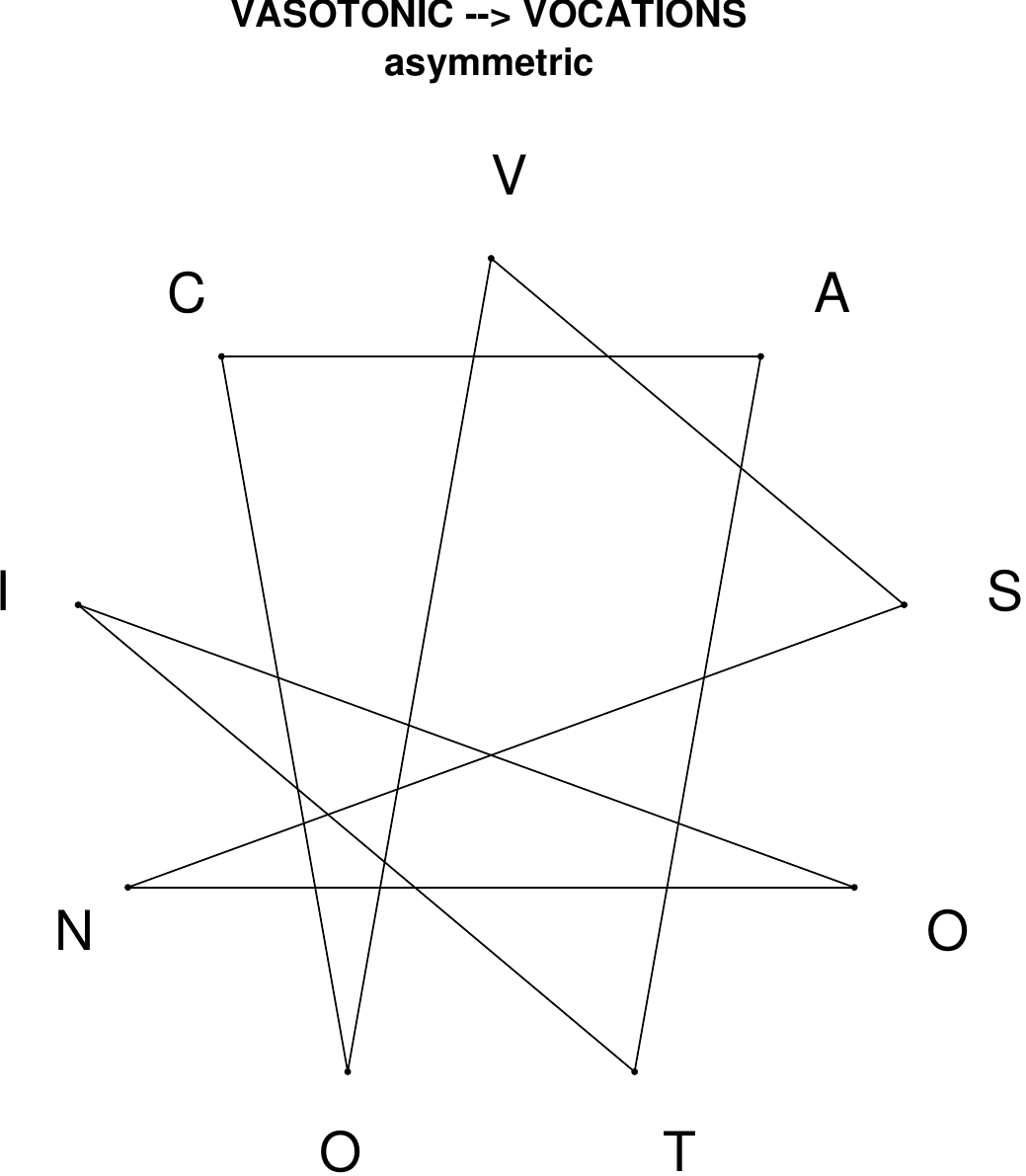}
\end{subfigure}
\hfill
\begin{subfigure}[T]{0.19\textwidth}
\centering
\includegraphics[width=\textwidth]{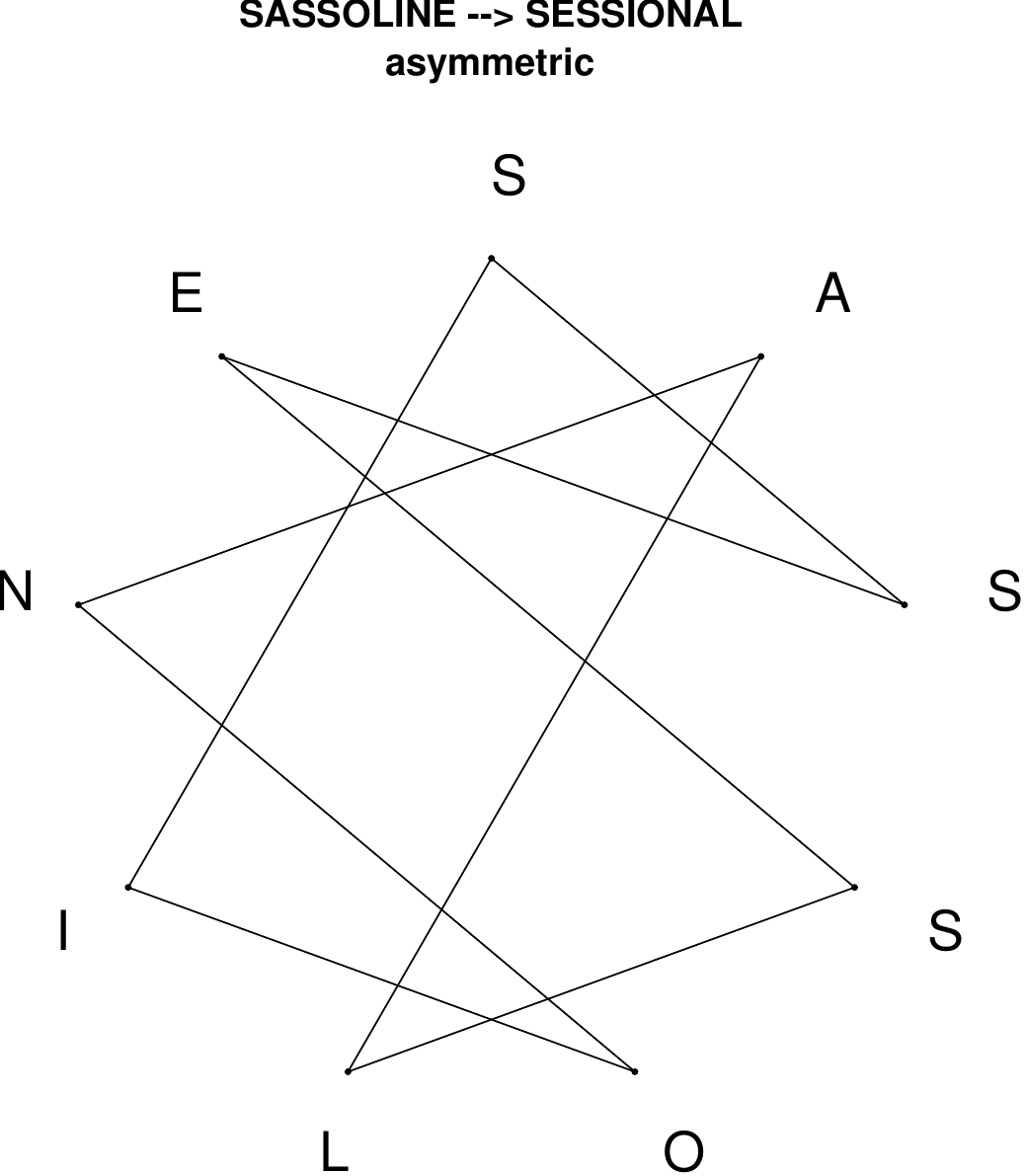}
\end{subfigure}
\hfill
\begin{subfigure}[T]{0.19\textwidth}
\centering
\includegraphics[width=\textwidth]{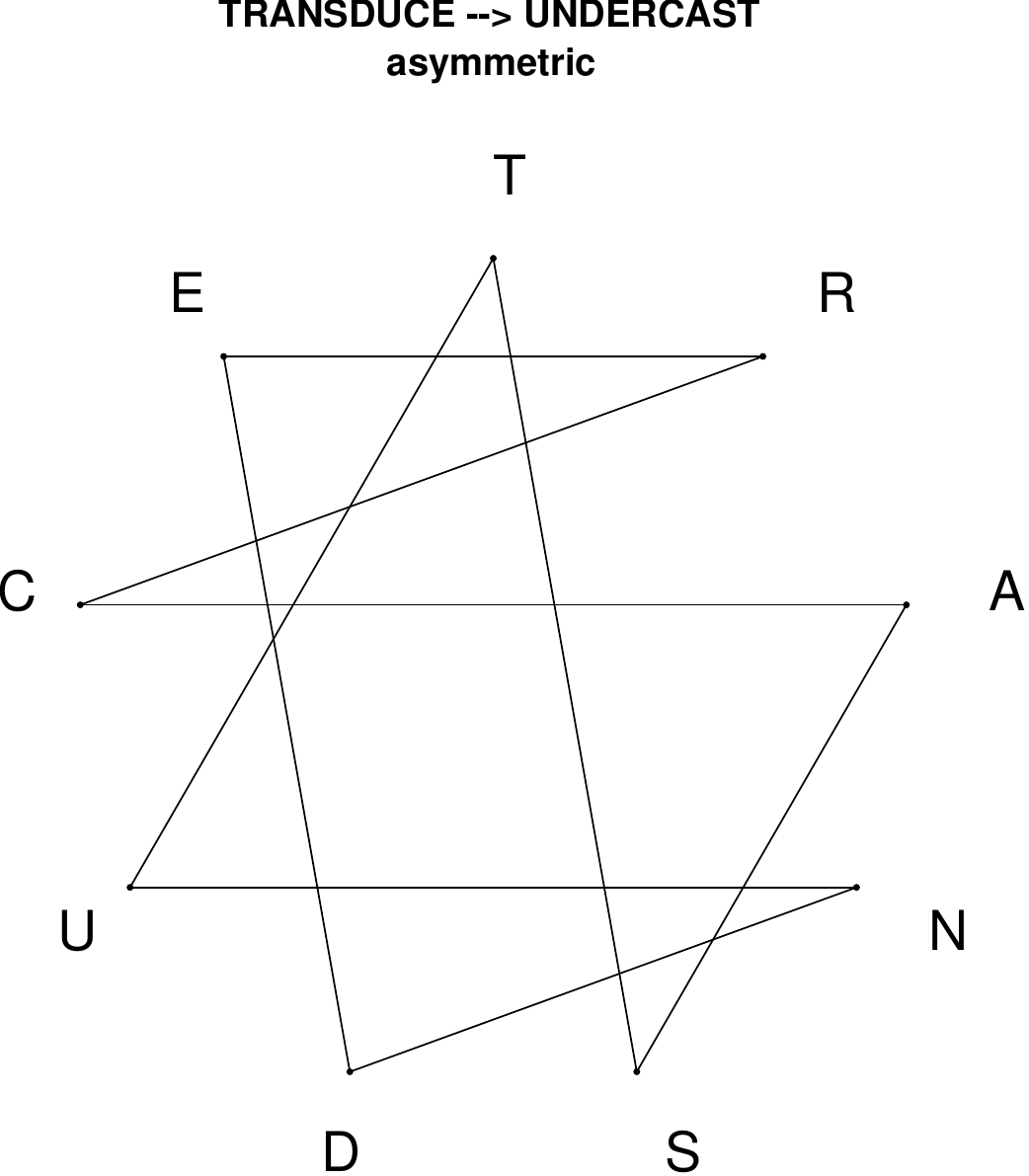}
\end{subfigure}
\hfill
\end{figure}

%%%%%%%%%%%%%%%%%%
\clearpage
\subsection{Star Anagrams $N = 8$}
The group of stars for $N=8$ is one of the richest, with numerous examples from all three classes. 

\subsubsection{Perfect Stars $N=8$}

\begin{figure}[H]
\centering
\begin{subfigure}[T]{0.19\textwidth}
\centering
\includegraphics[width=\textwidth]{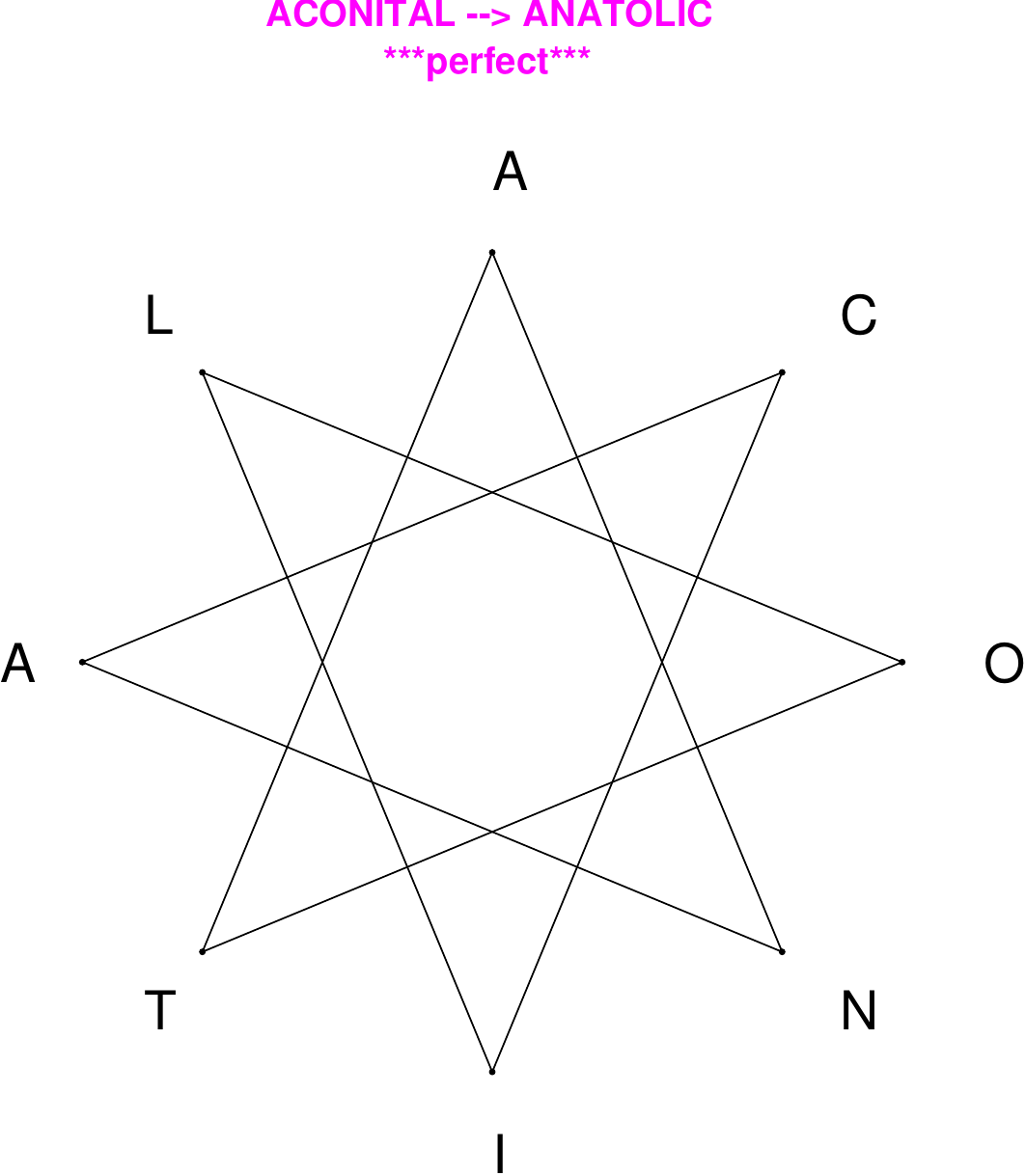}
\end{subfigure}
\hfill
\begin{subfigure}[T]{0.19\textwidth}
\centering
\includegraphics[width=\textwidth]{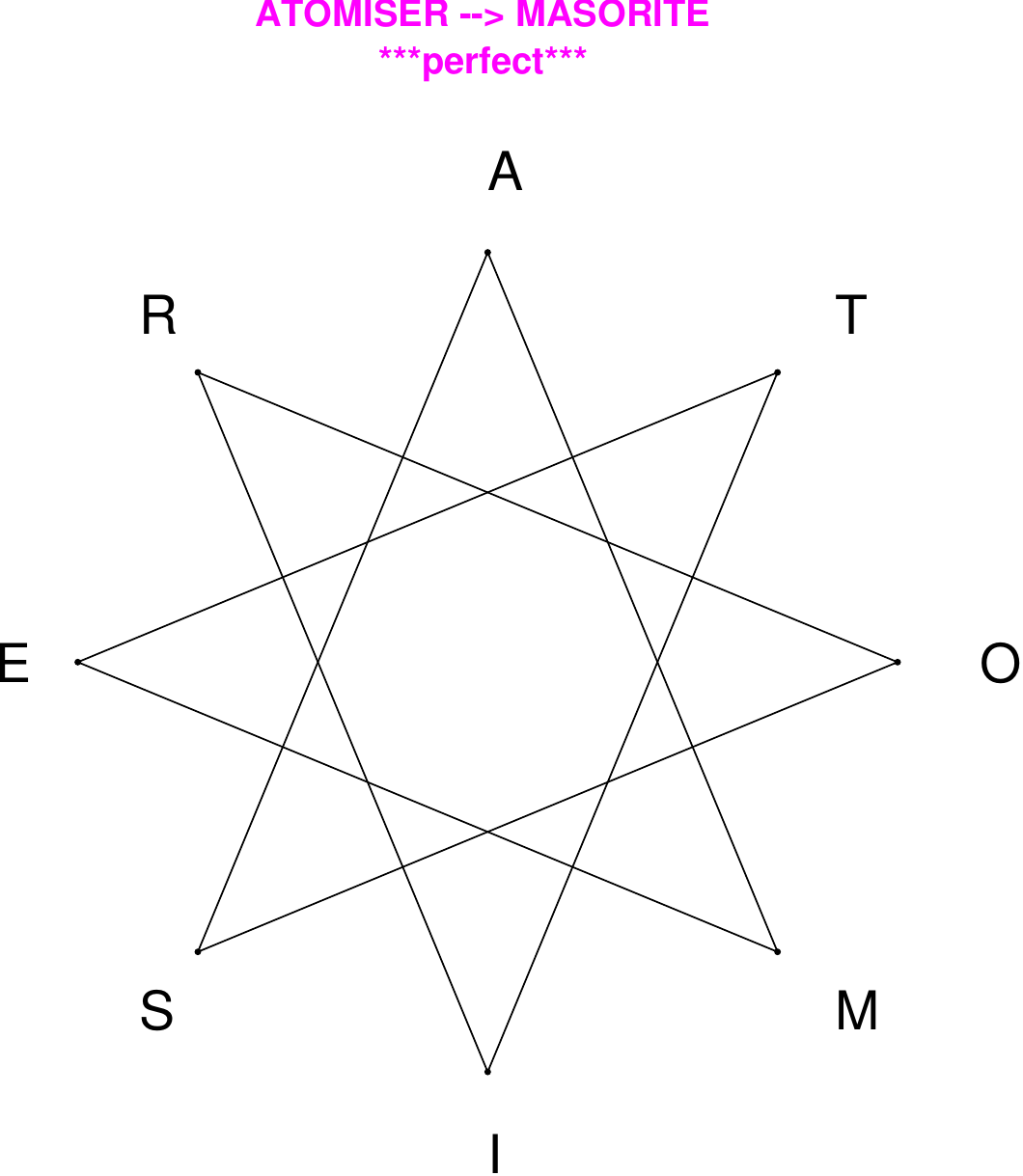}
\end{subfigure}
\hfill
\begin{subfigure}[T]{0.19\textwidth}
\centering
\includegraphics[width=\textwidth]{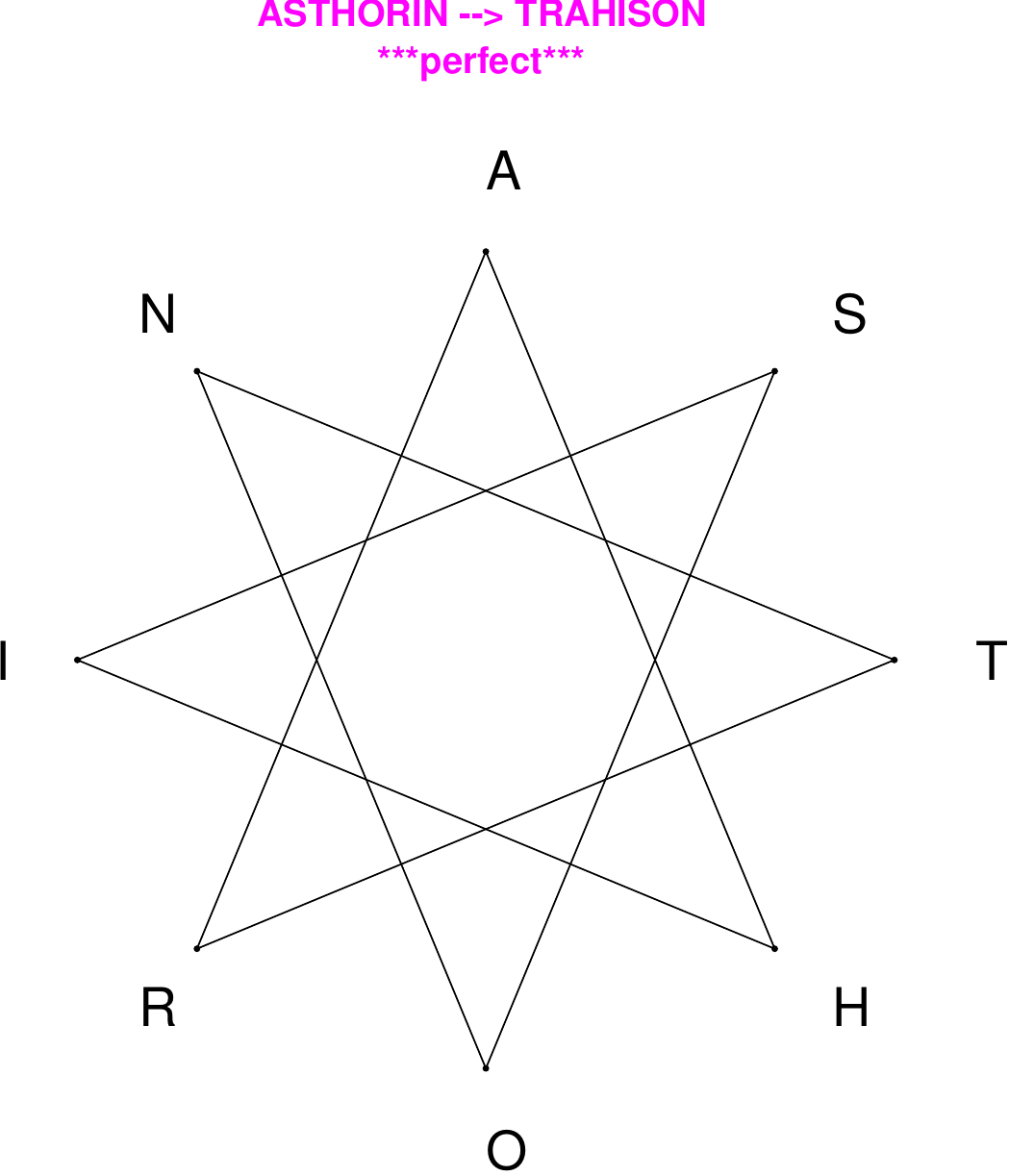}
\end{subfigure}
\hfill
\begin{subfigure}[T]{0.19\textwidth}
\centering
\includegraphics[width=\textwidth]{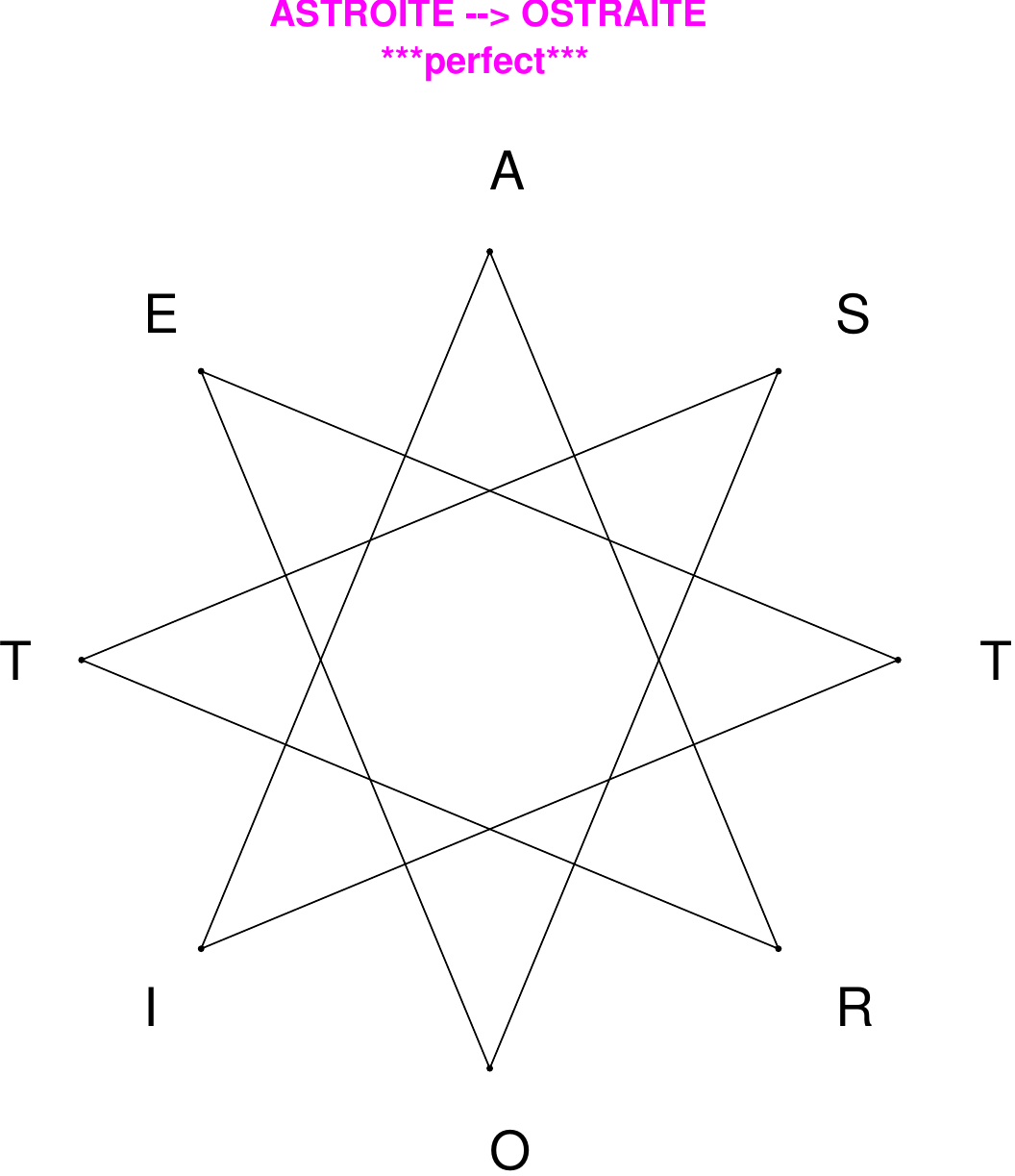}
\end{subfigure}
\hfill
\begin{subfigure}[T]{0.19\textwidth}
\centering
\includegraphics[width=\textwidth]{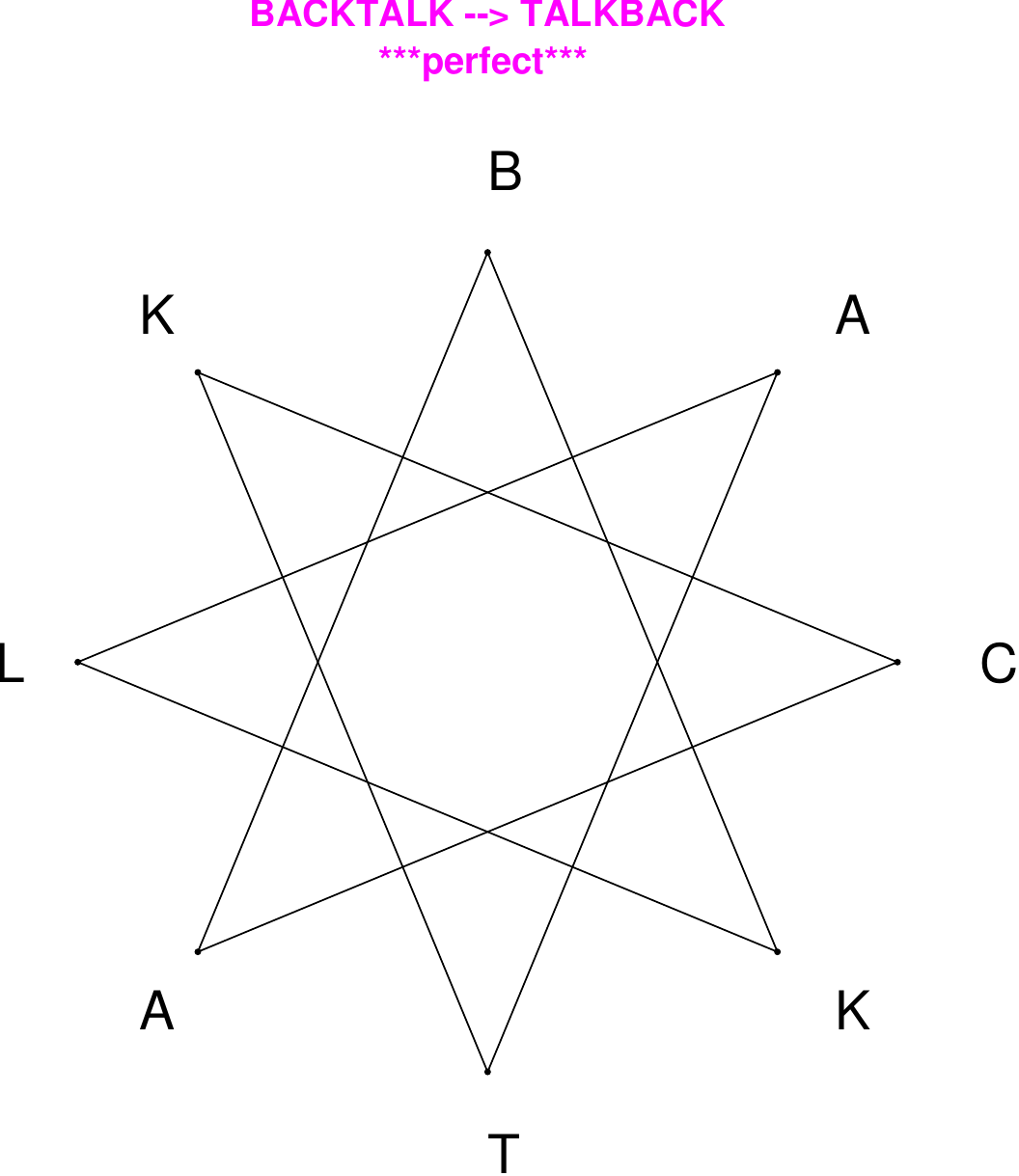}
\end{subfigure}
\end{figure}

\begin{figure}[H]
\centering
\begin{subfigure}[T]{0.19\textwidth}
\centering
\includegraphics[width=\textwidth]{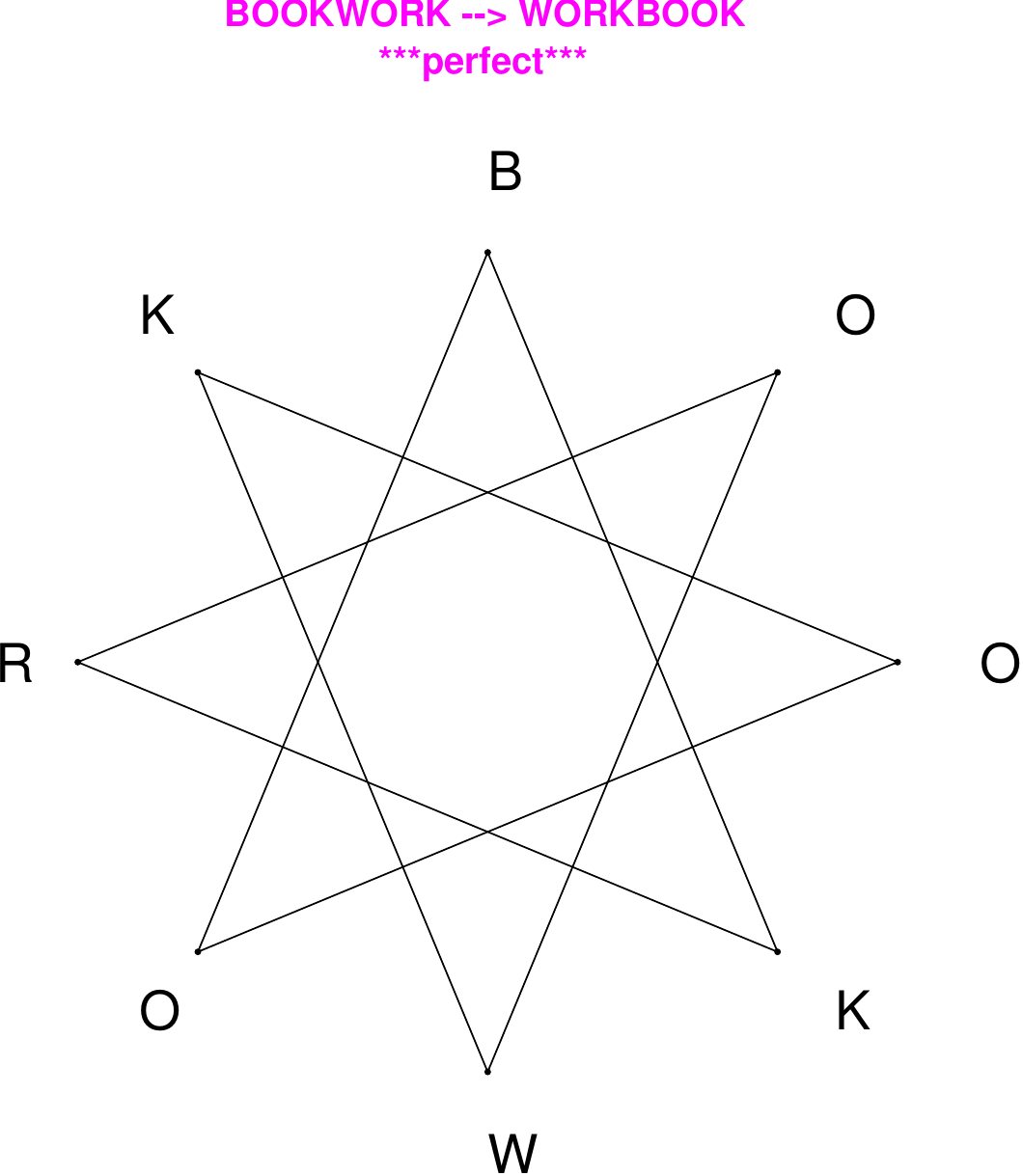}
\end{subfigure}
\hfill
\begin{subfigure}[T]{0.19\textwidth}
\centering
\includegraphics[width=\textwidth]{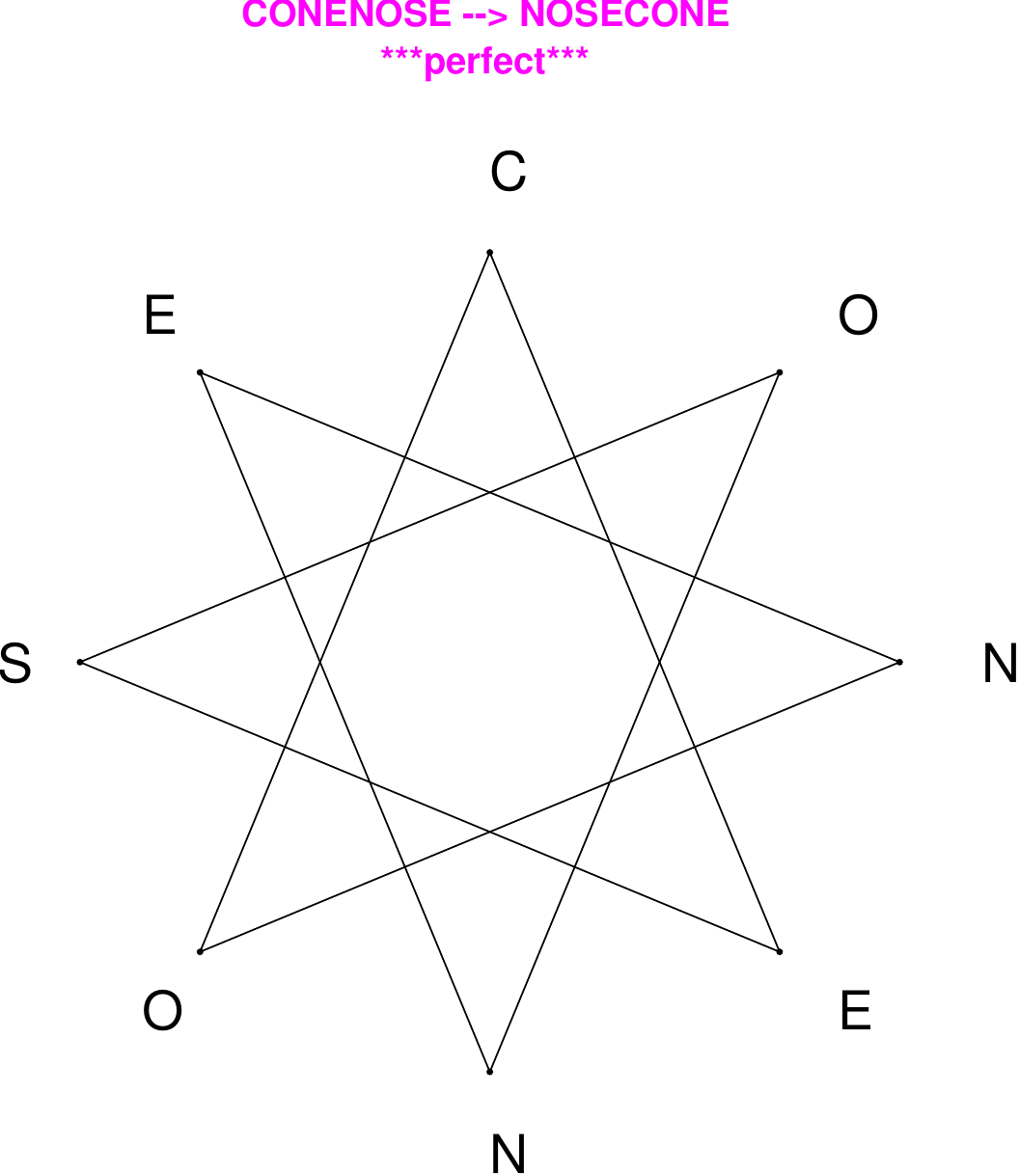}
\end{subfigure}
\hfill
\begin{subfigure}[T]{0.19\textwidth}
\centering
\includegraphics[width=\textwidth]{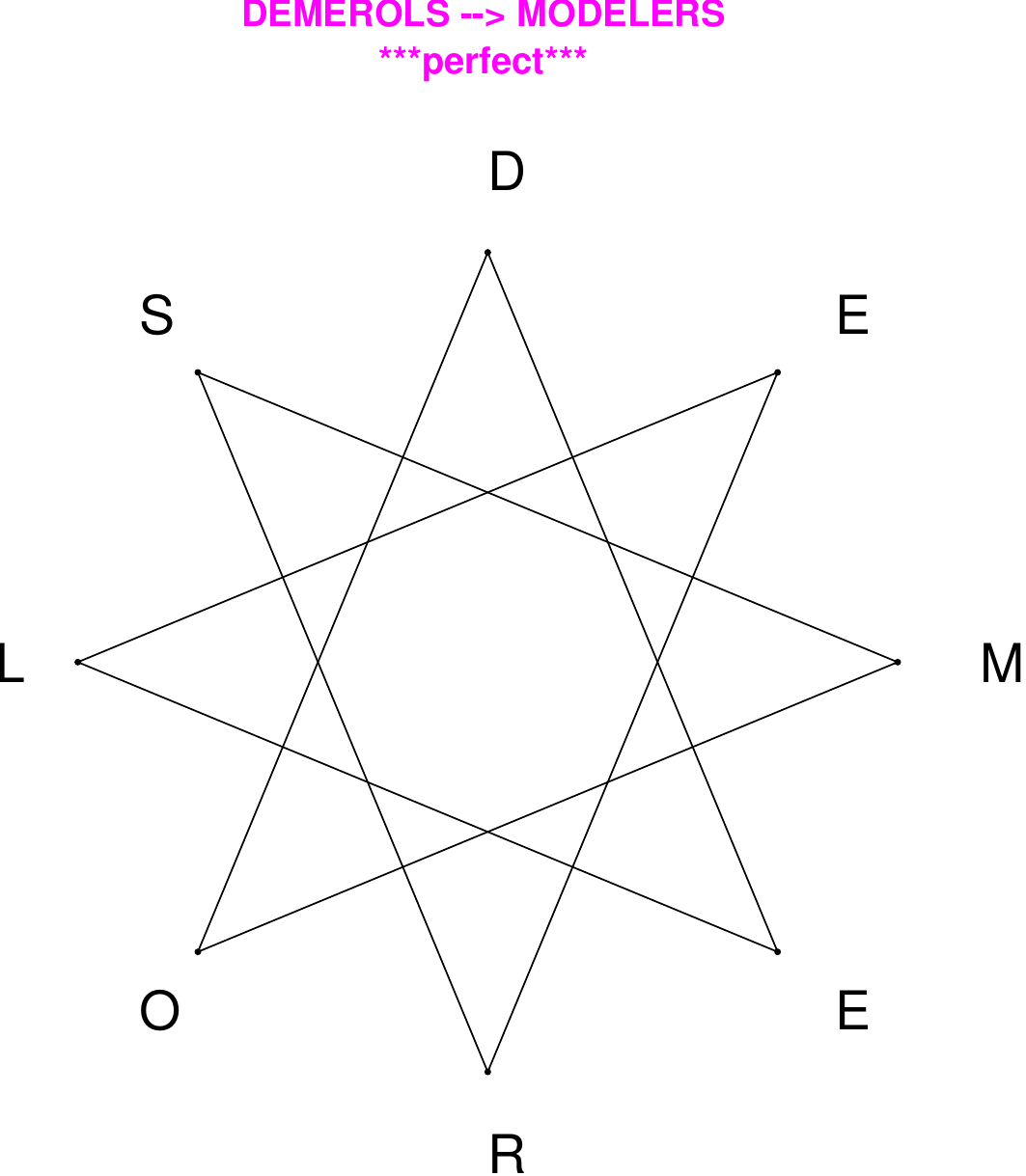}
\end{subfigure}
\hfill
\begin{subfigure}[T]{0.19\textwidth}
\centering
\includegraphics[width=\textwidth]{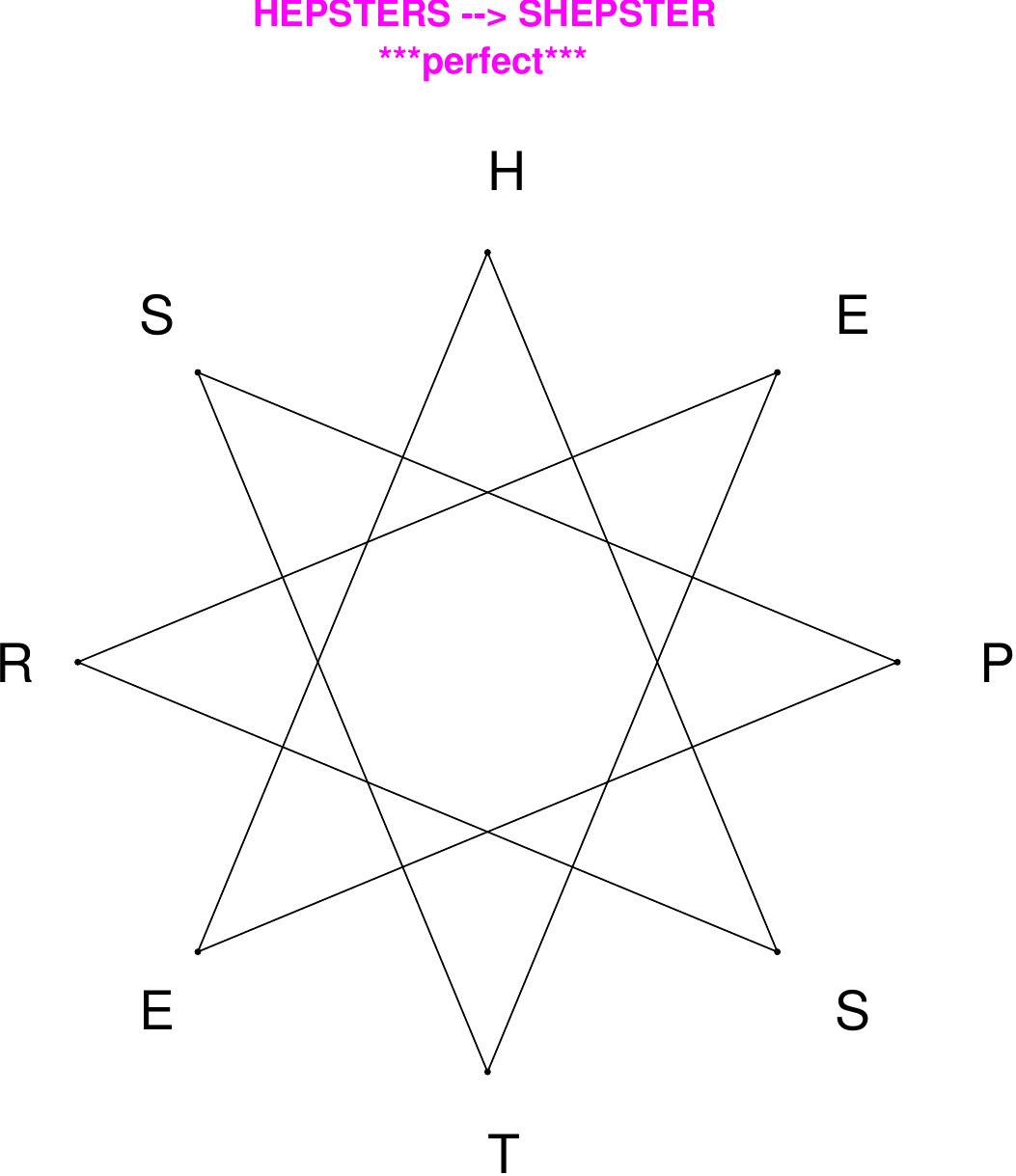}
\end{subfigure}
\hfill
\begin{subfigure}[T]{0.19\textwidth}
\centering
\includegraphics[width=\textwidth]{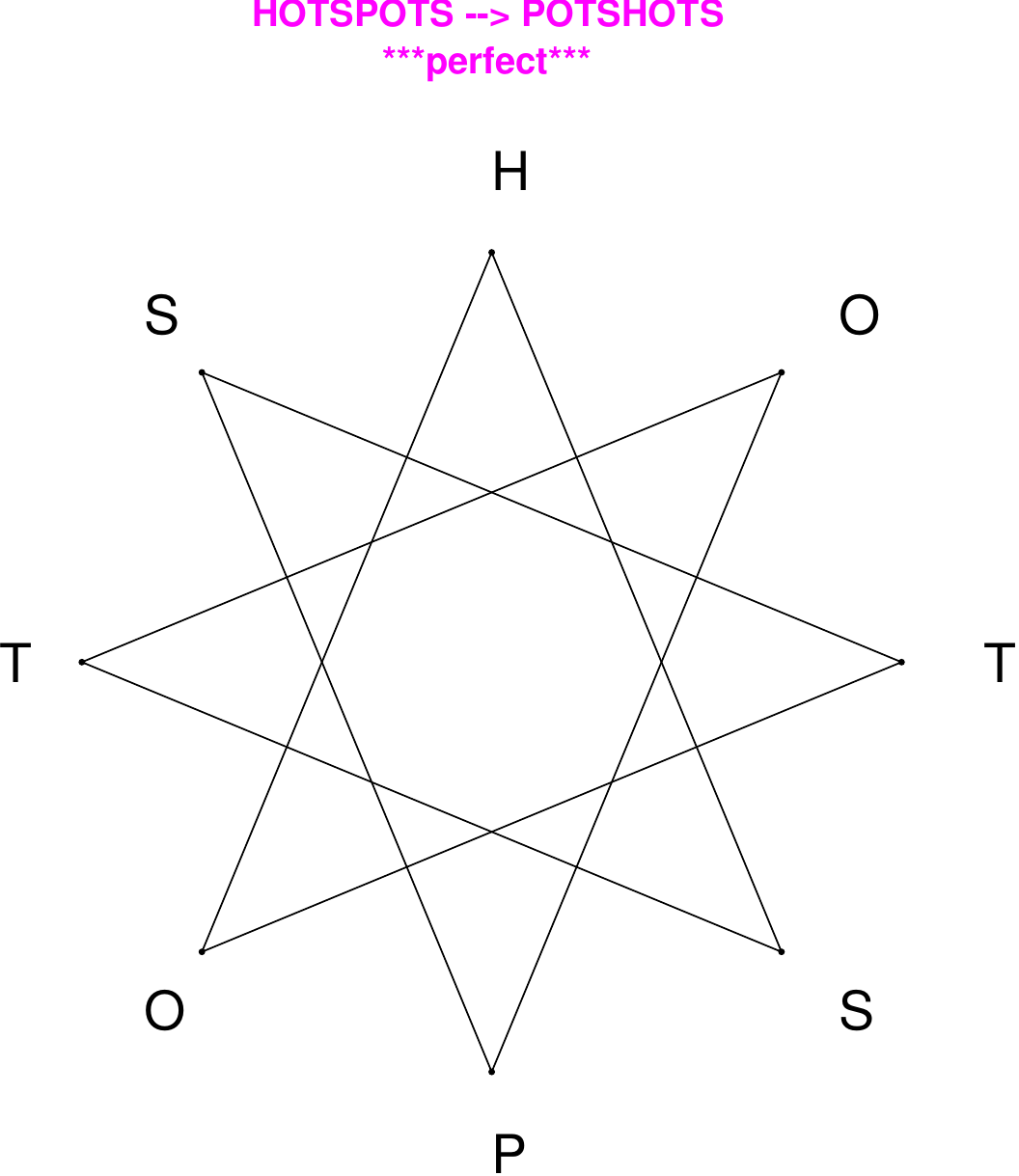}
\end{subfigure}
\end{figure}

\begin{figure}[H]
\centering
\begin{subfigure}[T]{0.19\textwidth}
\centering
\includegraphics[width=\textwidth]{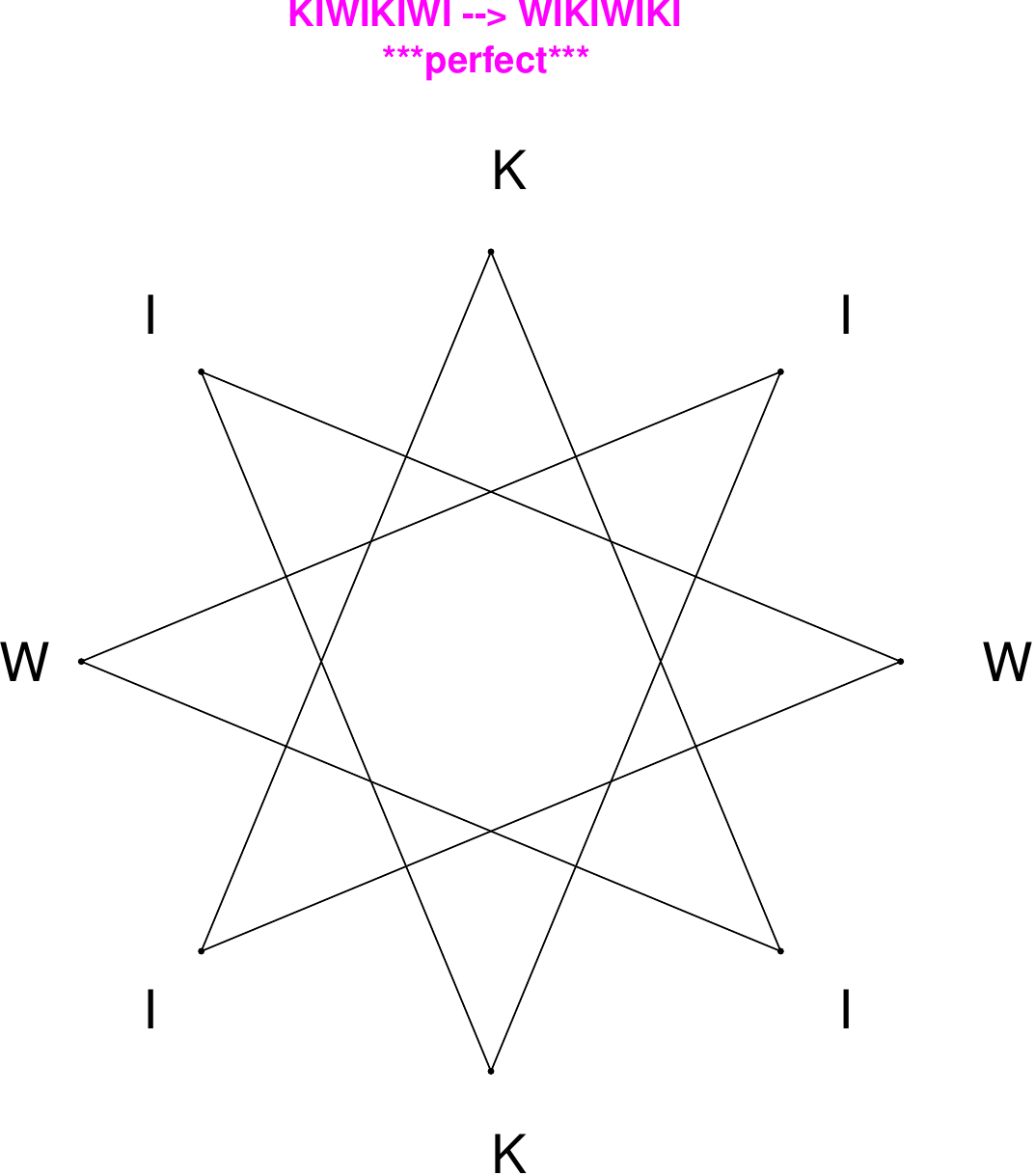}
\end{subfigure}
\hfill
\begin{subfigure}[T]{0.19\textwidth}
\centering
\includegraphics[width=\textwidth]{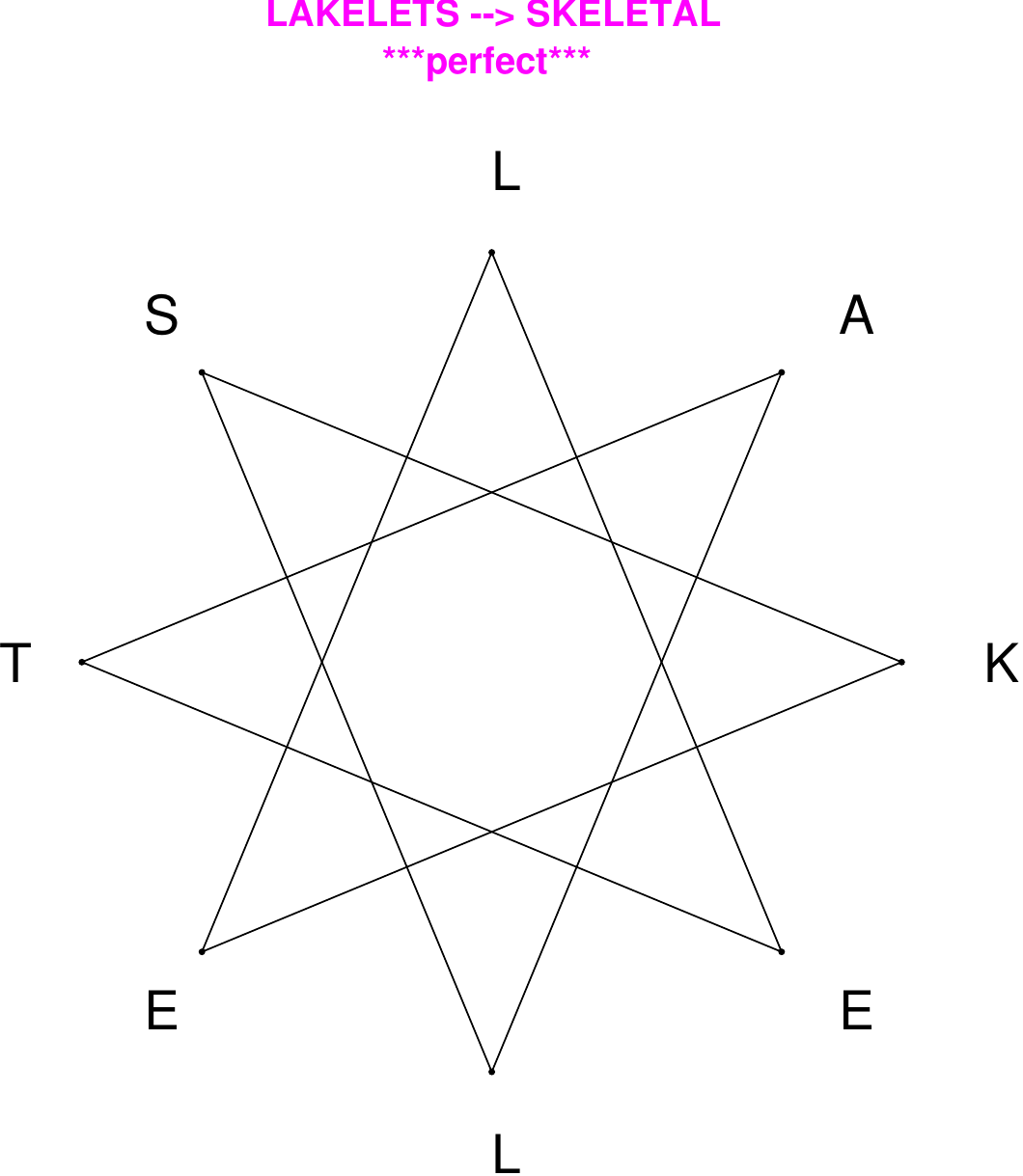}
\end{subfigure}
\hfill
\begin{subfigure}[T]{0.19\textwidth}
\centering
\includegraphics[width=\textwidth]{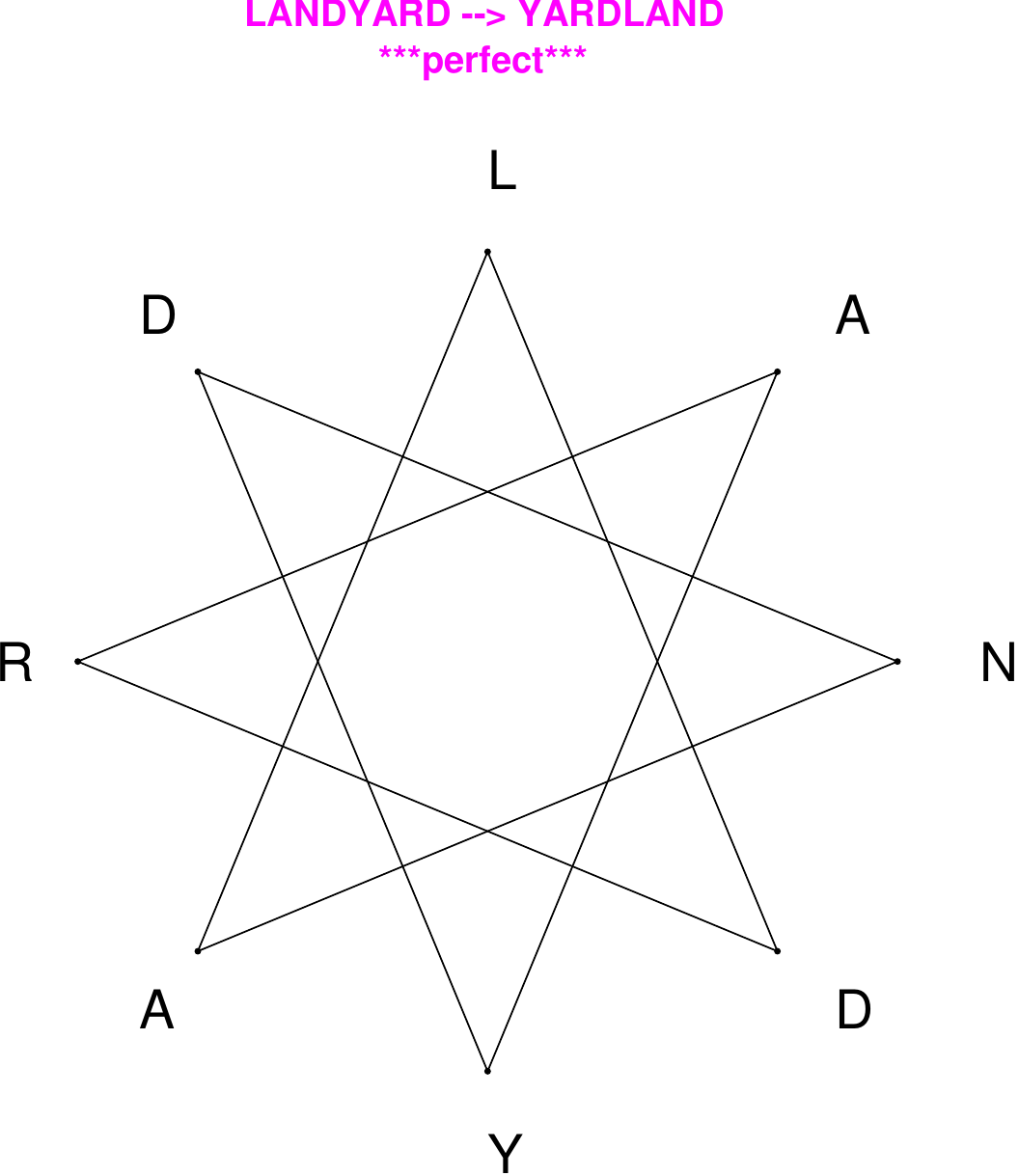}
\end{subfigure}
\hfill
\begin{subfigure}[T]{0.19\textwidth}
\centering
\includegraphics[width=\textwidth]{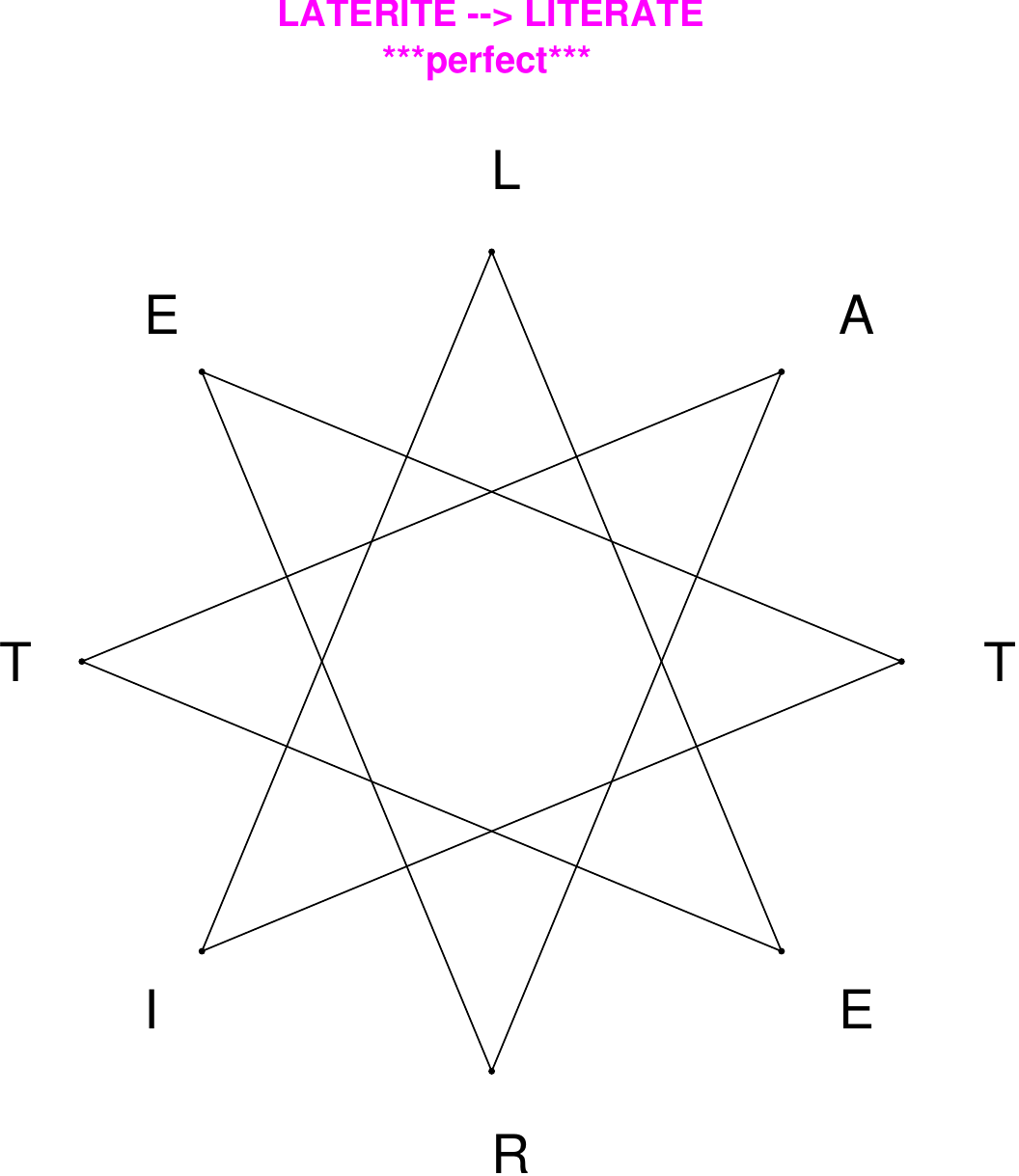}
\end{subfigure}
\hfill
\begin{subfigure}[T]{0.19\textwidth}
\centering
\includegraphics[width=\textwidth]{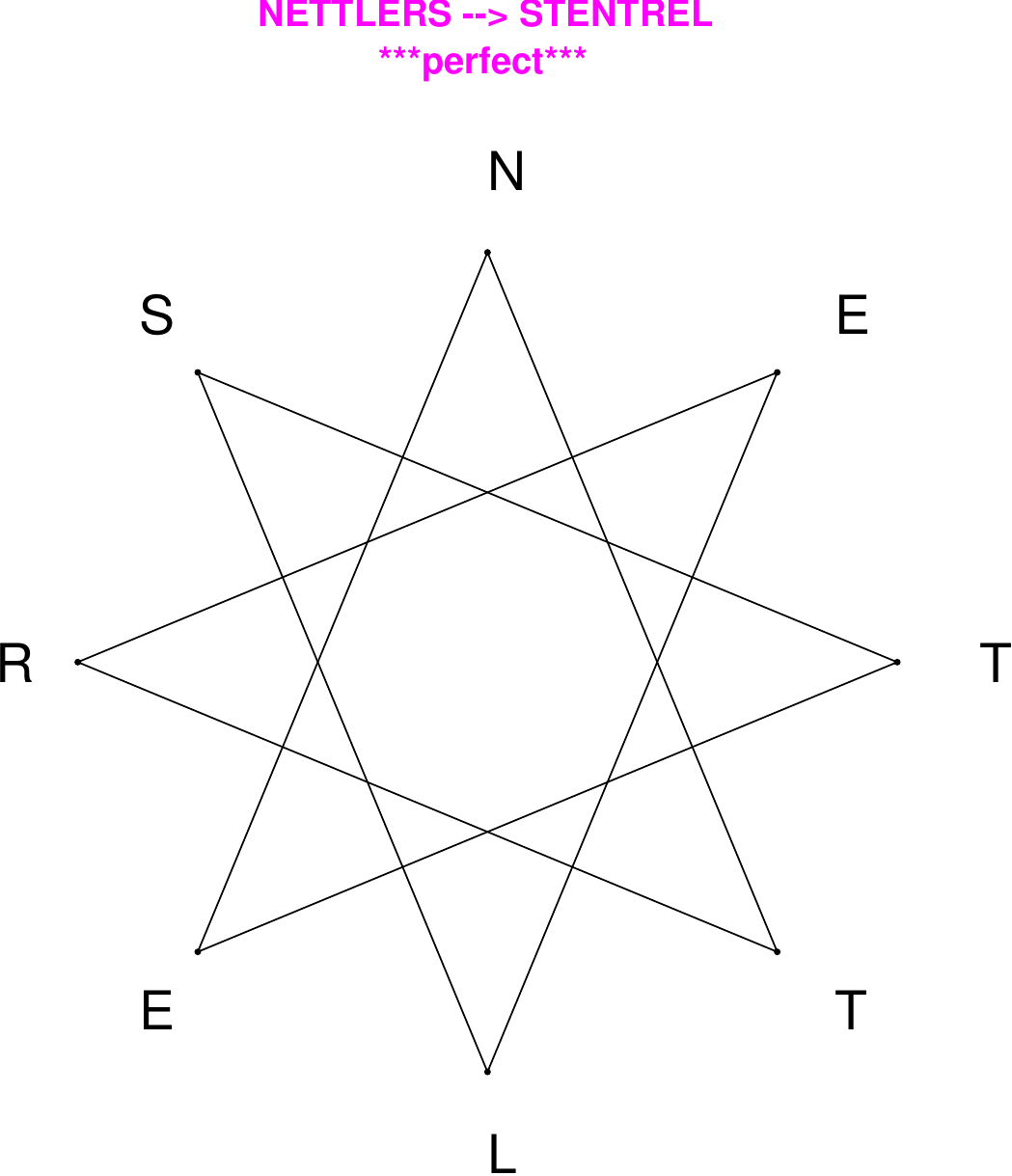}
\end{subfigure}
\end{figure}

\begin{figure}[H]
\centering
\begin{subfigure}[T]{0.19\textwidth}
\centering
\includegraphics[width=\textwidth]{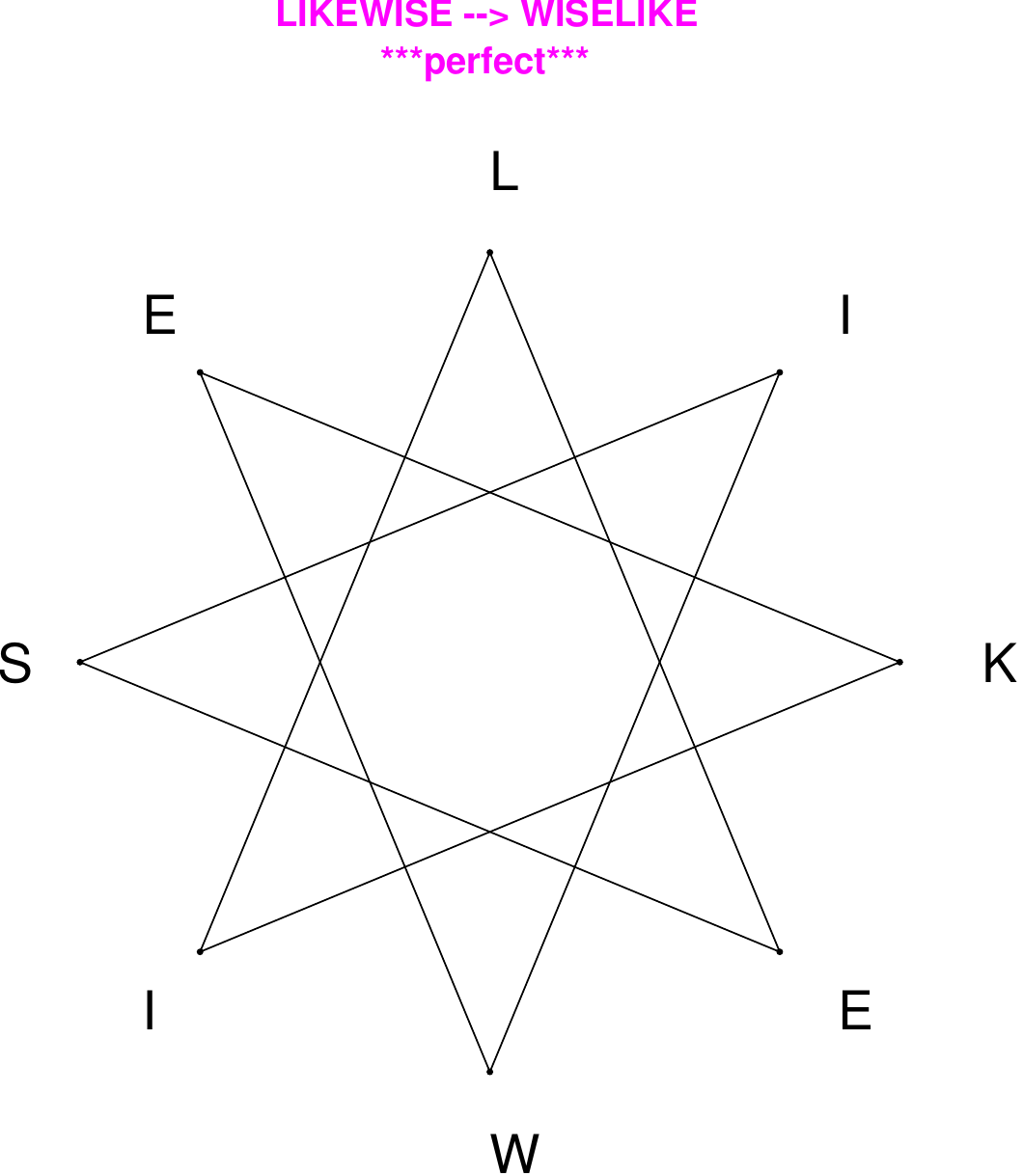}
\end{subfigure}
\hfill
\begin{subfigure}[T]{0.19\textwidth}
\centering
\includegraphics[width=\textwidth]{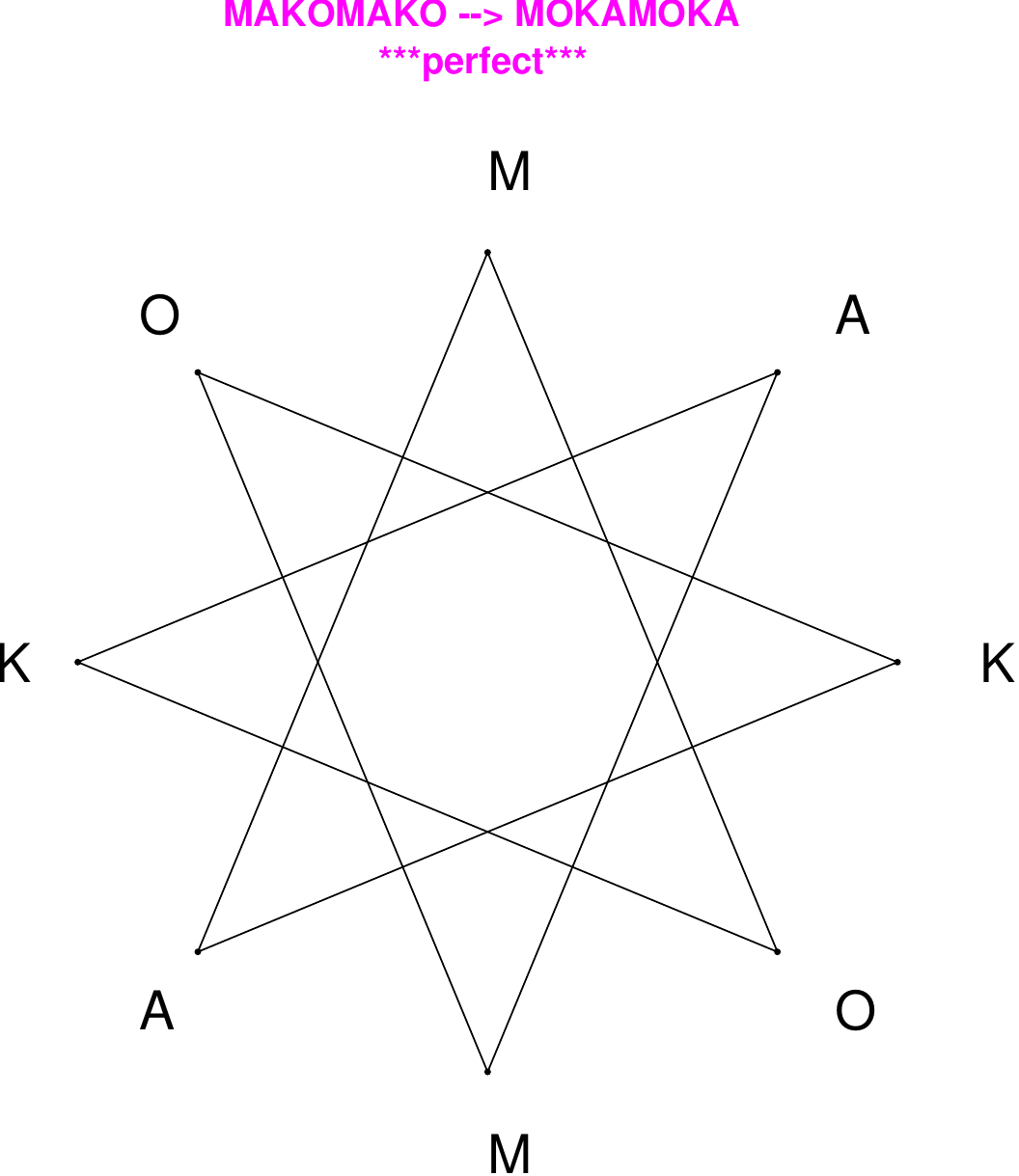}
\end{subfigure}
\hfill
\begin{subfigure}[T]{0.19\textwidth}
\centering
\includegraphics[width=\textwidth]{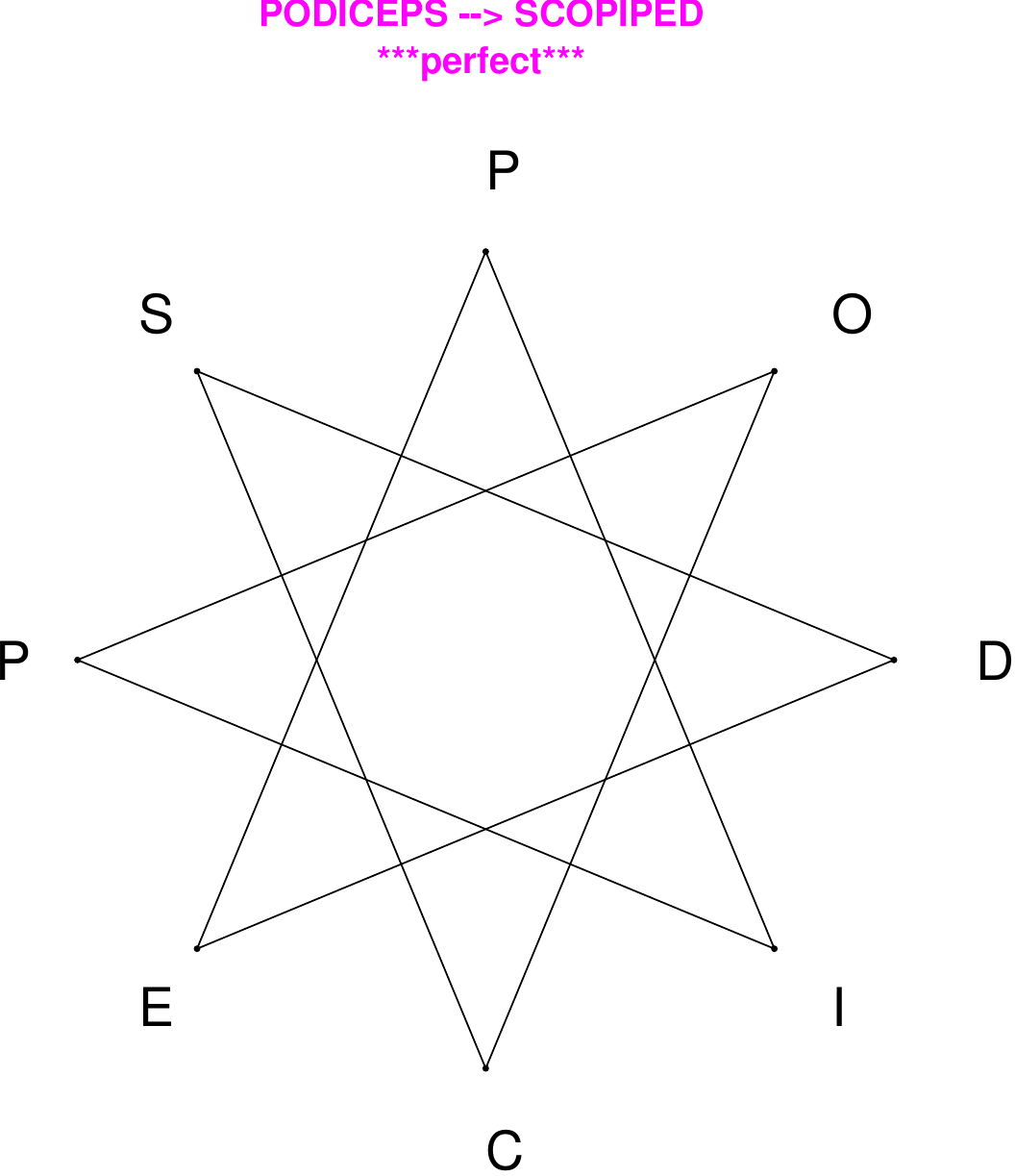}
\end{subfigure}
\hfill
\end{figure}

\subsubsection{Symmetric Stars $N=8$}

\begin{figure}[H]
\centering
\begin{subfigure}[T]{0.19\textwidth}
\centering
\includegraphics[width=\textwidth]{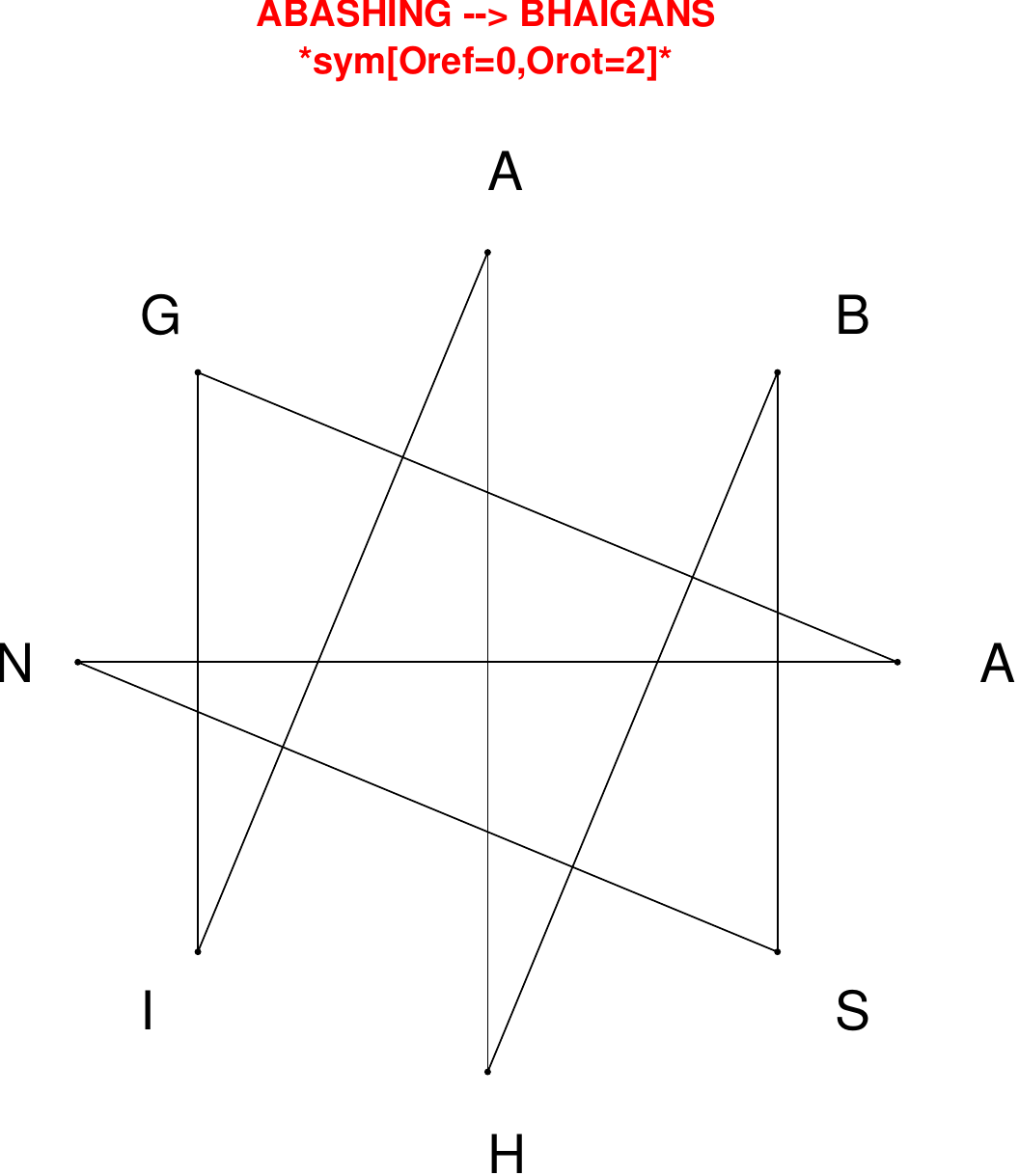}
\end{subfigure}
\hfill
\begin{subfigure}[T]{0.19\textwidth}
\centering
\includegraphics[width=\textwidth]{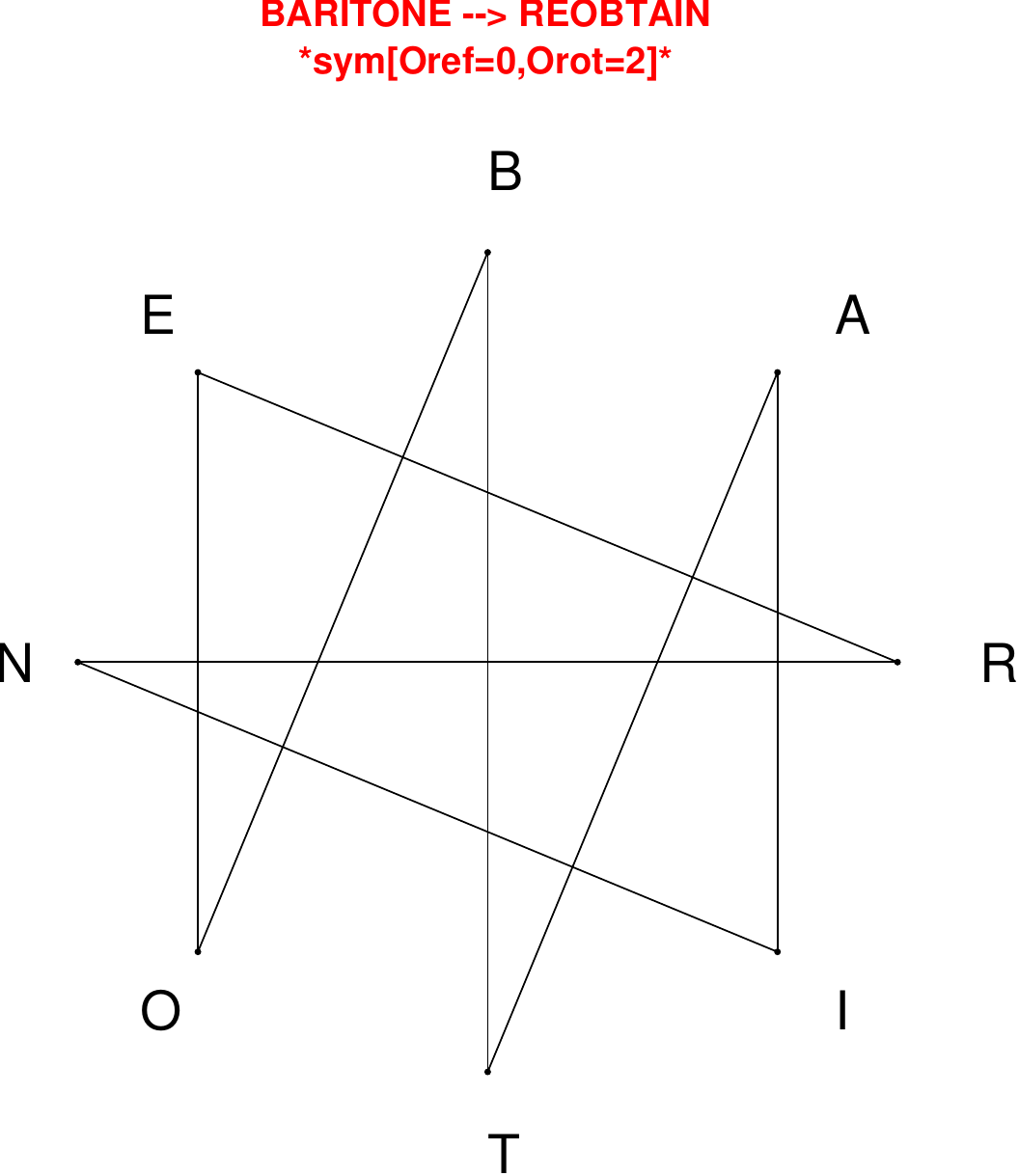}
\end{subfigure}
\hfill
\begin{subfigure}[T]{0.19\textwidth}
\centering
\includegraphics[width=\textwidth]{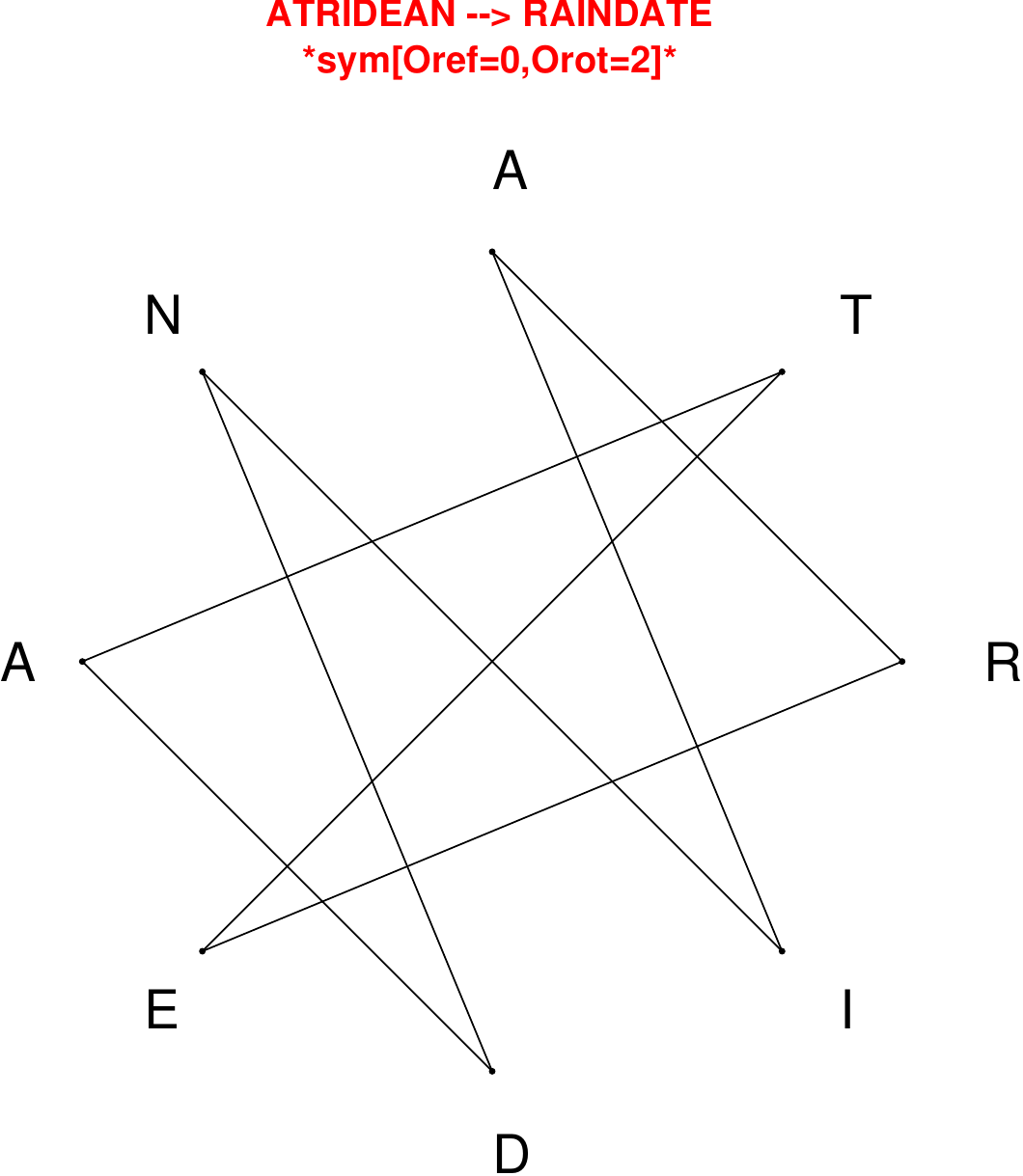}
\end{subfigure}
\hfill
\begin{subfigure}[T]{0.19\textwidth}
\centering
\includegraphics[width=\textwidth]{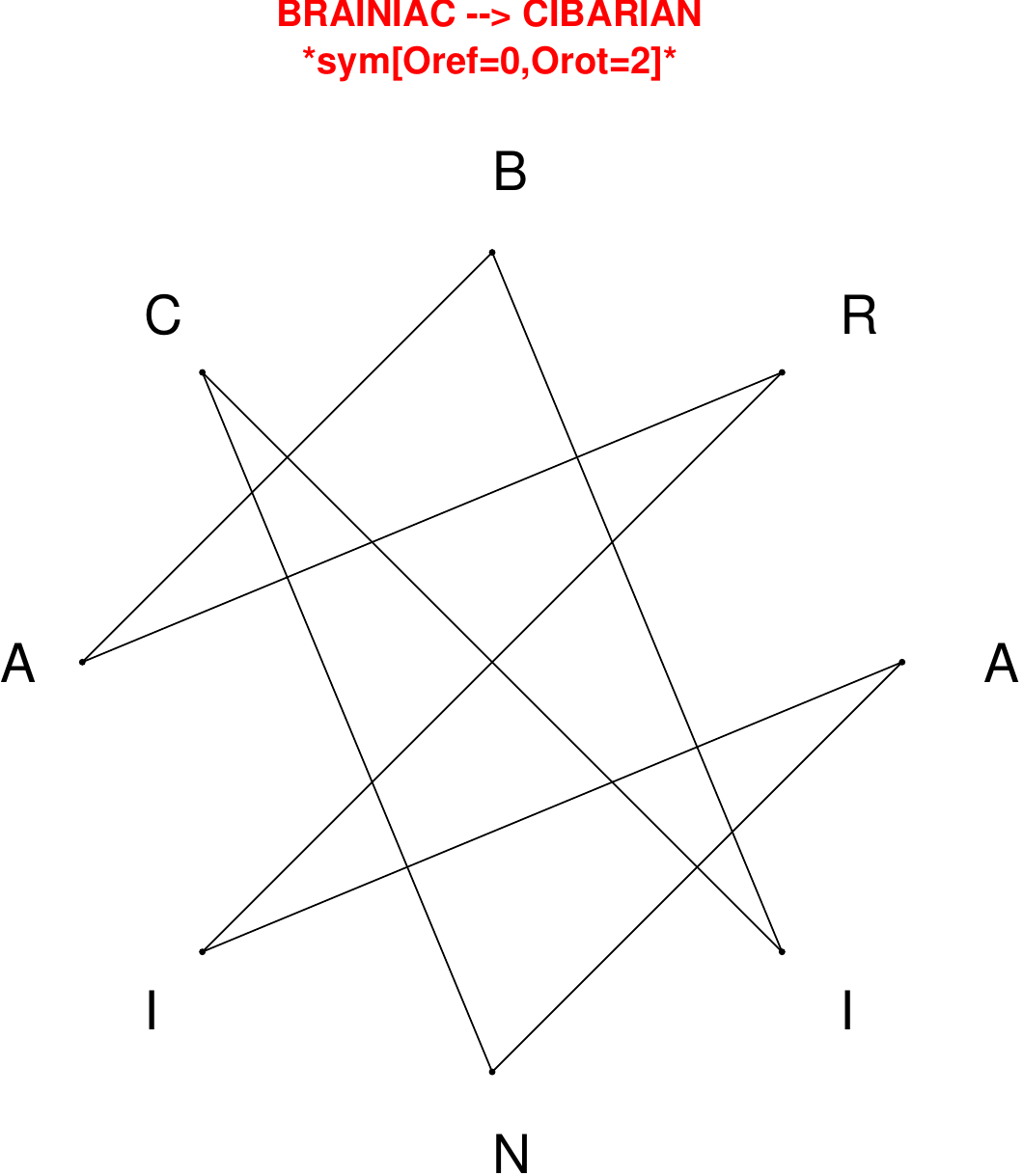}
\end{subfigure}
\hfill
\begin{subfigure}[T]{0.19\textwidth}
\centering
\includegraphics[width=\textwidth]{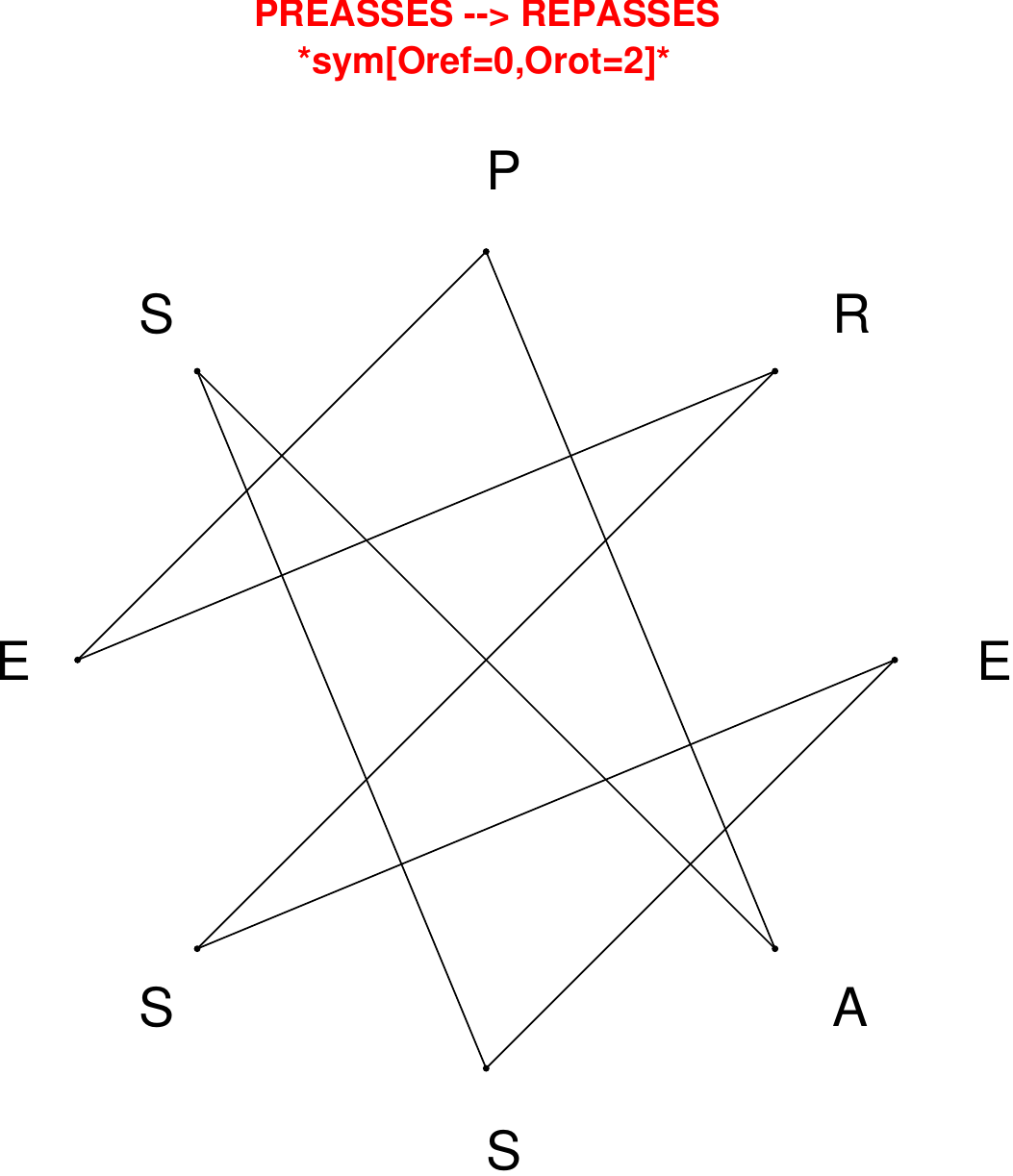}
\end{subfigure}
\end{figure}

\begin{figure}[H]
\centering
\begin{subfigure}[T]{0.19\textwidth}
\centering
\includegraphics[width=\textwidth]{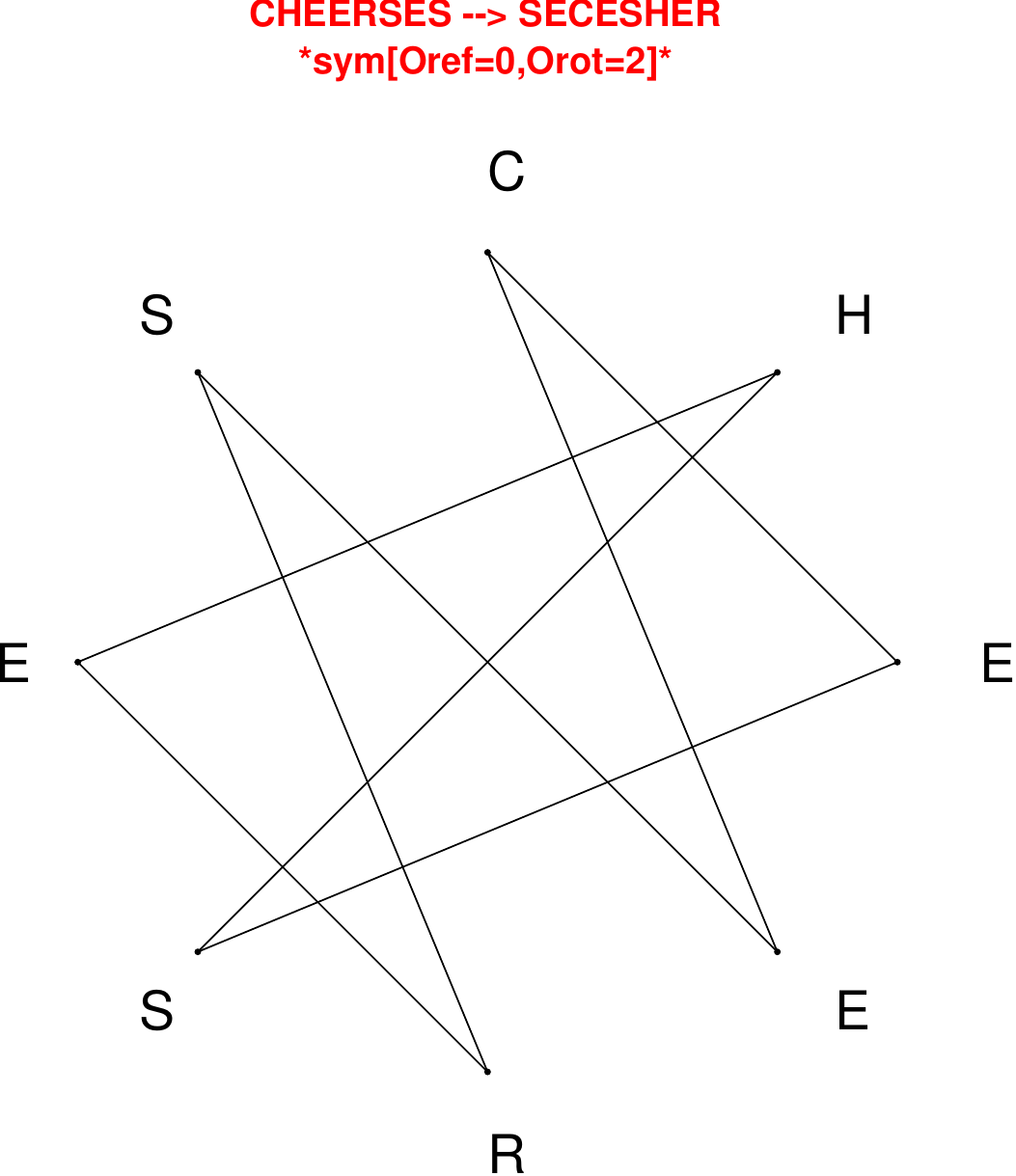}
\end{subfigure}
\hfill
\begin{subfigure}[T]{0.19\textwidth}
\centering
\includegraphics[width=\textwidth]{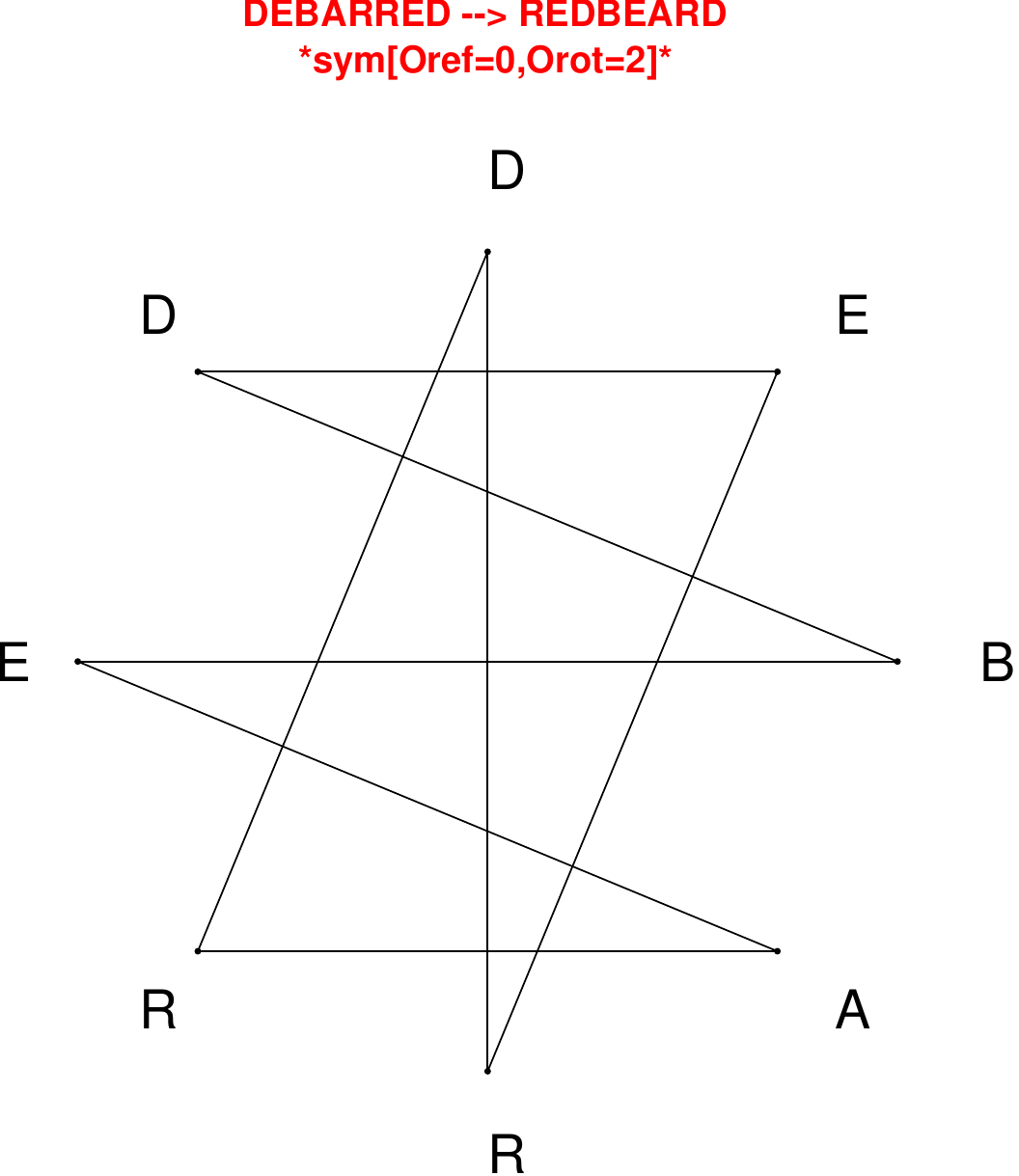}
\end{subfigure}
\hfill
\begin{subfigure}[T]{0.19\textwidth}
\centering
\includegraphics[width=\textwidth]{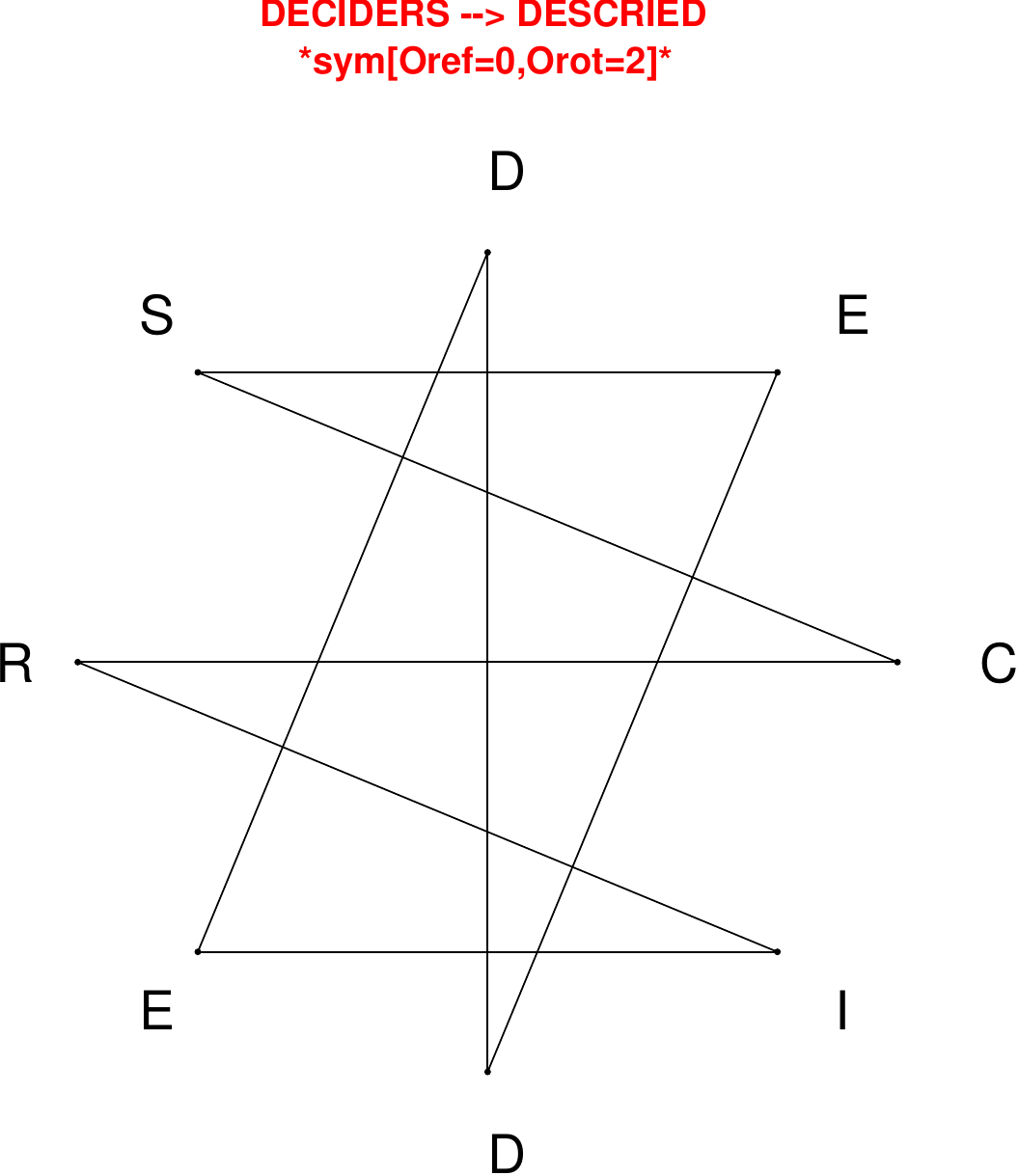}
\end{subfigure}
\hfill
\begin{subfigure}[T]{0.19\textwidth}
\centering
\includegraphics[width=\textwidth]{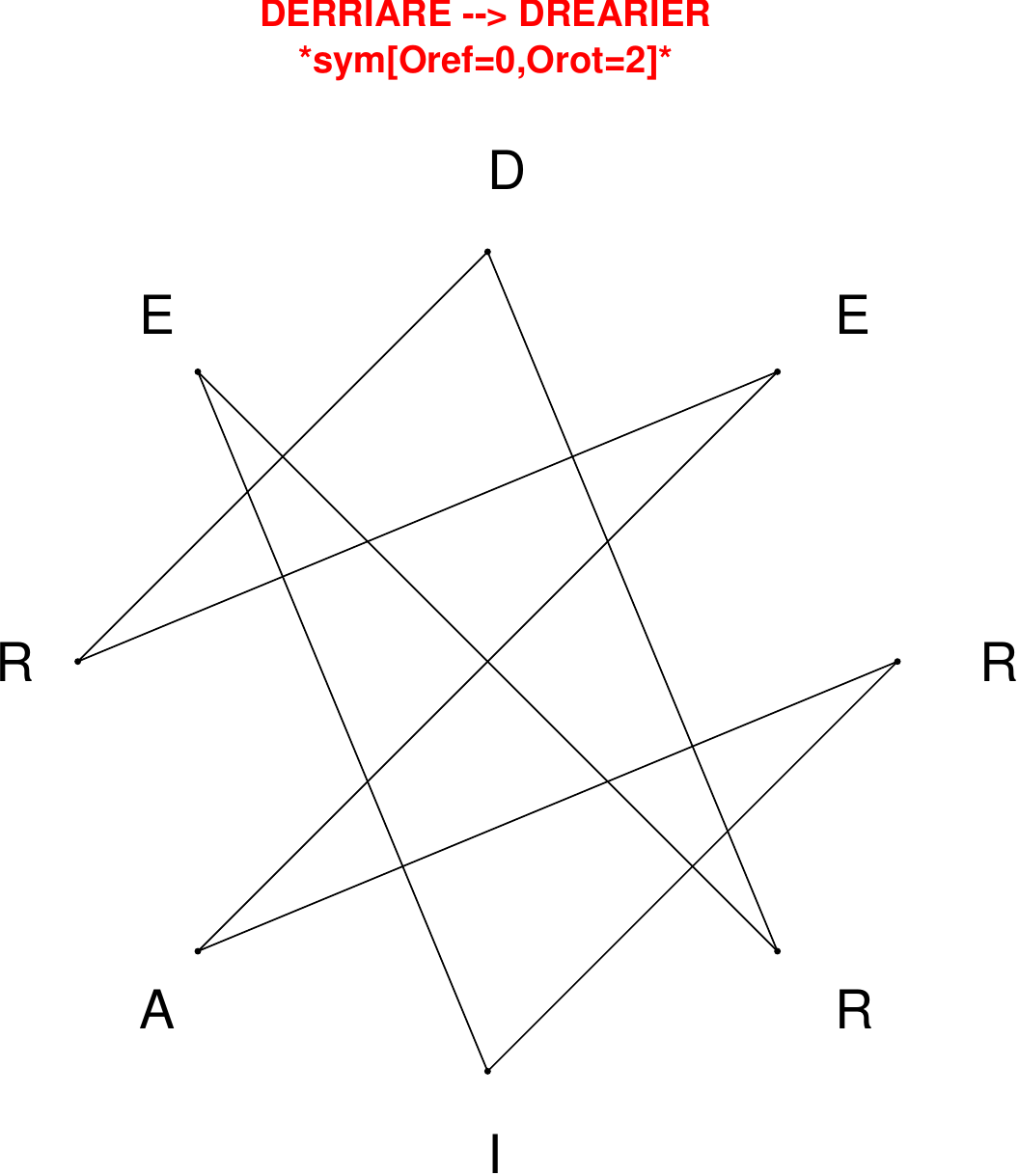}
\end{subfigure}
\hfill
\begin{subfigure}[T]{0.19\textwidth}
\centering
\includegraphics[width=\textwidth]{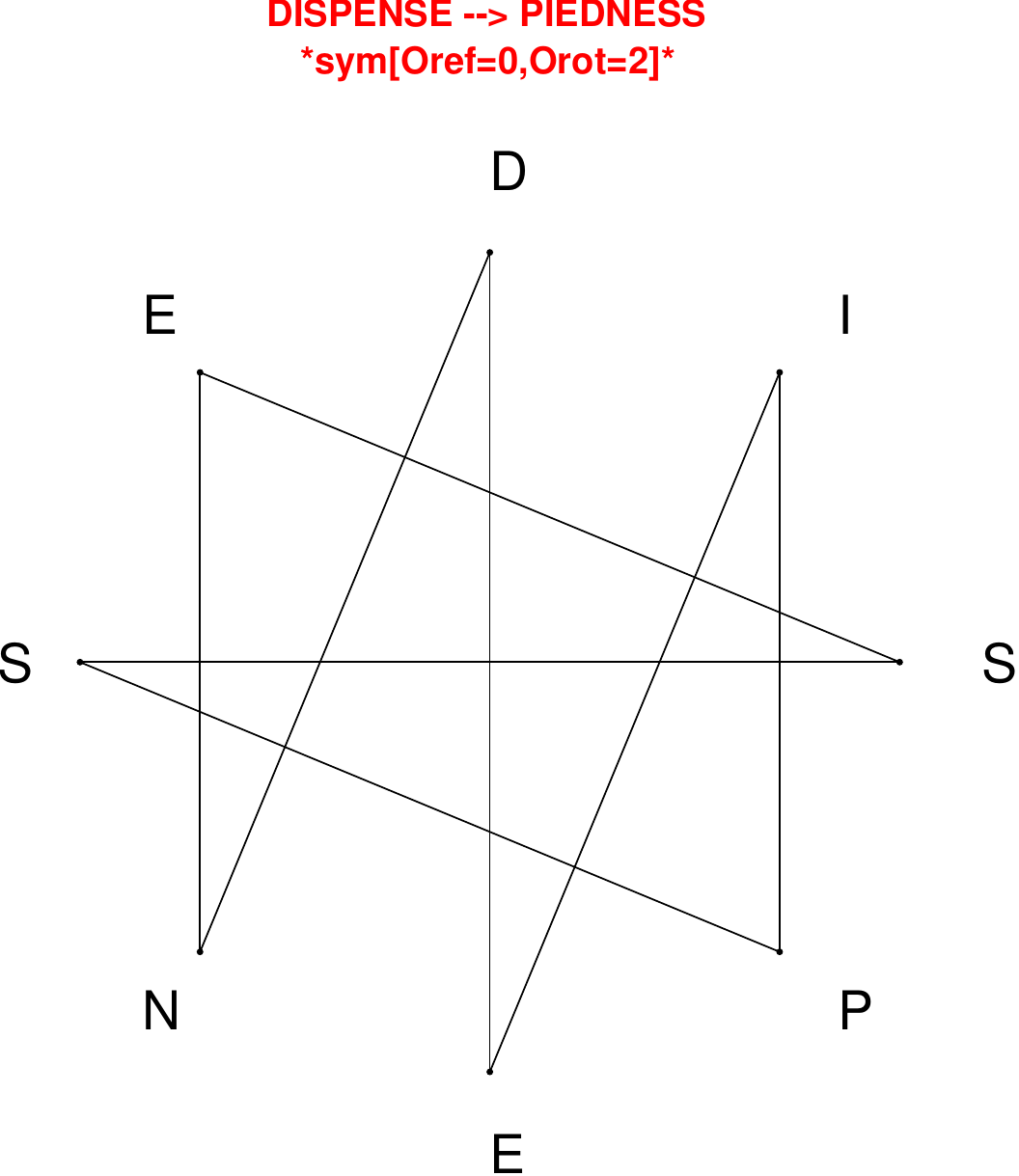}
\end{subfigure}
\end{figure}

\begin{figure}[H]
\centering
\begin{subfigure}[T]{0.19\textwidth}
\centering
\includegraphics[width=\textwidth]{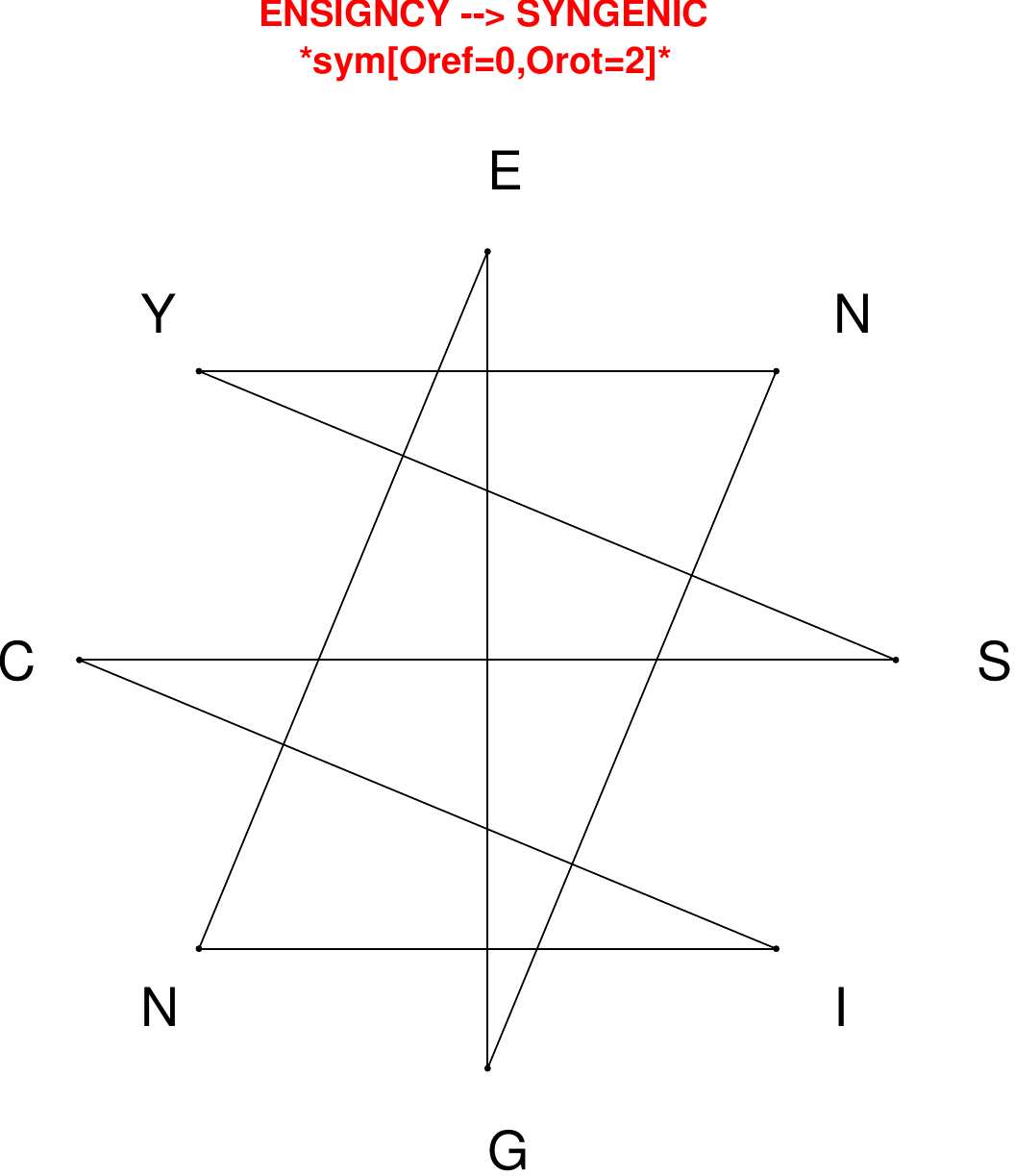}
\end{subfigure}
\hfill
\begin{subfigure}[T]{0.19\textwidth}
\centering
\includegraphics[width=\textwidth]{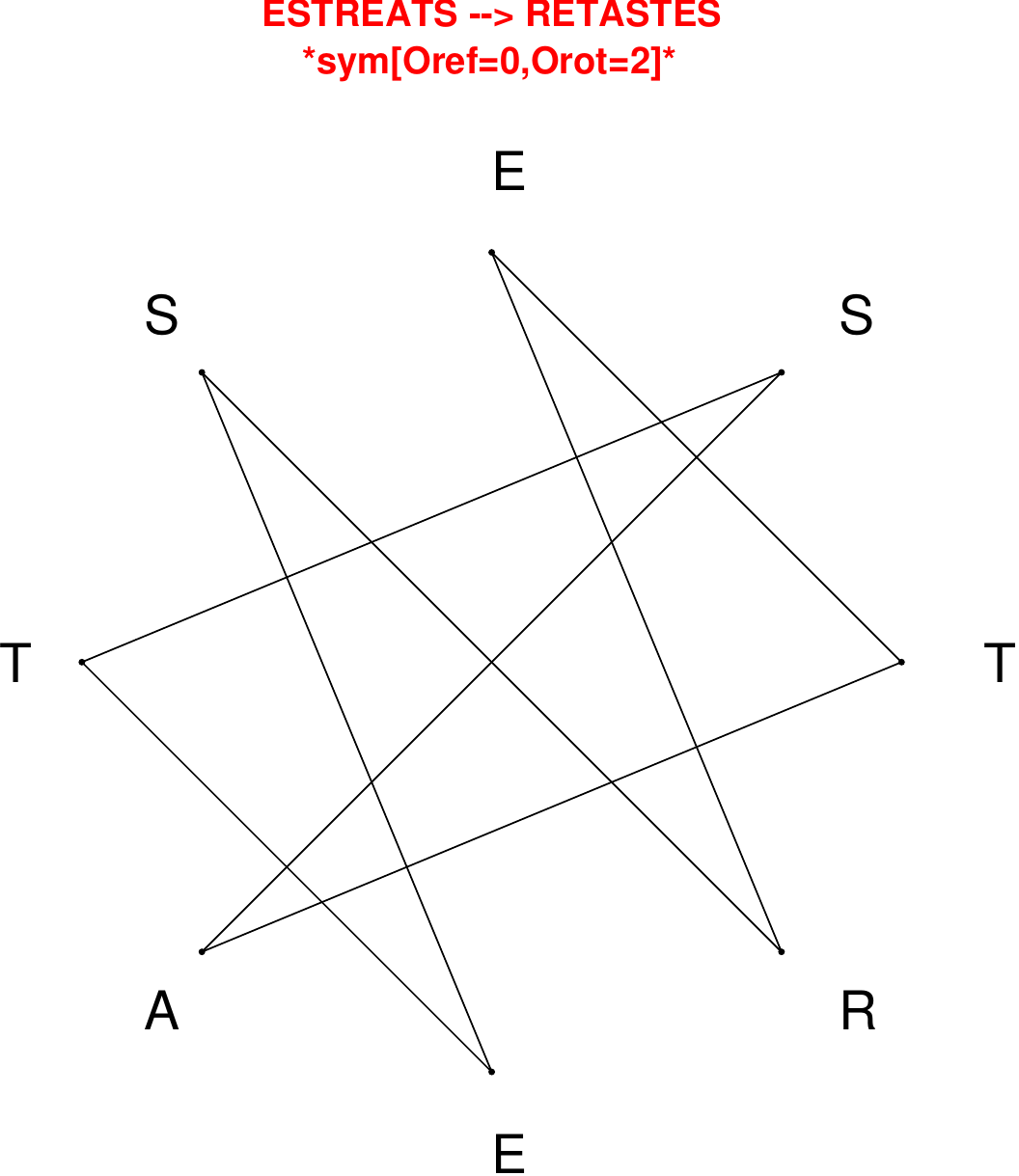}
\end{subfigure}
\hfill
\begin{subfigure}[T]{0.19\textwidth}
\centering
\includegraphics[width=\textwidth]{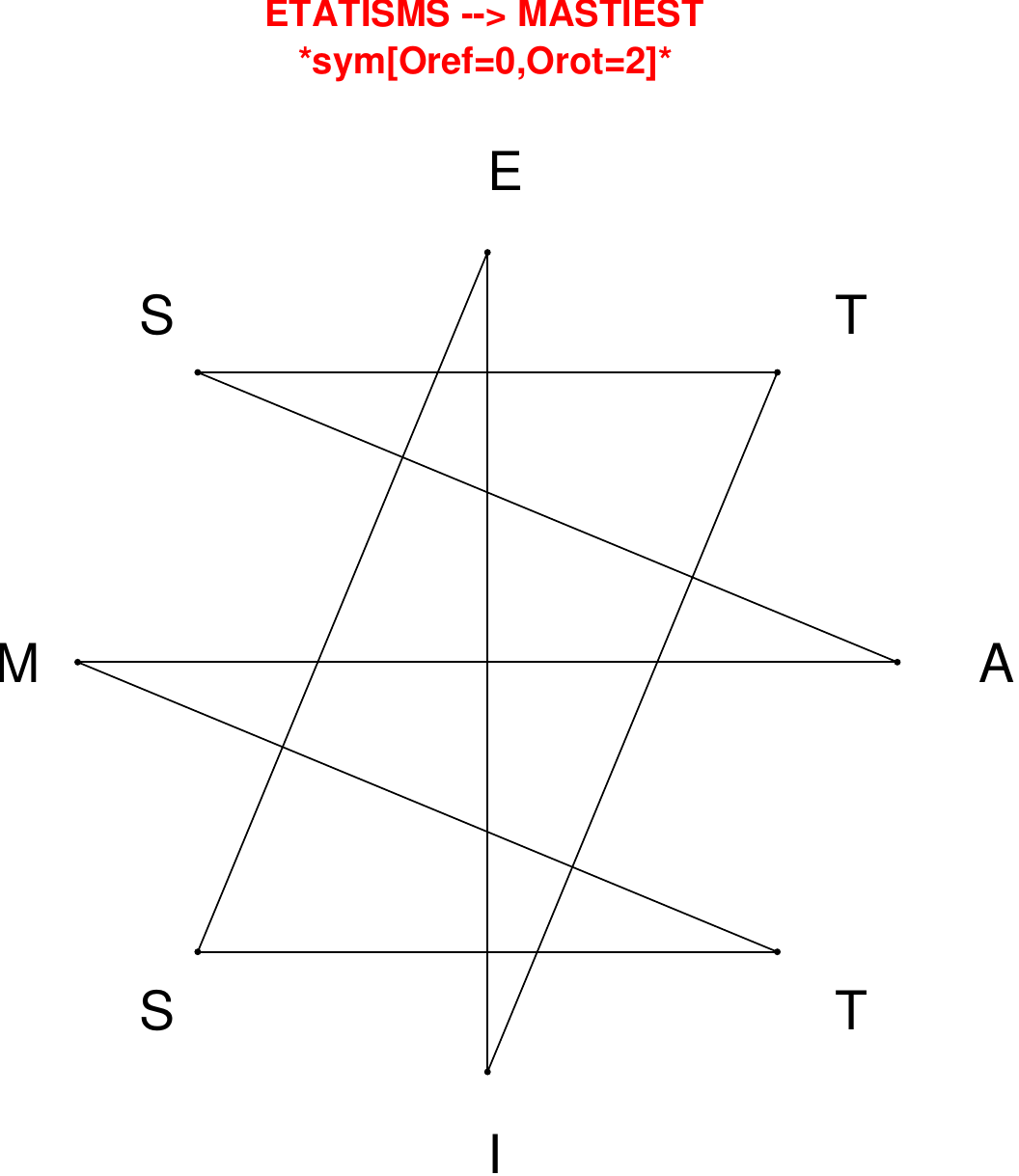}
\end{subfigure}
\hfill
\begin{subfigure}[T]{0.19\textwidth}
\centering
\includegraphics[width=\textwidth]{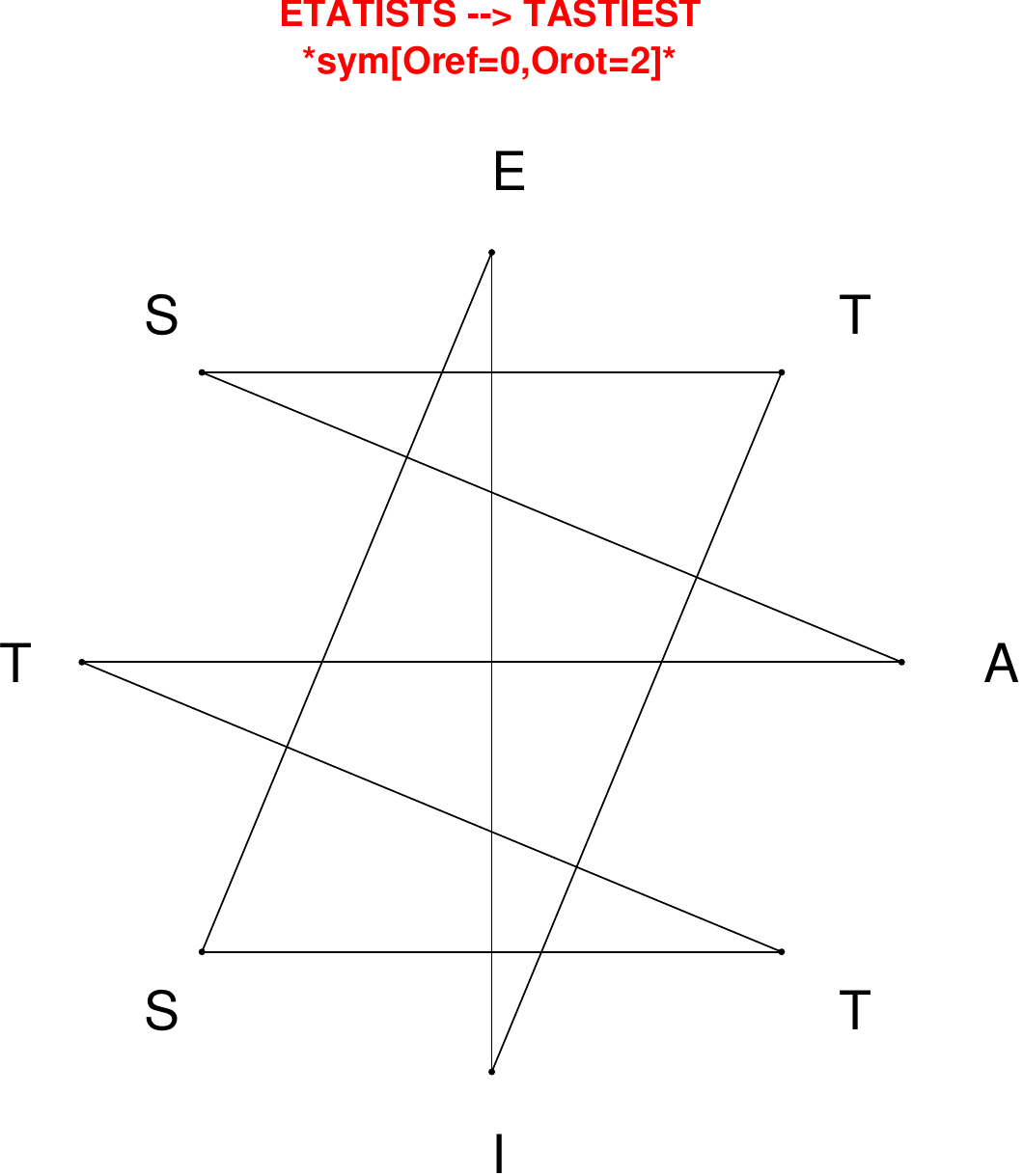}
\end{subfigure}
\hfill
\begin{subfigure}[T]{0.19\textwidth}
\centering
\includegraphics[width=\textwidth]{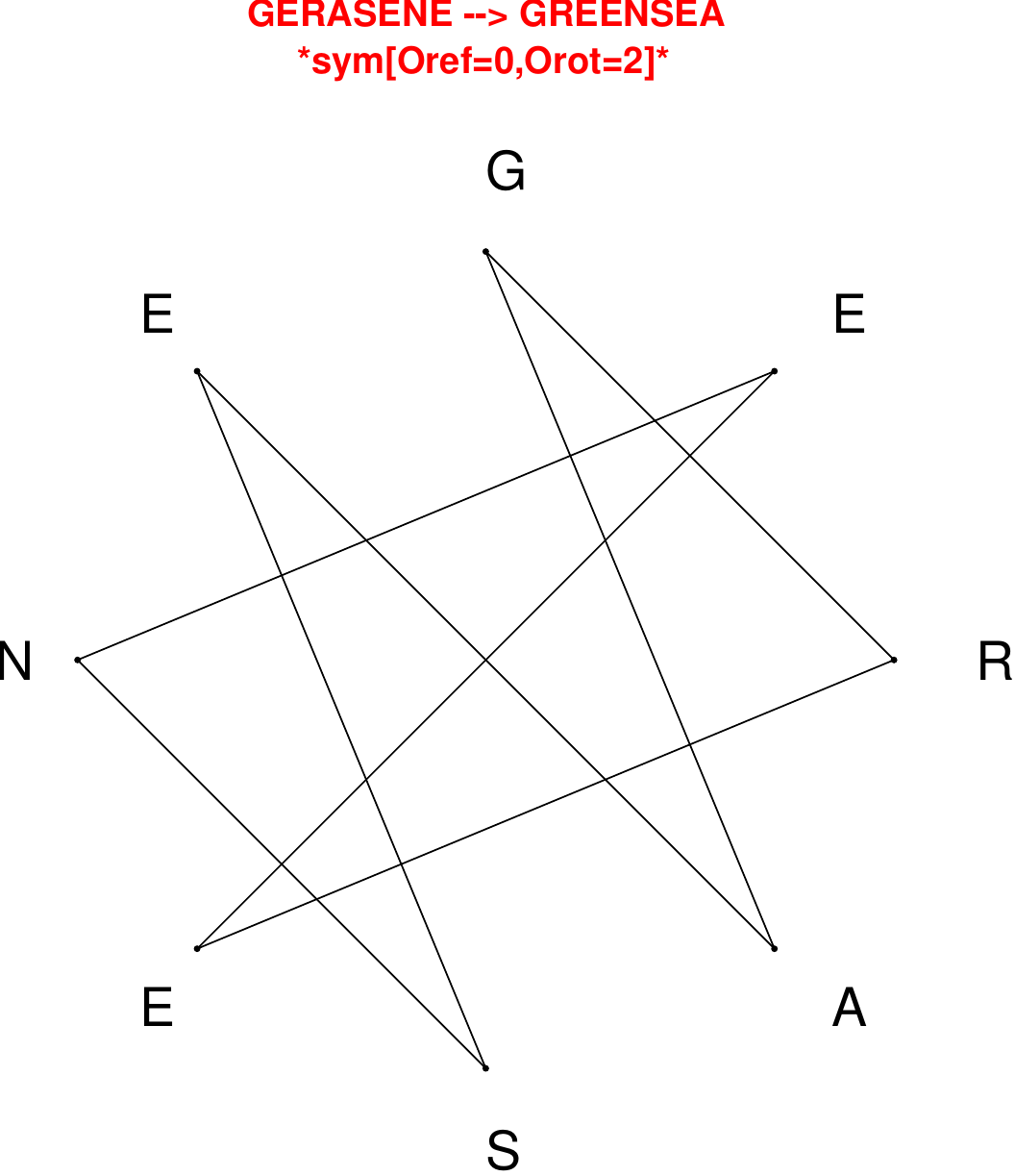}
\end{subfigure}
\end{figure}

\begin{figure}[H]
\centering
\begin{subfigure}[T]{0.19\textwidth}
\centering
\includegraphics[width=\textwidth]{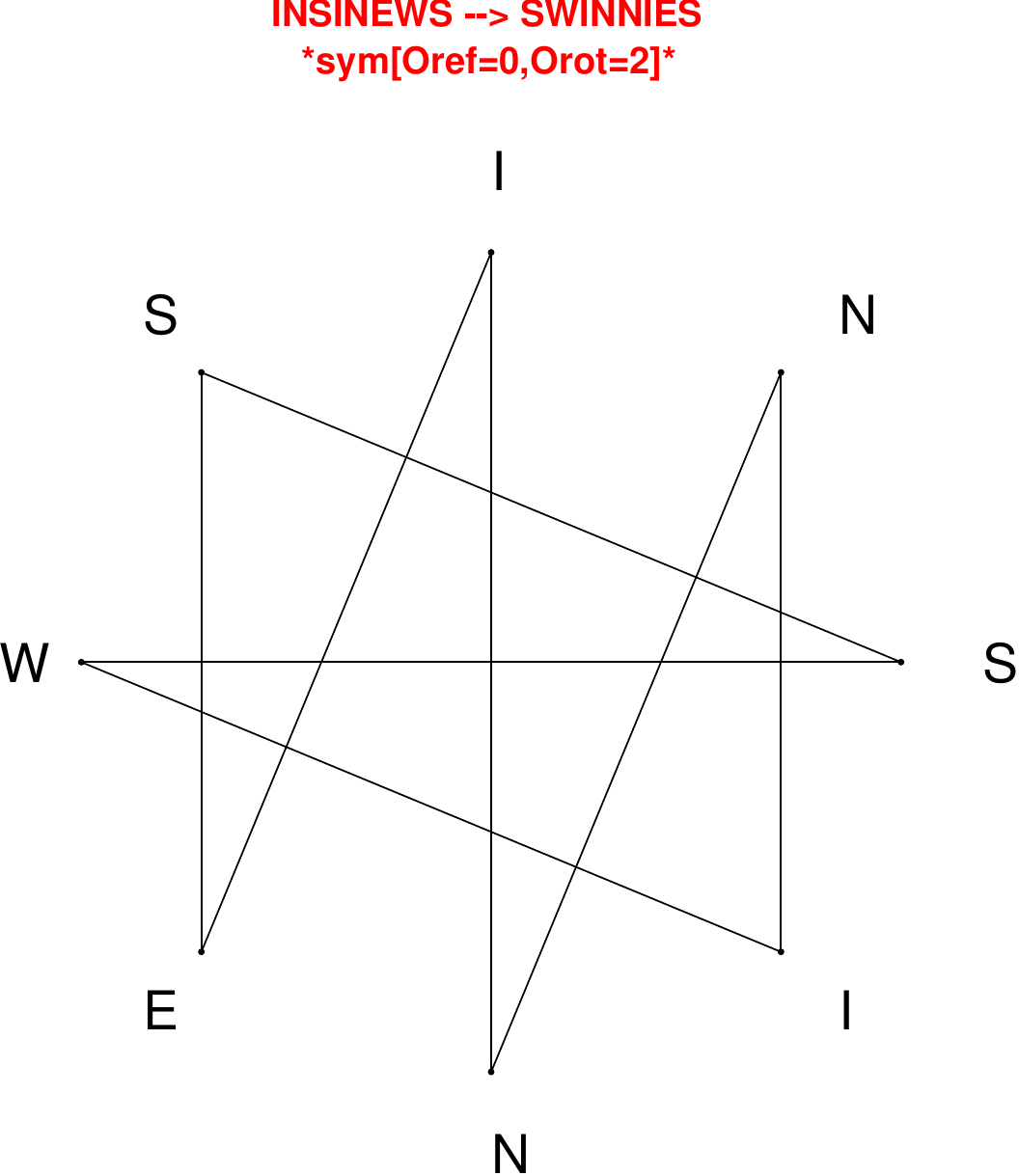}
\end{subfigure}
\hfill
\begin{subfigure}[T]{0.19\textwidth}
\centering
\includegraphics[width=\textwidth]{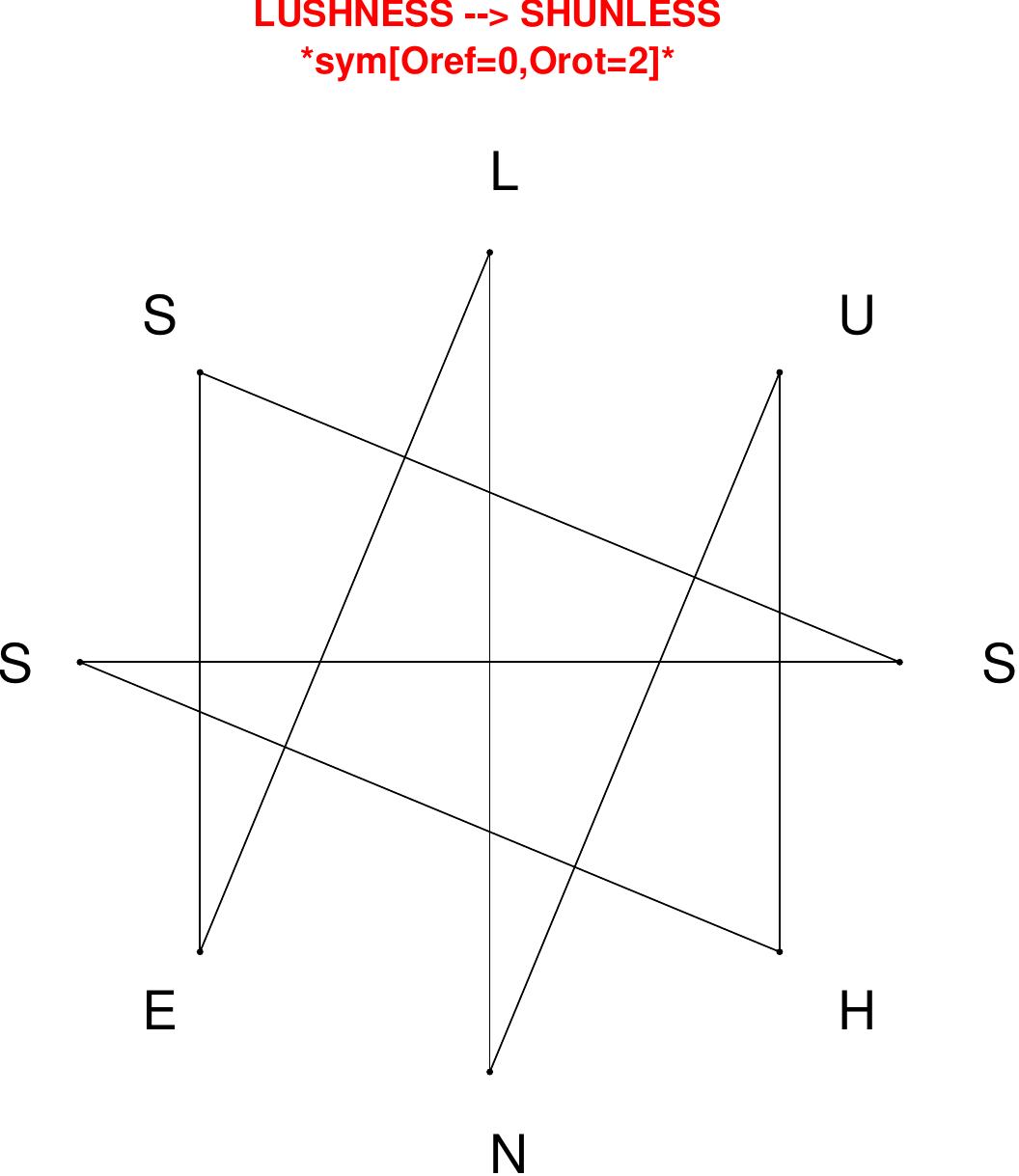}
\end{subfigure}
\hfill
\begin{subfigure}[T]{0.19\textwidth}
\centering
\includegraphics[width=\textwidth]{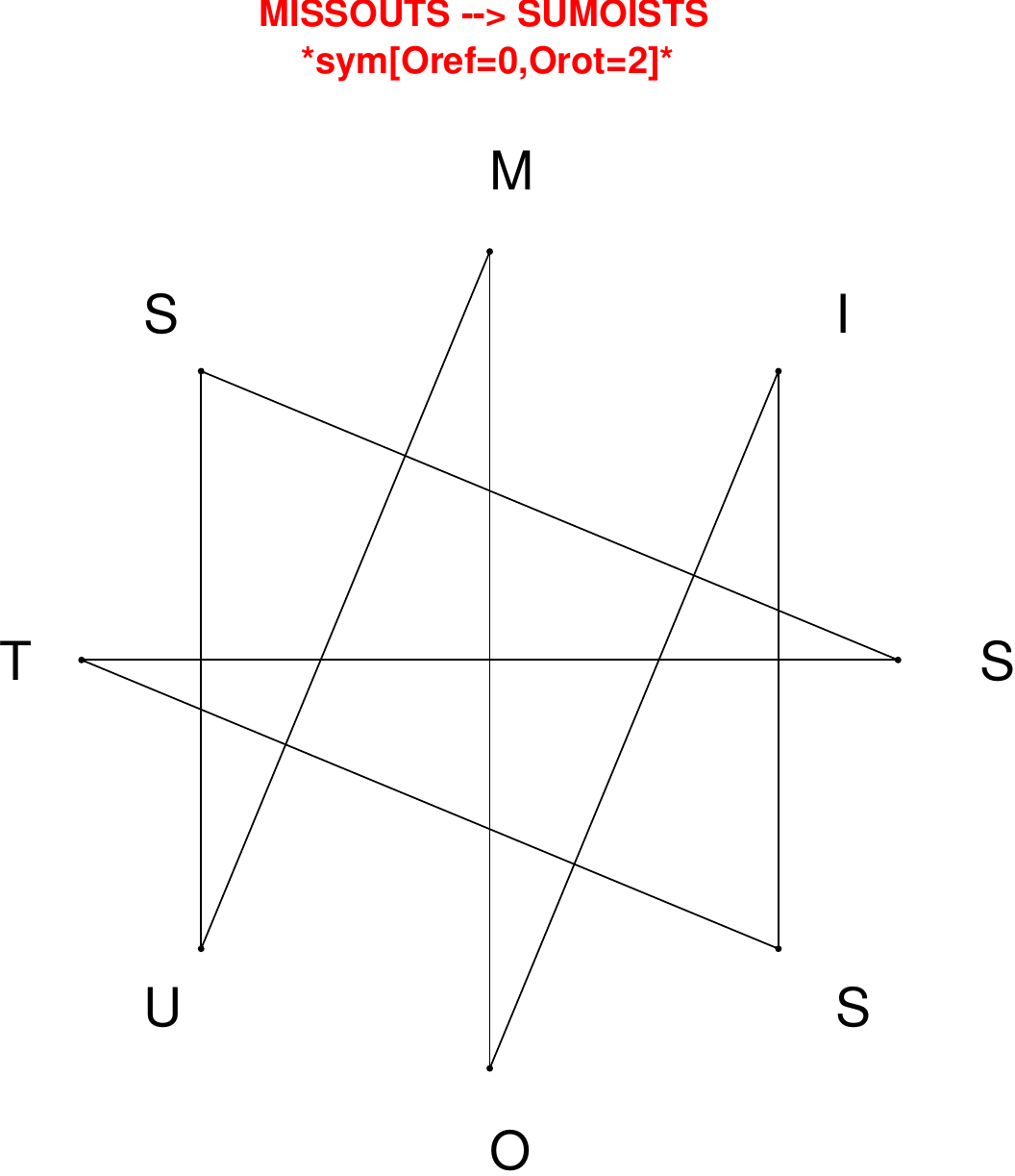}
\end{subfigure}
\hfill
\begin{subfigure}[T]{0.19\textwidth}
\centering
\includegraphics[width=\textwidth]{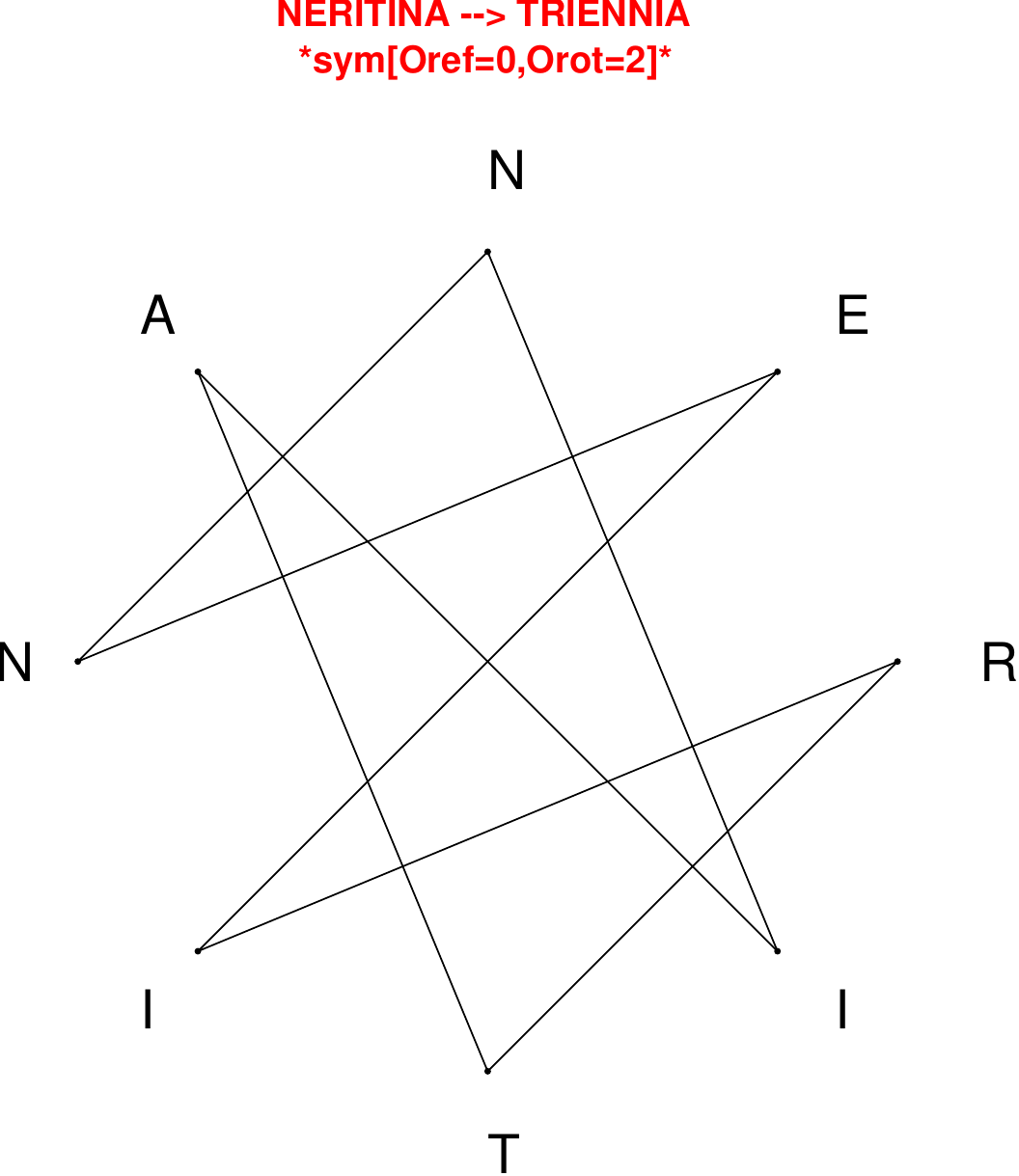}
\end{subfigure}
\hfill
\begin{subfigure}[T]{0.19\textwidth}
\centering
\includegraphics[width=\textwidth]{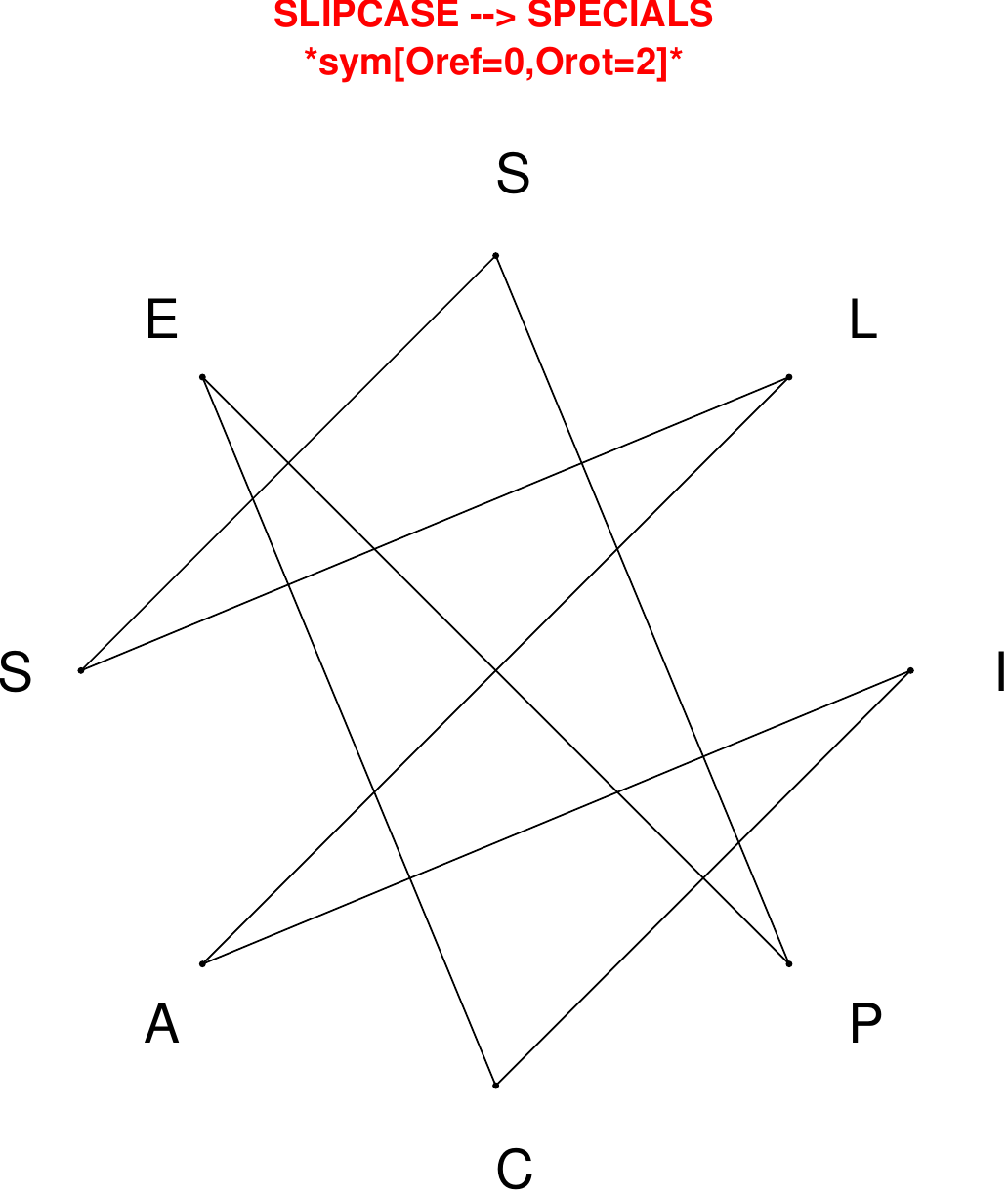}
\end{subfigure}
\end{figure}

\begin{figure}[H]
\centering
\begin{subfigure}[T]{0.19\textwidth}
\centering
\includegraphics[width=\textwidth]{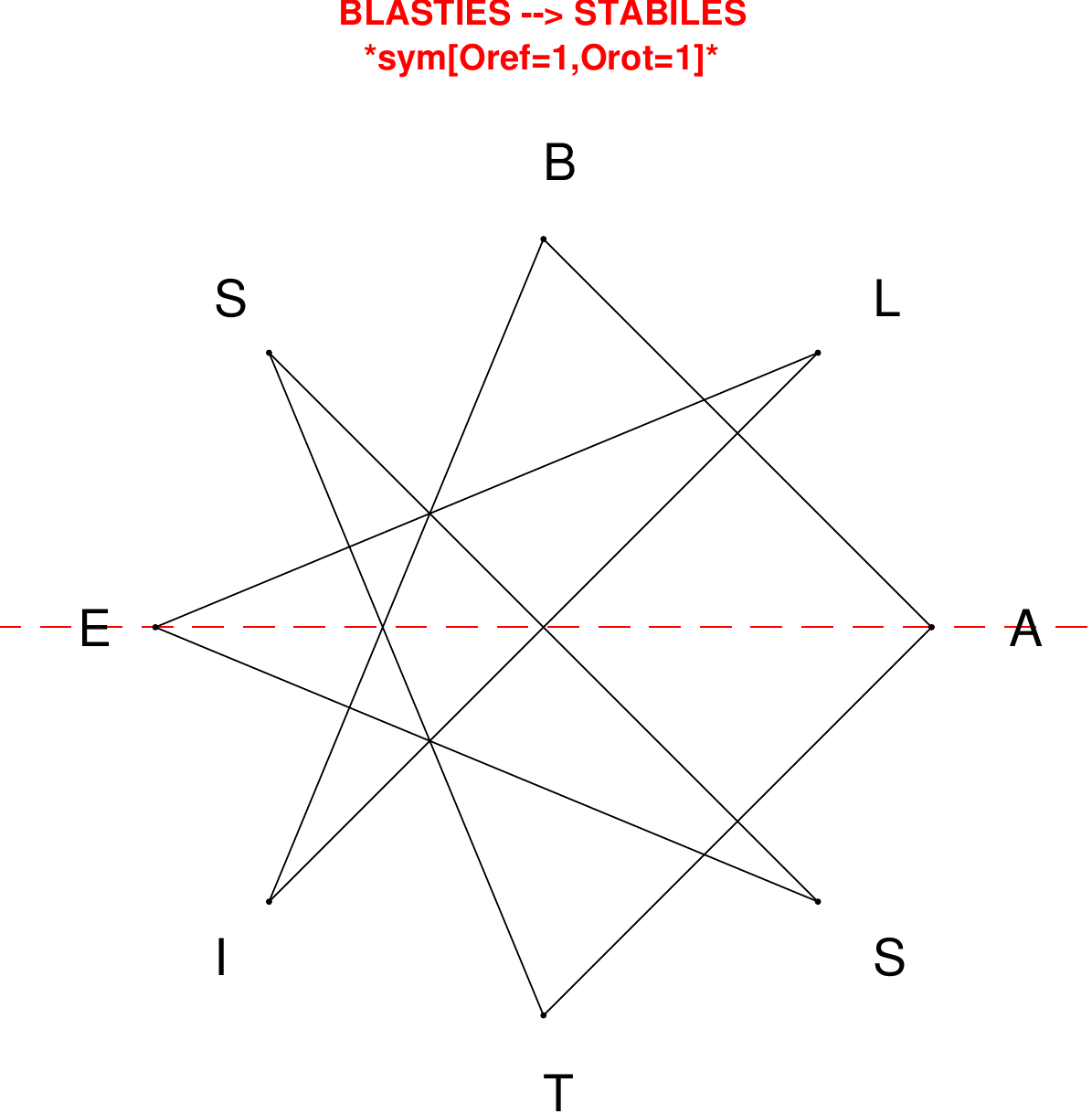}
\end{subfigure}
\hfill
\begin{subfigure}[T]{0.19\textwidth}
\centering
\includegraphics[width=\textwidth]{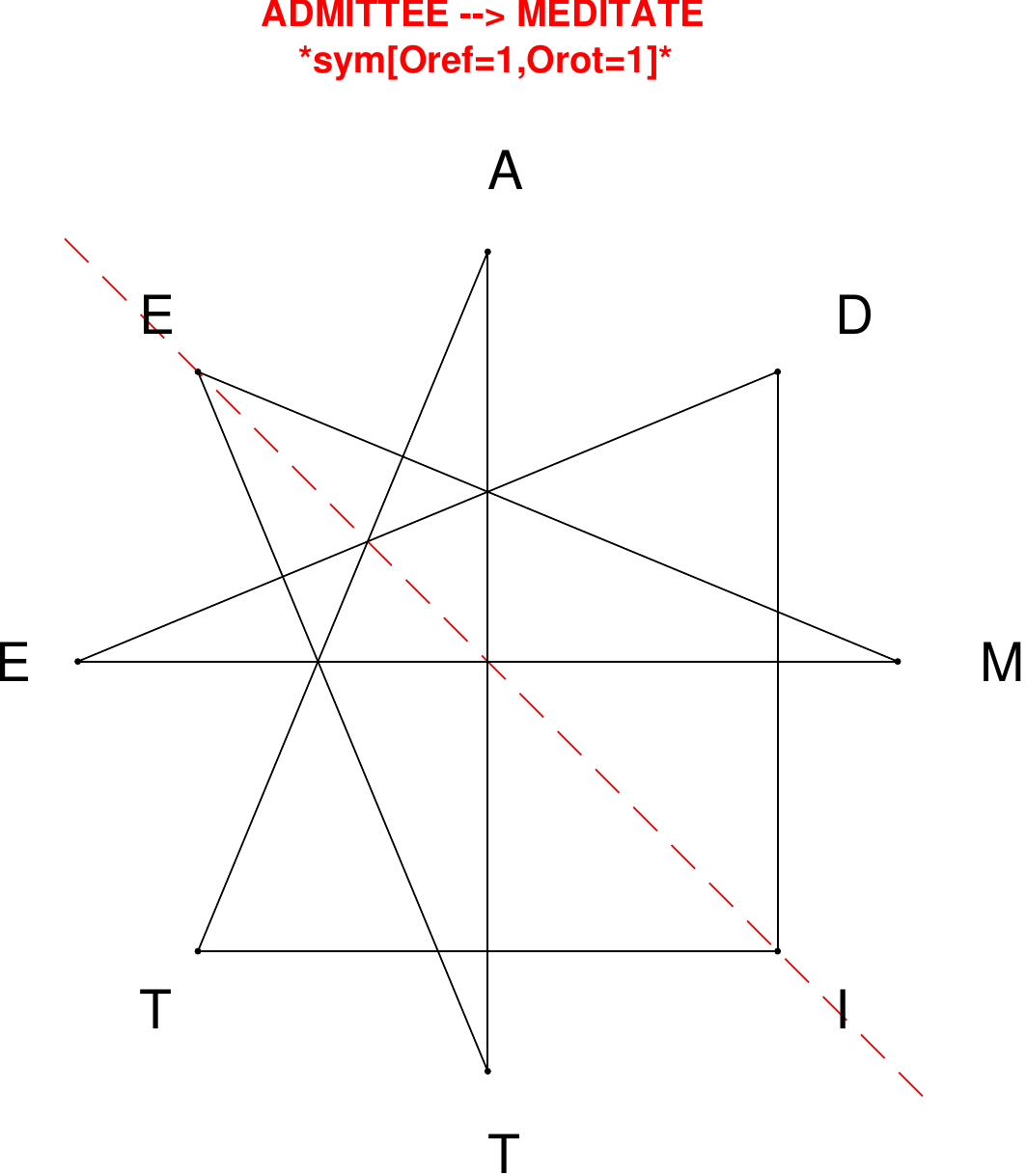}
\end{subfigure}
\hfill
\begin{subfigure}[T]{0.19\textwidth}
\centering
\includegraphics[width=\textwidth]{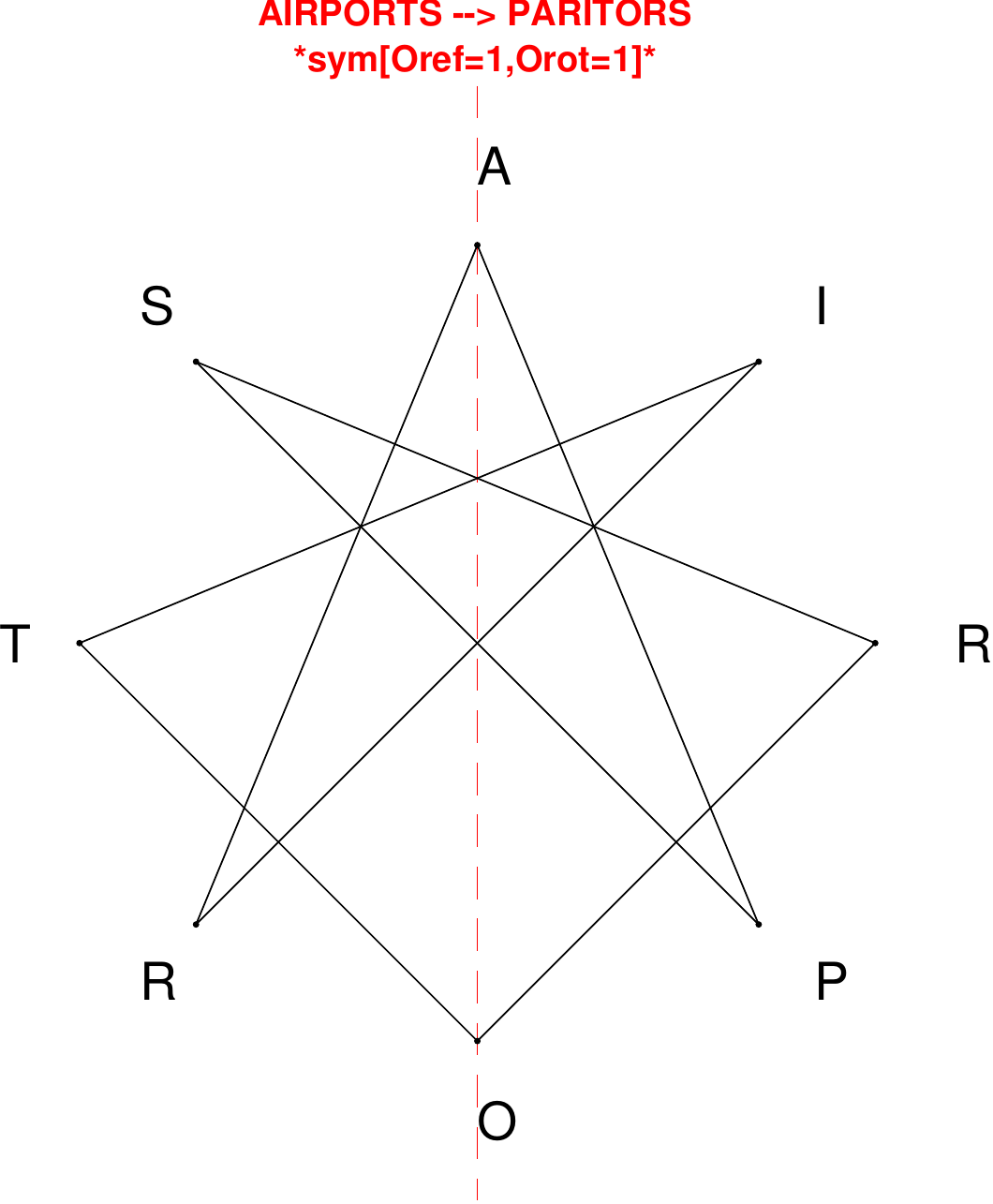}
\end{subfigure}
\hfill
\begin{subfigure}[T]{0.19\textwidth}
\centering
\includegraphics[width=\textwidth]{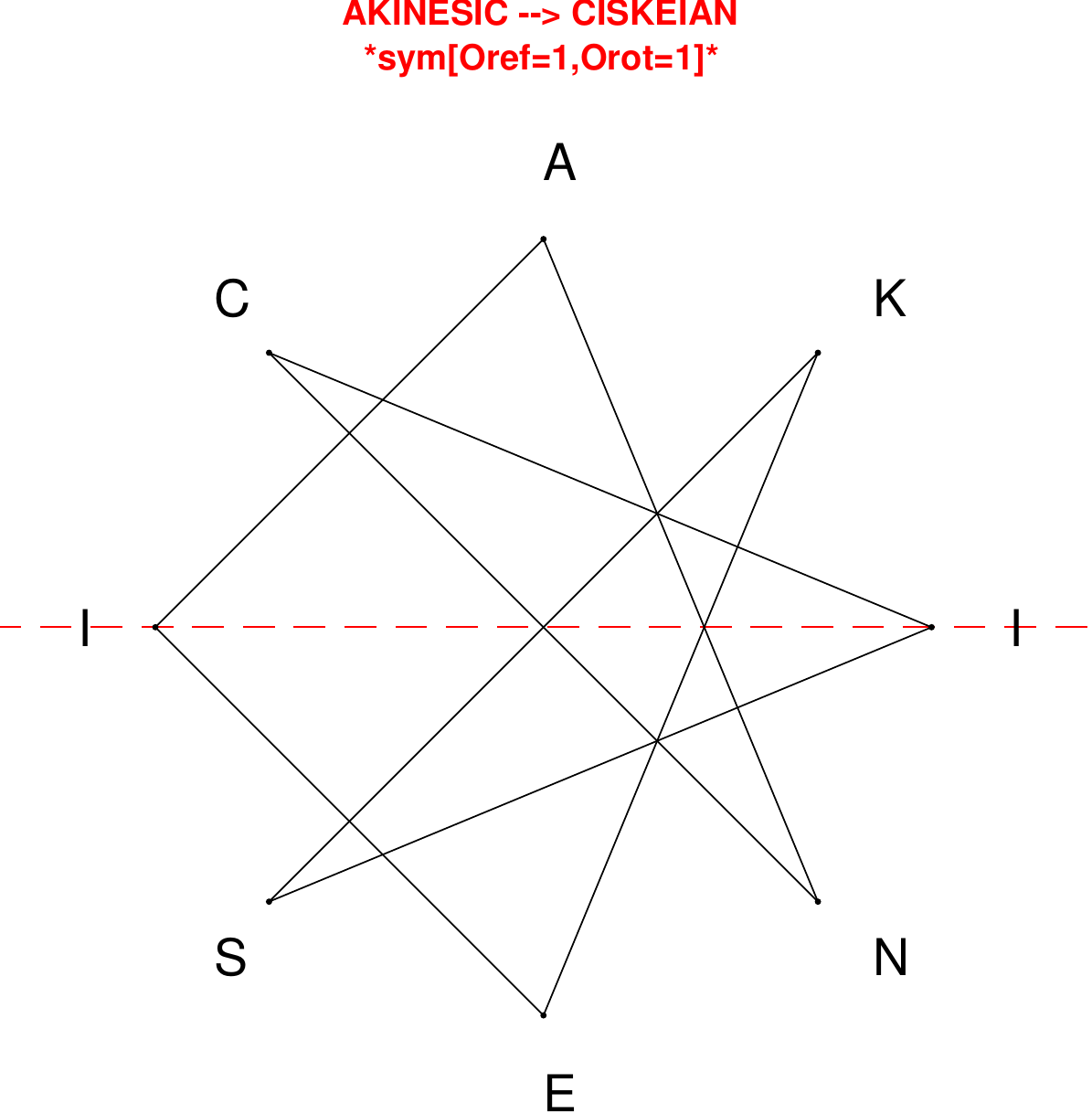}
\end{subfigure}
\hfill
\begin{subfigure}[T]{0.19\textwidth}
\centering
\includegraphics[width=\textwidth]{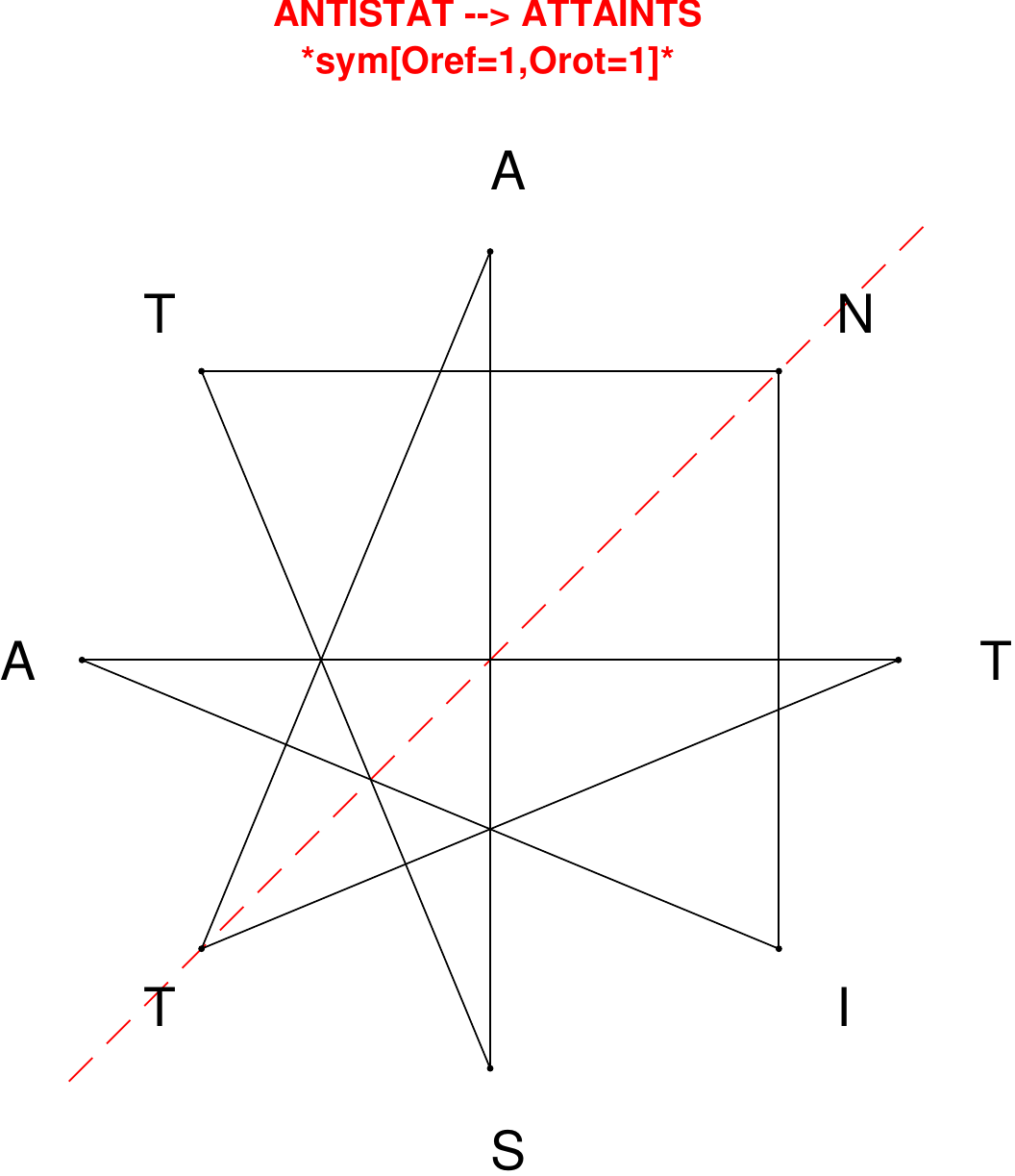}
\end{subfigure}
\end{figure}

\begin{figure}[H]
\centering
\begin{subfigure}[T]{0.19\textwidth}
\centering
\includegraphics[width=\textwidth]{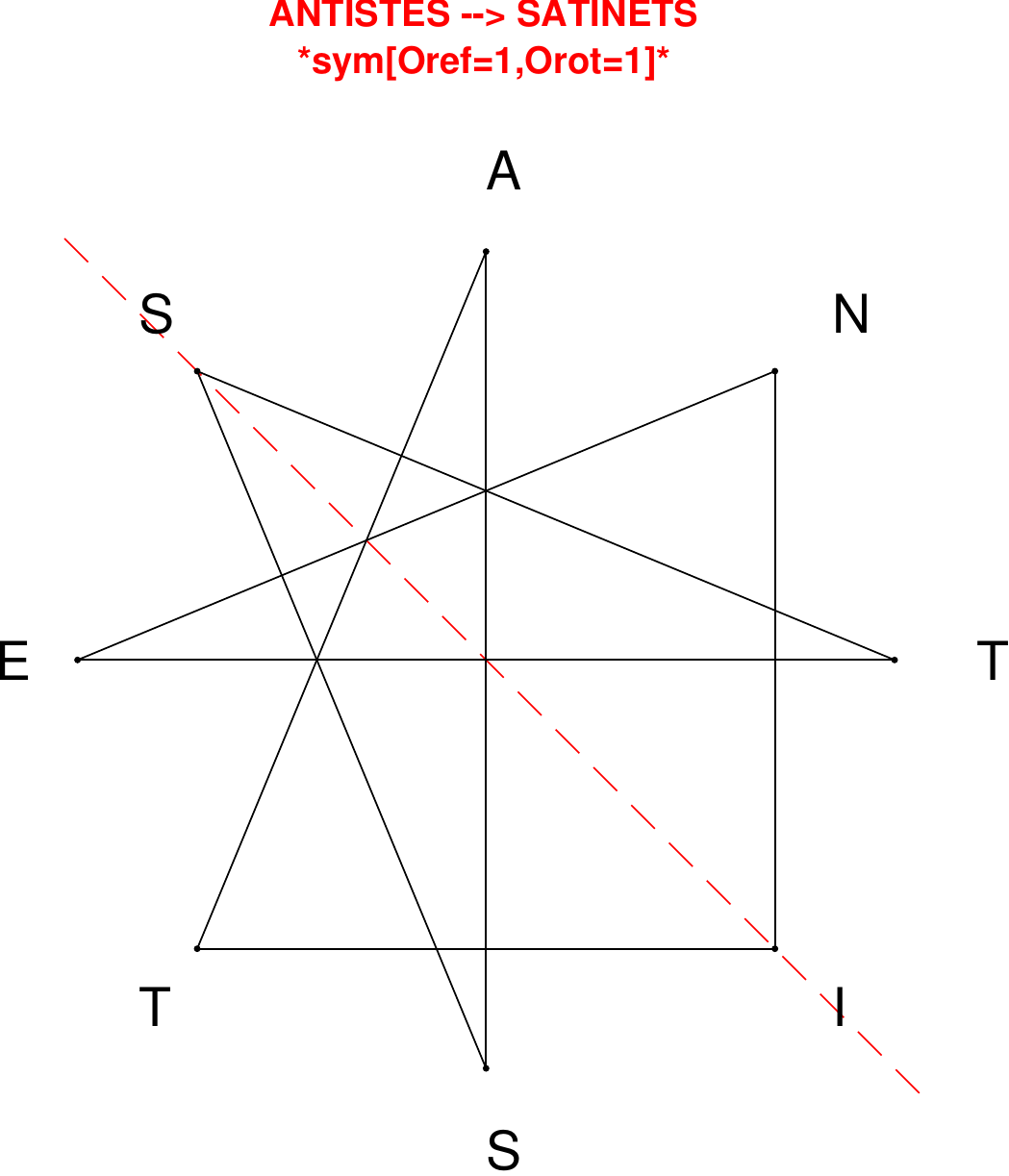}
\end{subfigure}
\hfill
\begin{subfigure}[T]{0.19\textwidth}
\centering
\includegraphics[width=\textwidth]{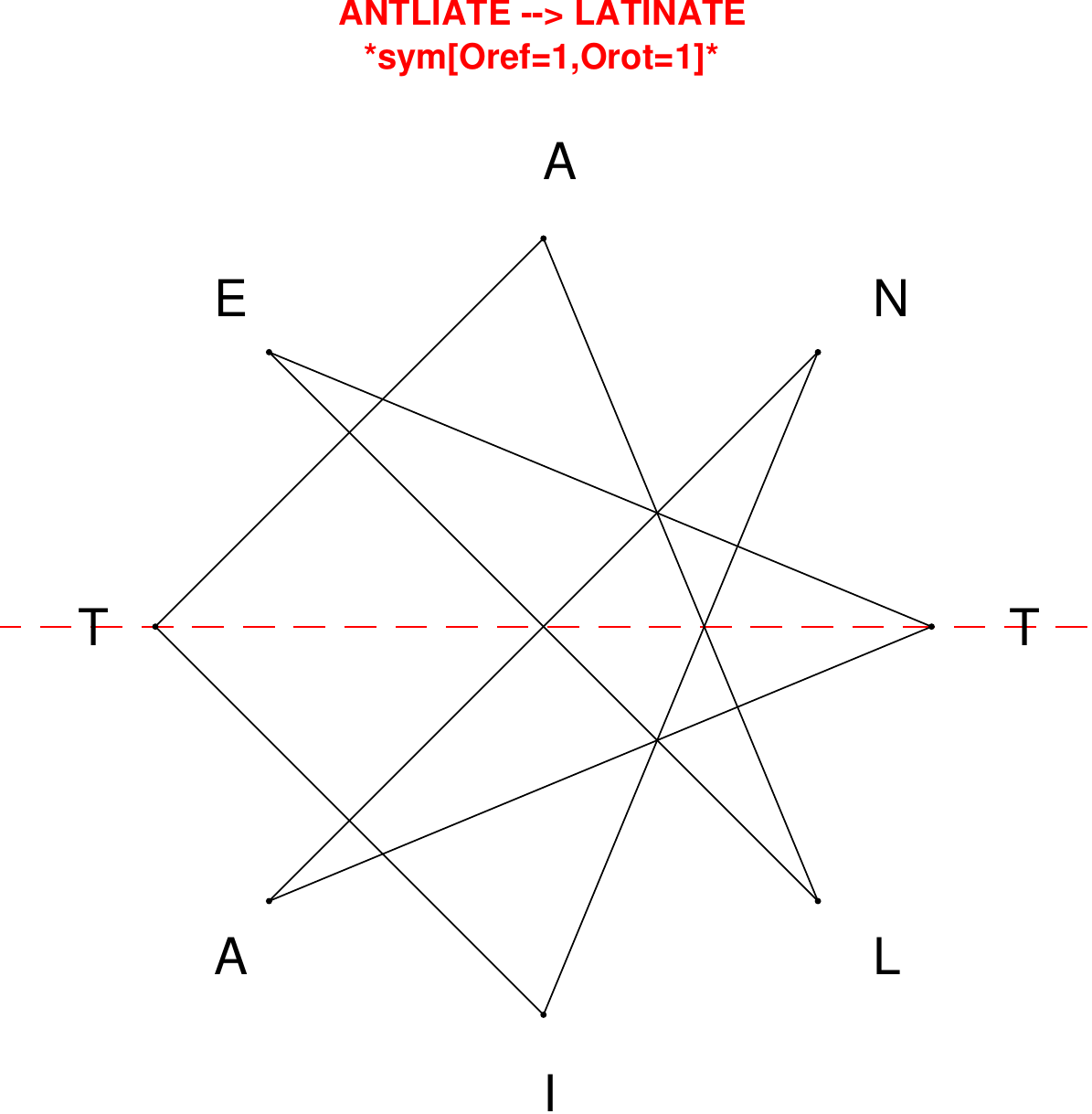}
\end{subfigure}
\hfill
\begin{subfigure}[T]{0.19\textwidth}
\centering
\includegraphics[width=\textwidth]{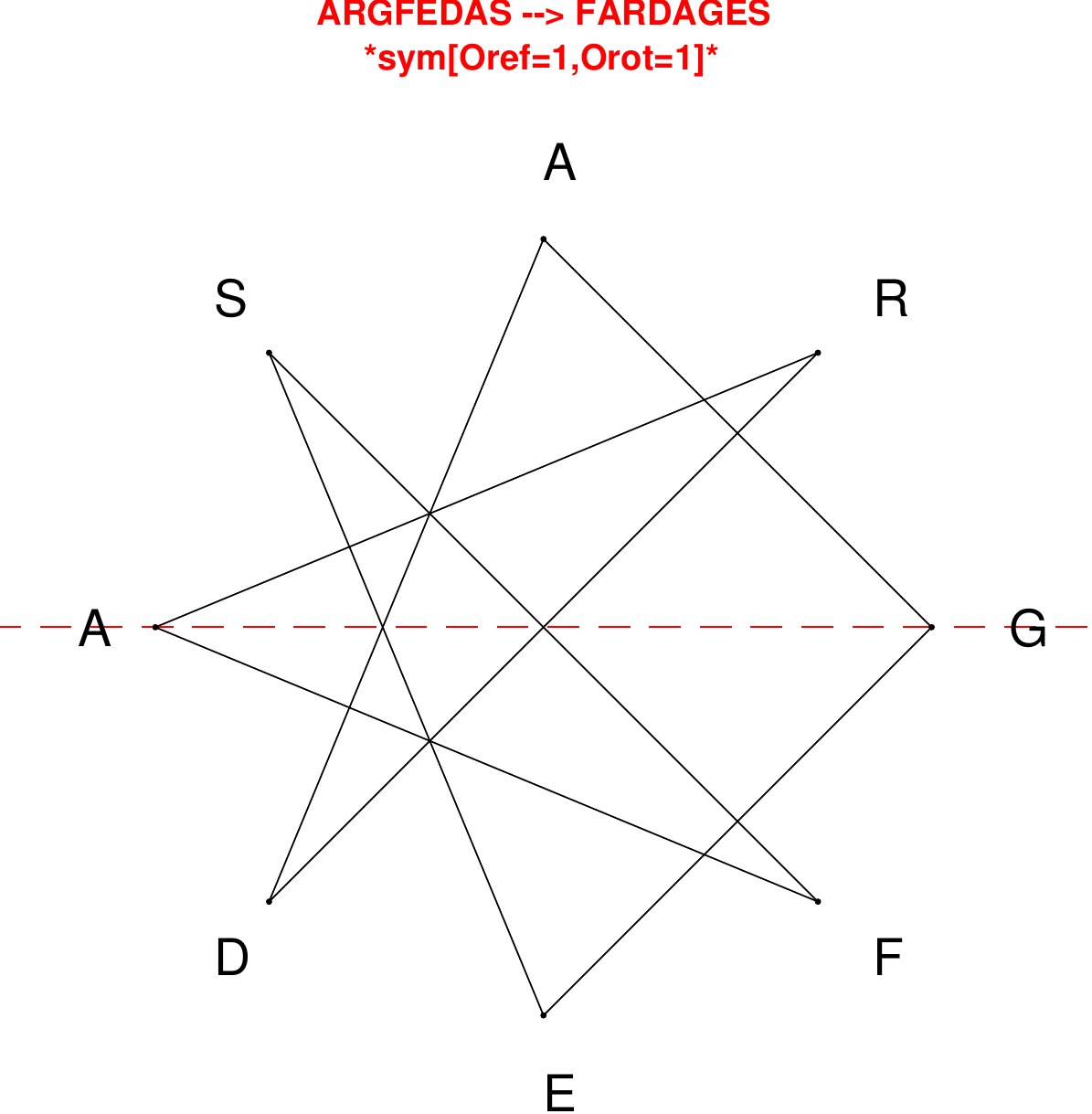}
\end{subfigure}
\hfill
\begin{subfigure}[T]{0.19\textwidth}
\centering
\includegraphics[width=\textwidth]{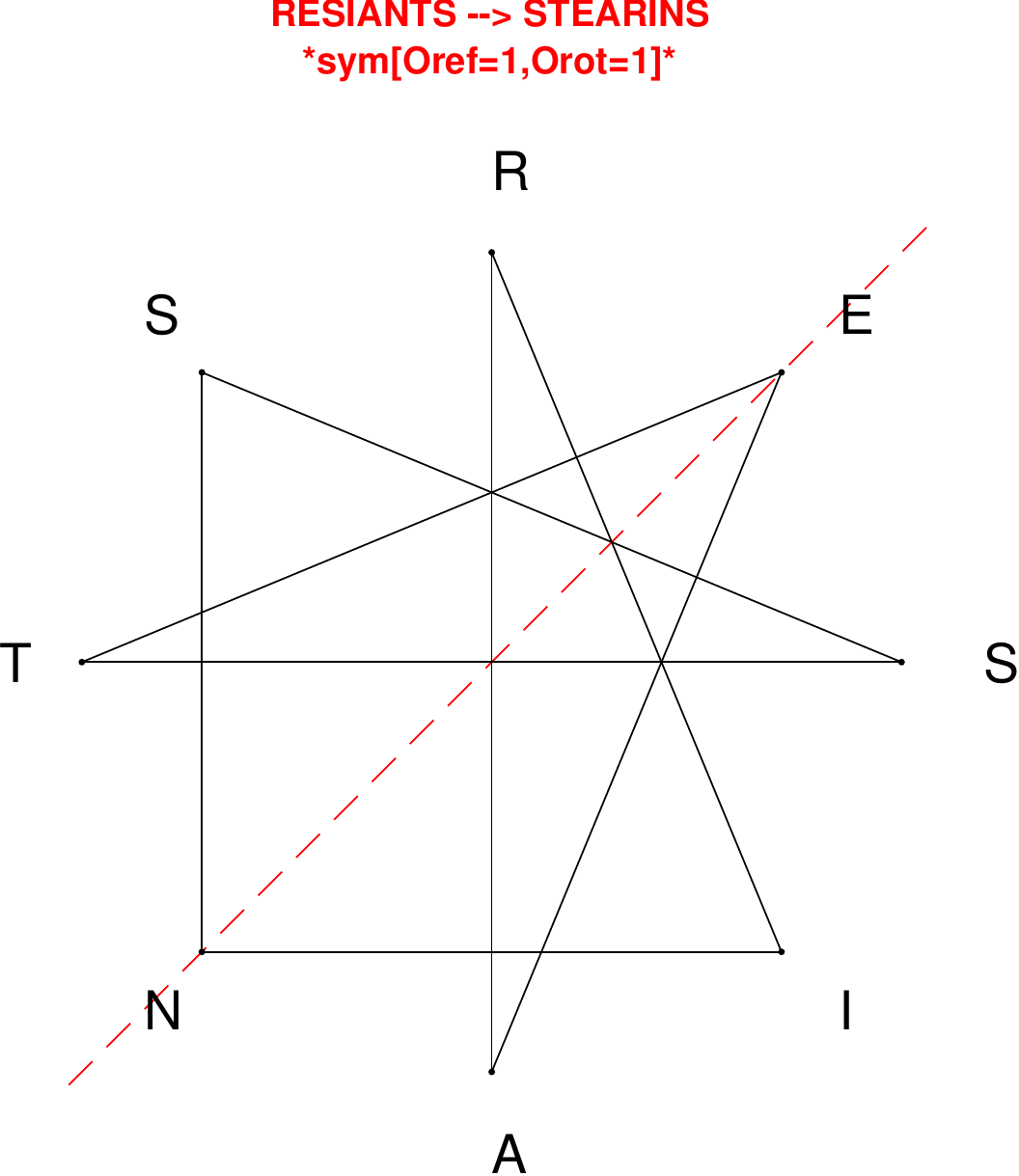}
\end{subfigure}
\hfill
\begin{subfigure}[T]{0.19\textwidth}
\centering
\includegraphics[width=\textwidth]{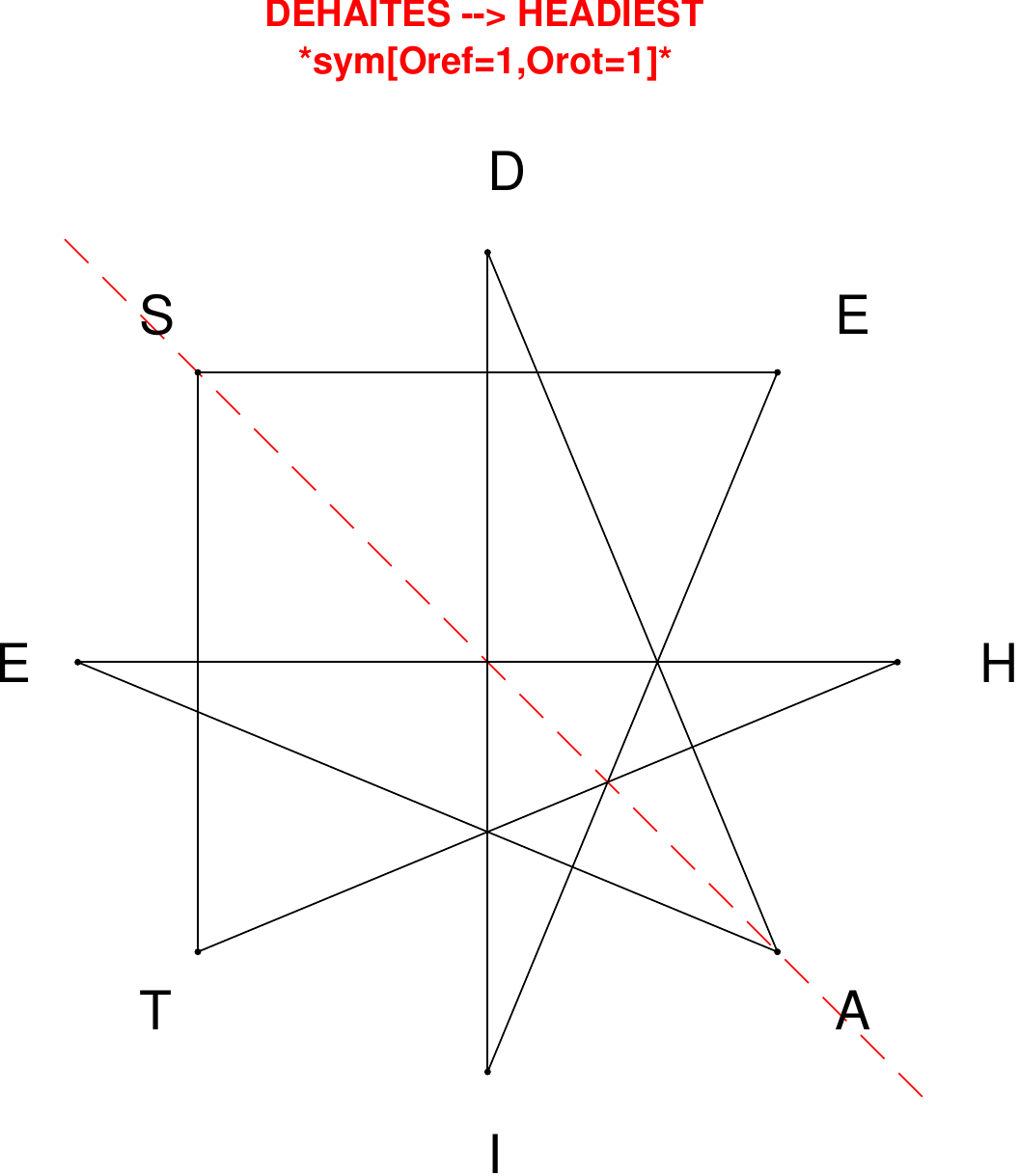}
\end{subfigure}
\end{figure}

\begin{figure}[H]
\centering
\begin{subfigure}[T]{0.19\textwidth}
\centering
\includegraphics[width=\textwidth]{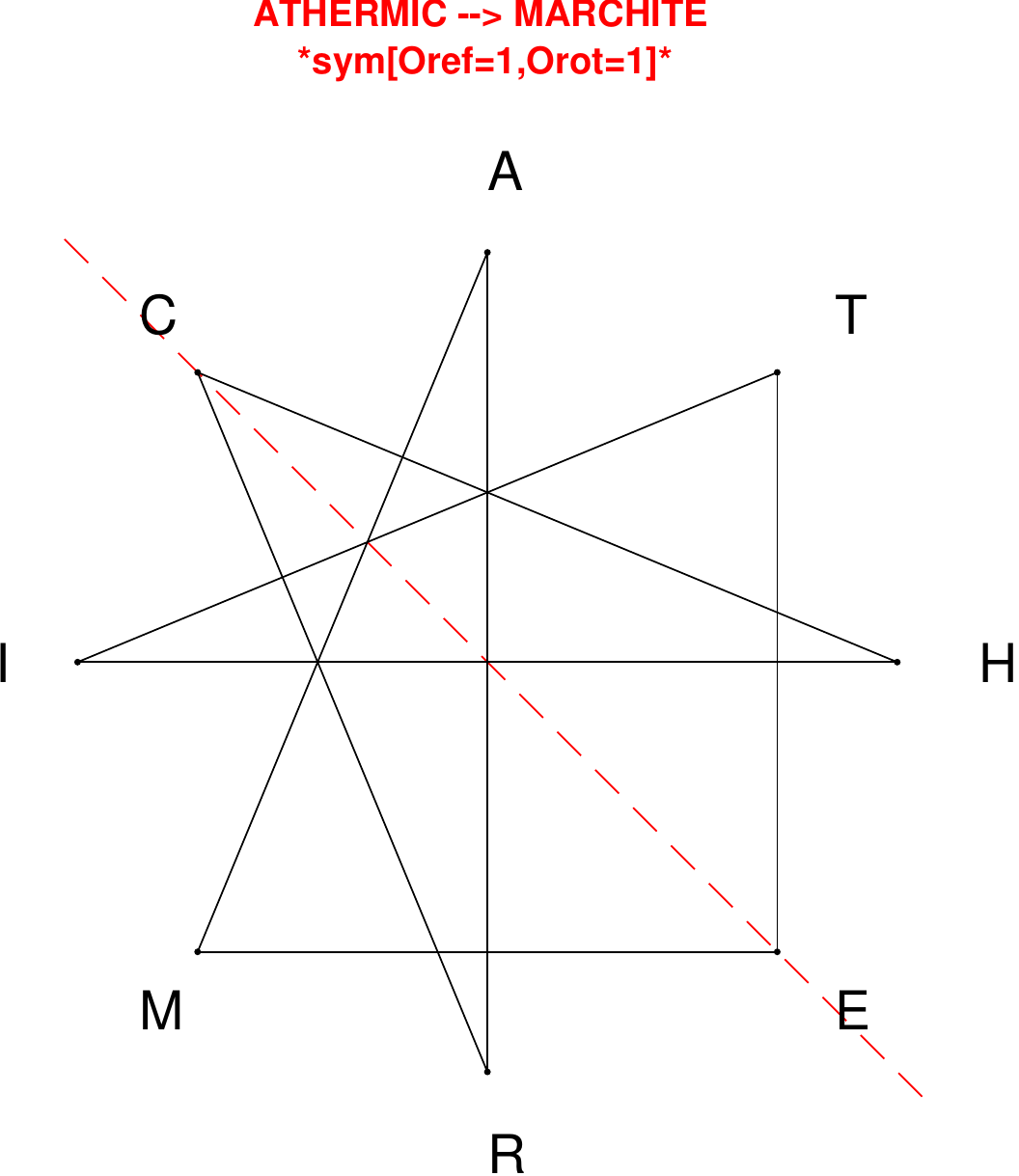}
\end{subfigure}
\hfill
\begin{subfigure}[T]{0.19\textwidth}
\centering
\includegraphics[width=\textwidth]{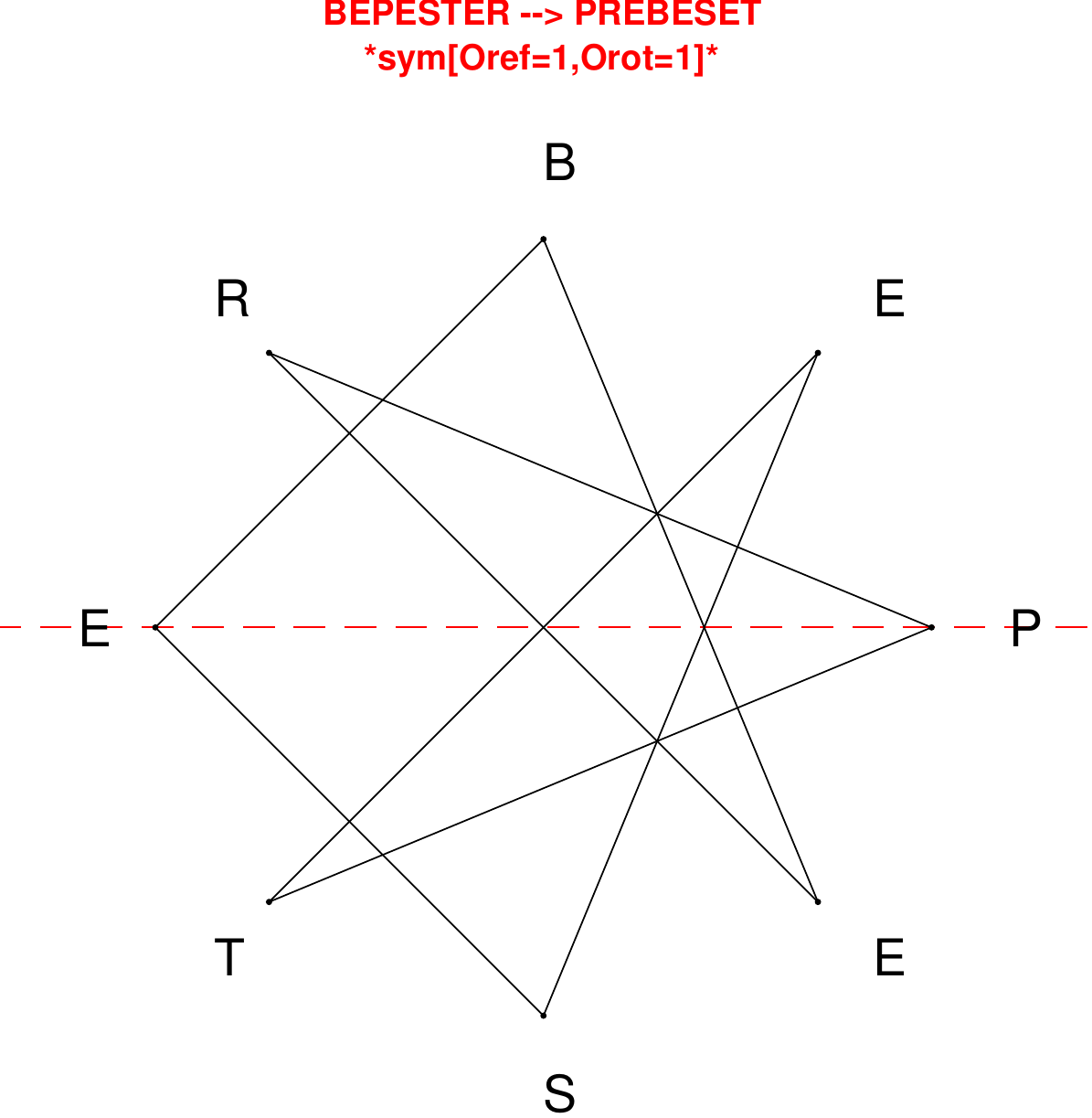}
\end{subfigure}
\hfill
\begin{subfigure}[T]{0.19\textwidth}
\centering
\includegraphics[width=\textwidth]{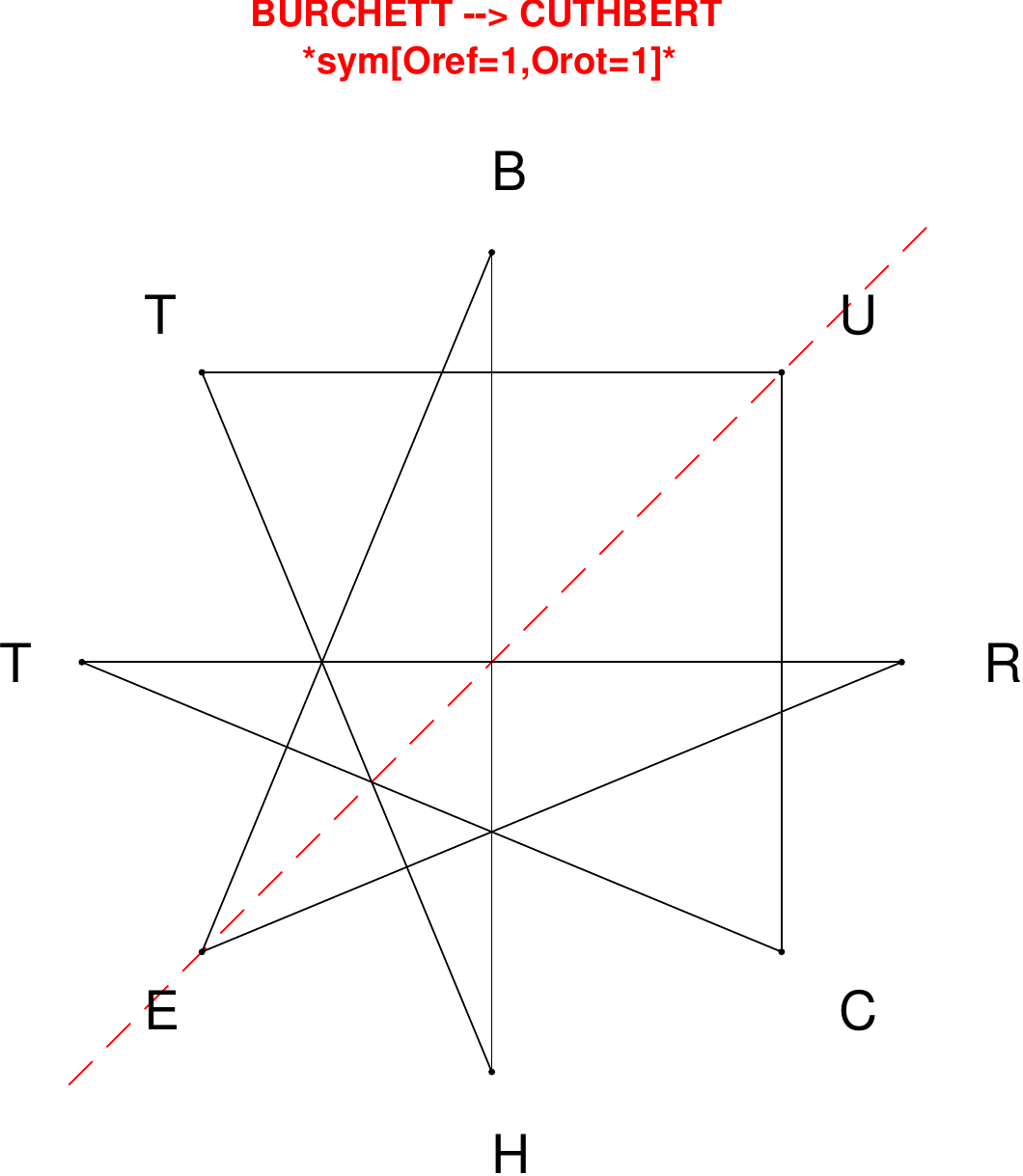}
\end{subfigure}
\hfill
\begin{subfigure}[T]{0.19\textwidth}
\centering
\includegraphics[width=\textwidth]{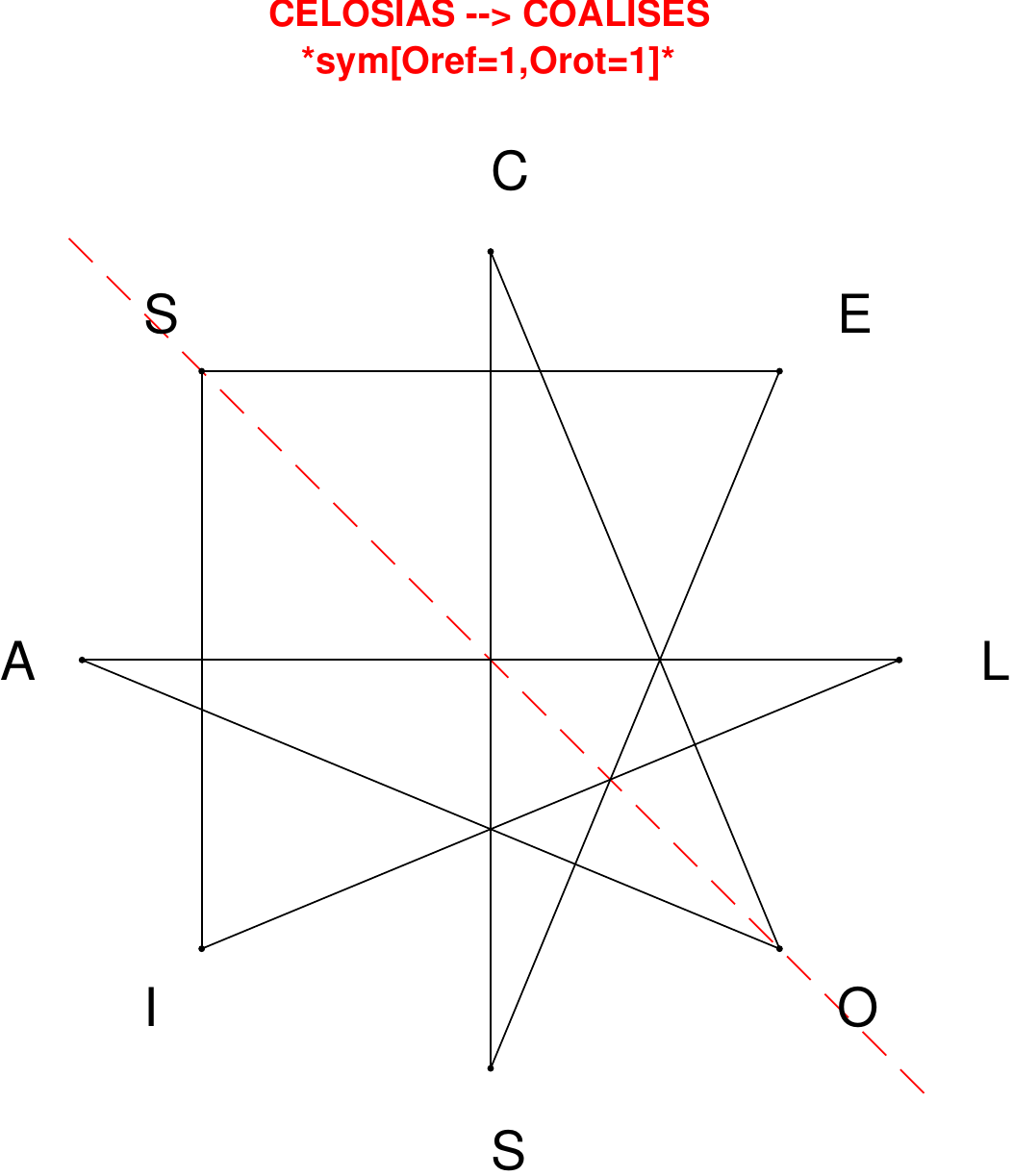}
\end{subfigure}
\hfill
\begin{subfigure}[T]{0.19\textwidth}
\centering
\includegraphics[width=\textwidth]{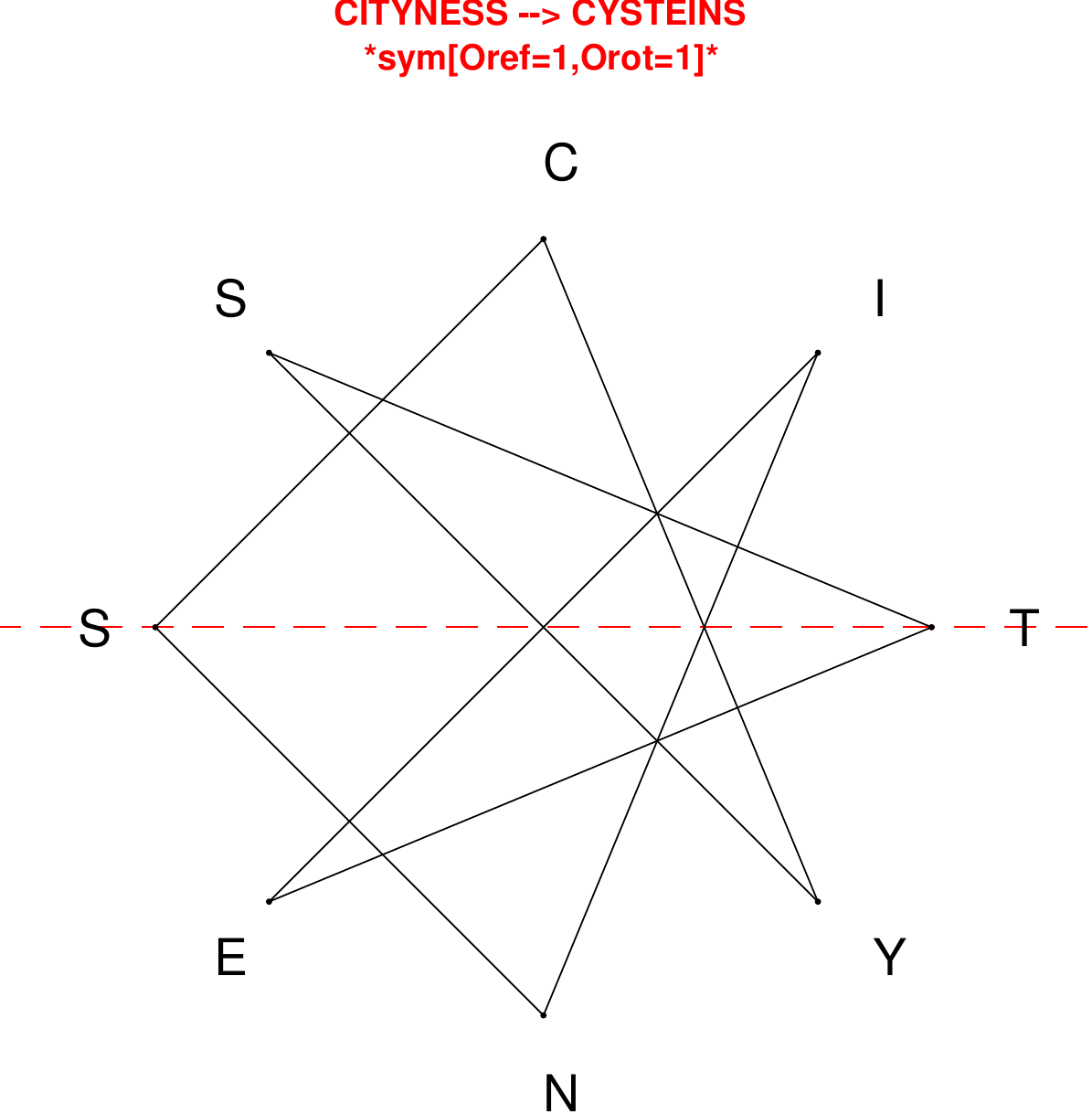}
\end{subfigure}
\end{figure}

\begin{figure}[H]
\centering
\begin{subfigure}[T]{0.19\textwidth}
\centering
\includegraphics[width=\textwidth]{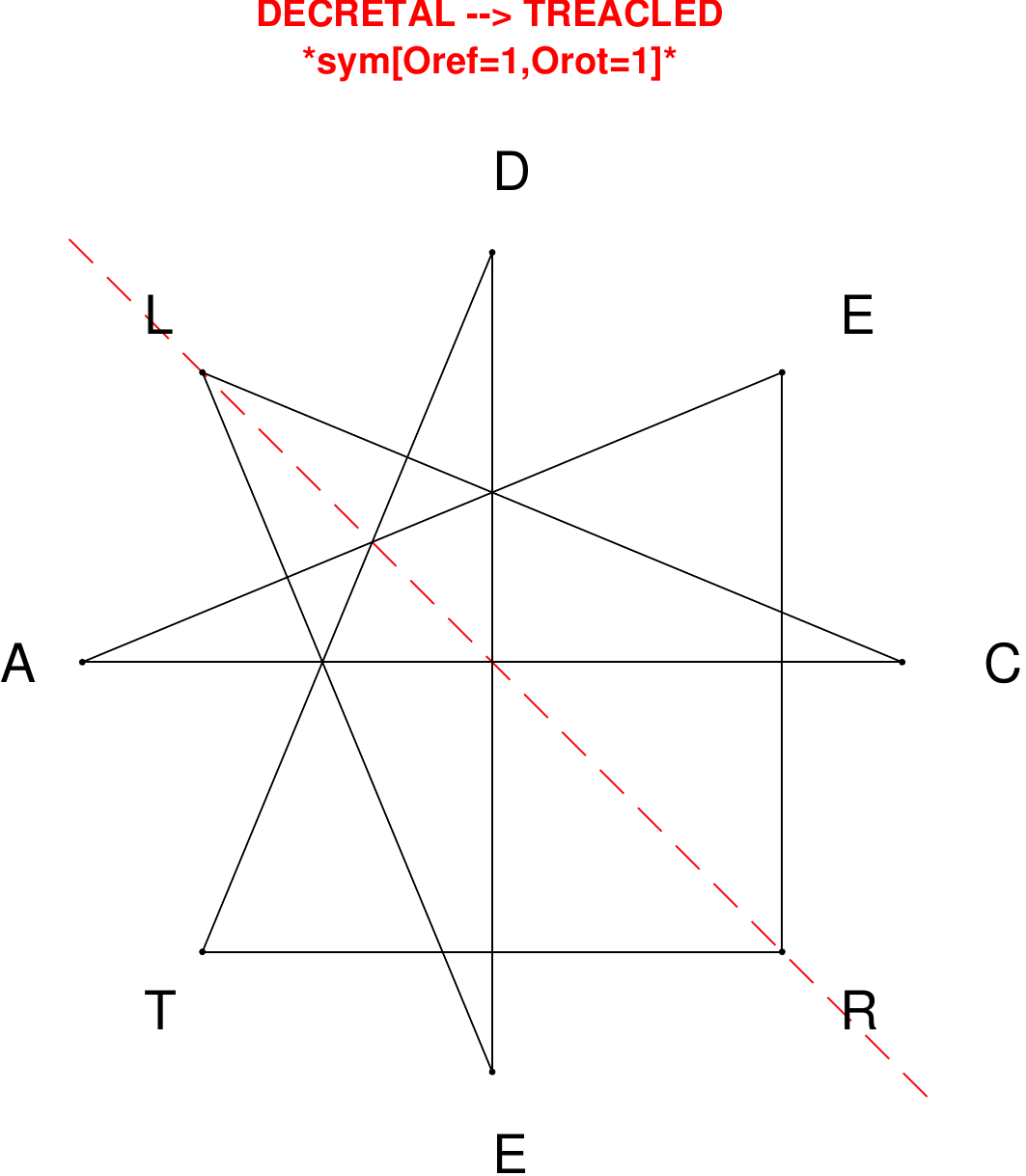}
\end{subfigure}
\hfill
\begin{subfigure}[T]{0.19\textwidth}
\centering
\includegraphics[width=\textwidth]{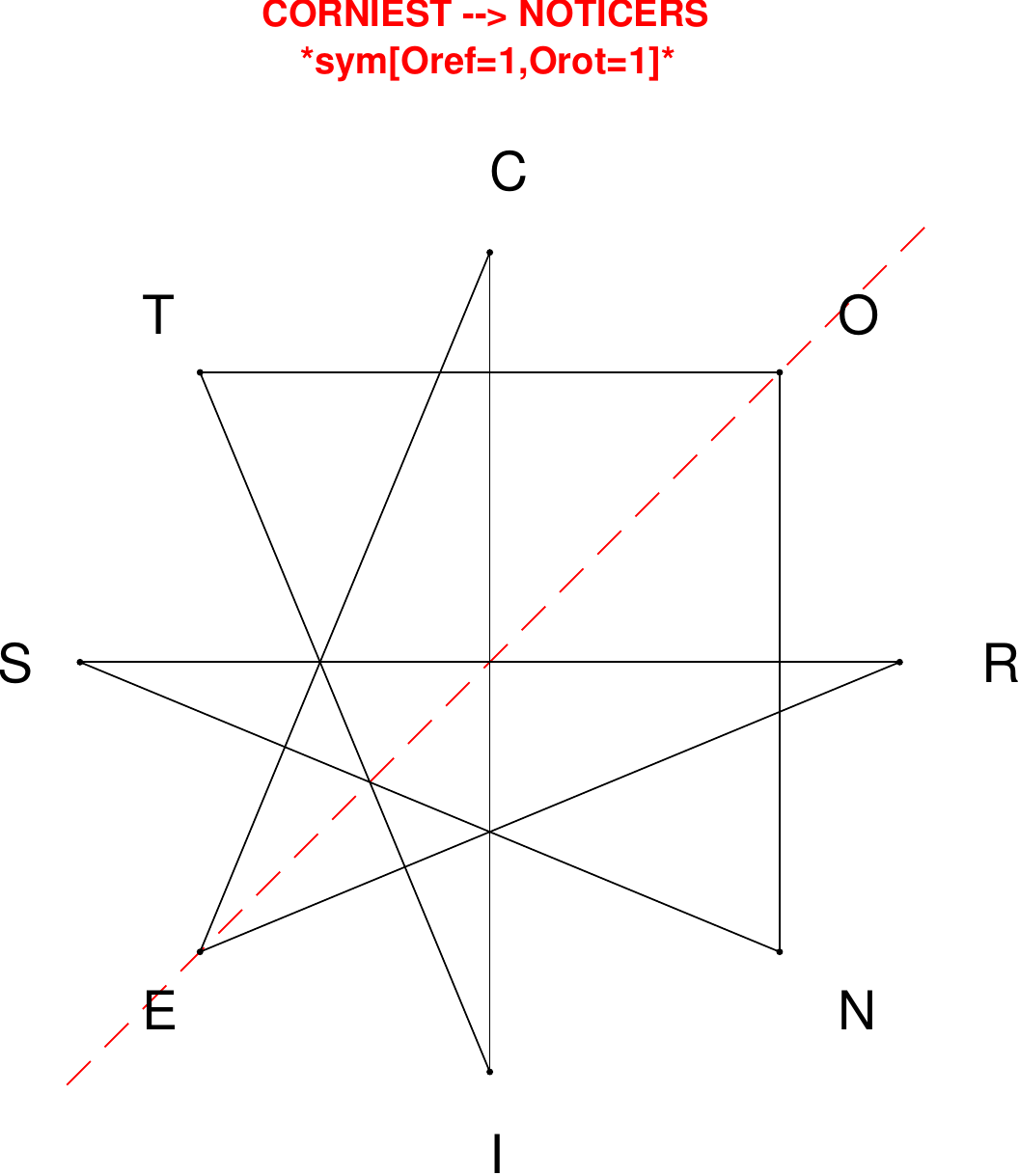}
\end{subfigure}
\hfill
\begin{subfigure}[T]{0.19\textwidth}
\centering
\includegraphics[width=\textwidth]{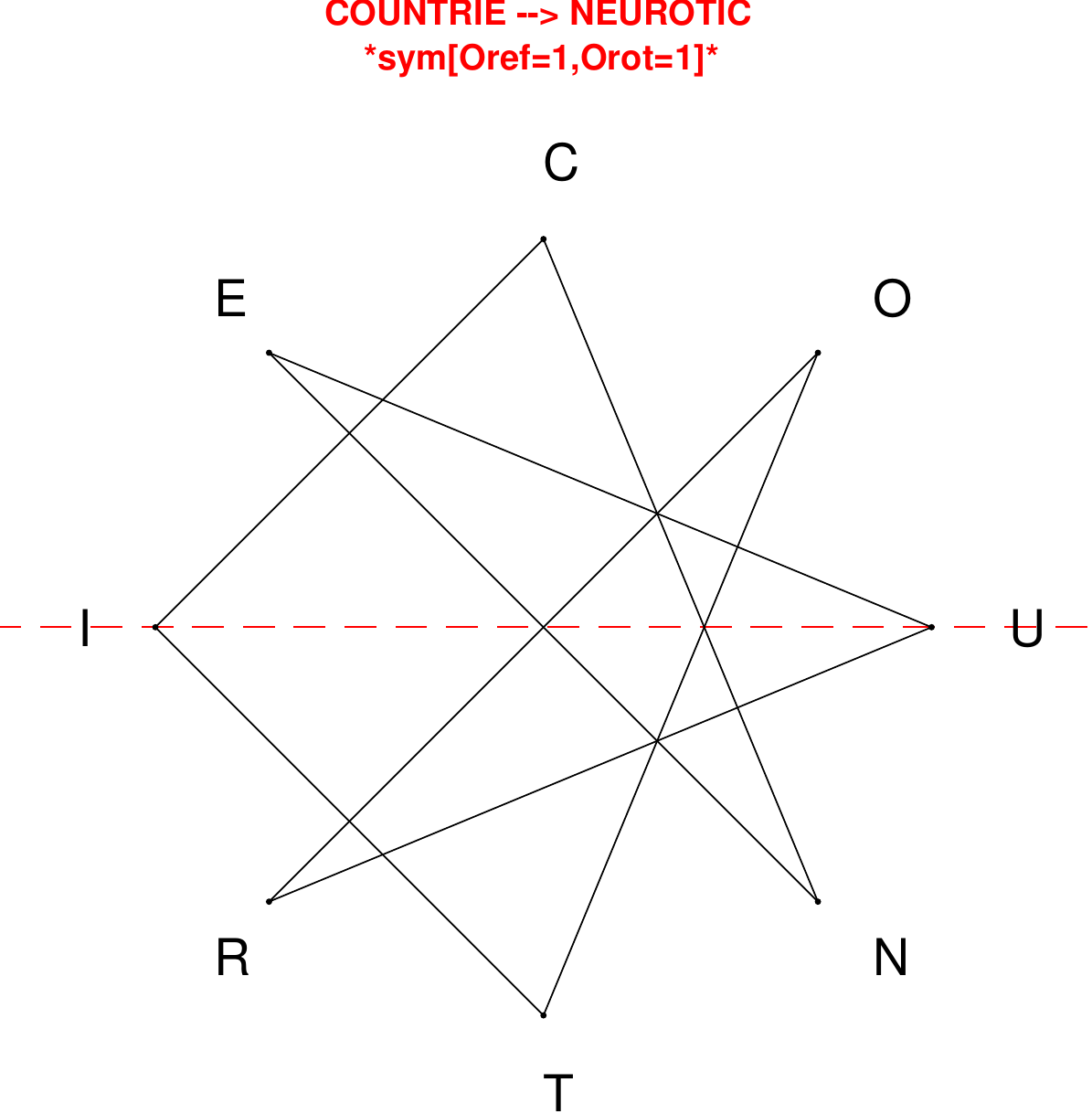}
\end{subfigure}
\hfill
\begin{subfigure}[T]{0.19\textwidth}
\centering
\includegraphics[width=\textwidth]{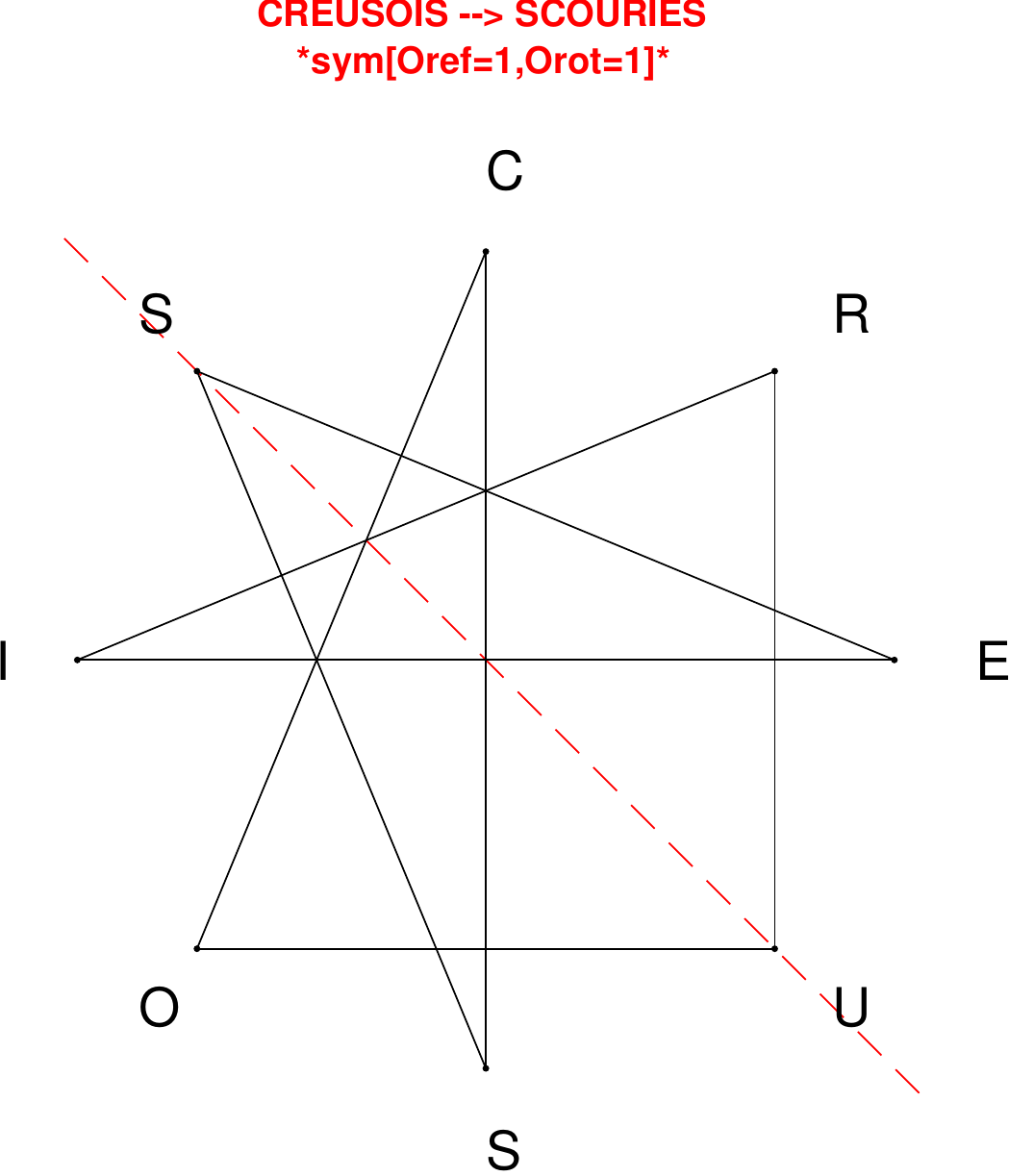}
\end{subfigure}
\hfill
\begin{subfigure}[T]{0.19\textwidth}
\centering
\includegraphics[width=\textwidth]{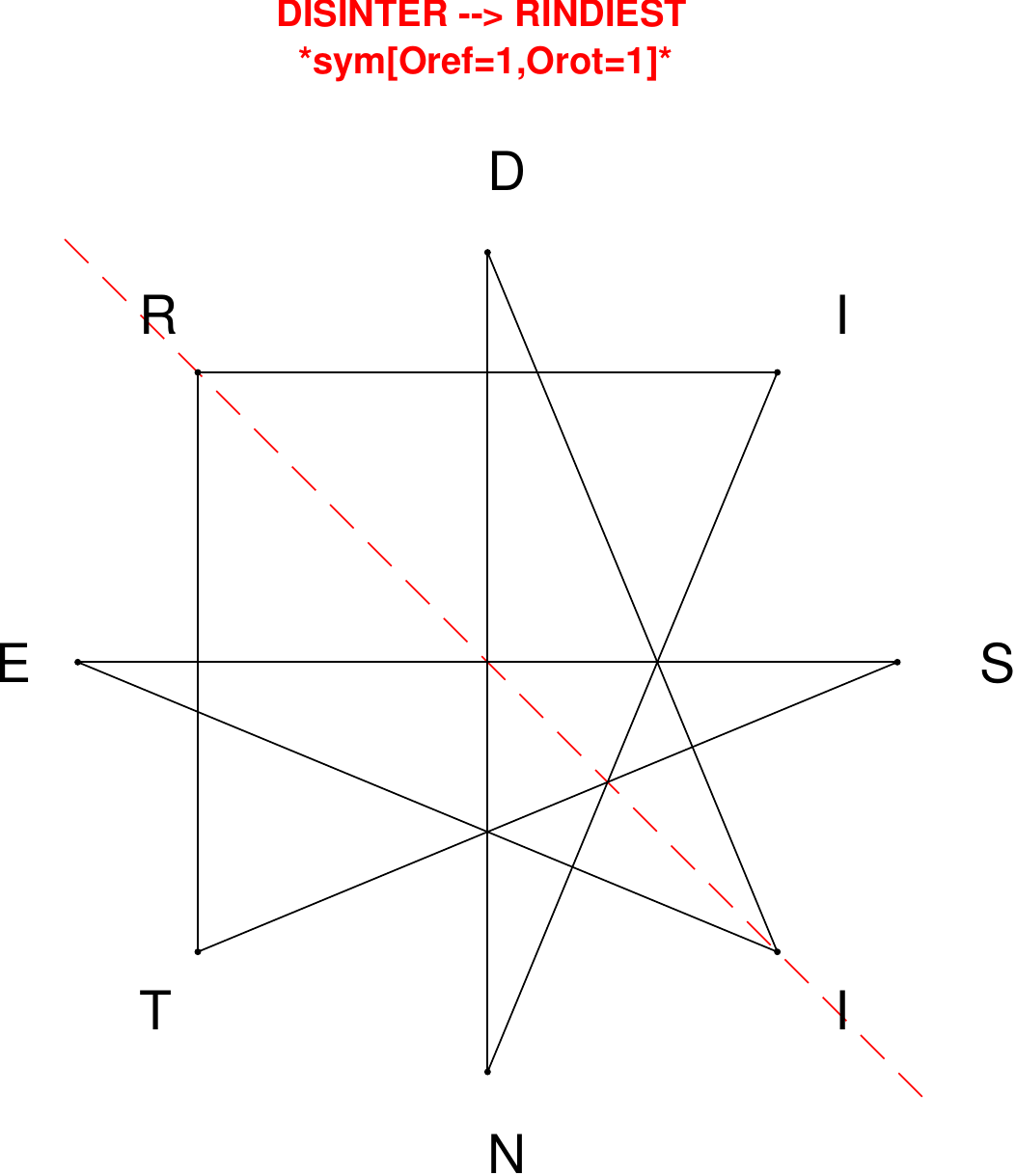}
\end{subfigure}
\end{figure}

\begin{figure}[H]
\centering
\begin{subfigure}[T]{0.19\textwidth}
\centering
\includegraphics[width=\textwidth]{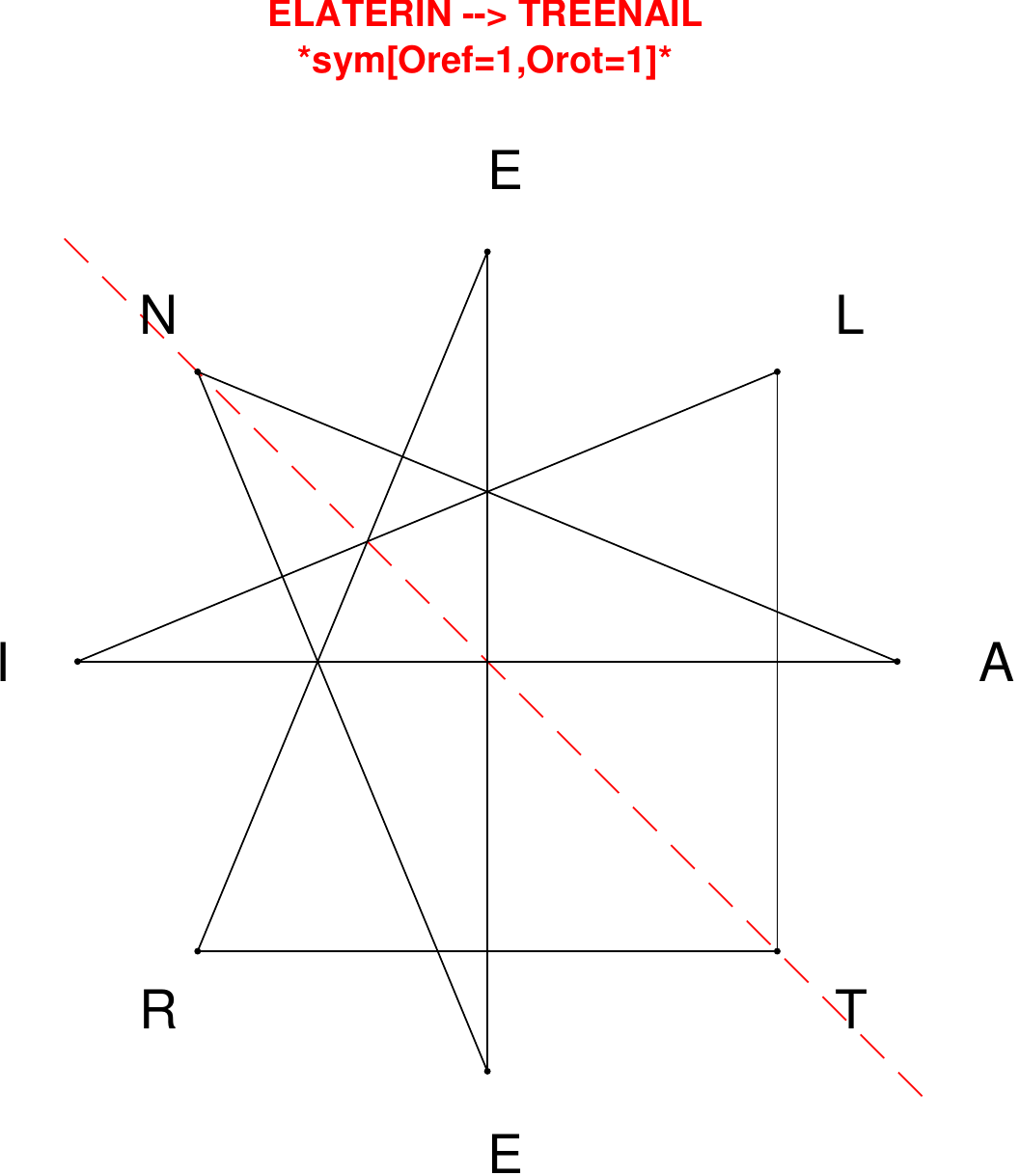}
\end{subfigure}
\hfill
\begin{subfigure}[T]{0.19\textwidth}
\centering
\includegraphics[width=\textwidth]{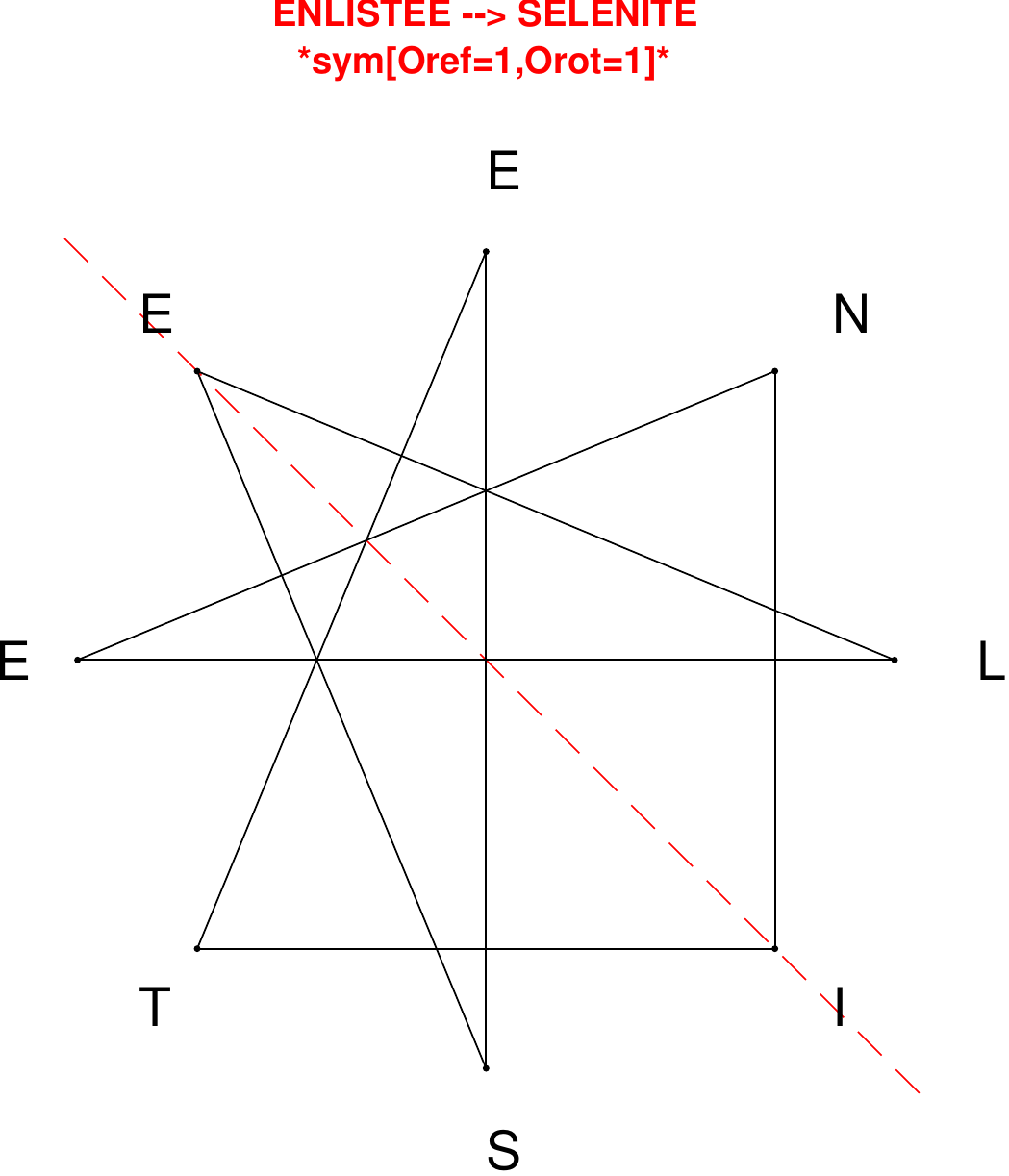}
\end{subfigure}
\hfill
\begin{subfigure}[T]{0.19\textwidth}
\centering
\includegraphics[width=\textwidth]{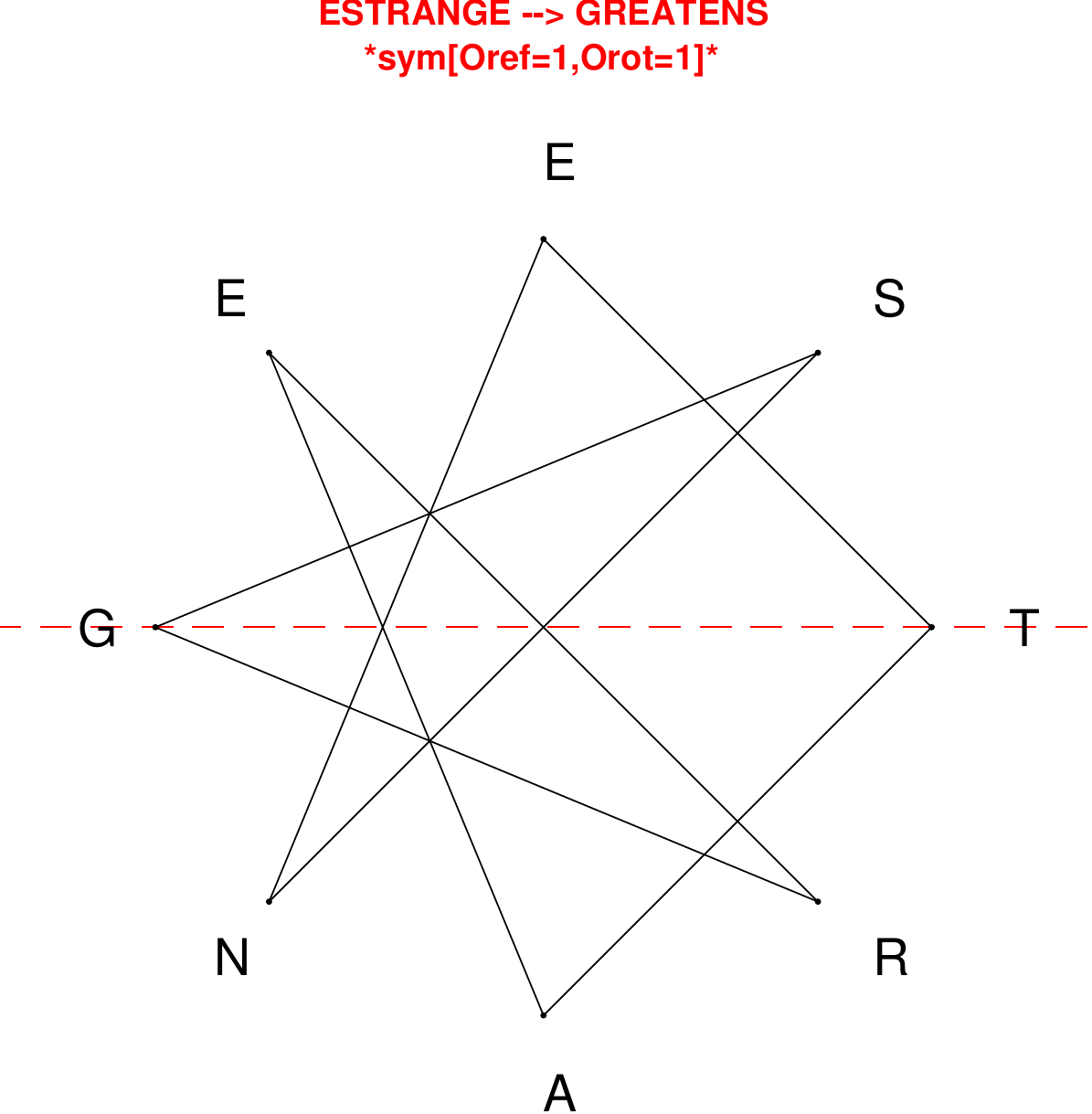}
\end{subfigure}
\hfill
\begin{subfigure}[T]{0.19\textwidth}
\centering
\includegraphics[width=\textwidth]{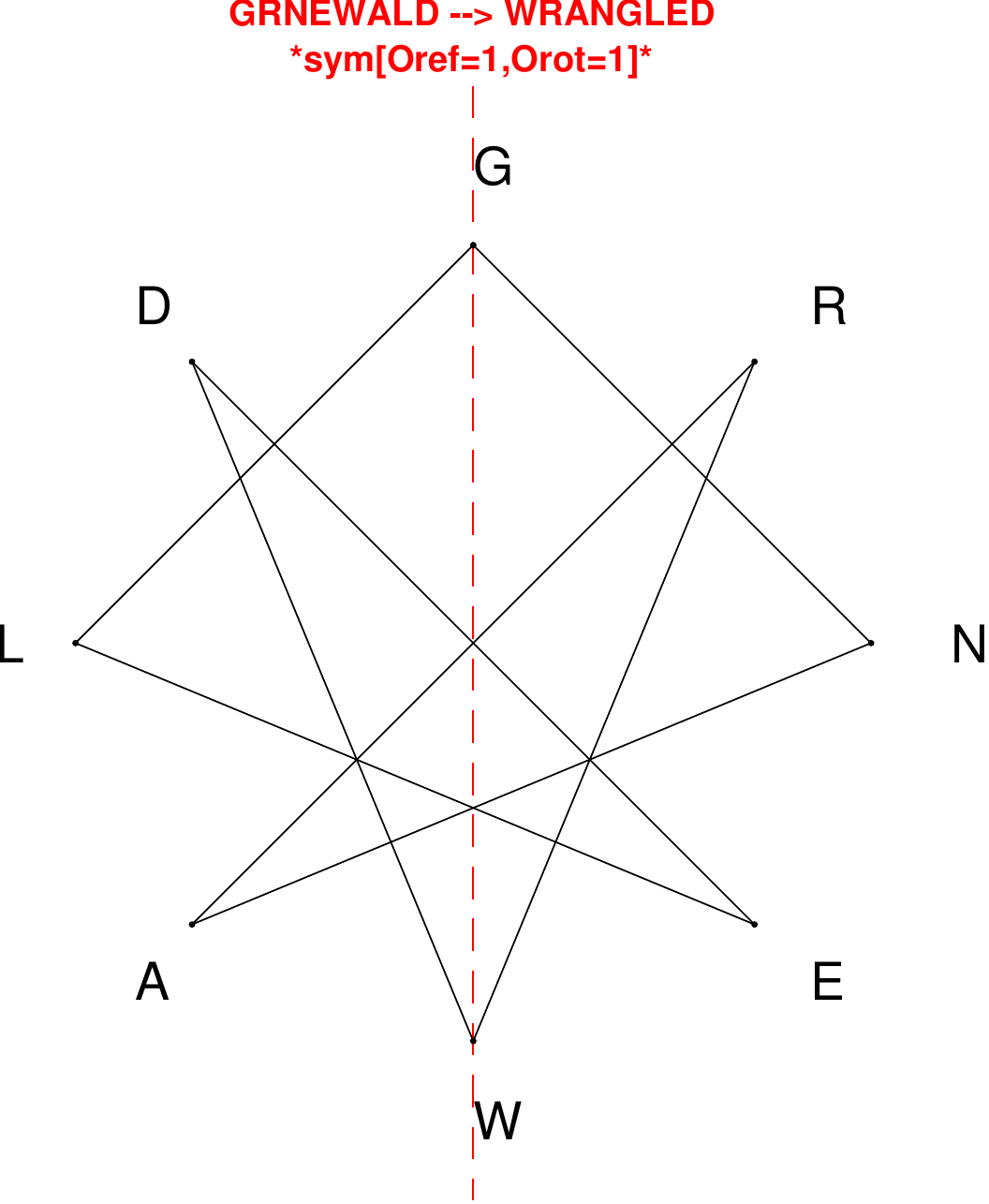}
\end{subfigure}
\hfill
\begin{subfigure}[T]{0.19\textwidth}
\centering
\includegraphics[width=\textwidth]{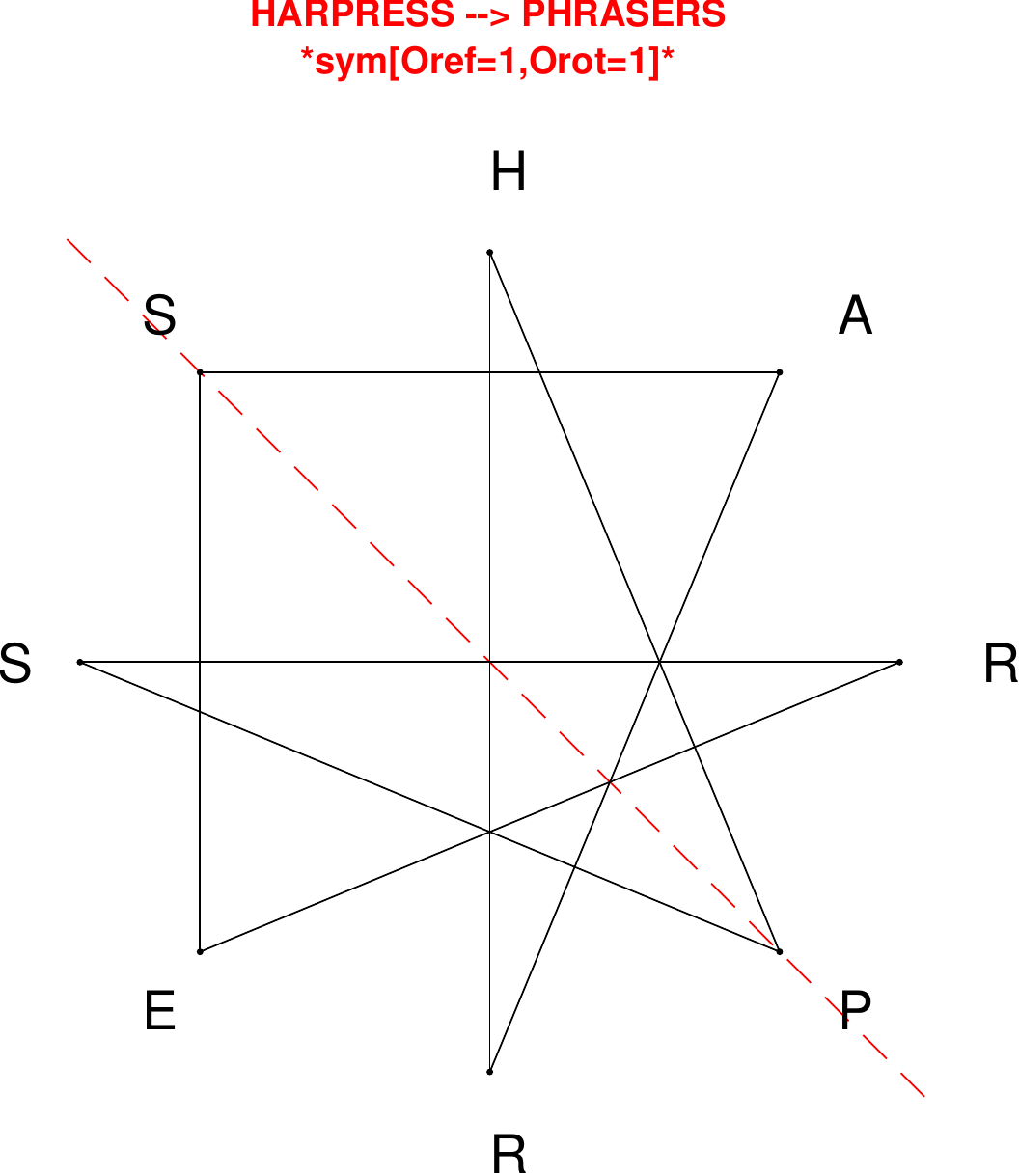}
\end{subfigure}
\end{figure}

\begin{figure}[H]
\centering
\begin{subfigure}[T]{0.19\textwidth}
\centering
\includegraphics[width=\textwidth]{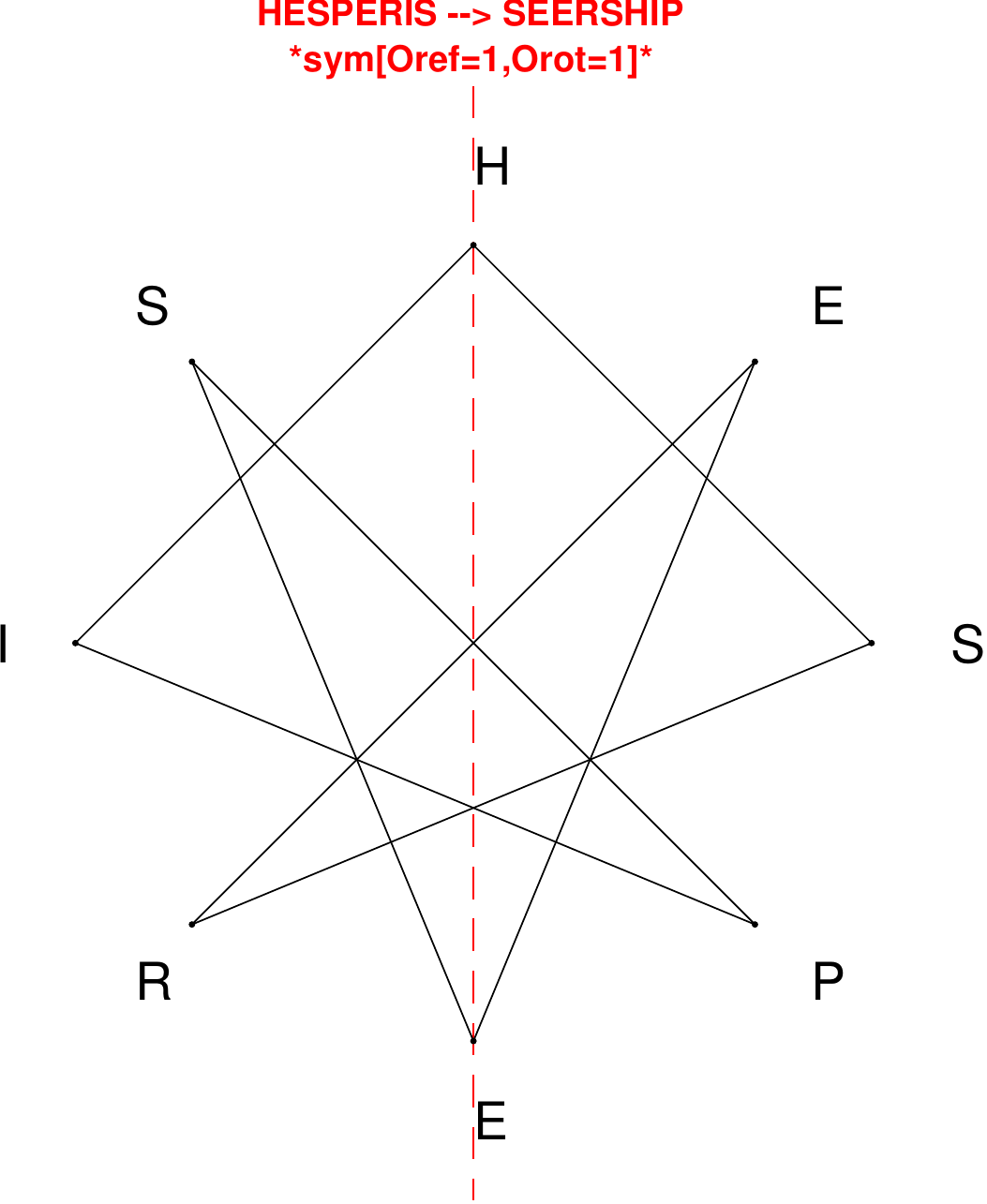}
\end{subfigure}
\hfill
\begin{subfigure}[T]{0.19\textwidth}
\centering
\includegraphics[width=\textwidth]{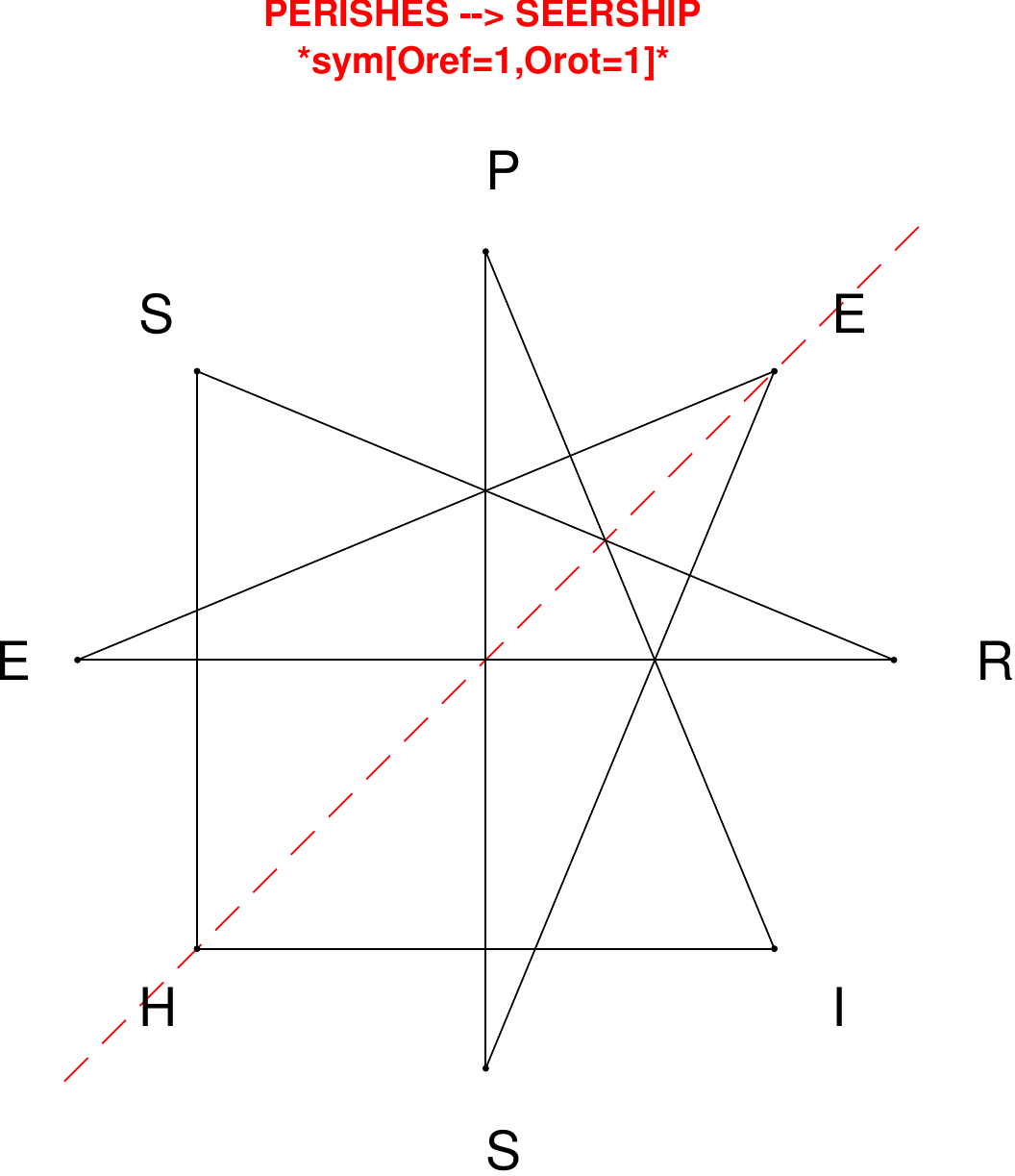}
\end{subfigure}
\hfill
\begin{subfigure}[T]{0.19\textwidth}
\centering
\includegraphics[width=\textwidth]{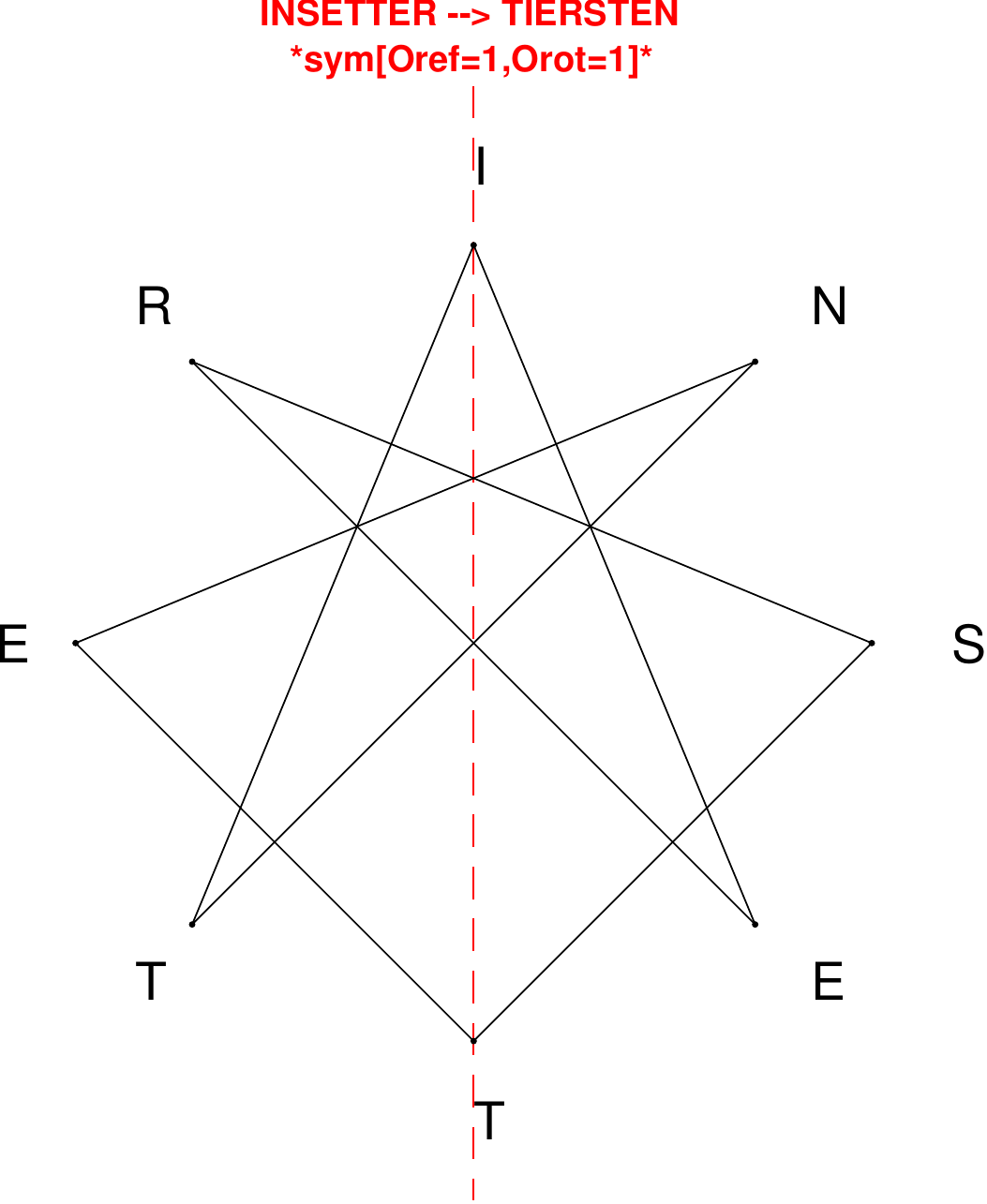}
\end{subfigure}
\hfill
\begin{subfigure}[T]{0.19\textwidth}
\centering
\includegraphics[width=\textwidth]{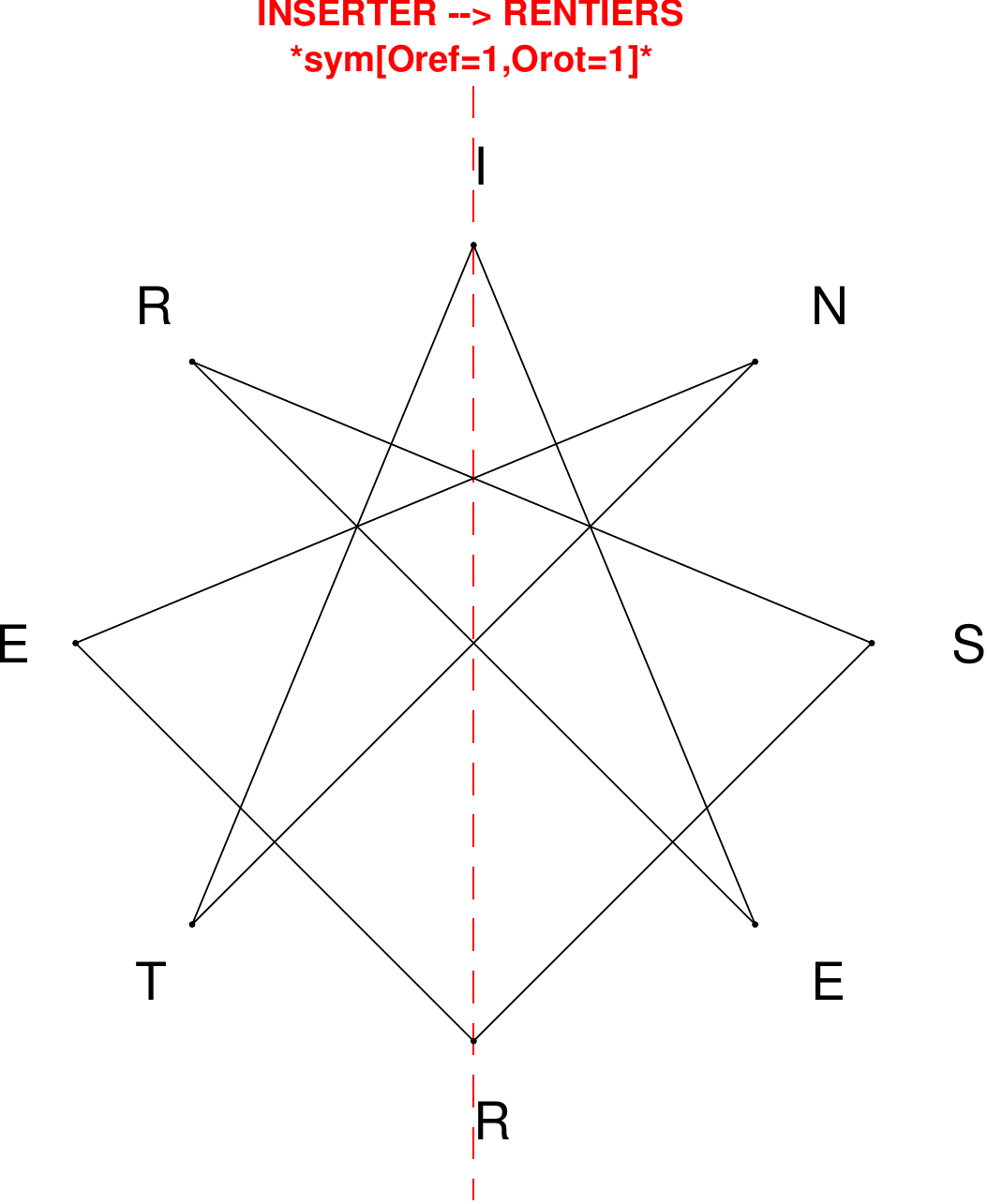}
\end{subfigure}
\hfill
\begin{subfigure}[T]{0.19\textwidth}
\centering
\includegraphics[width=\textwidth]{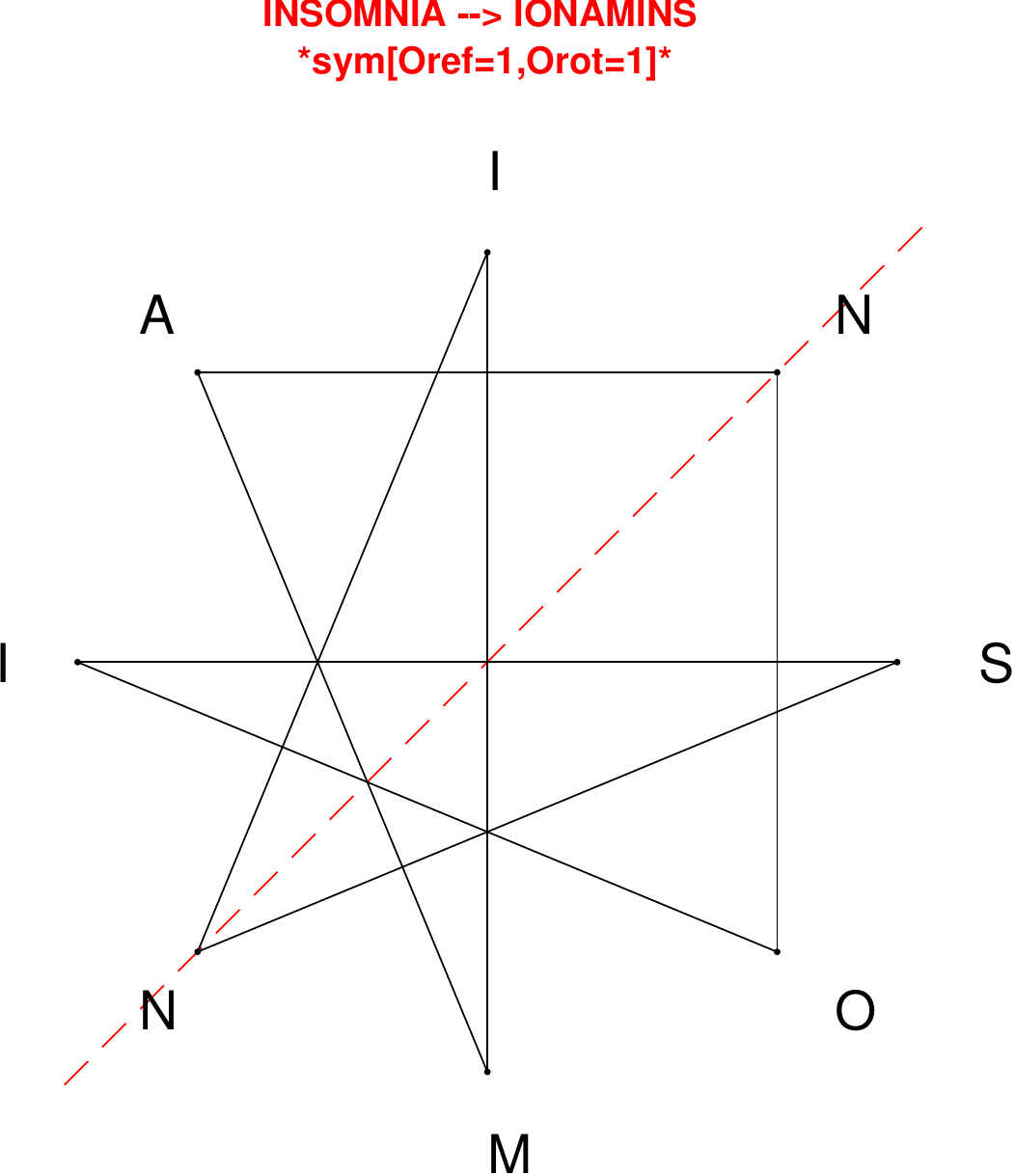}
\end{subfigure}
\end{figure}

\begin{figure}[H]
\centering
\begin{subfigure}[T]{0.19\textwidth}
\centering
\includegraphics[width=\textwidth]{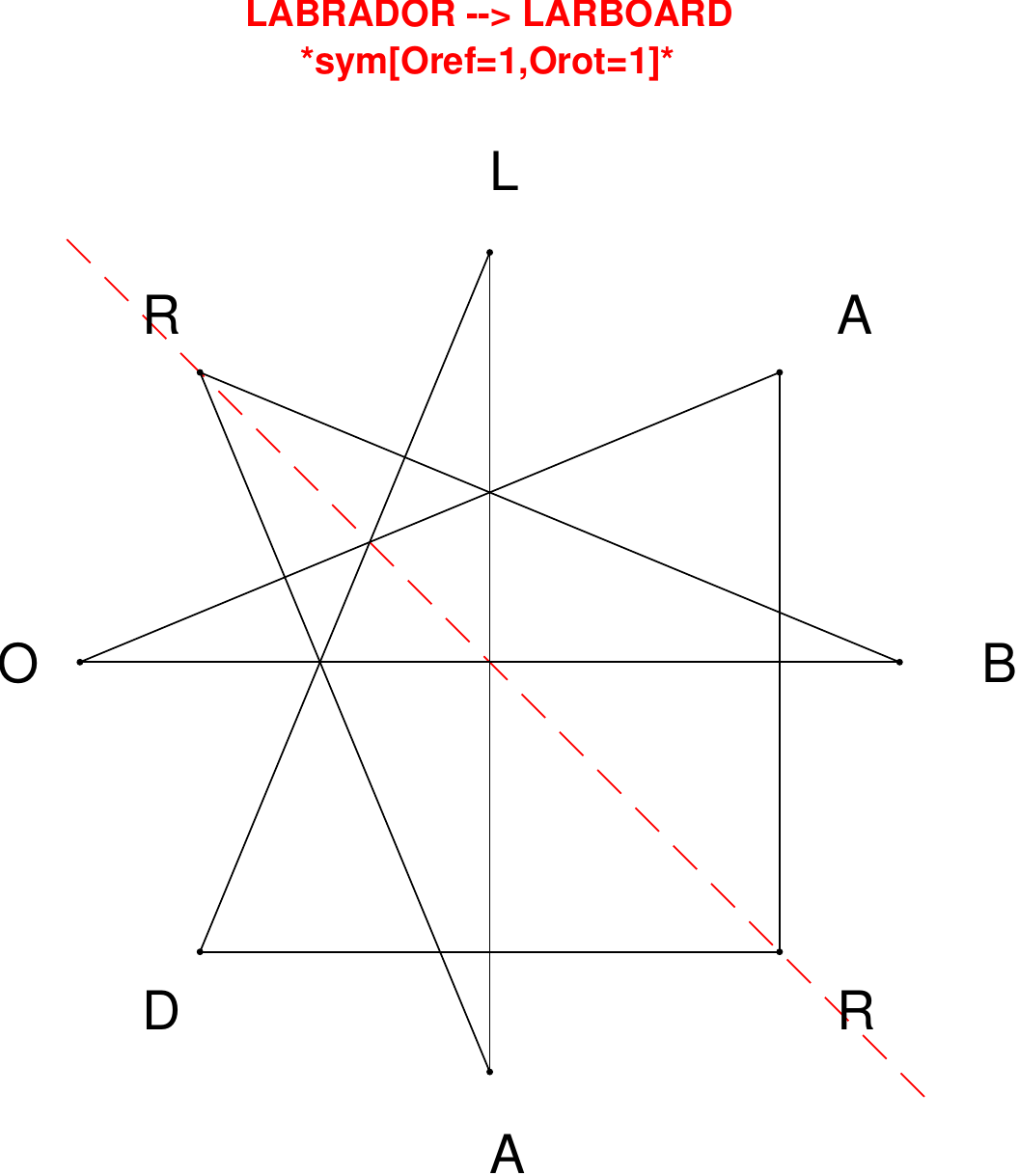}
\end{subfigure}
\hfill
\begin{subfigure}[T]{0.19\textwidth}
\centering
\includegraphics[width=\textwidth]{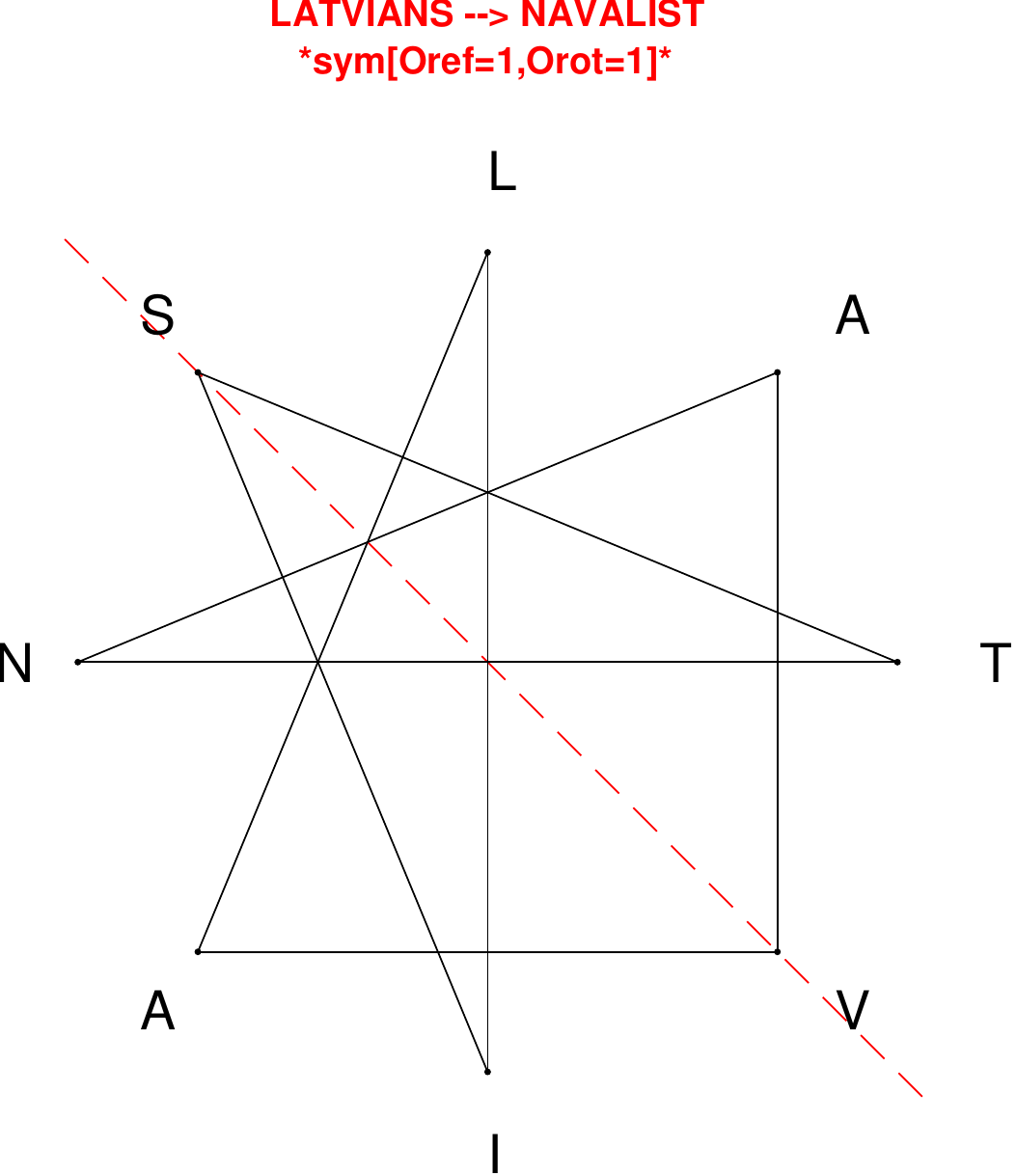}
\end{subfigure}
\hfill
\begin{subfigure}[T]{0.19\textwidth}
\centering
\includegraphics[width=\textwidth]{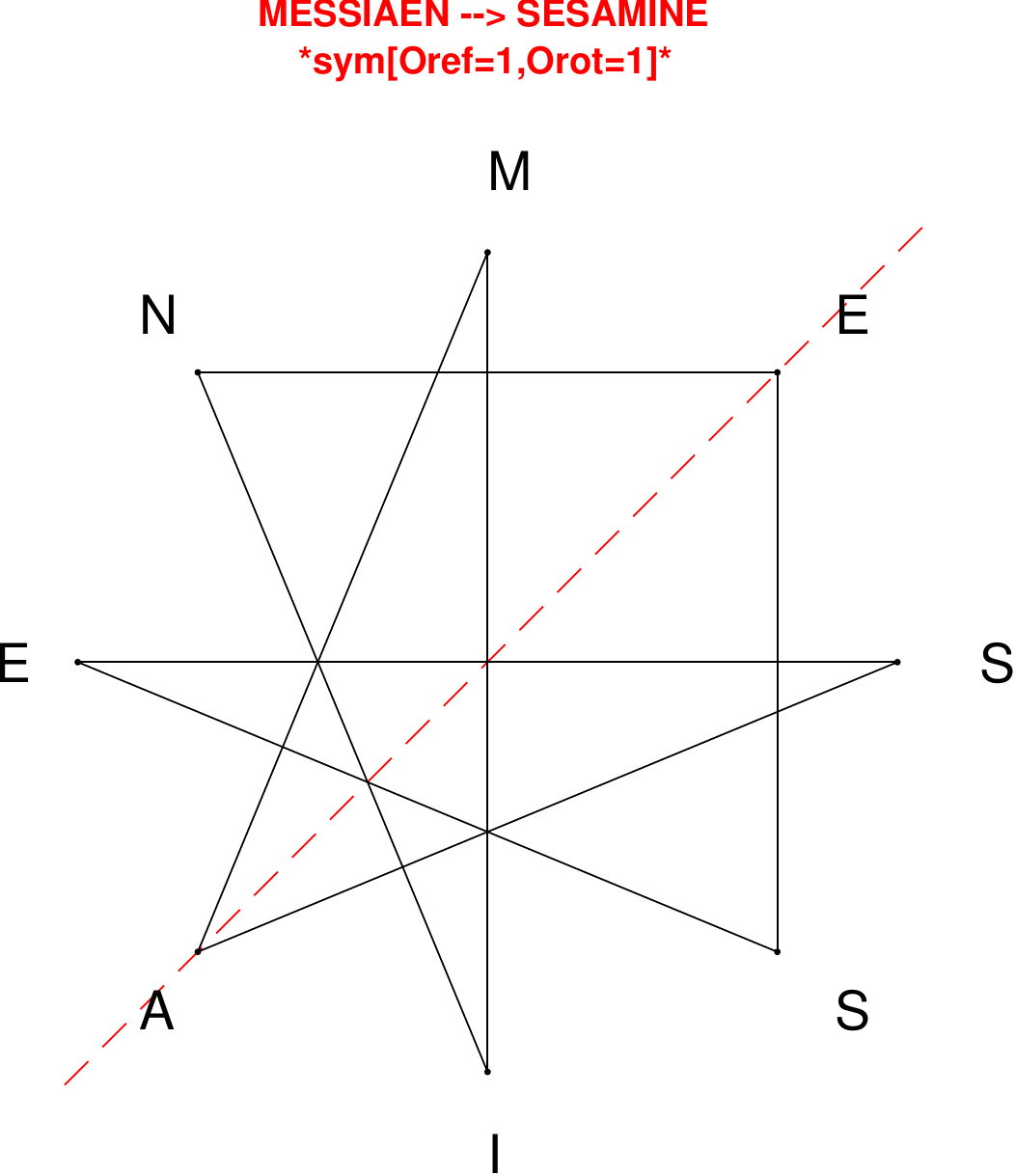}
\end{subfigure}
\hfill
\begin{subfigure}[T]{0.19\textwidth}
\centering
\includegraphics[width=\textwidth]{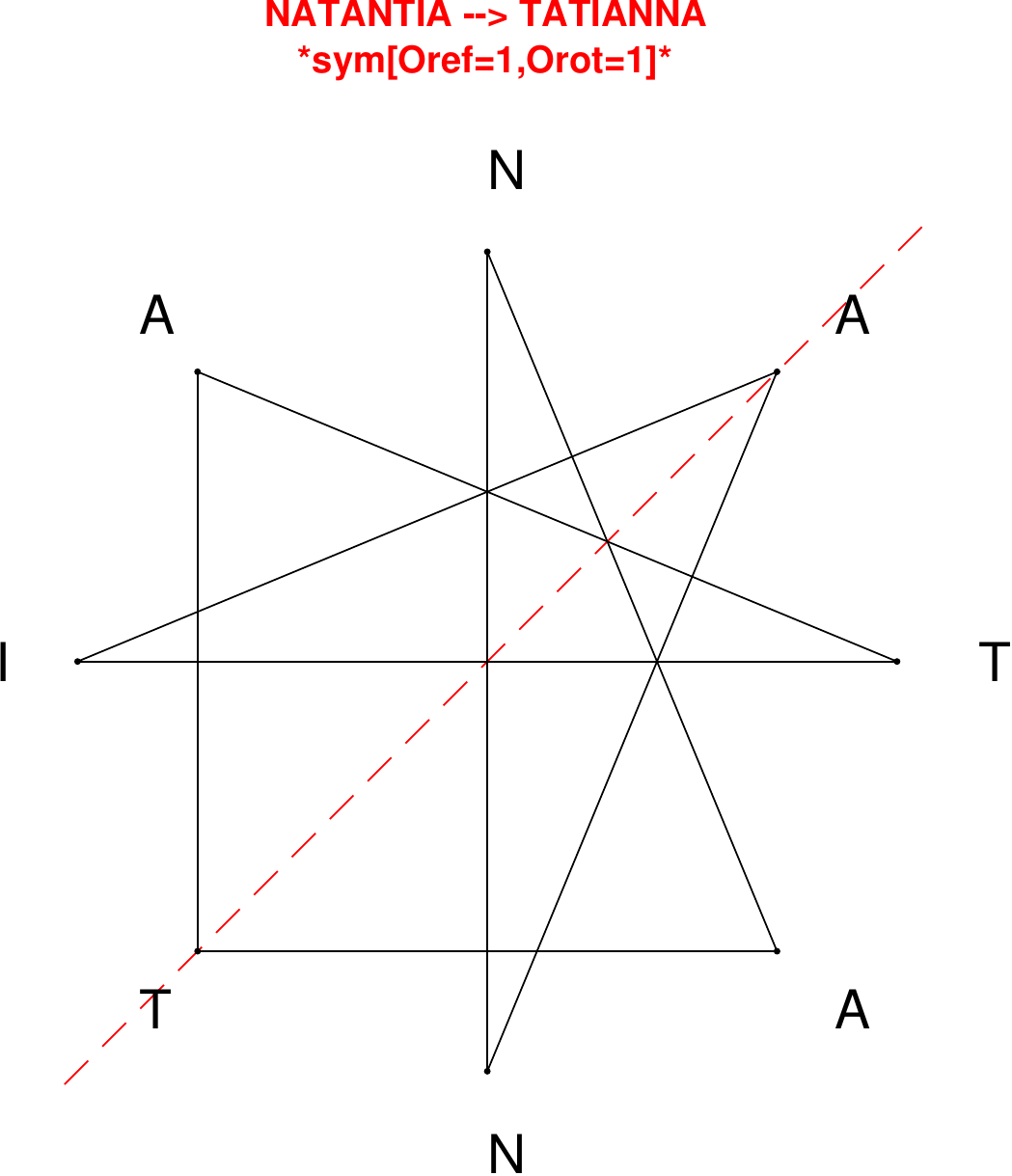}
\end{subfigure}
\hfill
\begin{subfigure}[T]{0.19\textwidth}
\centering
\includegraphics[width=\textwidth]{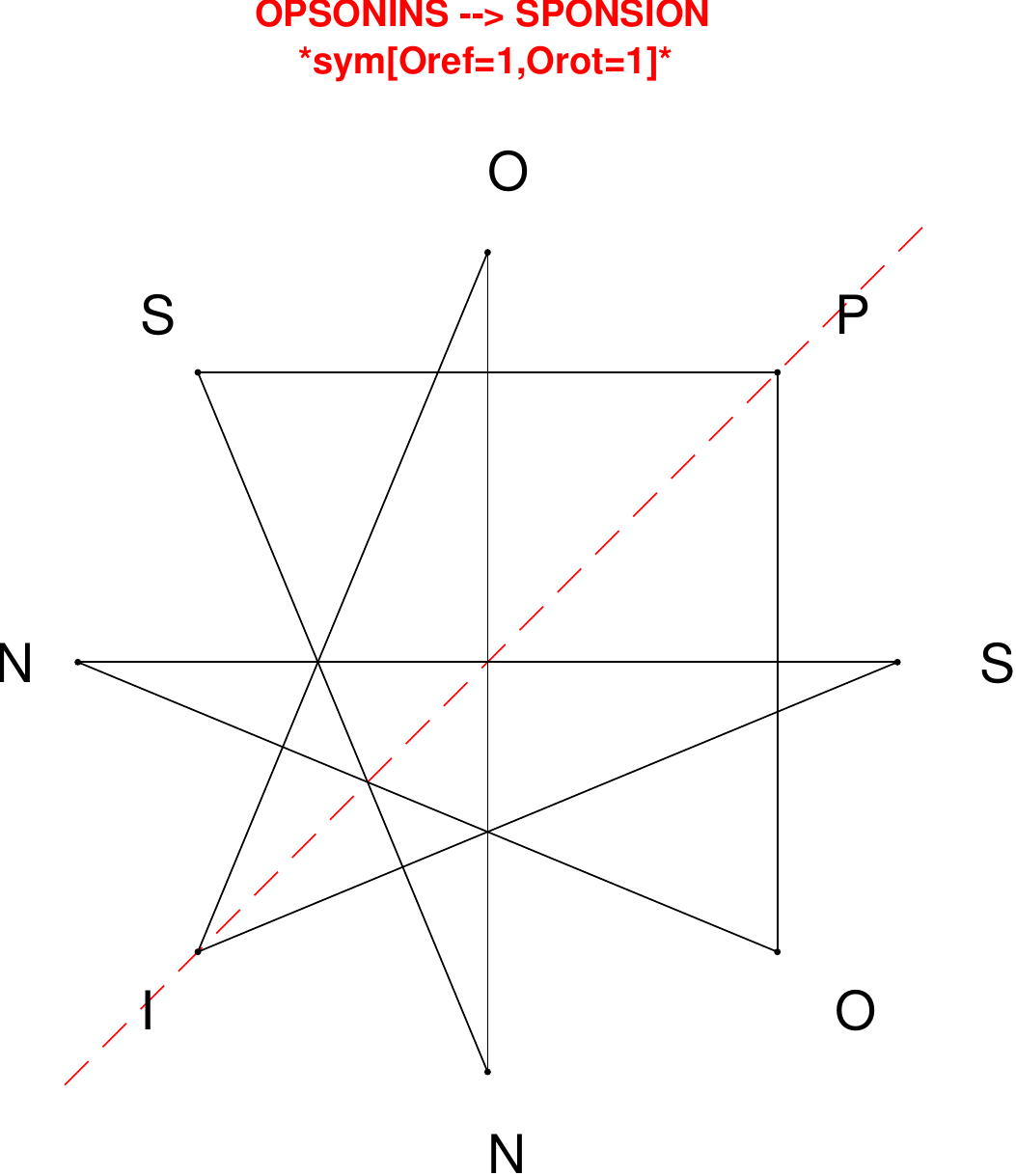}
\end{subfigure}
\end{figure}

\begin{figure}[H]
\centering
\begin{subfigure}[T]{0.19\textwidth}
\centering
\includegraphics[width=\textwidth]{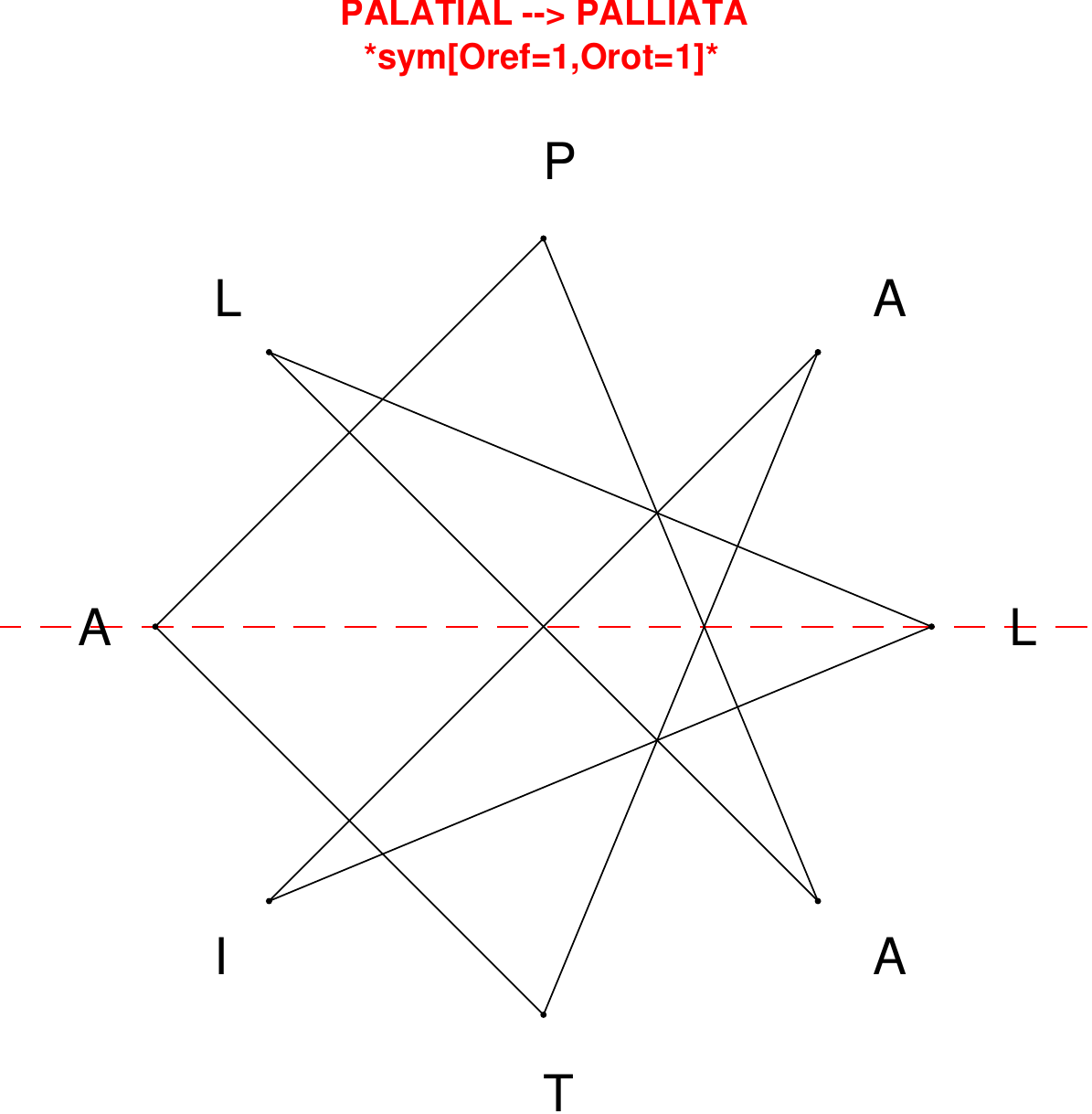}
\end{subfigure}
\hfill
\begin{subfigure}[T]{0.19\textwidth}
\centering
\includegraphics[width=\textwidth]{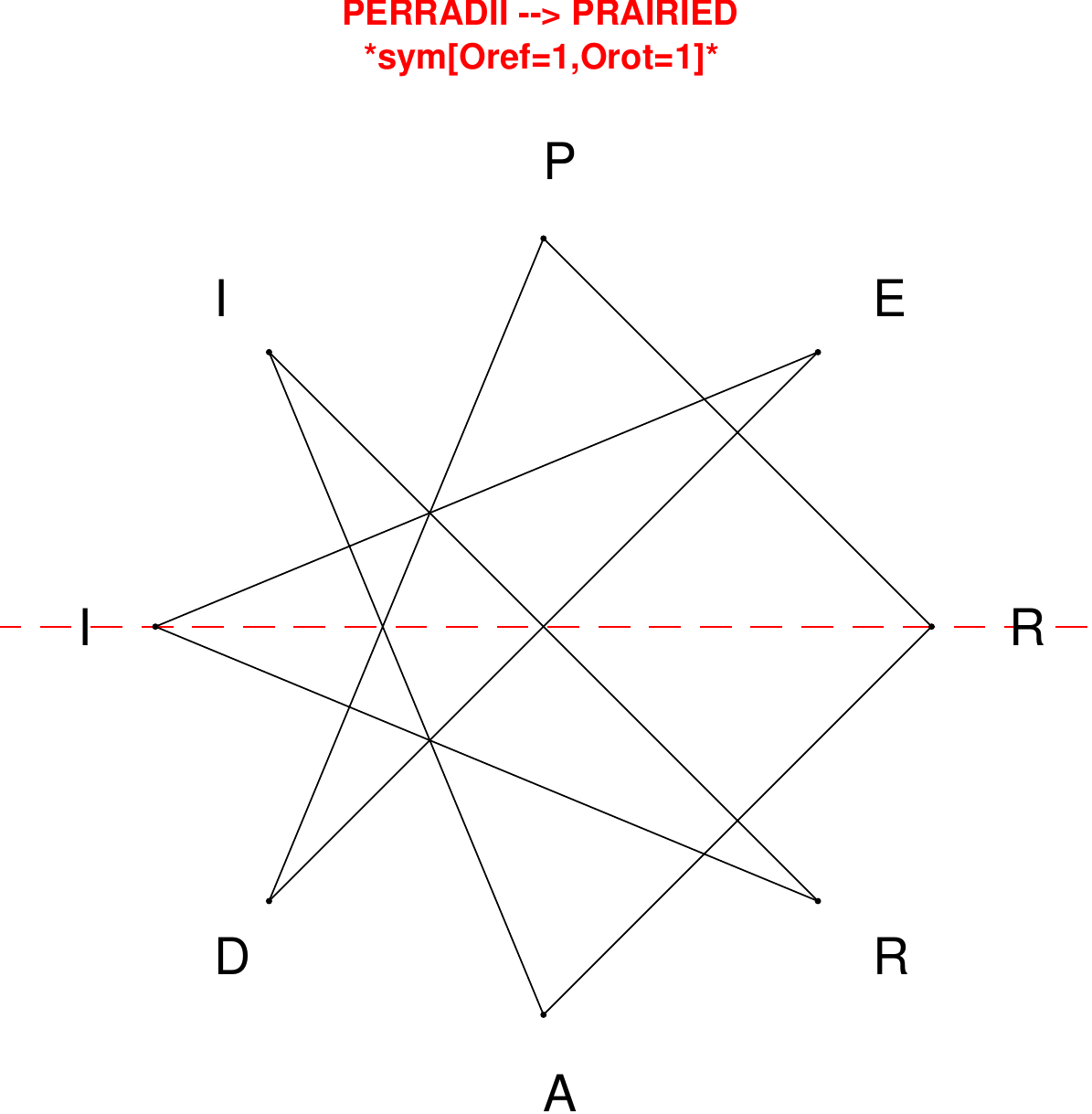}
\end{subfigure}
\hfill
\begin{subfigure}[T]{0.19\textwidth}
\centering
\includegraphics[width=\textwidth]{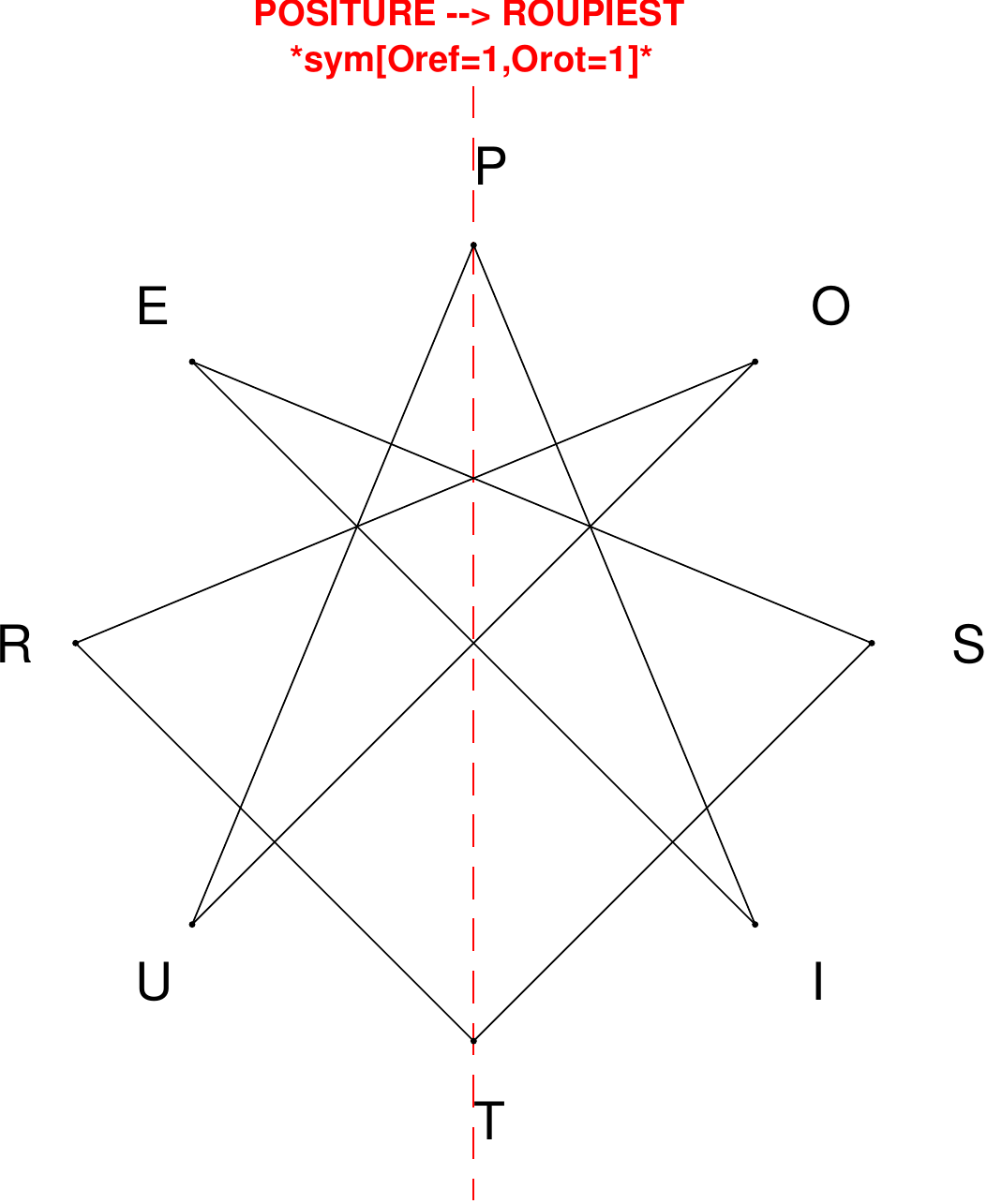}
\end{subfigure}
\hfill
\begin{subfigure}[T]{0.19\textwidth}
\centering
\includegraphics[width=\textwidth]{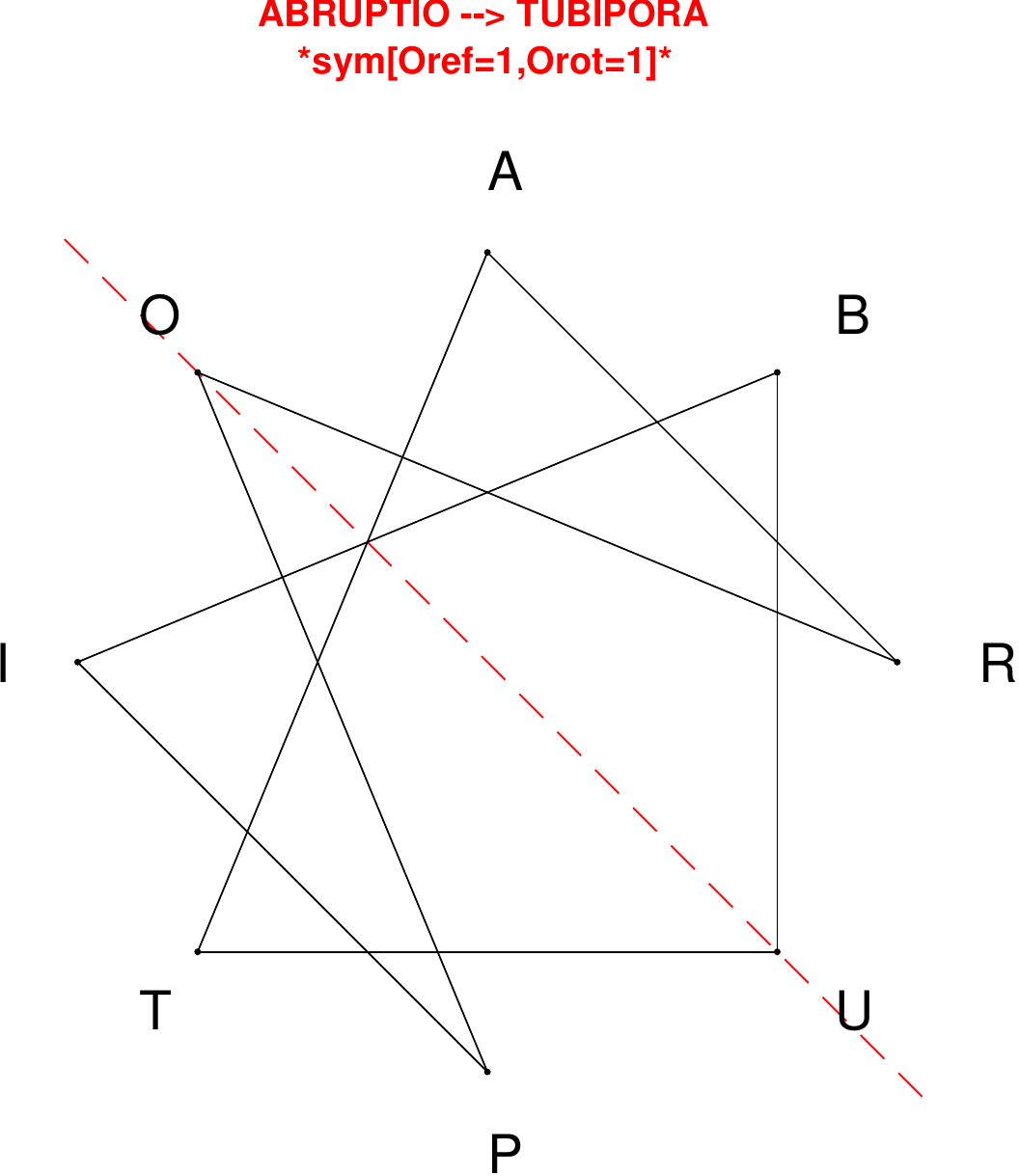}
\end{subfigure}
\hfill
\begin{subfigure}[T]{0.19\textwidth}
\centering
\includegraphics[width=\textwidth]{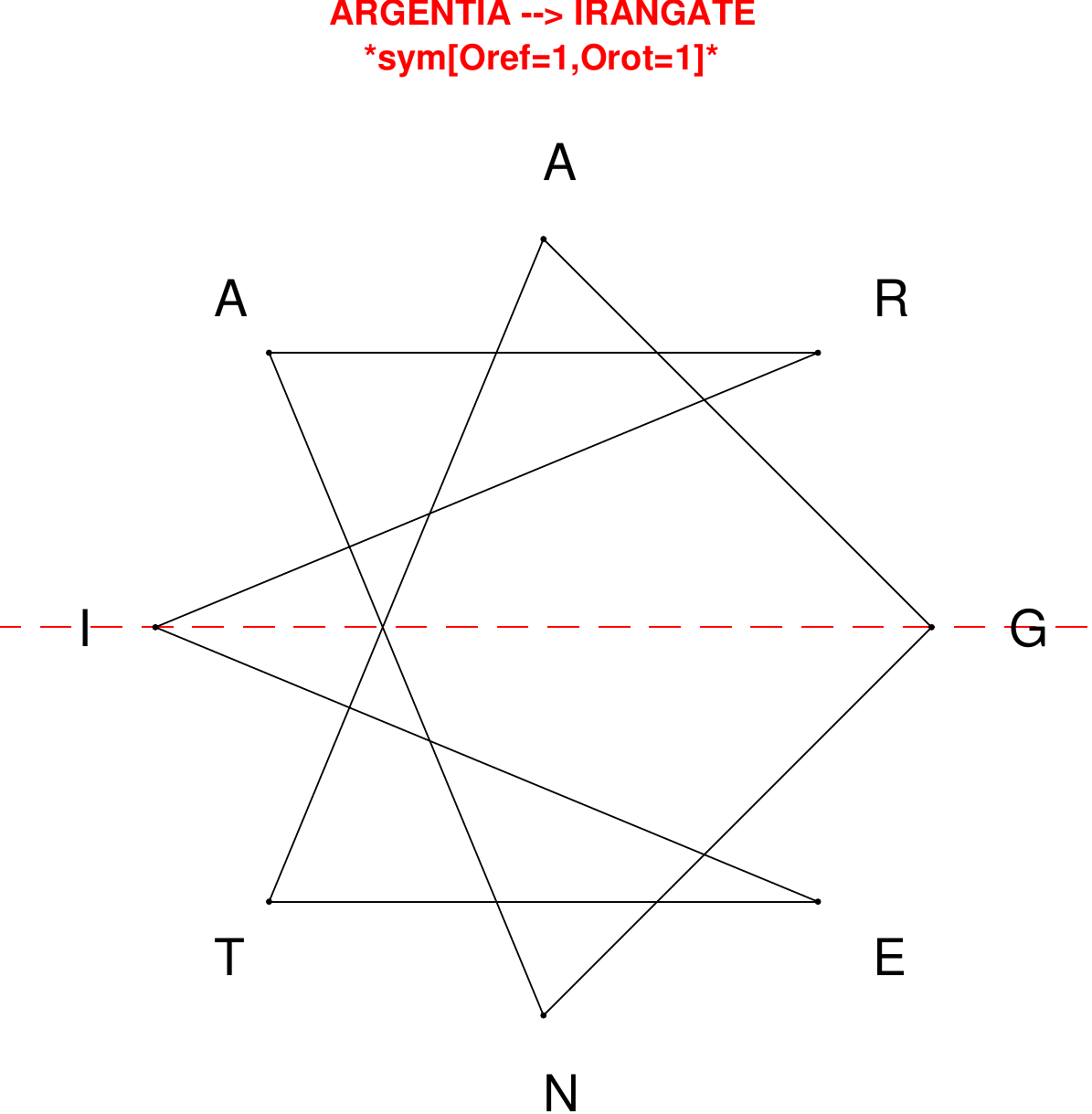}
\end{subfigure}
\end{figure}

\begin{figure}[H]
\centering
\begin{subfigure}[T]{0.19\textwidth}
\centering
\includegraphics[width=\textwidth]{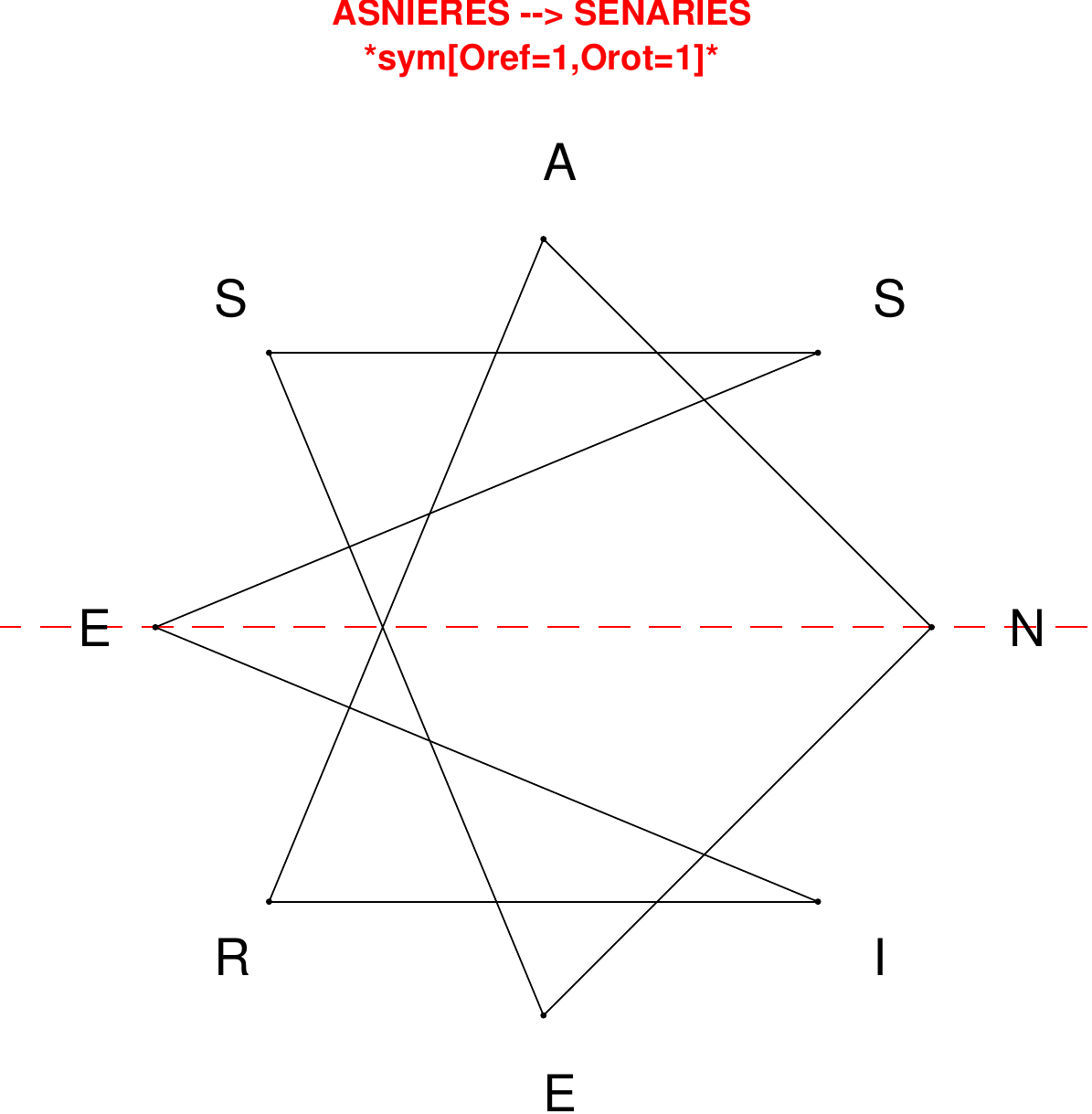}
\end{subfigure}
\hfill
\begin{subfigure}[T]{0.19\textwidth}
\centering
\includegraphics[width=\textwidth]{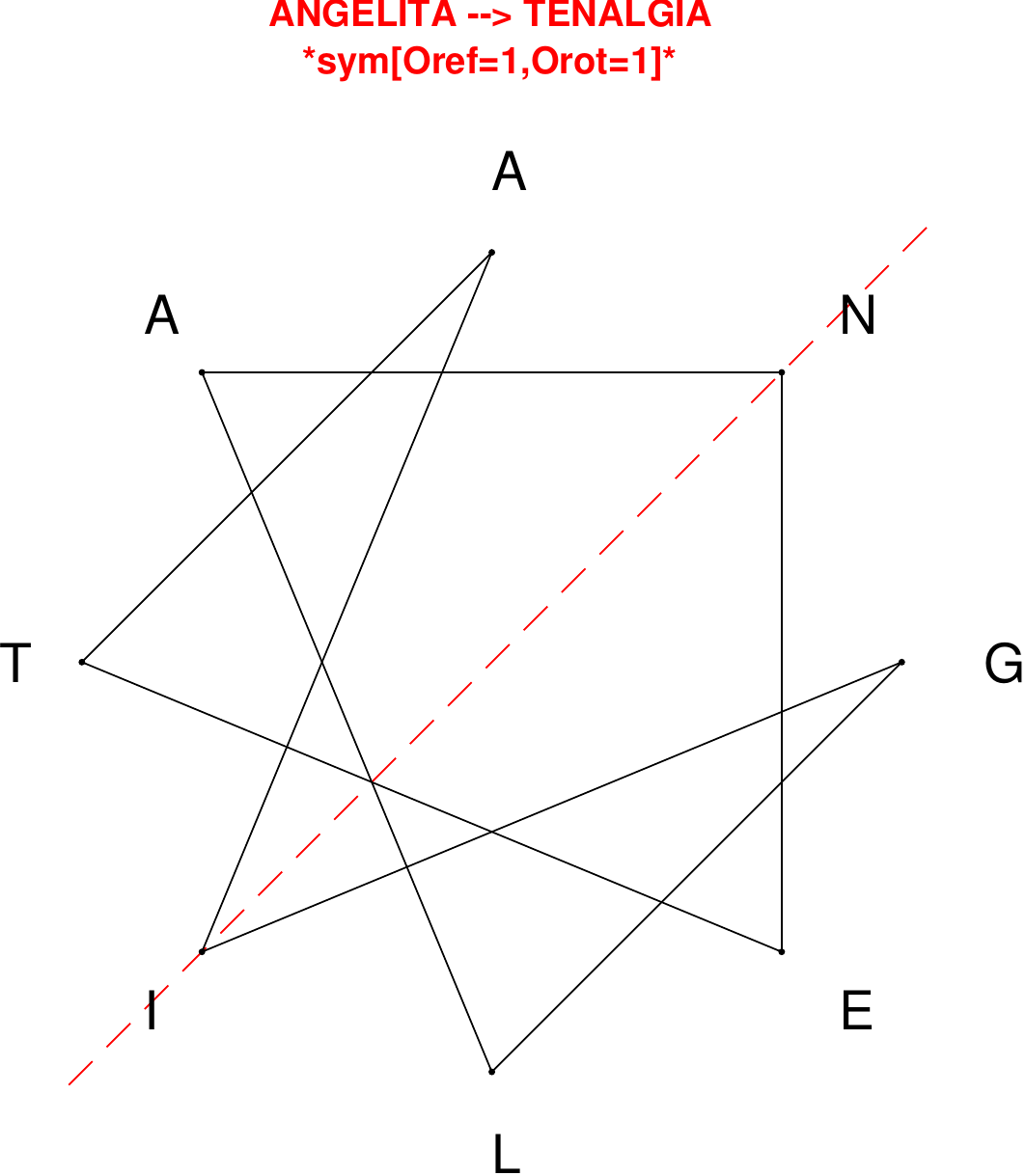}
\end{subfigure}
\hfill
\begin{subfigure}[T]{0.19\textwidth}
\centering
\includegraphics[width=\textwidth]{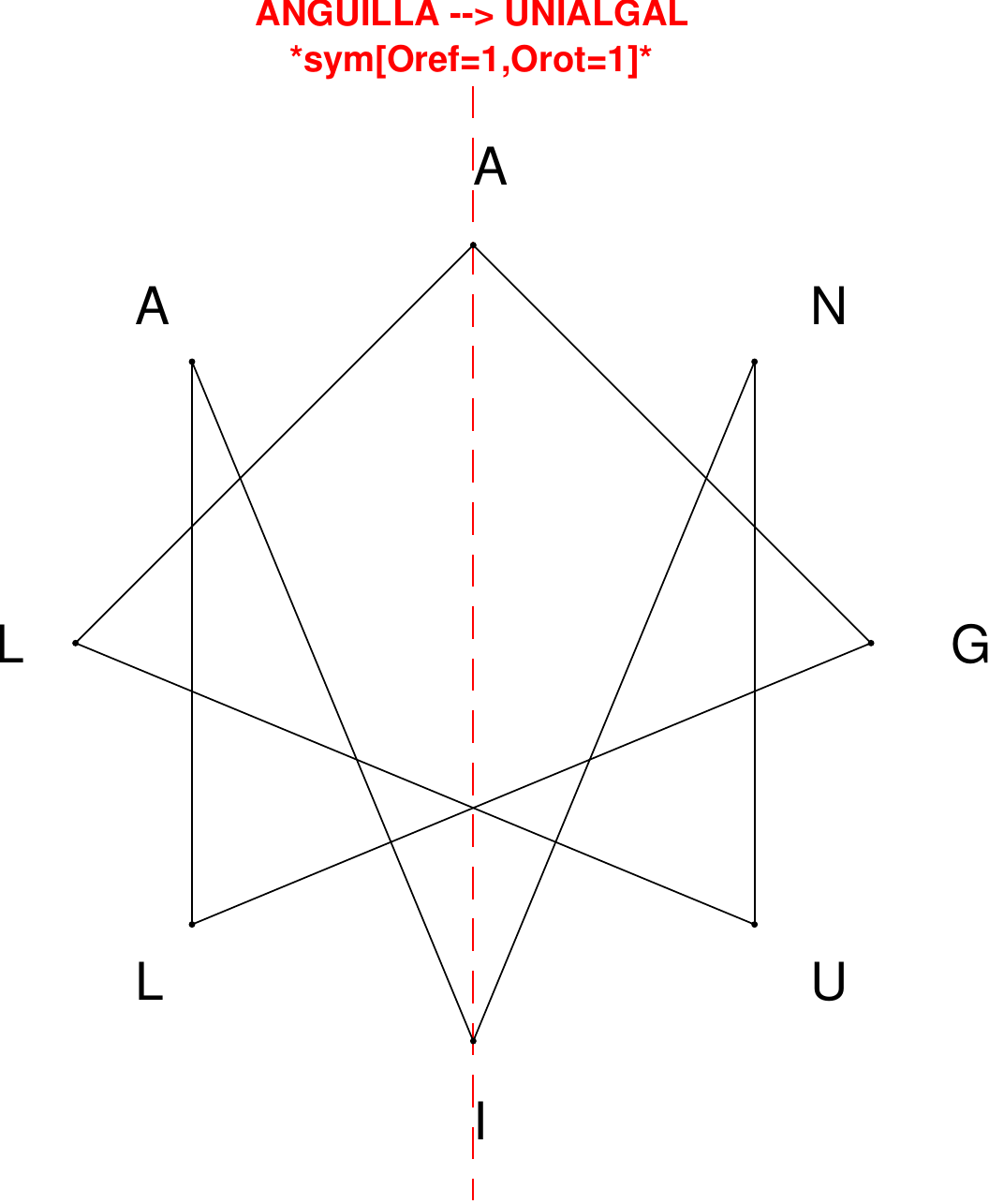}
\end{subfigure}
\hfill
\begin{subfigure}[T]{0.19\textwidth}
\centering
\includegraphics[width=\textwidth]{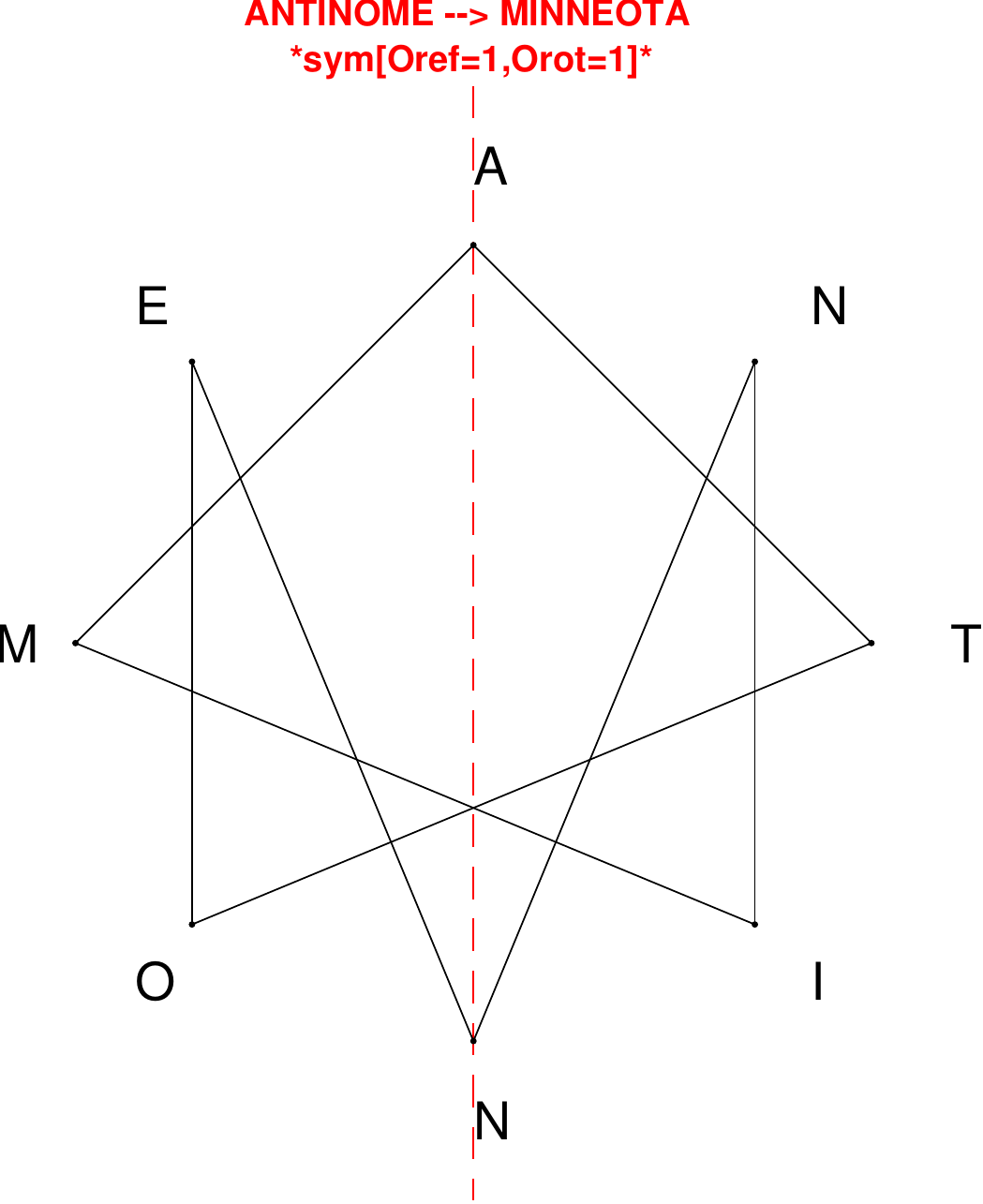}
\end{subfigure}
\hfill
\begin{subfigure}[T]{0.19\textwidth}
\centering
\includegraphics[width=\textwidth]{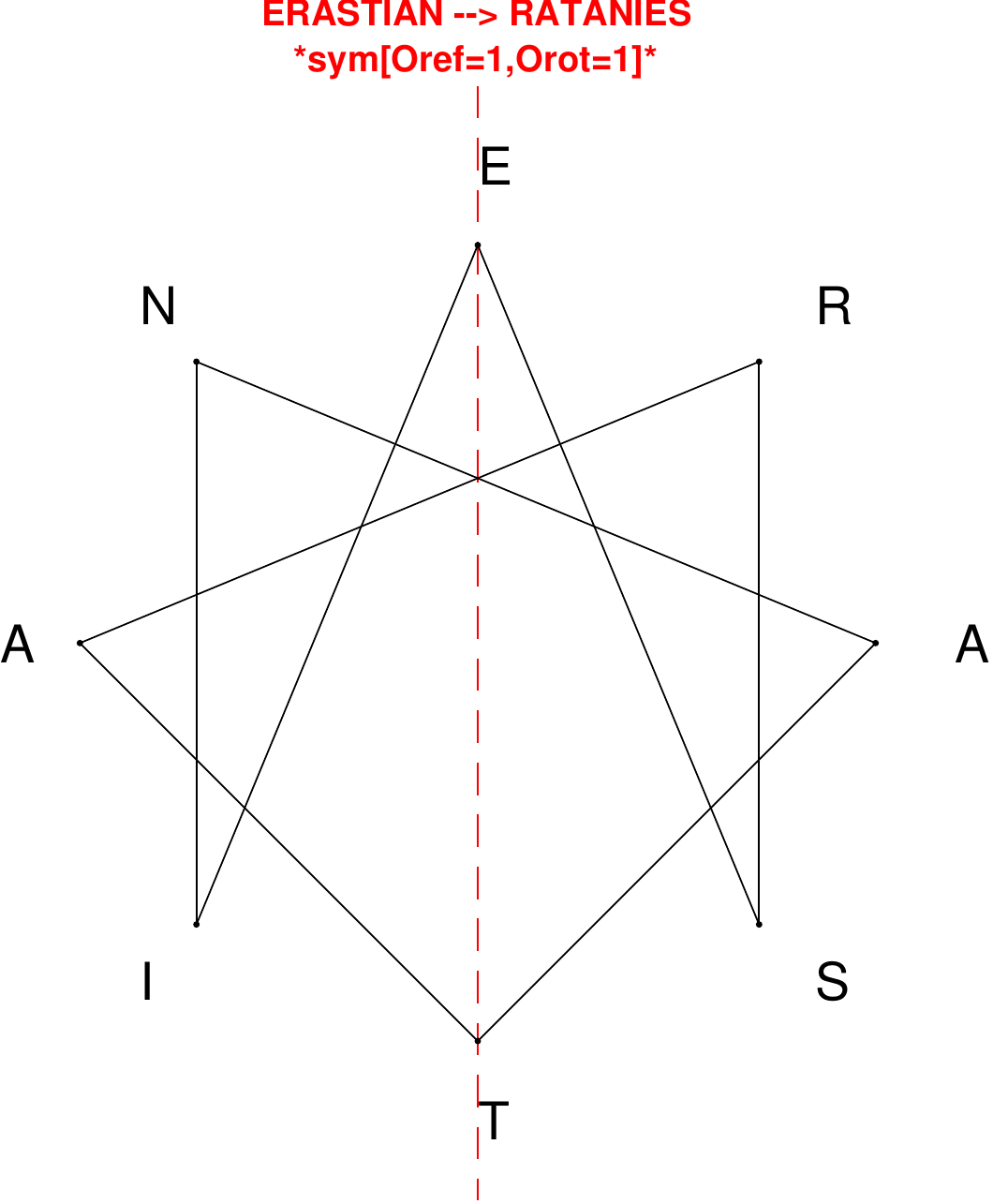}
\end{subfigure}
\end{figure}

\begin{figure}[H]
\centering
\begin{subfigure}[T]{0.19\textwidth}
\centering
\includegraphics[width=\textwidth]{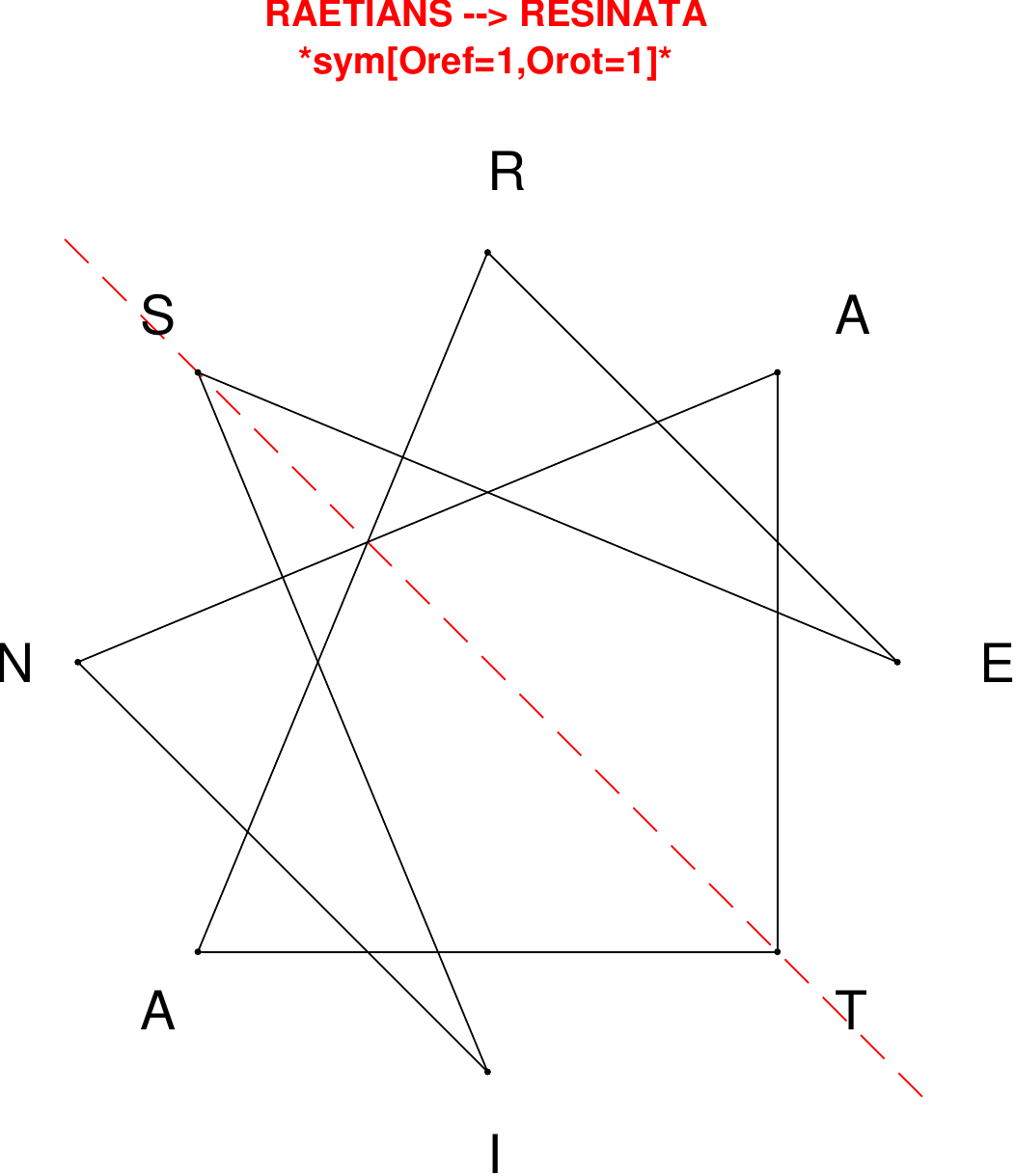}
\end{subfigure}
\hfill
\begin{subfigure}[T]{0.19\textwidth}
\centering
\includegraphics[width=\textwidth]{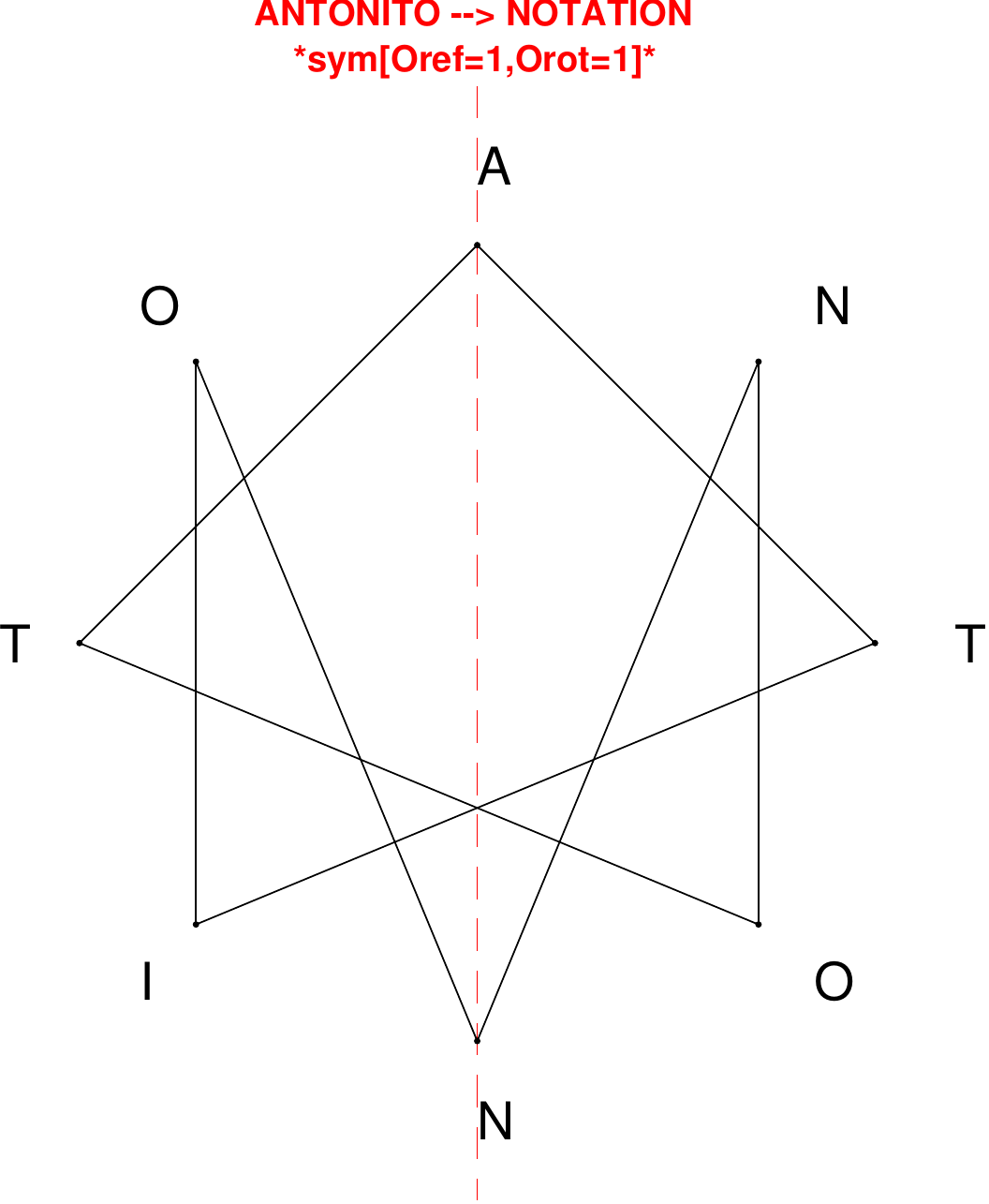}
\end{subfigure}
\hfill
\begin{subfigure}[T]{0.19\textwidth}
\centering
\includegraphics[width=\textwidth]{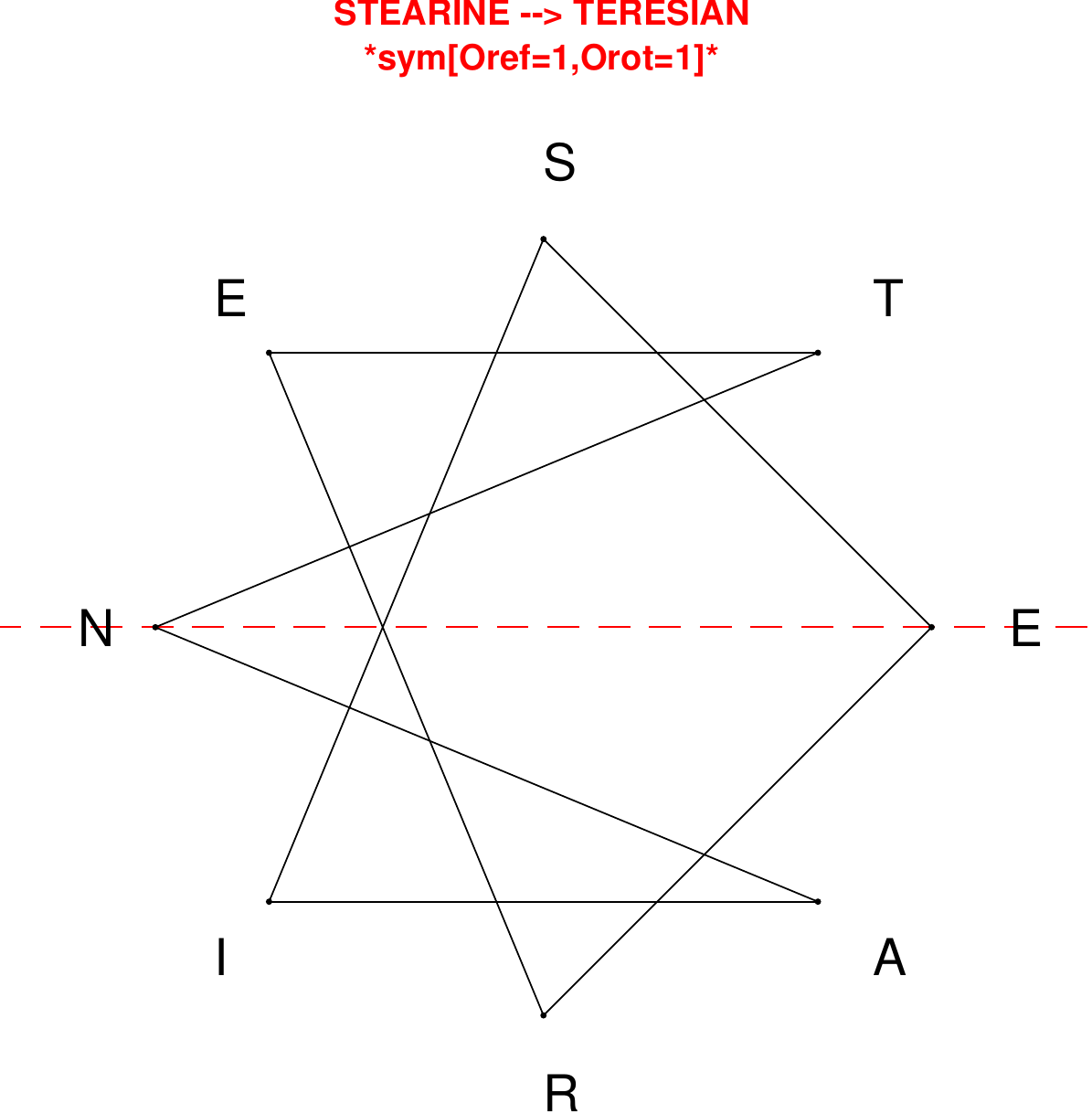}
\end{subfigure}
\hfill
\begin{subfigure}[T]{0.19\textwidth}
\centering
\includegraphics[width=\textwidth]{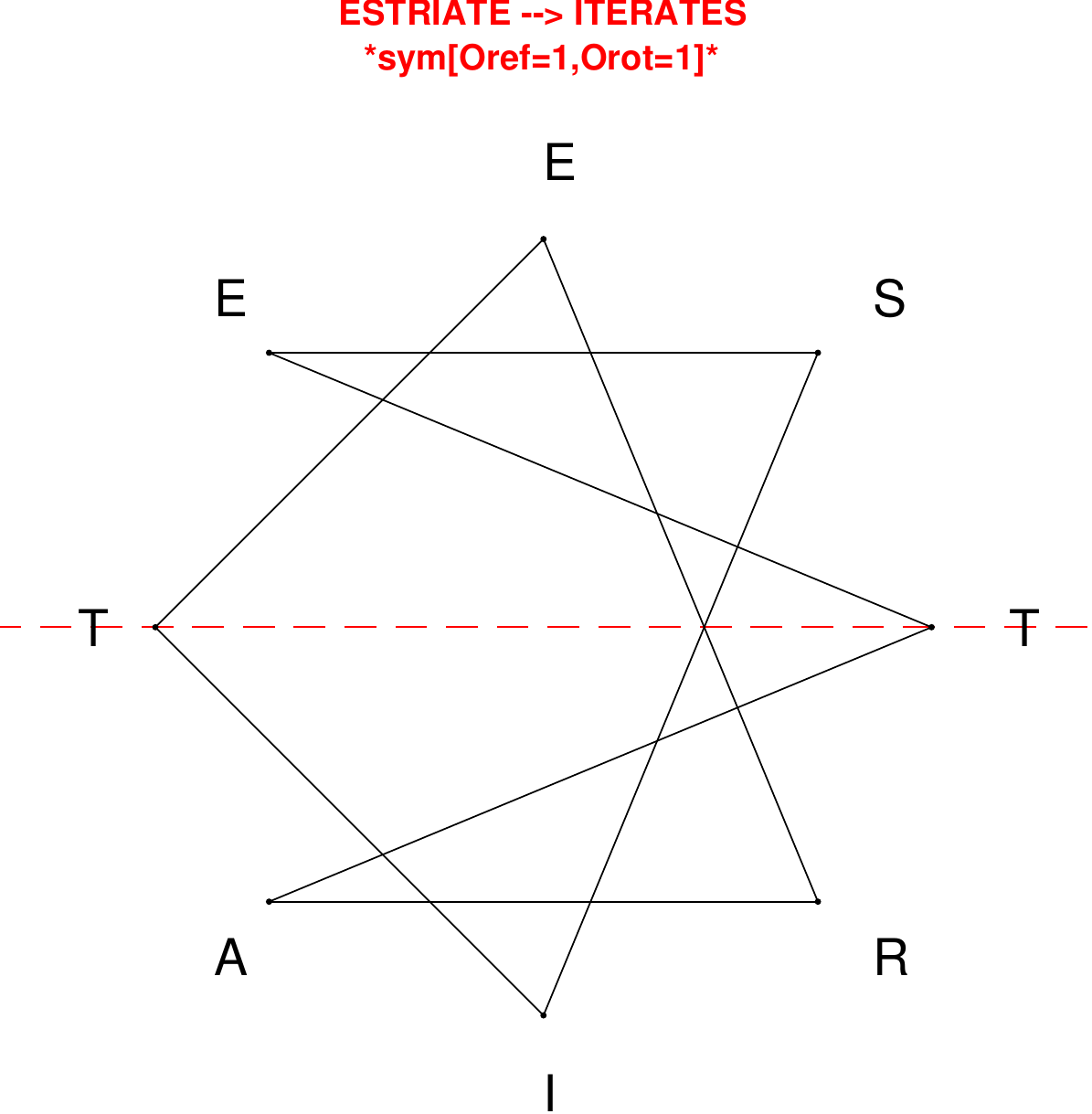}
\end{subfigure}
\hfill
\begin{subfigure}[T]{0.19\textwidth}
\centering
\includegraphics[width=\textwidth]{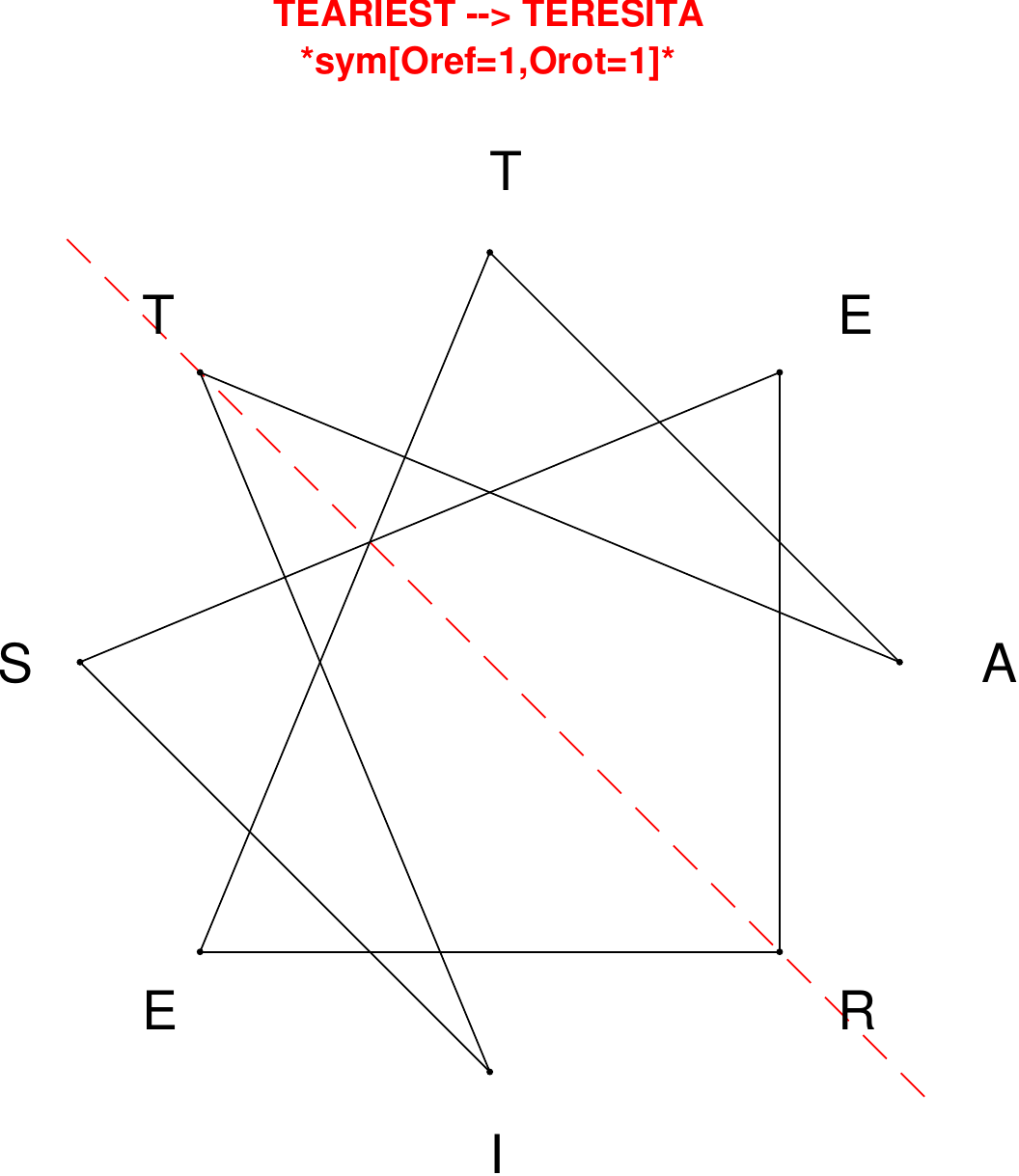}
\end{subfigure}
\end{figure}

\begin{figure}[H]
\centering
\begin{subfigure}[T]{0.19\textwidth}
\centering
\includegraphics[width=\textwidth]{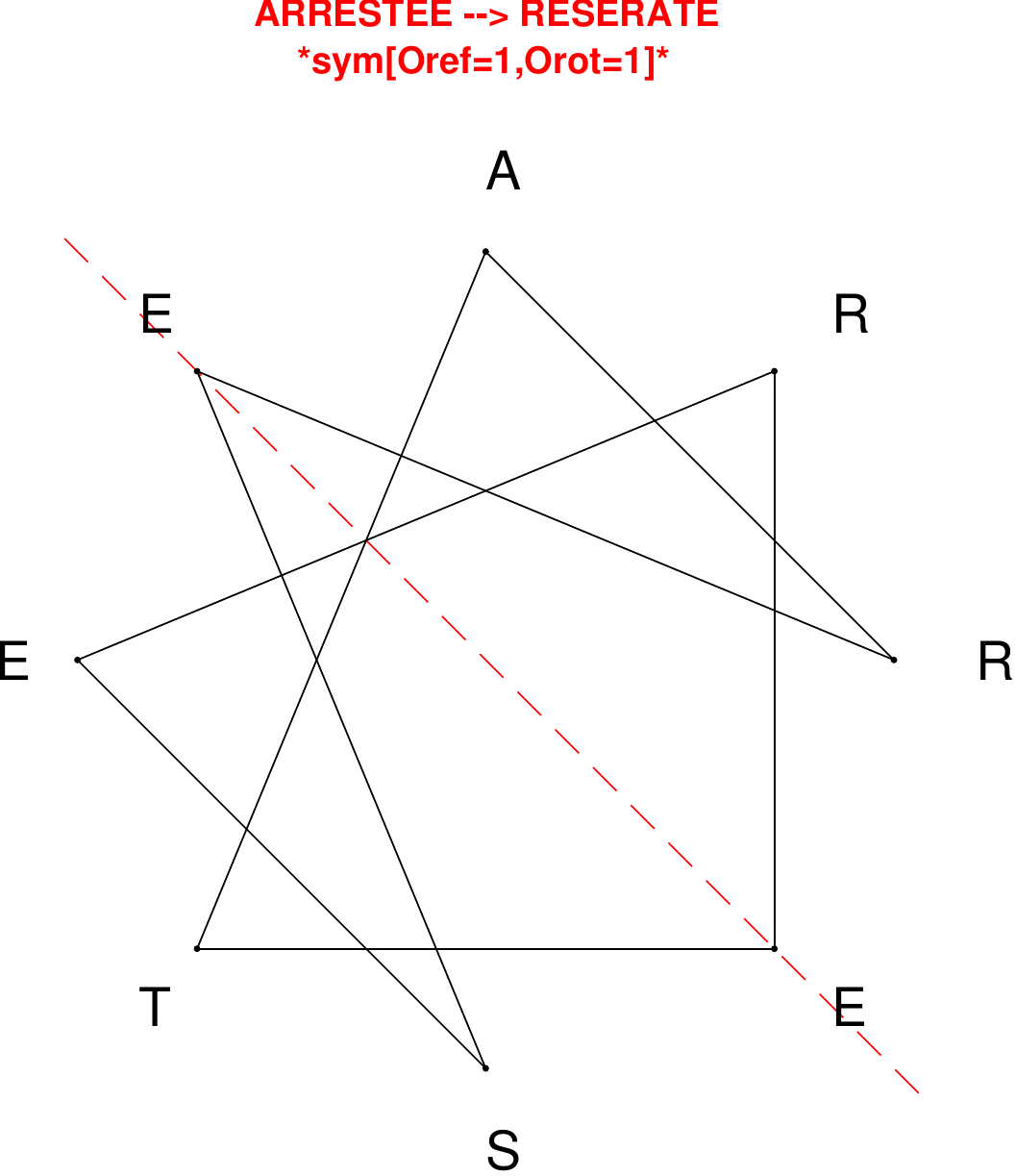}
\end{subfigure}
\hfill
\begin{subfigure}[T]{0.19\textwidth}
\centering
\includegraphics[width=\textwidth]{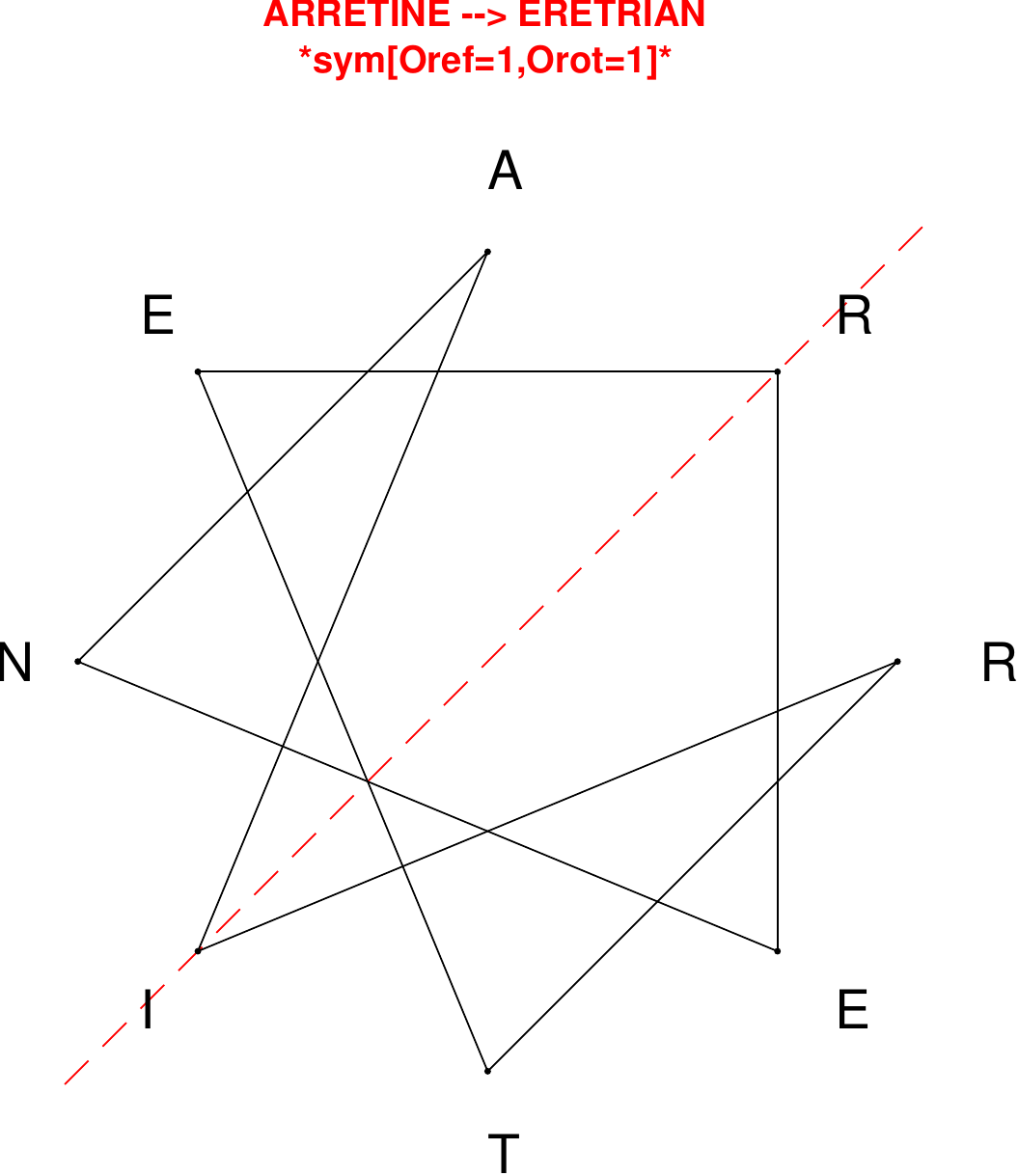}
\end{subfigure}
\hfill
\begin{subfigure}[T]{0.19\textwidth}
\centering
\includegraphics[width=\textwidth]{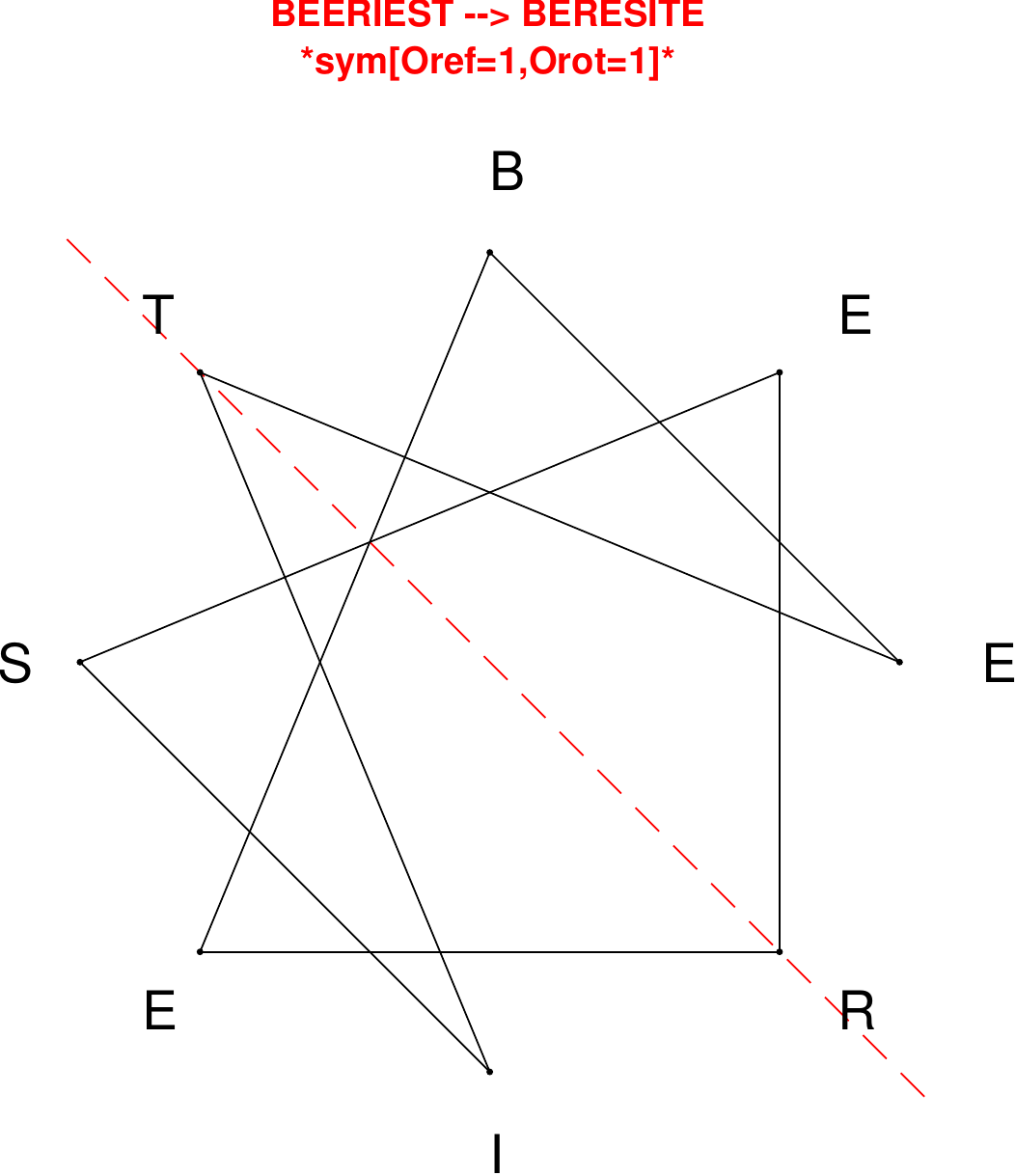}
\end{subfigure}
\hfill
\begin{subfigure}[T]{0.19\textwidth}
\centering
\includegraphics[width=\textwidth]{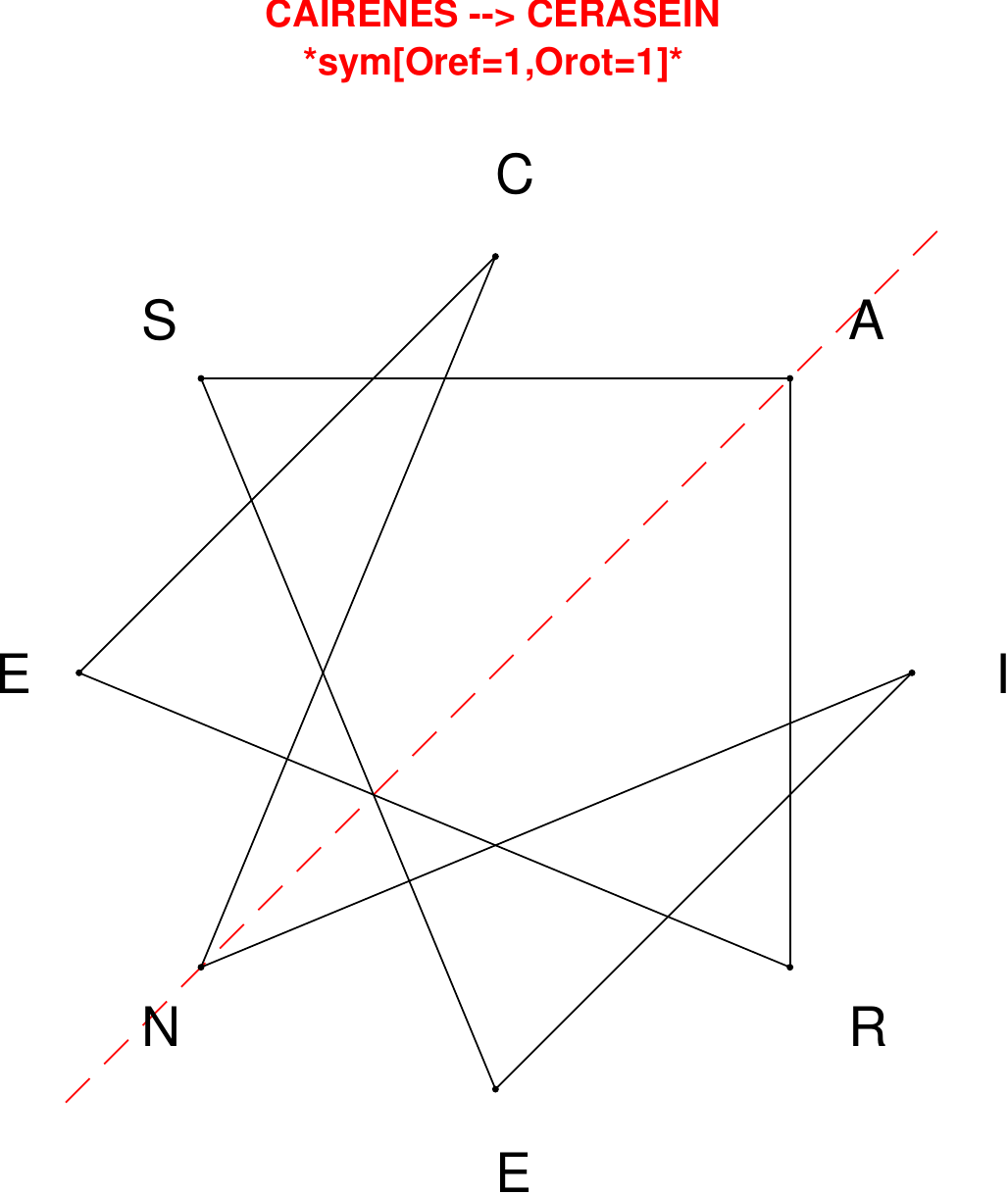}
\end{subfigure}
\hfill
\begin{subfigure}[T]{0.19\textwidth}
\centering
\includegraphics[width=\textwidth]{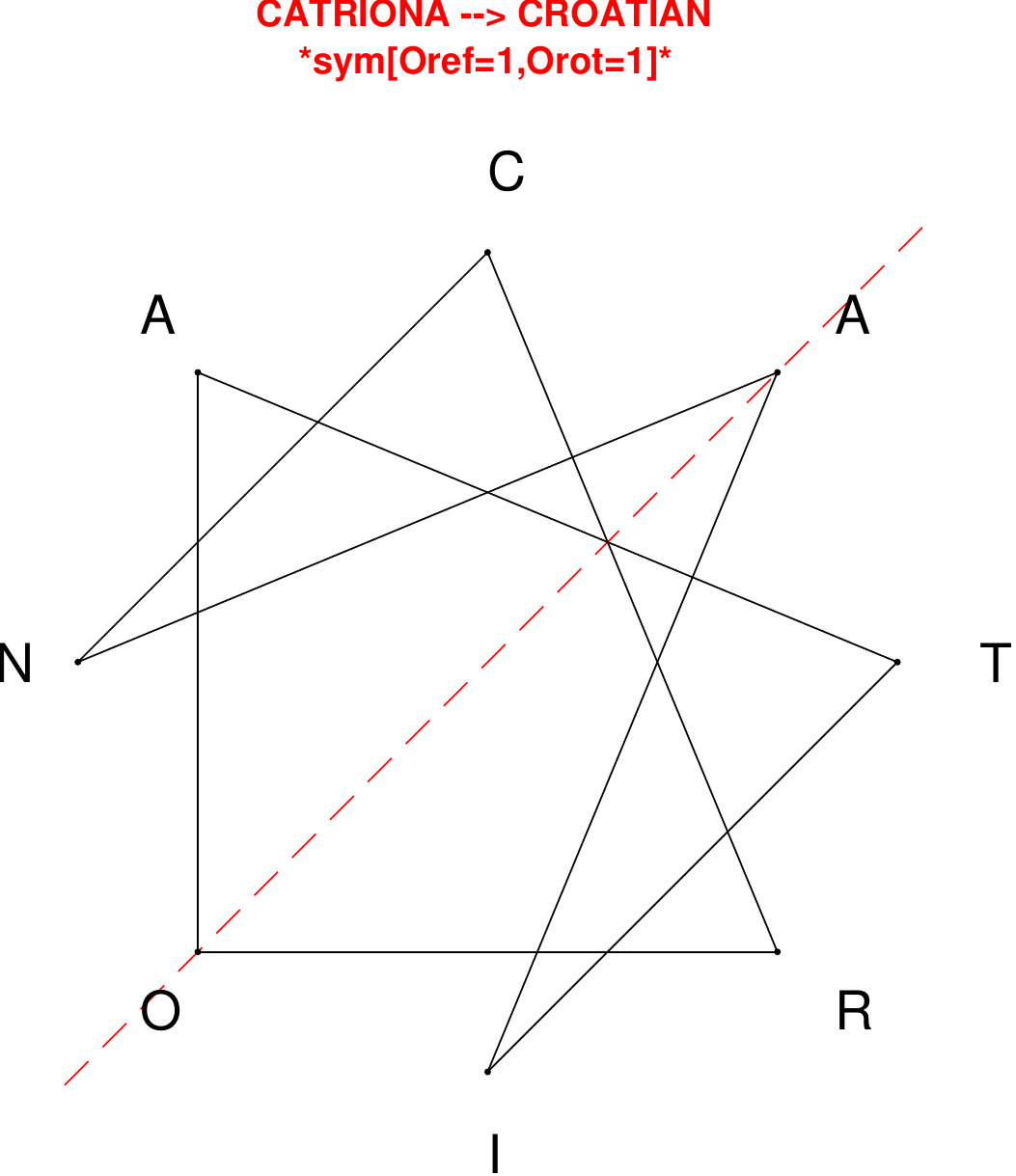}
\end{subfigure}
\end{figure}

\begin{figure}[H]
\centering
\begin{subfigure}[T]{0.19\textwidth}
\centering
\includegraphics[width=\textwidth]{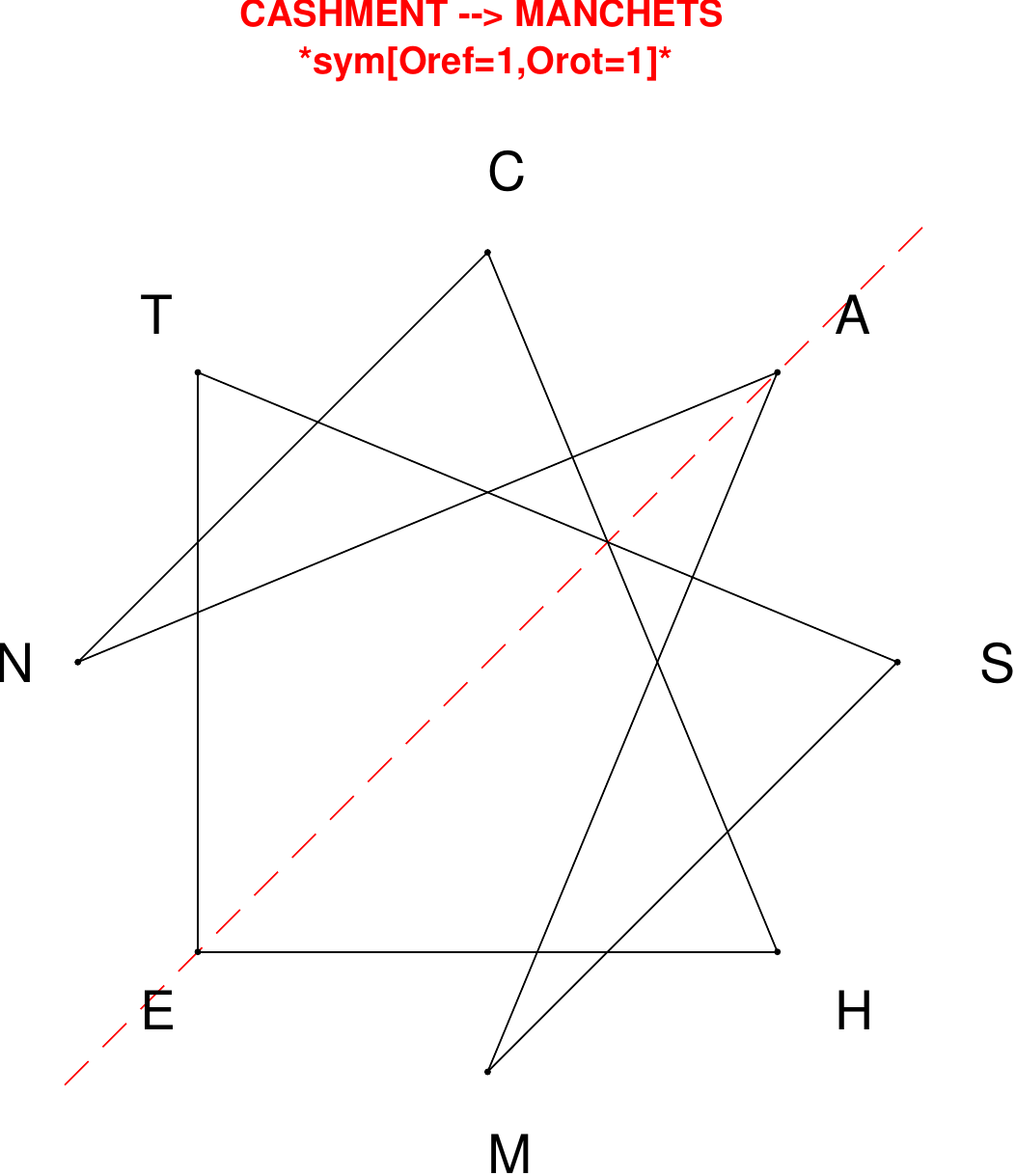}
\end{subfigure}
\hfill
\begin{subfigure}[T]{0.19\textwidth}
\centering
\includegraphics[width=\textwidth]{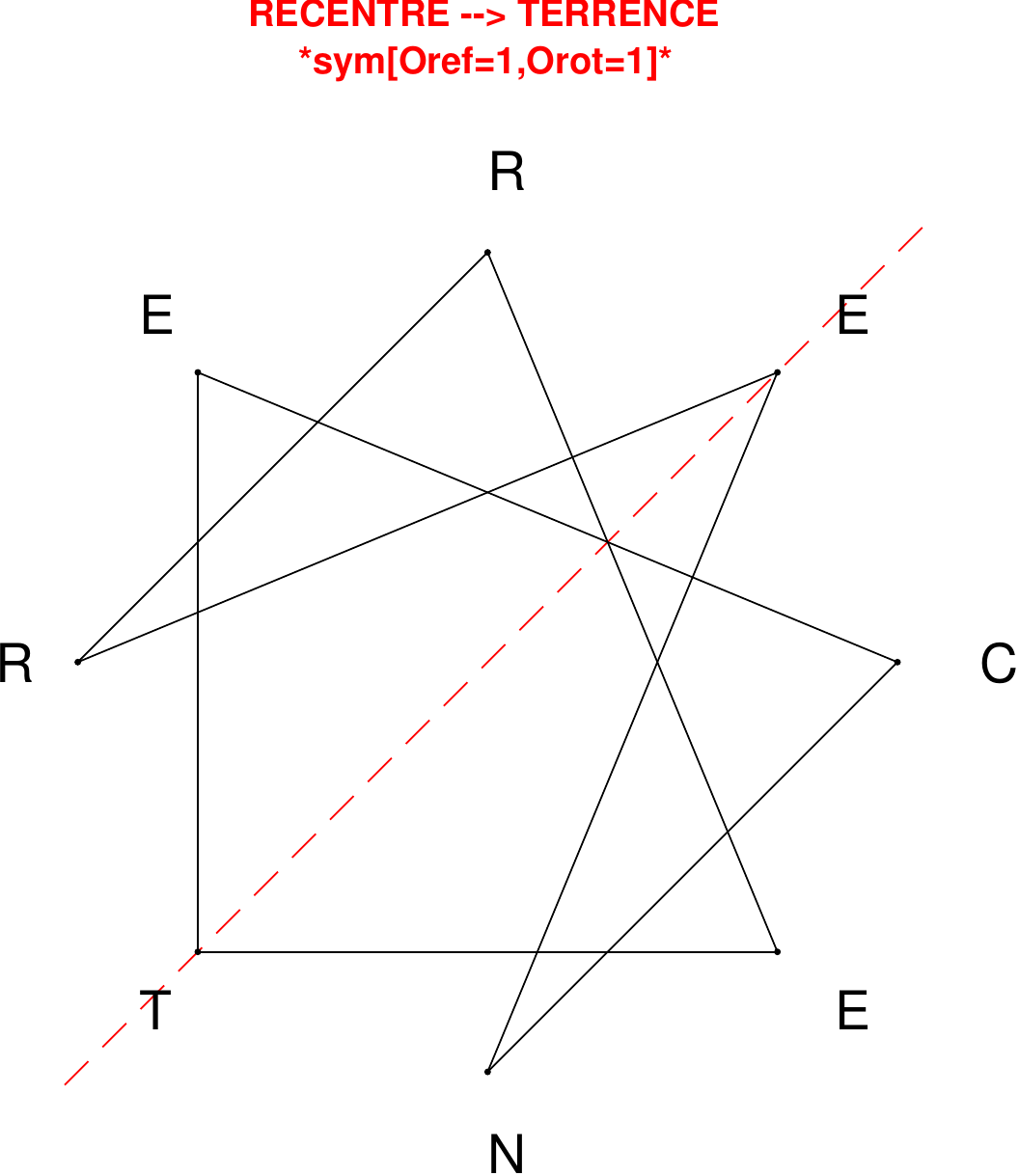}
\end{subfigure}
\hfill
\begin{subfigure}[T]{0.19\textwidth}
\centering
\includegraphics[width=\textwidth]{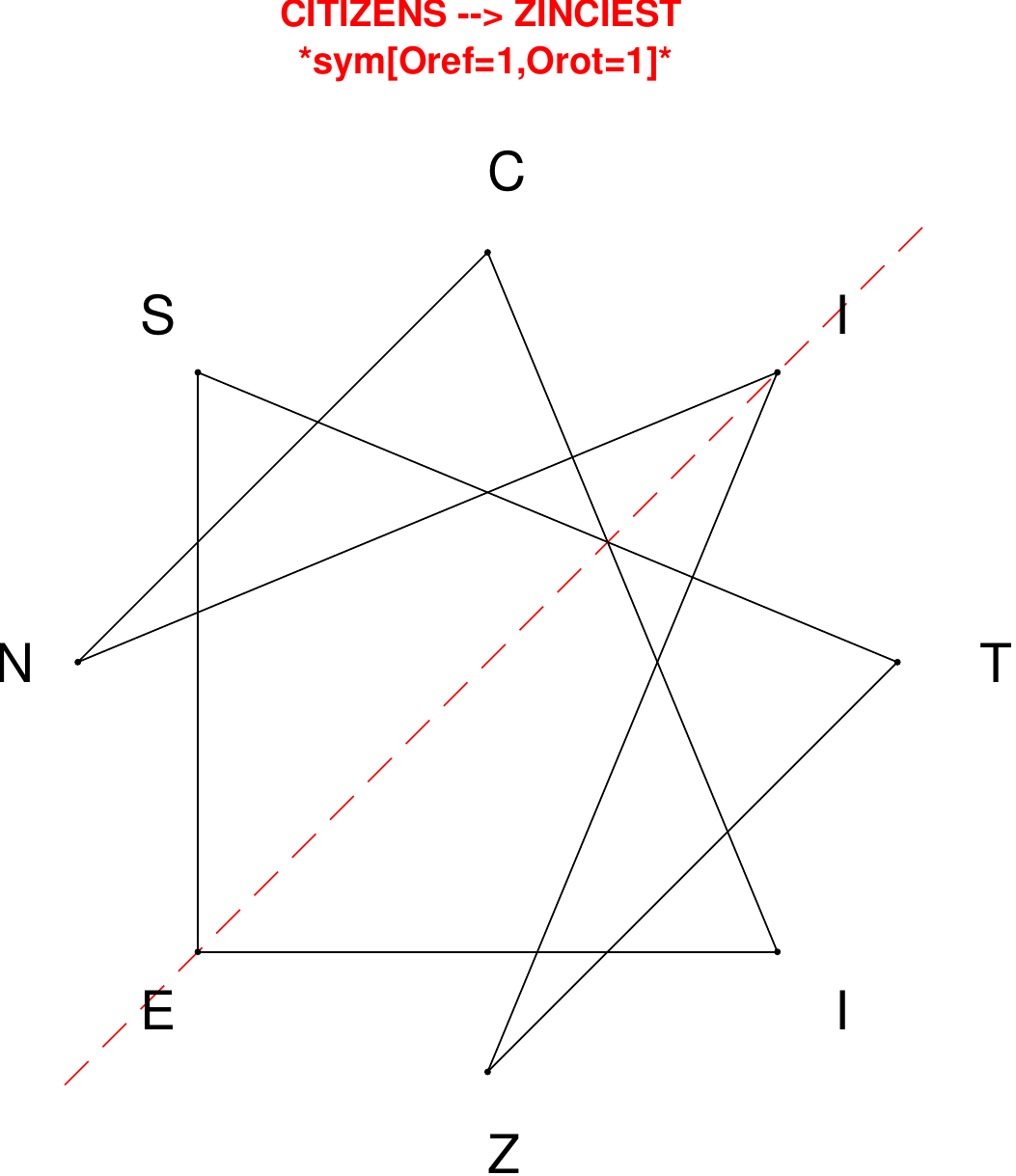}
\end{subfigure}
\hfill
\begin{subfigure}[T]{0.19\textwidth}
\centering
\includegraphics[width=\textwidth]{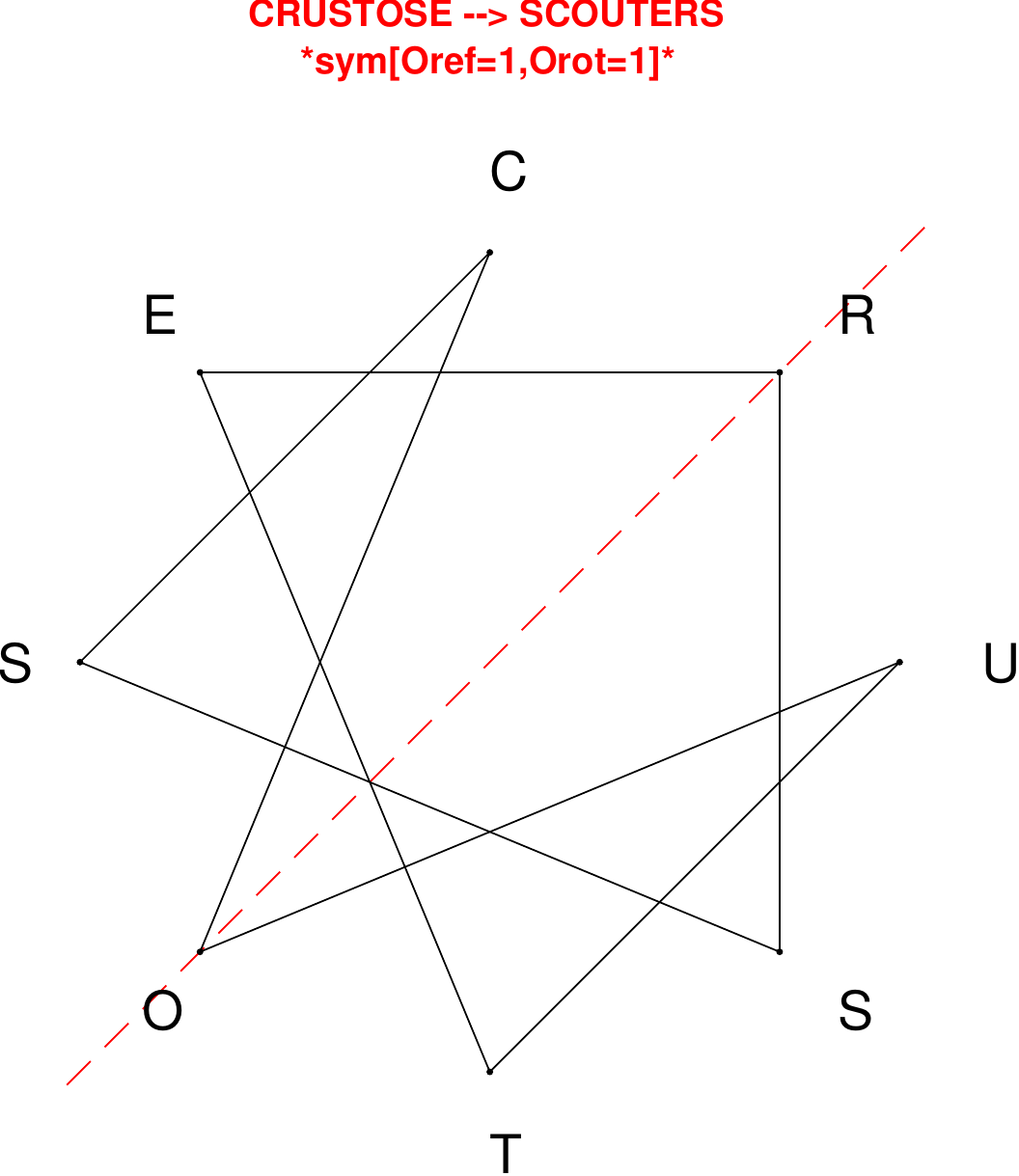}
\end{subfigure}
\hfill
\begin{subfigure}[T]{0.19\textwidth}
\centering
\includegraphics[width=\textwidth]{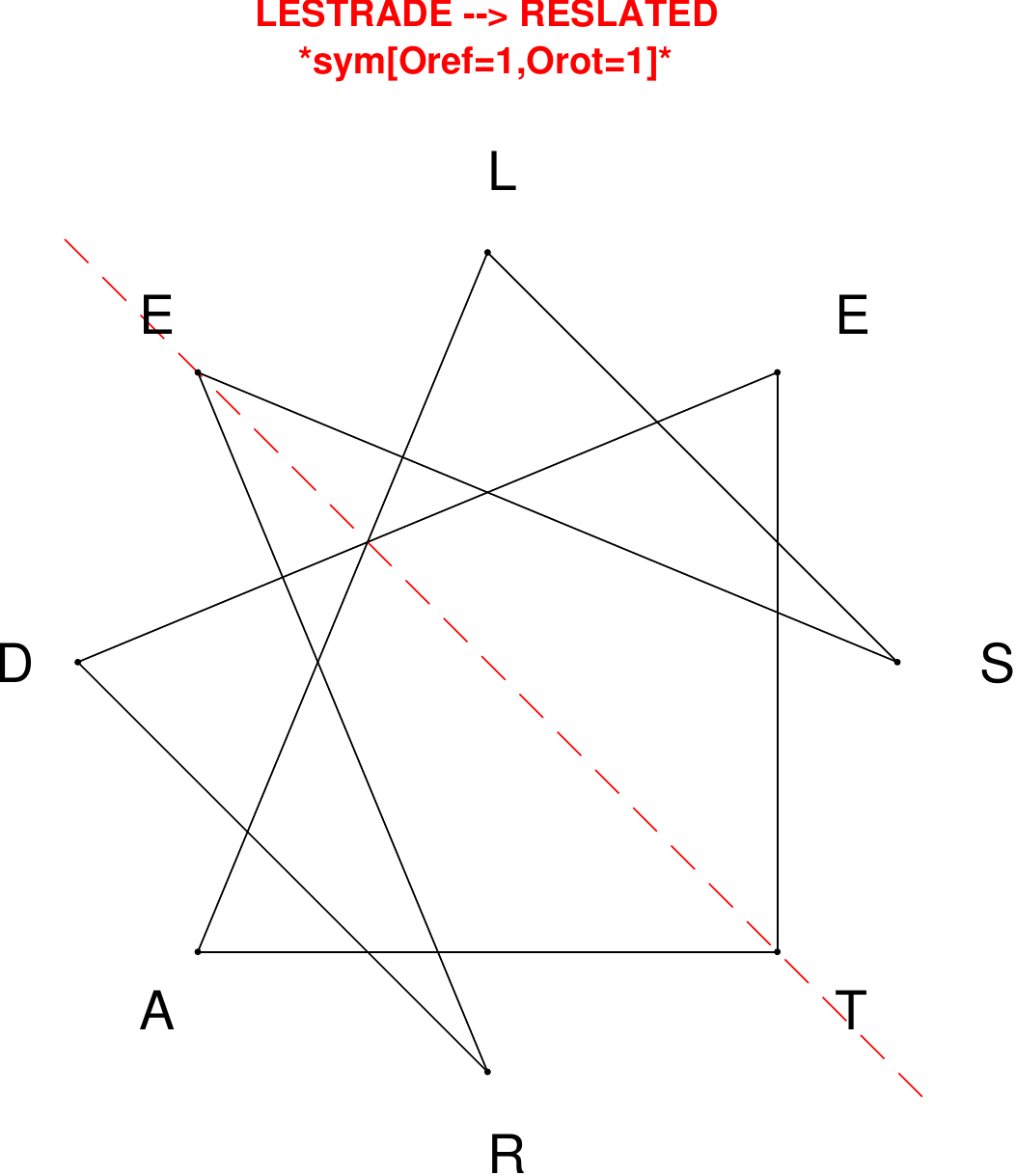}
\end{subfigure}
\end{figure}

\begin{figure}[H]
\centering
\begin{subfigure}[T]{0.19\textwidth}
\centering
\includegraphics[width=\textwidth]{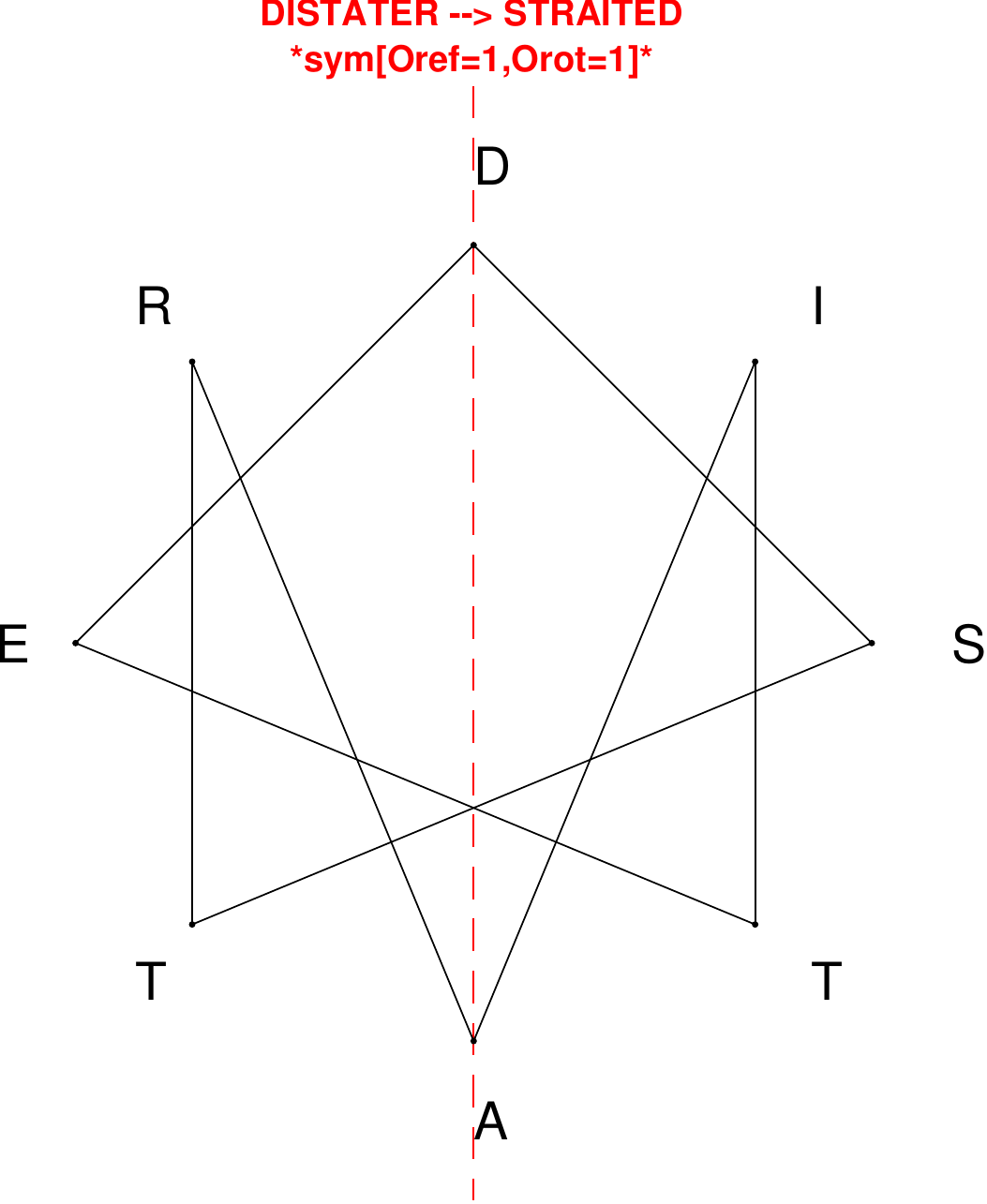}
\end{subfigure}
\hfill
\begin{subfigure}[T]{0.19\textwidth}
\centering
\includegraphics[width=\textwidth]{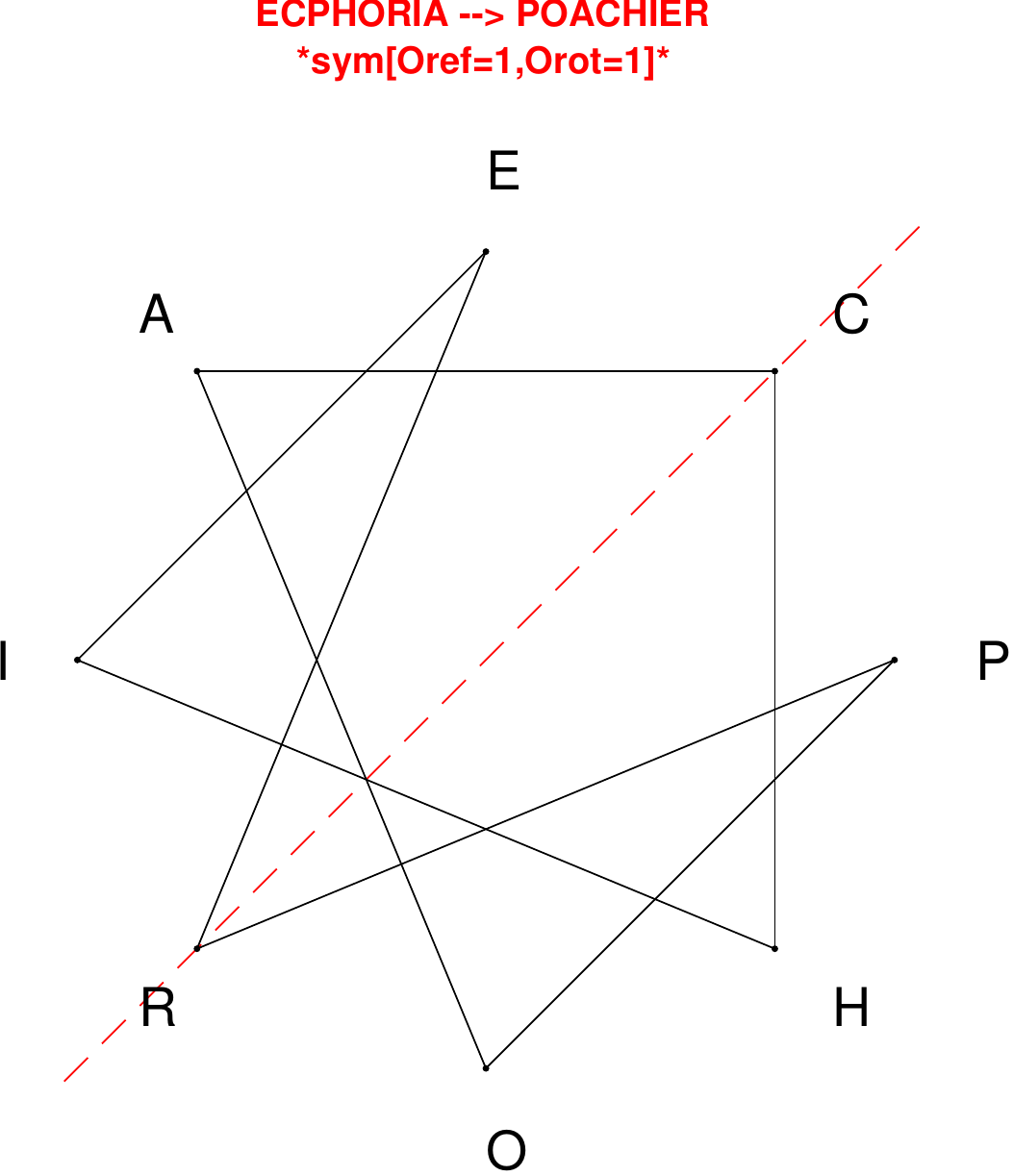}
\end{subfigure}
\hfill
\begin{subfigure}[T]{0.19\textwidth}
\centering
\includegraphics[width=\textwidth]{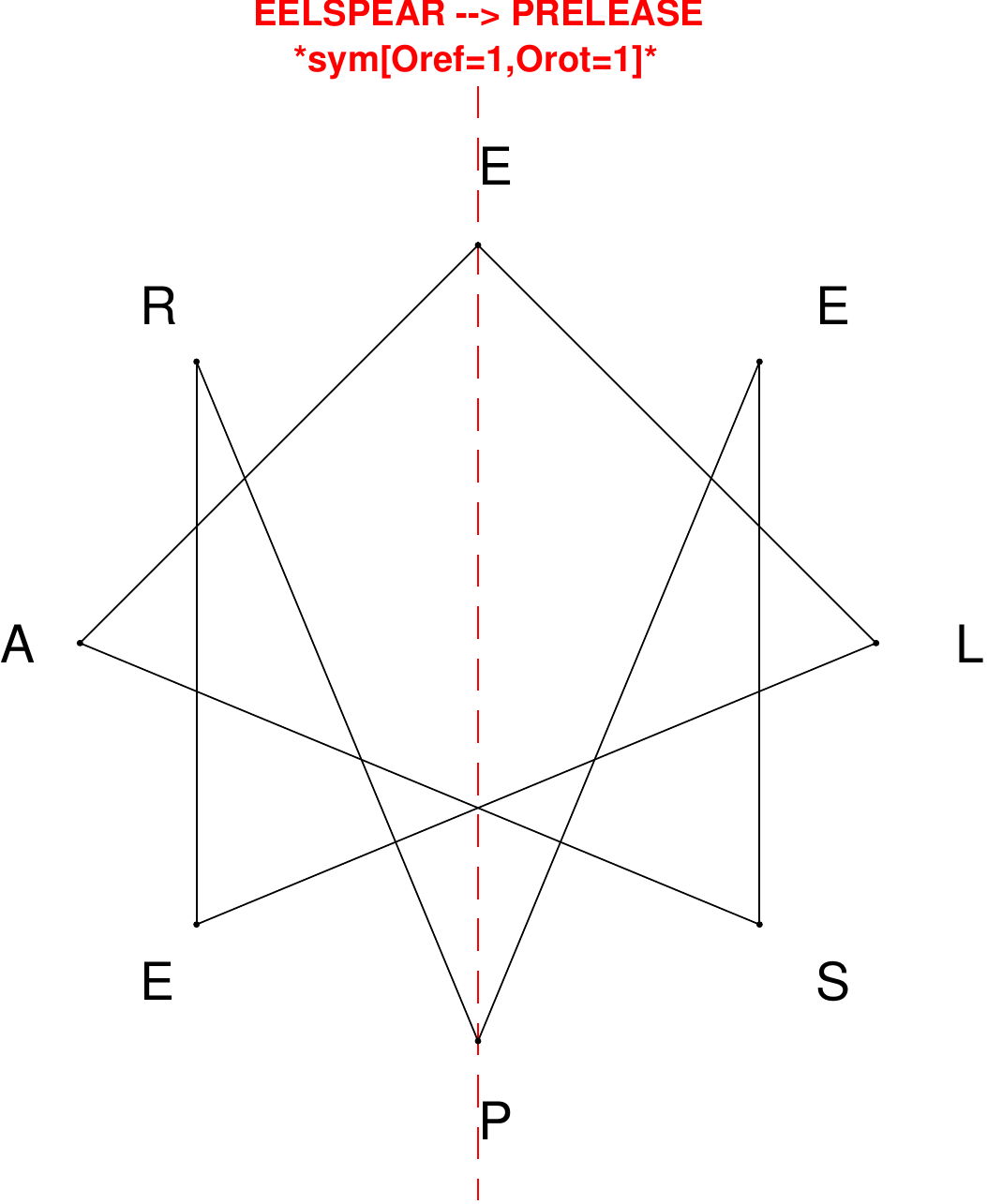}
\end{subfigure}
\hfill
\begin{subfigure}[T]{0.19\textwidth}
\centering
\includegraphics[width=\textwidth]{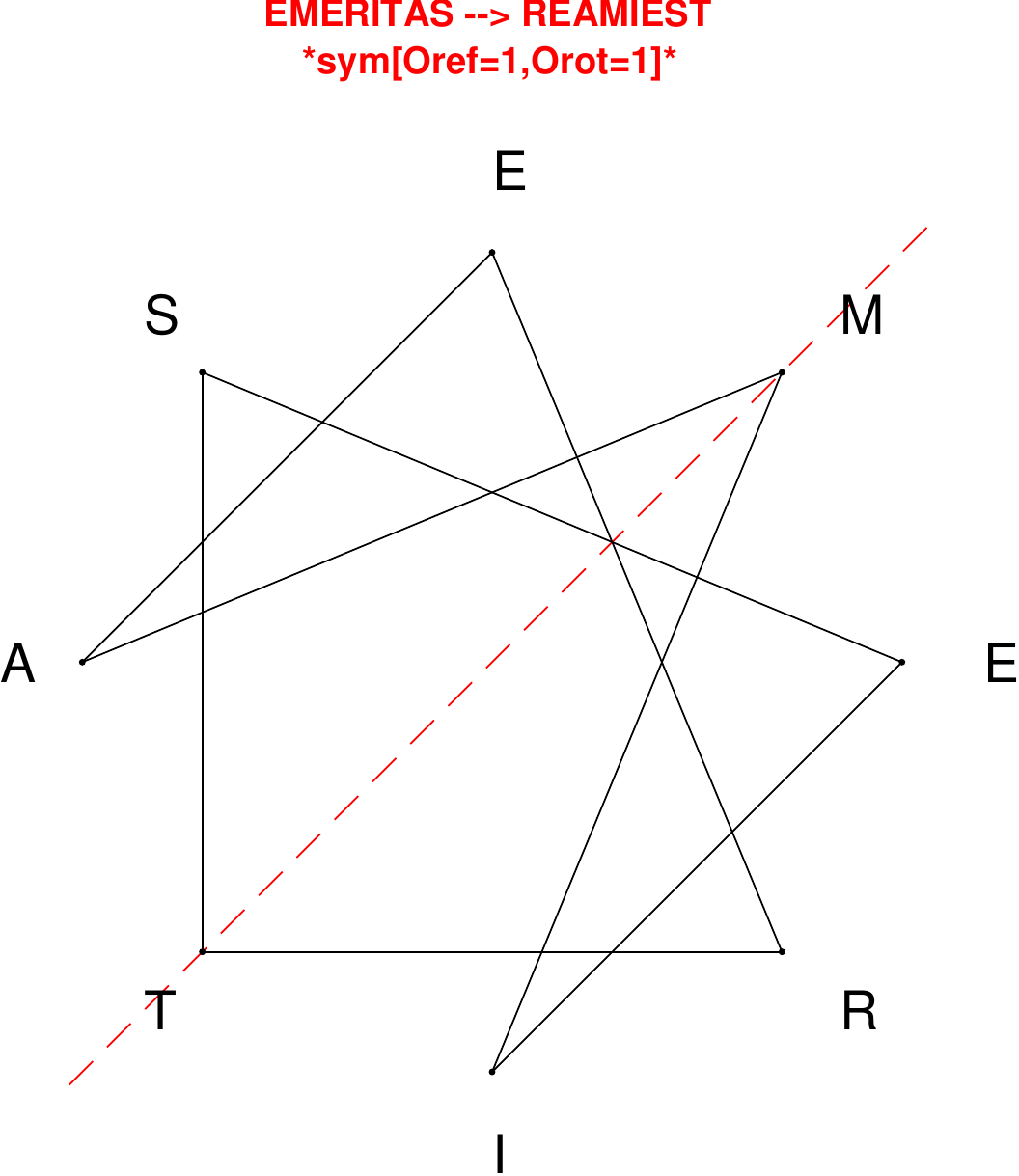}
\end{subfigure}
\hfill
\begin{subfigure}[T]{0.19\textwidth}
\centering
\includegraphics[width=\textwidth]{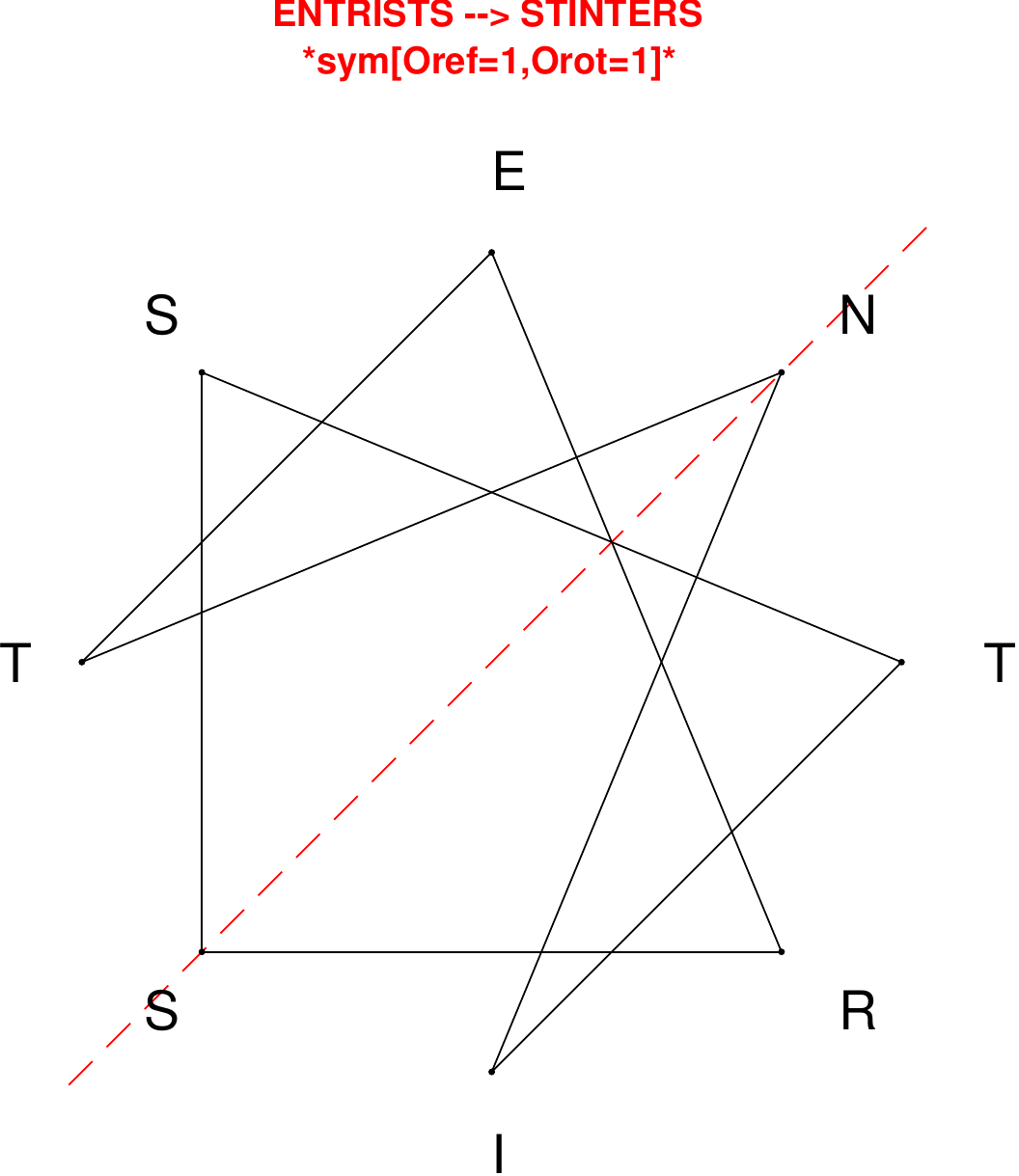}
\end{subfigure}
\end{figure}

\begin{figure}[H]
\centering
\begin{subfigure}[T]{0.19\textwidth}
\centering
\includegraphics[width=\textwidth]{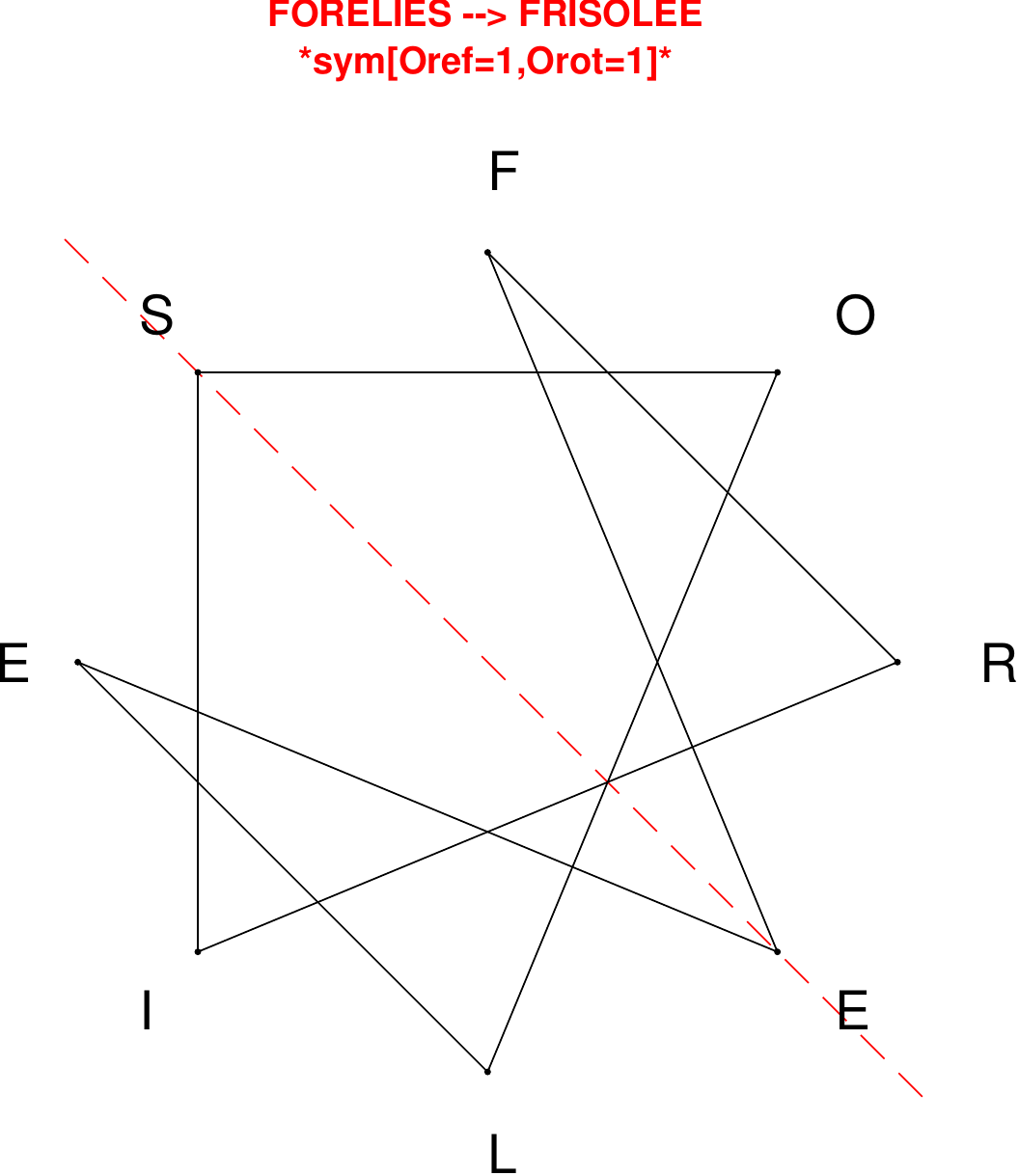}
\end{subfigure}
\hfill
\begin{subfigure}[T]{0.19\textwidth}
\centering
\includegraphics[width=\textwidth]{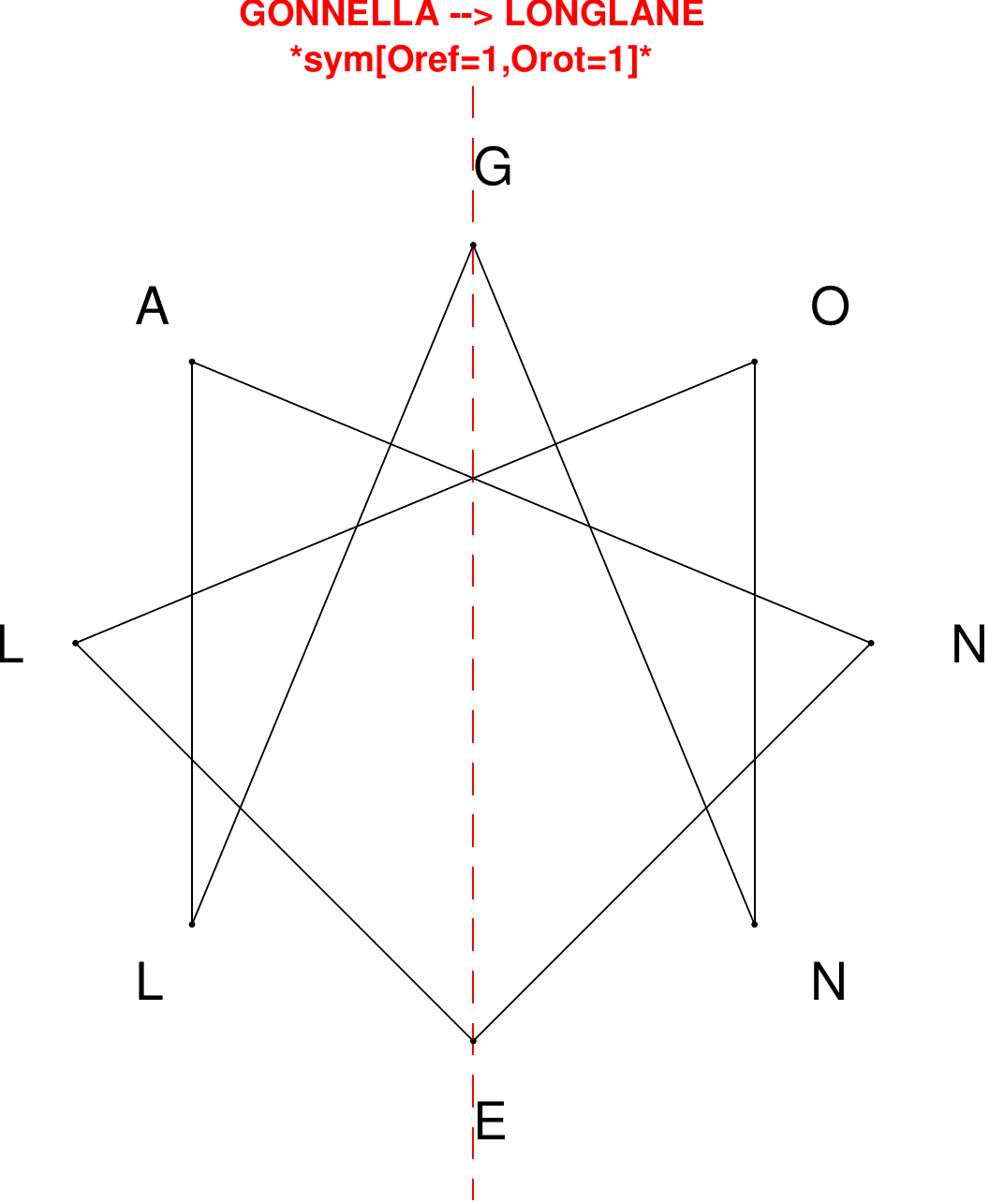}
\end{subfigure}
\hfill
\begin{subfigure}[T]{0.19\textwidth}
\centering
\includegraphics[width=\textwidth]{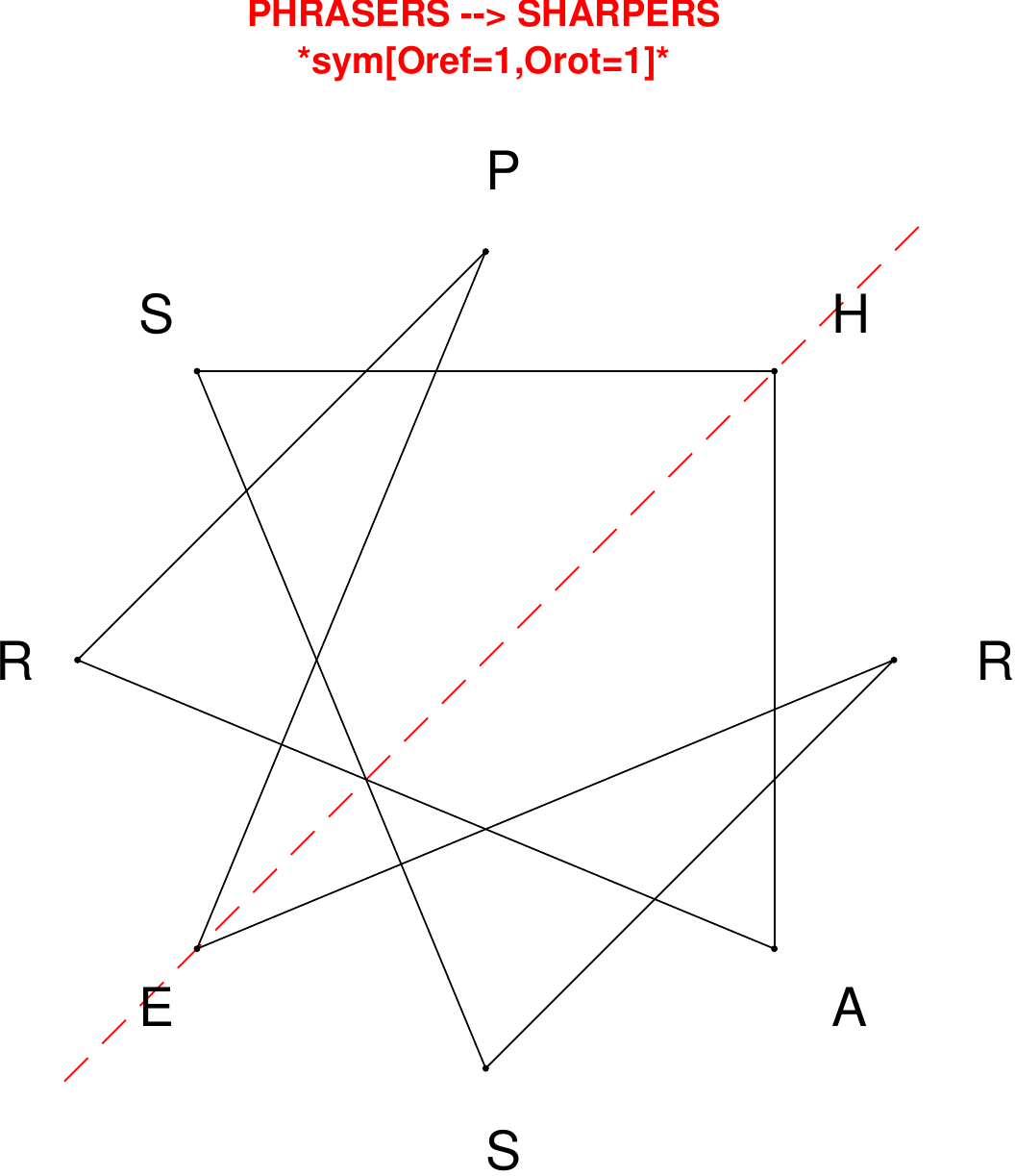}
\end{subfigure}
\hfill
\begin{subfigure}[T]{0.19\textwidth}
\centering
\includegraphics[width=\textwidth]{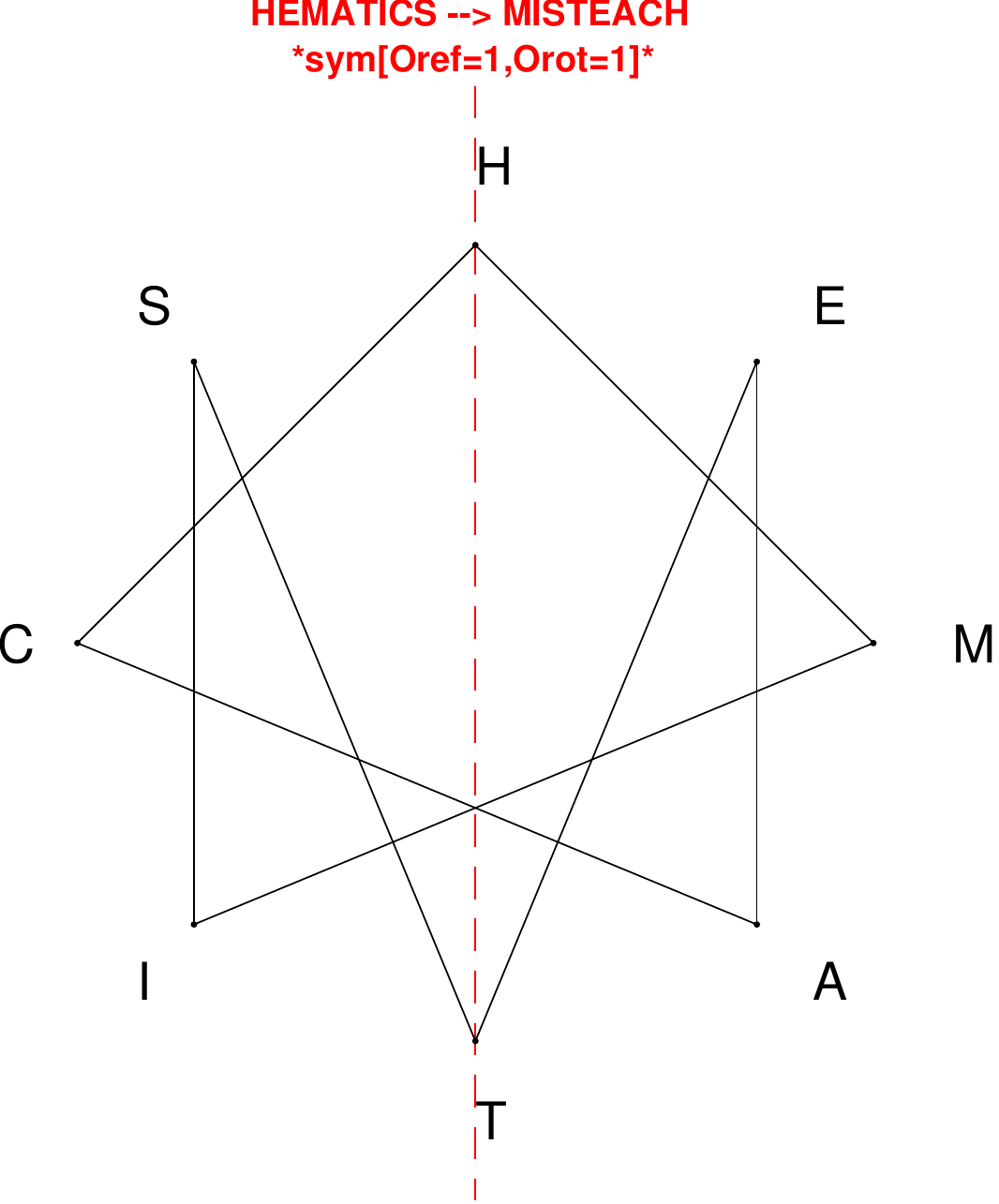}
\end{subfigure}
\hfill
\begin{subfigure}[T]{0.19\textwidth}
\centering
\includegraphics[width=\textwidth]{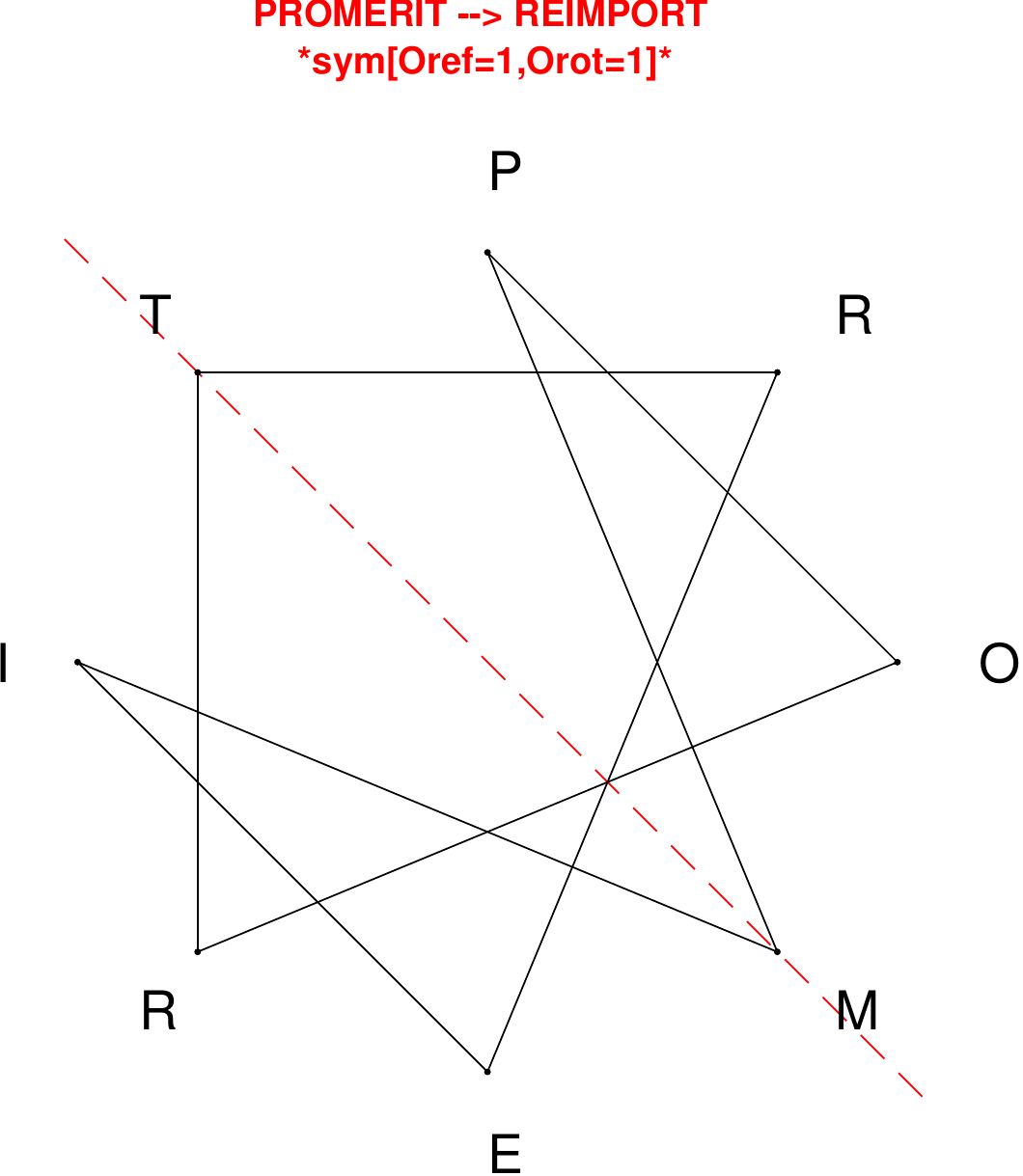}
\end{subfigure}
\end{figure}

\begin{figure}[H]
\centering
\begin{subfigure}[T]{0.19\textwidth}
\centering
\includegraphics[width=\textwidth]{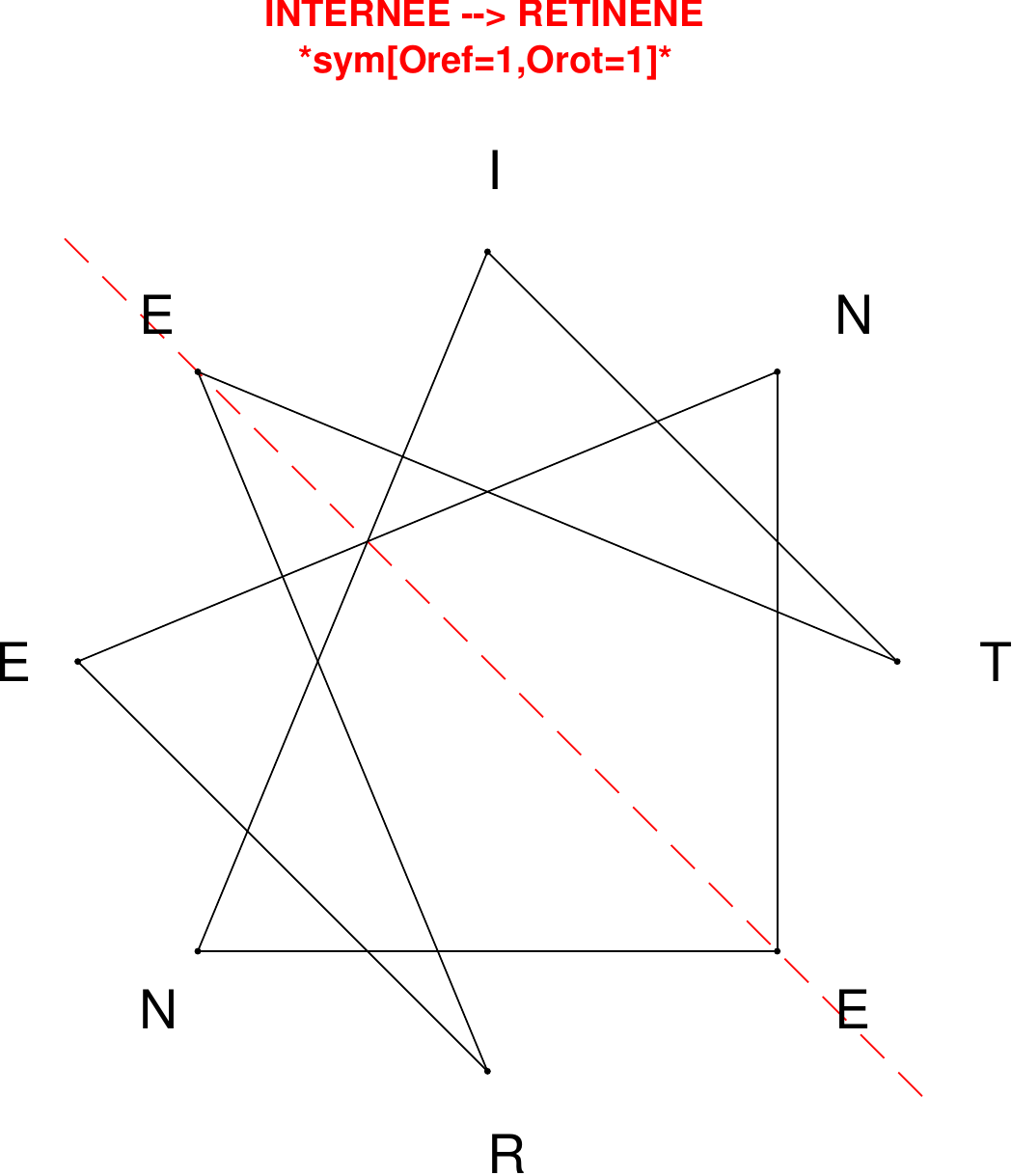}
\end{subfigure}
\hfill
\begin{subfigure}[T]{0.19\textwidth}
\centering
\includegraphics[width=\textwidth]{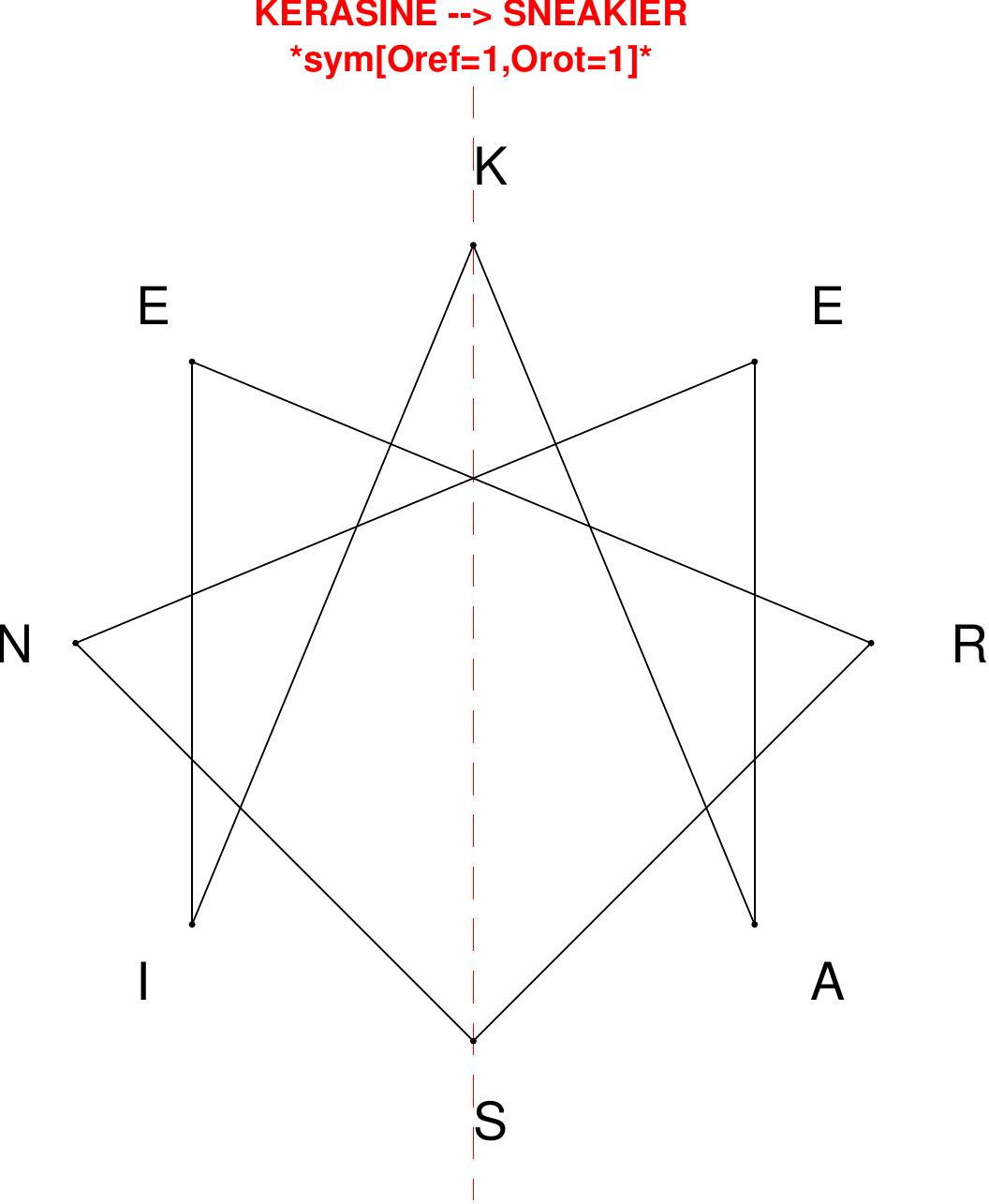}
\end{subfigure}
\hfill
\begin{subfigure}[T]{0.19\textwidth}
\centering
\includegraphics[width=\textwidth]{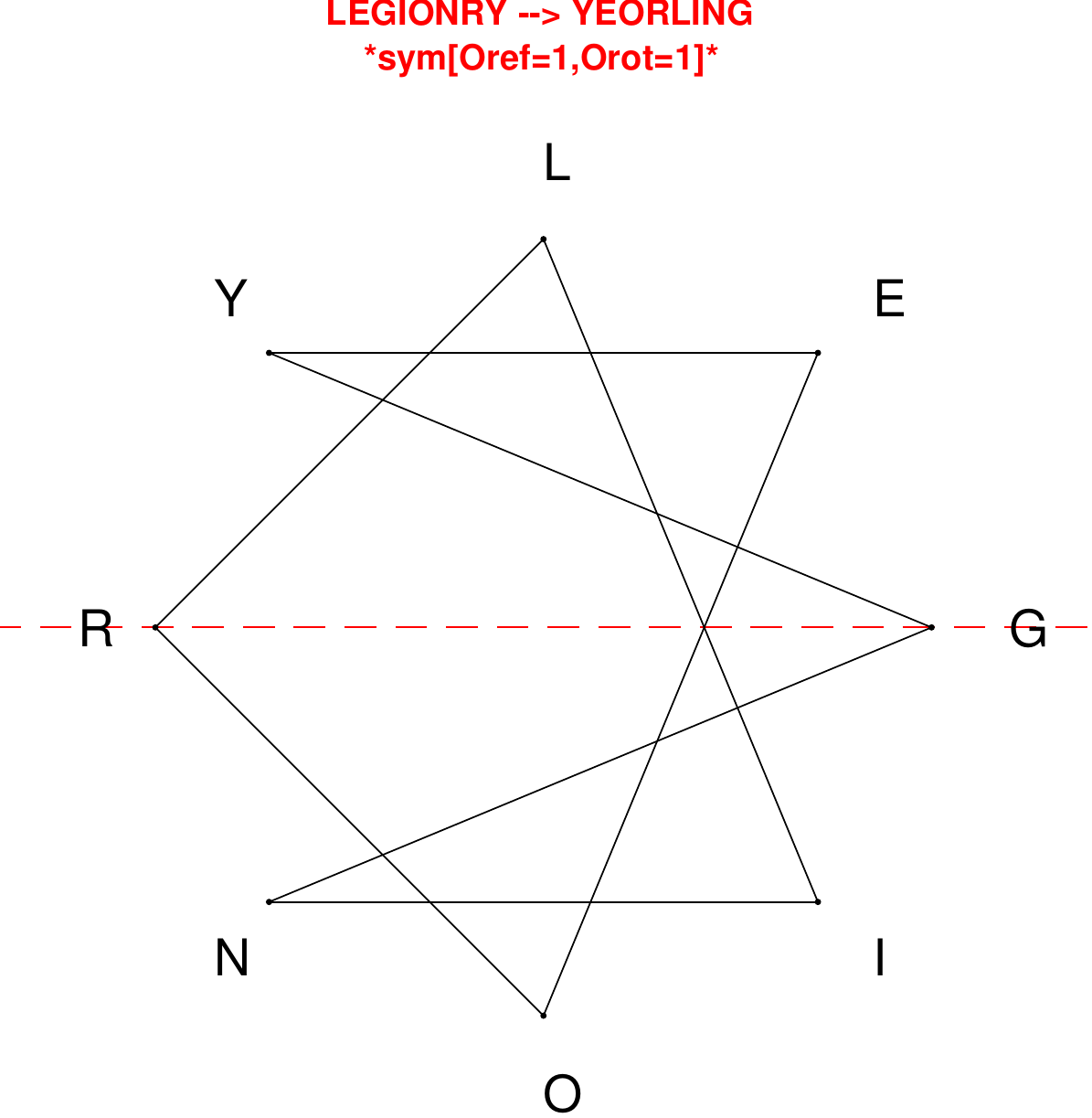}
\end{subfigure}
\hfill
\begin{subfigure}[T]{0.19\textwidth}
\centering
\includegraphics[width=\textwidth]{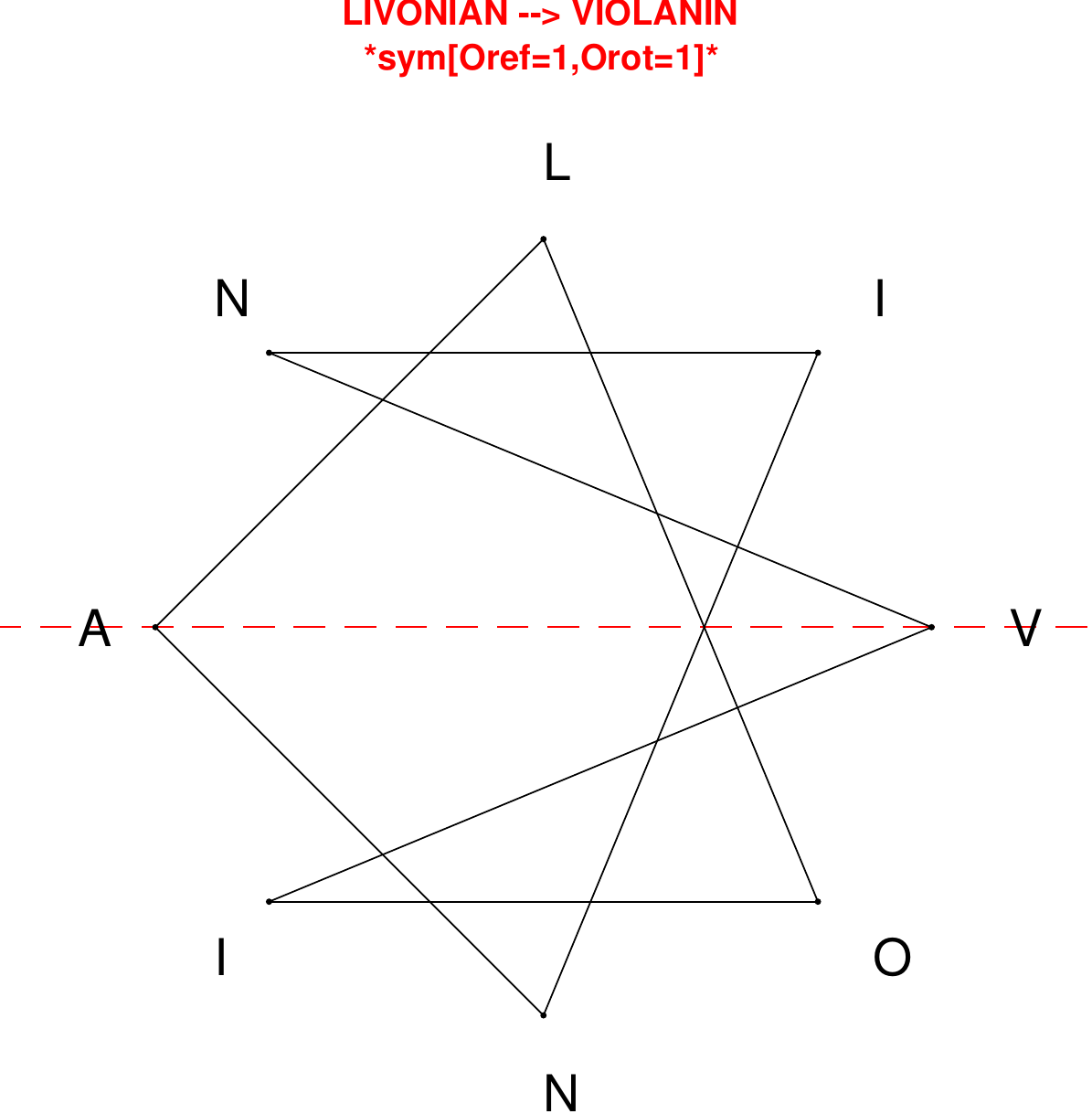}
\end{subfigure}
\hfill
\begin{subfigure}[T]{0.19\textwidth}
\centering
\includegraphics[width=\textwidth]{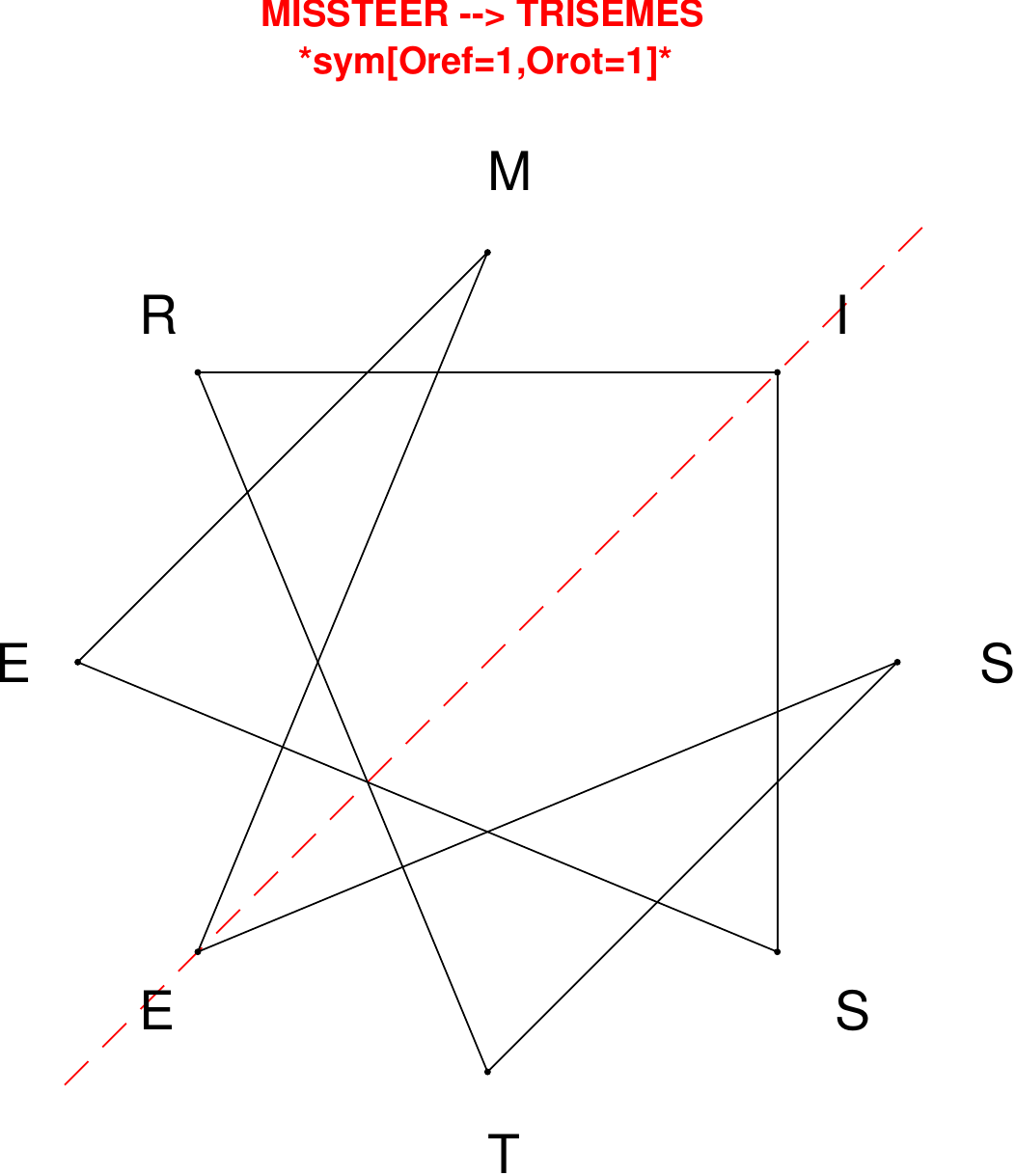}
\end{subfigure}
\end{figure}

\begin{figure}[H]
\centering
\begin{subfigure}[T]{0.19\textwidth}
\centering
\includegraphics[width=\textwidth]{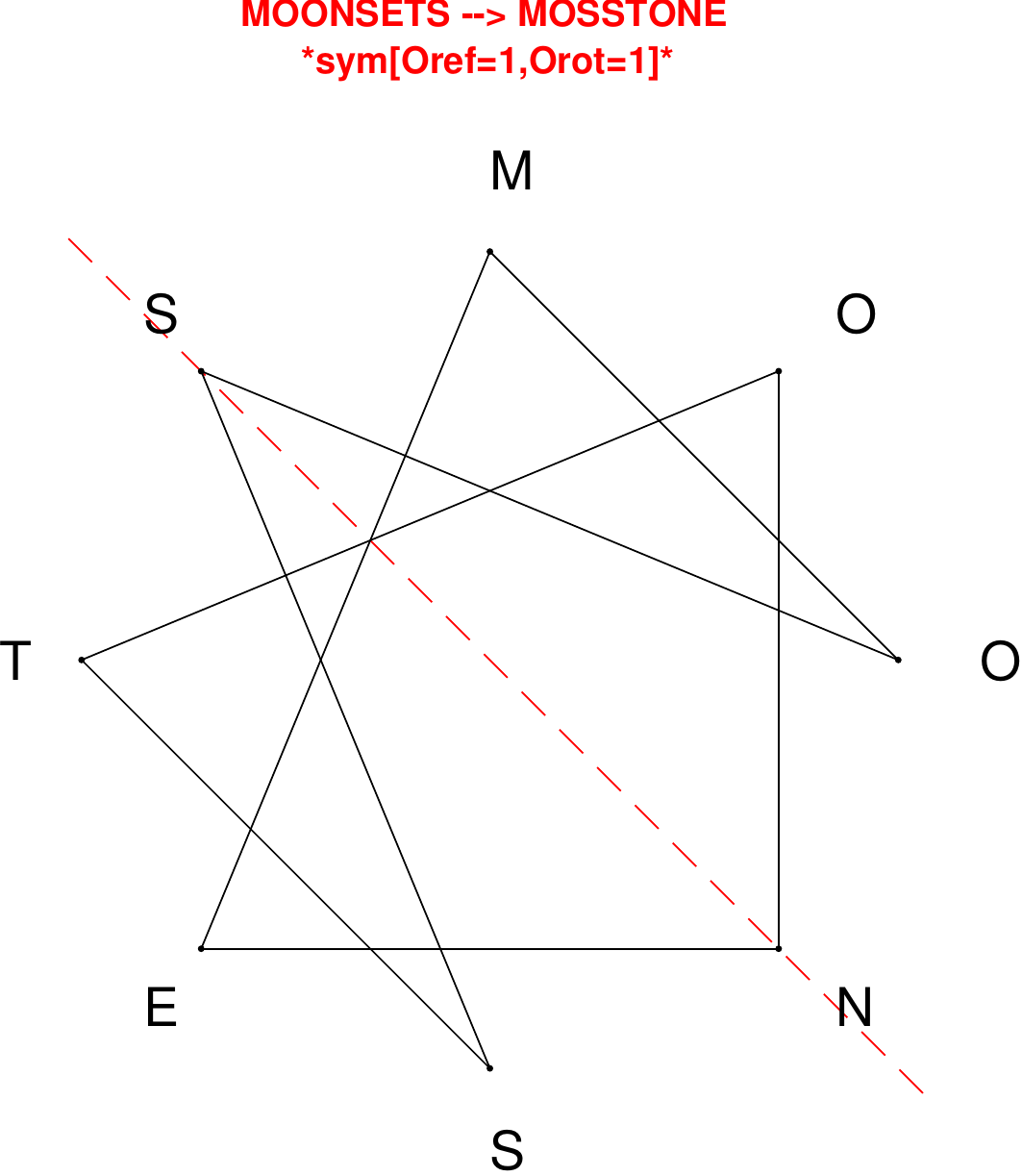}
\end{subfigure}
\hfill
\begin{subfigure}[T]{0.19\textwidth}
\centering
\includegraphics[width=\textwidth]{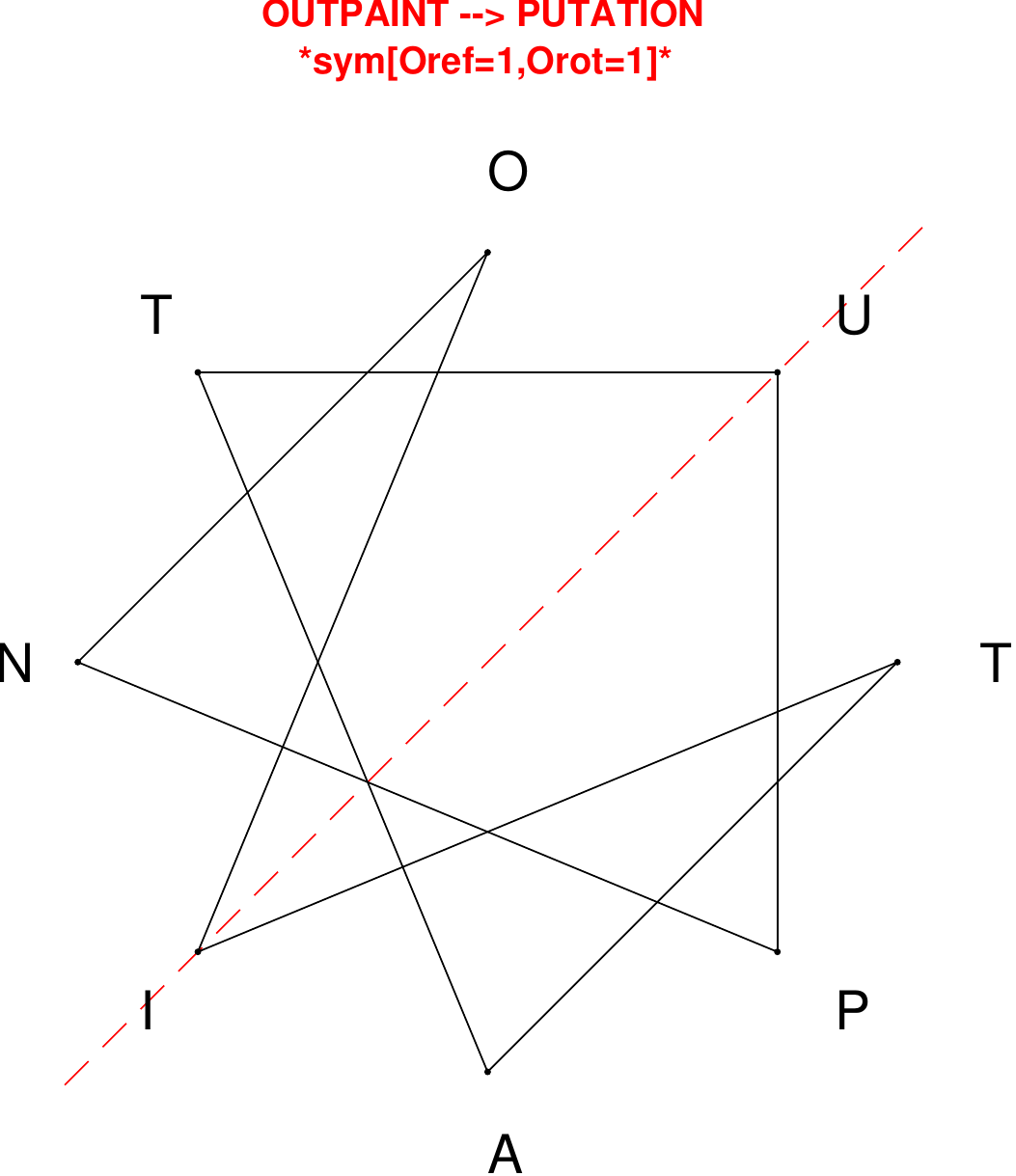}
\end{subfigure}
\hfill
\begin{subfigure}[T]{0.19\textwidth}
\centering
\includegraphics[width=\textwidth]{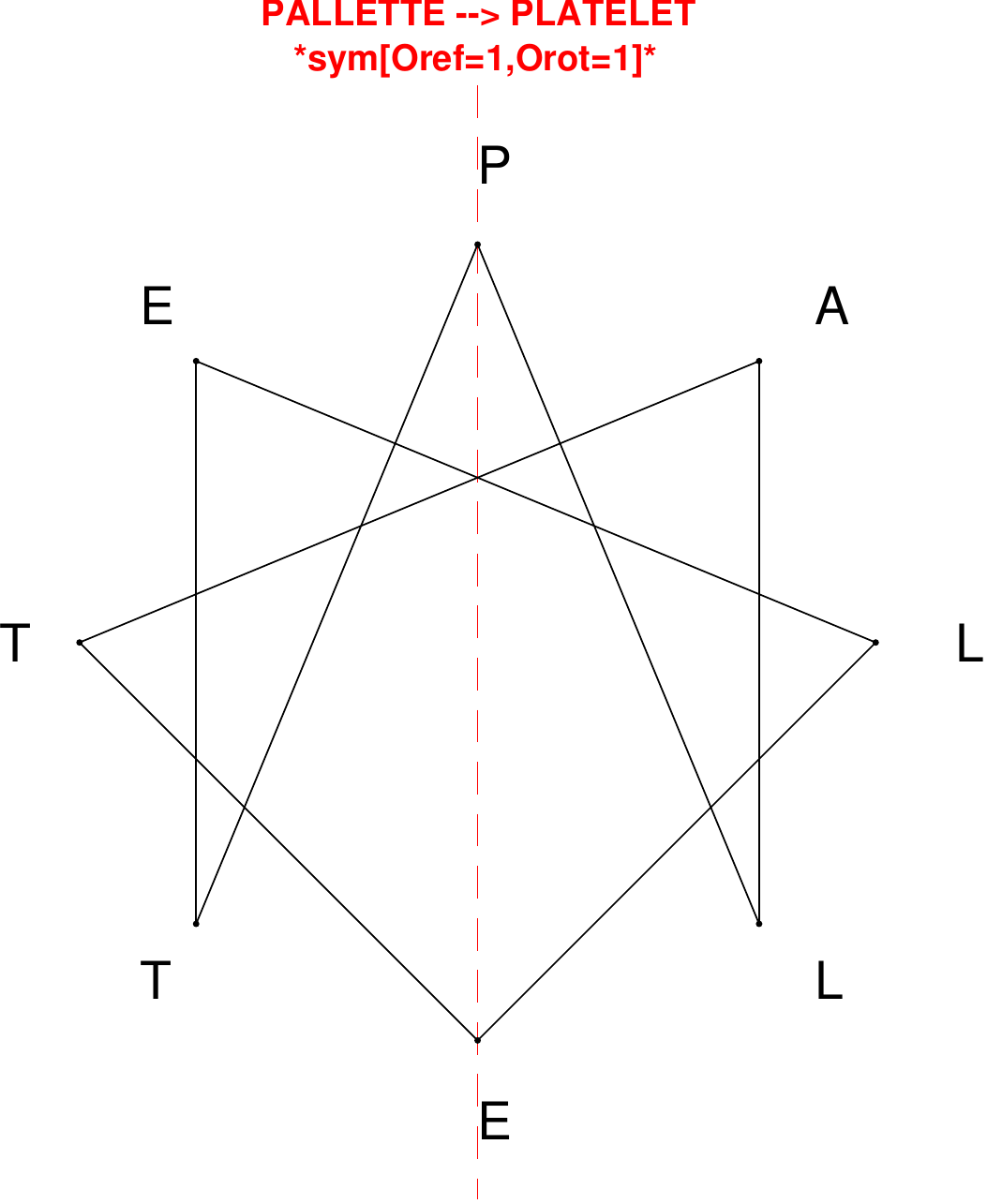}
\end{subfigure}
\hfill
\begin{subfigure}[T]{0.19\textwidth}
\centering
\includegraphics[width=\textwidth]{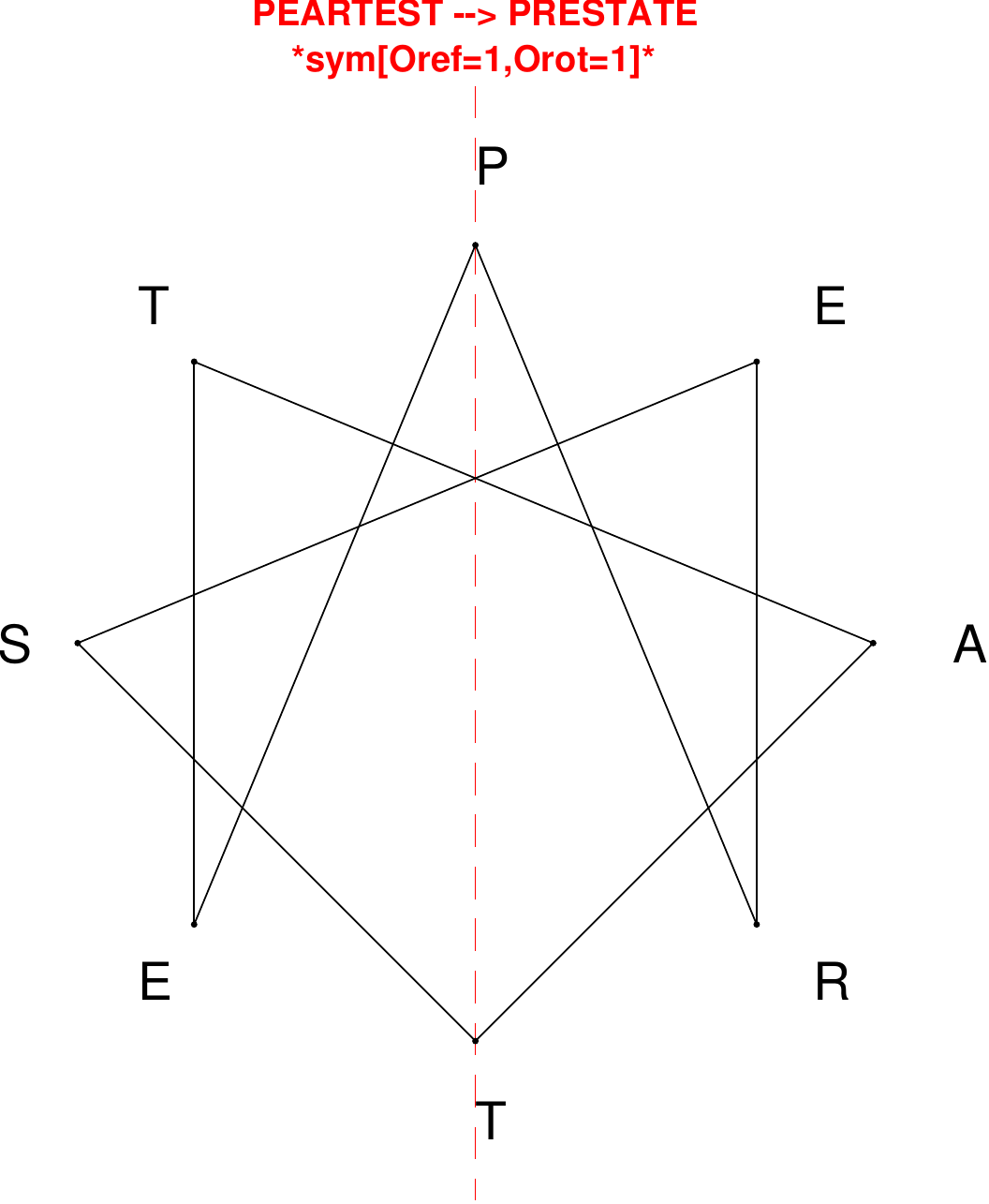}
\end{subfigure}
\hfill
\begin{subfigure}[T]{0.19\textwidth}
\centering
\includegraphics[width=\textwidth]{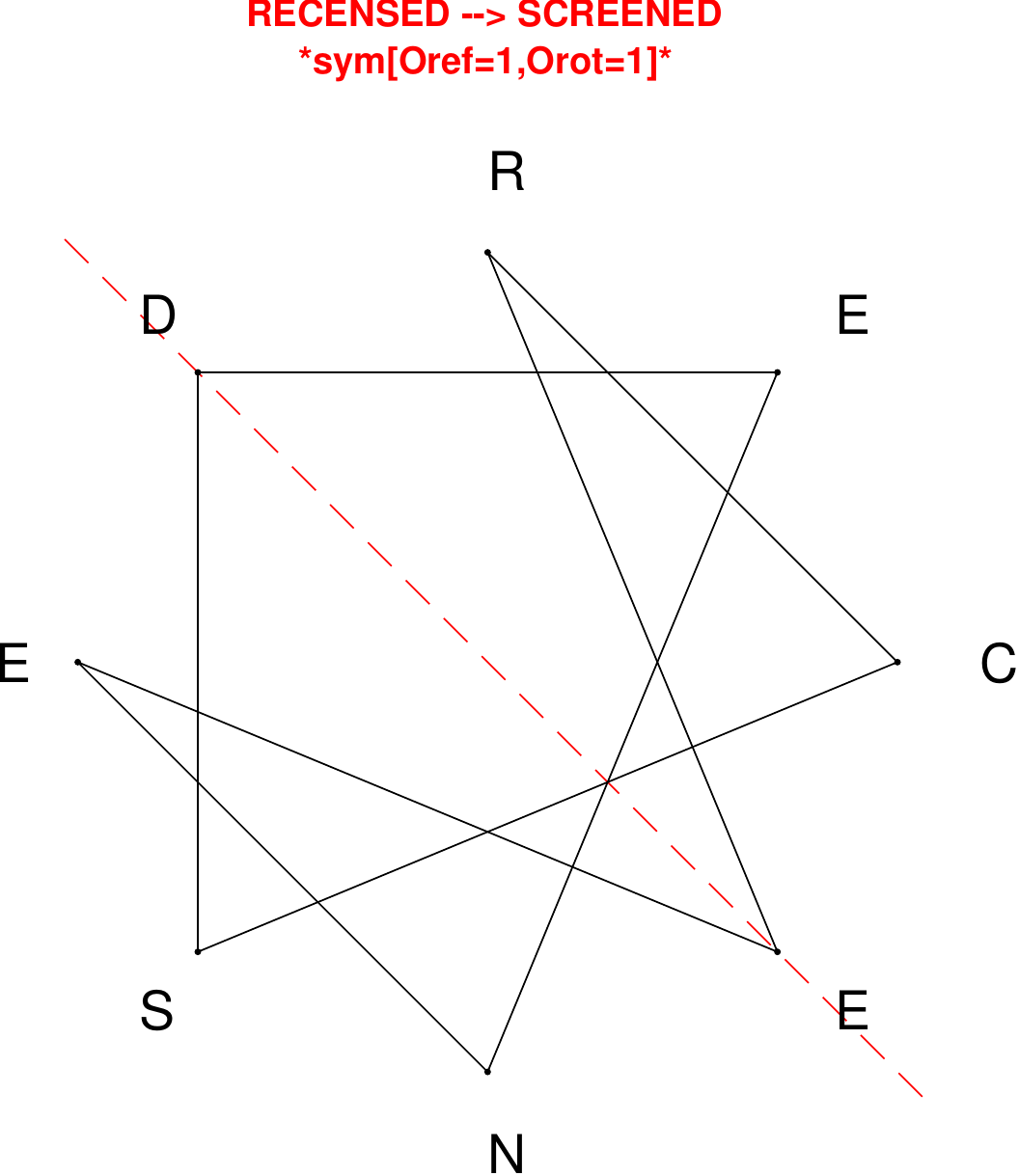}
\end{subfigure}
\end{figure}

\begin{figure}[H]
\centering
\begin{subfigure}[T]{0.19\textwidth}
\centering
\includegraphics[width=\textwidth]{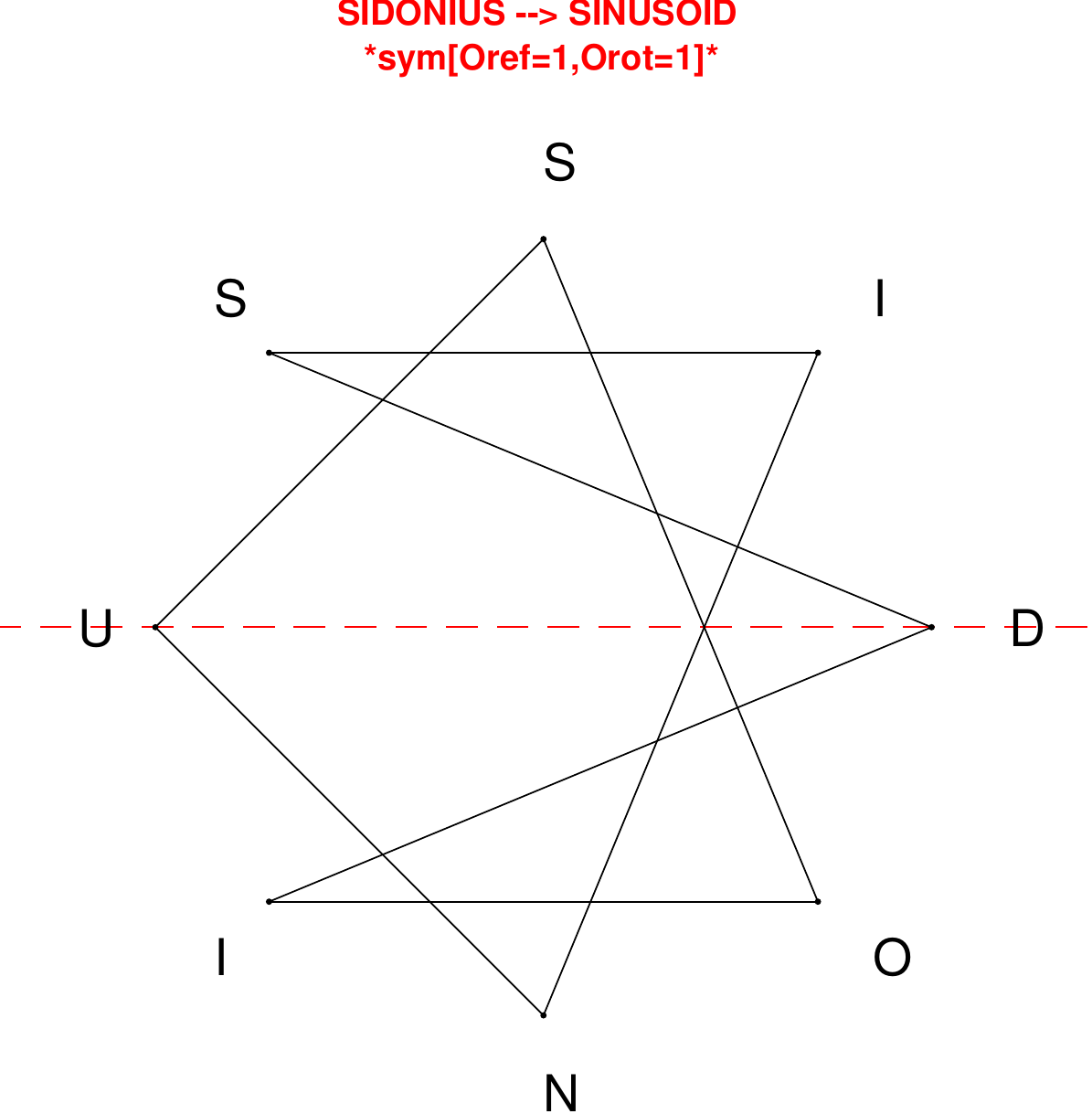}
\end{subfigure}
\hfill
\begin{subfigure}[T]{0.19\textwidth}
\centering
\includegraphics[width=\textwidth]{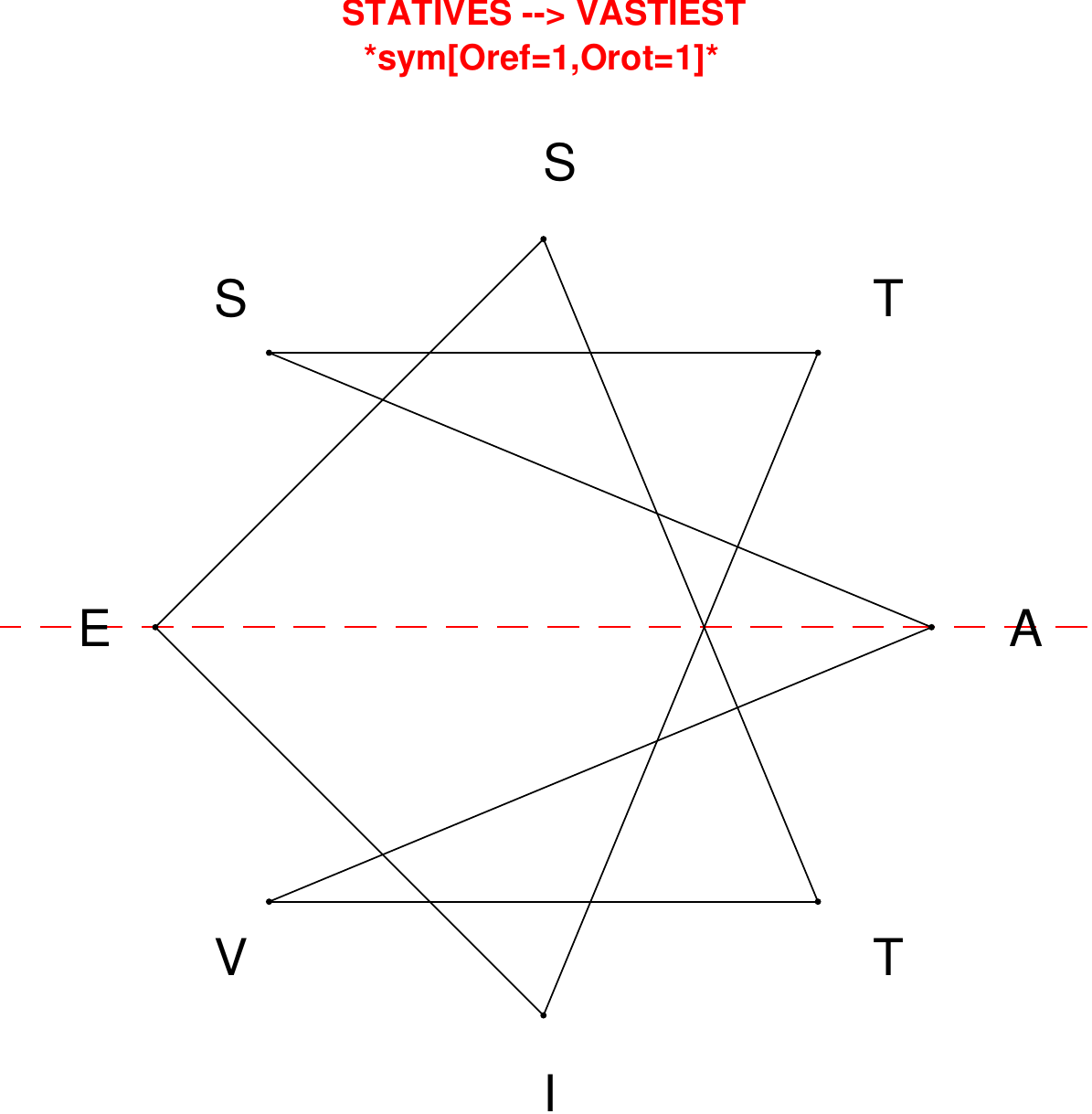}
\end{subfigure}
\hfill
\begin{subfigure}[T]{0.19\textwidth}
\centering
\includegraphics[width=\textwidth]{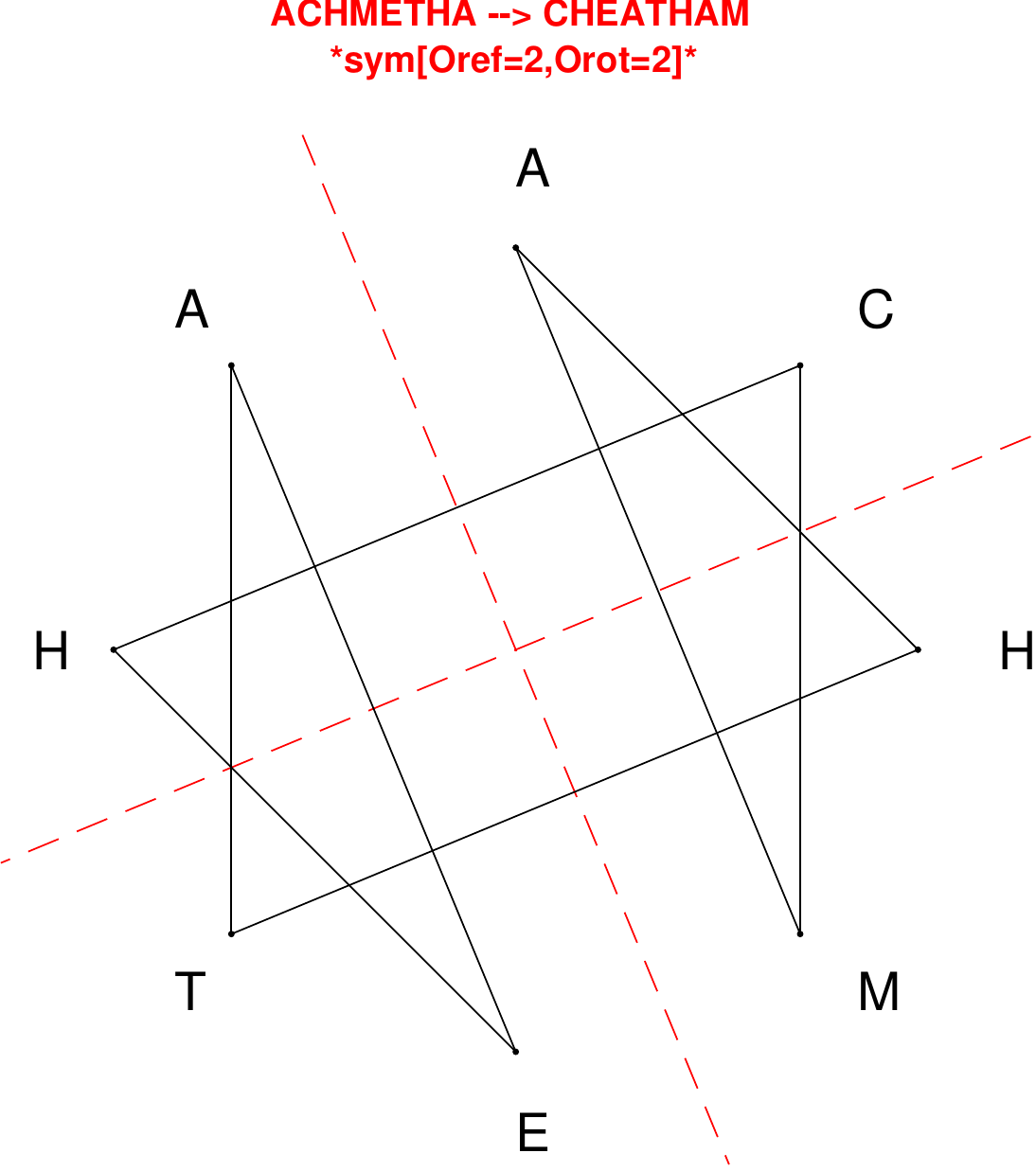}
\end{subfigure}
\hfill
\begin{subfigure}[T]{0.19\textwidth}
\centering
\includegraphics[width=\textwidth]{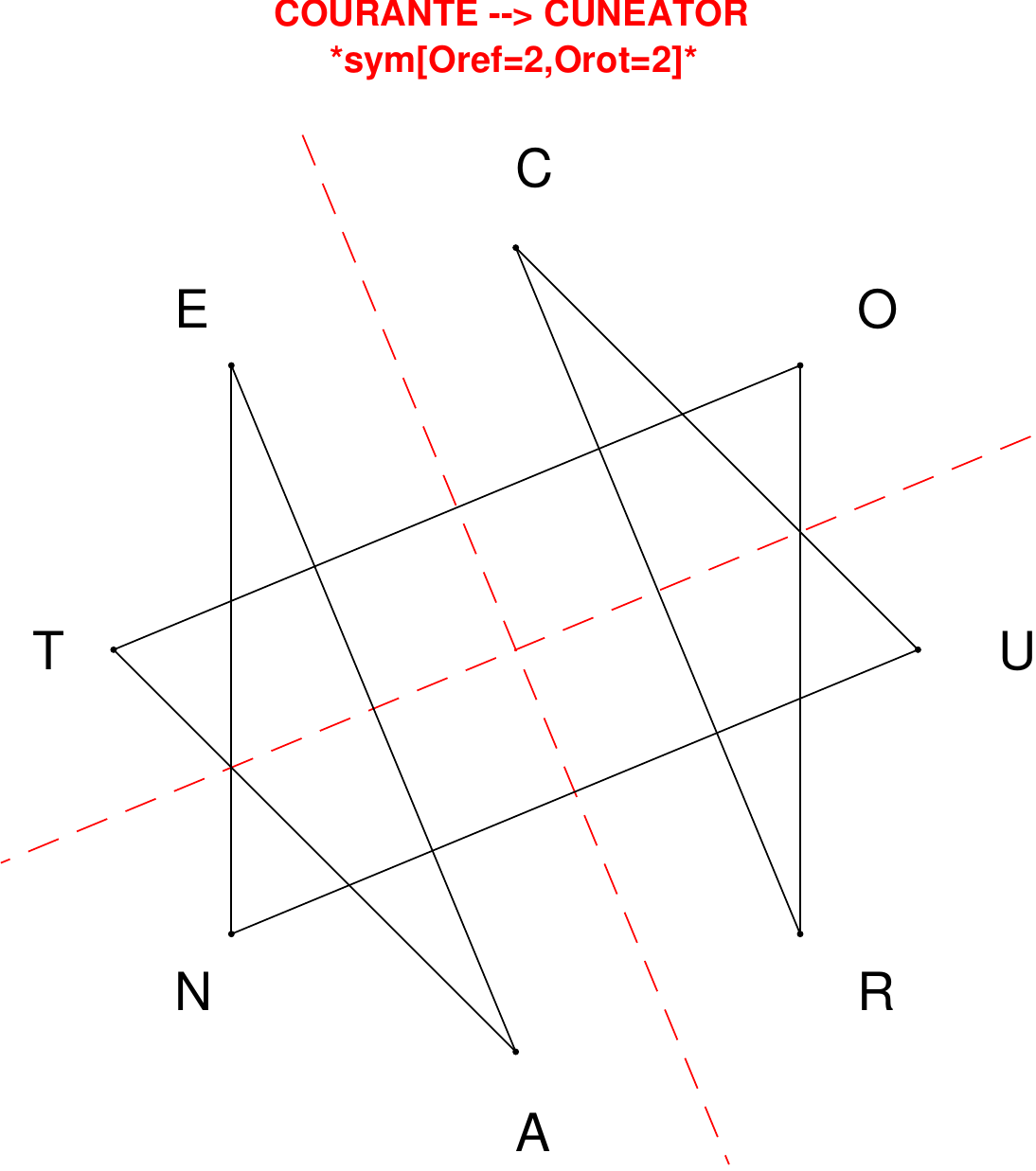}
\end{subfigure}
\hfill
\begin{subfigure}[T]{0.19\textwidth}
\centering
\includegraphics[width=\textwidth]{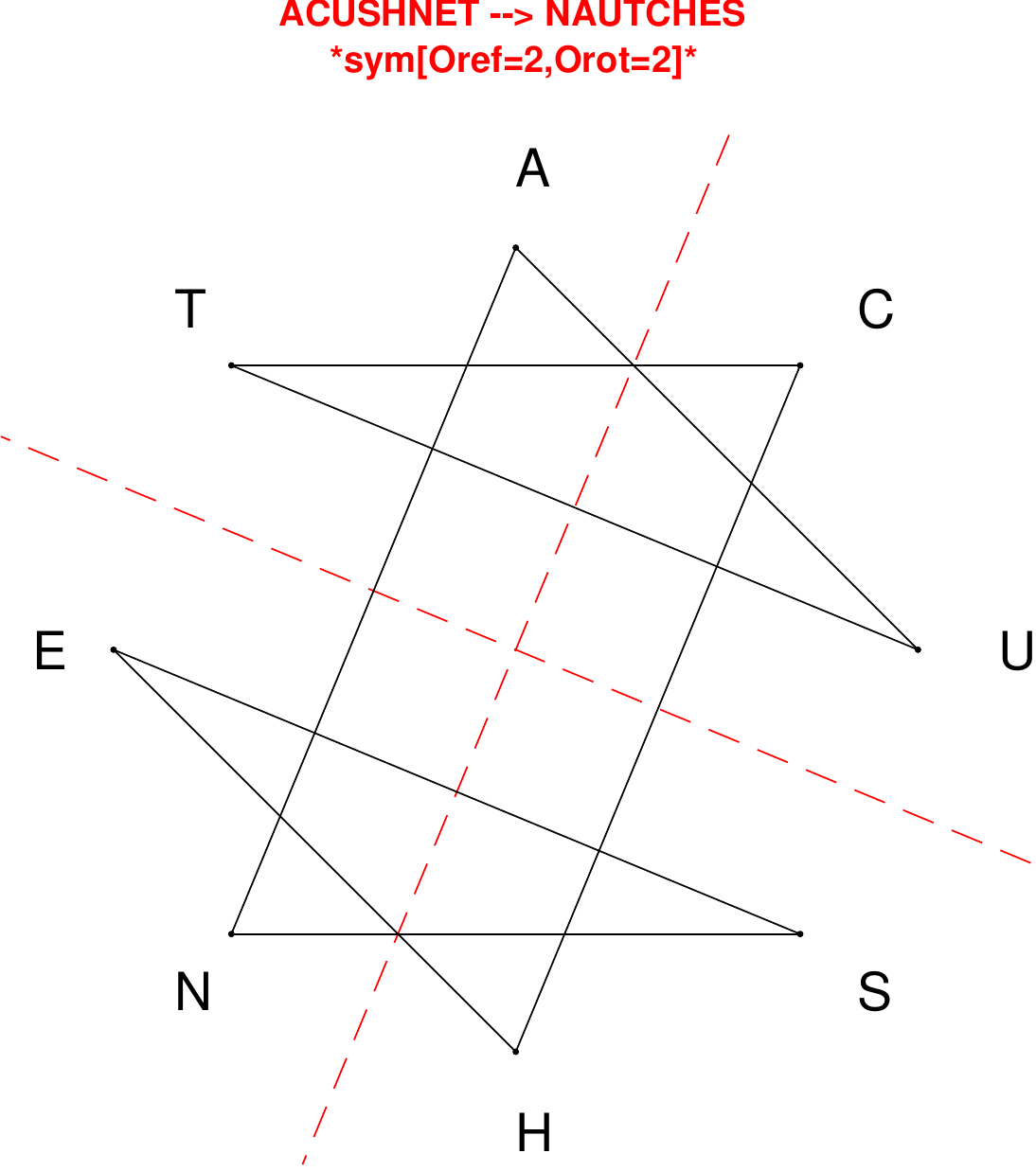}
\end{subfigure}
\end{figure}

\begin{figure}[H]
\centering
\begin{subfigure}[T]{0.19\textwidth}
\centering
\includegraphics[width=\textwidth]{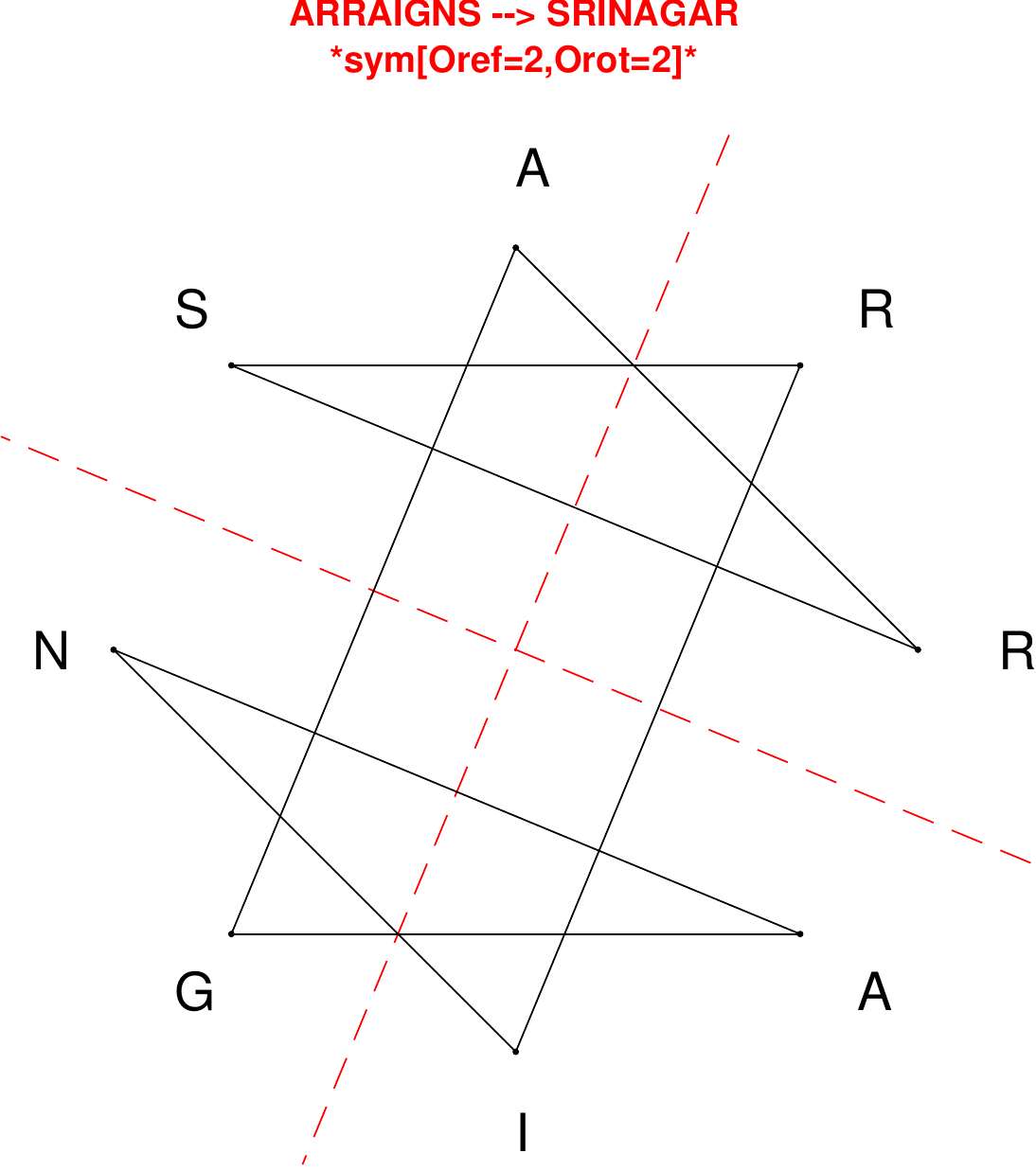}
\end{subfigure}
\hfill
\begin{subfigure}[T]{0.19\textwidth}
\centering
\includegraphics[width=\textwidth]{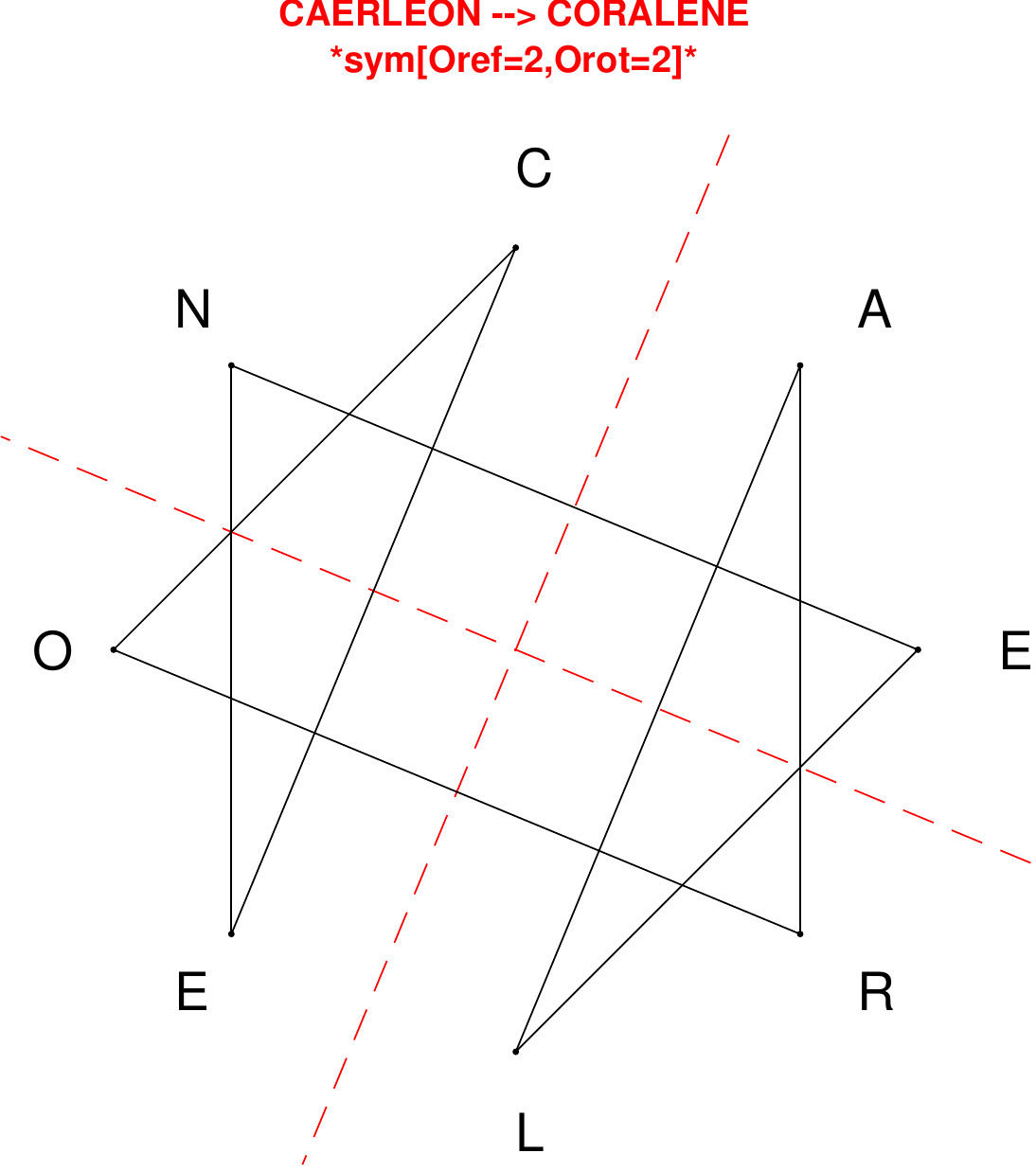}
\end{subfigure}
\hfill
\begin{subfigure}[T]{0.19\textwidth}
\centering
\includegraphics[width=\textwidth]{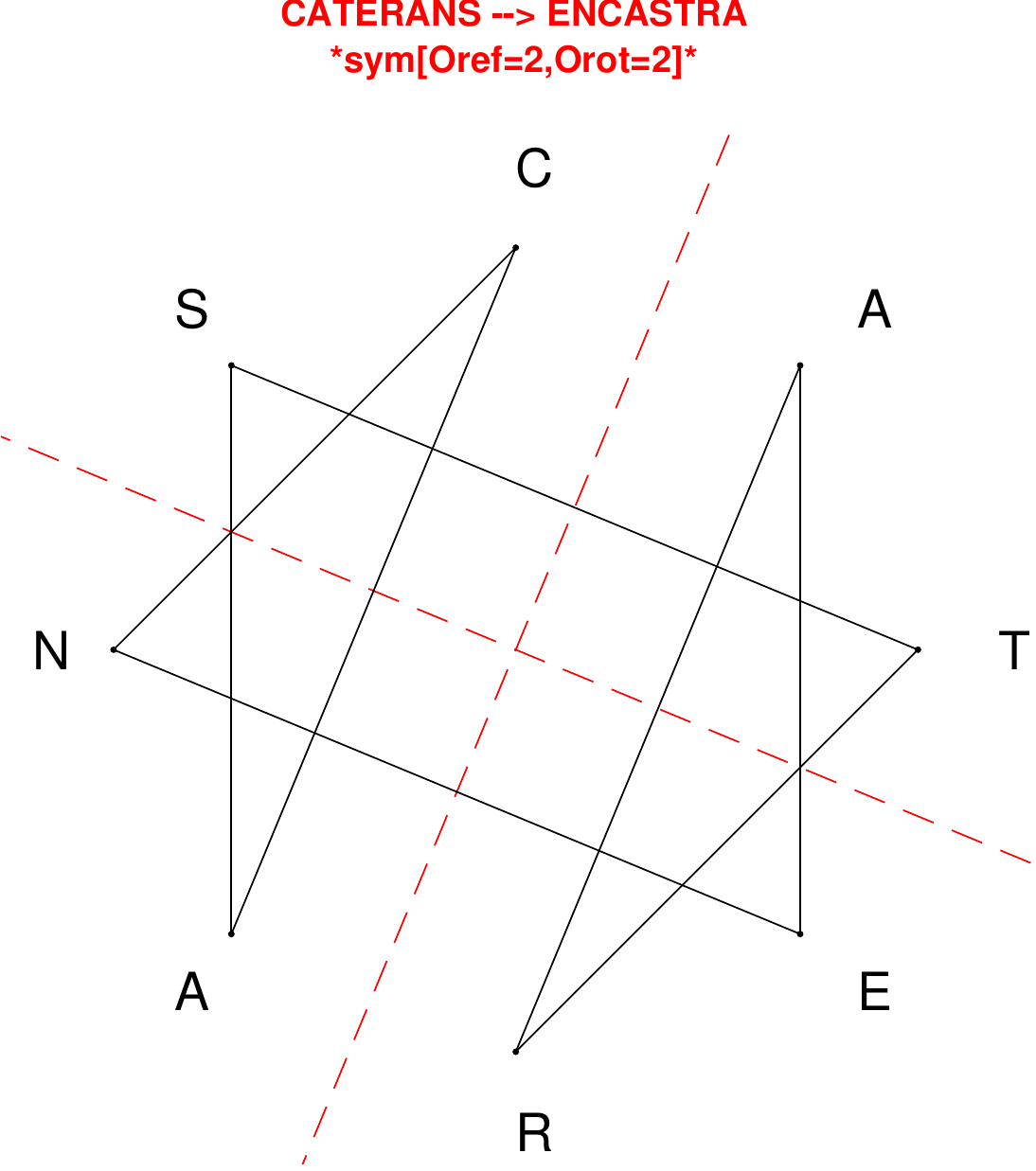}
\end{subfigure}
\hfill
\begin{subfigure}[T]{0.19\textwidth}
\centering
\includegraphics[width=\textwidth]{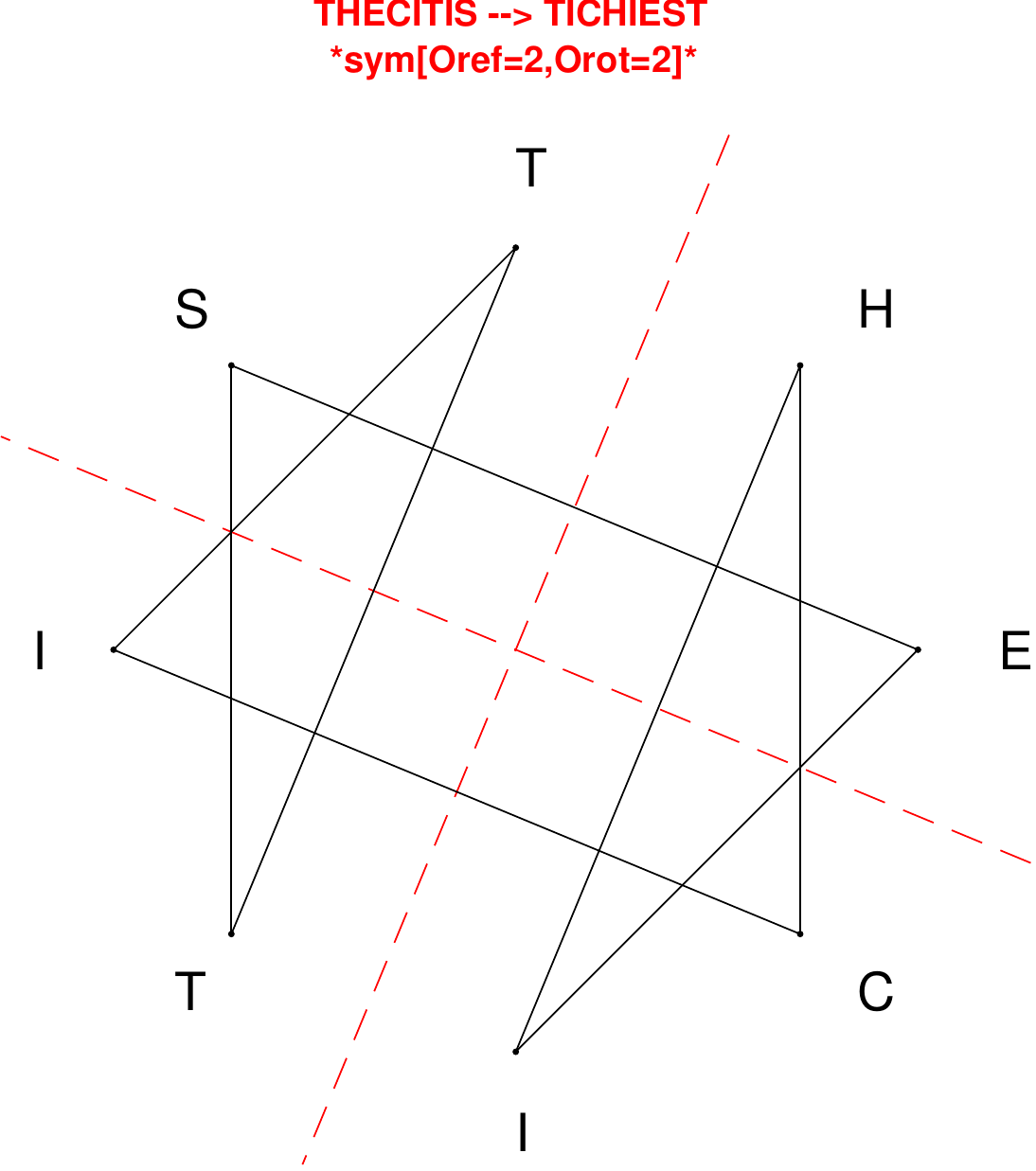}
\end{subfigure}
\hfill
\begin{subfigure}[T]{0.19\textwidth}
\centering
\includegraphics[width=\textwidth]{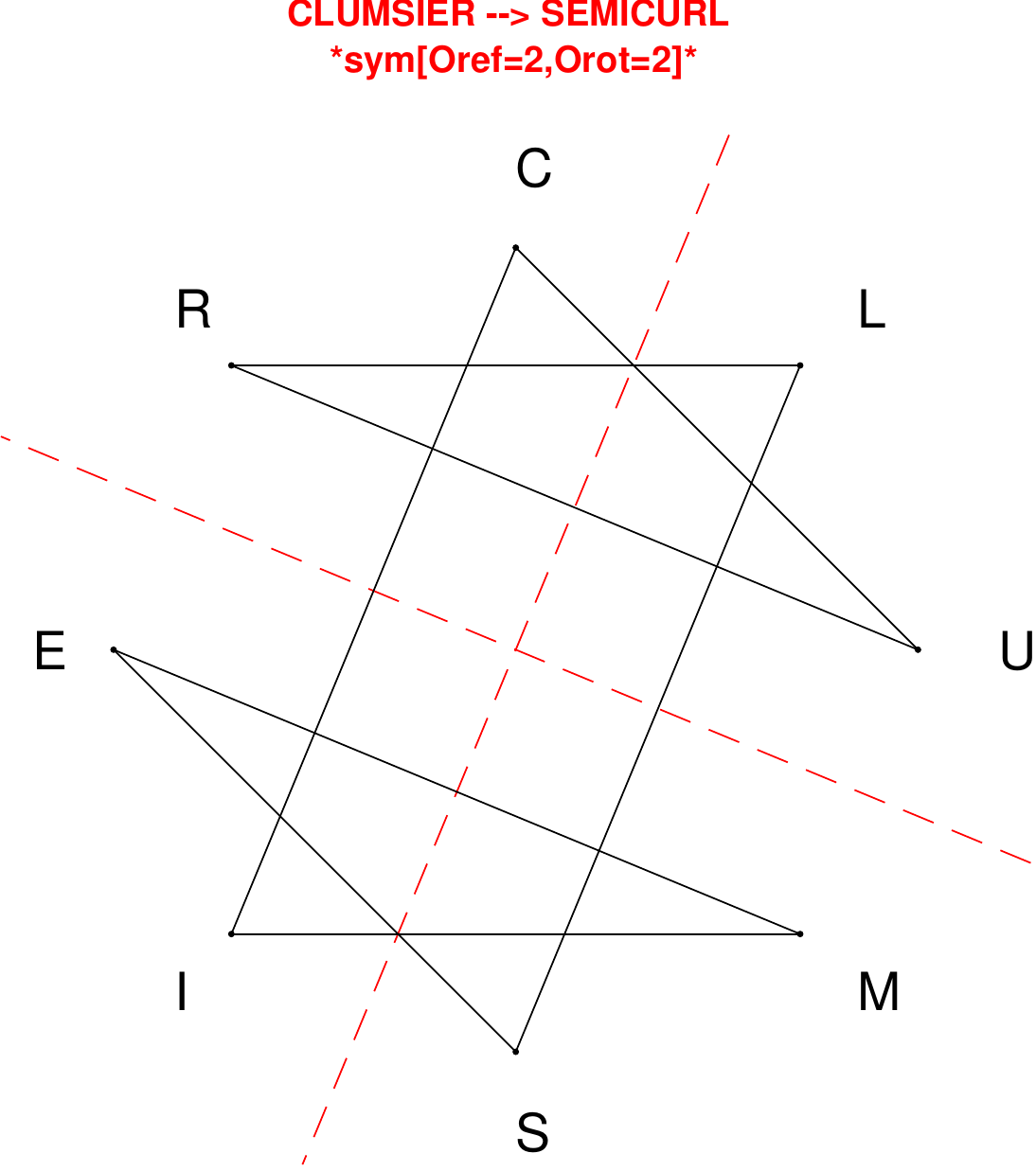}
\end{subfigure}
\end{figure}

\begin{figure}[H]
\centering
\begin{subfigure}[T]{0.19\textwidth}
\centering
\includegraphics[width=\textwidth]{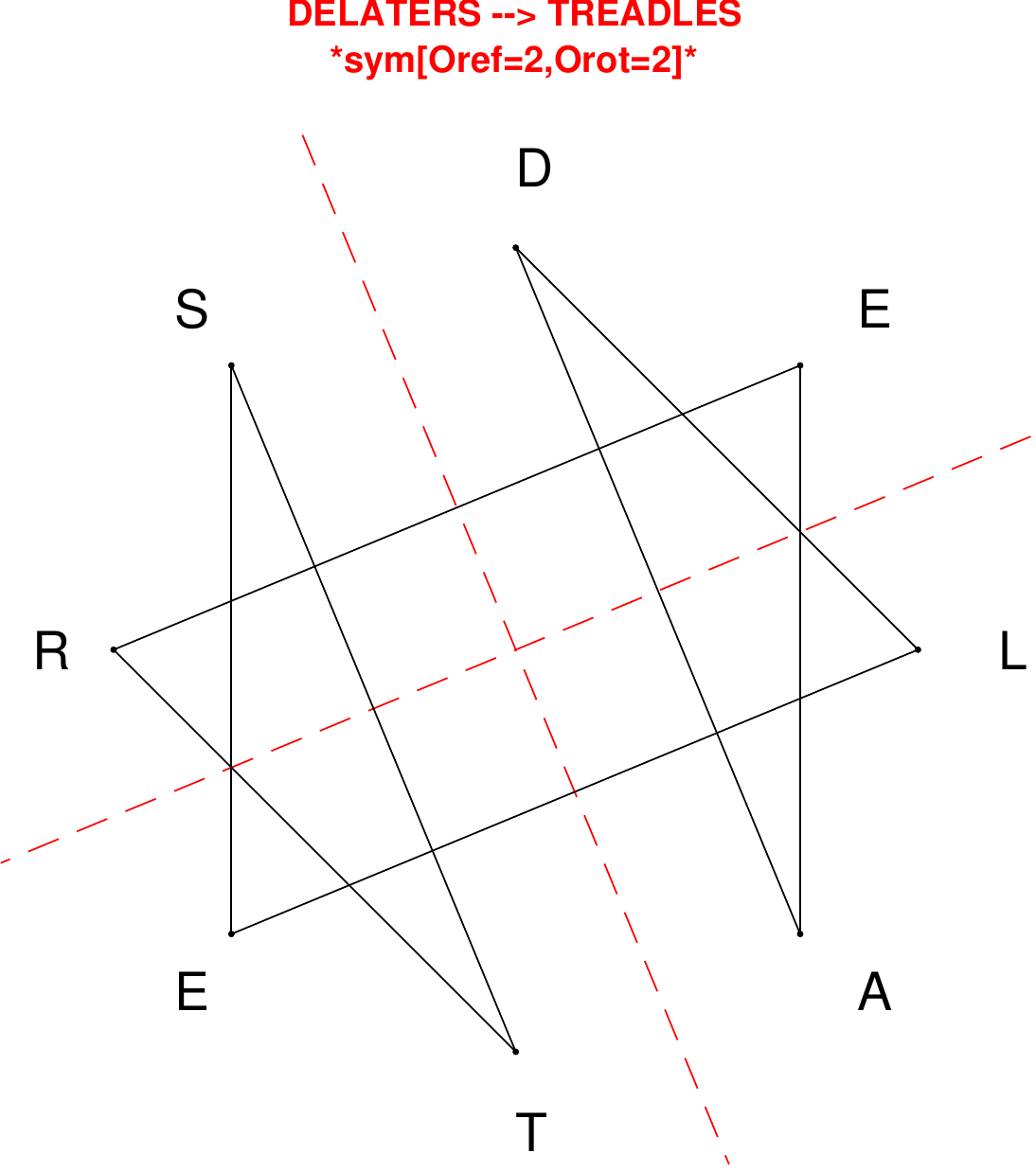}
\end{subfigure}
\hfill
\begin{subfigure}[T]{0.19\textwidth}
\centering
\includegraphics[width=\textwidth]{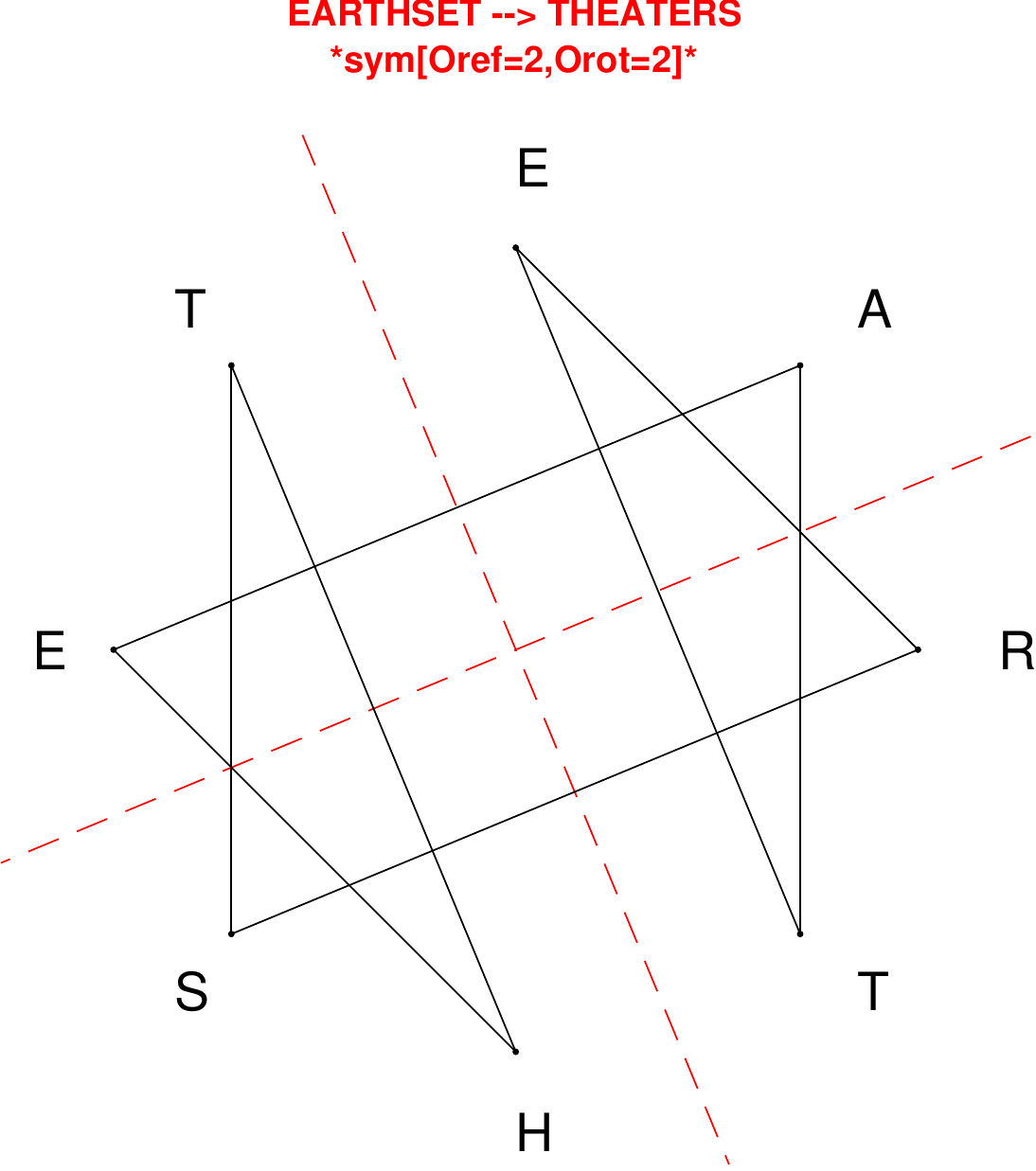}
\end{subfigure}
\hfill
\begin{subfigure}[T]{0.19\textwidth}
\centering
\includegraphics[width=\textwidth]{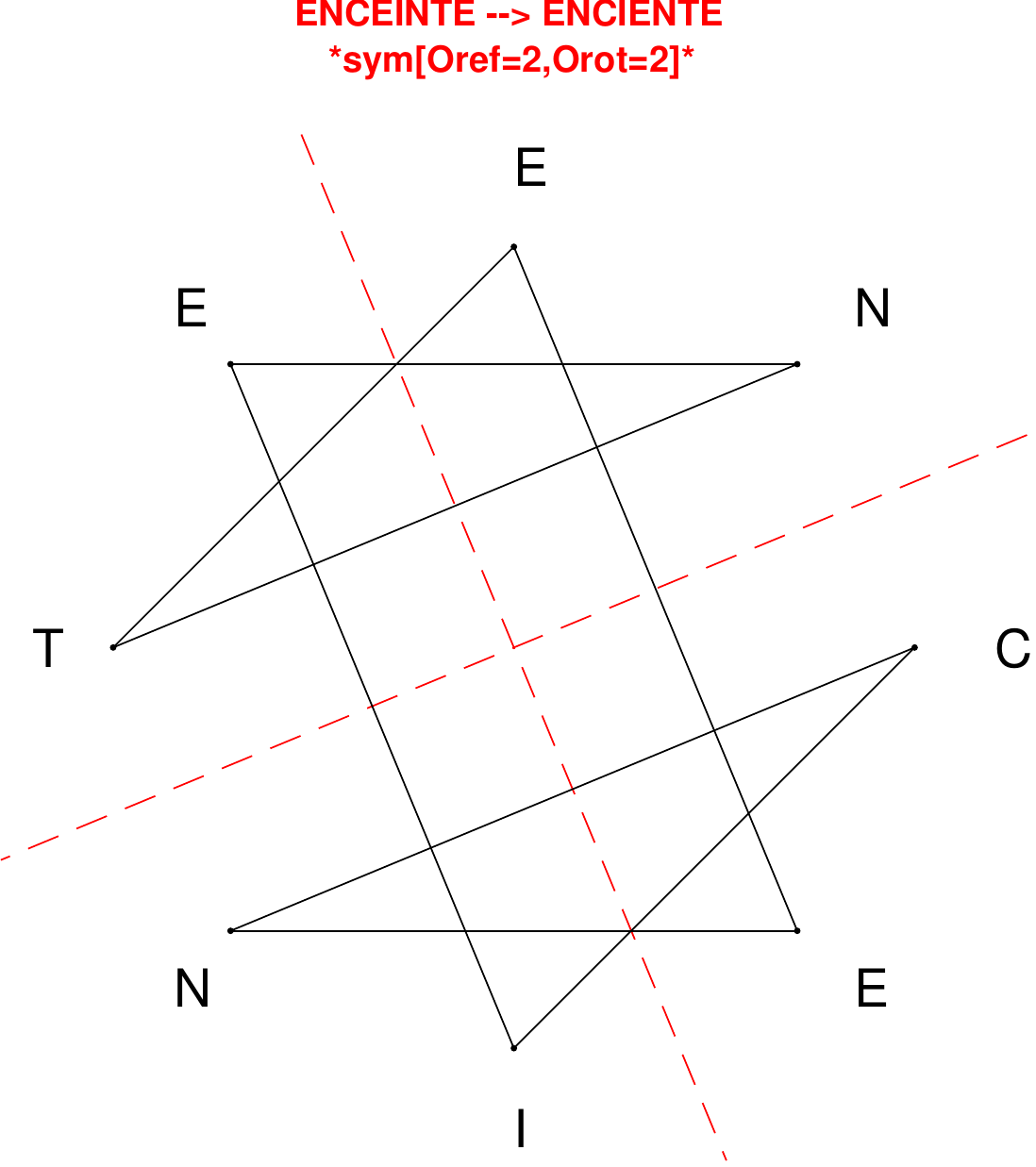}
\end{subfigure}
\hfill
\begin{subfigure}[T]{0.19\textwidth}
\centering
\includegraphics[width=\textwidth]{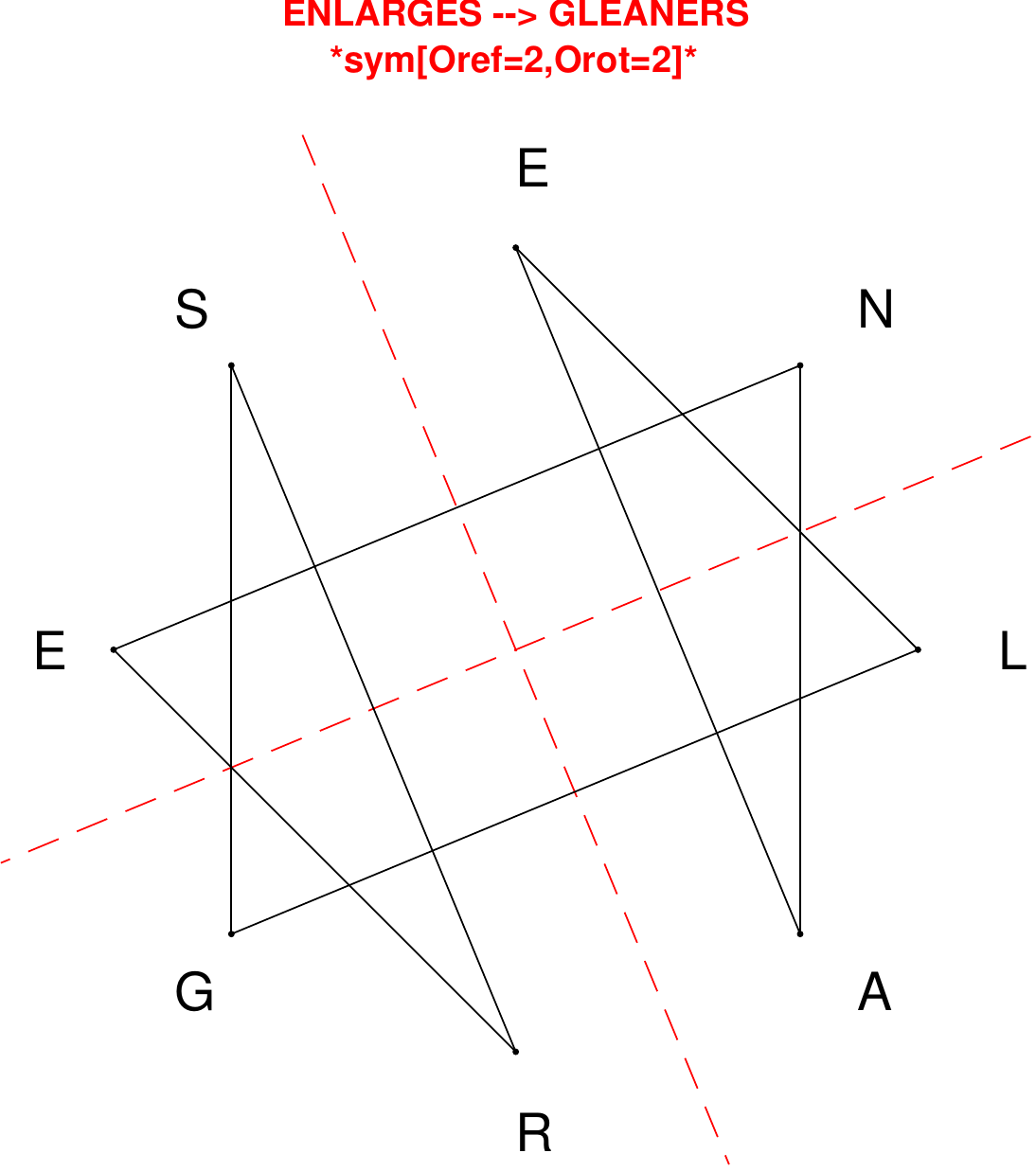}
\end{subfigure}
\hfill
\begin{subfigure}[T]{0.19\textwidth}
\centering
\includegraphics[width=\textwidth]{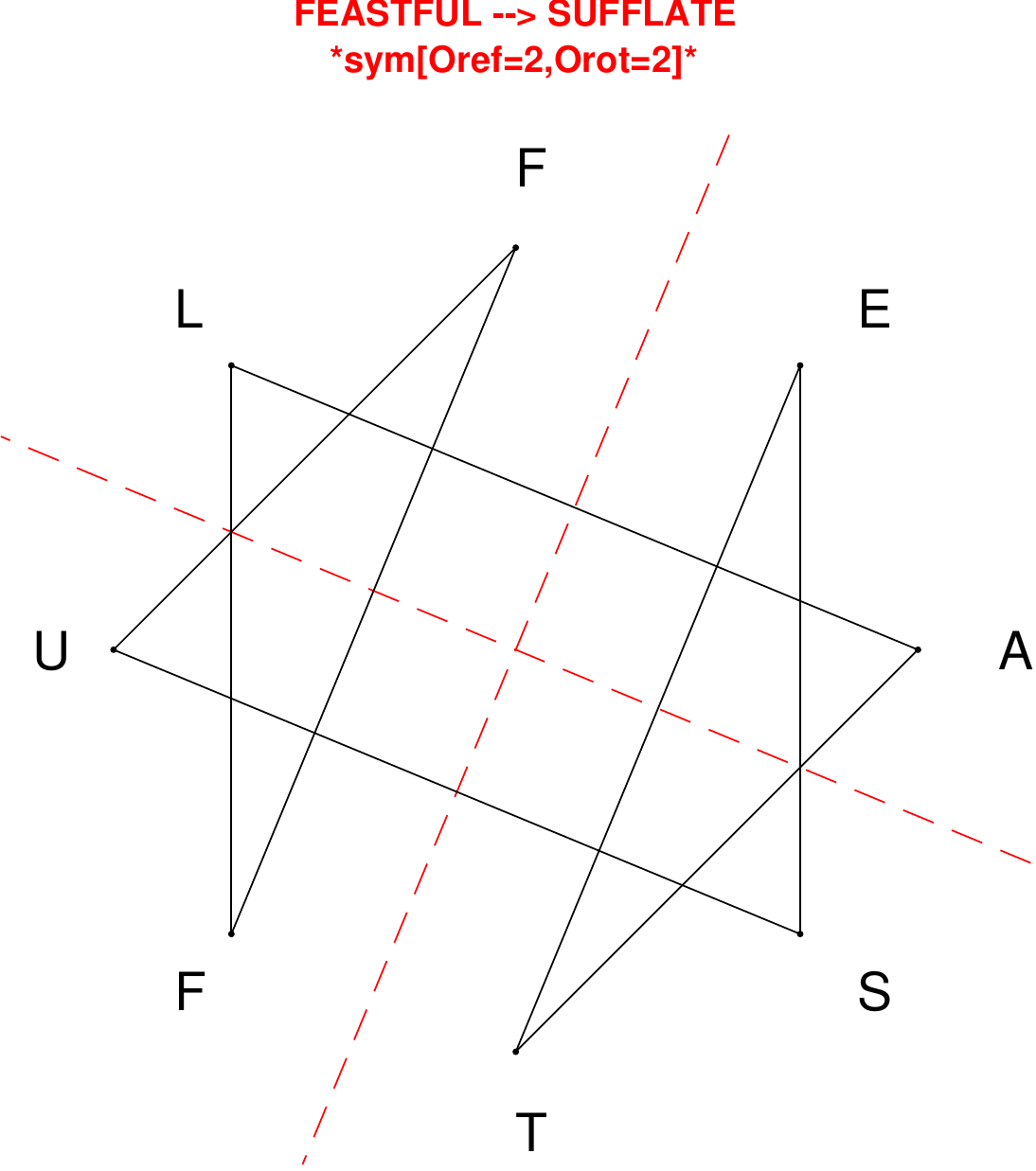}
\end{subfigure}
\end{figure}

\begin{figure}[H]
\centering
\begin{subfigure}[T]{0.19\textwidth}
\centering
\includegraphics[width=\textwidth]{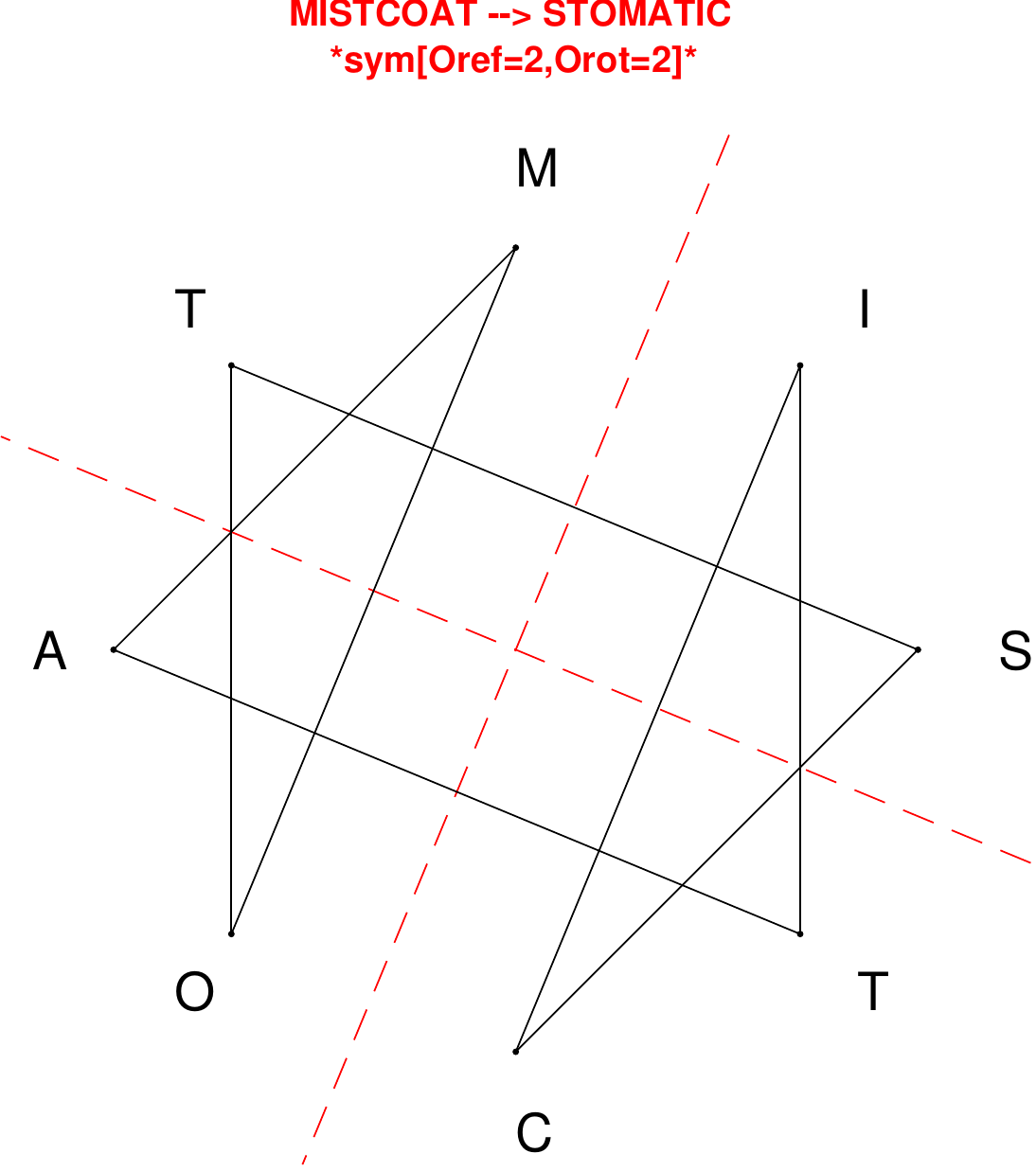}
\end{subfigure}
\hfill
\begin{subfigure}[T]{0.19\textwidth}
\centering
\includegraphics[width=\textwidth]{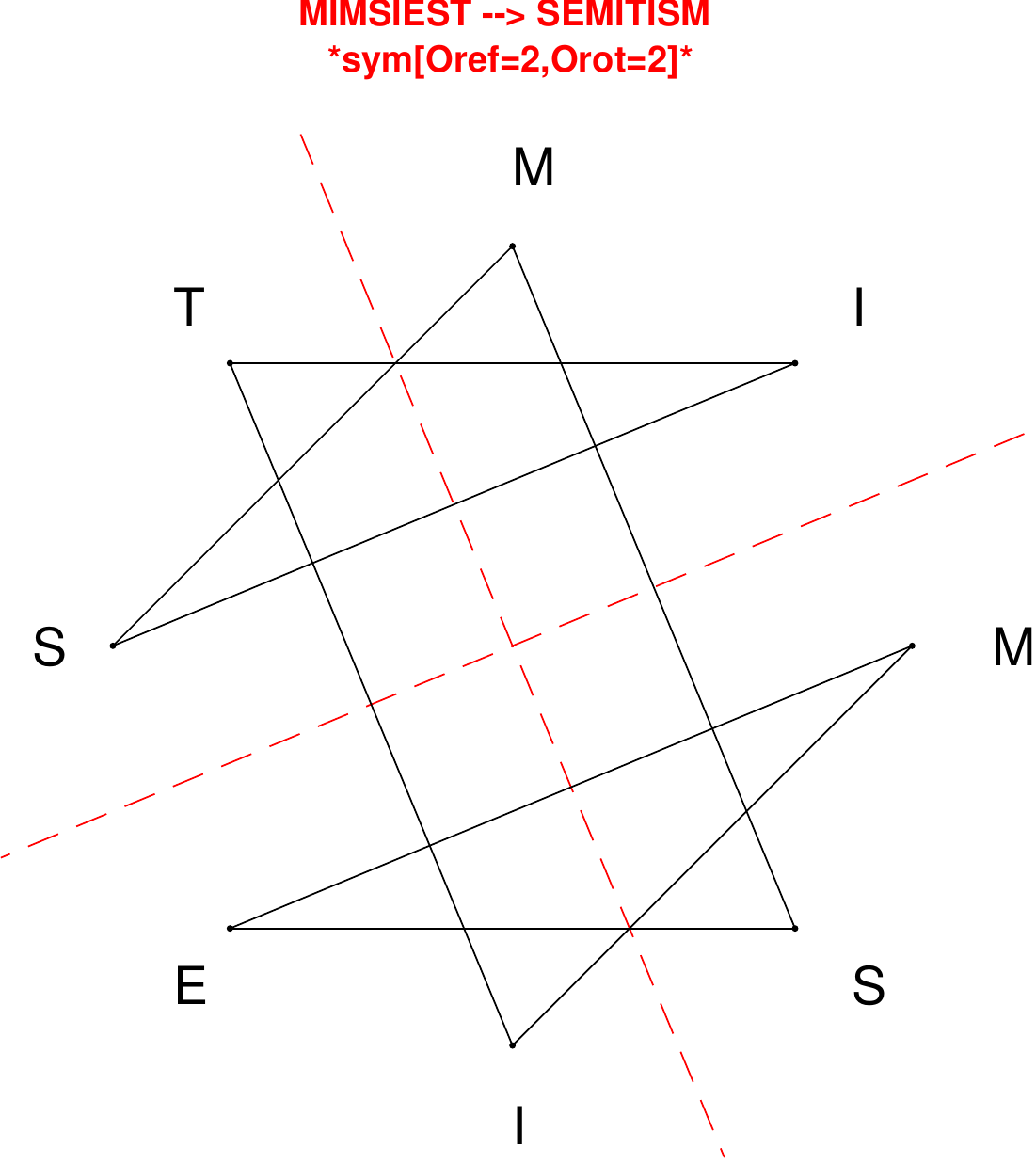}
\end{subfigure}
\hfill
\begin{subfigure}[T]{0.19\textwidth}
\centering
\includegraphics[width=\textwidth]{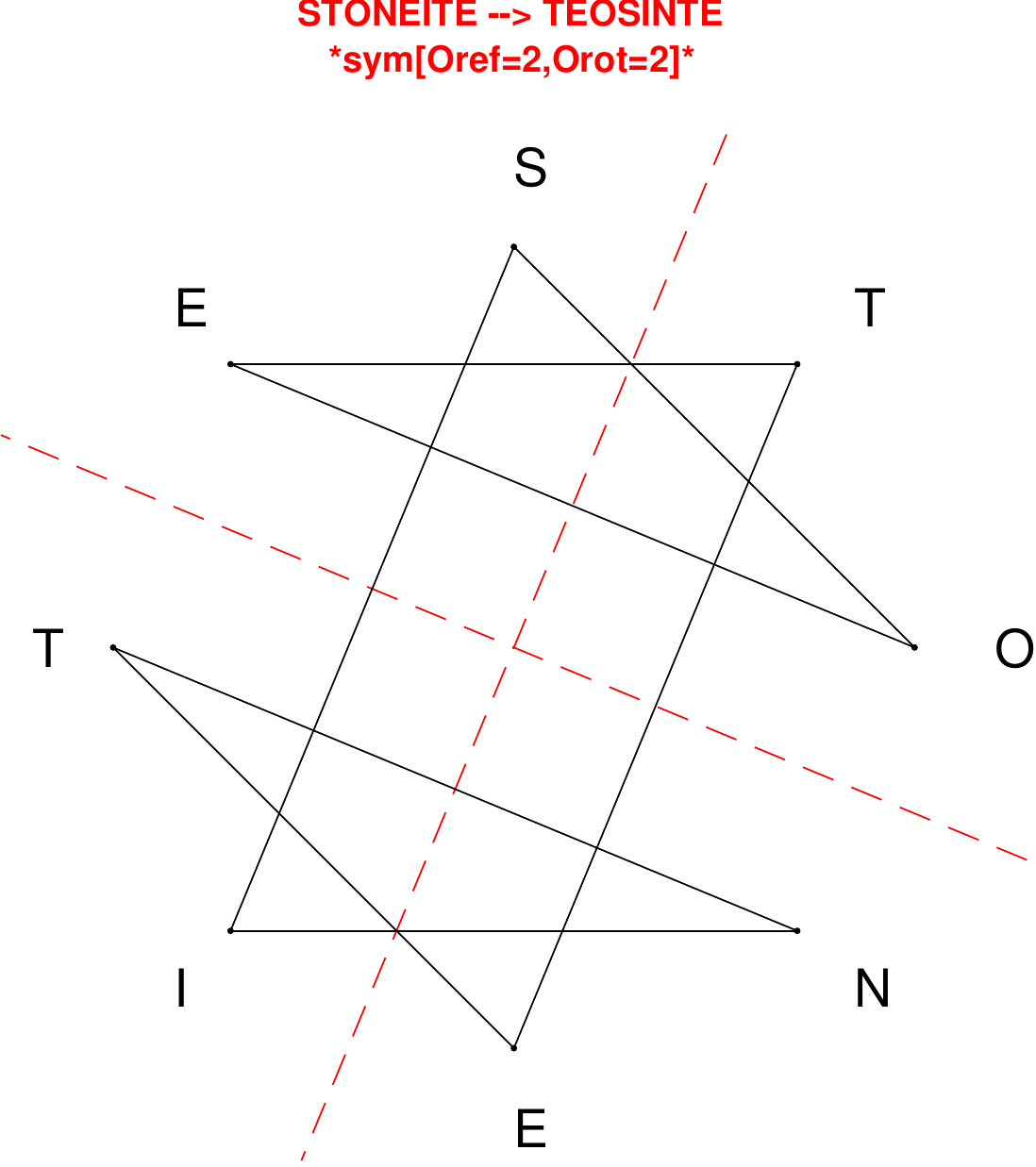}
\end{subfigure}
\hfill
\begin{subfigure}[T]{0.19\textwidth}
\centering
\includegraphics[width=\textwidth]{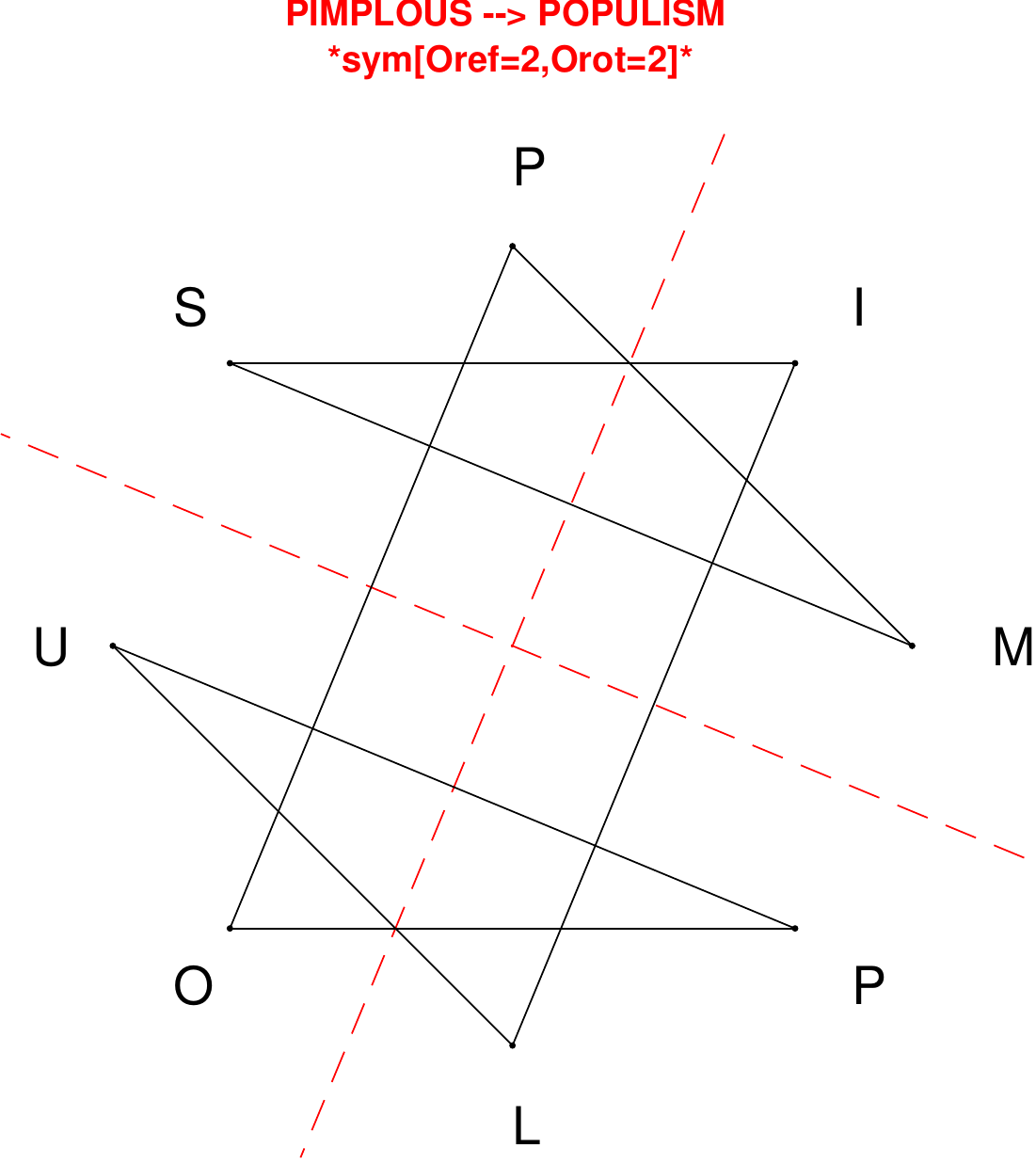}
\end{subfigure}
\hfill
\begin{subfigure}[T]{0.19\textwidth}
\centering
\includegraphics[width=\textwidth]{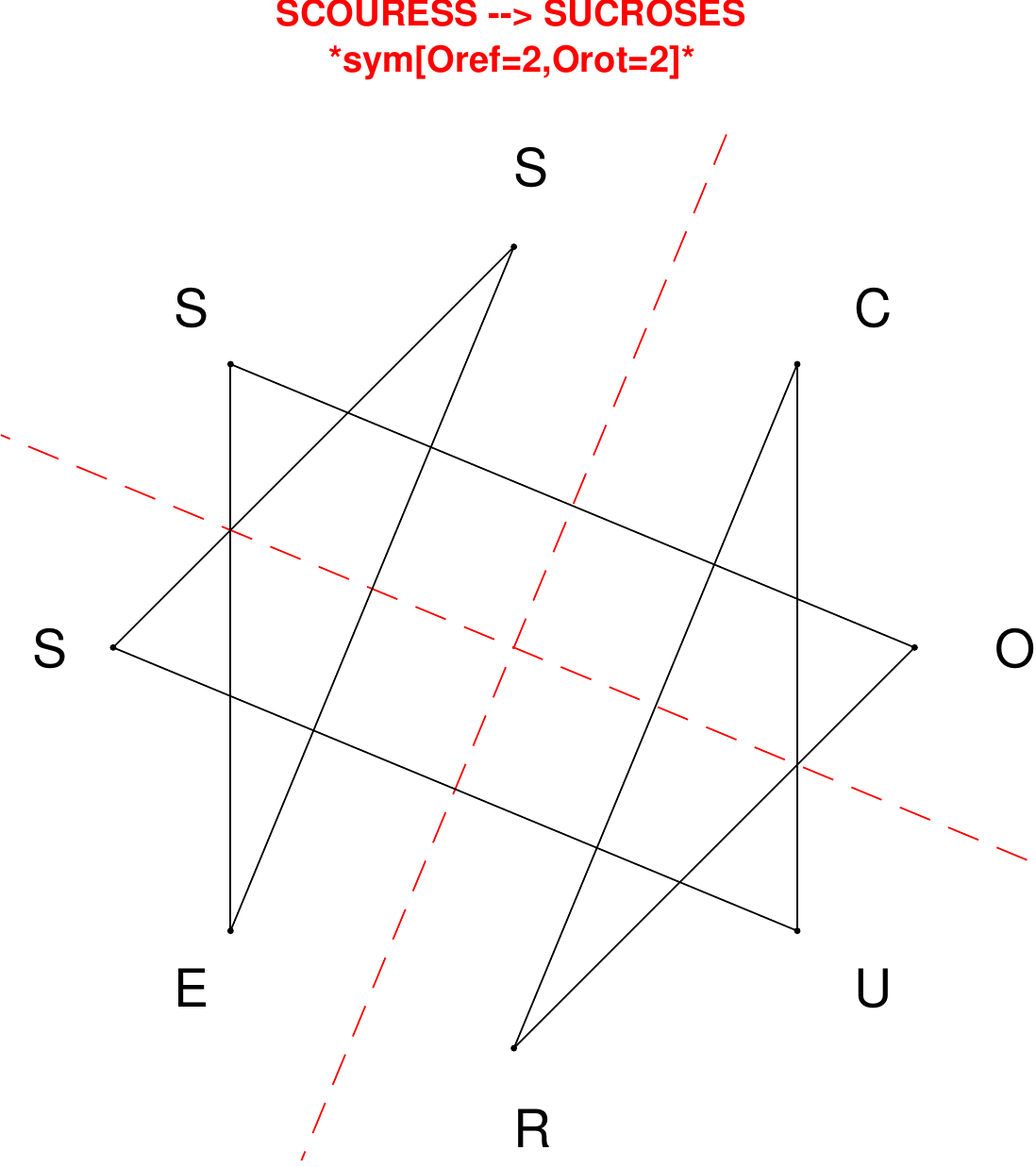}
\end{subfigure}
\end{figure}

\begin{figure}[H]
\centering
\begin{subfigure}[T]{0.19\textwidth}
\centering
\includegraphics[width=\textwidth]{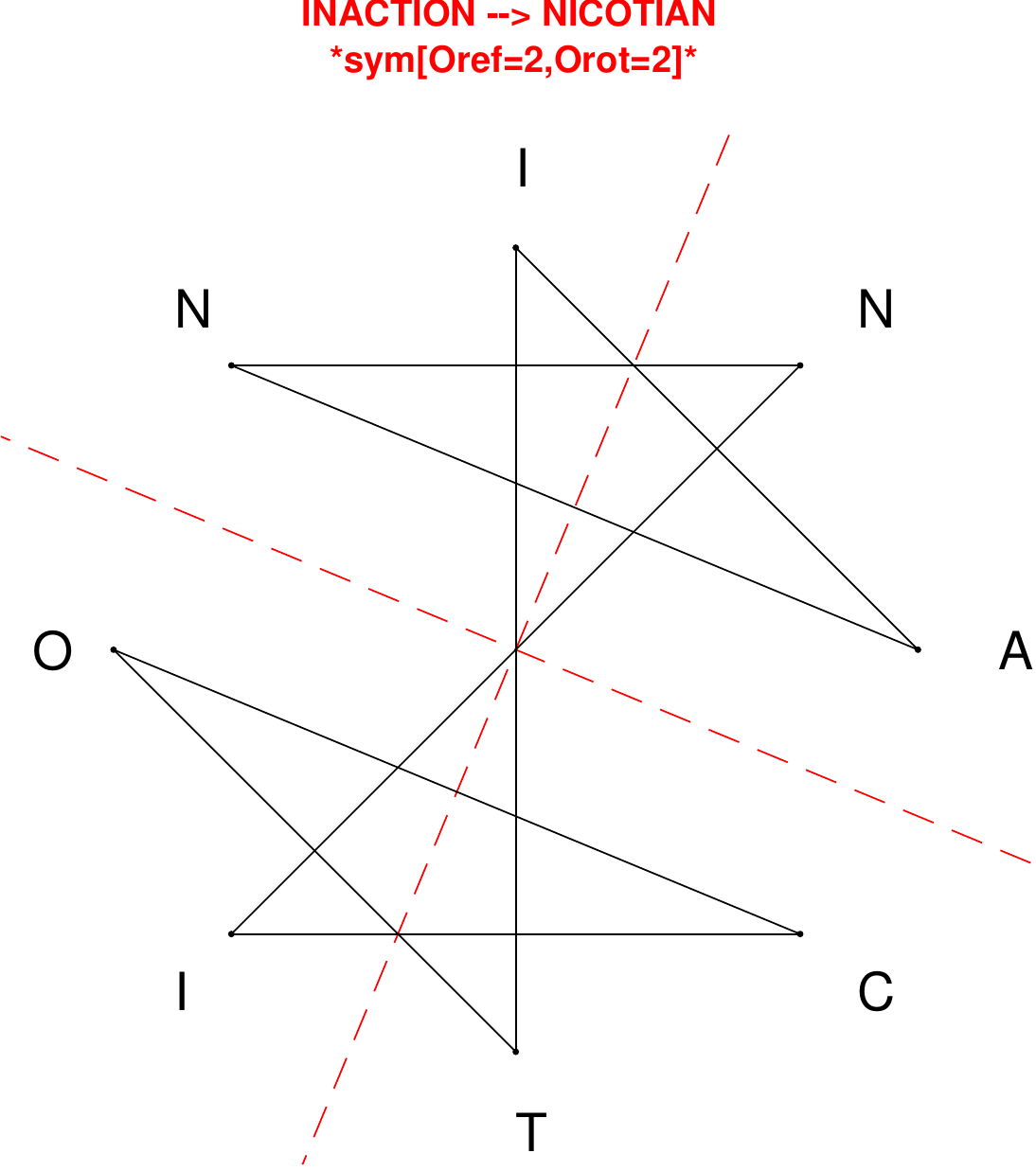}
\end{subfigure}
\hfill
\begin{subfigure}[T]{0.19\textwidth}
\centering
\includegraphics[width=\textwidth]{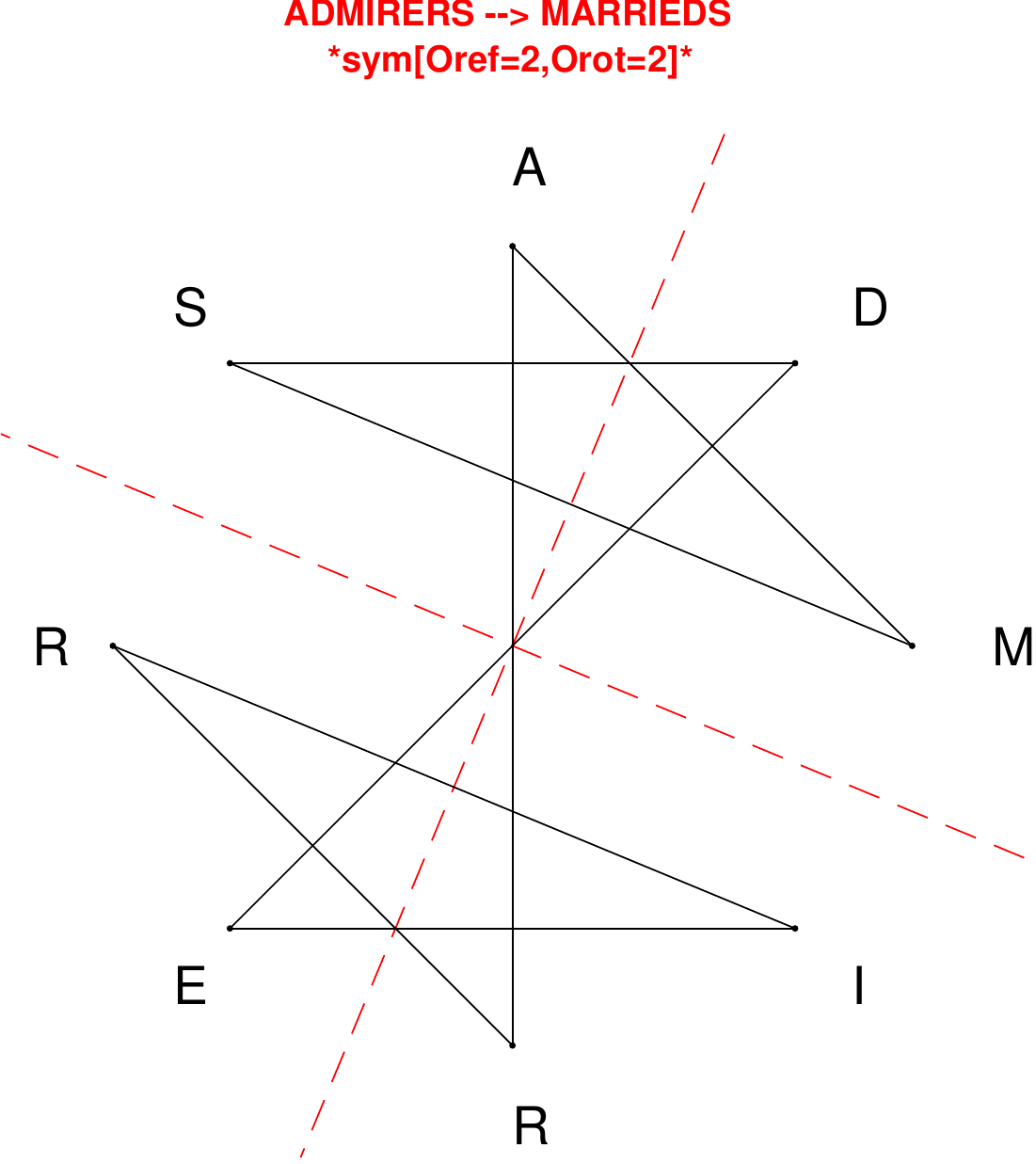}
\end{subfigure}
\hfill
\begin{subfigure}[T]{0.19\textwidth}
\centering
\includegraphics[width=\textwidth]{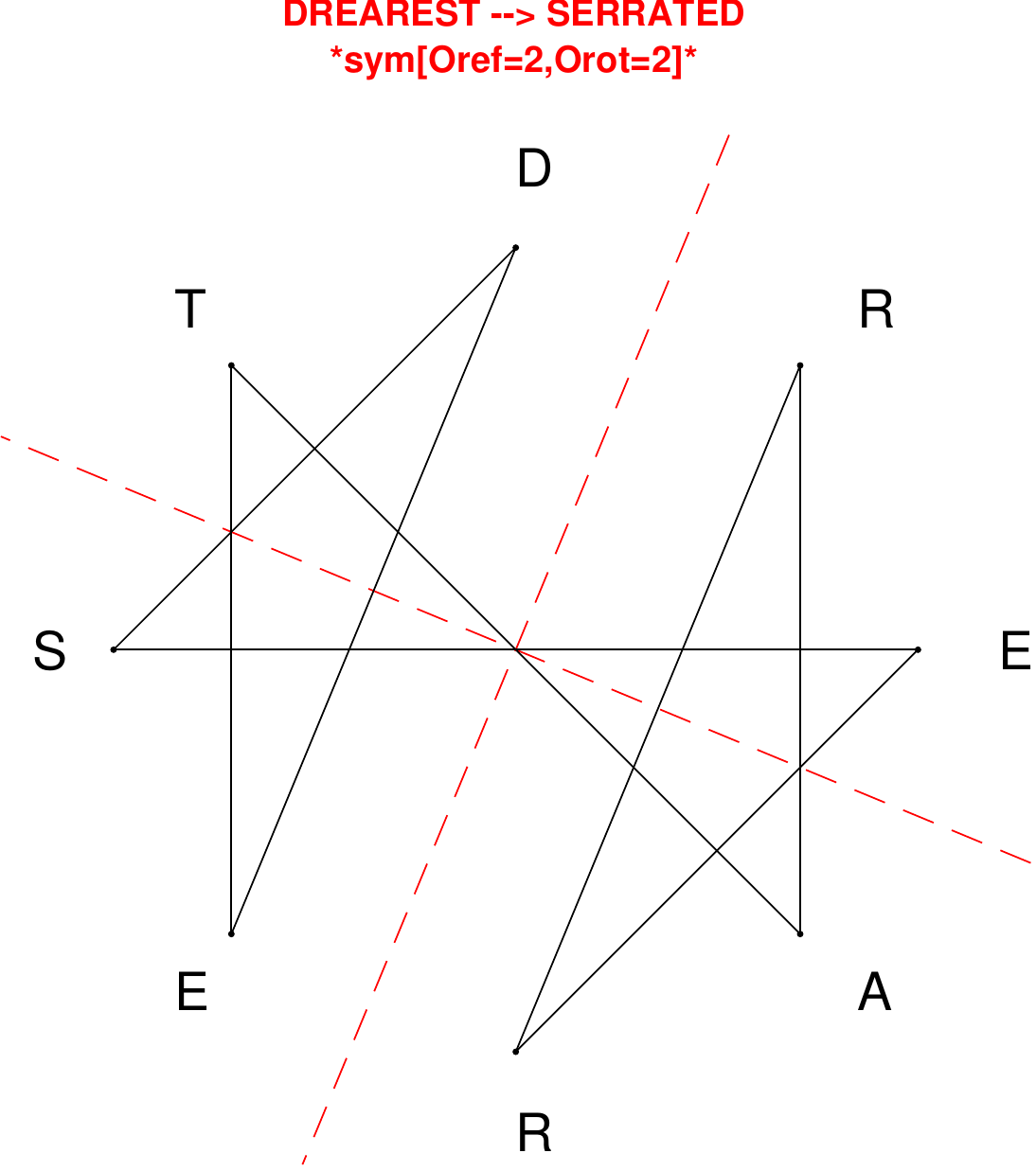}
\end{subfigure}
\hfill
\begin{subfigure}[T]{0.19\textwidth}
\centering
\includegraphics[width=\textwidth]{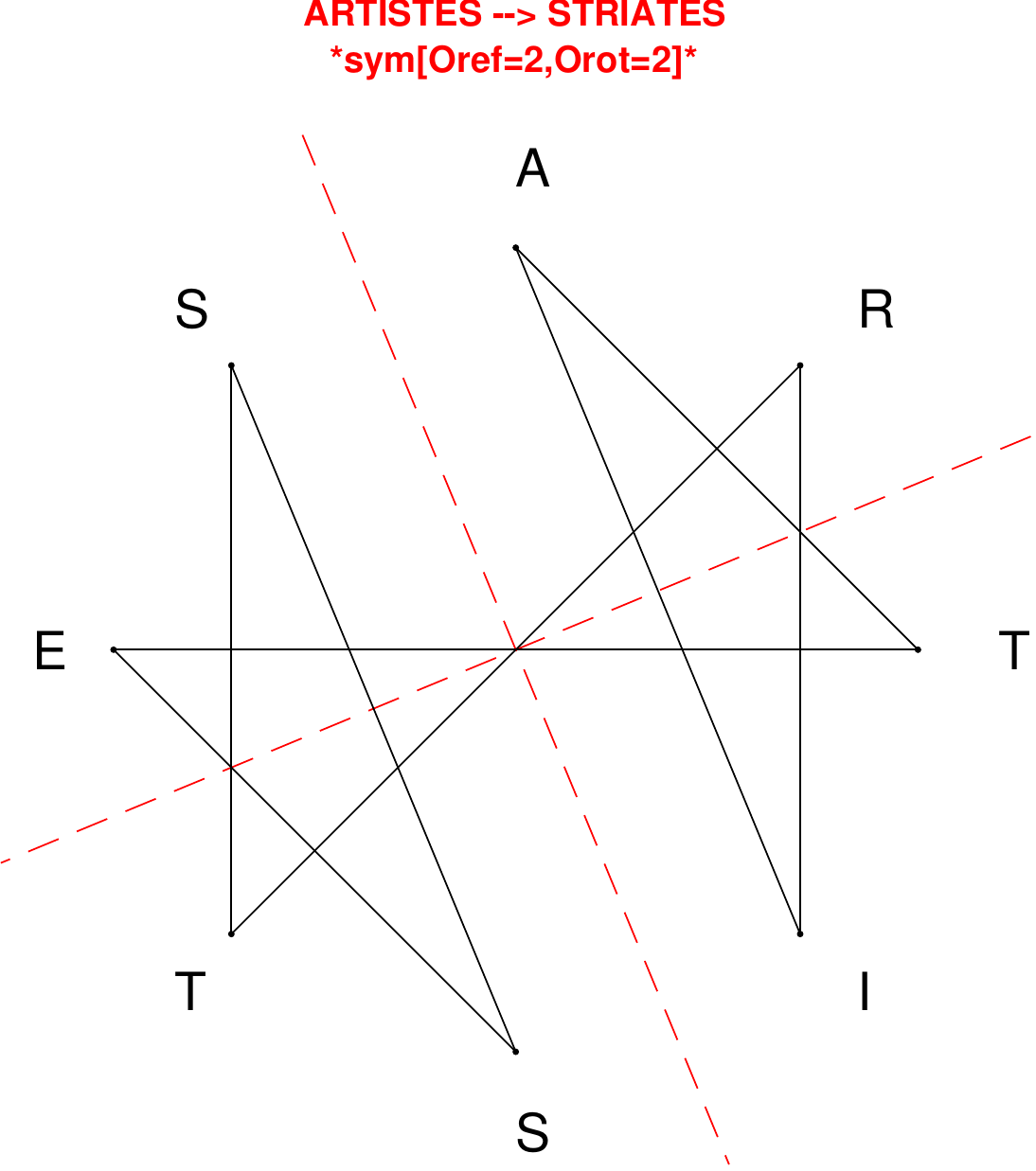}
\end{subfigure}
\hfill
\begin{subfigure}[T]{0.19\textwidth}
\centering
\includegraphics[width=\textwidth]{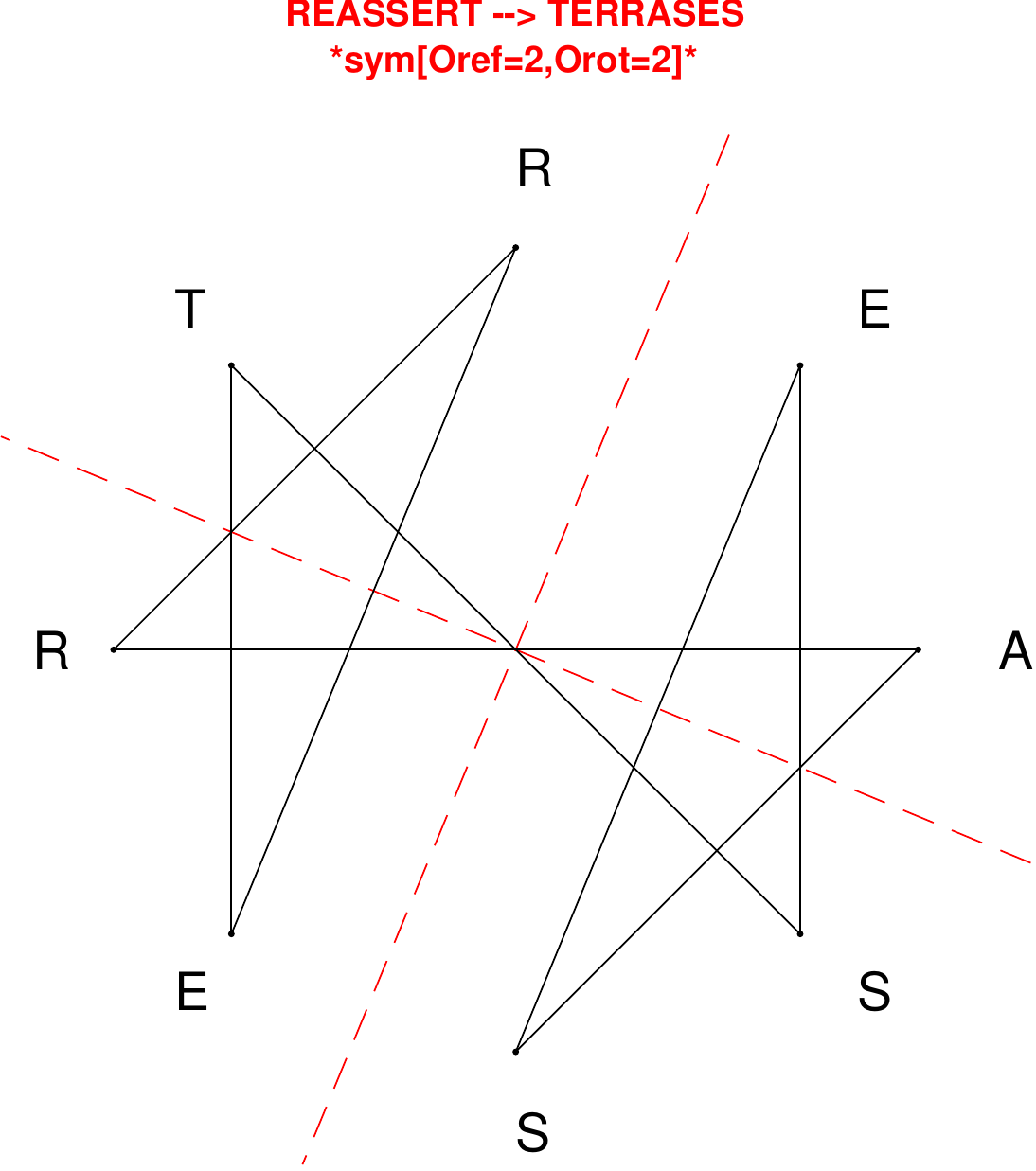}
\end{subfigure}
\end{figure}

\begin{figure}[H]
\centering
\begin{subfigure}[T]{0.19\textwidth}
\centering
\includegraphics[width=\textwidth]{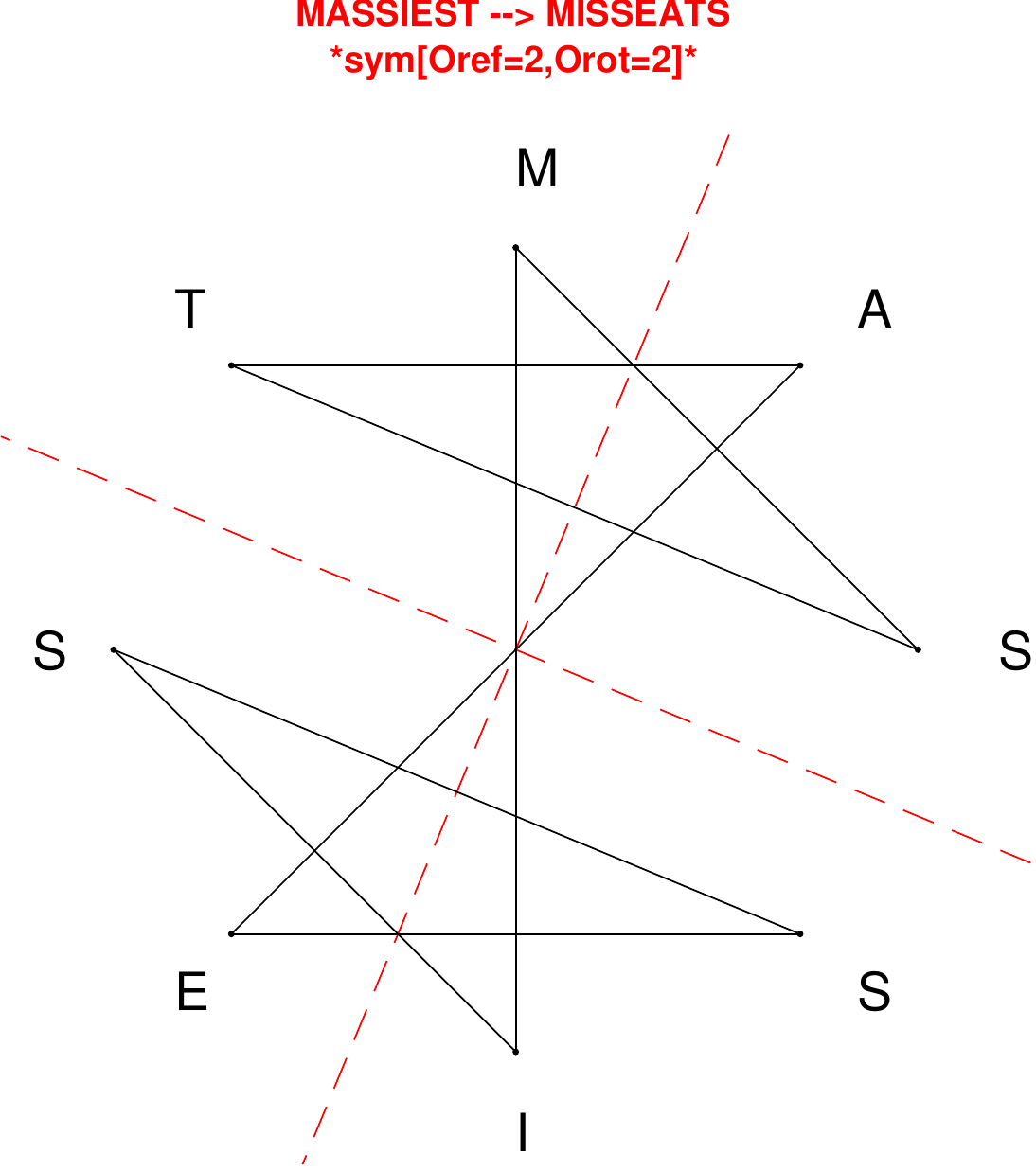}
\end{subfigure}
\hfill
\begin{subfigure}[T]{0.19\textwidth}
\centering
\includegraphics[width=\textwidth]{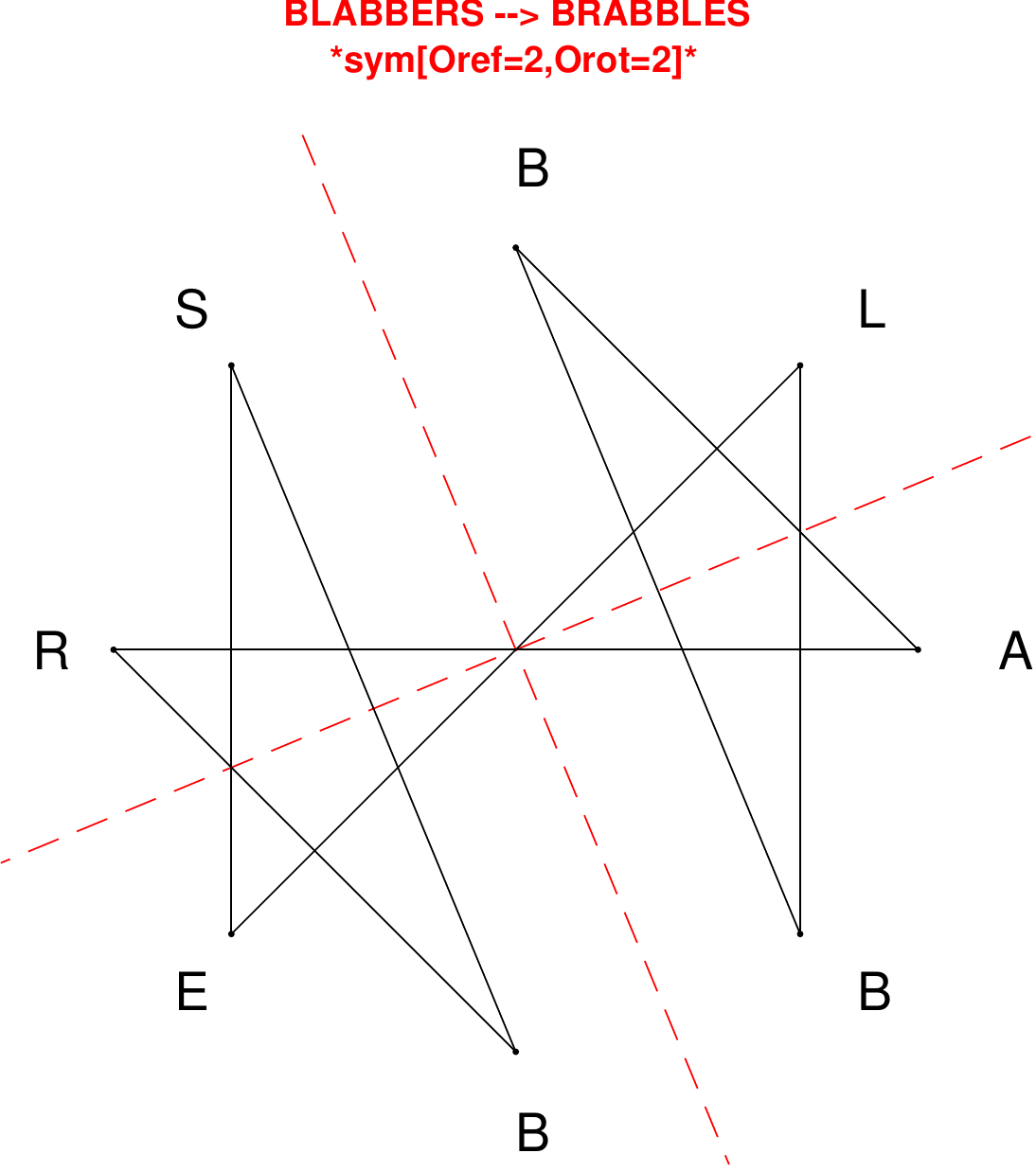}
\end{subfigure}
\hfill
\begin{subfigure}[T]{0.19\textwidth}
\centering
\includegraphics[width=\textwidth]{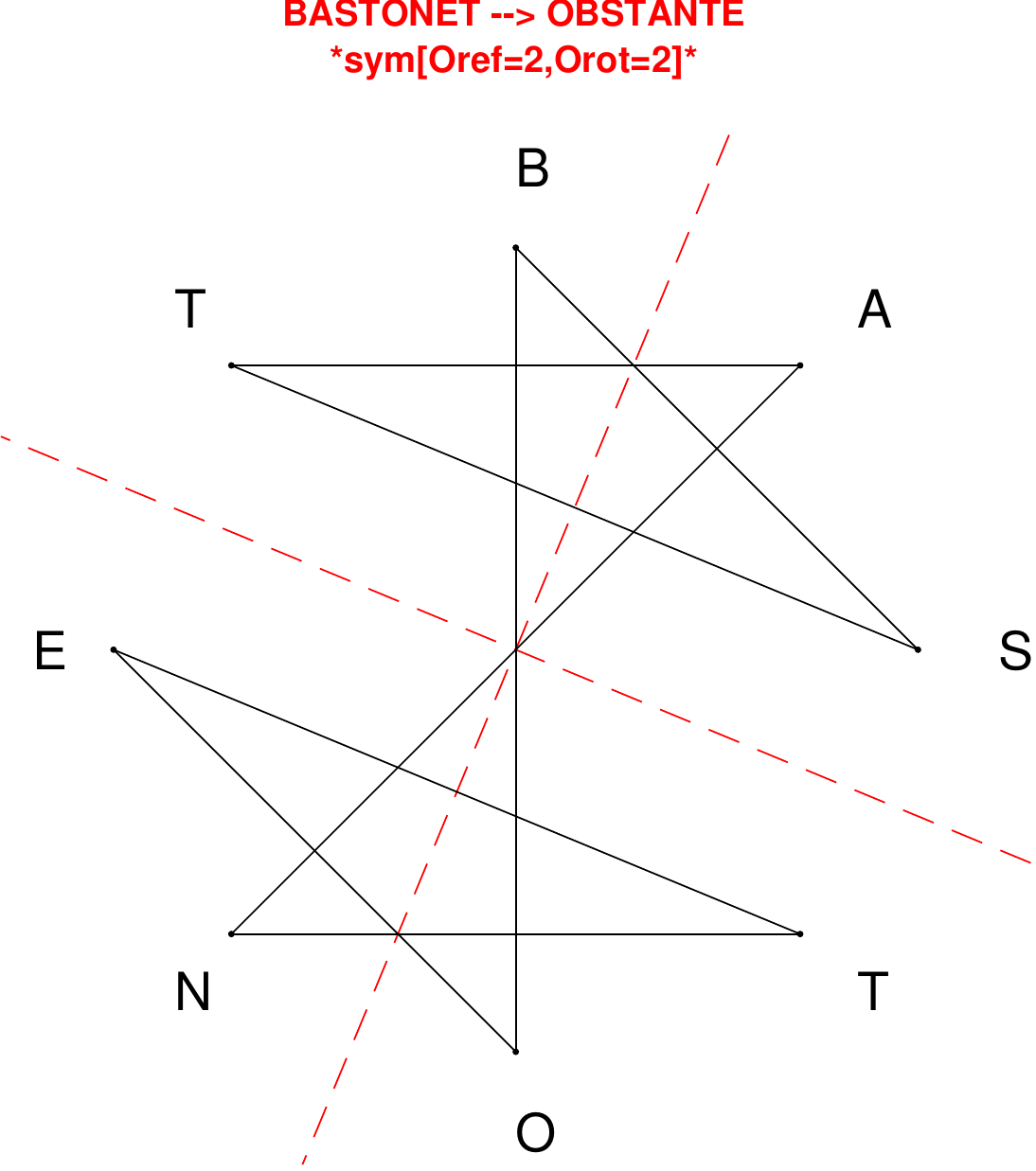}
\end{subfigure}
\hfill
\begin{subfigure}[T]{0.19\textwidth}
\centering
\includegraphics[width=\textwidth]{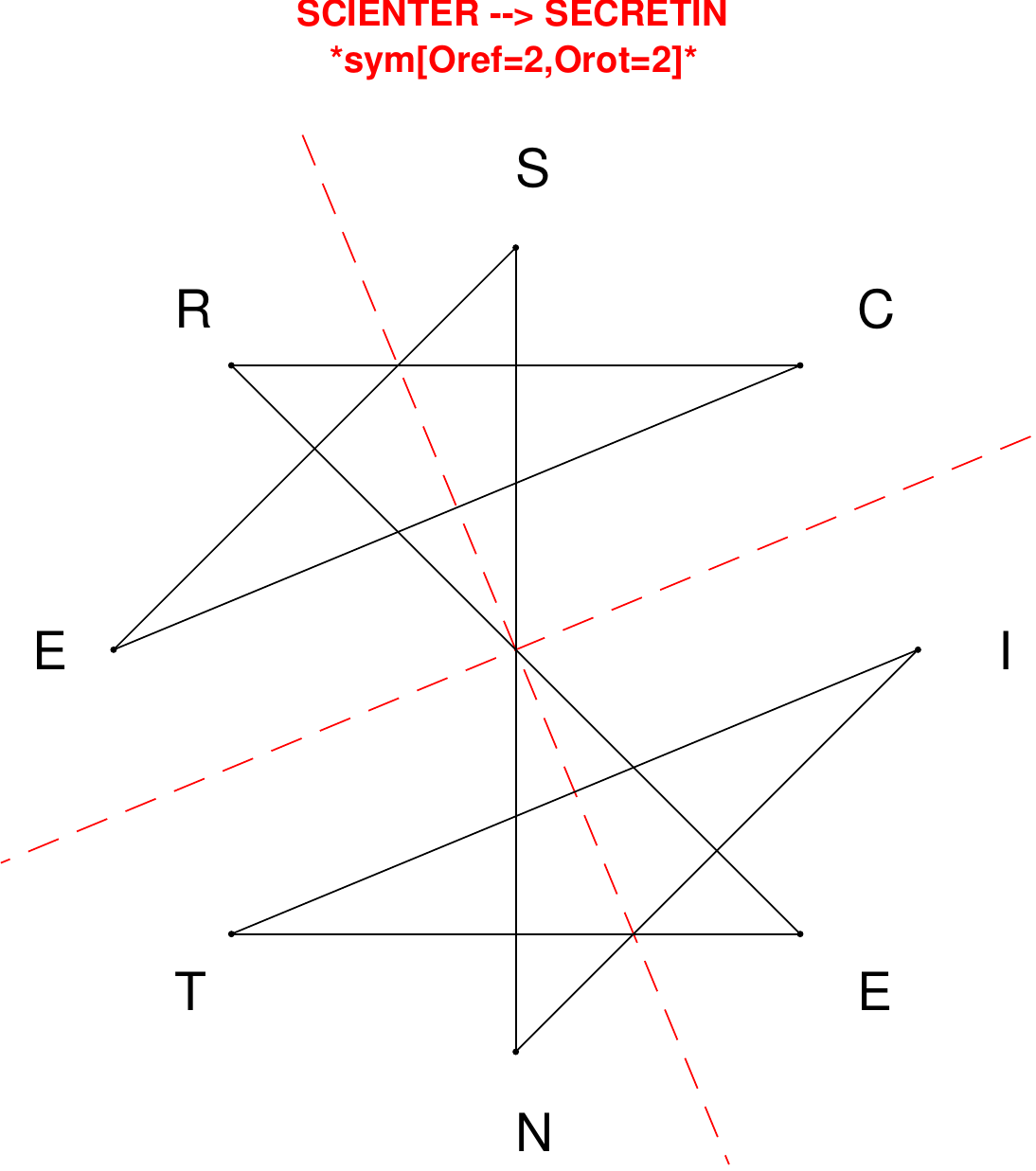}
\end{subfigure}
\hfill
\begin{subfigure}[T]{0.19\textwidth}
\centering
\includegraphics[width=\textwidth]{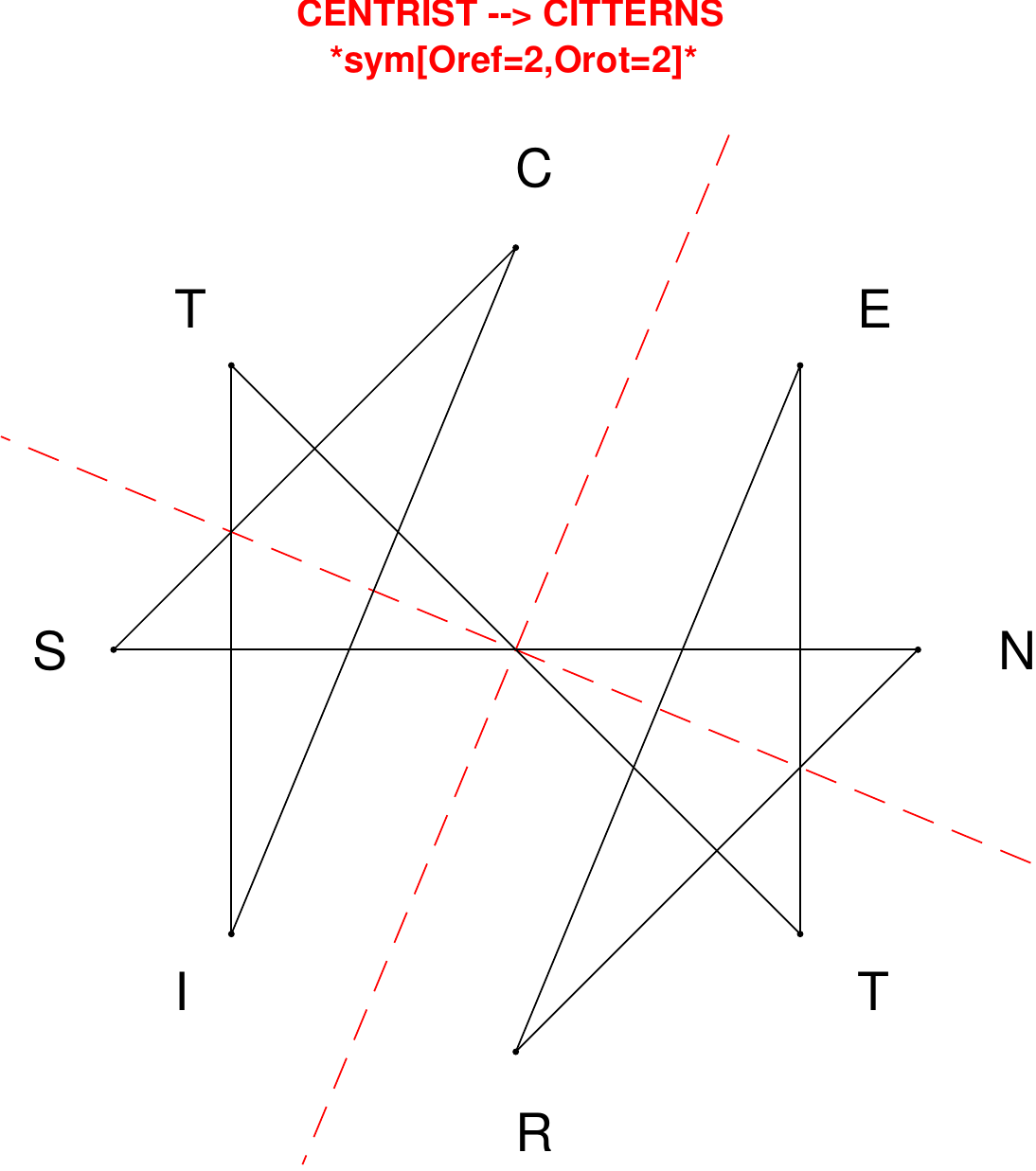}
\end{subfigure}
\end{figure}

\begin{figure}[H]
\centering
\begin{subfigure}[T]{0.19\textwidth}
\centering
\includegraphics[width=\textwidth]{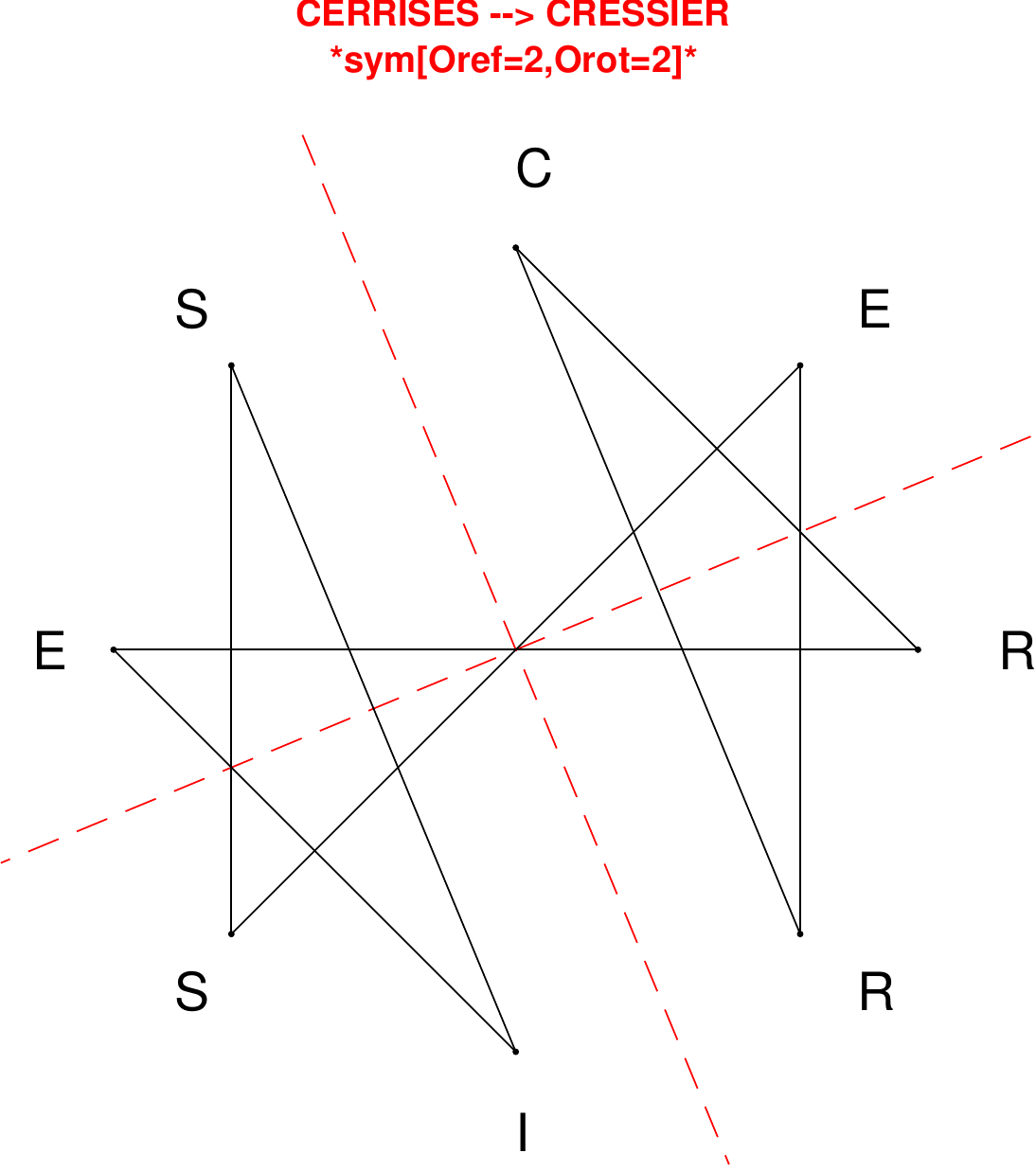}
\end{subfigure}
\hfill
\begin{subfigure}[T]{0.19\textwidth}
\centering
\includegraphics[width=\textwidth]{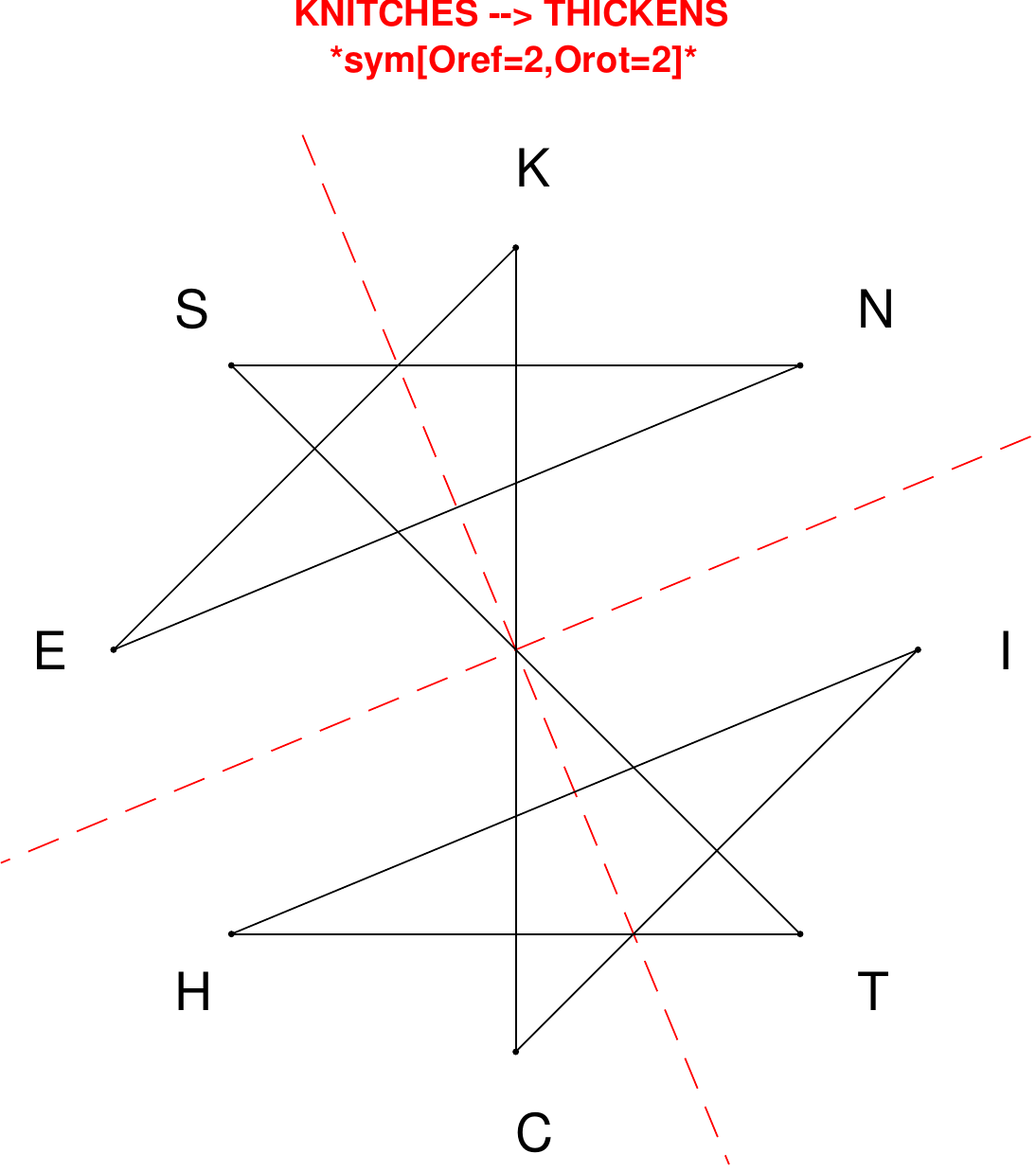}
\end{subfigure}
\hfill
\begin{subfigure}[T]{0.19\textwidth}
\centering
\includegraphics[width=\textwidth]{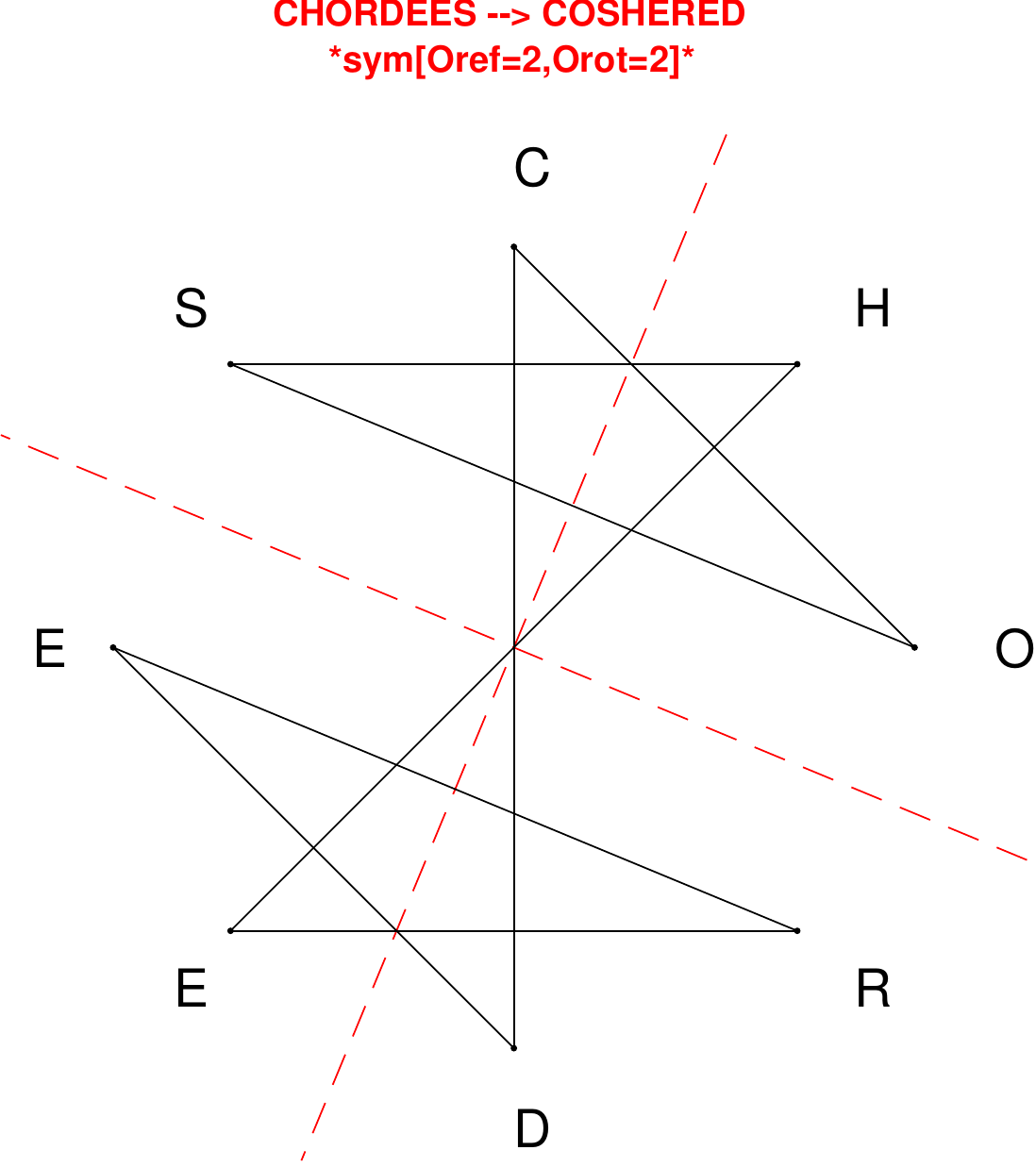}
\end{subfigure}
\hfill
\begin{subfigure}[T]{0.19\textwidth}
\centering
\includegraphics[width=\textwidth]{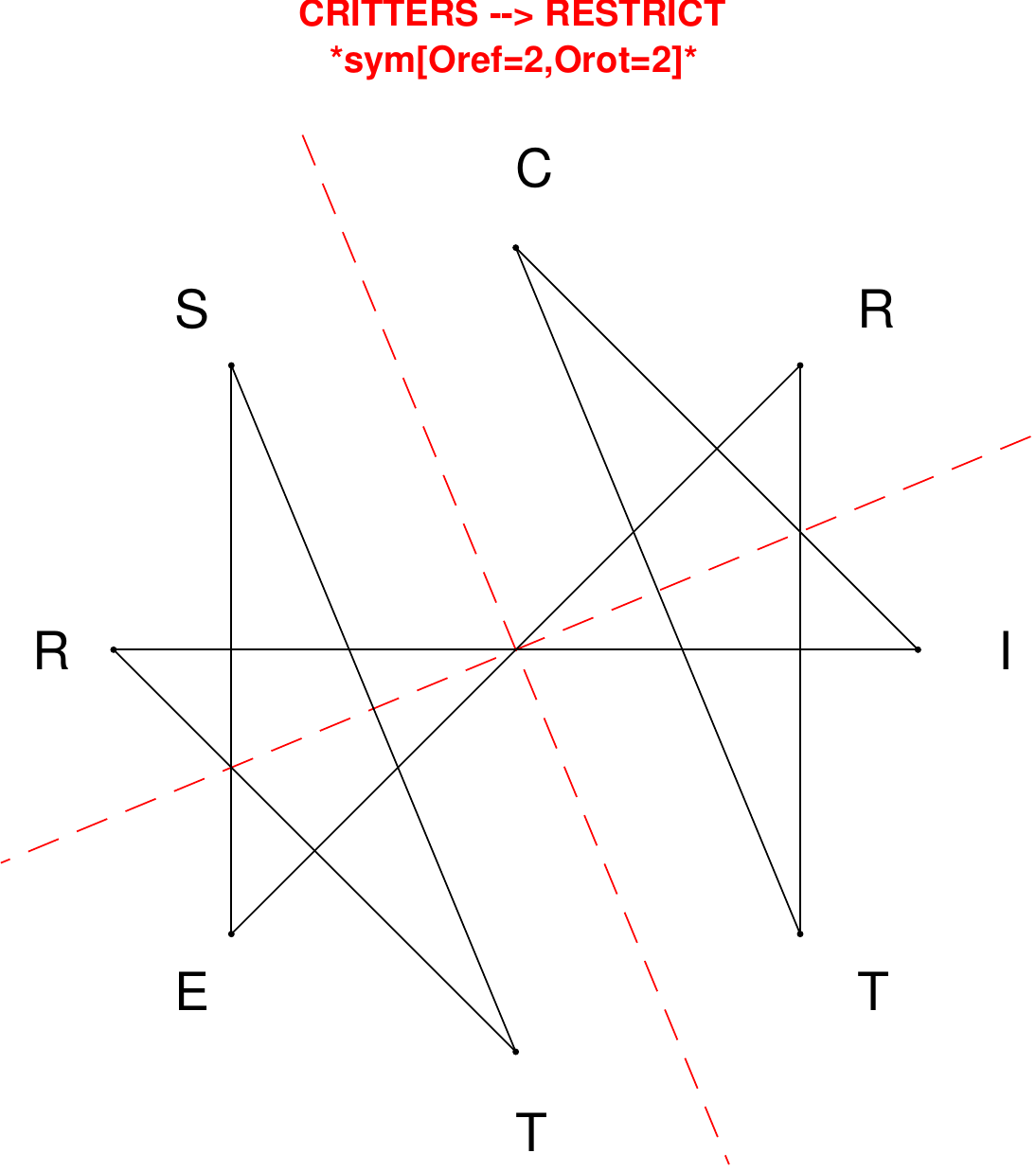}
\end{subfigure}
\hfill
\begin{subfigure}[T]{0.19\textwidth}
\centering
\includegraphics[width=\textwidth]{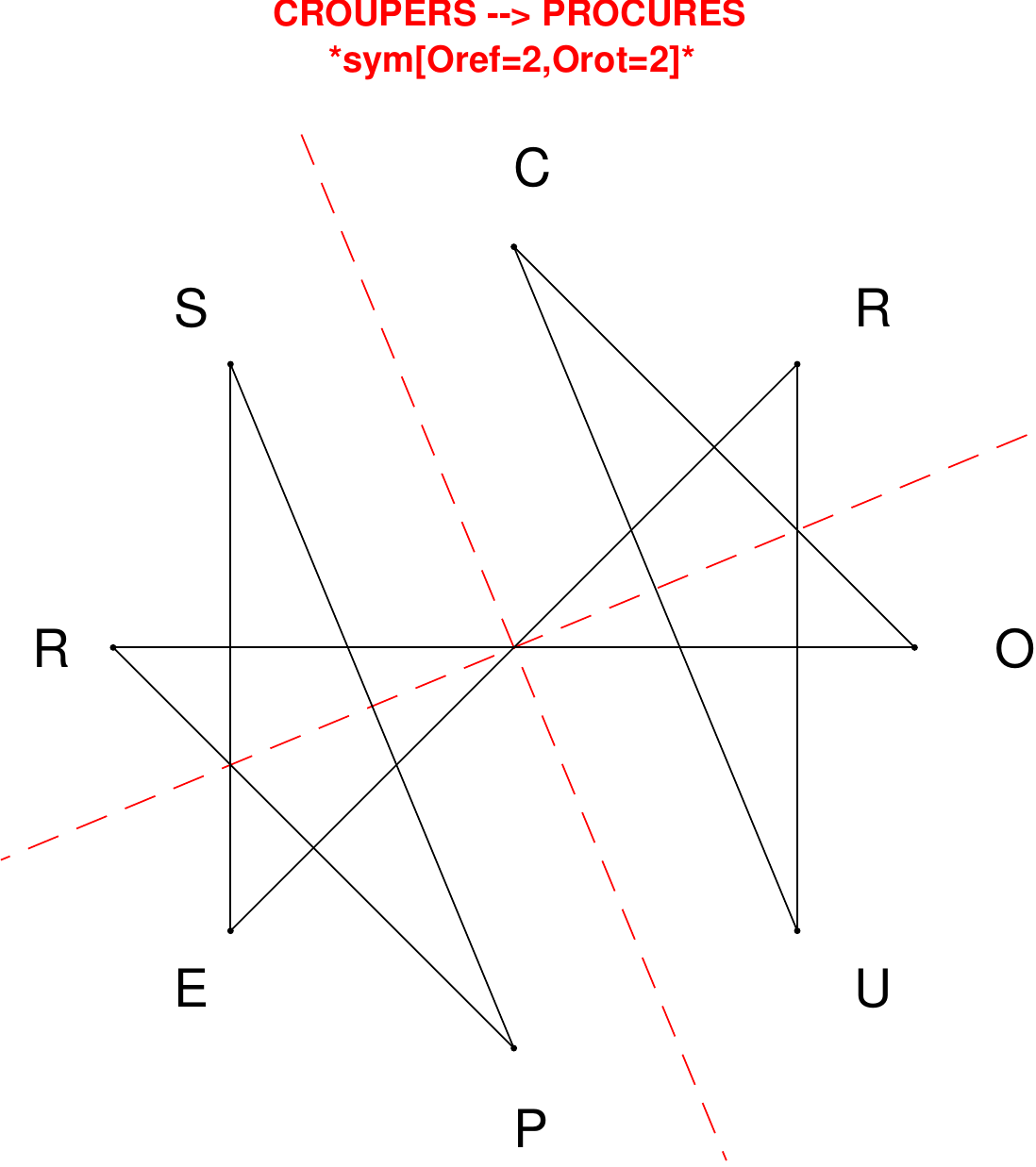}
\end{subfigure}
\end{figure}

\begin{figure}[H]
\centering
\begin{subfigure}[T]{0.19\textwidth}
\centering
\includegraphics[width=\textwidth]{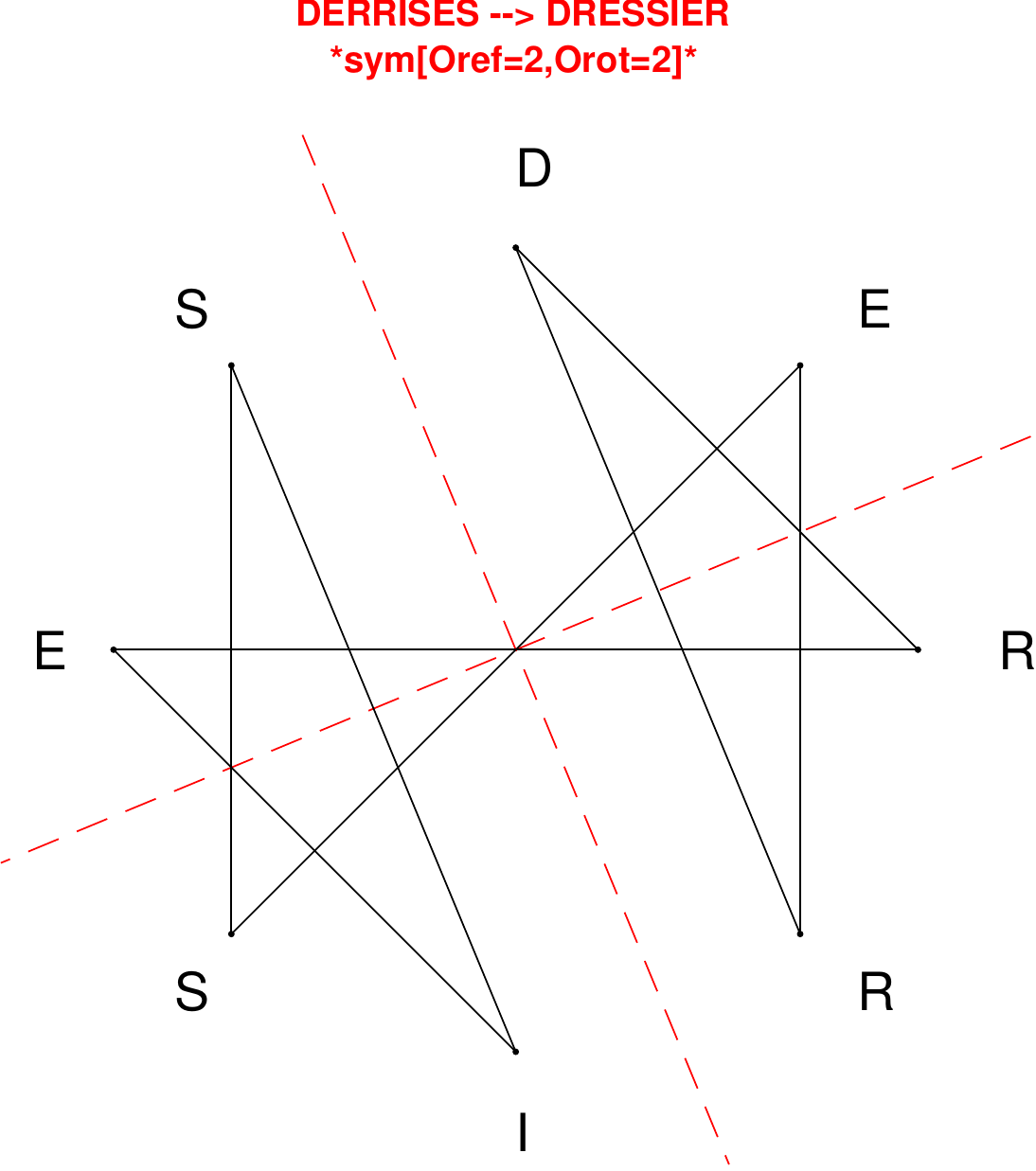}
\end{subfigure}
\hfill
\begin{subfigure}[T]{0.19\textwidth}
\centering
\includegraphics[width=\textwidth]{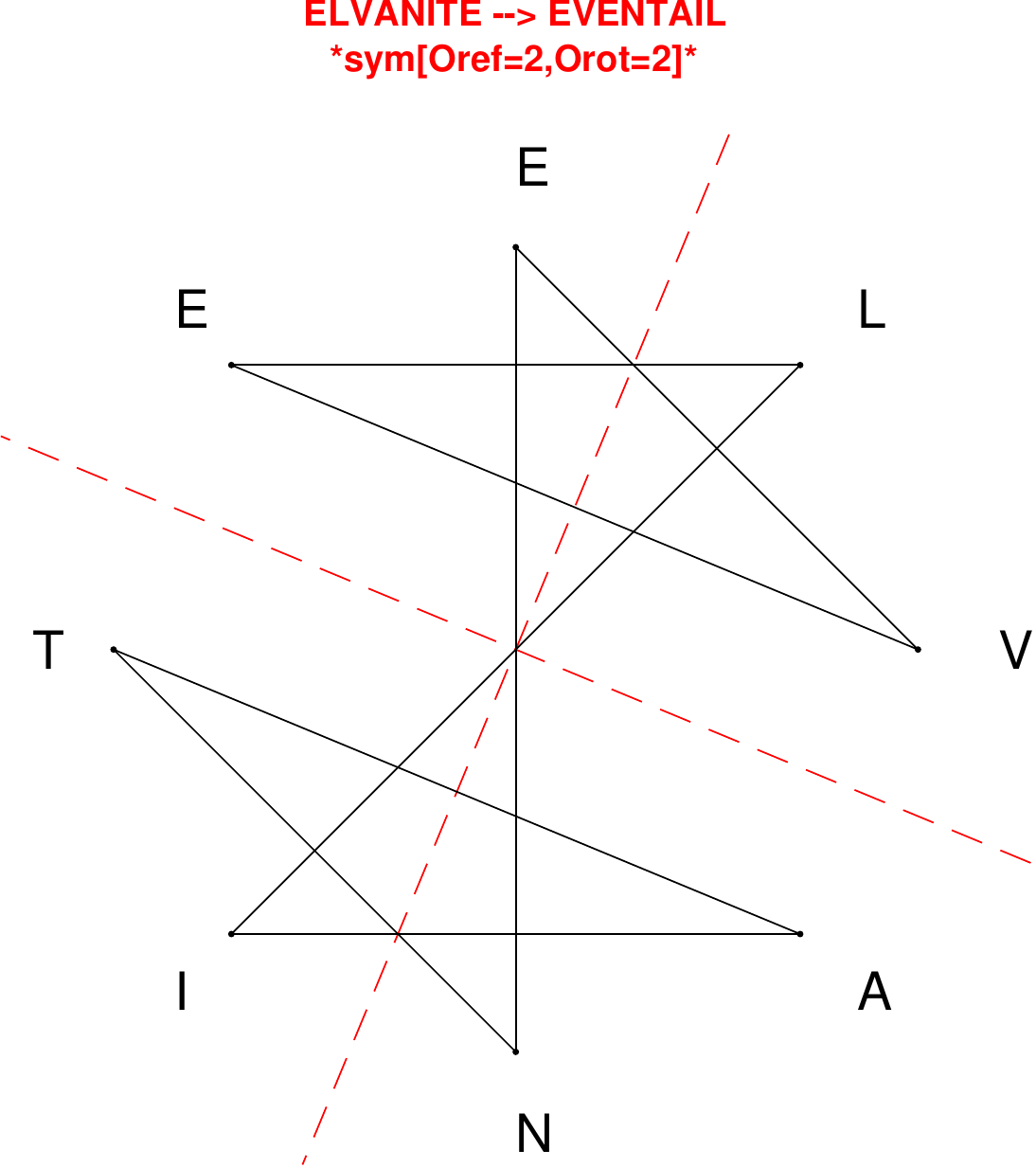}
\end{subfigure}
\hfill
\begin{subfigure}[T]{0.19\textwidth}
\centering
\includegraphics[width=\textwidth]{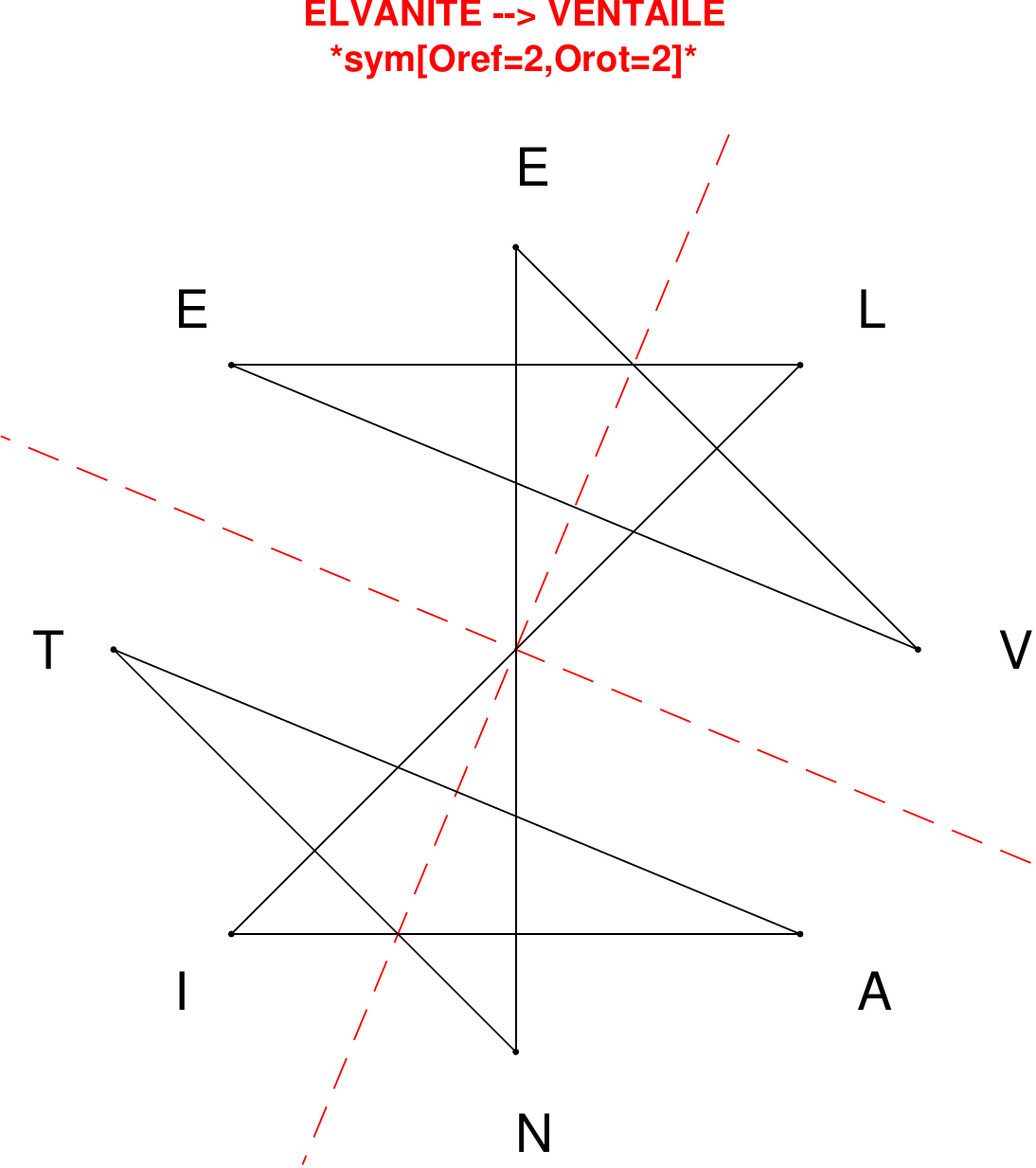}
\end{subfigure}
\hfill
\begin{subfigure}[T]{0.19\textwidth}
\centering
\includegraphics[width=\textwidth]{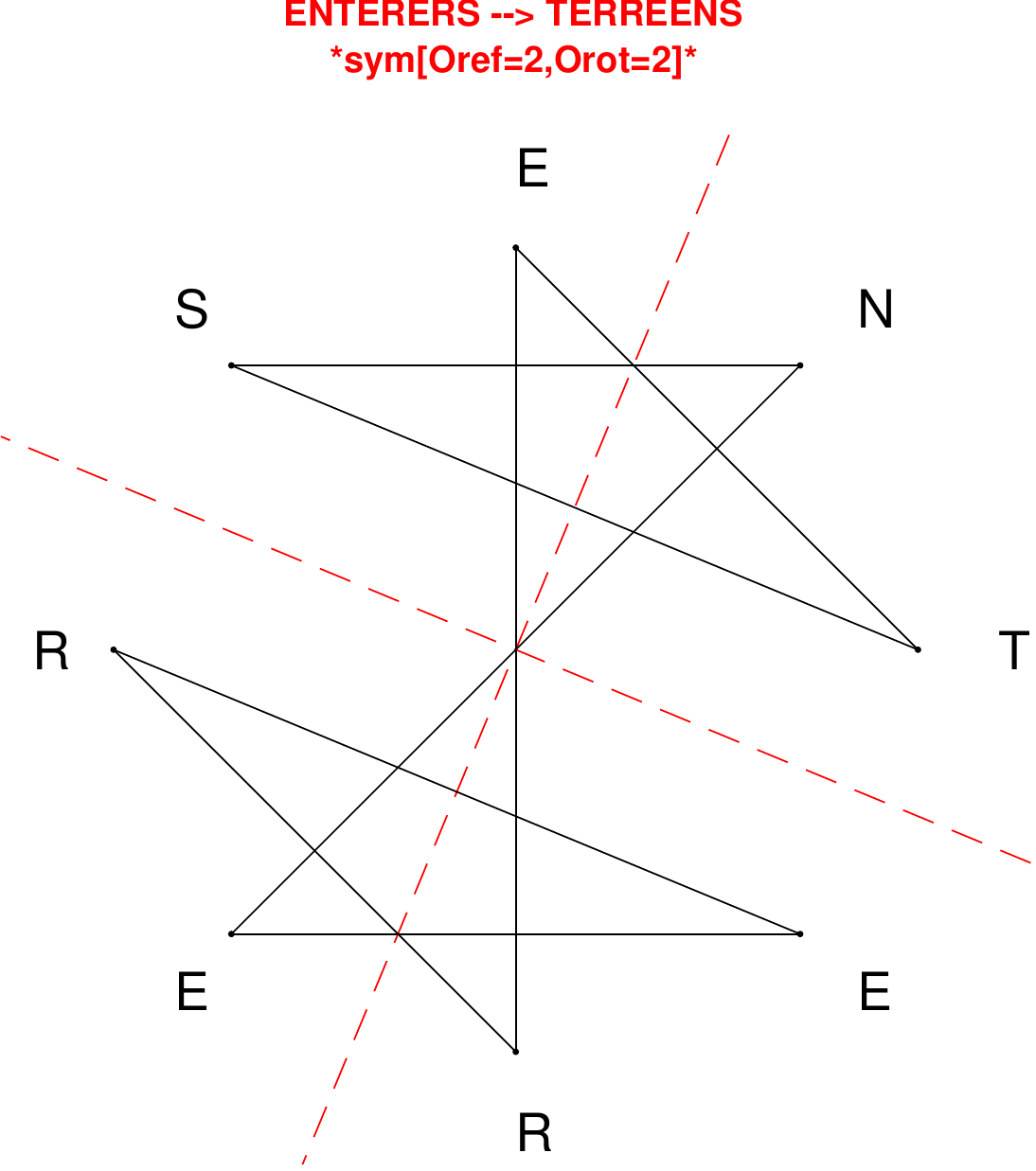}
\end{subfigure}
\hfill
\begin{subfigure}[T]{0.19\textwidth}
\centering
\includegraphics[width=\textwidth]{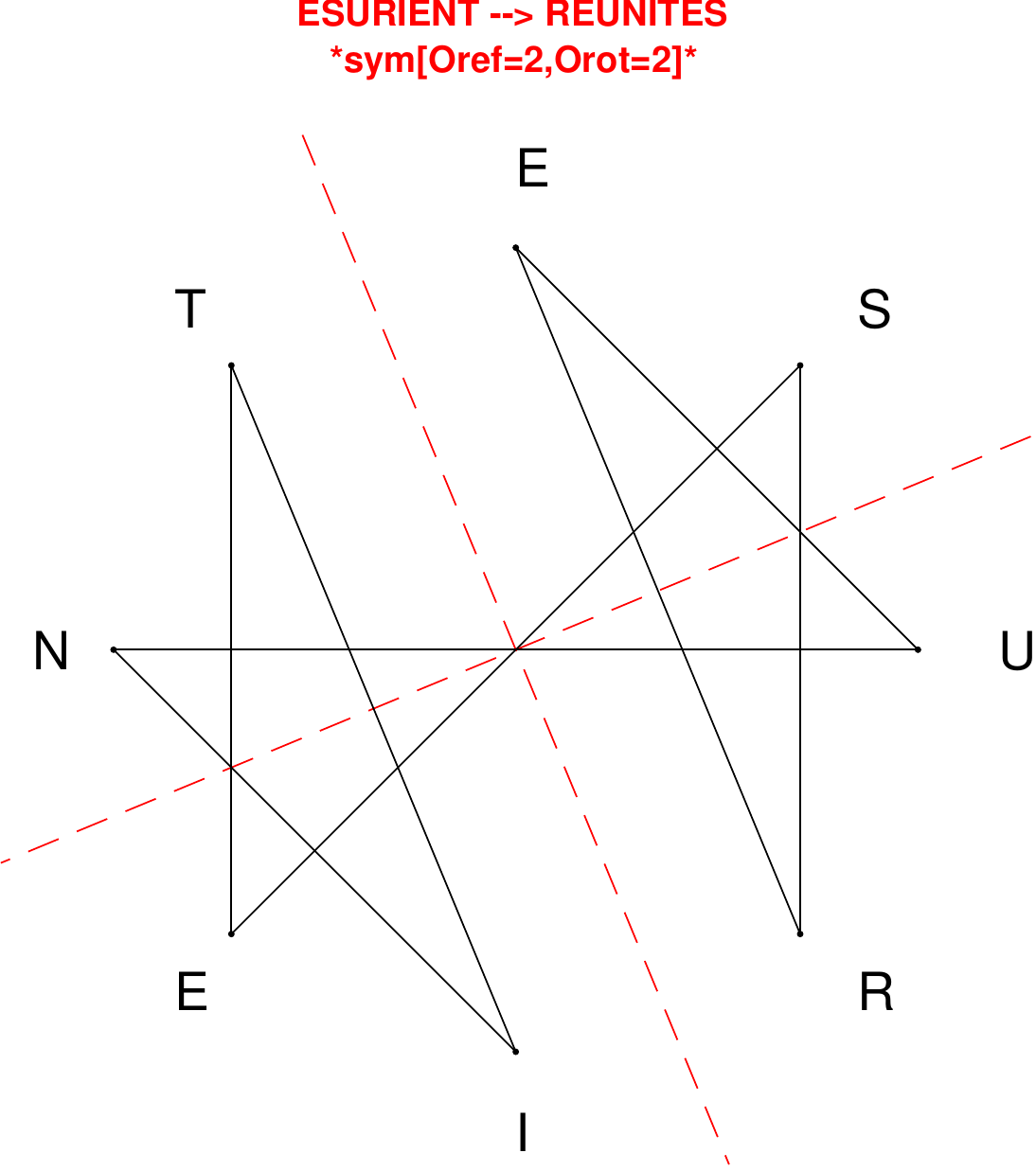}
\end{subfigure}
\end{figure}

\begin{figure}[H]
\centering
\begin{subfigure}[T]{0.19\textwidth}
\centering
\includegraphics[width=\textwidth]{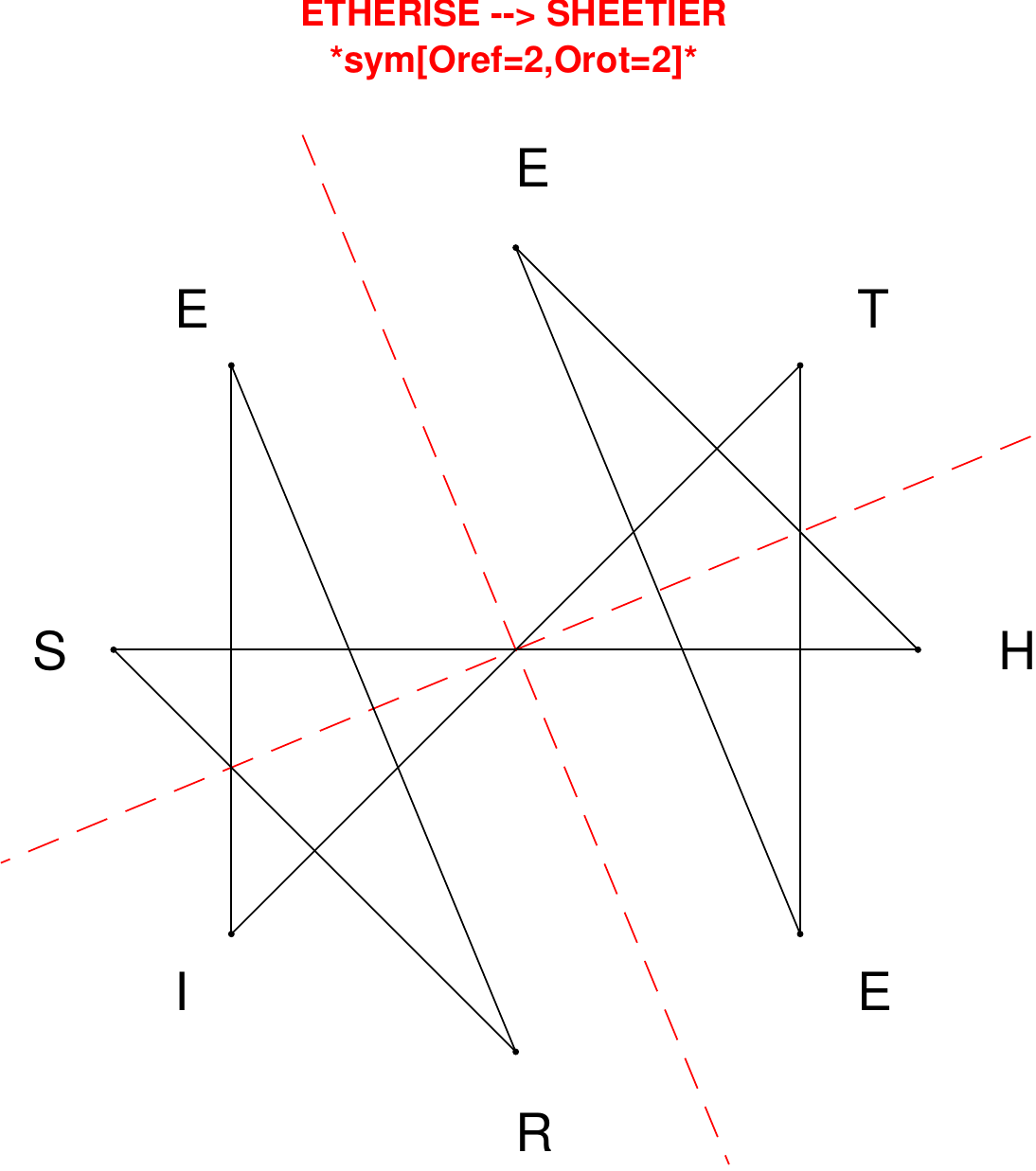}
\end{subfigure}
\hfill
\begin{subfigure}[T]{0.19\textwidth}
\centering
\includegraphics[width=\textwidth]{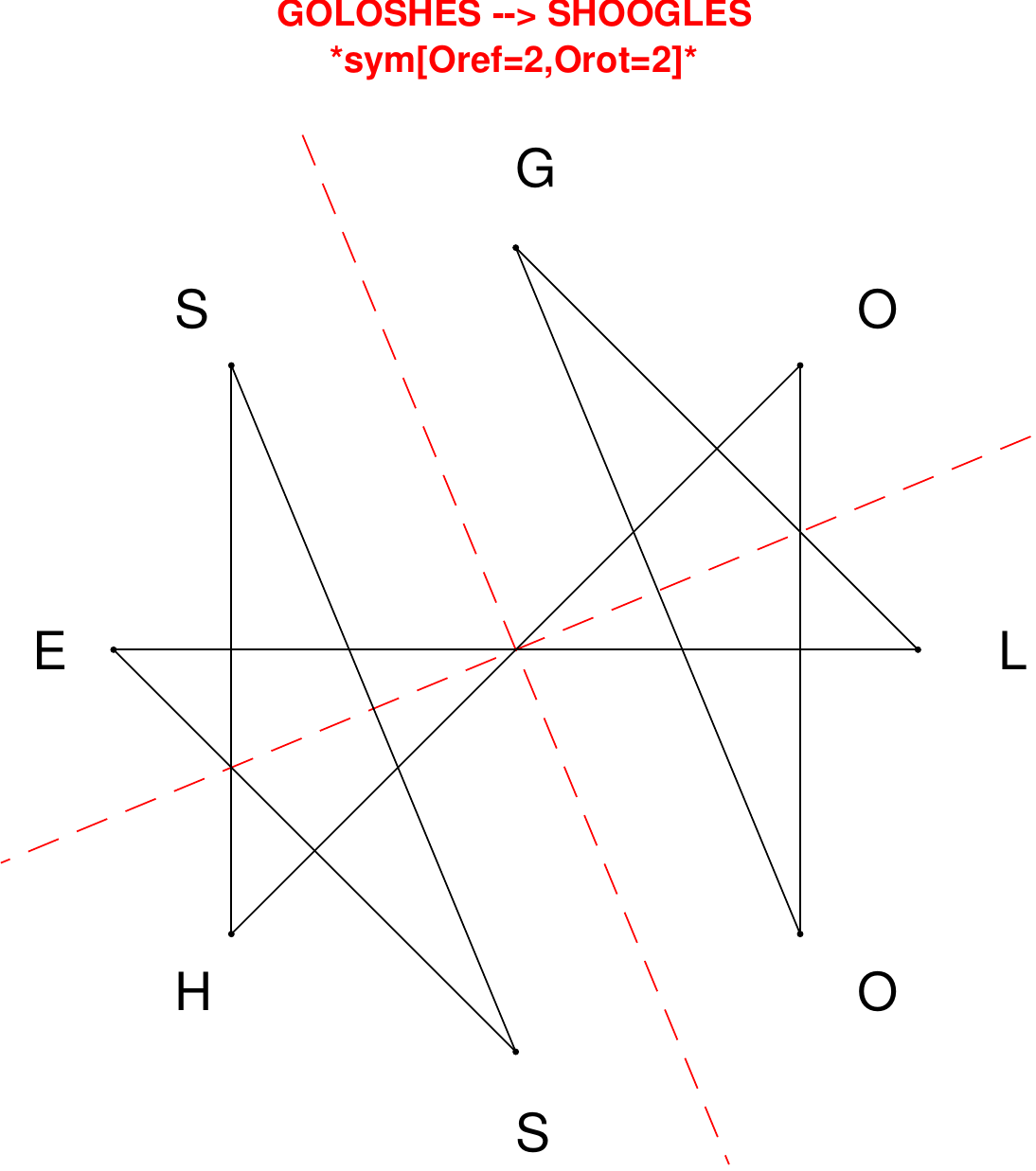}
\end{subfigure}
\hfill
\begin{subfigure}[T]{0.19\textwidth}
\centering
\includegraphics[width=\textwidth]{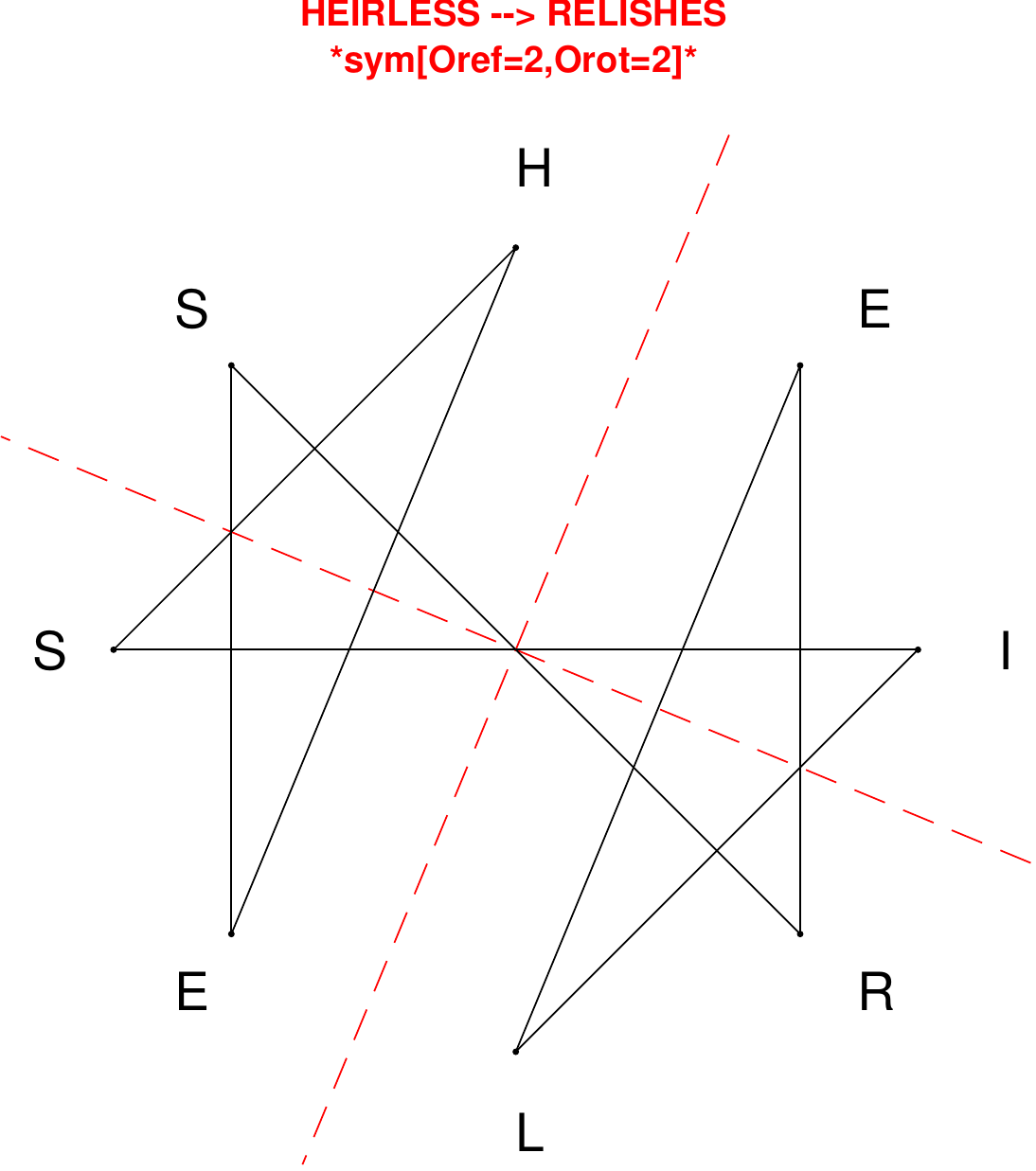}
\end{subfigure}
\hfill
\begin{subfigure}[T]{0.19\textwidth}
\centering
\includegraphics[width=\textwidth]{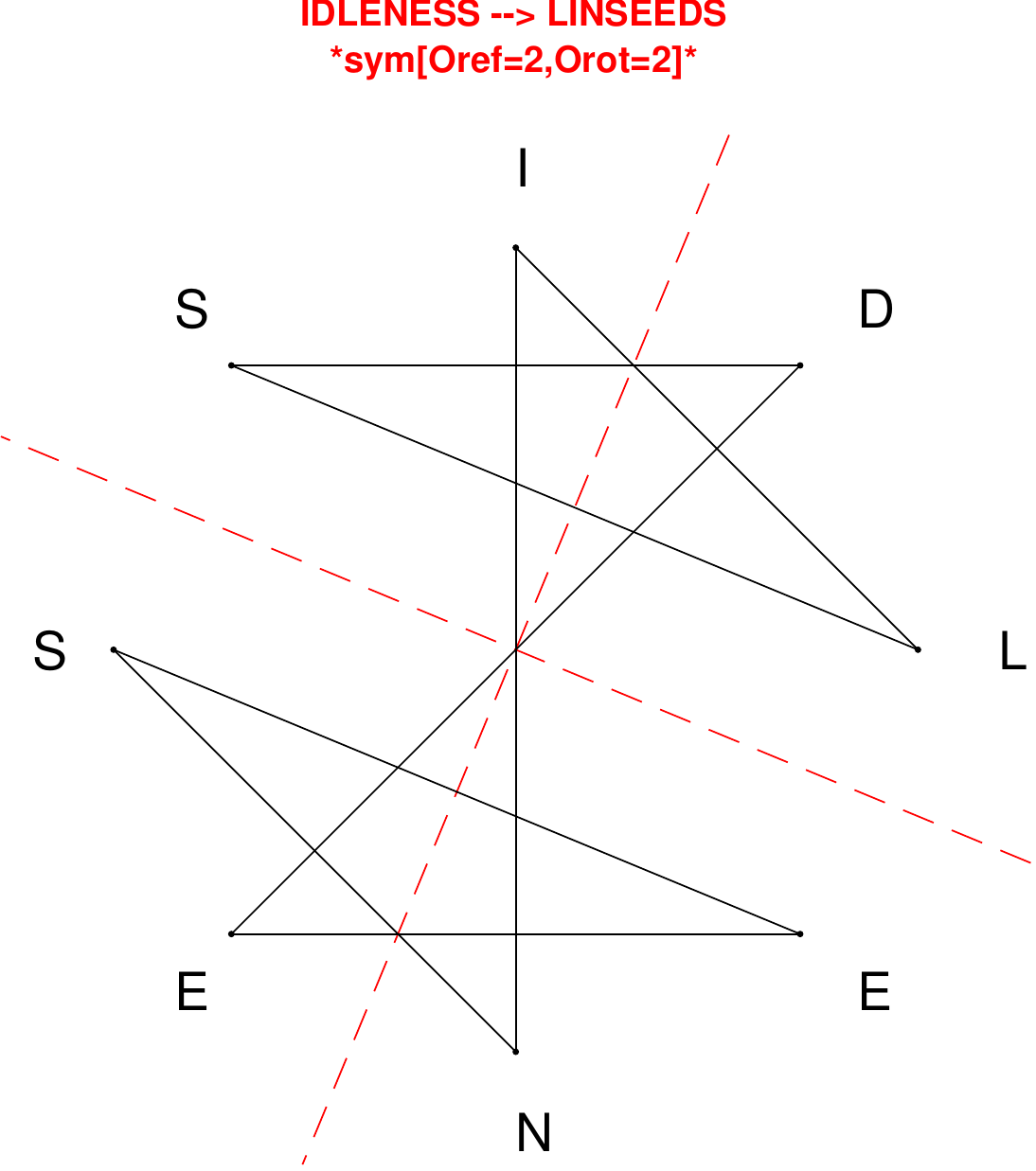}
\end{subfigure}
\hfill
\begin{subfigure}[T]{0.19\textwidth}
\centering
\includegraphics[width=\textwidth]{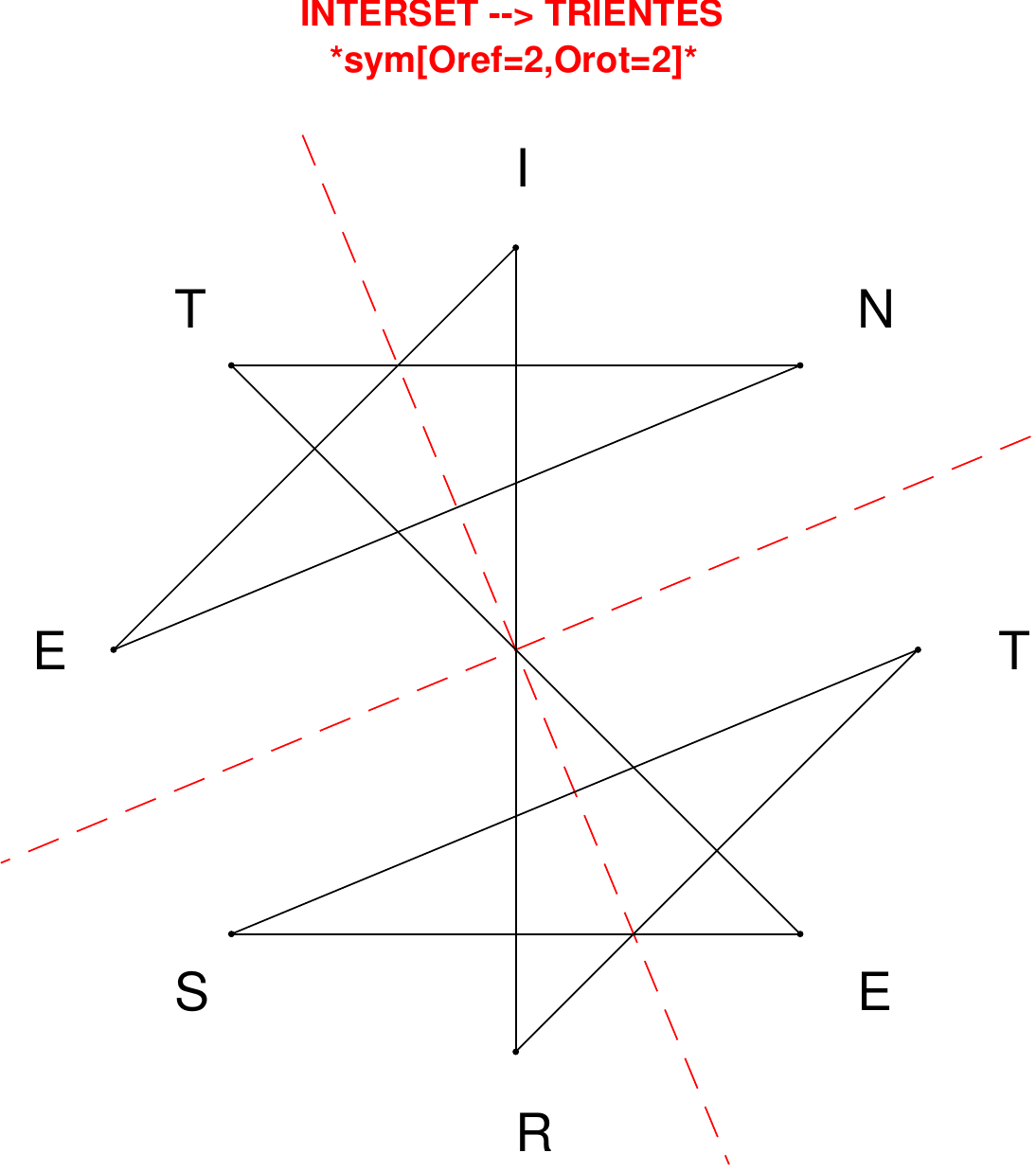}
\end{subfigure}
\end{figure}

\begin{figure}[H]
\centering
\begin{subfigure}[T]{0.19\textwidth}
\centering
\includegraphics[width=\textwidth]{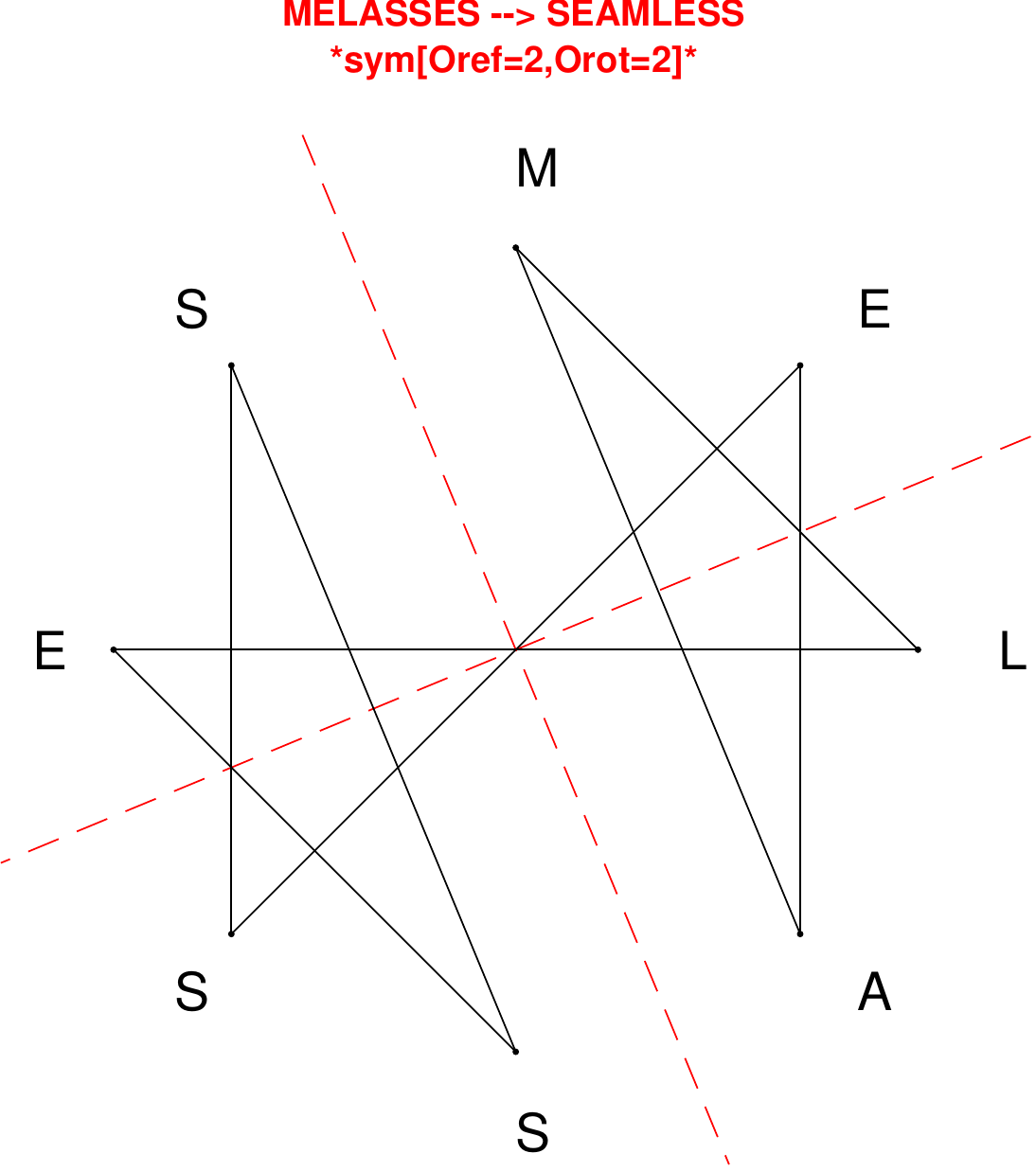}
\end{subfigure}
\hfill
\begin{subfigure}[T]{0.19\textwidth}
\centering
\includegraphics[width=\textwidth]{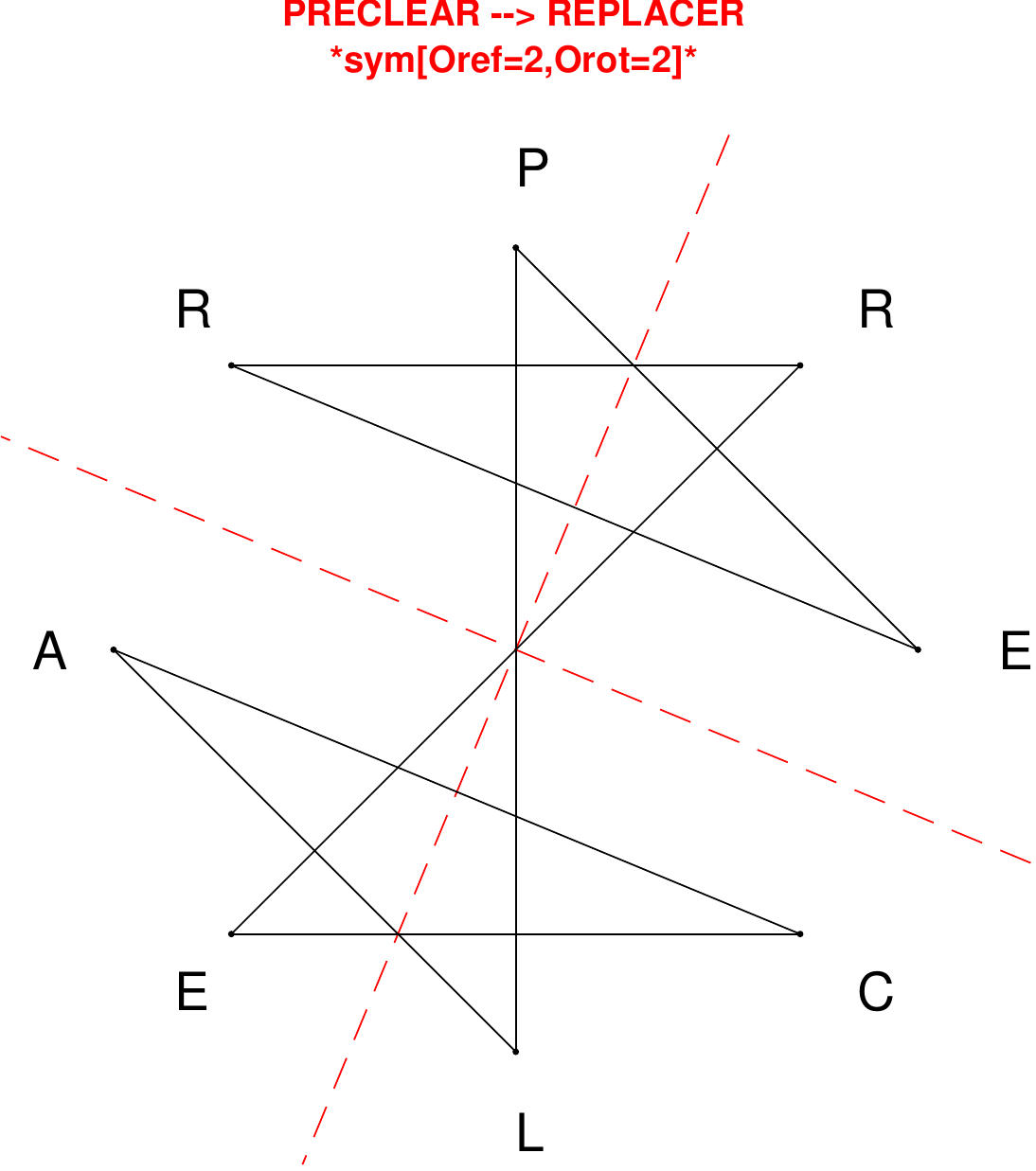}
\end{subfigure}
\hfill
\begin{subfigure}[T]{0.19\textwidth}
\centering
\includegraphics[width=\textwidth]{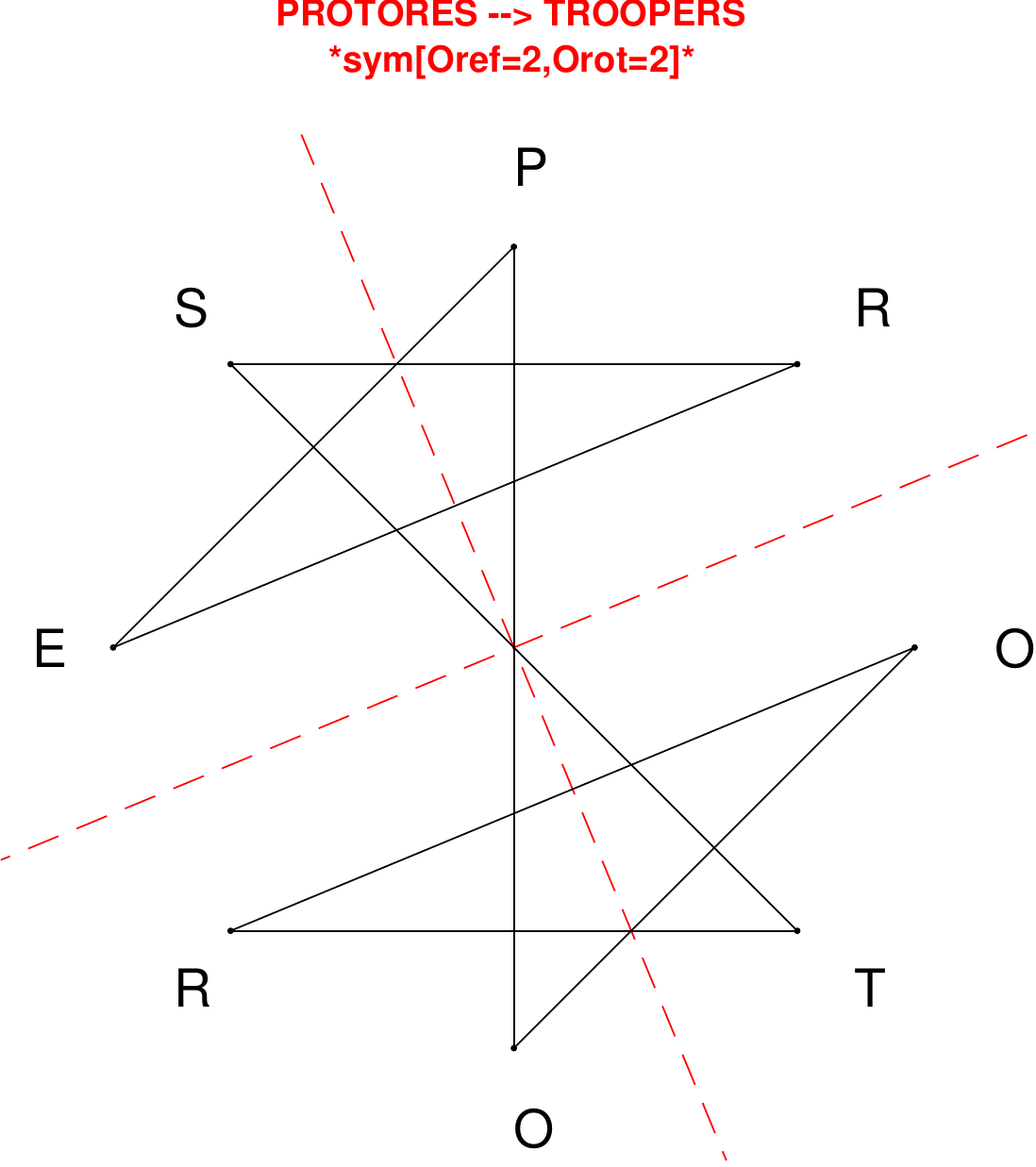}
\end{subfigure}
\hfill
\begin{subfigure}[T]{0.19\textwidth}
\centering
\includegraphics[width=\textwidth]{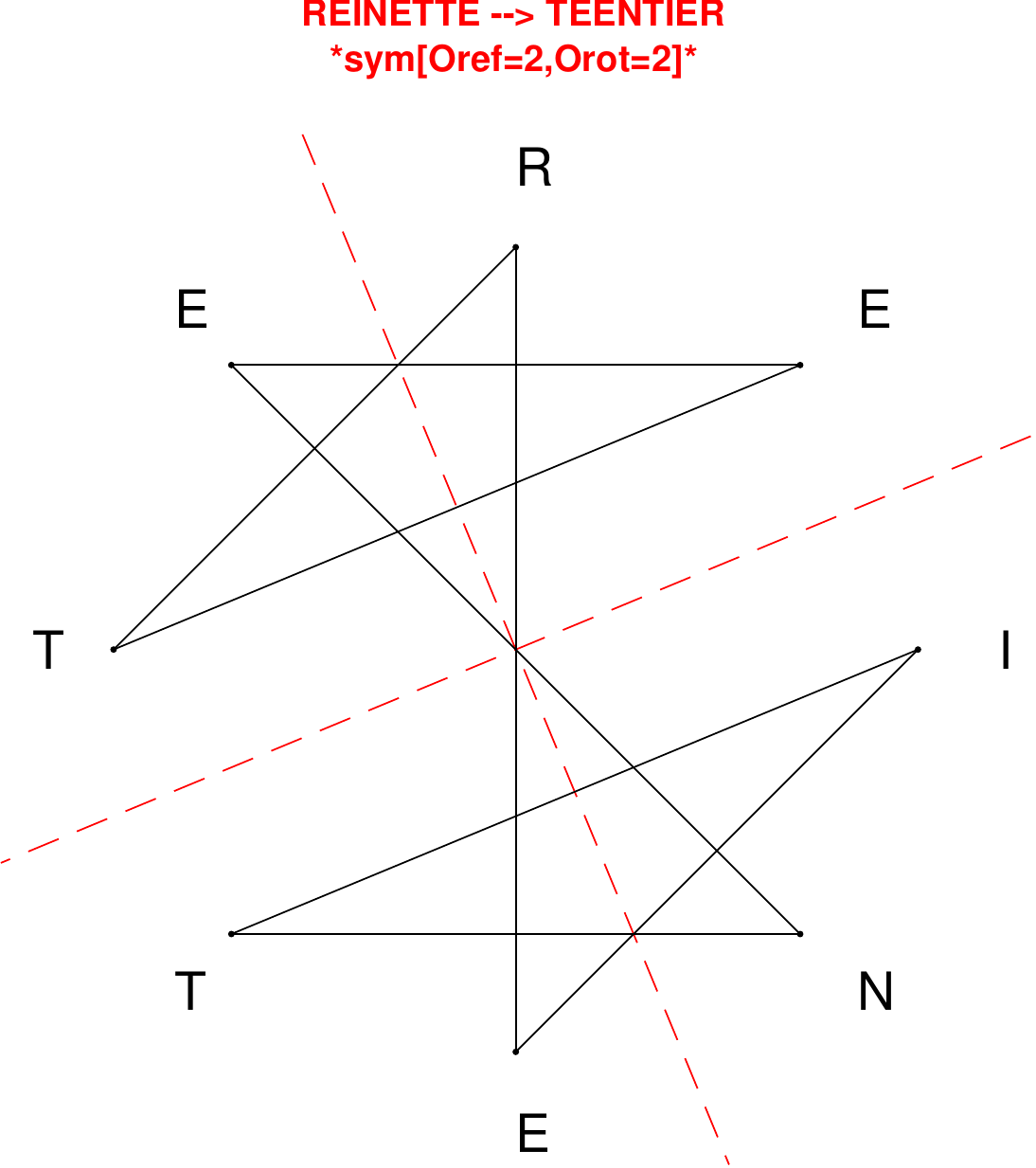}
\end{subfigure}
\hfill
\begin{subfigure}[T]{0.19\textwidth}
\centering
\includegraphics[width=\textwidth]{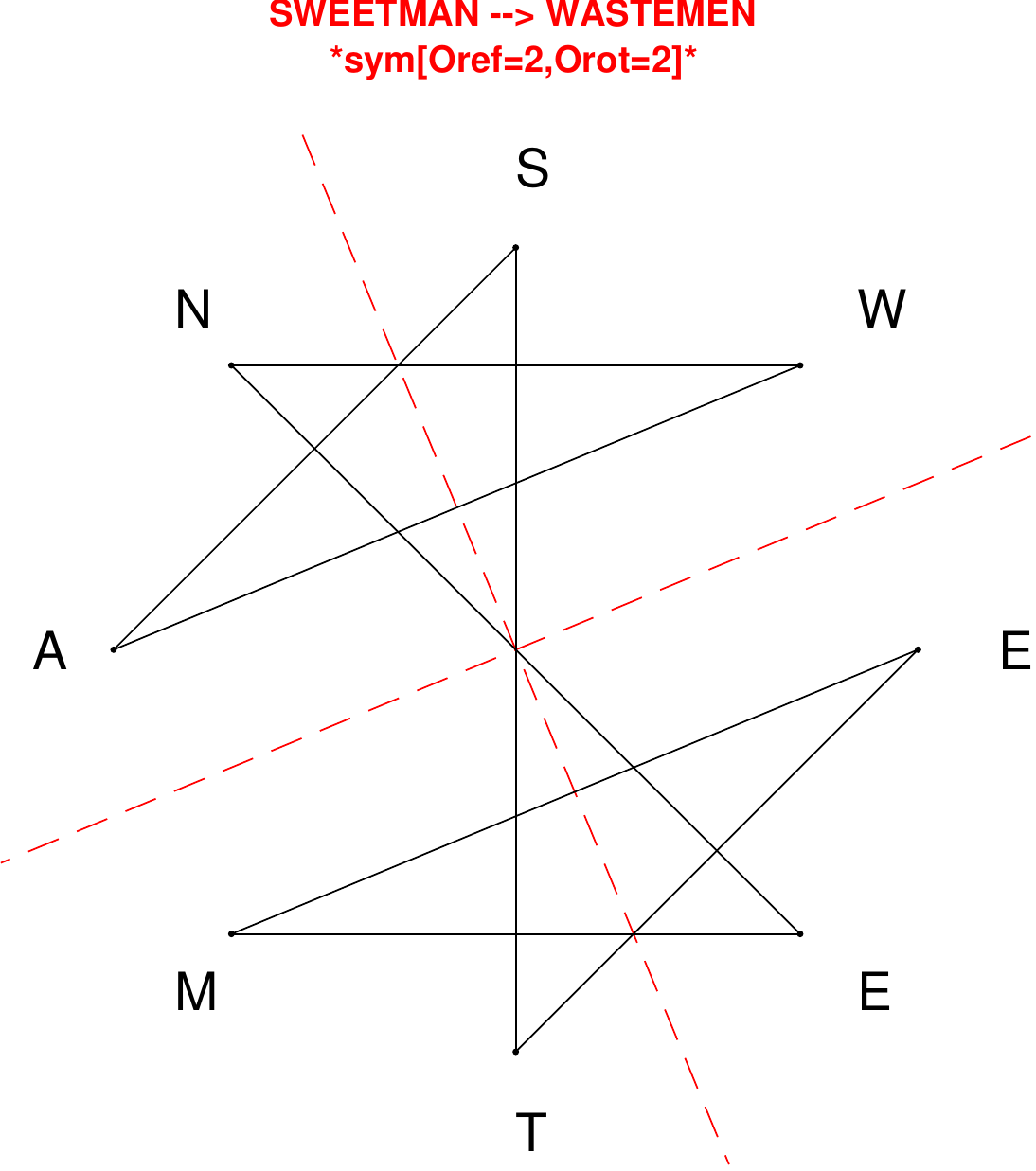}
\end{subfigure}
\end{figure}

\begin{figure}[H]
\centering
\begin{subfigure}[T]{0.19\textwidth}
\centering
\includegraphics[width=\textwidth]{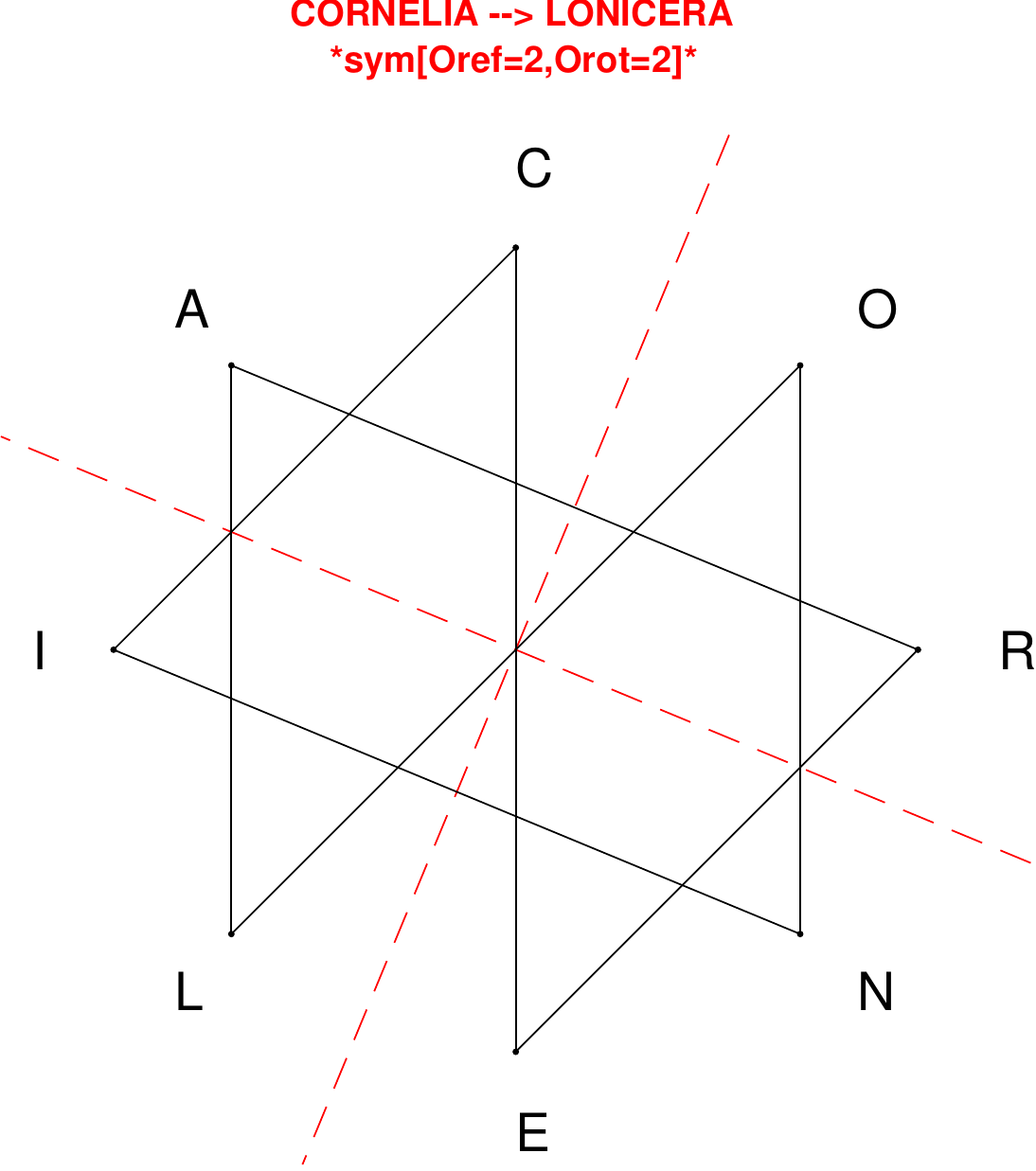}
\end{subfigure}
\hfill
\begin{subfigure}[T]{0.19\textwidth}
\centering
\includegraphics[width=\textwidth]{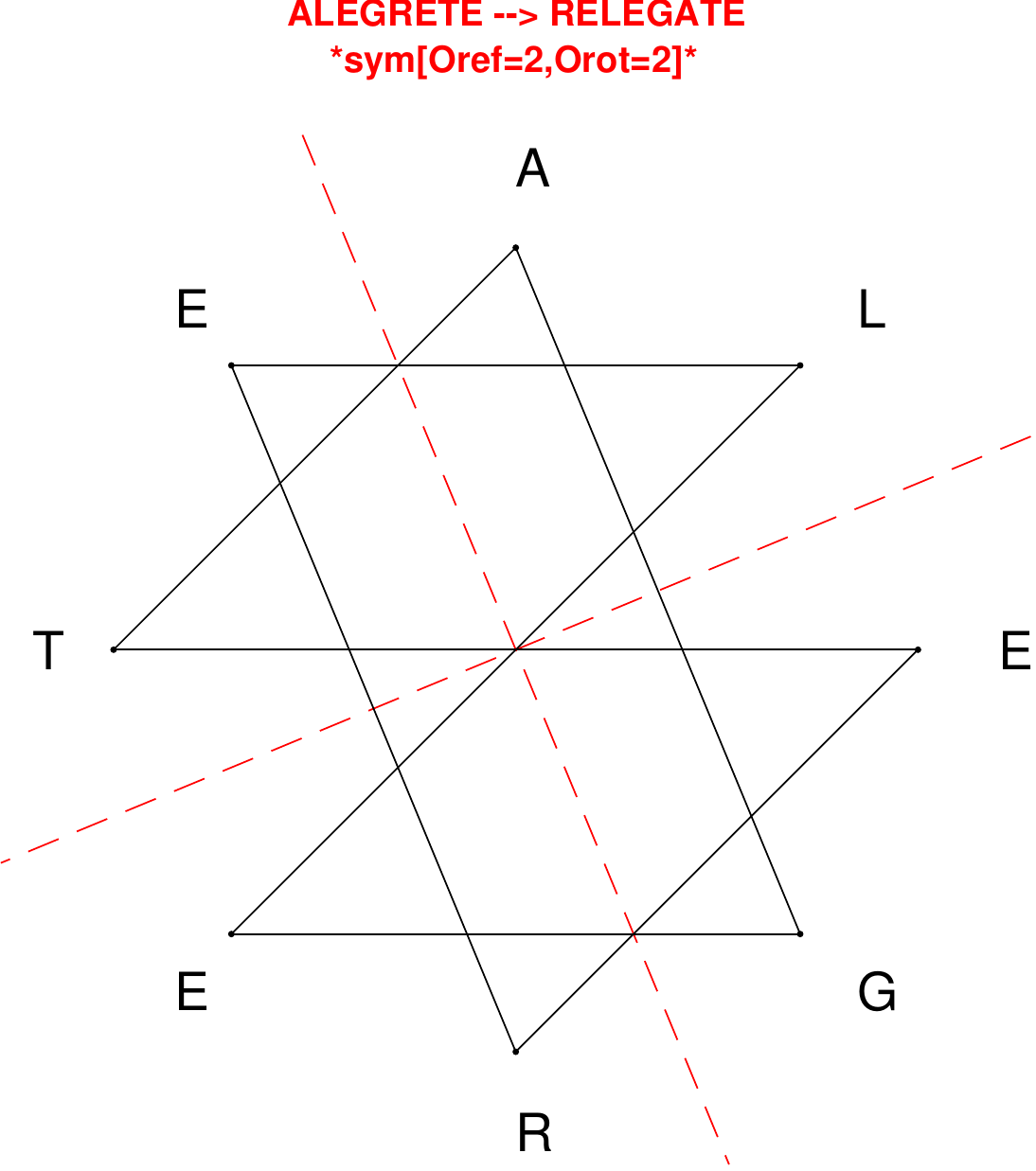}
\end{subfigure}
\hfill
\begin{subfigure}[T]{0.19\textwidth}
\centering
\includegraphics[width=\textwidth]{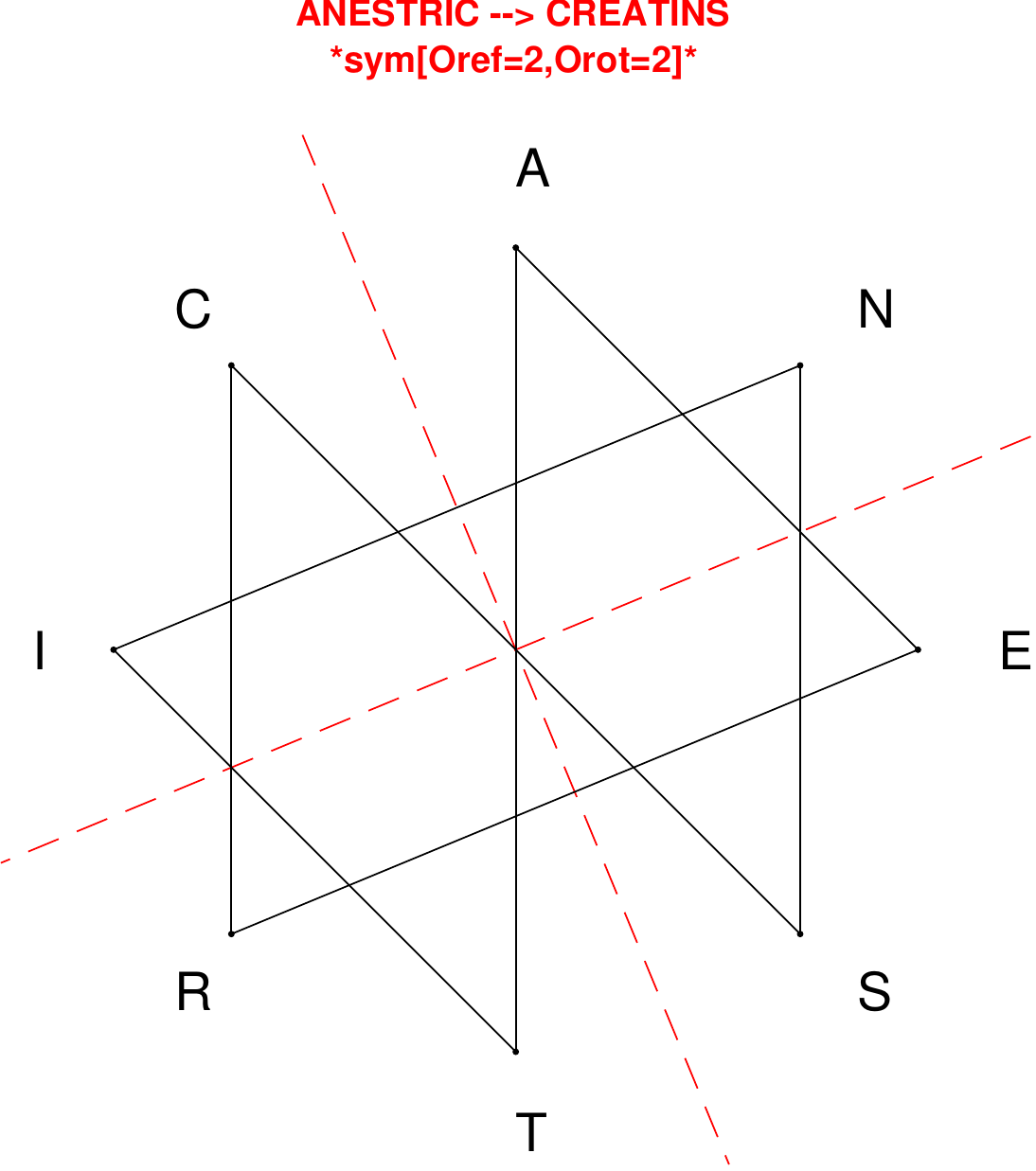}
\end{subfigure}
\hfill
\begin{subfigure}[T]{0.19\textwidth}
\centering
\includegraphics[width=\textwidth]{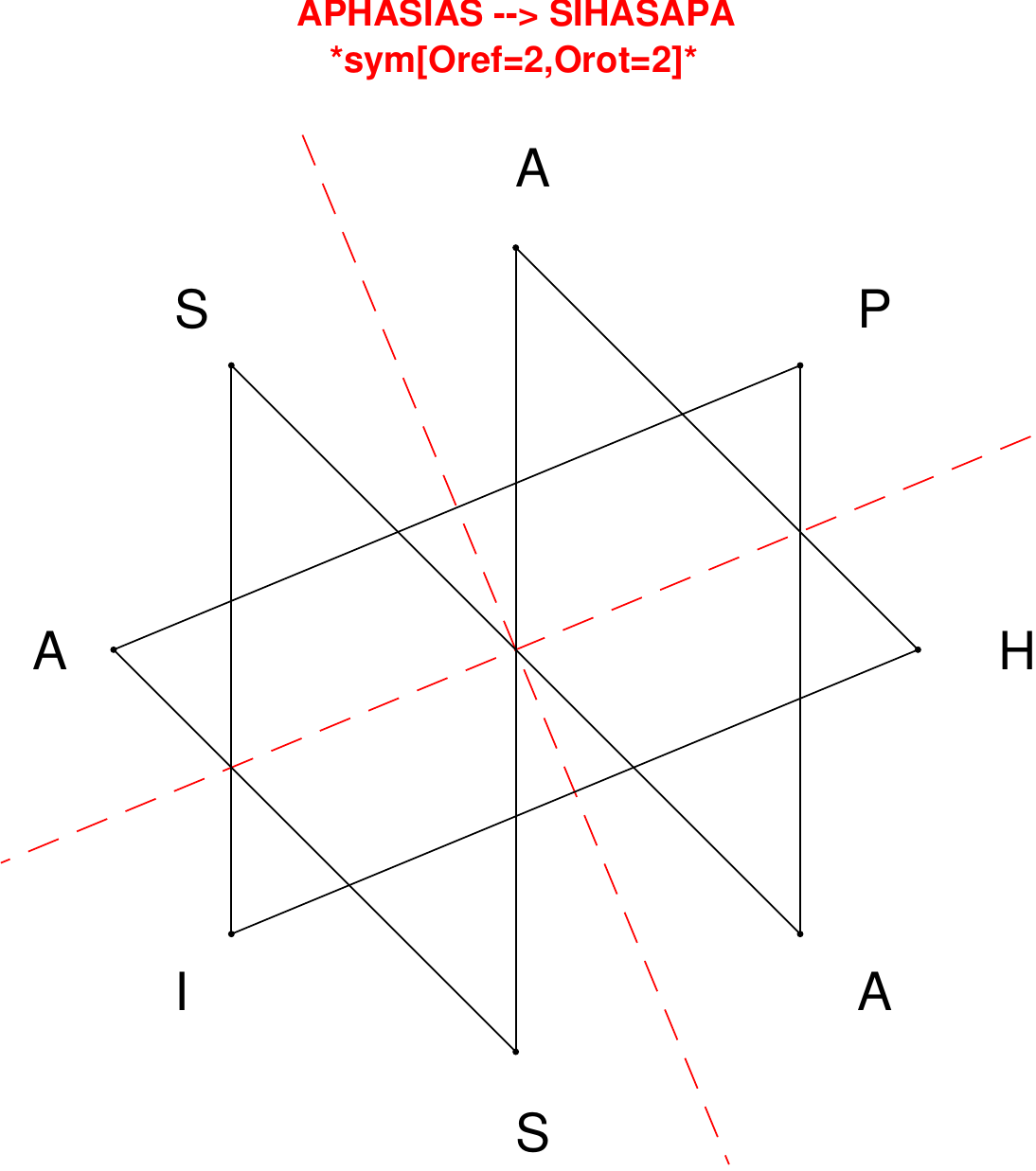}
\end{subfigure}
\hfill
\begin{subfigure}[T]{0.19\textwidth}
\centering
\includegraphics[width=\textwidth]{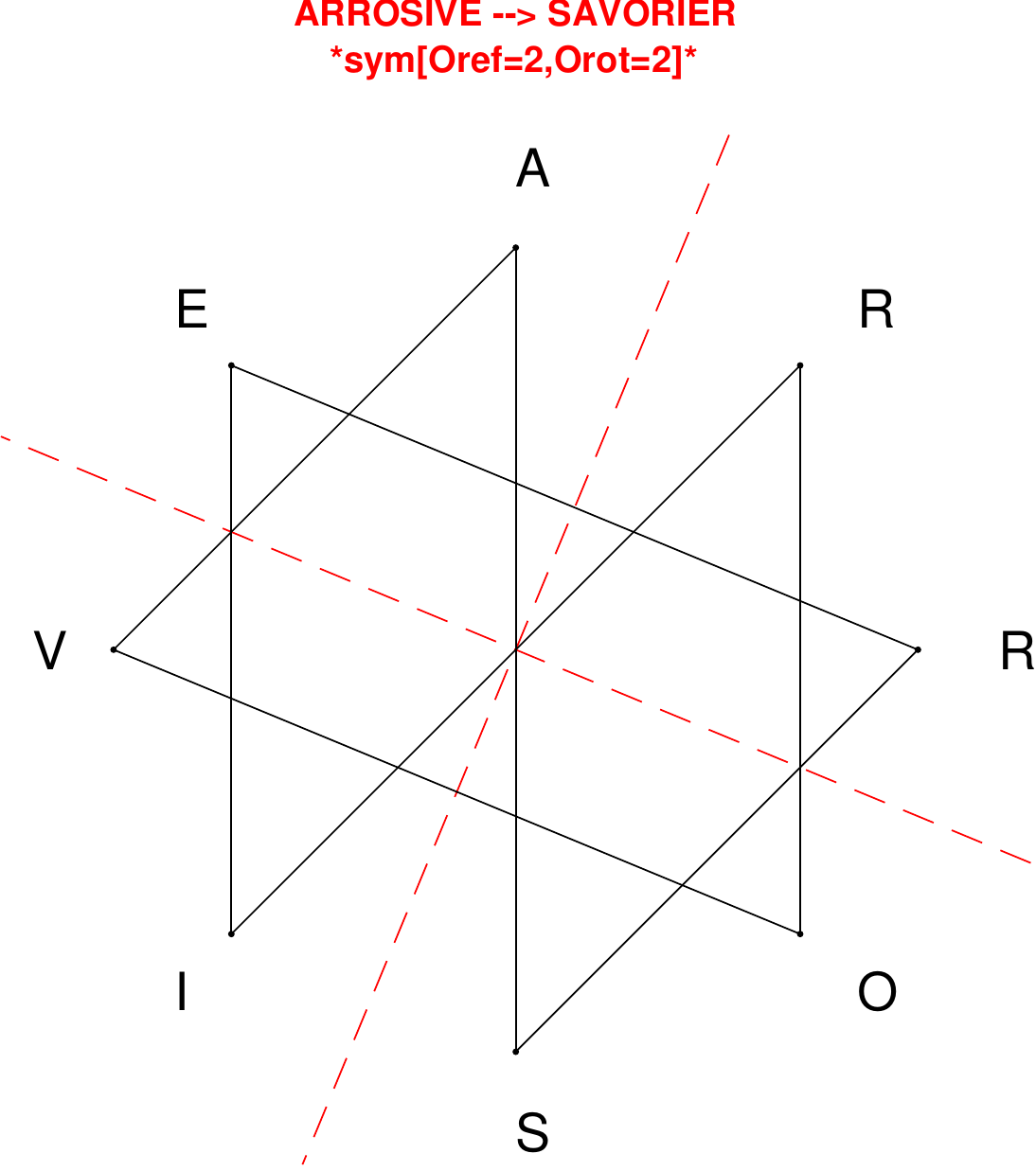}
\end{subfigure}
\end{figure}

\begin{figure}[H]
\centering
\begin{subfigure}[T]{0.19\textwidth}
\centering
\includegraphics[width=\textwidth]{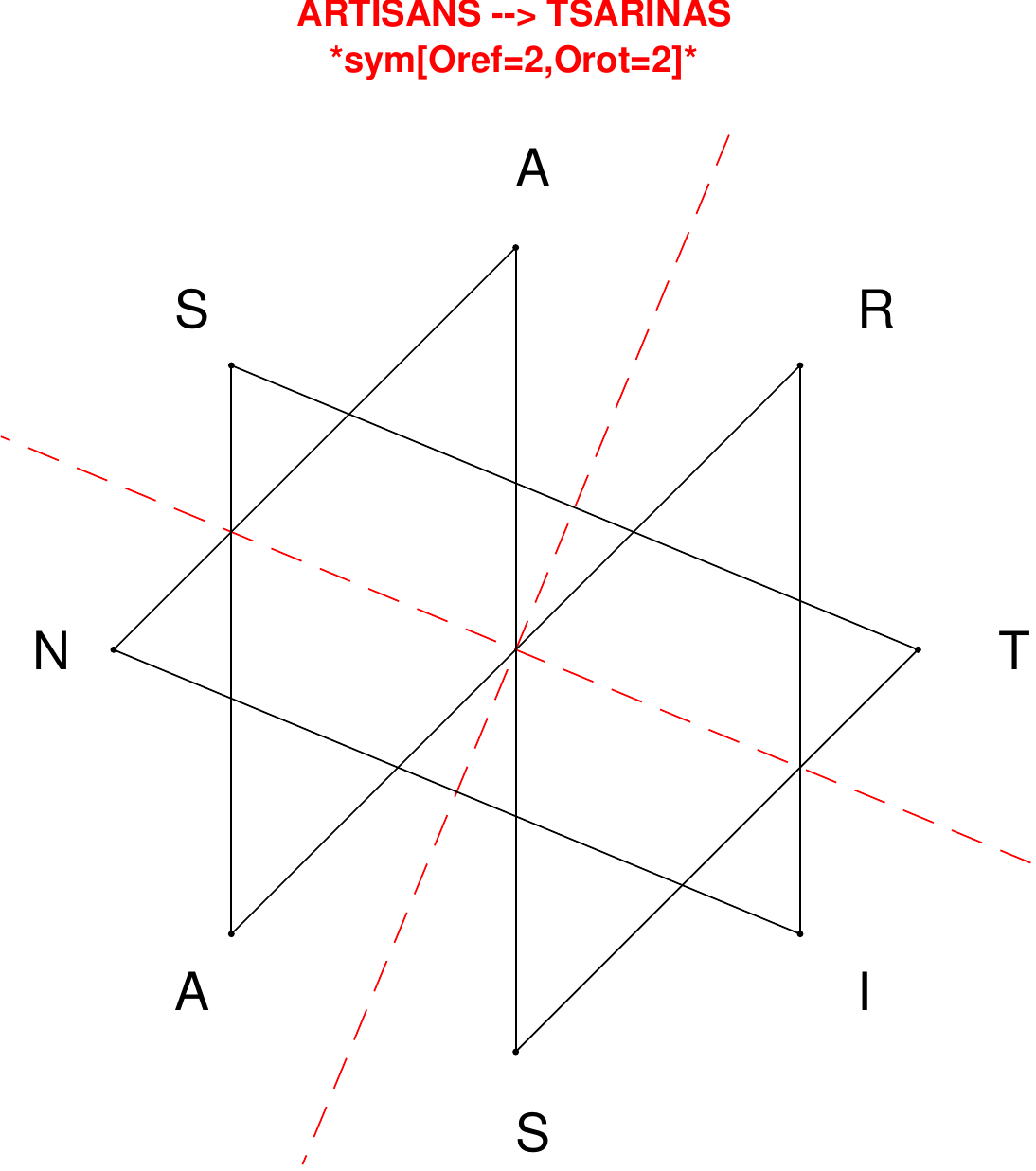}
\end{subfigure}
\hfill
\begin{subfigure}[T]{0.19\textwidth}
\centering
\includegraphics[width=\textwidth]{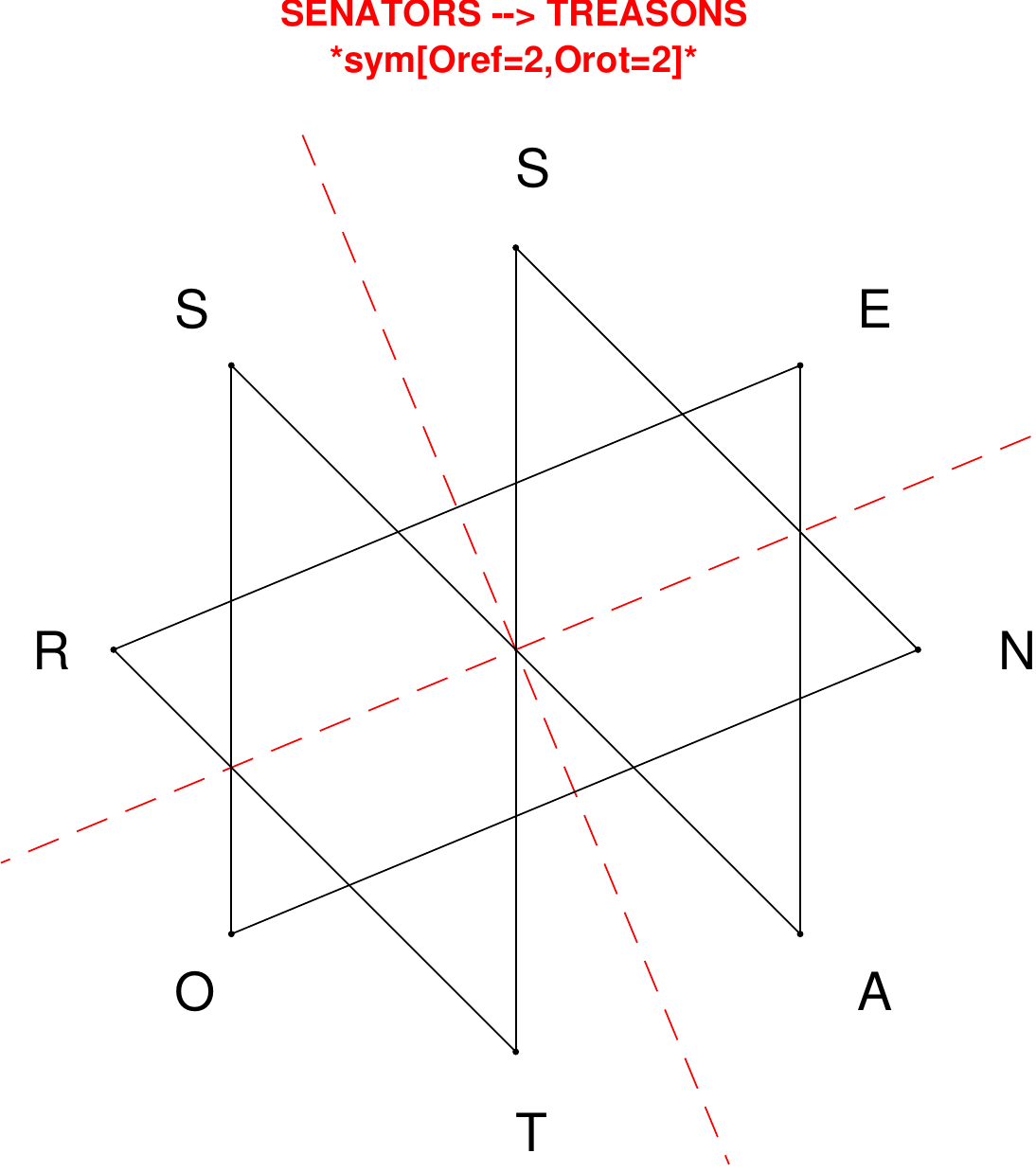}
\end{subfigure}
\hfill
\begin{subfigure}[T]{0.19\textwidth}
\centering
\includegraphics[width=\textwidth]{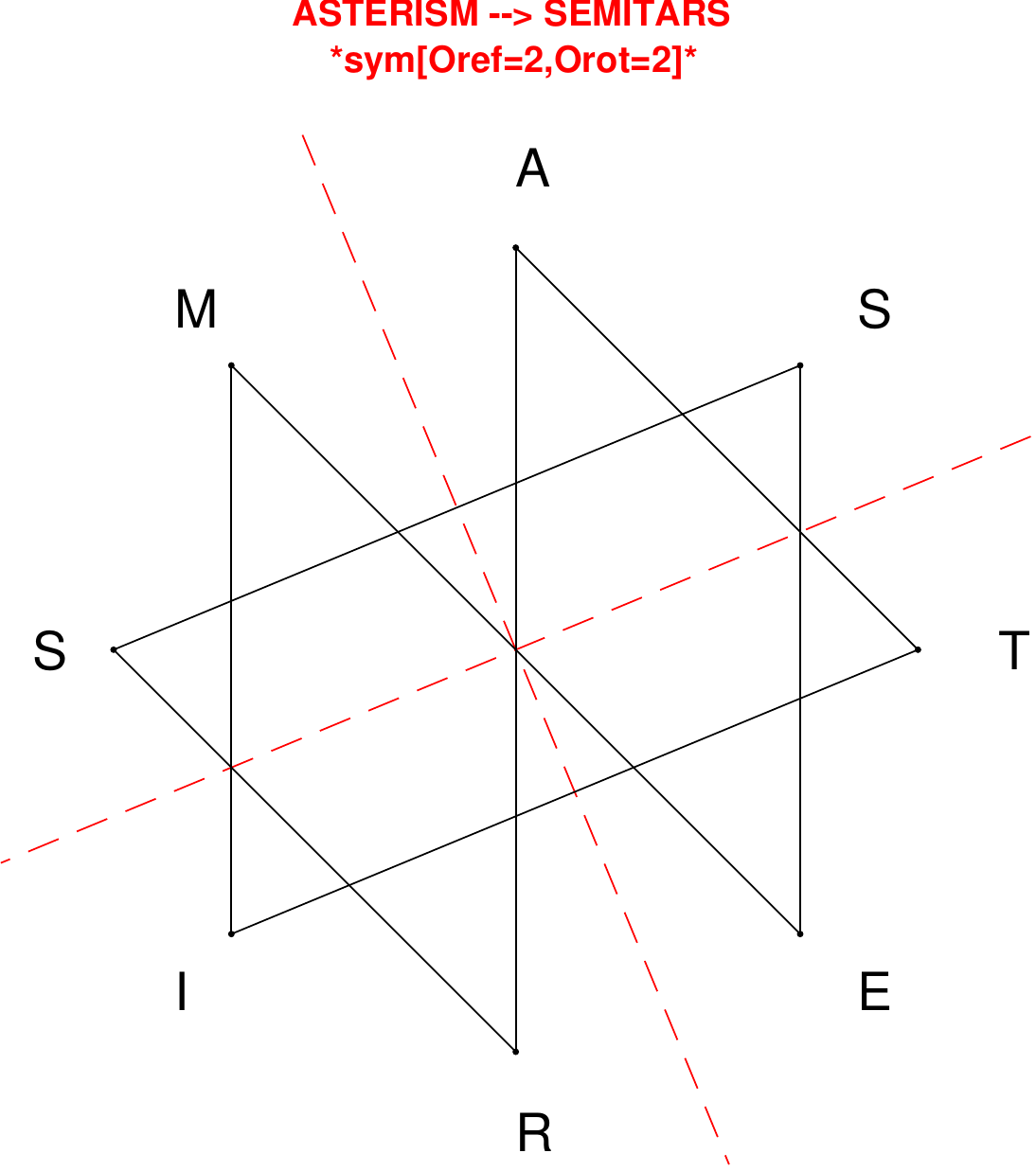}
\end{subfigure}
\hfill
\begin{subfigure}[T]{0.19\textwidth}
\centering
\includegraphics[width=\textwidth]{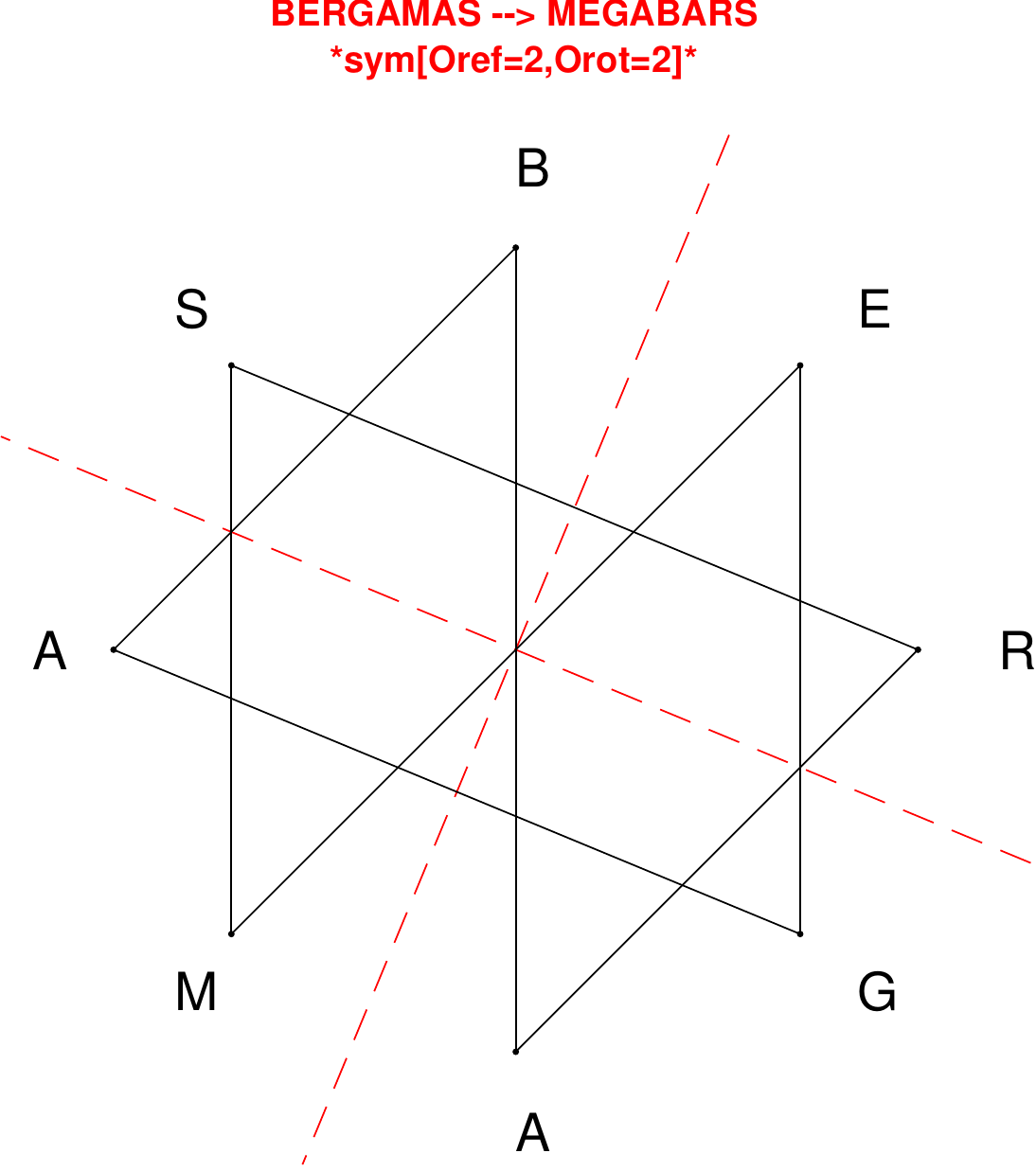}
\end{subfigure}
\hfill
\begin{subfigure}[T]{0.19\textwidth}
\centering
\includegraphics[width=\textwidth]{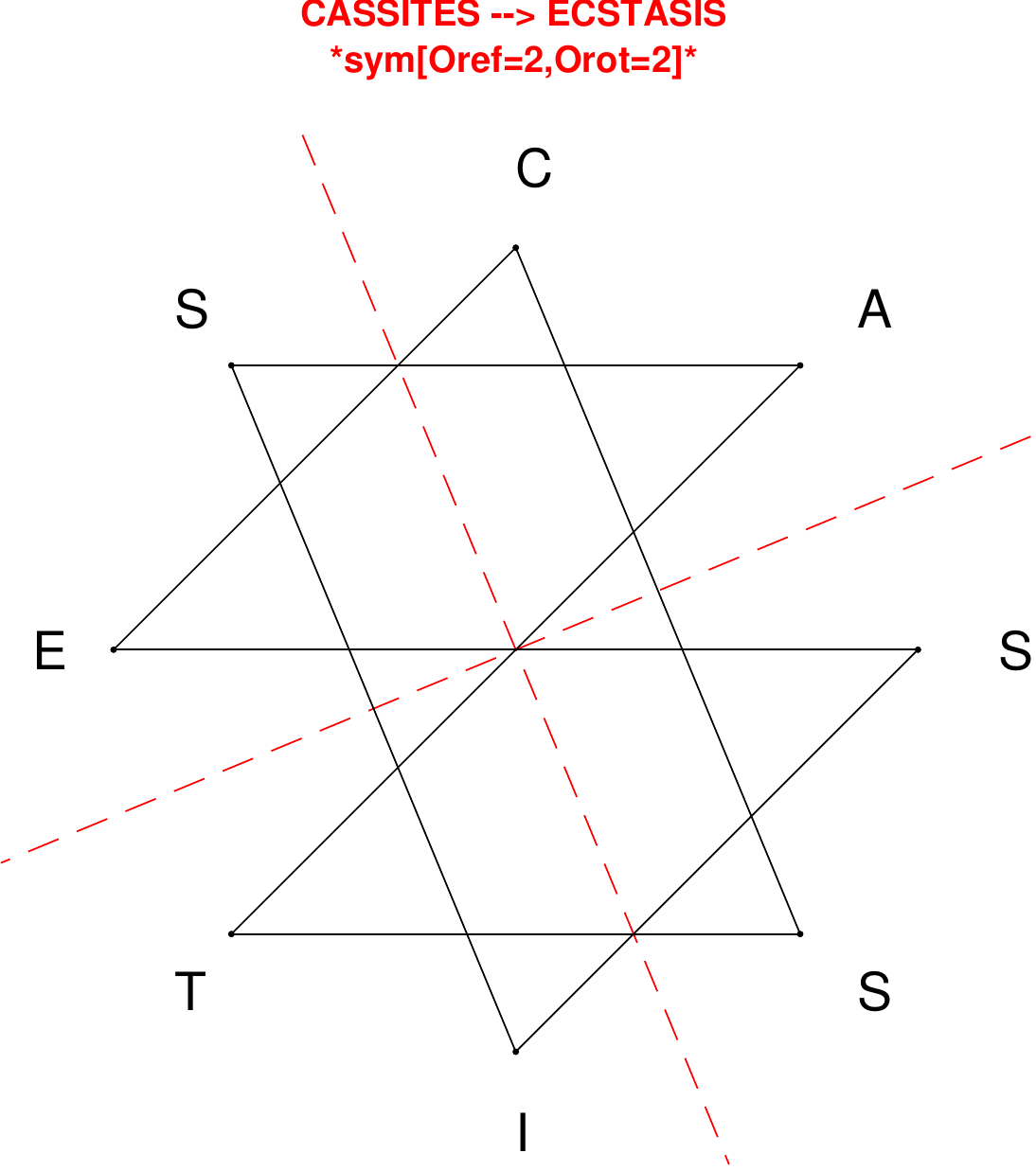}
\end{subfigure}
\end{figure}

\begin{figure}[H]
\centering
\begin{subfigure}[T]{0.19\textwidth}
\centering
\includegraphics[width=\textwidth]{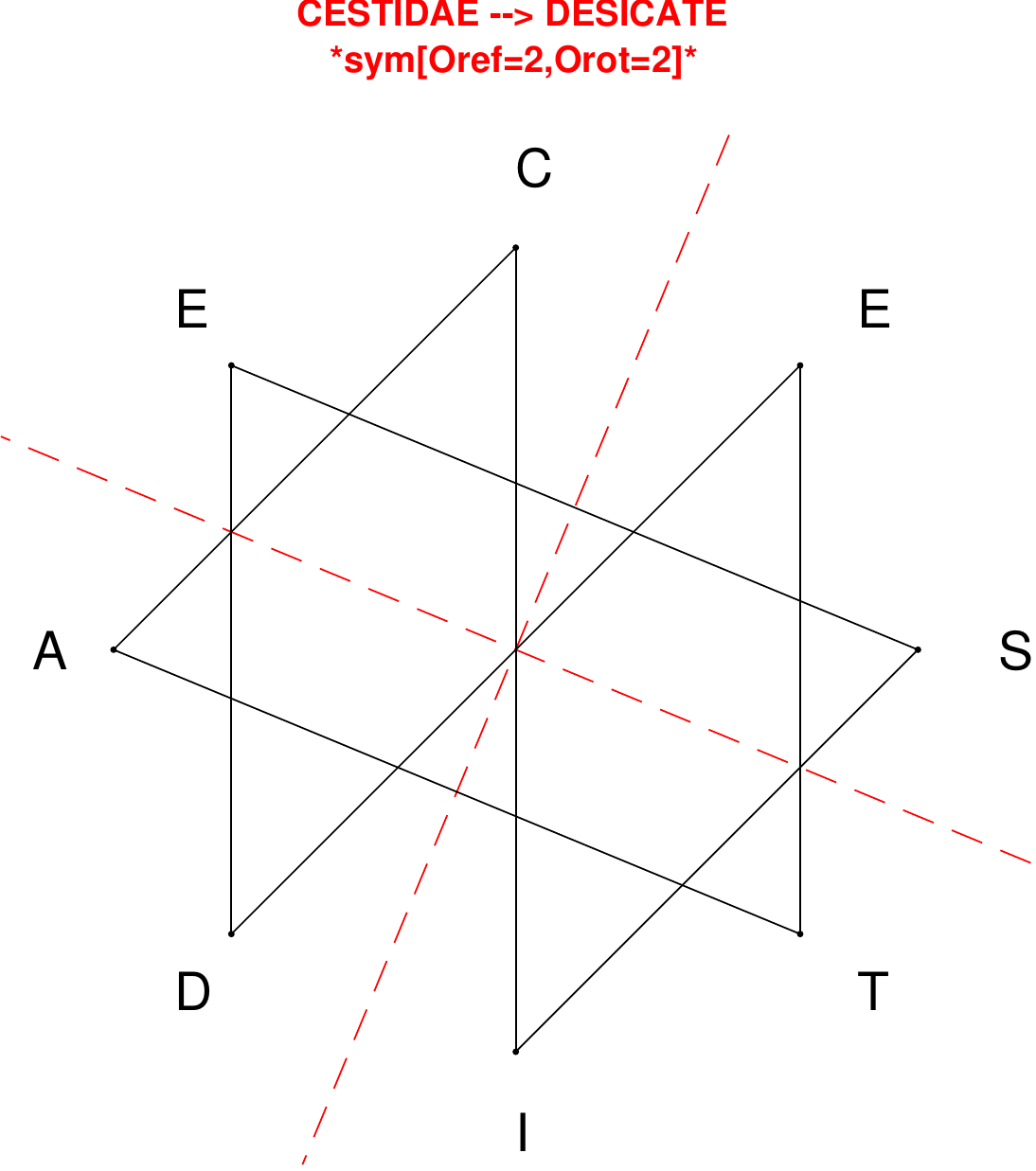}
\end{subfigure}
\hfill
\begin{subfigure}[T]{0.19\textwidth}
\centering
\includegraphics[width=\textwidth]{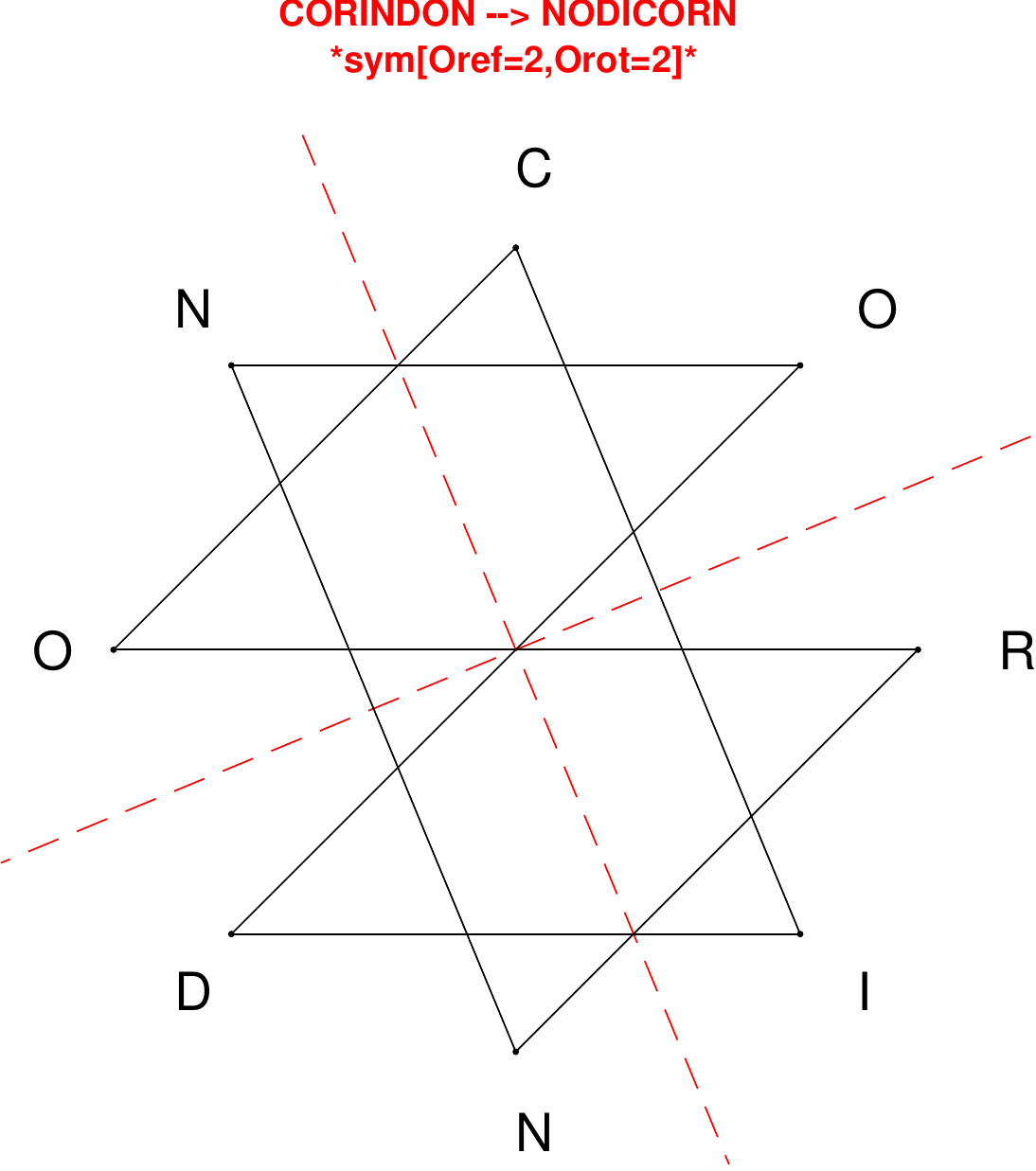}
\end{subfigure}
\hfill
\begin{subfigure}[T]{0.19\textwidth}
\centering
\includegraphics[width=\textwidth]{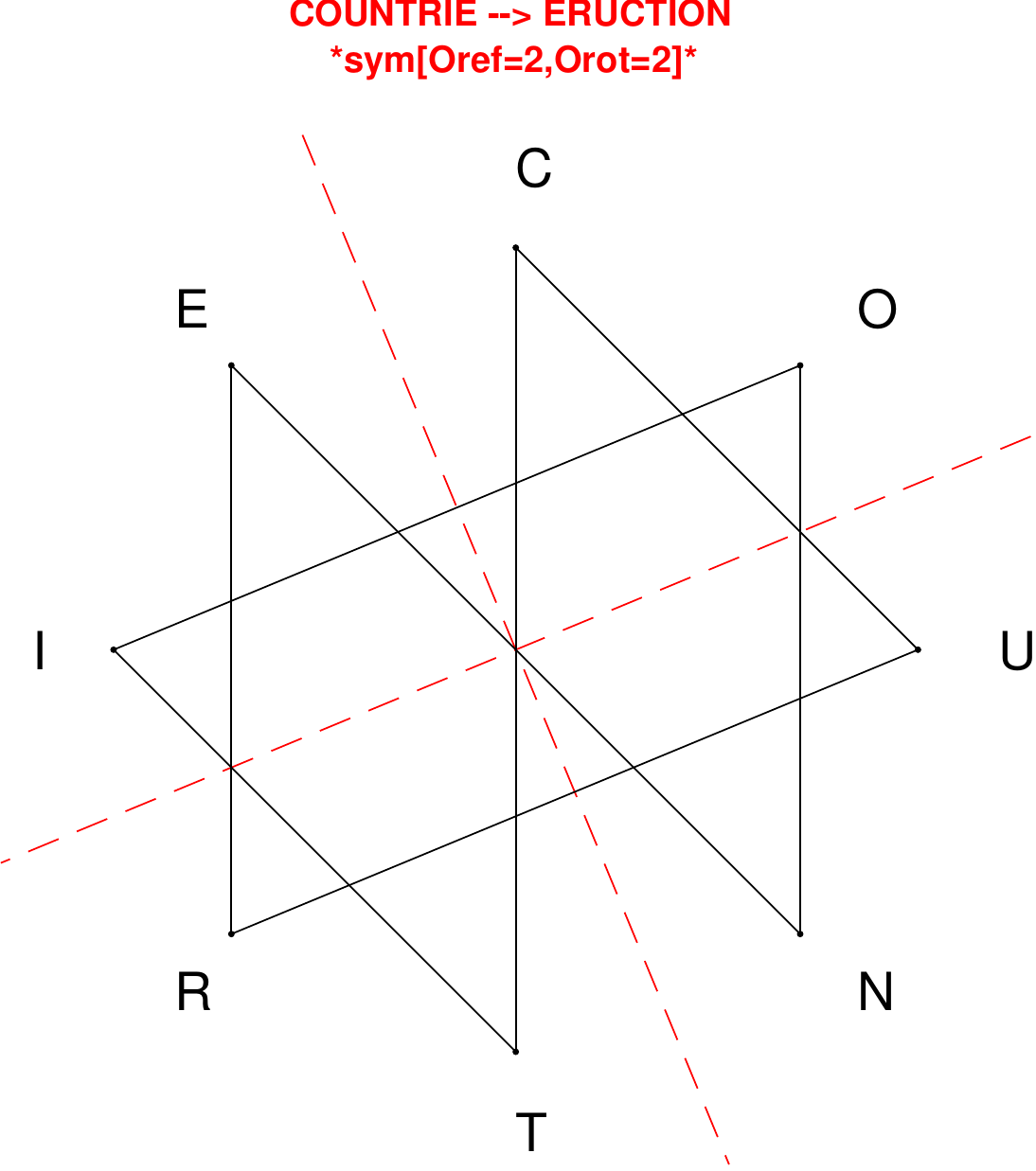}
\end{subfigure}
\hfill
\begin{subfigure}[T]{0.19\textwidth}
\centering
\includegraphics[width=\textwidth]{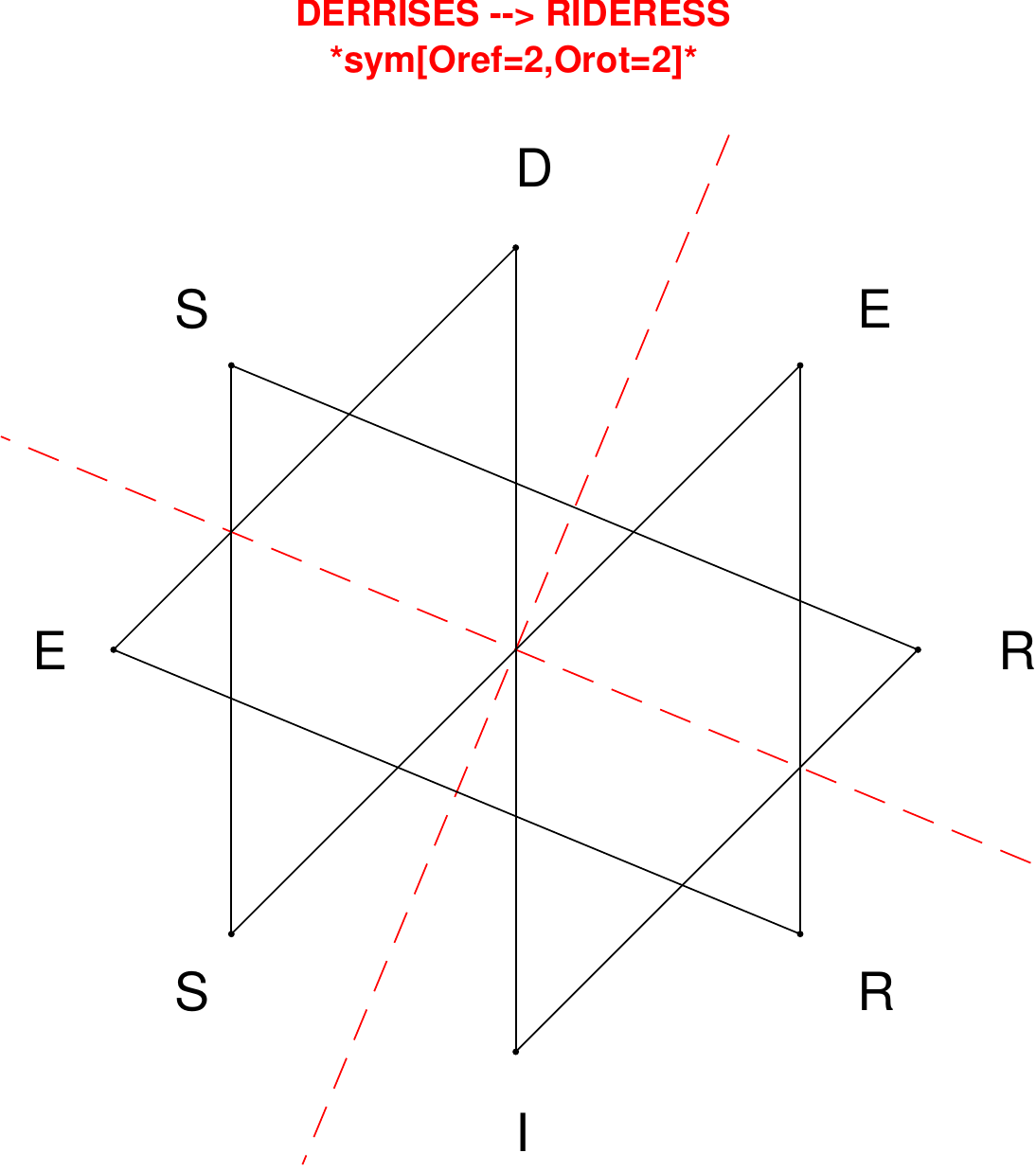}
\end{subfigure}
\hfill
\begin{subfigure}[T]{0.19\textwidth}
\centering
\includegraphics[width=\textwidth]{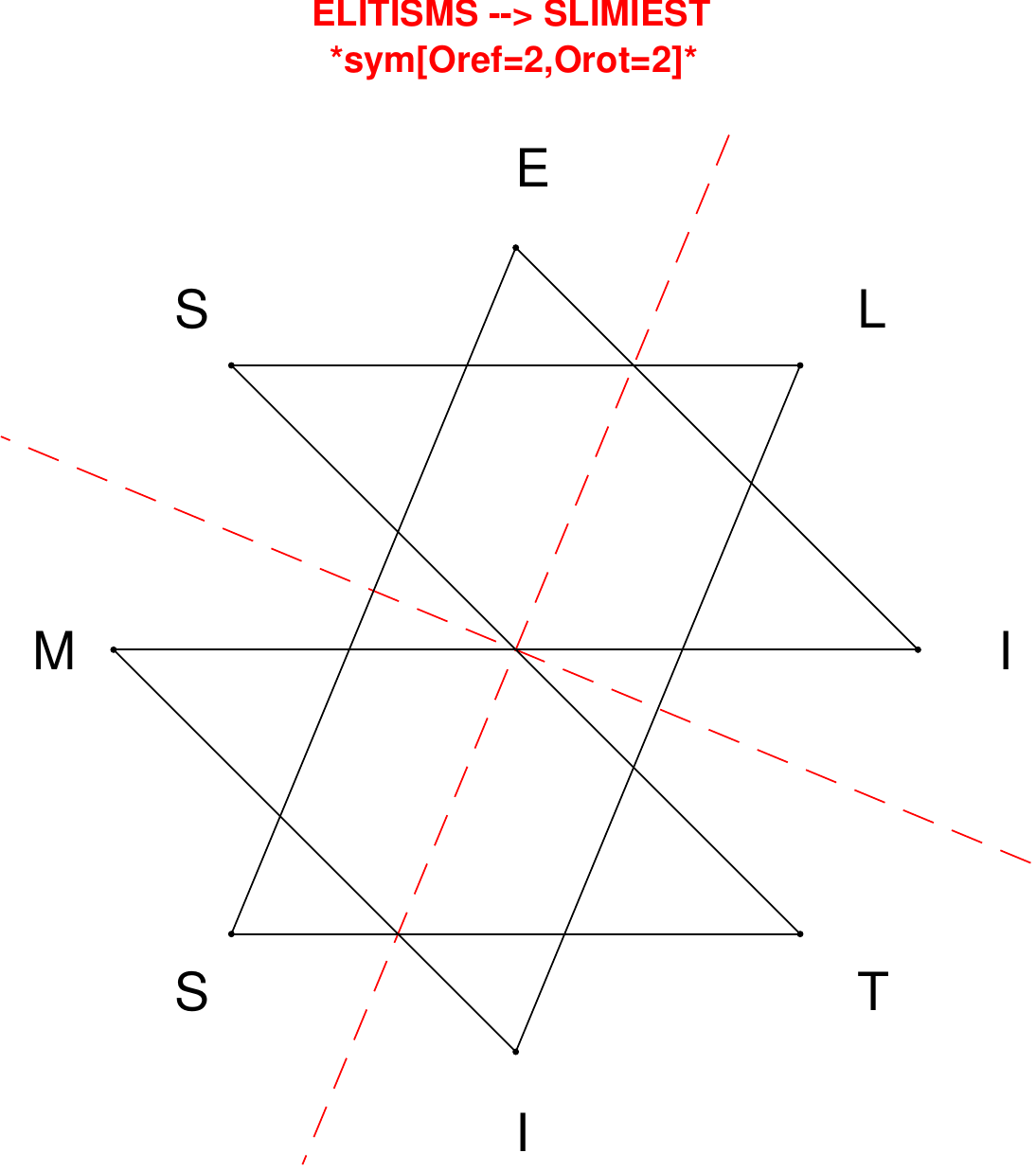}
\end{subfigure}
\end{figure}

\begin{figure}[H]
\centering
\begin{subfigure}[T]{0.19\textwidth}
\centering
\includegraphics[width=\textwidth]{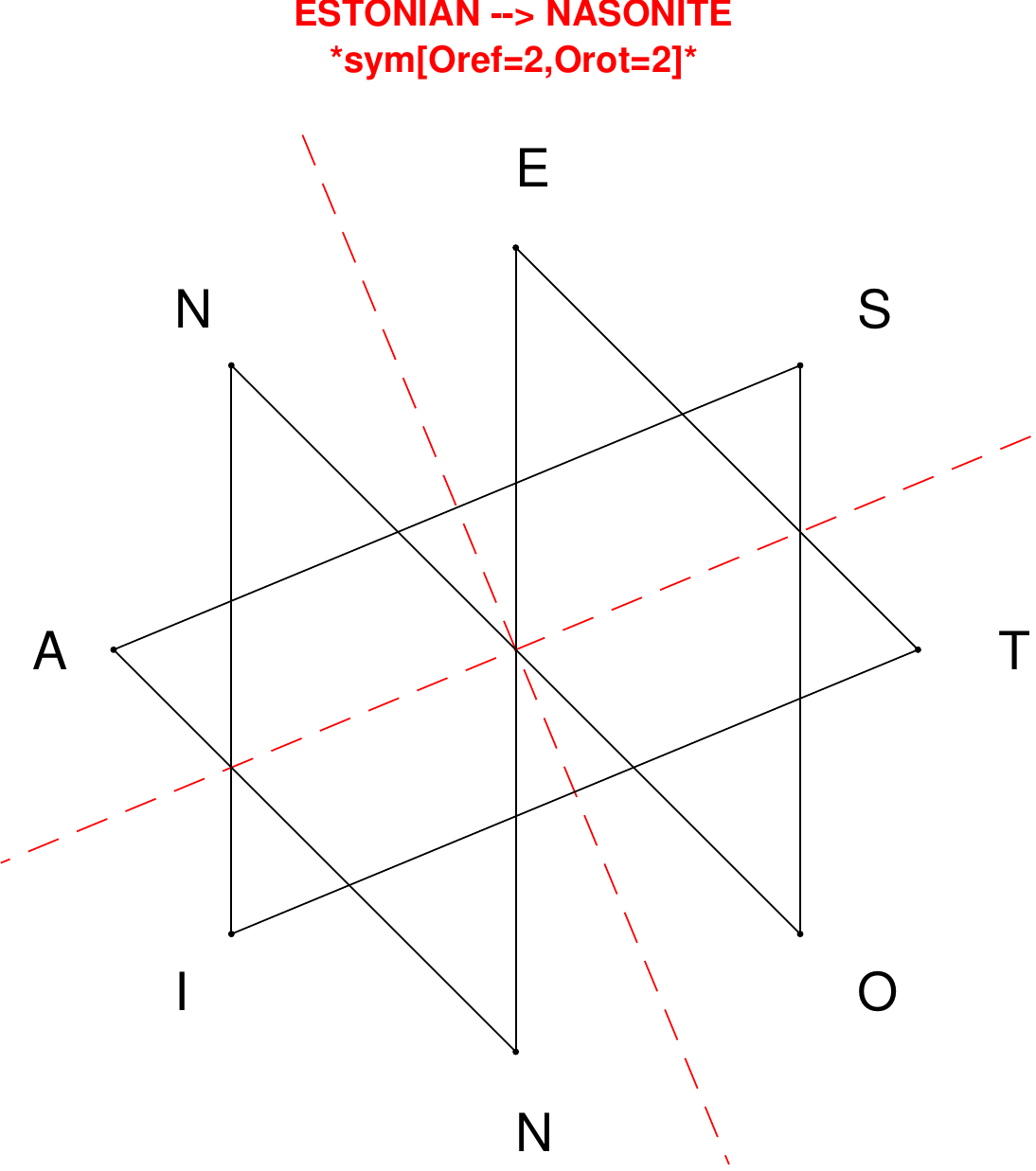}
\end{subfigure}
\hfill
\begin{subfigure}[T]{0.19\textwidth}
\centering
\includegraphics[width=\textwidth]{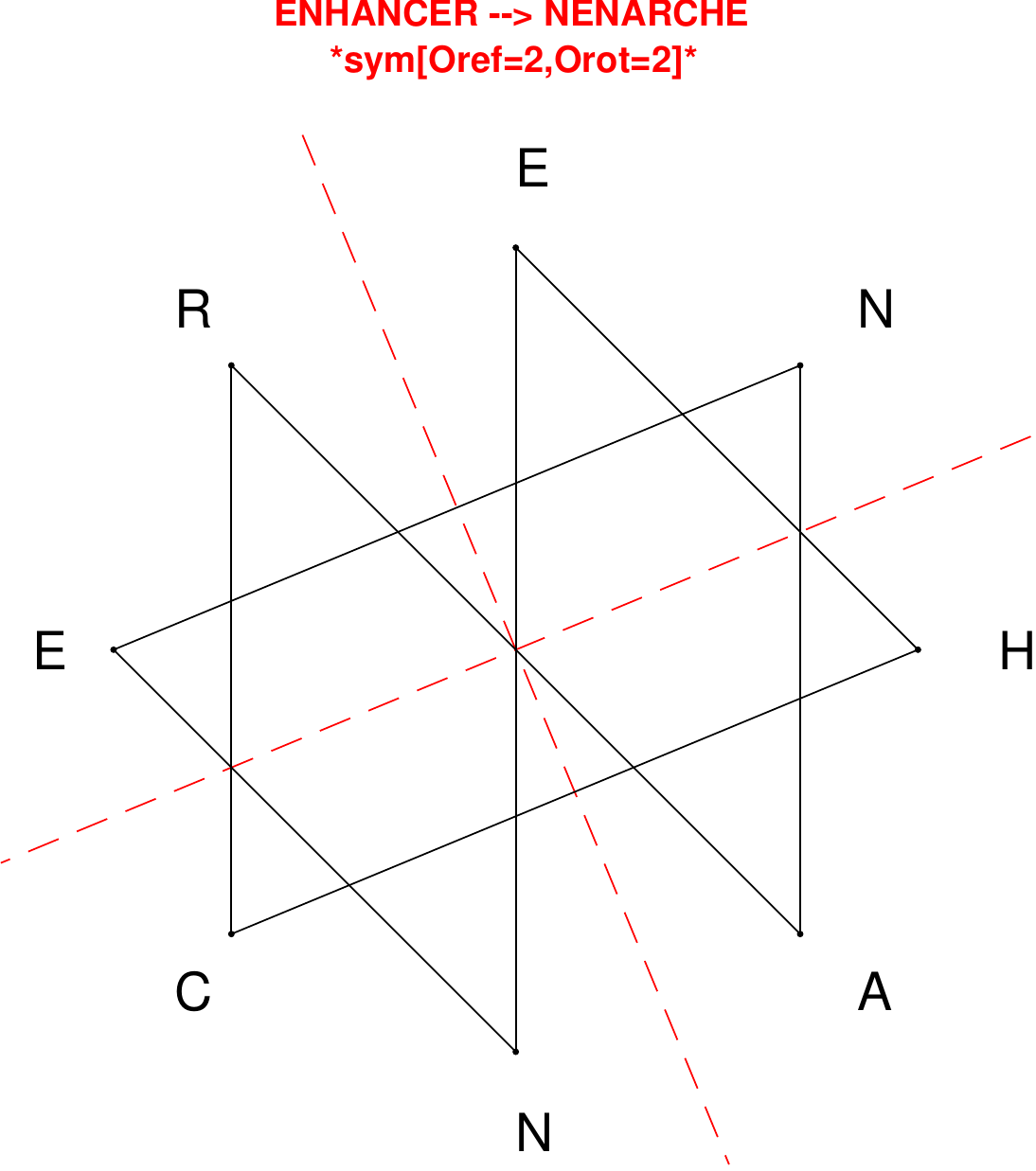}
\end{subfigure}
\hfill
\begin{subfigure}[T]{0.19\textwidth}
\centering
\includegraphics[width=\textwidth]{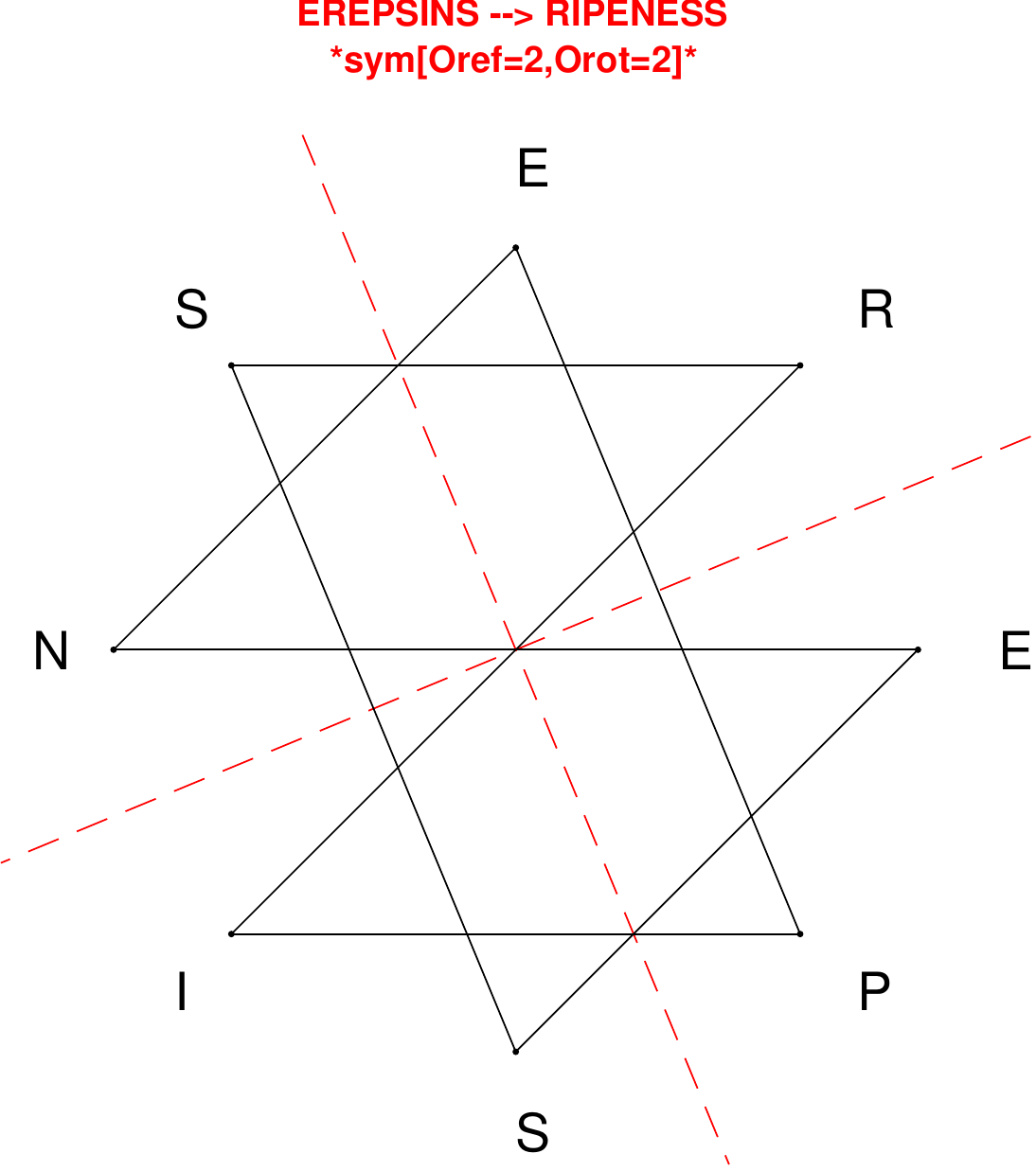}
\end{subfigure}
\hfill
\begin{subfigure}[T]{0.19\textwidth}
\centering
\includegraphics[width=\textwidth]{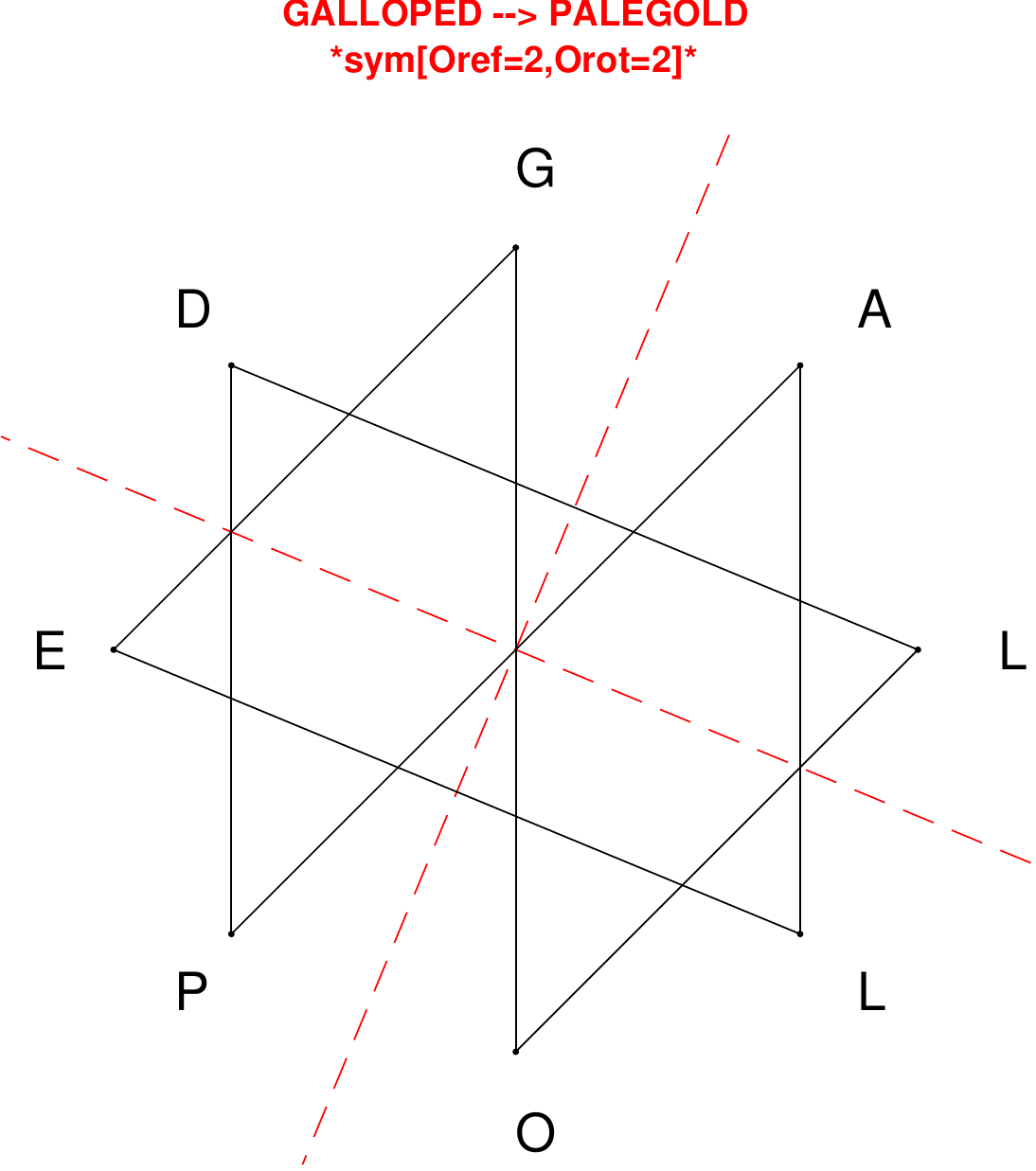}
\end{subfigure}
\hfill
\begin{subfigure}[T]{0.19\textwidth}
\centering
\includegraphics[width=\textwidth]{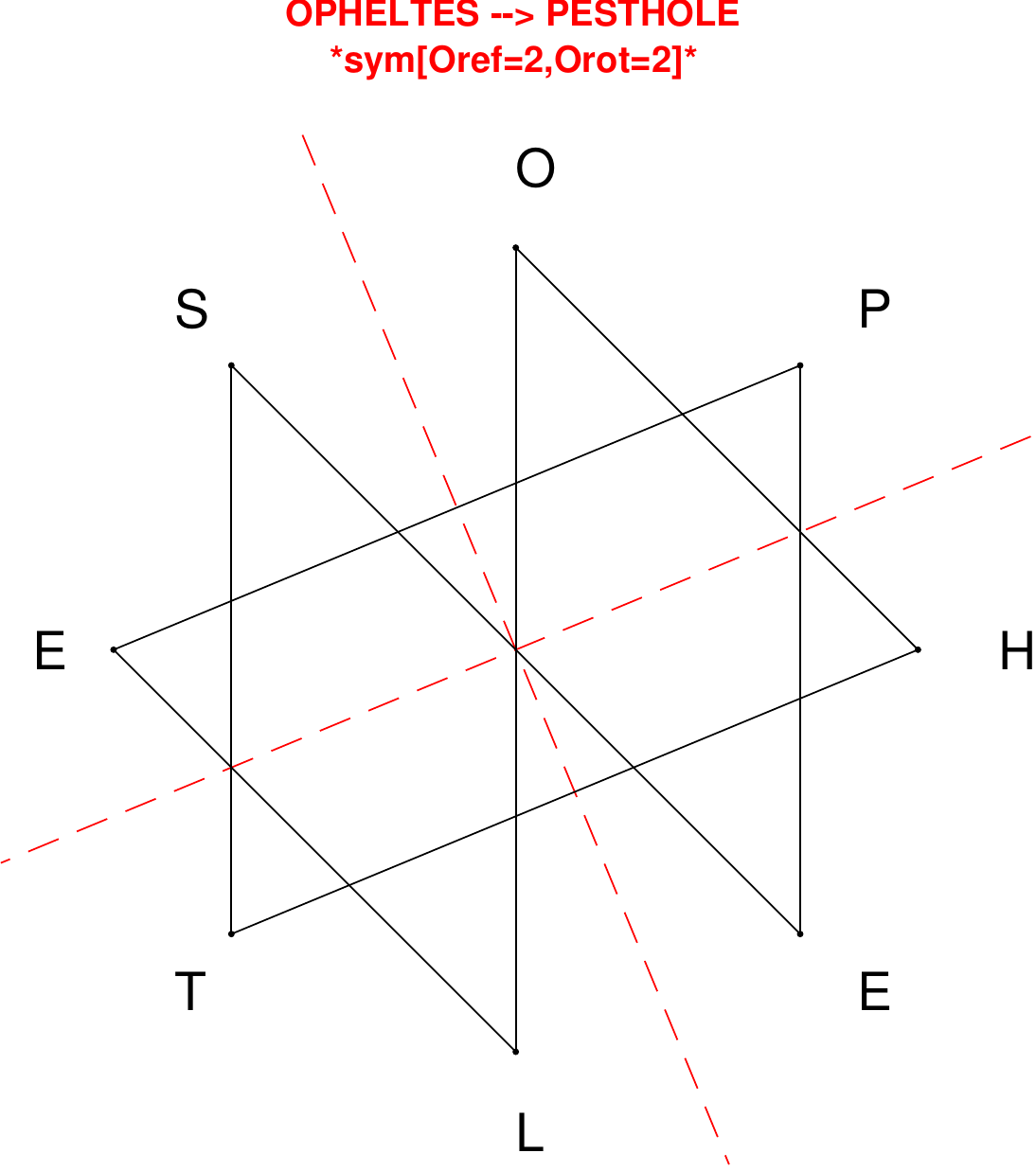}
\end{subfigure}
\end{figure}

\begin{figure}[H]
\centering
\begin{subfigure}[T]{0.19\textwidth}
\centering
\includegraphics[width=\textwidth]{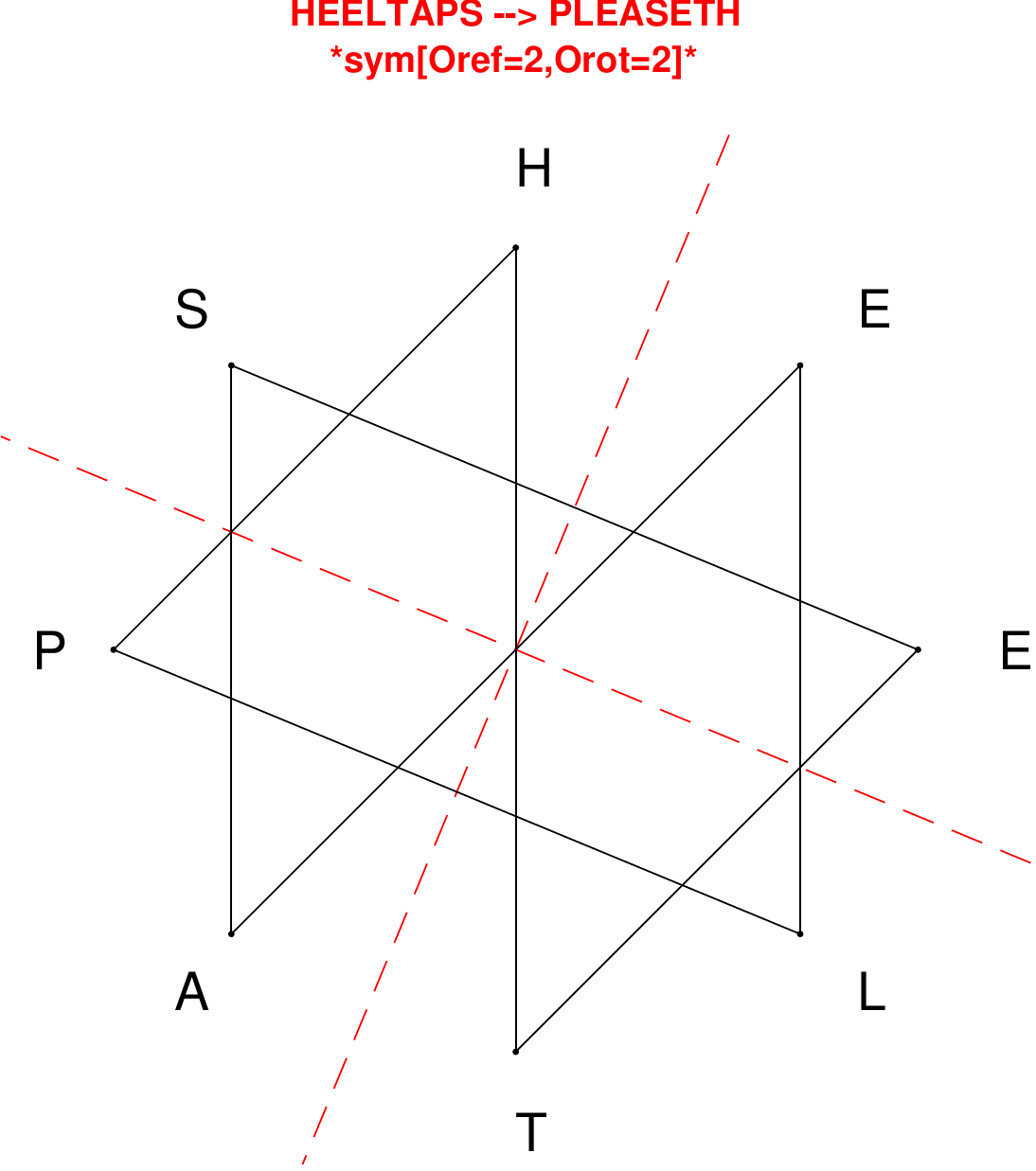}
\end{subfigure}
\hfill
\begin{subfigure}[T]{0.19\textwidth}
\centering
\includegraphics[width=\textwidth]{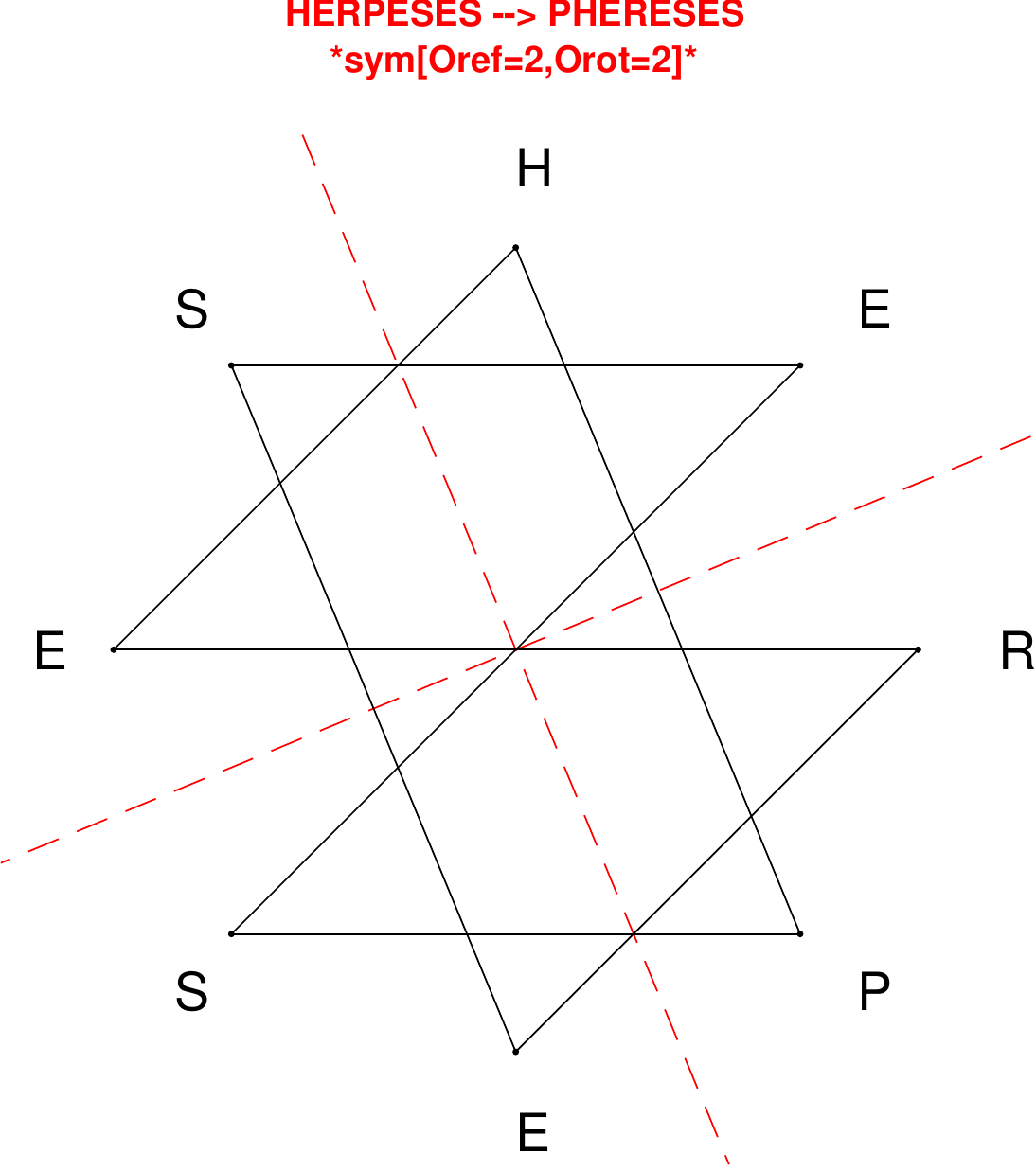}
\end{subfigure}
\hfill
\begin{subfigure}[T]{0.19\textwidth}
\centering
\includegraphics[width=\textwidth]{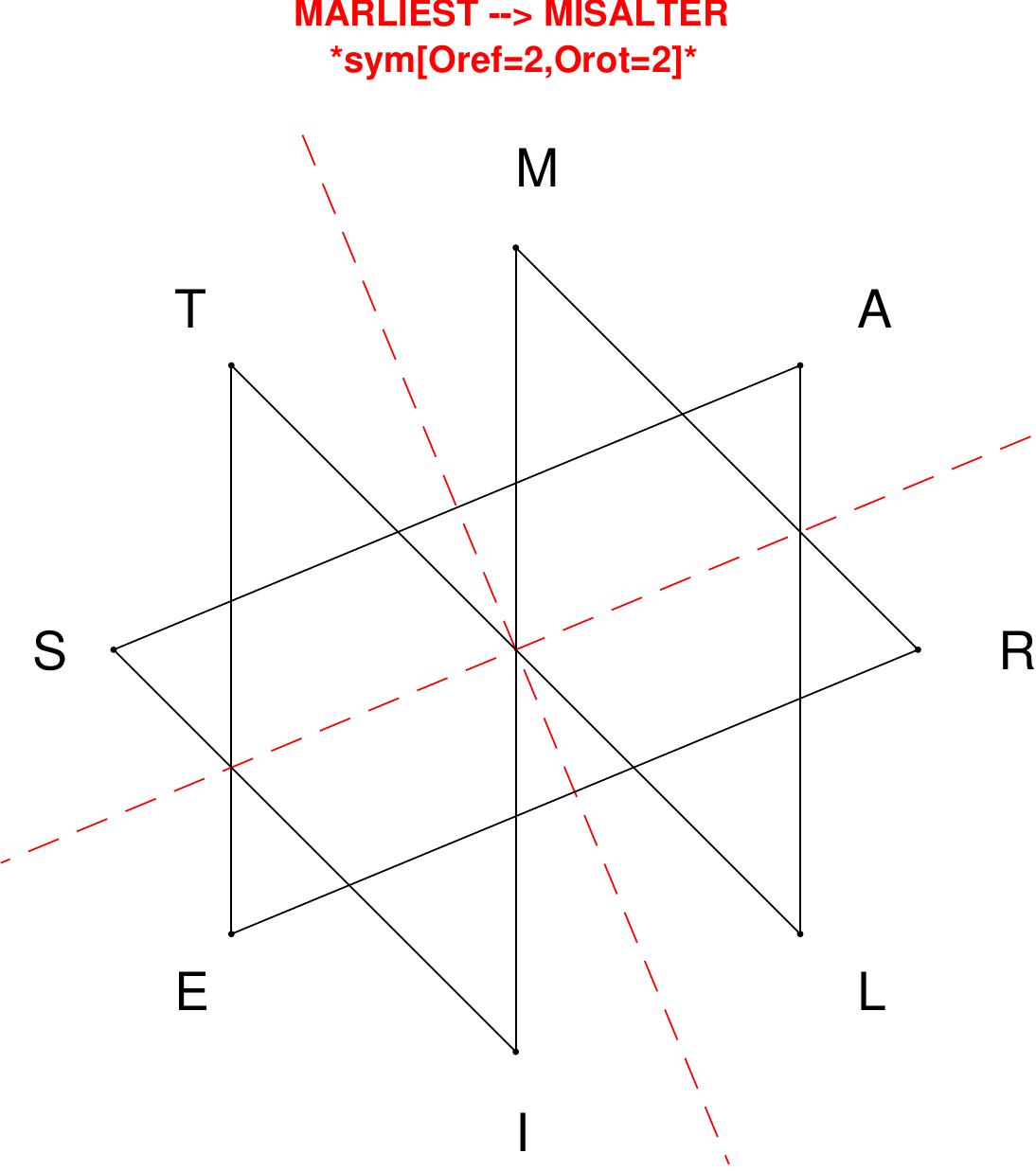}
\end{subfigure}
\hfill
\begin{subfigure}[T]{0.19\textwidth}
\centering
\includegraphics[width=\textwidth]{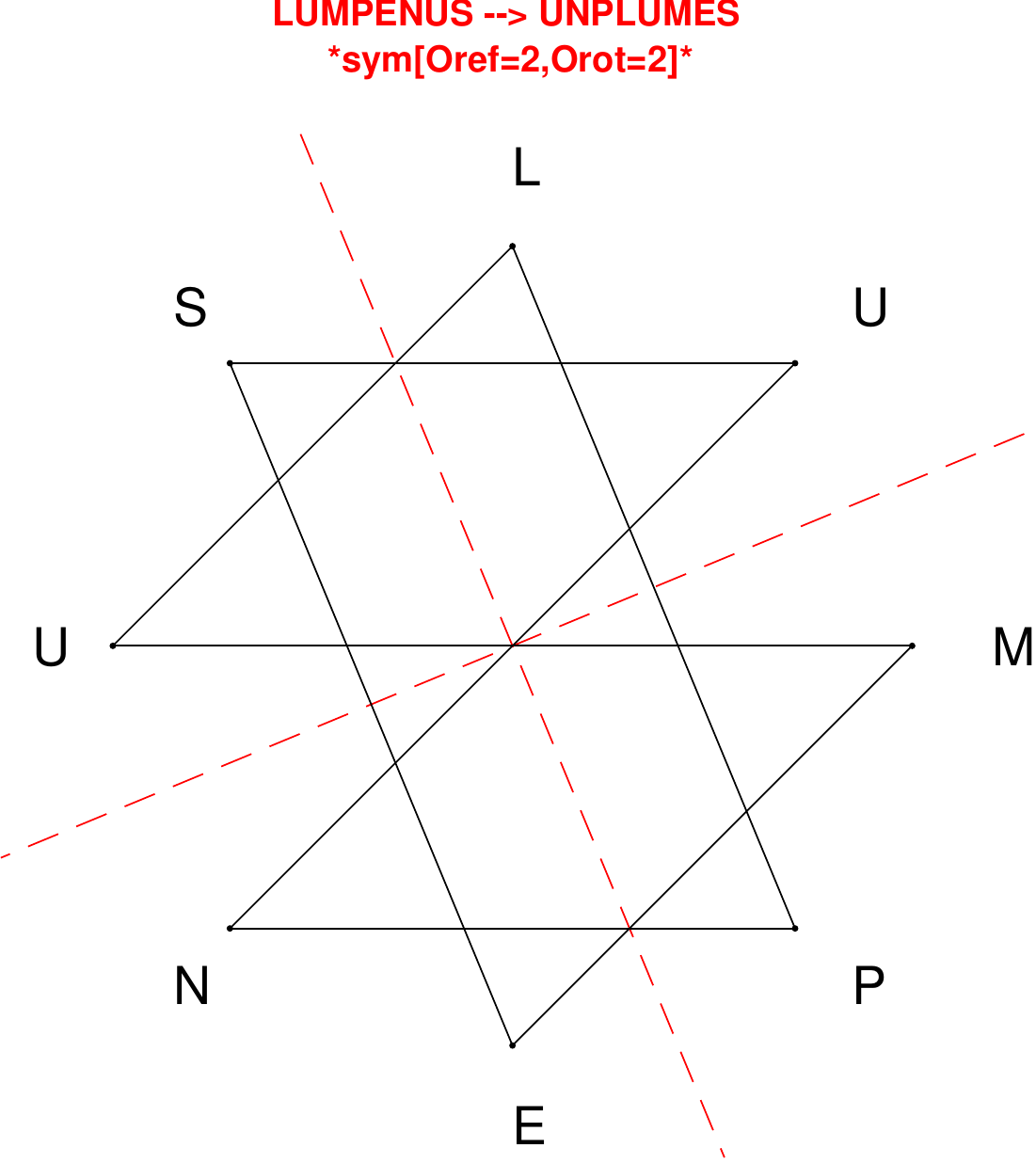}
\end{subfigure}
\hfill
\begin{subfigure}[T]{0.19\textwidth}
\centering
\includegraphics[width=\textwidth]{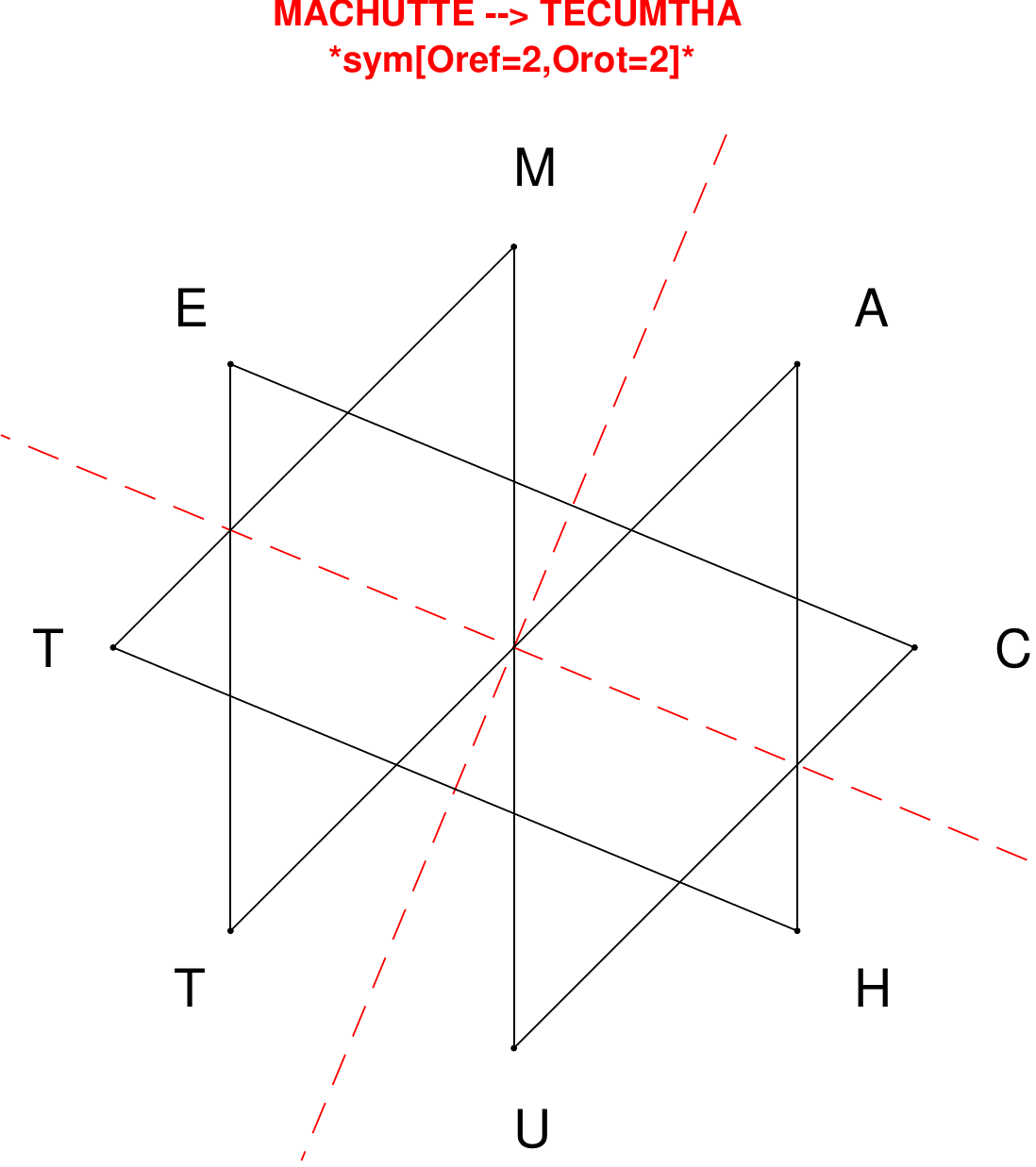}
\end{subfigure}
\end{figure}

\begin{figure}[H]
\centering
\begin{subfigure}[T]{0.19\textwidth}
\centering
\includegraphics[width=\textwidth]{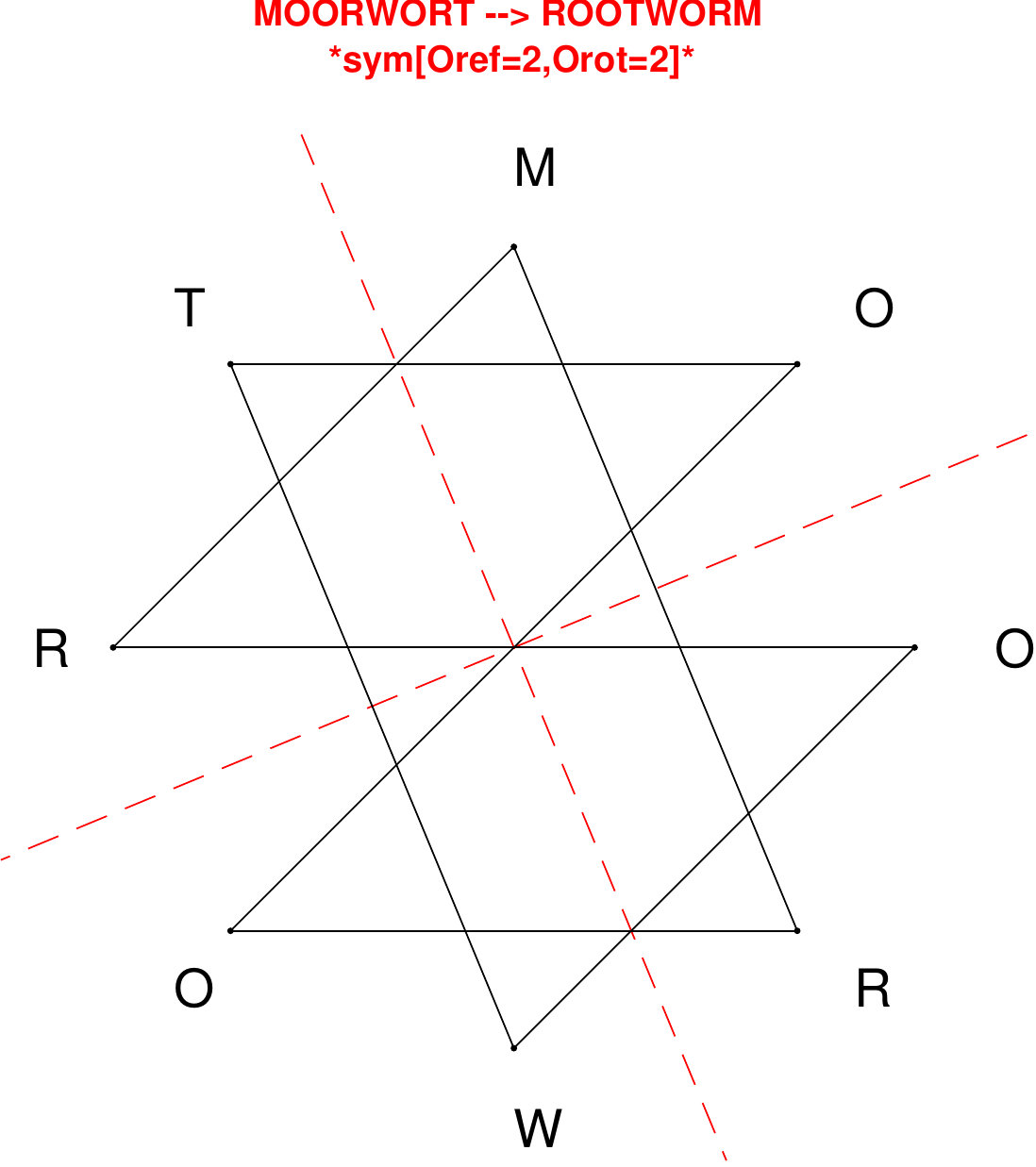}
\end{subfigure}
\hfill
\begin{subfigure}[T]{0.19\textwidth}
\centering
\includegraphics[width=\textwidth]{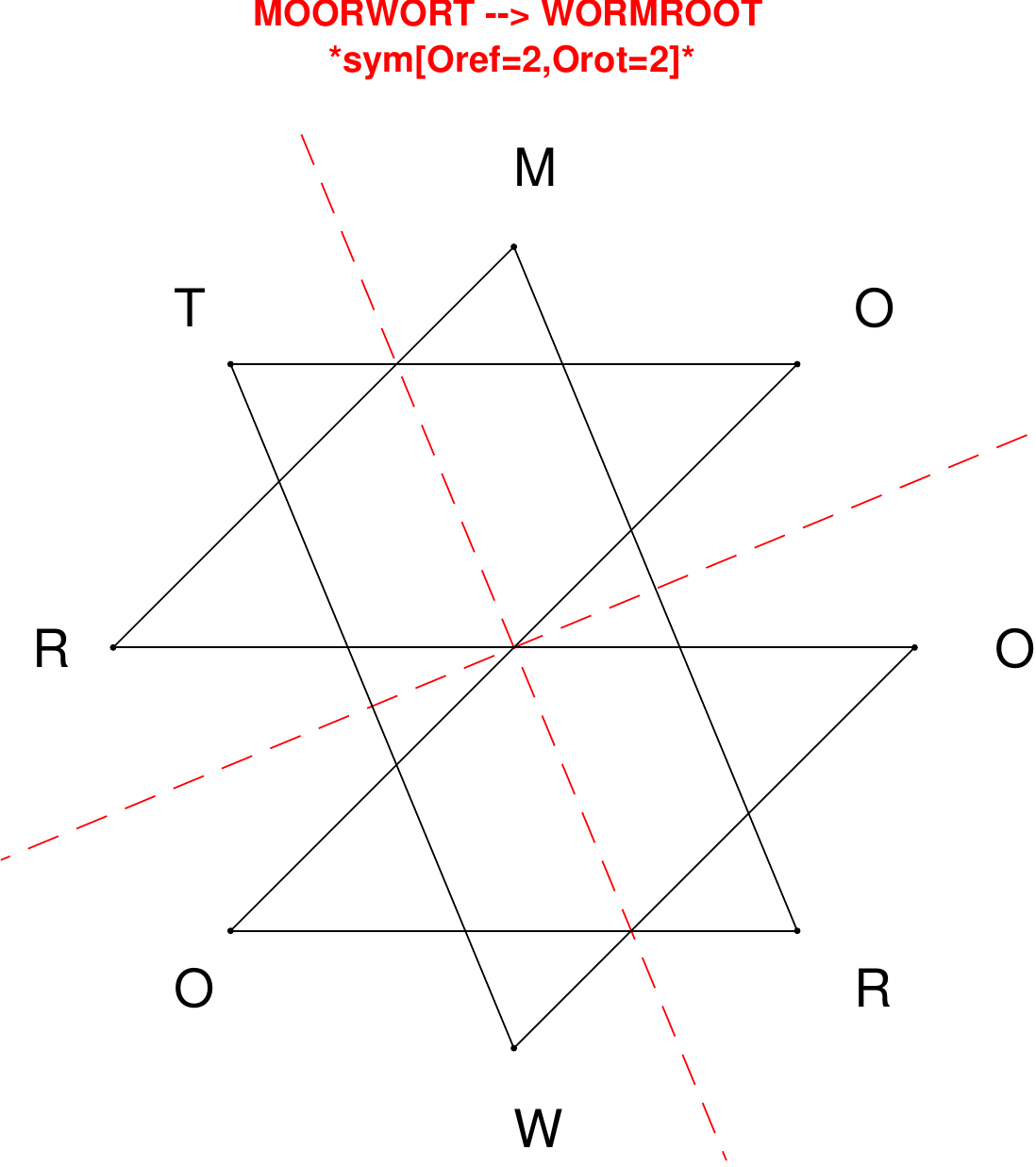}
\end{subfigure}
\hfill
\begin{subfigure}[T]{0.19\textwidth}
\centering
\includegraphics[width=\textwidth]{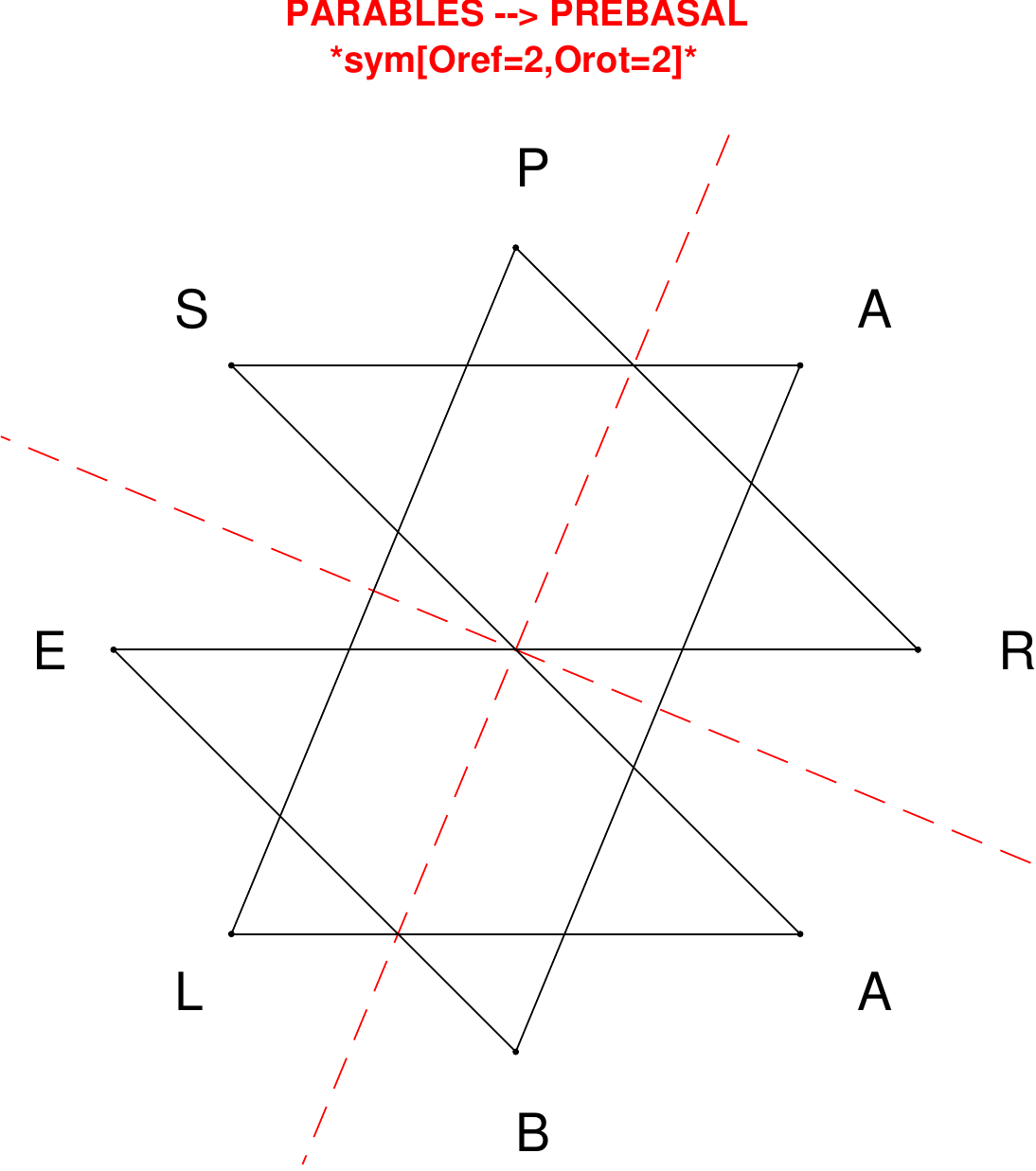}
\end{subfigure}
\hfill
\begin{subfigure}[T]{0.19\textwidth}
\centering
\includegraphics[width=\textwidth]{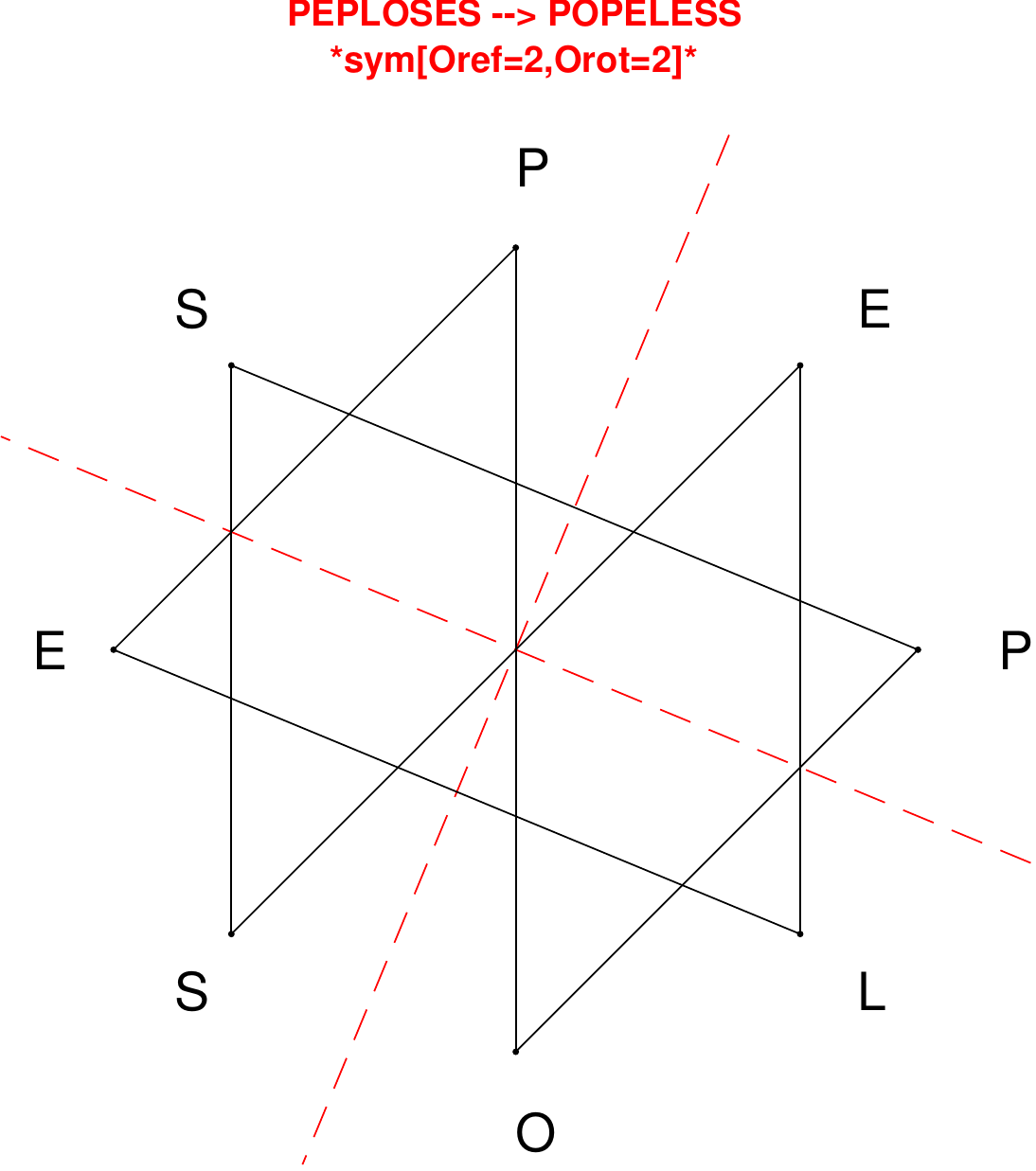}
\end{subfigure}
\hfill
\begin{subfigure}[T]{0.19\textwidth}
\centering
\includegraphics[width=\textwidth]{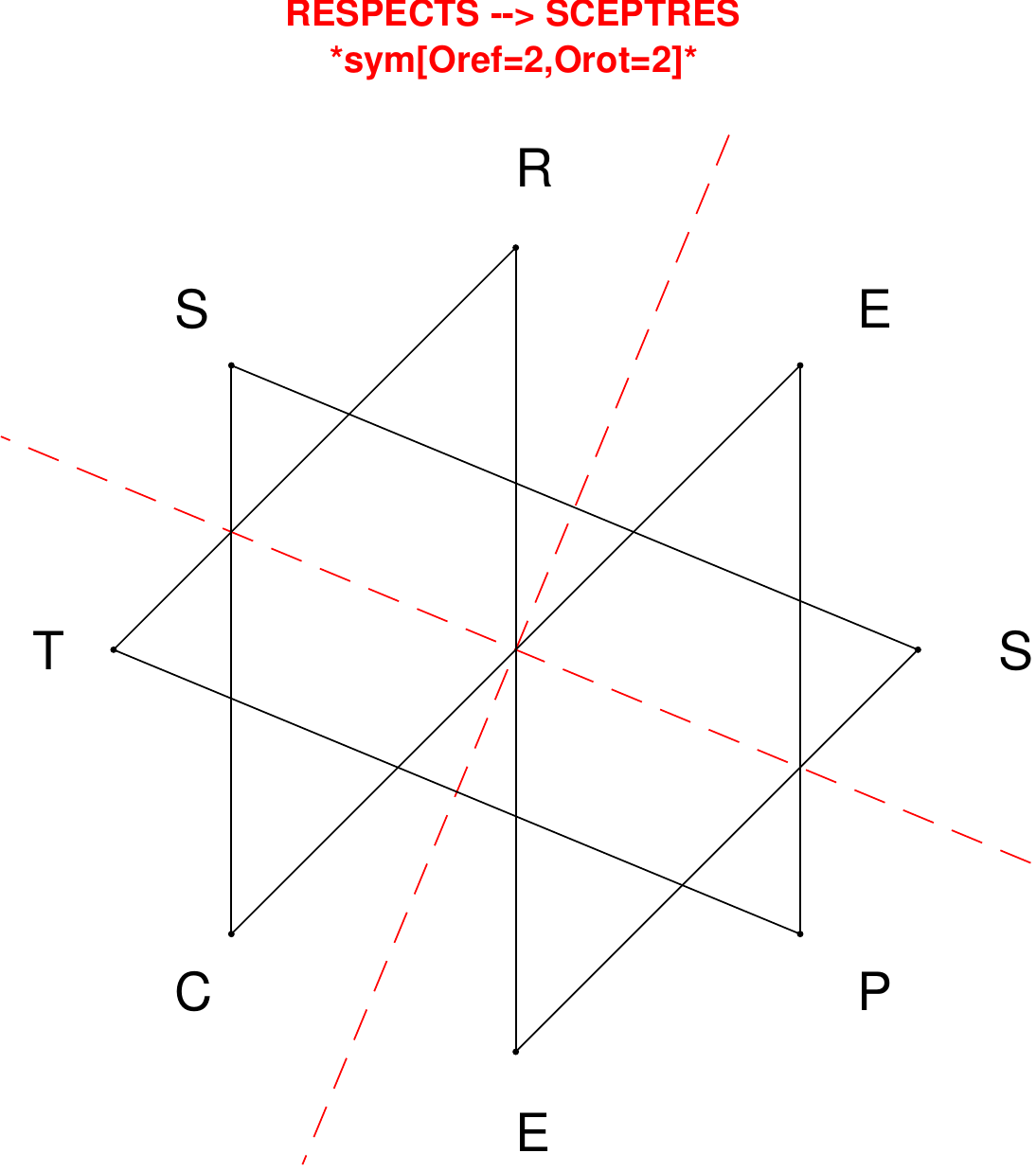}
\end{subfigure}
\end{figure}

\begin{figure}[H]
\centering
\begin{subfigure}[T]{0.19\textwidth}
\centering
\includegraphics[width=\textwidth]{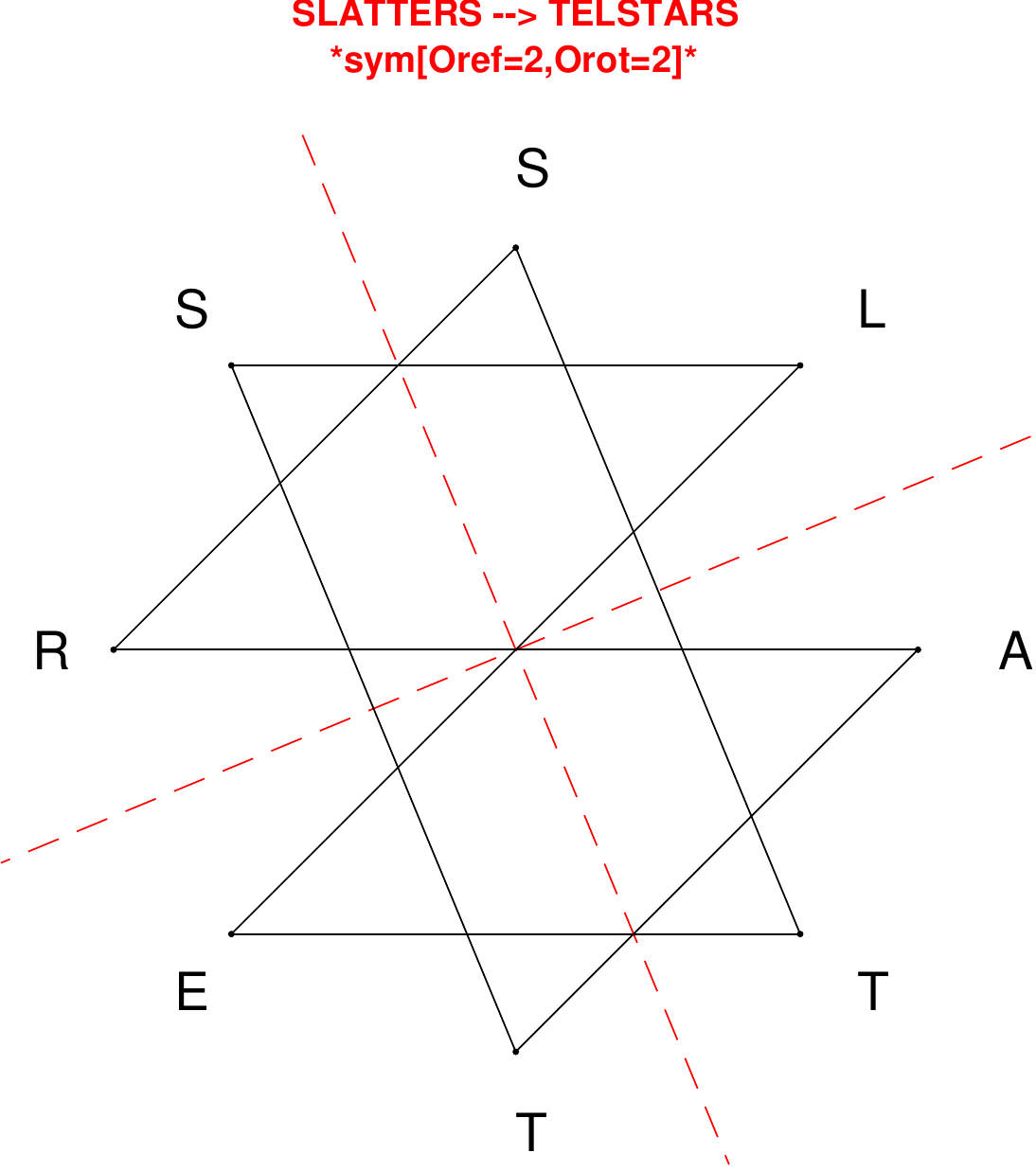}
\end{subfigure}
\hfill
\begin{subfigure}[T]{0.19\textwidth}
\centering
\includegraphics[width=\textwidth]{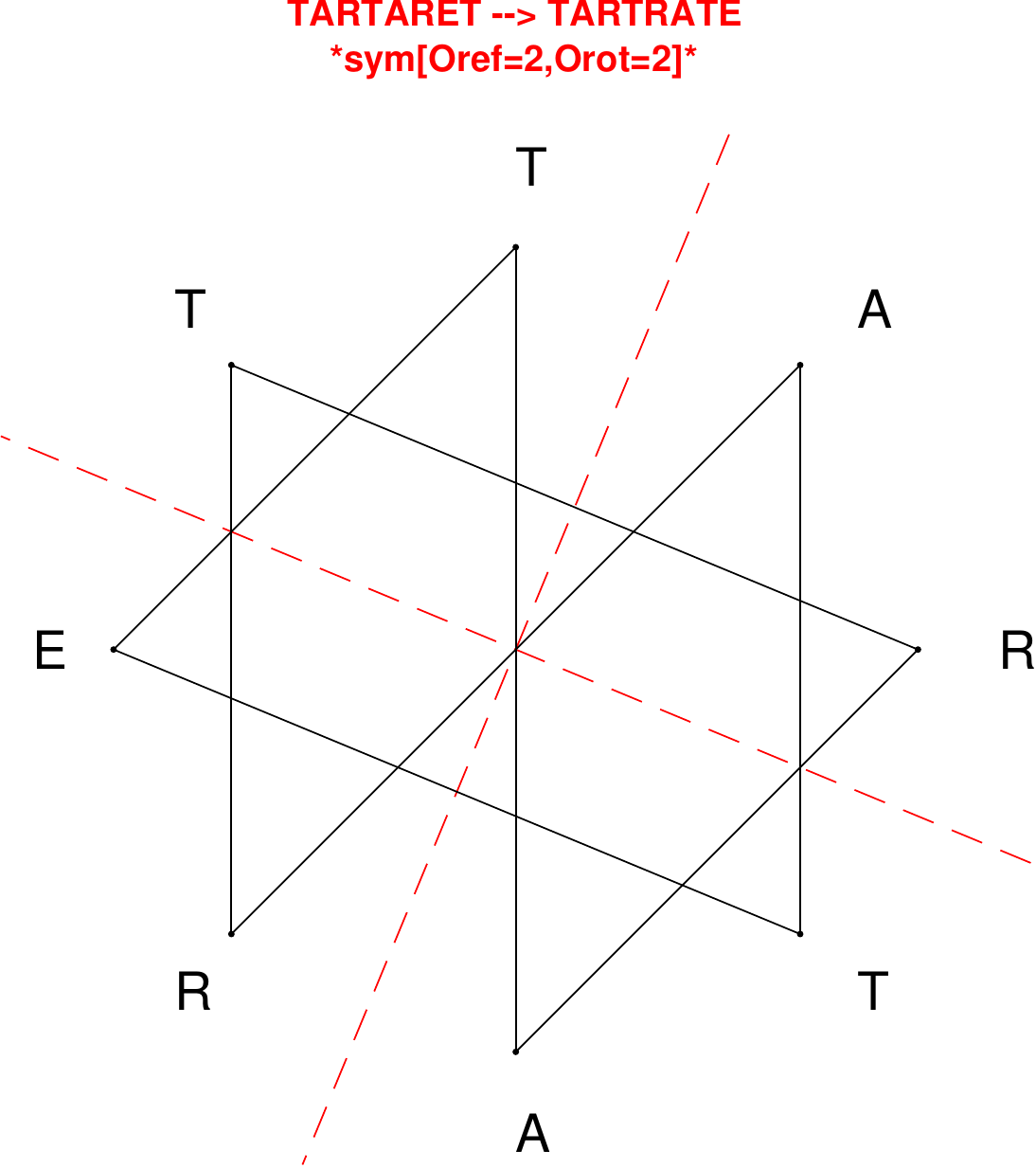}
\end{subfigure}
\hfill
\begin{subfigure}[T]{0.19\textwidth}
\centering
\includegraphics[width=\textwidth]{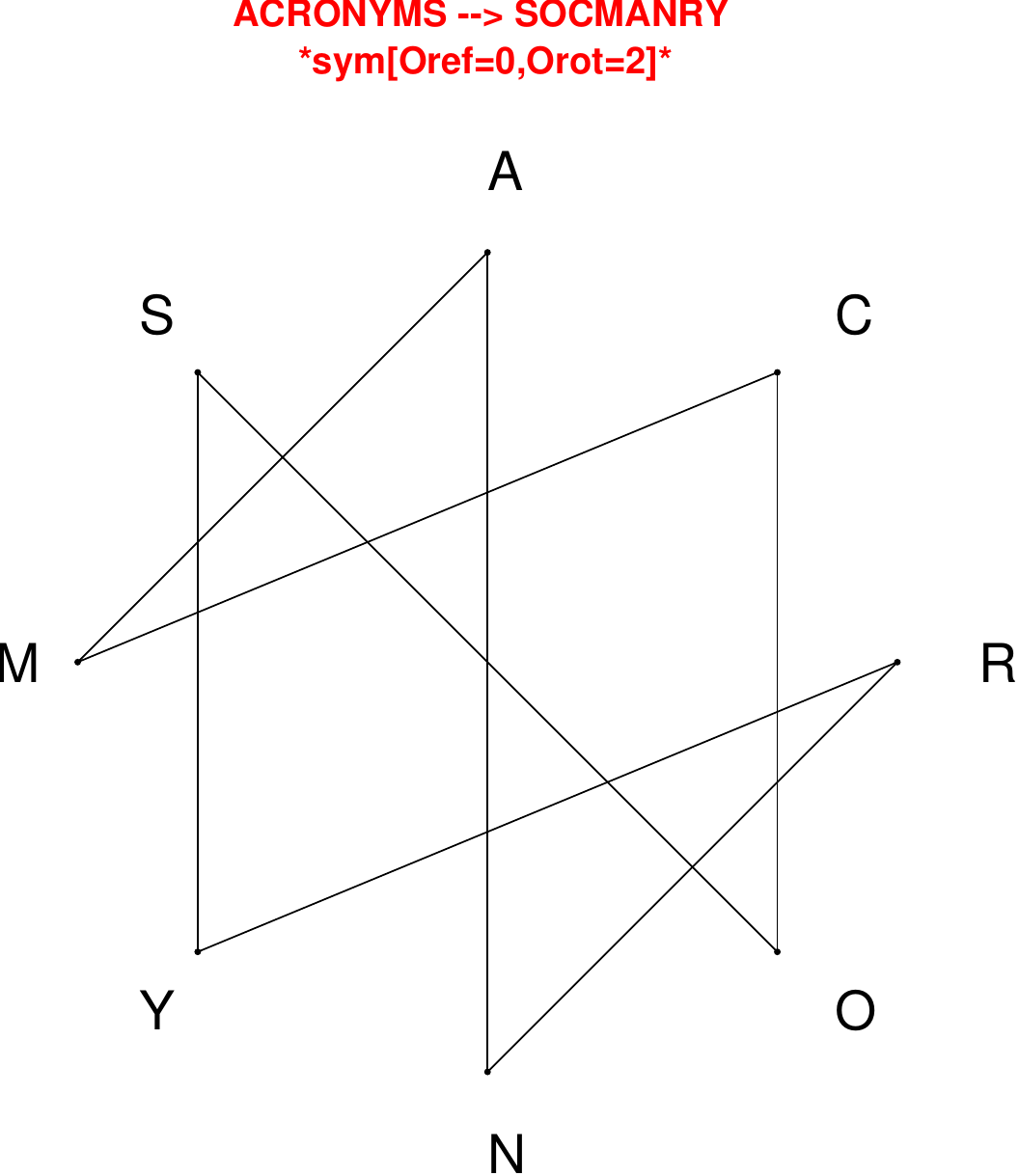}
\end{subfigure}
\hfill
\begin{subfigure}[T]{0.19\textwidth}
\centering
\includegraphics[width=\textwidth]{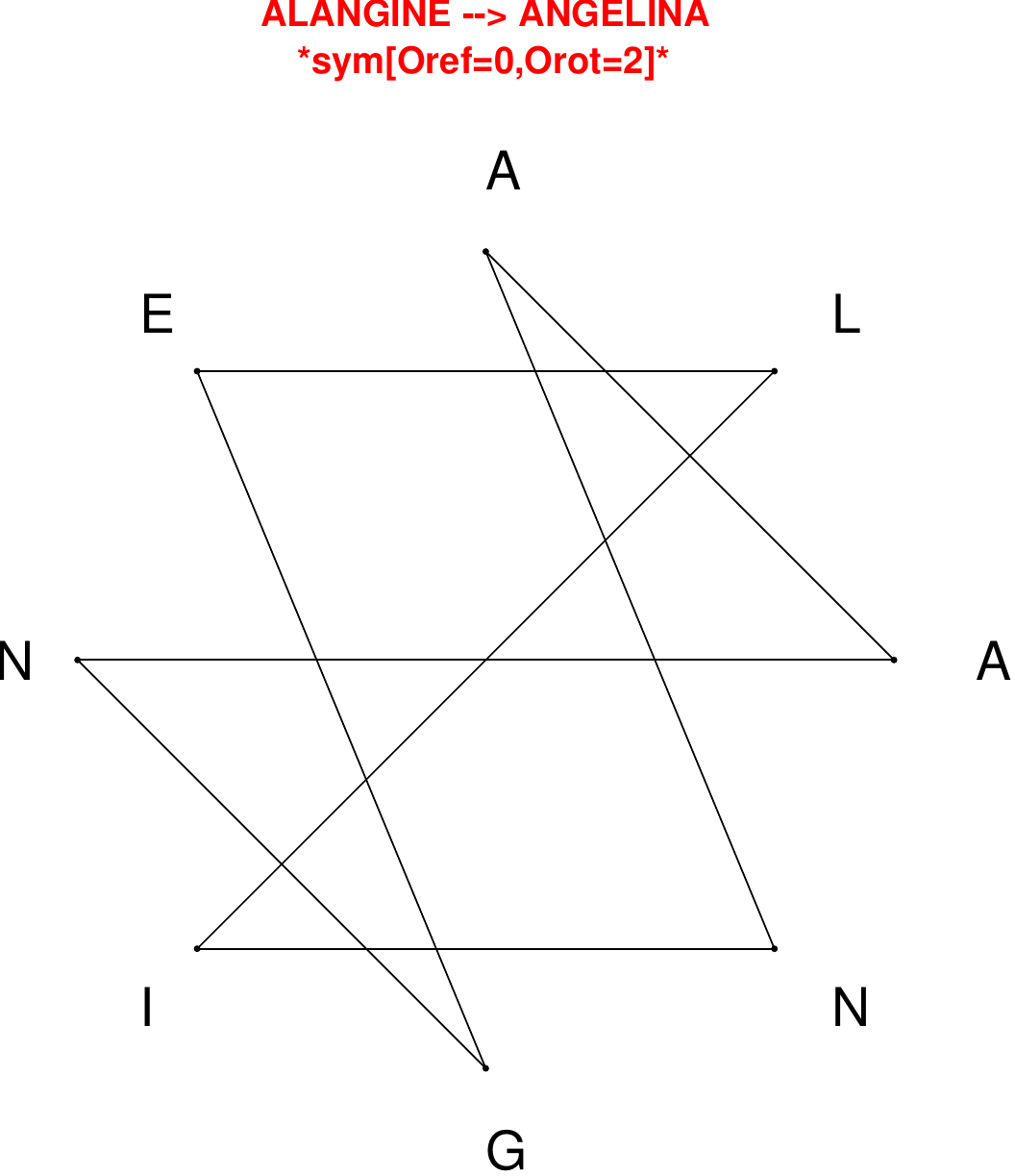}
\end{subfigure}
\hfill
\begin{subfigure}[T]{0.19\textwidth}
\centering
\includegraphics[width=\textwidth]{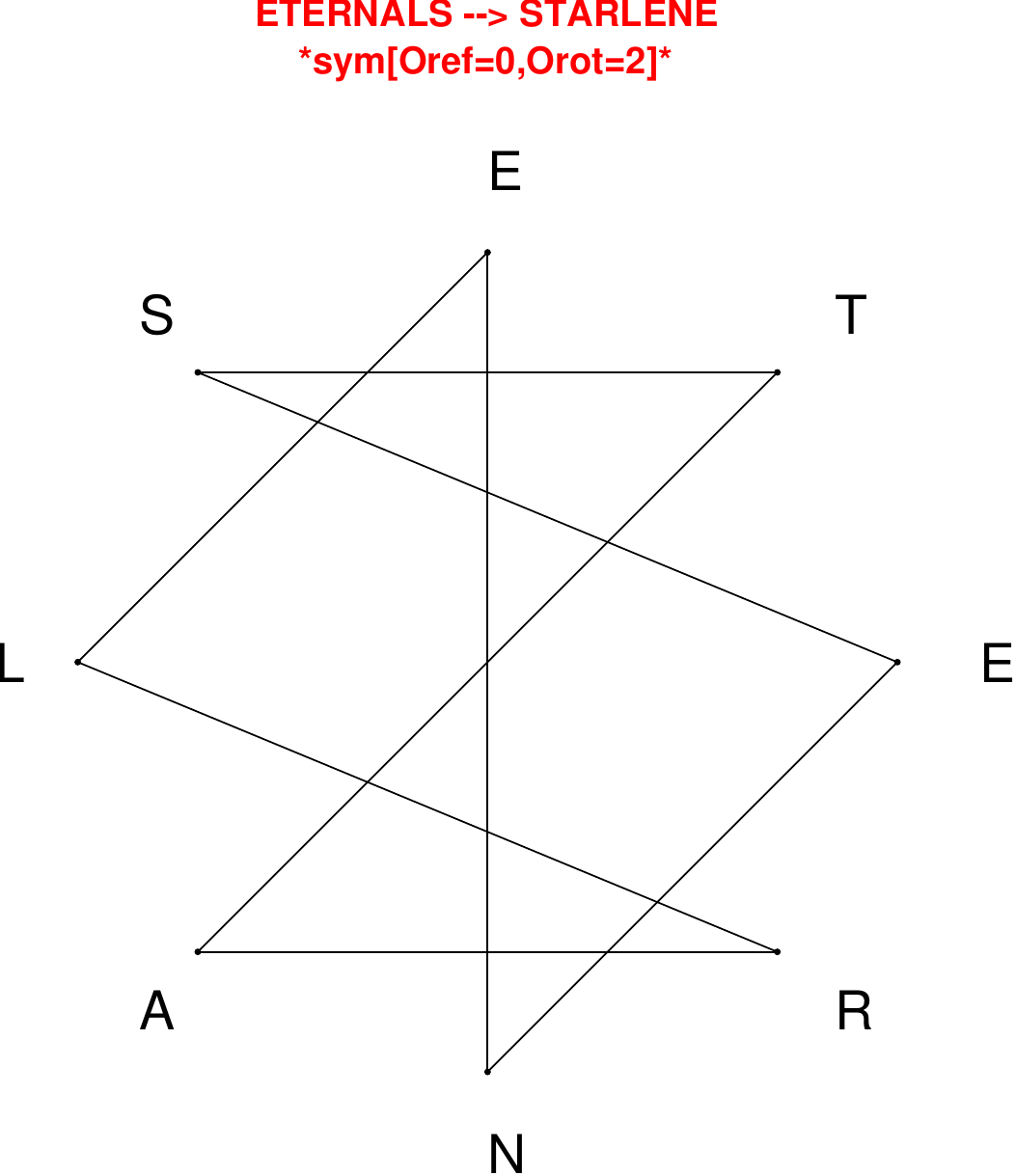}
\end{subfigure}
\end{figure}

\begin{figure}[H]
\centering
\begin{subfigure}[T]{0.19\textwidth}
\centering
\includegraphics[width=\textwidth]{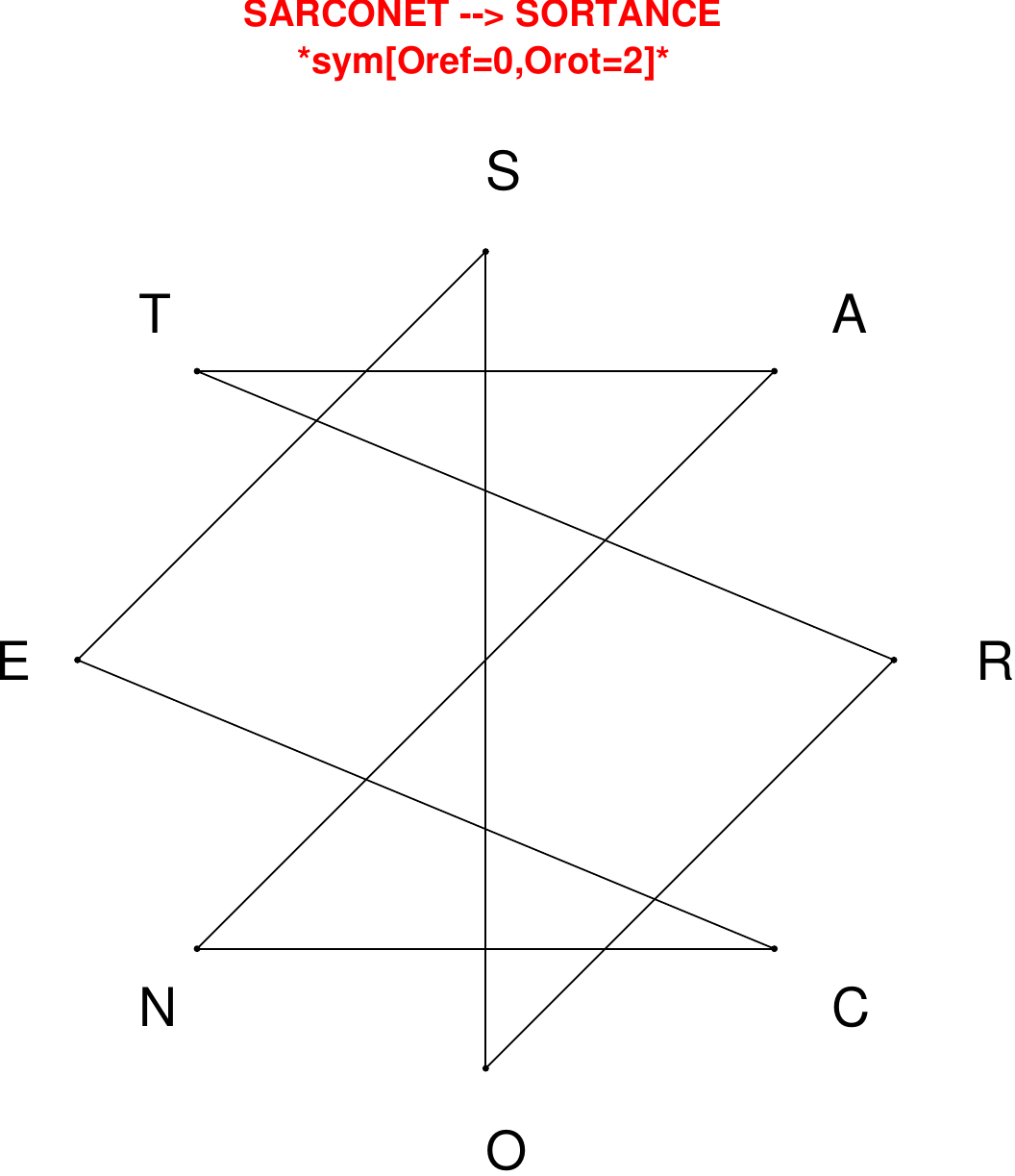}
\end{subfigure}
\hfill
\begin{subfigure}[T]{0.19\textwidth}
\centering
\includegraphics[width=\textwidth]{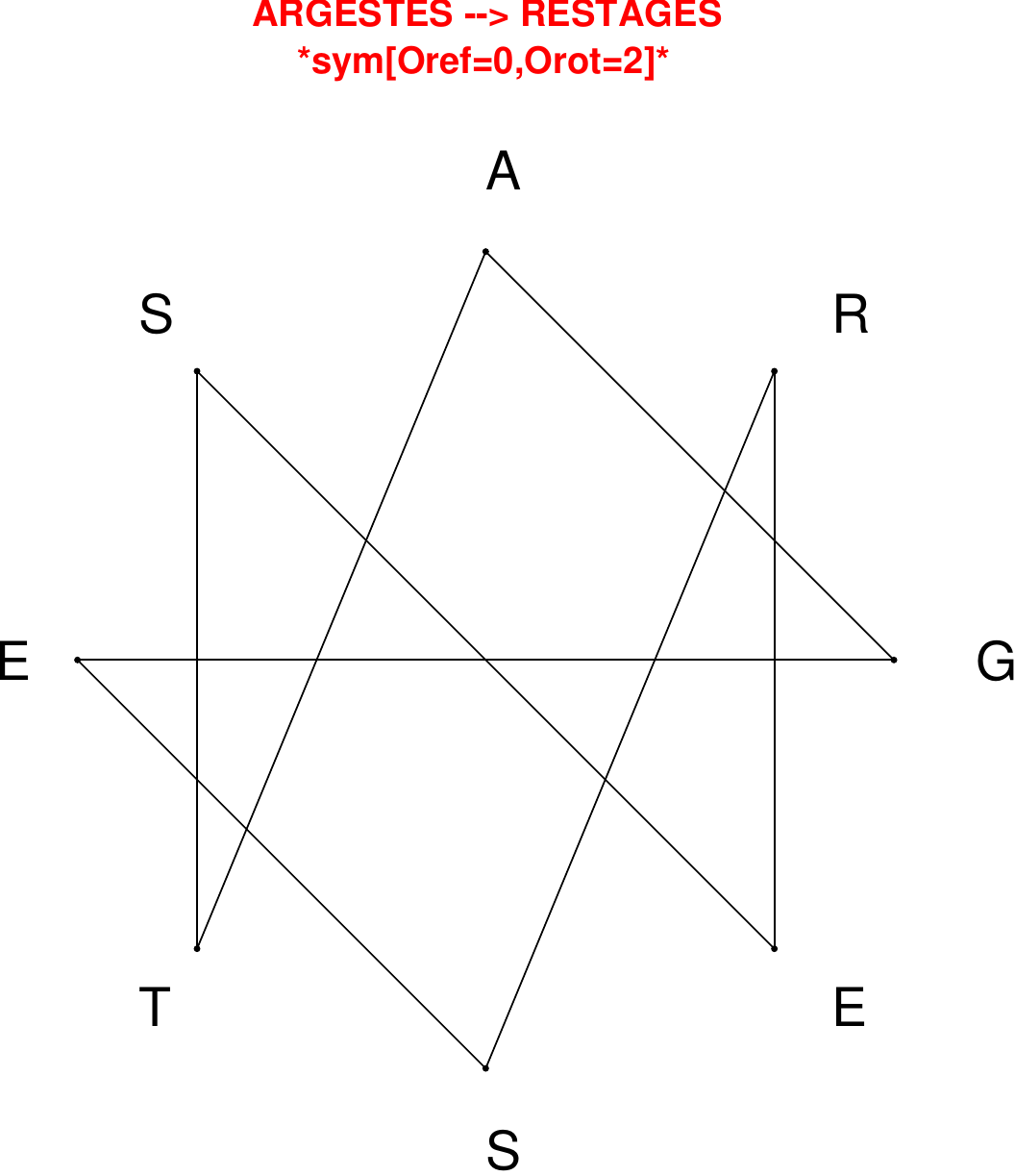}
\end{subfigure}
\hfill
\begin{subfigure}[T]{0.19\textwidth}
\centering
\includegraphics[width=\textwidth]{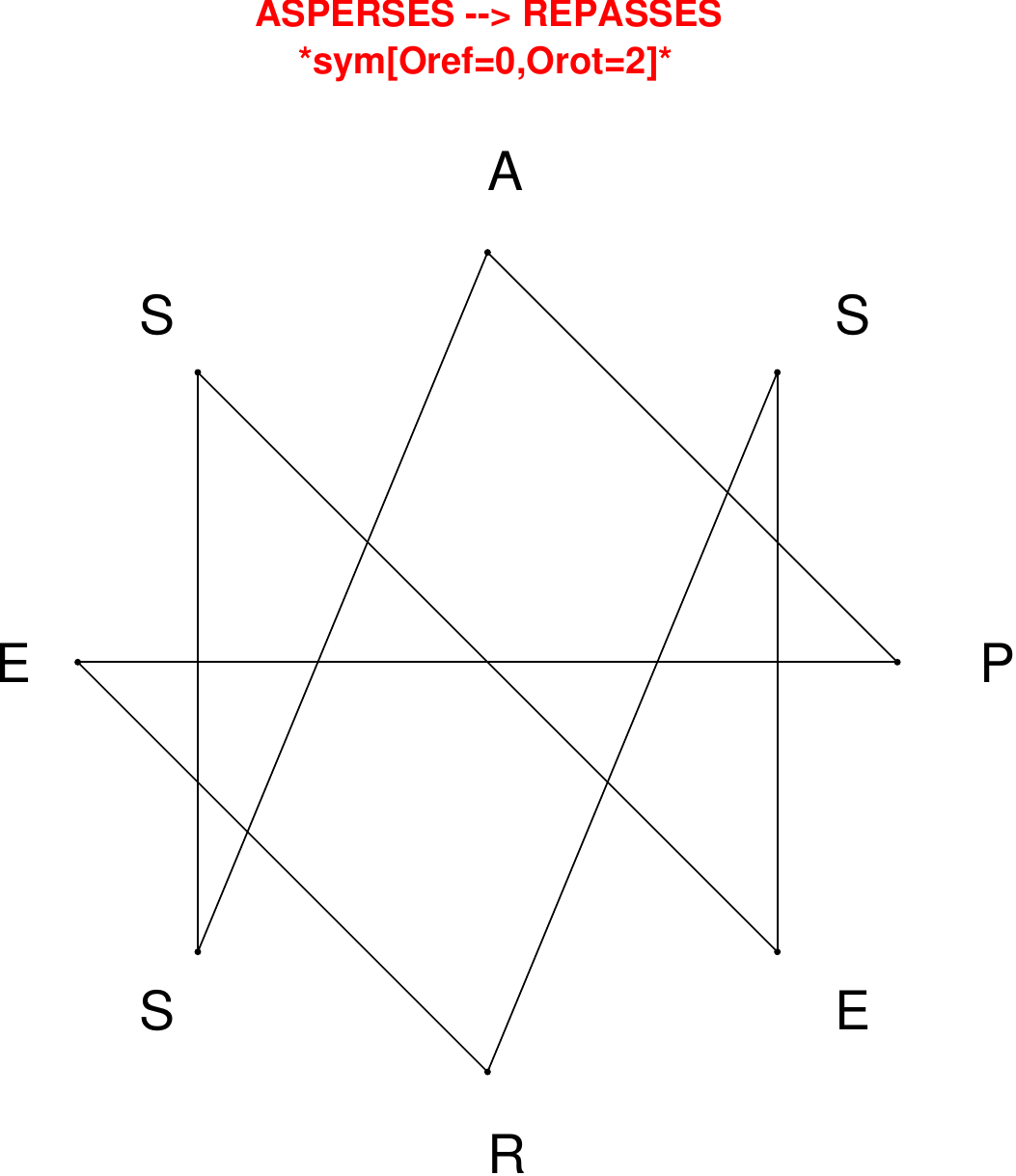}
\end{subfigure}
\hfill
\begin{subfigure}[T]{0.19\textwidth}
\centering
\includegraphics[width=\textwidth]{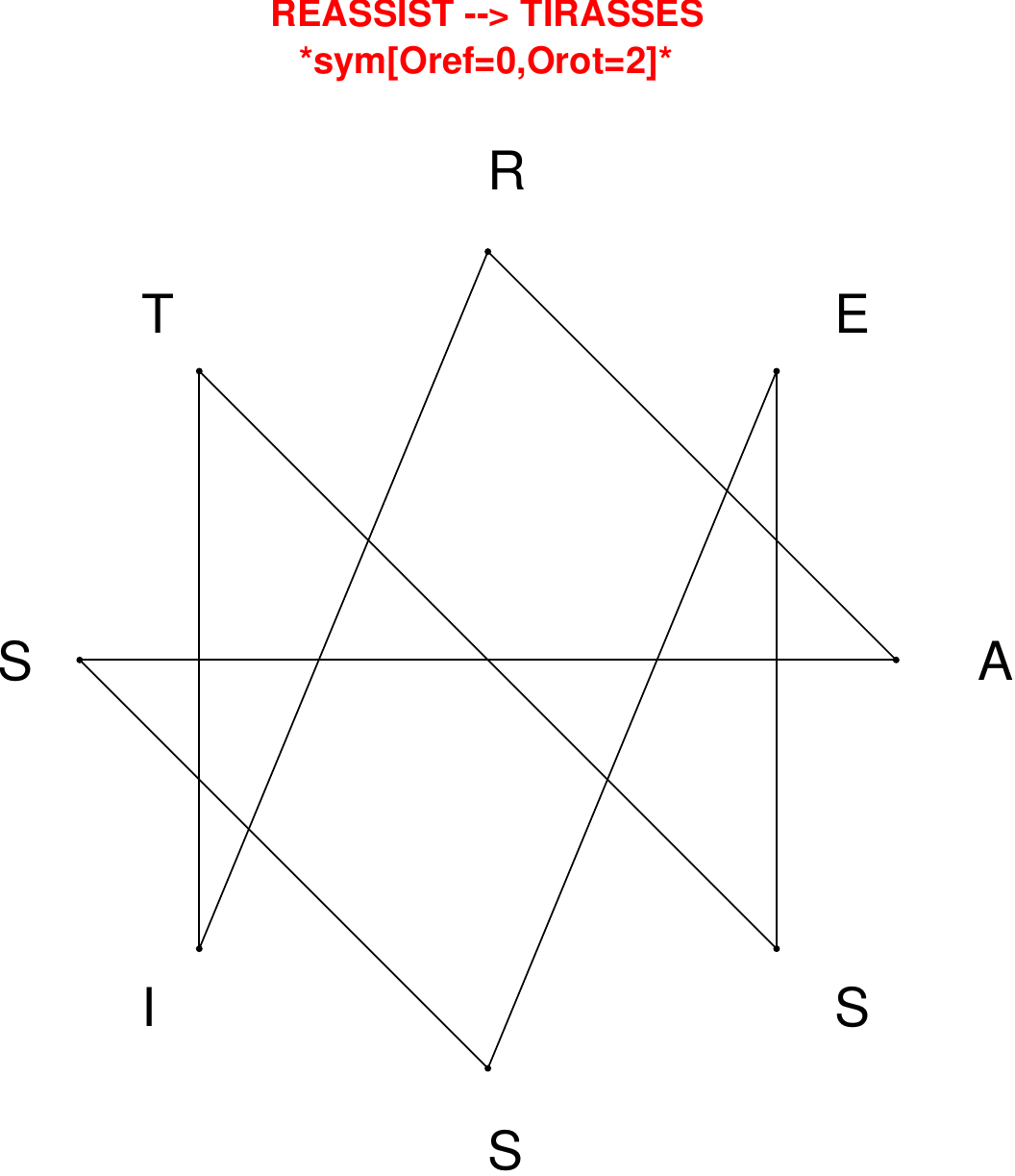}
\end{subfigure}
\hfill
\begin{subfigure}[T]{0.19\textwidth}
\centering
\includegraphics[width=\textwidth]{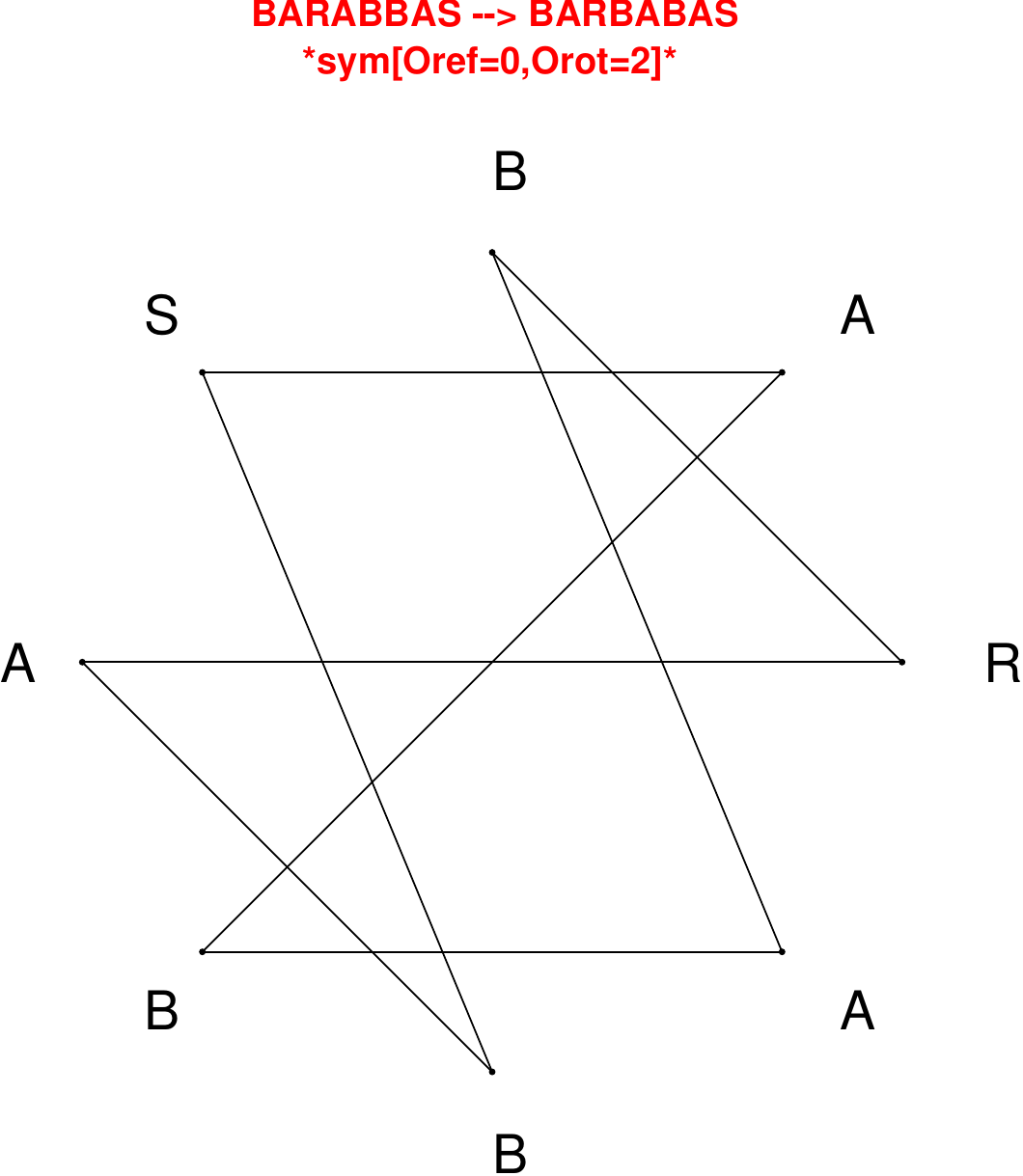}
\end{subfigure}
\end{figure}

\begin{figure}[H]
\centering
\begin{subfigure}[T]{0.19\textwidth}
\centering
\includegraphics[width=\textwidth]{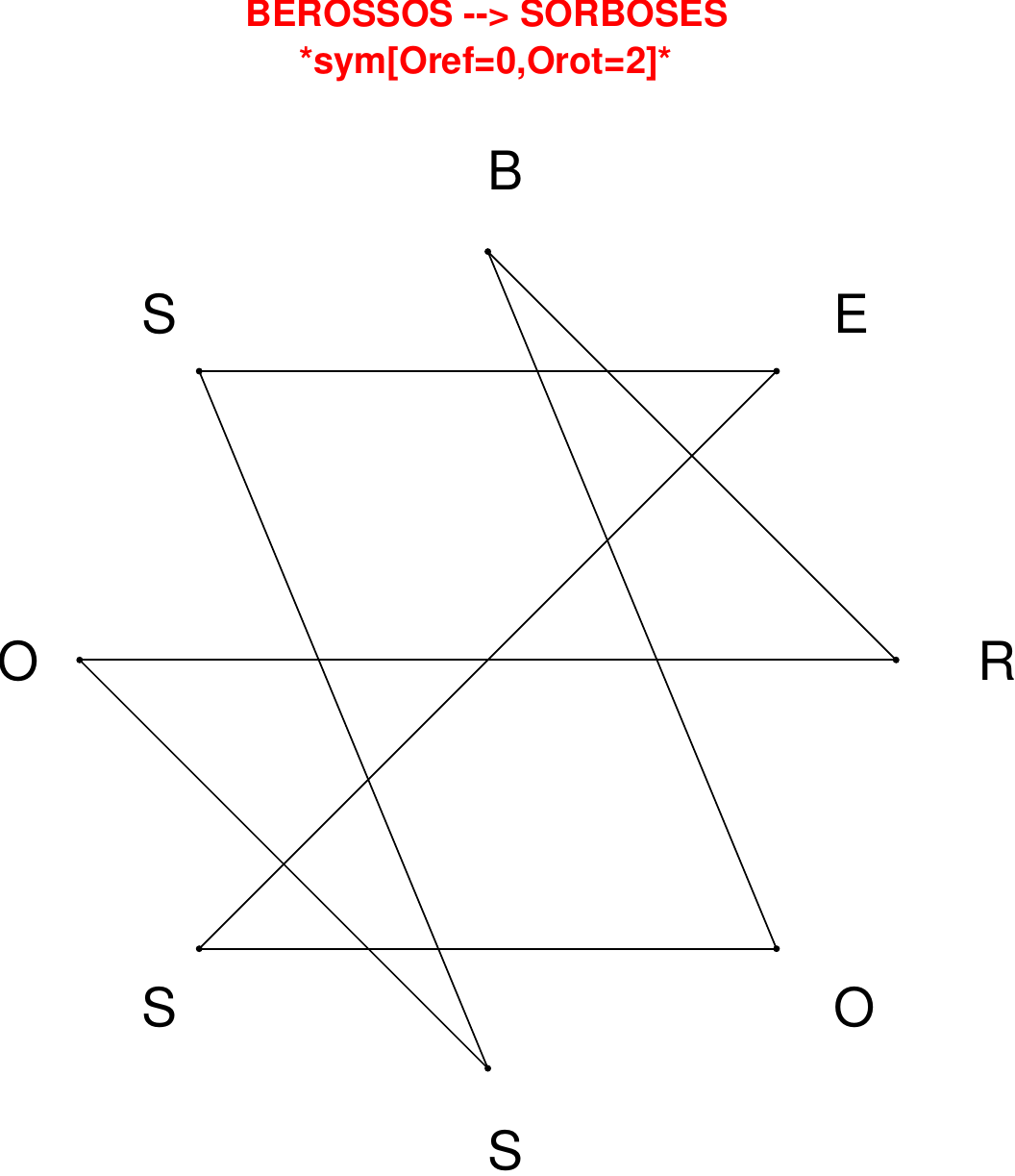}
\end{subfigure}
\hfill
\begin{subfigure}[T]{0.19\textwidth}
\centering
\includegraphics[width=\textwidth]{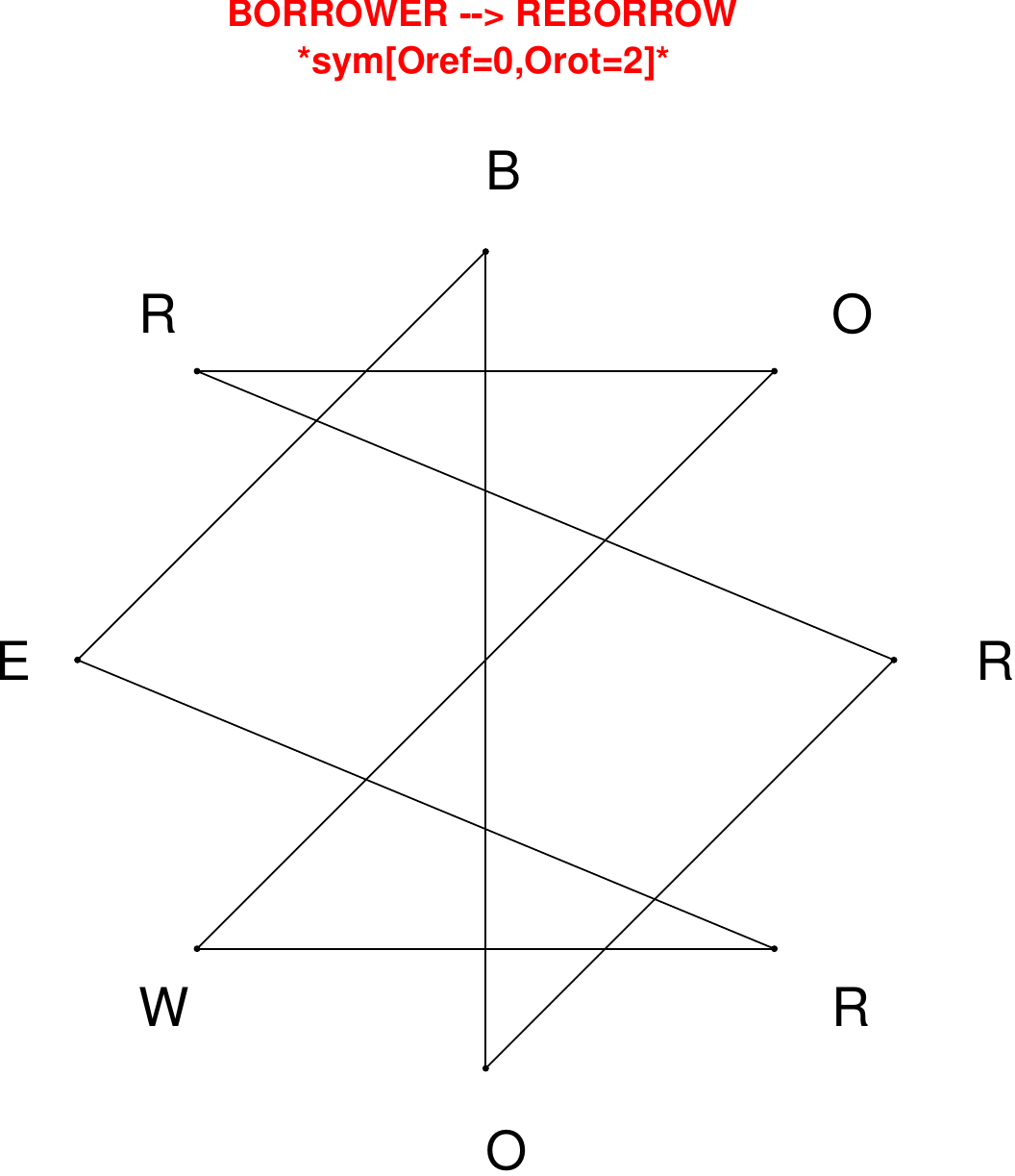}
\end{subfigure}
\hfill
\begin{subfigure}[T]{0.19\textwidth}
\centering
\includegraphics[width=\textwidth]{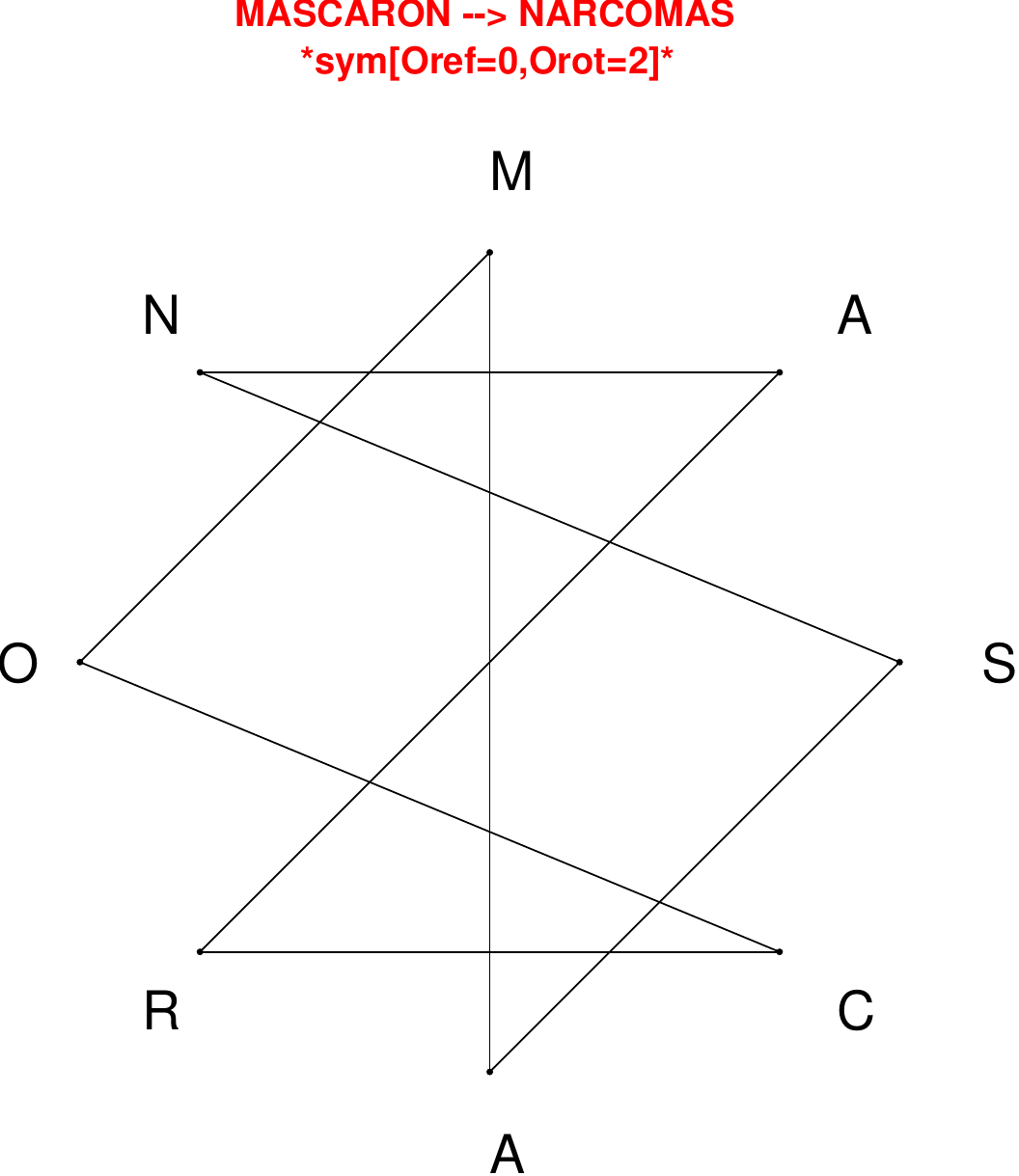}
\end{subfigure}
\hfill
\begin{subfigure}[T]{0.19\textwidth}
\centering
\includegraphics[width=\textwidth]{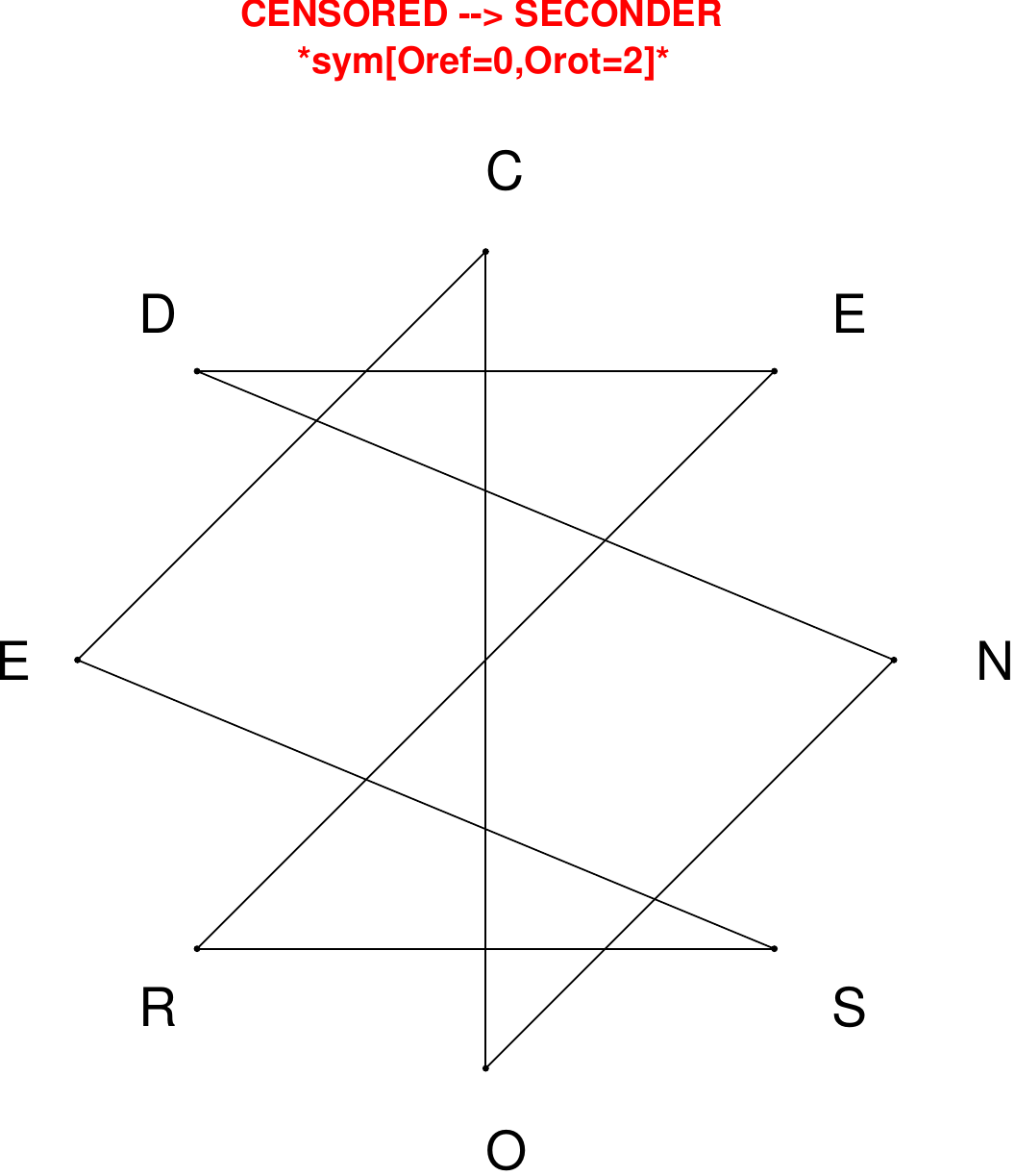}
\end{subfigure}
\hfill
\begin{subfigure}[T]{0.19\textwidth}
\centering
\includegraphics[width=\textwidth]{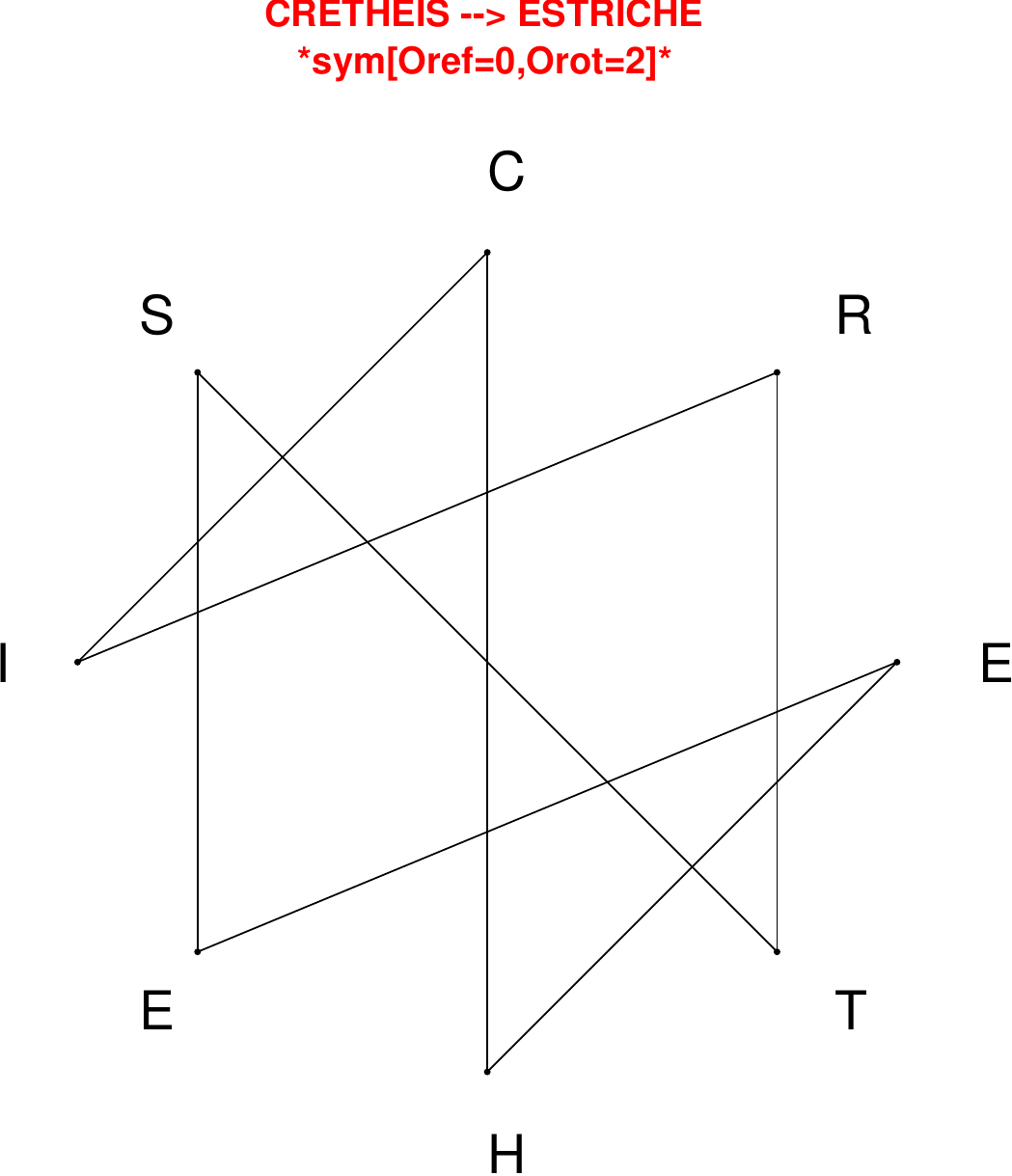}
\end{subfigure}
\end{figure}

\begin{figure}[H]
\centering
\begin{subfigure}[T]{0.19\textwidth}
\centering
\includegraphics[width=\textwidth]{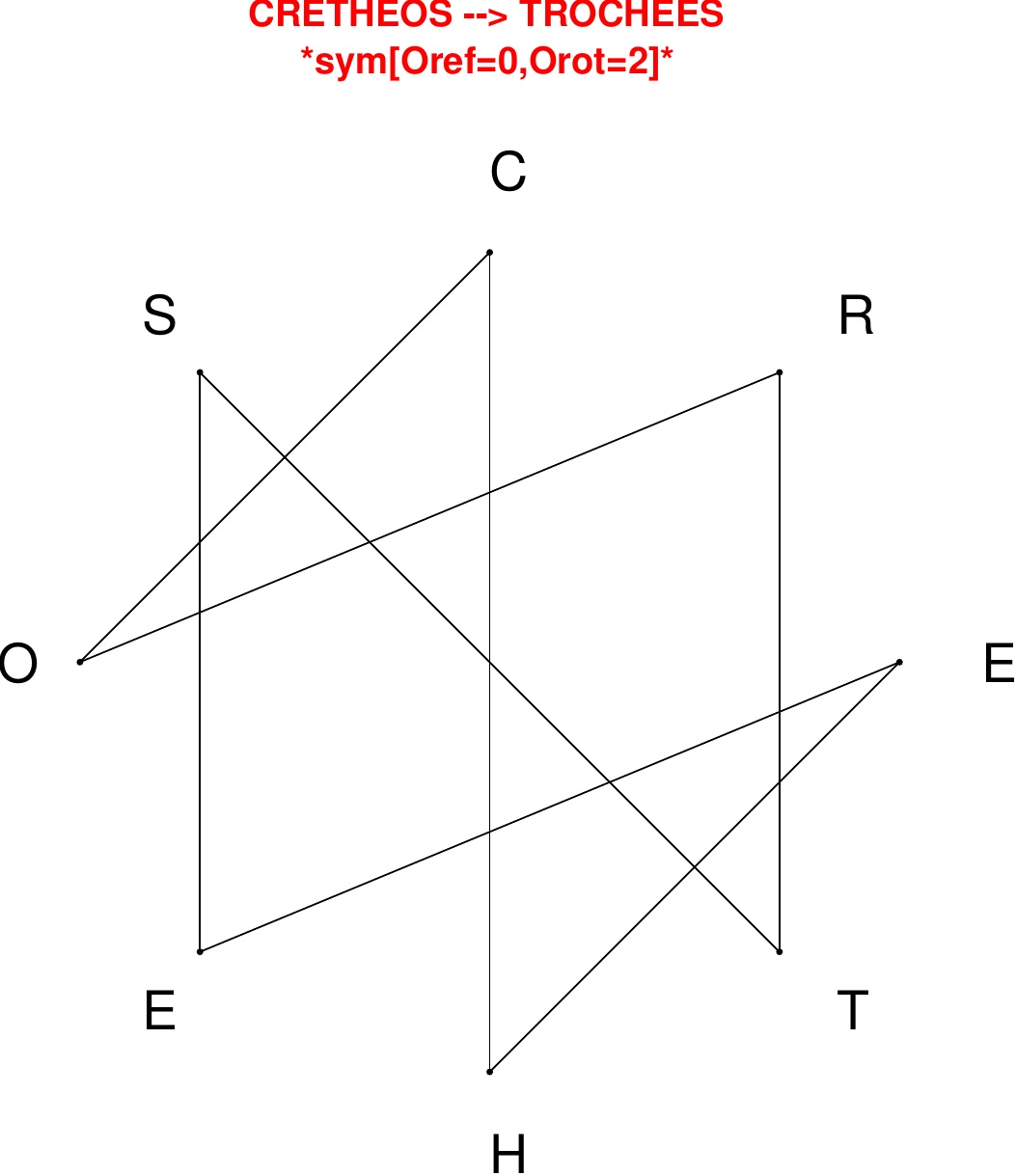}
\end{subfigure}
\hfill
\begin{subfigure}[T]{0.19\textwidth}
\centering
\includegraphics[width=\textwidth]{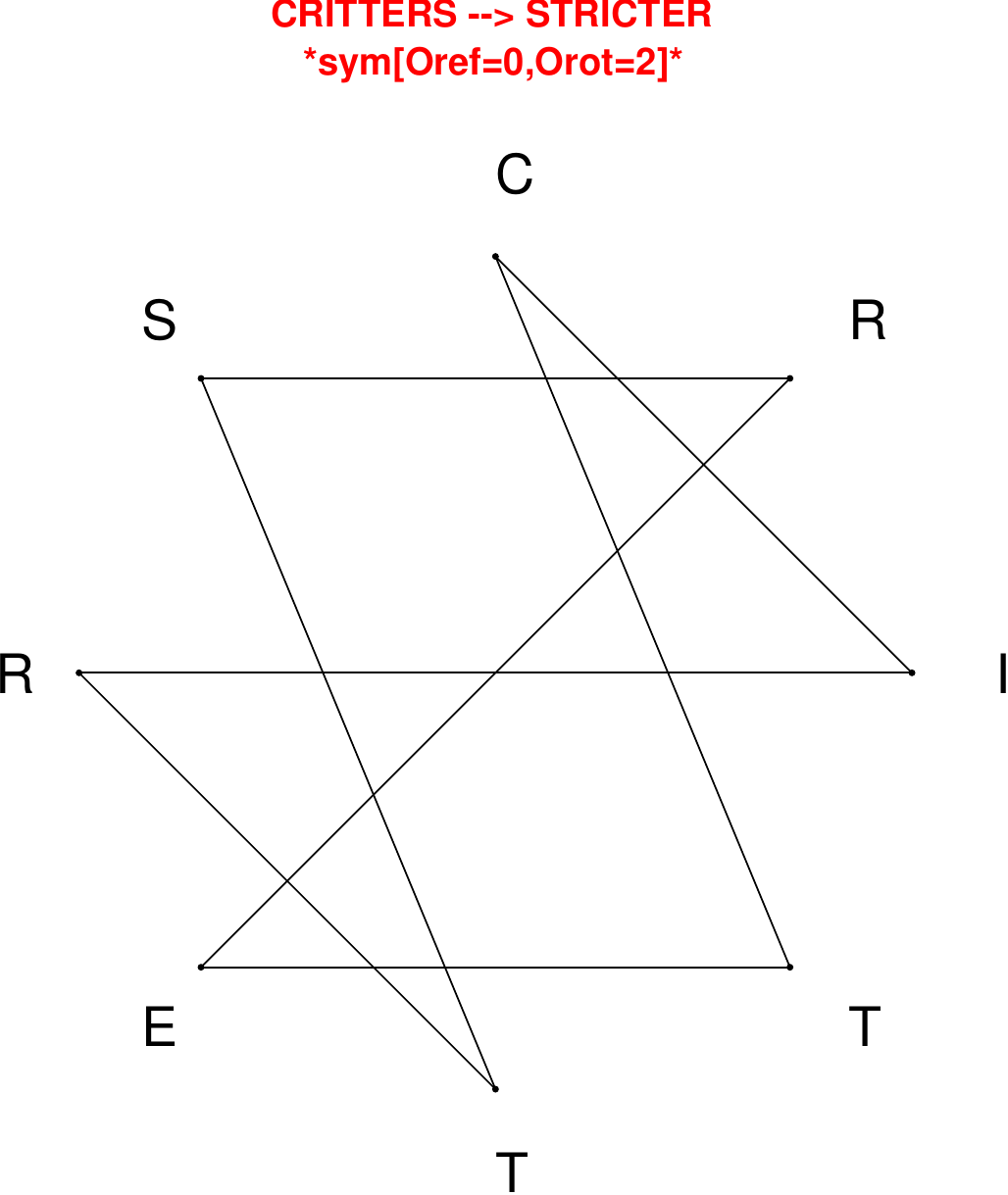}
\end{subfigure}
\hfill
\begin{subfigure}[T]{0.19\textwidth}
\centering
\includegraphics[width=\textwidth]{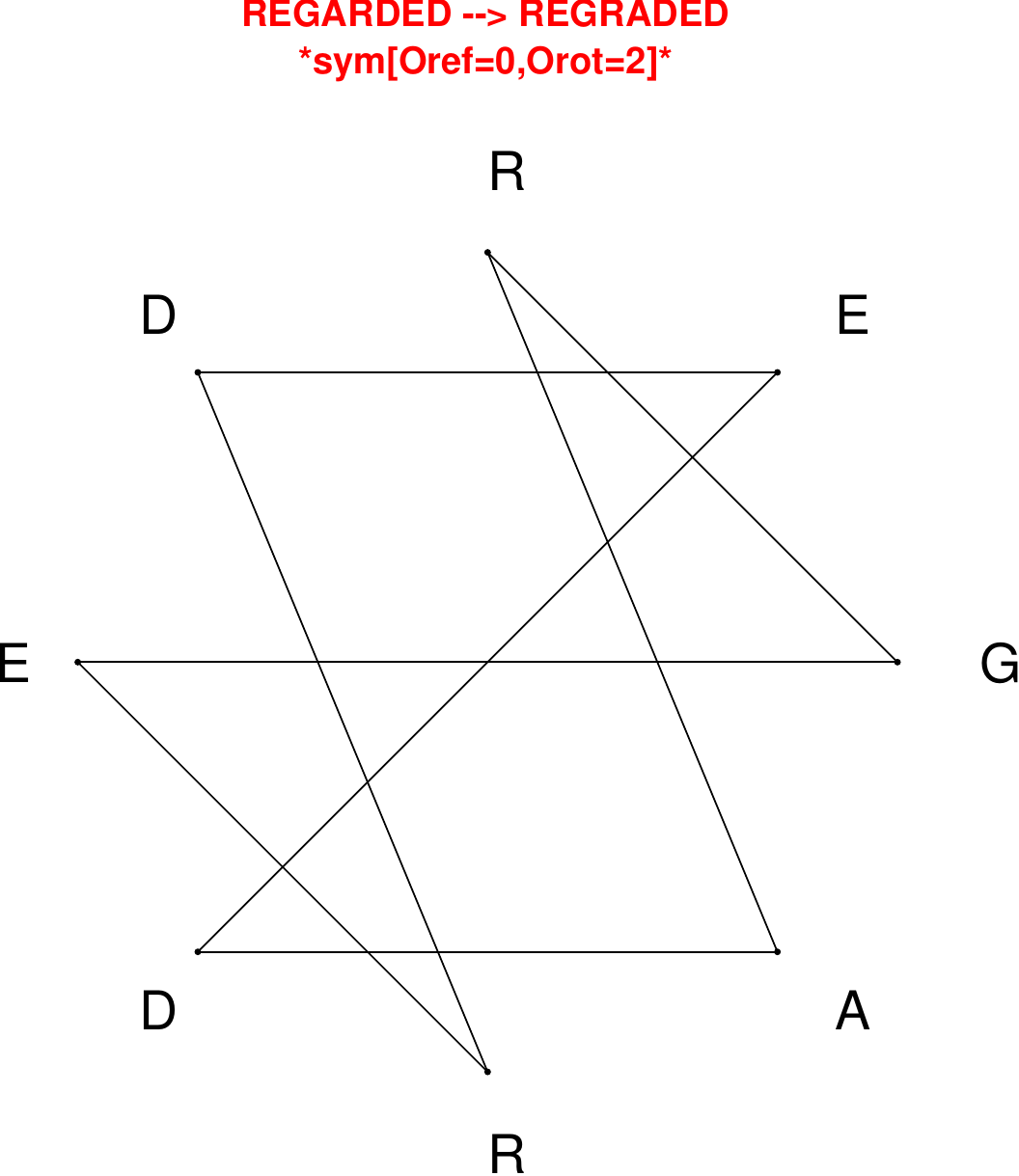}
\end{subfigure}
\hfill
\begin{subfigure}[T]{0.19\textwidth}
\centering
\includegraphics[width=\textwidth]{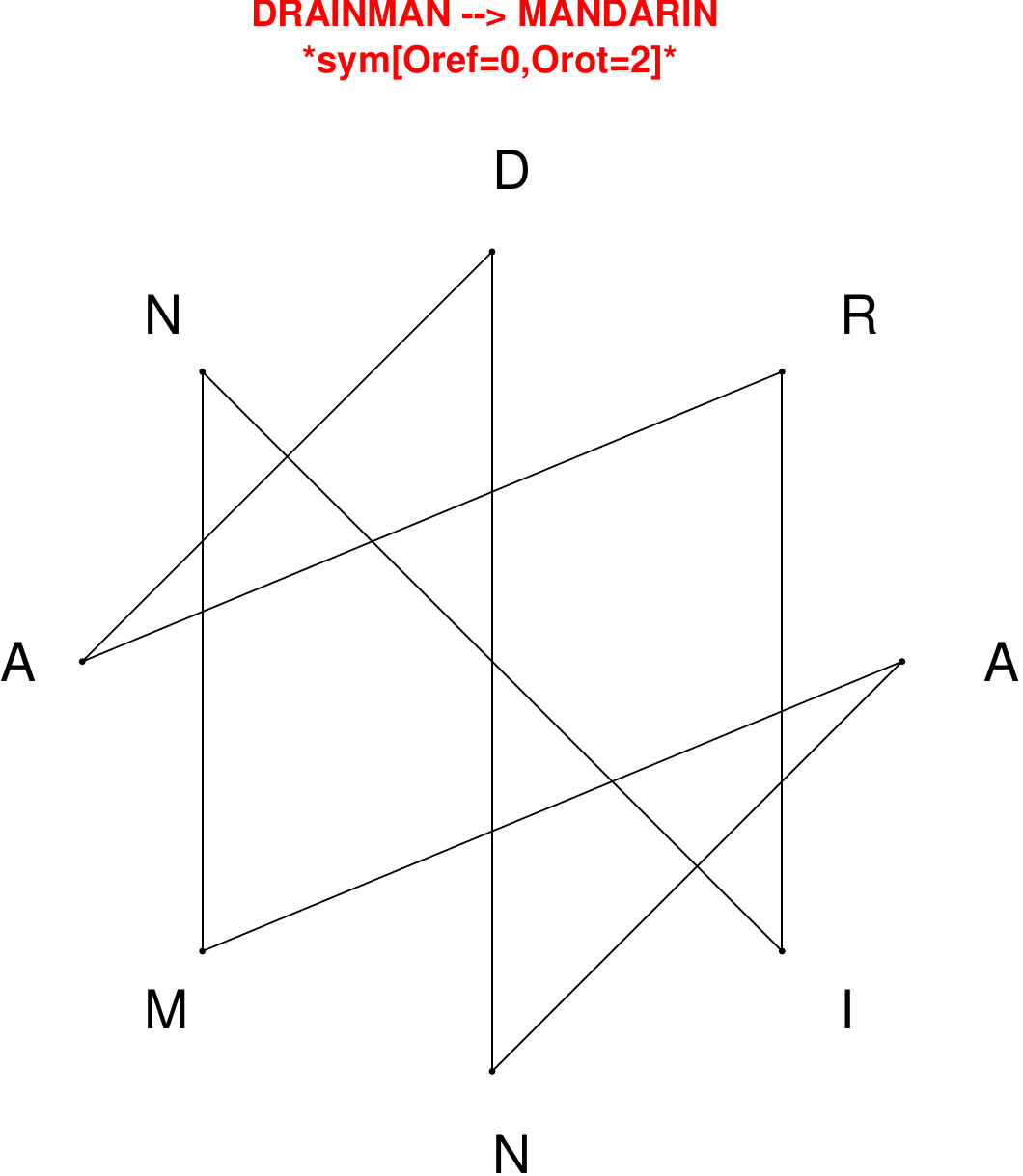}
\end{subfigure}
\hfill
\begin{subfigure}[T]{0.19\textwidth}
\centering
\includegraphics[width=\textwidth]{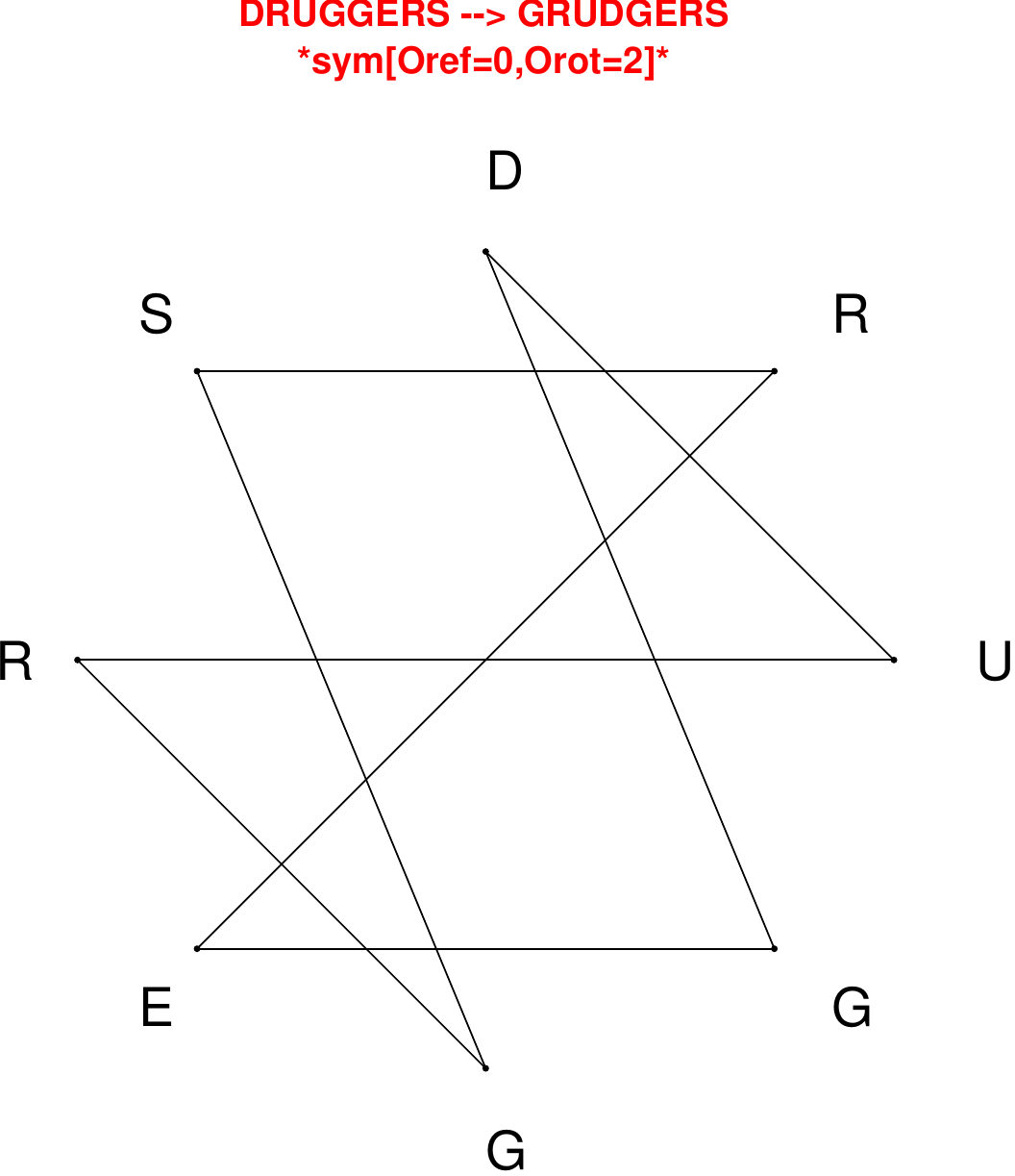}
\end{subfigure}
\end{figure}

\begin{figure}[H]
\centering
\begin{subfigure}[T]{0.19\textwidth}
\centering
\includegraphics[width=\textwidth]{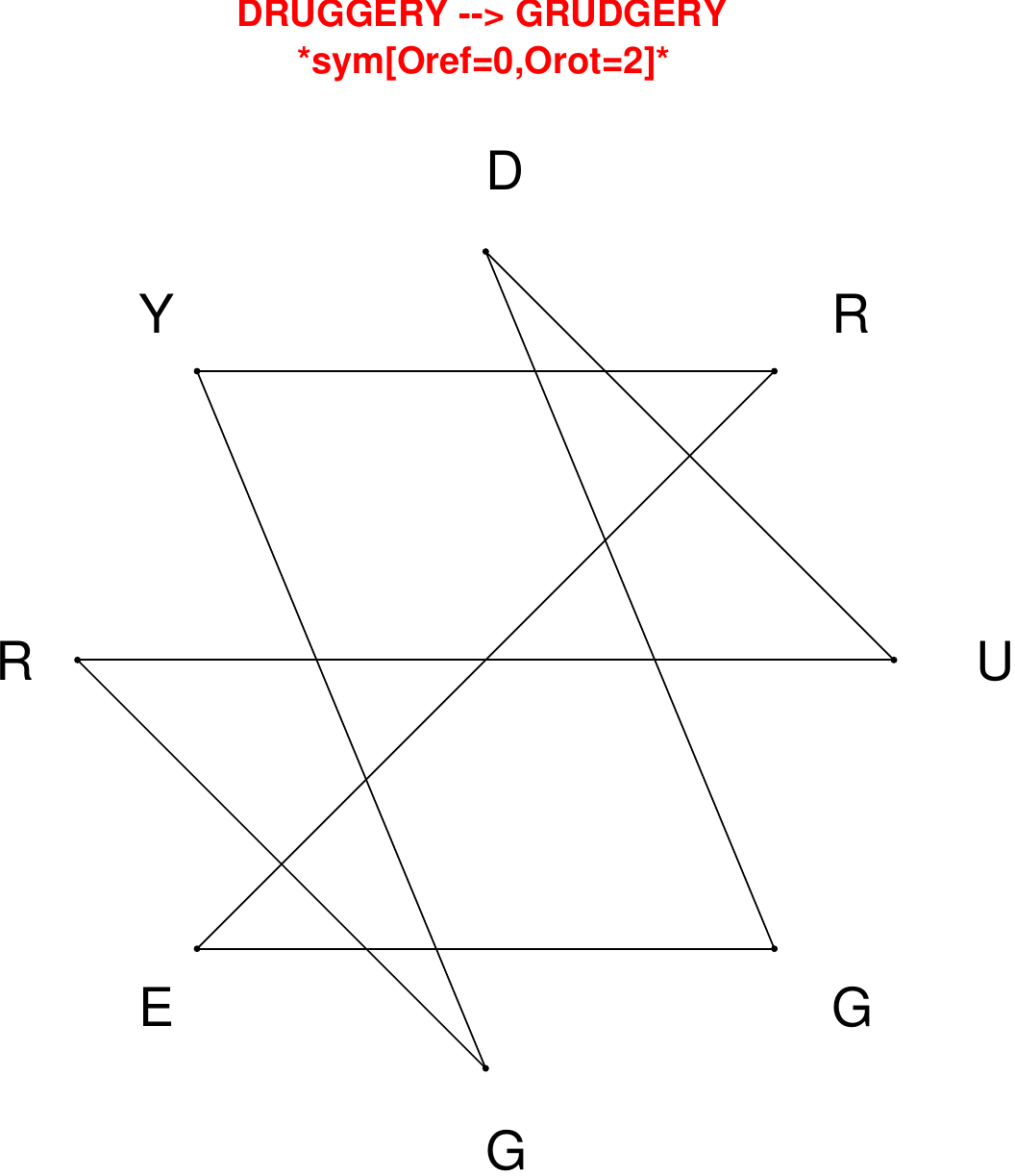}
\end{subfigure}
\hfill
\begin{subfigure}[T]{0.19\textwidth}
\centering
\includegraphics[width=\textwidth]{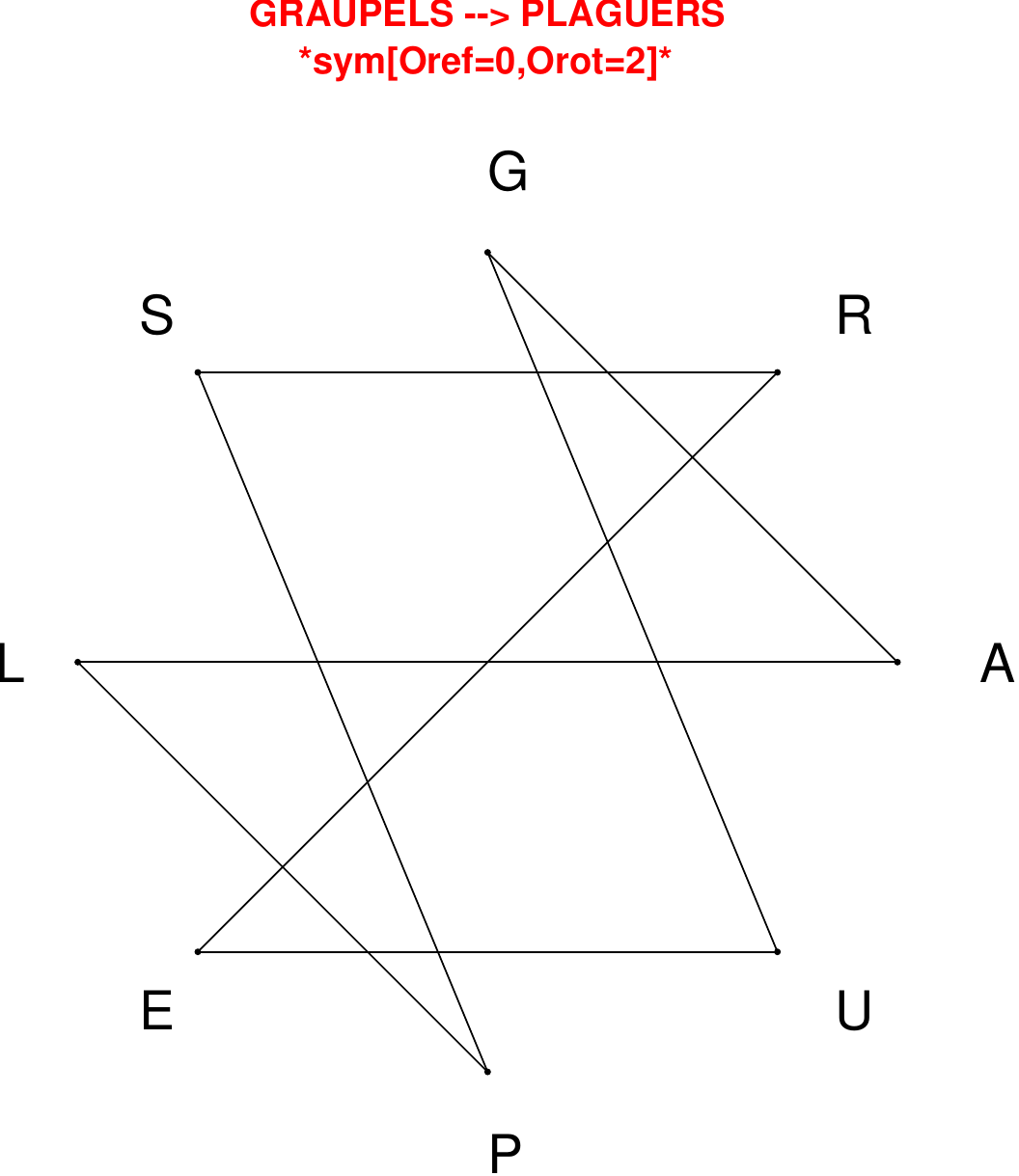}
\end{subfigure}
\hfill
\begin{subfigure}[T]{0.19\textwidth}
\centering
\includegraphics[width=\textwidth]{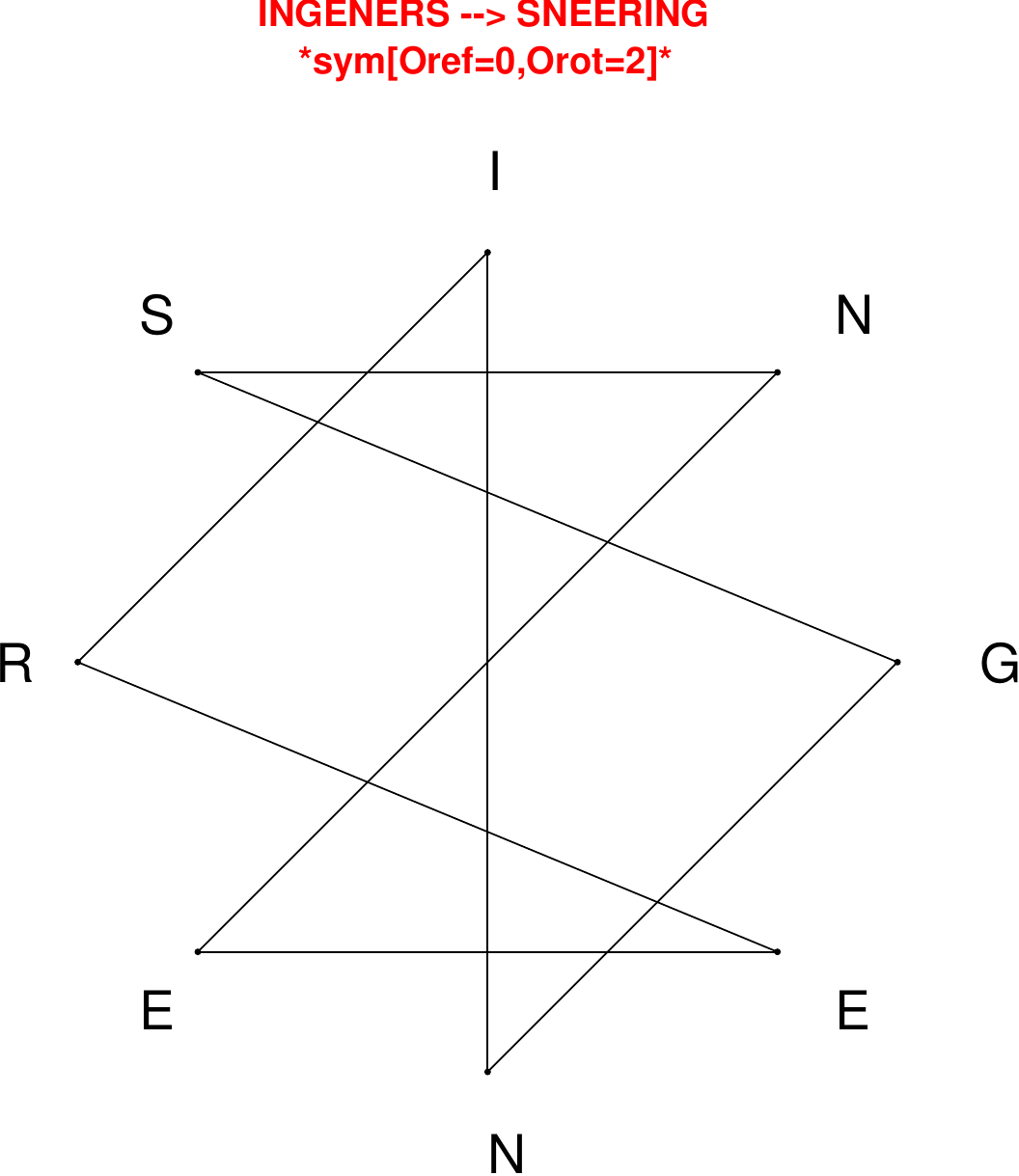}
\end{subfigure}
\hfill
\begin{subfigure}[T]{0.19\textwidth}
\centering
\includegraphics[width=\textwidth]{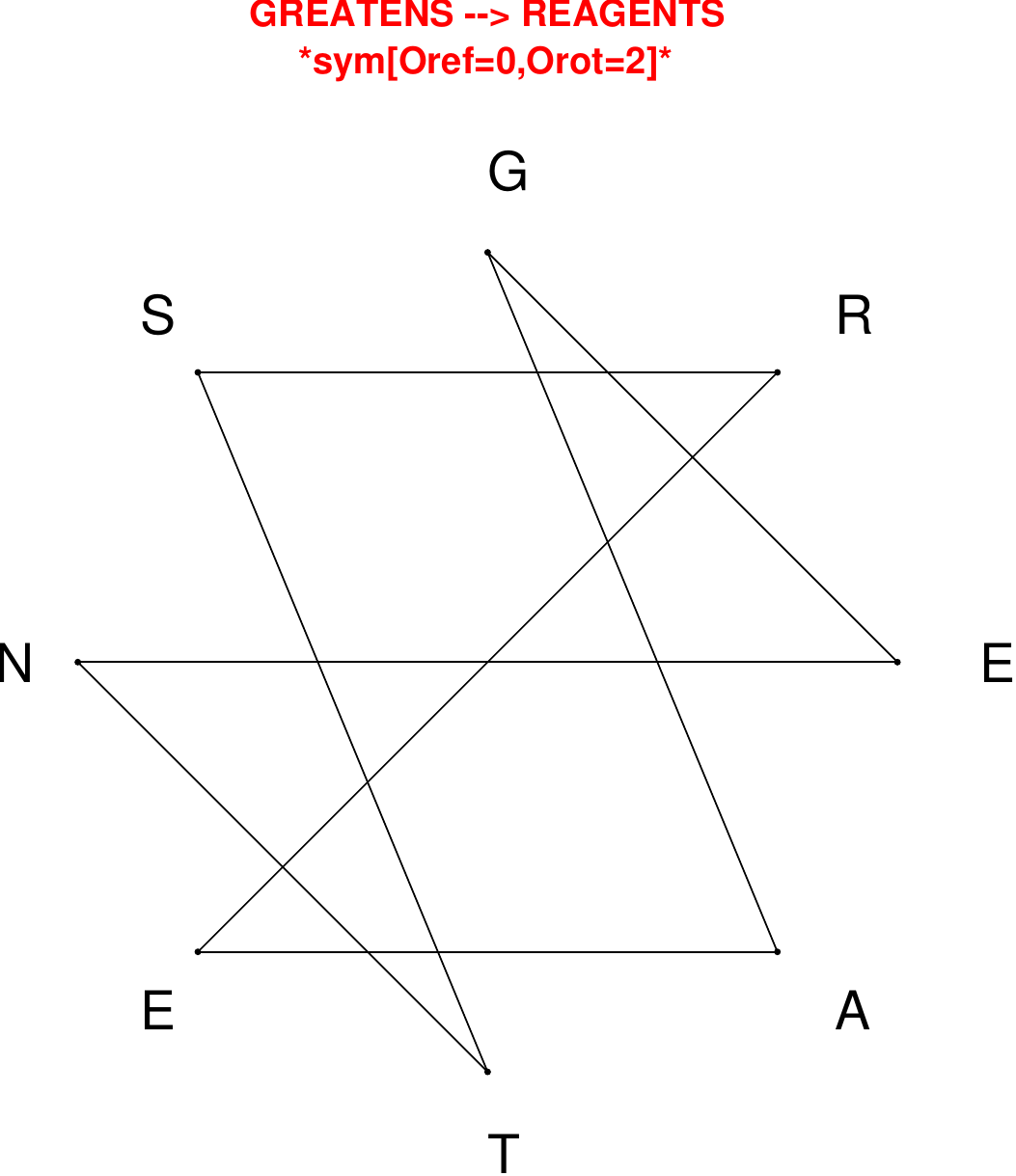}
\end{subfigure}
\hfill
\begin{subfigure}[T]{0.19\textwidth}
\centering
\includegraphics[width=\textwidth]{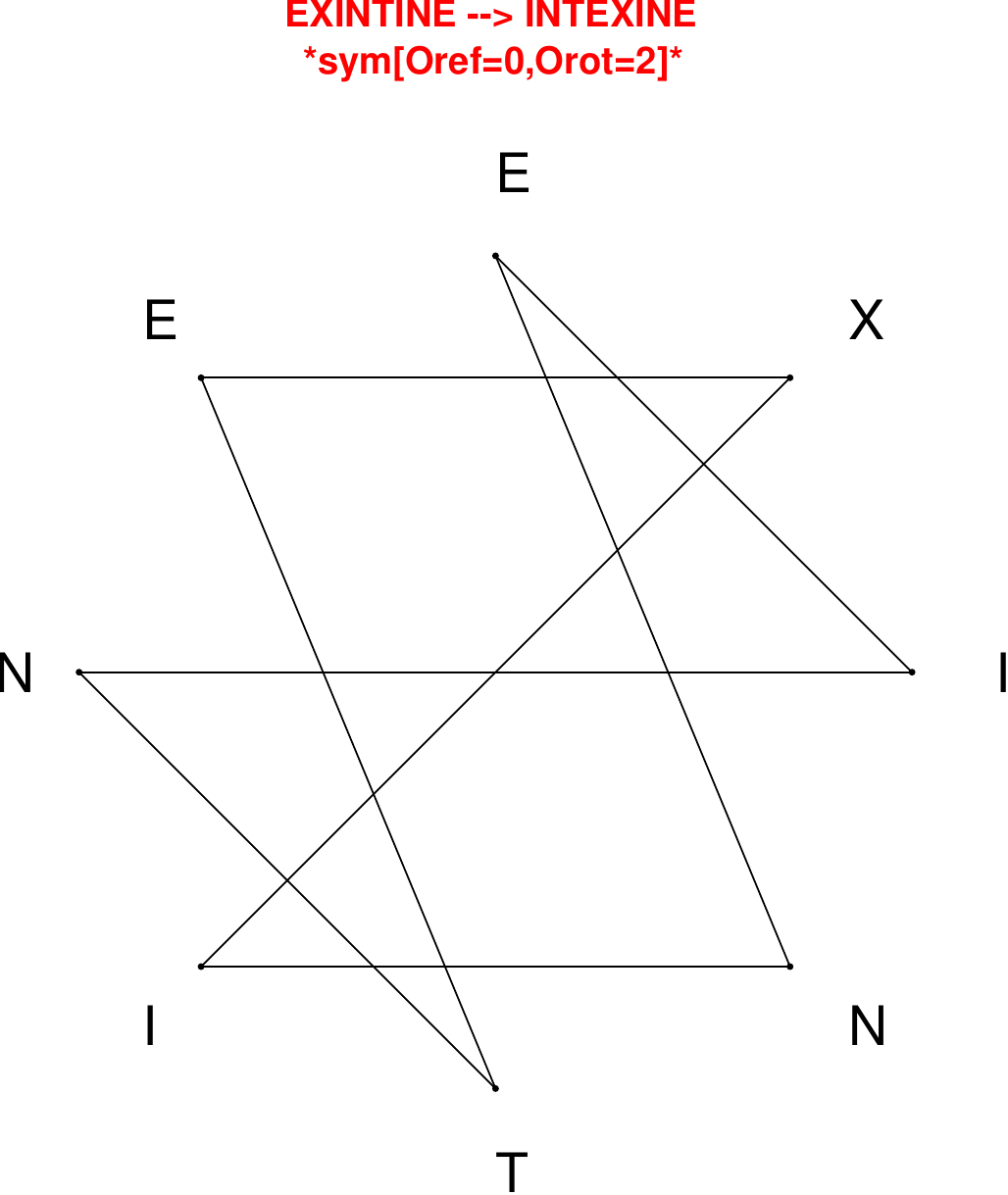}
\end{subfigure}
\end{figure}

\begin{figure}[H]
\centering
\begin{subfigure}[T]{0.19\textwidth}
\centering
\includegraphics[width=\textwidth]{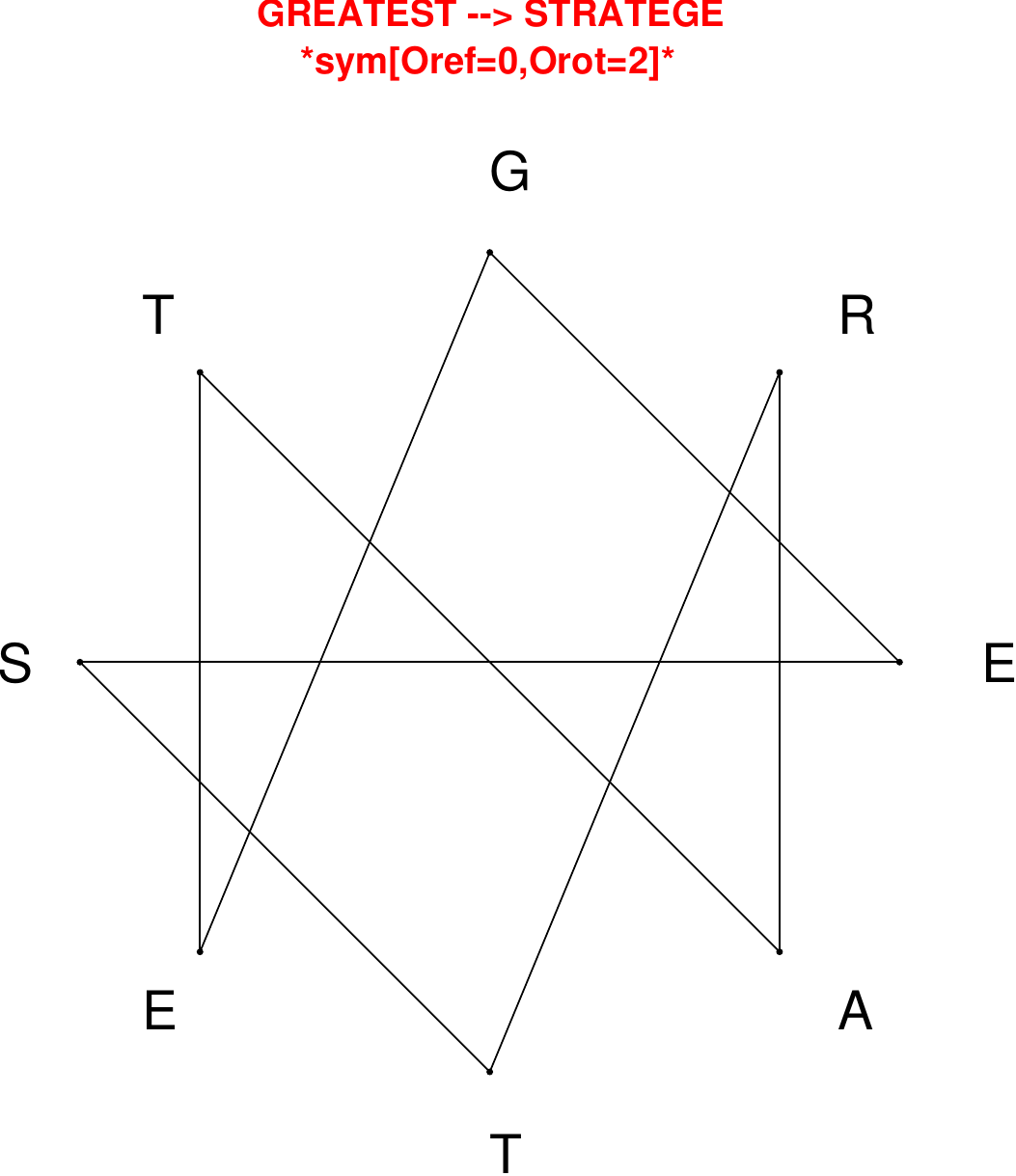}
\end{subfigure}
\hfill
\begin{subfigure}[T]{0.19\textwidth}
\centering
\includegraphics[width=\textwidth]{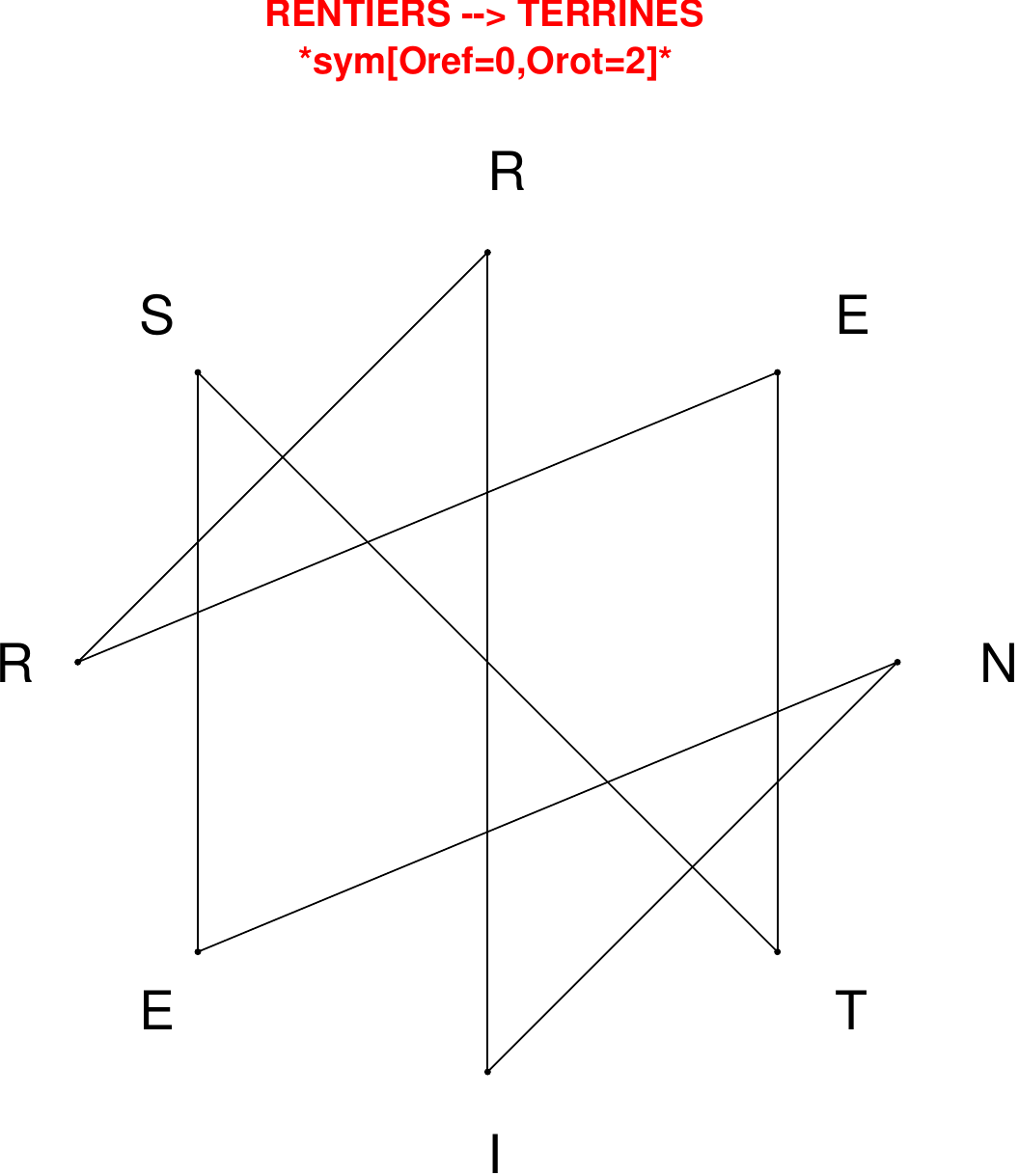}
\end{subfigure}
\hfill
\begin{subfigure}[T]{0.19\textwidth}
\centering
\includegraphics[width=\textwidth]{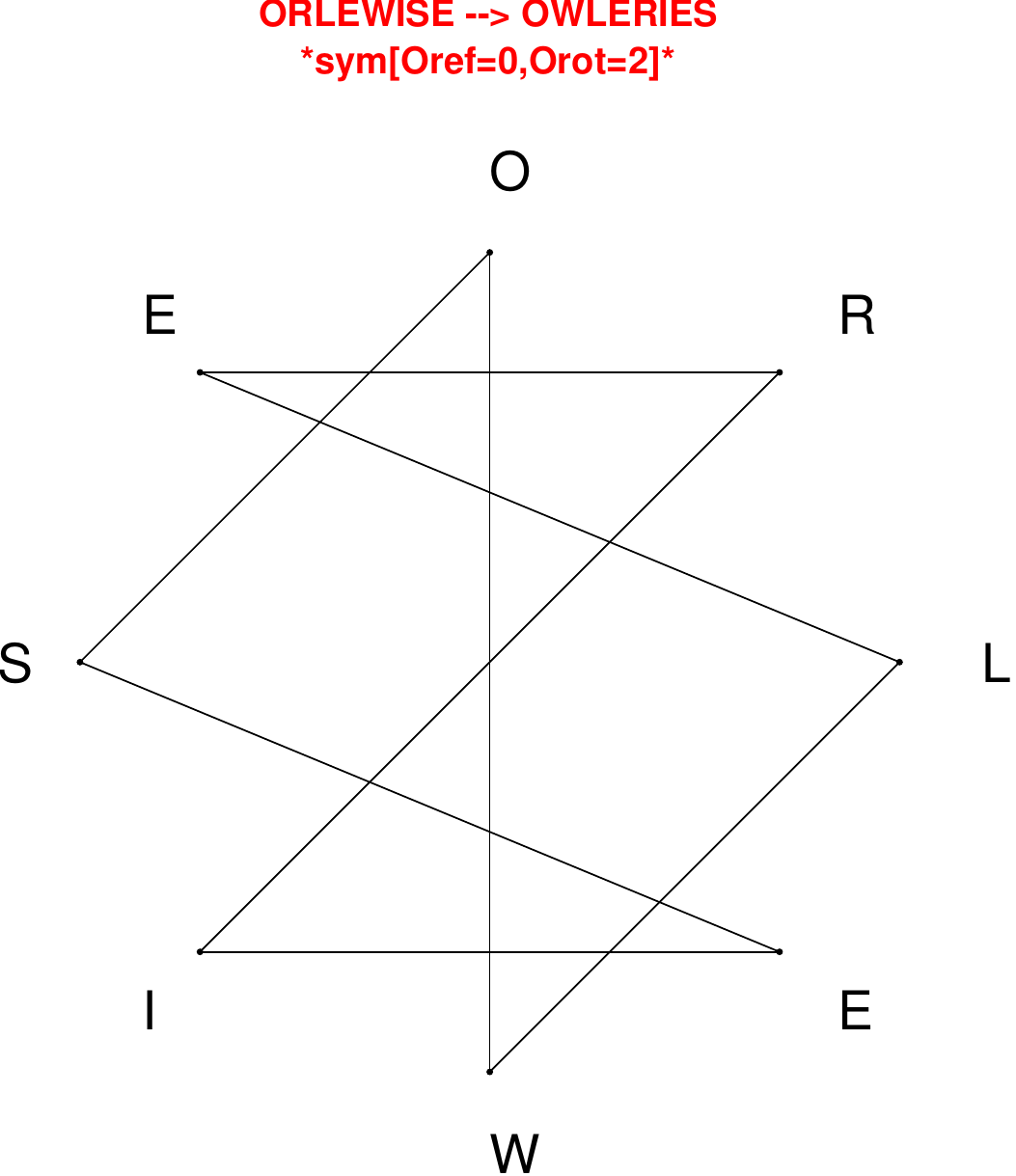}
\end{subfigure}
\hfill
\begin{subfigure}[T]{0.19\textwidth}
\centering
\includegraphics[width=\textwidth]{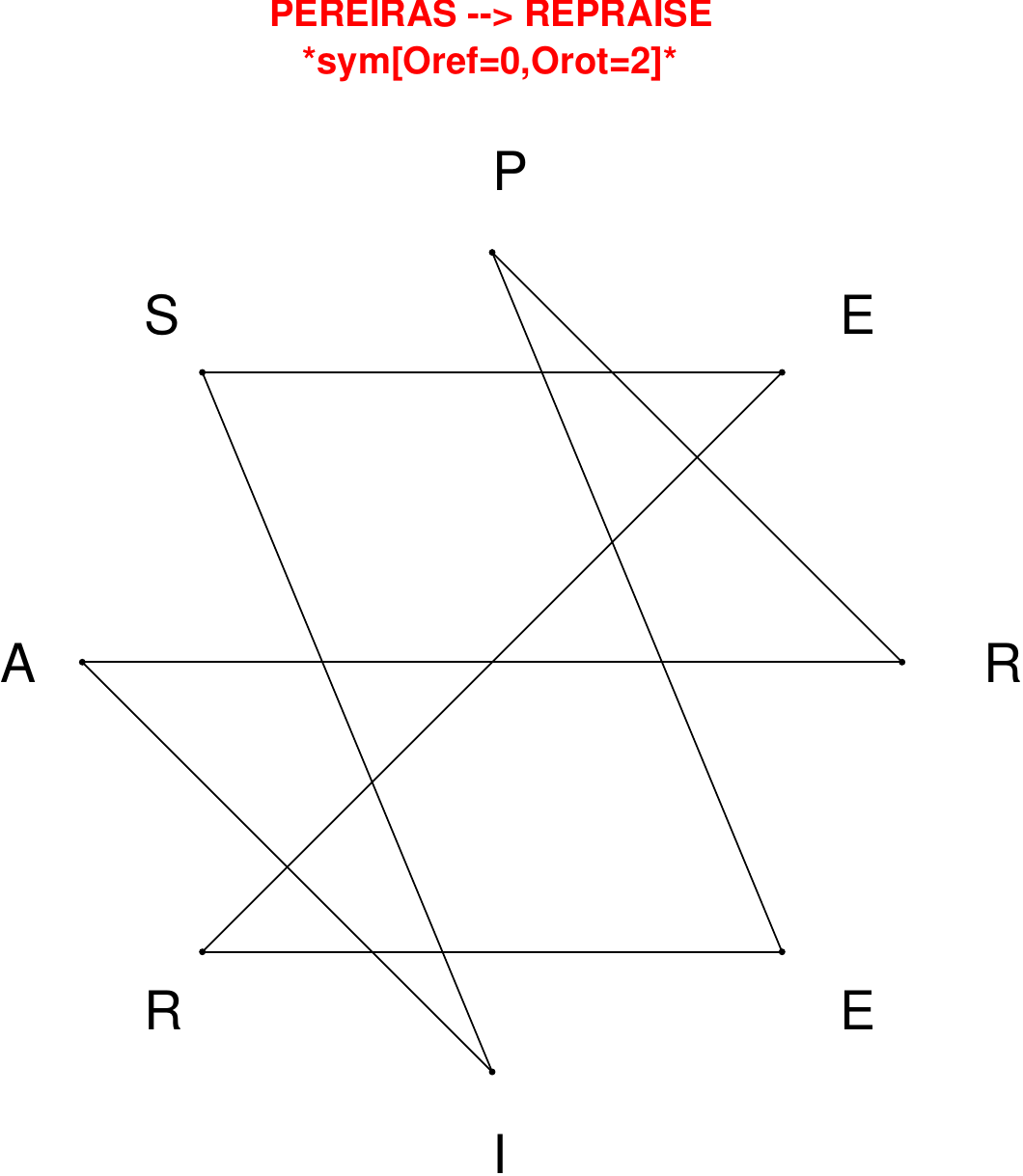}
\end{subfigure}
\hfill
\begin{subfigure}[T]{0.19\textwidth}
\centering
\includegraphics[width=\textwidth]{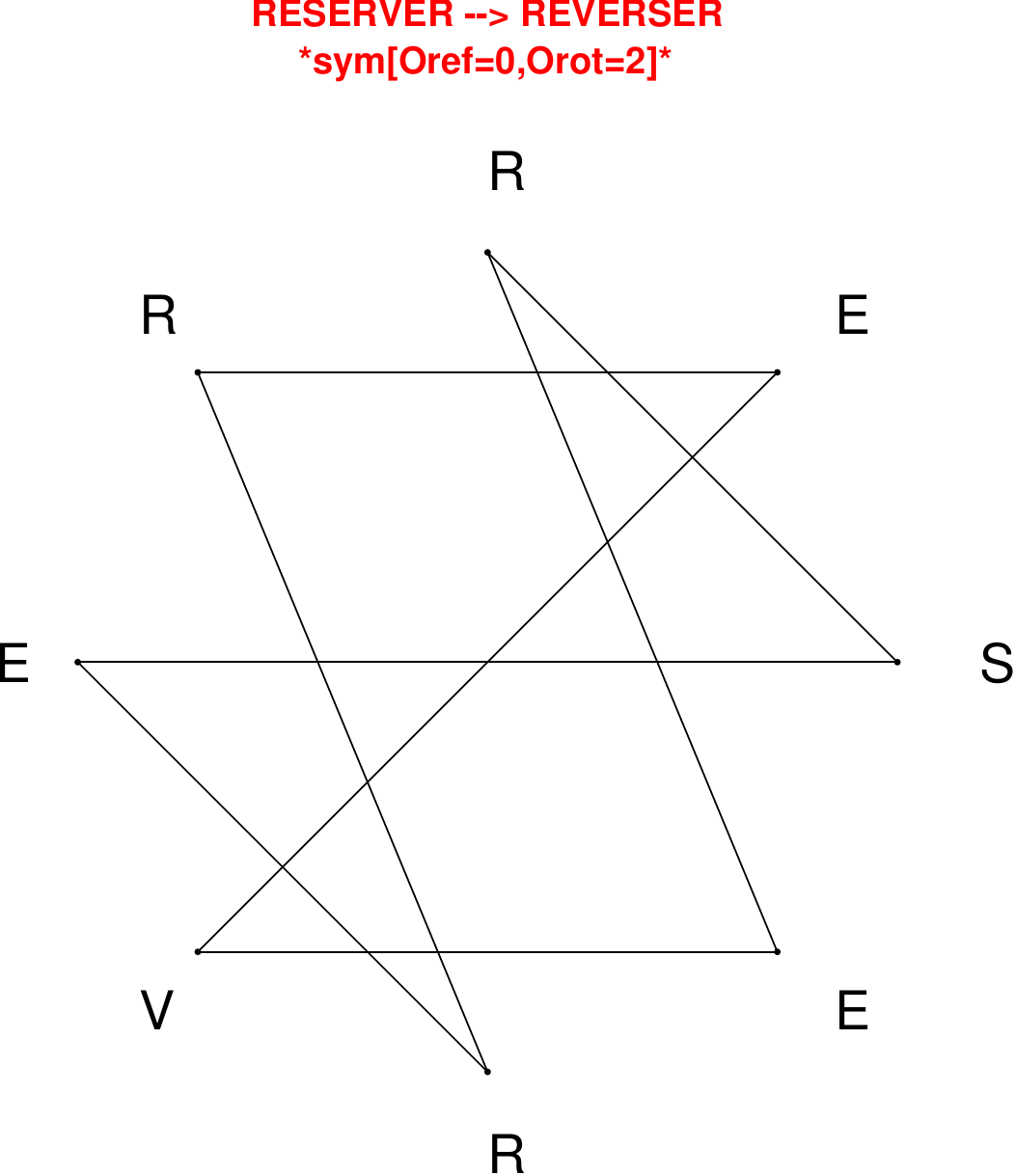}
\end{subfigure}
\end{figure}

\begin{figure}[H]
\centering
\begin{subfigure}[T]{0.19\textwidth}
\centering
\includegraphics[width=\textwidth]{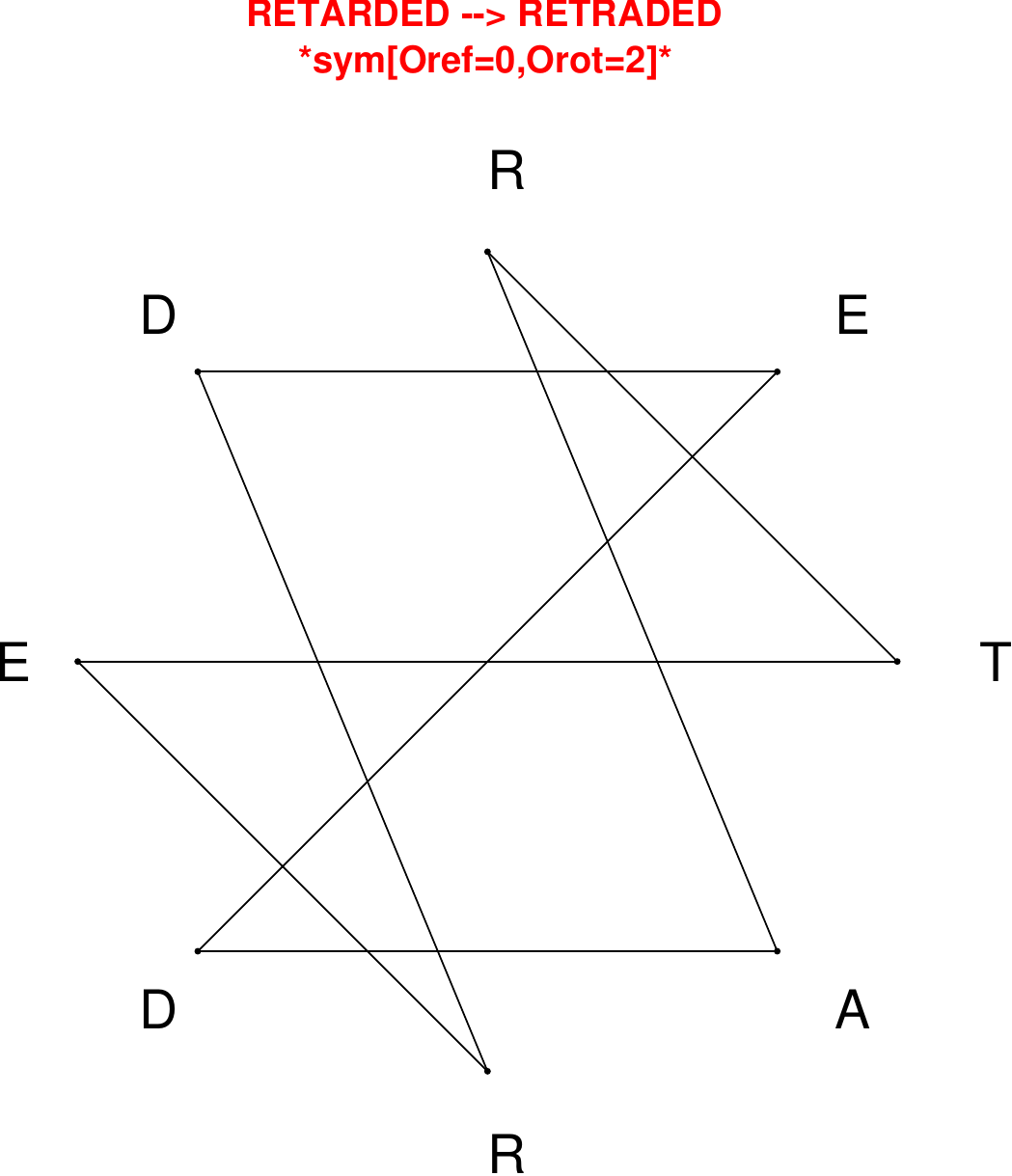}
\end{subfigure}
\hfill
\begin{subfigure}[T]{0.19\textwidth}
\centering
\includegraphics[width=\textwidth]{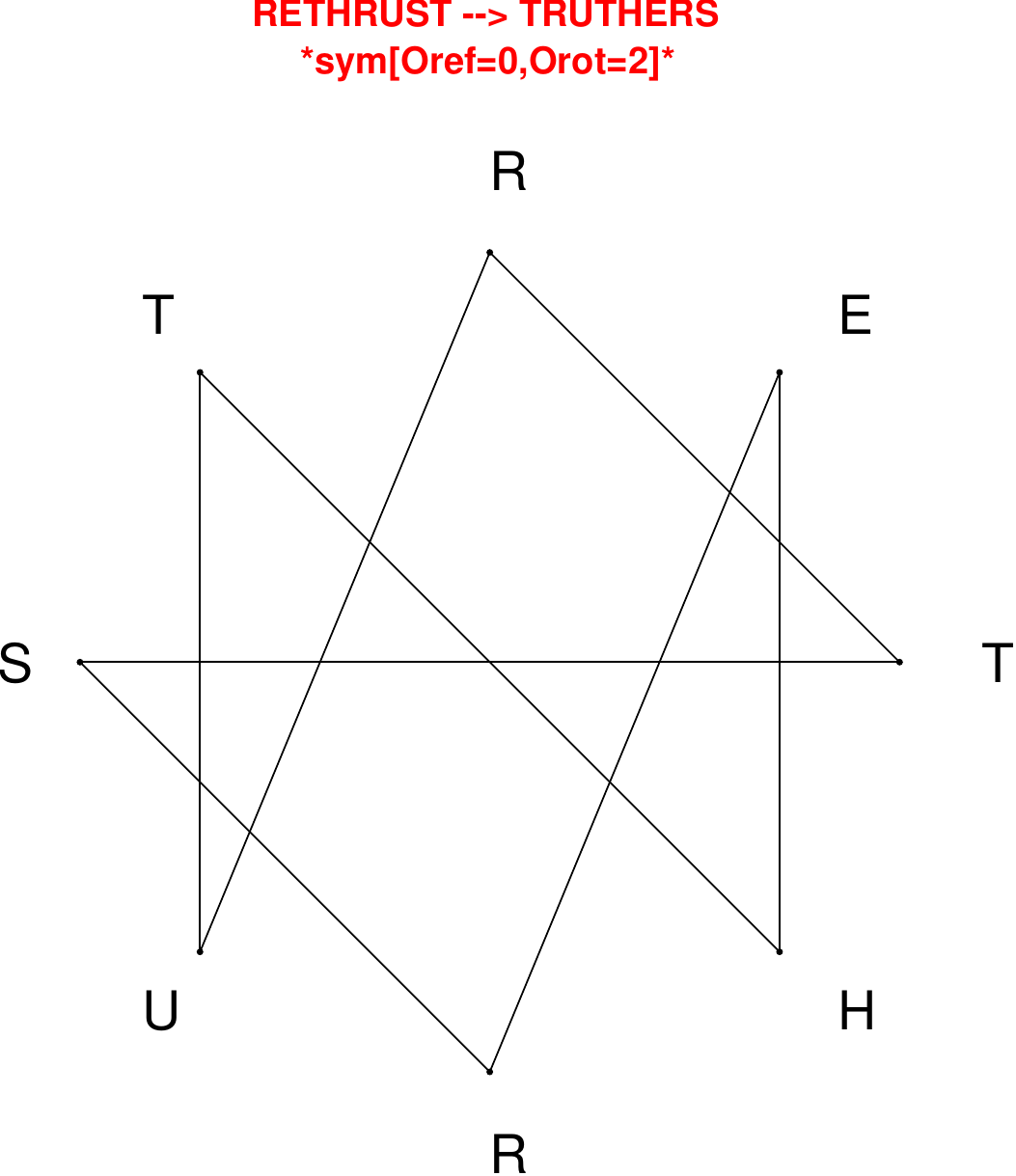}
\end{subfigure}
\hfill
\begin{subfigure}[T]{0.19\textwidth}
\centering
\includegraphics[width=\textwidth]{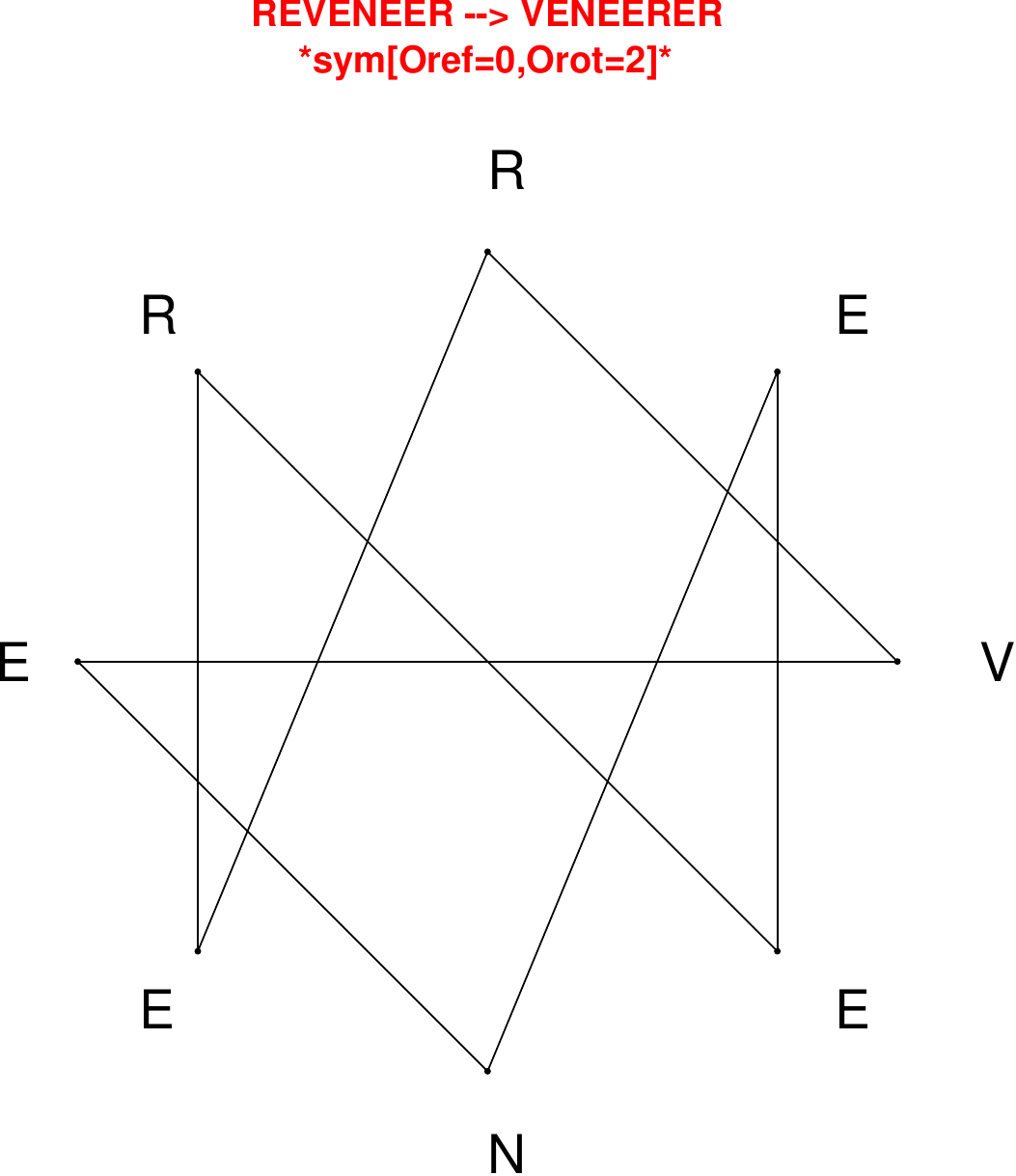}
\end{subfigure}
\hfill
\begin{subfigure}[T]{0.19\textwidth}
\centering
\includegraphics[width=\textwidth]{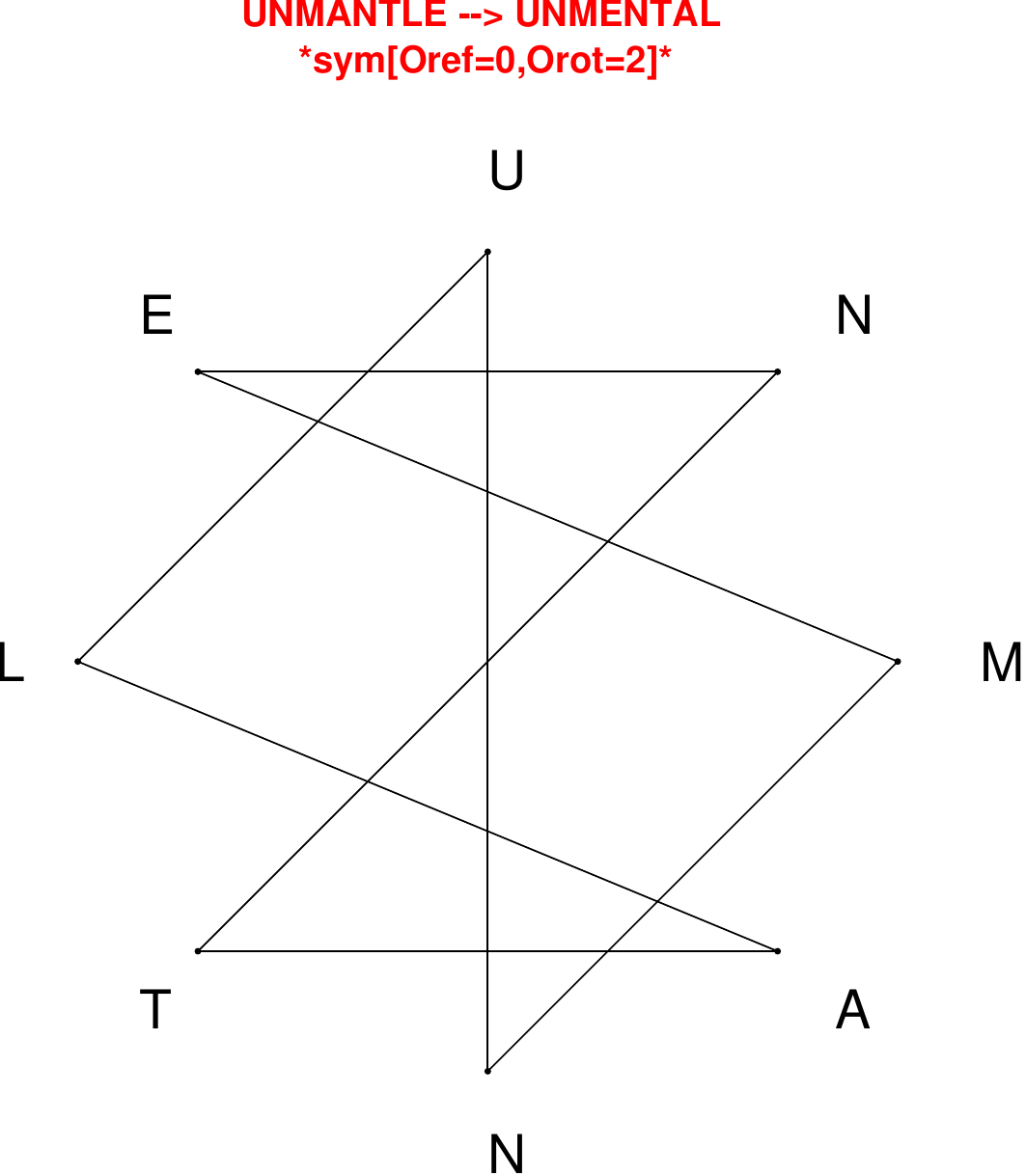}
\end{subfigure}
\hfill
\begin{subfigure}[T]{0.19\textwidth}
\centering
\includegraphics[width=\textwidth]{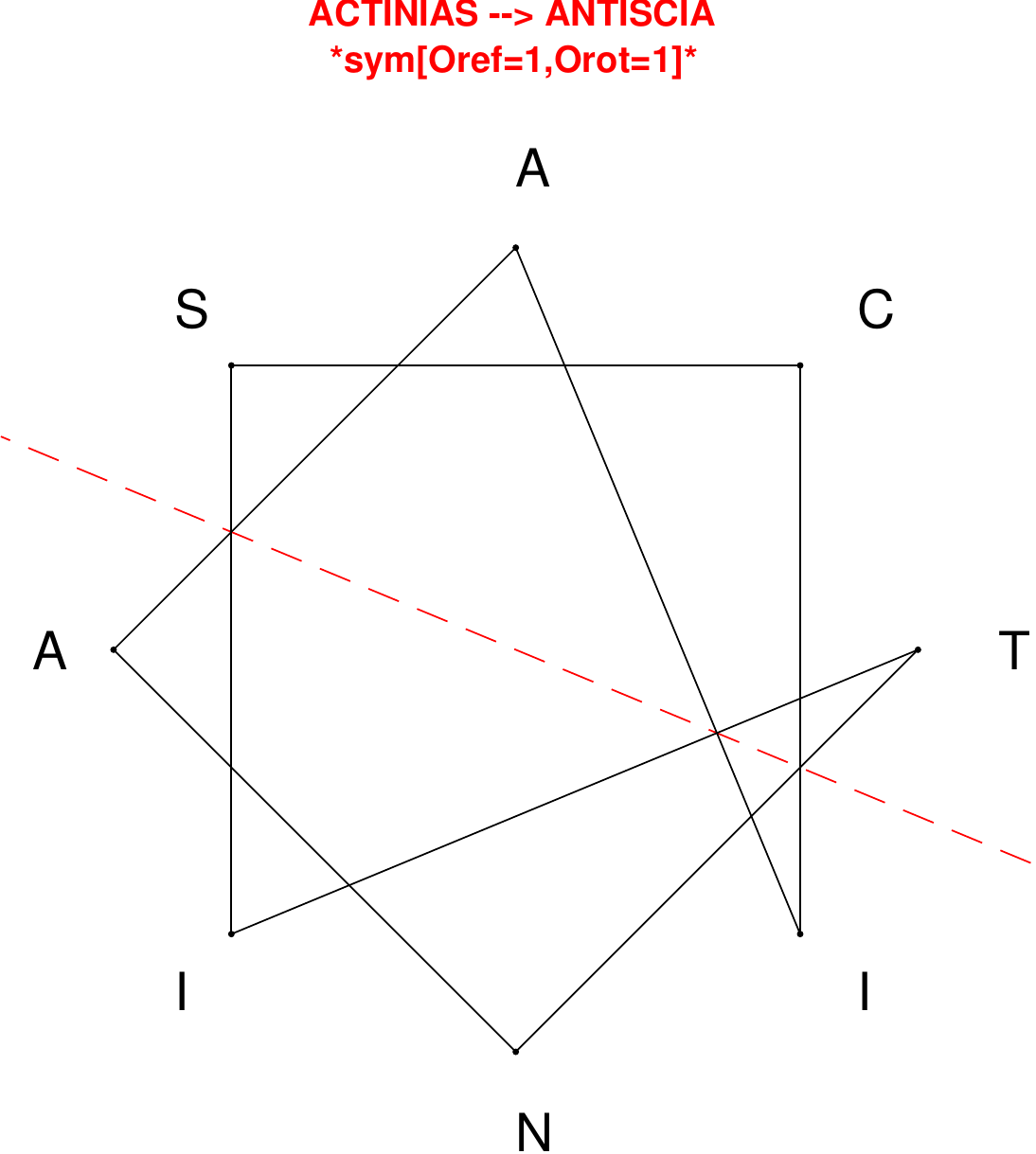}
\end{subfigure}
\end{figure}

\begin{figure}[H]
\centering
\begin{subfigure}[T]{0.19\textwidth}
\centering
\includegraphics[width=\textwidth]{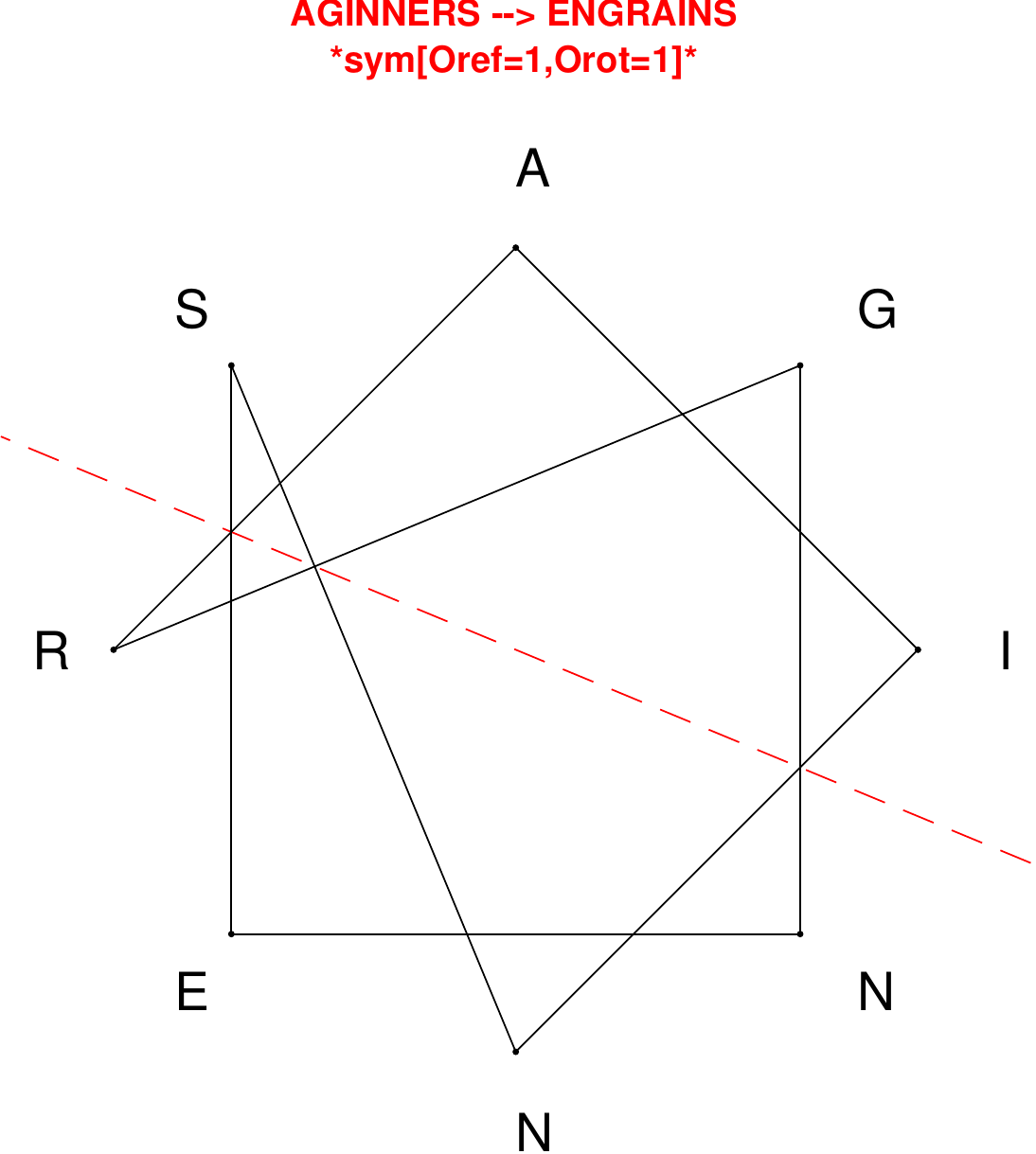}
\end{subfigure}
\hfill
\begin{subfigure}[T]{0.19\textwidth}
\centering
\includegraphics[width=\textwidth]{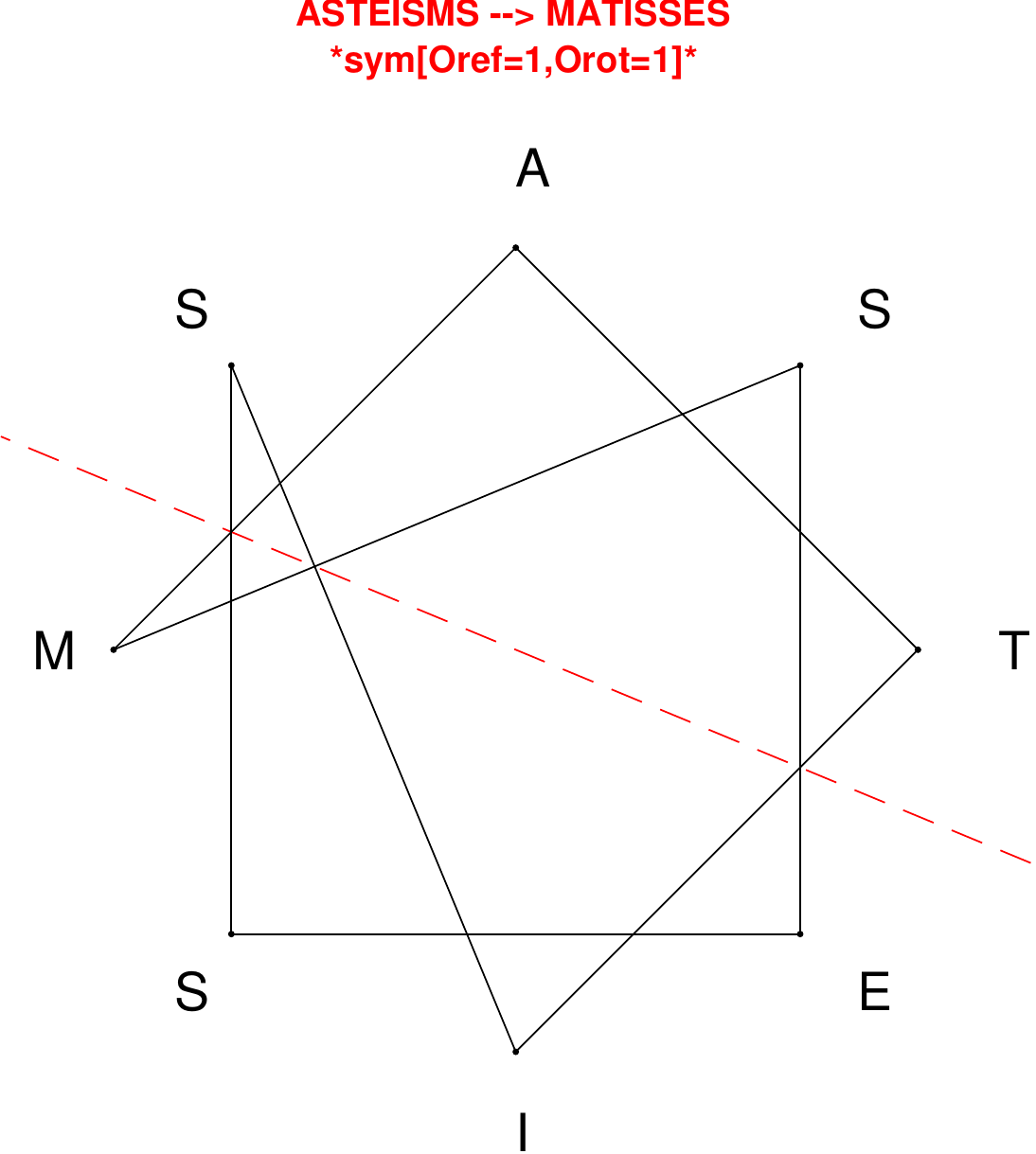}
\end{subfigure}
\hfill
\begin{subfigure}[T]{0.19\textwidth}
\centering
\includegraphics[width=\textwidth]{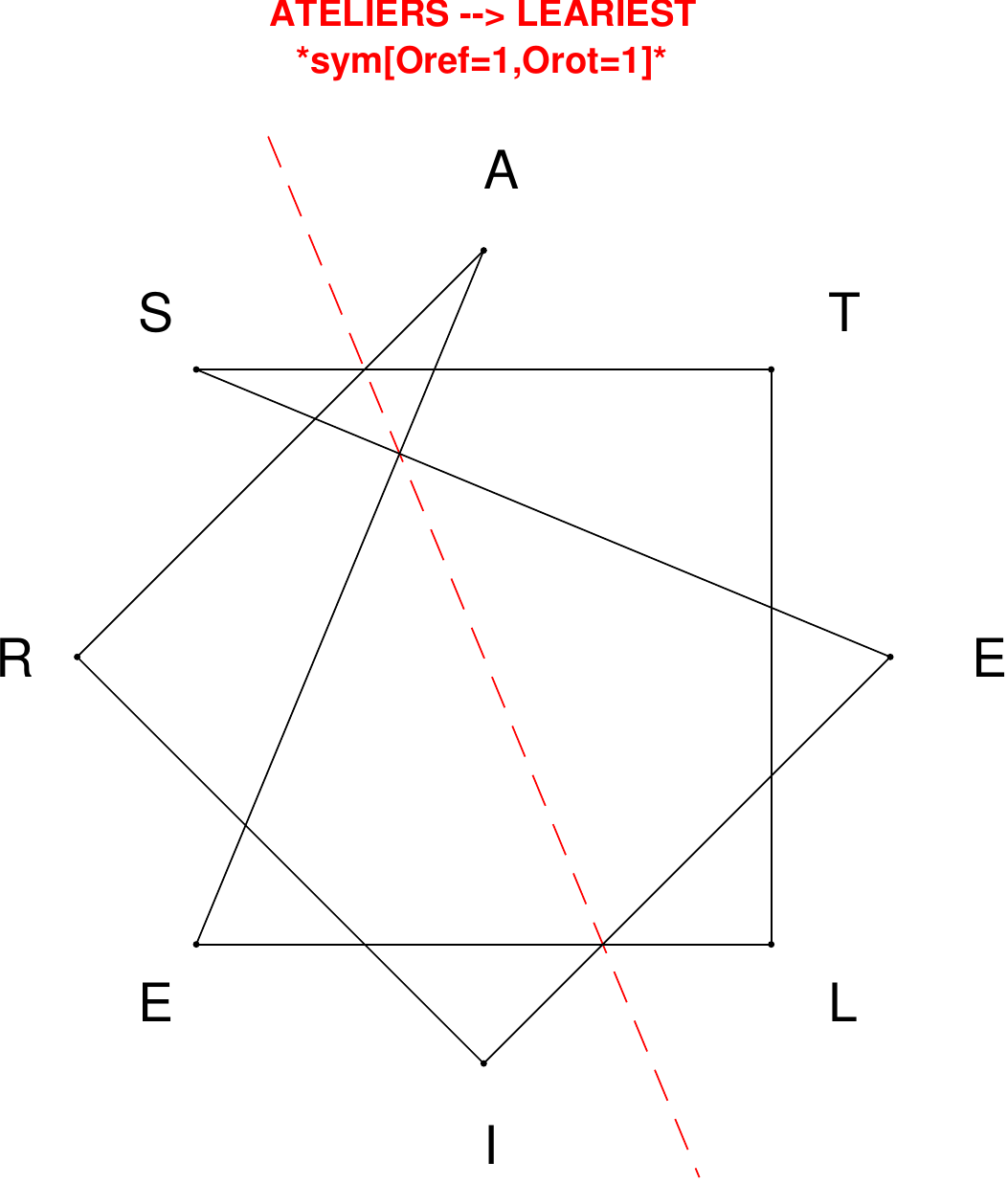}
\end{subfigure}
\hfill
\begin{subfigure}[T]{0.19\textwidth}
\centering
\includegraphics[width=\textwidth]{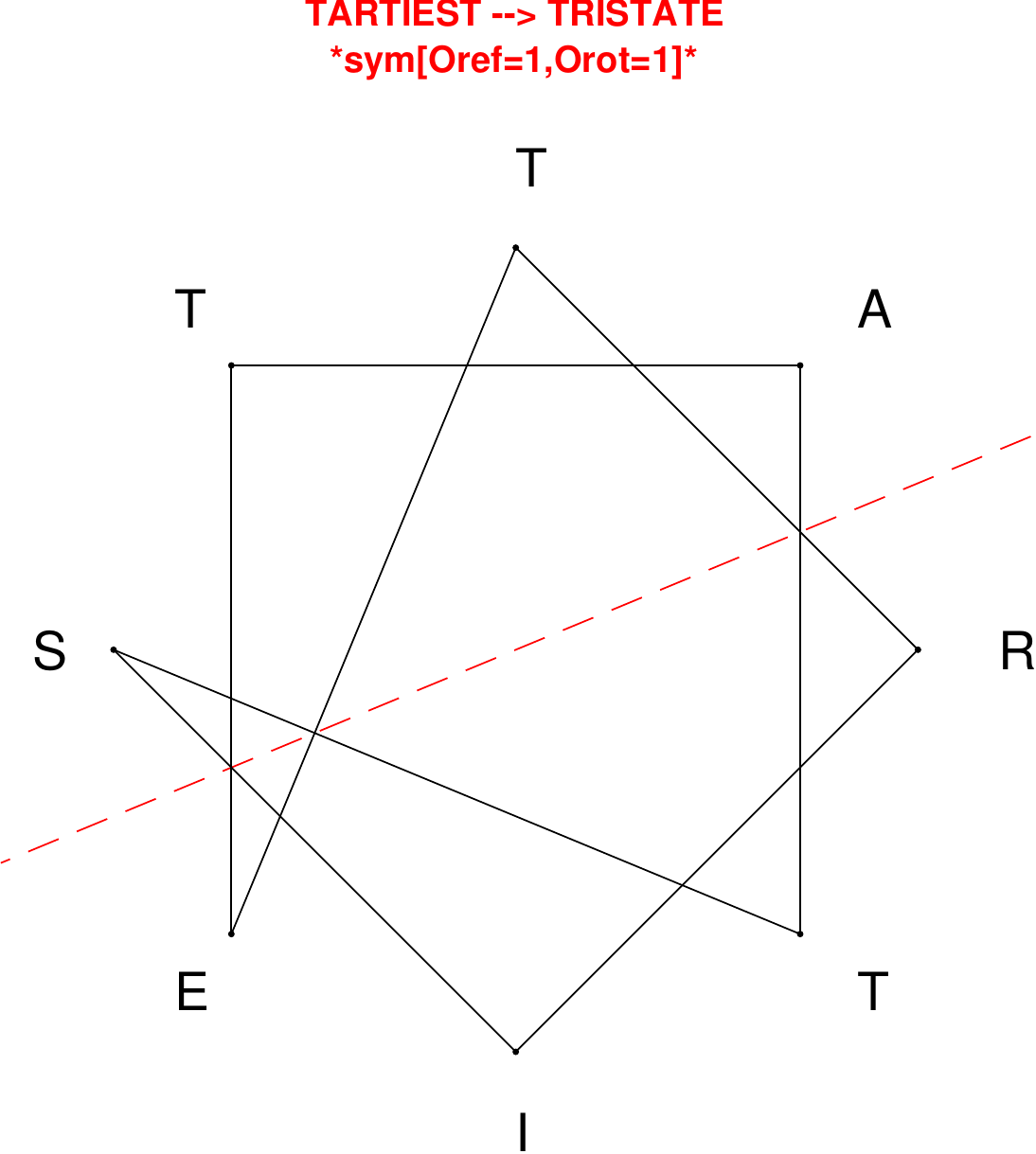}
\end{subfigure}
\hfill
\begin{subfigure}[T]{0.19\textwidth}
\centering
\includegraphics[width=\textwidth]{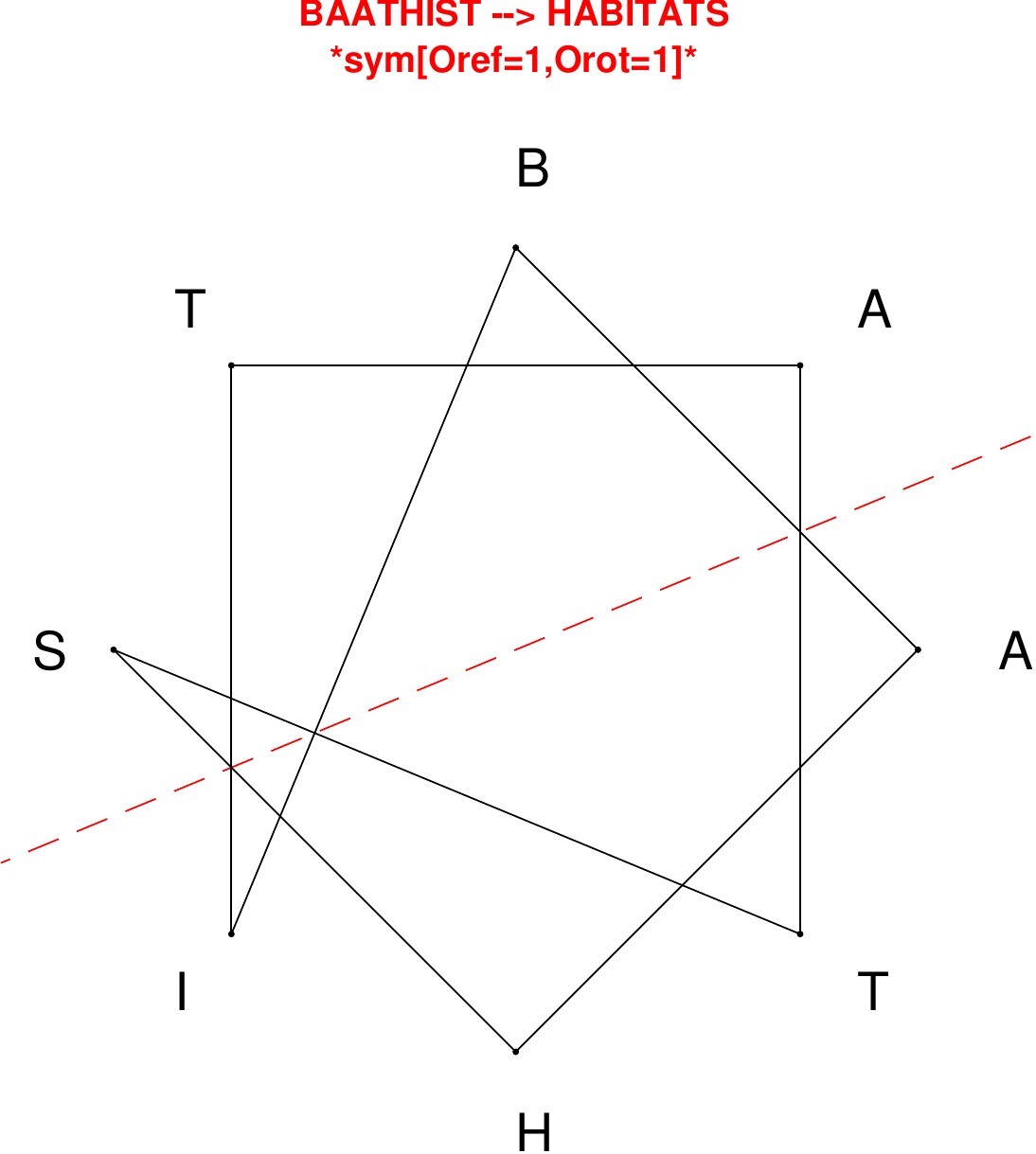}
\end{subfigure}
\end{figure}

\begin{figure}[H]
\centering
\begin{subfigure}[T]{0.19\textwidth}
\centering
\includegraphics[width=\textwidth]{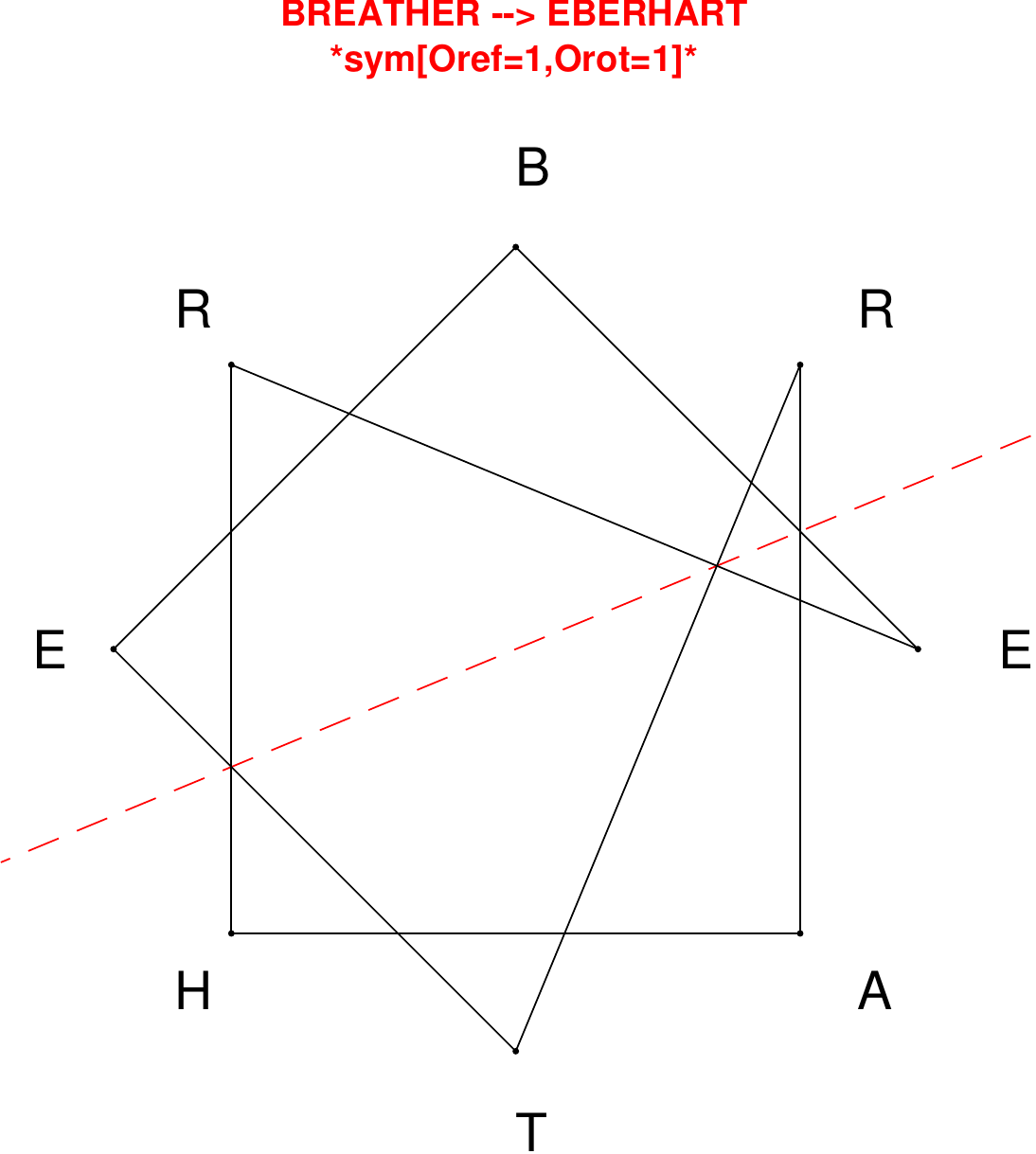}
\end{subfigure}
\hfill
\begin{subfigure}[T]{0.19\textwidth}
\centering
\includegraphics[width=\textwidth]{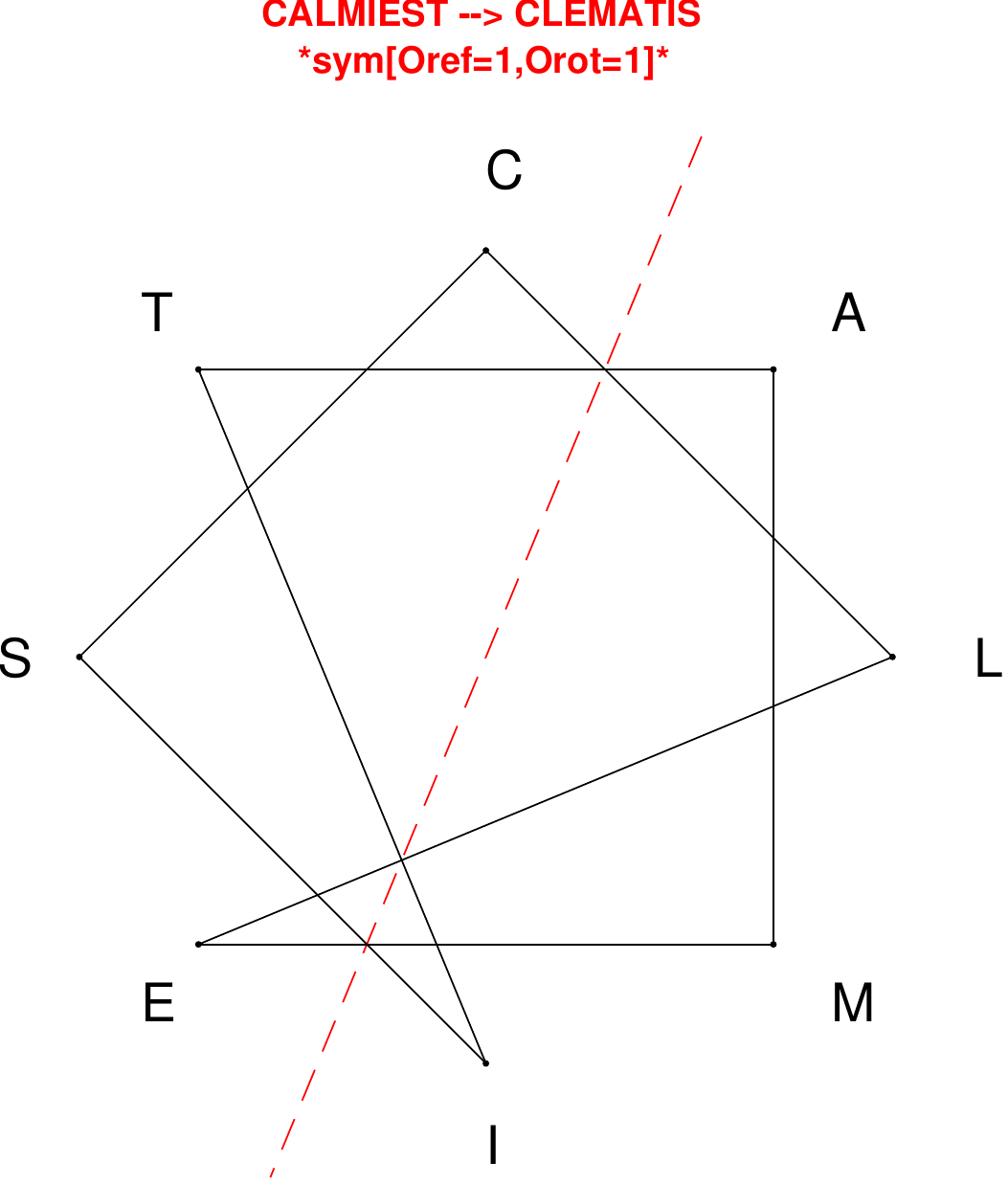}
\end{subfigure}
\hfill
\begin{subfigure}[T]{0.19\textwidth}
\centering
\includegraphics[width=\textwidth]{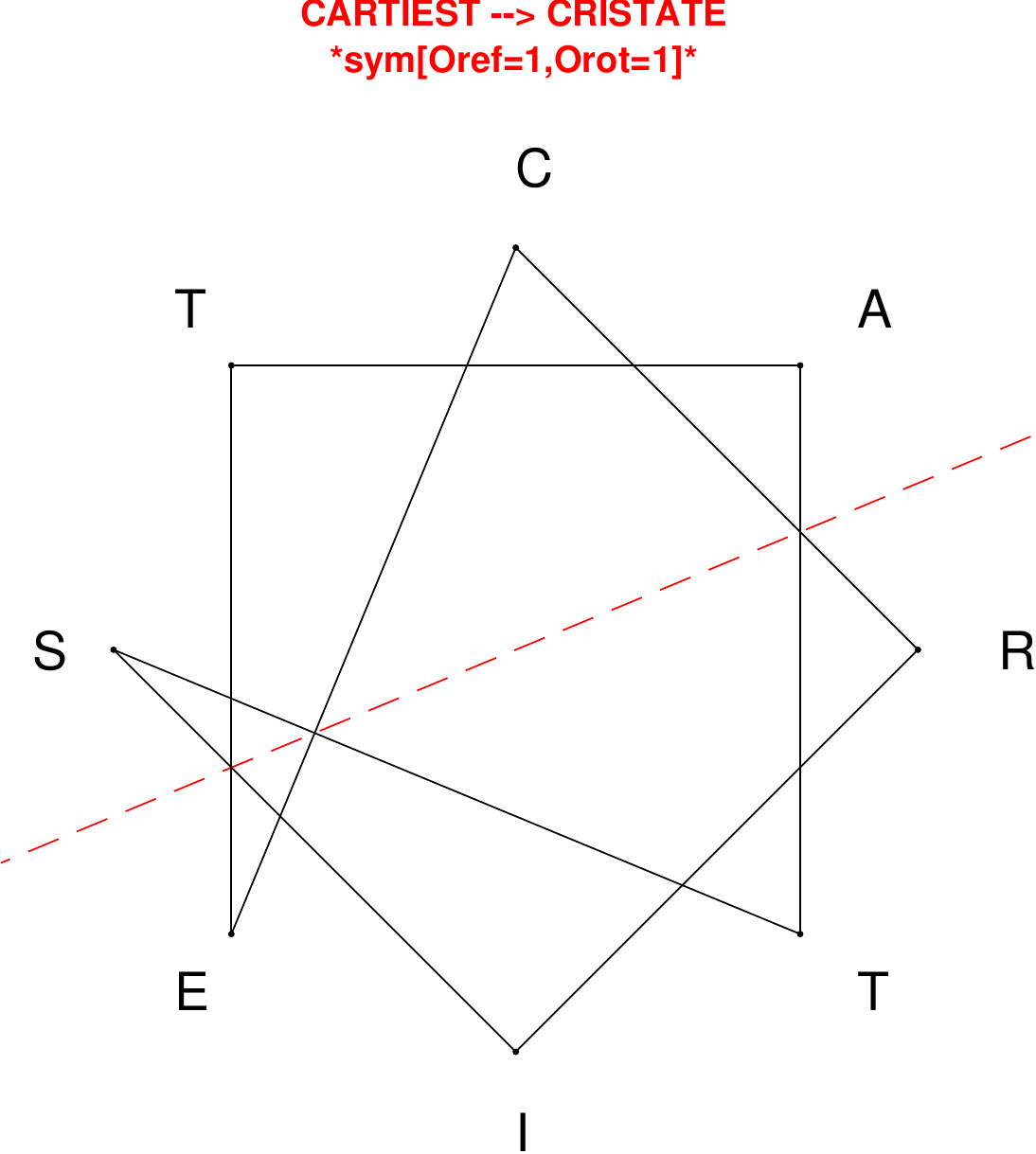}
\end{subfigure}
\hfill
\begin{subfigure}[T]{0.19\textwidth}
\centering
\includegraphics[width=\textwidth]{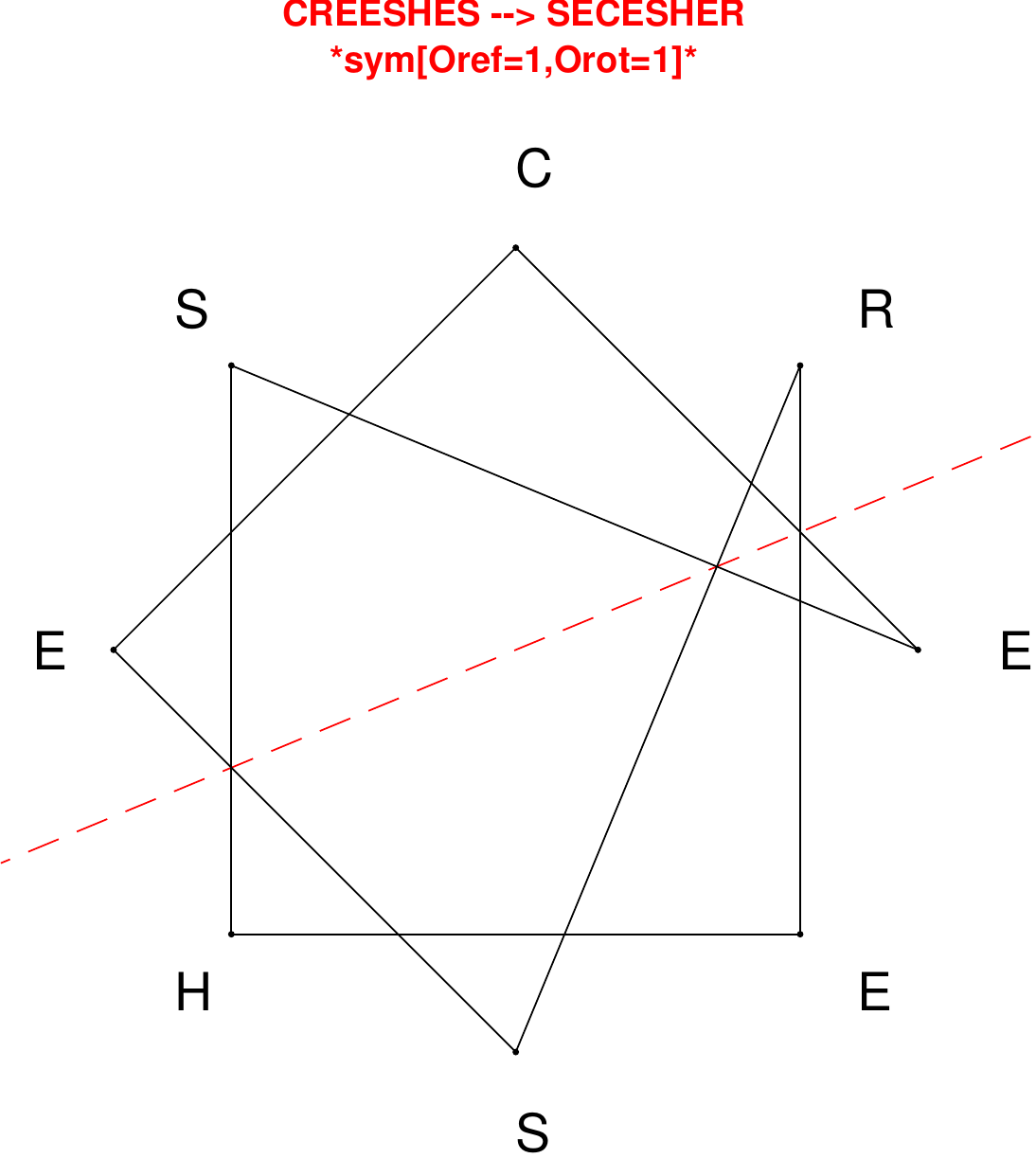}
\end{subfigure}
\hfill
\begin{subfigure}[T]{0.19\textwidth}
\centering
\includegraphics[width=\textwidth]{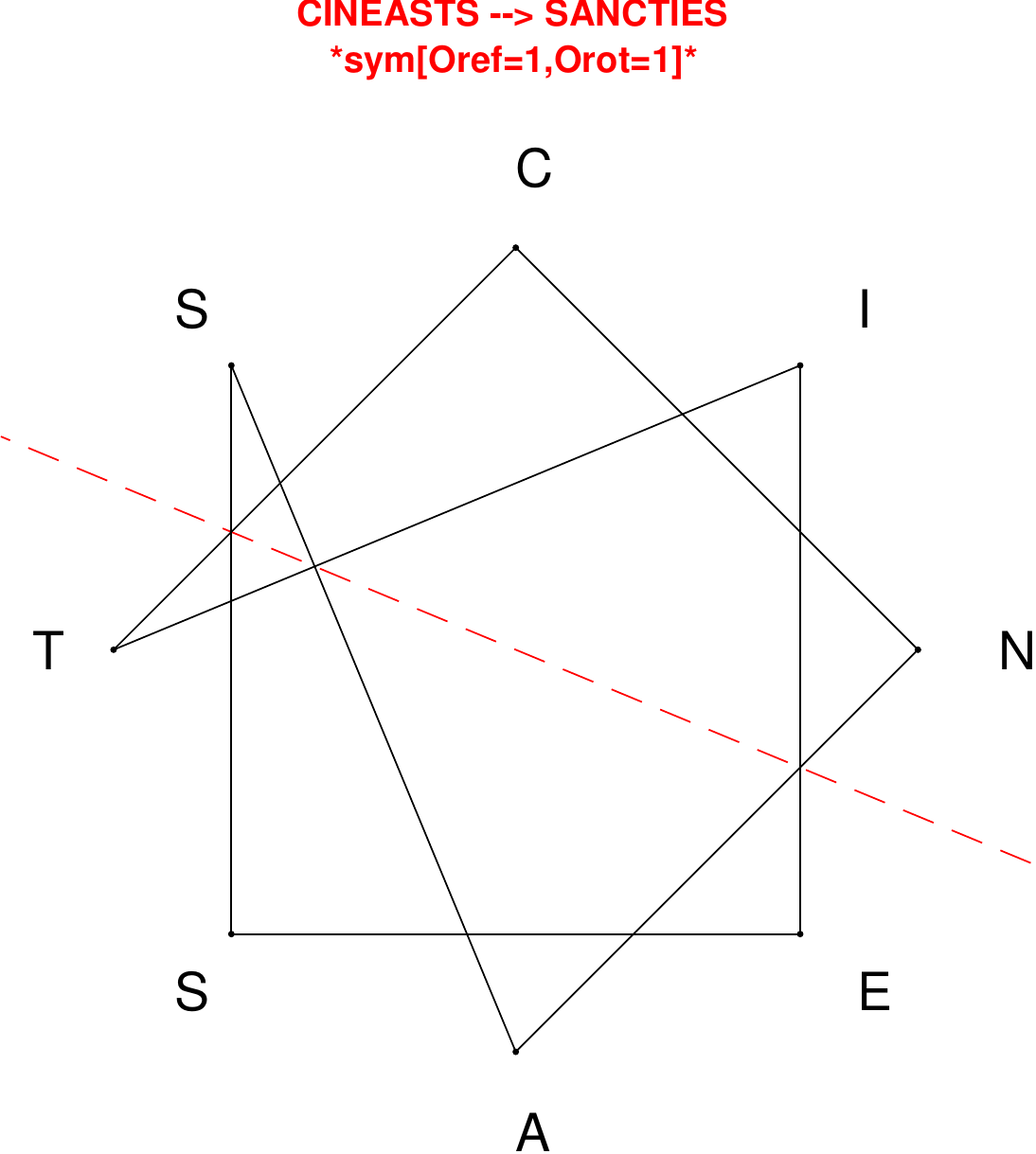}
\end{subfigure}
\end{figure}

\begin{figure}[H]
\centering
\begin{subfigure}[T]{0.19\textwidth}
\centering
\includegraphics[width=\textwidth]{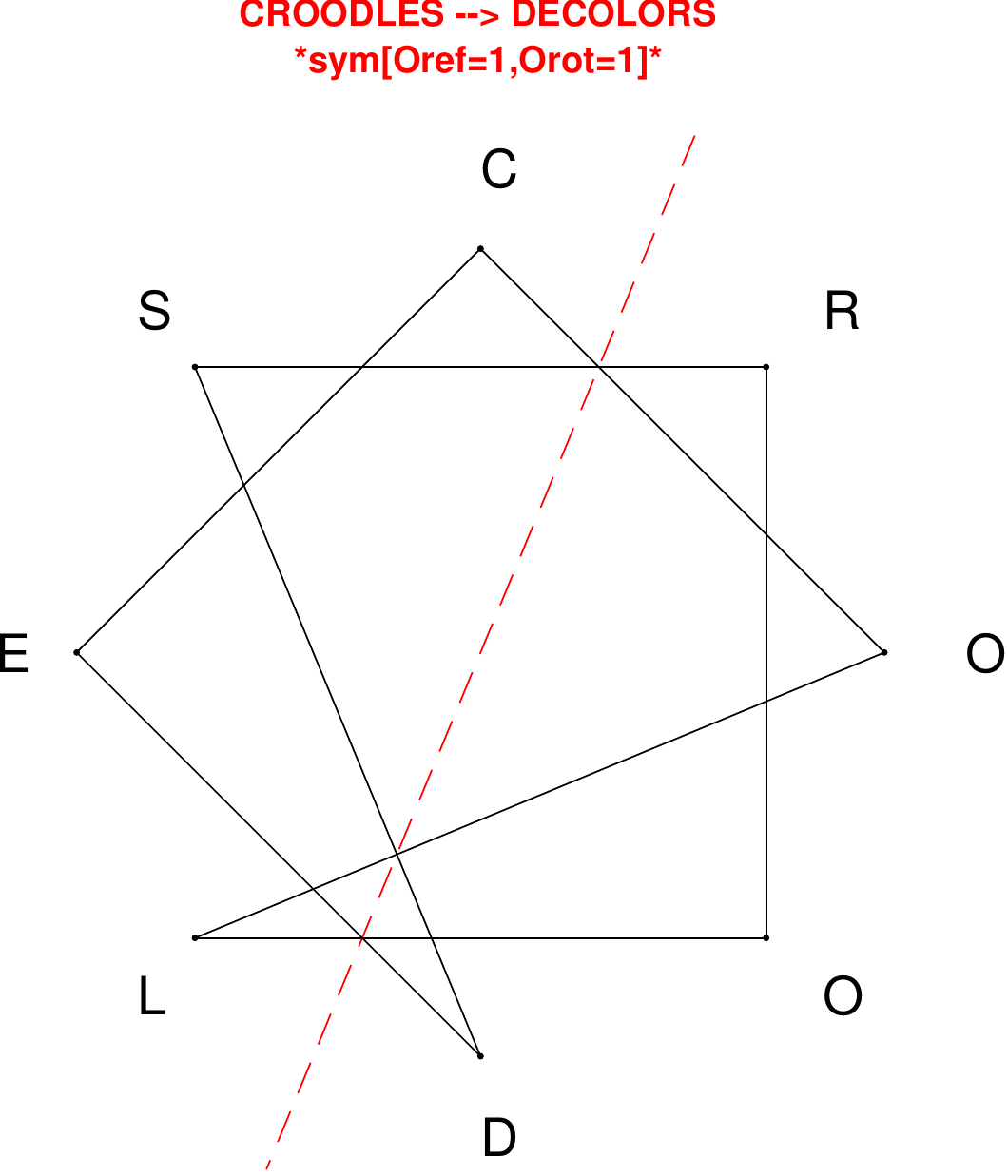}
\end{subfigure}
\hfill
\begin{subfigure}[T]{0.19\textwidth}
\centering
\includegraphics[width=\textwidth]{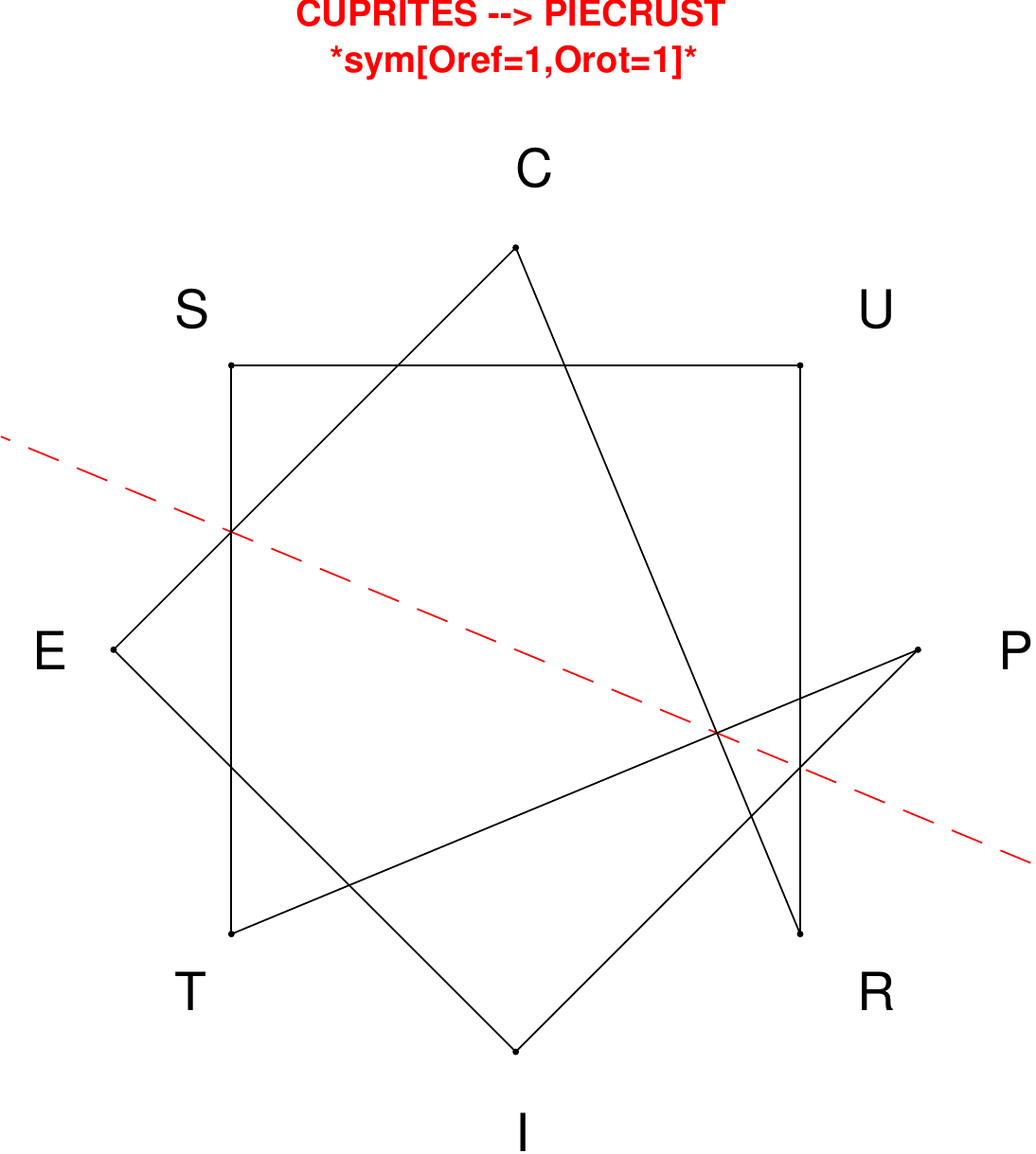}
\end{subfigure}
\hfill
\begin{subfigure}[T]{0.19\textwidth}
\centering
\includegraphics[width=\textwidth]{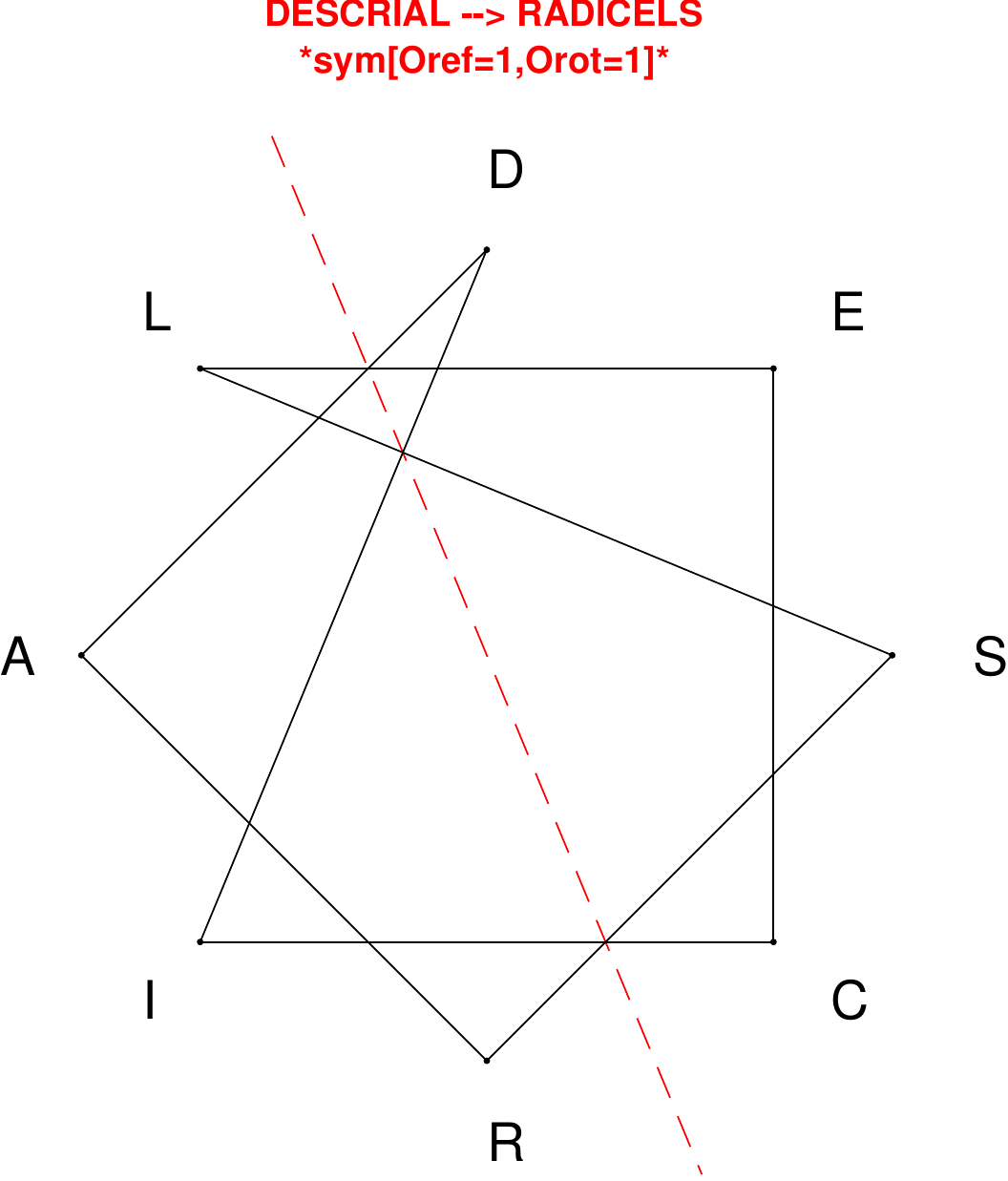}
\end{subfigure}
\hfill
\begin{subfigure}[T]{0.19\textwidth}
\centering
\includegraphics[width=\textwidth]{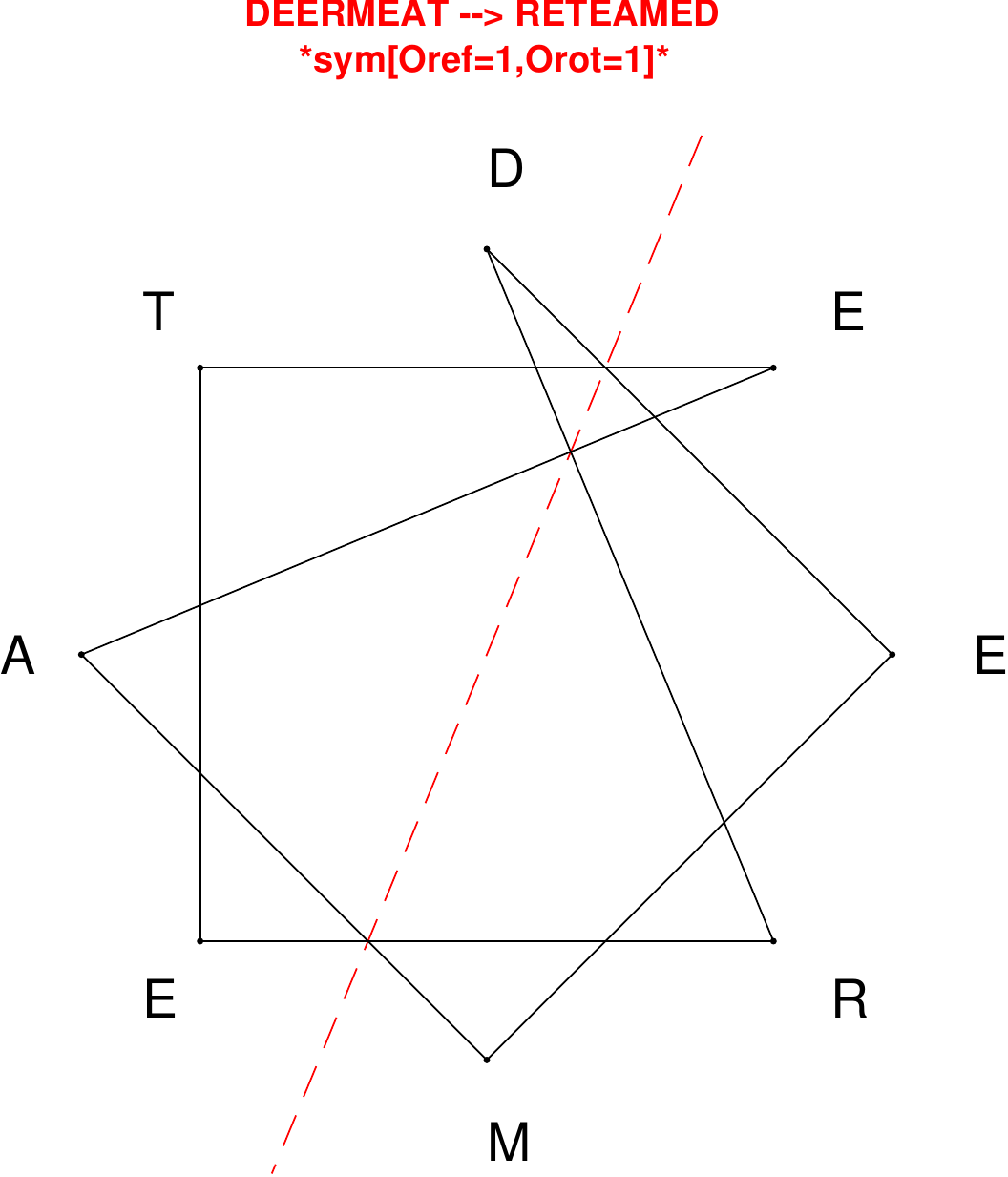}
\end{subfigure}
\hfill
\begin{subfigure}[T]{0.19\textwidth}
\centering
\includegraphics[width=\textwidth]{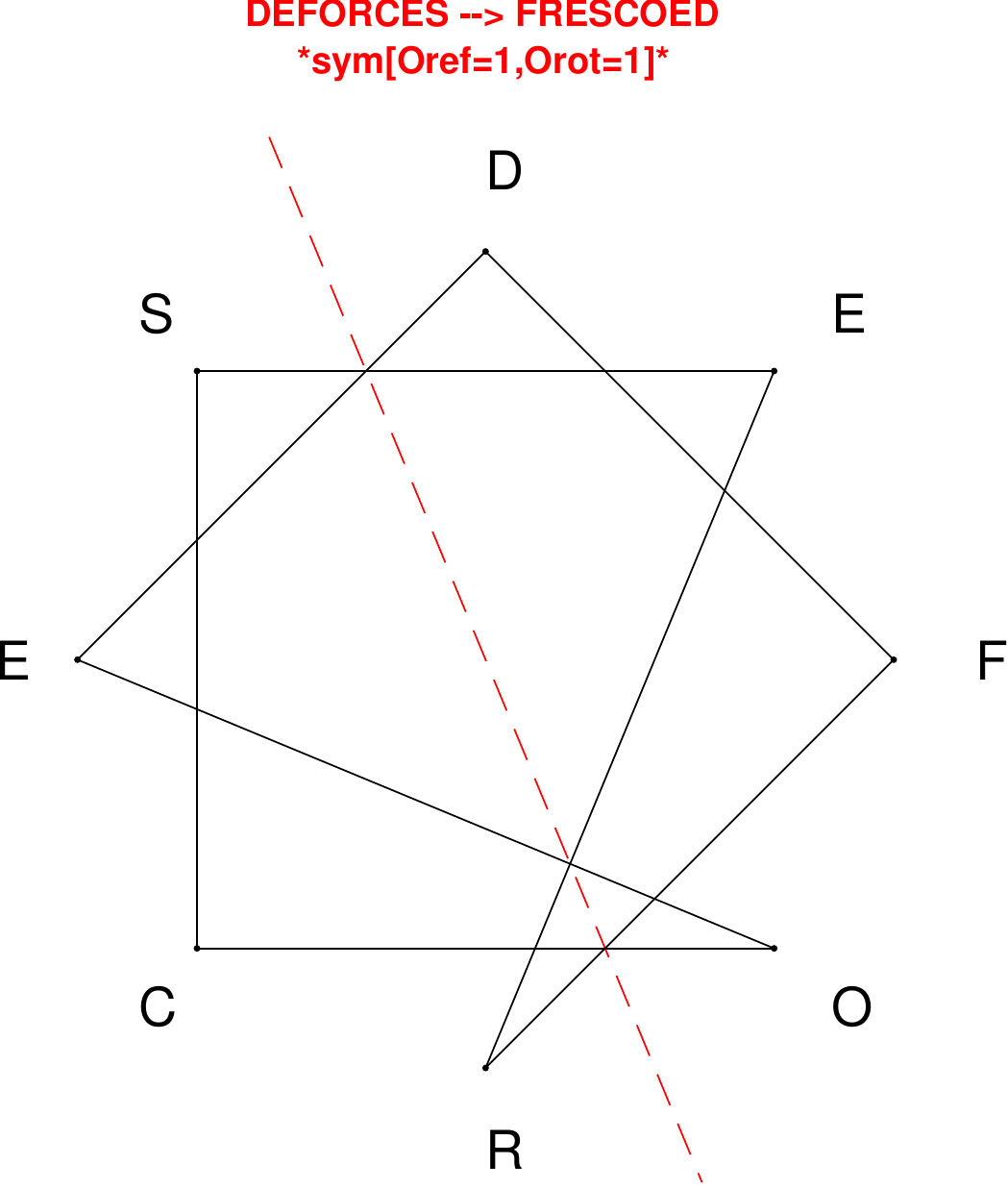}
\end{subfigure}
\end{figure}

\begin{figure}[H]
\centering
\begin{subfigure}[T]{0.19\textwidth}
\centering
\includegraphics[width=\textwidth]{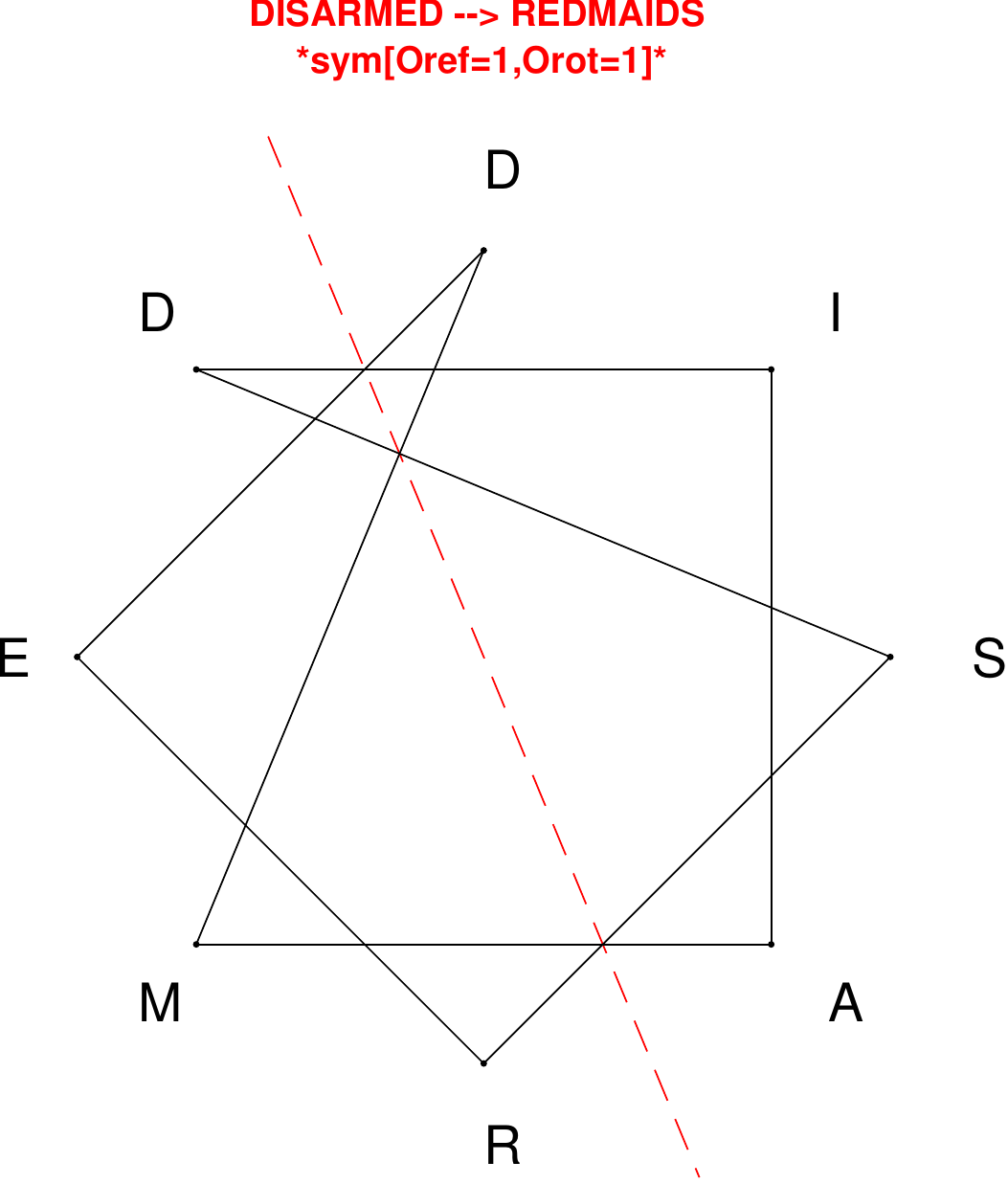}
\end{subfigure}
\hfill
\begin{subfigure}[T]{0.19\textwidth}
\centering
\includegraphics[width=\textwidth]{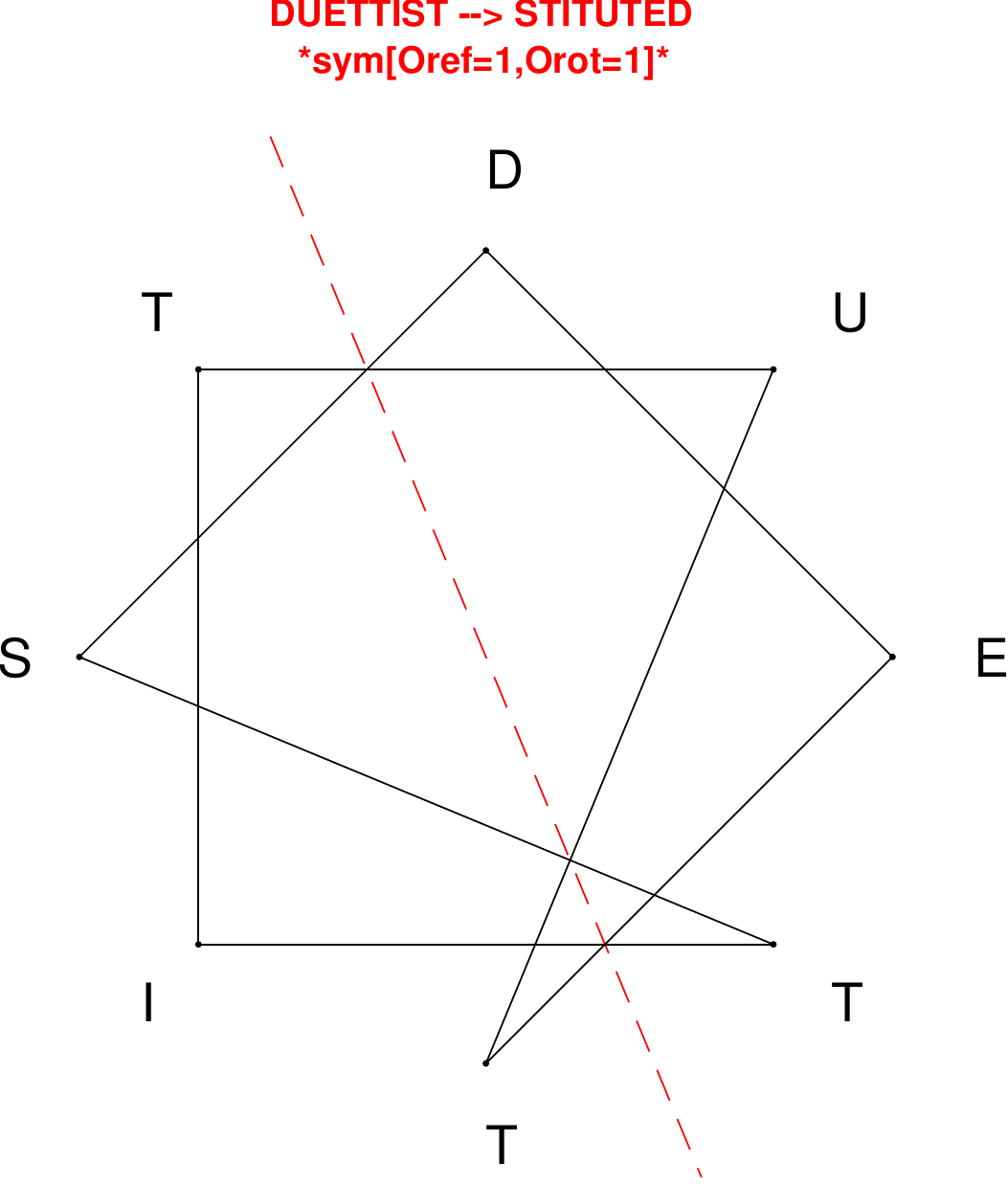}
\end{subfigure}
\hfill
\begin{subfigure}[T]{0.19\textwidth}
\centering
\includegraphics[width=\textwidth]{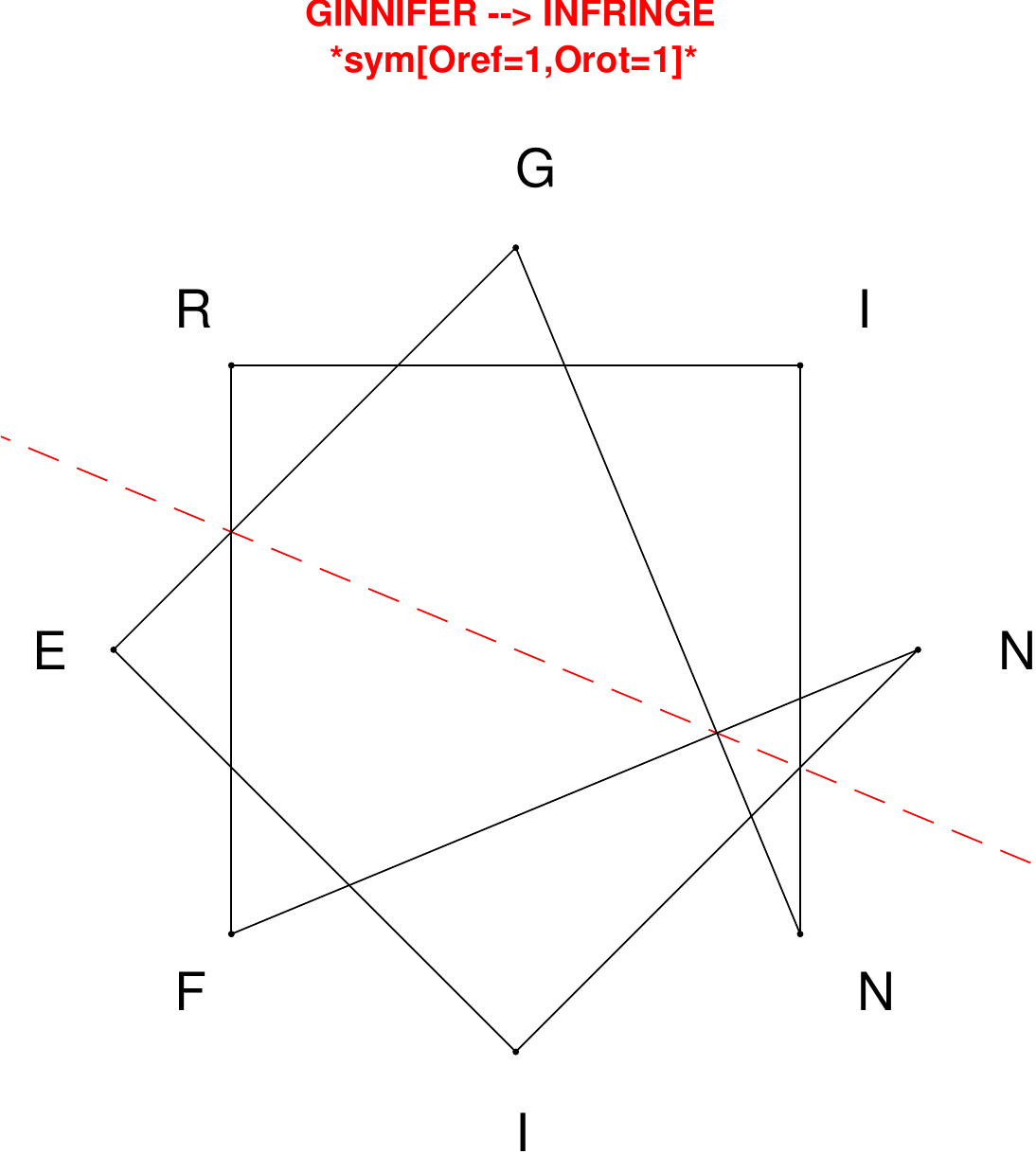}
\end{subfigure}
\hfill
\begin{subfigure}[T]{0.19\textwidth}
\centering
\includegraphics[width=\textwidth]{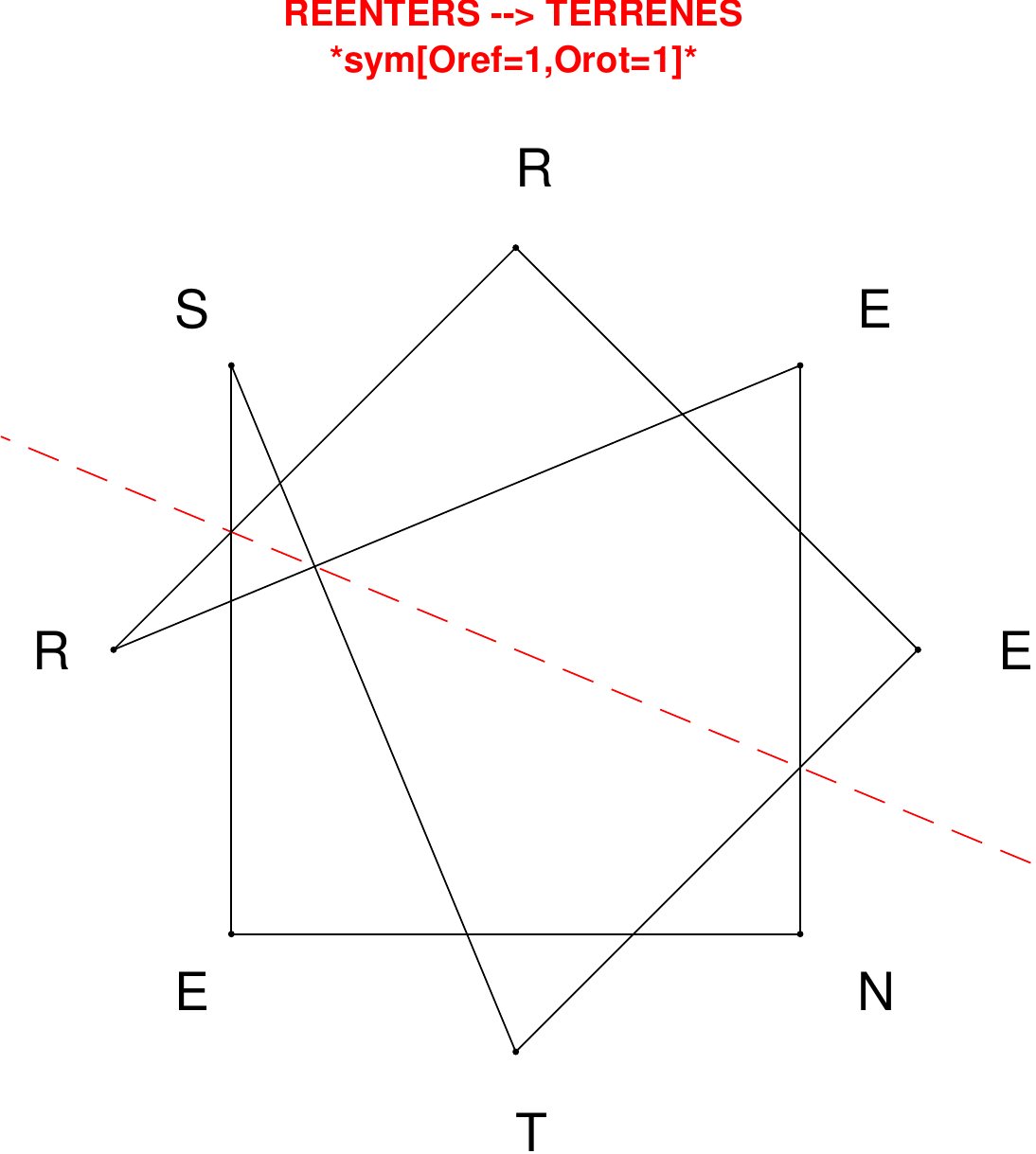}
\end{subfigure}
\hfill
\begin{subfigure}[T]{0.19\textwidth}
\centering
\includegraphics[width=\textwidth]{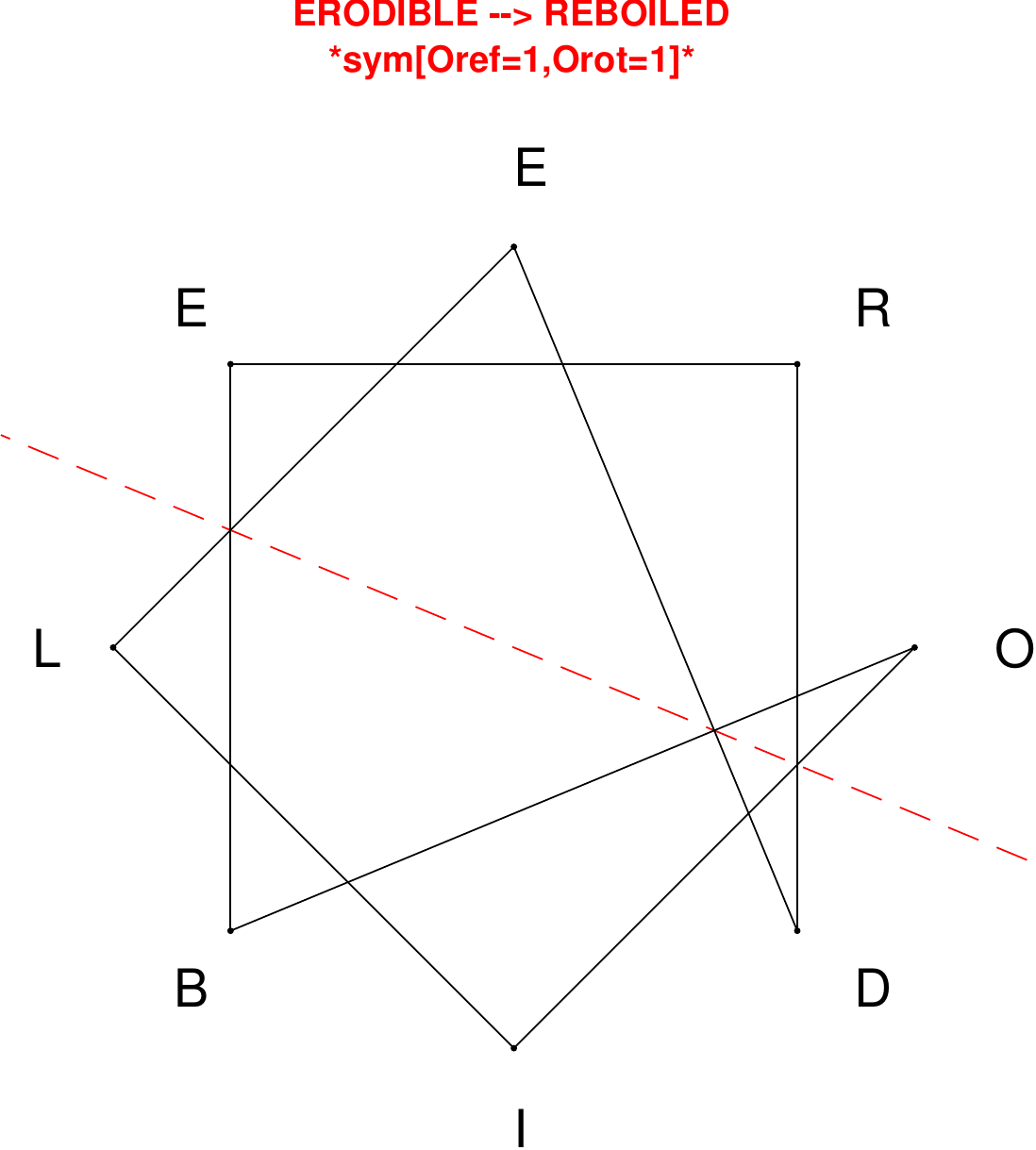}
\end{subfigure}
\end{figure}

\begin{figure}[H]
\centering
\begin{subfigure}[T]{0.19\textwidth}
\centering
\includegraphics[width=\textwidth]{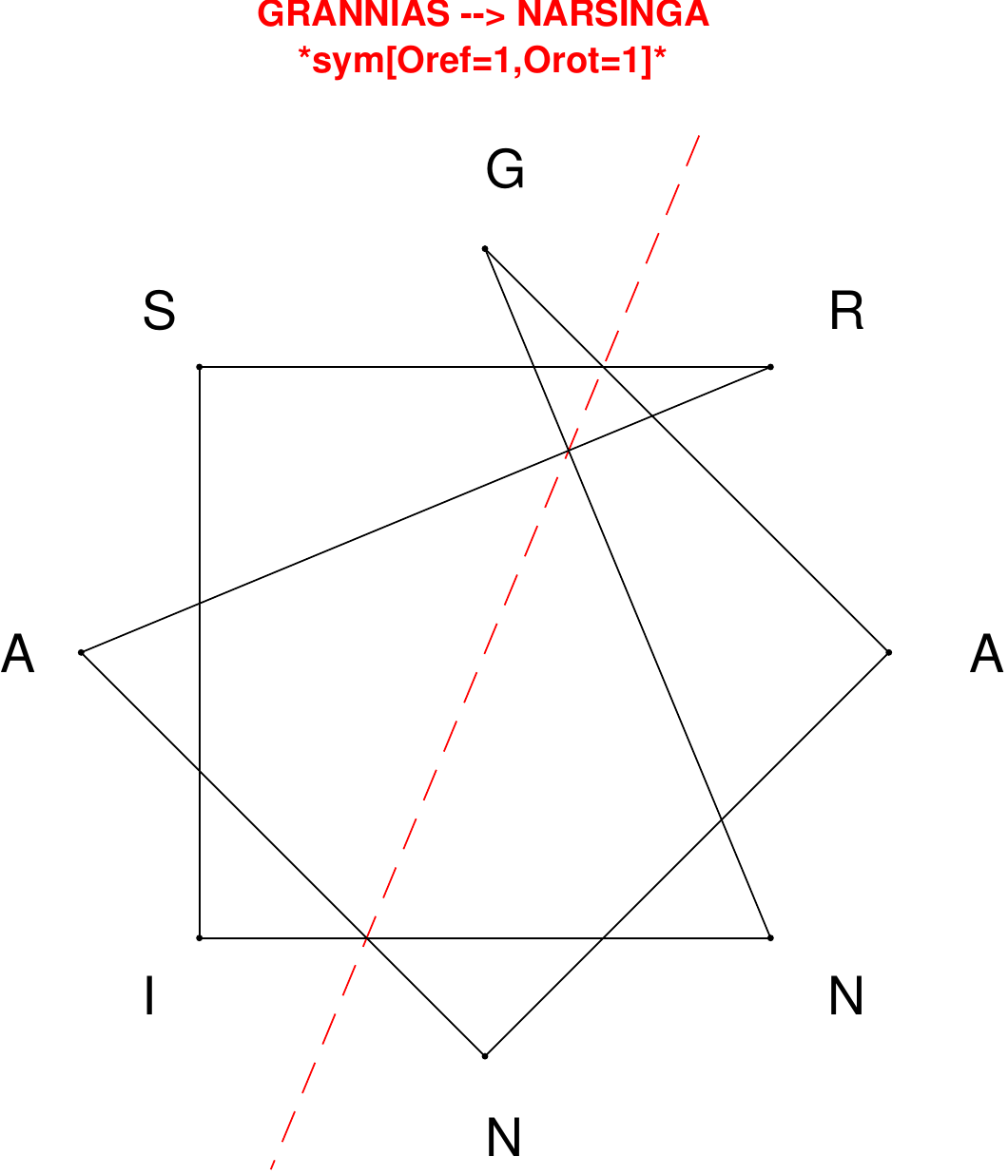}
\end{subfigure}
\hfill
\begin{subfigure}[T]{0.19\textwidth}
\centering
\includegraphics[width=\textwidth]{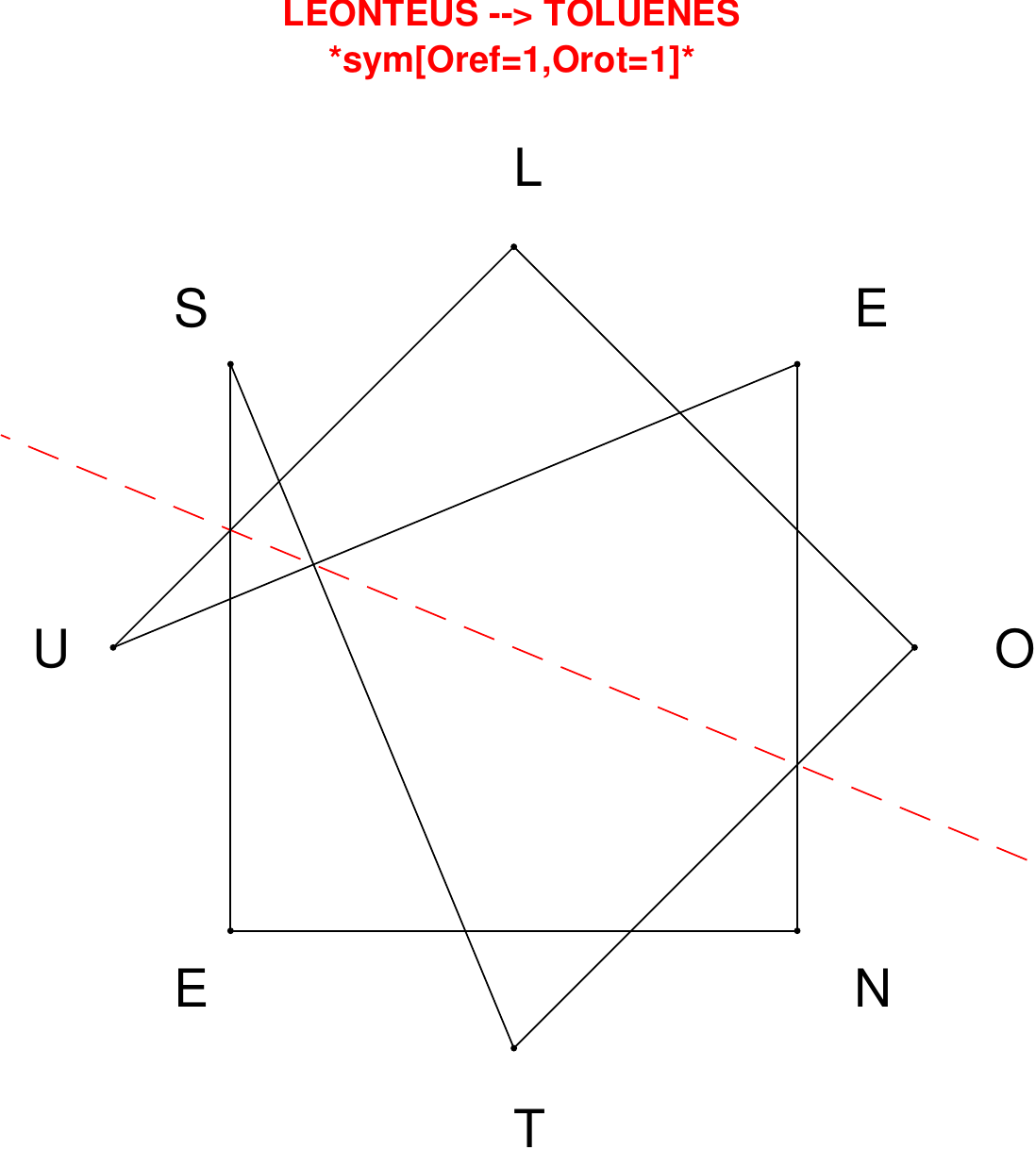}
\end{subfigure}
\hfill
\begin{subfigure}[T]{0.19\textwidth}
\centering
\includegraphics[width=\textwidth]{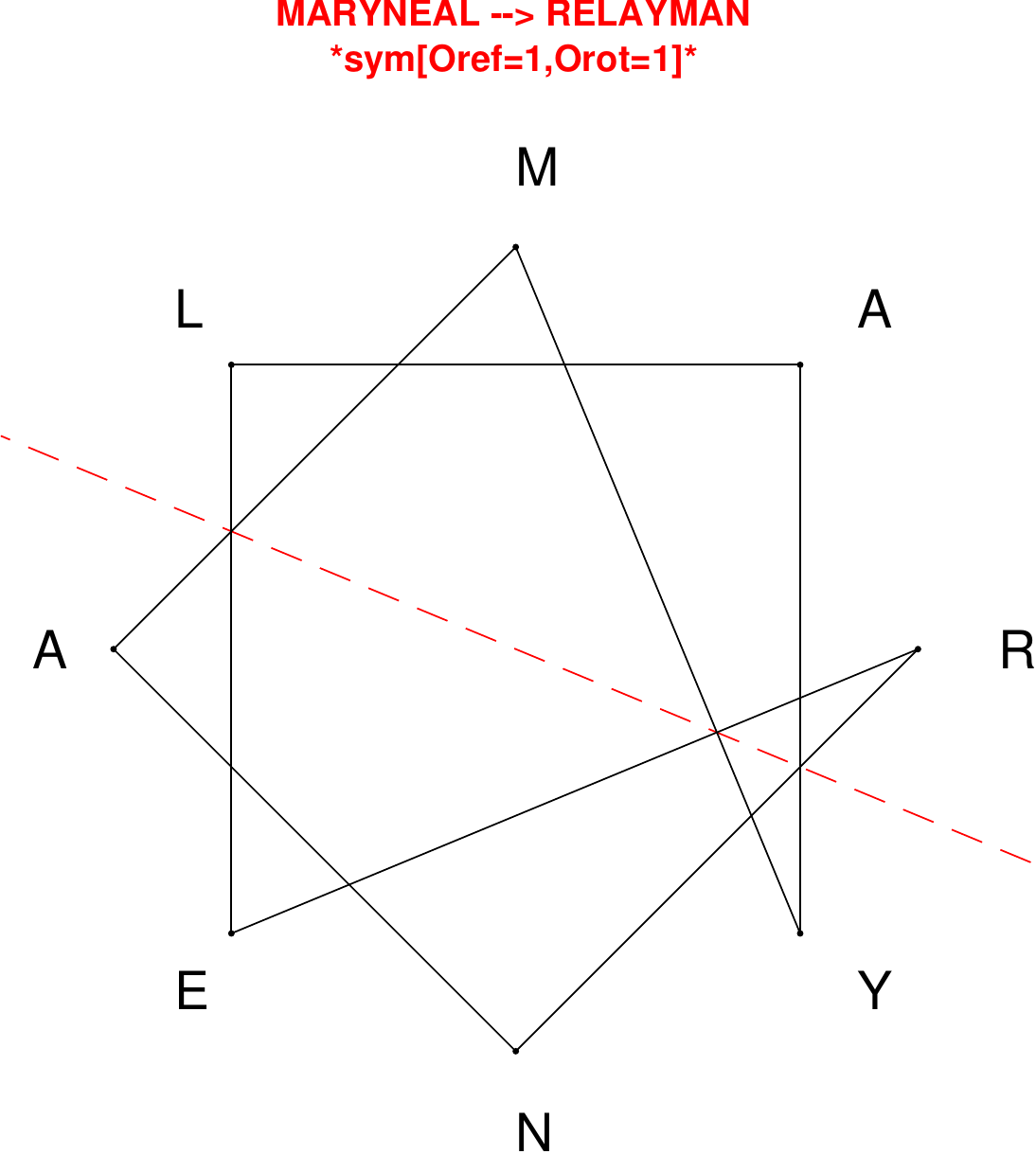}
\end{subfigure}
\hfill
\begin{subfigure}[T]{0.19\textwidth}
\centering
\includegraphics[width=\textwidth]{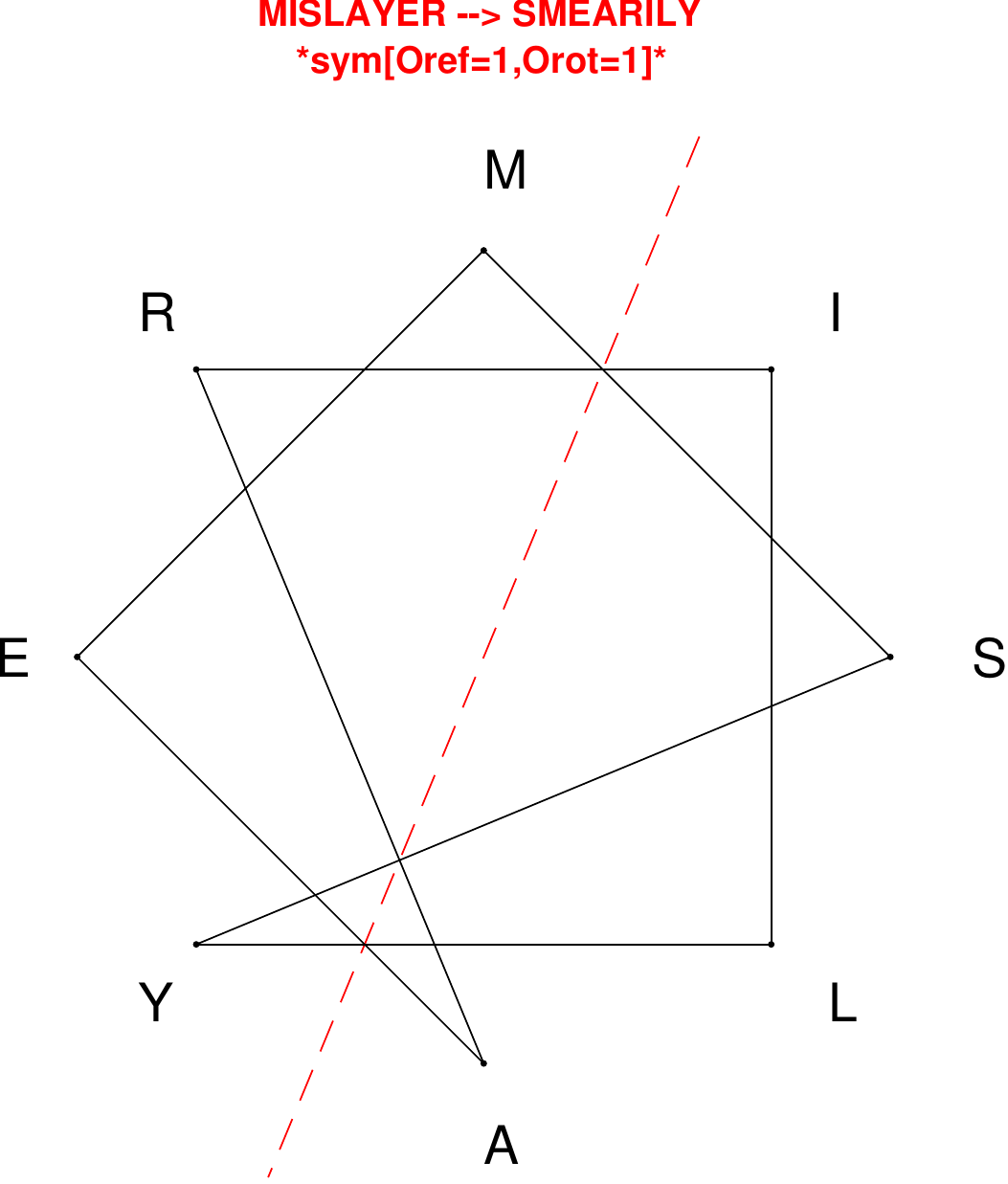}
\end{subfigure}
\hfill
\begin{subfigure}[T]{0.19\textwidth}
\centering
\includegraphics[width=\textwidth]{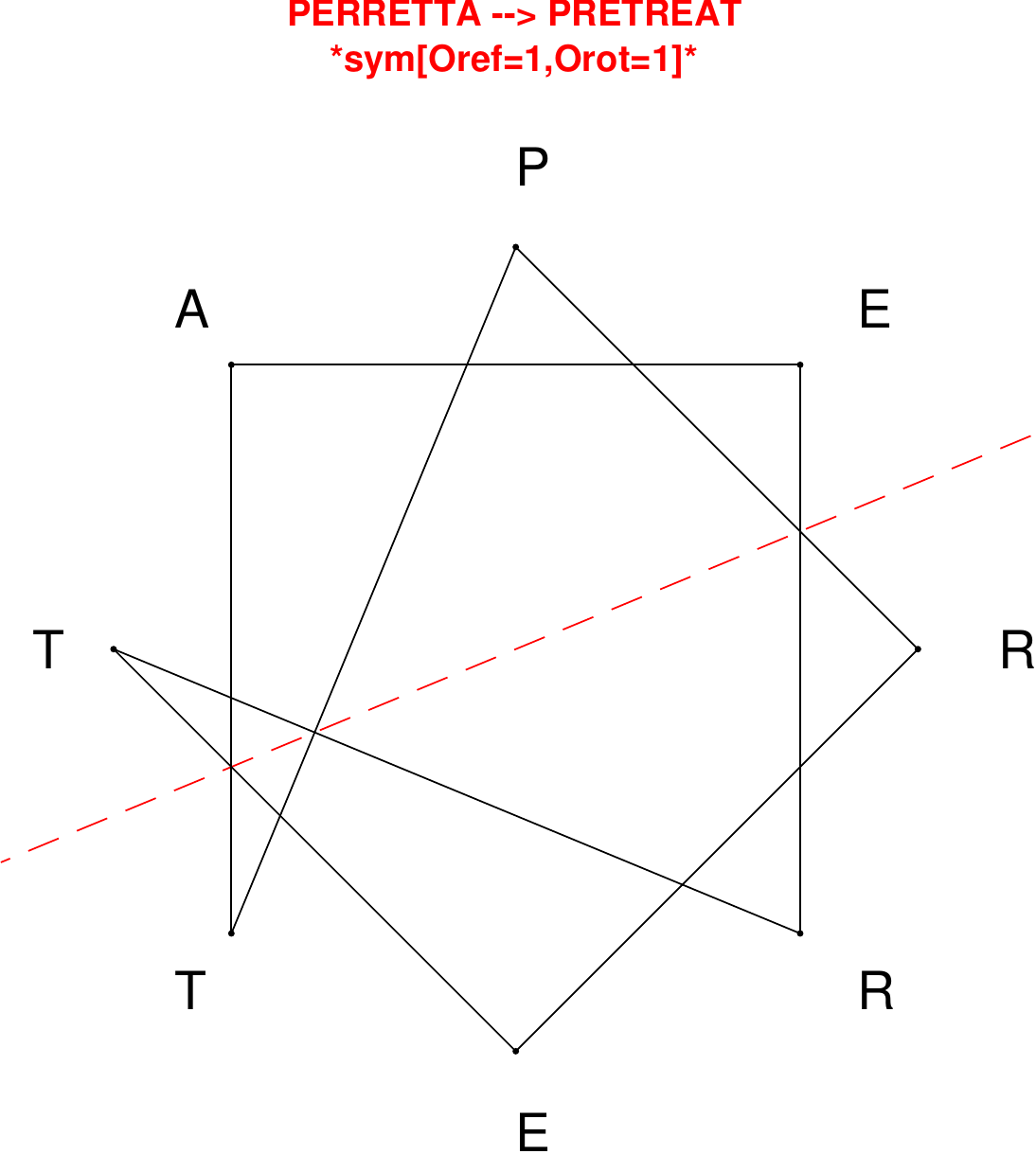}
\end{subfigure}
\end{figure}

\begin{figure}[H]
\centering
\begin{subfigure}[T]{0.19\textwidth}
\centering
\includegraphics[width=\textwidth]{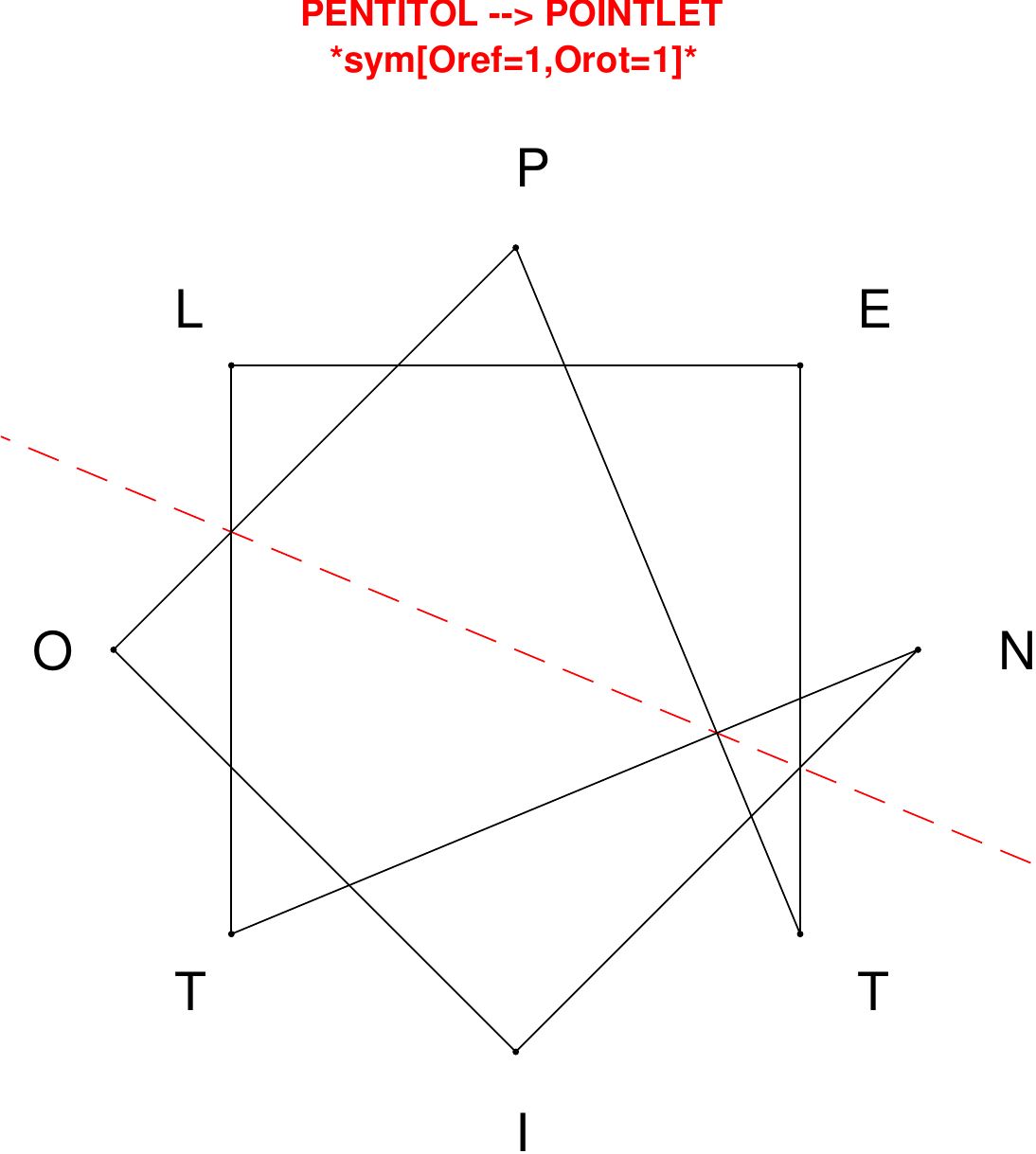}
\end{subfigure}
\hfill
\begin{subfigure}[T]{0.19\textwidth}
\centering
\includegraphics[width=\textwidth]{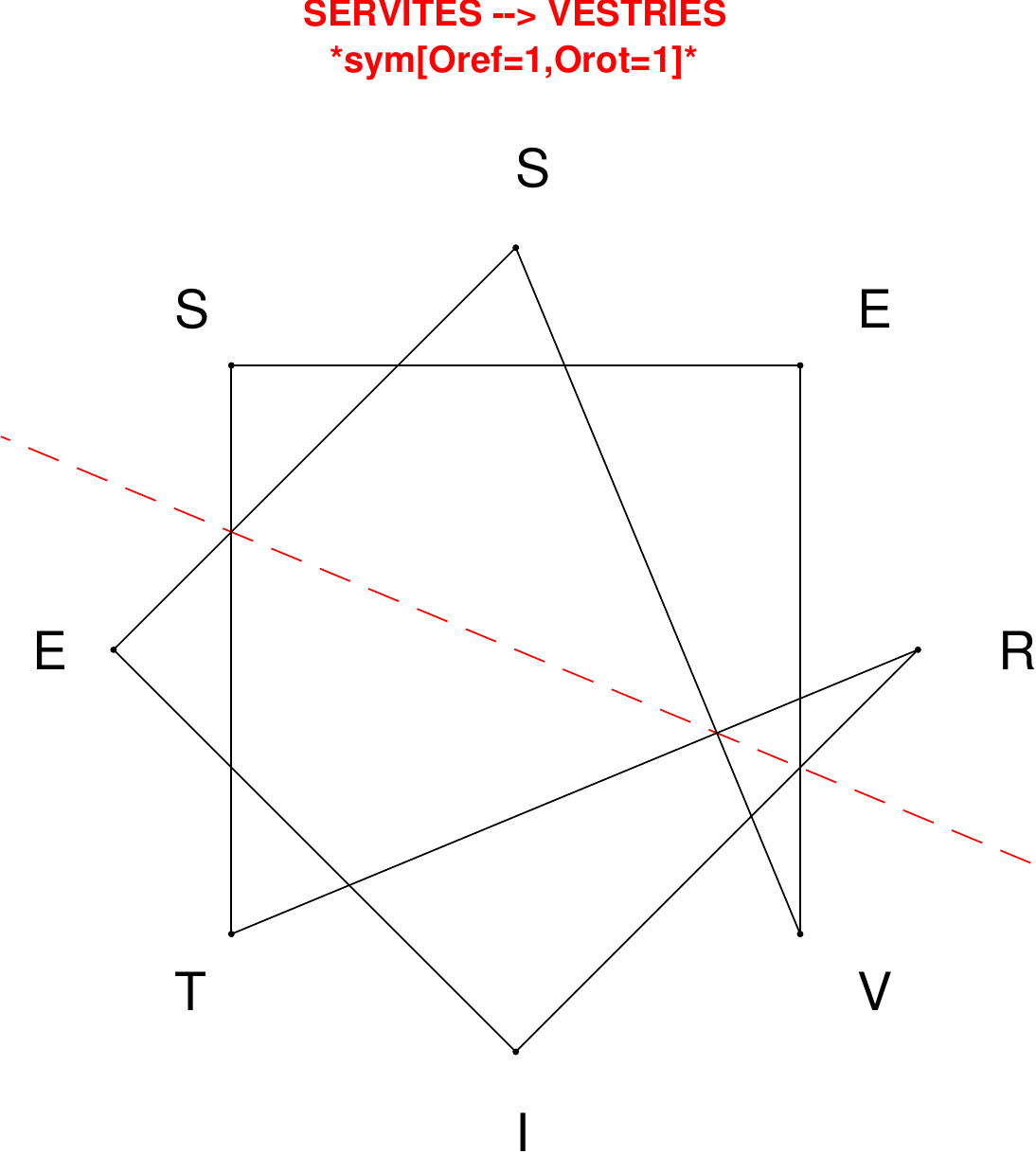}
\end{subfigure}
\hfill
\begin{subfigure}[T]{0.19\textwidth}
\centering
\includegraphics[width=\textwidth]{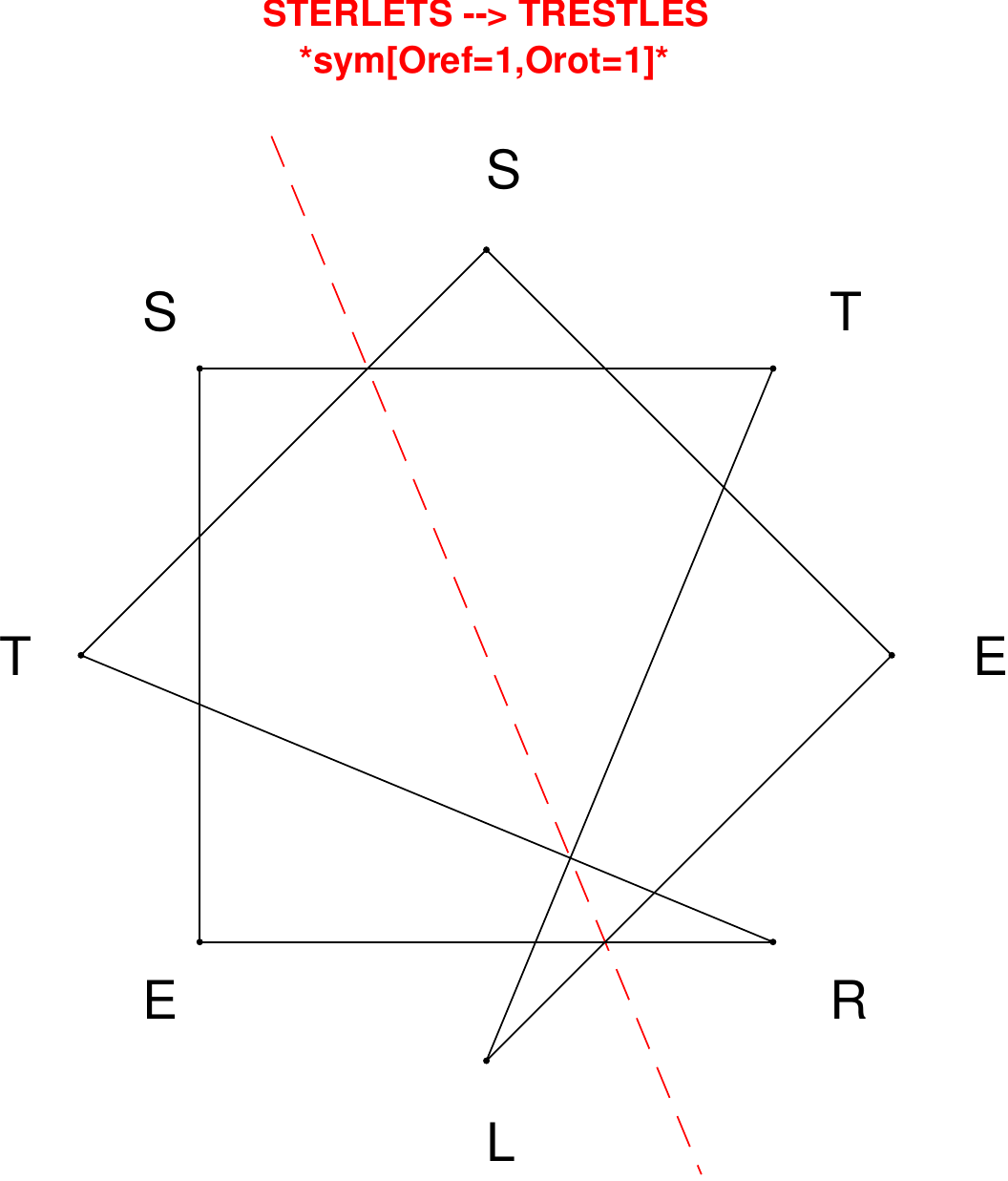}
\end{subfigure}
\hfill
\begin{subfigure}[T]{0.19\textwidth}
\centering
\includegraphics[width=\textwidth]{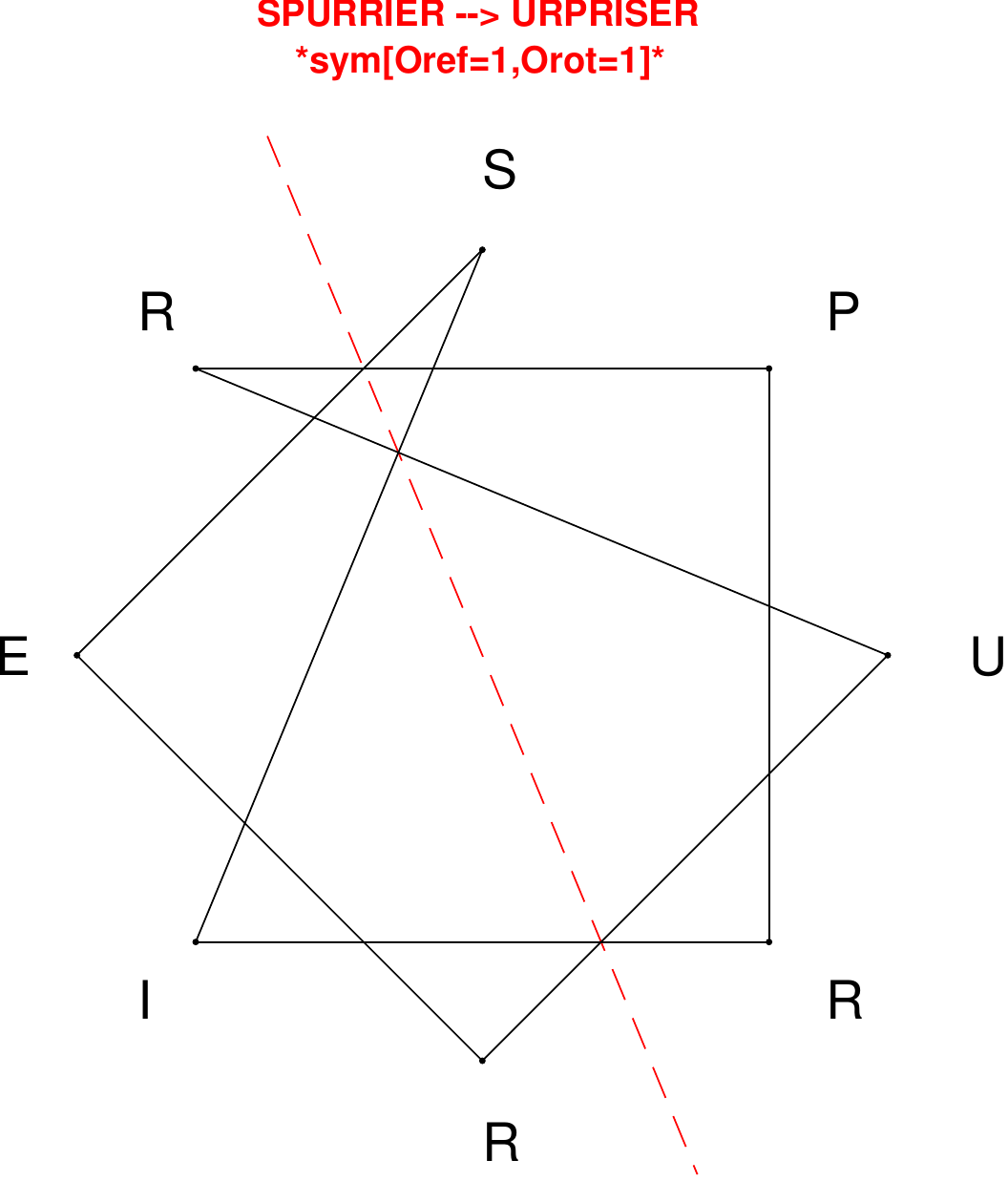}
\end{subfigure}
\hfill
\begin{subfigure}[T]{0.19\textwidth}
\centering
\includegraphics[width=\textwidth]{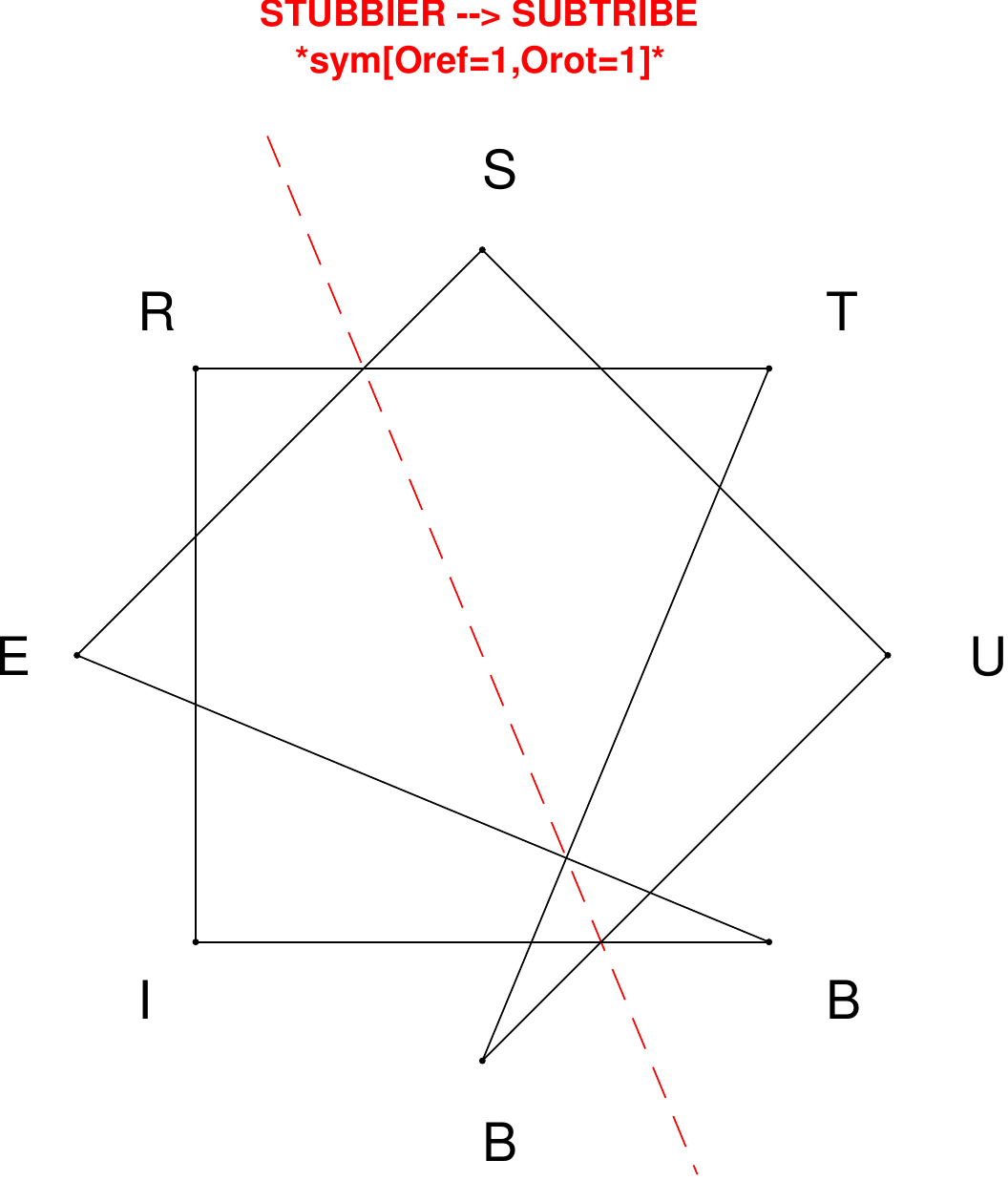}
\end{subfigure}
\end{figure}

\begin{figure}[H]
\centering
\begin{subfigure}[T]{0.19\textwidth}
\centering
\includegraphics[width=\textwidth]{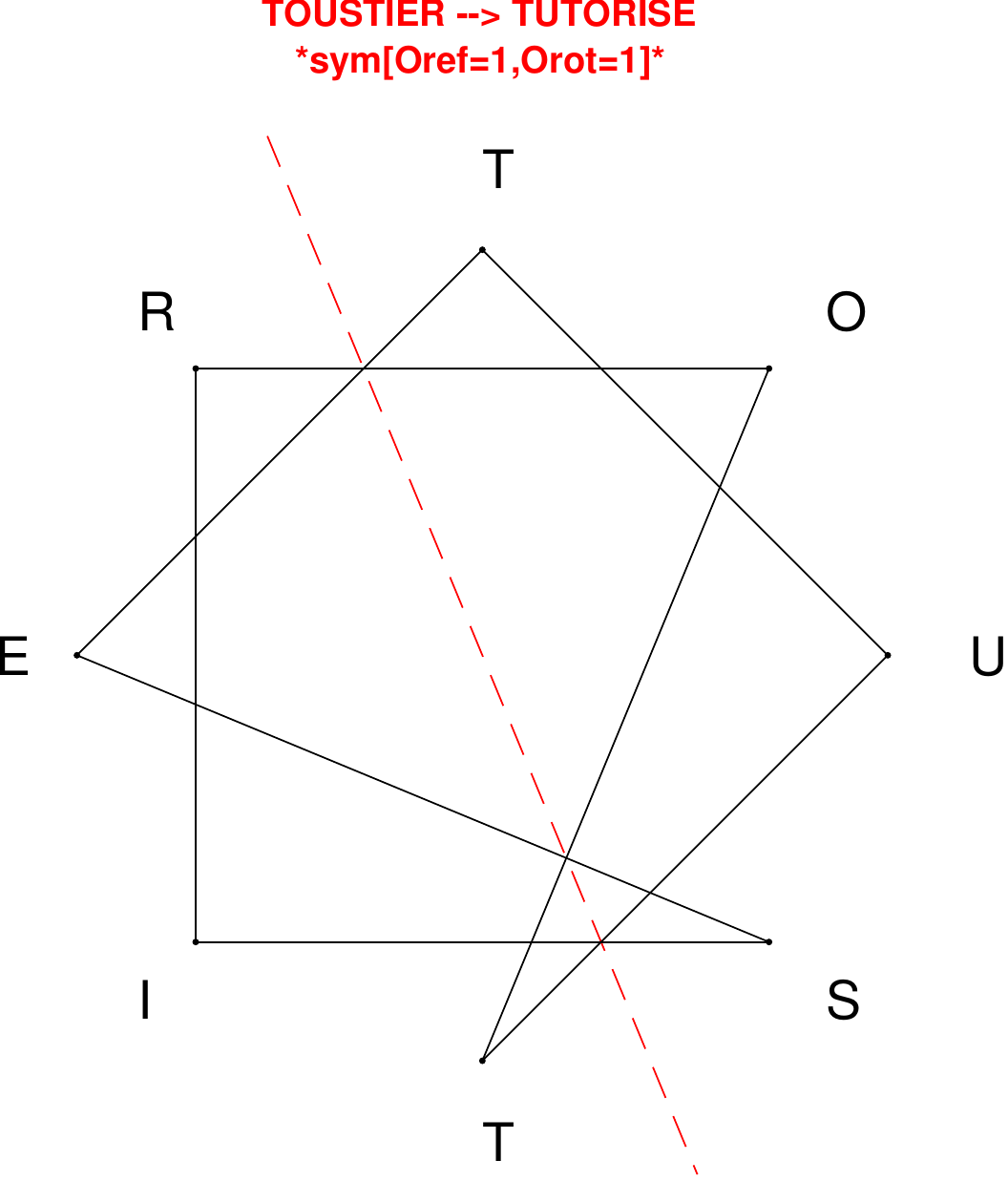}
\end{subfigure}
\hfill
\begin{subfigure}[T]{0.19\textwidth}
\centering
\includegraphics[width=\textwidth]{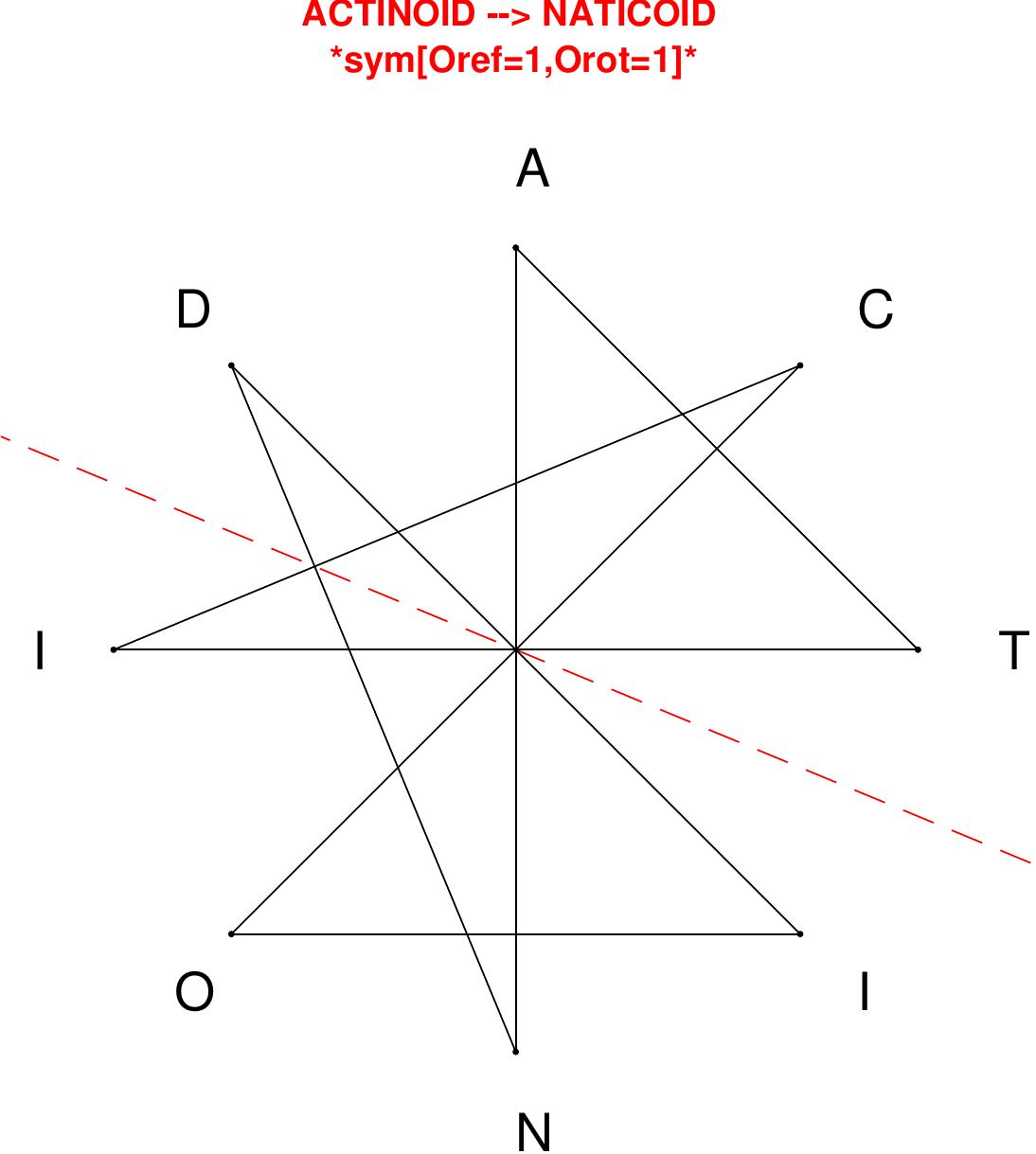}
\end{subfigure}
\hfill
\begin{subfigure}[T]{0.19\textwidth}
\centering
\includegraphics[width=\textwidth]{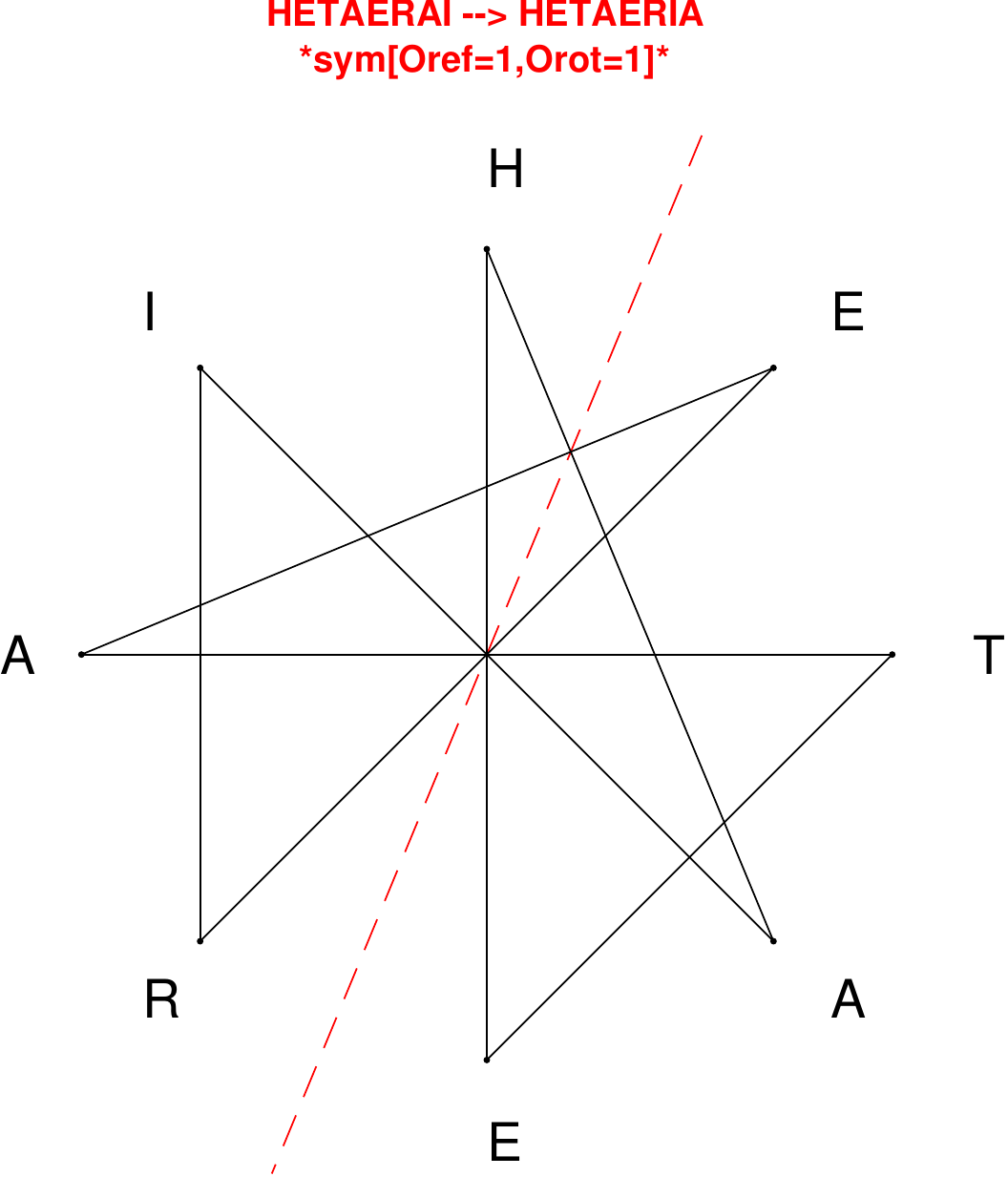}
\end{subfigure}
\hfill
\begin{subfigure}[T]{0.19\textwidth}
\centering
\includegraphics[width=\textwidth]{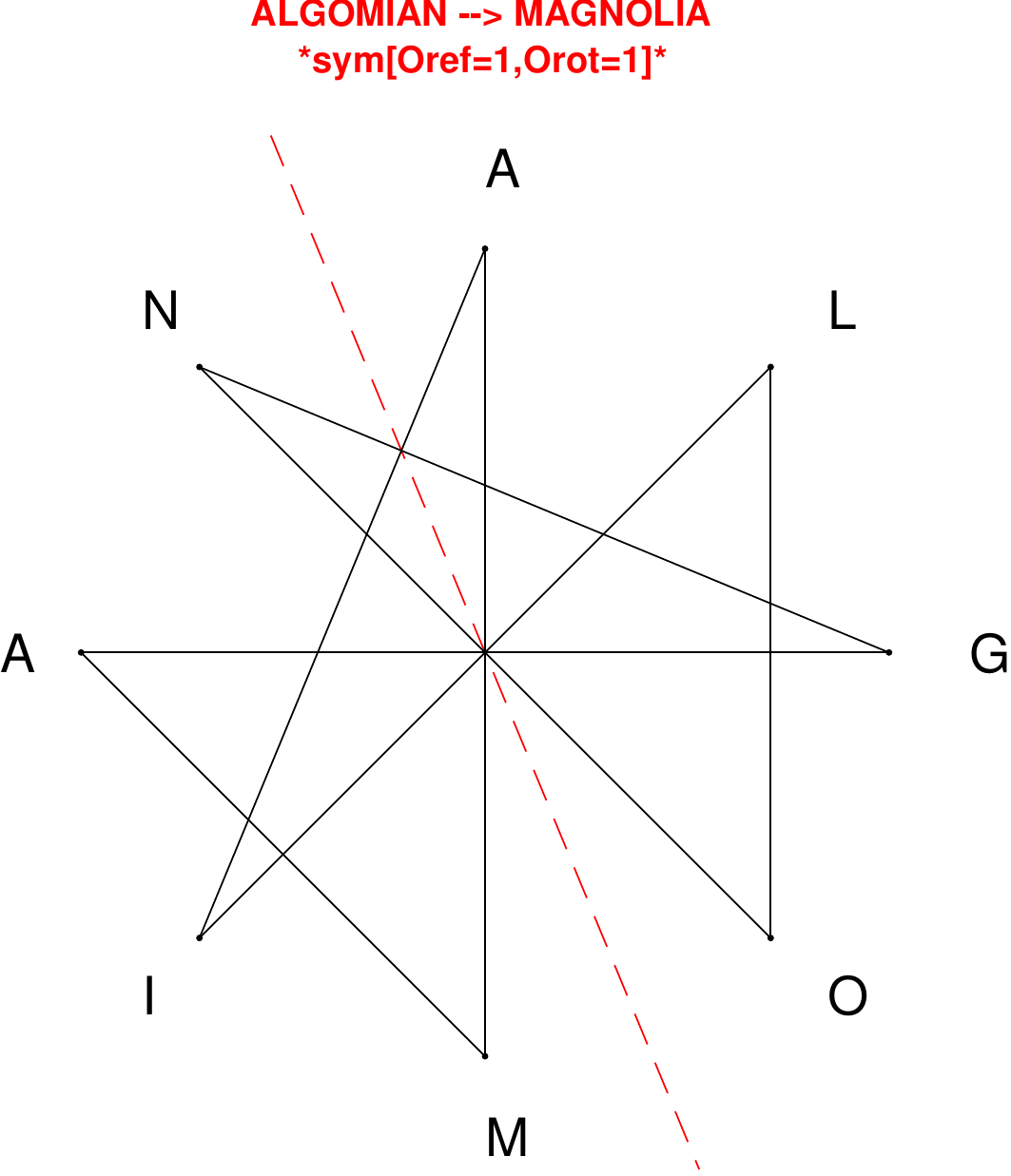}
\end{subfigure}
\hfill
\begin{subfigure}[T]{0.19\textwidth}
\centering
\includegraphics[width=\textwidth]{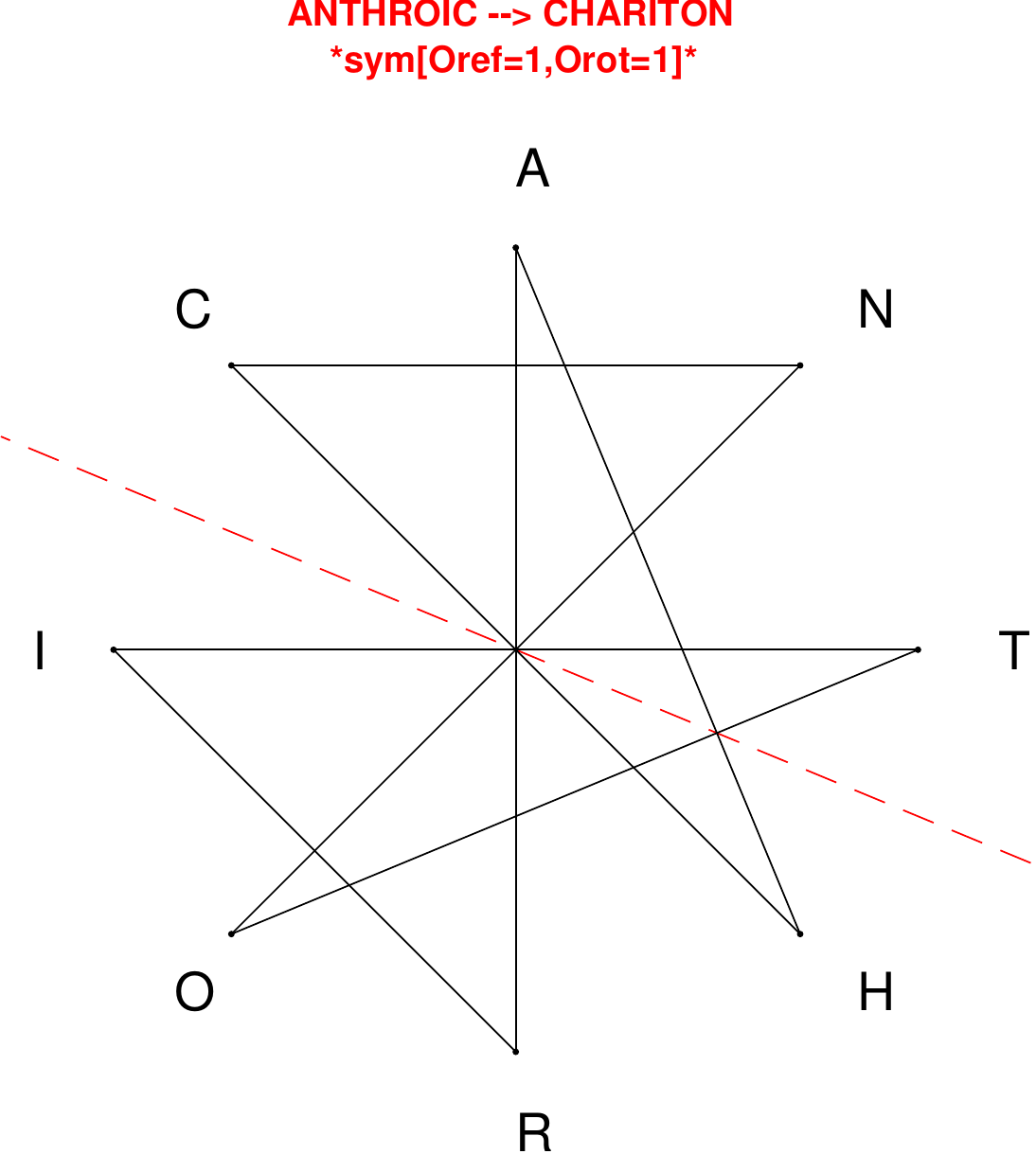}
\end{subfigure}
\end{figure}

\begin{figure}[H]
\centering
\begin{subfigure}[T]{0.19\textwidth}
\centering
\includegraphics[width=\textwidth]{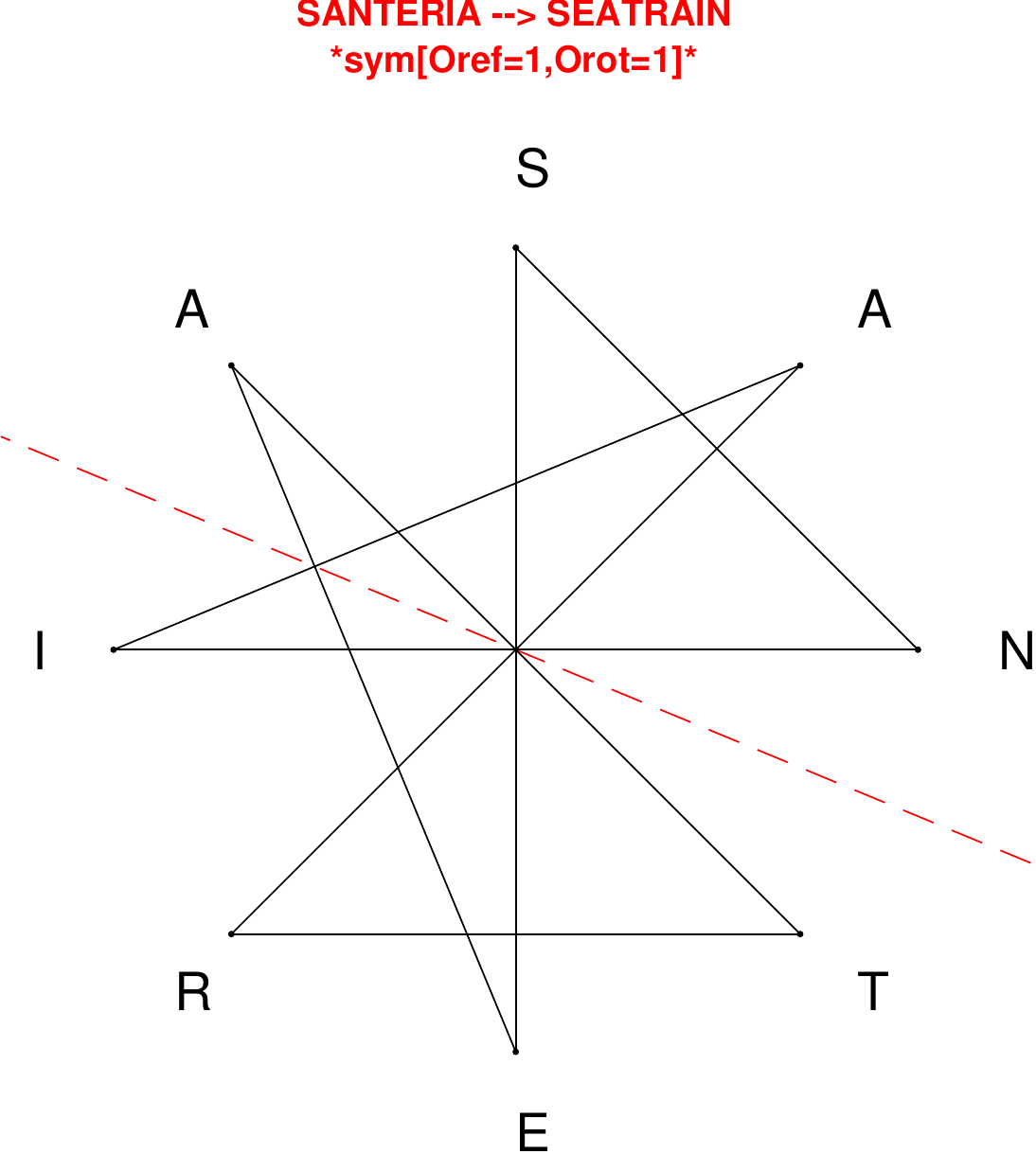}
\end{subfigure}
\hfill
\begin{subfigure}[T]{0.19\textwidth}
\centering
\includegraphics[width=\textwidth]{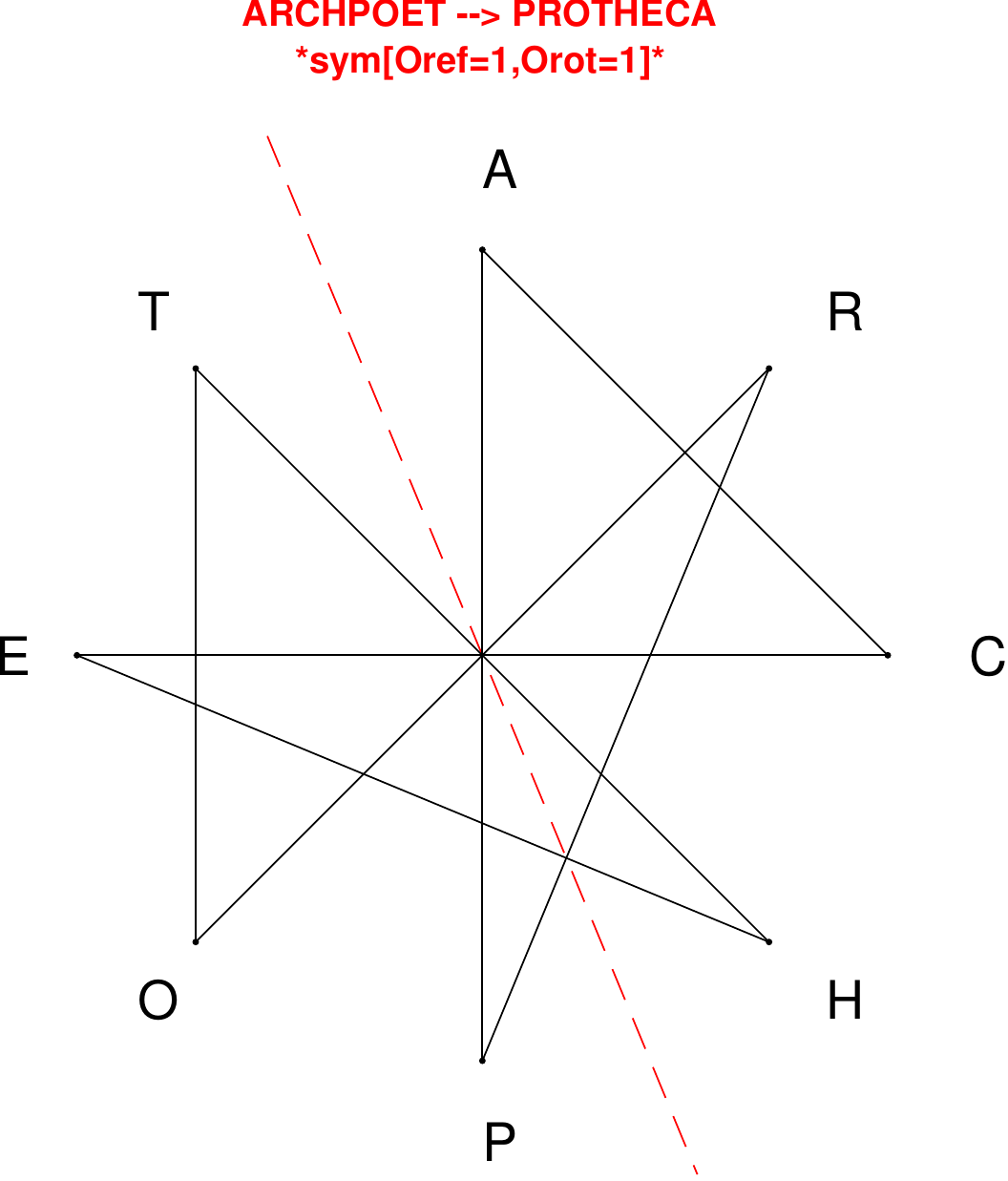}
\end{subfigure}
\hfill
\begin{subfigure}[T]{0.19\textwidth}
\centering
\includegraphics[width=\textwidth]{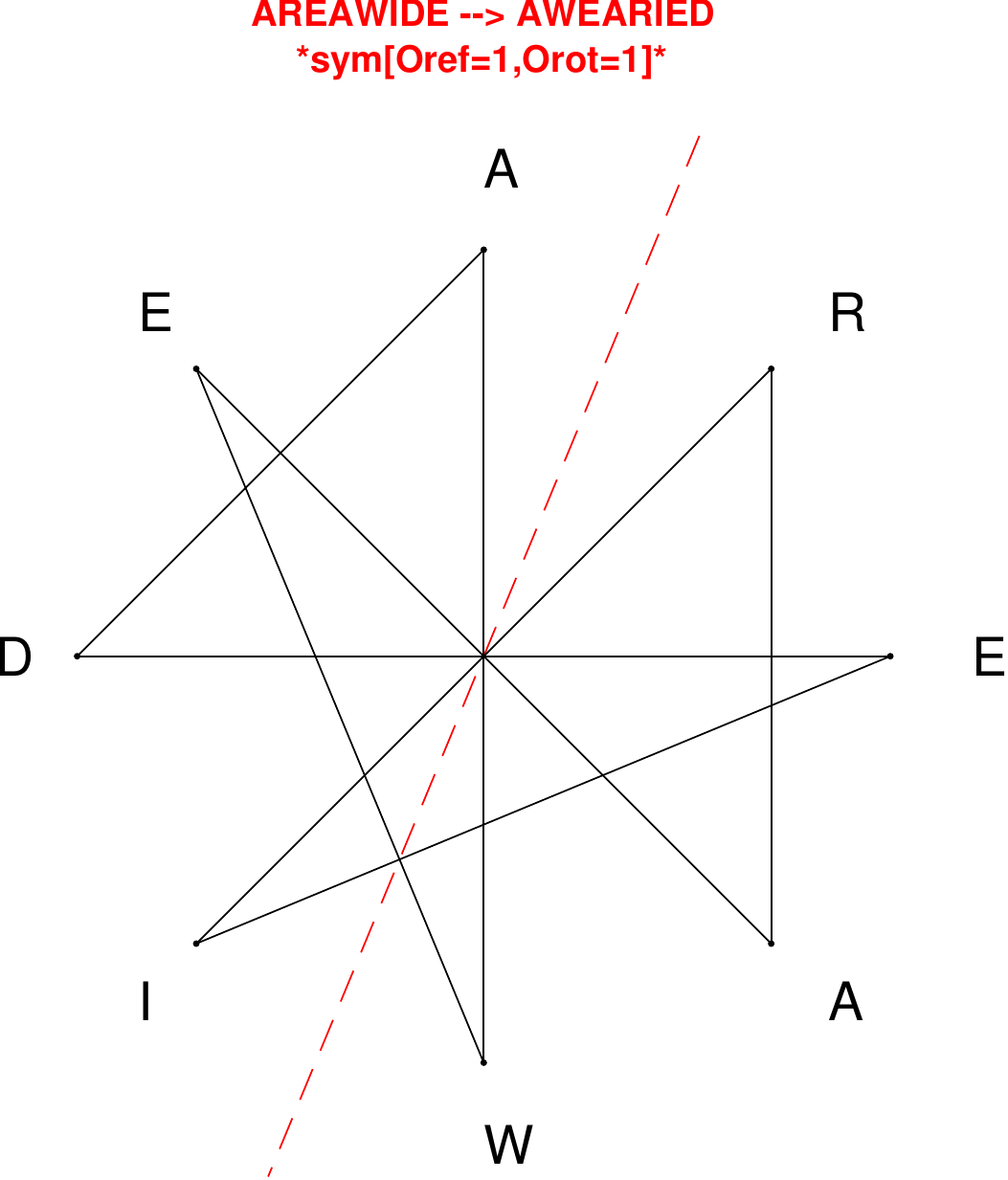}
\end{subfigure}
\hfill
\begin{subfigure}[T]{0.19\textwidth}
\centering
\includegraphics[width=\textwidth]{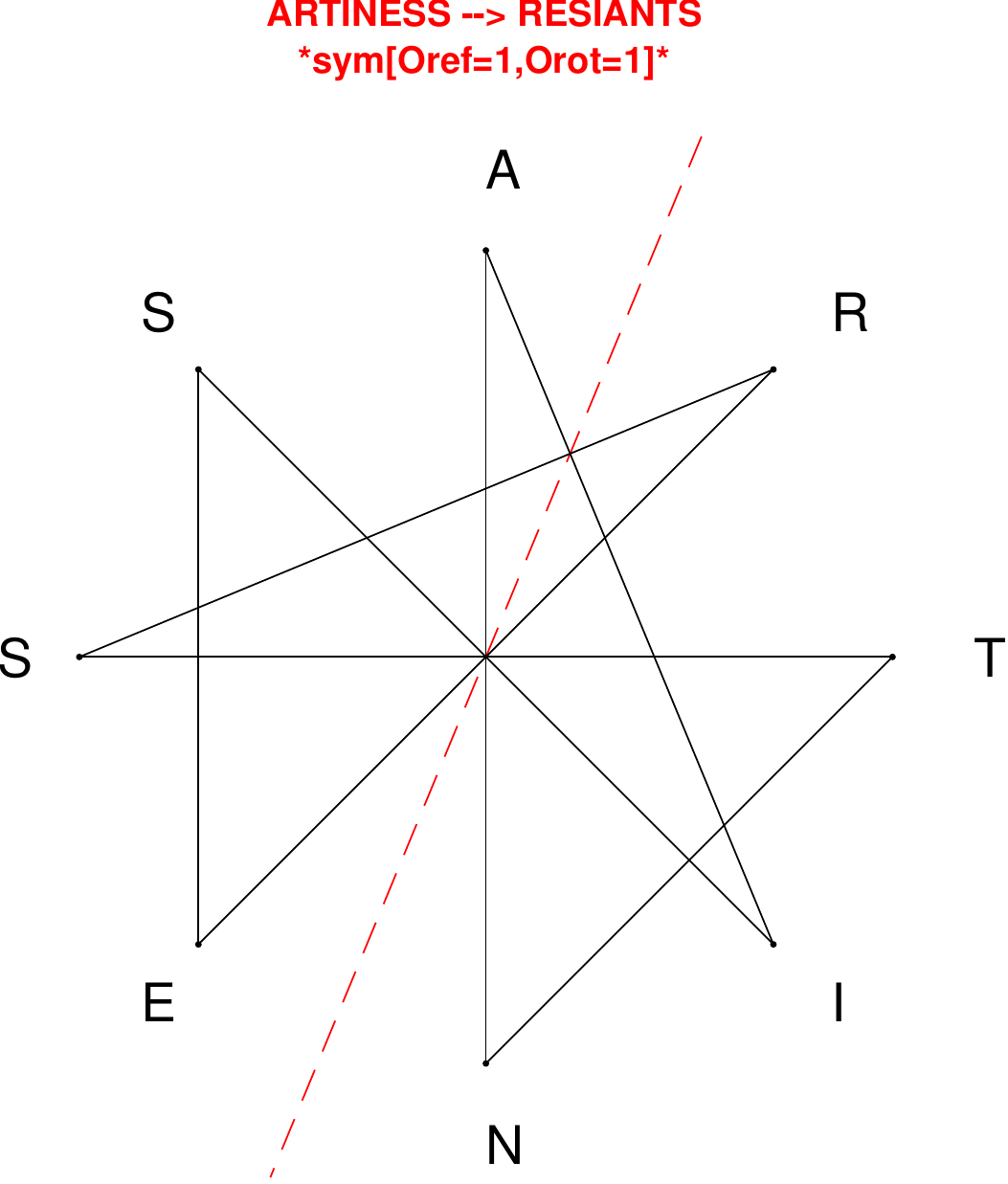}
\end{subfigure}
\hfill
\begin{subfigure}[T]{0.19\textwidth}
\centering
\includegraphics[width=\textwidth]{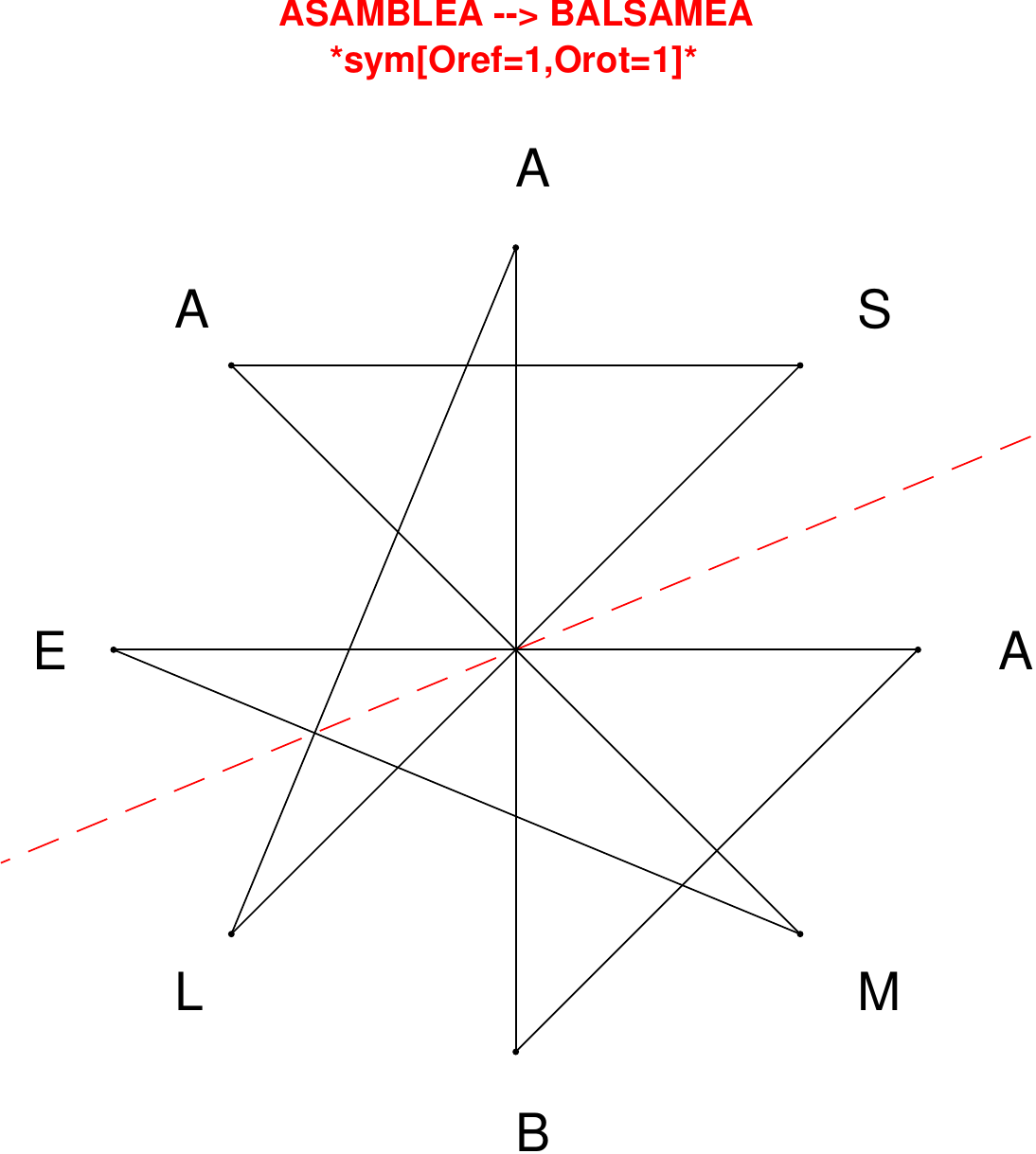}
\end{subfigure}
\end{figure}

\begin{figure}[H]
\centering
\begin{subfigure}[T]{0.19\textwidth}
\centering
\includegraphics[width=\textwidth]{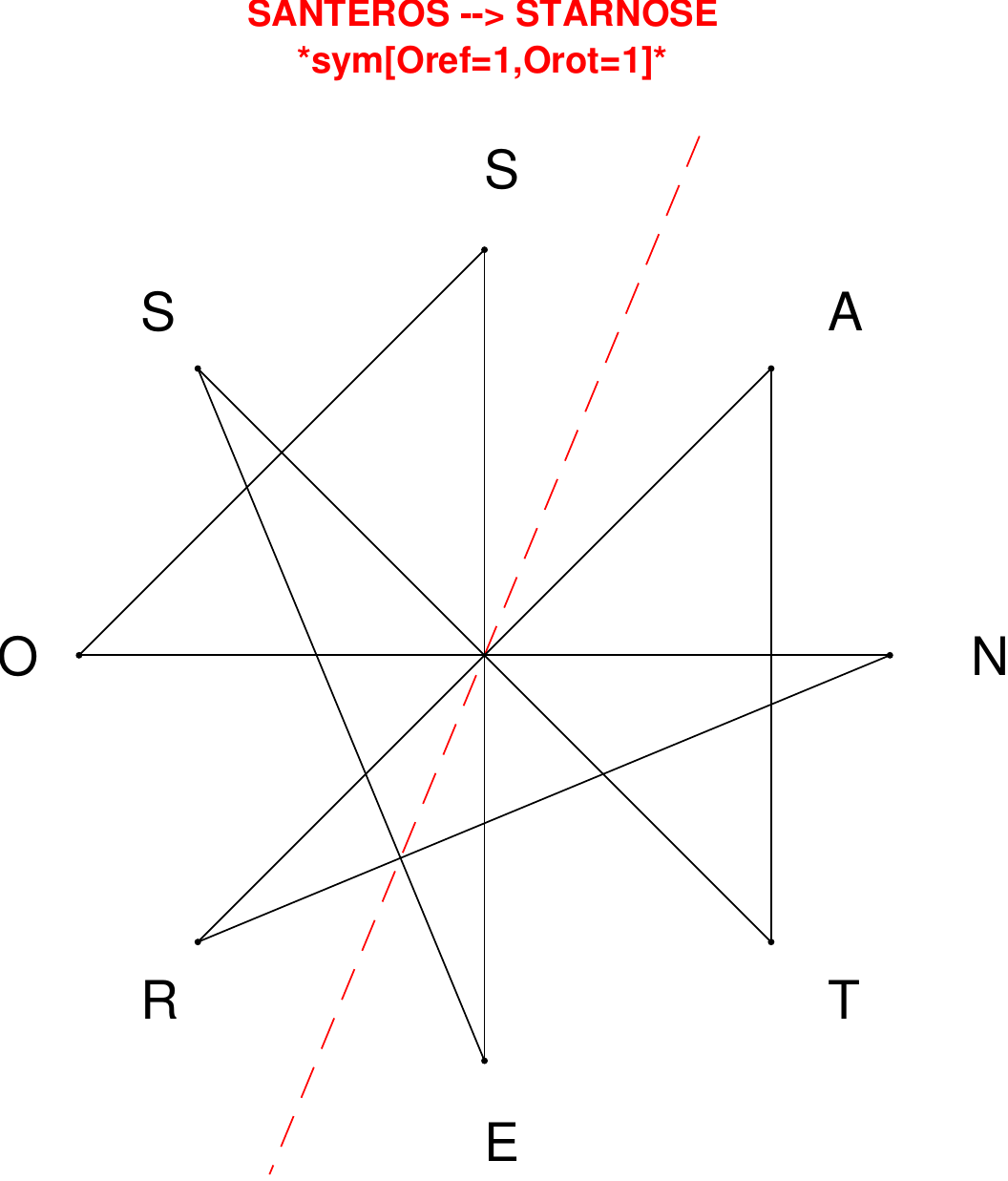}
\end{subfigure}
\hfill
\begin{subfigure}[T]{0.19\textwidth}
\centering
\includegraphics[width=\textwidth]{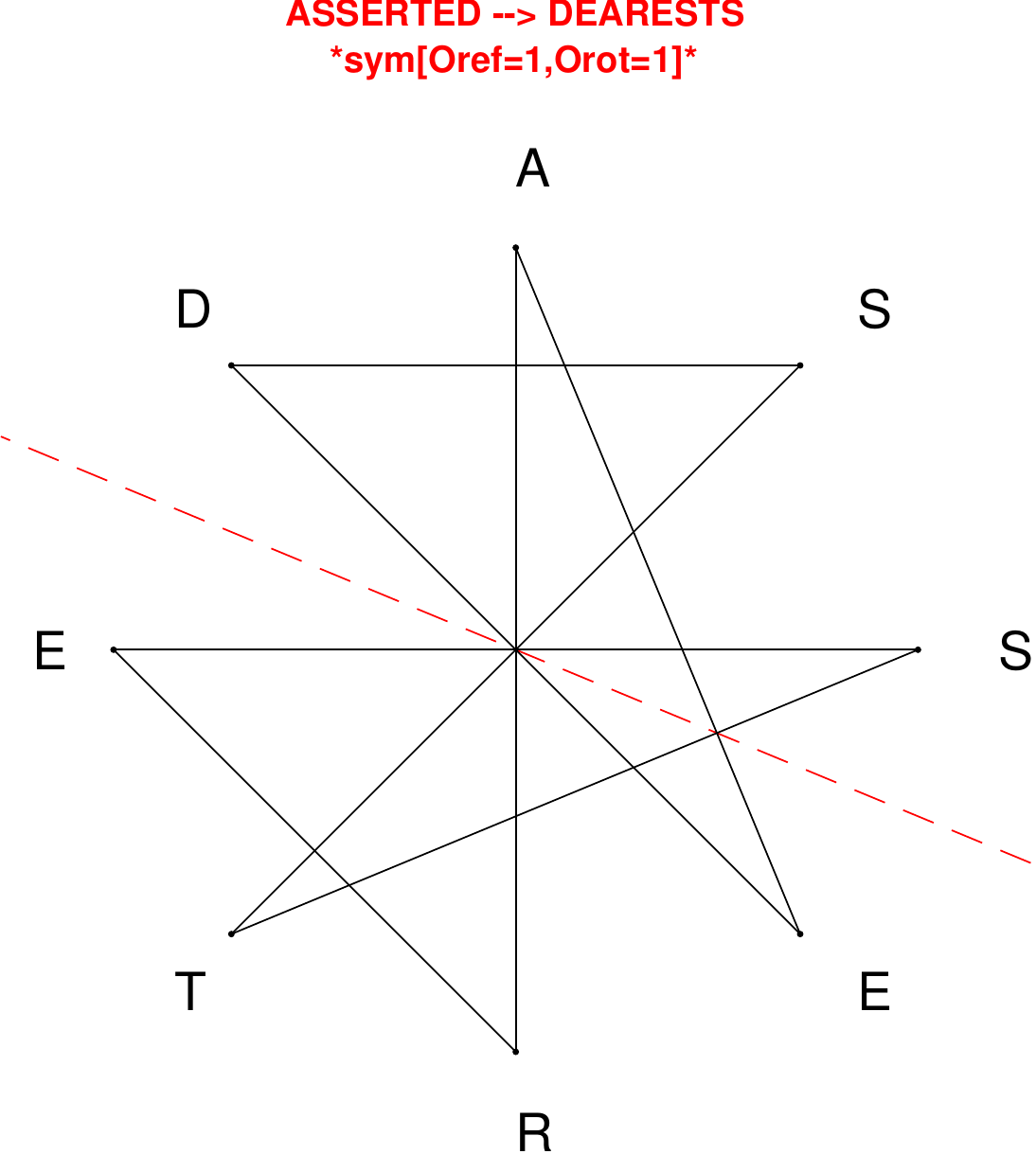}
\end{subfigure}
\hfill
\begin{subfigure}[T]{0.19\textwidth}
\centering
\includegraphics[width=\textwidth]{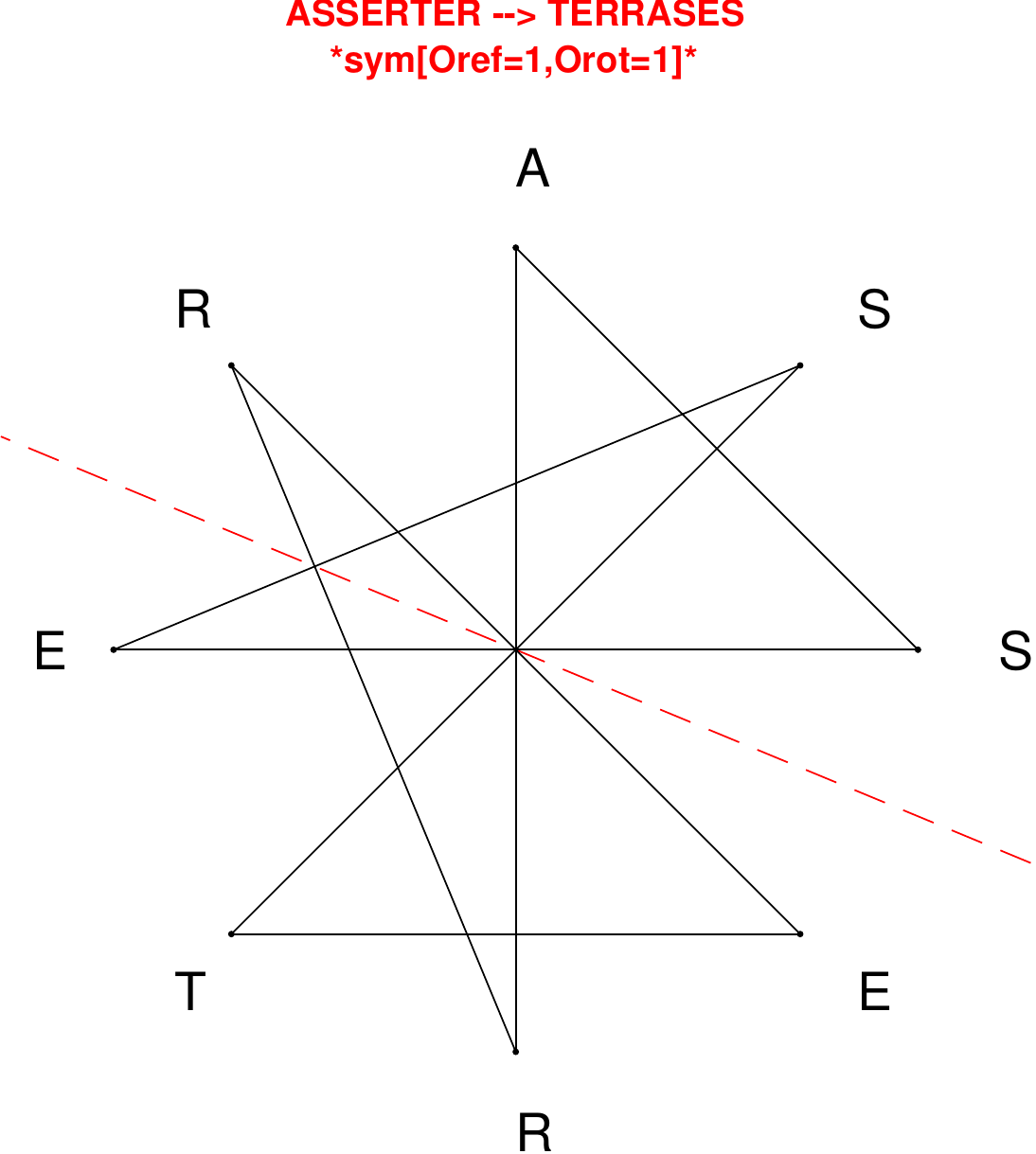}
\end{subfigure}
\hfill
\begin{subfigure}[T]{0.19\textwidth}
\centering
\includegraphics[width=\textwidth]{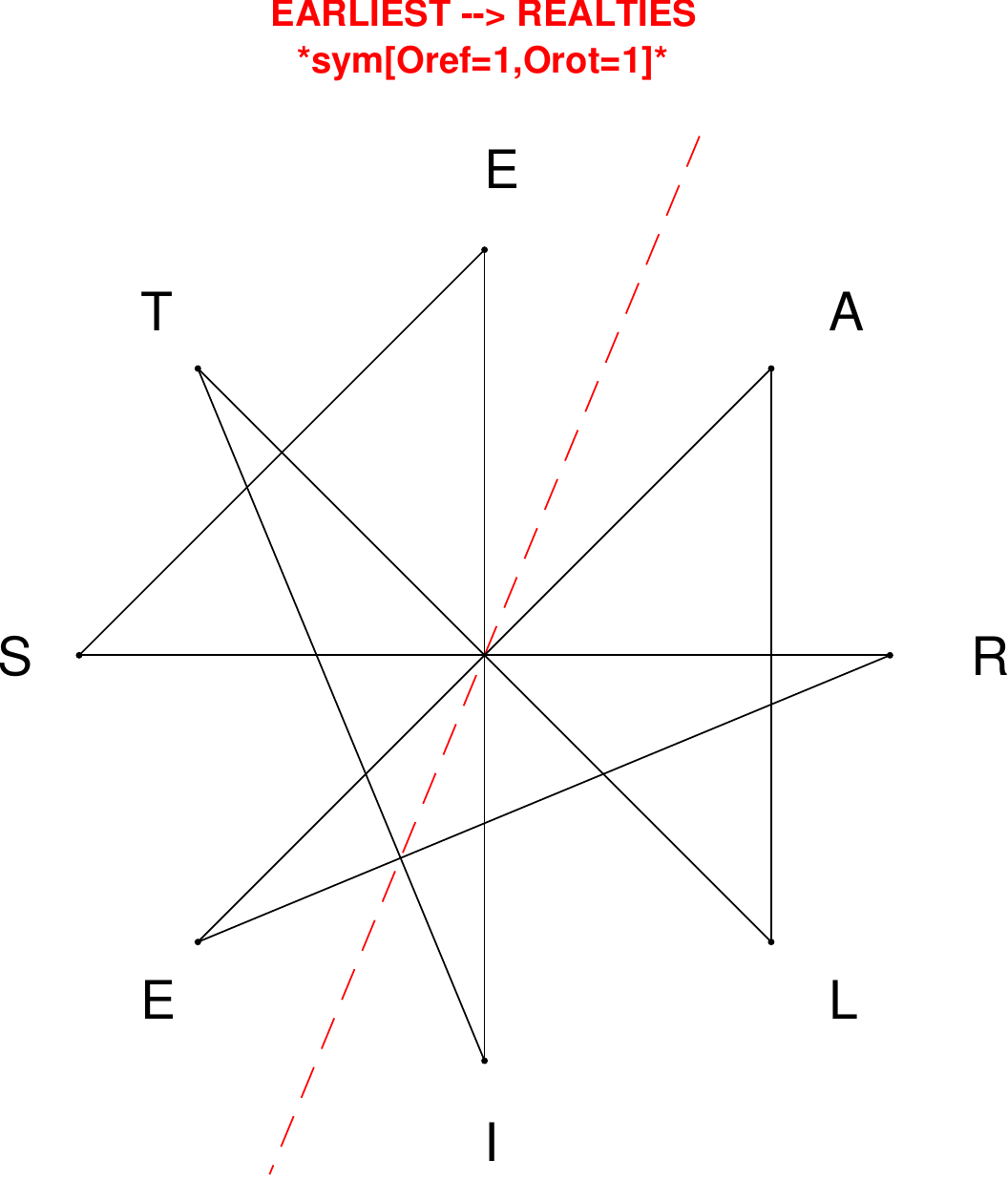}
\end{subfigure}
\hfill
\begin{subfigure}[T]{0.19\textwidth}
\centering
\includegraphics[width=\textwidth]{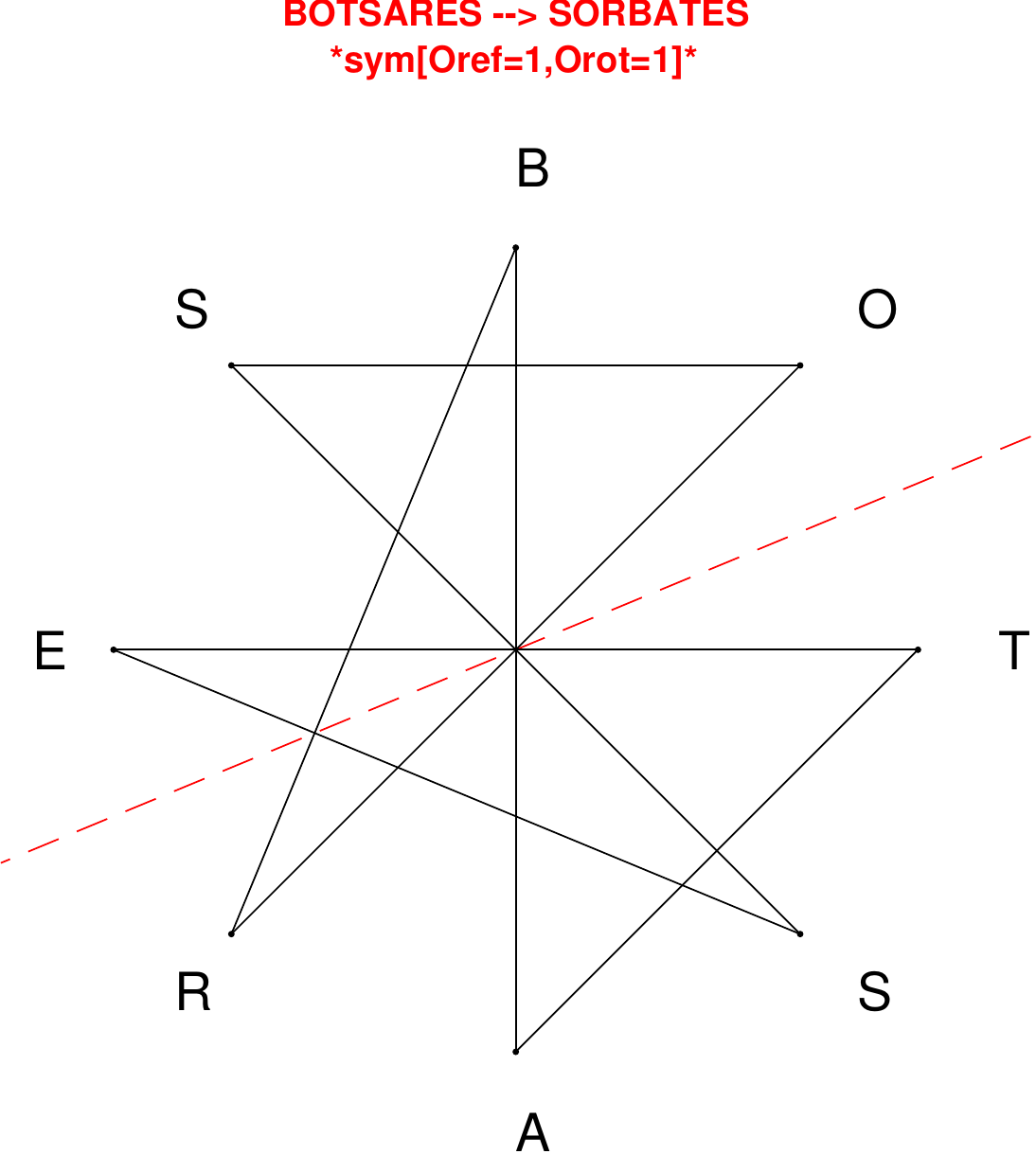}
\end{subfigure}
\end{figure}

\begin{figure}[H]
\centering
\begin{subfigure}[T]{0.19\textwidth}
\centering
\includegraphics[width=\textwidth]{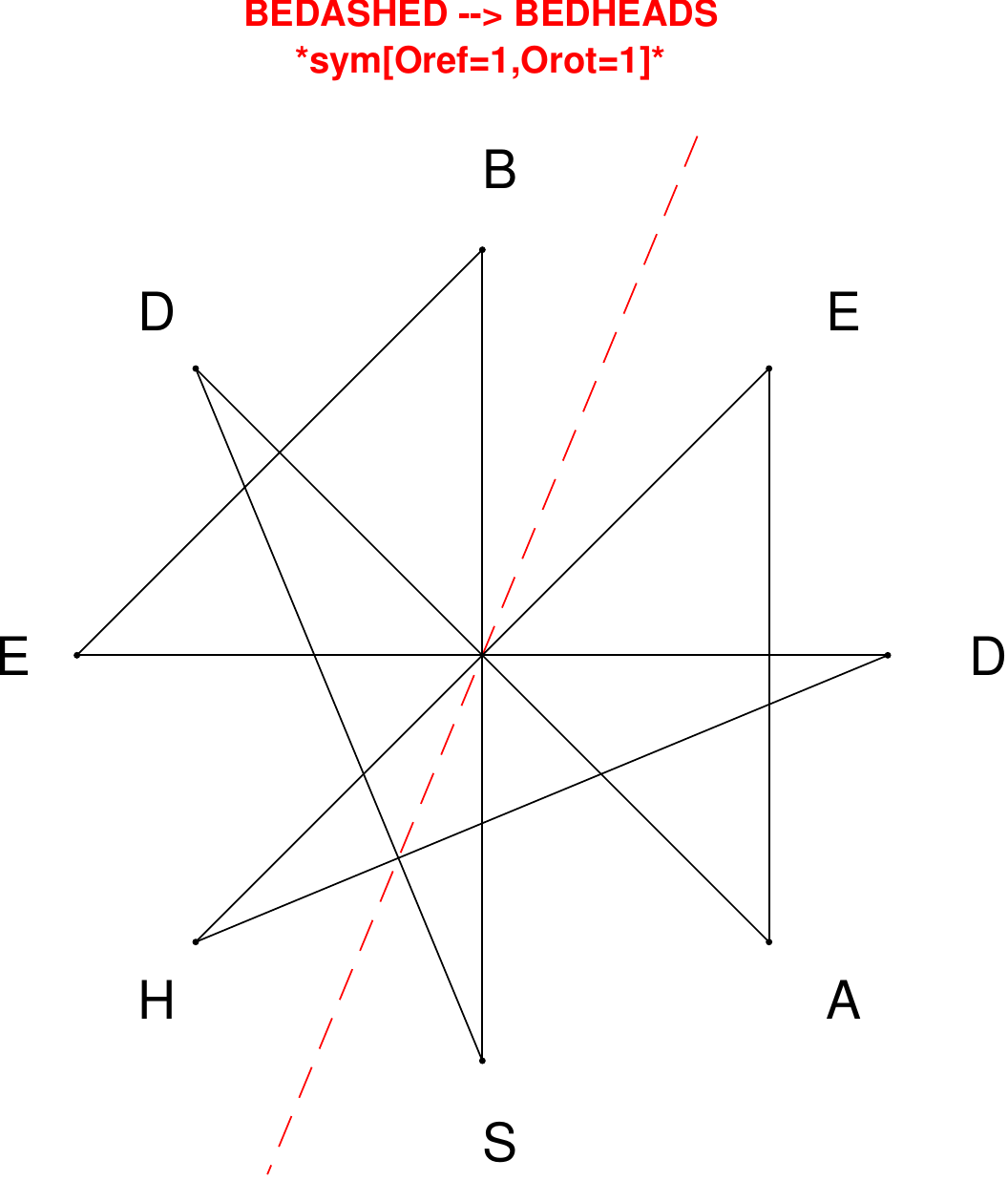}
\end{subfigure}
\hfill
\begin{subfigure}[T]{0.19\textwidth}
\centering
\includegraphics[width=\textwidth]{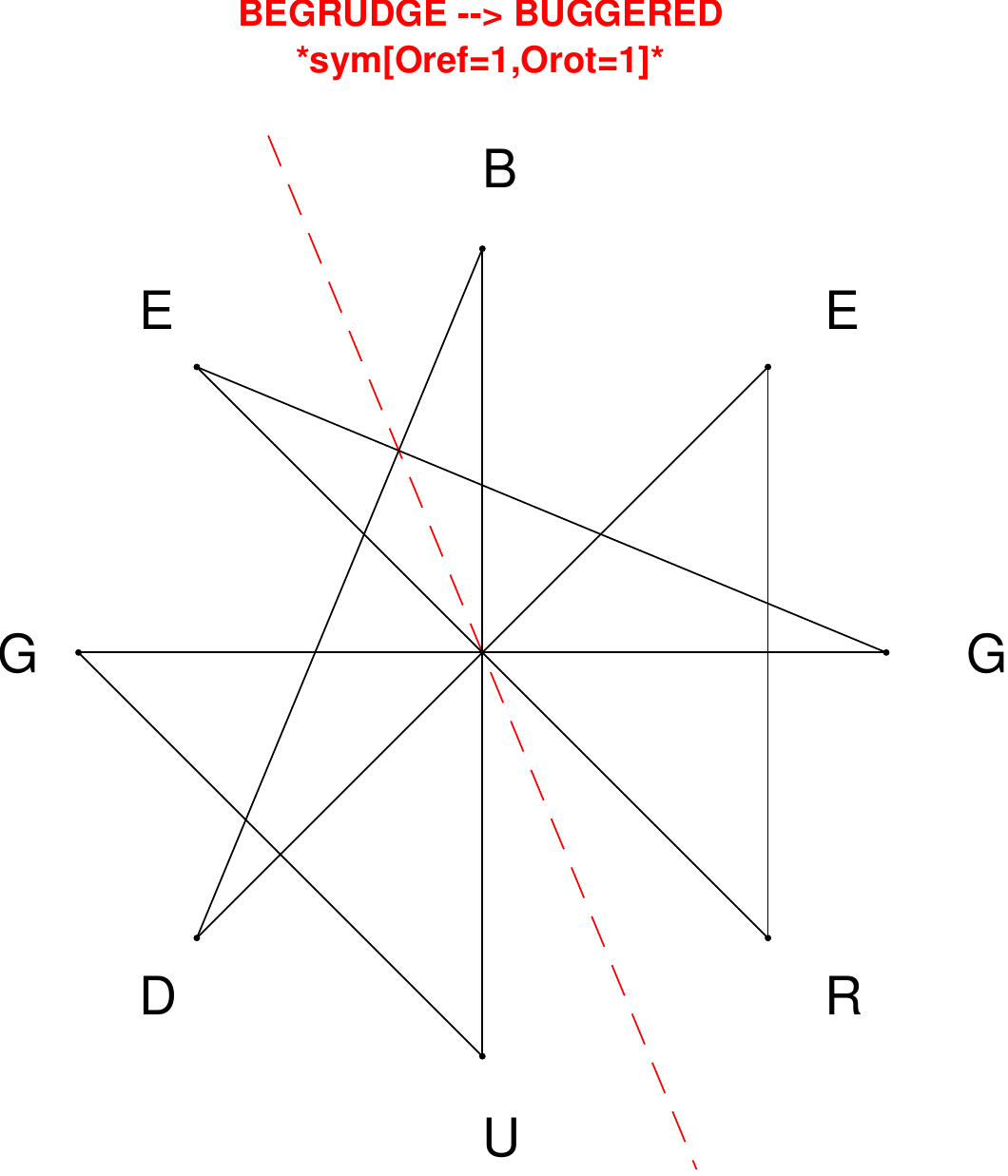}
\end{subfigure}
\hfill
\begin{subfigure}[T]{0.19\textwidth}
\centering
\includegraphics[width=\textwidth]{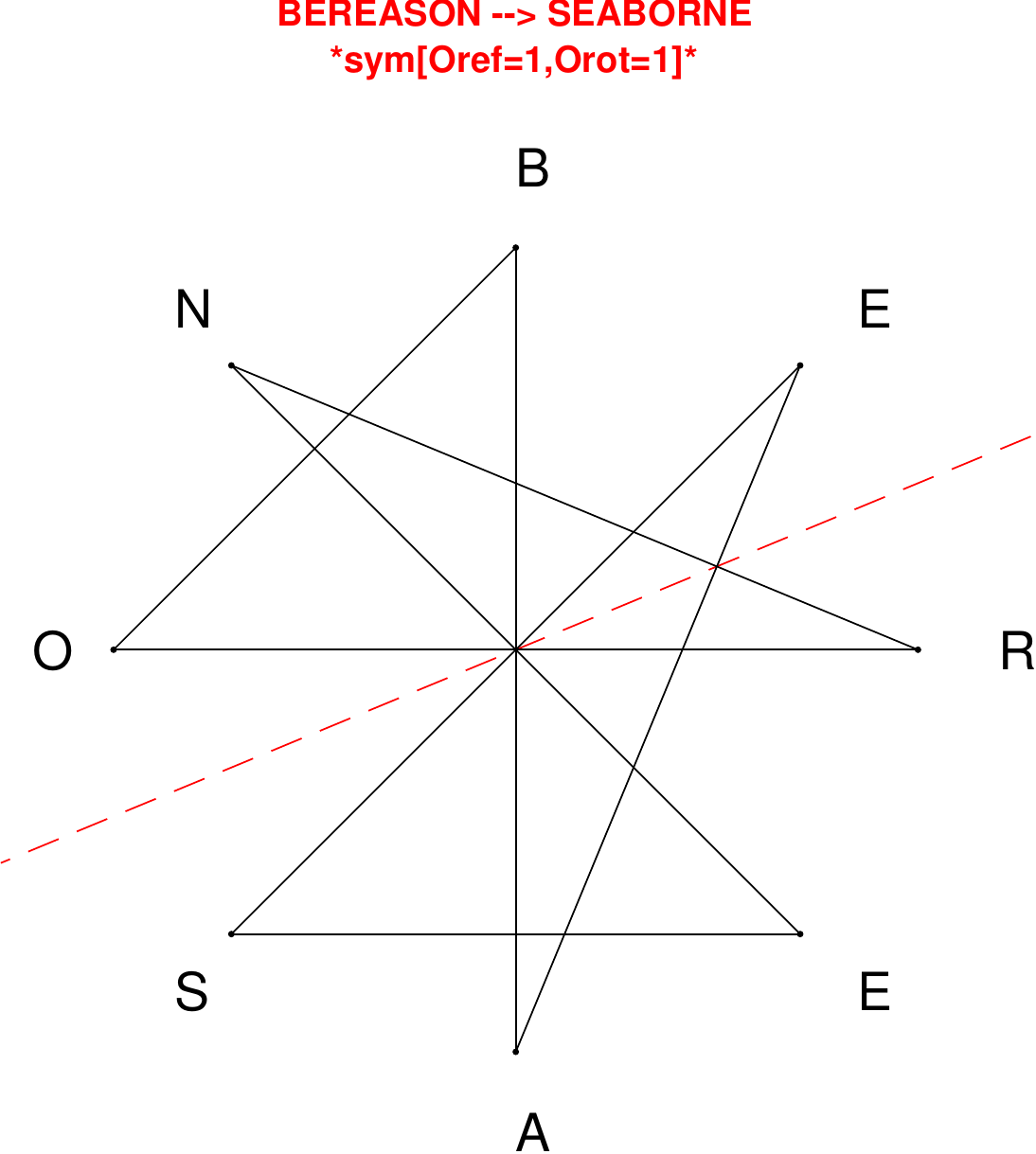}
\end{subfigure}
\hfill
\begin{subfigure}[T]{0.19\textwidth}
\centering
\includegraphics[width=\textwidth]{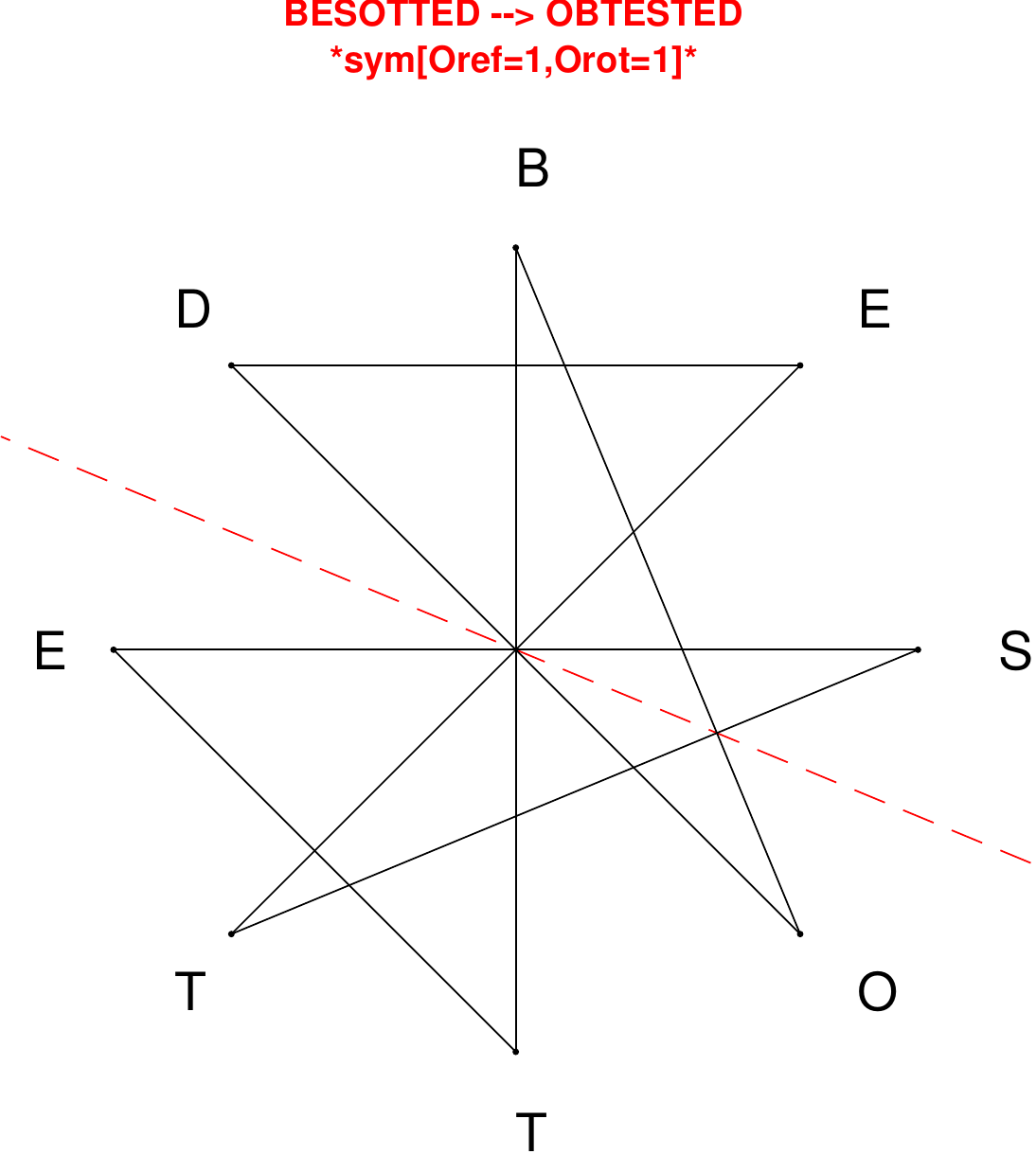}
\end{subfigure}
\hfill
\begin{subfigure}[T]{0.19\textwidth}
\centering
\includegraphics[width=\textwidth]{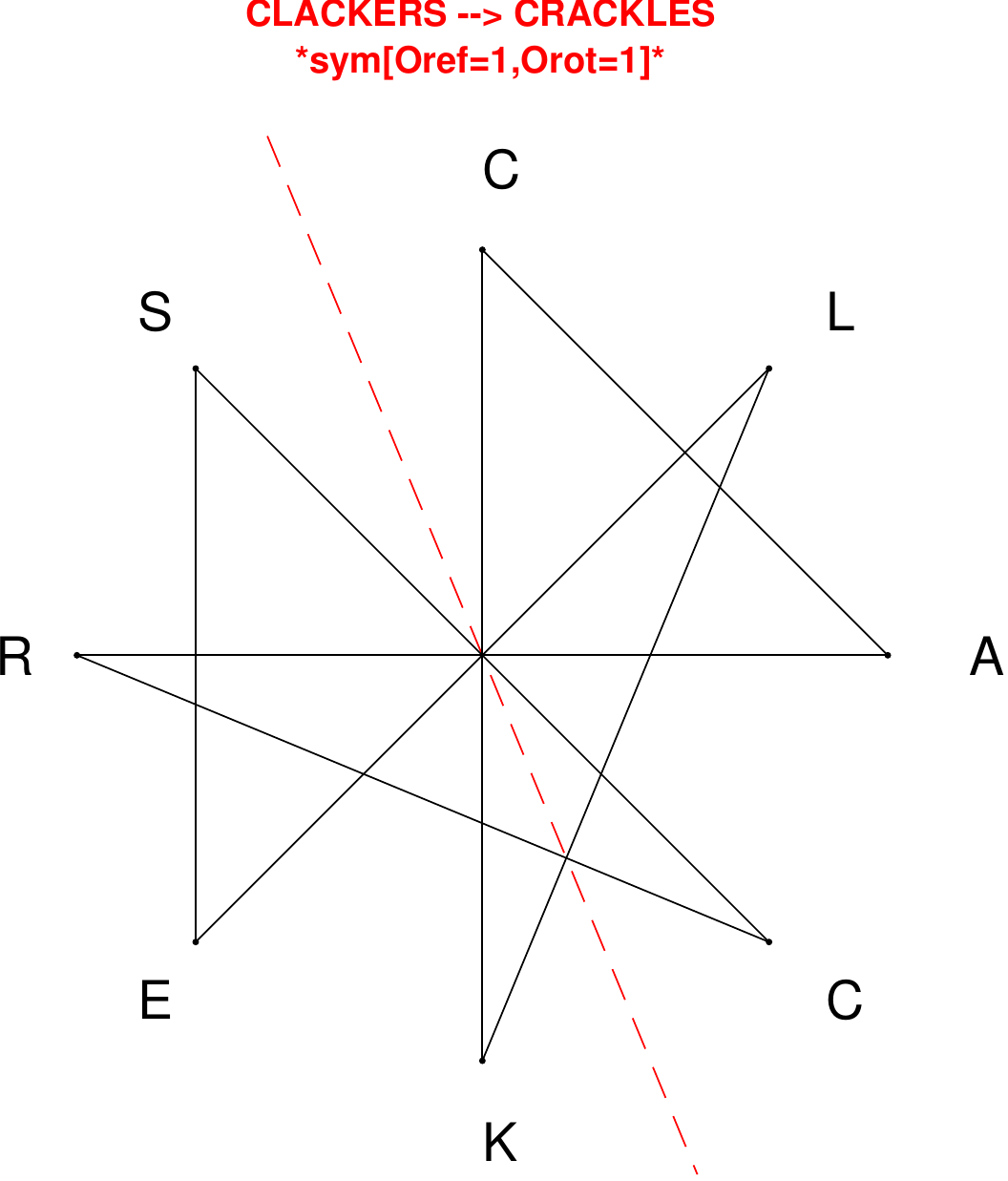}
\end{subfigure}
\end{figure}

\begin{figure}[H]
\centering
\begin{subfigure}[T]{0.19\textwidth}
\centering
\includegraphics[width=\textwidth]{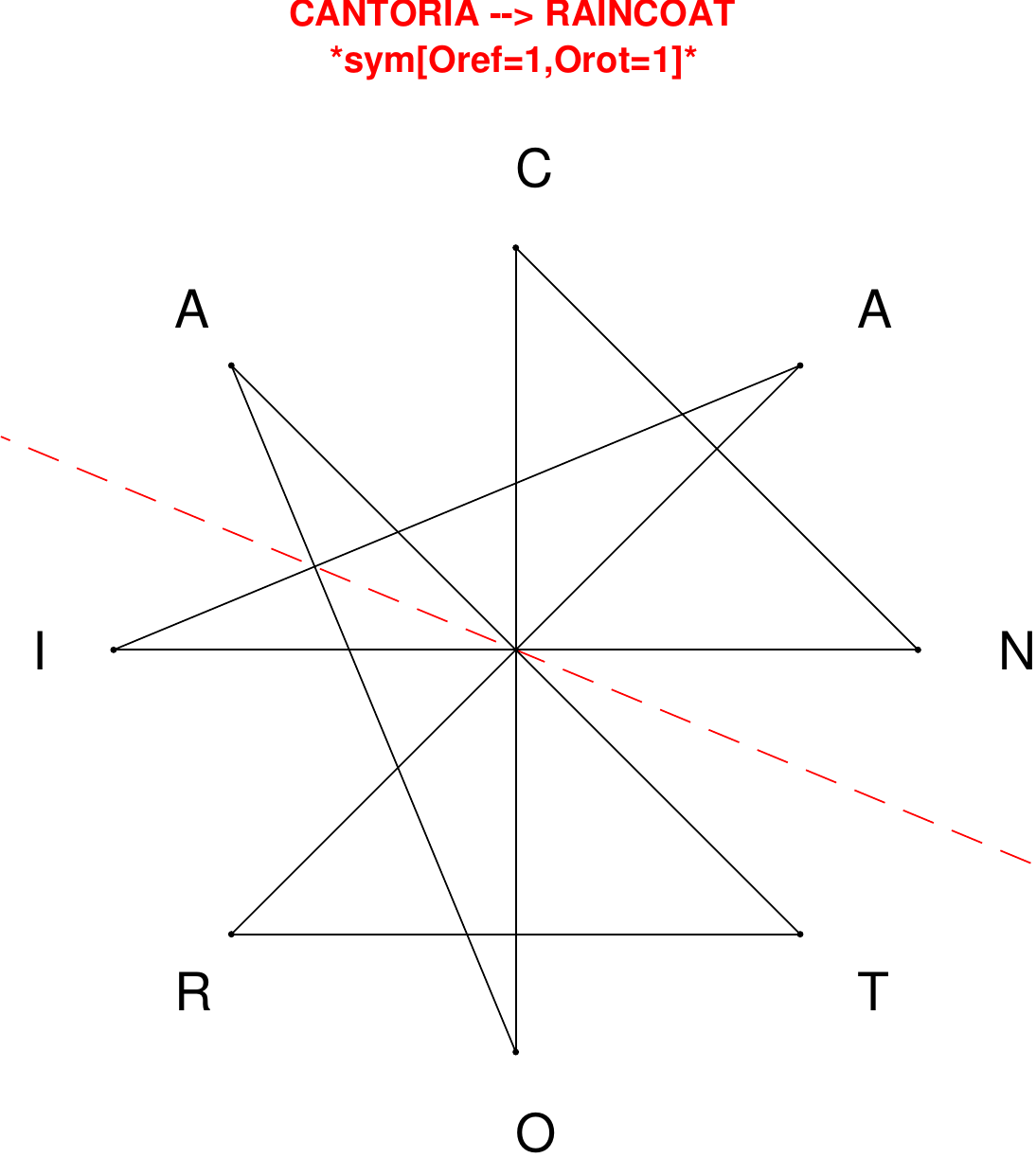}
\end{subfigure}
\hfill
\begin{subfigure}[T]{0.19\textwidth}
\centering
\includegraphics[width=\textwidth]{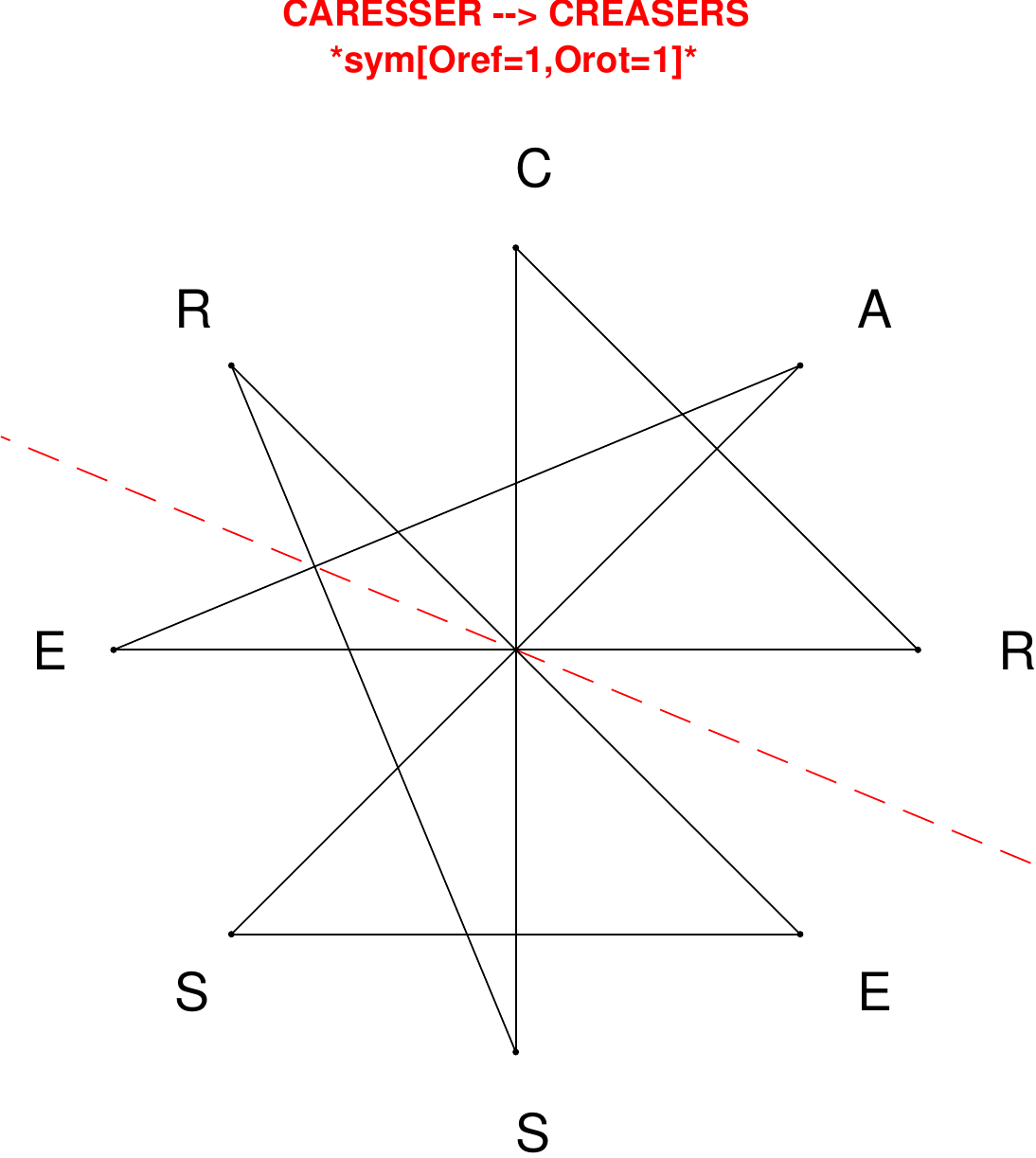}
\end{subfigure}
\hfill
\begin{subfigure}[T]{0.19\textwidth}
\centering
\includegraphics[width=\textwidth]{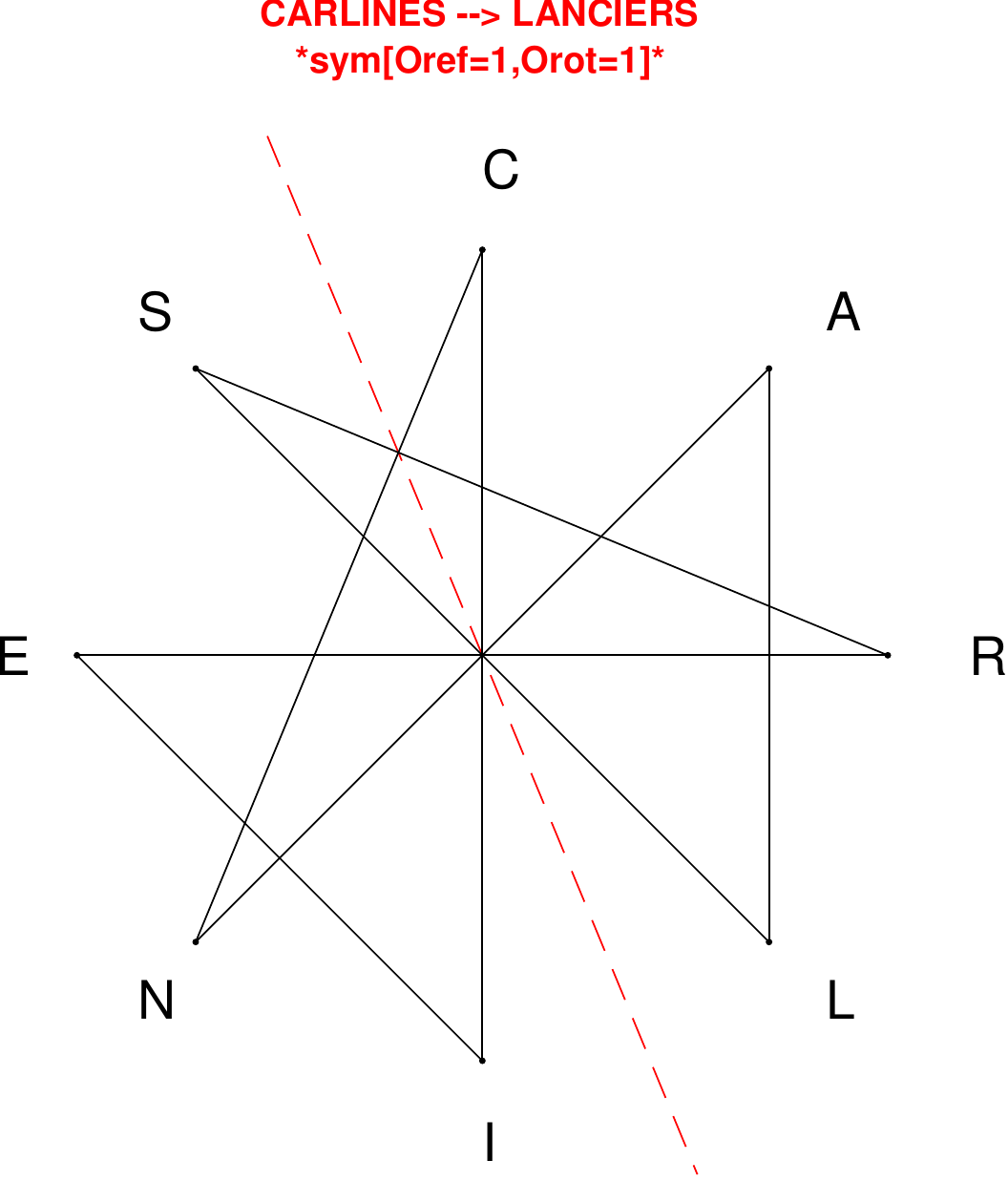}
\end{subfigure}
\hfill
\begin{subfigure}[T]{0.19\textwidth}
\centering
\includegraphics[width=\textwidth]{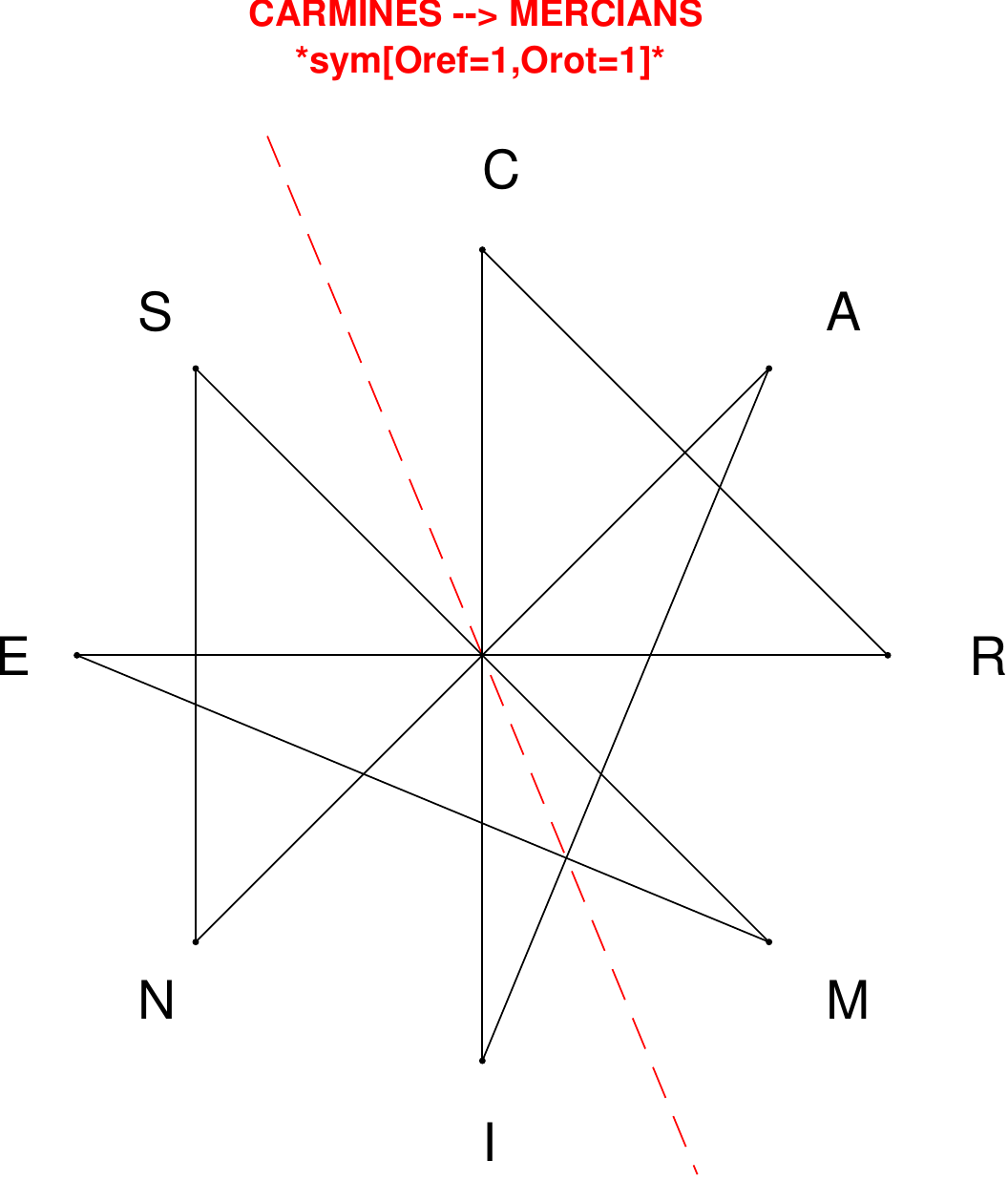}
\end{subfigure}
\hfill
\begin{subfigure}[T]{0.19\textwidth}
\centering
\includegraphics[width=\textwidth]{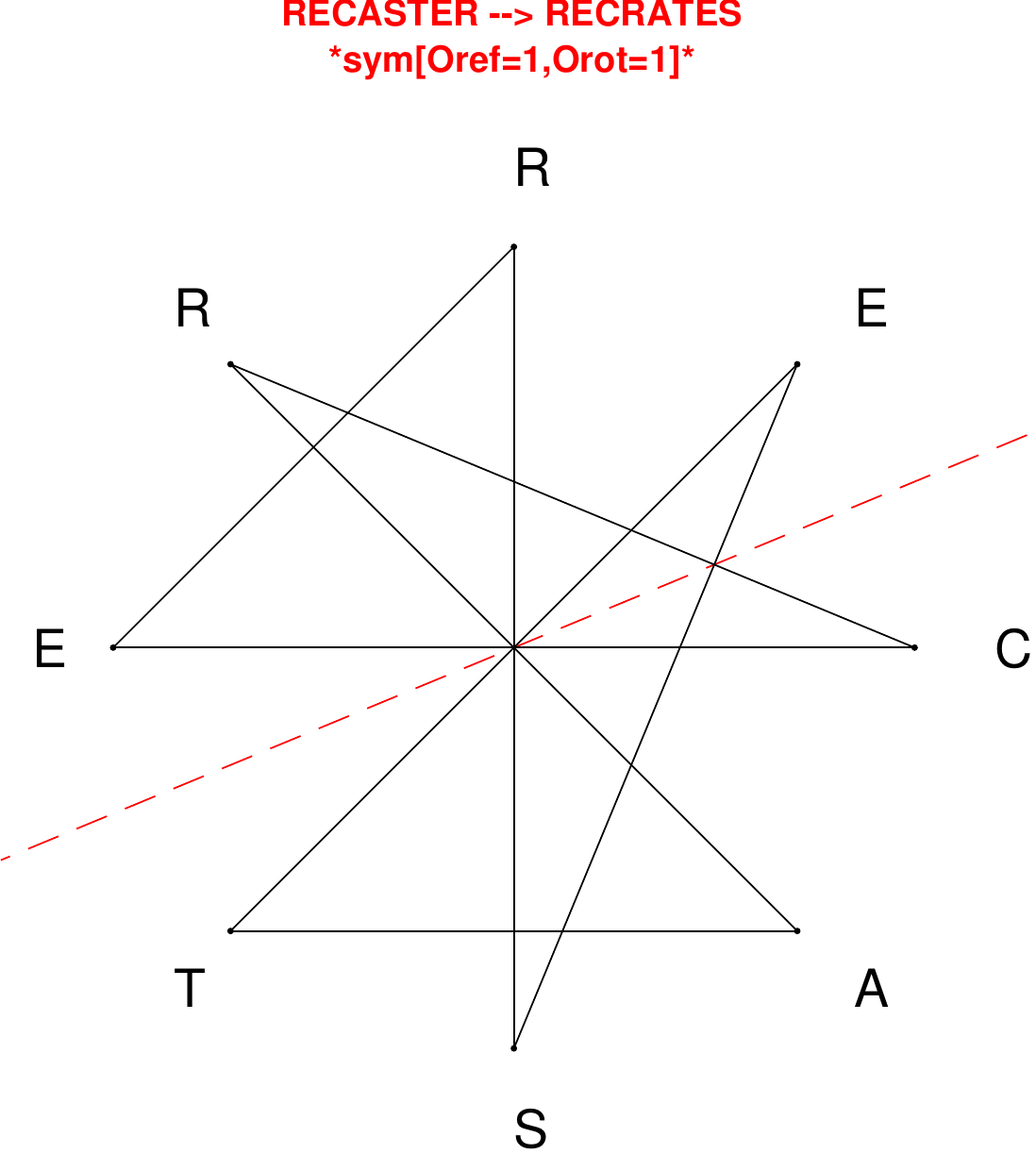}
\end{subfigure}
\end{figure}

\begin{figure}[H]
\centering
\begin{subfigure}[T]{0.19\textwidth}
\centering
\includegraphics[width=\textwidth]{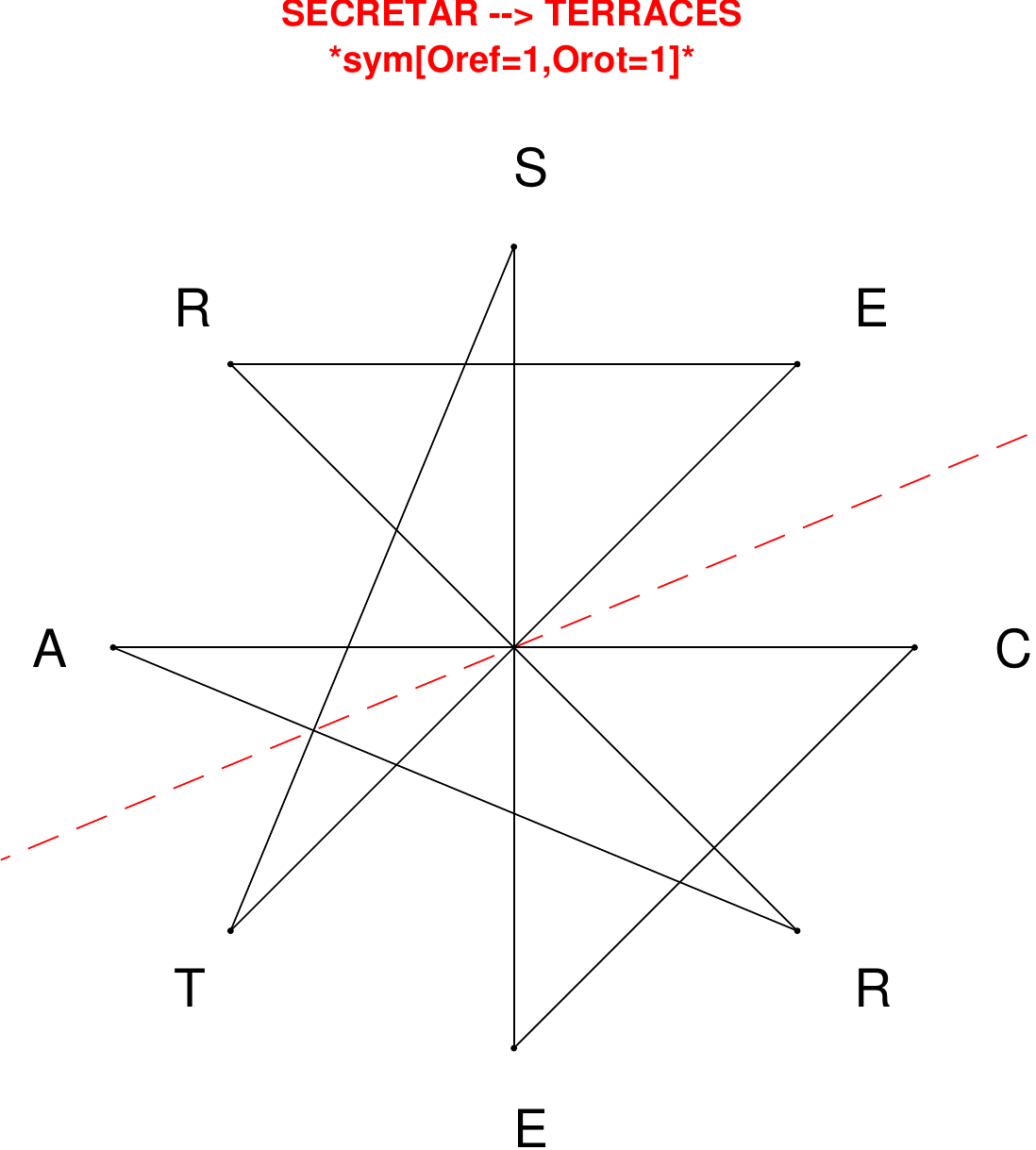}
\end{subfigure}
\hfill
\begin{subfigure}[T]{0.19\textwidth}
\centering
\includegraphics[width=\textwidth]{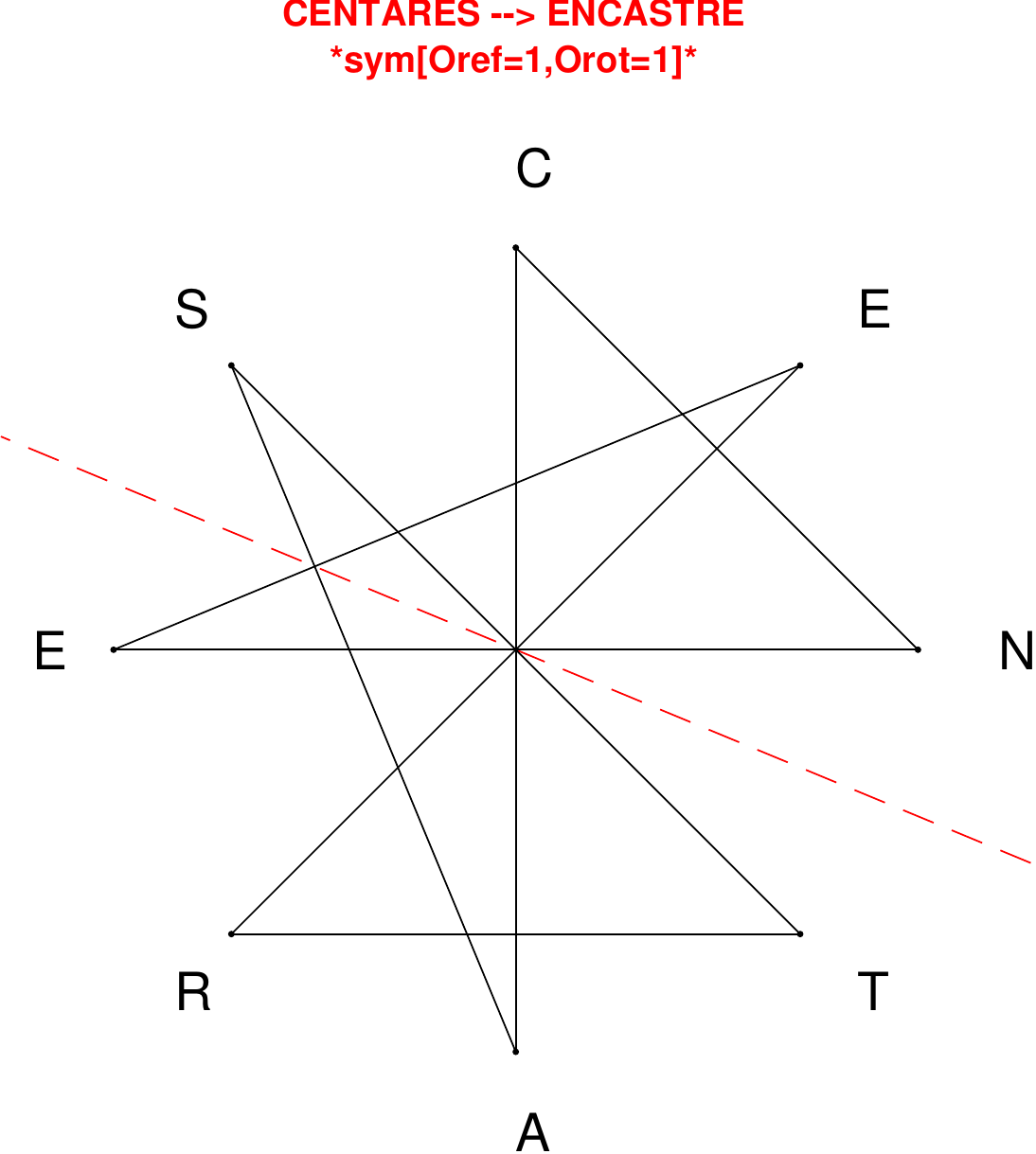}
\end{subfigure}
\hfill
\begin{subfigure}[T]{0.19\textwidth}
\centering
\includegraphics[width=\textwidth]{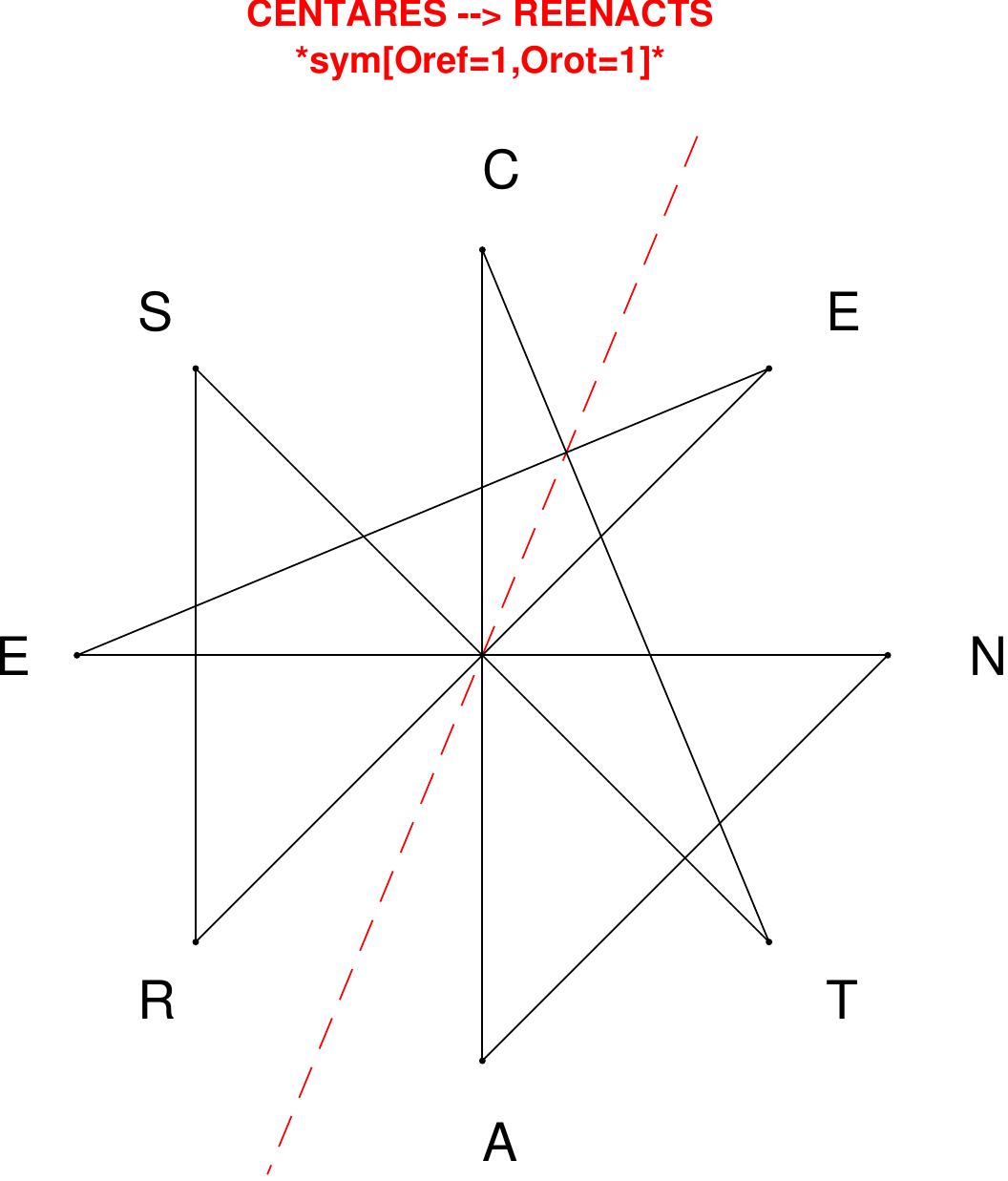}
\end{subfigure}
\hfill
\begin{subfigure}[T]{0.19\textwidth}
\centering
\includegraphics[width=\textwidth]{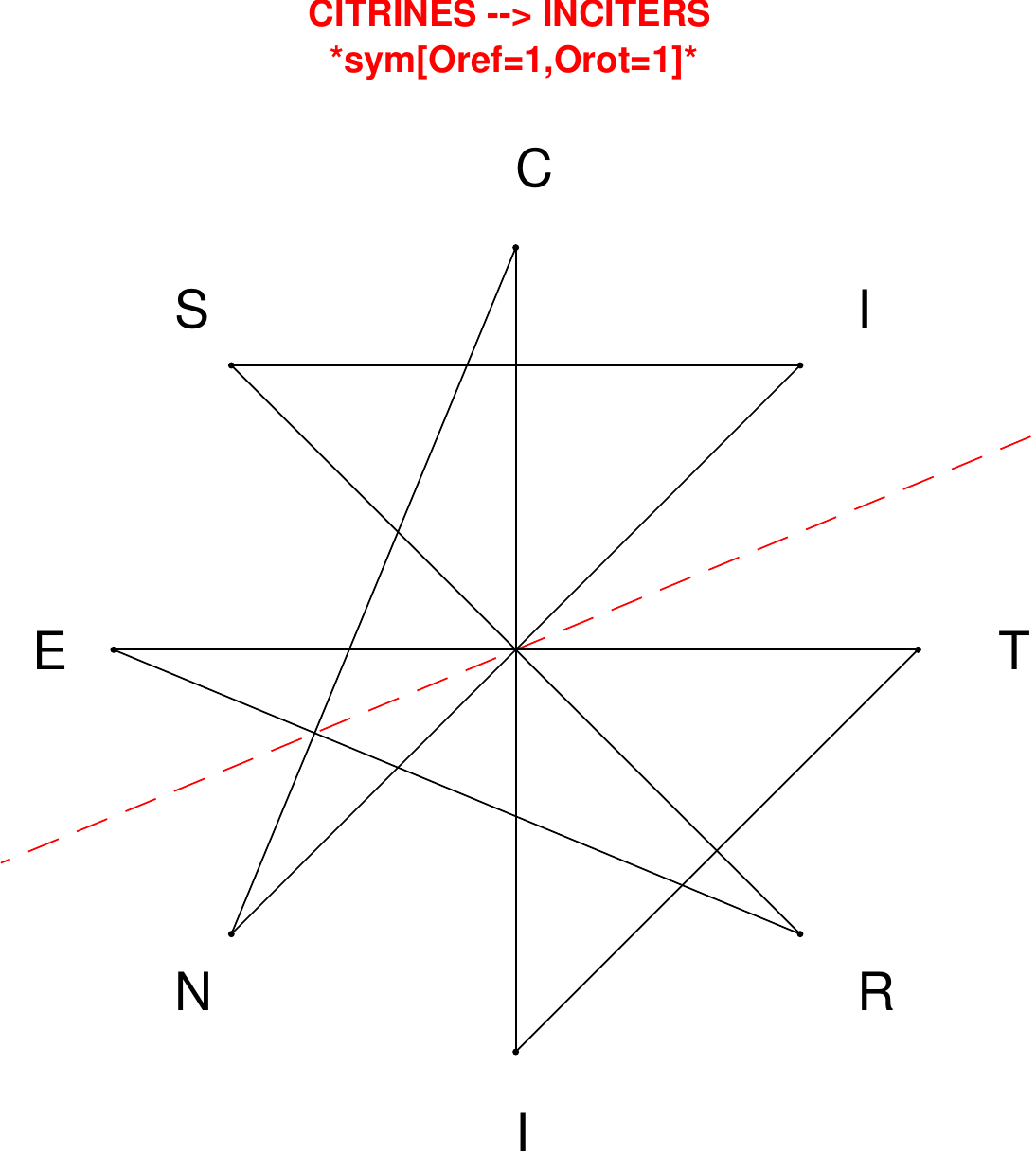}
\end{subfigure}
\hfill
\begin{subfigure}[T]{0.19\textwidth}
\centering
\includegraphics[width=\textwidth]{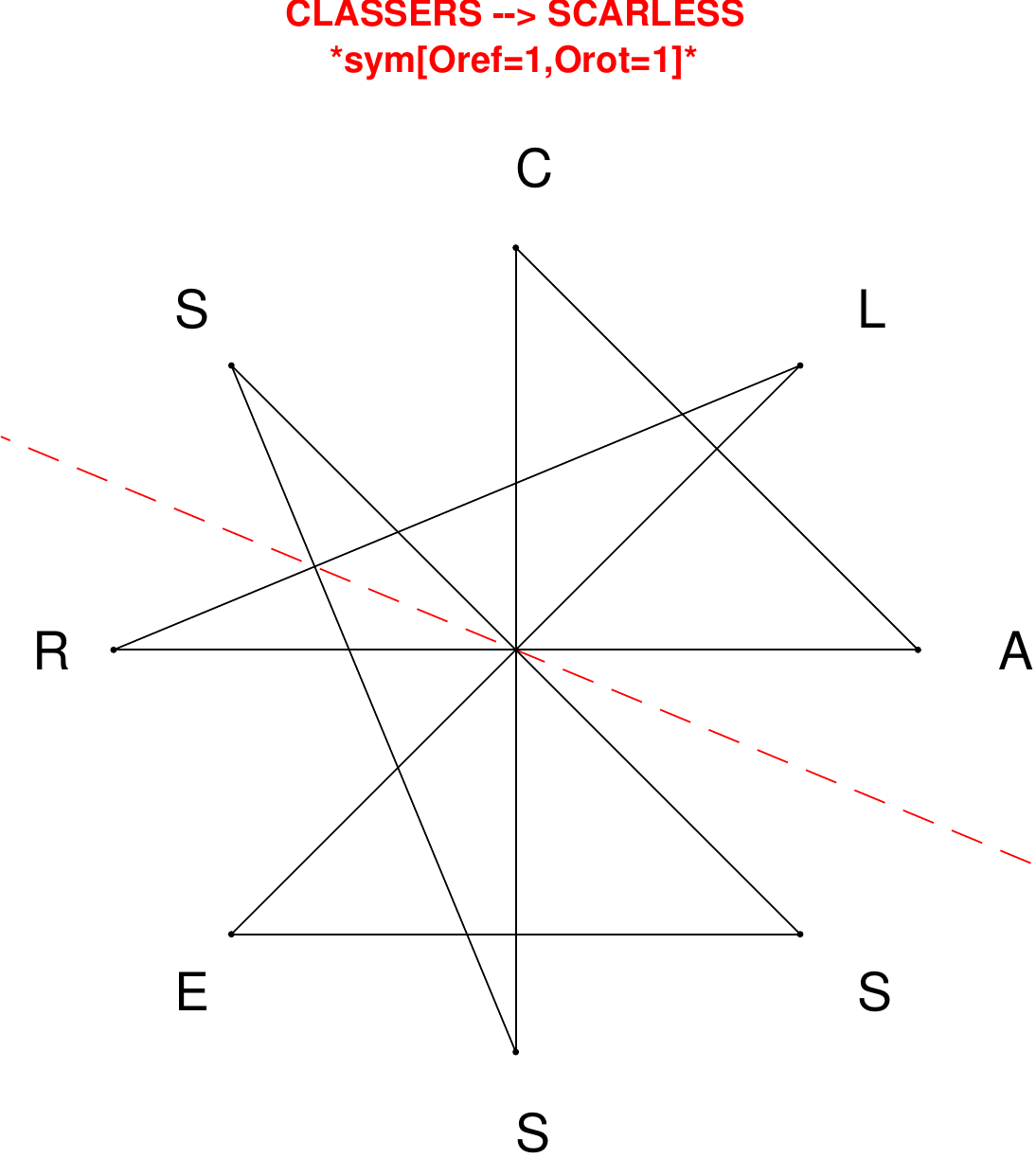}
\end{subfigure}
\end{figure}

\begin{figure}[H]
\centering
\begin{subfigure}[T]{0.19\textwidth}
\centering
\includegraphics[width=\textwidth]{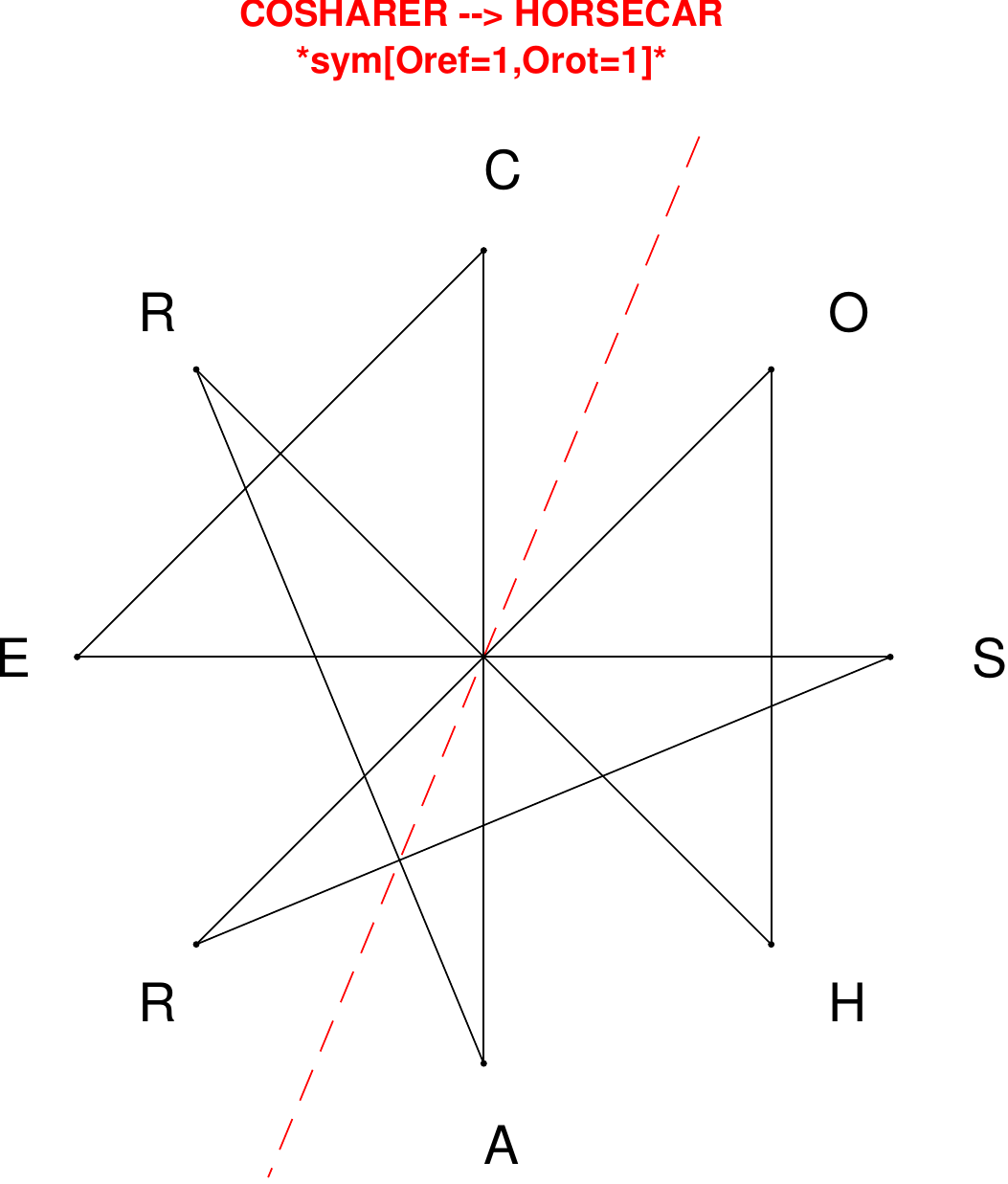}
\end{subfigure}
\hfill
\begin{subfigure}[T]{0.19\textwidth}
\centering
\includegraphics[width=\textwidth]{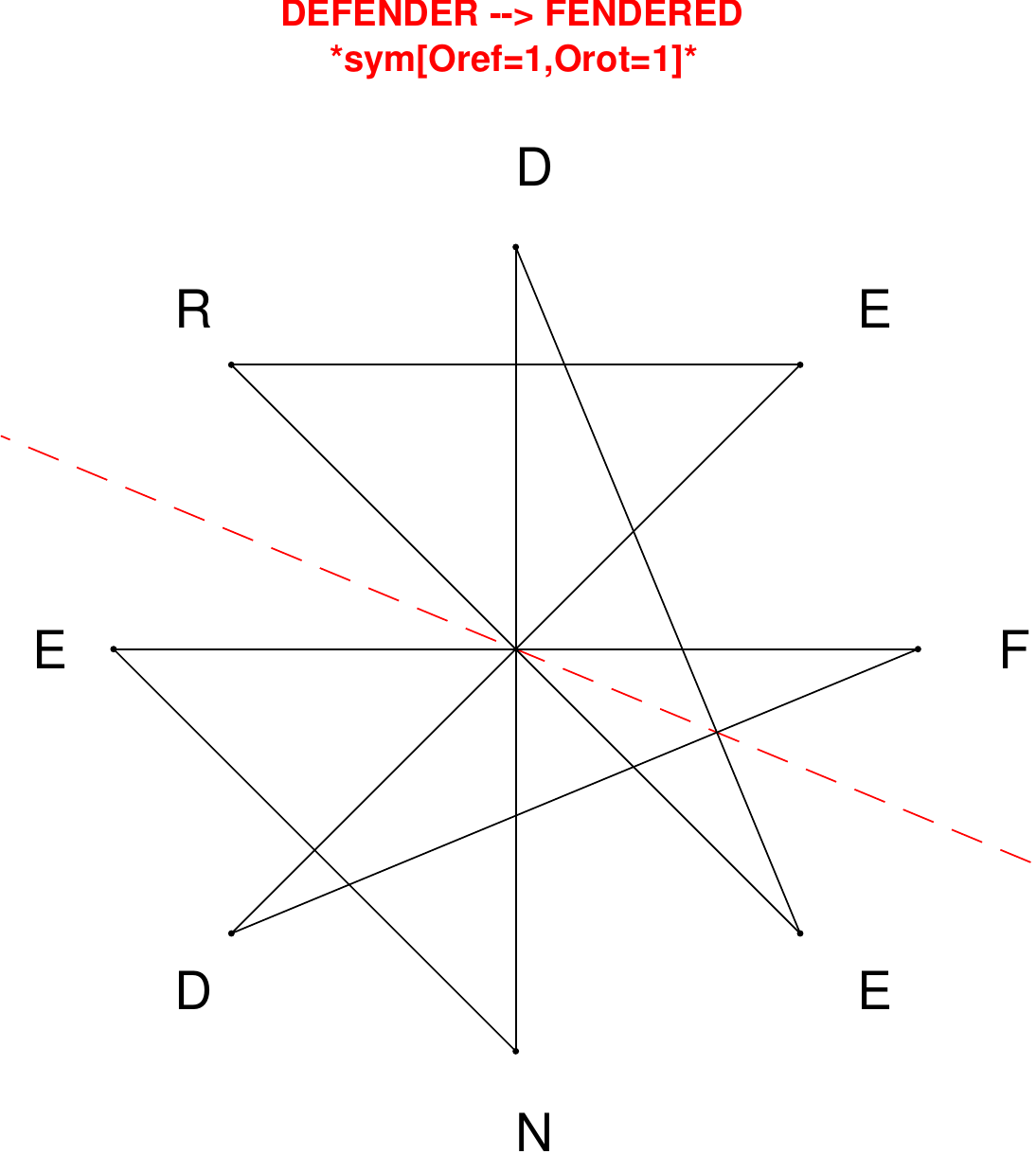}
\end{subfigure}
\hfill
\begin{subfigure}[T]{0.19\textwidth}
\centering
\includegraphics[width=\textwidth]{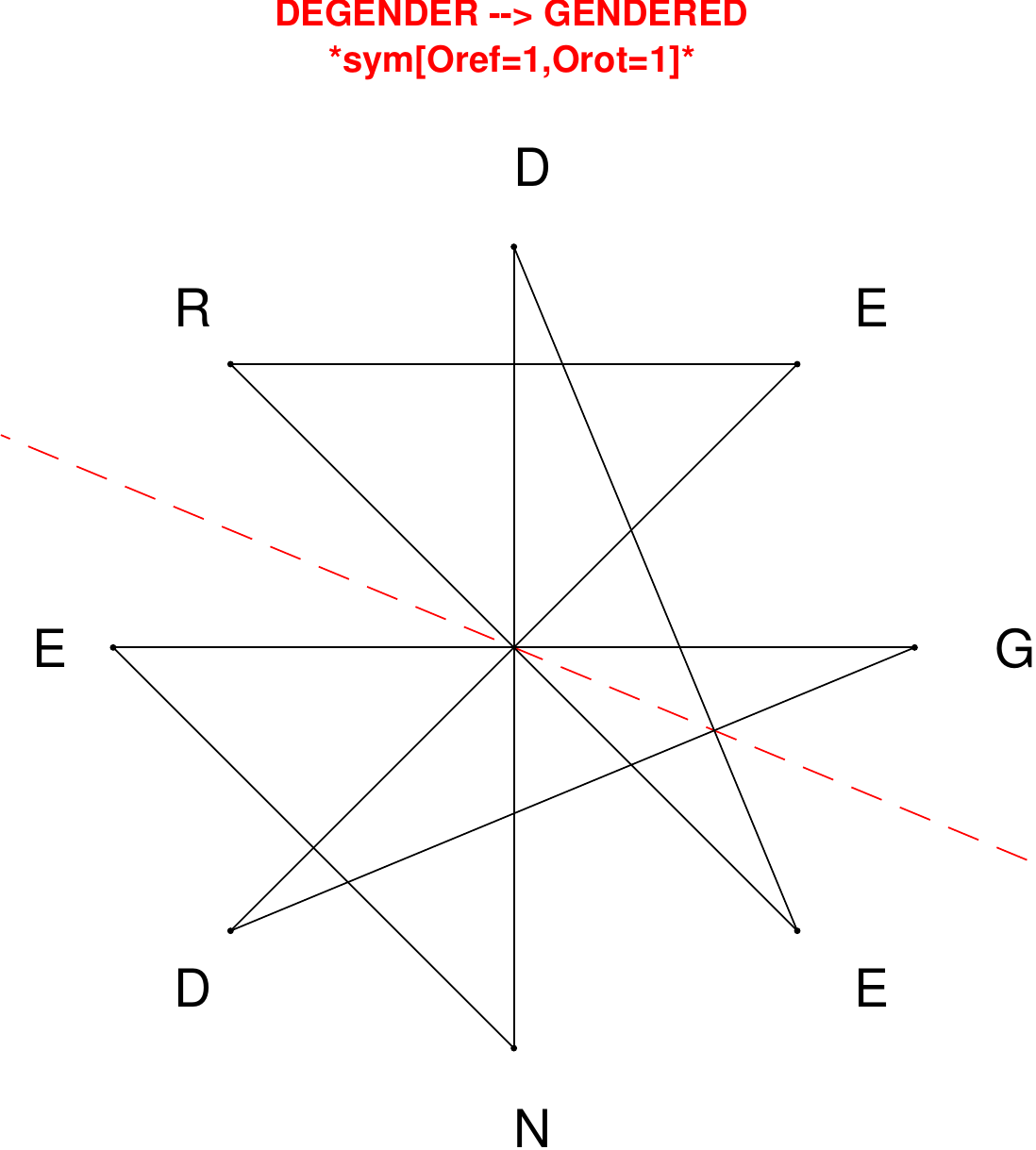}
\end{subfigure}
\hfill
\begin{subfigure}[T]{0.19\textwidth}
\centering
\includegraphics[width=\textwidth]{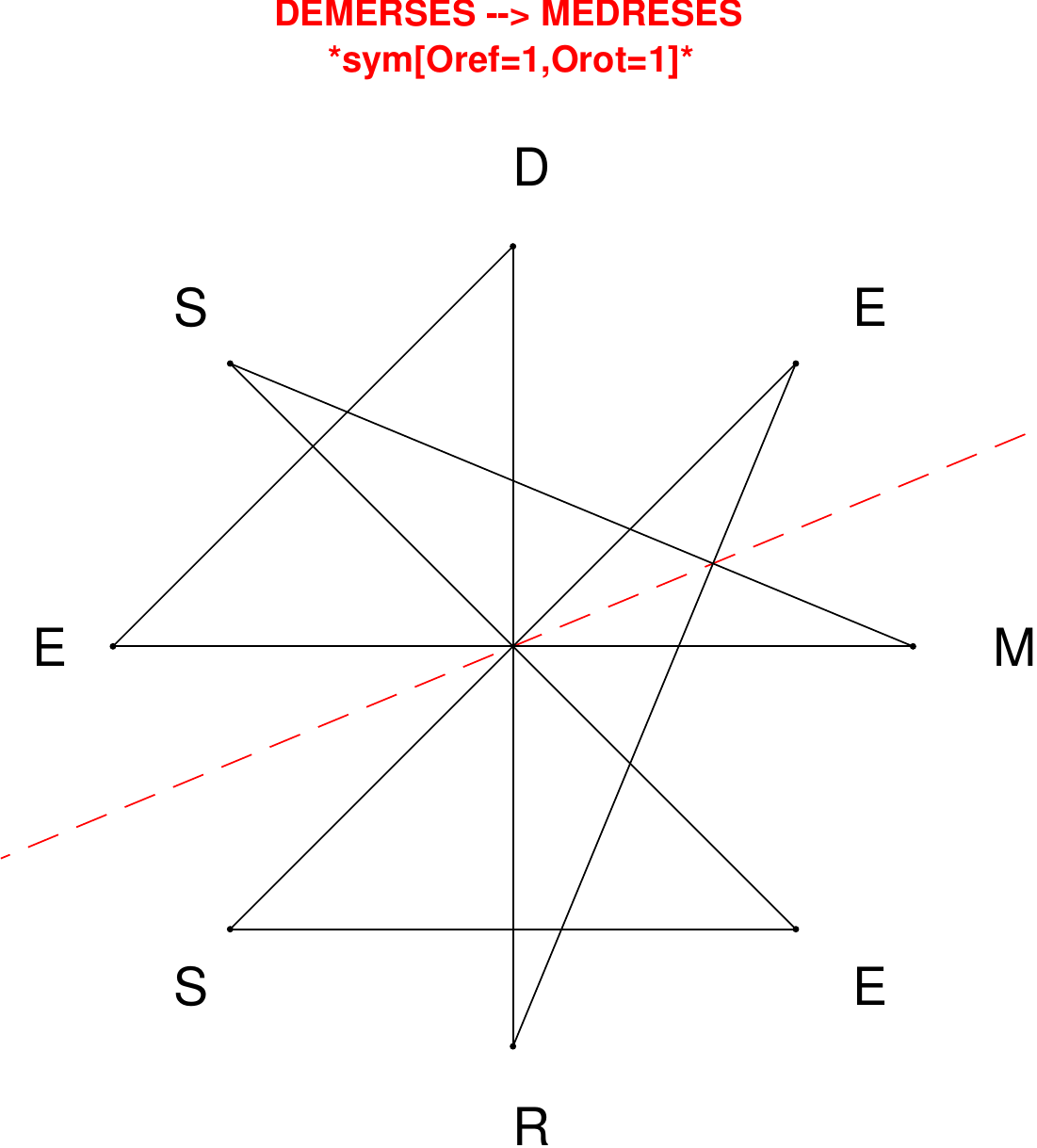}
\end{subfigure}
\hfill
\begin{subfigure}[T]{0.19\textwidth}
\centering
\includegraphics[width=\textwidth]{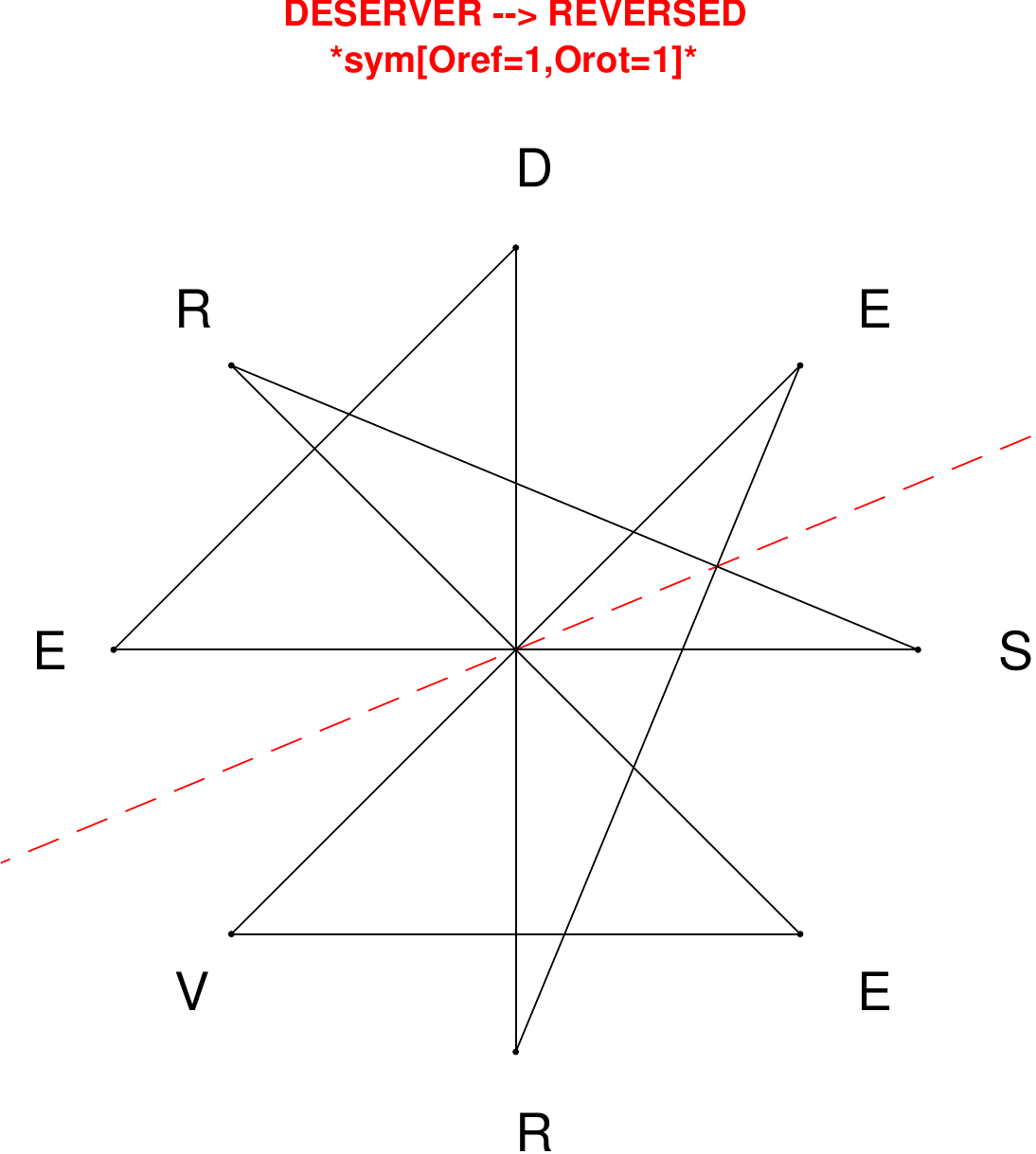}
\end{subfigure}
\end{figure}

\begin{figure}[H]
\centering
\begin{subfigure}[T]{0.19\textwidth}
\centering
\includegraphics[width=\textwidth]{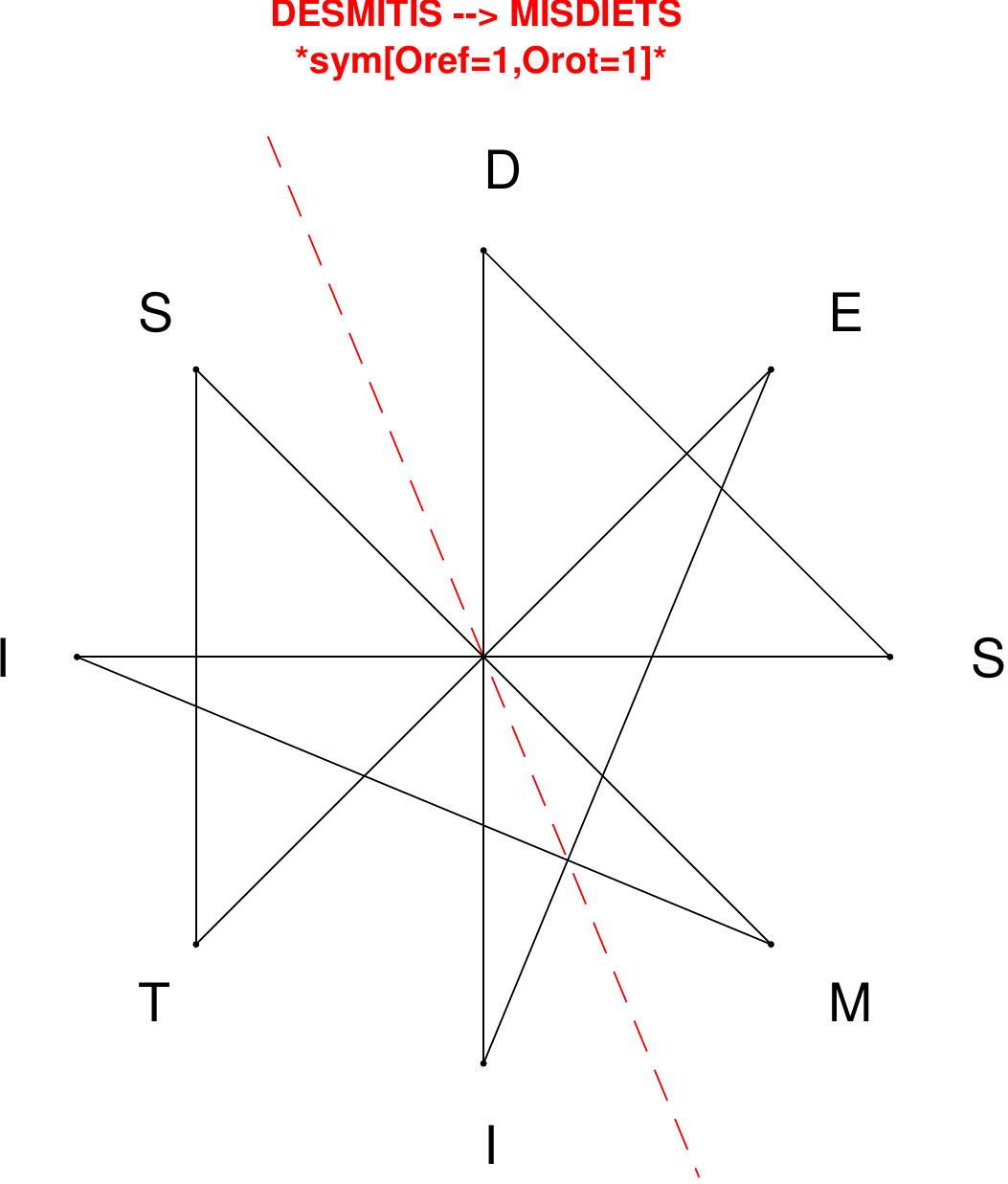}
\end{subfigure}
\hfill
\begin{subfigure}[T]{0.19\textwidth}
\centering
\includegraphics[width=\textwidth]{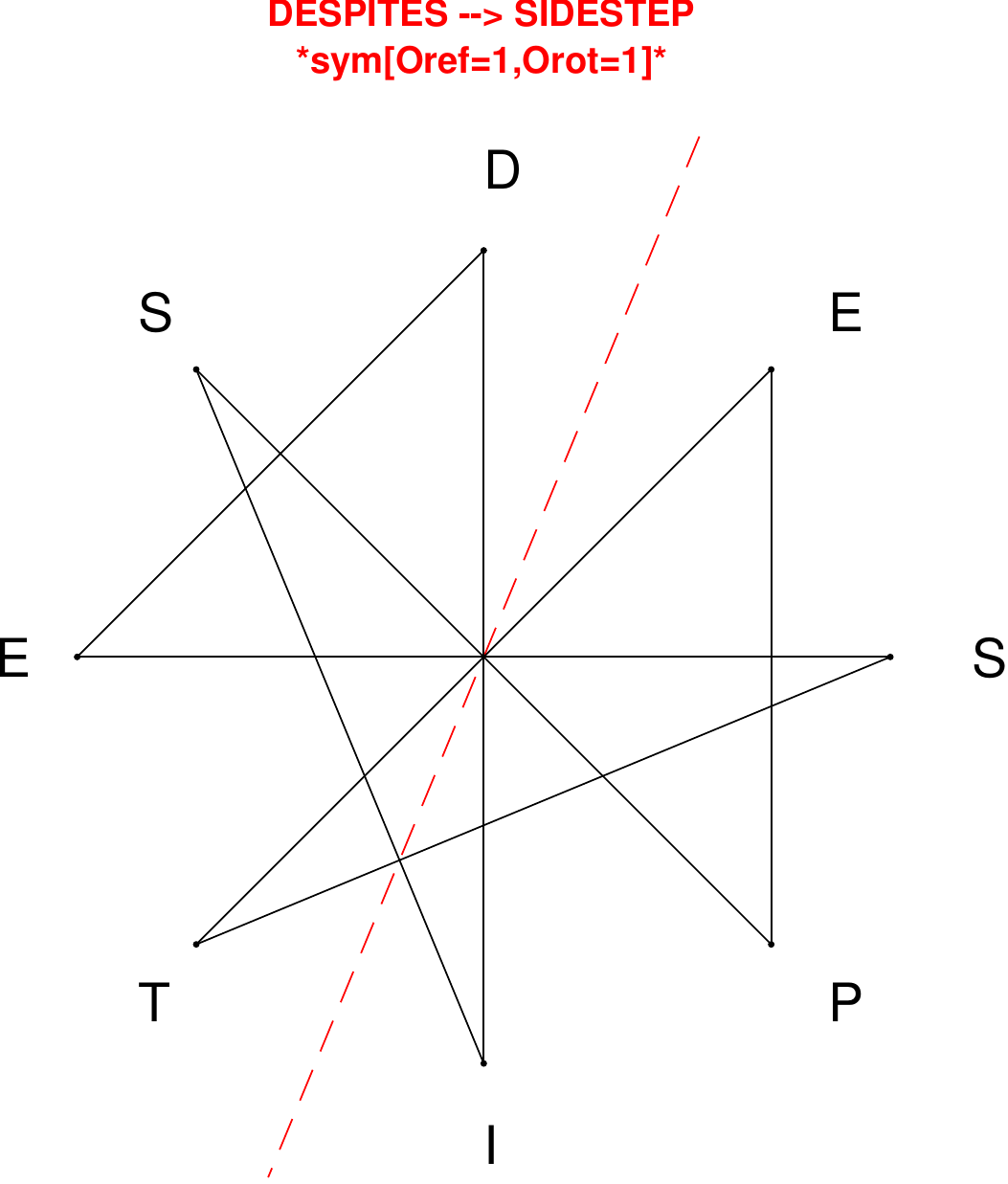}
\end{subfigure}
\hfill
\begin{subfigure}[T]{0.19\textwidth}
\centering
\includegraphics[width=\textwidth]{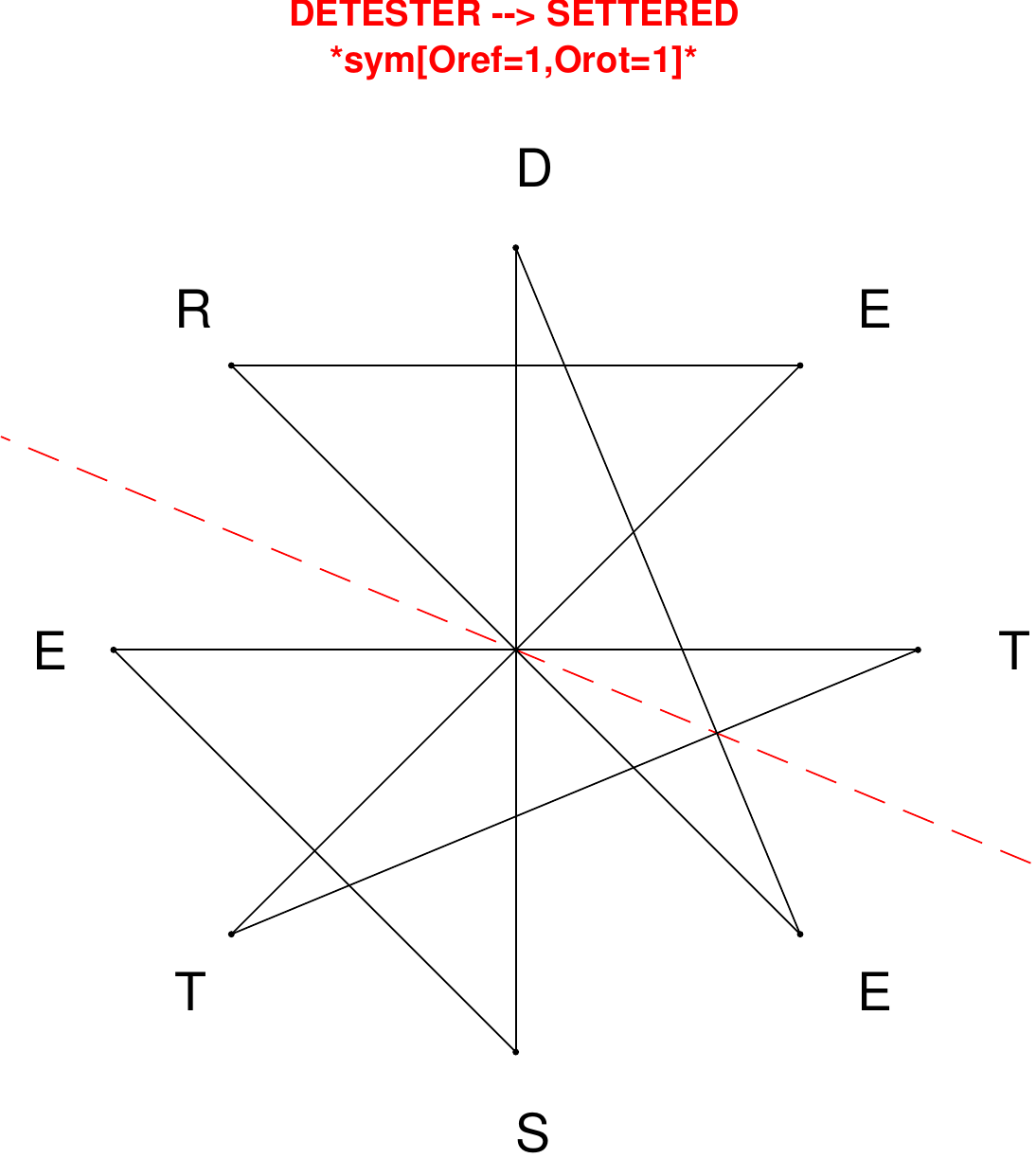}
\end{subfigure}
\hfill
\begin{subfigure}[T]{0.19\textwidth}
\centering
\includegraphics[width=\textwidth]{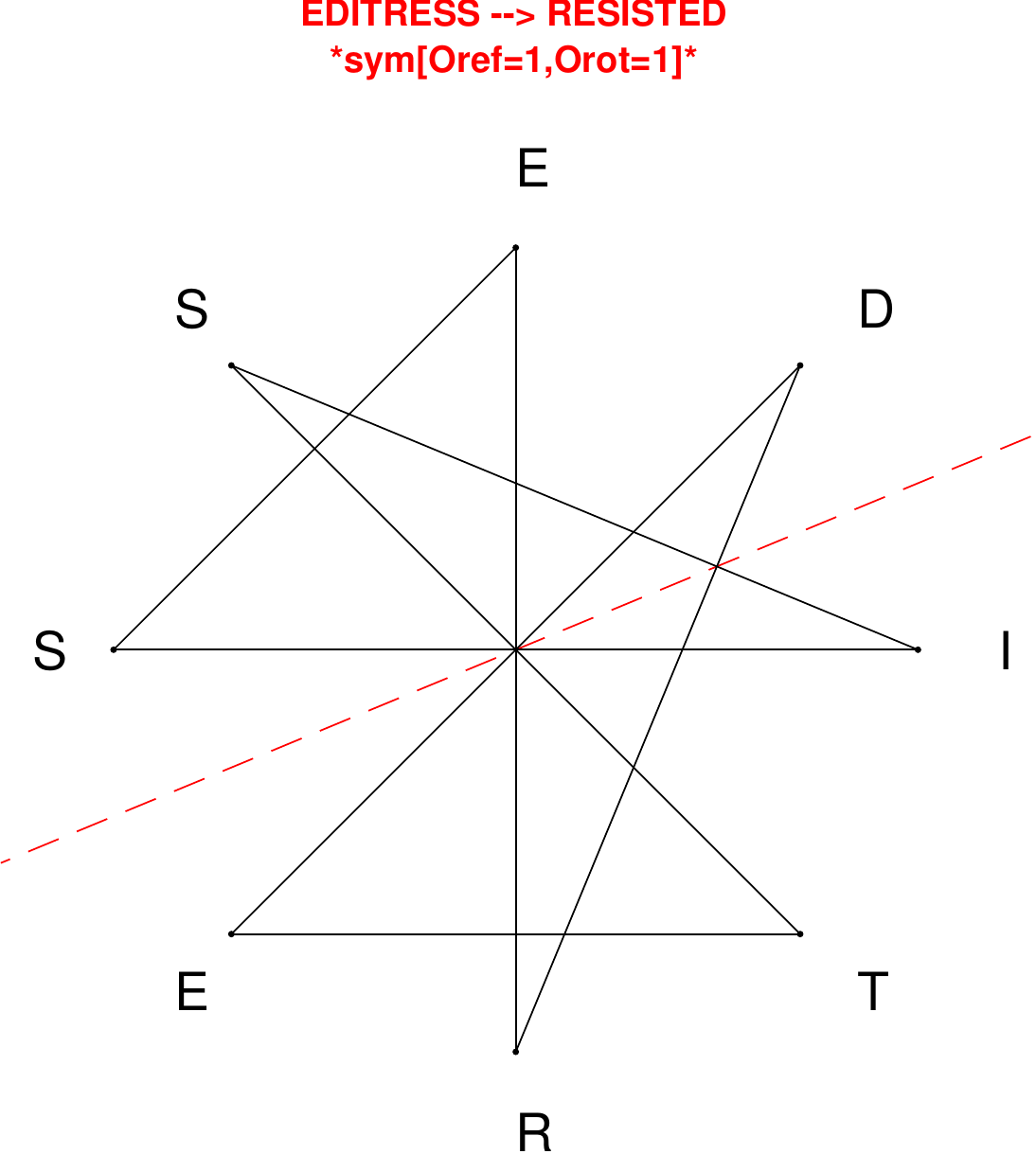}
\end{subfigure}
\hfill
\begin{subfigure}[T]{0.19\textwidth}
\centering
\includegraphics[width=\textwidth]{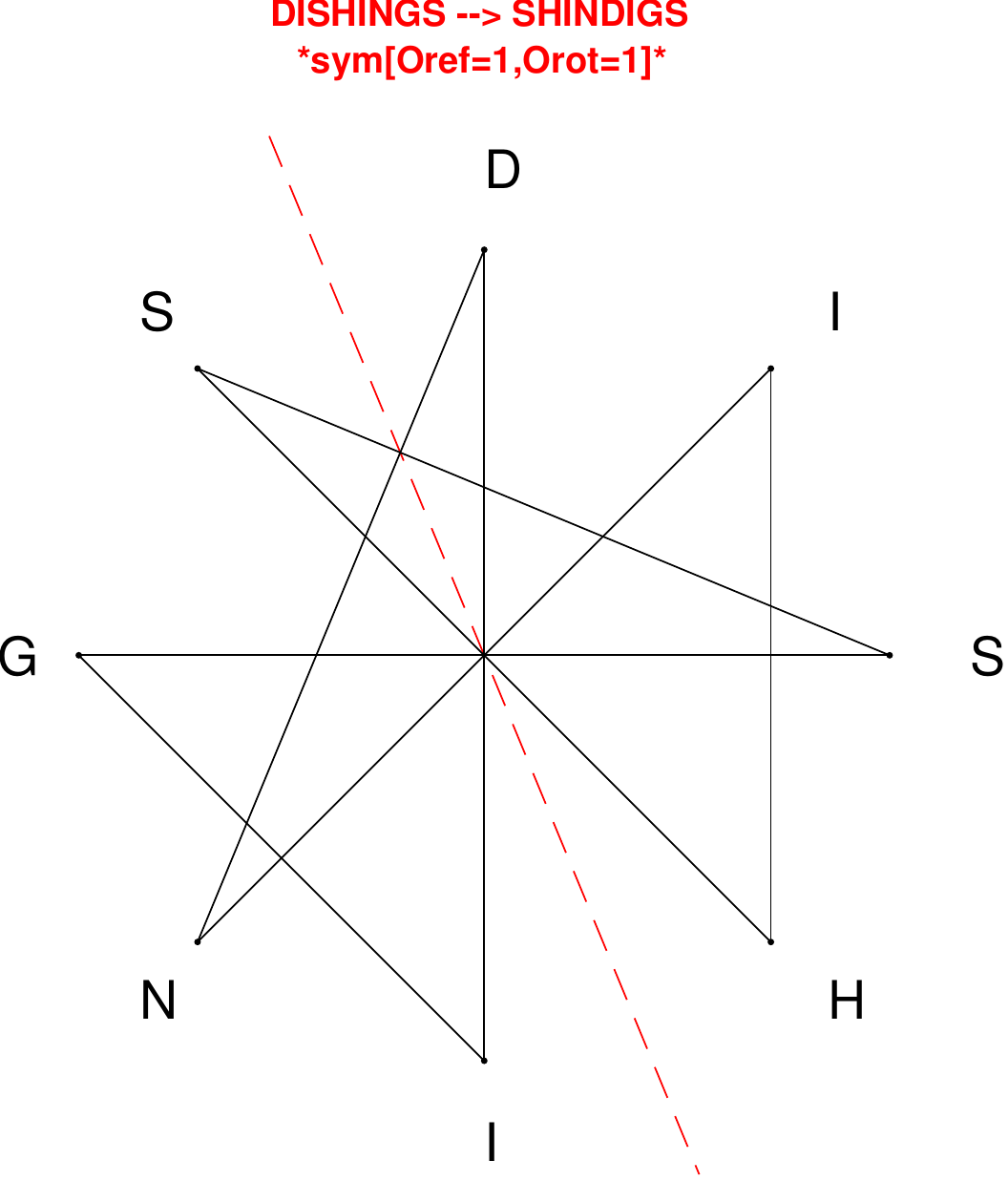}
\end{subfigure}
\end{figure}

\begin{figure}[H]
\centering
\begin{subfigure}[T]{0.19\textwidth}
\centering
\includegraphics[width=\textwidth]{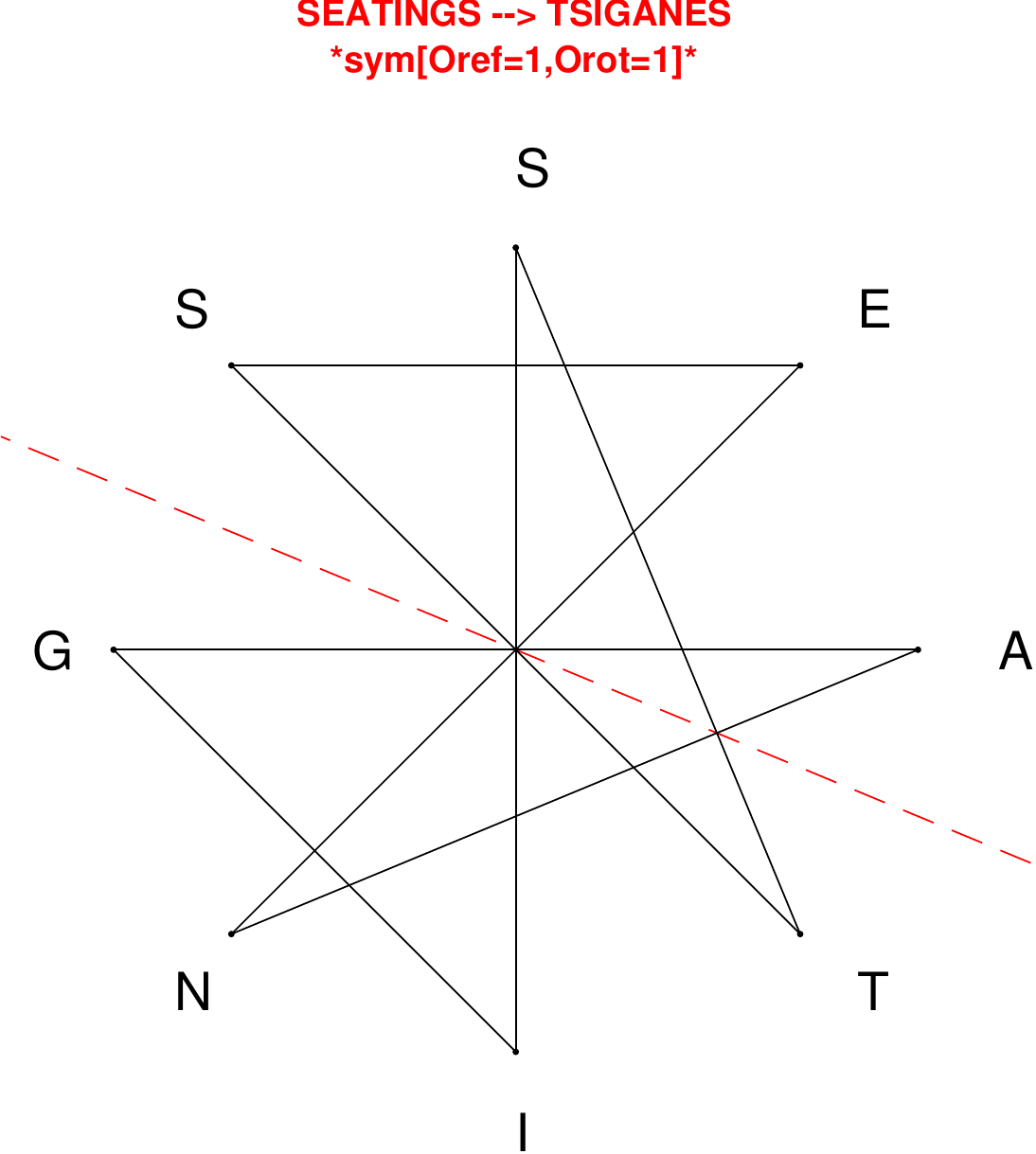}
\end{subfigure}
\hfill
\begin{subfigure}[T]{0.19\textwidth}
\centering
\includegraphics[width=\textwidth]{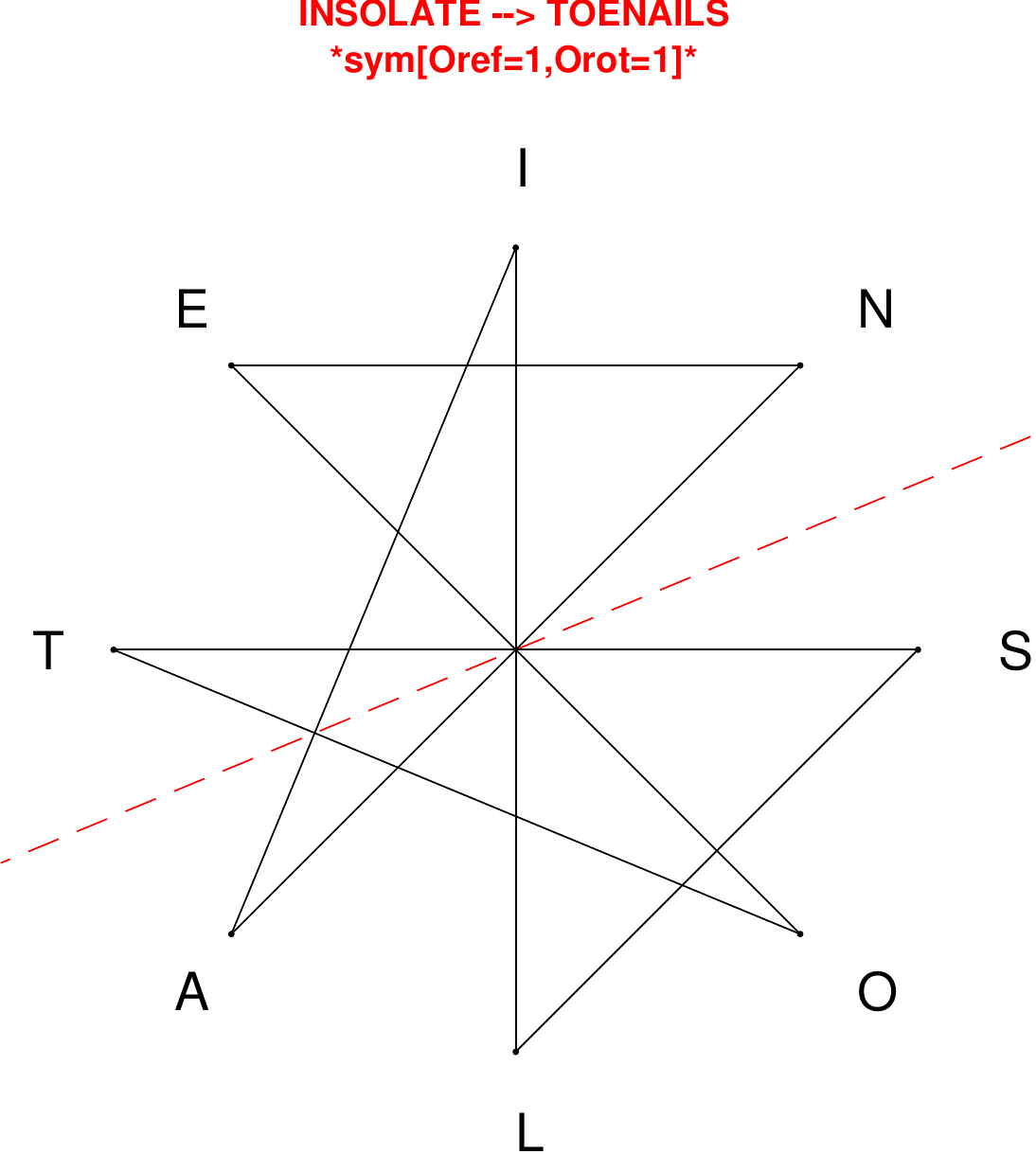}
\end{subfigure}
\hfill
\begin{subfigure}[T]{0.19\textwidth}
\centering
\includegraphics[width=\textwidth]{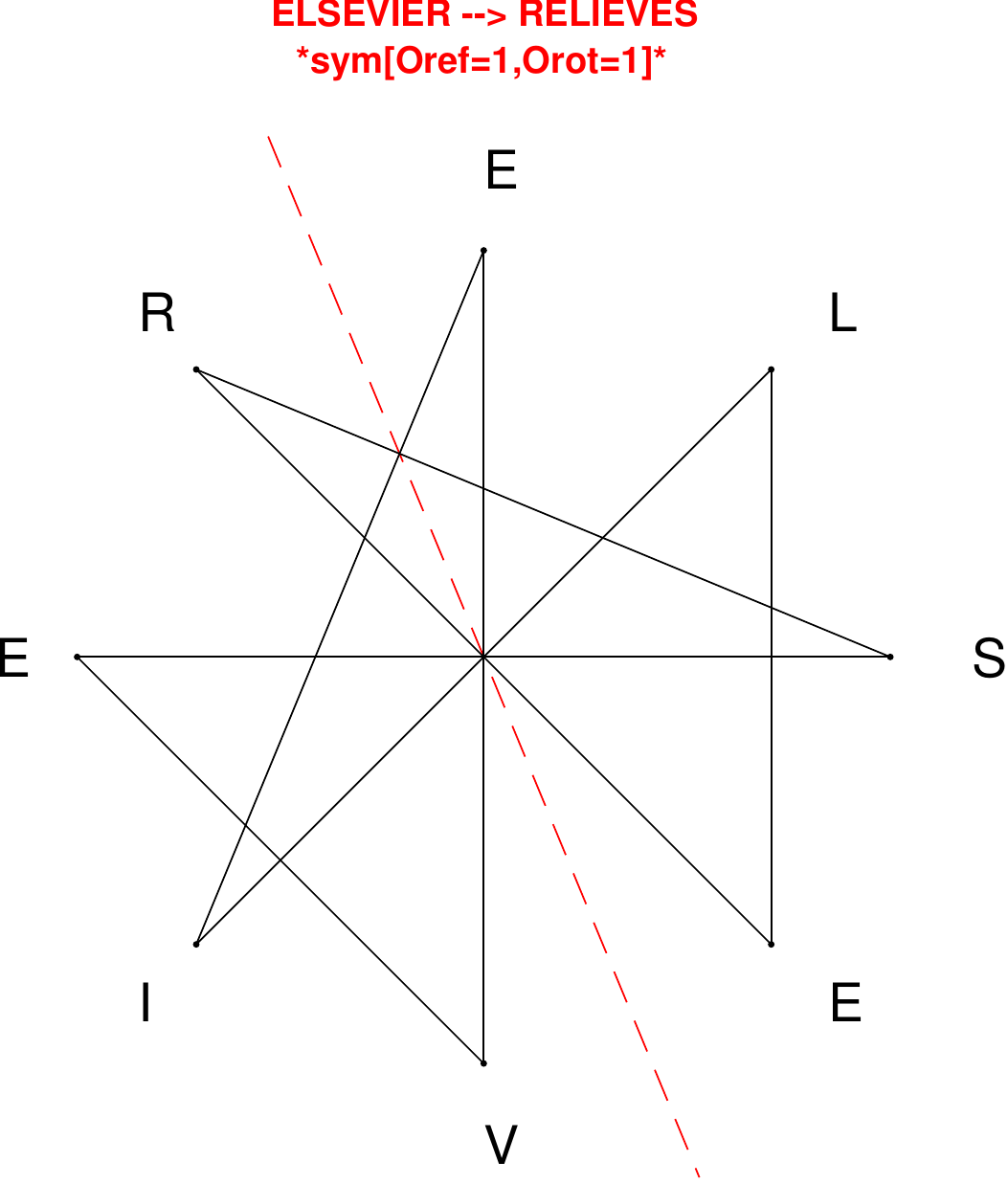}
\end{subfigure}
\hfill
\begin{subfigure}[T]{0.19\textwidth}
\centering
\includegraphics[width=\textwidth]{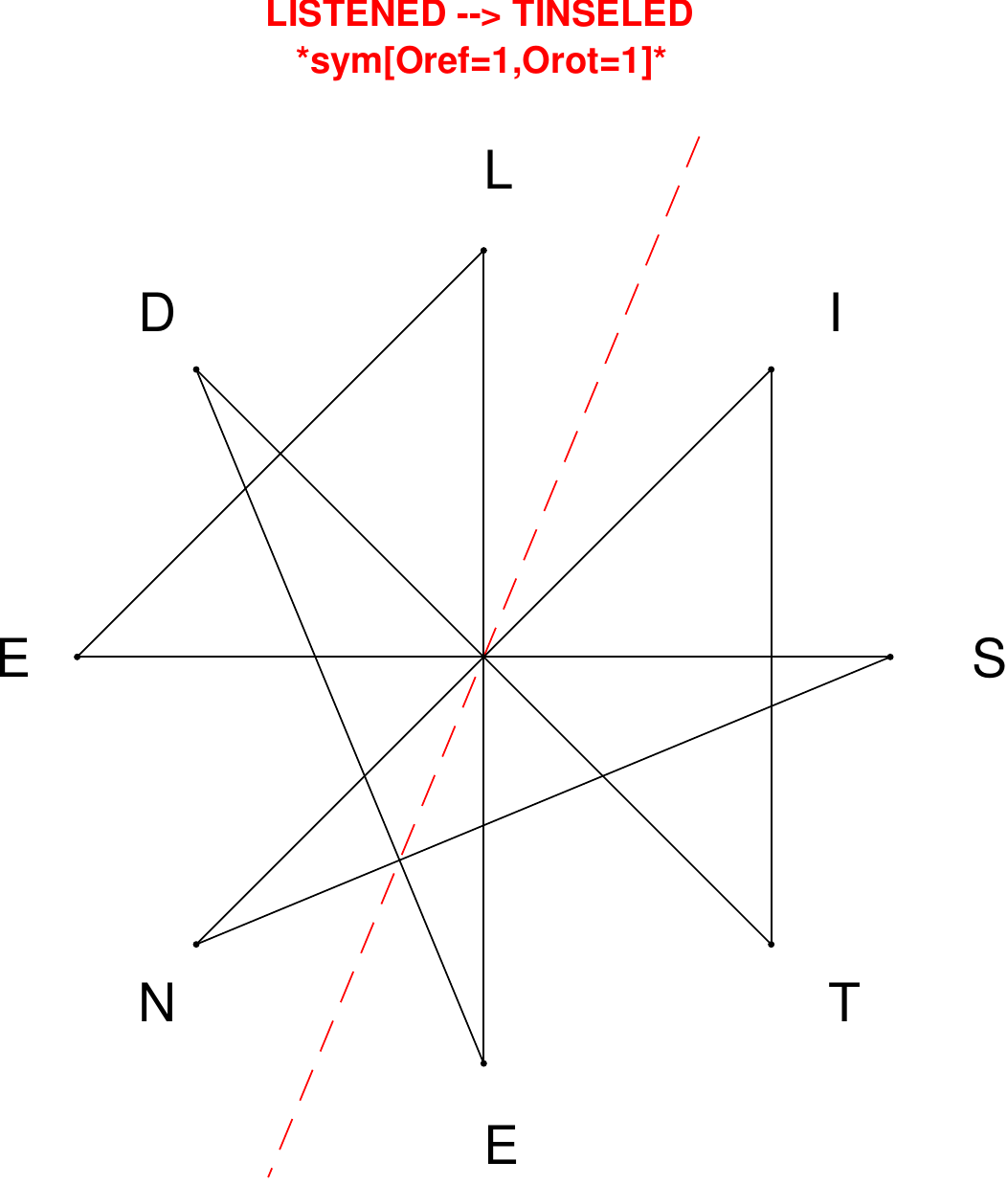}
\end{subfigure}
\hfill
\begin{subfigure}[T]{0.19\textwidth}
\centering
\includegraphics[width=\textwidth]{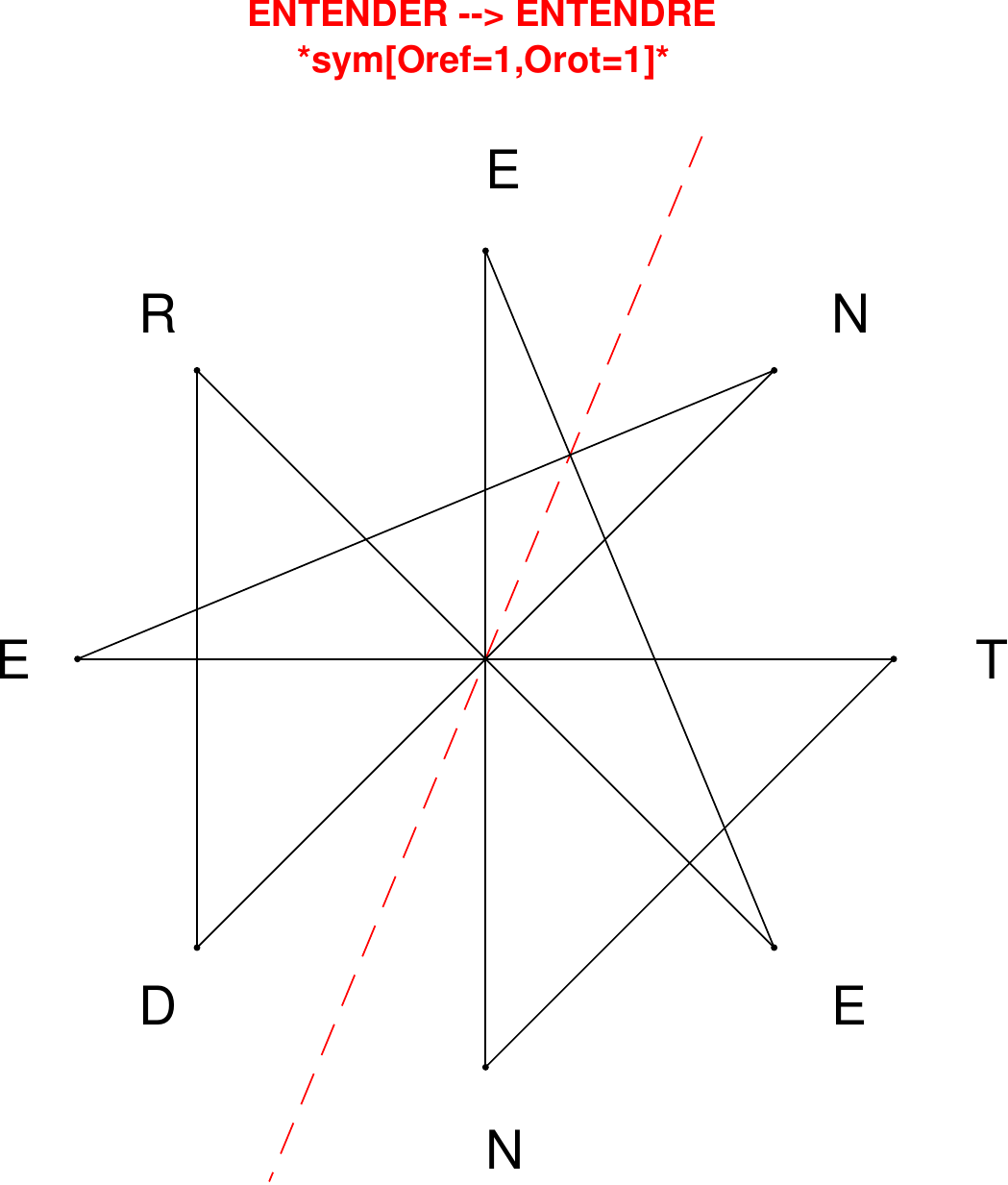}
\end{subfigure}
\end{figure}

\begin{figure}[H]
\centering
\begin{subfigure}[T]{0.19\textwidth}
\centering
\includegraphics[width=\textwidth]{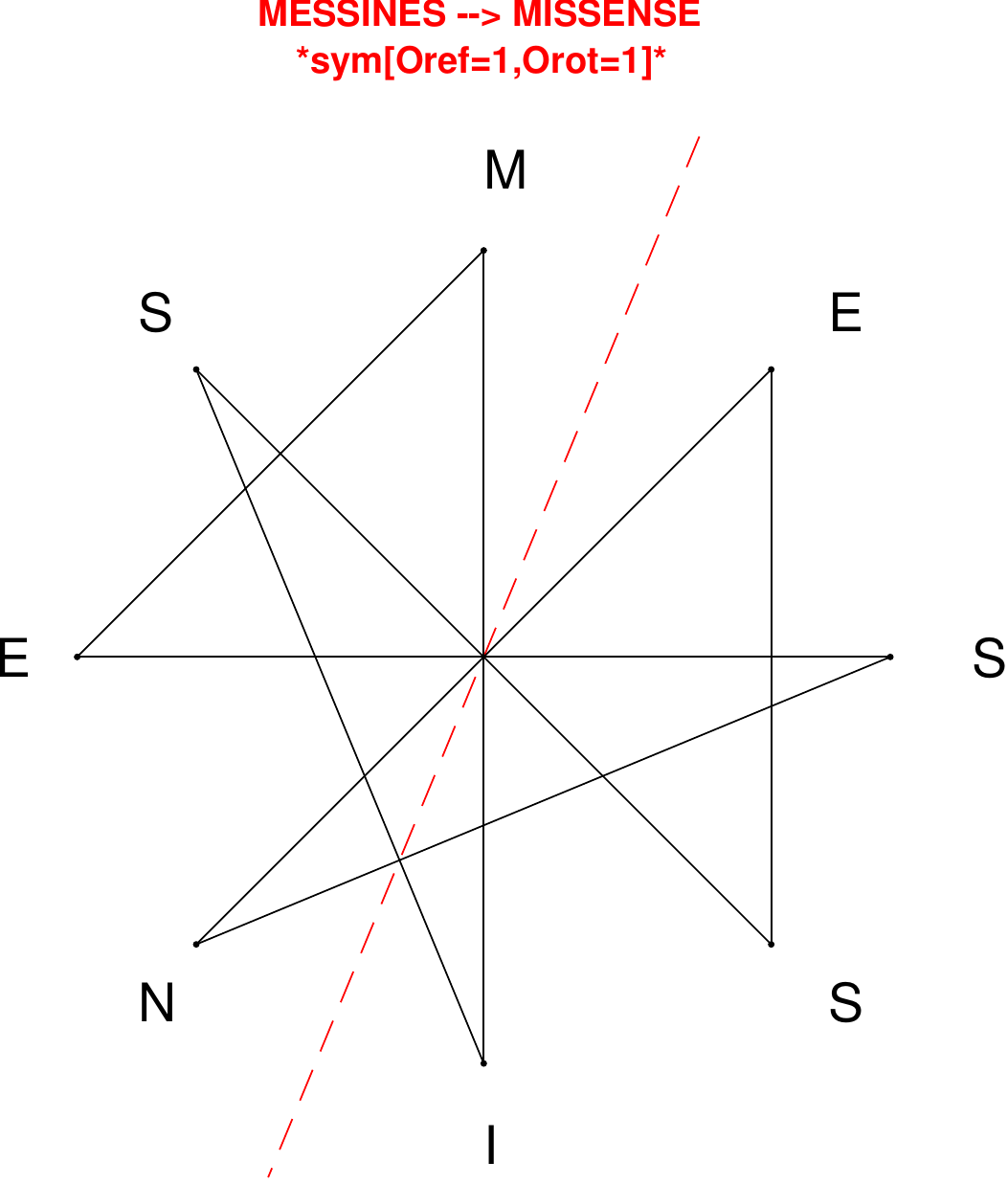}
\end{subfigure}
\hfill
\begin{subfigure}[T]{0.19\textwidth}
\centering
\includegraphics[width=\textwidth]{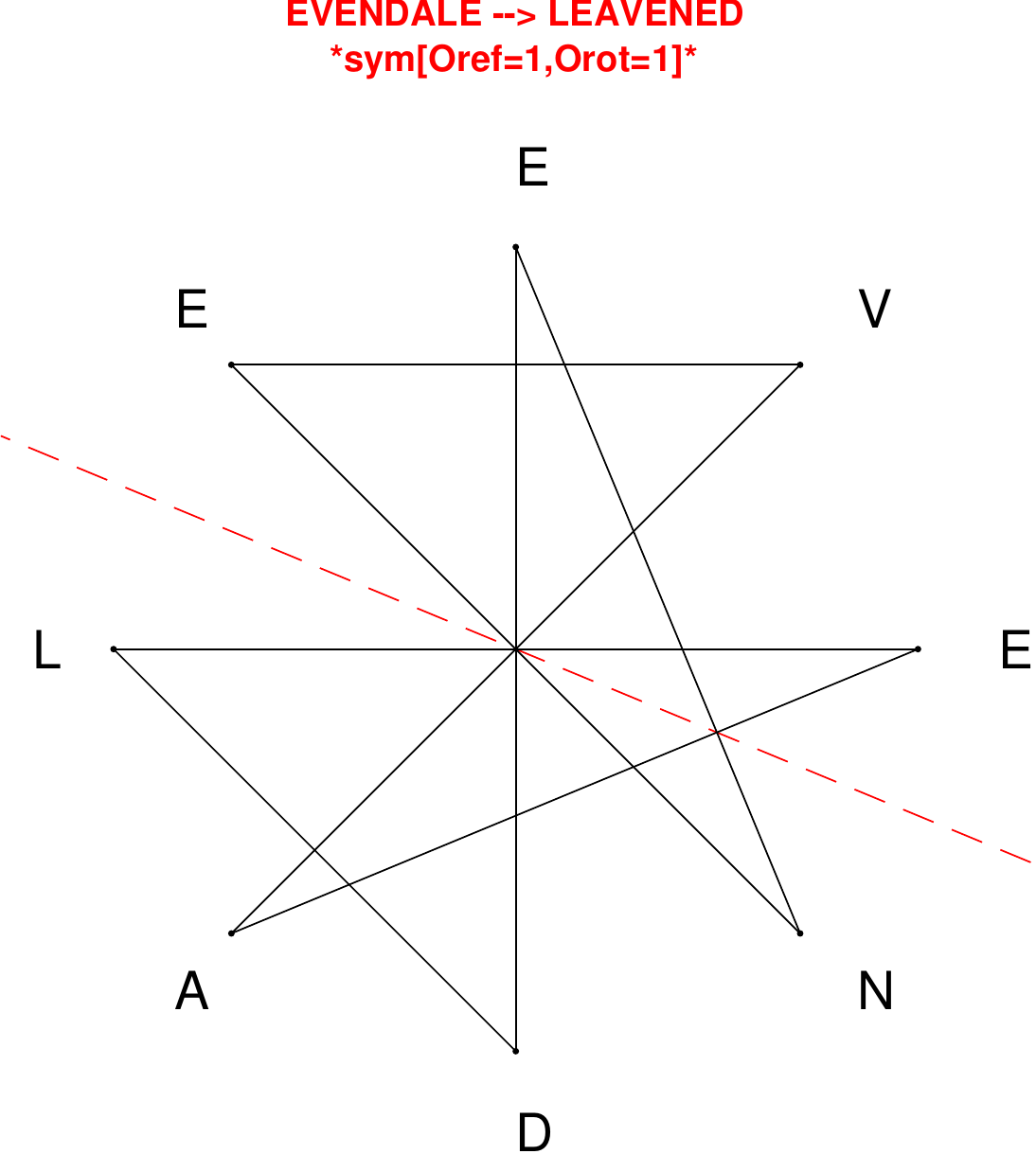}
\end{subfigure}
\hfill
\begin{subfigure}[T]{0.19\textwidth}
\centering
\includegraphics[width=\textwidth]{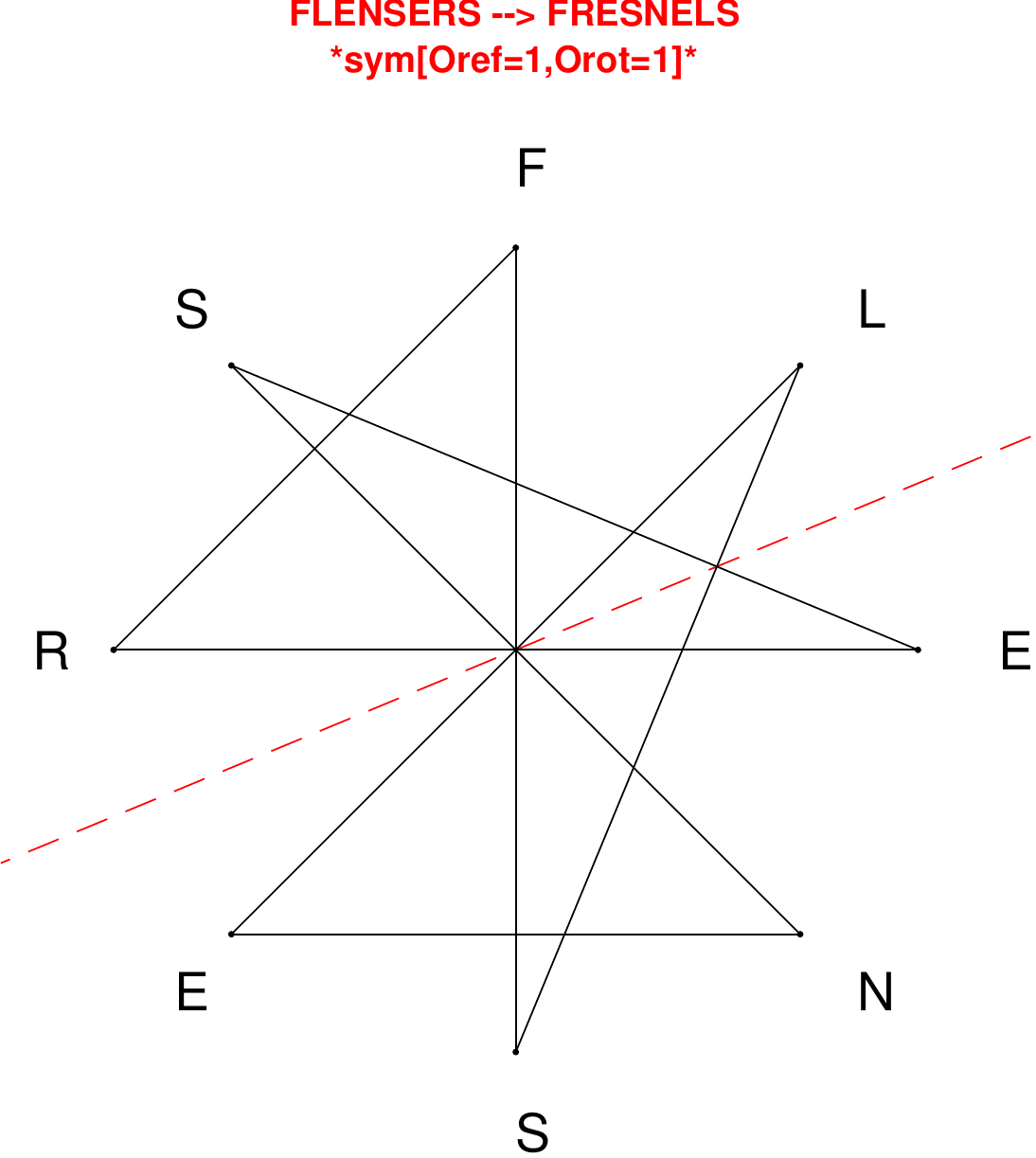}
\end{subfigure}
\hfill
\begin{subfigure}[T]{0.19\textwidth}
\centering
\includegraphics[width=\textwidth]{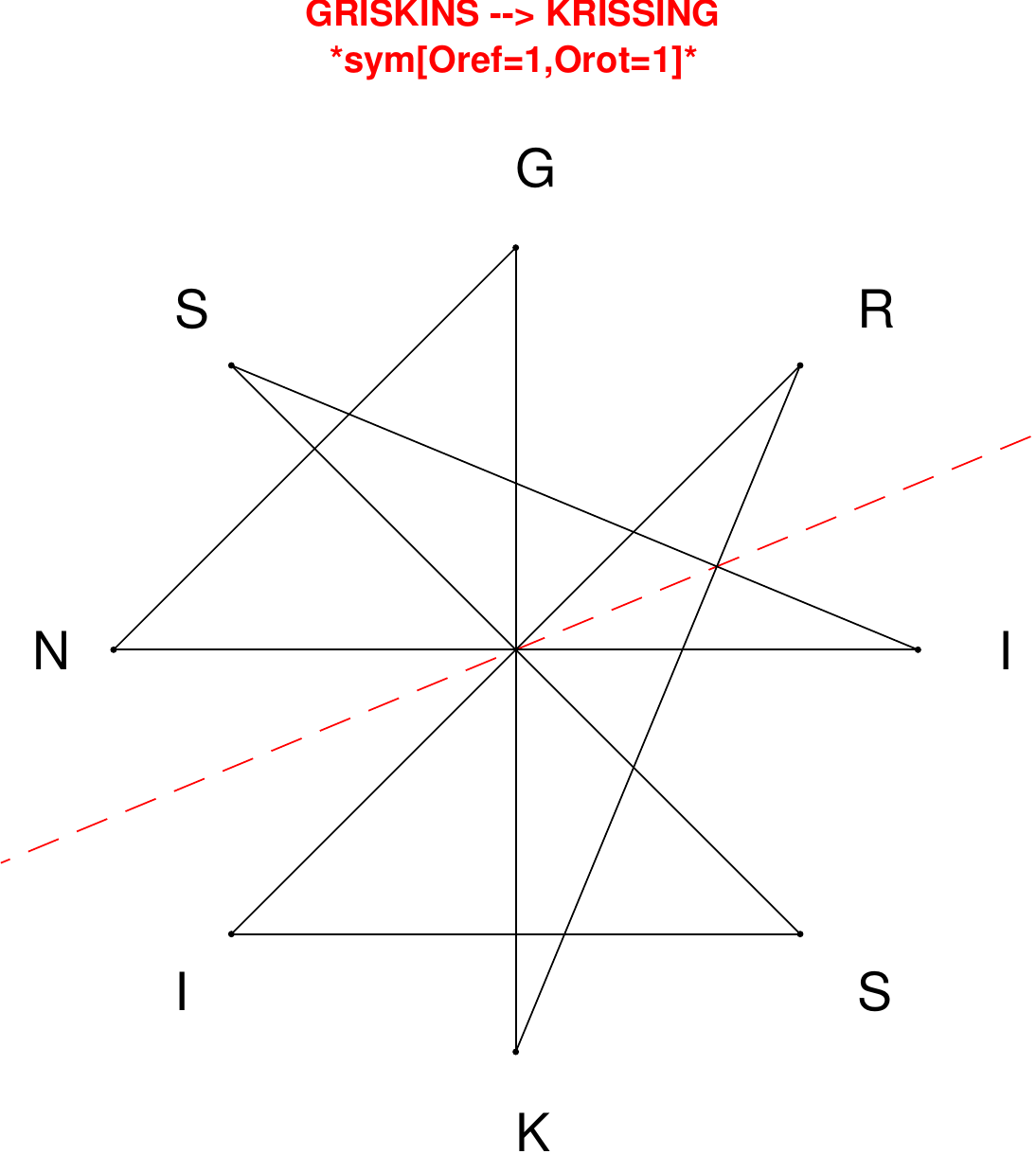}
\end{subfigure}
\hfill
\begin{subfigure}[T]{0.19\textwidth}
\centering
\includegraphics[width=\textwidth]{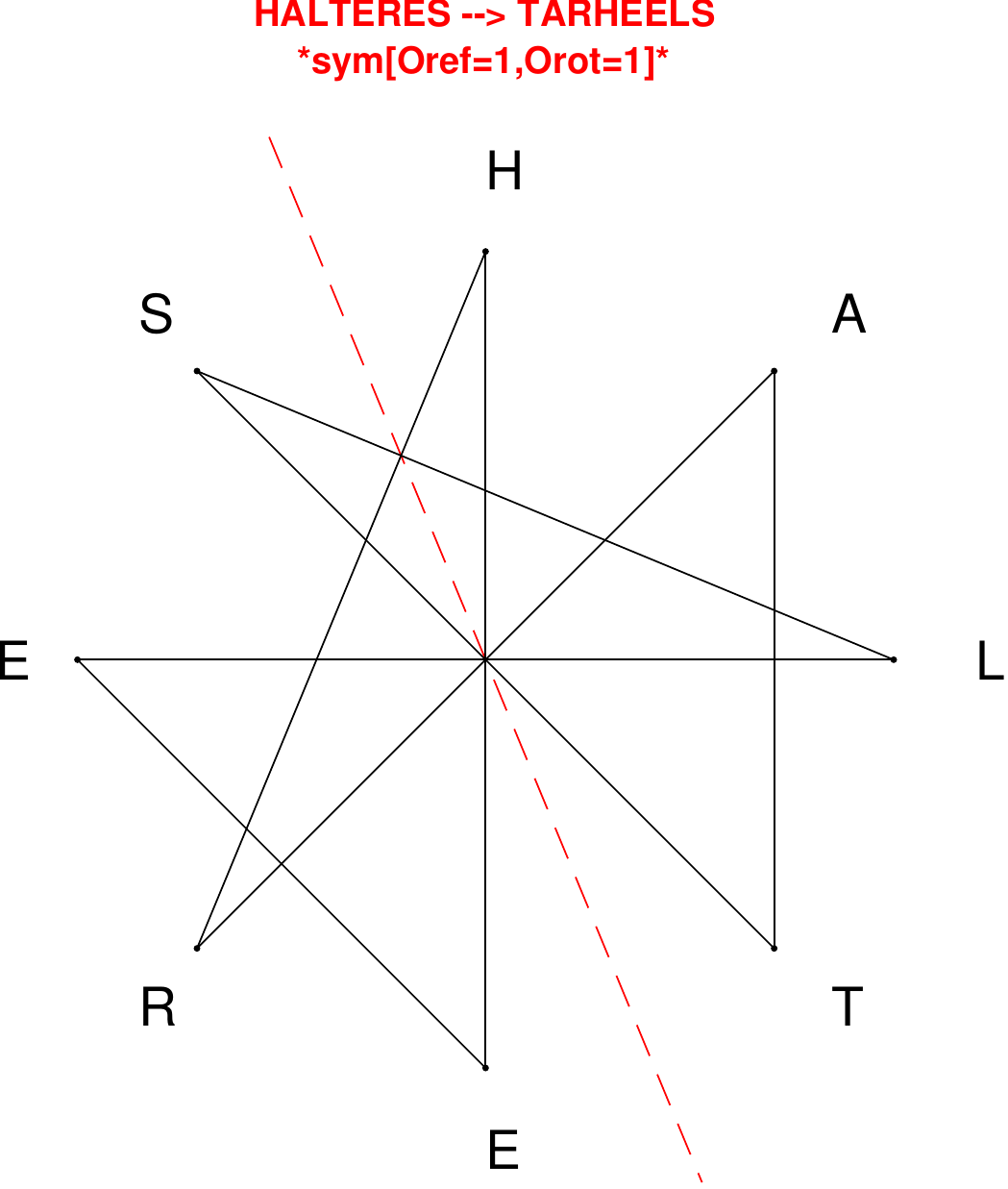}
\end{subfigure}
\end{figure}

\begin{figure}[H]
\centering
\begin{subfigure}[T]{0.19\textwidth}
\centering
\includegraphics[width=\textwidth]{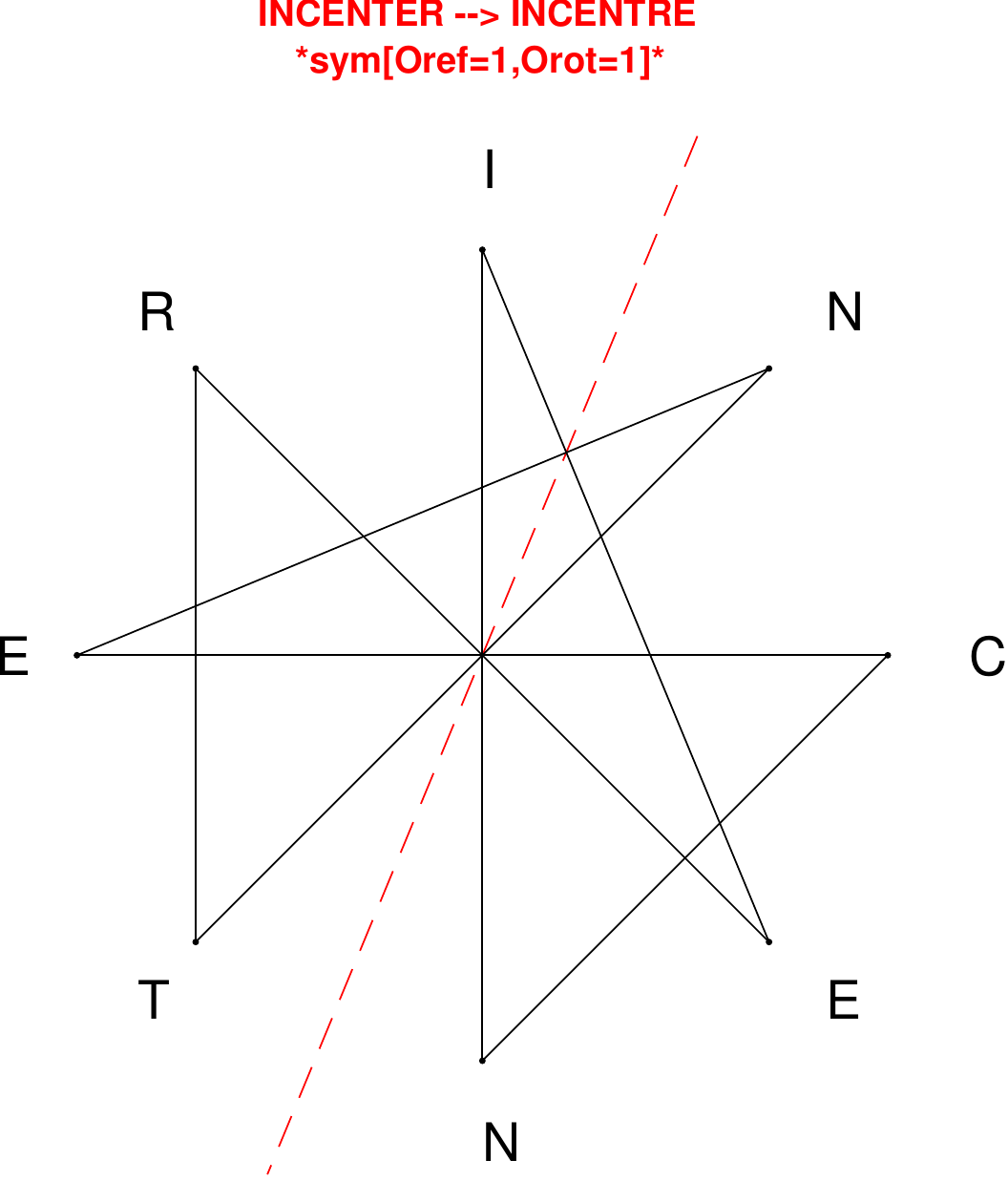}
\end{subfigure}
\hfill
\begin{subfigure}[T]{0.19\textwidth}
\centering
\includegraphics[width=\textwidth]{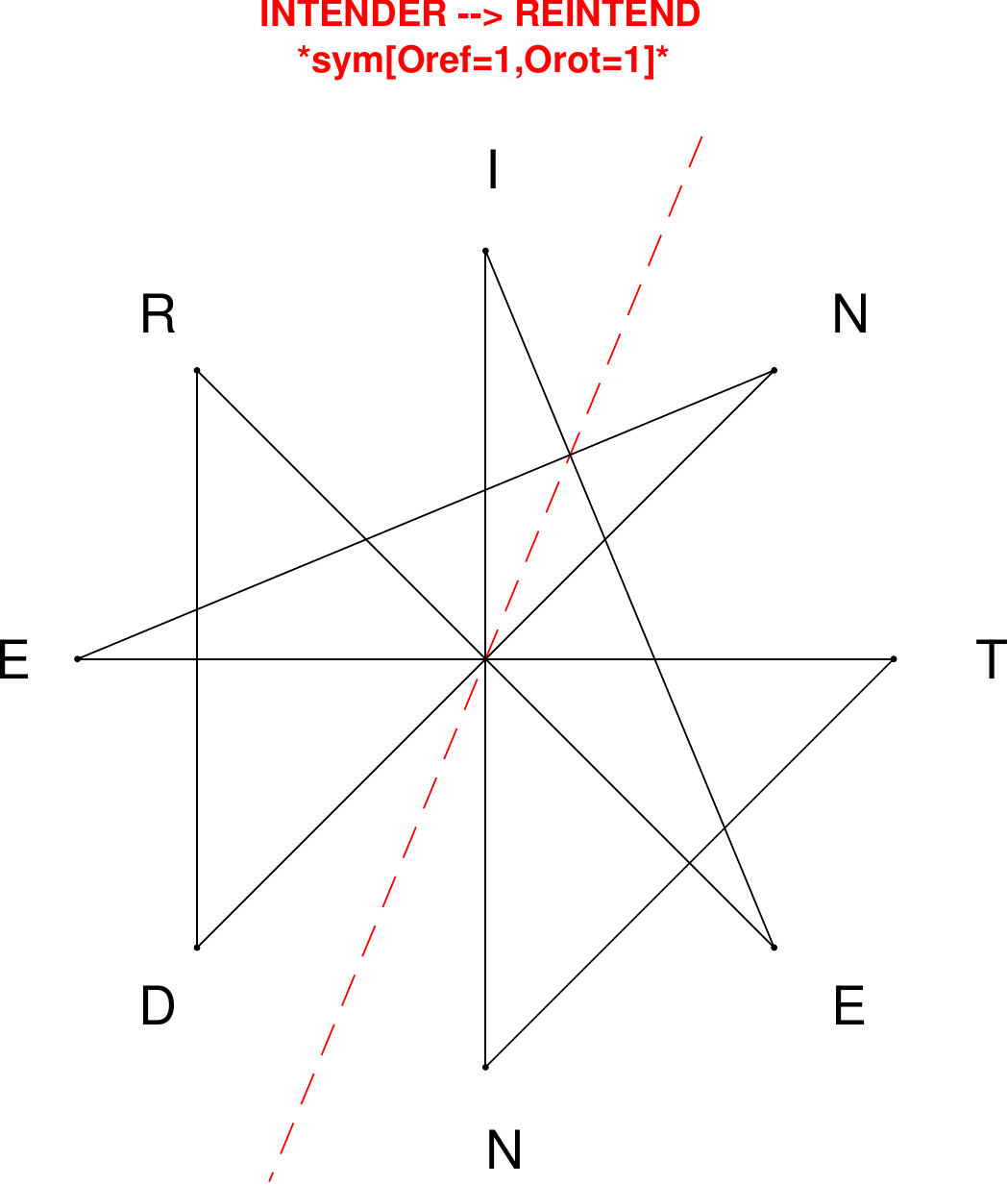}
\end{subfigure}
\hfill
\begin{subfigure}[T]{0.19\textwidth}
\centering
\includegraphics[width=\textwidth]{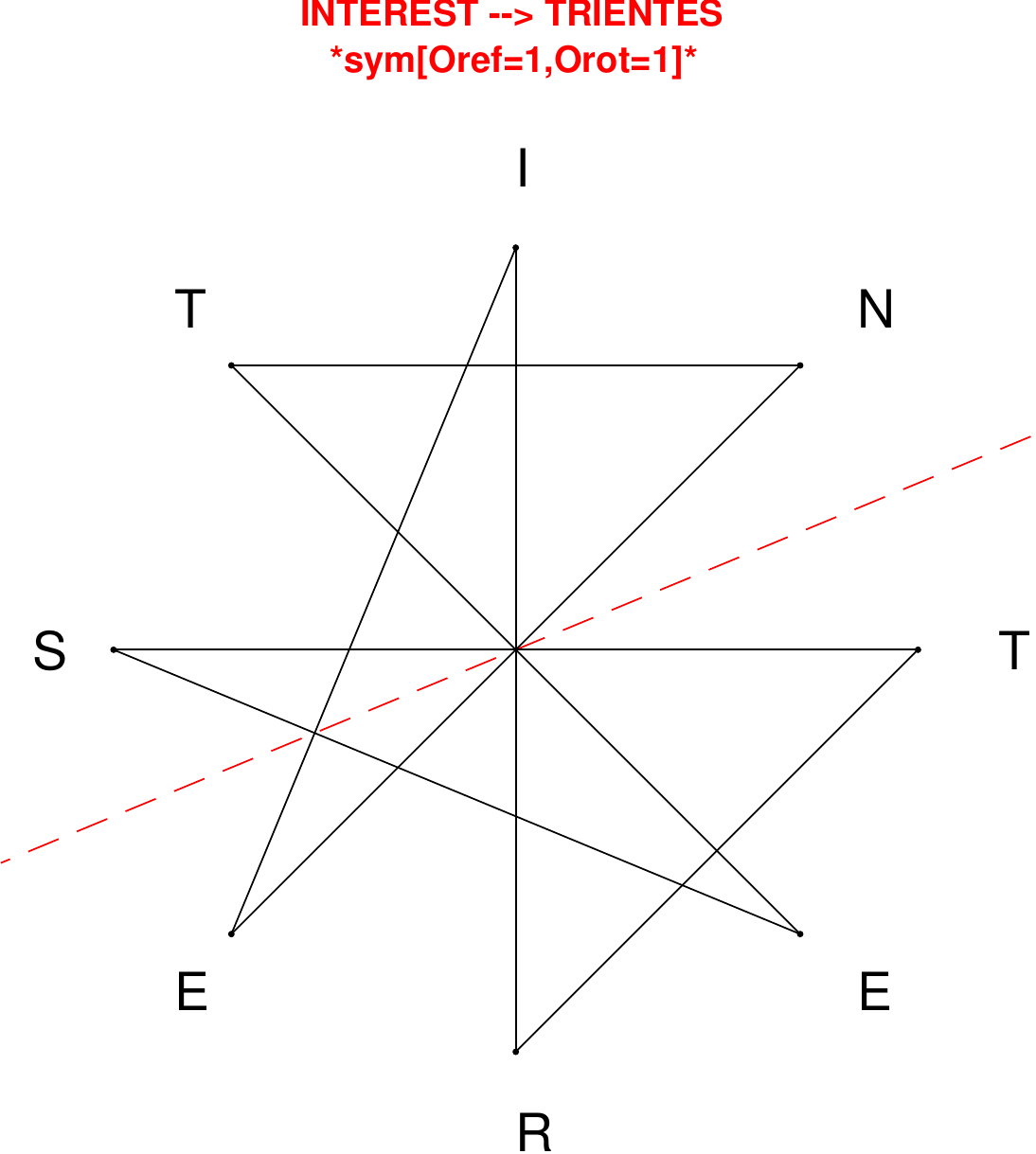}
\end{subfigure}
\hfill
\begin{subfigure}[T]{0.19\textwidth}
\centering
\includegraphics[width=\textwidth]{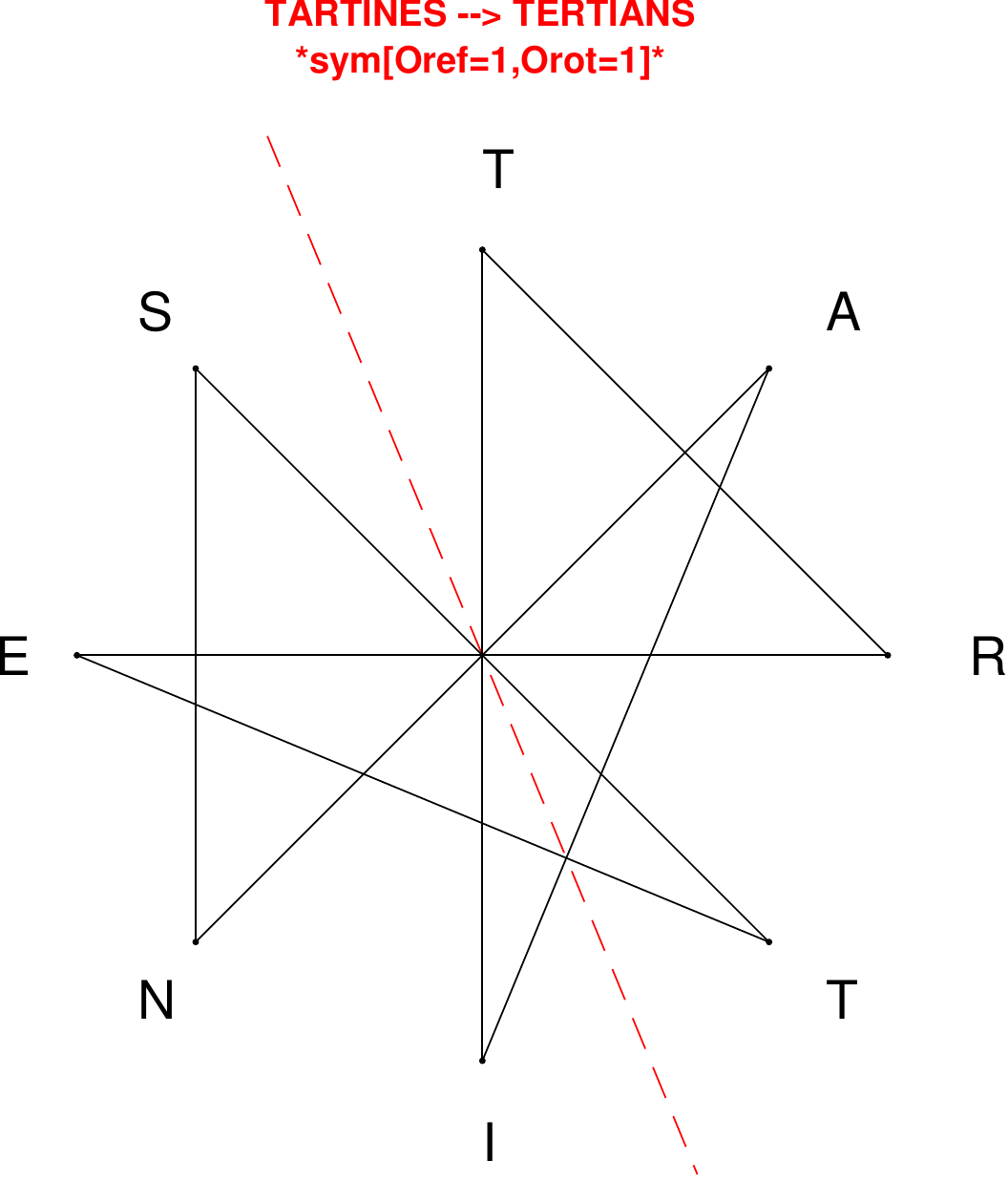}
\end{subfigure}
\hfill
\begin{subfigure}[T]{0.19\textwidth}
\centering
\includegraphics[width=\textwidth]{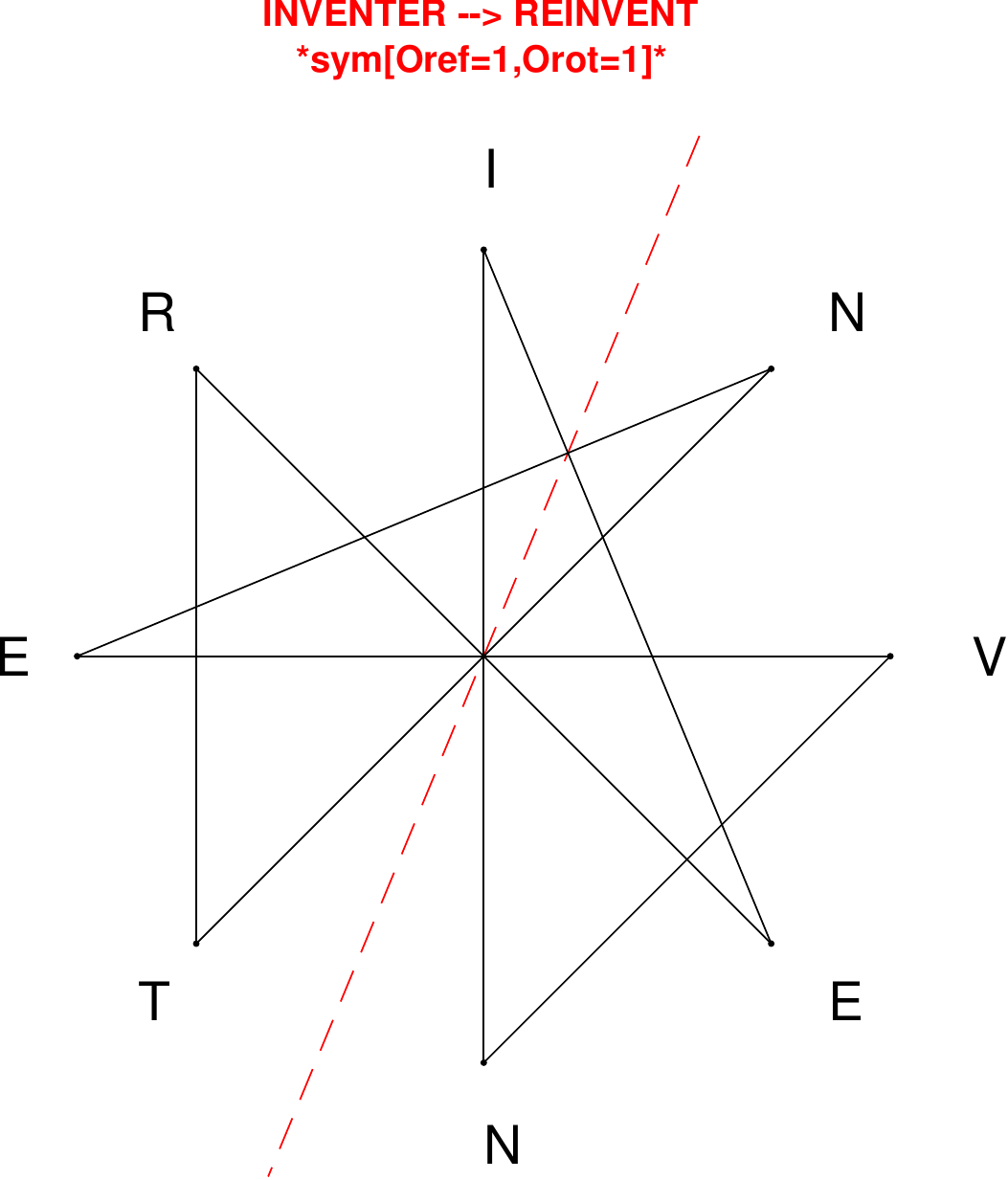}
\end{subfigure}
\end{figure}

\begin{figure}[H]
\centering
\begin{subfigure}[T]{0.19\textwidth}
\centering
\includegraphics[width=\textwidth]{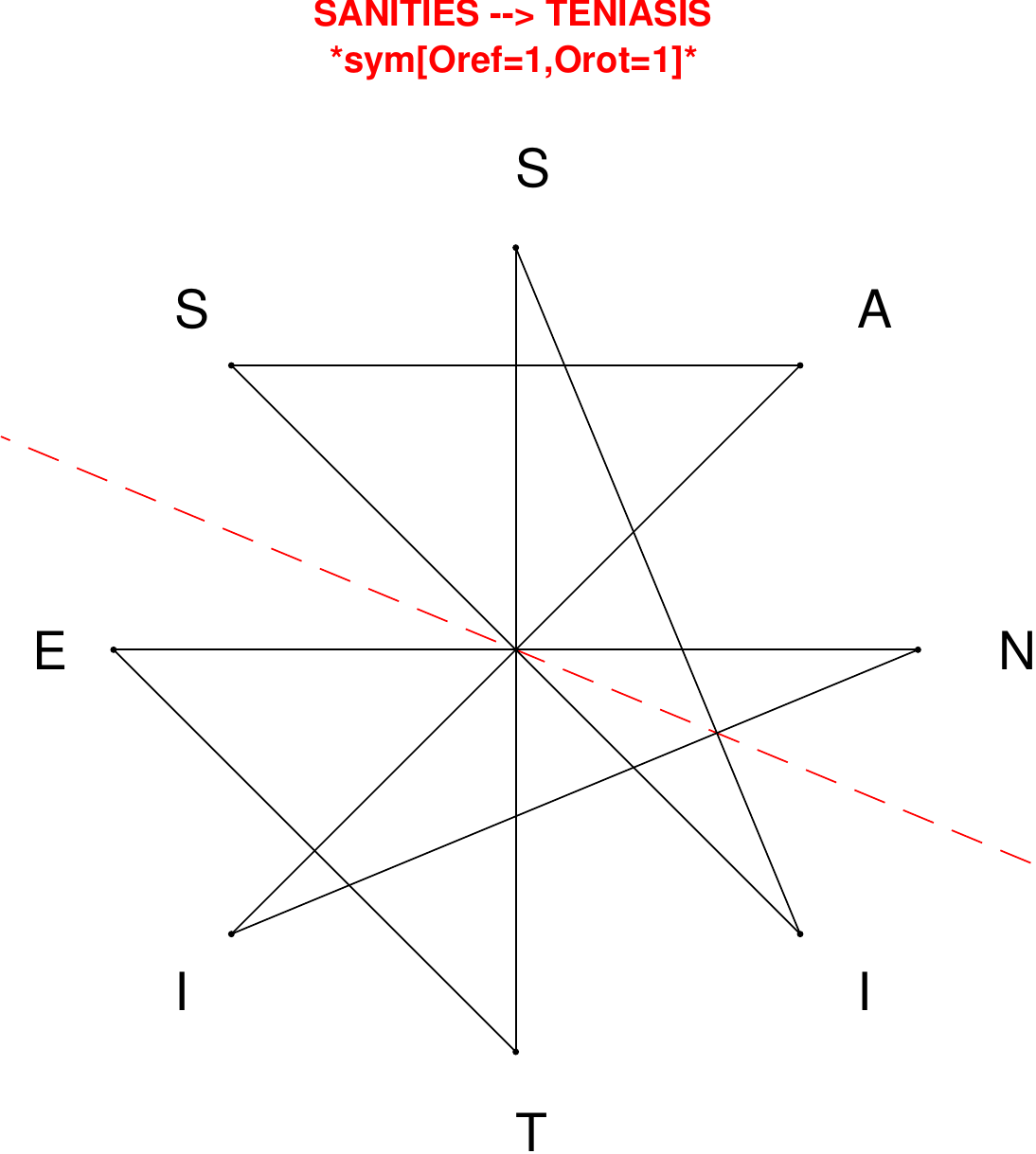}
\end{subfigure}
\hfill
\begin{subfigure}[T]{0.19\textwidth}
\centering
\includegraphics[width=\textwidth]{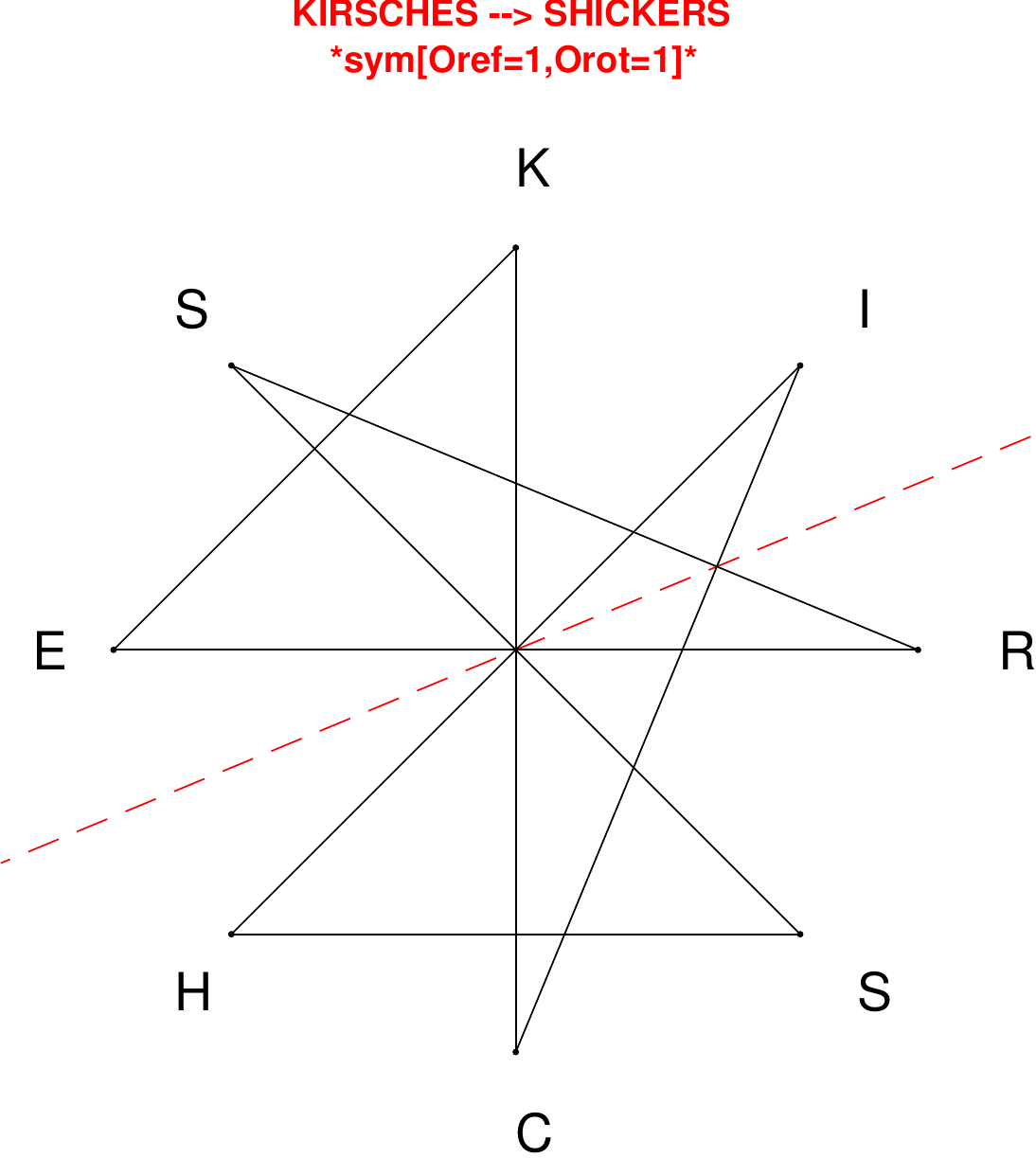}
\end{subfigure}
\hfill
\begin{subfigure}[T]{0.19\textwidth}
\centering
\includegraphics[width=\textwidth]{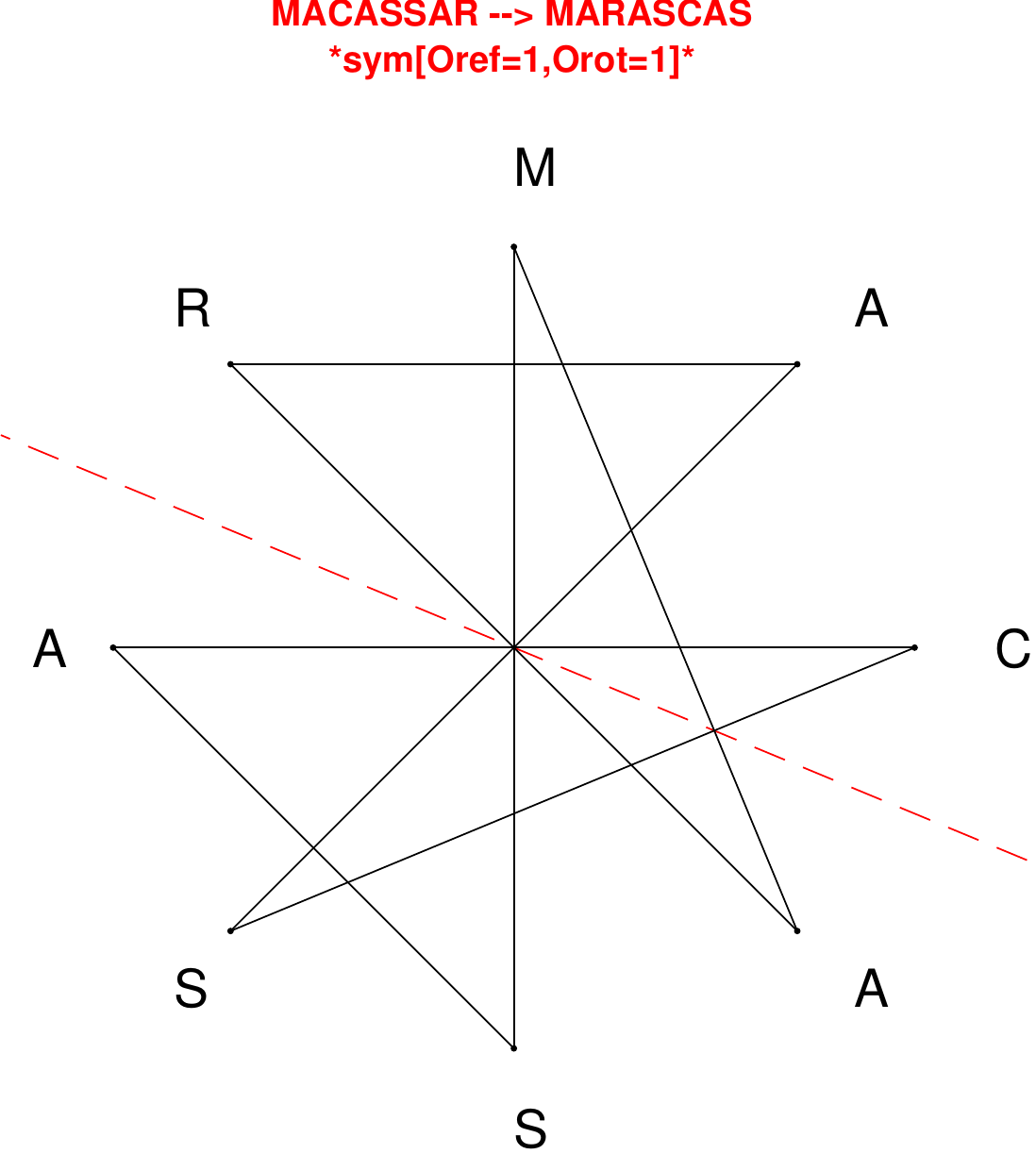}
\end{subfigure}
\hfill
\begin{subfigure}[T]{0.19\textwidth}
\centering
\includegraphics[width=\textwidth]{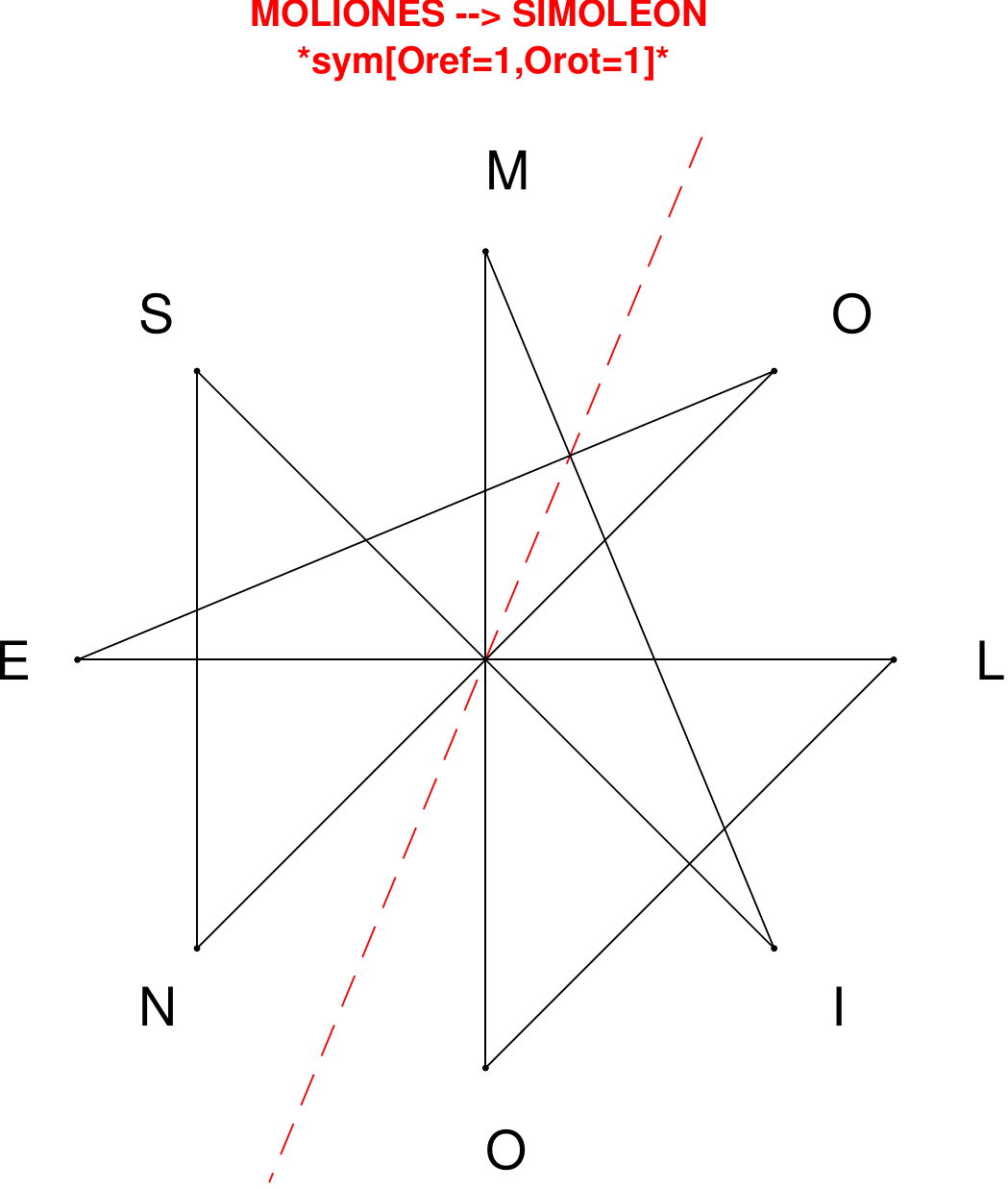}
\end{subfigure}
\hfill
\begin{subfigure}[T]{0.19\textwidth}
\centering
\includegraphics[width=\textwidth]{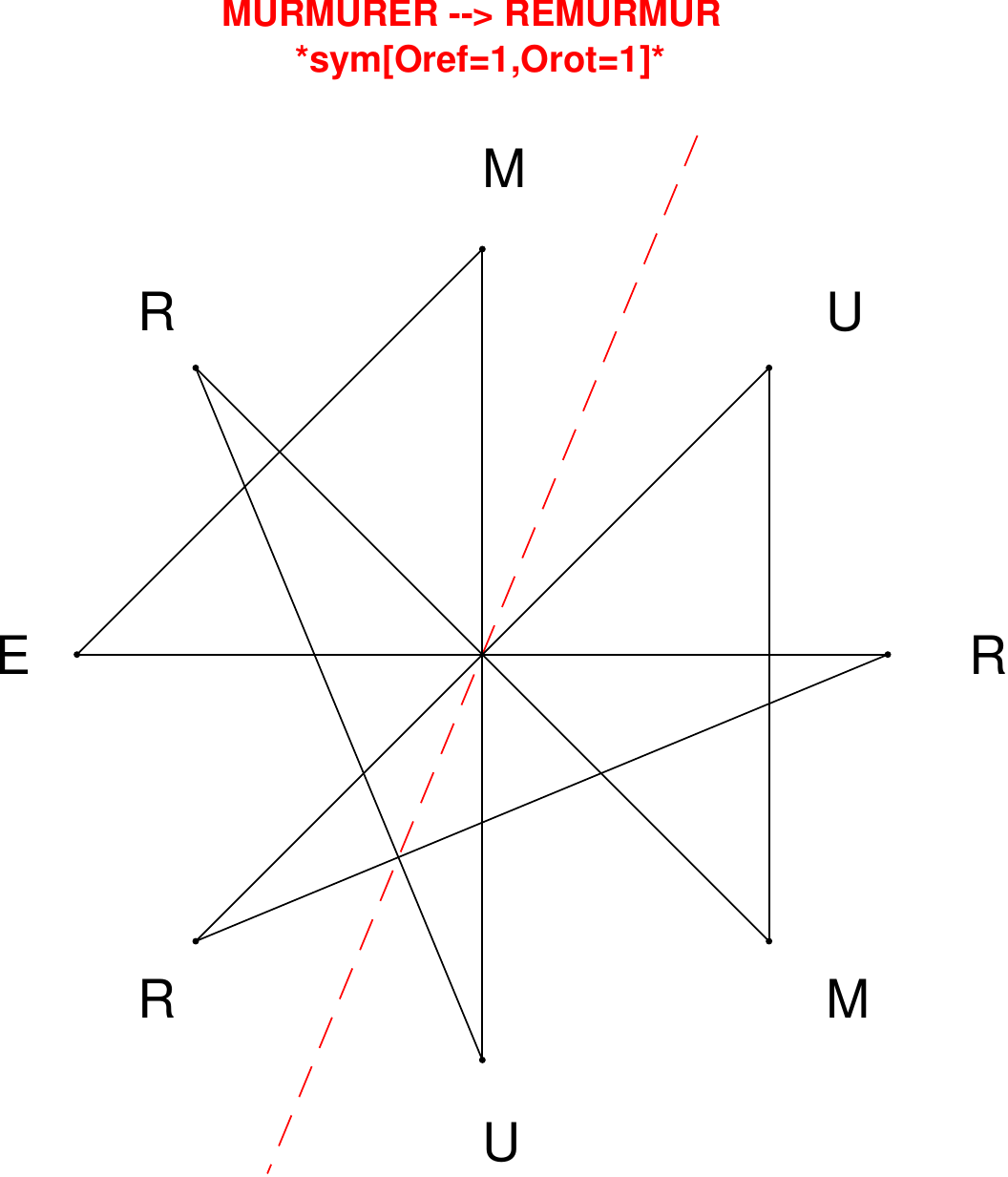}
\end{subfigure}
\end{figure}

\begin{figure}[H]
\centering
\begin{subfigure}[T]{0.19\textwidth}
\centering
\includegraphics[width=\textwidth]{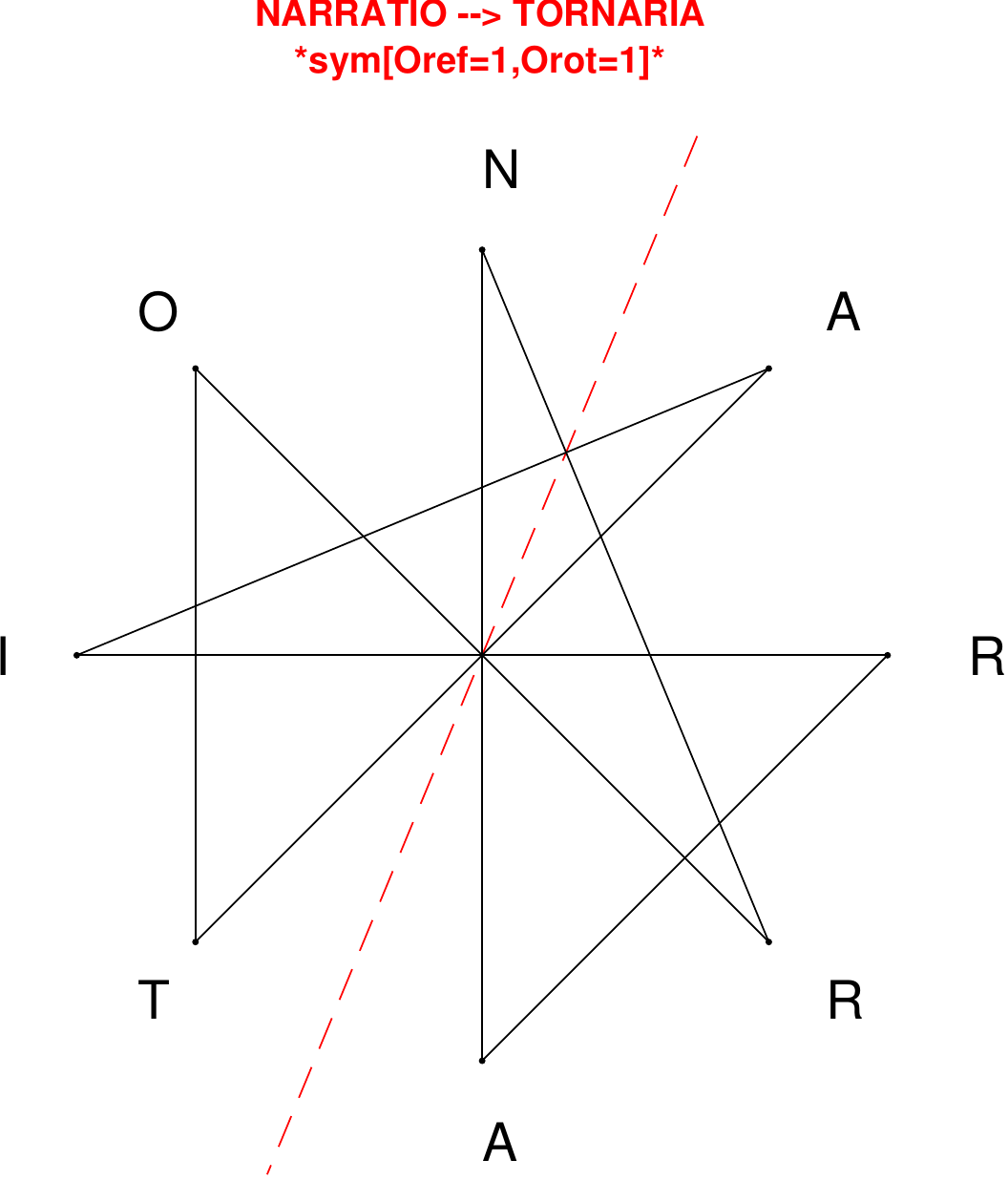}
\end{subfigure}
\hfill
\begin{subfigure}[T]{0.19\textwidth}
\centering
\includegraphics[width=\textwidth]{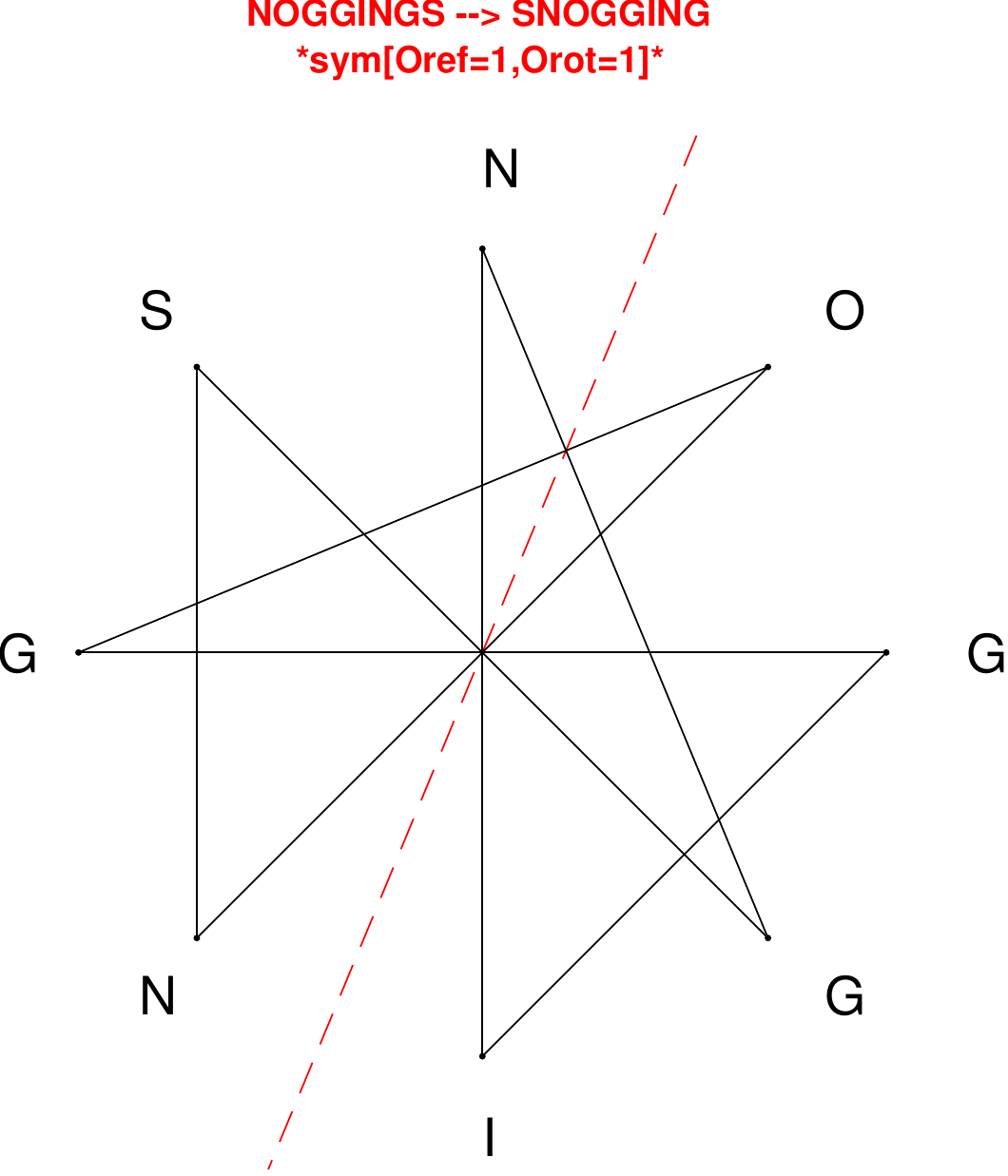}
\end{subfigure}
\hfill
\begin{subfigure}[T]{0.19\textwidth}
\centering
\includegraphics[width=\textwidth]{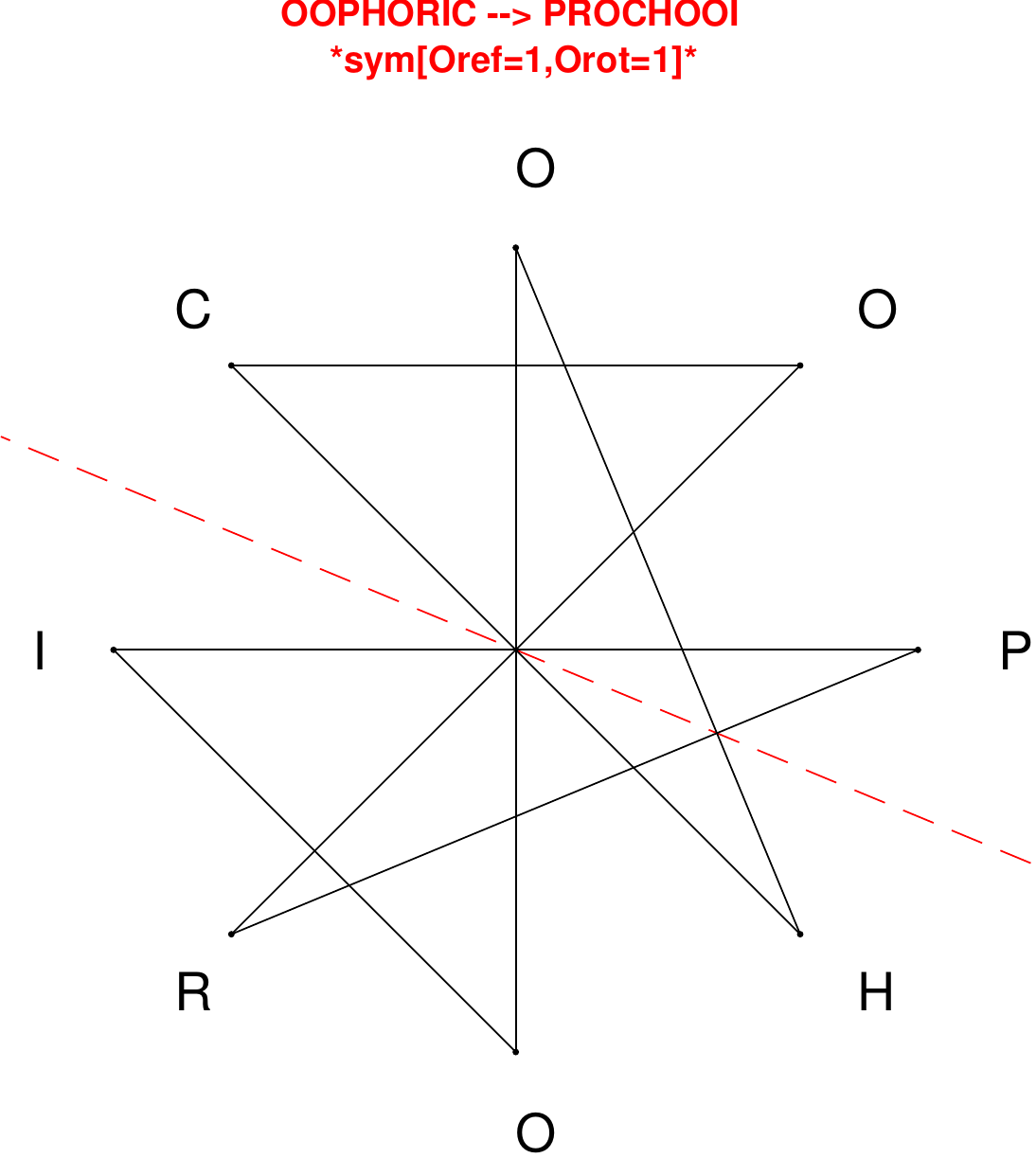}
\end{subfigure}
\hfill
\begin{subfigure}[T]{0.19\textwidth}
\centering
\includegraphics[width=\textwidth]{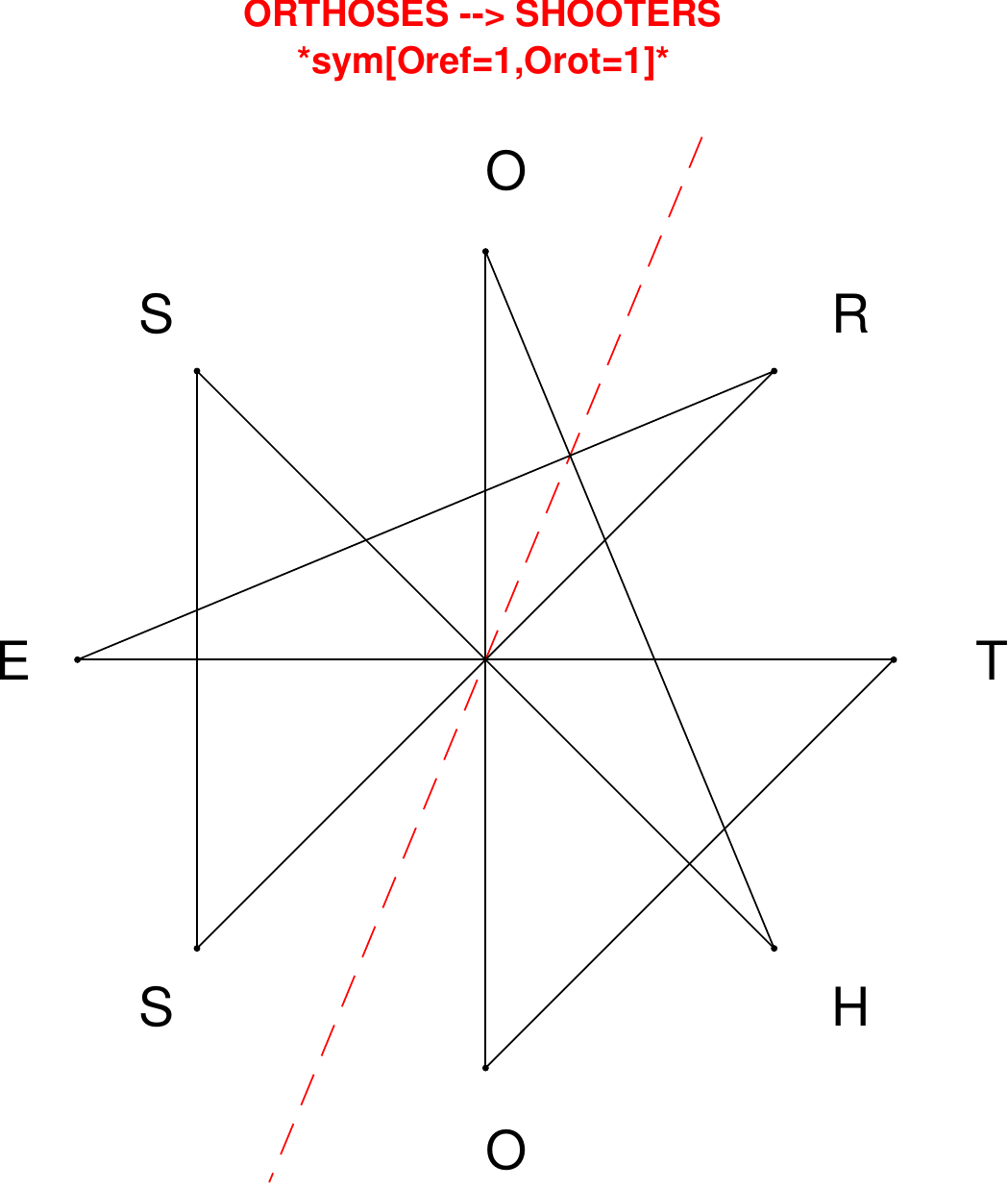}
\end{subfigure}
\hfill
\begin{subfigure}[T]{0.19\textwidth}
\centering
\includegraphics[width=\textwidth]{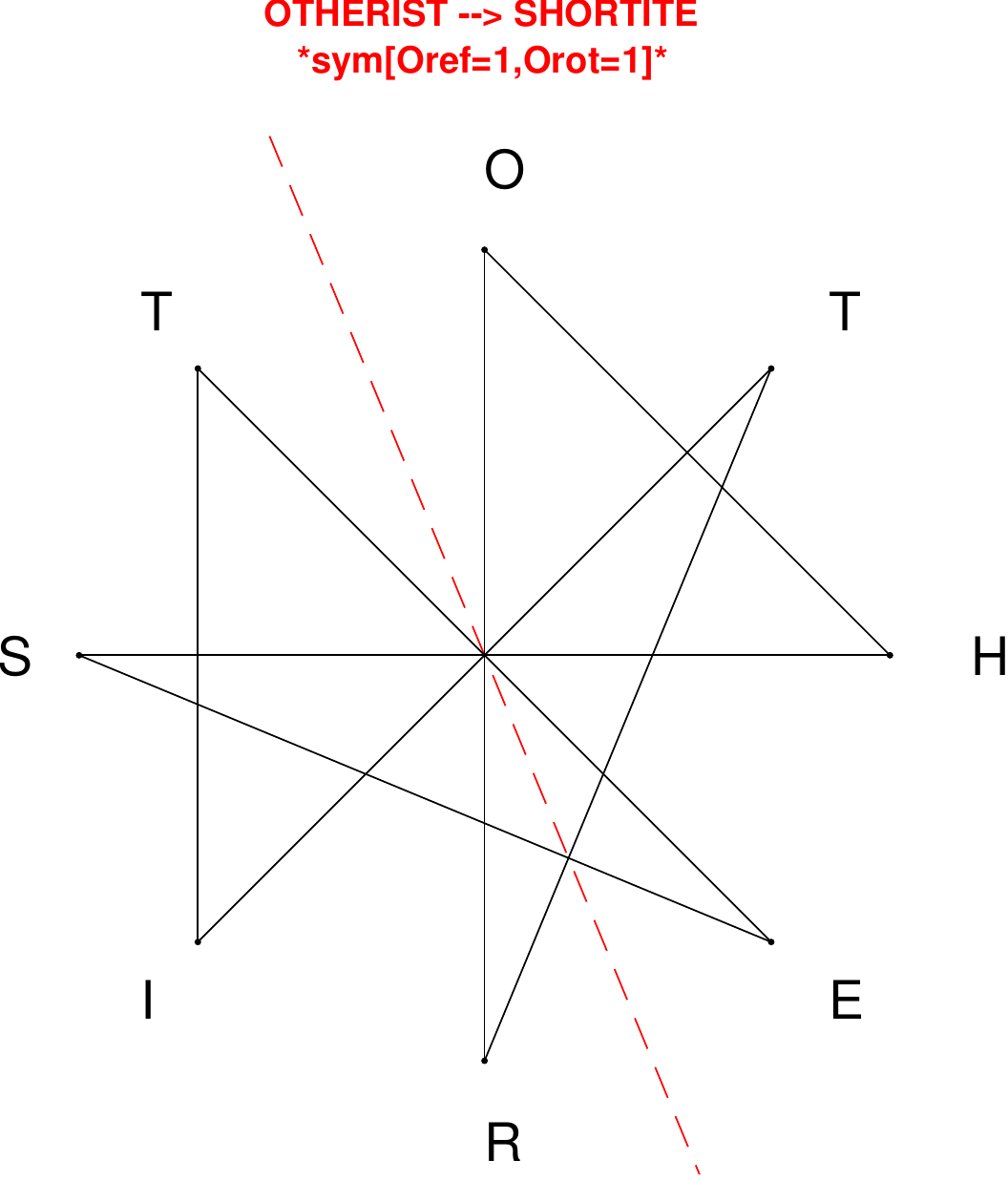}
\end{subfigure}
\end{figure}

\begin{figure}[H]
\centering
\begin{subfigure}[T]{0.19\textwidth}
\centering
\includegraphics[width=\textwidth]{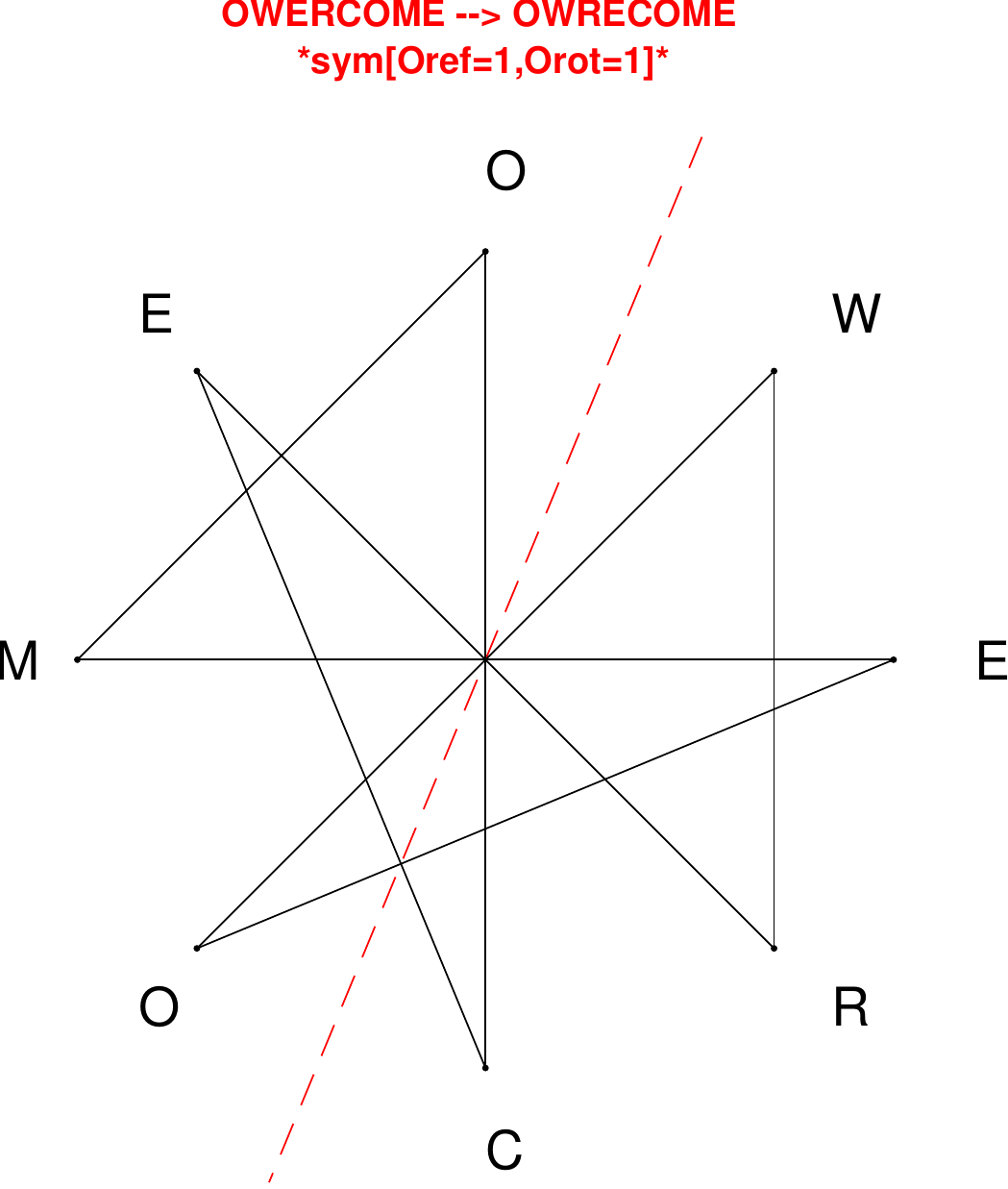}
\end{subfigure}
\hfill
\begin{subfigure}[T]{0.19\textwidth}
\centering
\includegraphics[width=\textwidth]{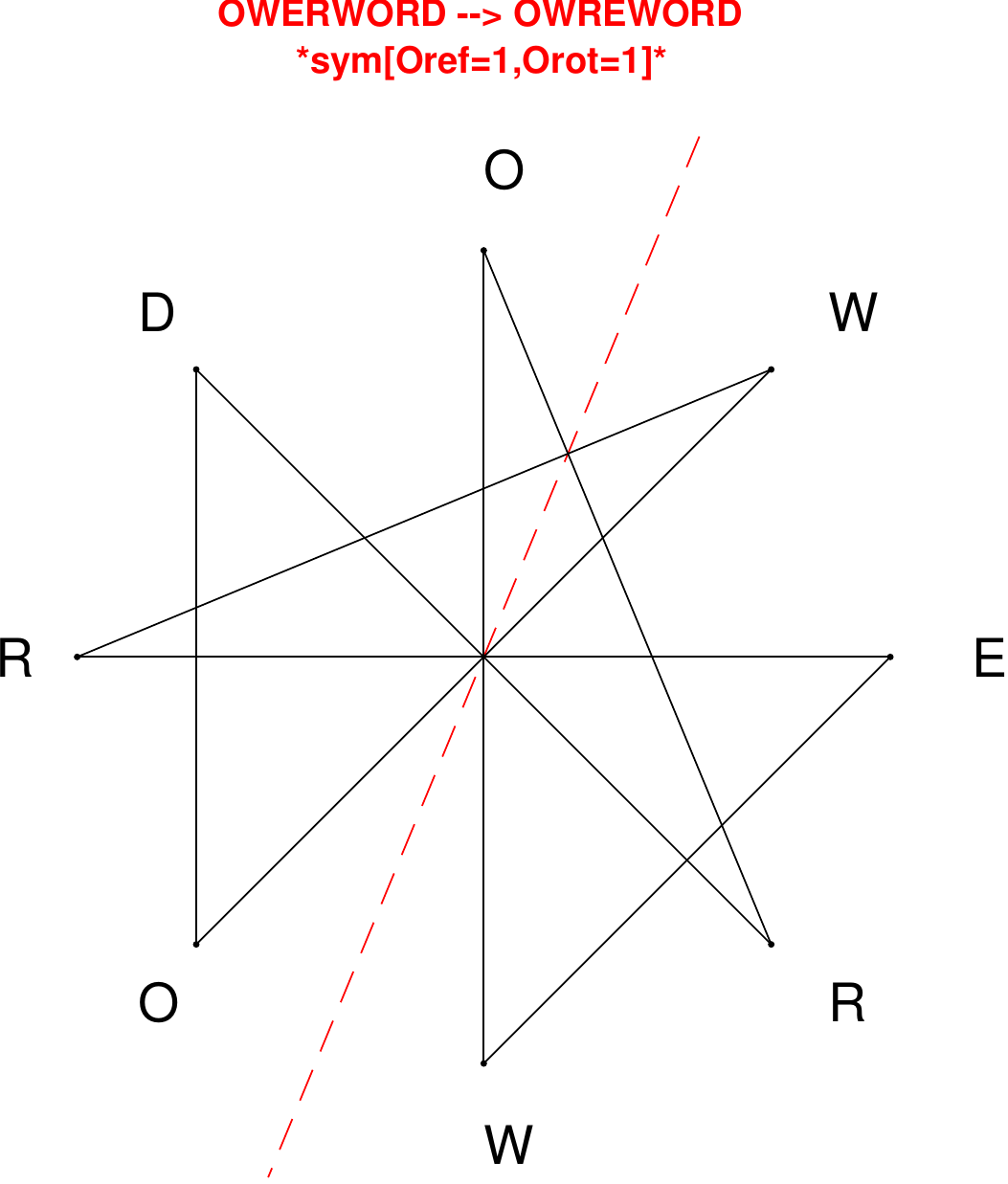}
\end{subfigure}
\hfill
\begin{subfigure}[T]{0.19\textwidth}
\centering
\includegraphics[width=\textwidth]{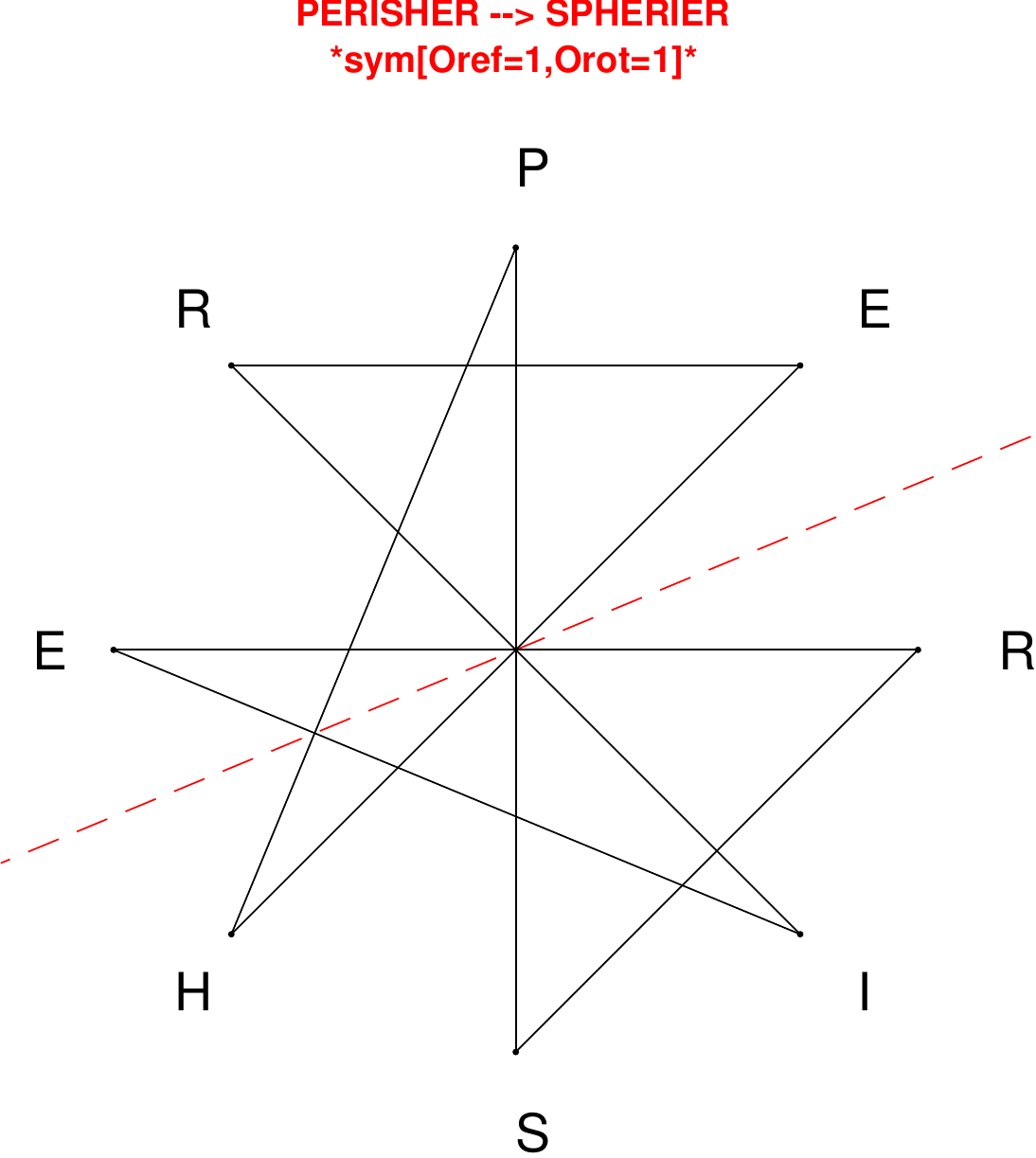}
\end{subfigure}
\hfill
\begin{subfigure}[T]{0.19\textwidth}
\centering
\includegraphics[width=\textwidth]{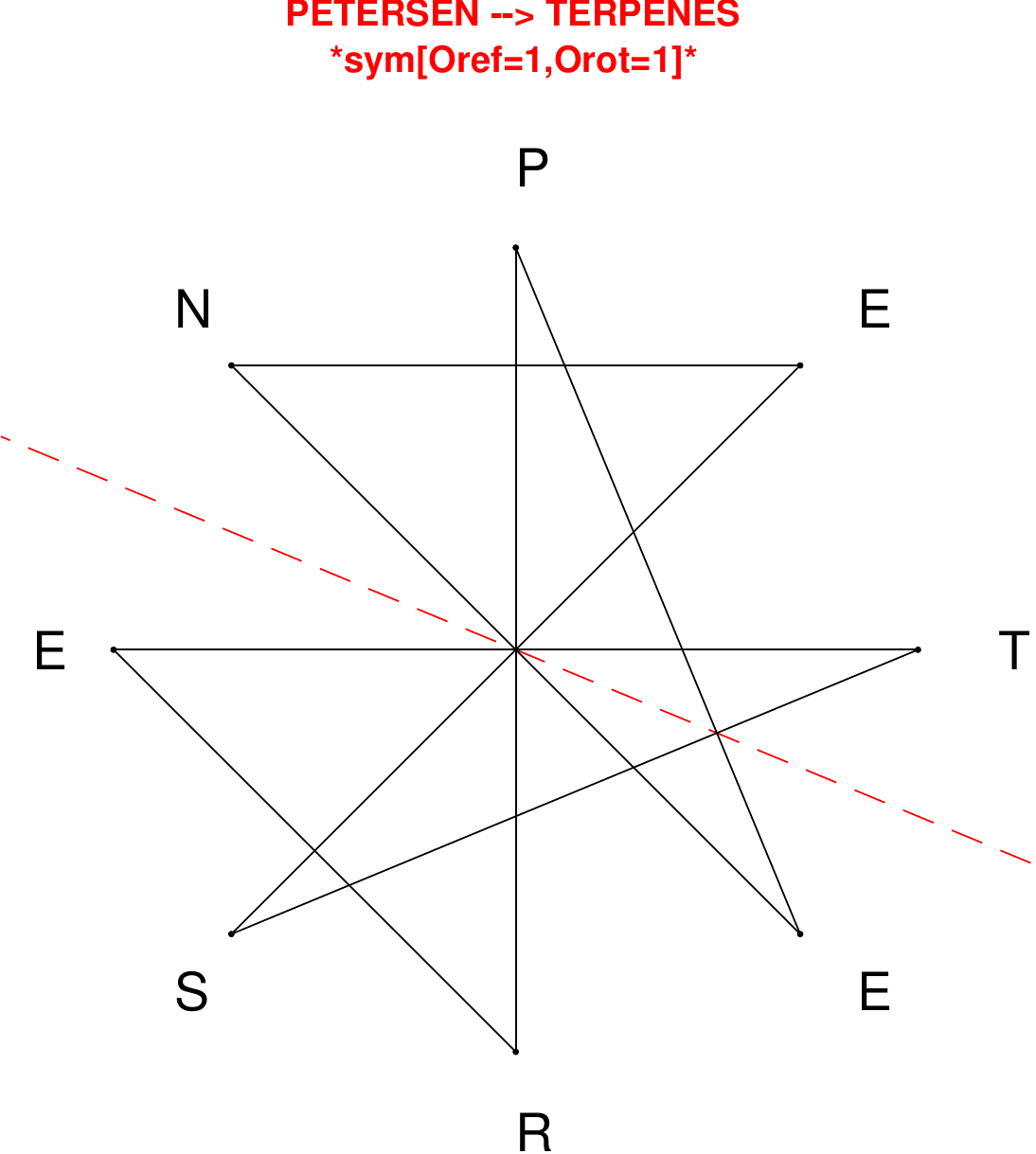}
\end{subfigure}
\hfill
\begin{subfigure}[T]{0.19\textwidth}
\centering
\includegraphics[width=\textwidth]{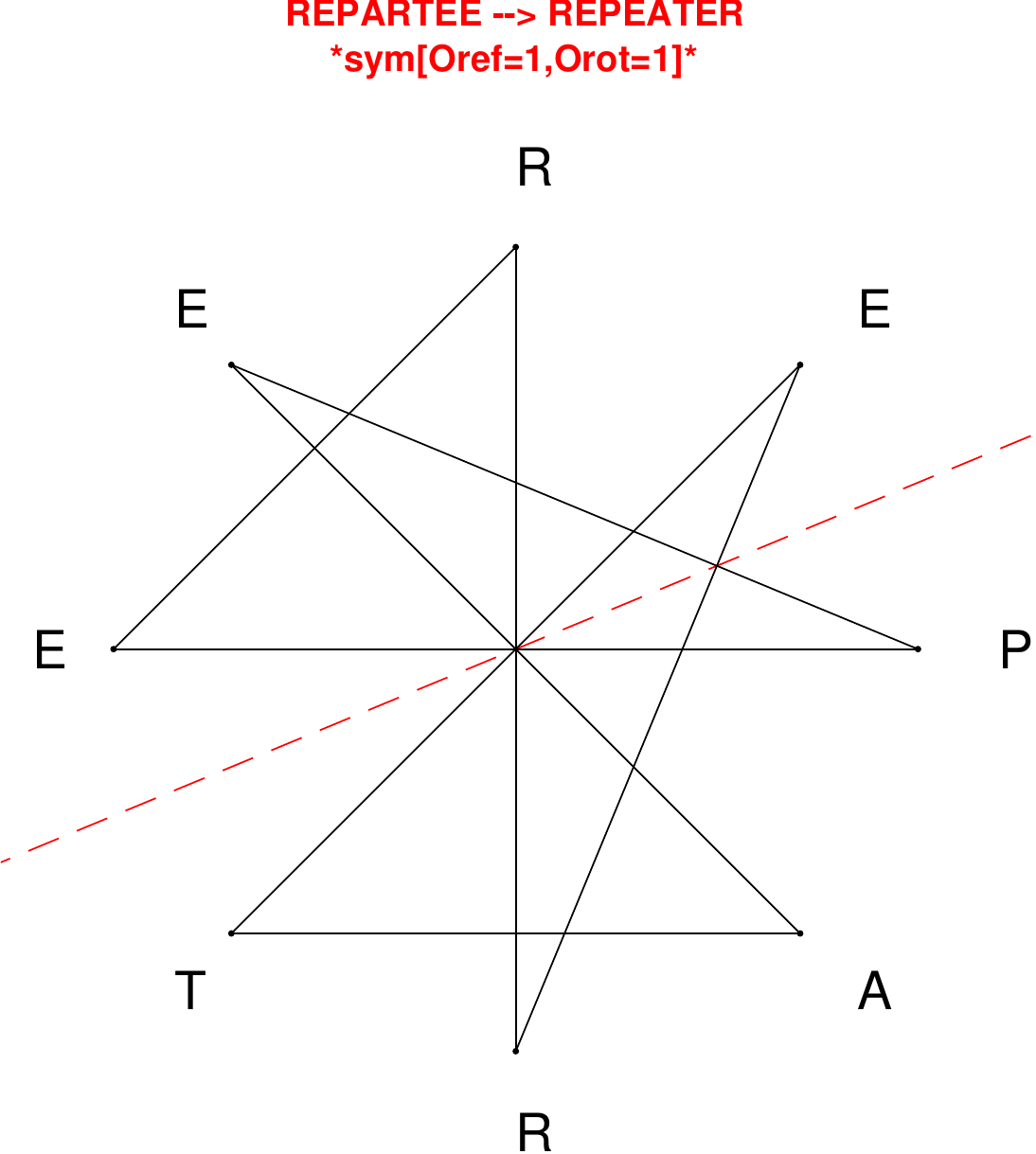}
\end{subfigure}
\end{figure}

\begin{figure}[H]
\centering
\begin{subfigure}[T]{0.19\textwidth}
\centering
\includegraphics[width=\textwidth]{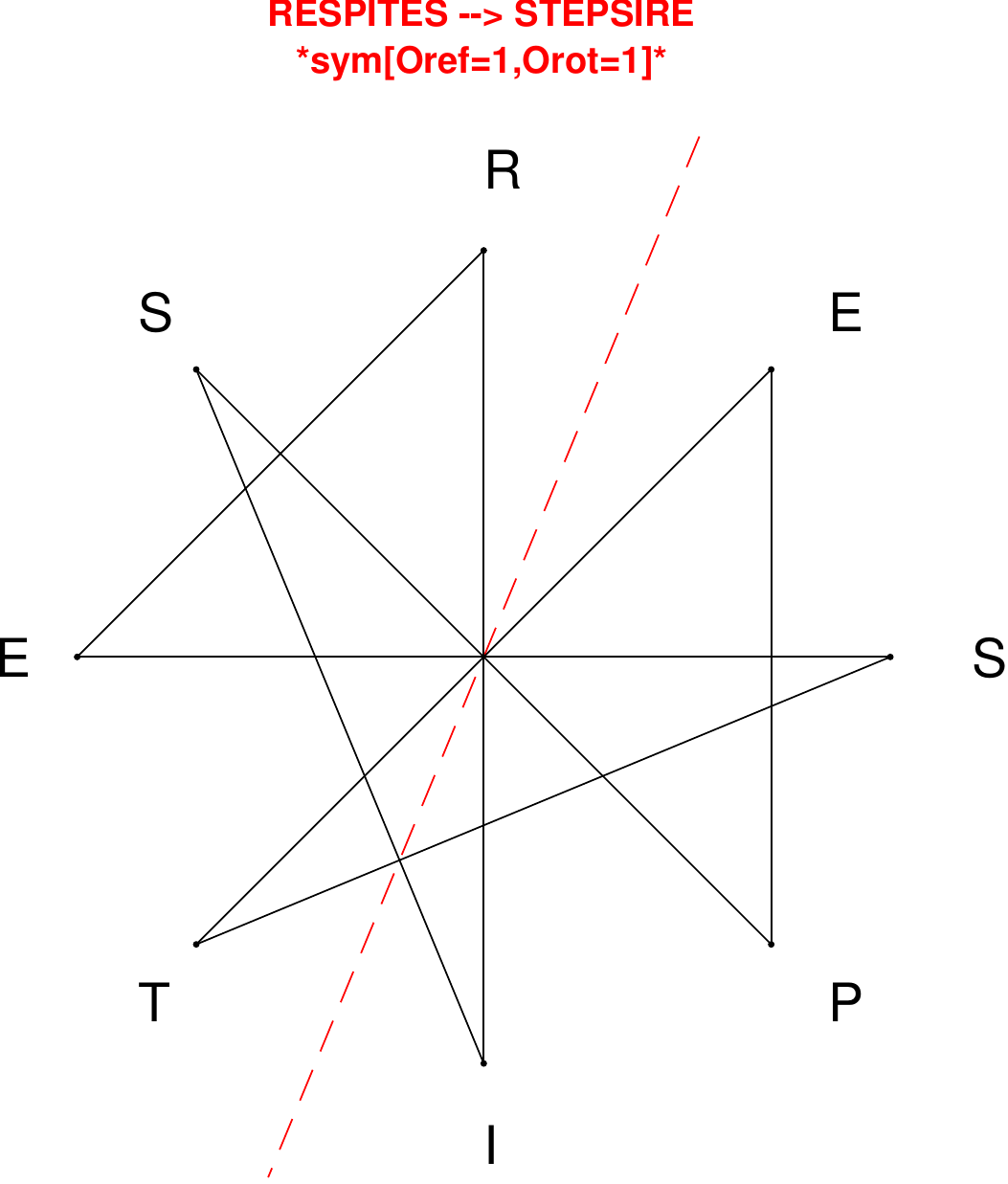}
\end{subfigure}
\hfill
\begin{subfigure}[T]{0.19\textwidth}
\centering
\includegraphics[width=\textwidth]{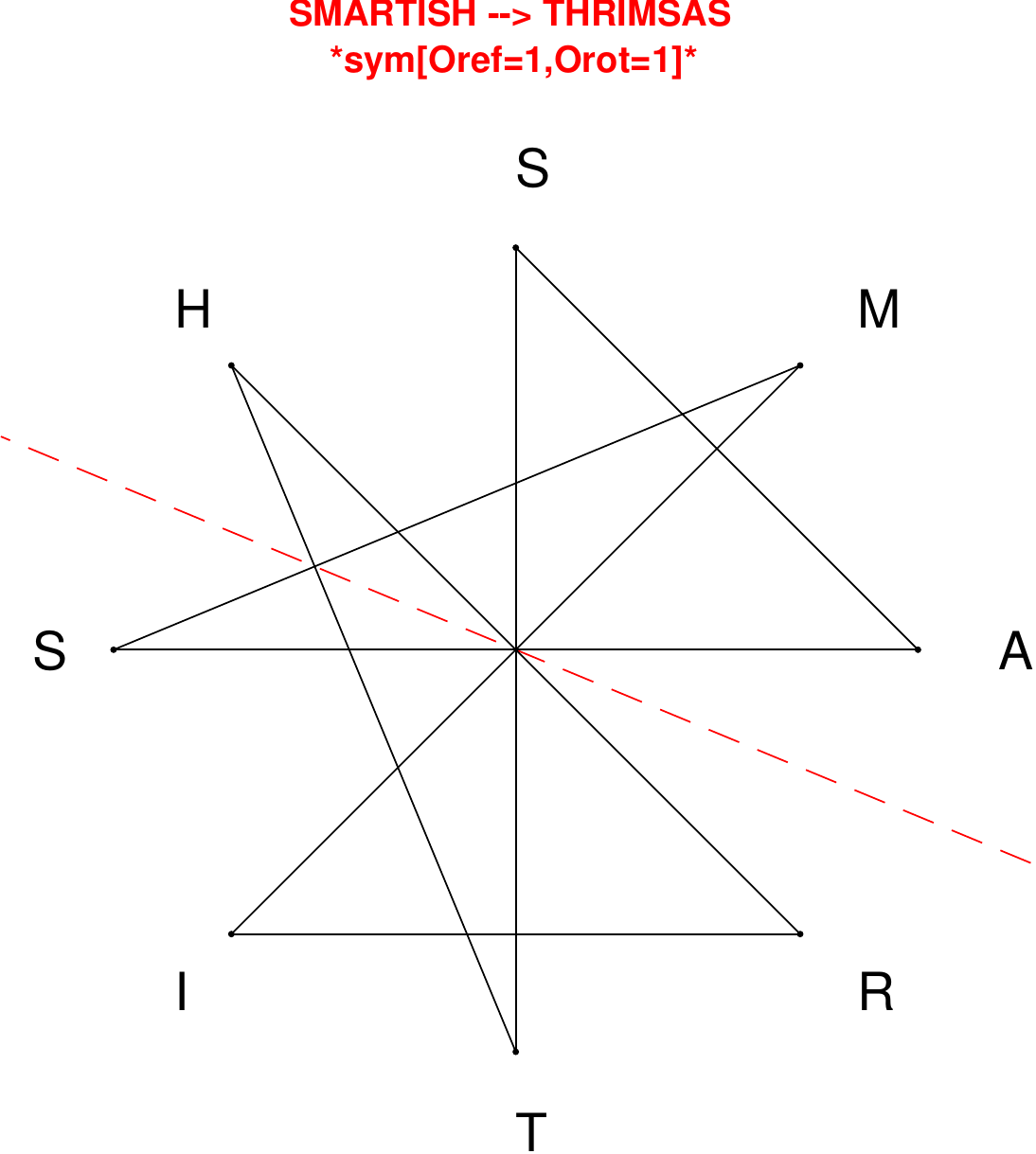}
\end{subfigure}
\hfill
\begin{subfigure}[T]{0.19\textwidth}
\centering
\includegraphics[width=\textwidth]{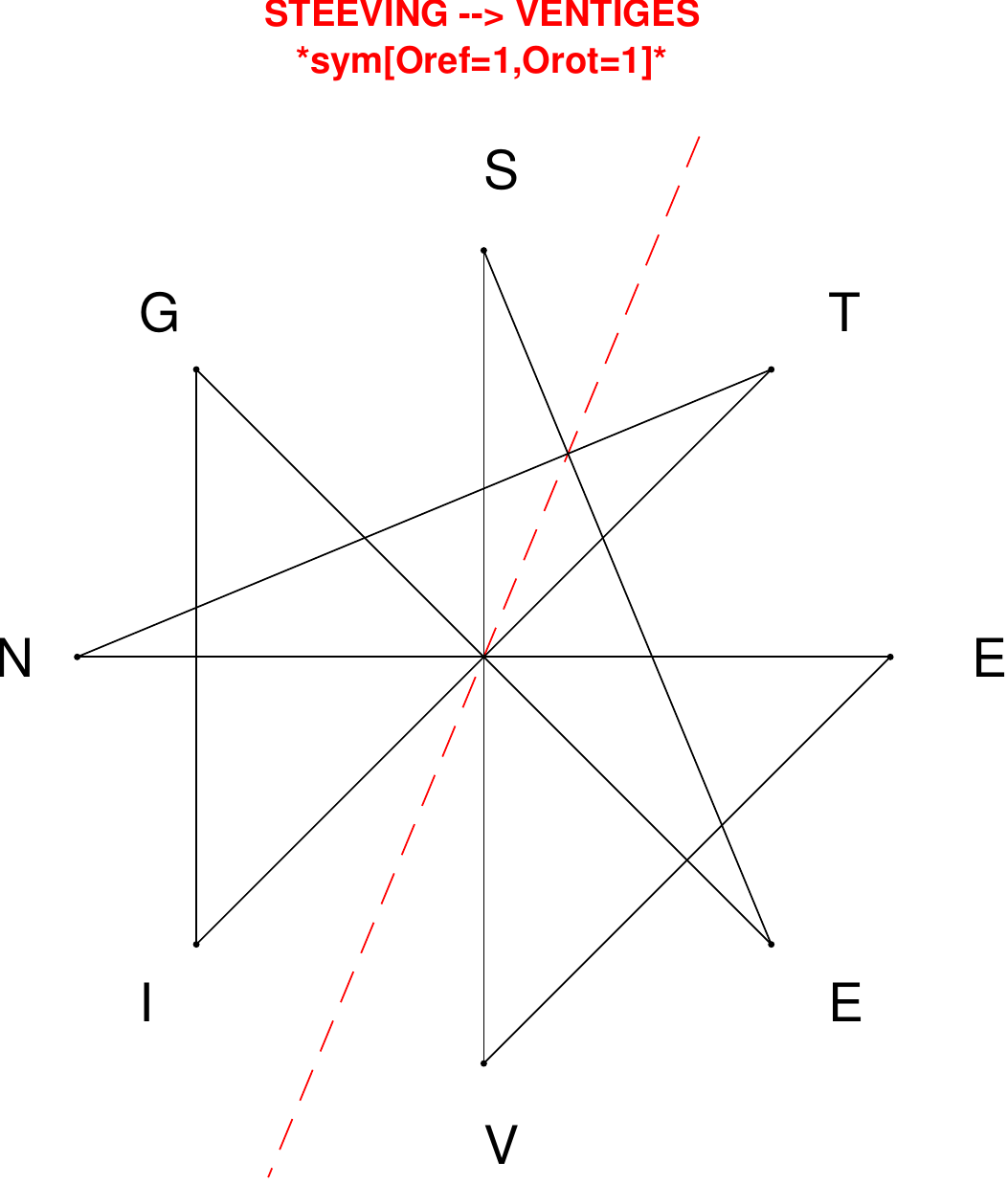}
\end{subfigure}
\hfill
\begin{subfigure}[T]{0.19\textwidth}
\centering
\includegraphics[width=\textwidth]{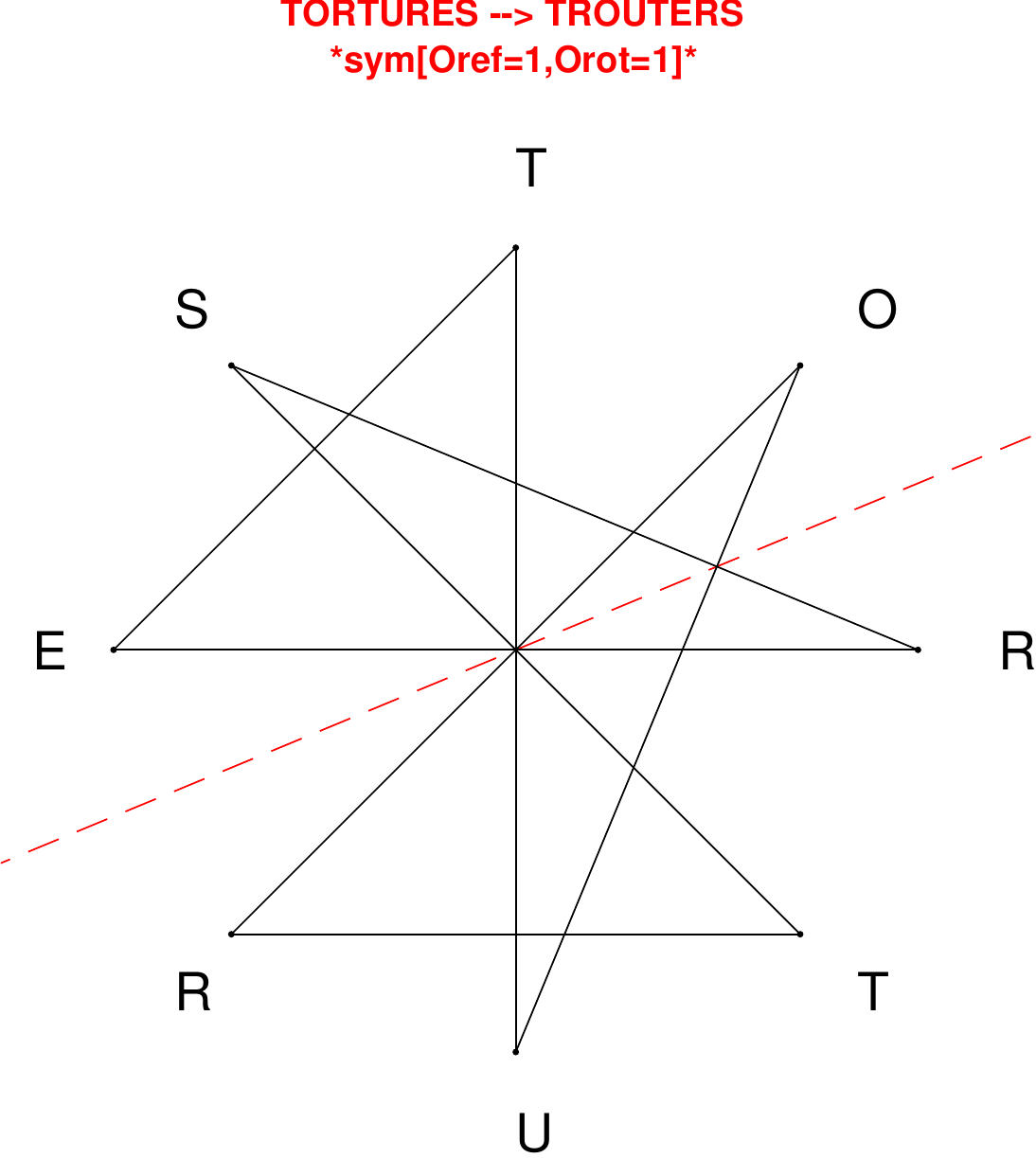}
\end{subfigure}
\hfill
\begin{subfigure}[T]{0.19\textwidth}
\centering
\includegraphics[width=\textwidth]{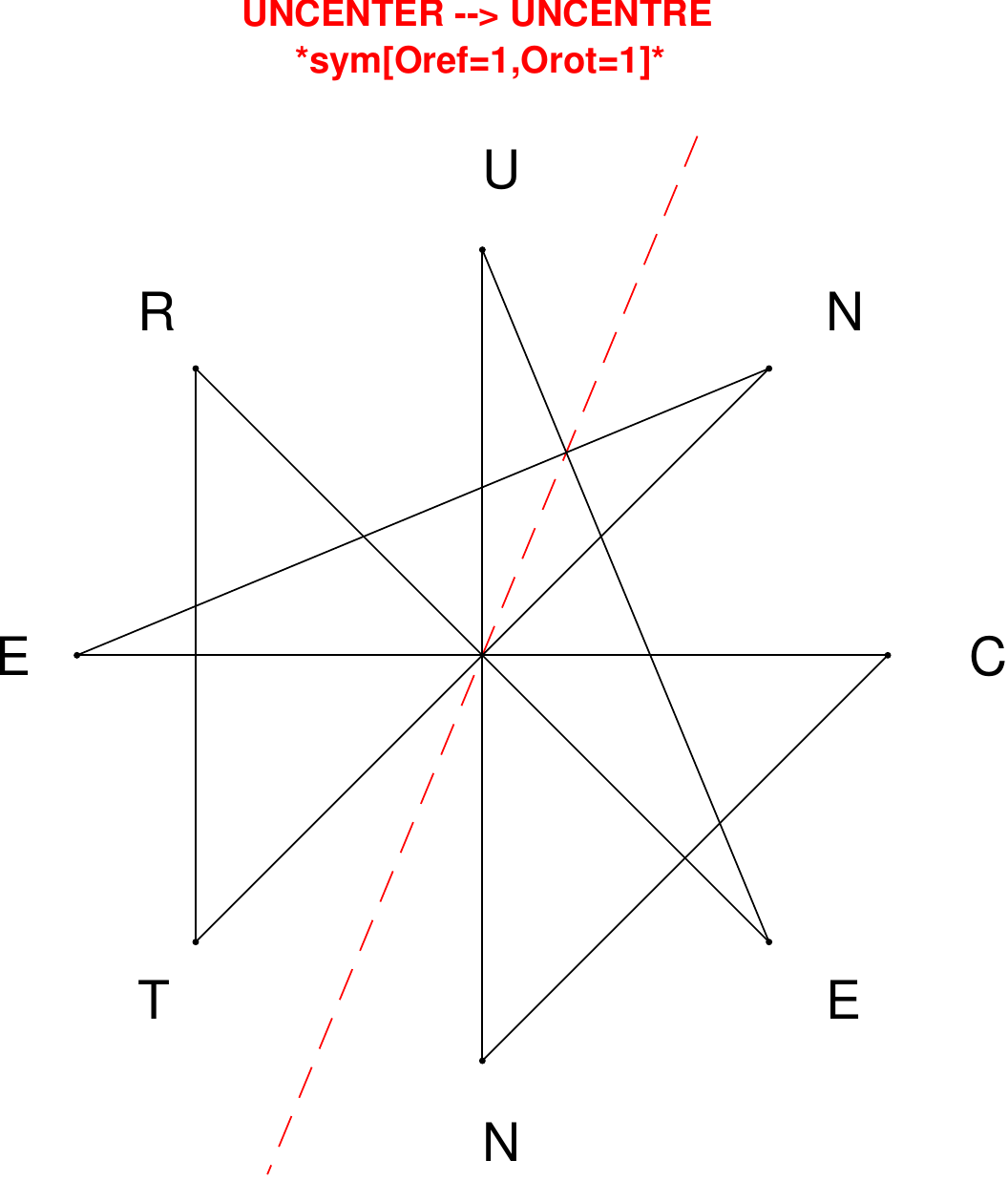}
\end{subfigure}
\end{figure}

\begin{figure}[H]
\centering
\begin{subfigure}[T]{0.19\textwidth}
\centering
\includegraphics[width=\textwidth]{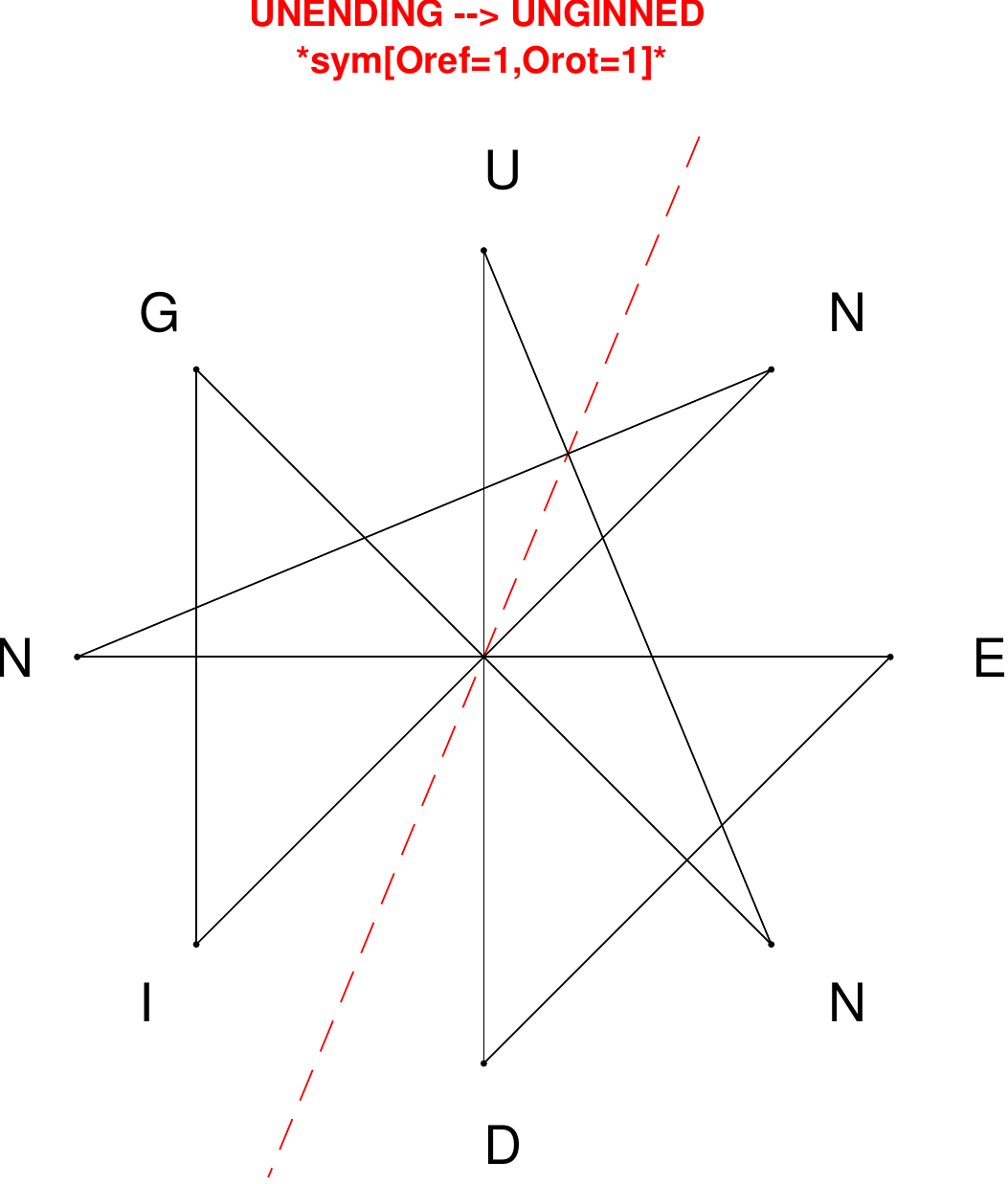}
\end{subfigure}
\hfill
\begin{subfigure}[T]{0.19\textwidth}
\centering
\includegraphics[width=\textwidth]{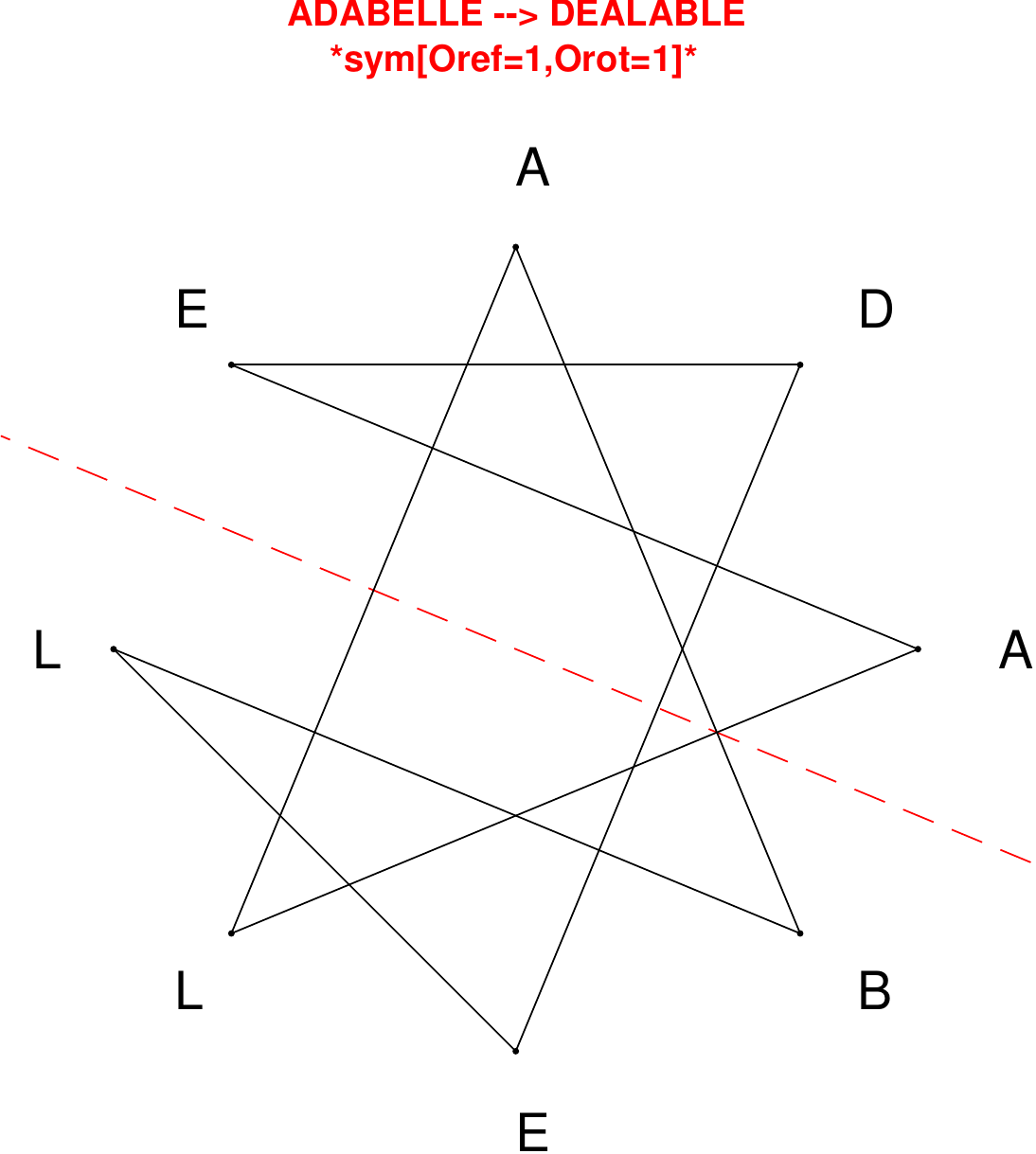}
\end{subfigure}
\hfill
\begin{subfigure}[T]{0.19\textwidth}
\centering
\includegraphics[width=\textwidth]{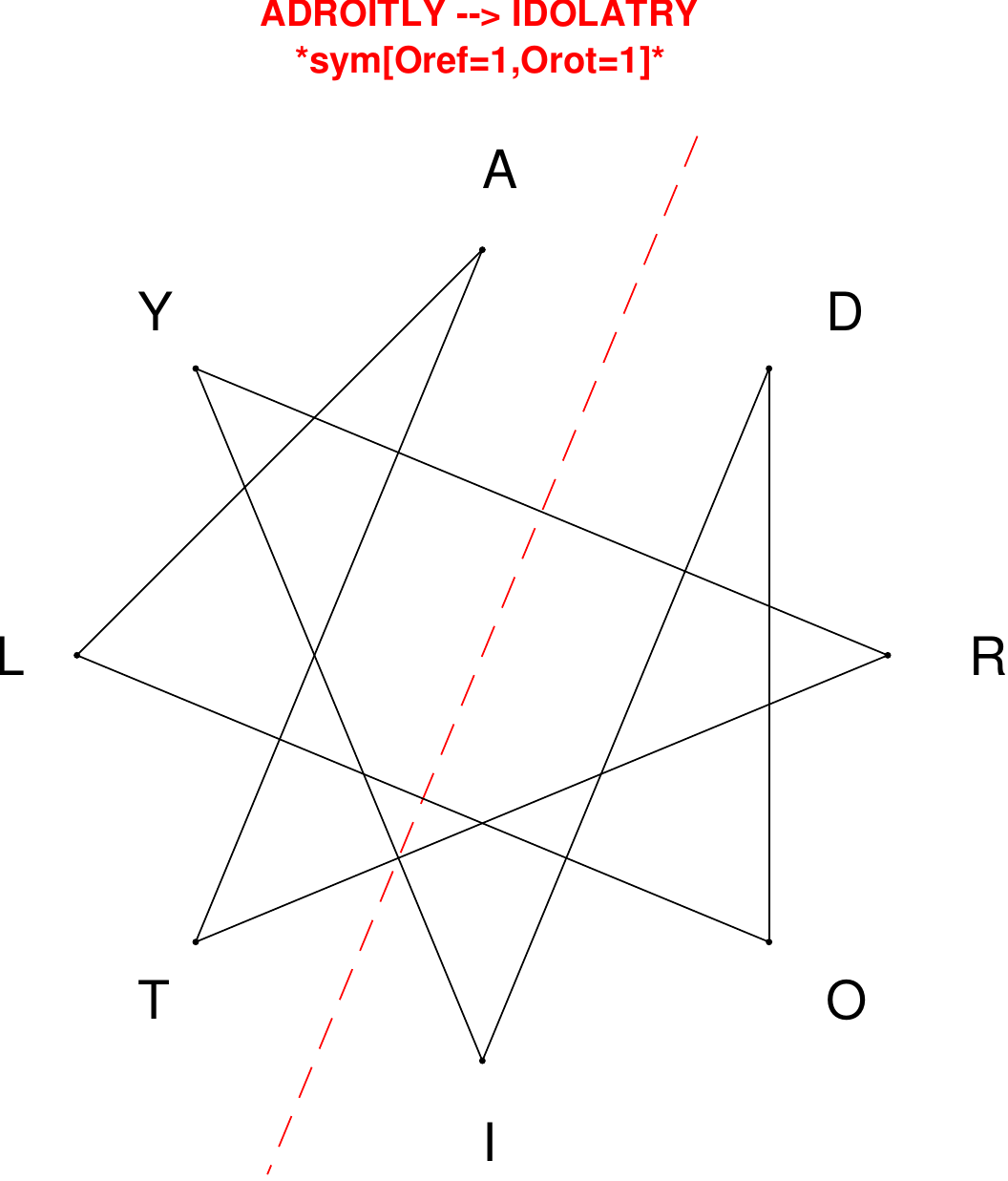}
\end{subfigure}
\hfill
\begin{subfigure}[T]{0.19\textwidth}
\centering
\includegraphics[width=\textwidth]{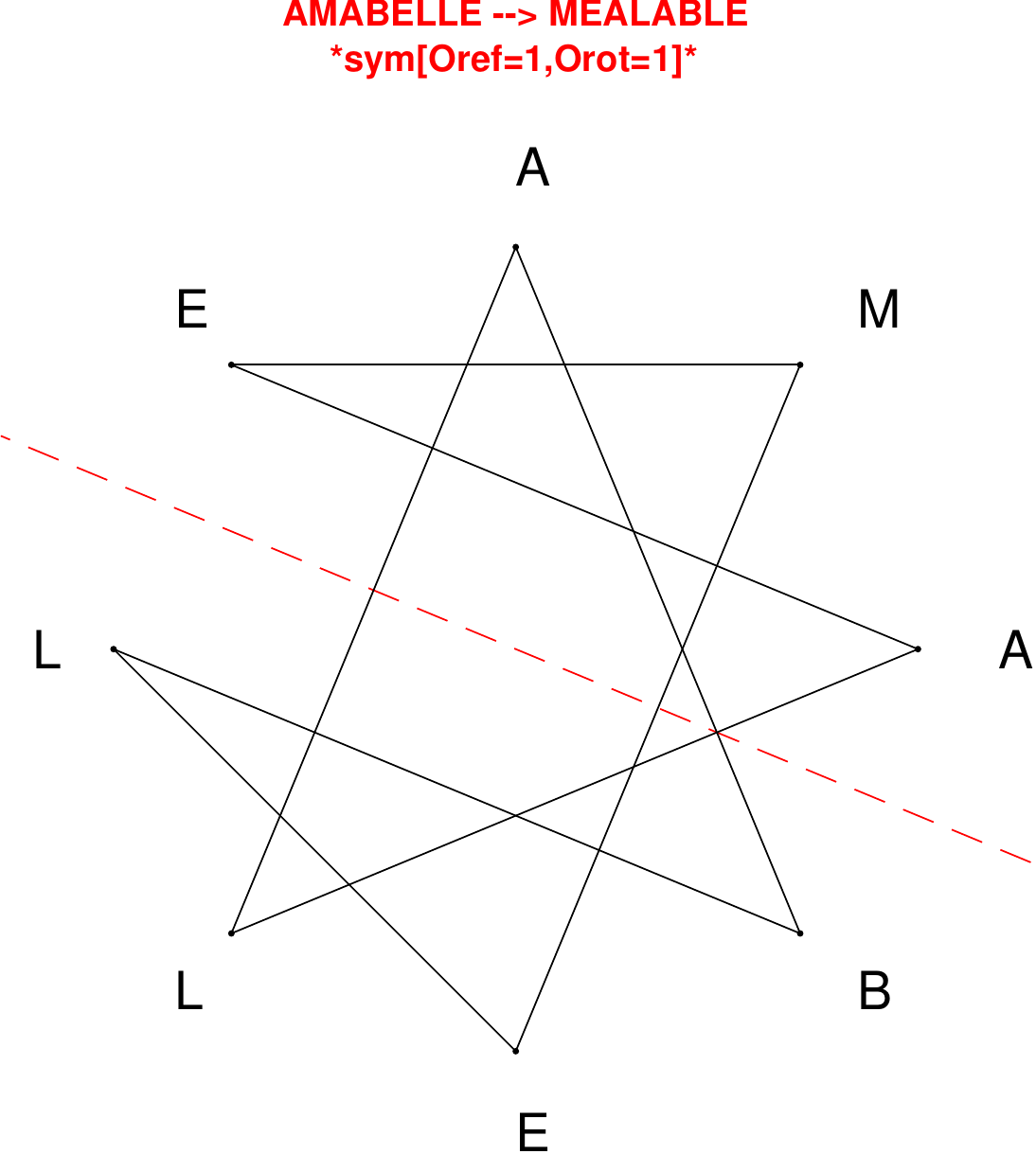}
\end{subfigure}
\hfill
\begin{subfigure}[T]{0.19\textwidth}
\centering
\includegraphics[width=\textwidth]{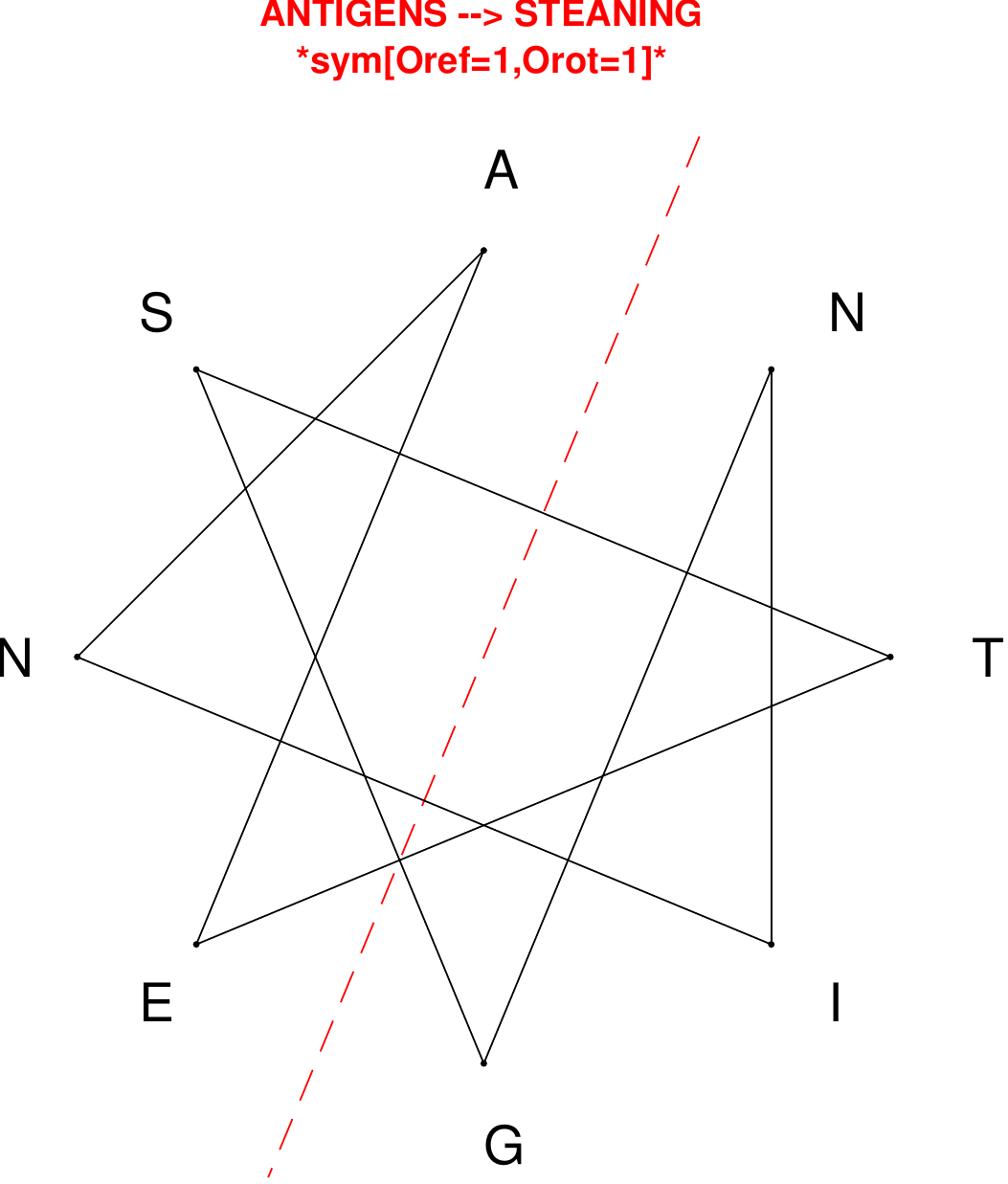}
\end{subfigure}
\end{figure}

\begin{figure}[H]
\centering
\begin{subfigure}[T]{0.19\textwidth}
\centering
\includegraphics[width=\textwidth]{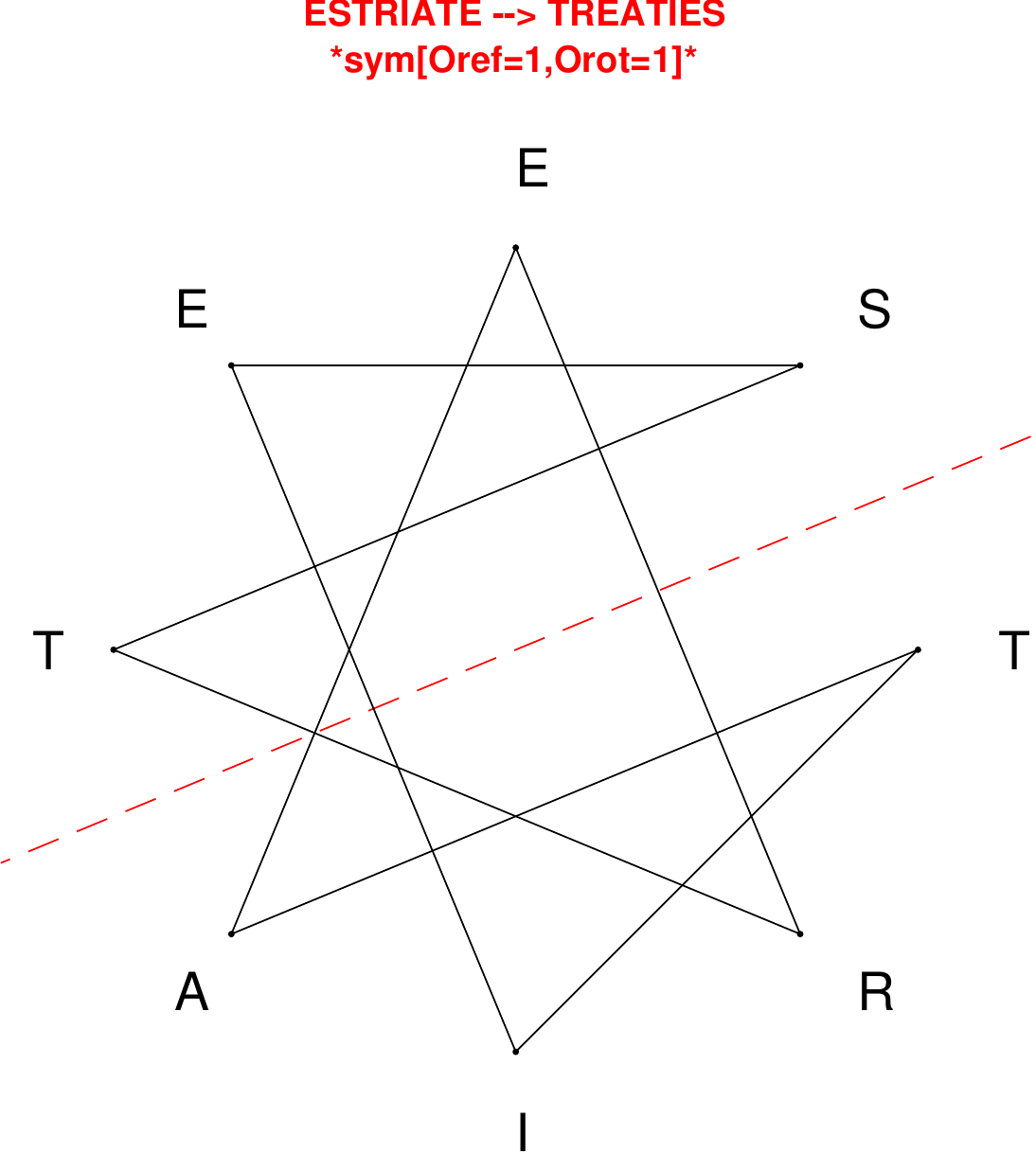}
\end{subfigure}
\hfill
\begin{subfigure}[T]{0.19\textwidth}
\centering
\includegraphics[width=\textwidth]{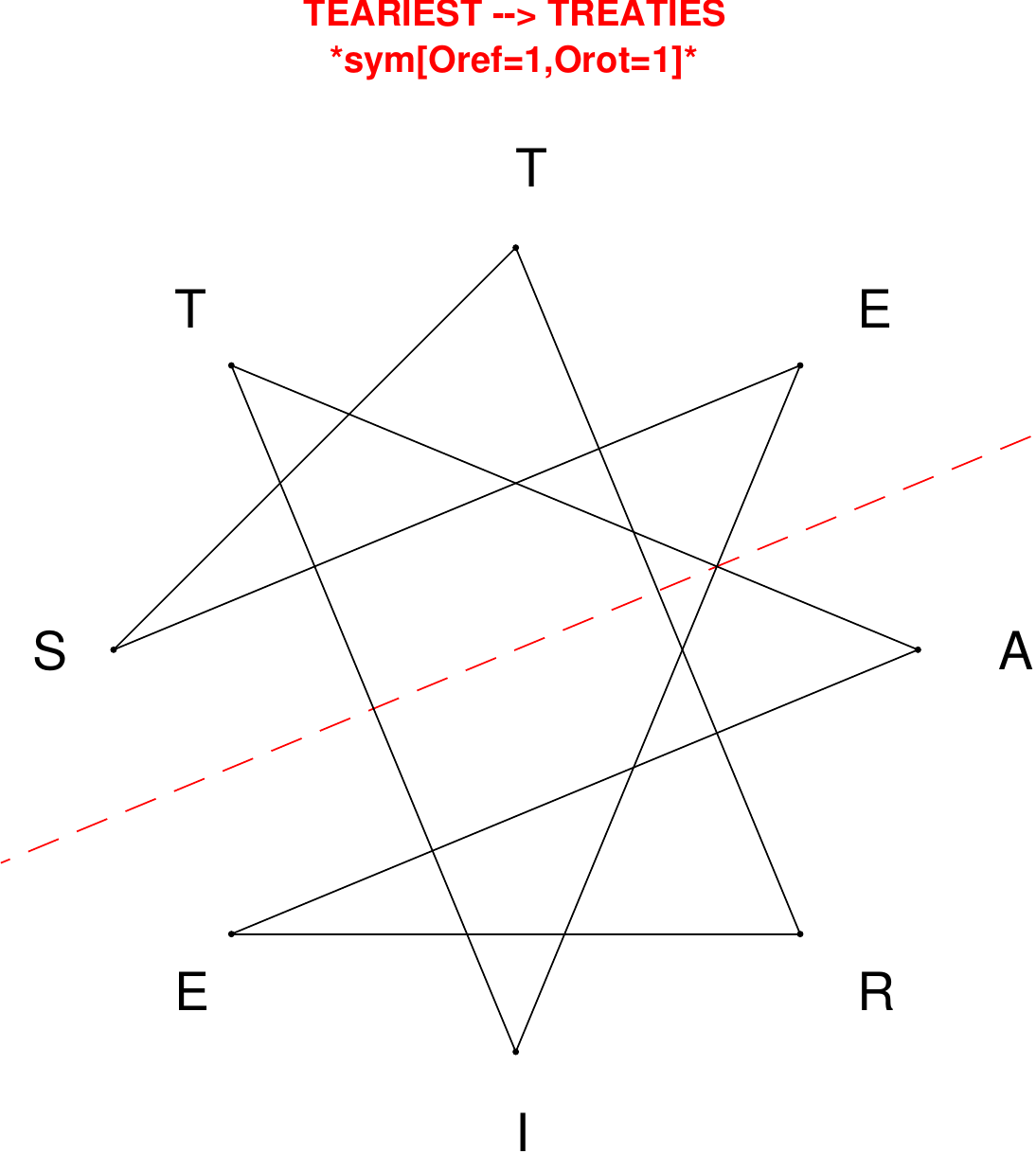}
\end{subfigure}
\hfill
\begin{subfigure}[T]{0.19\textwidth}
\centering
\includegraphics[width=\textwidth]{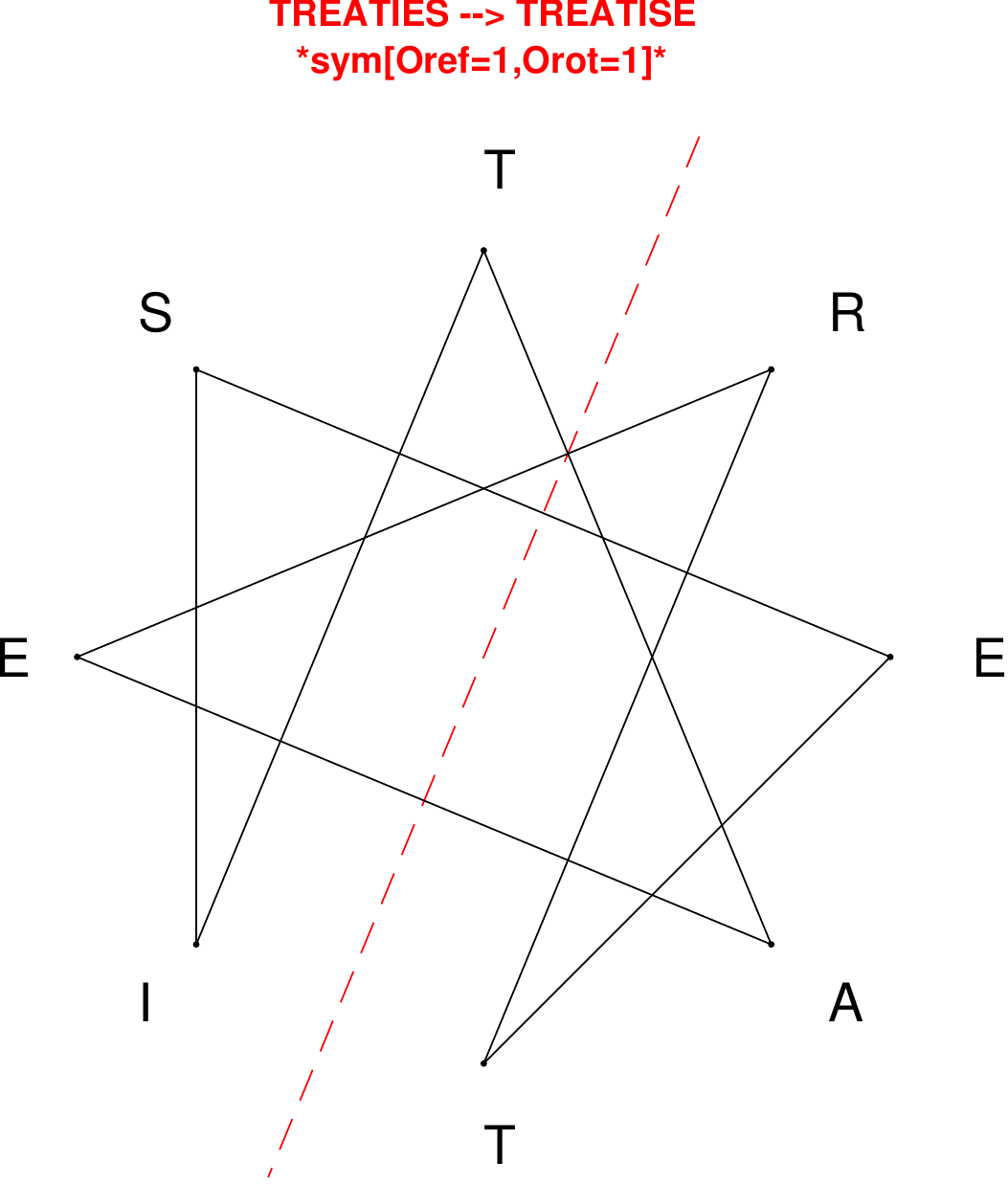}
\end{subfigure}
\hfill
\begin{subfigure}[T]{0.19\textwidth}
\centering
\includegraphics[width=\textwidth]{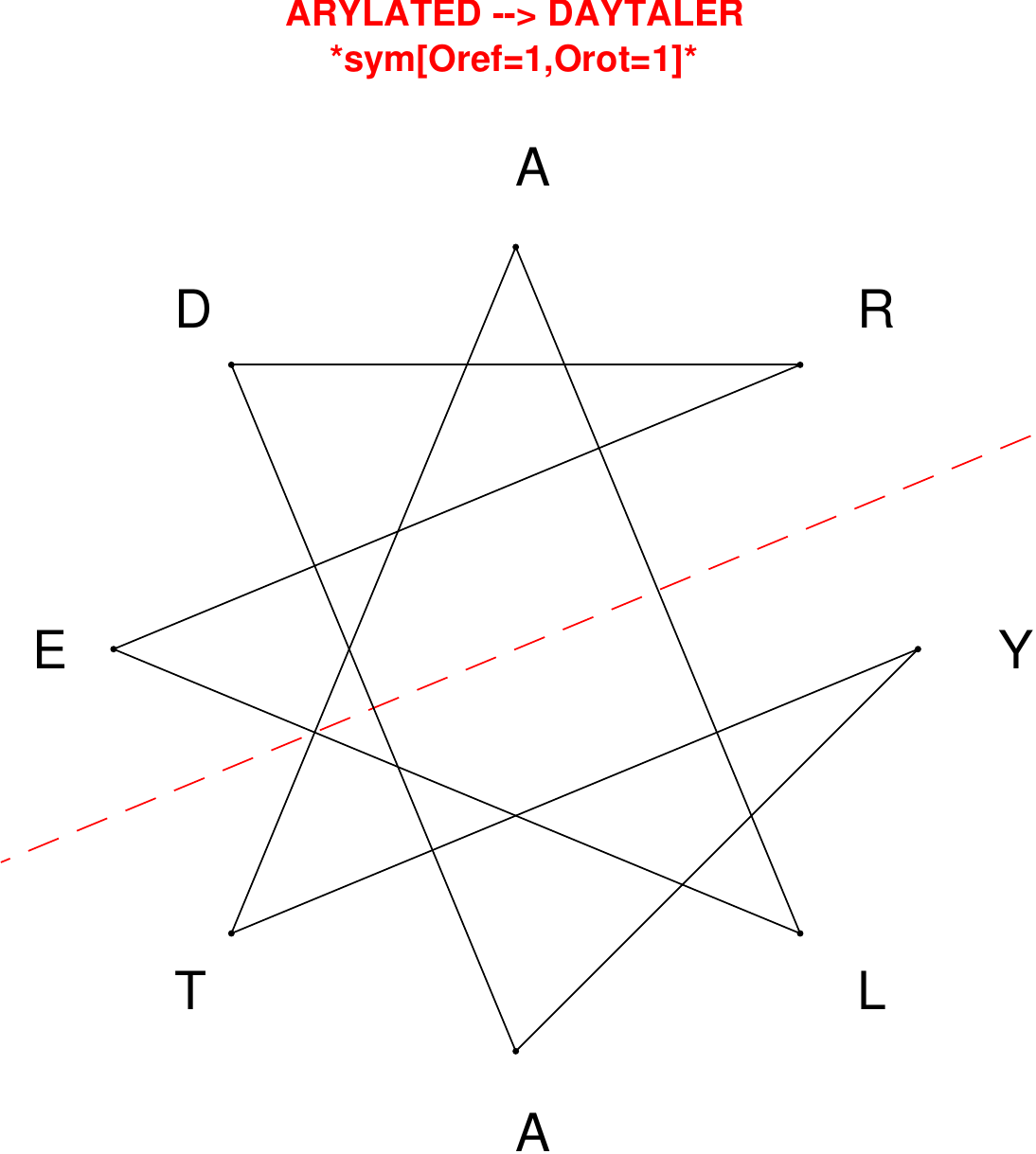}
\end{subfigure}
\hfill
\begin{subfigure}[T]{0.19\textwidth}
\centering
\includegraphics[width=\textwidth]{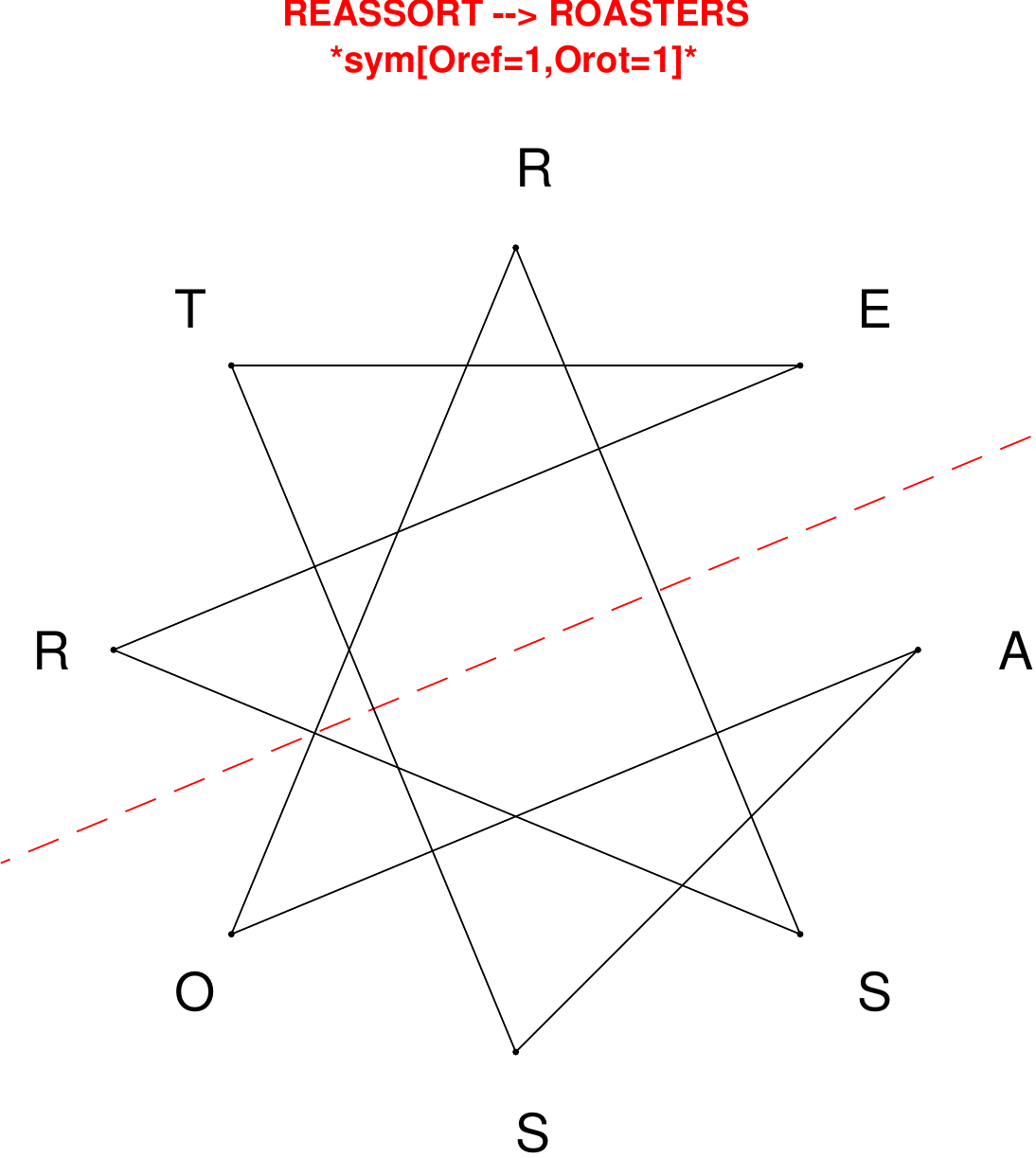}
\end{subfigure}
\end{figure}

\begin{figure}[H]
\centering
\begin{subfigure}[T]{0.19\textwidth}
\centering
\includegraphics[width=\textwidth]{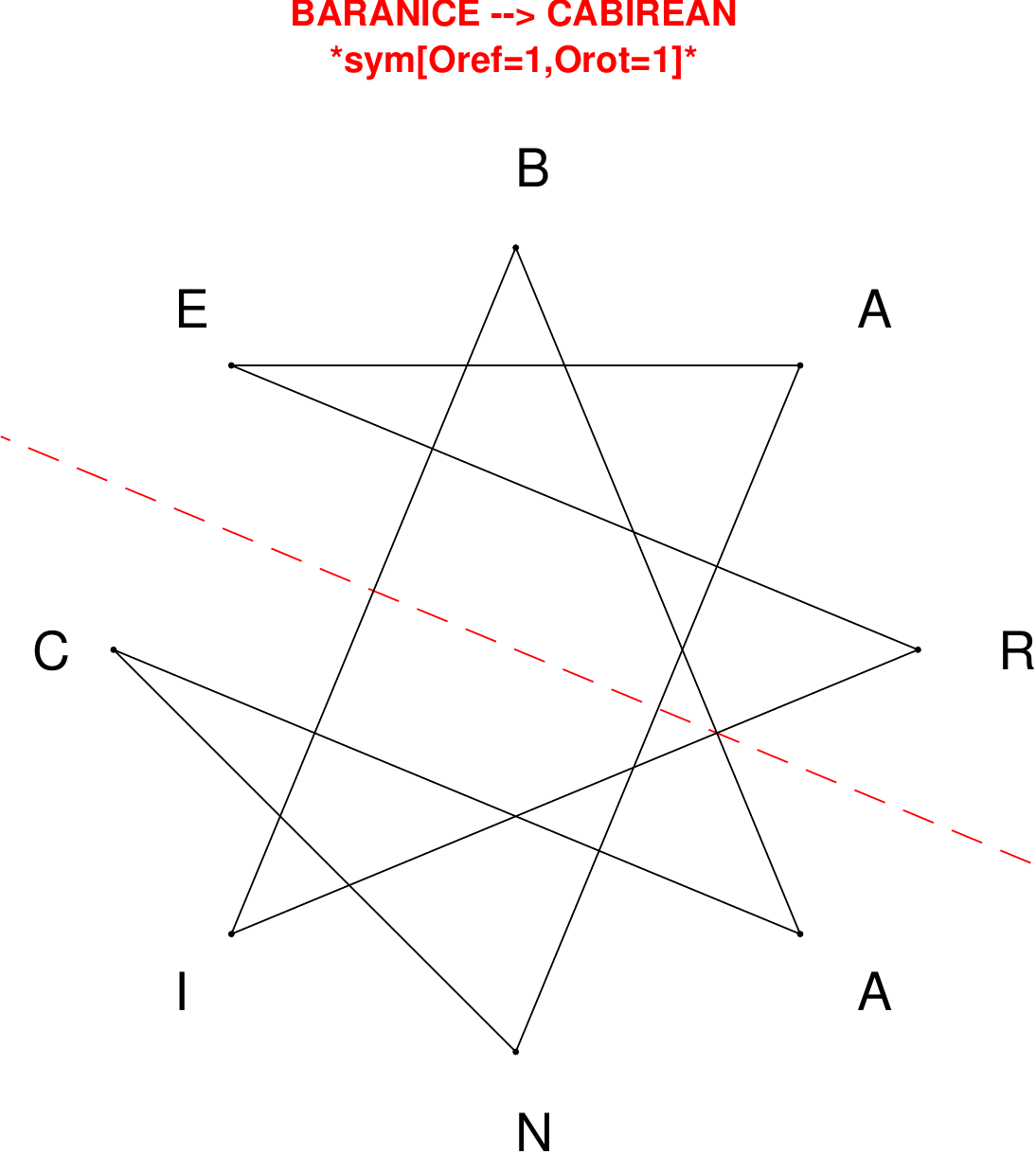}
\end{subfigure}
\hfill
\begin{subfigure}[T]{0.19\textwidth}
\centering
\includegraphics[width=\textwidth]{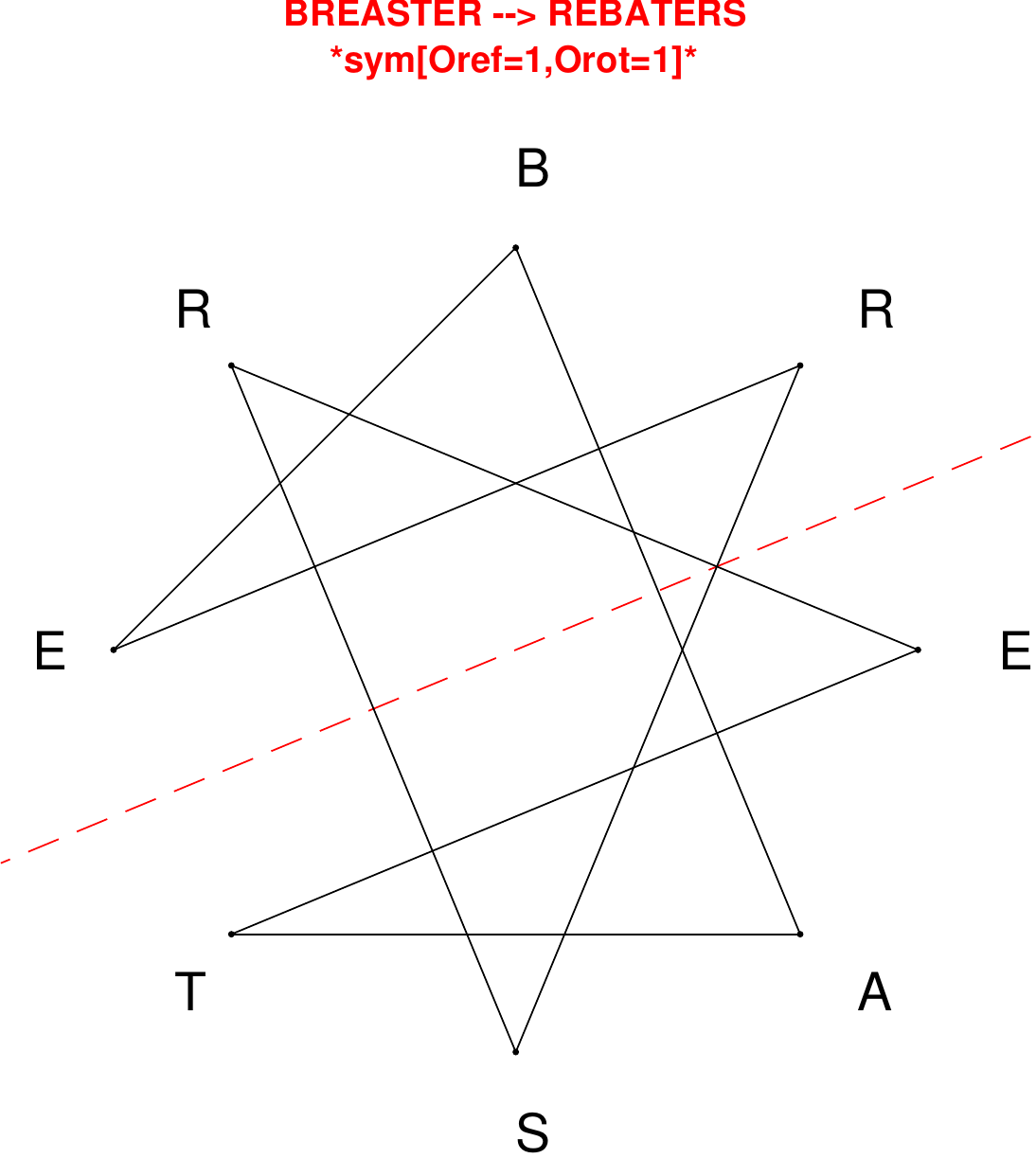}
\end{subfigure}
\hfill
\begin{subfigure}[T]{0.19\textwidth}
\centering
\includegraphics[width=\textwidth]{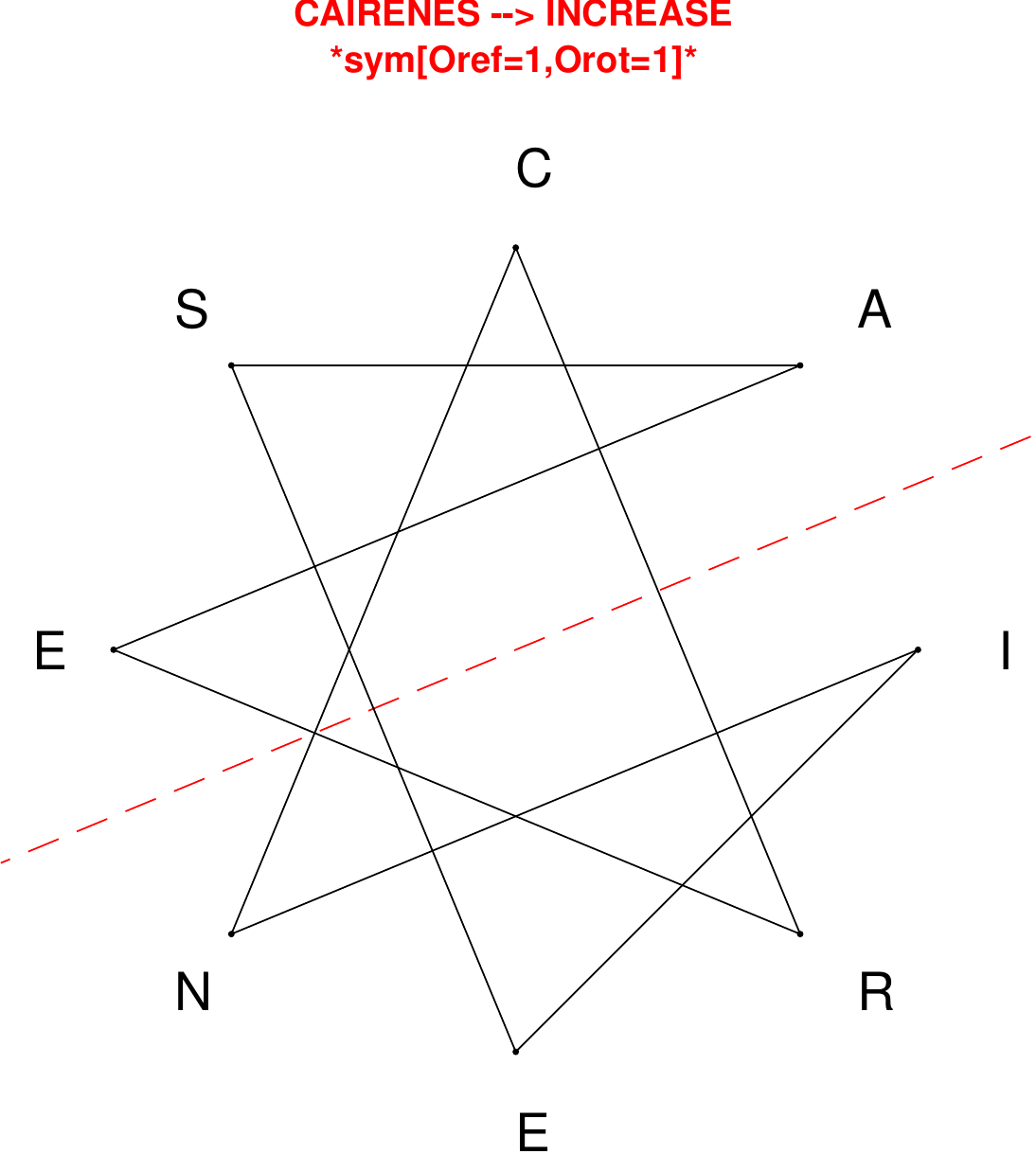}
\end{subfigure}
\hfill
\begin{subfigure}[T]{0.19\textwidth}
\centering
\includegraphics[width=\textwidth]{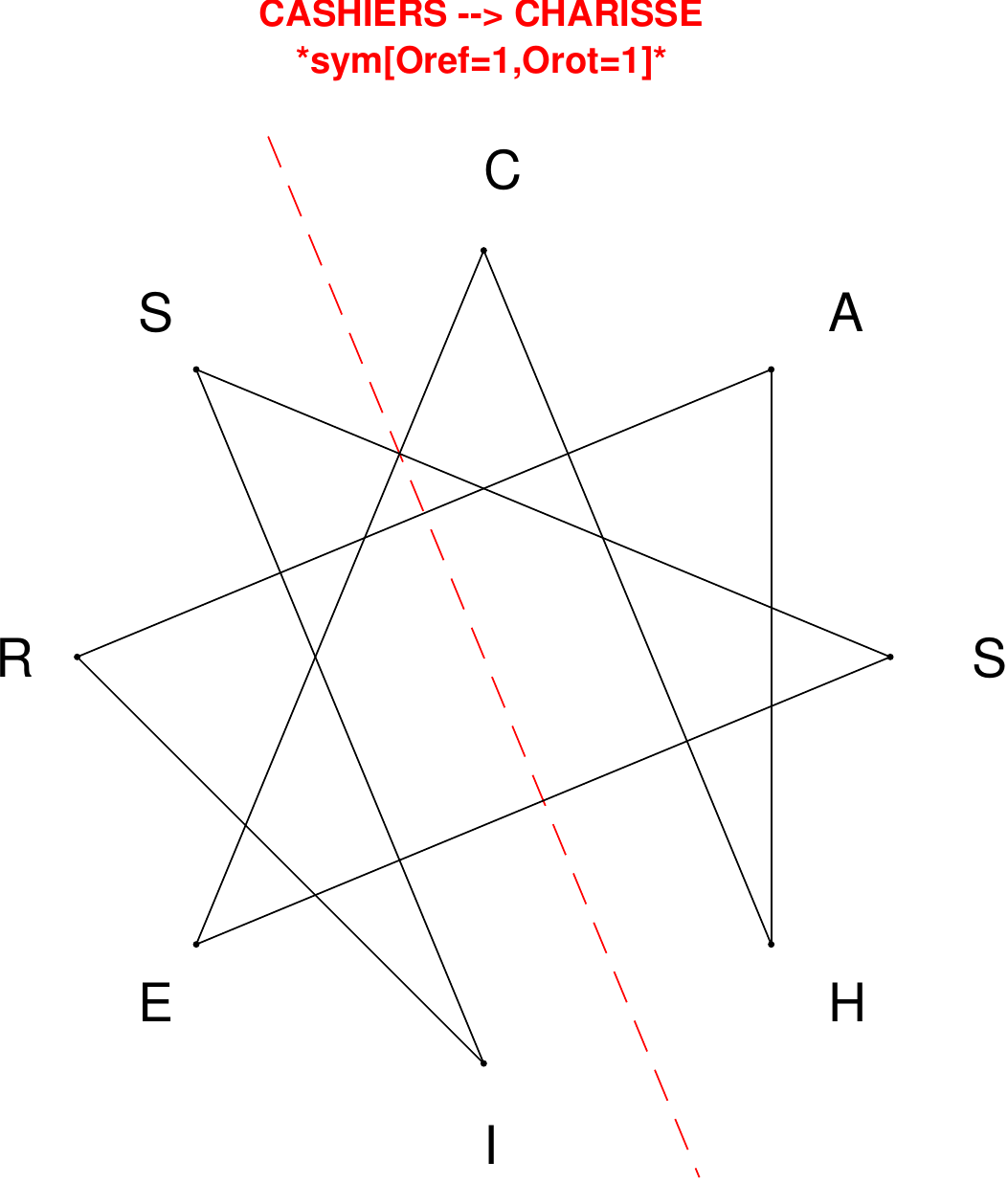}
\end{subfigure}
\hfill
\begin{subfigure}[T]{0.19\textwidth}
\centering
\includegraphics[width=\textwidth]{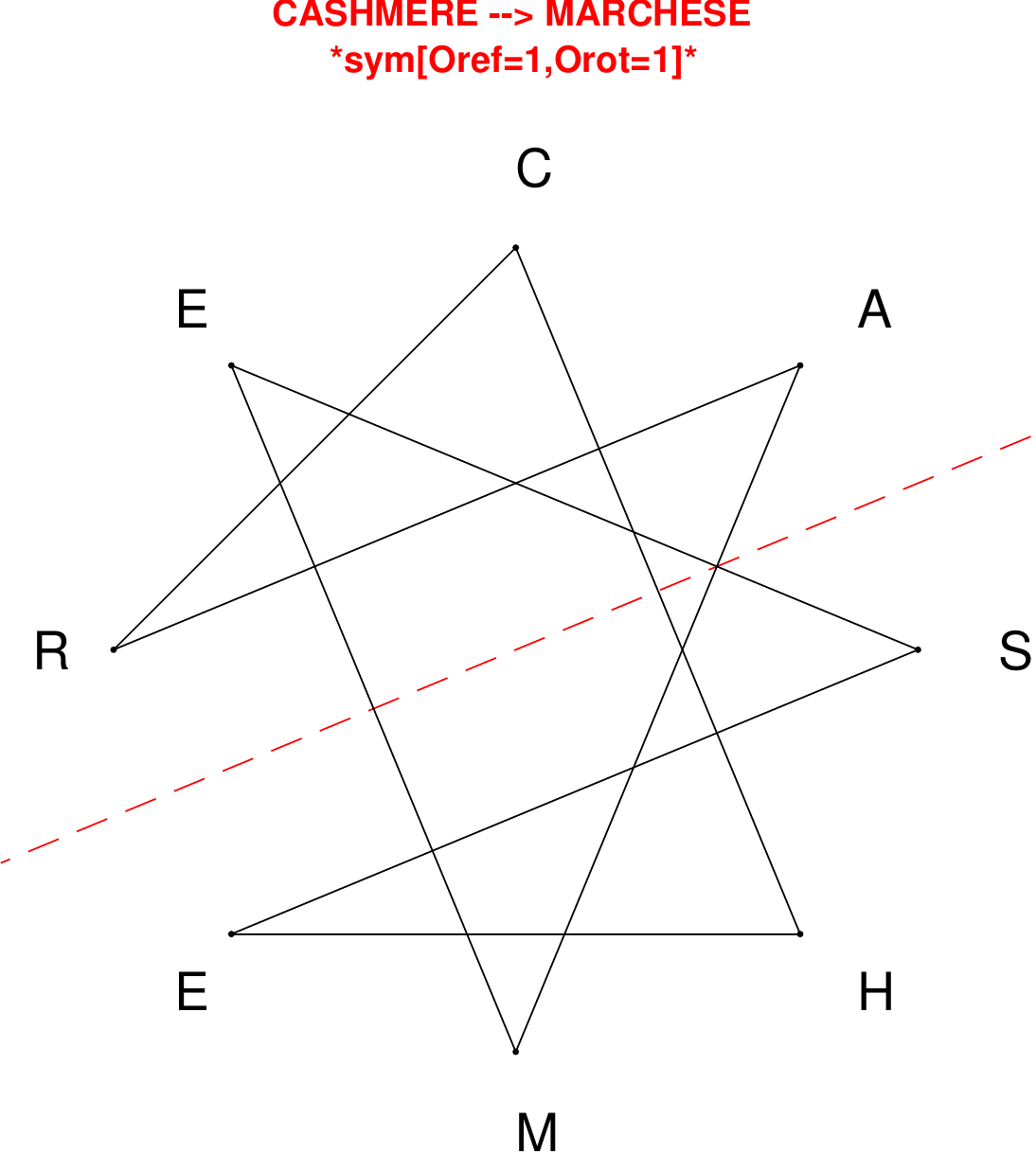}
\end{subfigure}
\end{figure}

\begin{figure}[H]
\centering
\begin{subfigure}[T]{0.19\textwidth}
\centering
\includegraphics[width=\textwidth]{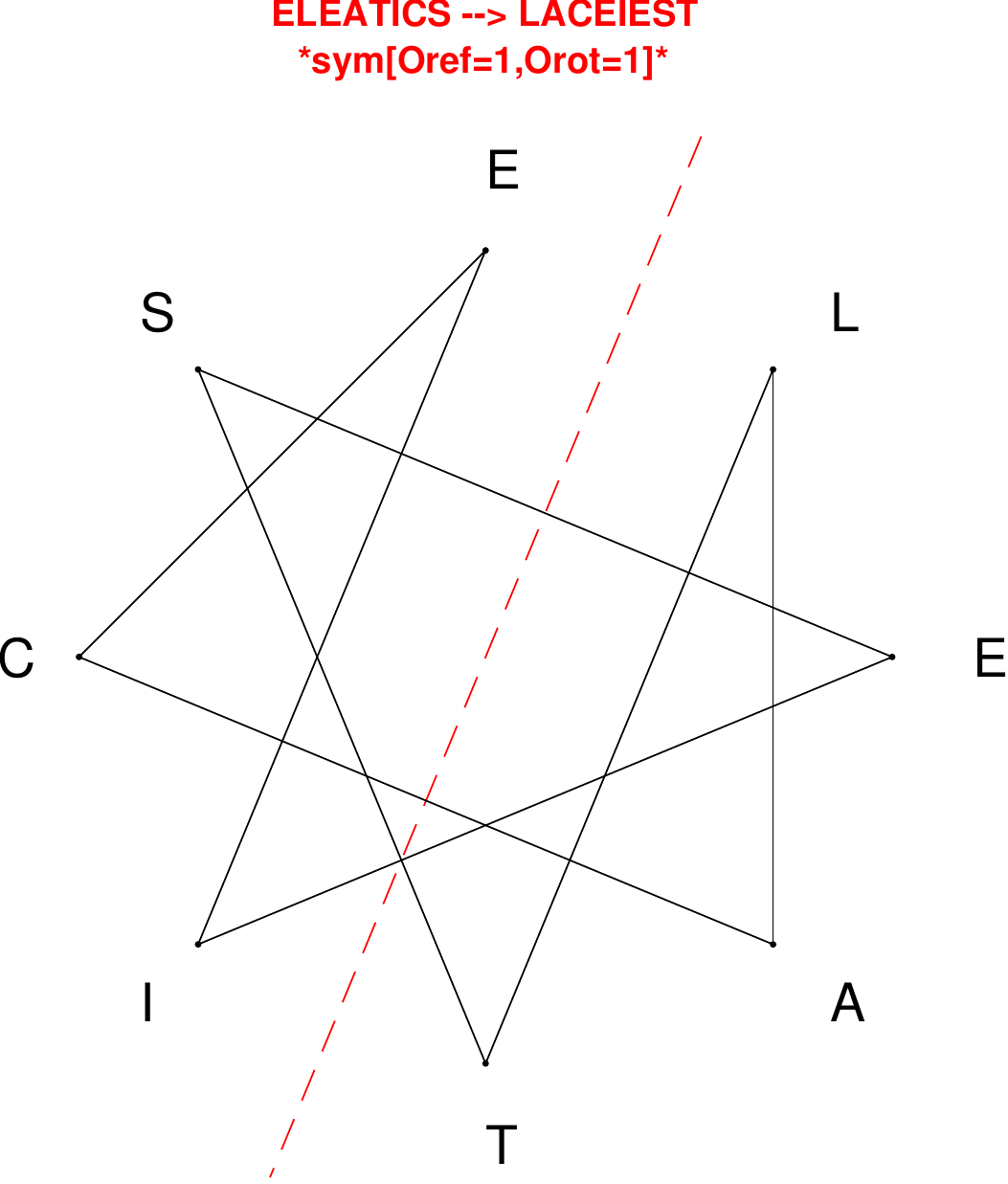}
\end{subfigure}
\hfill
\begin{subfigure}[T]{0.19\textwidth}
\centering
\includegraphics[width=\textwidth]{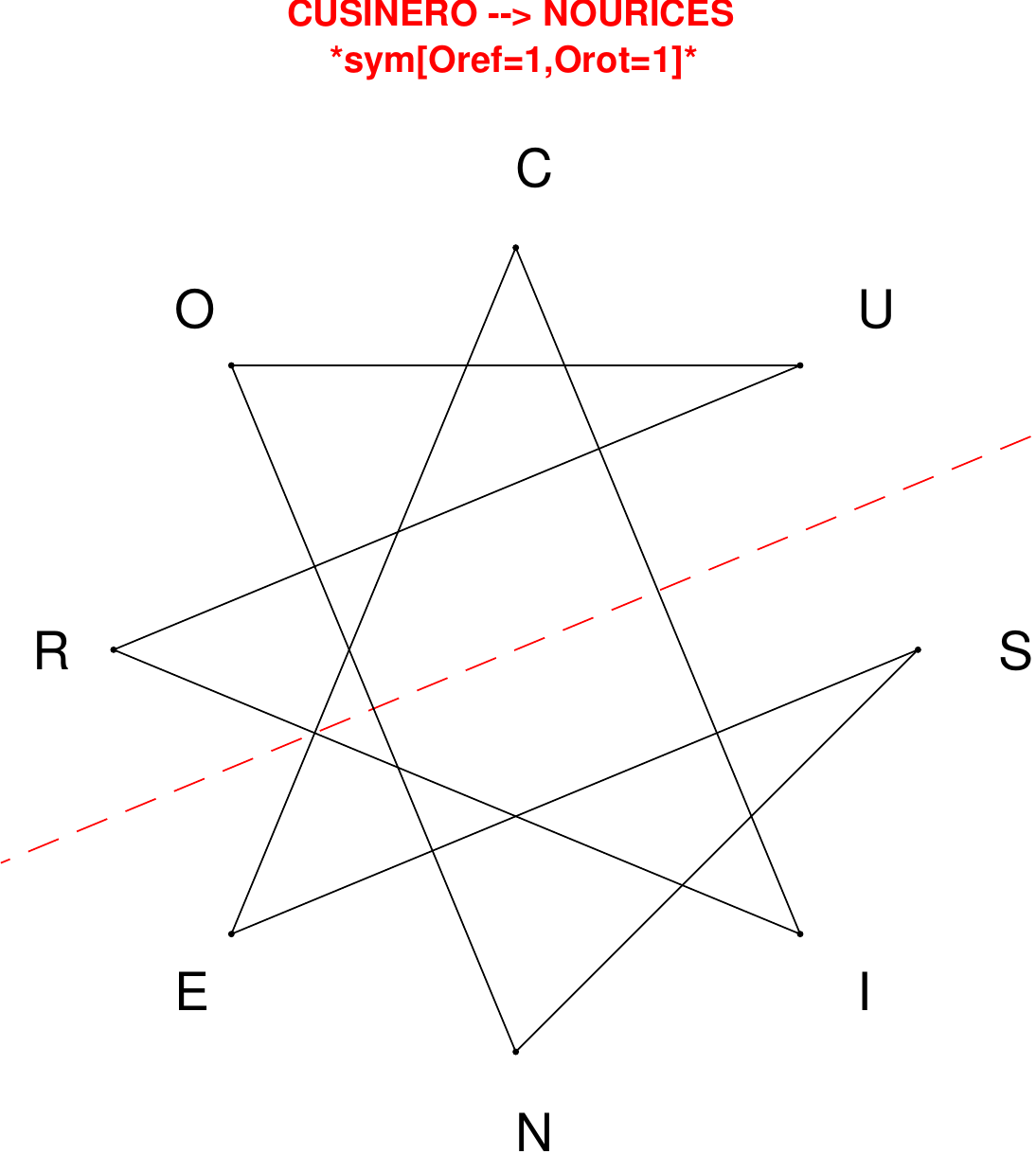}
\end{subfigure}
\hfill
\begin{subfigure}[T]{0.19\textwidth}
\centering
\includegraphics[width=\textwidth]{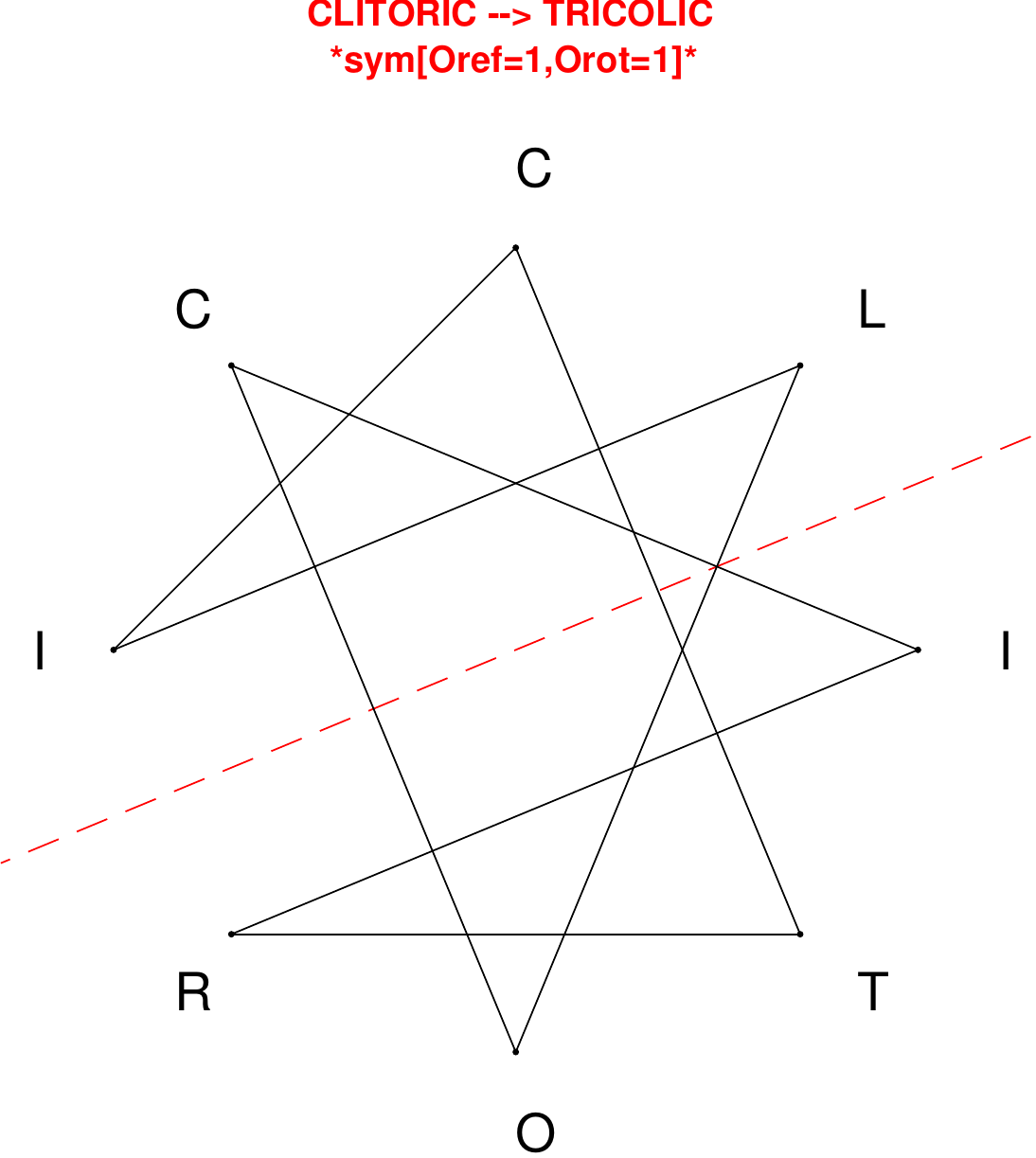}
\end{subfigure}
\hfill
\begin{subfigure}[T]{0.19\textwidth}
\centering
\includegraphics[width=\textwidth]{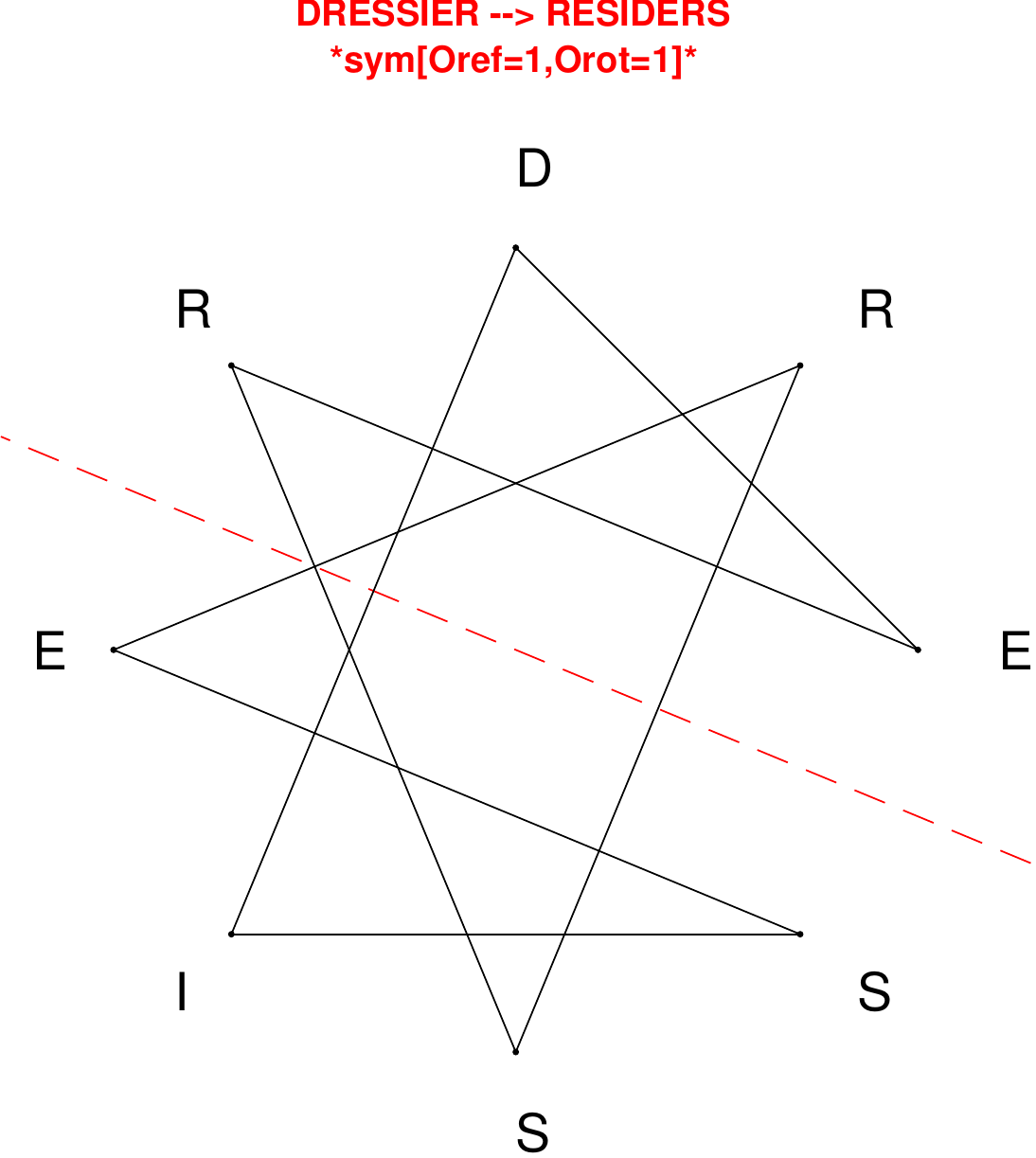}
\end{subfigure}
\hfill
\begin{subfigure}[T]{0.19\textwidth}
\centering
\includegraphics[width=\textwidth]{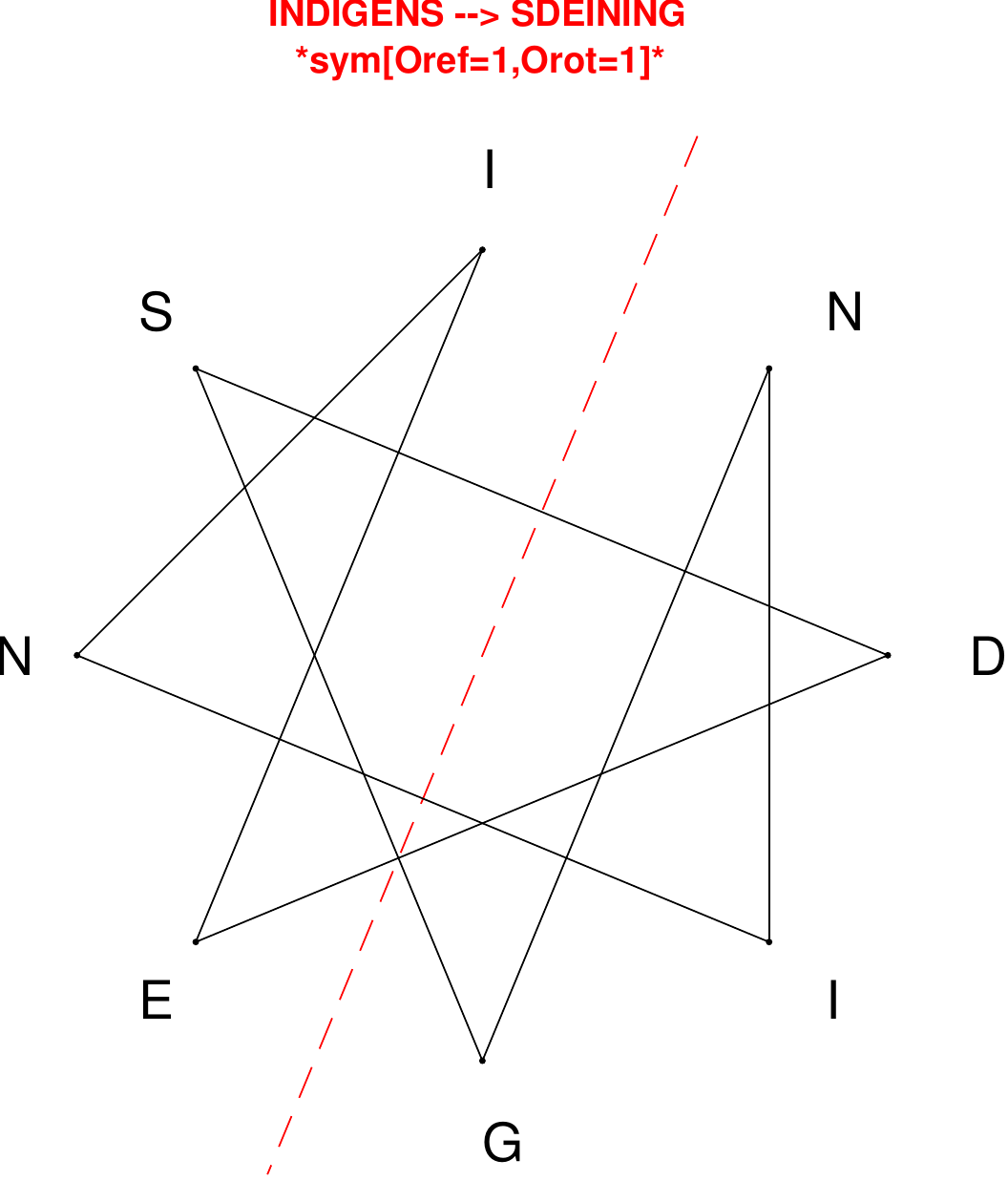}
\end{subfigure}
\end{figure}

\begin{figure}[H]
\centering
\begin{subfigure}[T]{0.19\textwidth}
\centering
\includegraphics[width=\textwidth]{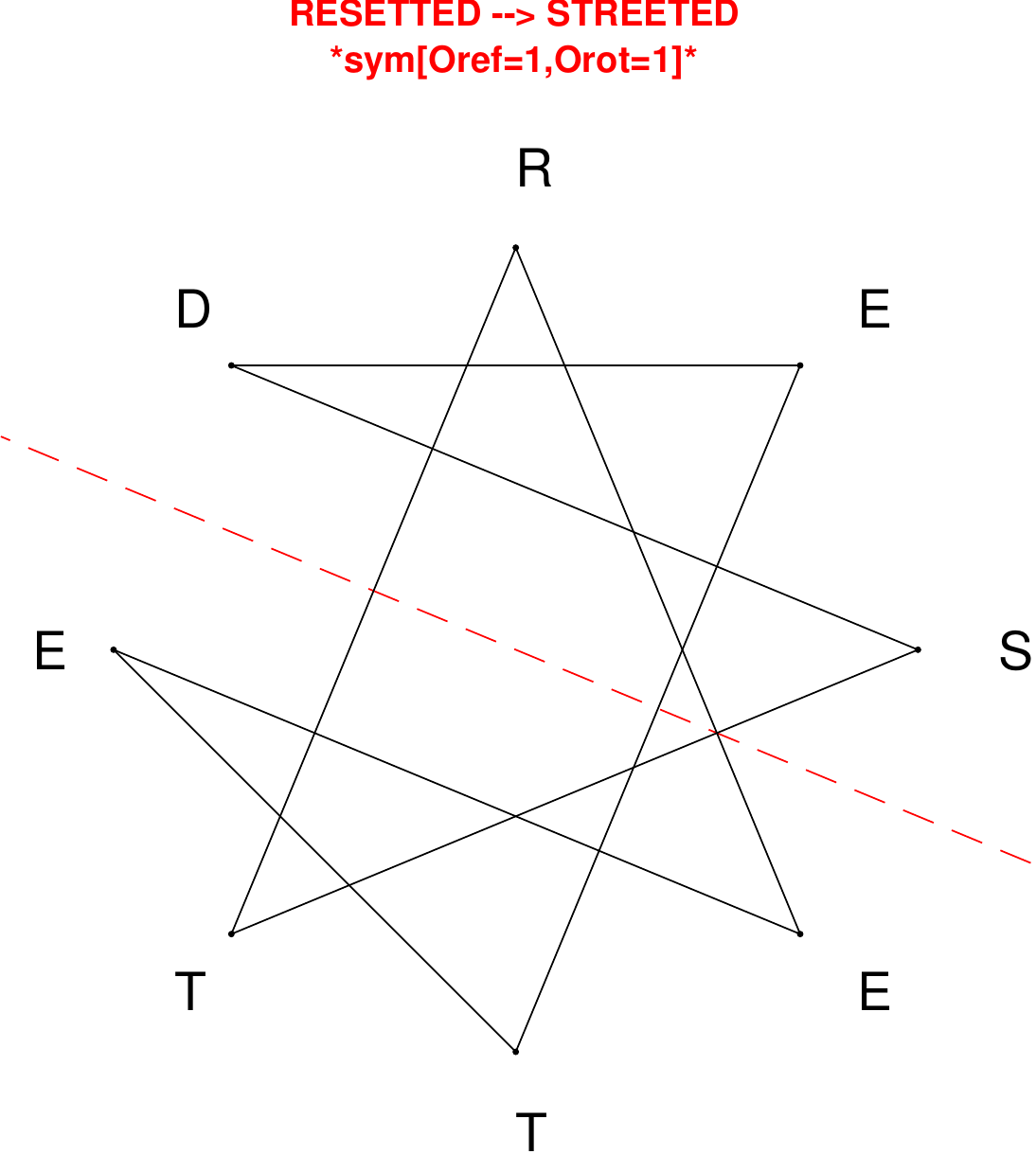}
\end{subfigure}
\hfill
\begin{subfigure}[T]{0.19\textwidth}
\centering
\includegraphics[width=\textwidth]{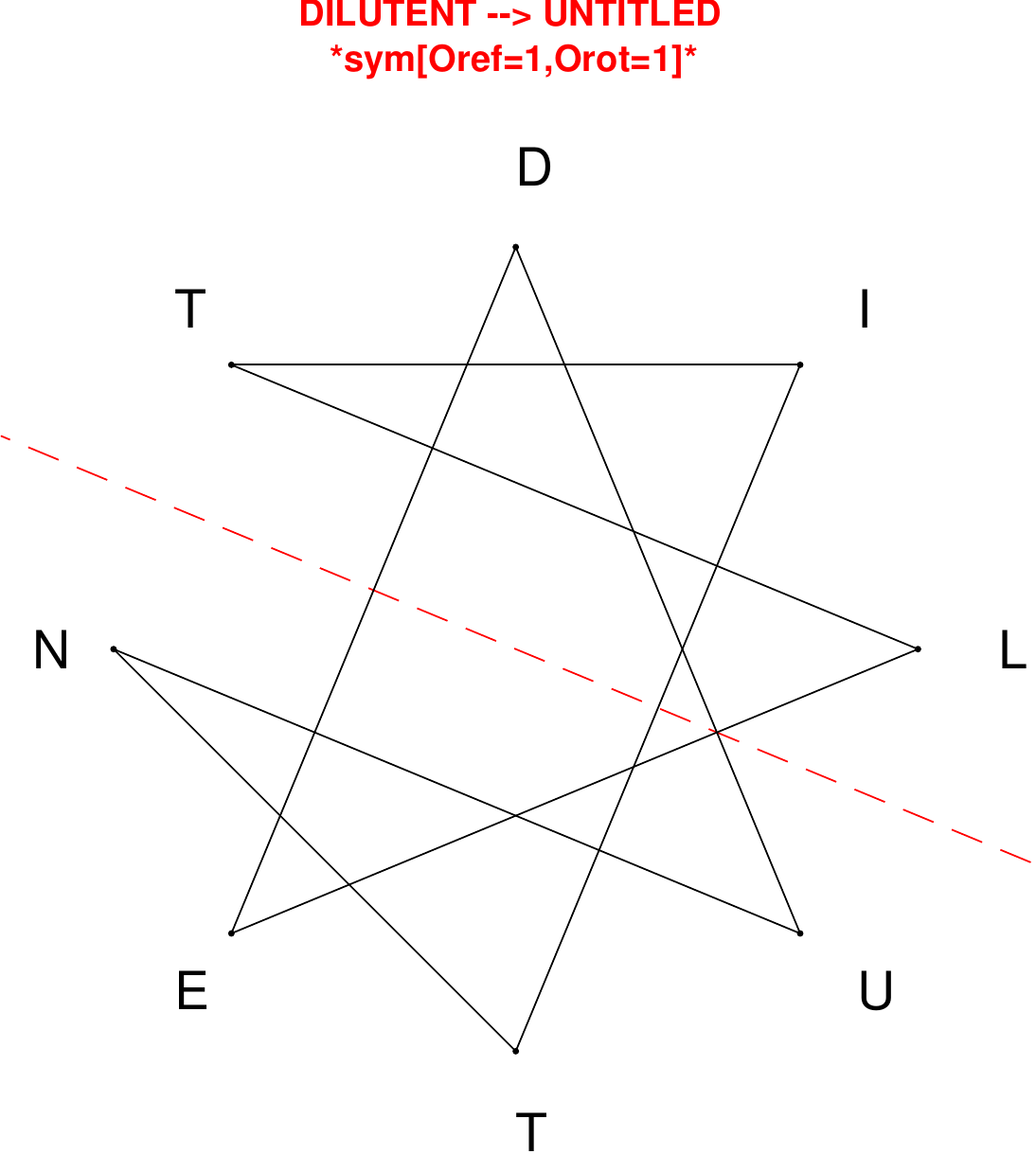}
\end{subfigure}
\hfill
\begin{subfigure}[T]{0.19\textwidth}
\centering
\includegraphics[width=\textwidth]{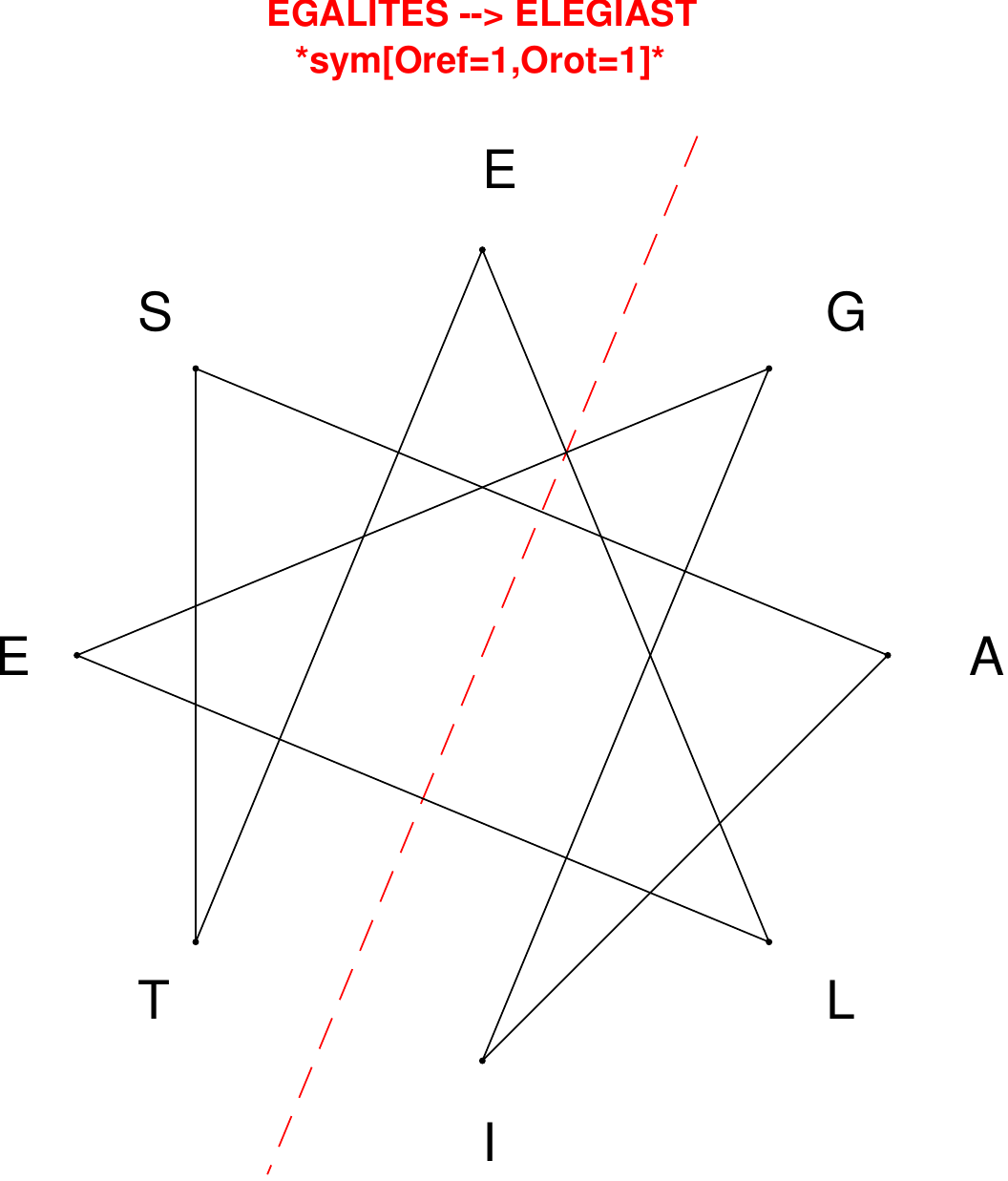}
\end{subfigure}
\hfill
\begin{subfigure}[T]{0.19\textwidth}
\centering
\includegraphics[width=\textwidth]{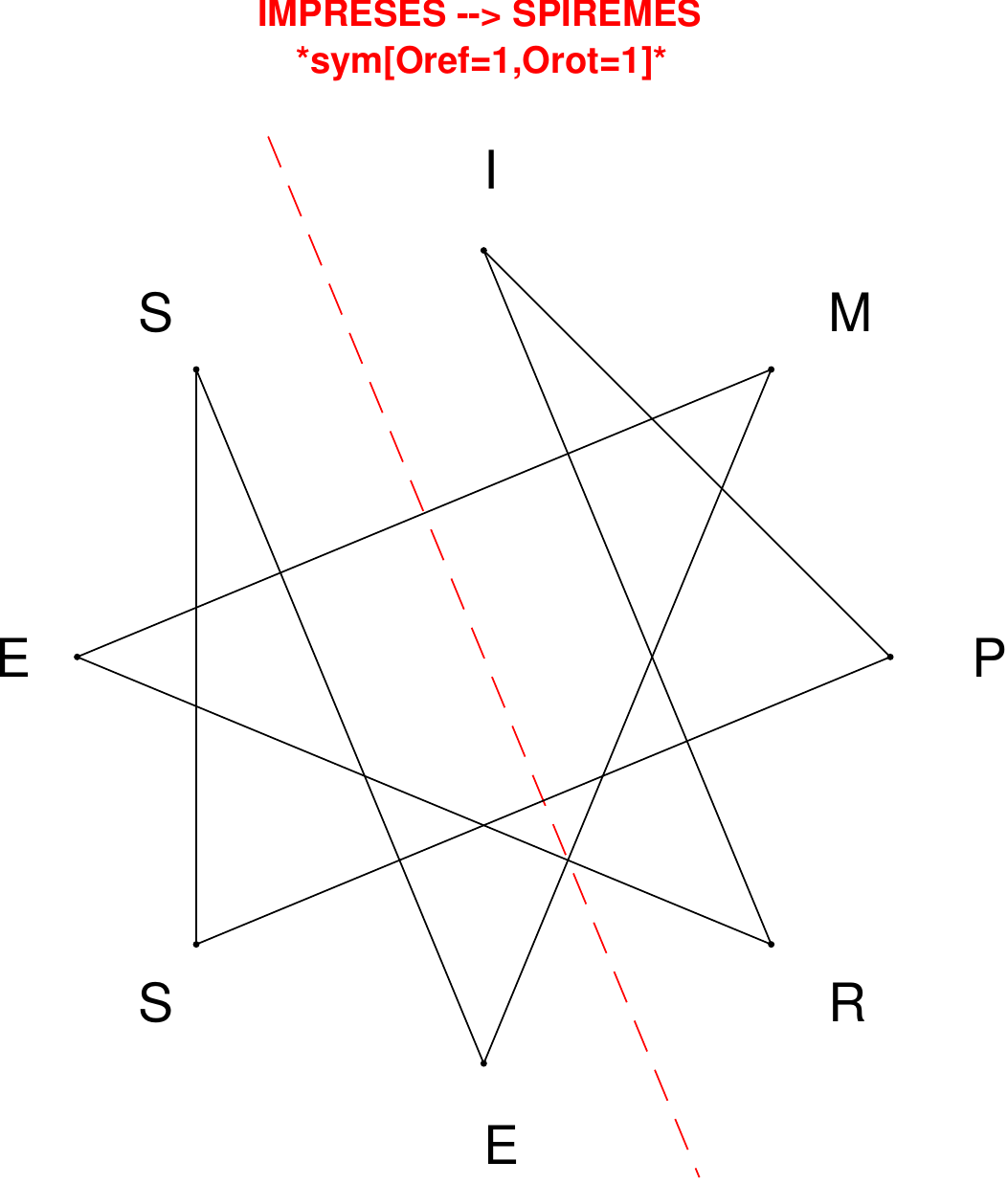}
\end{subfigure}
\hfill
\begin{subfigure}[T]{0.19\textwidth}
\centering
\includegraphics[width=\textwidth]{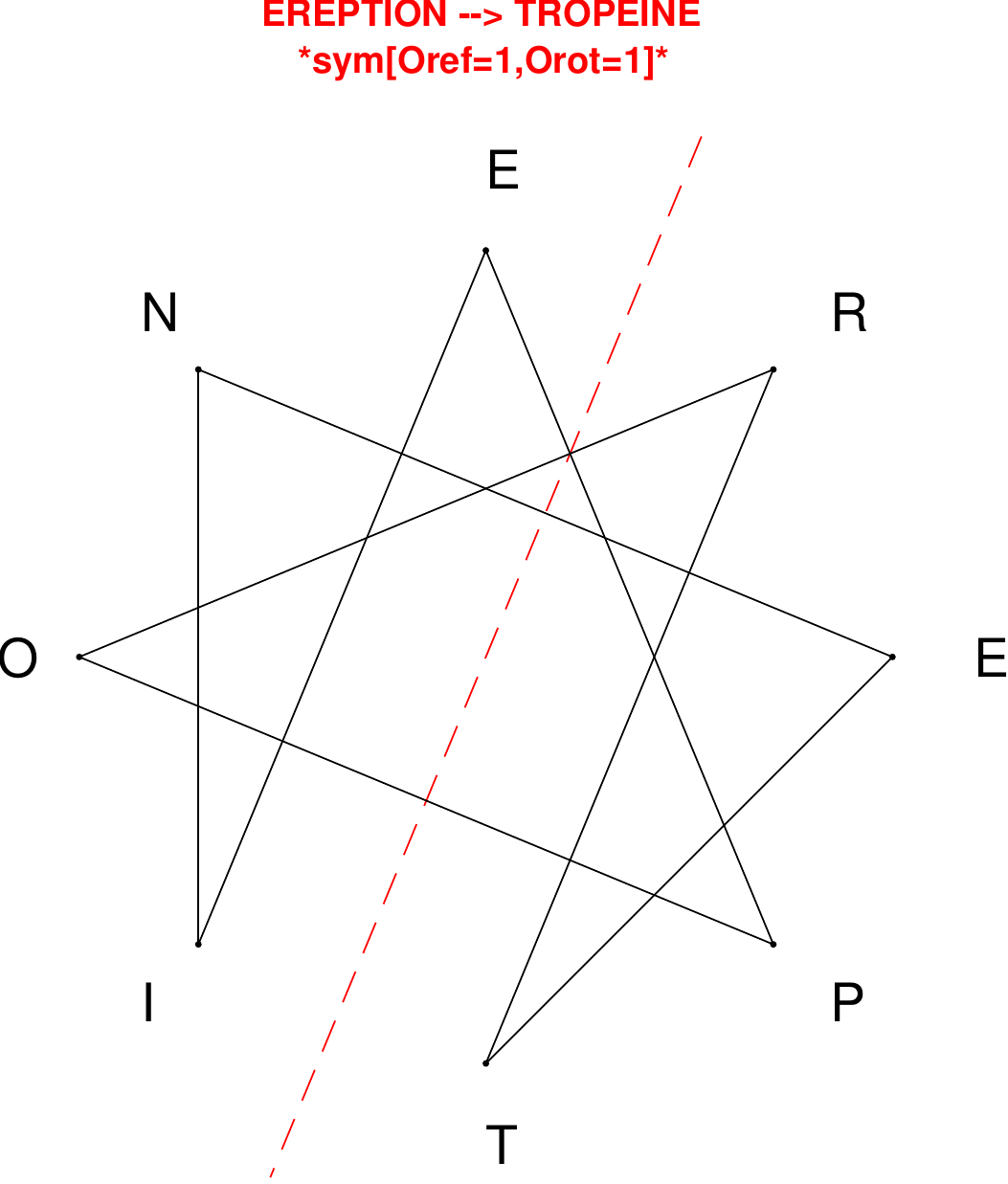}
\end{subfigure}
\end{figure}

\begin{figure}[H]
\centering
\begin{subfigure}[T]{0.19\textwidth}
\centering
\includegraphics[width=\textwidth]{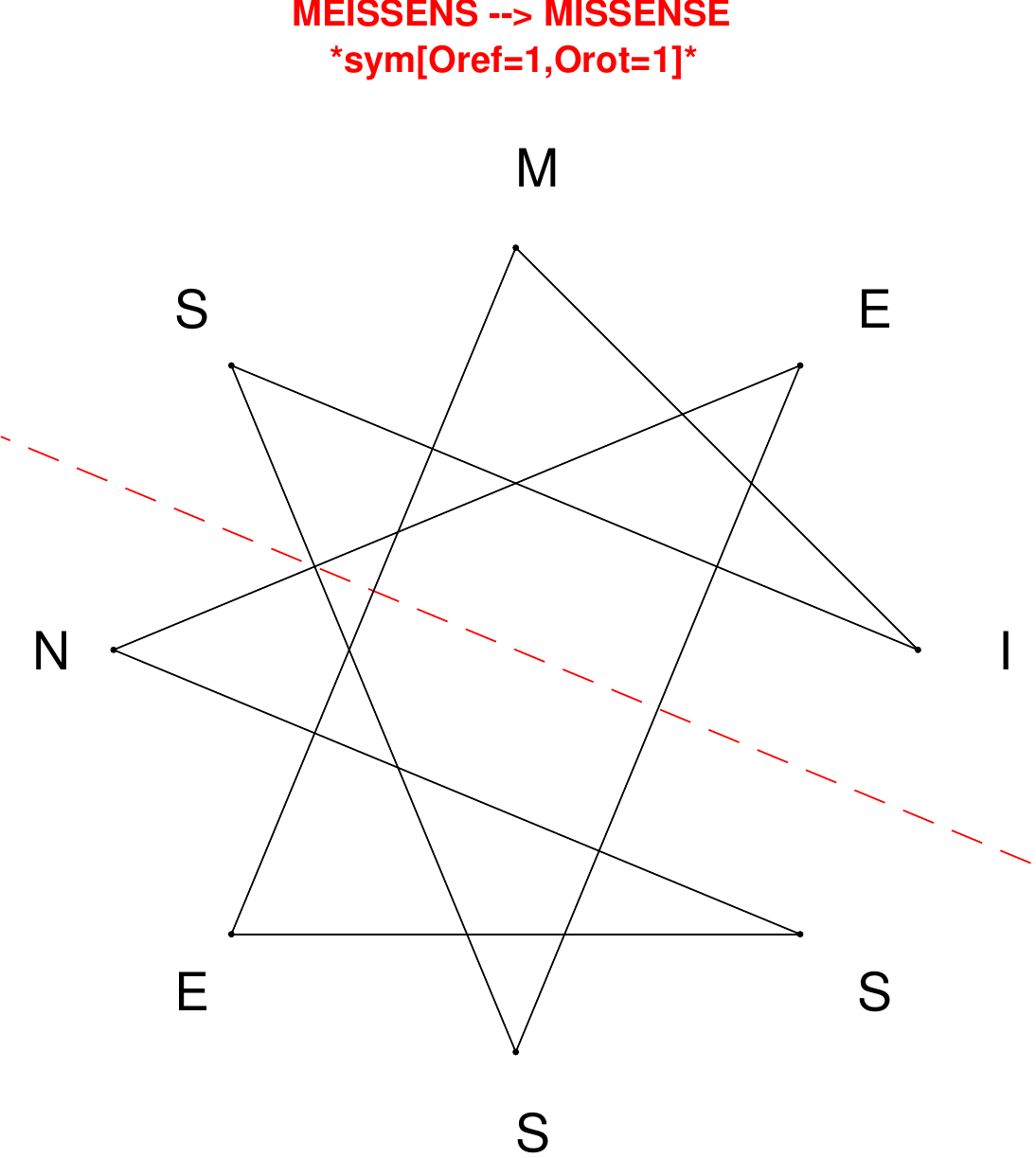}
\end{subfigure}
\hfill
\begin{subfigure}[T]{0.19\textwidth}
\centering
\includegraphics[width=\textwidth]{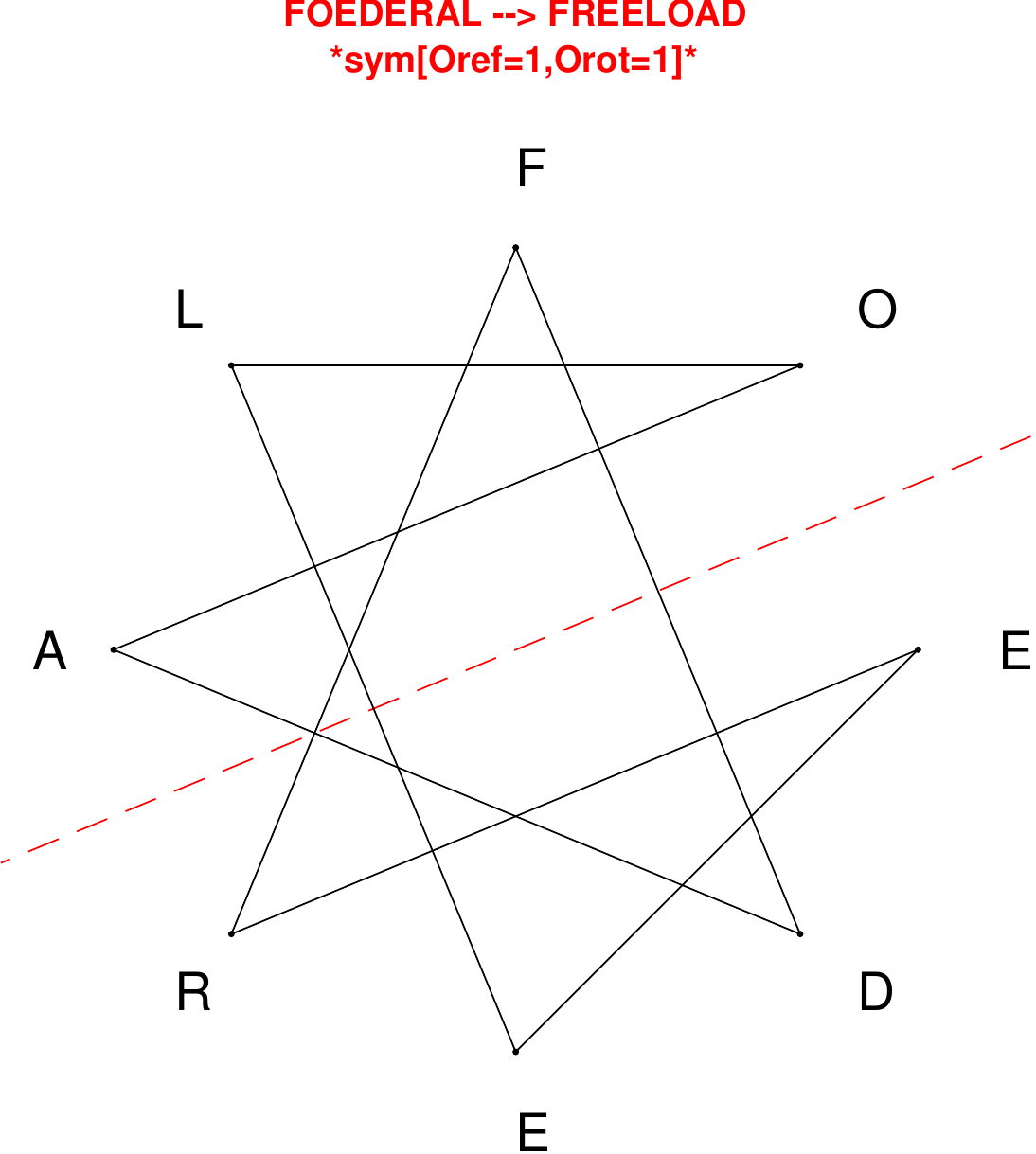}
\end{subfigure}
\hfill
\begin{subfigure}[T]{0.19\textwidth}
\centering
\includegraphics[width=\textwidth]{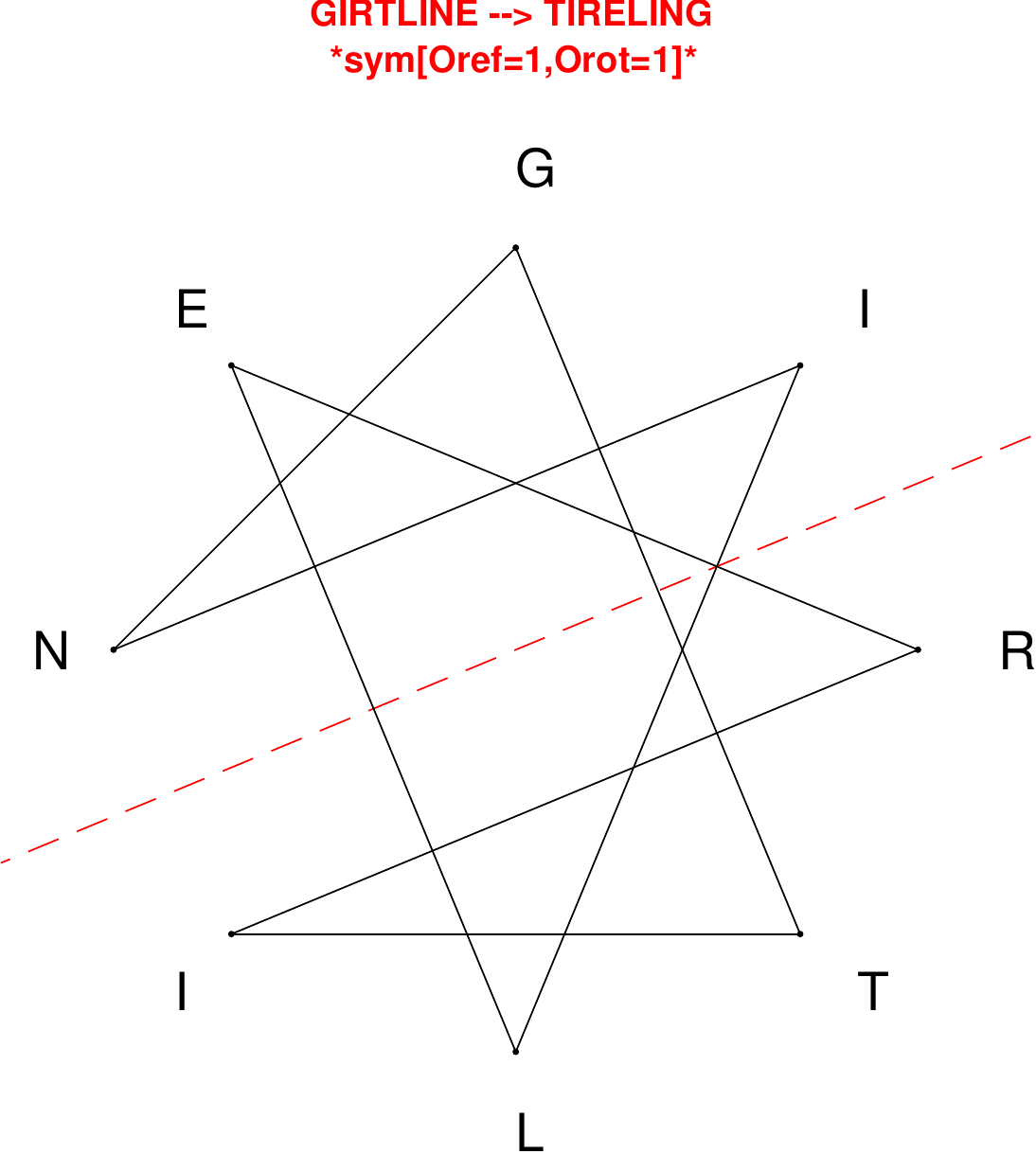}
\end{subfigure}
\hfill
\begin{subfigure}[T]{0.19\textwidth}
\centering
\includegraphics[width=\textwidth]{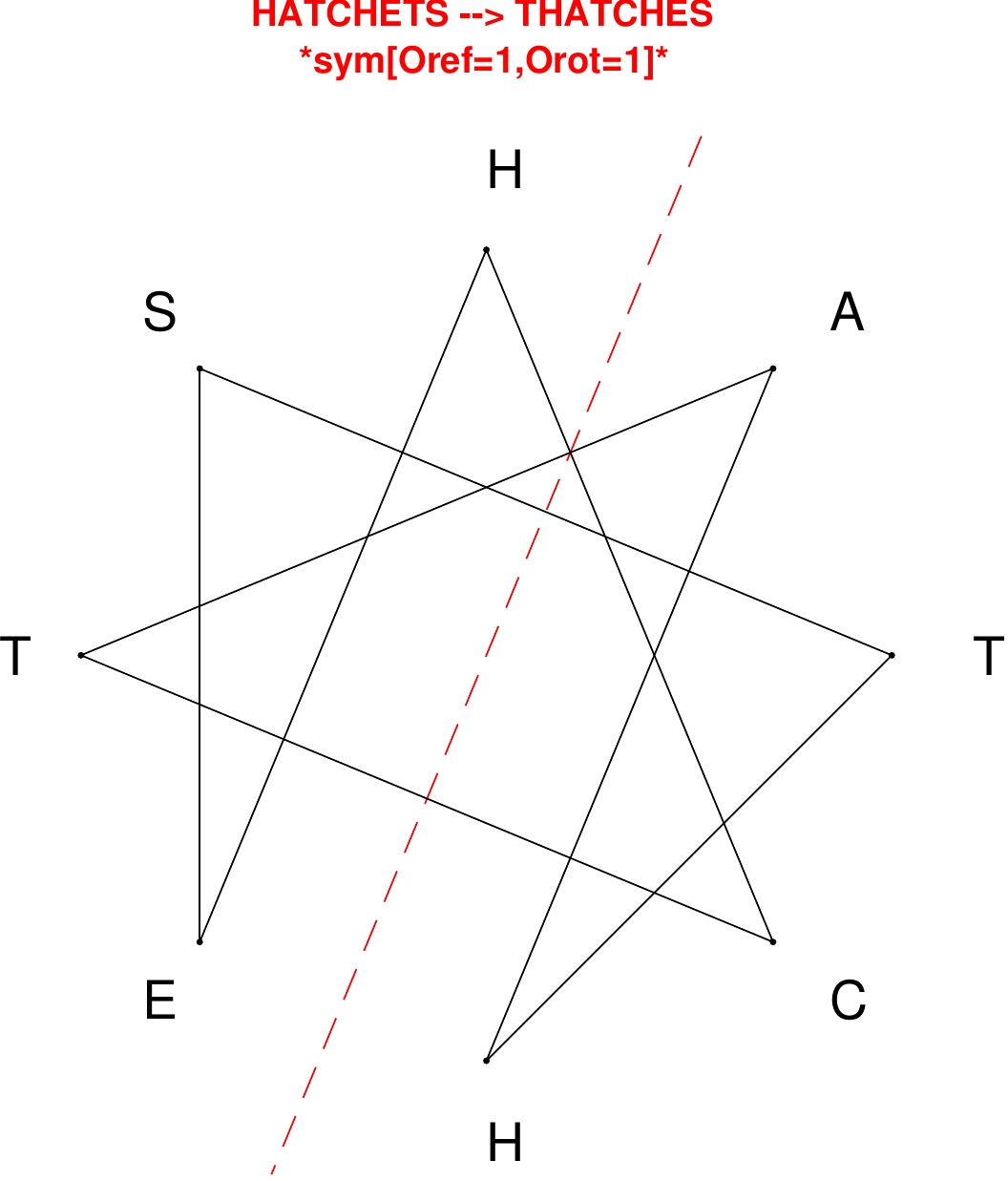}
\end{subfigure}
\hfill
\begin{subfigure}[T]{0.19\textwidth}
\centering
\includegraphics[width=\textwidth]{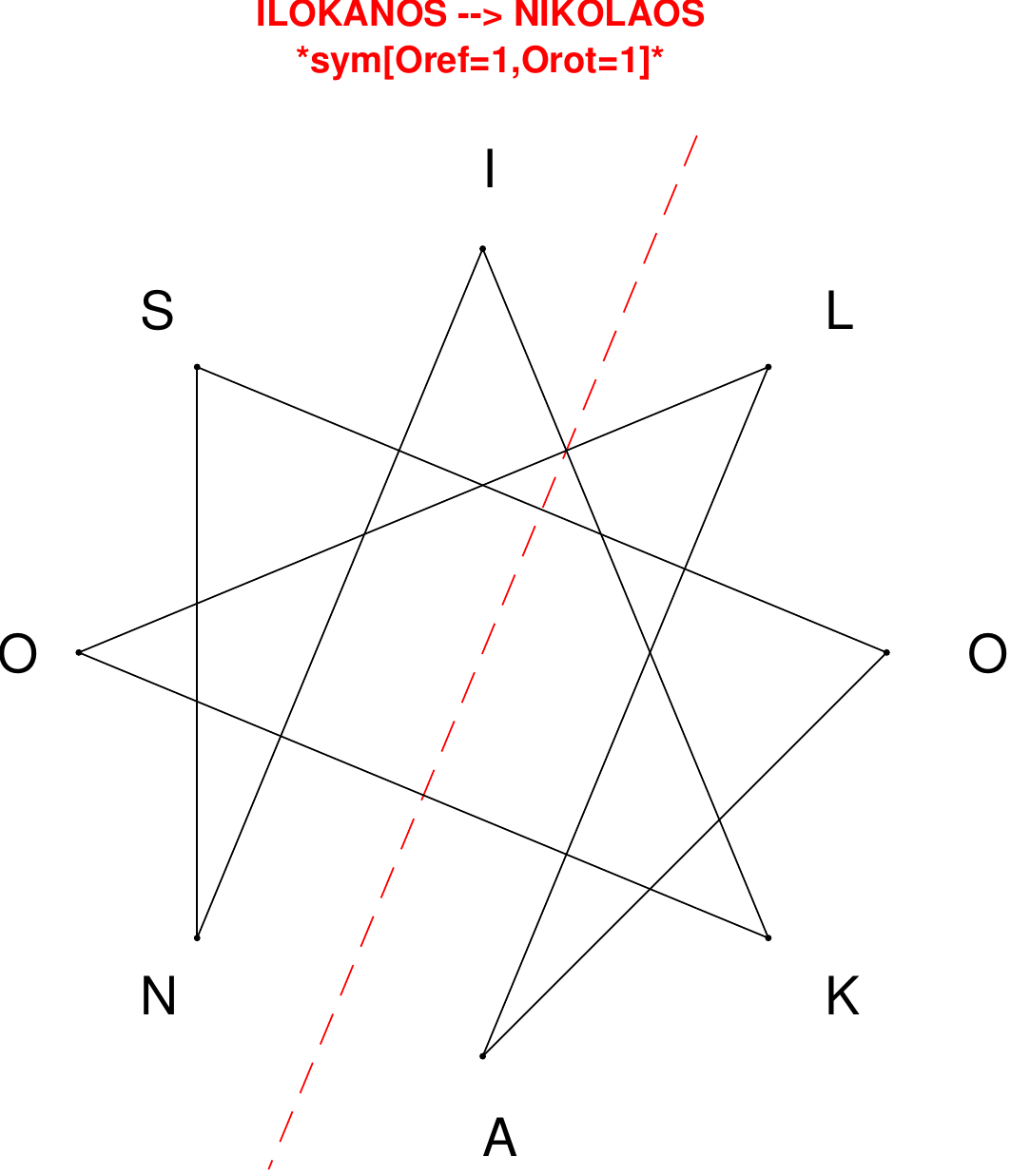}
\end{subfigure}
\end{figure}

\begin{figure}[H]
\centering
\begin{subfigure}[T]{0.19\textwidth}
\centering
\includegraphics[width=\textwidth]{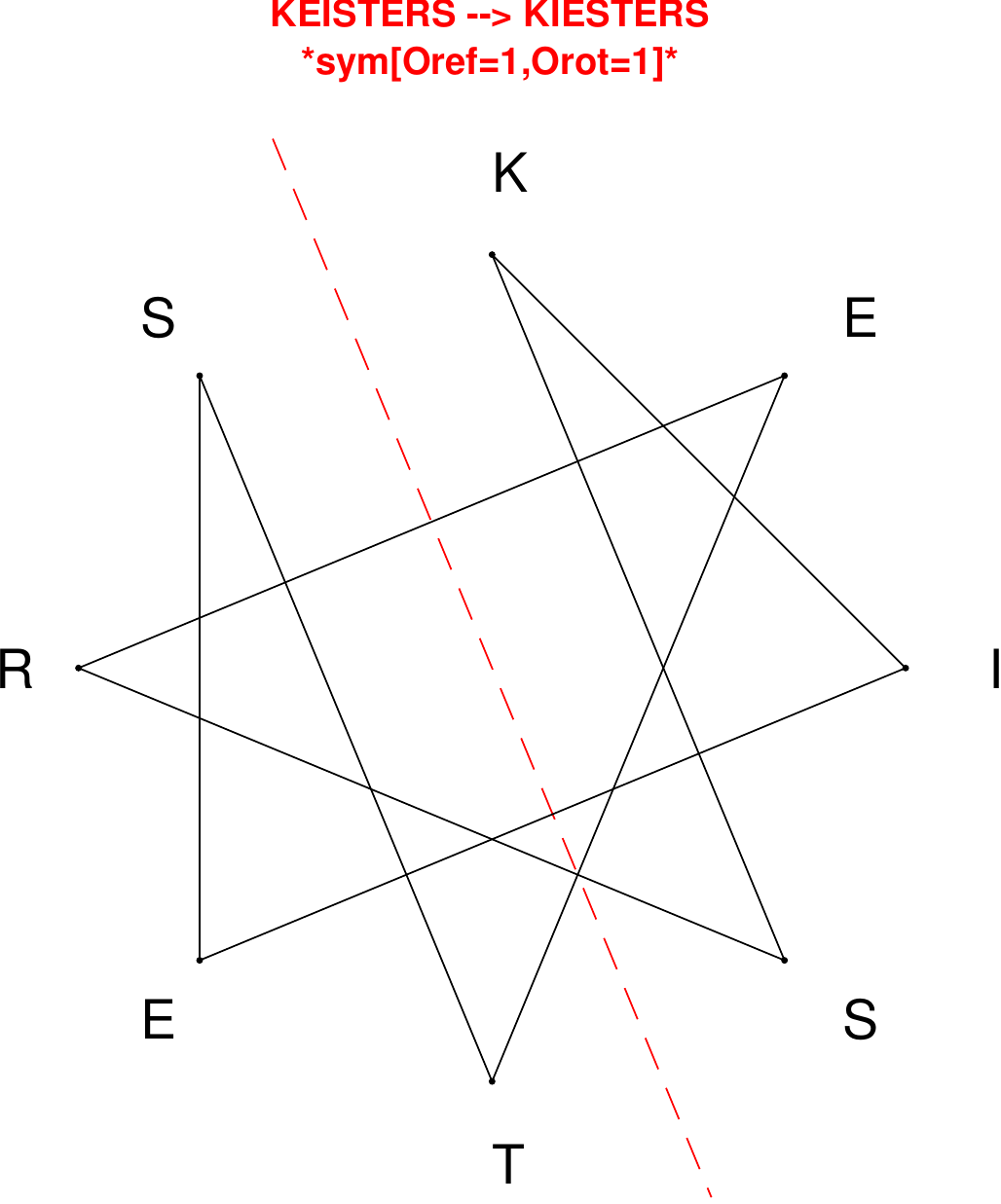}
\end{subfigure}
\hfill
\begin{subfigure}[T]{0.19\textwidth}
\centering
\includegraphics[width=\textwidth]{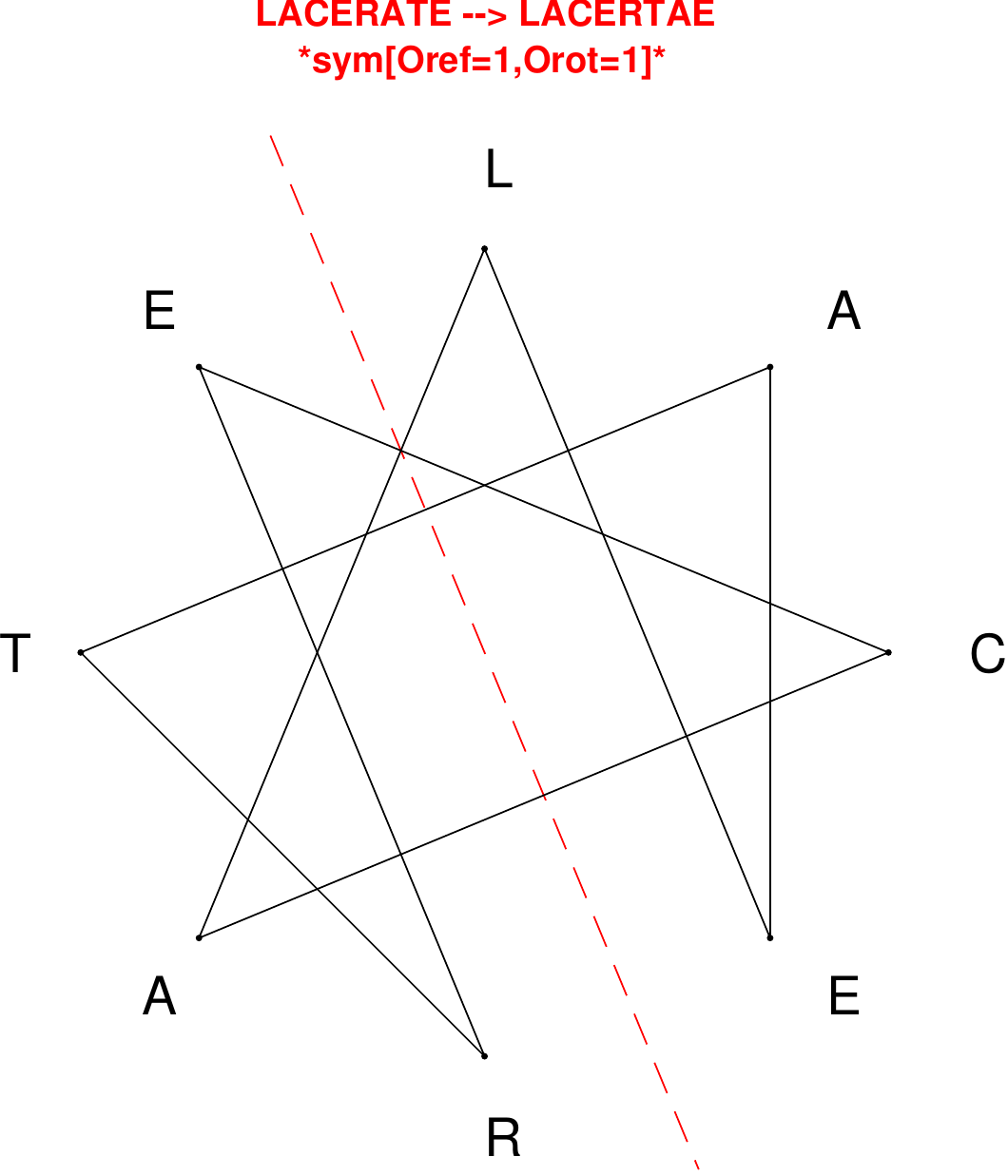}
\end{subfigure}
\hfill
\begin{subfigure}[T]{0.19\textwidth}
\centering
\includegraphics[width=\textwidth]{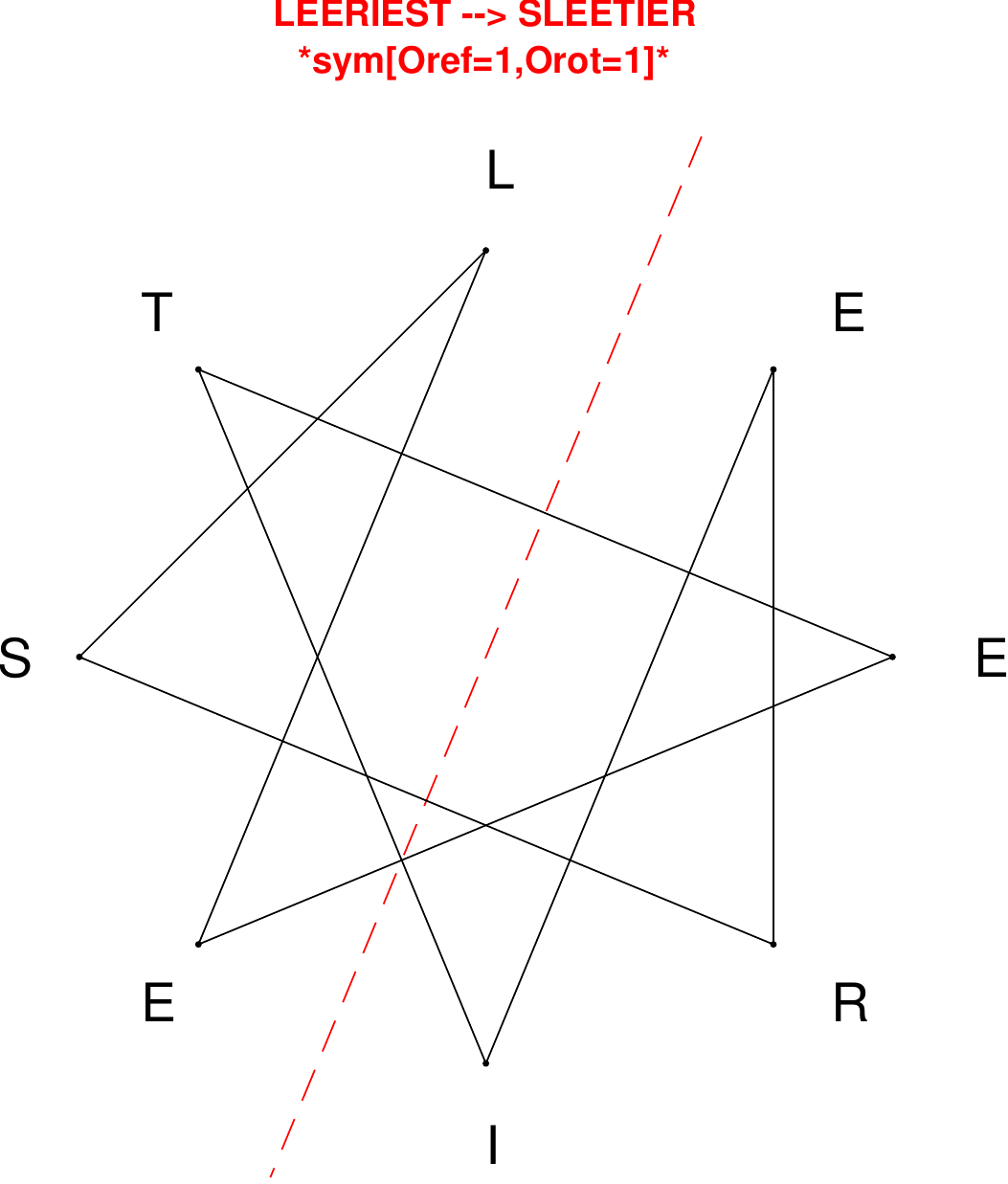}
\end{subfigure}
\hfill
\begin{subfigure}[T]{0.19\textwidth}
\centering
\includegraphics[width=\textwidth]{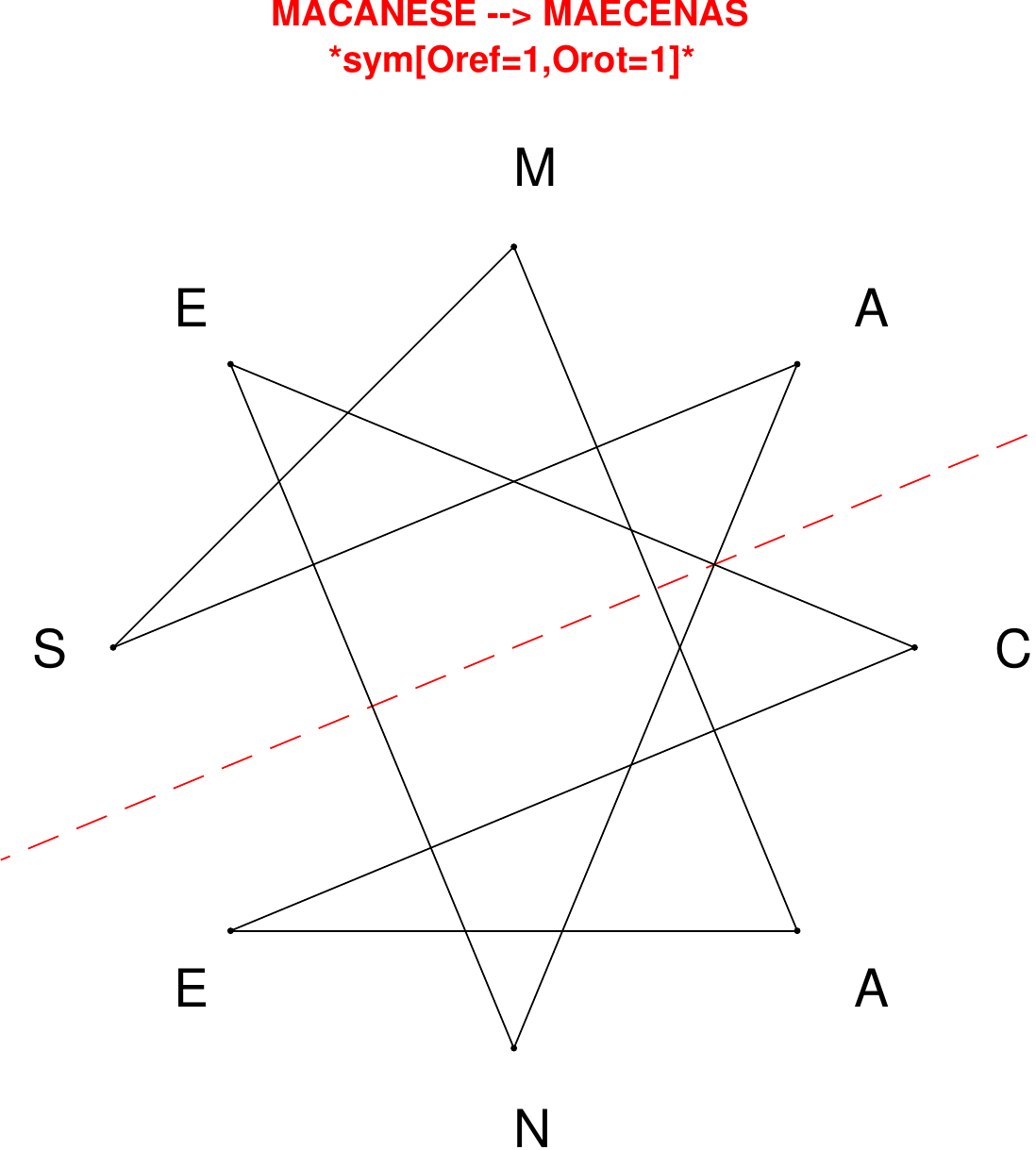}
\end{subfigure}
\hfill
\begin{subfigure}[T]{0.19\textwidth}
\centering
\includegraphics[width=\textwidth]{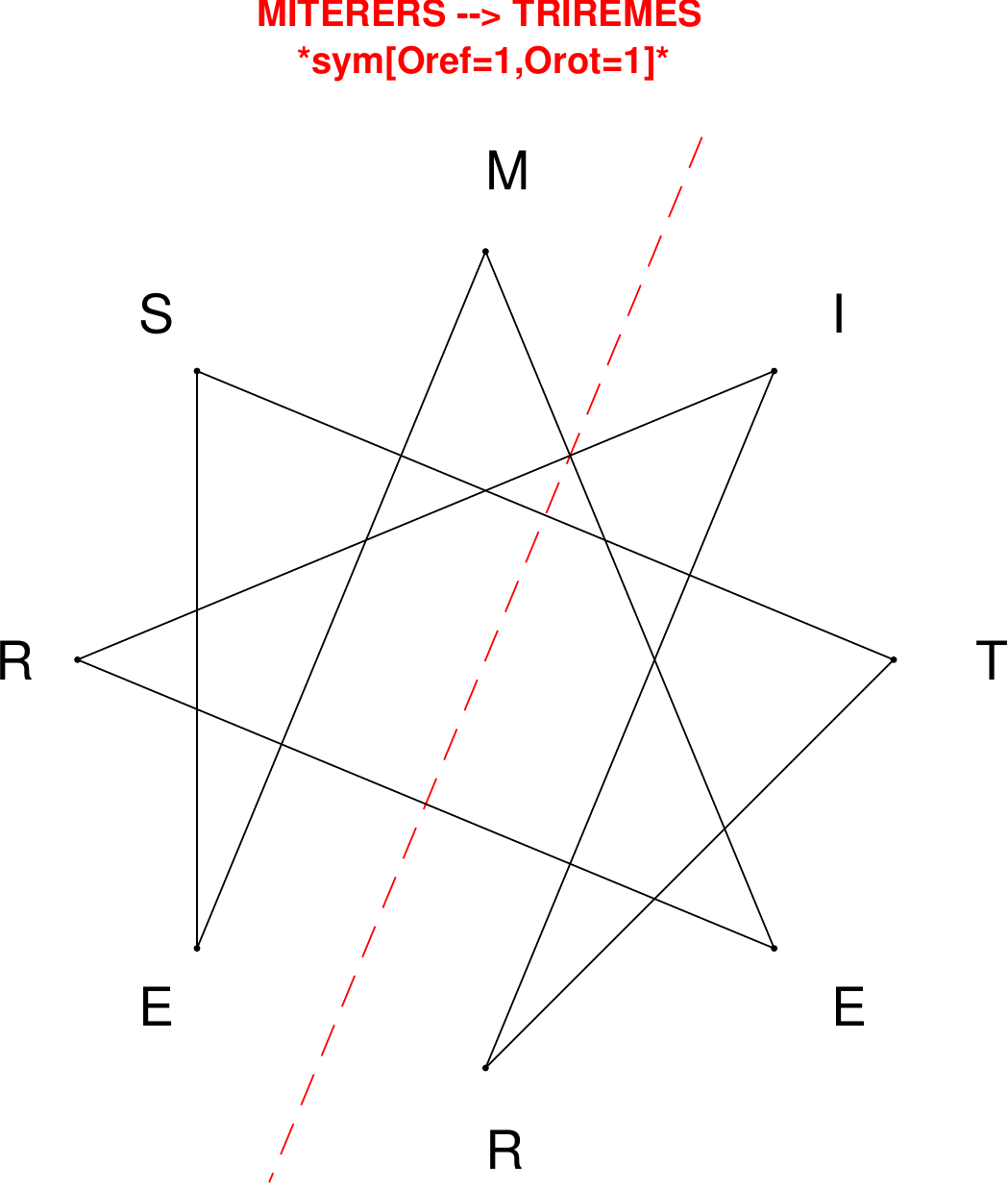}
\end{subfigure}
\end{figure}

\begin{figure}[H]
\centering
\begin{subfigure}[T]{0.19\textwidth}
\centering
\includegraphics[width=\textwidth]{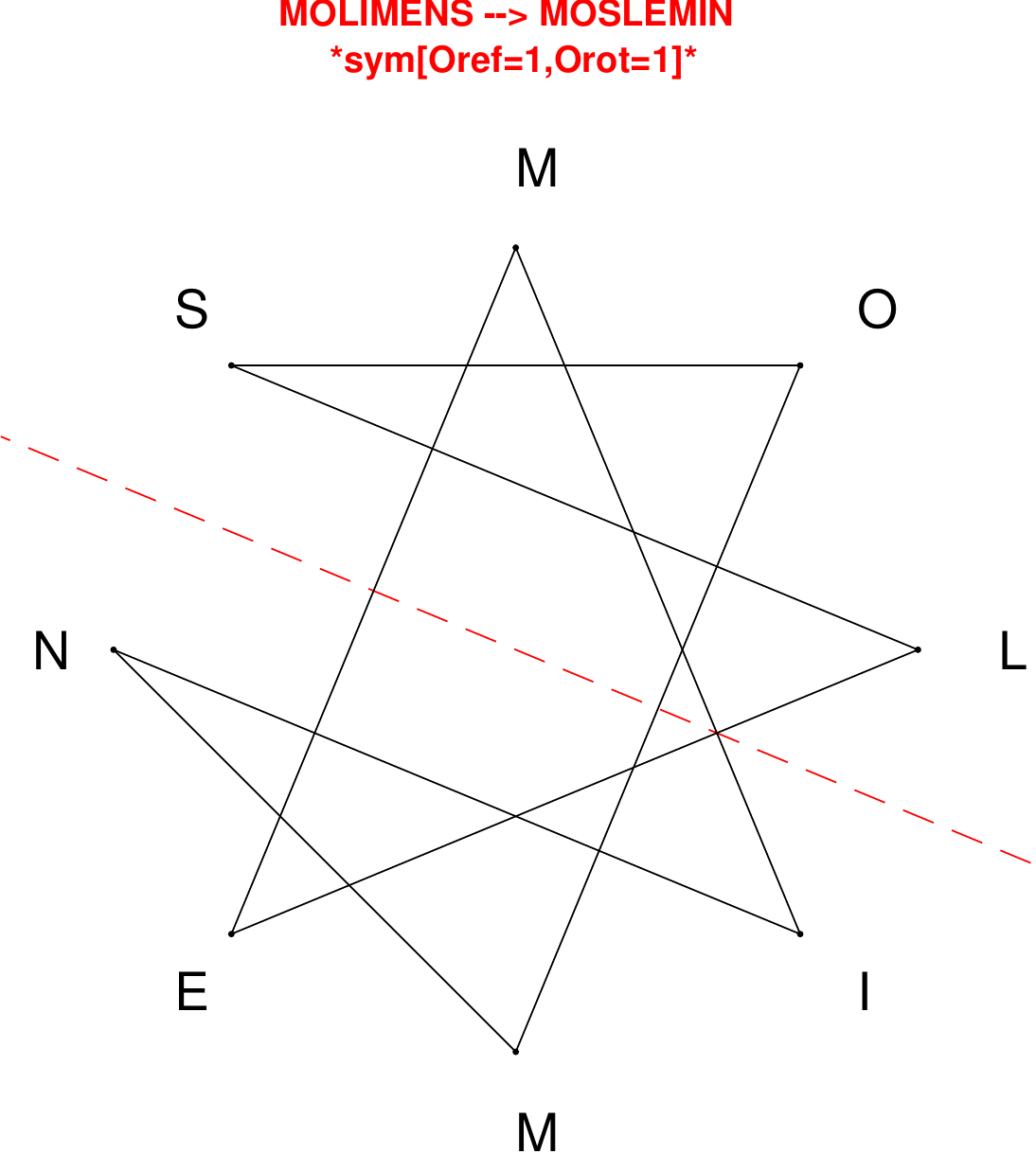}
\end{subfigure}
\hfill
\begin{subfigure}[T]{0.19\textwidth}
\centering
\includegraphics[width=\textwidth]{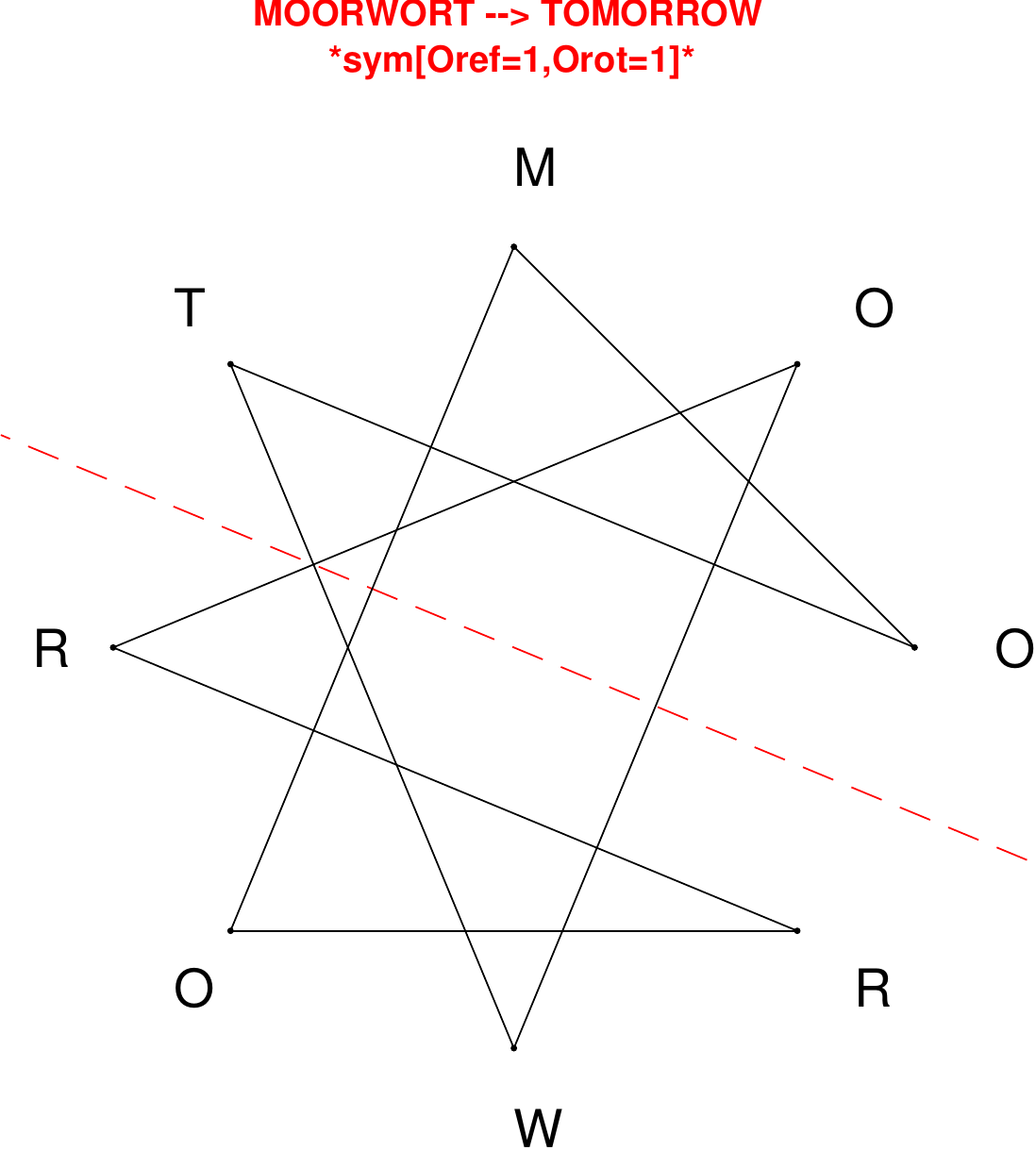}
\end{subfigure}
\hfill
\begin{subfigure}[T]{0.19\textwidth}
\centering
\includegraphics[width=\textwidth]{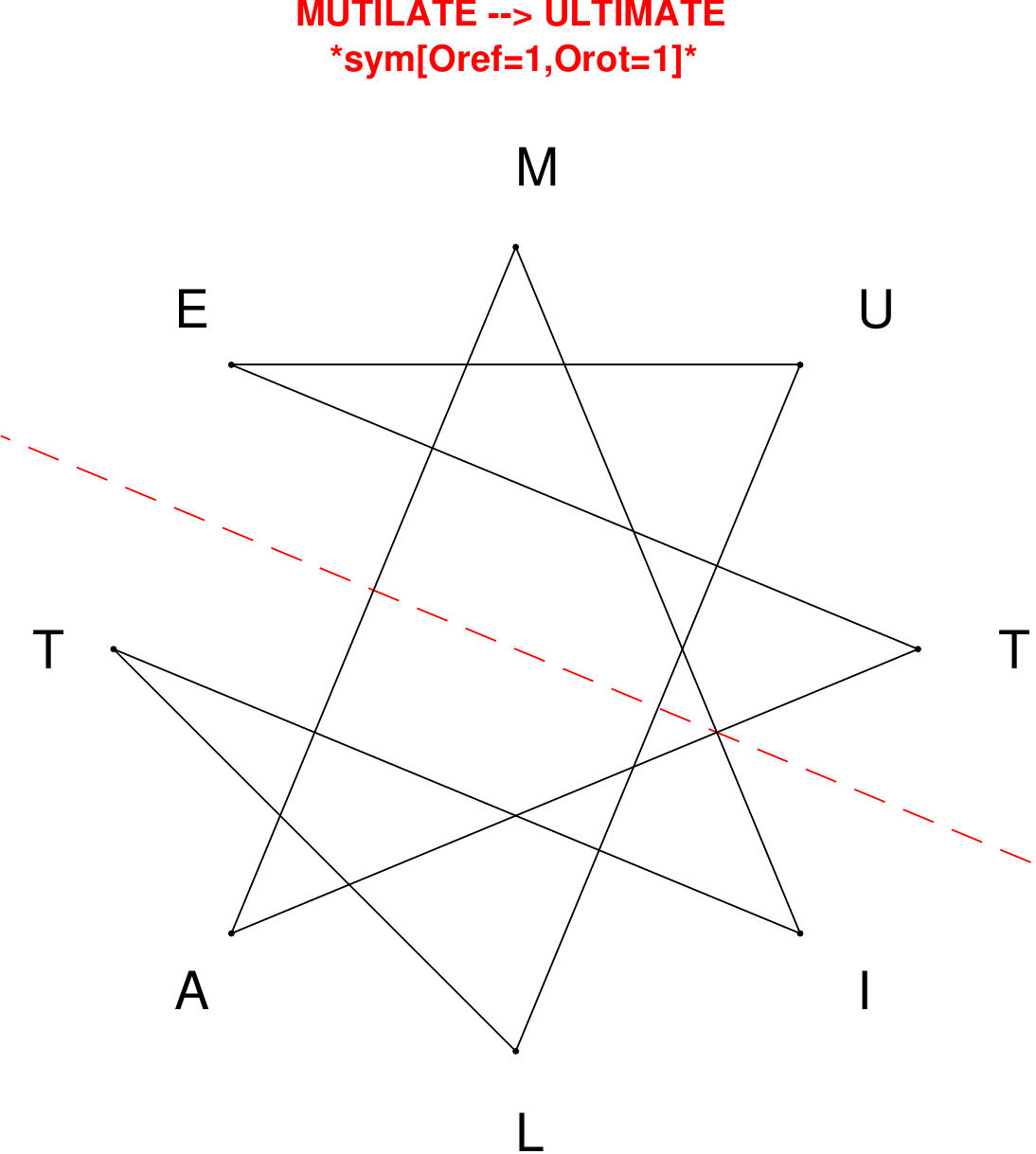}
\end{subfigure}
\hfill
\begin{subfigure}[T]{0.19\textwidth}
\centering
\includegraphics[width=\textwidth]{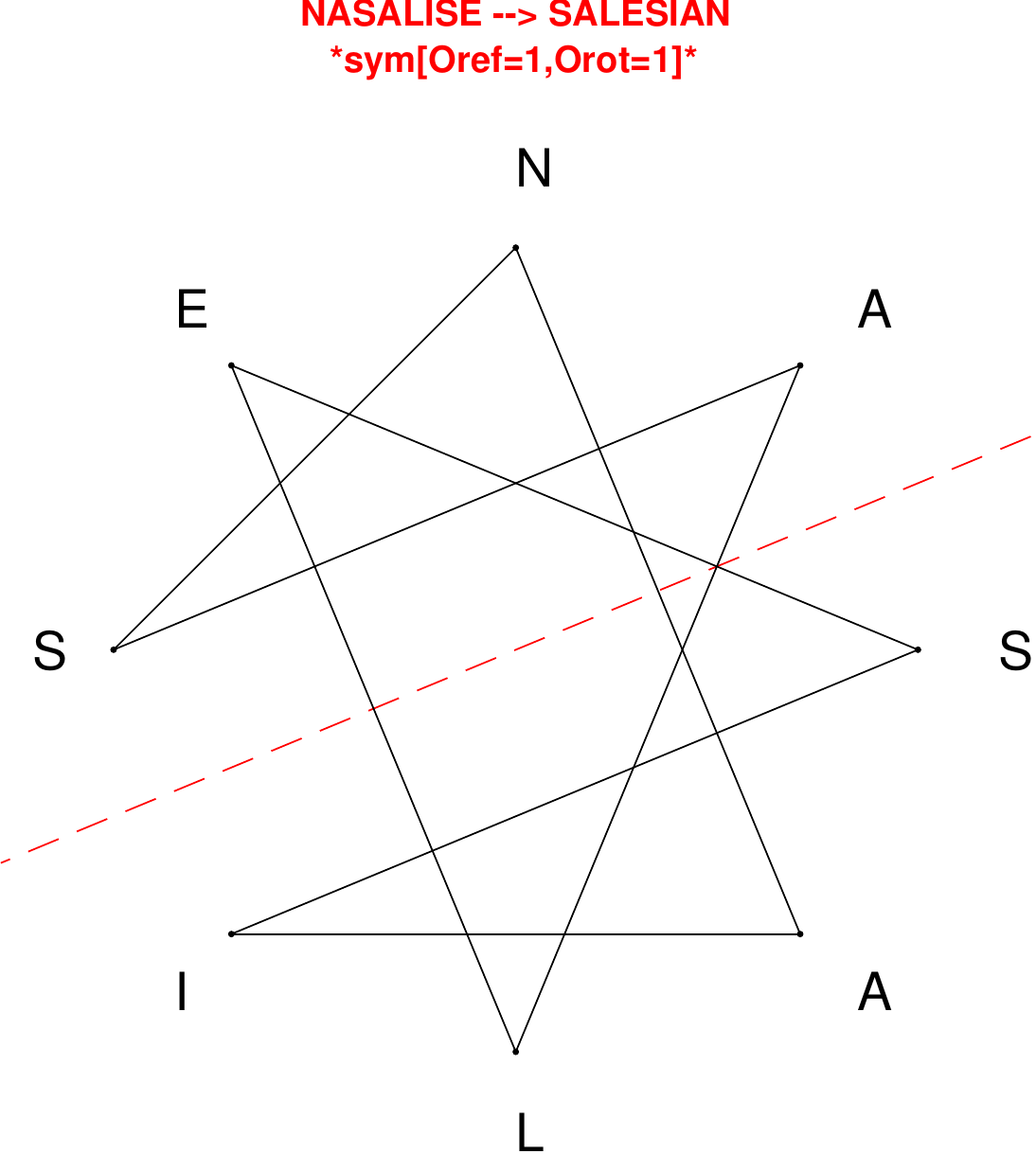}
\end{subfigure}
\hfill
\begin{subfigure}[T]{0.19\textwidth}
\centering
\includegraphics[width=\textwidth]{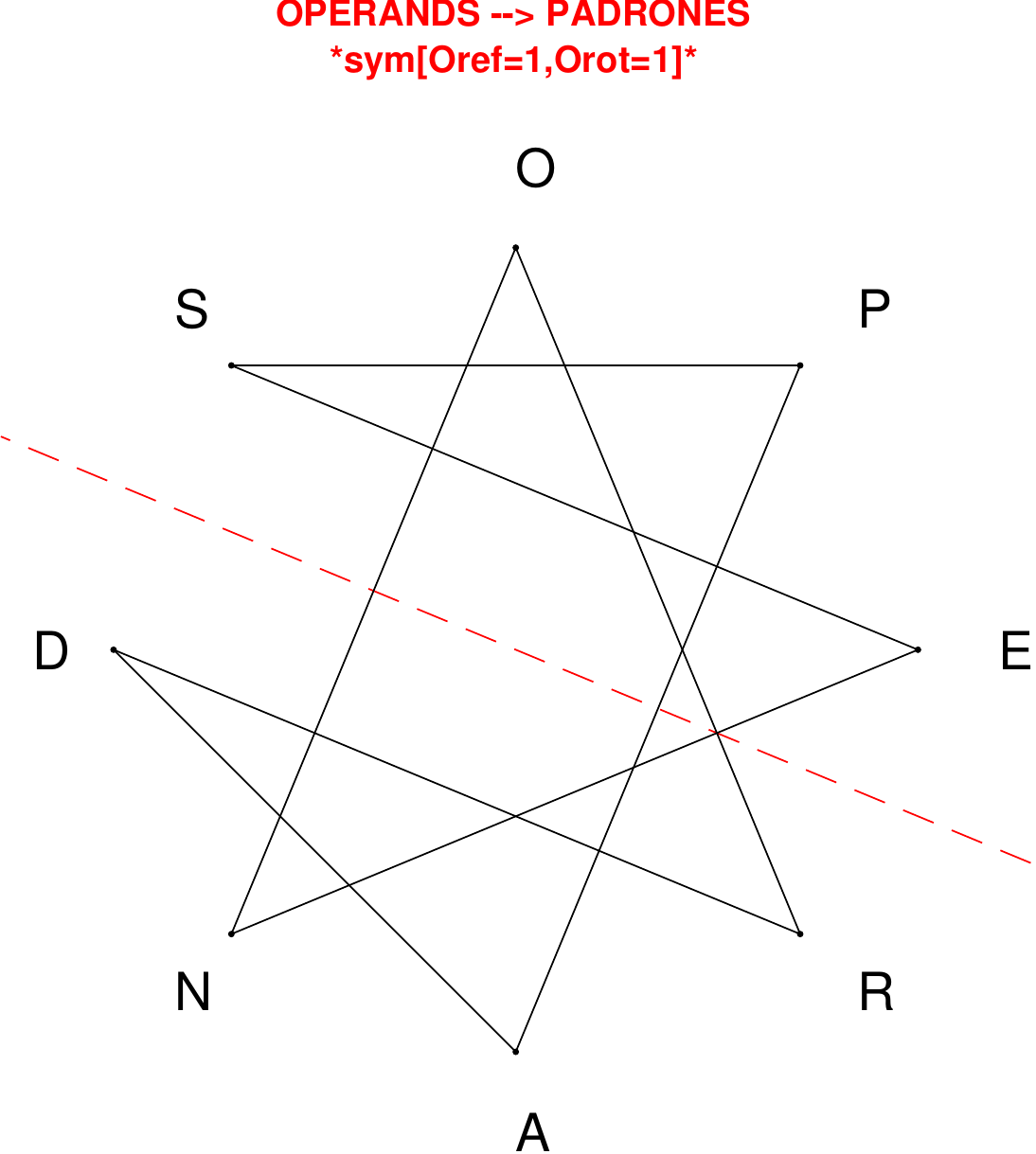}
\end{subfigure}
\end{figure}

\begin{figure}[H]
\centering
\begin{subfigure}[T]{0.19\textwidth}
\centering
\includegraphics[width=\textwidth]{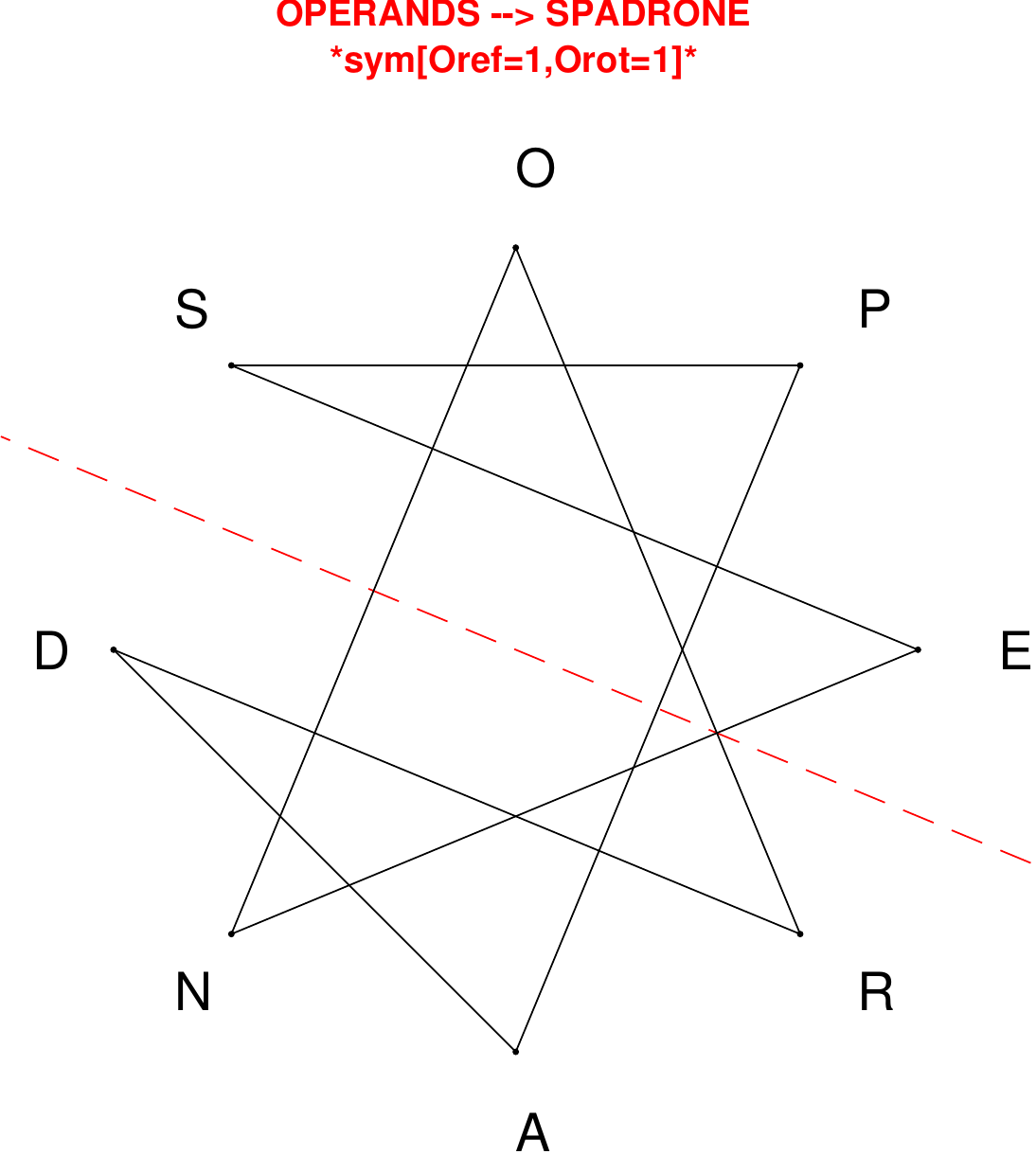}
\end{subfigure}
\hfill
\begin{subfigure}[T]{0.19\textwidth}
\centering
\includegraphics[width=\textwidth]{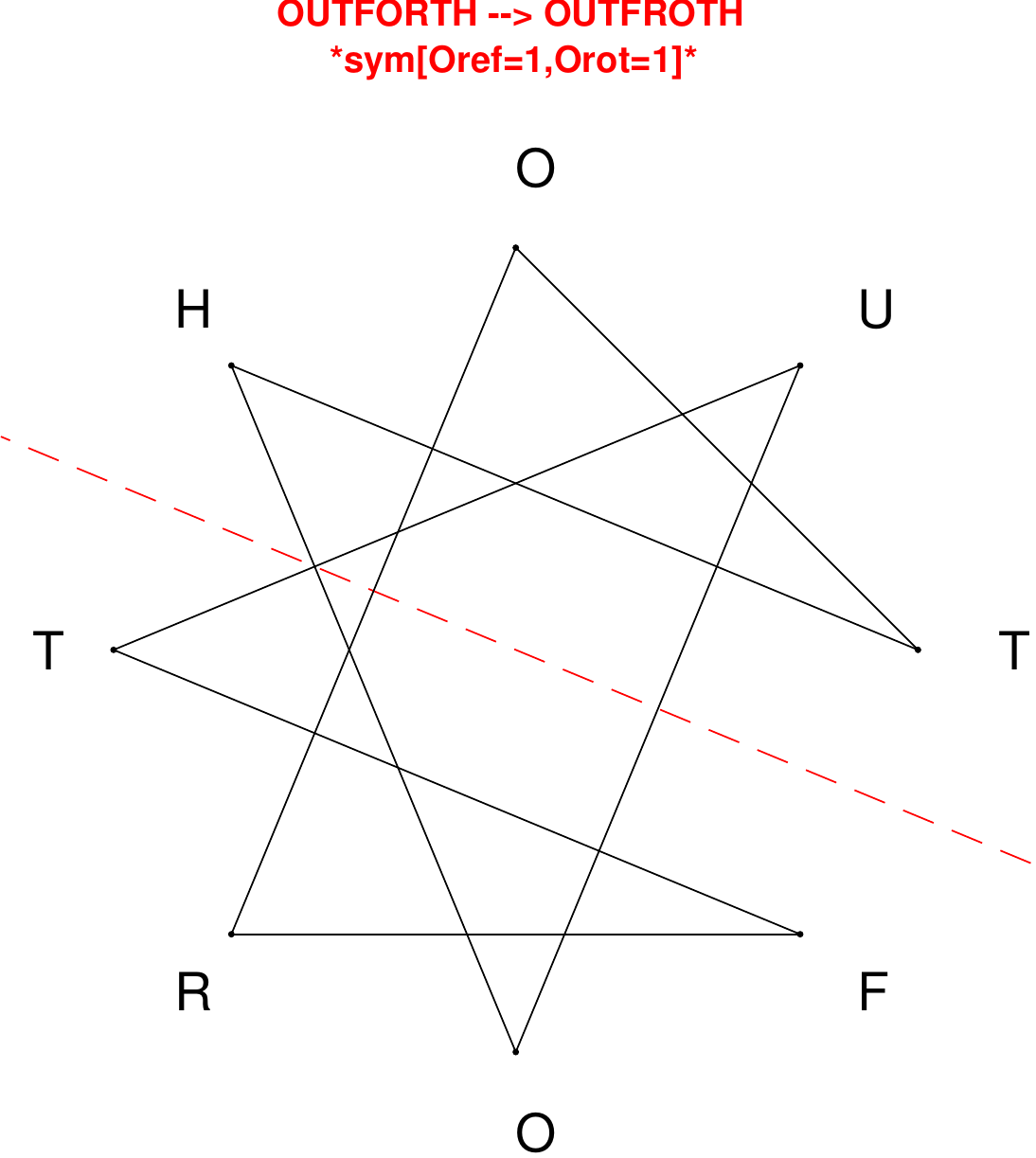}
\end{subfigure}
\hfill
\begin{subfigure}[T]{0.19\textwidth}
\centering
\includegraphics[width=\textwidth]{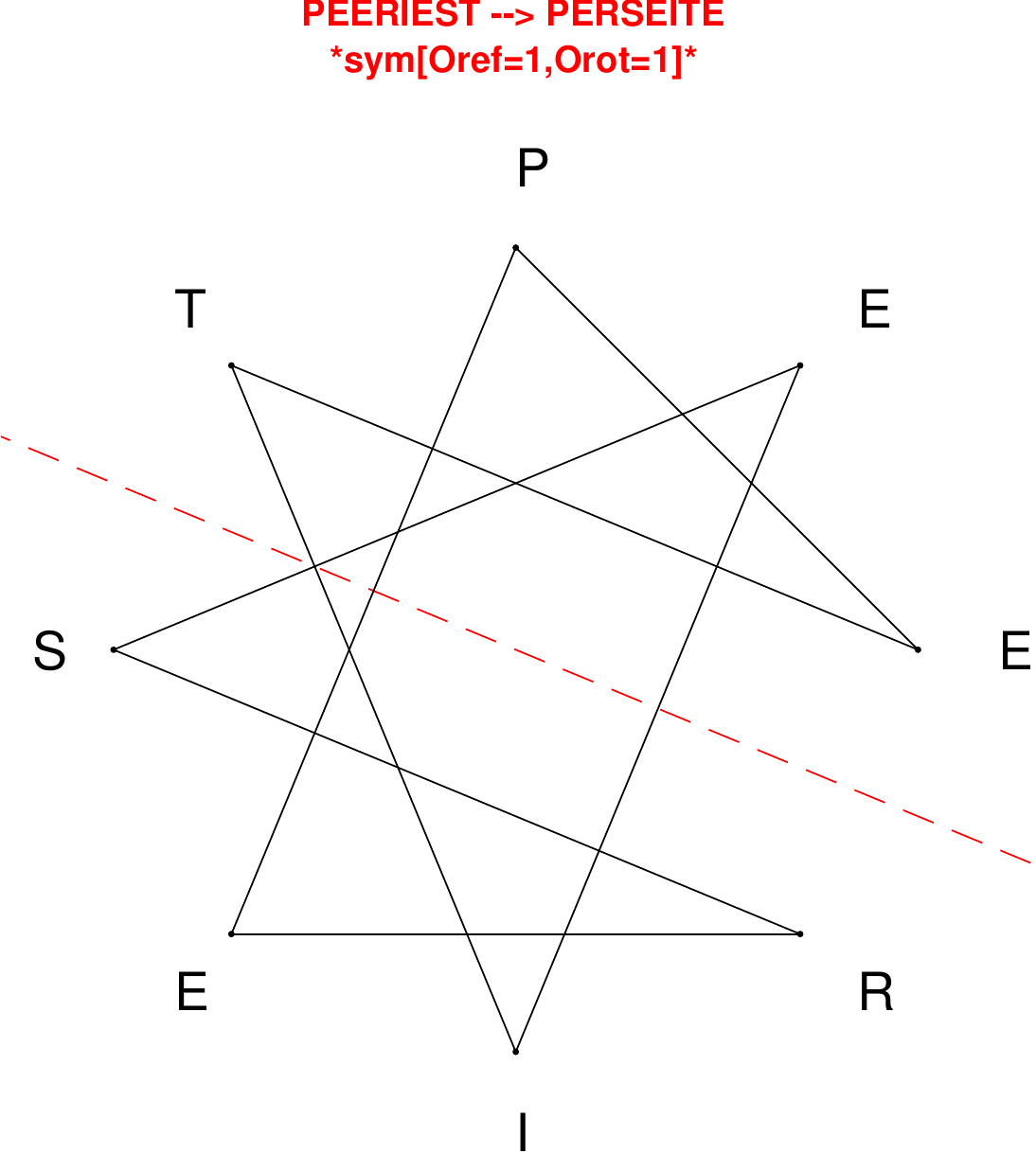}
\end{subfigure}
\hfill
\begin{subfigure}[T]{0.19\textwidth}
\centering
\includegraphics[width=\textwidth]{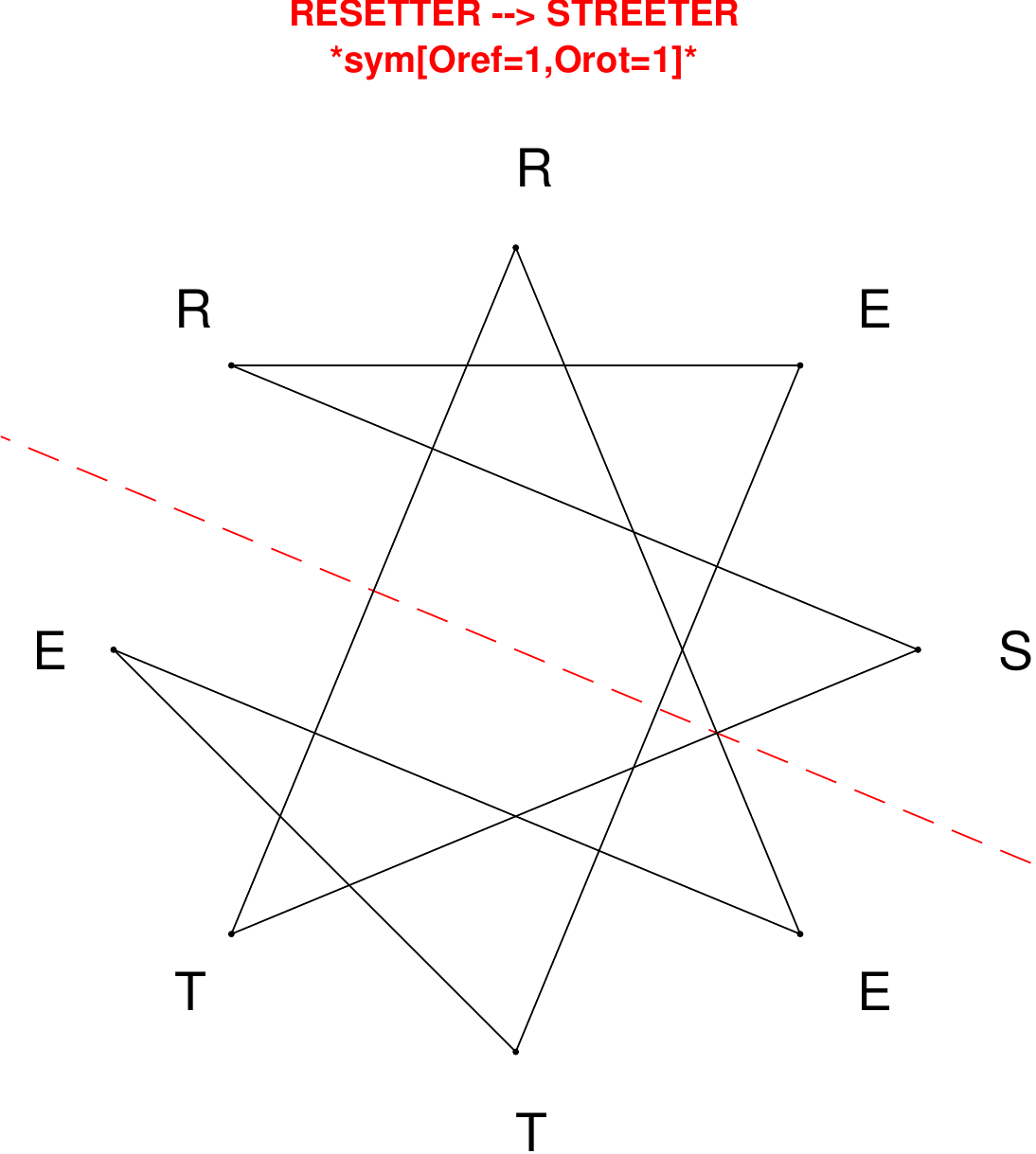}
\end{subfigure}
\hfill
\begin{subfigure}[T]{0.19\textwidth}
\centering
\includegraphics[width=\textwidth]{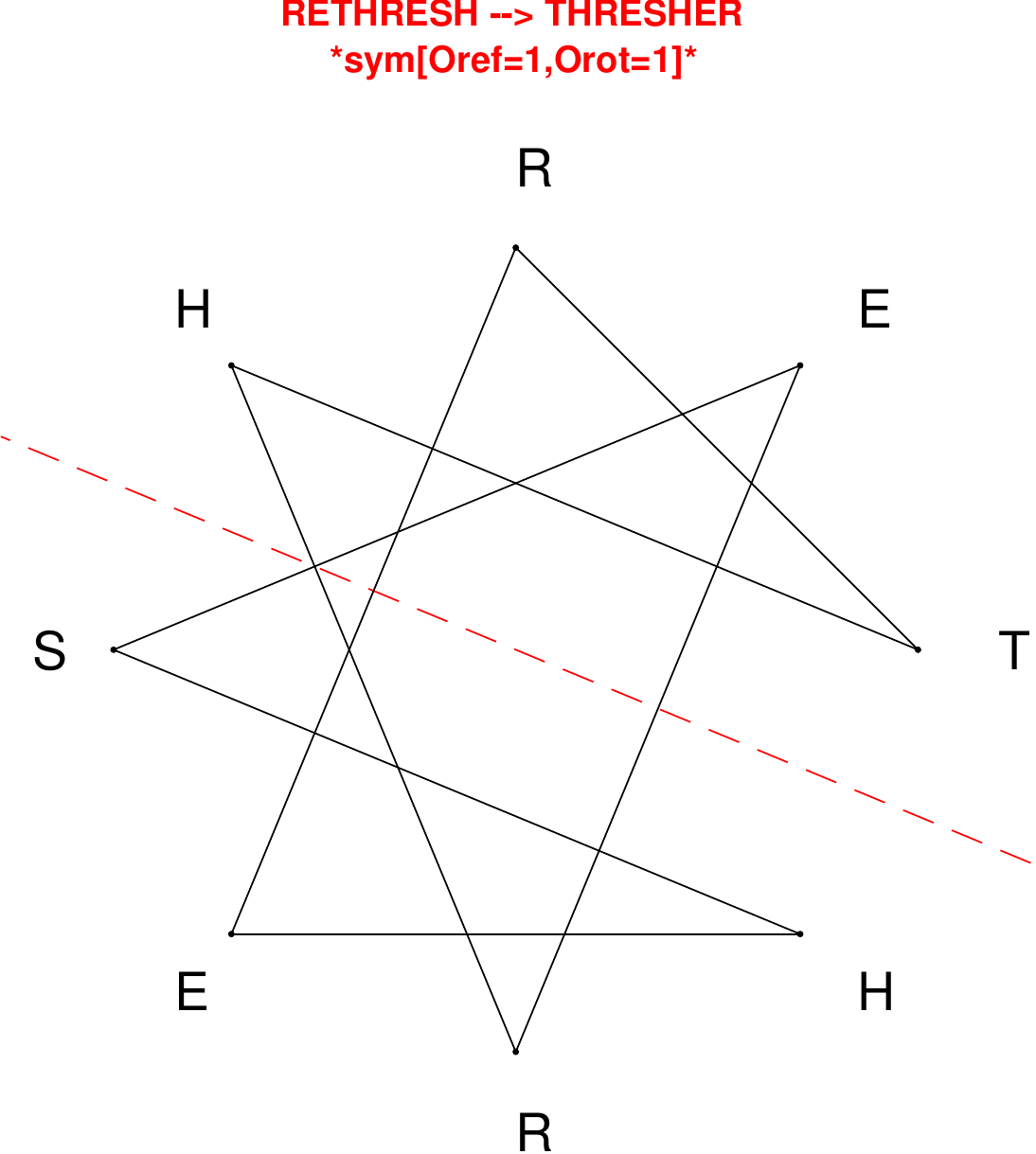}
\end{subfigure}
\end{figure}

\begin{figure}[H]
\centering
\begin{subfigure}[T]{0.19\textwidth}
\centering
\includegraphics[width=\textwidth]{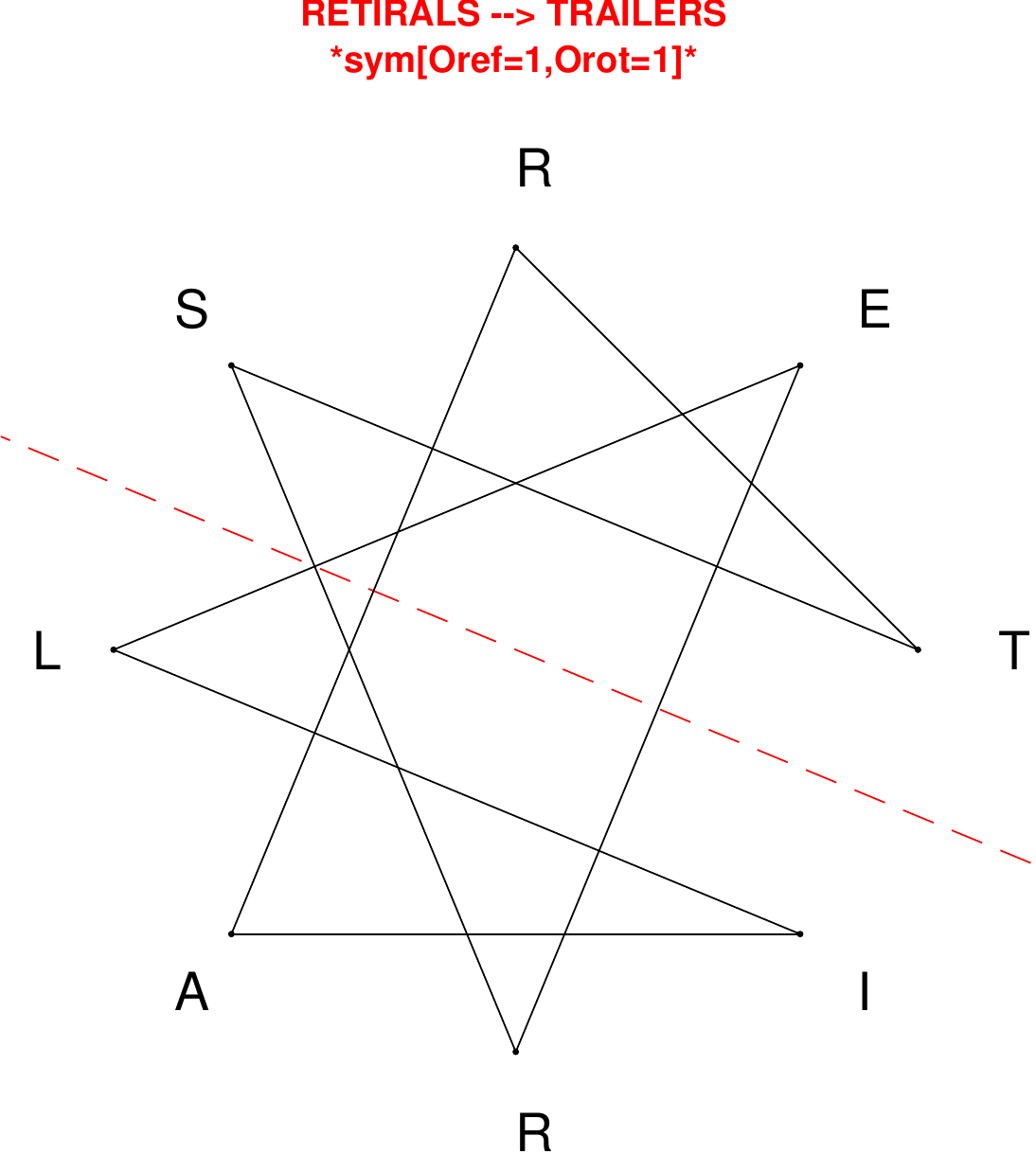}
\end{subfigure}
\hfill
\begin{subfigure}[T]{0.19\textwidth}
\centering
\includegraphics[width=\textwidth]{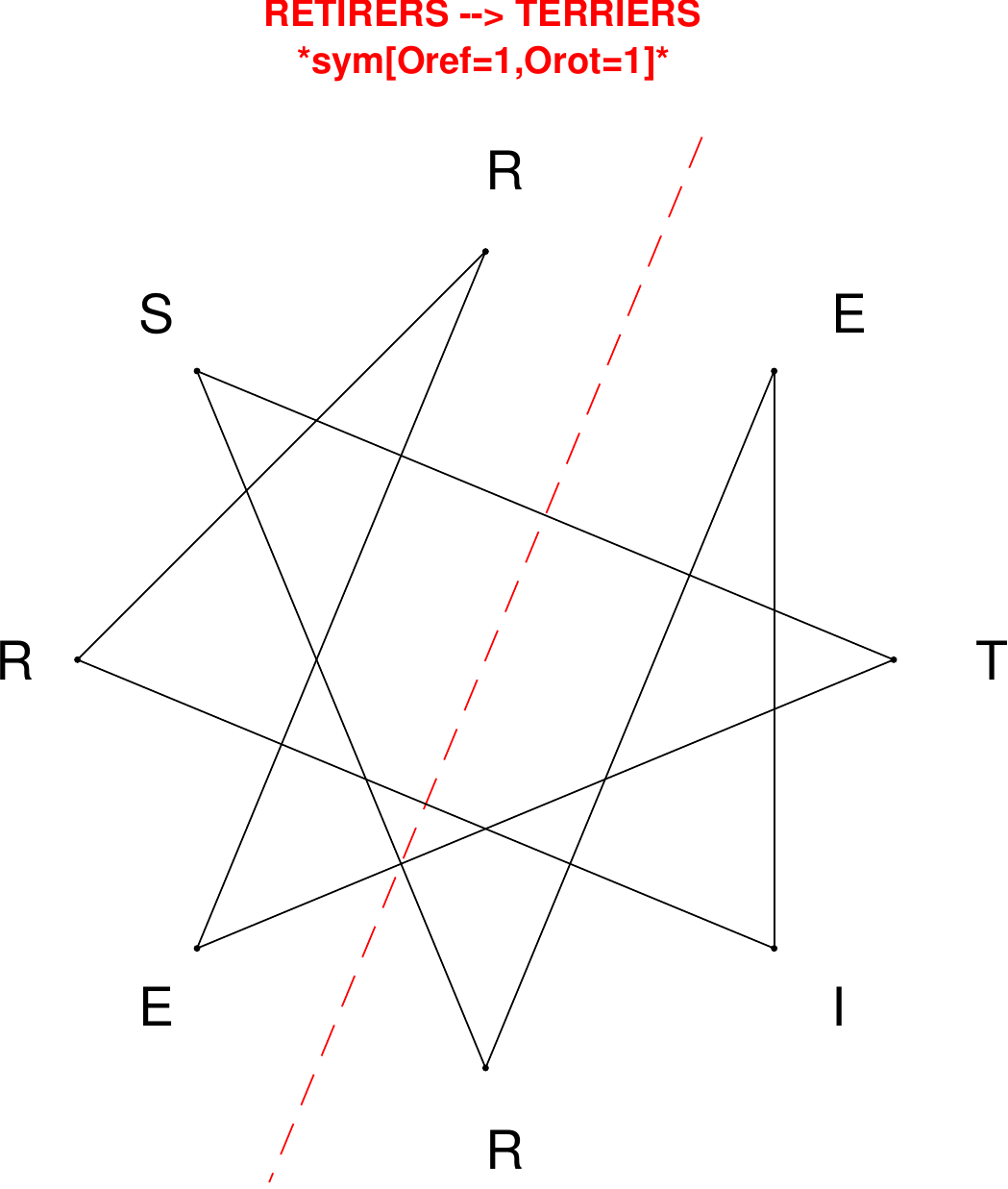}
\end{subfigure}
\hfill
\begin{subfigure}[T]{0.19\textwidth}
\centering
\includegraphics[width=\textwidth]{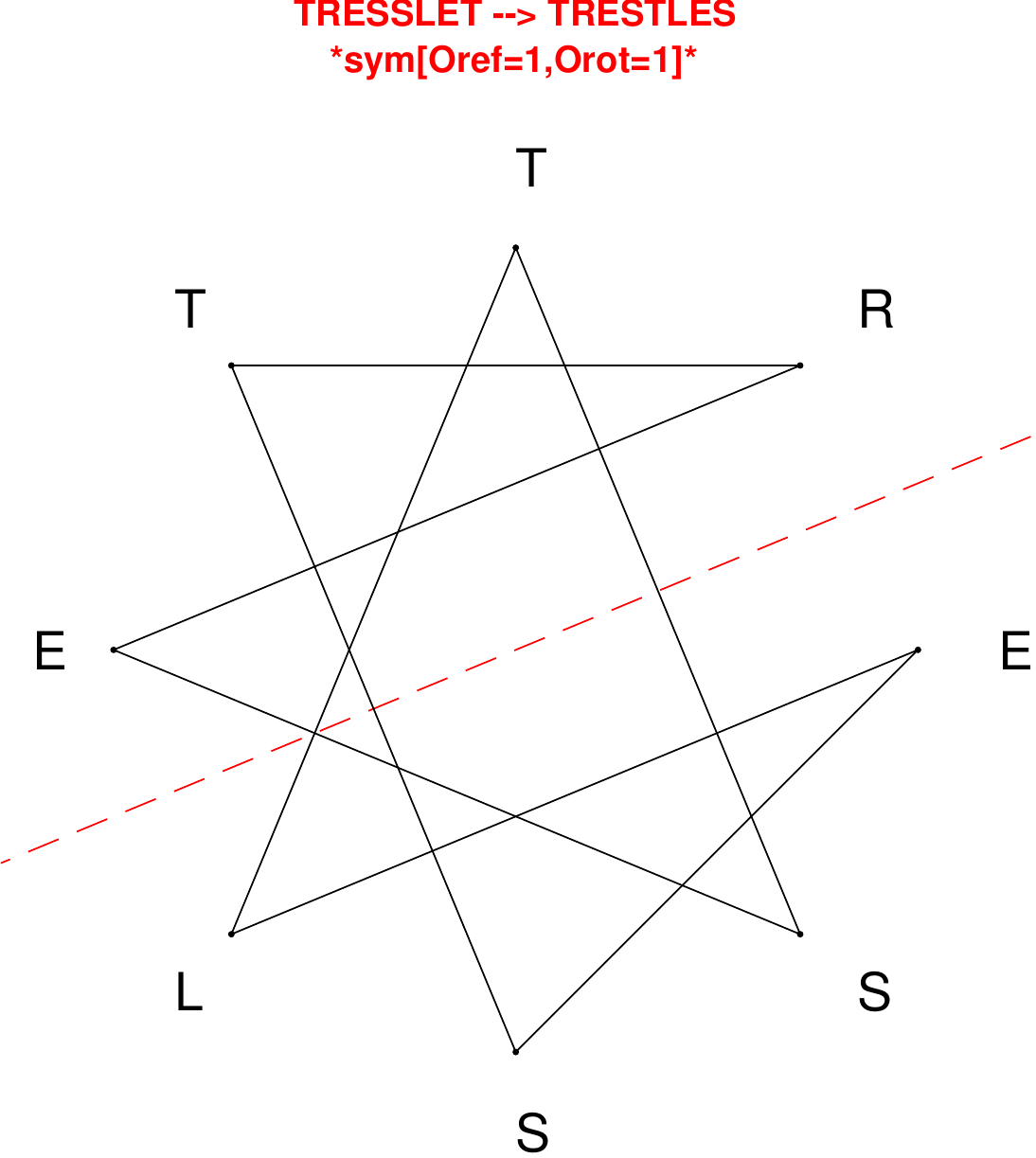}
\end{subfigure}
\hfill
\begin{subfigure}[T]{0.19\textwidth}
\centering
\includegraphics[width=\textwidth]{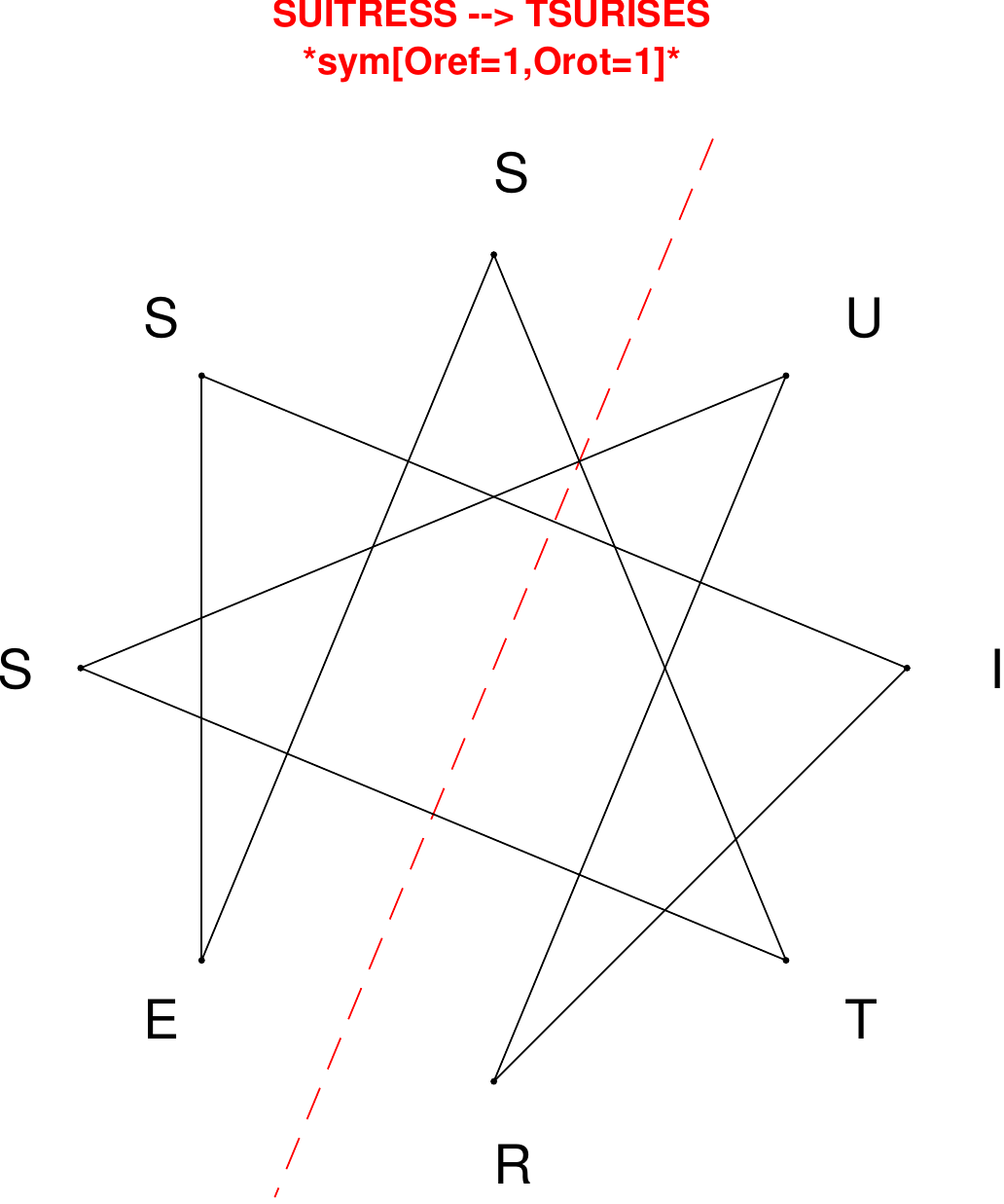}
\end{subfigure}
\hfill
\begin{subfigure}[T]{0.19\textwidth}
\centering
\includegraphics[width=\textwidth]{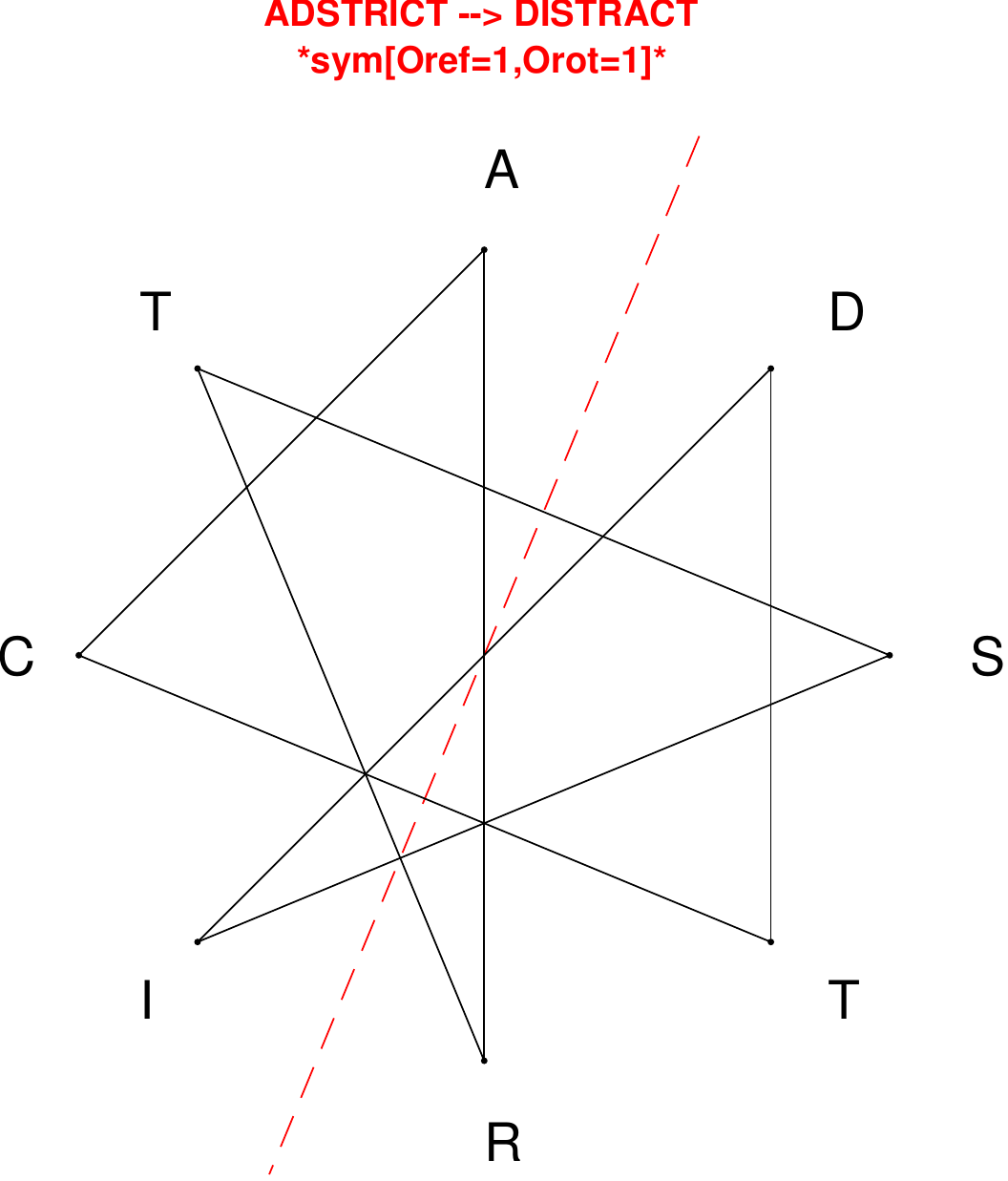}
\end{subfigure}
\end{figure}

\begin{figure}[H]
\centering
\begin{subfigure}[T]{0.19\textwidth}
\centering
\includegraphics[width=\textwidth]{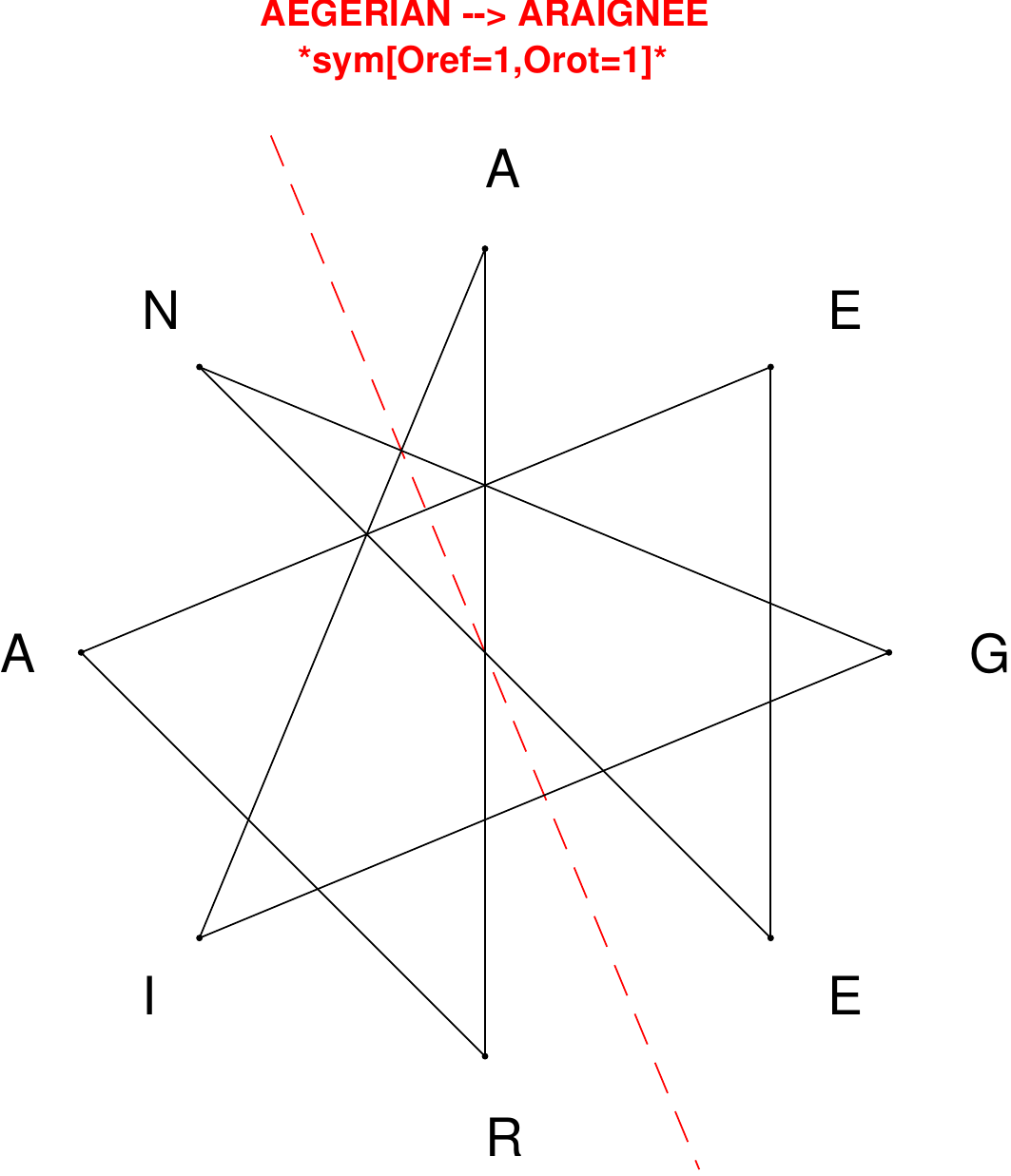}
\end{subfigure}
\hfill
\begin{subfigure}[T]{0.19\textwidth}
\centering
\includegraphics[width=\textwidth]{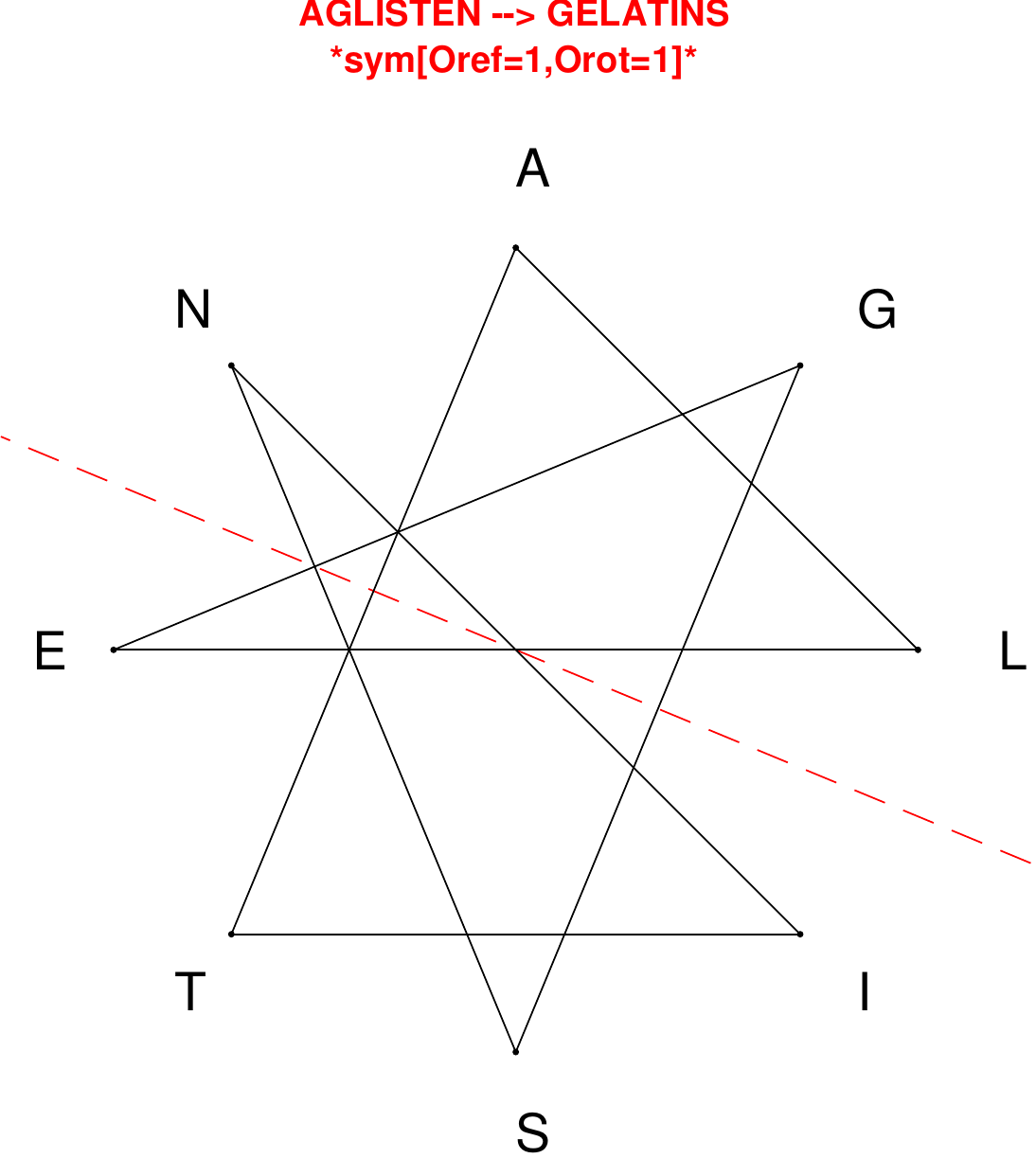}
\end{subfigure}
\hfill
\begin{subfigure}[T]{0.19\textwidth}
\centering
\includegraphics[width=\textwidth]{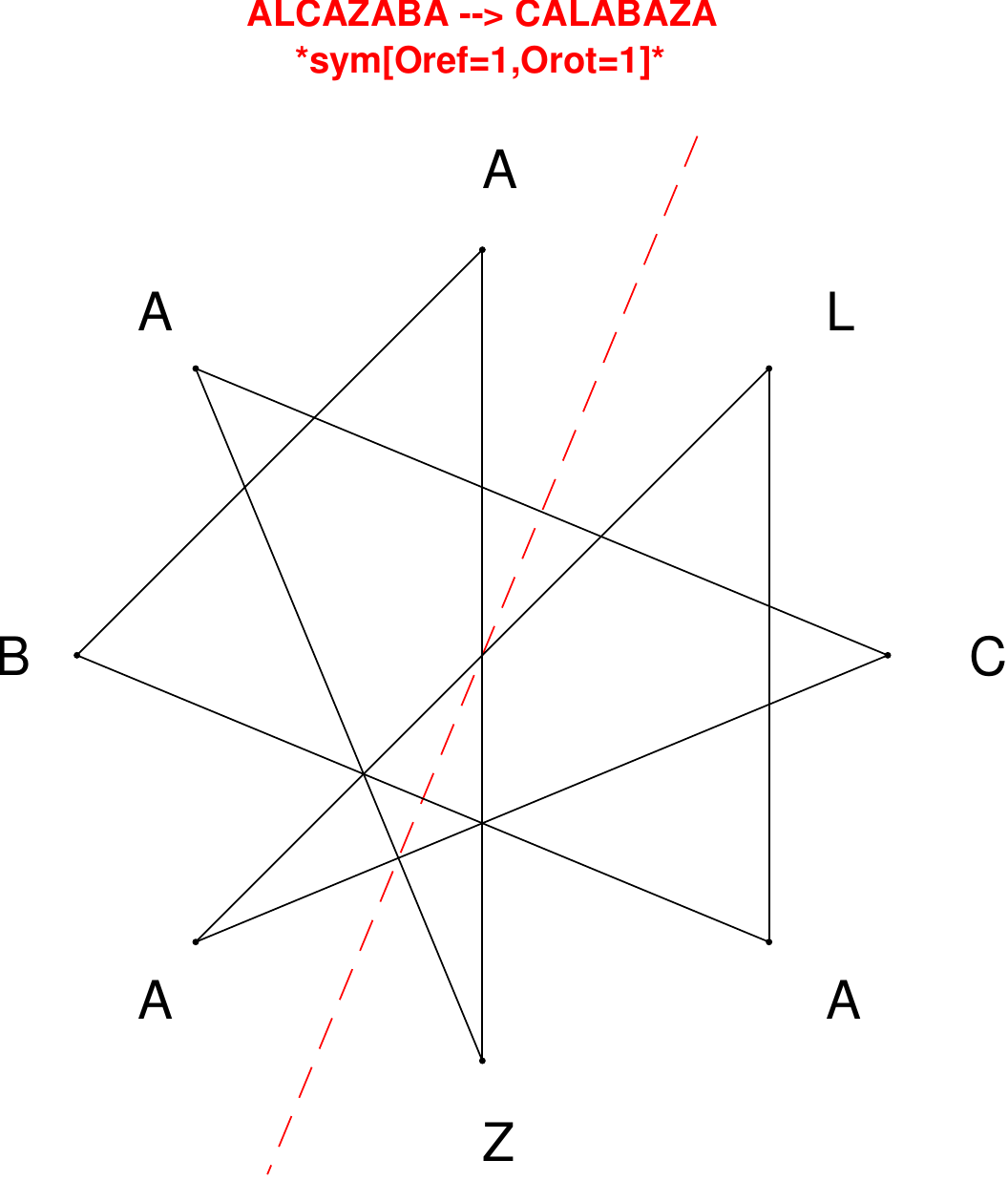}
\end{subfigure}
\hfill
\begin{subfigure}[T]{0.19\textwidth}
\centering
\includegraphics[width=\textwidth]{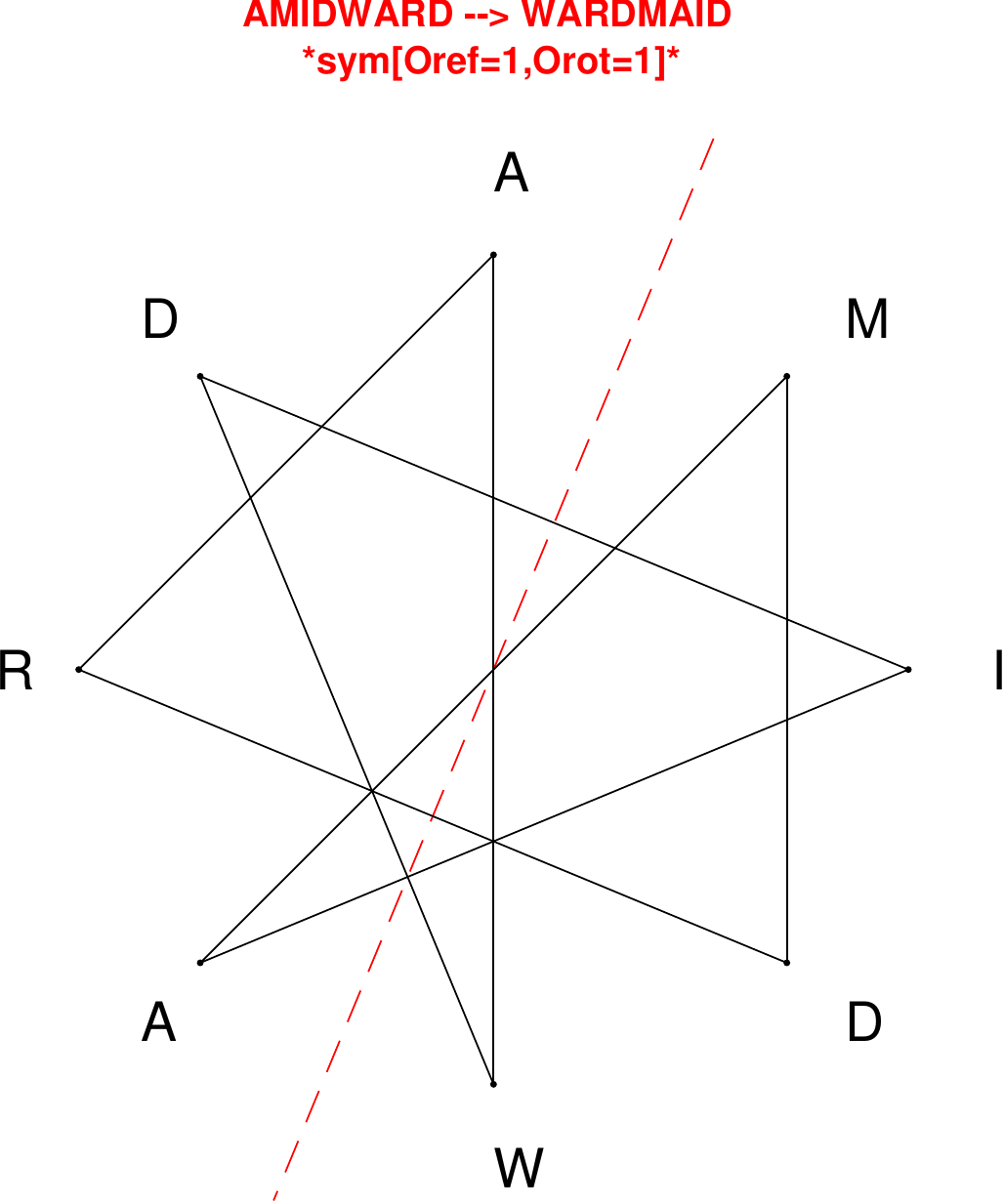}
\end{subfigure}
\hfill
\begin{subfigure}[T]{0.19\textwidth}
\centering
\includegraphics[width=\textwidth]{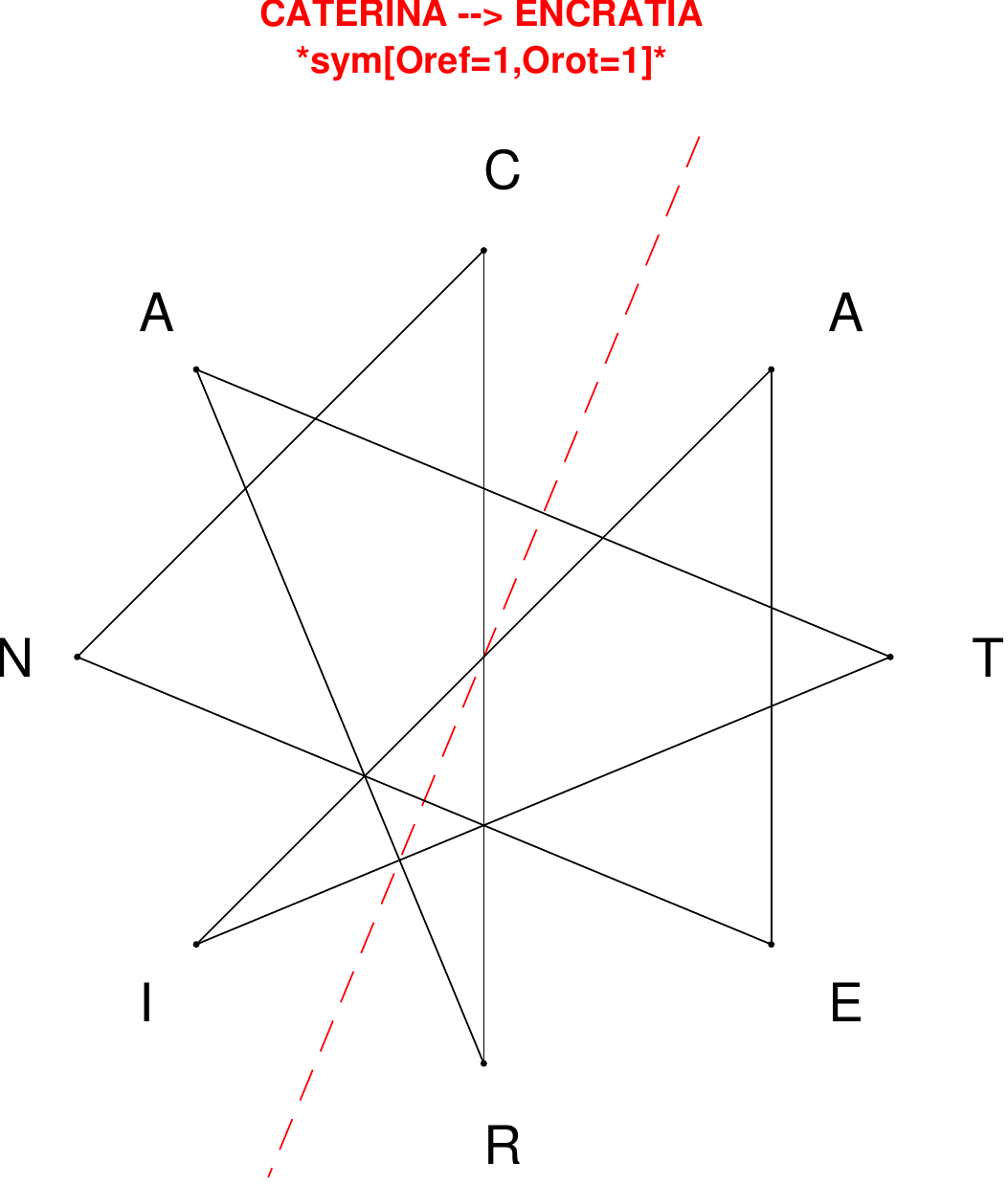}
\end{subfigure}
\end{figure}

\begin{figure}[H]
\centering
\begin{subfigure}[T]{0.19\textwidth}
\centering
\includegraphics[width=\textwidth]{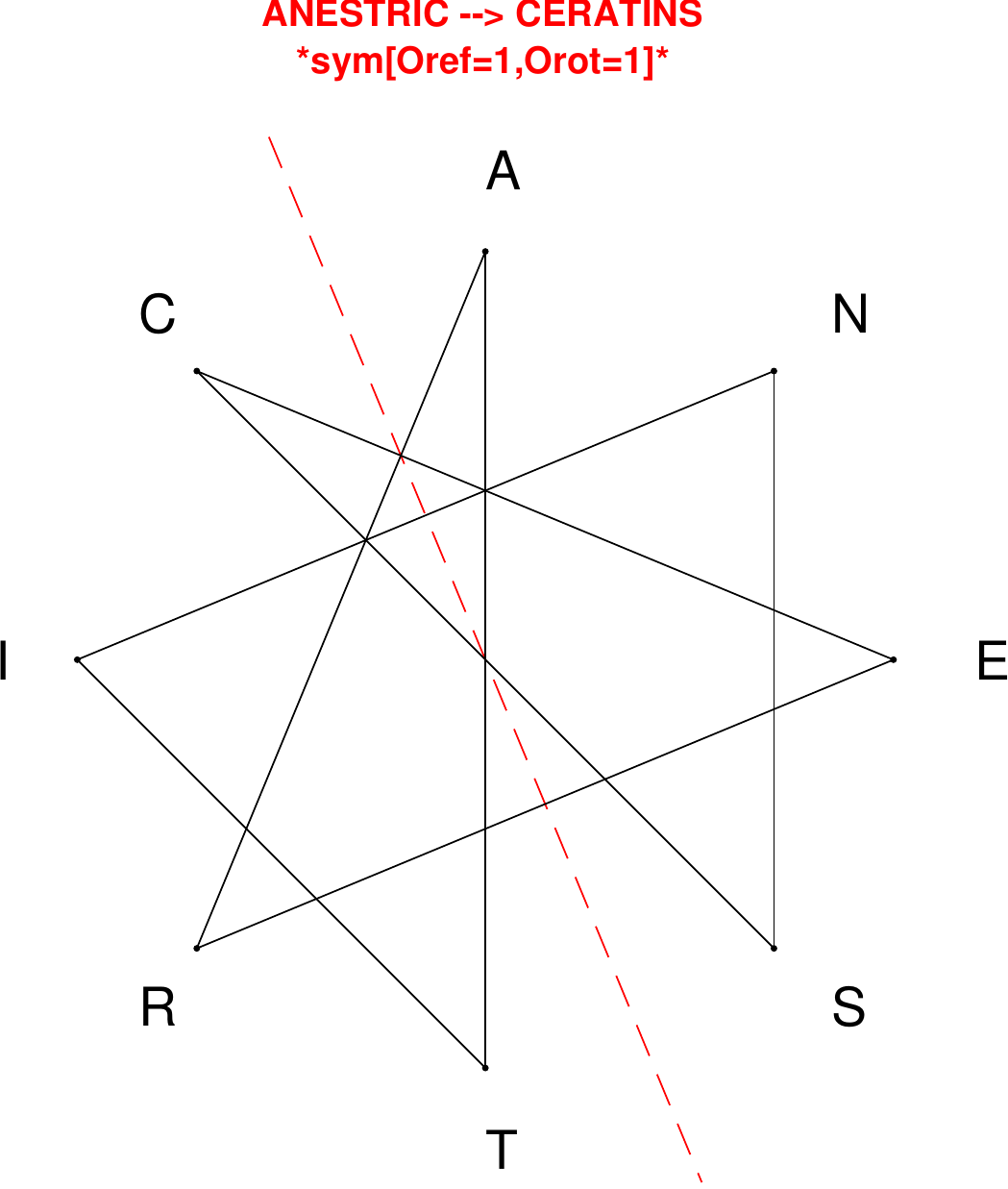}
\end{subfigure}
\hfill
\begin{subfigure}[T]{0.19\textwidth}
\centering
\includegraphics[width=\textwidth]{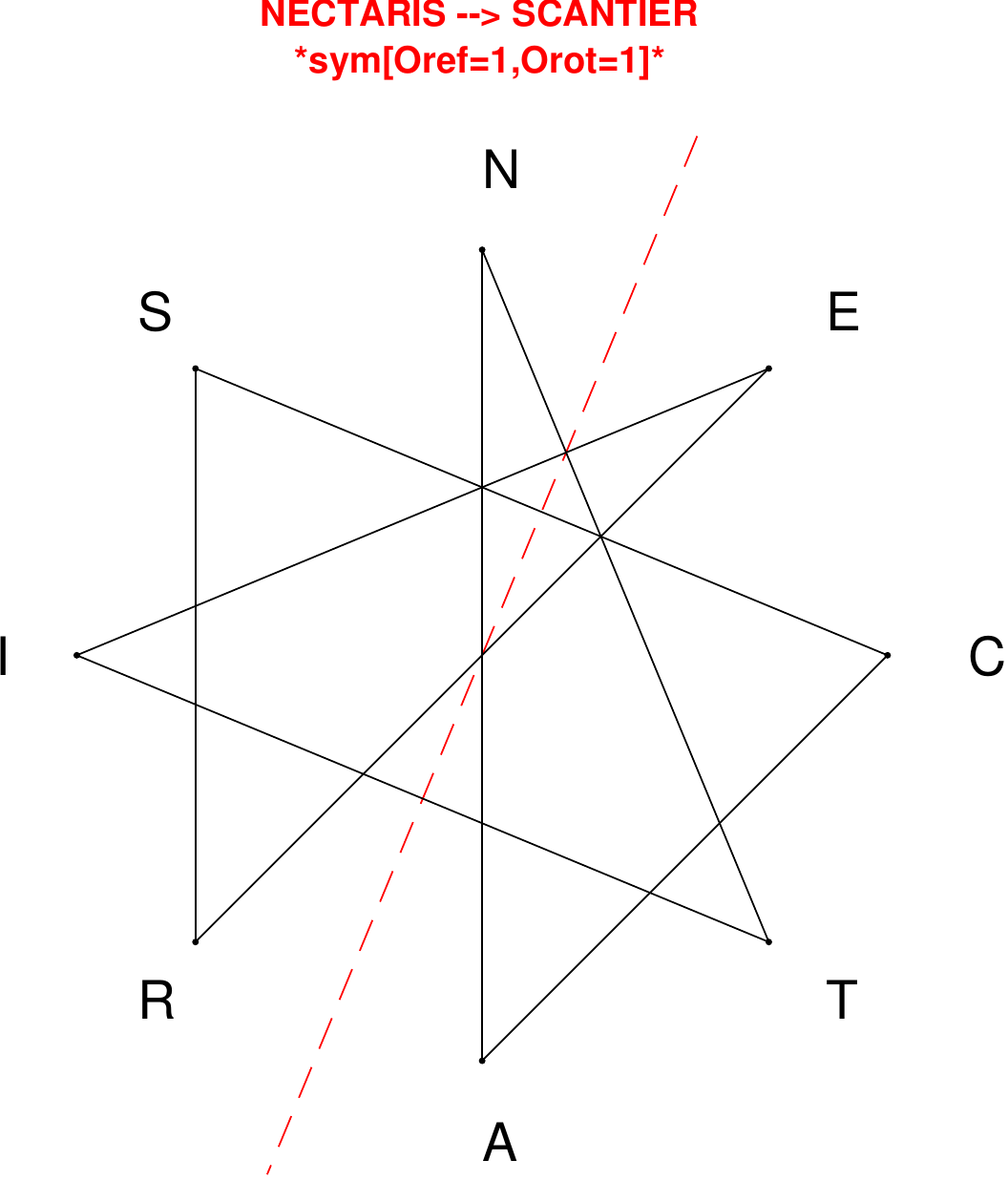}
\end{subfigure}
\hfill
\begin{subfigure}[T]{0.19\textwidth}
\centering
\includegraphics[width=\textwidth]{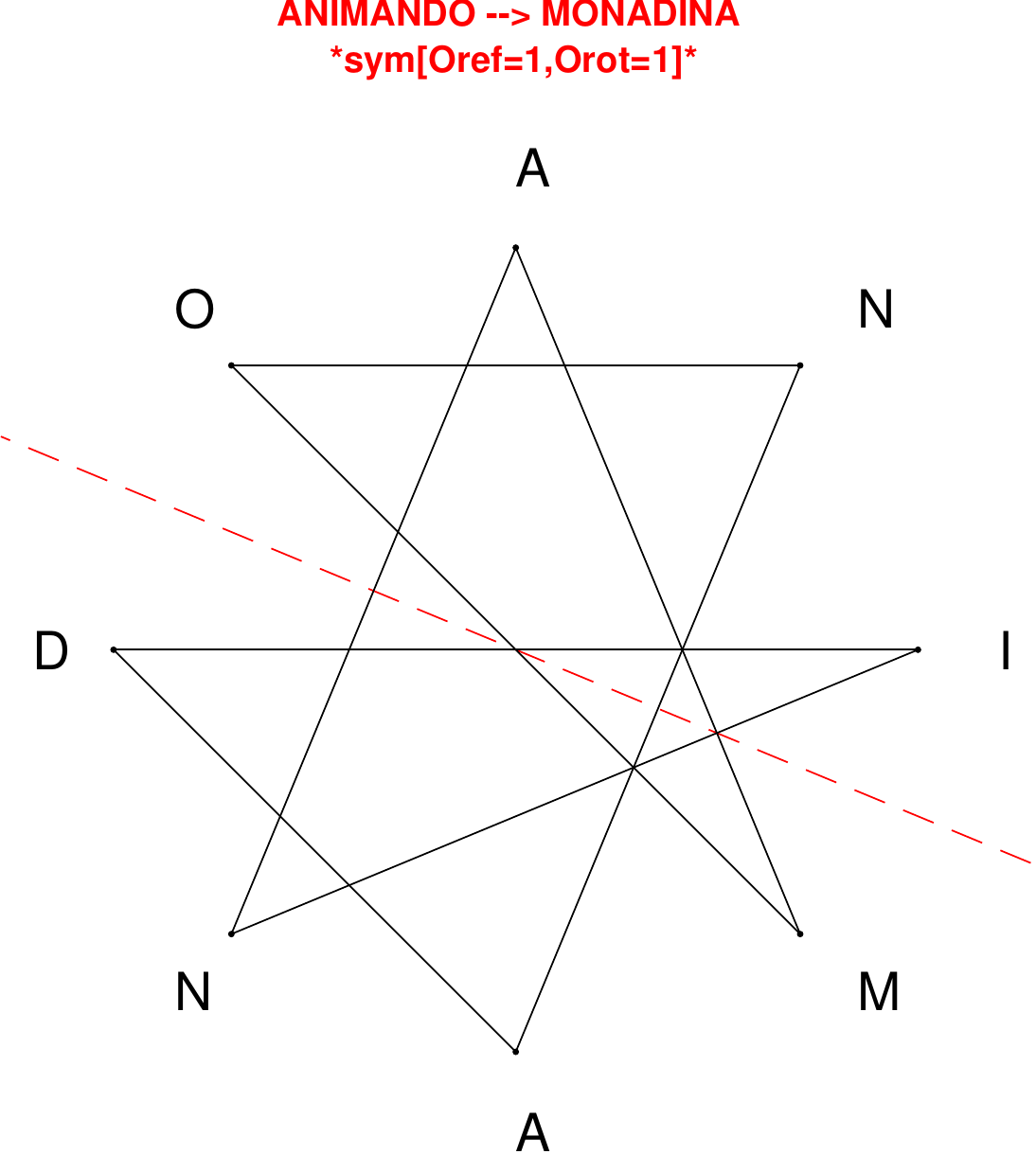}
\end{subfigure}
\hfill
\begin{subfigure}[T]{0.19\textwidth}
\centering
\includegraphics[width=\textwidth]{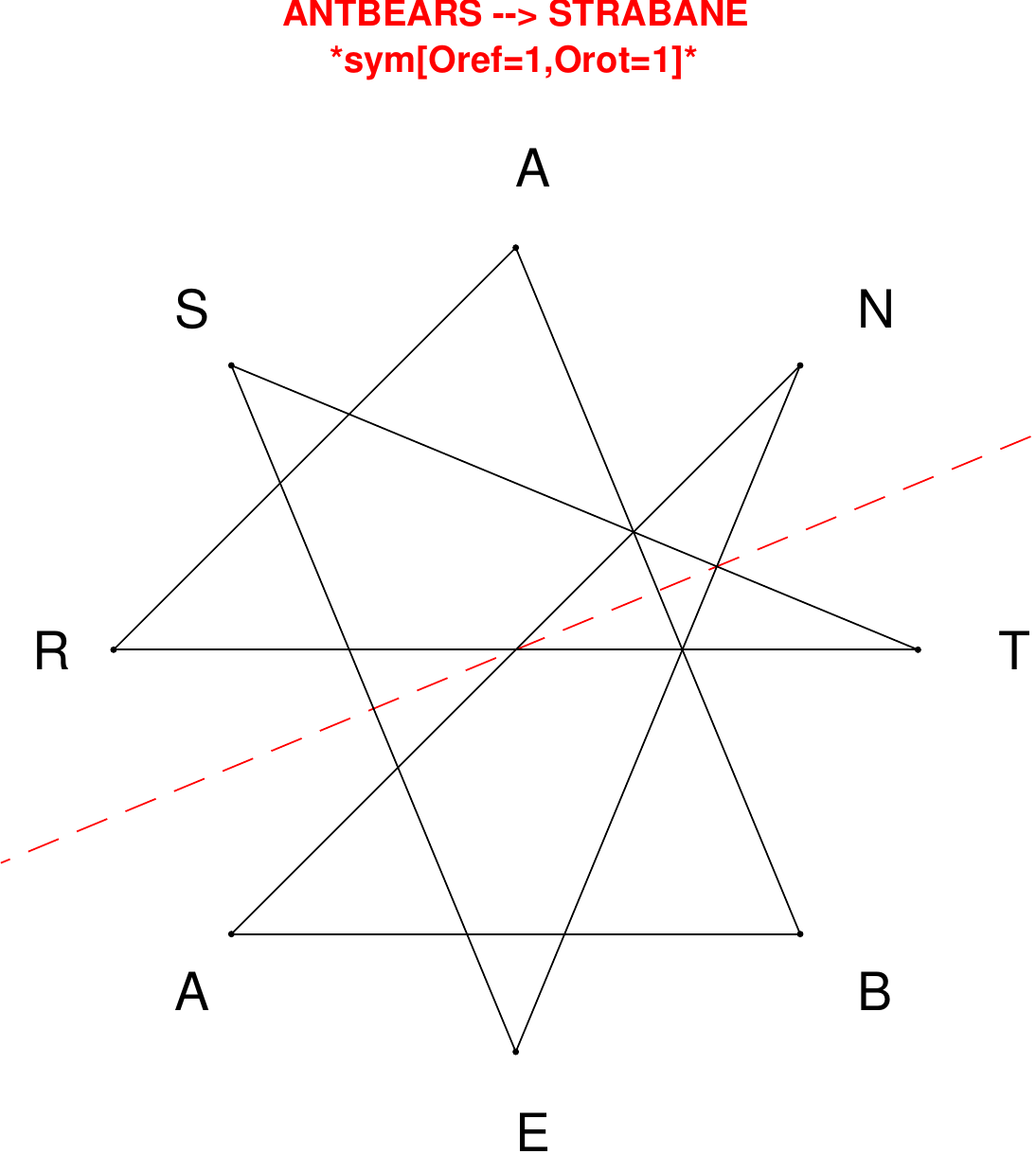}
\end{subfigure}
\hfill
\begin{subfigure}[T]{0.19\textwidth}
\centering
\includegraphics[width=\textwidth]{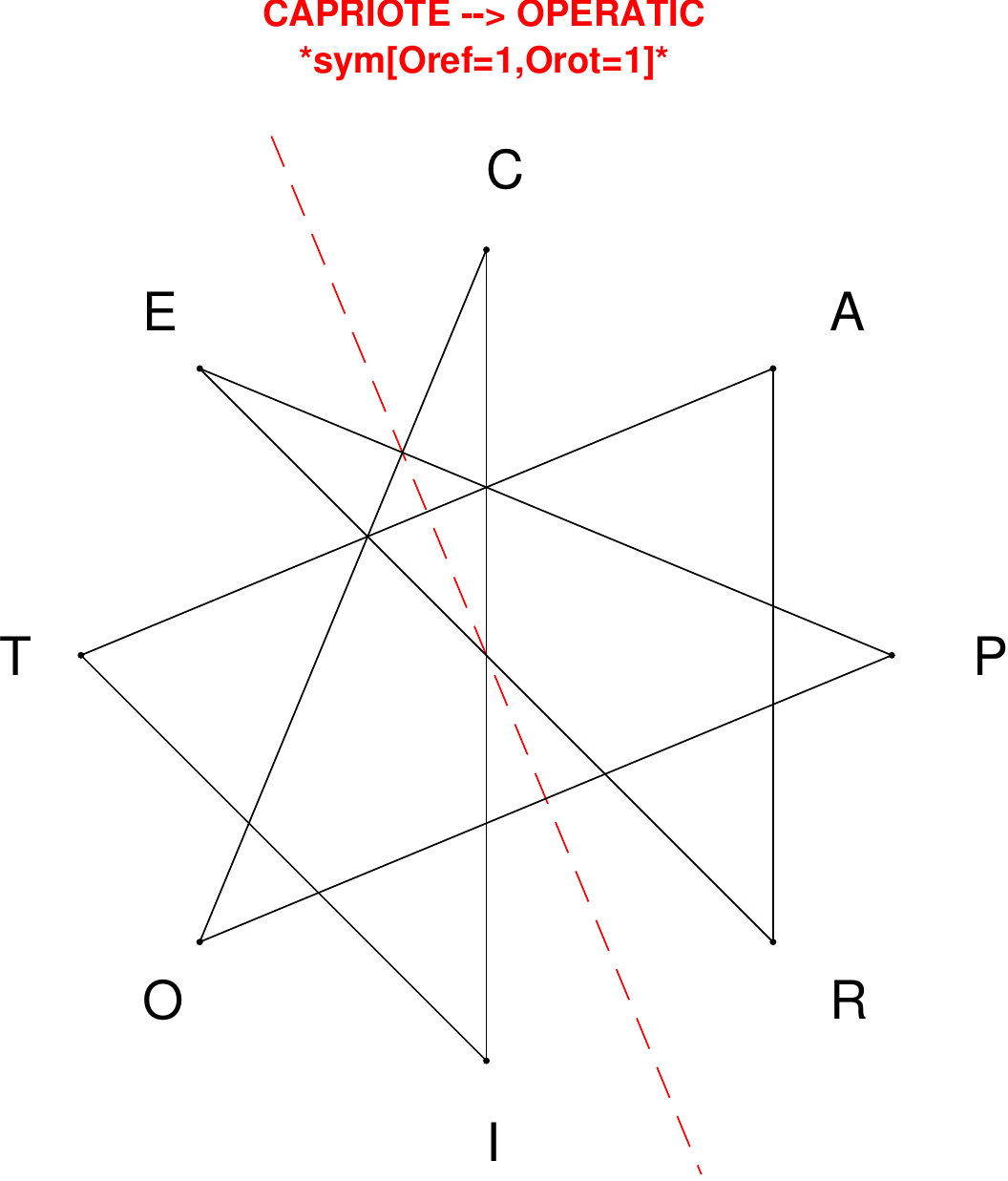}
\end{subfigure}
\end{figure}

\begin{figure}[H]
\centering
\begin{subfigure}[T]{0.19\textwidth}
\centering
\includegraphics[width=\textwidth]{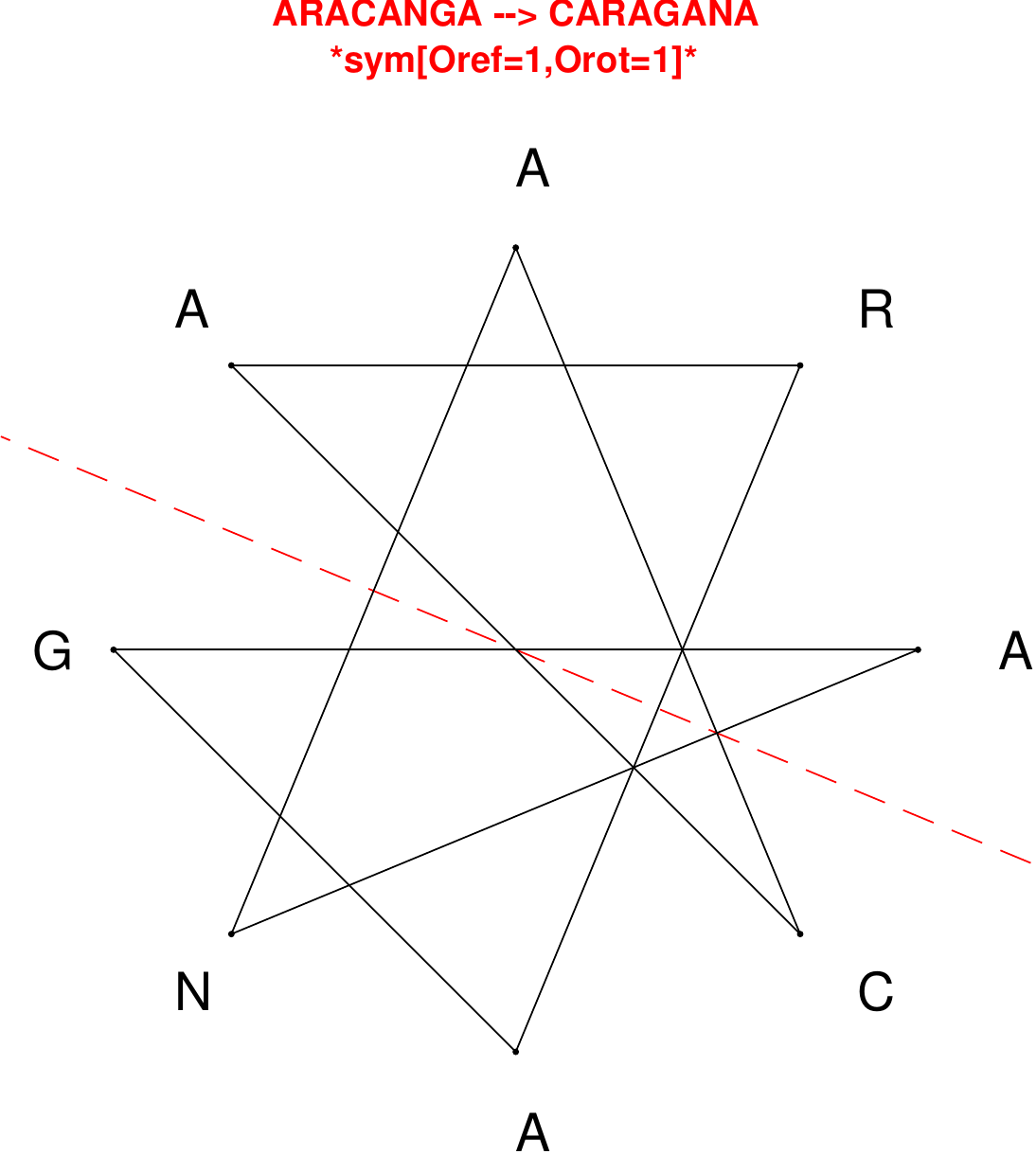}
\end{subfigure}
\hfill
\begin{subfigure}[T]{0.19\textwidth}
\centering
\includegraphics[width=\textwidth]{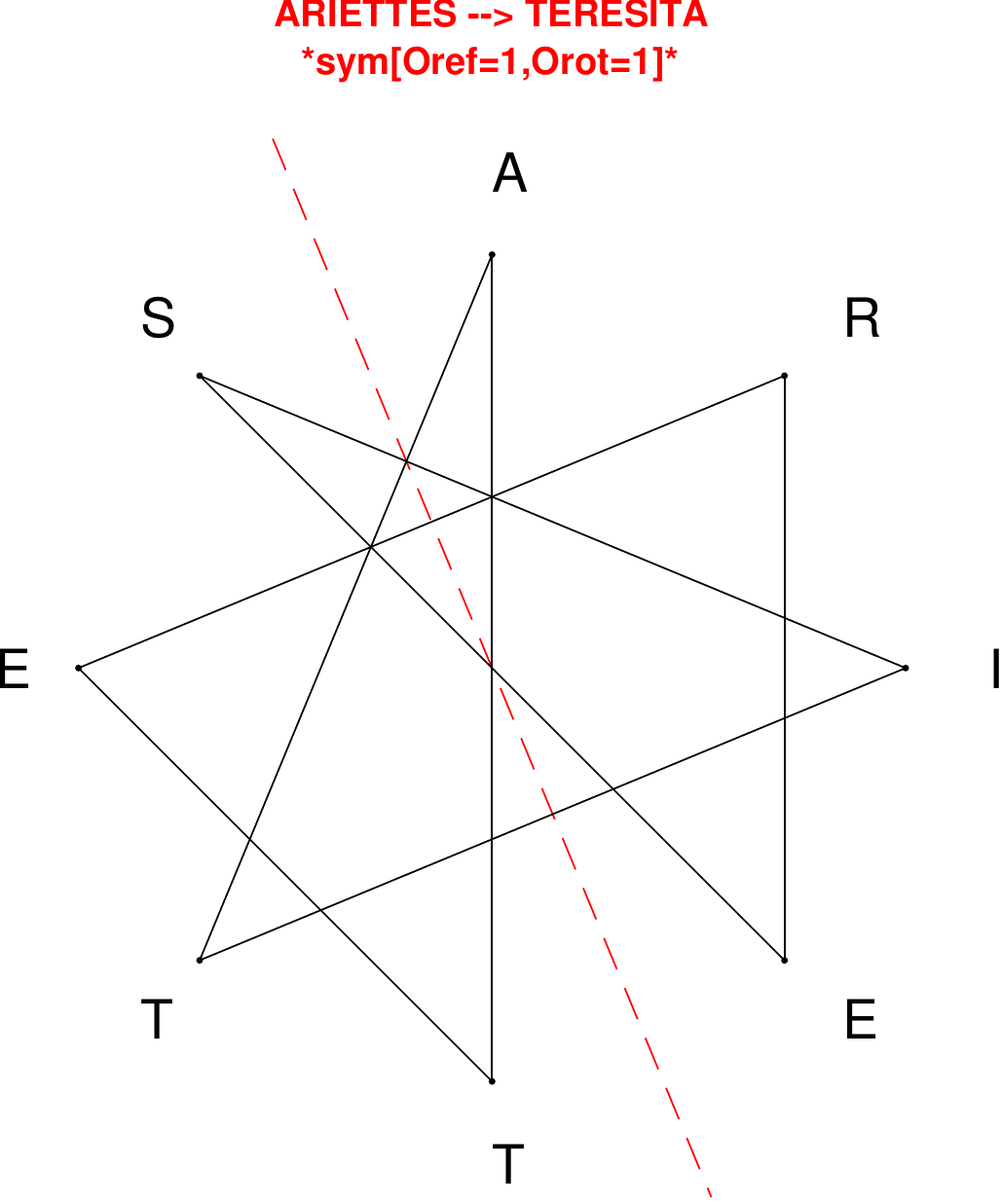}
\end{subfigure}
\hfill
\begin{subfigure}[T]{0.19\textwidth}
\centering
\includegraphics[width=\textwidth]{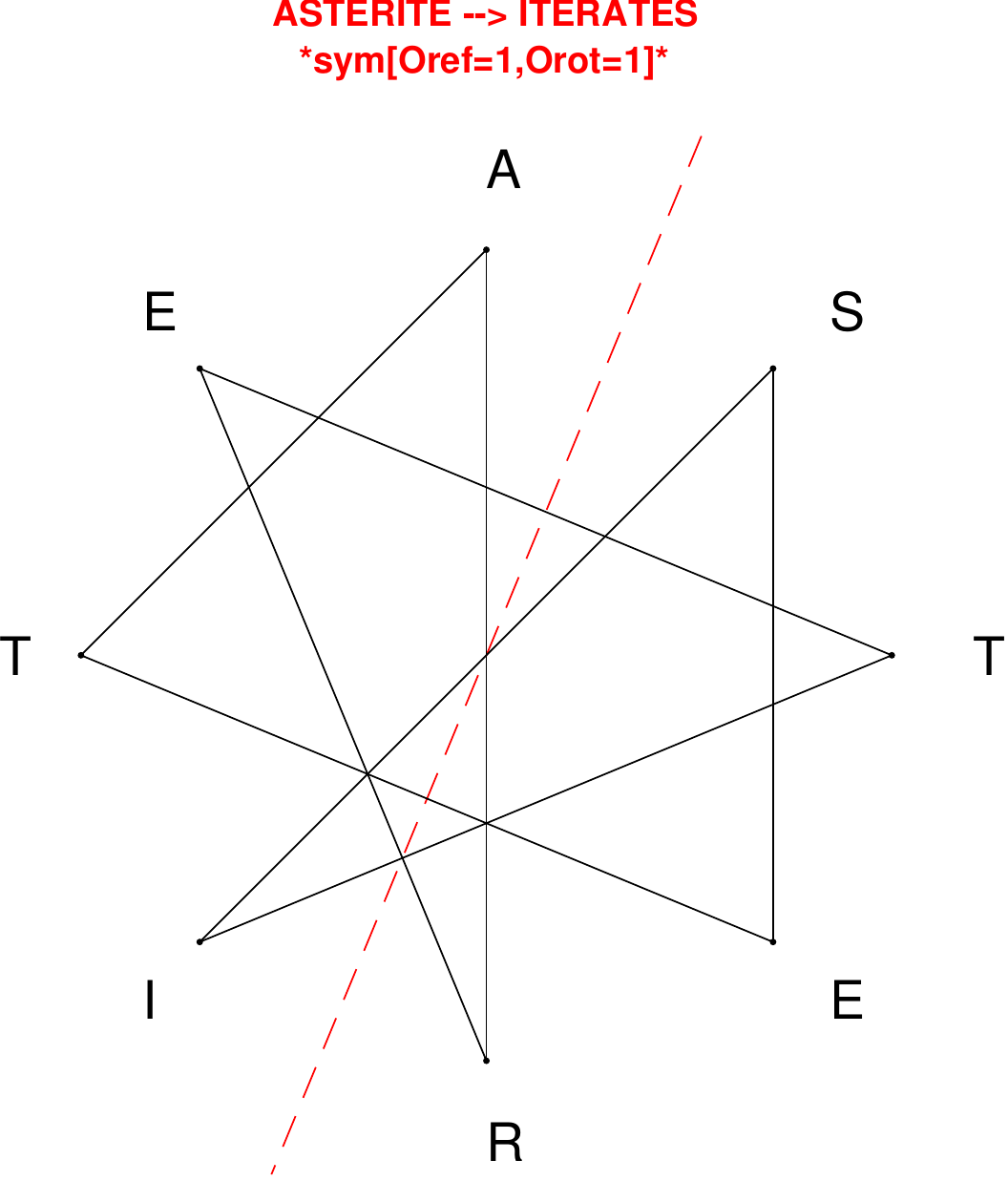}
\end{subfigure}
\hfill
\begin{subfigure}[T]{0.19\textwidth}
\centering
\includegraphics[width=\textwidth]{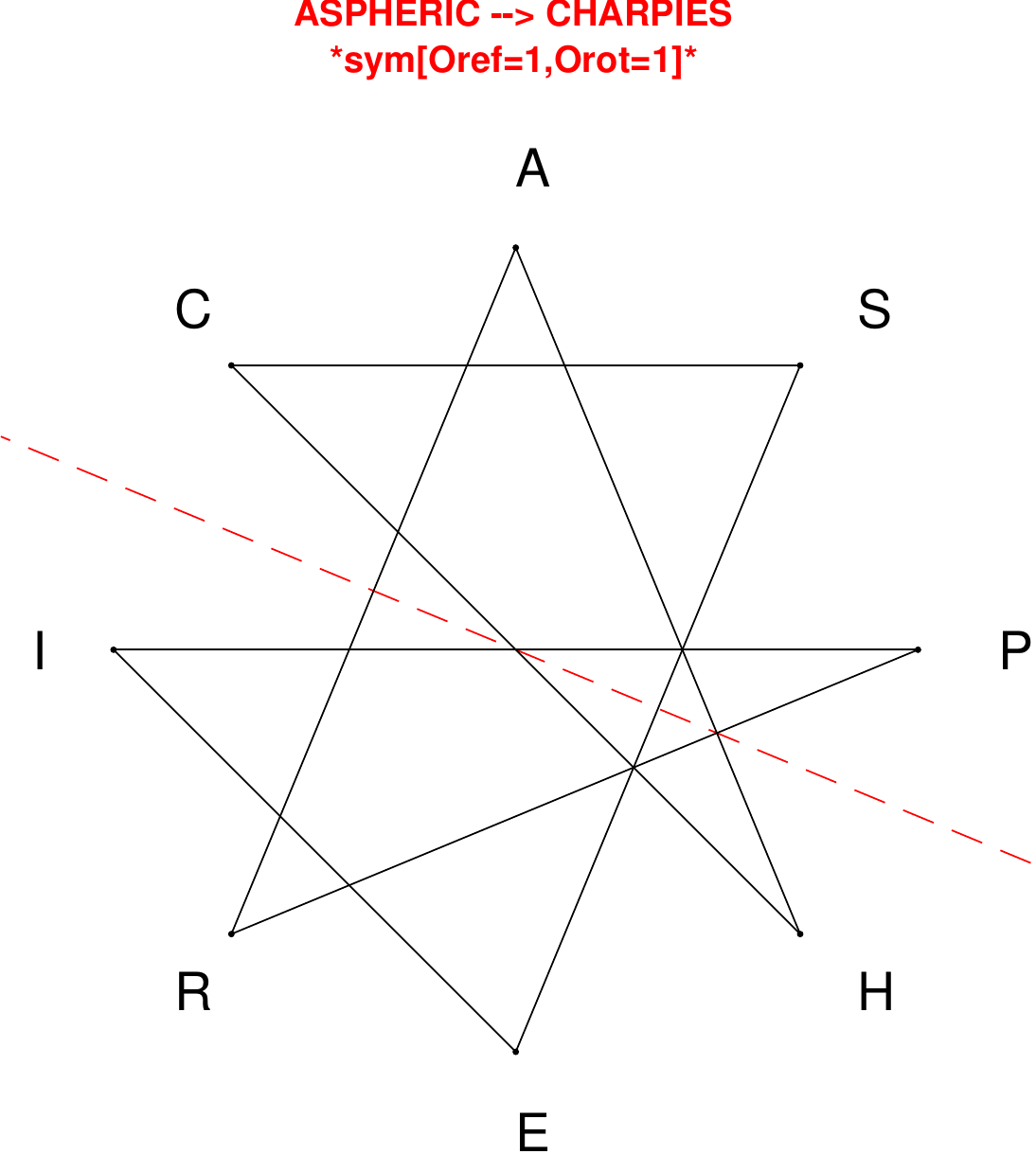}
\end{subfigure}
\hfill
\begin{subfigure}[T]{0.19\textwidth}
\centering
\includegraphics[width=\textwidth]{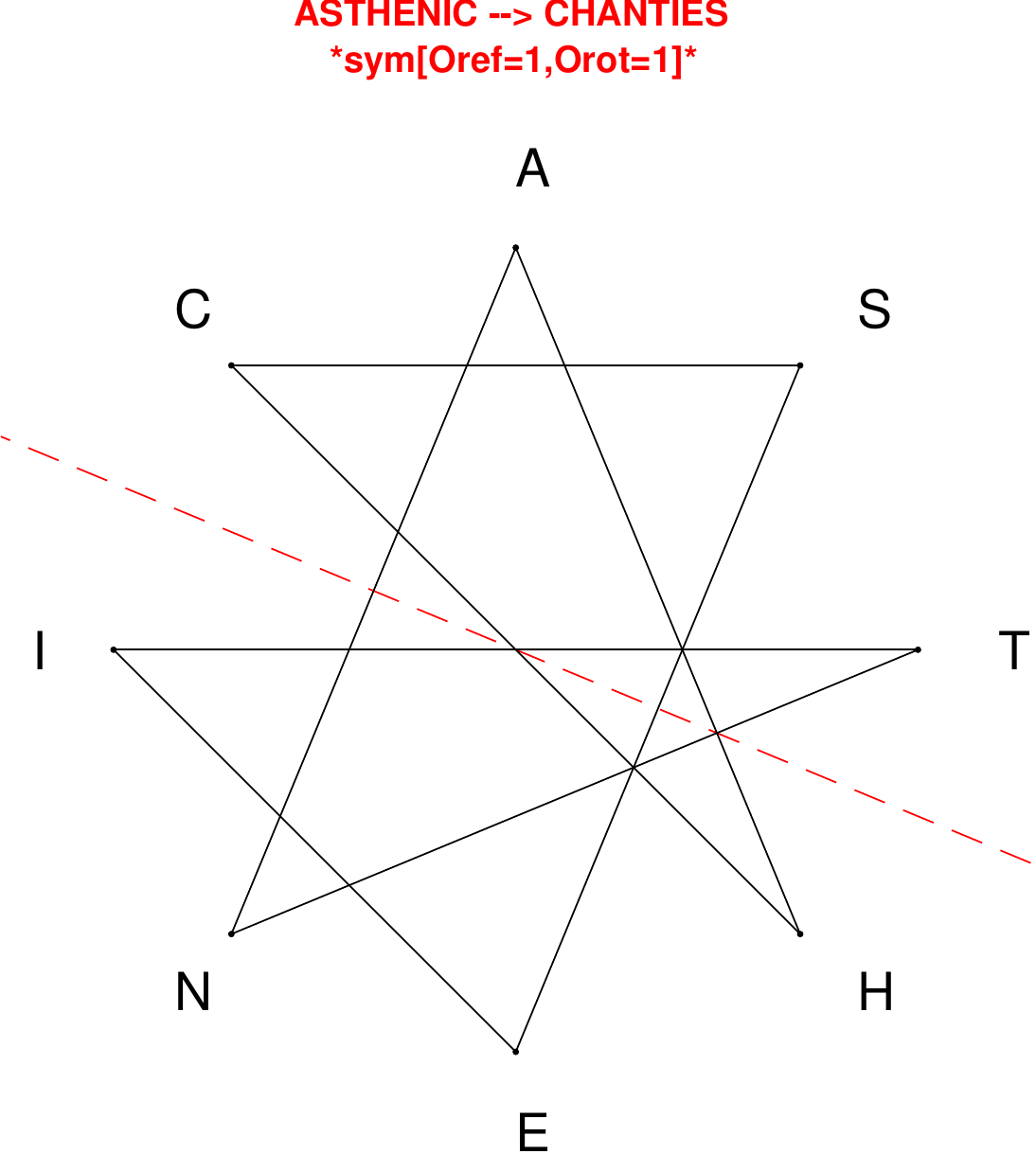}
\end{subfigure}
\end{figure}

\begin{figure}[H]
\centering
\begin{subfigure}[T]{0.19\textwidth}
\centering
\includegraphics[width=\textwidth]{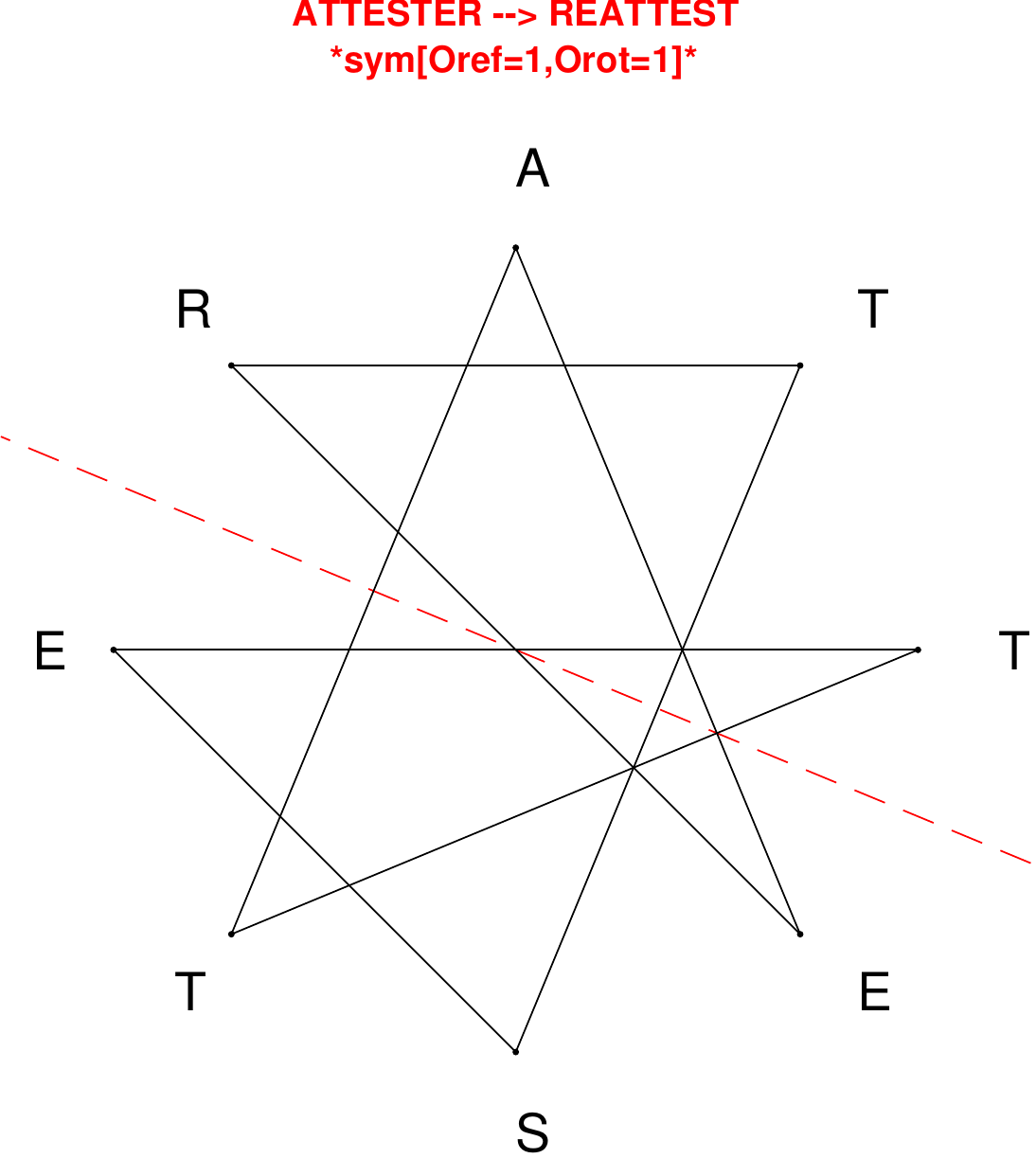}
\end{subfigure}
\hfill
\begin{subfigure}[T]{0.19\textwidth}
\centering
\includegraphics[width=\textwidth]{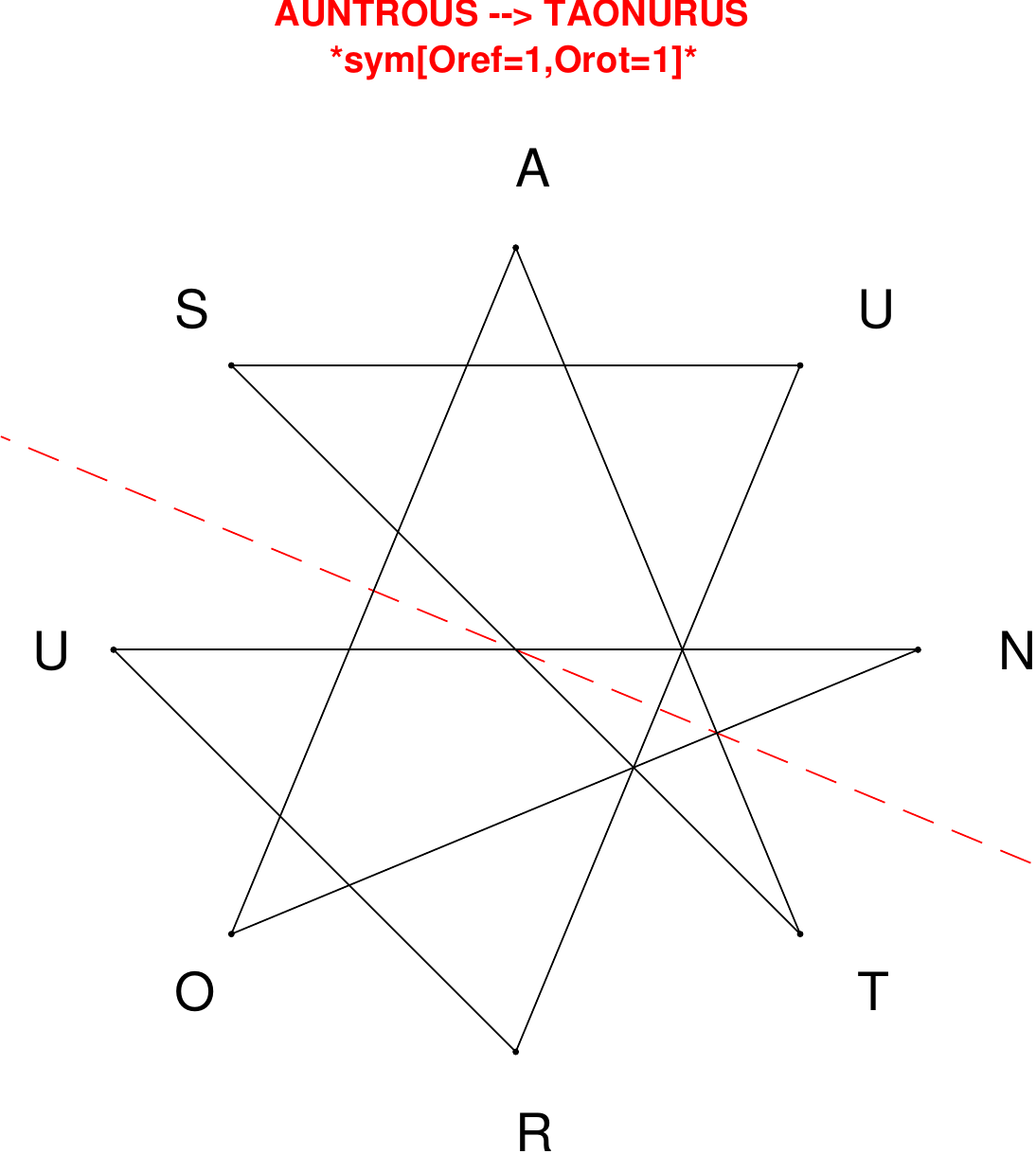}
\end{subfigure}
\hfill
\begin{subfigure}[T]{0.19\textwidth}
\centering
\includegraphics[width=\textwidth]{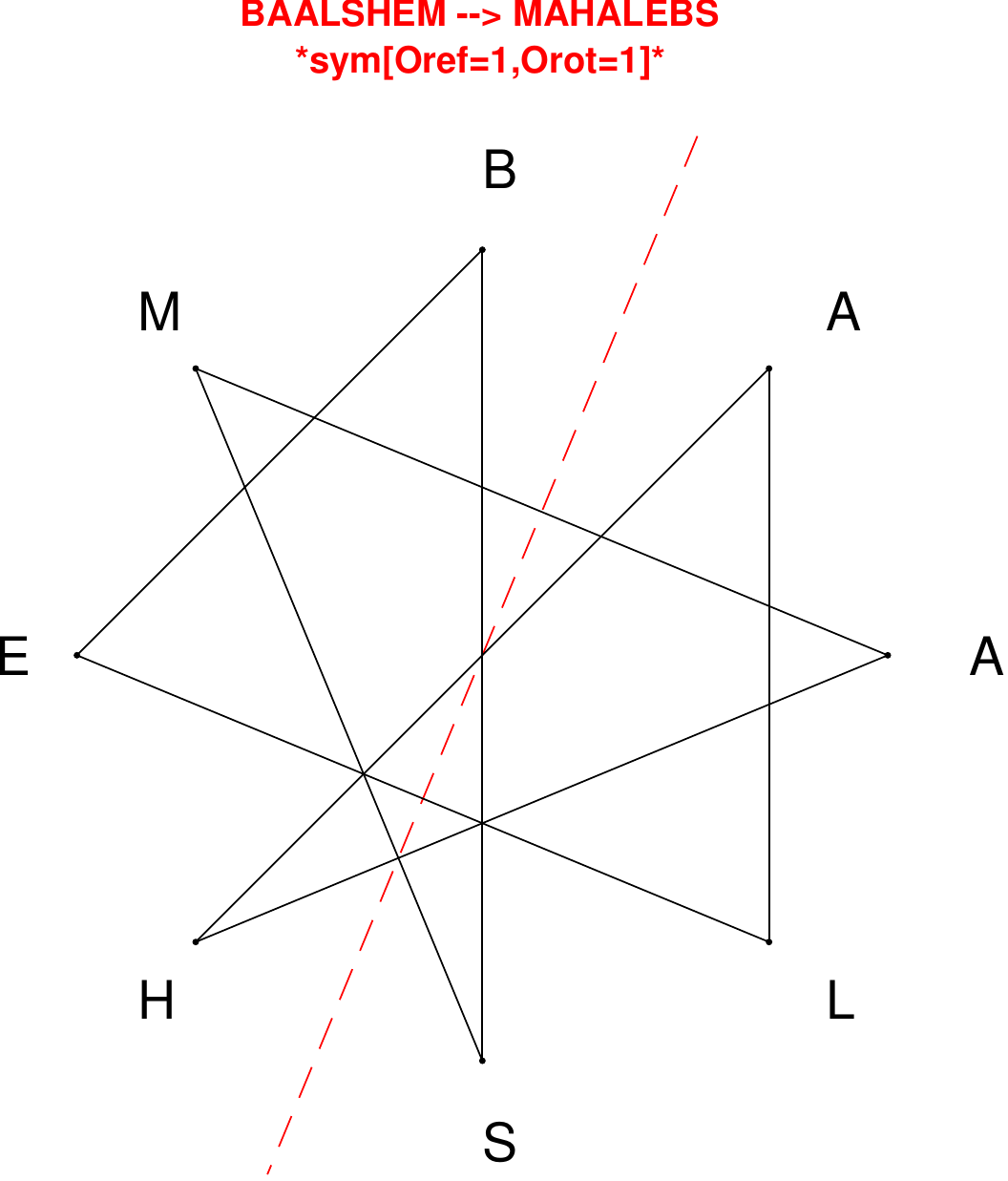}
\end{subfigure}
\hfill
\begin{subfigure}[T]{0.19\textwidth}
\centering
\includegraphics[width=\textwidth]{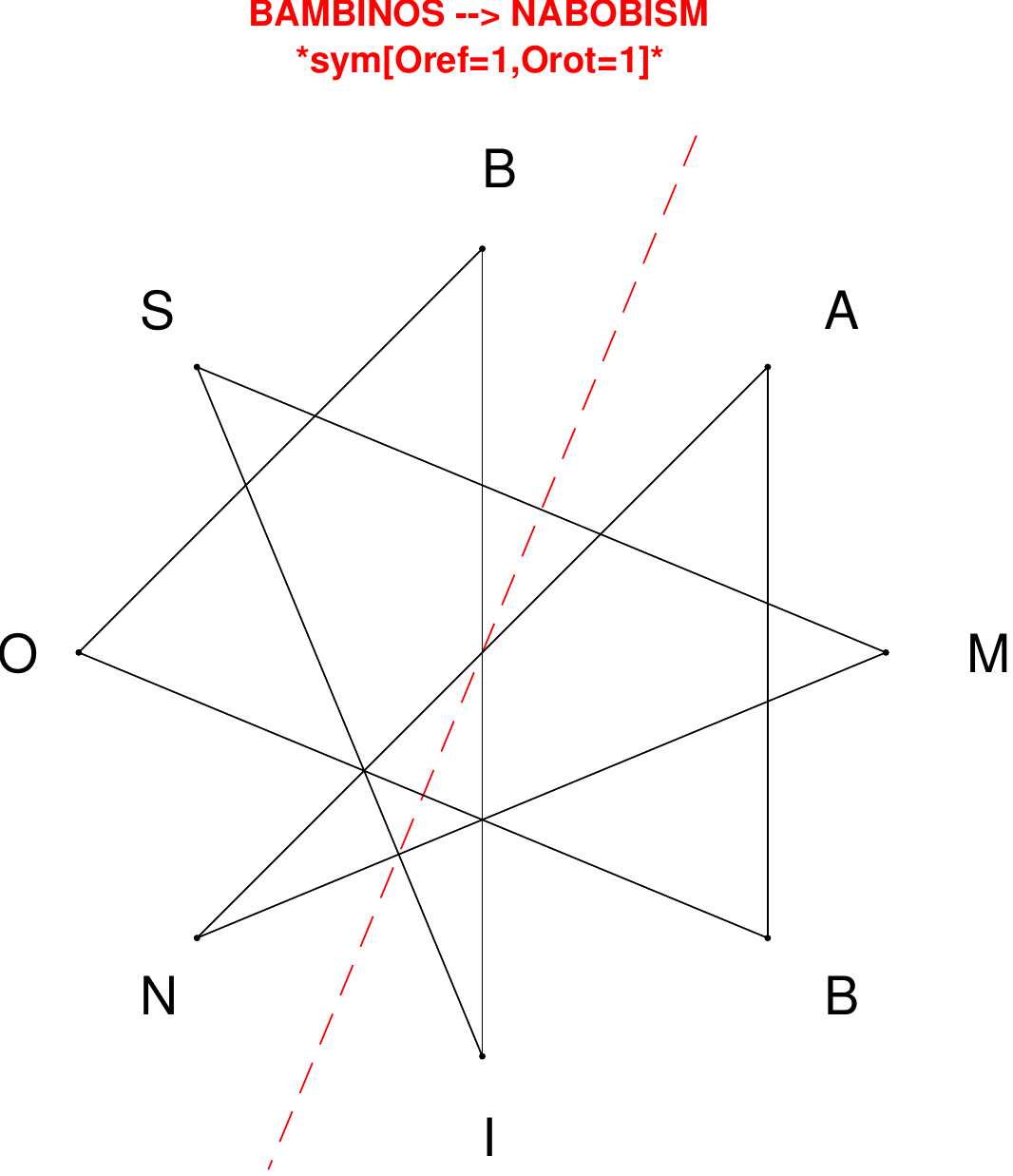}
\end{subfigure}
\hfill
\begin{subfigure}[T]{0.19\textwidth}
\centering
\includegraphics[width=\textwidth]{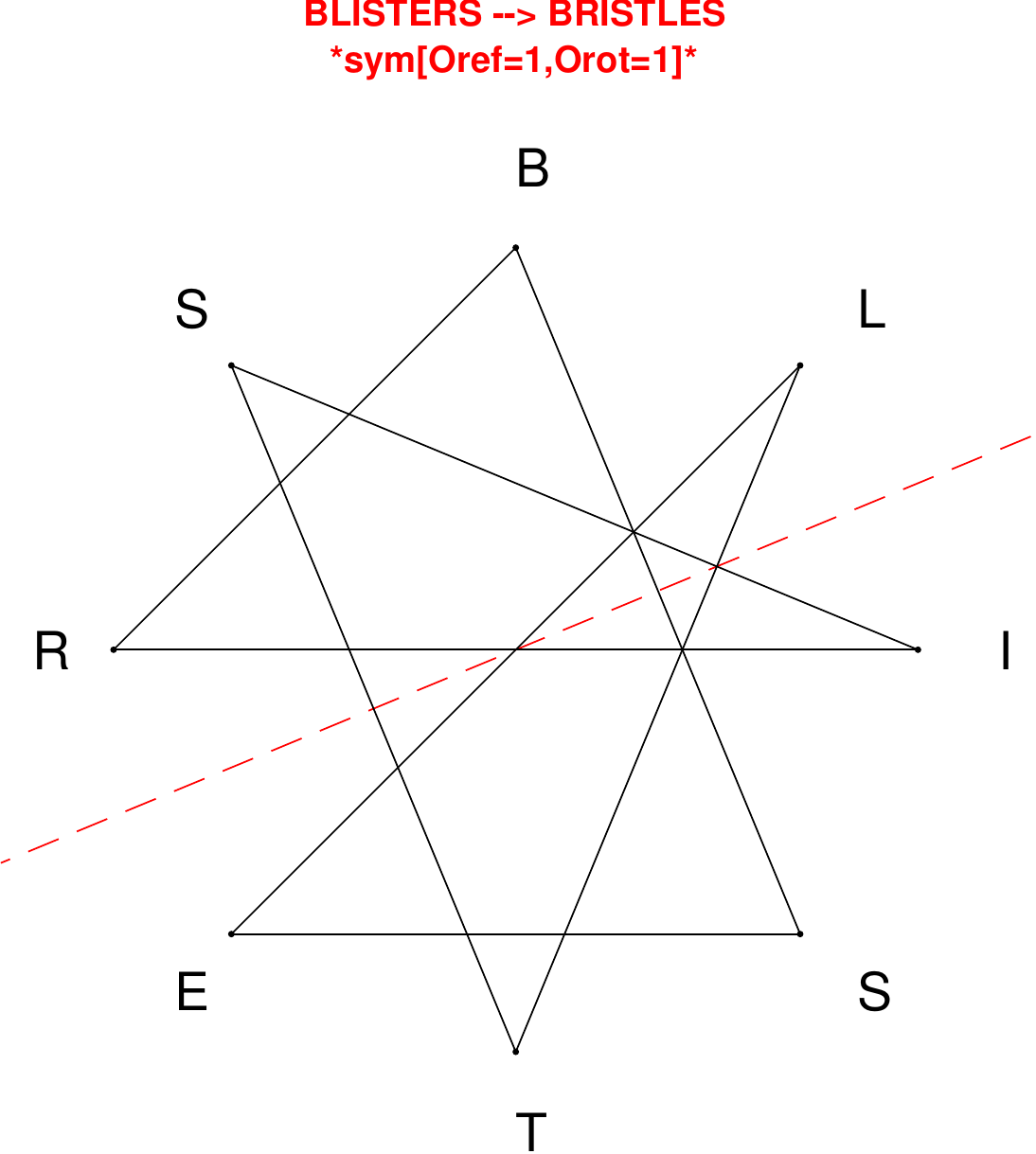}
\end{subfigure}
\end{figure}

\begin{figure}[H]
\centering
\begin{subfigure}[T]{0.19\textwidth}
\centering
\includegraphics[width=\textwidth]{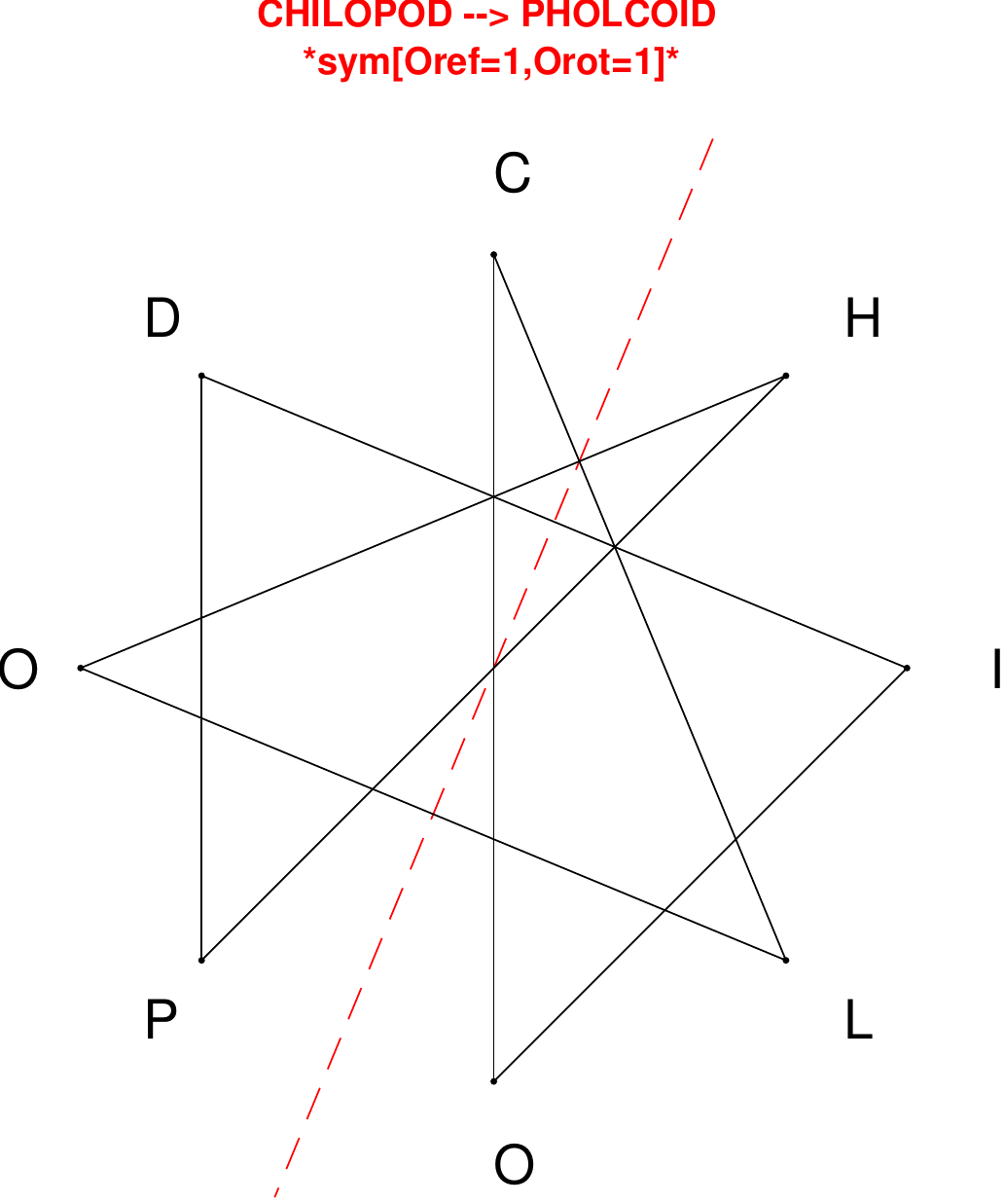}
\end{subfigure}
\hfill
\begin{subfigure}[T]{0.19\textwidth}
\centering
\includegraphics[width=\textwidth]{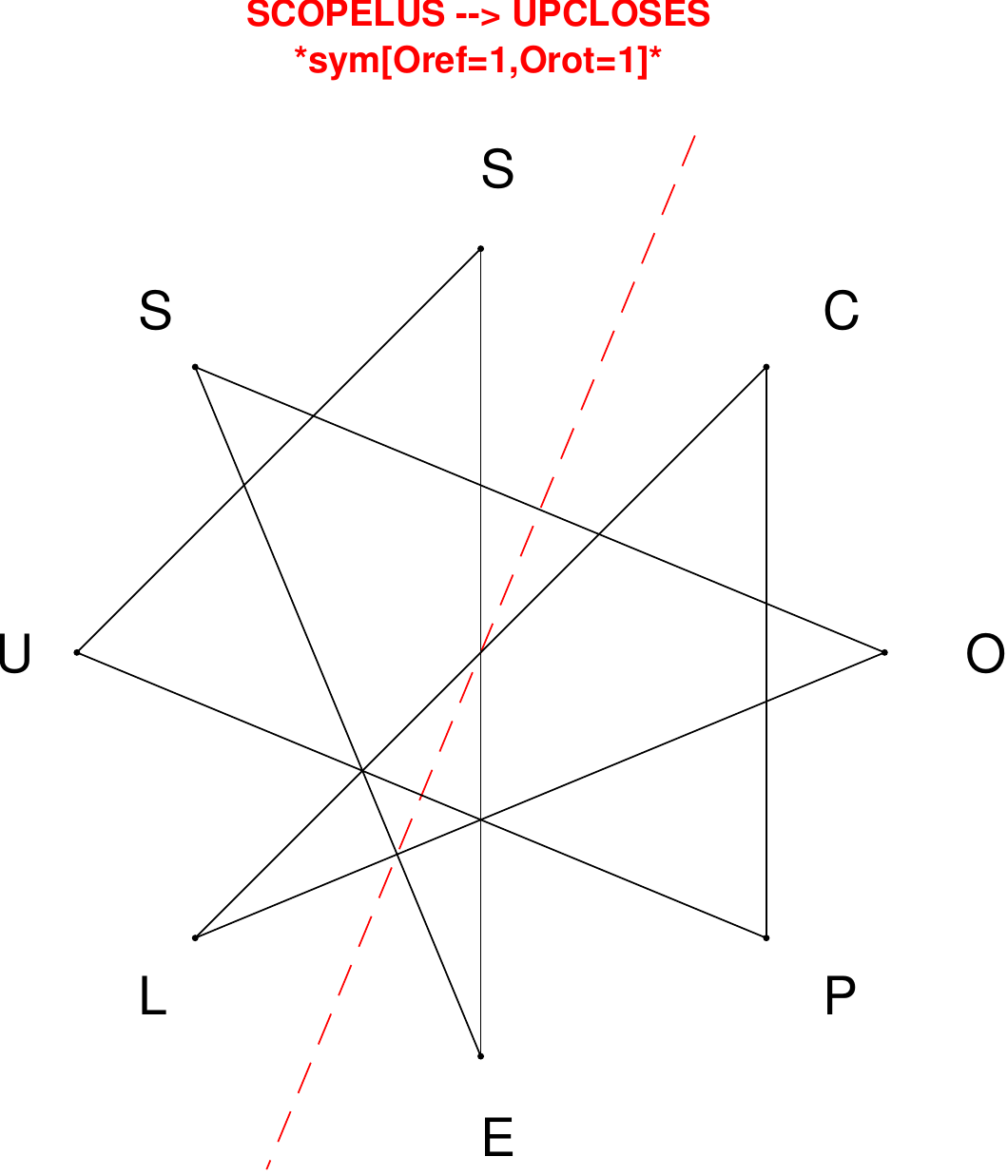}
\end{subfigure}
\hfill
\begin{subfigure}[T]{0.19\textwidth}
\centering
\includegraphics[width=\textwidth]{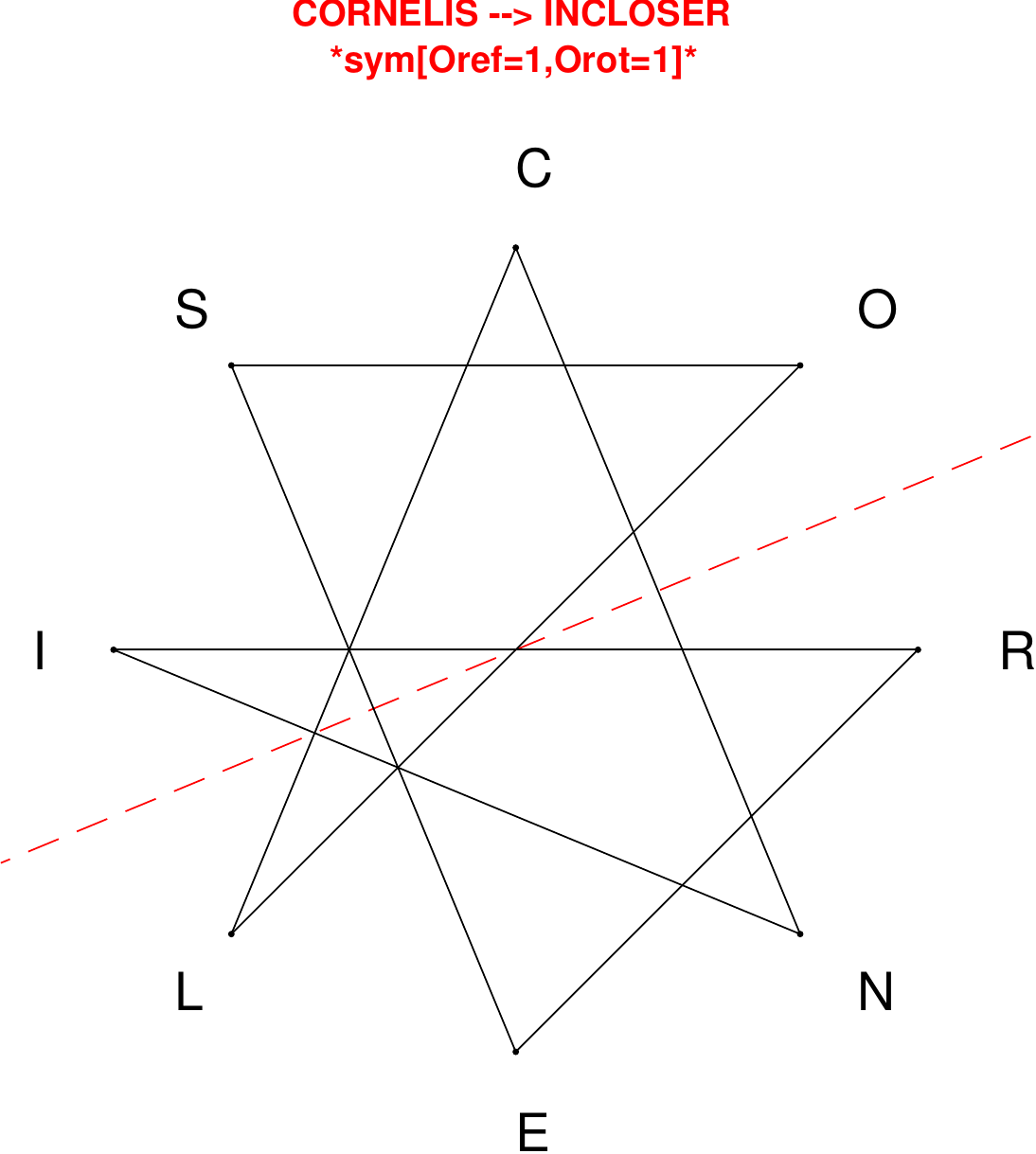}
\end{subfigure}
\hfill
\begin{subfigure}[T]{0.19\textwidth}
\centering
\includegraphics[width=\textwidth]{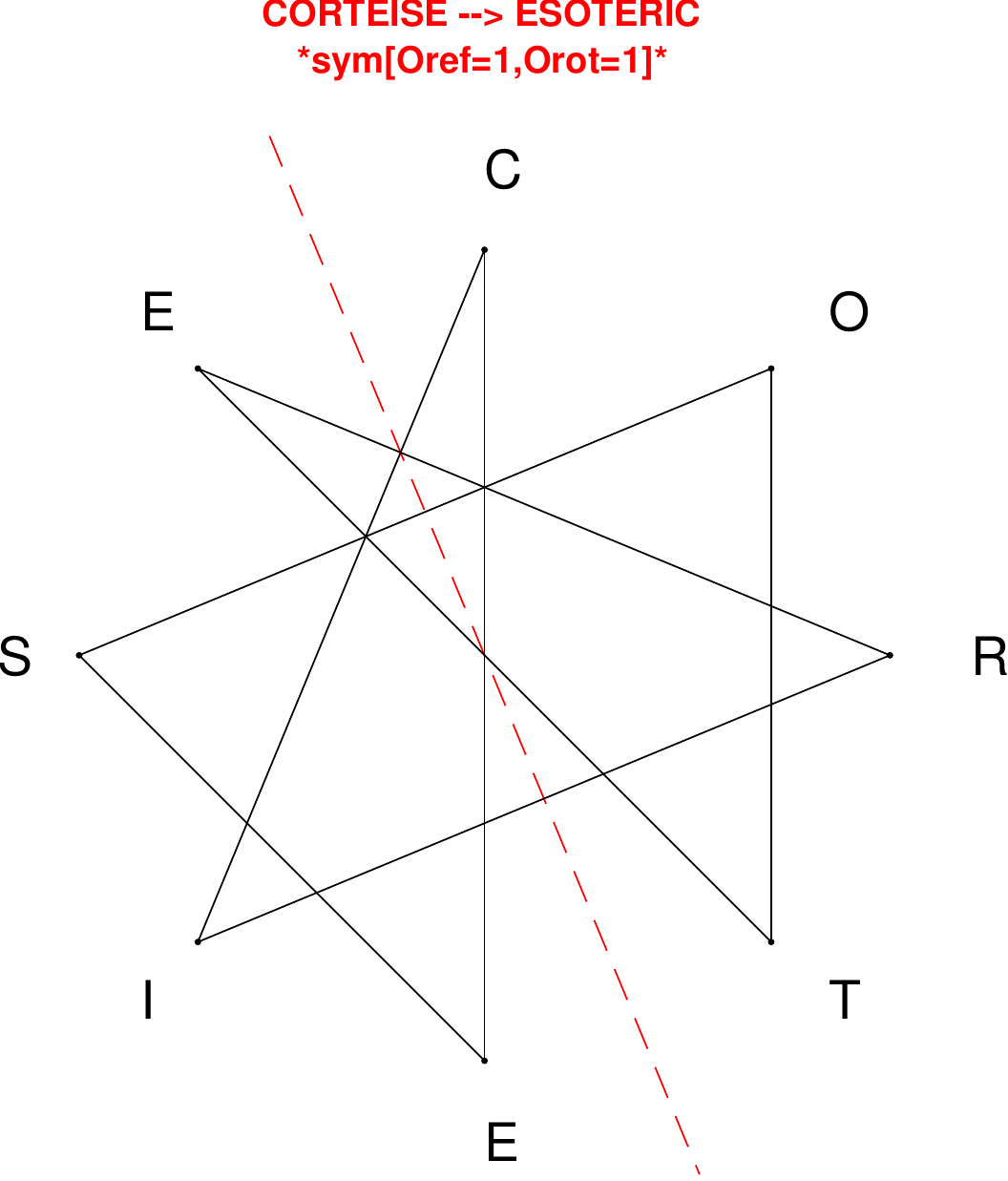}
\end{subfigure}
\hfill
\begin{subfigure}[T]{0.19\textwidth}
\centering
\includegraphics[width=\textwidth]{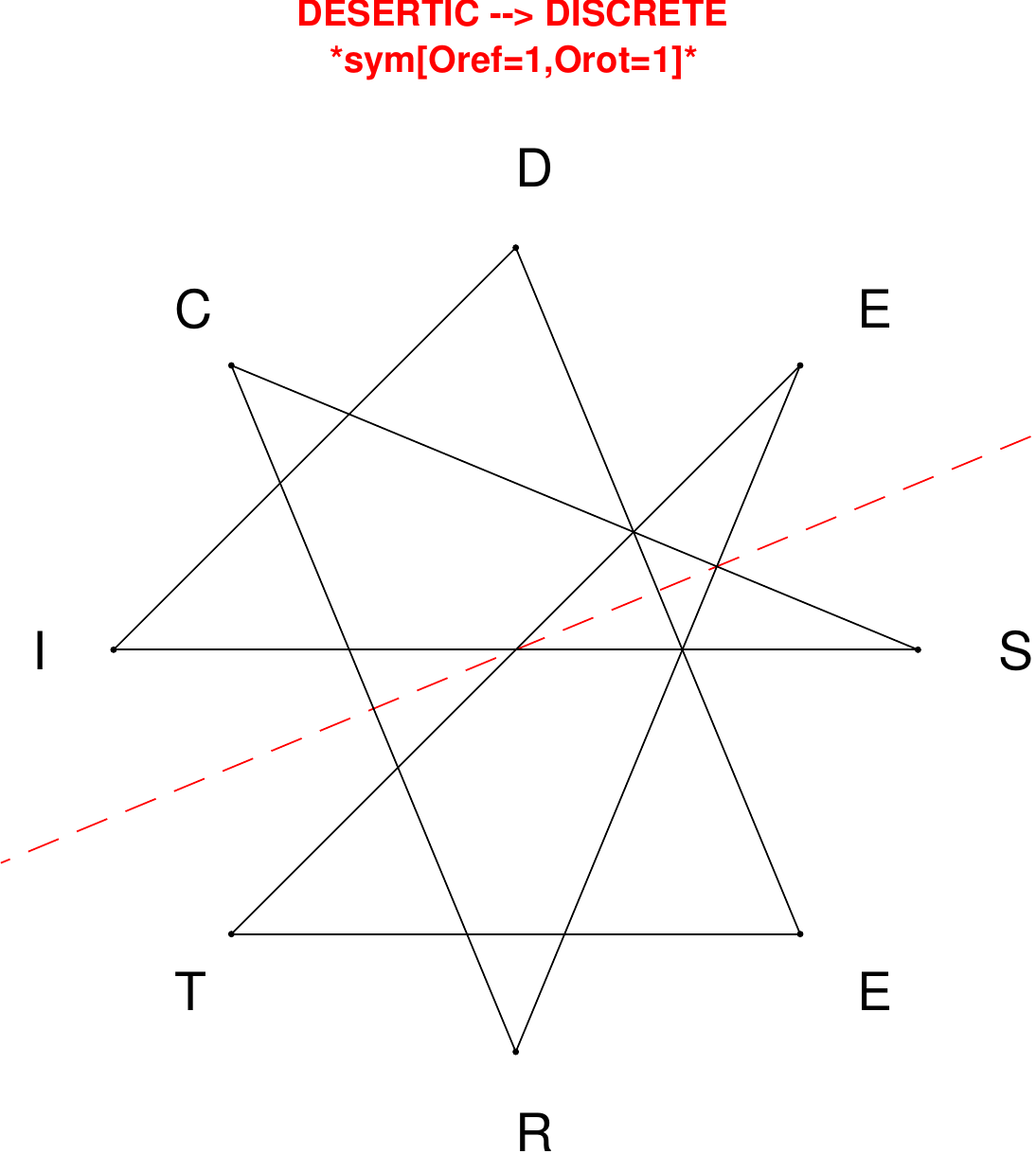}
\end{subfigure}
\end{figure}

\begin{figure}[H]
\centering
\begin{subfigure}[T]{0.19\textwidth}
\centering
\includegraphics[width=\textwidth]{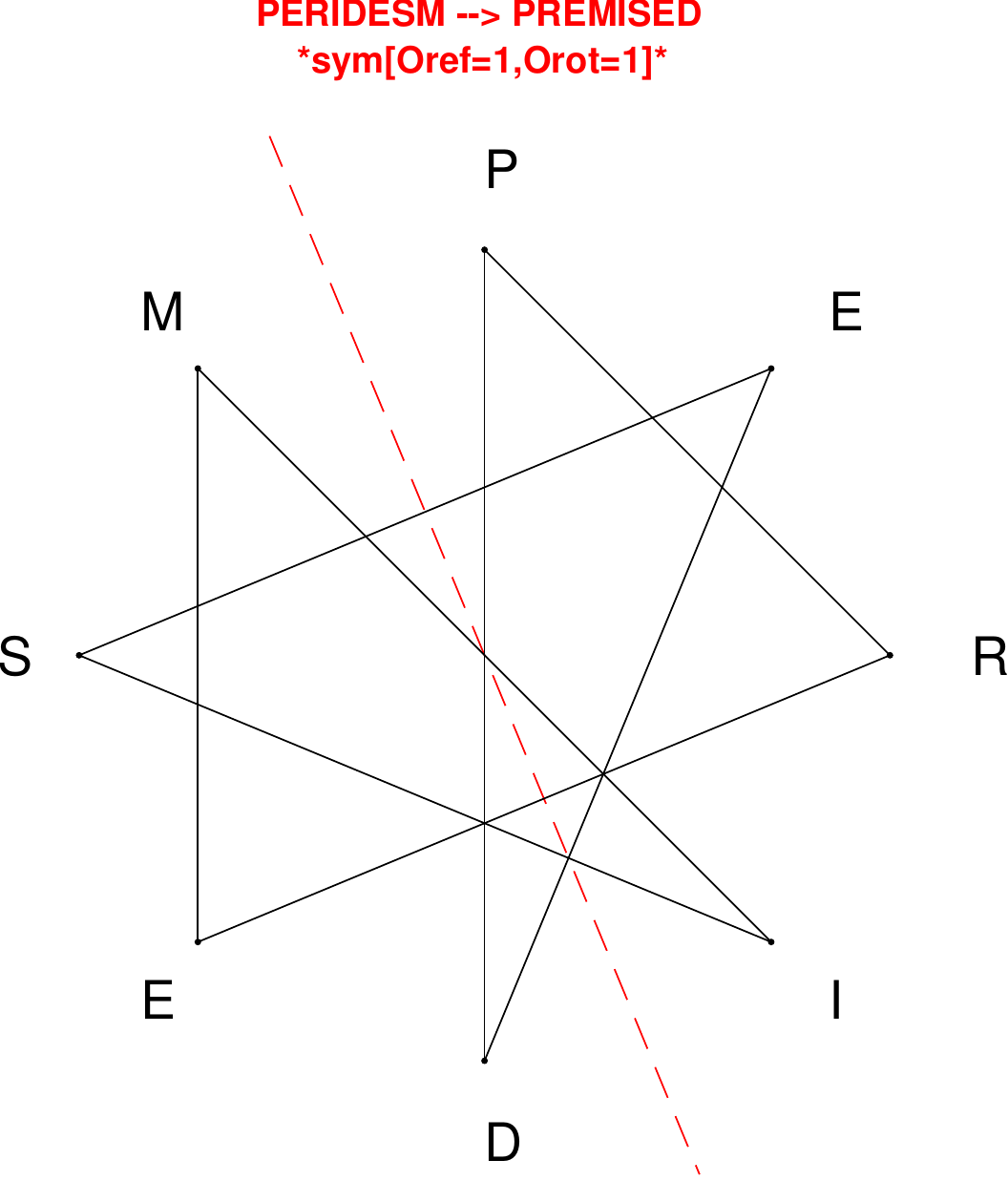}
\end{subfigure}
\hfill
\begin{subfigure}[T]{0.19\textwidth}
\centering
\includegraphics[width=\textwidth]{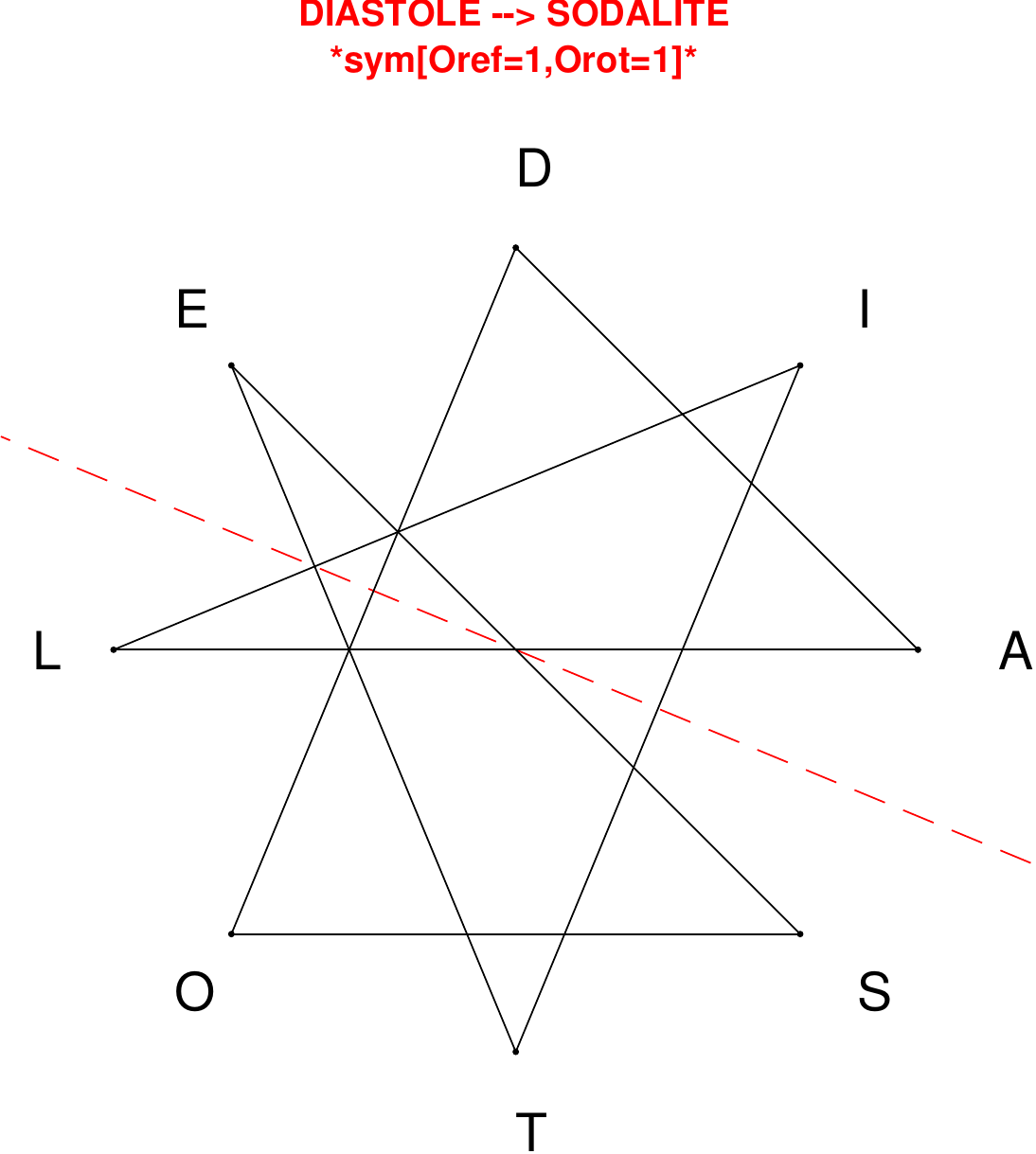}
\end{subfigure}
\hfill
\begin{subfigure}[T]{0.19\textwidth}
\centering
\includegraphics[width=\textwidth]{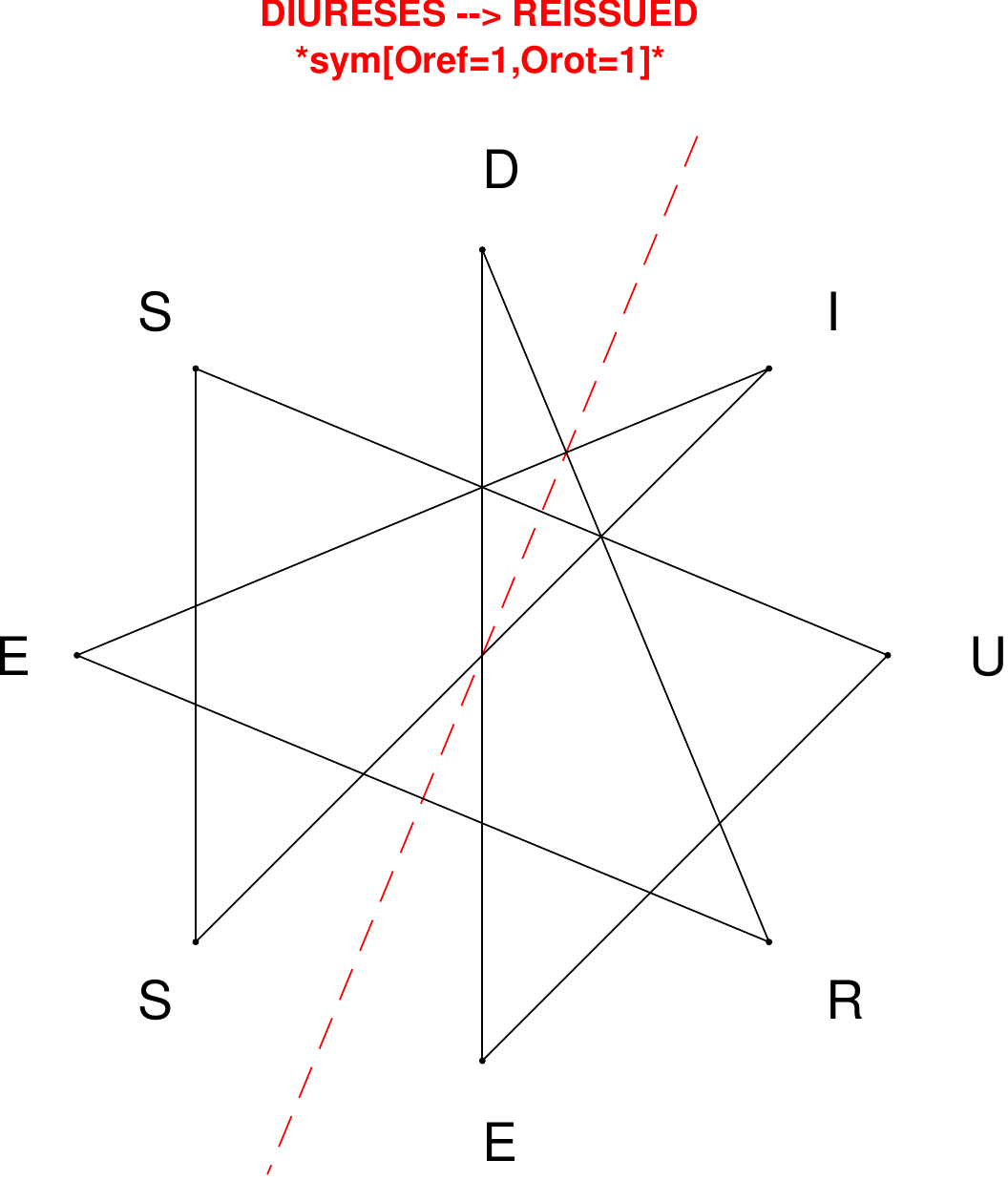}
\end{subfigure}
\hfill
\begin{subfigure}[T]{0.19\textwidth}
\centering
\includegraphics[width=\textwidth]{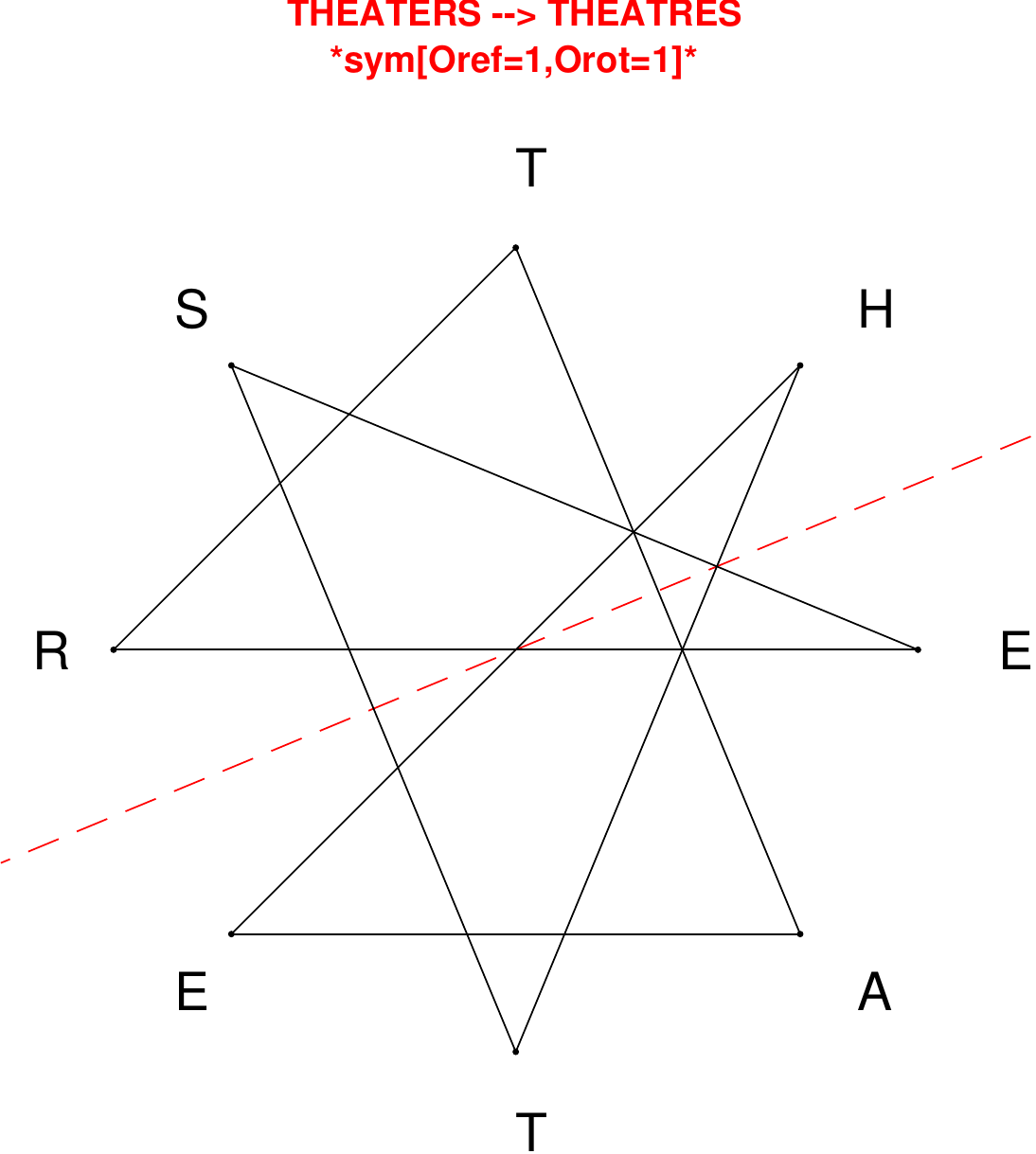}
\end{subfigure}
\hfill
\begin{subfigure}[T]{0.19\textwidth}
\centering
\includegraphics[width=\textwidth]{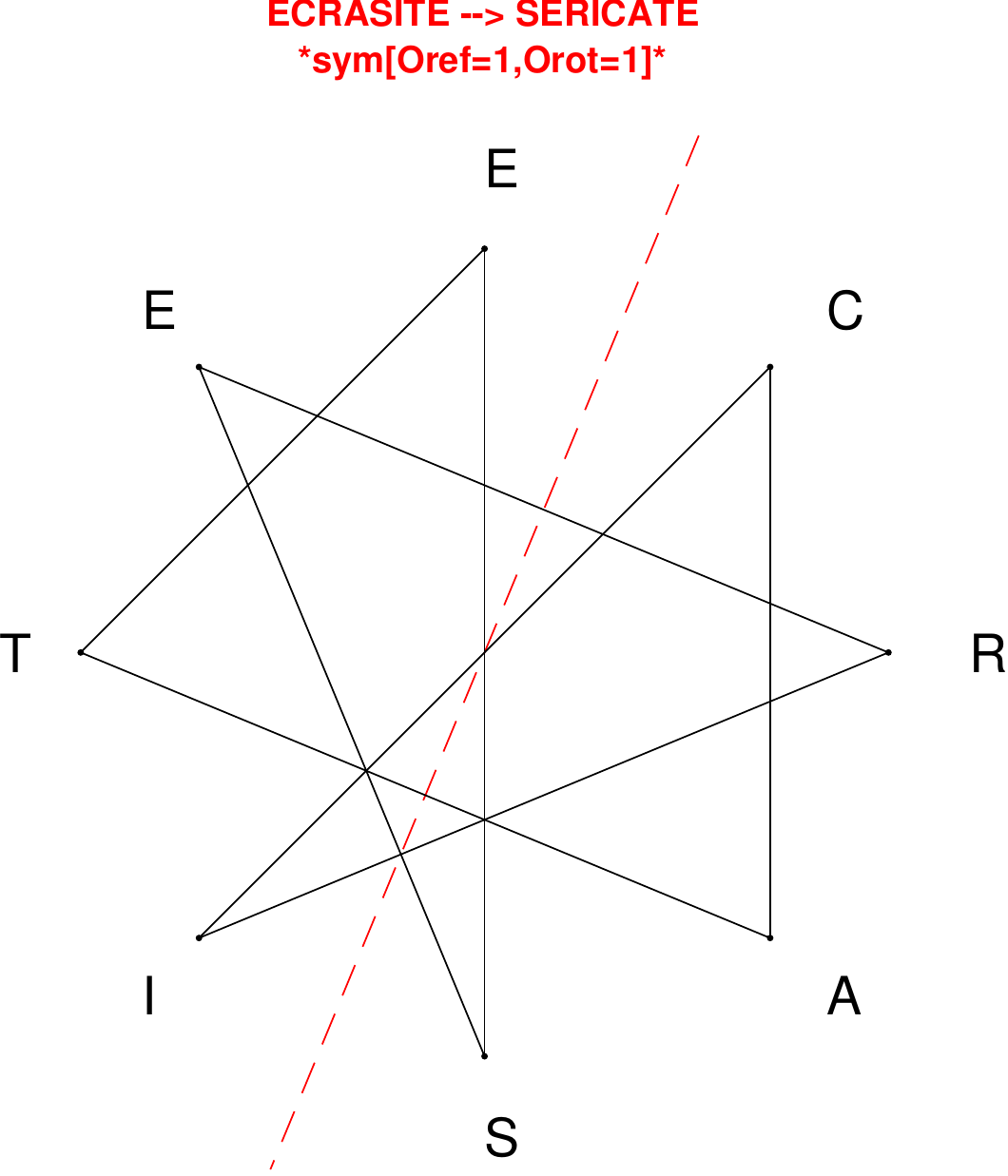}
\end{subfigure}
\end{figure}

\begin{figure}[H]
\centering
\begin{subfigure}[T]{0.19\textwidth}
\centering
\includegraphics[width=\textwidth]{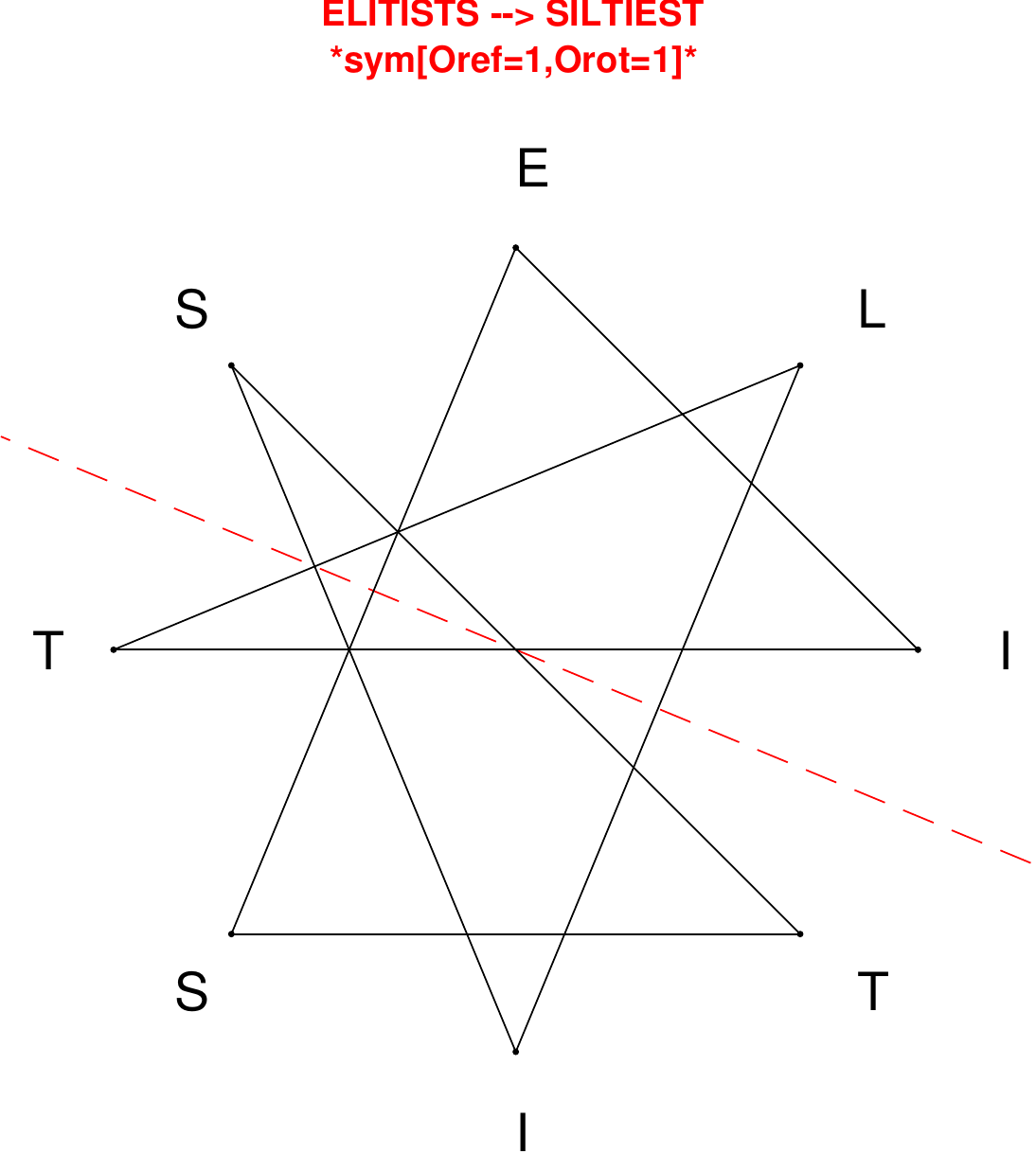}
\end{subfigure}
\hfill
\begin{subfigure}[T]{0.19\textwidth}
\centering
\includegraphics[width=\textwidth]{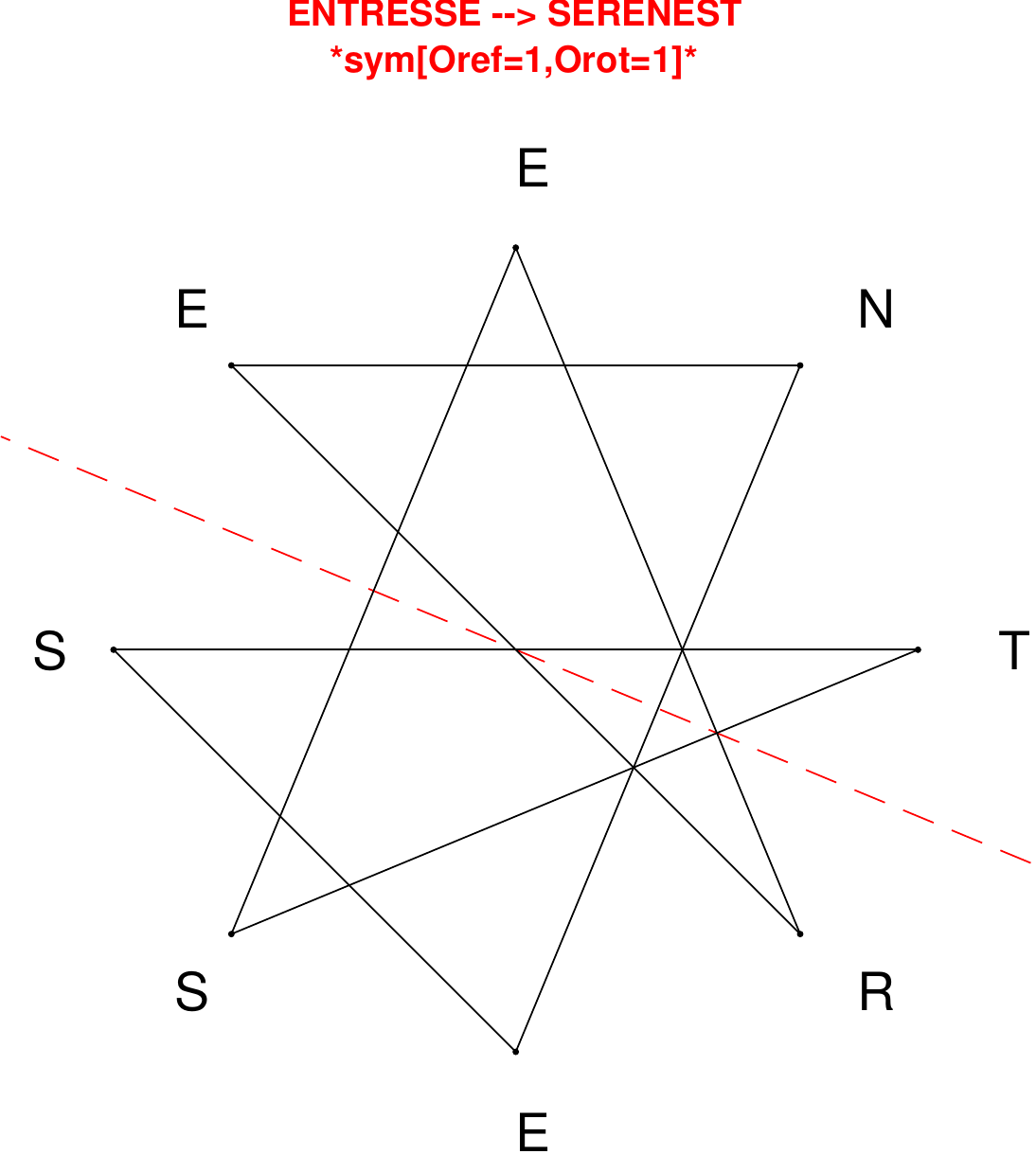}
\end{subfigure}
\hfill
\begin{subfigure}[T]{0.19\textwidth}
\centering
\includegraphics[width=\textwidth]{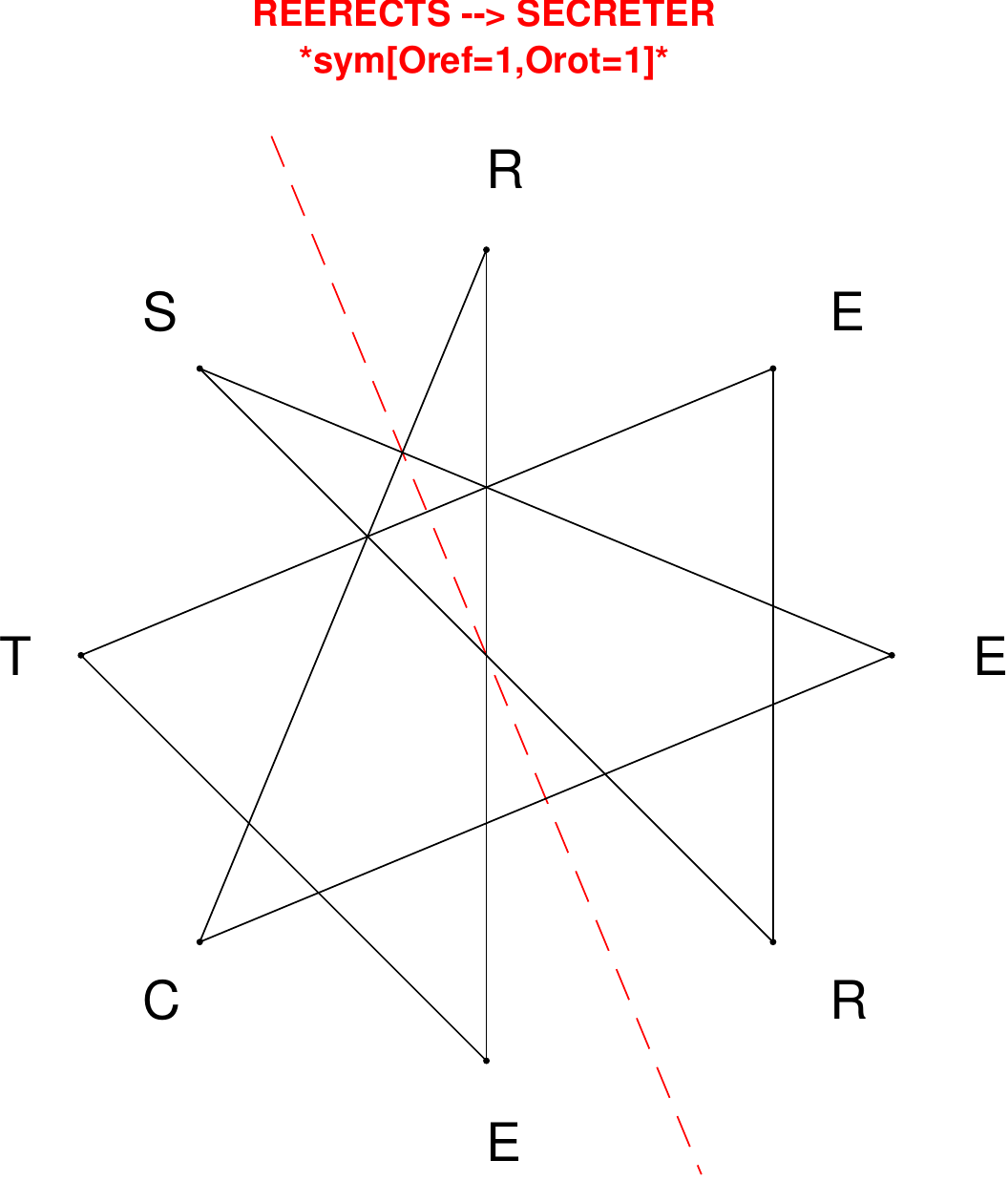}
\end{subfigure}
\hfill
\begin{subfigure}[T]{0.19\textwidth}
\centering
\includegraphics[width=\textwidth]{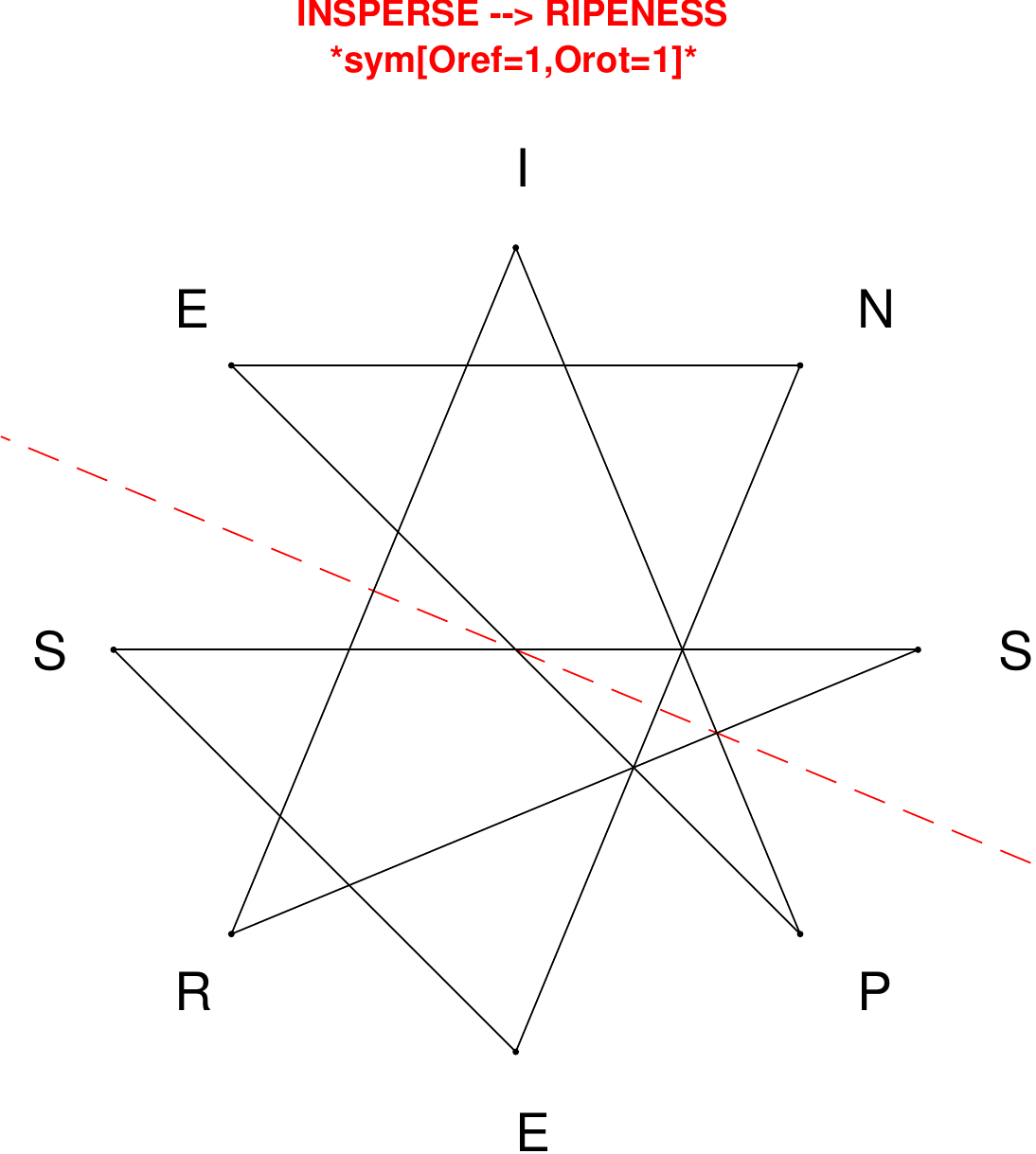}
\end{subfigure}
\hfill
\begin{subfigure}[T]{0.19\textwidth}
\centering
\includegraphics[width=\textwidth]{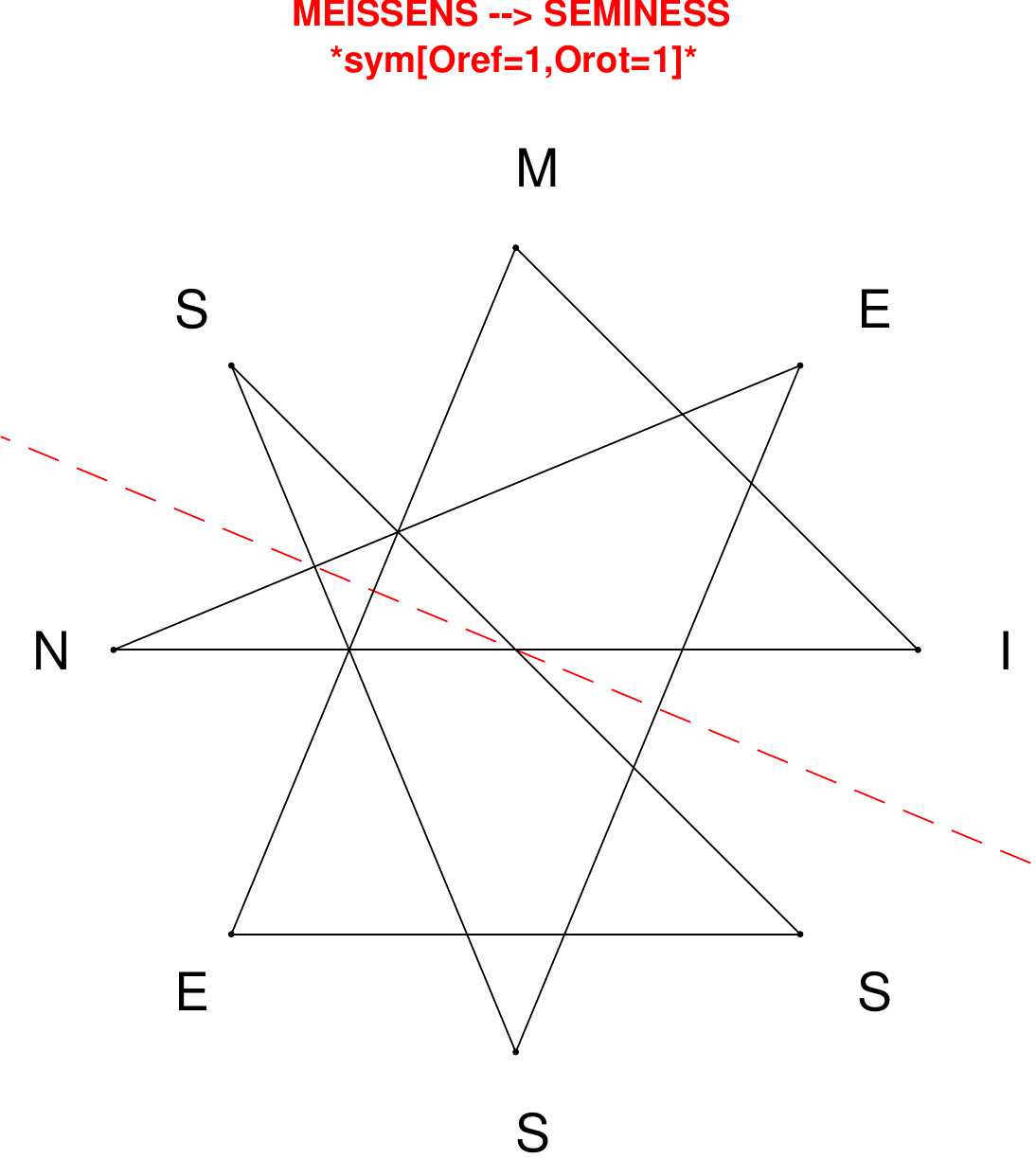}
\end{subfigure}
\end{figure}

\begin{figure}[H]
\centering
\begin{subfigure}[T]{0.19\textwidth}
\centering
\includegraphics[width=\textwidth]{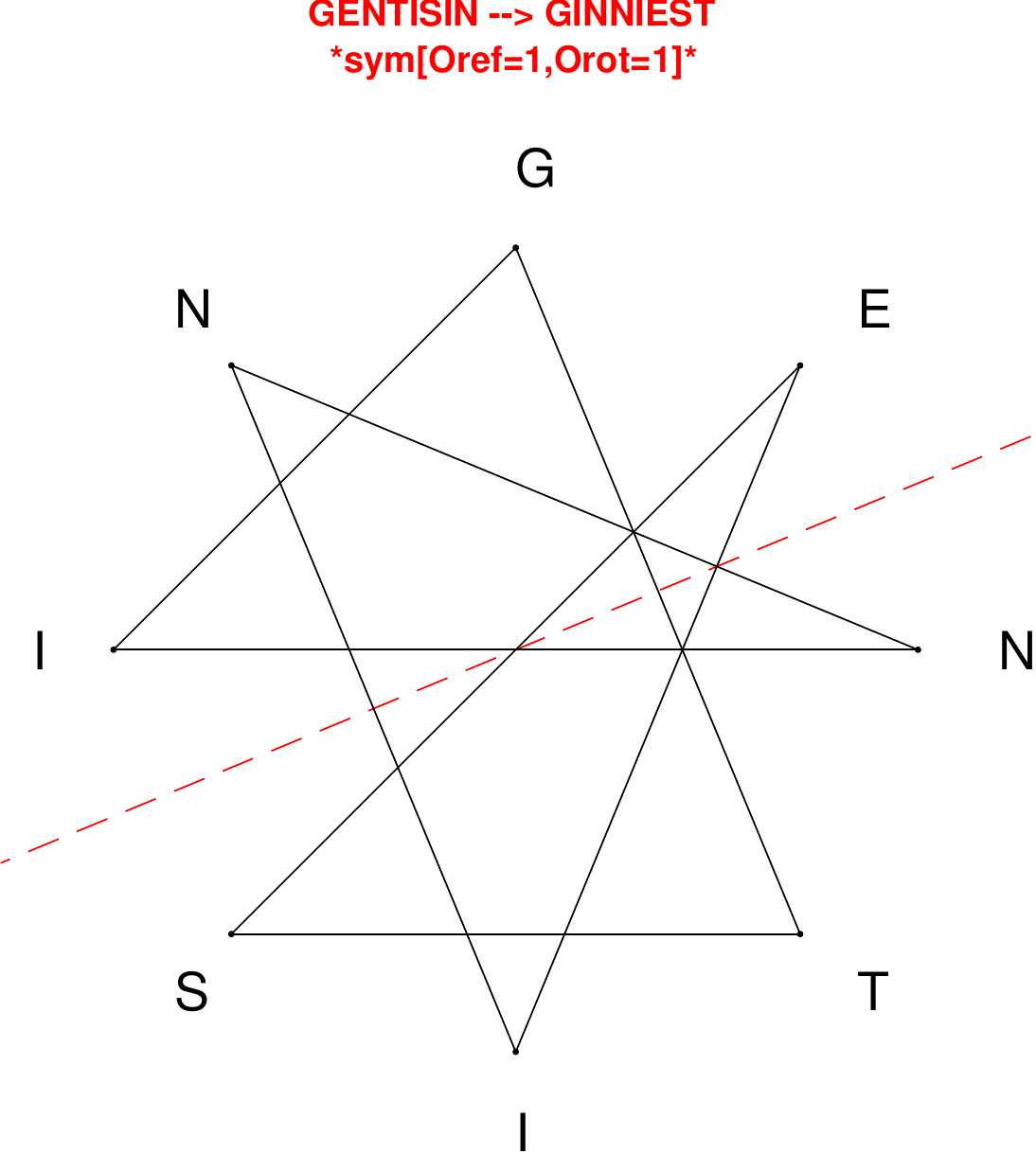}
\end{subfigure}
\hfill
\begin{subfigure}[T]{0.19\textwidth}
\centering
\includegraphics[width=\textwidth]{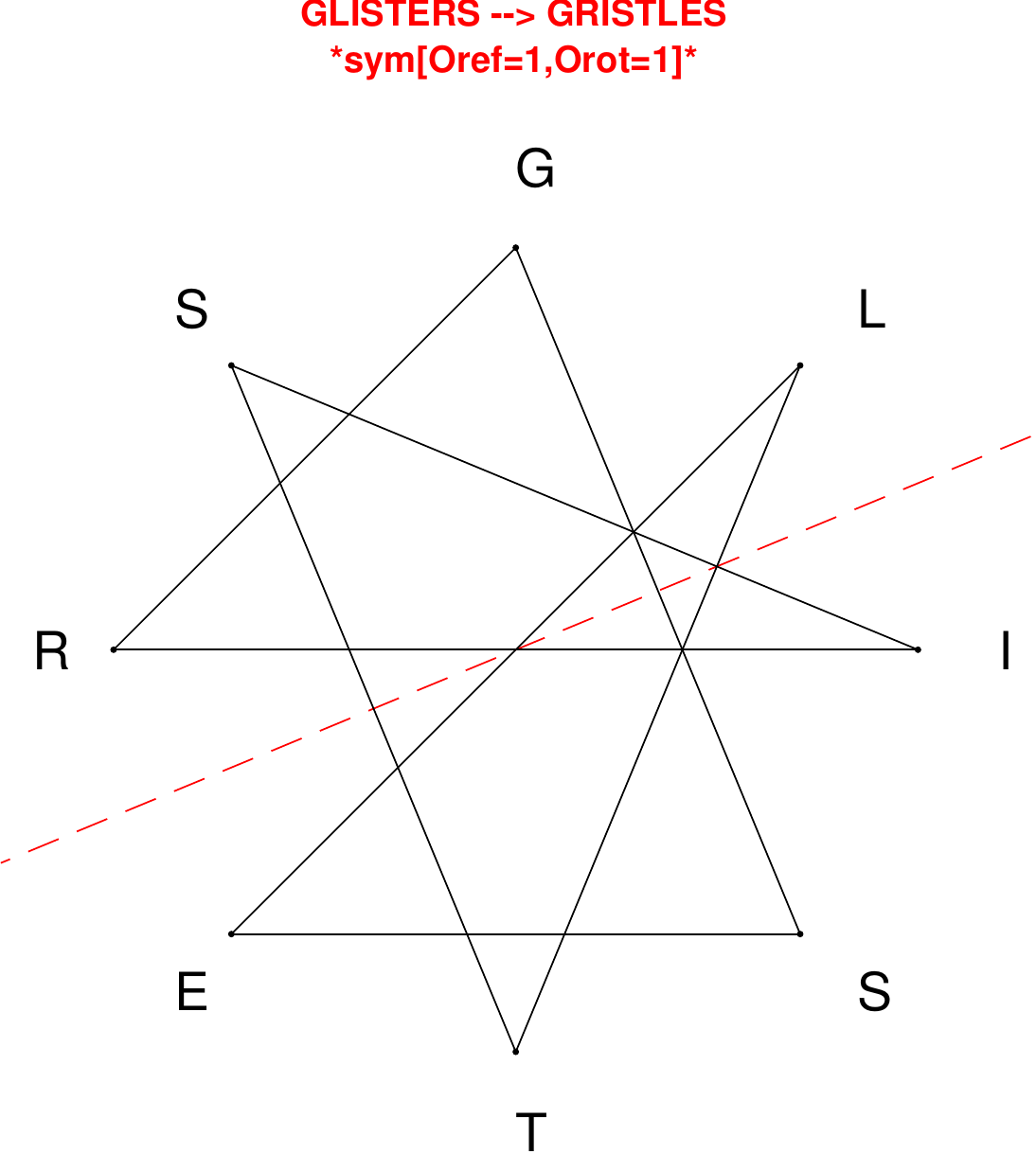}
\end{subfigure}
\hfill
\begin{subfigure}[T]{0.19\textwidth}
\centering
\includegraphics[width=\textwidth]{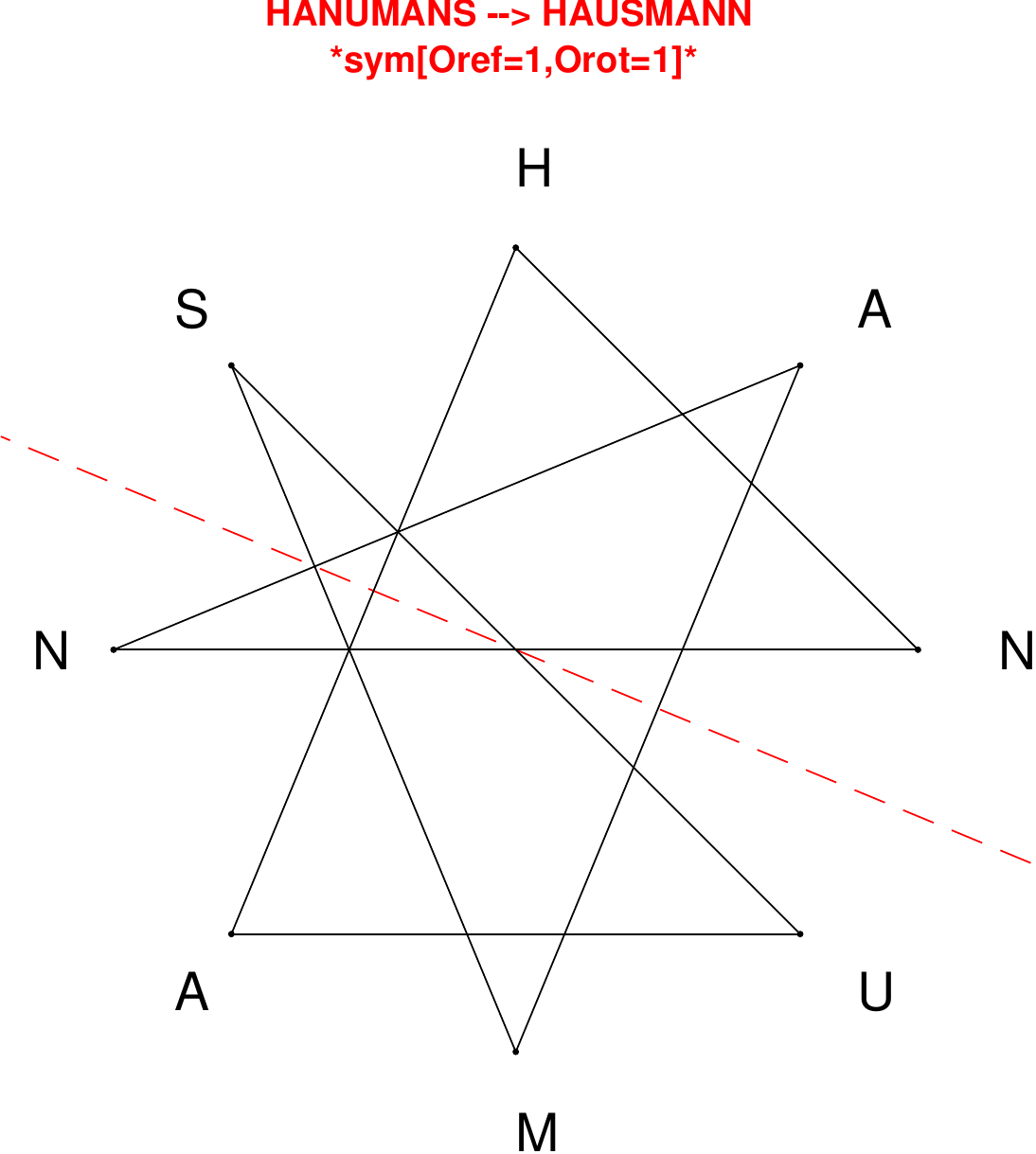}
\end{subfigure}
\hfill
\begin{subfigure}[T]{0.19\textwidth}
\centering
\includegraphics[width=\textwidth]{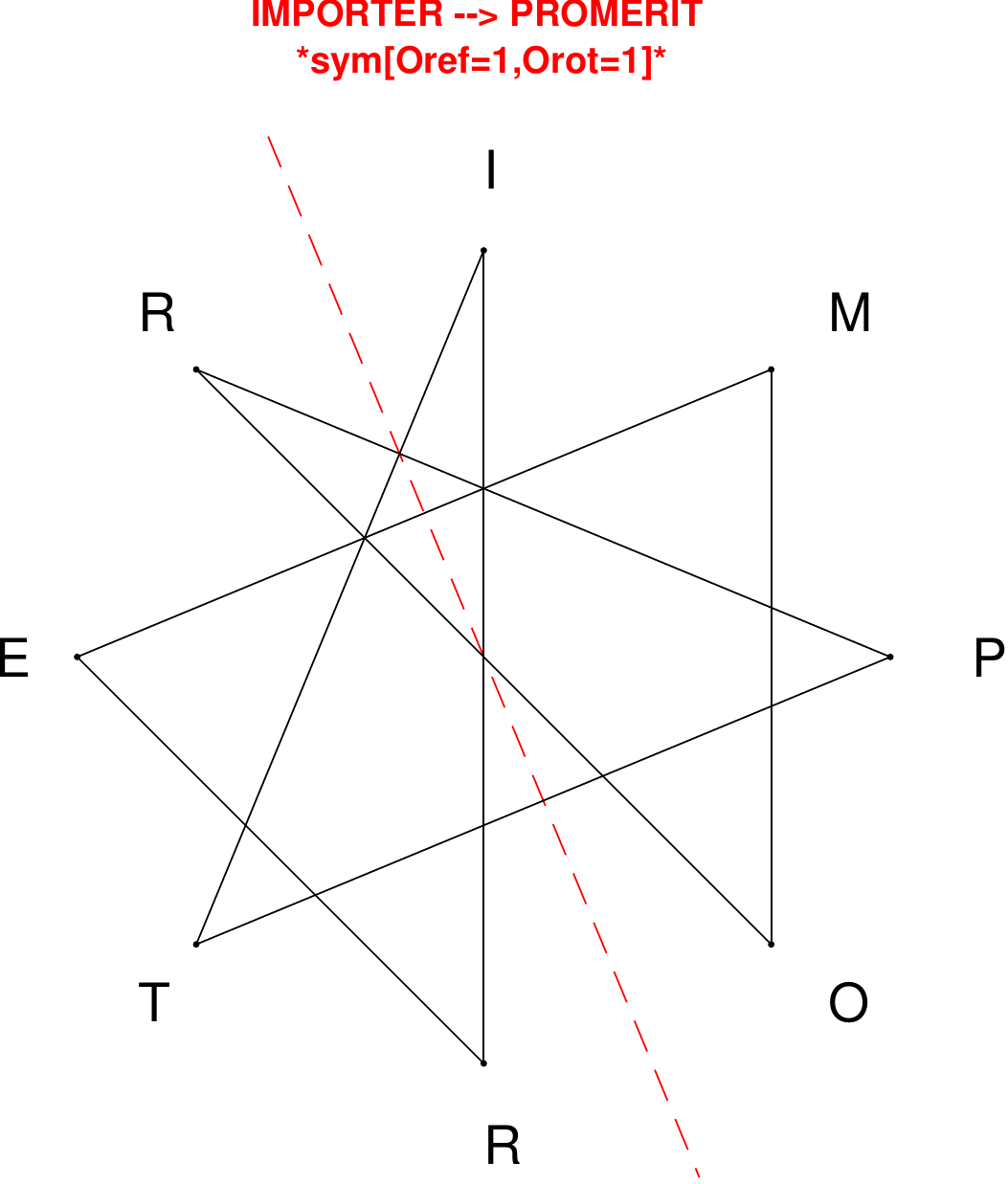}
\end{subfigure}
\hfill
\begin{subfigure}[T]{0.19\textwidth}
\centering
\includegraphics[width=\textwidth]{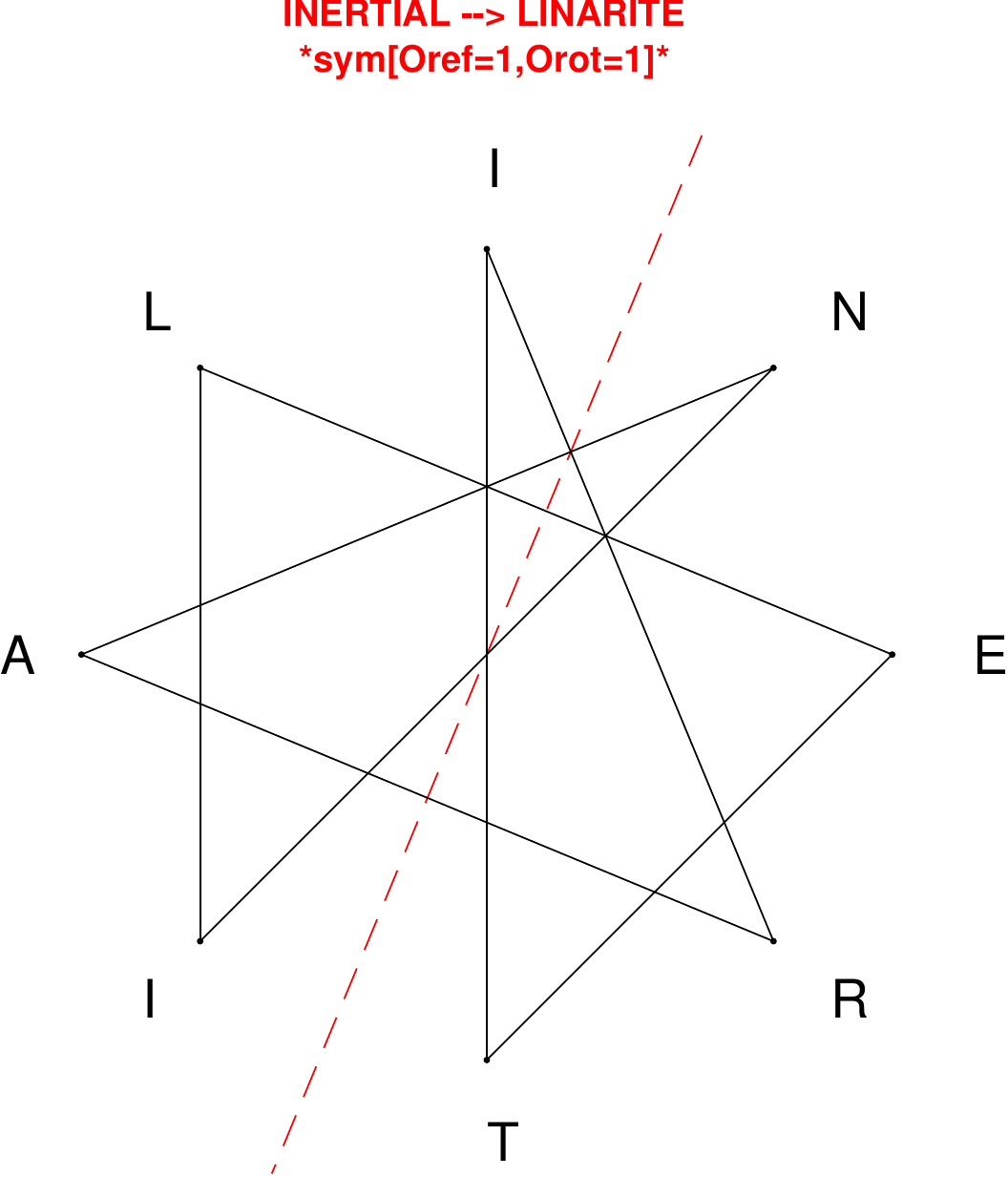}
\end{subfigure}
\end{figure}

\begin{figure}[H]
\centering
\begin{subfigure}[T]{0.19\textwidth}
\centering
\includegraphics[width=\textwidth]{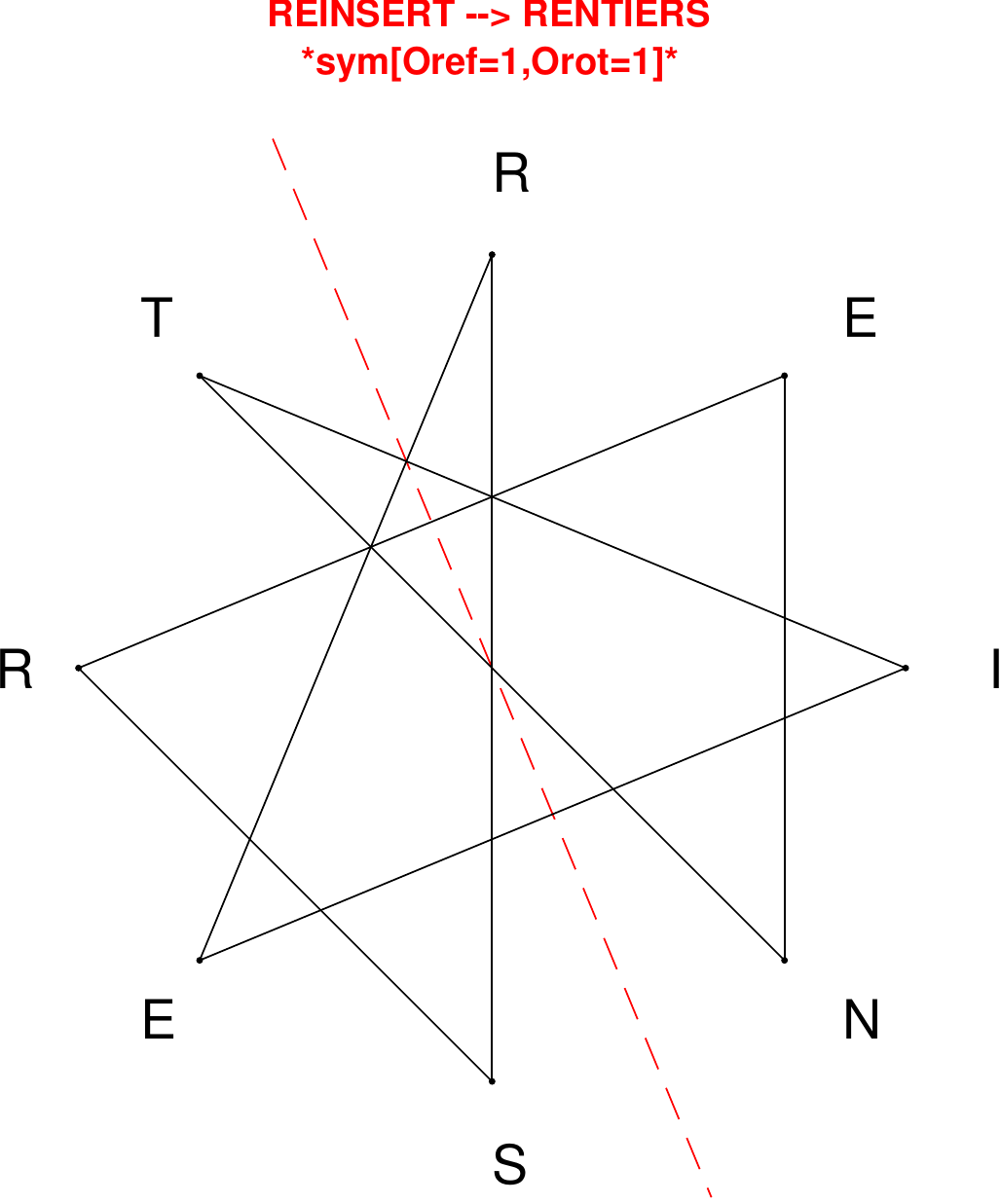}
\end{subfigure}
\hfill
\begin{subfigure}[T]{0.19\textwidth}
\centering
\includegraphics[width=\textwidth]{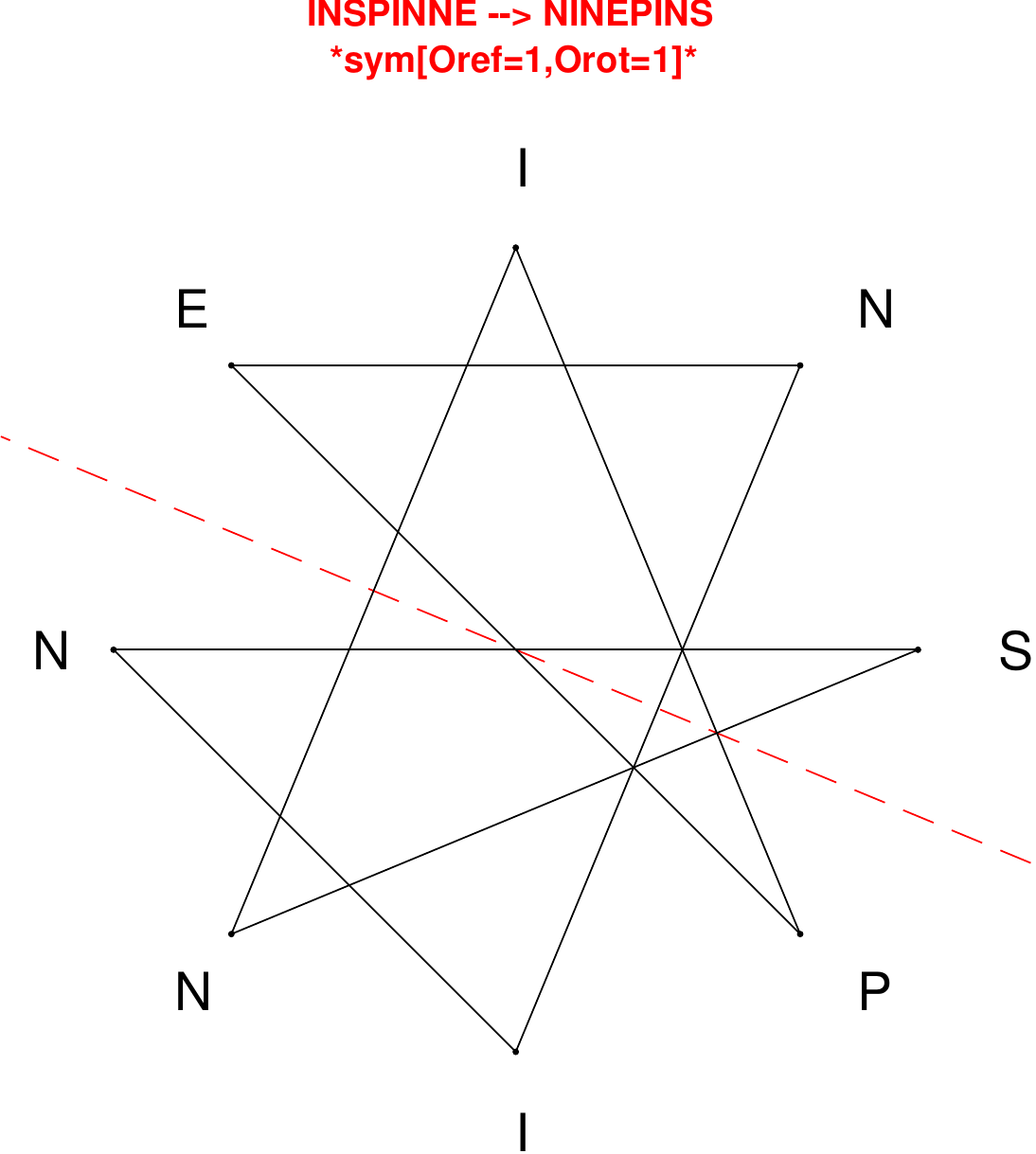}
\end{subfigure}
\hfill
\begin{subfigure}[T]{0.19\textwidth}
\centering
\includegraphics[width=\textwidth]{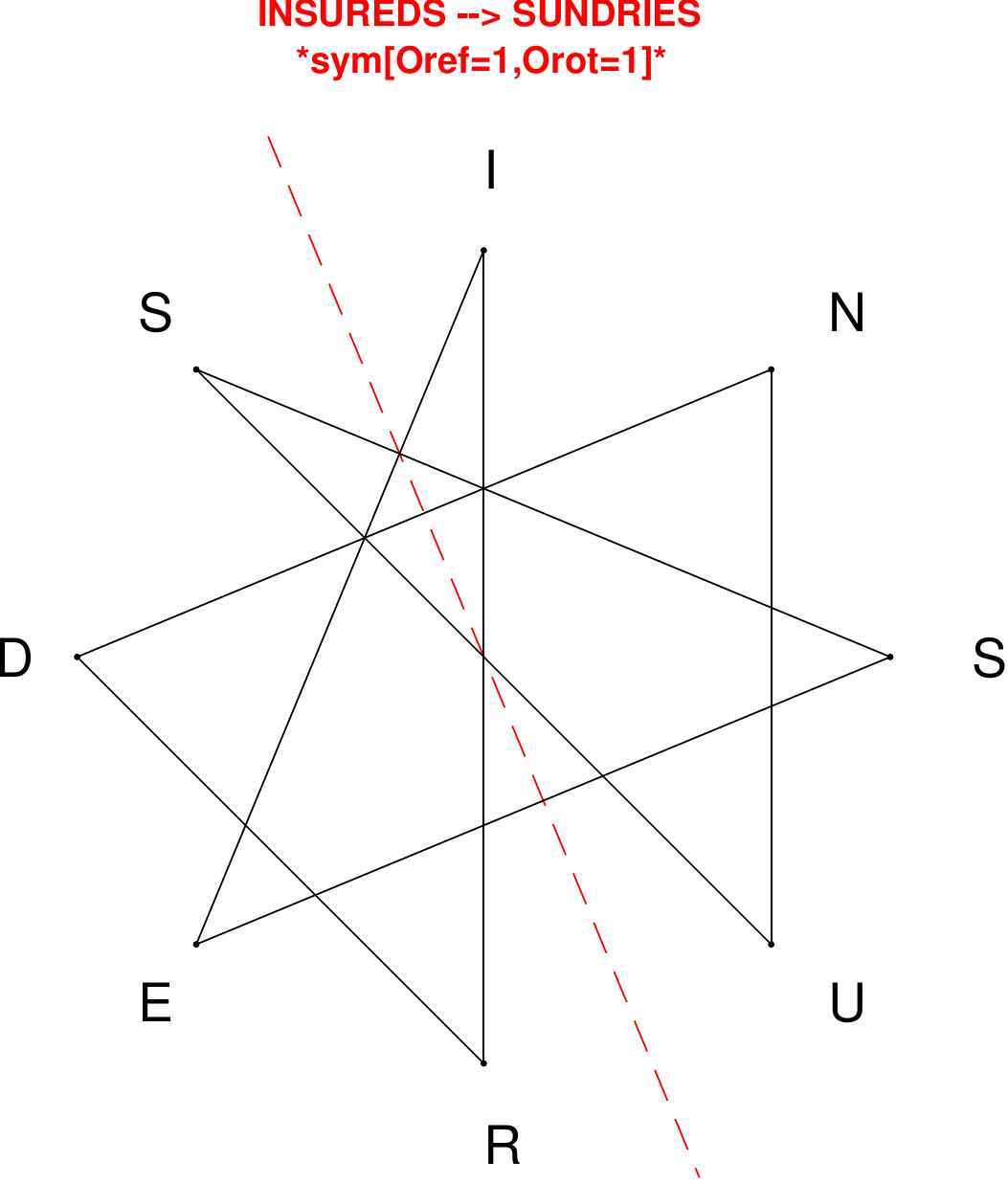}
\end{subfigure}
\hfill
\begin{subfigure}[T]{0.19\textwidth}
\centering
\includegraphics[width=\textwidth]{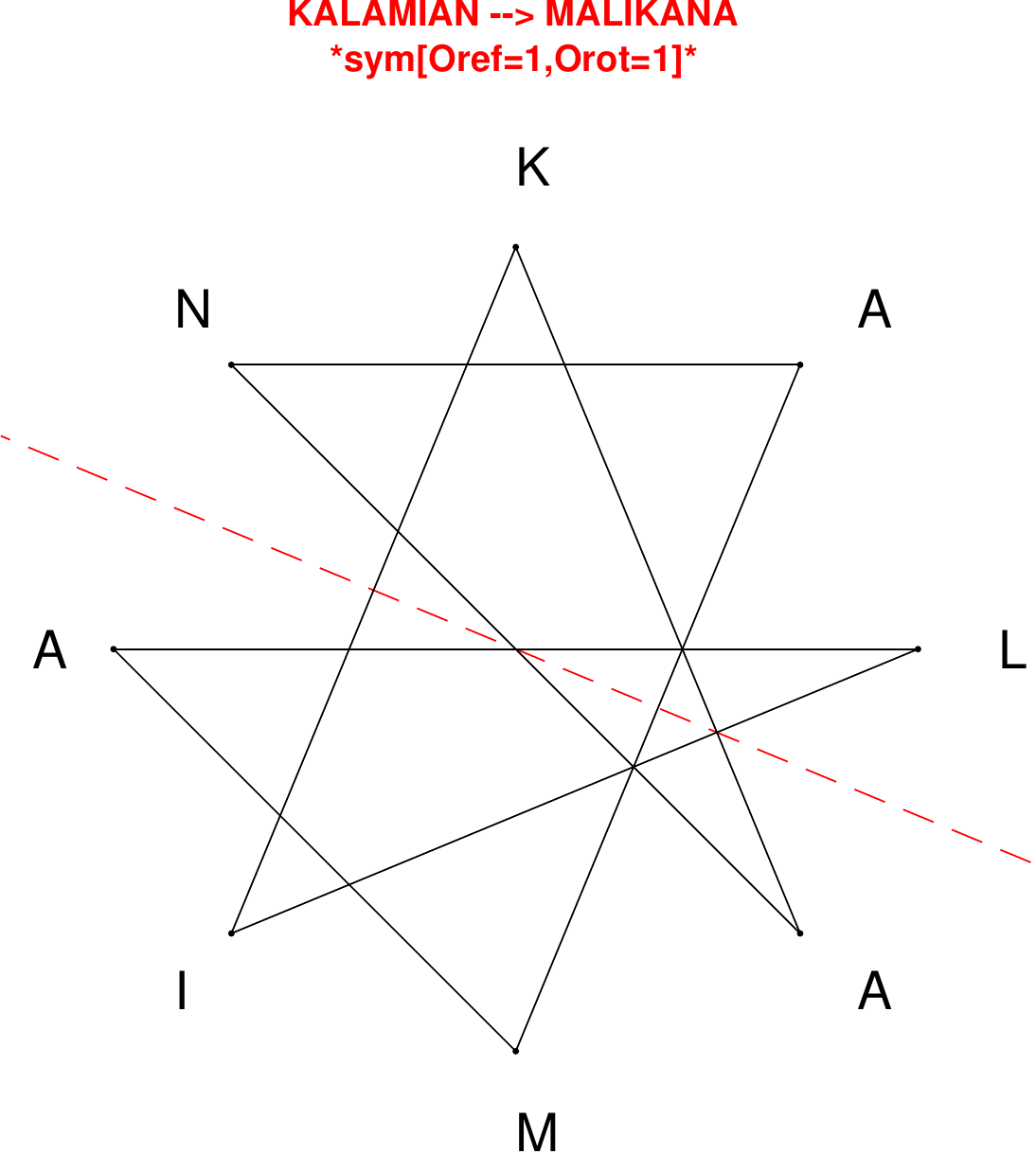}
\end{subfigure}
\hfill
\begin{subfigure}[T]{0.19\textwidth}
\centering
\includegraphics[width=\textwidth]{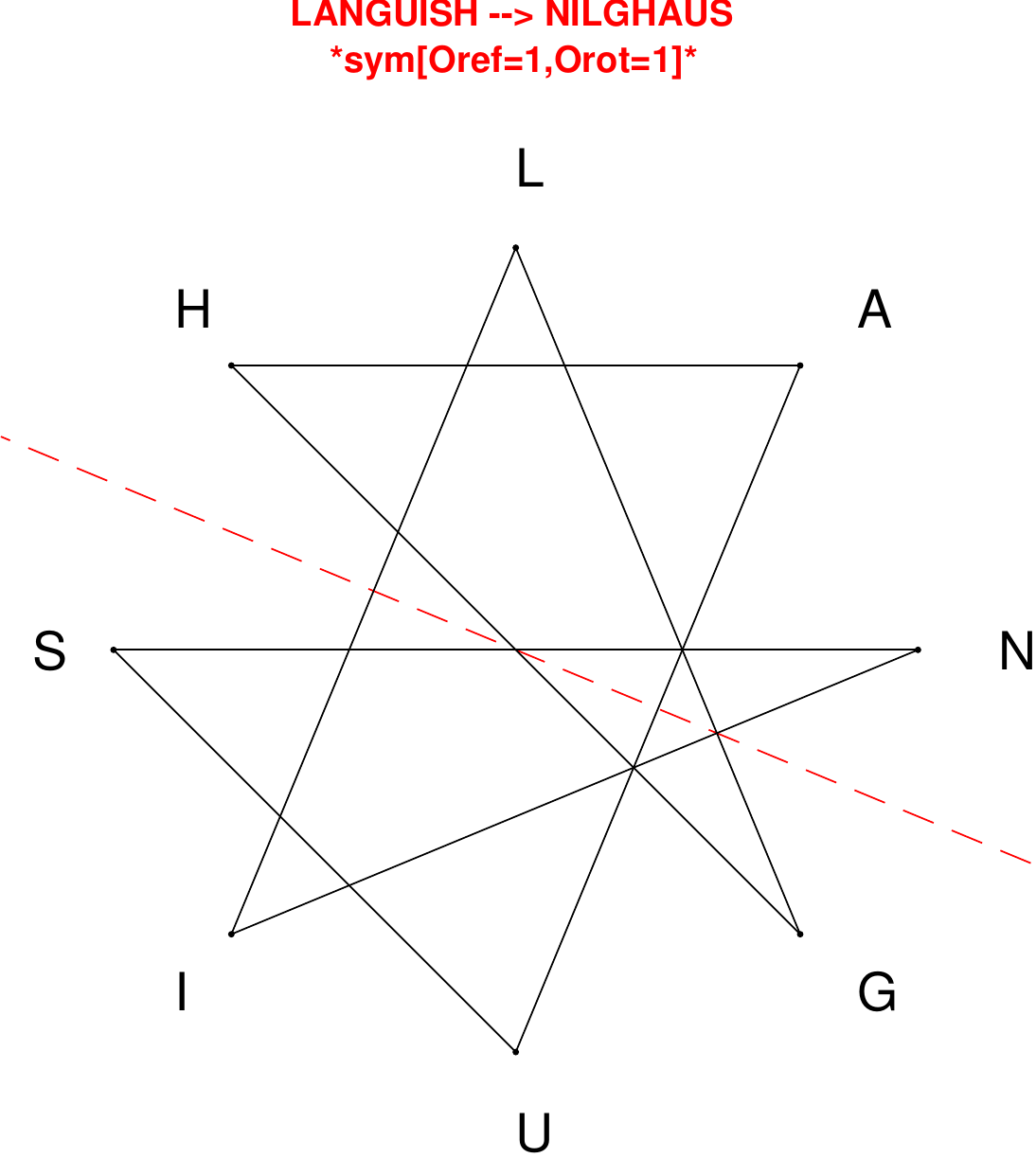}
\end{subfigure}
\end{figure}

\begin{figure}[H]
\centering
\begin{subfigure}[T]{0.19\textwidth}
\centering
\includegraphics[width=\textwidth]{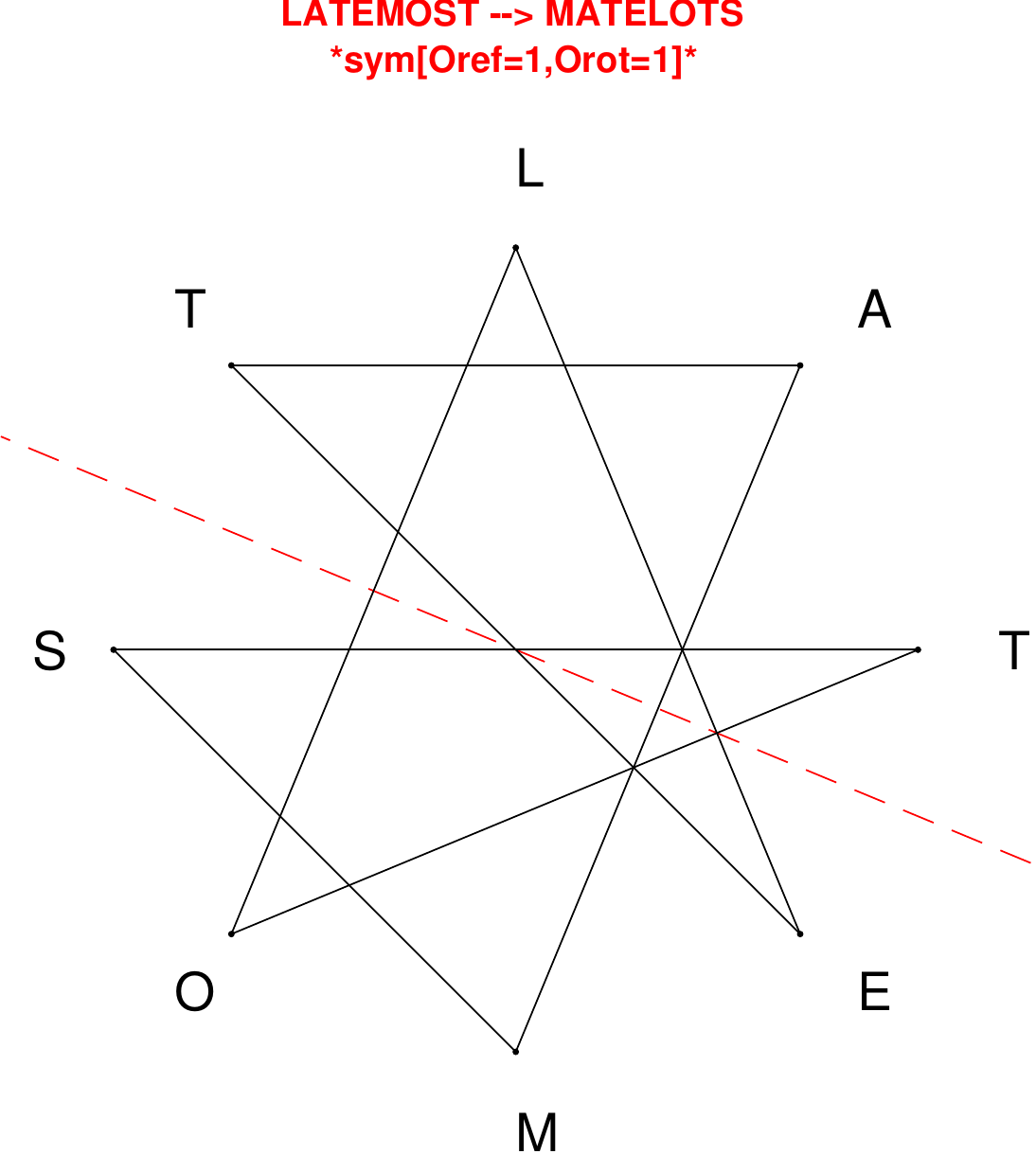}
\end{subfigure}
\hfill
\begin{subfigure}[T]{0.19\textwidth}
\centering
\includegraphics[width=\textwidth]{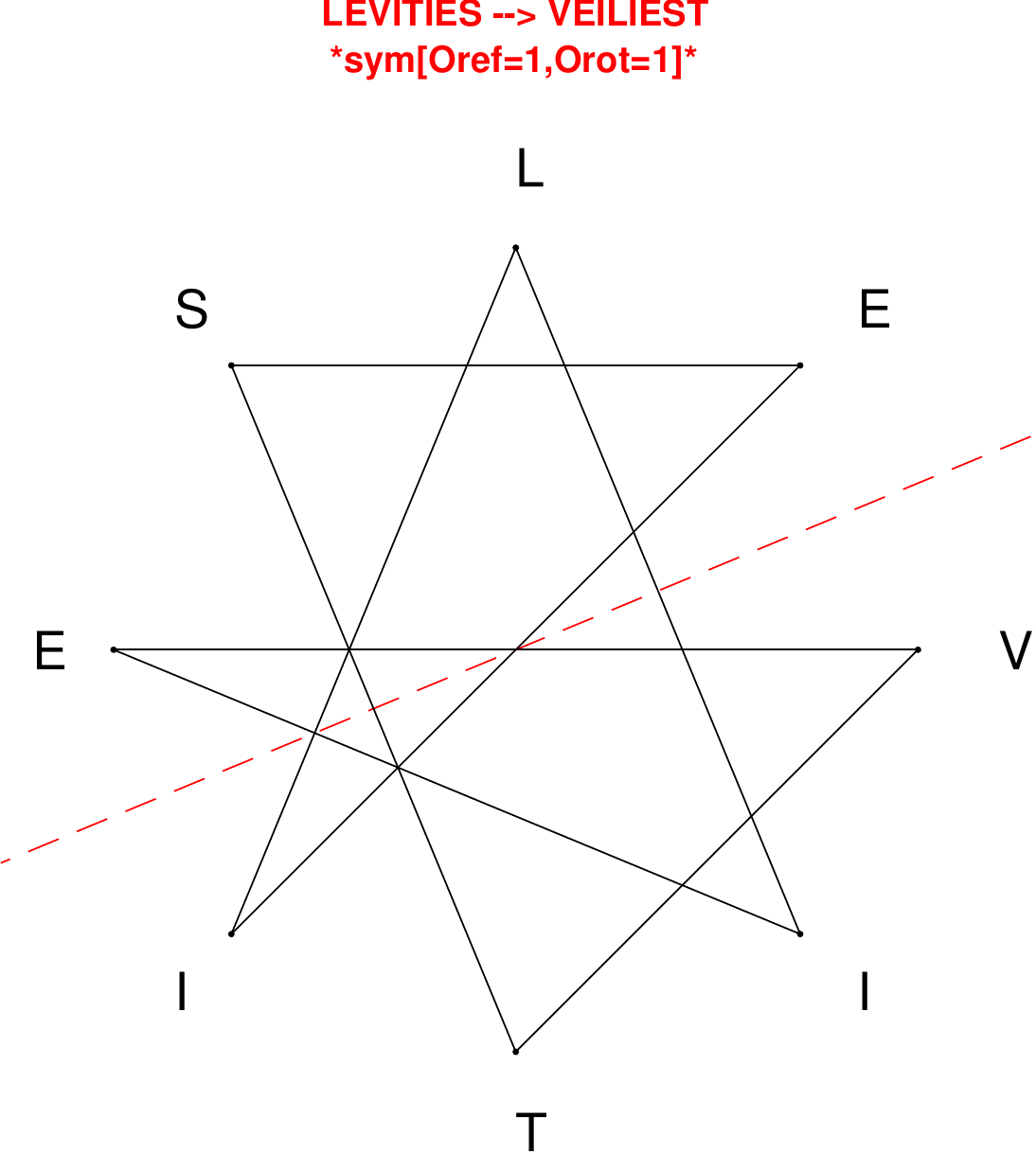}
\end{subfigure}
\hfill
\begin{subfigure}[T]{0.19\textwidth}
\centering
\includegraphics[width=\textwidth]{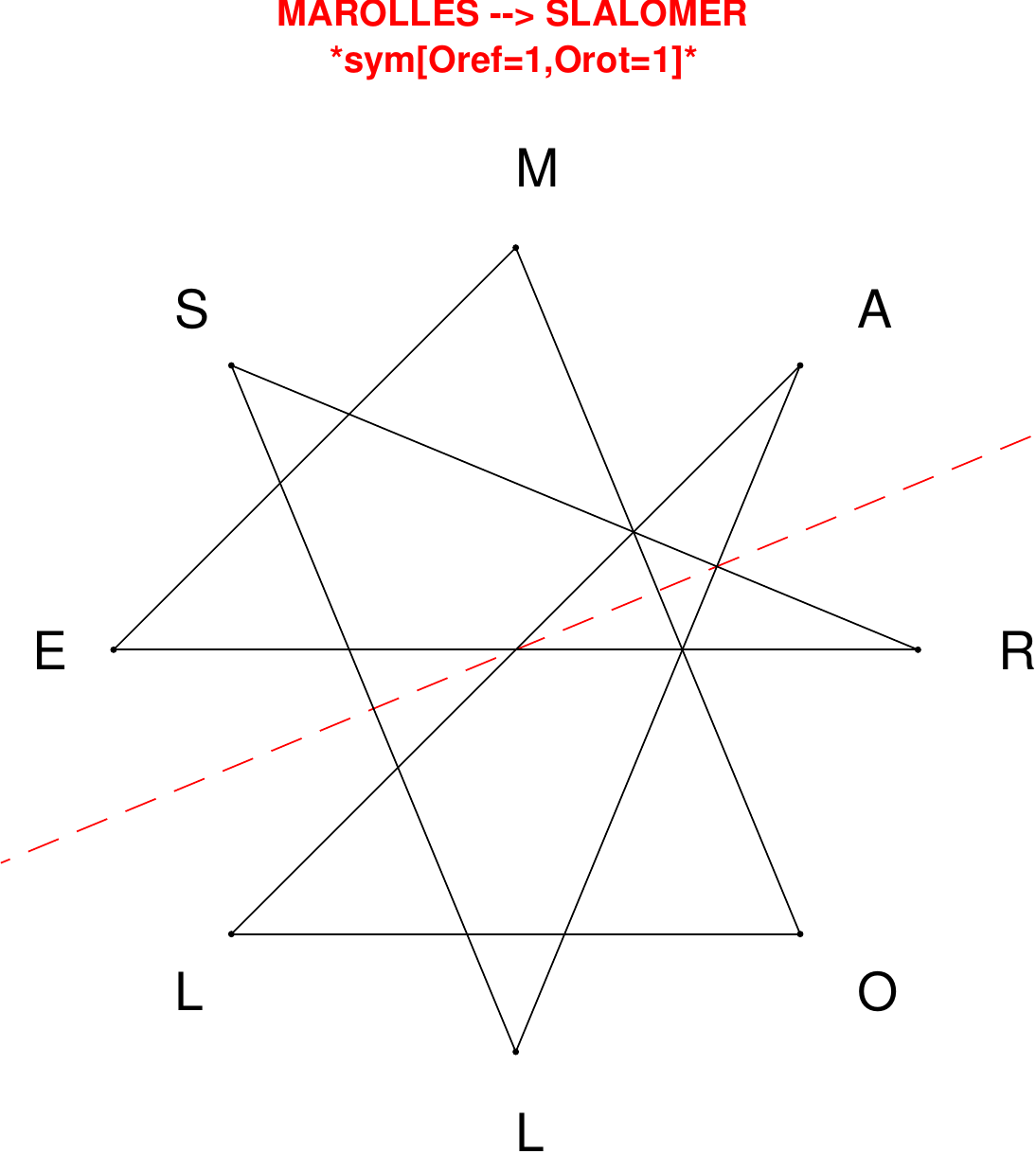}
\end{subfigure}
\hfill
\begin{subfigure}[T]{0.19\textwidth}
\centering
\includegraphics[width=\textwidth]{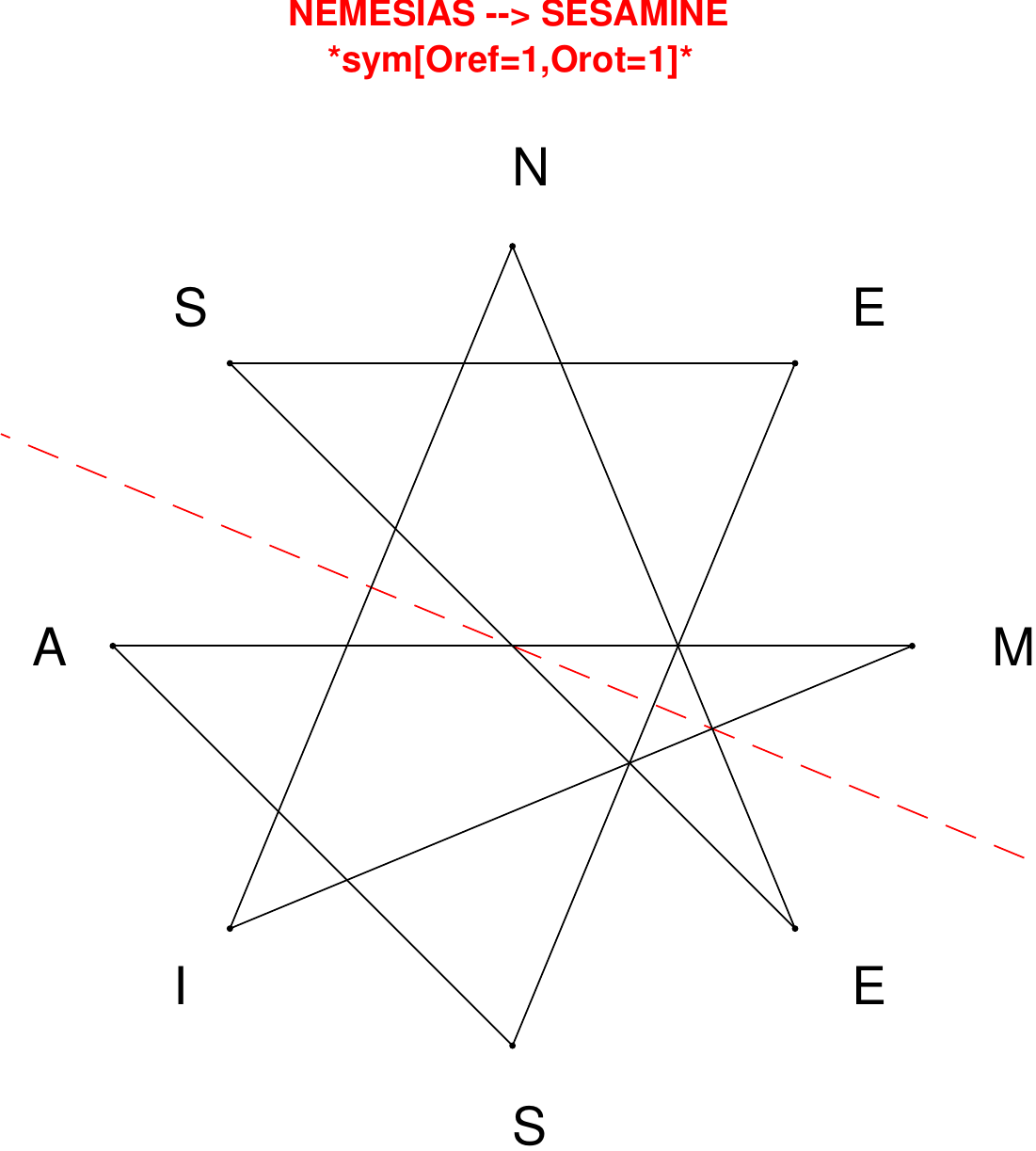}
\end{subfigure}
\hfill
\begin{subfigure}[T]{0.19\textwidth}
\centering
\includegraphics[width=\textwidth]{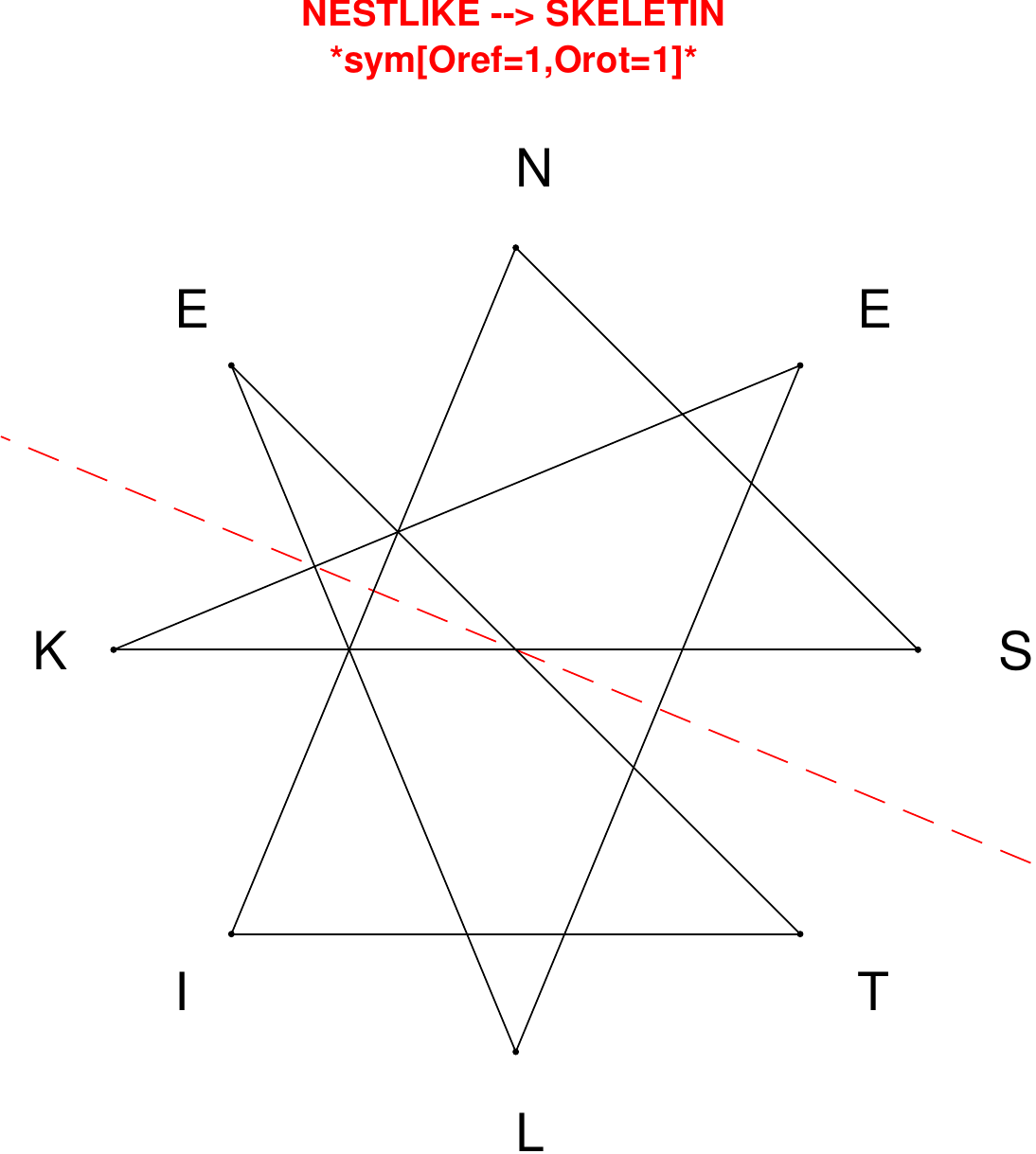}
\end{subfigure}
\end{figure}

\begin{figure}[H]
\centering
\begin{subfigure}[T]{0.19\textwidth}
\centering
\includegraphics[width=\textwidth]{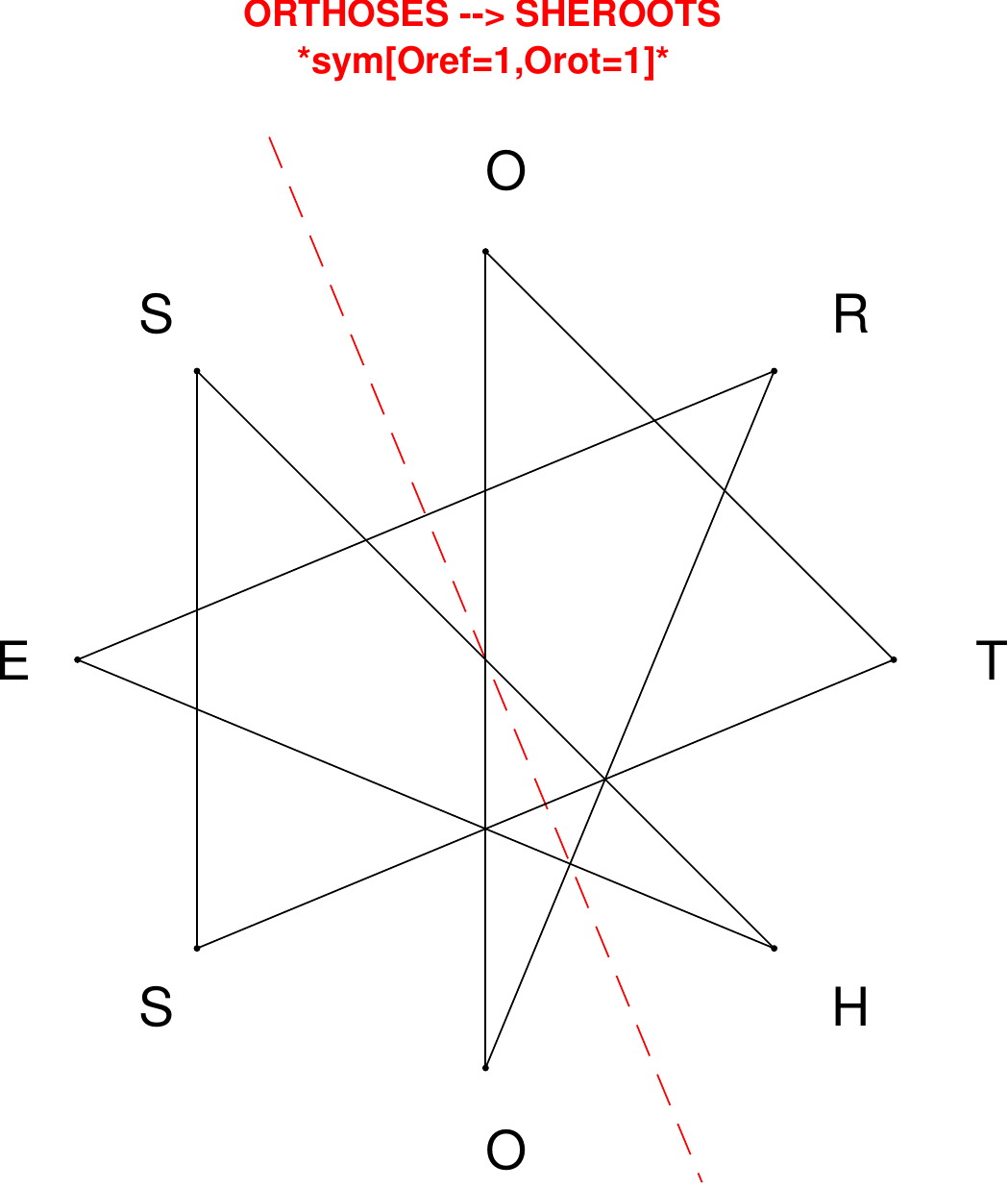}
\end{subfigure}
\hfill
\begin{subfigure}[T]{0.19\textwidth}
\centering
\includegraphics[width=\textwidth]{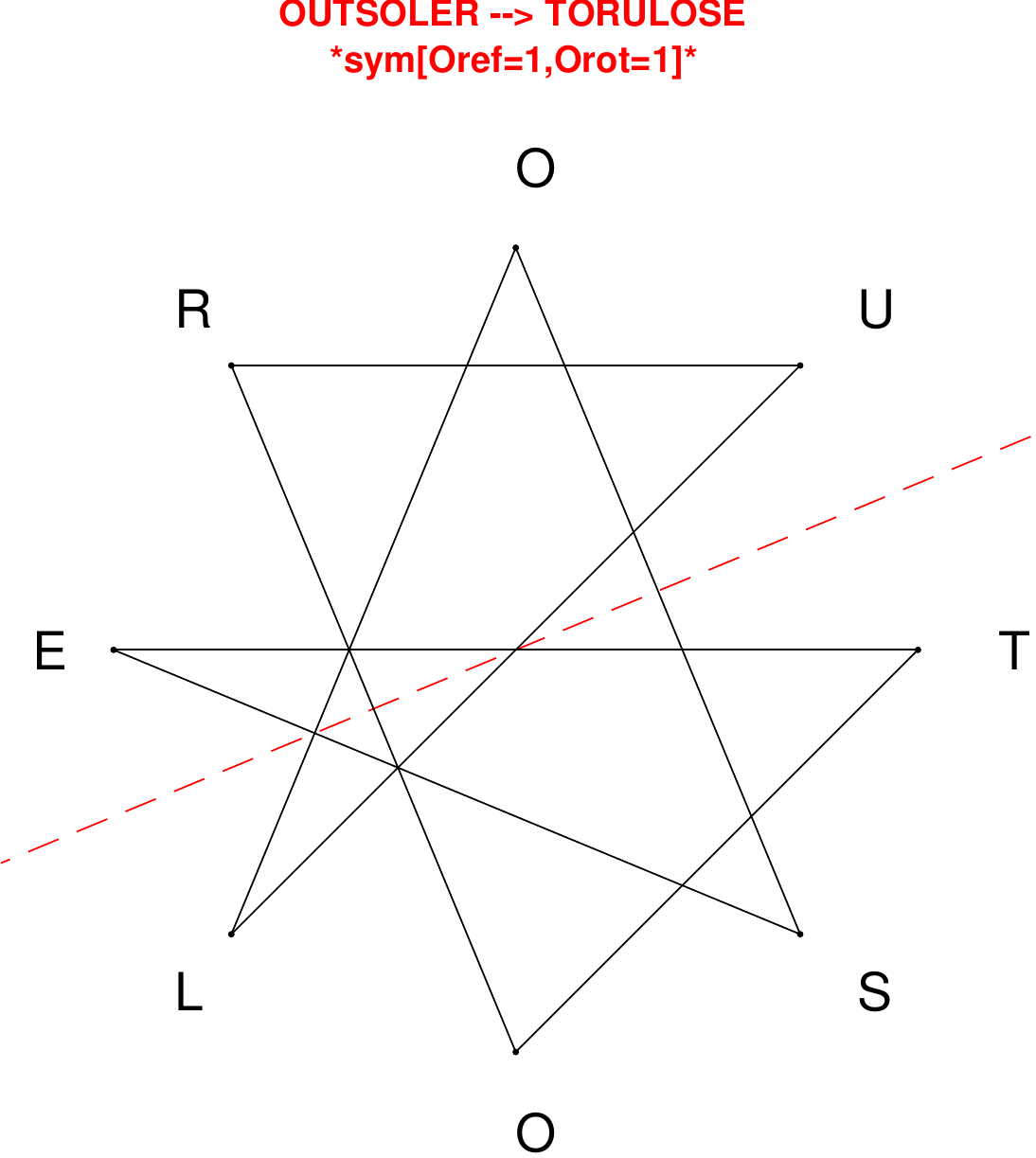}
\end{subfigure}
\hfill
\begin{subfigure}[T]{0.19\textwidth}
\centering
\includegraphics[width=\textwidth]{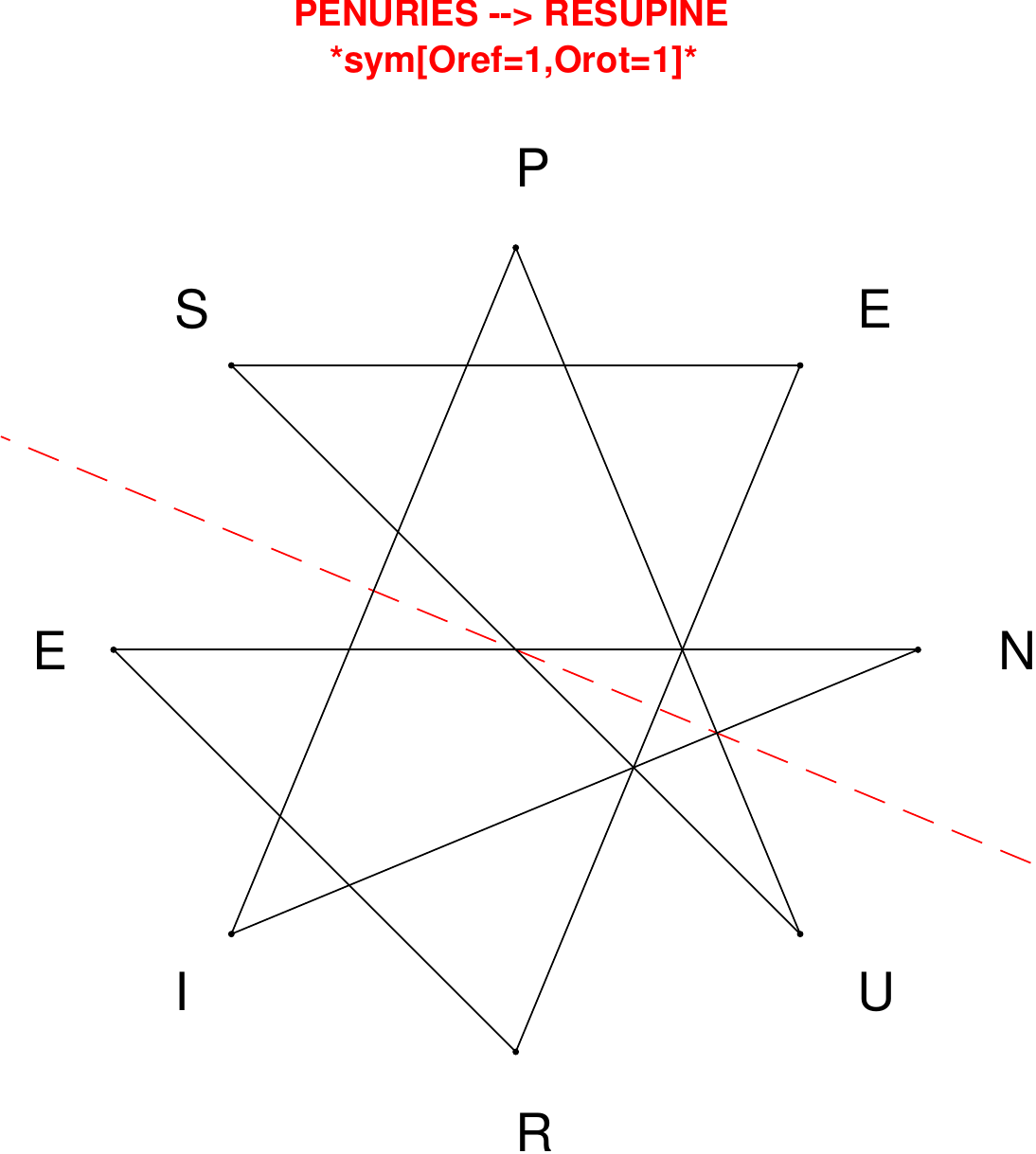}
\end{subfigure}
\hfill
\begin{subfigure}[T]{0.19\textwidth}
\centering
\includegraphics[width=\textwidth]{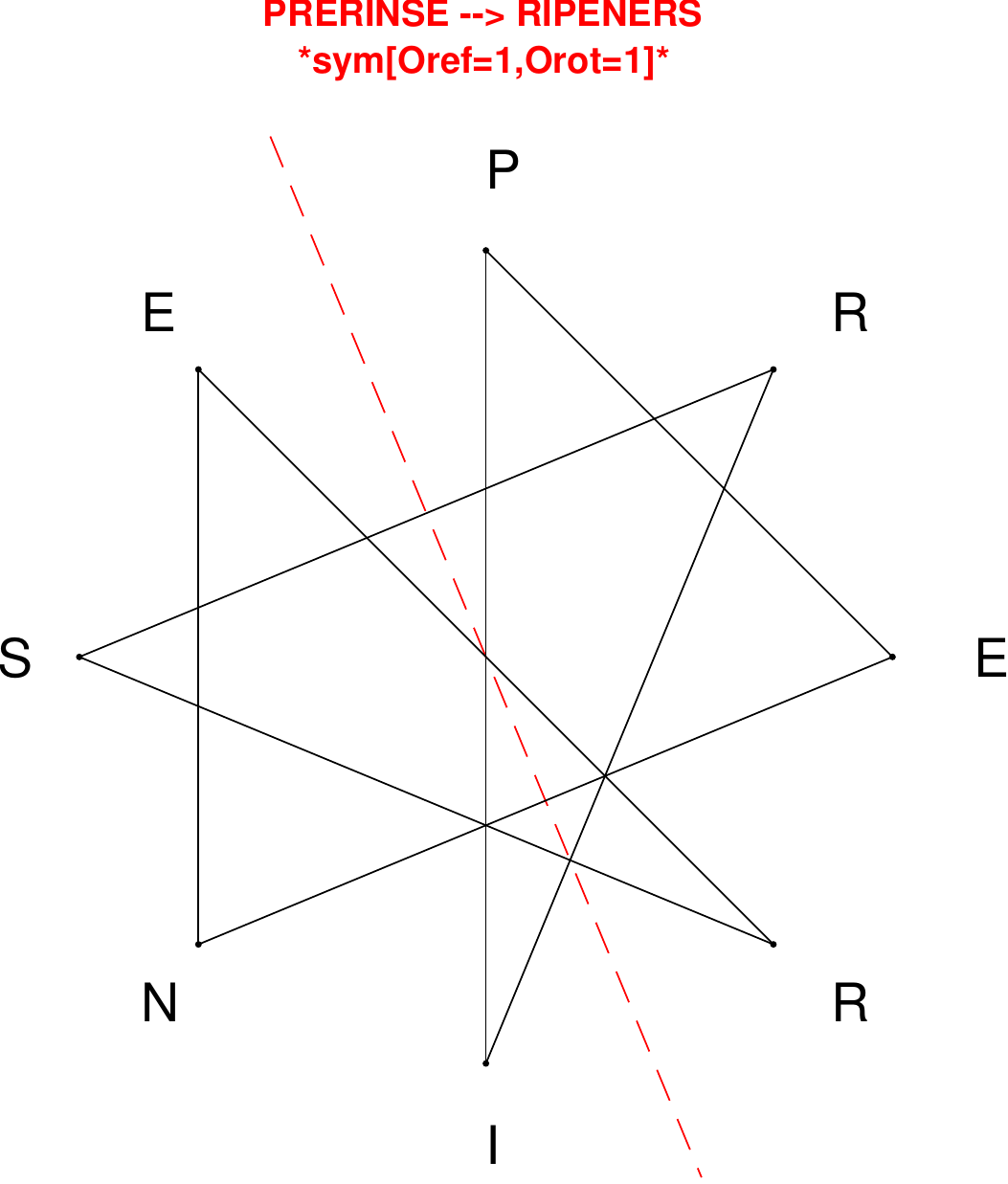}
\end{subfigure}
\hfill
\begin{subfigure}[T]{0.19\textwidth}
\centering
\includegraphics[width=\textwidth]{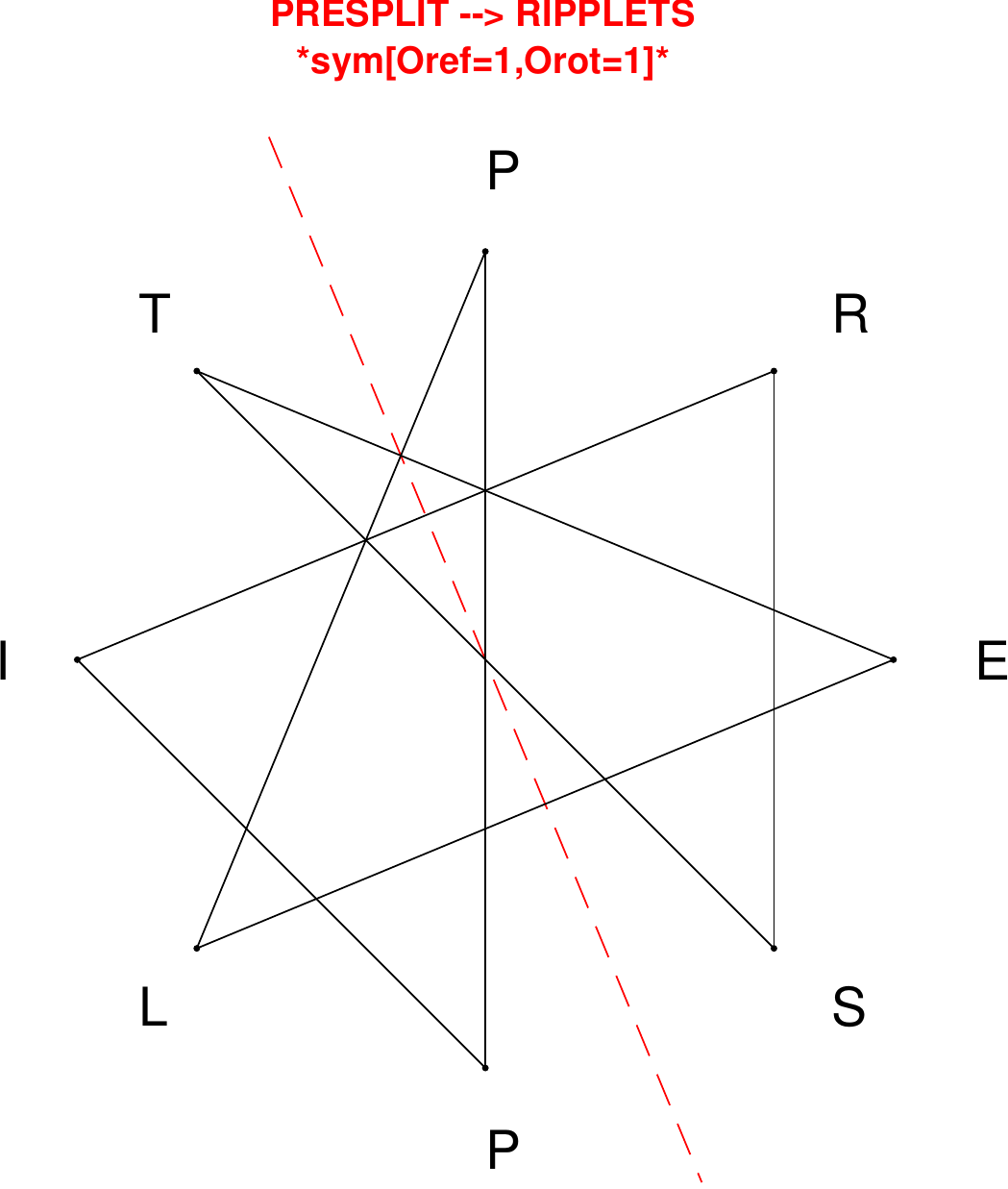}
\end{subfigure}
\end{figure}

\begin{figure}[H]
\centering
\begin{subfigure}[T]{0.19\textwidth}
\centering
\includegraphics[width=\textwidth]{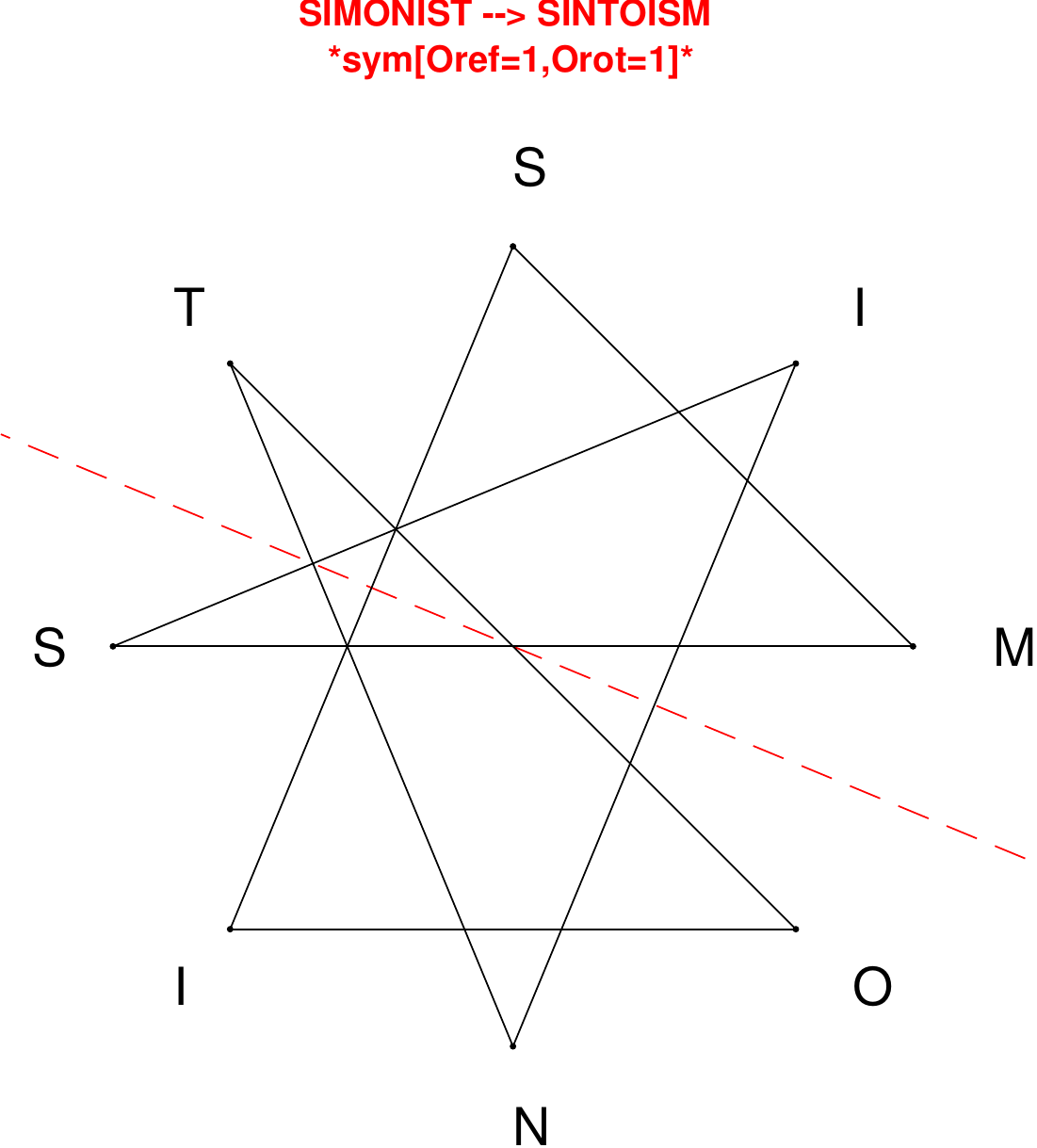}
\end{subfigure}
\hfill
\begin{subfigure}[T]{0.19\textwidth}
\centering
\includegraphics[width=\textwidth]{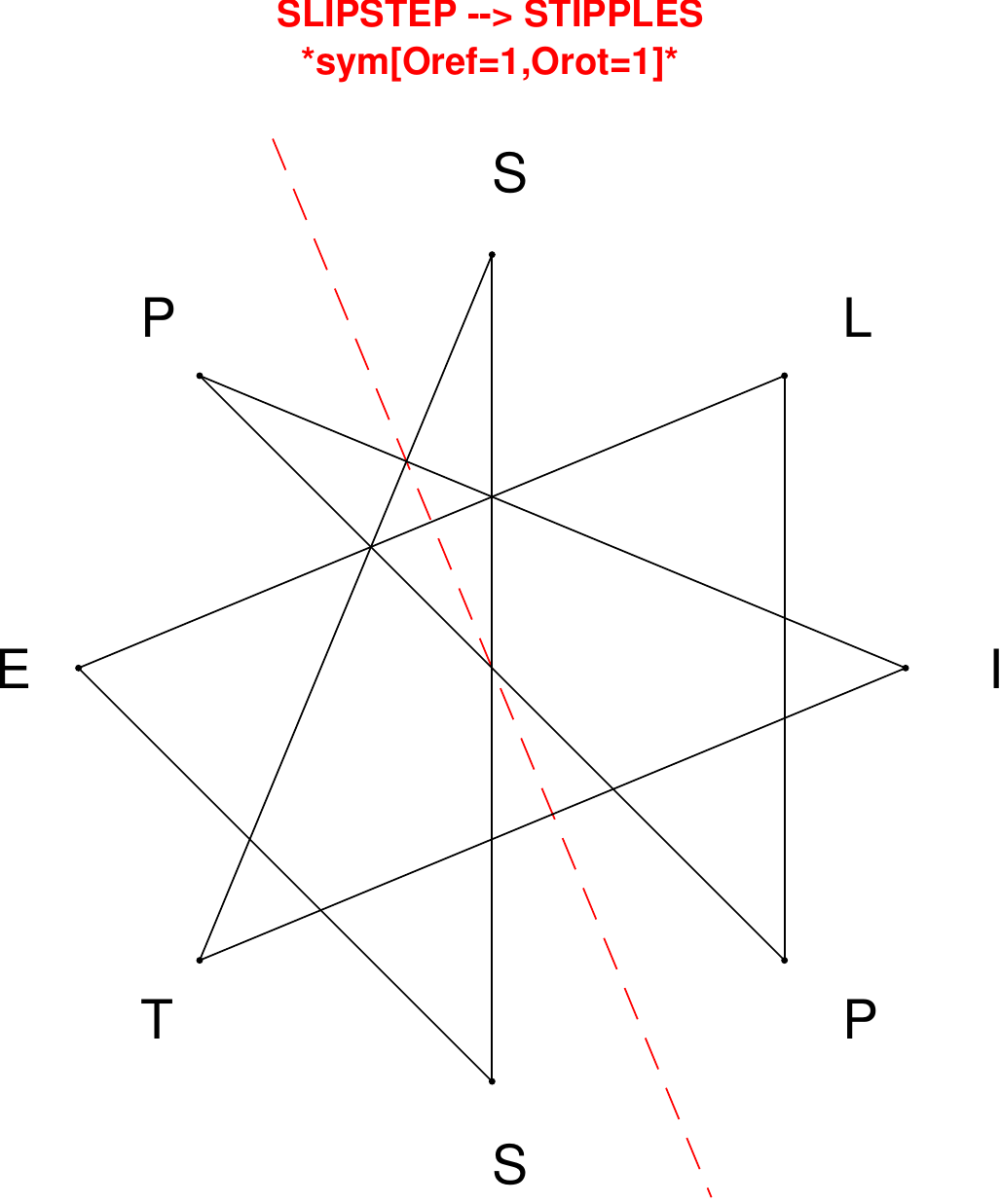}
\end{subfigure}
\hfill
\begin{subfigure}[T]{0.19\textwidth}
\centering
\includegraphics[width=\textwidth]{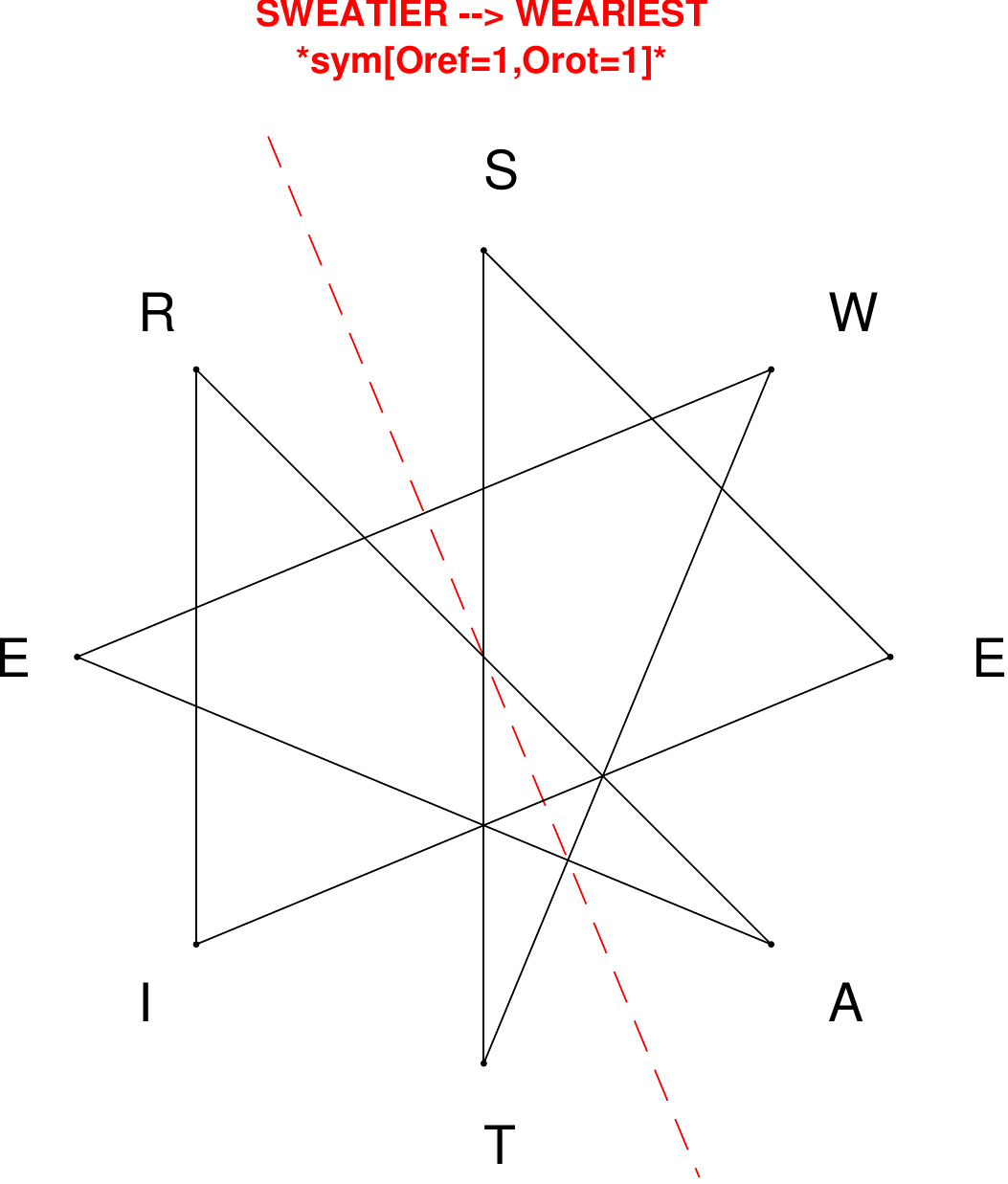}
\end{subfigure}
\hfill
\begin{subfigure}[T]{0.19\textwidth}
\centering
\includegraphics[width=\textwidth]{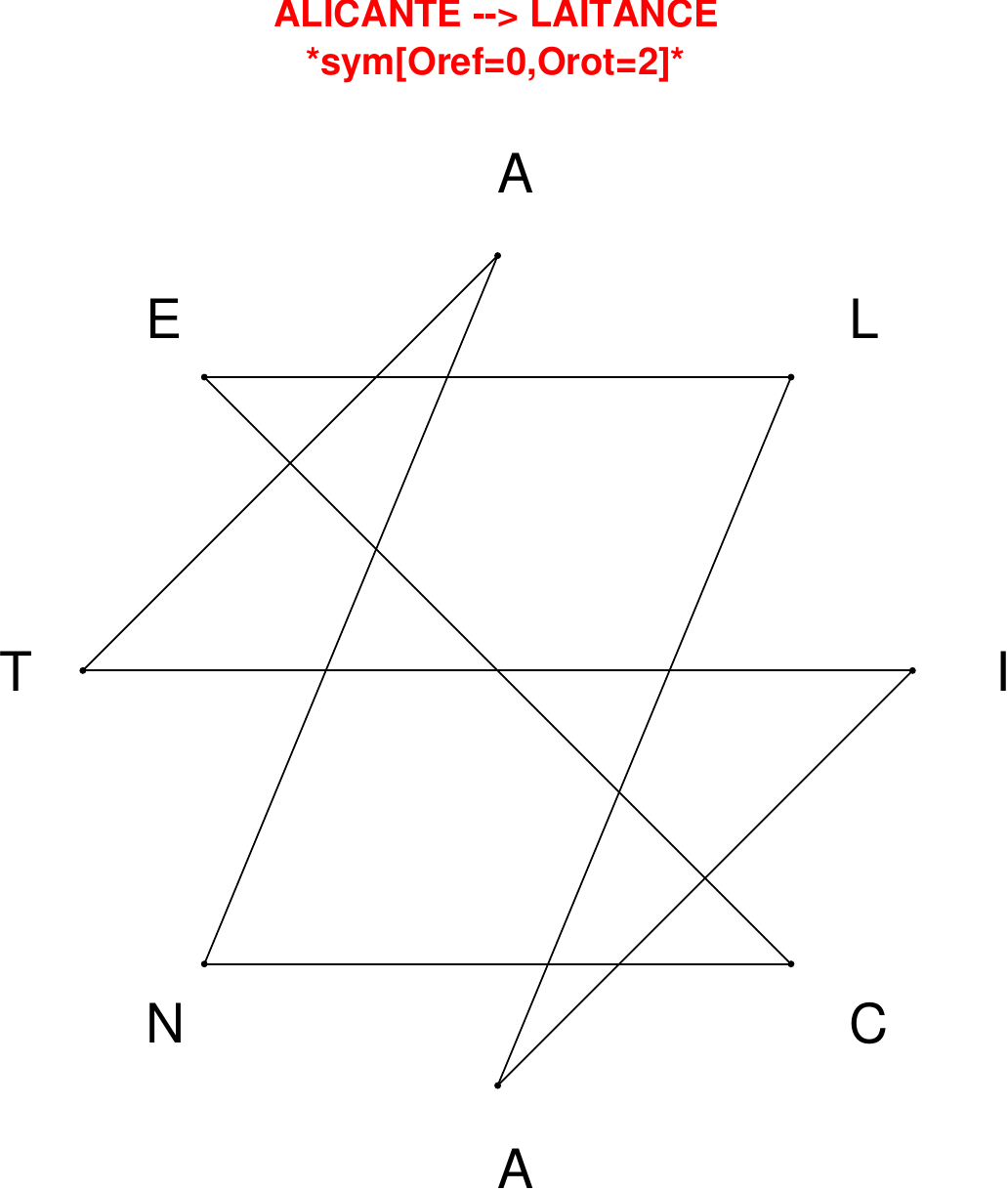}
\end{subfigure}
\hfill
\begin{subfigure}[T]{0.19\textwidth}
\centering
\includegraphics[width=\textwidth]{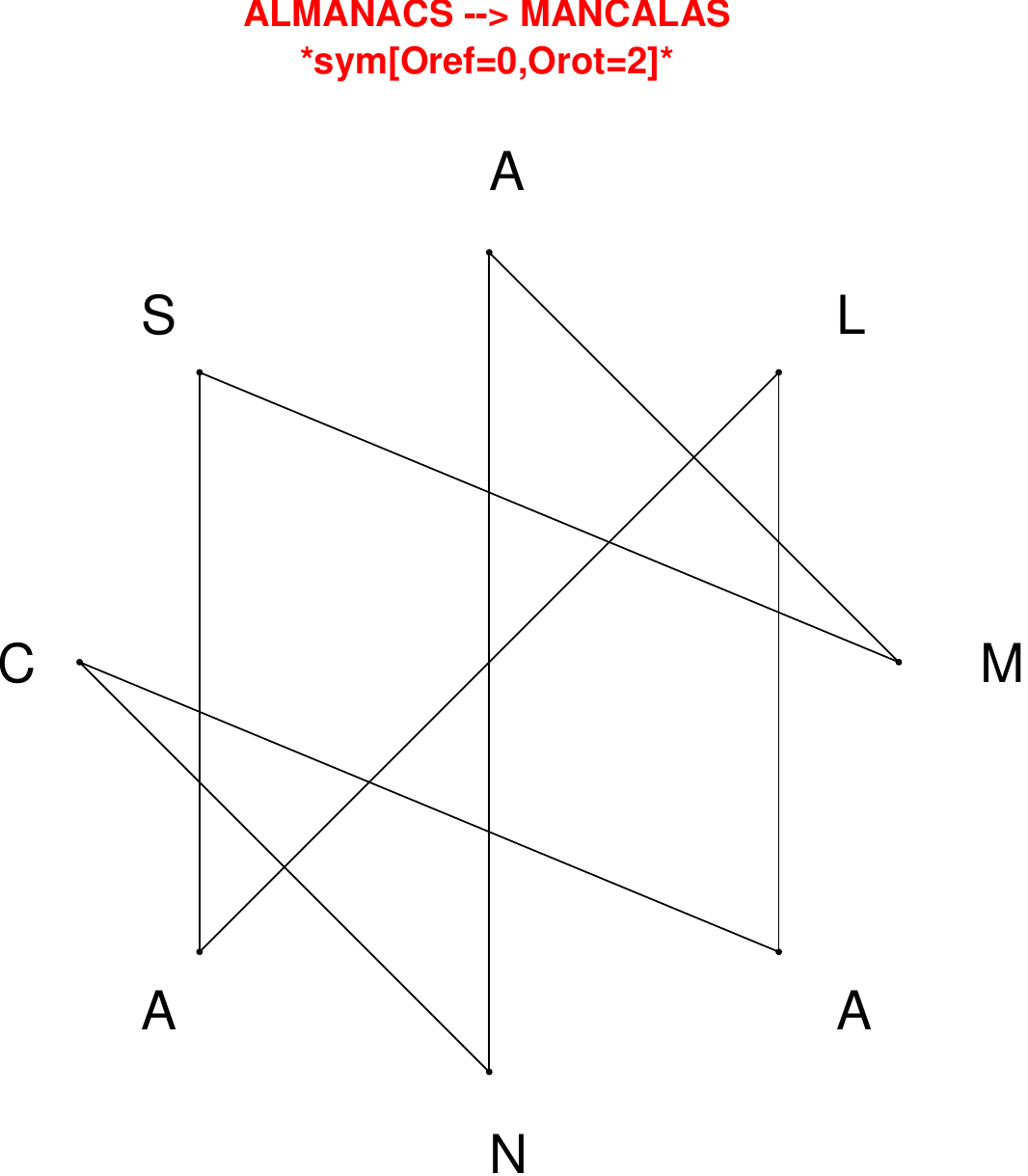}
\end{subfigure}
\end{figure}

\begin{figure}[H]
\centering
\begin{subfigure}[T]{0.19\textwidth}
\centering
\includegraphics[width=\textwidth]{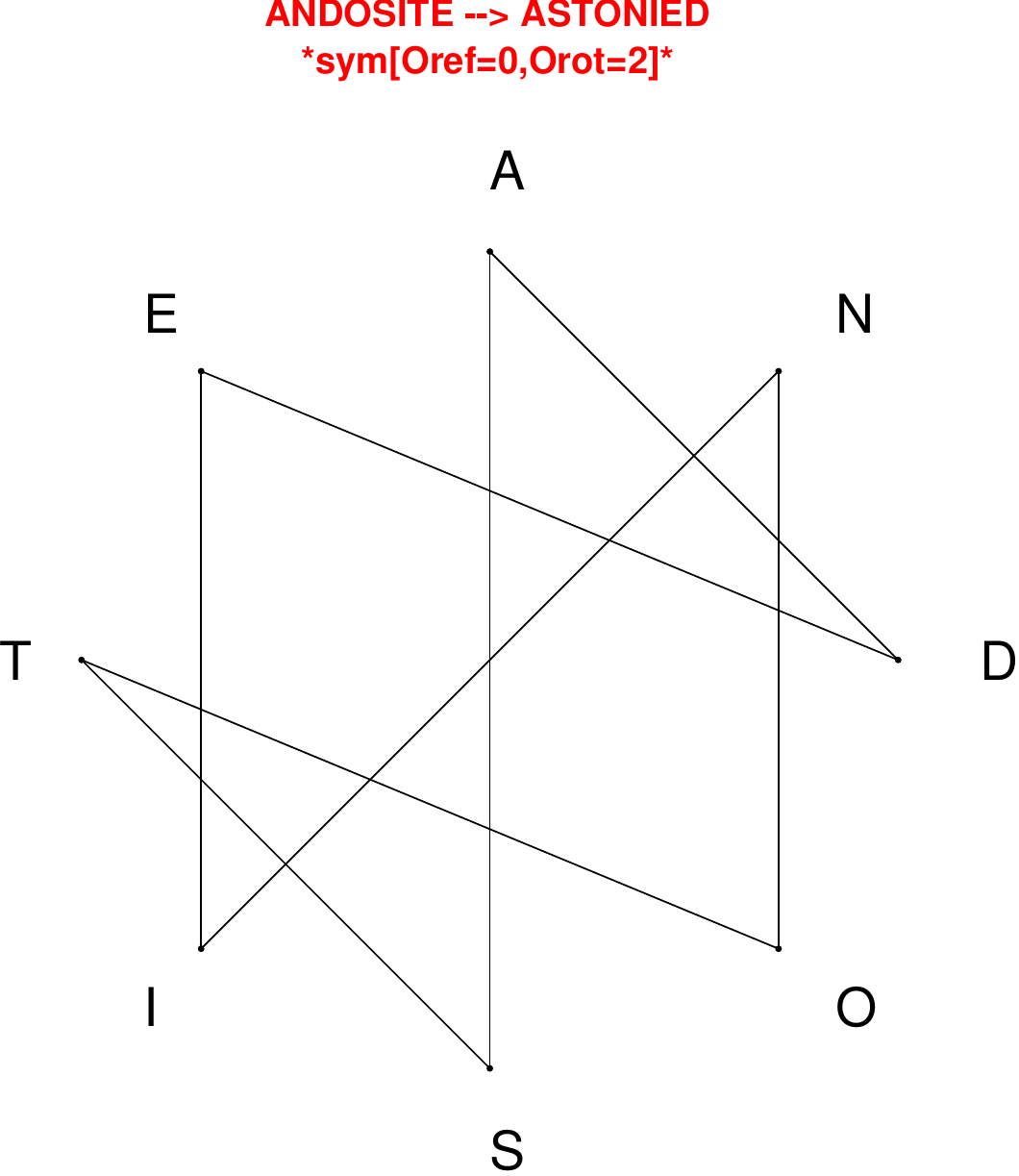}
\end{subfigure}
\hfill
\begin{subfigure}[T]{0.19\textwidth}
\centering
\includegraphics[width=\textwidth]{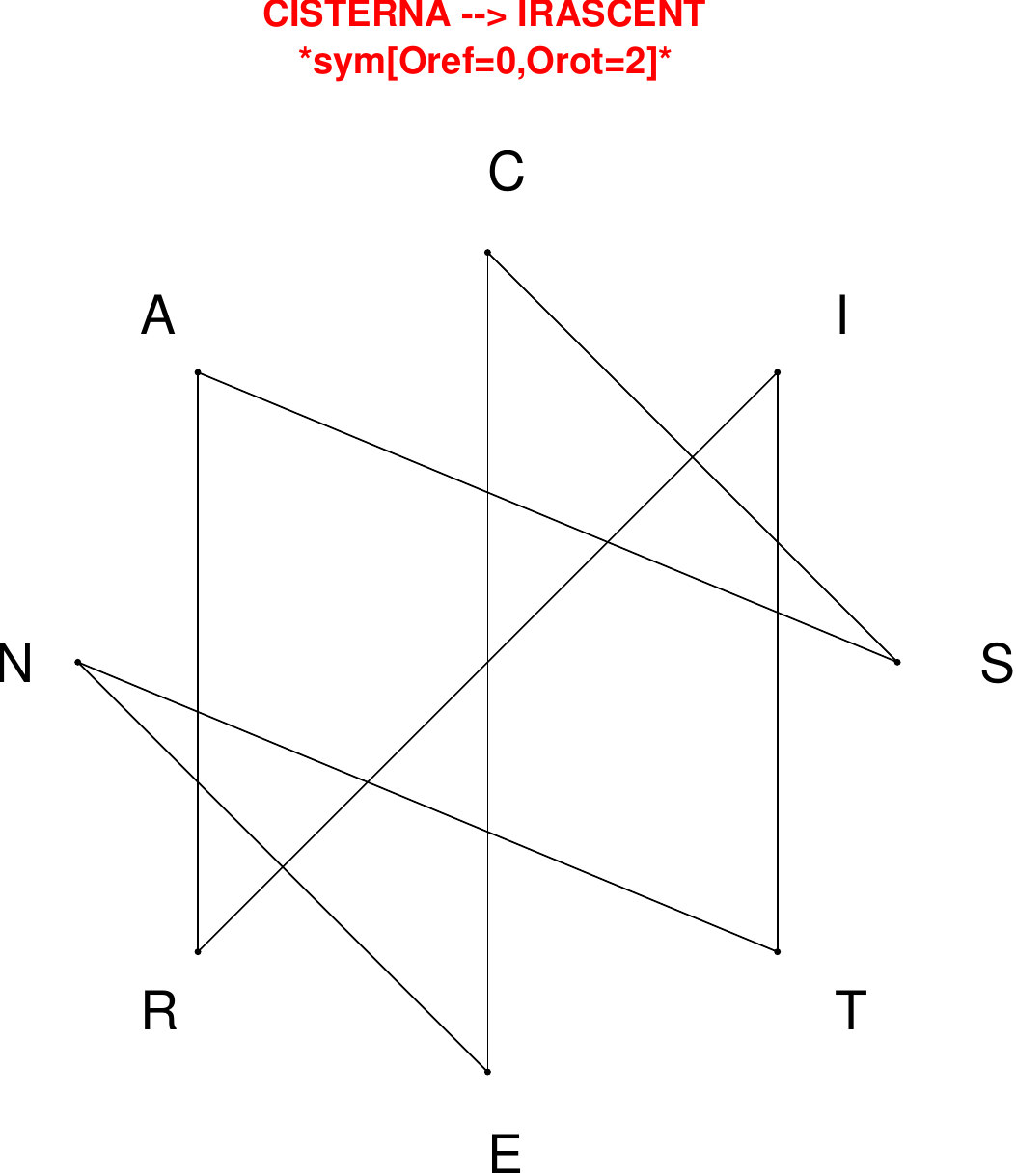}
\end{subfigure}
\hfill
\begin{subfigure}[T]{0.19\textwidth}
\centering
\includegraphics[width=\textwidth]{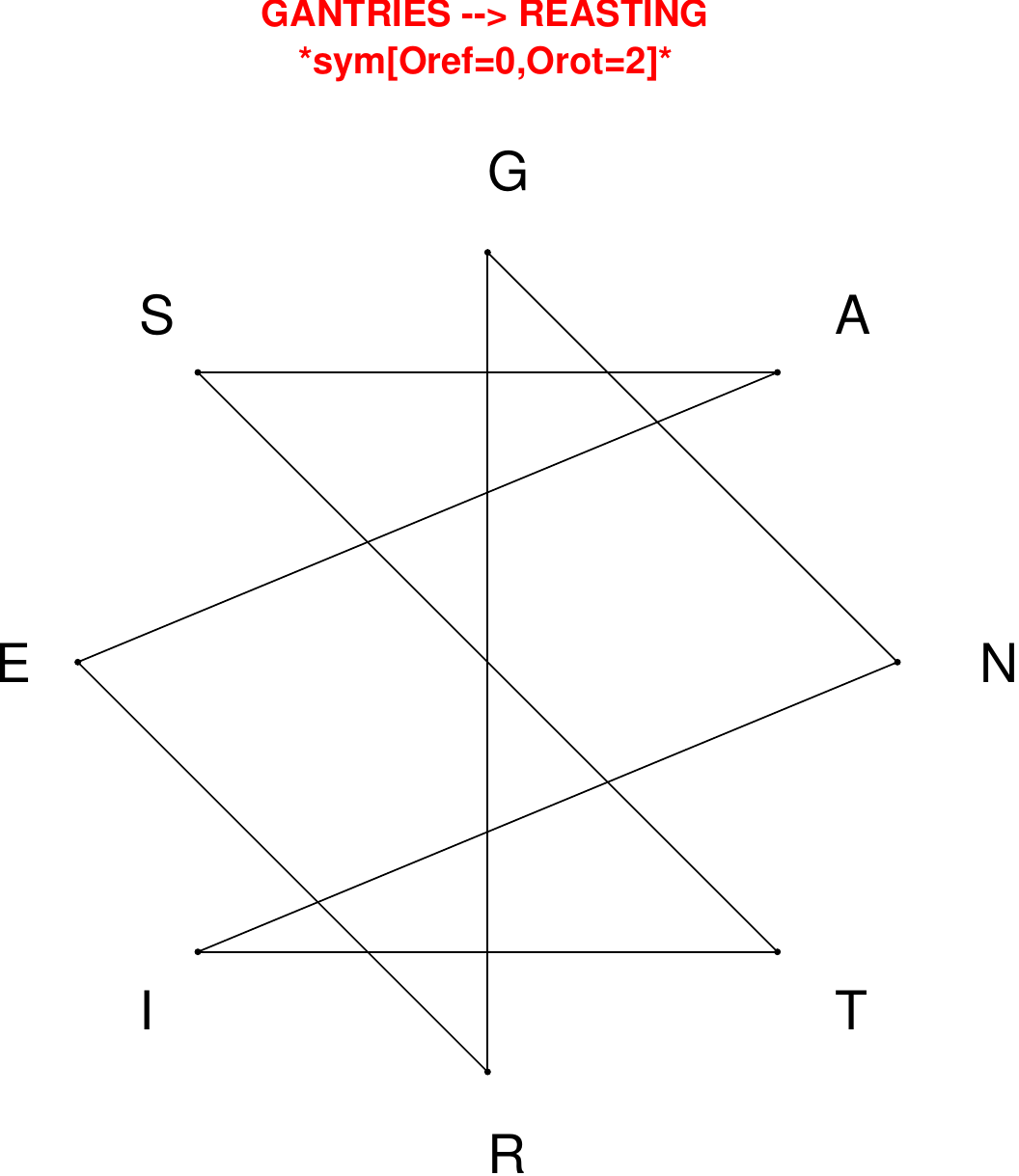}
\end{subfigure}
\hfill
\begin{subfigure}[T]{0.19\textwidth}
\centering
\includegraphics[width=\textwidth]{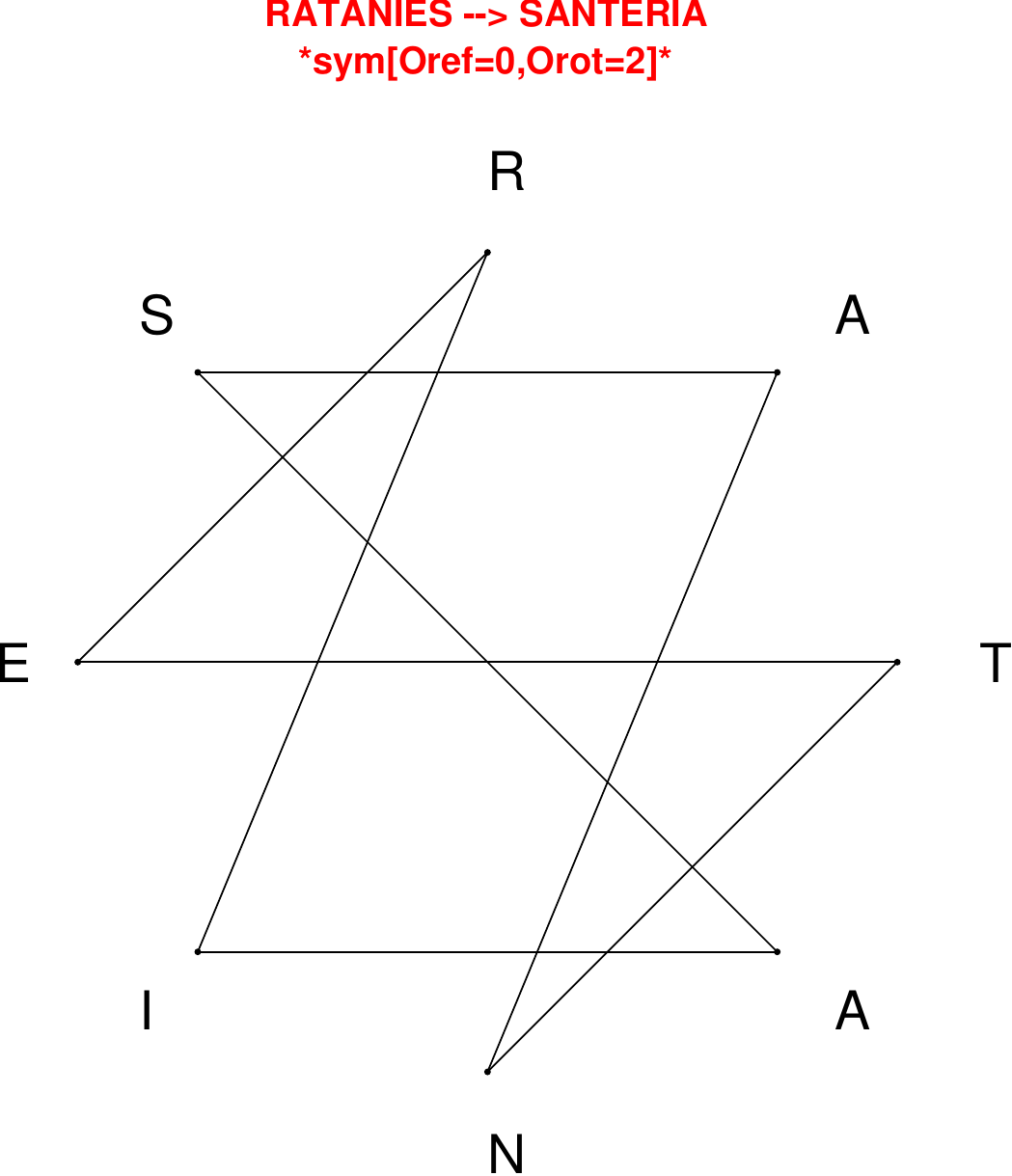}
\end{subfigure}
\hfill
\begin{subfigure}[T]{0.19\textwidth}
\centering
\includegraphics[width=\textwidth]{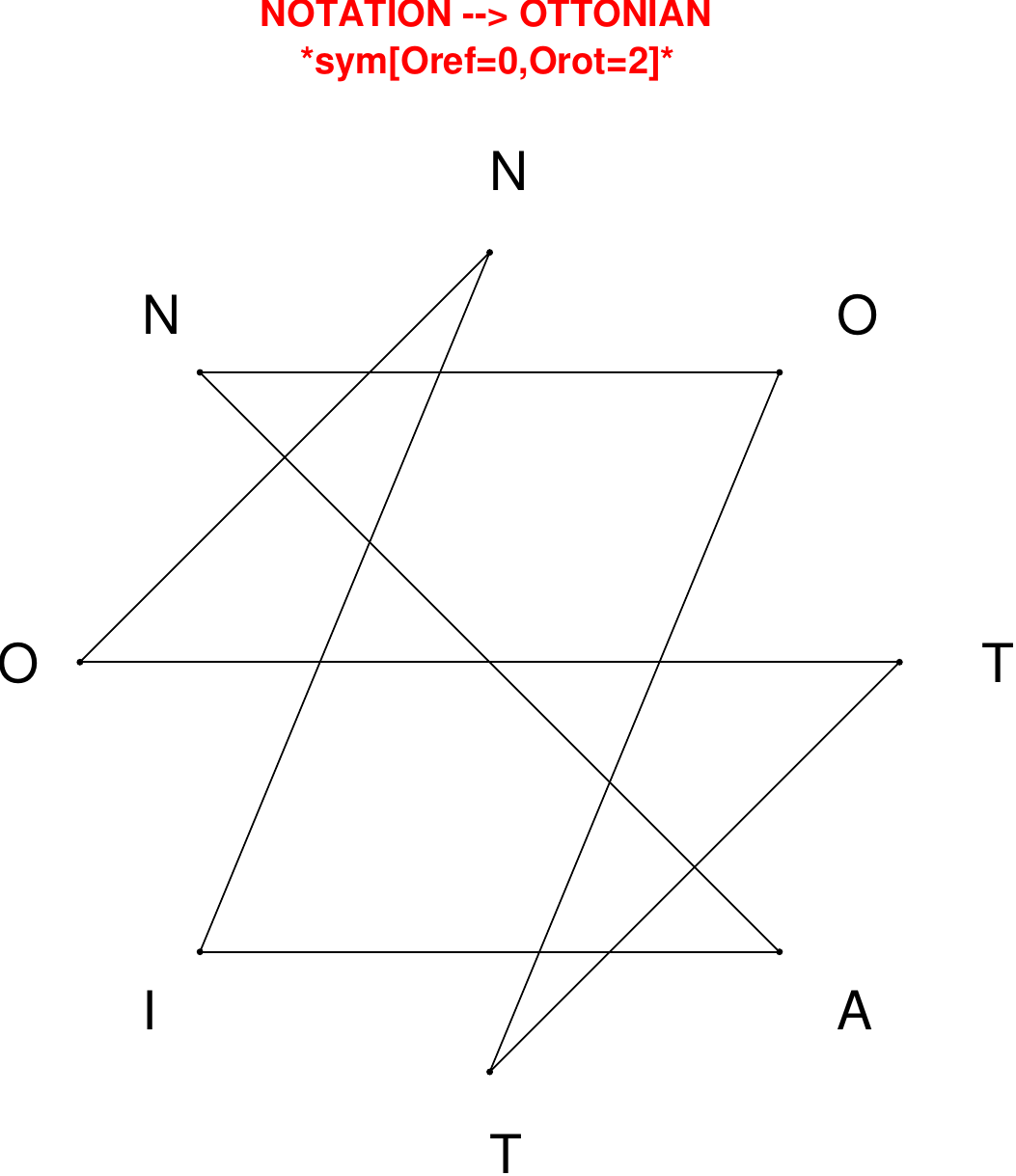}
\end{subfigure}
\end{figure}

\begin{figure}[H]
\centering
\begin{subfigure}[T]{0.19\textwidth}
\centering
\includegraphics[width=\textwidth]{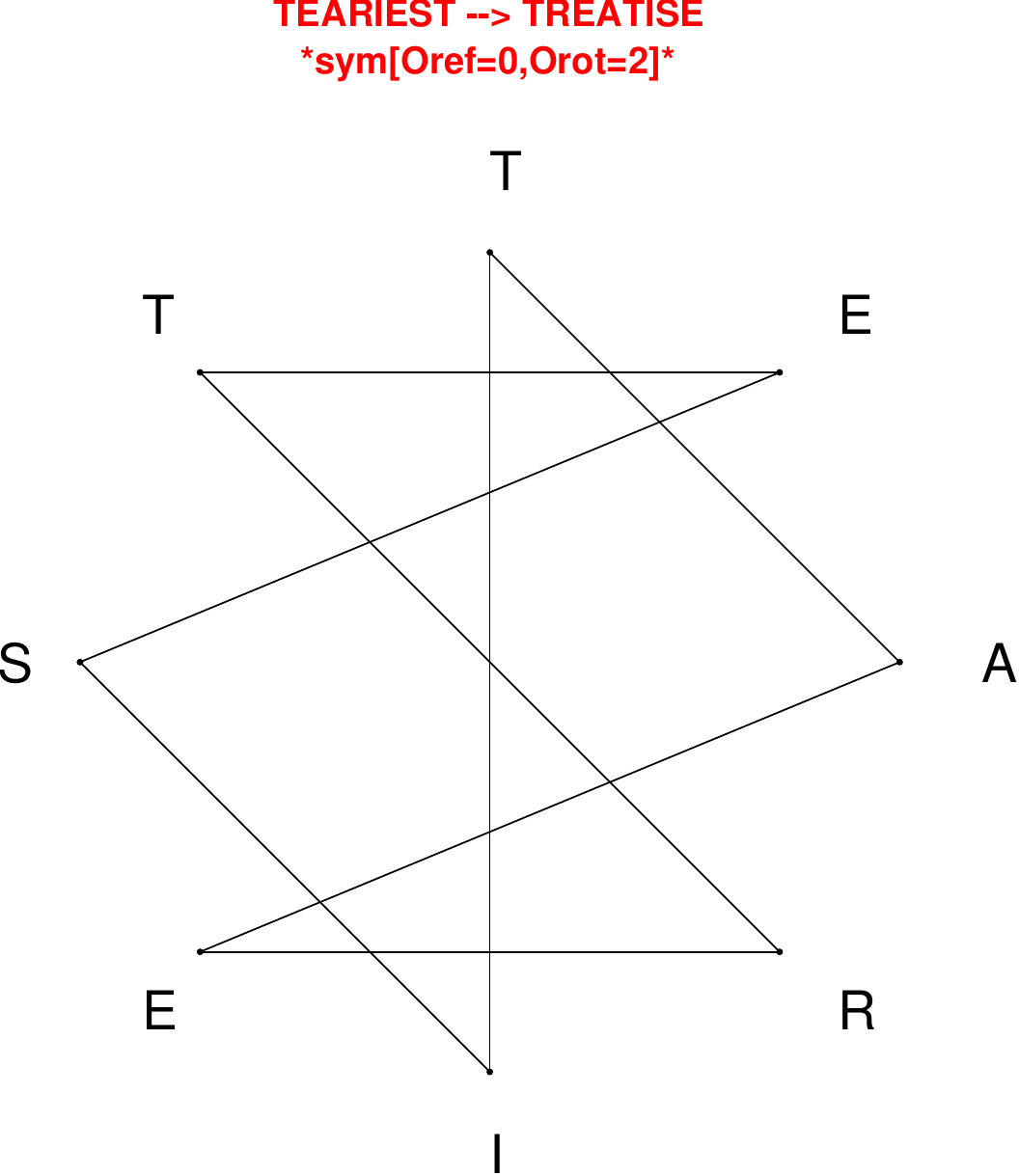}
\end{subfigure}
\hfill
\begin{subfigure}[T]{0.19\textwidth}
\centering
\includegraphics[width=\textwidth]{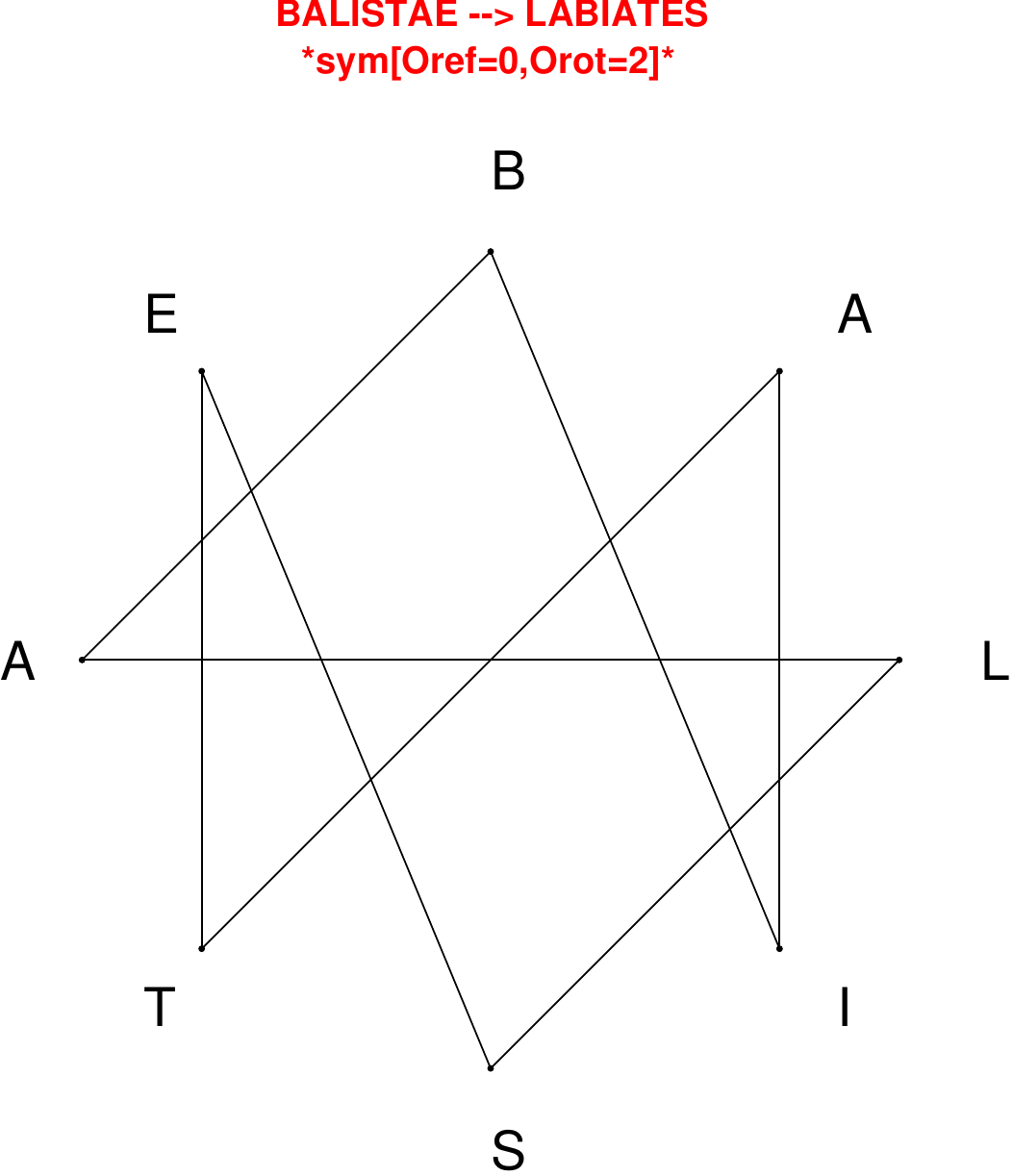}
\end{subfigure}
\hfill
\begin{subfigure}[T]{0.19\textwidth}
\centering
\includegraphics[width=\textwidth]{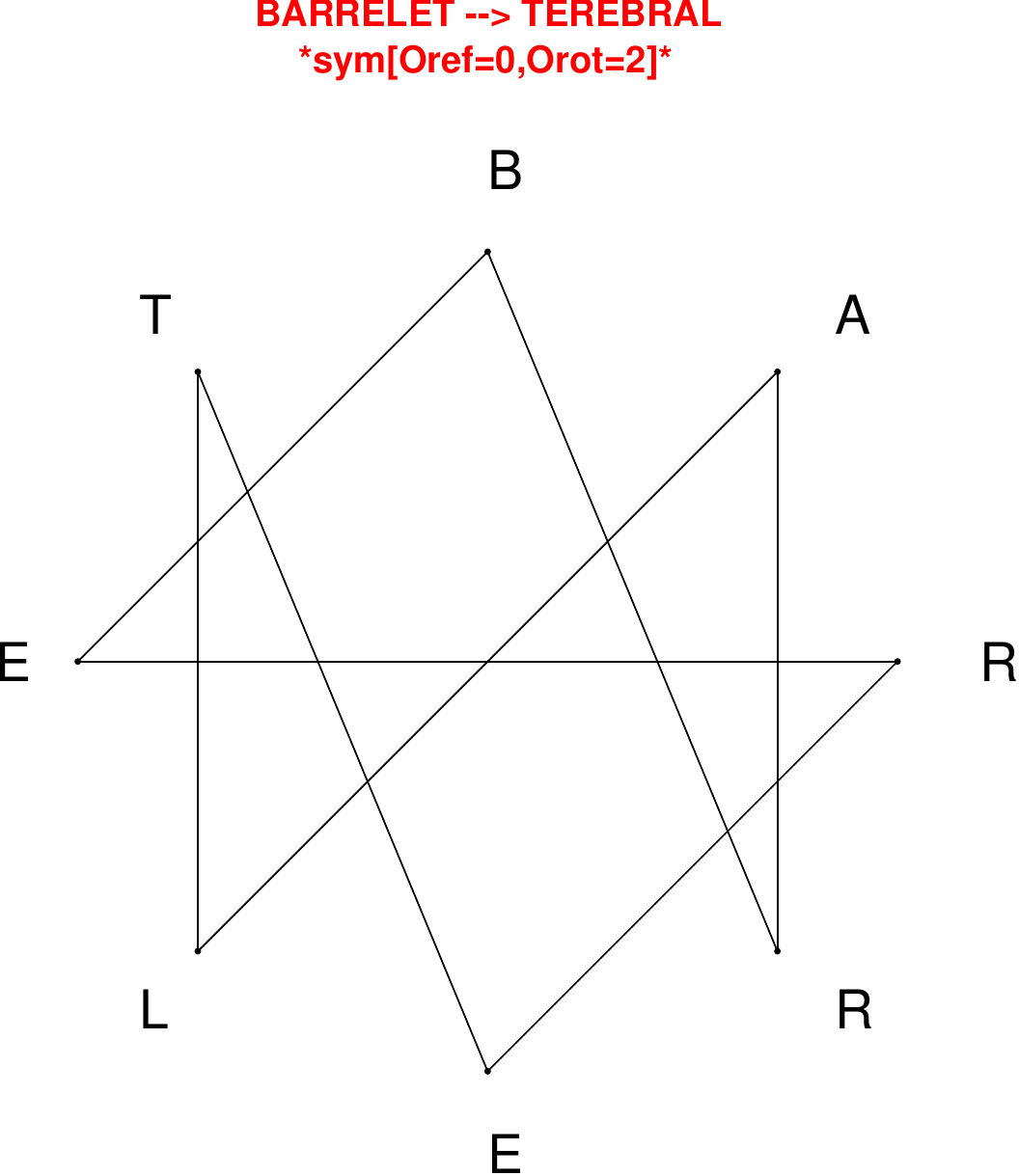}
\end{subfigure}
\hfill
\begin{subfigure}[T]{0.19\textwidth}
\centering
\includegraphics[width=\textwidth]{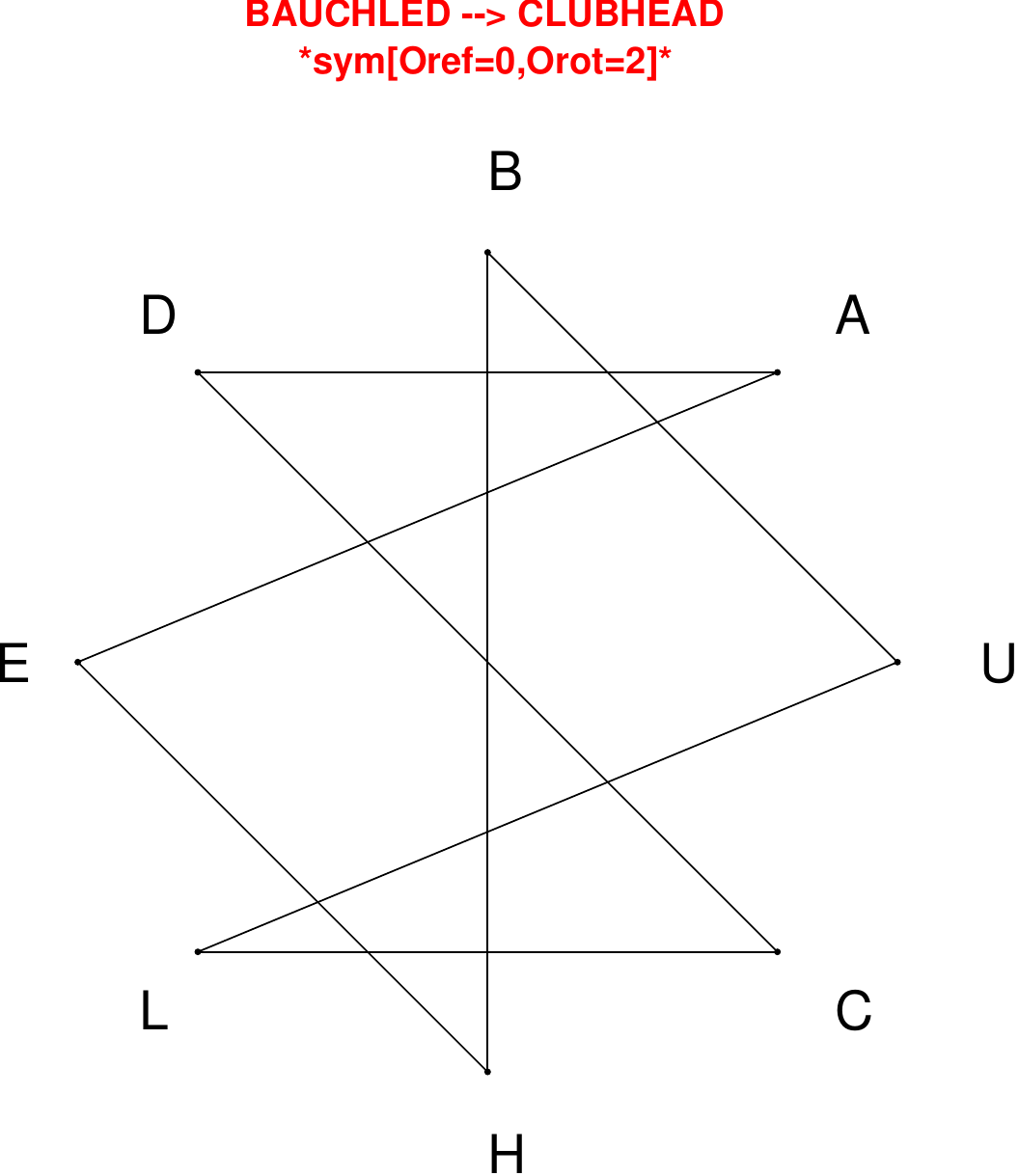}
\end{subfigure}
\hfill
\begin{subfigure}[T]{0.19\textwidth}
\centering
\includegraphics[width=\textwidth]{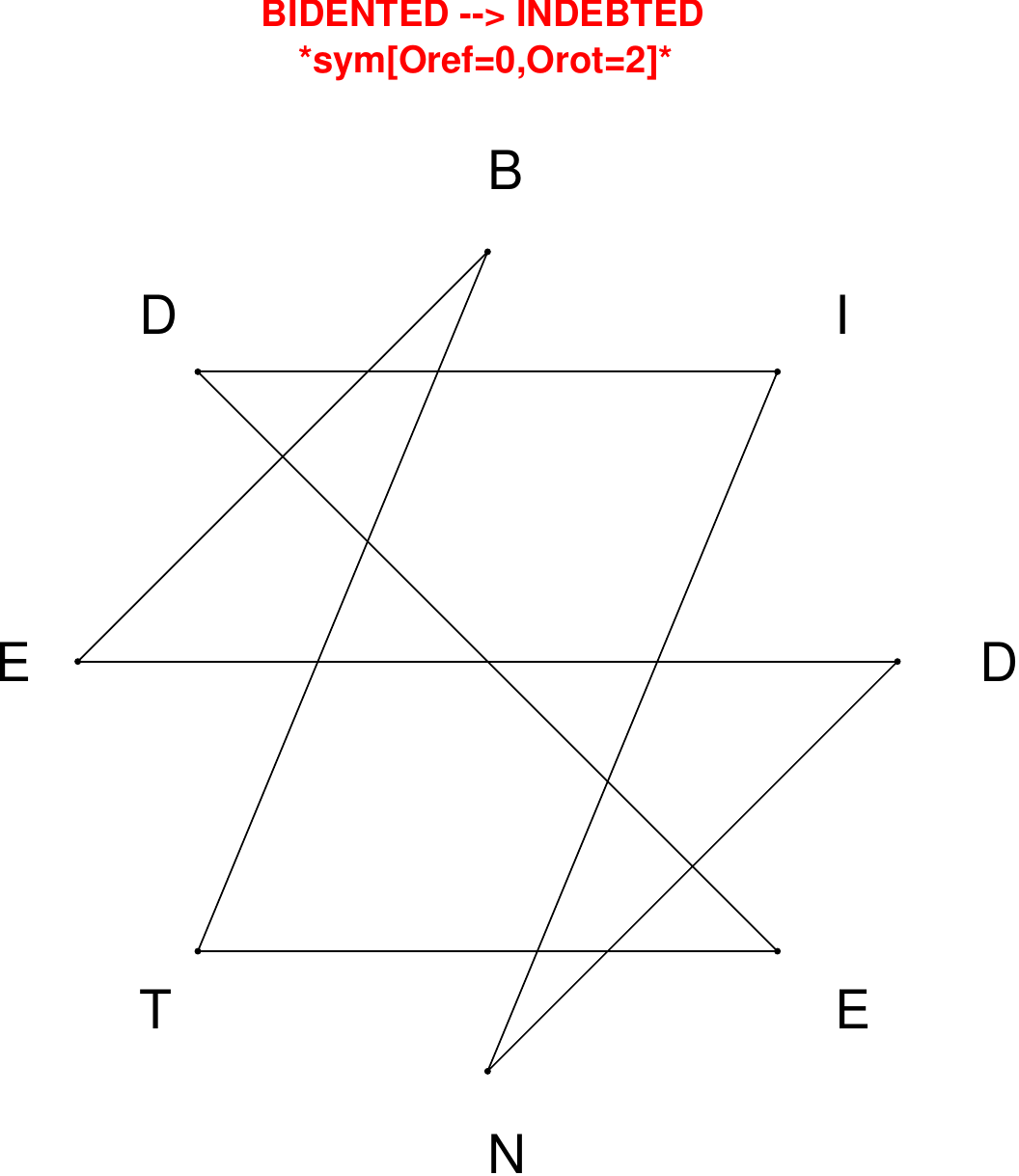}
\end{subfigure}
\end{figure}

\begin{figure}[H]
\centering
\begin{subfigure}[T]{0.19\textwidth}
\centering
\includegraphics[width=\textwidth]{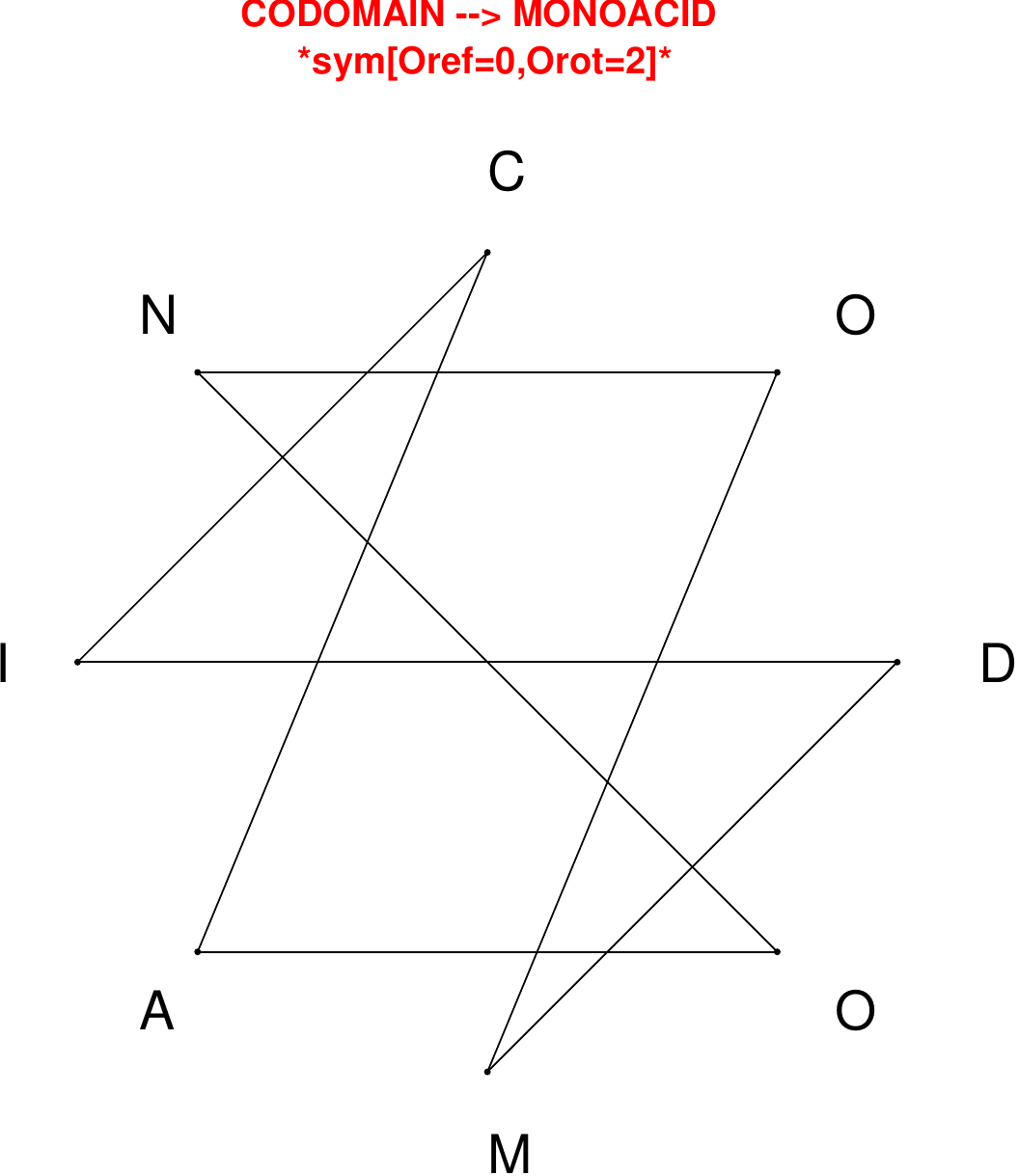}
\end{subfigure}
\hfill
\begin{subfigure}[T]{0.19\textwidth}
\centering
\includegraphics[width=\textwidth]{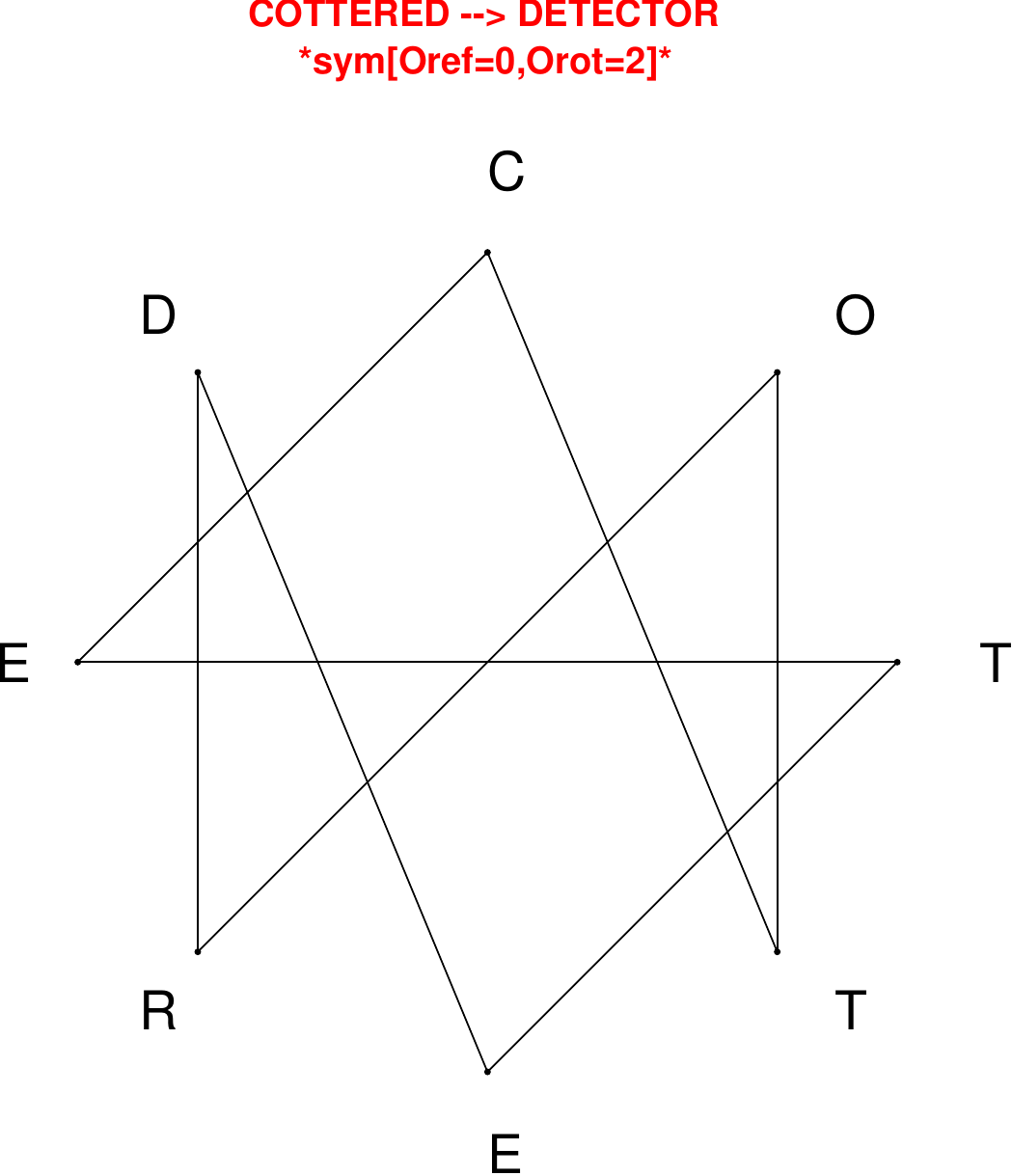}
\end{subfigure}
\hfill
\begin{subfigure}[T]{0.19\textwidth}
\centering
\includegraphics[width=\textwidth]{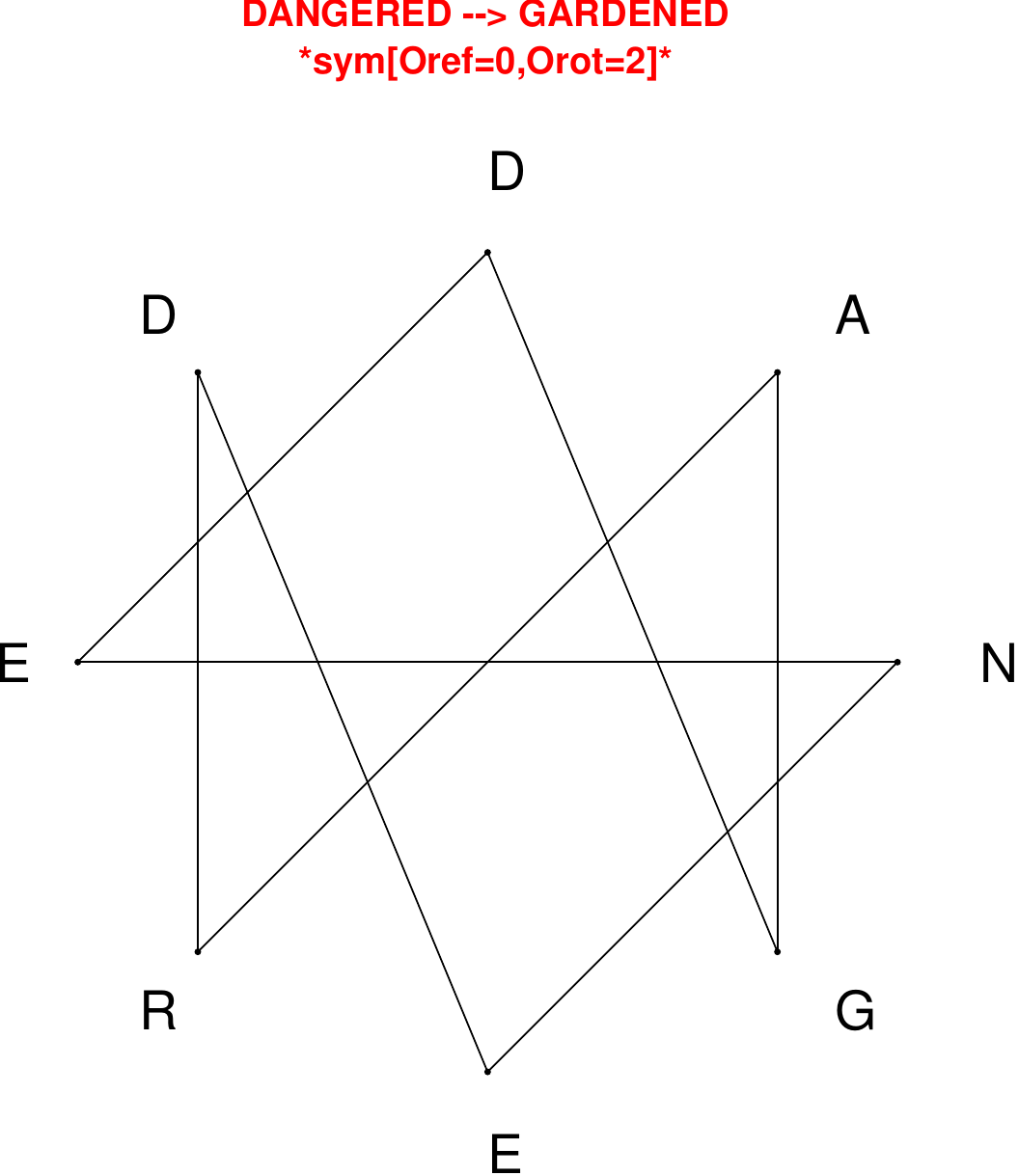}
\end{subfigure}
\hfill
\begin{subfigure}[T]{0.19\textwidth}
\centering
\includegraphics[width=\textwidth]{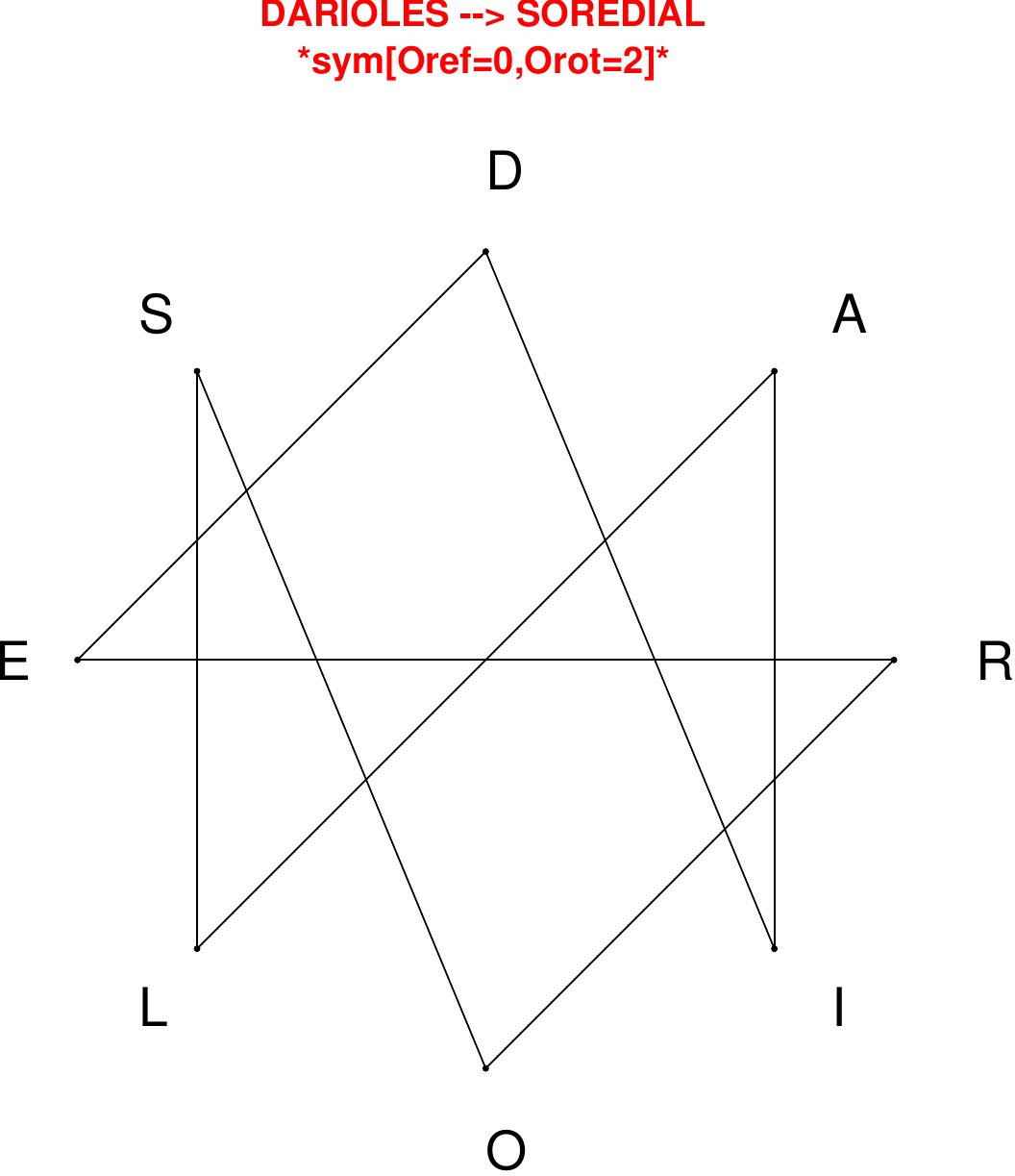}
\end{subfigure}
\hfill
\begin{subfigure}[T]{0.19\textwidth}
\centering
\includegraphics[width=\textwidth]{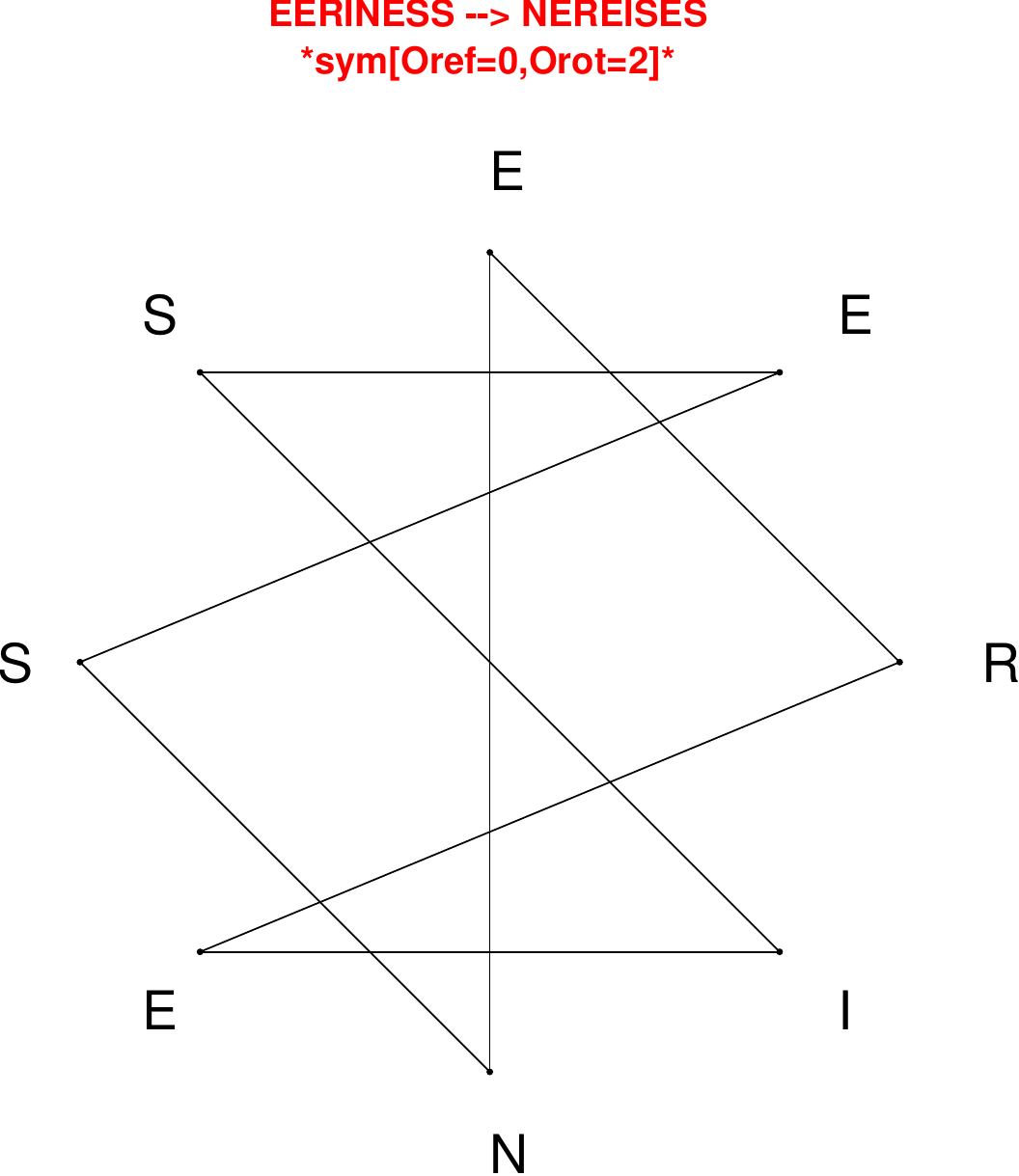}
\end{subfigure}
\end{figure}

\begin{figure}[H]
\centering
\begin{subfigure}[T]{0.19\textwidth}
\centering
\includegraphics[width=\textwidth]{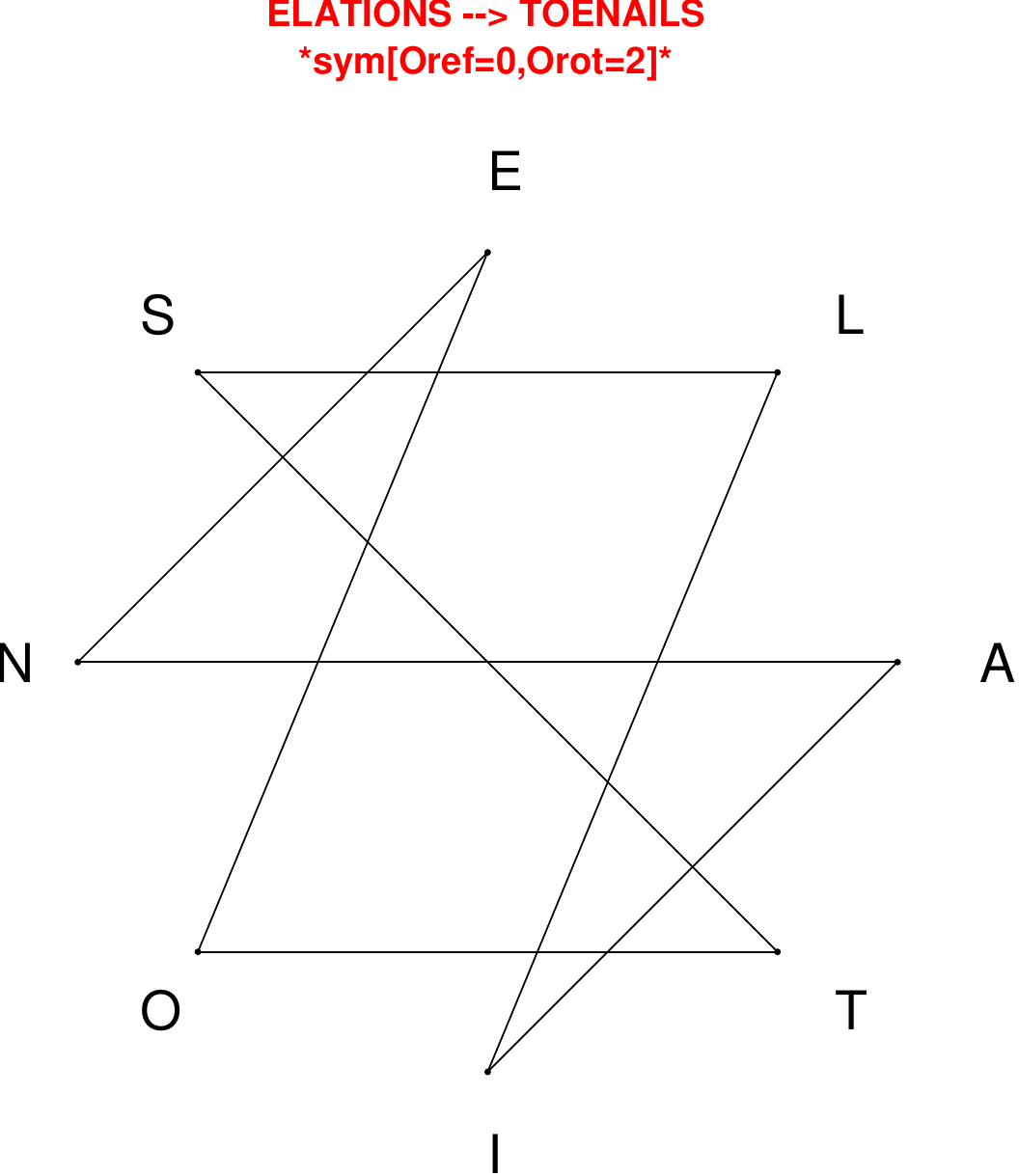}
\end{subfigure}
\hfill
\begin{subfigure}[T]{0.19\textwidth}
\centering
\includegraphics[width=\textwidth]{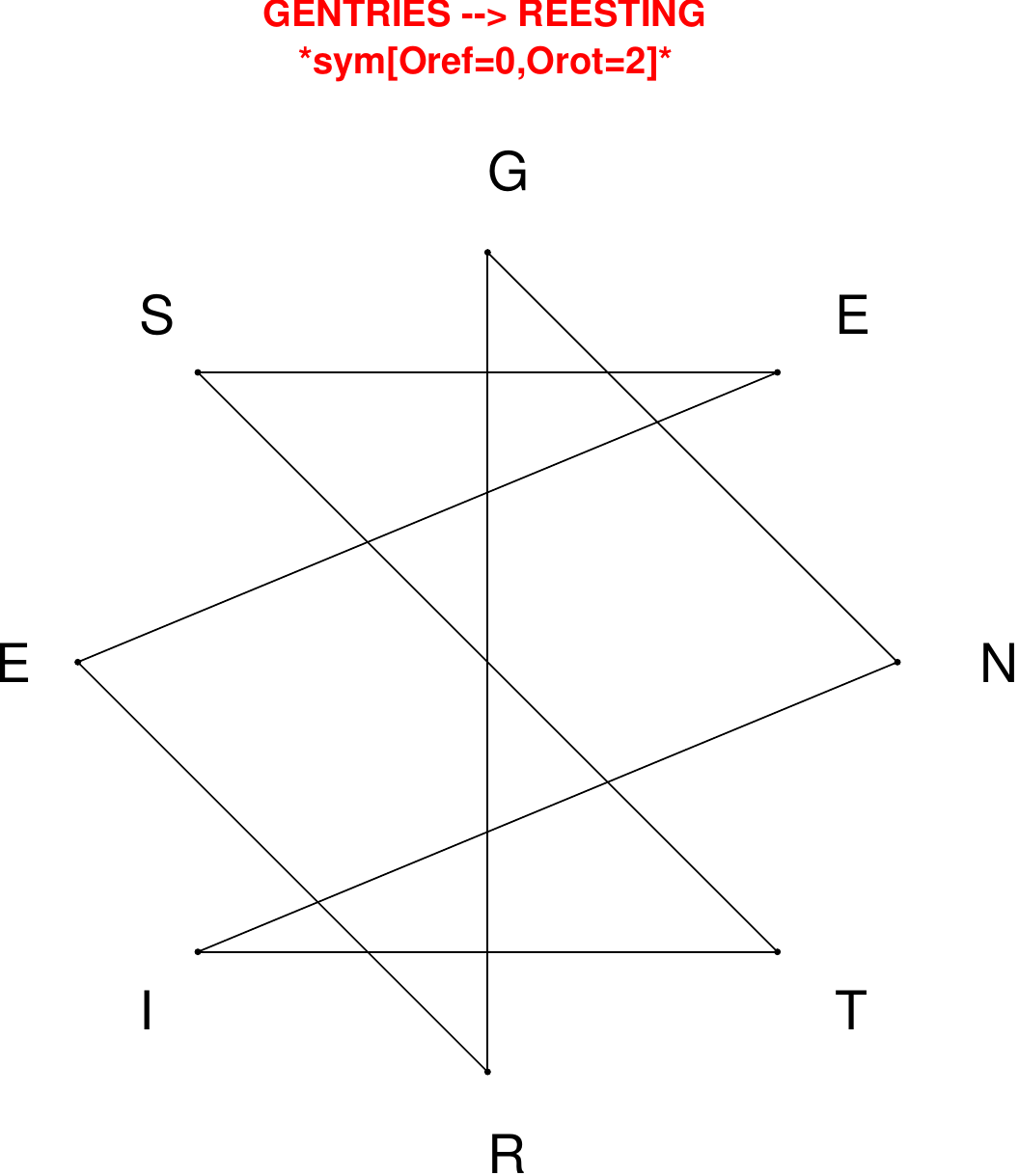}
\end{subfigure}
\hfill
\begin{subfigure}[T]{0.19\textwidth}
\centering
\includegraphics[width=\textwidth]{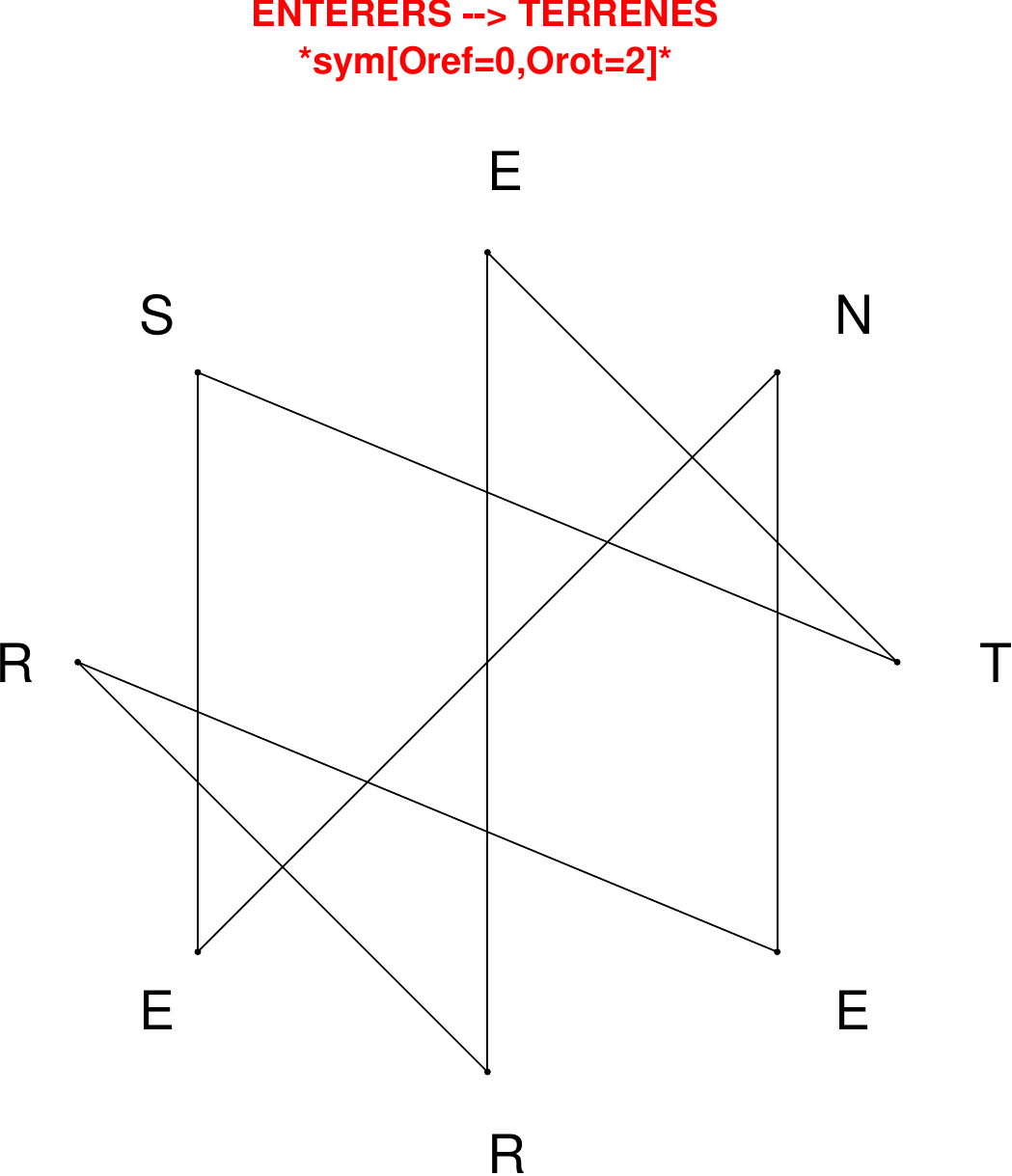}
\end{subfigure}
\hfill
\begin{subfigure}[T]{0.19\textwidth}
\centering
\includegraphics[width=\textwidth]{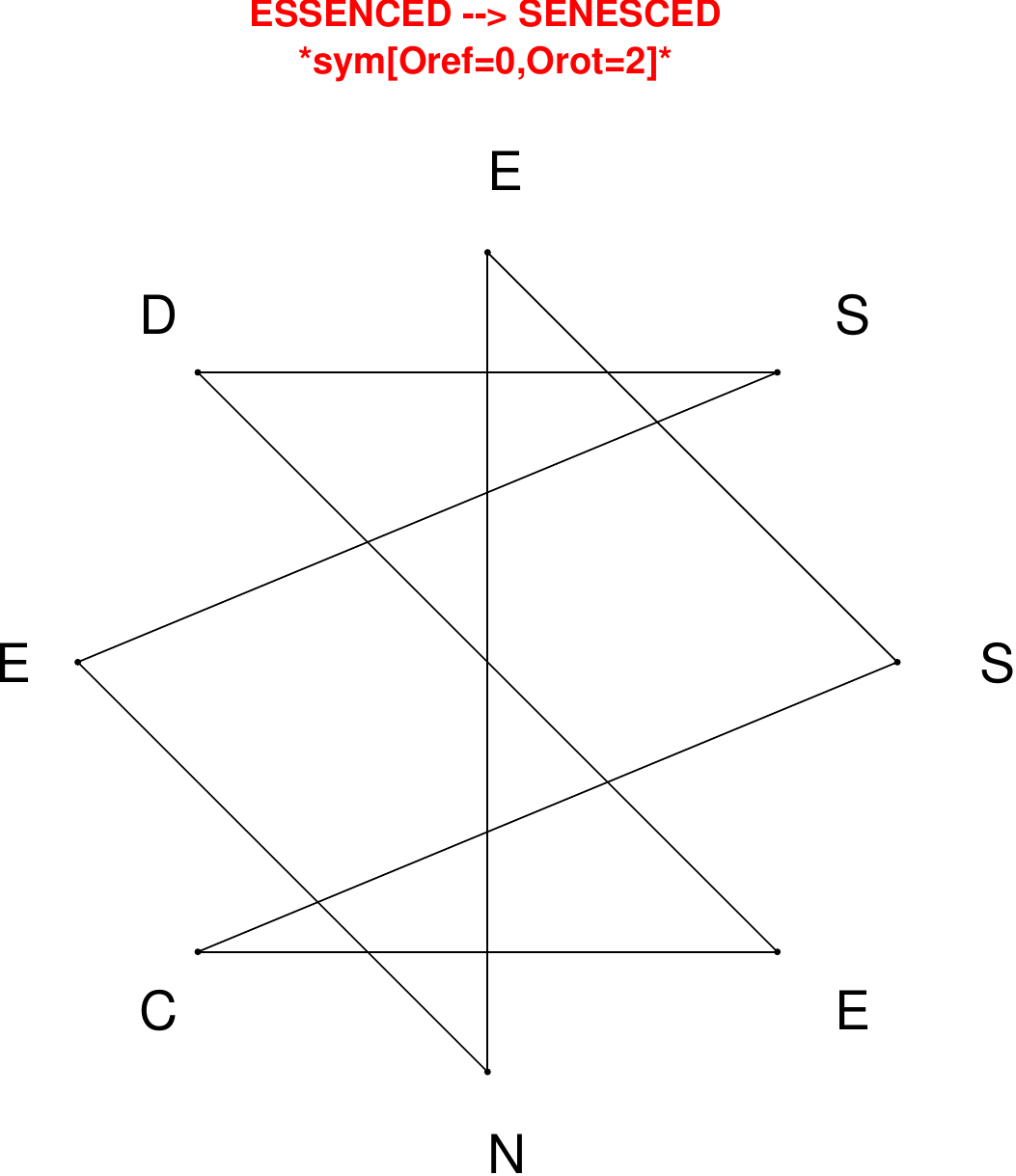}
\end{subfigure}
\hfill
\begin{subfigure}[T]{0.19\textwidth}
\centering
\includegraphics[width=\textwidth]{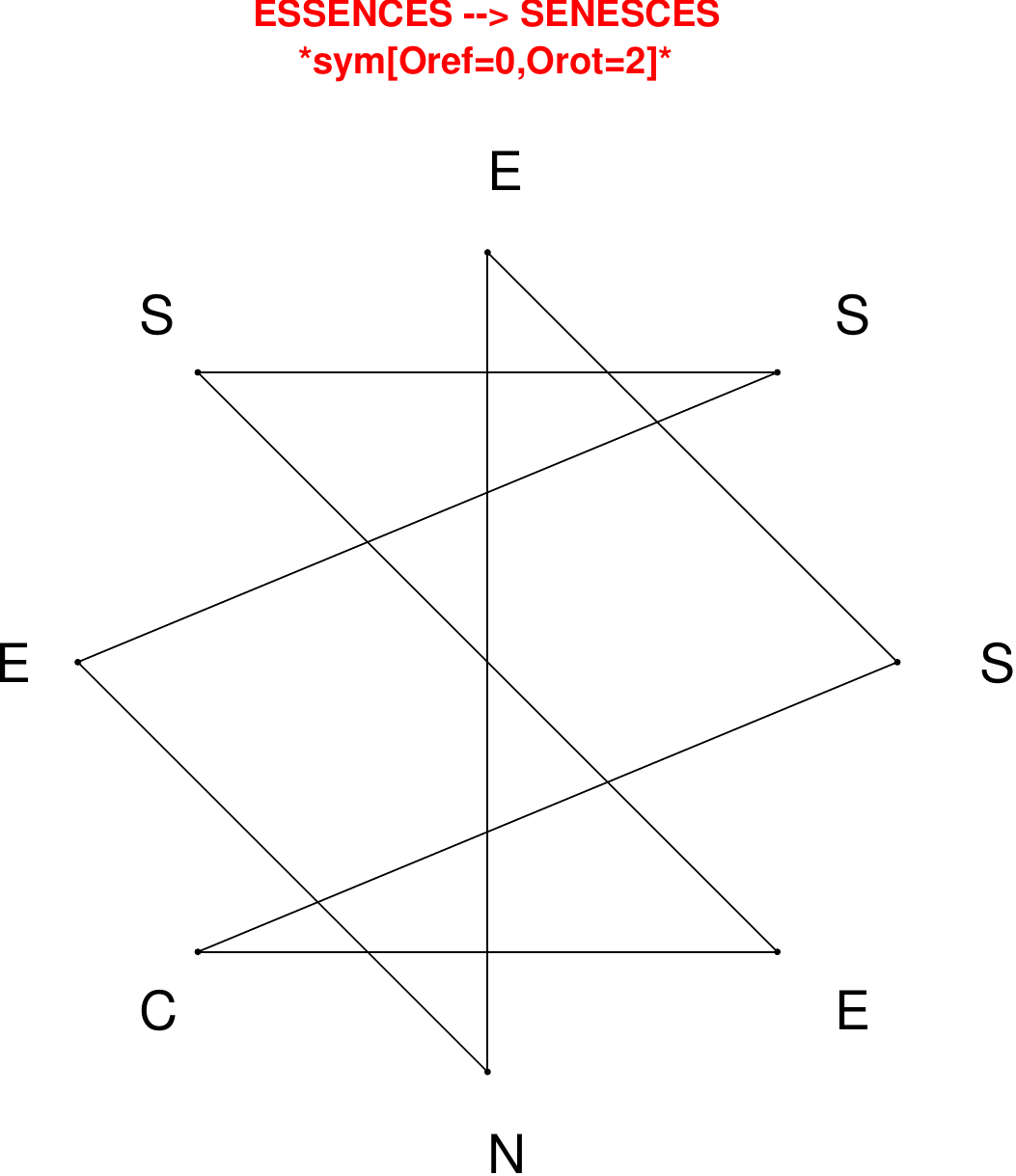}
\end{subfigure}
\end{figure}

\begin{figure}[H]
\centering
\begin{subfigure}[T]{0.19\textwidth}
\centering
\includegraphics[width=\textwidth]{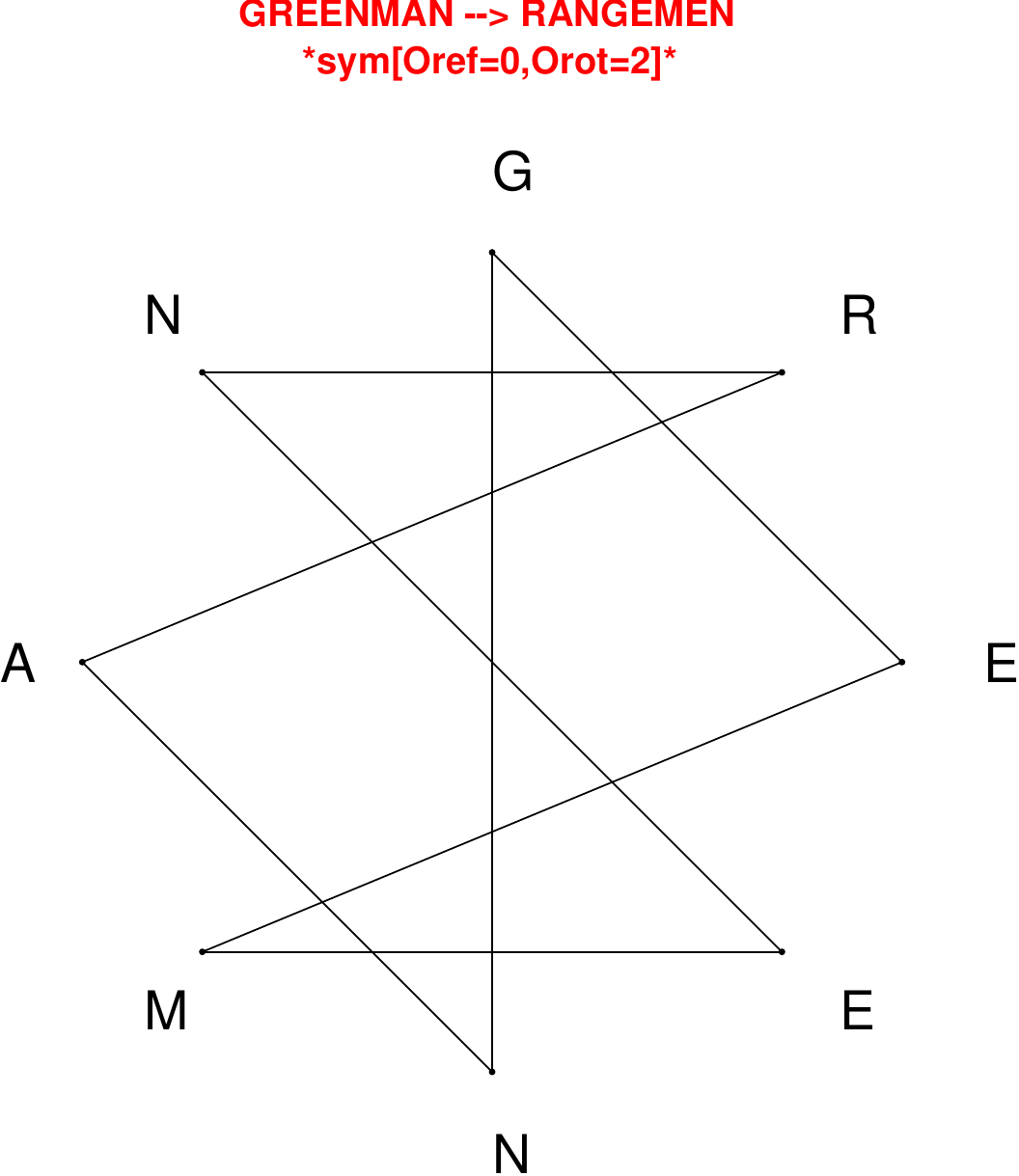}
\end{subfigure}
\hfill
\begin{subfigure}[T]{0.19\textwidth}
\centering
\includegraphics[width=\textwidth]{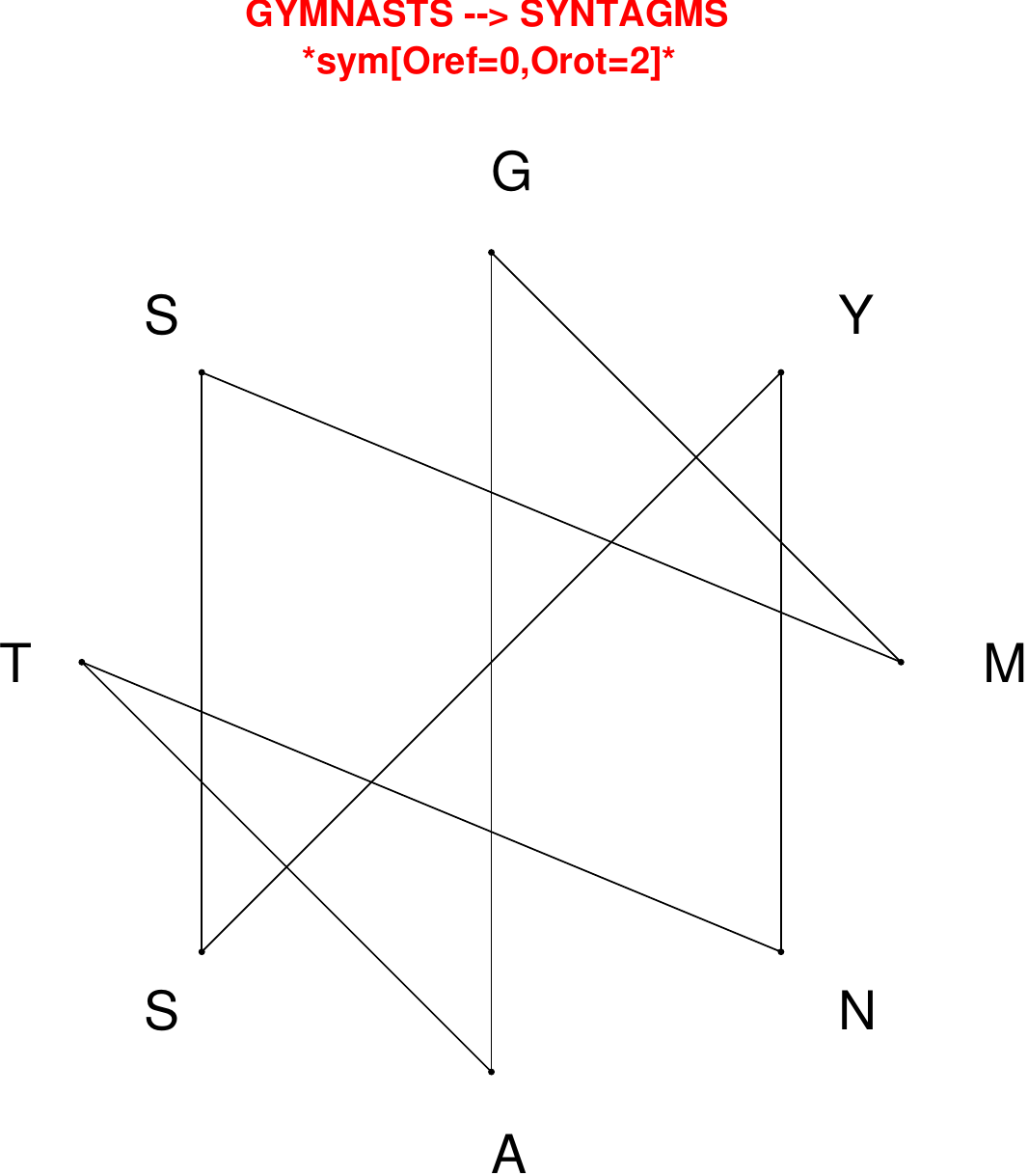}
\end{subfigure}
\hfill
\begin{subfigure}[T]{0.19\textwidth}
\centering
\includegraphics[width=\textwidth]{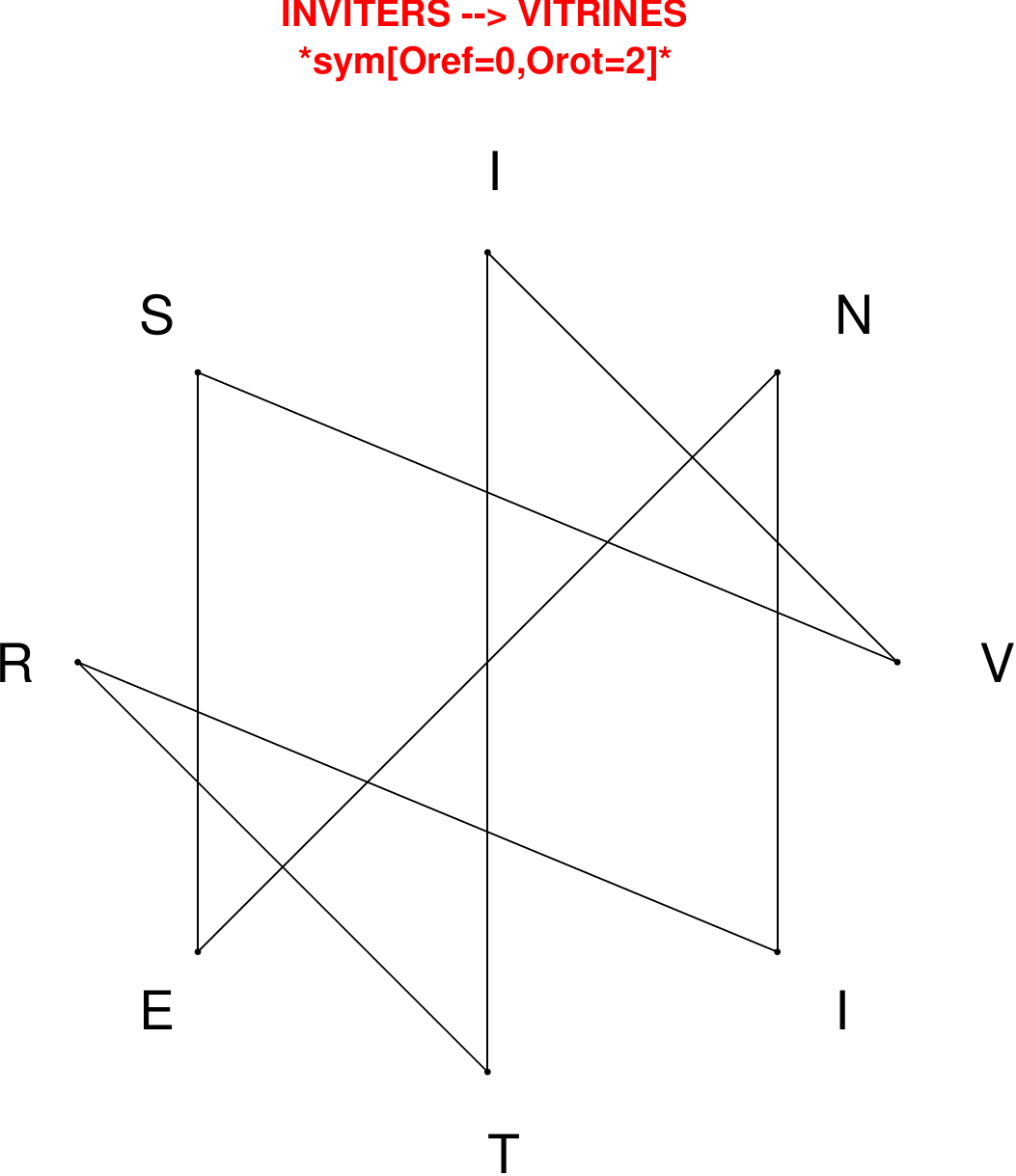}
\end{subfigure}
\hfill
\begin{subfigure}[T]{0.19\textwidth}
\centering
\includegraphics[width=\textwidth]{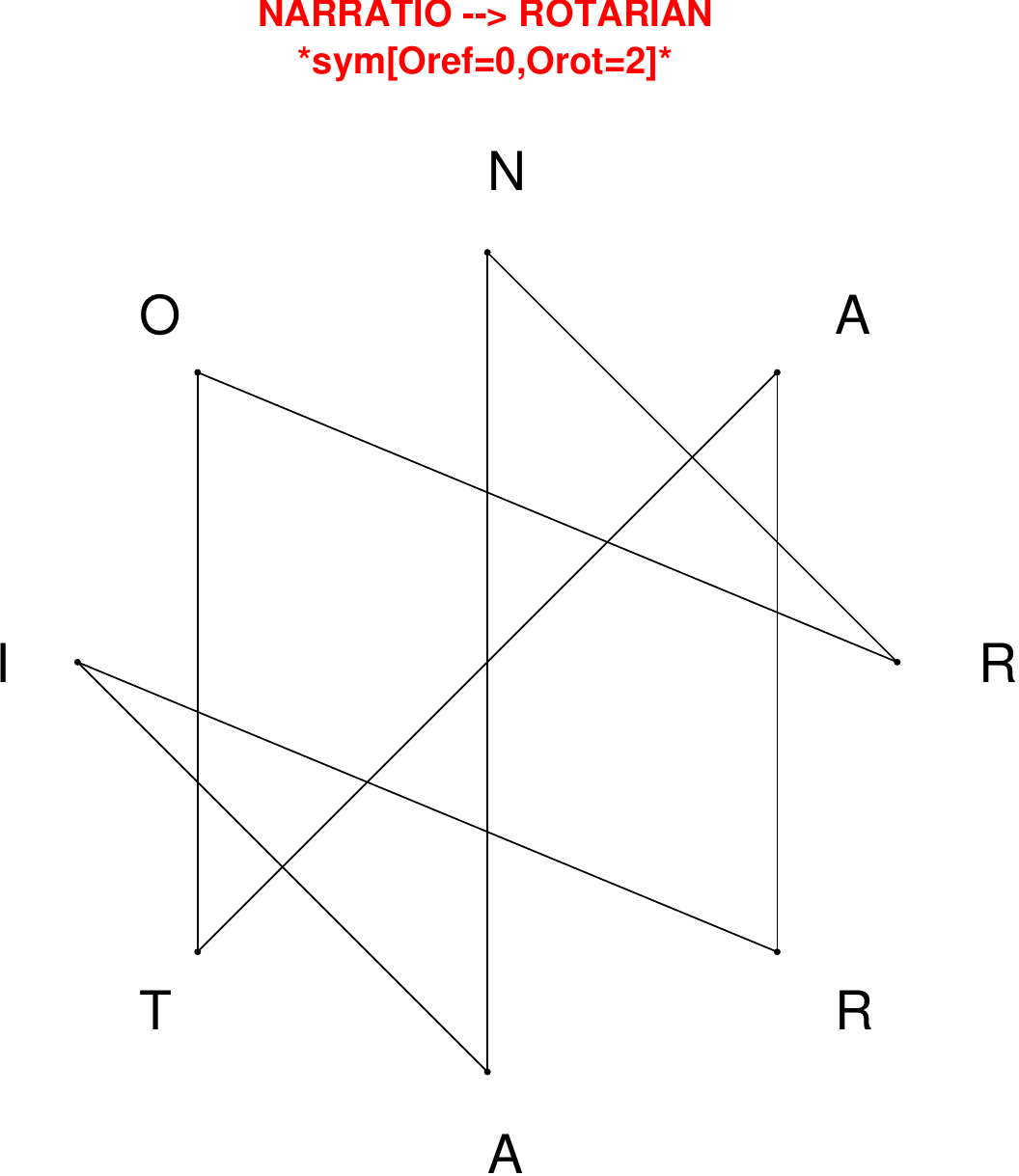}
\end{subfigure}
\hfill
\begin{subfigure}[T]{0.19\textwidth}
\centering
\includegraphics[width=\textwidth]{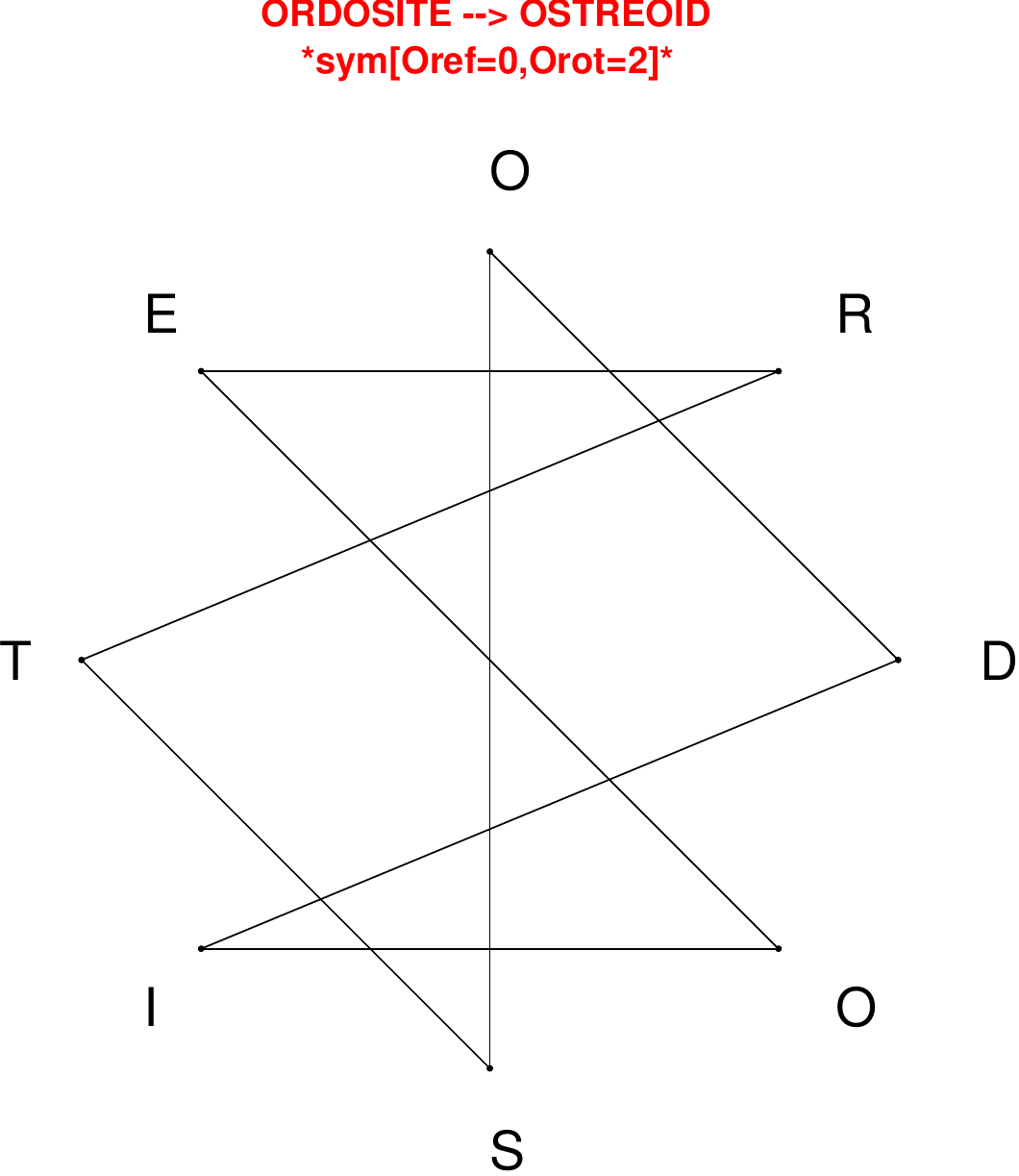}
\end{subfigure}
\end{figure}

\begin{figure}[H]
\centering
\begin{subfigure}[T]{0.19\textwidth}
\centering
\includegraphics[width=\textwidth]{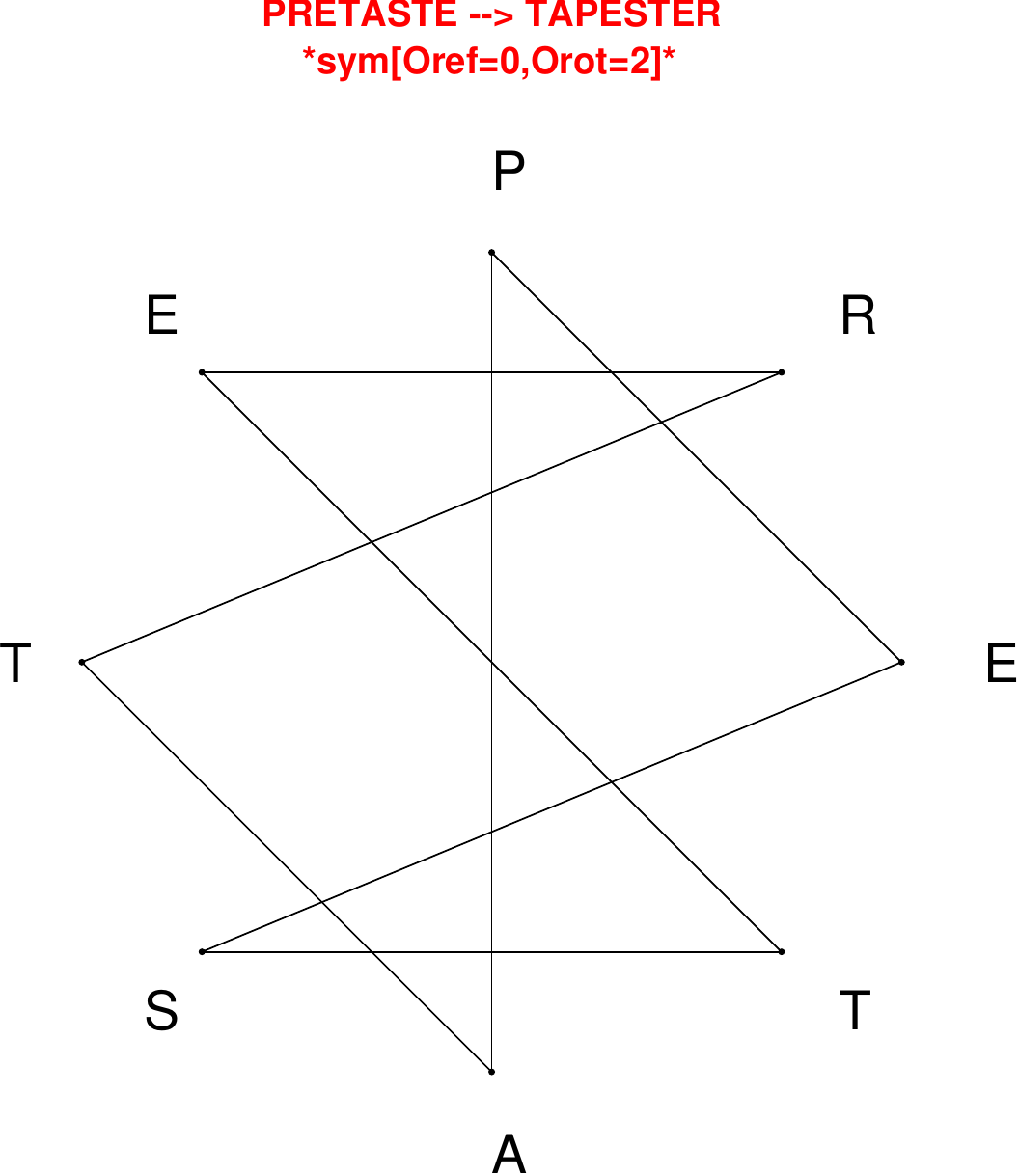}
\end{subfigure}
\hfill
\begin{subfigure}[T]{0.19\textwidth}
\centering
\includegraphics[width=\textwidth]{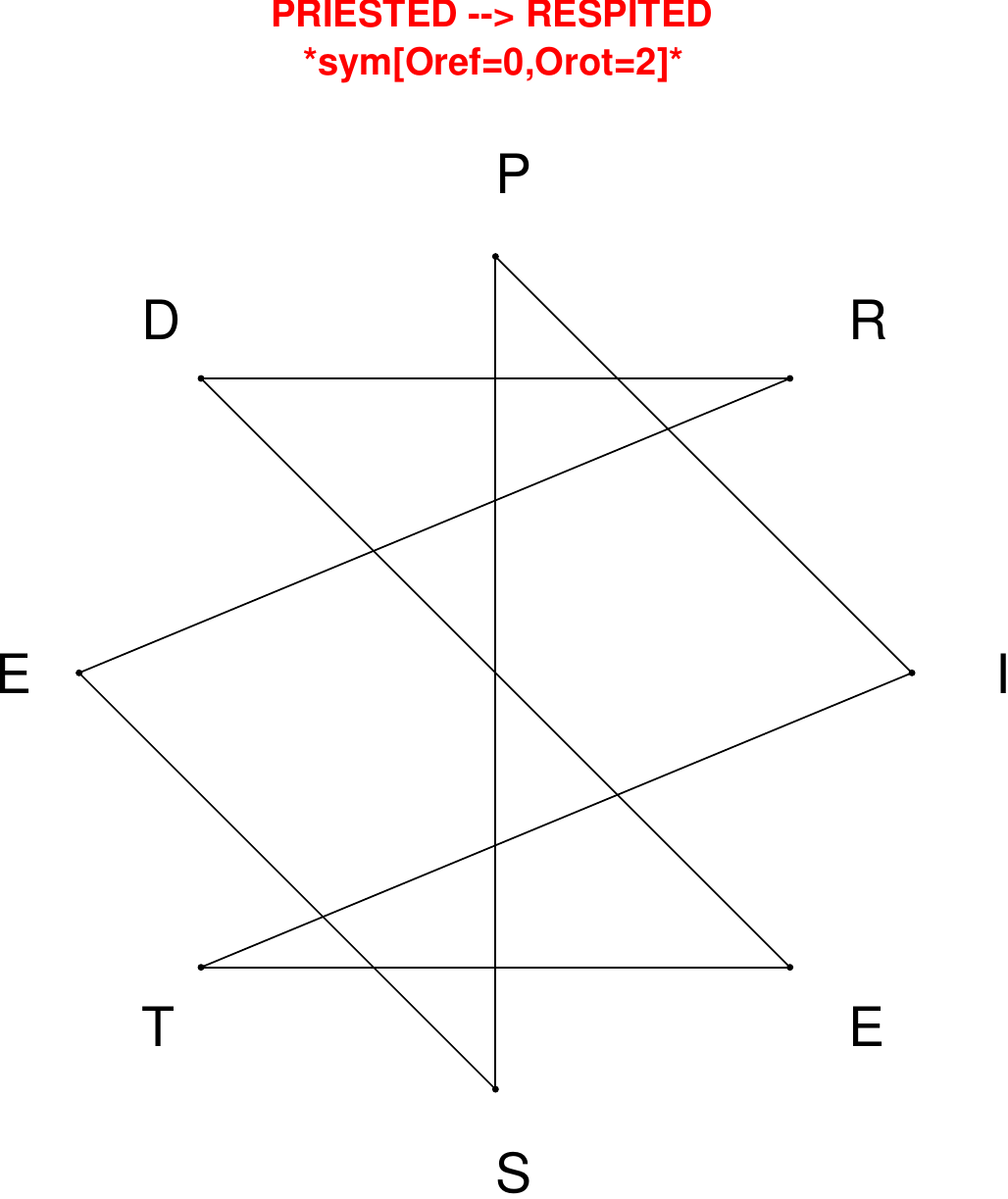}
\end{subfigure}
\hfill
\begin{subfigure}[T]{0.19\textwidth}
\centering
\includegraphics[width=\textwidth]{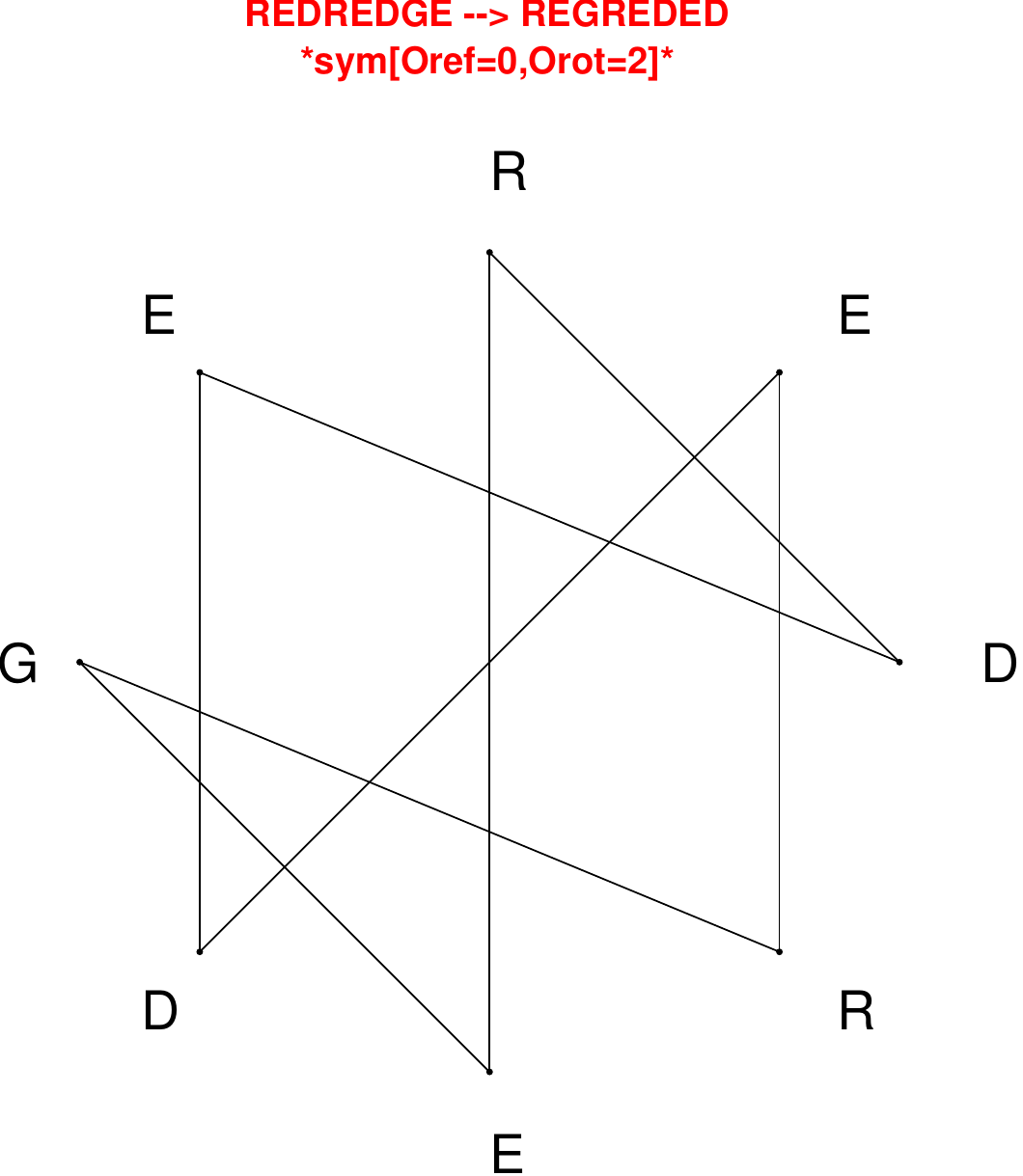}
\end{subfigure}
\hfill
\begin{subfigure}[T]{0.19\textwidth}
\centering
\includegraphics[width=\textwidth]{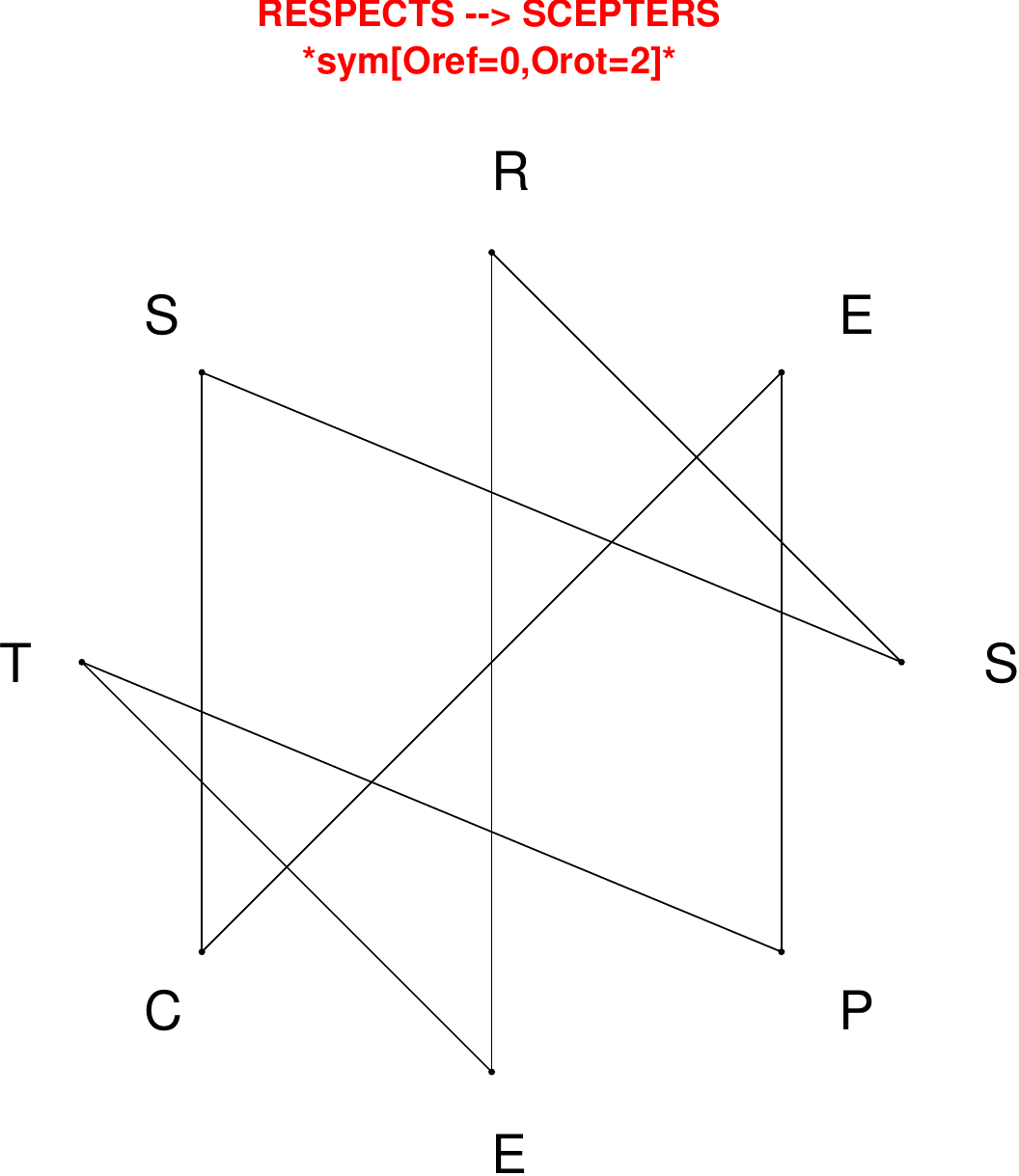}
\end{subfigure}
\hfill
\begin{subfigure}[T]{0.19\textwidth}
\centering
\includegraphics[width=\textwidth]{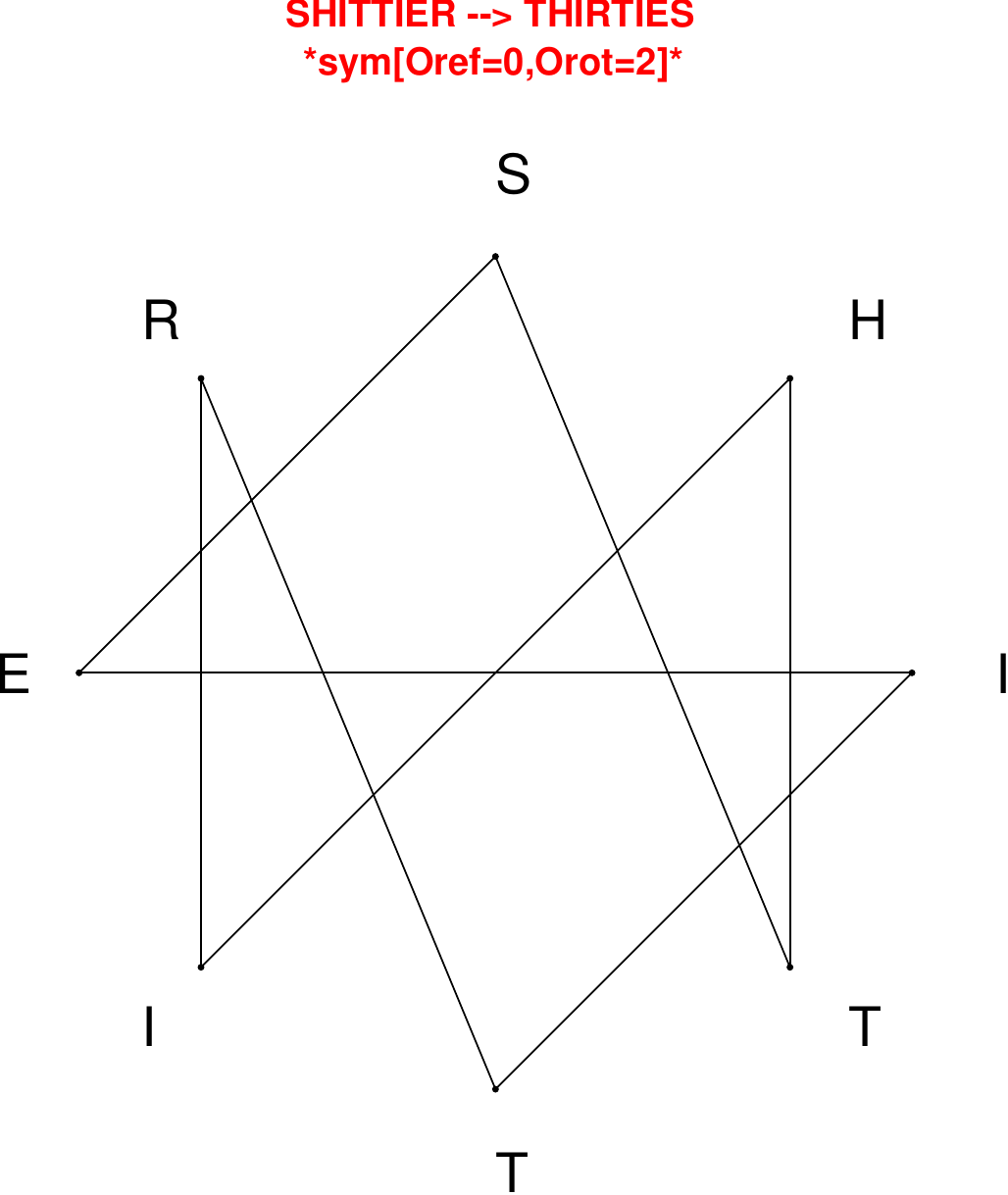}
\end{subfigure}
\end{figure}

\begin{figure}[H]
\centering
\begin{subfigure}[T]{0.19\textwidth}
\centering
\includegraphics[width=\textwidth]{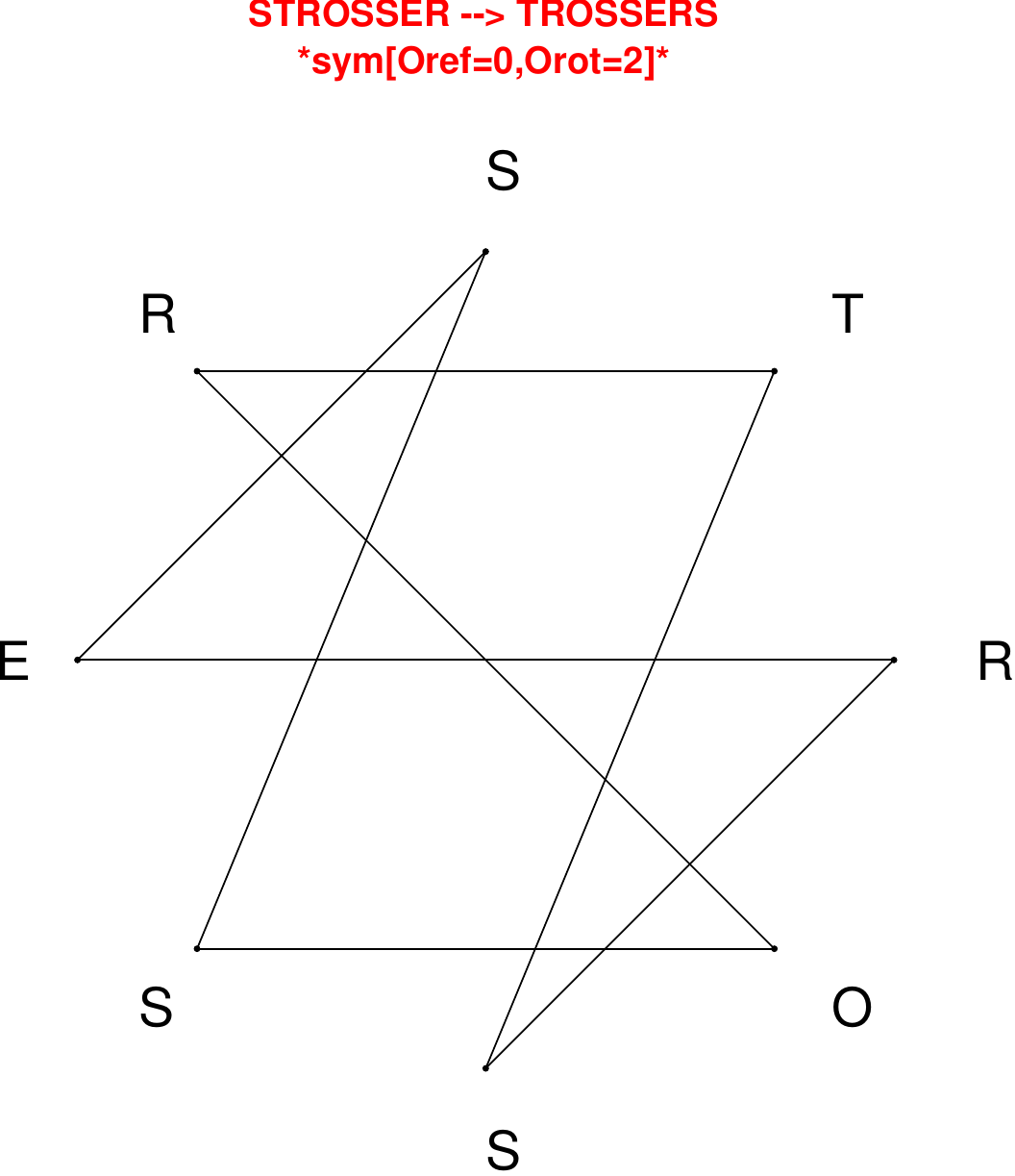}
\end{subfigure}
\hfill
\begin{subfigure}[T]{0.19\textwidth}
\centering
\includegraphics[width=\textwidth]{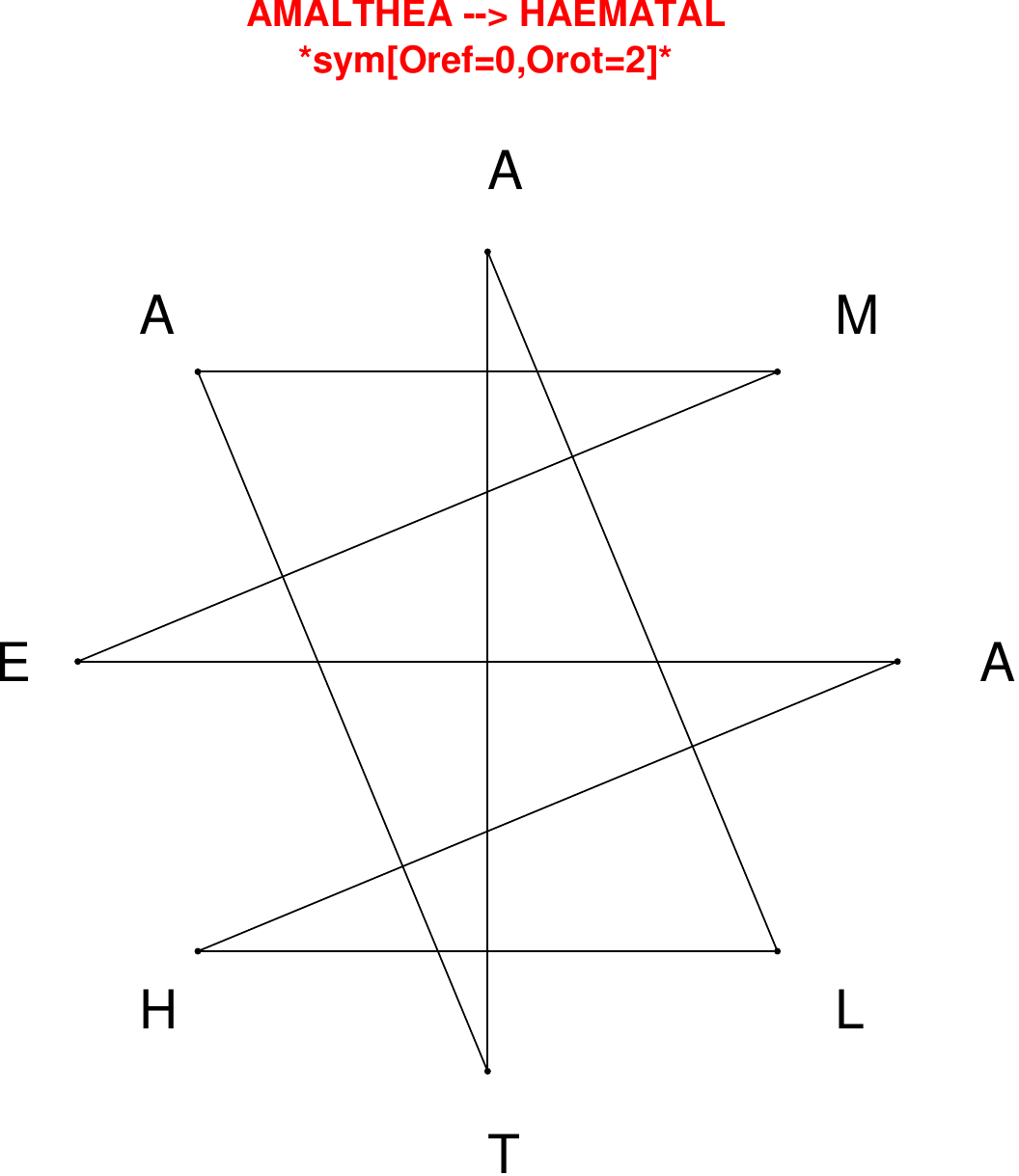}
\end{subfigure}
\hfill
\begin{subfigure}[T]{0.19\textwidth}
\centering
\includegraphics[width=\textwidth]{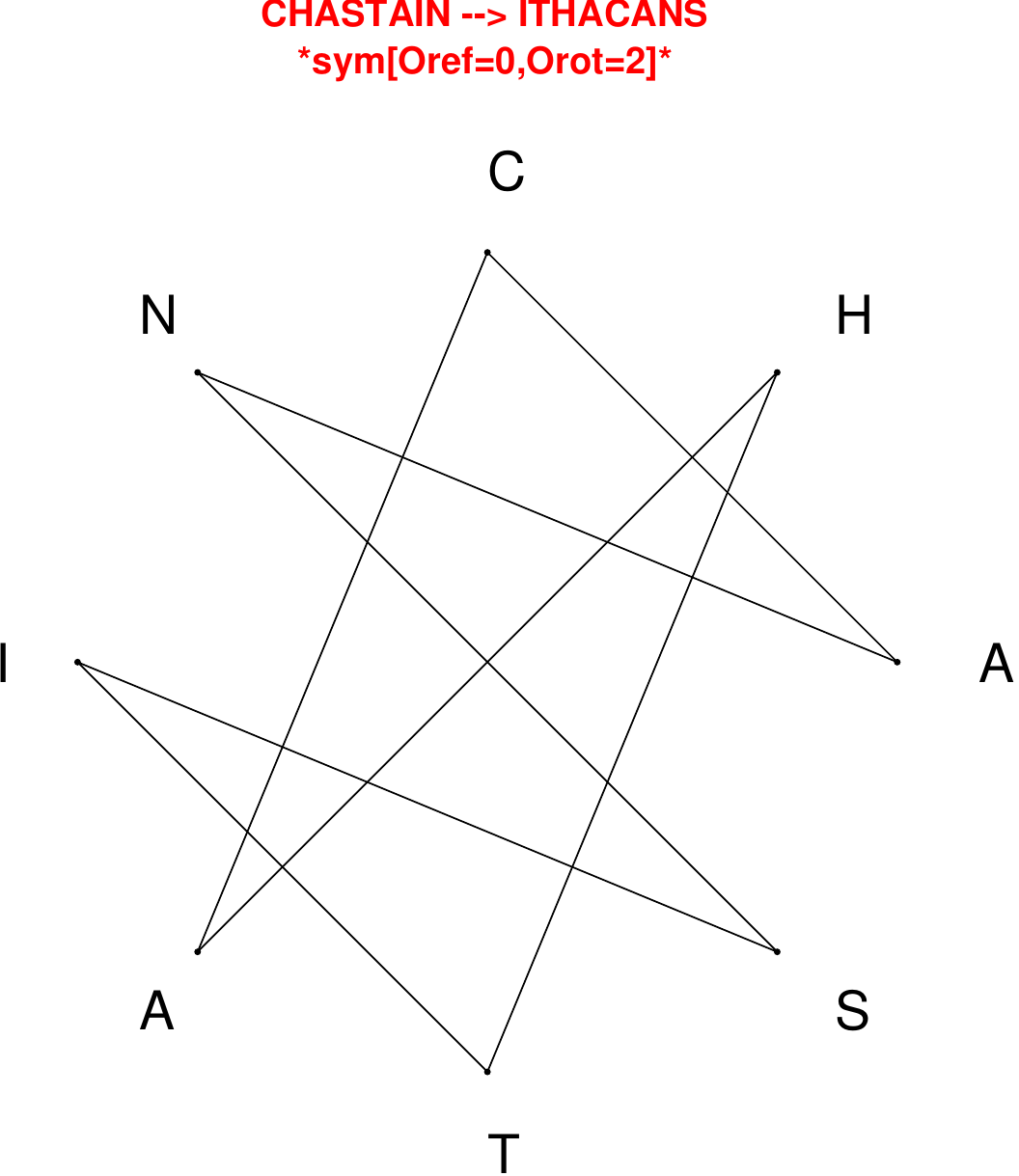}
\end{subfigure}
\hfill
\begin{subfigure}[T]{0.19\textwidth}
\centering
\includegraphics[width=\textwidth]{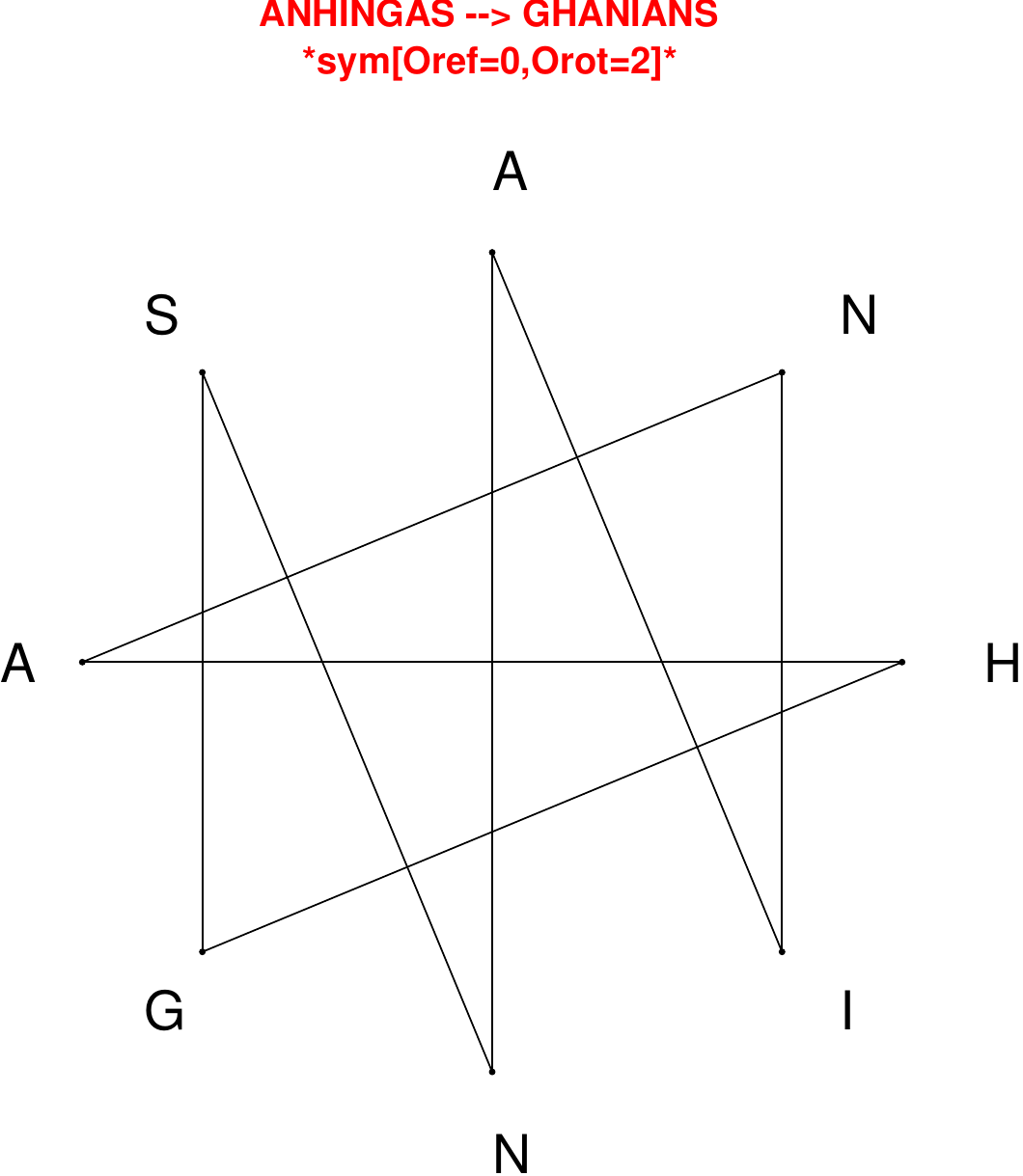}
\end{subfigure}
\hfill
\begin{subfigure}[T]{0.19\textwidth}
\centering
\includegraphics[width=\textwidth]{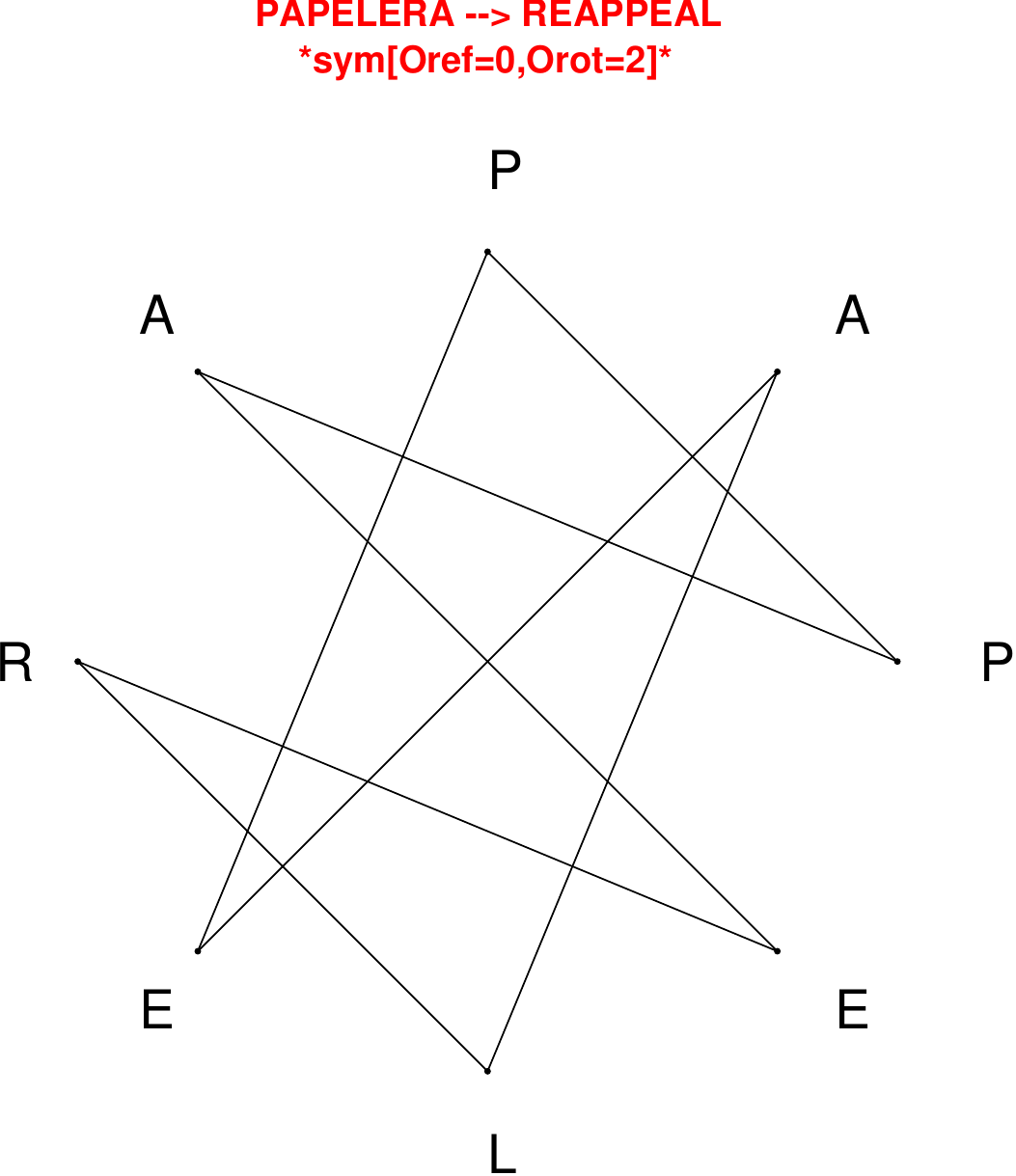}
\end{subfigure}
\end{figure}

\begin{figure}[H]
\centering
\begin{subfigure}[T]{0.19\textwidth}
\centering
\includegraphics[width=\textwidth]{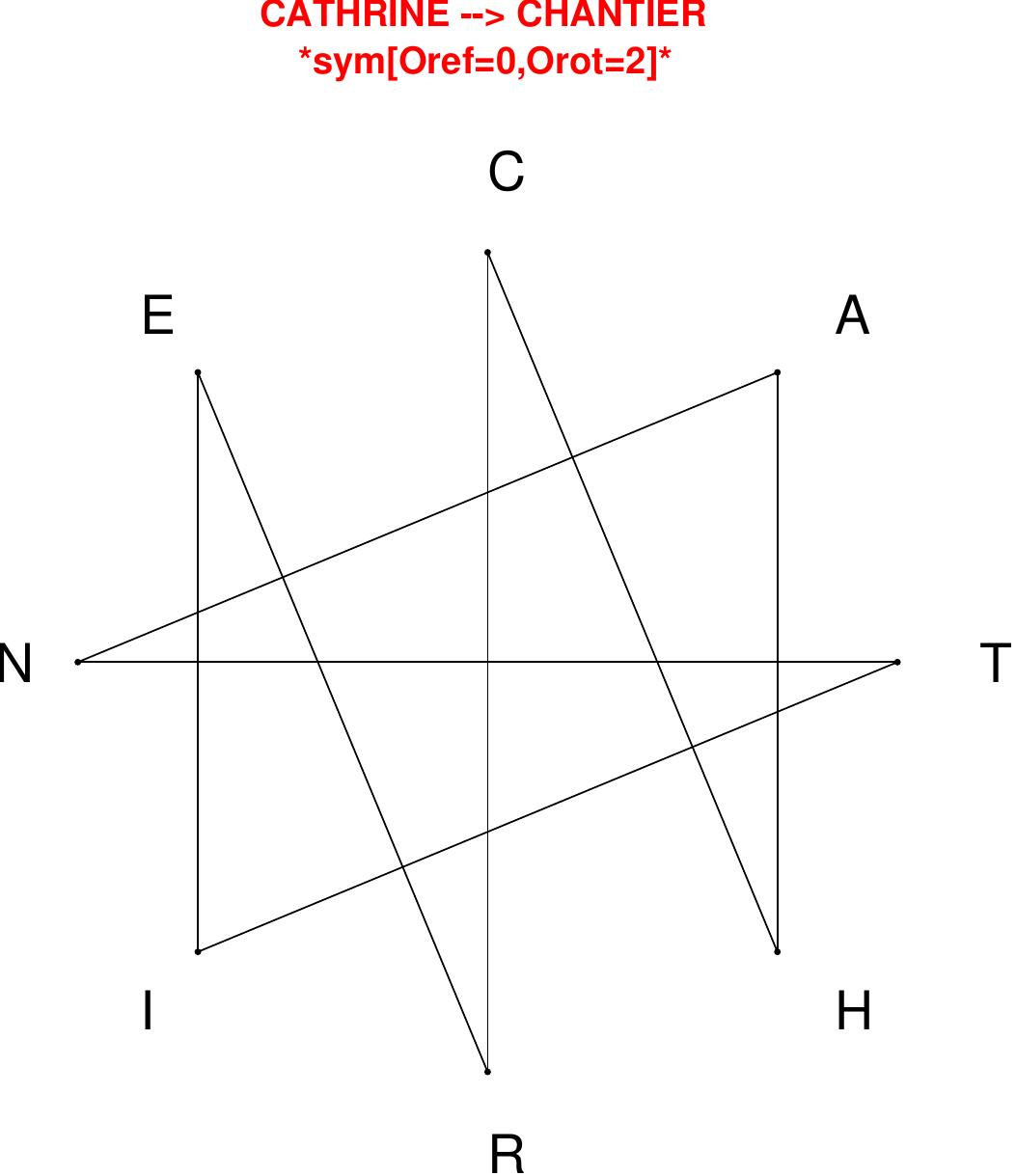}
\end{subfigure}
\hfill
\begin{subfigure}[T]{0.19\textwidth}
\centering
\includegraphics[width=\textwidth]{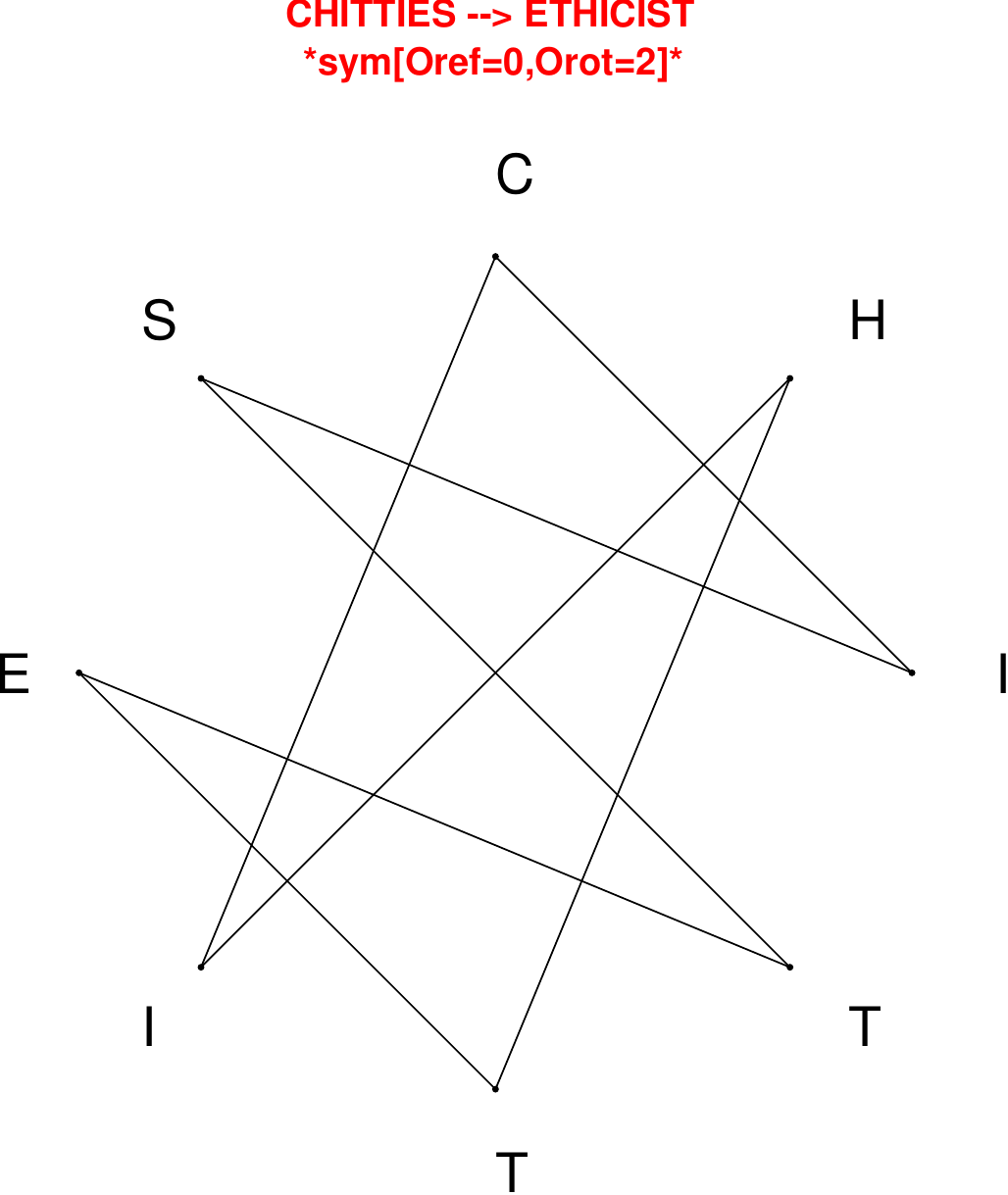}
\end{subfigure}
\hfill
\begin{subfigure}[T]{0.19\textwidth}
\centering
\includegraphics[width=\textwidth]{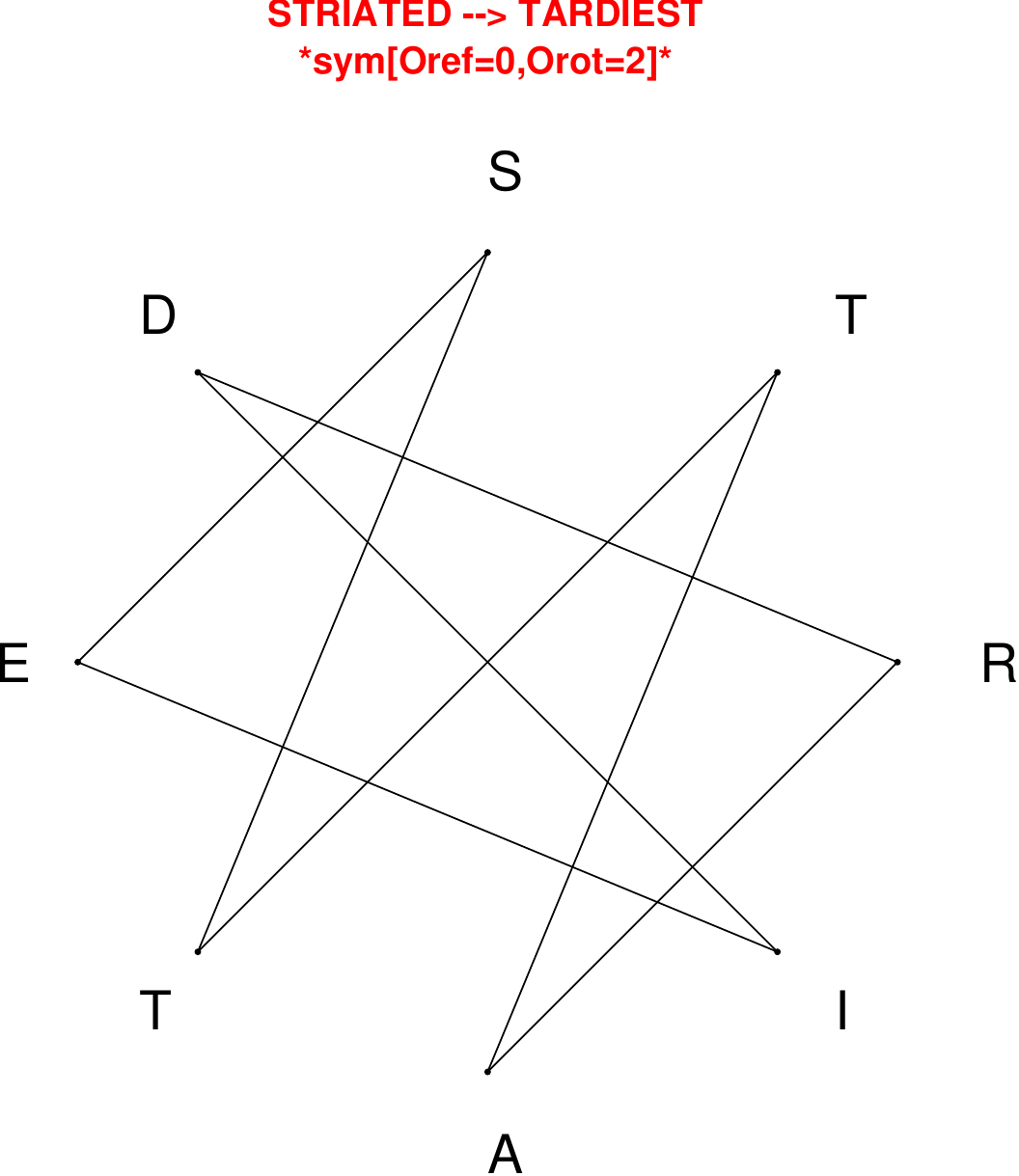}
\end{subfigure}
\hfill
\begin{subfigure}[T]{0.19\textwidth}
\centering
\includegraphics[width=\textwidth]{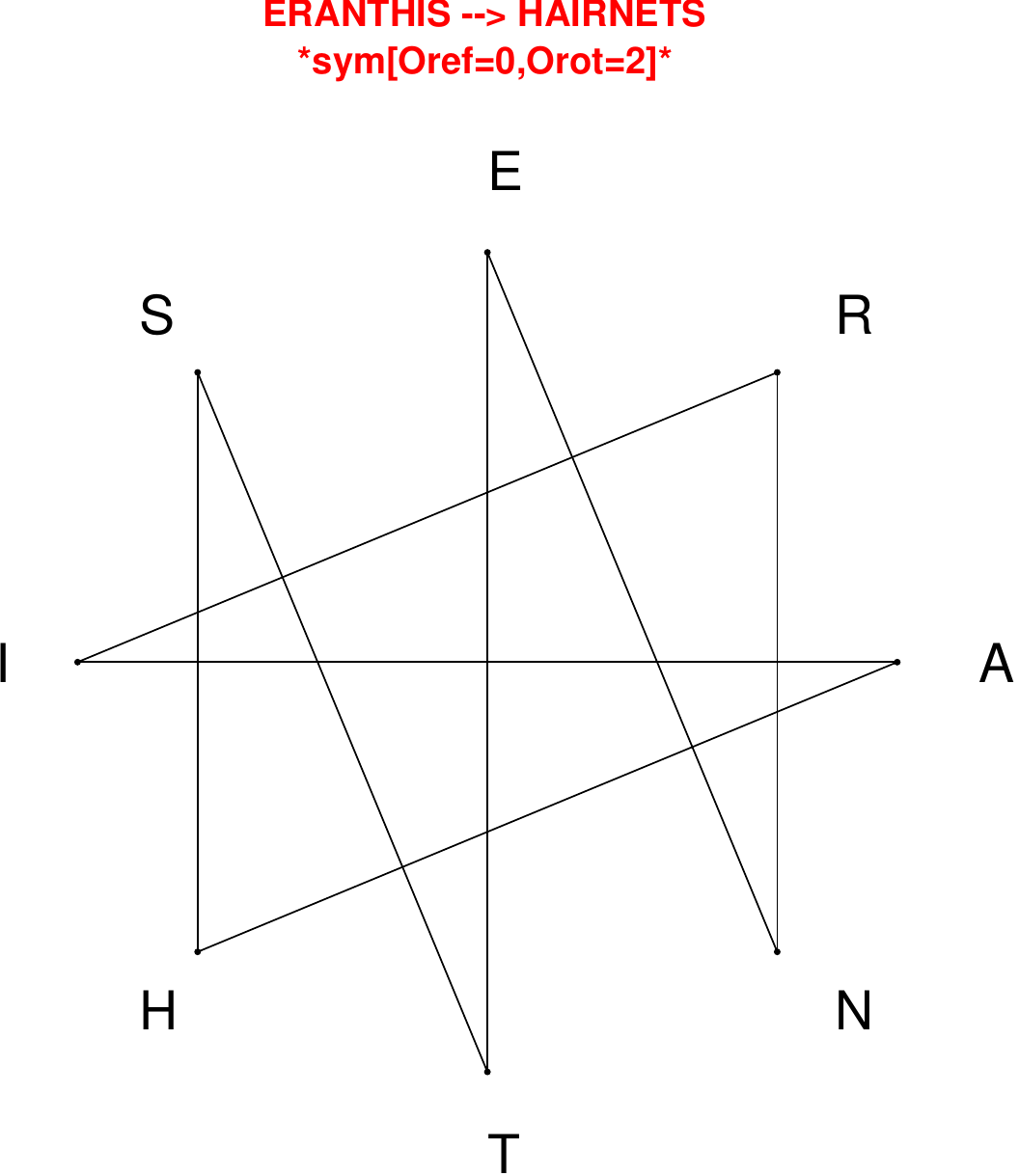}
\end{subfigure}
\hfill
\begin{subfigure}[T]{0.19\textwidth}
\centering
\includegraphics[width=\textwidth]{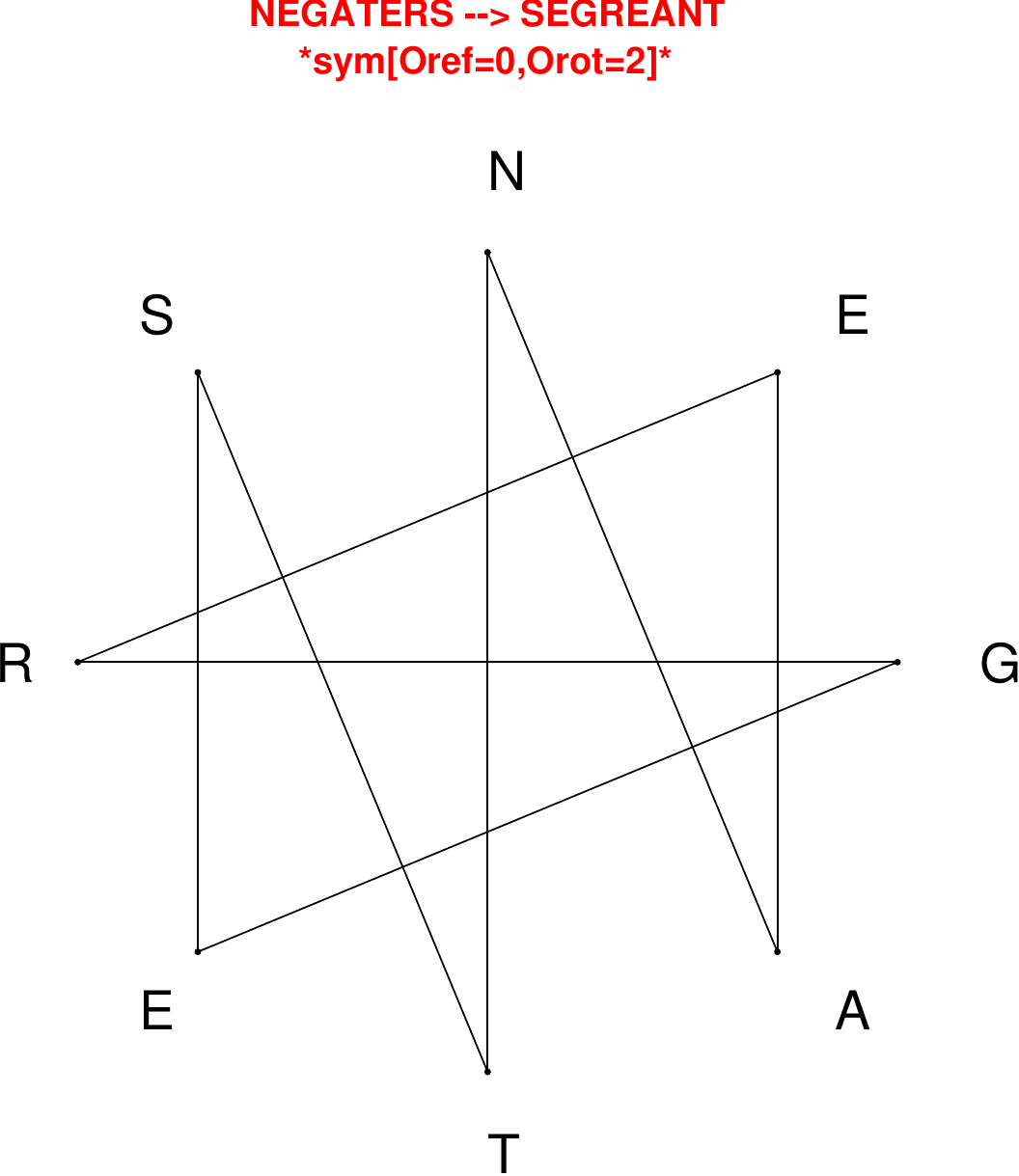}
\end{subfigure}
\end{figure}

\begin{figure}[H]
\centering
\begin{subfigure}[T]{0.19\textwidth}
\centering
\includegraphics[width=\textwidth]{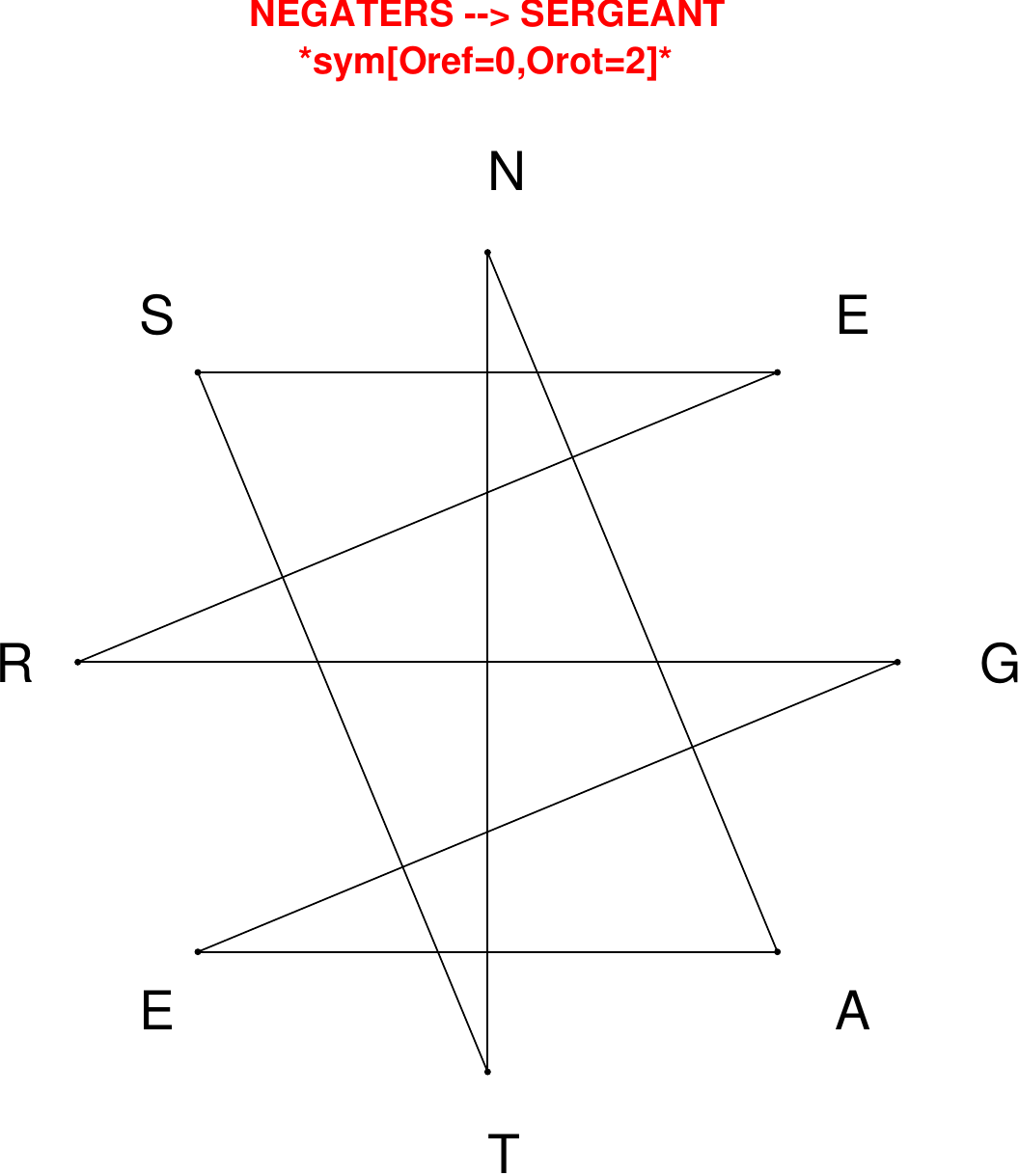}
\end{subfigure}
\hfill
\begin{subfigure}[T]{0.19\textwidth}
\centering
\includegraphics[width=\textwidth]{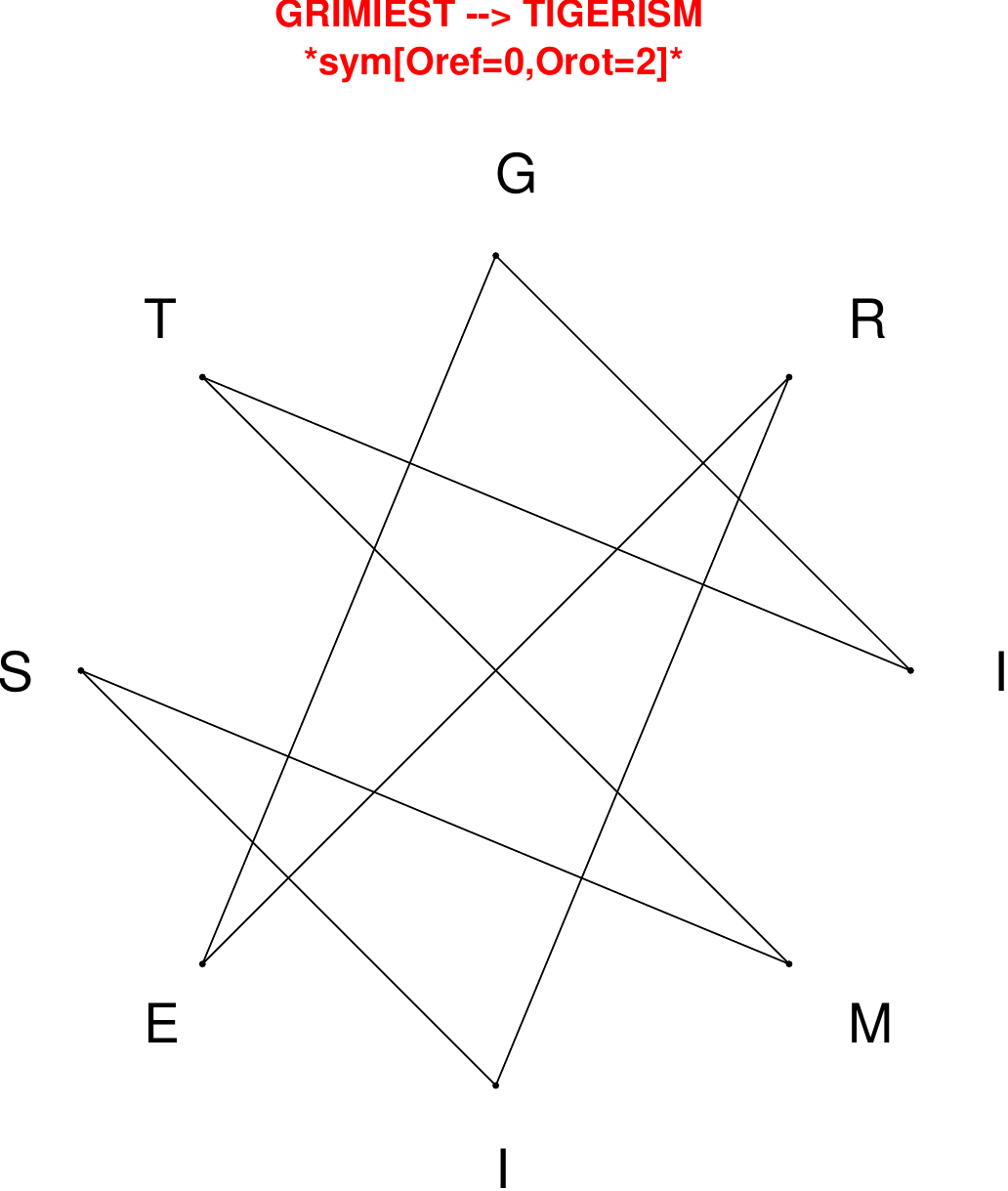}
\end{subfigure}
\hfill
\begin{subfigure}[T]{0.19\textwidth}
\centering
\includegraphics[width=\textwidth]{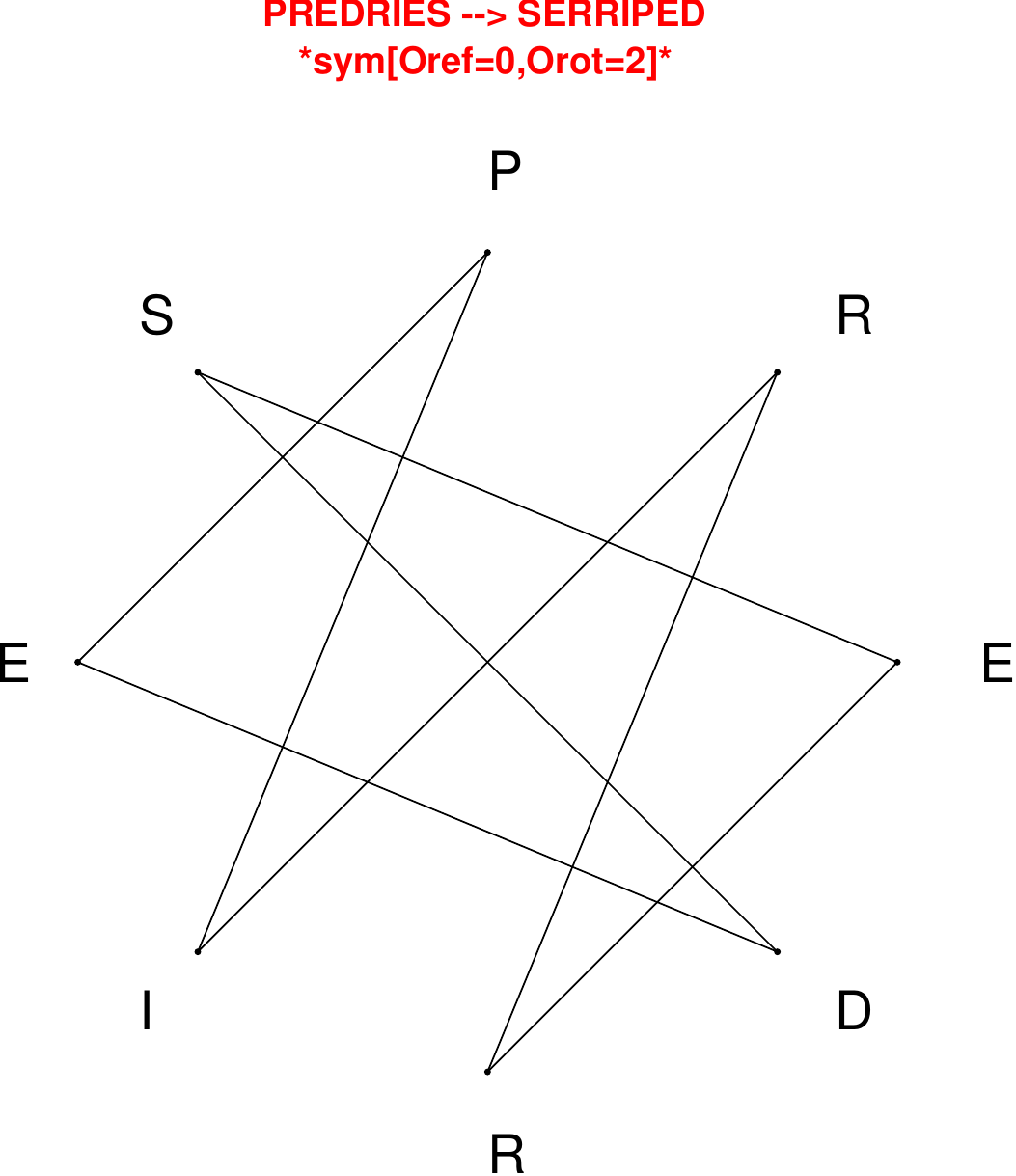}
\end{subfigure}
\hfill
\begin{subfigure}[T]{0.19\textwidth}
\centering
\includegraphics[width=\textwidth]{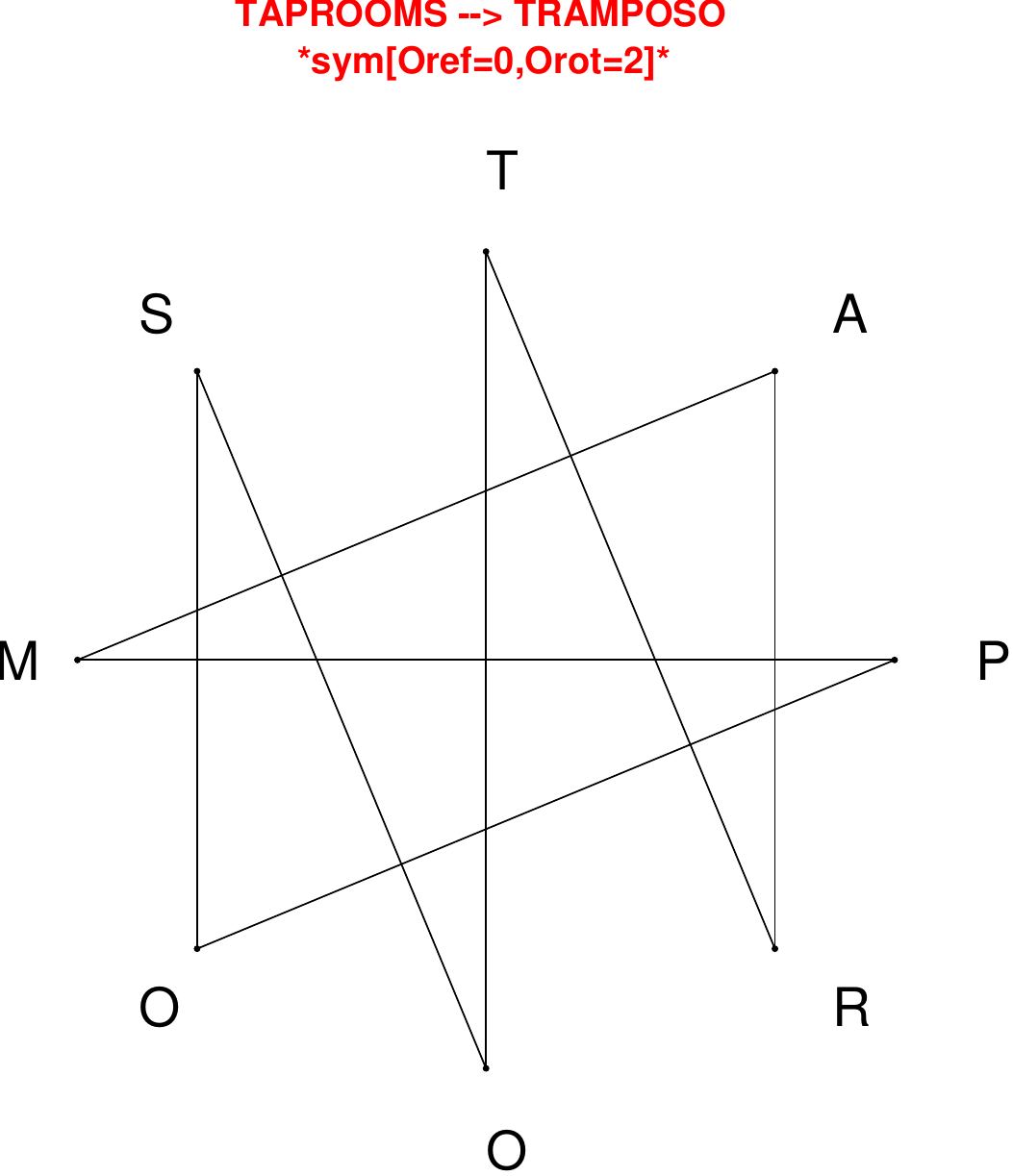}
\end{subfigure}
\hfill
\begin{subfigure}[T]{0.19\textwidth}
\centering
\includegraphics[width=\textwidth]{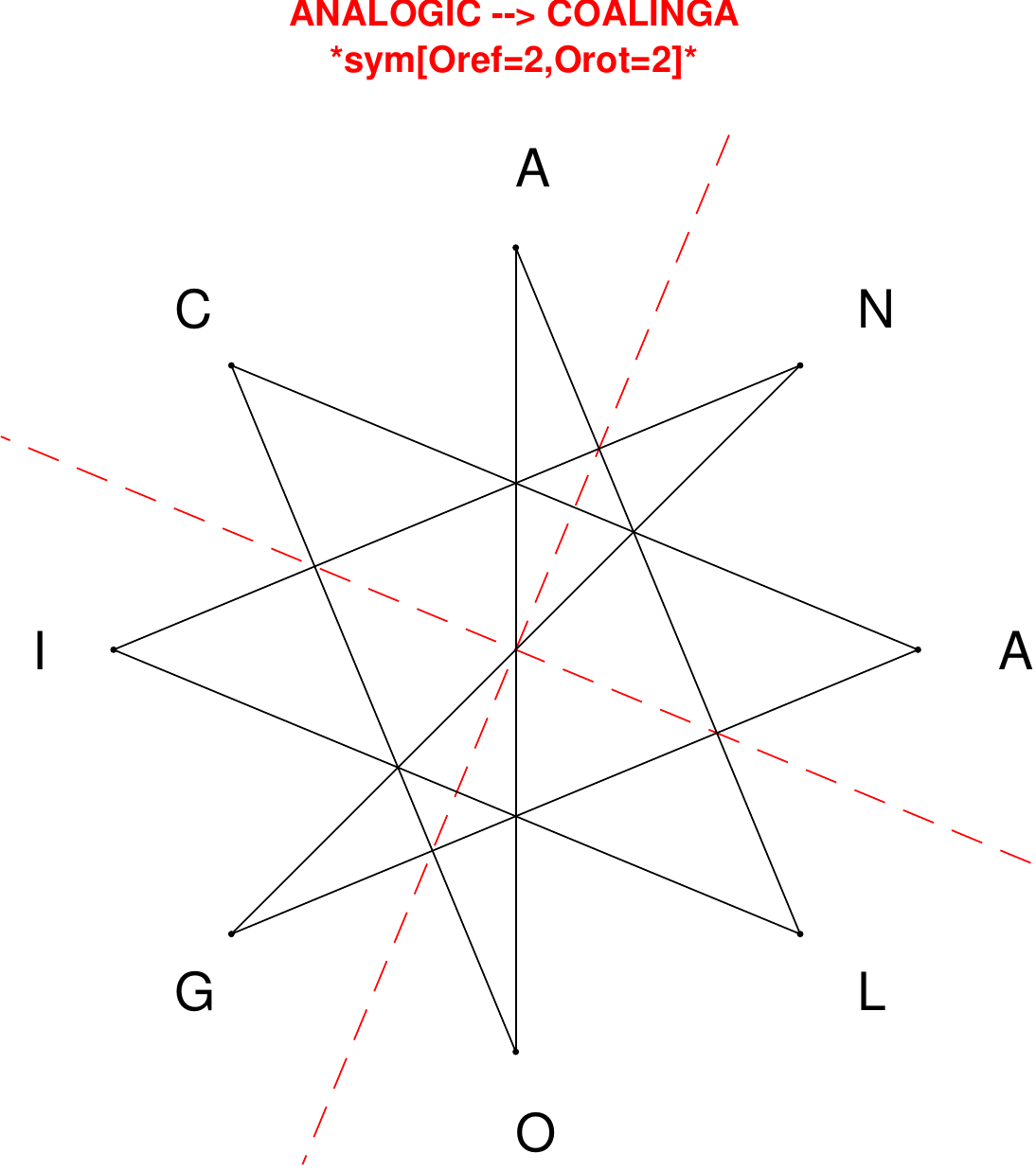}
\end{subfigure}
\end{figure}

\begin{figure}[H]
\centering
\begin{subfigure}[T]{0.19\textwidth}
\centering
\includegraphics[width=\textwidth]{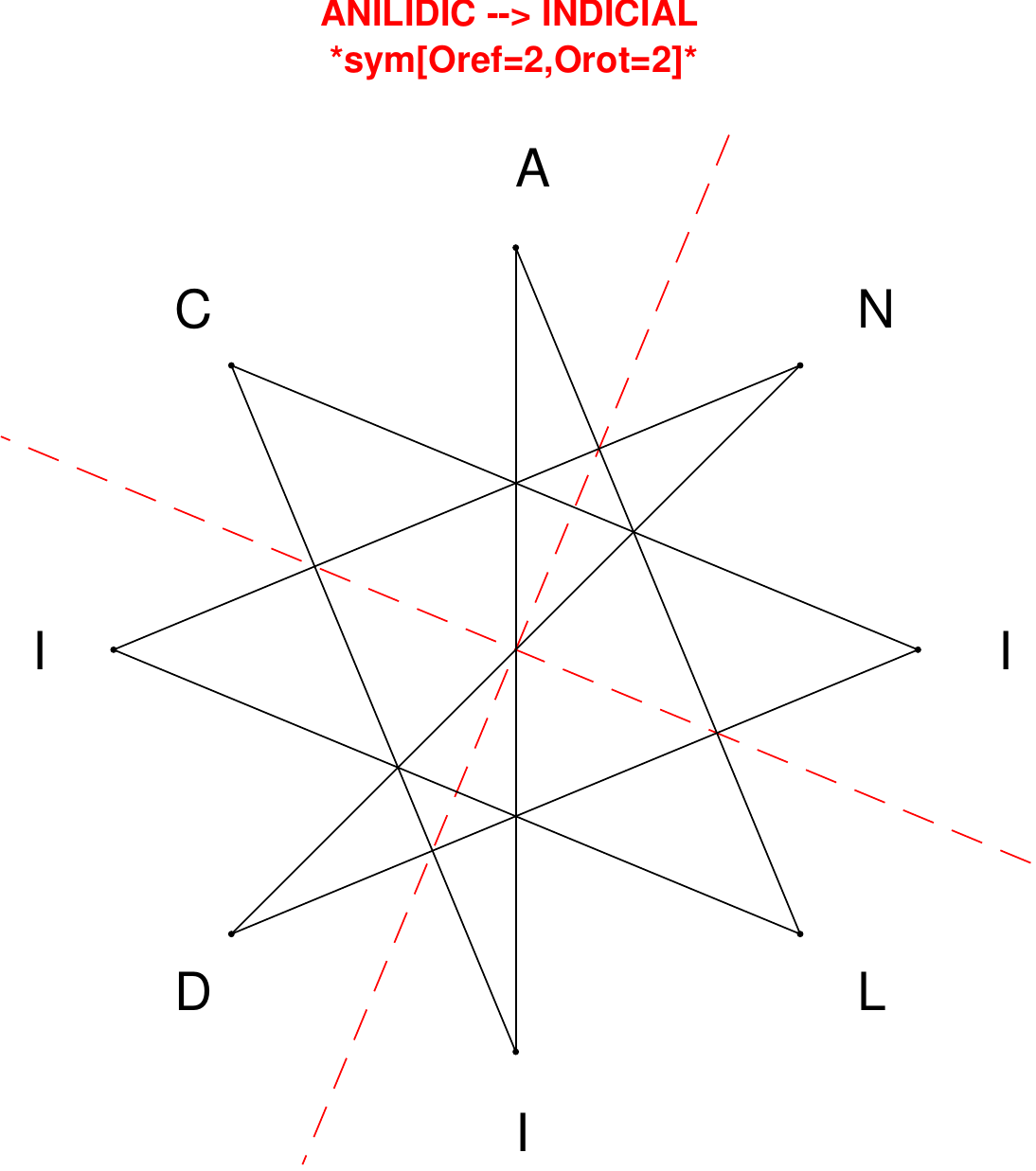}
\end{subfigure}
\hfill
\begin{subfigure}[T]{0.19\textwidth}
\centering
\includegraphics[width=\textwidth]{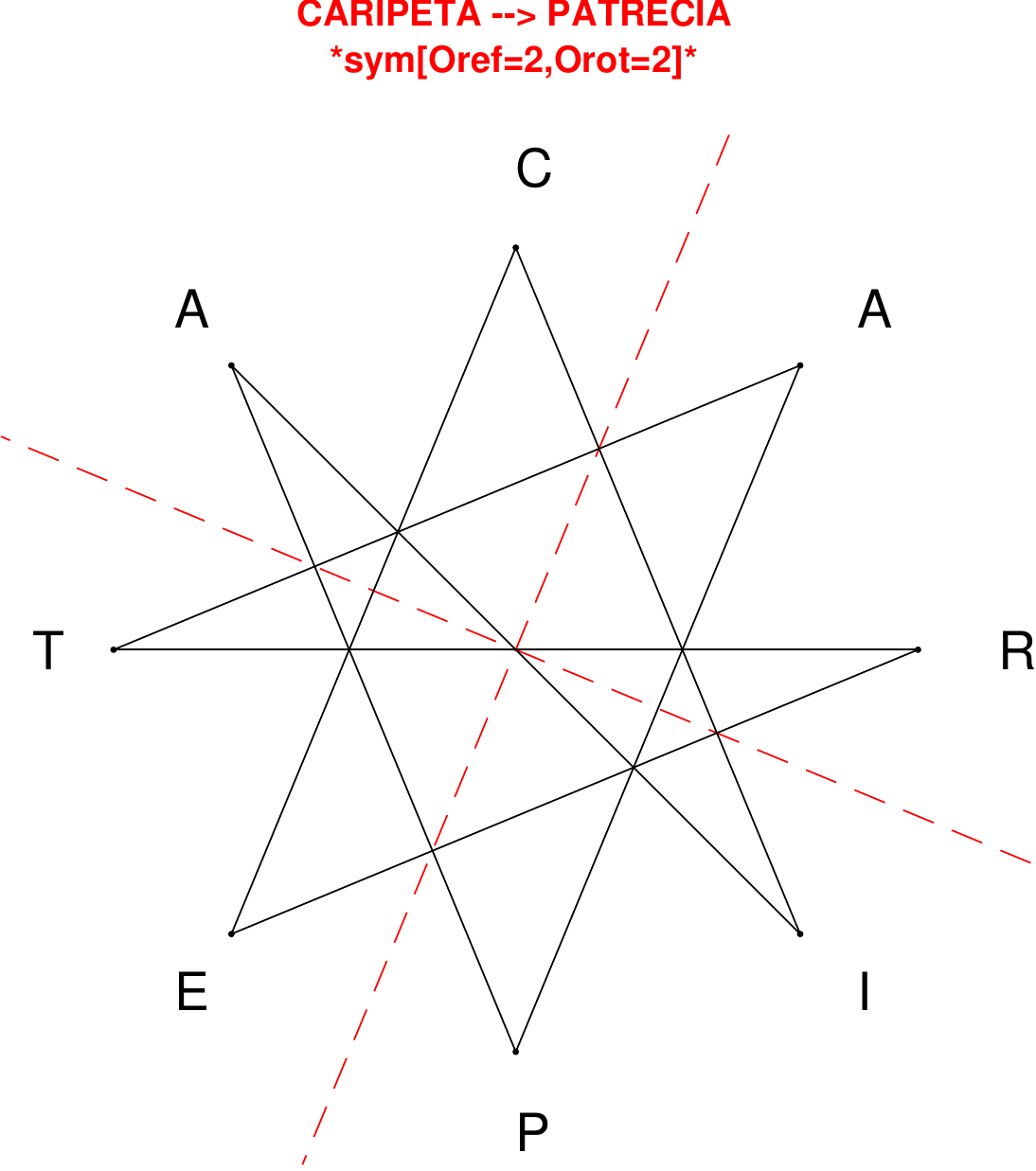}
\end{subfigure}
\hfill
\begin{subfigure}[T]{0.19\textwidth}
\centering
\includegraphics[width=\textwidth]{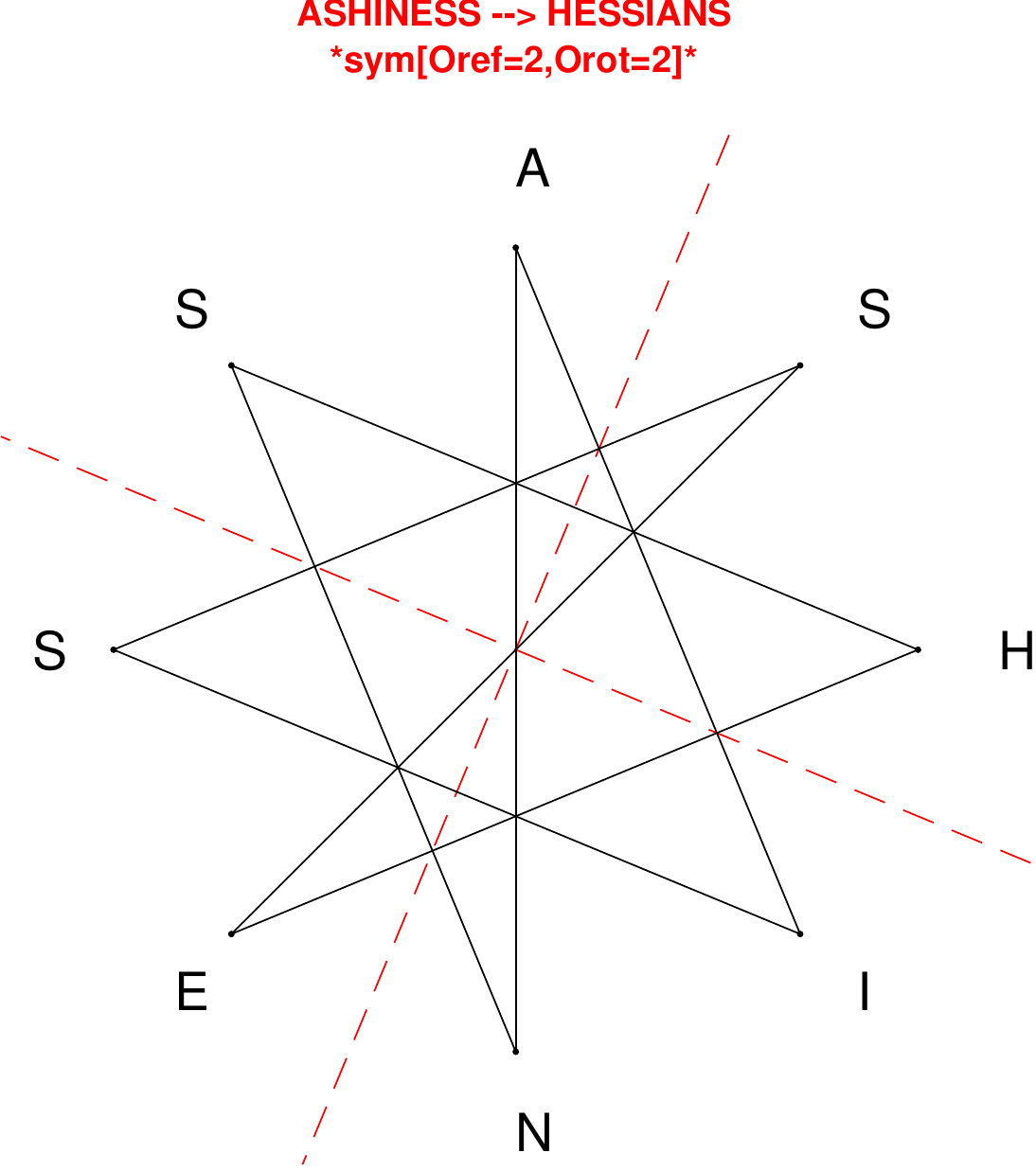}
\end{subfigure}
\hfill
\begin{subfigure}[T]{0.19\textwidth}
\centering
\includegraphics[width=\textwidth]{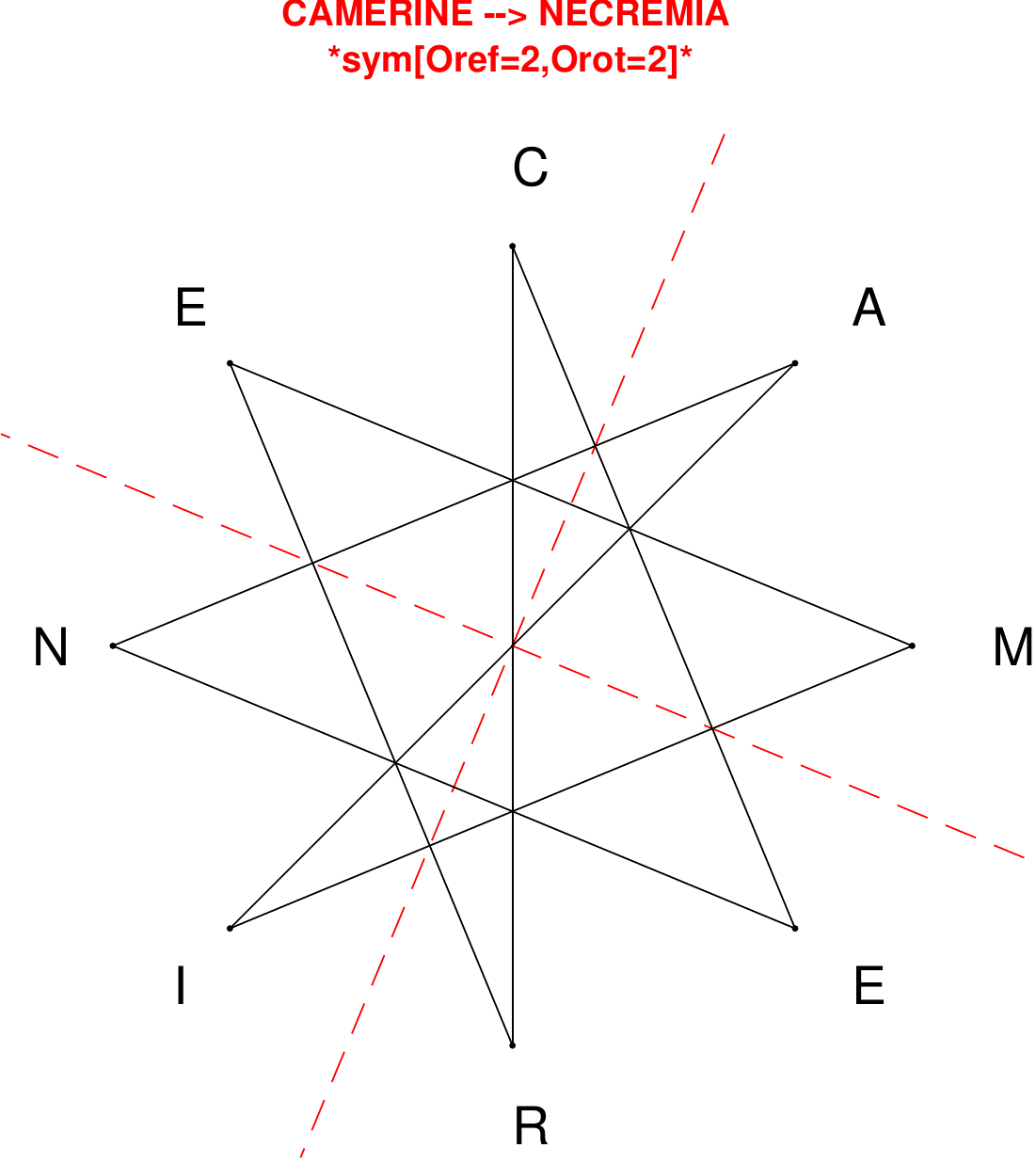}
\end{subfigure}
\hfill
\begin{subfigure}[T]{0.19\textwidth}
\centering
\includegraphics[width=\textwidth]{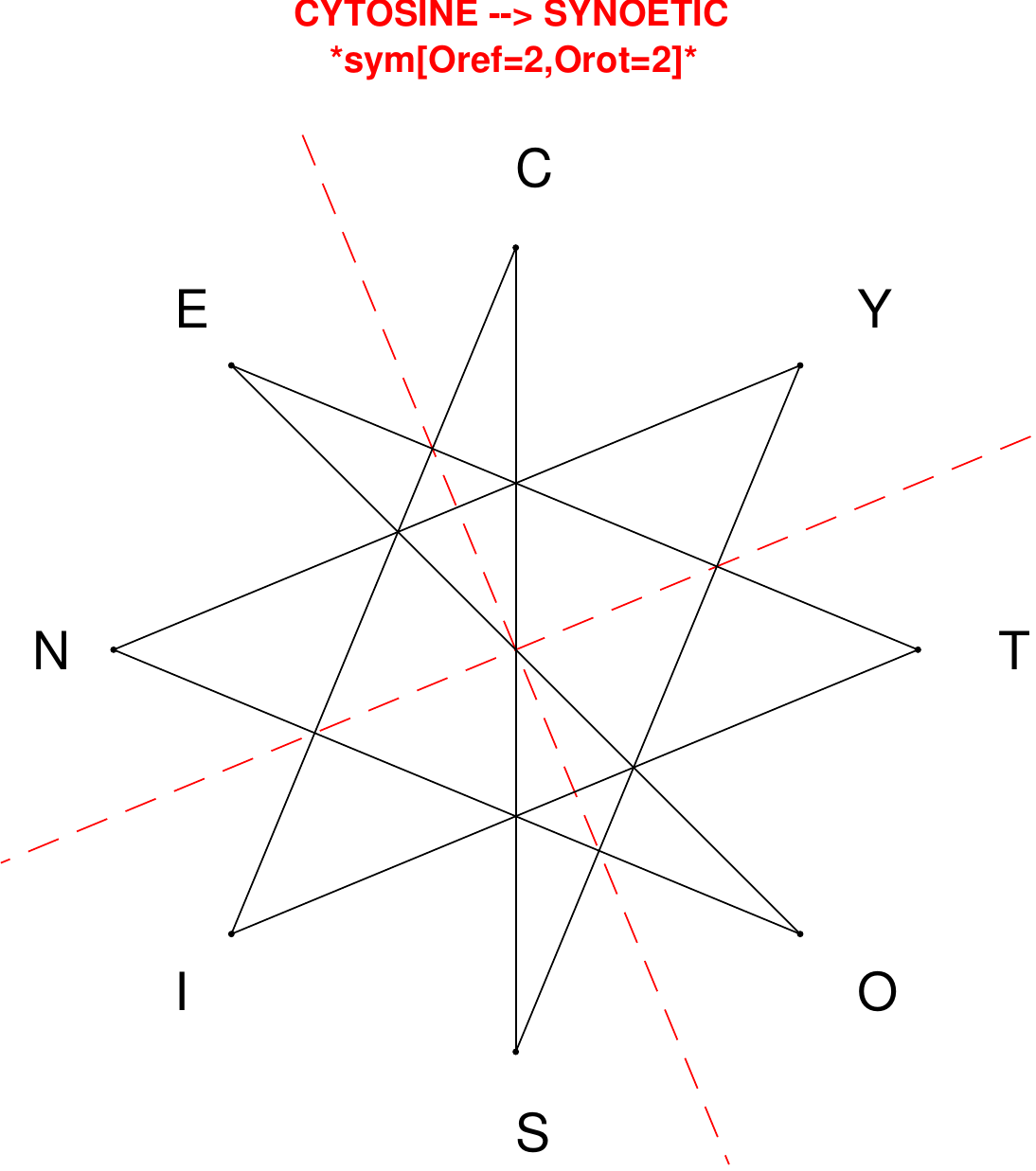}
\end{subfigure}
\end{figure}

\begin{figure}[H]
\centering
\begin{subfigure}[T]{0.19\textwidth}
\centering
\includegraphics[width=\textwidth]{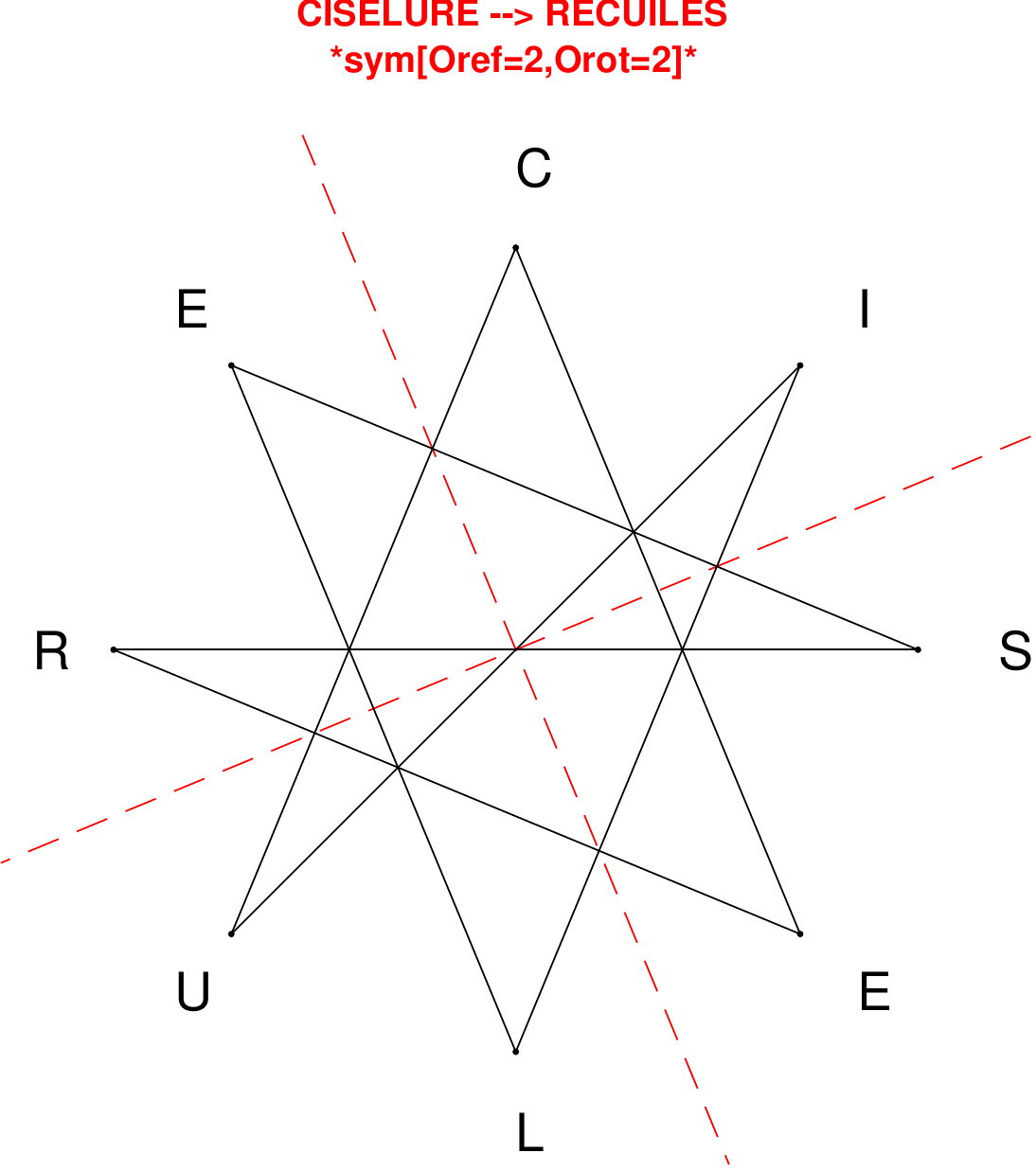}
\end{subfigure}
\hfill
\begin{subfigure}[T]{0.19\textwidth}
\centering
\includegraphics[width=\textwidth]{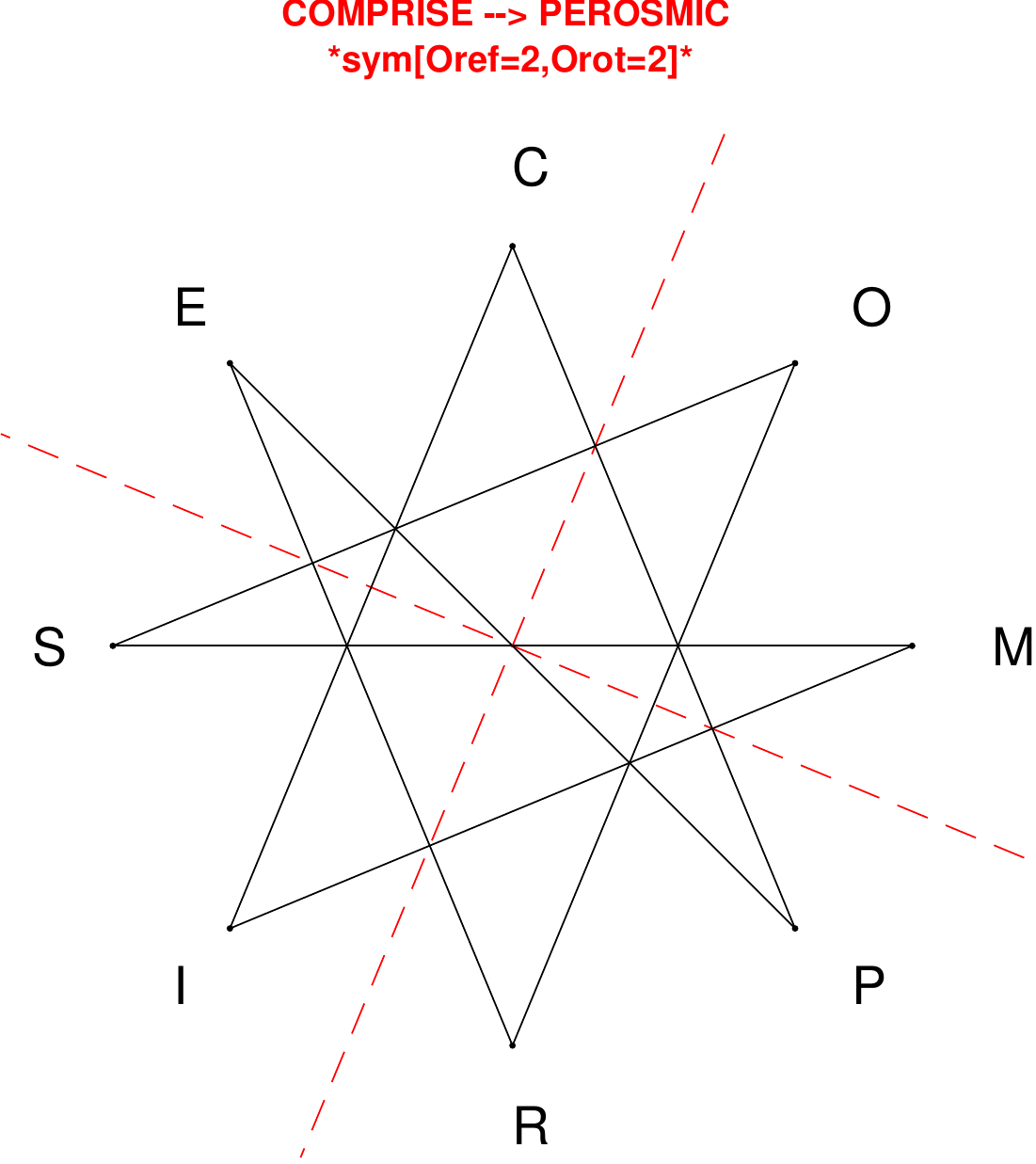}
\end{subfigure}
\hfill
\begin{subfigure}[T]{0.19\textwidth}
\centering
\includegraphics[width=\textwidth]{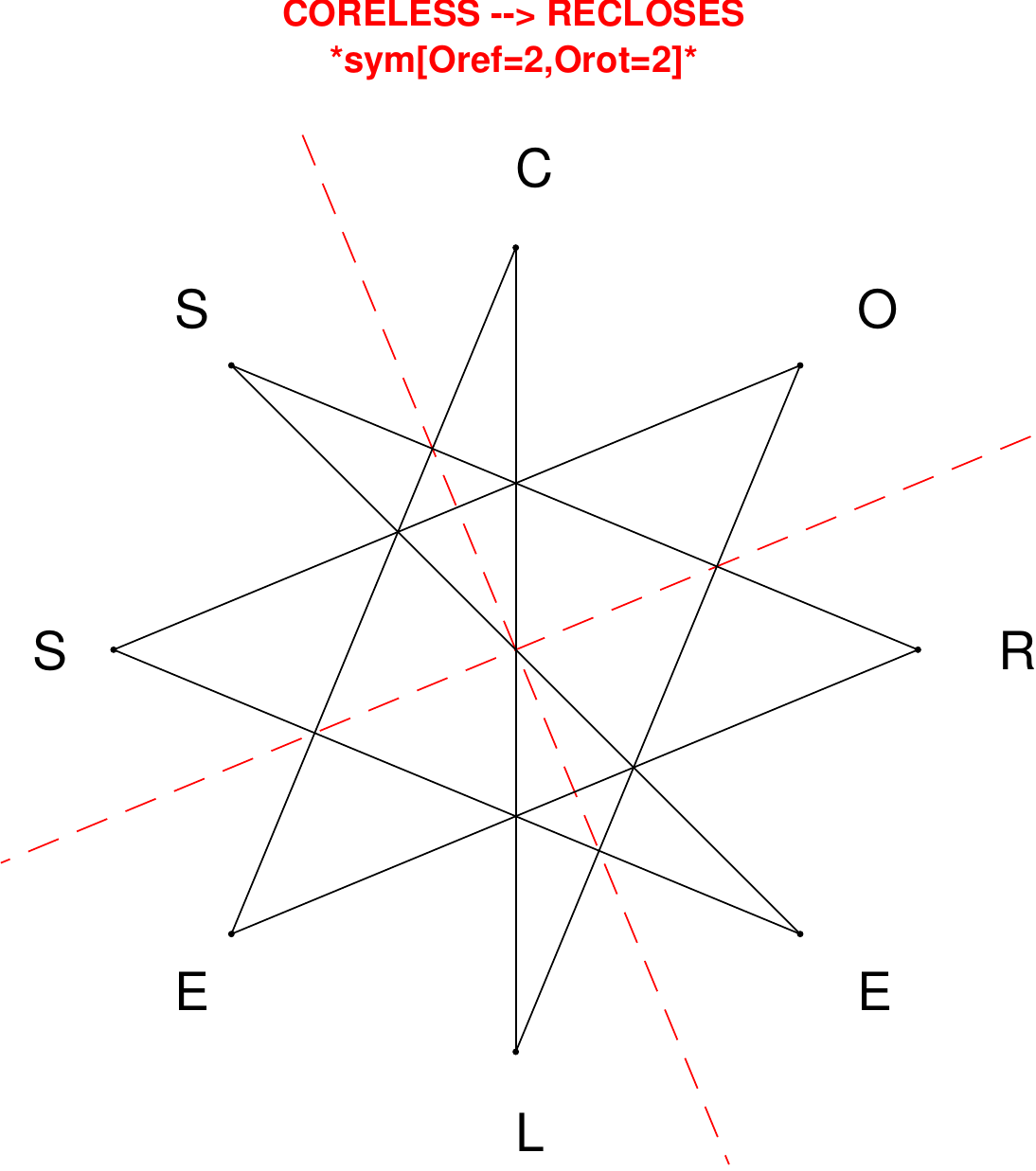}
\end{subfigure}
\hfill
\begin{subfigure}[T]{0.19\textwidth}
\centering
\includegraphics[width=\textwidth]{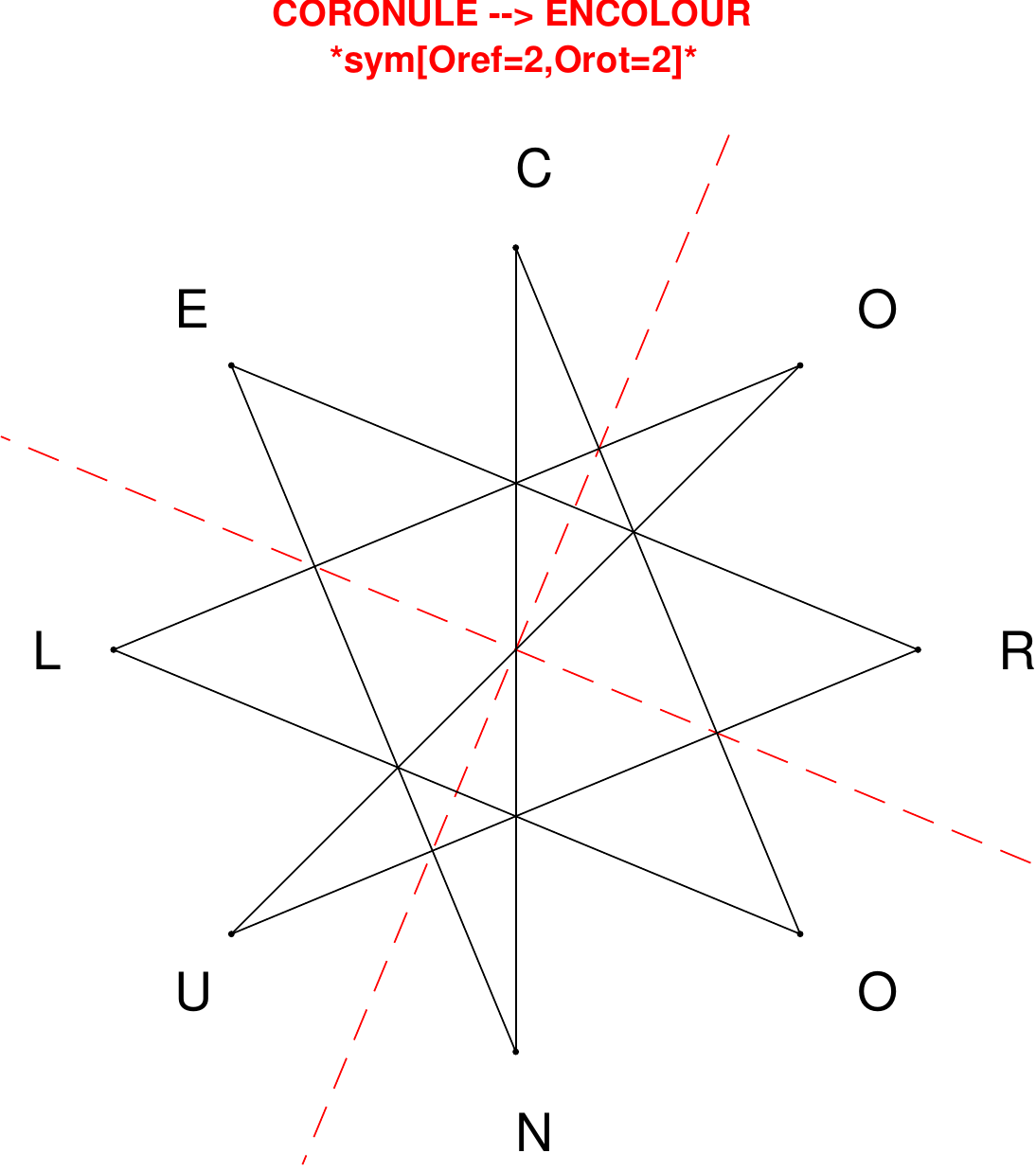}
\end{subfigure}
\hfill
\begin{subfigure}[T]{0.19\textwidth}
\centering
\includegraphics[width=\textwidth]{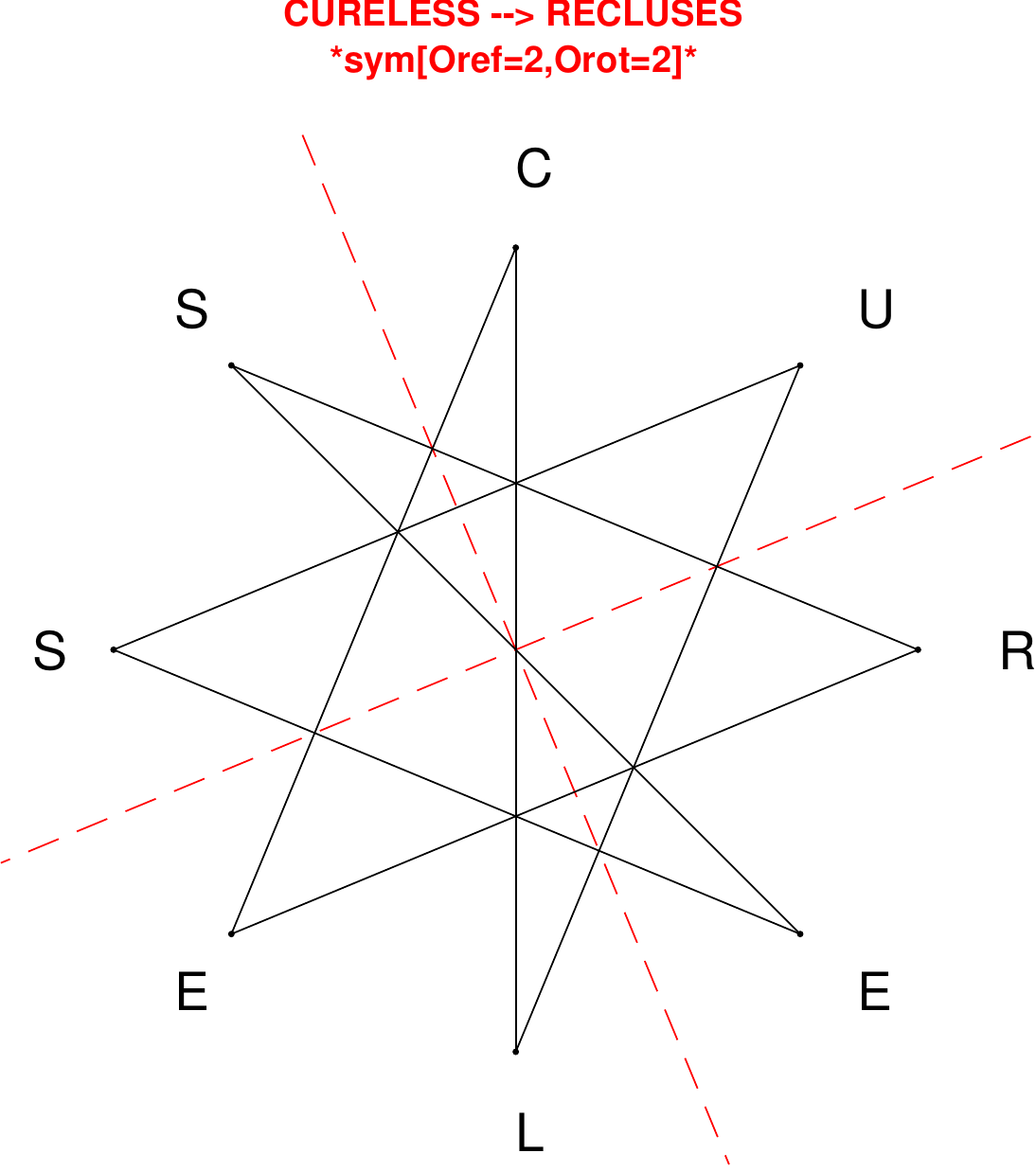}
\end{subfigure}
\end{figure}

\begin{figure}[H]
\centering
\begin{subfigure}[T]{0.19\textwidth}
\centering
\includegraphics[width=\textwidth]{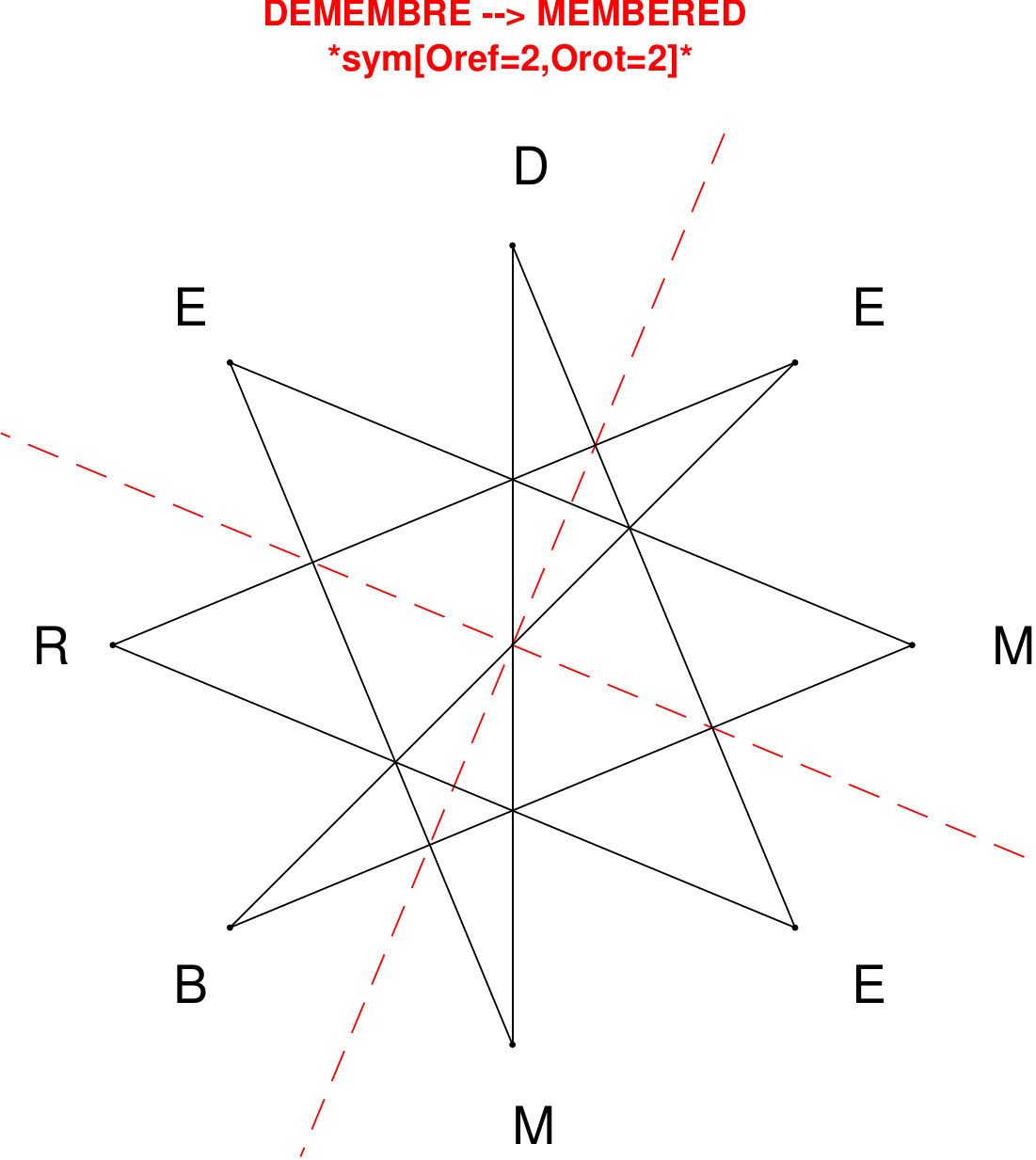}
\end{subfigure}
\hfill
\begin{subfigure}[T]{0.19\textwidth}
\centering
\includegraphics[width=\textwidth]{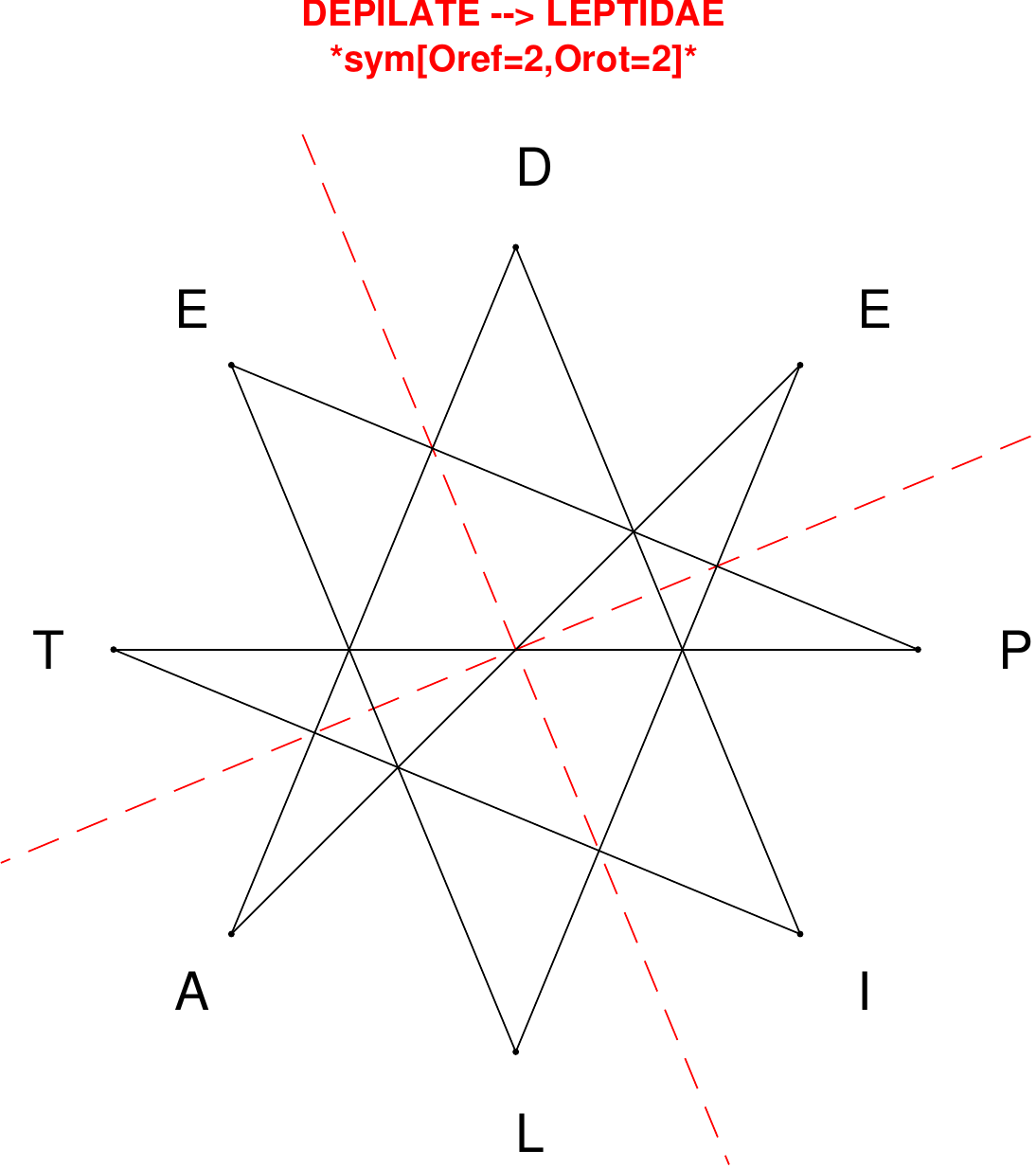}
\end{subfigure}
\hfill
\begin{subfigure}[T]{0.19\textwidth}
\centering
\includegraphics[width=\textwidth]{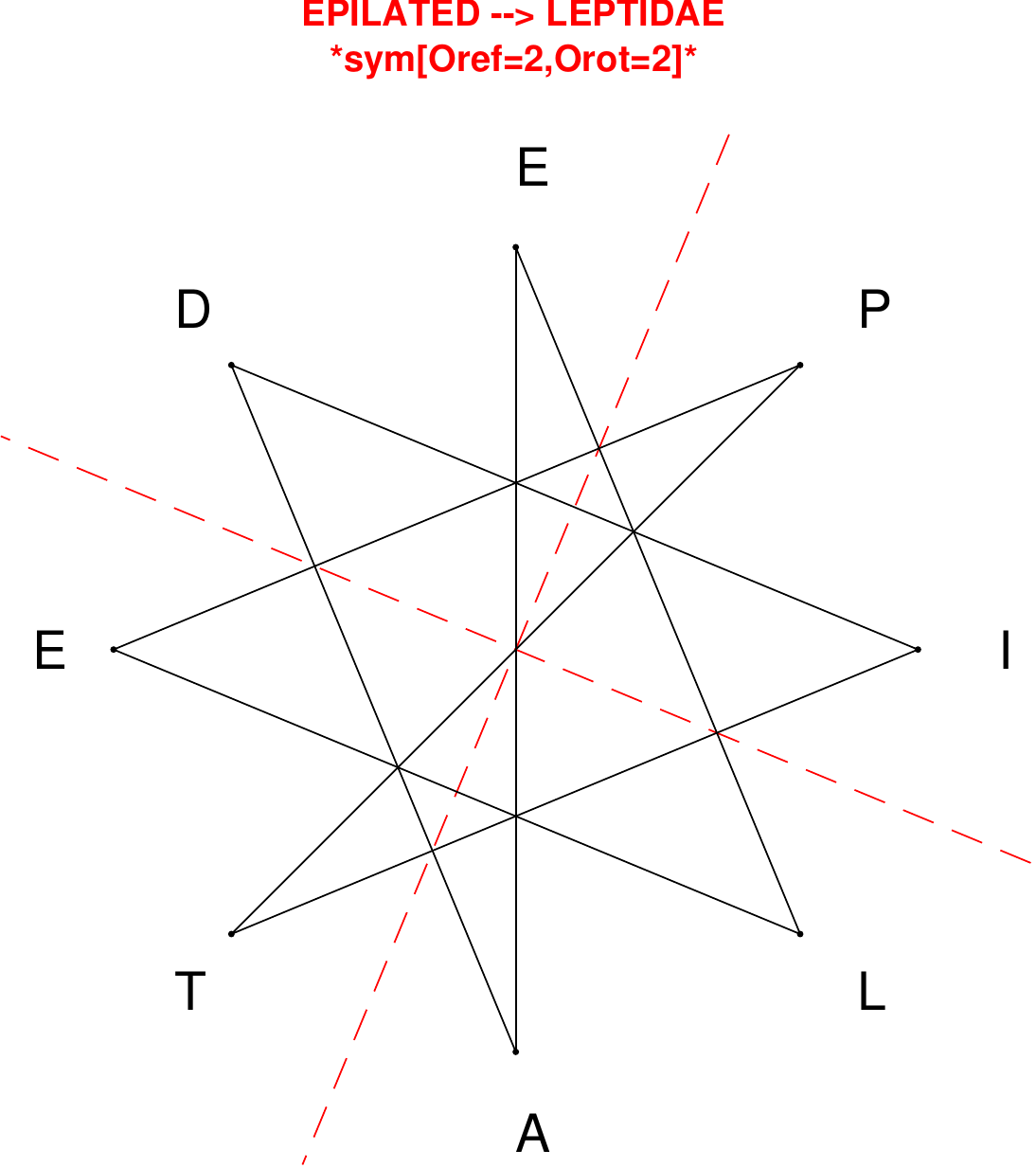}
\end{subfigure}
\hfill
\begin{subfigure}[T]{0.19\textwidth}
\centering
\includegraphics[width=\textwidth]{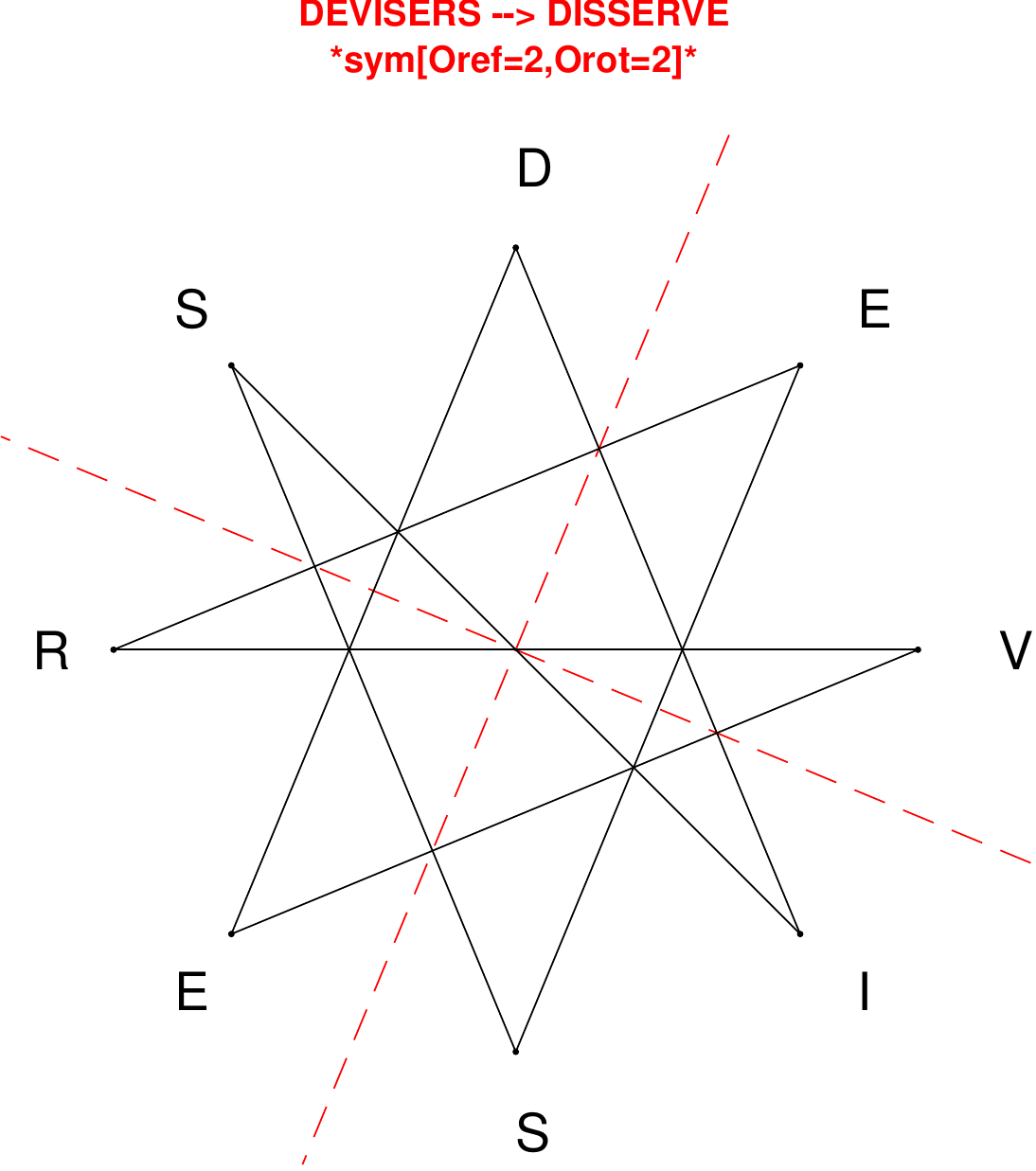}
\end{subfigure}
\hfill
\begin{subfigure}[T]{0.19\textwidth}
\centering
\includegraphics[width=\textwidth]{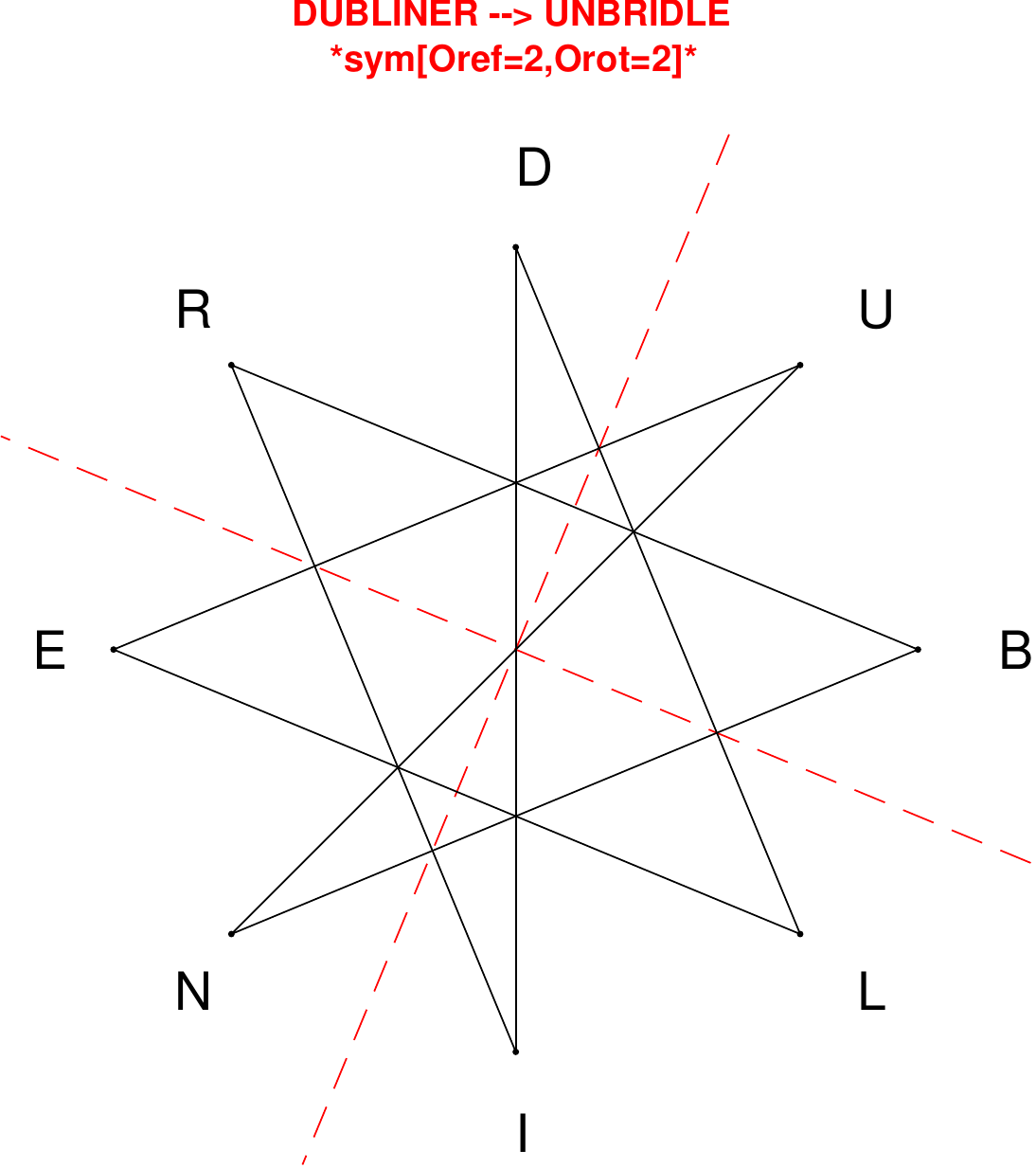}
\end{subfigure}
\end{figure}

\begin{figure}[H]
\centering
\begin{subfigure}[T]{0.19\textwidth}
\centering
\includegraphics[width=\textwidth]{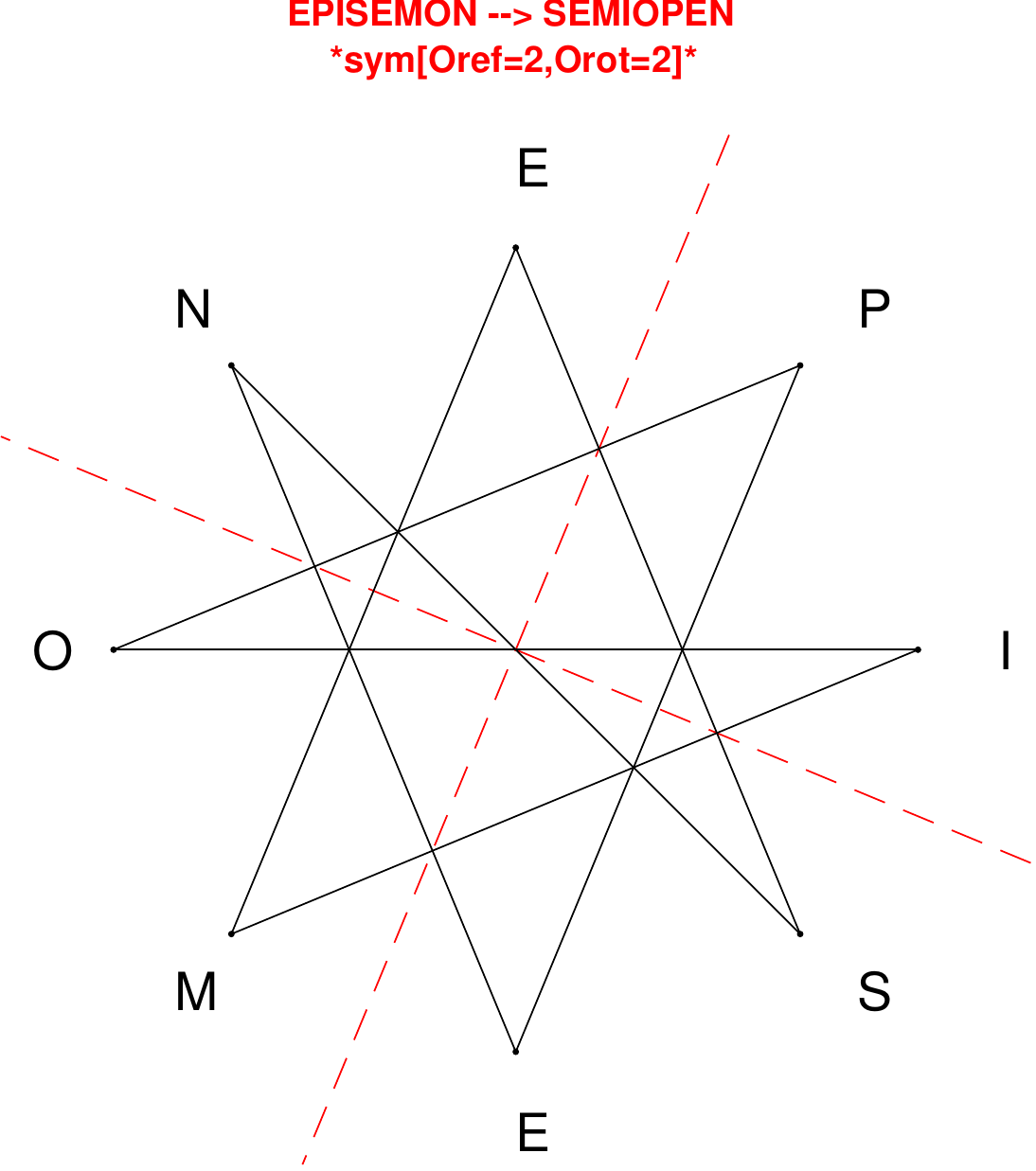}
\end{subfigure}
\hfill
\begin{subfigure}[T]{0.19\textwidth}
\centering
\includegraphics[width=\textwidth]{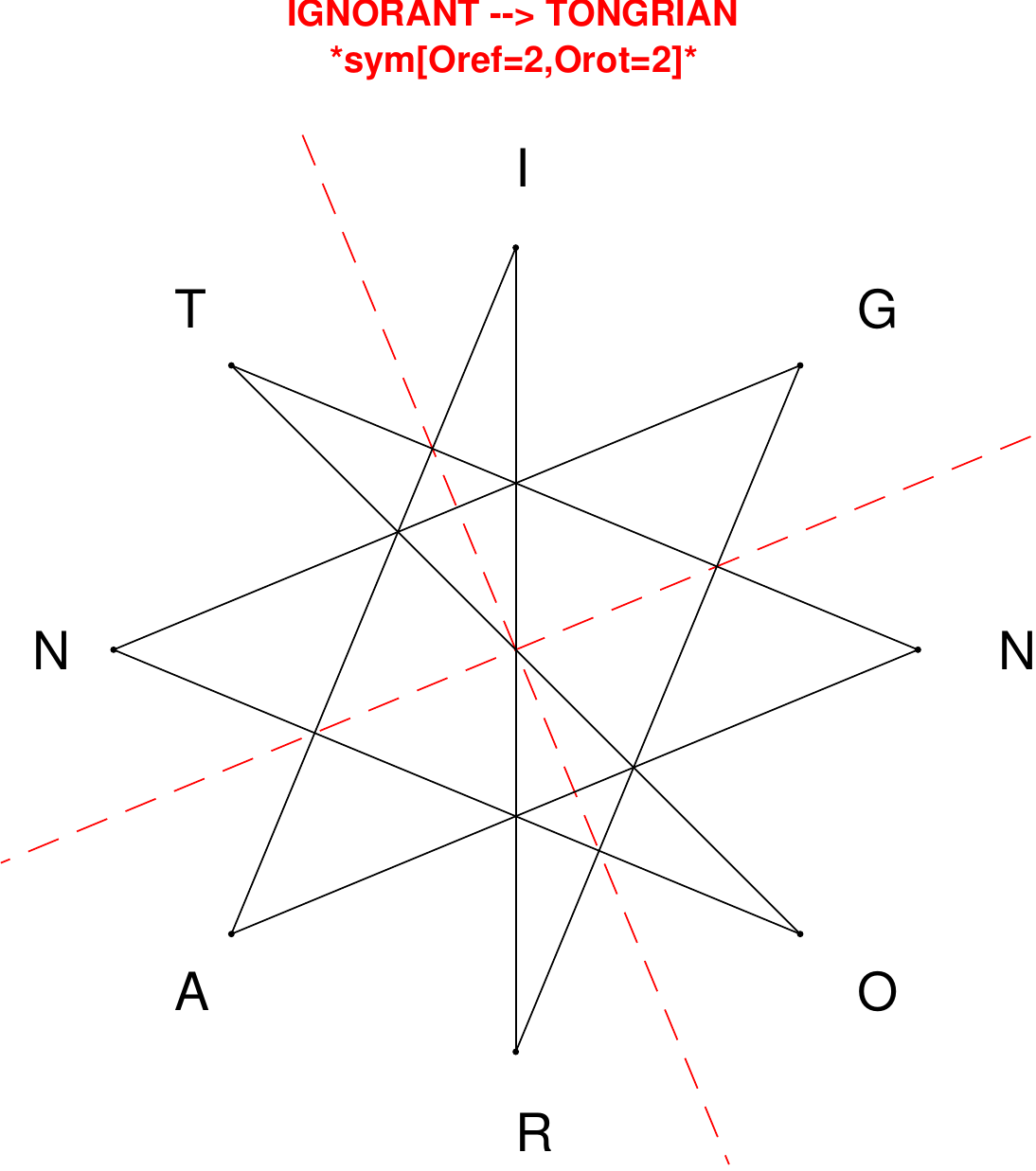}
\end{subfigure}
\hfill
\begin{subfigure}[T]{0.19\textwidth}
\centering
\includegraphics[width=\textwidth]{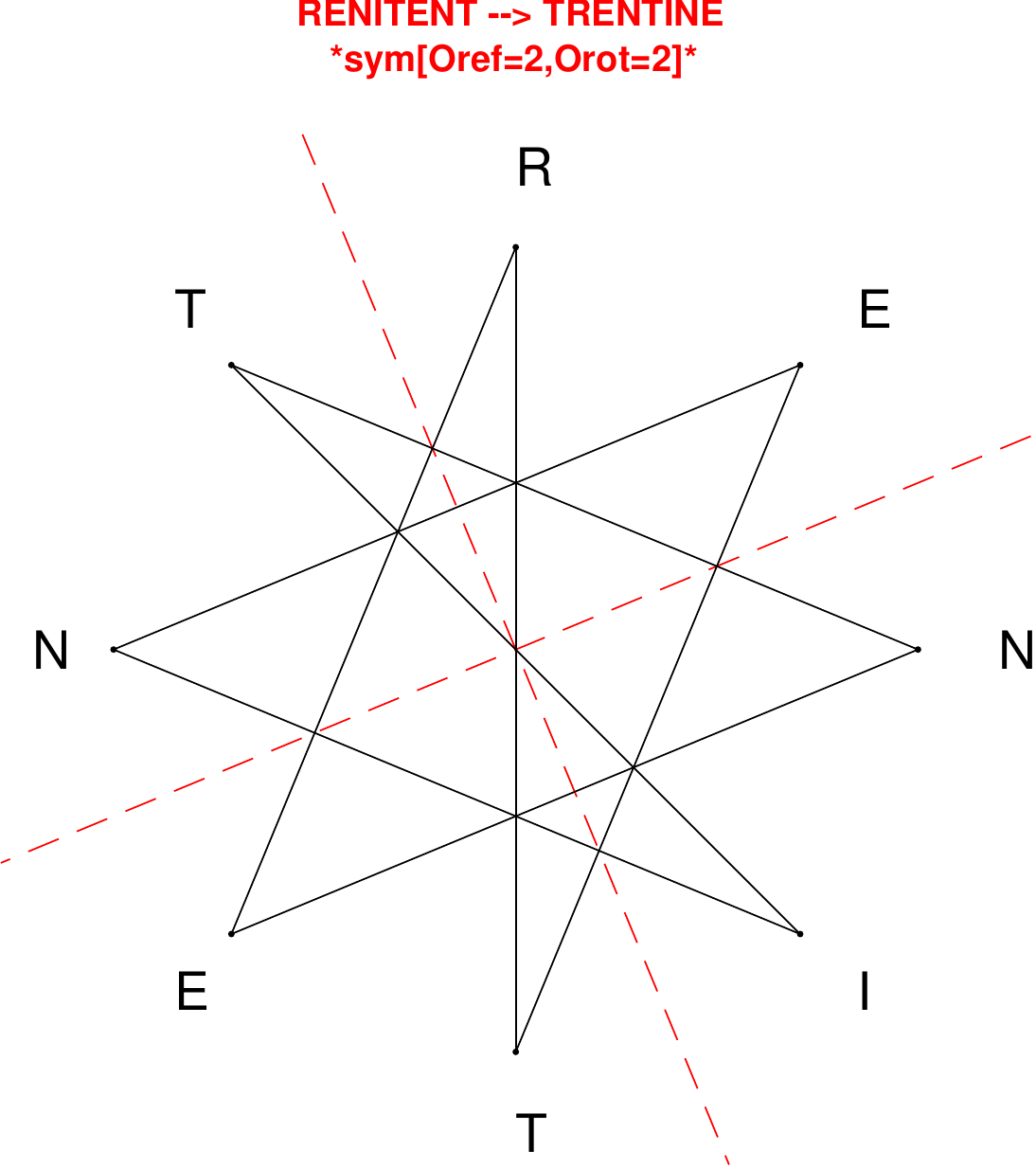}
\end{subfigure}
\hfill
\begin{subfigure}[T]{0.19\textwidth}
\centering
\includegraphics[width=\textwidth]{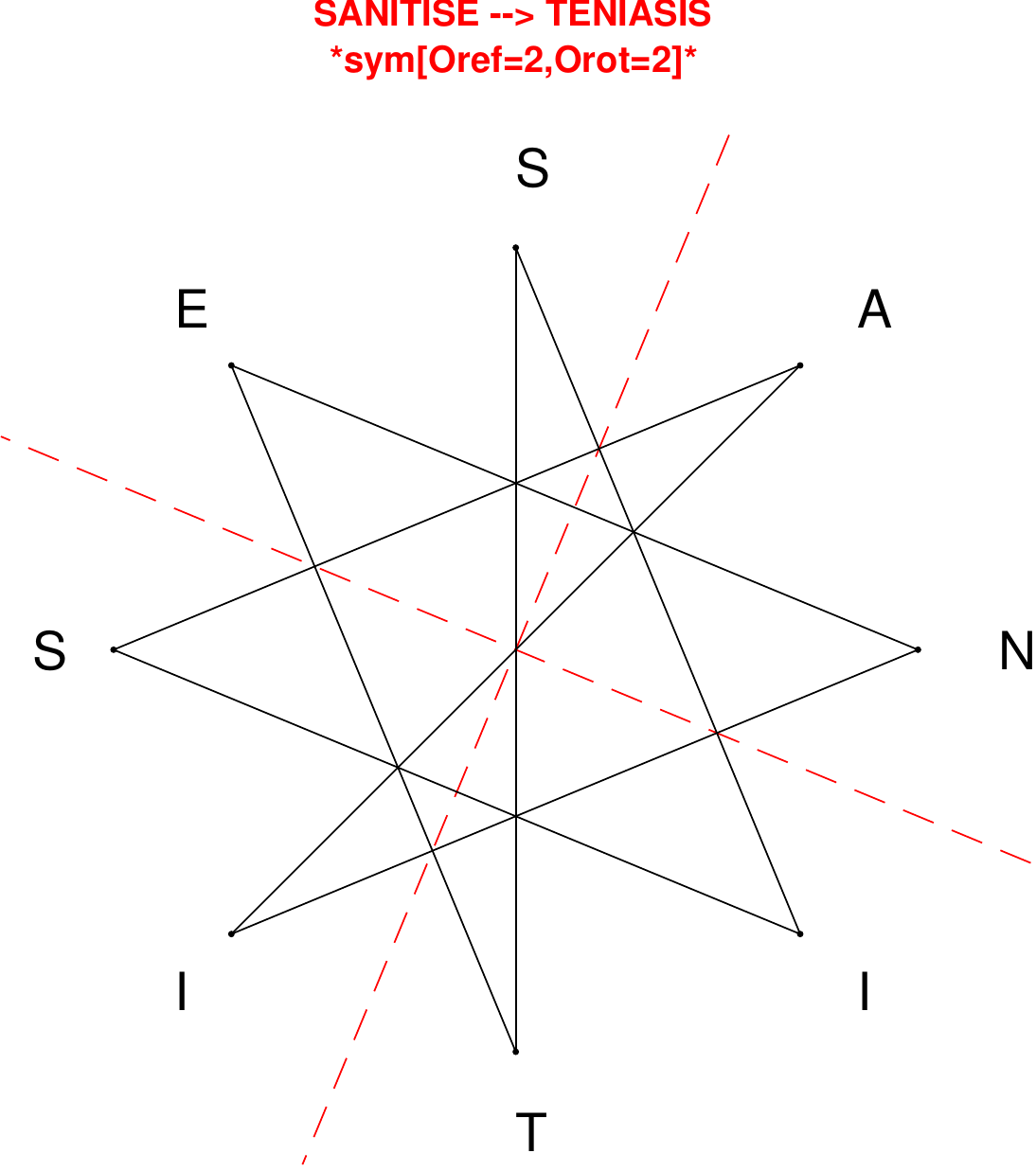}
\end{subfigure}
\hfill
\begin{subfigure}[T]{0.19\textwidth}
\centering
\includegraphics[width=\textwidth]{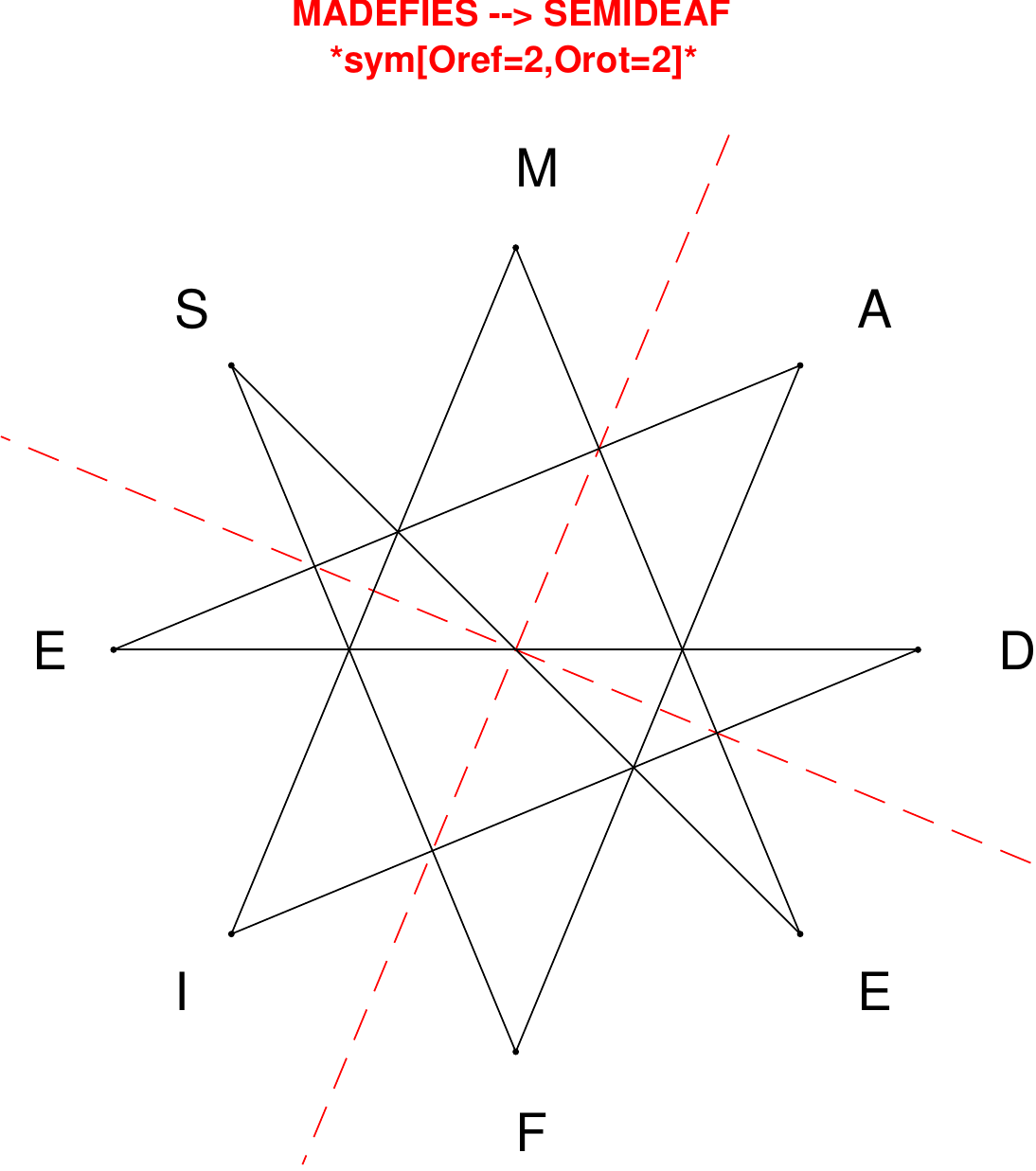}
\end{subfigure}
\end{figure}

\begin{figure}[H]
\centering
\begin{subfigure}[T]{0.19\textwidth}
\centering
\includegraphics[width=\textwidth]{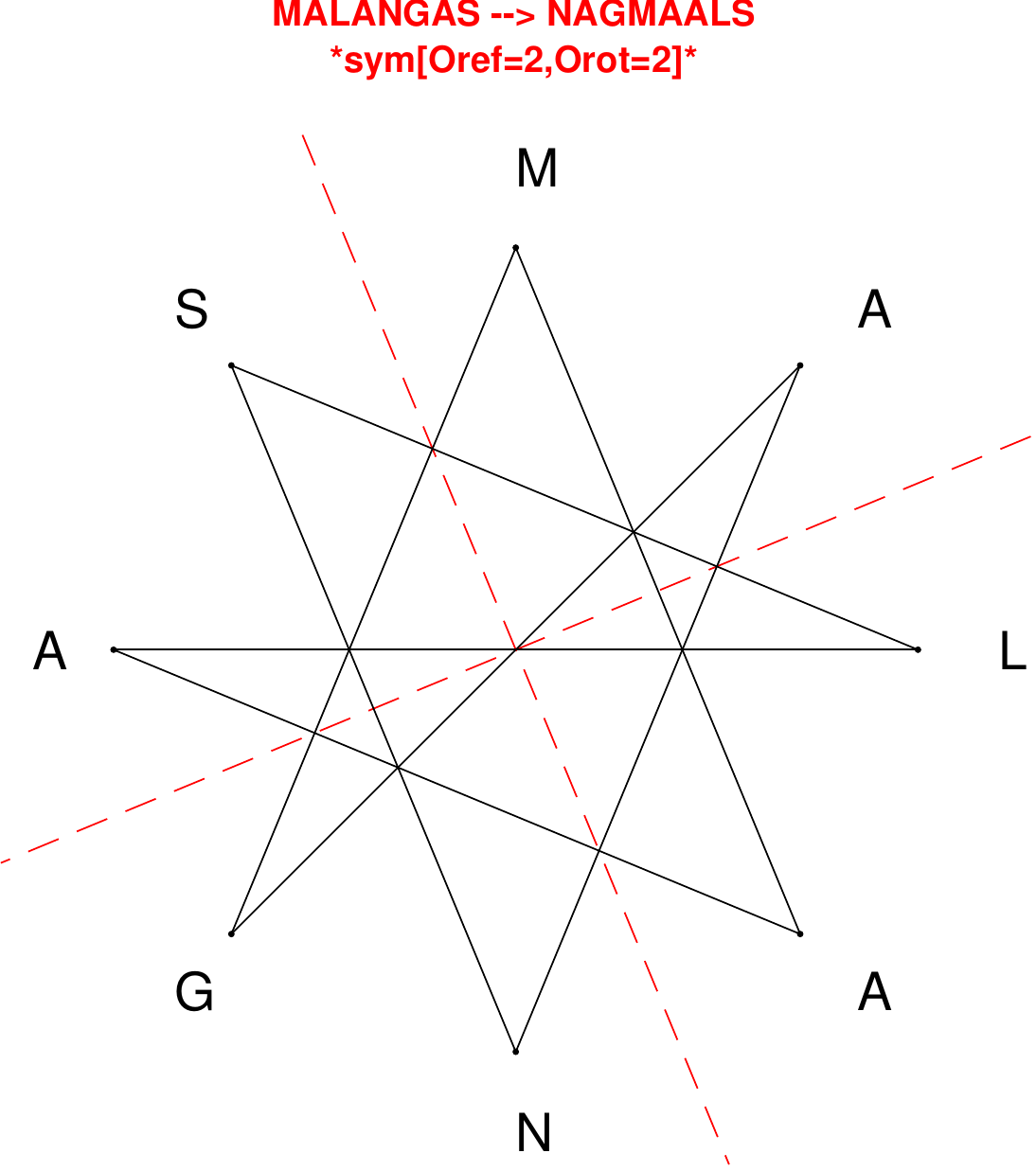}
\end{subfigure}
\hfill
\begin{subfigure}[T]{0.19\textwidth}
\centering
\includegraphics[width=\textwidth]{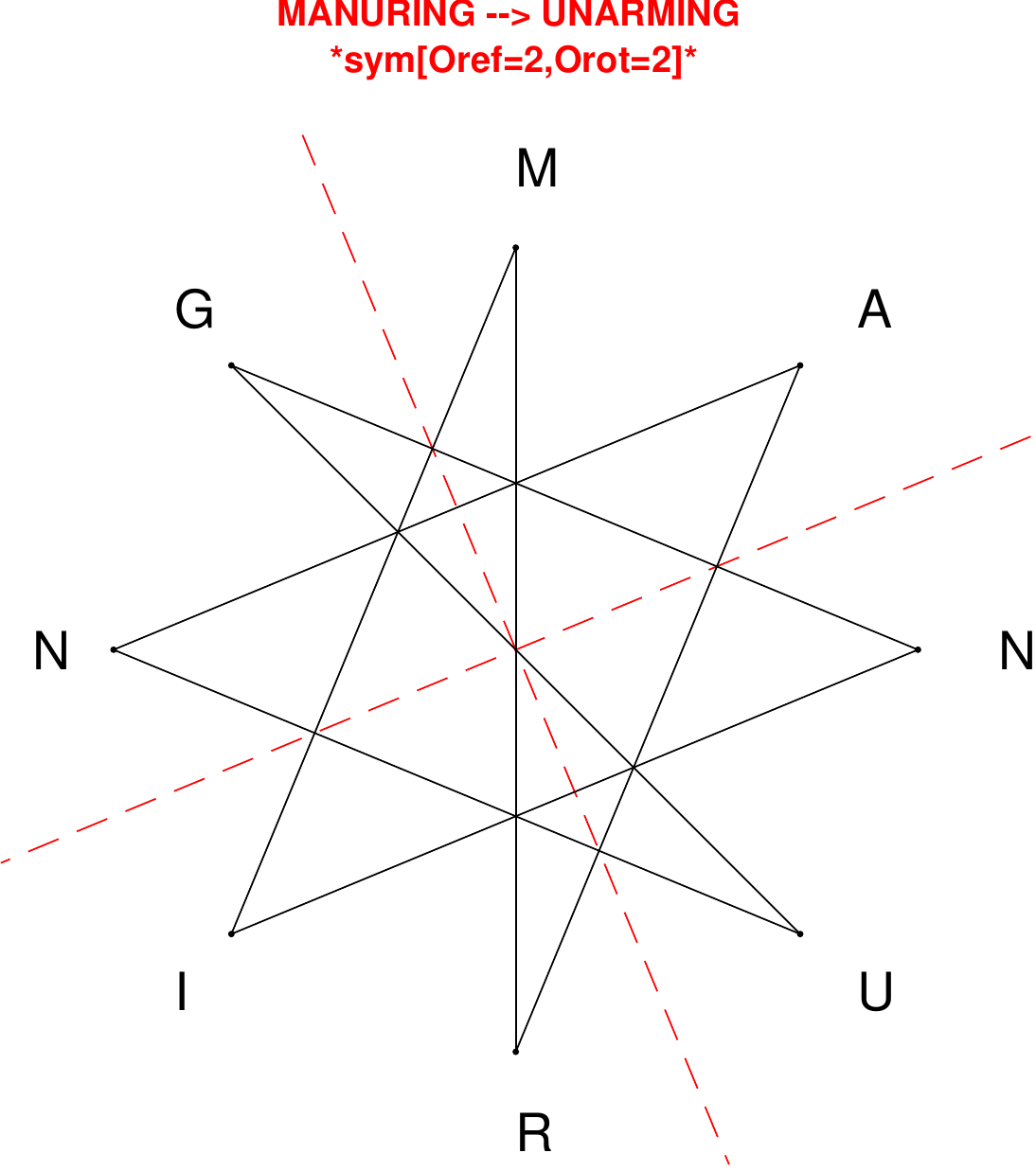}
\end{subfigure}
\hfill
\begin{subfigure}[T]{0.19\textwidth}
\centering
\includegraphics[width=\textwidth]{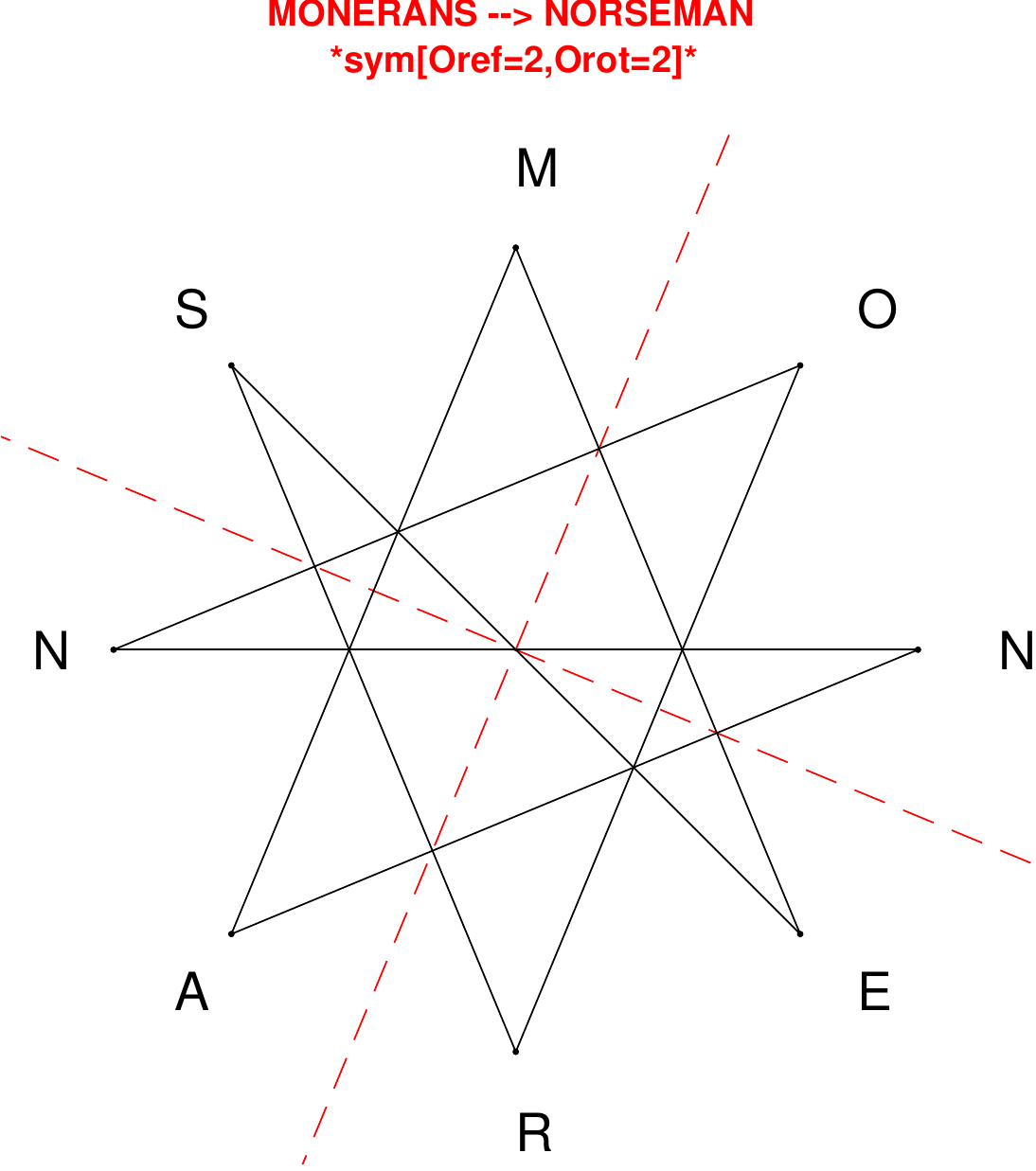}
\end{subfigure}
\hfill
\begin{subfigure}[T]{0.19\textwidth}
\centering
\includegraphics[width=\textwidth]{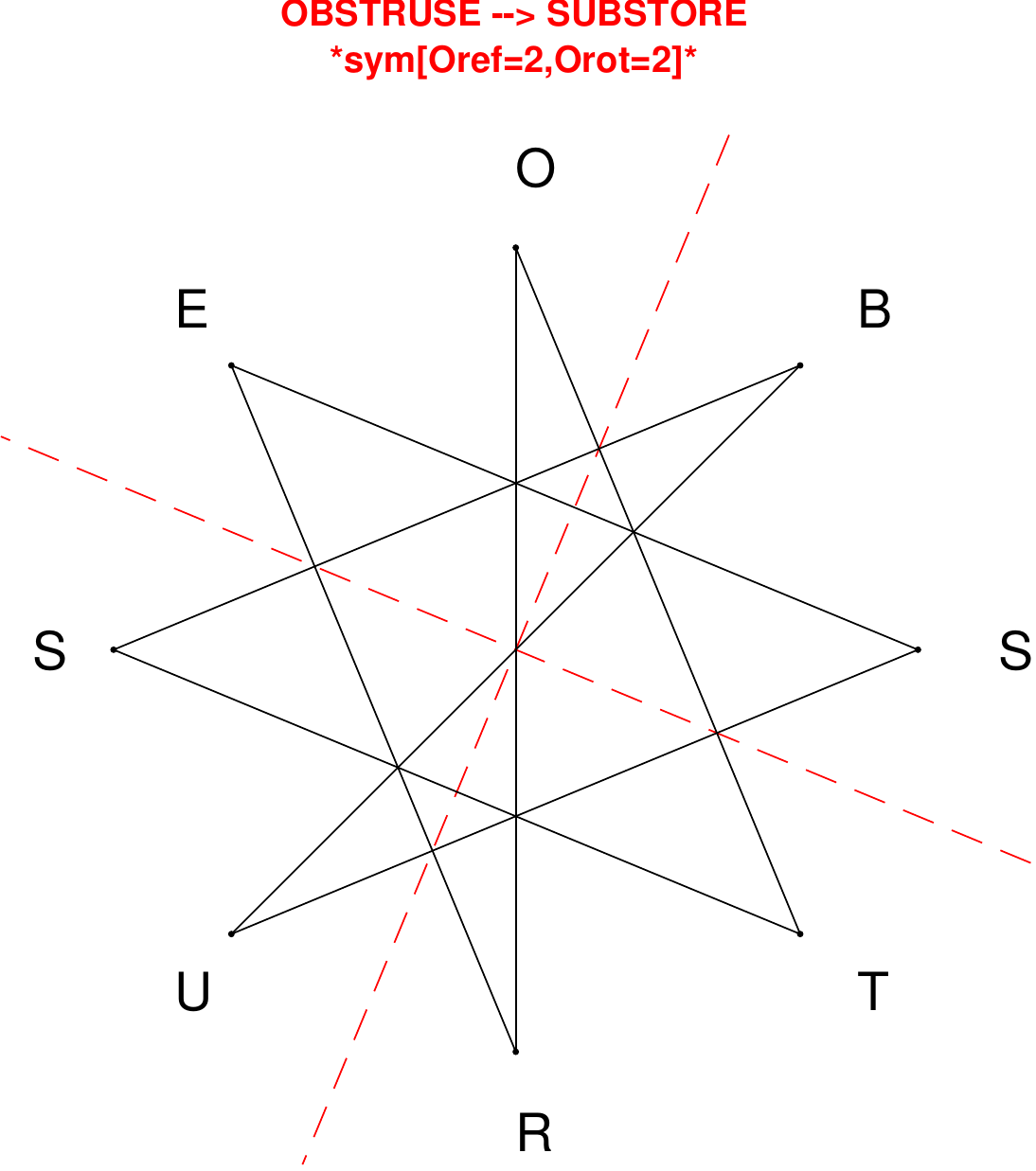}
\end{subfigure}
\hfill
\begin{subfigure}[T]{0.19\textwidth}
\centering
\includegraphics[width=\textwidth]{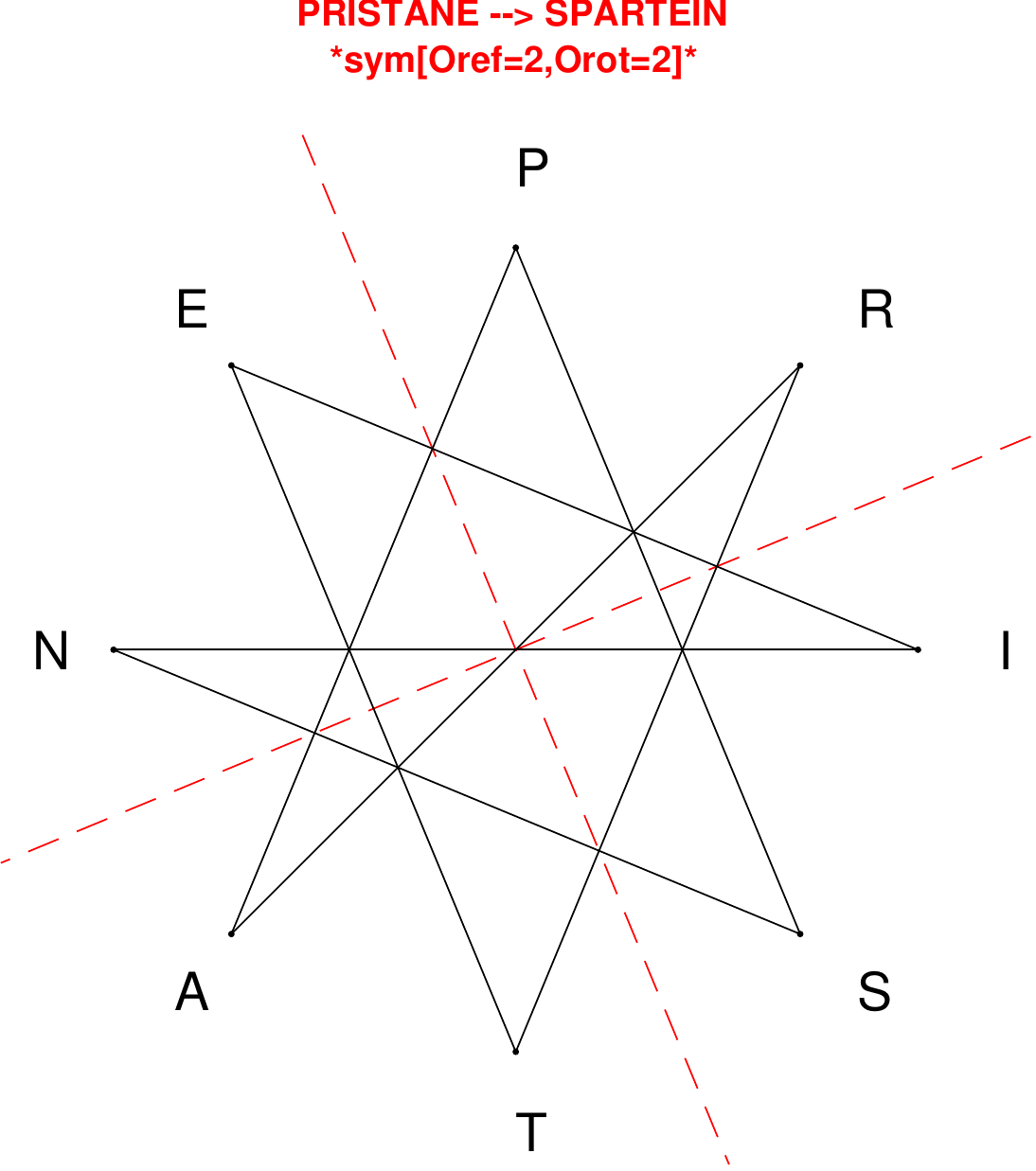}
\end{subfigure}
\end{figure}

\begin{figure}[H]
\centering
\begin{subfigure}[T]{0.19\textwidth}
\centering
\includegraphics[width=\textwidth]{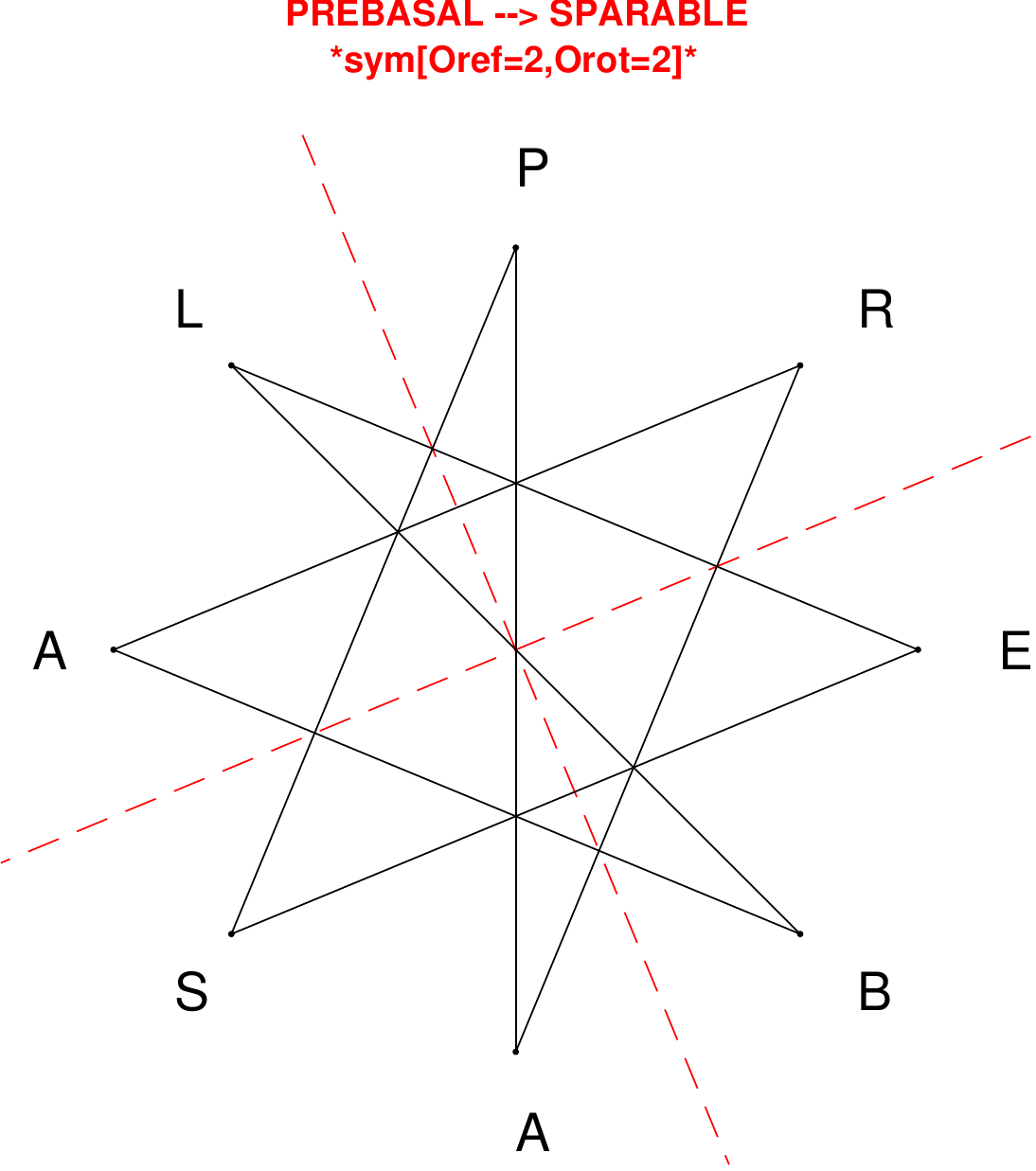}
\end{subfigure}
\hfill
\begin{subfigure}[T]{0.19\textwidth}
\centering
\includegraphics[width=\textwidth]{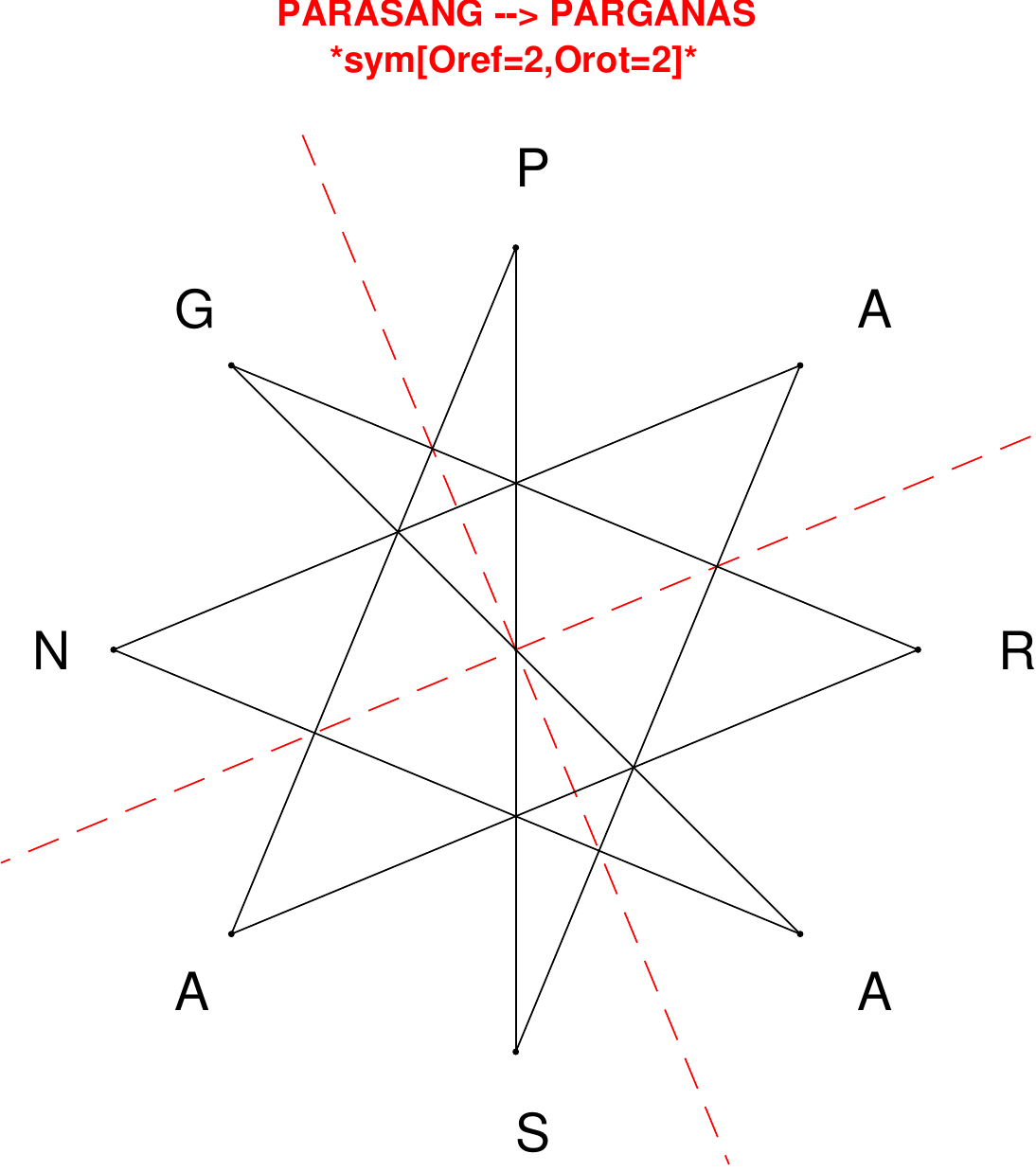}
\end{subfigure}
\hfill
\begin{subfigure}[T]{0.19\textwidth}
\centering
\includegraphics[width=\textwidth]{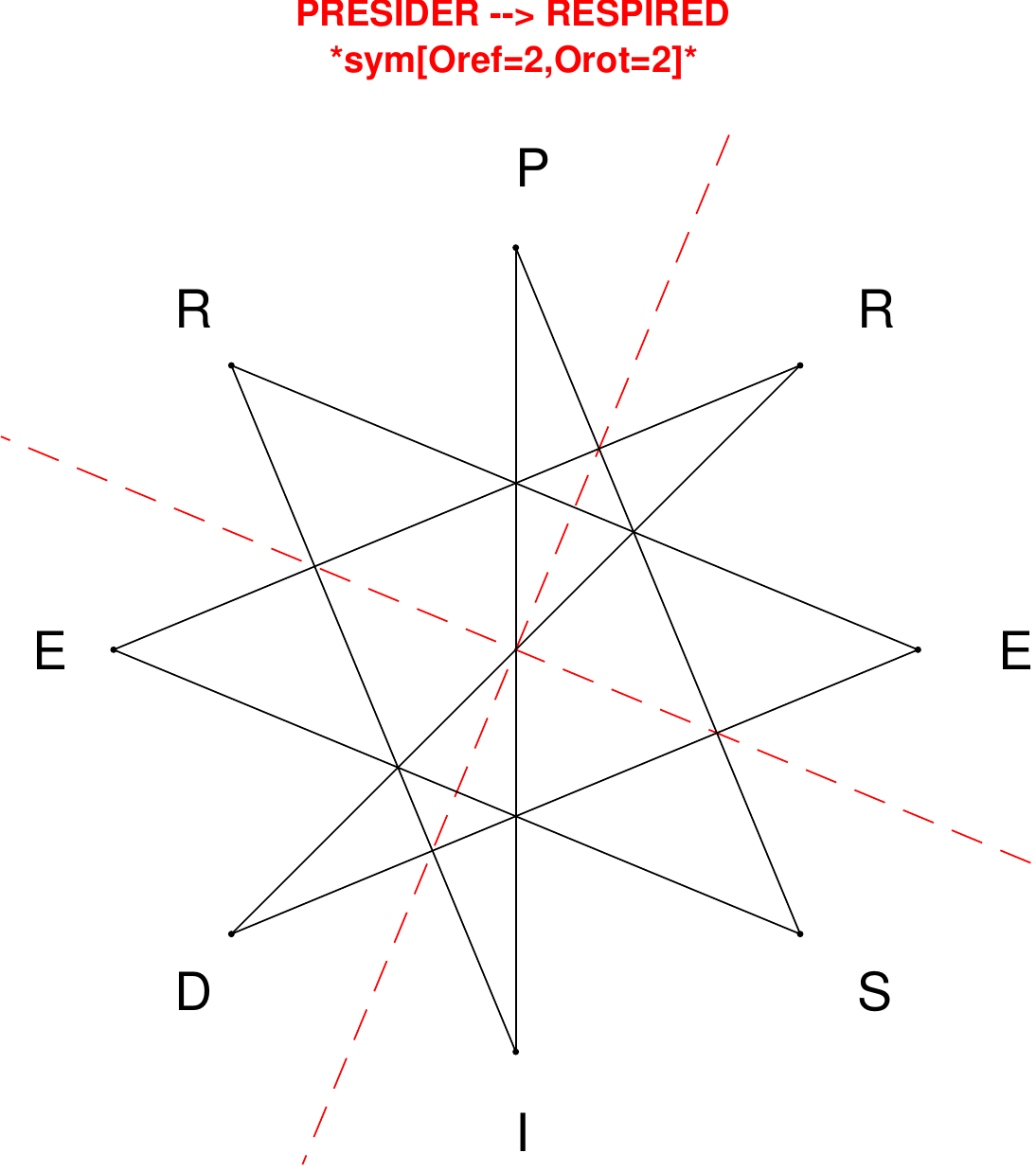}
\end{subfigure}
\hfill
\begin{subfigure}[T]{0.19\textwidth}
\centering
\includegraphics[width=\textwidth]{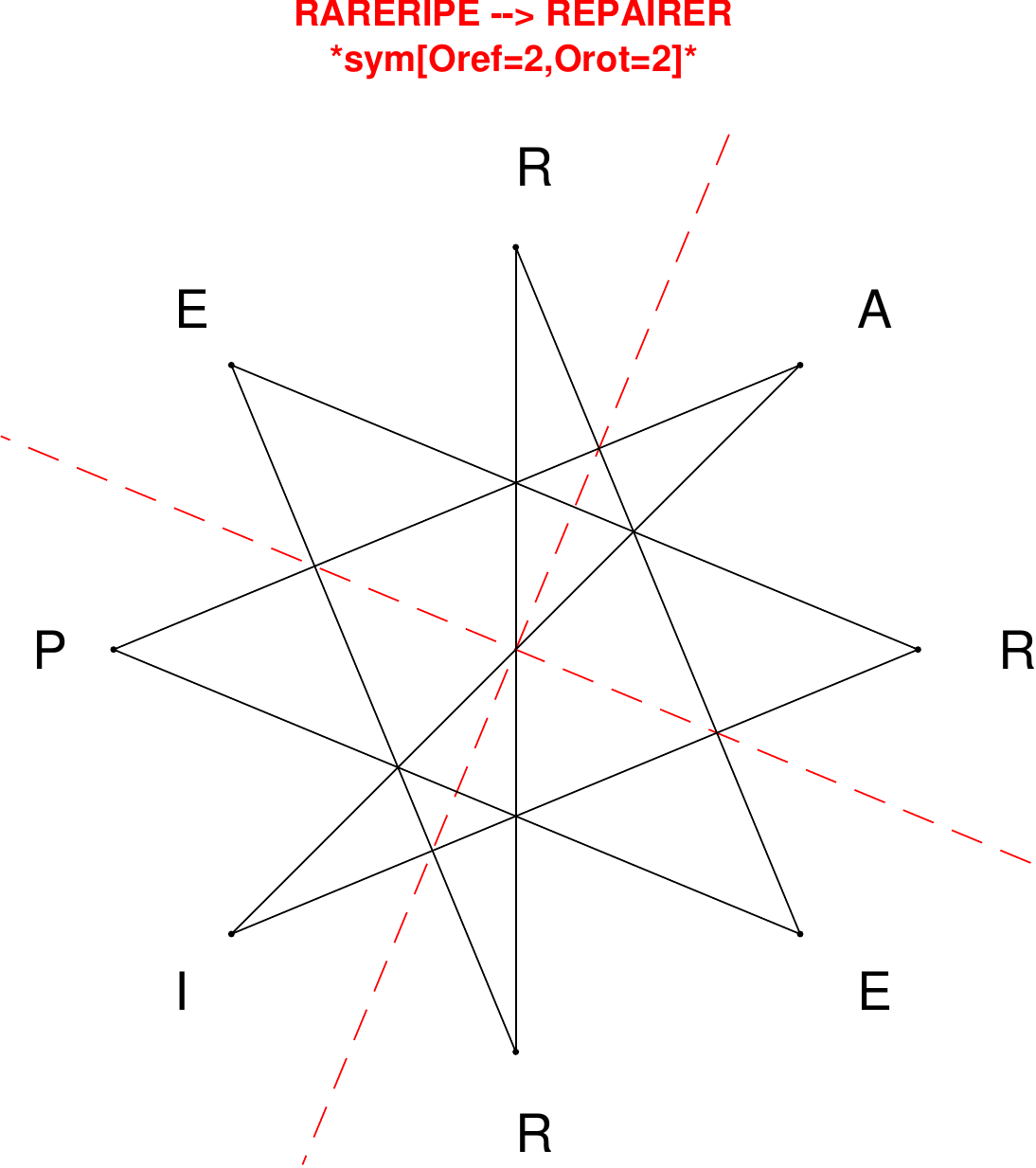}
\end{subfigure}
\hfill
\begin{subfigure}[T]{0.19\textwidth}
\centering
\includegraphics[width=\textwidth]{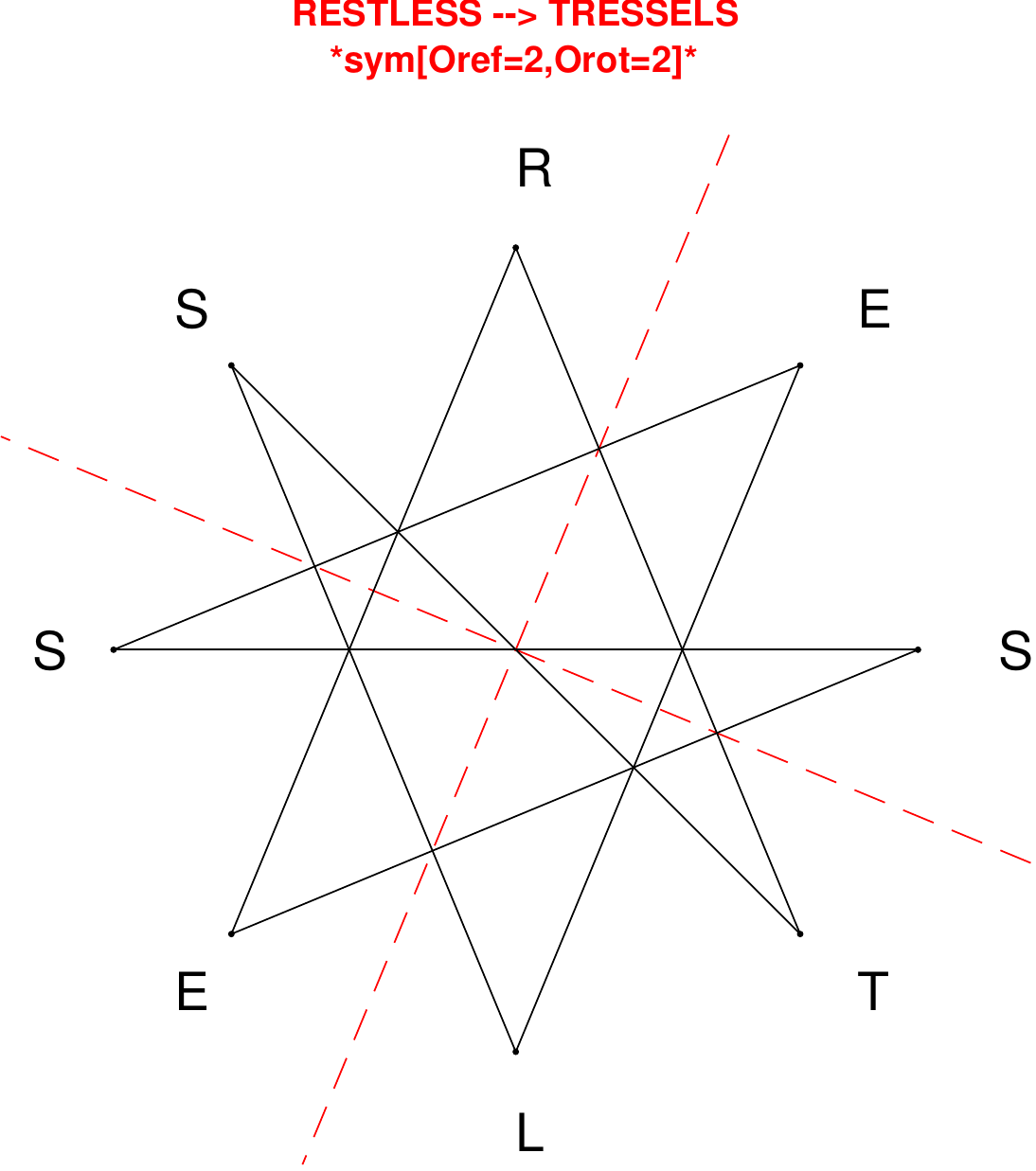}
\end{subfigure}
\end{figure}

\begin{figure}[H]
\centering
\begin{subfigure}[T]{0.19\textwidth}
\centering
\includegraphics[width=\textwidth]{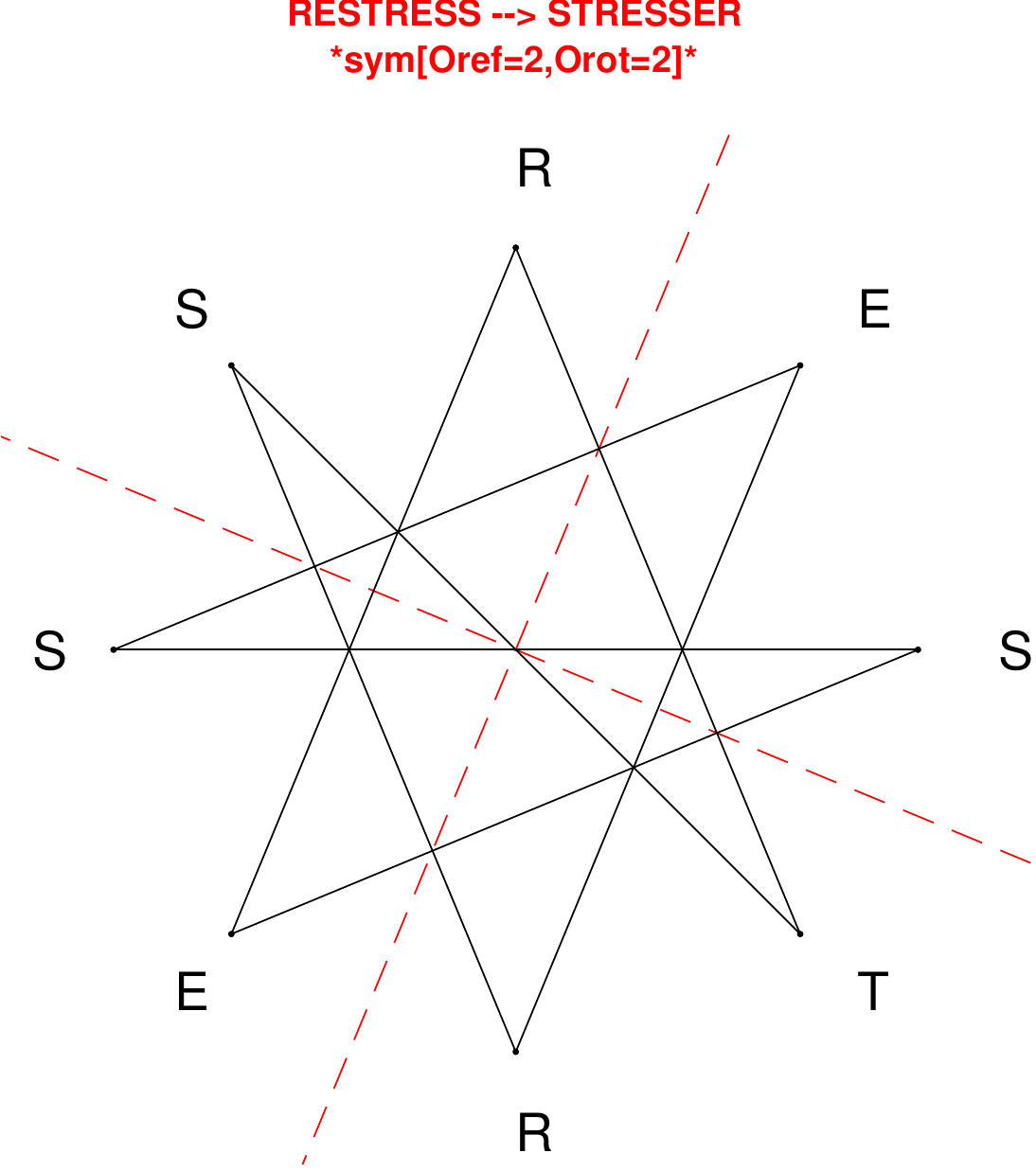}
\end{subfigure}
\hfill
\end{figure}

\subsubsection{Asymmetric Stars $N=8$}

\begin{figure}[H]
\centering
\begin{subfigure}[T]{0.19\textwidth}
\centering
\includegraphics[width=\textwidth]{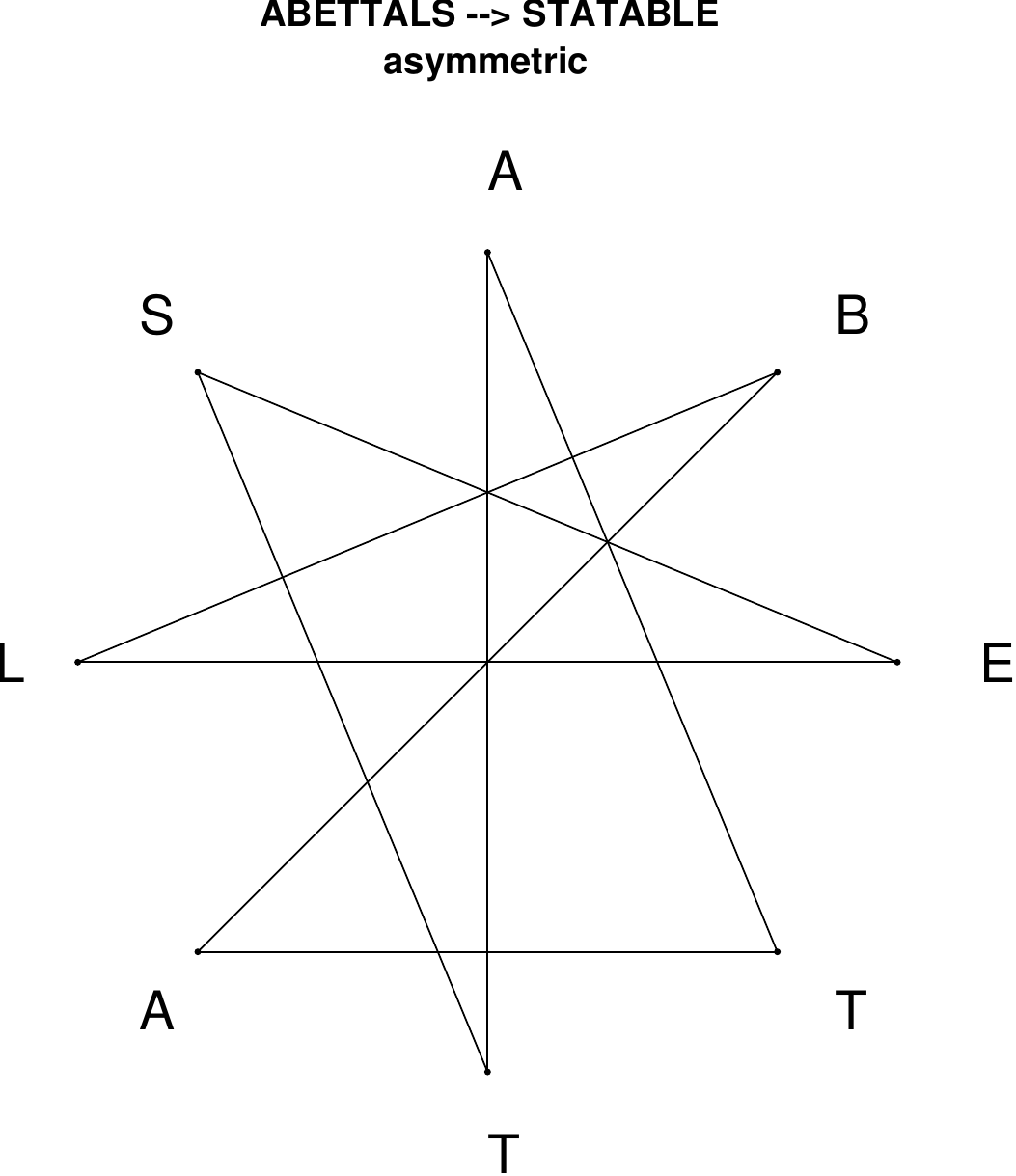}
\end{subfigure}
\hfill
\begin{subfigure}[T]{0.19\textwidth}
\centering
\includegraphics[width=\textwidth]{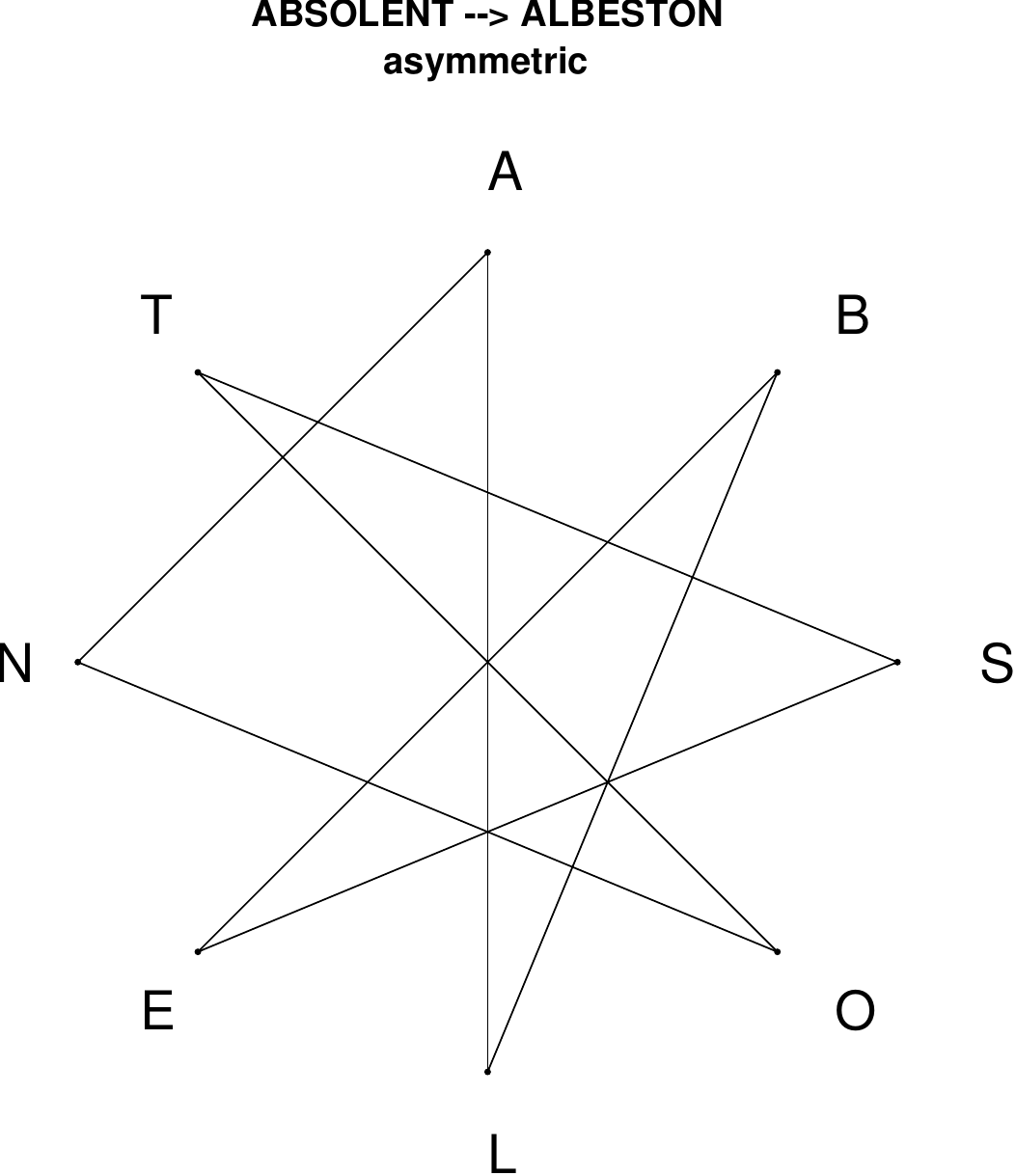}
\end{subfigure}
\hfill
\begin{subfigure}[T]{0.19\textwidth}
\centering
\includegraphics[width=\textwidth]{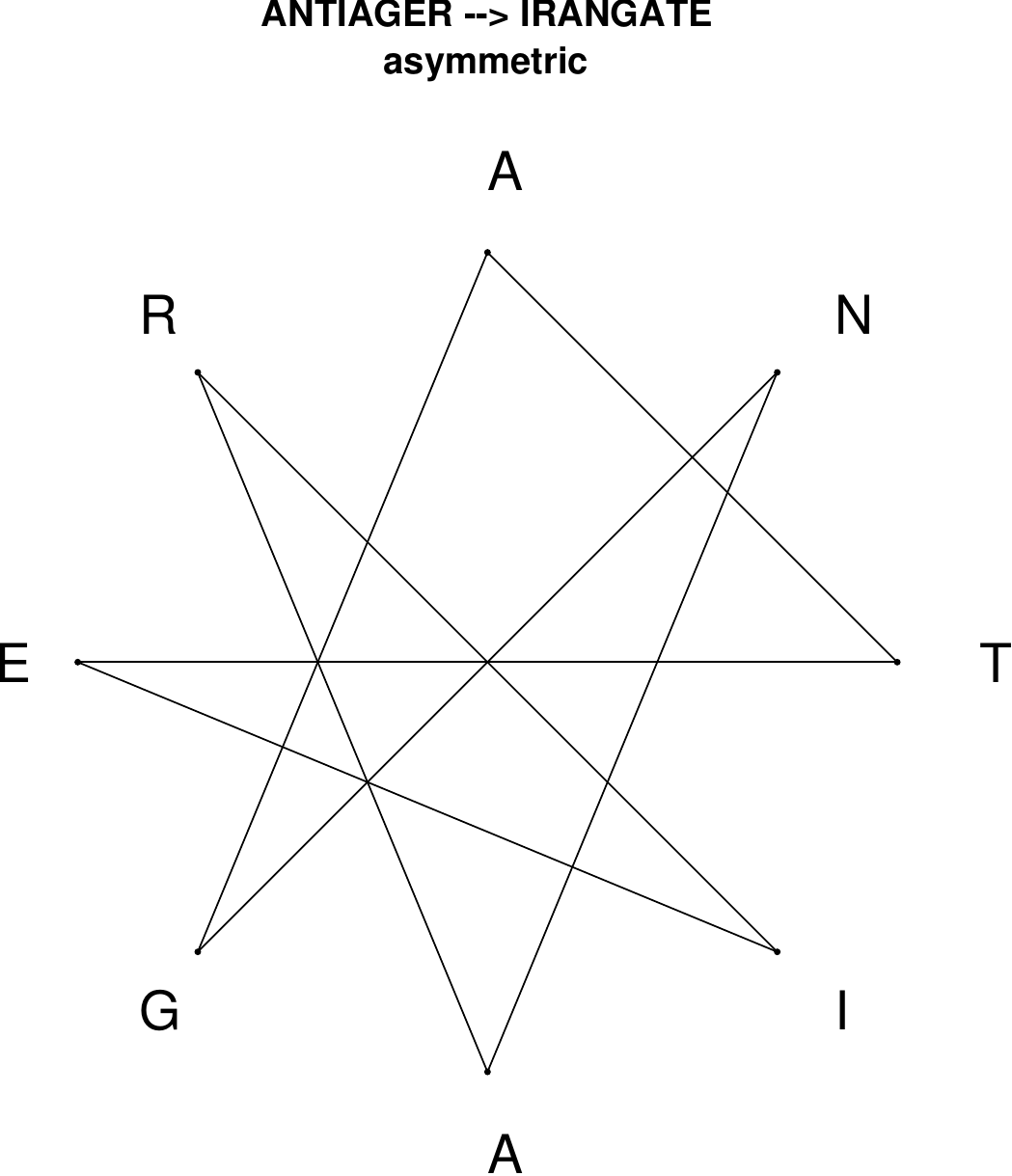}
\end{subfigure}
\hfill
\begin{subfigure}[T]{0.19\textwidth}
\centering
\includegraphics[width=\textwidth]{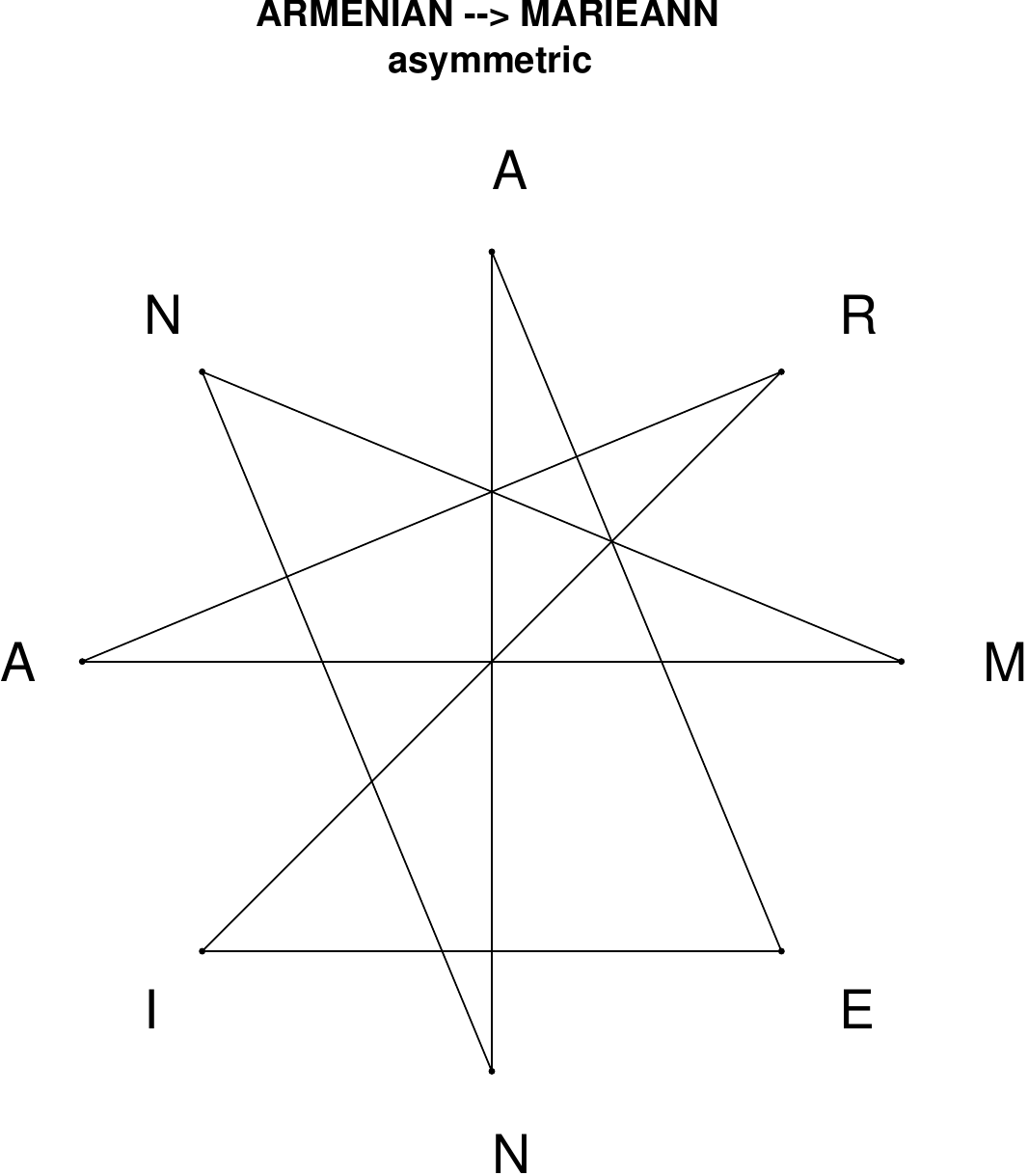}
\end{subfigure}
\hfill
\begin{subfigure}[T]{0.19\textwidth}
\centering
\includegraphics[width=\textwidth]{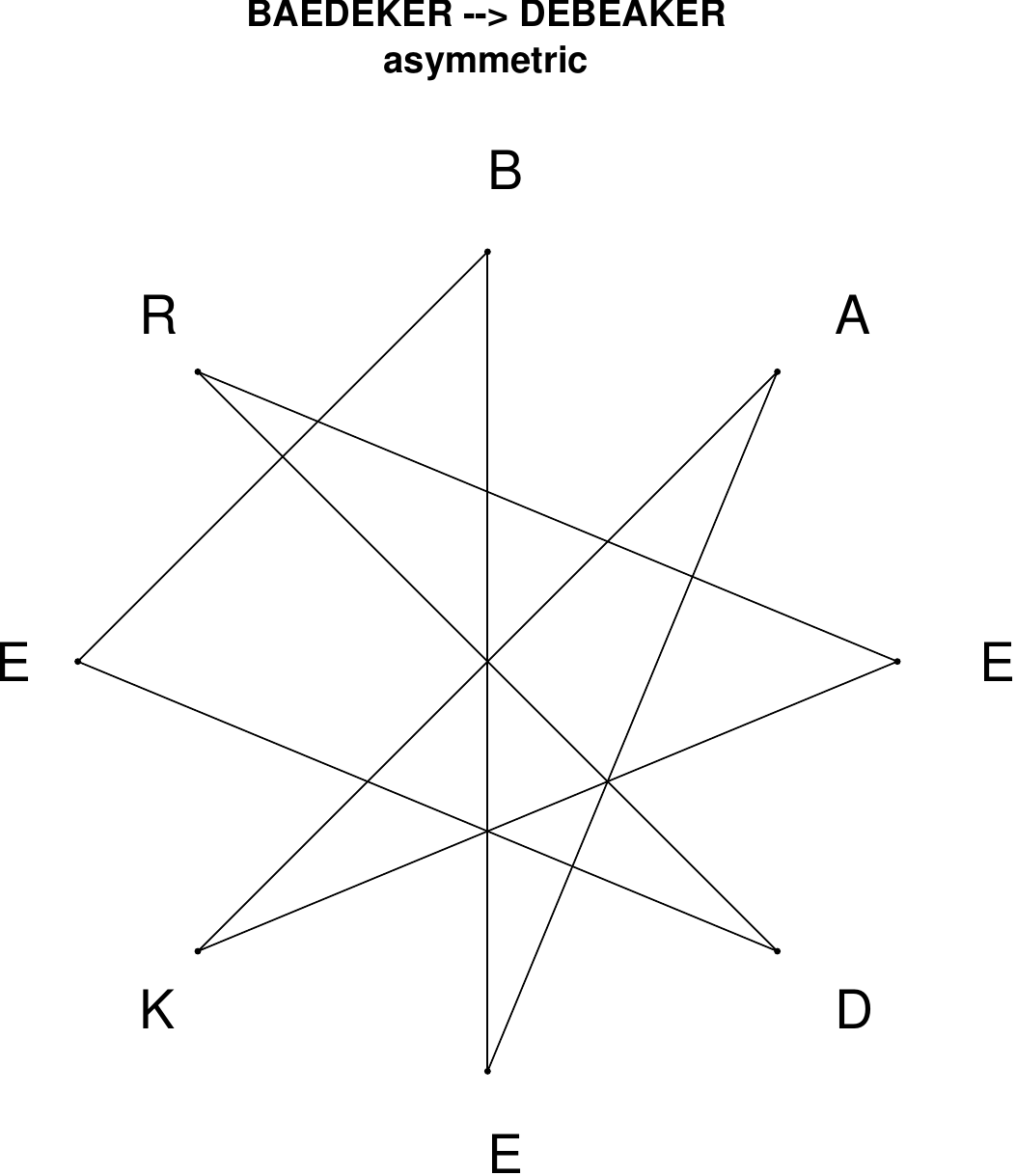}
\end{subfigure}
\end{figure}

\begin{figure}[H]
\centering
\begin{subfigure}[T]{0.19\textwidth}
\centering
\includegraphics[width=\textwidth]{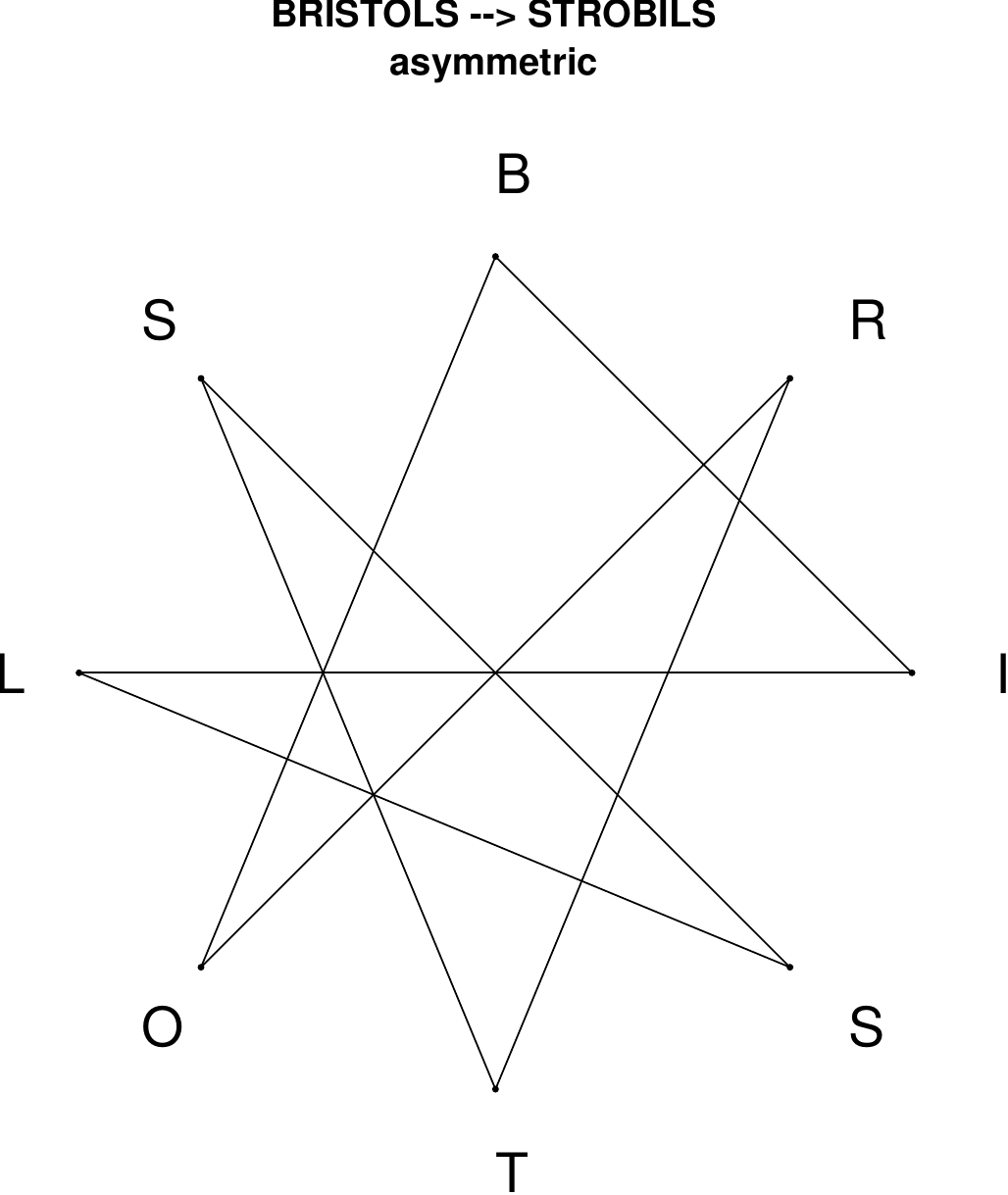}
\end{subfigure}
\hfill
\begin{subfigure}[T]{0.19\textwidth}
\centering
\includegraphics[width=\textwidth]{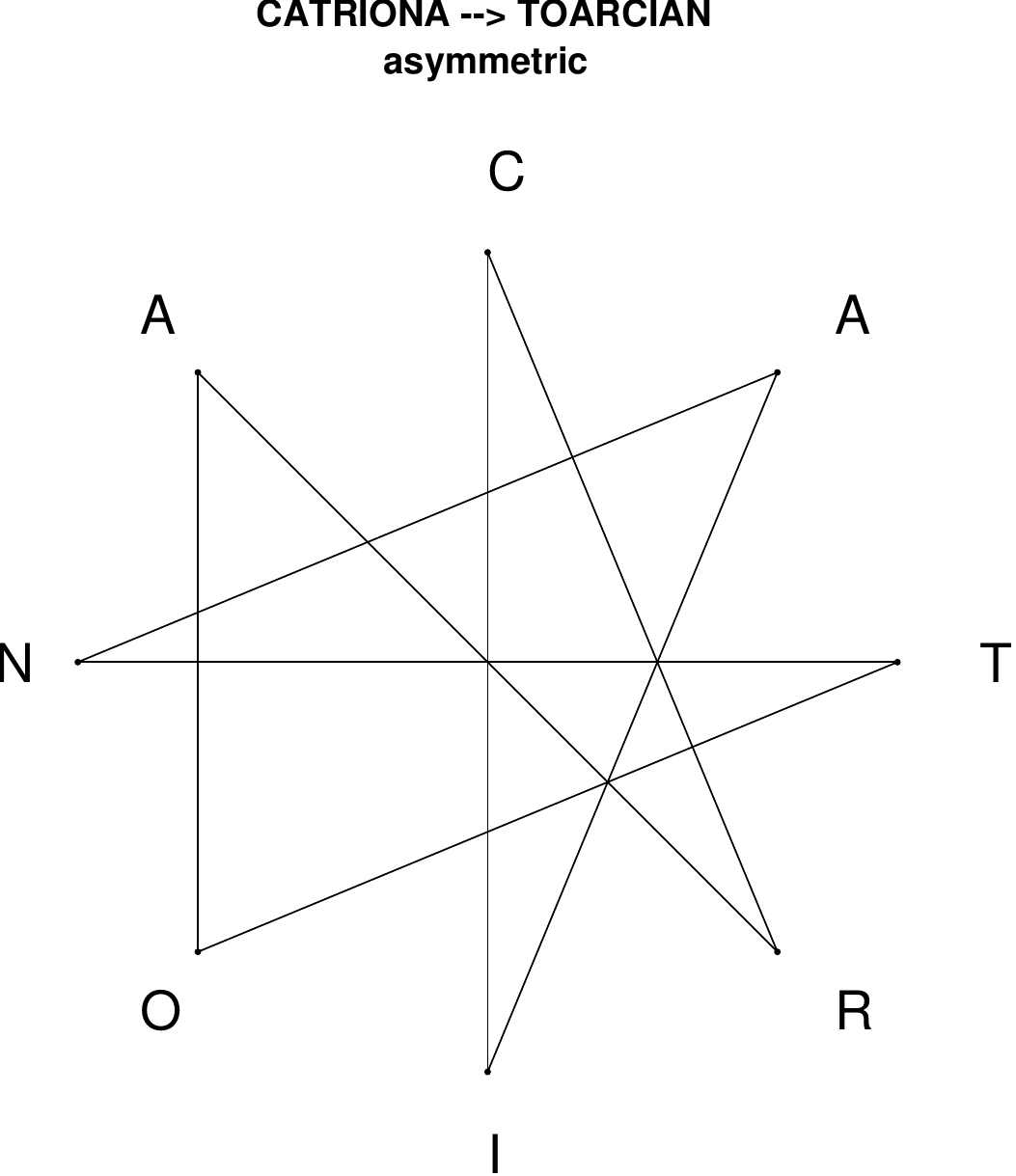}
\end{subfigure}
\hfill
\begin{subfigure}[T]{0.19\textwidth}
\centering
\includegraphics[width=\textwidth]{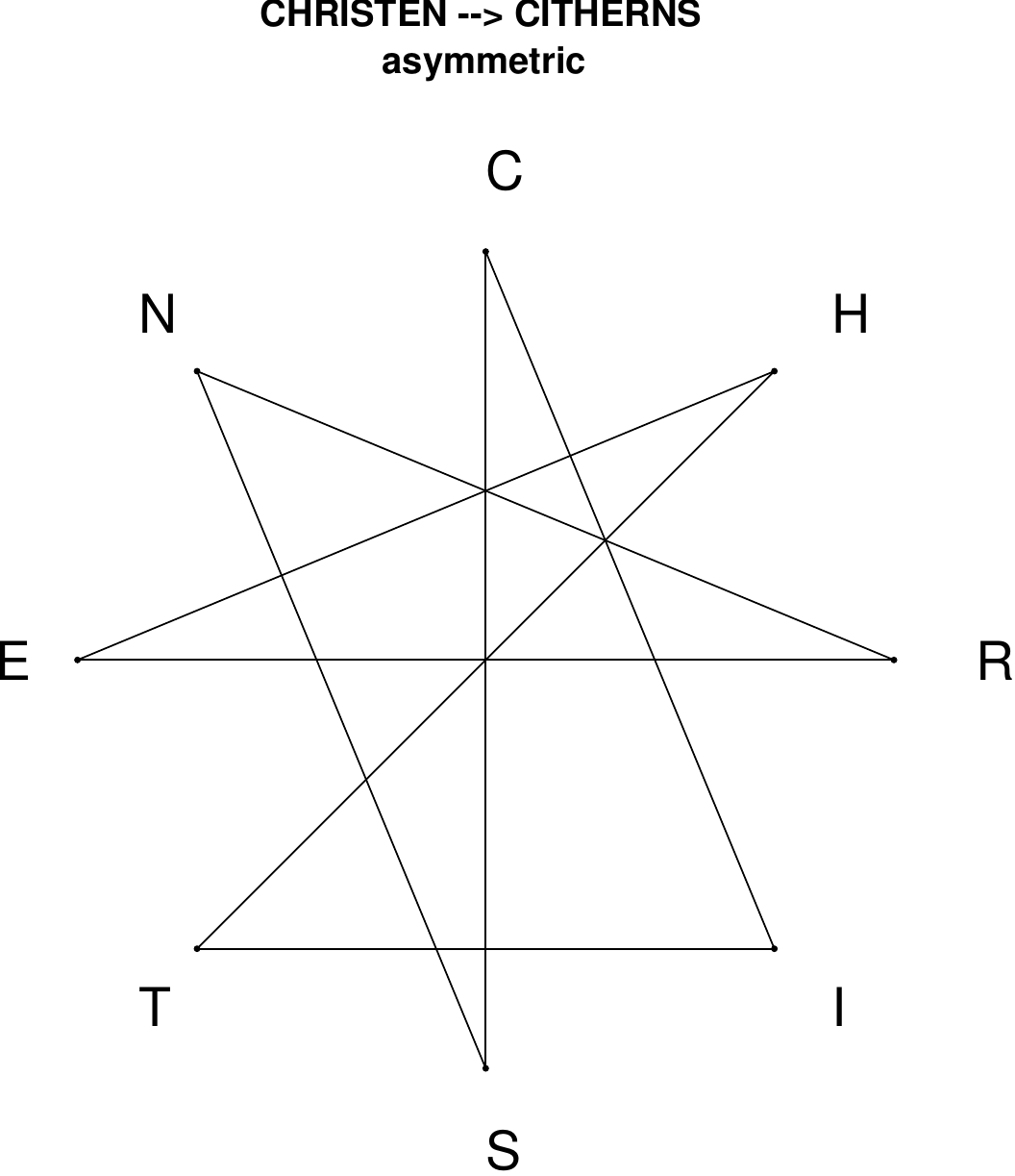}
\end{subfigure}
\hfill
\begin{subfigure}[T]{0.19\textwidth}
\centering
\includegraphics[width=\textwidth]{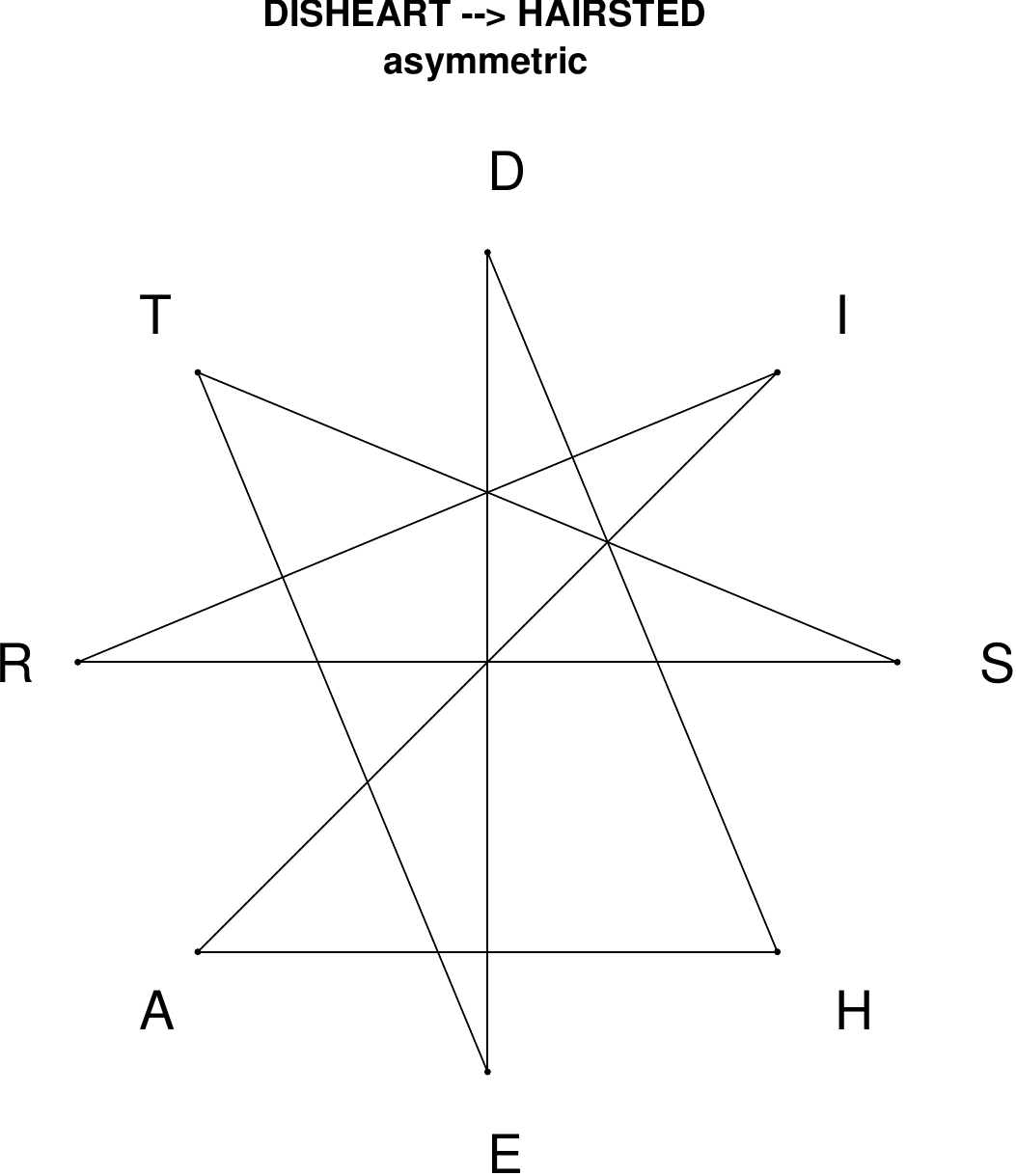}
\end{subfigure}
\hfill
\begin{subfigure}[T]{0.19\textwidth}
\centering
\includegraphics[width=\textwidth]{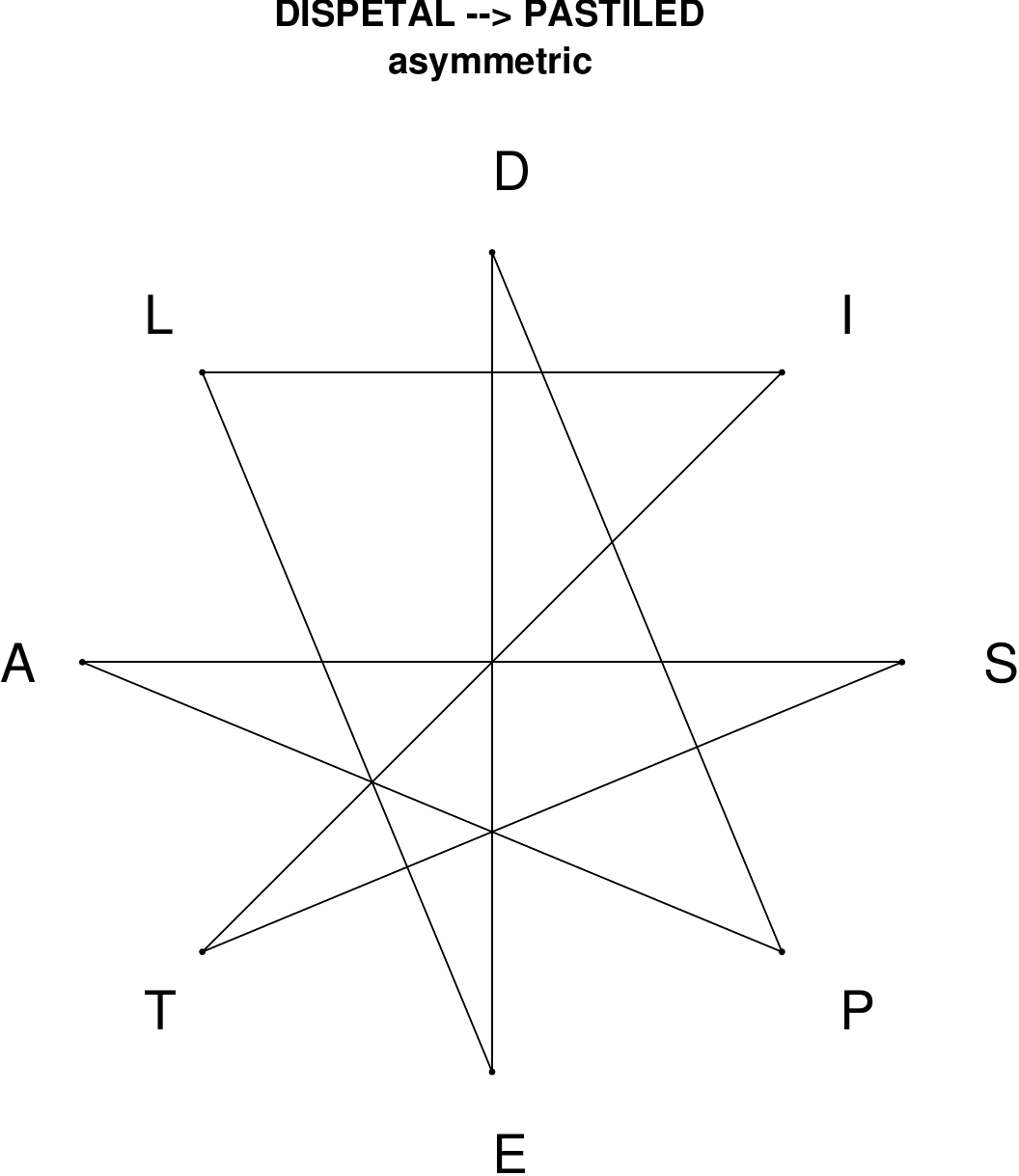}
\end{subfigure}
\end{figure}

\begin{figure}[H]
\centering
\begin{subfigure}[T]{0.19\textwidth}
\centering
\includegraphics[width=\textwidth]{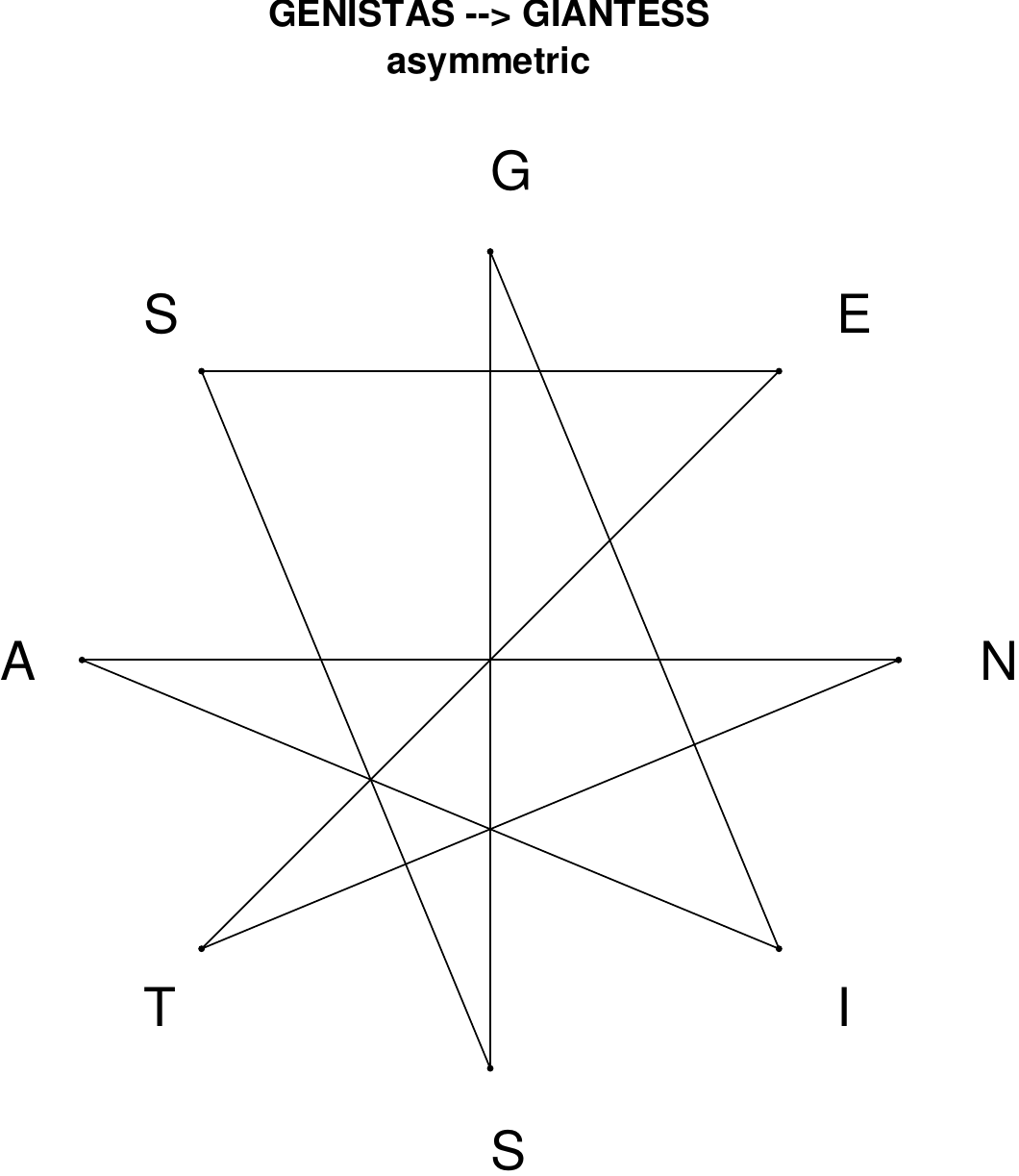}
\end{subfigure}
\hfill
\begin{subfigure}[T]{0.19\textwidth}
\centering
\includegraphics[width=\textwidth]{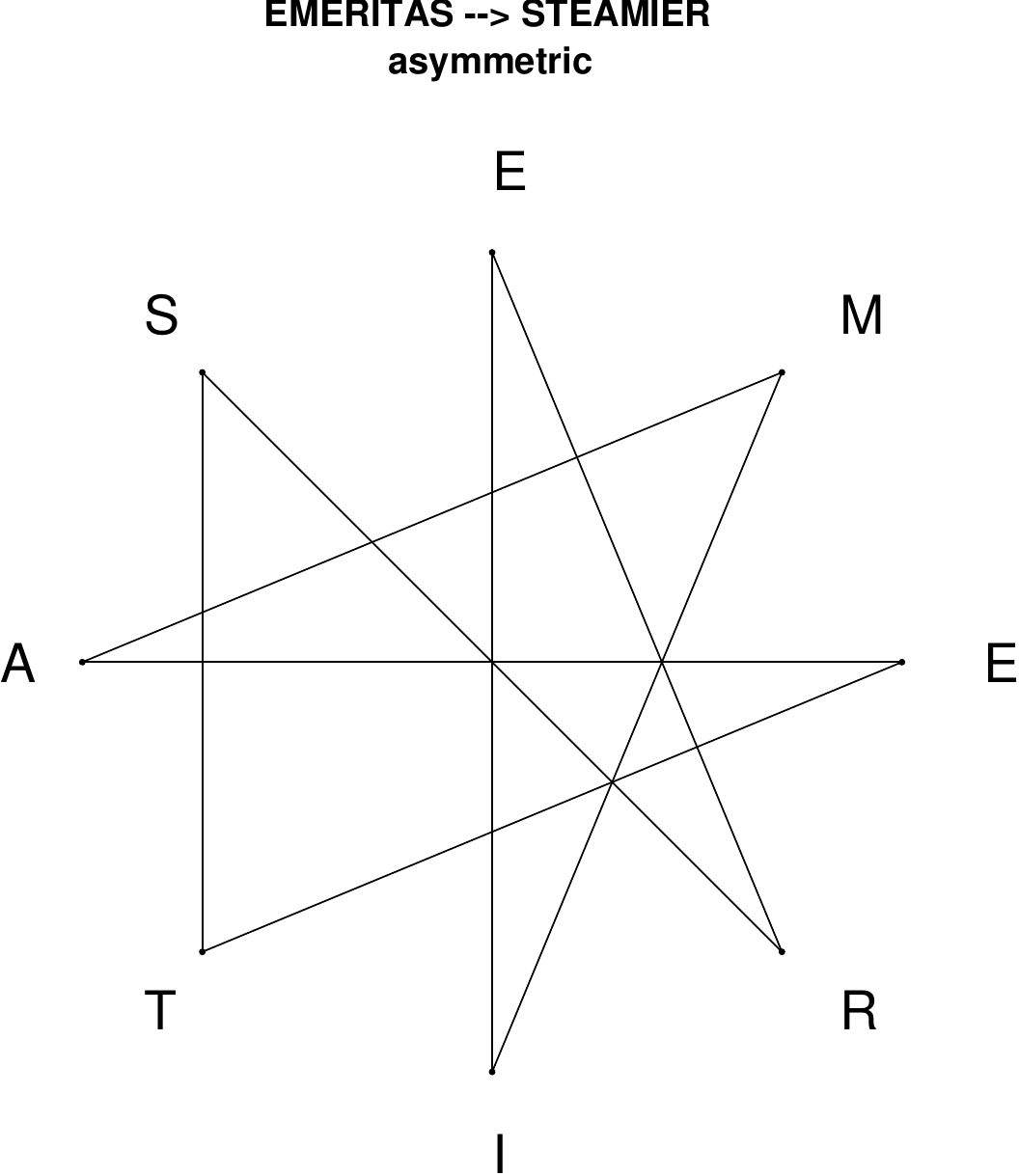}
\end{subfigure}
\hfill
\begin{subfigure}[T]{0.19\textwidth}
\centering
\includegraphics[width=\textwidth]{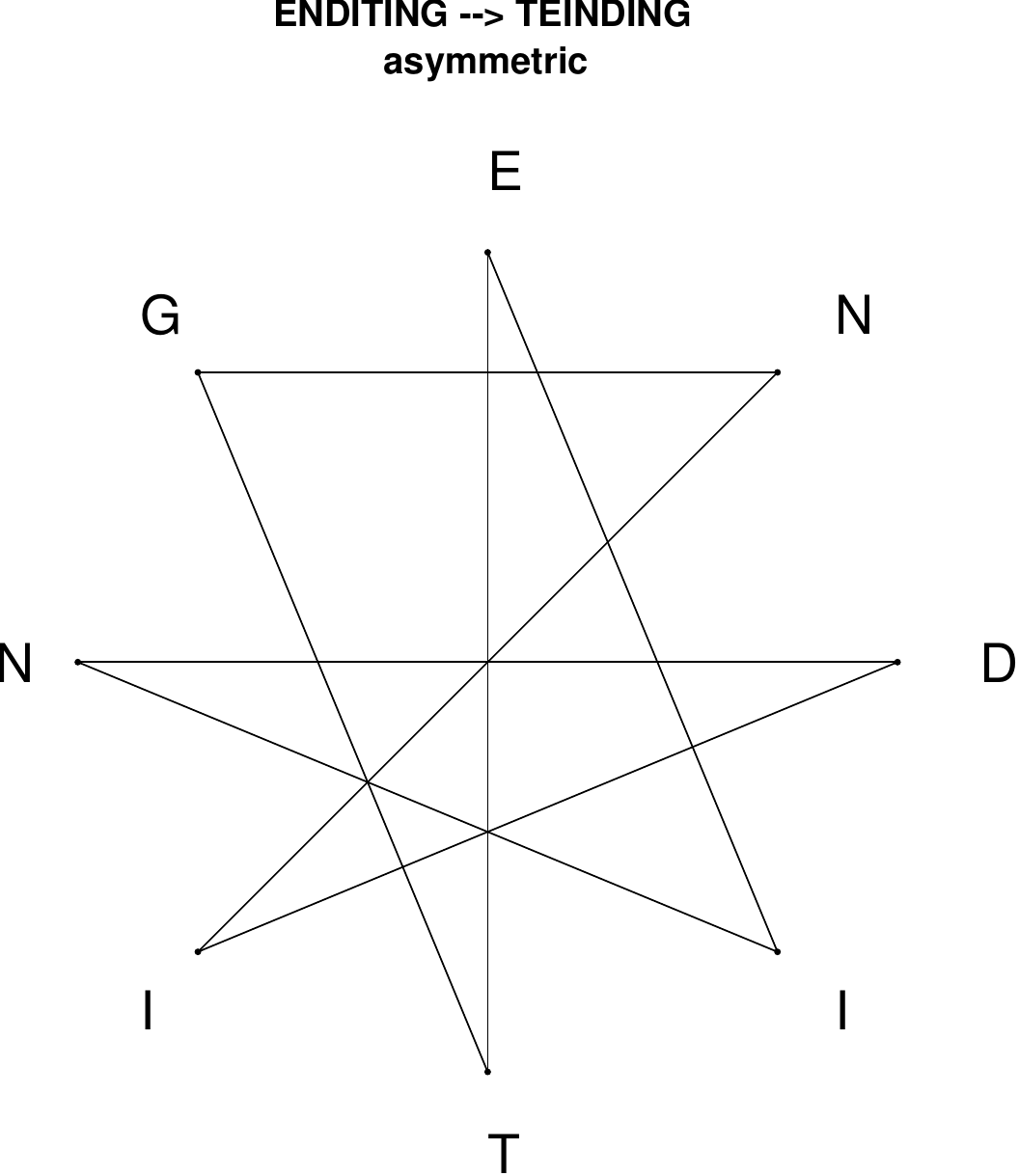}
\end{subfigure}
\hfill
\begin{subfigure}[T]{0.19\textwidth}
\centering
\includegraphics[width=\textwidth]{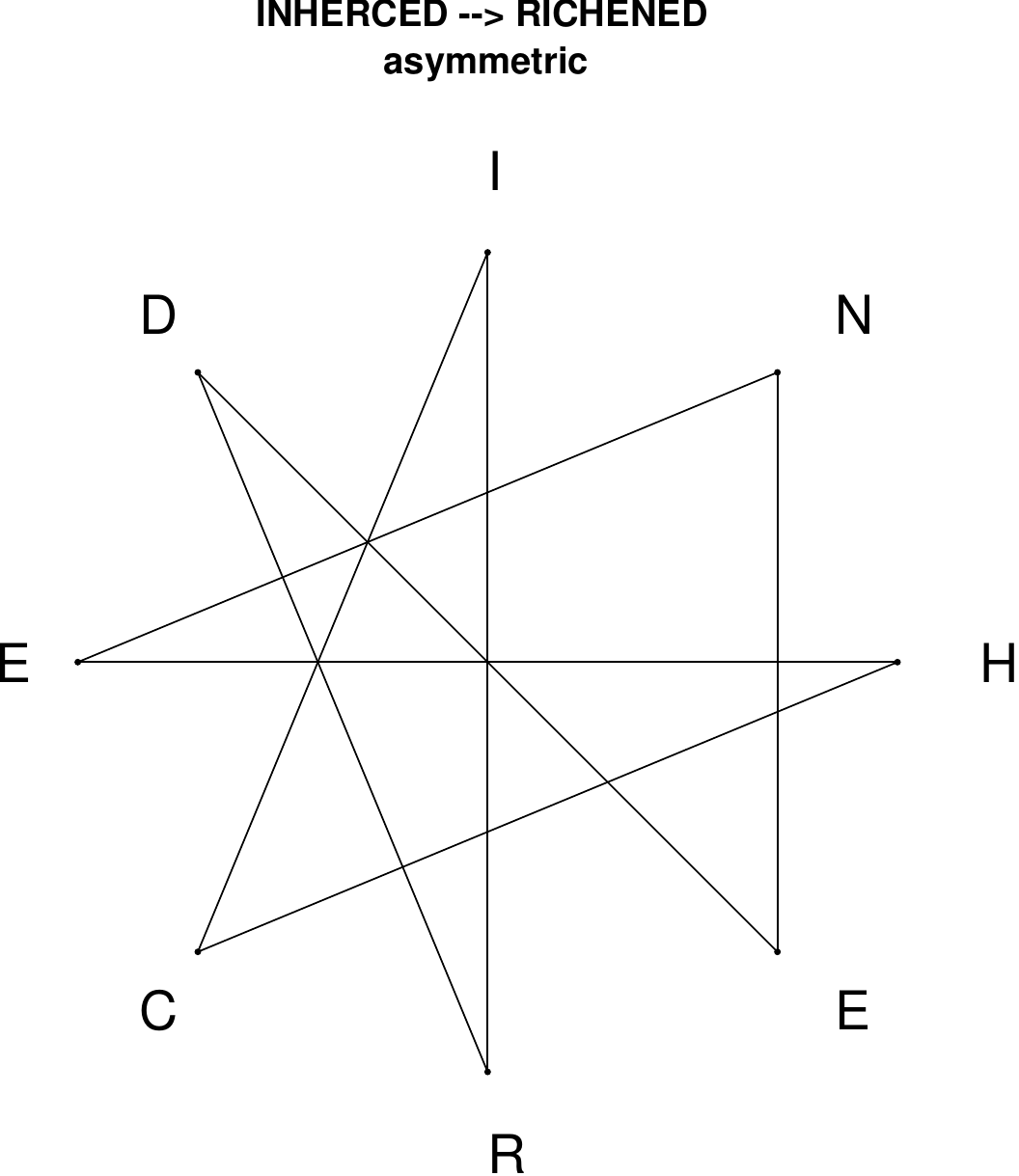}
\end{subfigure}
\hfill
\begin{subfigure}[T]{0.19\textwidth}
\centering
\includegraphics[width=\textwidth]{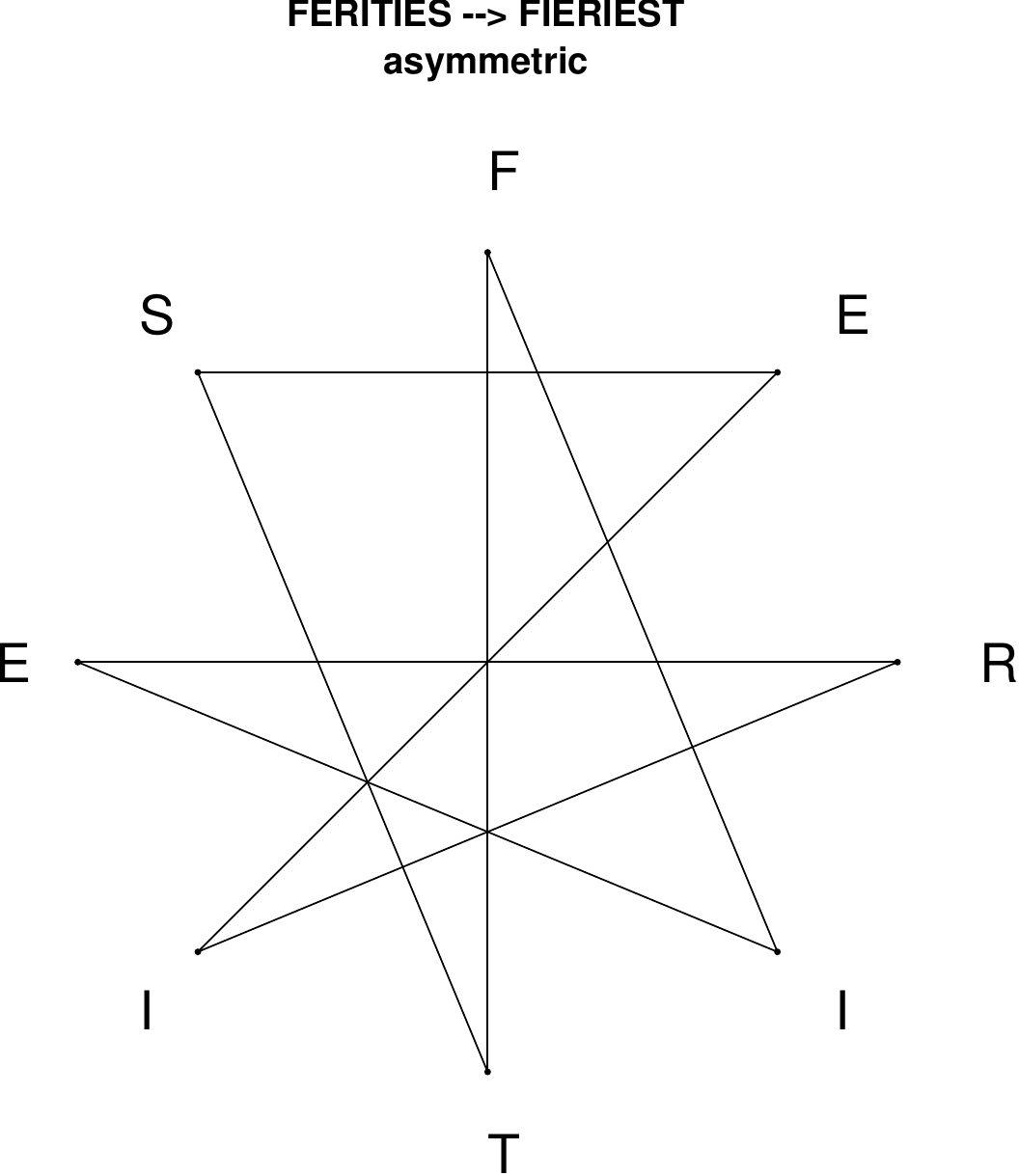}
\end{subfigure}
\end{figure}

\begin{figure}[H]
\centering
\begin{subfigure}[T]{0.19\textwidth}
\centering
\includegraphics[width=\textwidth]{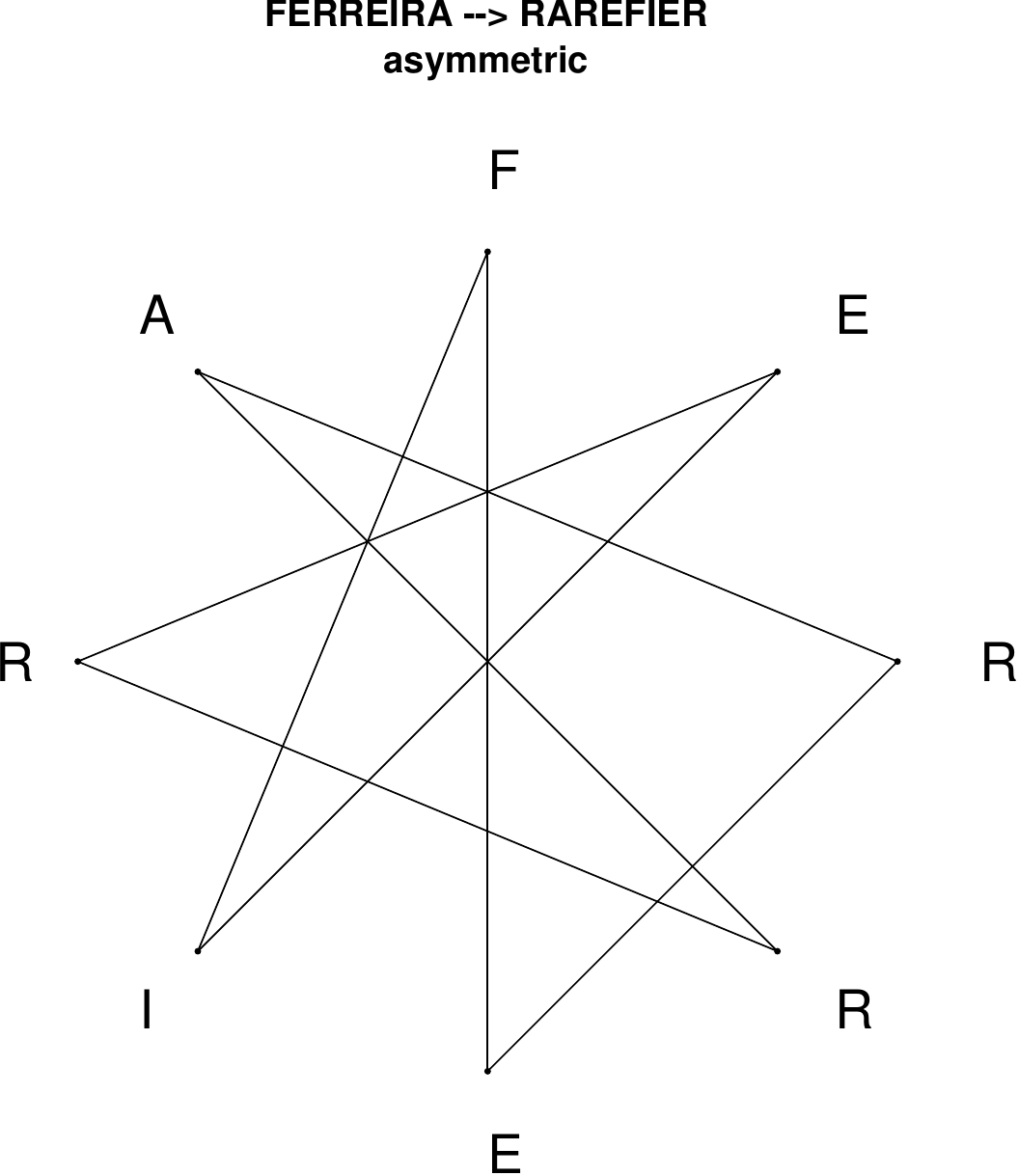}
\end{subfigure}
\hfill
\begin{subfigure}[T]{0.19\textwidth}
\centering
\includegraphics[width=\textwidth]{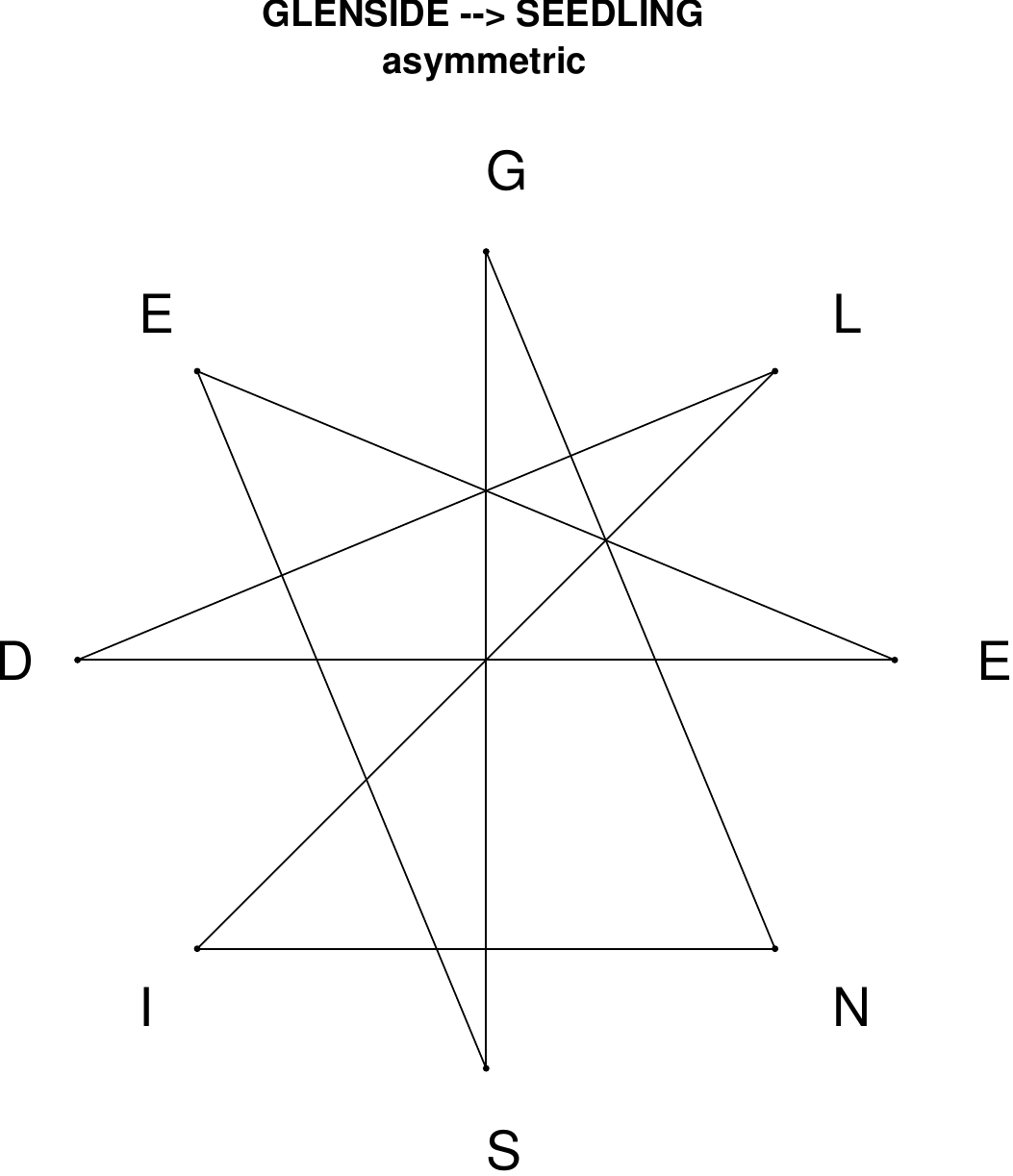}
\end{subfigure}
\hfill
\begin{subfigure}[T]{0.19\textwidth}
\centering
\includegraphics[width=\textwidth]{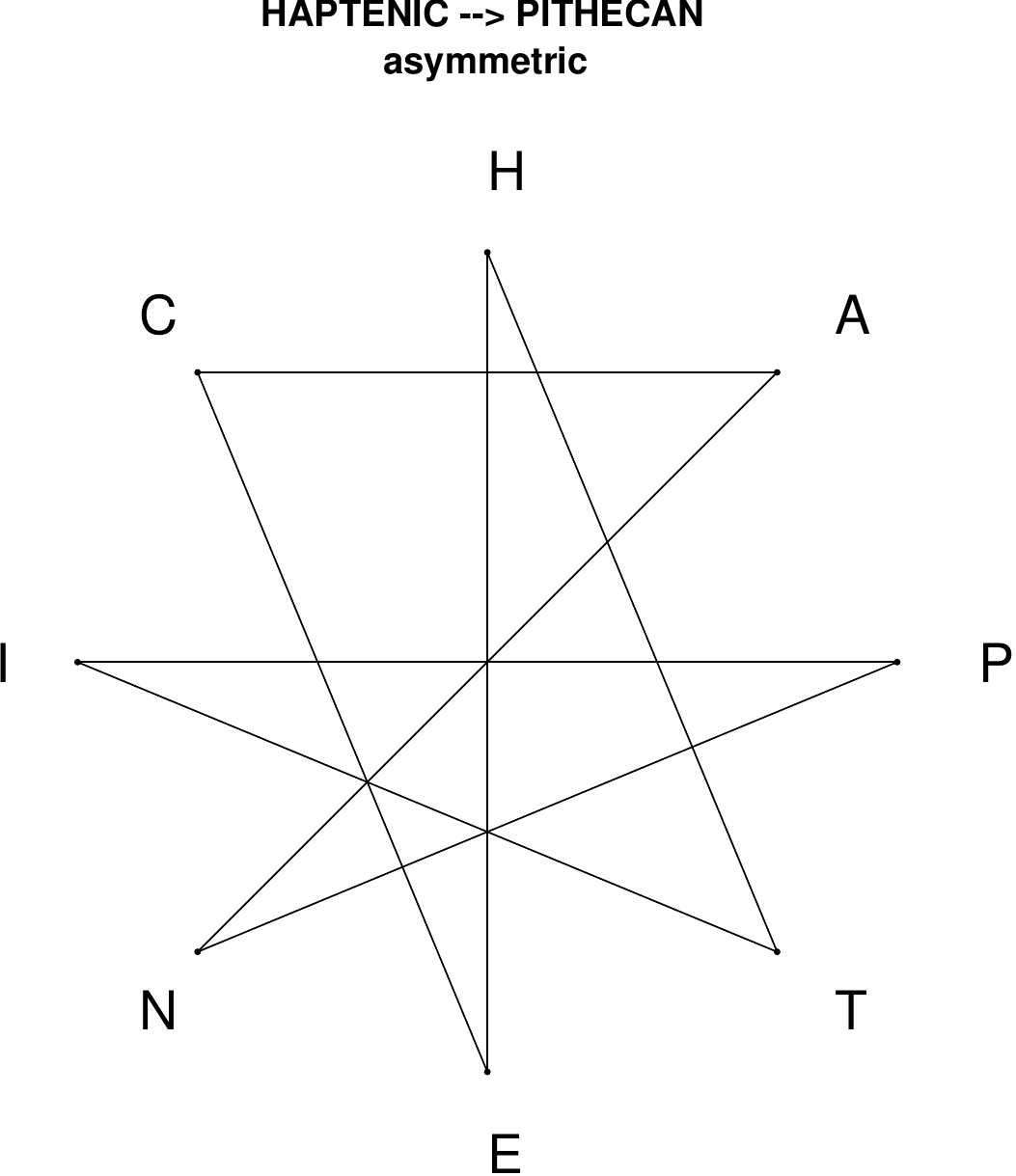}
\end{subfigure}
\hfill
\begin{subfigure}[T]{0.19\textwidth}
\centering
\includegraphics[width=\textwidth]{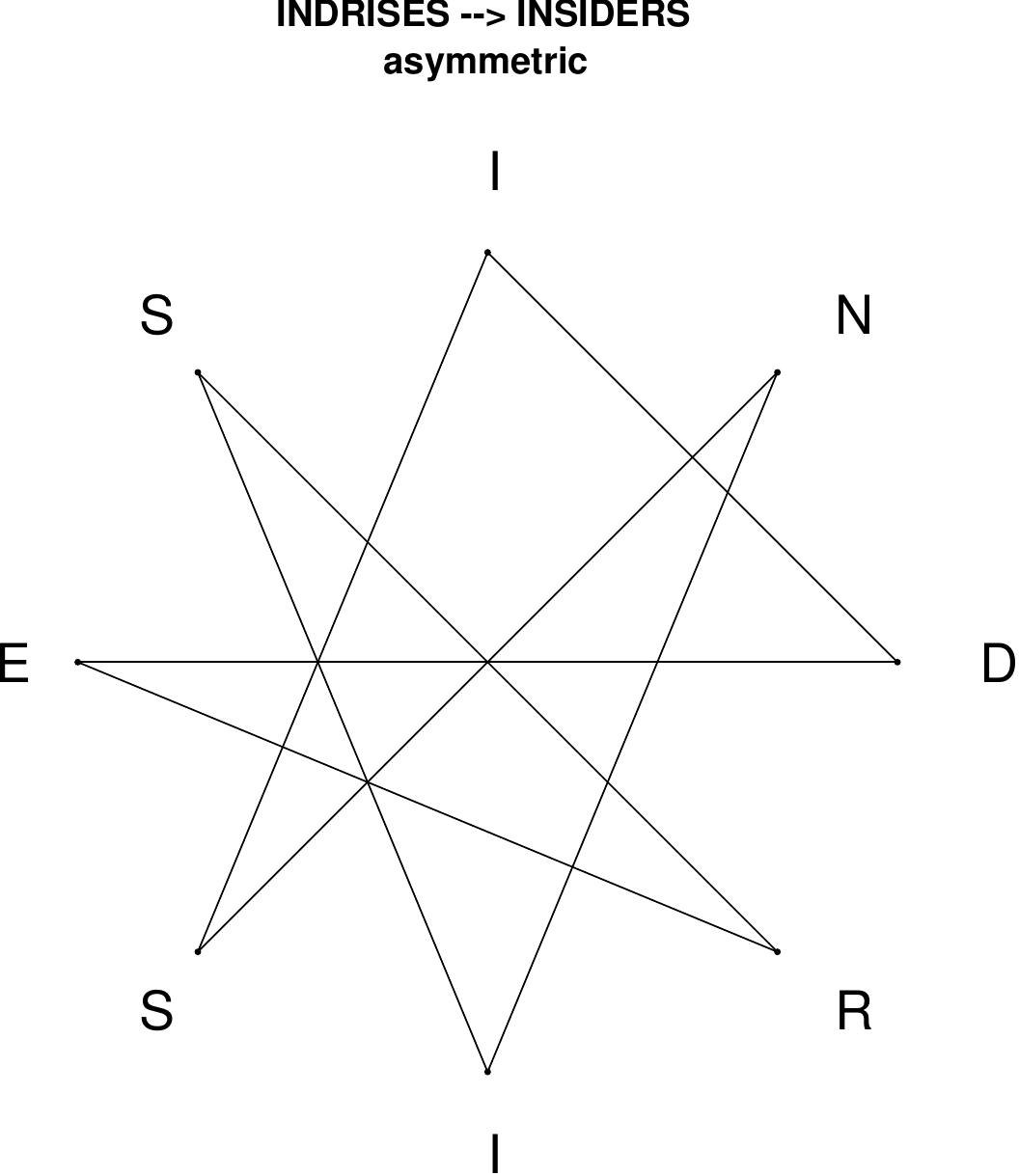}
\end{subfigure}
\hfill
\begin{subfigure}[T]{0.19\textwidth}
\centering
\includegraphics[width=\textwidth]{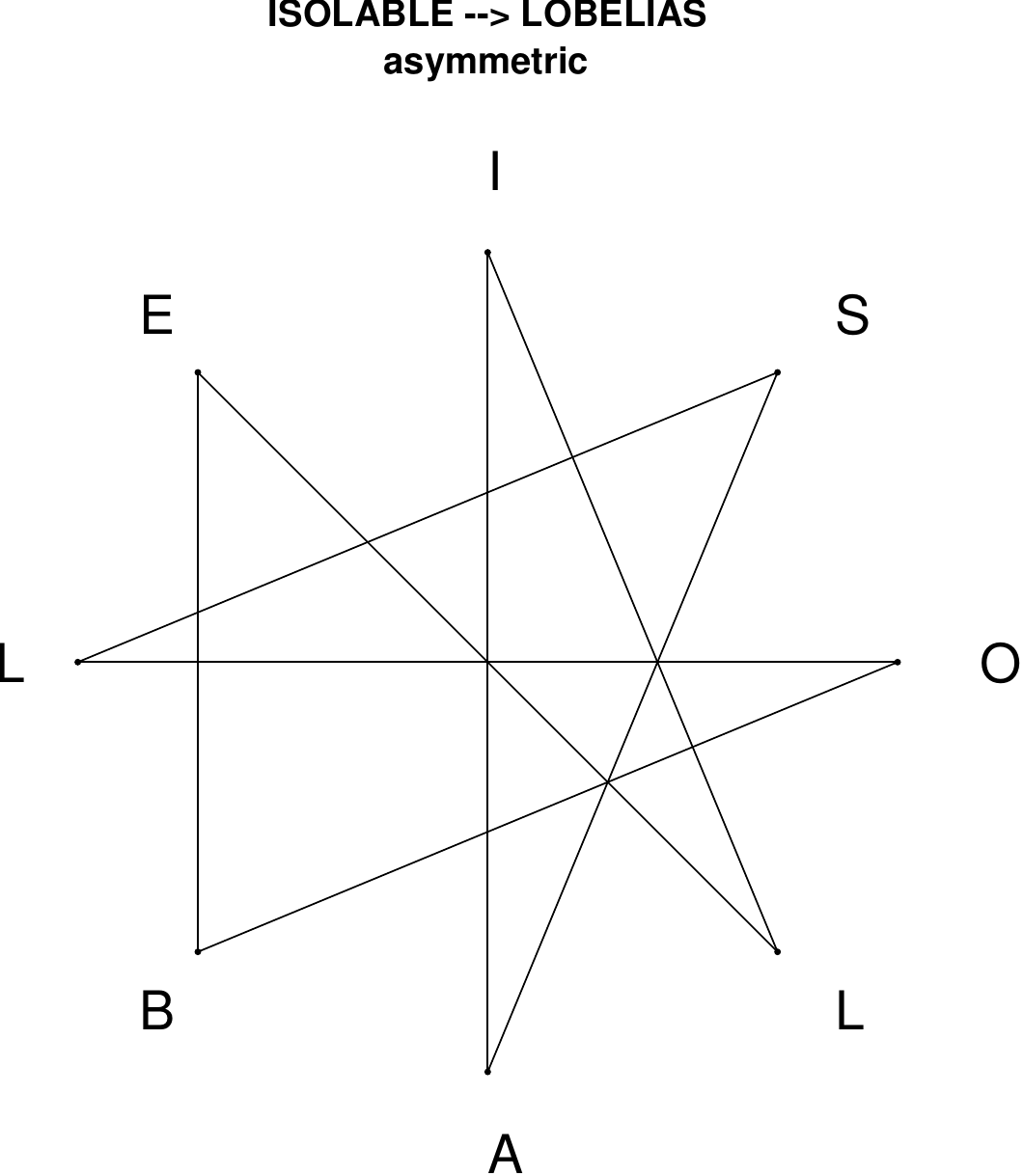}
\end{subfigure}
\end{figure}

\begin{figure}[H]
\centering
\begin{subfigure}[T]{0.19\textwidth}
\centering
\includegraphics[width=\textwidth]{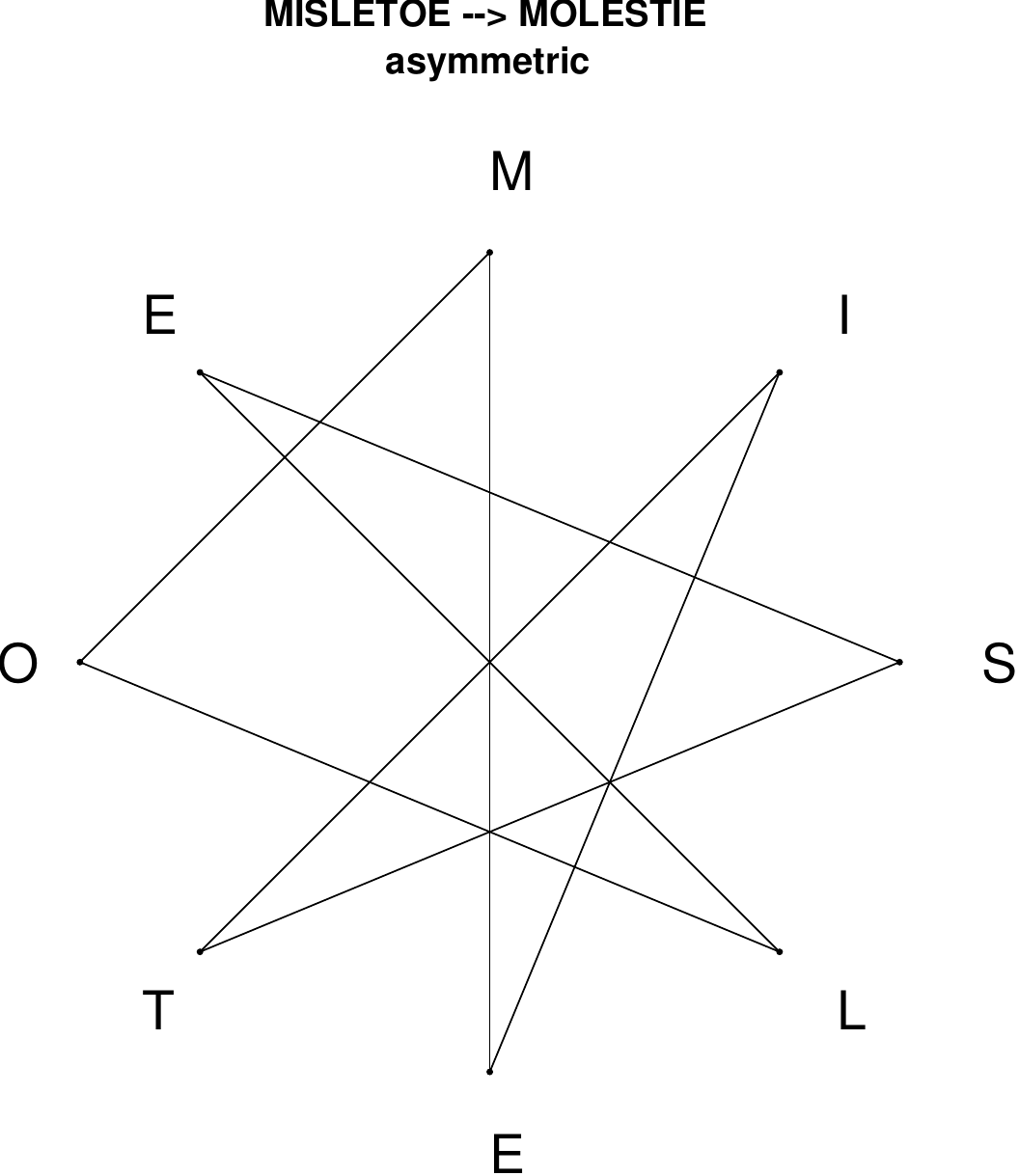}
\end{subfigure}
\hfill
\begin{subfigure}[T]{0.19\textwidth}
\centering
\includegraphics[width=\textwidth]{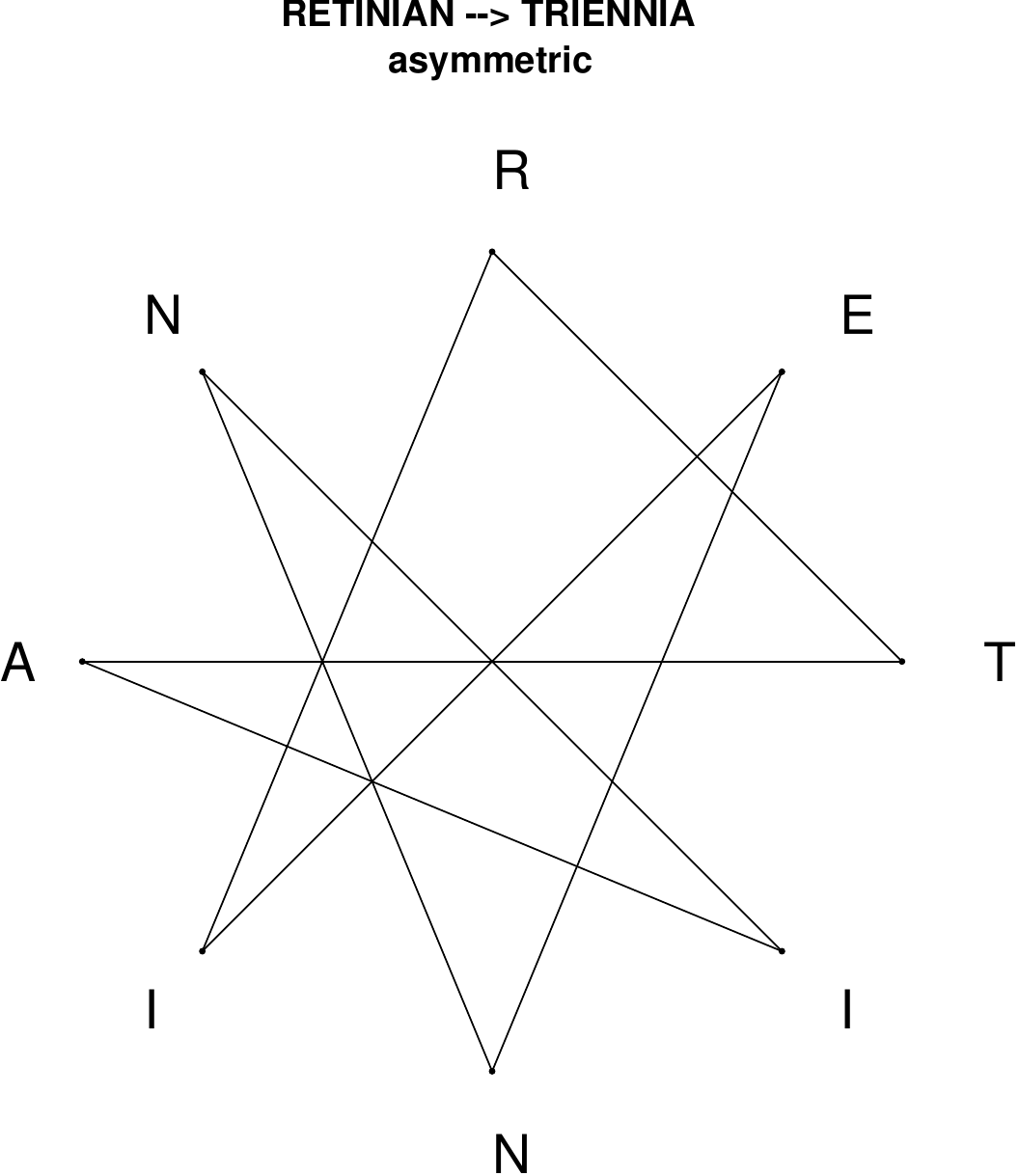}
\end{subfigure}
\hfill
\begin{subfigure}[T]{0.19\textwidth}
\centering
\includegraphics[width=\textwidth]{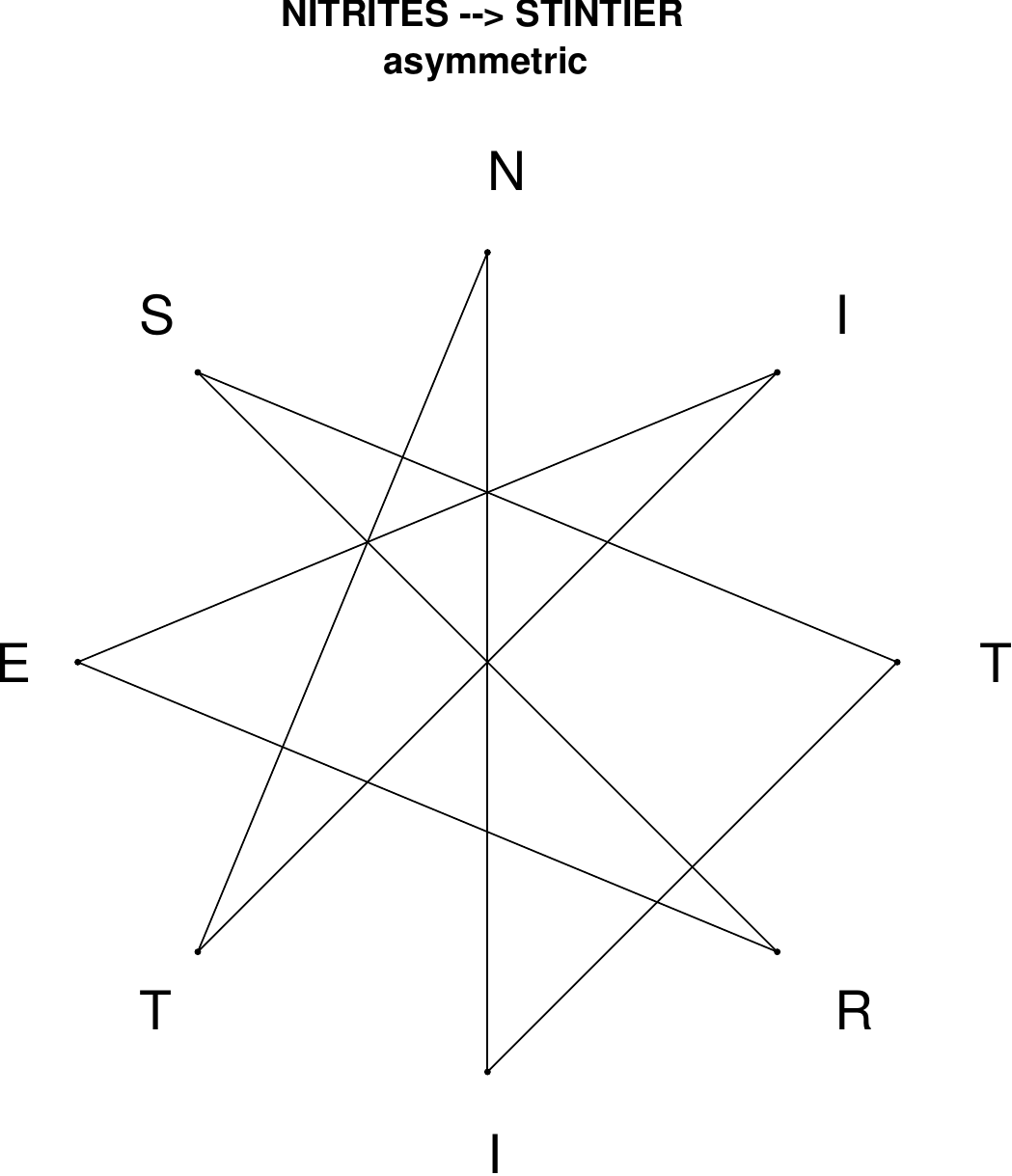}
\end{subfigure}
\hfill
\begin{subfigure}[T]{0.19\textwidth}
\centering
\includegraphics[width=\textwidth]{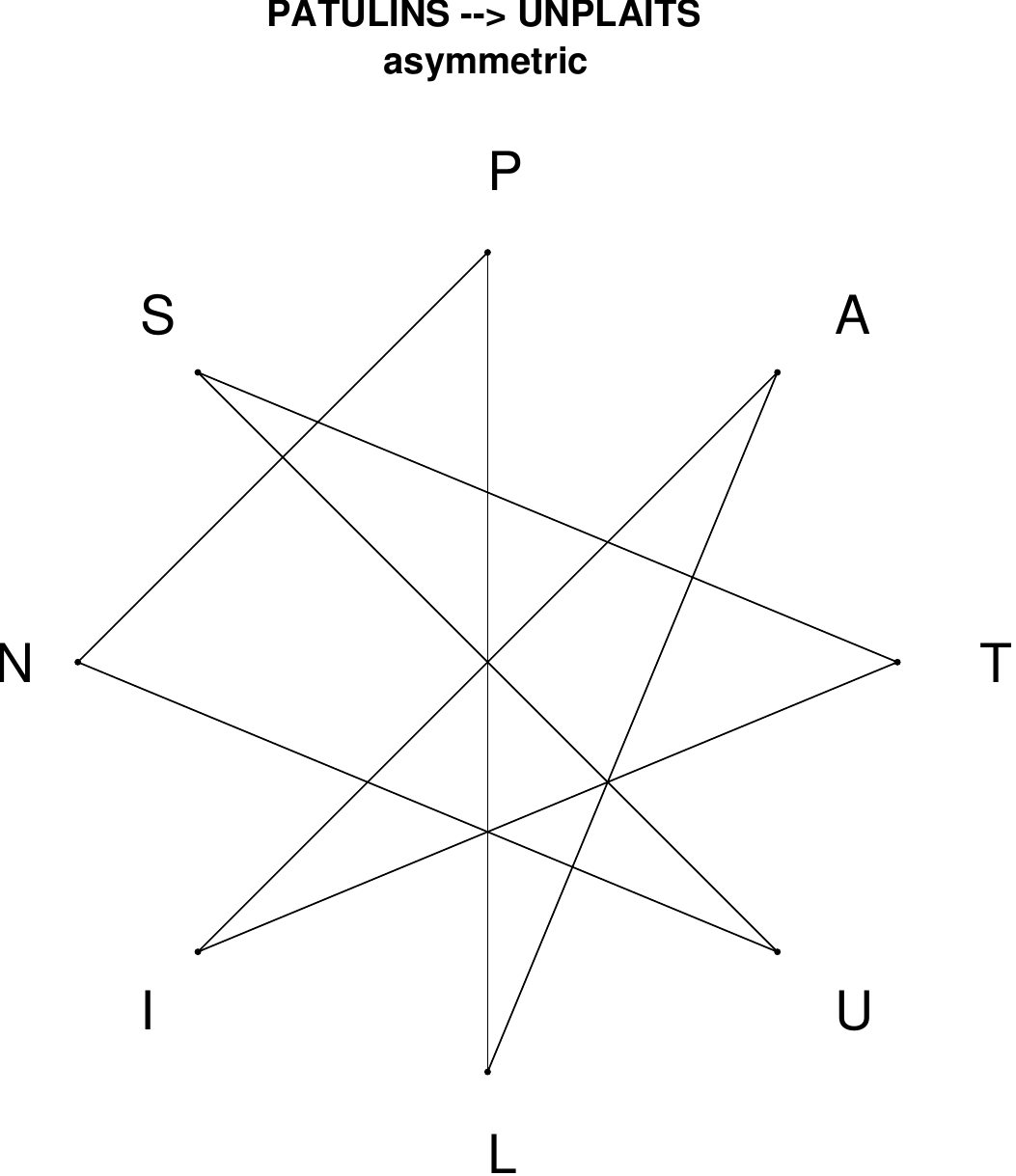}
\end{subfigure}
\hfill
\begin{subfigure}[T]{0.19\textwidth}
\centering
\includegraphics[width=\textwidth]{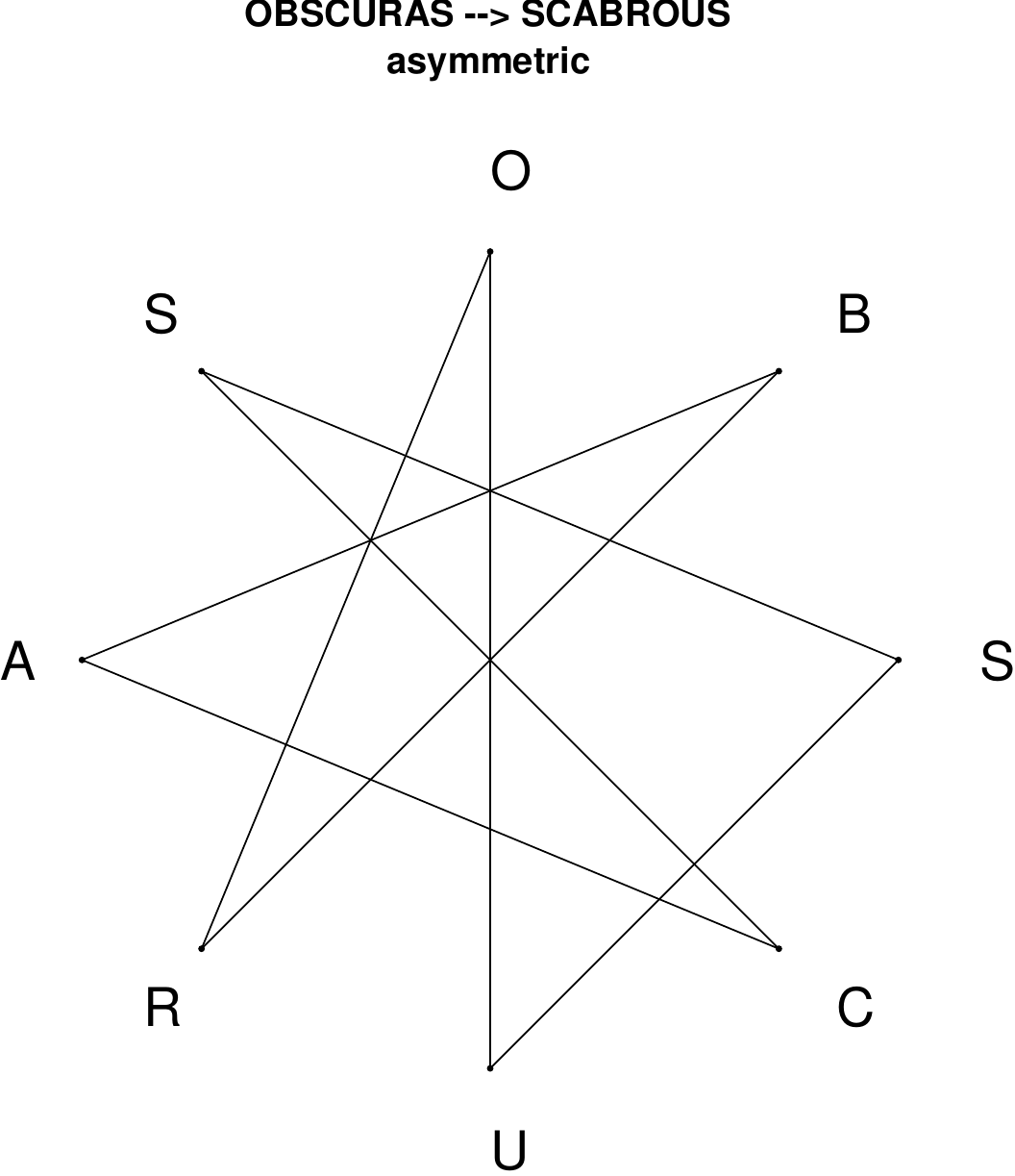}
\end{subfigure}
\end{figure}

\begin{figure}[H]
\centering
\begin{subfigure}[T]{0.19\textwidth}
\centering
\includegraphics[width=\textwidth]{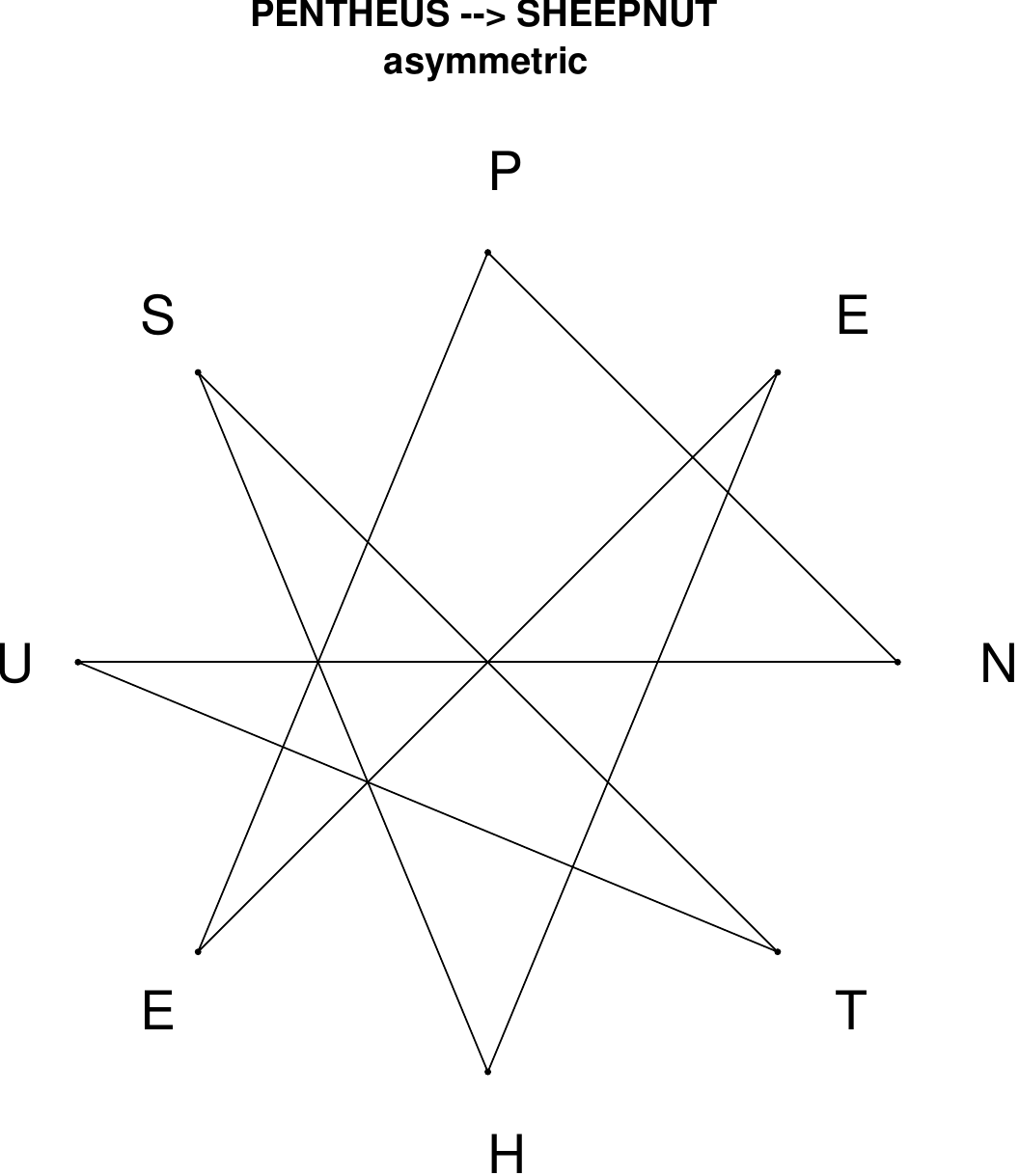}
\end{subfigure}
\hfill
\begin{subfigure}[T]{0.19\textwidth}
\centering
\includegraphics[width=\textwidth]{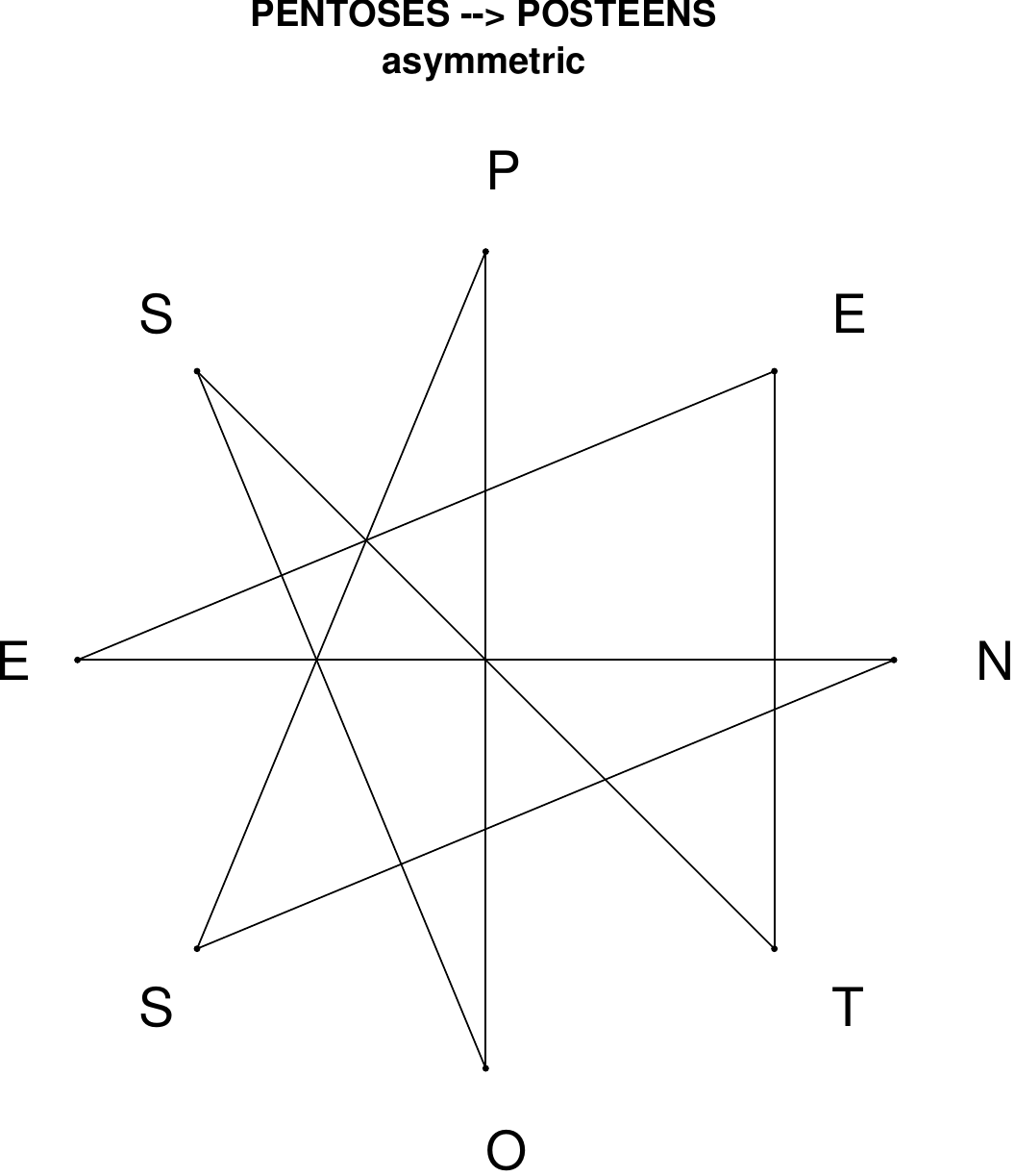}
\end{subfigure}
\hfill
\begin{subfigure}[T]{0.19\textwidth}
\centering
\includegraphics[width=\textwidth]{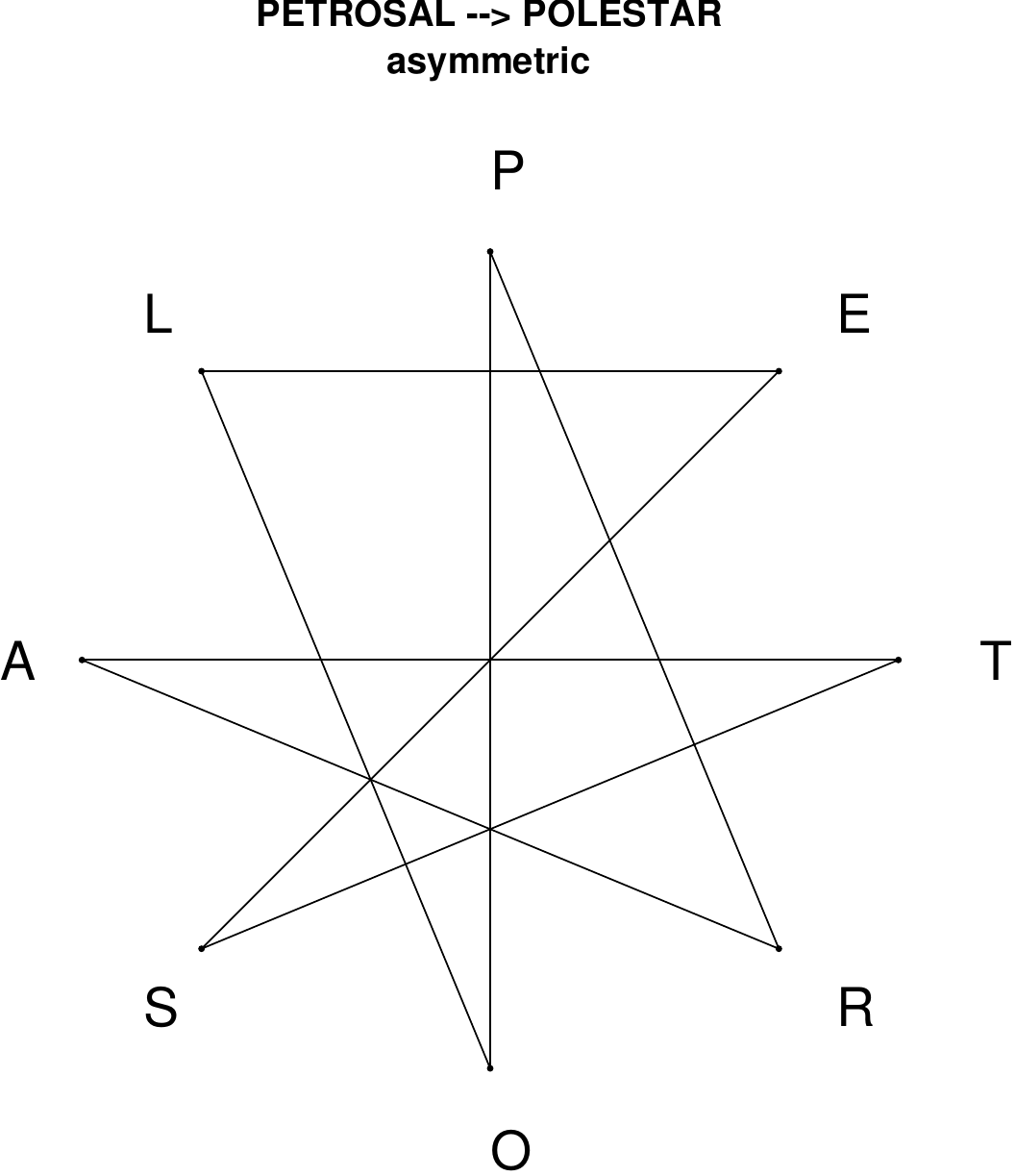}
\end{subfigure}
\hfill
\begin{subfigure}[T]{0.19\textwidth}
\centering
\includegraphics[width=\textwidth]{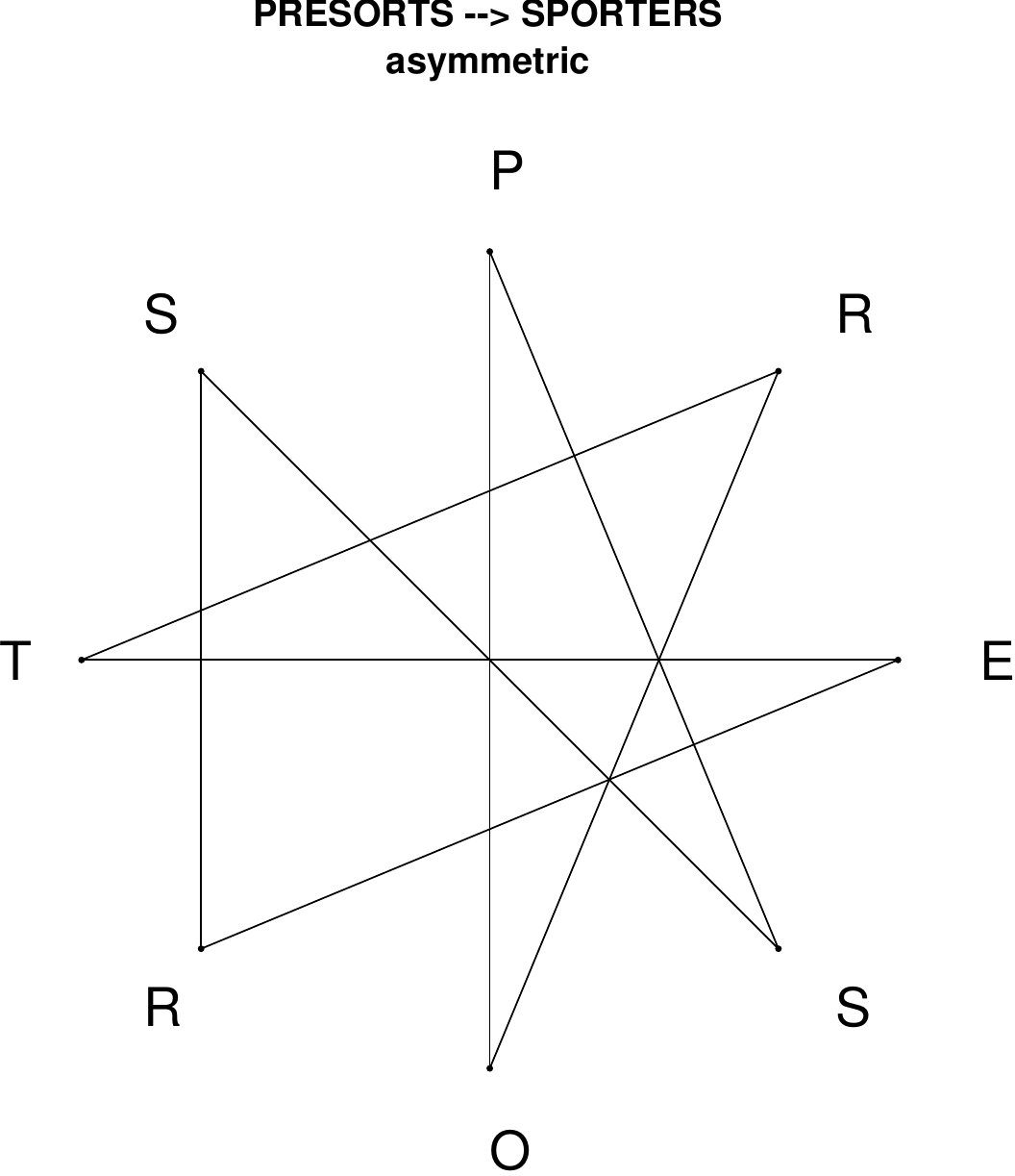}
\end{subfigure}
\hfill
\begin{subfigure}[T]{0.19\textwidth}
\centering
\includegraphics[width=\textwidth]{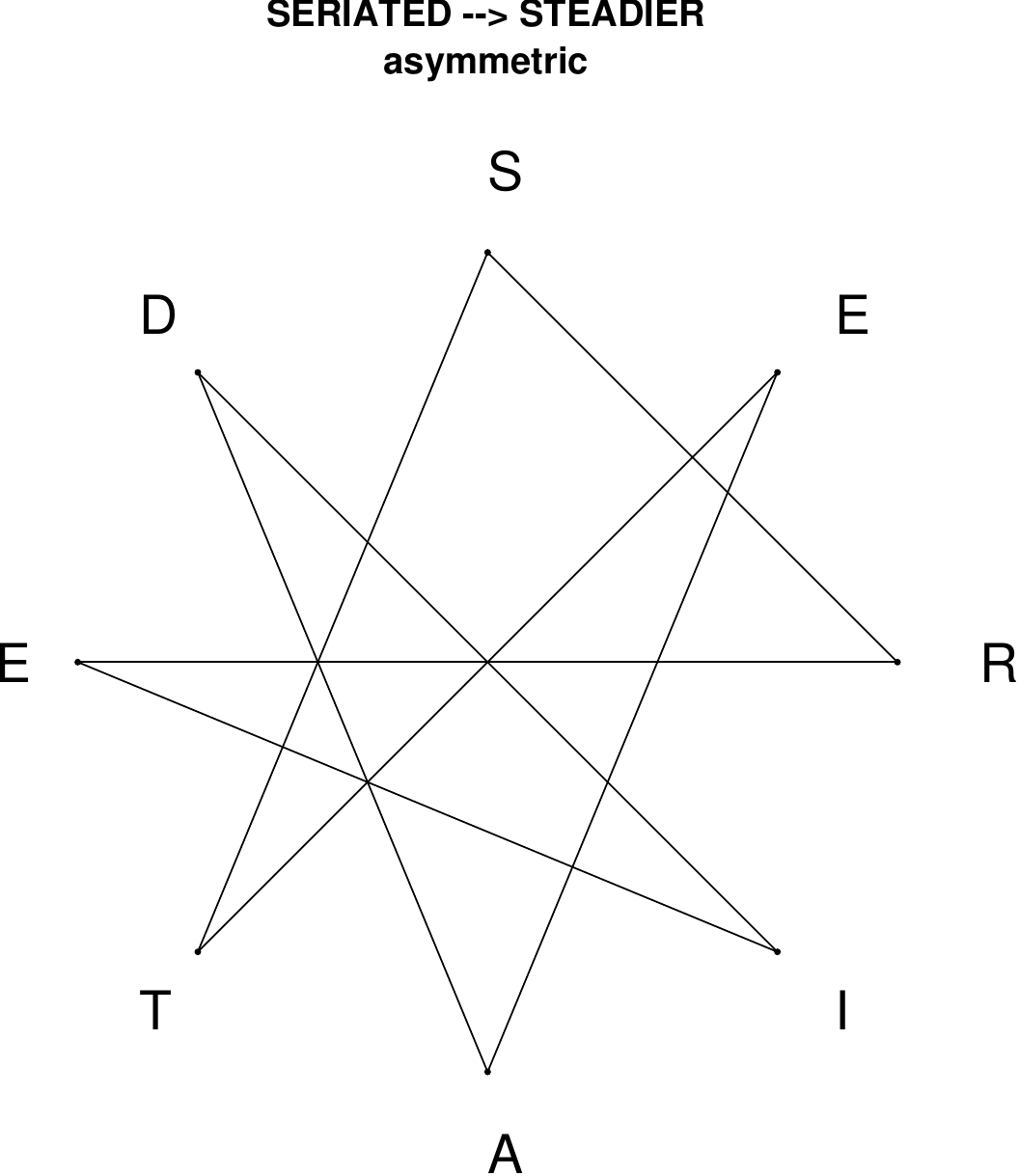}
\end{subfigure}
\end{figure}

\begin{figure}[H]
\centering
\begin{subfigure}[T]{0.19\textwidth}
\centering
\includegraphics[width=\textwidth]{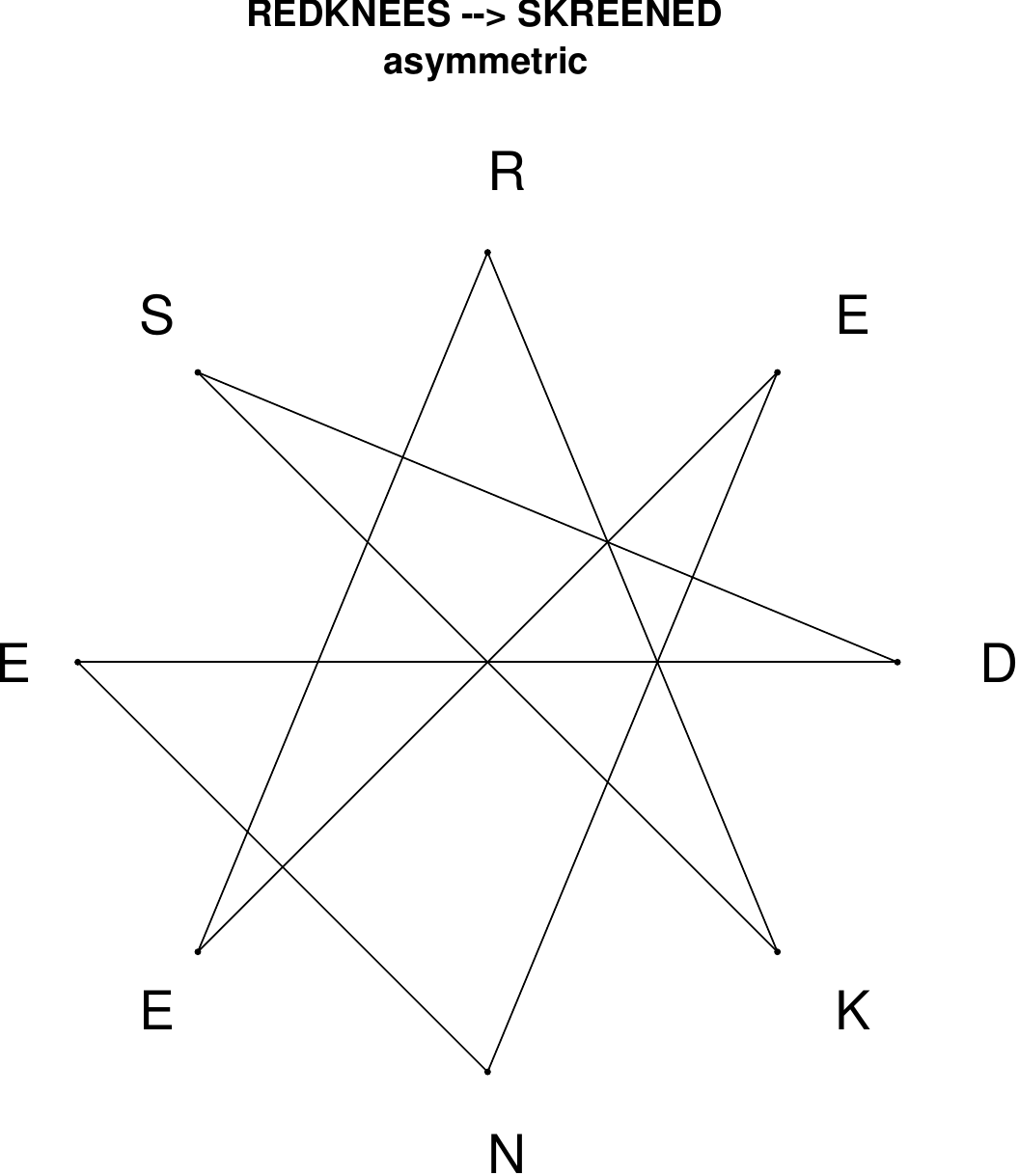}
\end{subfigure}
\hfill
\begin{subfigure}[T]{0.19\textwidth}
\centering
\includegraphics[width=\textwidth]{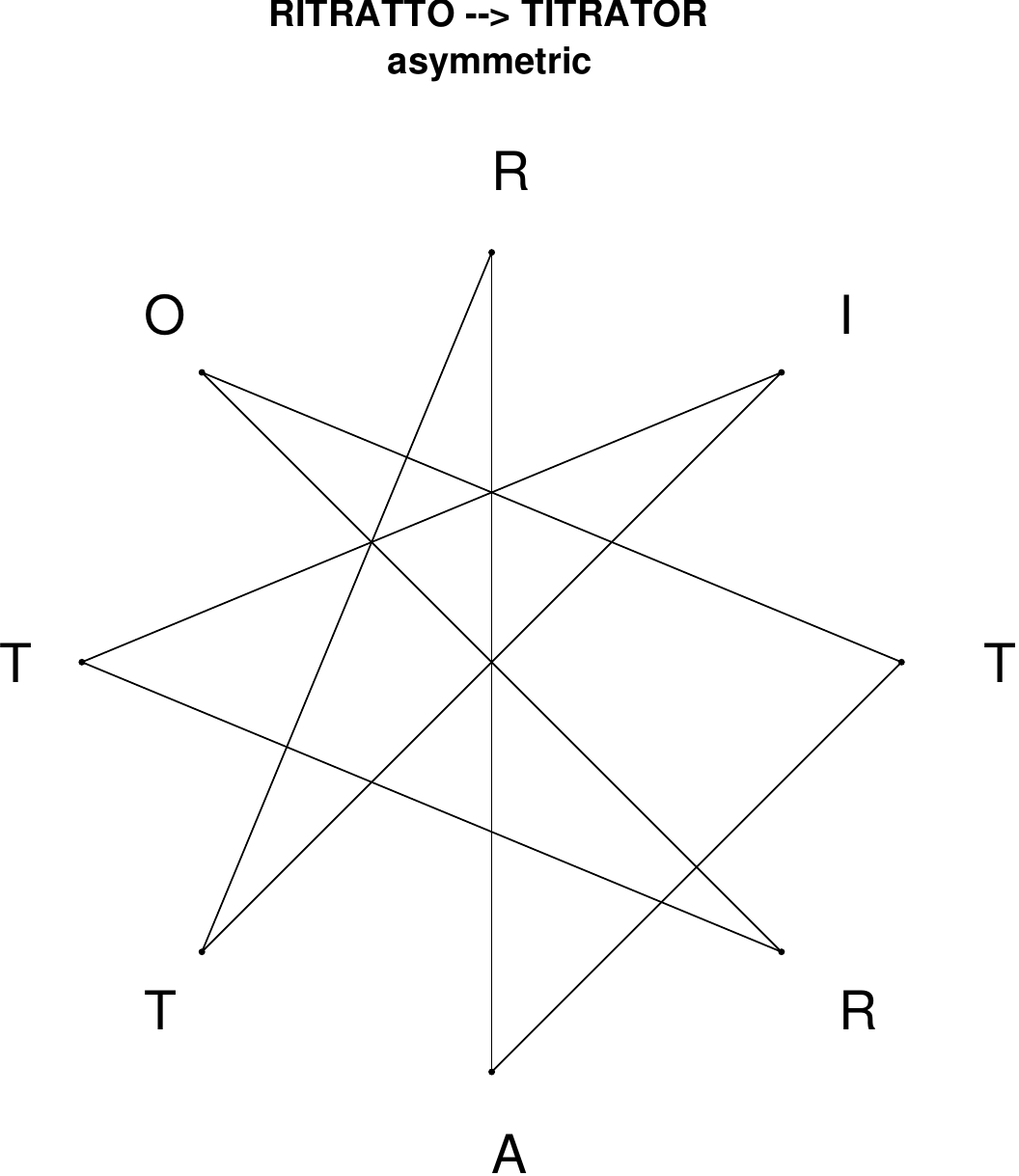}
\end{subfigure}
\hfill
\begin{subfigure}[T]{0.19\textwidth}
\centering
\includegraphics[width=\textwidth]{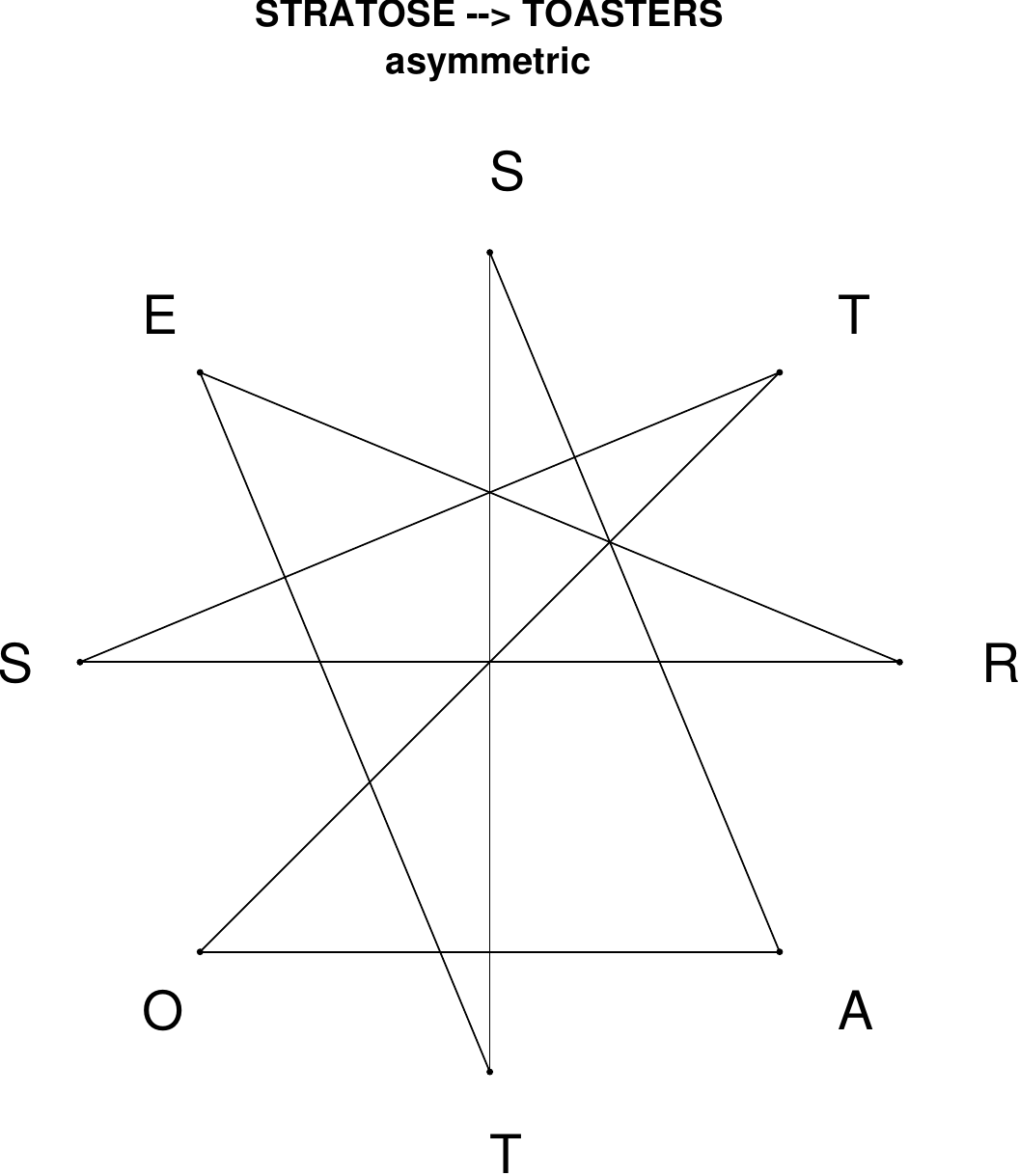}
\end{subfigure}
\hfill
\begin{subfigure}[T]{0.19\textwidth}
\centering
\includegraphics[width=\textwidth]{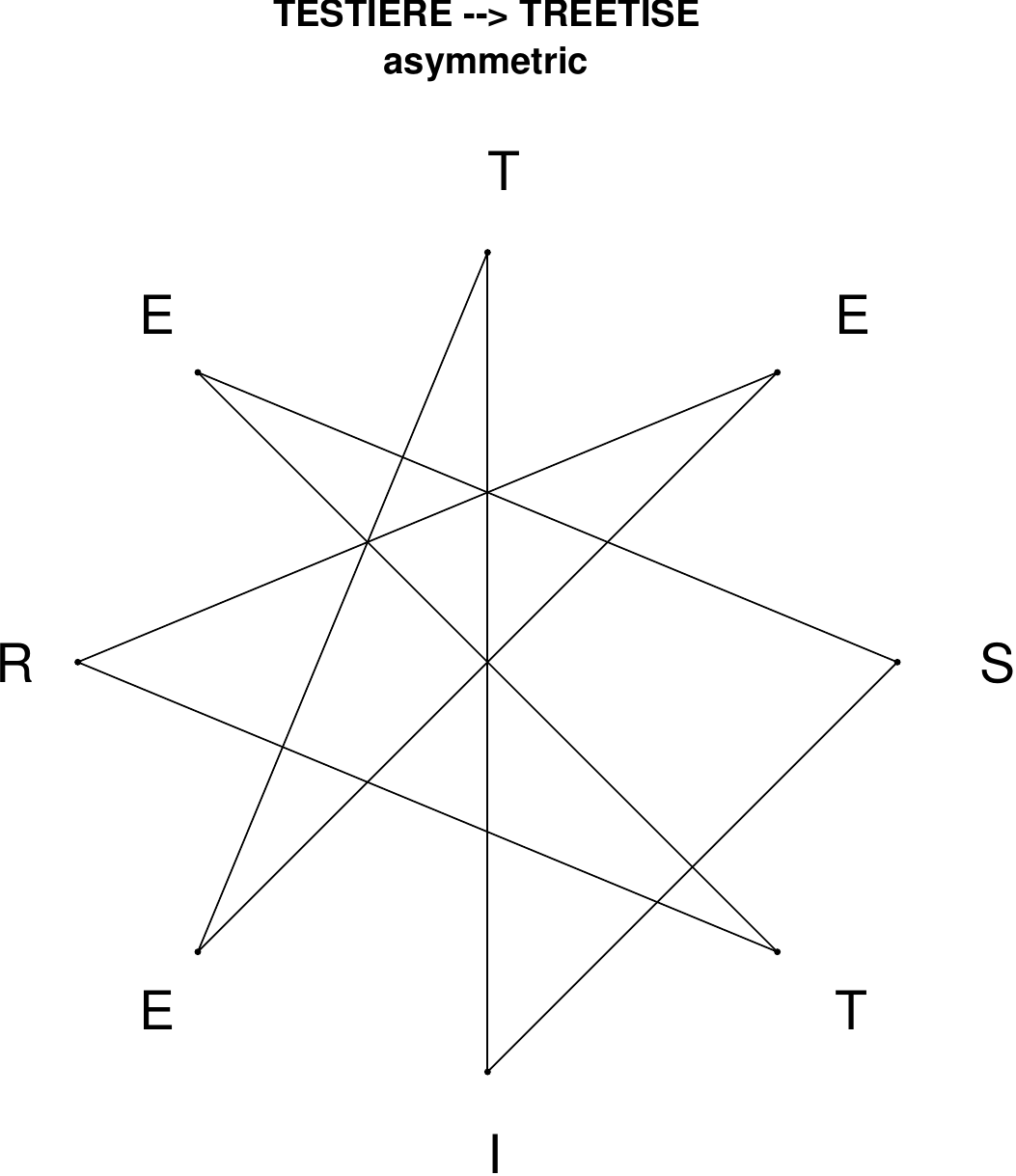}
\end{subfigure}
\hfill
\begin{subfigure}[T]{0.19\textwidth}
\centering
\includegraphics[width=\textwidth]{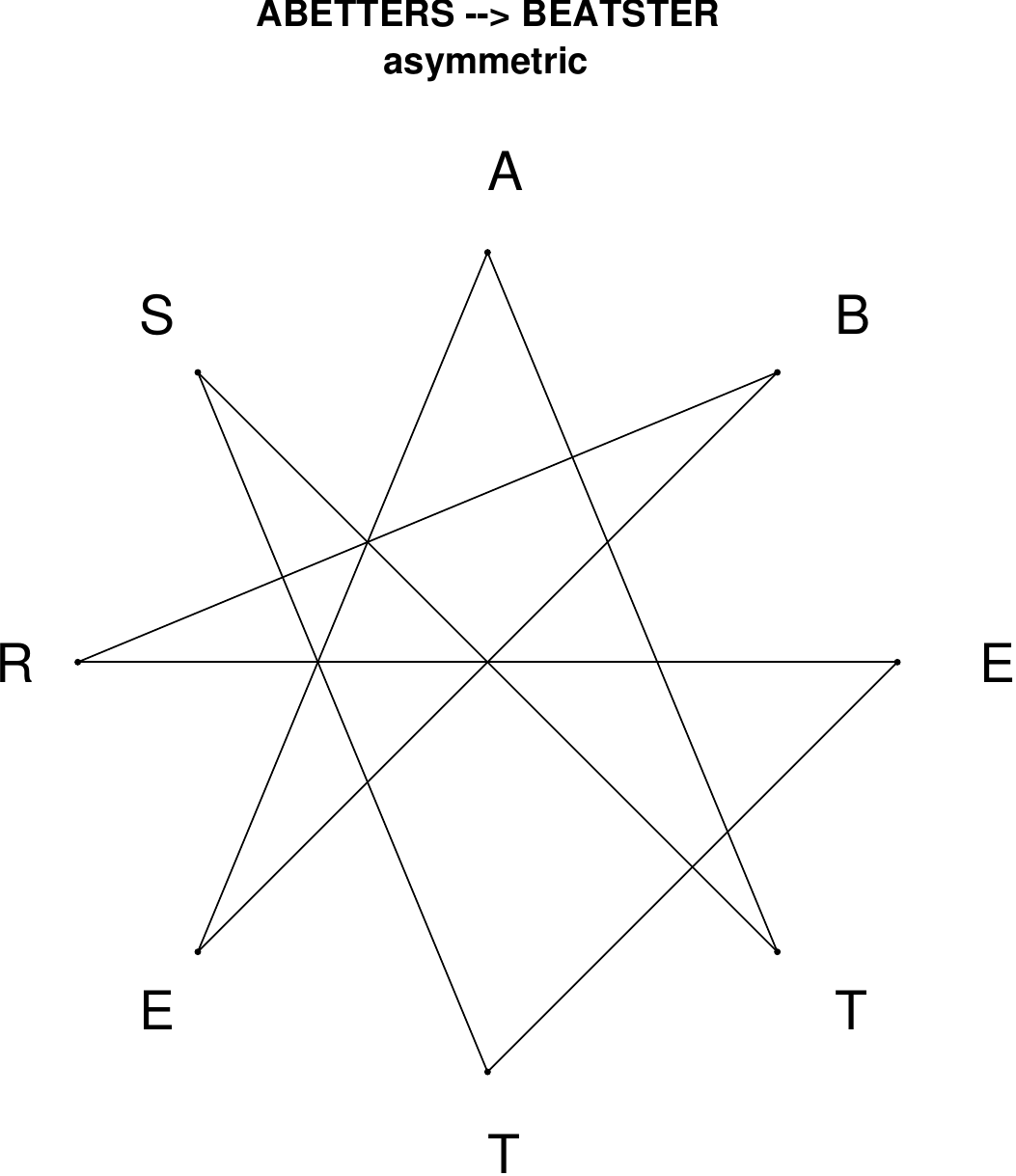}
\end{subfigure}
\end{figure}

\begin{figure}[H]
\centering
\begin{subfigure}[T]{0.19\textwidth}
\centering
\includegraphics[width=\textwidth]{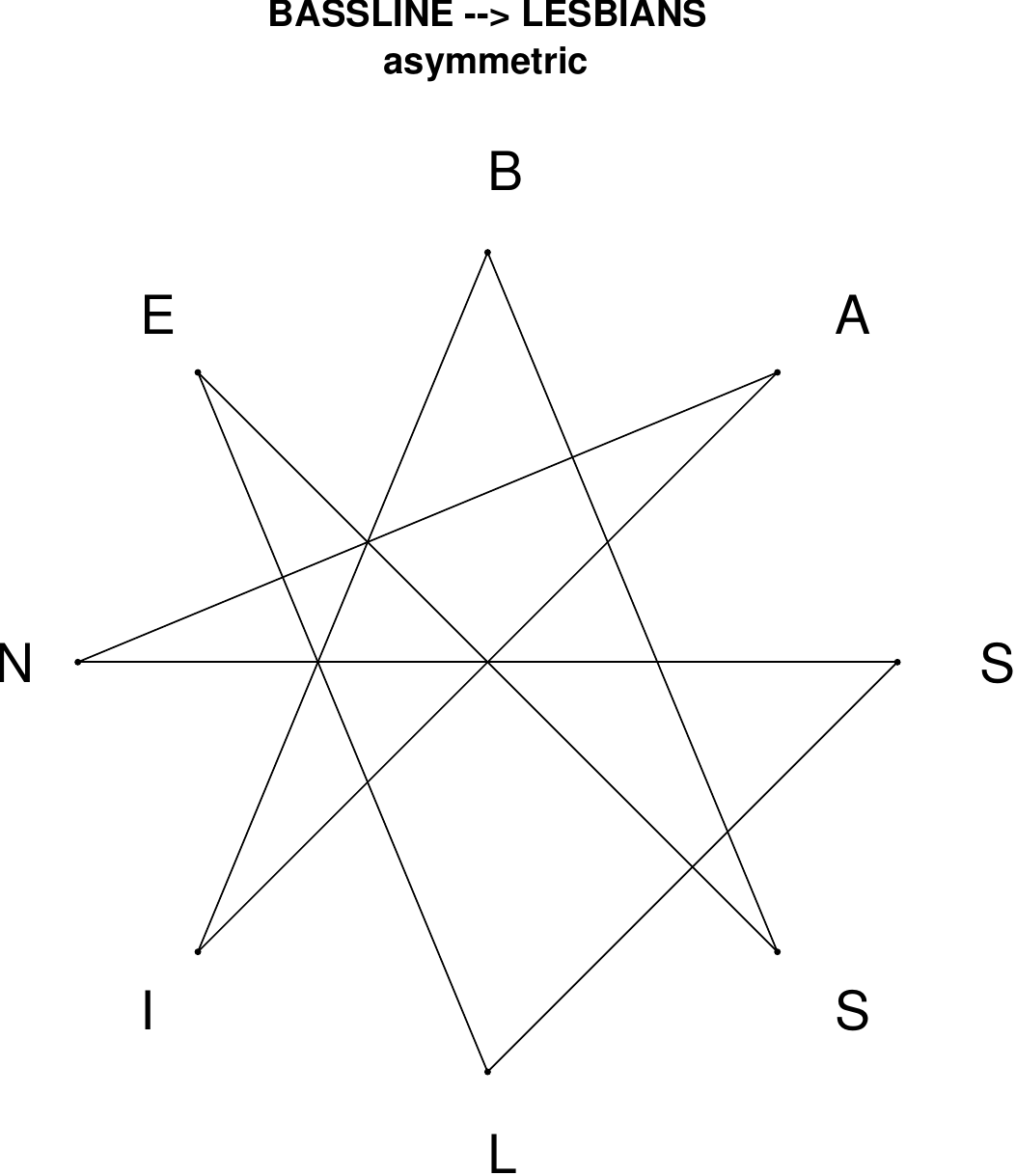}
\end{subfigure}
\hfill
\begin{subfigure}[T]{0.19\textwidth}
\centering
\includegraphics[width=\textwidth]{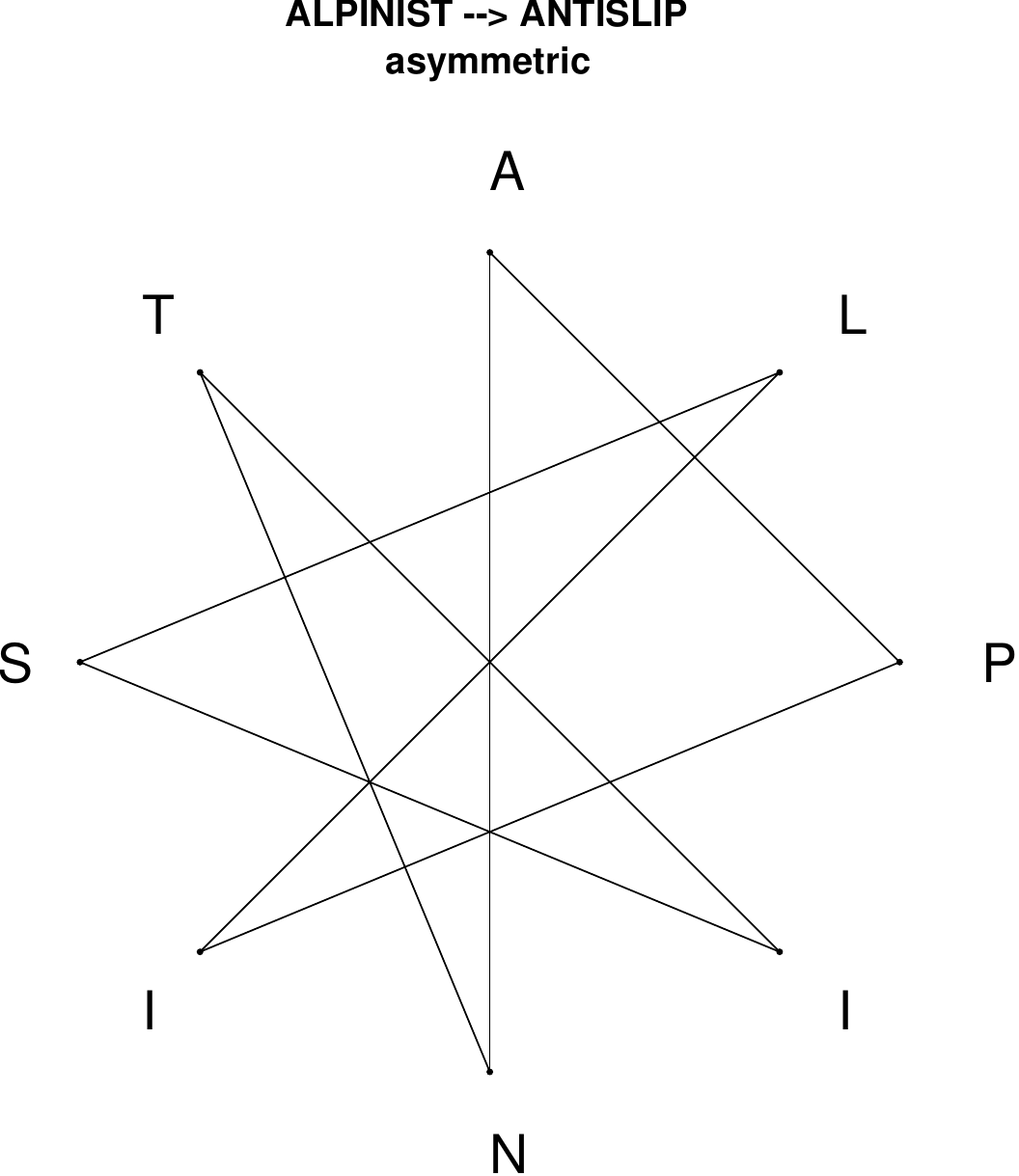}
\end{subfigure}
\hfill
\begin{subfigure}[T]{0.19\textwidth}
\centering
\includegraphics[width=\textwidth]{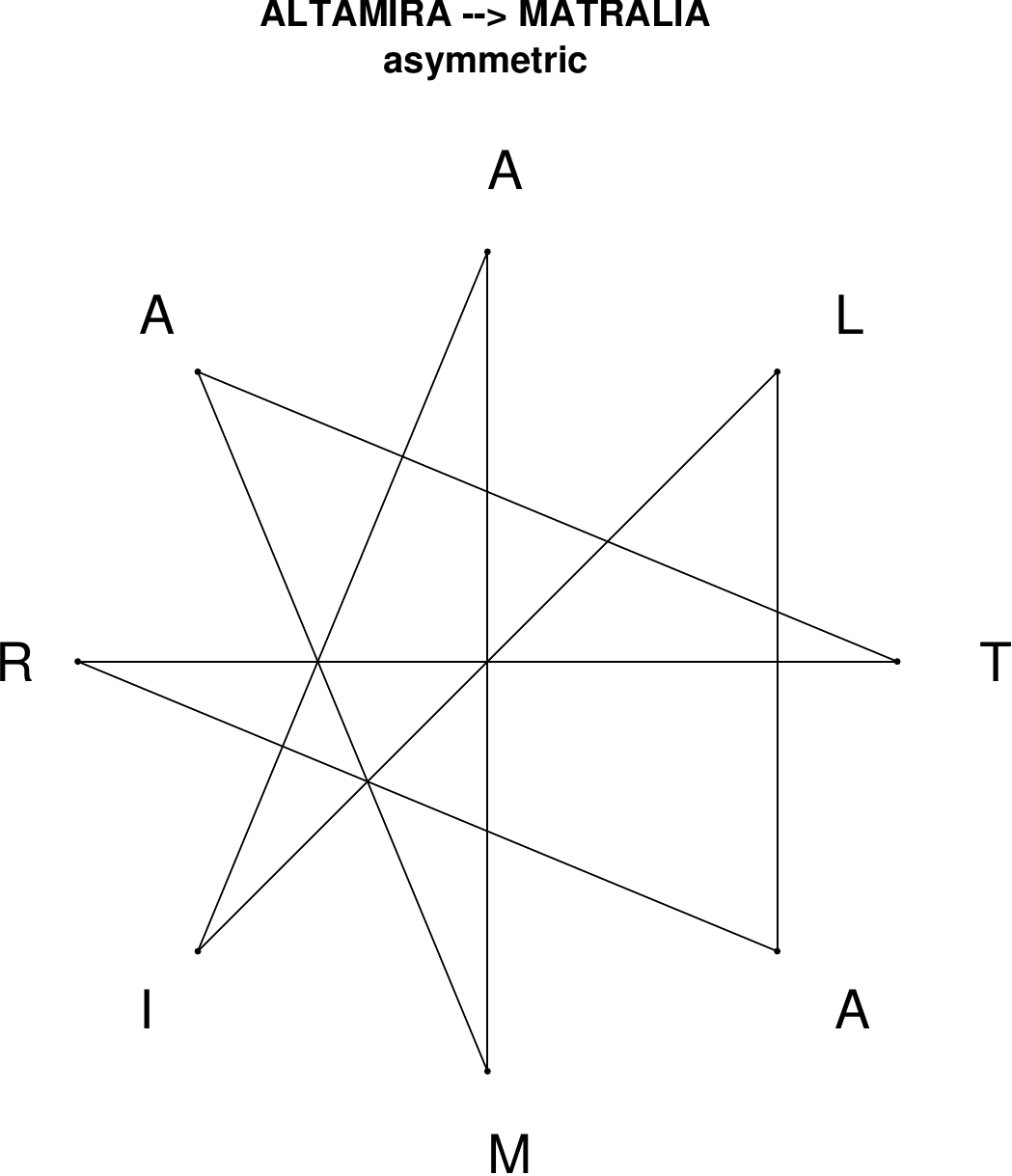}
\end{subfigure}
\hfill
\begin{subfigure}[T]{0.19\textwidth}
\centering
\includegraphics[width=\textwidth]{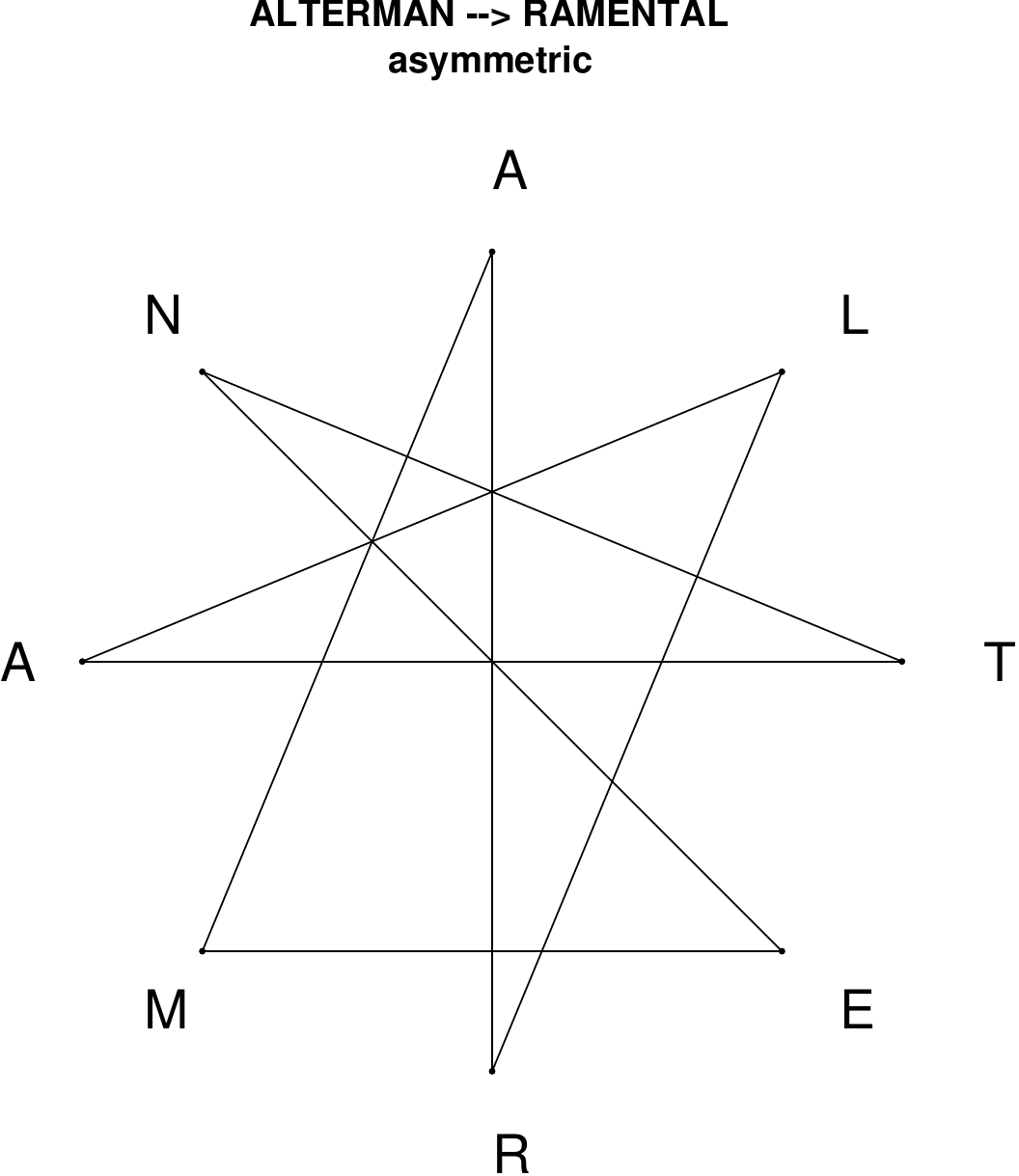}
\end{subfigure}
\hfill
\begin{subfigure}[T]{0.19\textwidth}
\centering
\includegraphics[width=\textwidth]{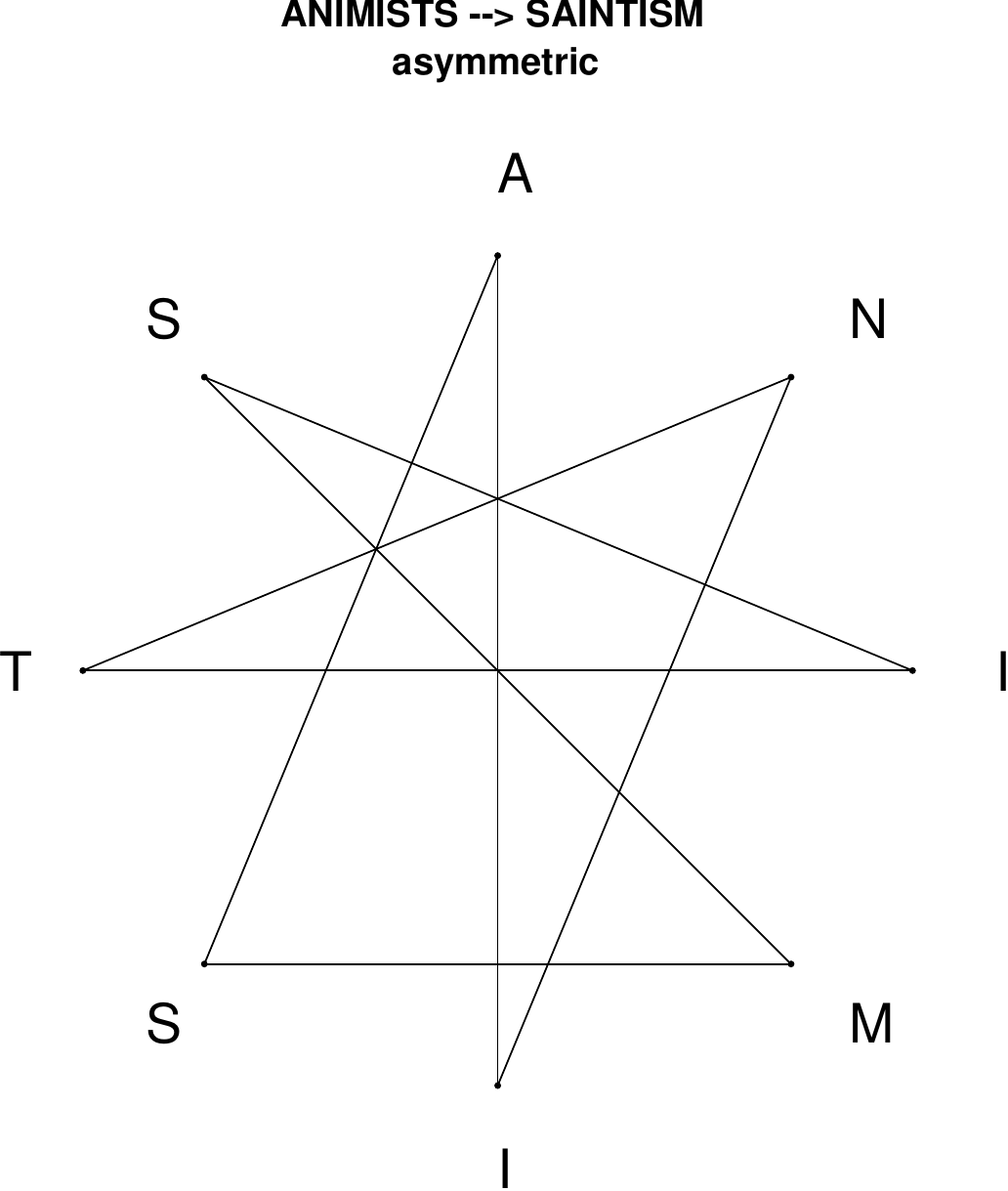}
\end{subfigure}
\end{figure}

\begin{figure}[H]
\centering
\begin{subfigure}[T]{0.19\textwidth}
\centering
\includegraphics[width=\textwidth]{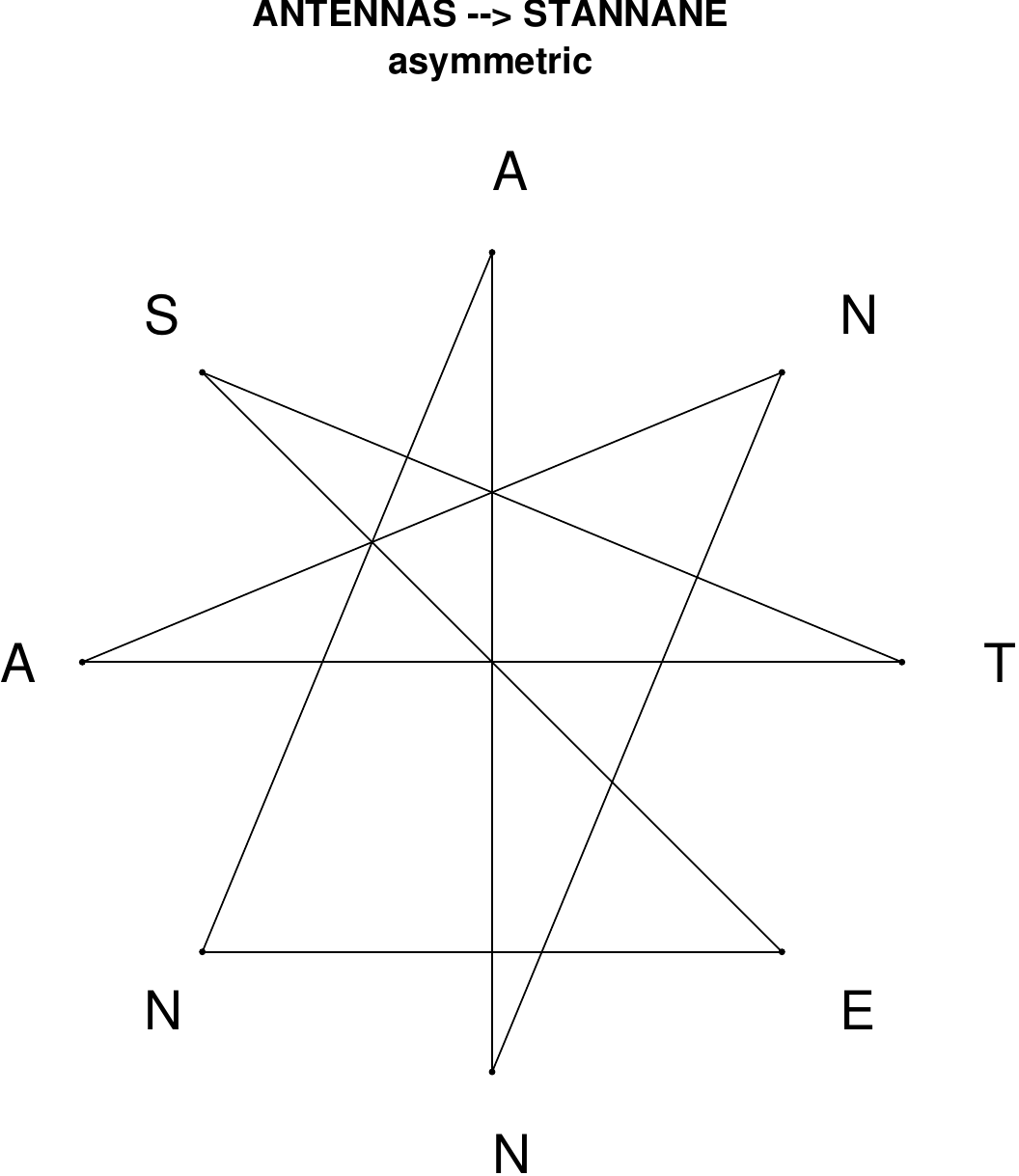}
\end{subfigure}
\hfill
\begin{subfigure}[T]{0.19\textwidth}
\centering
\includegraphics[width=\textwidth]{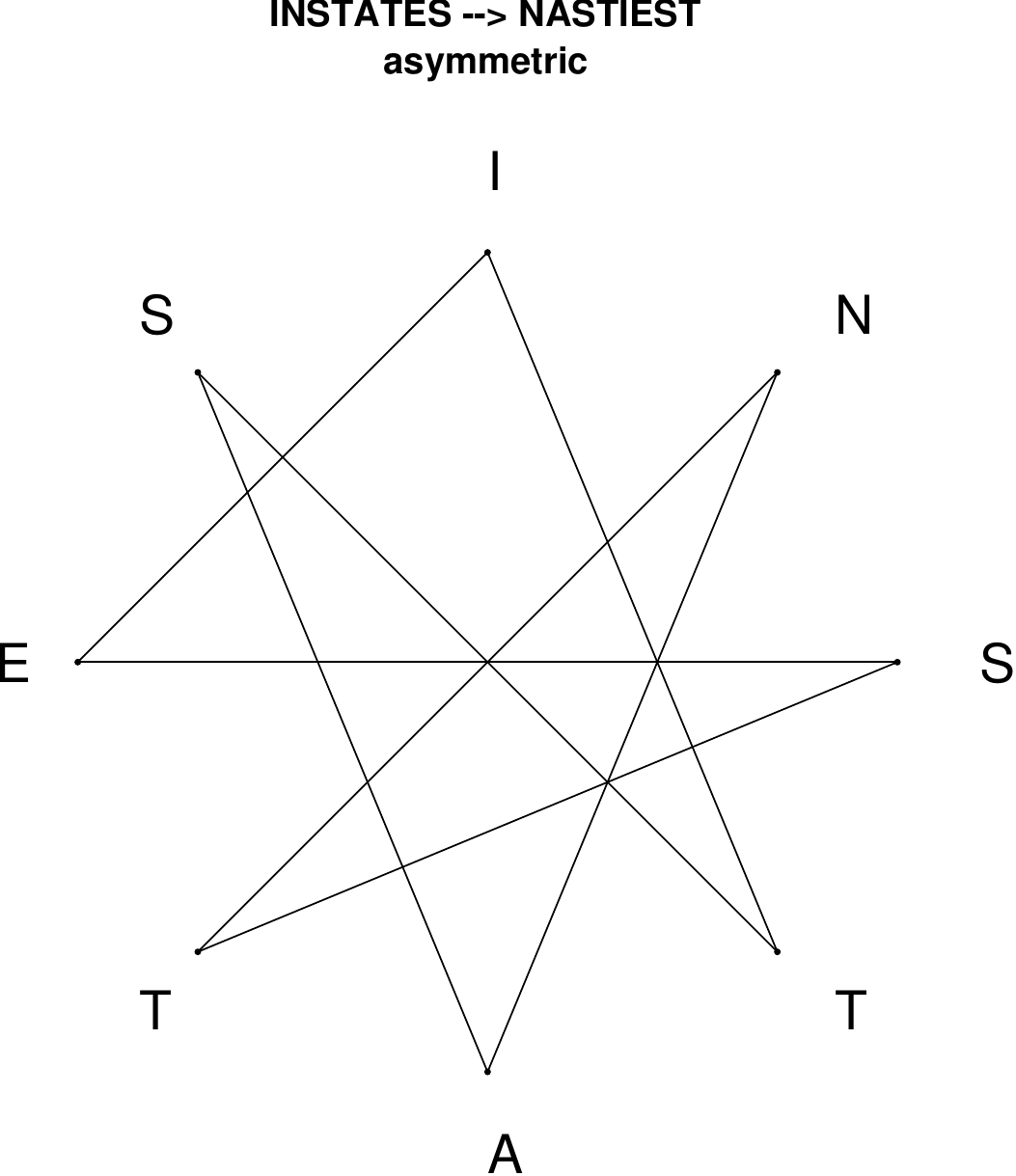}
\end{subfigure}
\hfill
\begin{subfigure}[T]{0.19\textwidth}
\centering
\includegraphics[width=\textwidth]{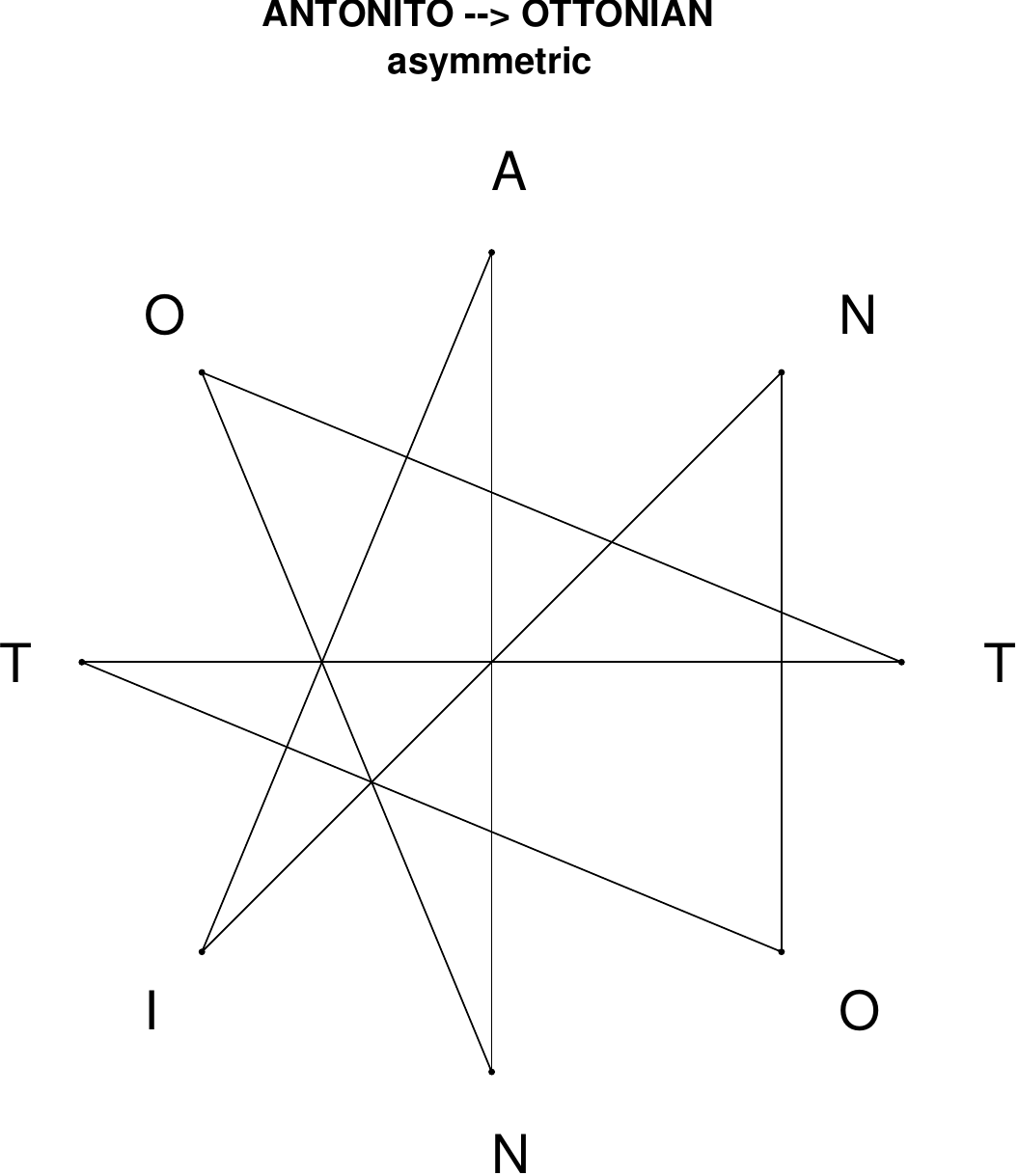}
\end{subfigure}
\hfill
\begin{subfigure}[T]{0.19\textwidth}
\centering
\includegraphics[width=\textwidth]{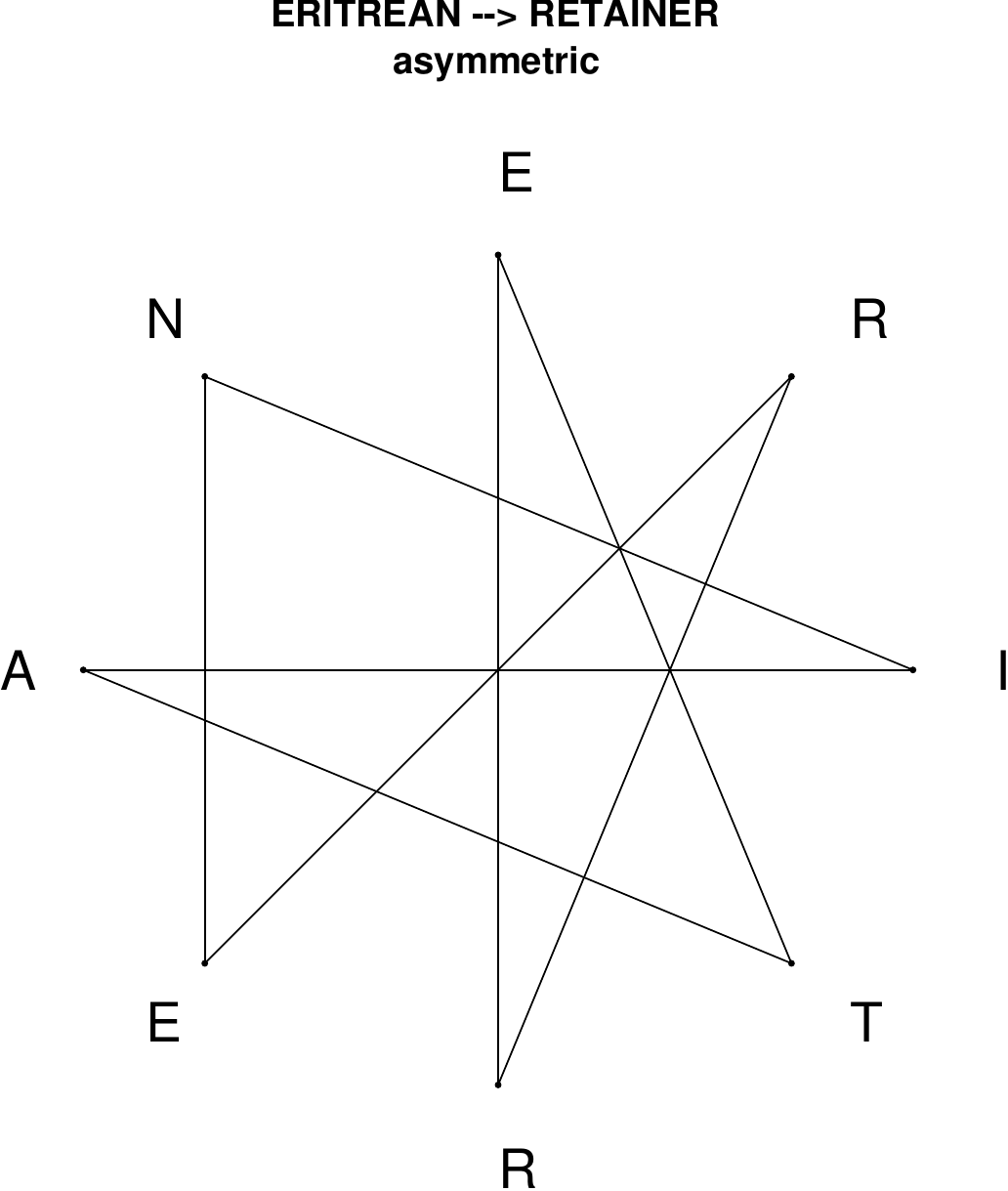}
\end{subfigure}
\hfill
\begin{subfigure}[T]{0.19\textwidth}
\centering
\includegraphics[width=\textwidth]{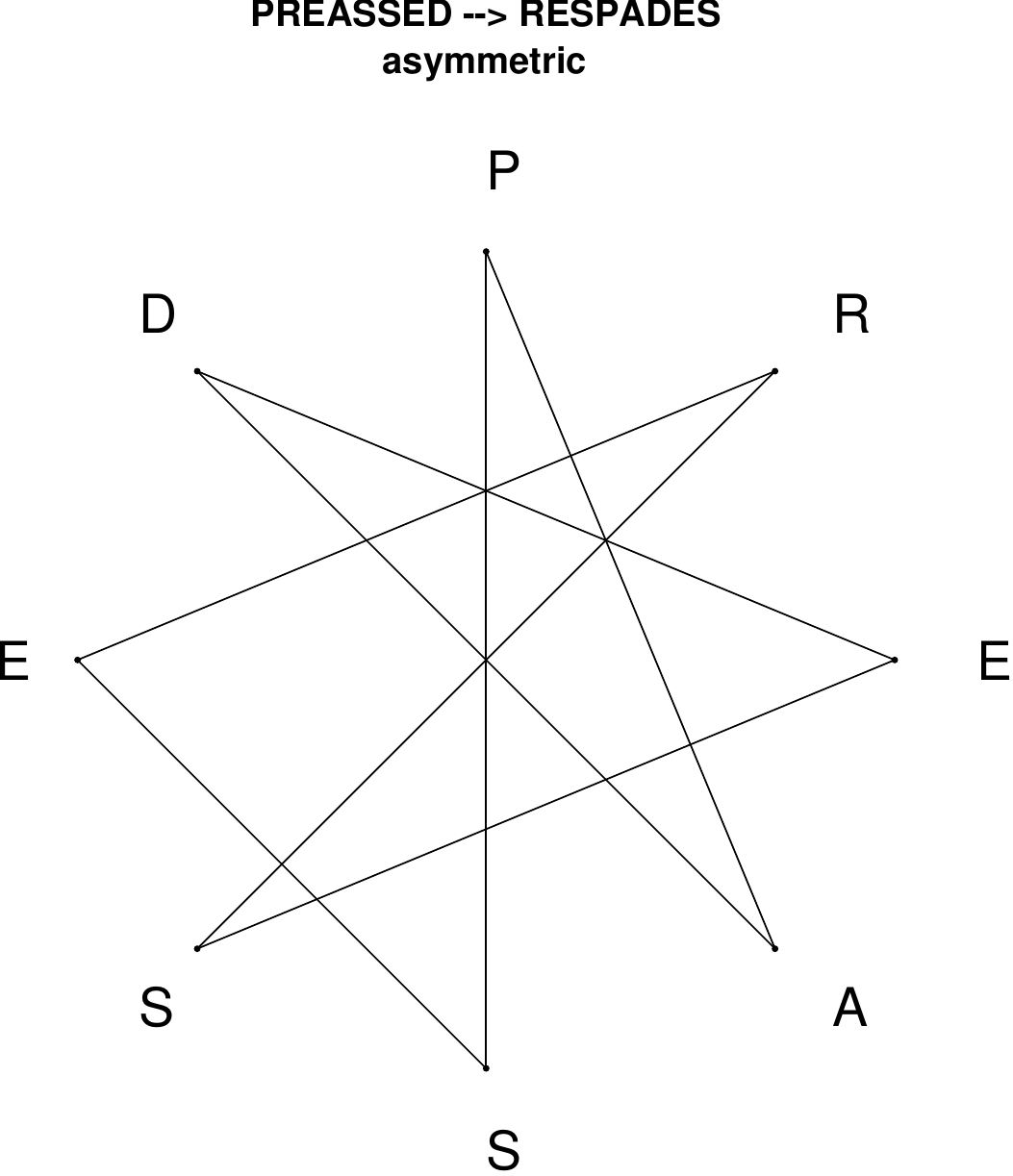}
\end{subfigure}
\end{figure}

\begin{figure}[H]
\centering
\begin{subfigure}[T]{0.19\textwidth}
\centering
\includegraphics[width=\textwidth]{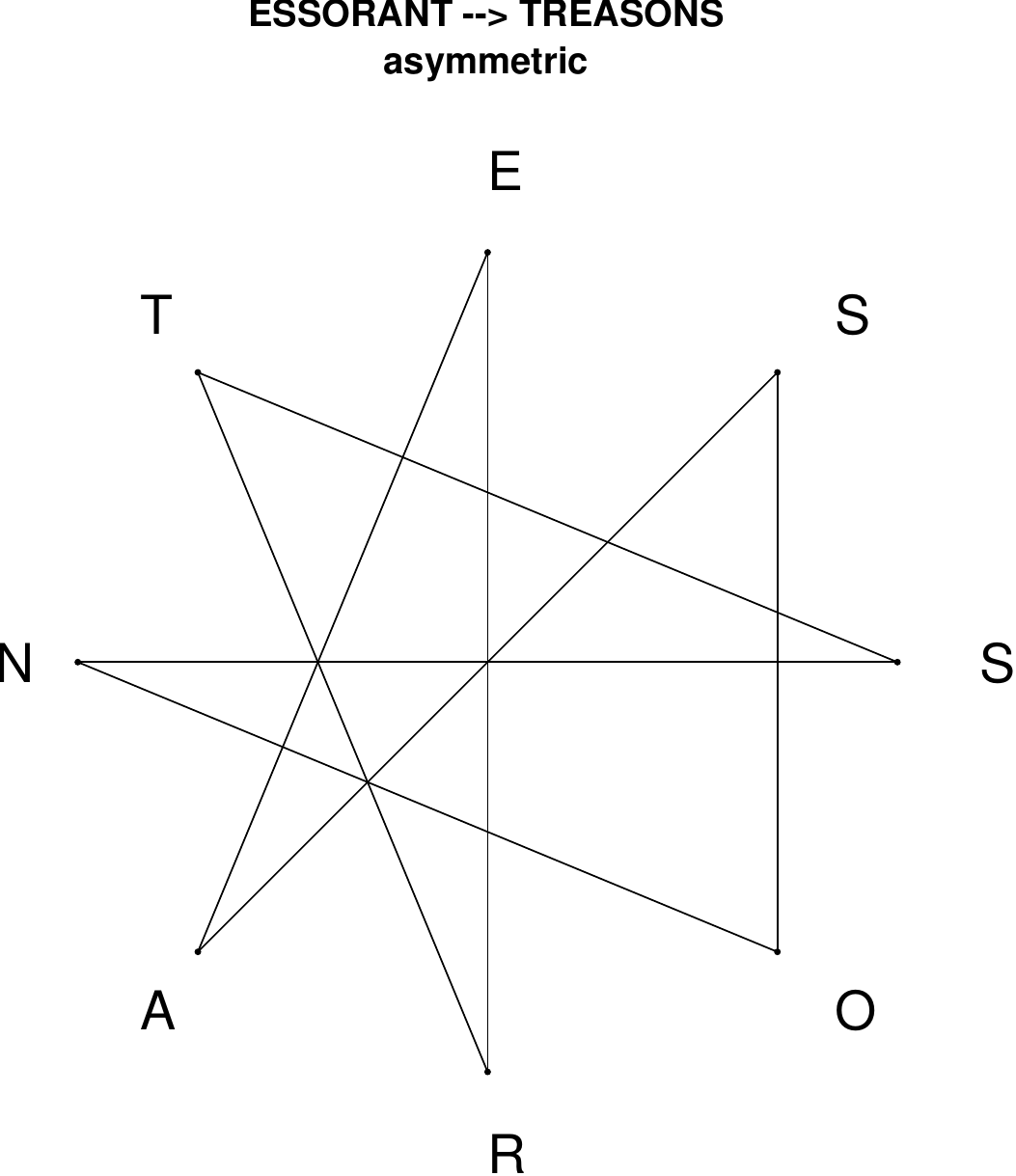}
\end{subfigure}
\hfill
\begin{subfigure}[T]{0.19\textwidth}
\centering
\includegraphics[width=\textwidth]{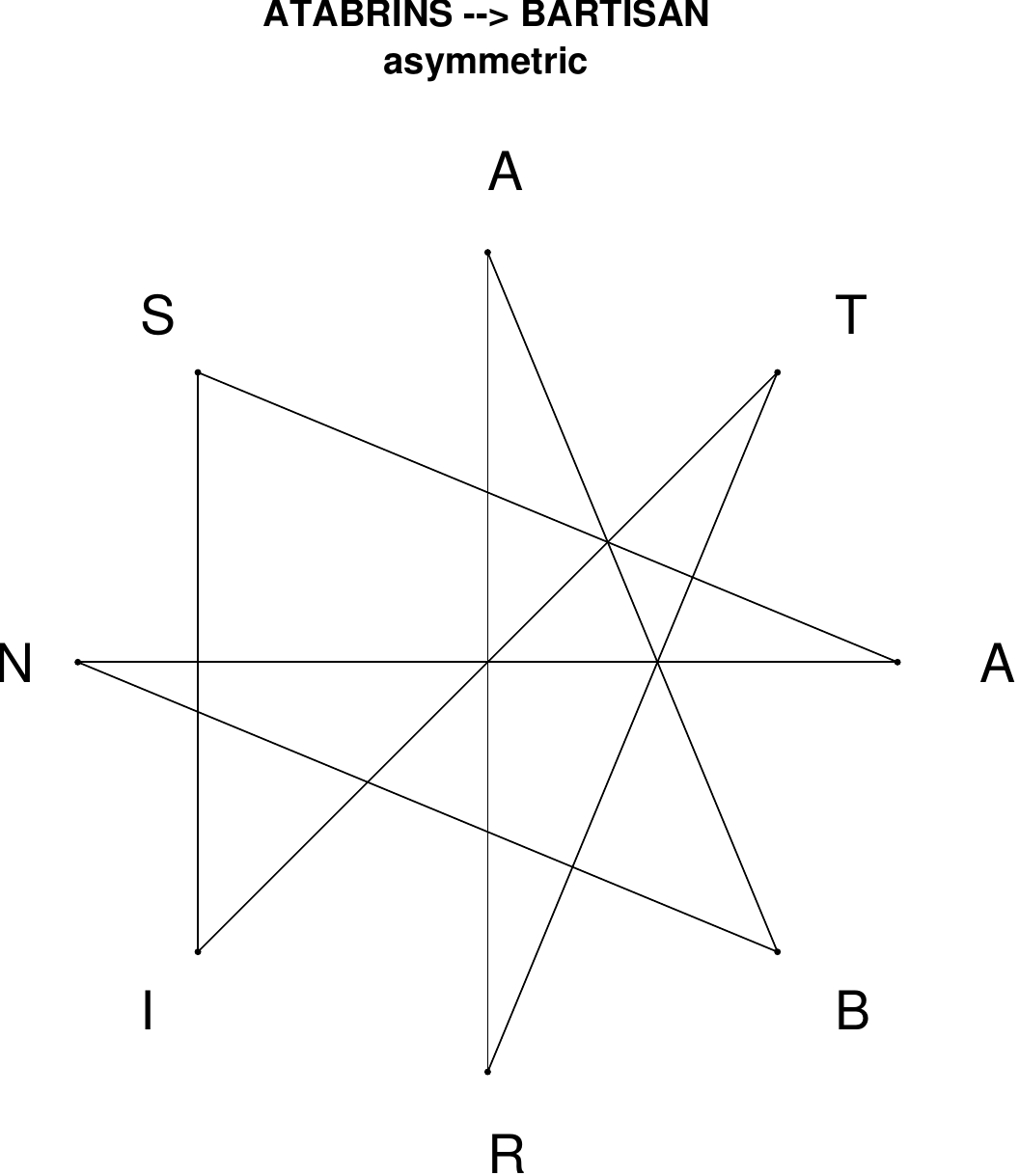}
\end{subfigure}
\hfill
\begin{subfigure}[T]{0.19\textwidth}
\centering
\includegraphics[width=\textwidth]{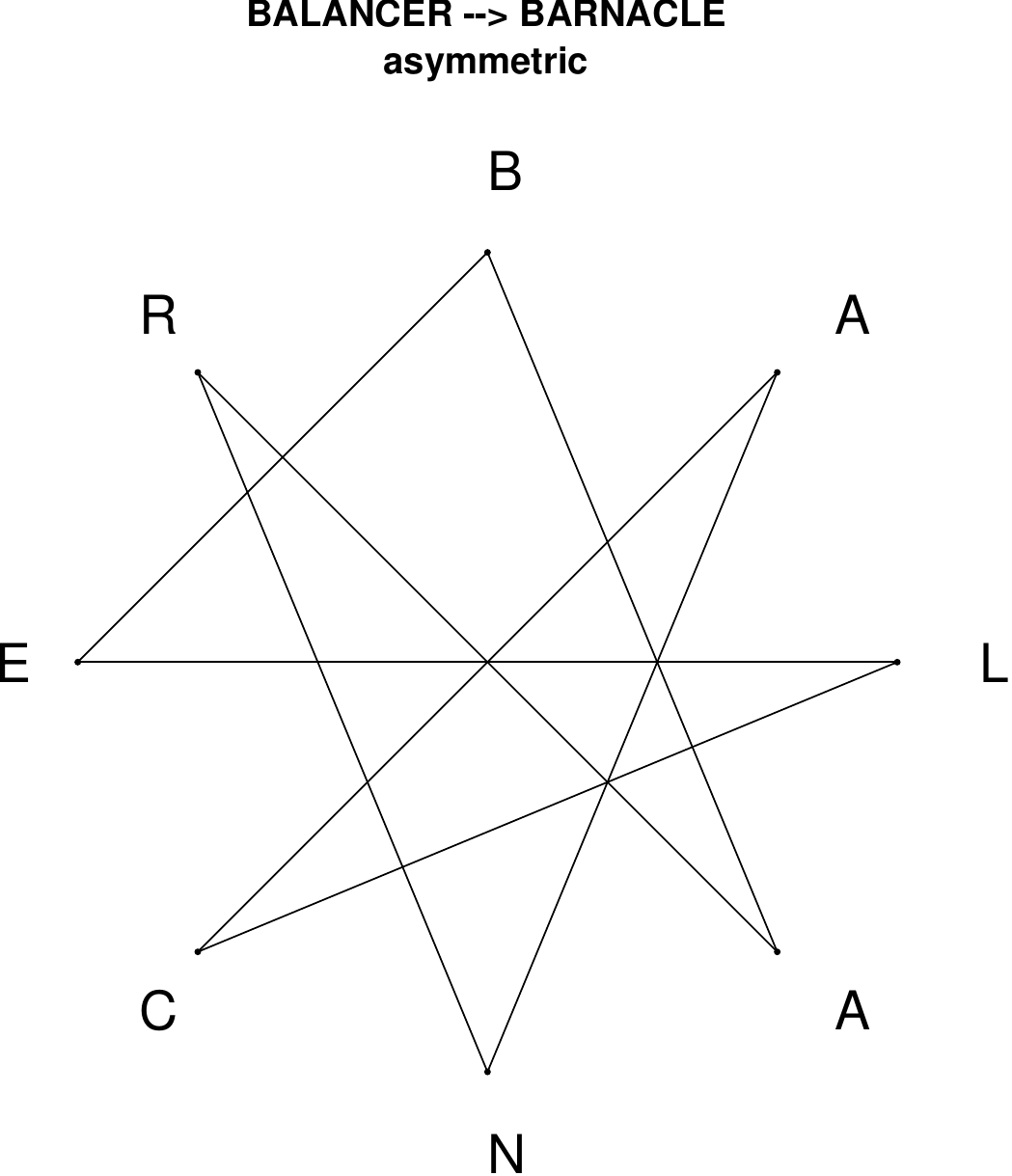}
\end{subfigure}
\hfill
\begin{subfigure}[T]{0.19\textwidth}
\centering
\includegraphics[width=\textwidth]{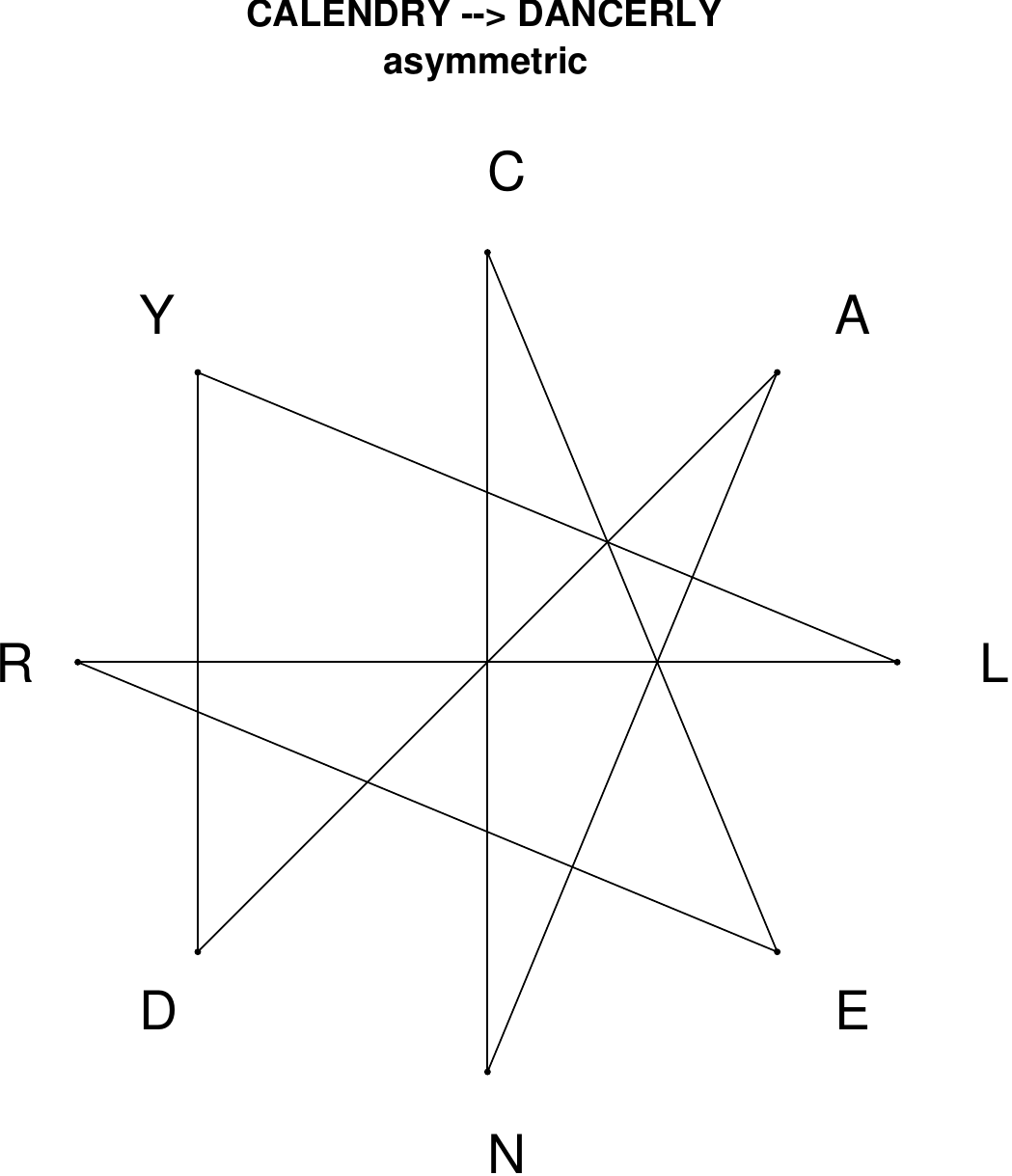}
\end{subfigure}
\hfill
\begin{subfigure}[T]{0.19\textwidth}
\centering
\includegraphics[width=\textwidth]{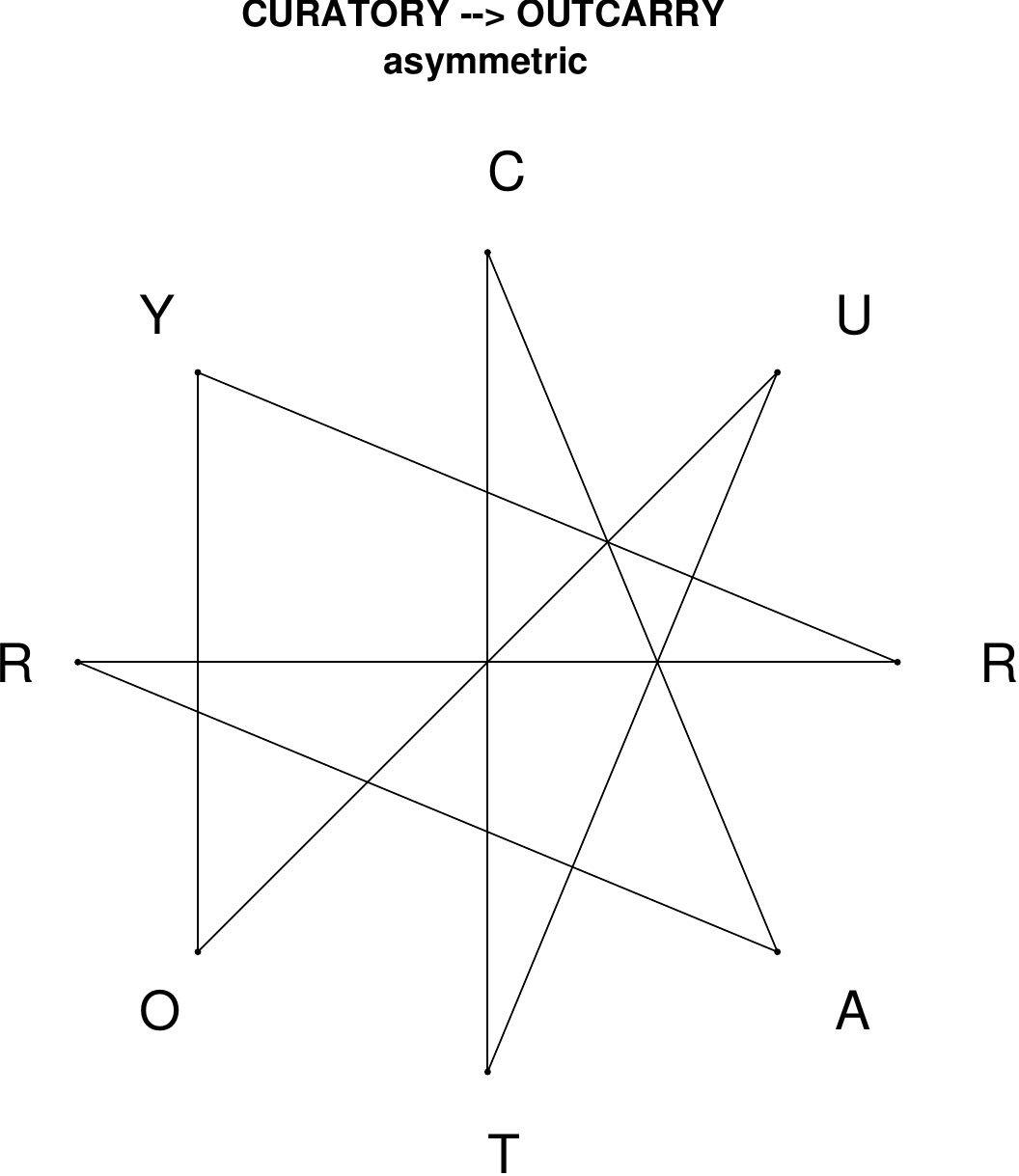}
\end{subfigure}
\end{figure}

\begin{figure}[H]
\centering
\begin{subfigure}[T]{0.19\textwidth}
\centering
\includegraphics[width=\textwidth]{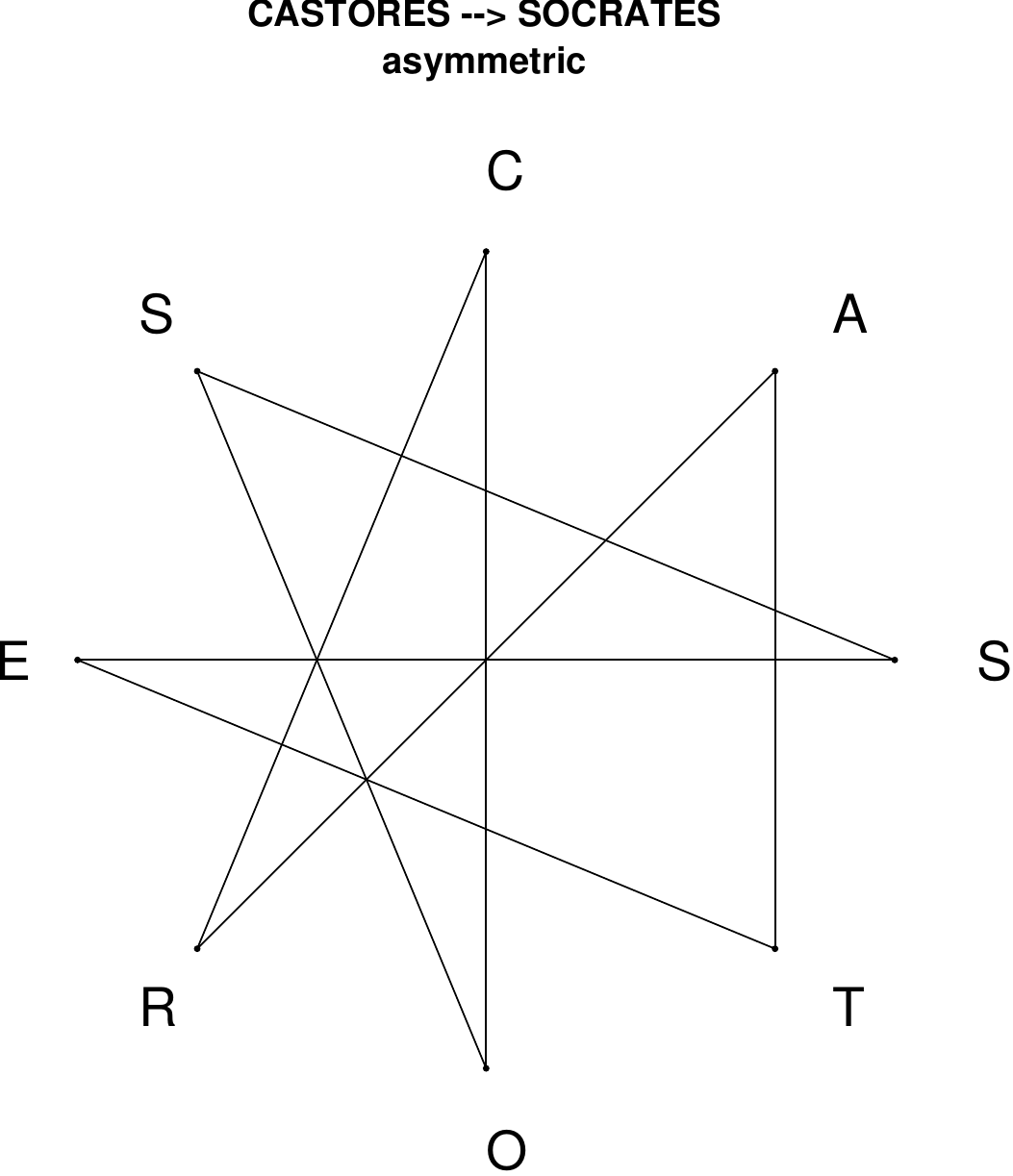}
\end{subfigure}
\hfill
\begin{subfigure}[T]{0.19\textwidth}
\centering
\includegraphics[width=\textwidth]{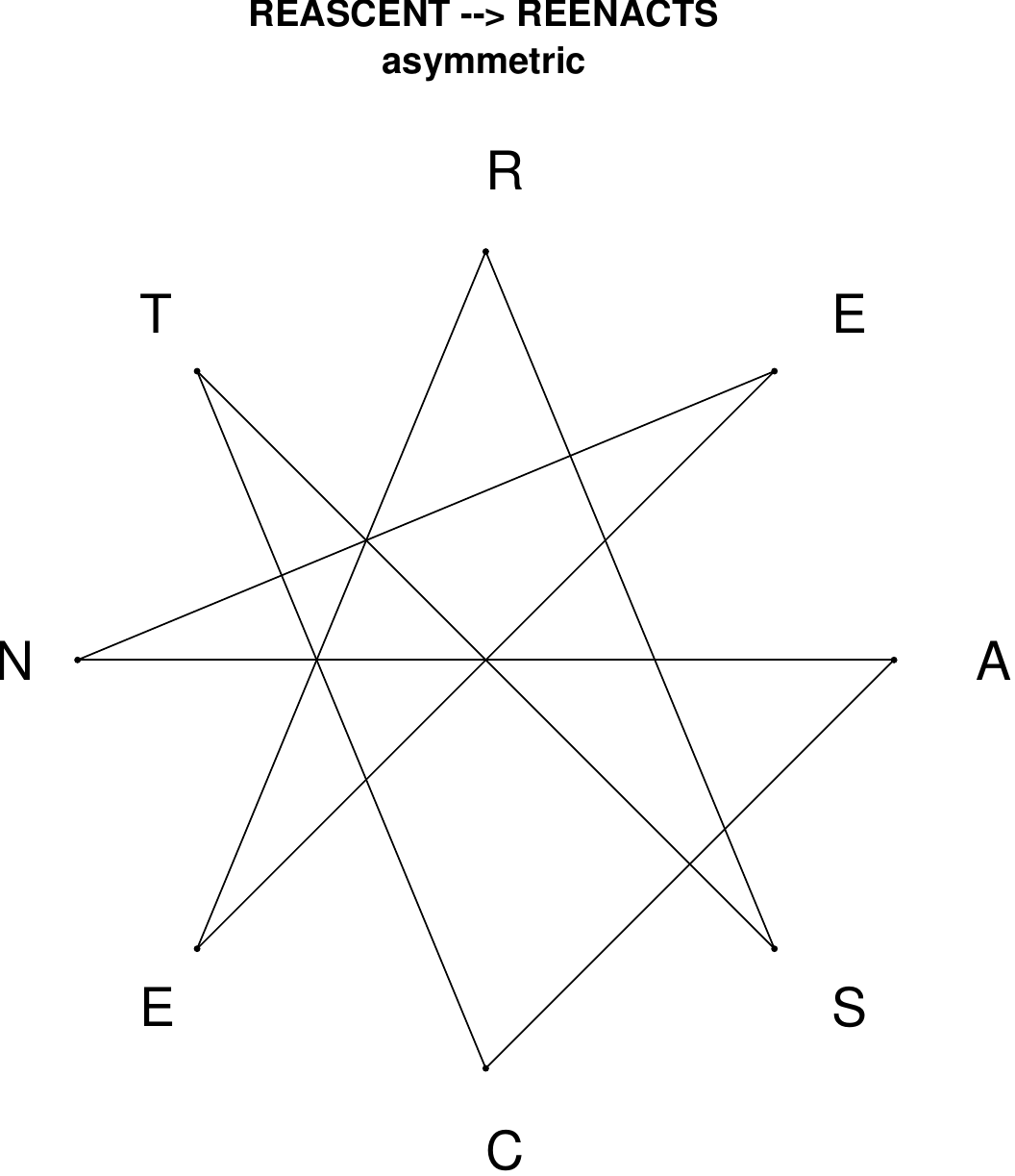}
\end{subfigure}
\hfill
\begin{subfigure}[T]{0.19\textwidth}
\centering
\includegraphics[width=\textwidth]{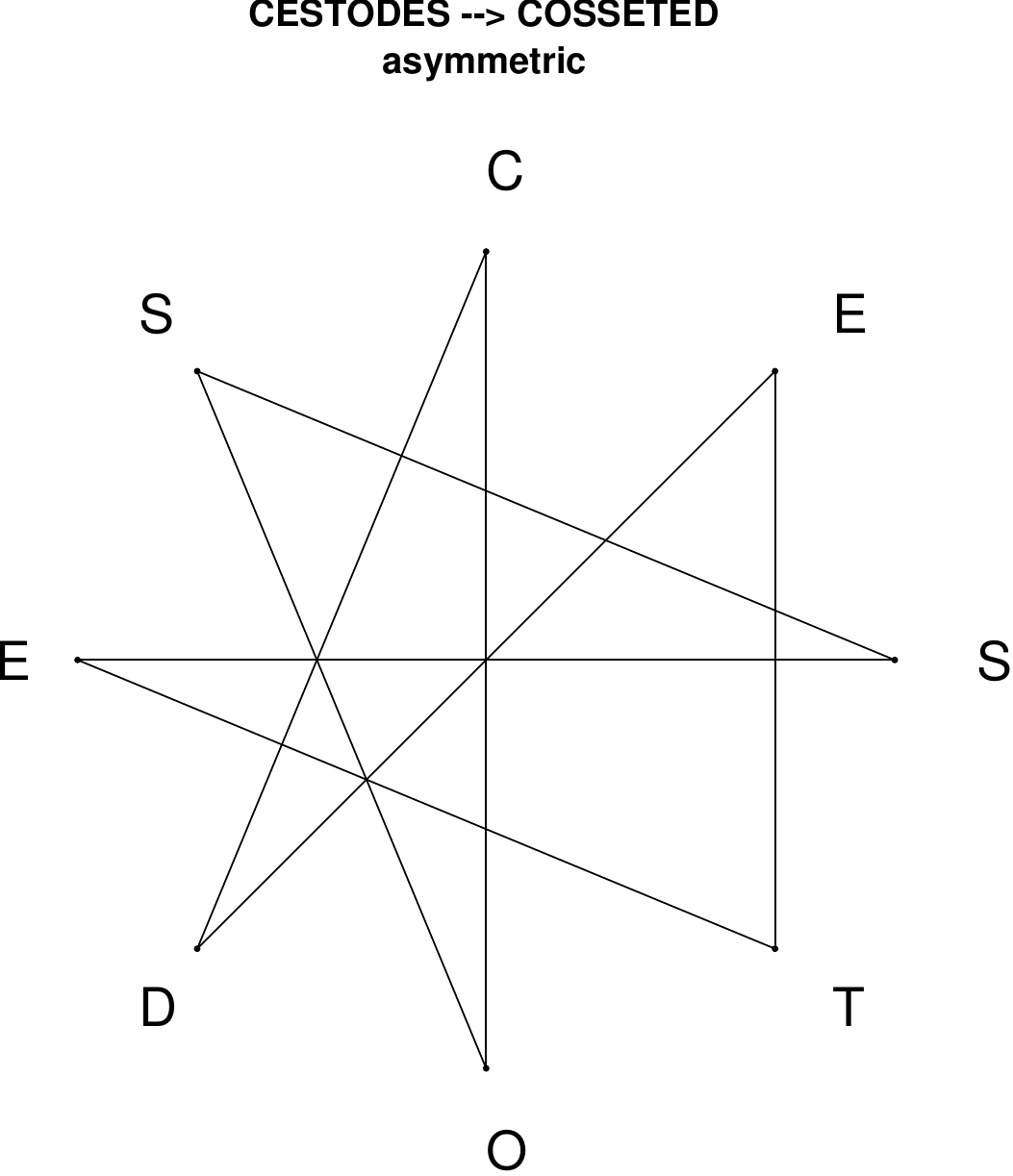}
\end{subfigure}
\hfill
\begin{subfigure}[T]{0.19\textwidth}
\centering
\includegraphics[width=\textwidth]{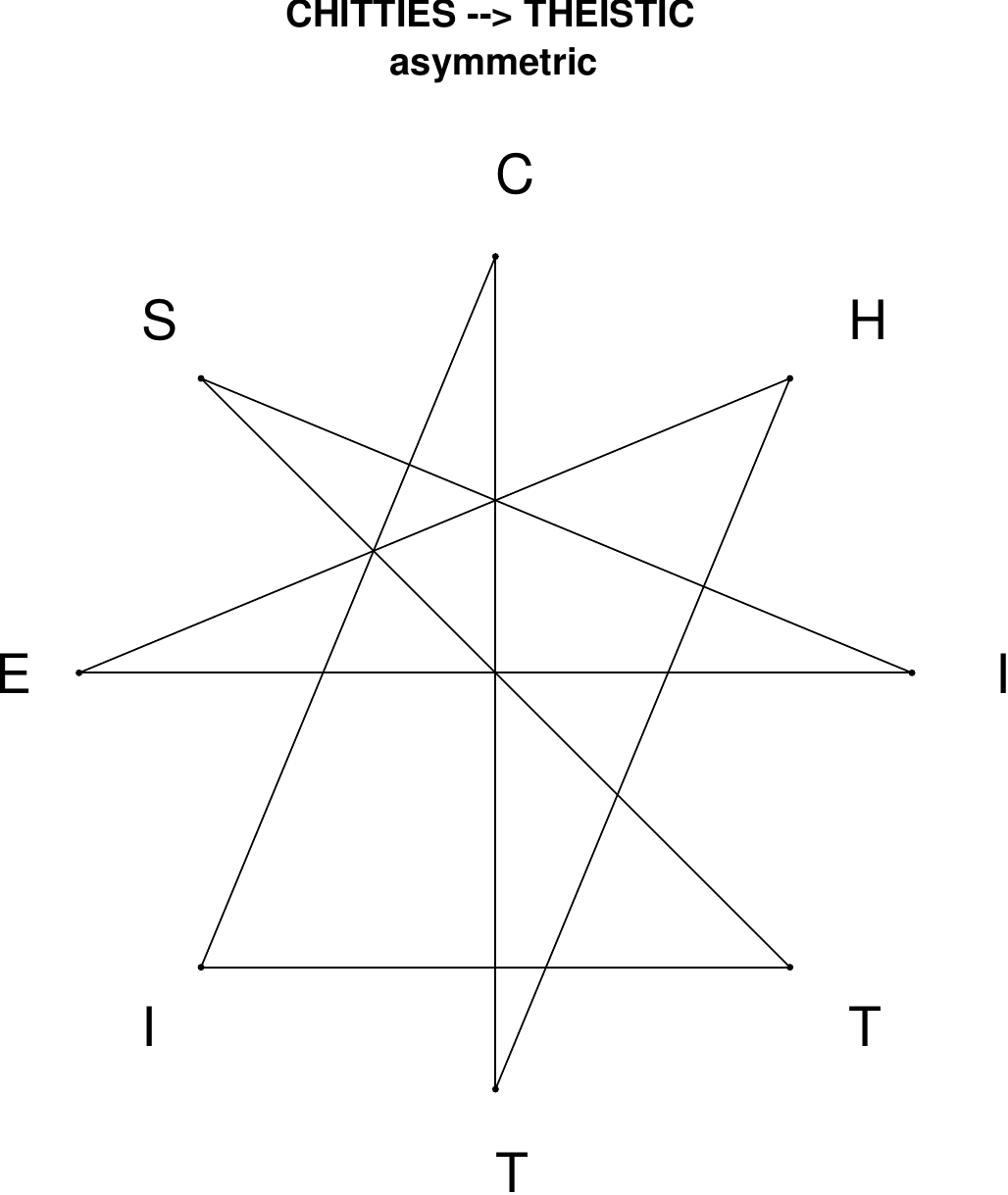}
\end{subfigure}
\hfill
\begin{subfigure}[T]{0.19\textwidth}
\centering
\includegraphics[width=\textwidth]{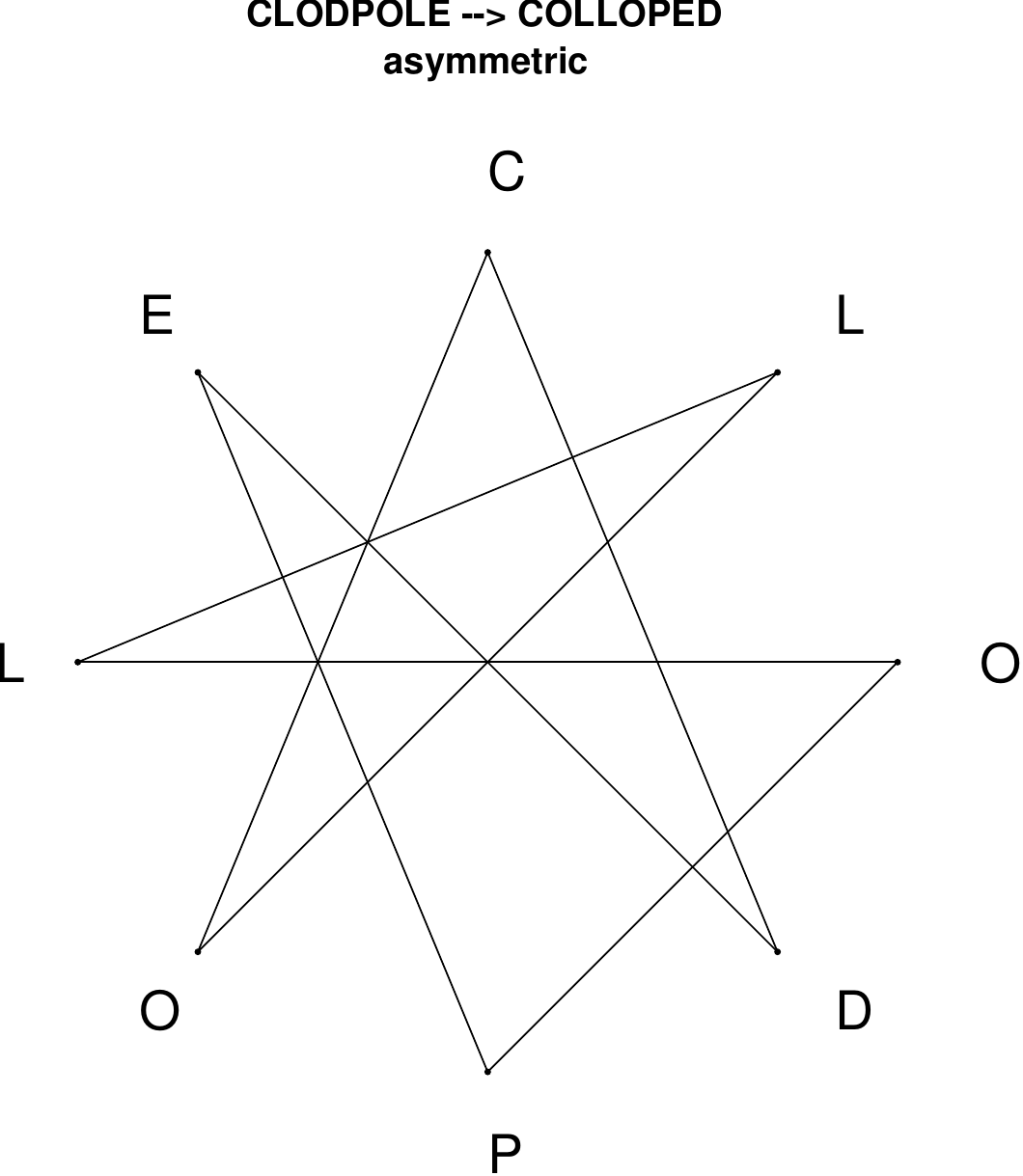}
\end{subfigure}
\end{figure}

\begin{figure}[H]
\centering
\begin{subfigure}[T]{0.19\textwidth}
\centering
\includegraphics[width=\textwidth]{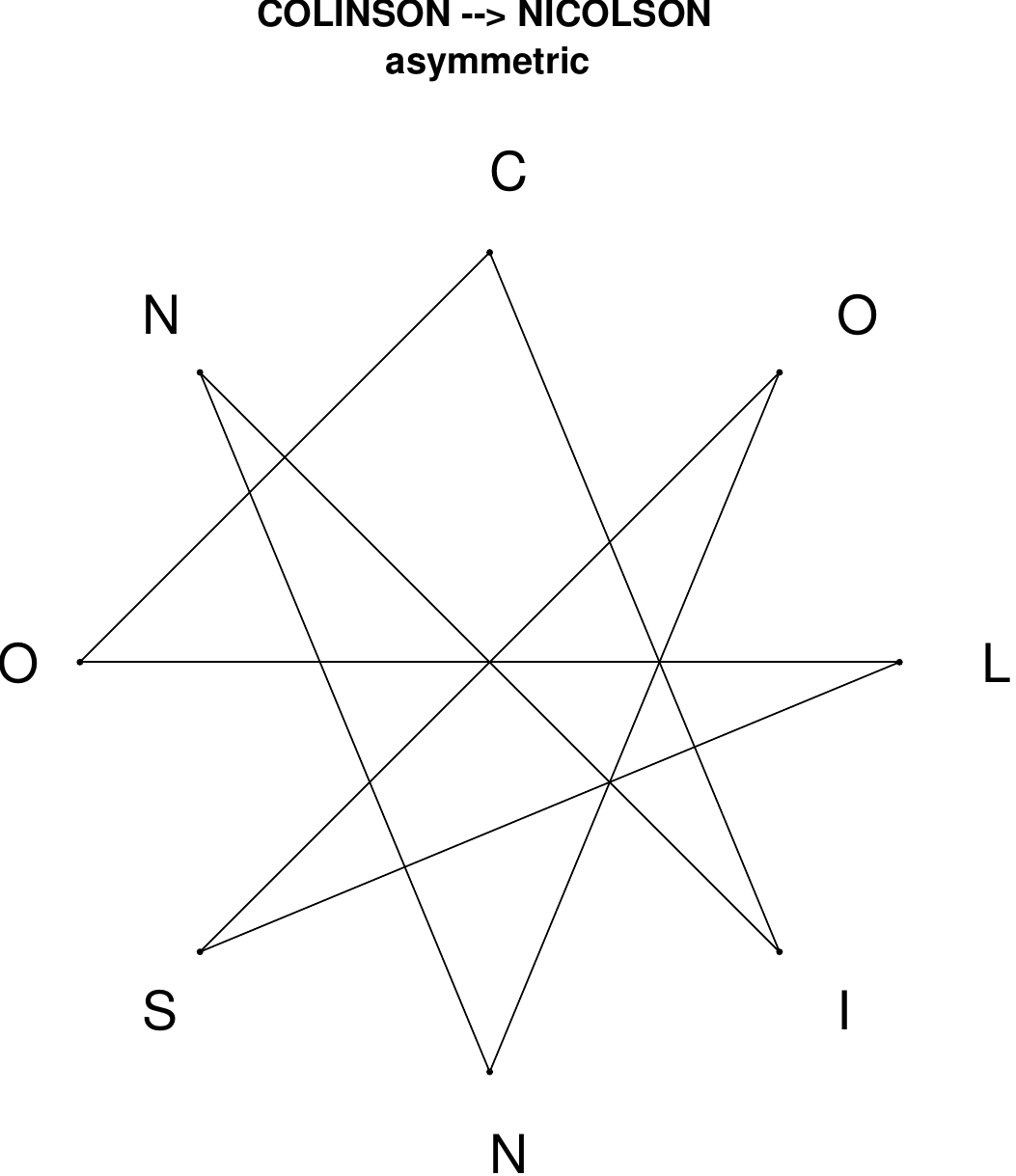}
\end{subfigure}
\hfill
\begin{subfigure}[T]{0.19\textwidth}
\centering
\includegraphics[width=\textwidth]{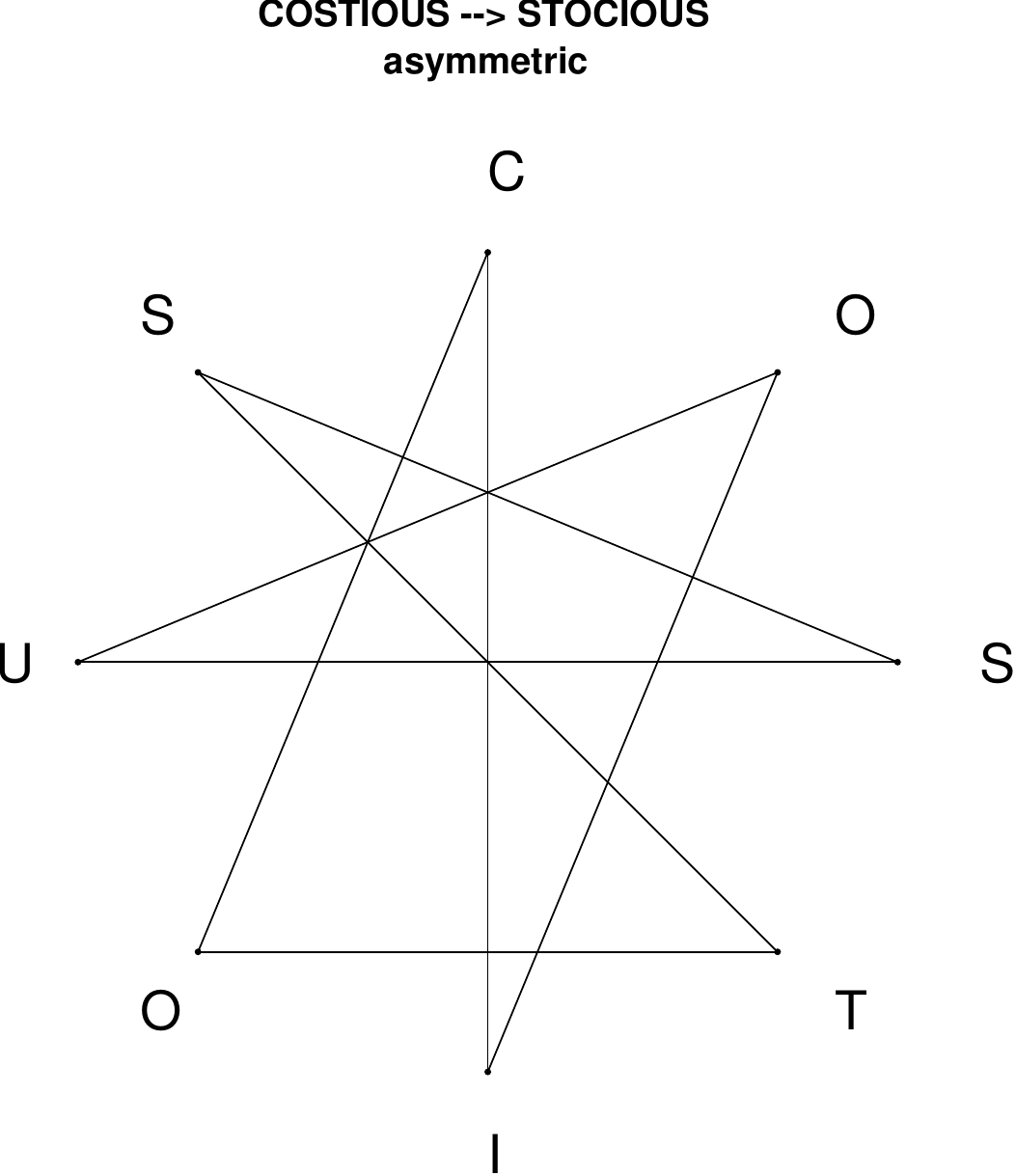}
\end{subfigure}
\hfill
\begin{subfigure}[T]{0.19\textwidth}
\centering
\includegraphics[width=\textwidth]{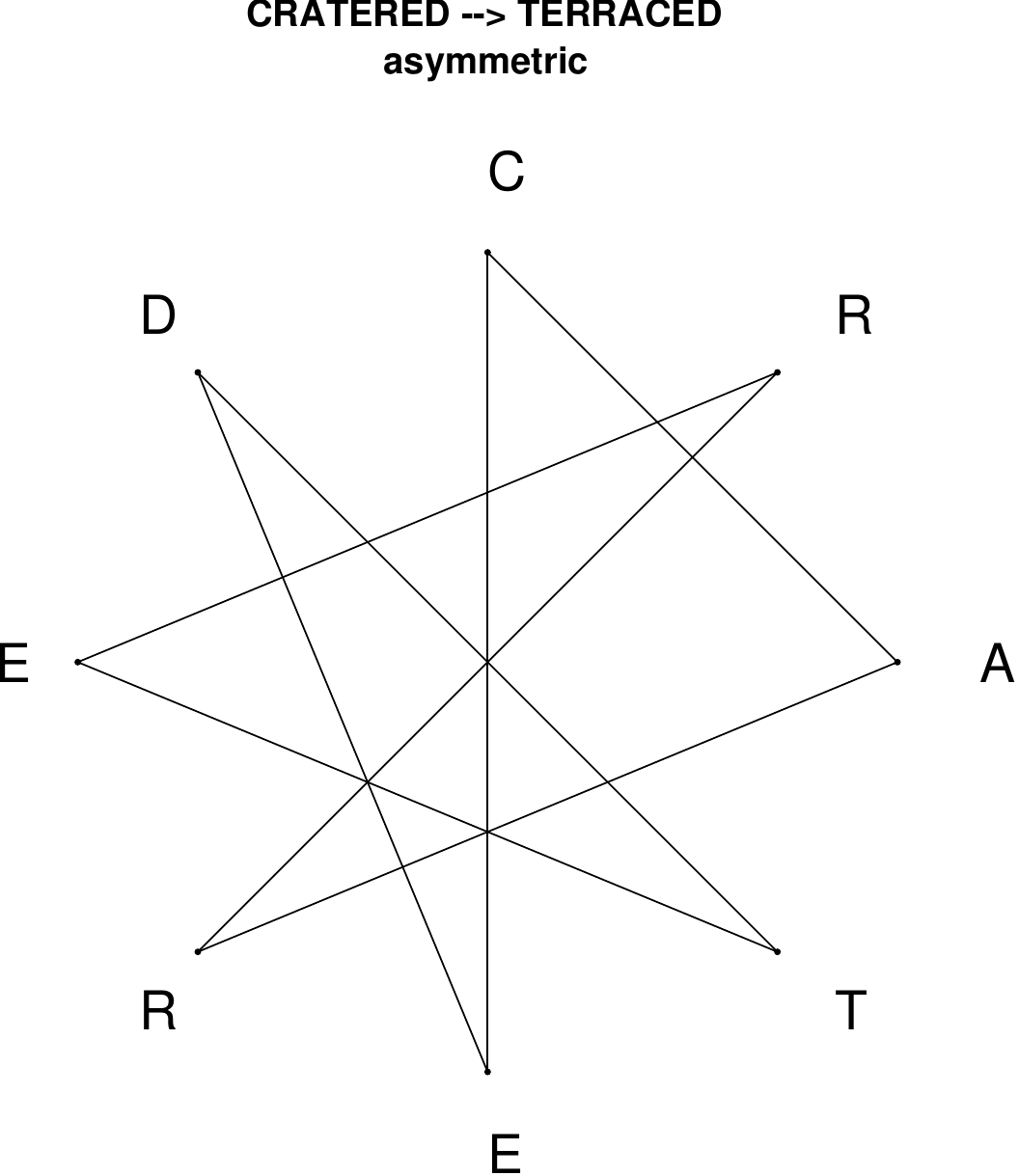}
\end{subfigure}
\hfill
\begin{subfigure}[T]{0.19\textwidth}
\centering
\includegraphics[width=\textwidth]{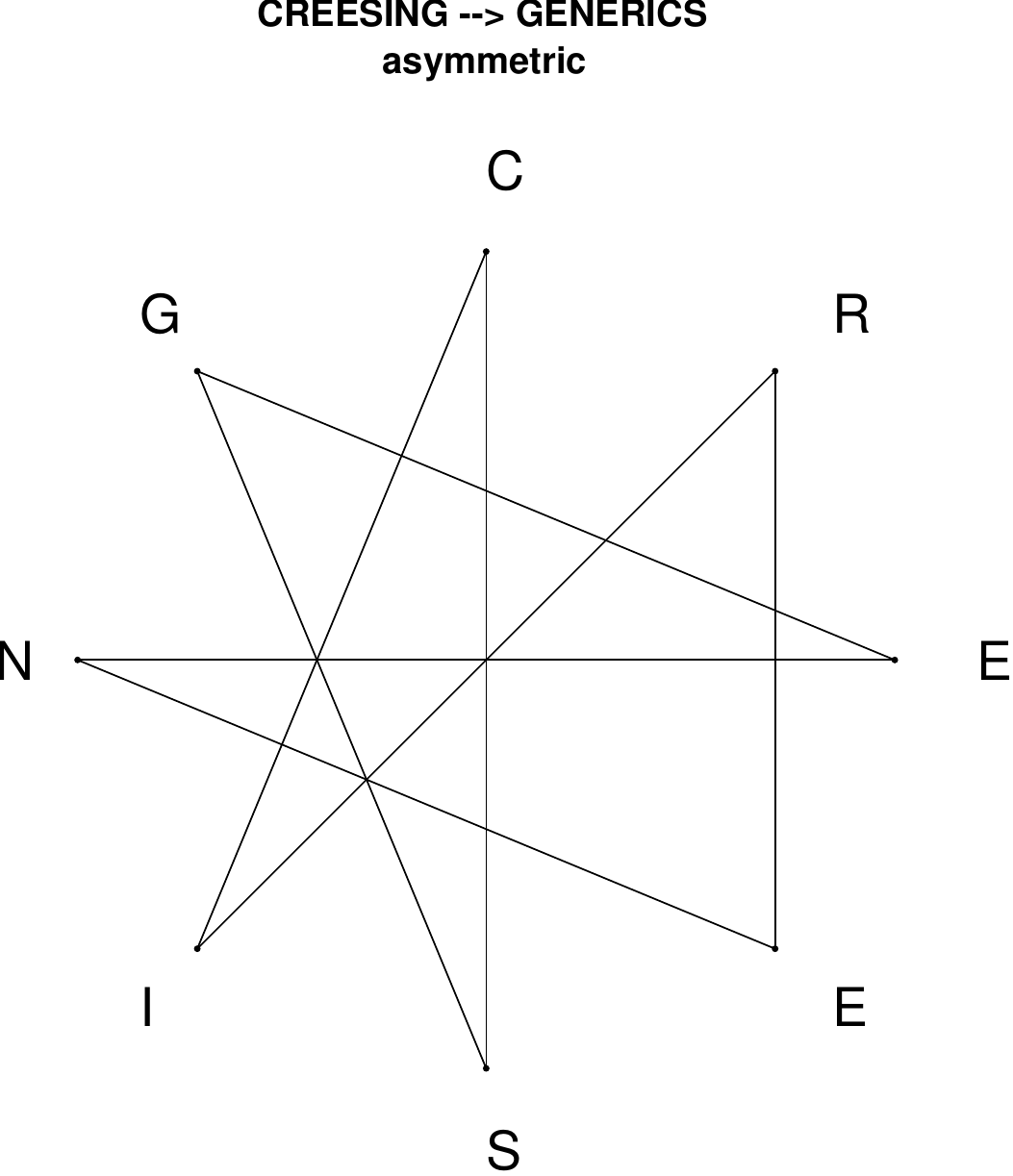}
\end{subfigure}
\hfill
\begin{subfigure}[T]{0.19\textwidth}
\centering
\includegraphics[width=\textwidth]{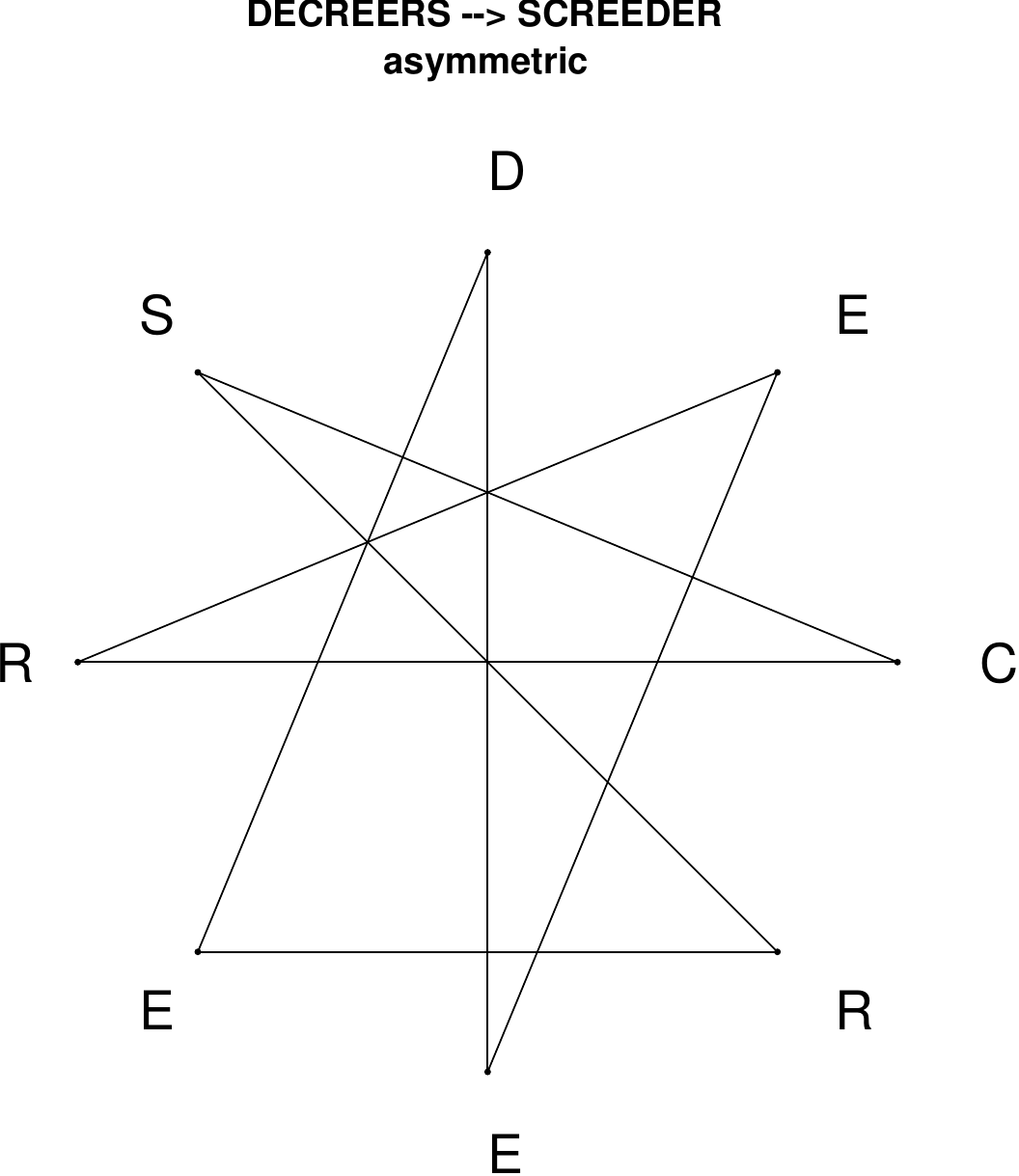}
\end{subfigure}
\end{figure}

\begin{figure}[H]
\centering
\begin{subfigure}[T]{0.19\textwidth}
\centering
\includegraphics[width=\textwidth]{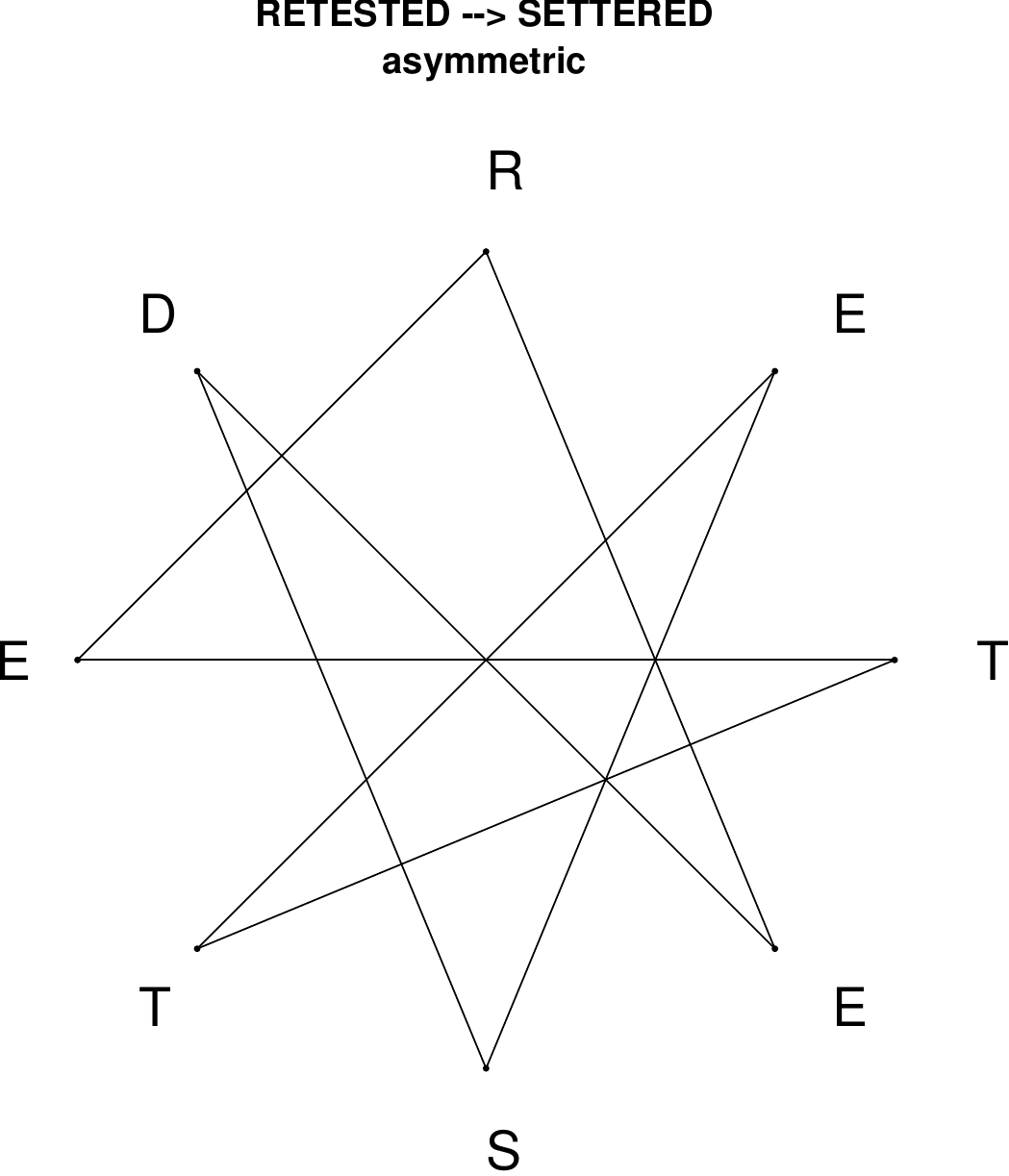}
\end{subfigure}
\hfill
\begin{subfigure}[T]{0.19\textwidth}
\centering
\includegraphics[width=\textwidth]{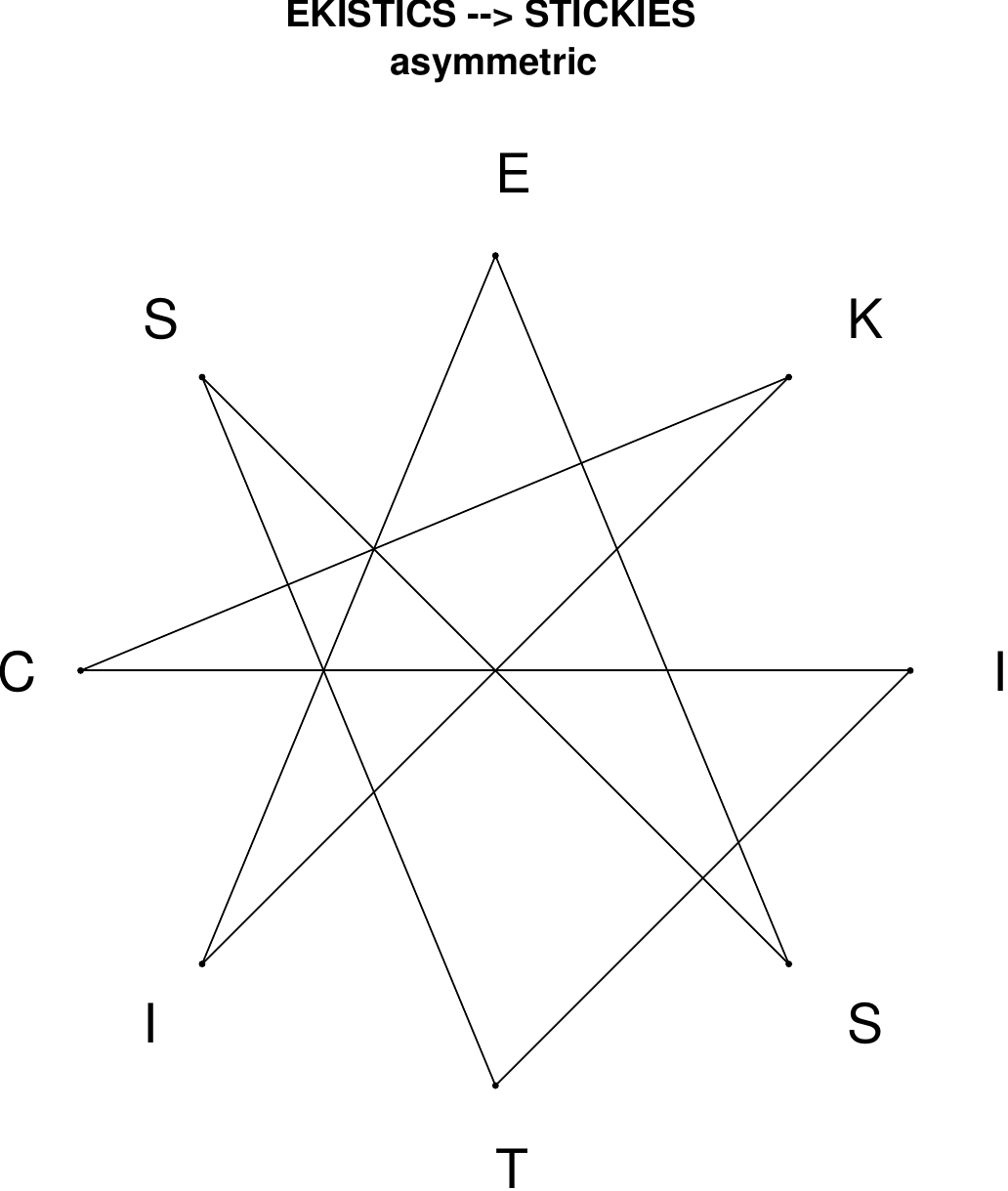}
\end{subfigure}
\hfill
\begin{subfigure}[T]{0.19\textwidth}
\centering
\includegraphics[width=\textwidth]{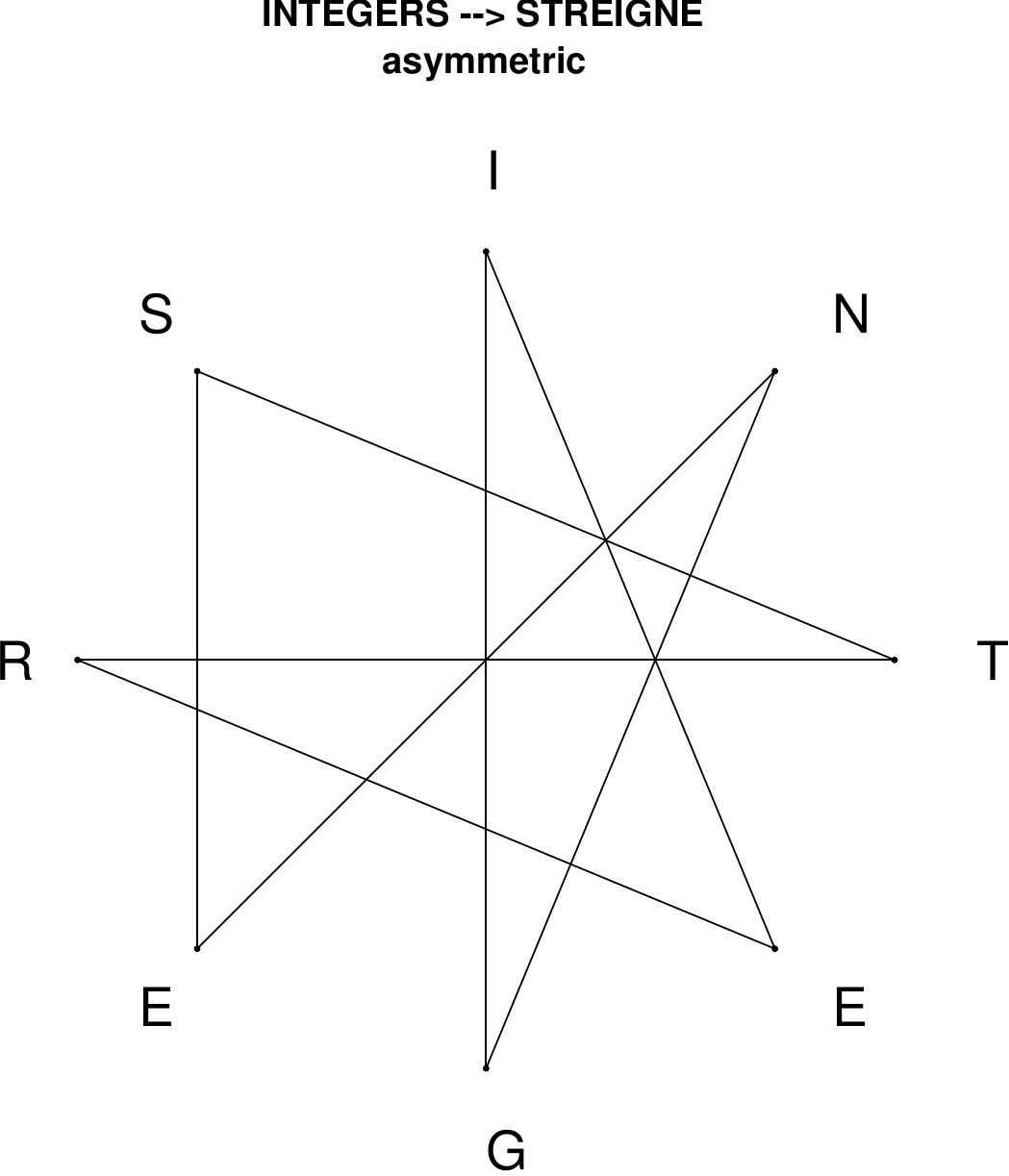}
\end{subfigure}
\hfill
\begin{subfigure}[T]{0.19\textwidth}
\centering
\includegraphics[width=\textwidth]{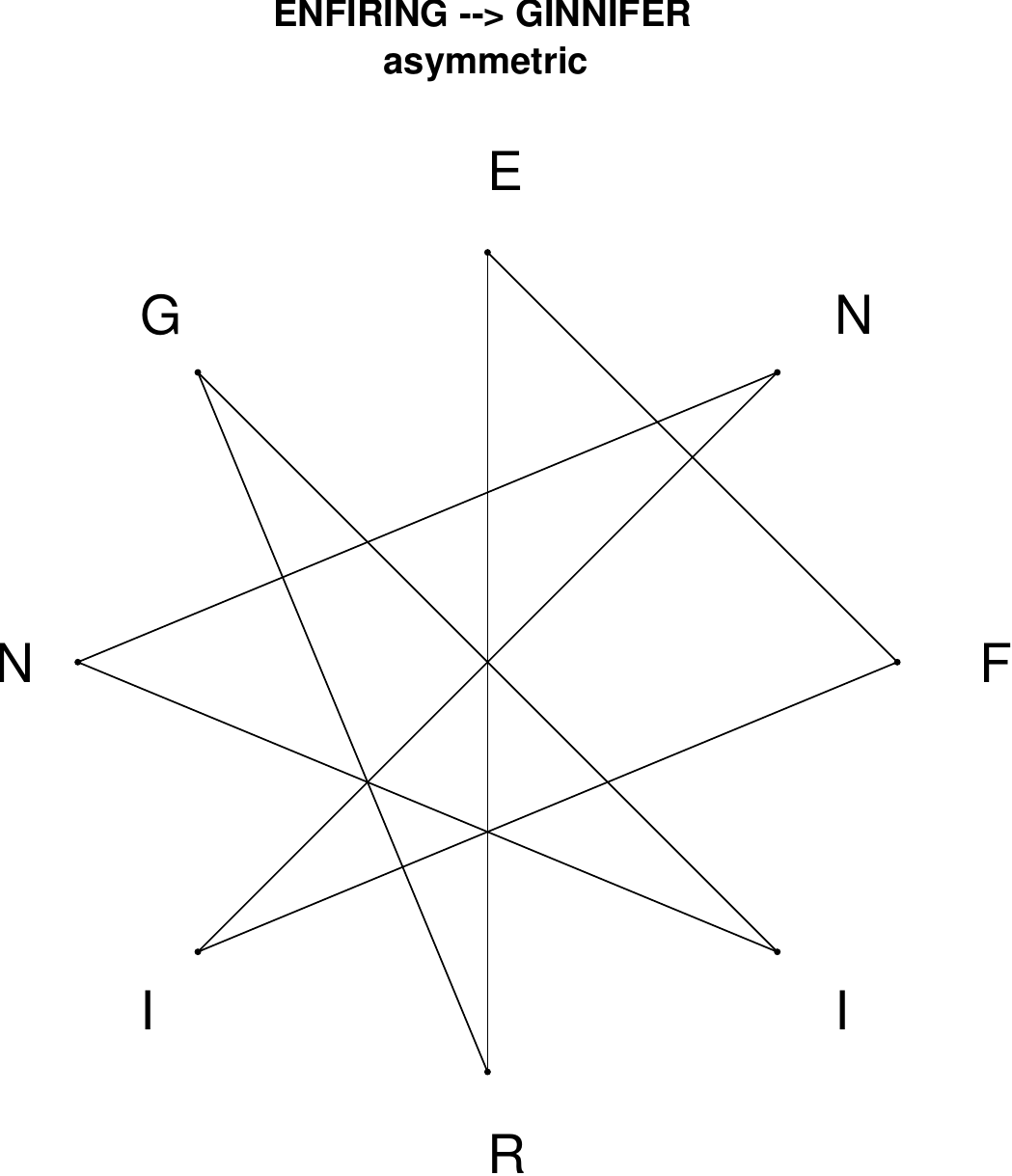}
\end{subfigure}
\hfill
\begin{subfigure}[T]{0.19\textwidth}
\centering
\includegraphics[width=\textwidth]{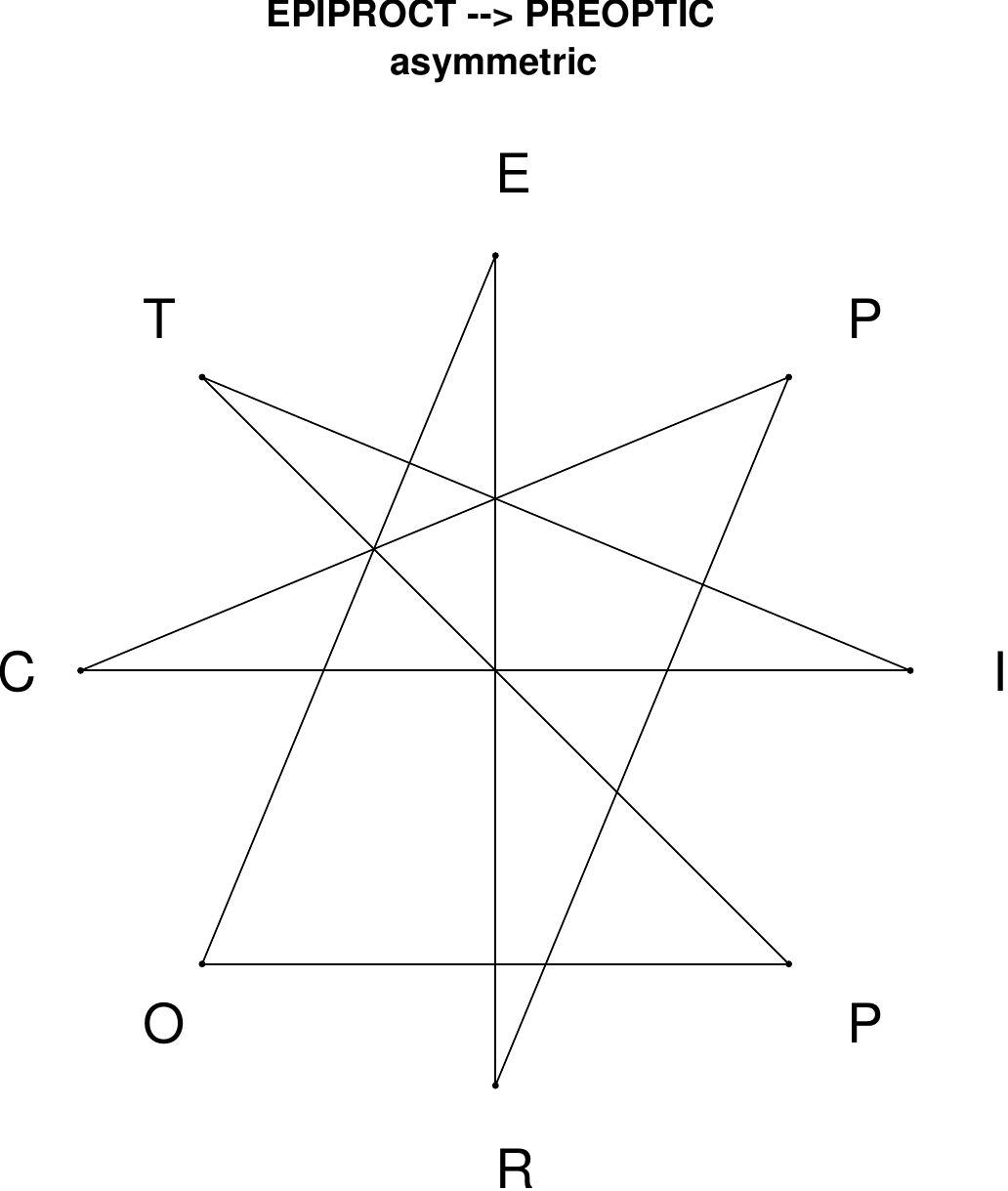}
\end{subfigure}
\end{figure}

\begin{figure}[H]
\centering
\begin{subfigure}[T]{0.19\textwidth}
\centering
\includegraphics[width=\textwidth]{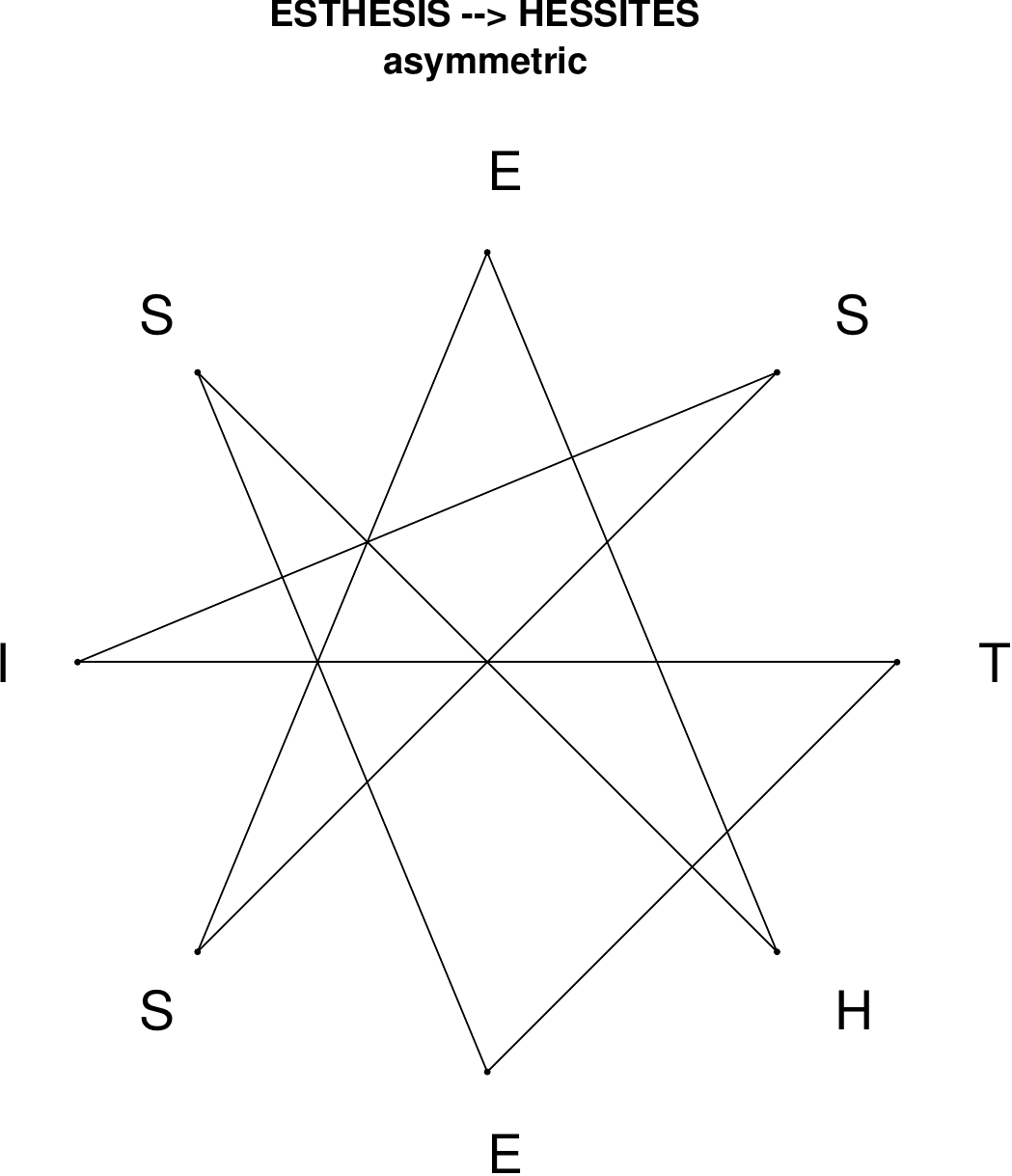}
\end{subfigure}
\hfill
\begin{subfigure}[T]{0.19\textwidth}
\centering
\includegraphics[width=\textwidth]{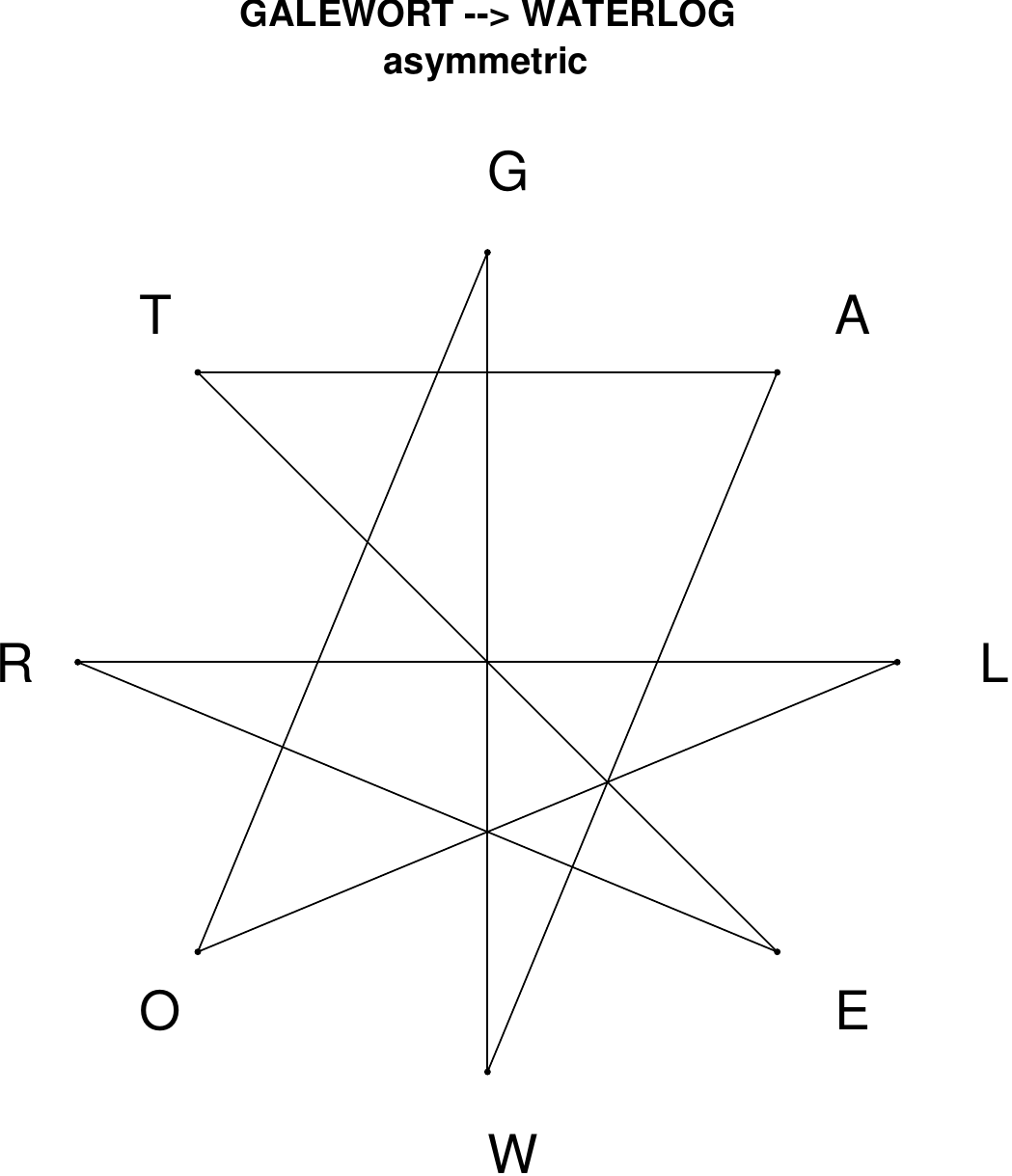}
\end{subfigure}
\hfill
\begin{subfigure}[T]{0.19\textwidth}
\centering
\includegraphics[width=\textwidth]{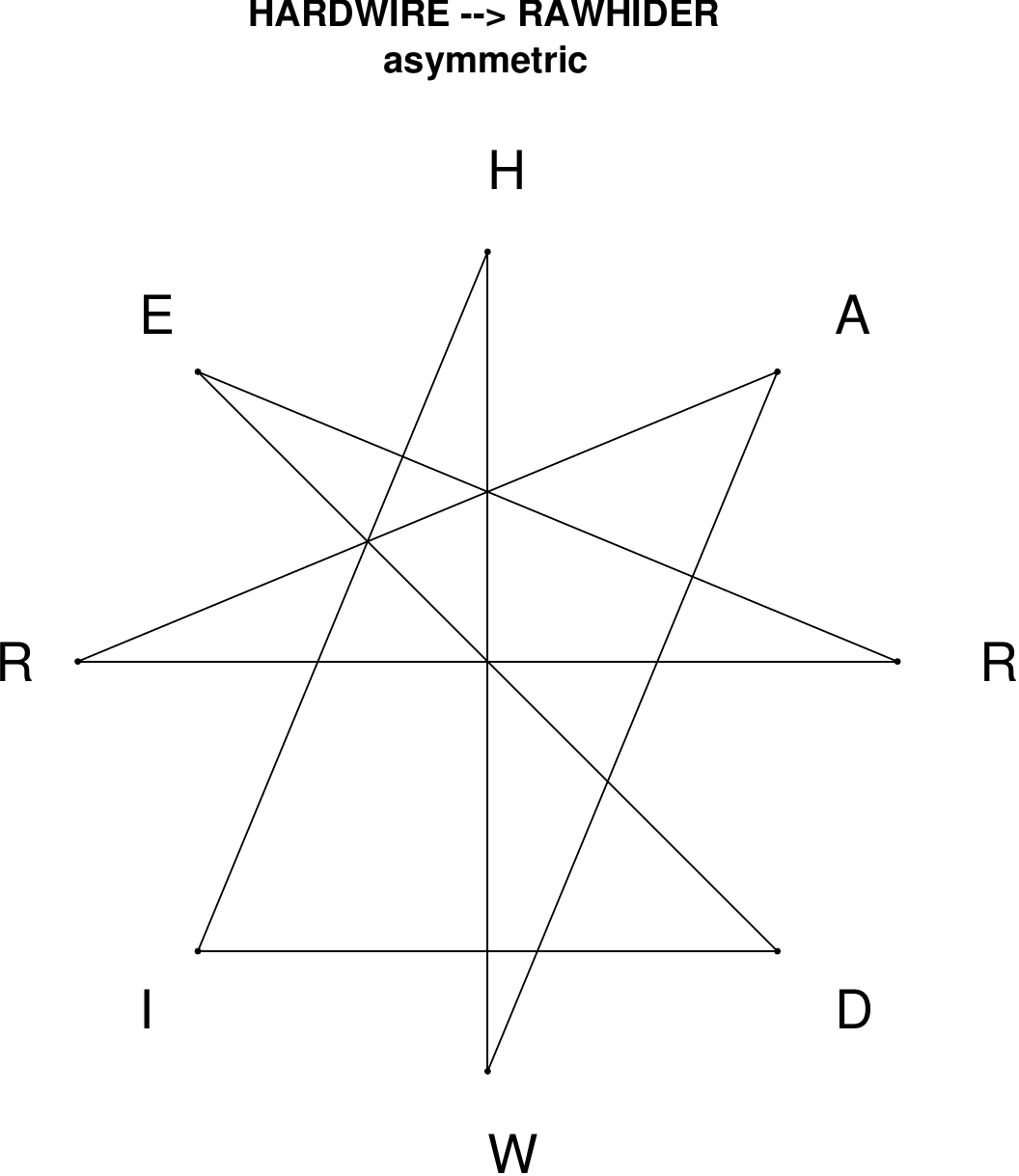}
\end{subfigure}
\hfill
\begin{subfigure}[T]{0.19\textwidth}
\centering
\includegraphics[width=\textwidth]{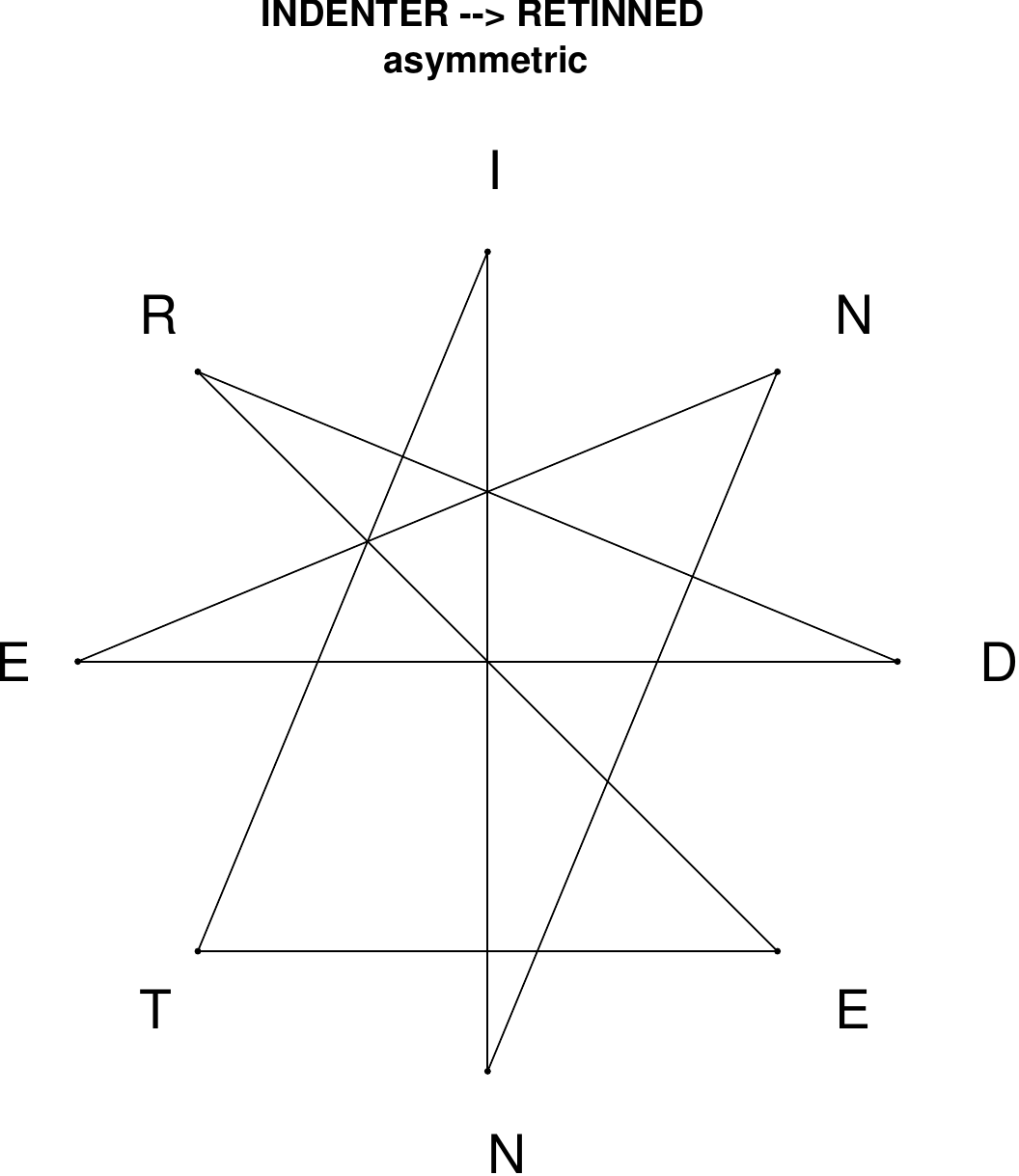}
\end{subfigure}
\hfill
\begin{subfigure}[T]{0.19\textwidth}
\centering
\includegraphics[width=\textwidth]{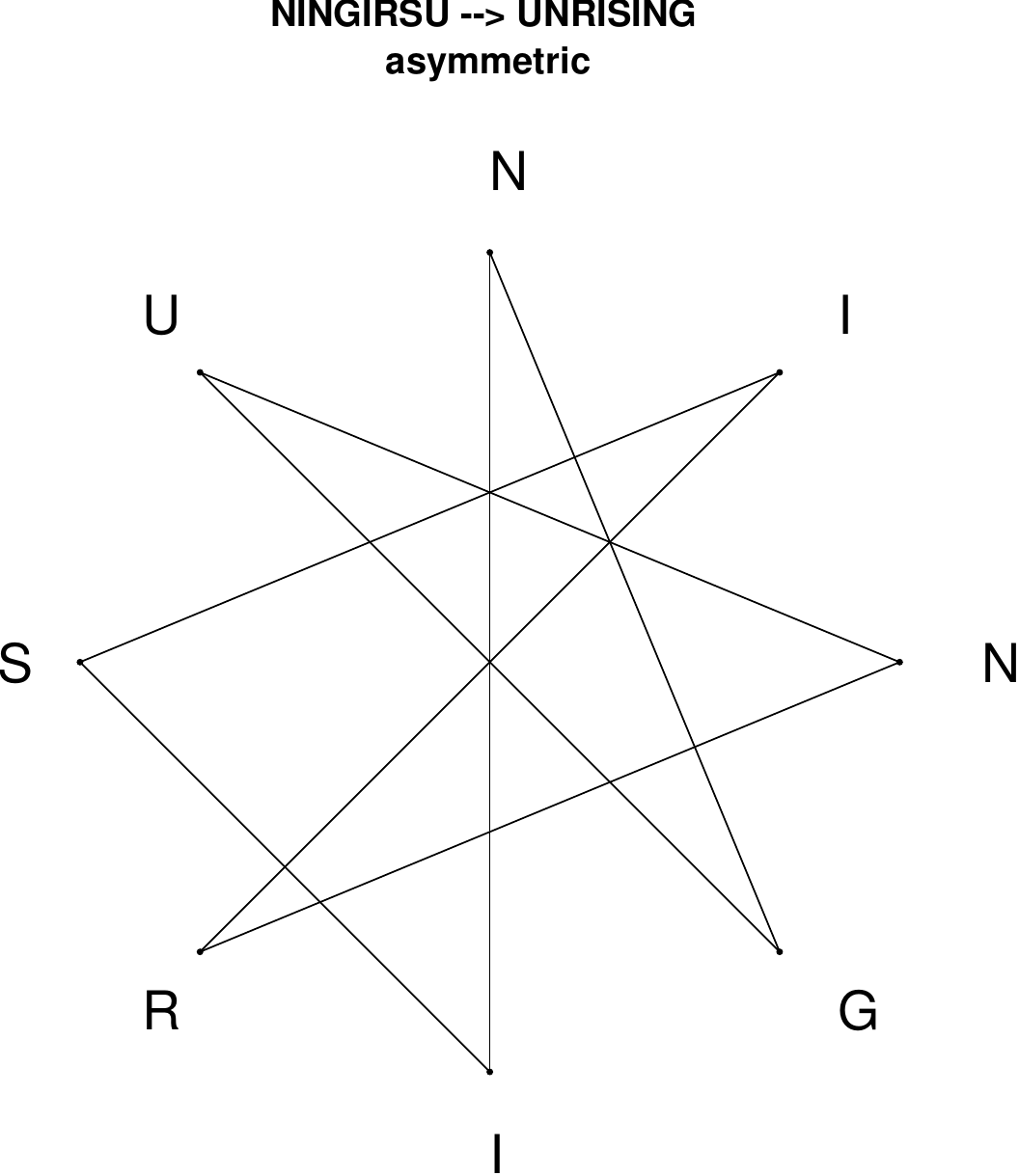}
\end{subfigure}
\end{figure}

\begin{figure}[H]
\centering
\begin{subfigure}[T]{0.19\textwidth}
\centering
\includegraphics[width=\textwidth]{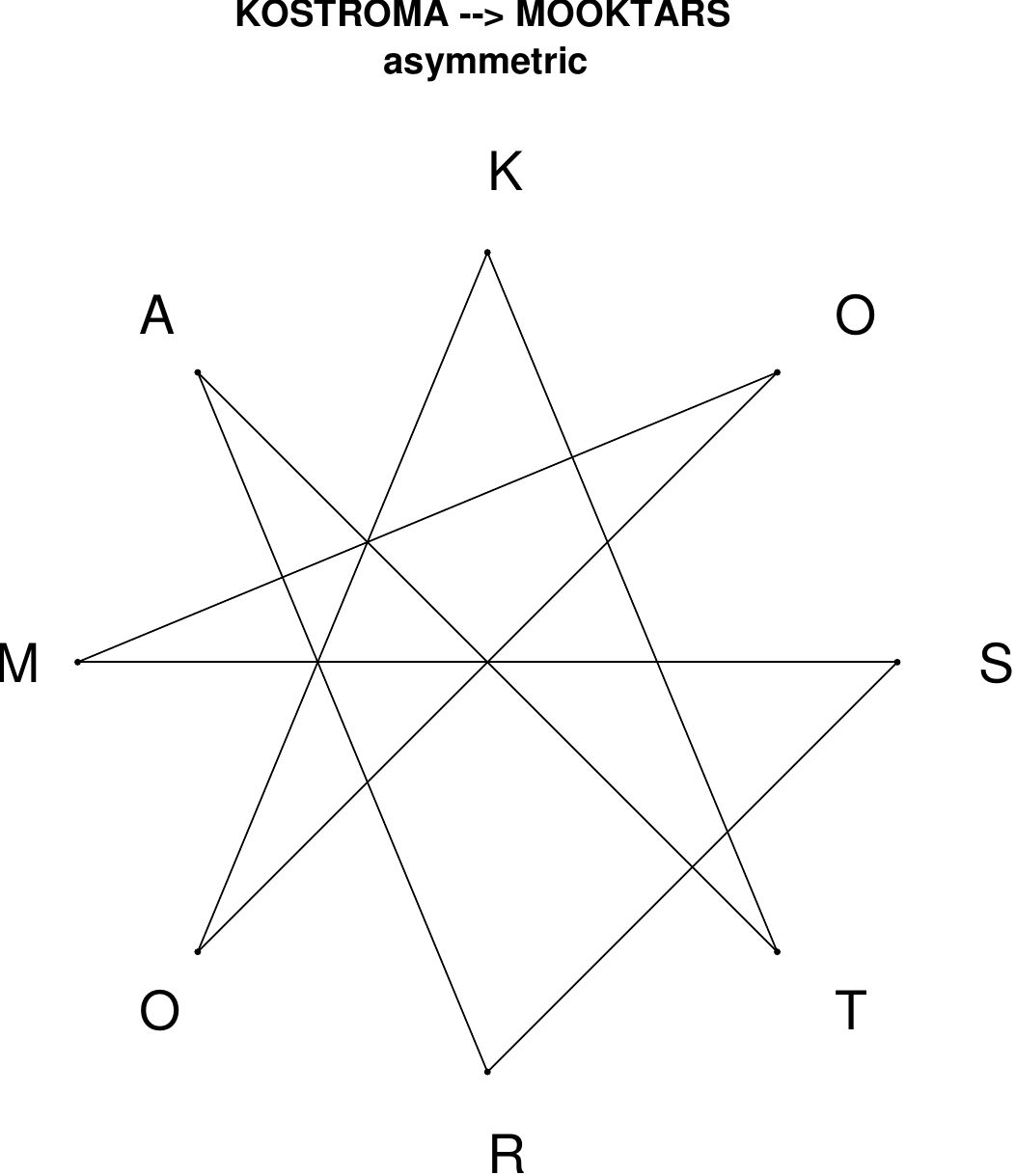}
\end{subfigure}
\hfill
\begin{subfigure}[T]{0.19\textwidth}
\centering
\includegraphics[width=\textwidth]{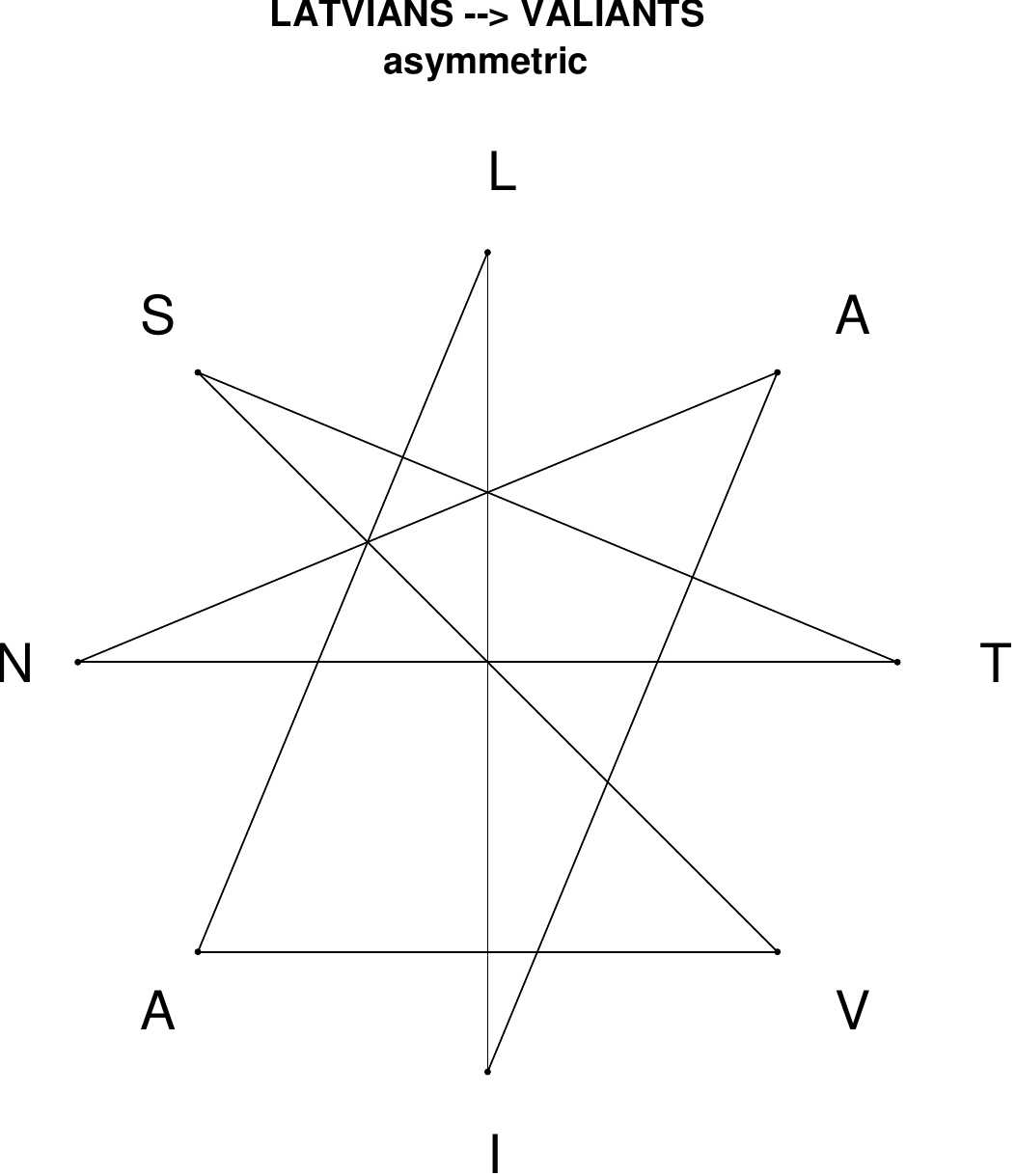}
\end{subfigure}
\hfill
\begin{subfigure}[T]{0.19\textwidth}
\centering
\includegraphics[width=\textwidth]{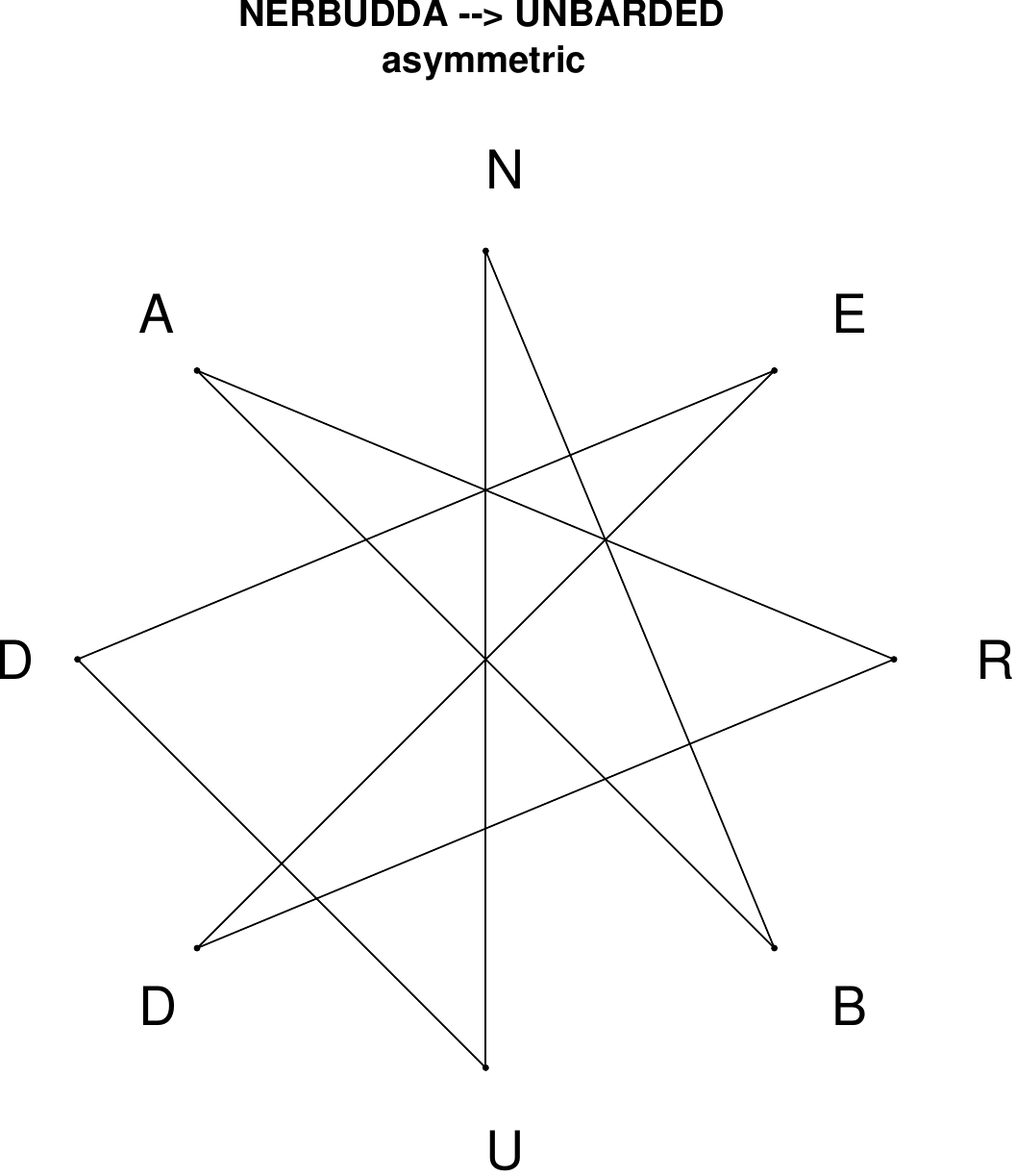}
\end{subfigure}
\hfill
\begin{subfigure}[T]{0.19\textwidth}
\centering
\includegraphics[width=\textwidth]{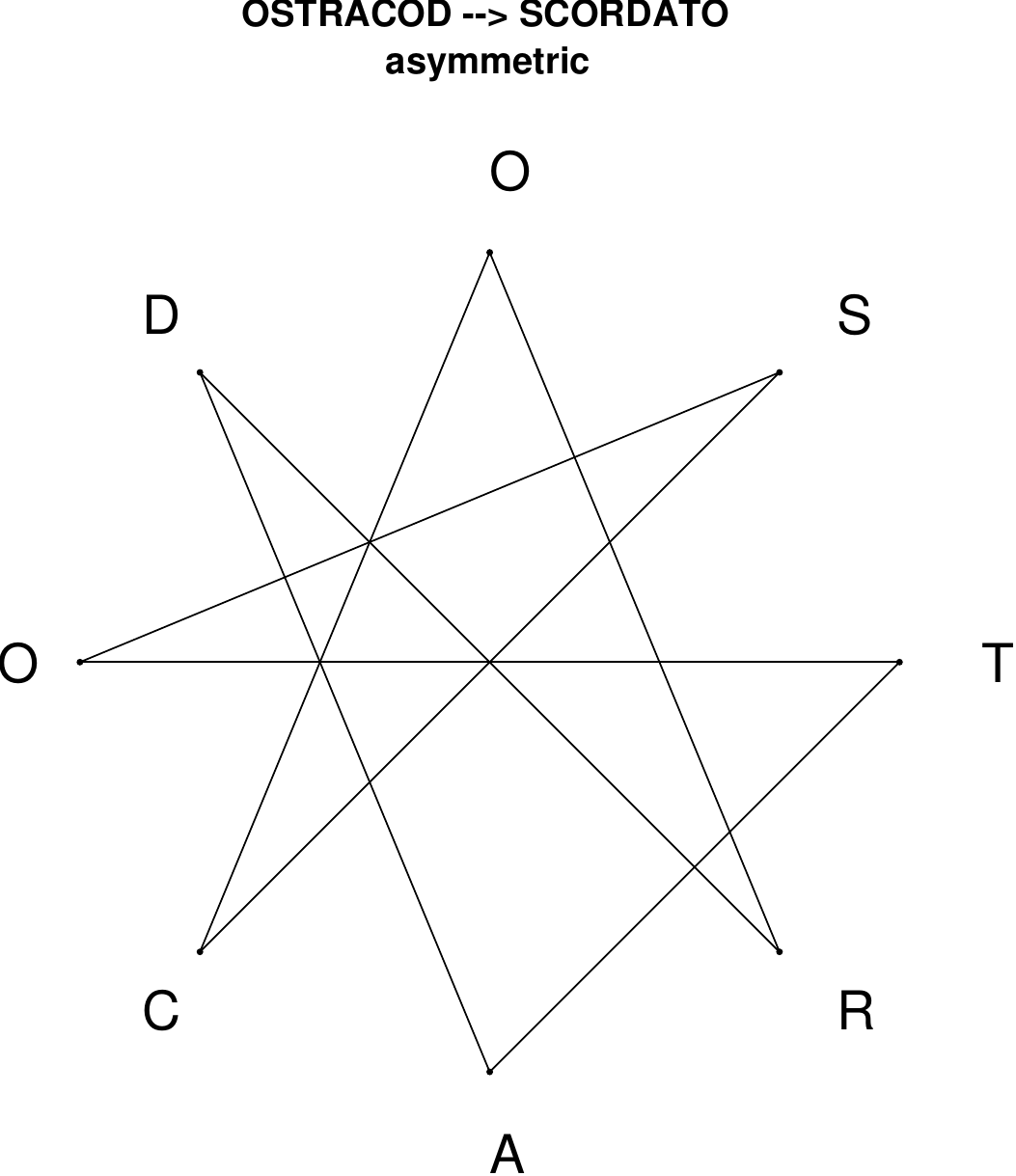}
\end{subfigure}
\hfill
\begin{subfigure}[T]{0.19\textwidth}
\centering
\includegraphics[width=\textwidth]{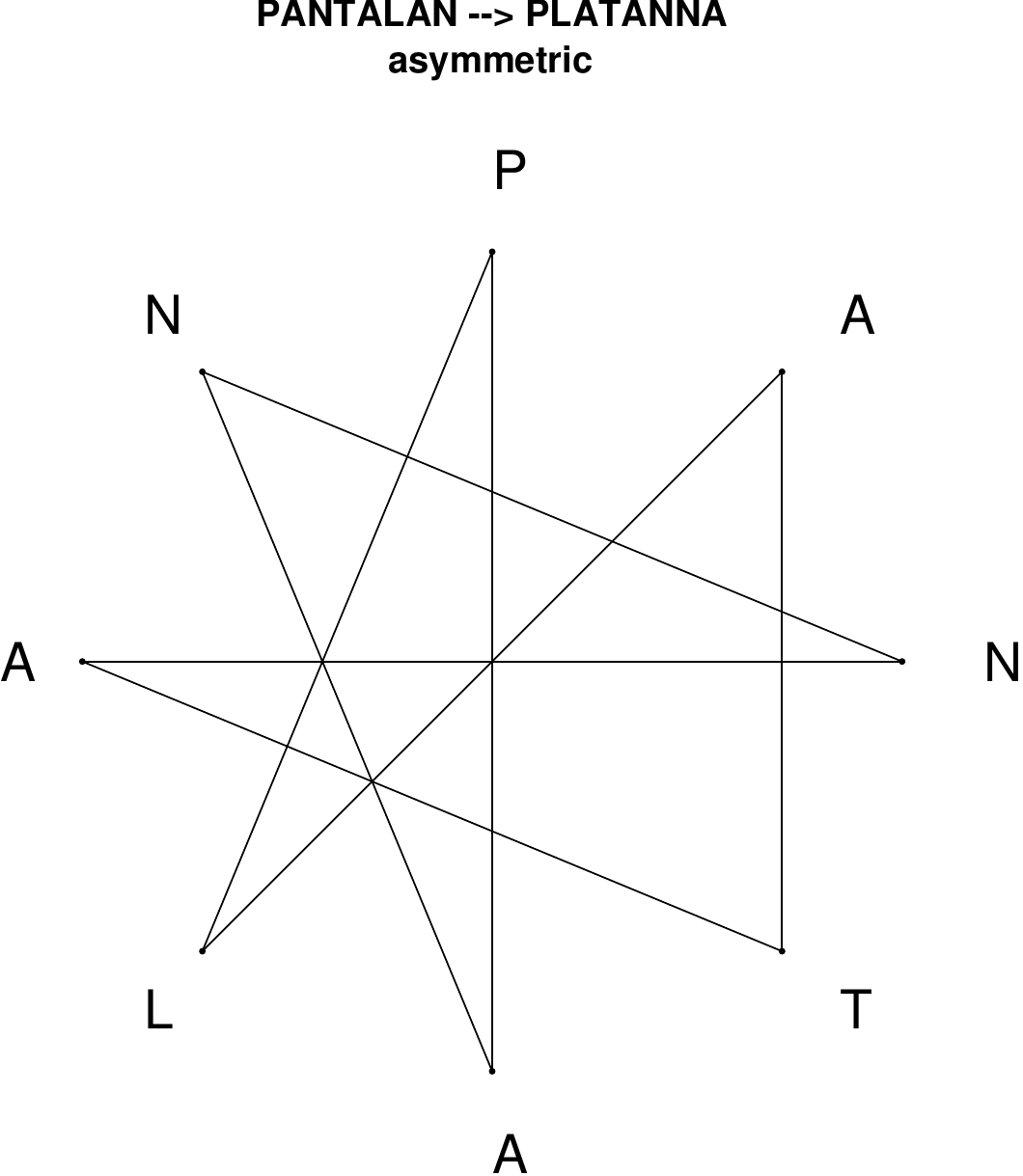}
\end{subfigure}
\end{figure}

\begin{figure}[H]
\centering
\begin{subfigure}[T]{0.19\textwidth}
\centering
\includegraphics[width=\textwidth]{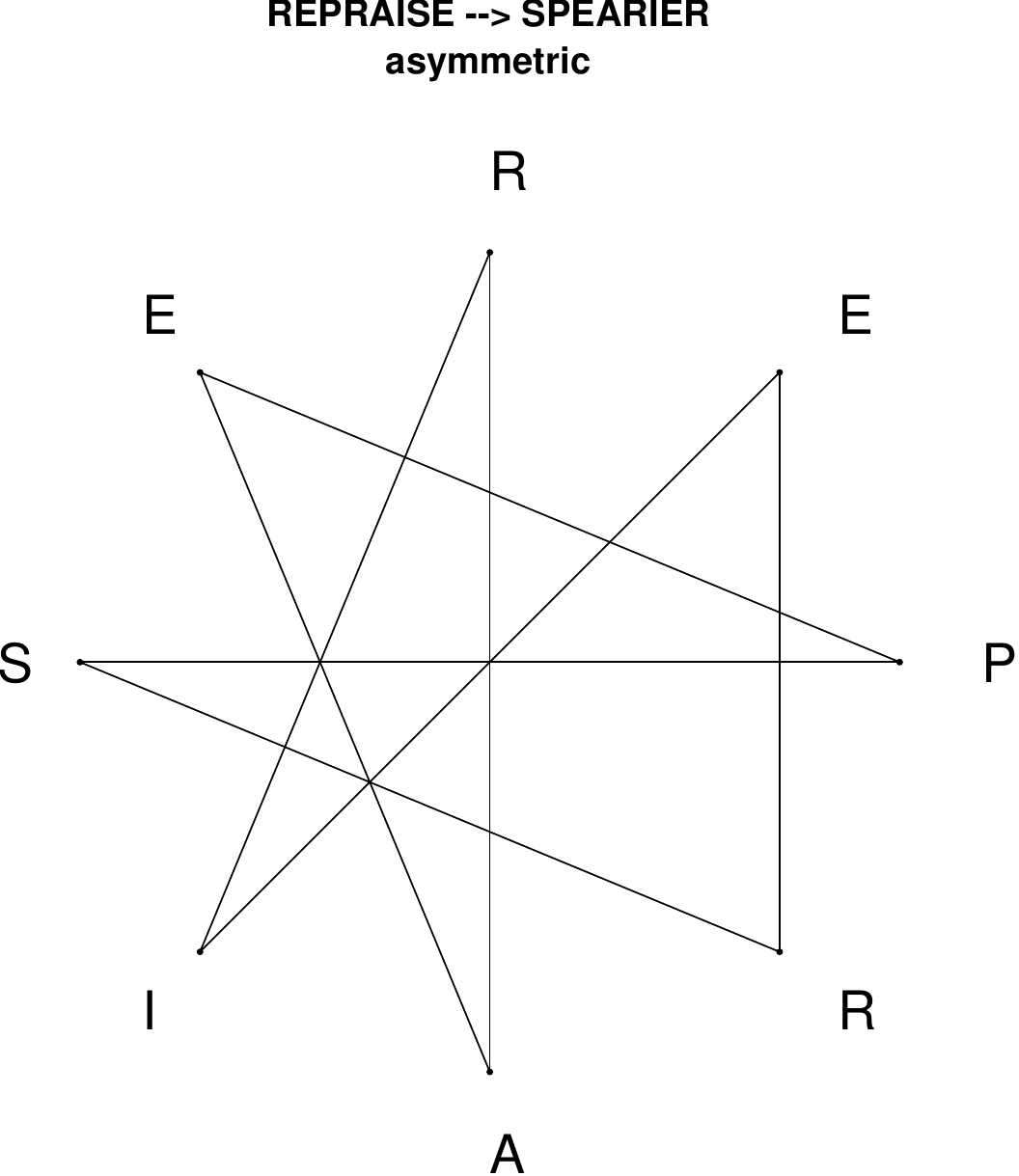}
\end{subfigure}
\hfill
\begin{subfigure}[T]{0.19\textwidth}
\centering
\includegraphics[width=\textwidth]{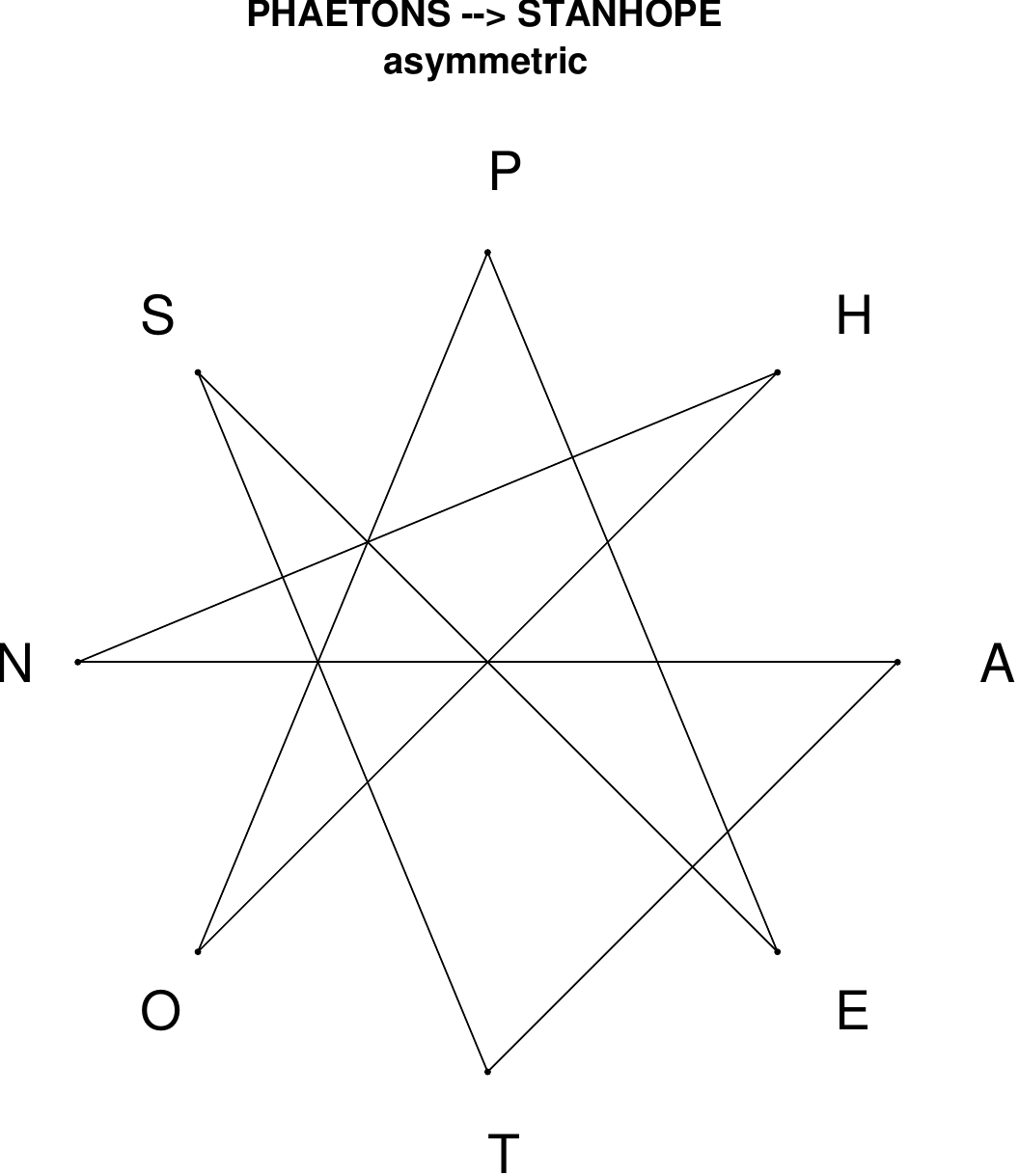}
\end{subfigure}
\hfill
\begin{subfigure}[T]{0.19\textwidth}
\centering
\includegraphics[width=\textwidth]{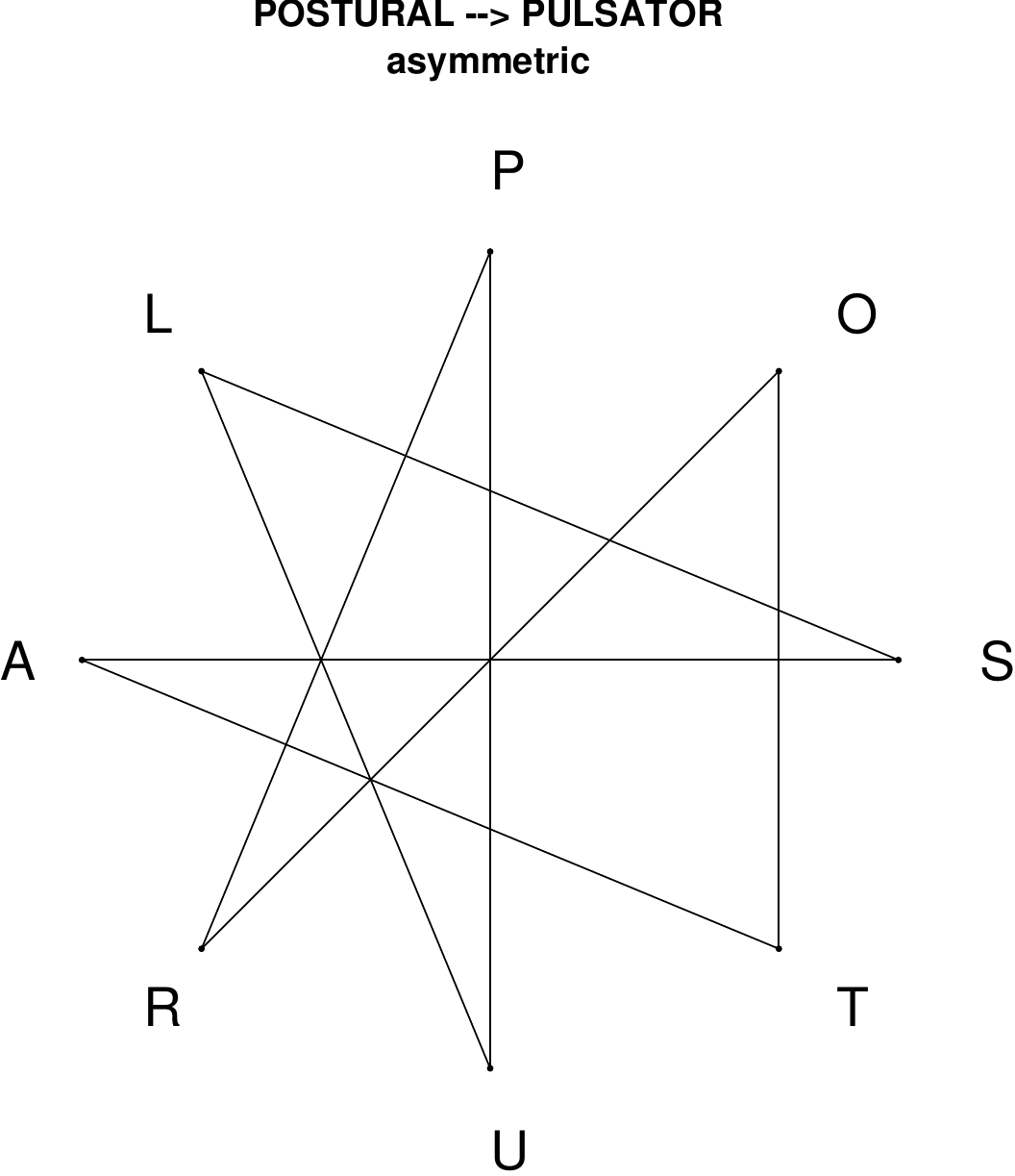}
\end{subfigure}
\hfill
\begin{subfigure}[T]{0.19\textwidth}
\centering
\includegraphics[width=\textwidth]{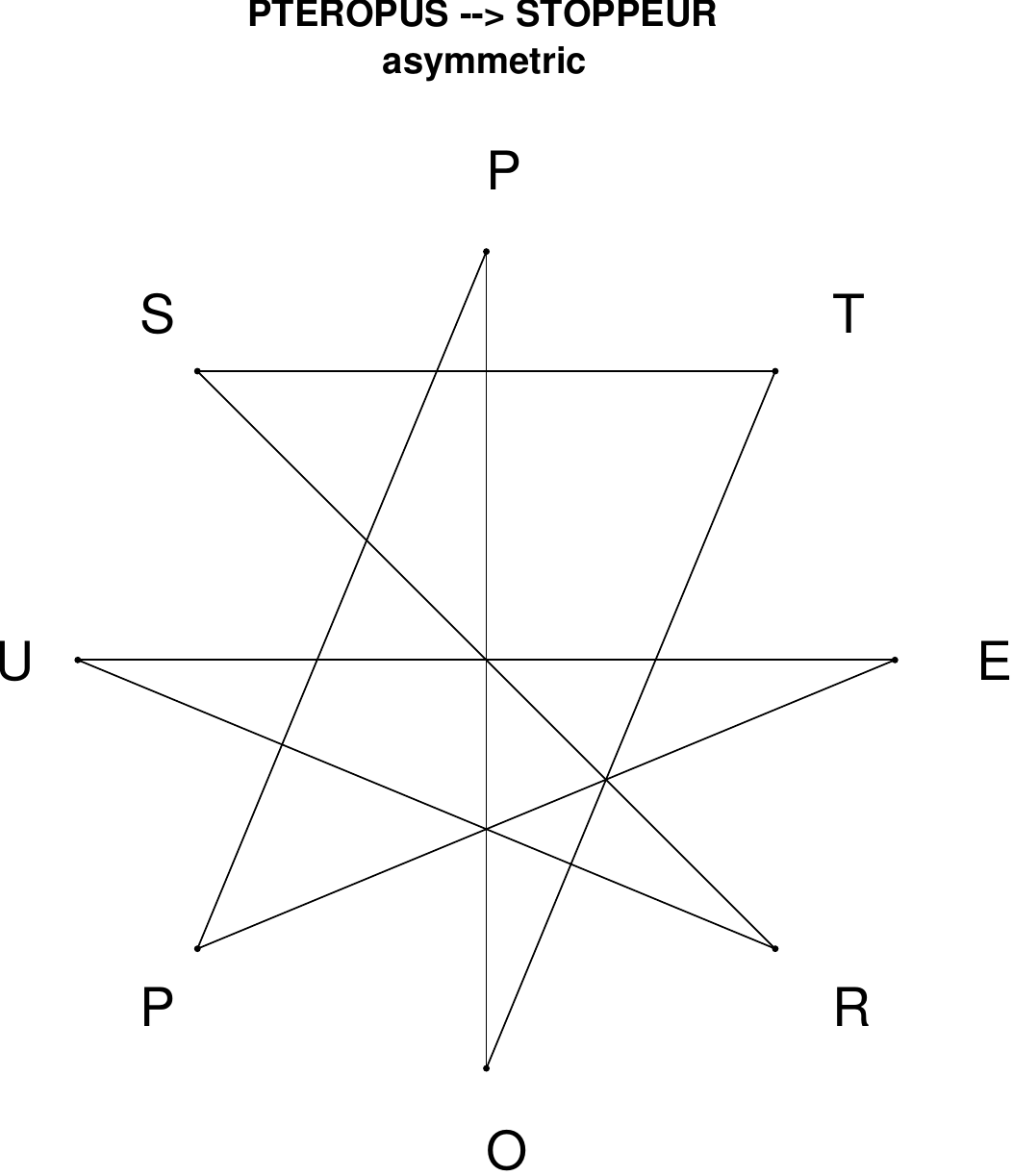}
\end{subfigure}
\hfill
\begin{subfigure}[T]{0.19\textwidth}
\centering
\includegraphics[width=\textwidth]{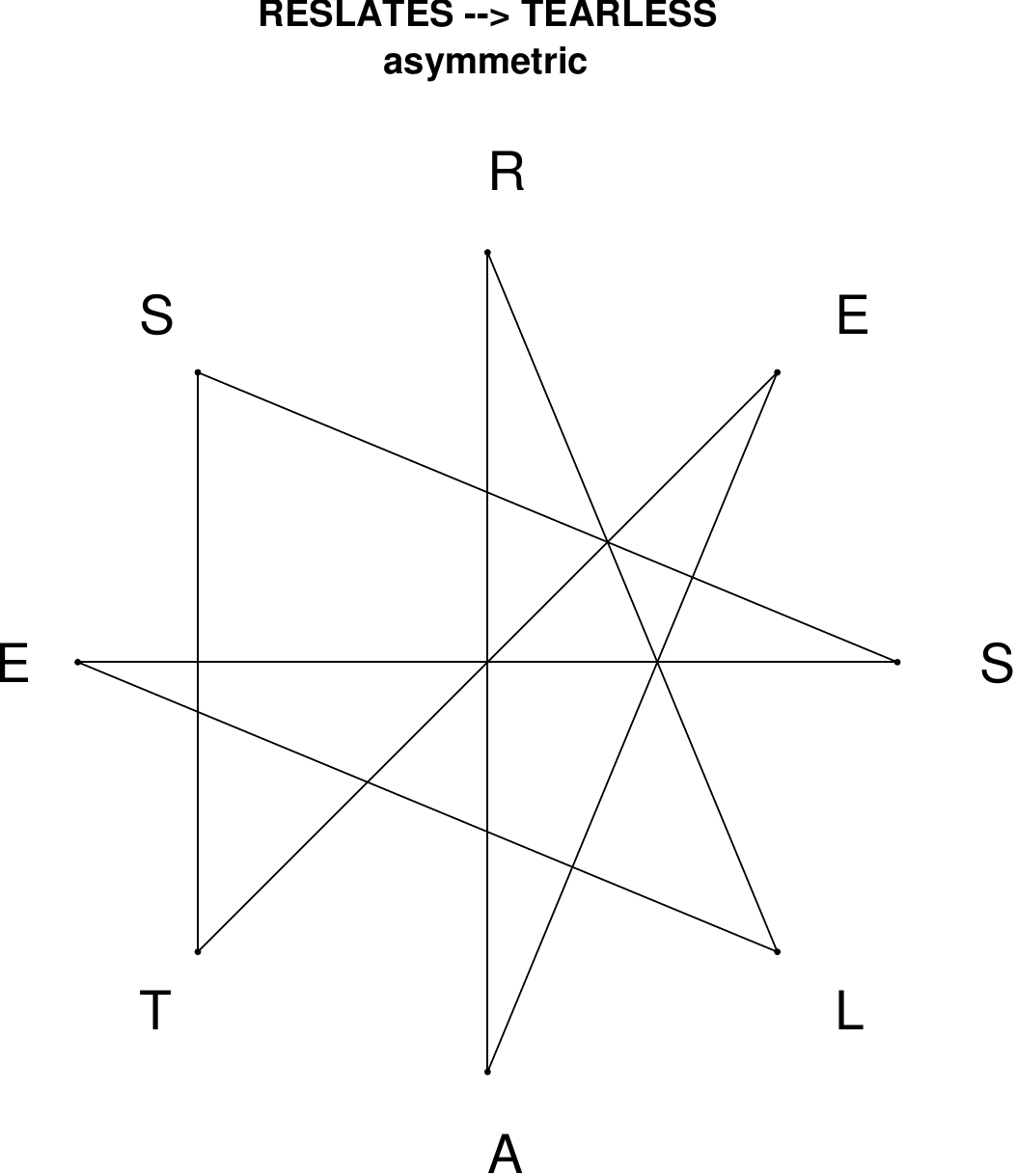}
\end{subfigure}
\end{figure}

\begin{figure}[H]
\centering
\begin{subfigure}[T]{0.19\textwidth}
\centering
\includegraphics[width=\textwidth]{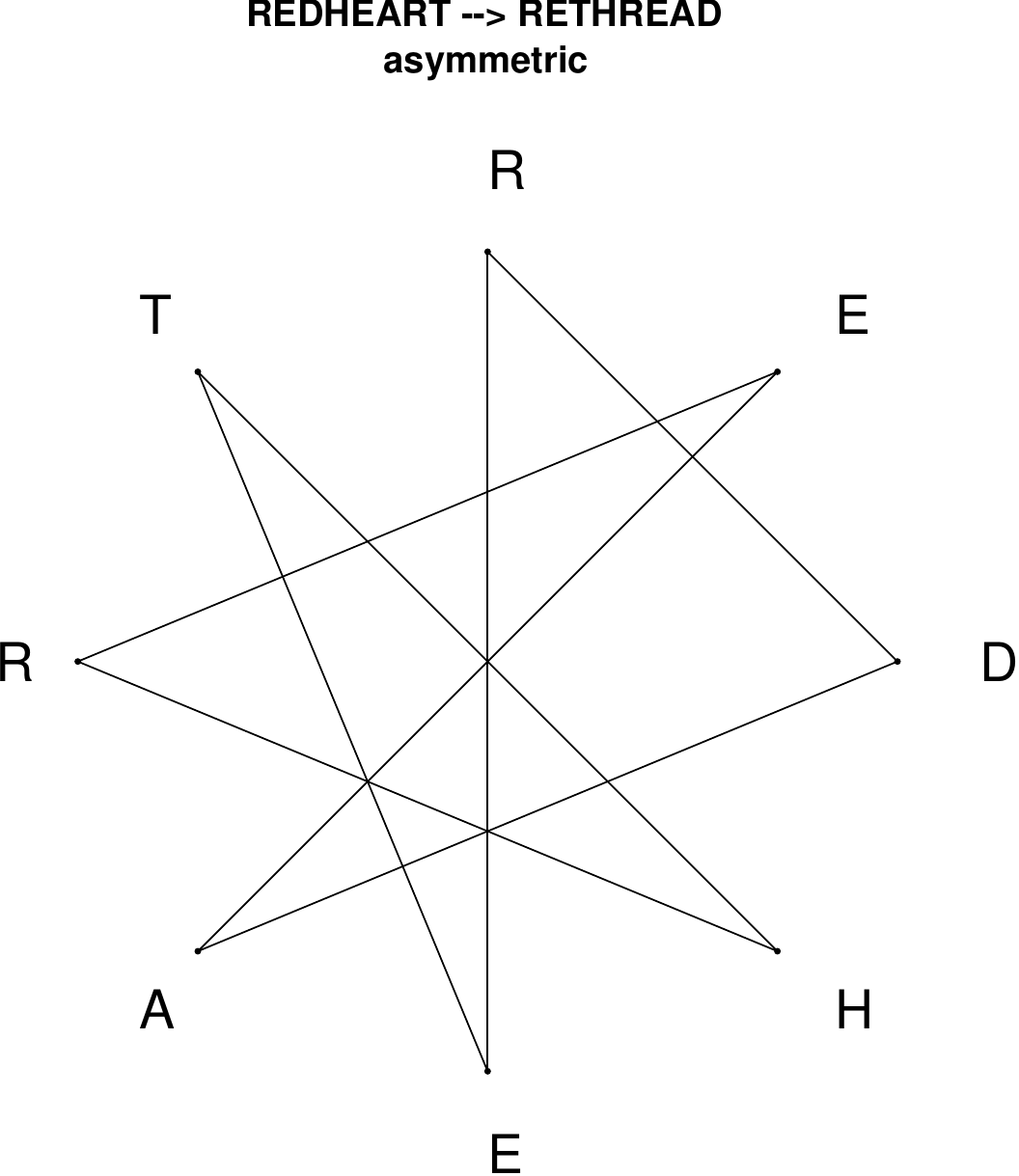}
\end{subfigure}
\hfill
\begin{subfigure}[T]{0.19\textwidth}
\centering
\includegraphics[width=\textwidth]{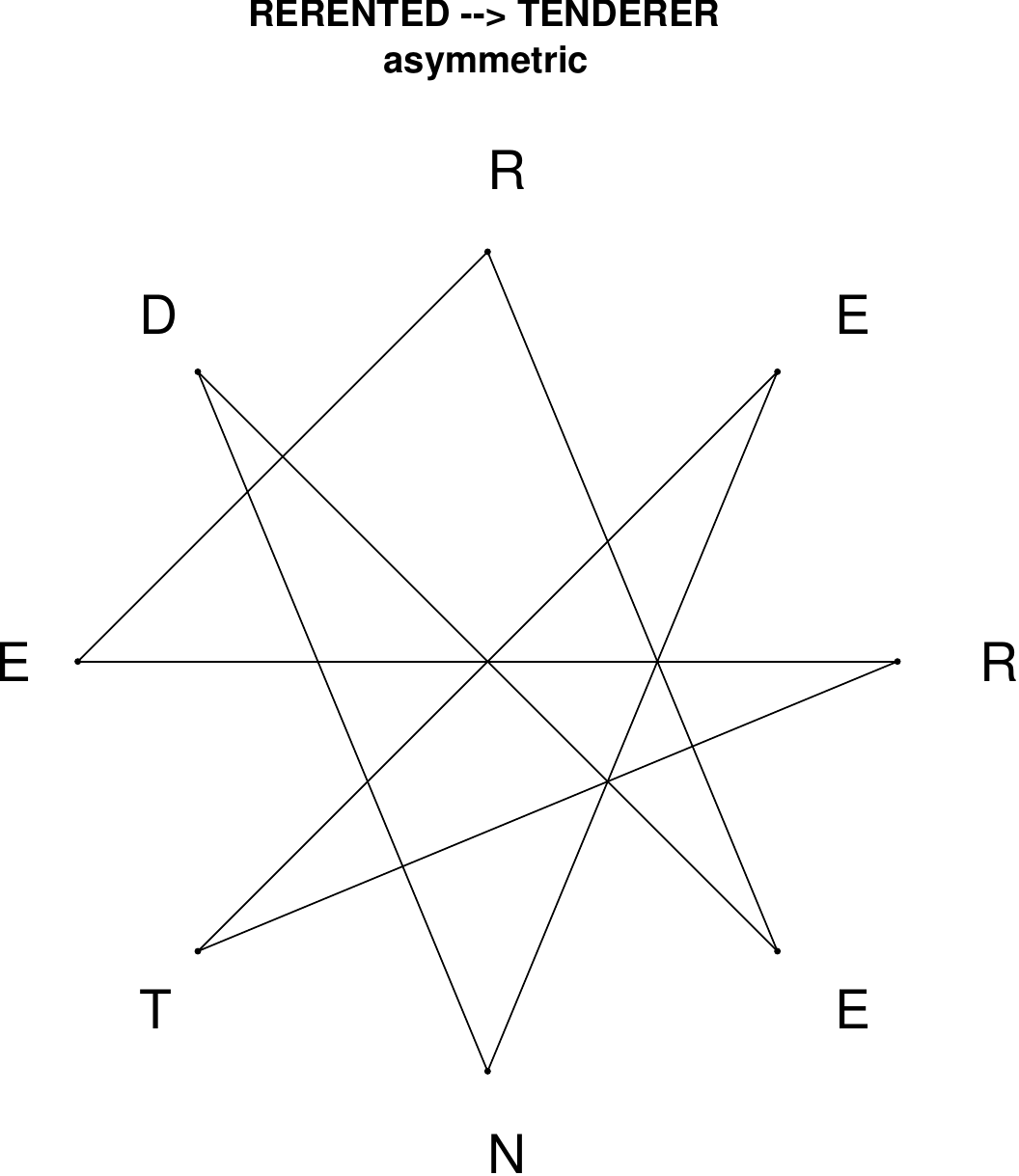}
\end{subfigure}
\hfill
\begin{subfigure}[T]{0.19\textwidth}
\centering
\includegraphics[width=\textwidth]{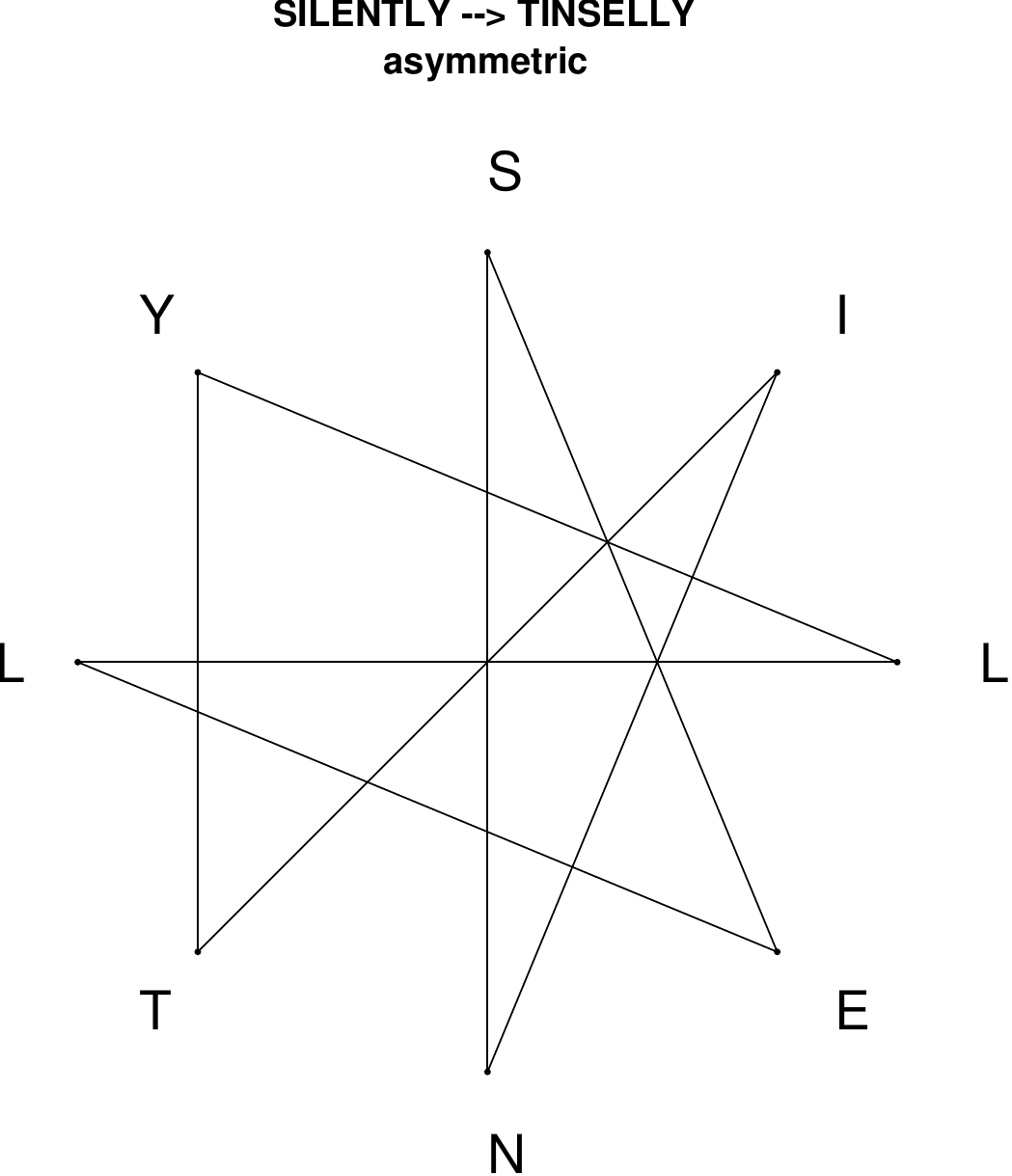}
\end{subfigure}
\hfill
\begin{subfigure}[T]{0.19\textwidth}
\centering
\includegraphics[width=\textwidth]{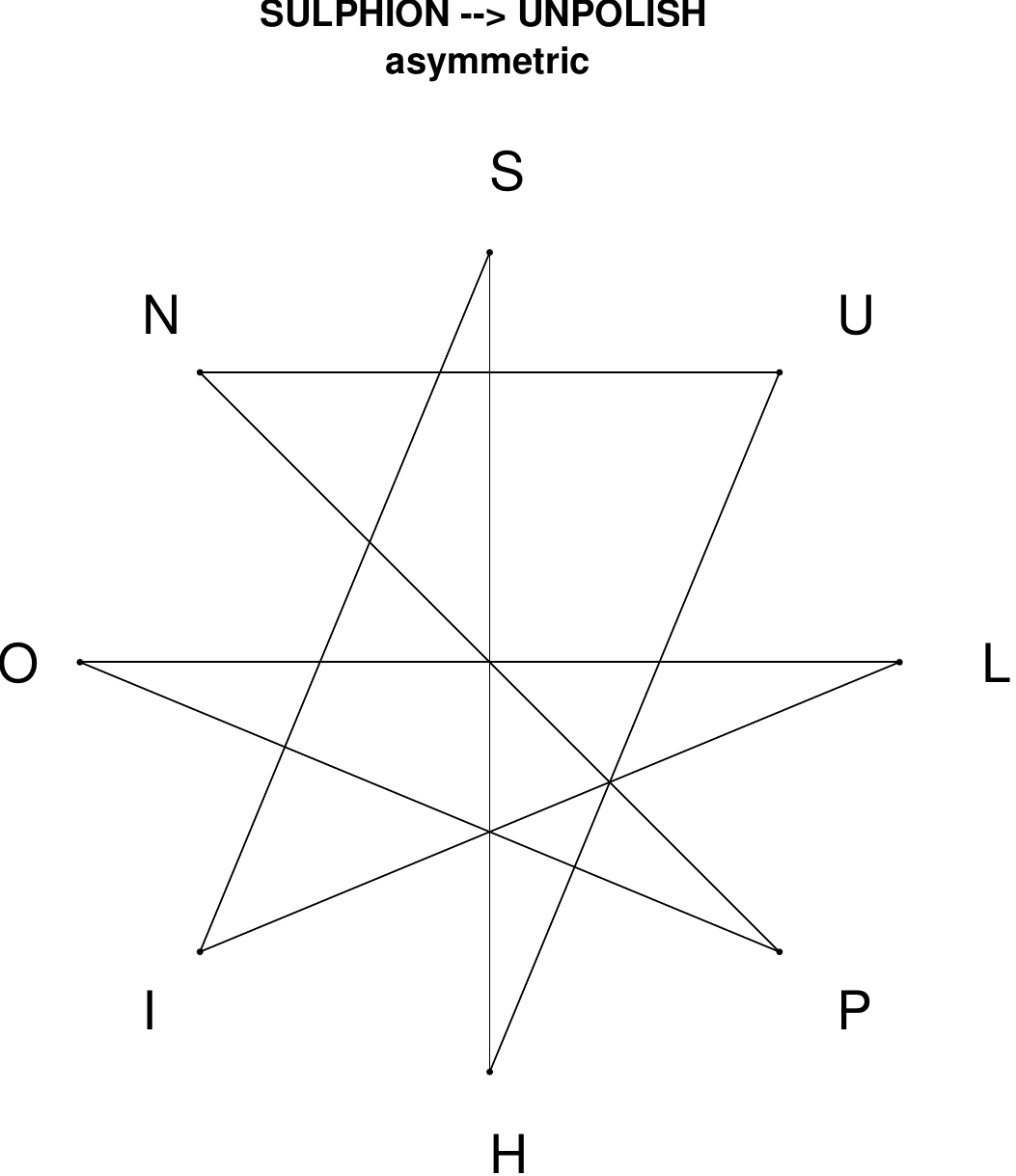}
\end{subfigure}
\hfill
\begin{subfigure}[T]{0.19\textwidth}
\centering
\includegraphics[width=\textwidth]{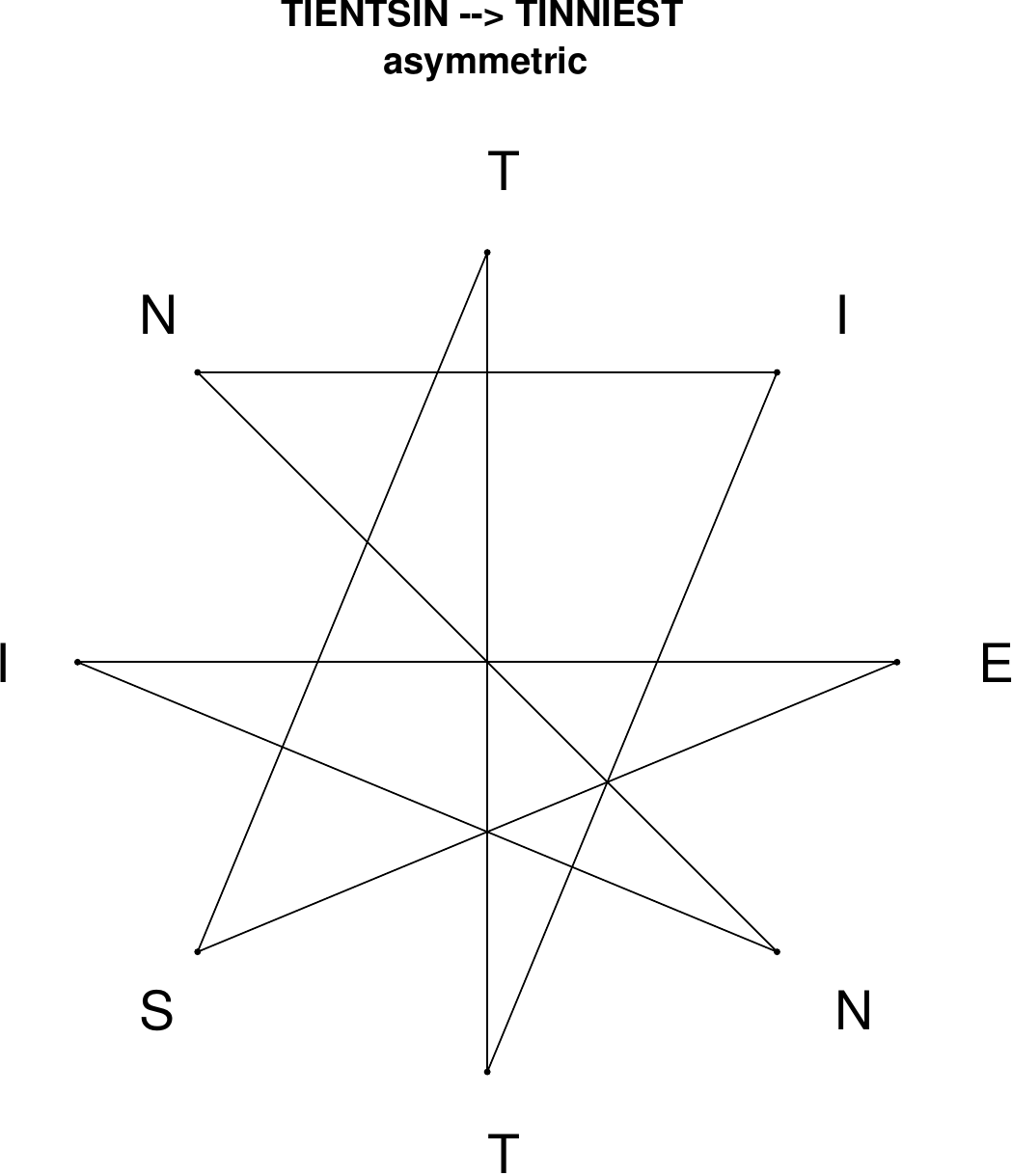}
\end{subfigure}
\end{figure}

\begin{figure}[H]
\centering
\begin{subfigure}[T]{0.19\textwidth}
\centering
\includegraphics[width=\textwidth]{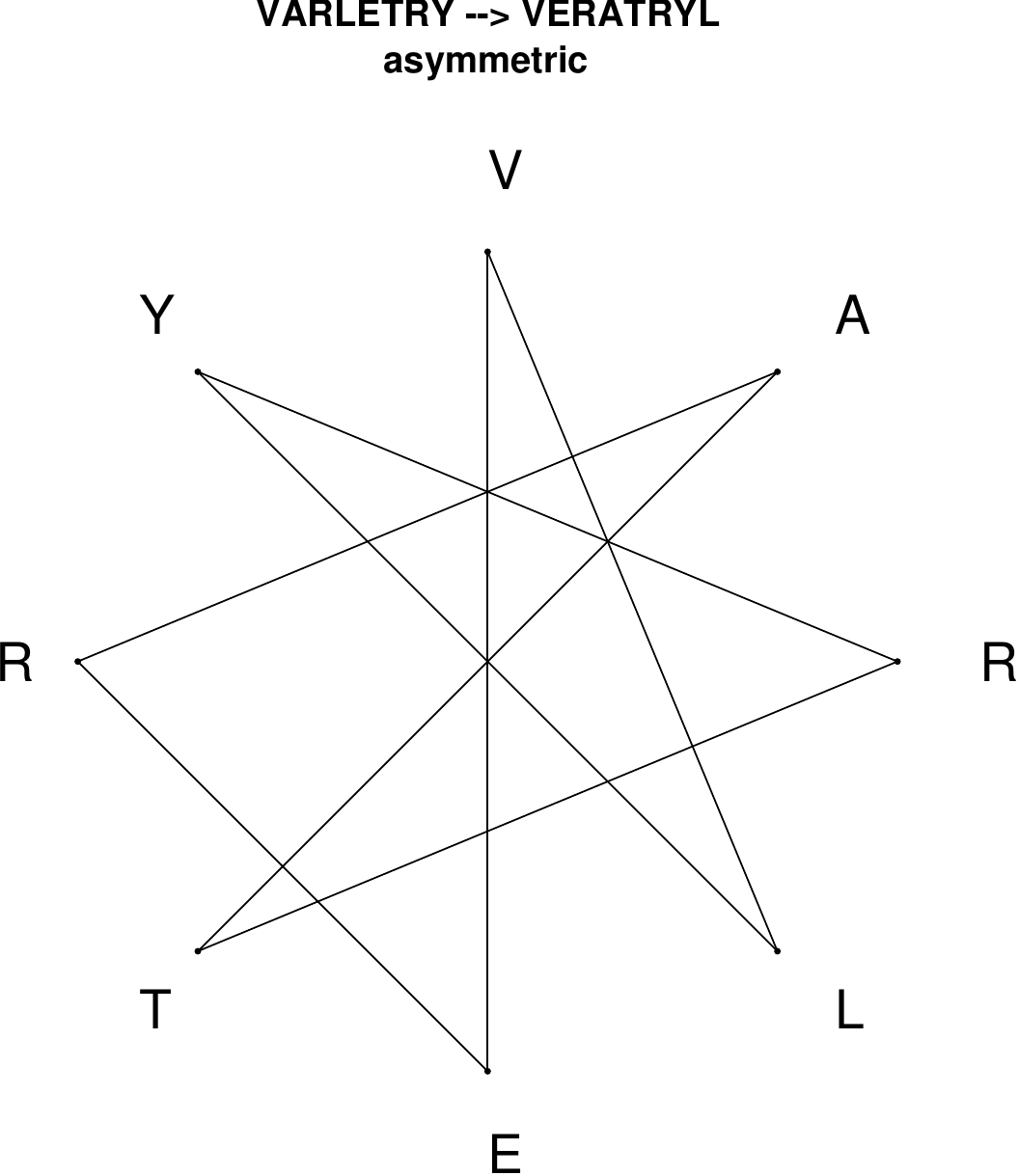}
\end{subfigure}
\hfill
\begin{subfigure}[T]{0.19\textwidth}
\centering
\includegraphics[width=\textwidth]{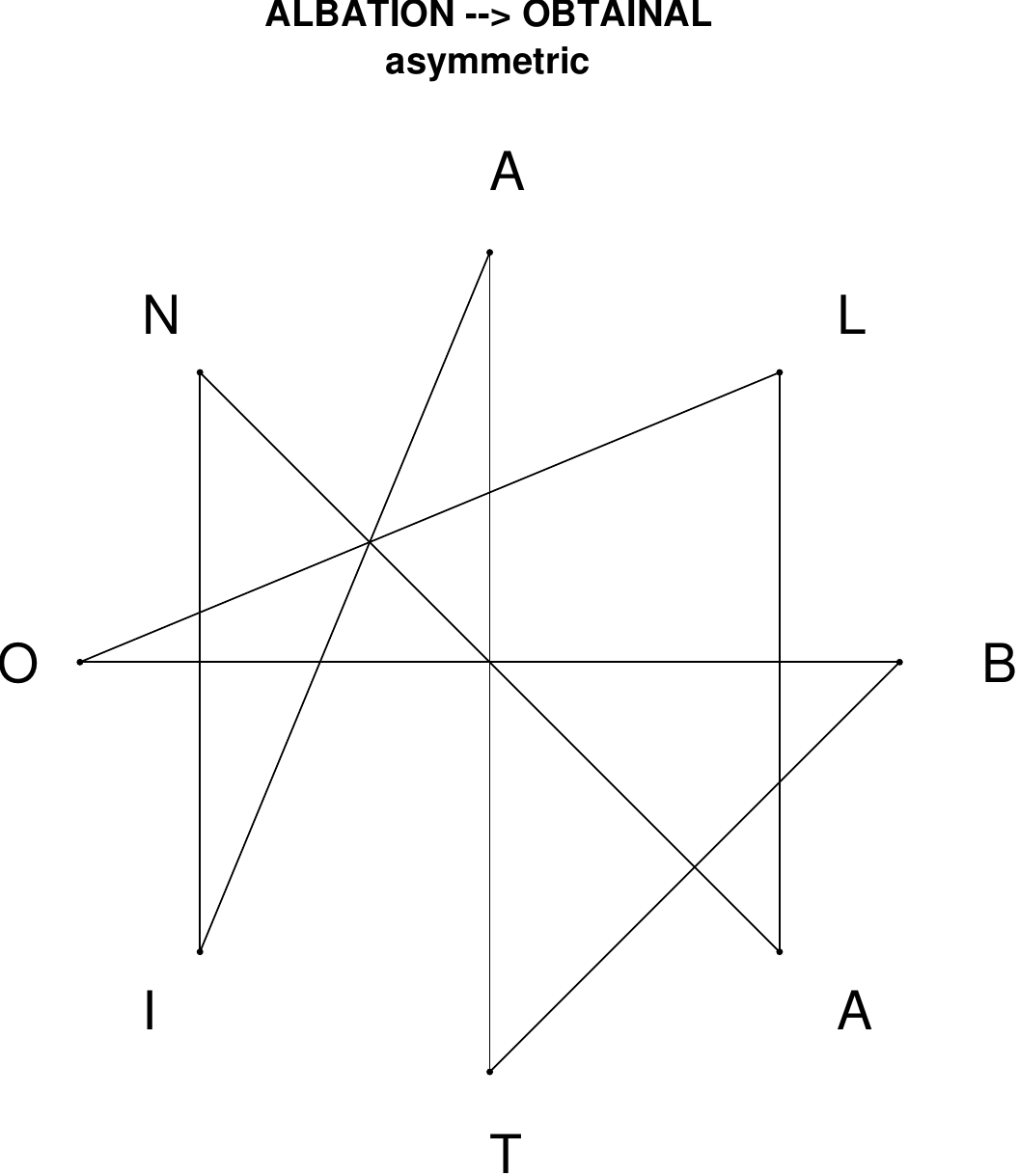}
\end{subfigure}
\hfill
\begin{subfigure}[T]{0.19\textwidth}
\centering
\includegraphics[width=\textwidth]{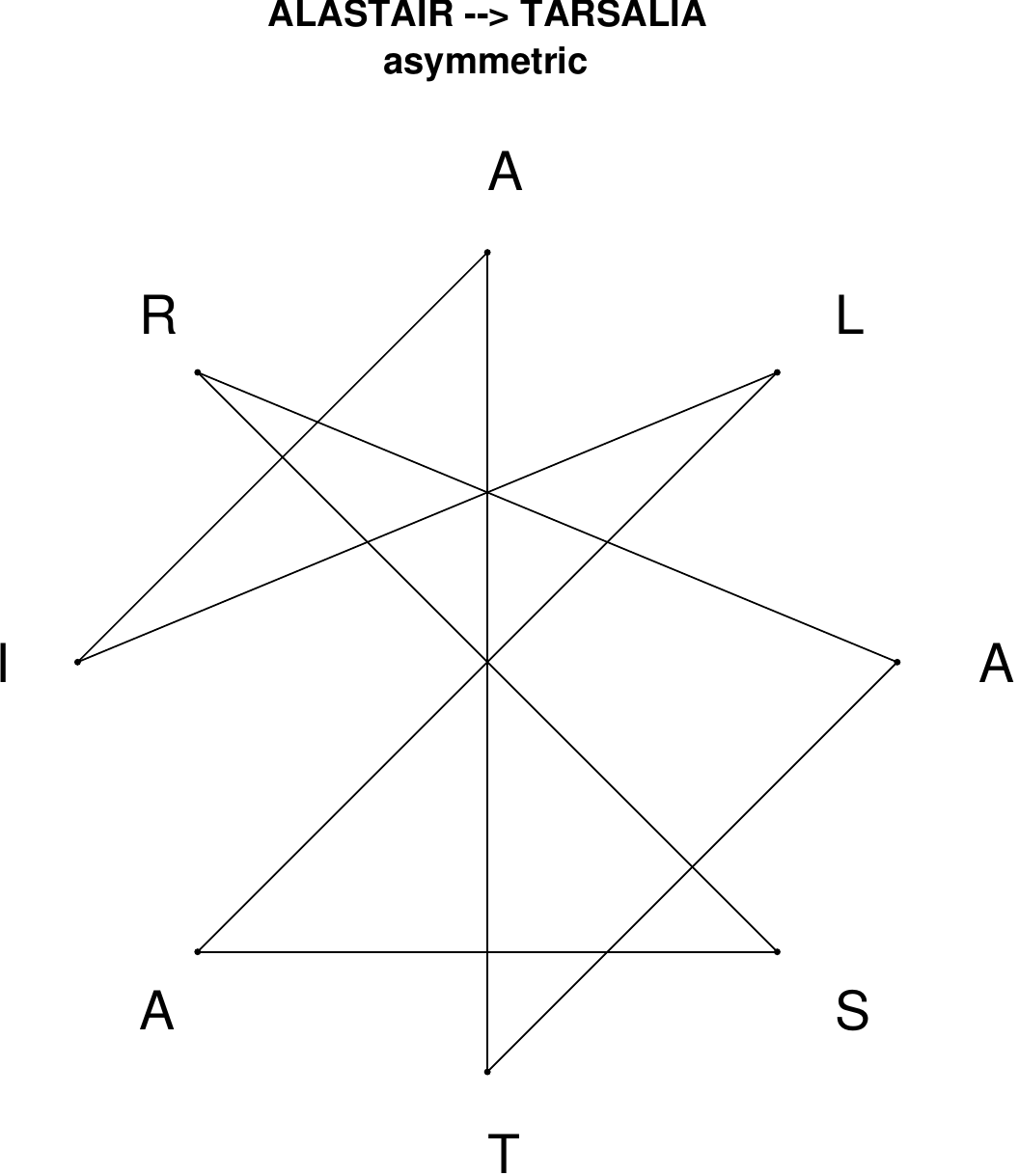}
\end{subfigure}
\hfill
\begin{subfigure}[T]{0.19\textwidth}
\centering
\includegraphics[width=\textwidth]{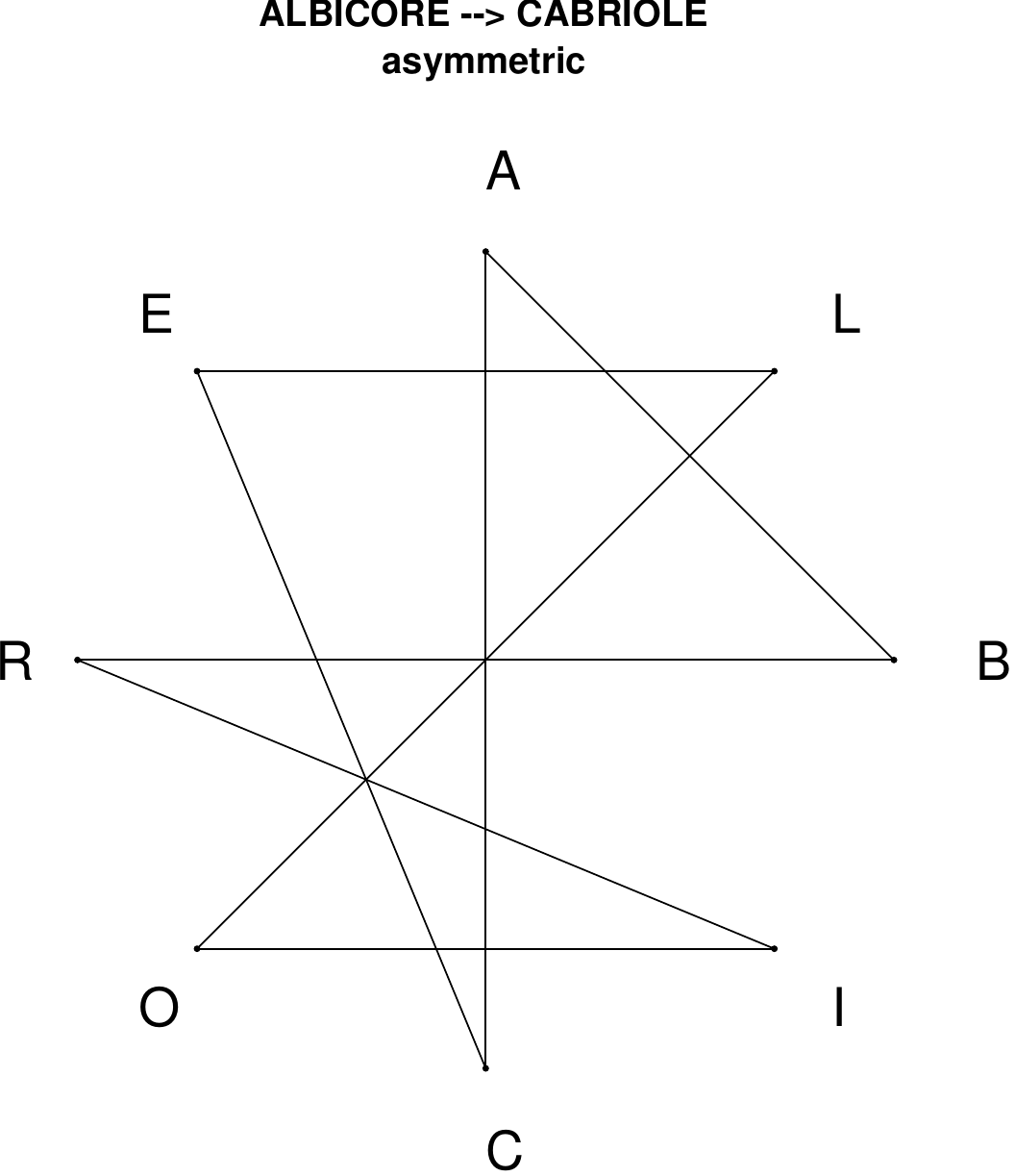}
\end{subfigure}
\hfill
\begin{subfigure}[T]{0.19\textwidth}
\centering
\includegraphics[width=\textwidth]{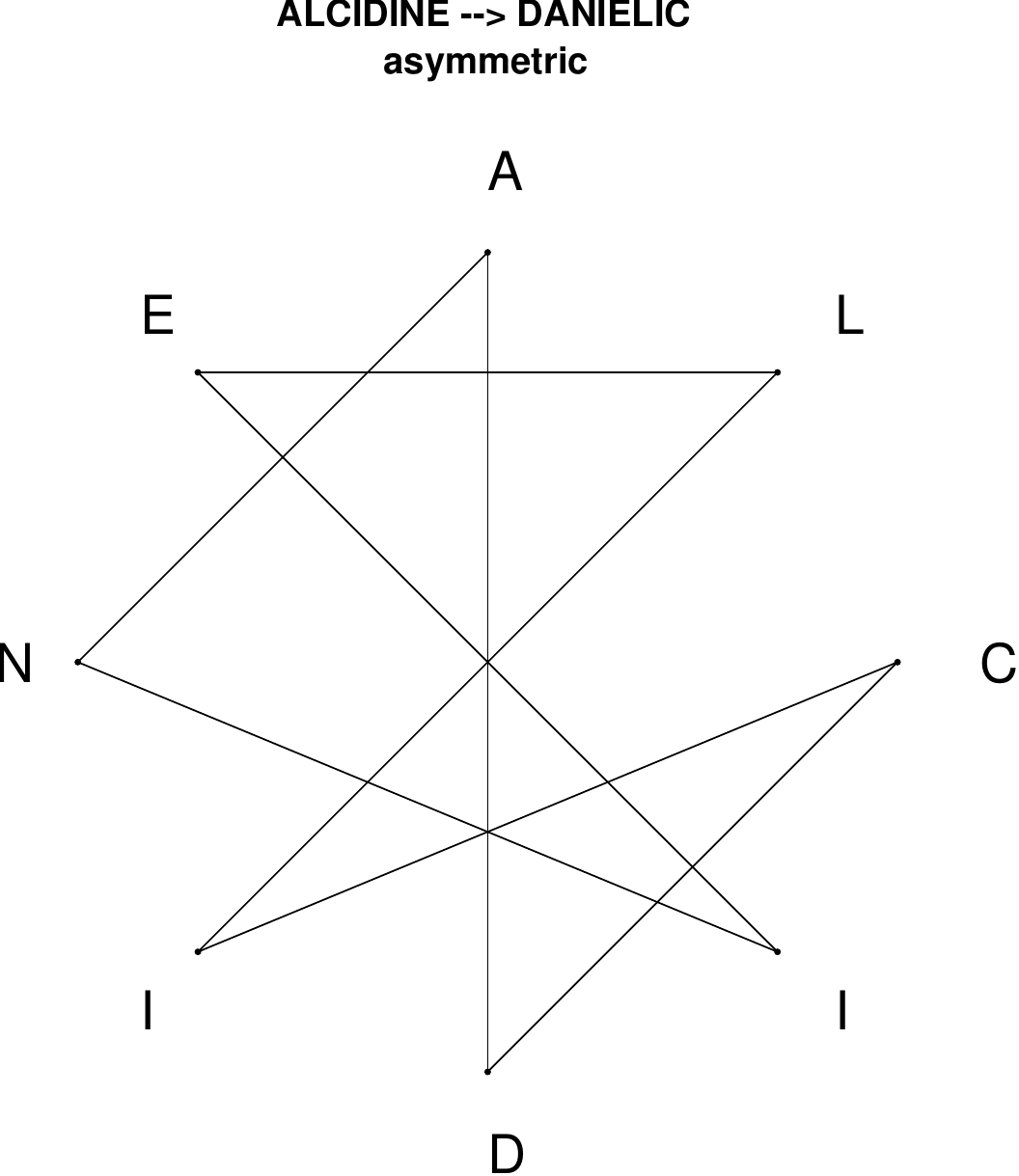}
\end{subfigure}
\end{figure}

\begin{figure}[H]
\centering
\begin{subfigure}[T]{0.19\textwidth}
\centering
\includegraphics[width=\textwidth]{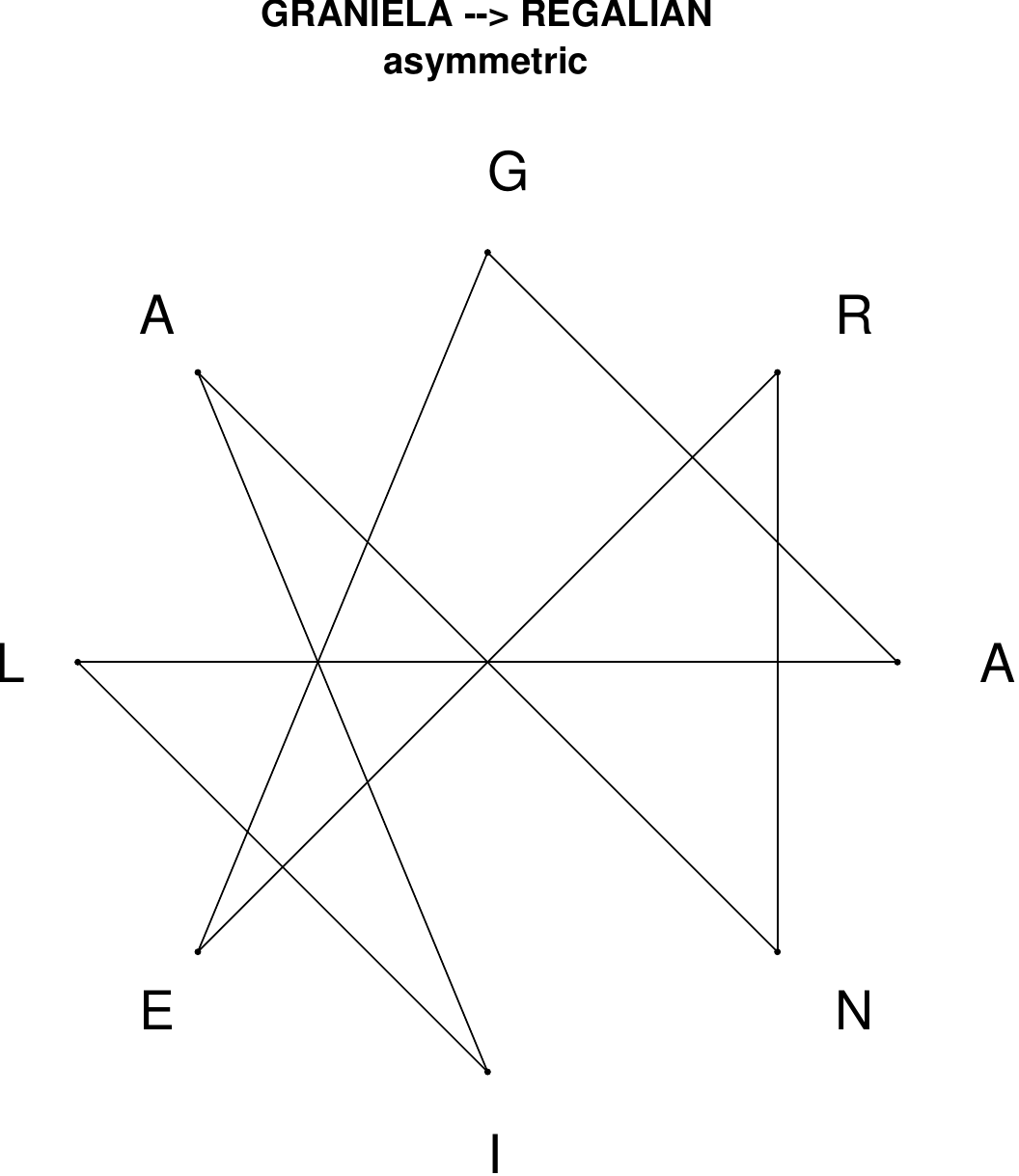}
\end{subfigure}
\hfill
\begin{subfigure}[T]{0.19\textwidth}
\centering
\includegraphics[width=\textwidth]{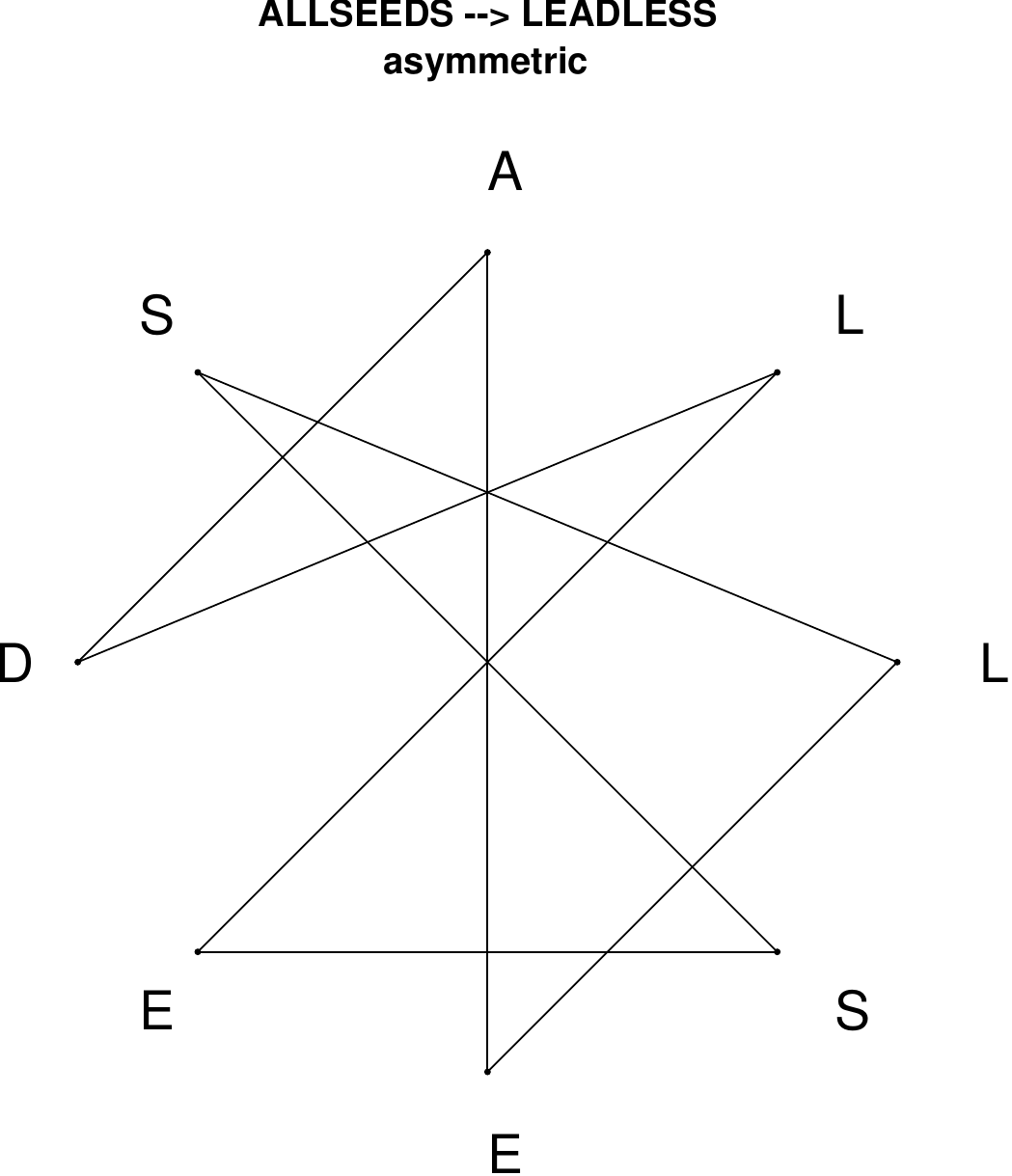}
\end{subfigure}
\hfill
\begin{subfigure}[T]{0.19\textwidth}
\centering
\includegraphics[width=\textwidth]{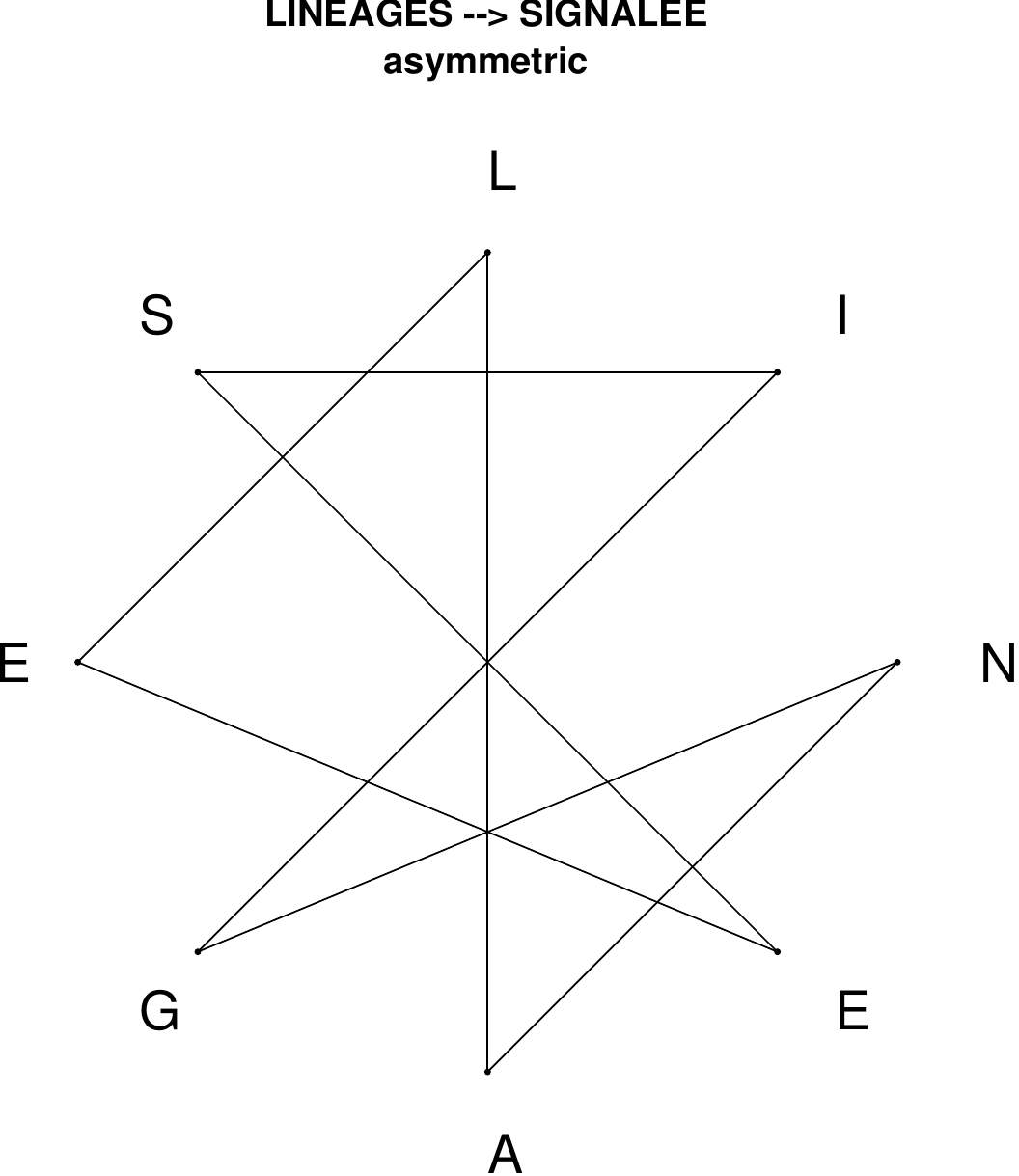}
\end{subfigure}
\hfill
\begin{subfigure}[T]{0.19\textwidth}
\centering
\includegraphics[width=\textwidth]{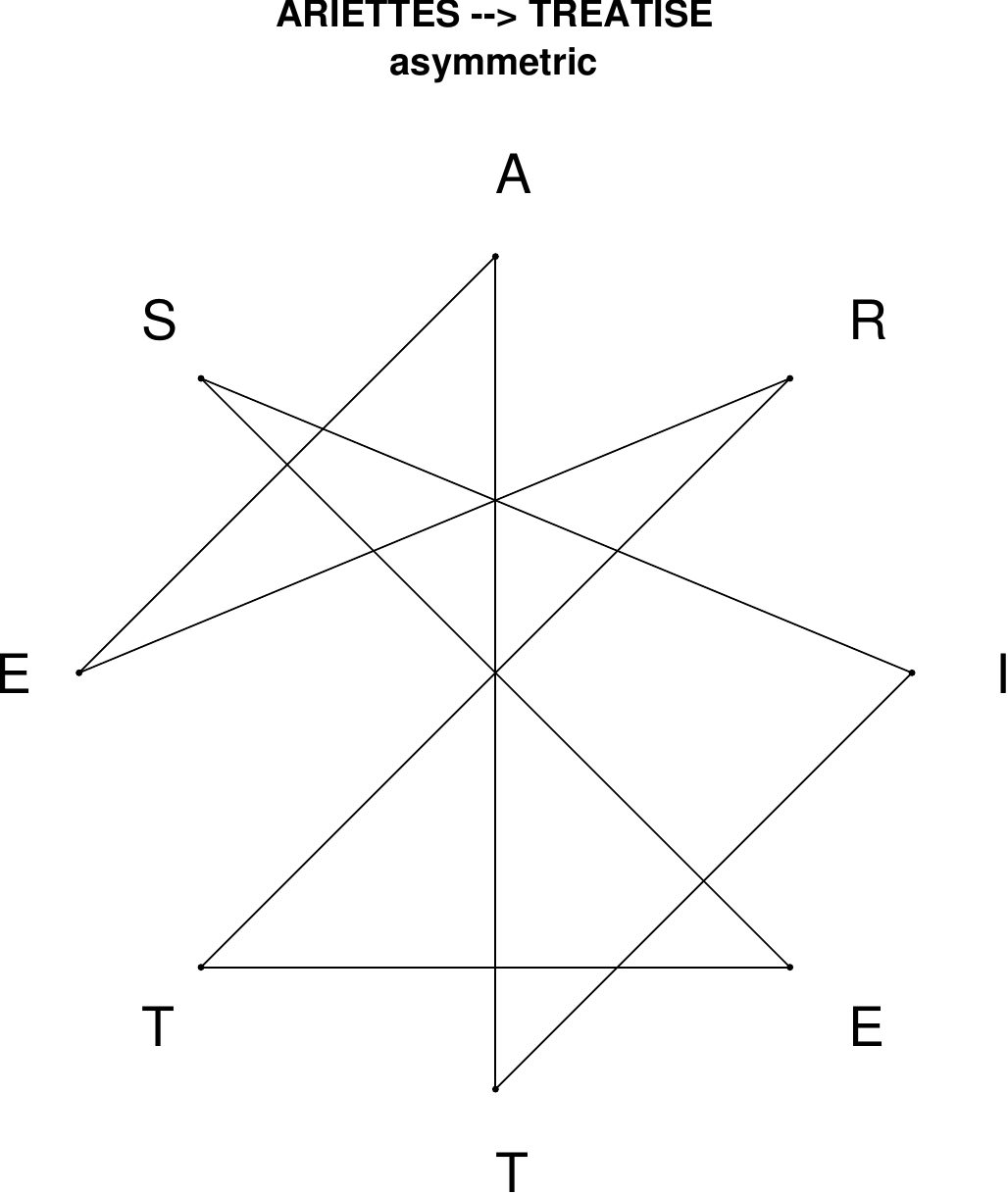}
\end{subfigure}
\hfill
\begin{subfigure}[T]{0.19\textwidth}
\centering
\includegraphics[width=\textwidth]{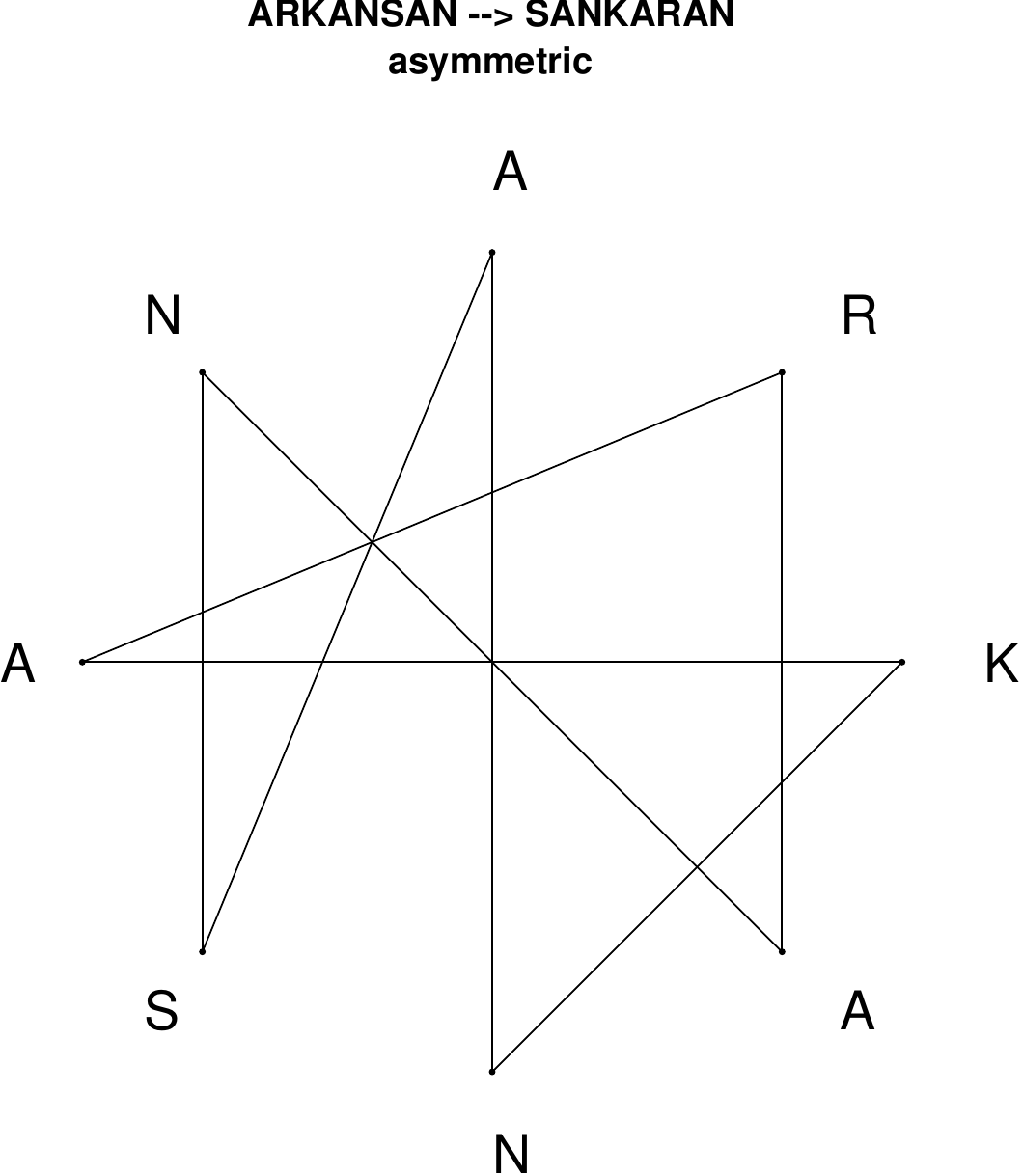}
\end{subfigure}
\end{figure}

\begin{figure}[H]
\centering
\begin{subfigure}[T]{0.19\textwidth}
\centering
\includegraphics[width=\textwidth]{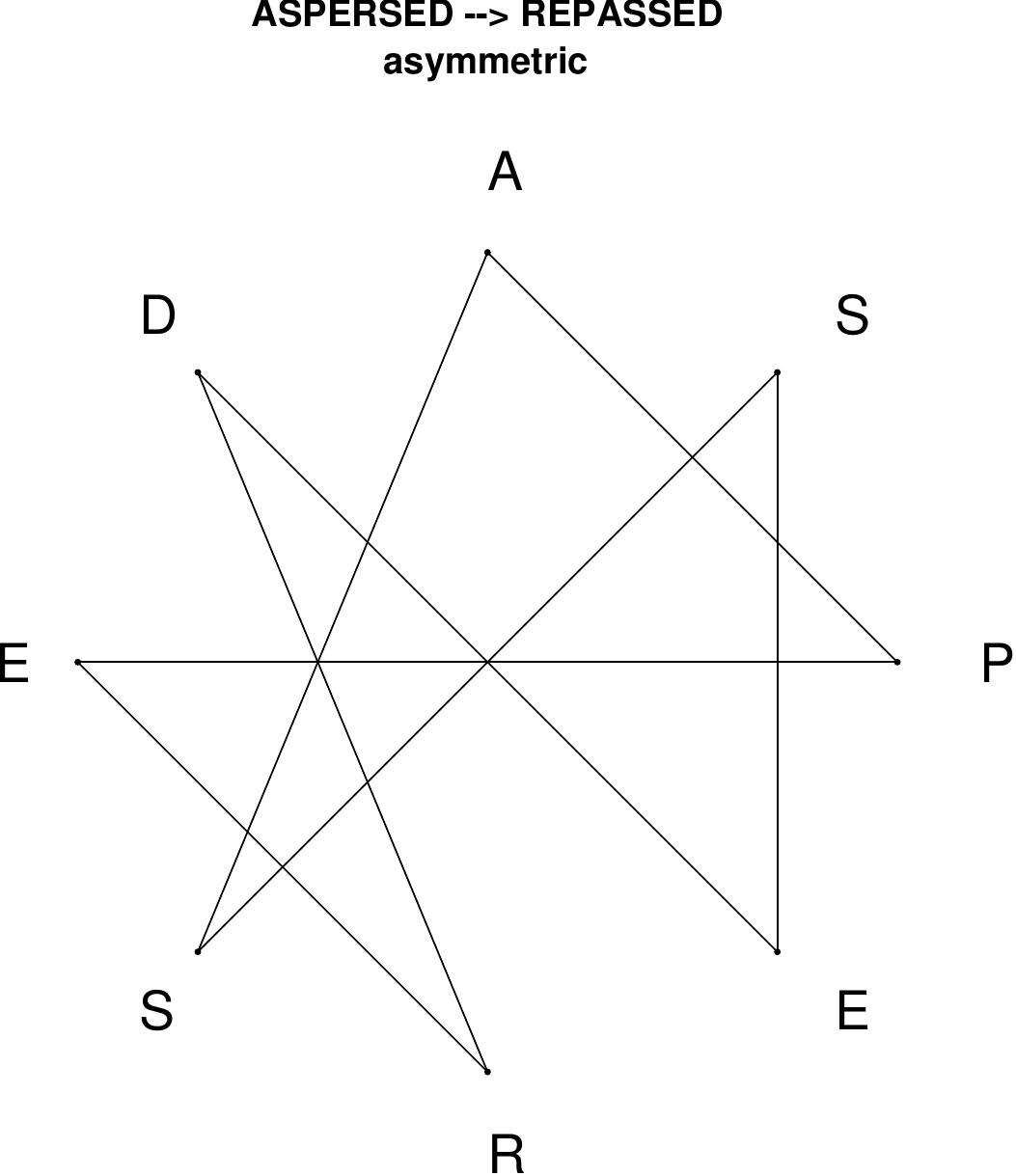}
\end{subfigure}
\hfill
\begin{subfigure}[T]{0.19\textwidth}
\centering
\includegraphics[width=\textwidth]{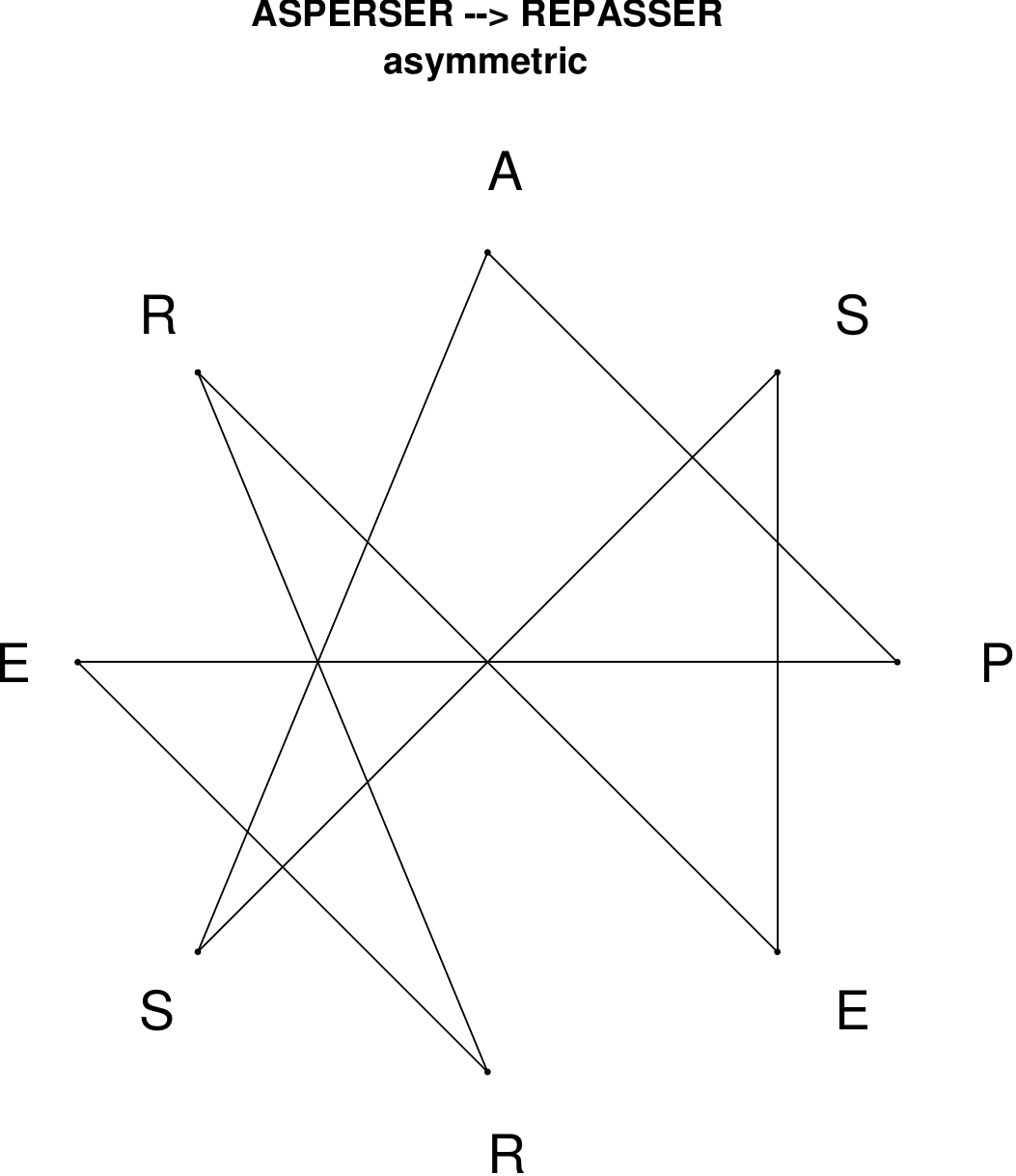}
\end{subfigure}
\hfill
\begin{subfigure}[T]{0.19\textwidth}
\centering
\includegraphics[width=\textwidth]{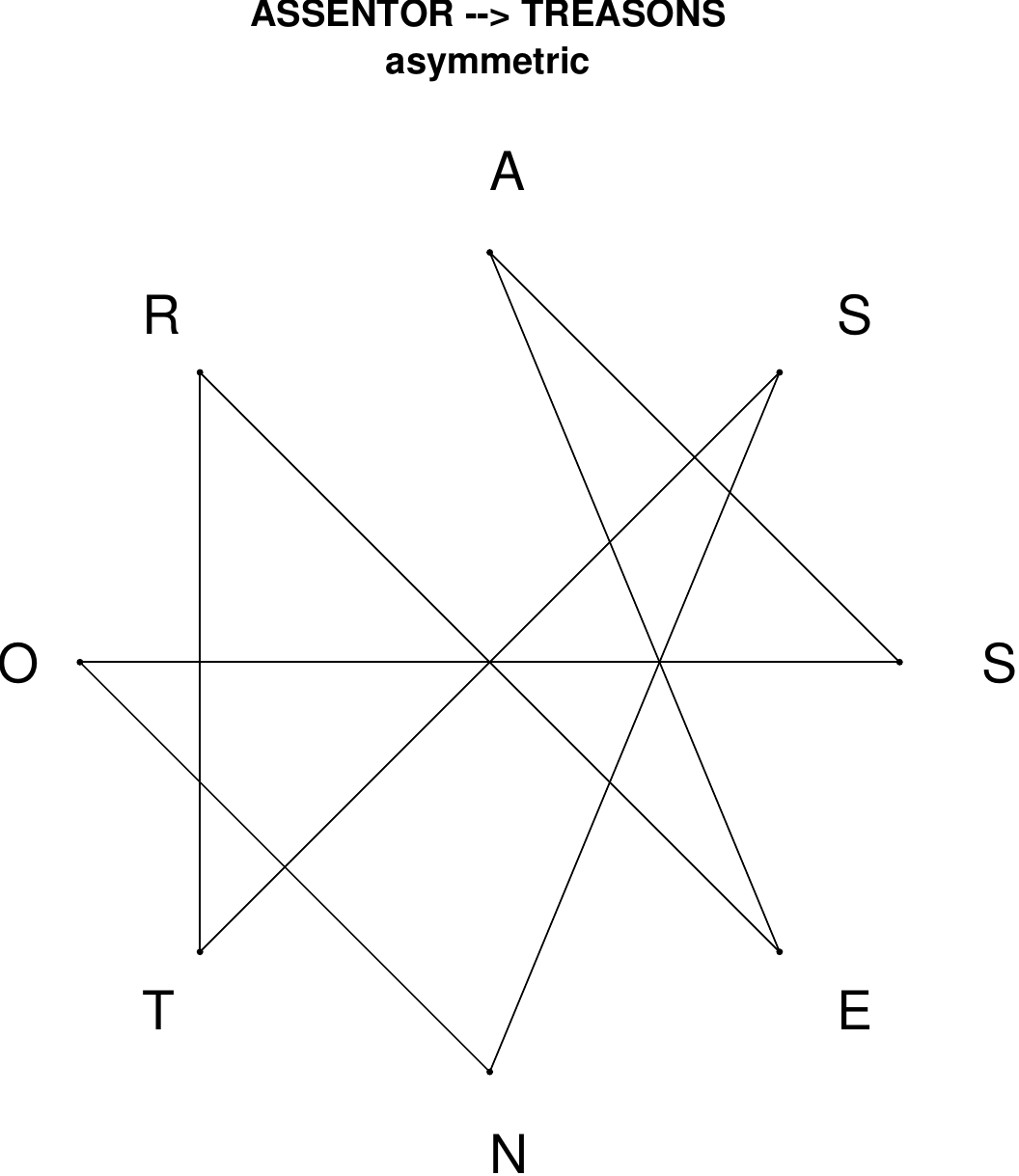}
\end{subfigure}
\hfill
\begin{subfigure}[T]{0.19\textwidth}
\centering
\includegraphics[width=\textwidth]{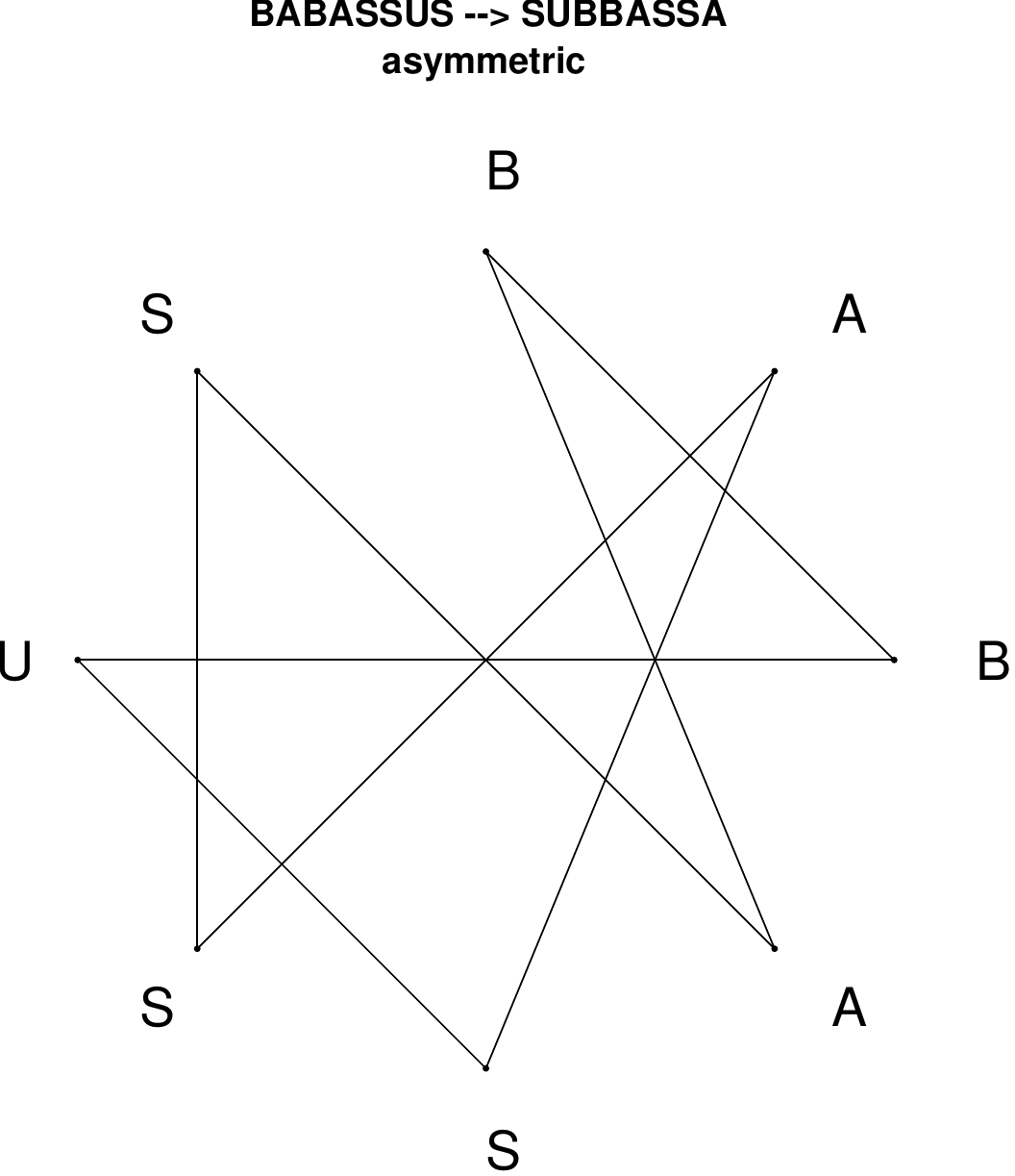}
\end{subfigure}
\hfill
\begin{subfigure}[T]{0.19\textwidth}
\centering
\includegraphics[width=\textwidth]{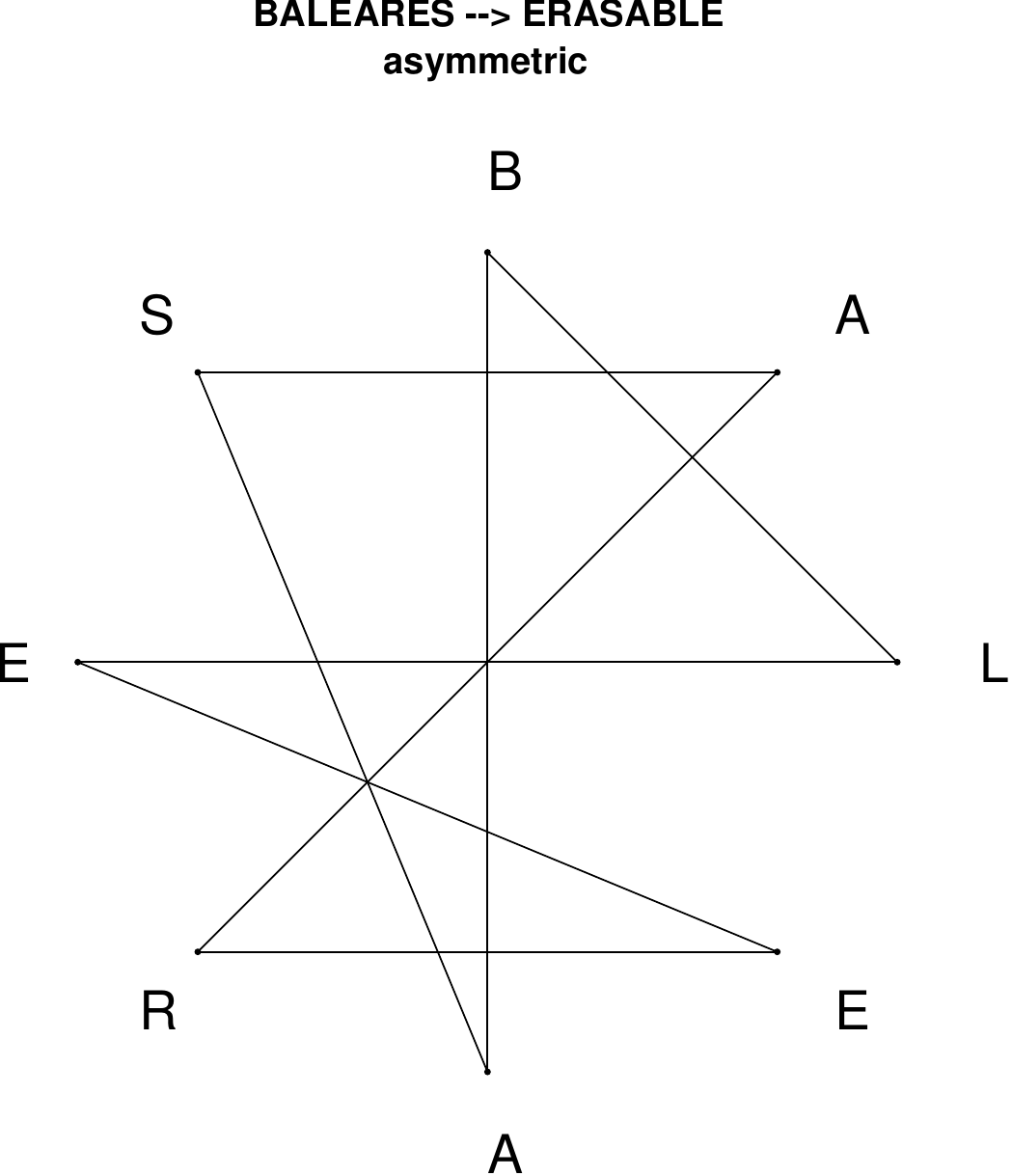}
\end{subfigure}
\end{figure}

\begin{figure}[H]
\centering
\begin{subfigure}[T]{0.19\textwidth}
\centering
\includegraphics[width=\textwidth]{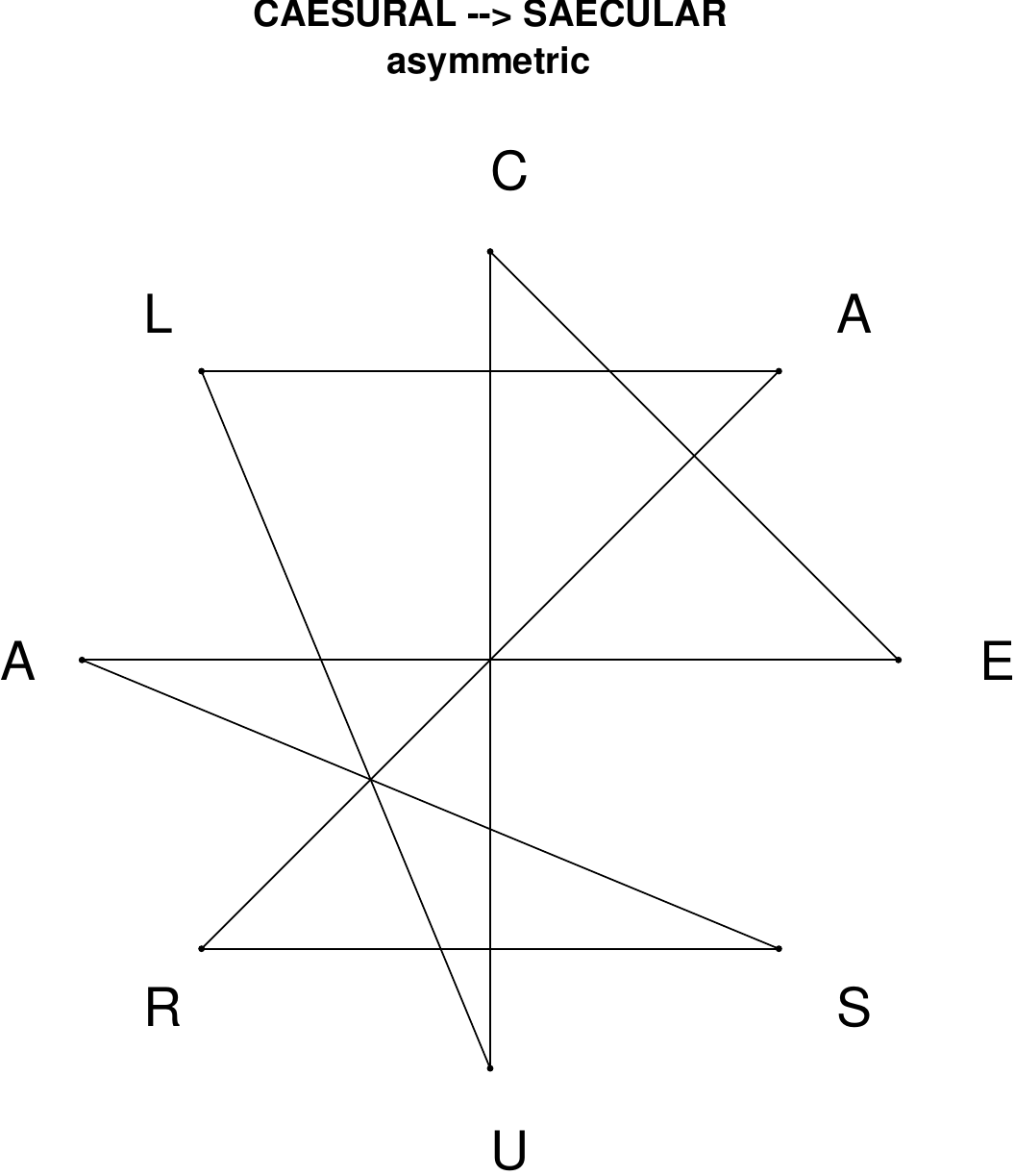}
\end{subfigure}
\hfill
\begin{subfigure}[T]{0.19\textwidth}
\centering
\includegraphics[width=\textwidth]{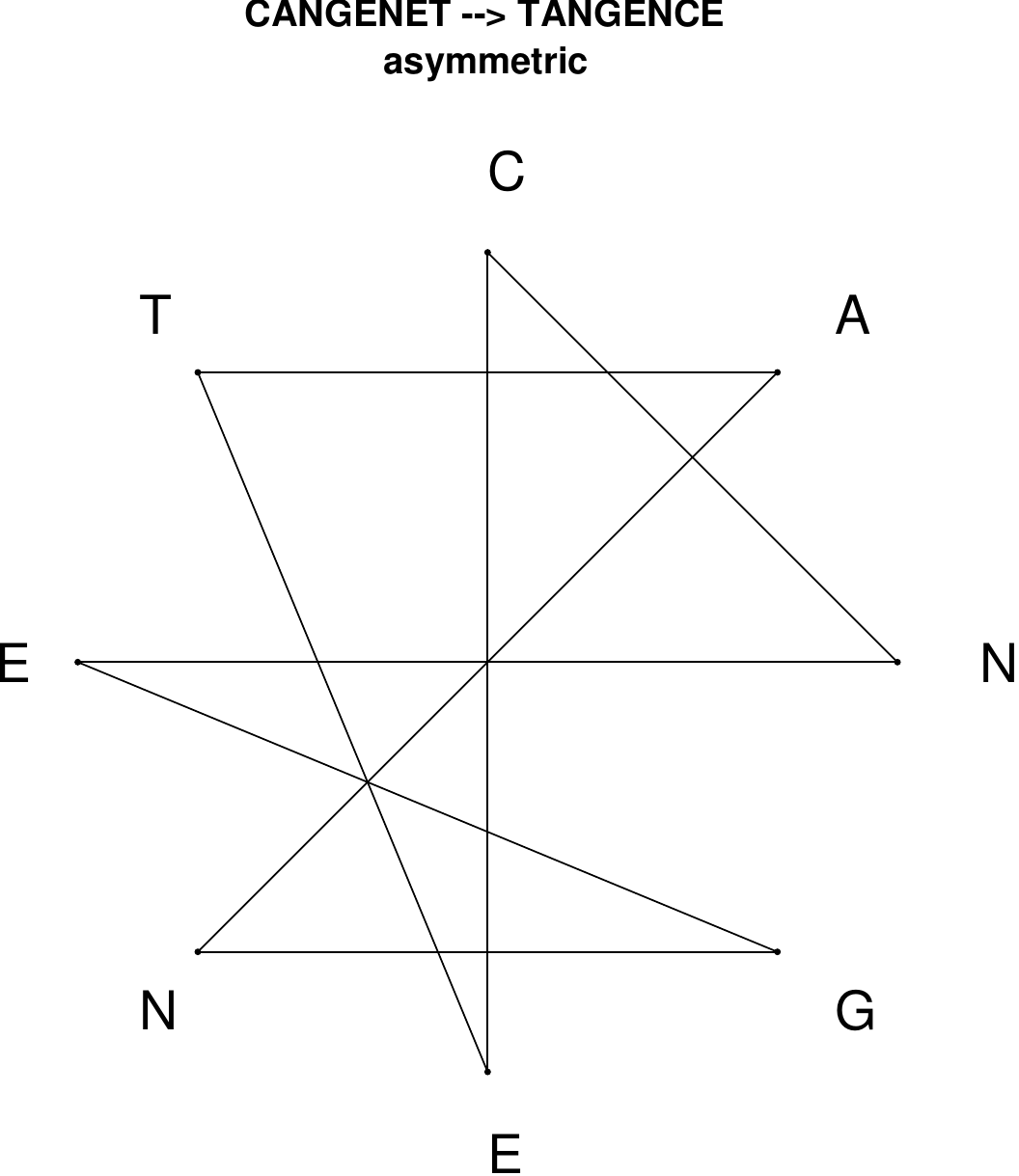}
\end{subfigure}
\hfill
\begin{subfigure}[T]{0.19\textwidth}
\centering
\includegraphics[width=\textwidth]{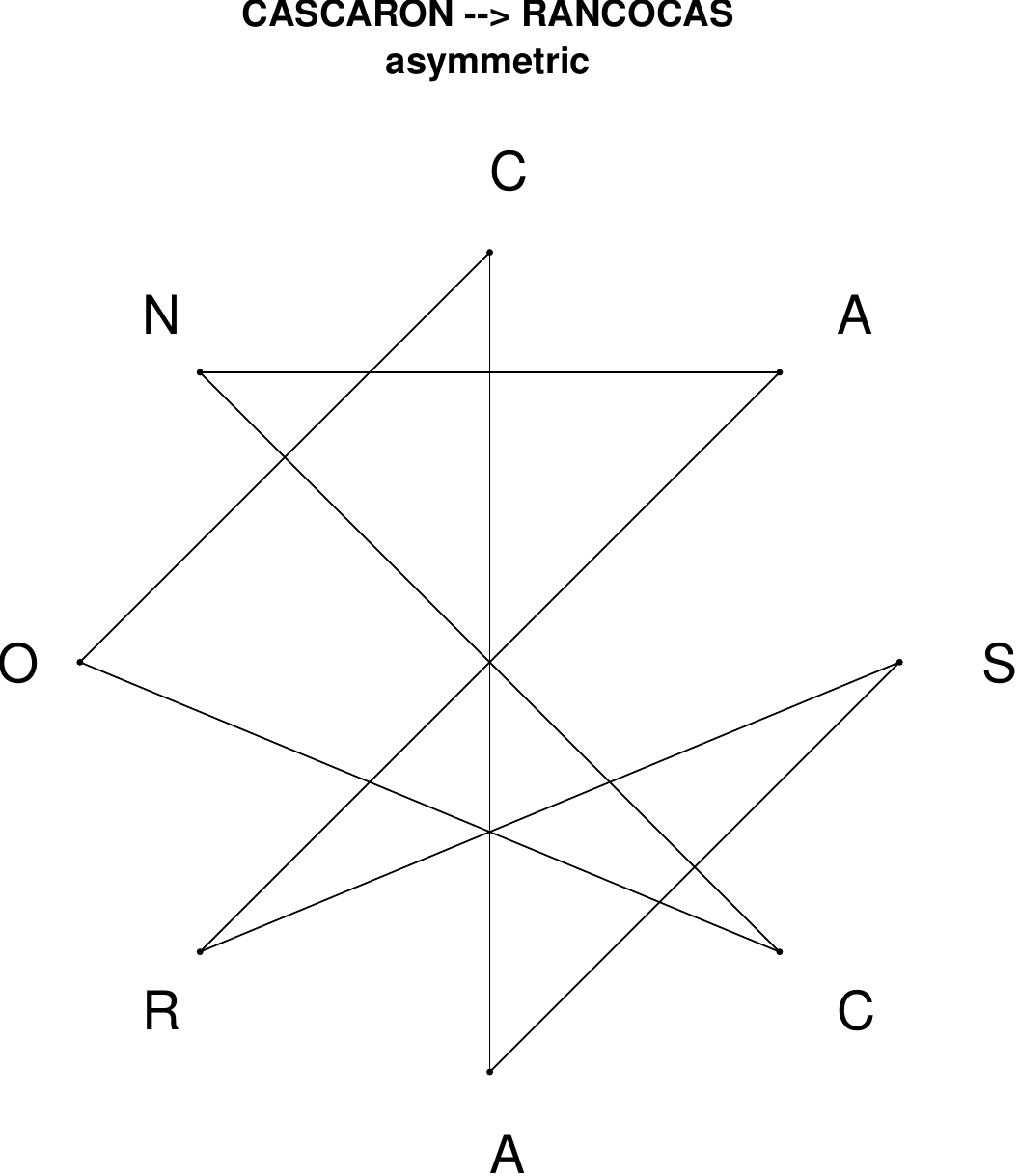}
\end{subfigure}
\hfill
\begin{subfigure}[T]{0.19\textwidth}
\centering
\includegraphics[width=\textwidth]{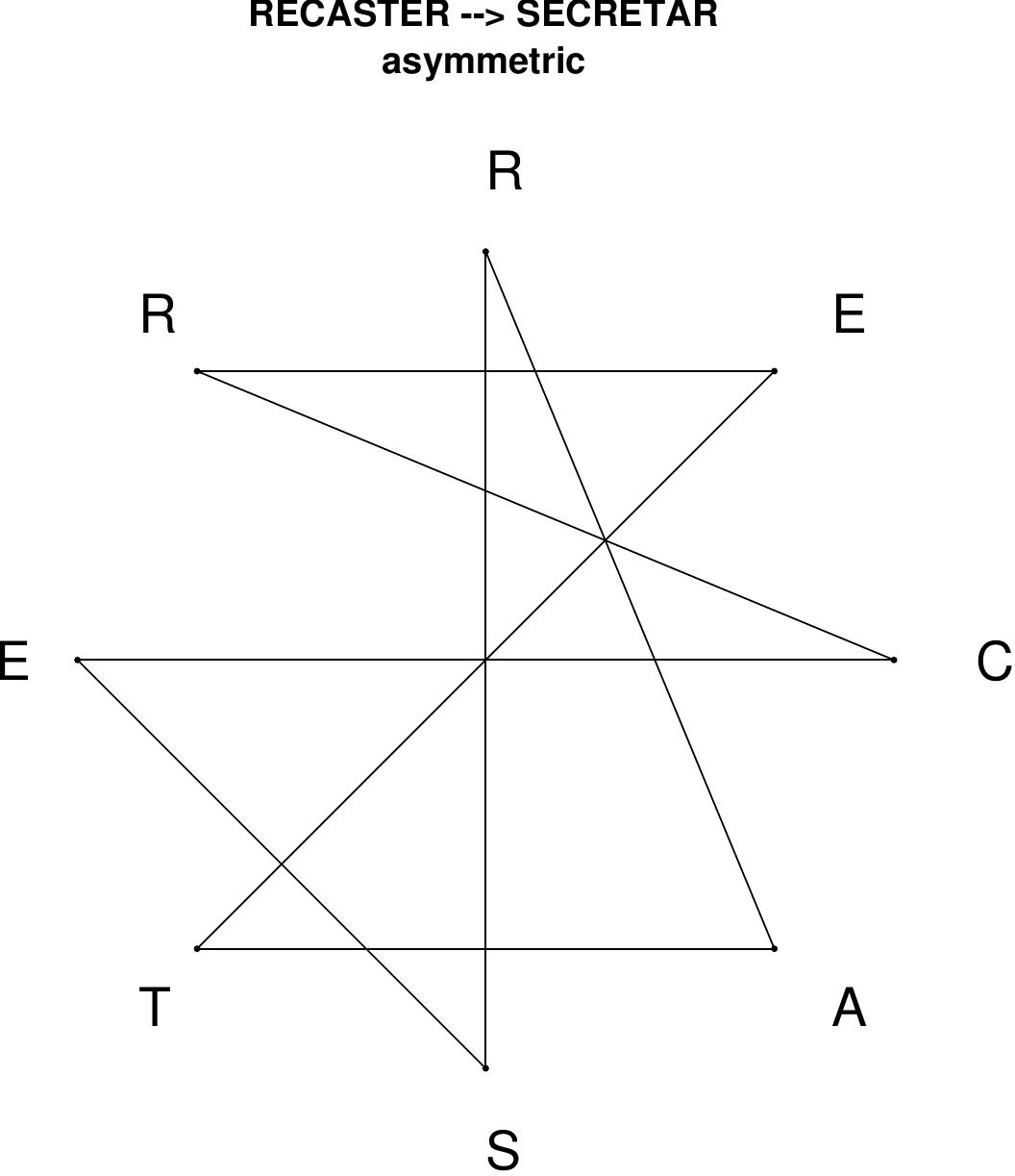}
\end{subfigure}
\hfill
\begin{subfigure}[T]{0.19\textwidth}
\centering
\includegraphics[width=\textwidth]{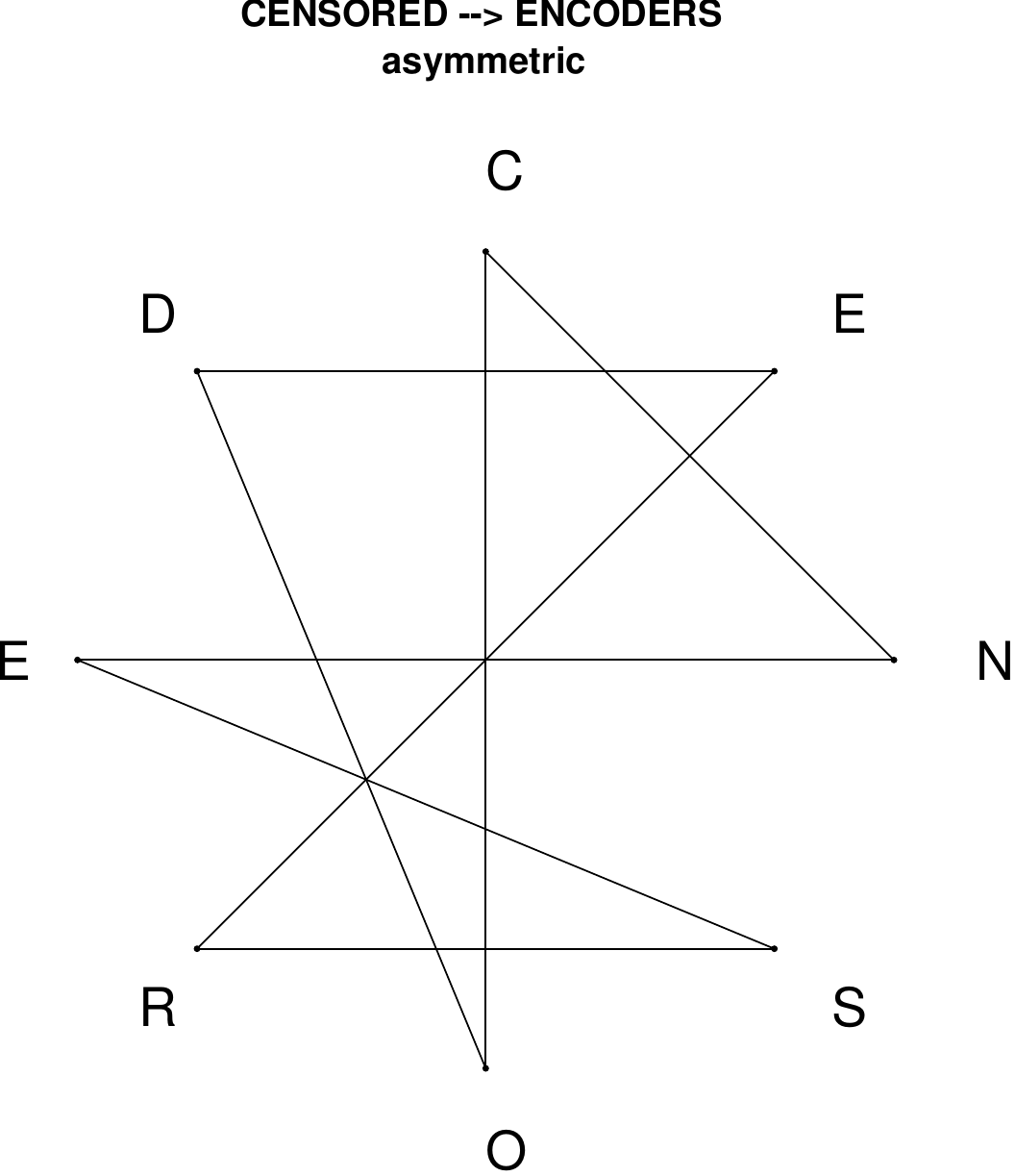}
\end{subfigure}
\end{figure}

\begin{figure}[H]
\centering
\begin{subfigure}[T]{0.19\textwidth}
\centering
\includegraphics[width=\textwidth]{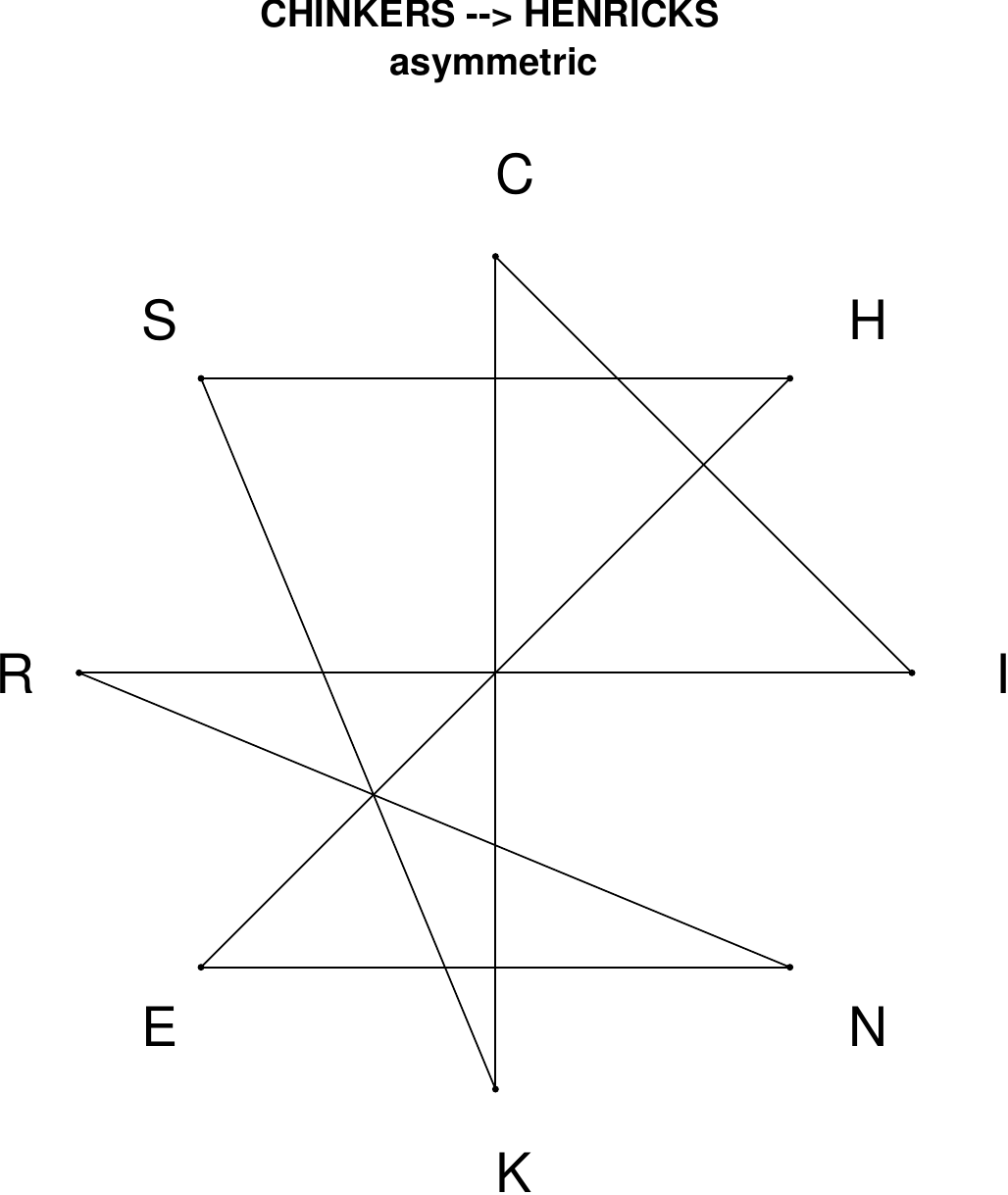}
\end{subfigure}
\hfill
\begin{subfigure}[T]{0.19\textwidth}
\centering
\includegraphics[width=\textwidth]{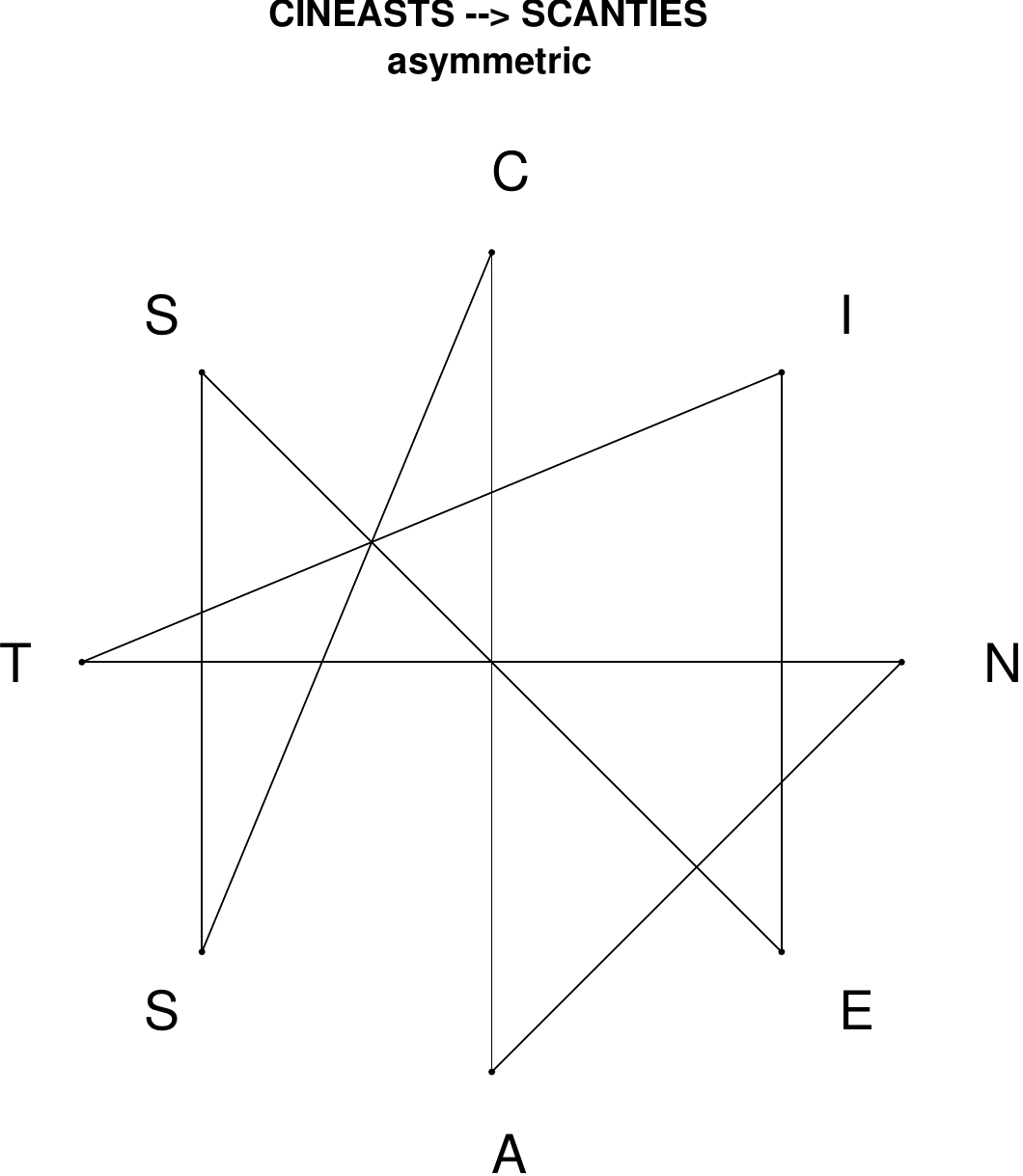}
\end{subfigure}
\hfill
\begin{subfigure}[T]{0.19\textwidth}
\centering
\includegraphics[width=\textwidth]{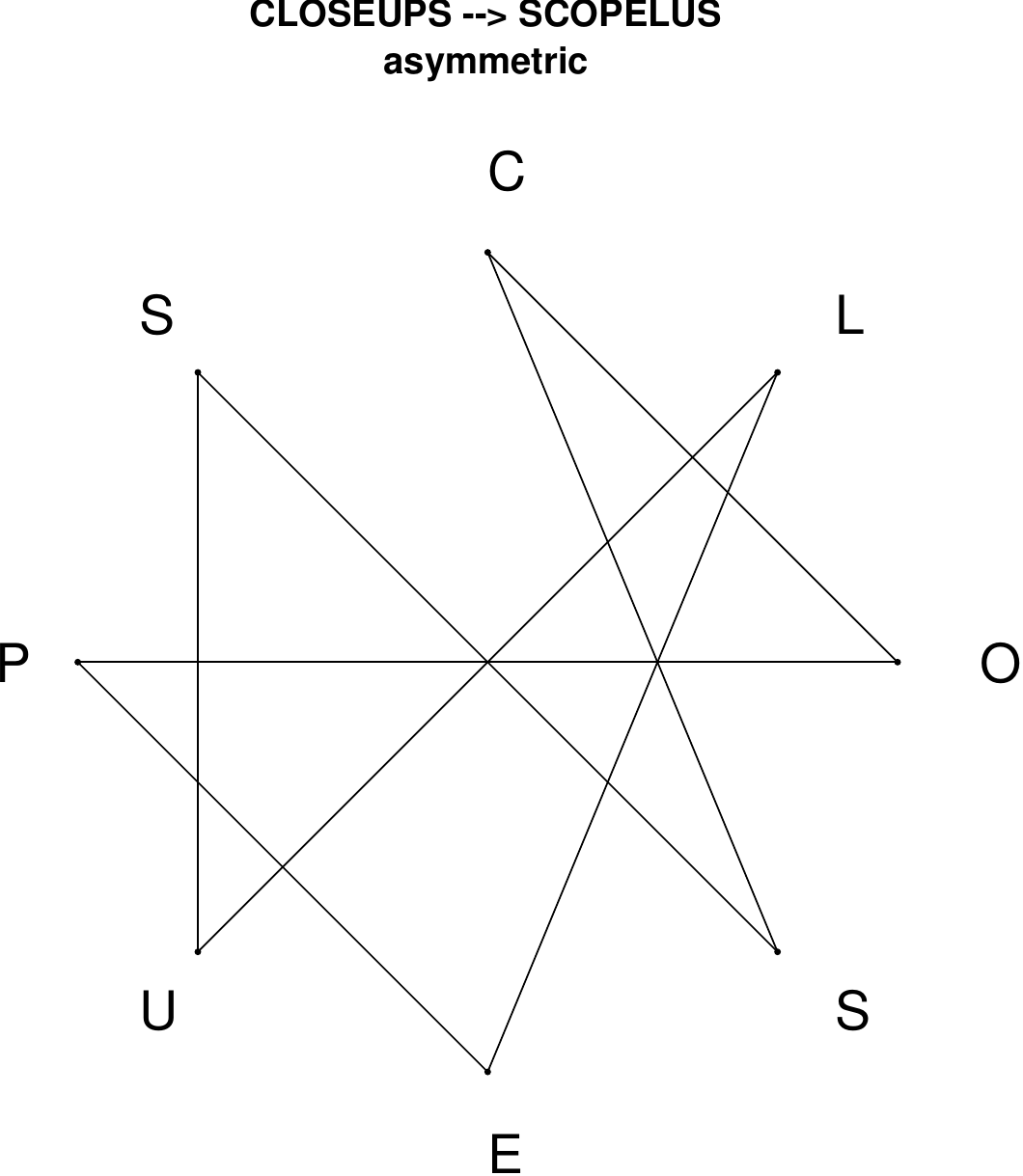}
\end{subfigure}
\hfill
\begin{subfigure}[T]{0.19\textwidth}
\centering
\includegraphics[width=\textwidth]{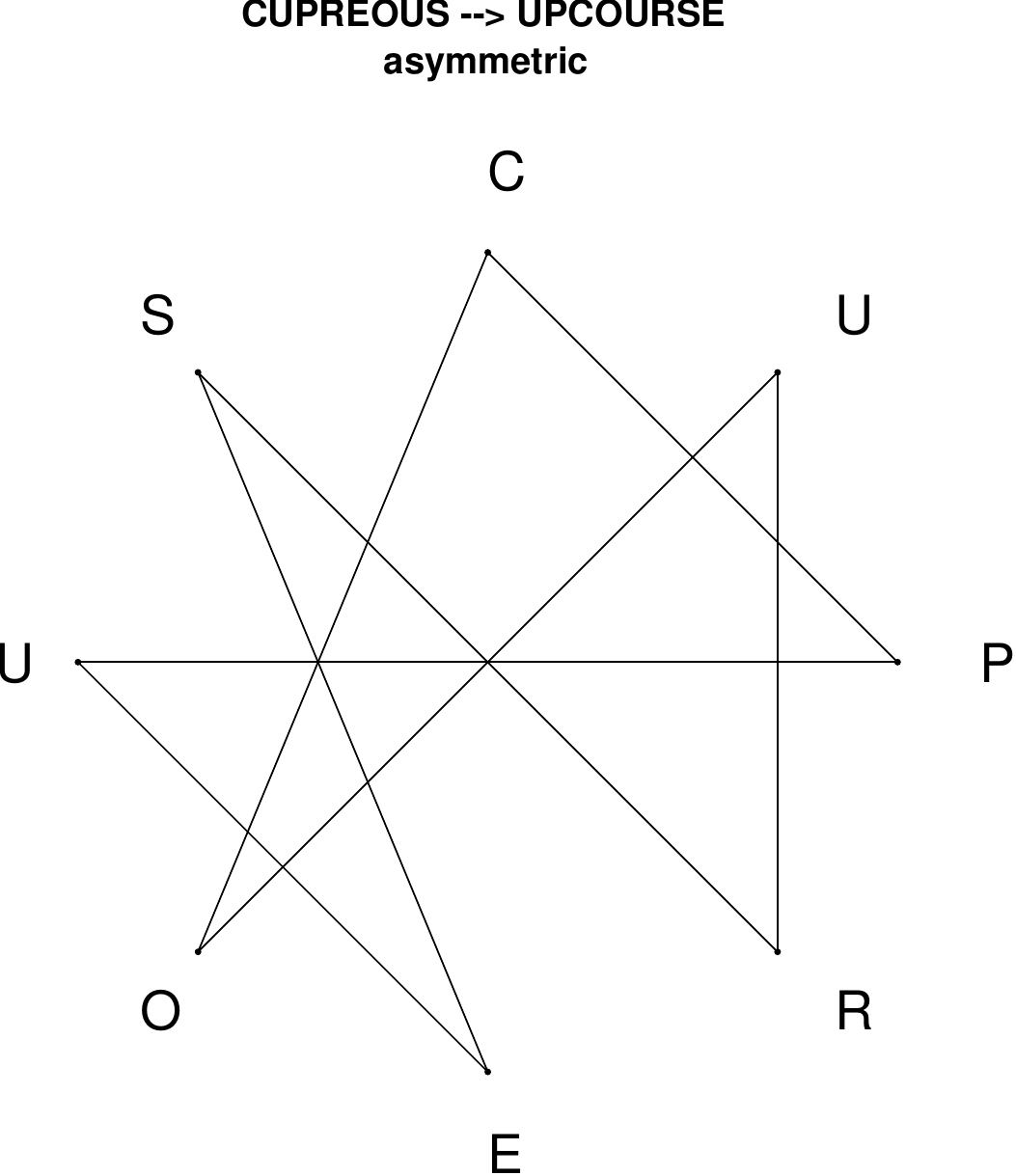}
\end{subfigure}
\hfill
\begin{subfigure}[T]{0.19\textwidth}
\centering
\includegraphics[width=\textwidth]{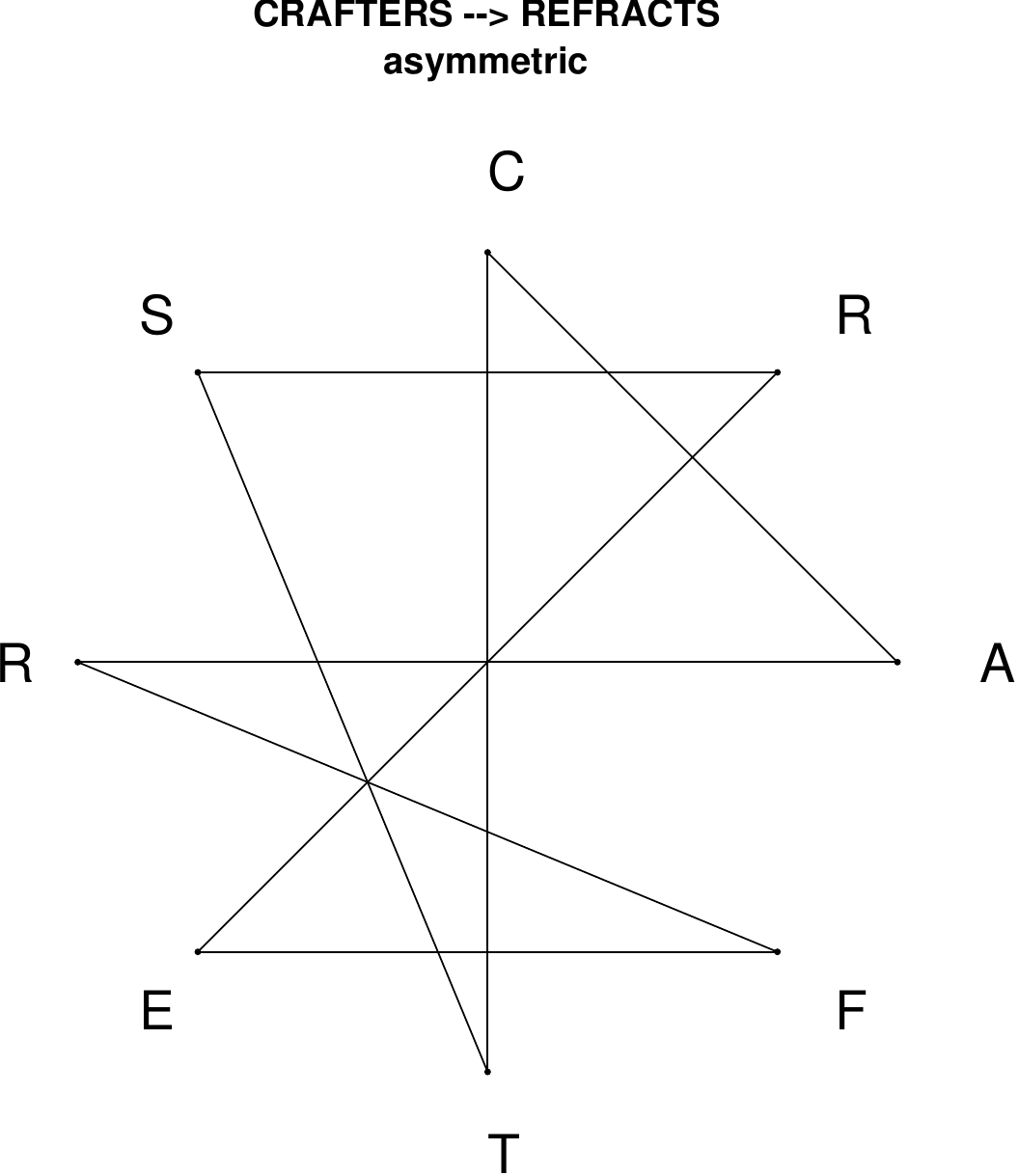}
\end{subfigure}
\end{figure}

\begin{figure}[H]
\centering
\begin{subfigure}[T]{0.19\textwidth}
\centering
\includegraphics[width=\textwidth]{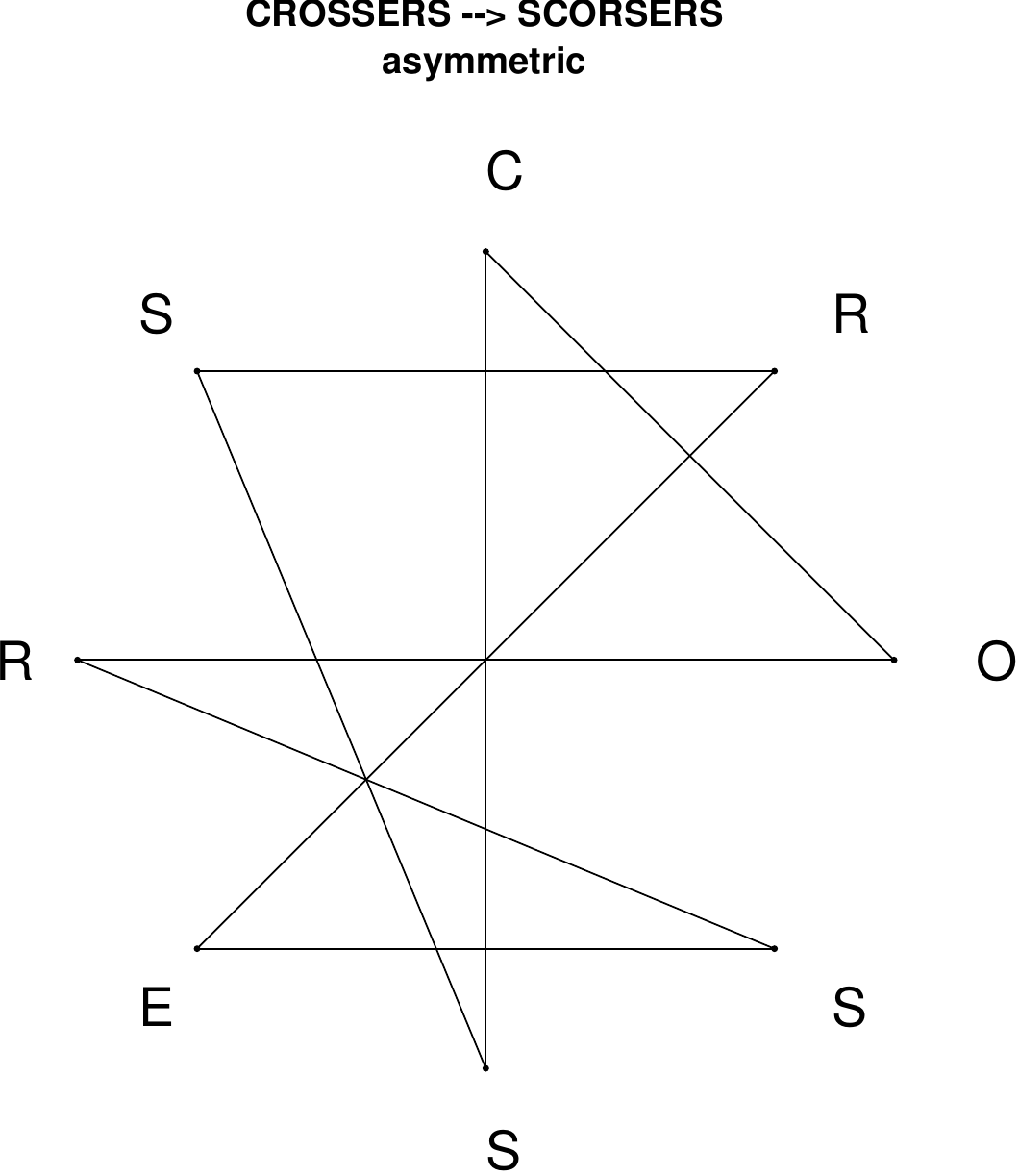}
\end{subfigure}
\hfill
\begin{subfigure}[T]{0.19\textwidth}
\centering
\includegraphics[width=\textwidth]{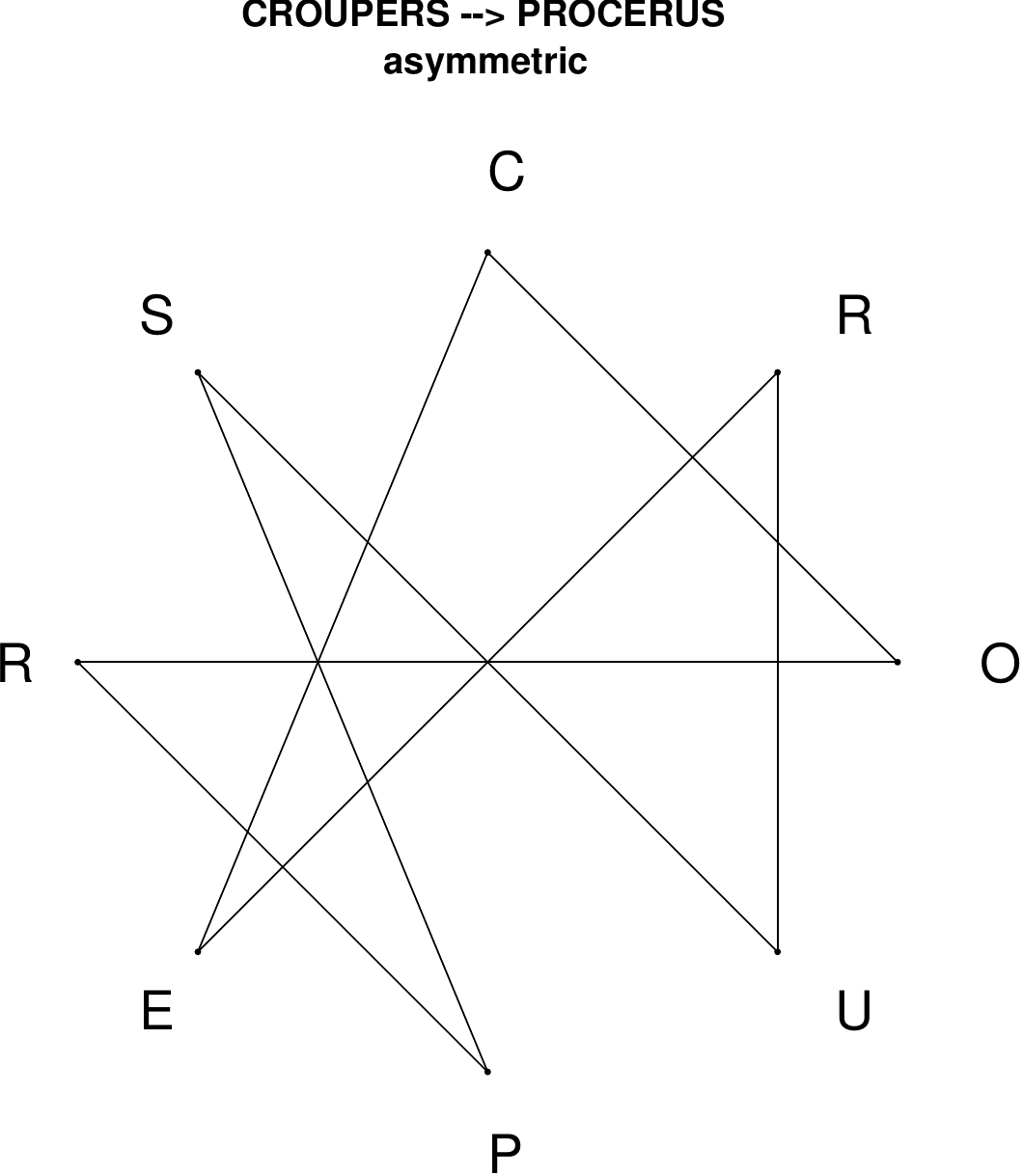}
\end{subfigure}
\hfill
\begin{subfigure}[T]{0.19\textwidth}
\centering
\includegraphics[width=\textwidth]{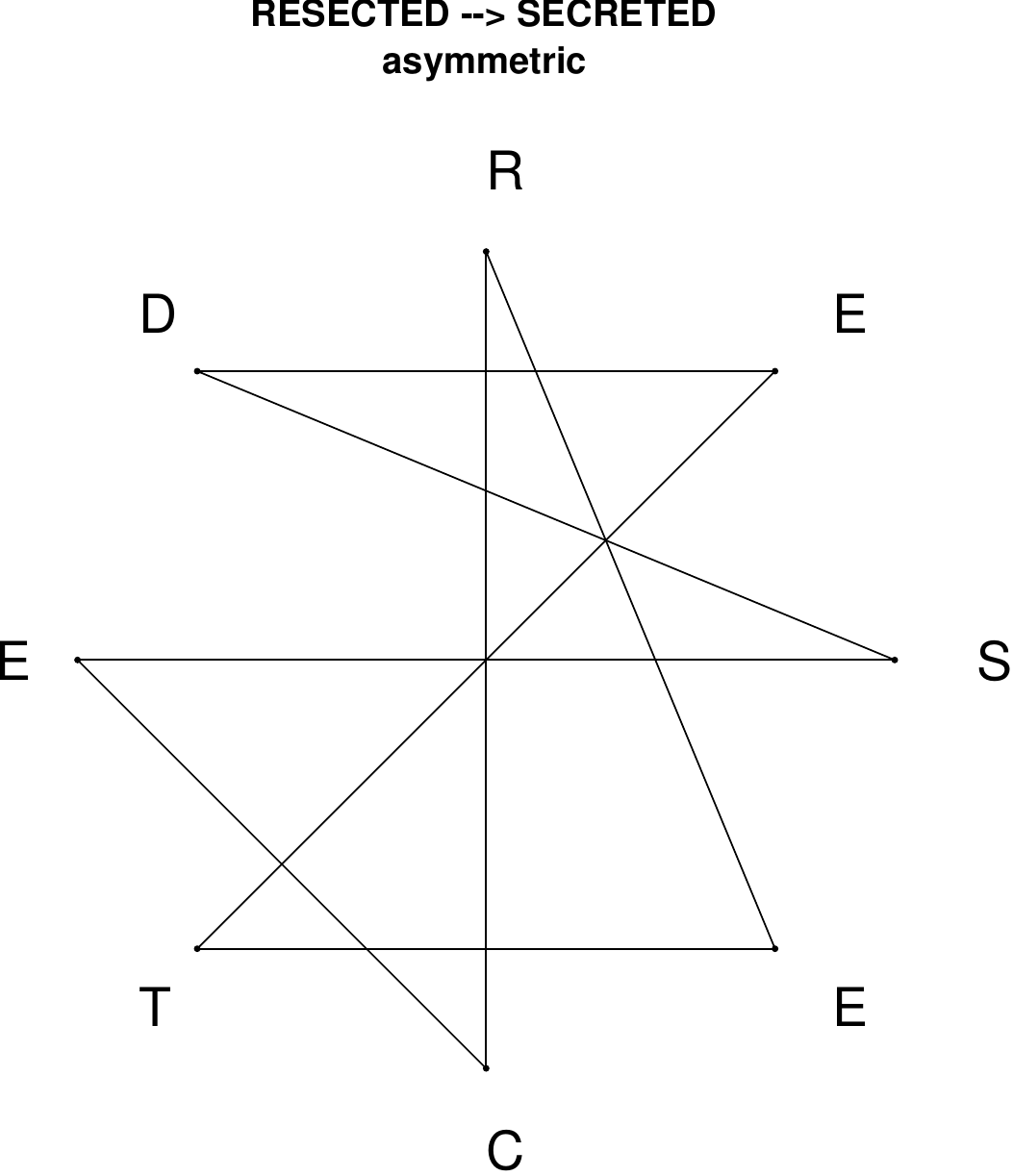}
\end{subfigure}
\hfill
\begin{subfigure}[T]{0.19\textwidth}
\centering
\includegraphics[width=\textwidth]{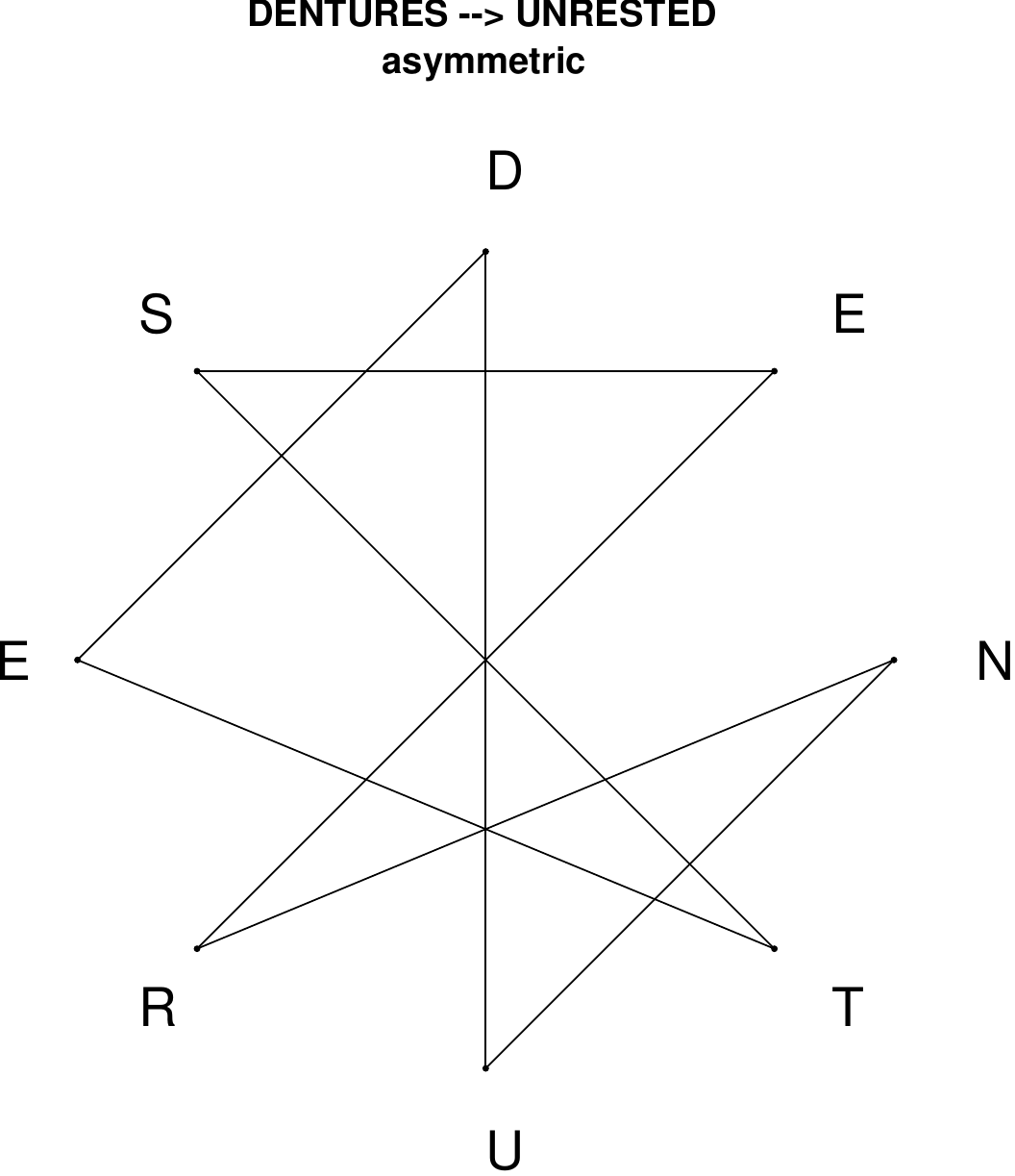}
\end{subfigure}
\hfill
\begin{subfigure}[T]{0.19\textwidth}
\centering
\includegraphics[width=\textwidth]{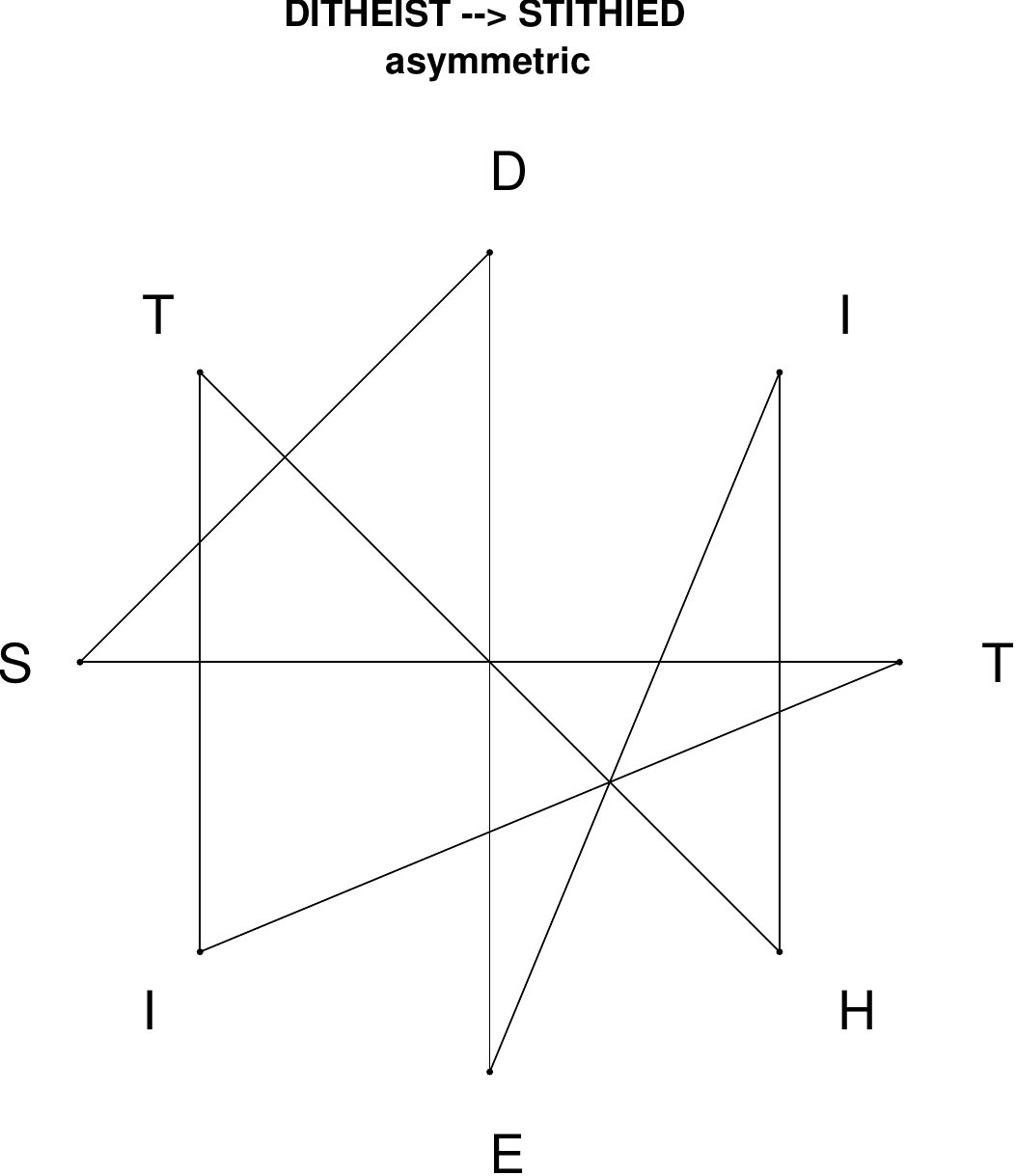}
\end{subfigure}
\end{figure}

\begin{figure}[H]
\centering
\begin{subfigure}[T]{0.19\textwidth}
\centering
\includegraphics[width=\textwidth]{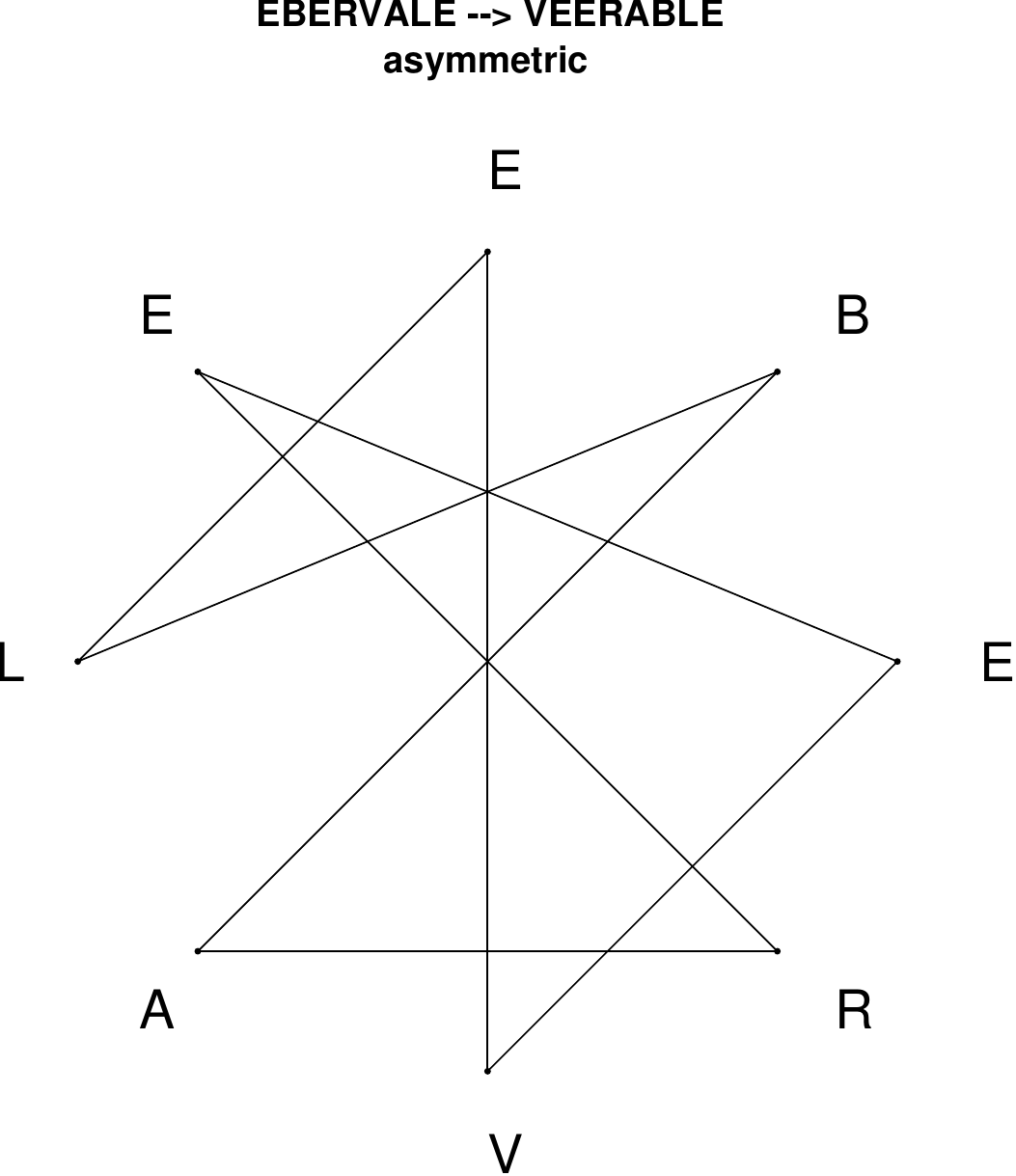}
\end{subfigure}
\hfill
\begin{subfigure}[T]{0.19\textwidth}
\centering
\includegraphics[width=\textwidth]{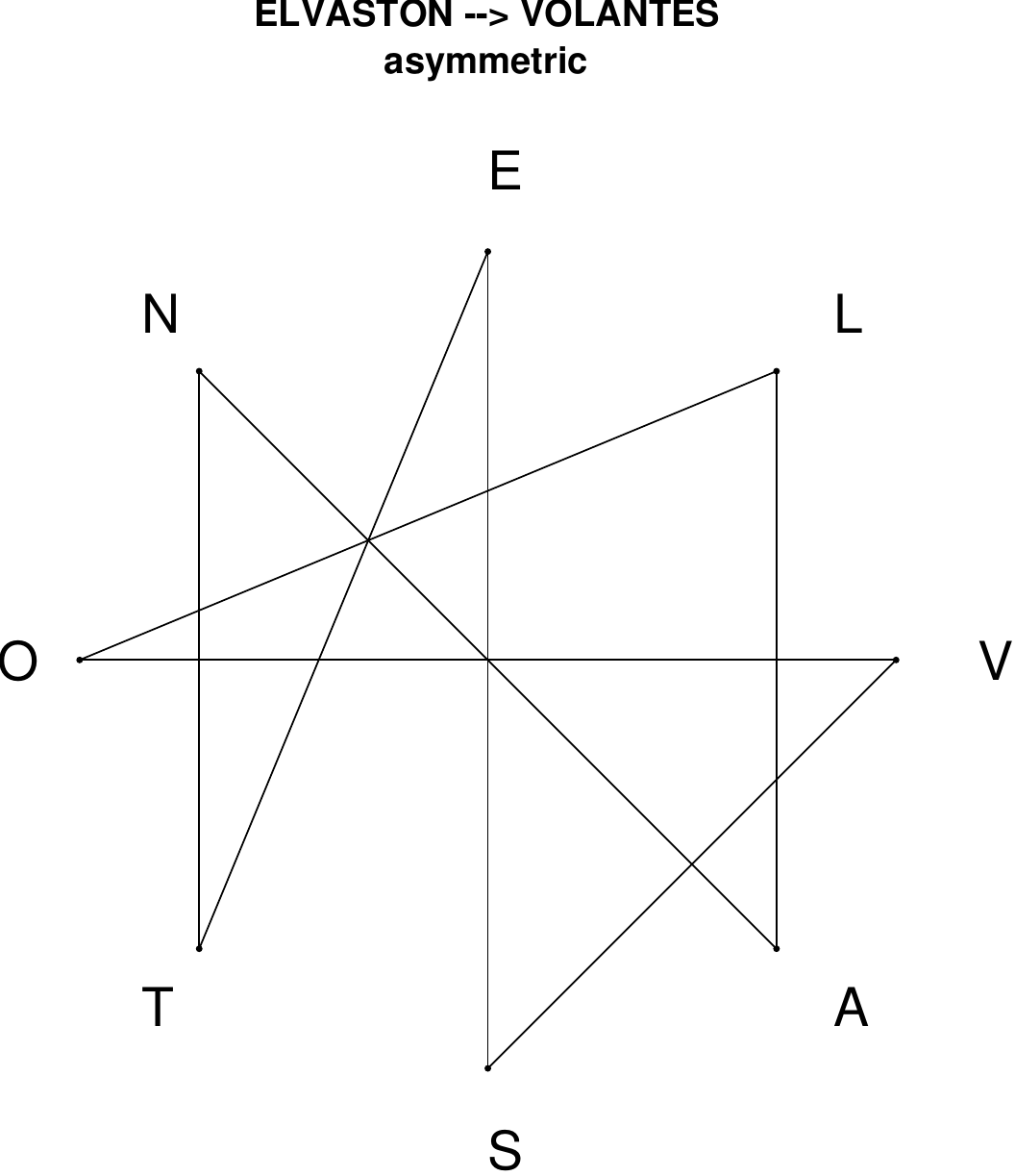}
\end{subfigure}
\hfill
\begin{subfigure}[T]{0.19\textwidth}
\centering
\includegraphics[width=\textwidth]{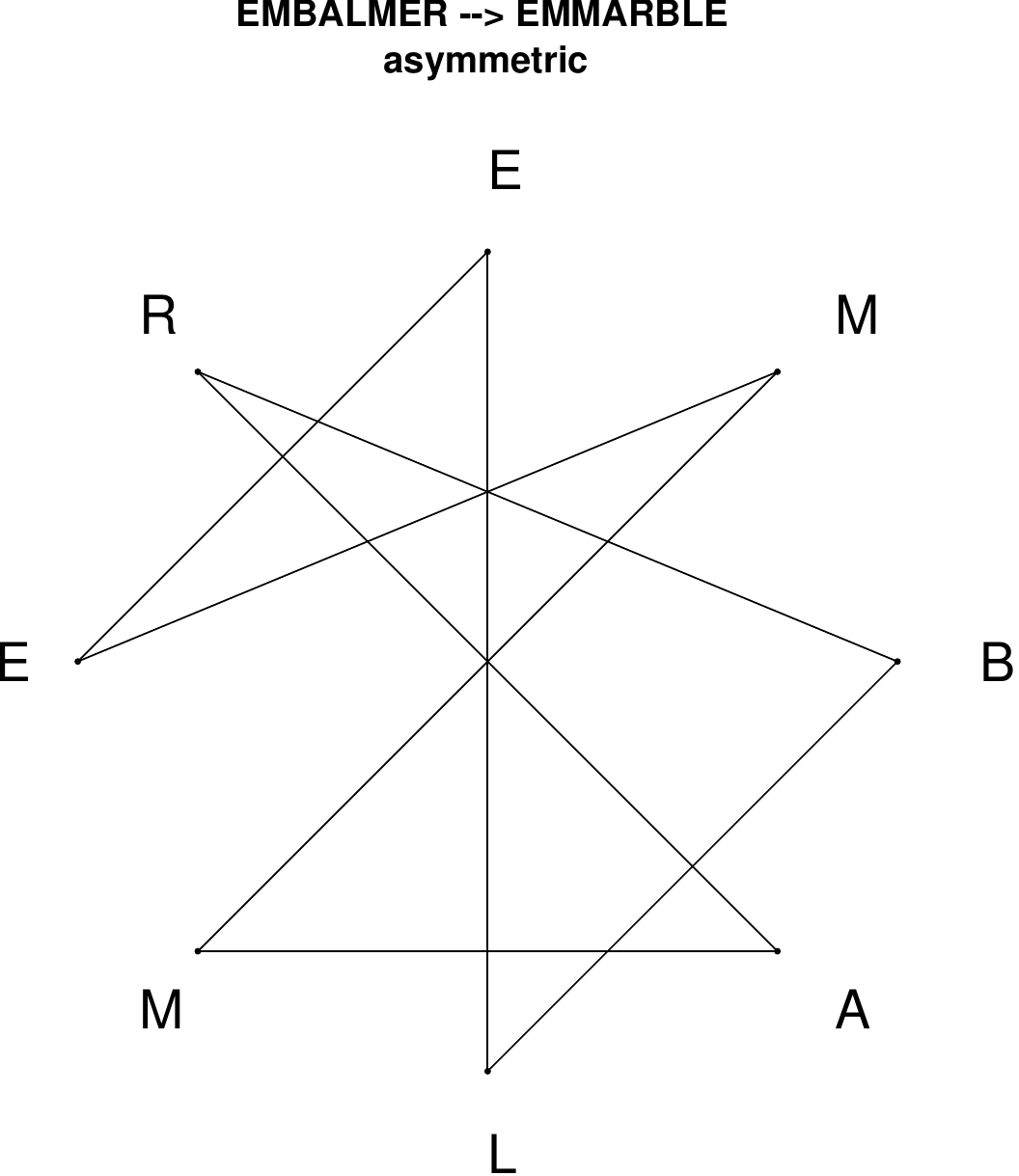}
\end{subfigure}
\hfill
\begin{subfigure}[T]{0.19\textwidth}
\centering
\includegraphics[width=\textwidth]{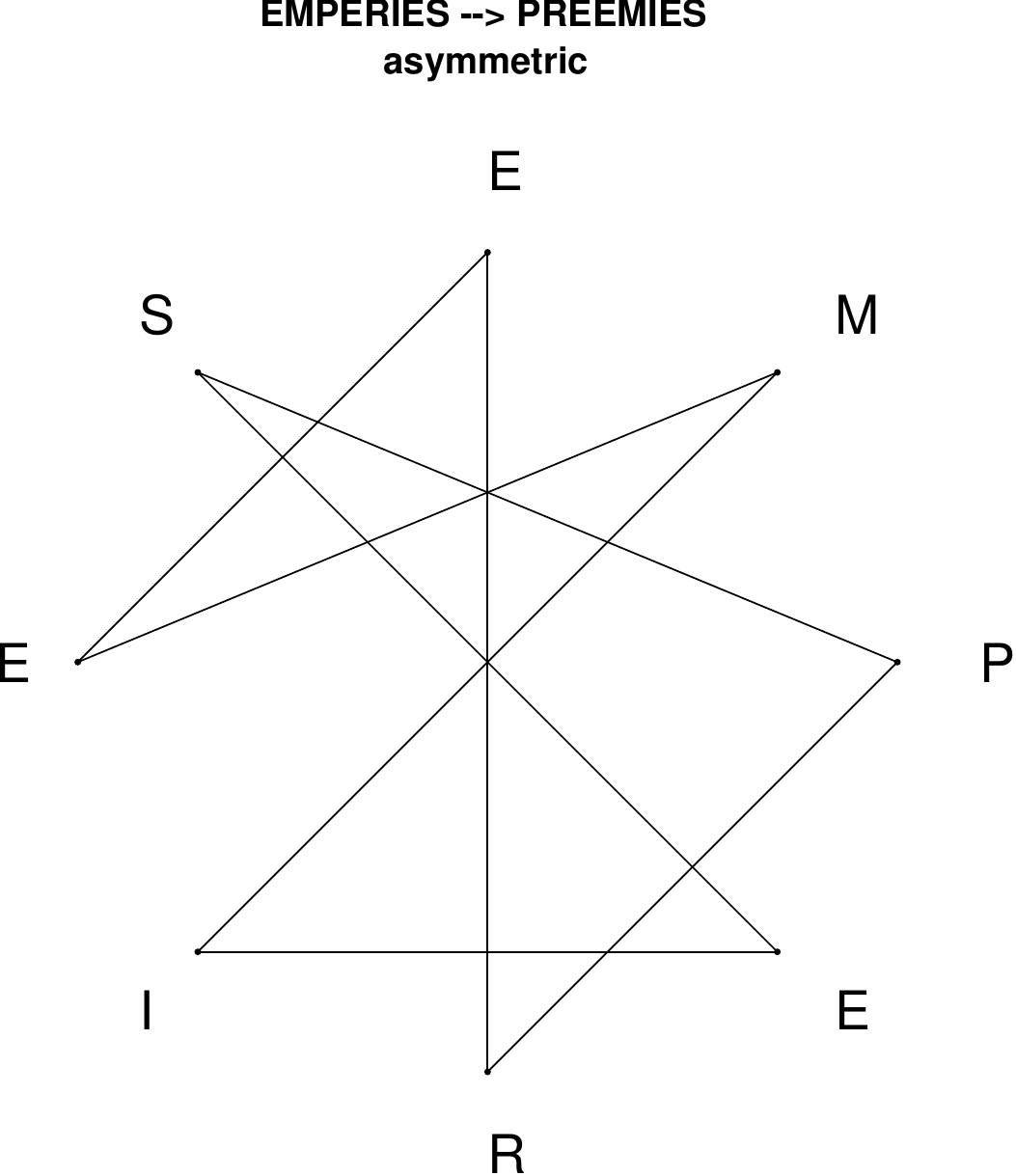}
\end{subfigure}
\hfill
\begin{subfigure}[T]{0.19\textwidth}
\centering
\includegraphics[width=\textwidth]{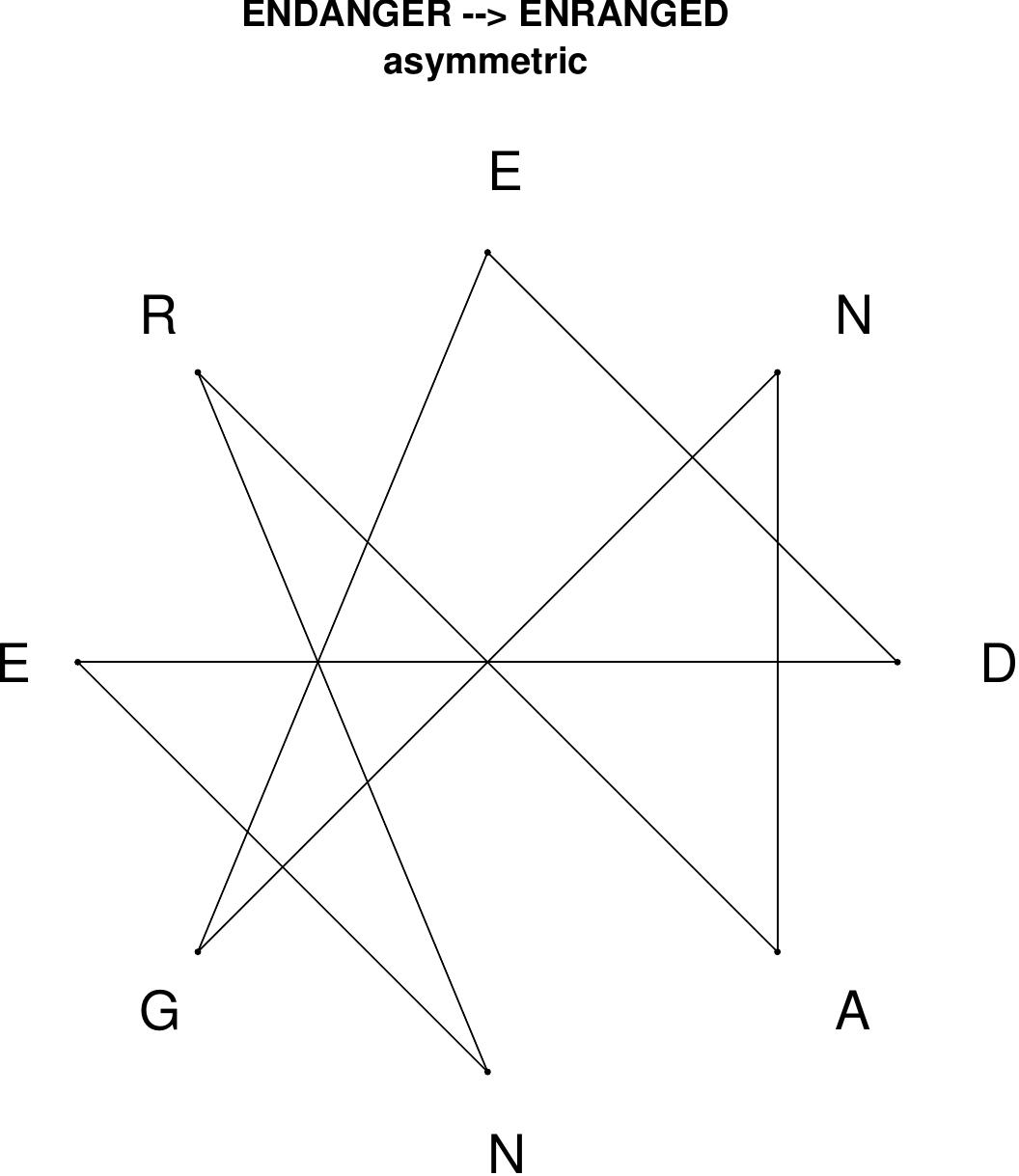}
\end{subfigure}
\end{figure}

\begin{figure}[H]
\centering
\begin{subfigure}[T]{0.19\textwidth}
\centering
\includegraphics[width=\textwidth]{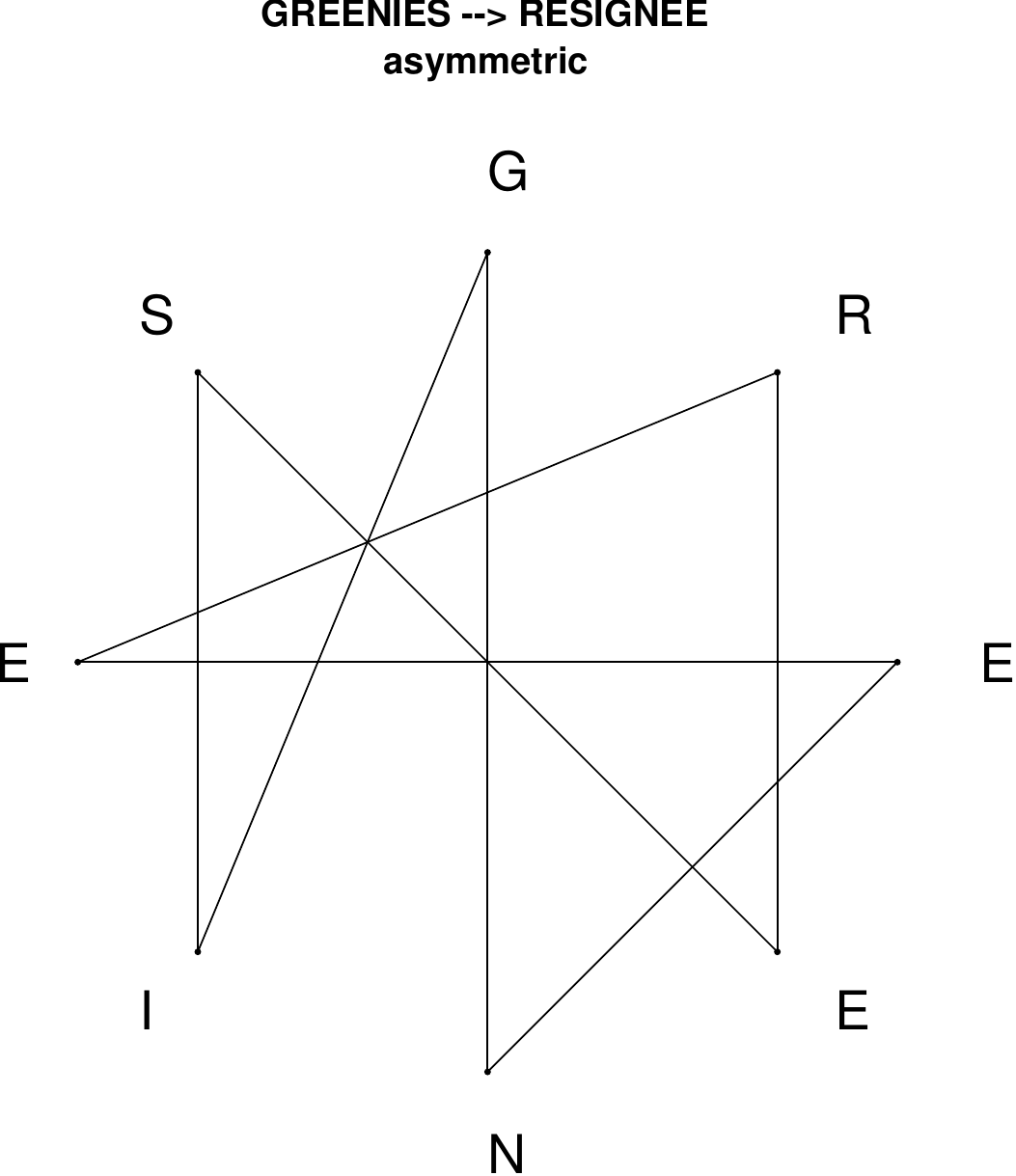}
\end{subfigure}
\hfill
\begin{subfigure}[T]{0.19\textwidth}
\centering
\includegraphics[width=\textwidth]{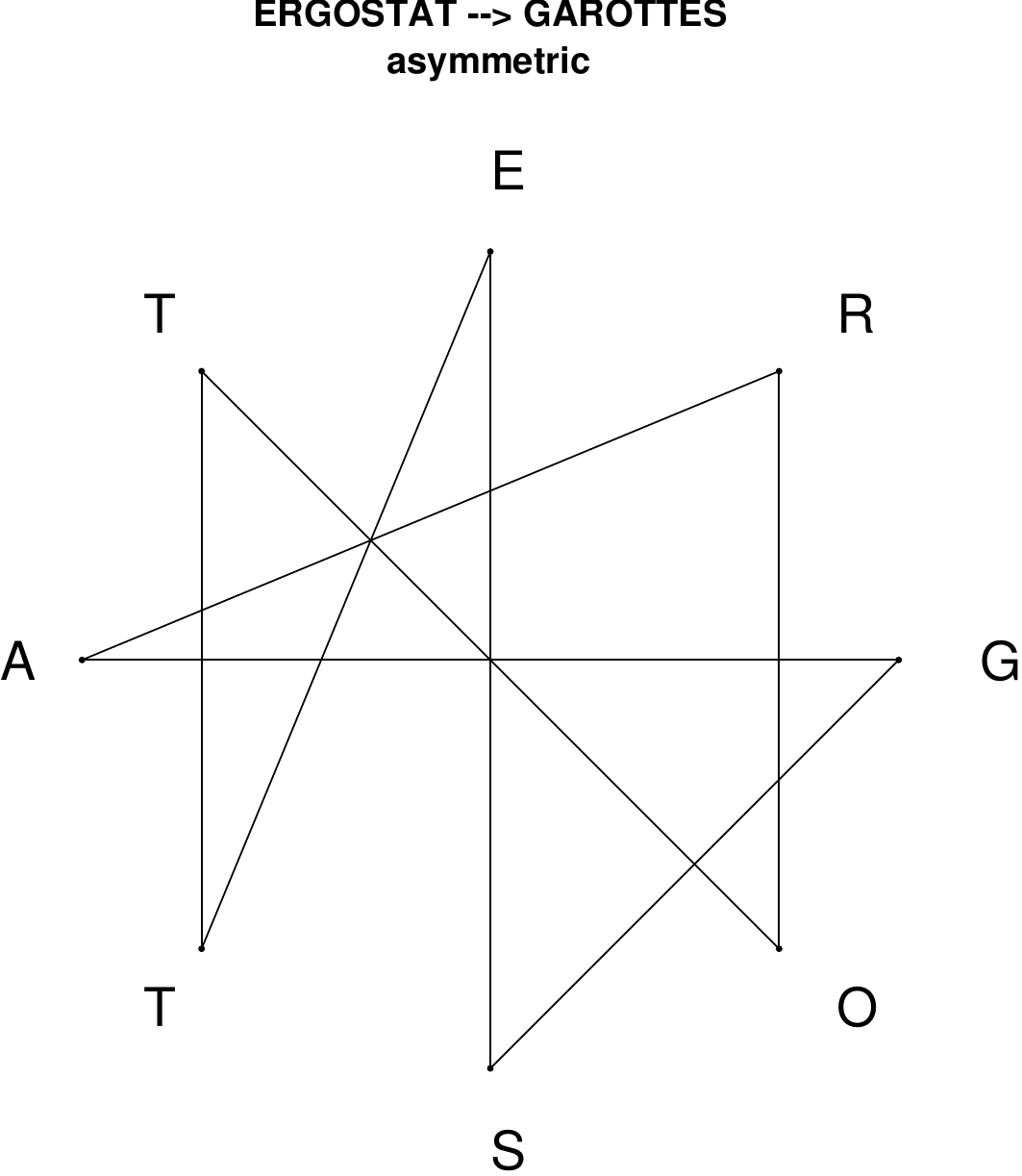}
\end{subfigure}
\hfill
\begin{subfigure}[T]{0.19\textwidth}
\centering
\includegraphics[width=\textwidth]{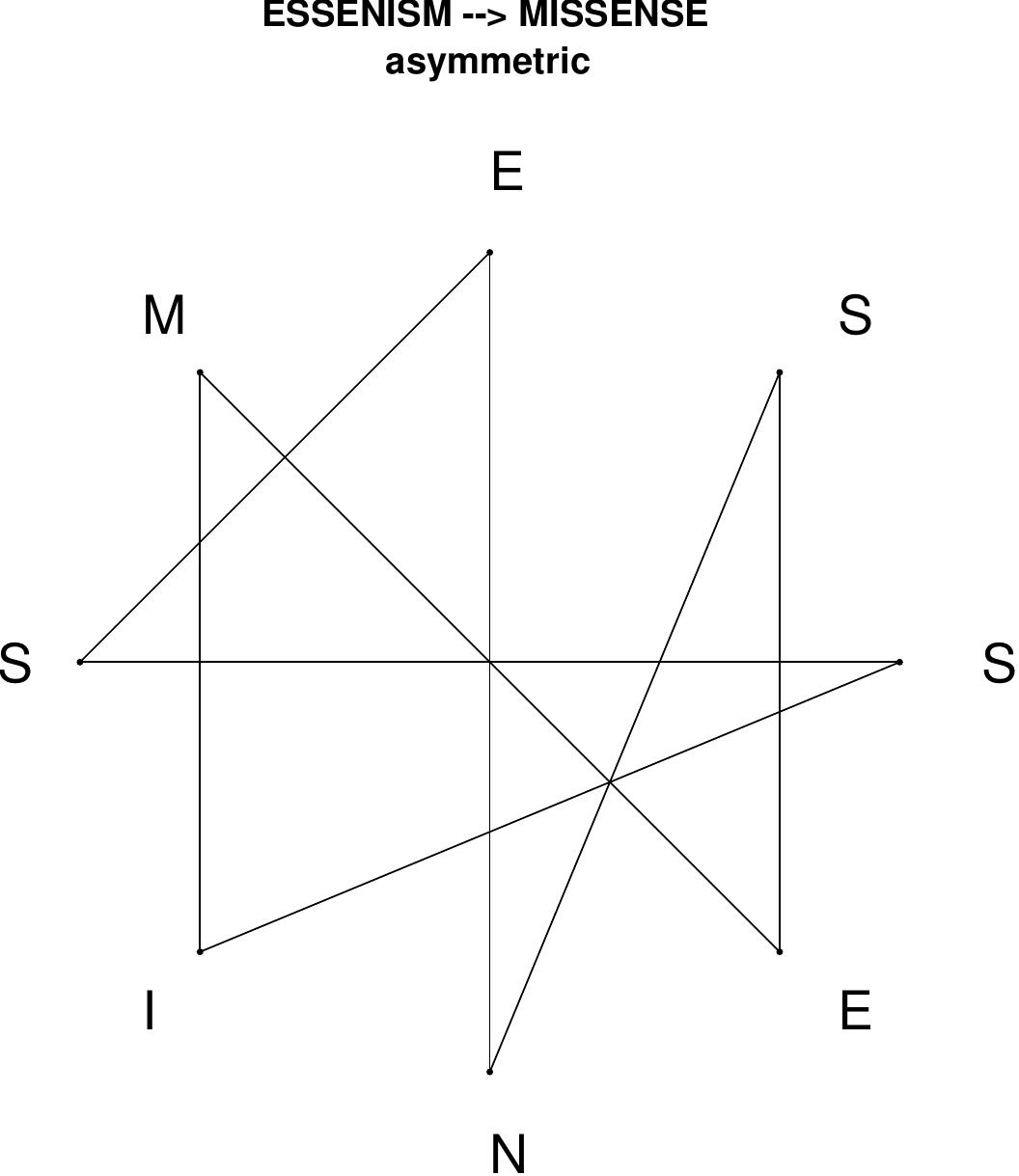}
\end{subfigure}
\hfill
\begin{subfigure}[T]{0.19\textwidth}
\centering
\includegraphics[width=\textwidth]{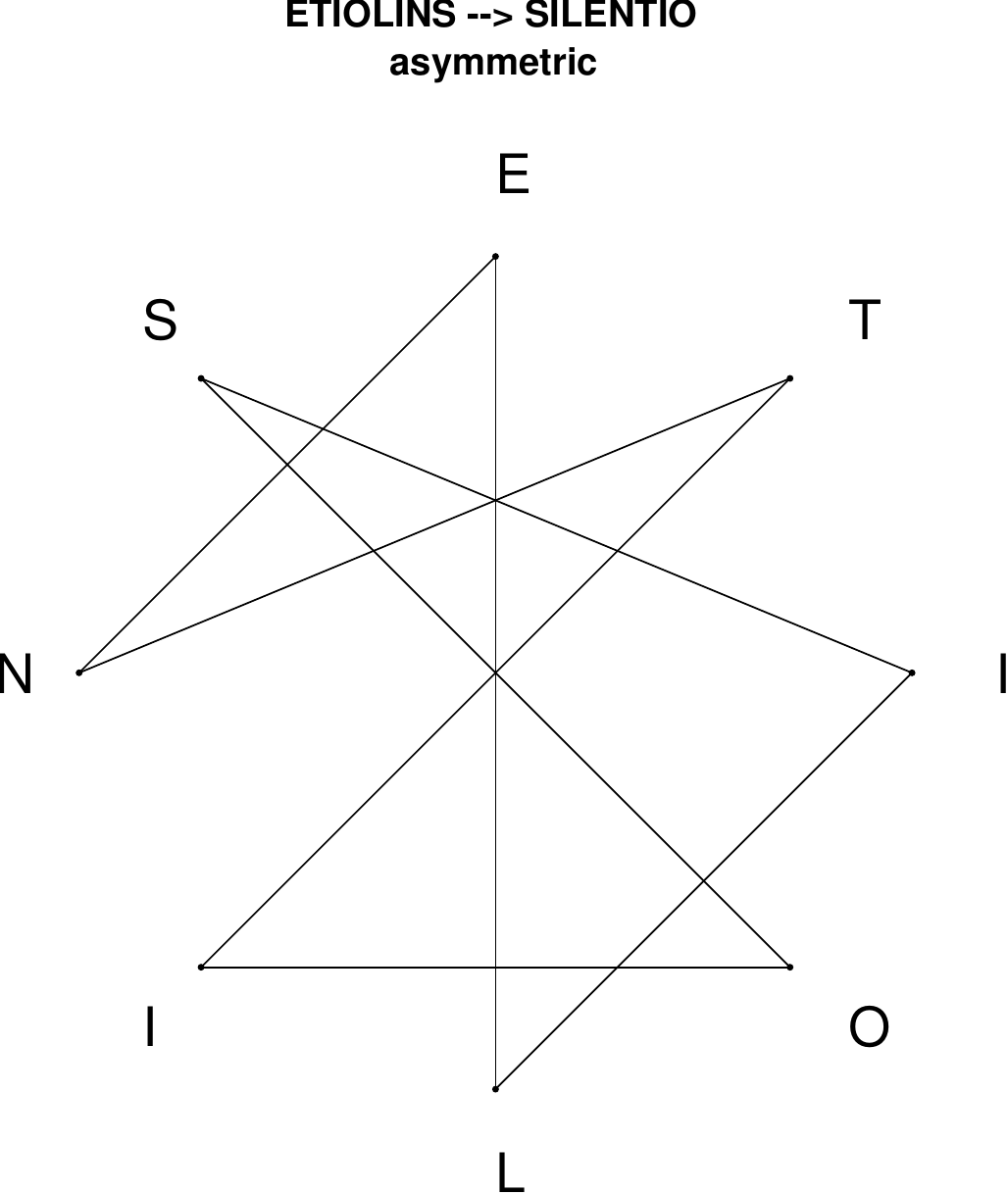}
\end{subfigure}
\hfill
\begin{subfigure}[T]{0.19\textwidth}
\centering
\includegraphics[width=\textwidth]{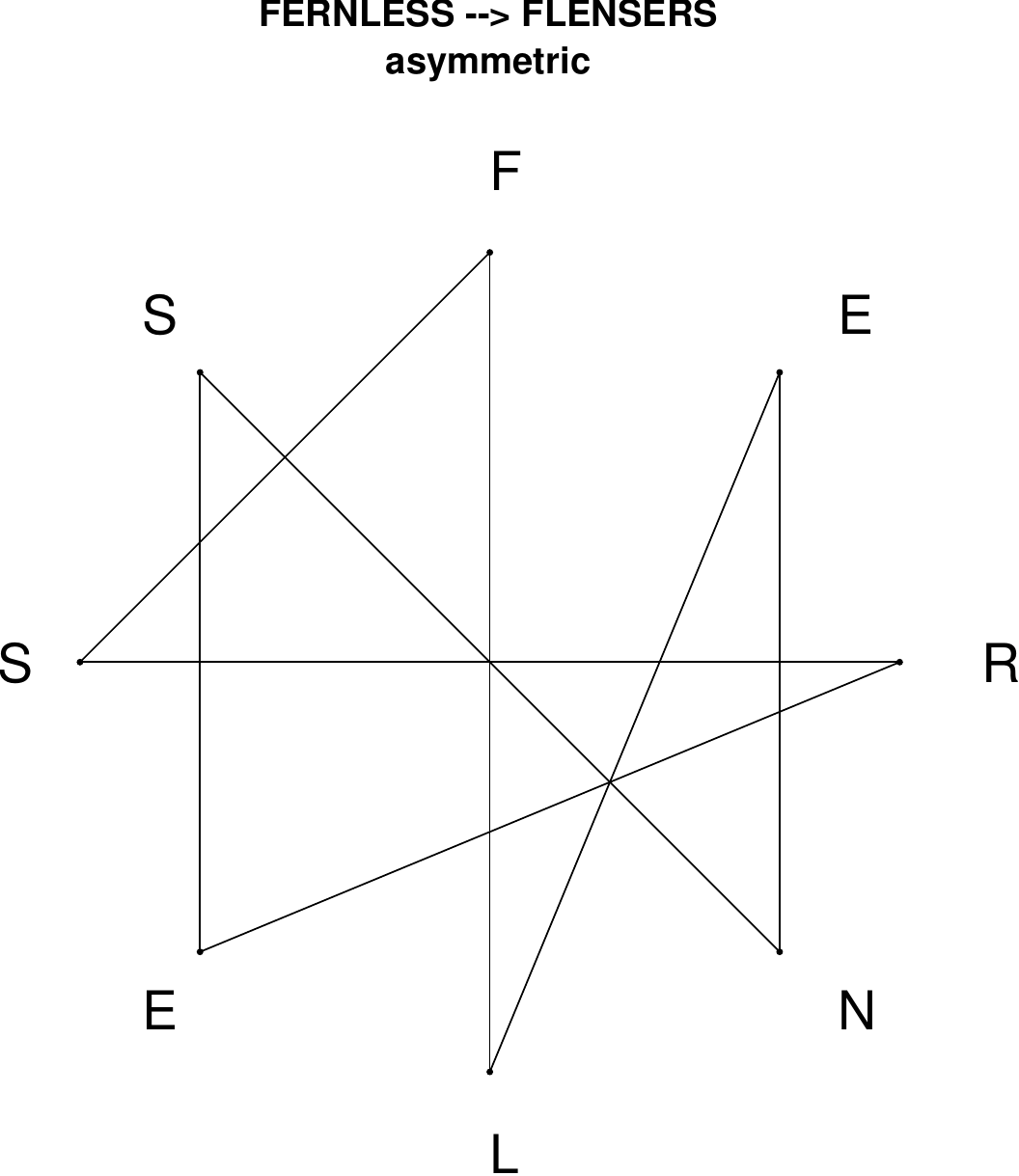}
\end{subfigure}
\end{figure}

\begin{figure}[H]
\centering
\begin{subfigure}[T]{0.19\textwidth}
\centering
\includegraphics[width=\textwidth]{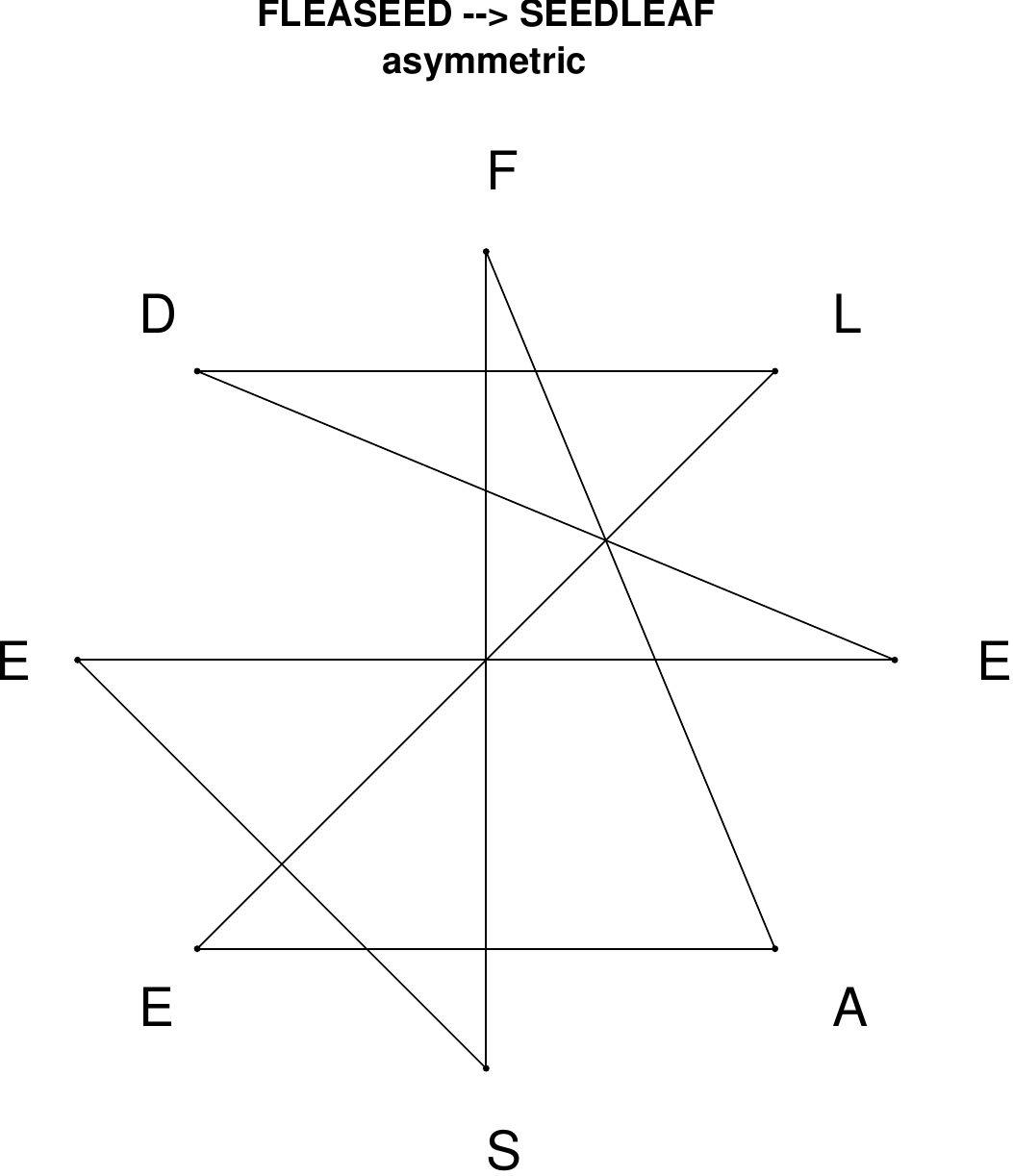}
\end{subfigure}
\hfill
\begin{subfigure}[T]{0.19\textwidth}
\centering
\includegraphics[width=\textwidth]{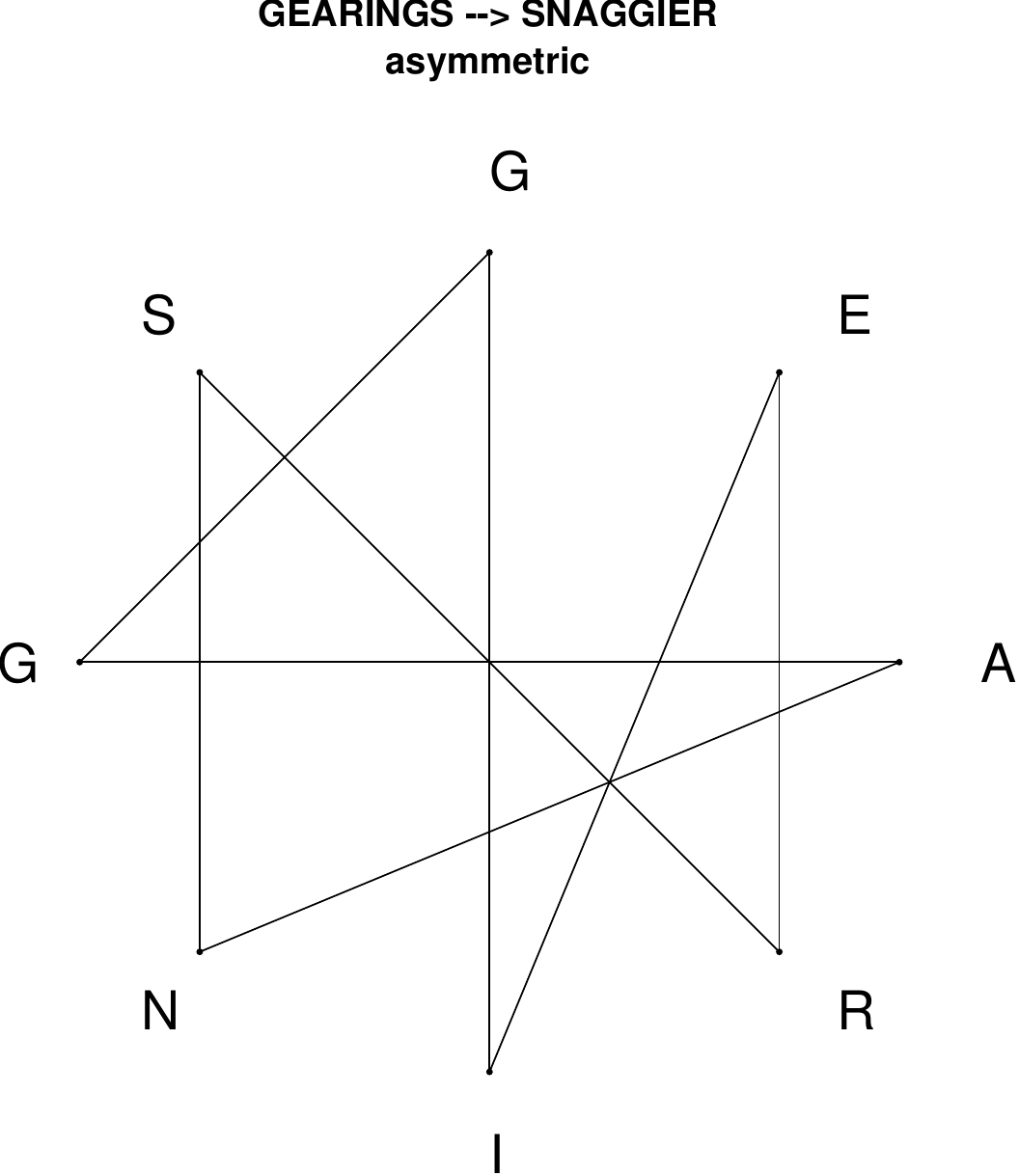}
\end{subfigure}
\hfill
\begin{subfigure}[T]{0.19\textwidth}
\centering
\includegraphics[width=\textwidth]{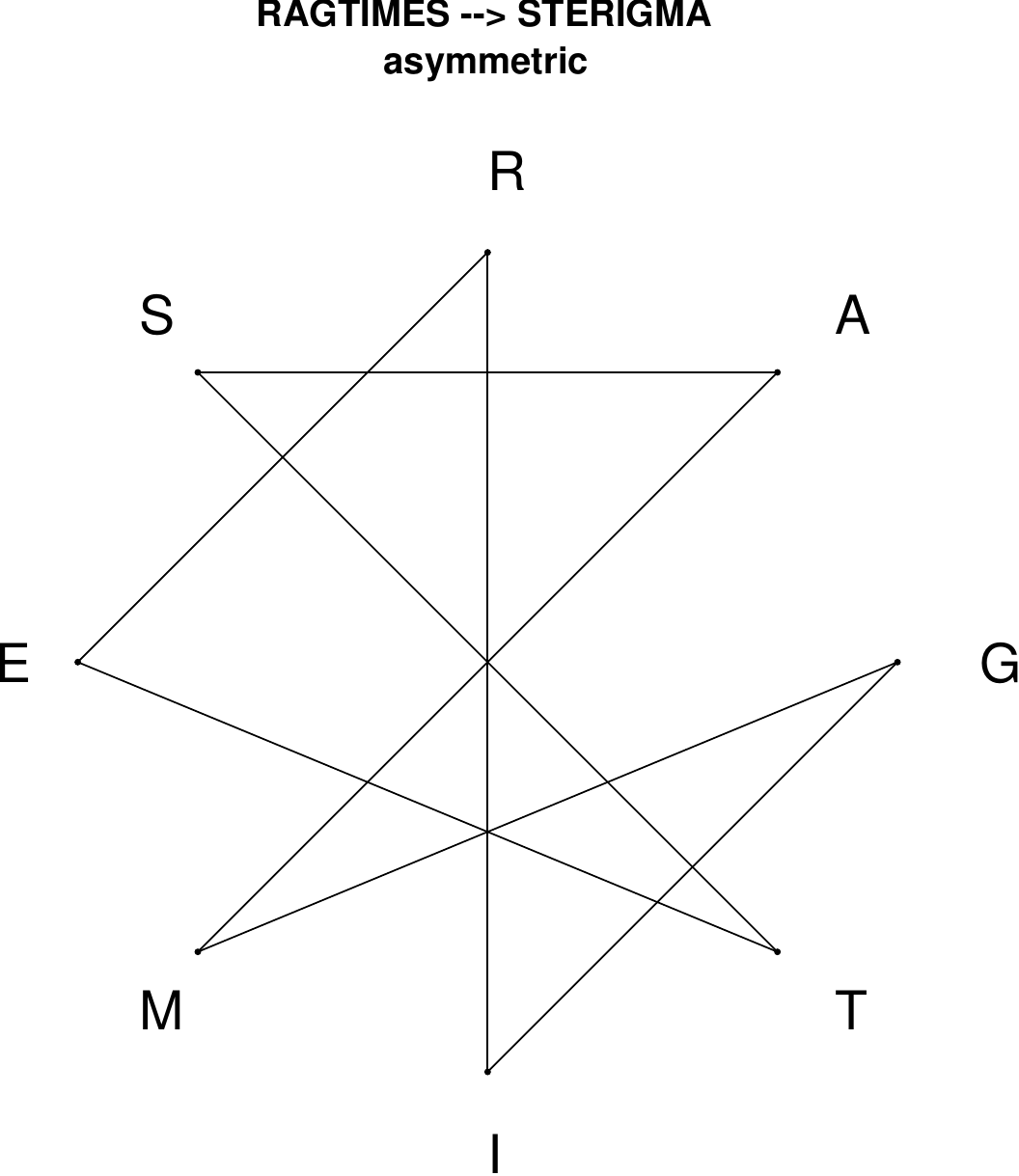}
\end{subfigure}
\hfill
\begin{subfigure}[T]{0.19\textwidth}
\centering
\includegraphics[width=\textwidth]{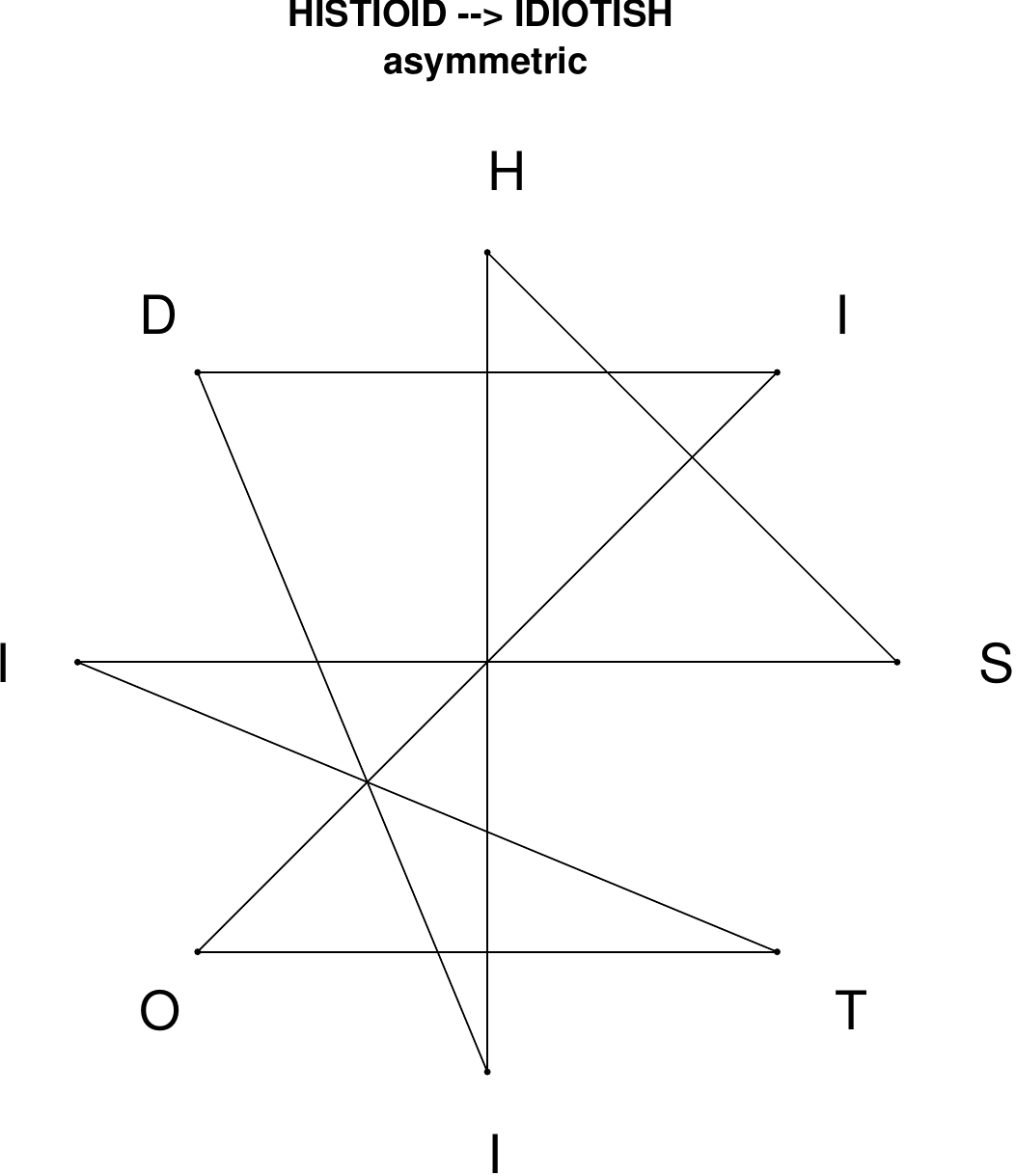}
\end{subfigure}
\hfill
\begin{subfigure}[T]{0.19\textwidth}
\centering
\includegraphics[width=\textwidth]{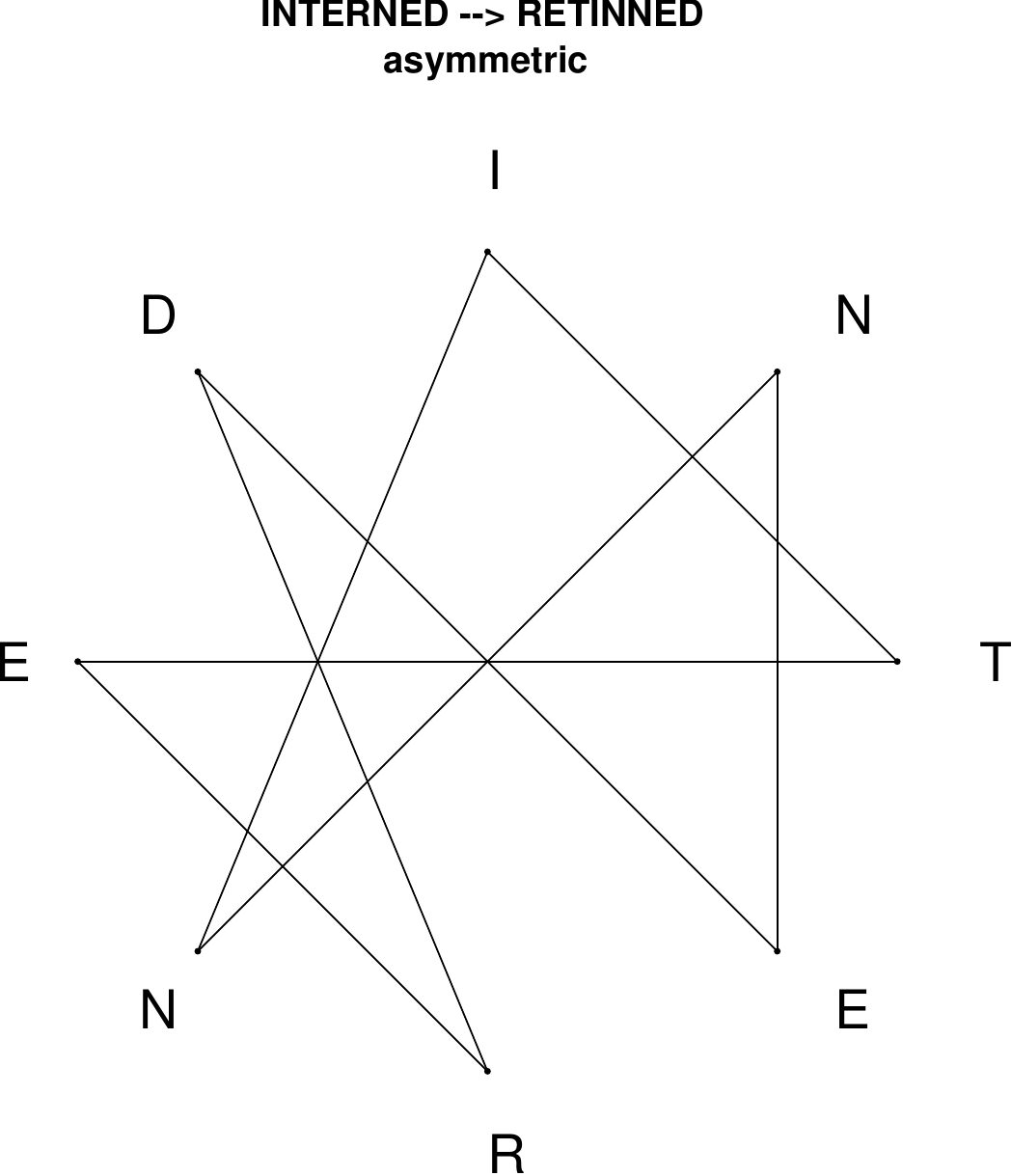}
\end{subfigure}
\end{figure}

\begin{figure}[H]
\centering
\begin{subfigure}[T]{0.19\textwidth}
\centering
\includegraphics[width=\textwidth]{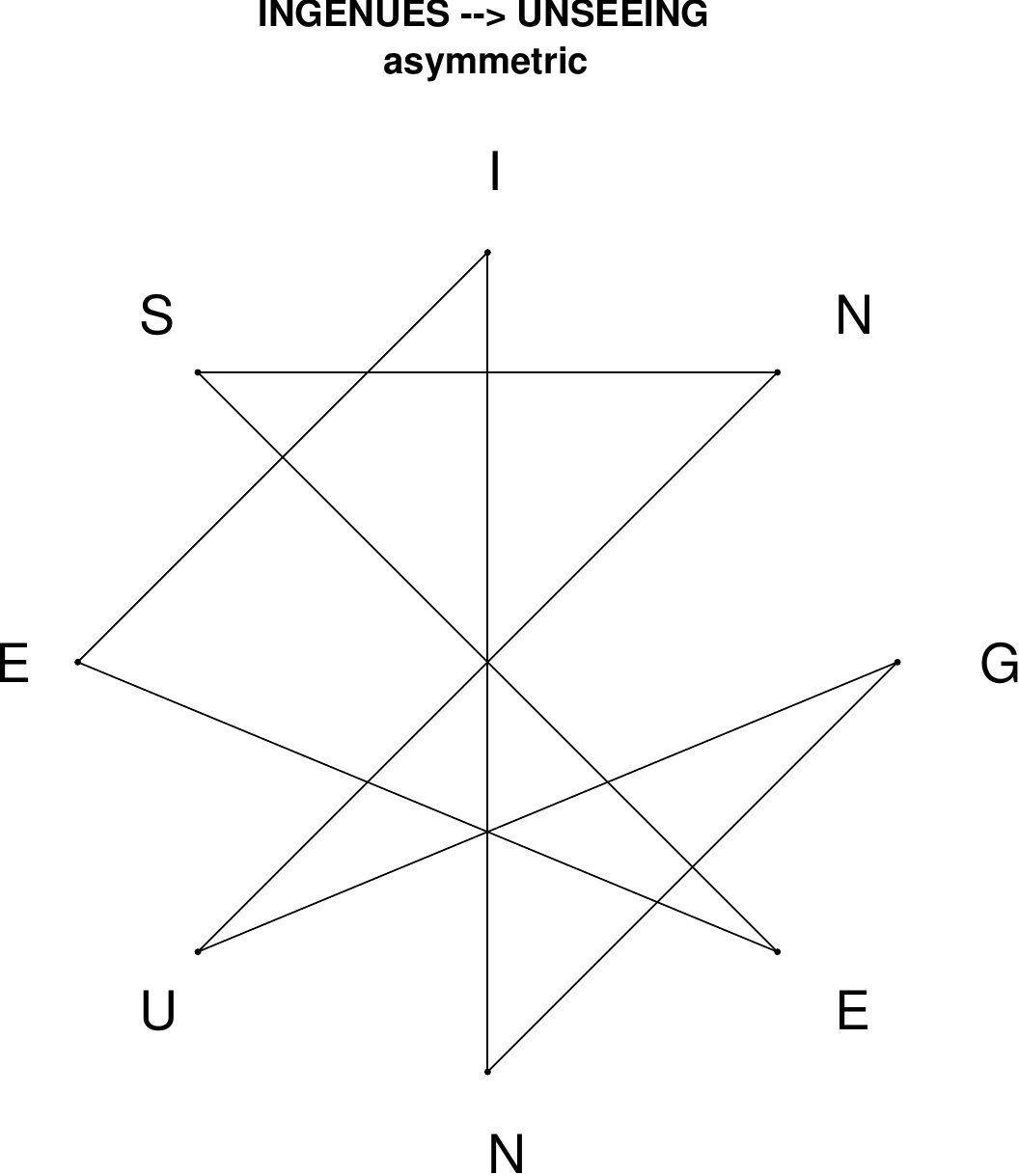}
\end{subfigure}
\hfill
\begin{subfigure}[T]{0.19\textwidth}
\centering
\includegraphics[width=\textwidth]{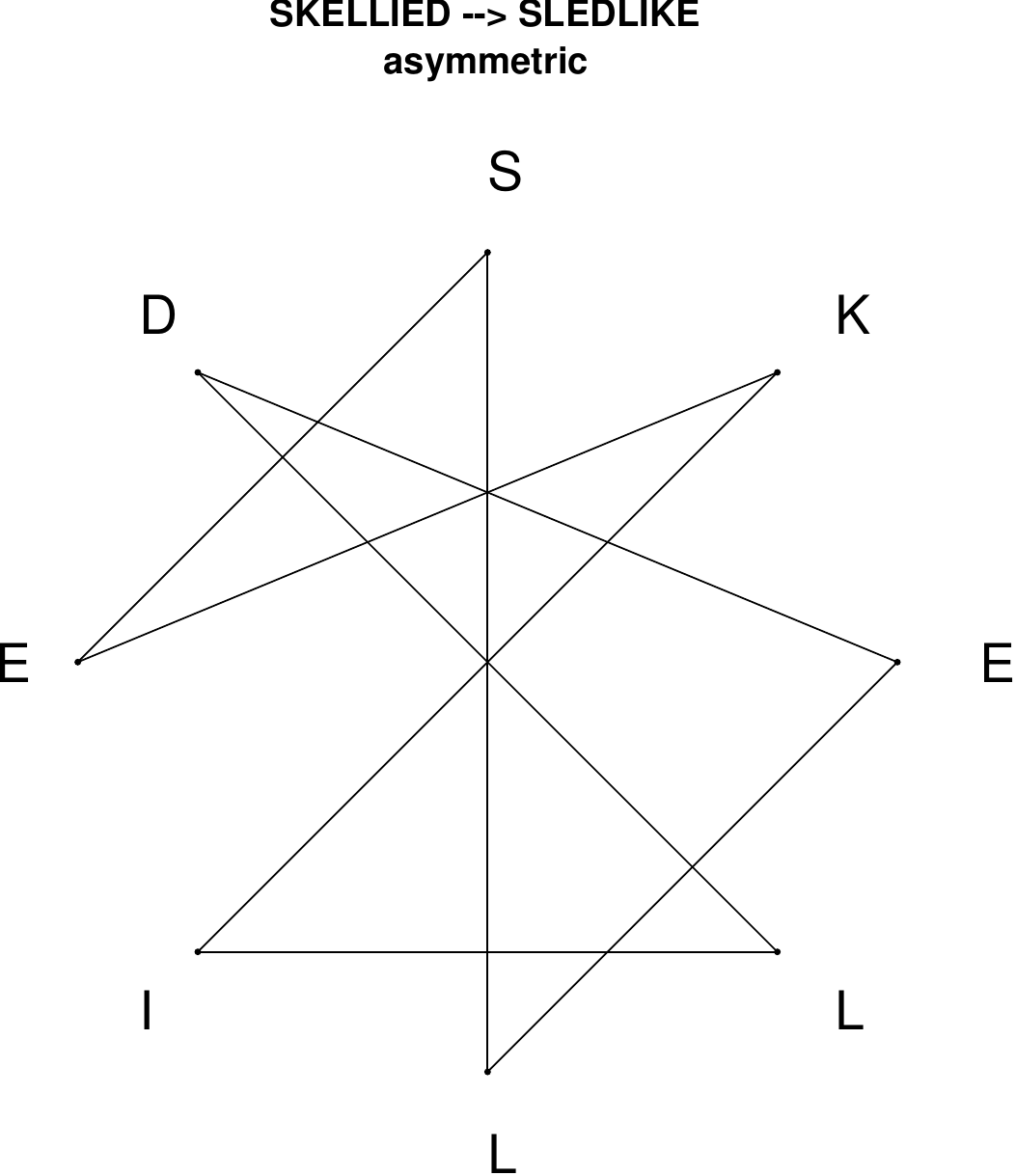}
\end{subfigure}
\hfill
\begin{subfigure}[T]{0.19\textwidth}
\centering
\includegraphics[width=\textwidth]{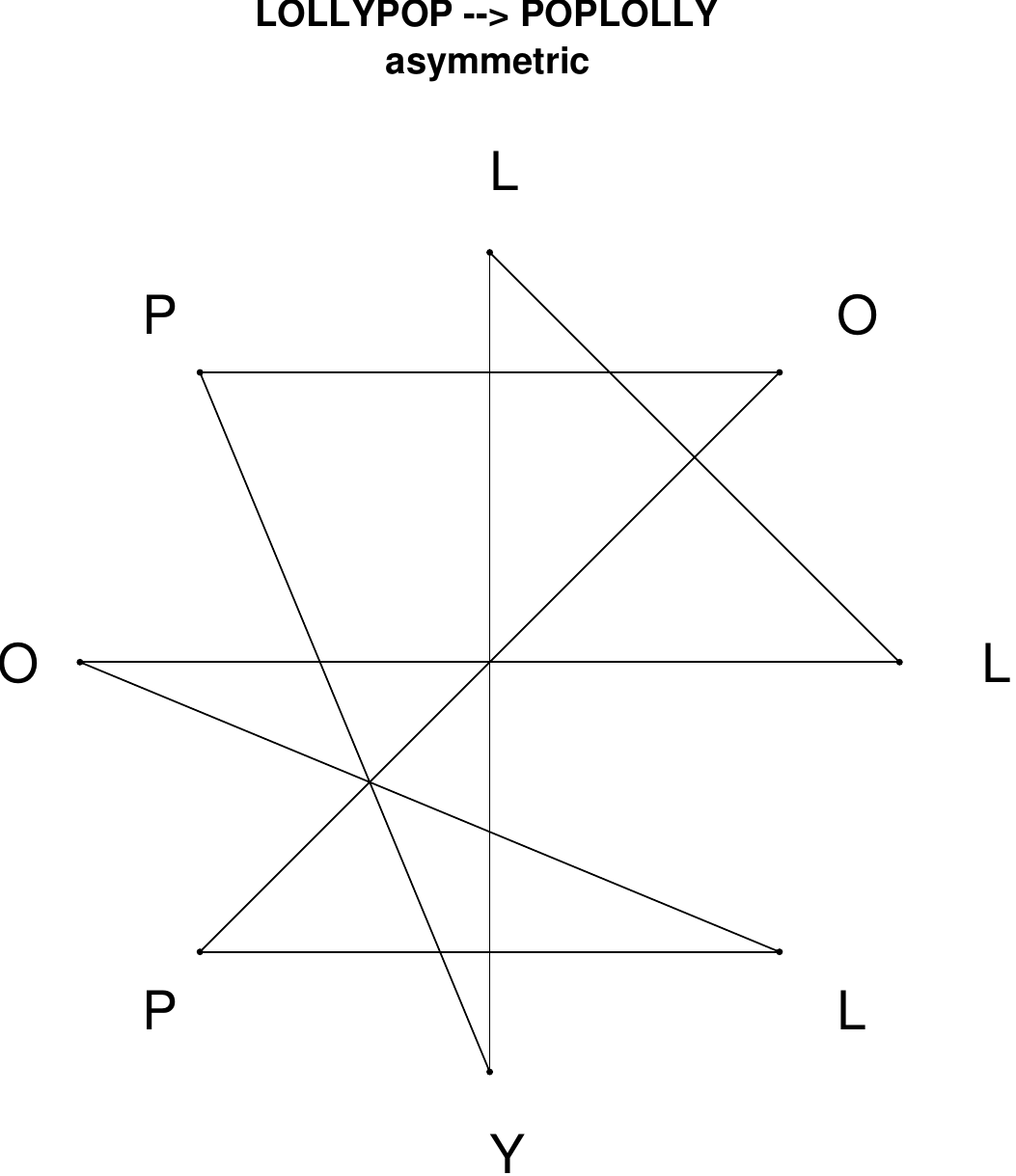}
\end{subfigure}
\hfill
\begin{subfigure}[T]{0.19\textwidth}
\centering
\includegraphics[width=\textwidth]{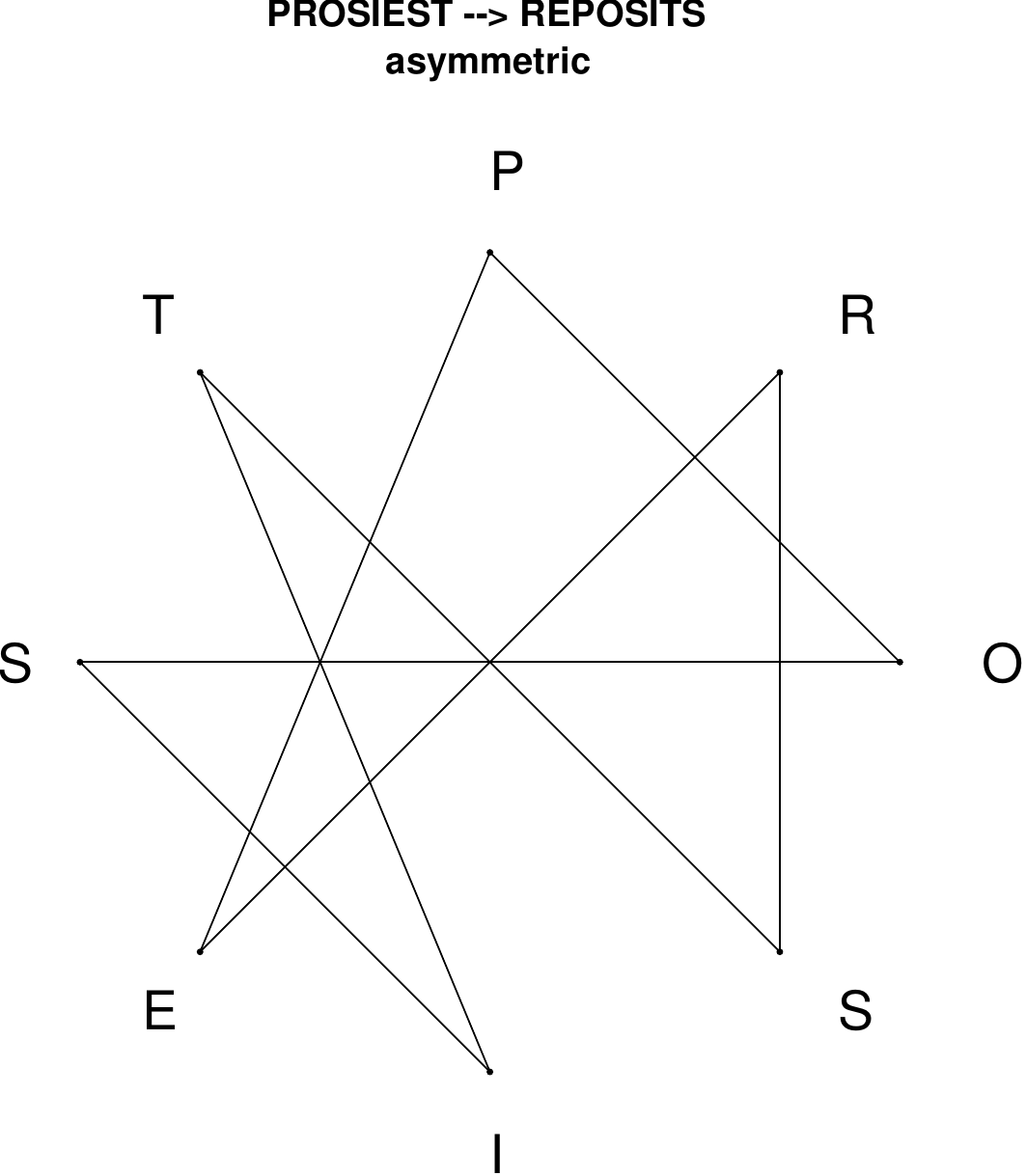}
\end{subfigure}
\hfill
\begin{subfigure}[T]{0.19\textwidth}
\centering
\includegraphics[width=\textwidth]{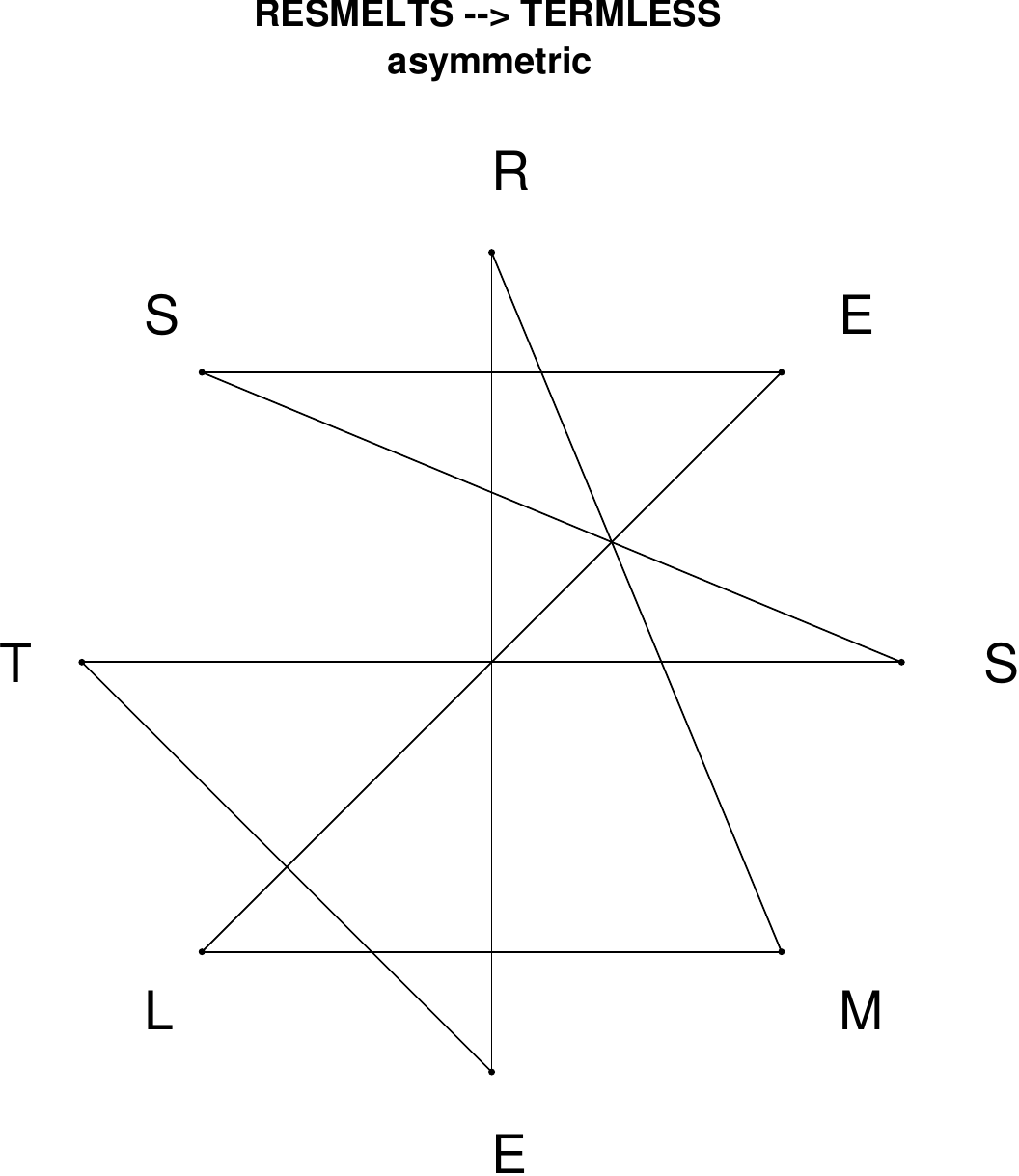}
\end{subfigure}
\end{figure}

\begin{figure}[H]
\centering
\begin{subfigure}[T]{0.19\textwidth}
\centering
\includegraphics[width=\textwidth]{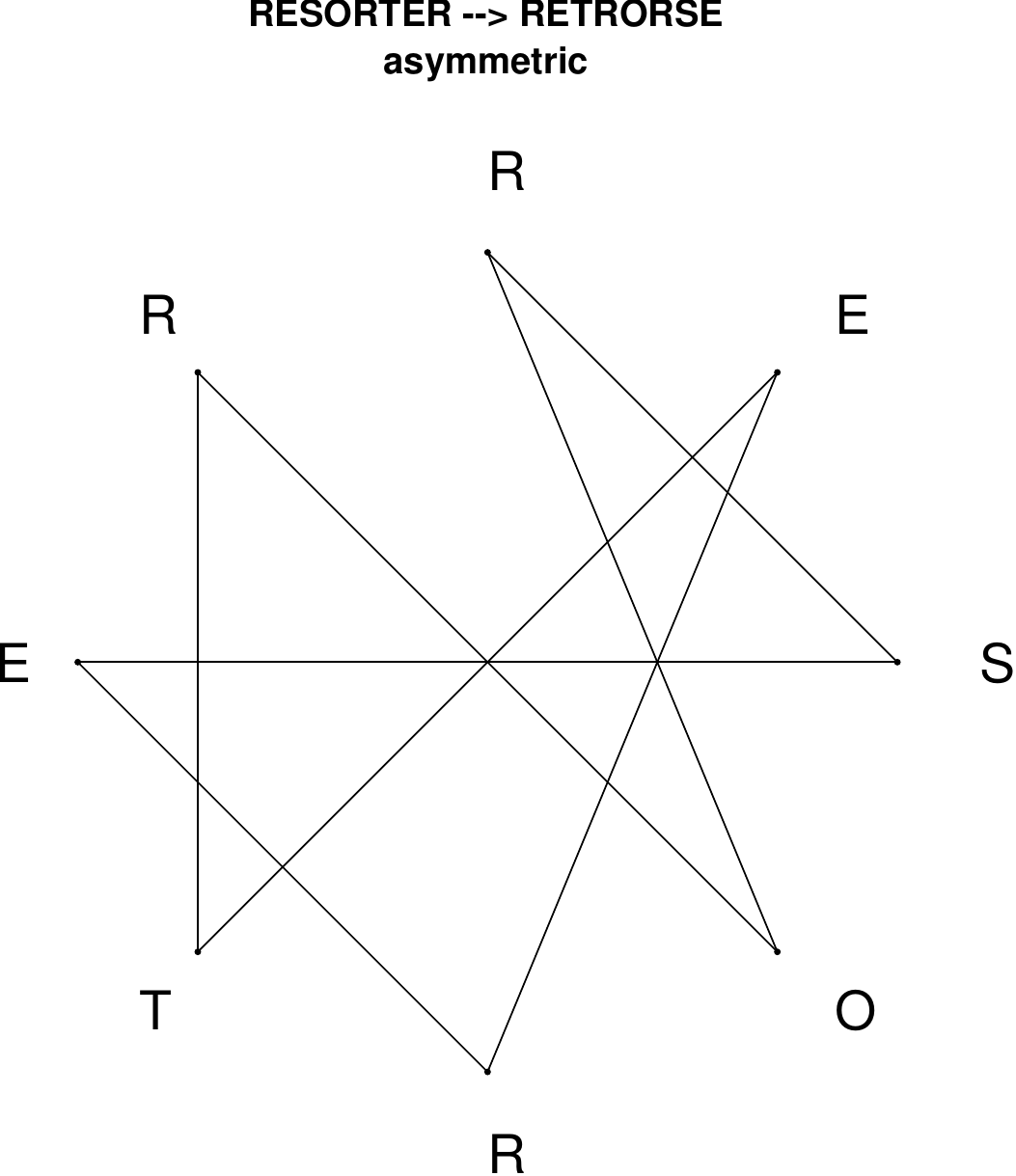}
\end{subfigure}
\hfill
\begin{subfigure}[T]{0.19\textwidth}
\centering
\includegraphics[width=\textwidth]{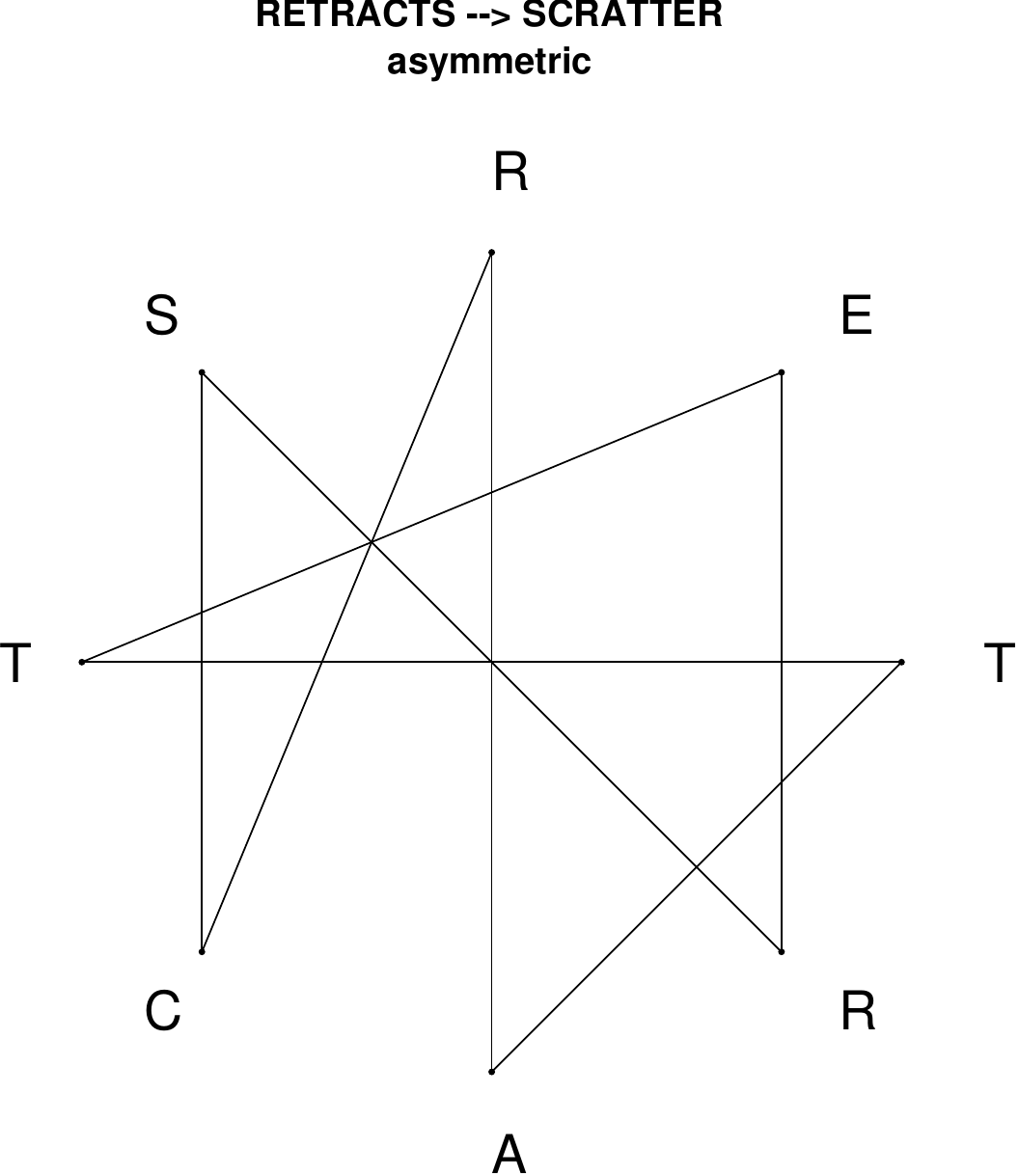}
\end{subfigure}
\hfill
\begin{subfigure}[T]{0.19\textwidth}
\centering
\includegraphics[width=\textwidth]{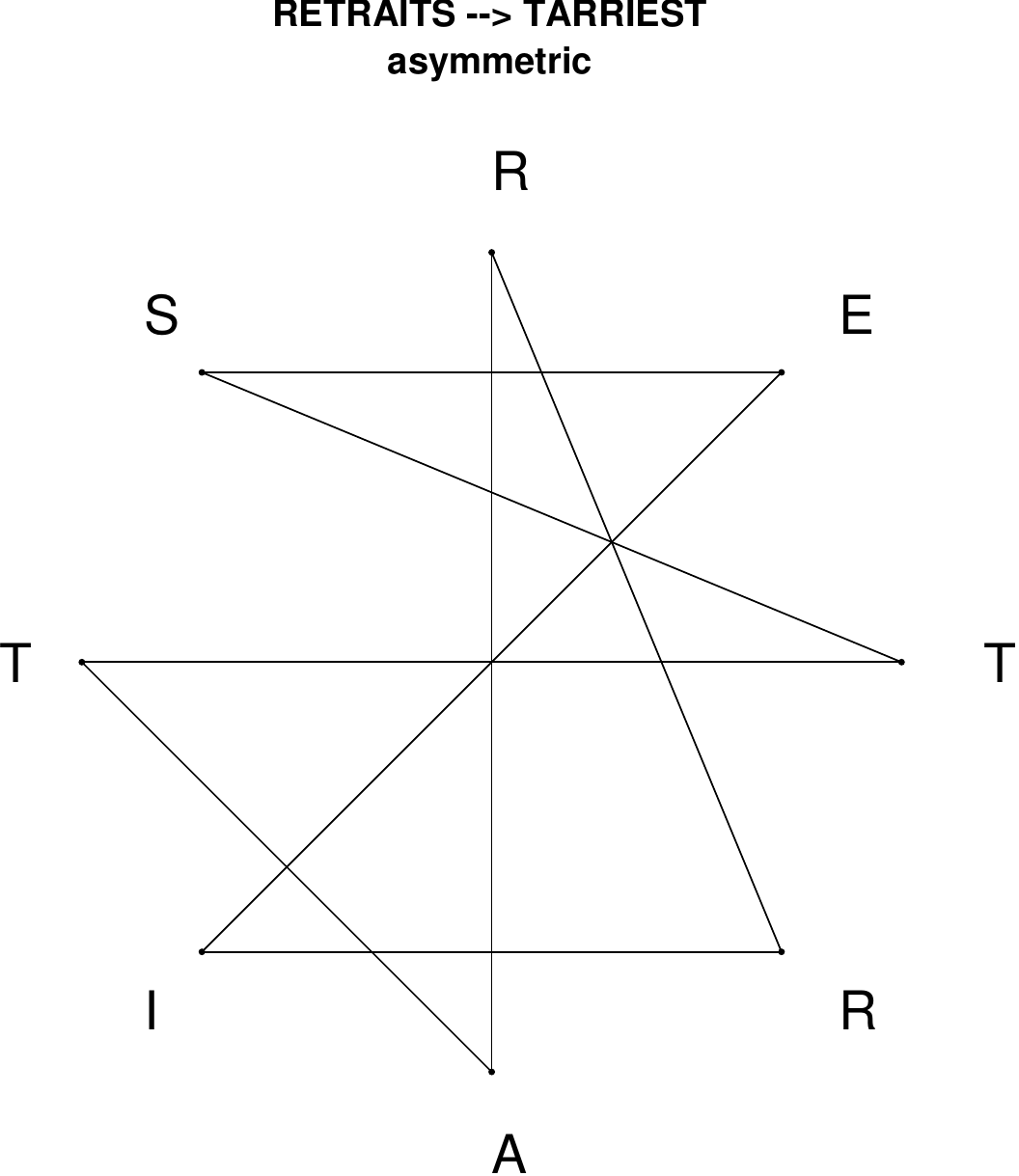}
\end{subfigure}
\hfill
\begin{subfigure}[T]{0.19\textwidth}
\centering
\includegraphics[width=\textwidth]{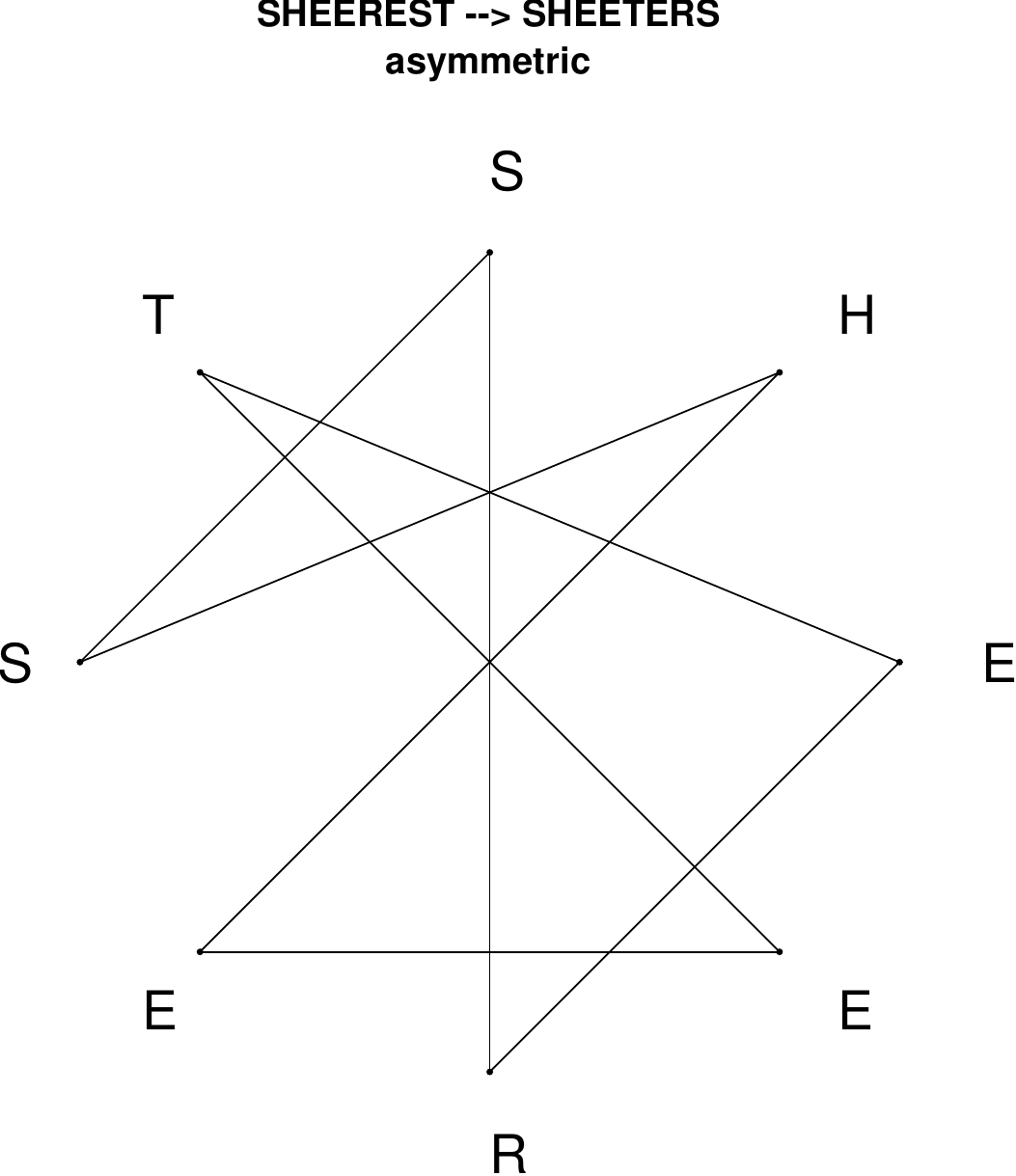}
\end{subfigure}
\hfill
\begin{subfigure}[T]{0.19\textwidth}
\centering
\includegraphics[width=\textwidth]{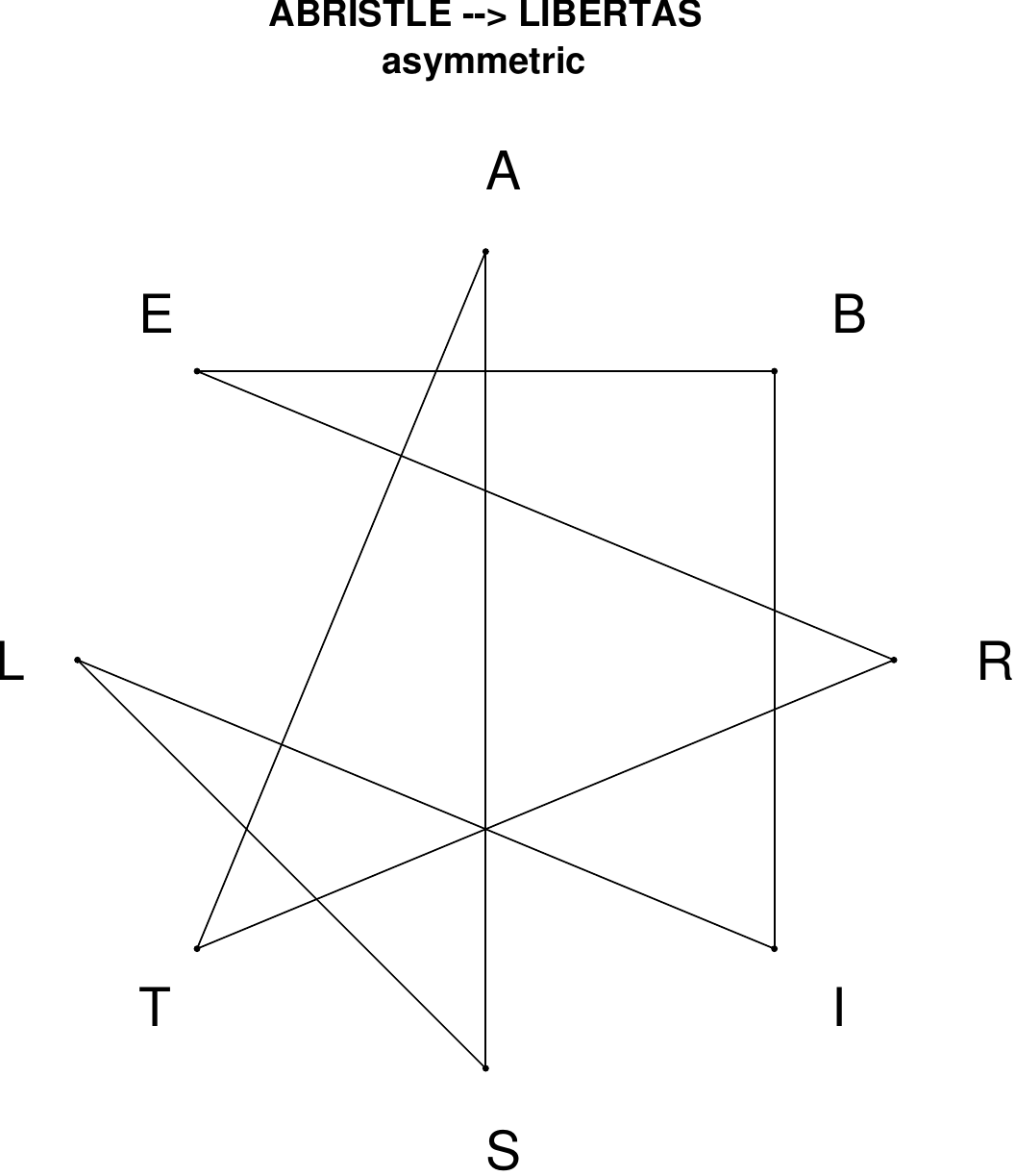}
\end{subfigure}
\end{figure}

\begin{figure}[H]
\centering
\begin{subfigure}[T]{0.19\textwidth}
\centering
\includegraphics[width=\textwidth]{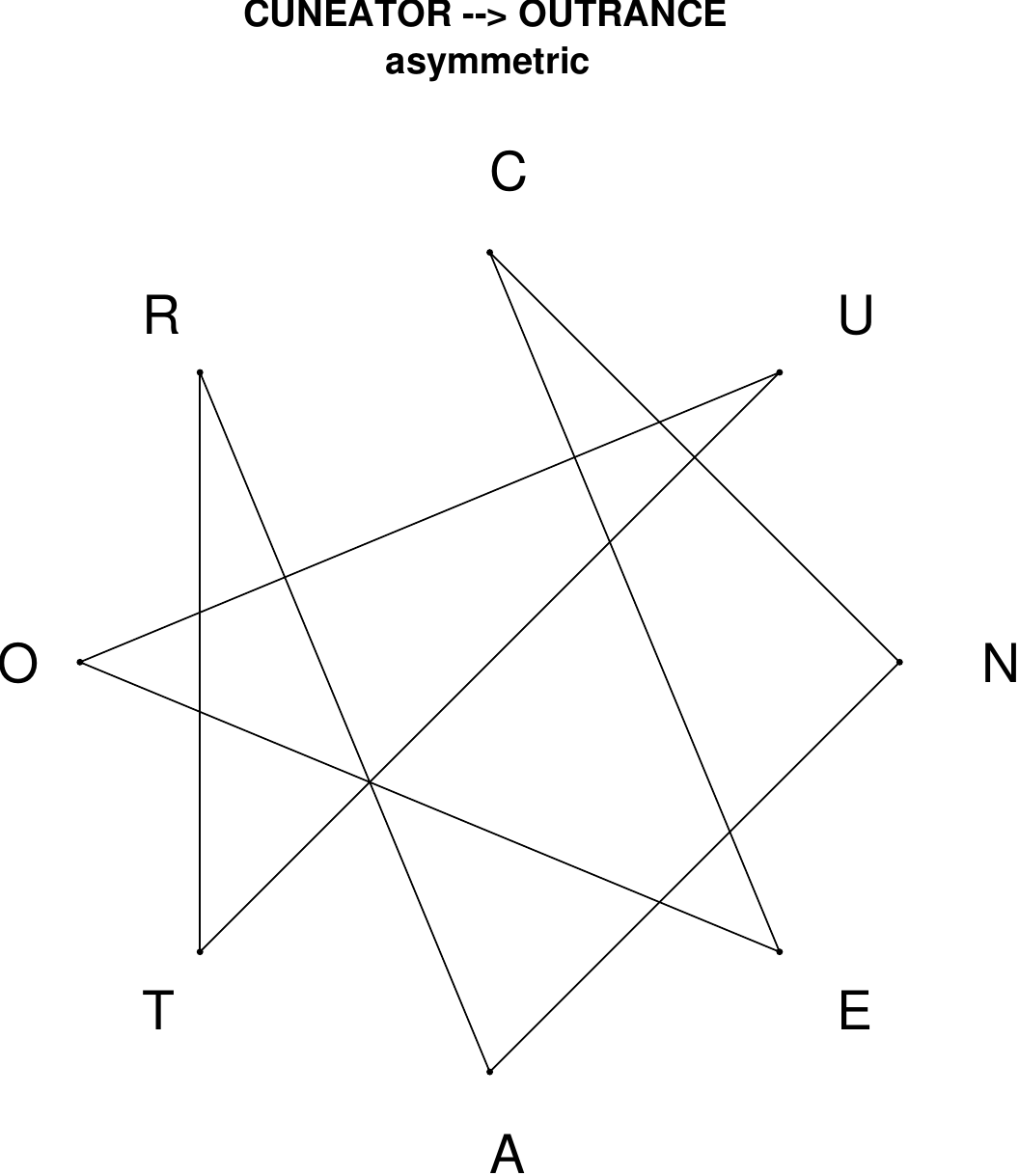}
\end{subfigure}
\hfill
\begin{subfigure}[T]{0.19\textwidth}
\centering
\includegraphics[width=\textwidth]{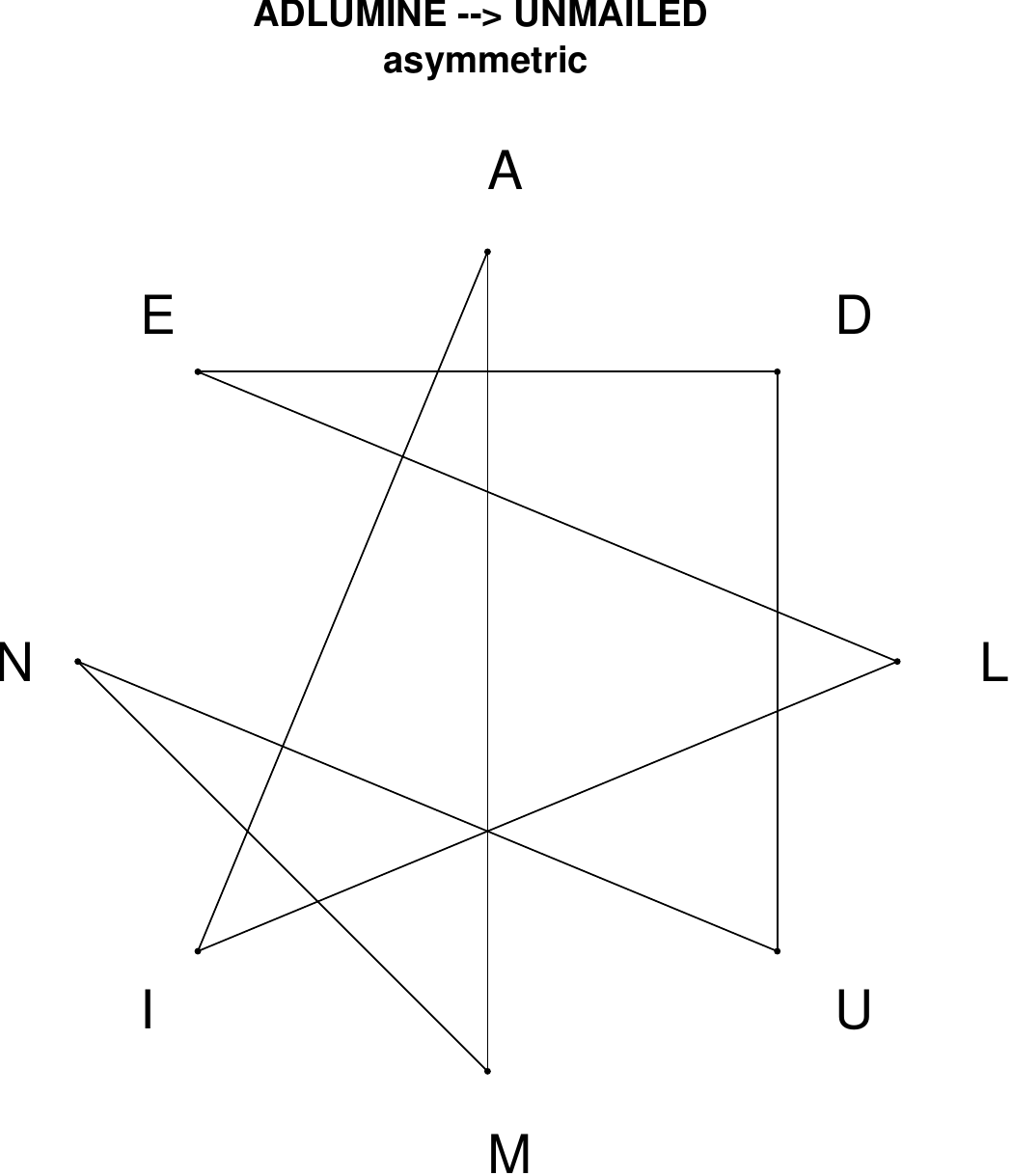}
\end{subfigure}
\hfill
\begin{subfigure}[T]{0.19\textwidth}
\centering
\includegraphics[width=\textwidth]{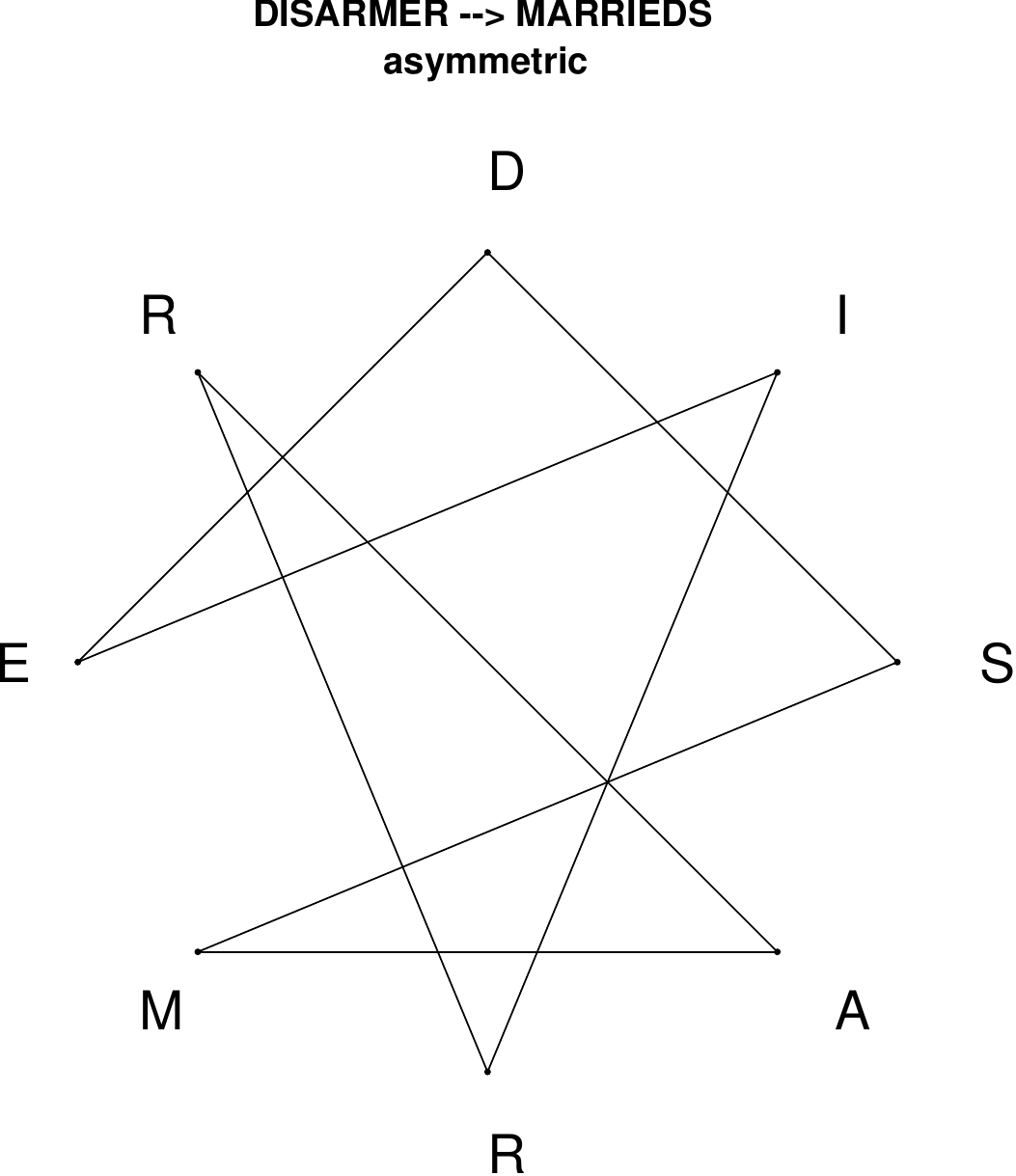}
\end{subfigure}
\hfill
\begin{subfigure}[T]{0.19\textwidth}
\centering
\includegraphics[width=\textwidth]{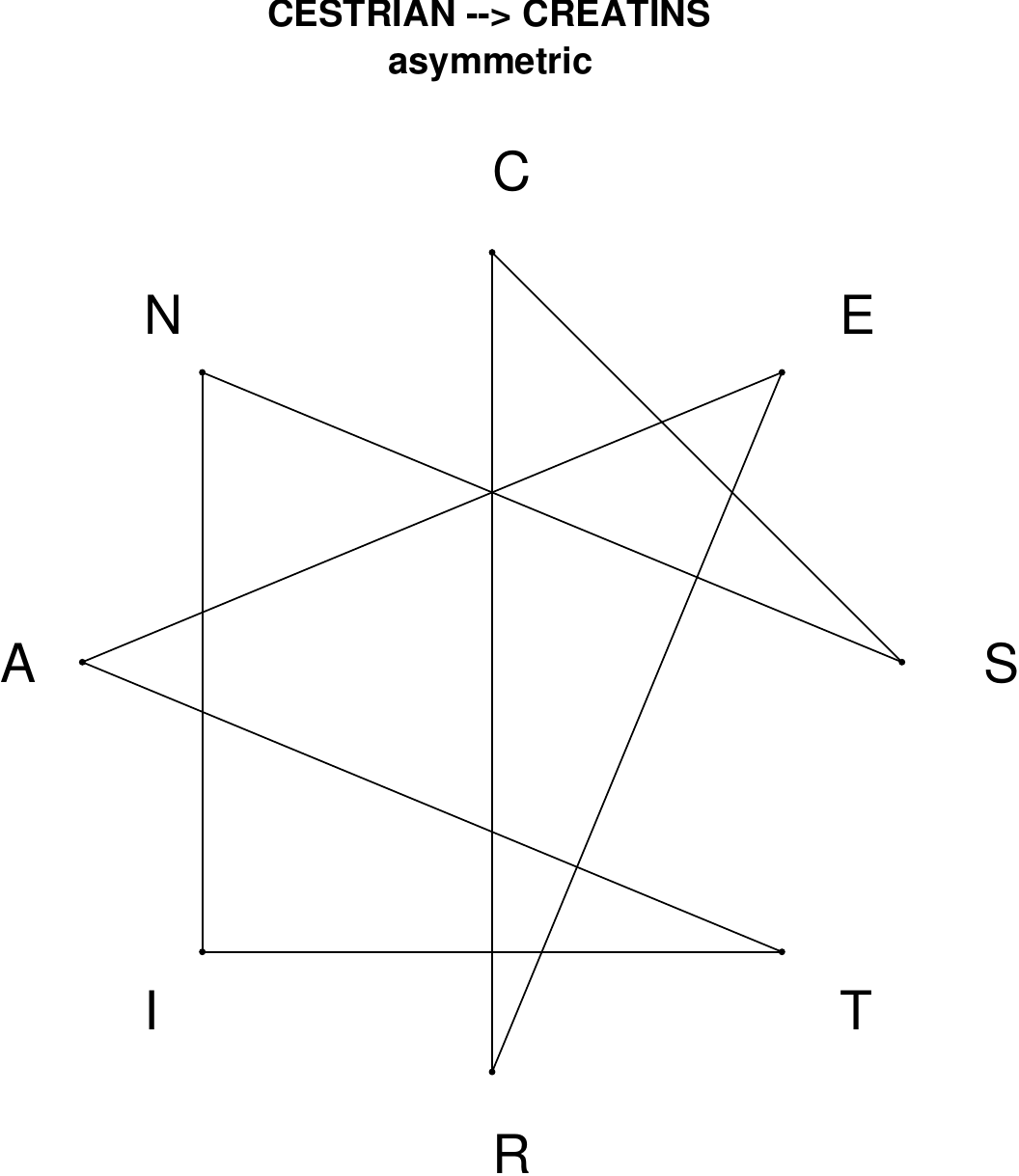}
\end{subfigure}
\hfill
\begin{subfigure}[T]{0.19\textwidth}
\centering
\includegraphics[width=\textwidth]{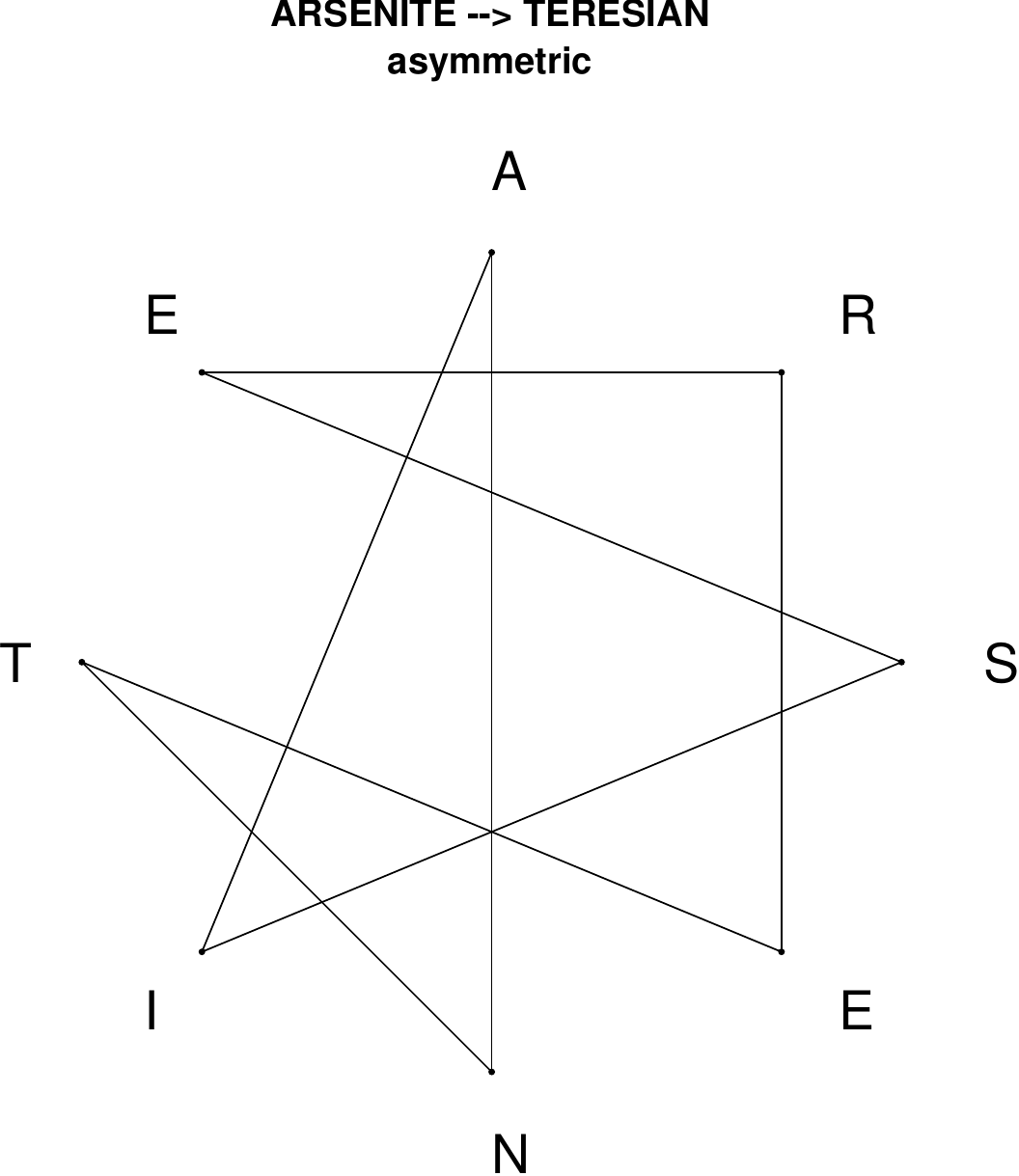}
\end{subfigure}
\end{figure}

\begin{figure}[H]
\centering
\begin{subfigure}[T]{0.19\textwidth}
\centering
\includegraphics[width=\textwidth]{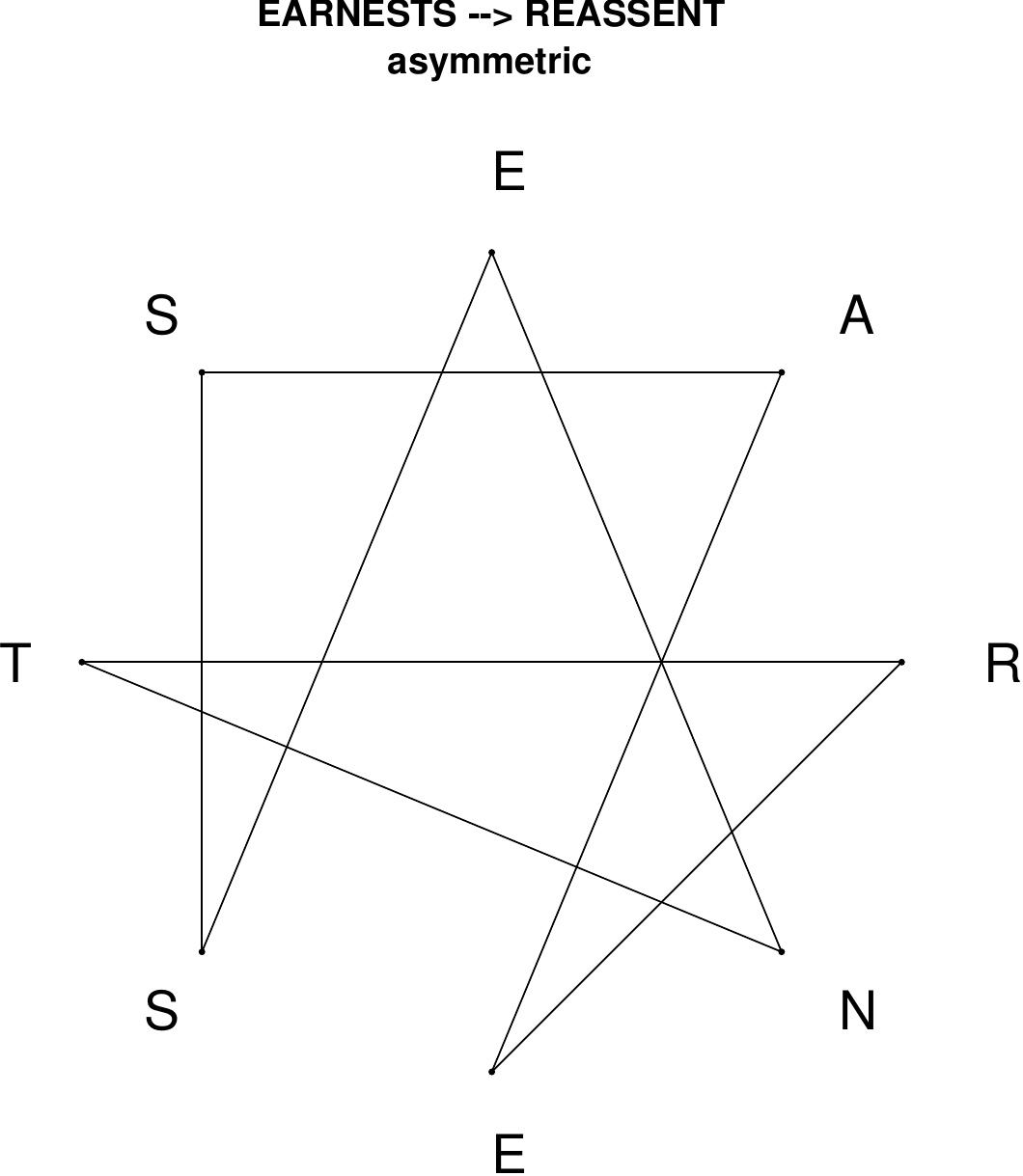}
\end{subfigure}
\hfill
\begin{subfigure}[T]{0.19\textwidth}
\centering
\includegraphics[width=\textwidth]{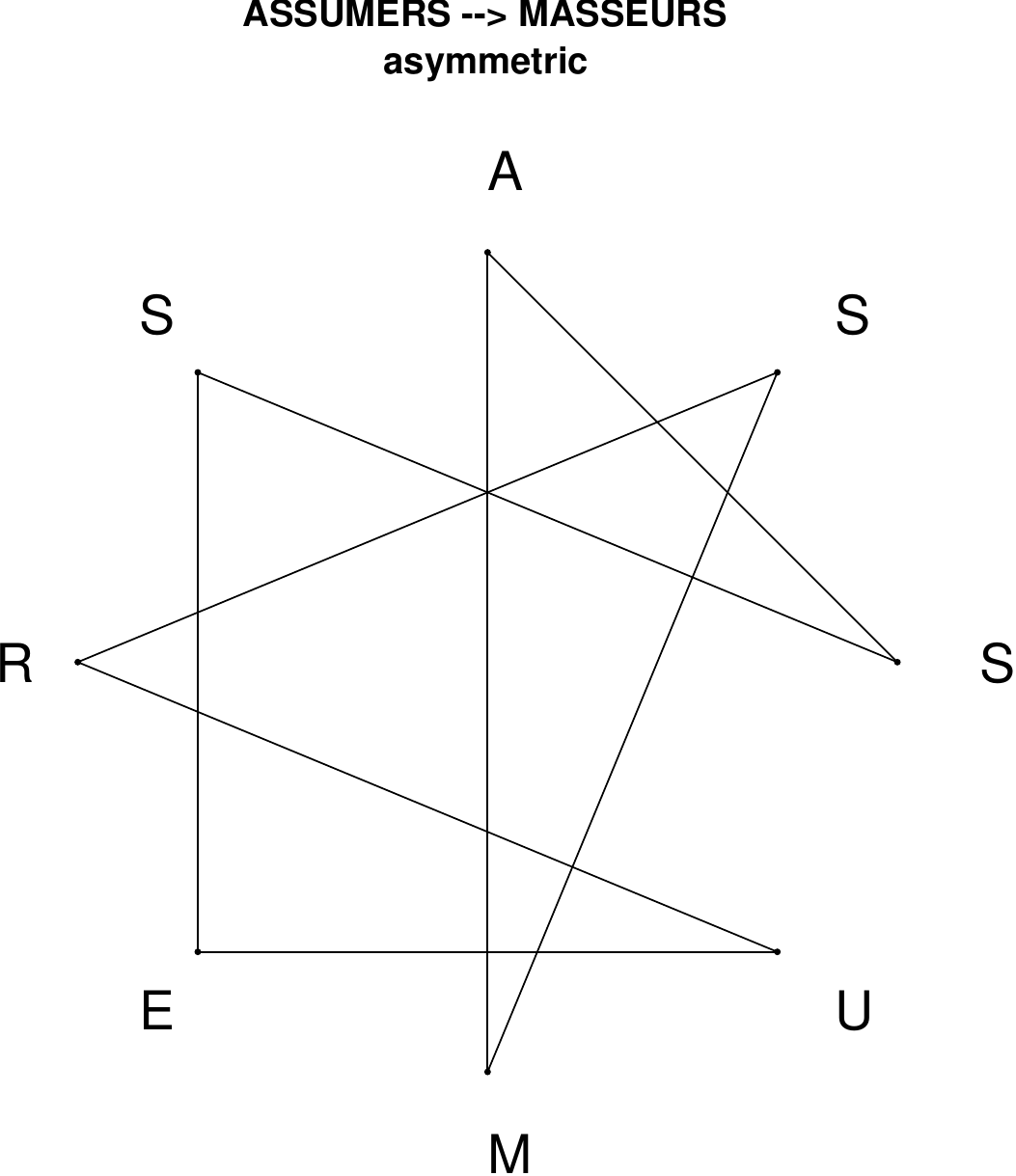}
\end{subfigure}
\hfill
\begin{subfigure}[T]{0.19\textwidth}
\centering
\includegraphics[width=\textwidth]{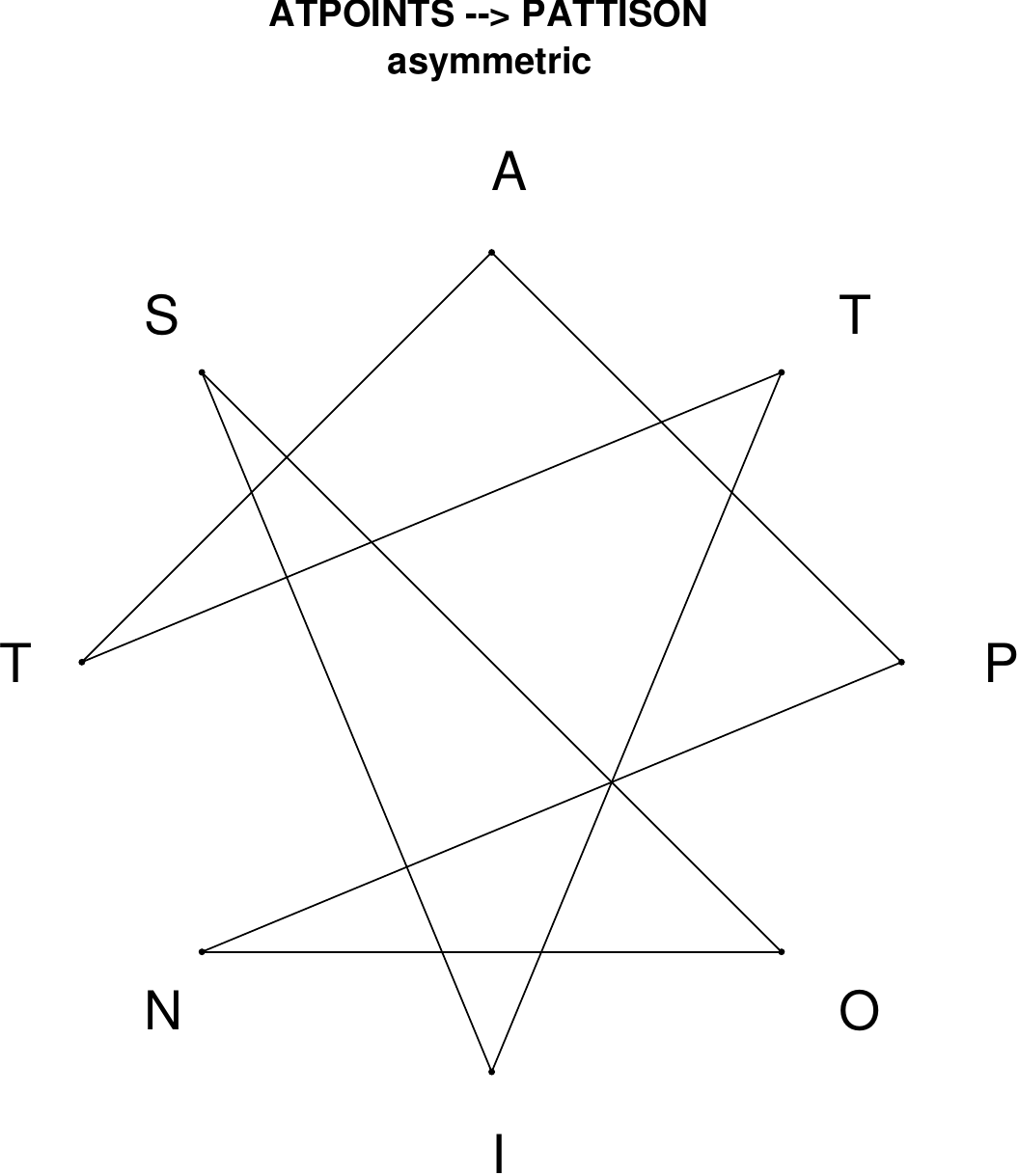}
\end{subfigure}
\hfill
\begin{subfigure}[T]{0.19\textwidth}
\centering
\includegraphics[width=\textwidth]{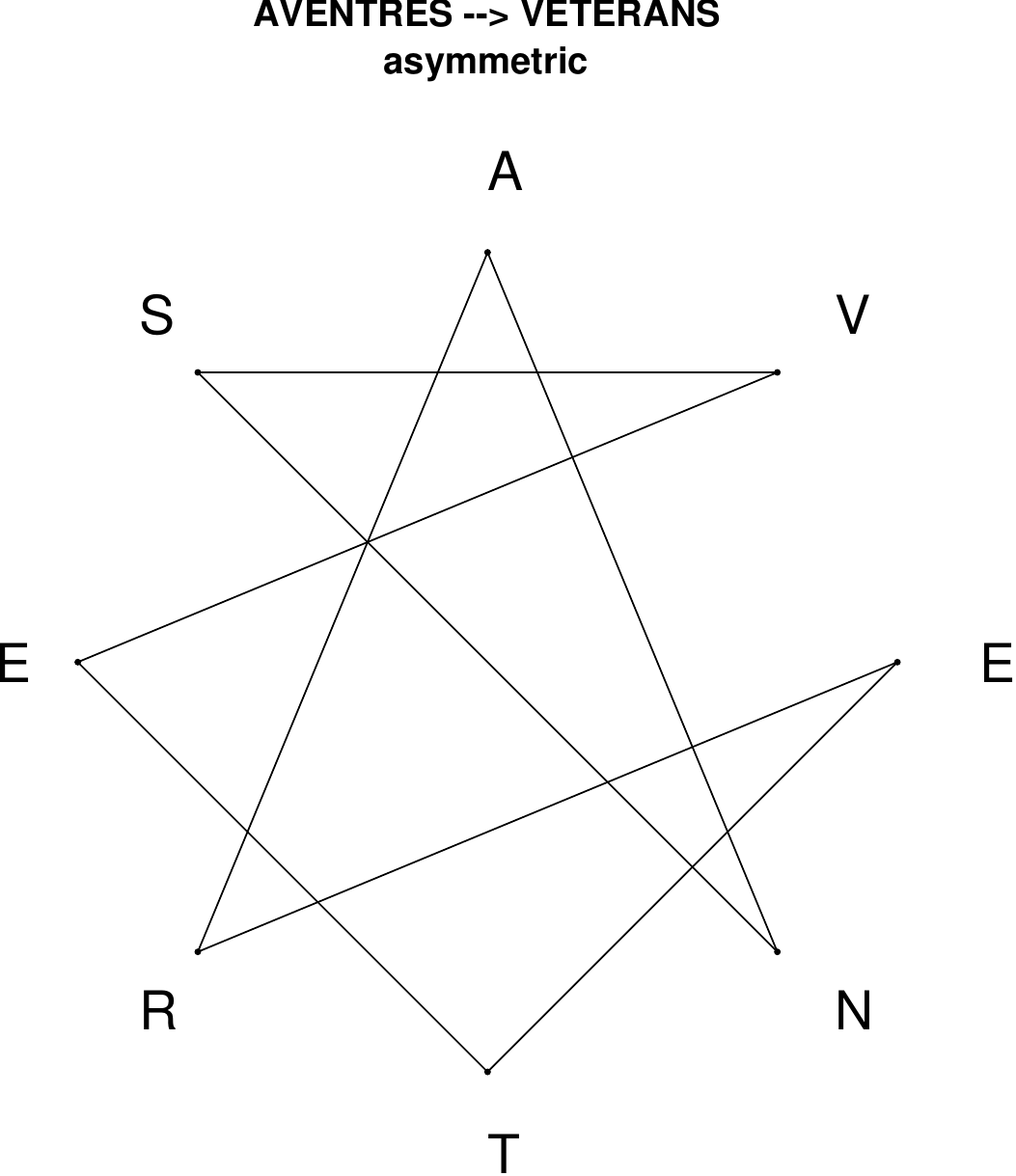}
\end{subfigure}
\hfill
\begin{subfigure}[T]{0.19\textwidth}
\centering
\includegraphics[width=\textwidth]{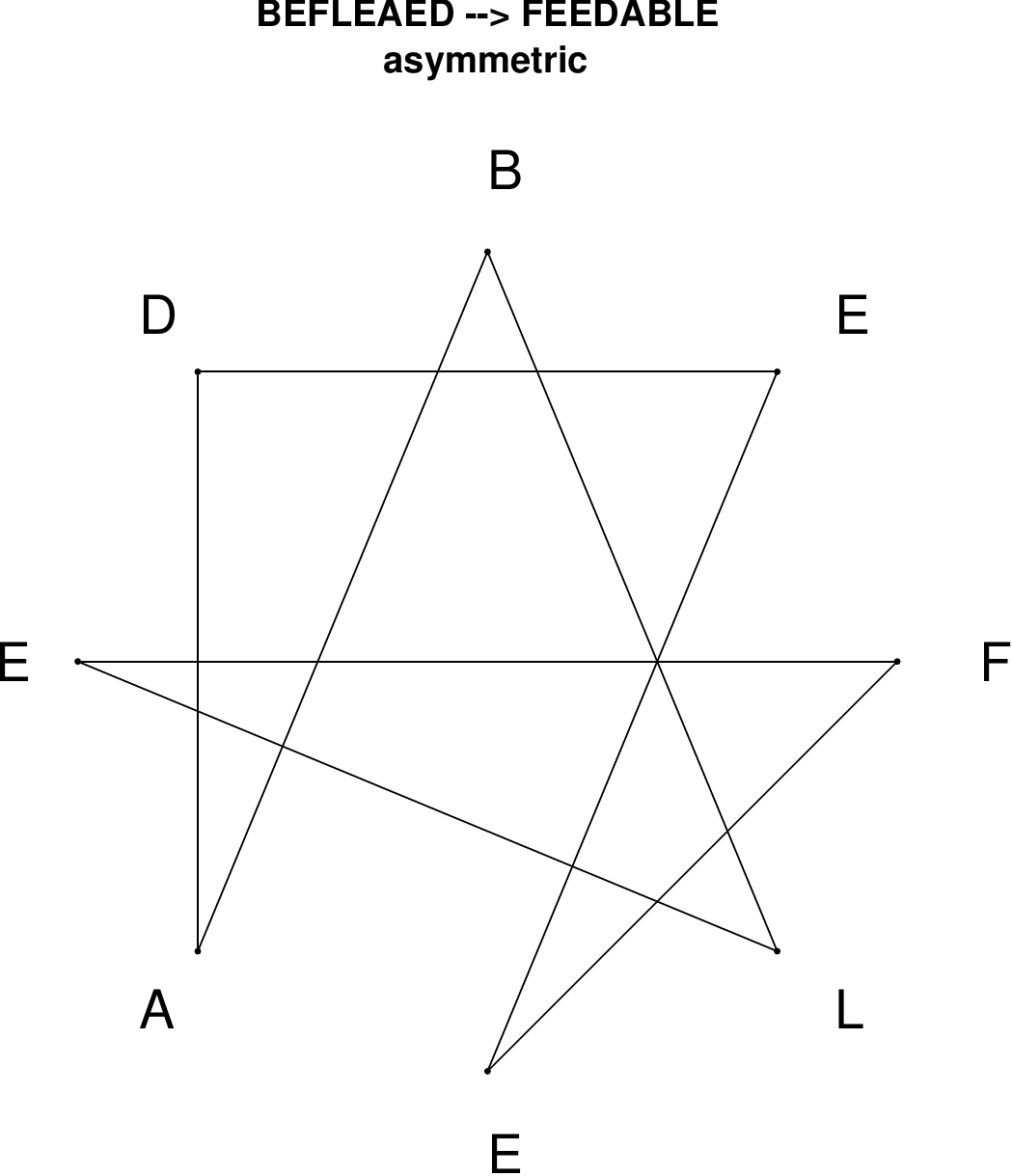}
\end{subfigure}
\end{figure}

\begin{figure}[H]
\centering
\begin{subfigure}[T]{0.19\textwidth}
\centering
\includegraphics[width=\textwidth]{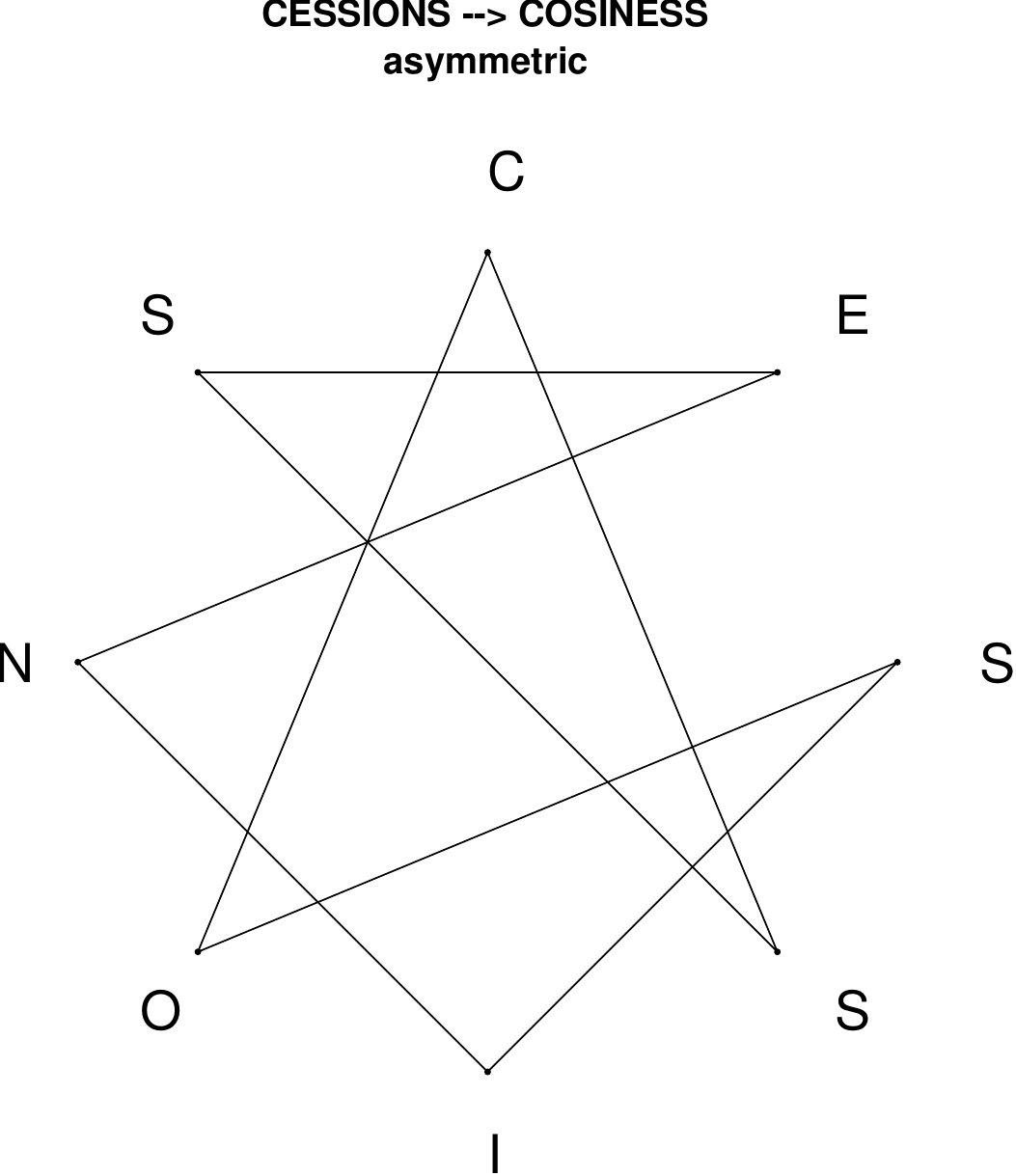}
\end{subfigure}
\hfill
\begin{subfigure}[T]{0.19\textwidth}
\centering
\includegraphics[width=\textwidth]{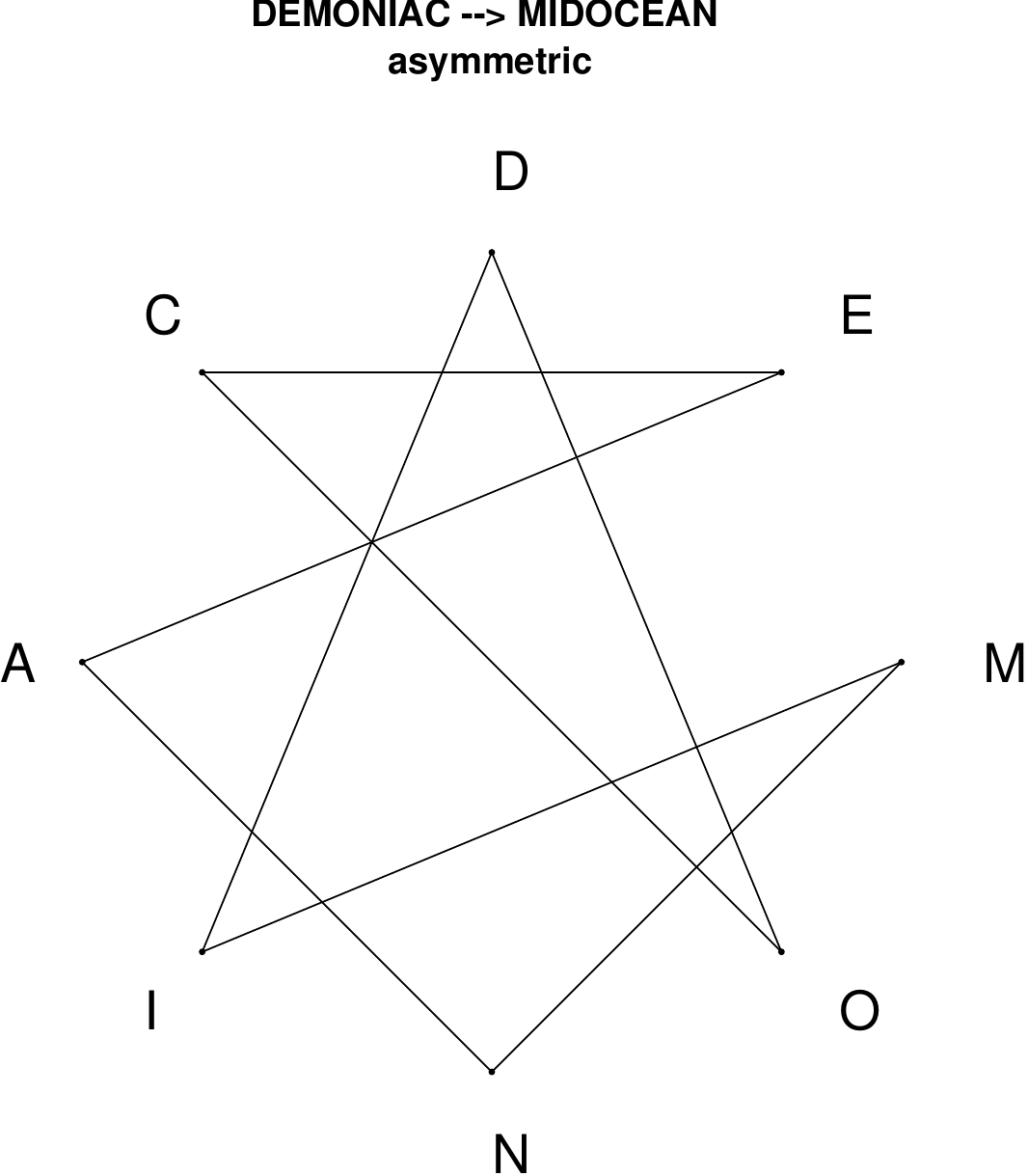}
\end{subfigure}
\hfill
\begin{subfigure}[T]{0.19\textwidth}
\centering
\includegraphics[width=\textwidth]{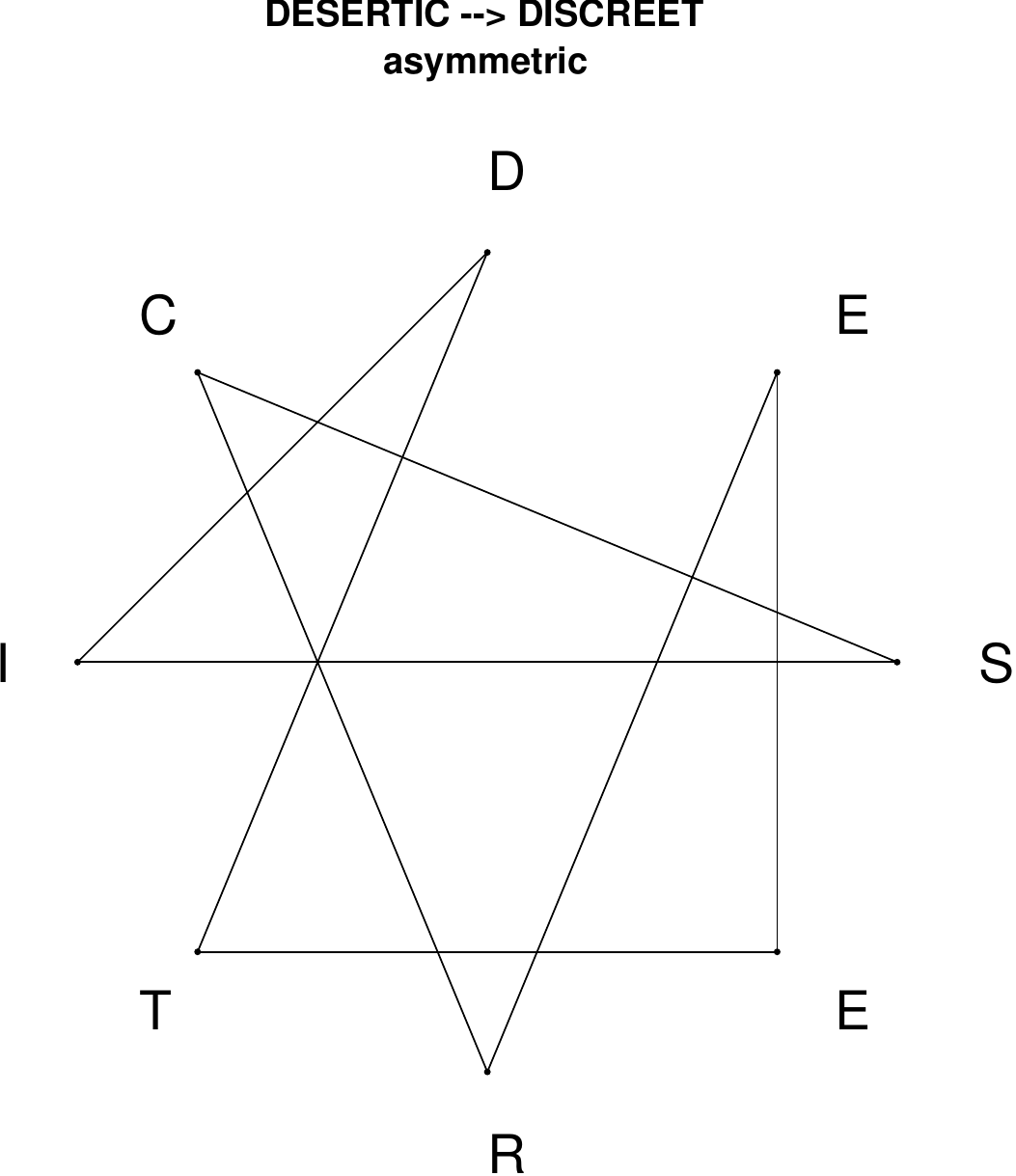}
\end{subfigure}
\hfill
\begin{subfigure}[T]{0.19\textwidth}
\centering
\includegraphics[width=\textwidth]{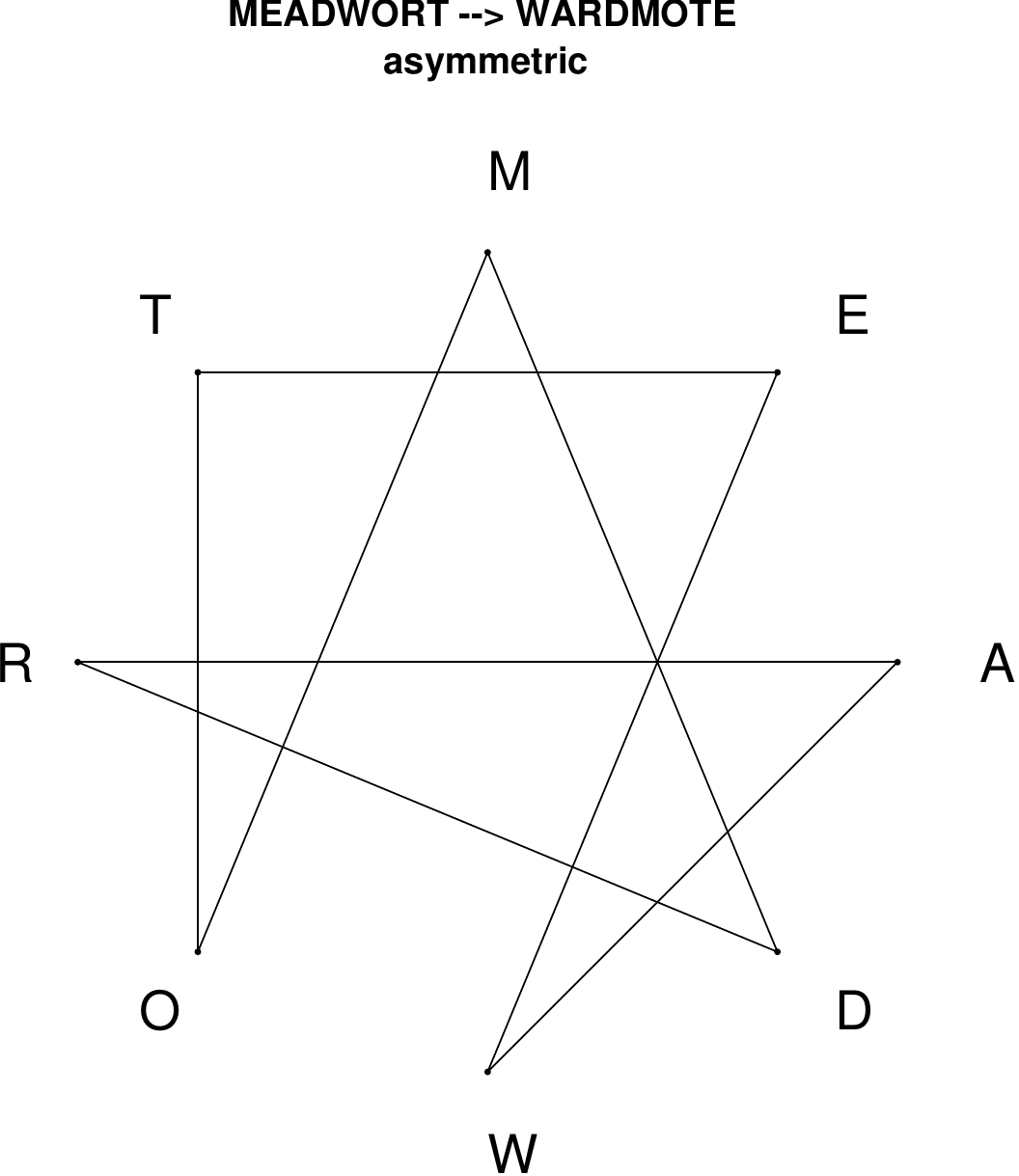}
\end{subfigure}
\hfill
\begin{subfigure}[T]{0.19\textwidth}
\centering
\includegraphics[width=\textwidth]{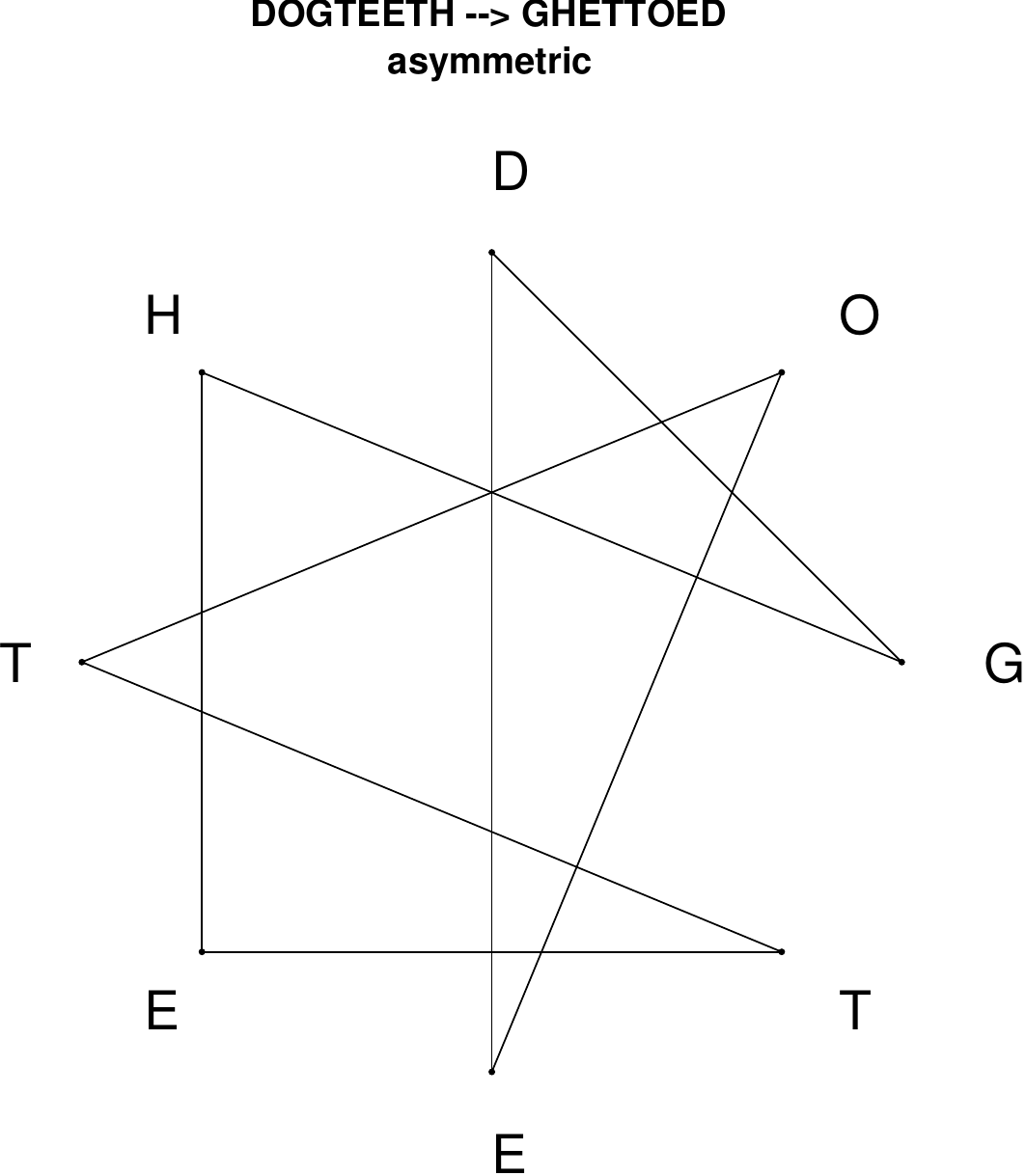}
\end{subfigure}
\end{figure}

\begin{figure}[H]
\centering
\begin{subfigure}[T]{0.19\textwidth}
\centering
\includegraphics[width=\textwidth]{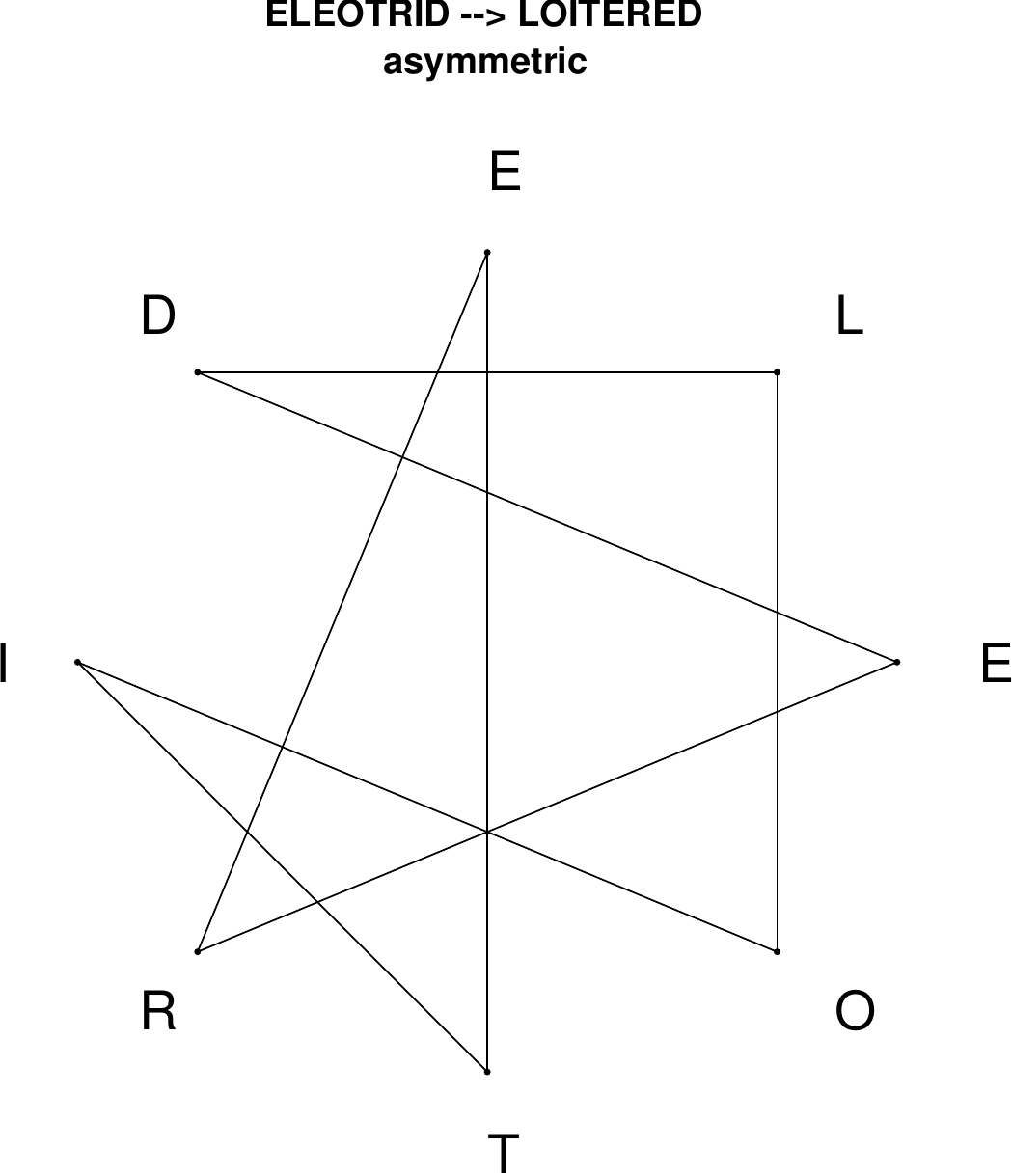}
\end{subfigure}
\hfill
\begin{subfigure}[T]{0.19\textwidth}
\centering
\includegraphics[width=\textwidth]{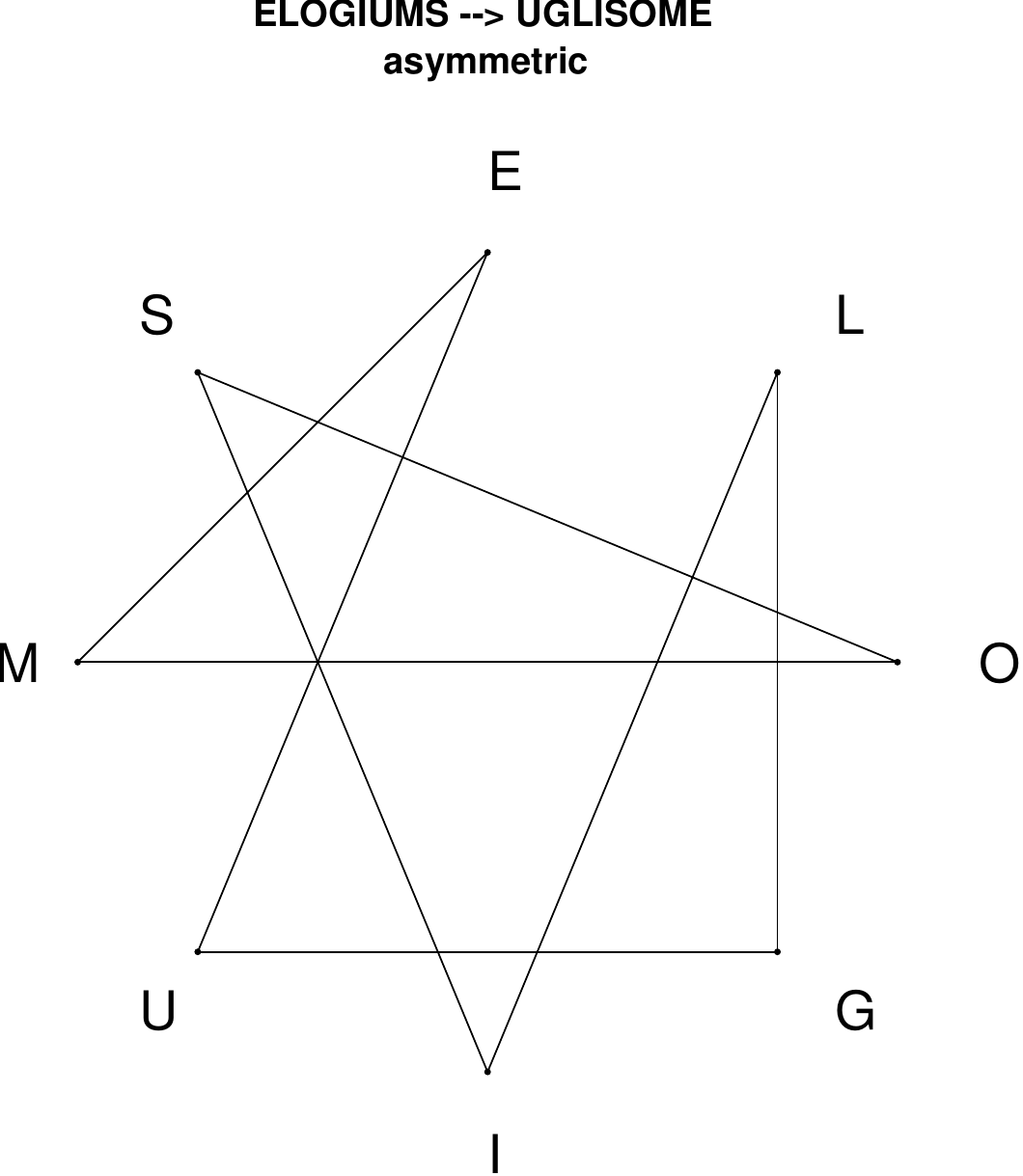}
\end{subfigure}
\hfill
\begin{subfigure}[T]{0.19\textwidth}
\centering
\includegraphics[width=\textwidth]{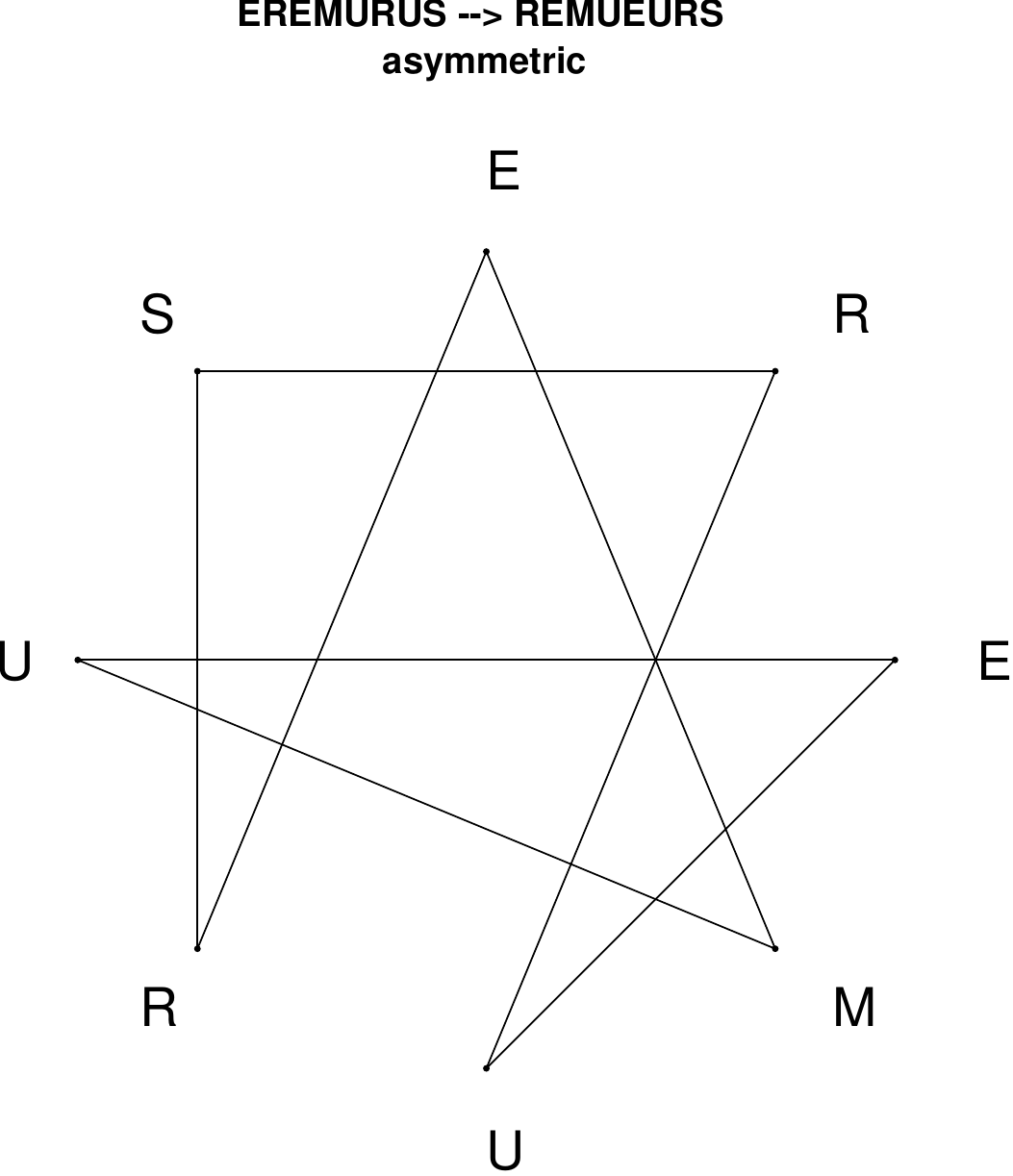}
\end{subfigure}
\hfill
\begin{subfigure}[T]{0.19\textwidth}
\centering
\includegraphics[width=\textwidth]{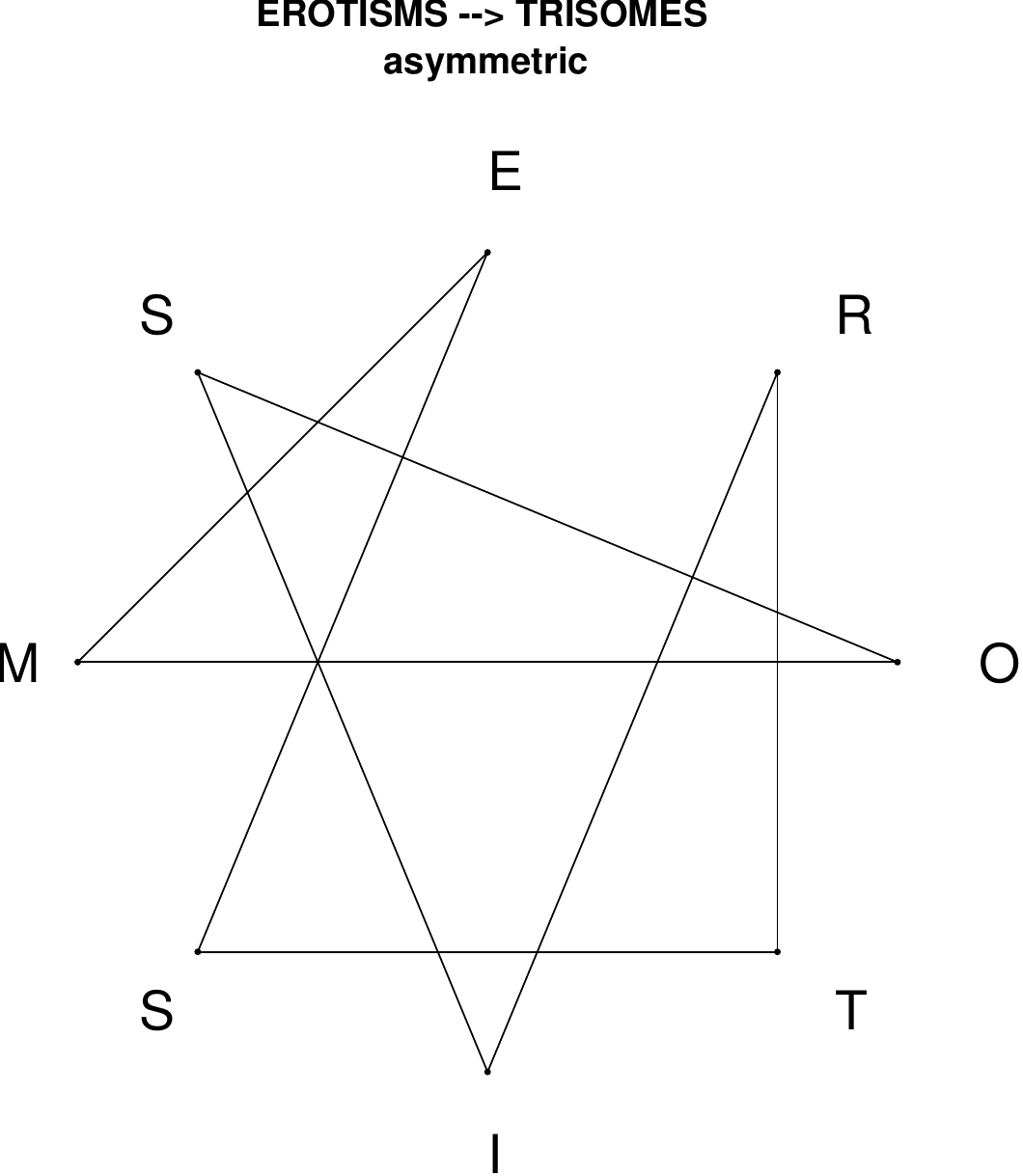}
\end{subfigure}
\hfill
\begin{subfigure}[T]{0.19\textwidth}
\centering
\includegraphics[width=\textwidth]{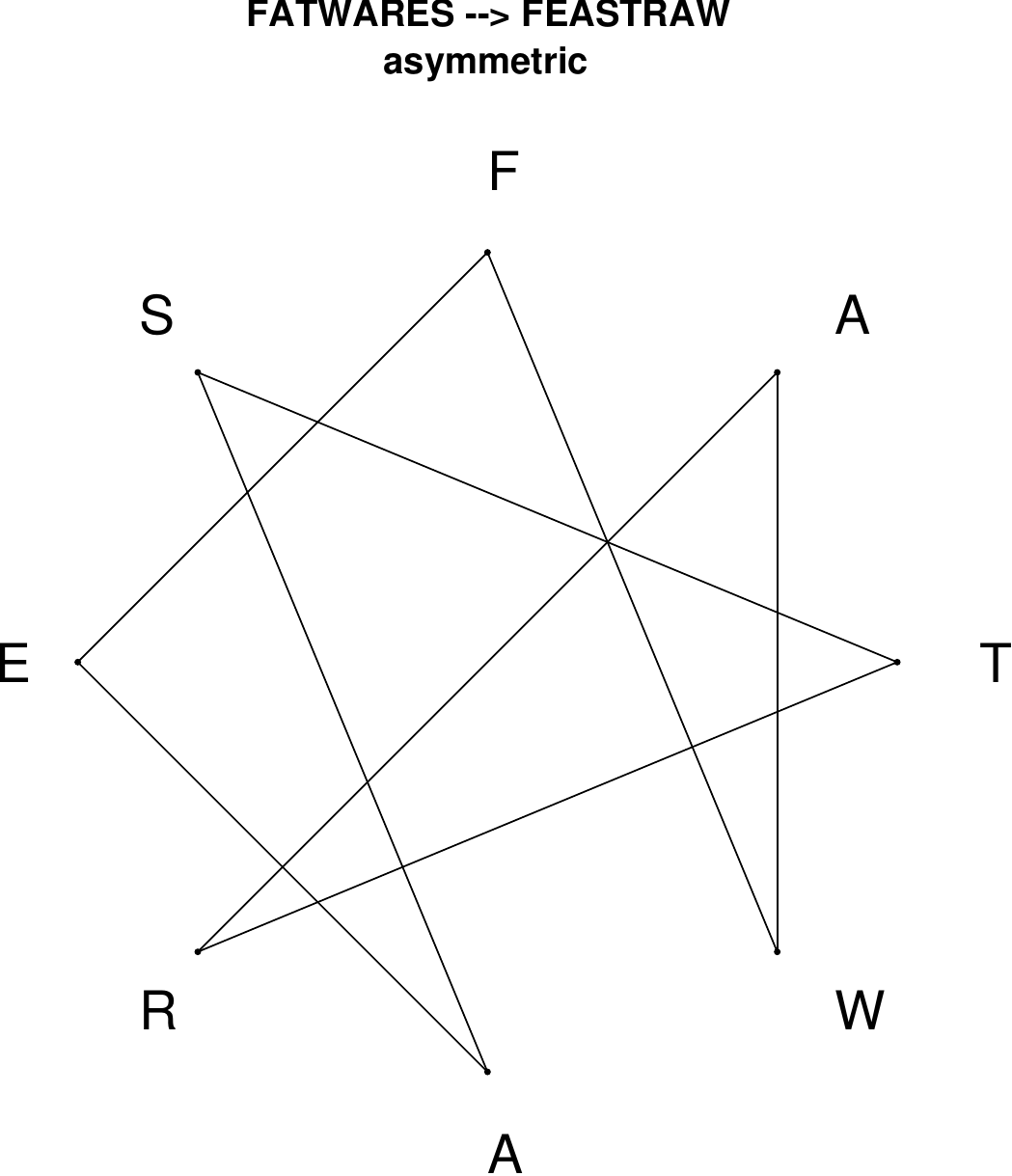}
\end{subfigure}
\end{figure}

\begin{figure}[H]
\centering
\begin{subfigure}[T]{0.19\textwidth}
\centering
\includegraphics[width=\textwidth]{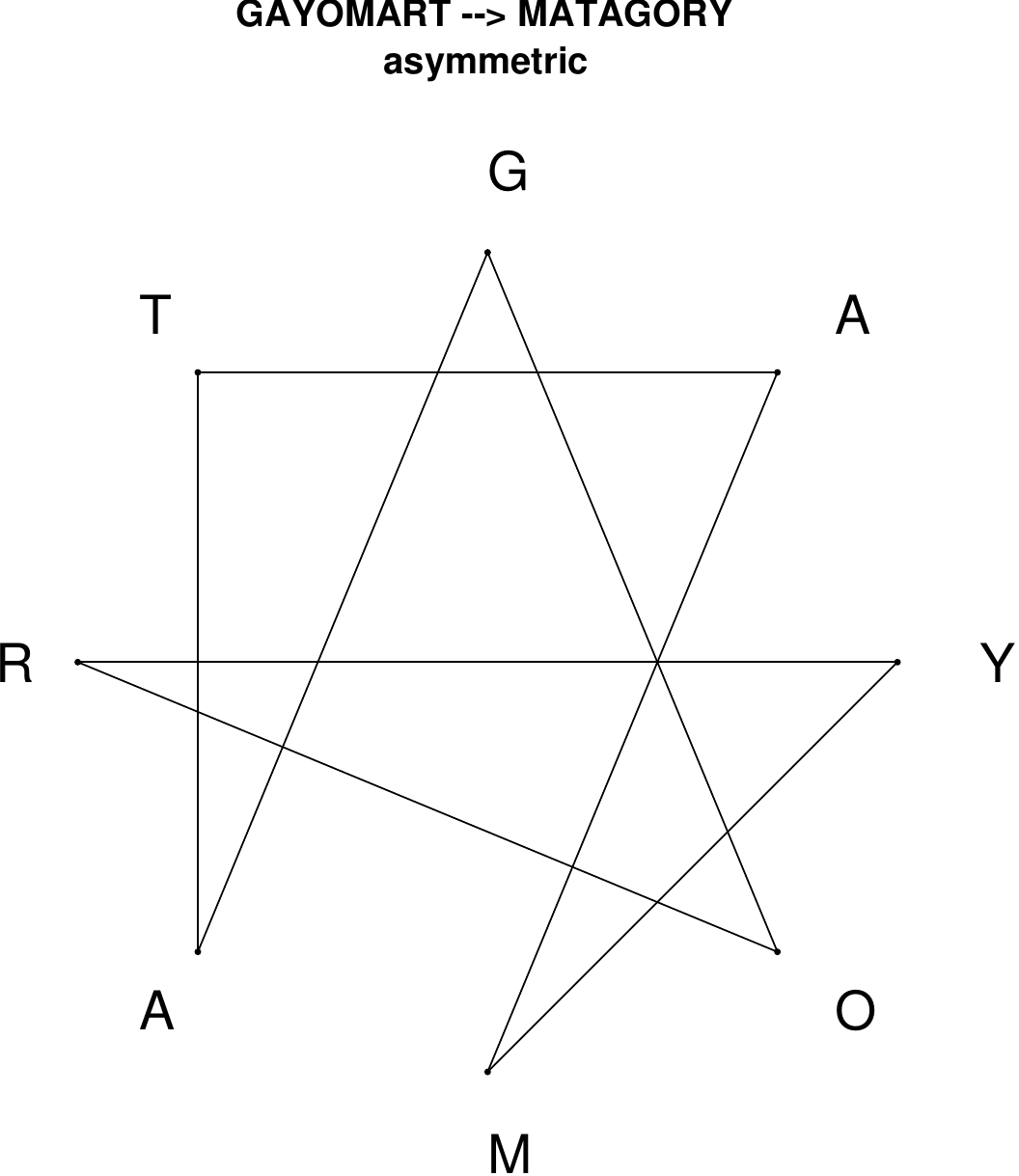}
\end{subfigure}
\hfill
\begin{subfigure}[T]{0.19\textwidth}
\centering
\includegraphics[width=\textwidth]{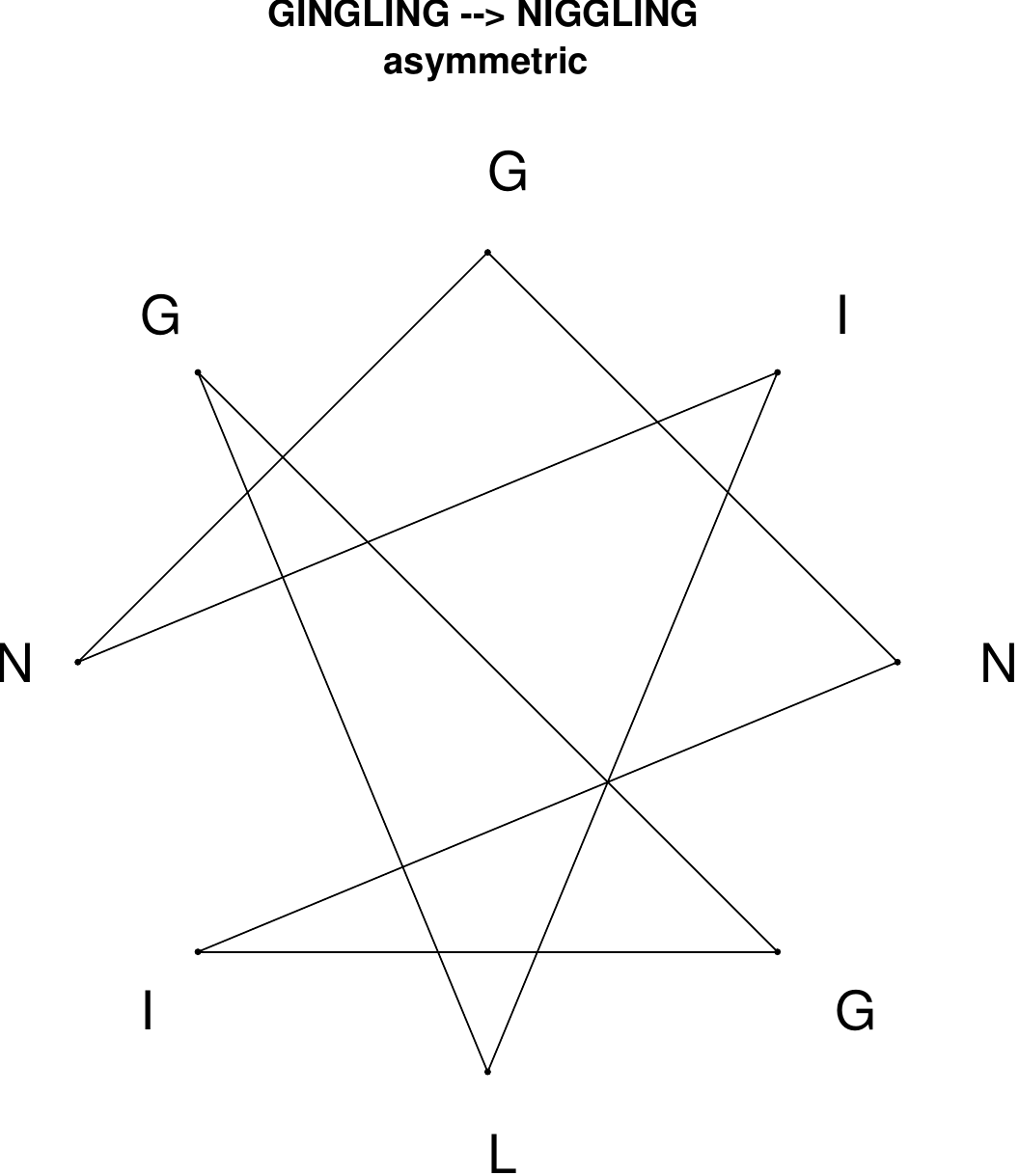}
\end{subfigure}
\hfill
\begin{subfigure}[T]{0.19\textwidth}
\centering
\includegraphics[width=\textwidth]{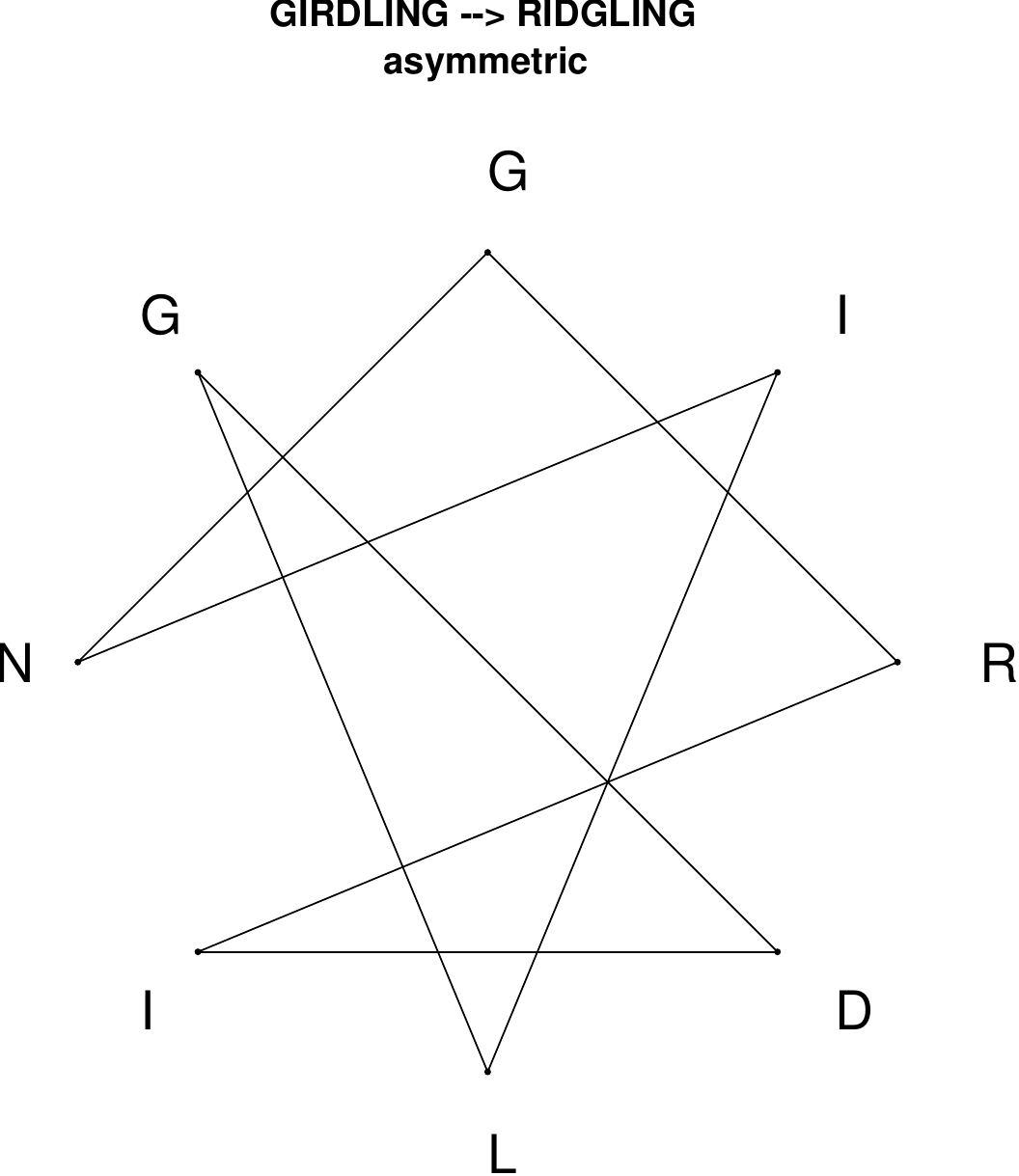}
\end{subfigure}
\hfill
\begin{subfigure}[T]{0.19\textwidth}
\centering
\includegraphics[width=\textwidth]{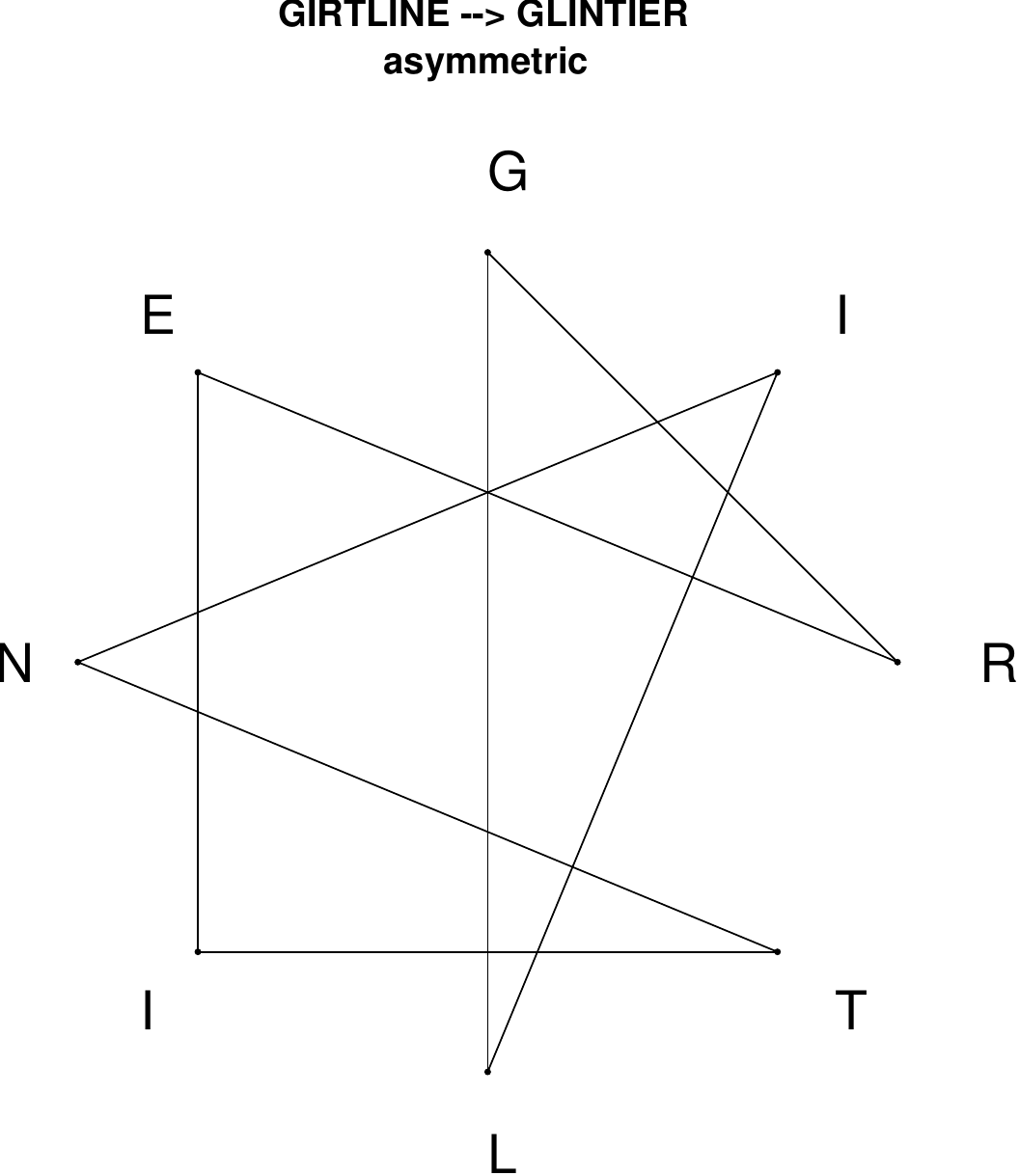}
\end{subfigure}
\hfill
\begin{subfigure}[T]{0.19\textwidth}
\centering
\includegraphics[width=\textwidth]{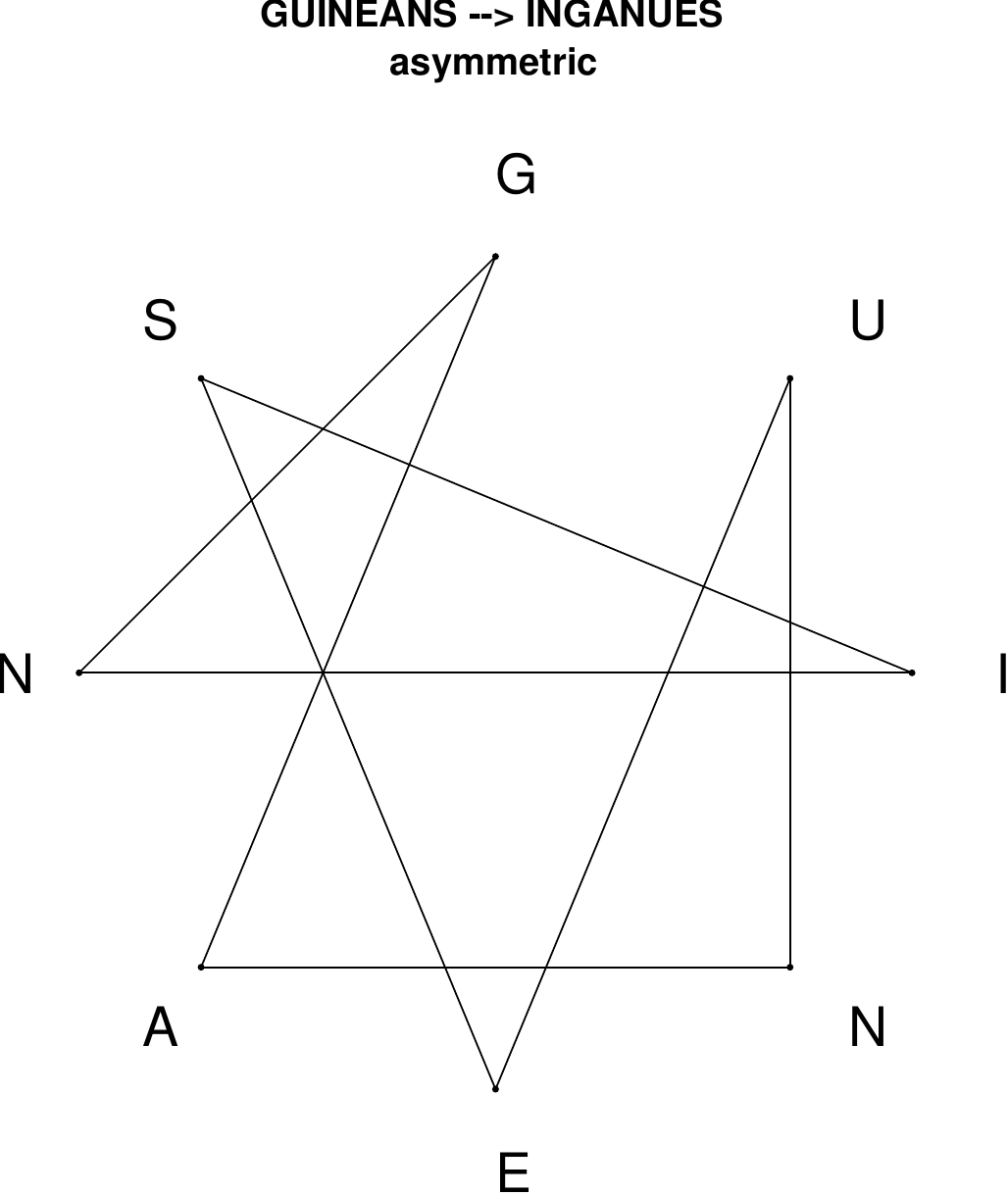}
\end{subfigure}
\end{figure}

\begin{figure}[H]
\centering
\begin{subfigure}[T]{0.19\textwidth}
\centering
\includegraphics[width=\textwidth]{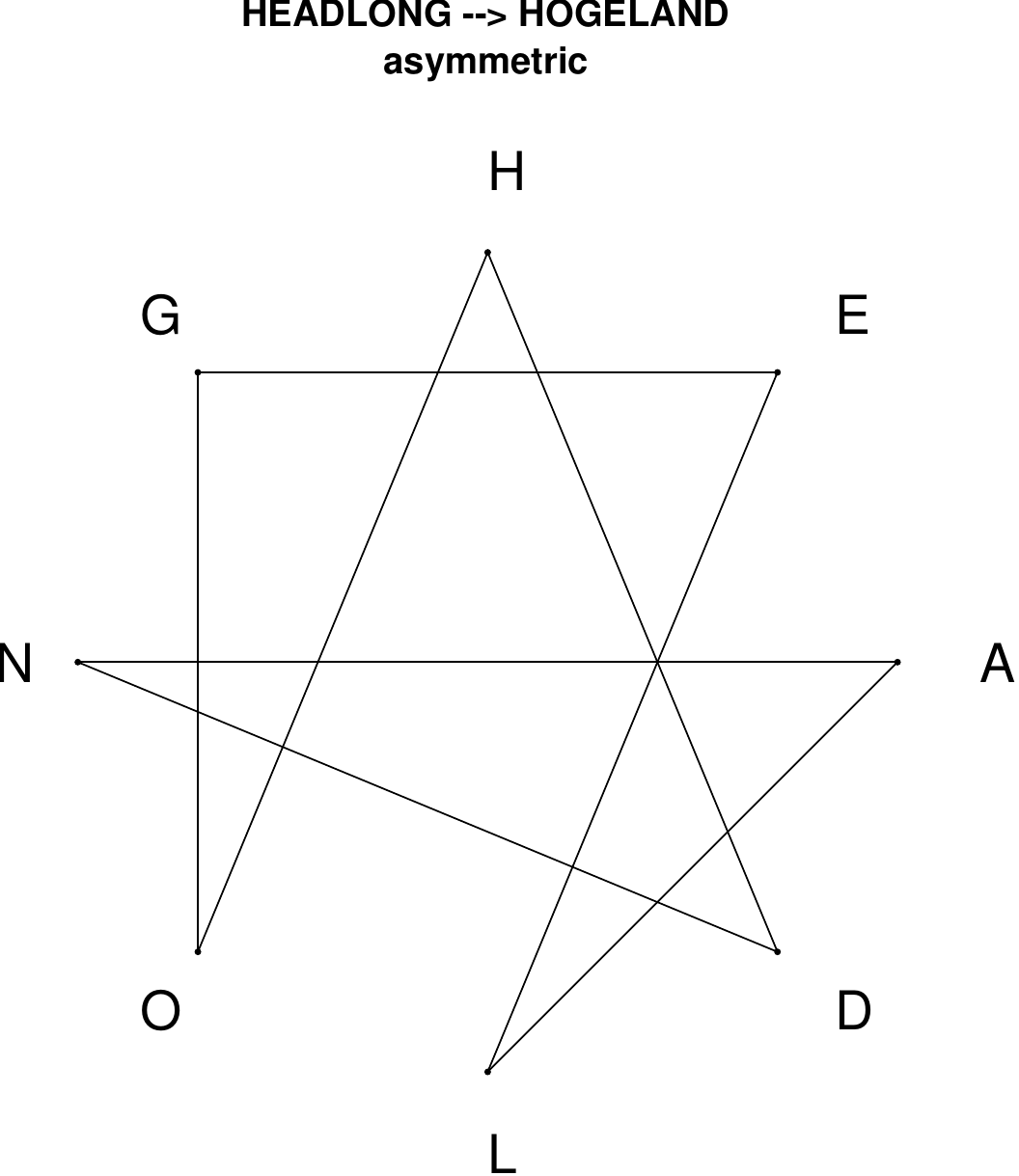}
\end{subfigure}
\hfill
\begin{subfigure}[T]{0.19\textwidth}
\centering
\includegraphics[width=\textwidth]{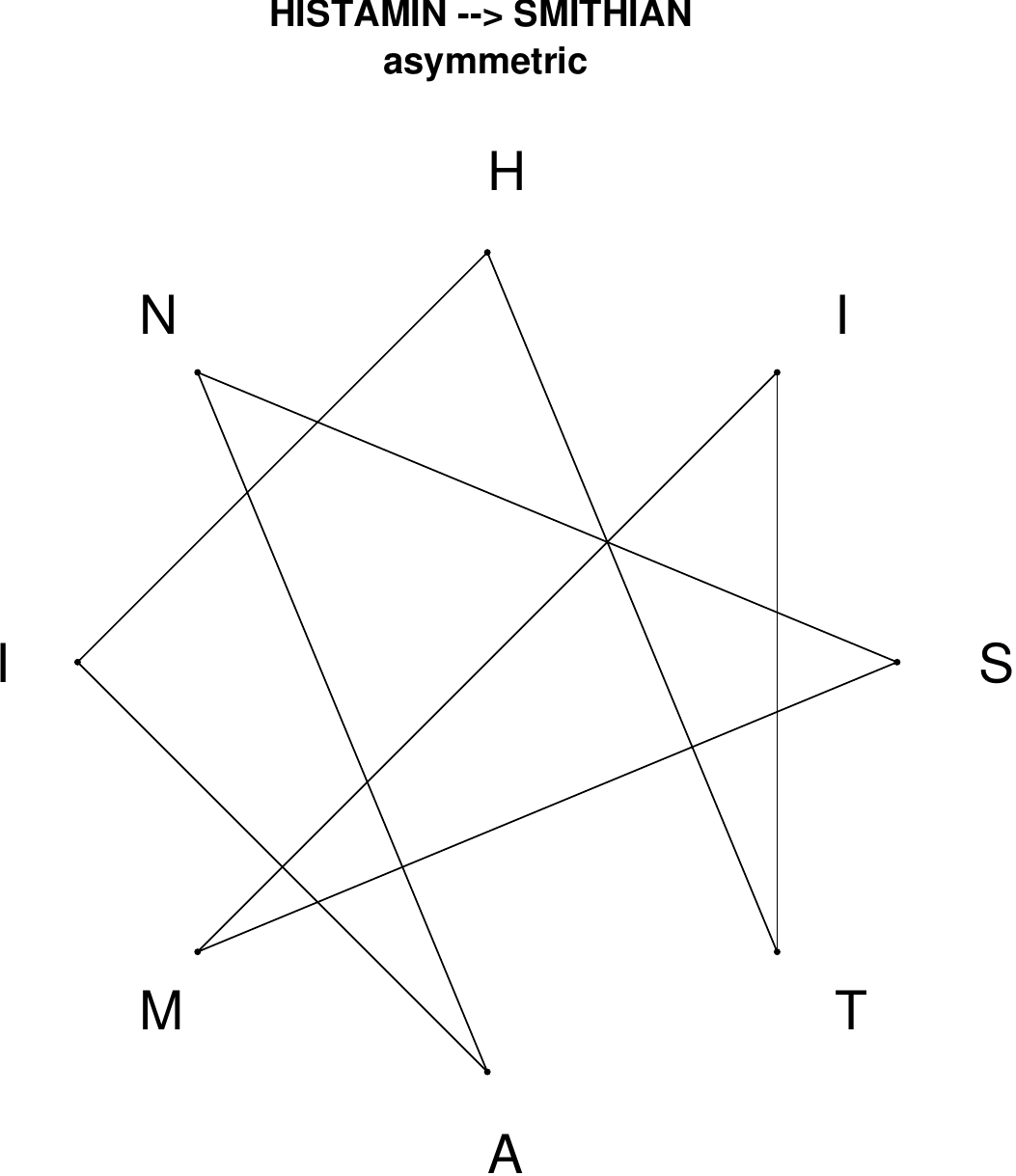}
\end{subfigure}
\hfill
\begin{subfigure}[T]{0.19\textwidth}
\centering
\includegraphics[width=\textwidth]{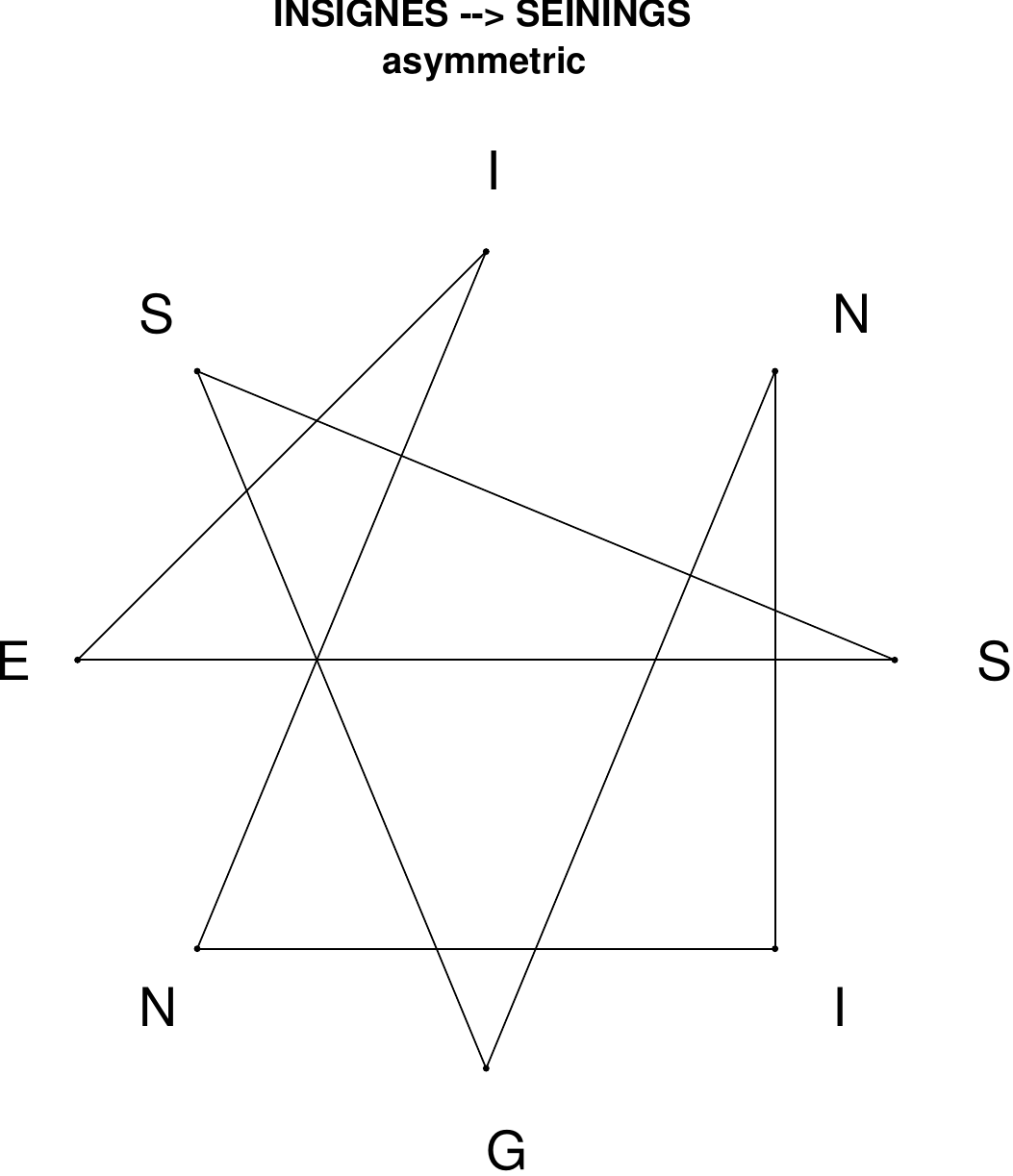}
\end{subfigure}
\hfill
\begin{subfigure}[T]{0.19\textwidth}
\centering
\includegraphics[width=\textwidth]{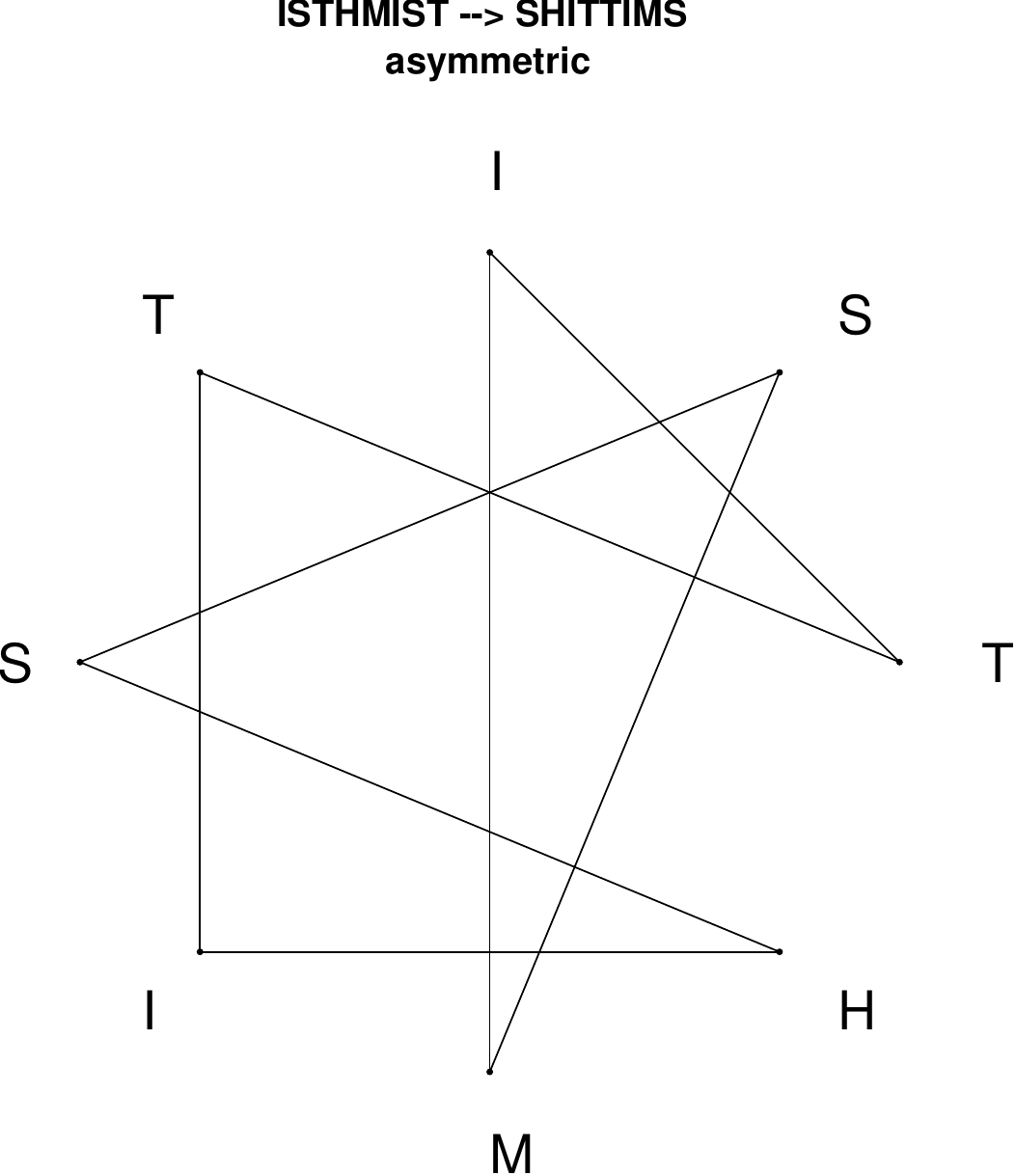}
\end{subfigure}
\hfill
\begin{subfigure}[T]{0.19\textwidth}
\centering
\includegraphics[width=\textwidth]{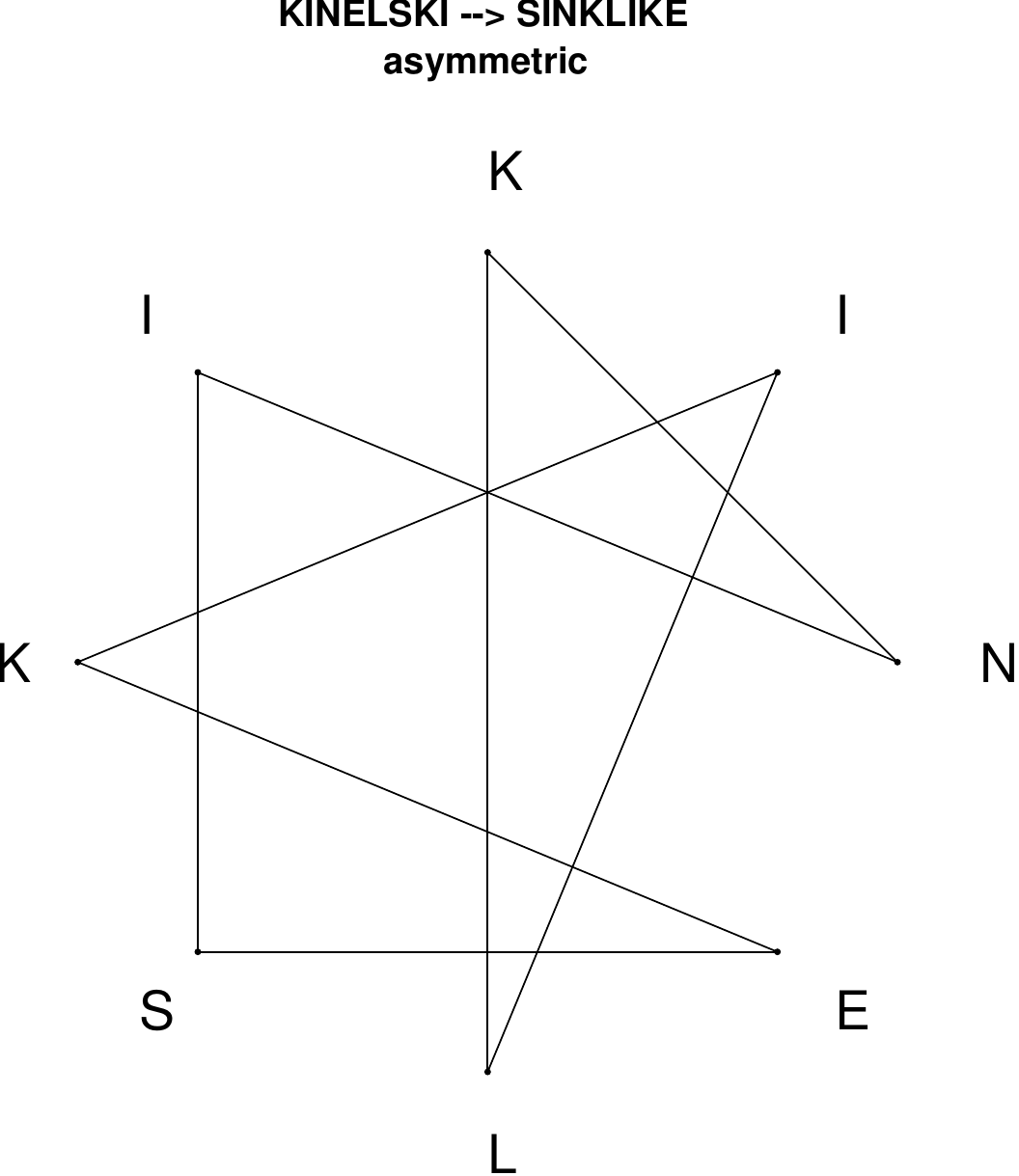}
\end{subfigure}
\end{figure}

\begin{figure}[H]
\centering
\begin{subfigure}[T]{0.19\textwidth}
\centering
\includegraphics[width=\textwidth]{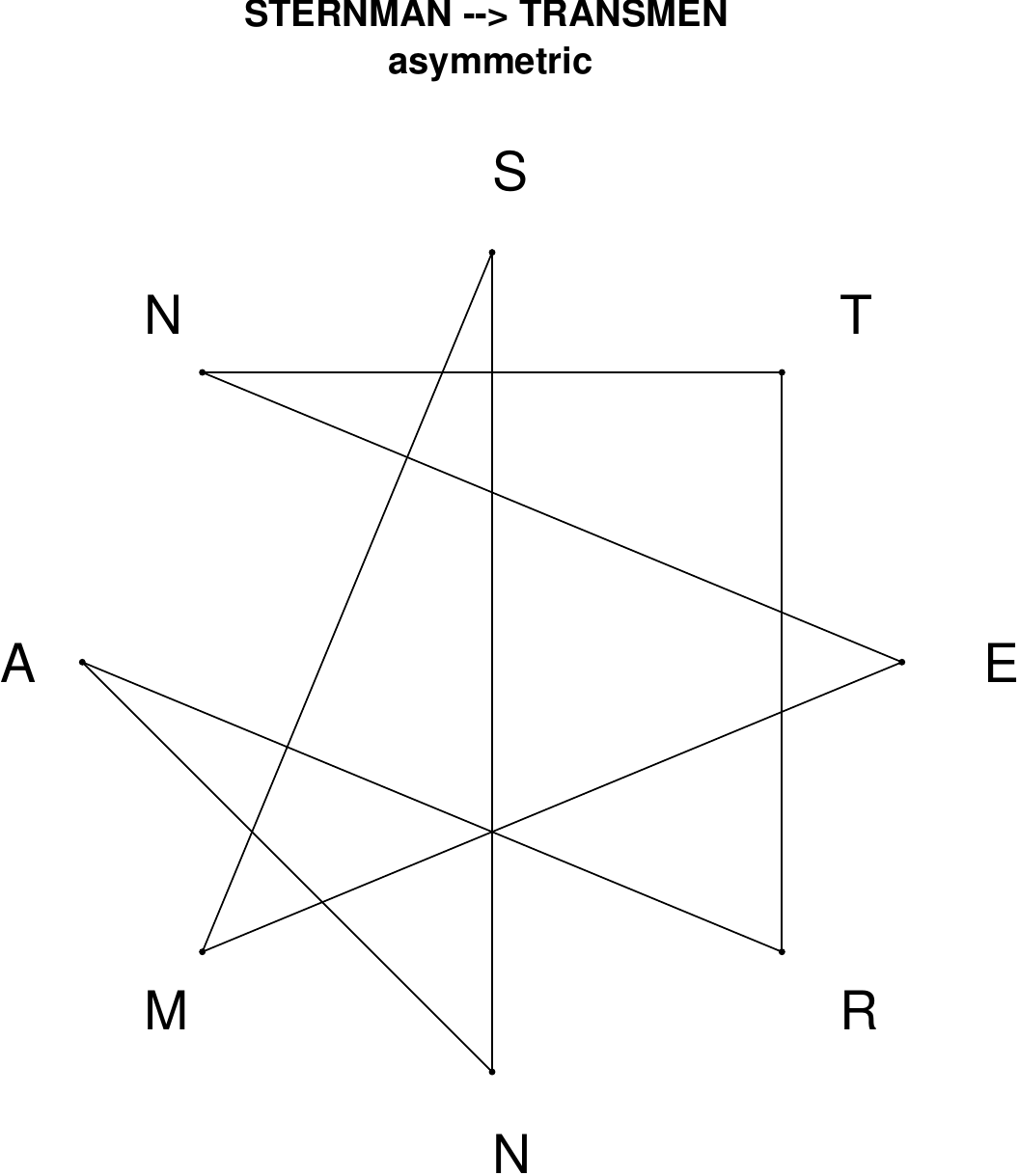}
\end{subfigure}
\hfill
\begin{subfigure}[T]{0.19\textwidth}
\centering
\includegraphics[width=\textwidth]{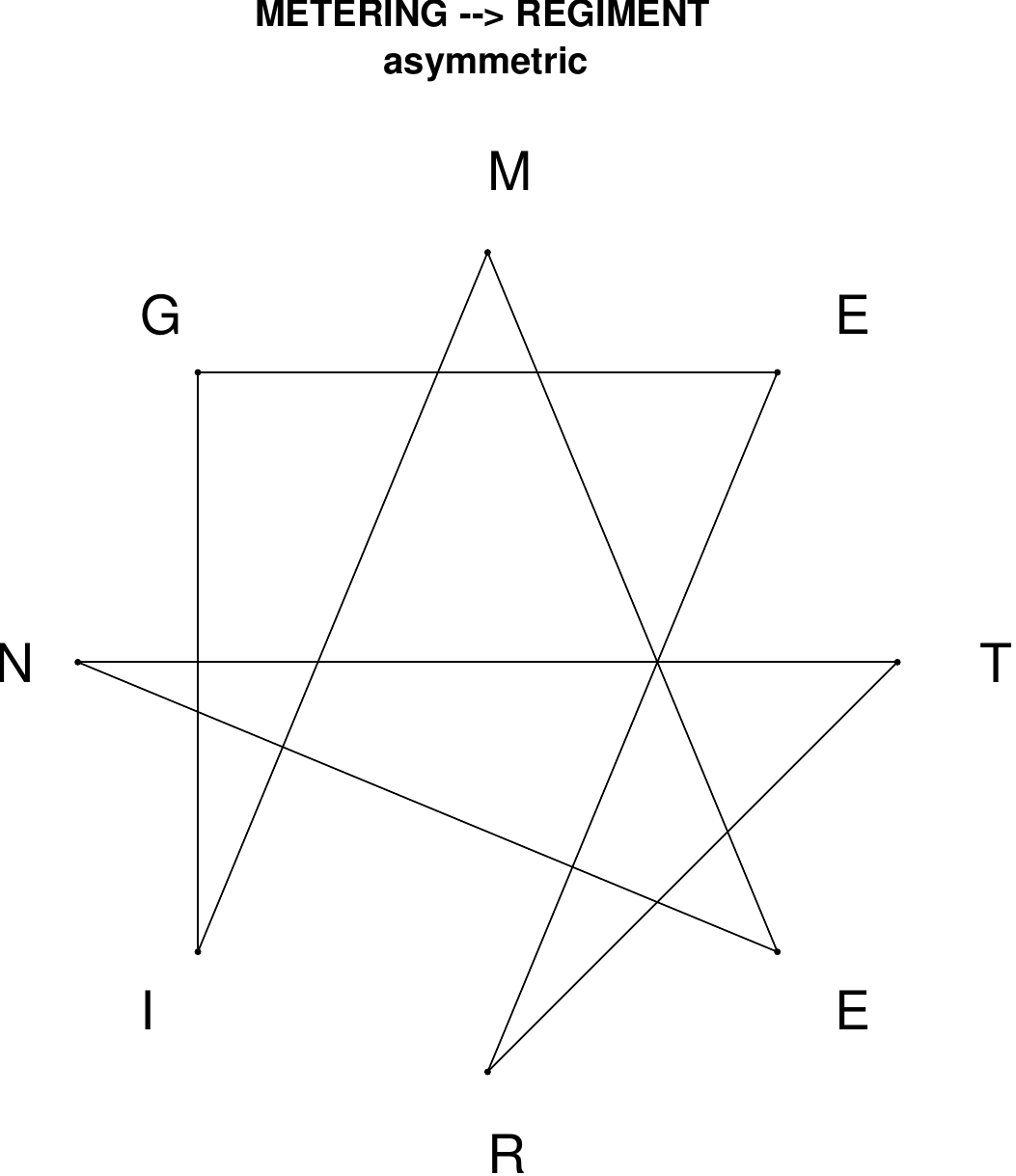}
\end{subfigure}
\hfill
\begin{subfigure}[T]{0.19\textwidth}
\centering
\includegraphics[width=\textwidth]{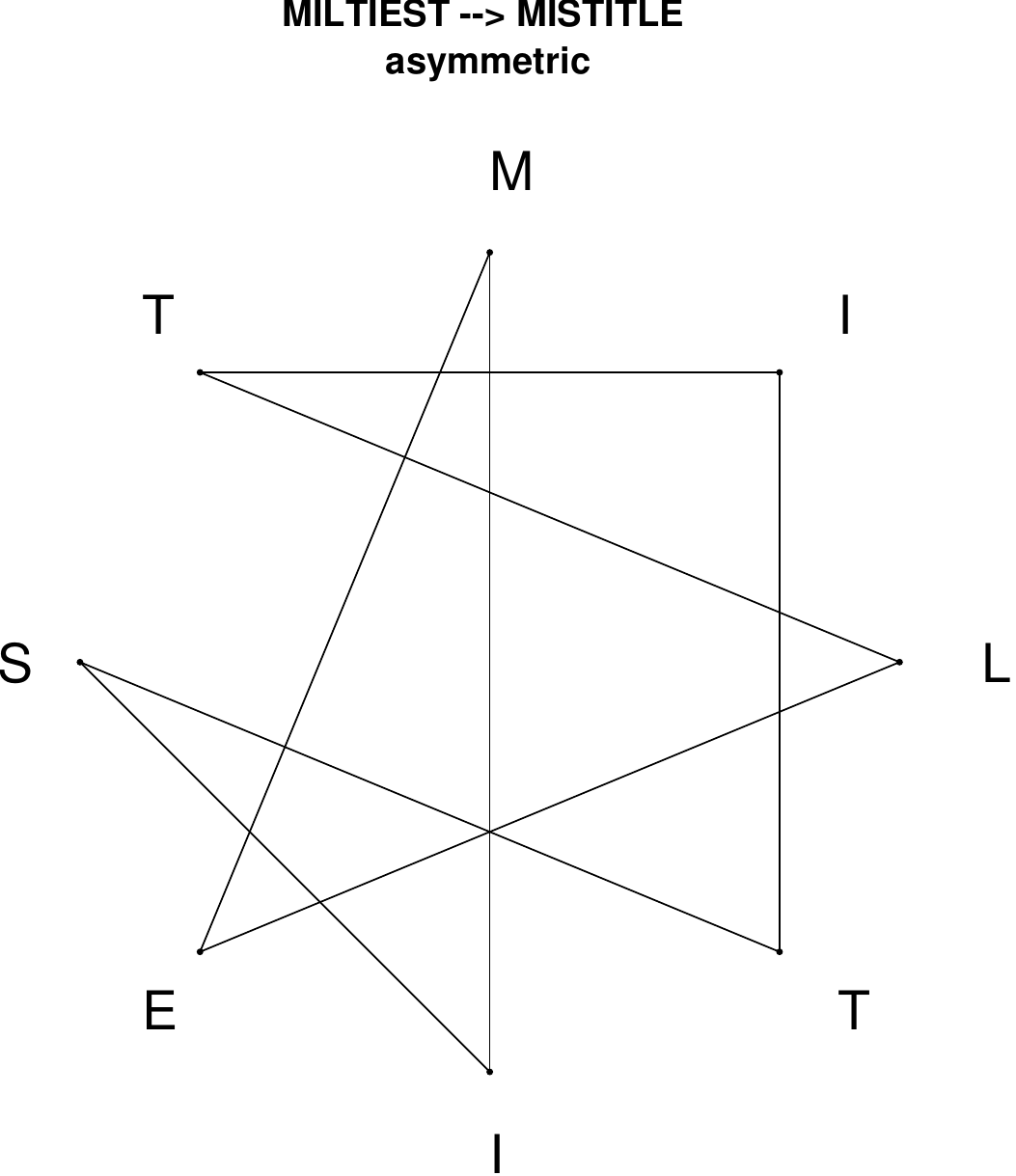}
\end{subfigure}
\hfill
\begin{subfigure}[T]{0.19\textwidth}
\centering
\includegraphics[width=\textwidth]{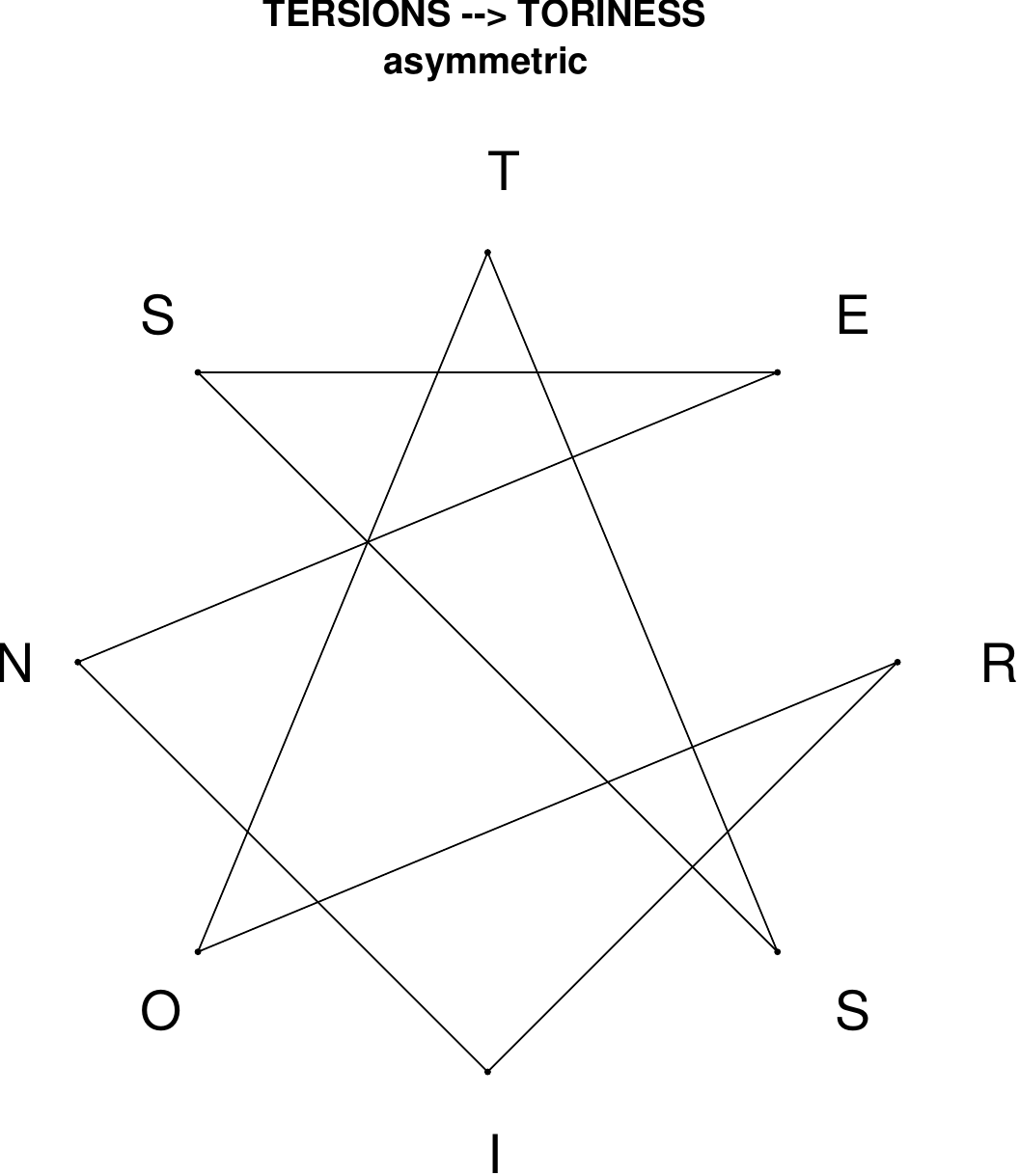}
\end{subfigure}
\hfill
\begin{subfigure}[T]{0.19\textwidth}
\centering
\includegraphics[width=\textwidth]{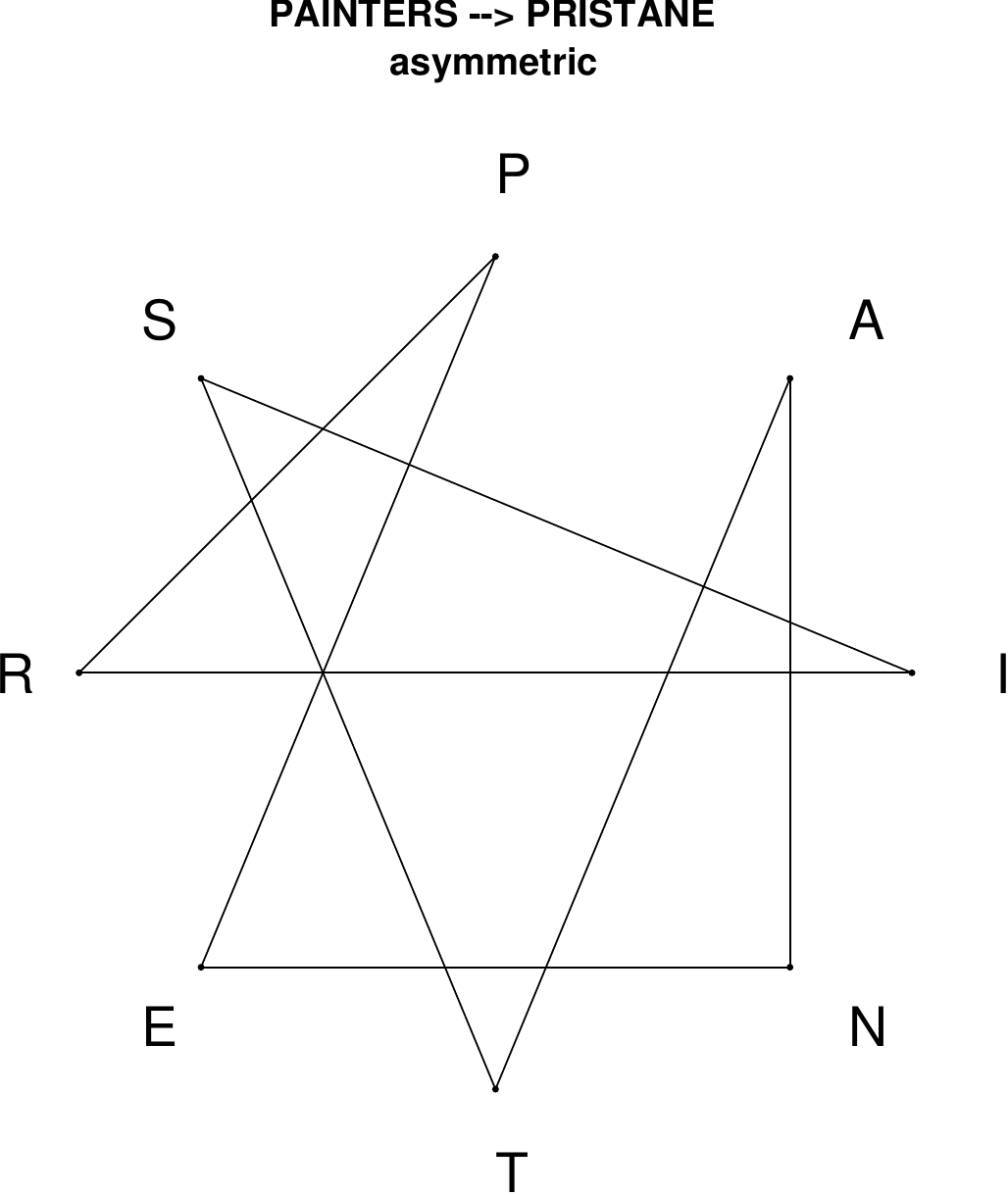}
\end{subfigure}
\end{figure}

\begin{figure}[H]
\centering
\begin{subfigure}[T]{0.19\textwidth}
\centering
\includegraphics[width=\textwidth]{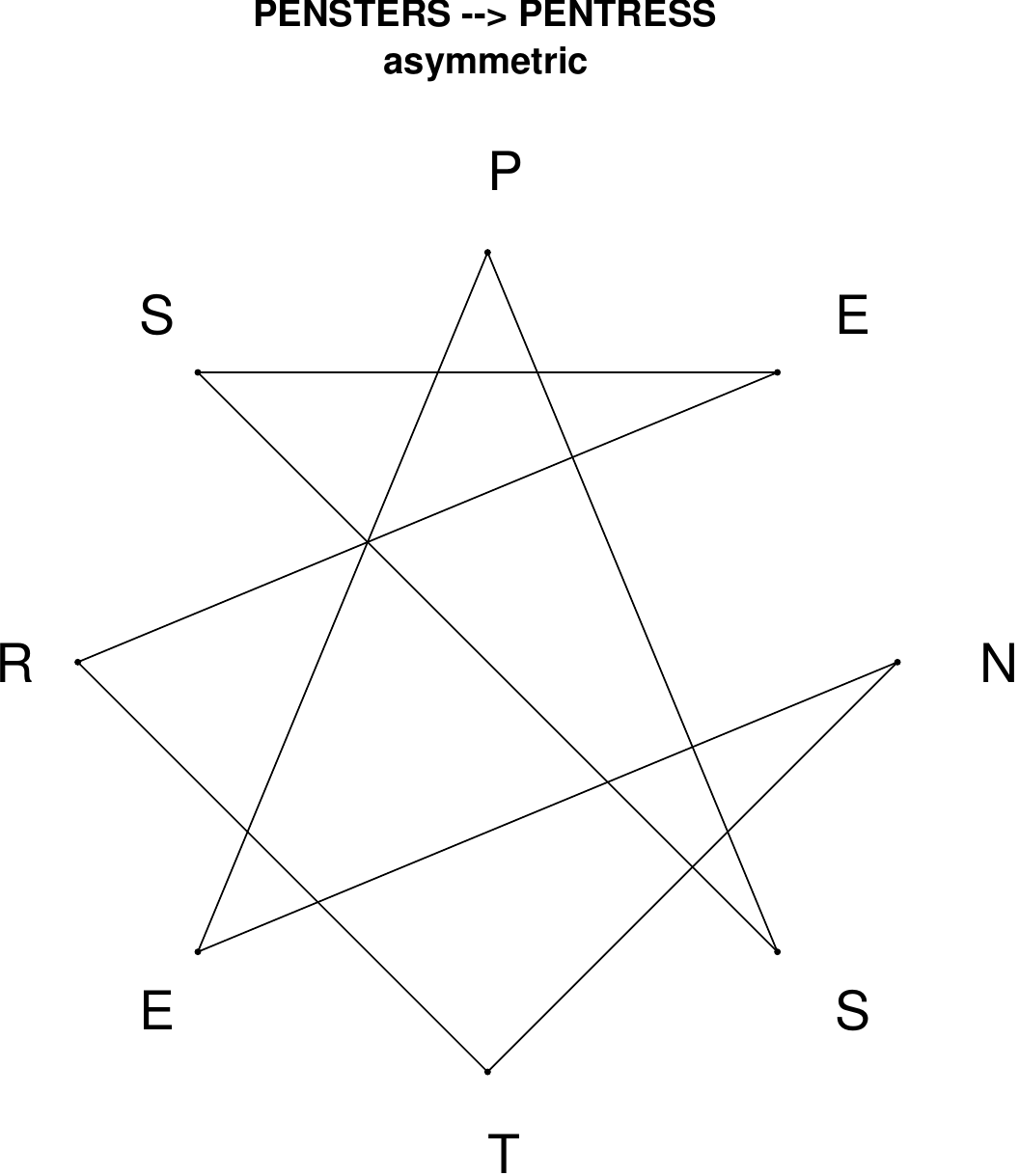}
\end{subfigure}
\hfill
\begin{subfigure}[T]{0.19\textwidth}
\centering
\includegraphics[width=\textwidth]{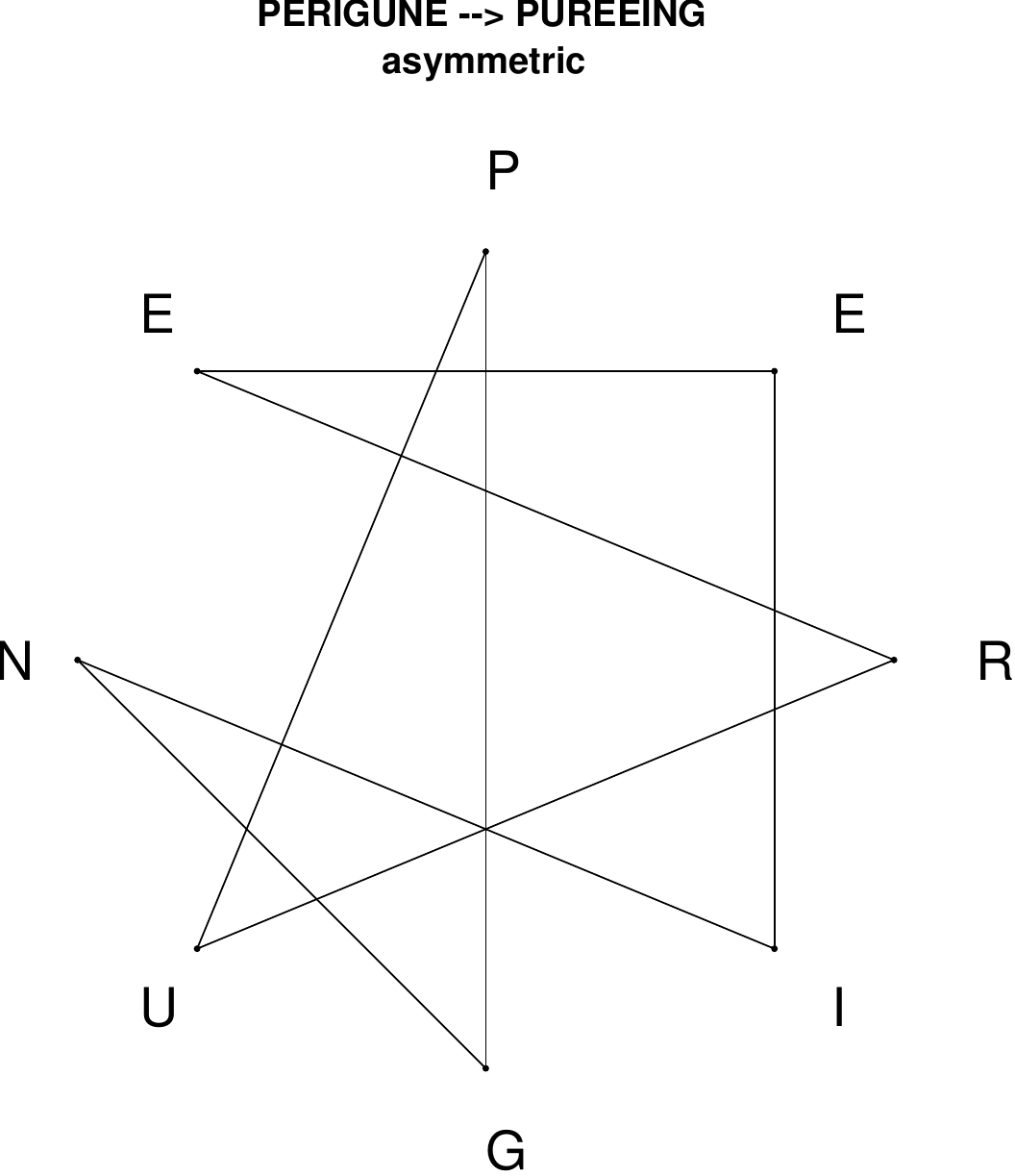}
\end{subfigure}
\hfill
\begin{subfigure}[T]{0.19\textwidth}
\centering
\includegraphics[width=\textwidth]{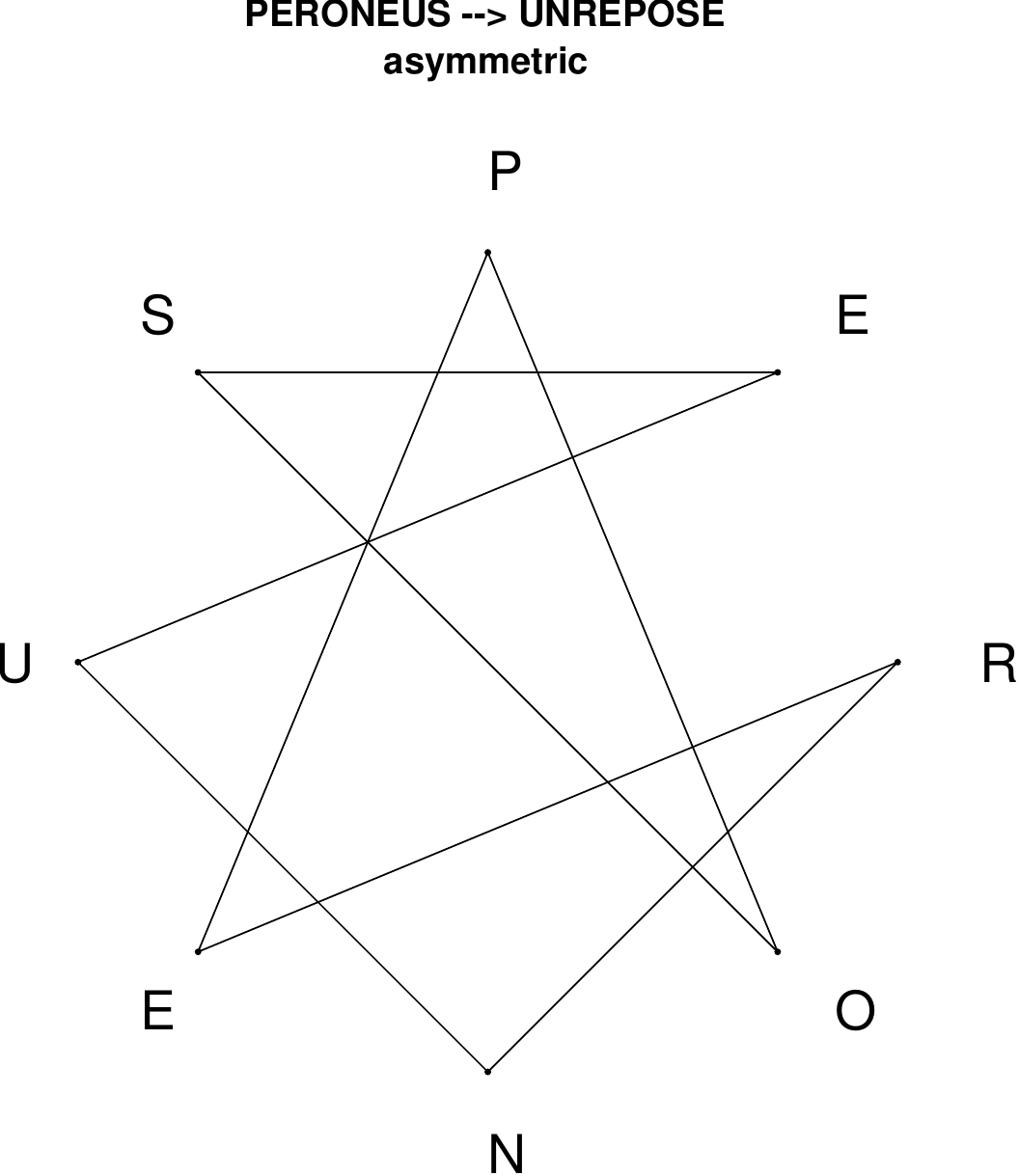}
\end{subfigure}
\hfill
\begin{subfigure}[T]{0.19\textwidth}
\centering
\includegraphics[width=\textwidth]{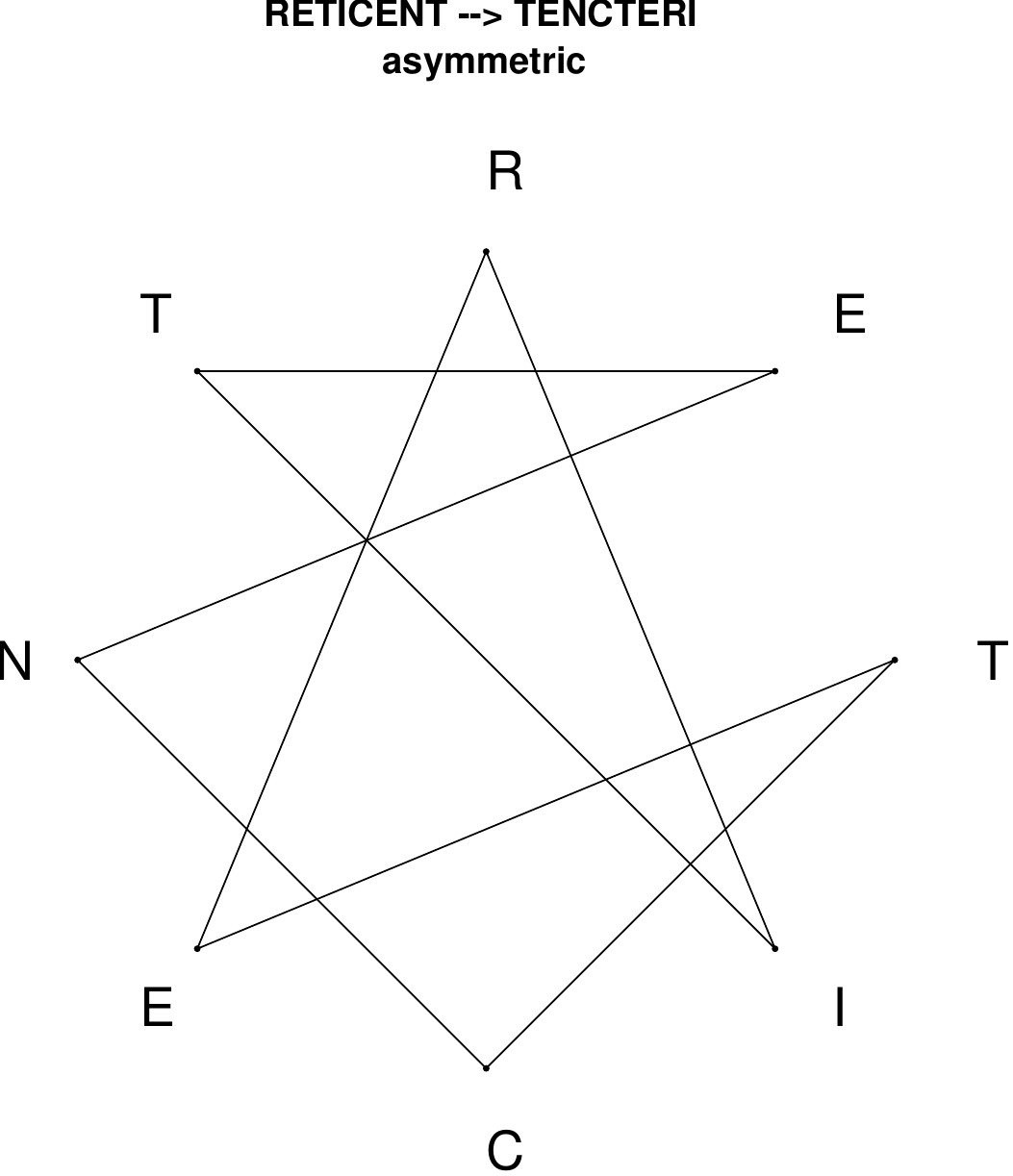}
\end{subfigure}
\hfill
\begin{subfigure}[T]{0.19\textwidth}
\centering
\includegraphics[width=\textwidth]{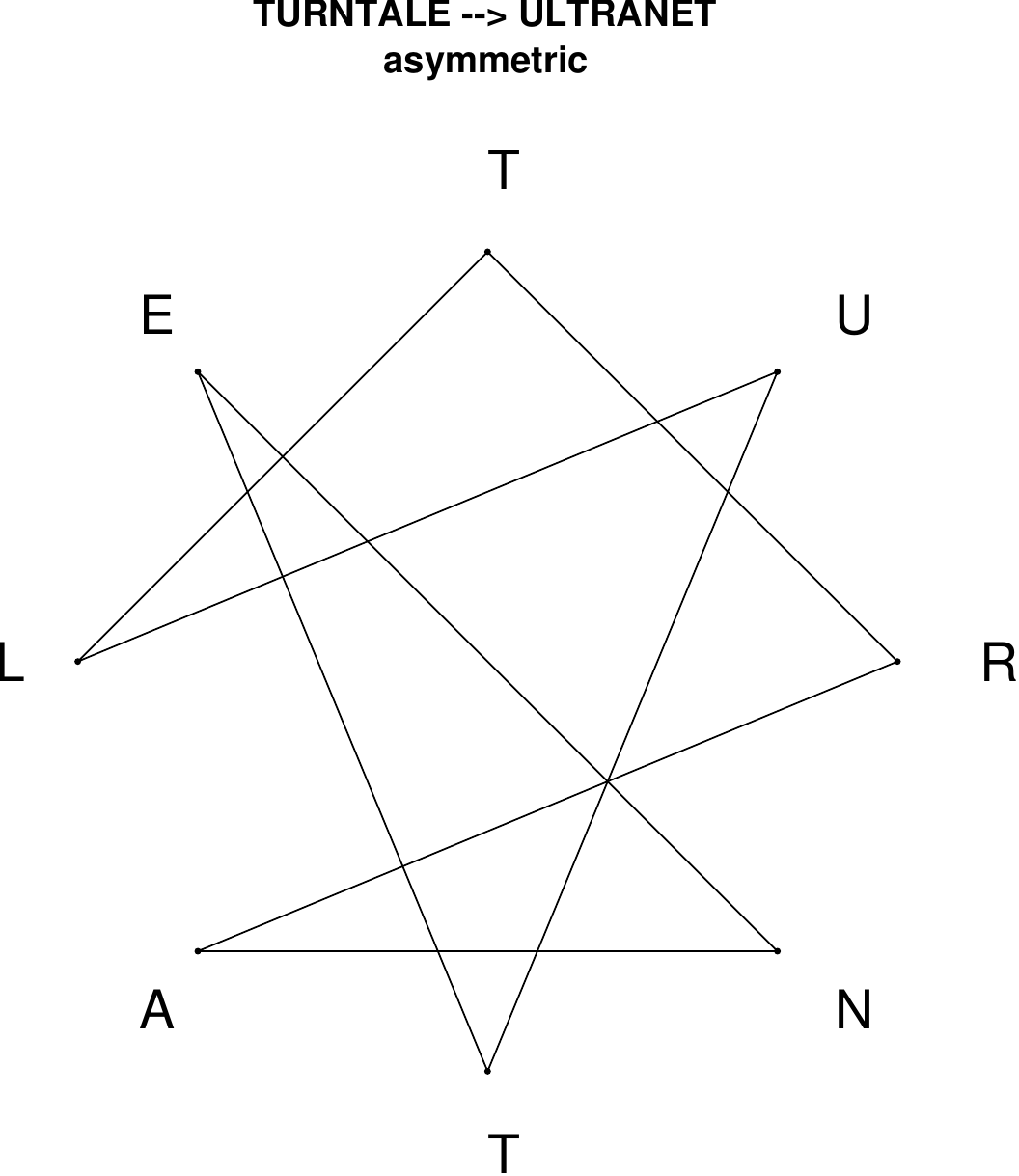}
\end{subfigure}
\end{figure}

\begin{figure}[H]
\centering
\begin{subfigure}[T]{0.19\textwidth}
\centering
\includegraphics[width=\textwidth]{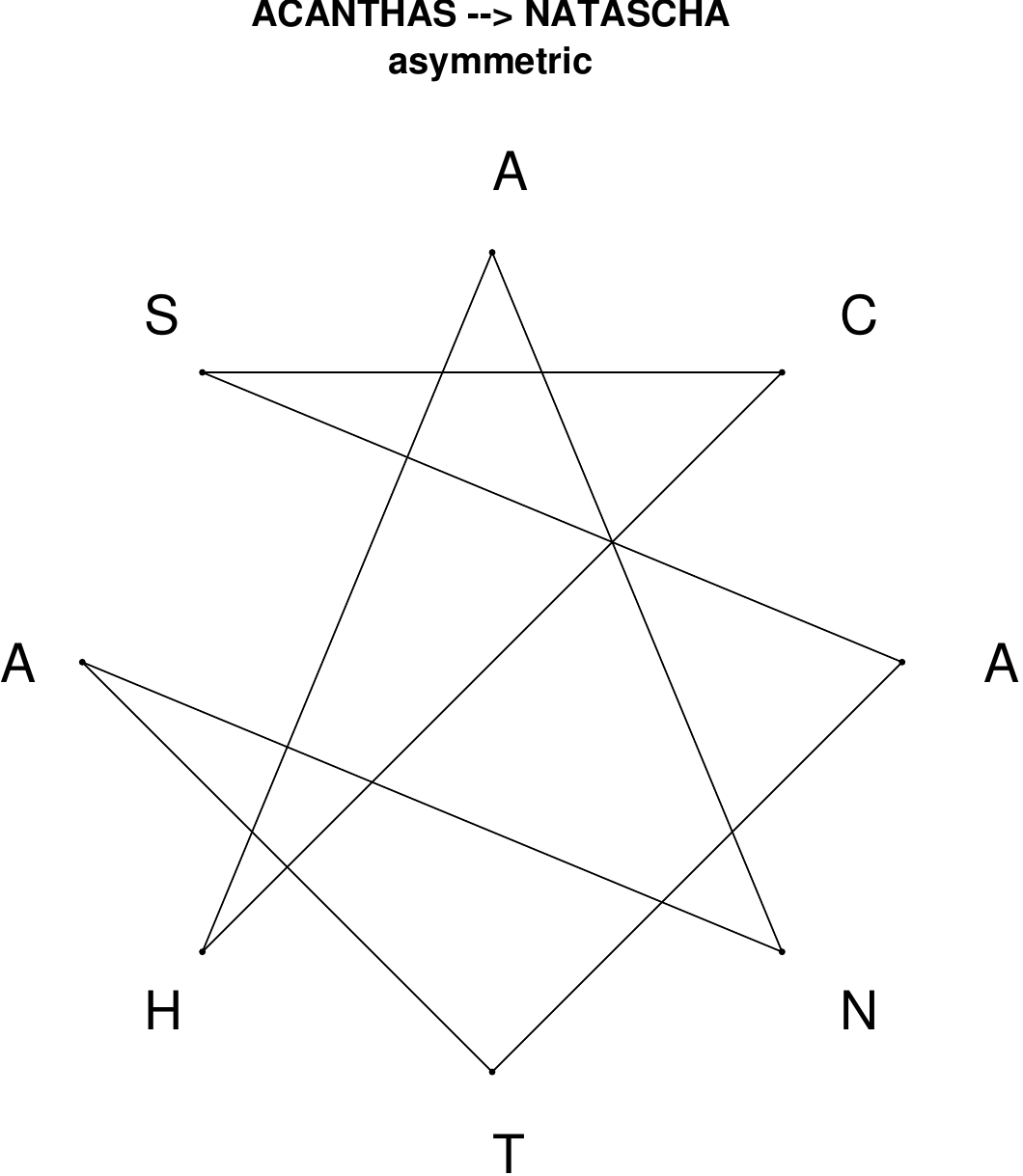}
\end{subfigure}
\hfill
\begin{subfigure}[T]{0.19\textwidth}
\centering
\includegraphics[width=\textwidth]{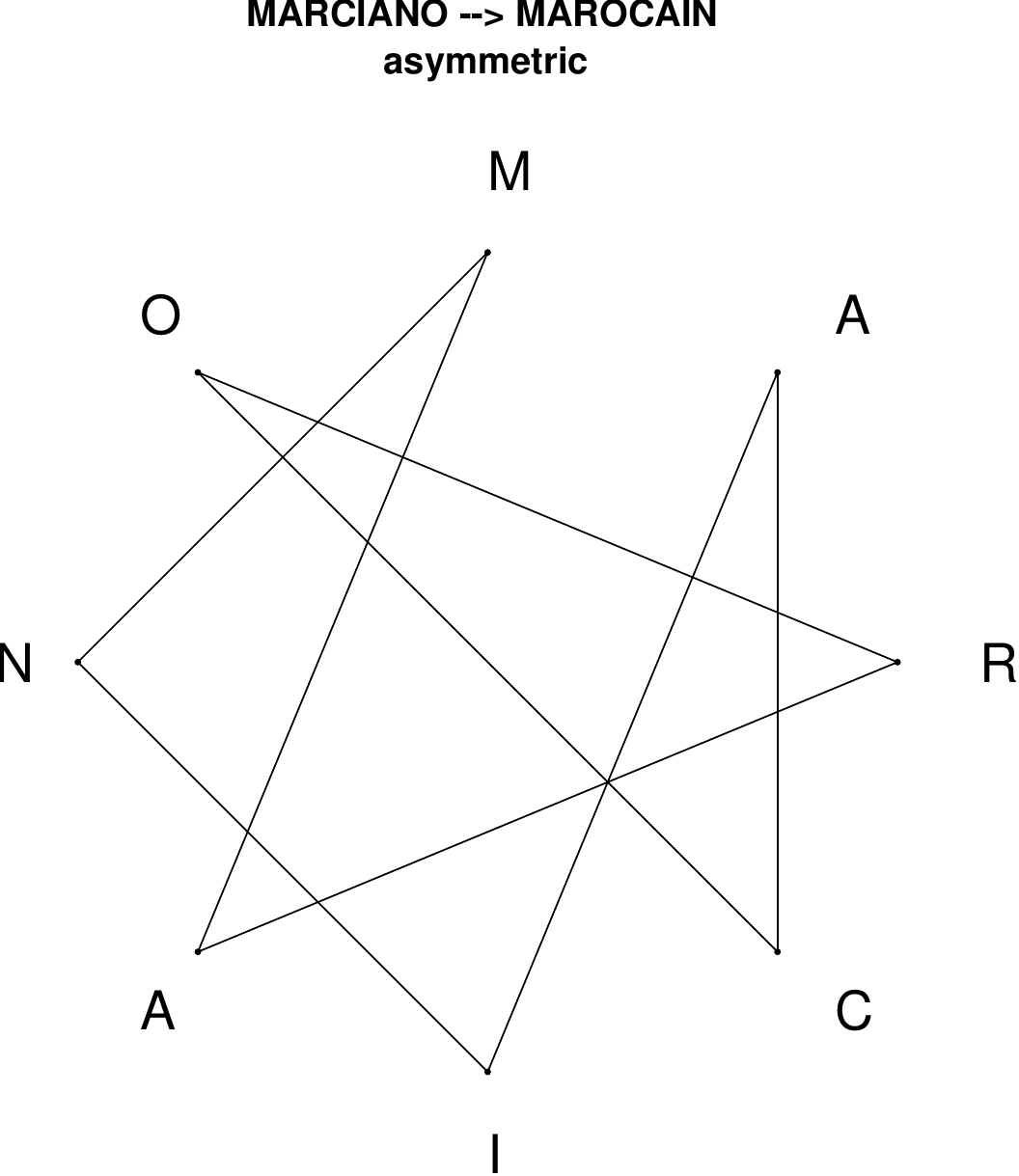}
\end{subfigure}
\hfill
\begin{subfigure}[T]{0.19\textwidth}
\centering
\includegraphics[width=\textwidth]{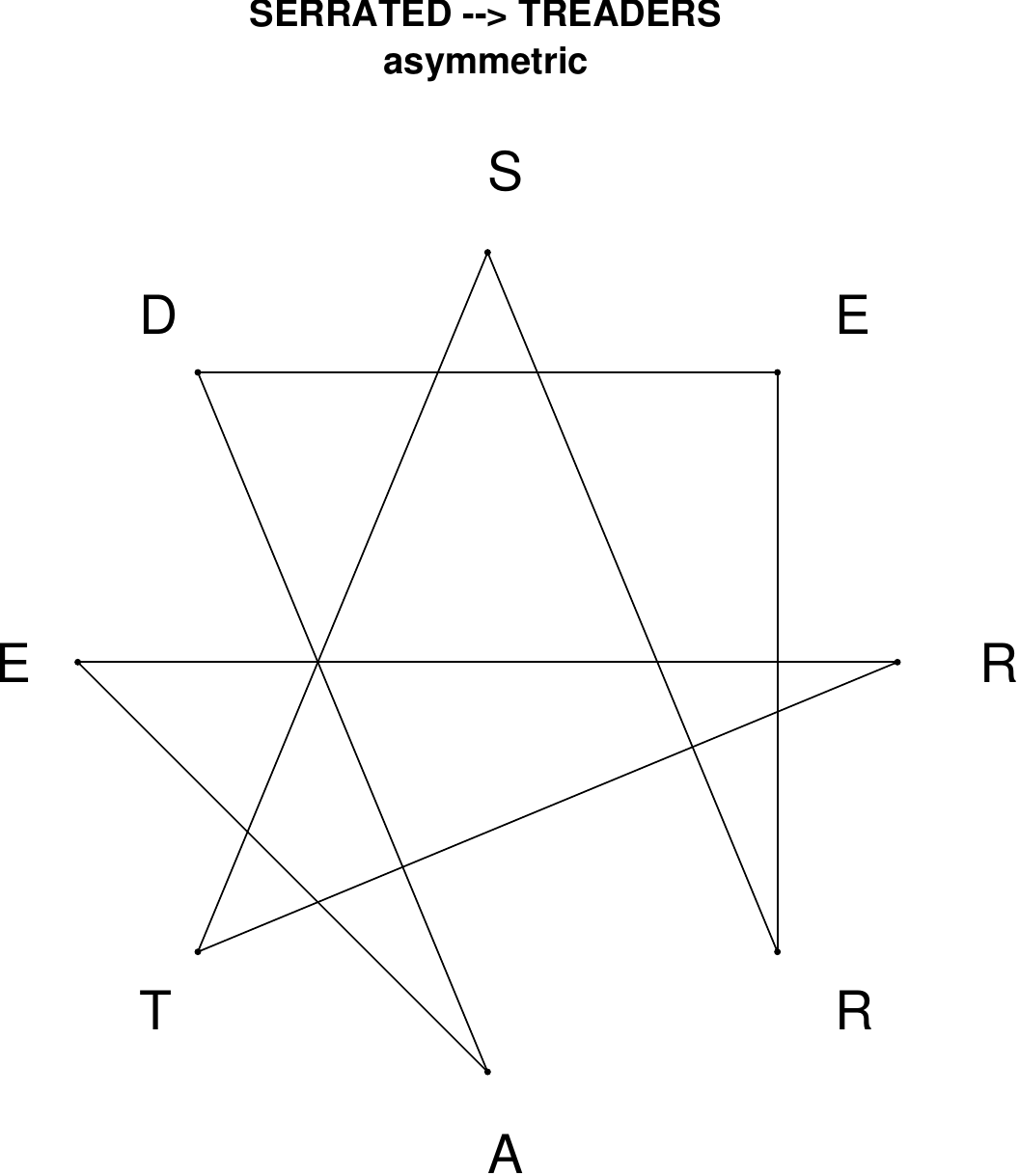}
\end{subfigure}
\hfill
\begin{subfigure}[T]{0.19\textwidth}
\centering
\includegraphics[width=\textwidth]{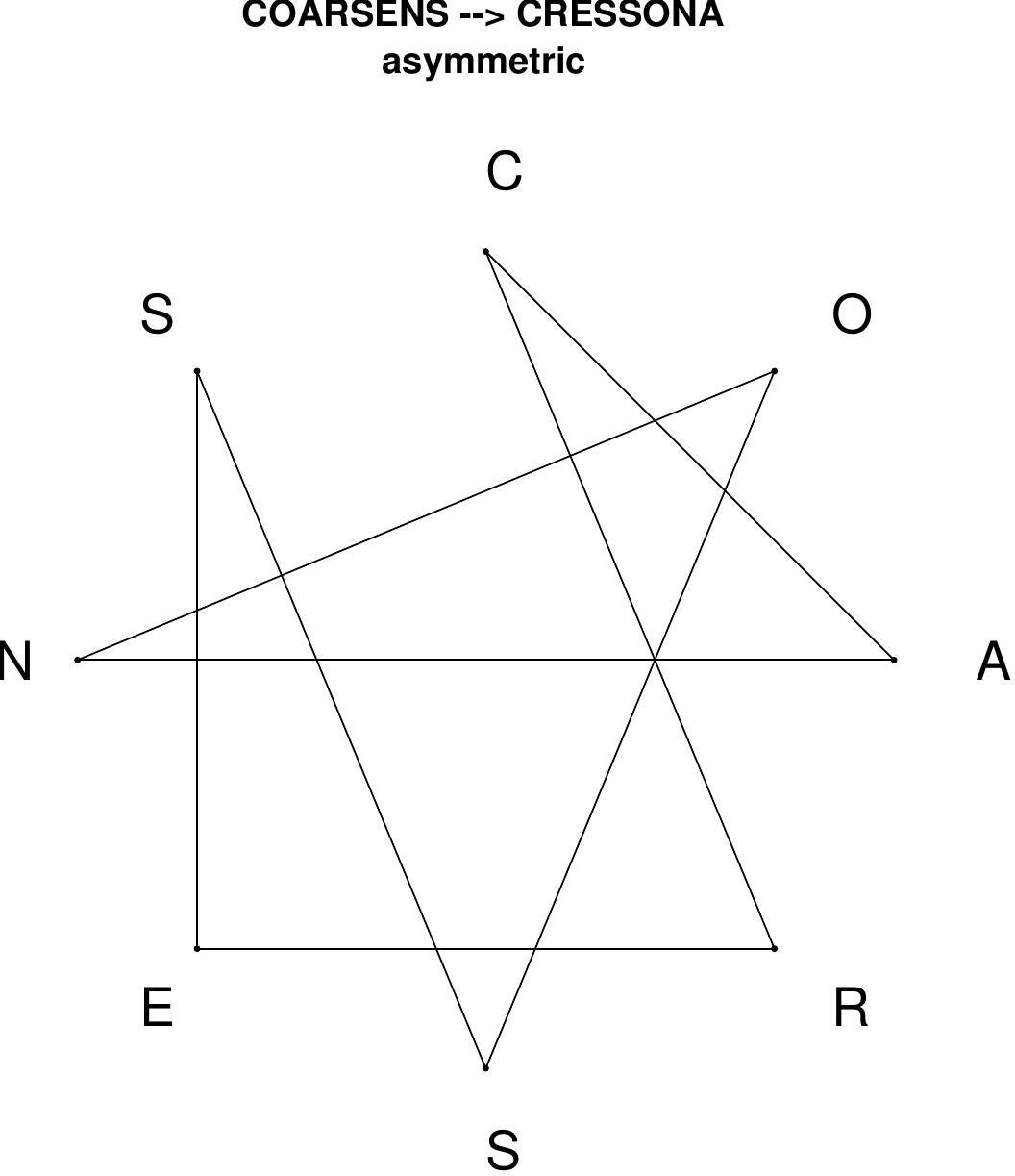}
\end{subfigure}
\hfill
\begin{subfigure}[T]{0.19\textwidth}
\centering
\includegraphics[width=\textwidth]{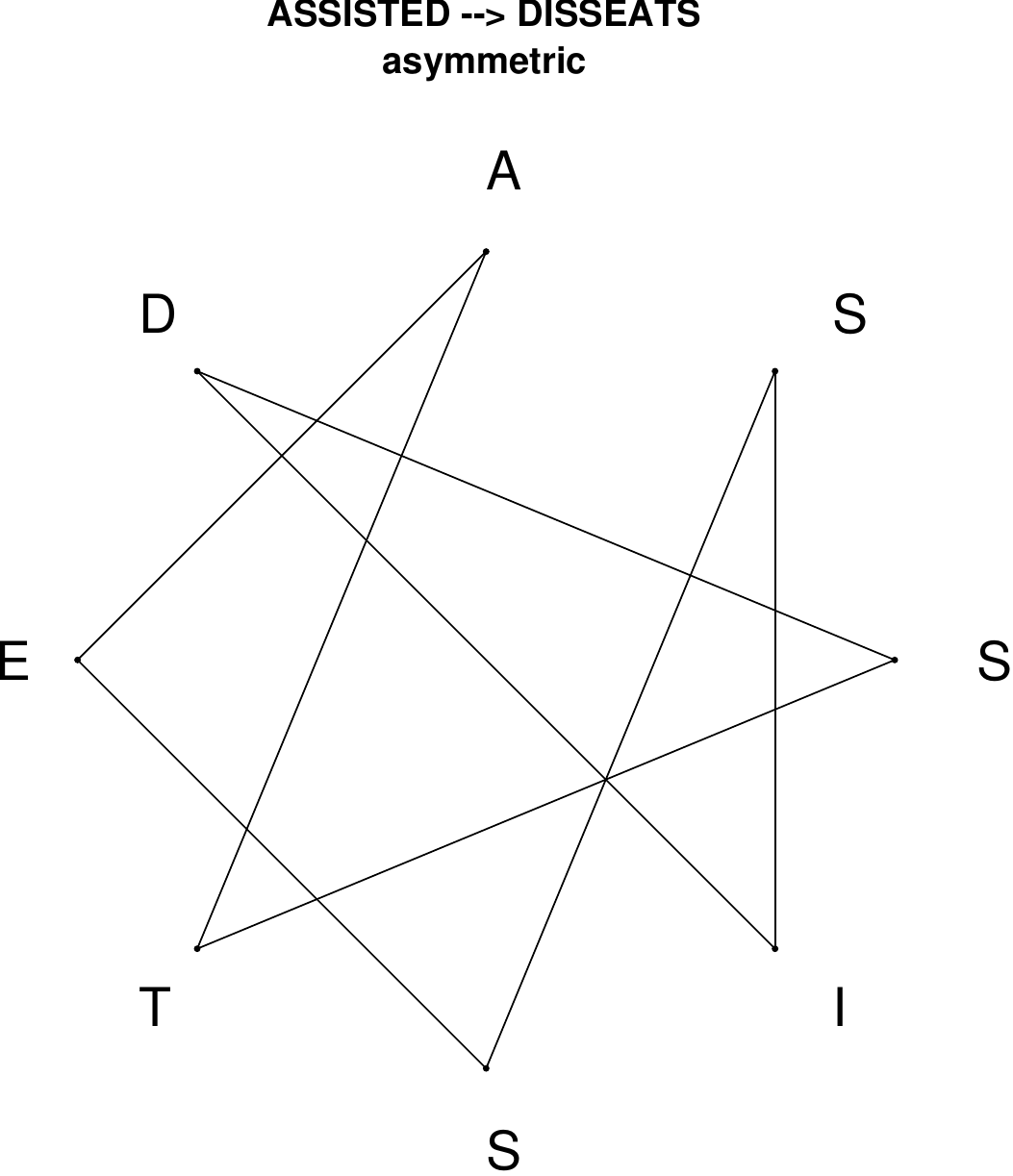}
\end{subfigure}
\end{figure}

\begin{figure}[H]
\centering
\begin{subfigure}[T]{0.19\textwidth}
\centering
\includegraphics[width=\textwidth]{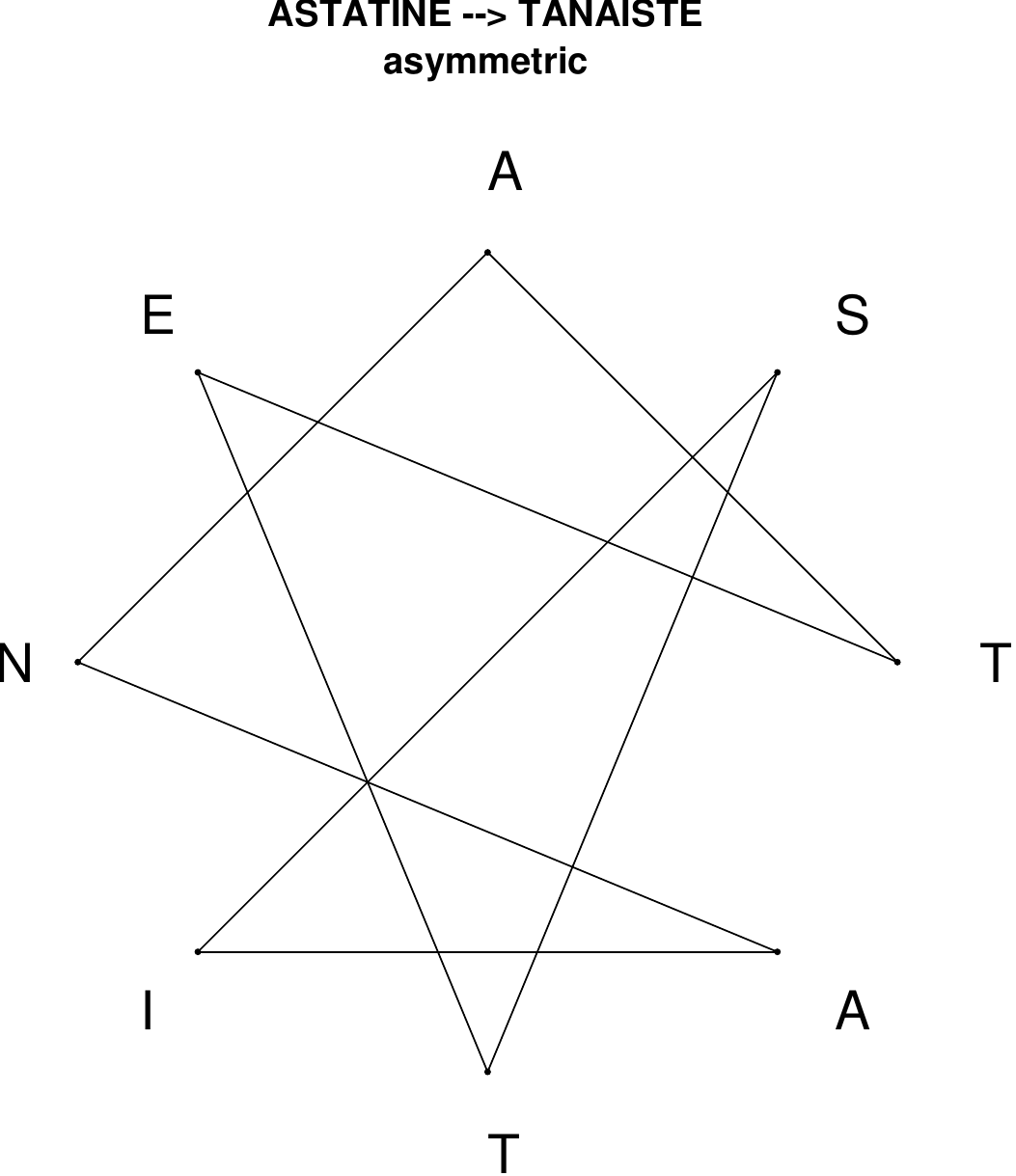}
\end{subfigure}
\hfill
\begin{subfigure}[T]{0.19\textwidth}
\centering
\includegraphics[width=\textwidth]{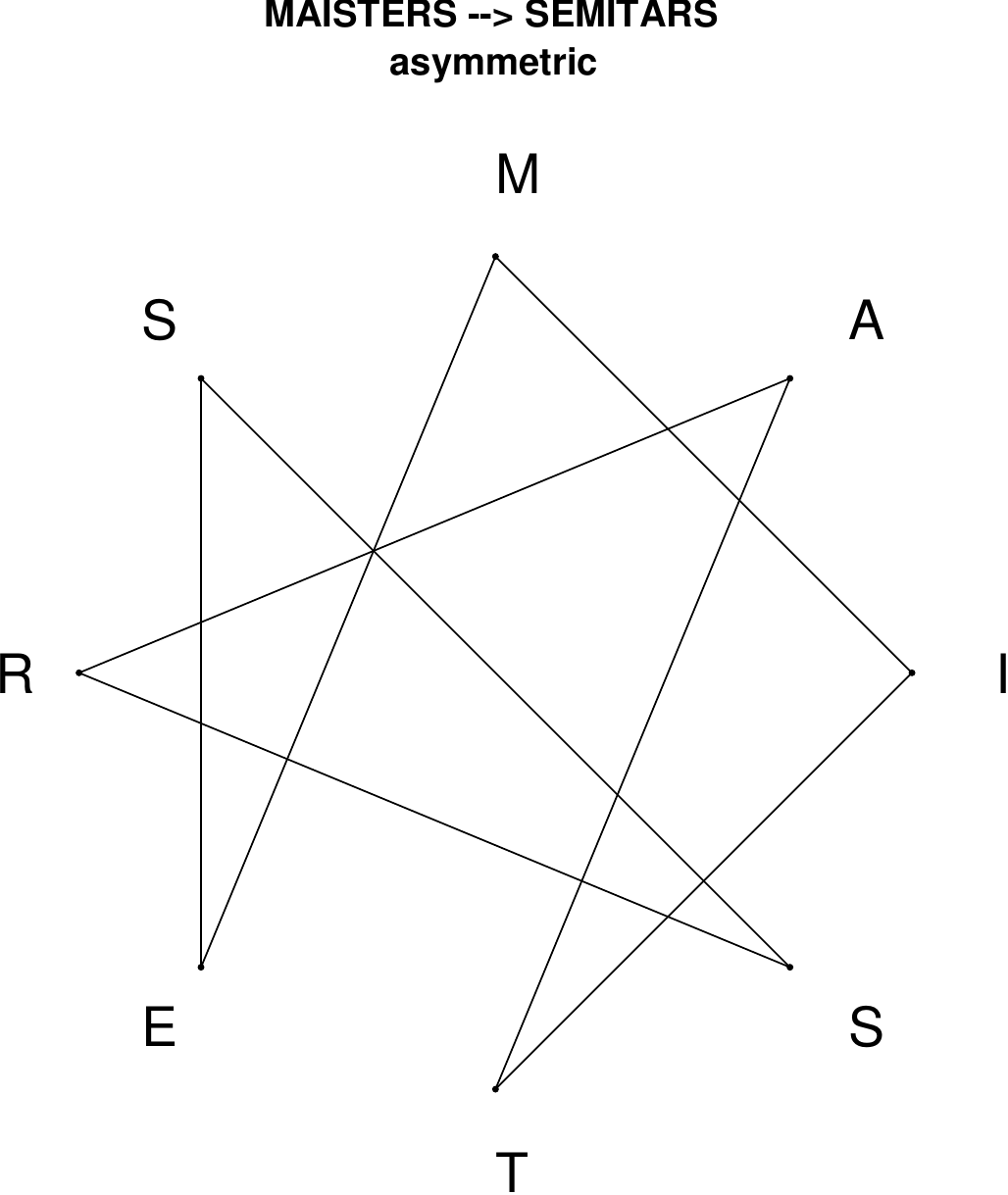}
\end{subfigure}
\hfill
\begin{subfigure}[T]{0.19\textwidth}
\centering
\includegraphics[width=\textwidth]{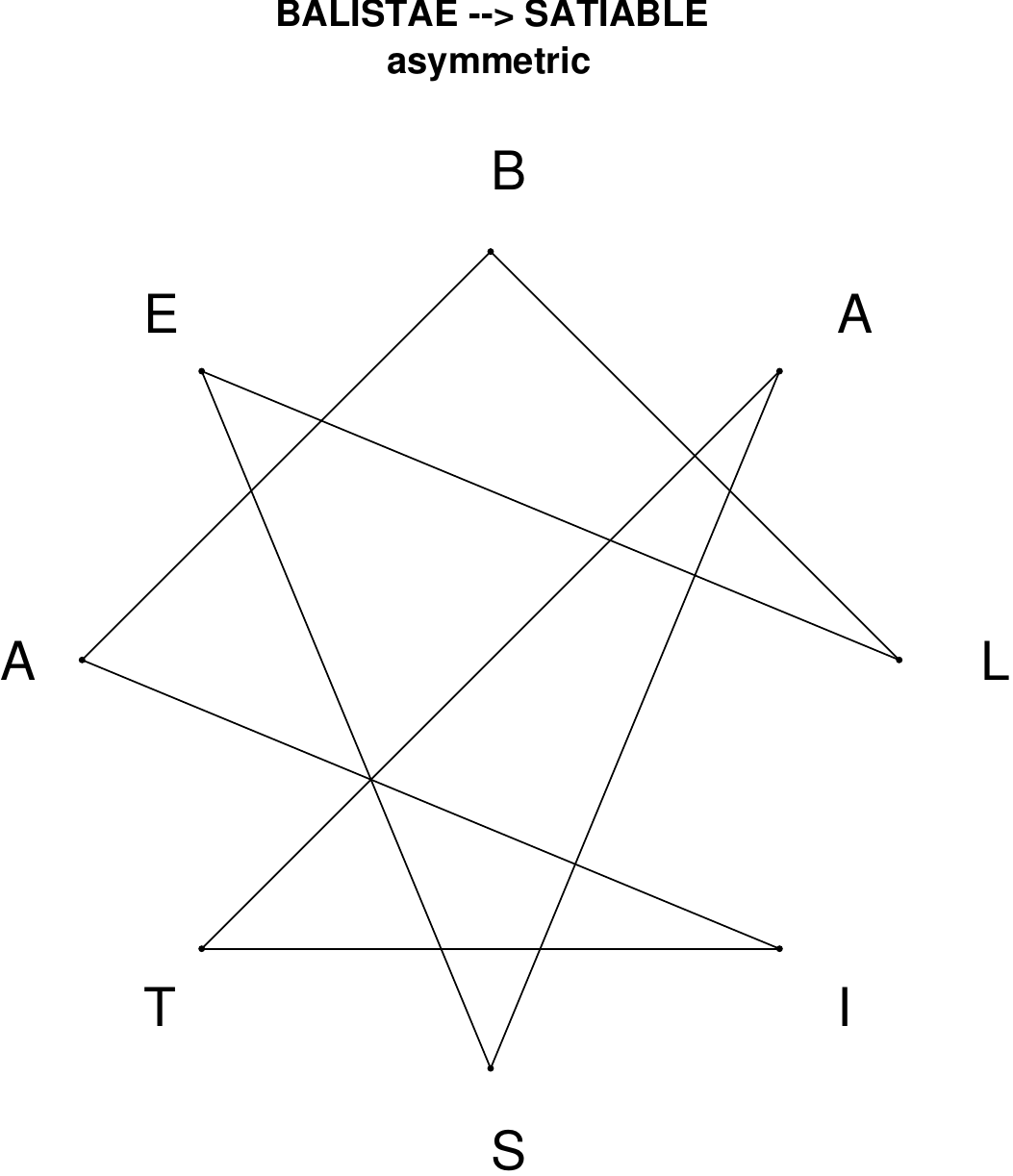}
\end{subfigure}
\hfill
\begin{subfigure}[T]{0.19\textwidth}
\centering
\includegraphics[width=\textwidth]{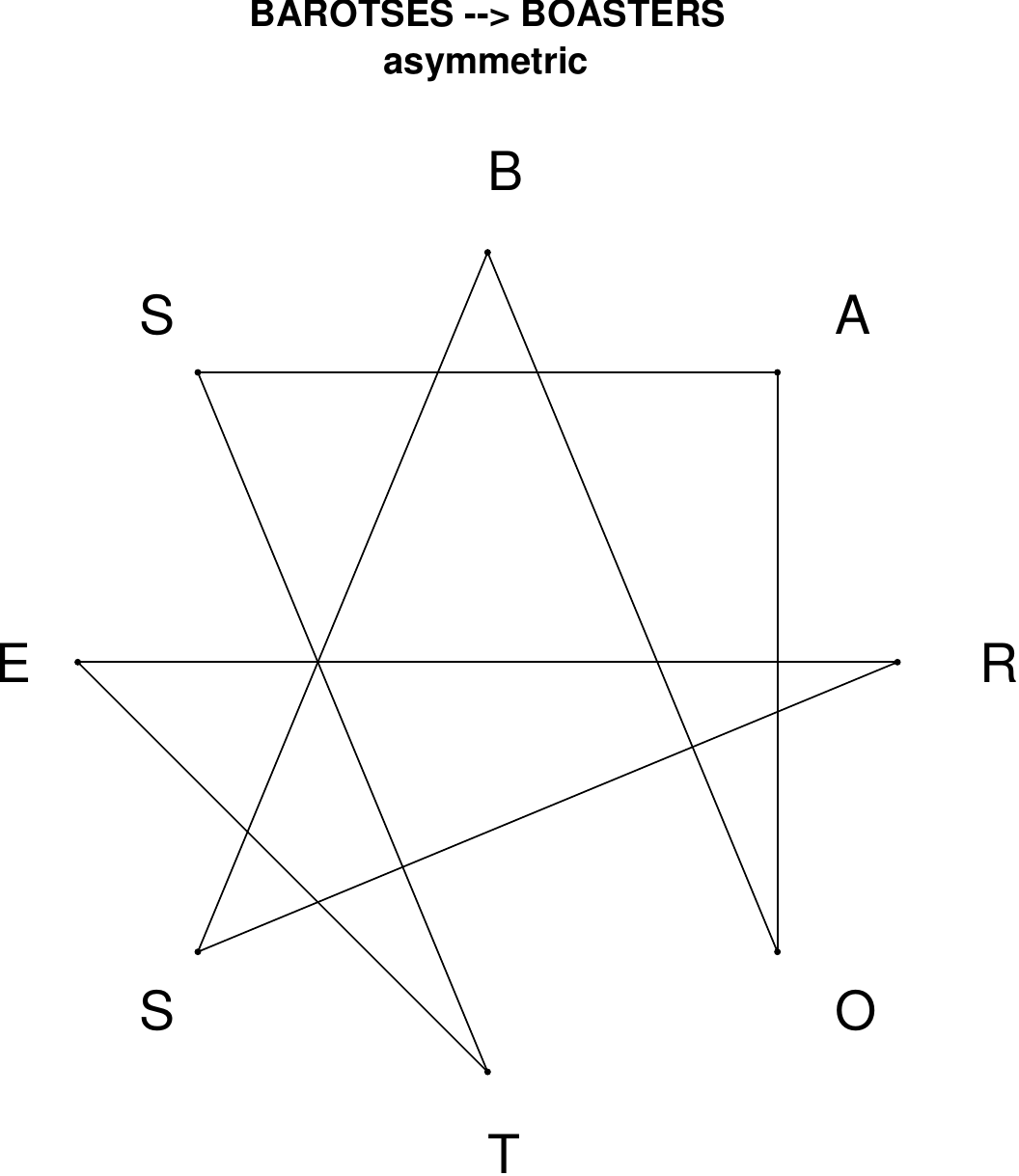}
\end{subfigure}
\hfill
\begin{subfigure}[T]{0.19\textwidth}
\centering
\includegraphics[width=\textwidth]{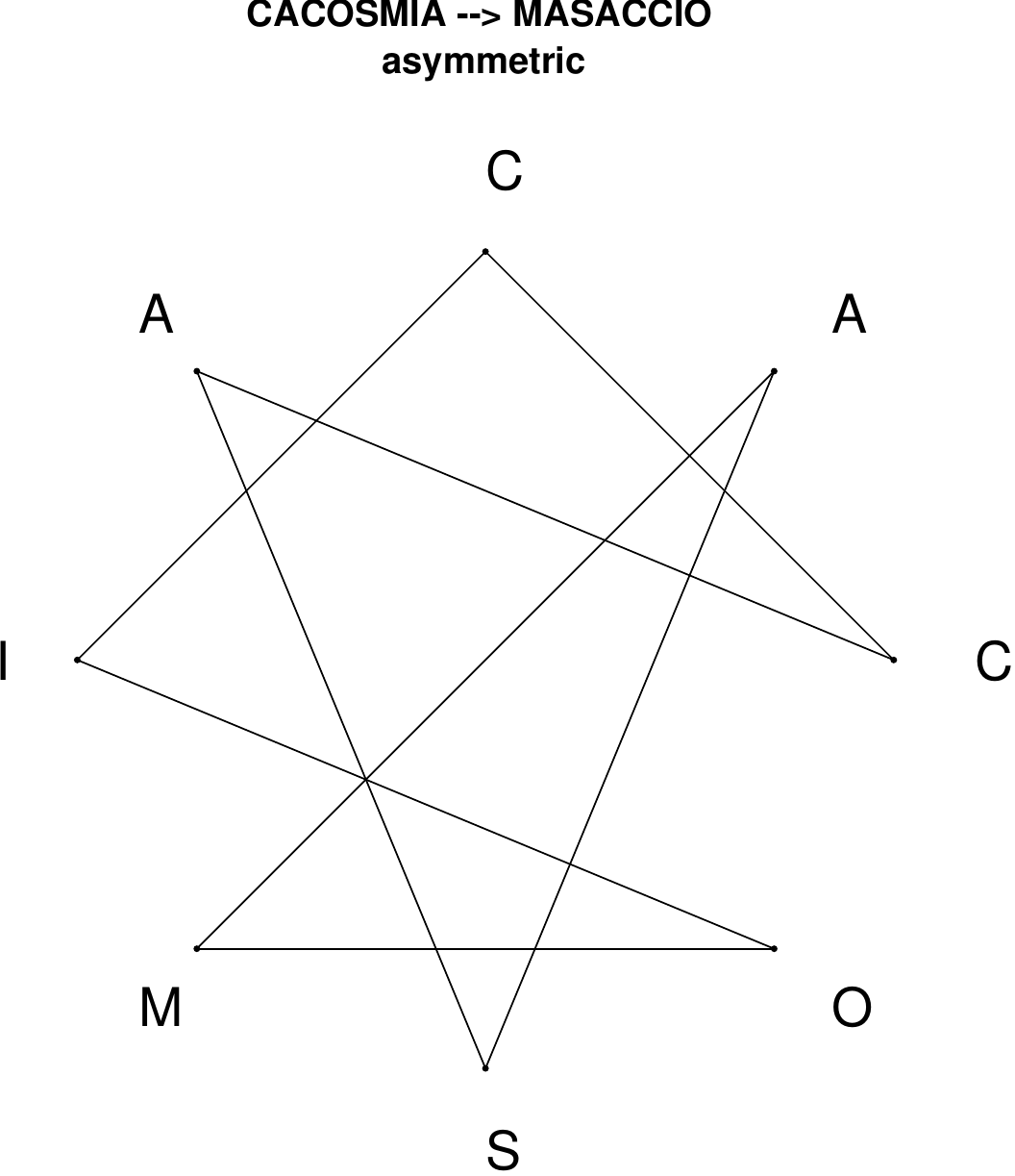}
\end{subfigure}
\end{figure}

\begin{figure}[H]
\centering
\begin{subfigure}[T]{0.19\textwidth}
\centering
\includegraphics[width=\textwidth]{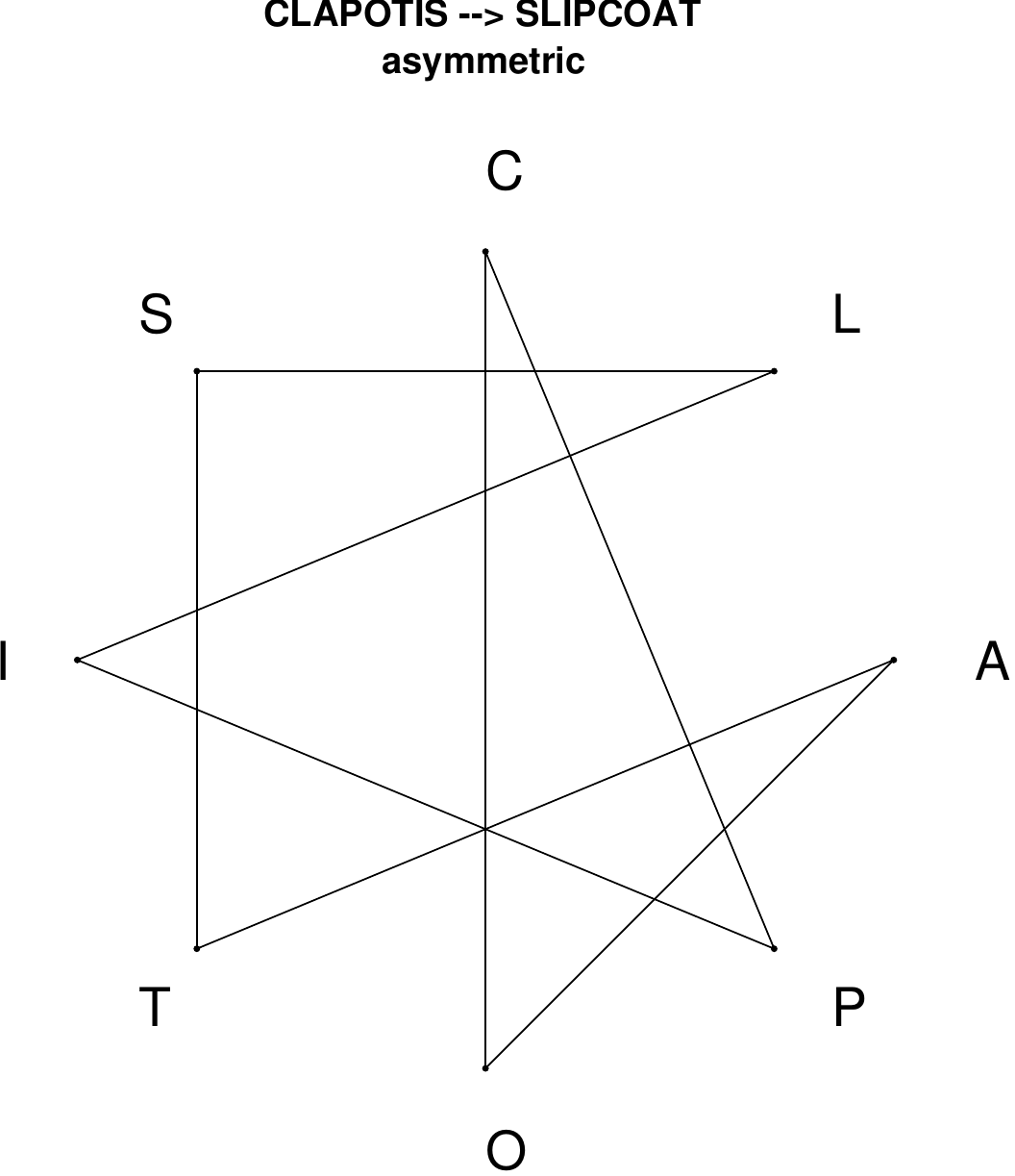}
\end{subfigure}
\hfill
\begin{subfigure}[T]{0.19\textwidth}
\centering
\includegraphics[width=\textwidth]{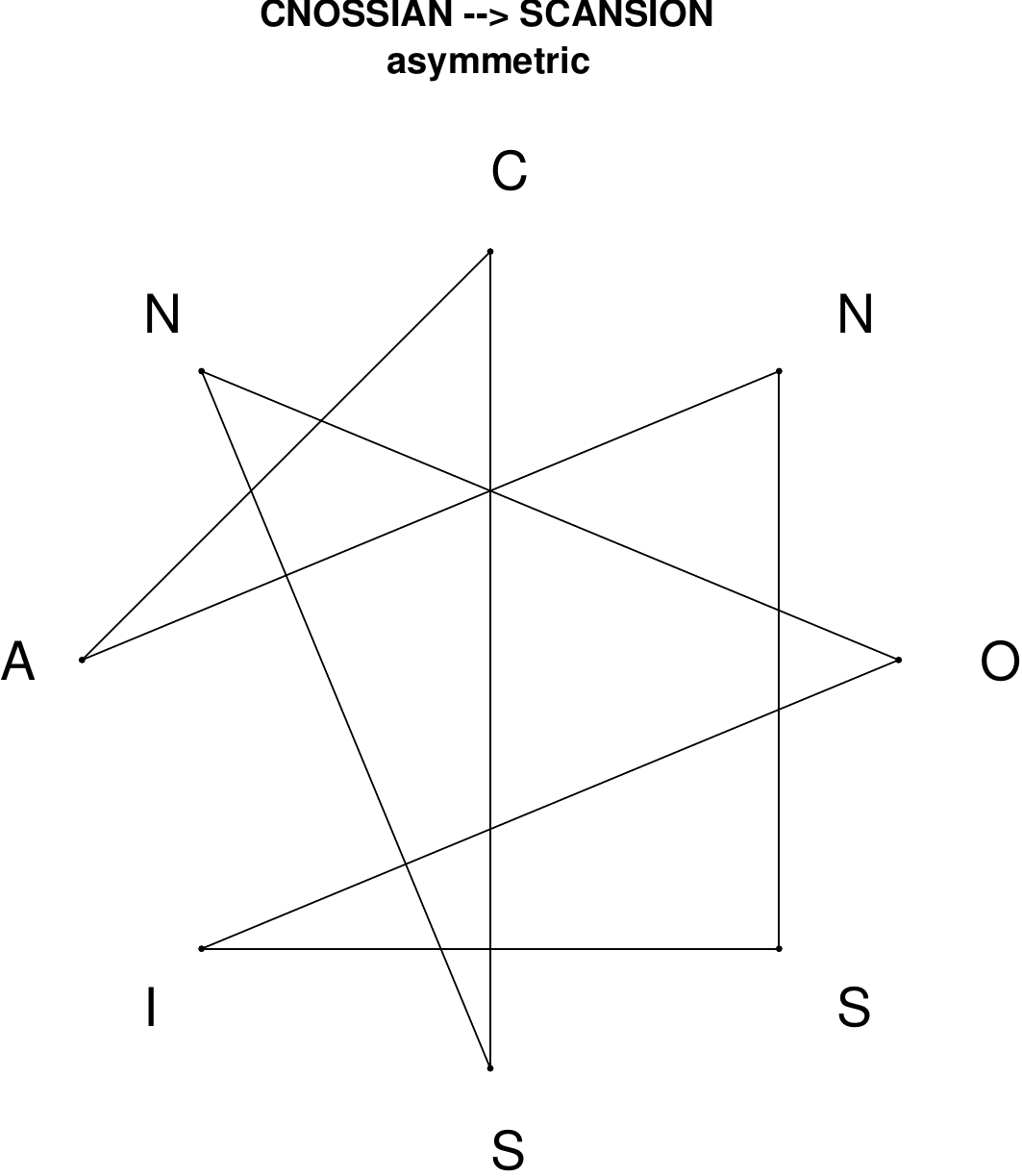}
\end{subfigure}
\hfill
\begin{subfigure}[T]{0.19\textwidth}
\centering
\includegraphics[width=\textwidth]{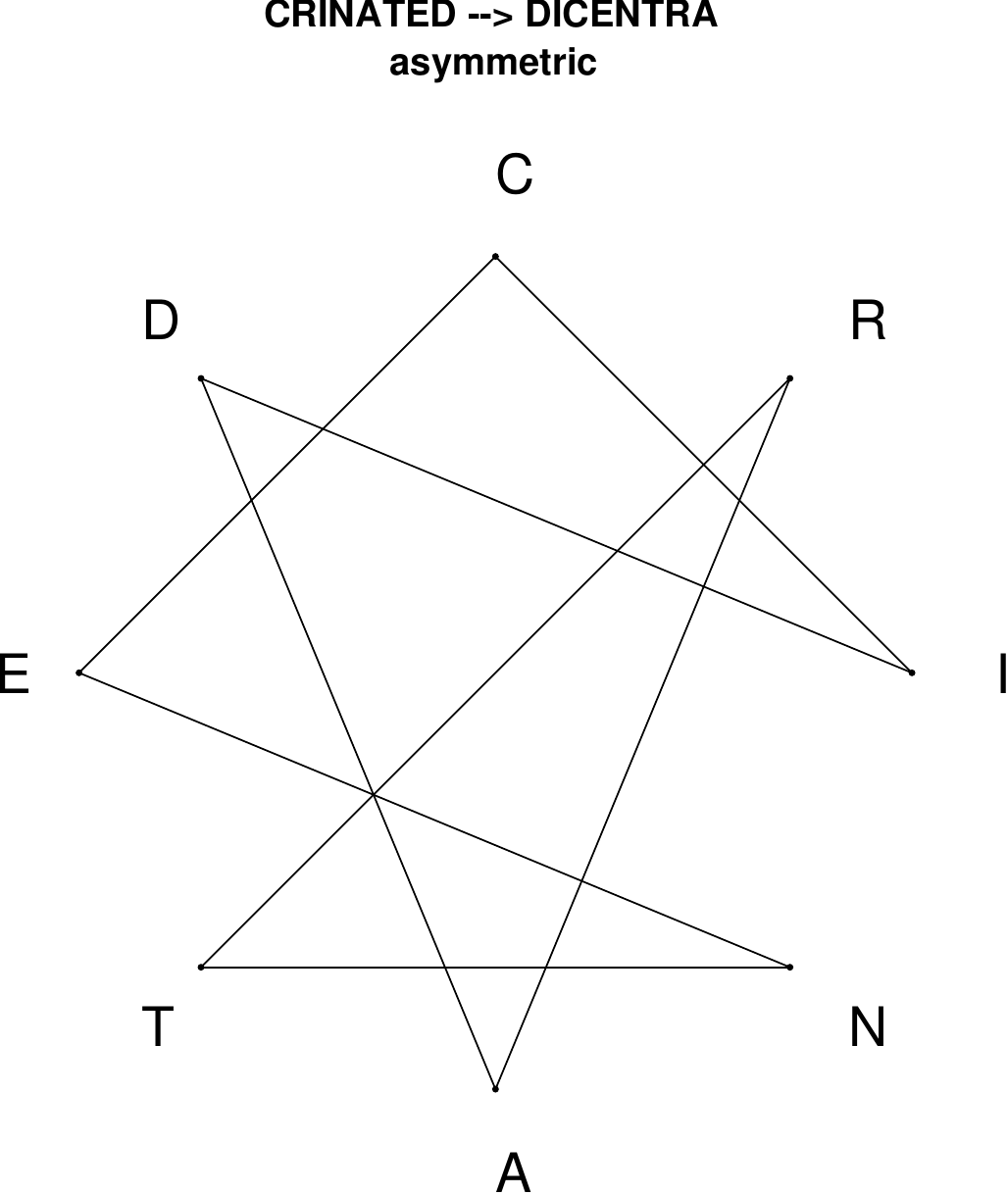}
\end{subfigure}
\hfill
\begin{subfigure}[T]{0.19\textwidth}
\centering
\includegraphics[width=\textwidth]{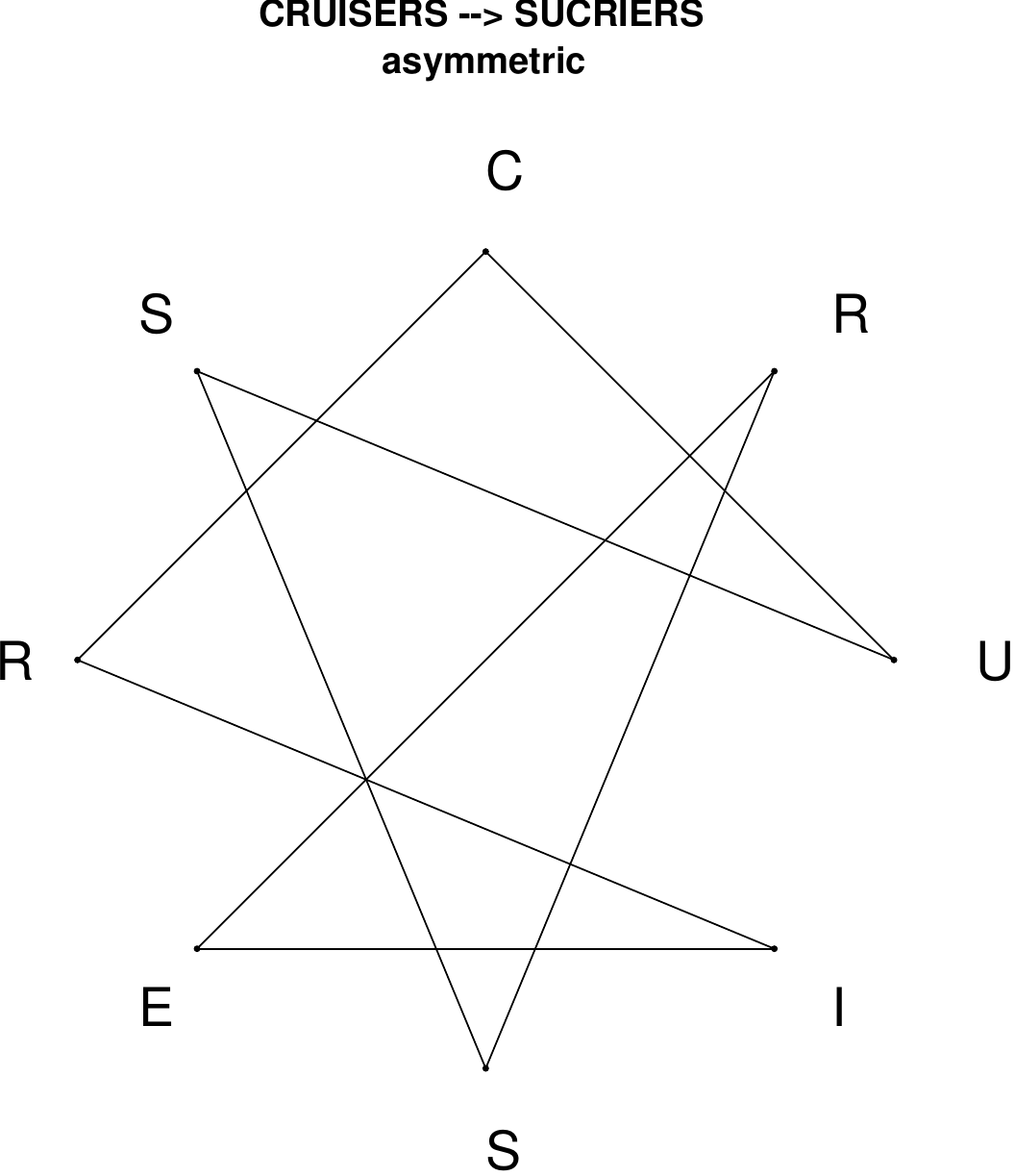}
\end{subfigure}
\hfill
\begin{subfigure}[T]{0.19\textwidth}
\centering
\includegraphics[width=\textwidth]{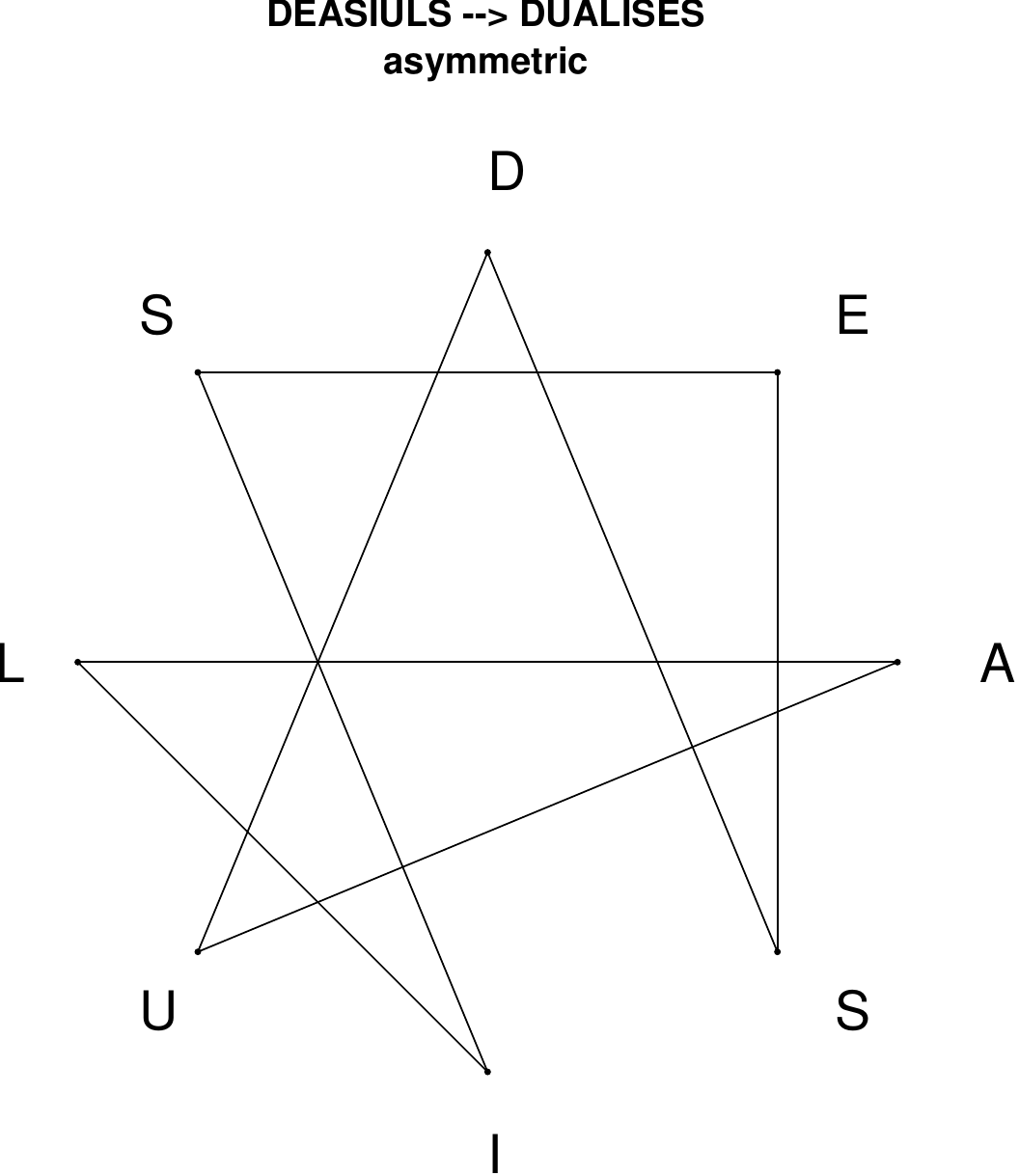}
\end{subfigure}
\end{figure}

\begin{figure}[H]
\centering
\begin{subfigure}[T]{0.19\textwidth}
\centering
\includegraphics[width=\textwidth]{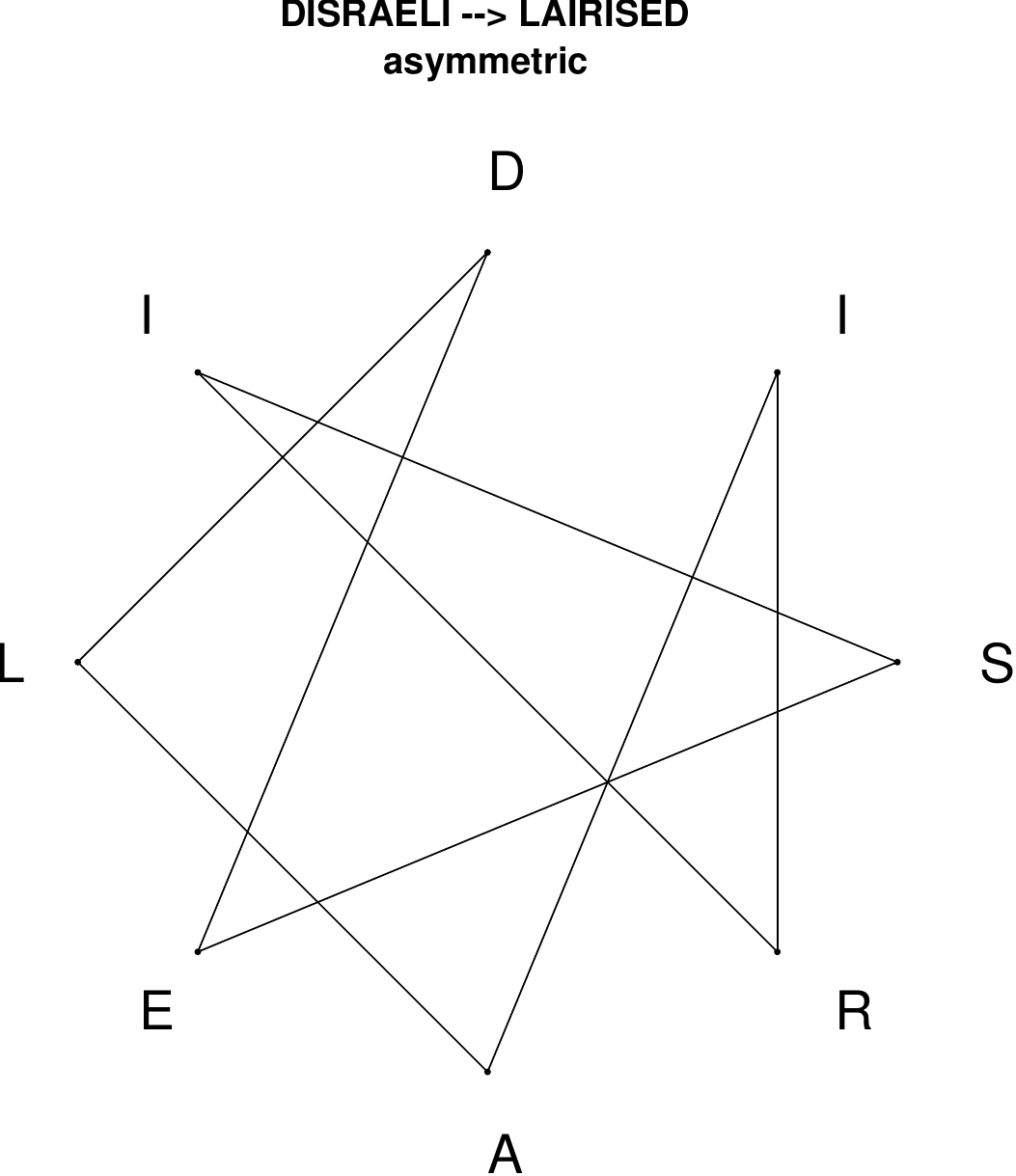}
\end{subfigure}
\hfill
\begin{subfigure}[T]{0.19\textwidth}
\centering
\includegraphics[width=\textwidth]{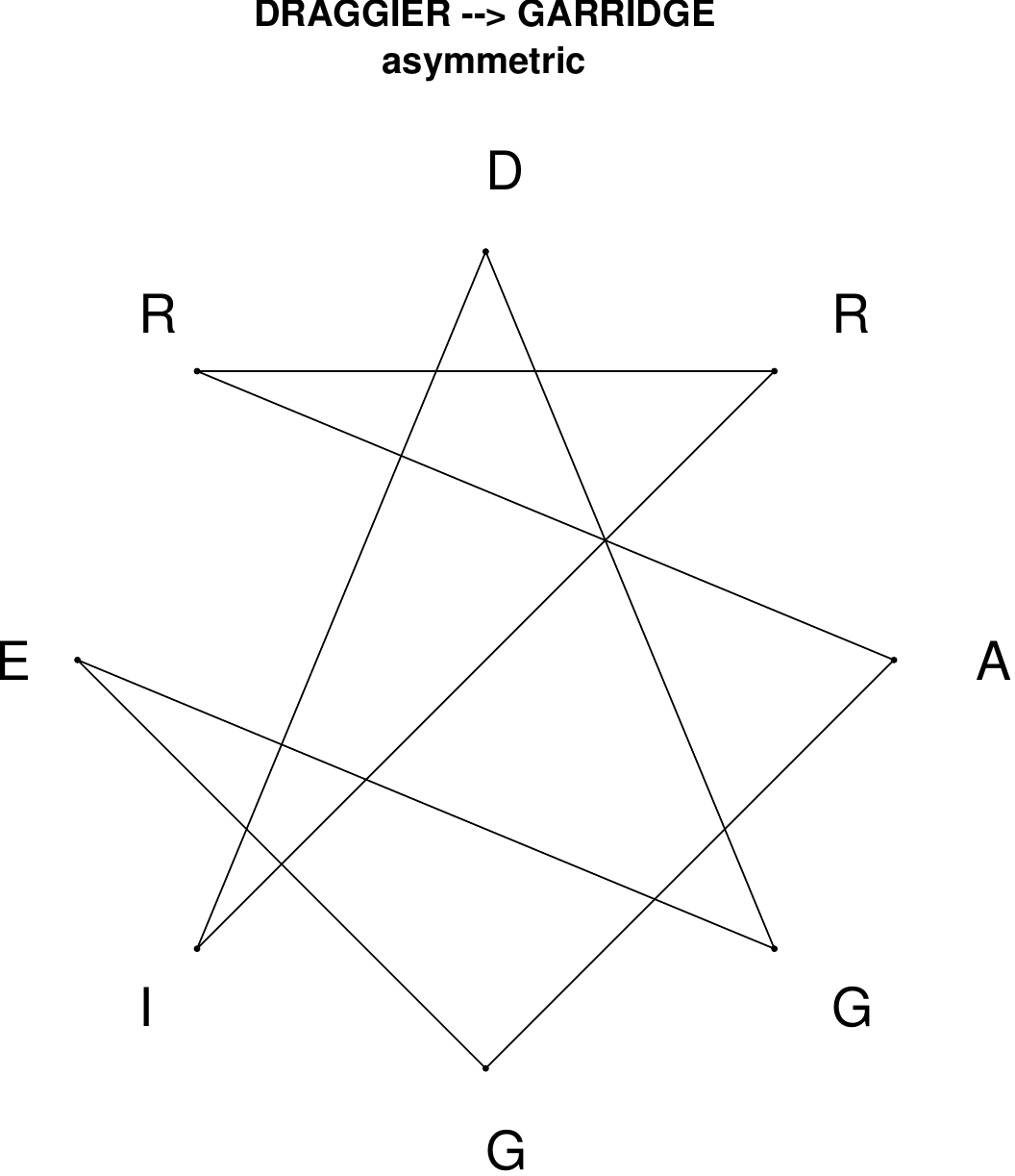}
\end{subfigure}
\hfill
\begin{subfigure}[T]{0.19\textwidth}
\centering
\includegraphics[width=\textwidth]{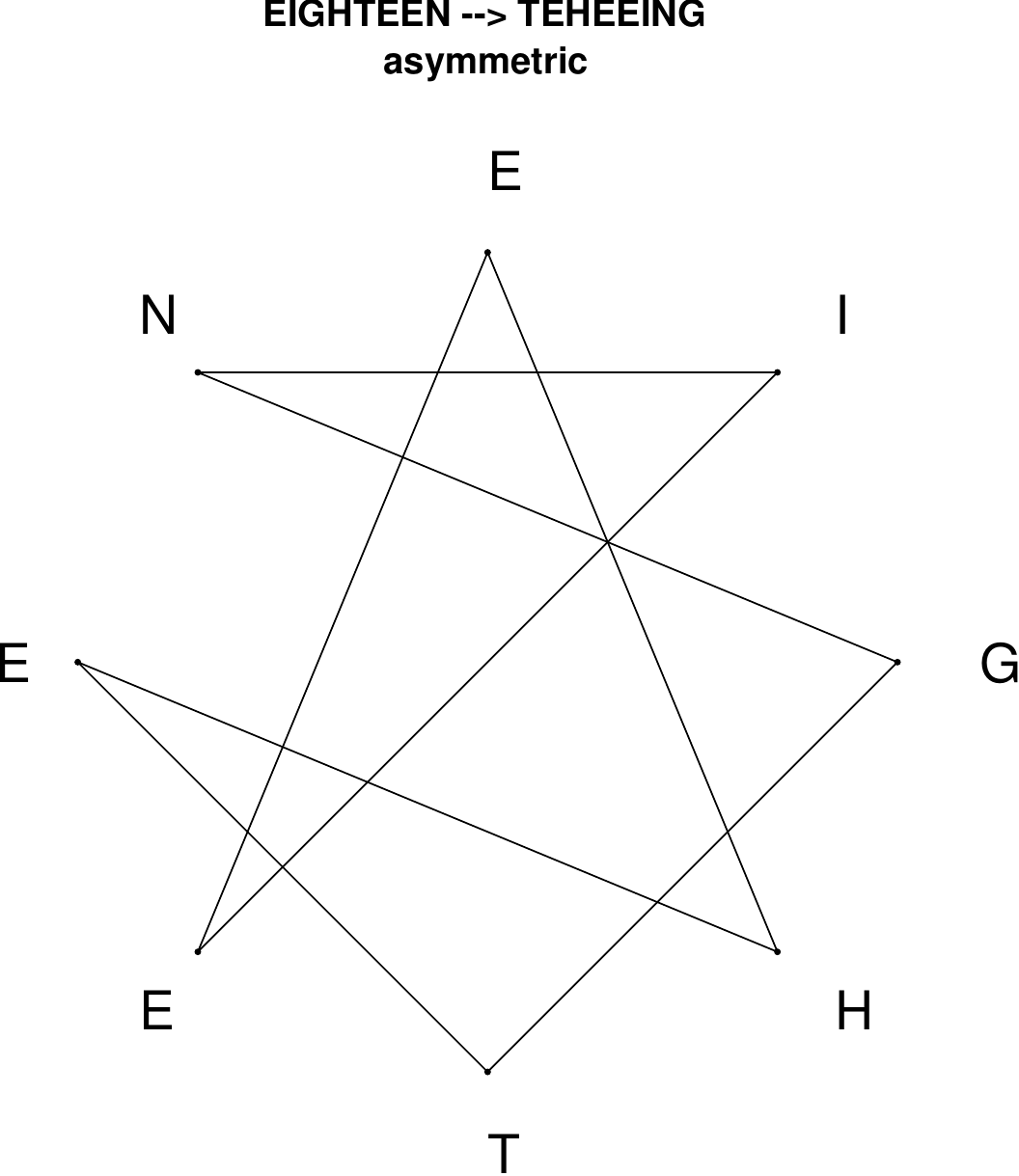}
\end{subfigure}
\hfill
\begin{subfigure}[T]{0.19\textwidth}
\centering
\includegraphics[width=\textwidth]{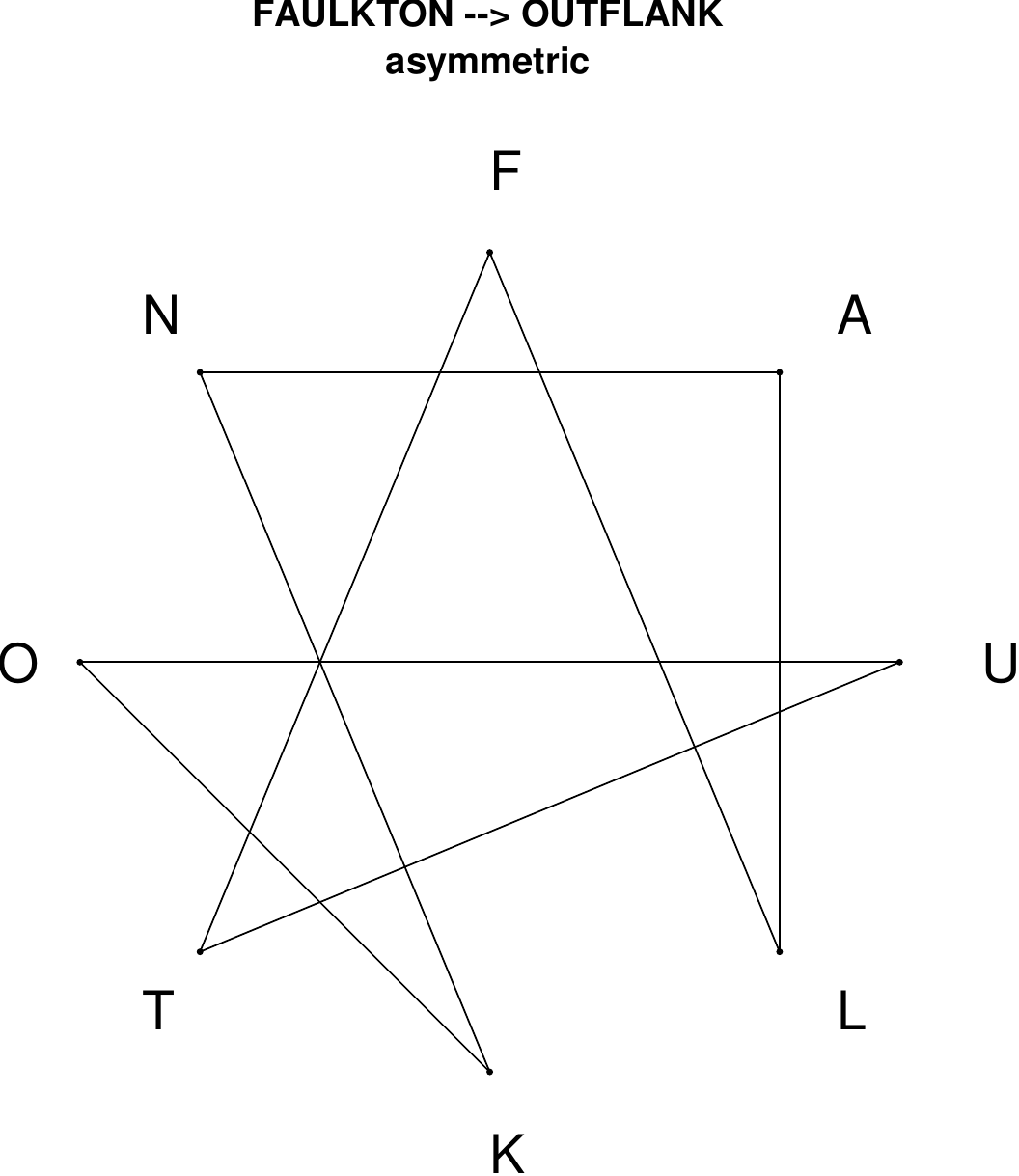}
\end{subfigure}
\hfill
\begin{subfigure}[T]{0.19\textwidth}
\centering
\includegraphics[width=\textwidth]{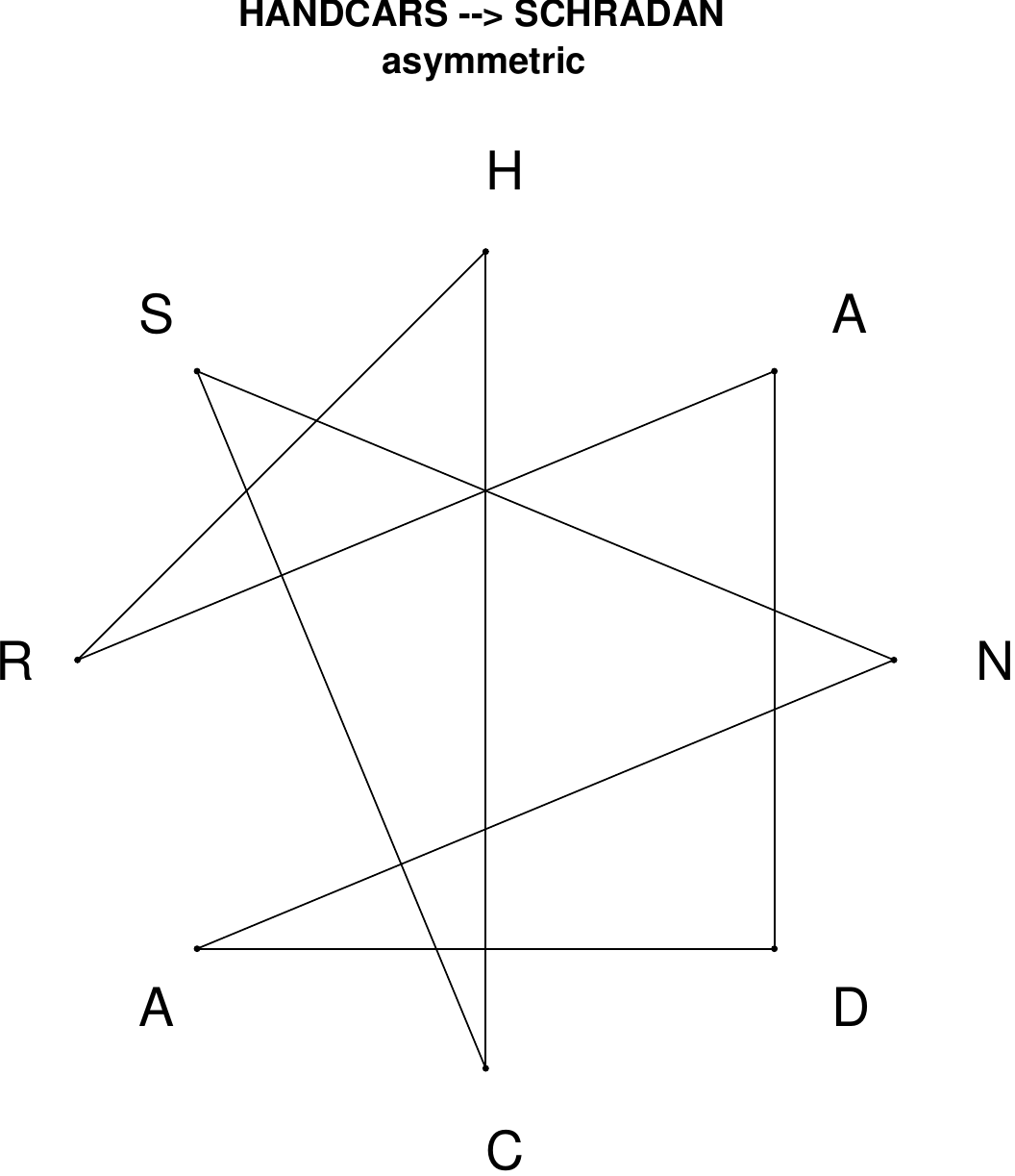}
\end{subfigure}
\end{figure}

\begin{figure}[H]
\centering
\begin{subfigure}[T]{0.19\textwidth}
\centering
\includegraphics[width=\textwidth]{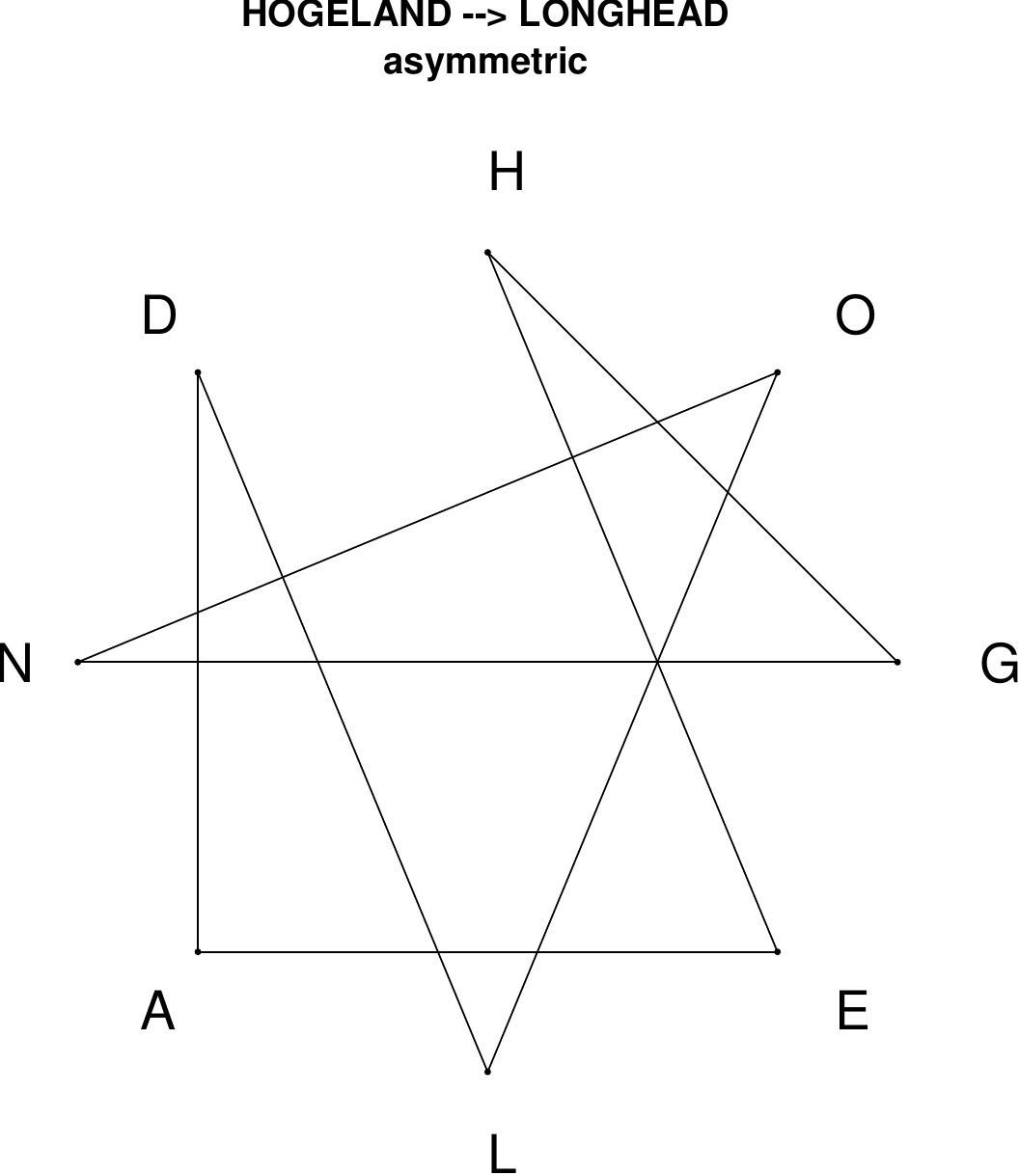}
\end{subfigure}
\hfill
\begin{subfigure}[T]{0.19\textwidth}
\centering
\includegraphics[width=\textwidth]{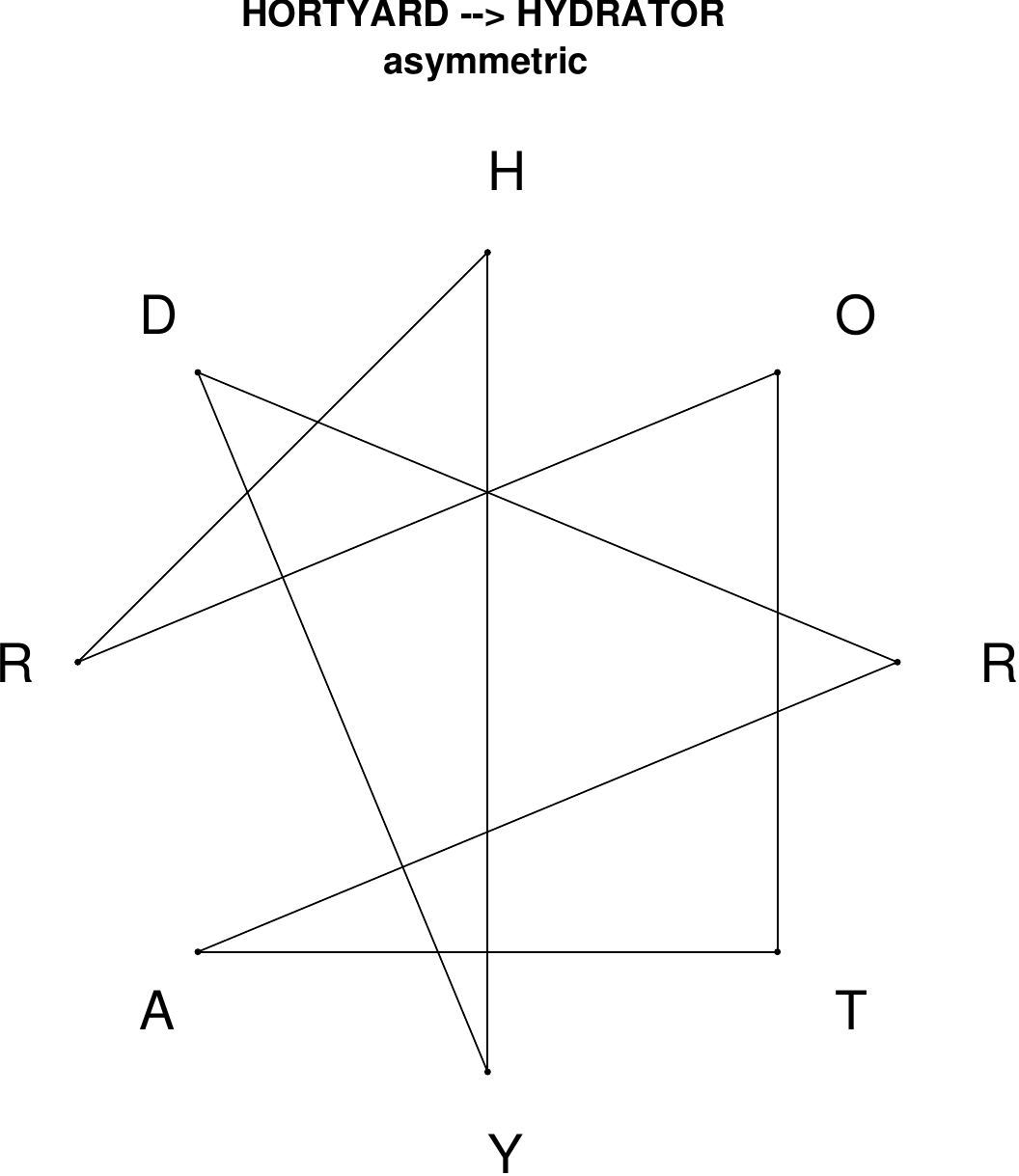}
\end{subfigure}
\hfill
\begin{subfigure}[T]{0.19\textwidth}
\centering
\includegraphics[width=\textwidth]{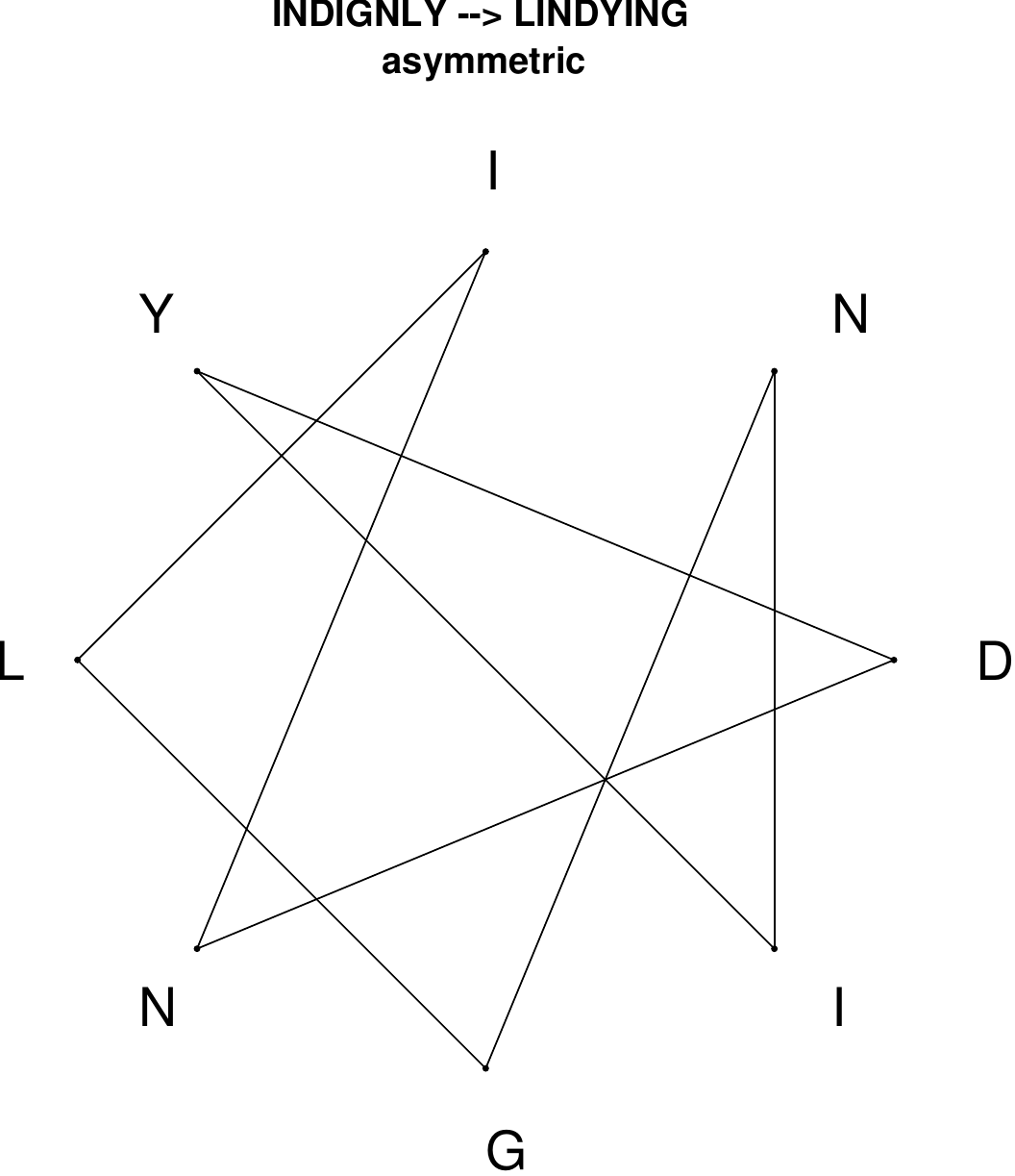}
\end{subfigure}
\hfill
\begin{subfigure}[T]{0.19\textwidth}
\centering
\includegraphics[width=\textwidth]{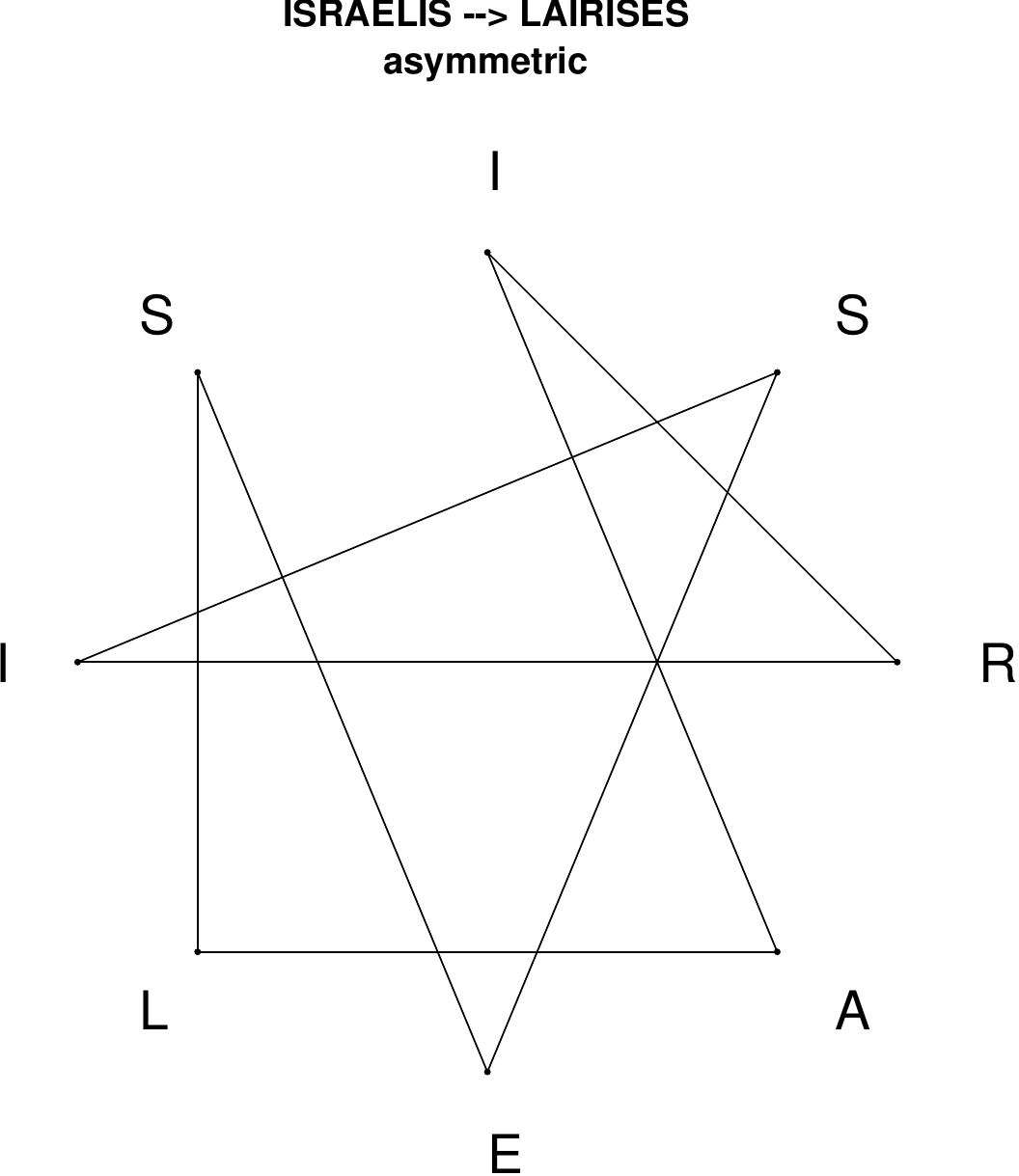}
\end{subfigure}
\hfill
\begin{subfigure}[T]{0.19\textwidth}
\centering
\includegraphics[width=\textwidth]{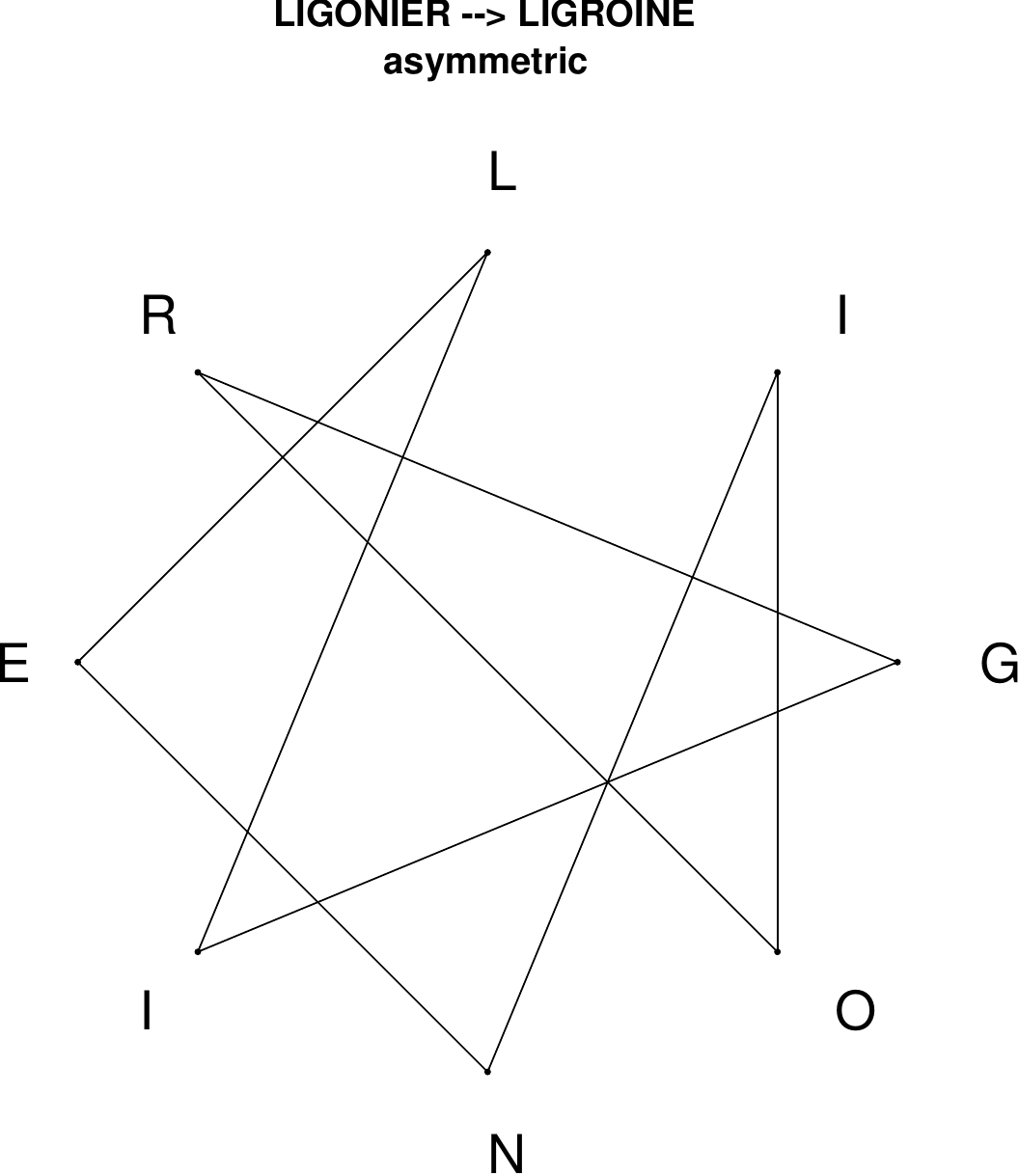}
\end{subfigure}
\end{figure}

\begin{figure}[H]
\centering
\begin{subfigure}[T]{0.19\textwidth}
\centering
\includegraphics[width=\textwidth]{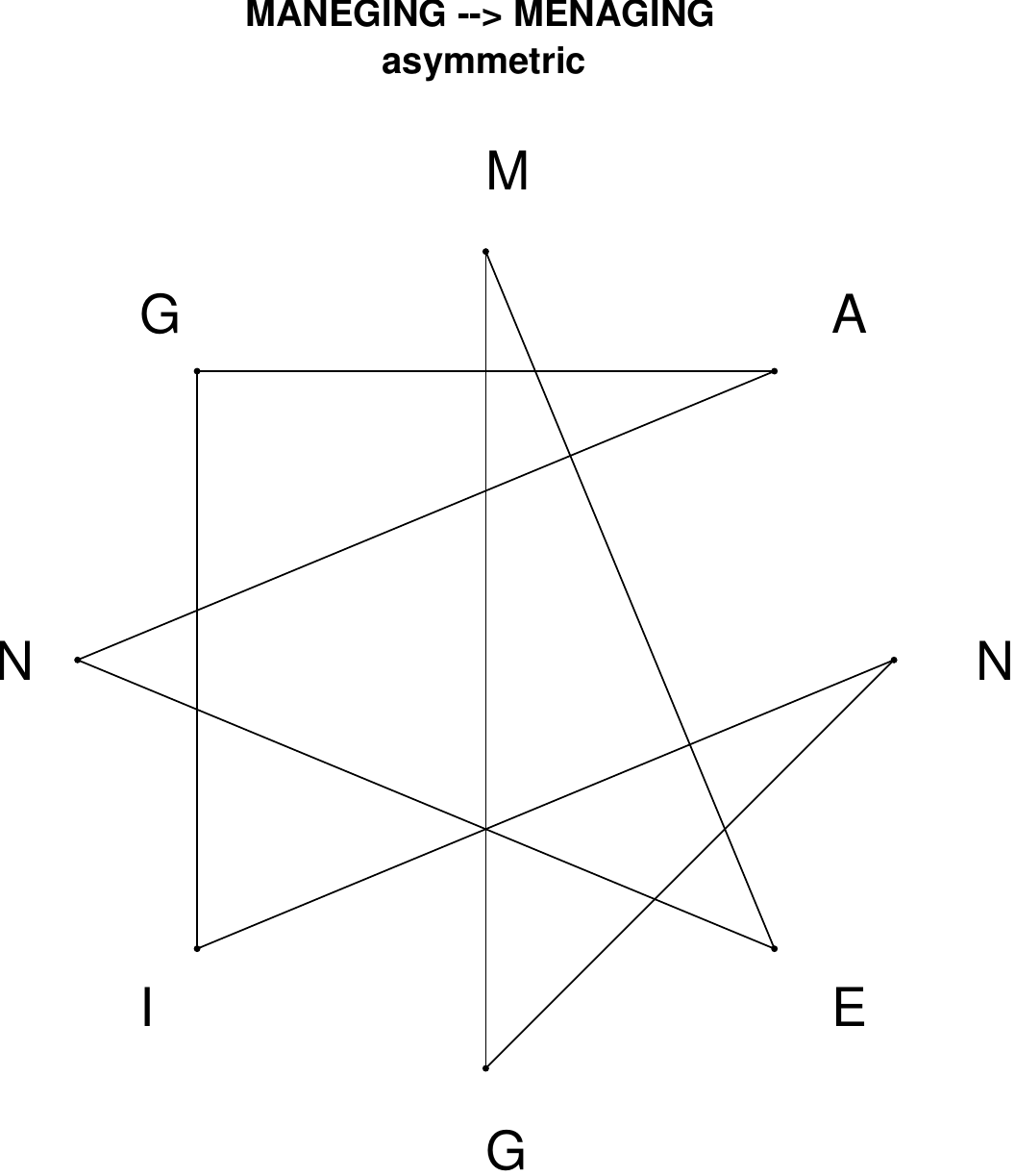}
\end{subfigure}
\hfill
\begin{subfigure}[T]{0.19\textwidth}
\centering
\includegraphics[width=\textwidth]{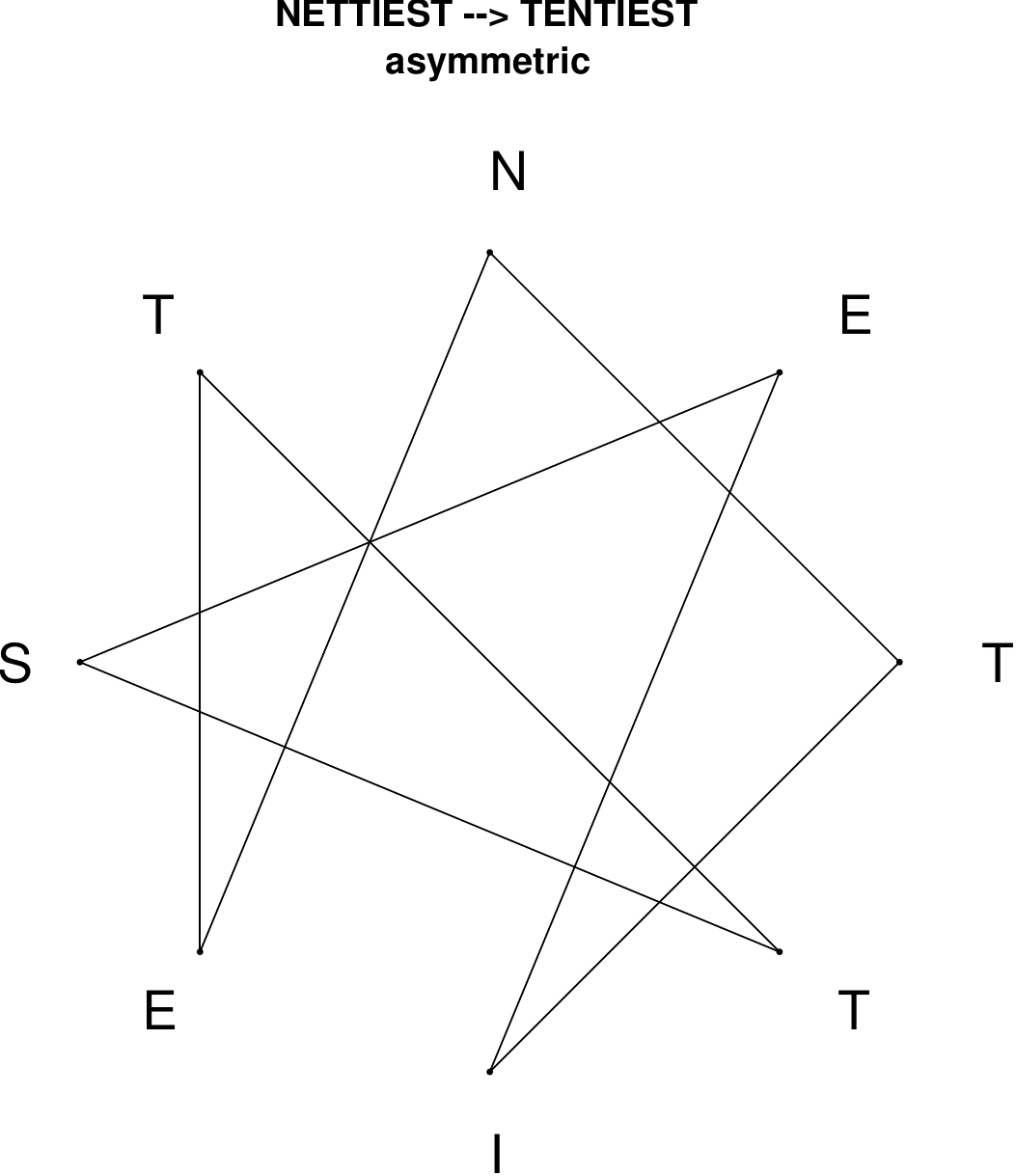}
\end{subfigure}
\hfill
\begin{subfigure}[T]{0.19\textwidth}
\centering
\includegraphics[width=\textwidth]{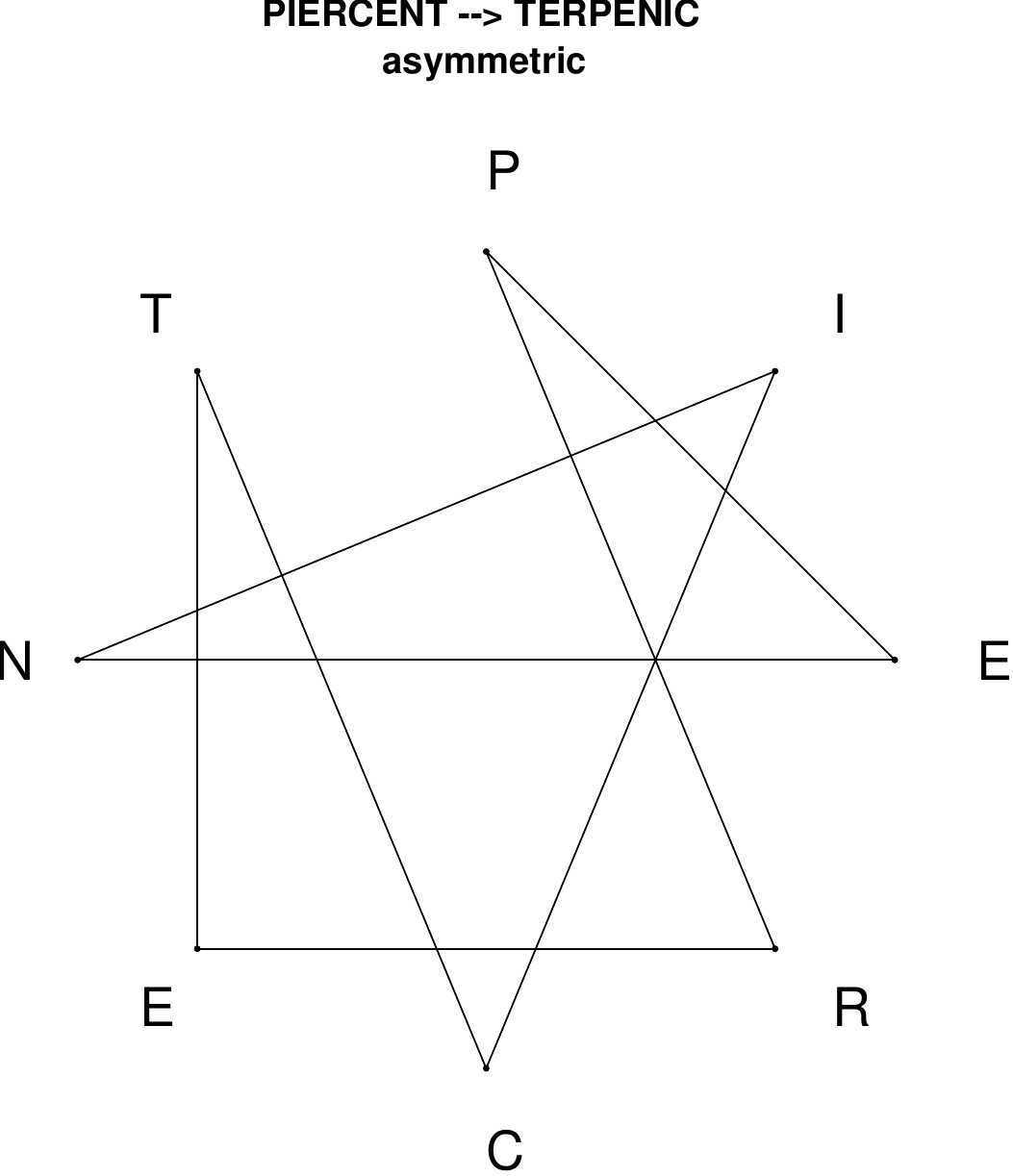}
\end{subfigure}
\hfill
\begin{subfigure}[T]{0.19\textwidth}
\centering
\includegraphics[width=\textwidth]{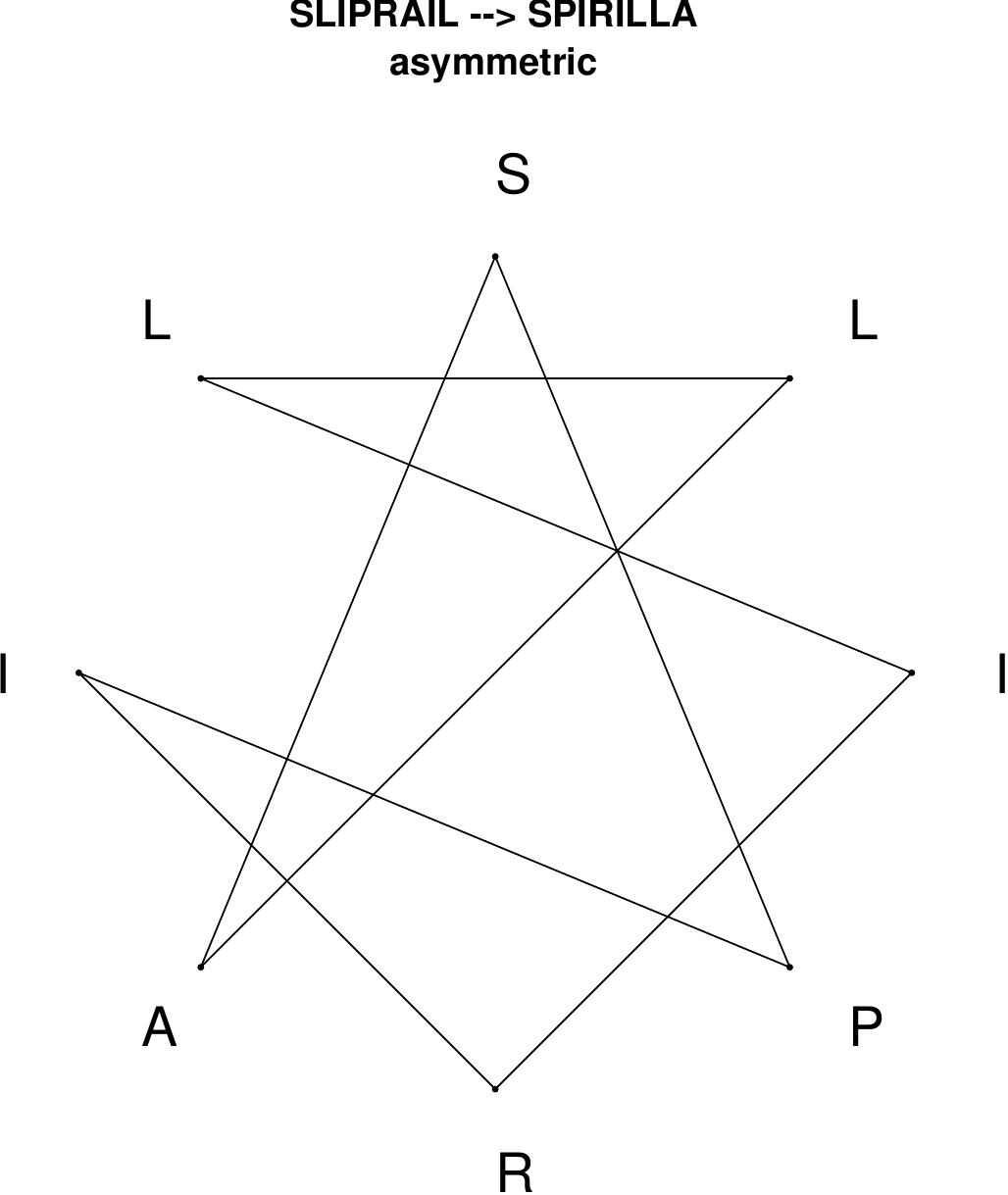}
\end{subfigure}
\hfill
\begin{subfigure}[T]{0.19\textwidth}
\centering
\includegraphics[width=\textwidth]{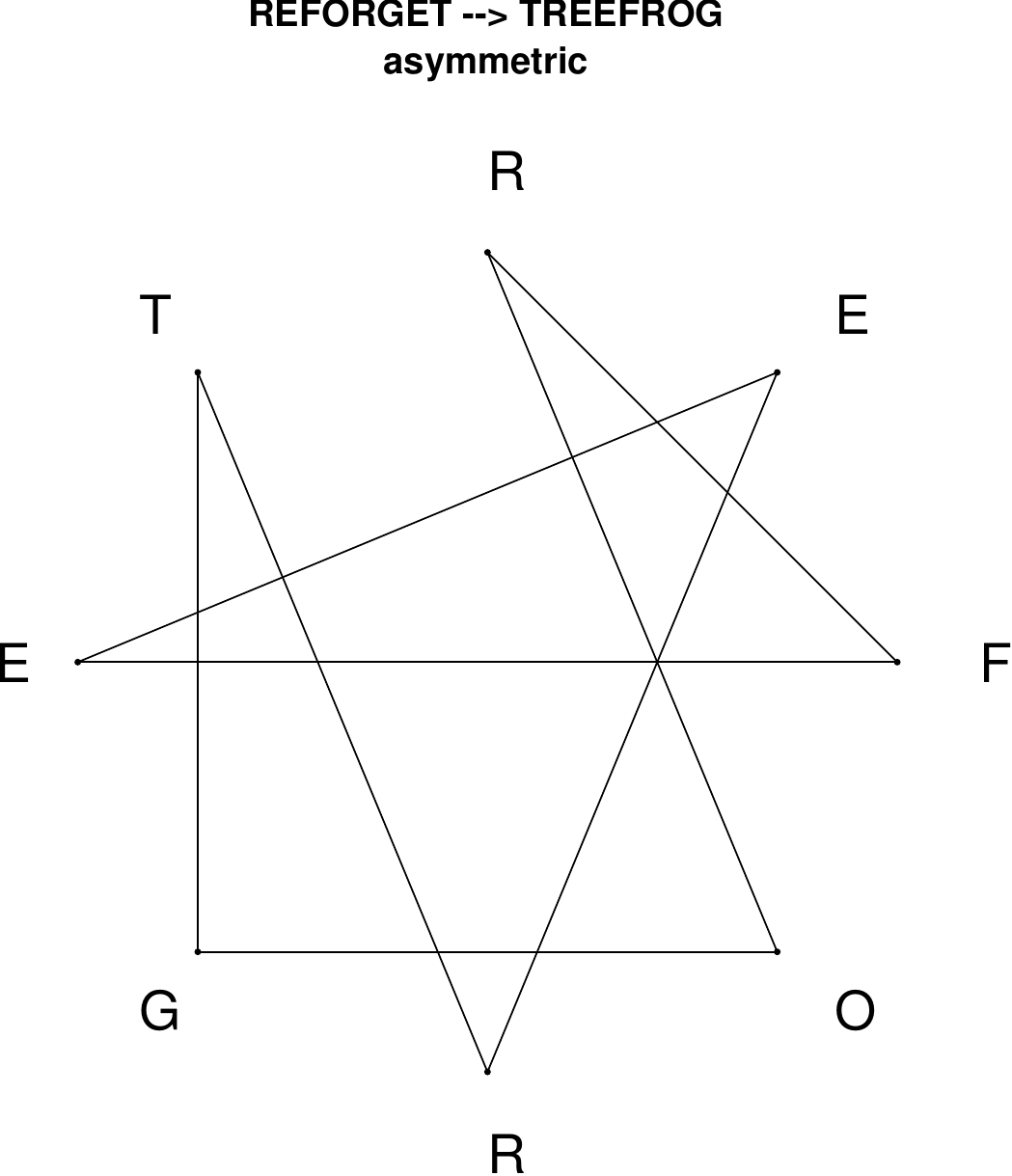}
\end{subfigure}
\end{figure}

\begin{figure}[H]
\centering
\begin{subfigure}[T]{0.19\textwidth}
\centering
\includegraphics[width=\textwidth]{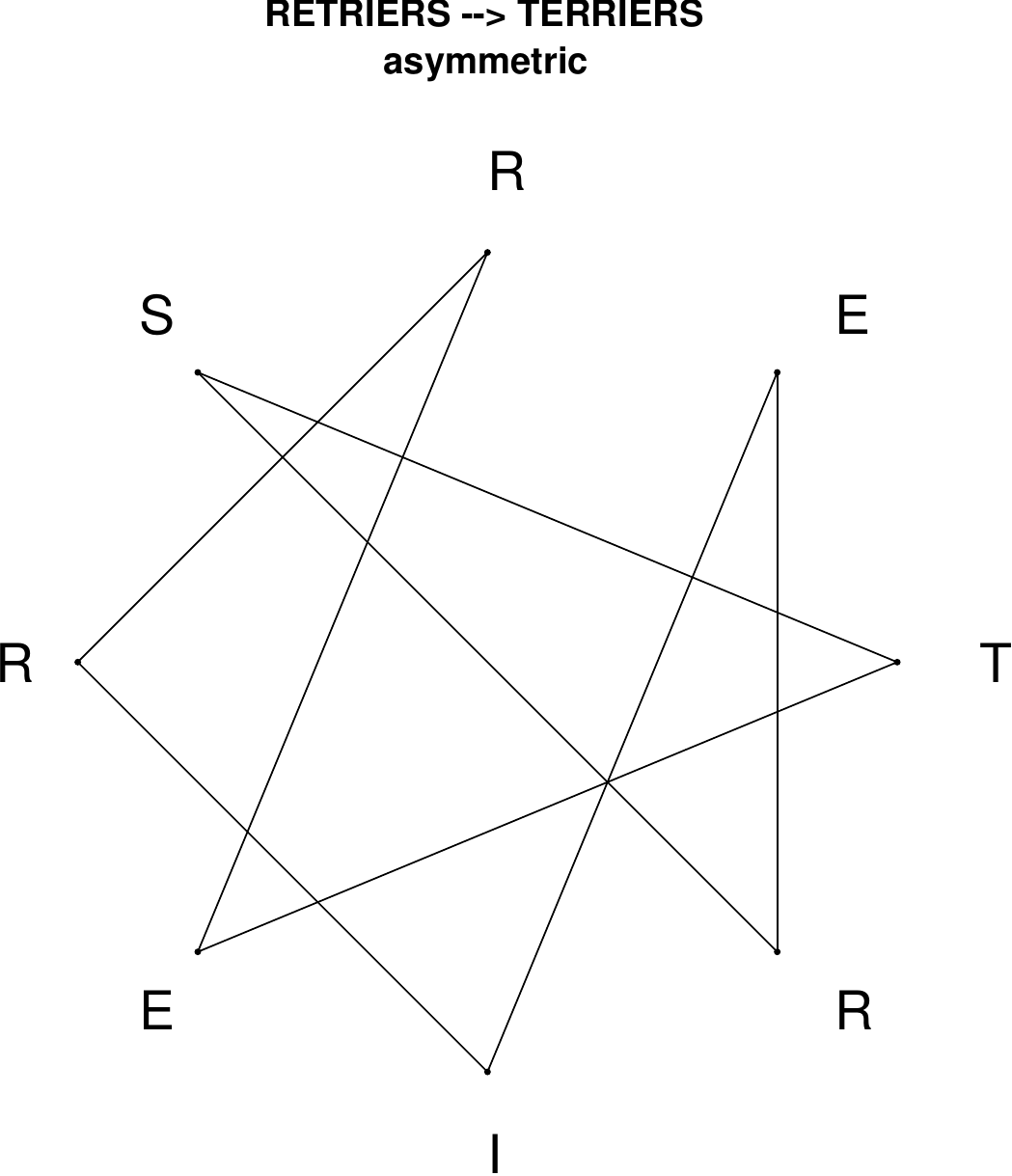}
\end{subfigure}
\hfill
\begin{subfigure}[T]{0.19\textwidth}
\centering
\includegraphics[width=\textwidth]{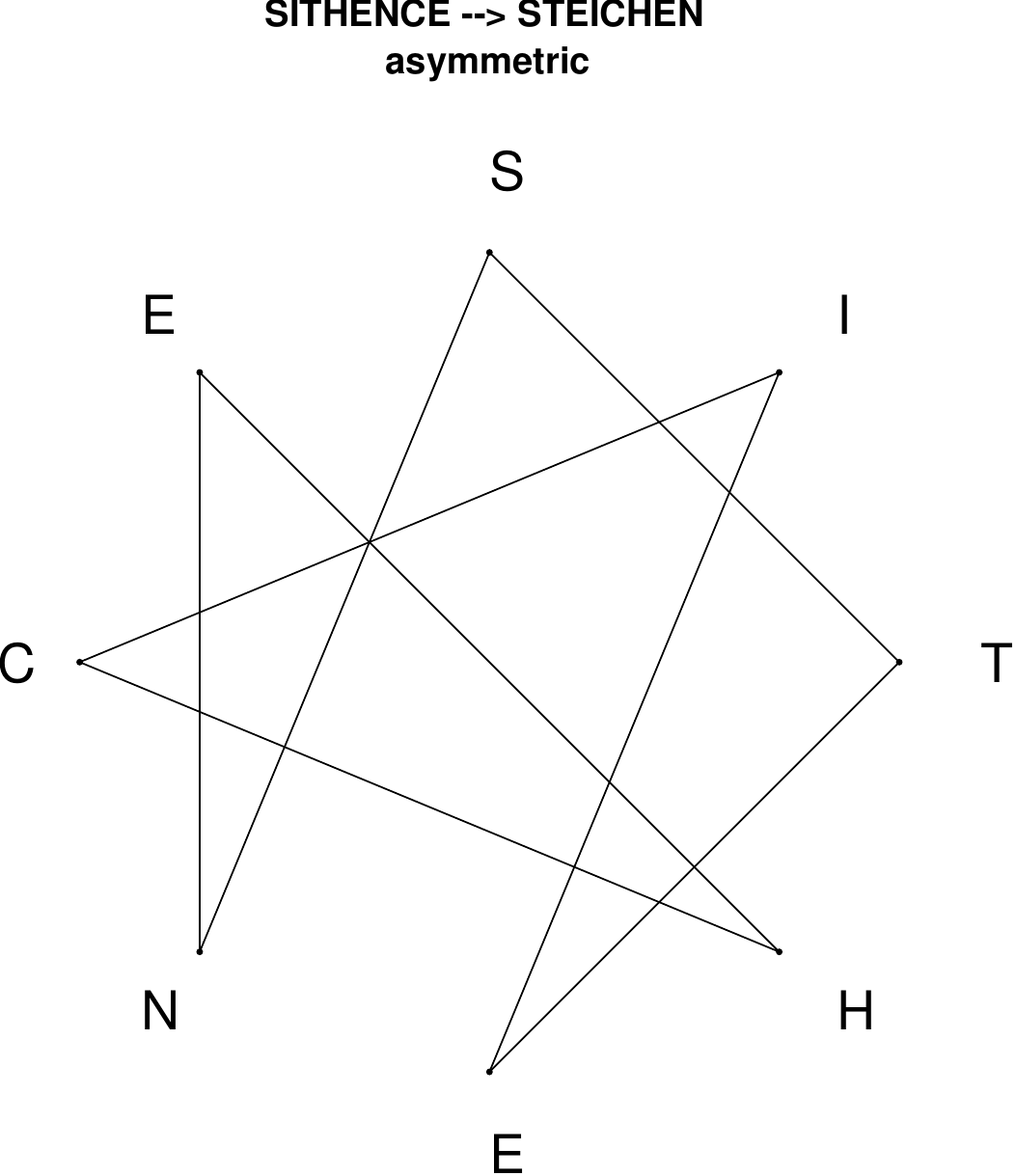}
\end{subfigure}
\hfill
\begin{subfigure}[T]{0.19\textwidth}
\centering
\includegraphics[width=\textwidth]{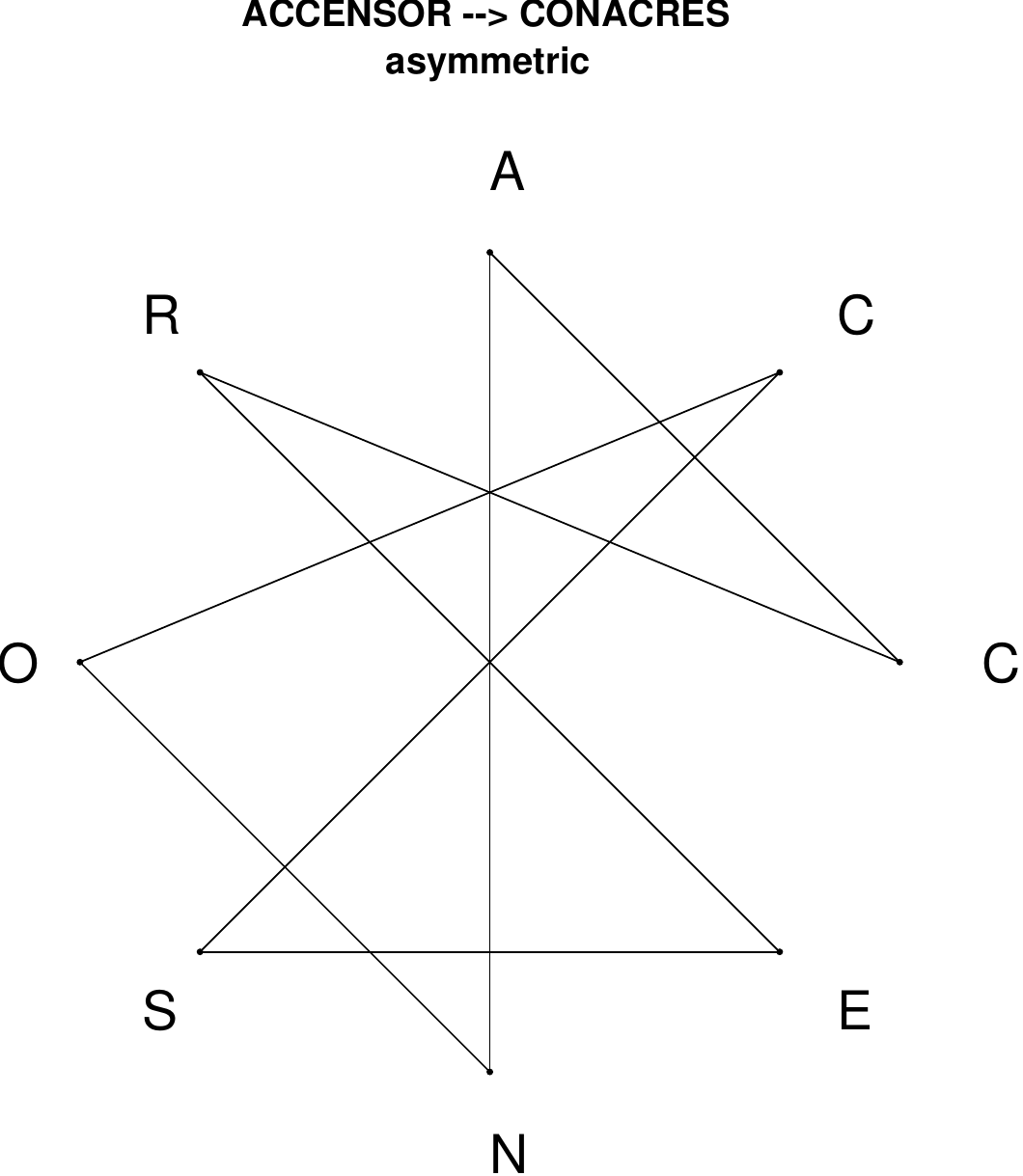}
\end{subfigure}
\hfill
\begin{subfigure}[T]{0.19\textwidth}
\centering
\includegraphics[width=\textwidth]{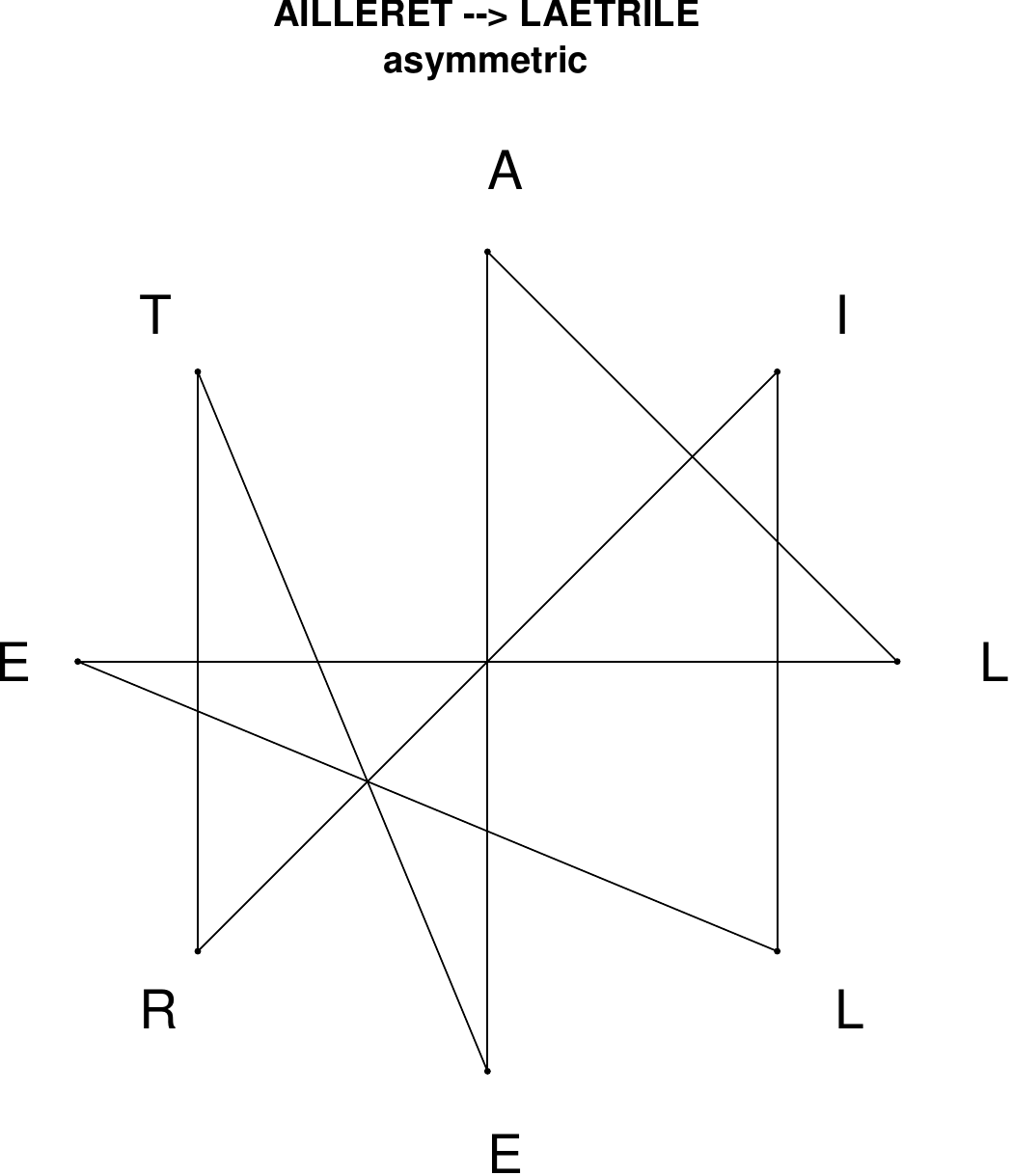}
\end{subfigure}
\hfill
\begin{subfigure}[T]{0.19\textwidth}
\centering
\includegraphics[width=\textwidth]{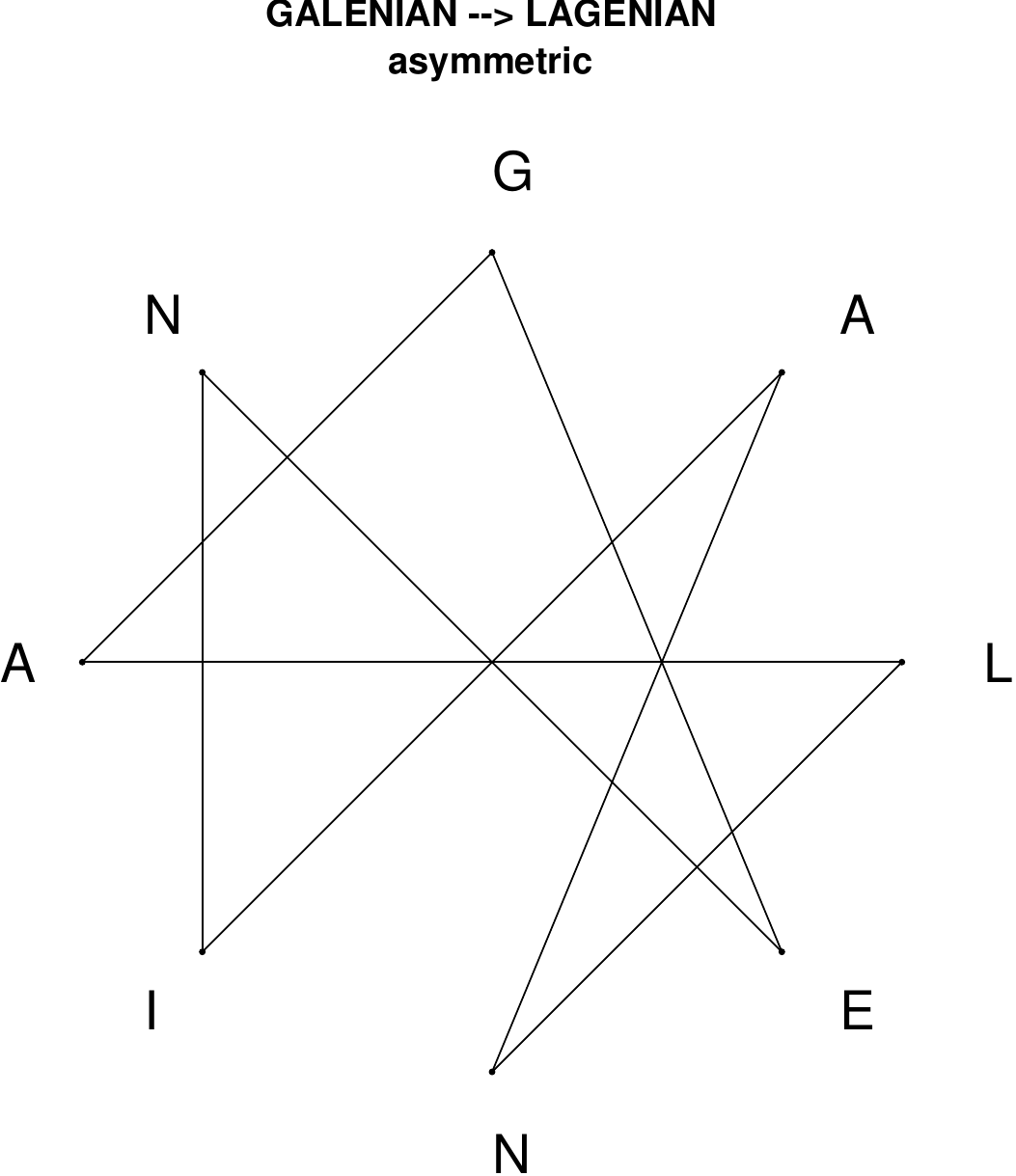}
\end{subfigure}
\end{figure}

\begin{figure}[H]
\centering
\begin{subfigure}[T]{0.19\textwidth}
\centering
\includegraphics[width=\textwidth]{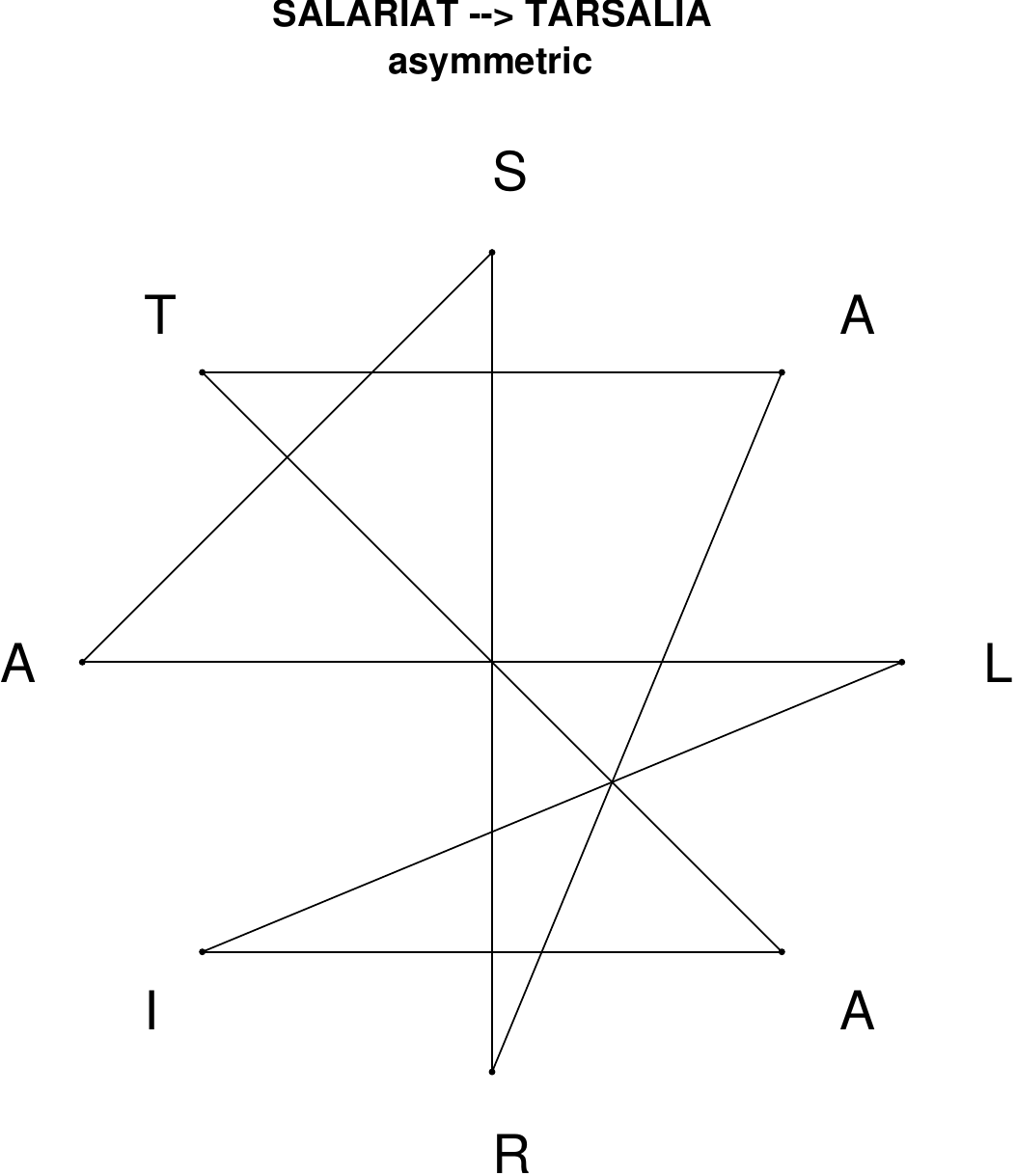}
\end{subfigure}
\hfill
\begin{subfigure}[T]{0.19\textwidth}
\centering
\includegraphics[width=\textwidth]{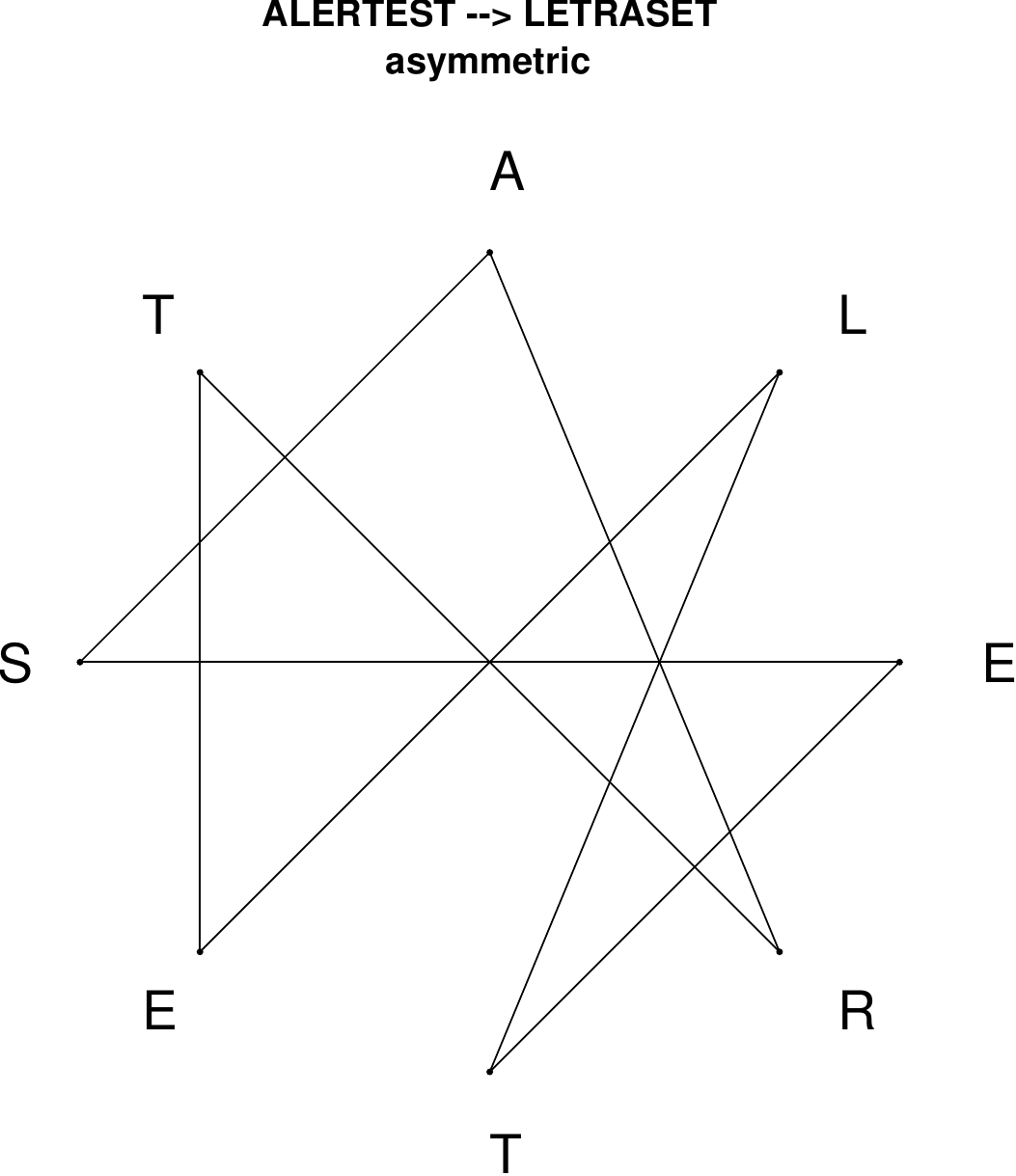}
\end{subfigure}
\hfill
\begin{subfigure}[T]{0.19\textwidth}
\centering
\includegraphics[width=\textwidth]{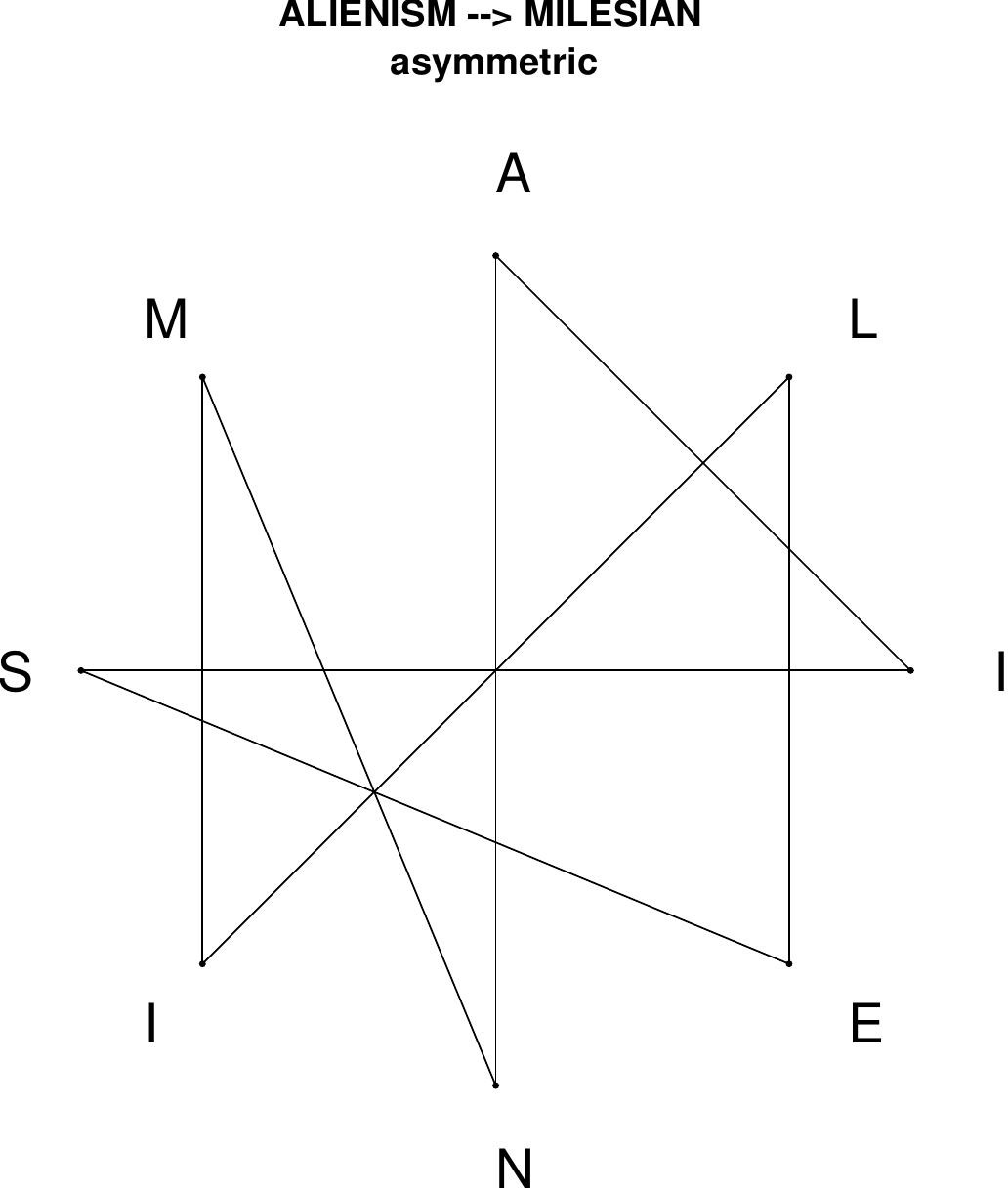}
\end{subfigure}
\hfill
\begin{subfigure}[T]{0.19\textwidth}
\centering
\includegraphics[width=\textwidth]{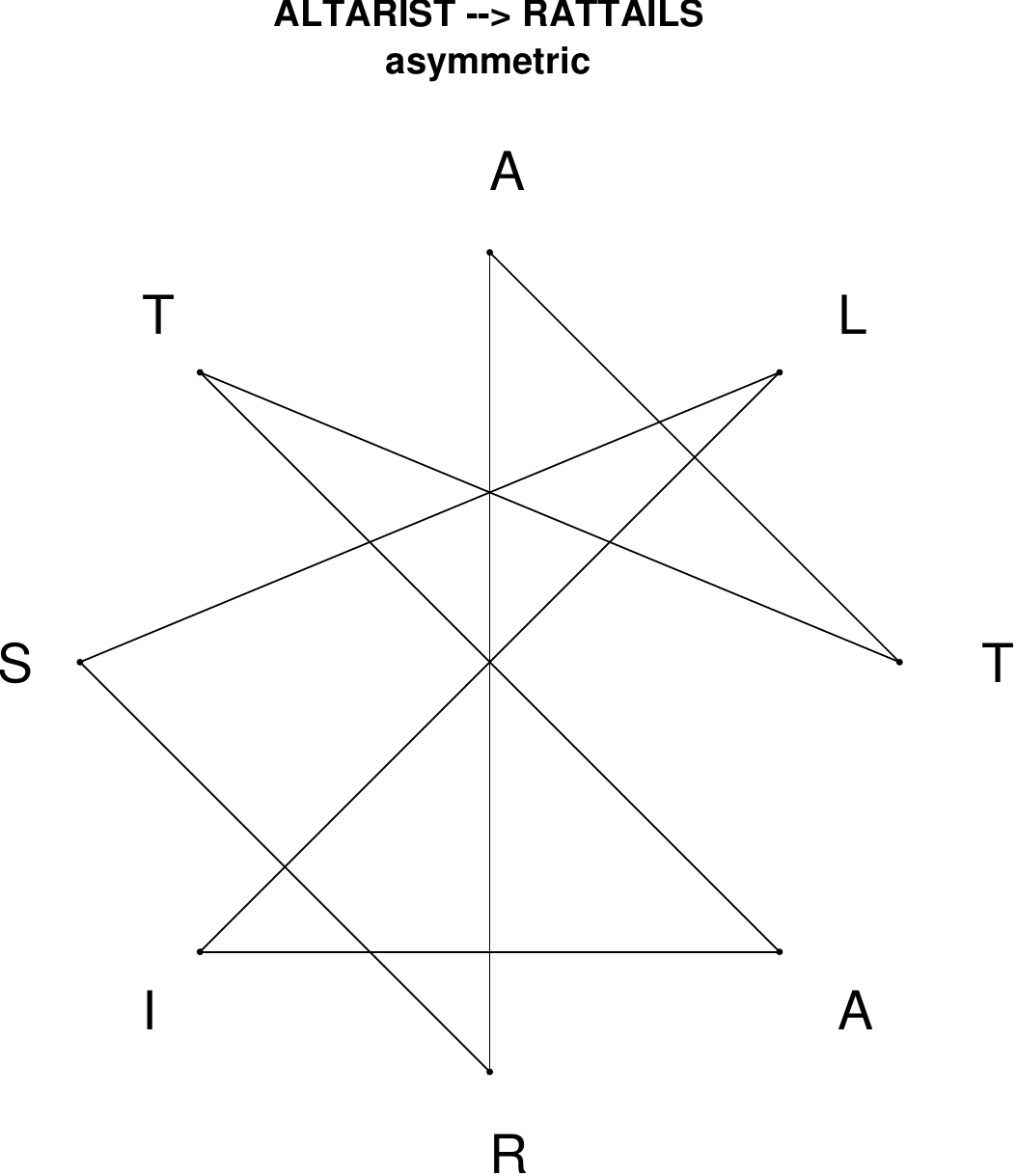}
\end{subfigure}
\hfill
\begin{subfigure}[T]{0.19\textwidth}
\centering
\includegraphics[width=\textwidth]{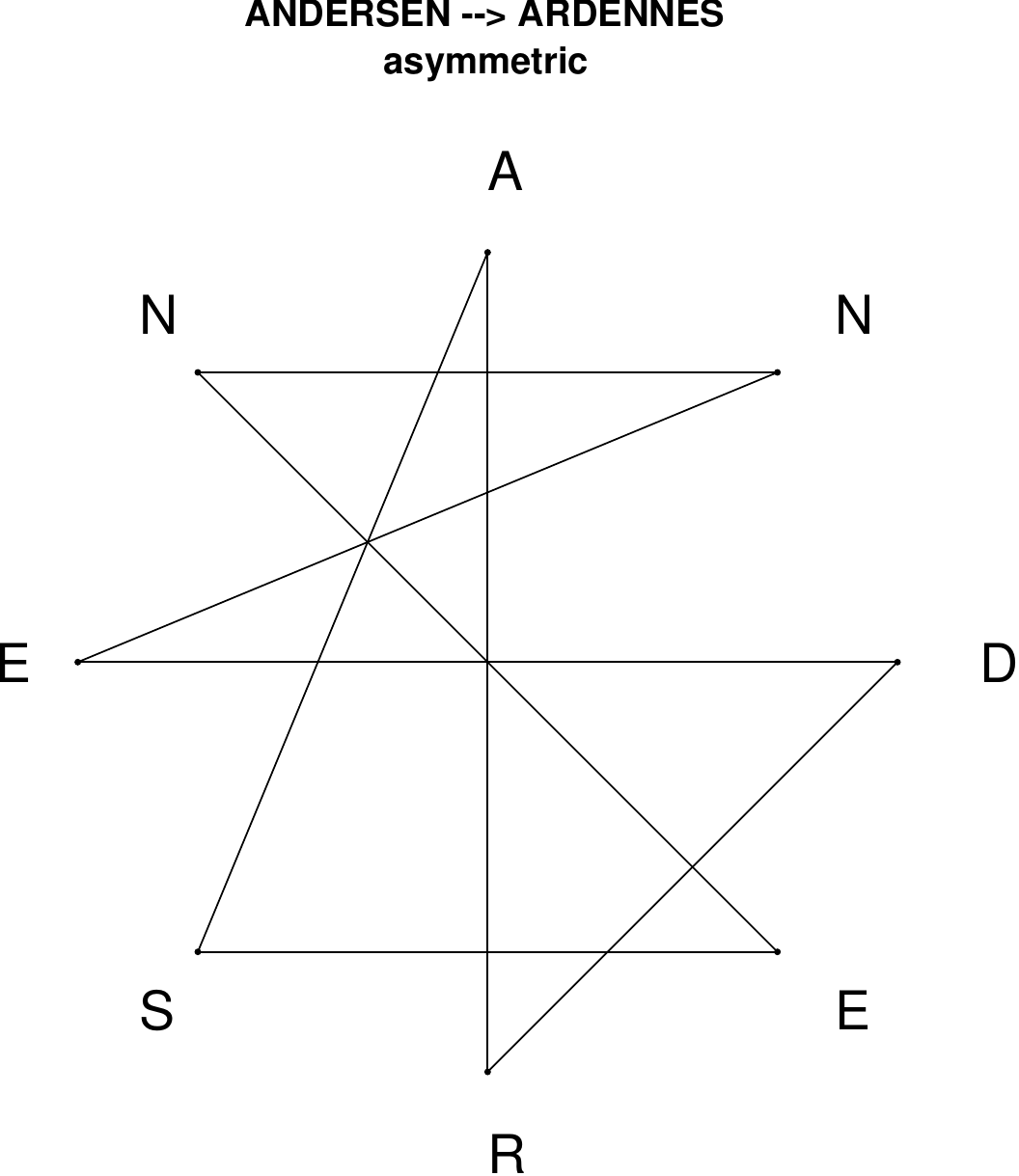}
\end{subfigure}
\end{figure}

\begin{figure}[H]
\centering
\begin{subfigure}[T]{0.19\textwidth}
\centering
\includegraphics[width=\textwidth]{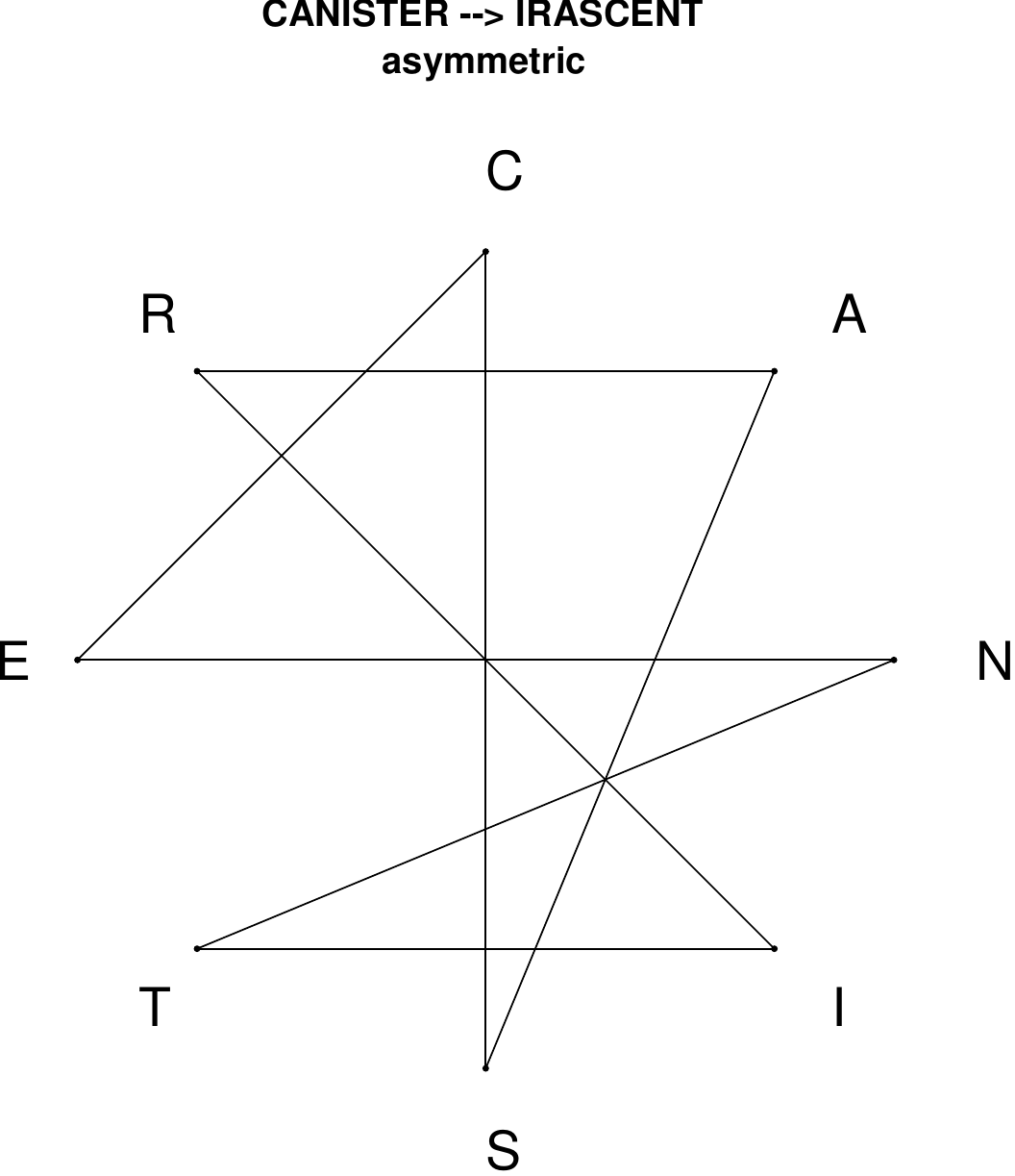}
\end{subfigure}
\hfill
\begin{subfigure}[T]{0.19\textwidth}
\centering
\includegraphics[width=\textwidth]{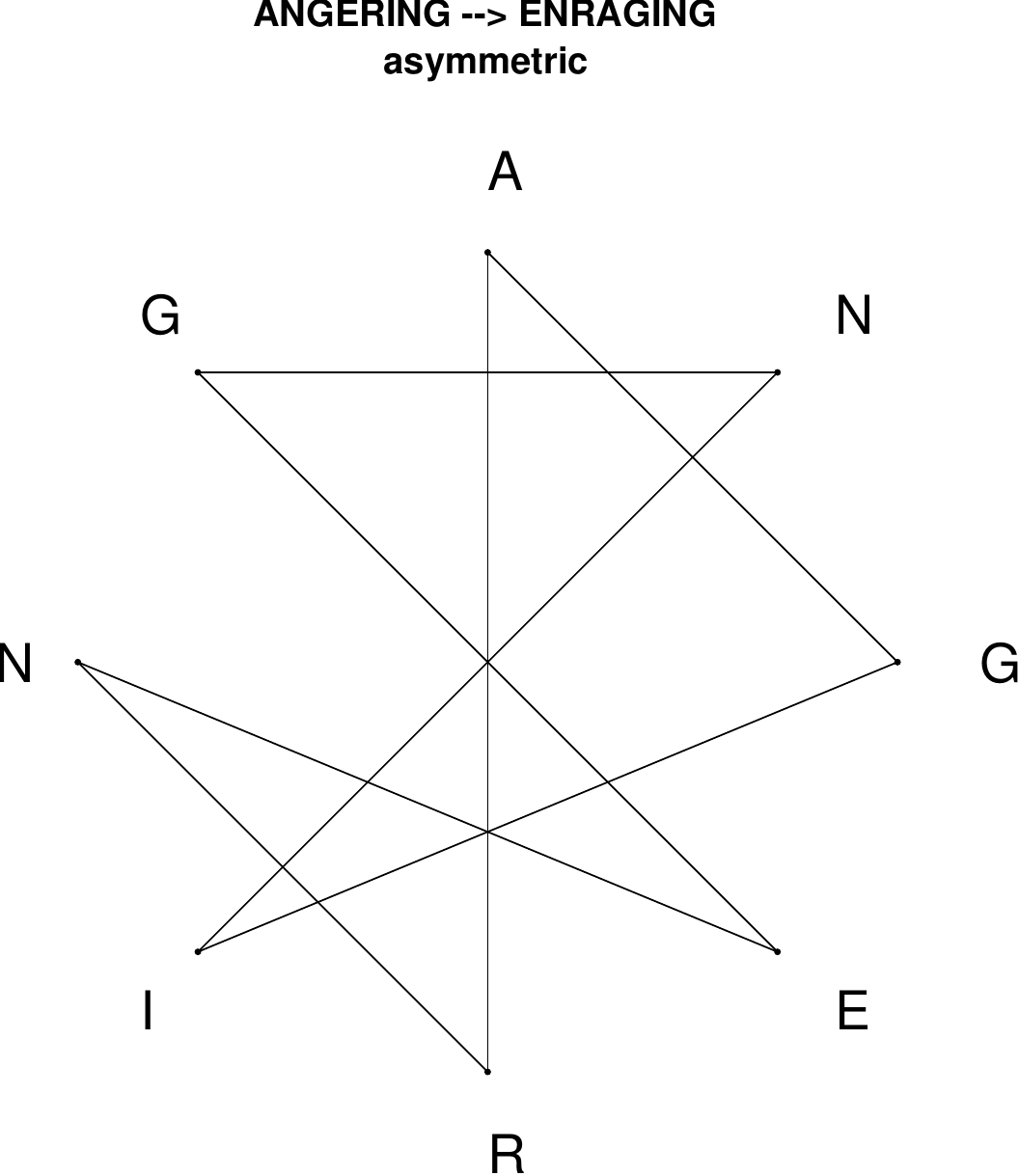}
\end{subfigure}
\hfill
\begin{subfigure}[T]{0.19\textwidth}
\centering
\includegraphics[width=\textwidth]{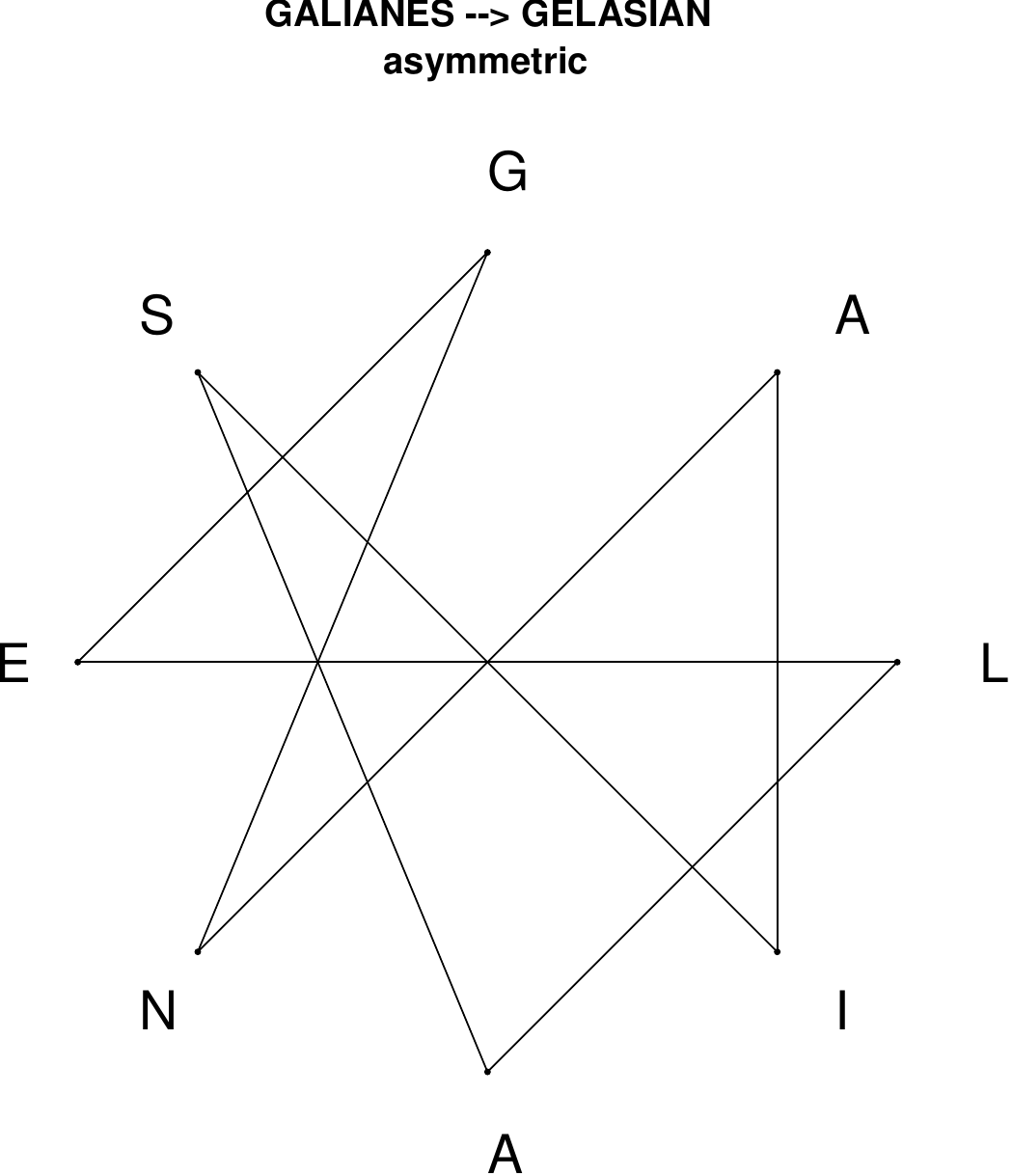}
\end{subfigure}
\hfill
\begin{subfigure}[T]{0.19\textwidth}
\centering
\includegraphics[width=\textwidth]{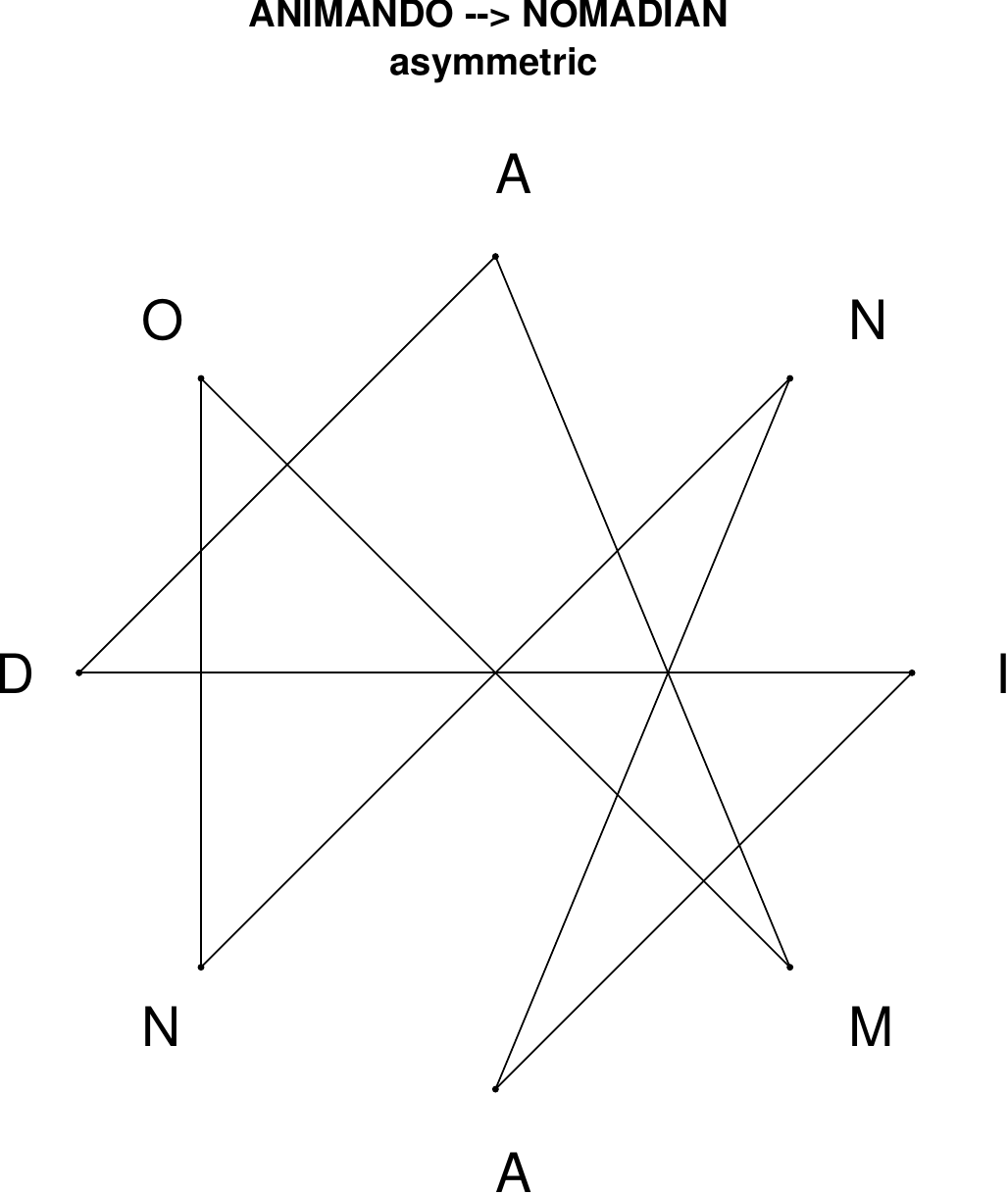}
\end{subfigure}
\hfill
\begin{subfigure}[T]{0.19\textwidth}
\centering
\includegraphics[width=\textwidth]{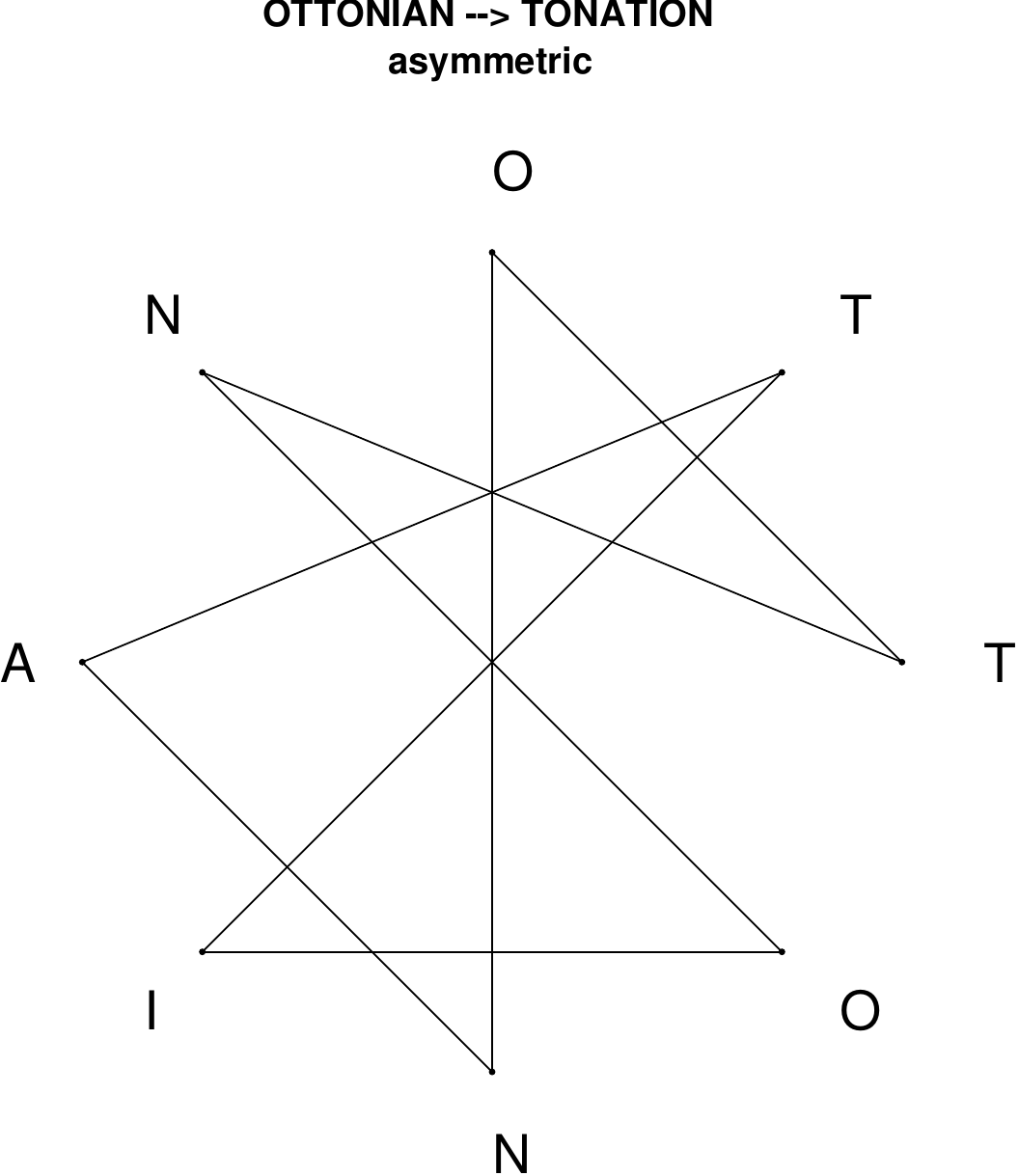}
\end{subfigure}
\end{figure}

\begin{figure}[H]
\centering
\begin{subfigure}[T]{0.19\textwidth}
\centering
\includegraphics[width=\textwidth]{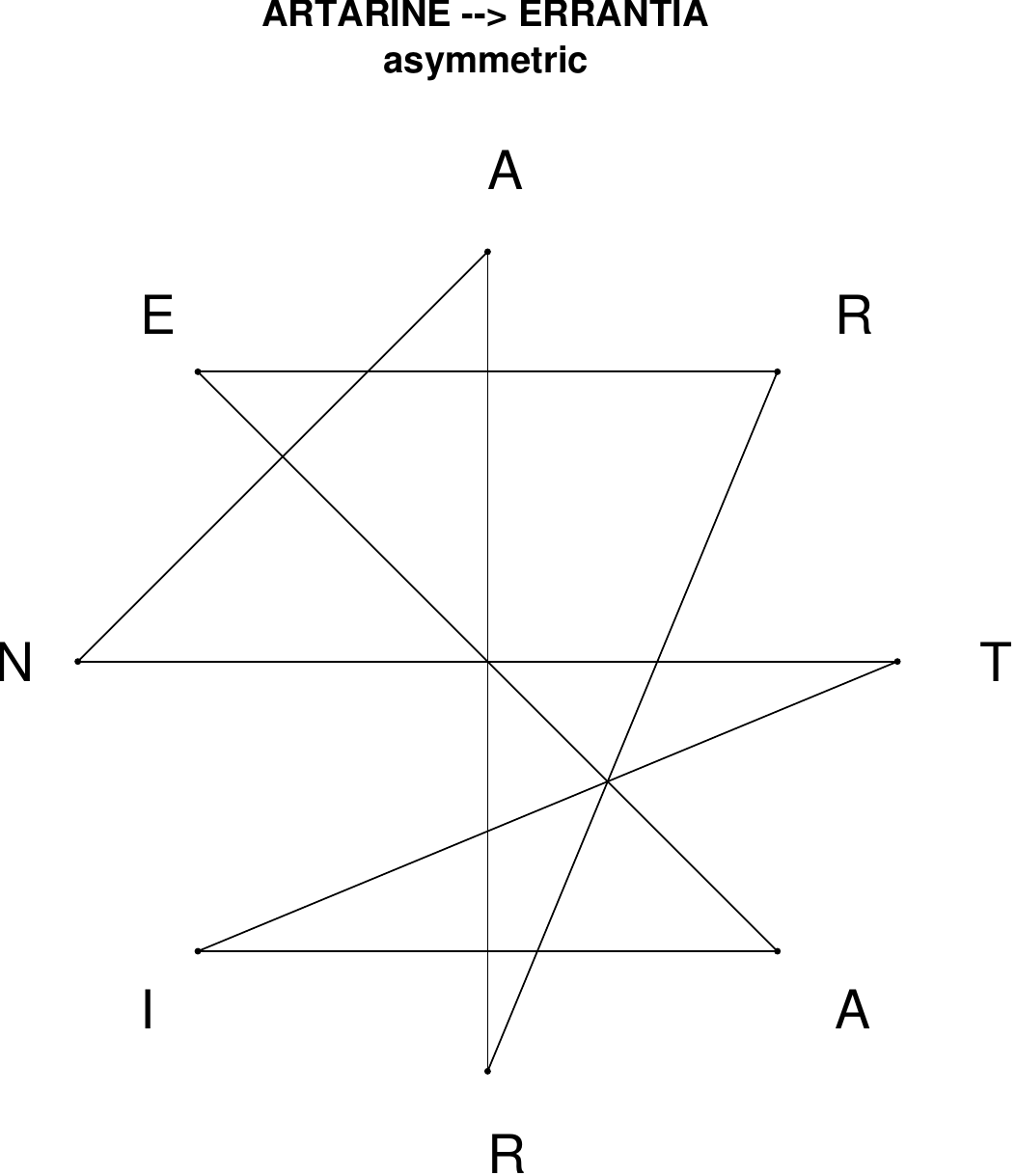}
\end{subfigure}
\hfill
\begin{subfigure}[T]{0.19\textwidth}
\centering
\includegraphics[width=\textwidth]{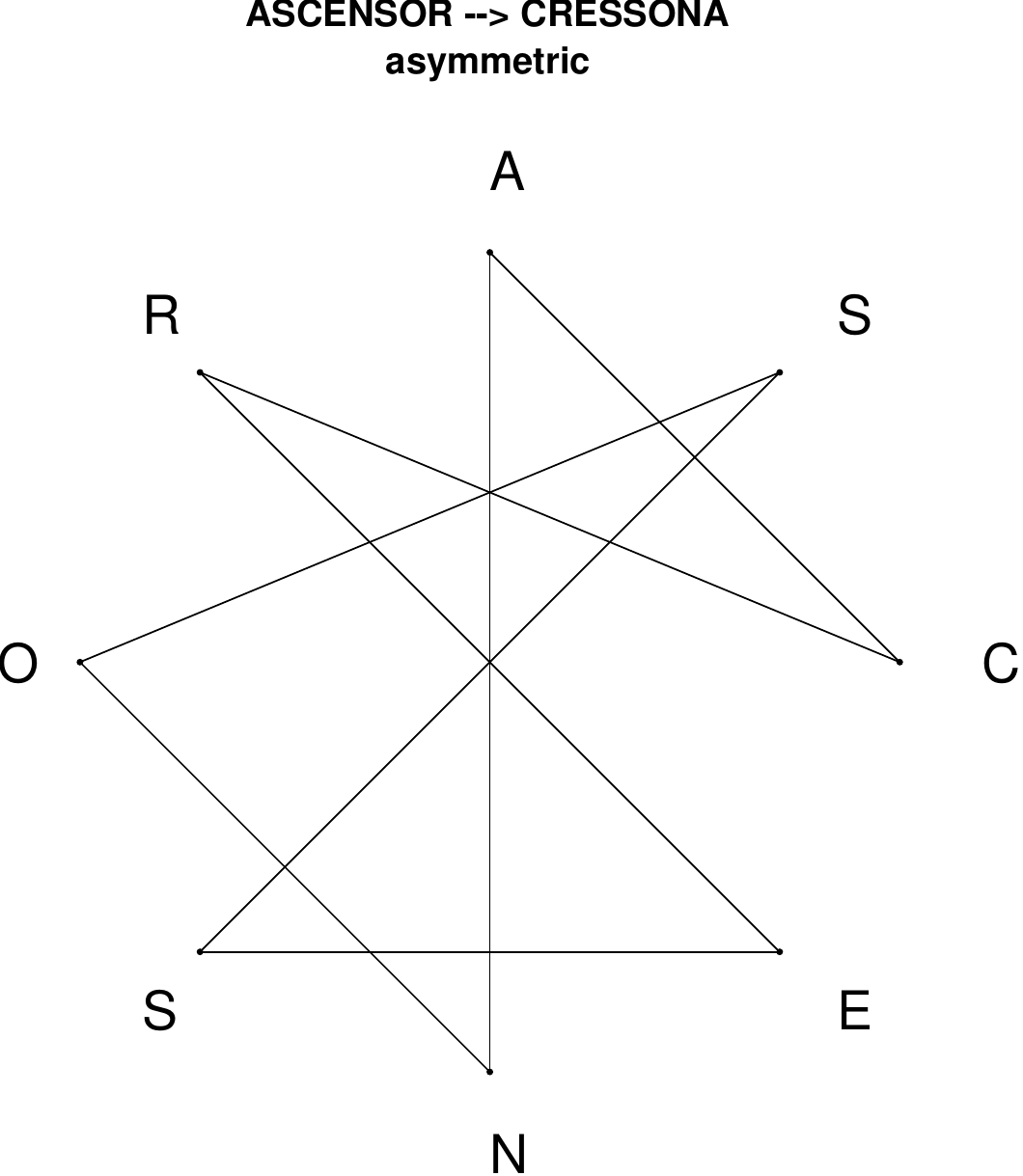}
\end{subfigure}
\hfill
\begin{subfigure}[T]{0.19\textwidth}
\centering
\includegraphics[width=\textwidth]{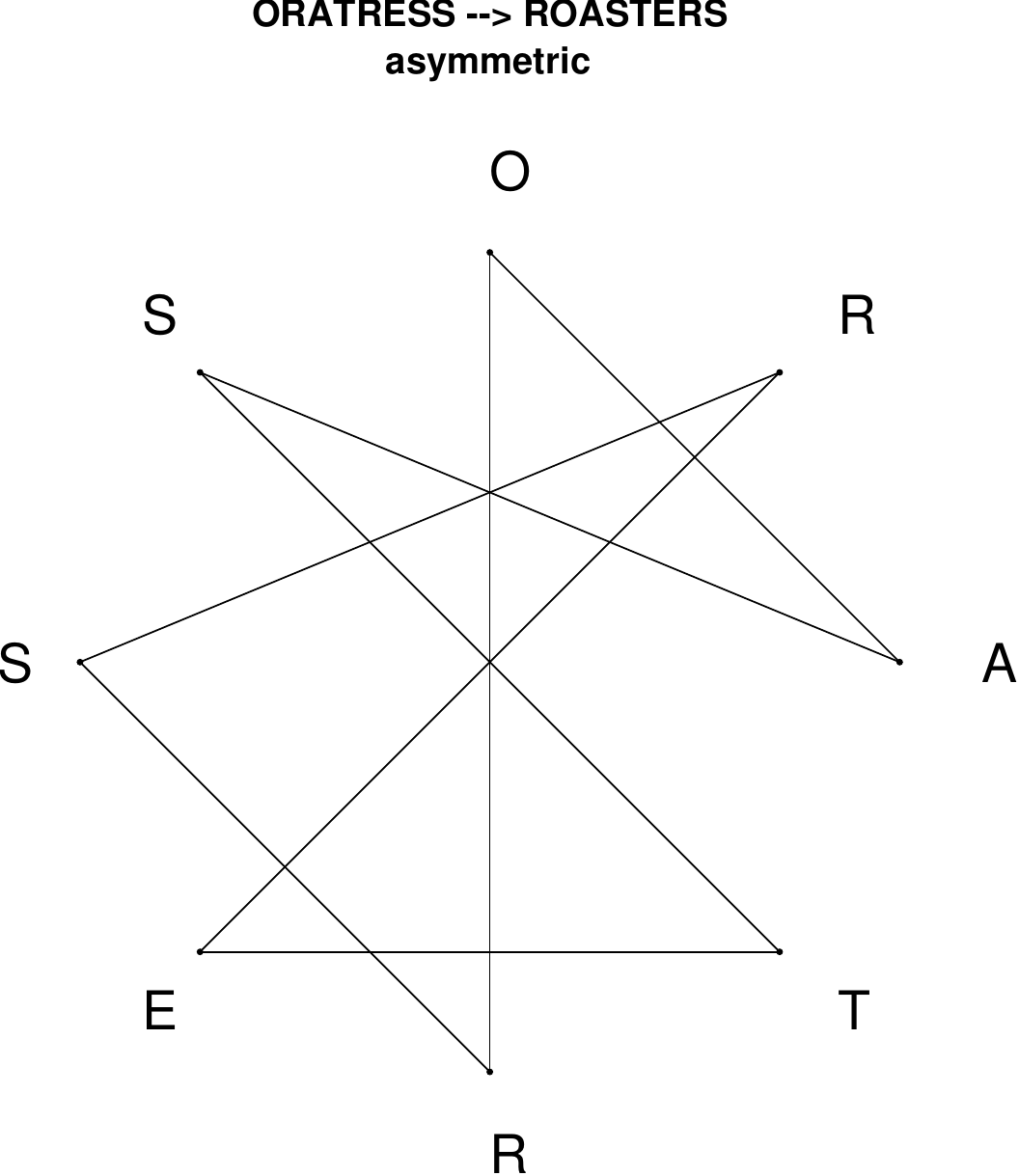}
\end{subfigure}
\hfill
\begin{subfigure}[T]{0.19\textwidth}
\centering
\includegraphics[width=\textwidth]{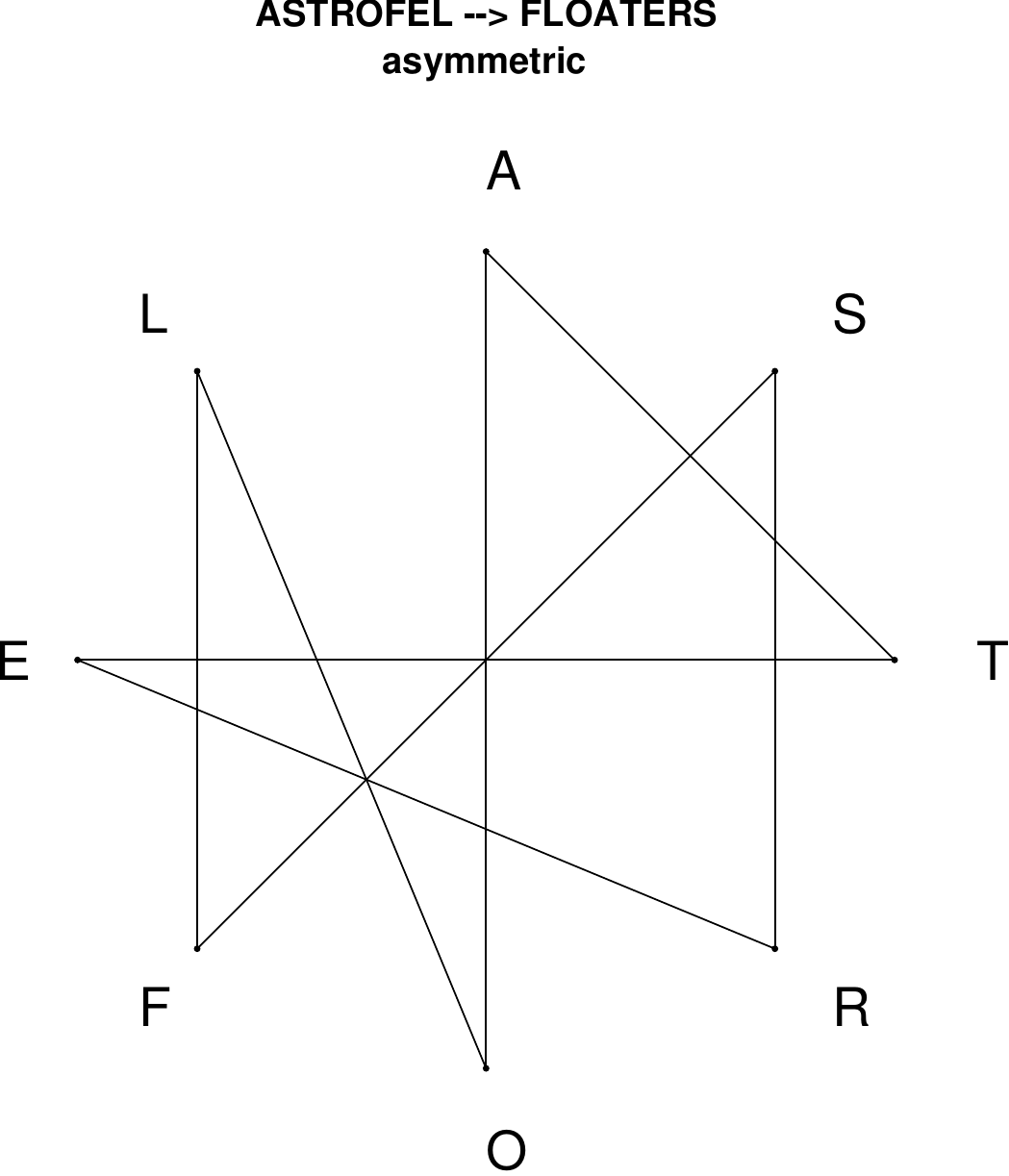}
\end{subfigure}
\hfill
\begin{subfigure}[T]{0.19\textwidth}
\centering
\includegraphics[width=\textwidth]{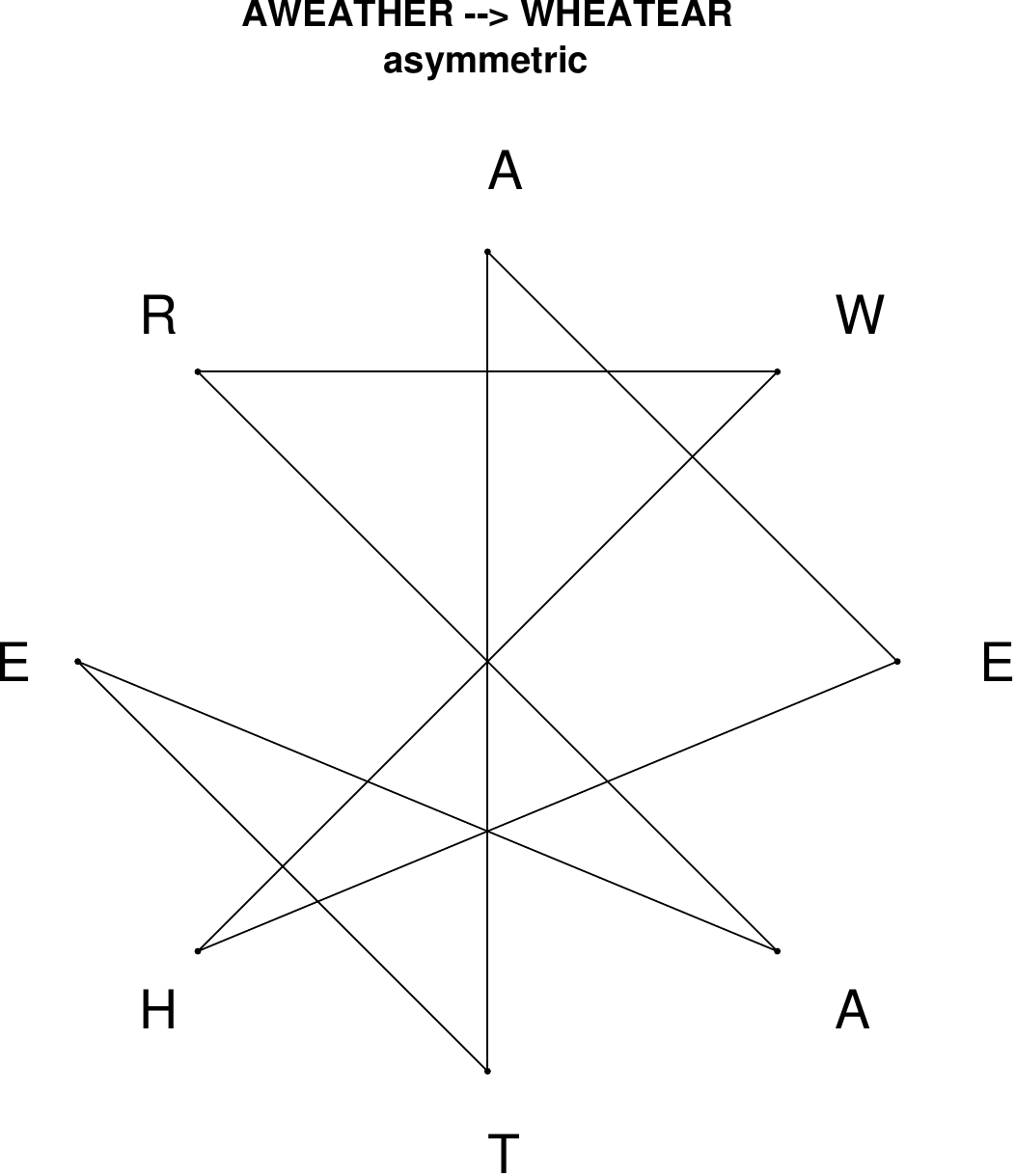}
\end{subfigure}
\end{figure}

\begin{figure}[H]
\centering
\begin{subfigure}[T]{0.19\textwidth}
\centering
\includegraphics[width=\textwidth]{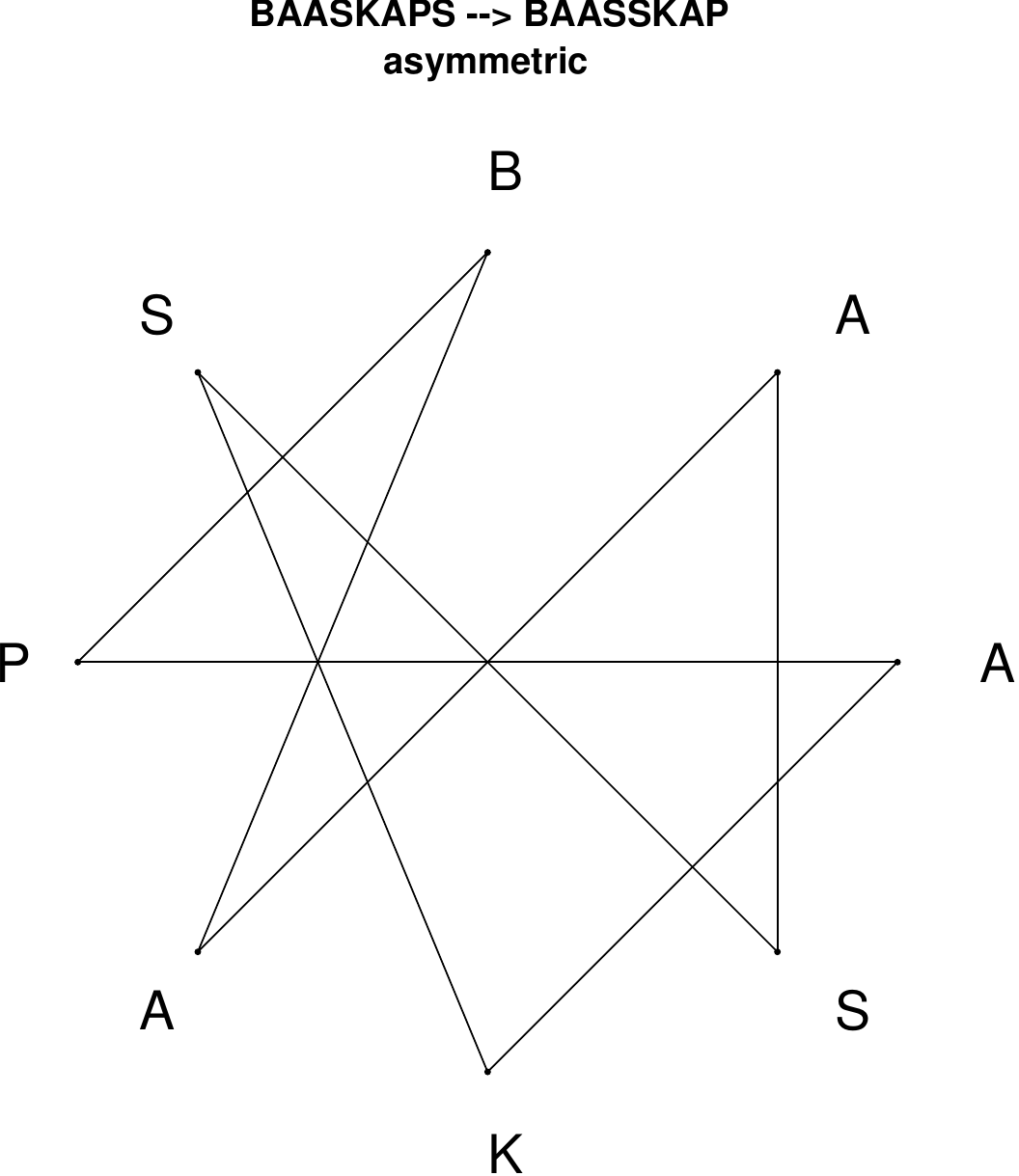}
\end{subfigure}
\hfill
\begin{subfigure}[T]{0.19\textwidth}
\centering
\includegraphics[width=\textwidth]{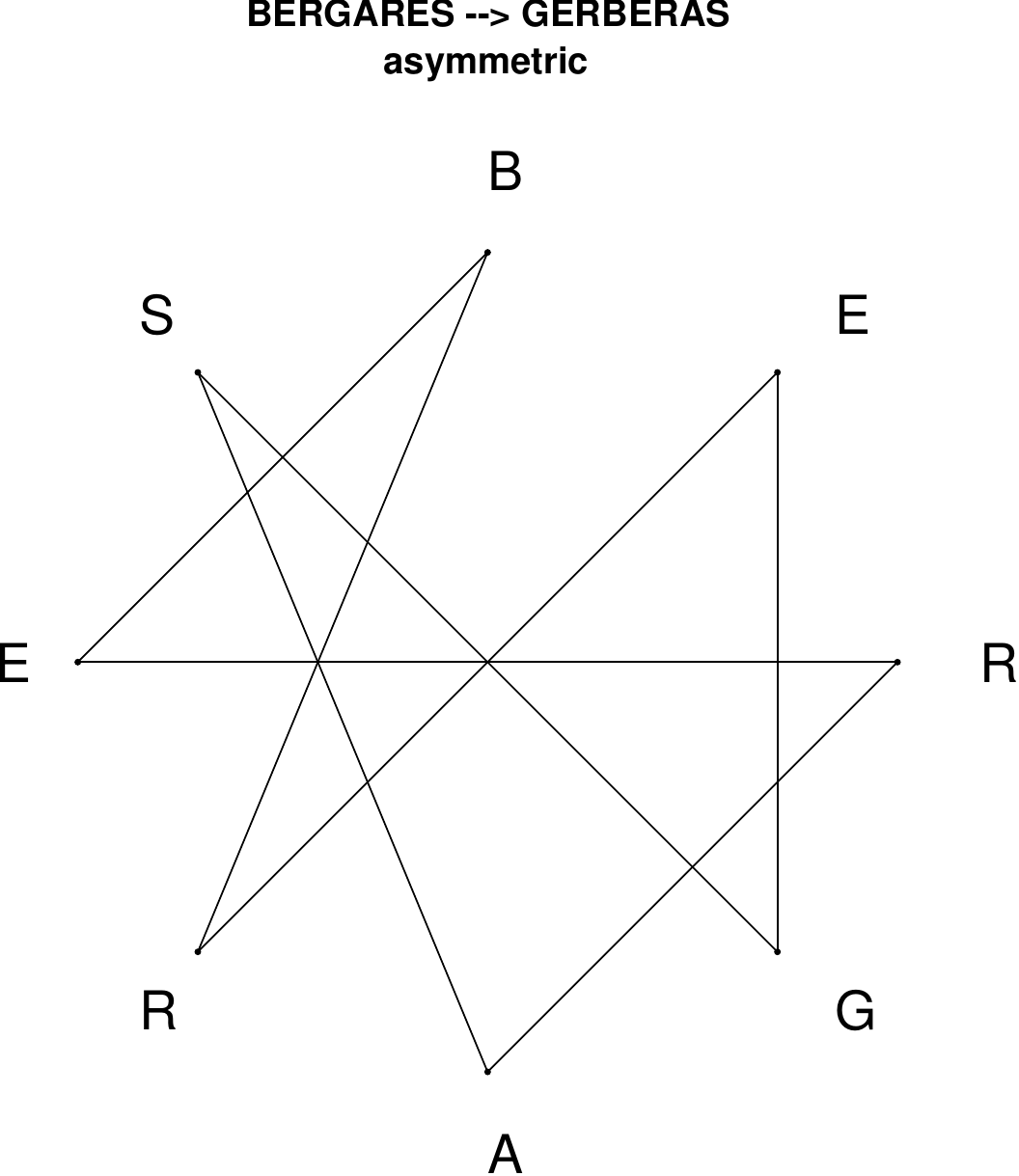}
\end{subfigure}
\hfill
\begin{subfigure}[T]{0.19\textwidth}
\centering
\includegraphics[width=\textwidth]{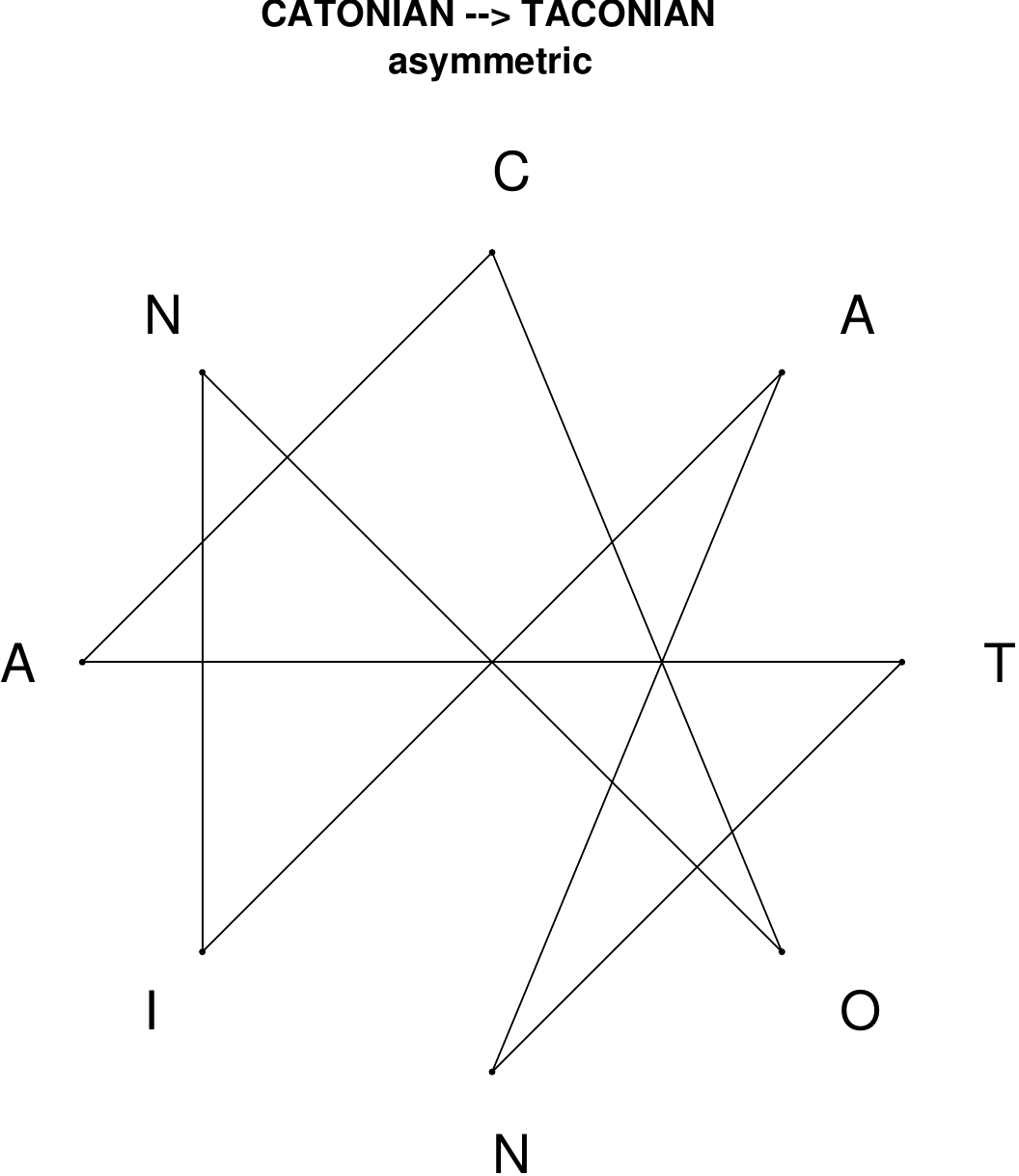}
\end{subfigure}
\hfill
\begin{subfigure}[T]{0.19\textwidth}
\centering
\includegraphics[width=\textwidth]{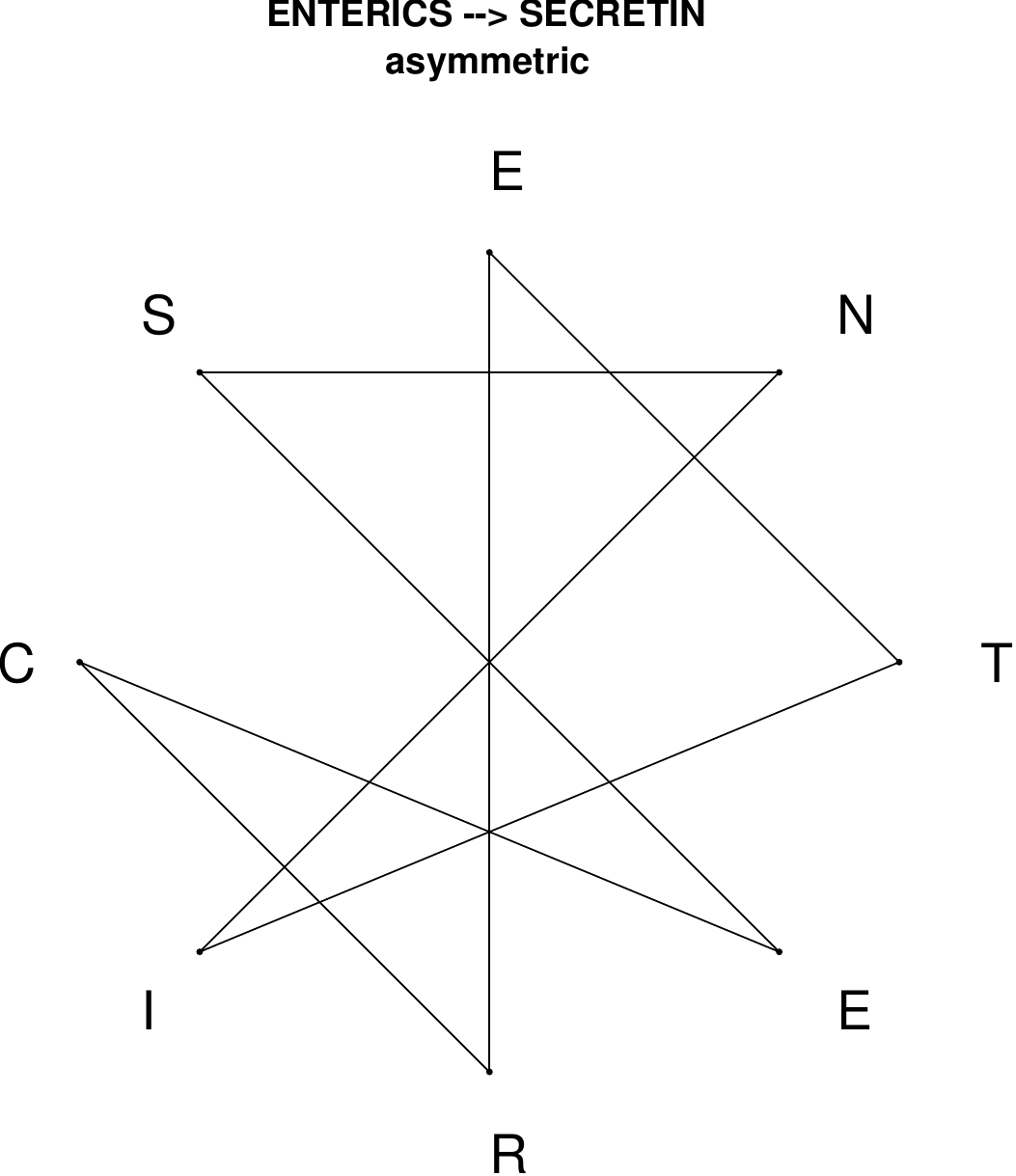}
\end{subfigure}
\hfill
\begin{subfigure}[T]{0.19\textwidth}
\centering
\includegraphics[width=\textwidth]{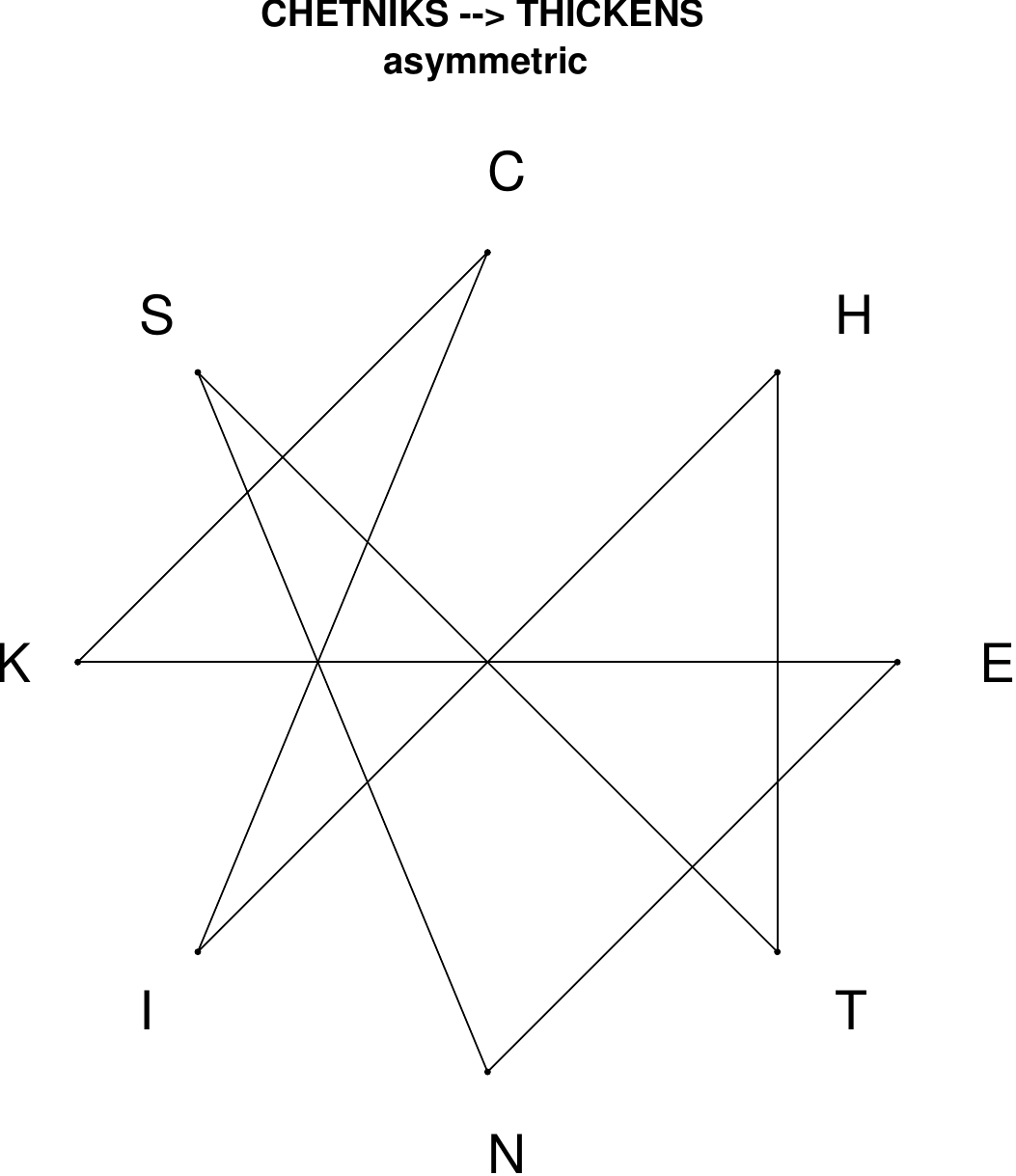}
\end{subfigure}
\end{figure}

\begin{figure}[H]
\centering
\begin{subfigure}[T]{0.19\textwidth}
\centering
\includegraphics[width=\textwidth]{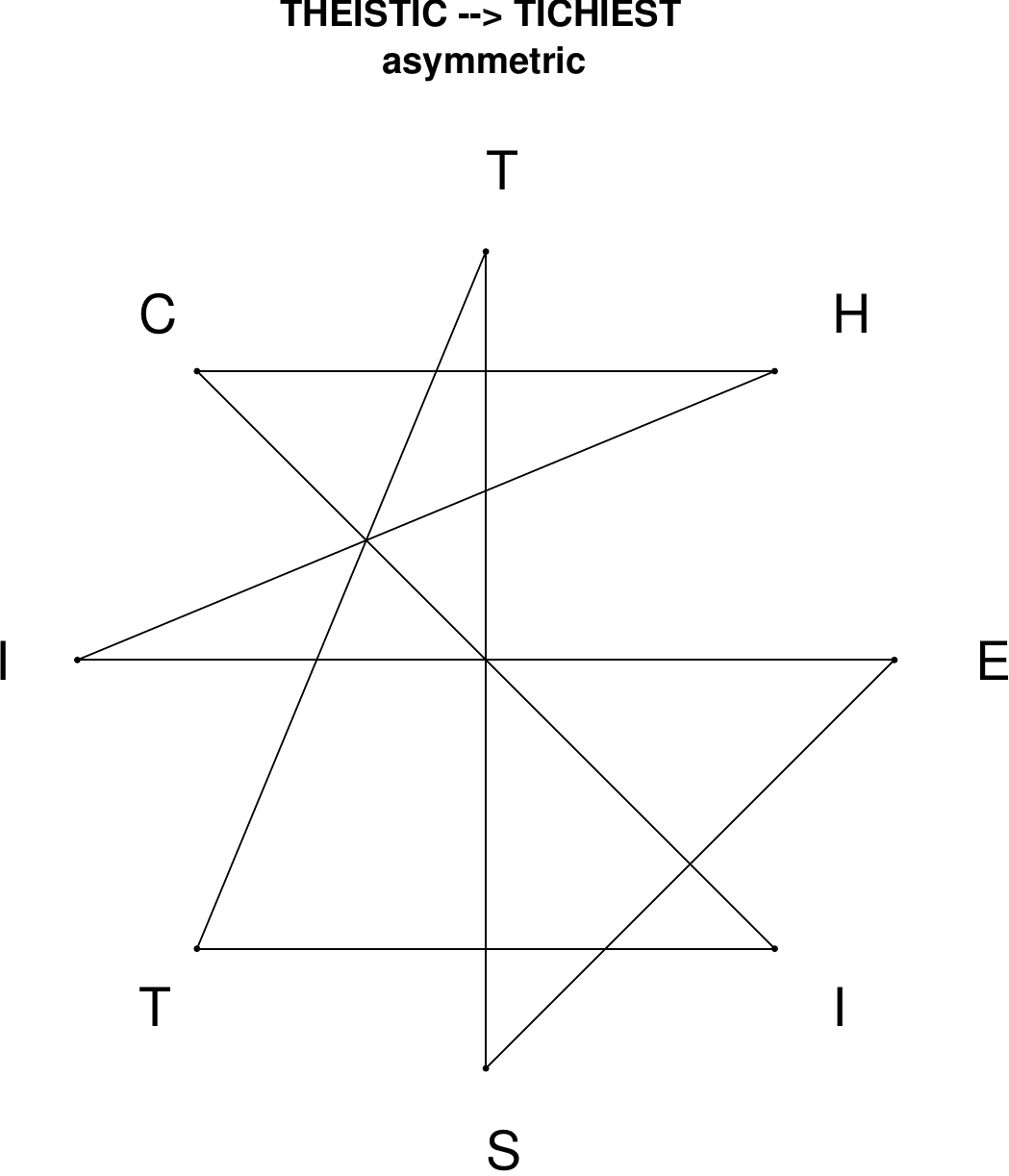}
\end{subfigure}
\hfill
\begin{subfigure}[T]{0.19\textwidth}
\centering
\includegraphics[width=\textwidth]{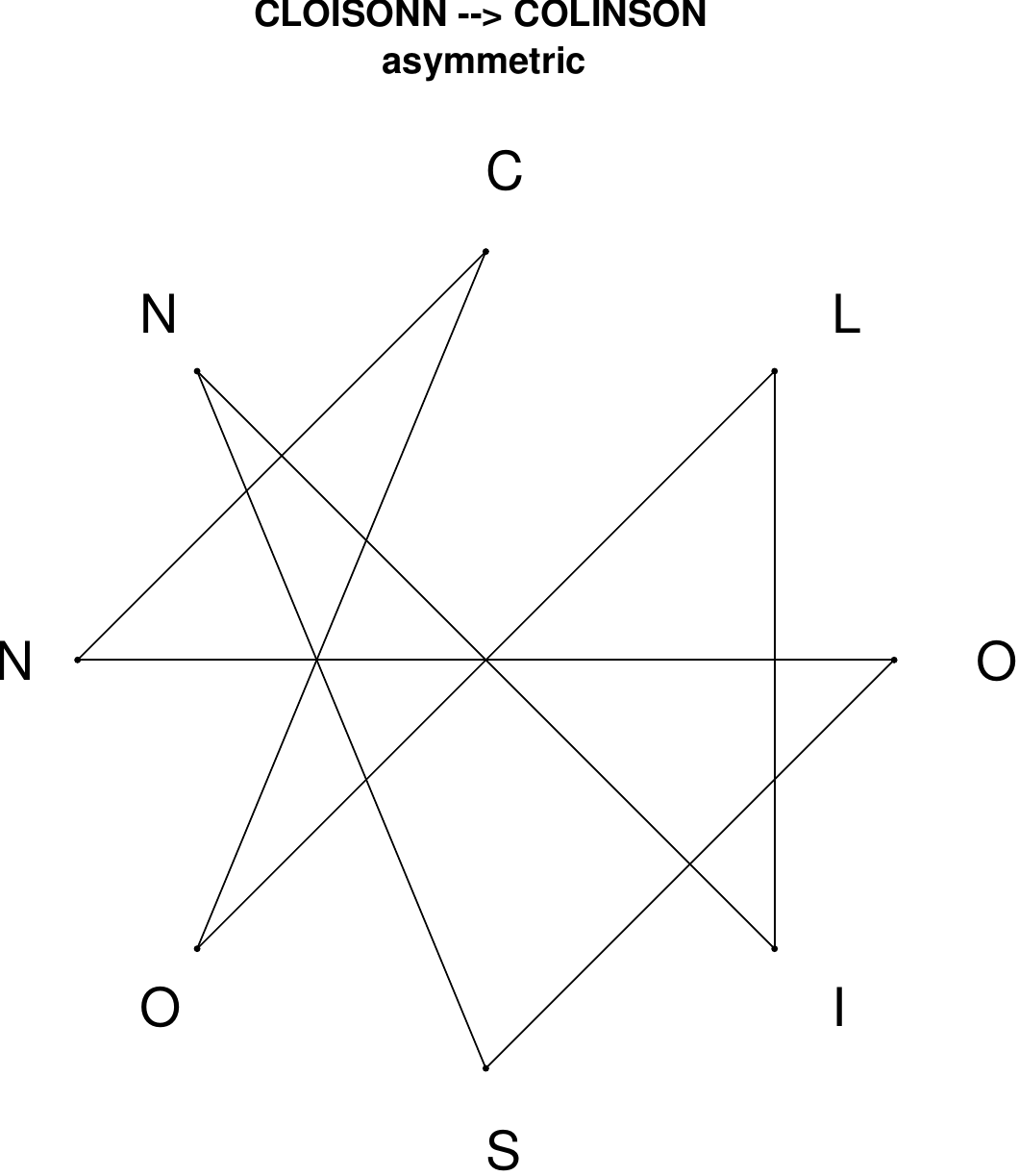}
\end{subfigure}
\hfill
\begin{subfigure}[T]{0.19\textwidth}
\centering
\includegraphics[width=\textwidth]{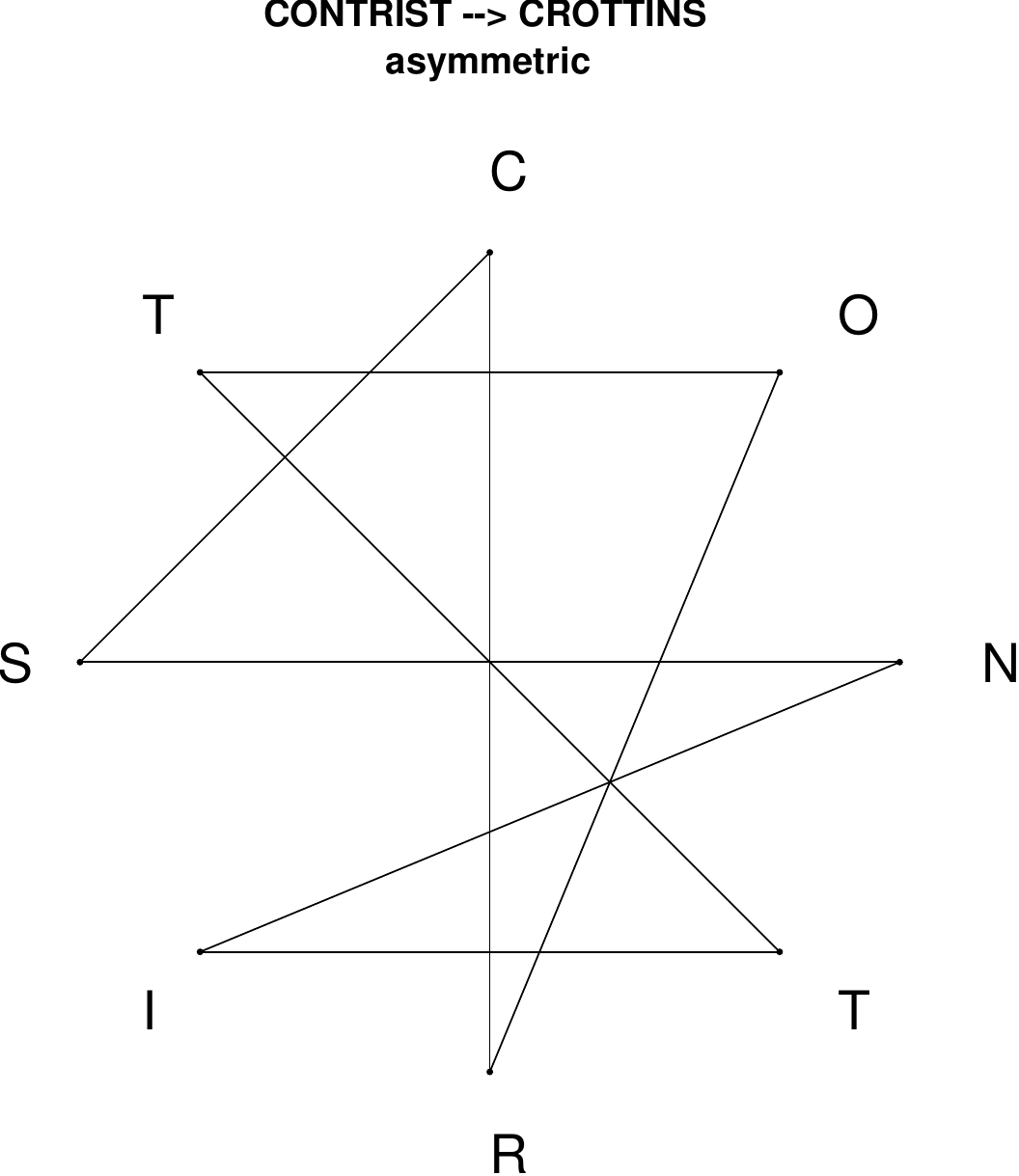}
\end{subfigure}
\hfill
\begin{subfigure}[T]{0.19\textwidth}
\centering
\includegraphics[width=\textwidth]{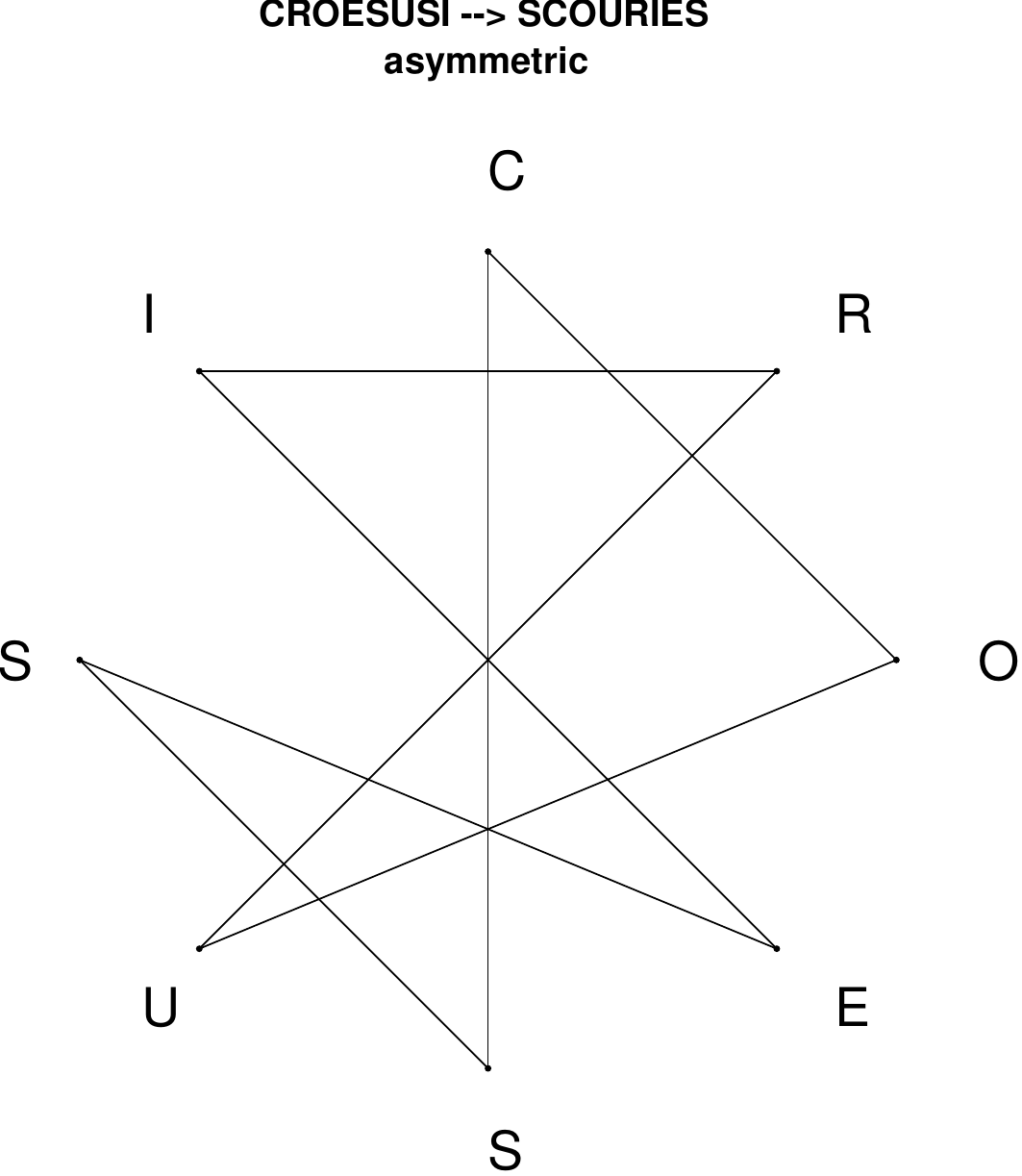}
\end{subfigure}
\hfill
\begin{subfigure}[T]{0.19\textwidth}
\centering
\includegraphics[width=\textwidth]{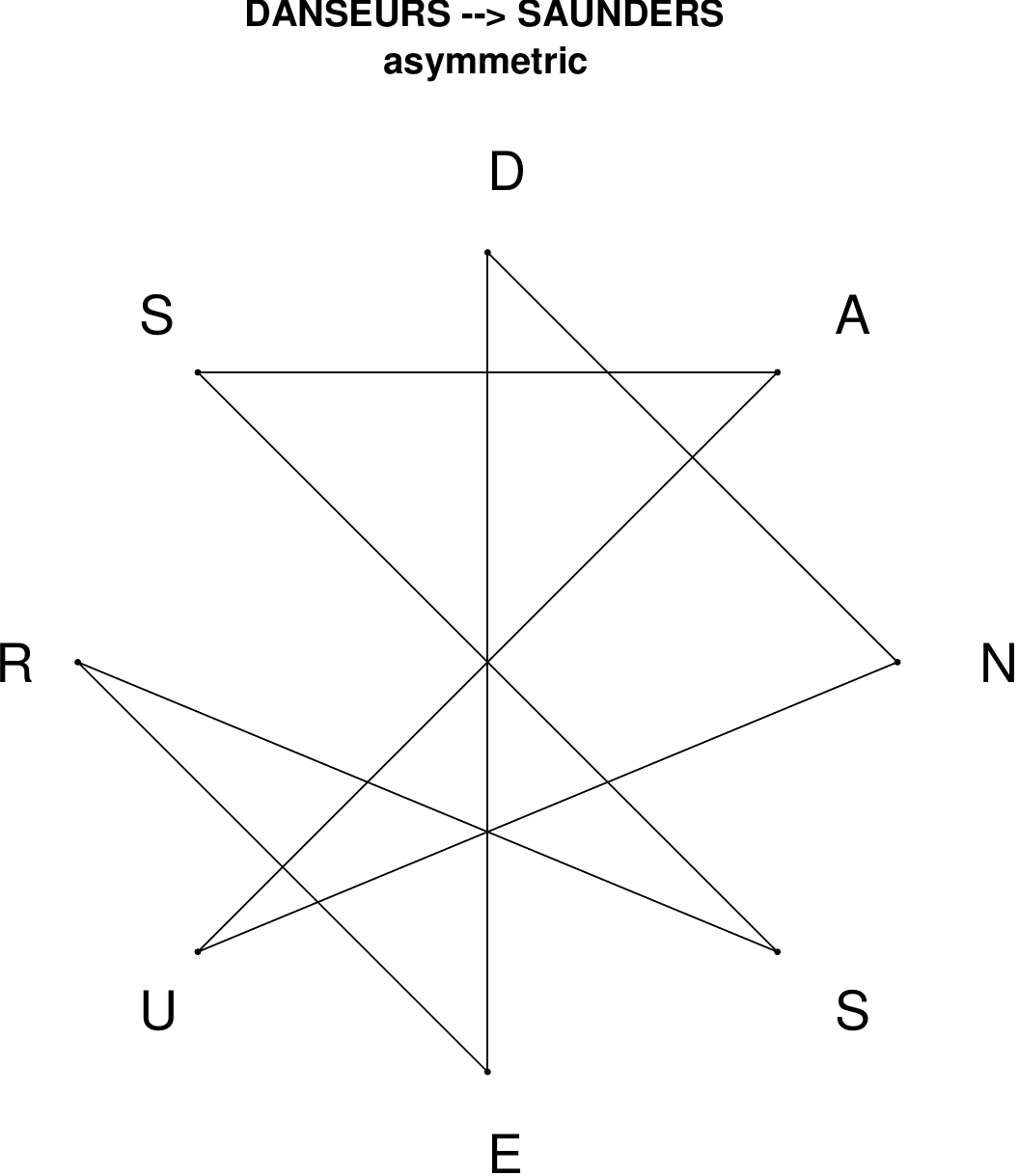}
\end{subfigure}
\end{figure}

\begin{figure}[H]
\centering
\begin{subfigure}[T]{0.19\textwidth}
\centering
\includegraphics[width=\textwidth]{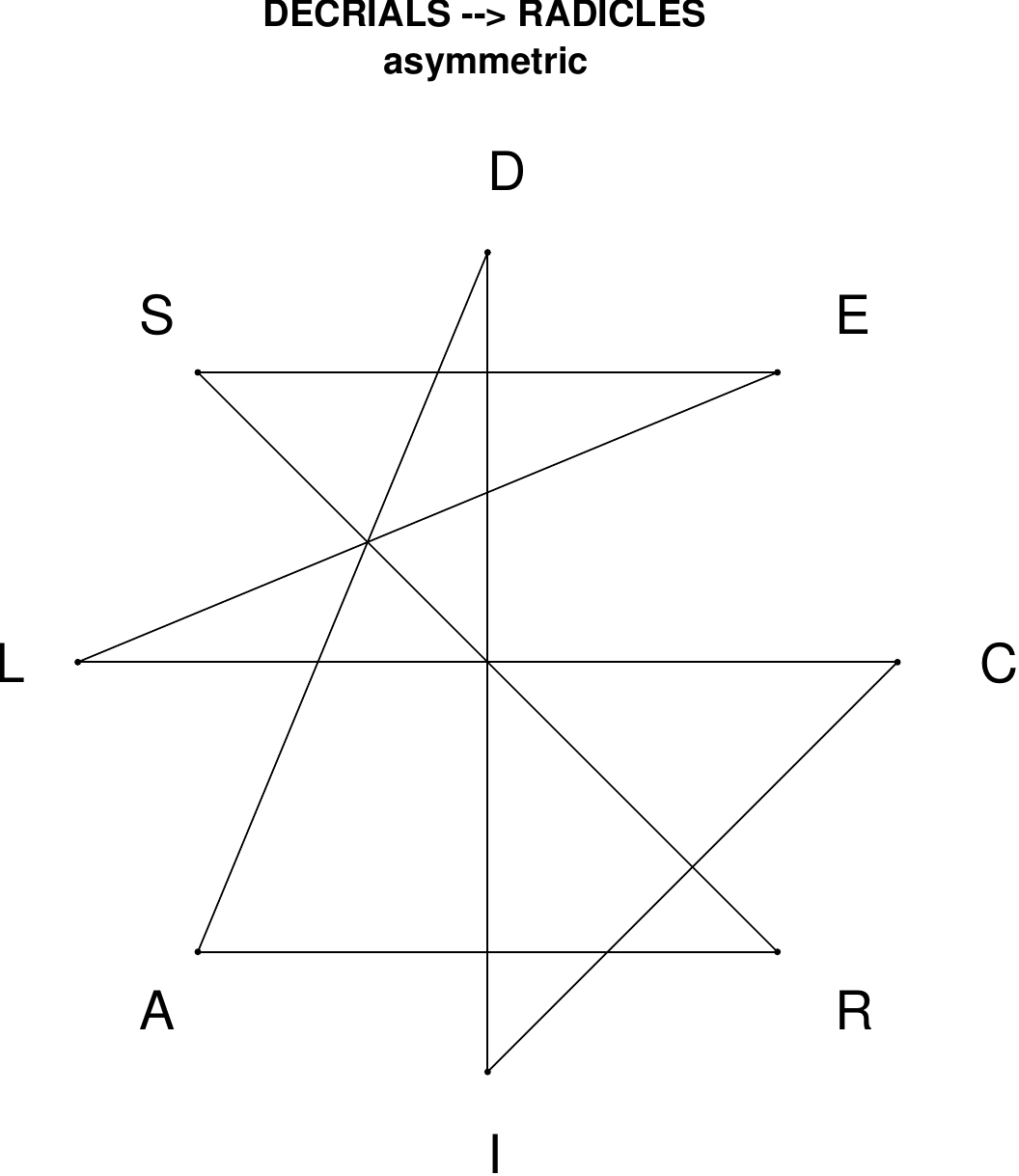}
\end{subfigure}
\hfill
\begin{subfigure}[T]{0.19\textwidth}
\centering
\includegraphics[width=\textwidth]{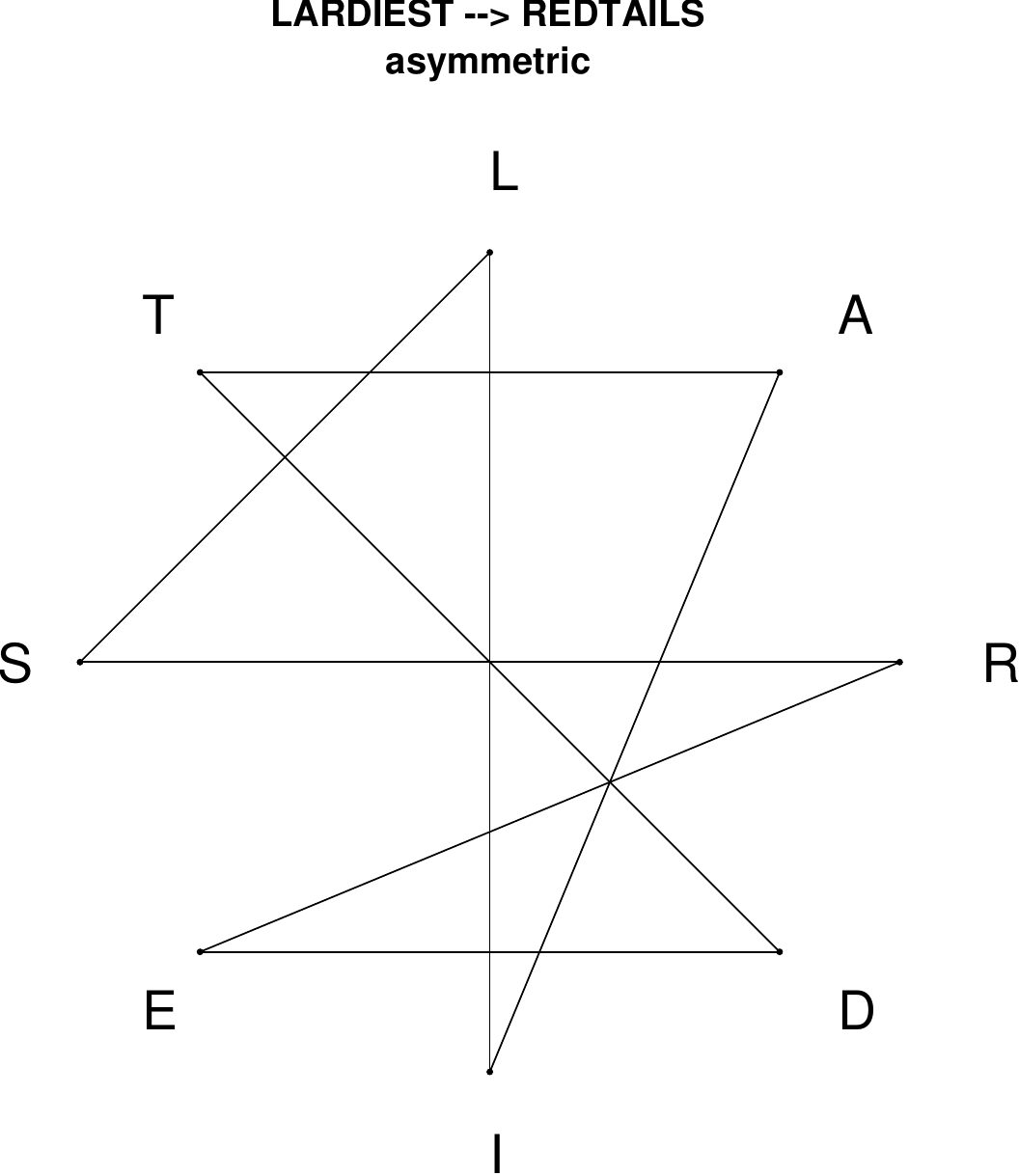}
\end{subfigure}
\hfill
\begin{subfigure}[T]{0.19\textwidth}
\centering
\includegraphics[width=\textwidth]{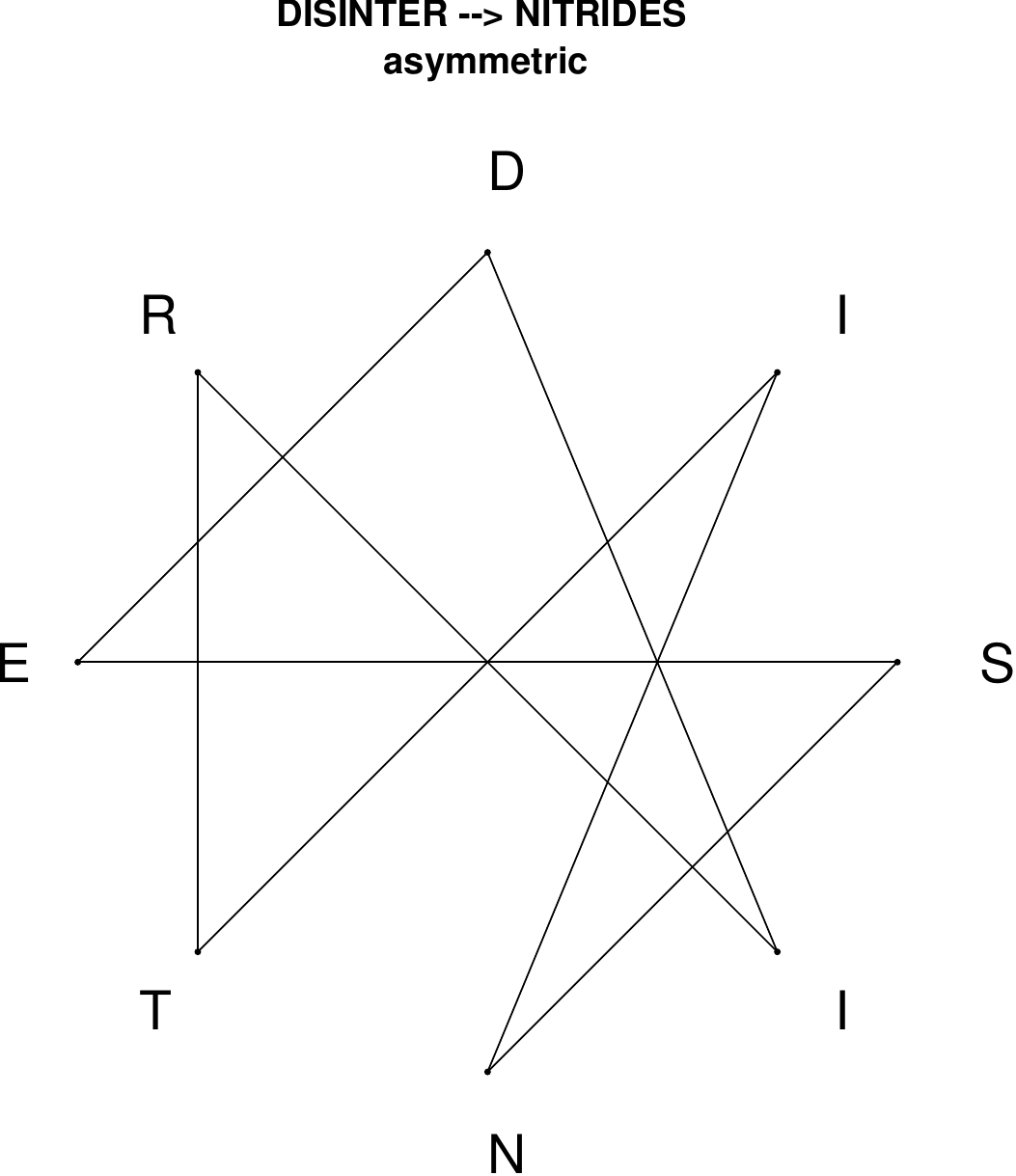}
\end{subfigure}
\hfill
\begin{subfigure}[T]{0.19\textwidth}
\centering
\includegraphics[width=\textwidth]{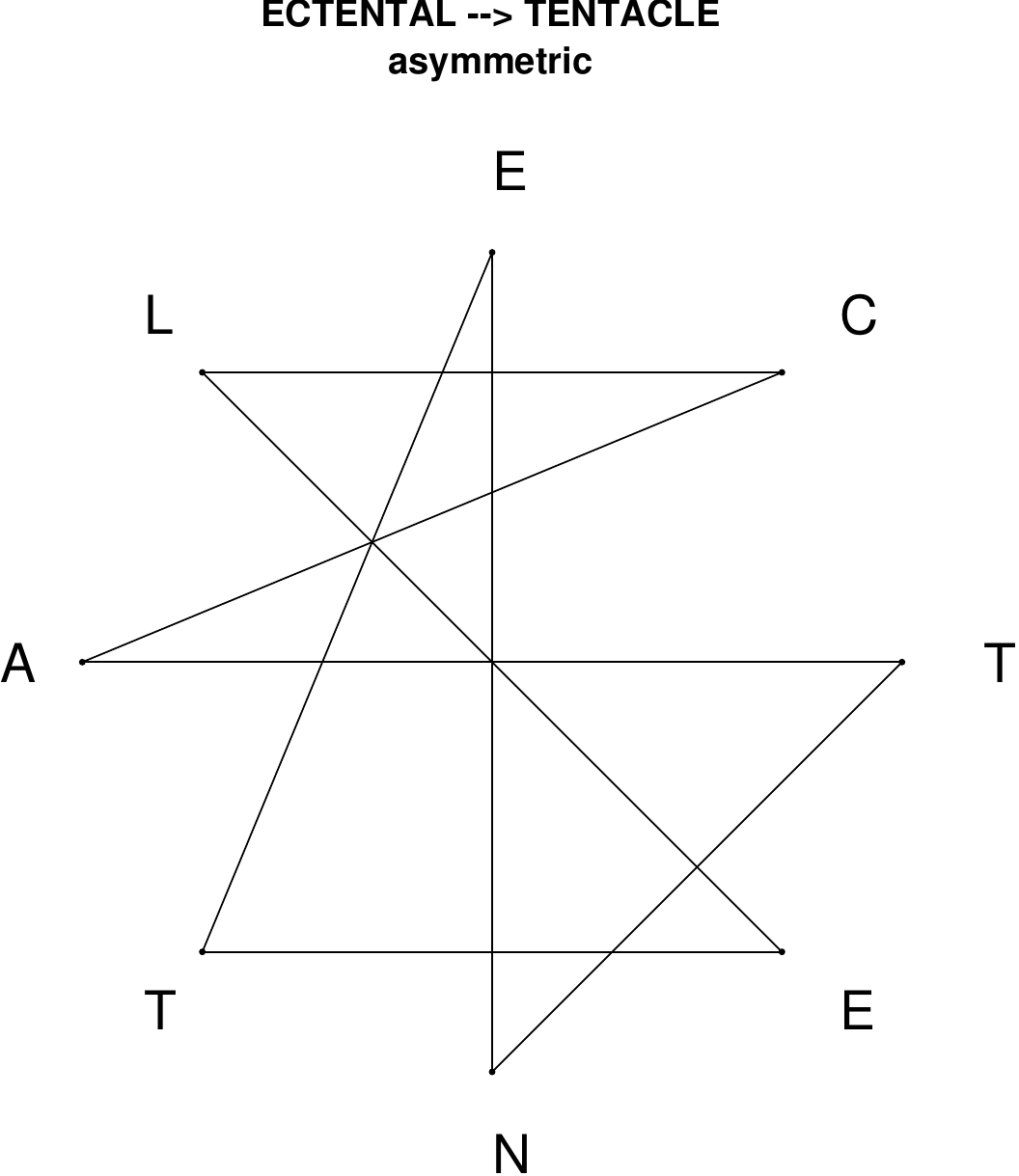}
\end{subfigure}
\hfill
\begin{subfigure}[T]{0.19\textwidth}
\centering
\includegraphics[width=\textwidth]{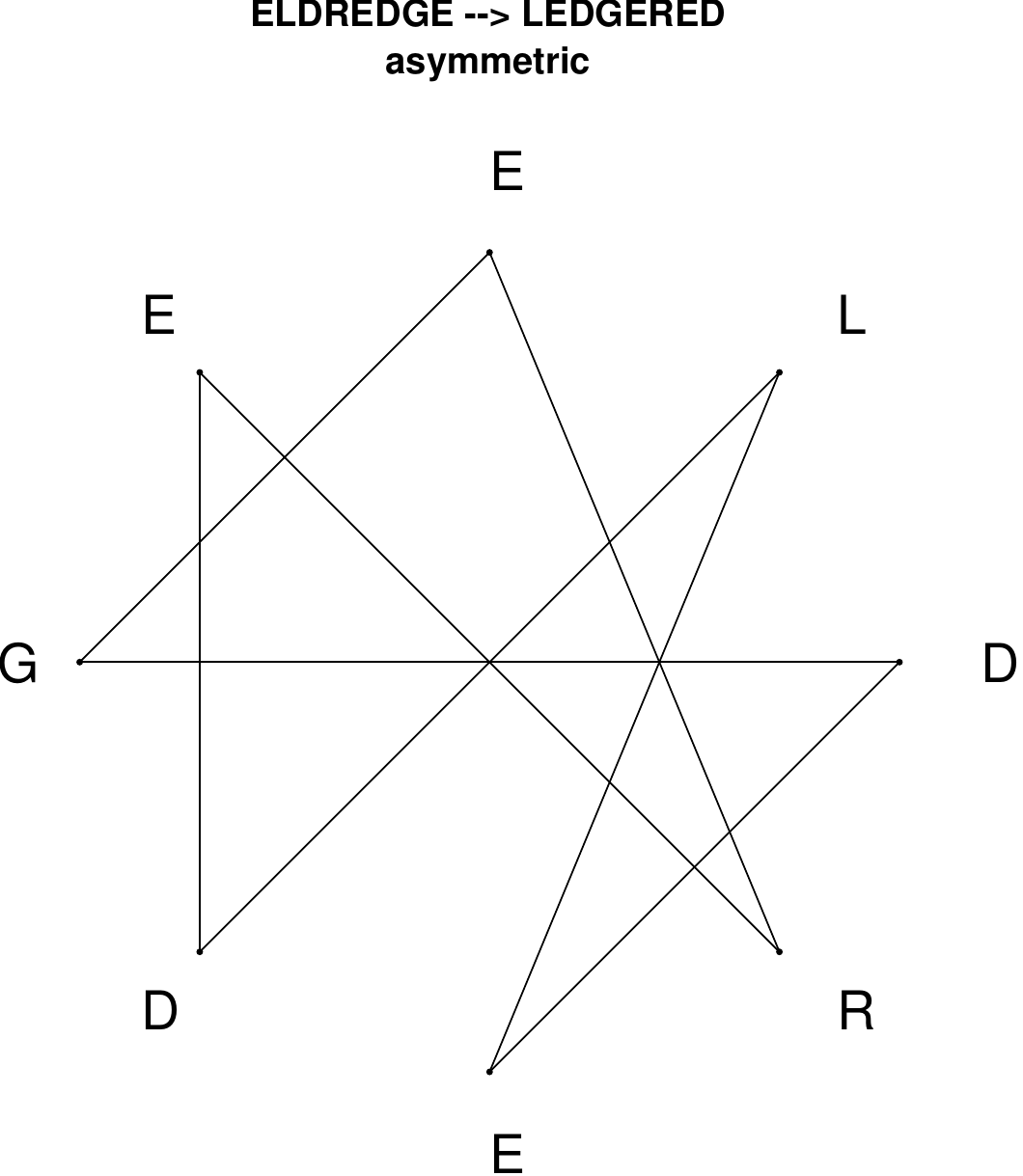}
\end{subfigure}
\end{figure}

\begin{figure}[H]
\centering
\begin{subfigure}[T]{0.19\textwidth}
\centering
\includegraphics[width=\textwidth]{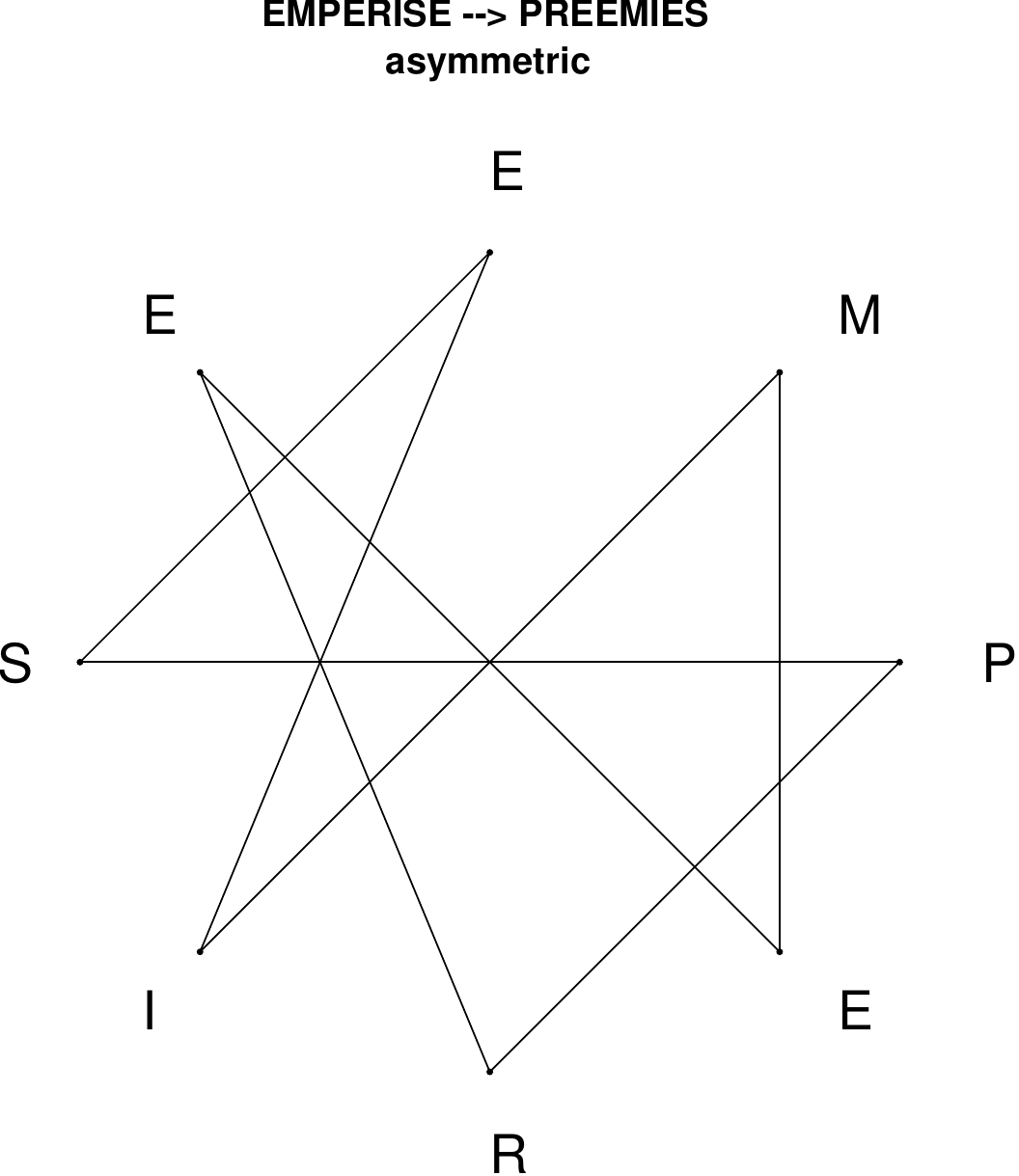}
\end{subfigure}
\hfill
\begin{subfigure}[T]{0.19\textwidth}
\centering
\includegraphics[width=\textwidth]{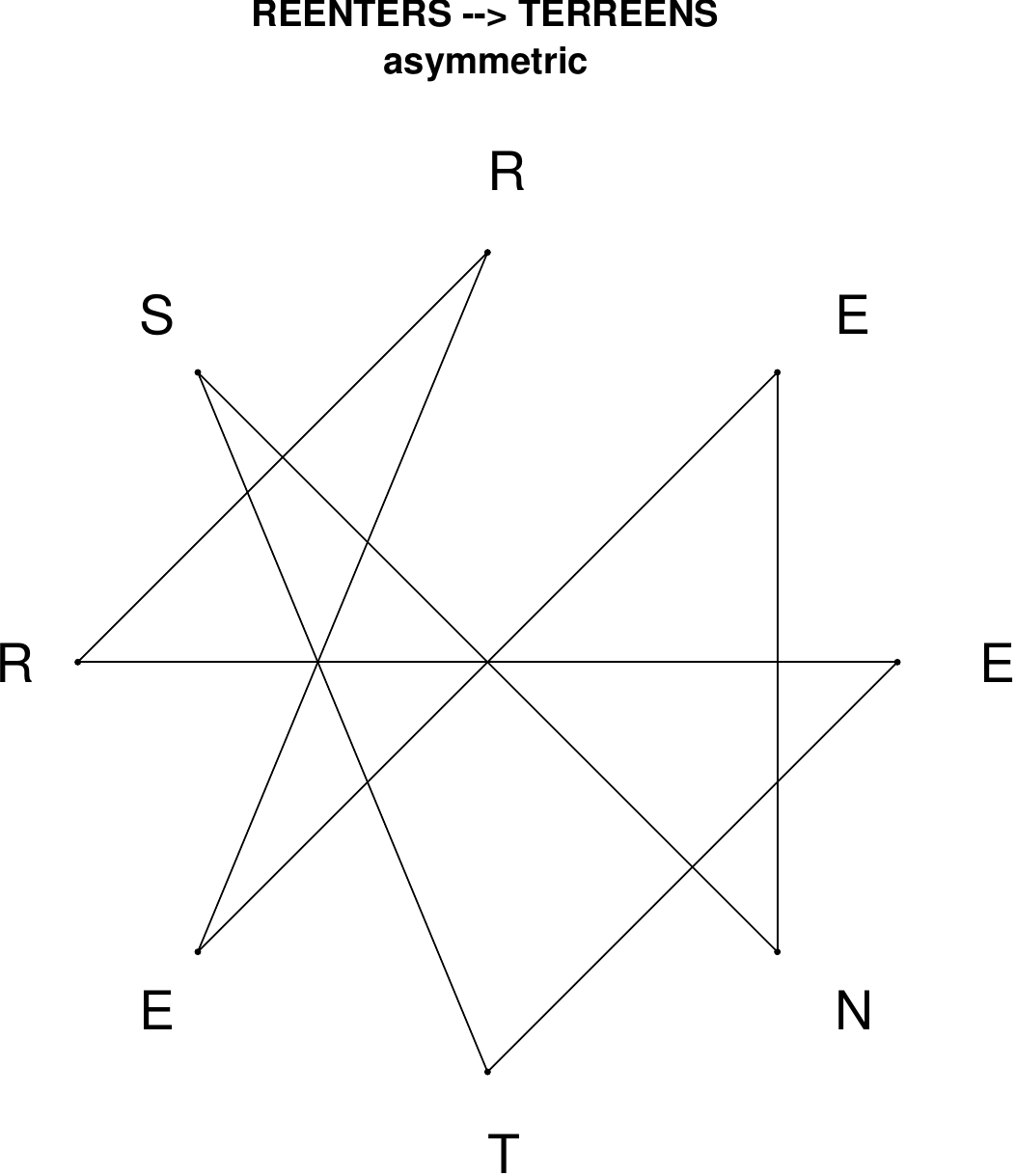}
\end{subfigure}
\hfill
\begin{subfigure}[T]{0.19\textwidth}
\centering
\includegraphics[width=\textwidth]{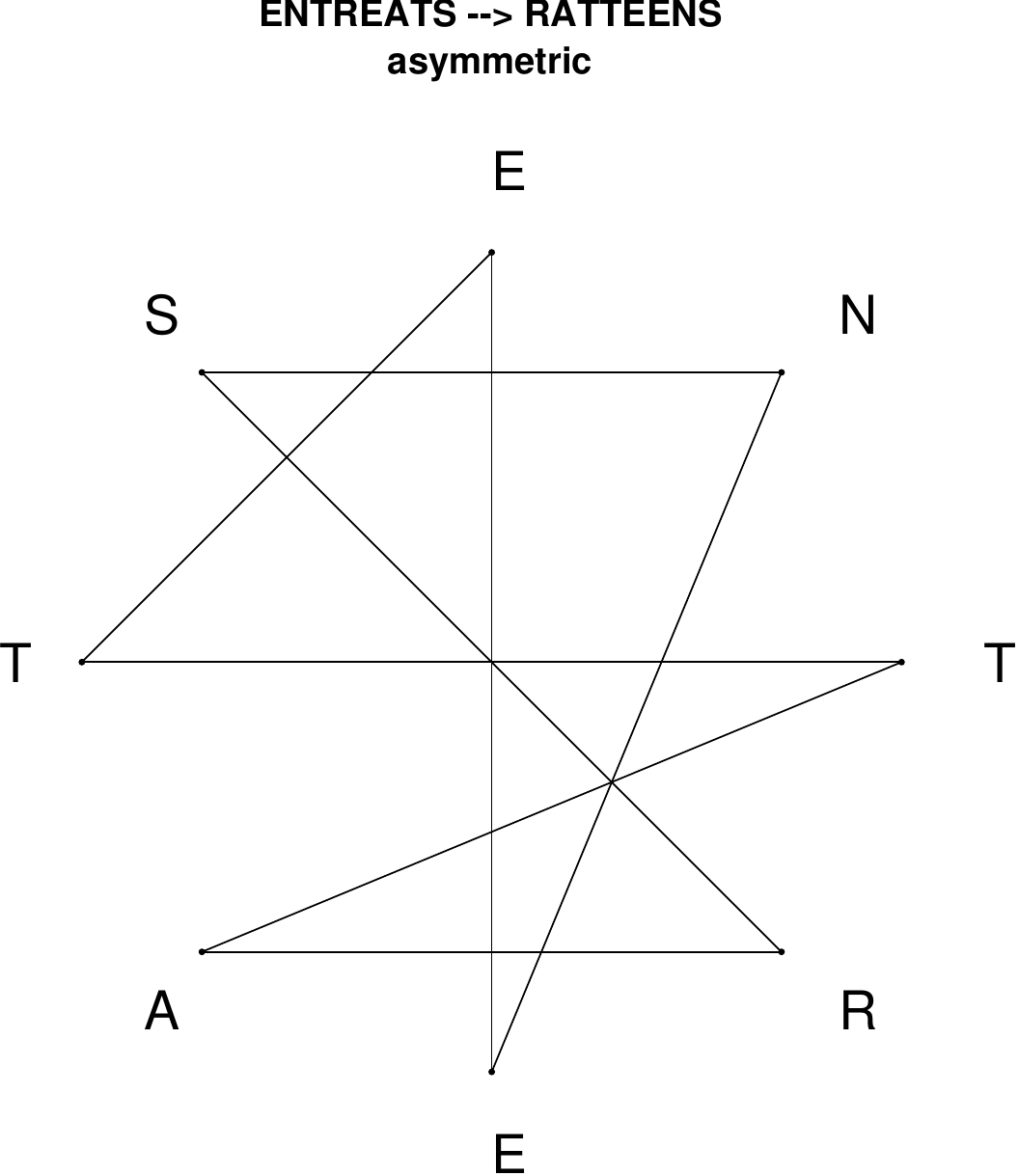}
\end{subfigure}
\hfill
\begin{subfigure}[T]{0.19\textwidth}
\centering
\includegraphics[width=\textwidth]{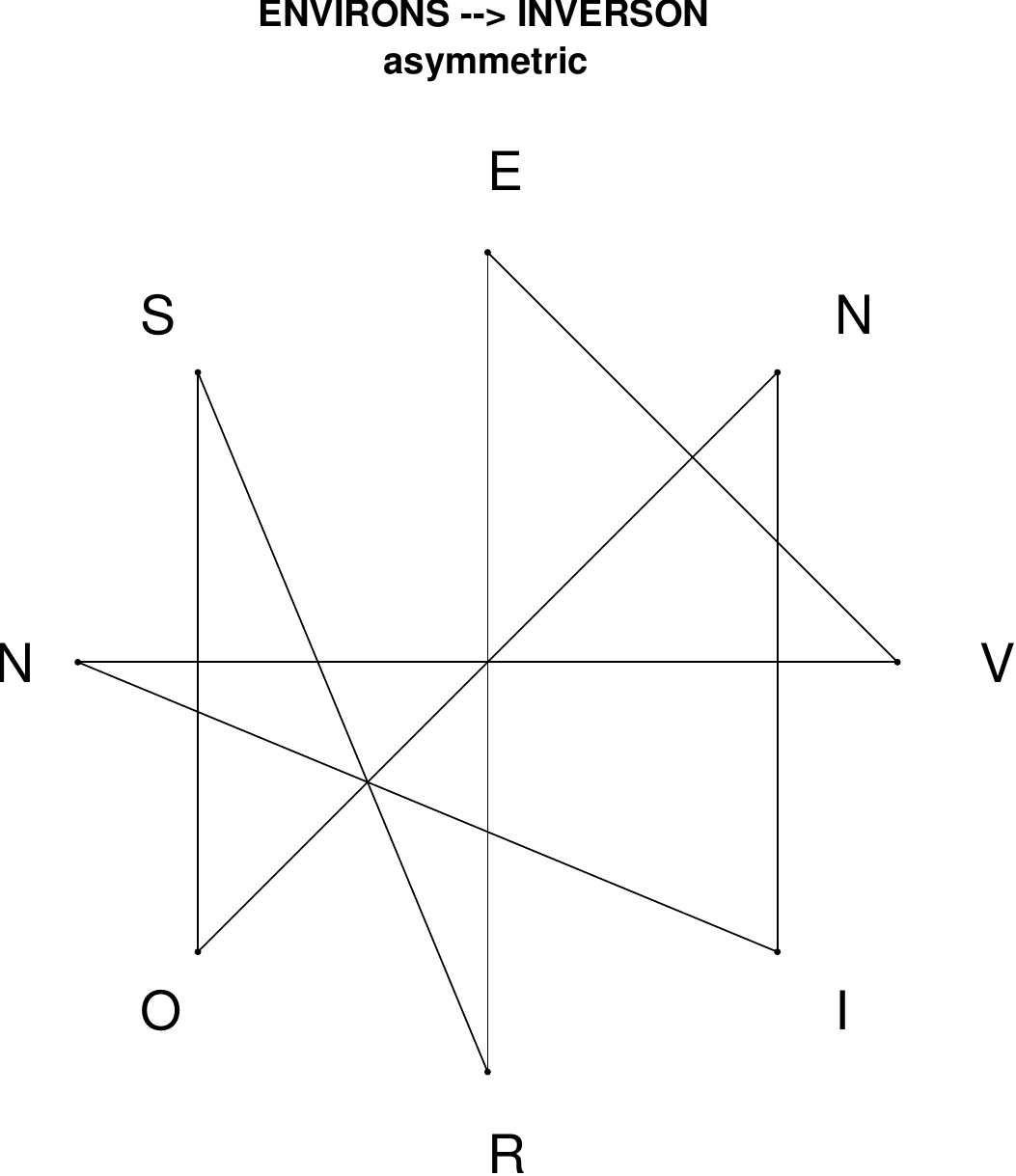}
\end{subfigure}
\hfill
\begin{subfigure}[T]{0.19\textwidth}
\centering
\includegraphics[width=\textwidth]{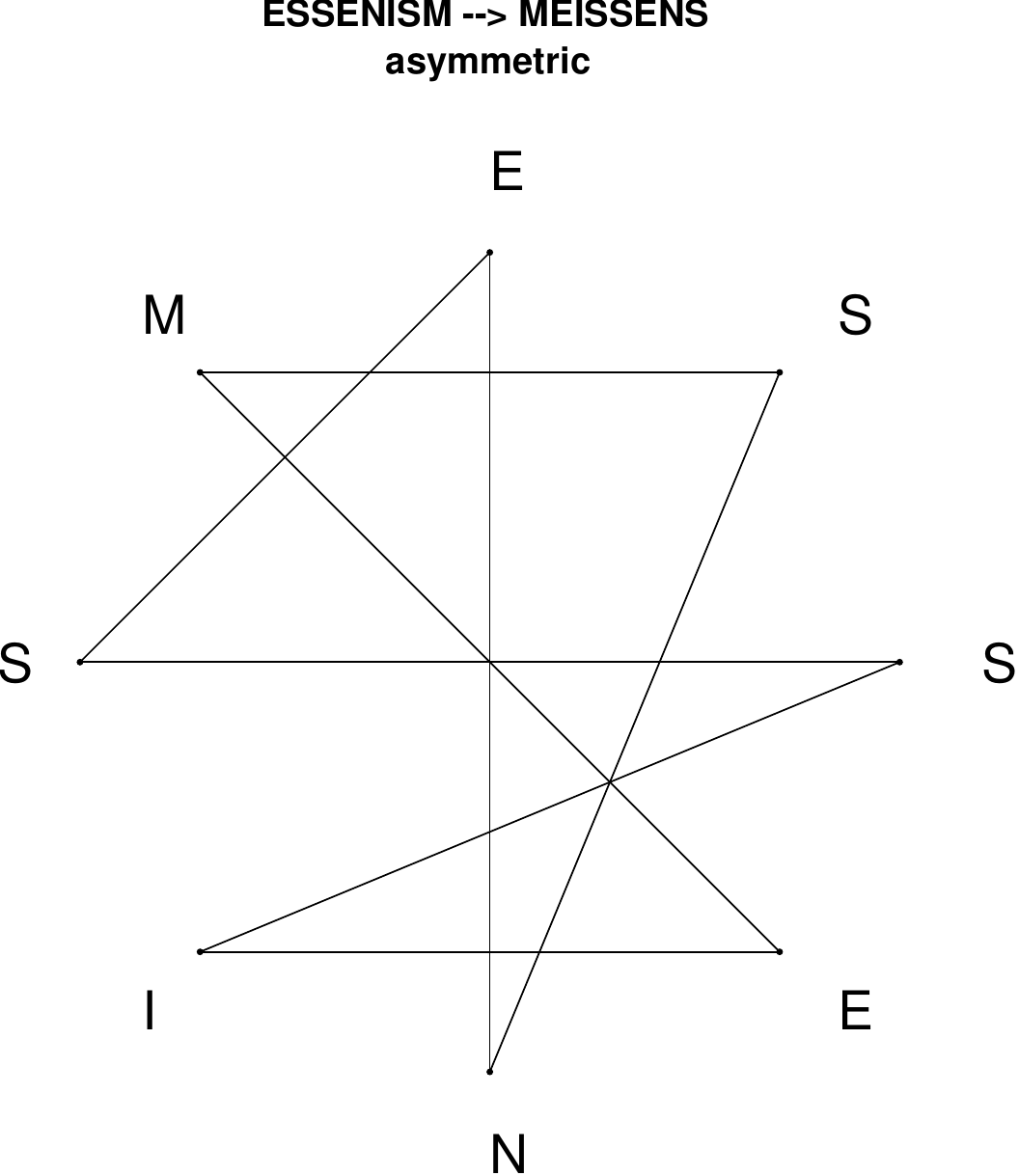}
\end{subfigure}
\end{figure}

\begin{figure}[H]
\centering
\begin{subfigure}[T]{0.19\textwidth}
\centering
\includegraphics[width=\textwidth]{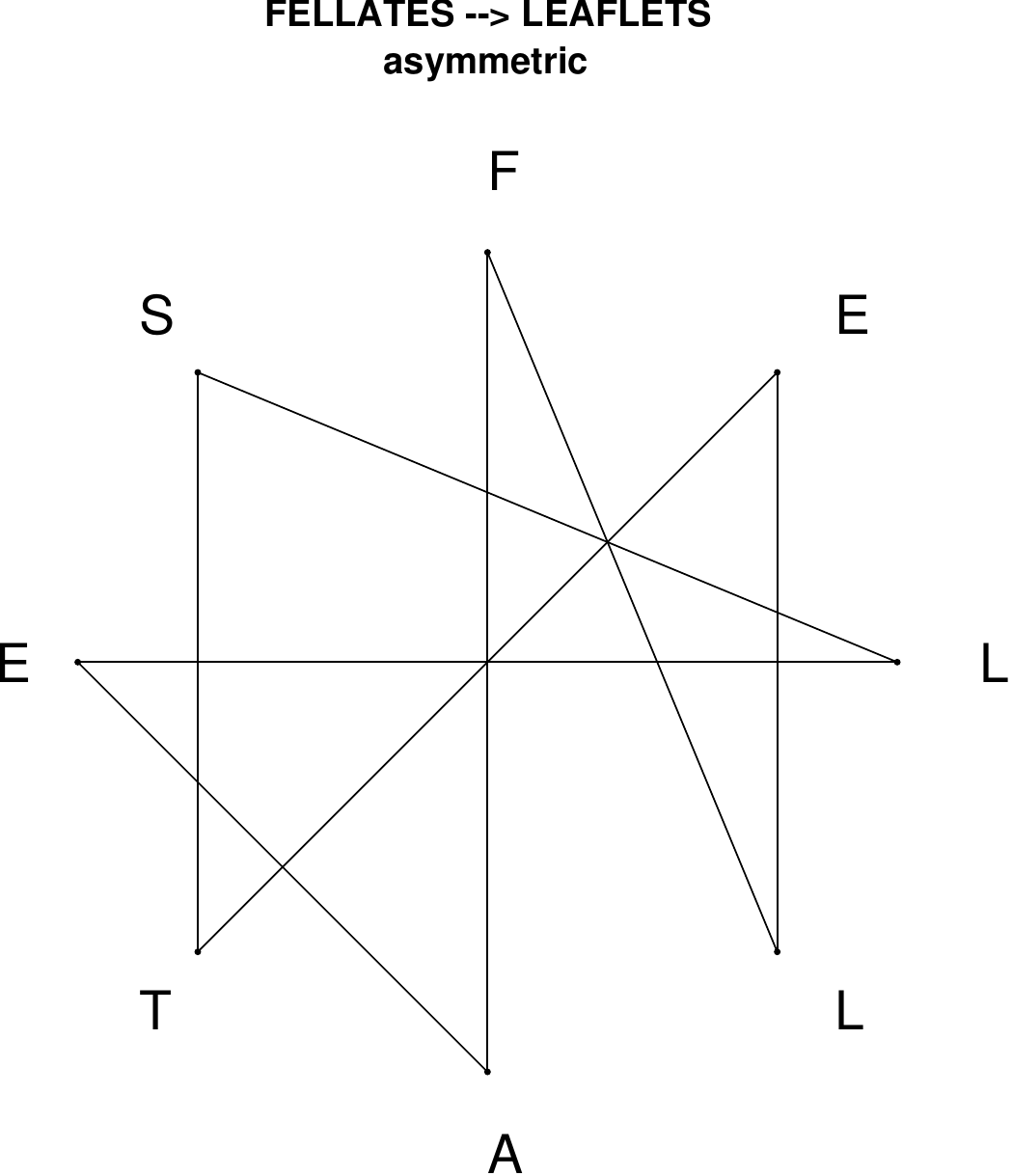}
\end{subfigure}
\hfill
\begin{subfigure}[T]{0.19\textwidth}
\centering
\includegraphics[width=\textwidth]{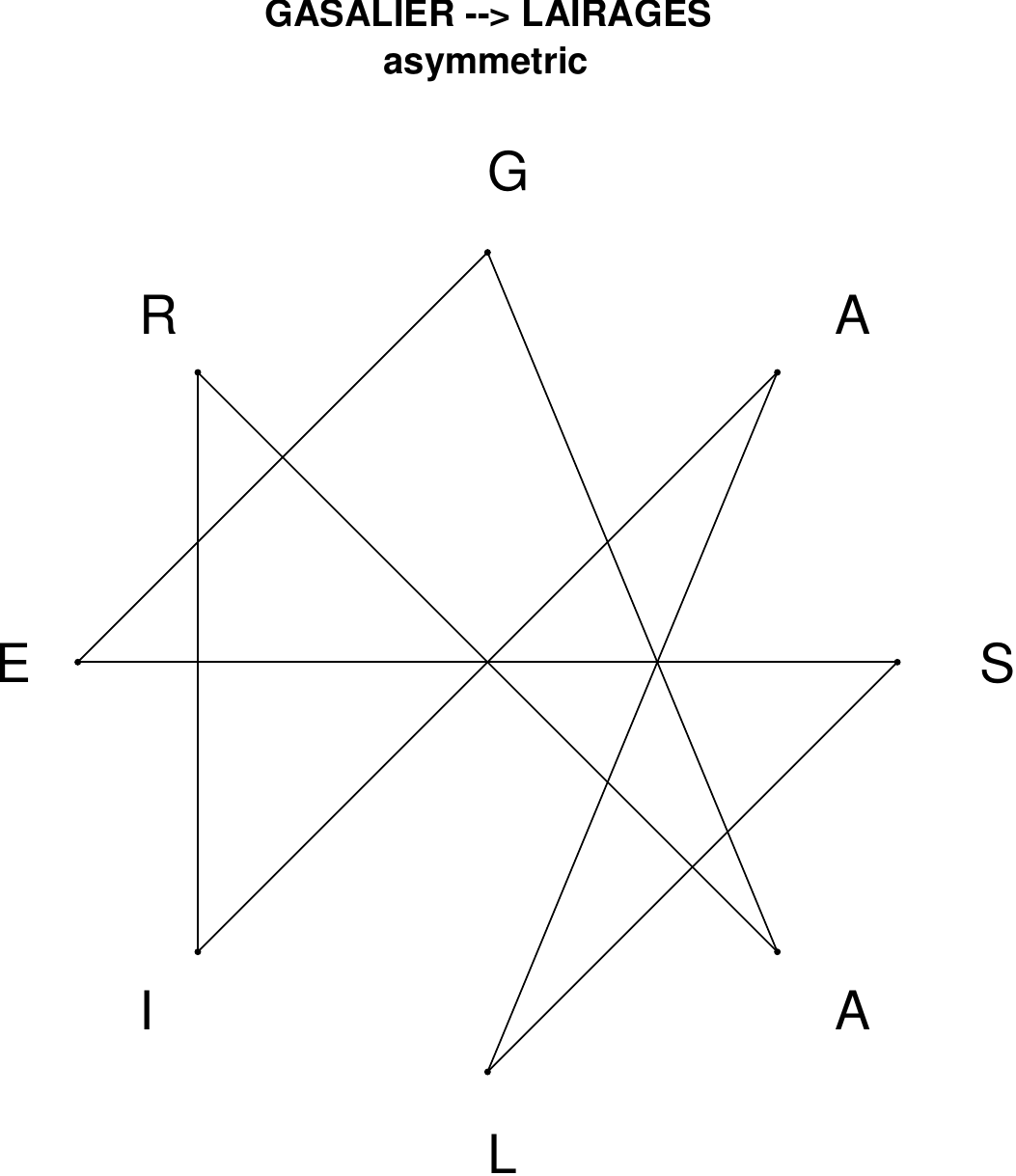}
\end{subfigure}
\hfill
\begin{subfigure}[T]{0.19\textwidth}
\centering
\includegraphics[width=\textwidth]{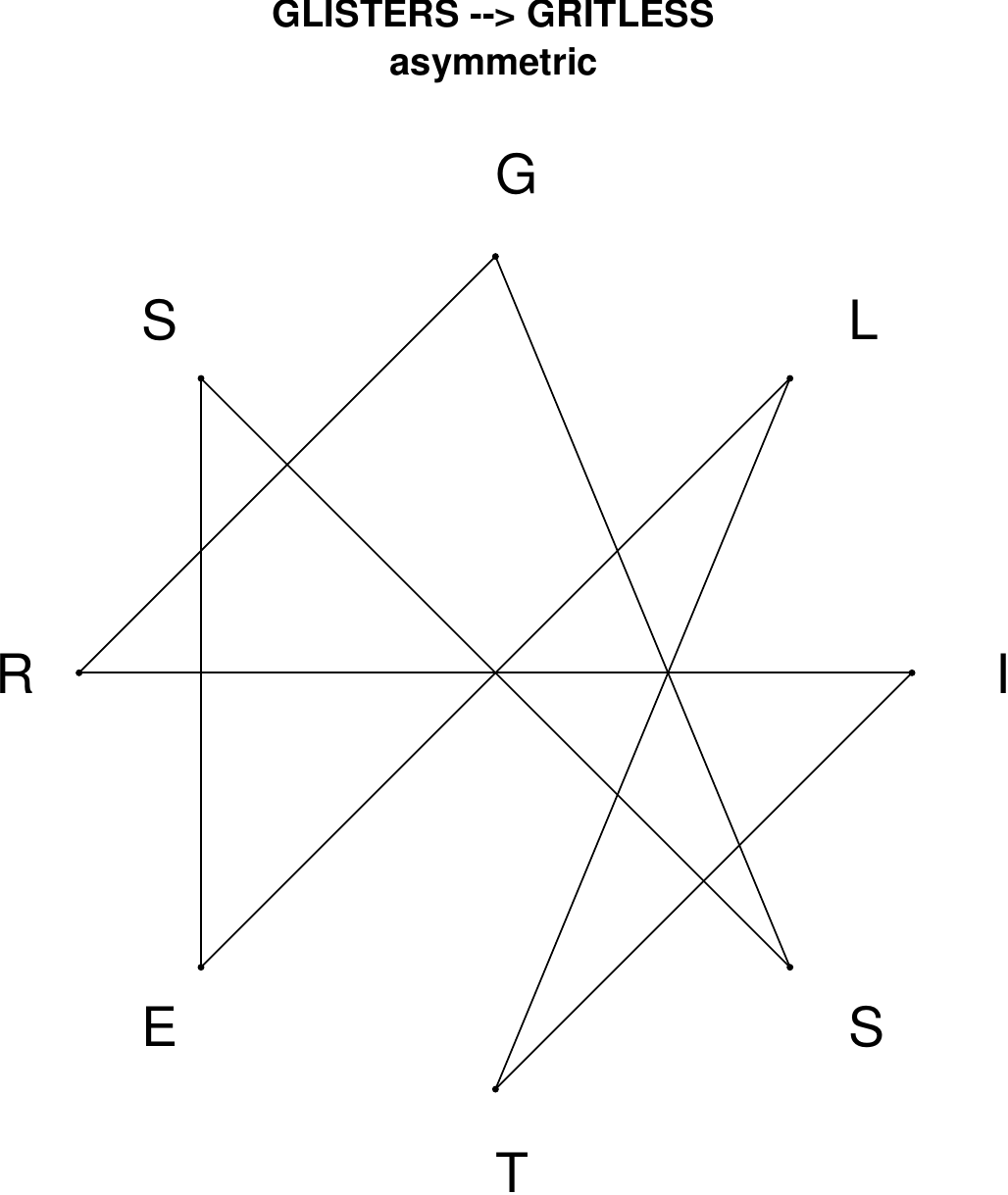}
\end{subfigure}
\hfill
\begin{subfigure}[T]{0.19\textwidth}
\centering
\includegraphics[width=\textwidth]{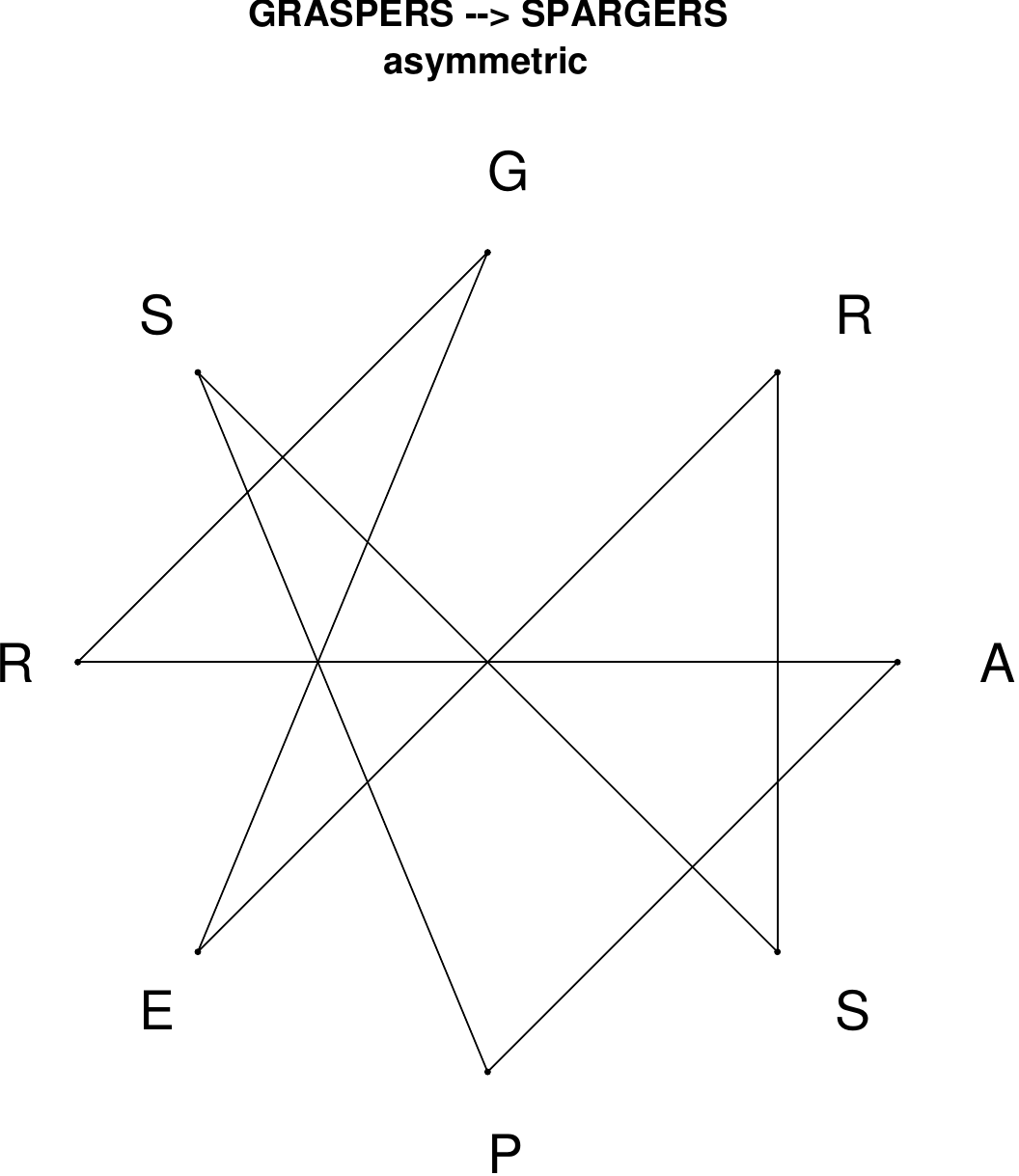}
\end{subfigure}
\hfill
\begin{subfigure}[T]{0.19\textwidth}
\centering
\includegraphics[width=\textwidth]{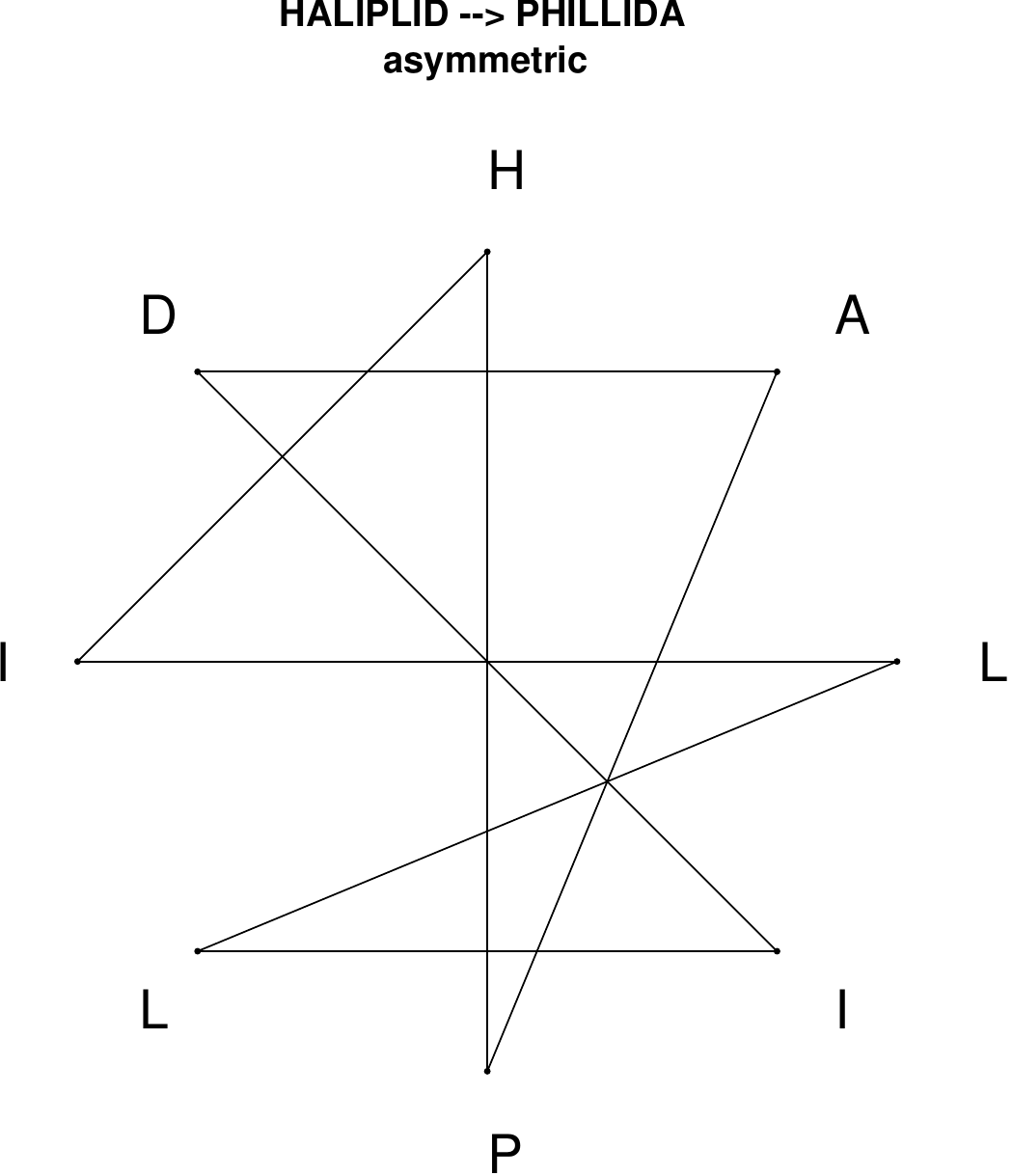}
\end{subfigure}
\end{figure}

\begin{figure}[H]
\centering
\begin{subfigure}[T]{0.19\textwidth}
\centering
\includegraphics[width=\textwidth]{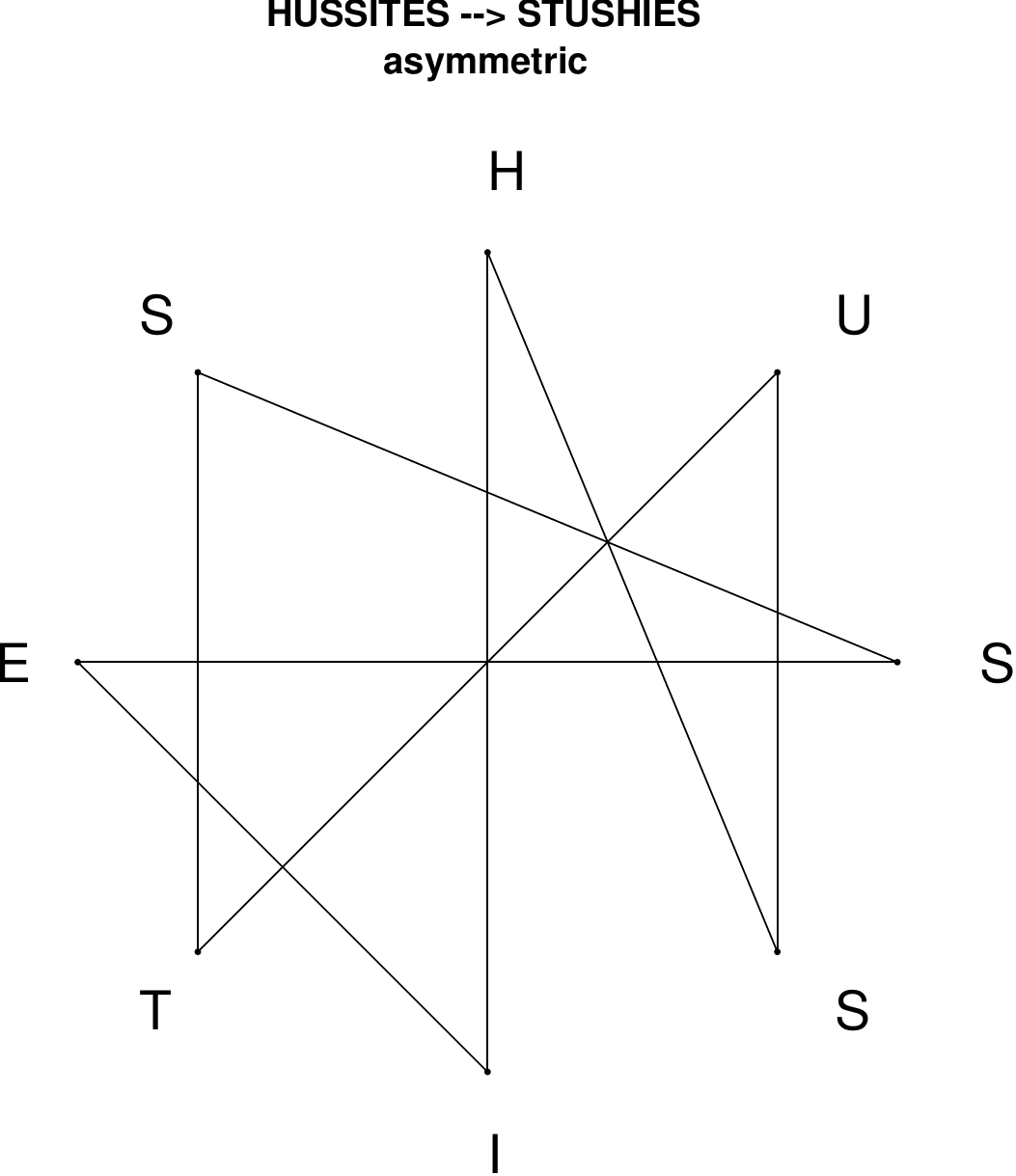}
\end{subfigure}
\hfill
\begin{subfigure}[T]{0.19\textwidth}
\centering
\includegraphics[width=\textwidth]{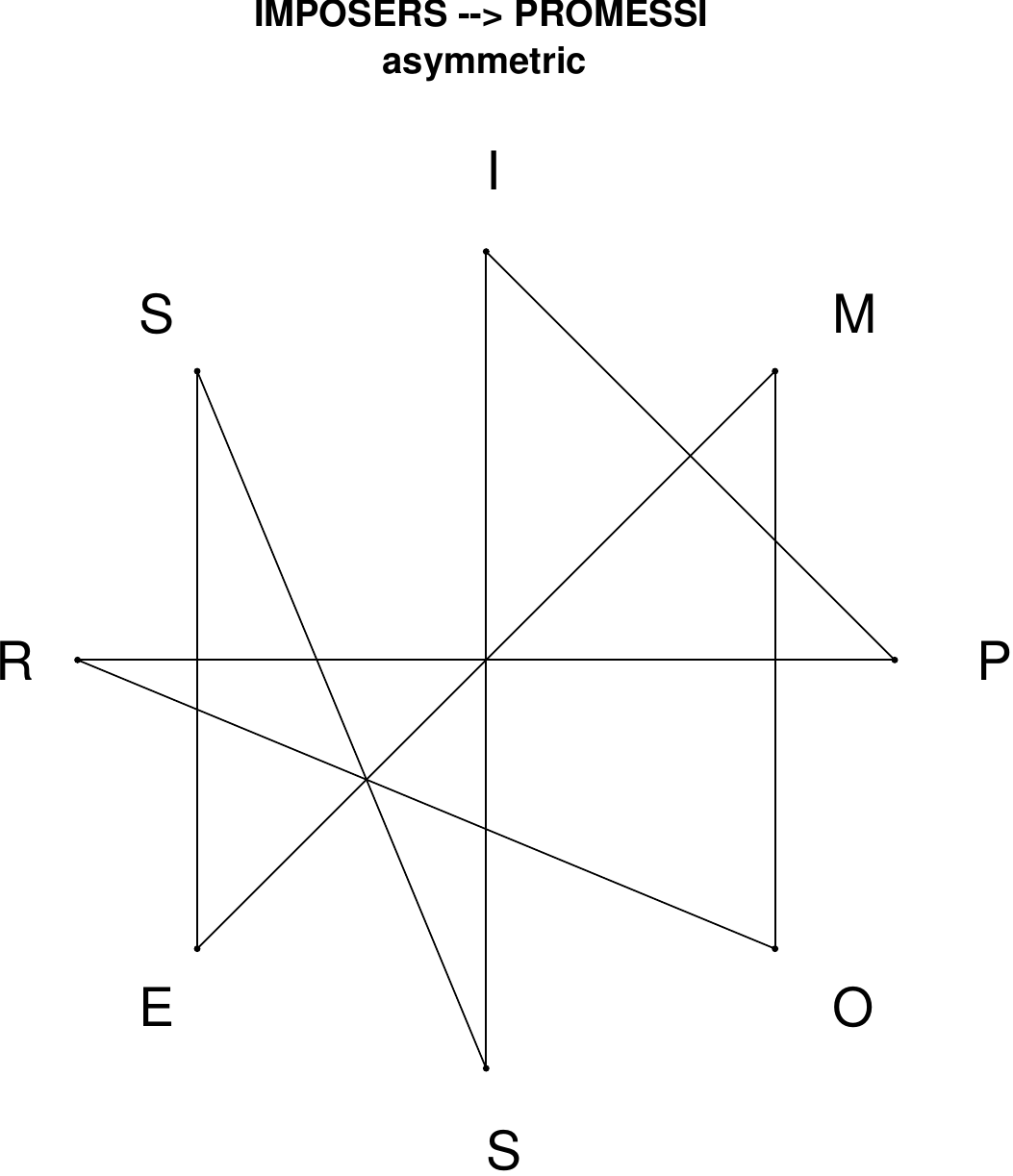}
\end{subfigure}
\hfill
\begin{subfigure}[T]{0.19\textwidth}
\centering
\includegraphics[width=\textwidth]{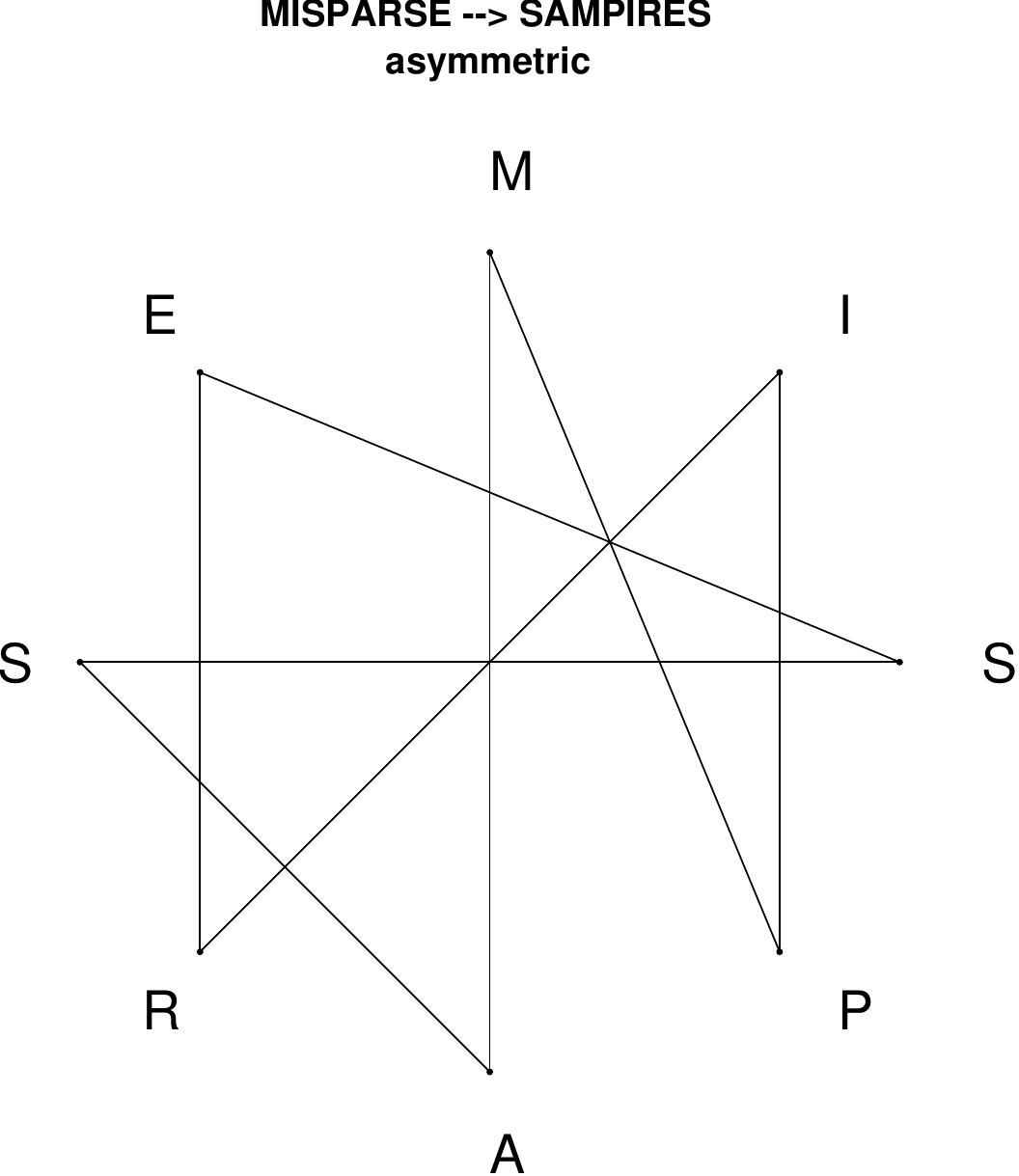}
\end{subfigure}
\hfill
\begin{subfigure}[T]{0.19\textwidth}
\centering
\includegraphics[width=\textwidth]{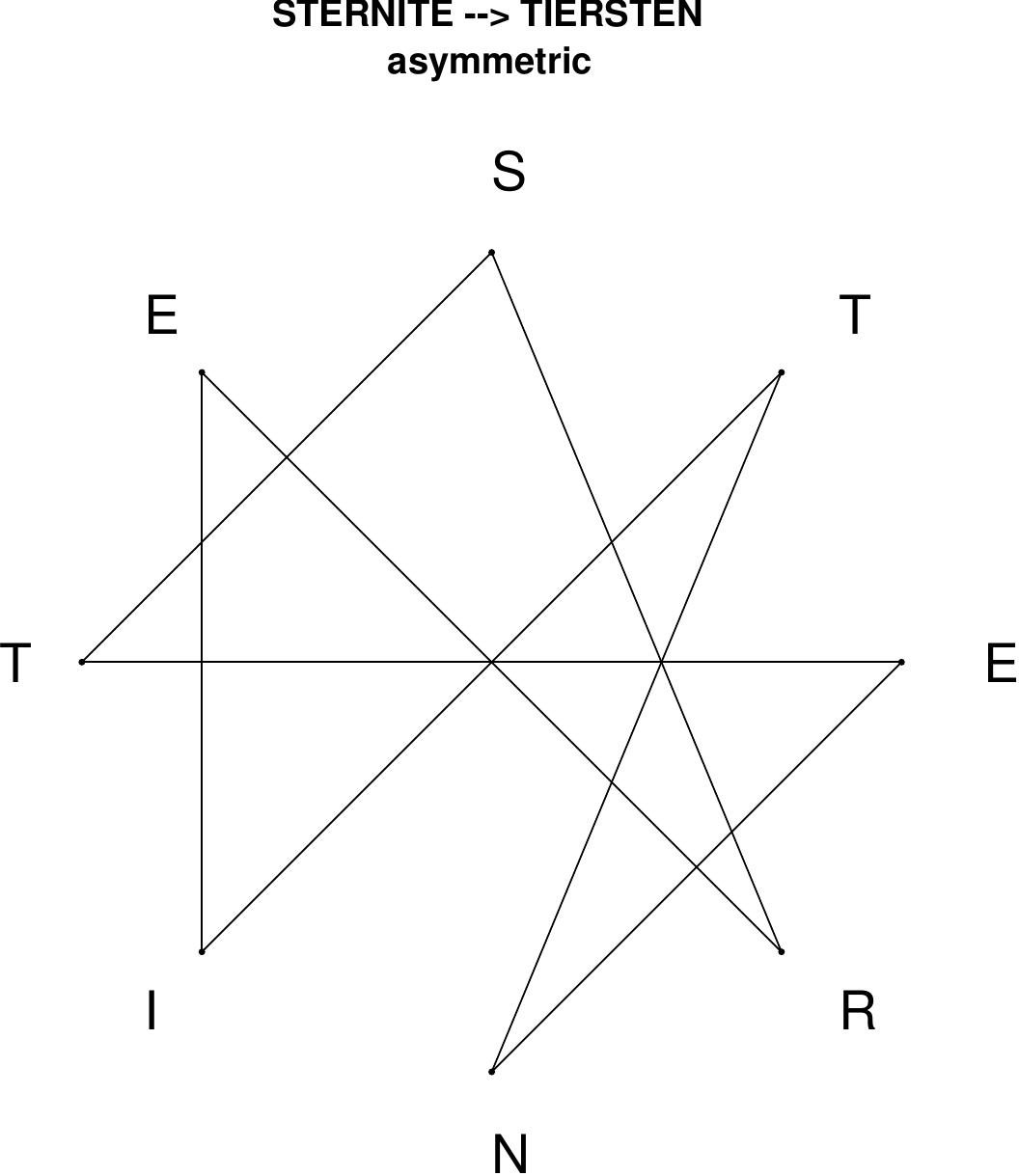}
\end{subfigure}
\hfill
\begin{subfigure}[T]{0.19\textwidth}
\centering
\includegraphics[width=\textwidth]{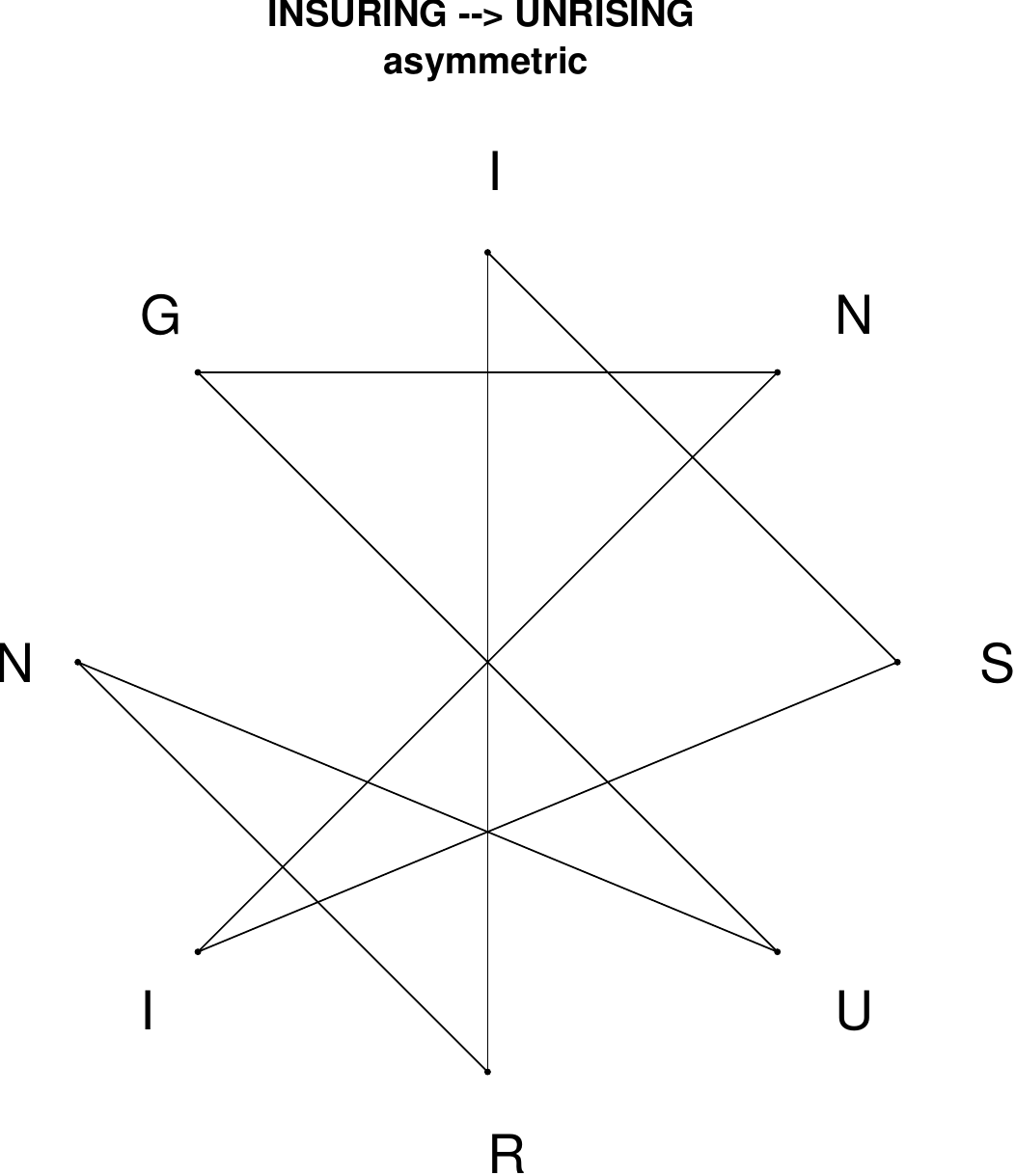}
\end{subfigure}
\end{figure}

\begin{figure}[H]
\centering
\begin{subfigure}[T]{0.19\textwidth}
\centering
\includegraphics[width=\textwidth]{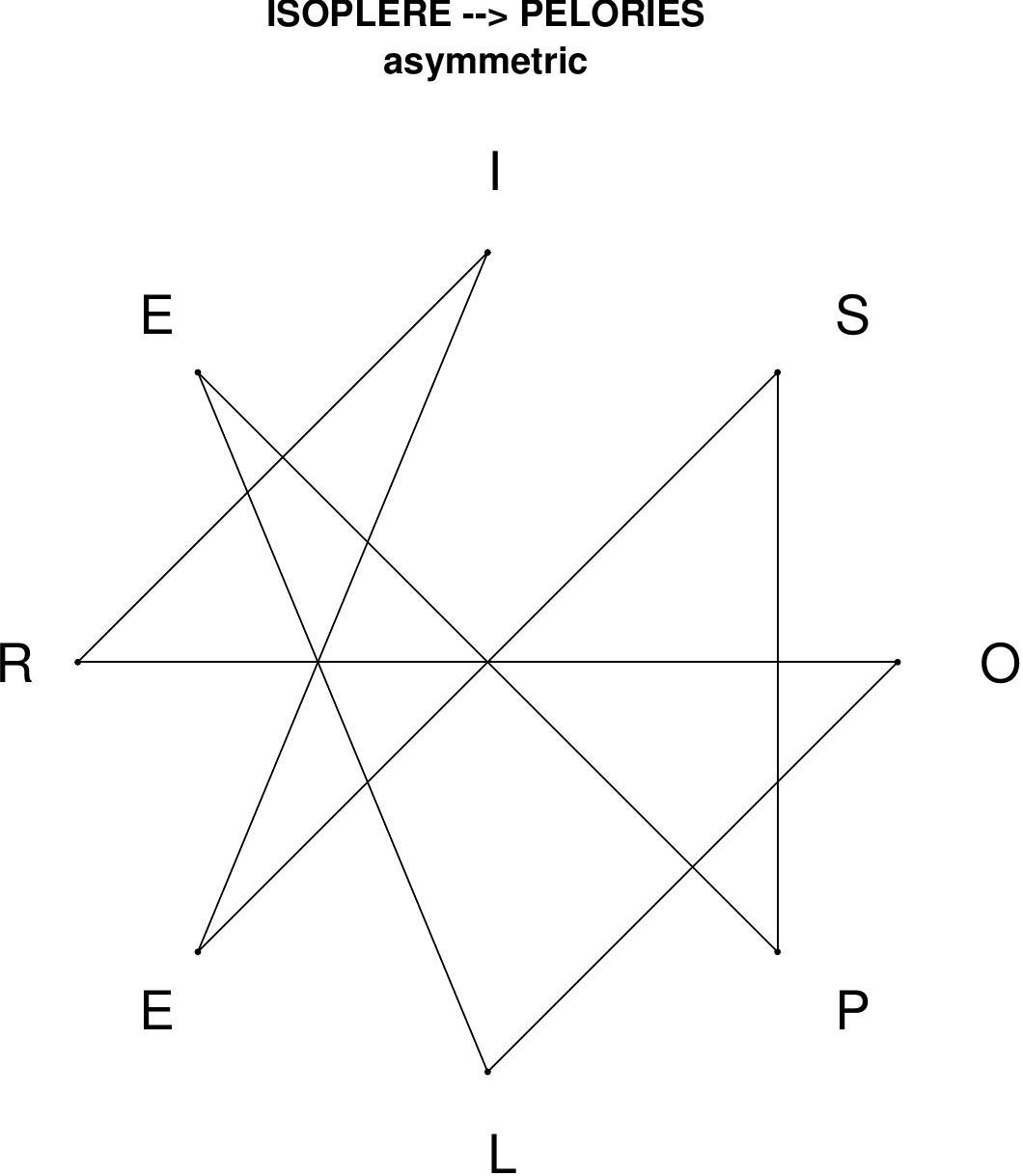}
\end{subfigure}
\hfill
\begin{subfigure}[T]{0.19\textwidth}
\centering
\includegraphics[width=\textwidth]{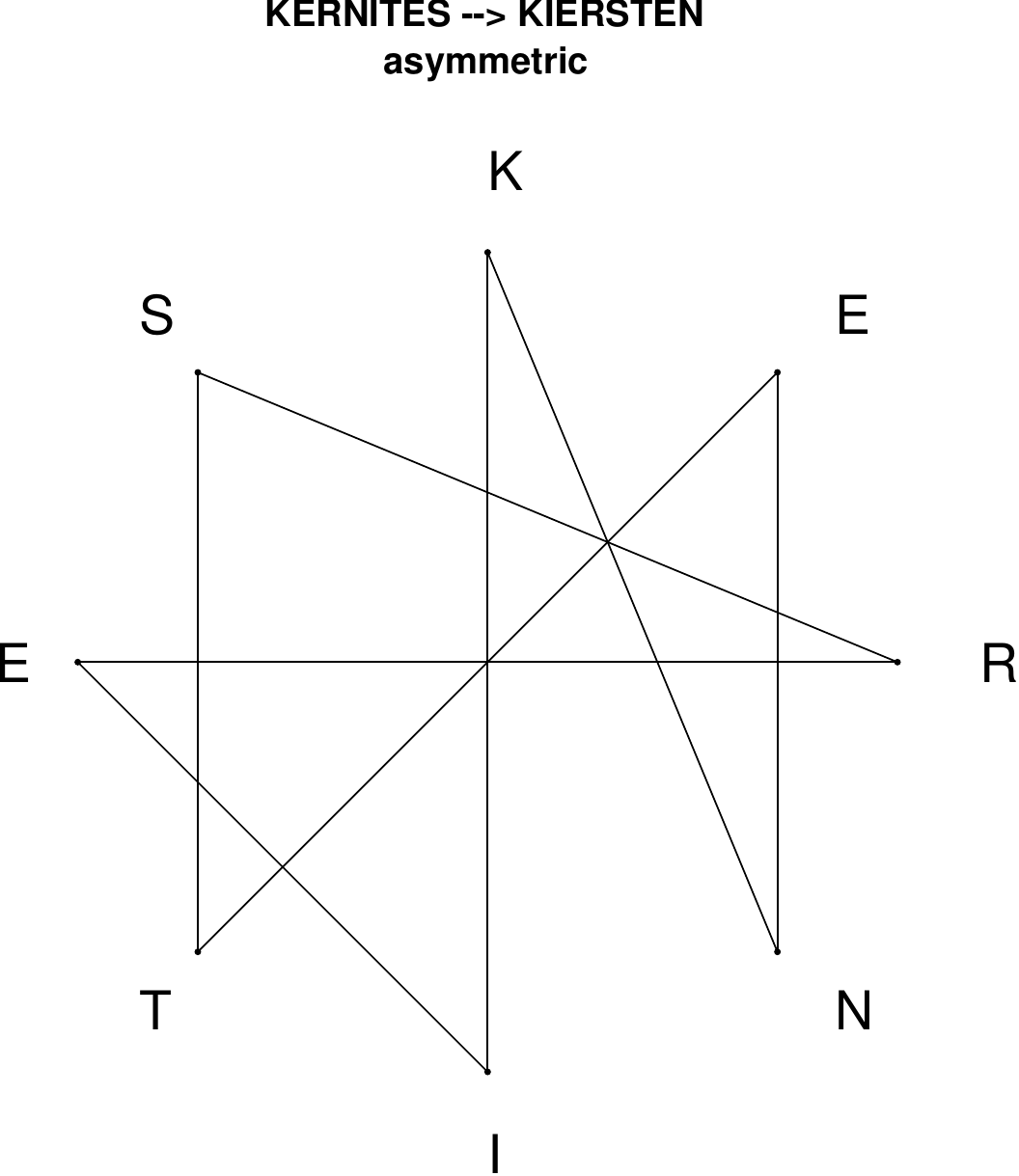}
\end{subfigure}
\hfill
\begin{subfigure}[T]{0.19\textwidth}
\centering
\includegraphics[width=\textwidth]{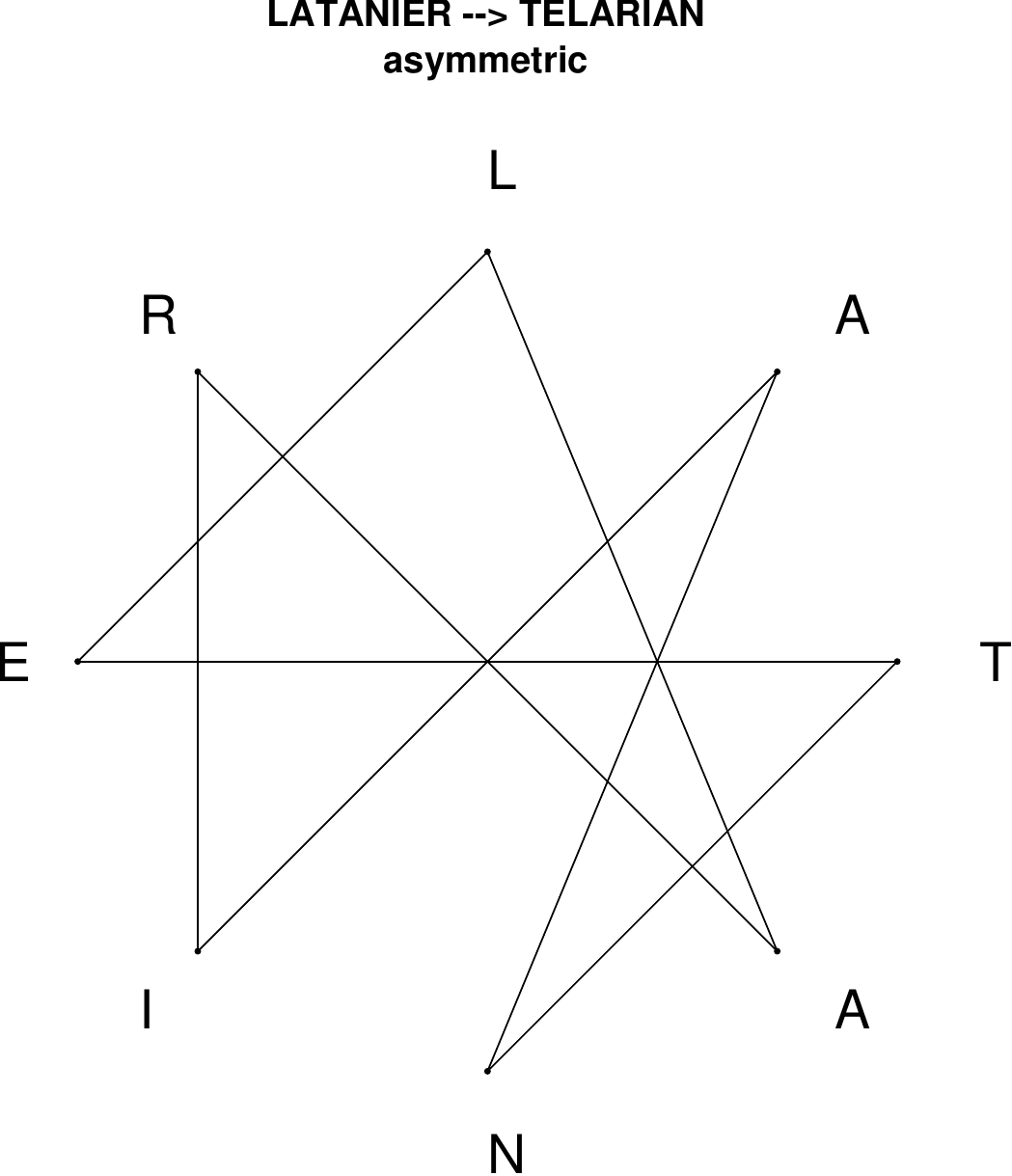}
\end{subfigure}
\hfill
\begin{subfigure}[T]{0.19\textwidth}
\centering
\includegraphics[width=\textwidth]{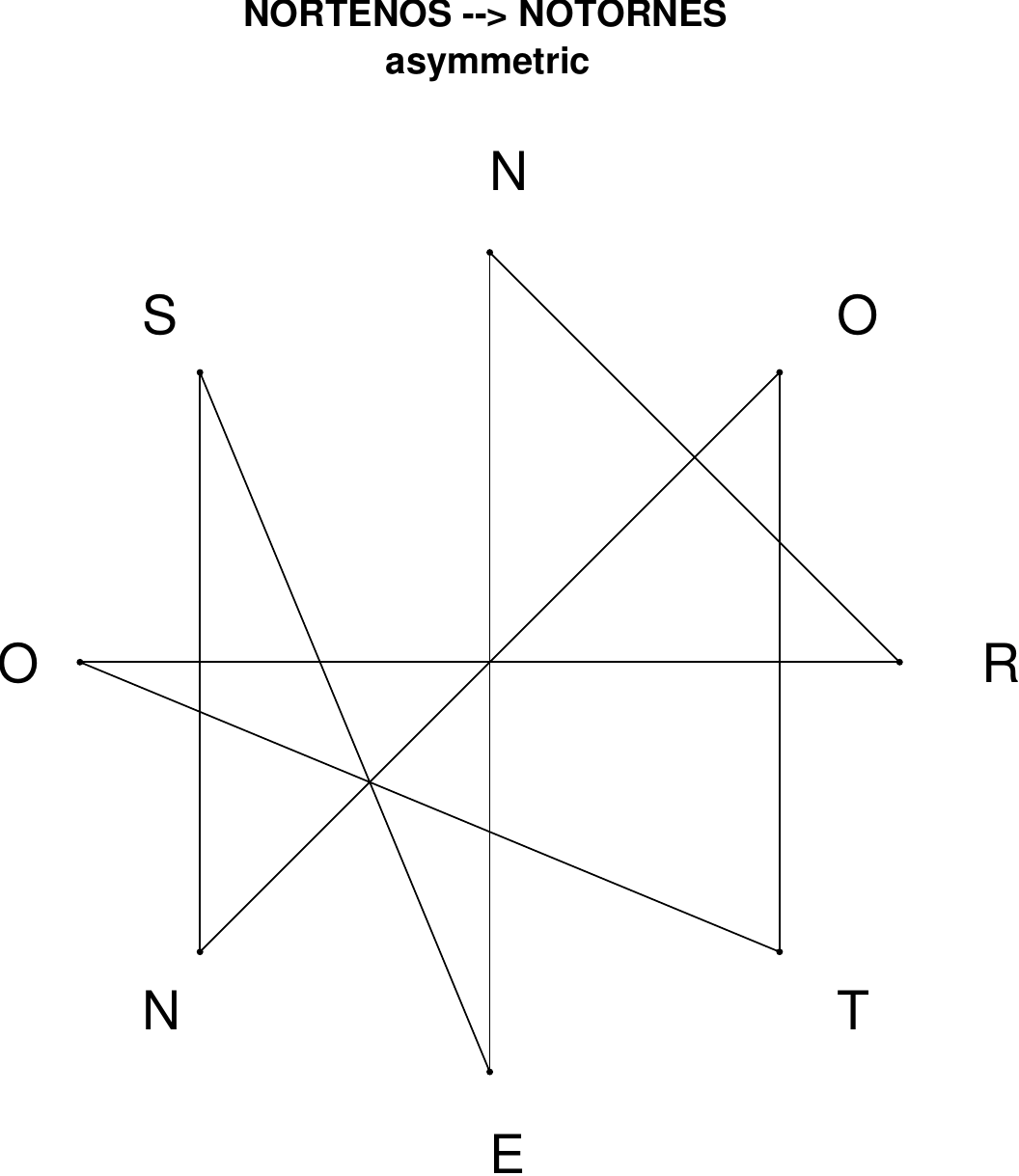}
\end{subfigure}
\hfill
\begin{subfigure}[T]{0.19\textwidth}
\centering
\includegraphics[width=\textwidth]{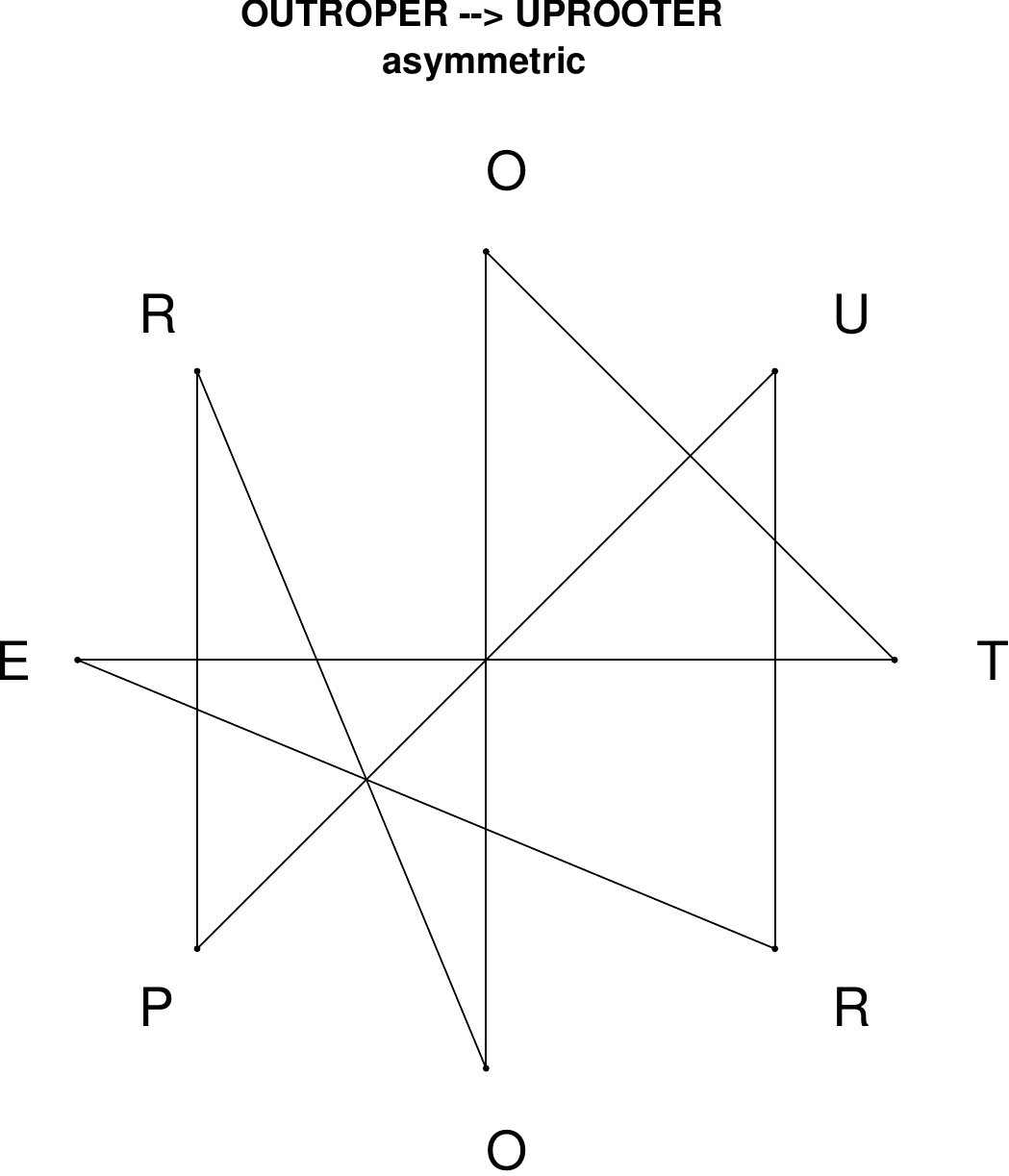}
\end{subfigure}
\end{figure}

\begin{figure}[H]
\centering
\begin{subfigure}[T]{0.19\textwidth}
\centering
\includegraphics[width=\textwidth]{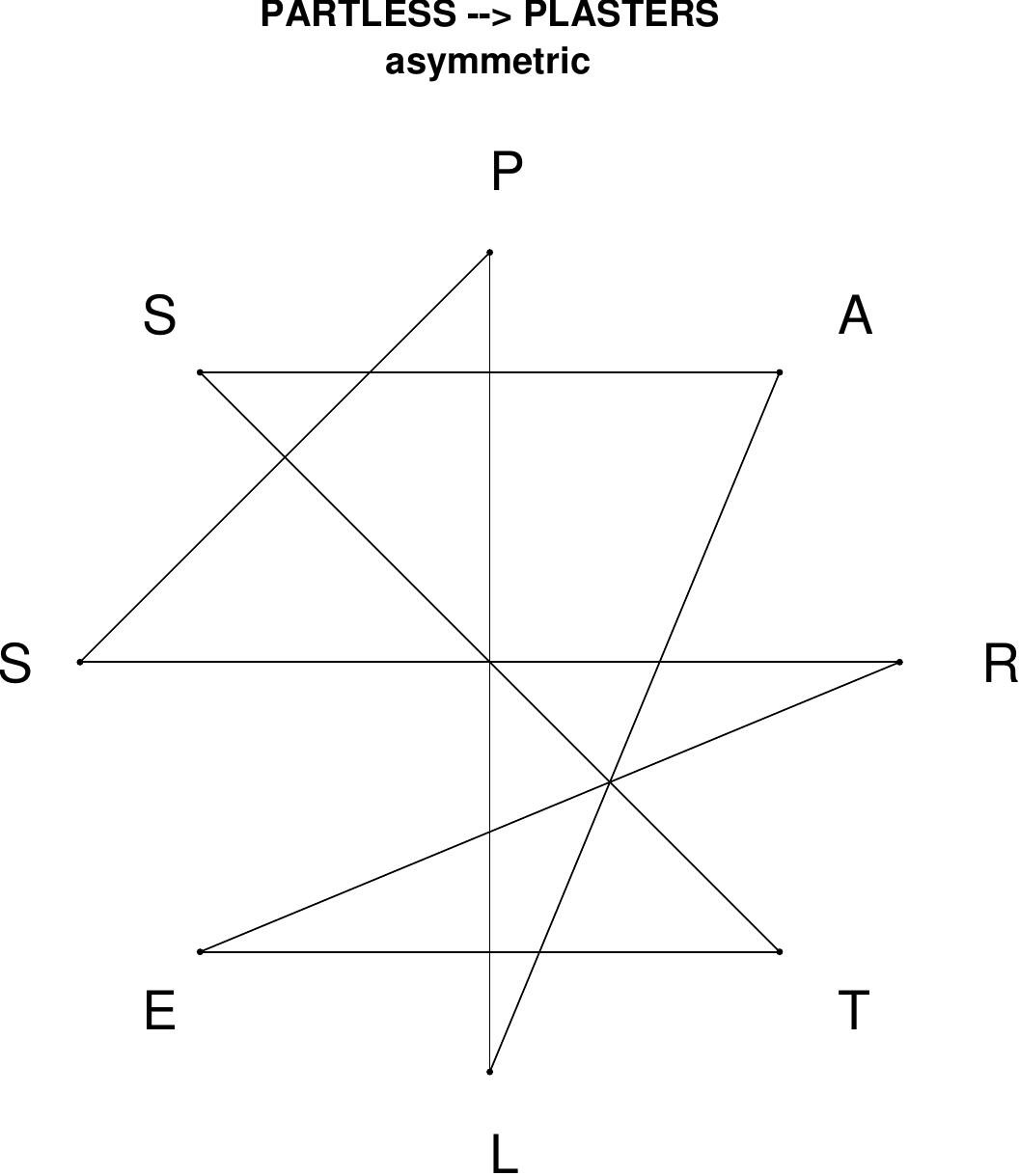}
\end{subfigure}
\hfill
\begin{subfigure}[T]{0.19\textwidth}
\centering
\includegraphics[width=\textwidth]{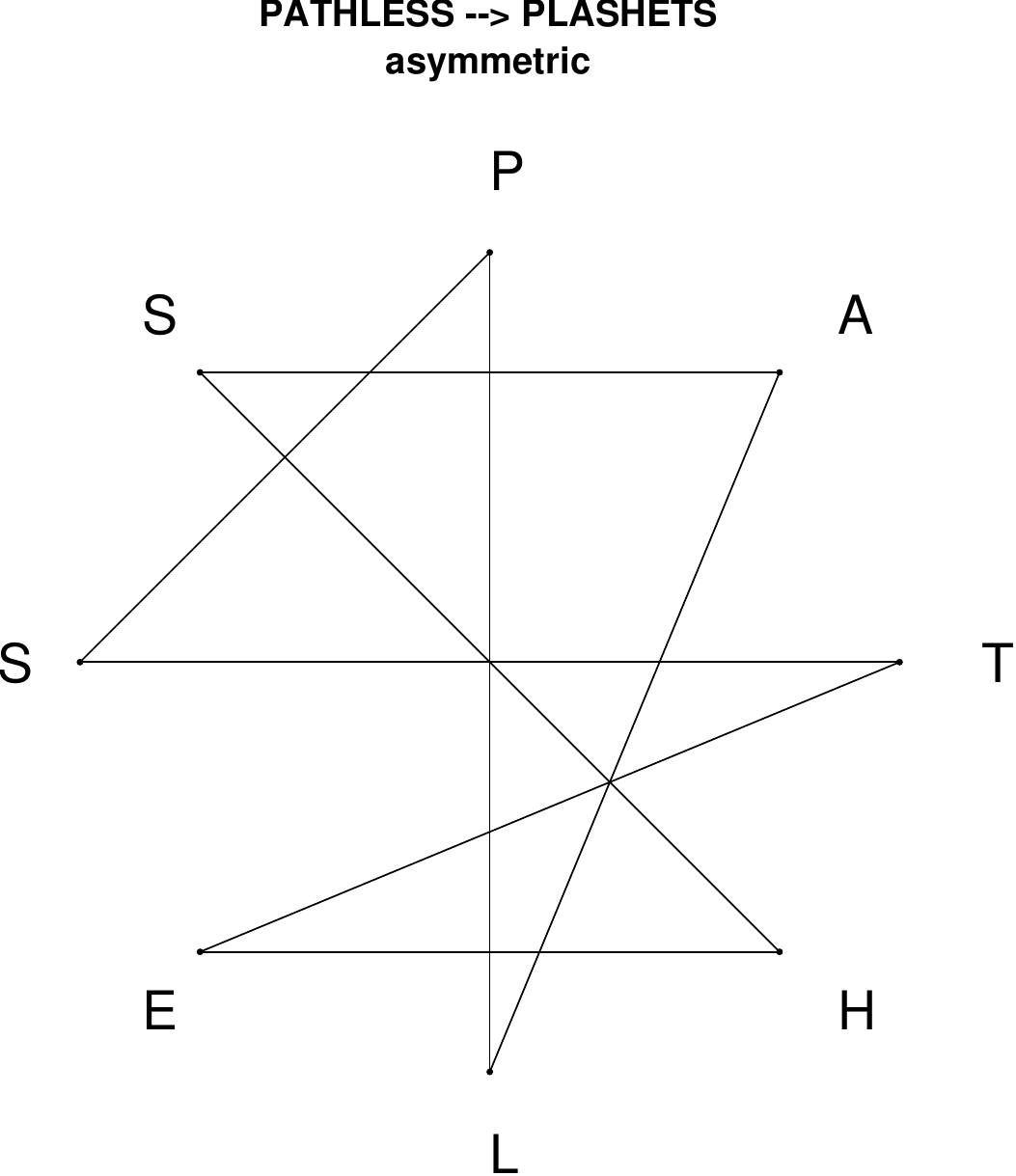}
\end{subfigure}
\hfill
\begin{subfigure}[T]{0.19\textwidth}
\centering
\includegraphics[width=\textwidth]{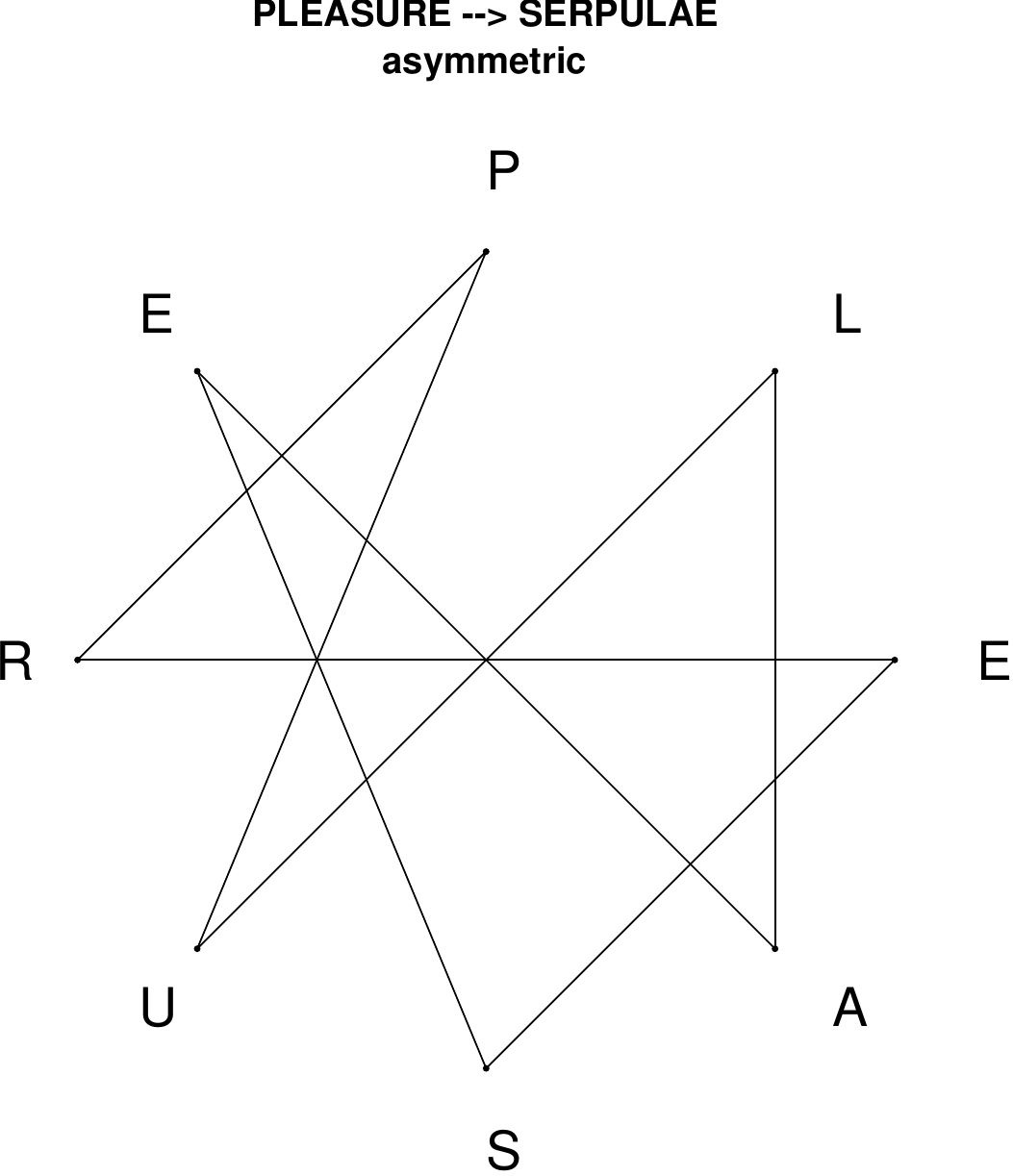}
\end{subfigure}
\hfill
\begin{subfigure}[T]{0.19\textwidth}
\centering
\includegraphics[width=\textwidth]{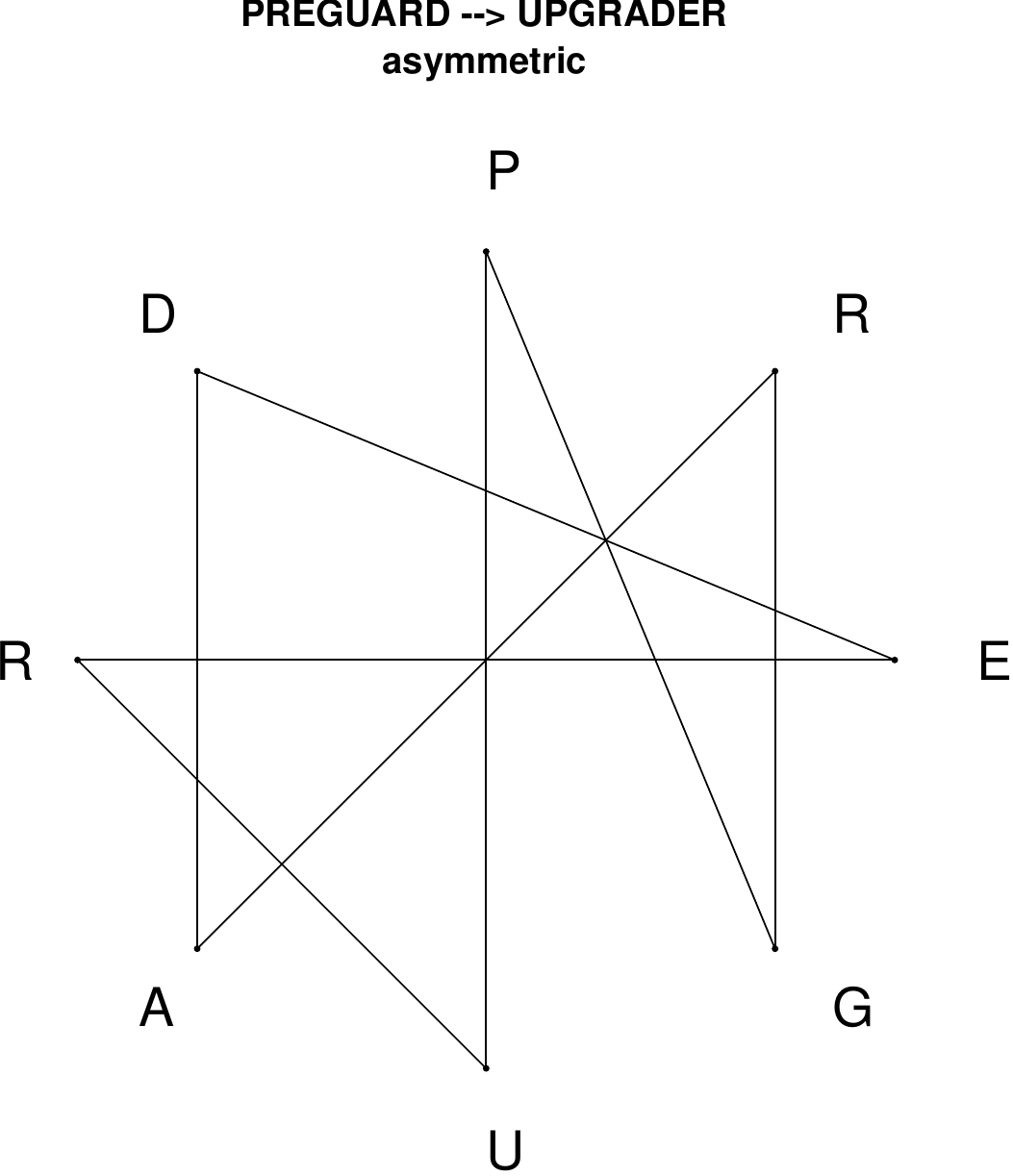}
\end{subfigure}
\hfill
\begin{subfigure}[T]{0.19\textwidth}
\centering
\includegraphics[width=\textwidth]{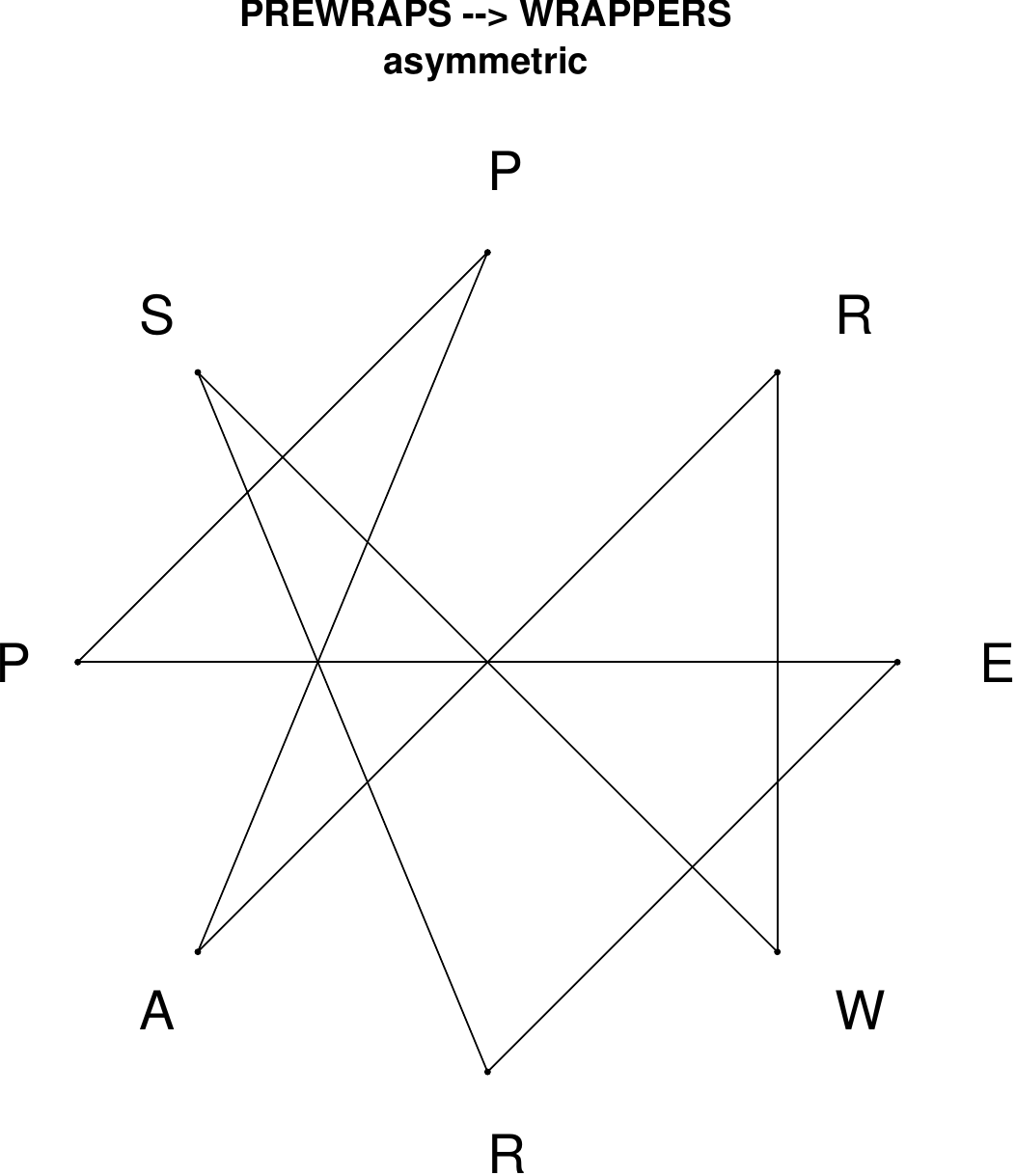}
\end{subfigure}
\end{figure}

\begin{figure}[H]
\centering
\begin{subfigure}[T]{0.19\textwidth}
\centering
\includegraphics[width=\textwidth]{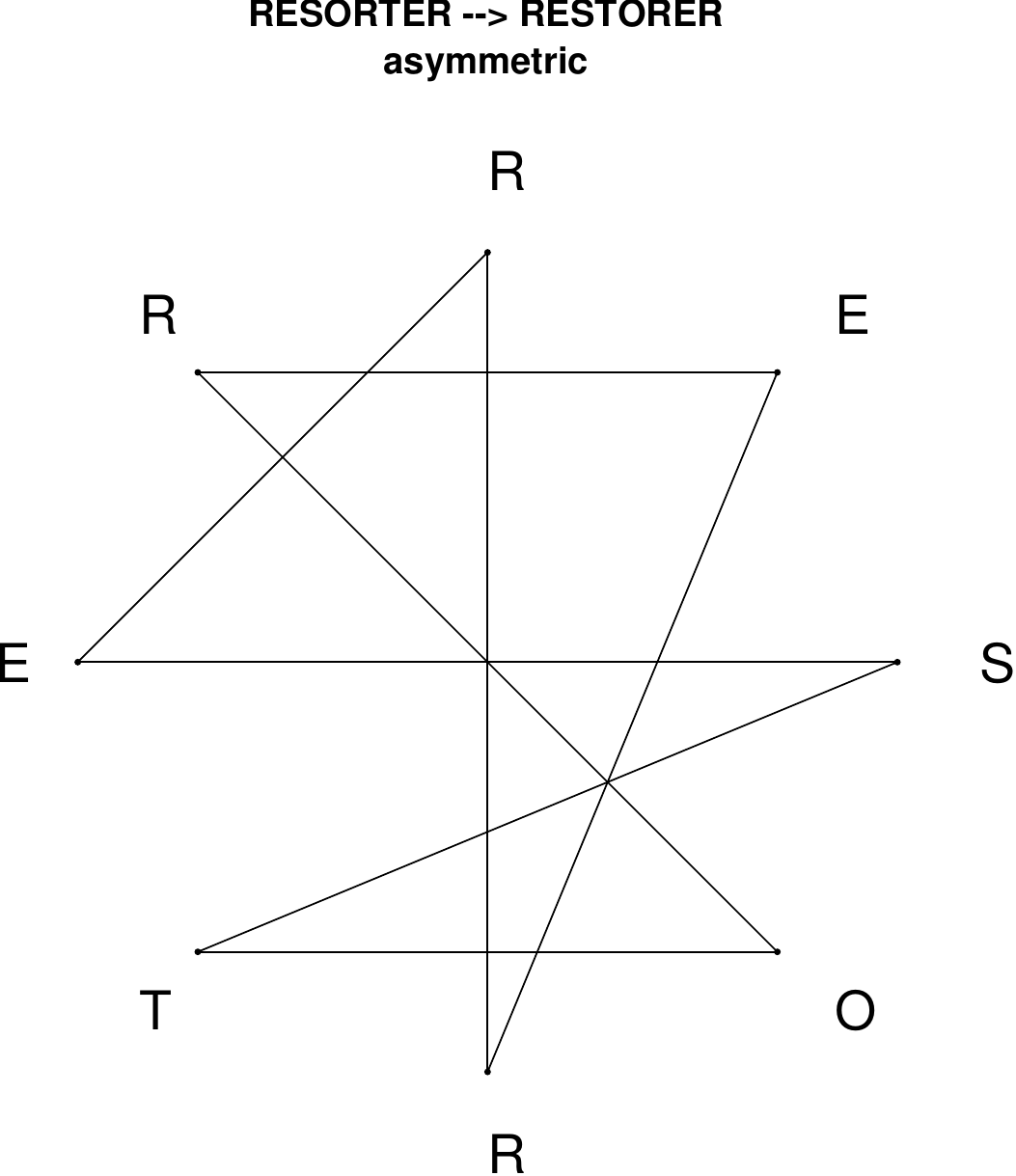}
\end{subfigure}
\hfill
\begin{subfigure}[T]{0.19\textwidth}
\centering
\includegraphics[width=\textwidth]{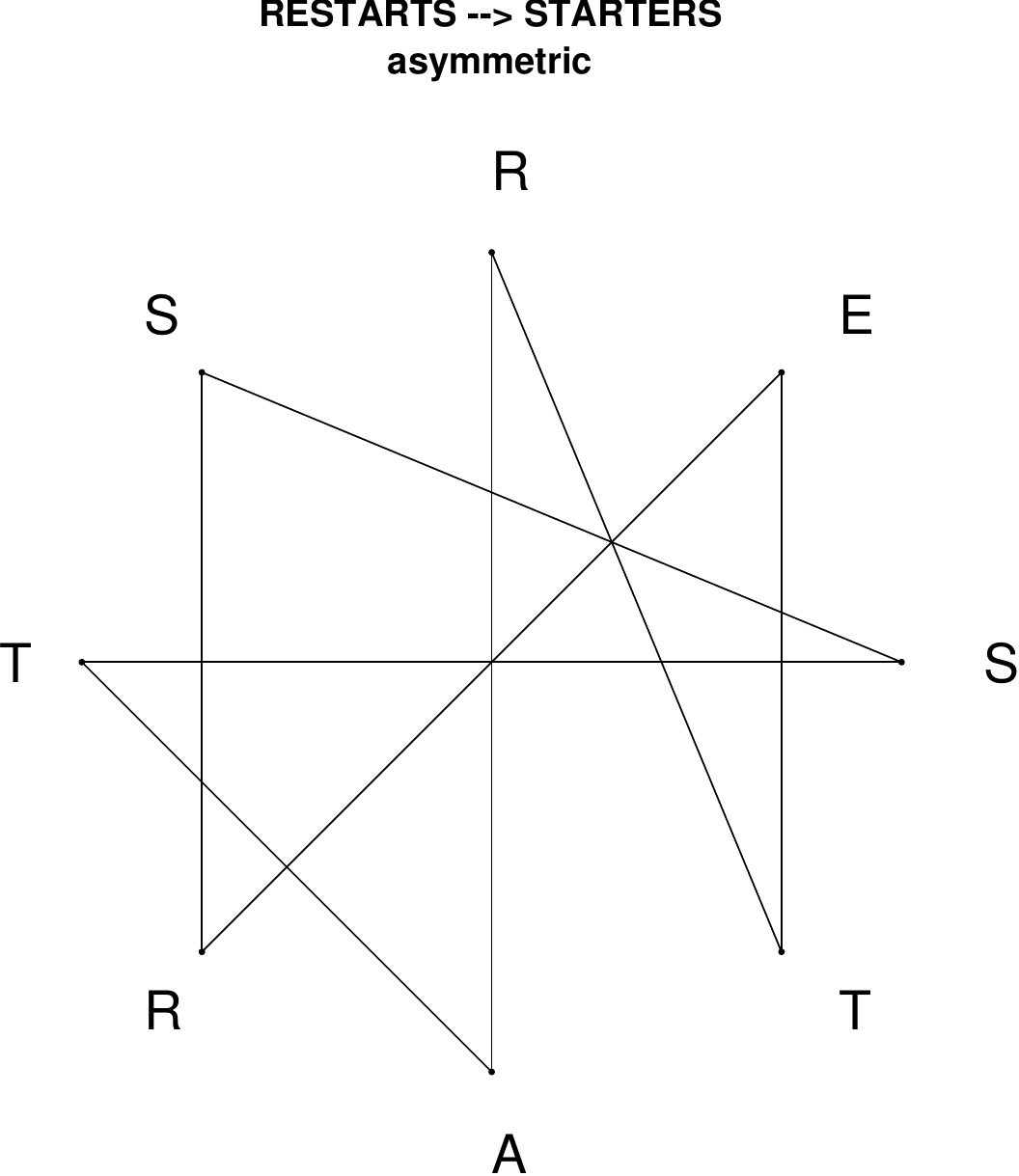}
\end{subfigure}
\hfill
\begin{subfigure}[T]{0.19\textwidth}
\centering
\includegraphics[width=\textwidth]{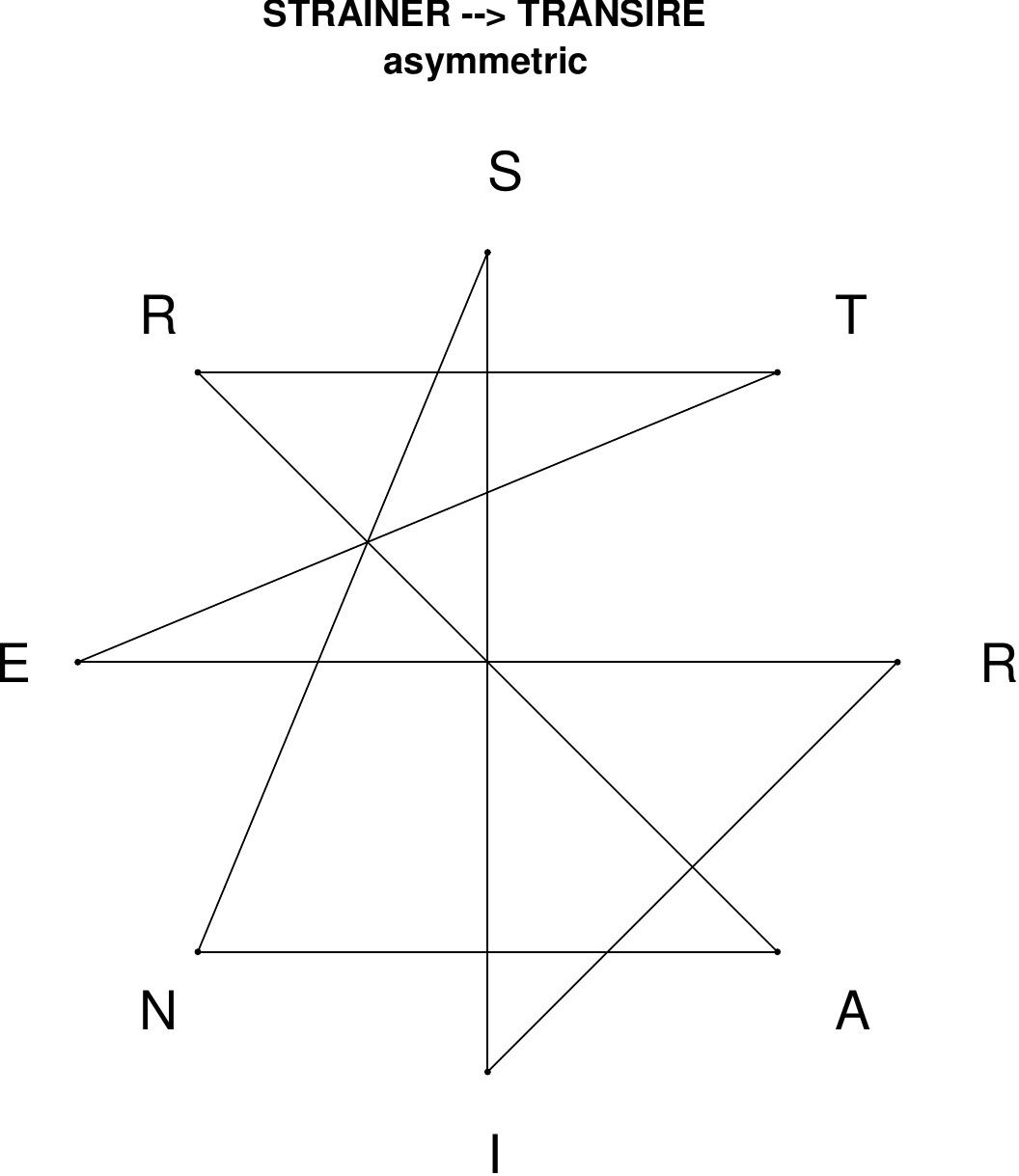}
\end{subfigure}
\hfill
\begin{subfigure}[T]{0.19\textwidth}
\centering
\includegraphics[width=\textwidth]{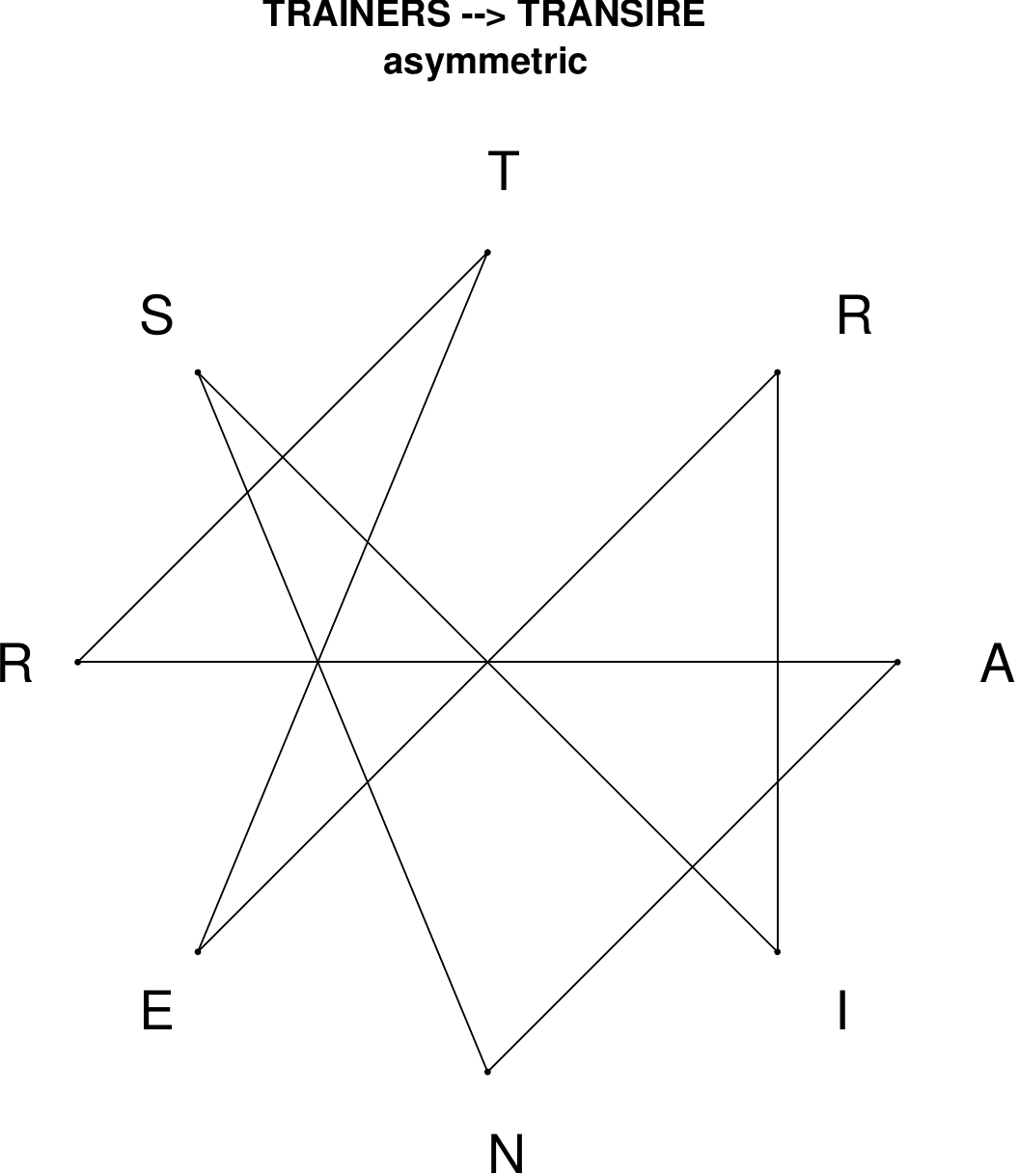}
\end{subfigure}
\hfill
\begin{subfigure}[T]{0.19\textwidth}
\centering
\includegraphics[width=\textwidth]{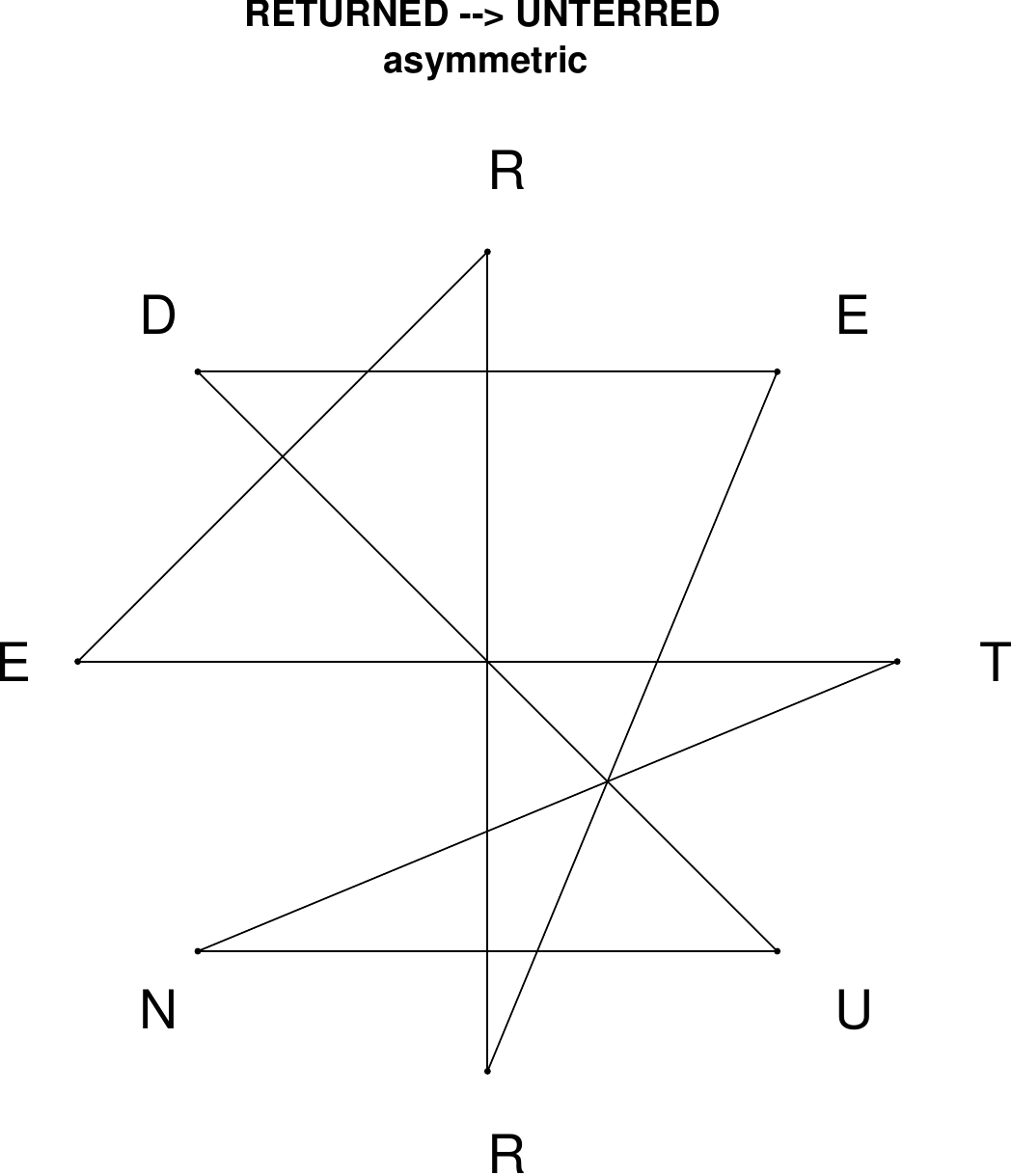}
\end{subfigure}
\end{figure}

\begin{figure}[H]
\centering
\begin{subfigure}[T]{0.19\textwidth}
\centering
\includegraphics[width=\textwidth]{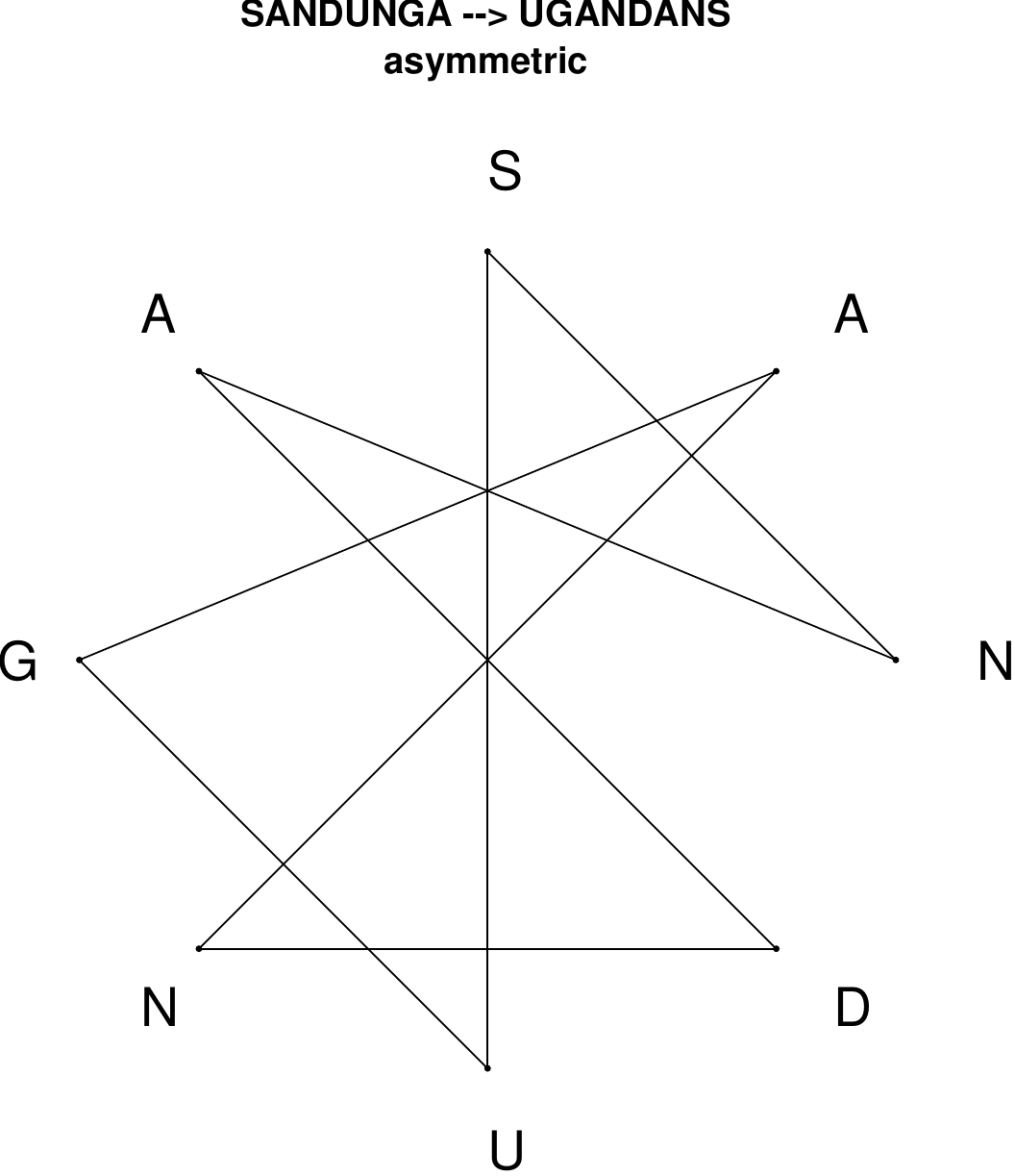}
\end{subfigure}
\hfill
\begin{subfigure}[T]{0.19\textwidth}
\centering
\includegraphics[width=\textwidth]{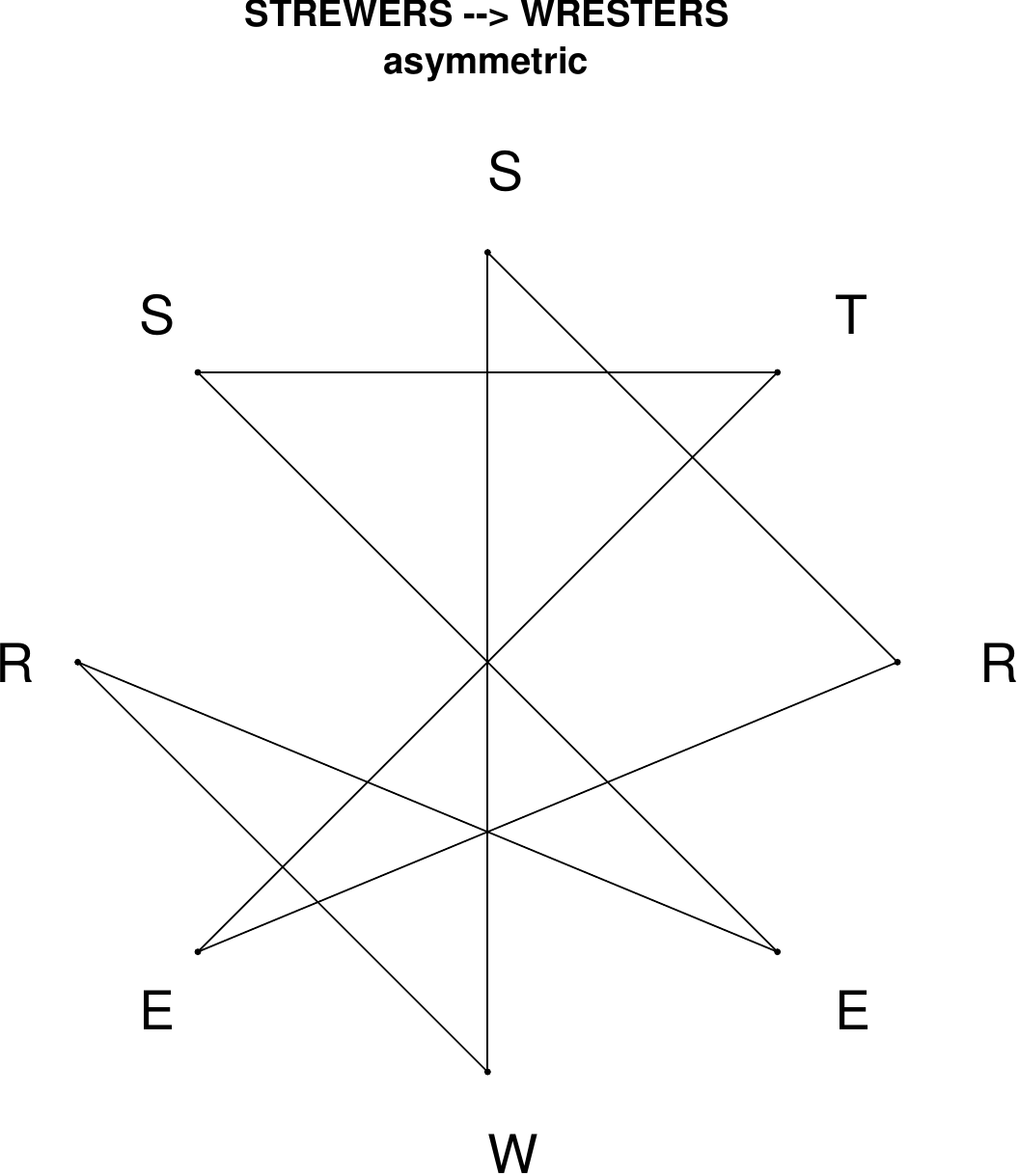}
\end{subfigure}
\hfill
\begin{subfigure}[T]{0.19\textwidth}
\centering
\includegraphics[width=\textwidth]{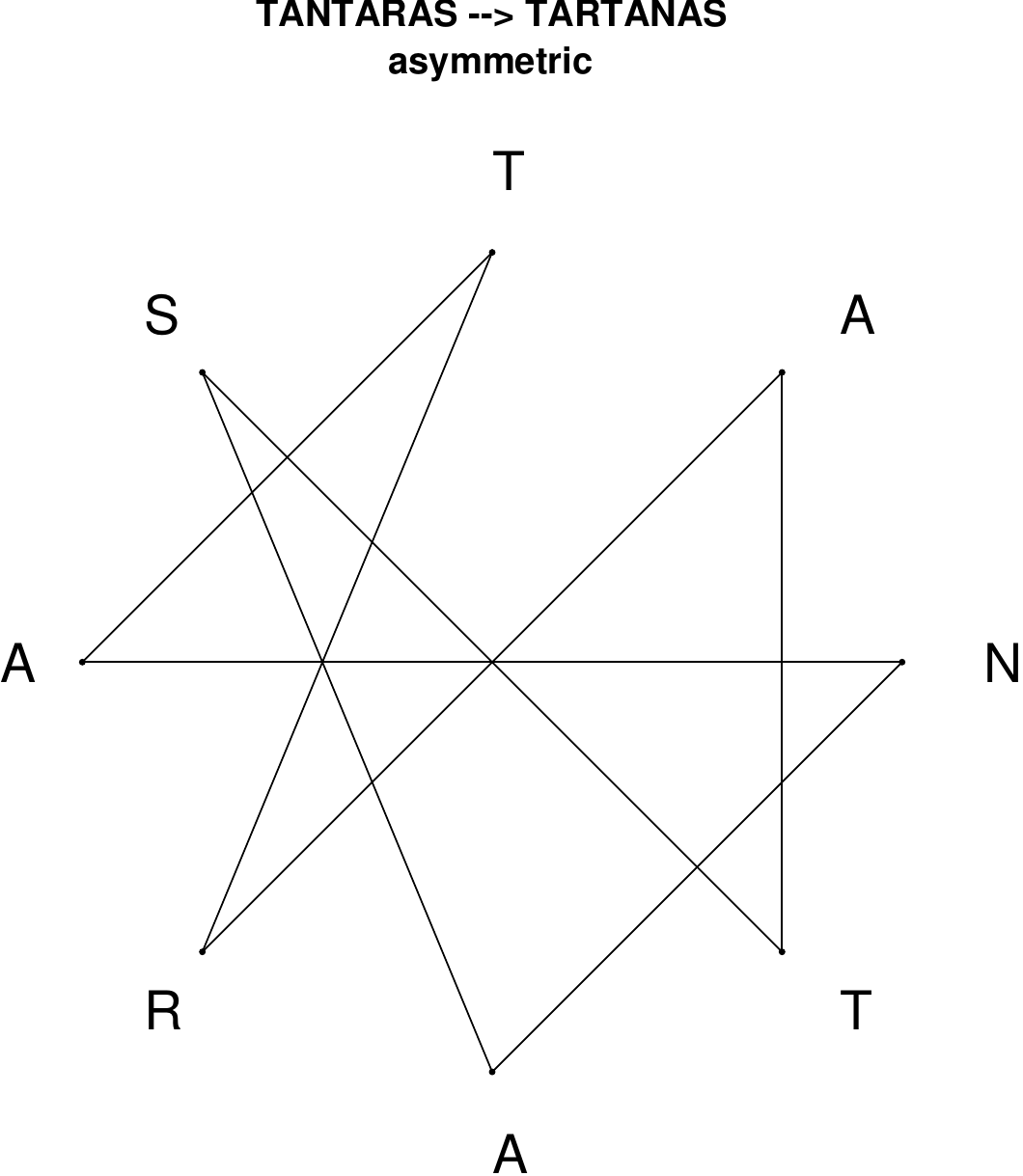}
\end{subfigure}
\hfill
\begin{subfigure}[T]{0.19\textwidth}
\centering
\includegraphics[width=\textwidth]{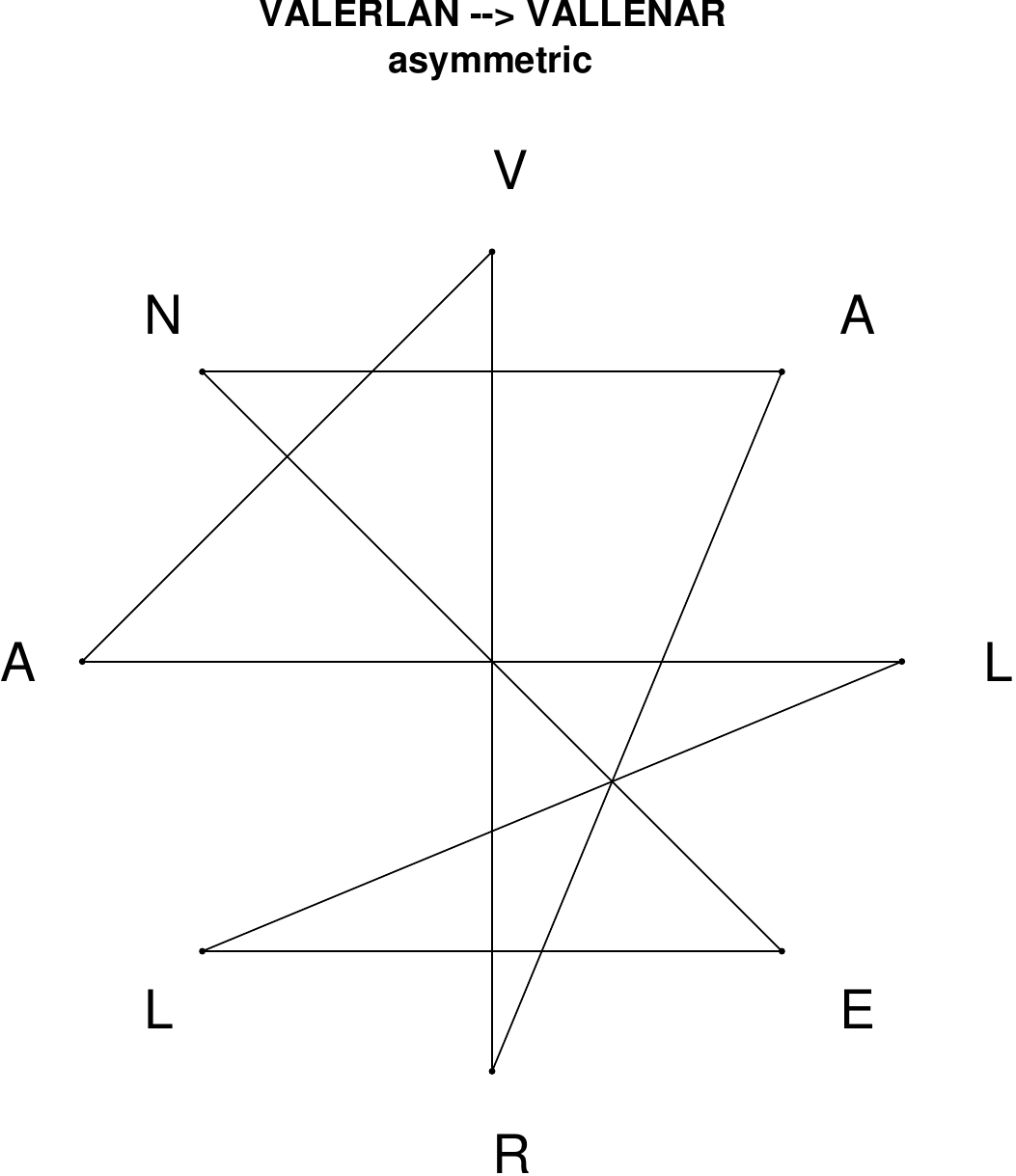}
\end{subfigure}
\hfill
\begin{subfigure}[T]{0.19\textwidth}
\centering
\includegraphics[width=\textwidth]{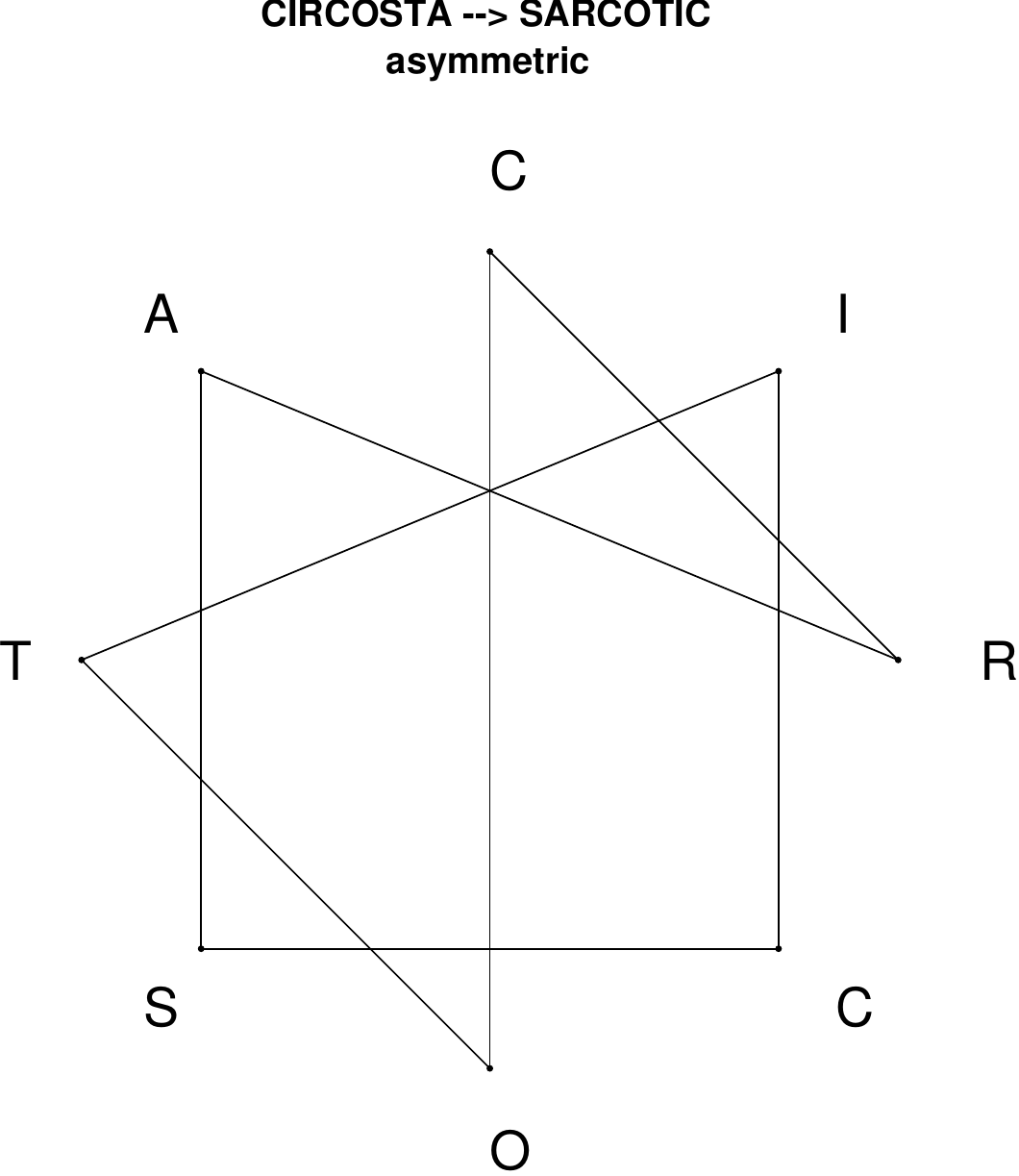}
\end{subfigure}
\end{figure}

\begin{figure}[H]
\centering
\begin{subfigure}[T]{0.19\textwidth}
\centering
\includegraphics[width=\textwidth]{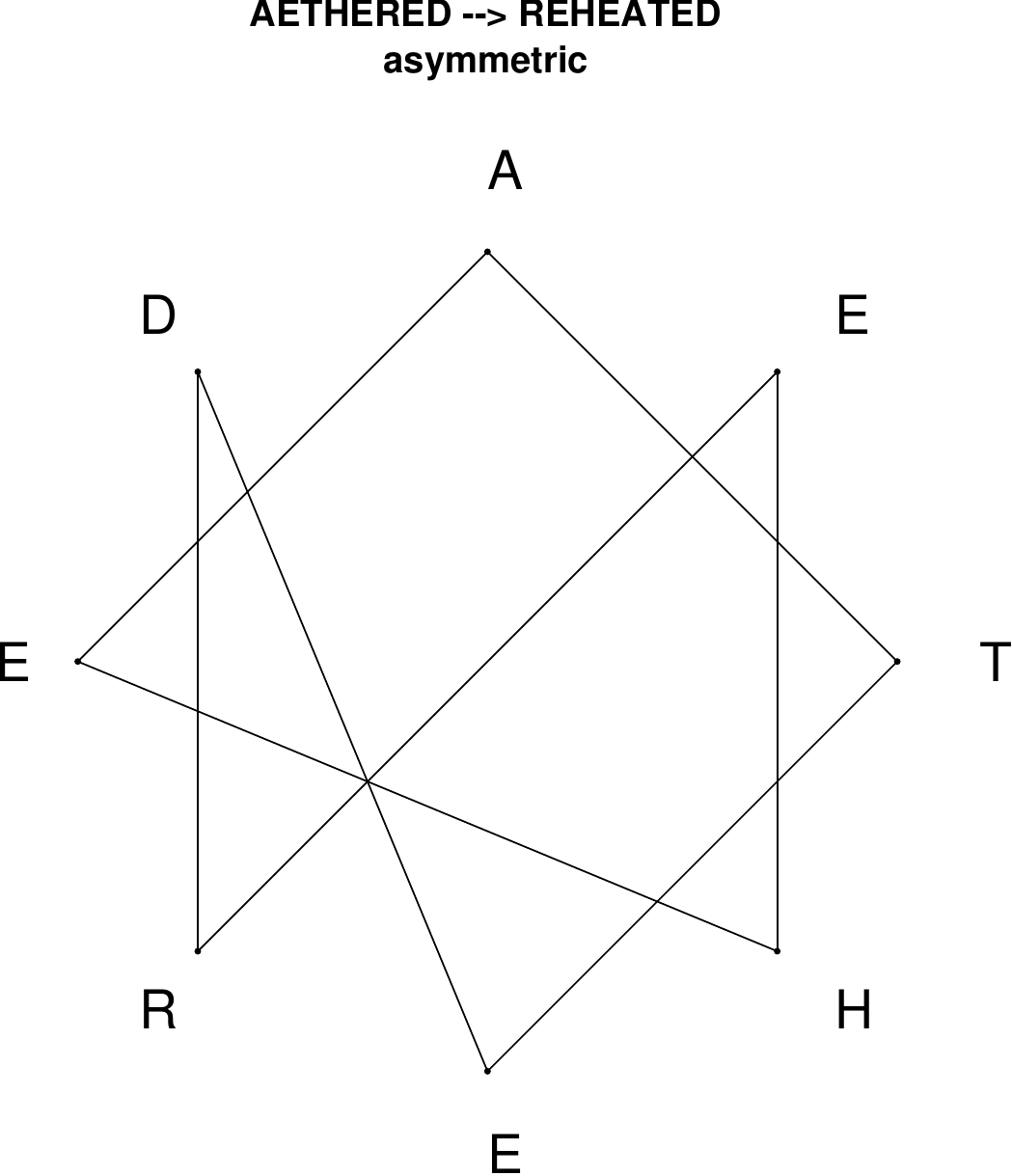}
\end{subfigure}
\hfill
\begin{subfigure}[T]{0.19\textwidth}
\centering
\includegraphics[width=\textwidth]{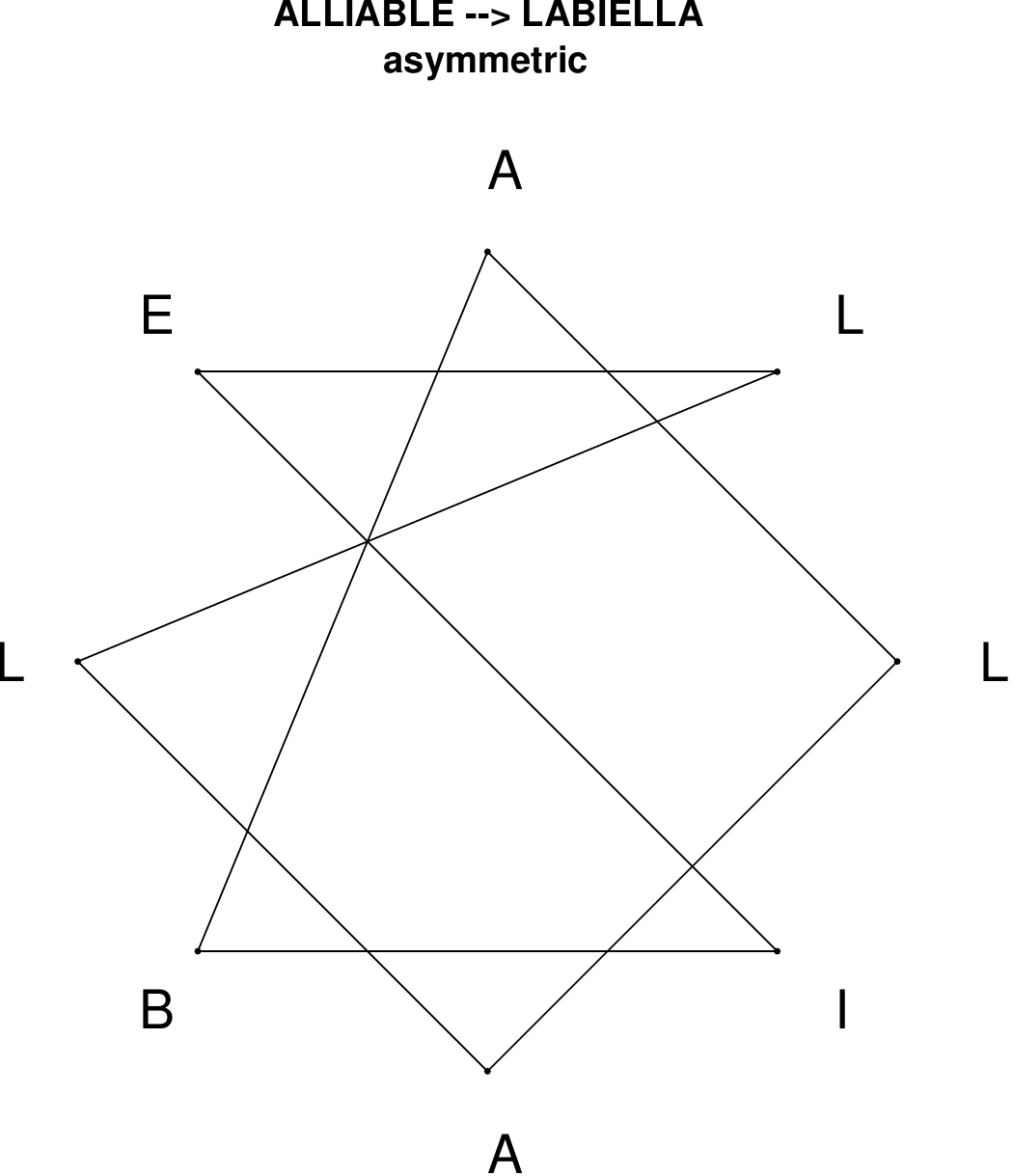}
\end{subfigure}
\hfill
\begin{subfigure}[T]{0.19\textwidth}
\centering
\includegraphics[width=\textwidth]{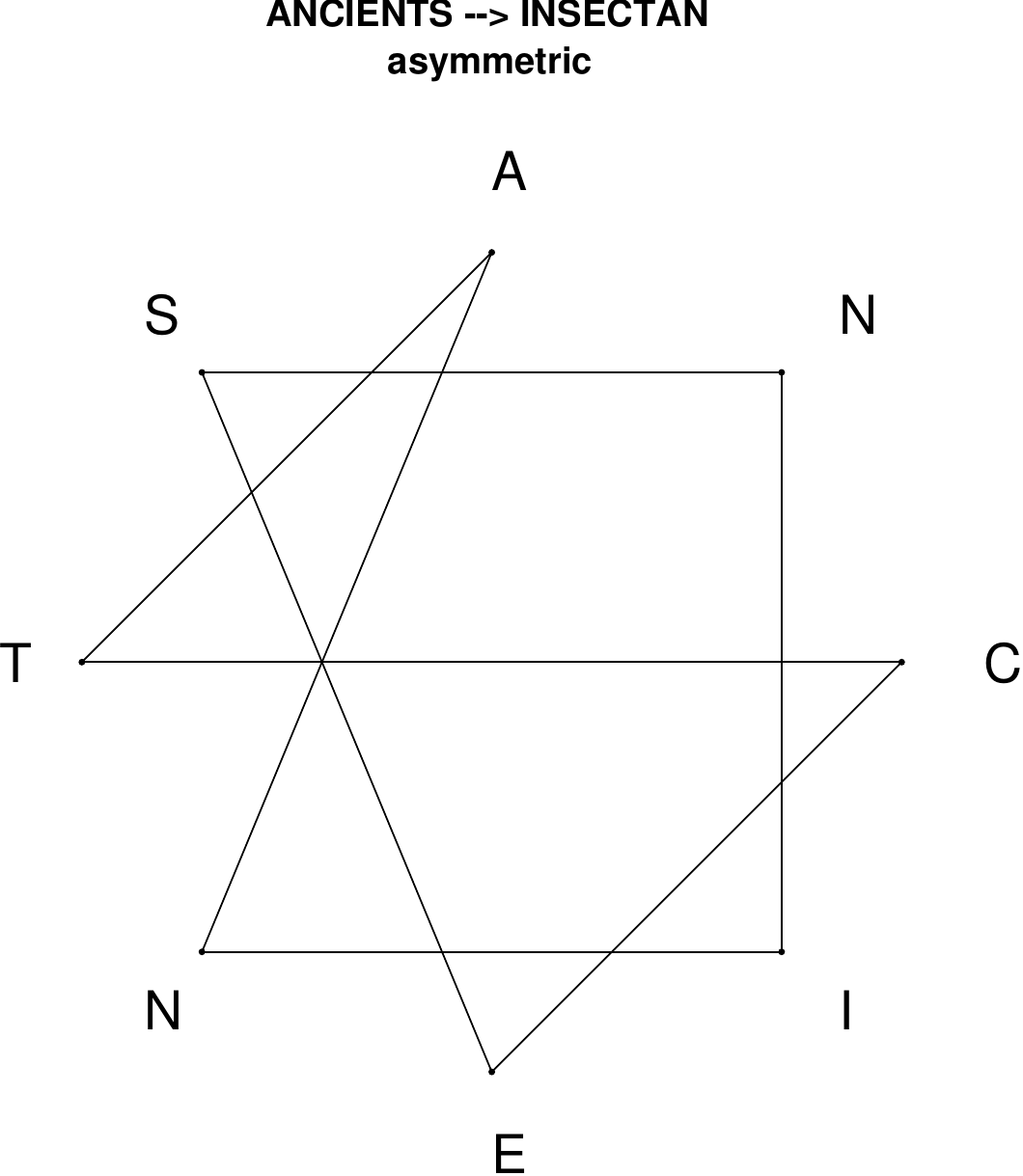}
\end{subfigure}
\hfill
\begin{subfigure}[T]{0.19\textwidth}
\centering
\includegraphics[width=\textwidth]{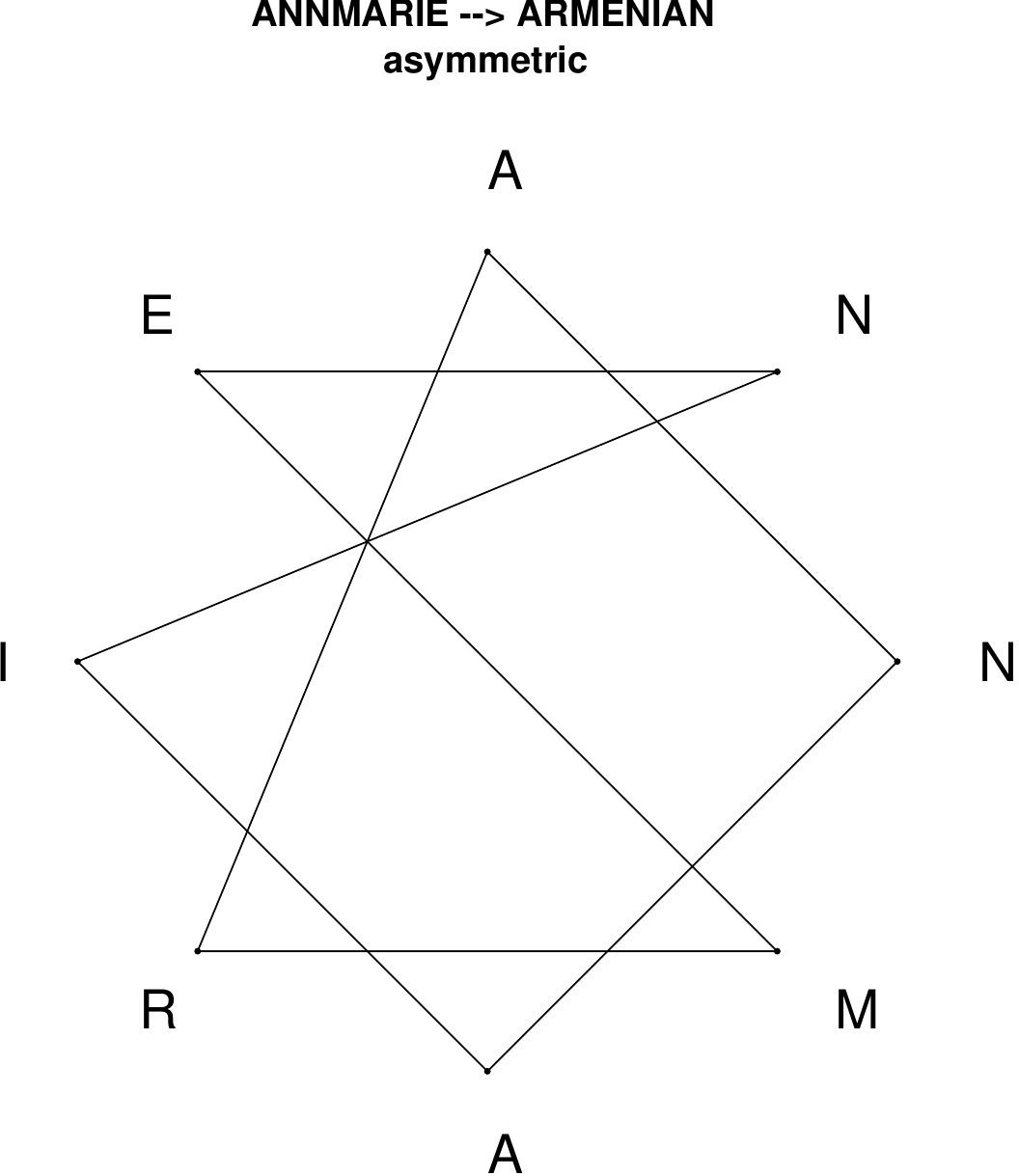}
\end{subfigure}
\hfill
\begin{subfigure}[T]{0.19\textwidth}
\centering
\includegraphics[width=\textwidth]{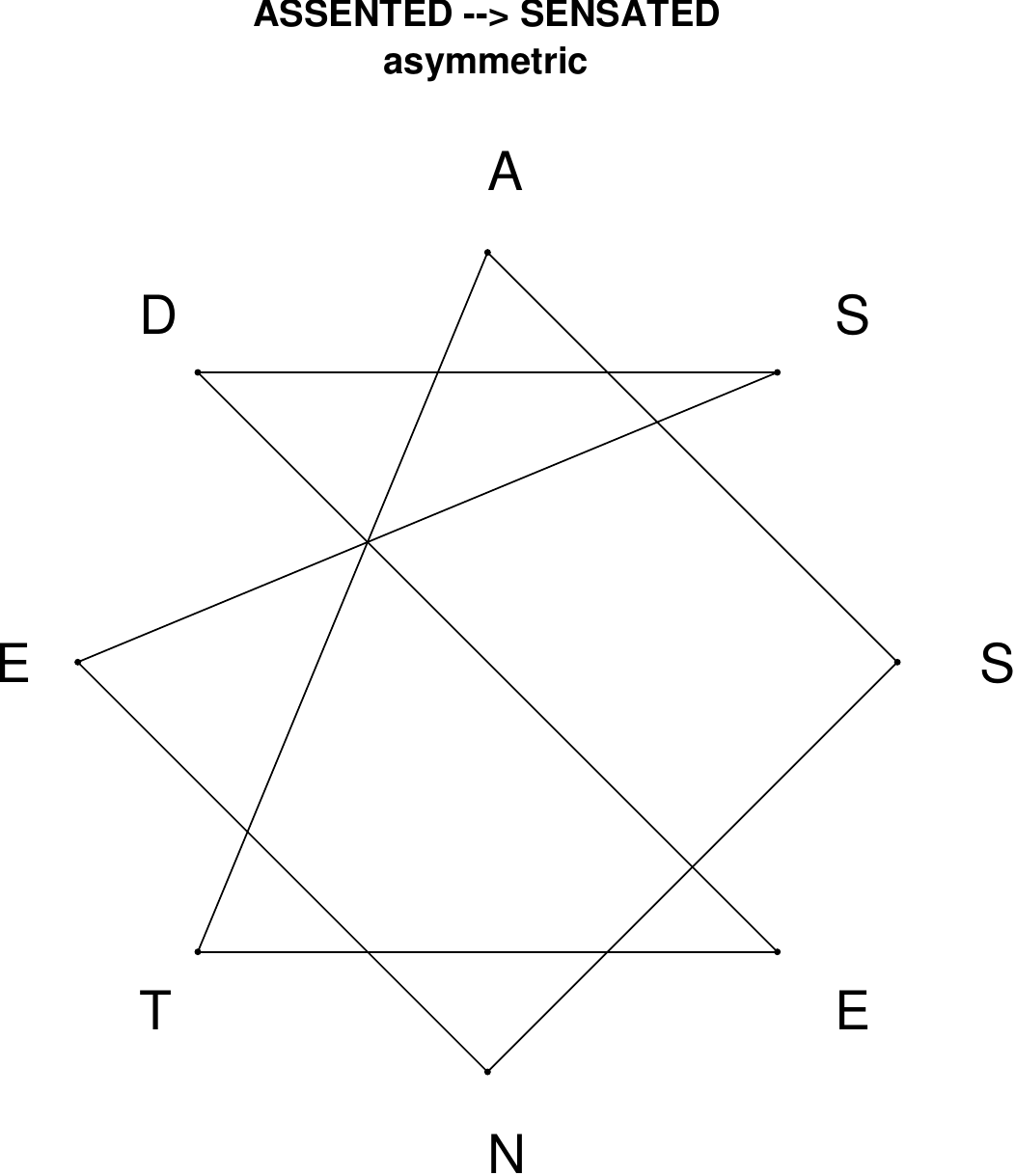}
\end{subfigure}
\end{figure}

\begin{figure}[H]
\centering
\begin{subfigure}[T]{0.19\textwidth}
\centering
\includegraphics[width=\textwidth]{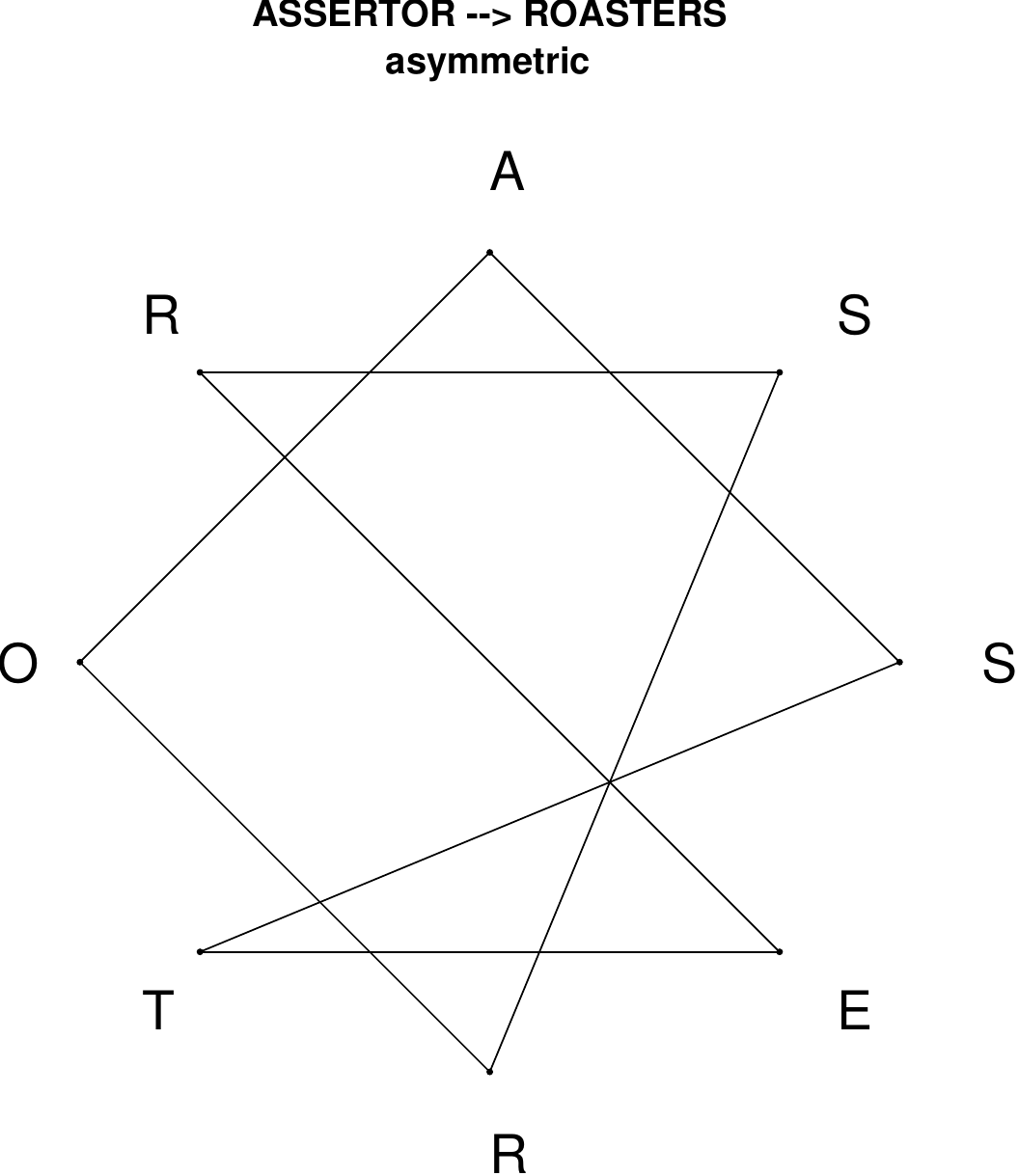}
\end{subfigure}
\hfill
\begin{subfigure}[T]{0.19\textwidth}
\centering
\includegraphics[width=\textwidth]{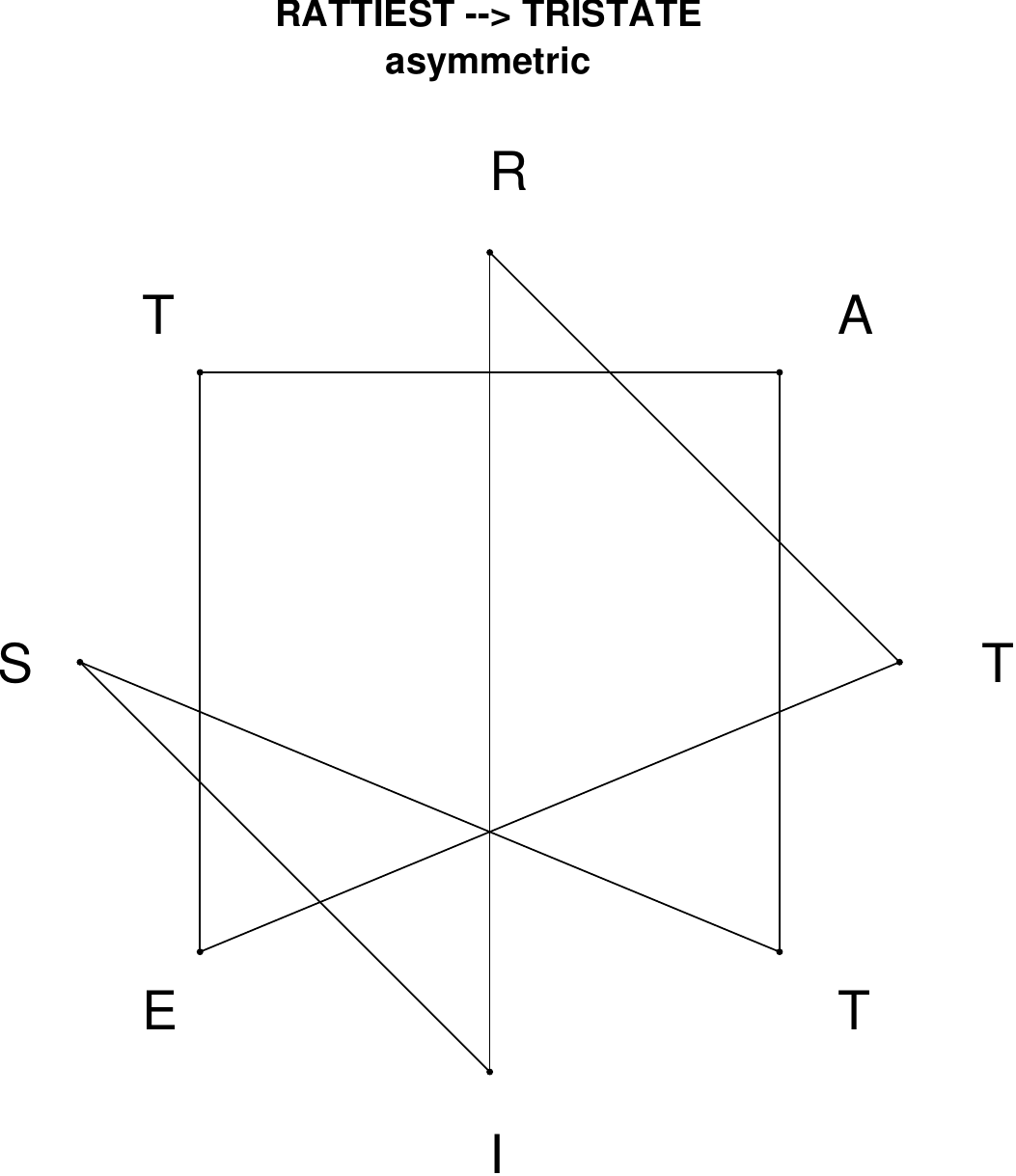}
\end{subfigure}
\hfill
\begin{subfigure}[T]{0.19\textwidth}
\centering
\includegraphics[width=\textwidth]{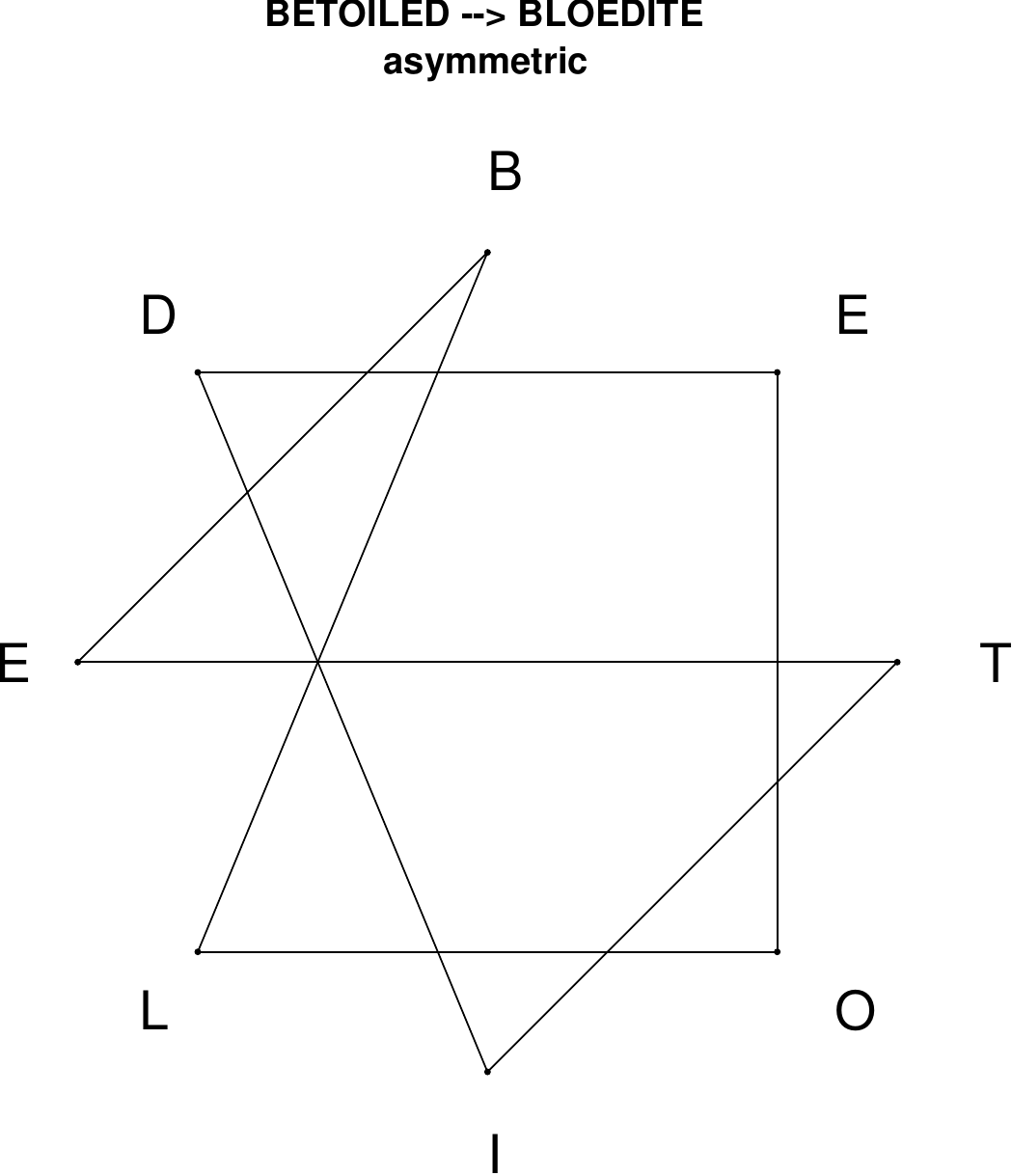}
\end{subfigure}
\hfill
\begin{subfigure}[T]{0.19\textwidth}
\centering
\includegraphics[width=\textwidth]{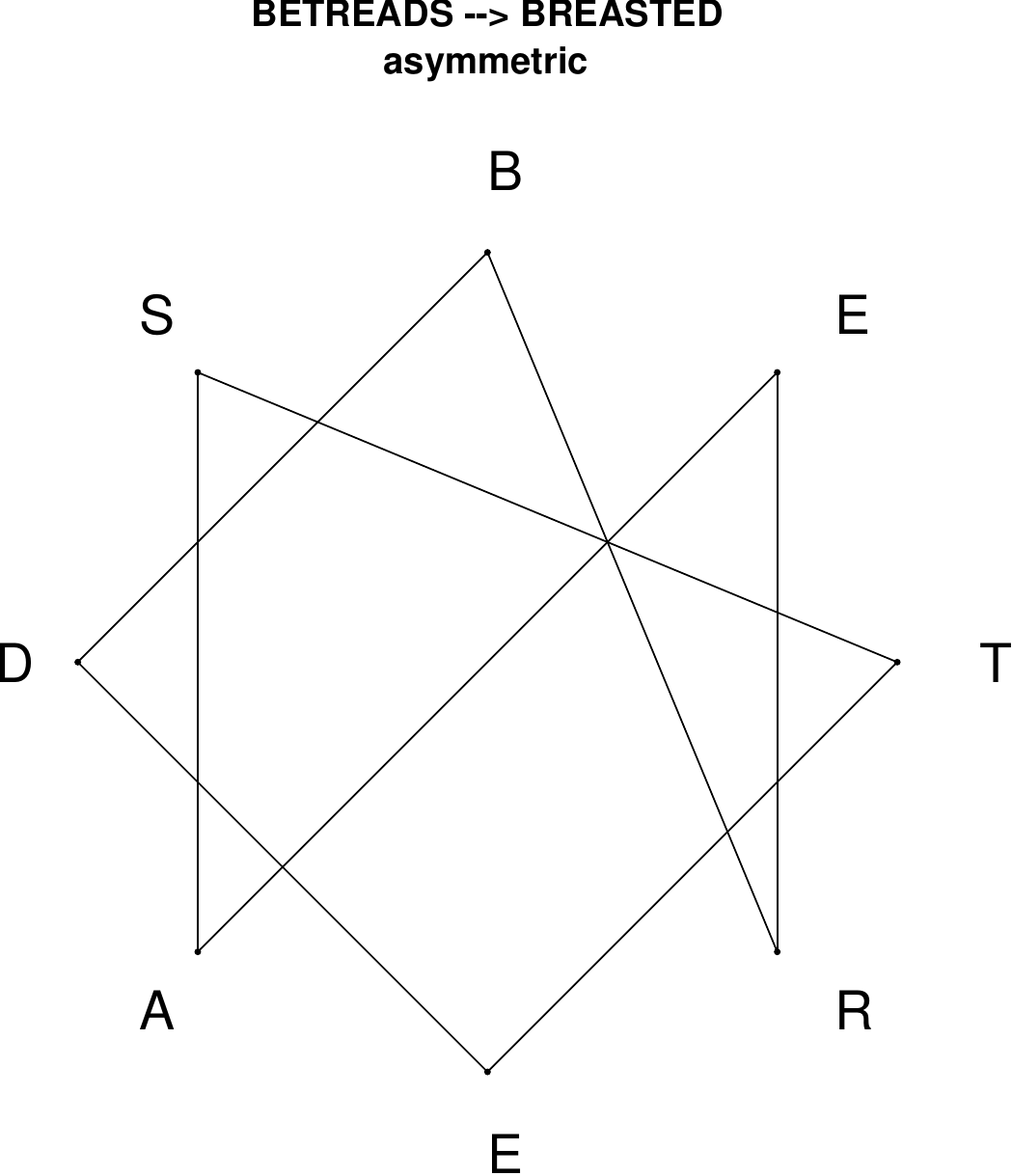}
\end{subfigure}
\hfill
\begin{subfigure}[T]{0.19\textwidth}
\centering
\includegraphics[width=\textwidth]{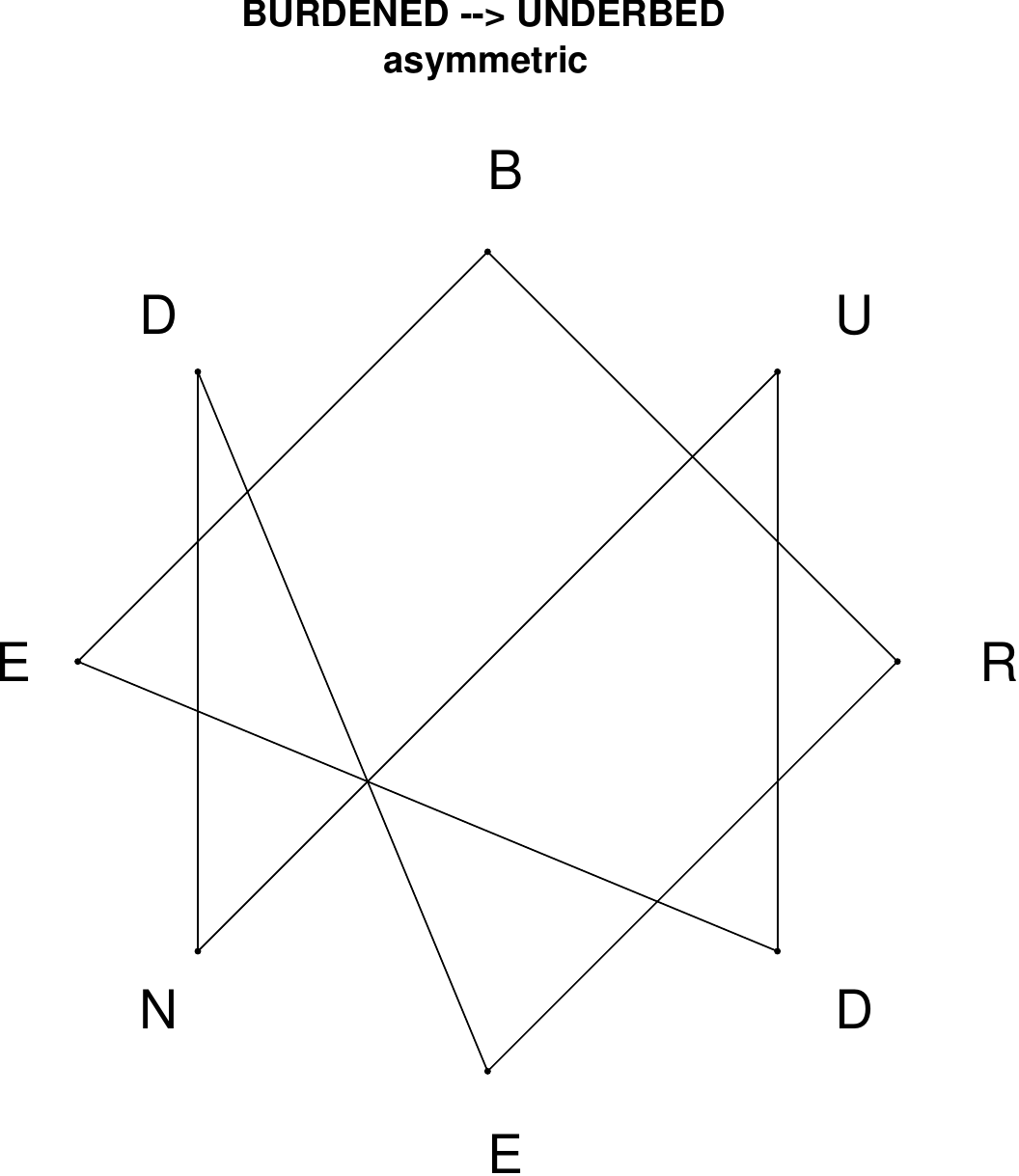}
\end{subfigure}
\end{figure}

\begin{figure}[H]
\centering
\begin{subfigure}[T]{0.19\textwidth}
\centering
\includegraphics[width=\textwidth]{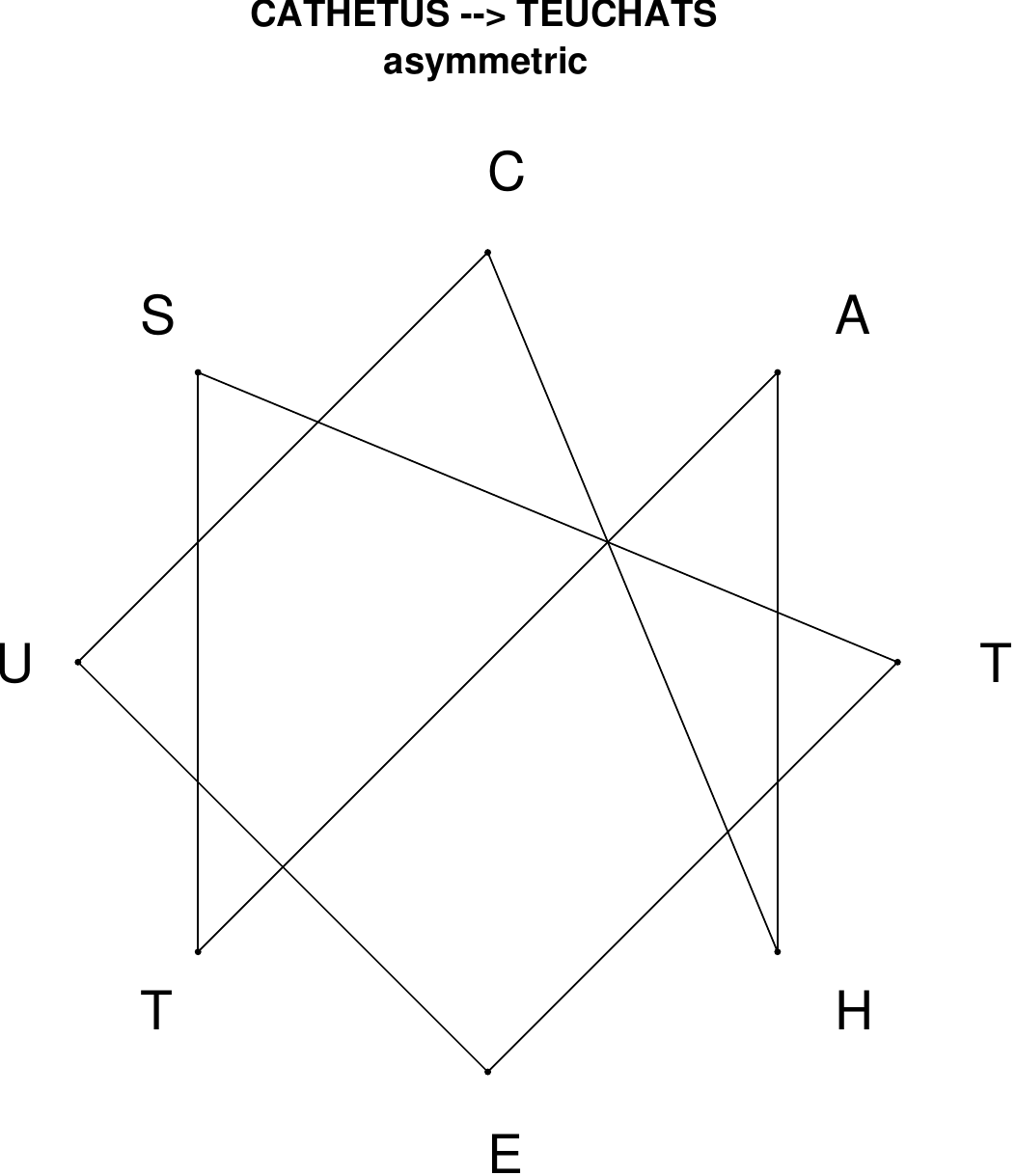}
\end{subfigure}
\hfill
\begin{subfigure}[T]{0.19\textwidth}
\centering
\includegraphics[width=\textwidth]{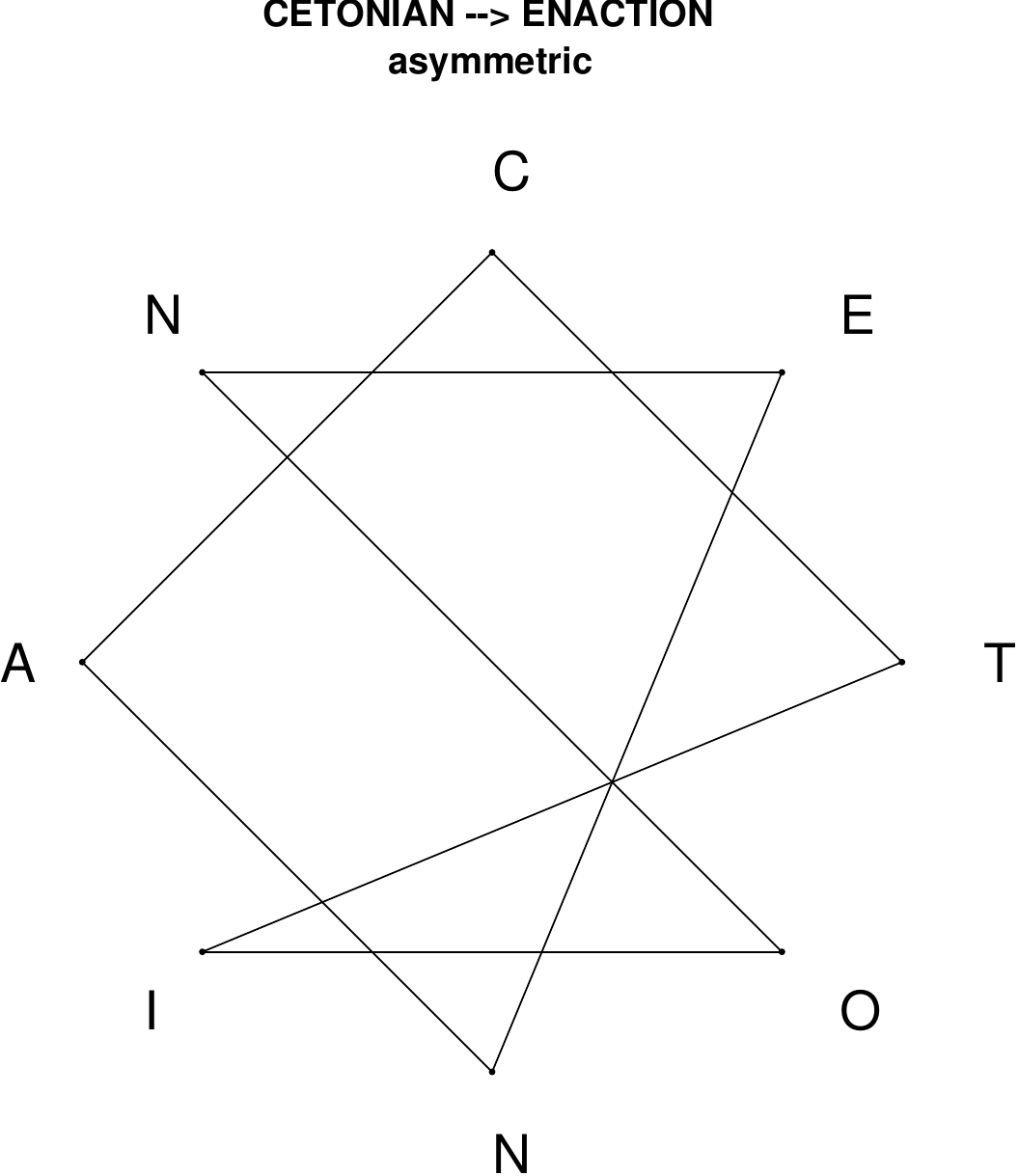}
\end{subfigure}
\hfill
\begin{subfigure}[T]{0.19\textwidth}
\centering
\includegraphics[width=\textwidth]{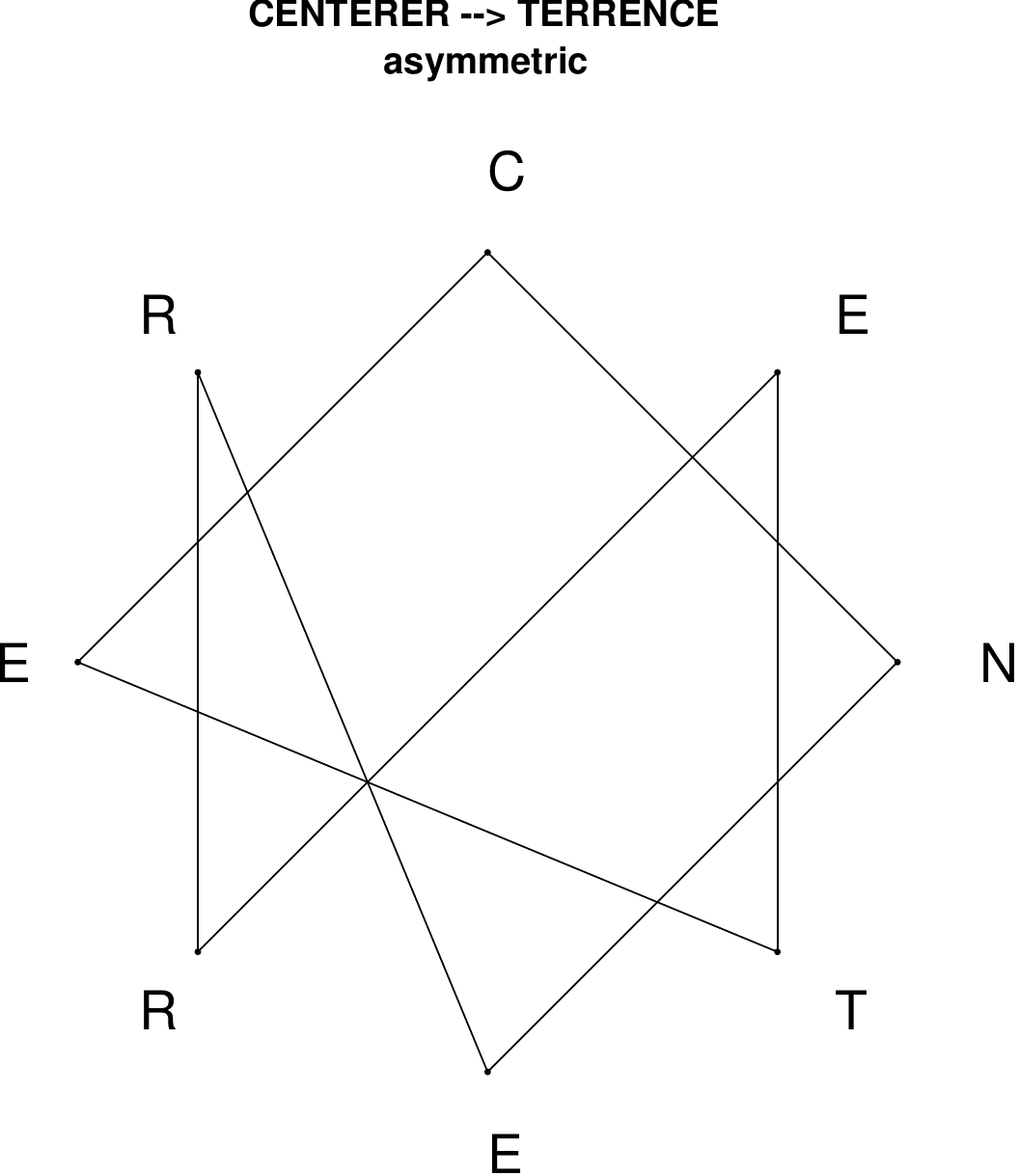}
\end{subfigure}
\hfill
\begin{subfigure}[T]{0.19\textwidth}
\centering
\includegraphics[width=\textwidth]{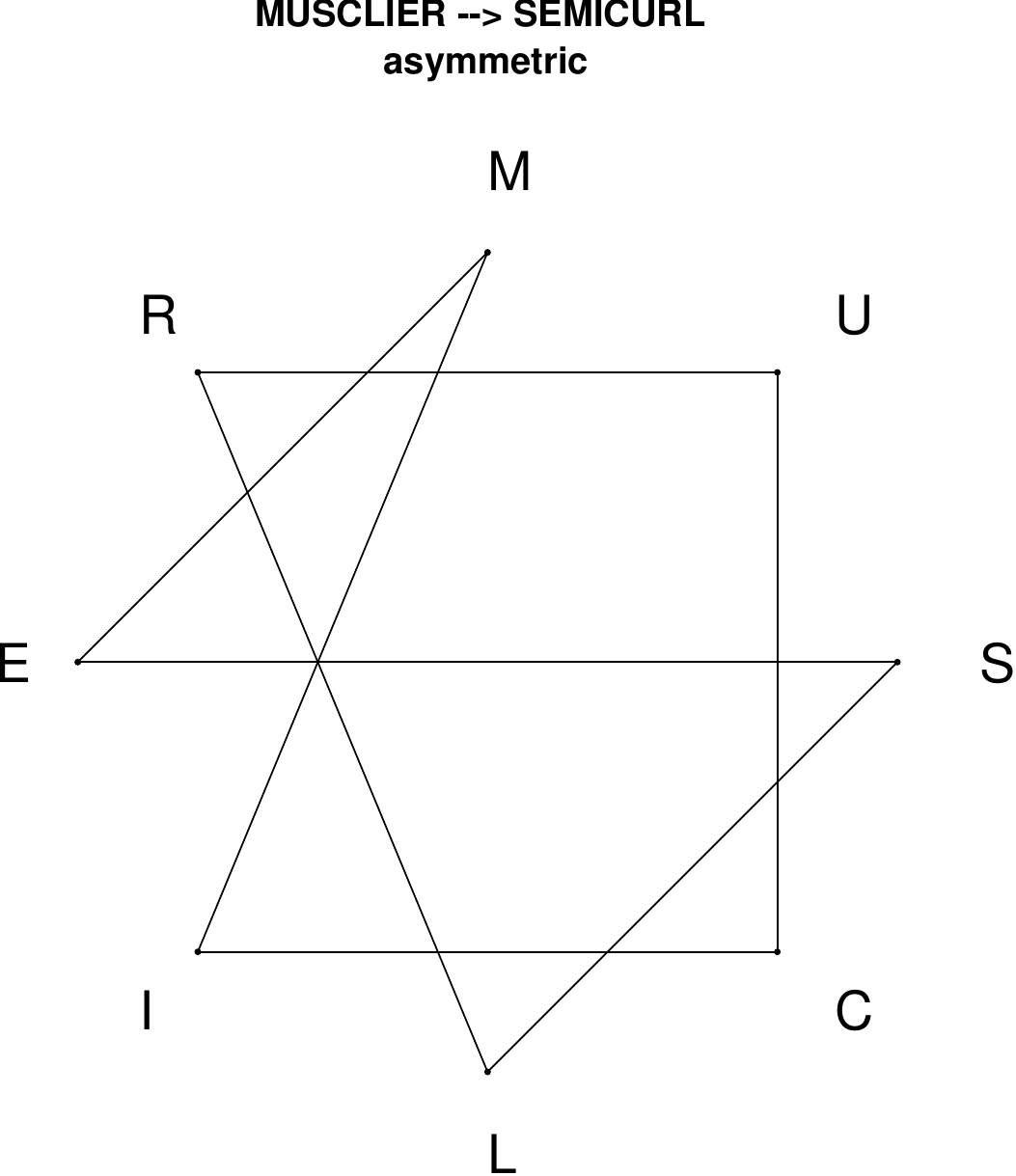}
\end{subfigure}
\hfill
\begin{subfigure}[T]{0.19\textwidth}
\centering
\includegraphics[width=\textwidth]{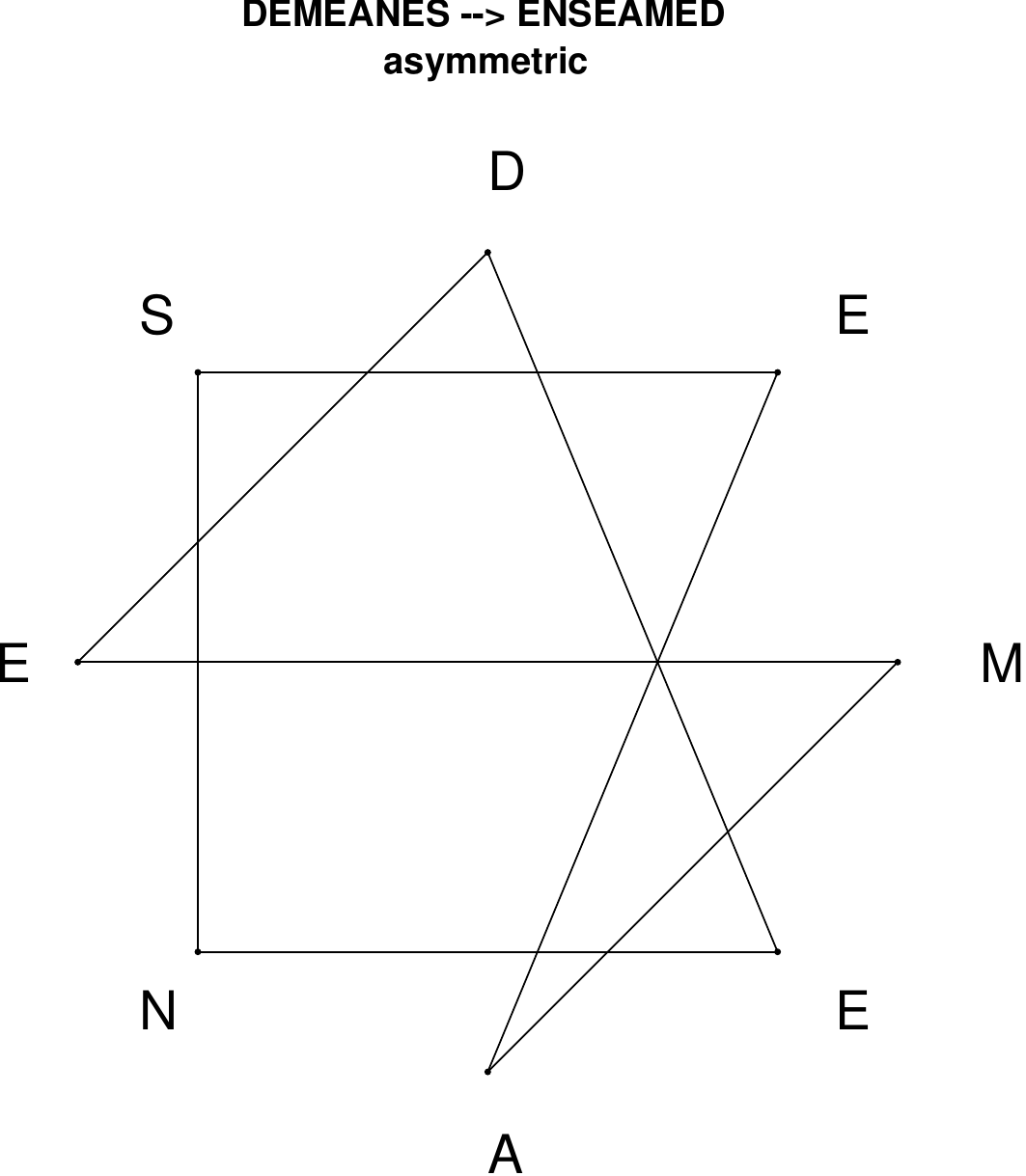}
\end{subfigure}
\end{figure}

\begin{figure}[H]
\centering
\begin{subfigure}[T]{0.19\textwidth}
\centering
\includegraphics[width=\textwidth]{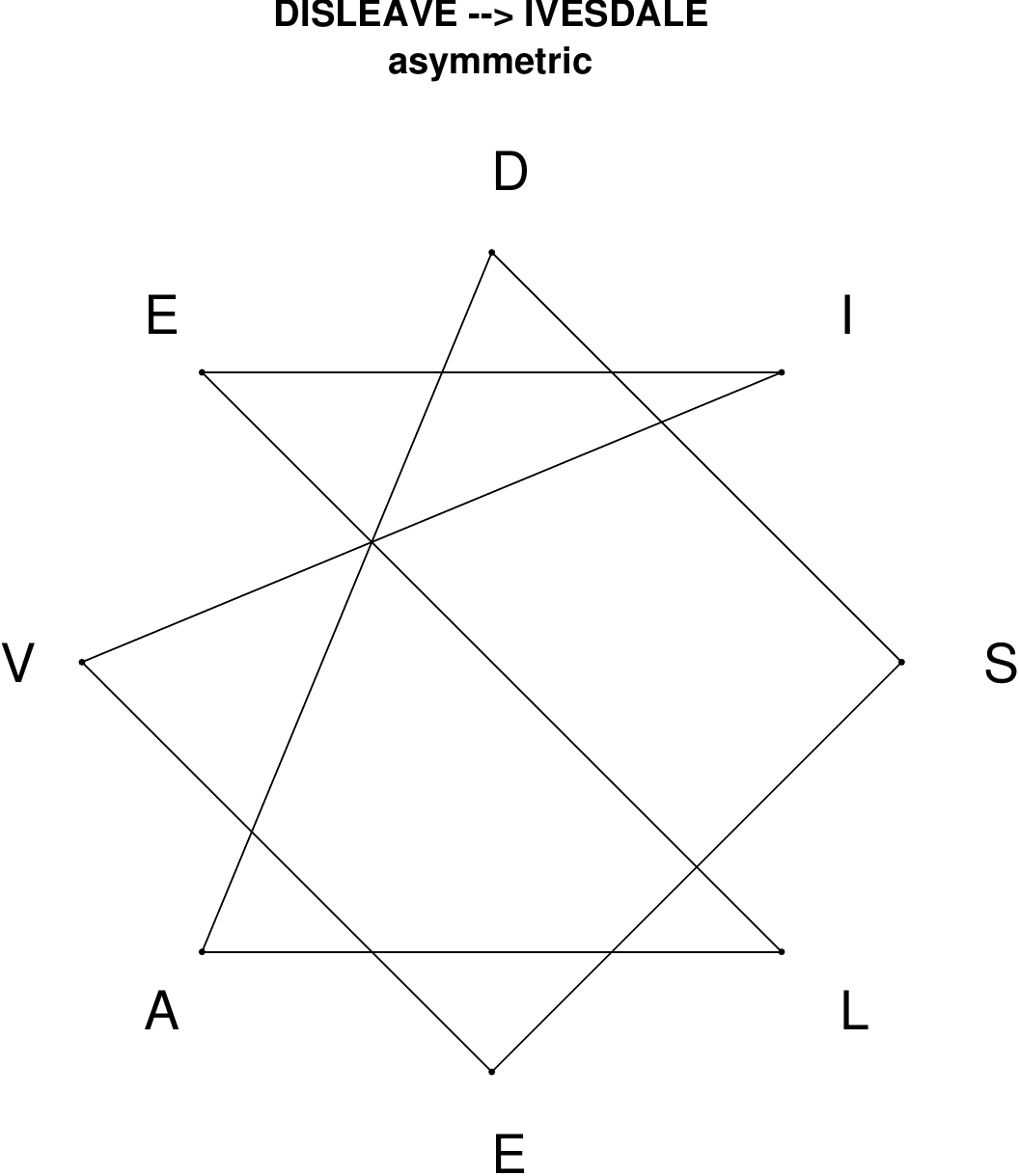}
\end{subfigure}
\hfill
\begin{subfigure}[T]{0.19\textwidth}
\centering
\includegraphics[width=\textwidth]{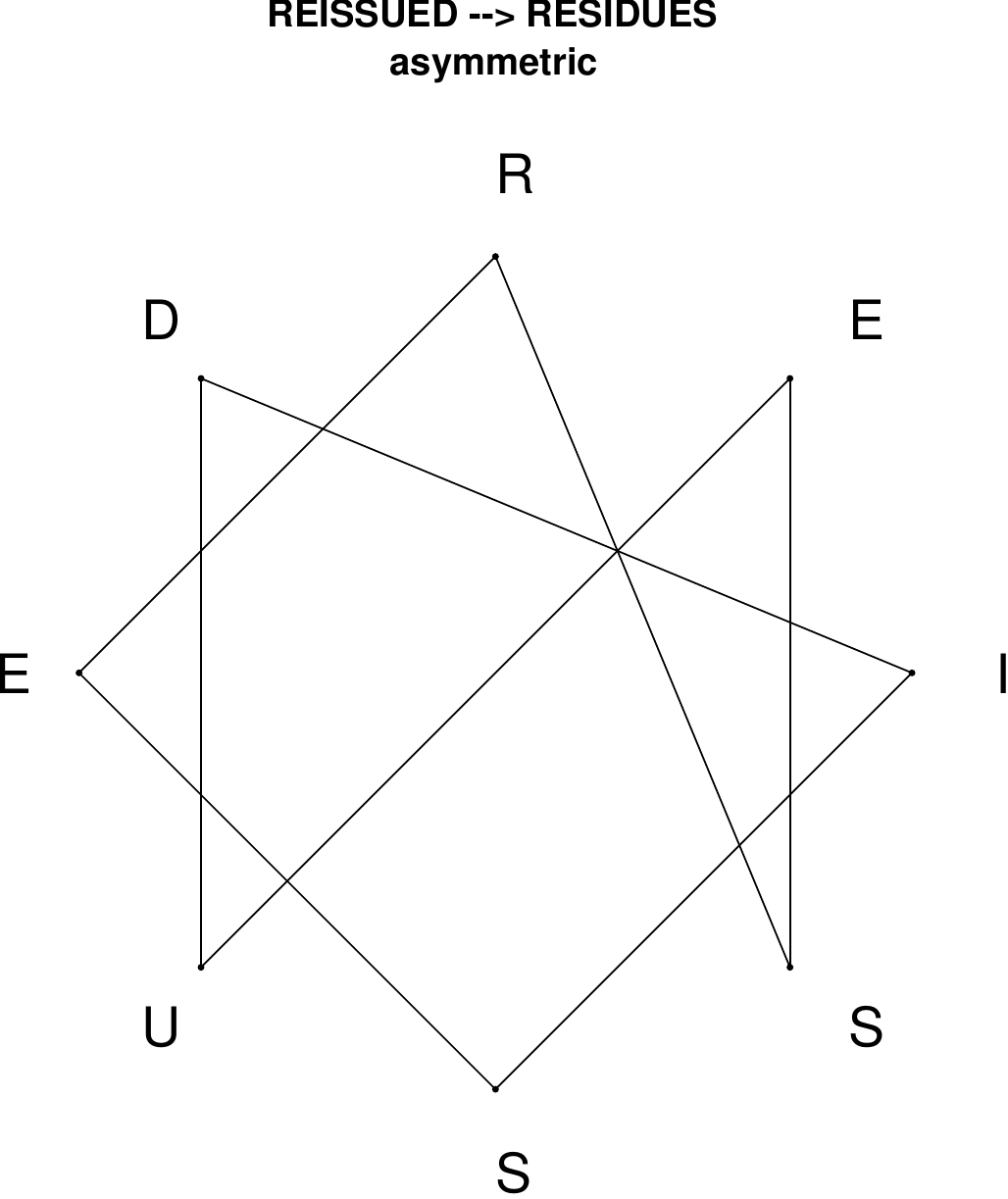}
\end{subfigure}
\hfill
\begin{subfigure}[T]{0.19\textwidth}
\centering
\includegraphics[width=\textwidth]{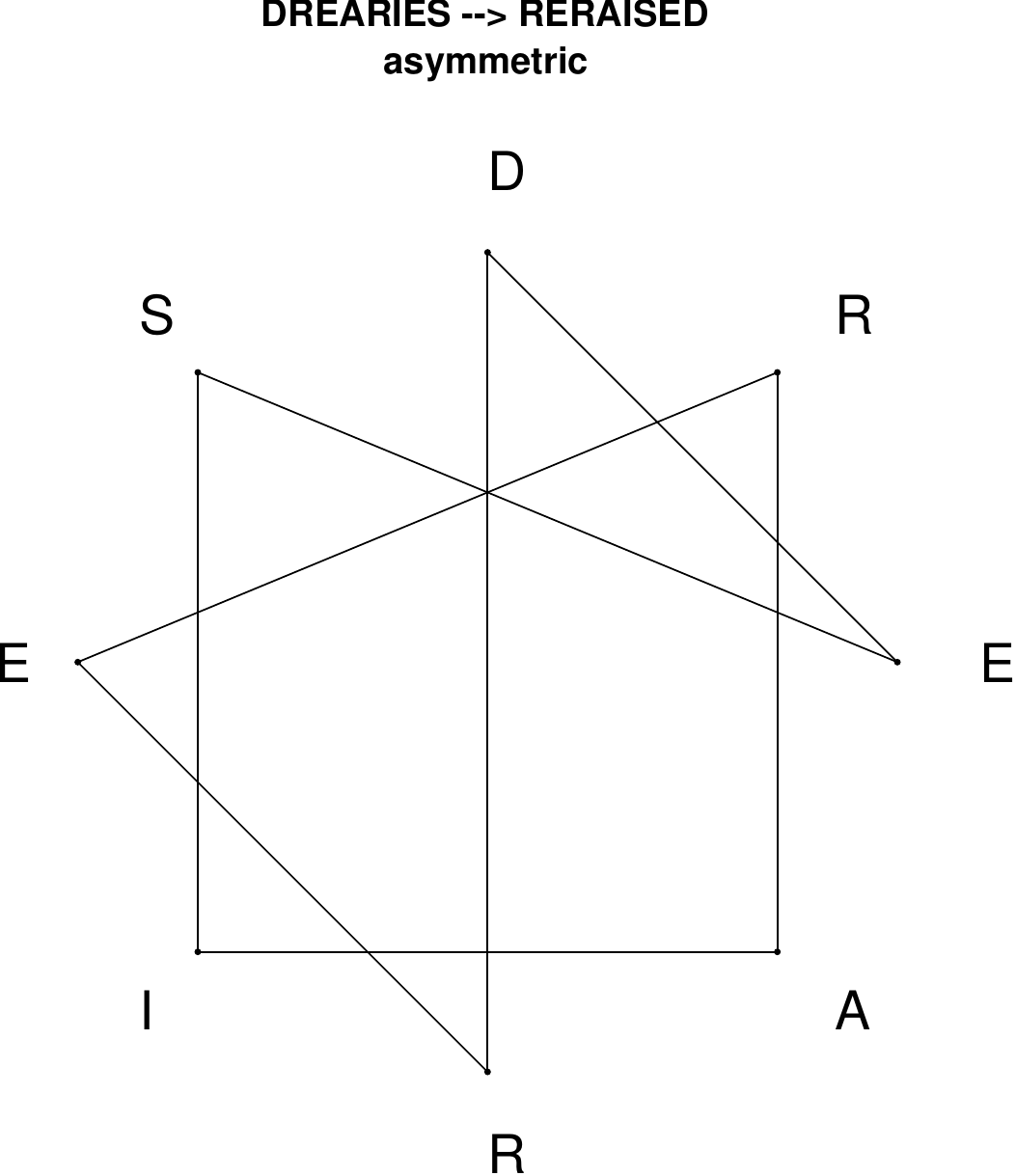}
\end{subfigure}
\hfill
\begin{subfigure}[T]{0.19\textwidth}
\centering
\includegraphics[width=\textwidth]{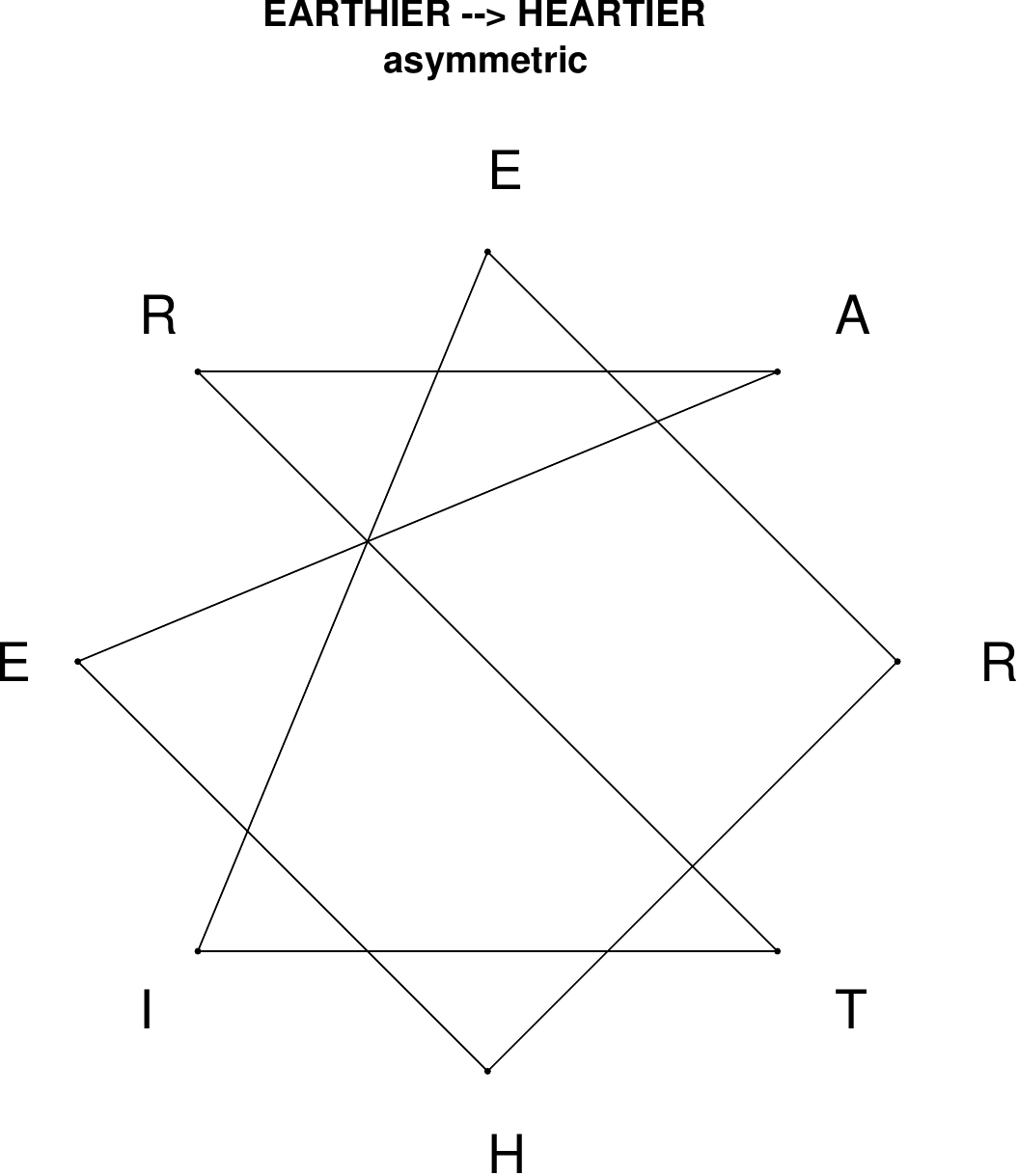}
\end{subfigure}
\hfill
\begin{subfigure}[T]{0.19\textwidth}
\centering
\includegraphics[width=\textwidth]{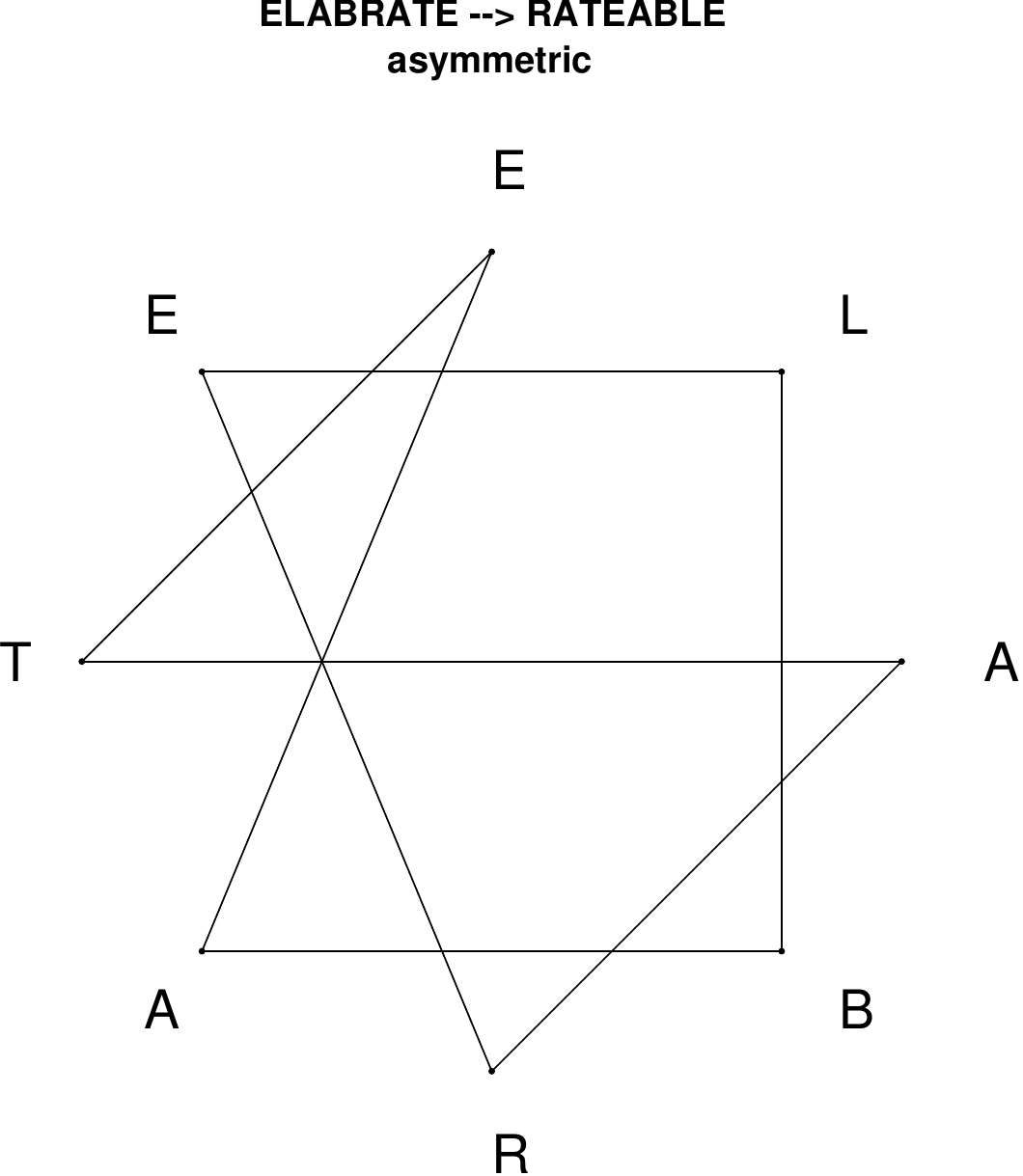}
\end{subfigure}
\end{figure}

\begin{figure}[H]
\centering
\begin{subfigure}[T]{0.19\textwidth}
\centering
\includegraphics[width=\textwidth]{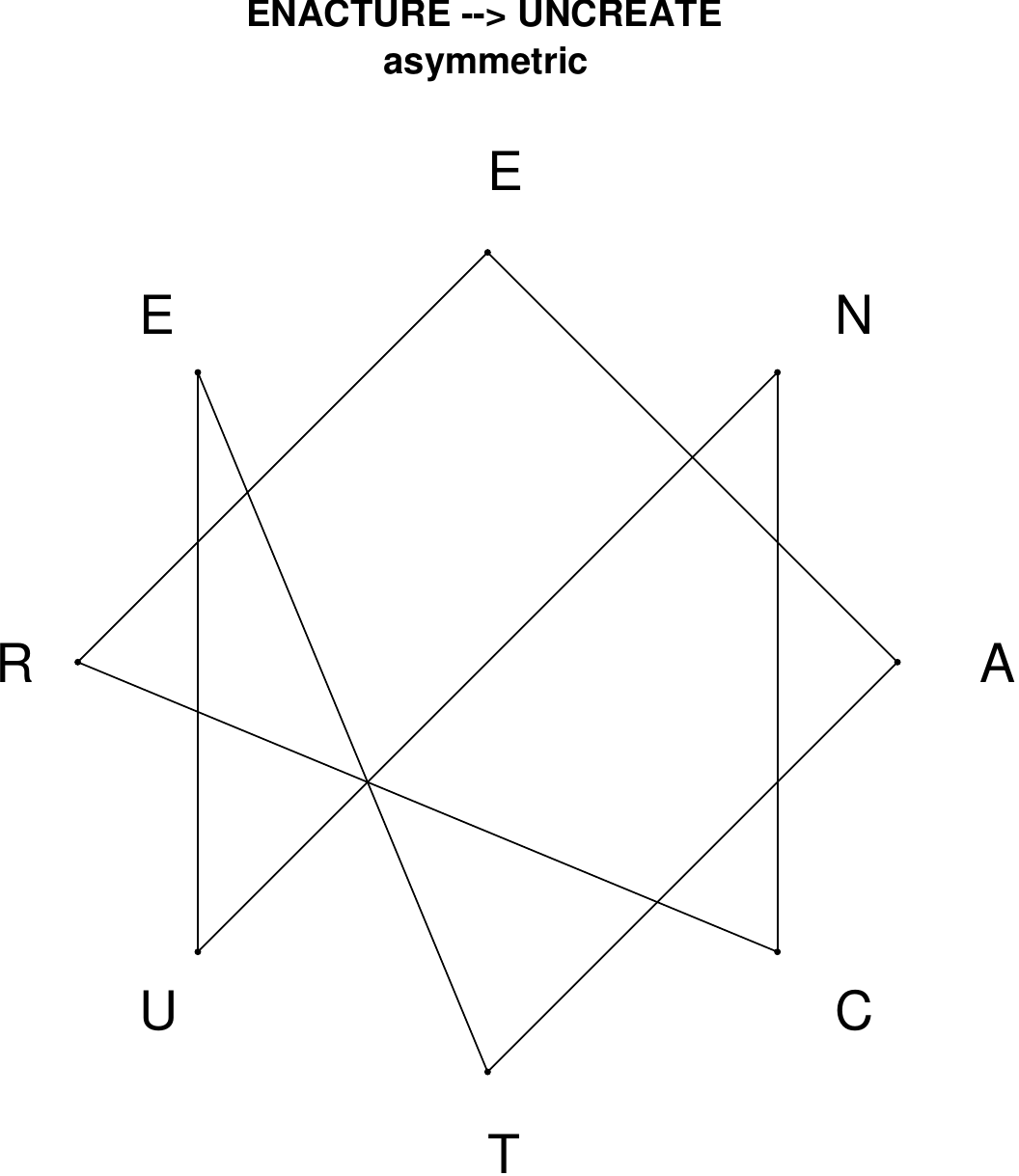}
\end{subfigure}
\hfill
\begin{subfigure}[T]{0.19\textwidth}
\centering
\includegraphics[width=\textwidth]{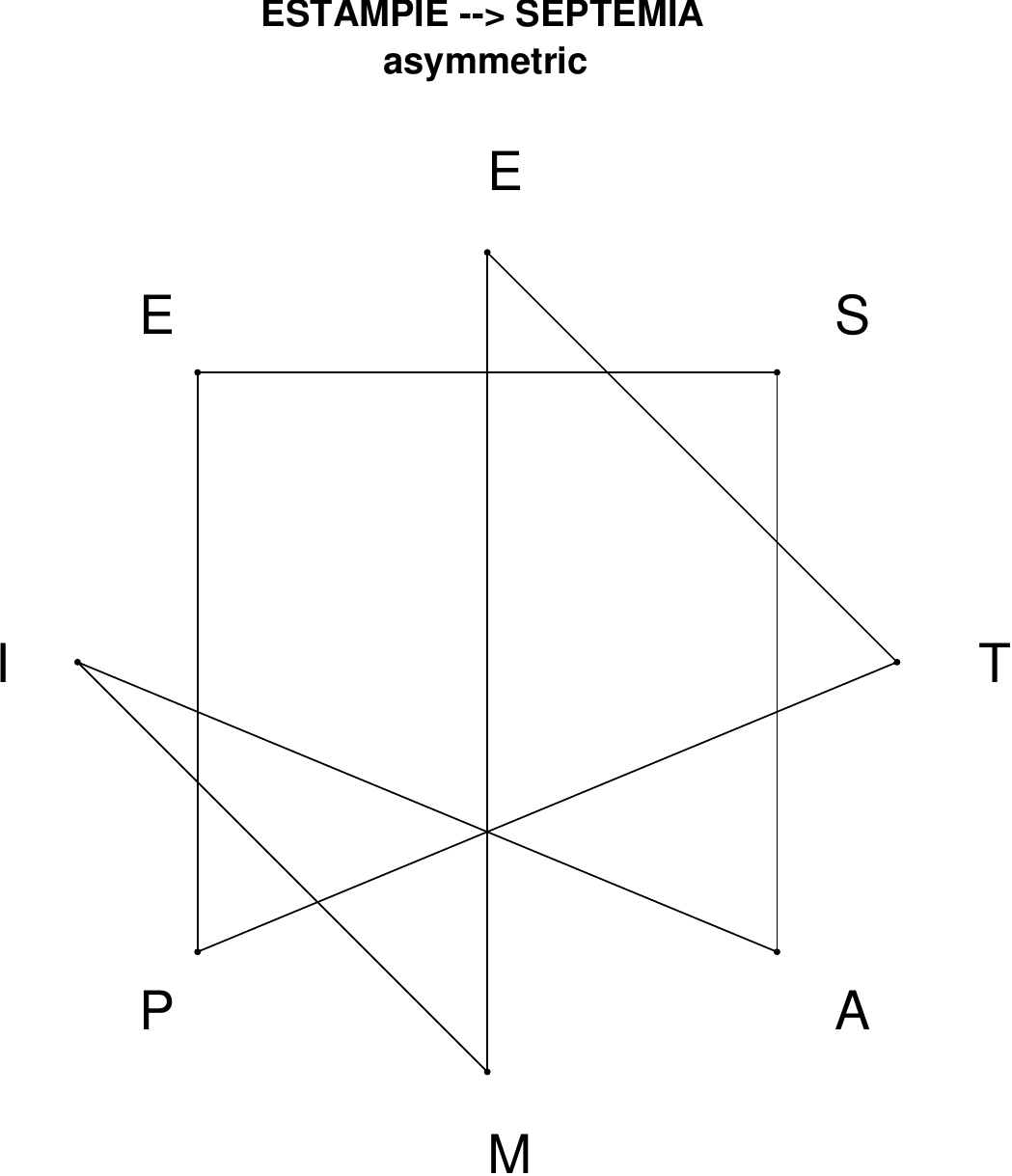}
\end{subfigure}
\hfill
\begin{subfigure}[T]{0.19\textwidth}
\centering
\includegraphics[width=\textwidth]{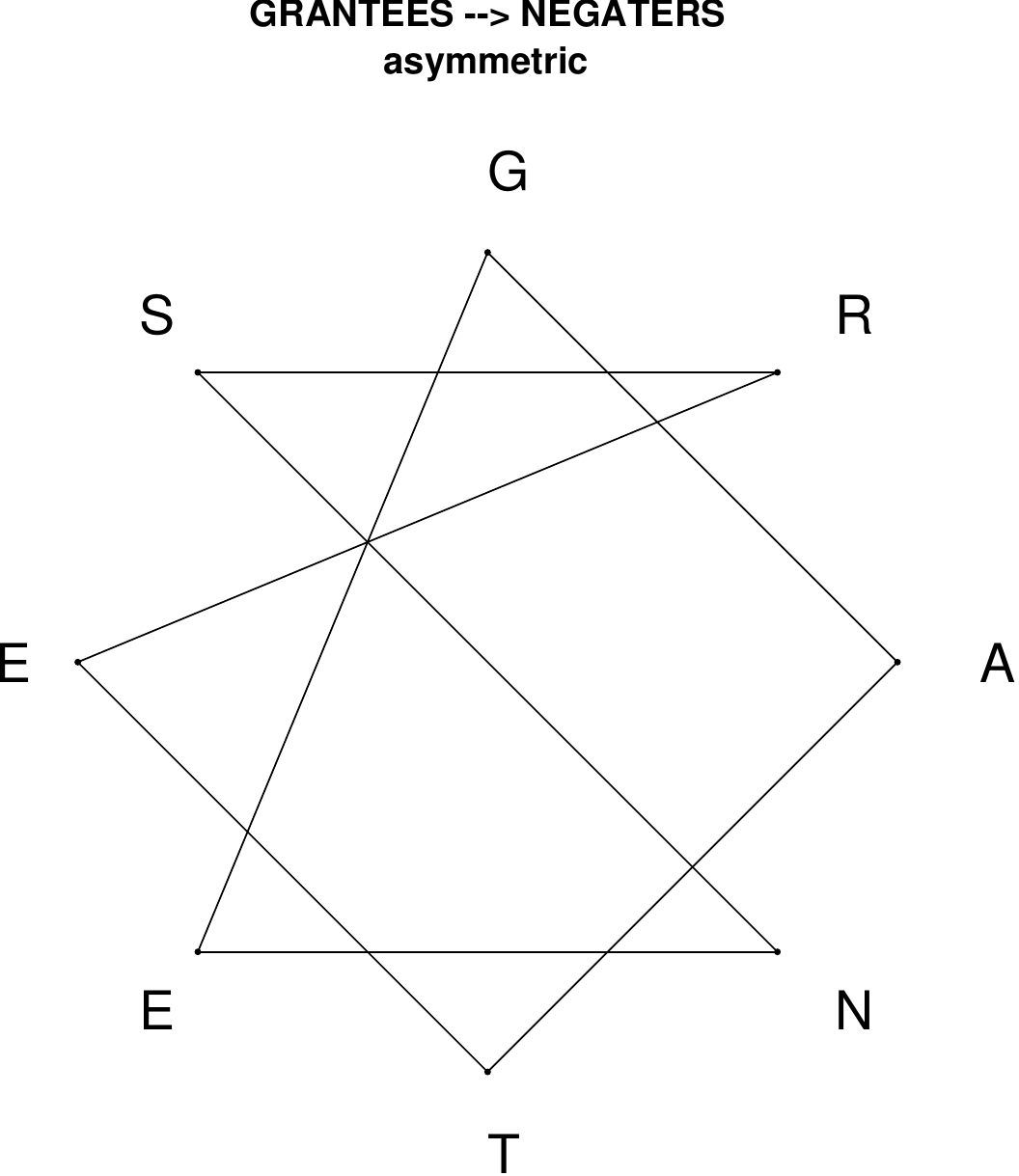}
\end{subfigure}
\hfill
\begin{subfigure}[T]{0.19\textwidth}
\centering
\includegraphics[width=\textwidth]{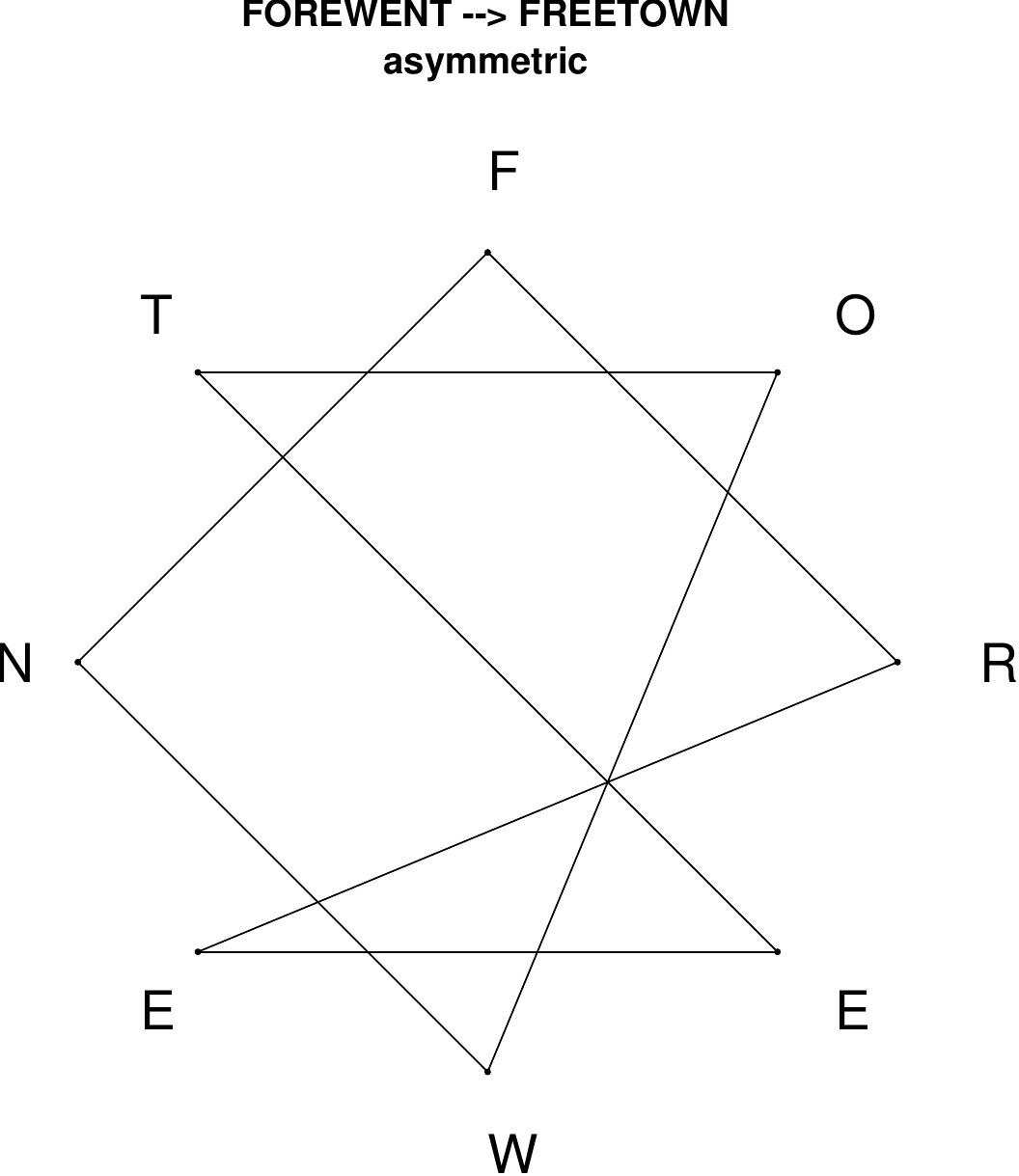}
\end{subfigure}
\hfill
\begin{subfigure}[T]{0.19\textwidth}
\centering
\includegraphics[width=\textwidth]{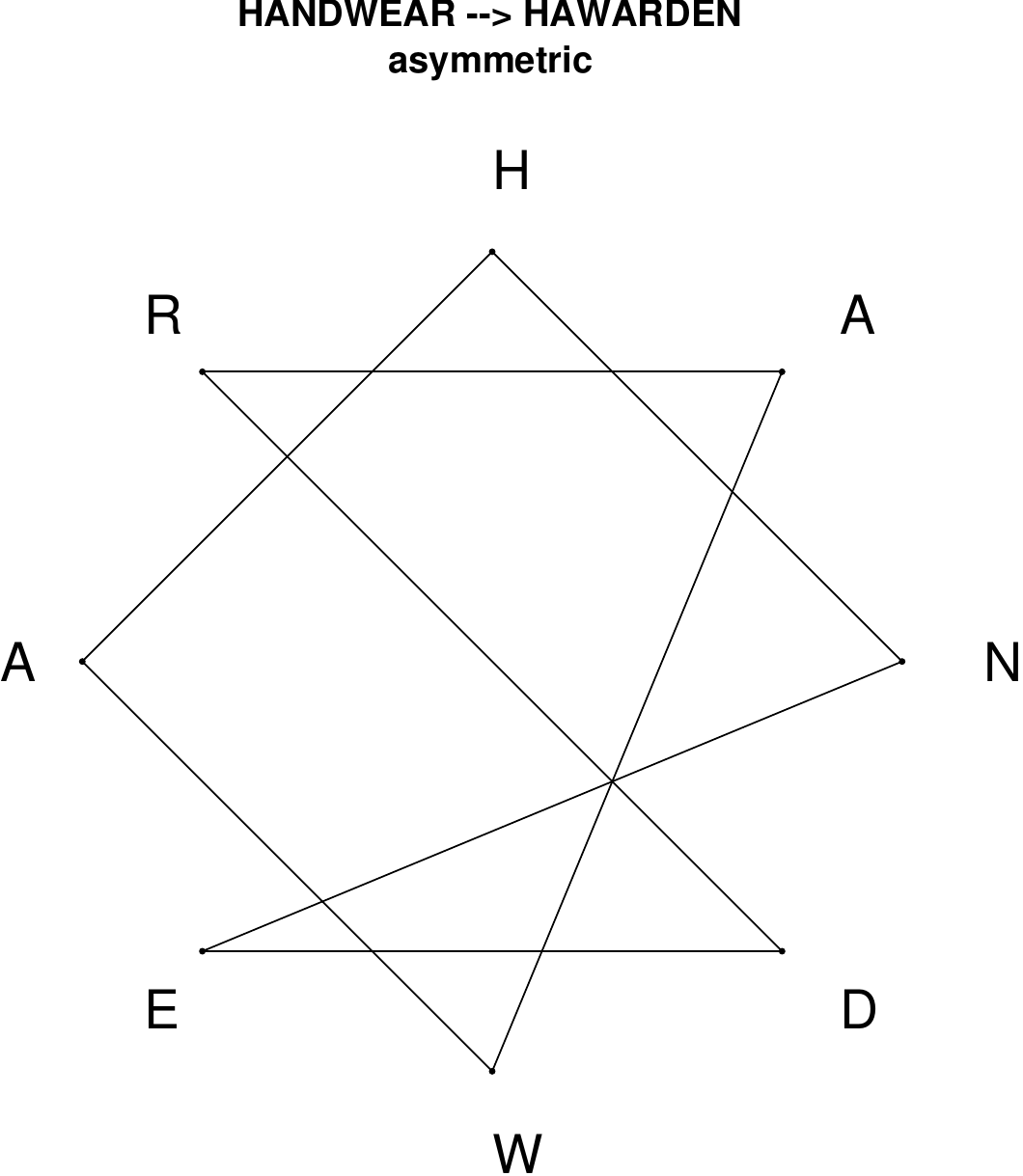}
\end{subfigure}
\end{figure}

\begin{figure}[H]
\centering
\begin{subfigure}[T]{0.19\textwidth}
\centering
\includegraphics[width=\textwidth]{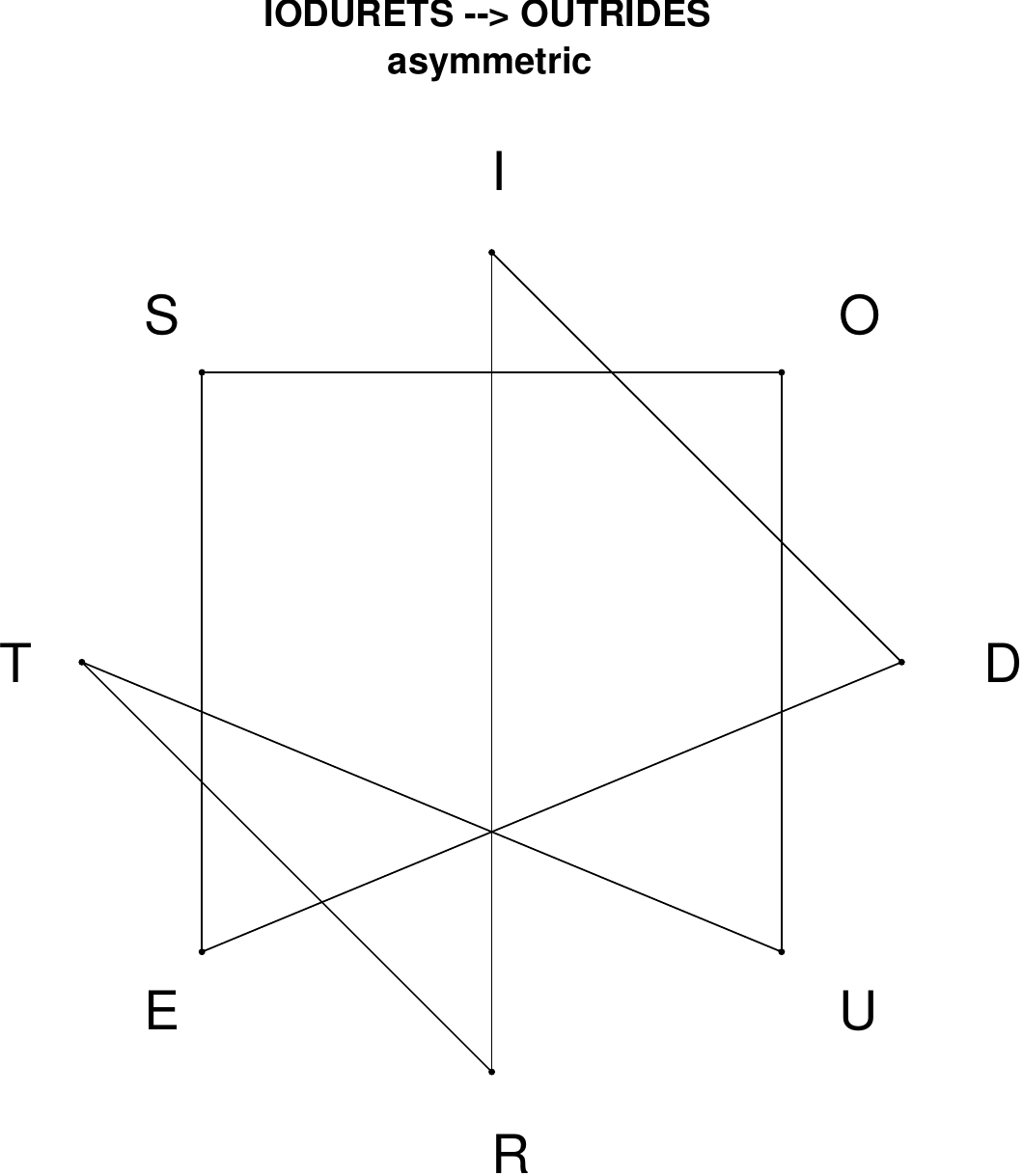}
\end{subfigure}
\hfill
\begin{subfigure}[T]{0.19\textwidth}
\centering
\includegraphics[width=\textwidth]{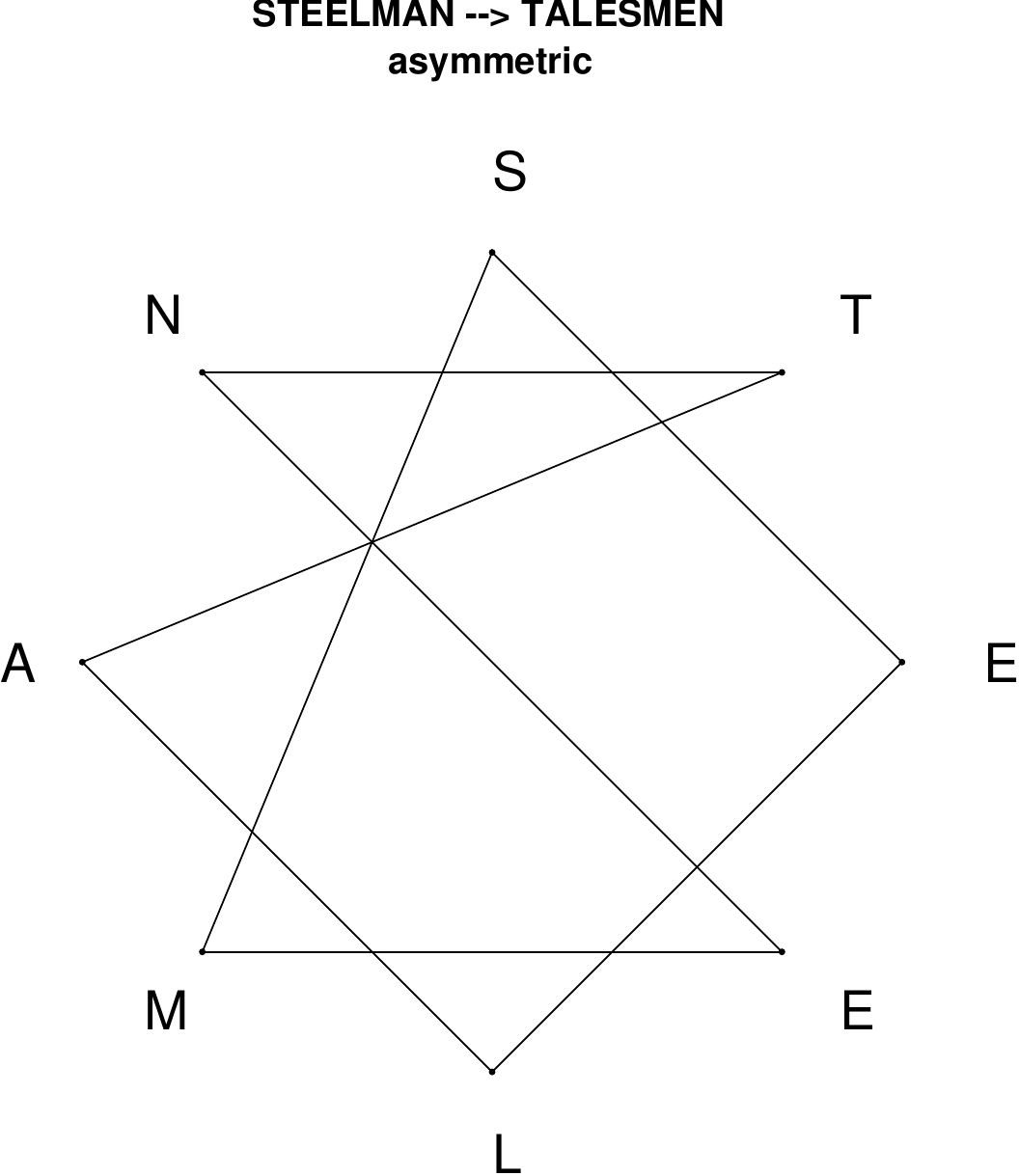}
\end{subfigure}
\hfill
\begin{subfigure}[T]{0.19\textwidth}
\centering
\includegraphics[width=\textwidth]{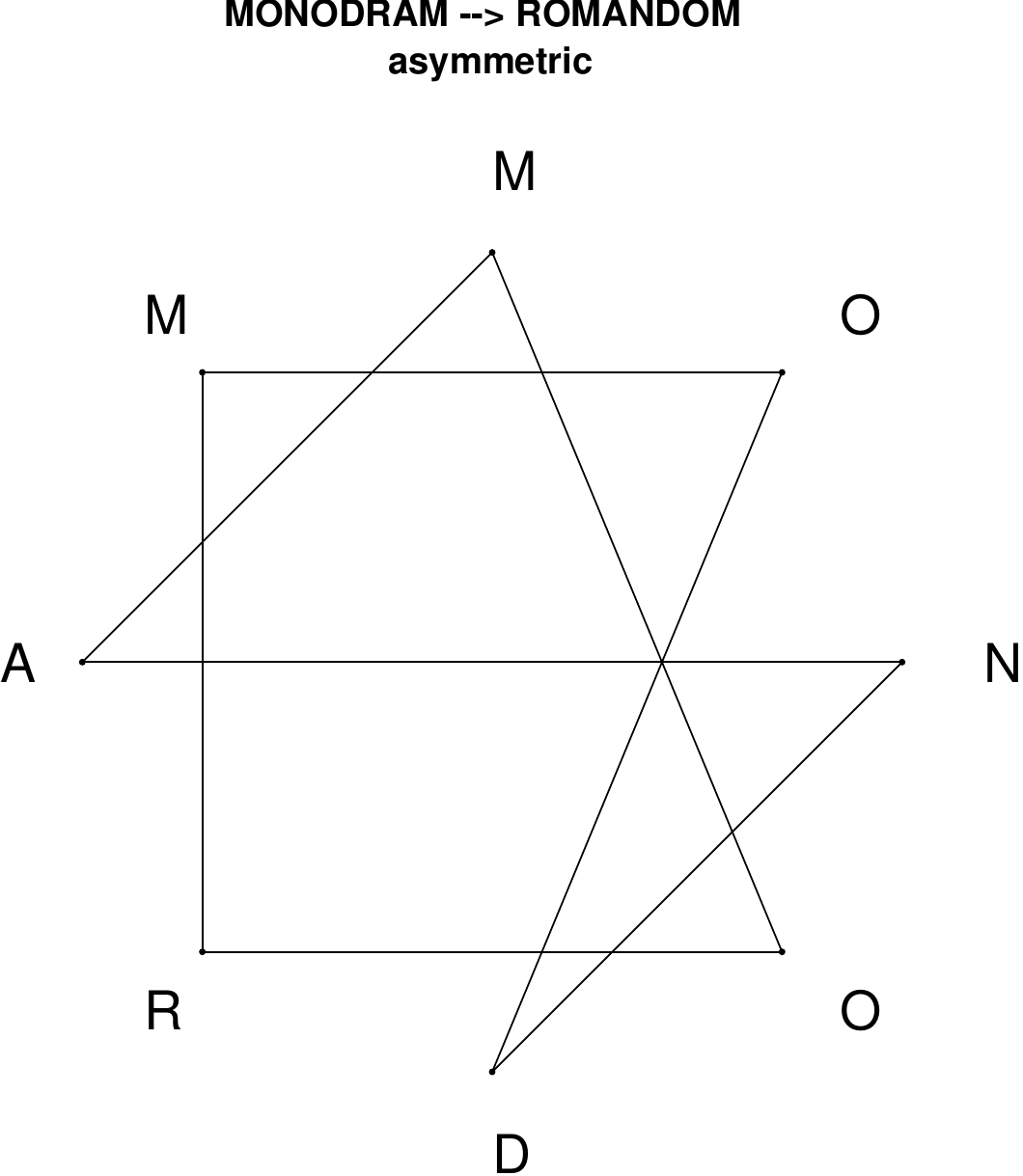}
\end{subfigure}
\hfill
\begin{subfigure}[T]{0.19\textwidth}
\centering
\includegraphics[width=\textwidth]{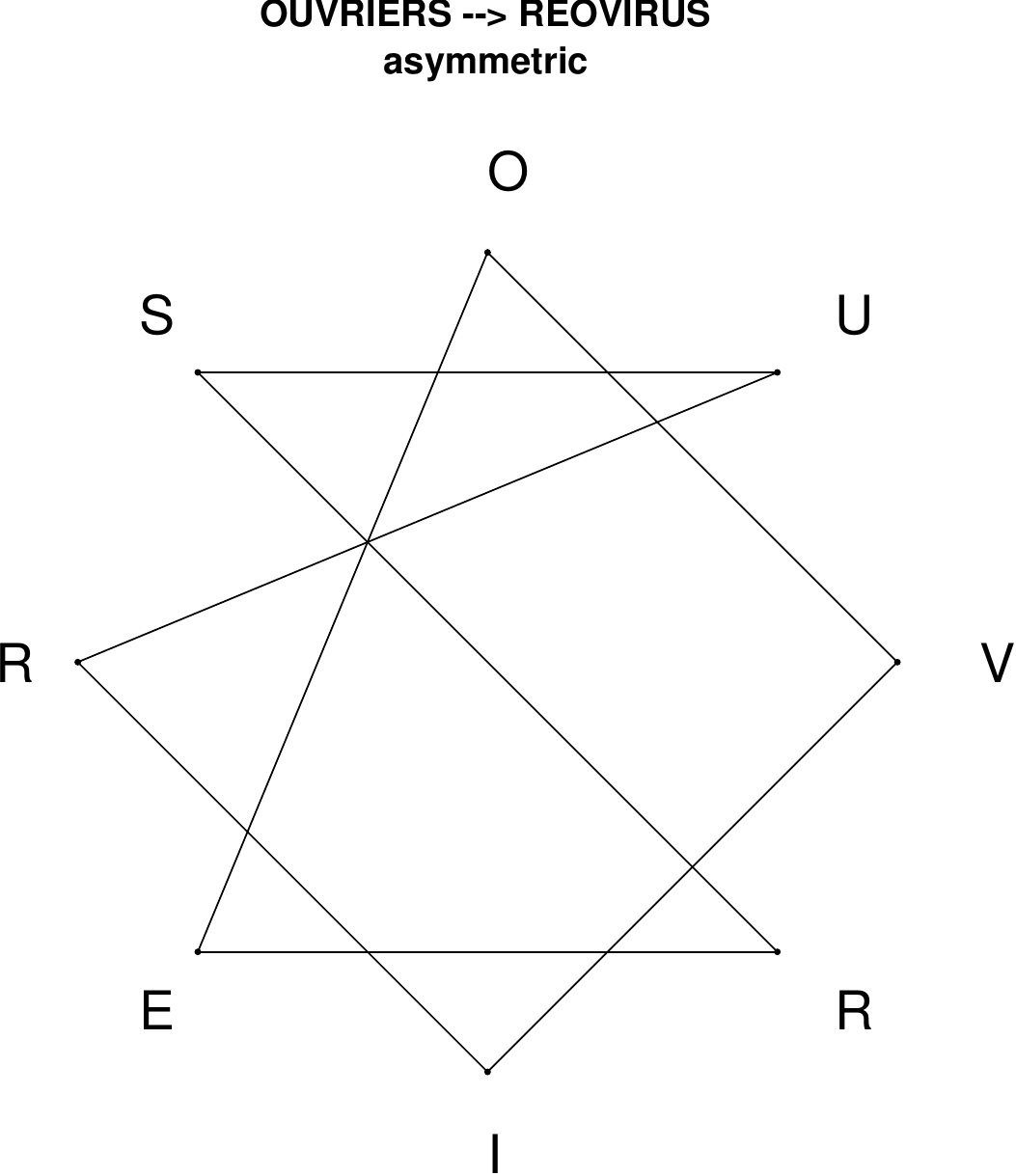}
\end{subfigure}
\hfill
\begin{subfigure}[T]{0.19\textwidth}
\centering
\includegraphics[width=\textwidth]{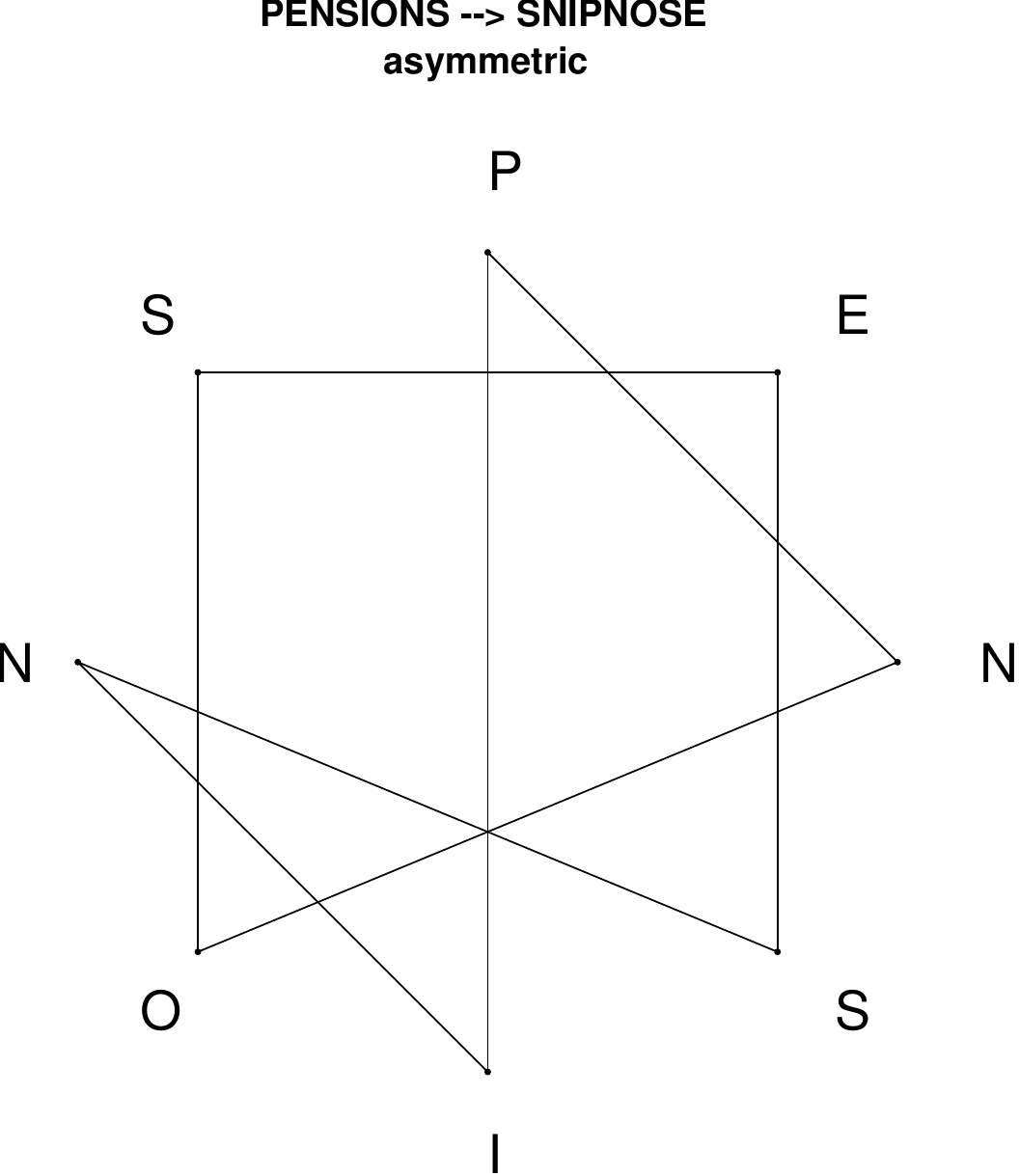}
\end{subfigure}
\end{figure}

\begin{figure}[H]
\centering
\begin{subfigure}[T]{0.19\textwidth}
\centering
\includegraphics[width=\textwidth]{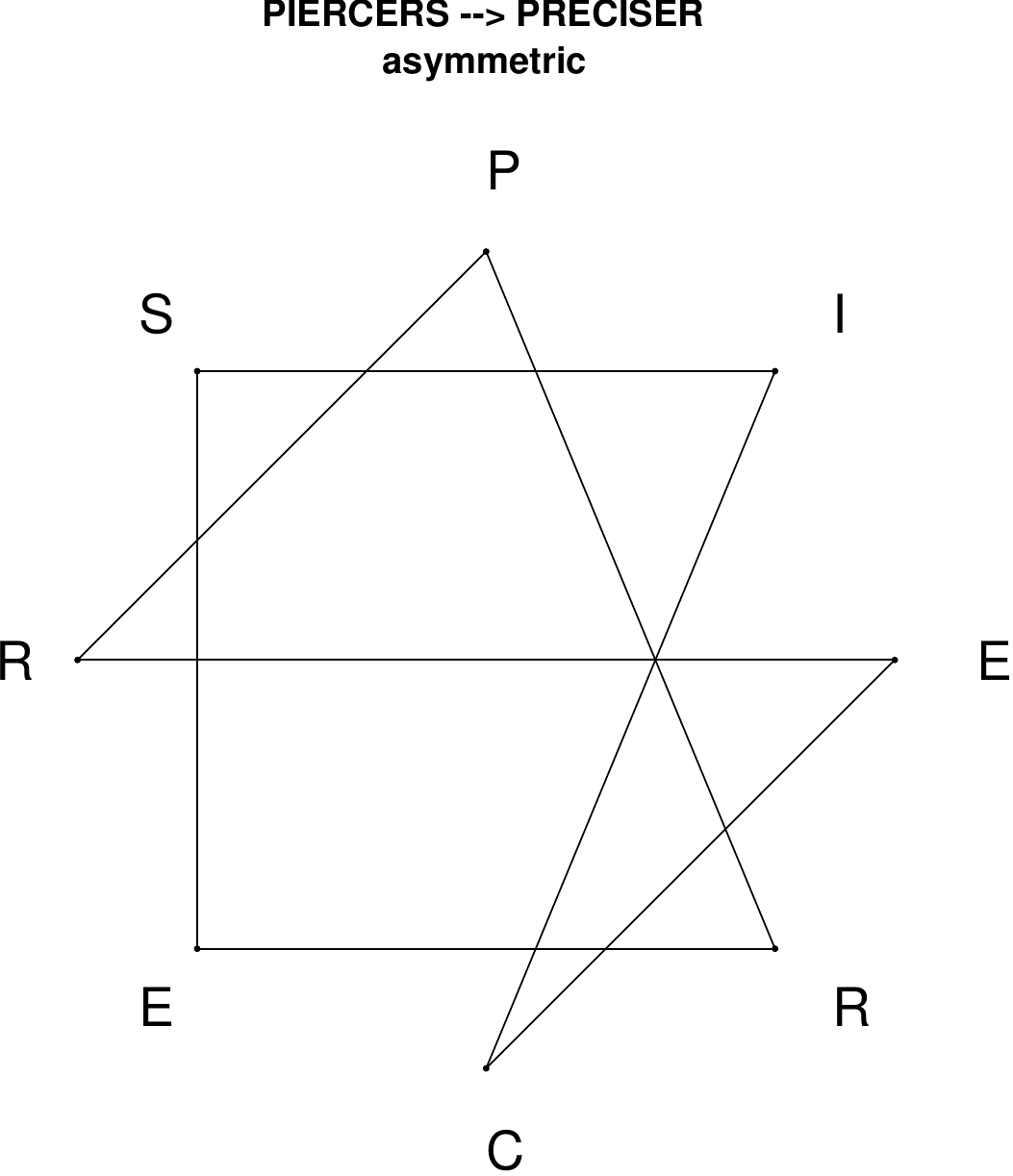}
\end{subfigure}
\hfill
\begin{subfigure}[T]{0.19\textwidth}
\centering
\includegraphics[width=\textwidth]{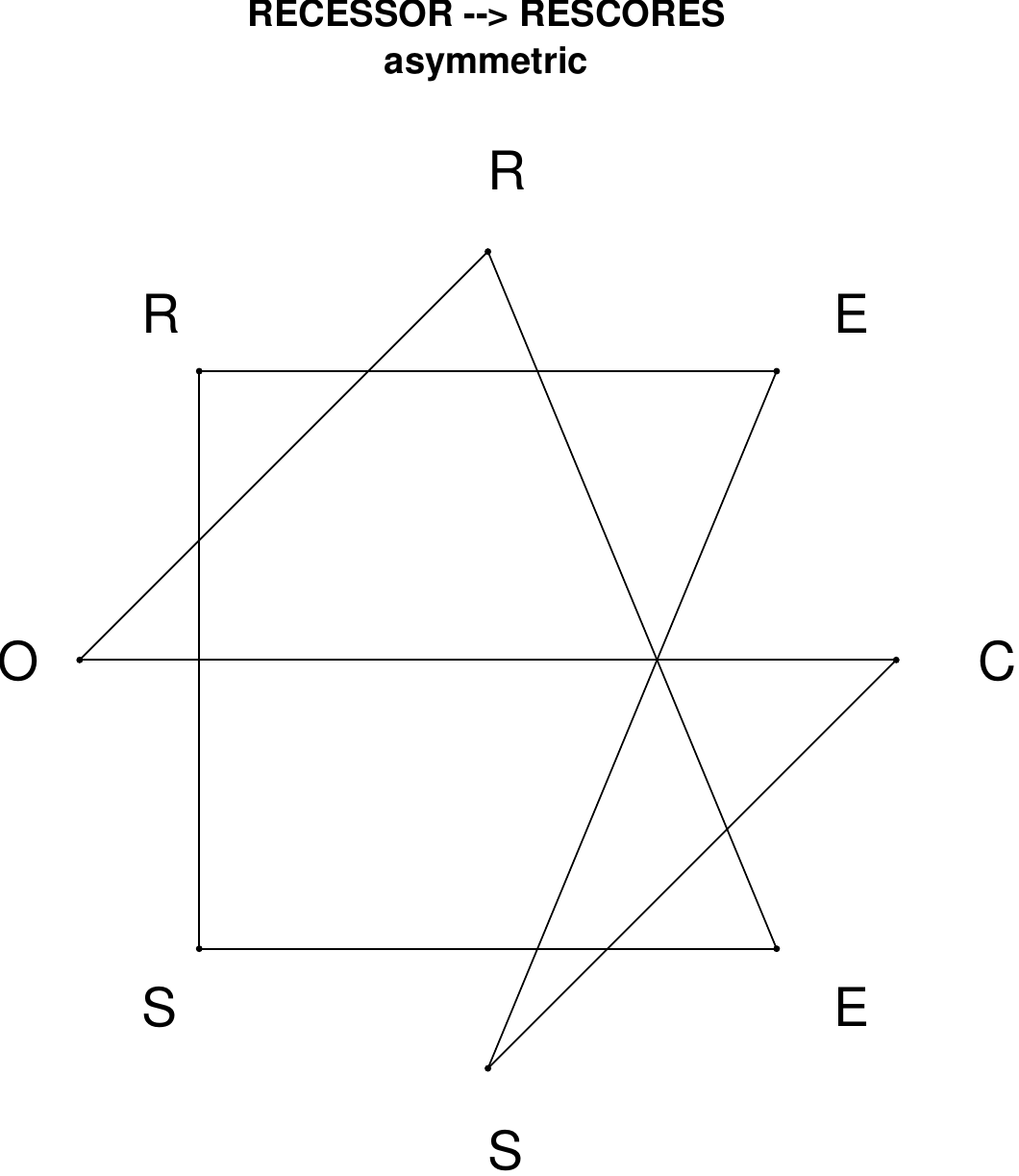}
\end{subfigure}
\hfill
\begin{subfigure}[T]{0.19\textwidth}
\centering
\includegraphics[width=\textwidth]{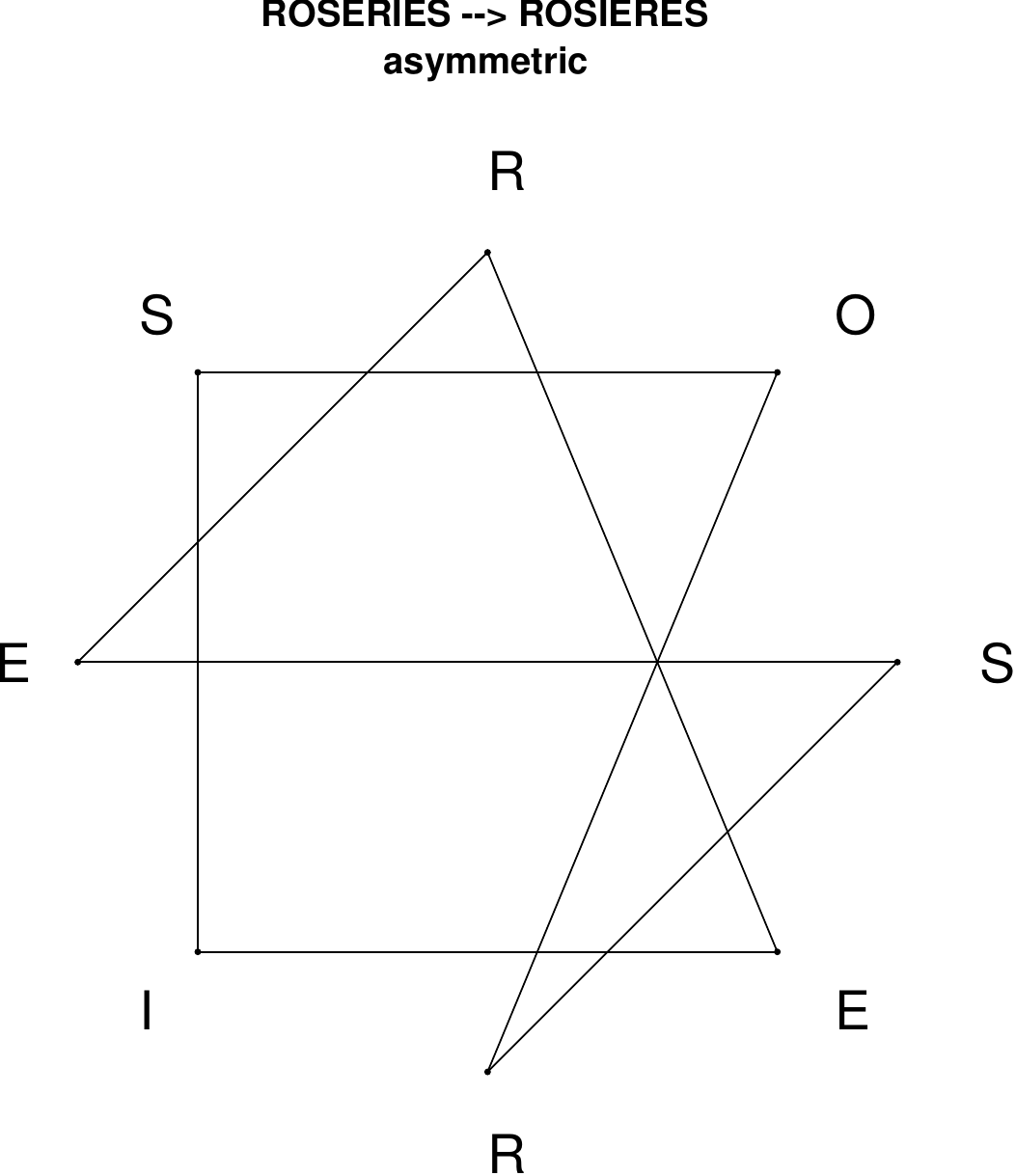}
\end{subfigure}
\hfill
\begin{subfigure}[T]{0.19\textwidth}
\centering
\includegraphics[width=\textwidth]{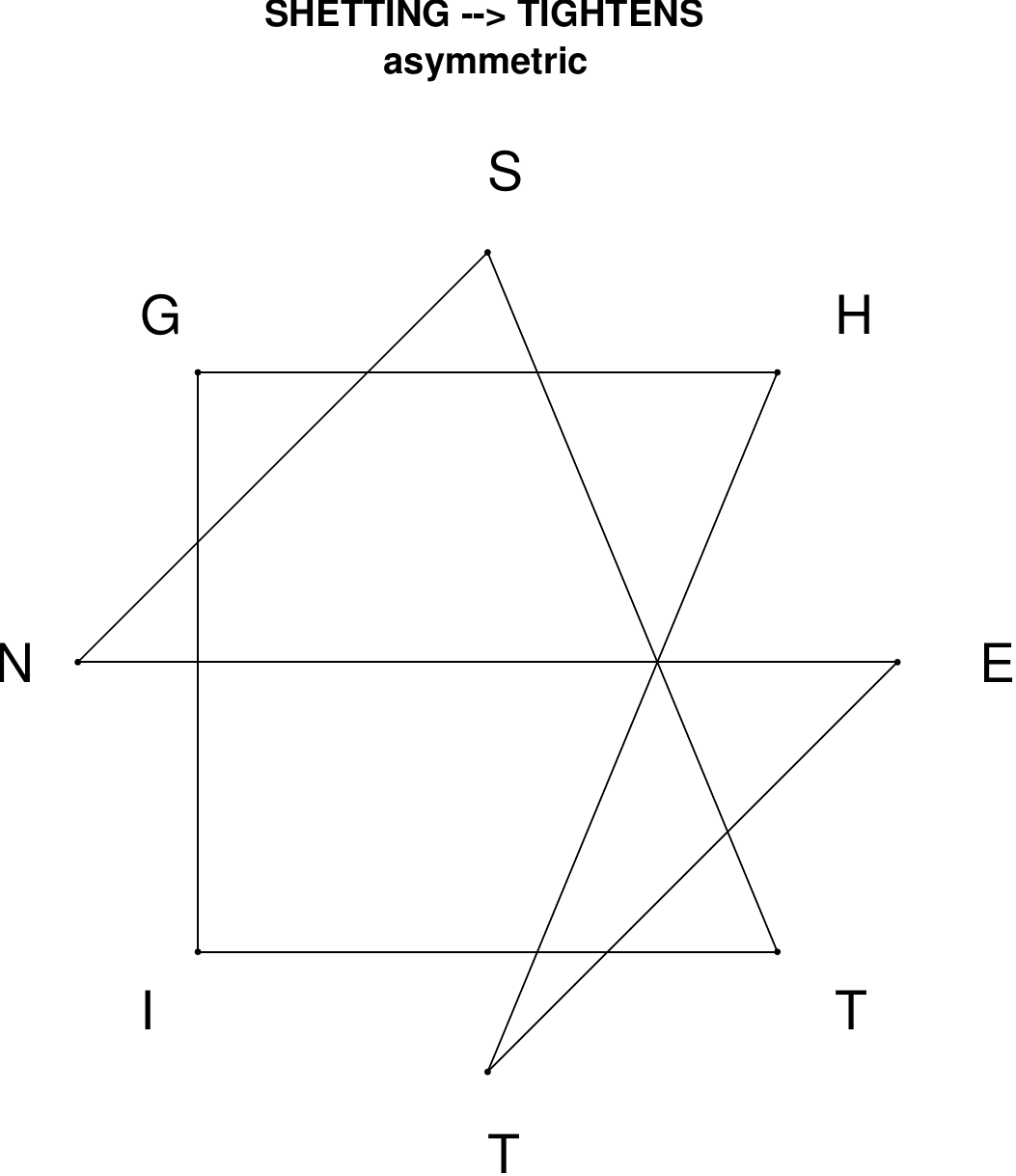}
\end{subfigure}
\hfill
\begin{subfigure}[T]{0.19\textwidth}
\centering
\includegraphics[width=\textwidth]{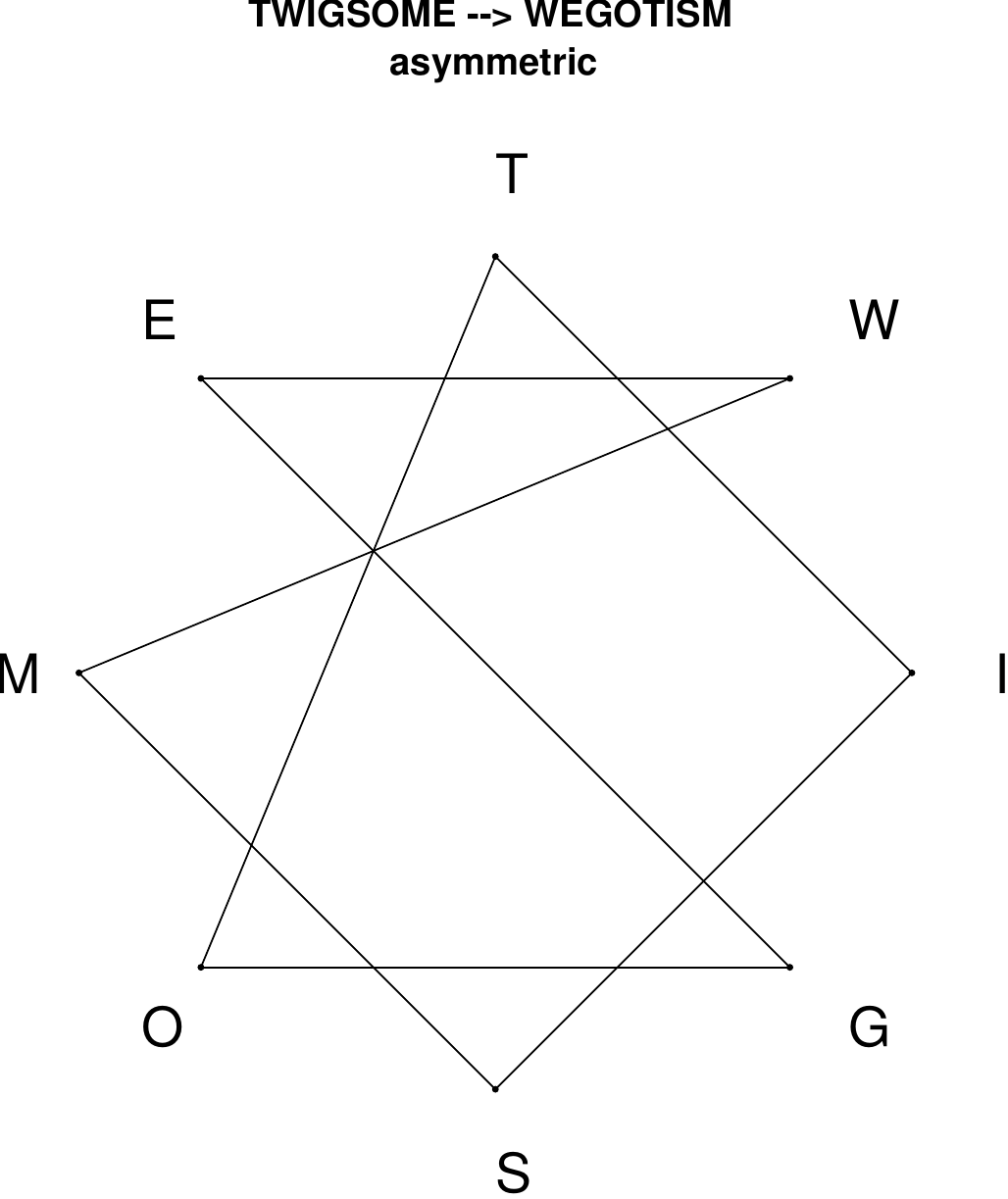}
\end{subfigure}
\end{figure}

\begin{figure}[H]
\centering
\begin{subfigure}[T]{0.19\textwidth}
\centering
\includegraphics[width=\textwidth]{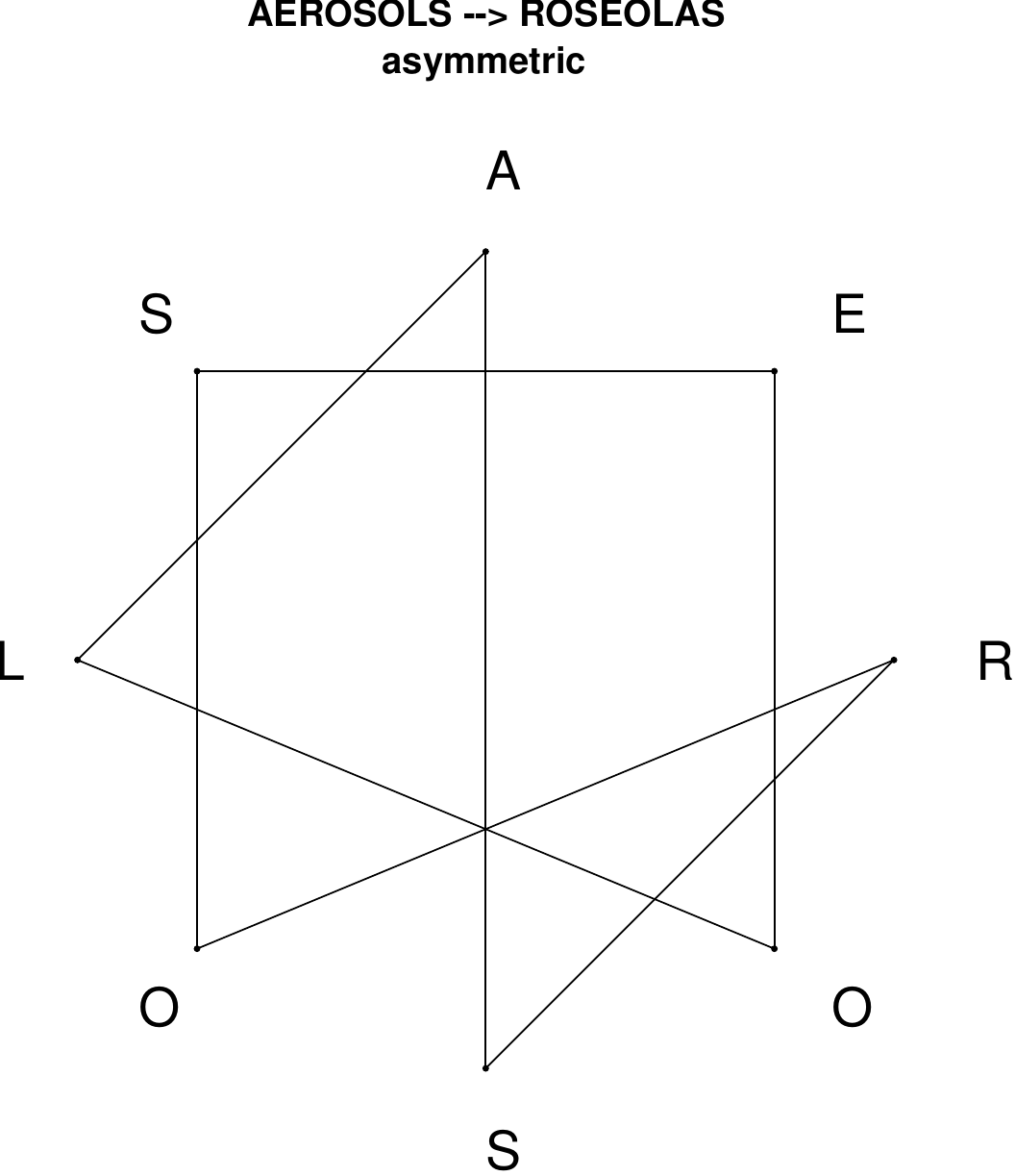}
\end{subfigure}
\hfill
\begin{subfigure}[T]{0.19\textwidth}
\centering
\includegraphics[width=\textwidth]{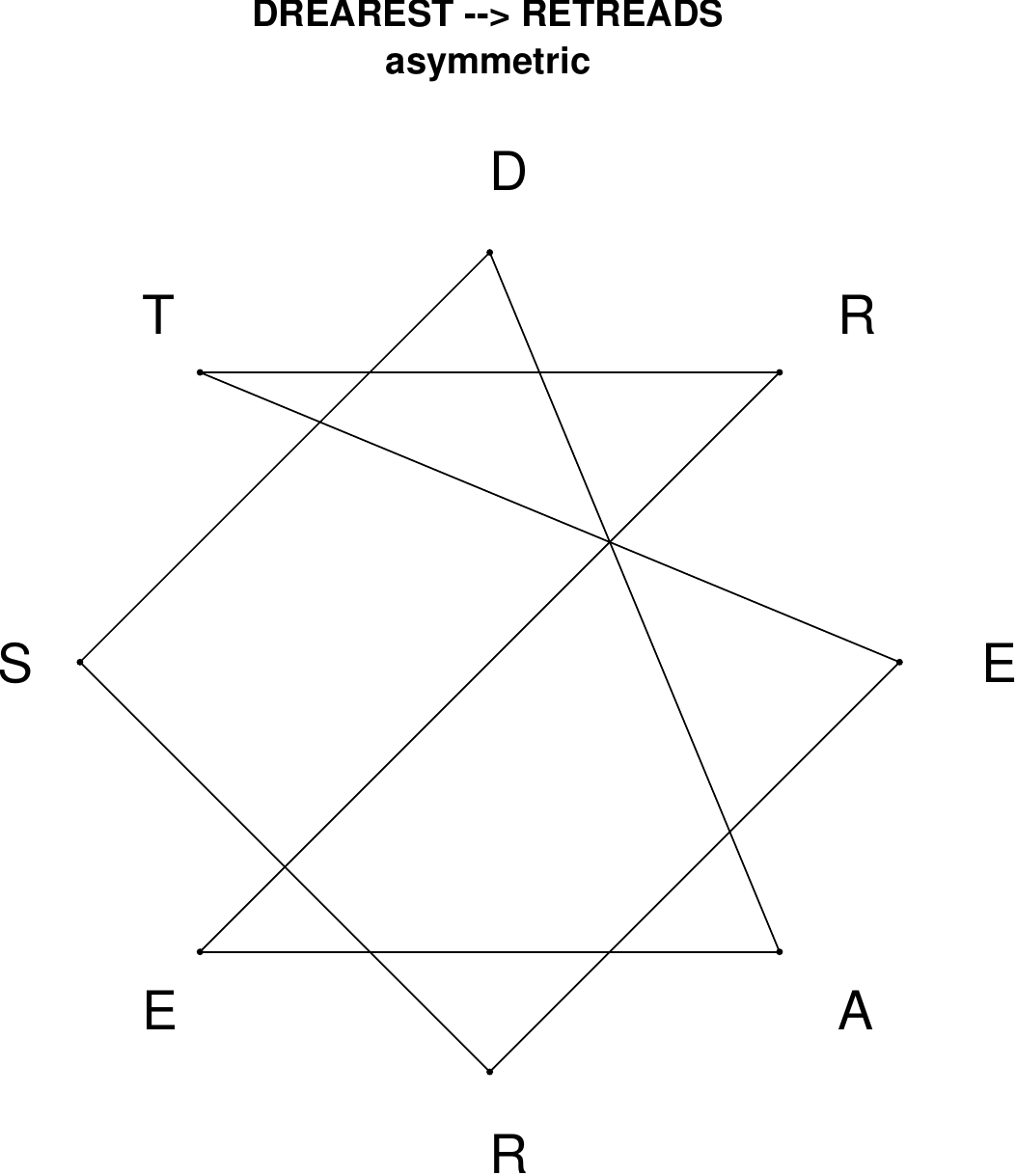}
\end{subfigure}
\hfill
\begin{subfigure}[T]{0.19\textwidth}
\centering
\includegraphics[width=\textwidth]{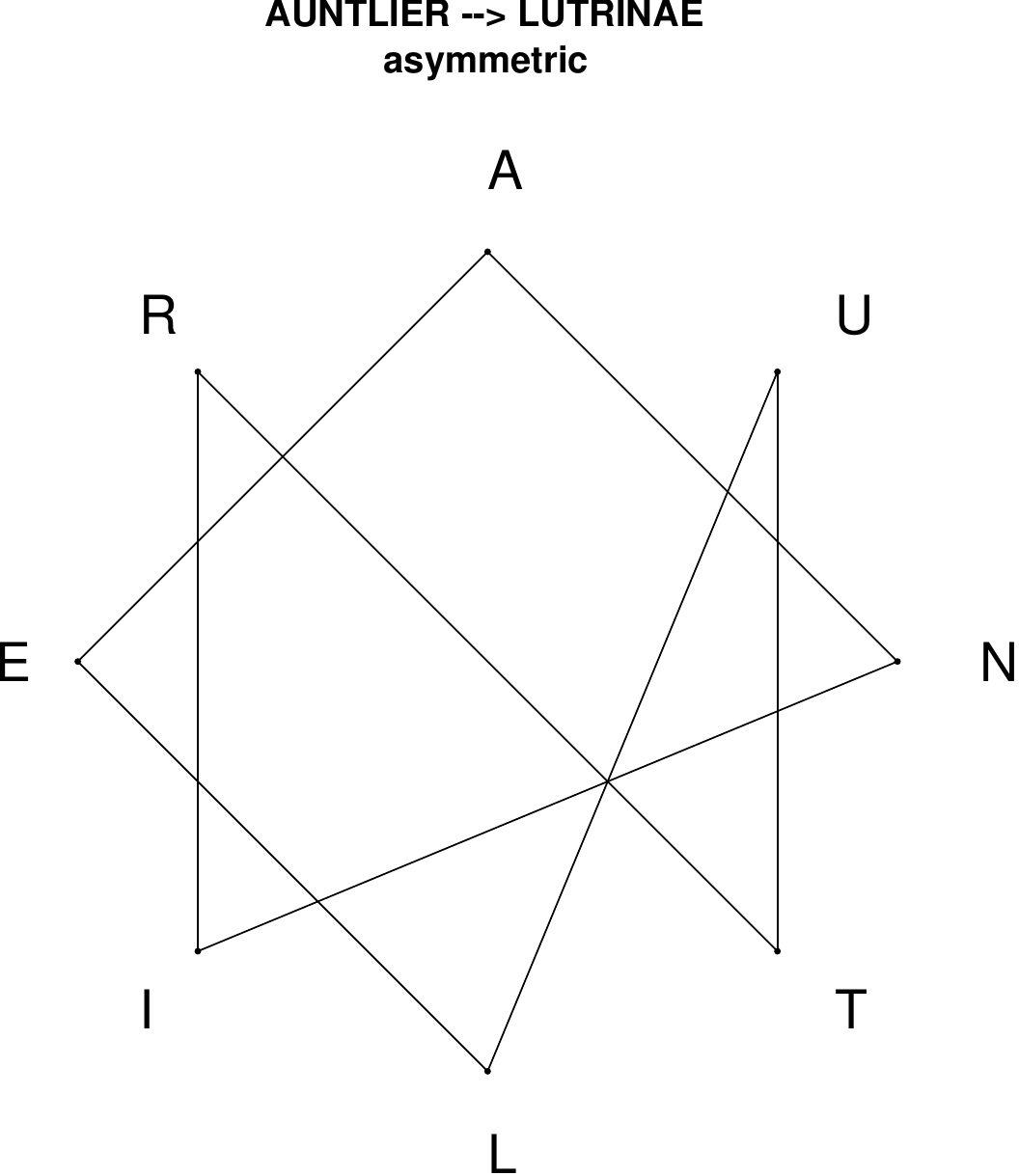}
\end{subfigure}
\hfill
\begin{subfigure}[T]{0.19\textwidth}
\centering
\includegraphics[width=\textwidth]{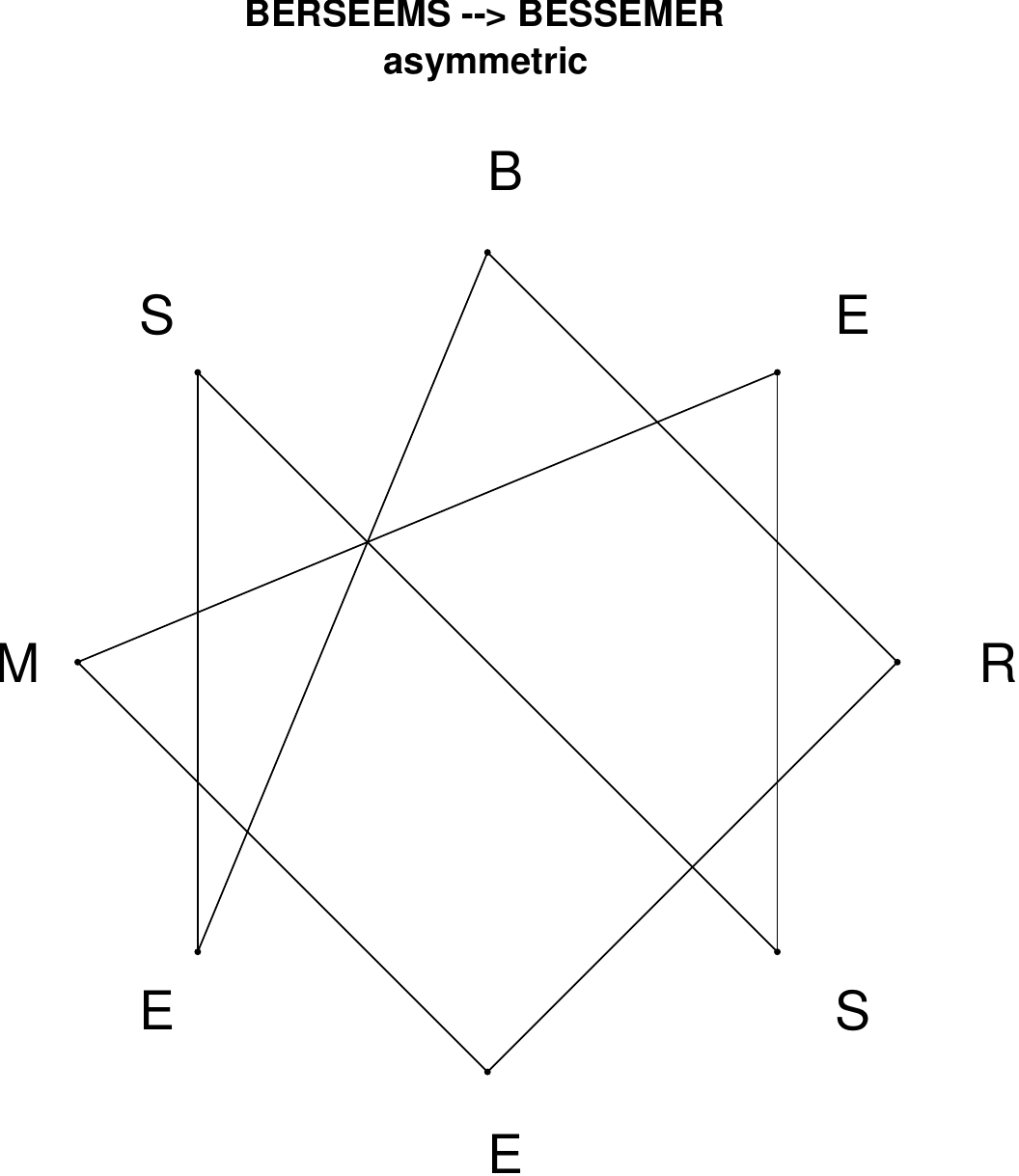}
\end{subfigure}
\hfill
\begin{subfigure}[T]{0.19\textwidth}
\centering
\includegraphics[width=\textwidth]{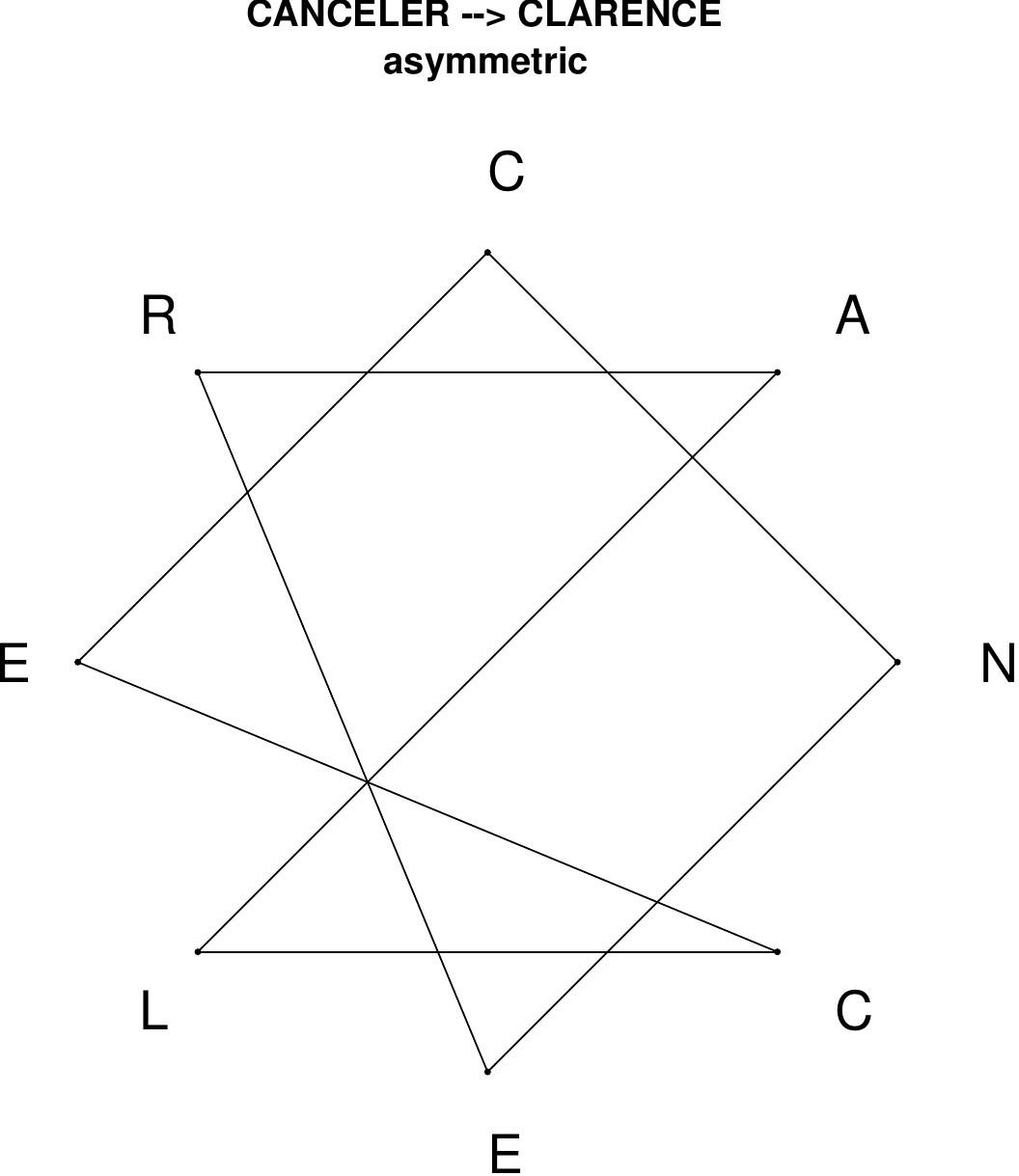}
\end{subfigure}
\end{figure}

\begin{figure}[H]
\centering
\begin{subfigure}[T]{0.19\textwidth}
\centering
\includegraphics[width=\textwidth]{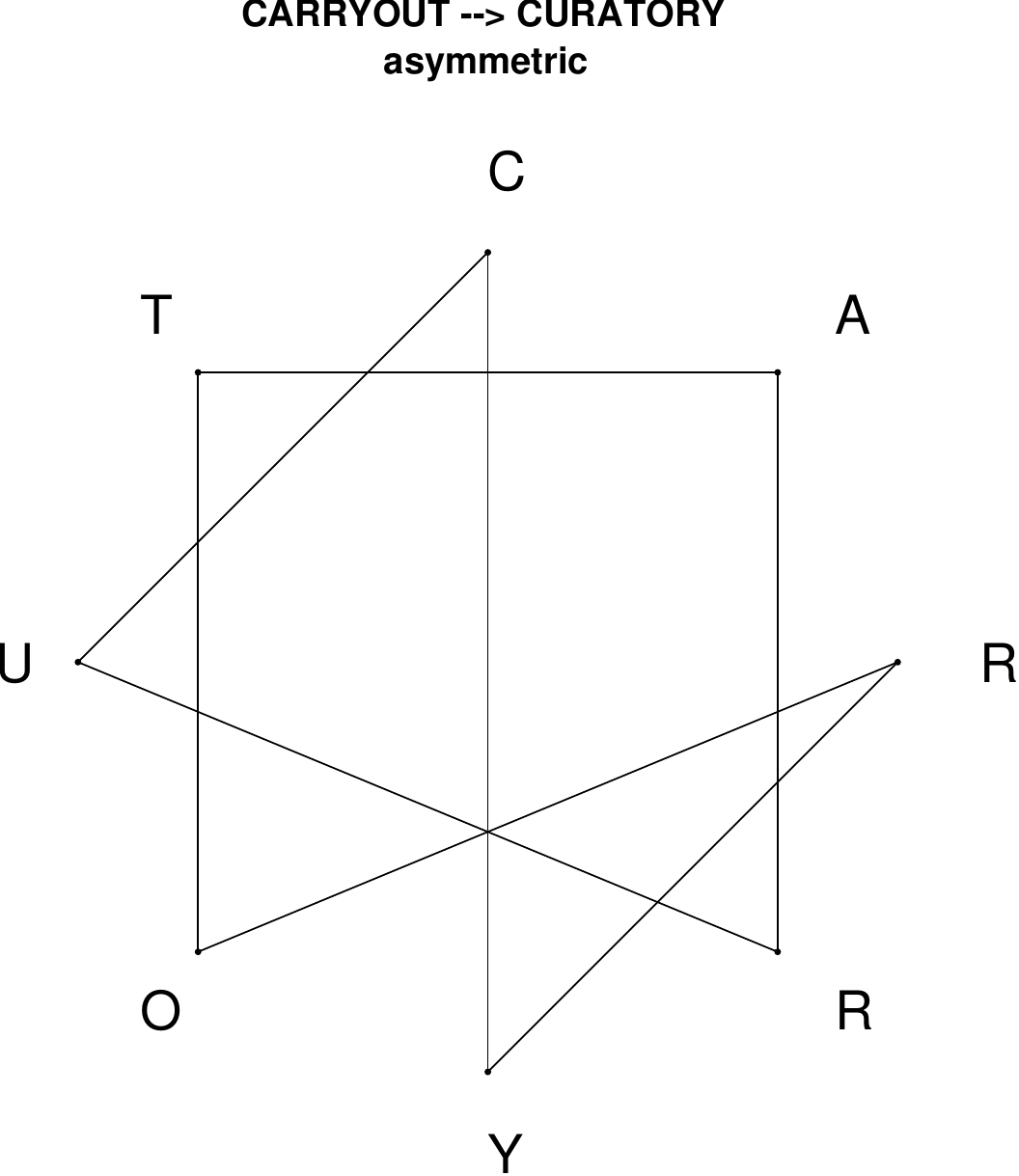}
\end{subfigure}
\hfill
\begin{subfigure}[T]{0.19\textwidth}
\centering
\includegraphics[width=\textwidth]{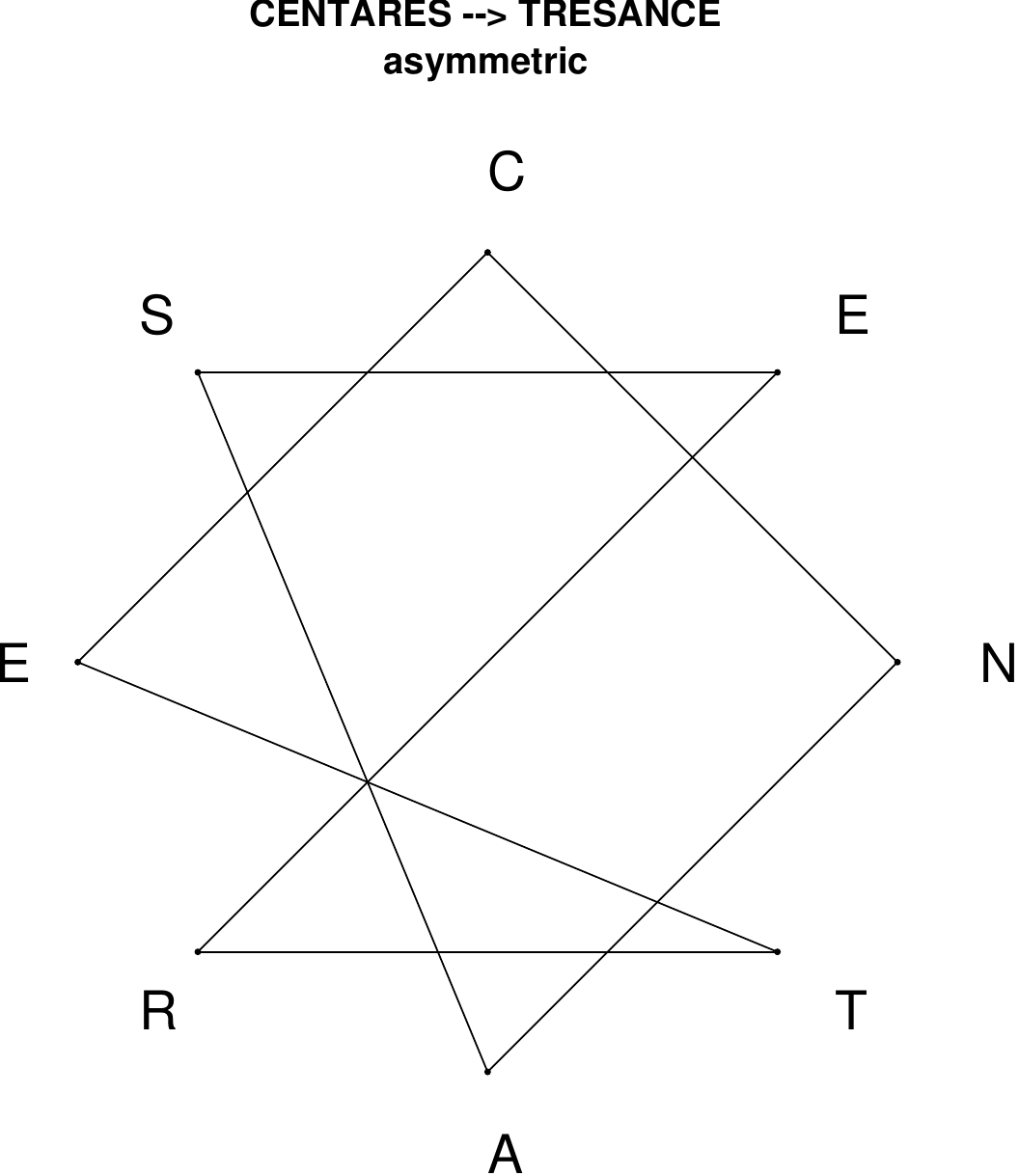}
\end{subfigure}
\hfill
\begin{subfigure}[T]{0.19\textwidth}
\centering
\includegraphics[width=\textwidth]{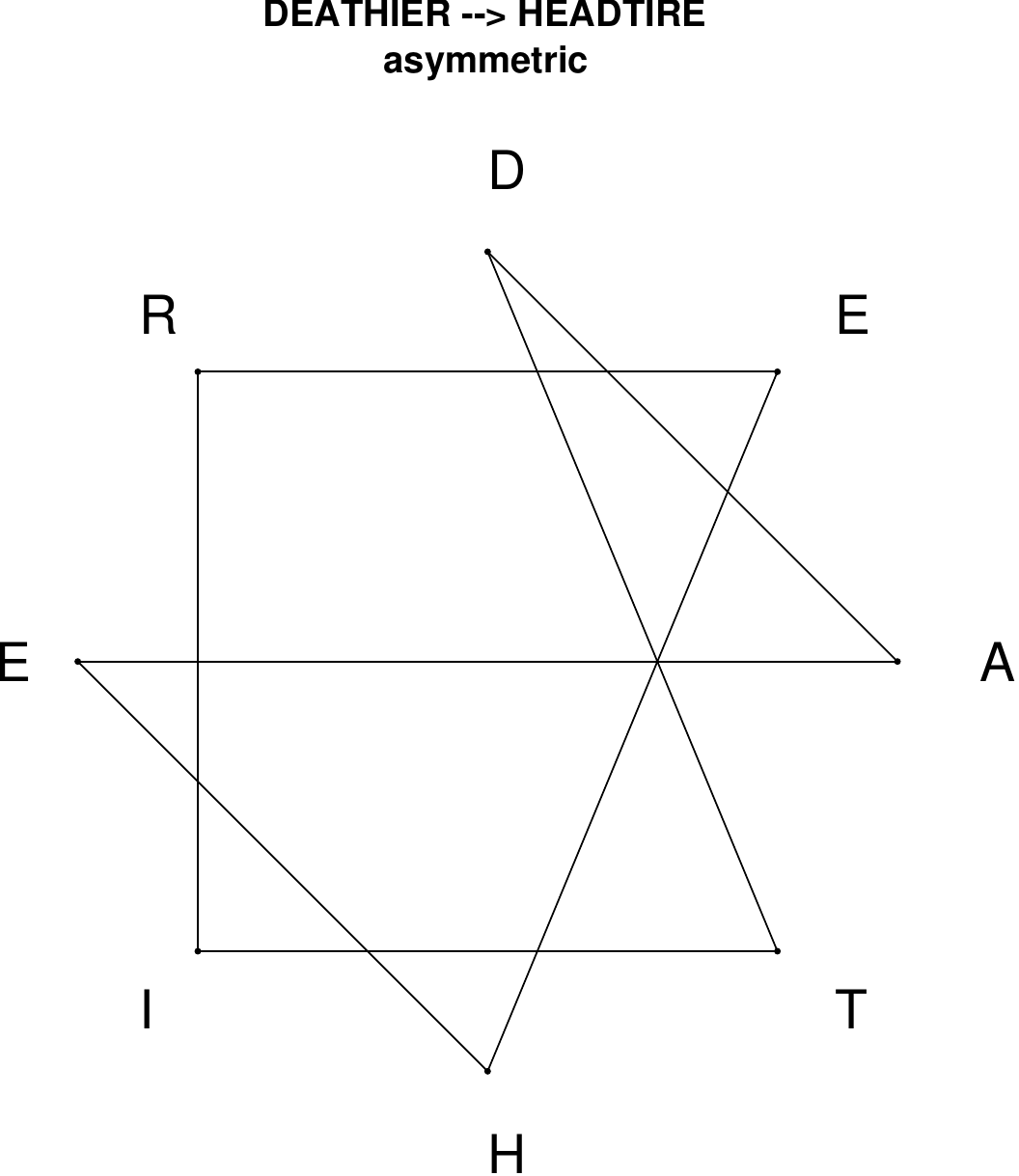}
\end{subfigure}
\hfill
\begin{subfigure}[T]{0.19\textwidth}
\centering
\includegraphics[width=\textwidth]{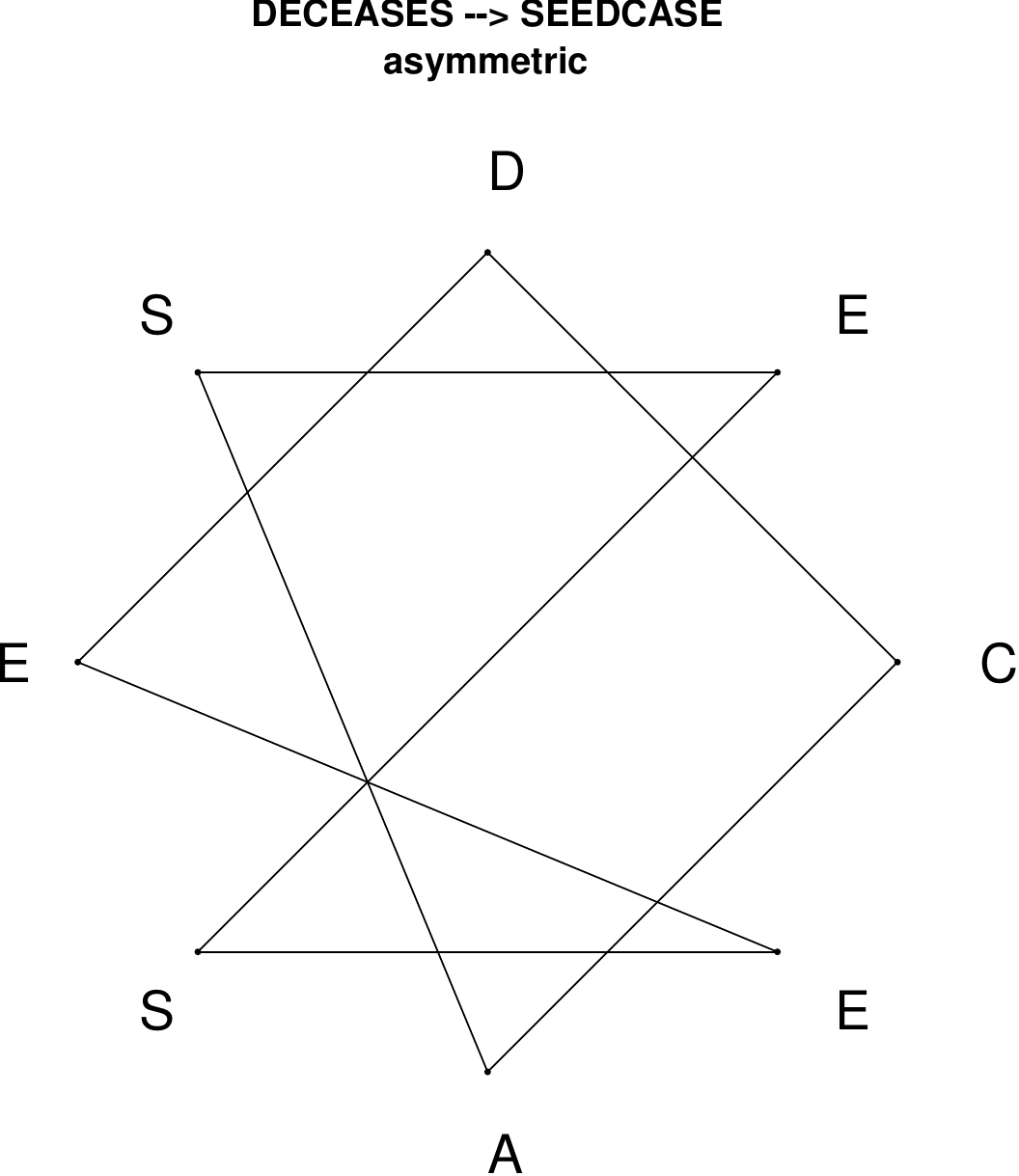}
\end{subfigure}
\hfill
\begin{subfigure}[T]{0.19\textwidth}
\centering
\includegraphics[width=\textwidth]{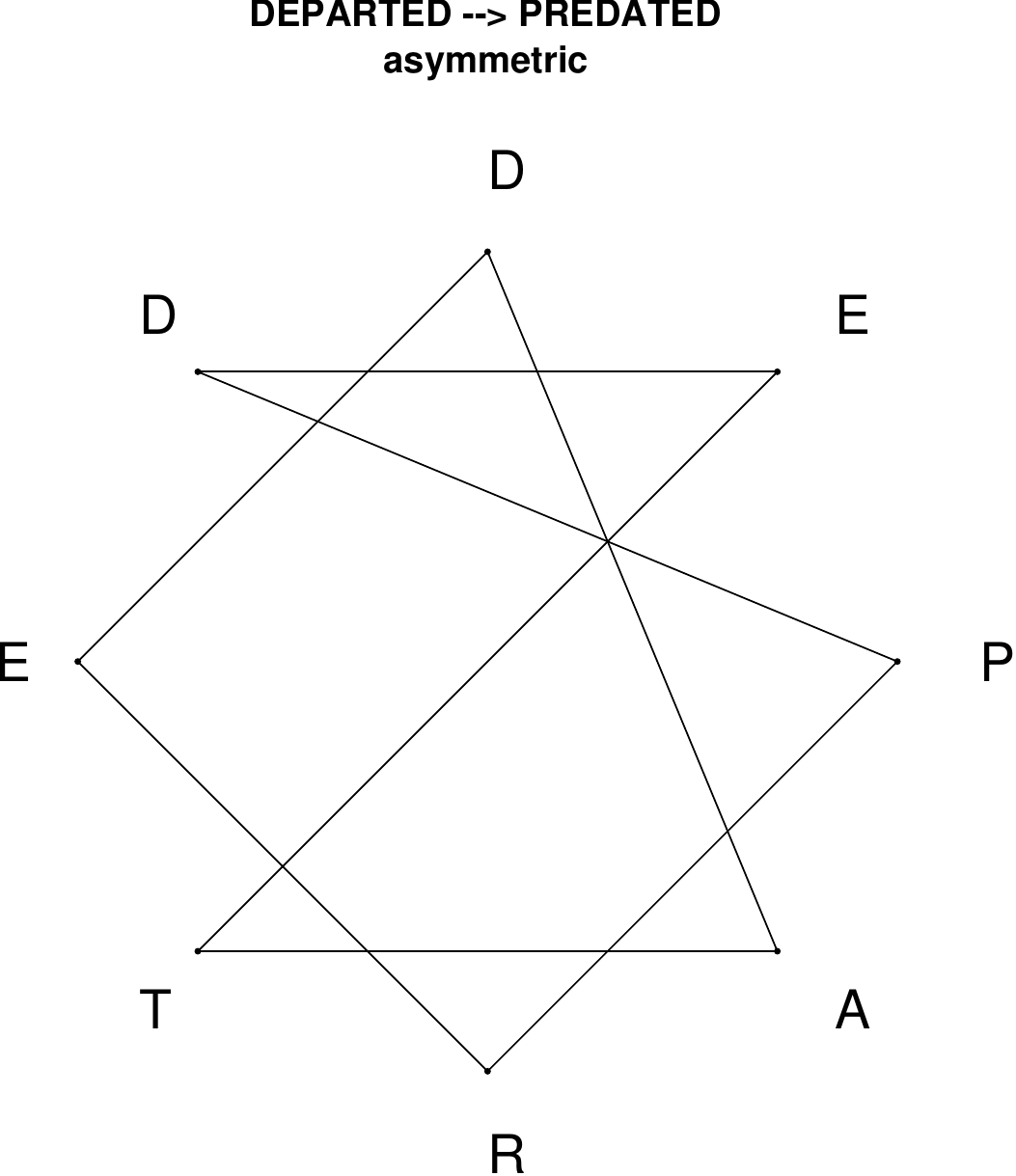}
\end{subfigure}
\end{figure}

\begin{figure}[H]
\centering
\begin{subfigure}[T]{0.19\textwidth}
\centering
\includegraphics[width=\textwidth]{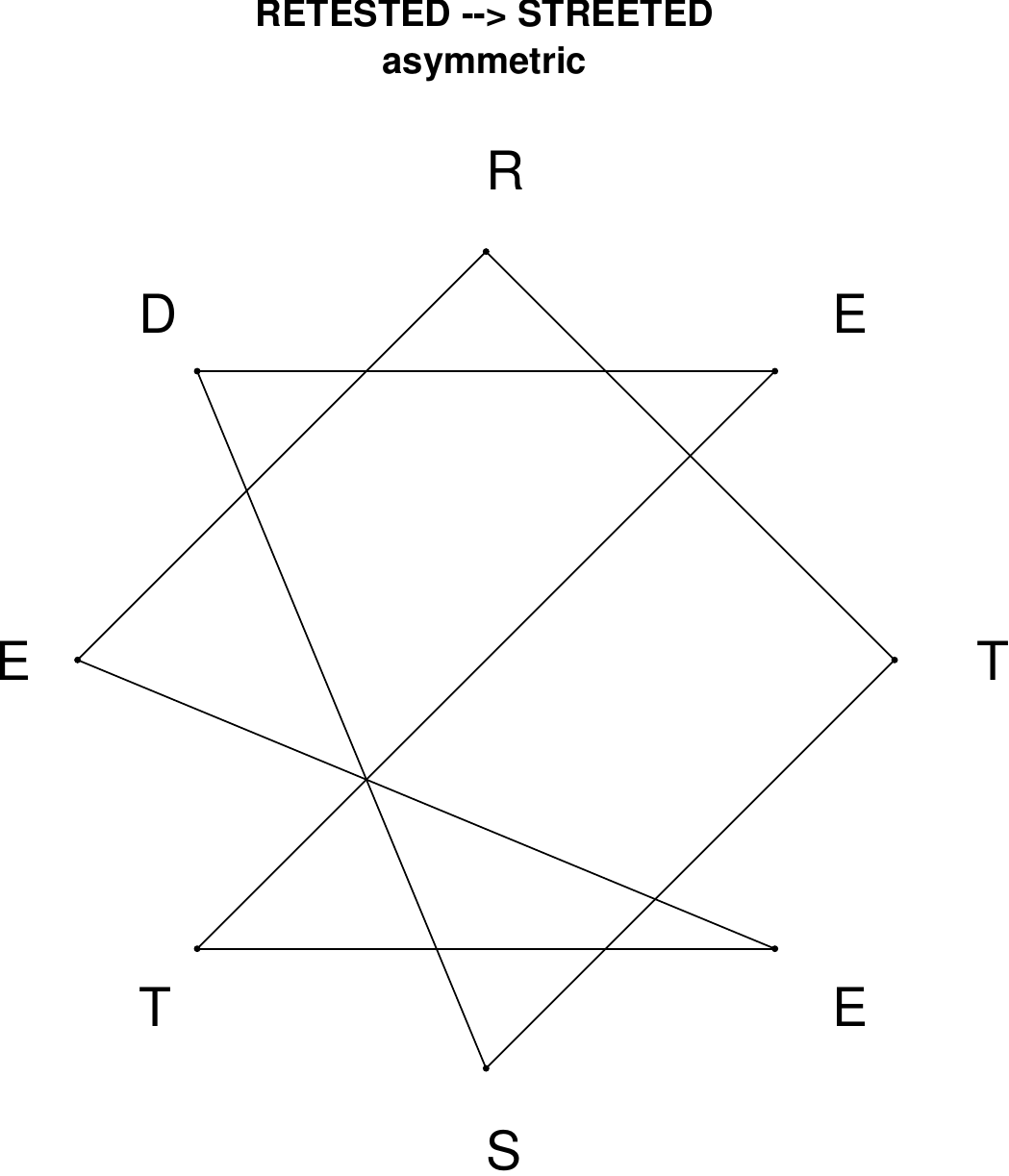}
\end{subfigure}
\hfill
\begin{subfigure}[T]{0.19\textwidth}
\centering
\includegraphics[width=\textwidth]{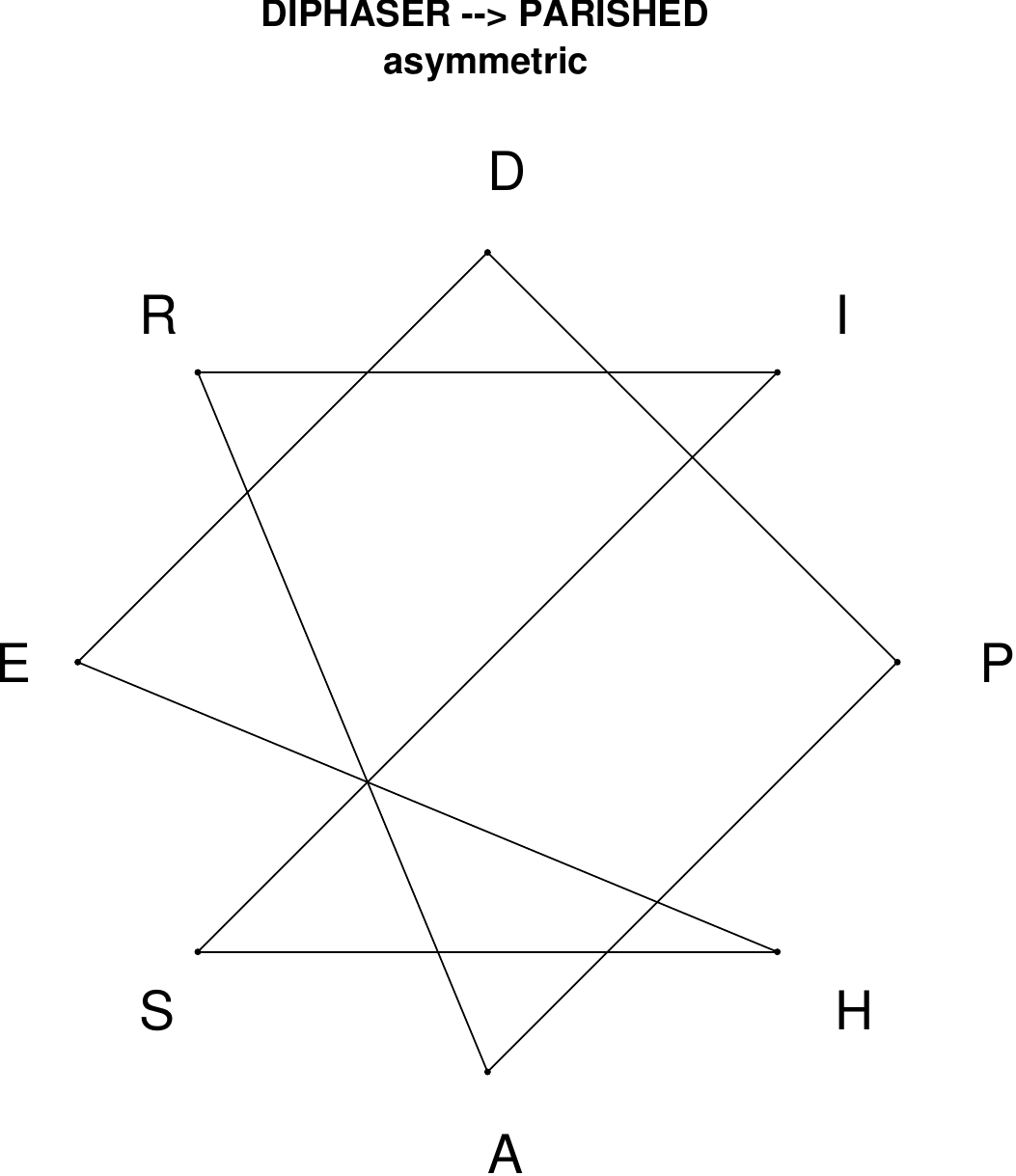}
\end{subfigure}
\hfill
\begin{subfigure}[T]{0.19\textwidth}
\centering
\includegraphics[width=\textwidth]{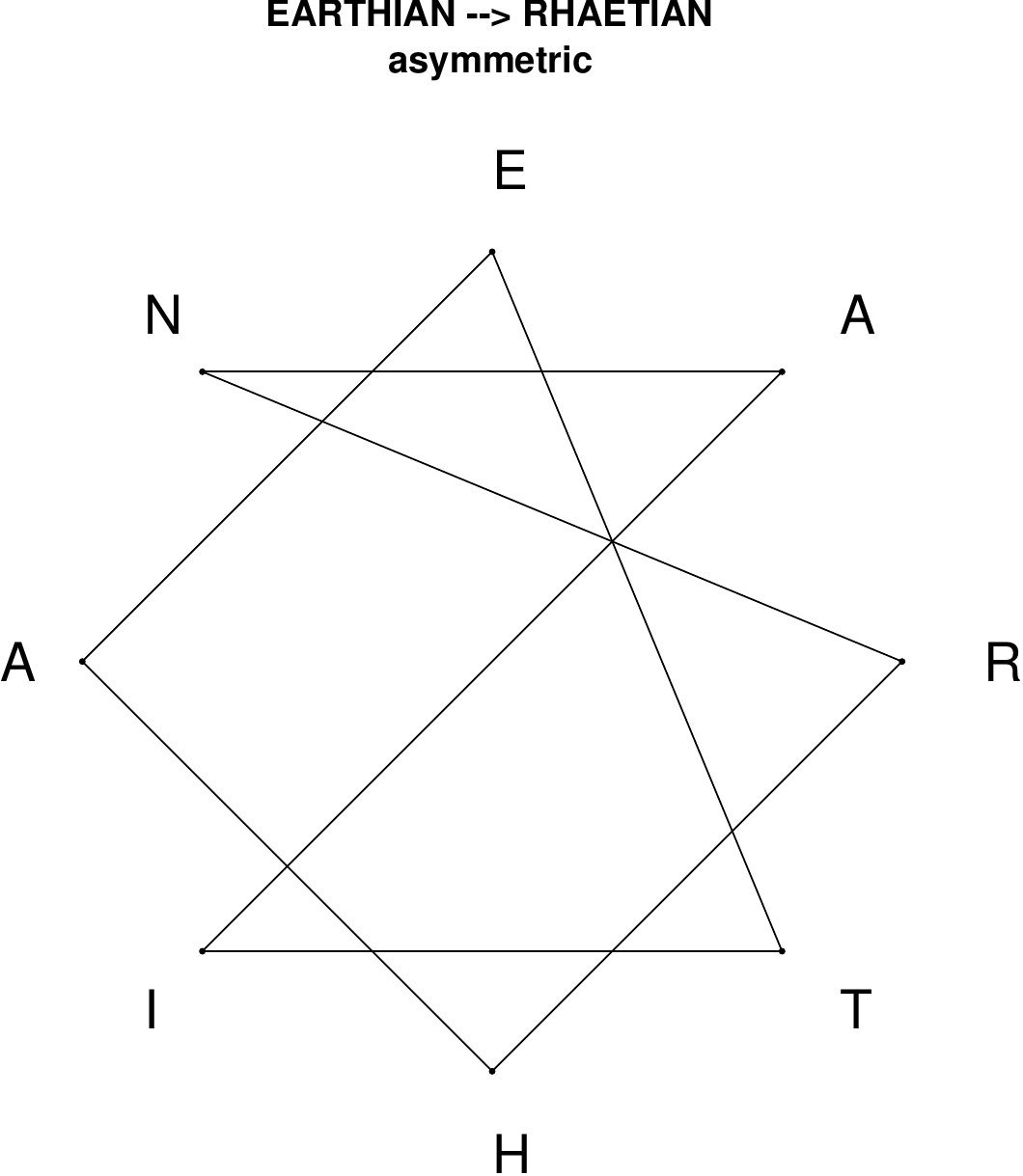}
\end{subfigure}
\hfill
\begin{subfigure}[T]{0.19\textwidth}
\centering
\includegraphics[width=\textwidth]{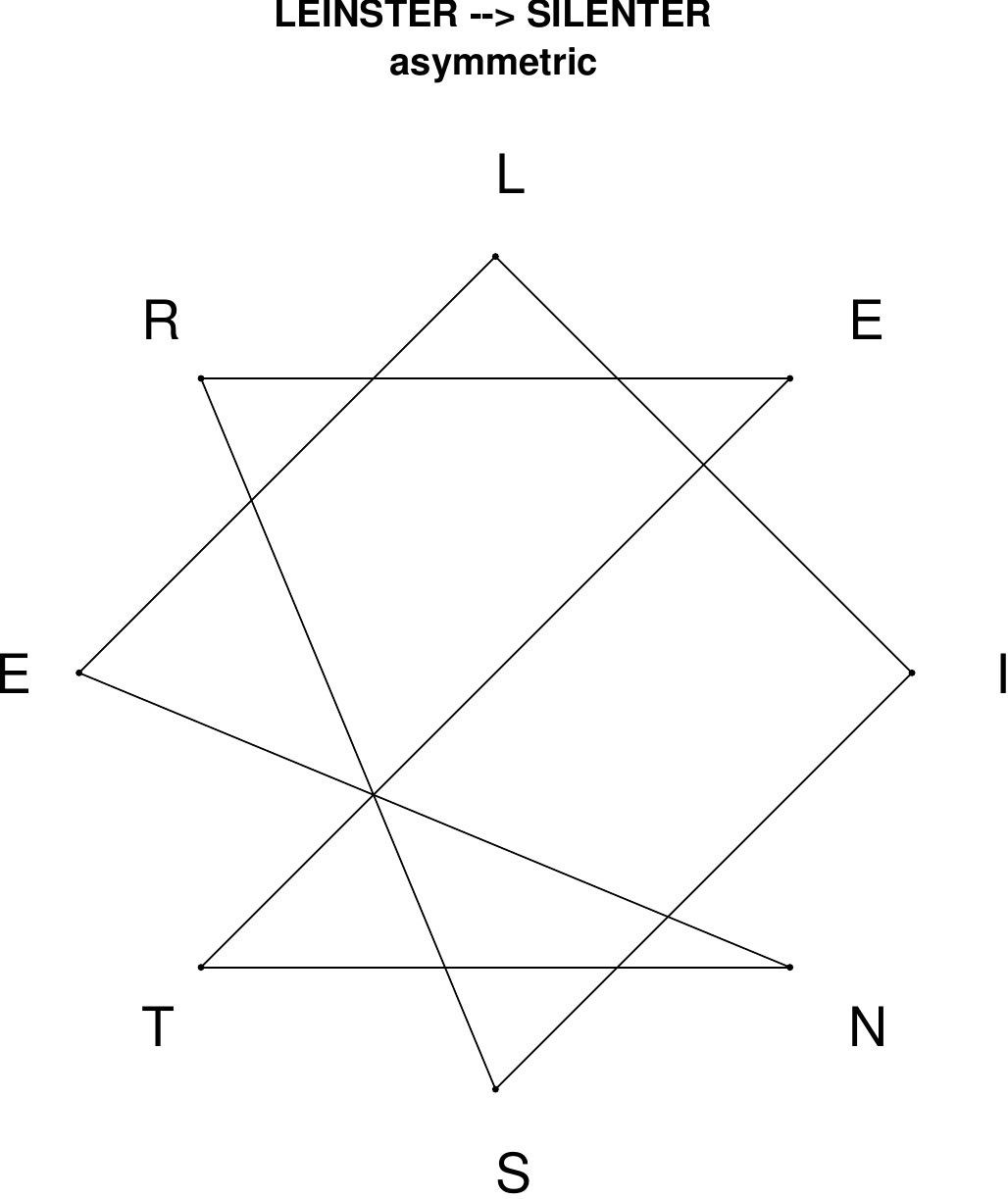}
\end{subfigure}
\hfill
\begin{subfigure}[T]{0.19\textwidth}
\centering
\includegraphics[width=\textwidth]{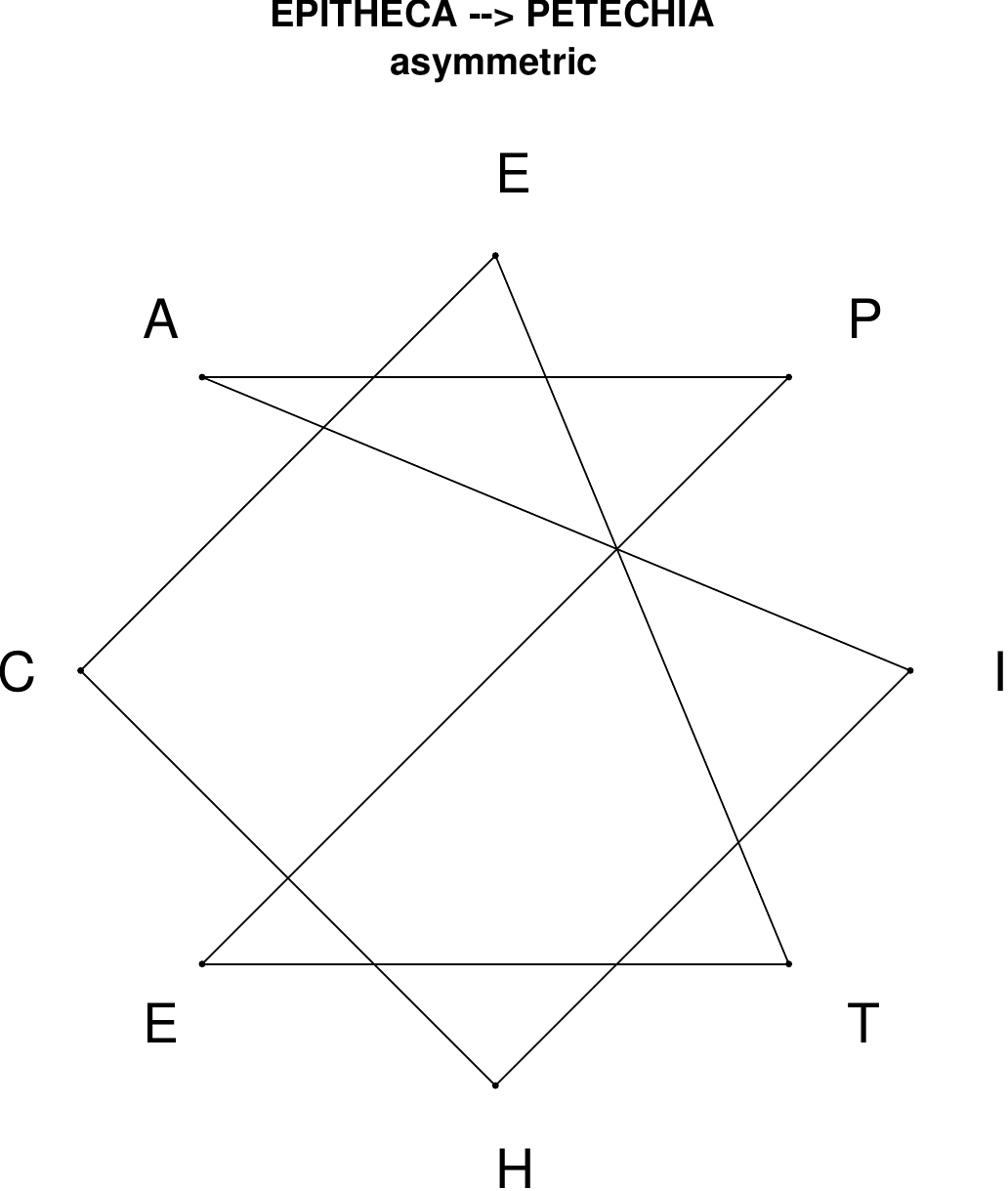}
\end{subfigure}
\end{figure}

\begin{figure}[H]
\centering
\begin{subfigure}[T]{0.19\textwidth}
\centering
\includegraphics[width=\textwidth]{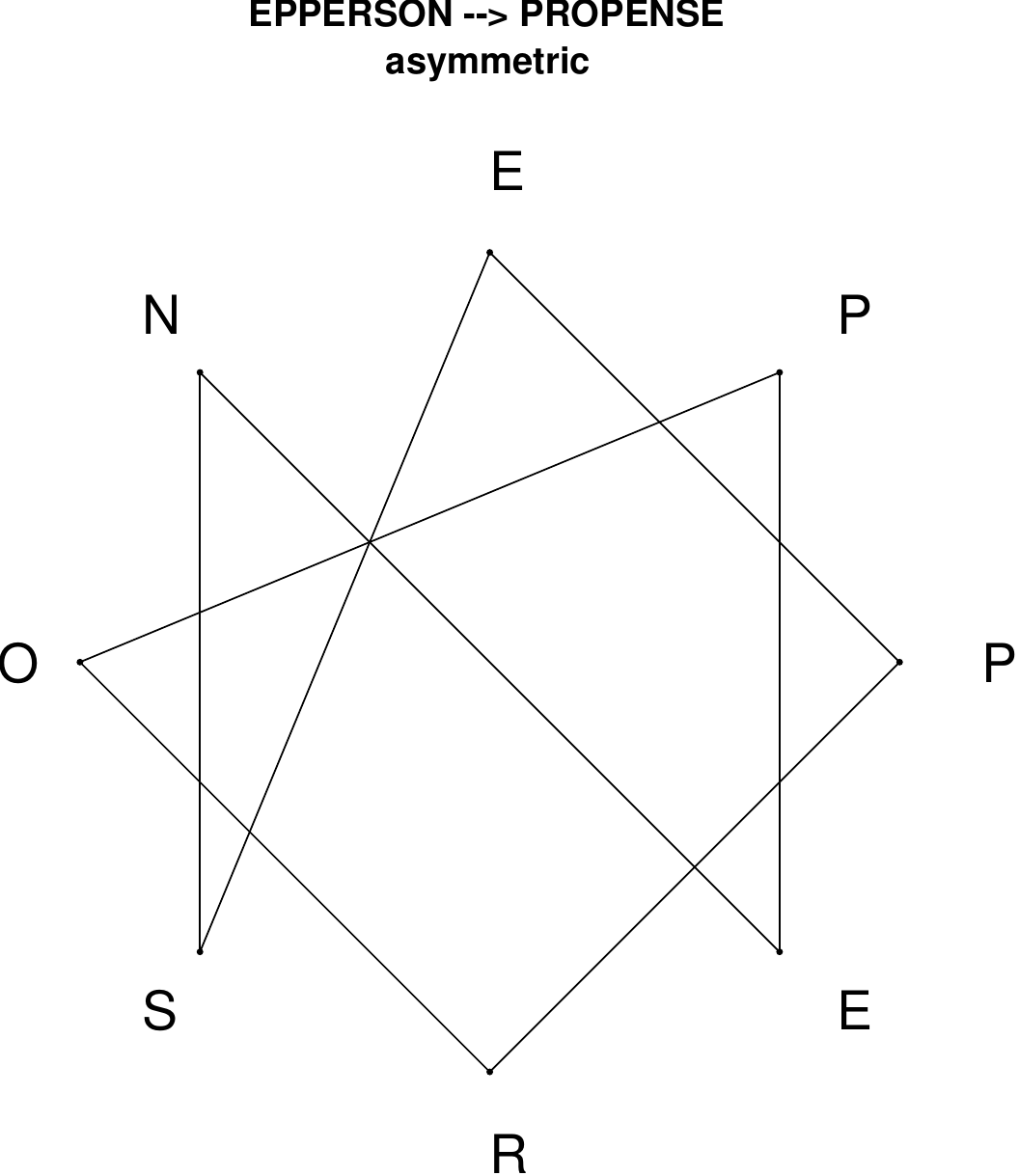}
\end{subfigure}
\hfill
\begin{subfigure}[T]{0.19\textwidth}
\centering
\includegraphics[width=\textwidth]{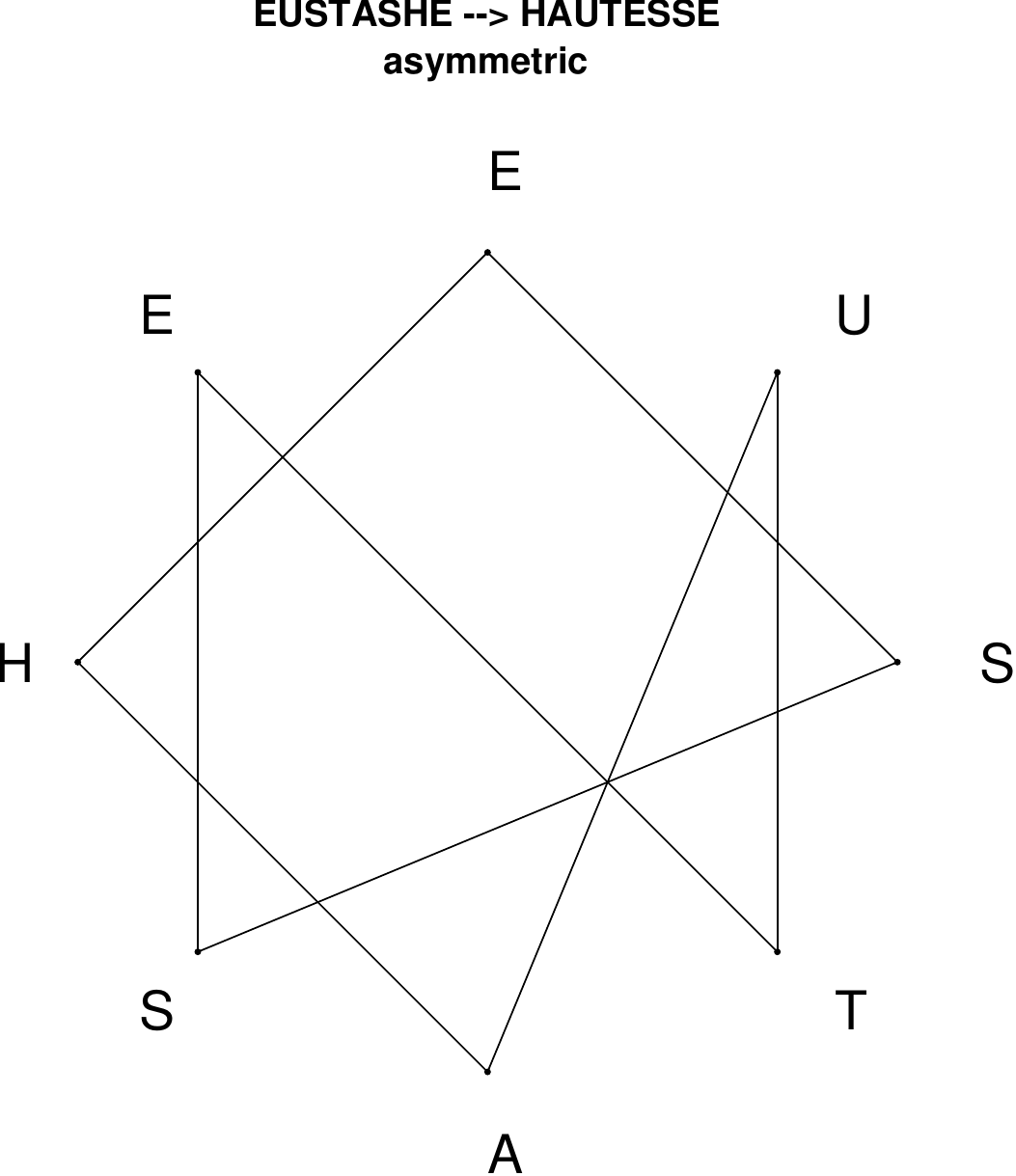}
\end{subfigure}
\hfill
\begin{subfigure}[T]{0.19\textwidth}
\centering
\includegraphics[width=\textwidth]{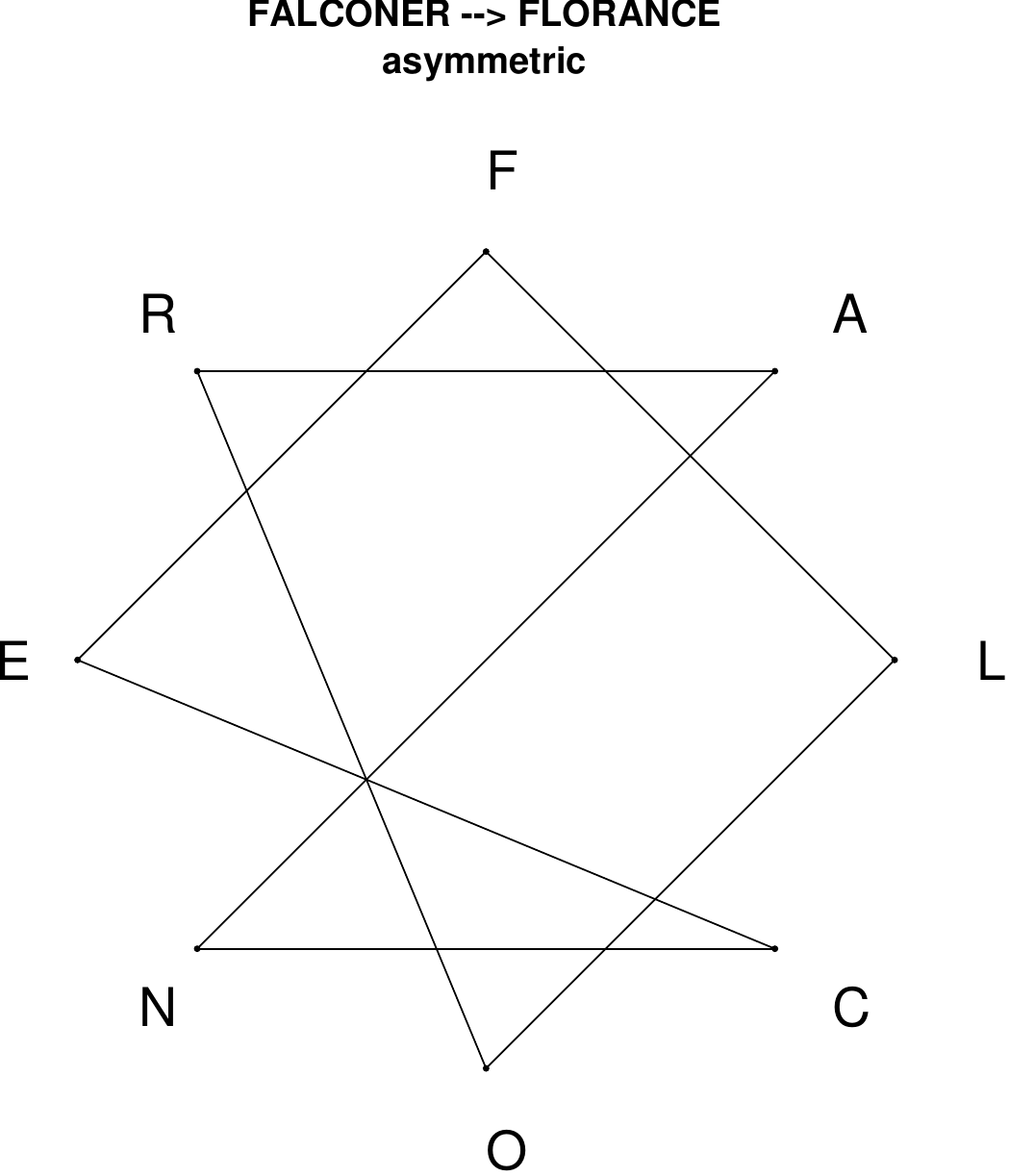}
\end{subfigure}
\hfill
\begin{subfigure}[T]{0.19\textwidth}
\centering
\includegraphics[width=\textwidth]{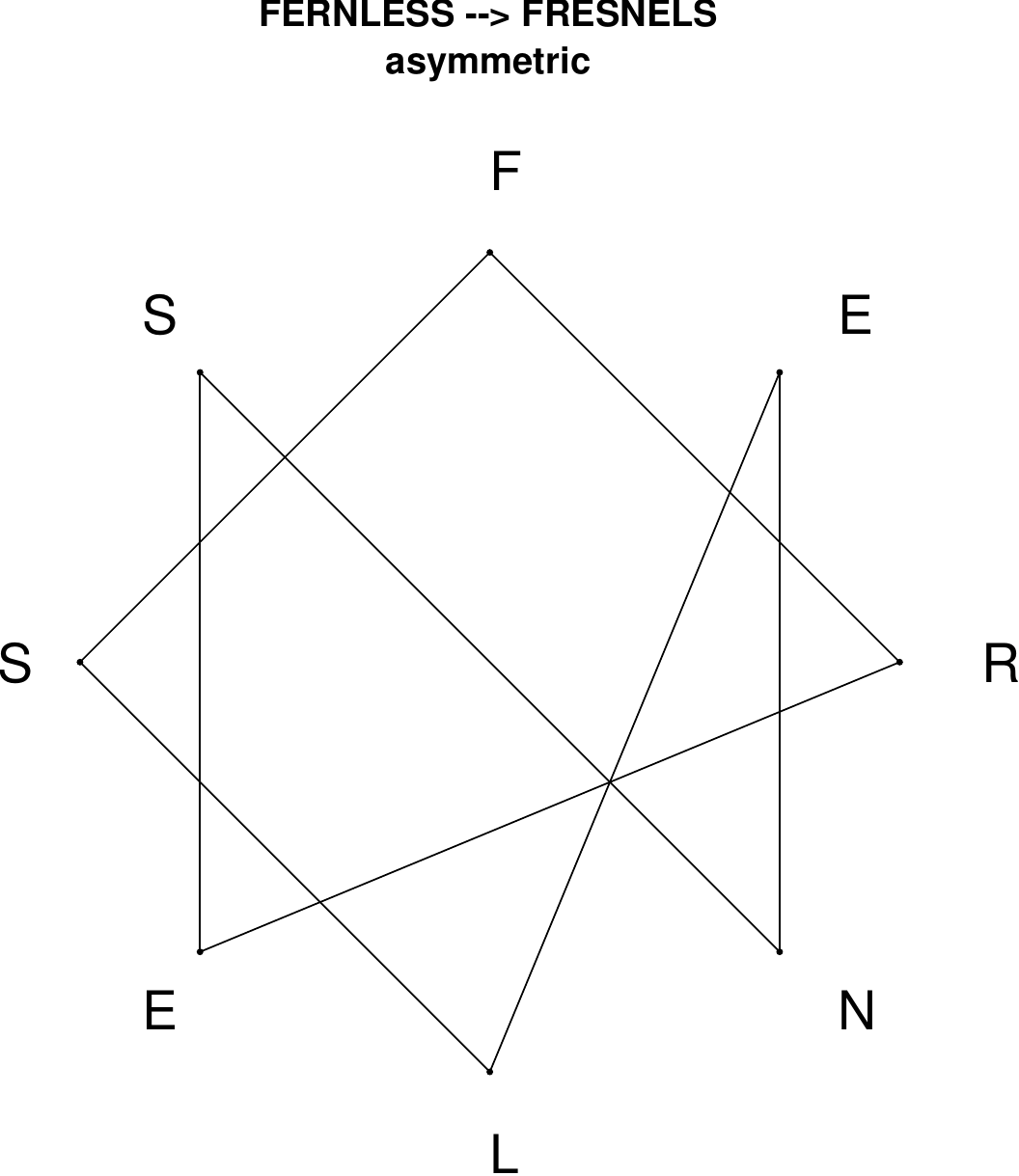}
\end{subfigure}
\hfill
\begin{subfigure}[T]{0.19\textwidth}
\centering
\includegraphics[width=\textwidth]{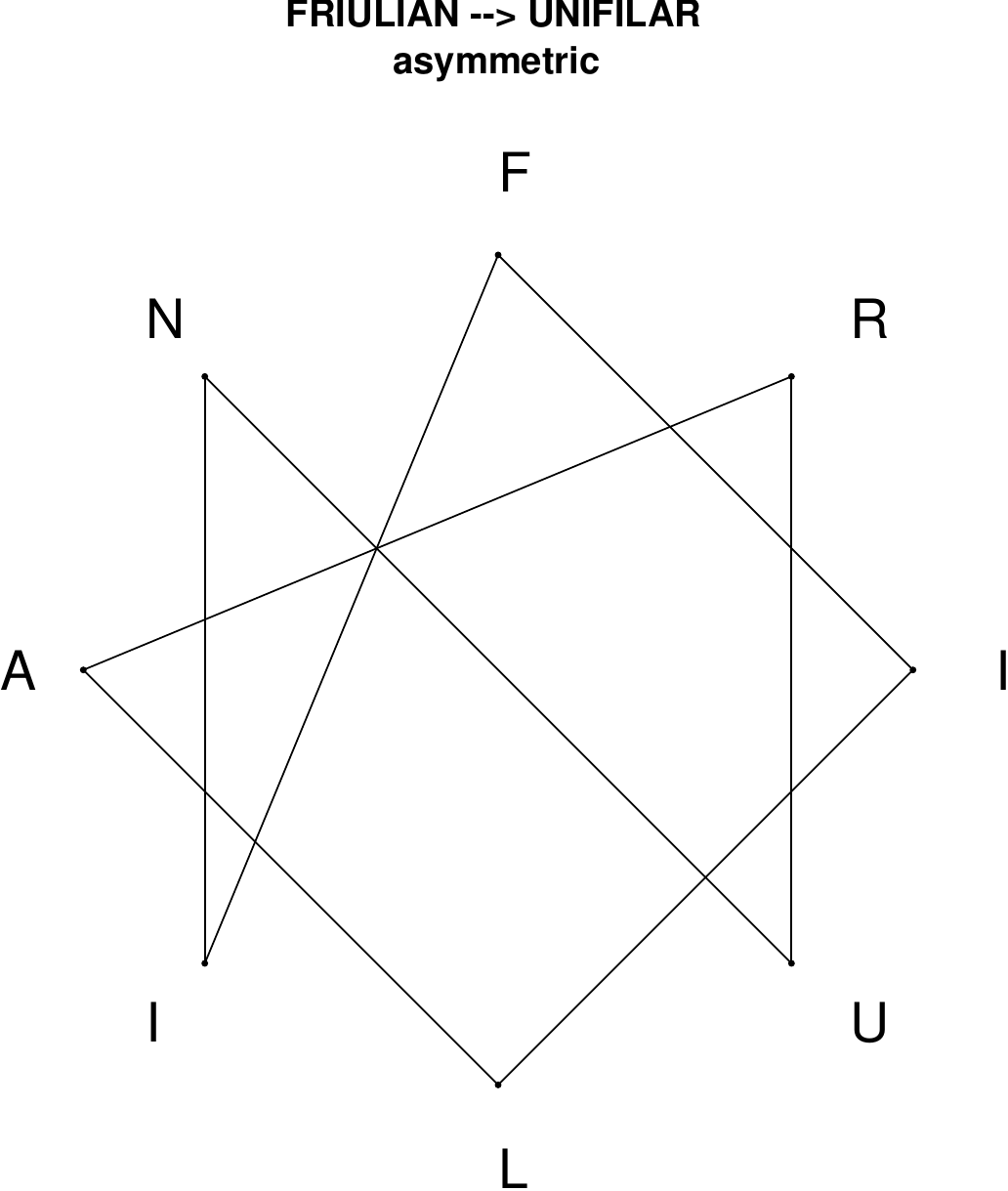}
\end{subfigure}
\end{figure}

\begin{figure}[H]
\centering
\begin{subfigure}[T]{0.19\textwidth}
\centering
\includegraphics[width=\textwidth]{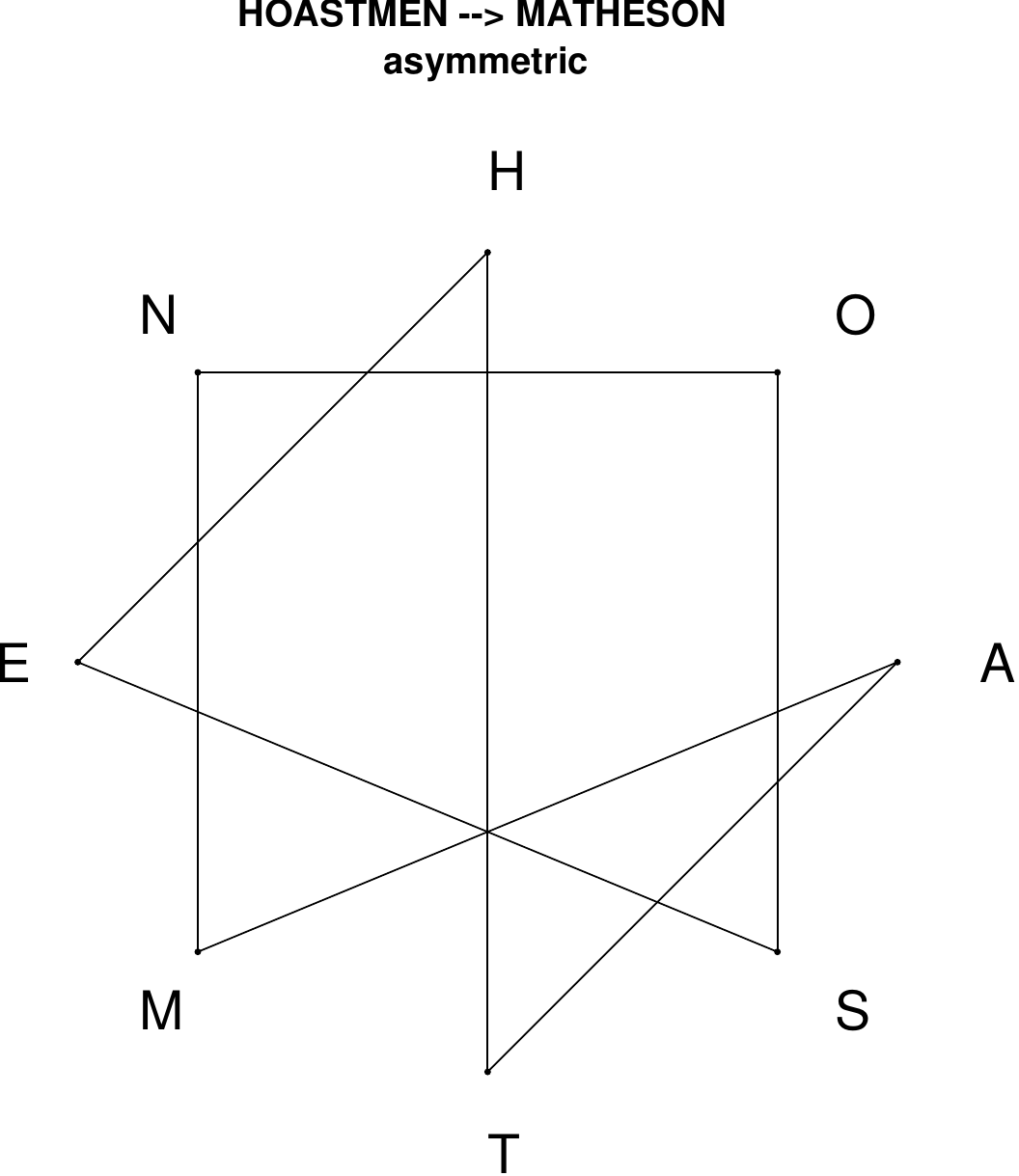}
\end{subfigure}
\hfill
\begin{subfigure}[T]{0.19\textwidth}
\centering
\includegraphics[width=\textwidth]{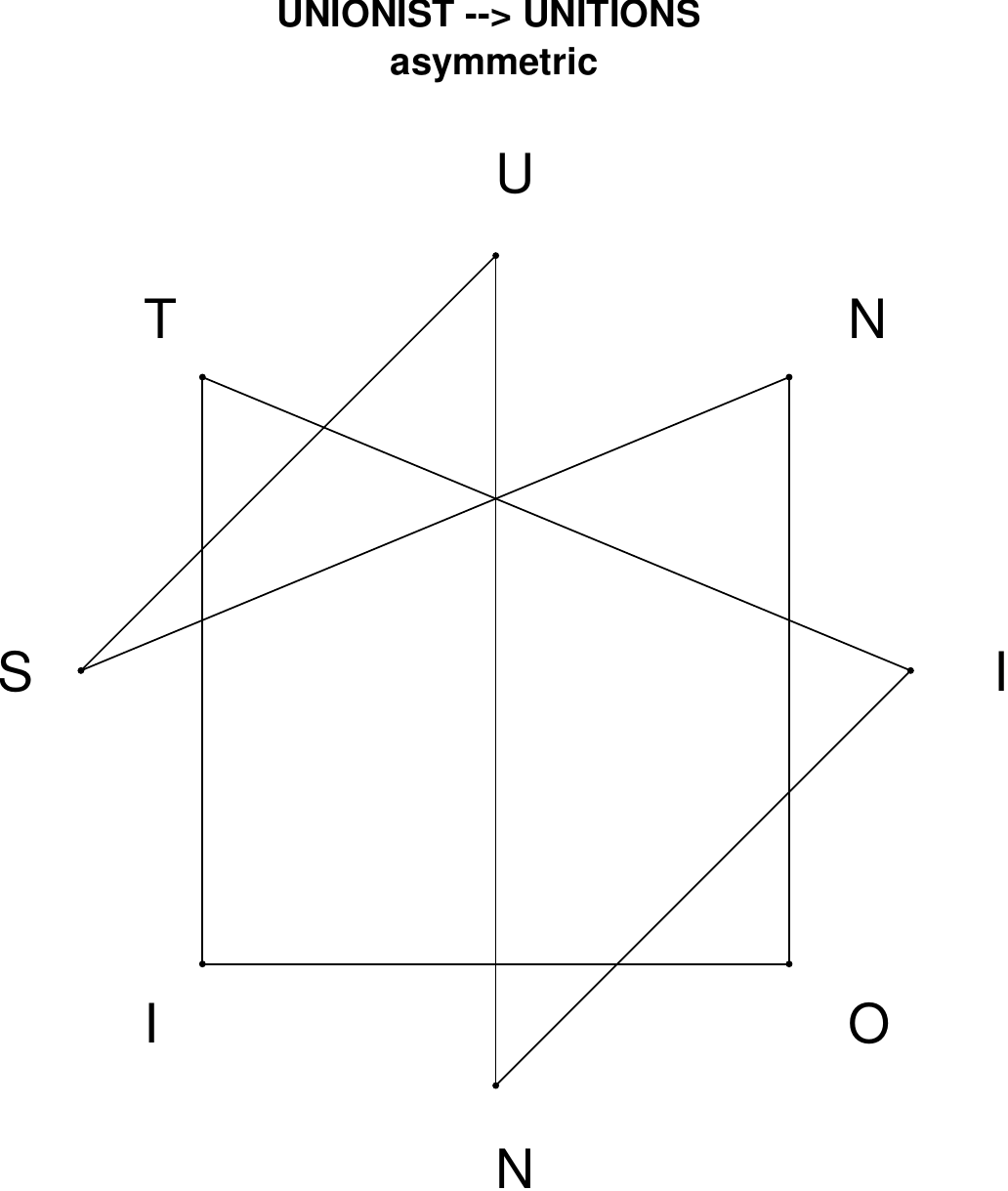}
\end{subfigure}
\hfill
\begin{subfigure}[T]{0.19\textwidth}
\centering
\includegraphics[width=\textwidth]{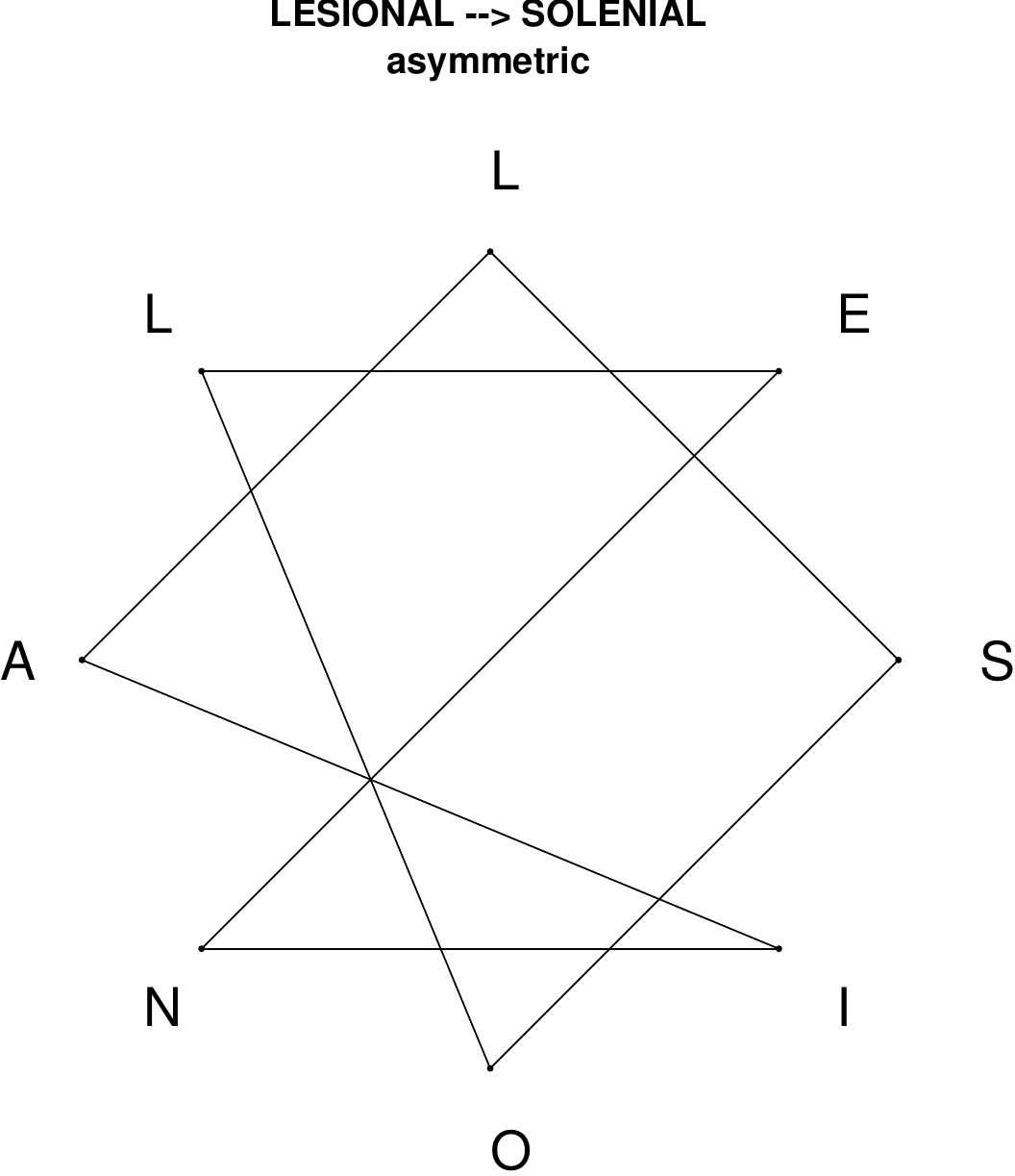}
\end{subfigure}
\hfill
\begin{subfigure}[T]{0.19\textwidth}
\centering
\includegraphics[width=\textwidth]{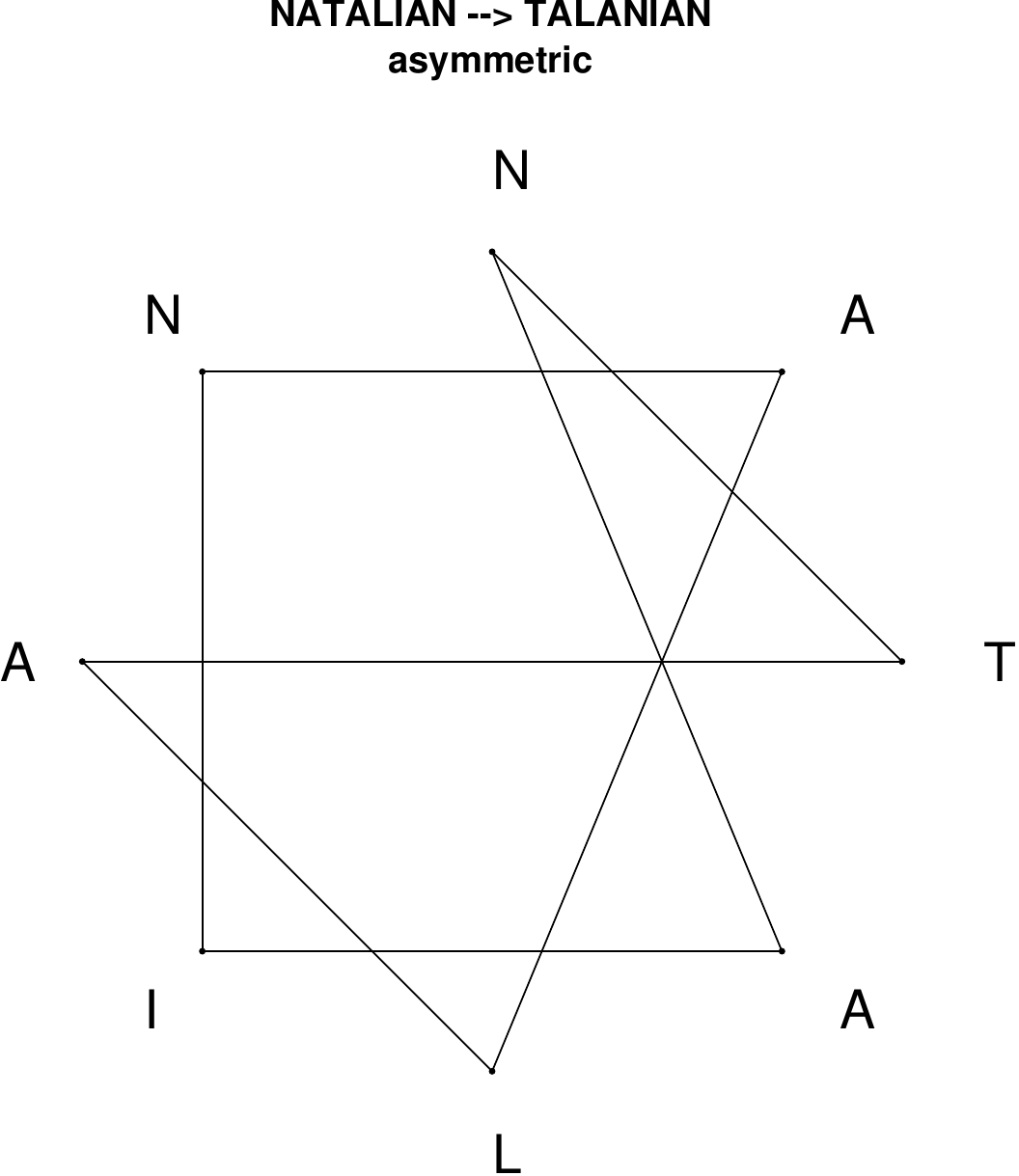}
\end{subfigure}
\hfill
\begin{subfigure}[T]{0.19\textwidth}
\centering
\includegraphics[width=\textwidth]{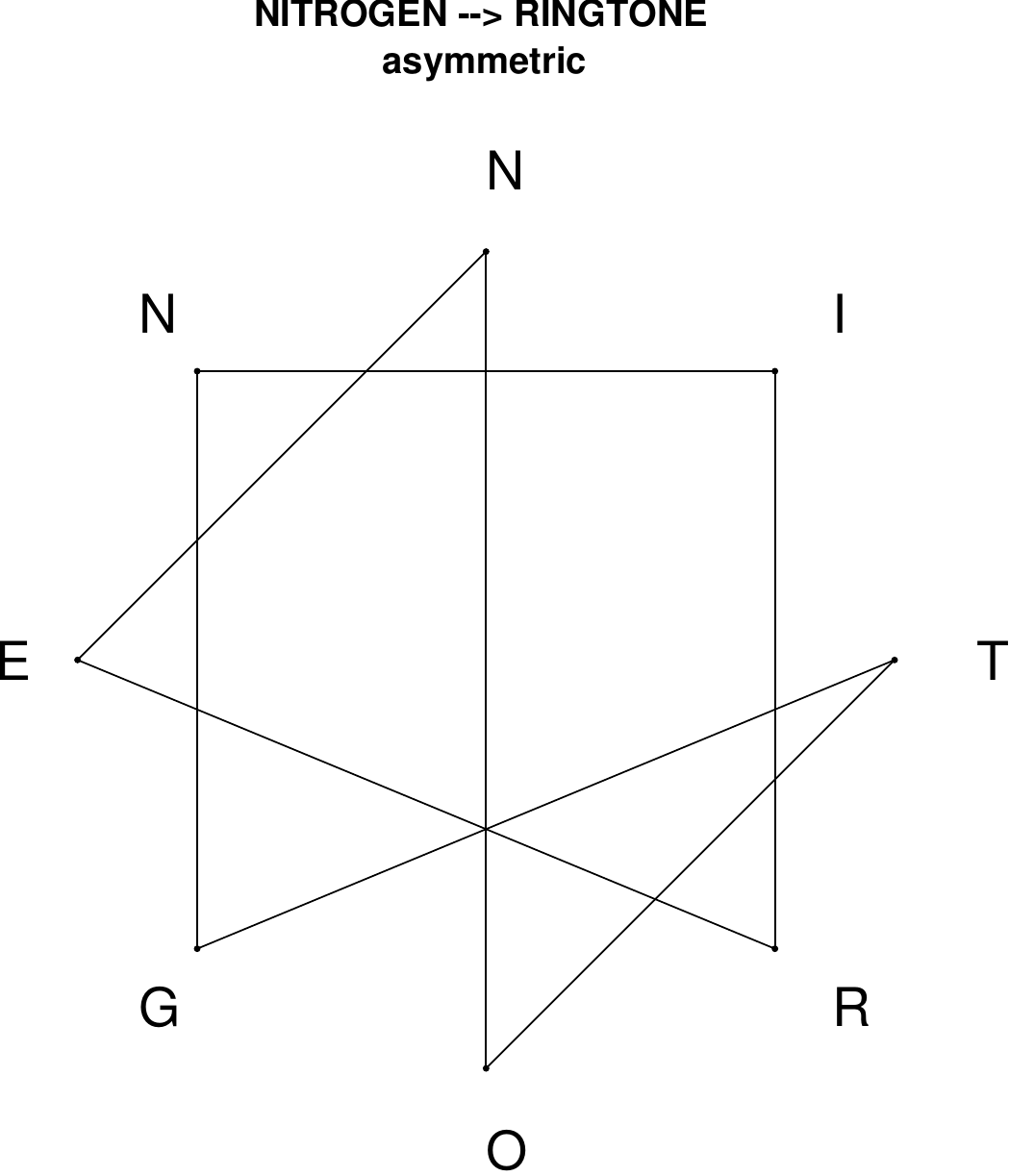}
\end{subfigure}
\end{figure}

\begin{figure}[H]
\centering
\begin{subfigure}[T]{0.19\textwidth}
\centering
\includegraphics[width=\textwidth]{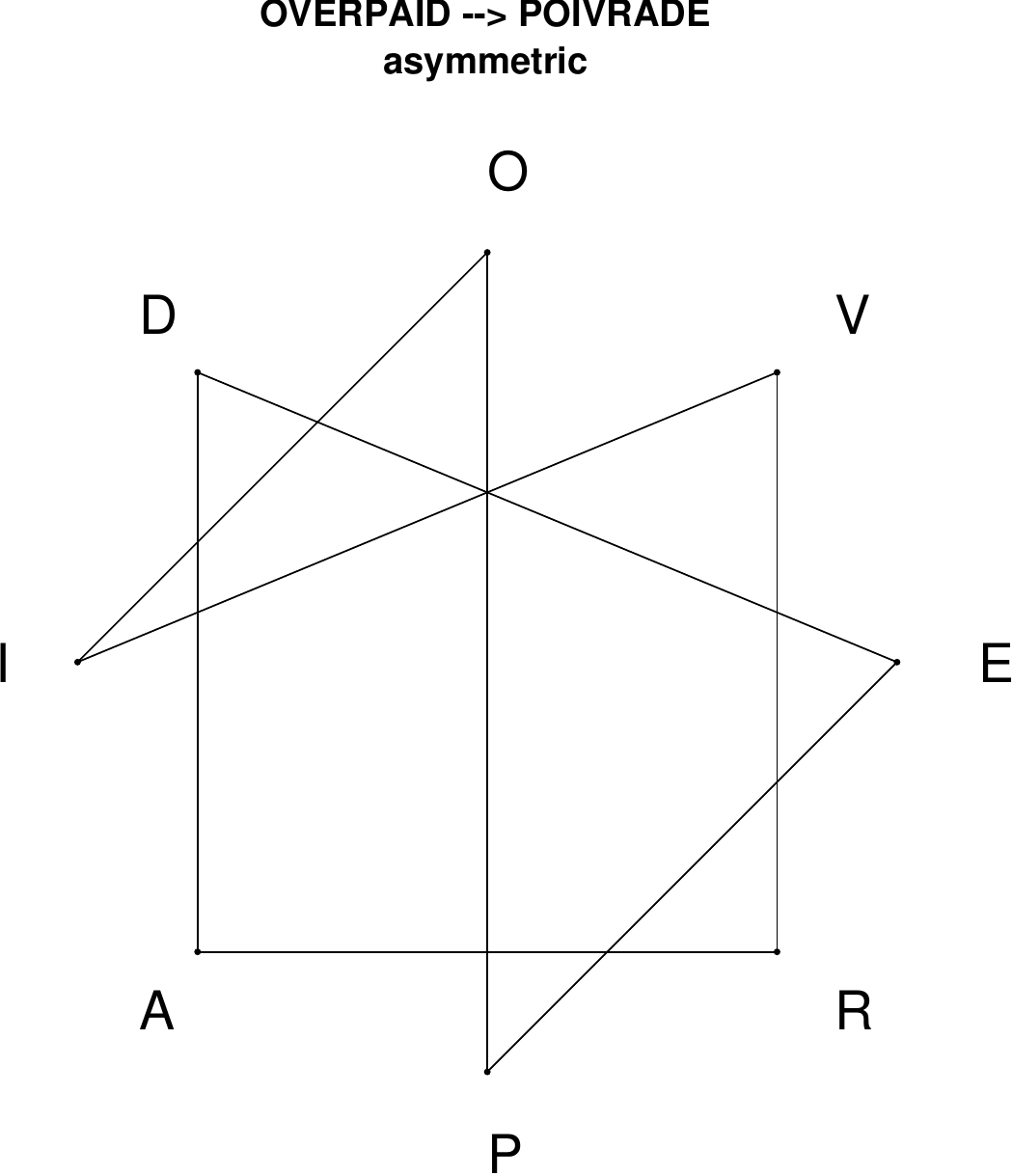}
\end{subfigure}
\hfill
\begin{subfigure}[T]{0.19\textwidth}
\centering
\includegraphics[width=\textwidth]{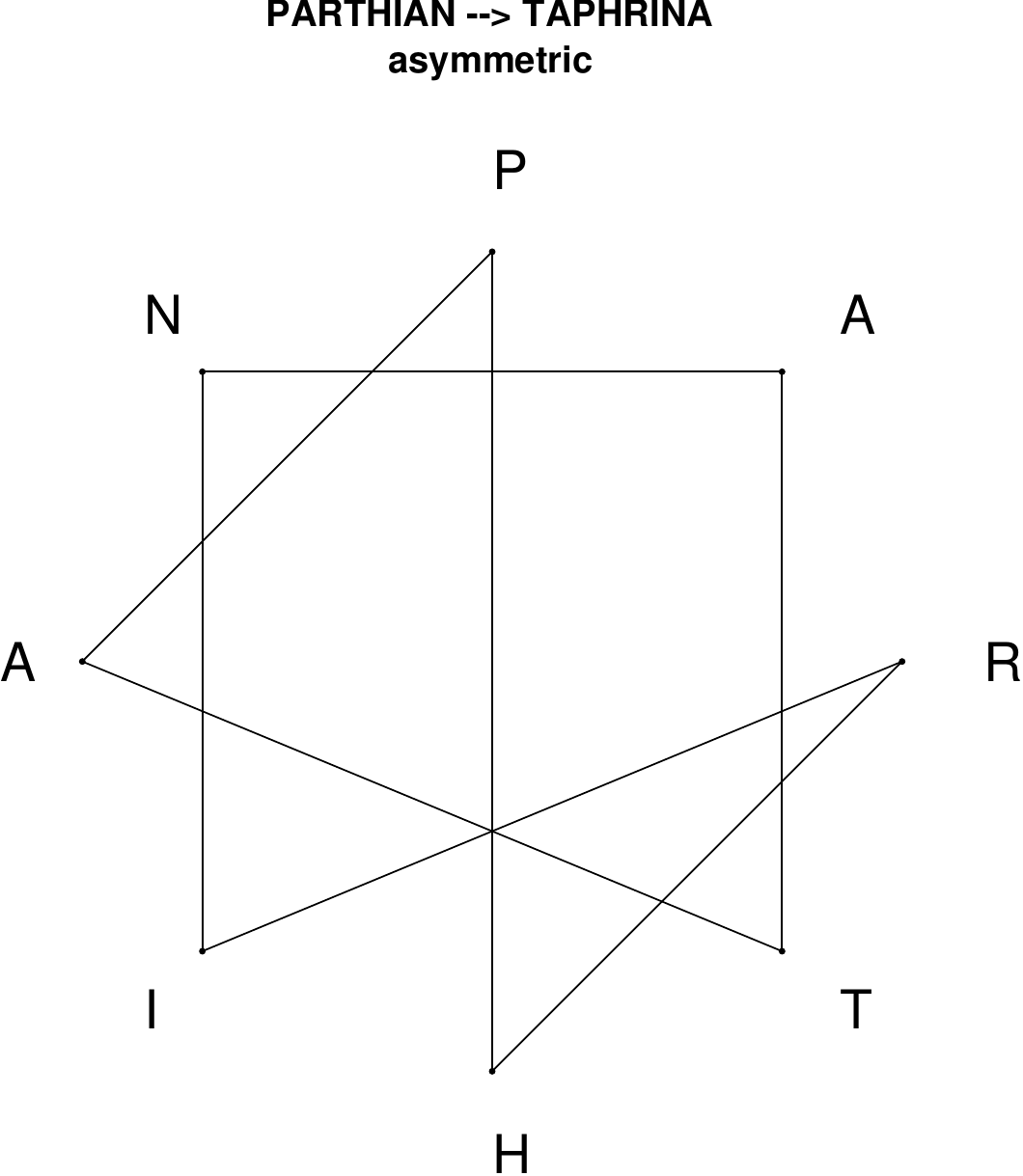}
\end{subfigure}
\hfill
\begin{subfigure}[T]{0.19\textwidth}
\centering
\includegraphics[width=\textwidth]{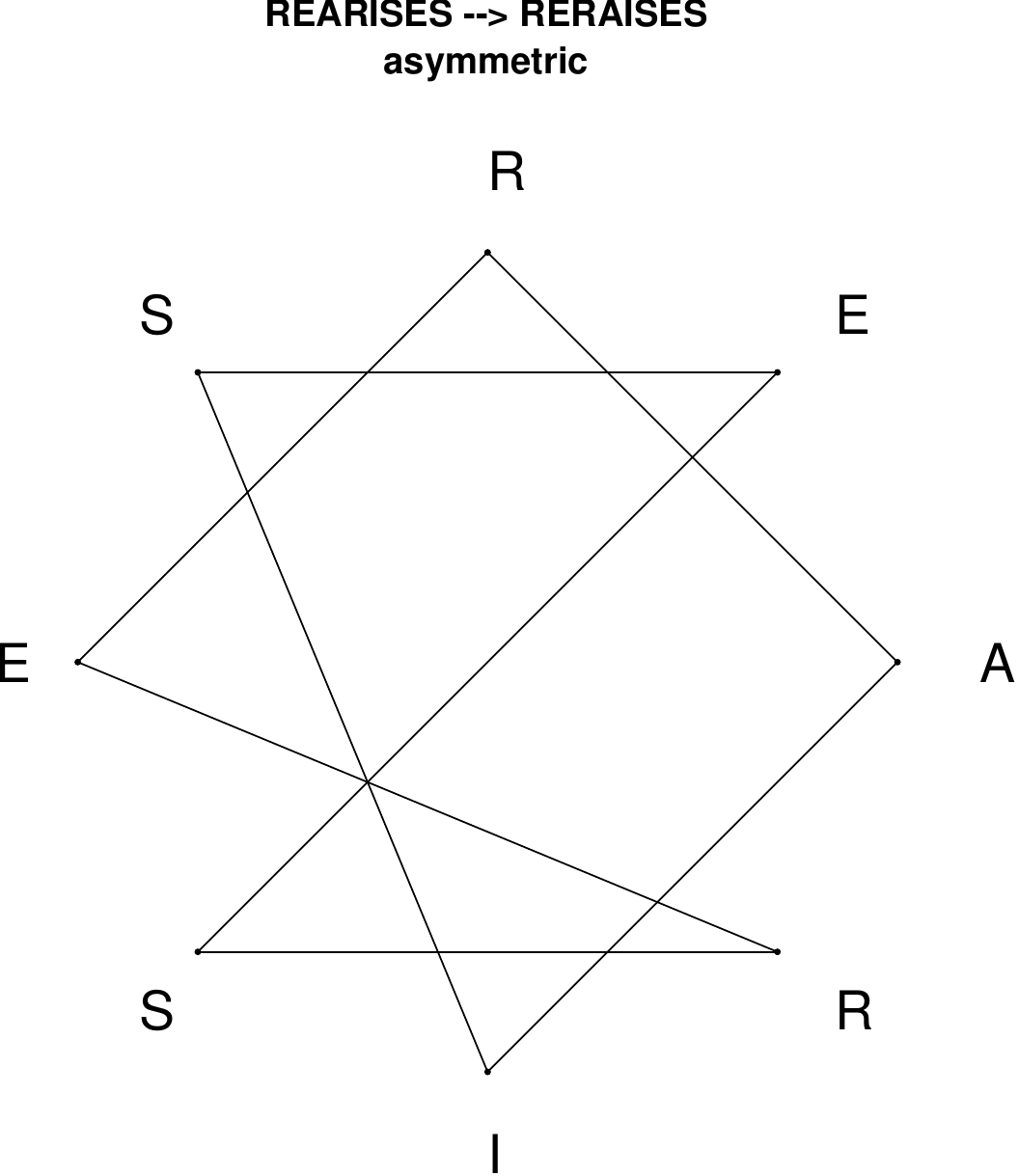}
\end{subfigure}
\hfill
\begin{subfigure}[T]{0.19\textwidth}
\centering
\includegraphics[width=\textwidth]{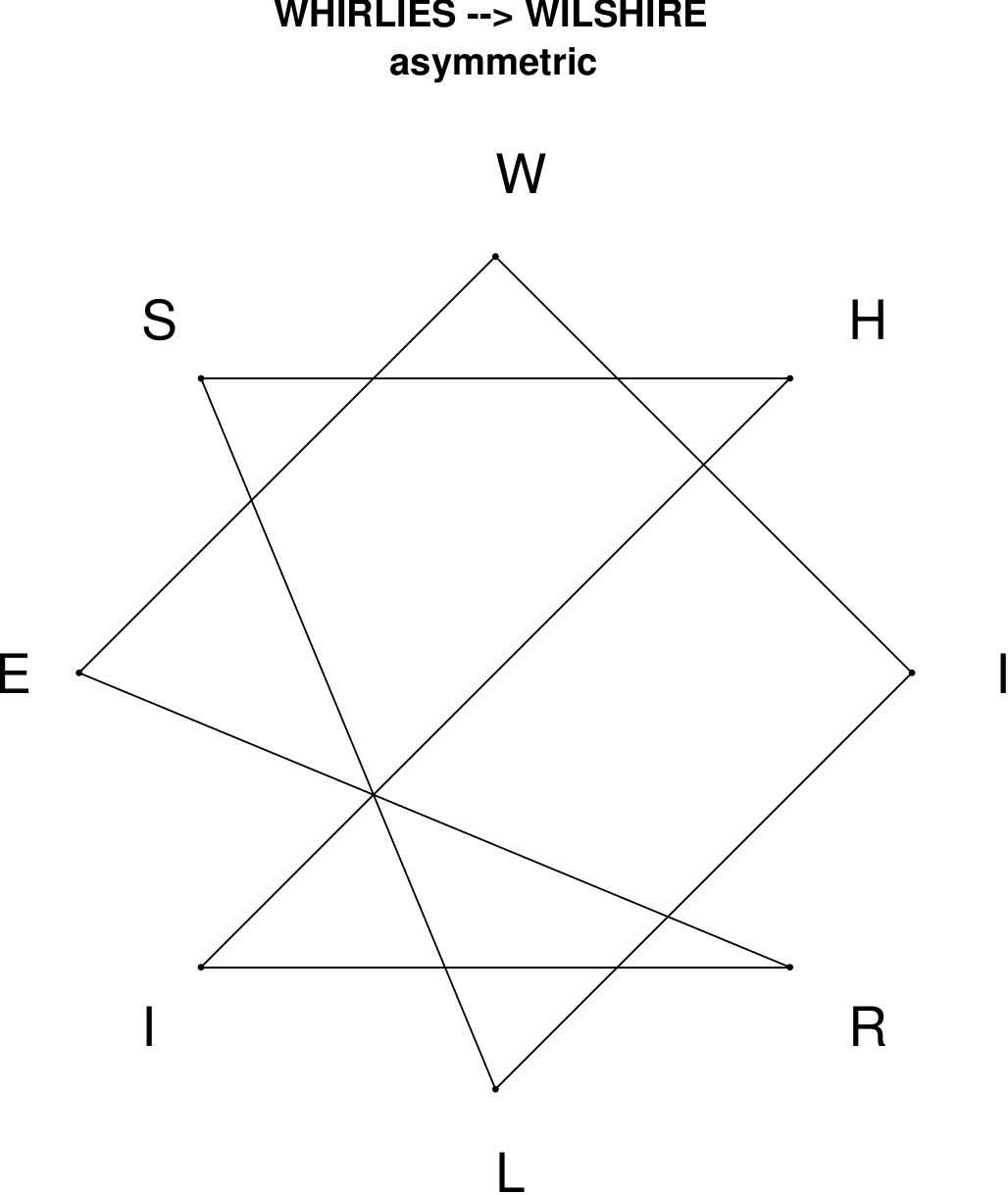}
\end{subfigure}
\hfill
\begin{subfigure}[T]{0.19\textwidth}
\centering
\includegraphics[width=\textwidth]{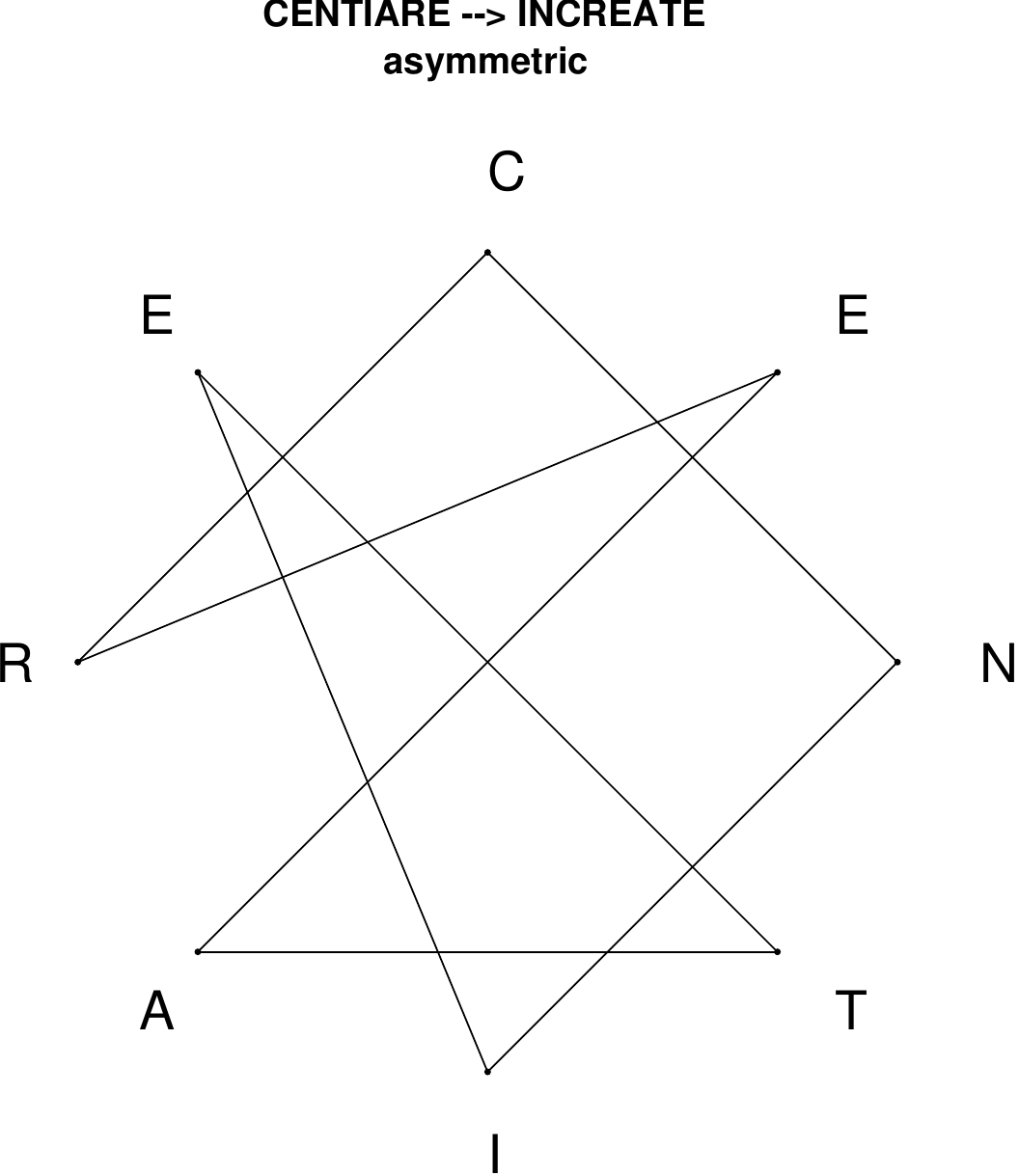}
\end{subfigure}
\end{figure}

\begin{figure}[H]
\centering
\begin{subfigure}[T]{0.19\textwidth}
\centering
\includegraphics[width=\textwidth]{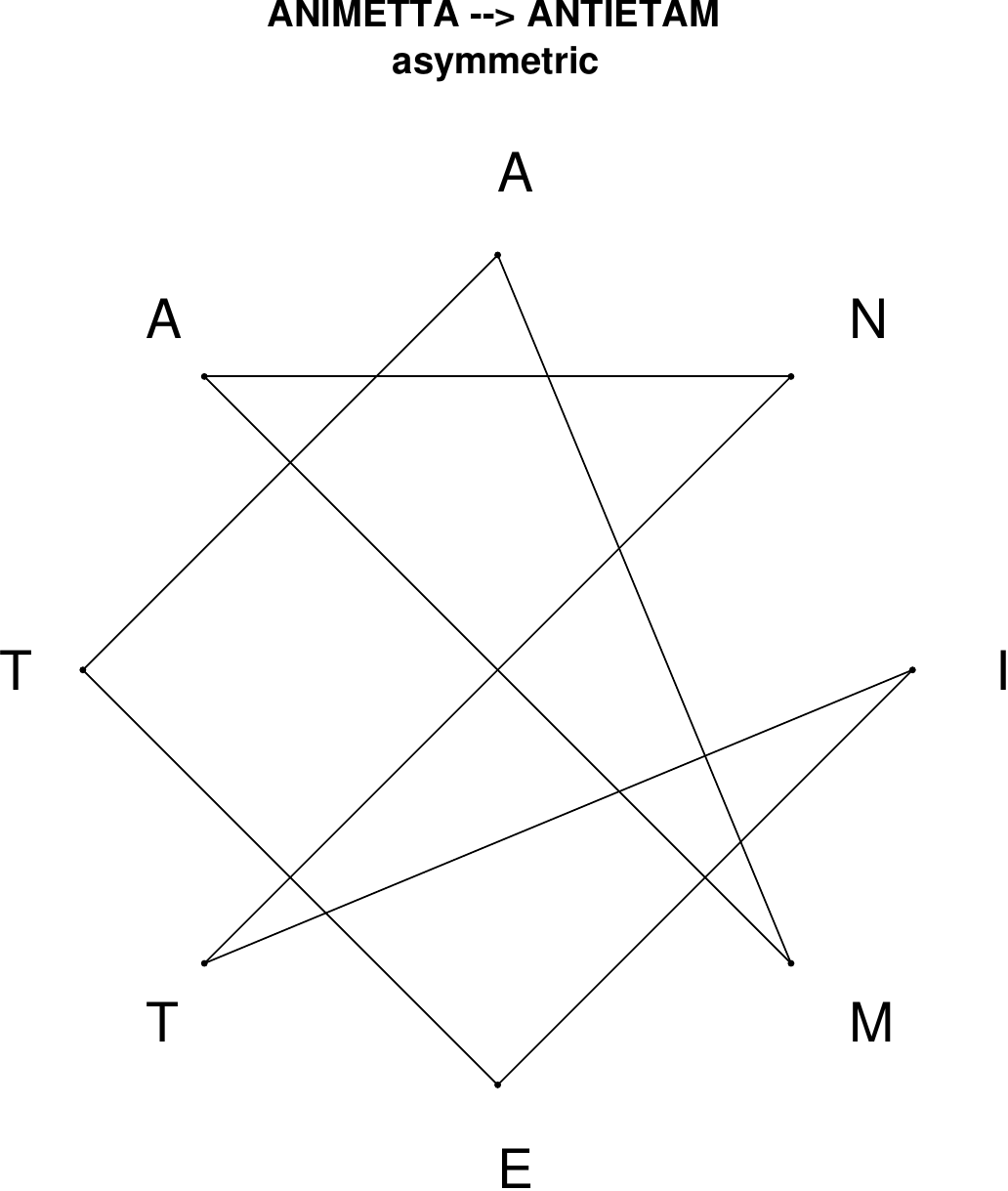}
\end{subfigure}
\hfill
\begin{subfigure}[T]{0.19\textwidth}
\centering
\includegraphics[width=\textwidth]{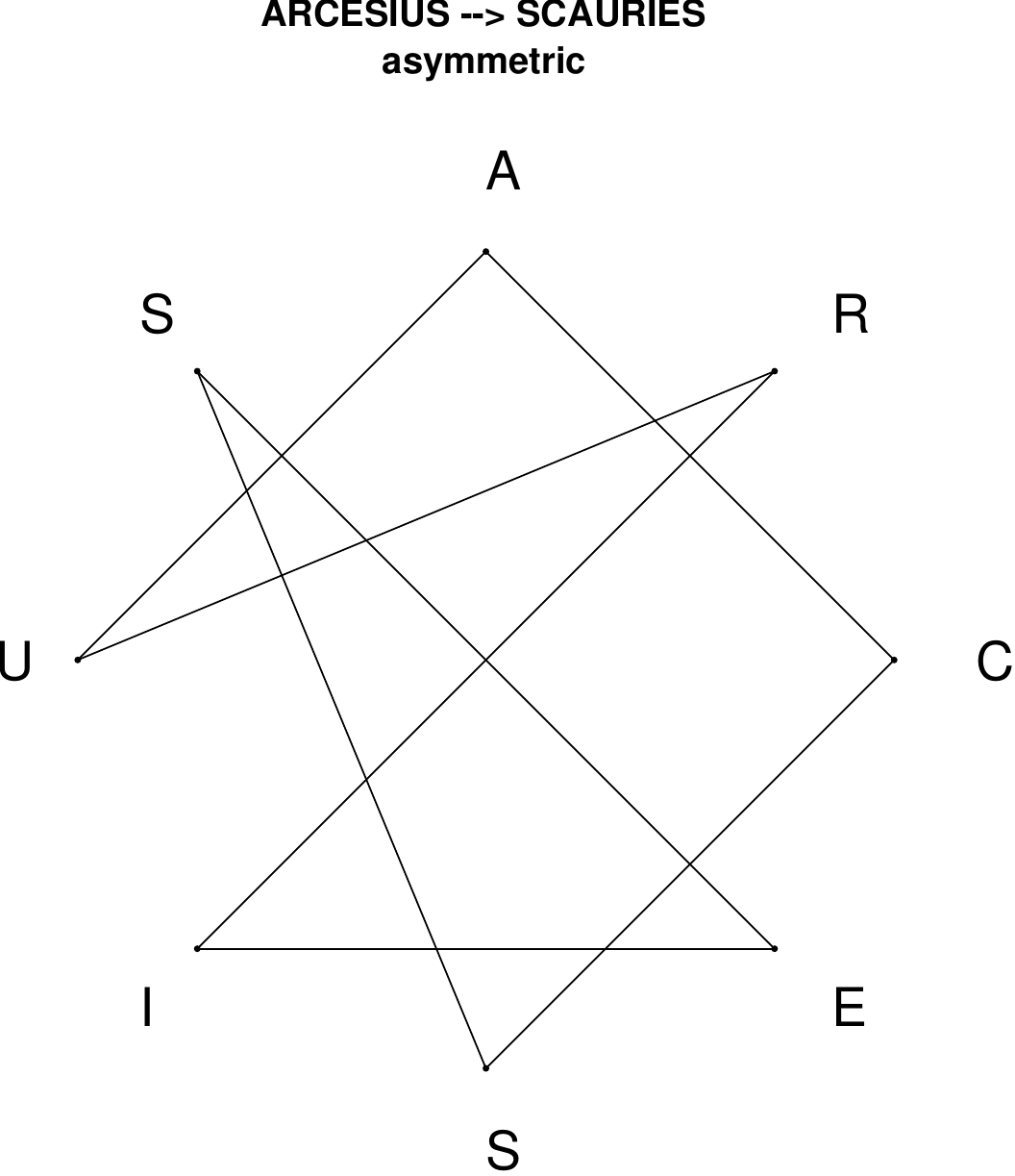}
\end{subfigure}
\hfill
\begin{subfigure}[T]{0.19\textwidth}
\centering
\includegraphics[width=\textwidth]{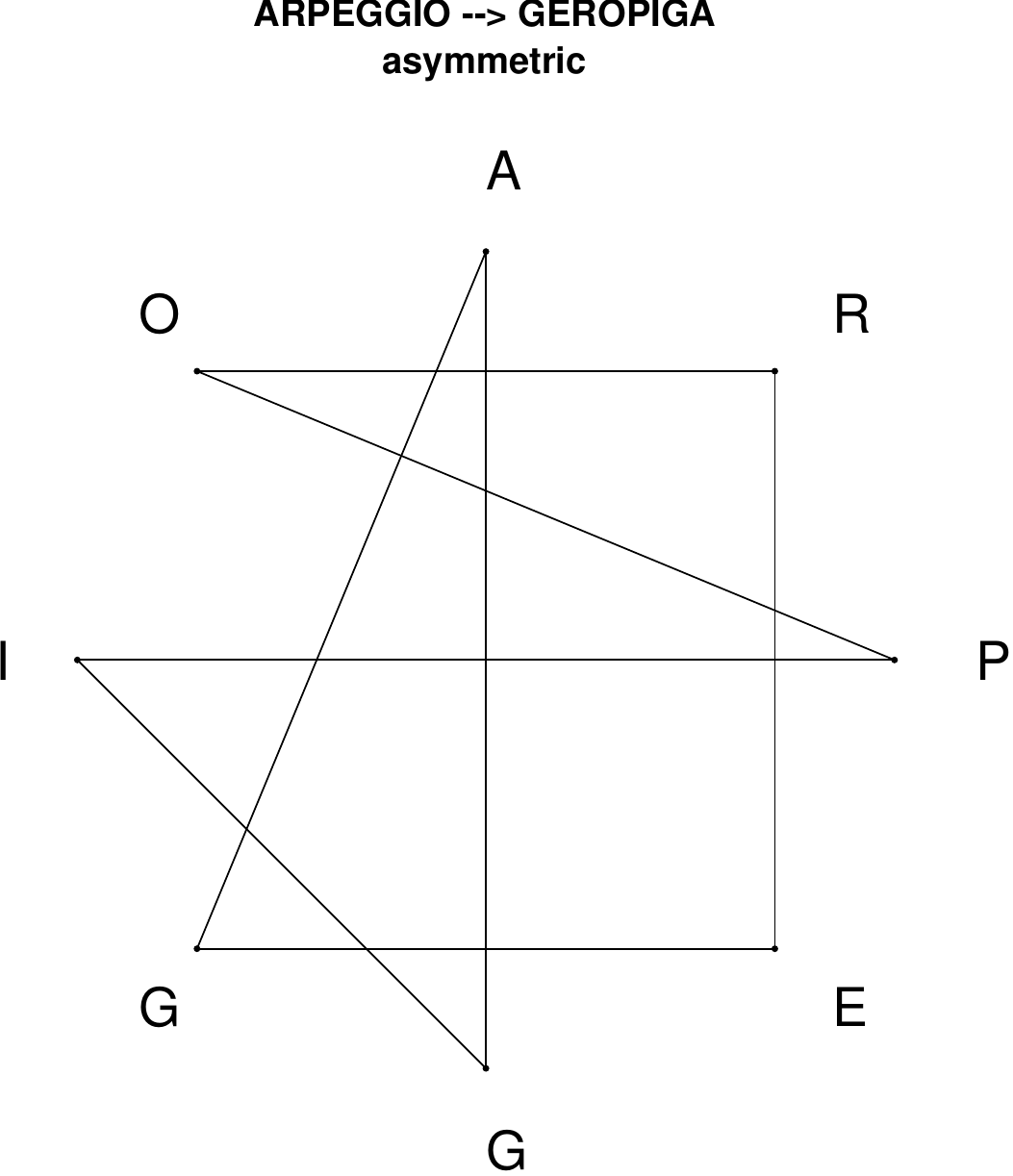}
\end{subfigure}
\hfill
\begin{subfigure}[T]{0.19\textwidth}
\centering
\includegraphics[width=\textwidth]{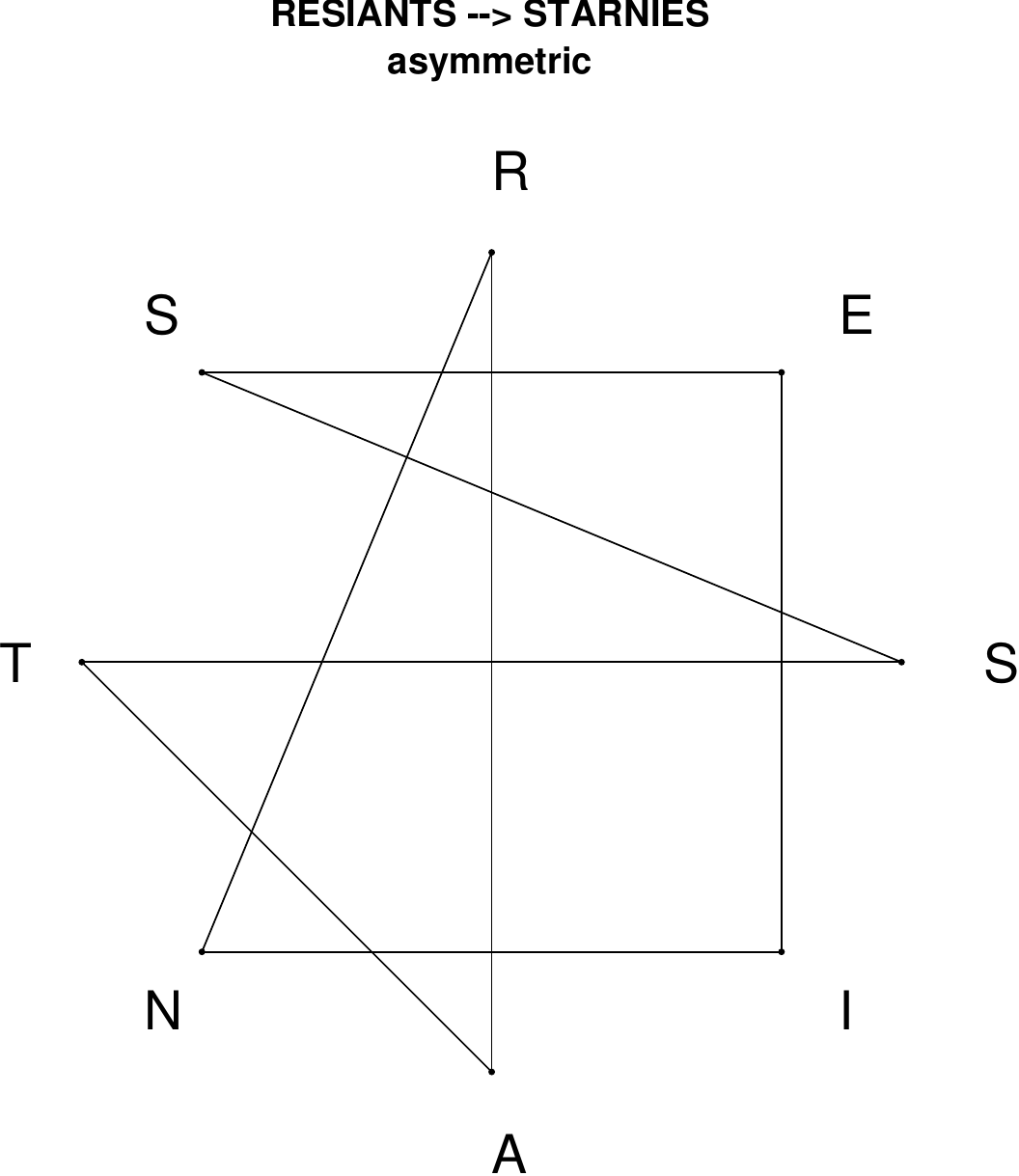}
\end{subfigure}
\hfill
\begin{subfigure}[T]{0.19\textwidth}
\centering
\includegraphics[width=\textwidth]{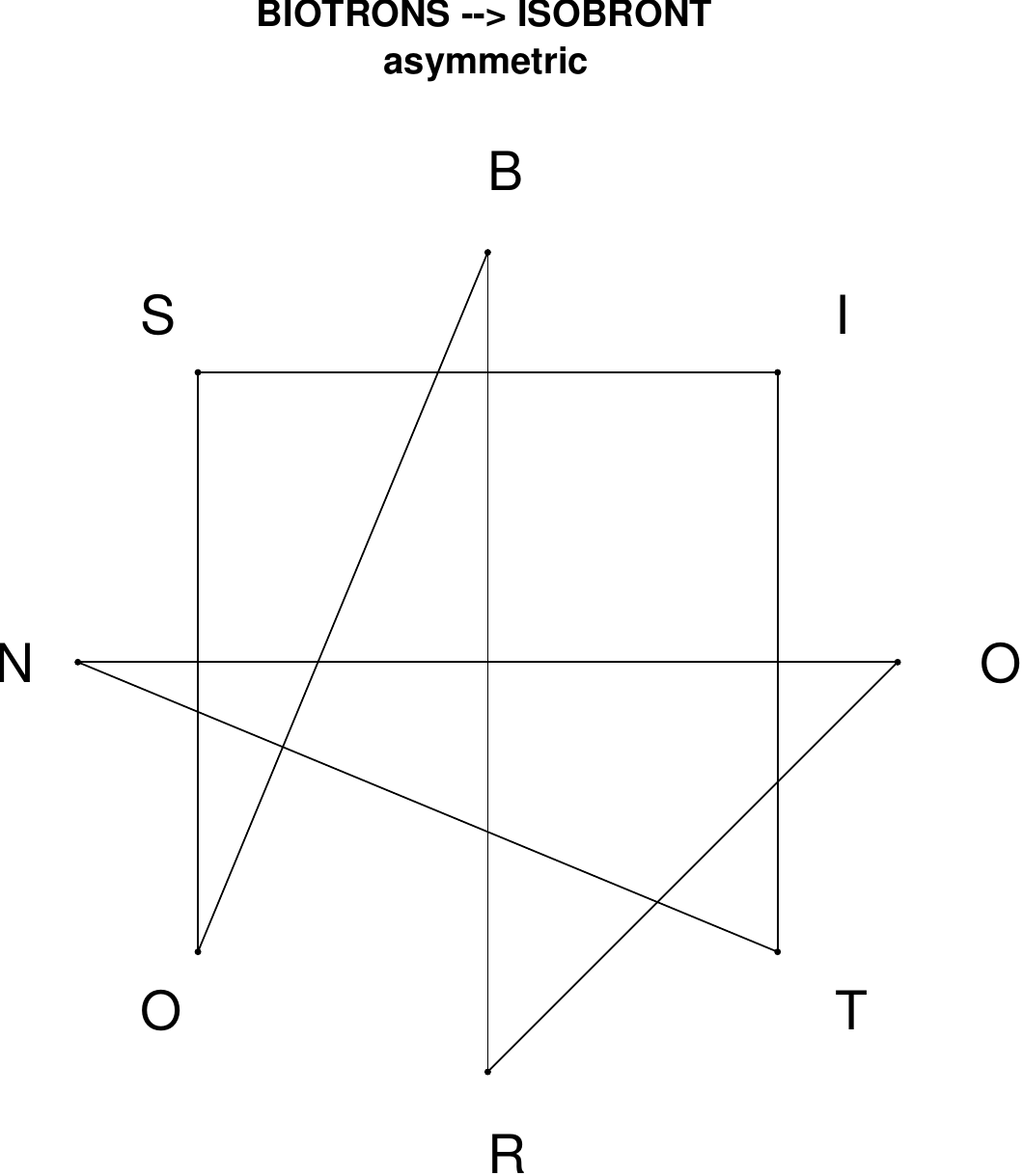}
\end{subfigure}
\end{figure}

\begin{figure}[H]
\centering
\begin{subfigure}[T]{0.19\textwidth}
\centering
\includegraphics[width=\textwidth]{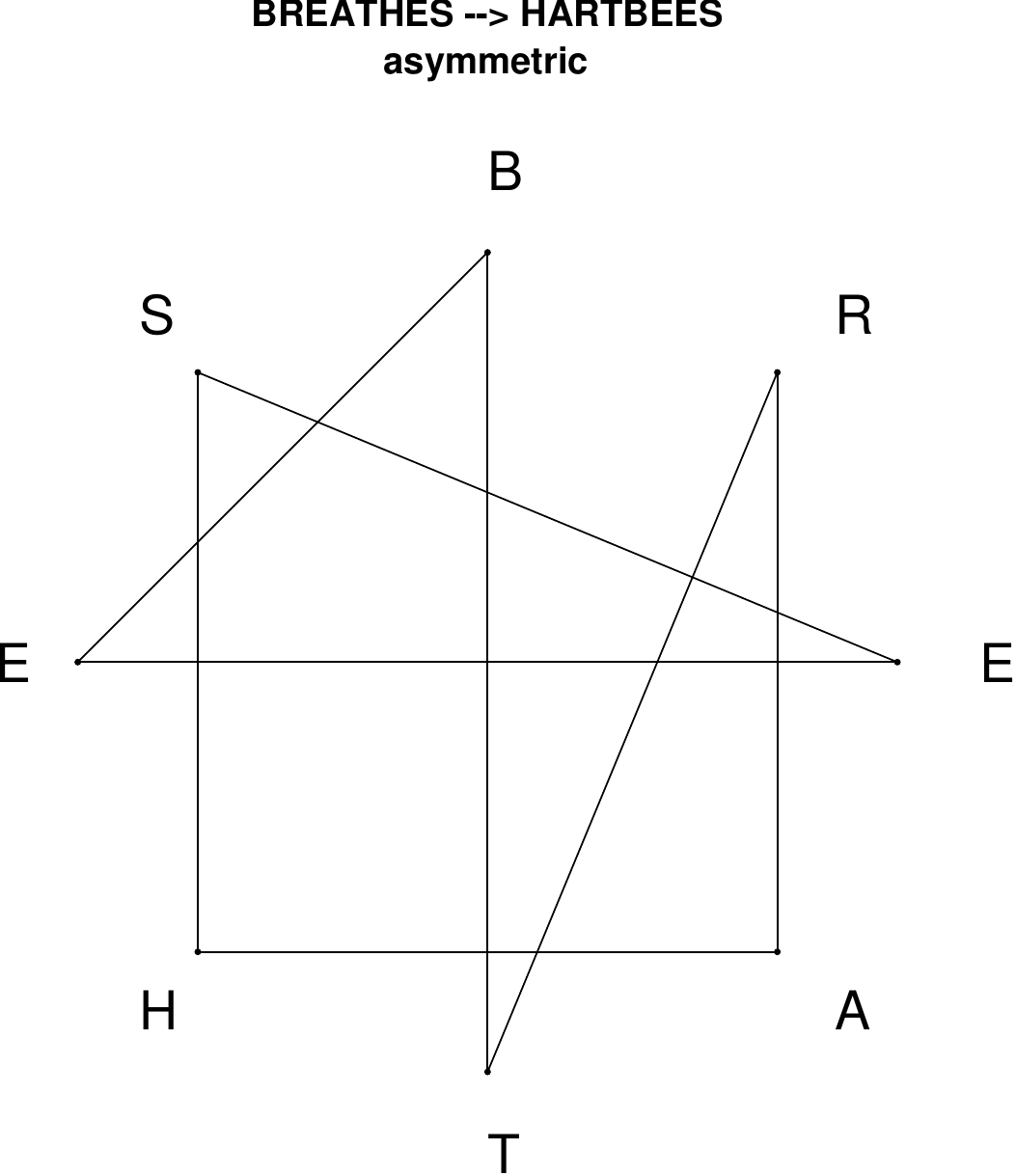}
\end{subfigure}
\hfill
\begin{subfigure}[T]{0.19\textwidth}
\centering
\includegraphics[width=\textwidth]{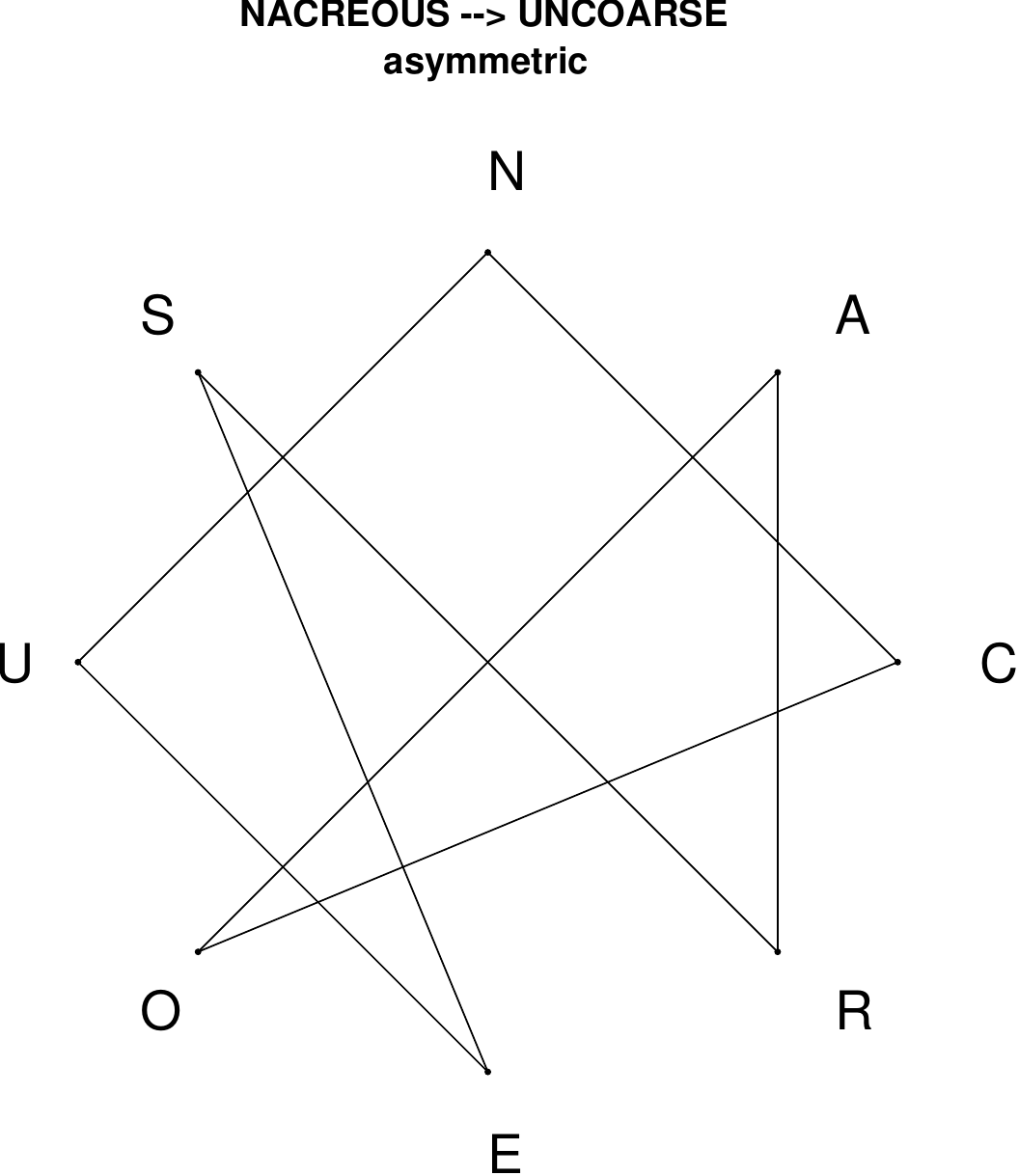}
\end{subfigure}
\hfill
\begin{subfigure}[T]{0.19\textwidth}
\centering
\includegraphics[width=\textwidth]{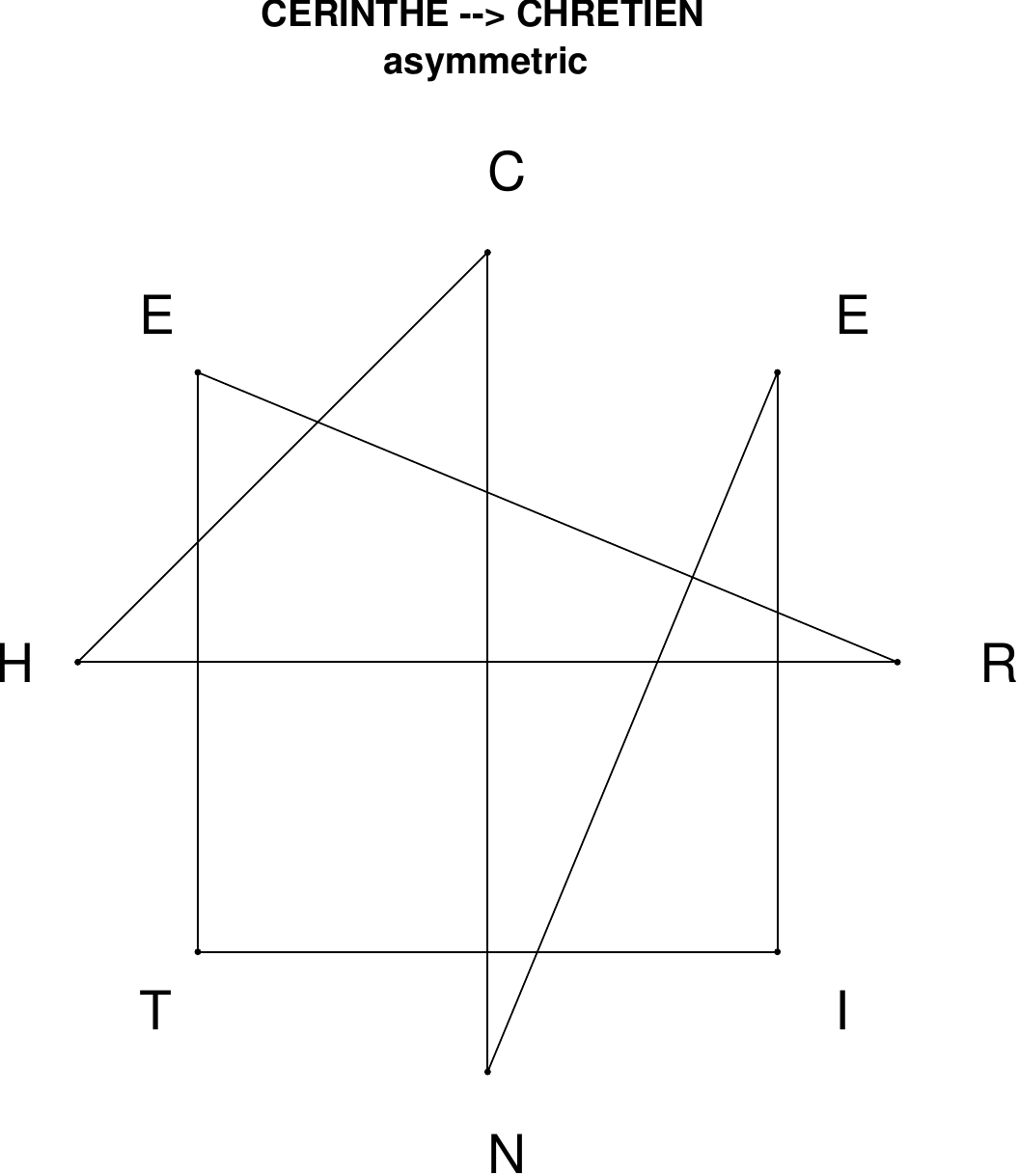}
\end{subfigure}
\hfill
\begin{subfigure}[T]{0.19\textwidth}
\centering
\includegraphics[width=\textwidth]{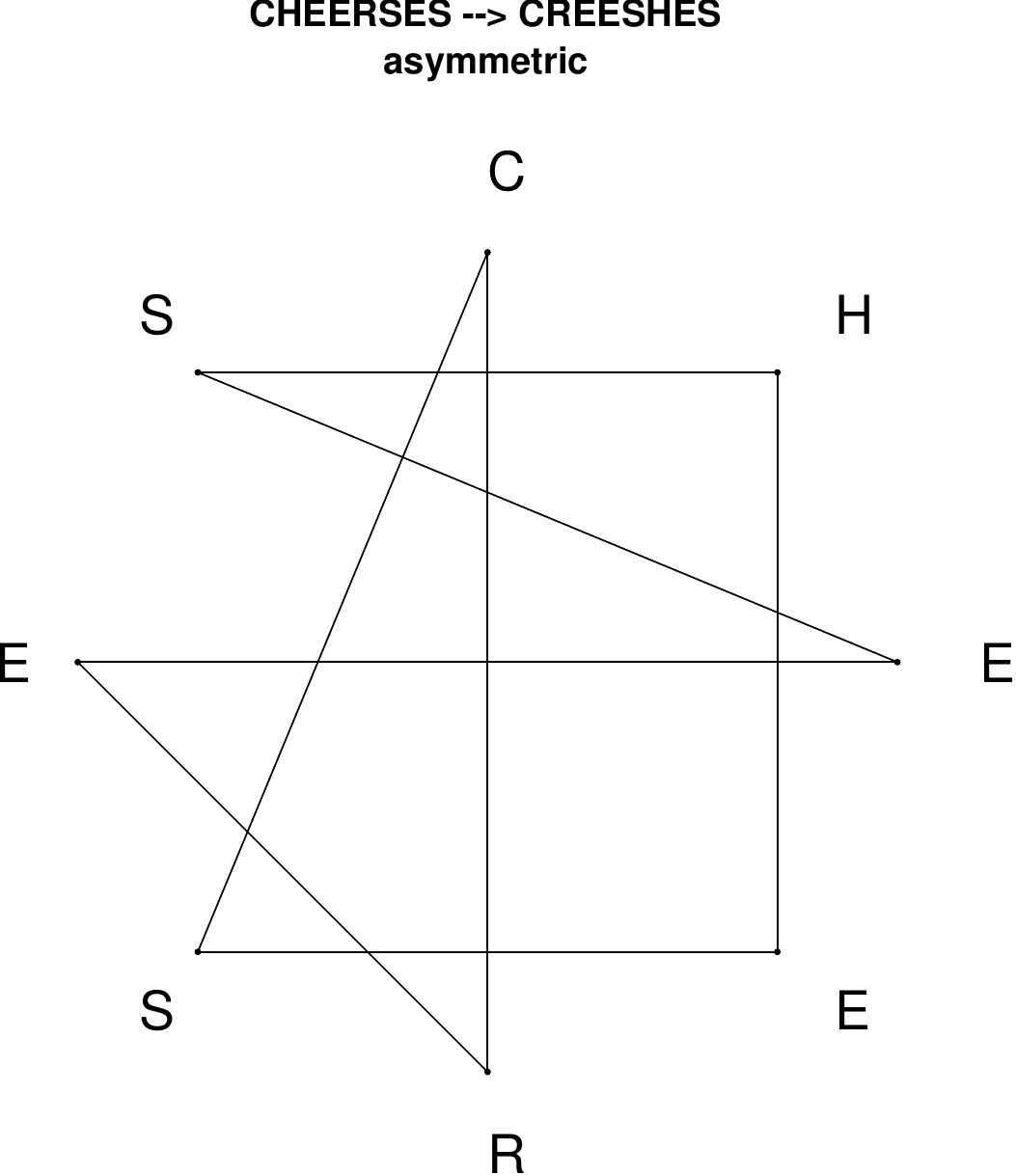}
\end{subfigure}
\hfill
\begin{subfigure}[T]{0.19\textwidth}
\centering
\includegraphics[width=\textwidth]{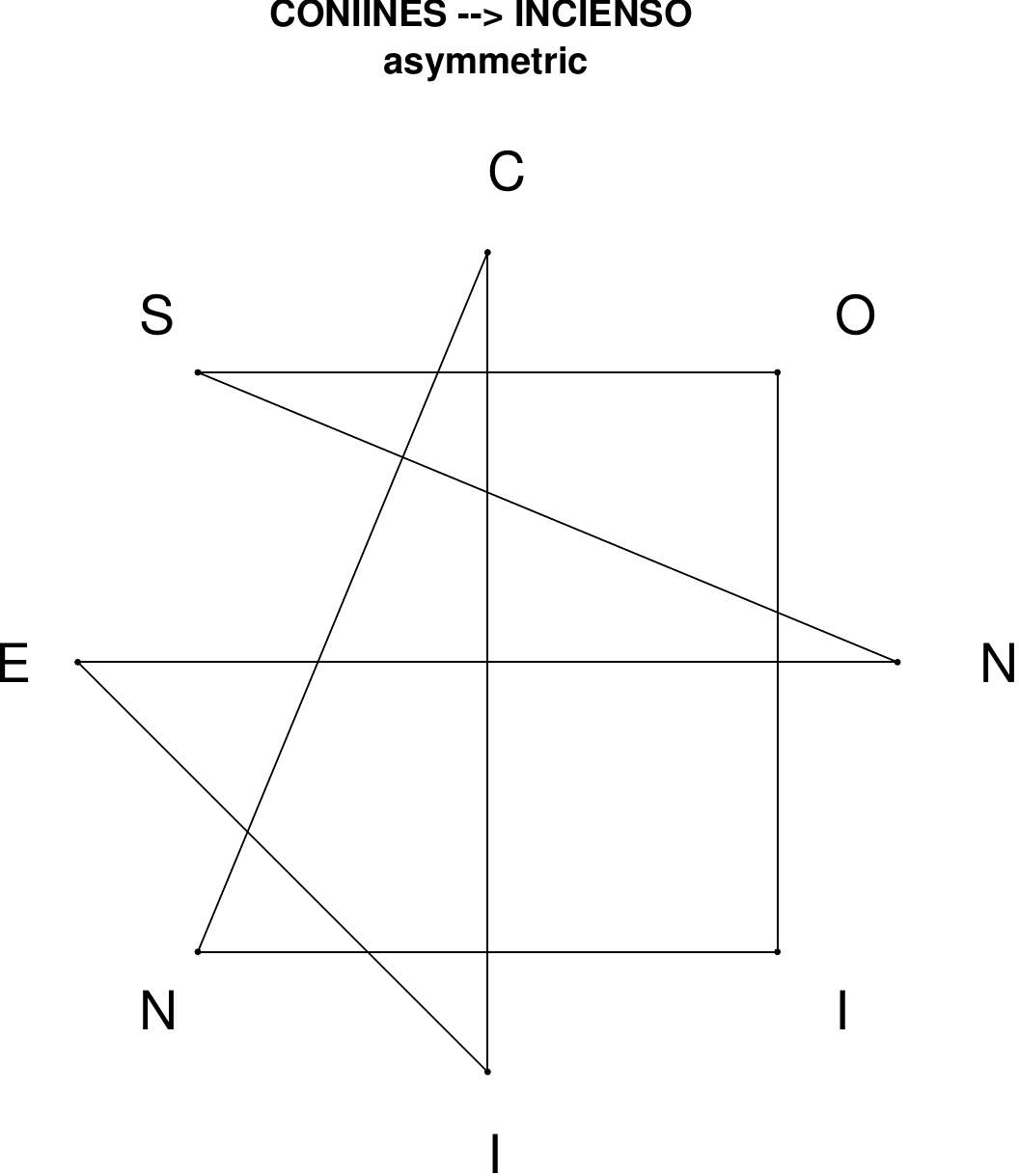}
\end{subfigure}
\end{figure}

\begin{figure}[H]
\centering
\begin{subfigure}[T]{0.19\textwidth}
\centering
\includegraphics[width=\textwidth]{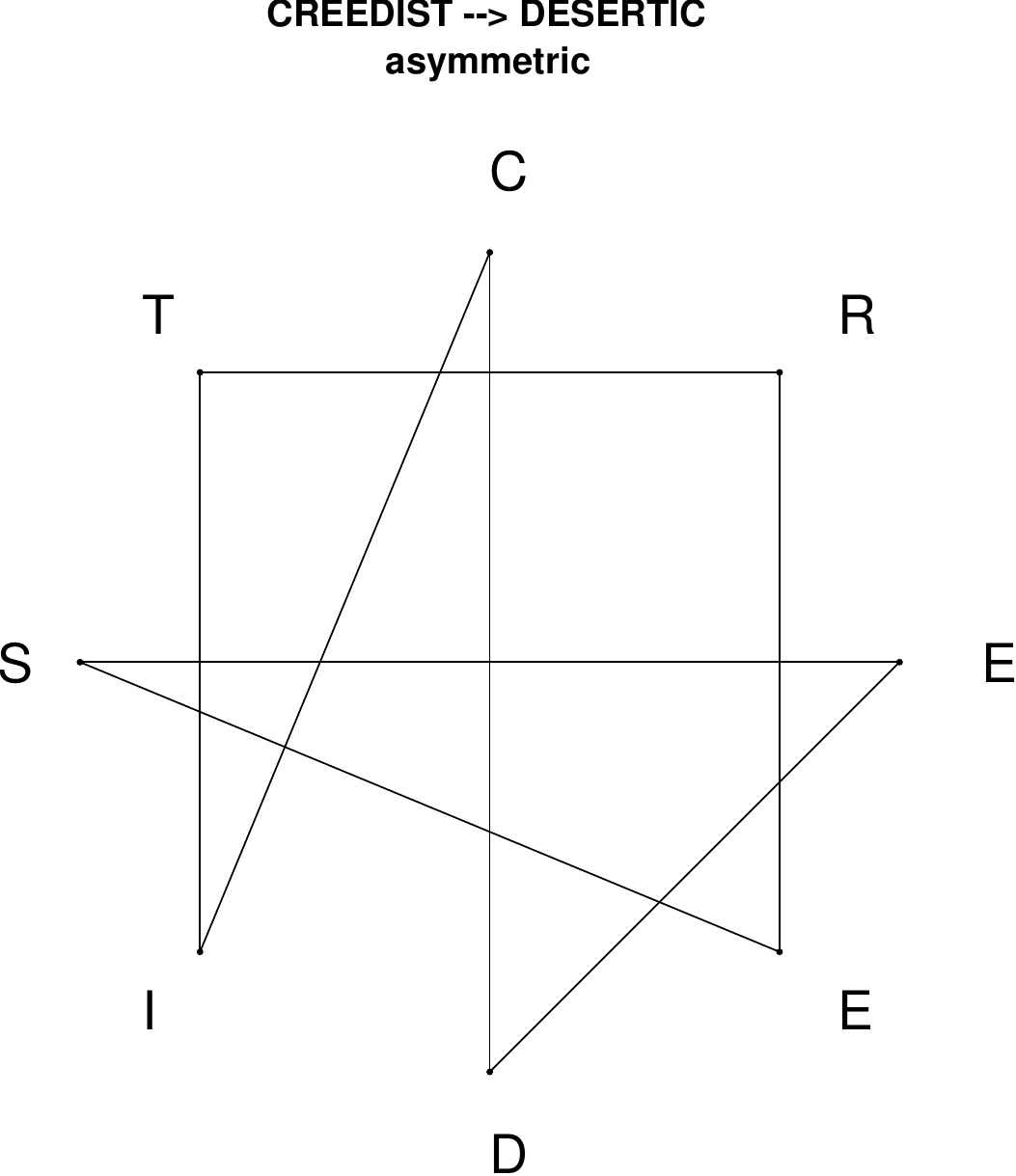}
\end{subfigure}
\hfill
\begin{subfigure}[T]{0.19\textwidth}
\centering
\includegraphics[width=\textwidth]{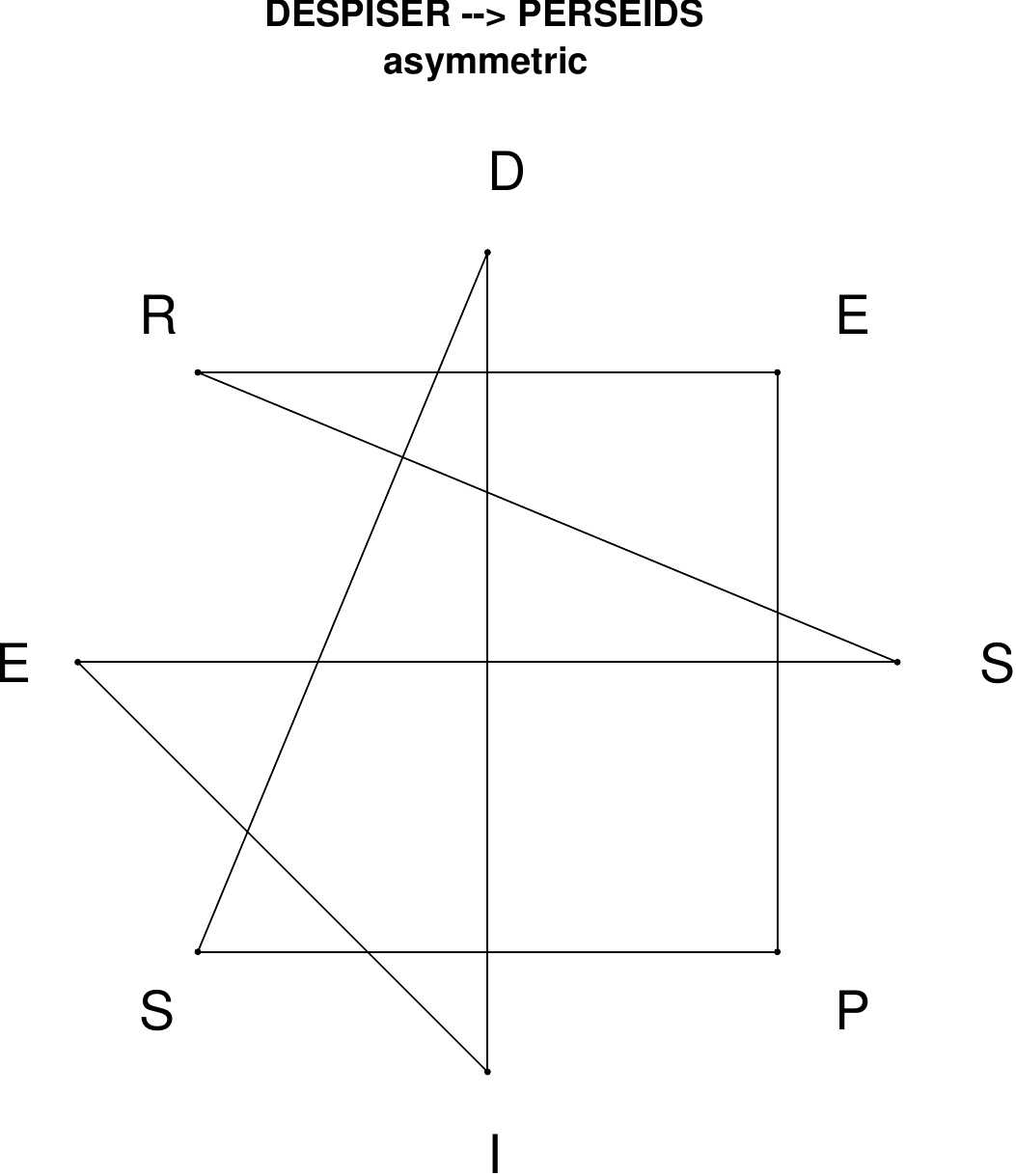}
\end{subfigure}
\hfill
\begin{subfigure}[T]{0.19\textwidth}
\centering
\includegraphics[width=\textwidth]{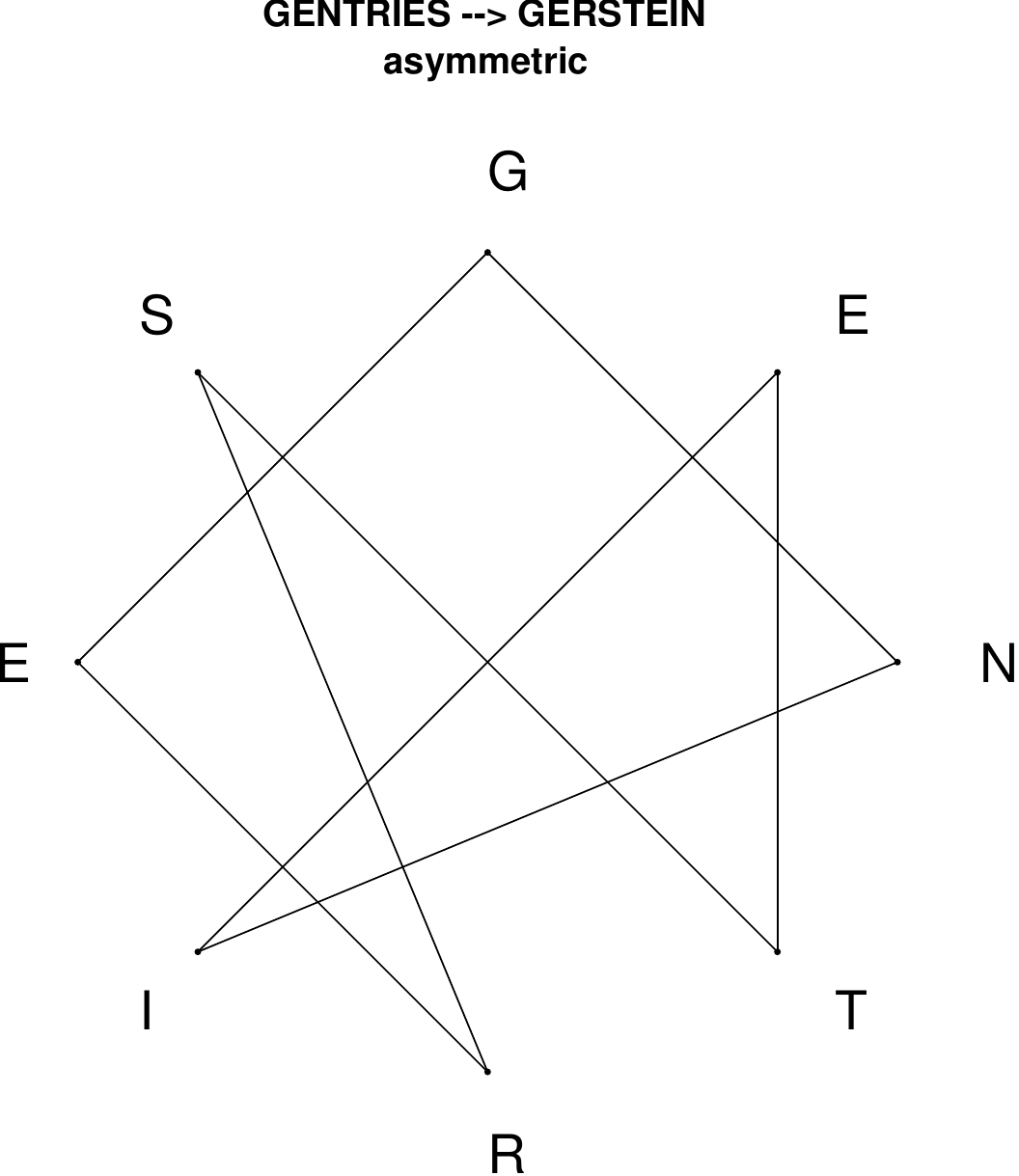}
\end{subfigure}
\hfill
\begin{subfigure}[T]{0.19\textwidth}
\centering
\includegraphics[width=\textwidth]{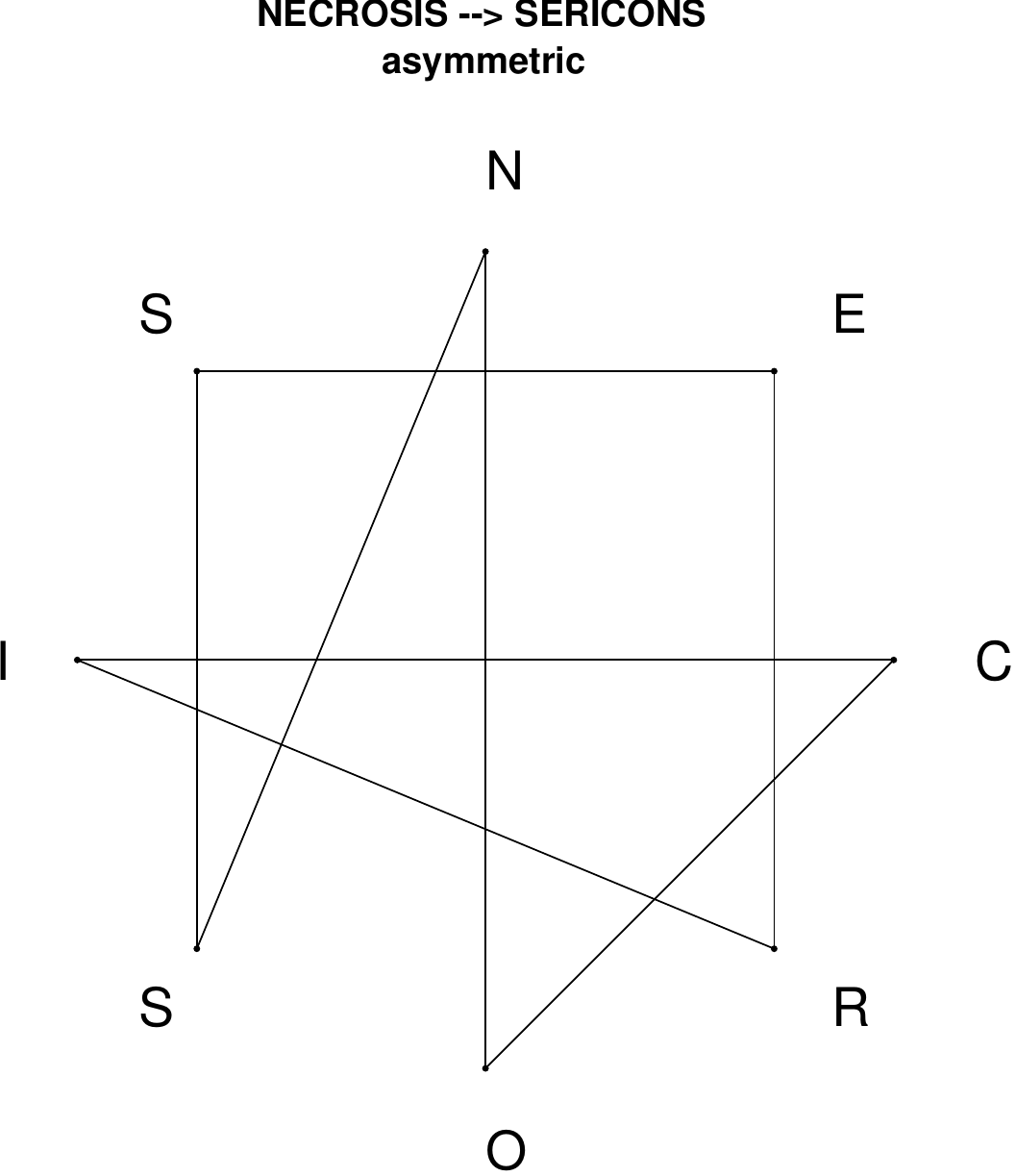}
\end{subfigure}
\hfill
\begin{subfigure}[T]{0.19\textwidth}
\centering
\includegraphics[width=\textwidth]{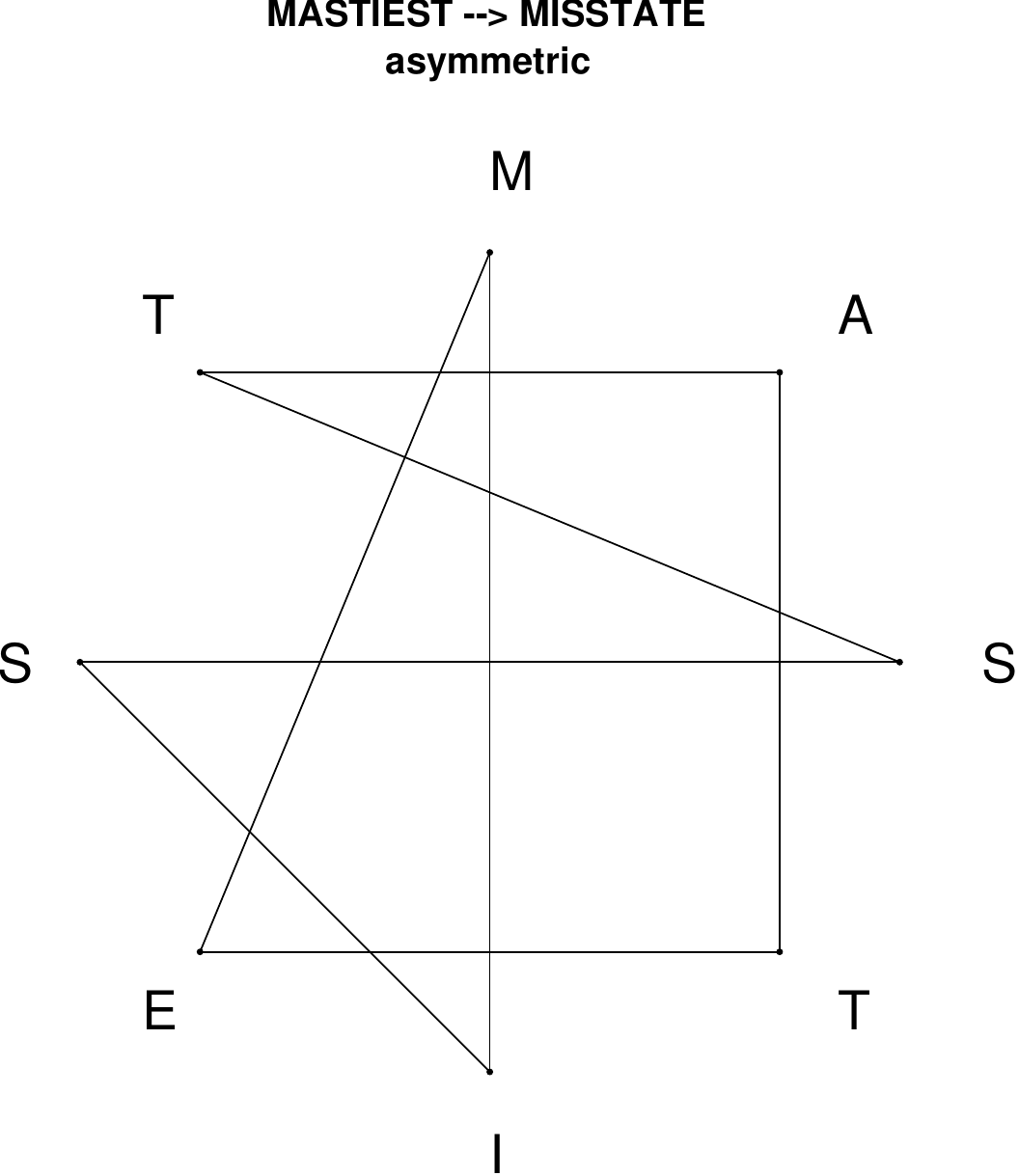}
\end{subfigure}
\end{figure}

\begin{figure}[H]
\centering
\begin{subfigure}[T]{0.19\textwidth}
\centering
\includegraphics[width=\textwidth]{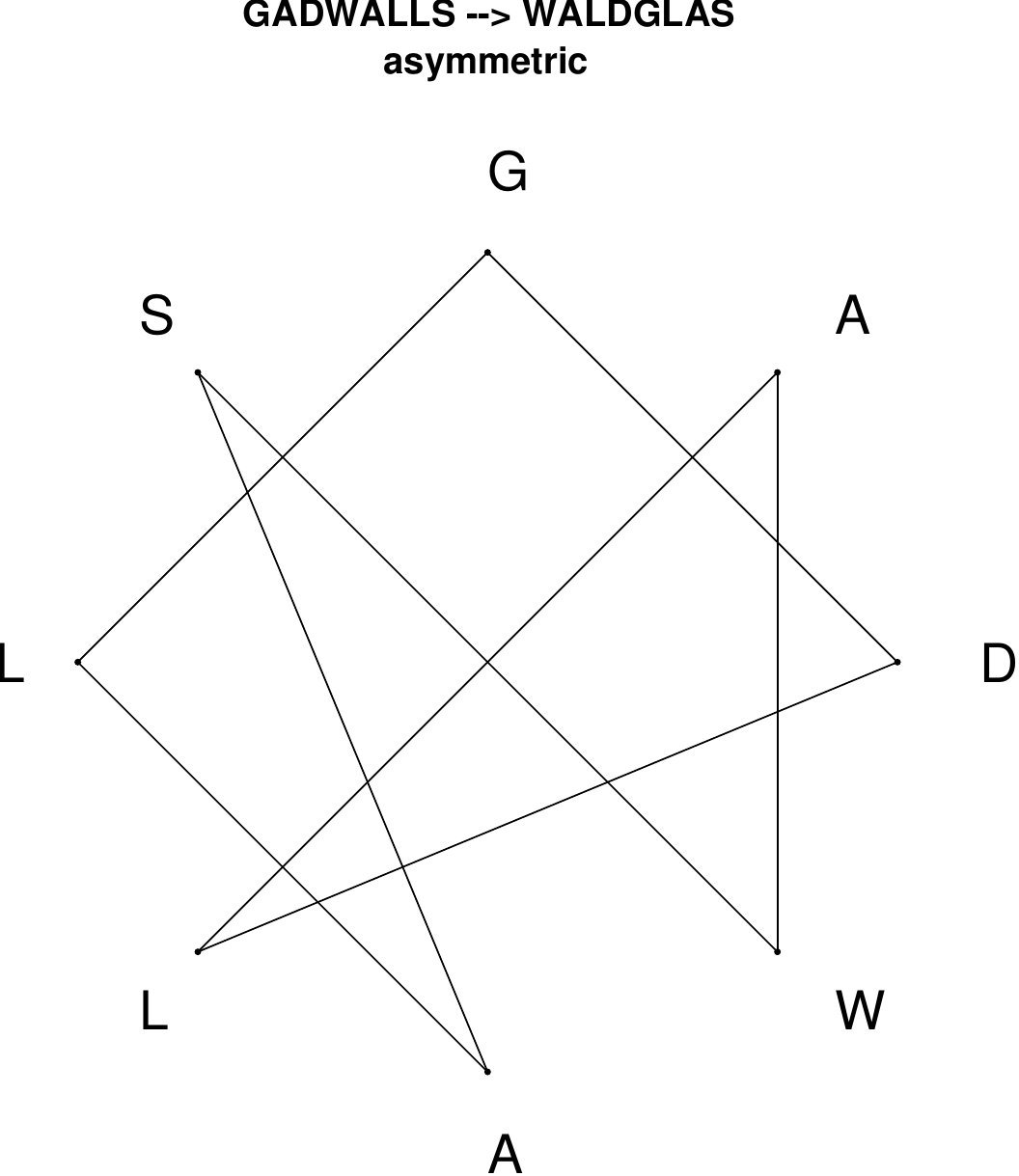}
\end{subfigure}
\hfill
\begin{subfigure}[T]{0.19\textwidth}
\centering
\includegraphics[width=\textwidth]{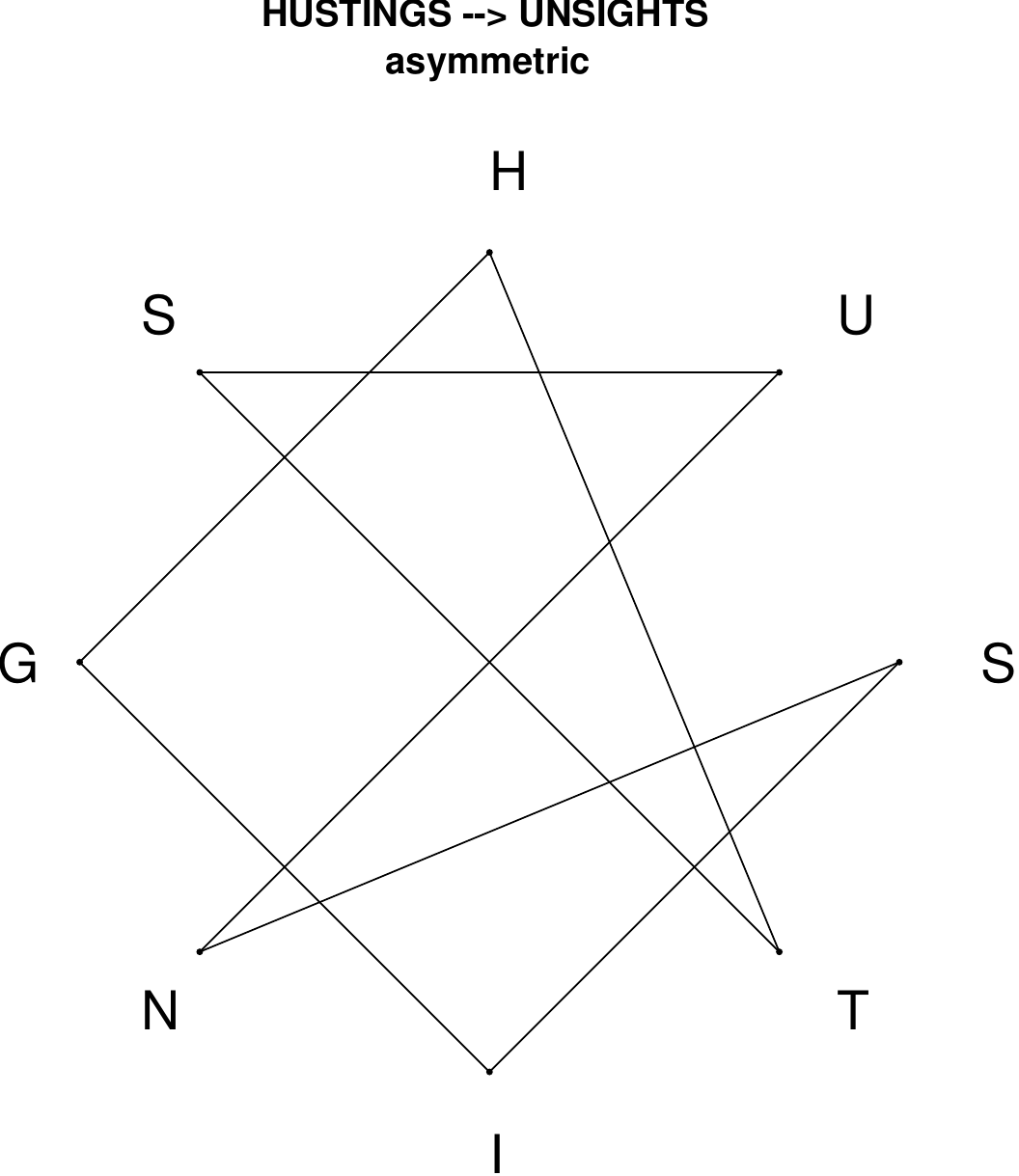}
\end{subfigure}
\hfill
\begin{subfigure}[T]{0.19\textwidth}
\centering
\includegraphics[width=\textwidth]{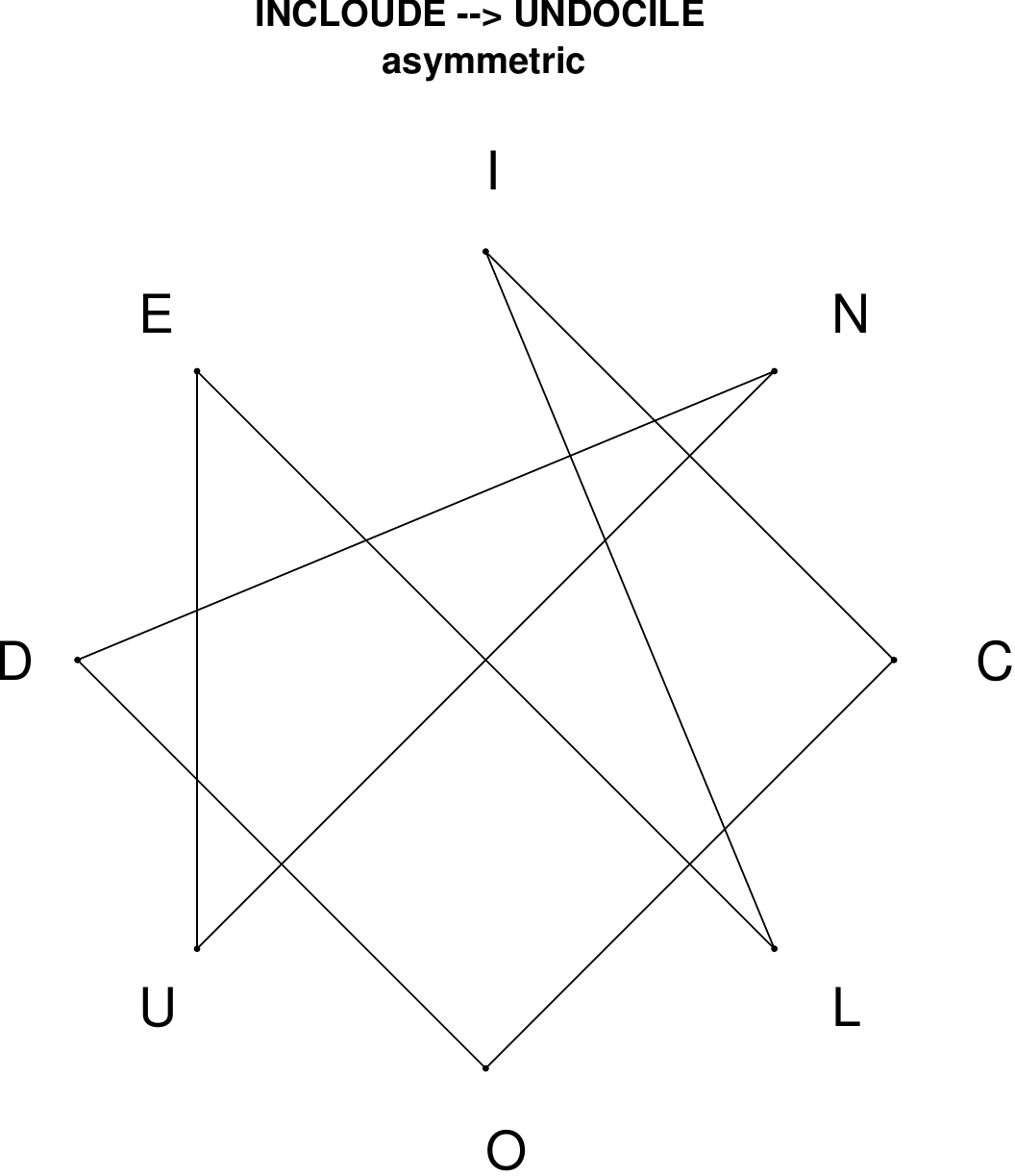}
\end{subfigure}
\hfill
\begin{subfigure}[T]{0.19\textwidth}
\centering
\includegraphics[width=\textwidth]{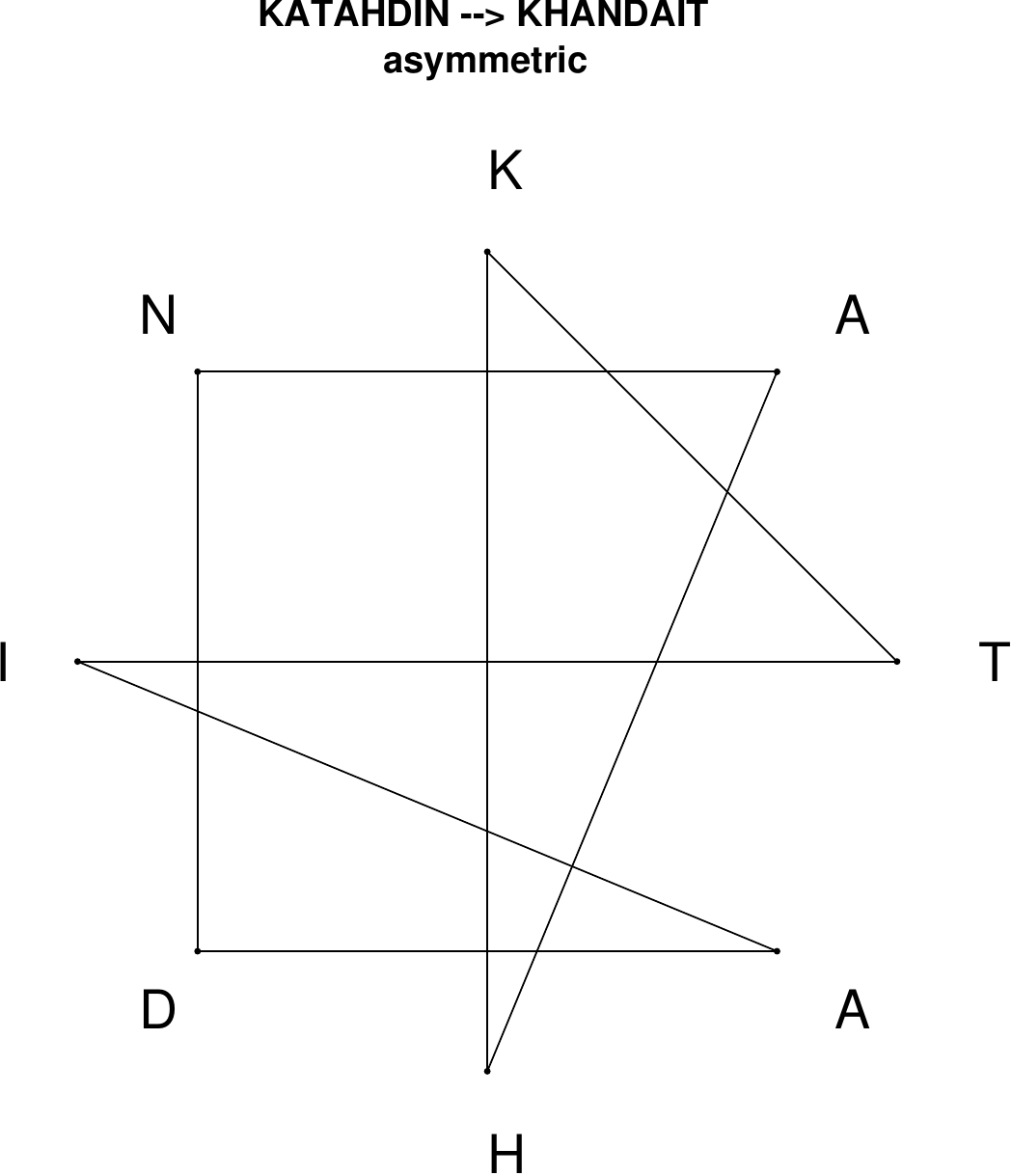}
\end{subfigure}
\hfill
\begin{subfigure}[T]{0.19\textwidth}
\centering
\includegraphics[width=\textwidth]{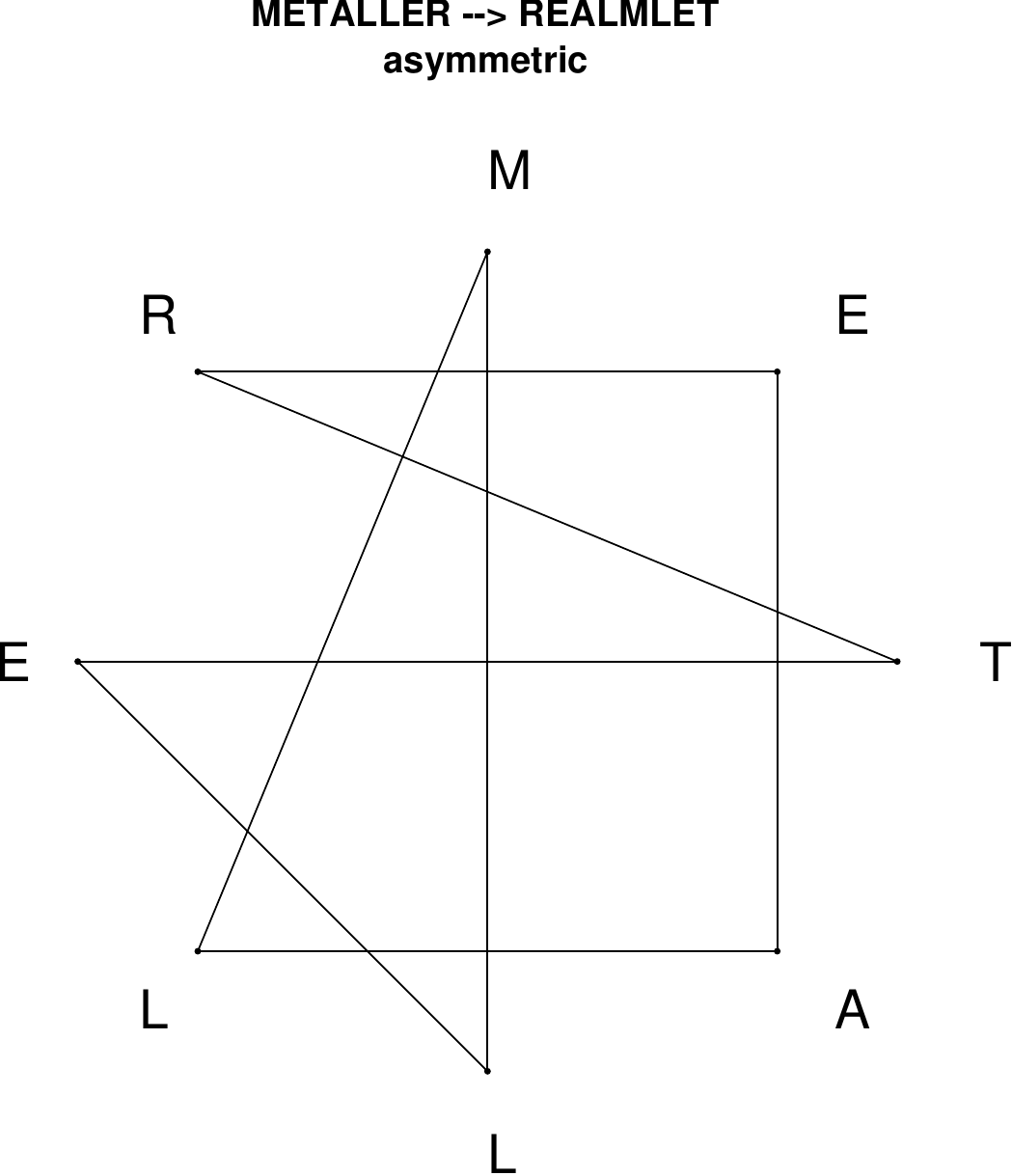}
\end{subfigure}
\end{figure}

\begin{figure}[H]
\centering
\begin{subfigure}[T]{0.19\textwidth}
\centering
\includegraphics[width=\textwidth]{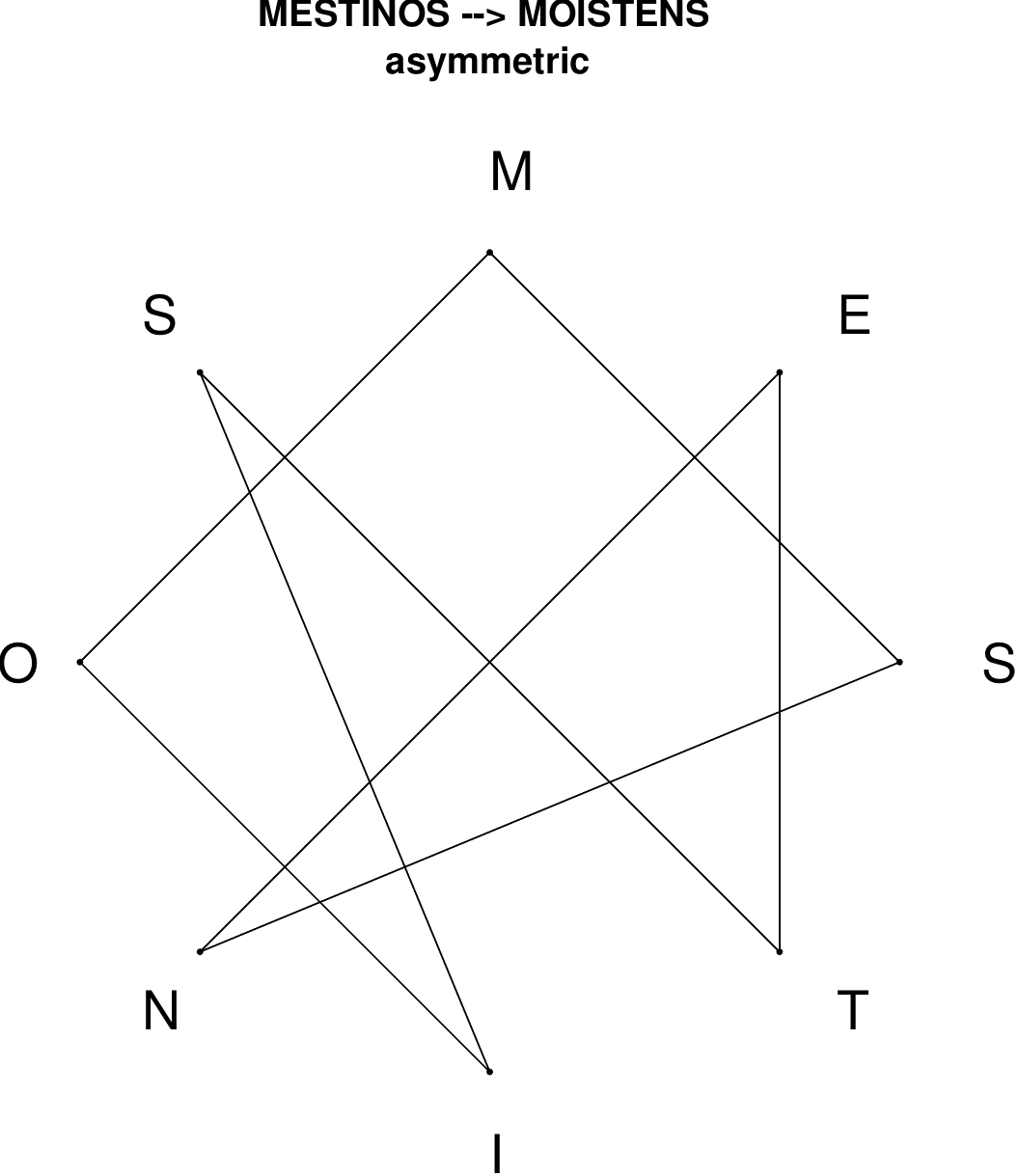}
\end{subfigure}
\hfill
\begin{subfigure}[T]{0.19\textwidth}
\centering
\includegraphics[width=\textwidth]{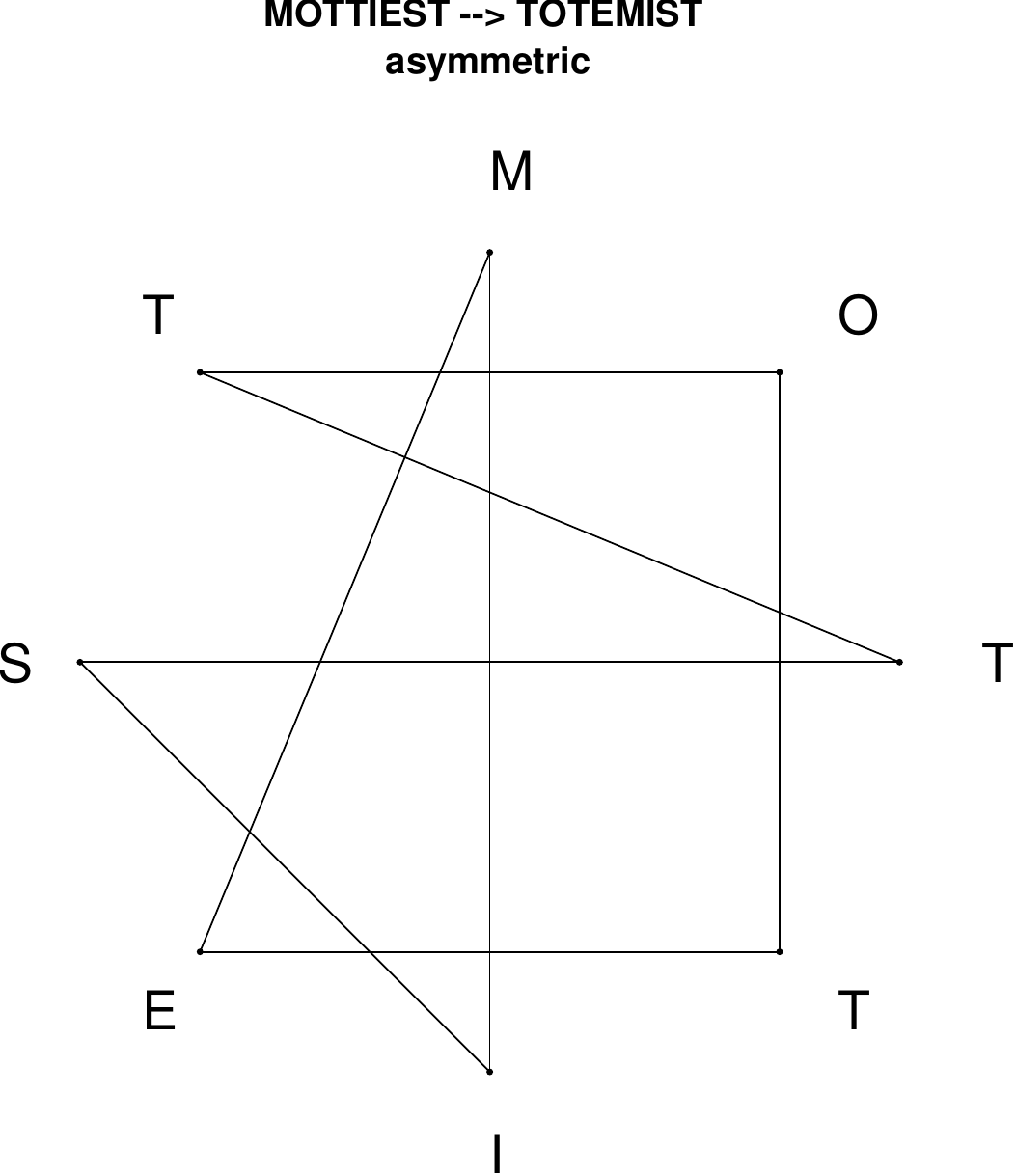}
\end{subfigure}
\hfill
\begin{subfigure}[T]{0.19\textwidth}
\centering
\includegraphics[width=\textwidth]{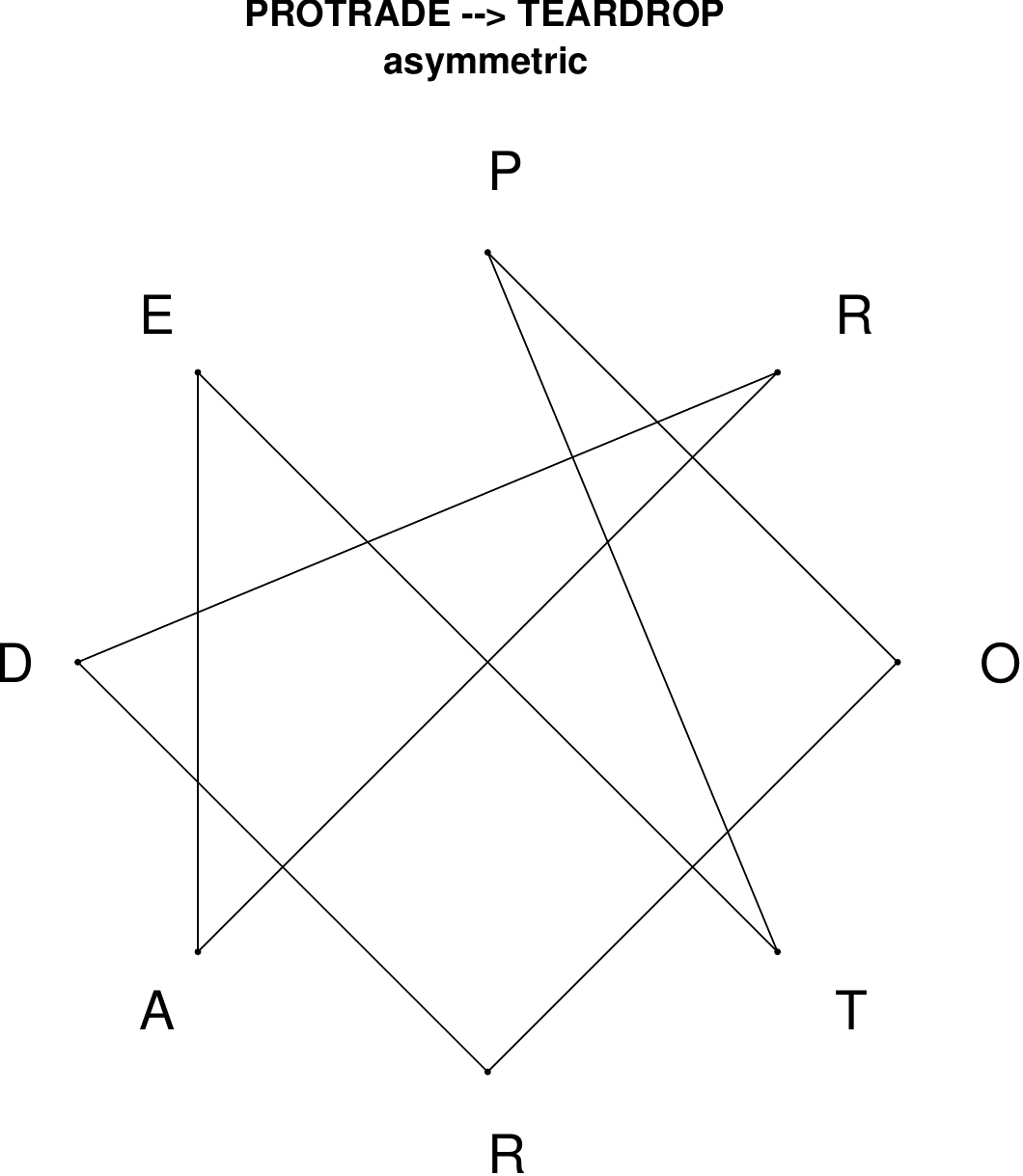}
\end{subfigure}
\hfill
\begin{subfigure}[T]{0.19\textwidth}
\centering
\includegraphics[width=\textwidth]{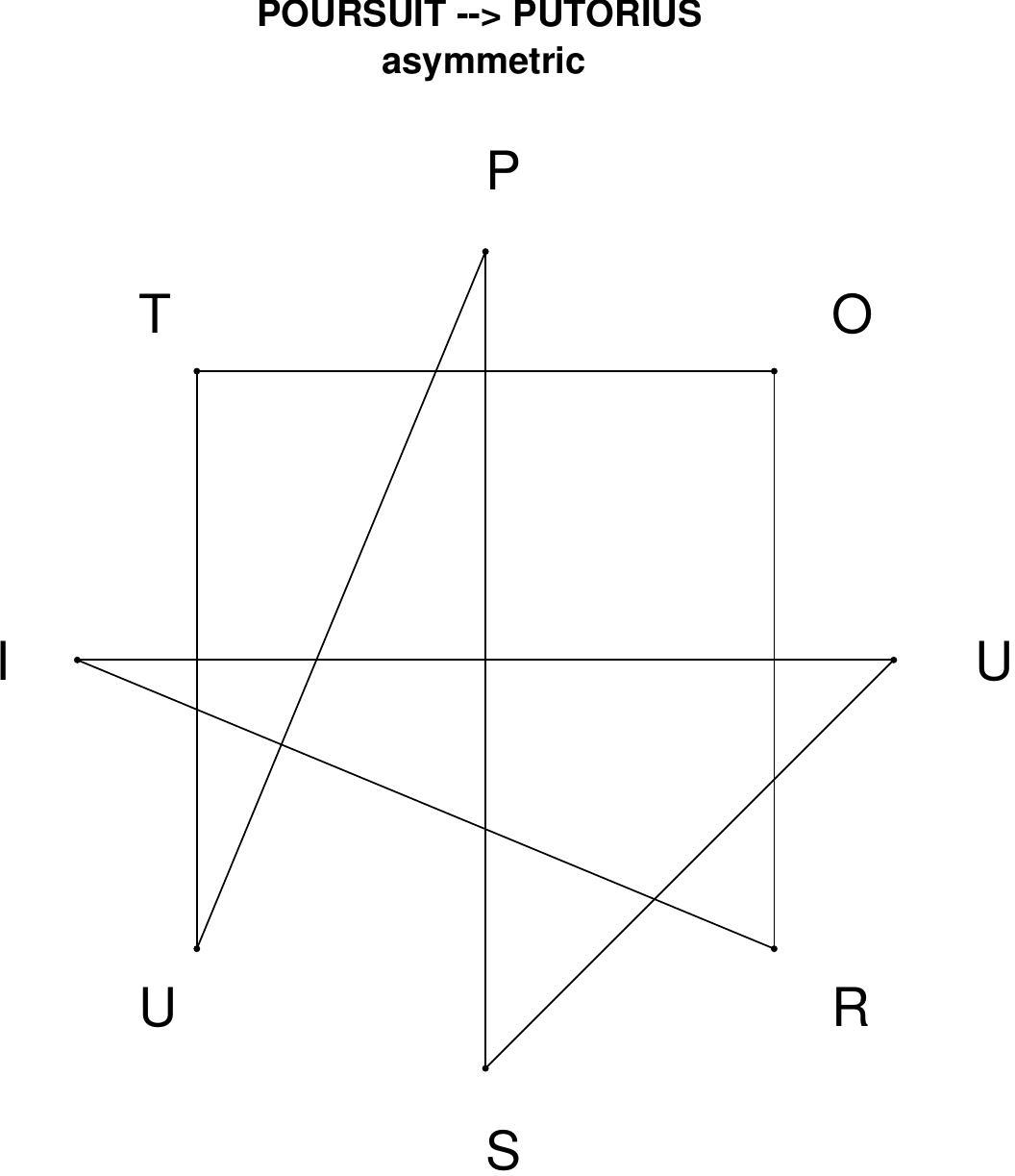}
\end{subfigure}
\hfill
\begin{subfigure}[T]{0.19\textwidth}
\centering
\includegraphics[width=\textwidth]{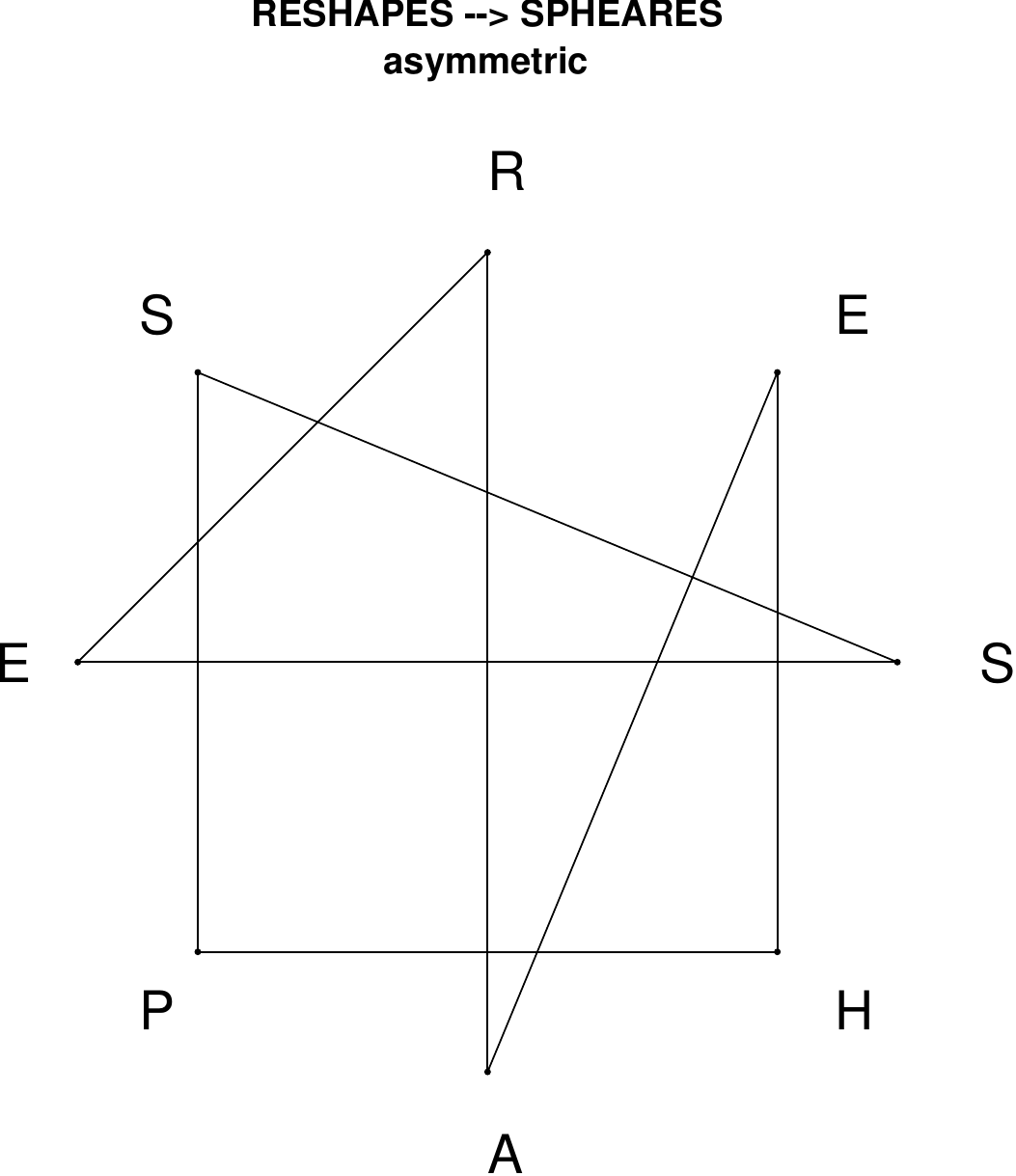}
\end{subfigure}
\end{figure}

\begin{figure}[H]
\centering
\begin{subfigure}[T]{0.19\textwidth}
\centering
\includegraphics[width=\textwidth]{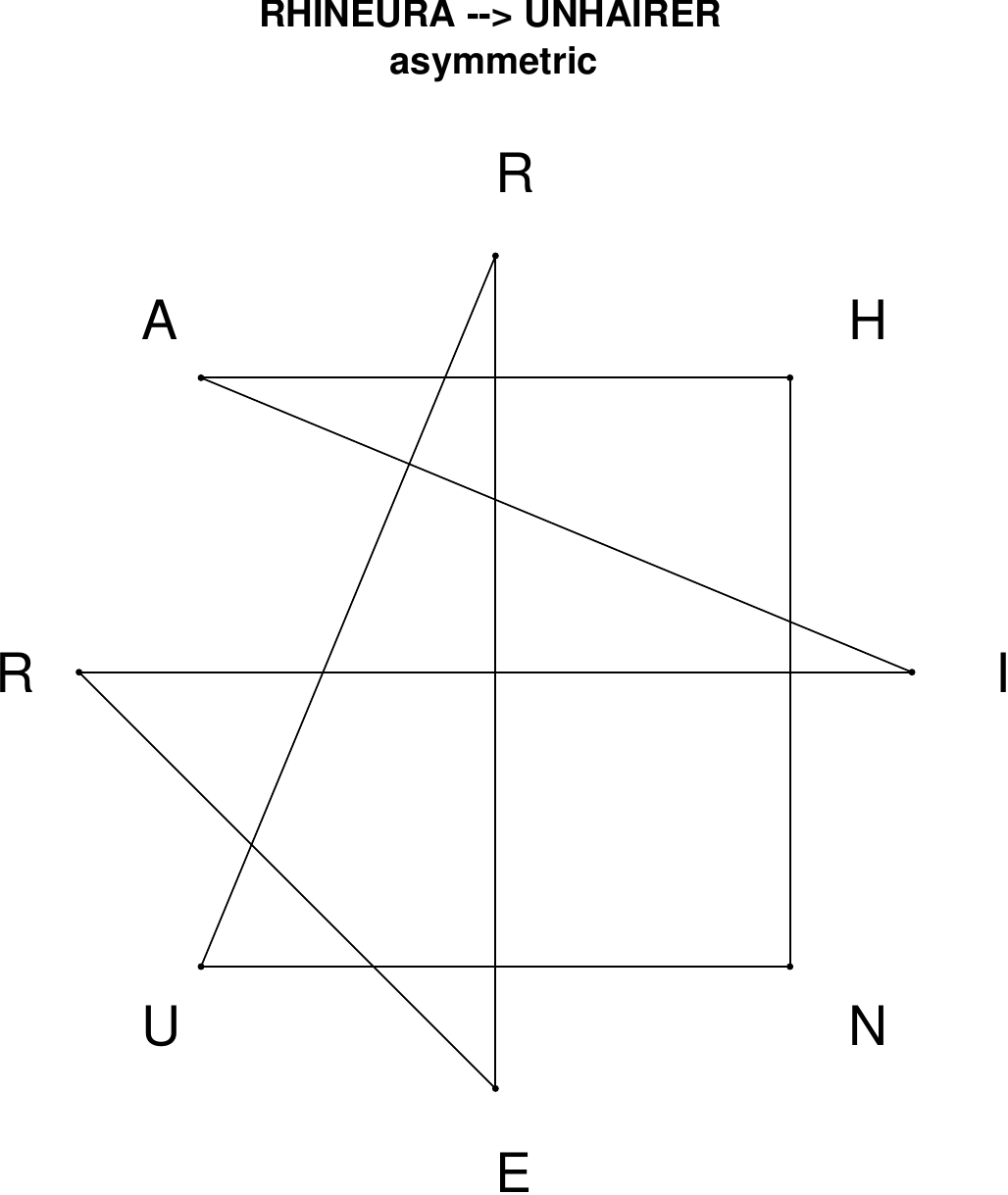}
\end{subfigure}
\hfill
\begin{subfigure}[T]{0.19\textwidth}
\centering
\includegraphics[width=\textwidth]{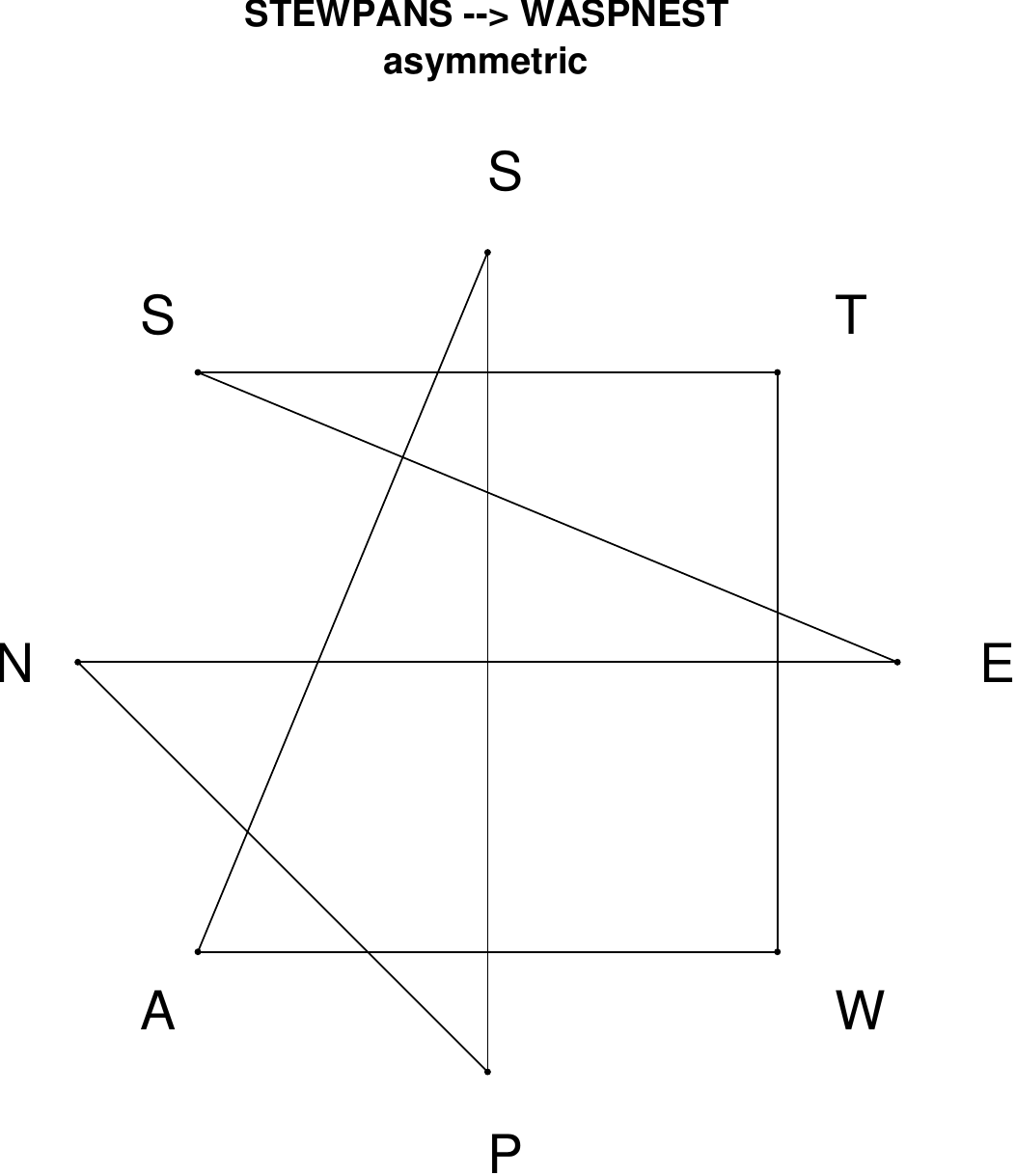}
\end{subfigure}
\hfill
\begin{subfigure}[T]{0.19\textwidth}
\centering
\includegraphics[width=\textwidth]{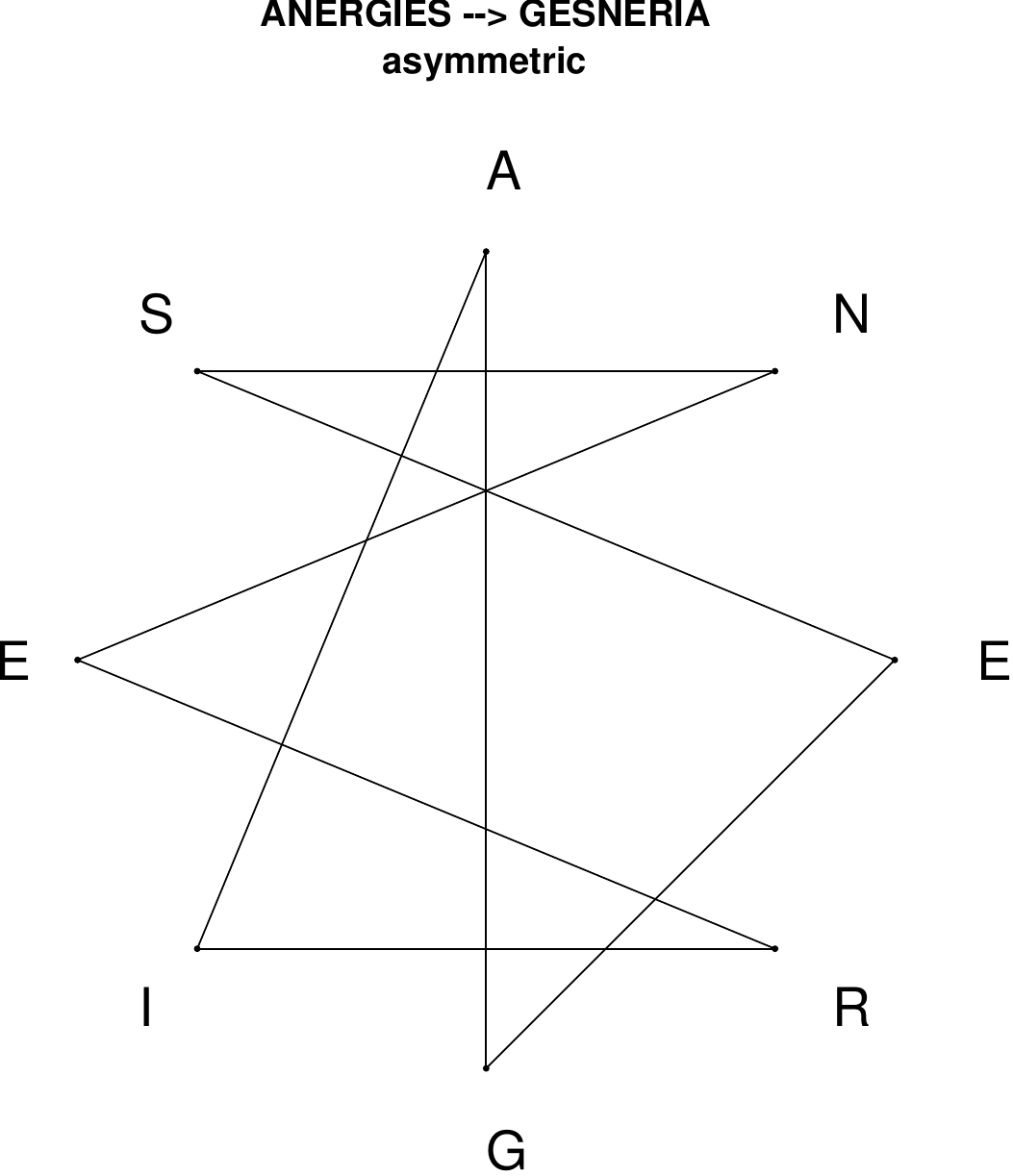}
\end{subfigure}
\hfill
\begin{subfigure}[T]{0.19\textwidth}
\centering
\includegraphics[width=\textwidth]{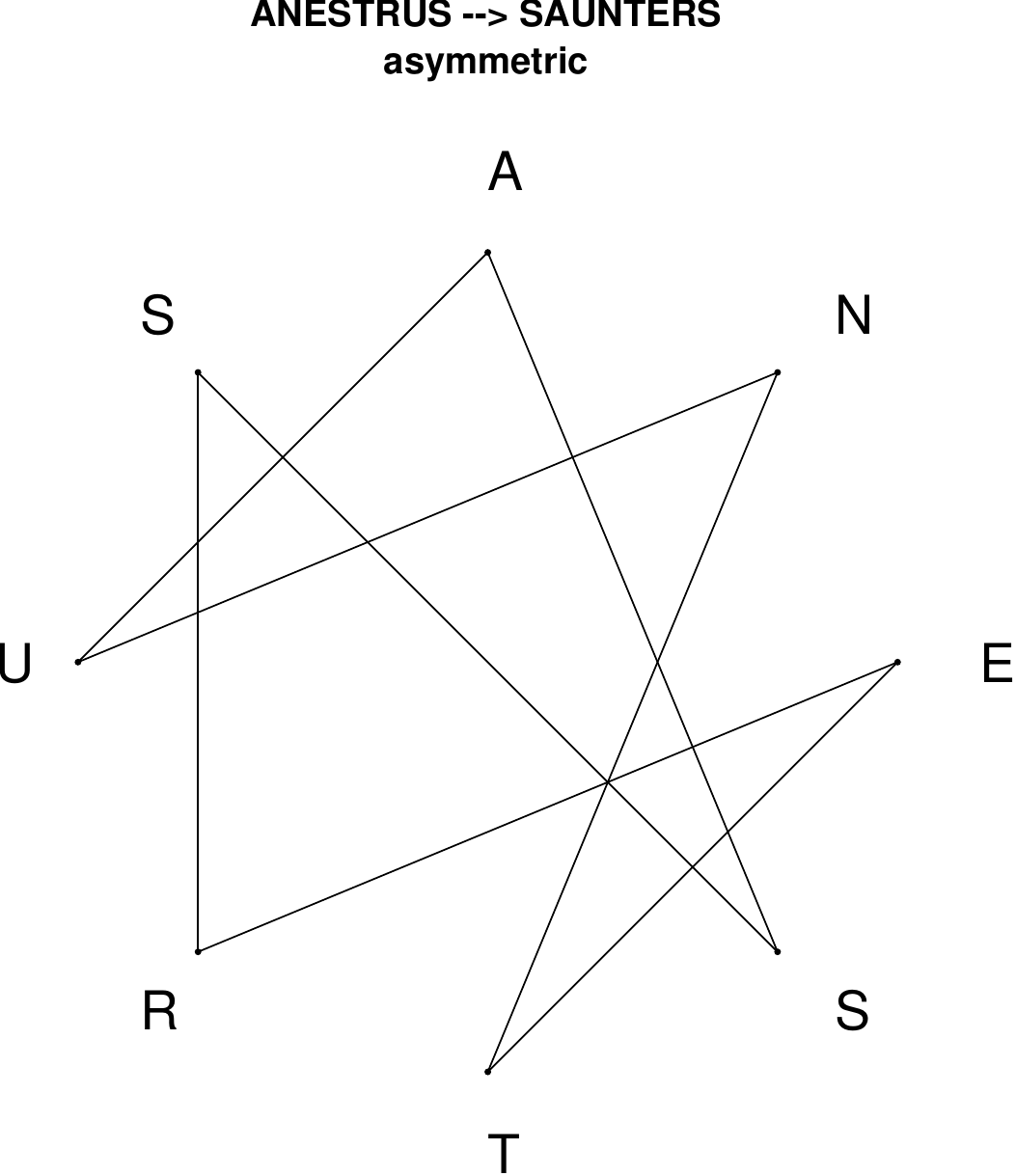}
\end{subfigure}
\hfill
\begin{subfigure}[T]{0.19\textwidth}
\centering
\includegraphics[width=\textwidth]{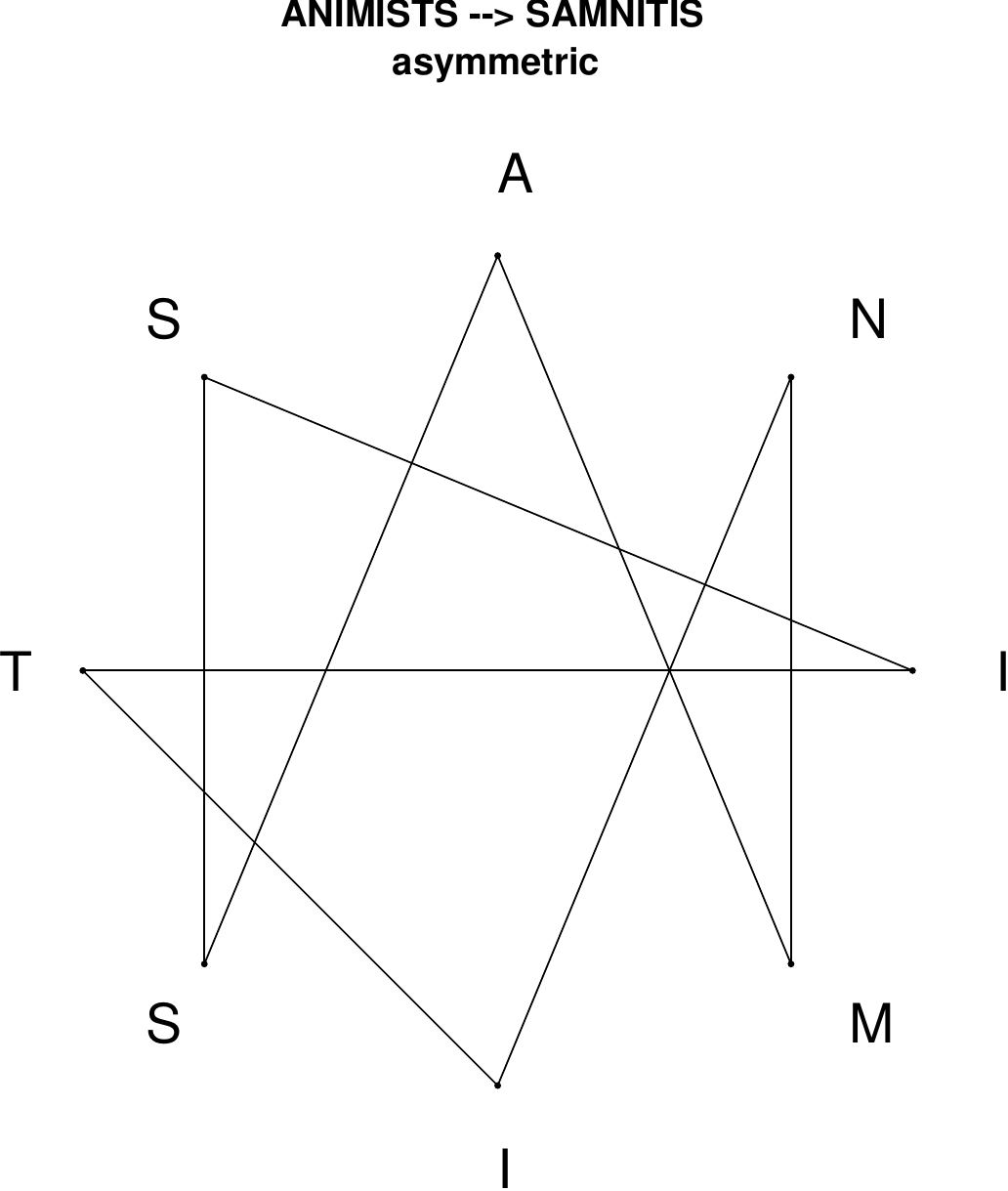}
\end{subfigure}
\end{figure}

\begin{figure}[H]
\centering
\begin{subfigure}[T]{0.19\textwidth}
\centering
\includegraphics[width=\textwidth]{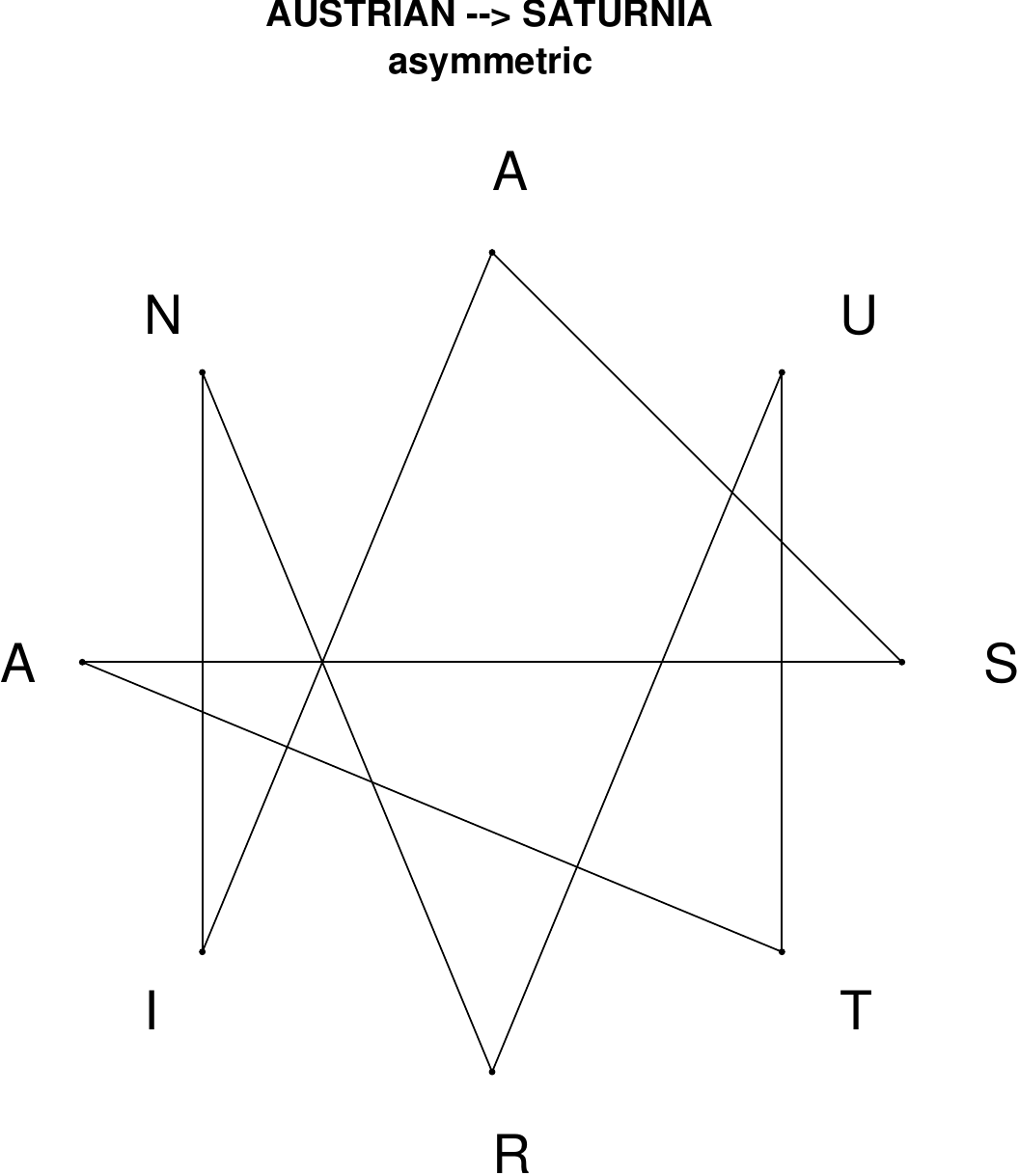}
\end{subfigure}
\hfill
\begin{subfigure}[T]{0.19\textwidth}
\centering
\includegraphics[width=\textwidth]{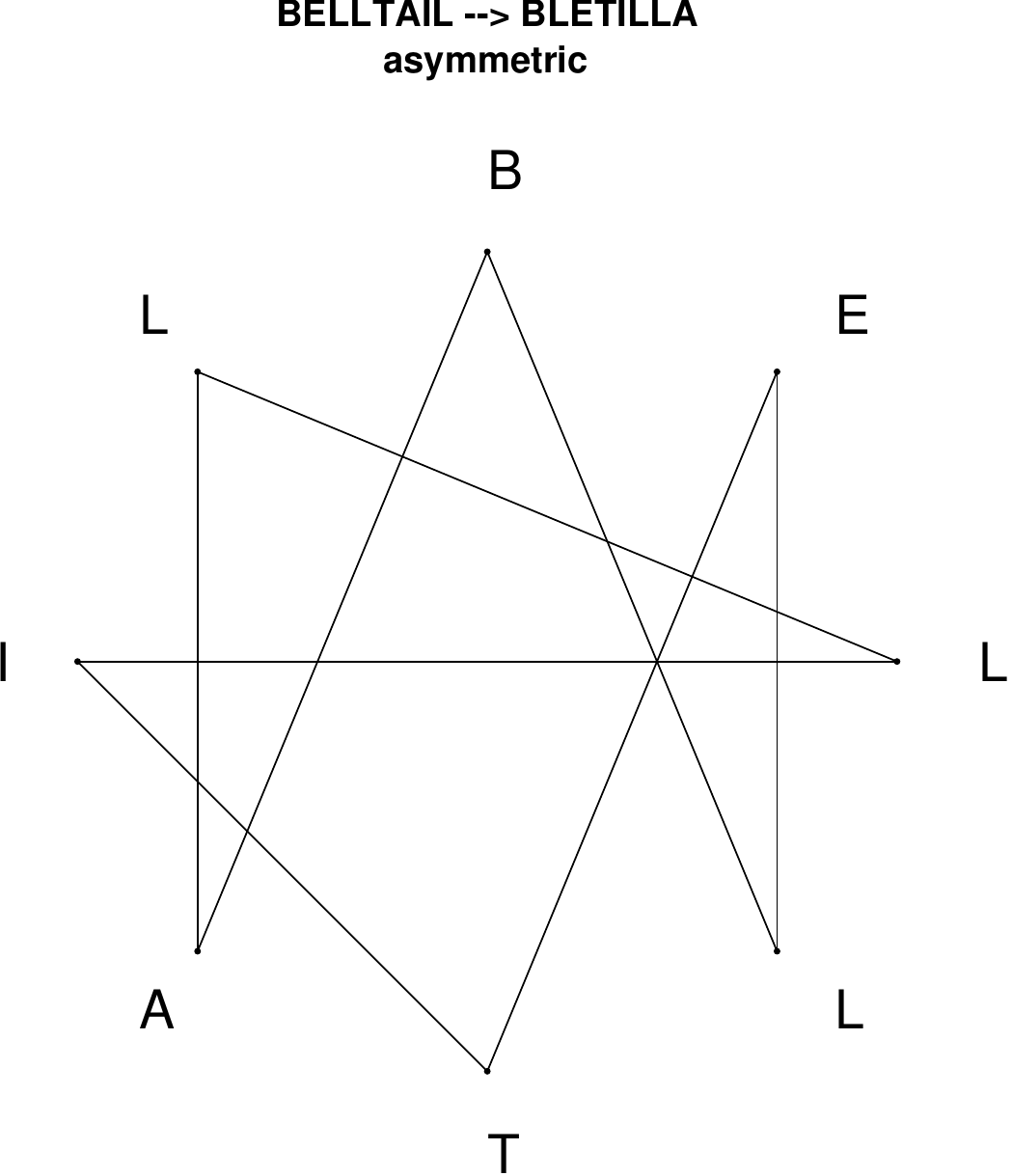}
\end{subfigure}
\hfill
\begin{subfigure}[T]{0.19\textwidth}
\centering
\includegraphics[width=\textwidth]{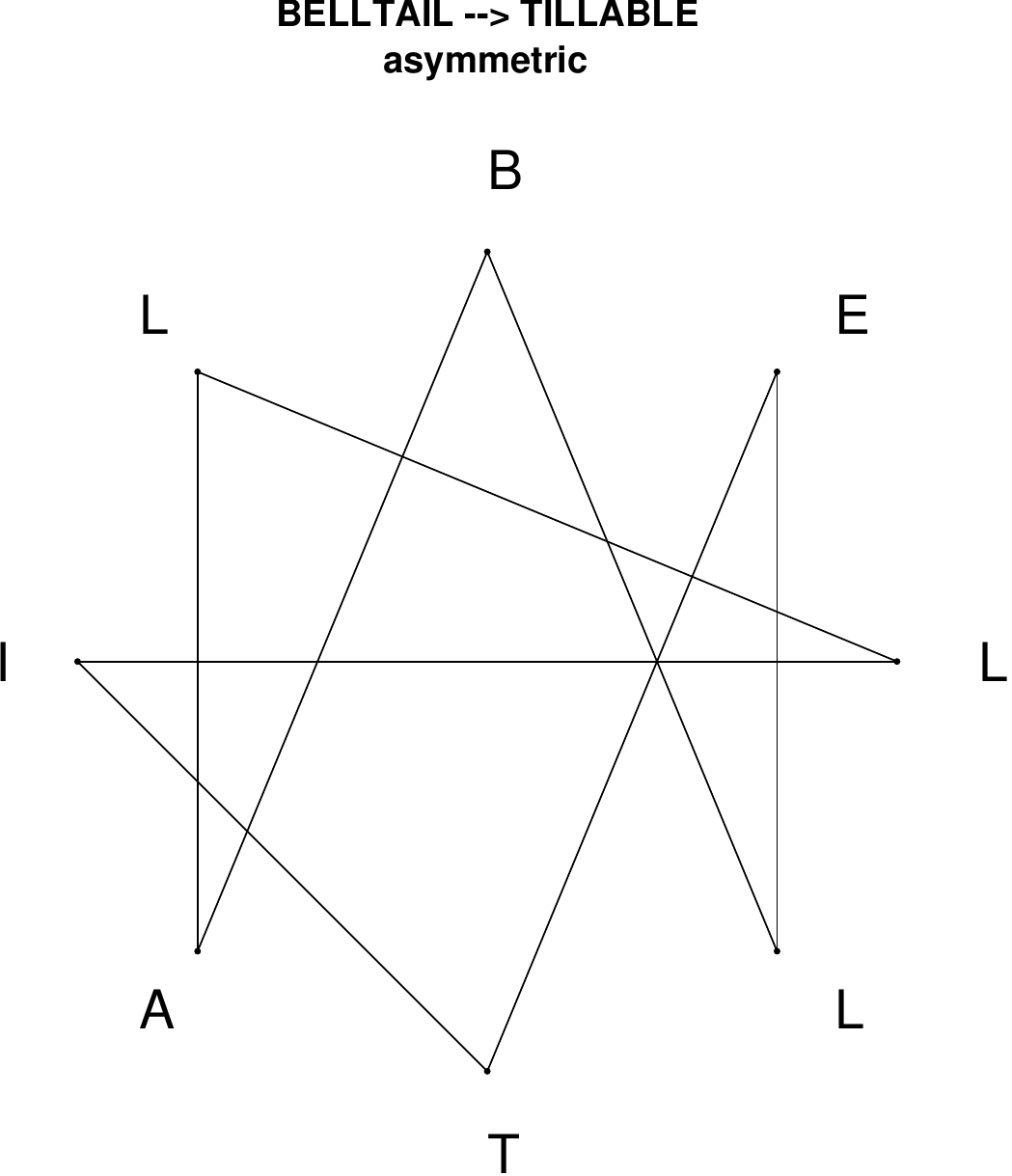}
\end{subfigure}
\hfill
\begin{subfigure}[T]{0.19\textwidth}
\centering
\includegraphics[width=\textwidth]{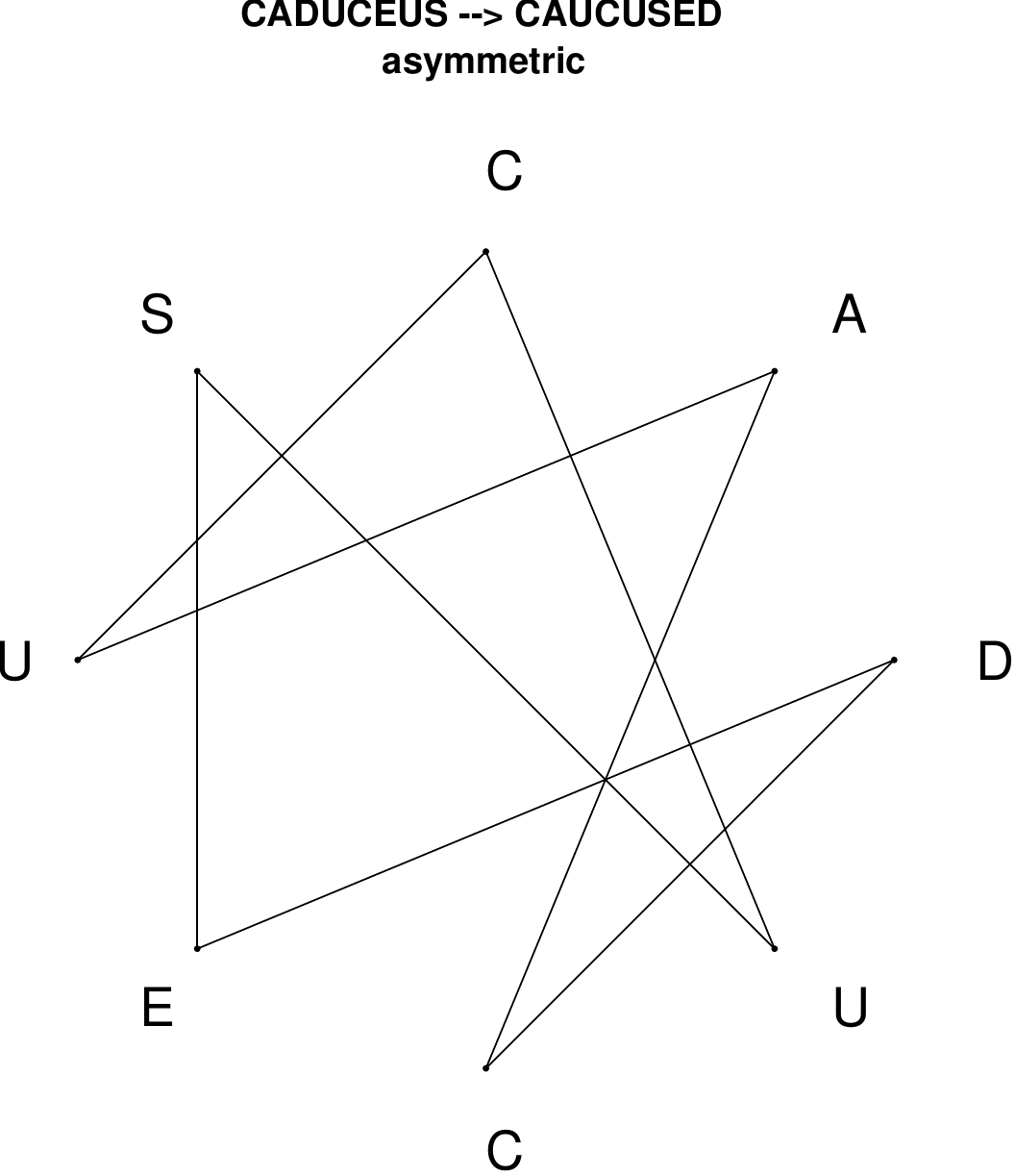}
\end{subfigure}
\hfill
\begin{subfigure}[T]{0.19\textwidth}
\centering
\includegraphics[width=\textwidth]{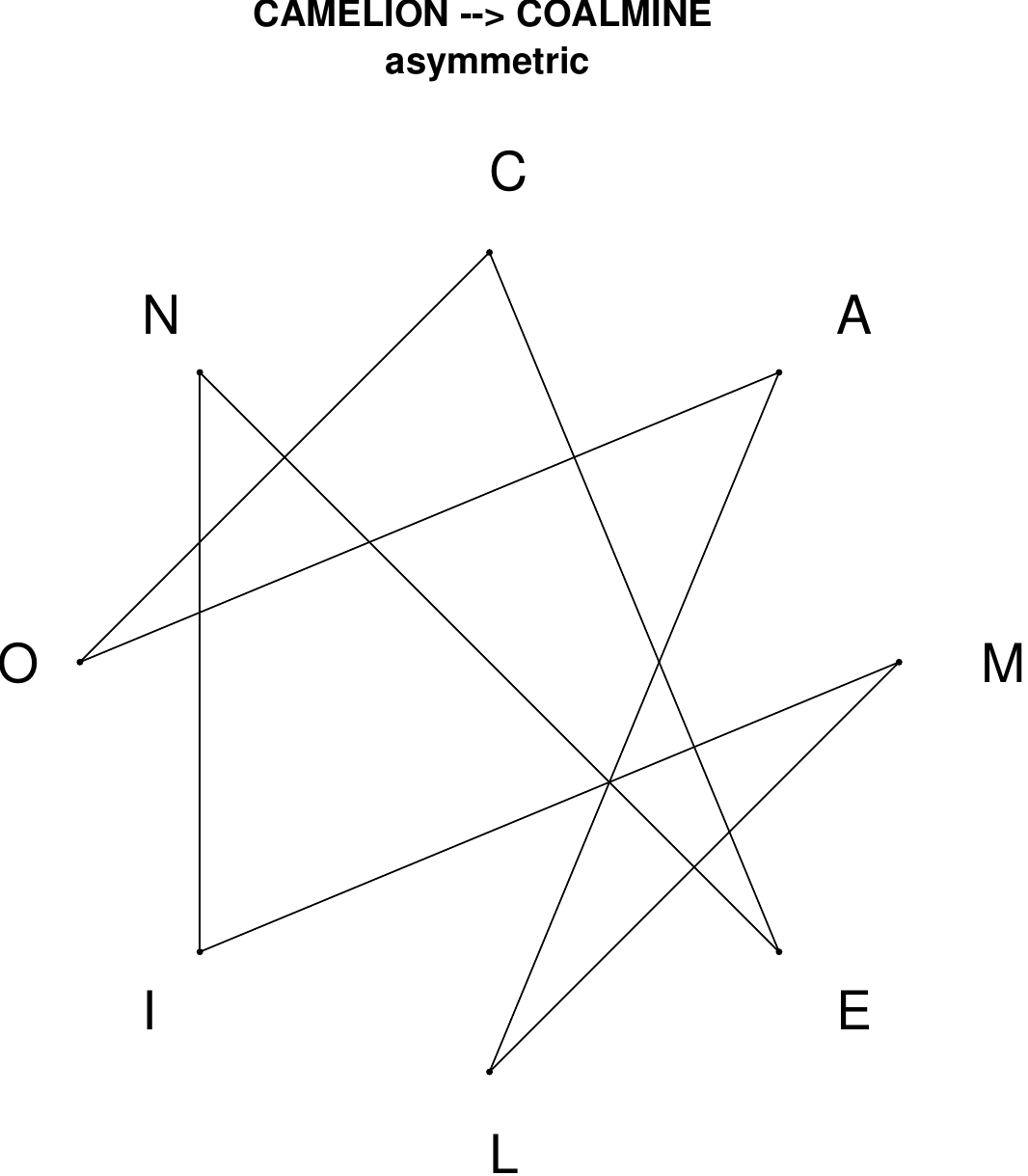}
\end{subfigure}
\end{figure}

\begin{figure}[H]
\centering
\begin{subfigure}[T]{0.19\textwidth}
\centering
\includegraphics[width=\textwidth]{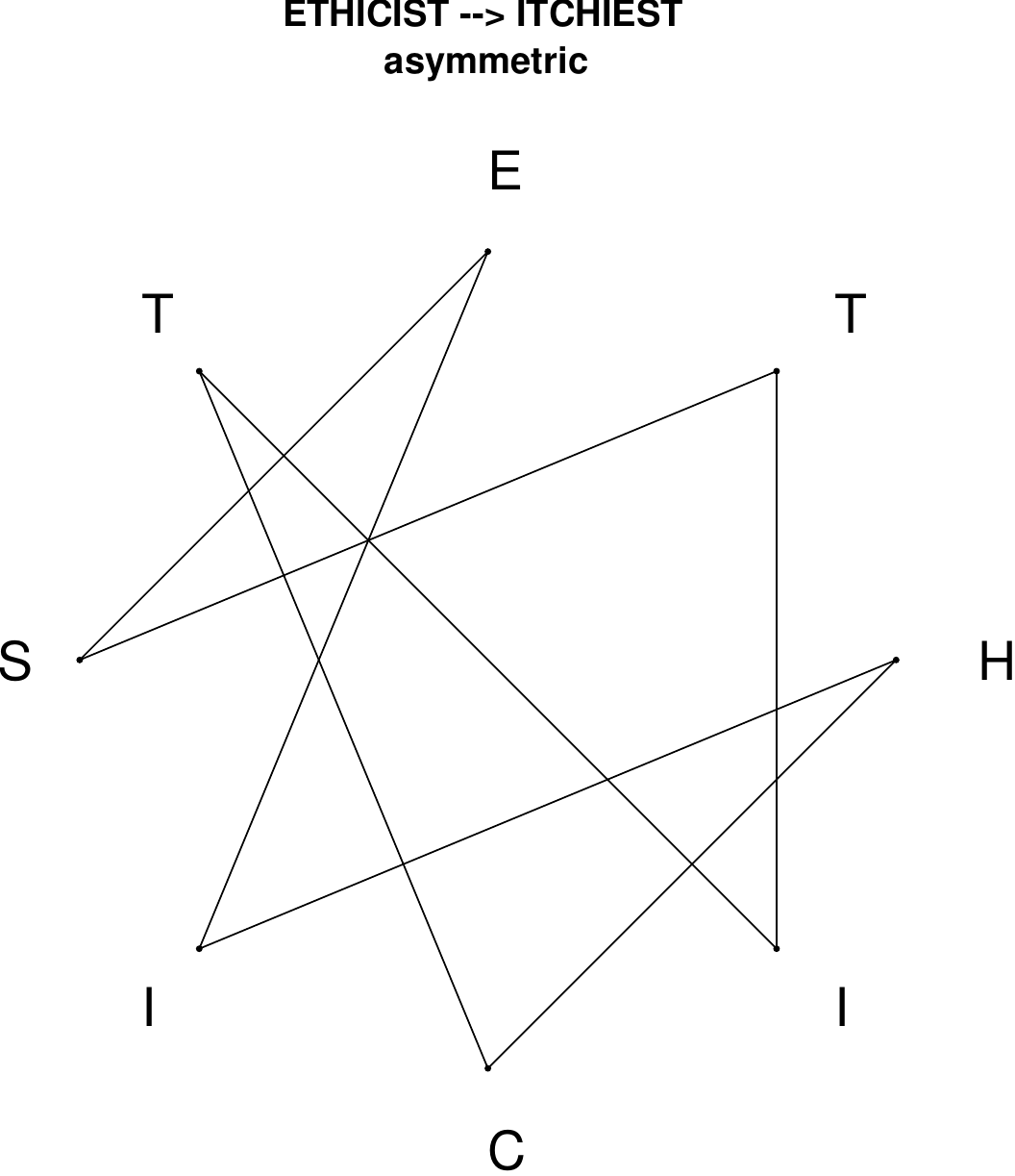}
\end{subfigure}
\hfill
\begin{subfigure}[T]{0.19\textwidth}
\centering
\includegraphics[width=\textwidth]{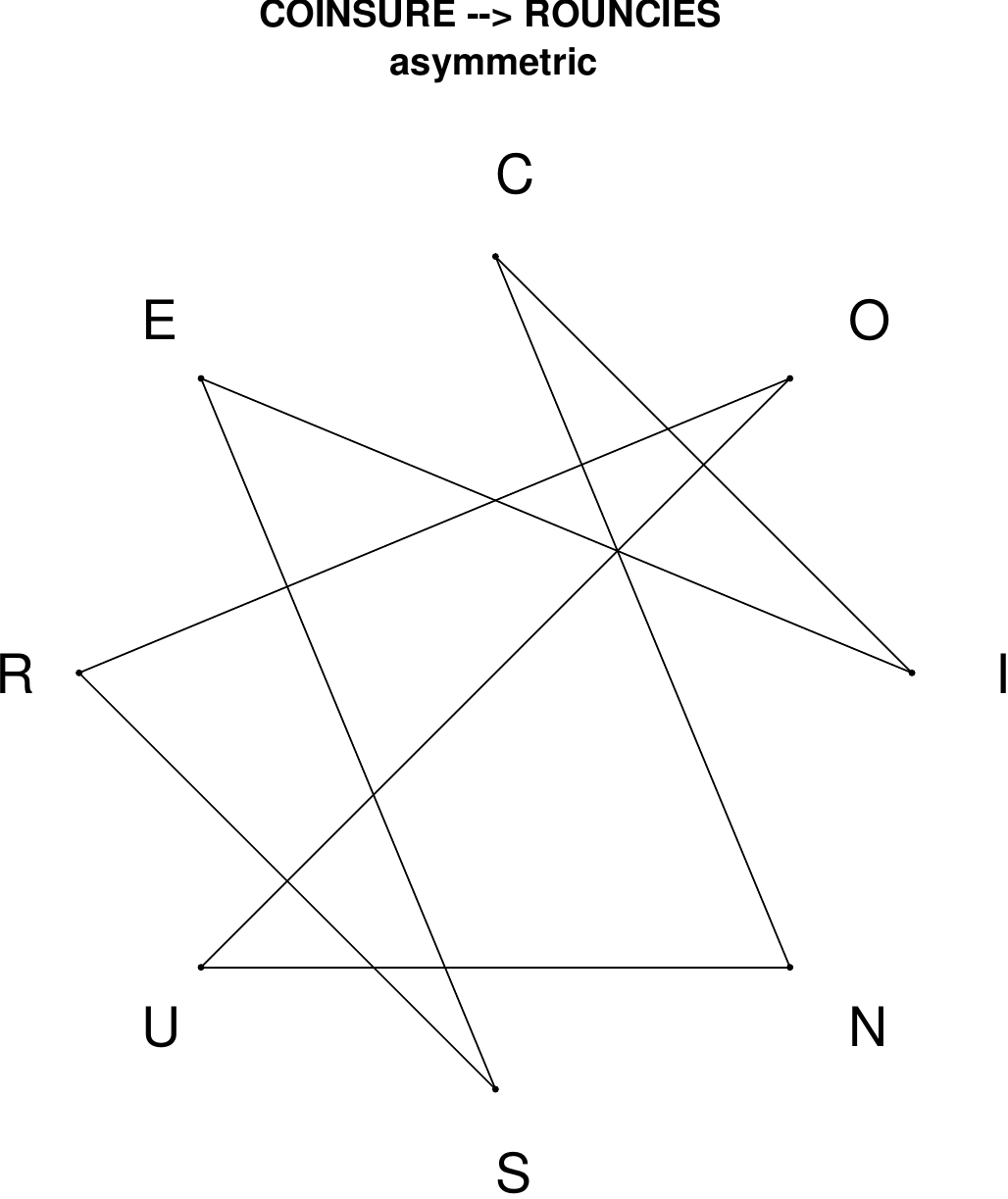}
\end{subfigure}
\hfill
\begin{subfigure}[T]{0.19\textwidth}
\centering
\includegraphics[width=\textwidth]{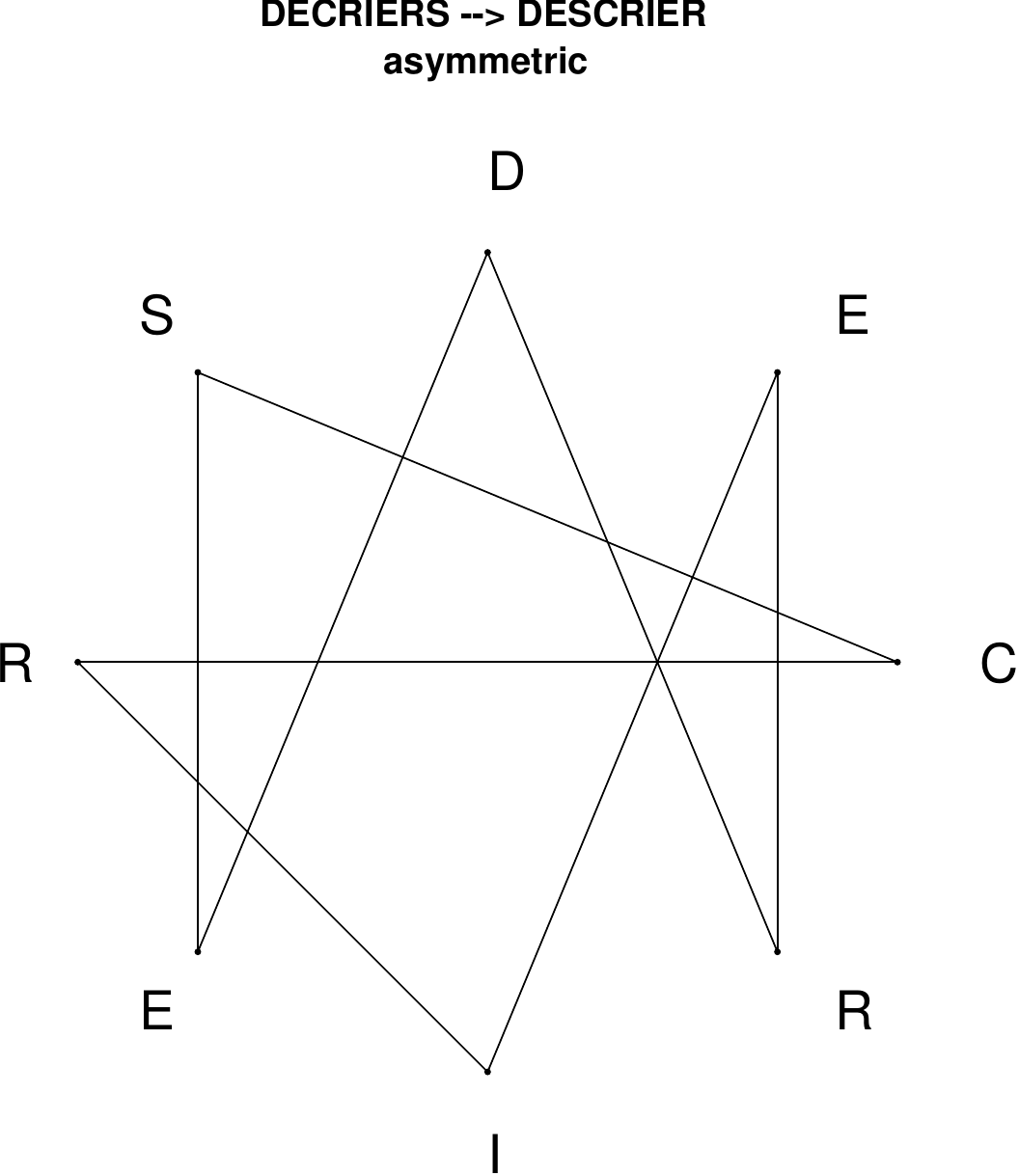}
\end{subfigure}
\hfill
\begin{subfigure}[T]{0.19\textwidth}
\centering
\includegraphics[width=\textwidth]{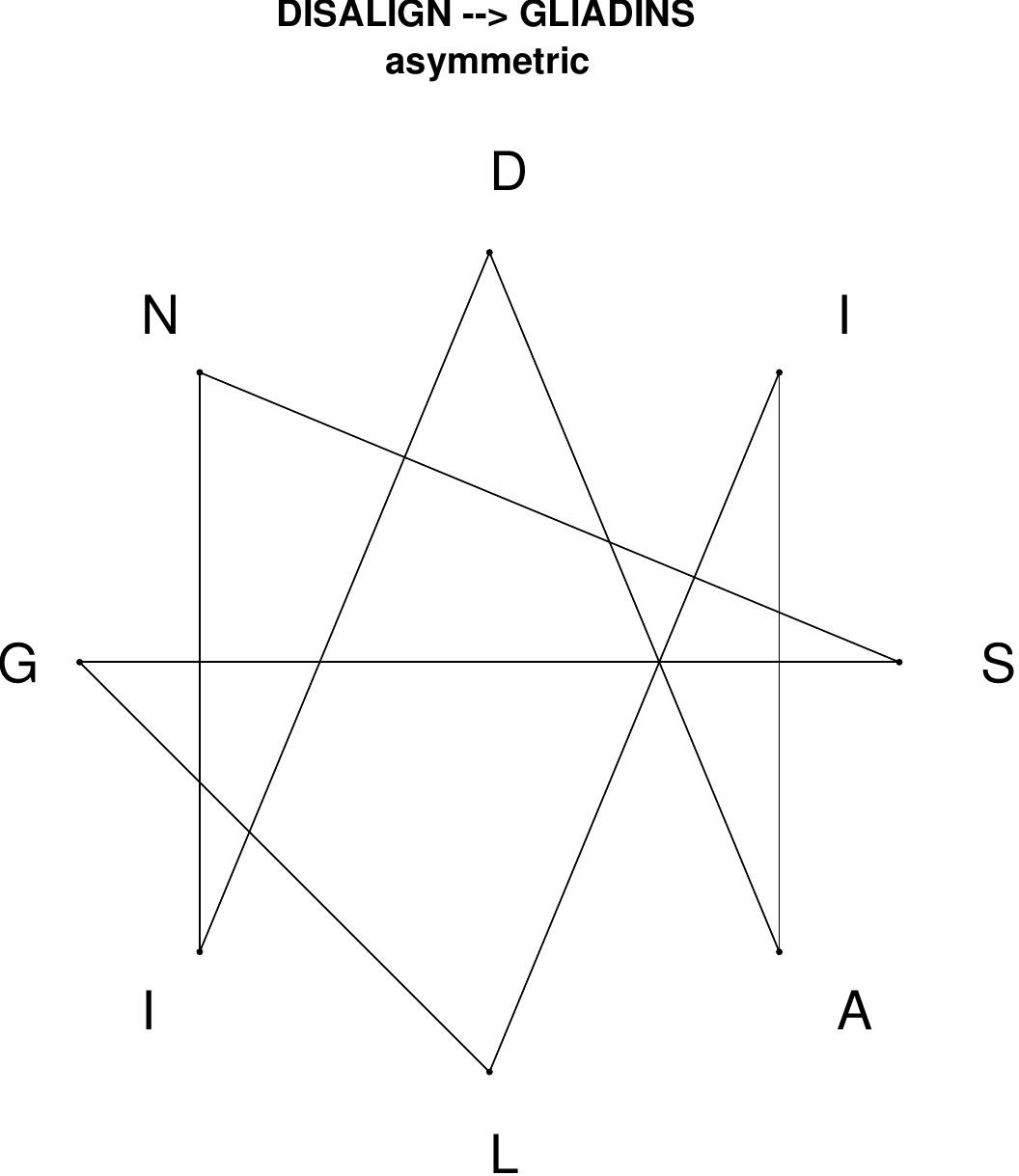}
\end{subfigure}
\hfill
\begin{subfigure}[T]{0.19\textwidth}
\centering
\includegraphics[width=\textwidth]{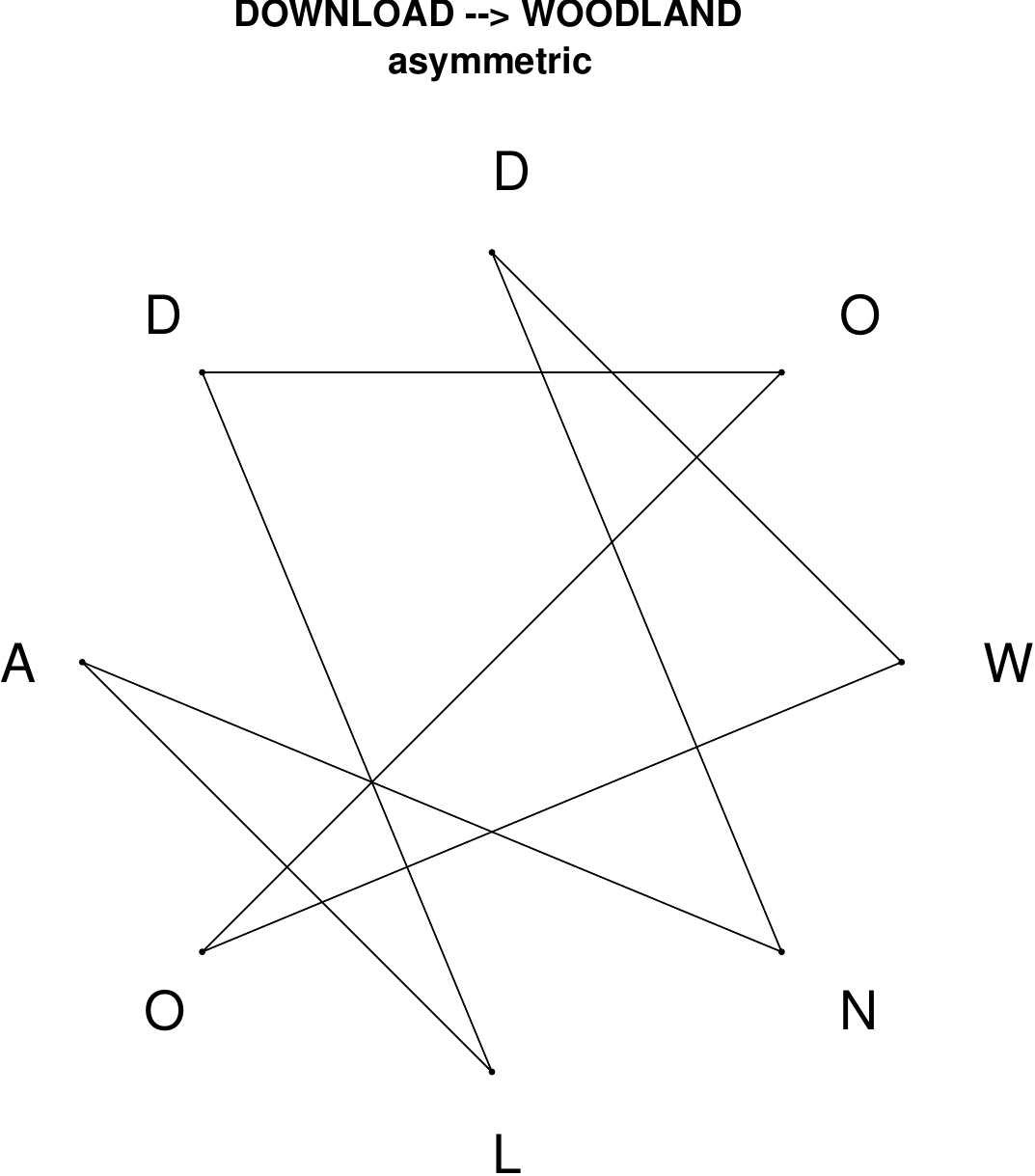}
\end{subfigure}
\end{figure}

\begin{figure}[H]
\centering
\begin{subfigure}[T]{0.19\textwidth}
\centering
\includegraphics[width=\textwidth]{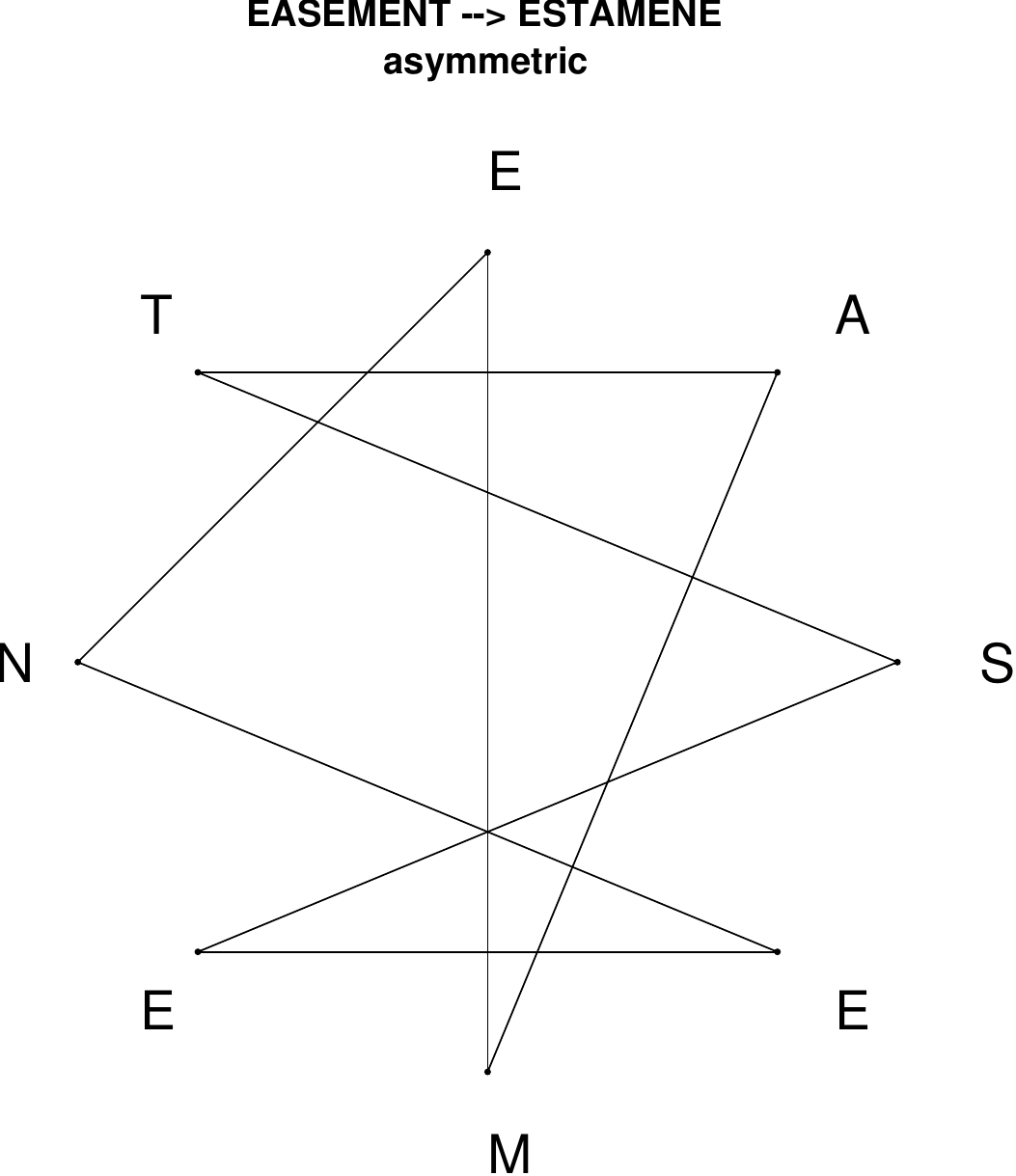}
\end{subfigure}
\hfill
\begin{subfigure}[T]{0.19\textwidth}
\centering
\includegraphics[width=\textwidth]{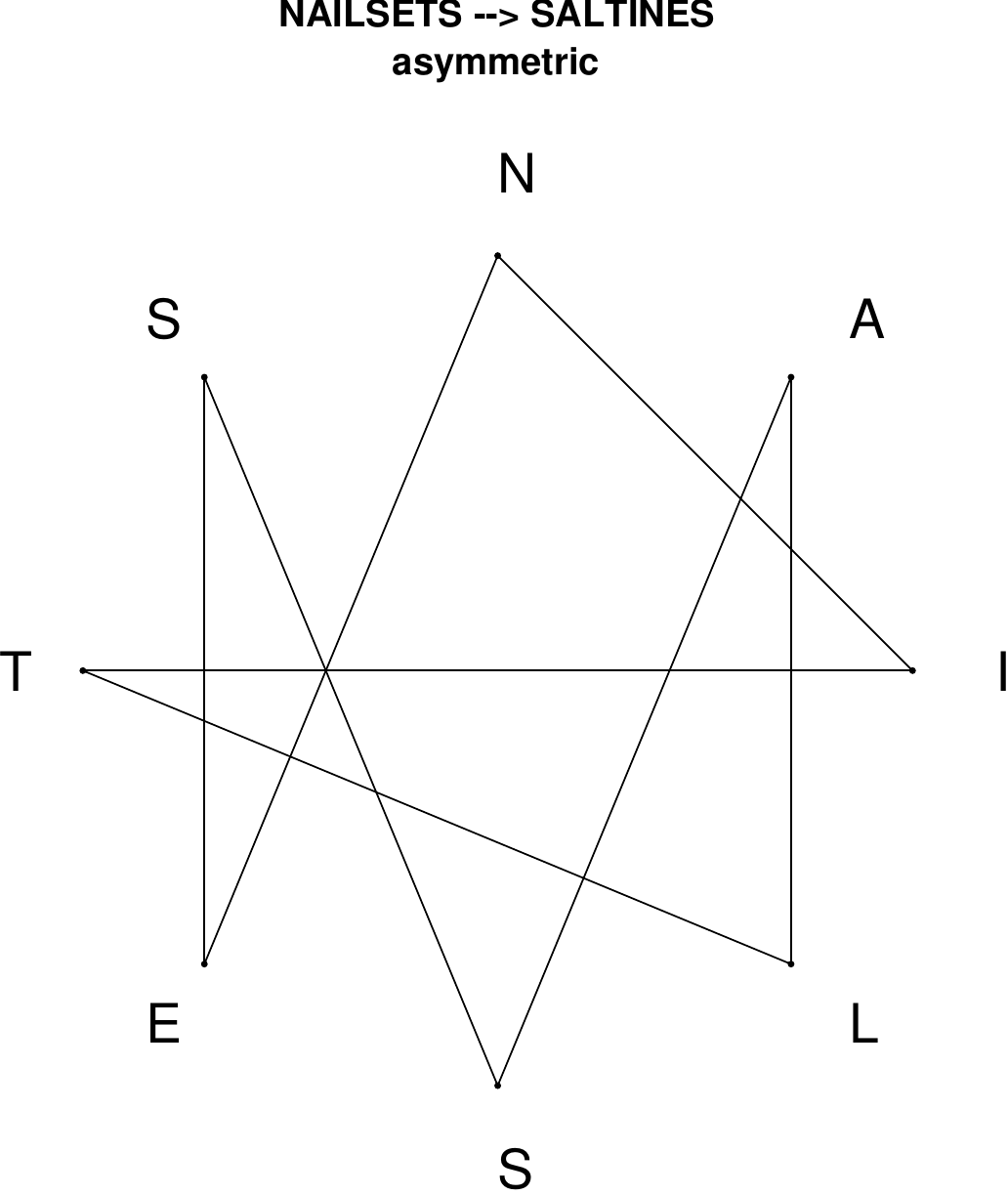}
\end{subfigure}
\hfill
\begin{subfigure}[T]{0.19\textwidth}
\centering
\includegraphics[width=\textwidth]{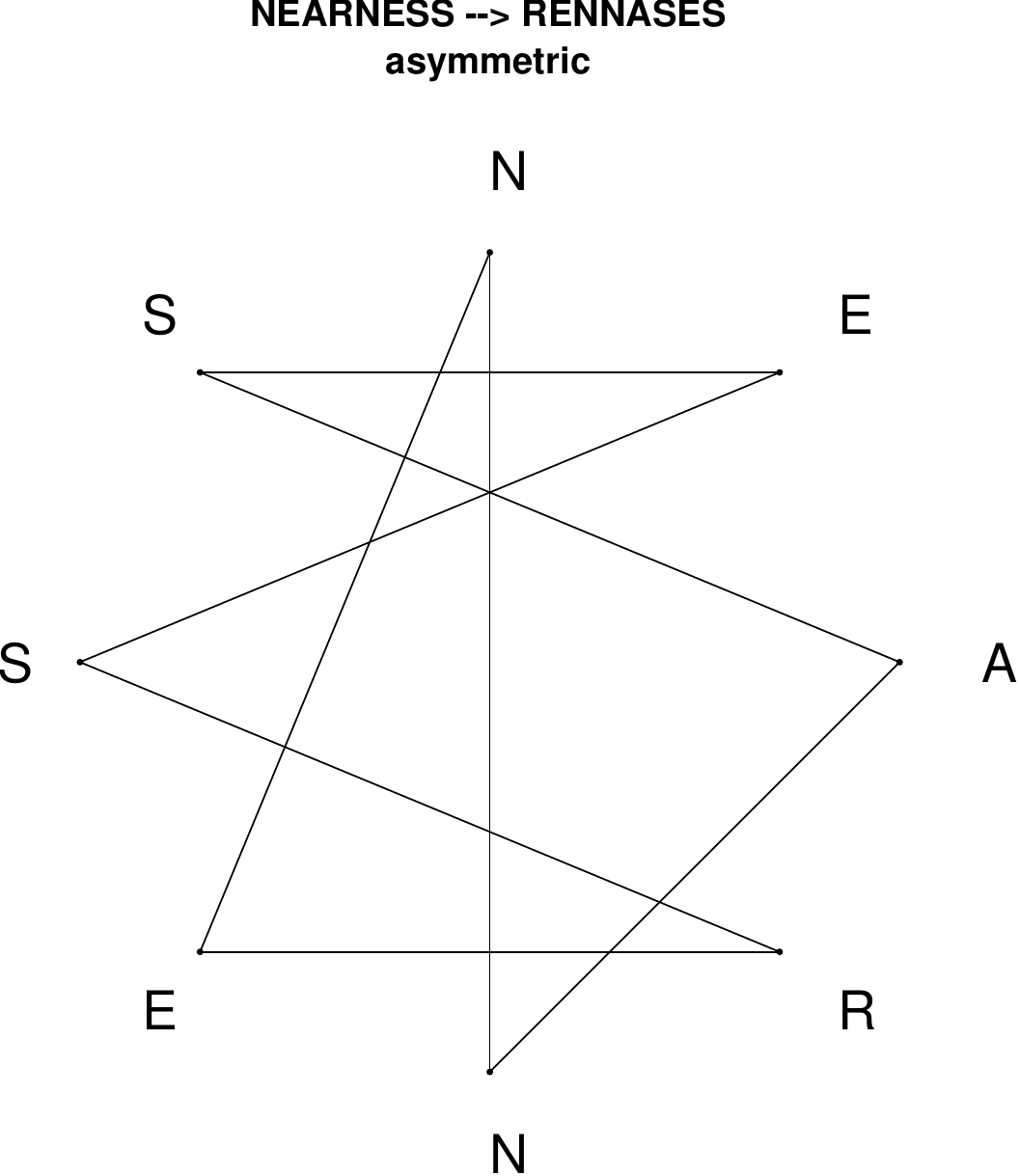}
\end{subfigure}
\hfill
\begin{subfigure}[T]{0.19\textwidth}
\centering
\includegraphics[width=\textwidth]{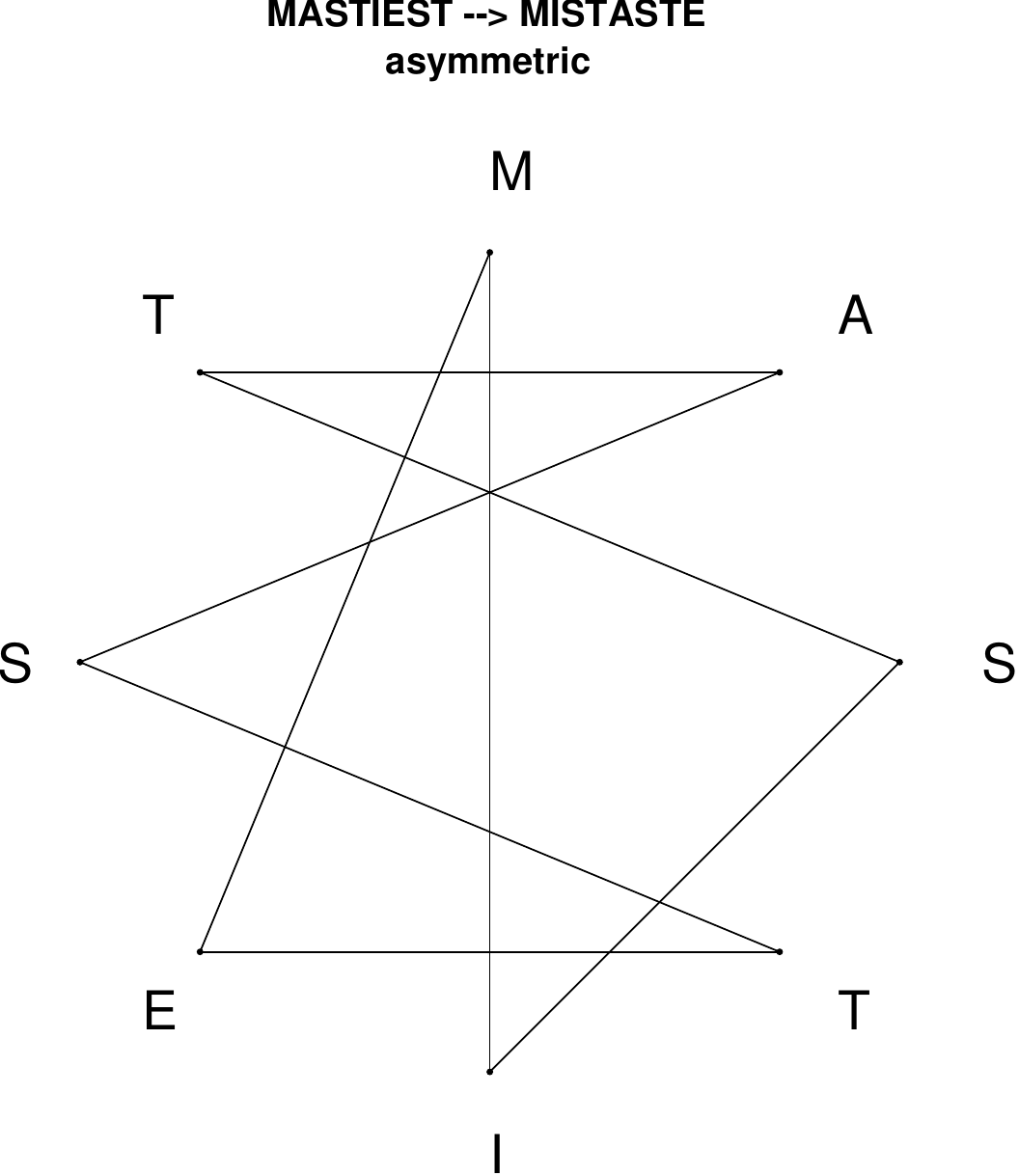}
\end{subfigure}
\hfill
\begin{subfigure}[T]{0.19\textwidth}
\centering
\includegraphics[width=\textwidth]{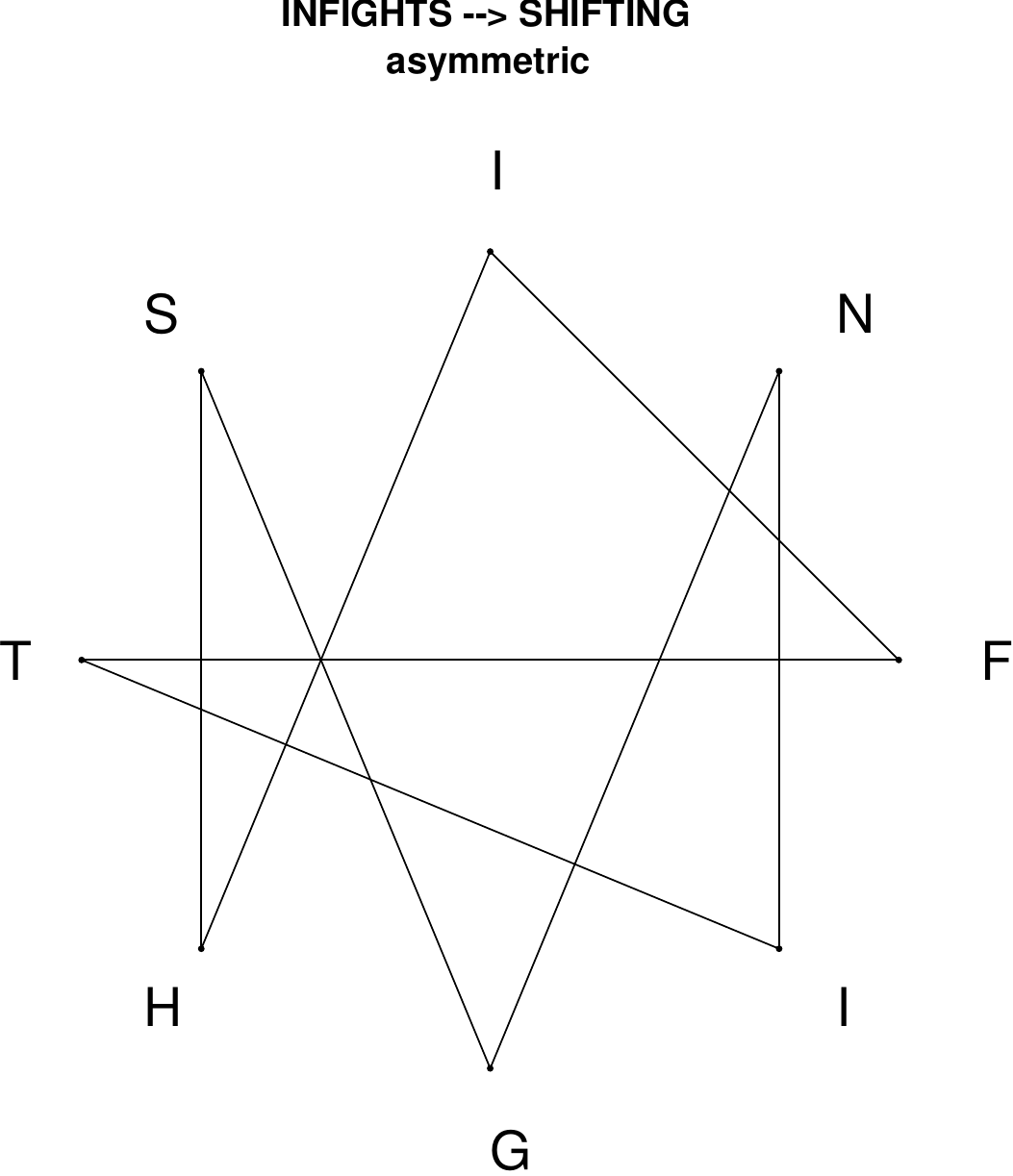}
\end{subfigure}
\end{figure}

\begin{figure}[H]
\centering
\begin{subfigure}[T]{0.19\textwidth}
\centering
\includegraphics[width=\textwidth]{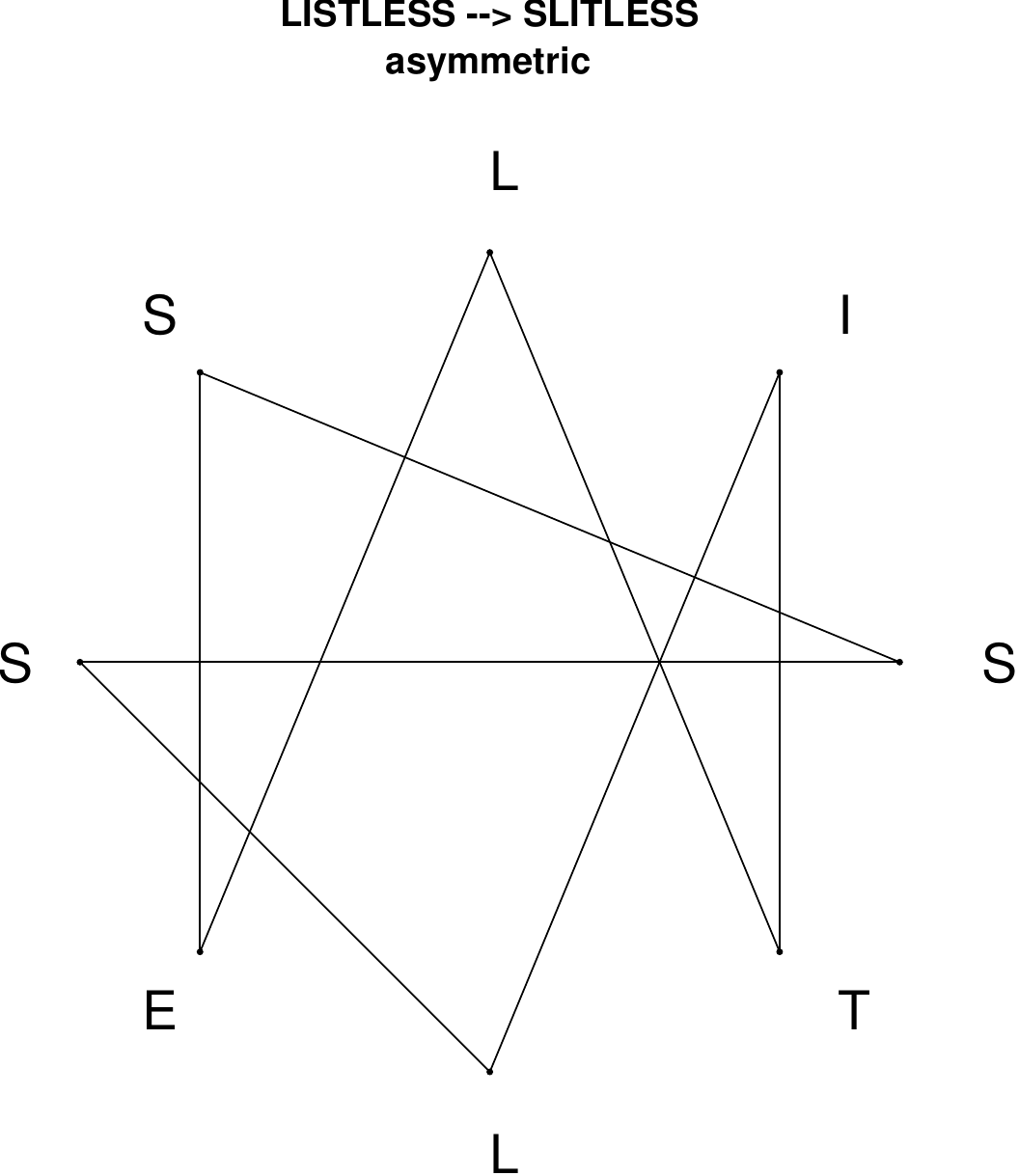}
\end{subfigure}
\hfill
\begin{subfigure}[T]{0.19\textwidth}
\centering
\includegraphics[width=\textwidth]{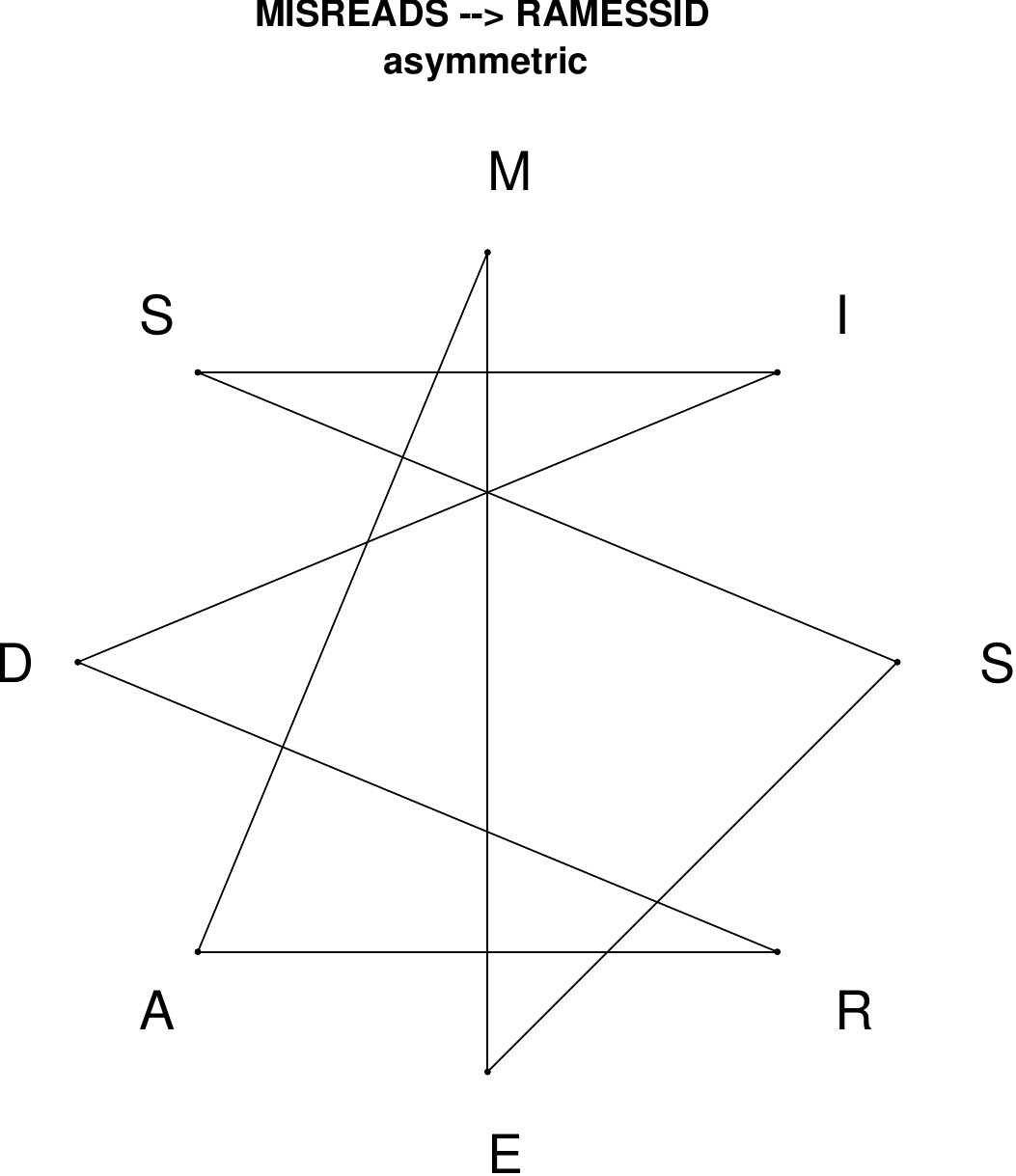}
\end{subfigure}
\hfill
\begin{subfigure}[T]{0.19\textwidth}
\centering
\includegraphics[width=\textwidth]{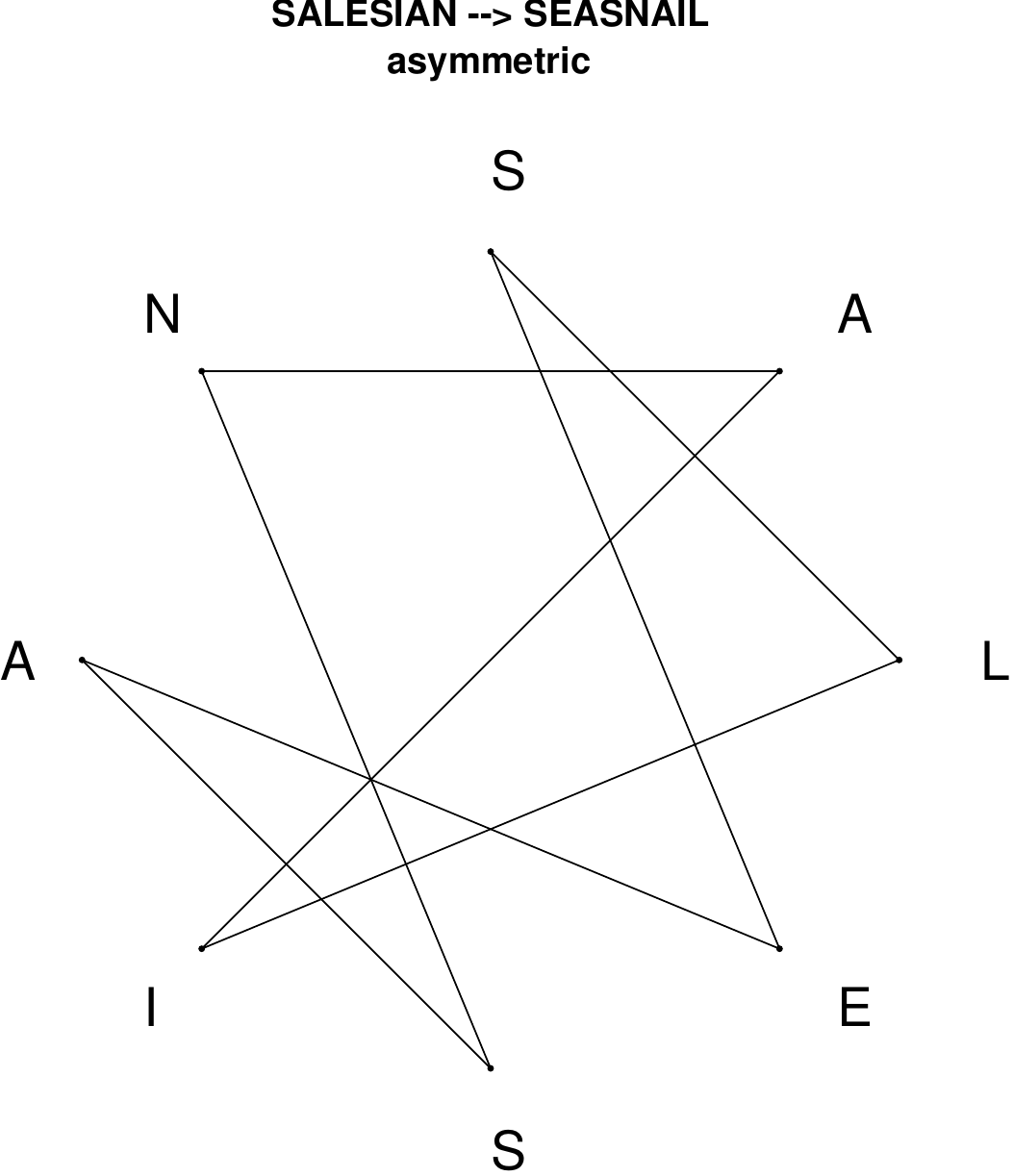}
\end{subfigure}
\hfill
\begin{subfigure}[T]{0.19\textwidth}
\centering
\includegraphics[width=\textwidth]{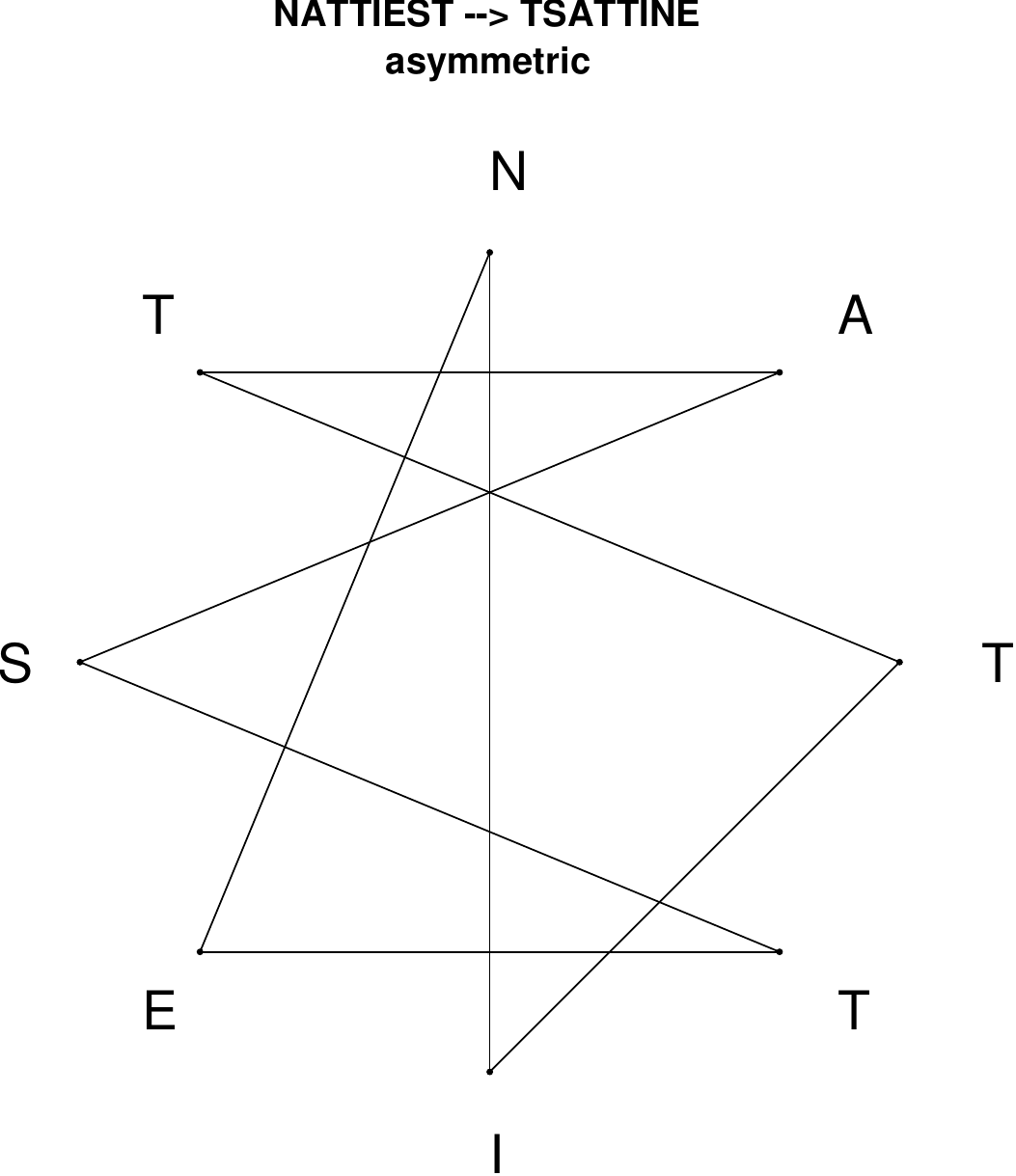}
\end{subfigure}
\hfill
\begin{subfigure}[T]{0.19\textwidth}
\centering
\includegraphics[width=\textwidth]{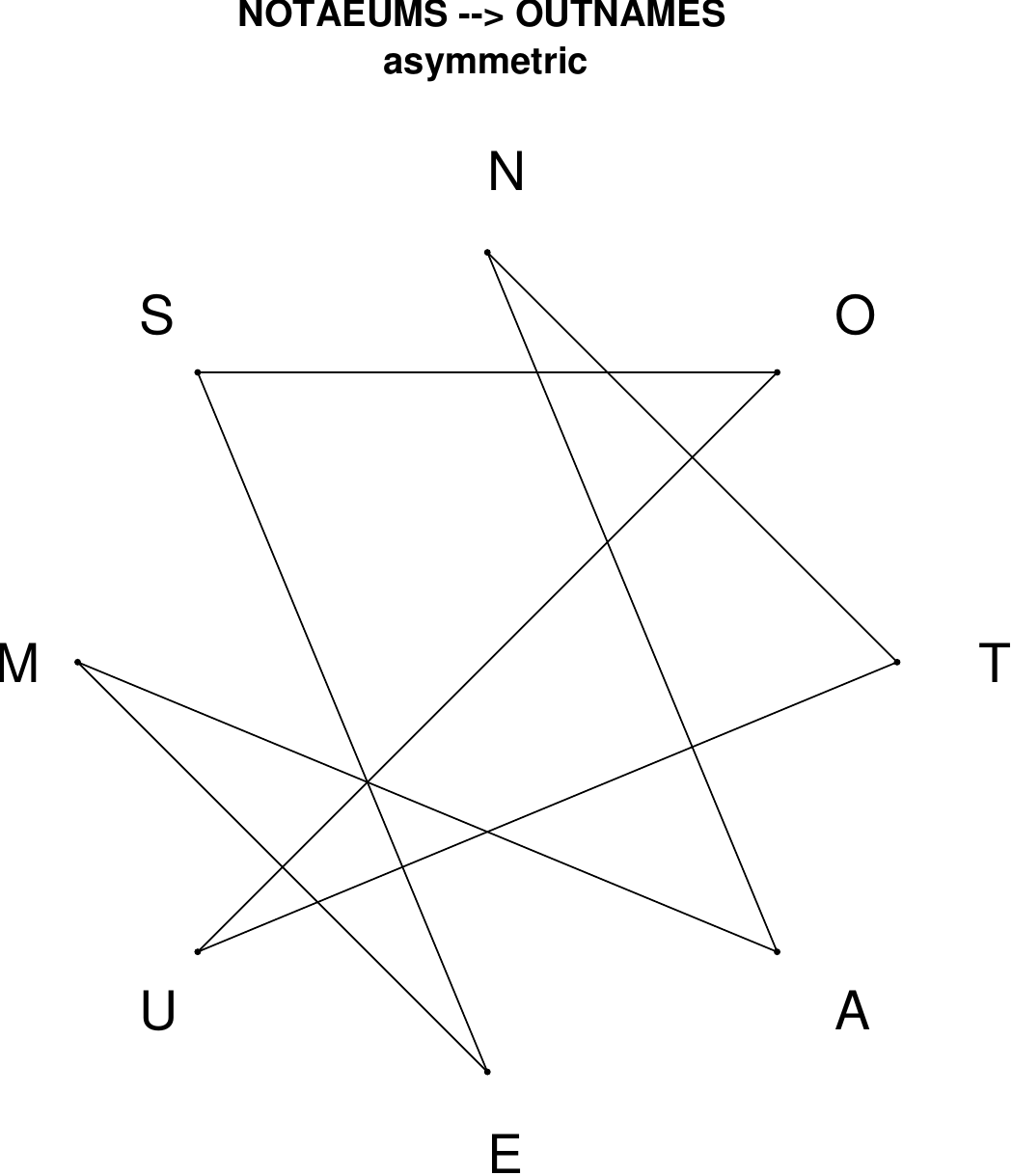}
\end{subfigure}
\end{figure}

\begin{figure}[H]
\centering
\begin{subfigure}[T]{0.19\textwidth}
\centering
\includegraphics[width=\textwidth]{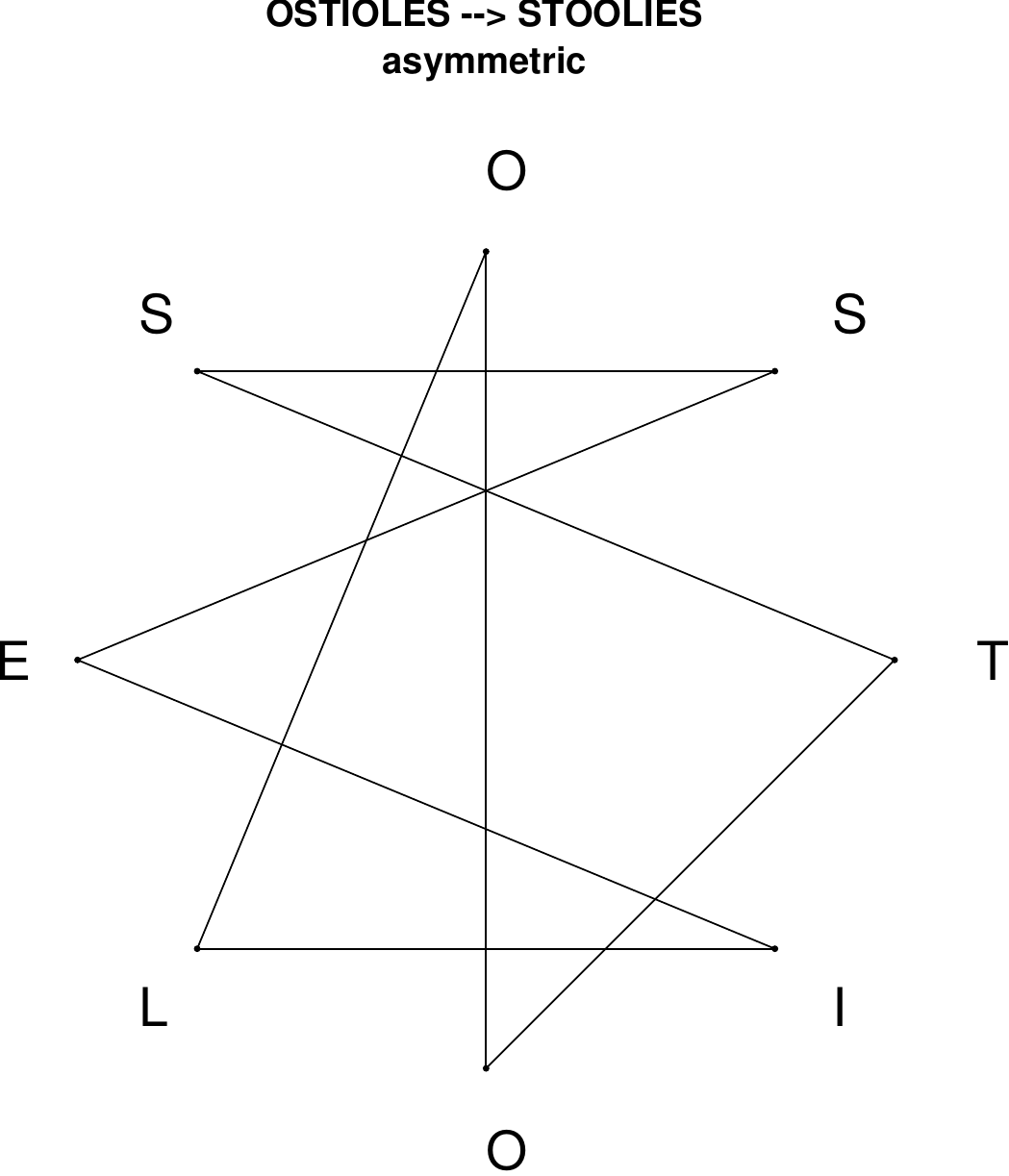}
\end{subfigure}
\hfill
\begin{subfigure}[T]{0.19\textwidth}
\centering
\includegraphics[width=\textwidth]{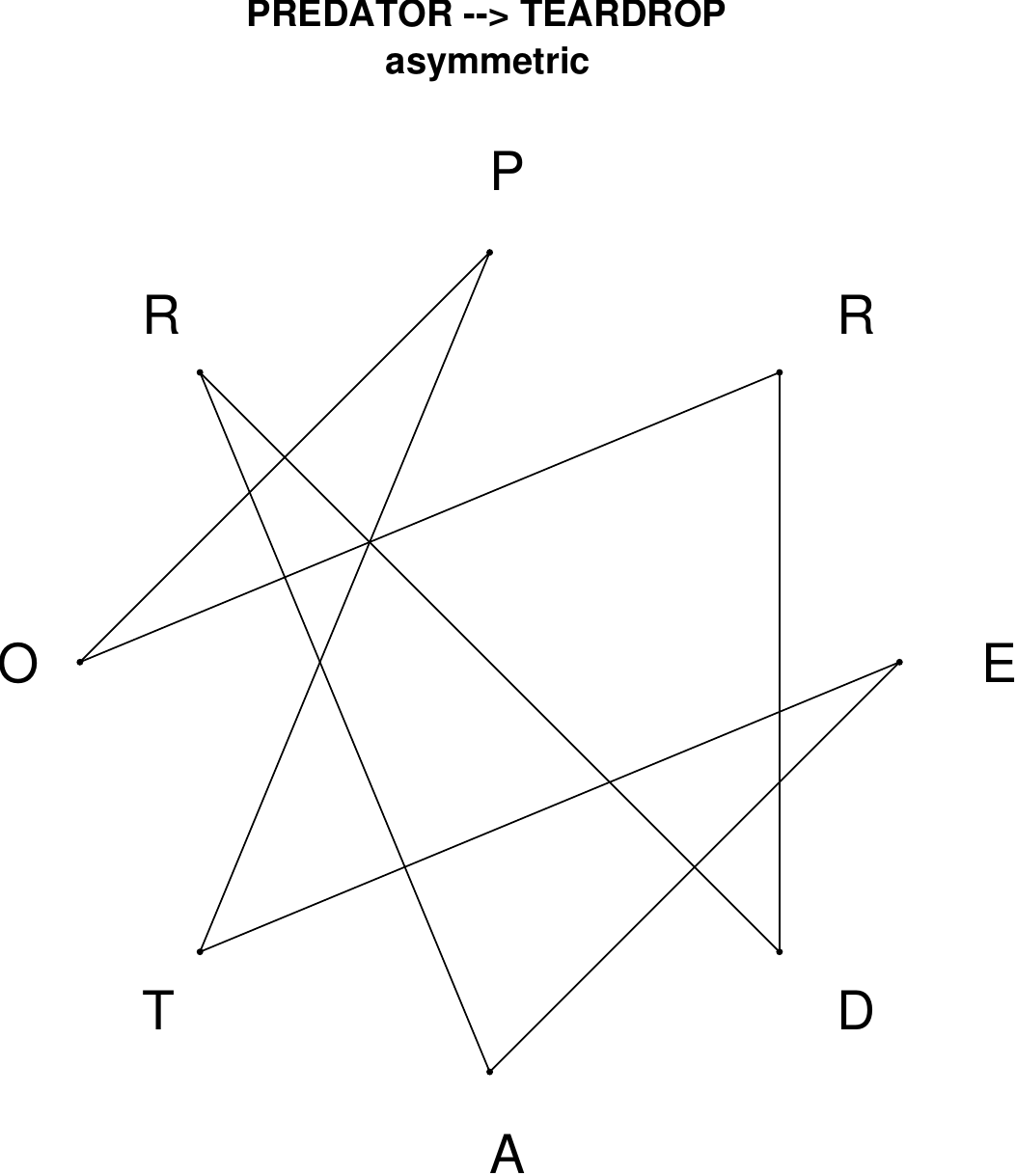}
\end{subfigure}
\hfill
\begin{subfigure}[T]{0.19\textwidth}
\centering
\includegraphics[width=\textwidth]{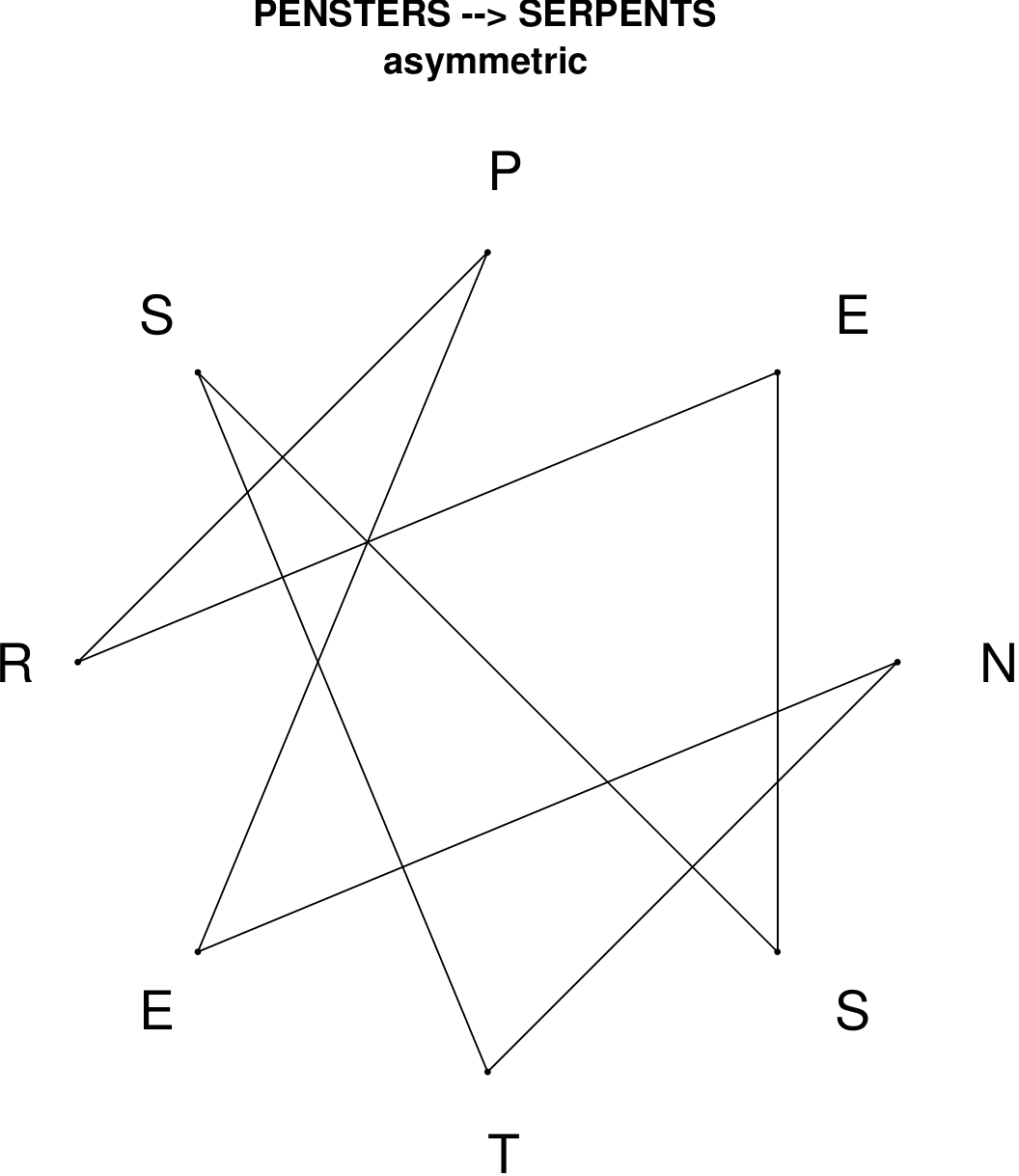}
\end{subfigure}
\hfill
\begin{subfigure}[T]{0.19\textwidth}
\centering
\includegraphics[width=\textwidth]{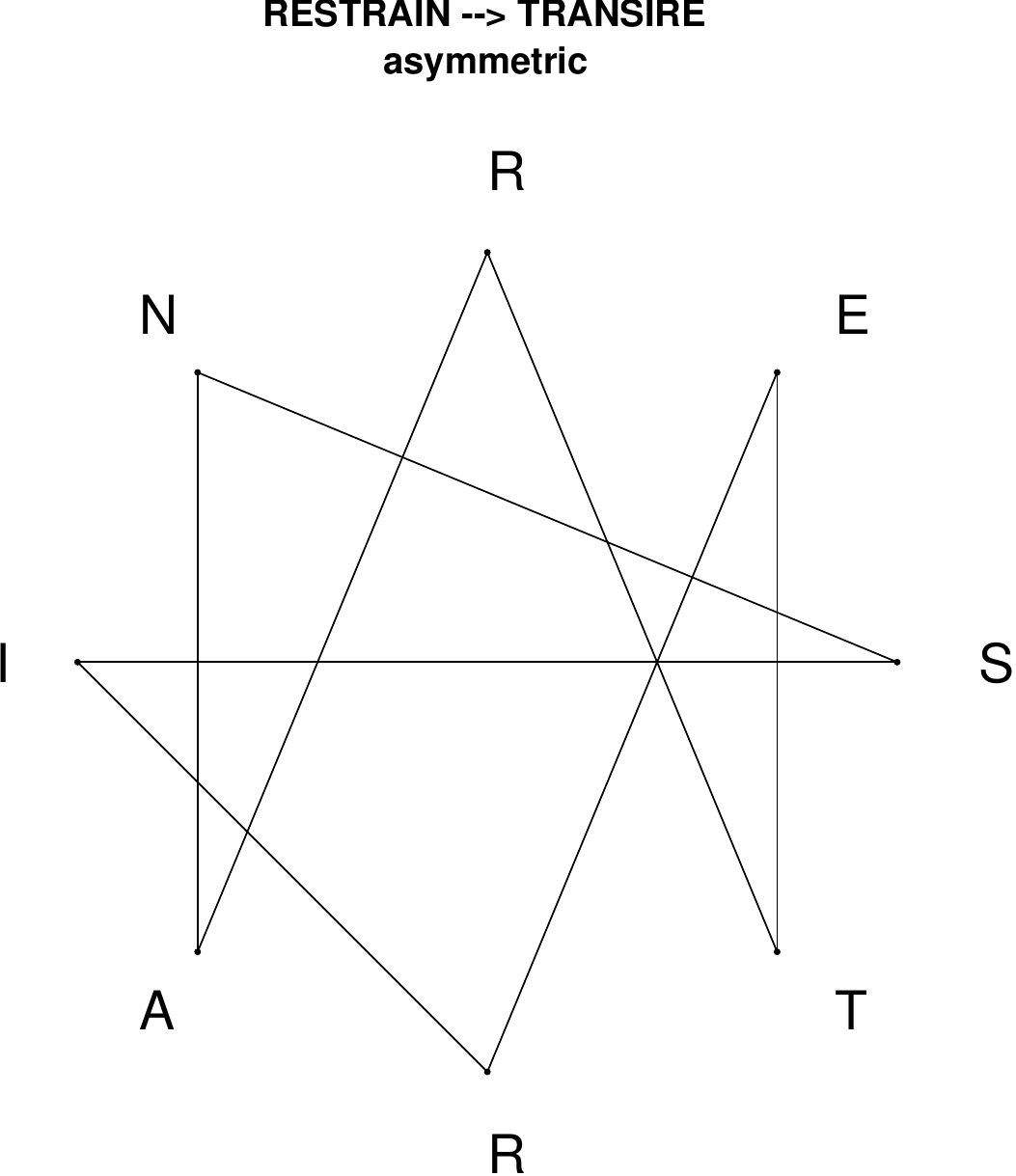}
\end{subfigure}
\hfill
\begin{subfigure}[T]{0.19\textwidth}
\centering
\includegraphics[width=\textwidth]{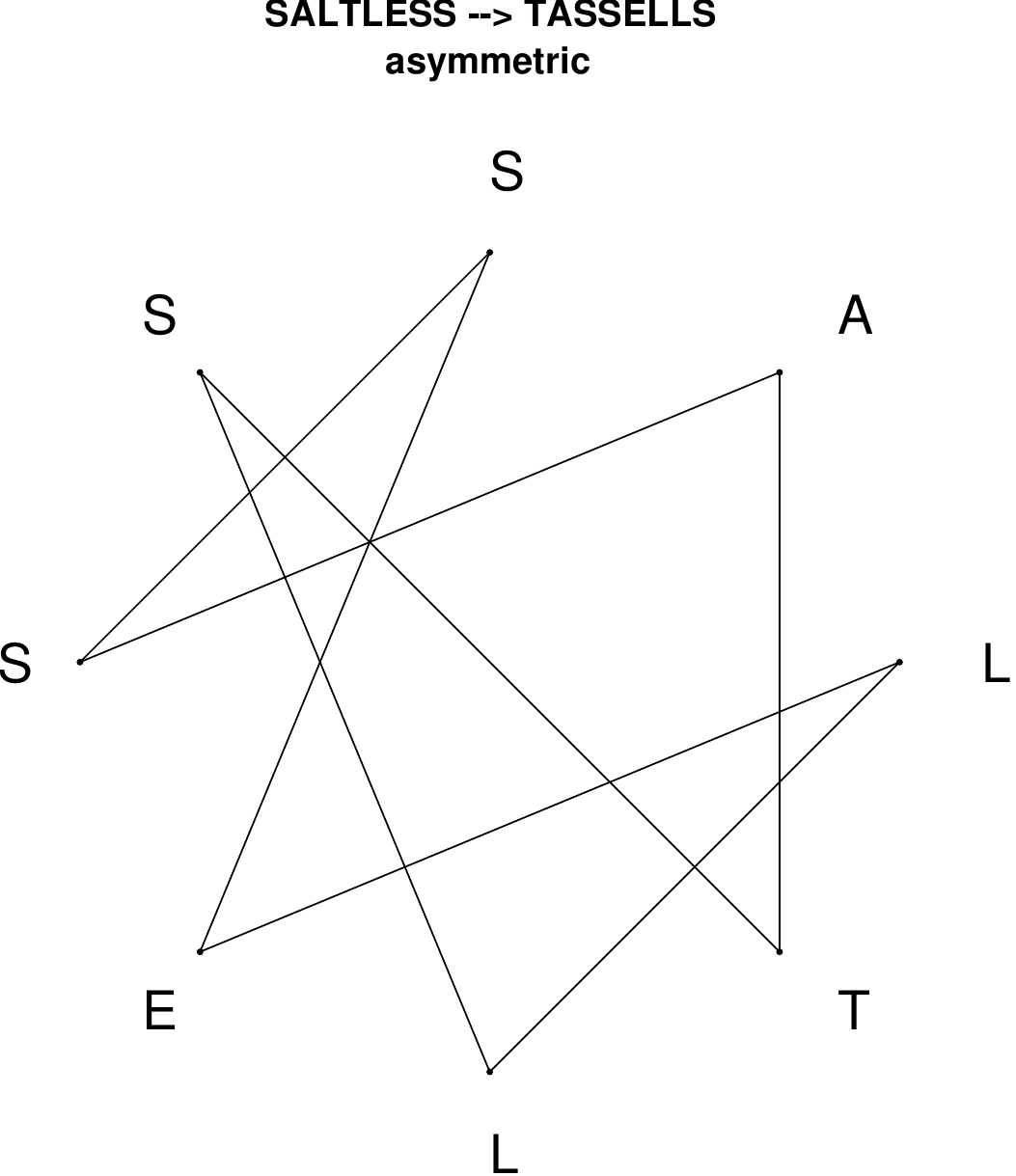}
\end{subfigure}
\end{figure}

\begin{figure}[H]
\centering
\begin{subfigure}[T]{0.19\textwidth}
\centering
\includegraphics[width=\textwidth]{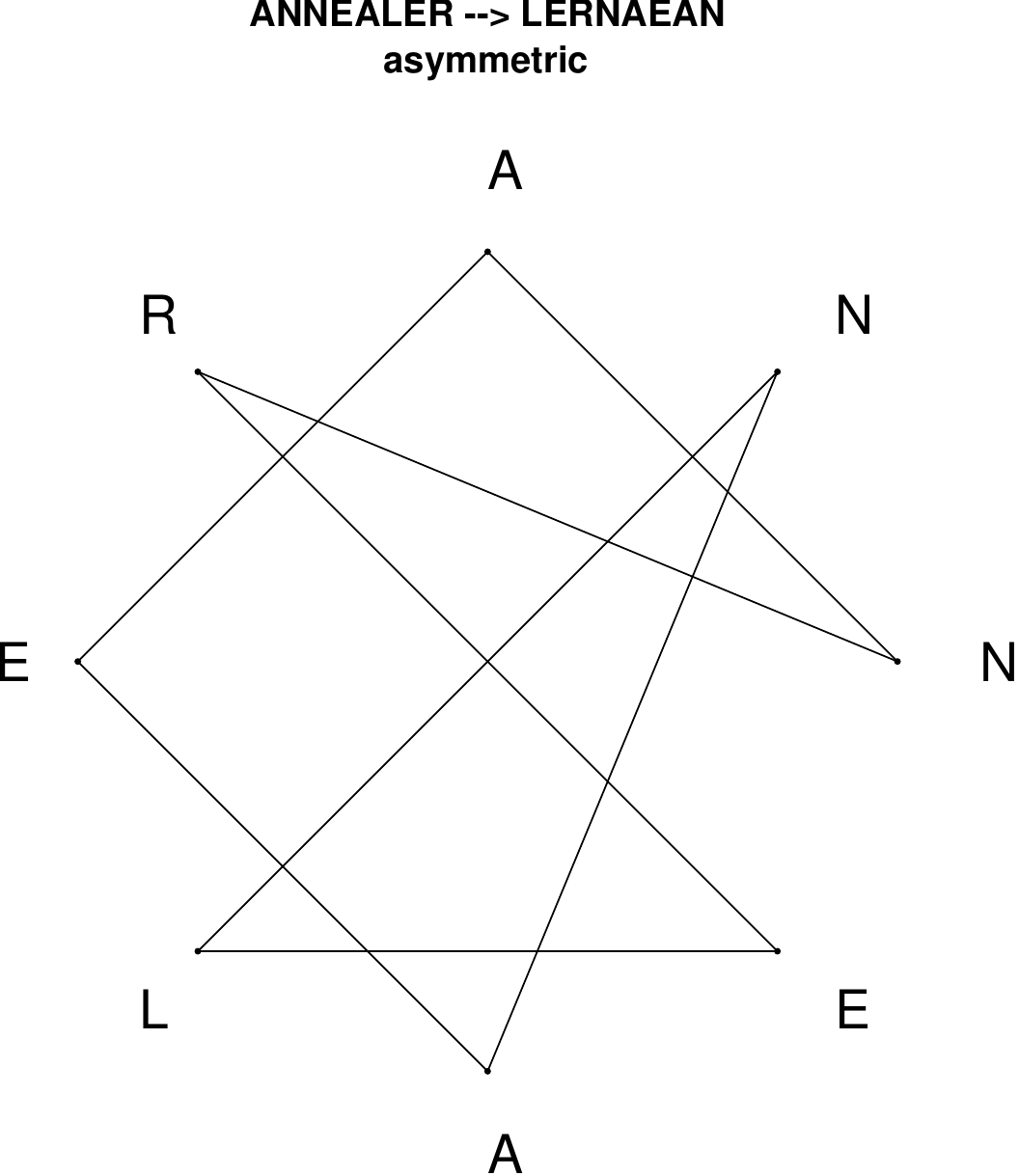}
\end{subfigure}
\hfill
\begin{subfigure}[T]{0.19\textwidth}
\centering
\includegraphics[width=\textwidth]{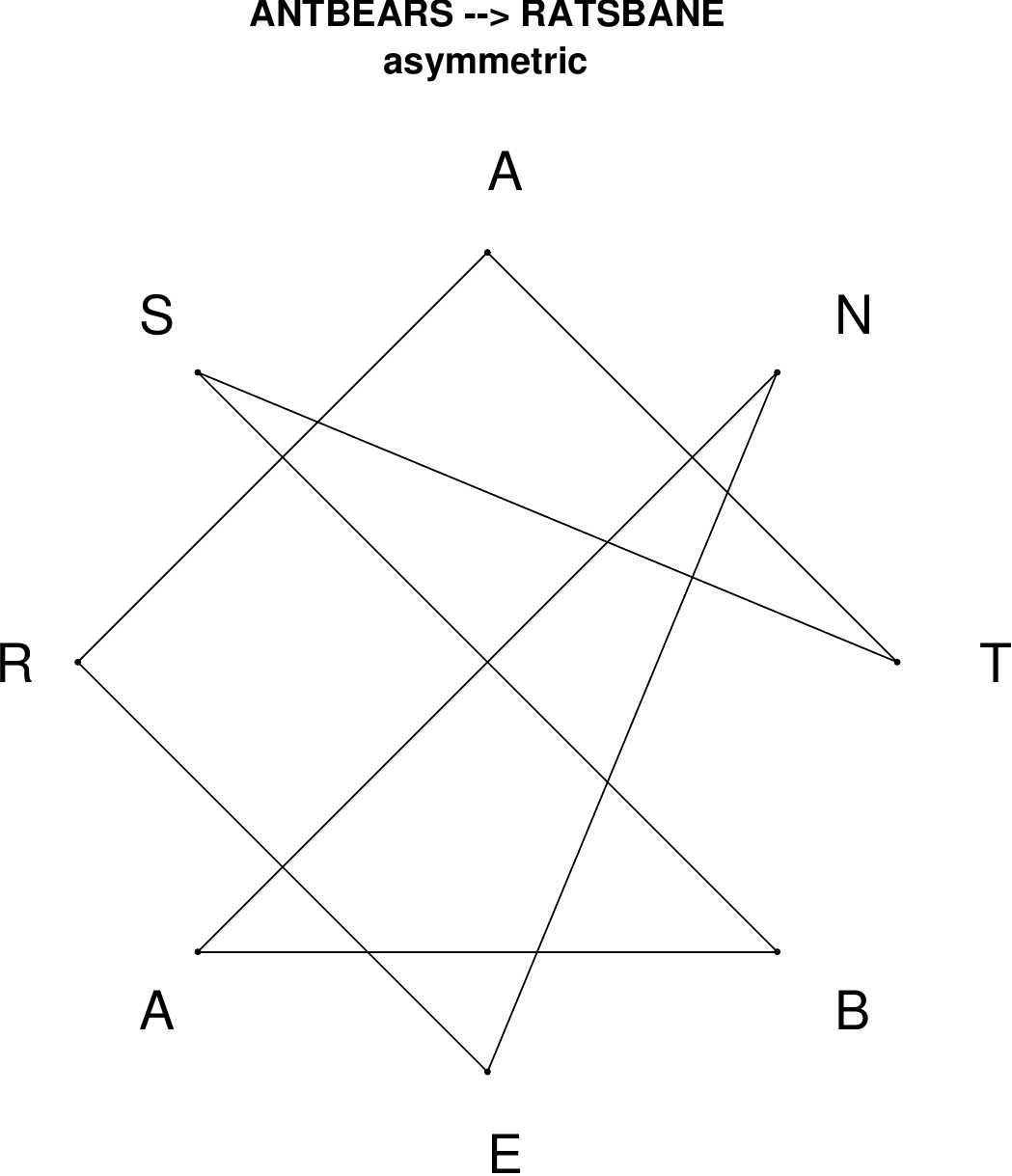}
\end{subfigure}
\hfill
\begin{subfigure}[T]{0.19\textwidth}
\centering
\includegraphics[width=\textwidth]{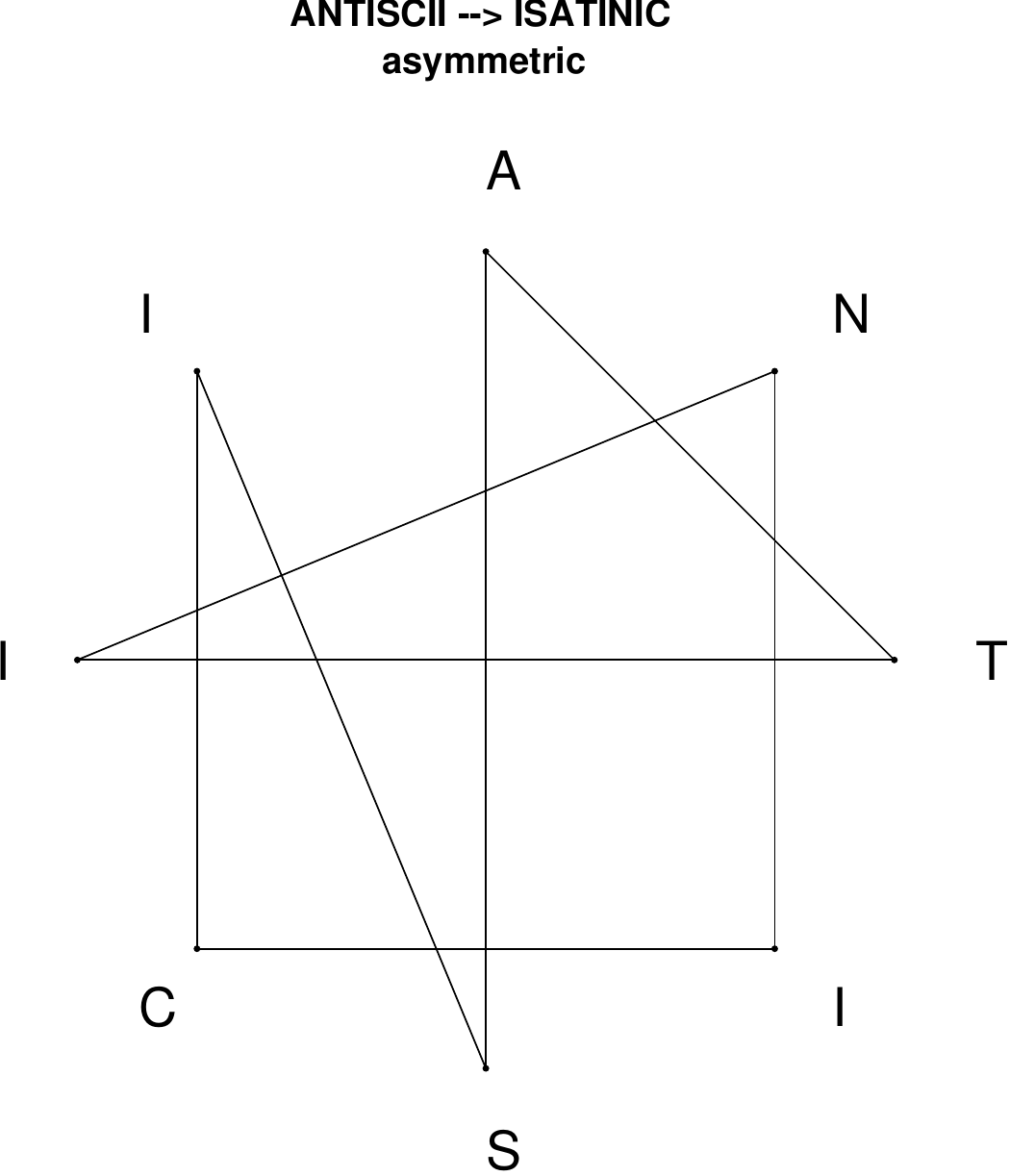}
\end{subfigure}
\hfill
\begin{subfigure}[T]{0.19\textwidth}
\centering
\includegraphics[width=\textwidth]{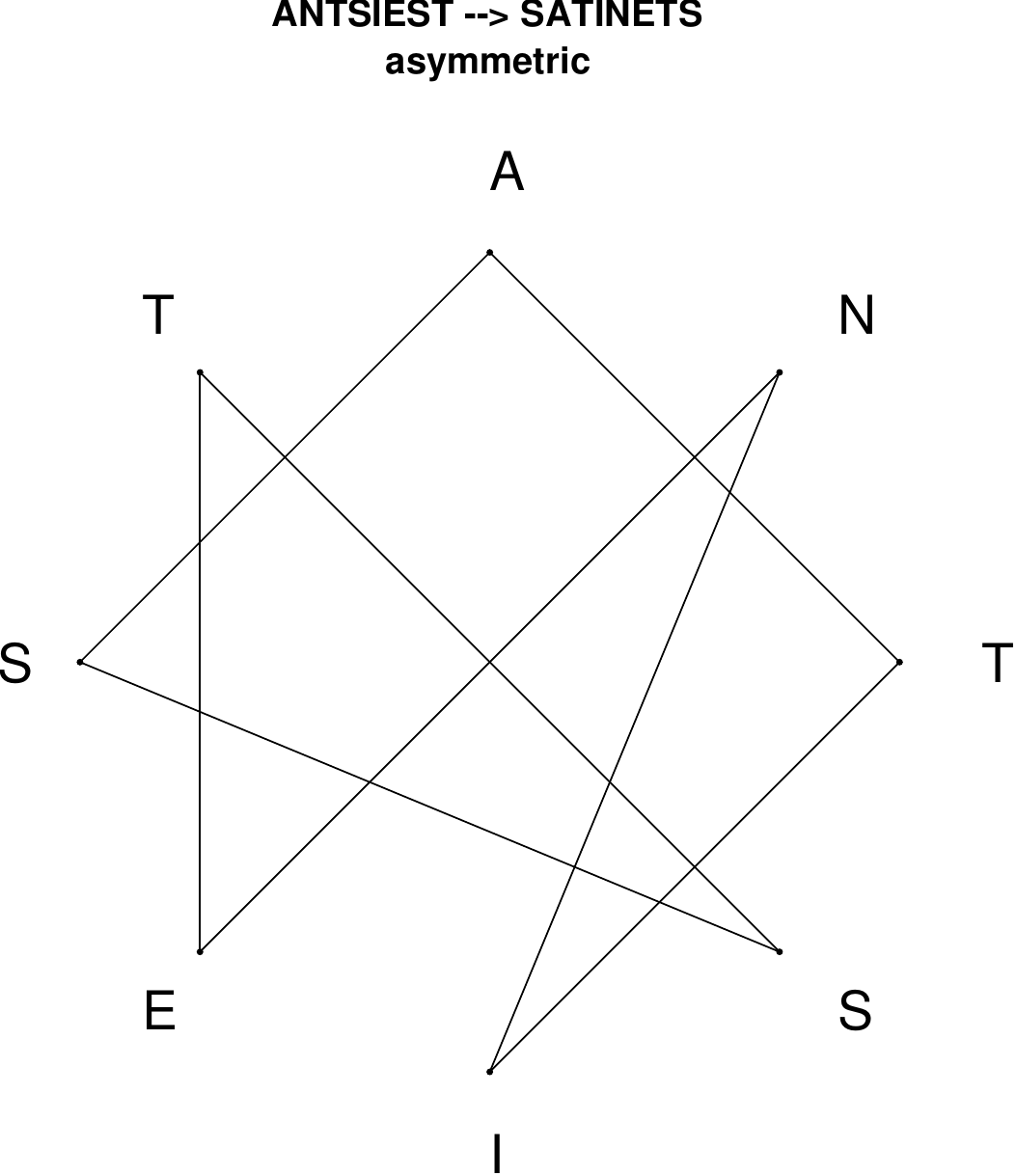}
\end{subfigure}
\hfill
\begin{subfigure}[T]{0.19\textwidth}
\centering
\includegraphics[width=\textwidth]{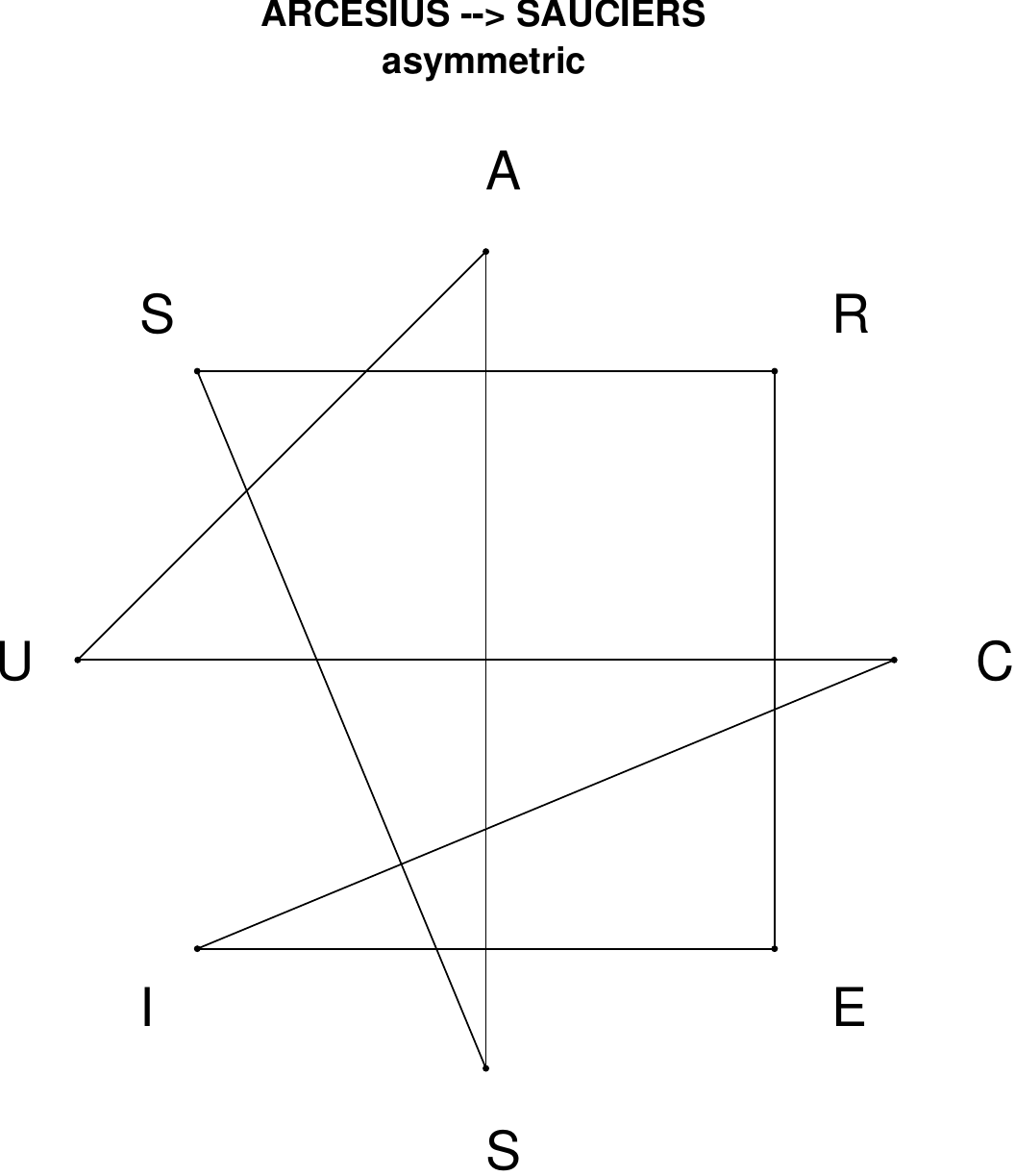}
\end{subfigure}
\end{figure}

\begin{figure}[H]
\centering
\begin{subfigure}[T]{0.19\textwidth}
\centering
\includegraphics[width=\textwidth]{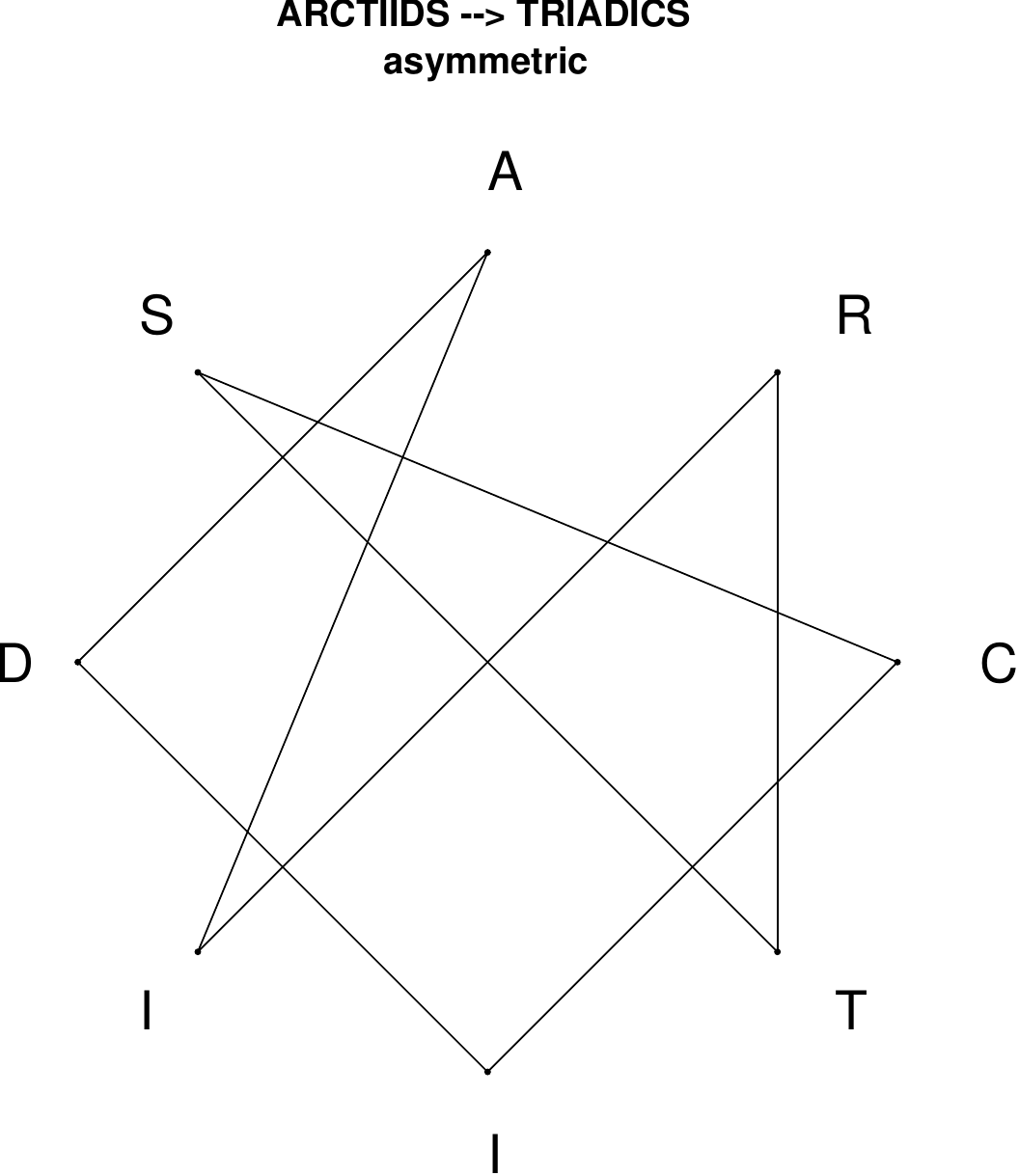}
\end{subfigure}
\hfill
\begin{subfigure}[T]{0.19\textwidth}
\centering
\includegraphics[width=\textwidth]{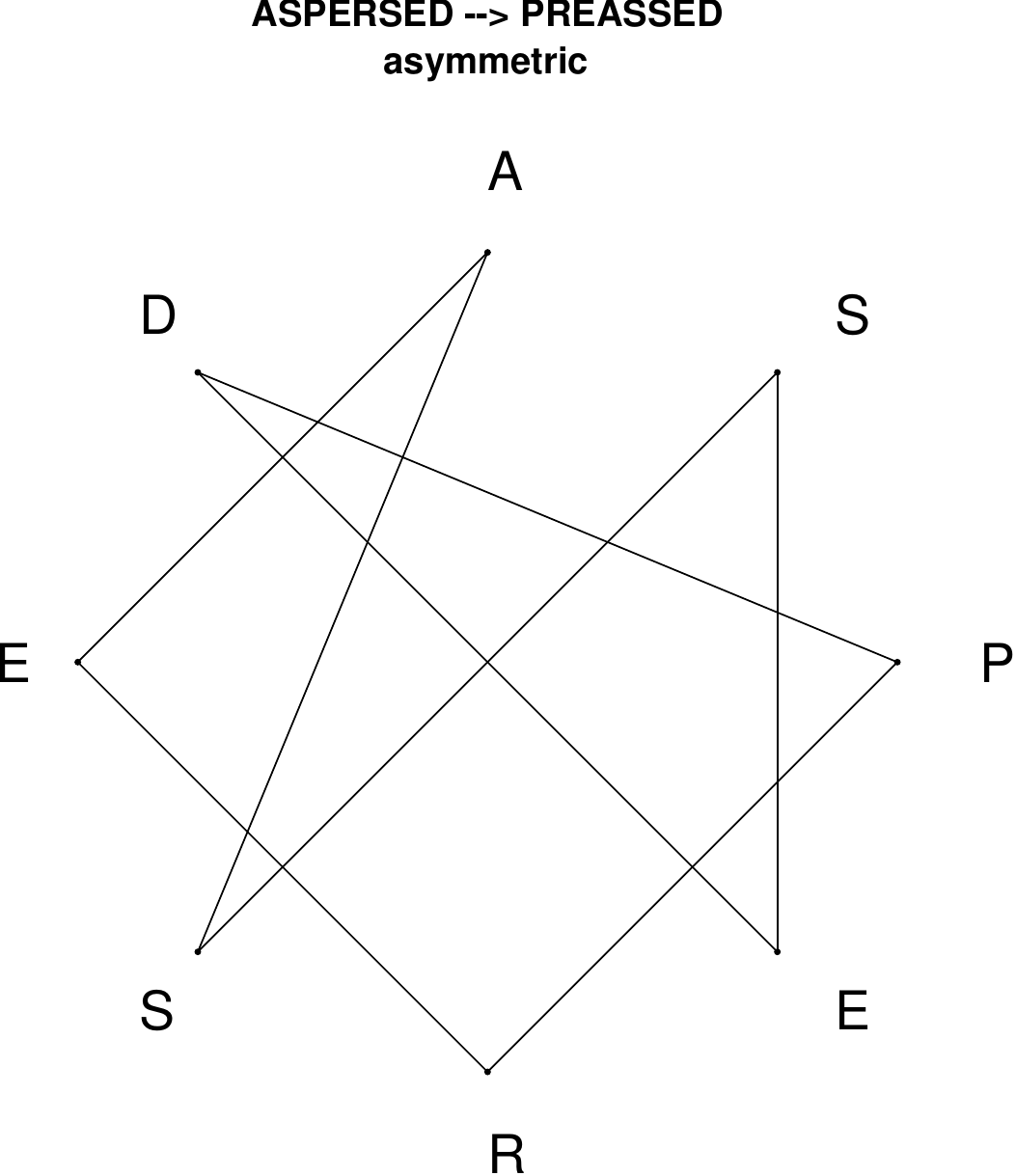}
\end{subfigure}
\hfill
\begin{subfigure}[T]{0.19\textwidth}
\centering
\includegraphics[width=\textwidth]{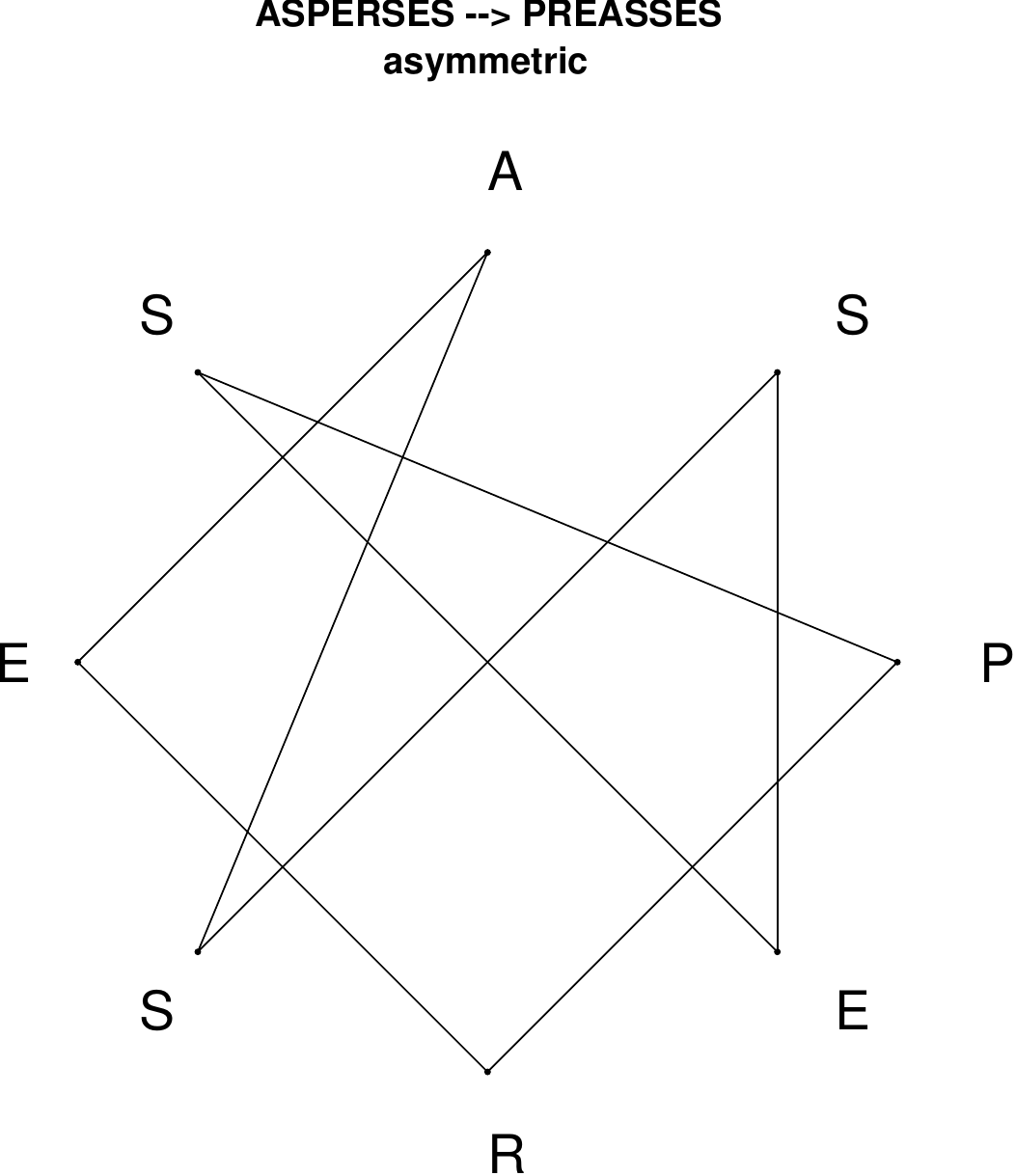}
\end{subfigure}
\hfill
\begin{subfigure}[T]{0.19\textwidth}
\centering
\includegraphics[width=\textwidth]{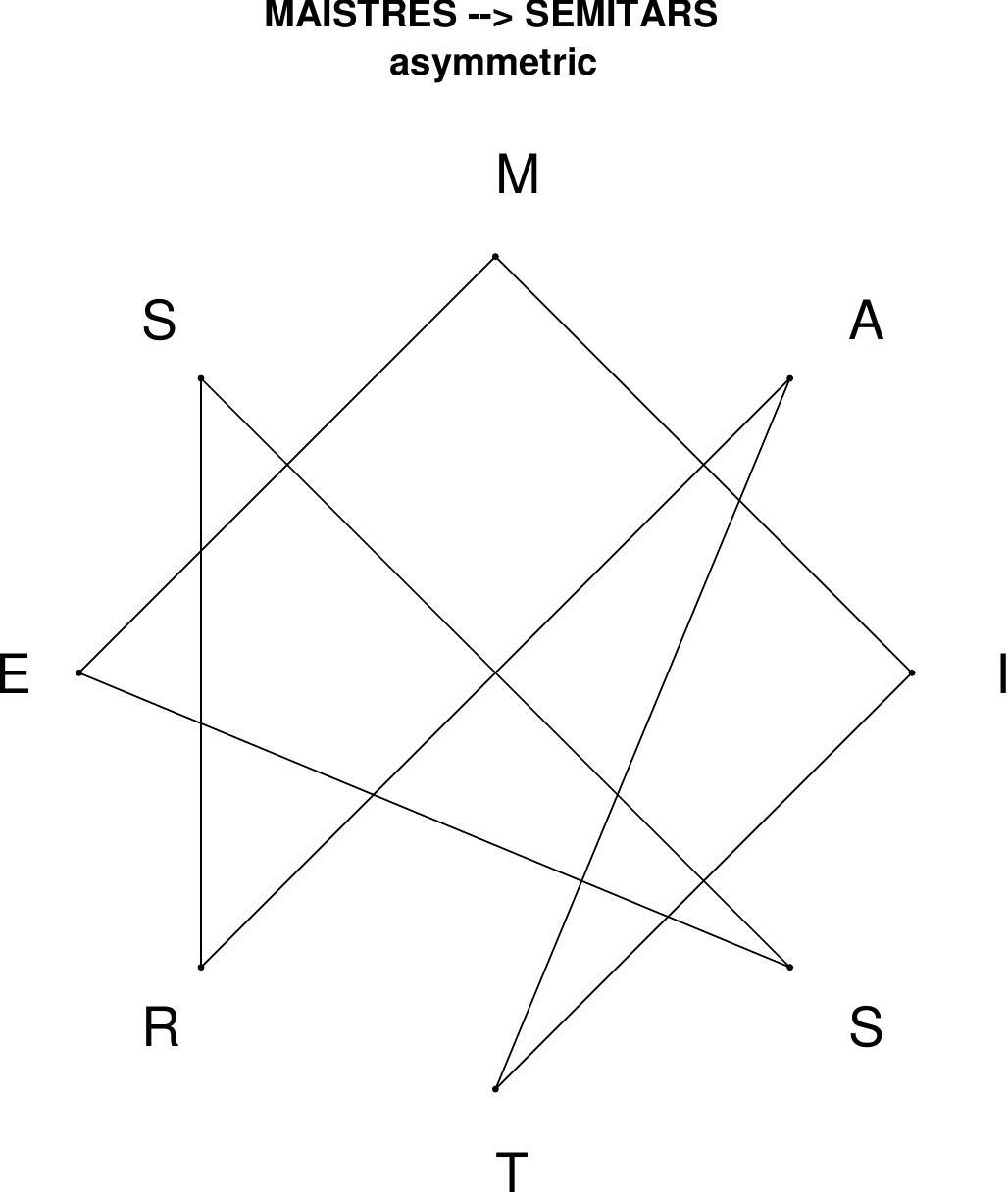}
\end{subfigure}
\hfill
\begin{subfigure}[T]{0.19\textwidth}
\centering
\includegraphics[width=\textwidth]{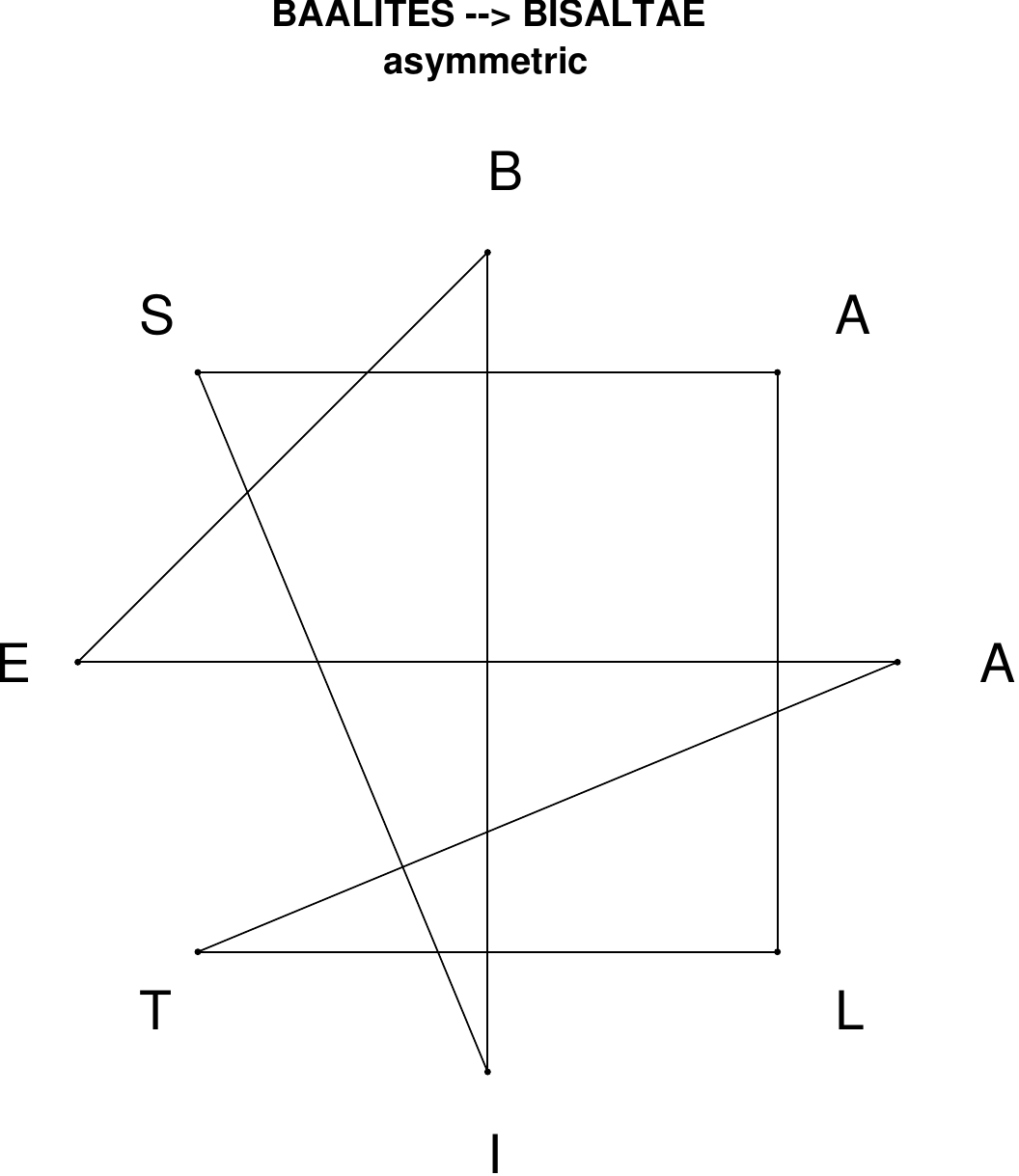}
\end{subfigure}
\end{figure}

\begin{figure}[H]
\centering
\begin{subfigure}[T]{0.19\textwidth}
\centering
\includegraphics[width=\textwidth]{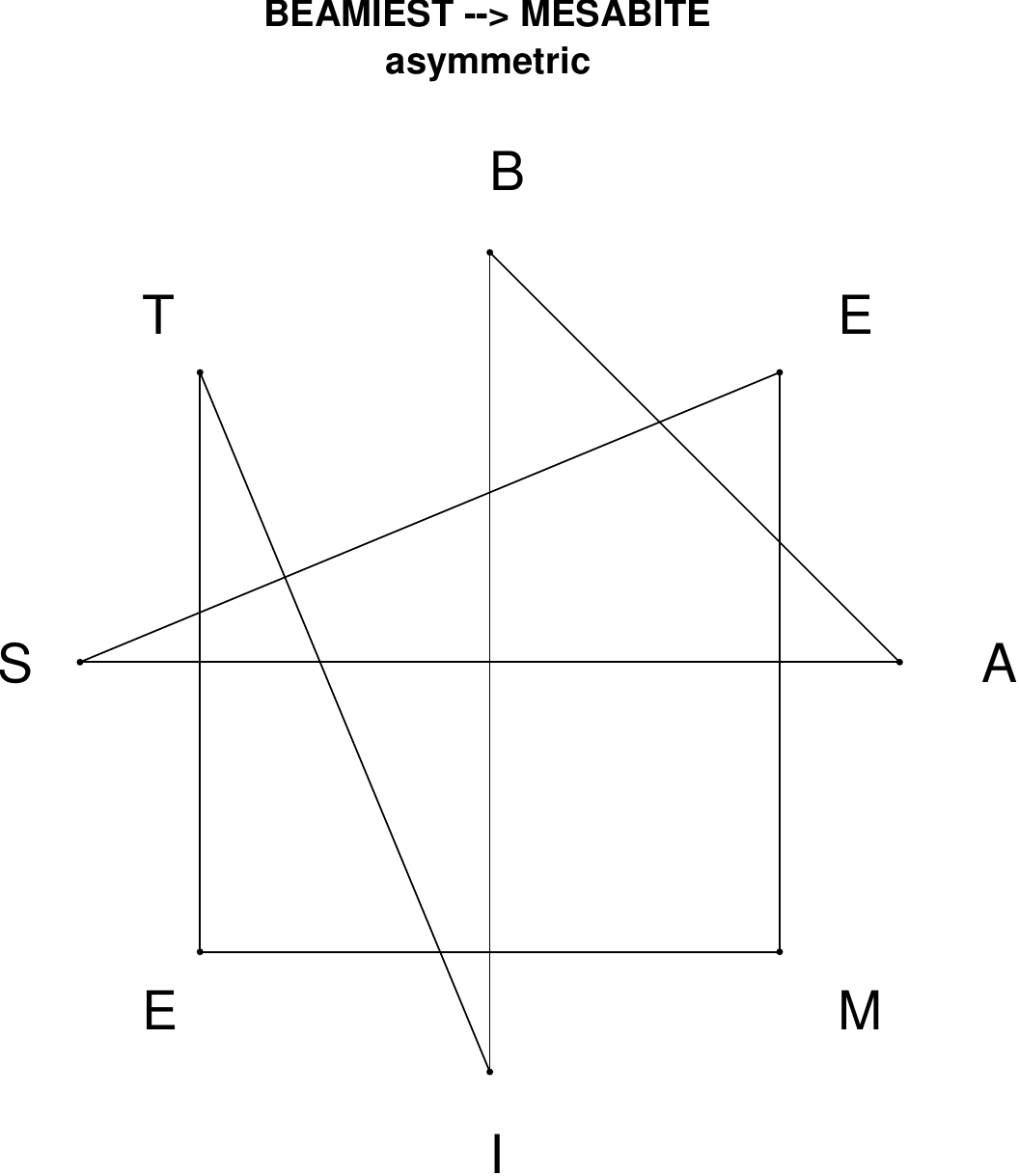}
\end{subfigure}
\hfill
\begin{subfigure}[T]{0.19\textwidth}
\centering
\includegraphics[width=\textwidth]{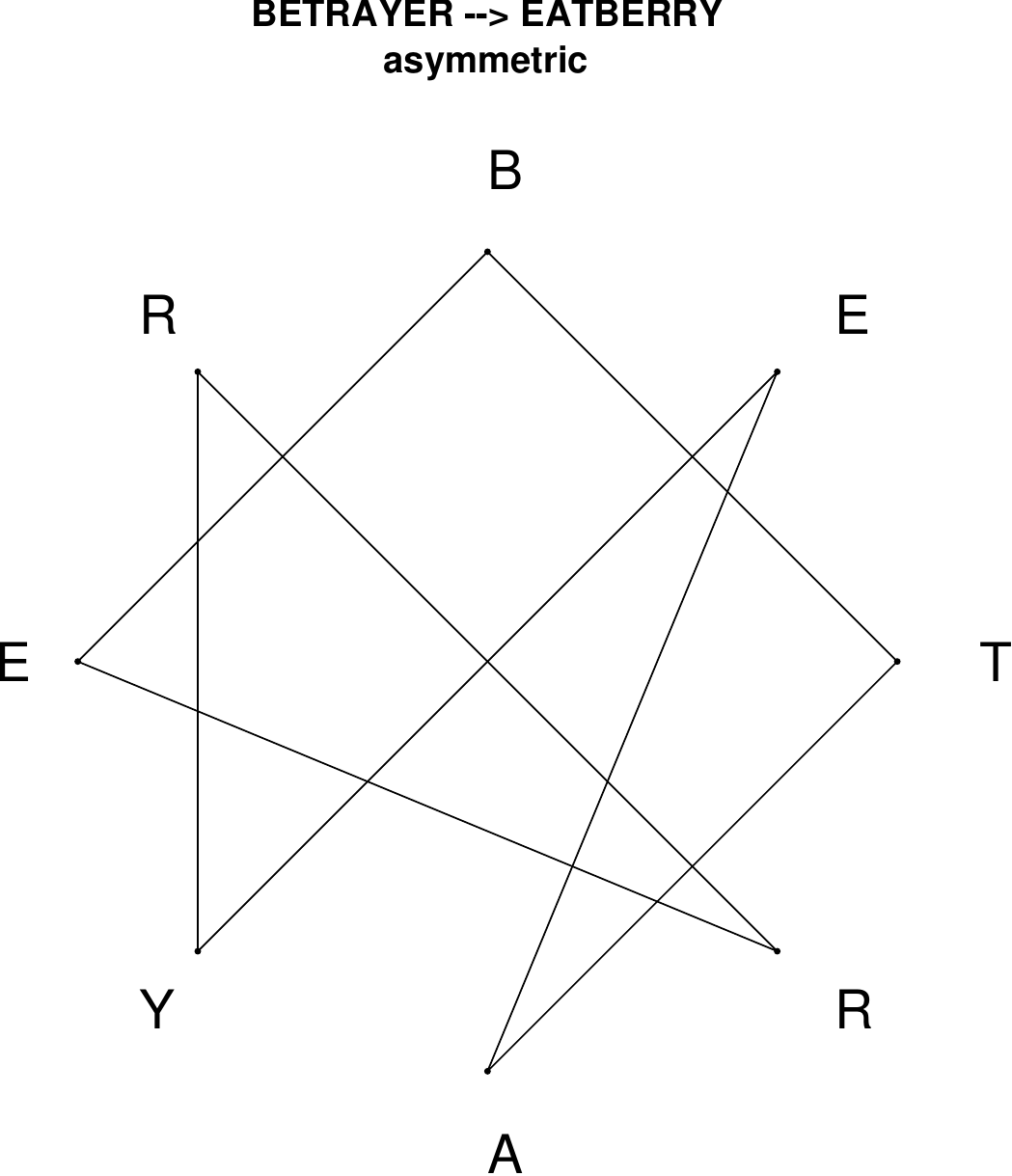}
\end{subfigure}
\hfill
\begin{subfigure}[T]{0.19\textwidth}
\centering
\includegraphics[width=\textwidth]{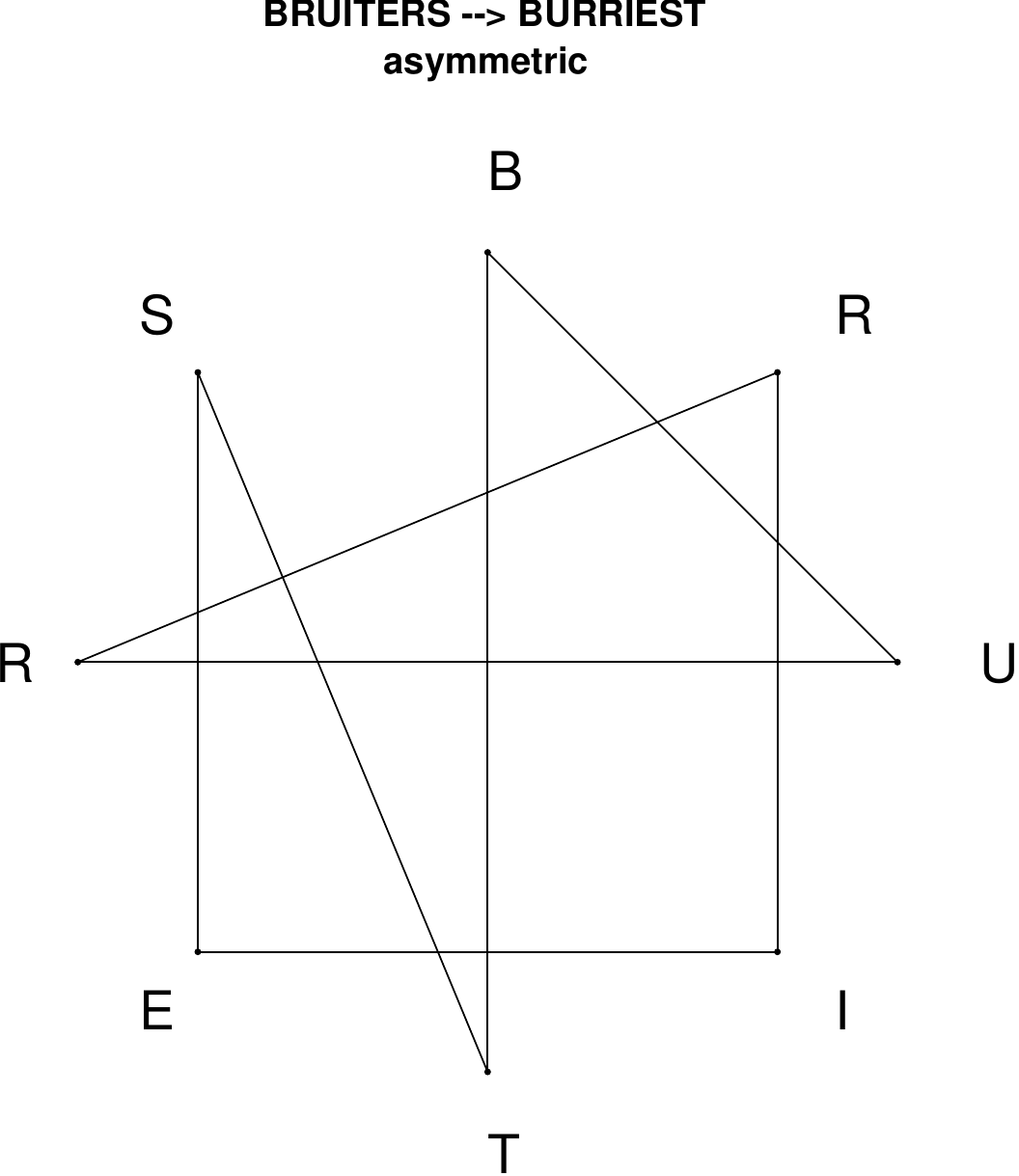}
\end{subfigure}
\hfill
\begin{subfigure}[T]{0.19\textwidth}
\centering
\includegraphics[width=\textwidth]{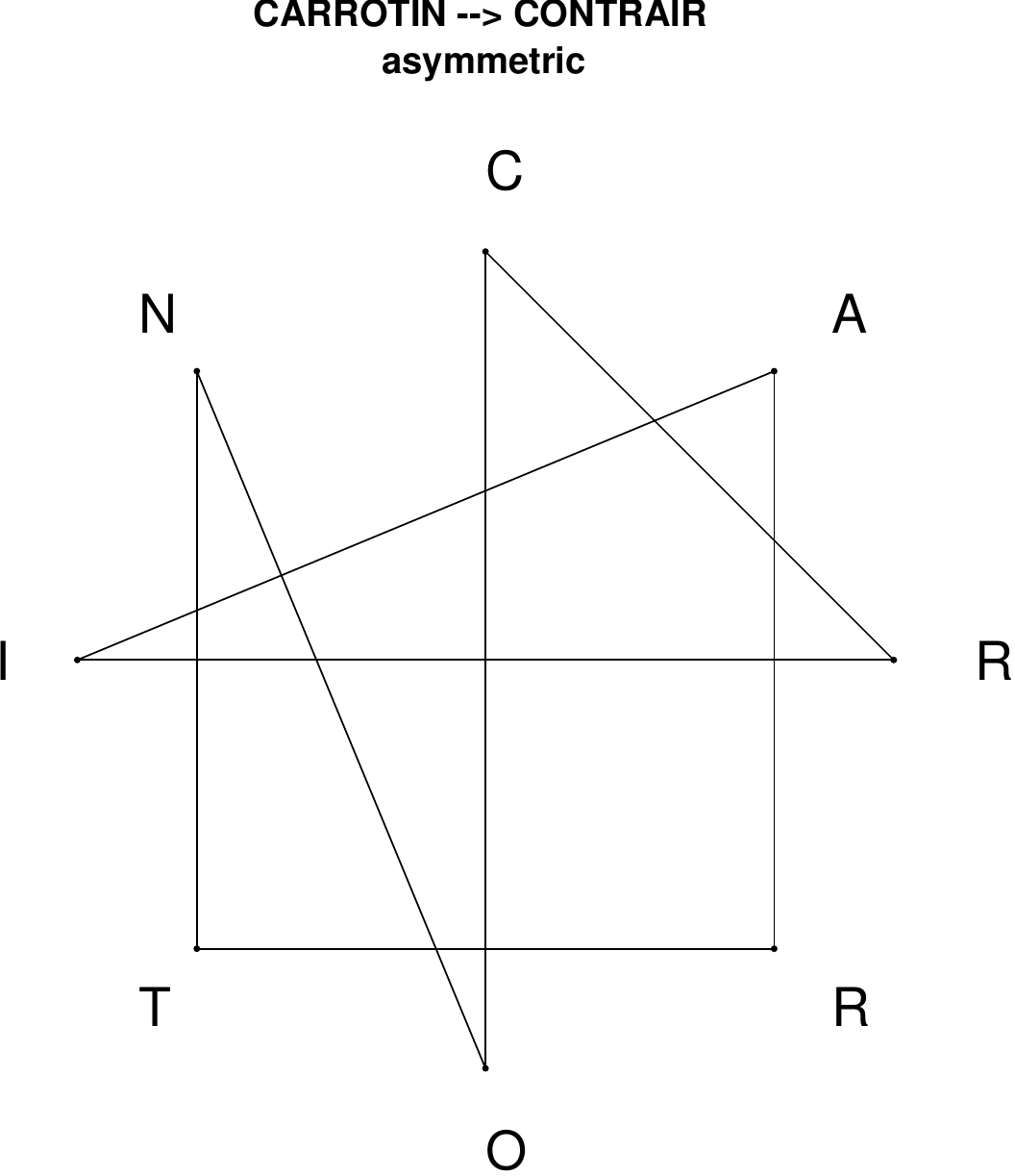}
\end{subfigure}
\hfill
\begin{subfigure}[T]{0.19\textwidth}
\centering
\includegraphics[width=\textwidth]{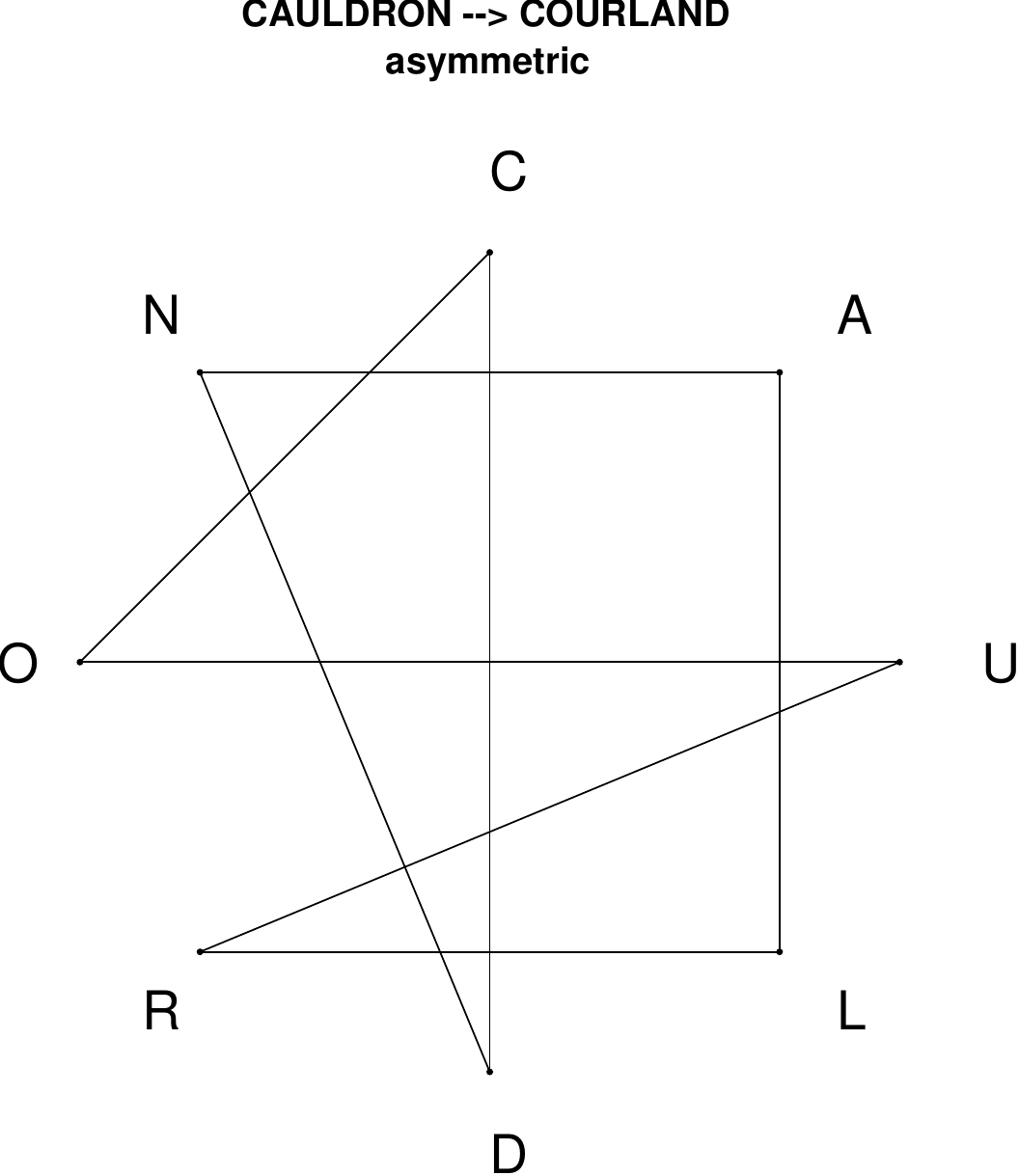}
\end{subfigure}
\end{figure}

\begin{figure}[H]
\centering
\begin{subfigure}[T]{0.19\textwidth}
\centering
\includegraphics[width=\textwidth]{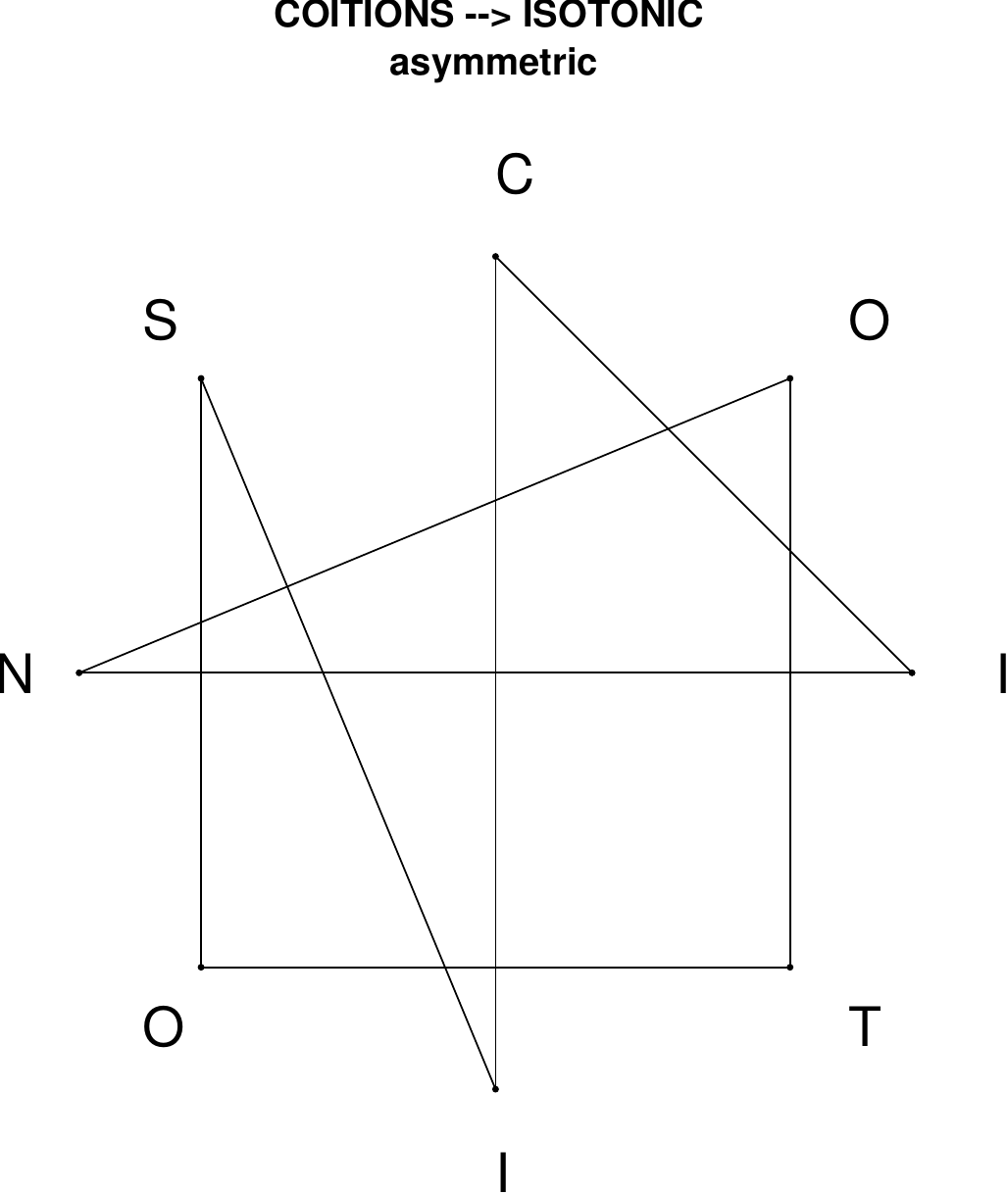}
\end{subfigure}
\hfill
\begin{subfigure}[T]{0.19\textwidth}
\centering
\includegraphics[width=\textwidth]{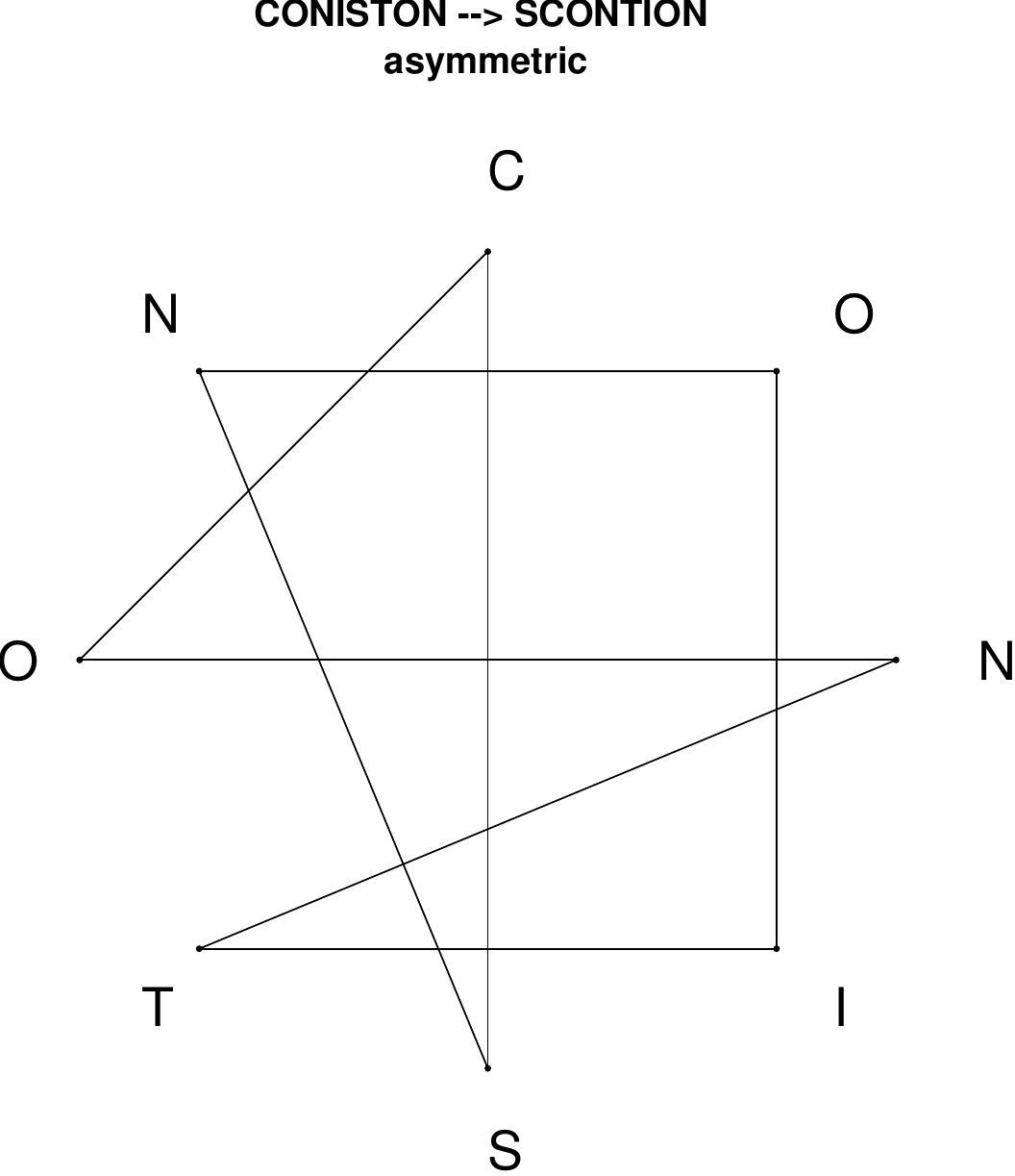}
\end{subfigure}
\hfill
\begin{subfigure}[T]{0.19\textwidth}
\centering
\includegraphics[width=\textwidth]{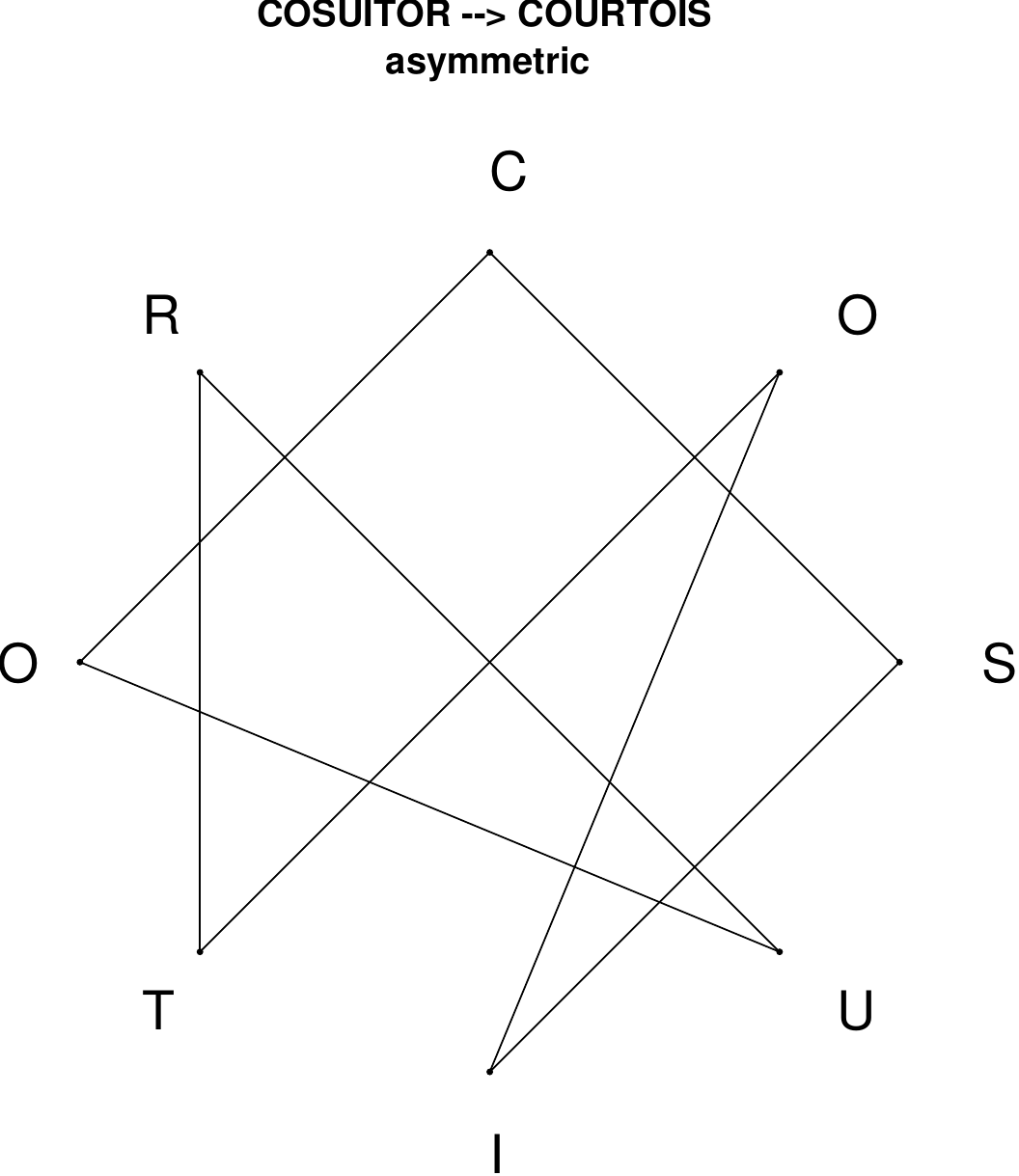}
\end{subfigure}
\hfill
\begin{subfigure}[T]{0.19\textwidth}
\centering
\includegraphics[width=\textwidth]{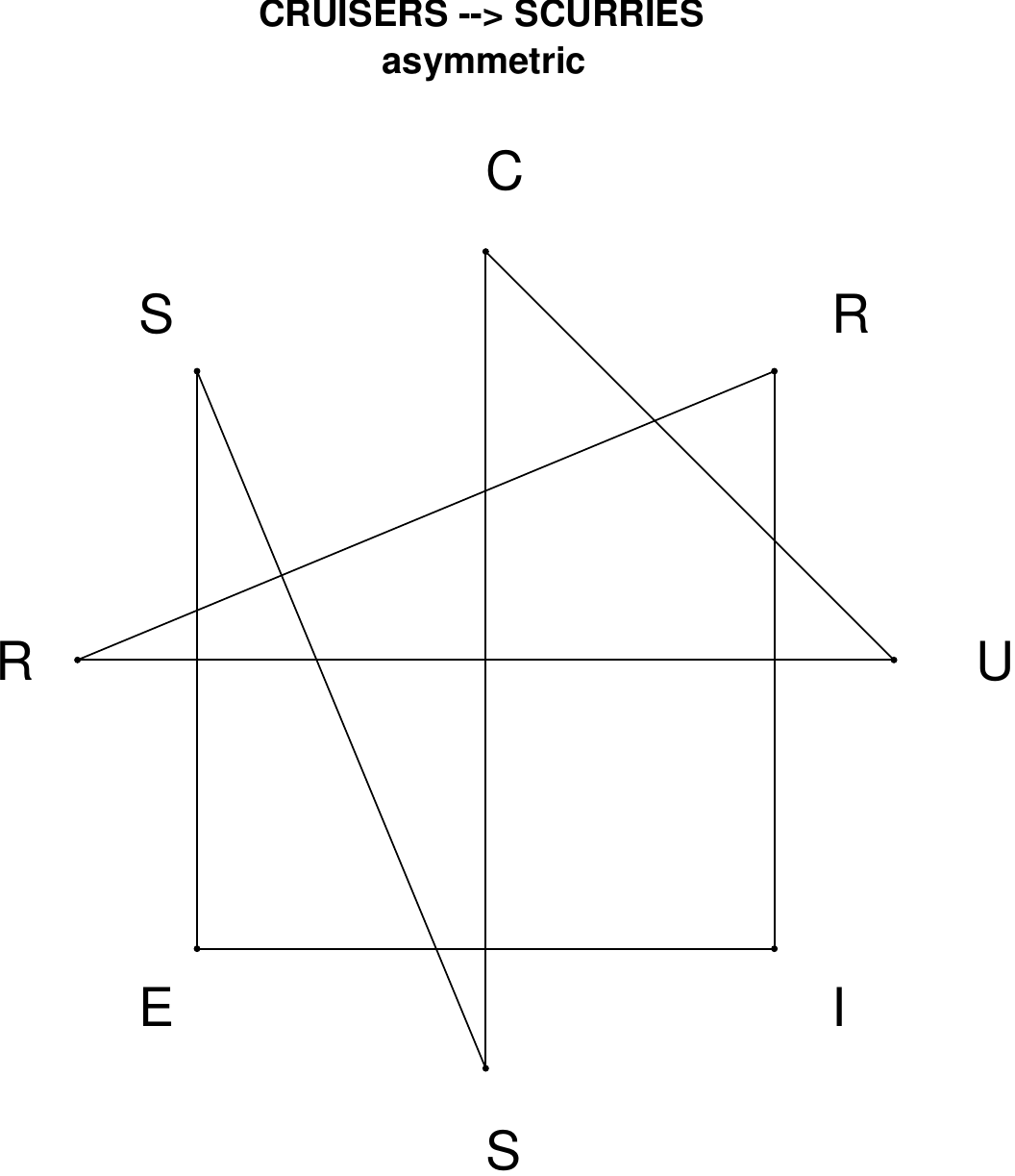}
\end{subfigure}
\hfill
\begin{subfigure}[T]{0.19\textwidth}
\centering
\includegraphics[width=\textwidth]{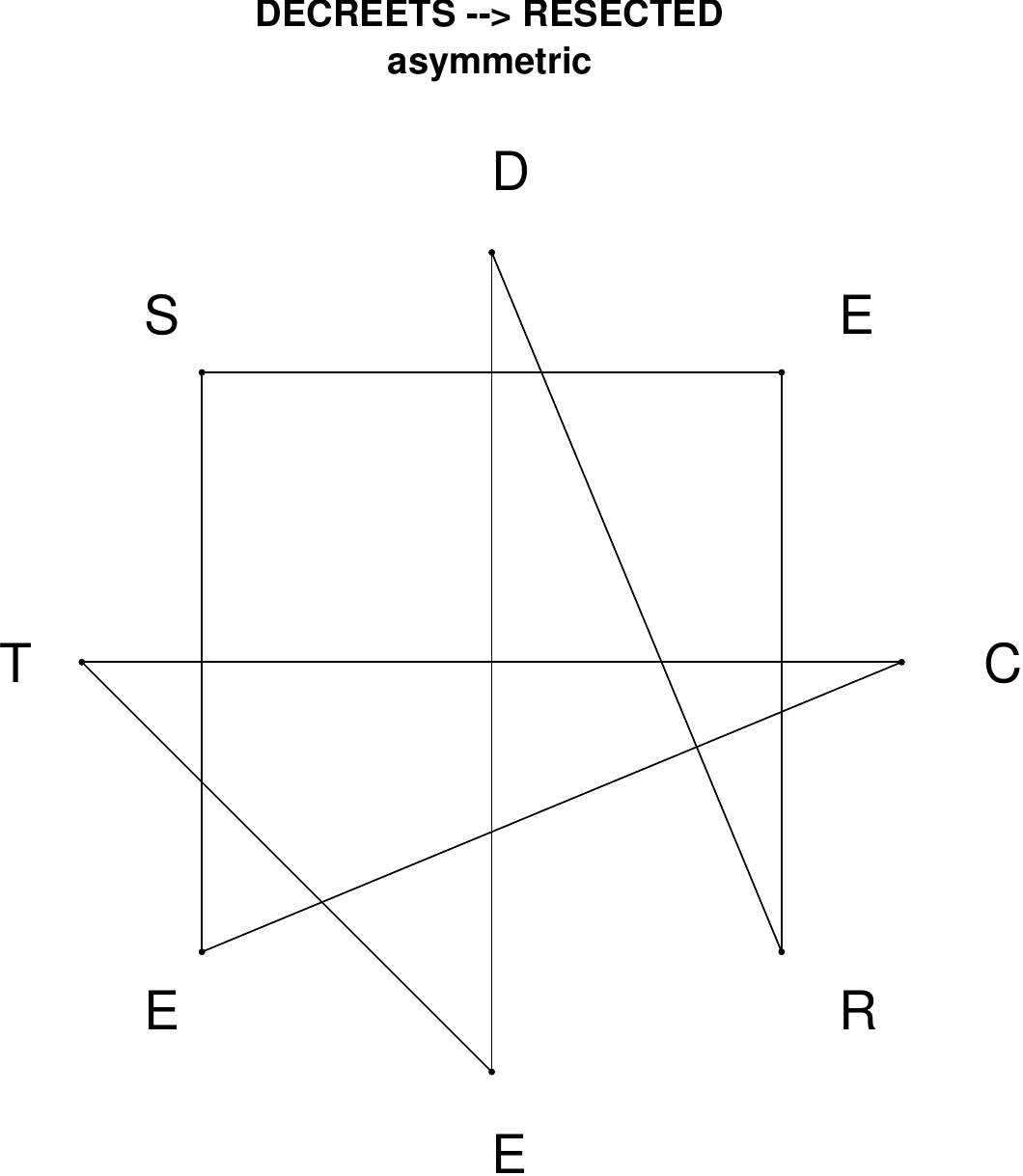}
\end{subfigure}
\end{figure}

\begin{figure}[H]
\centering
\begin{subfigure}[T]{0.19\textwidth}
\centering
\includegraphics[width=\textwidth]{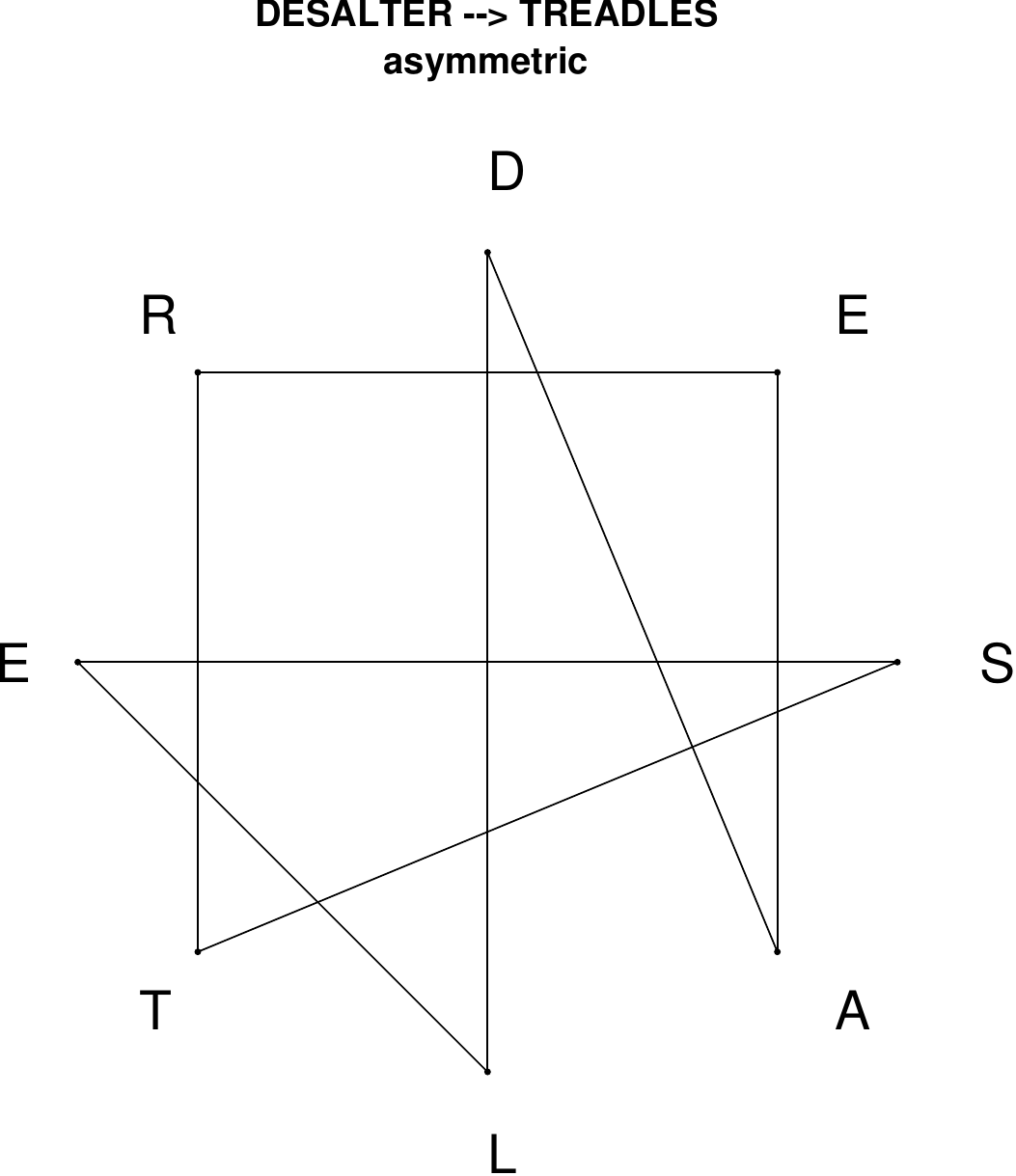}
\end{subfigure}
\hfill
\begin{subfigure}[T]{0.19\textwidth}
\centering
\includegraphics[width=\textwidth]{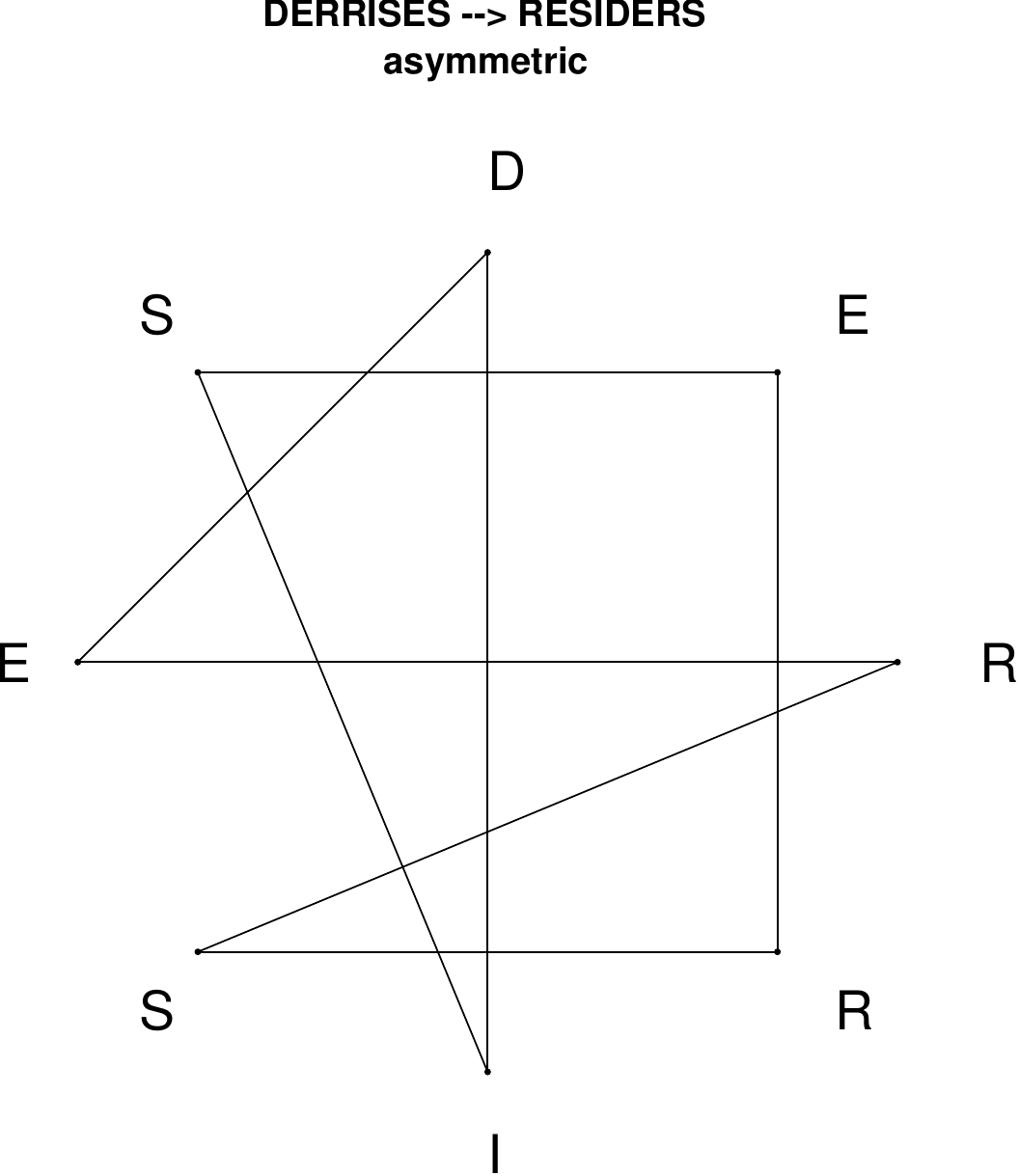}
\end{subfigure}
\hfill
\begin{subfigure}[T]{0.19\textwidth}
\centering
\includegraphics[width=\textwidth]{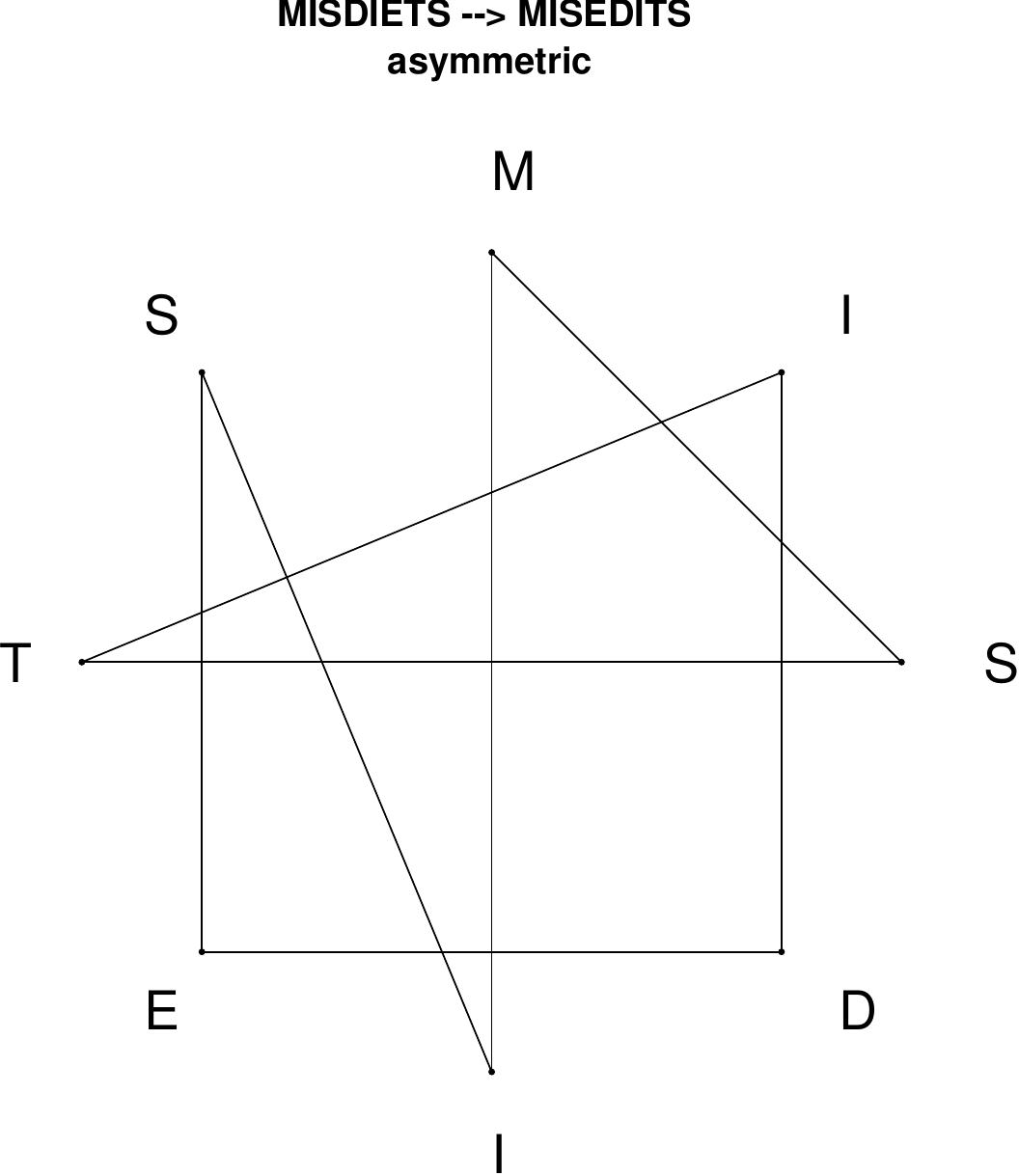}
\end{subfigure}
\hfill
\begin{subfigure}[T]{0.19\textwidth}
\centering
\includegraphics[width=\textwidth]{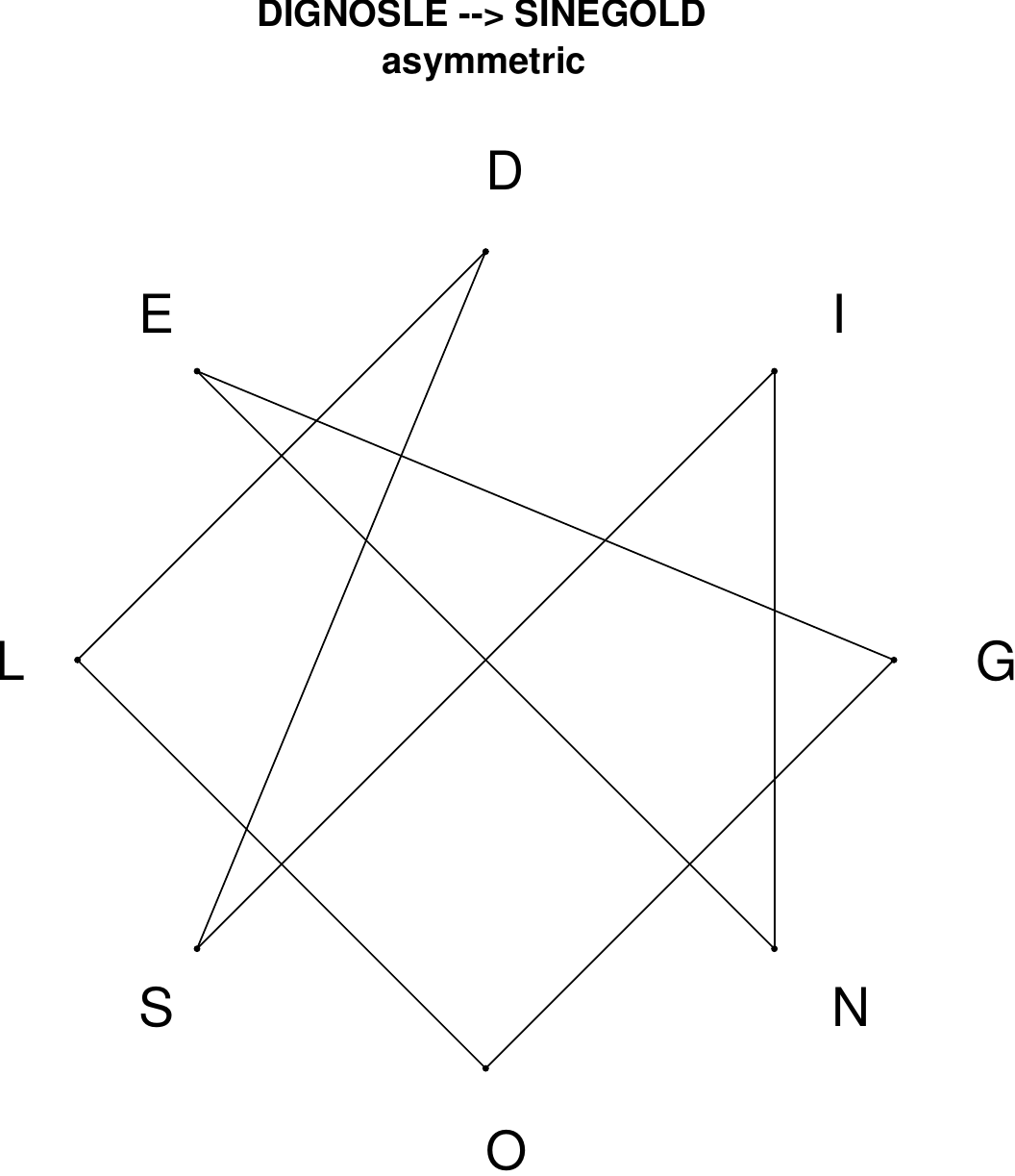}
\end{subfigure}
\hfill
\begin{subfigure}[T]{0.19\textwidth}
\centering
\includegraphics[width=\textwidth]{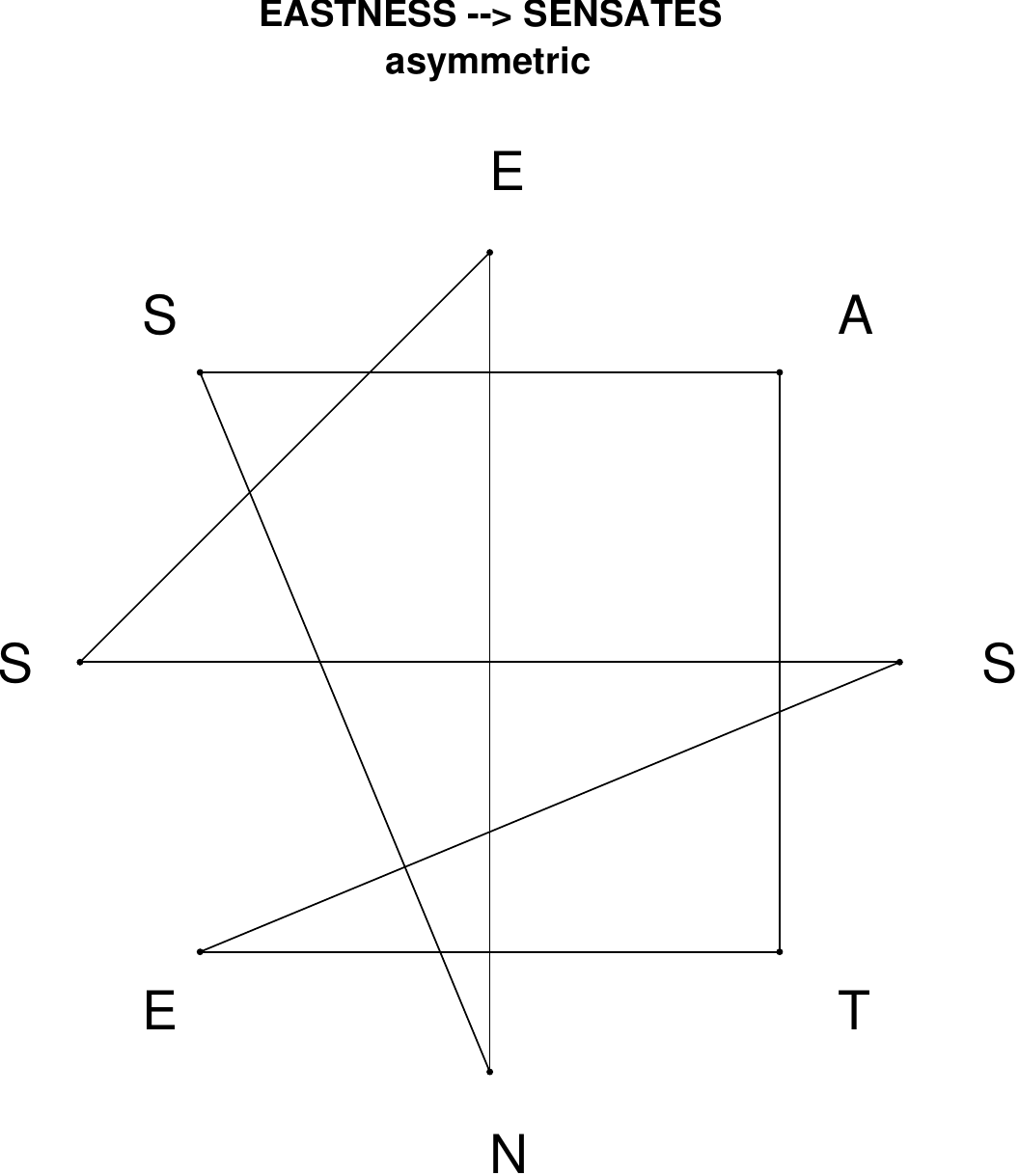}
\end{subfigure}
\end{figure}

\begin{figure}[H]
\centering
\begin{subfigure}[T]{0.19\textwidth}
\centering
\includegraphics[width=\textwidth]{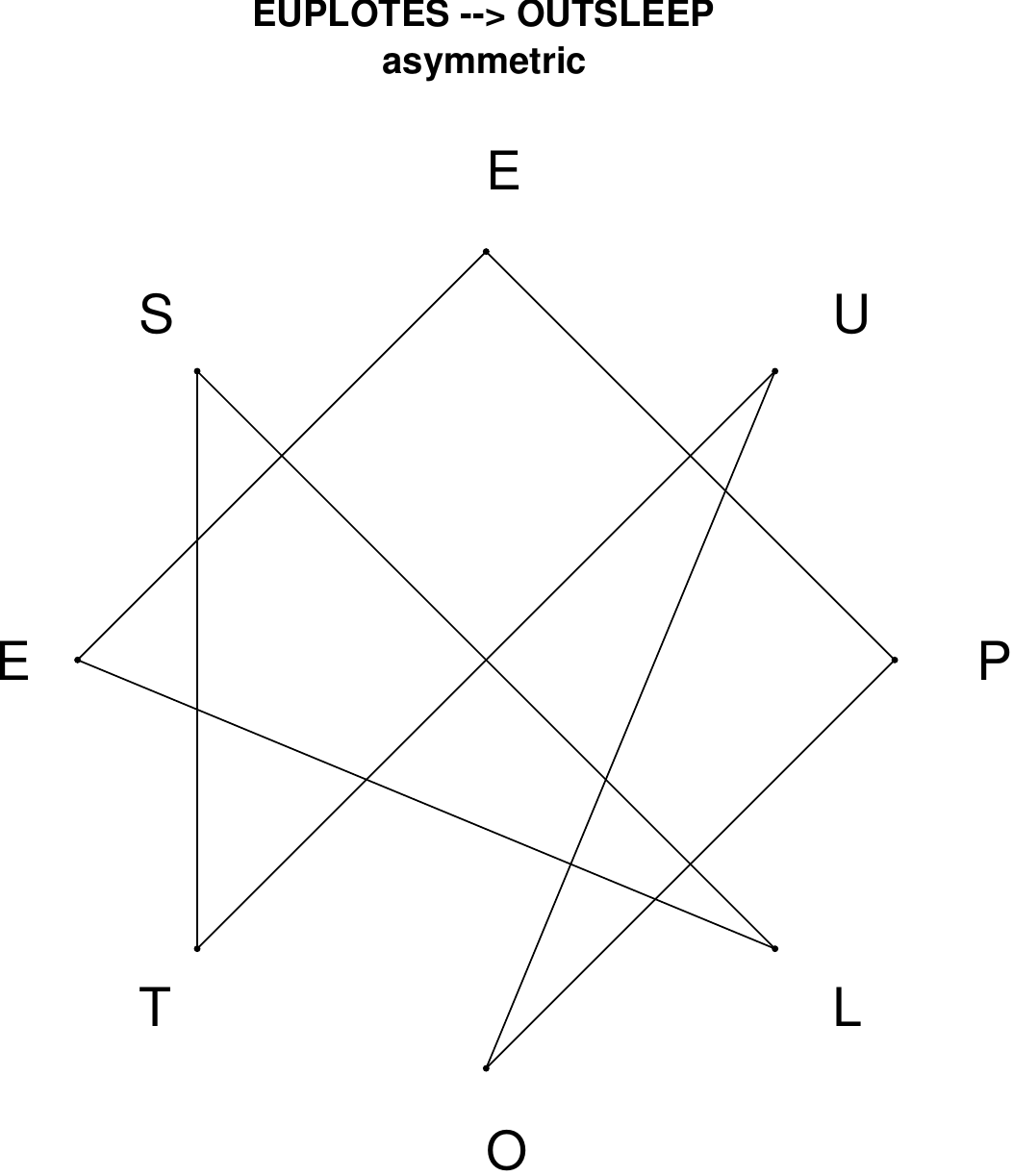}
\end{subfigure}
\hfill
\begin{subfigure}[T]{0.19\textwidth}
\centering
\includegraphics[width=\textwidth]{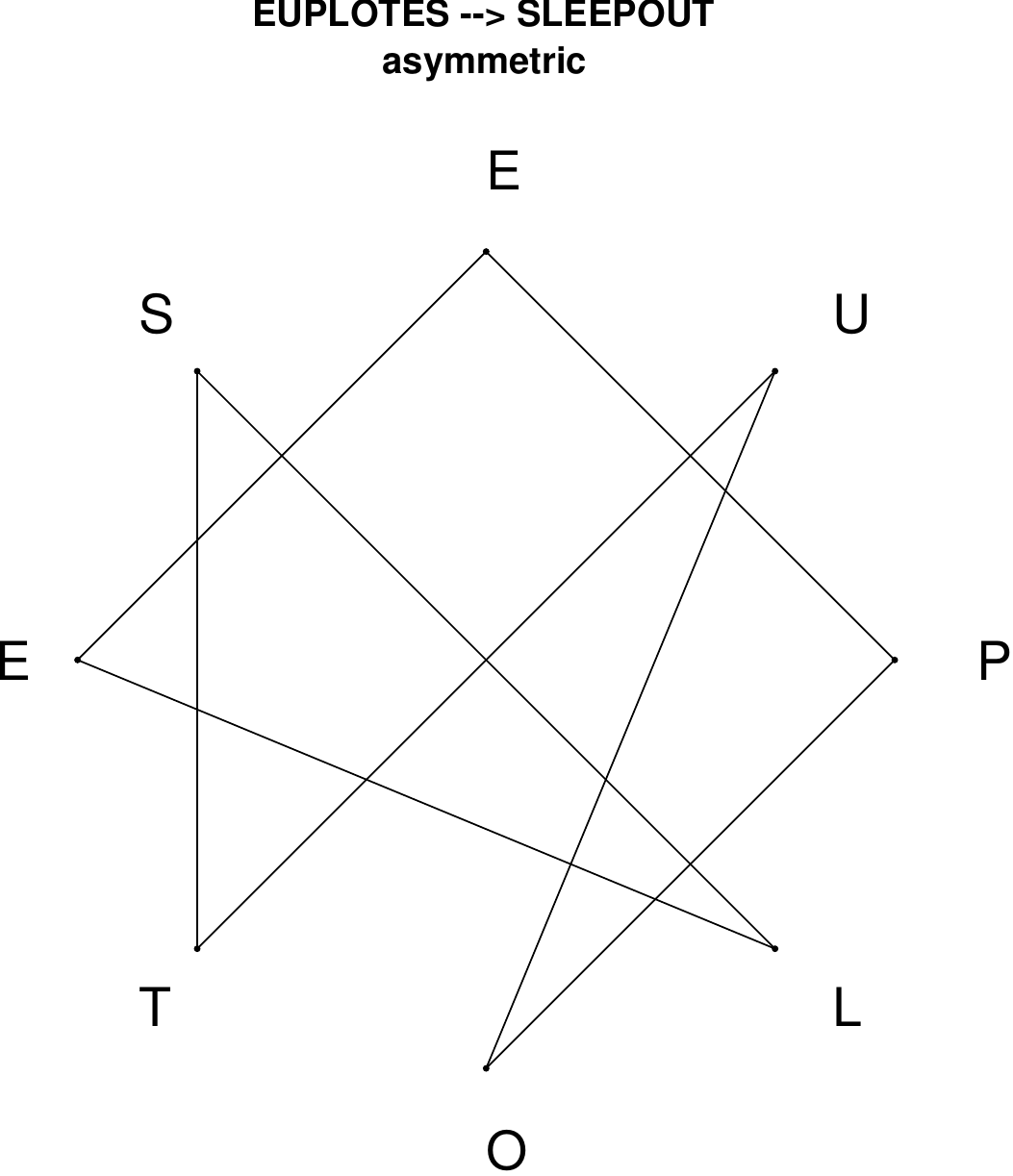}
\end{subfigure}
\hfill
\begin{subfigure}[T]{0.19\textwidth}
\centering
\includegraphics[width=\textwidth]{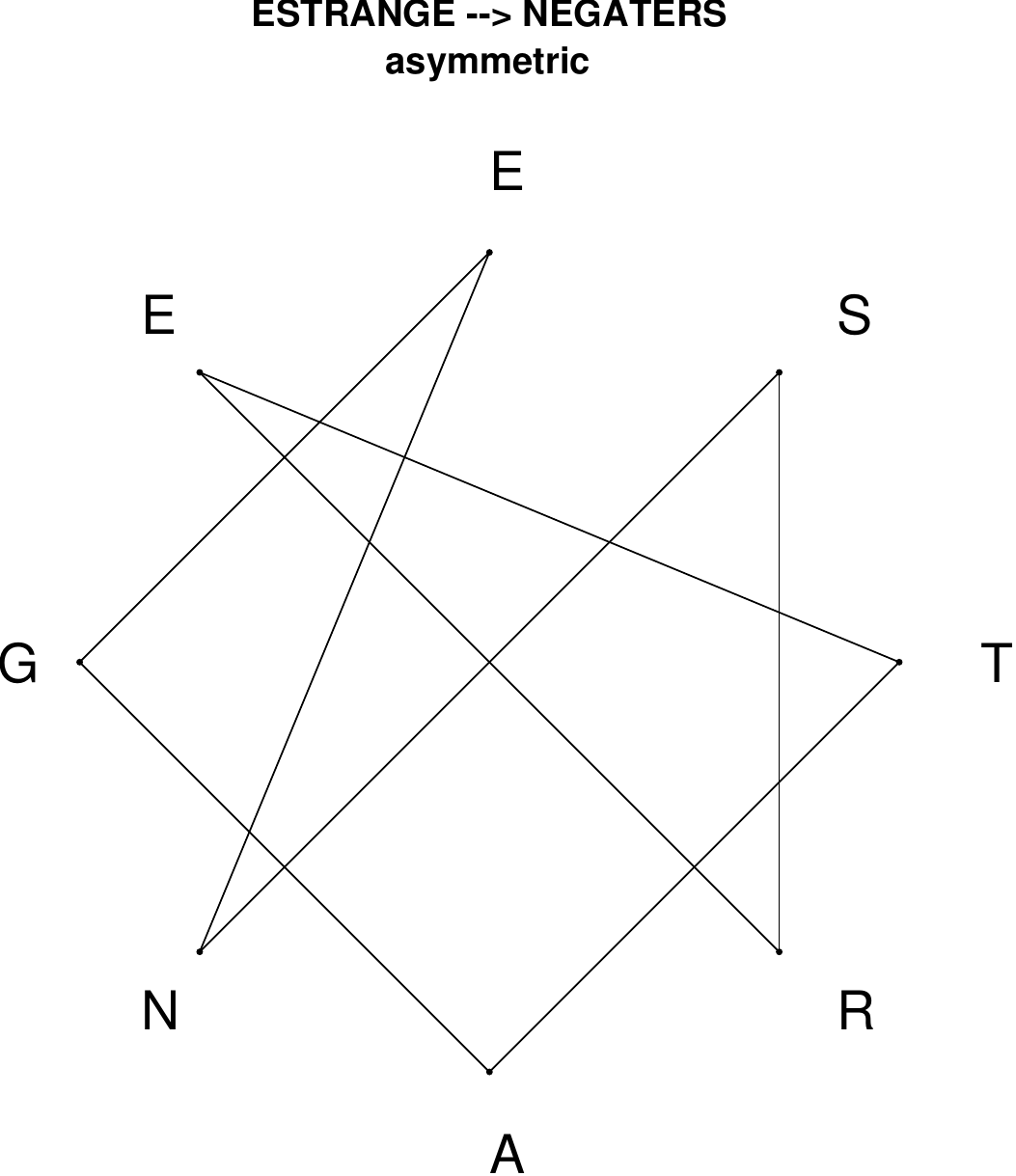}
\end{subfigure}
\hfill
\begin{subfigure}[T]{0.19\textwidth}
\centering
\includegraphics[width=\textwidth]{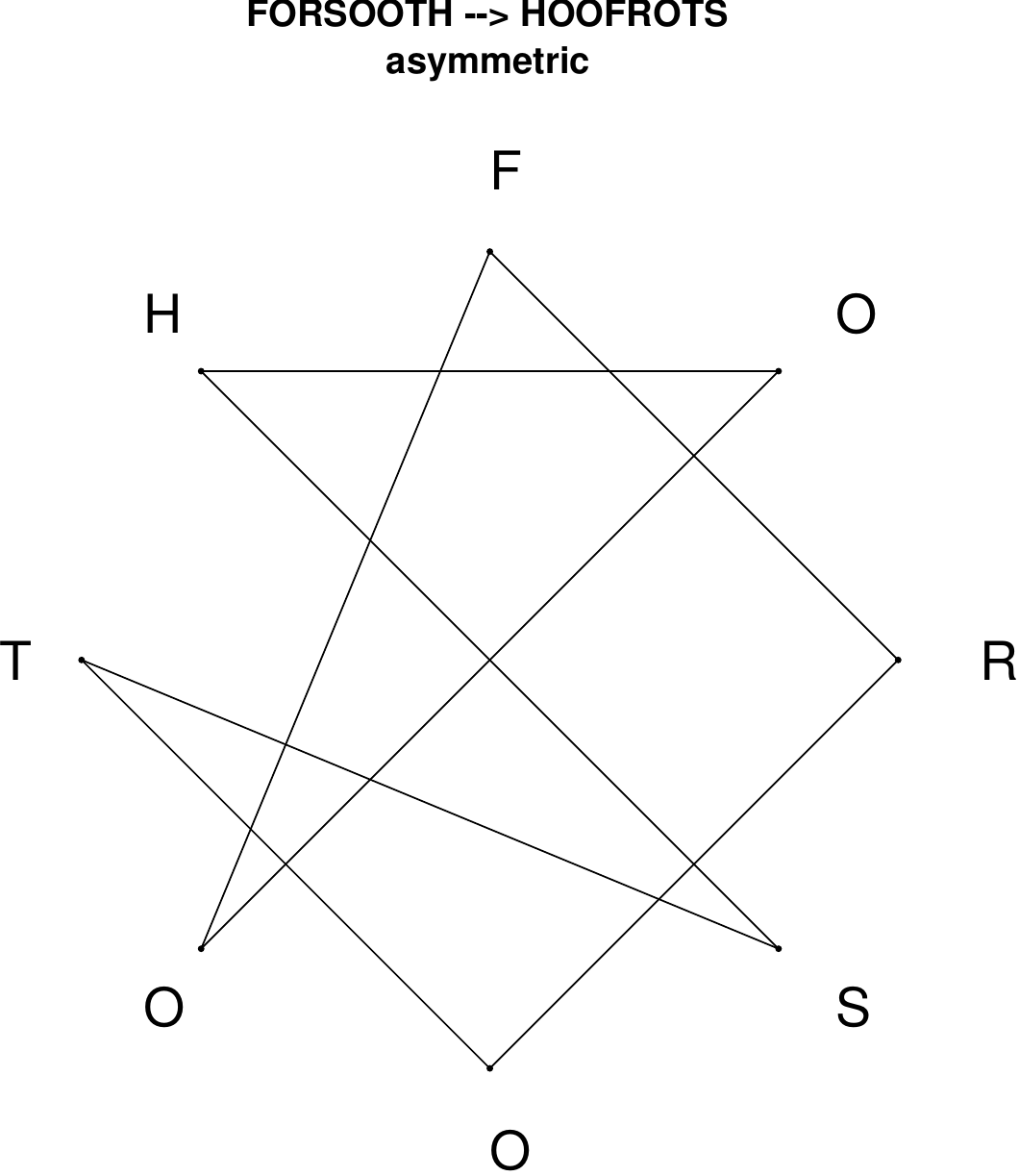}
\end{subfigure}
\hfill
\begin{subfigure}[T]{0.19\textwidth}
\centering
\includegraphics[width=\textwidth]{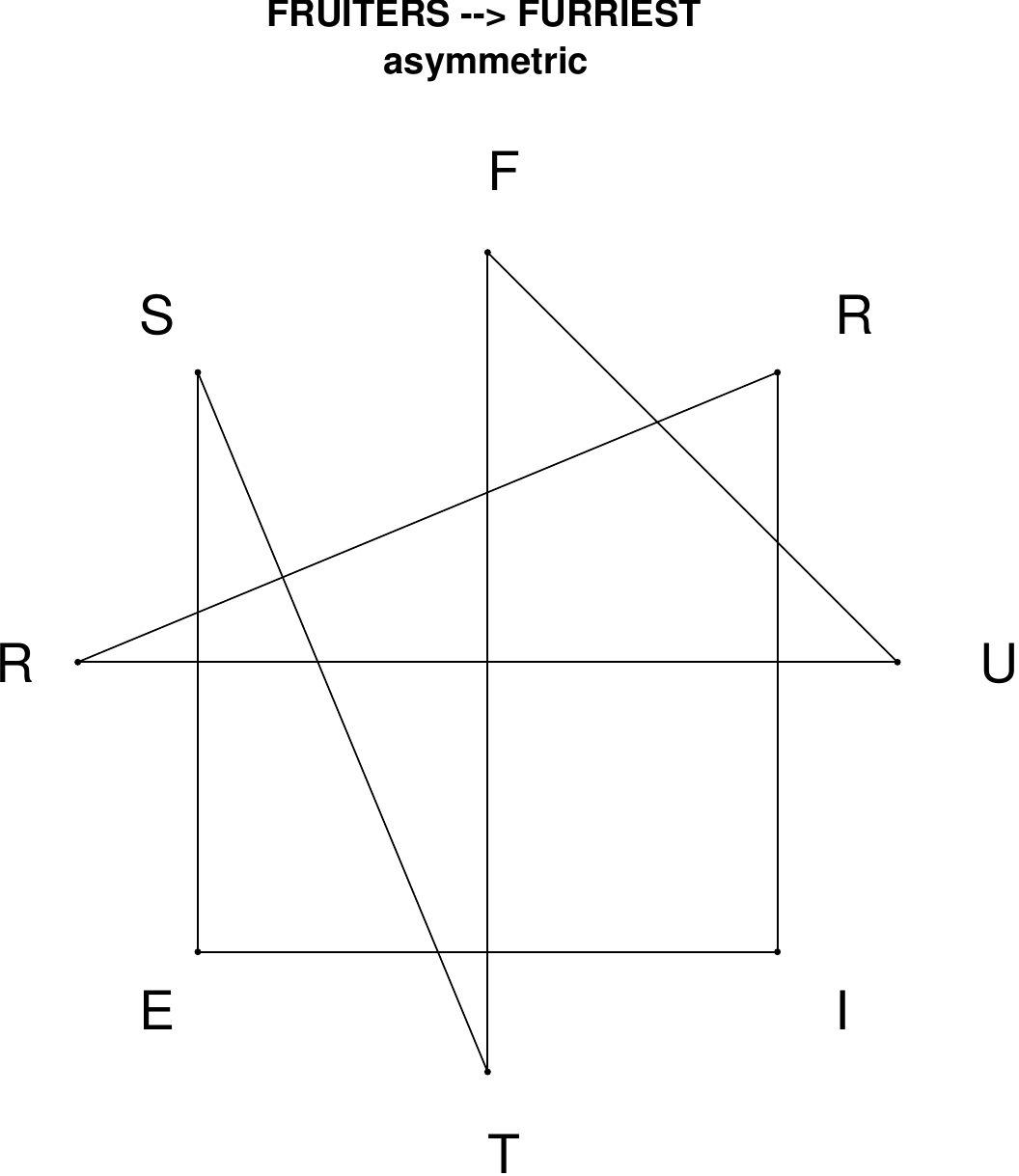}
\end{subfigure}
\end{figure}

\begin{figure}[H]
\centering
\begin{subfigure}[T]{0.19\textwidth}
\centering
\includegraphics[width=\textwidth]{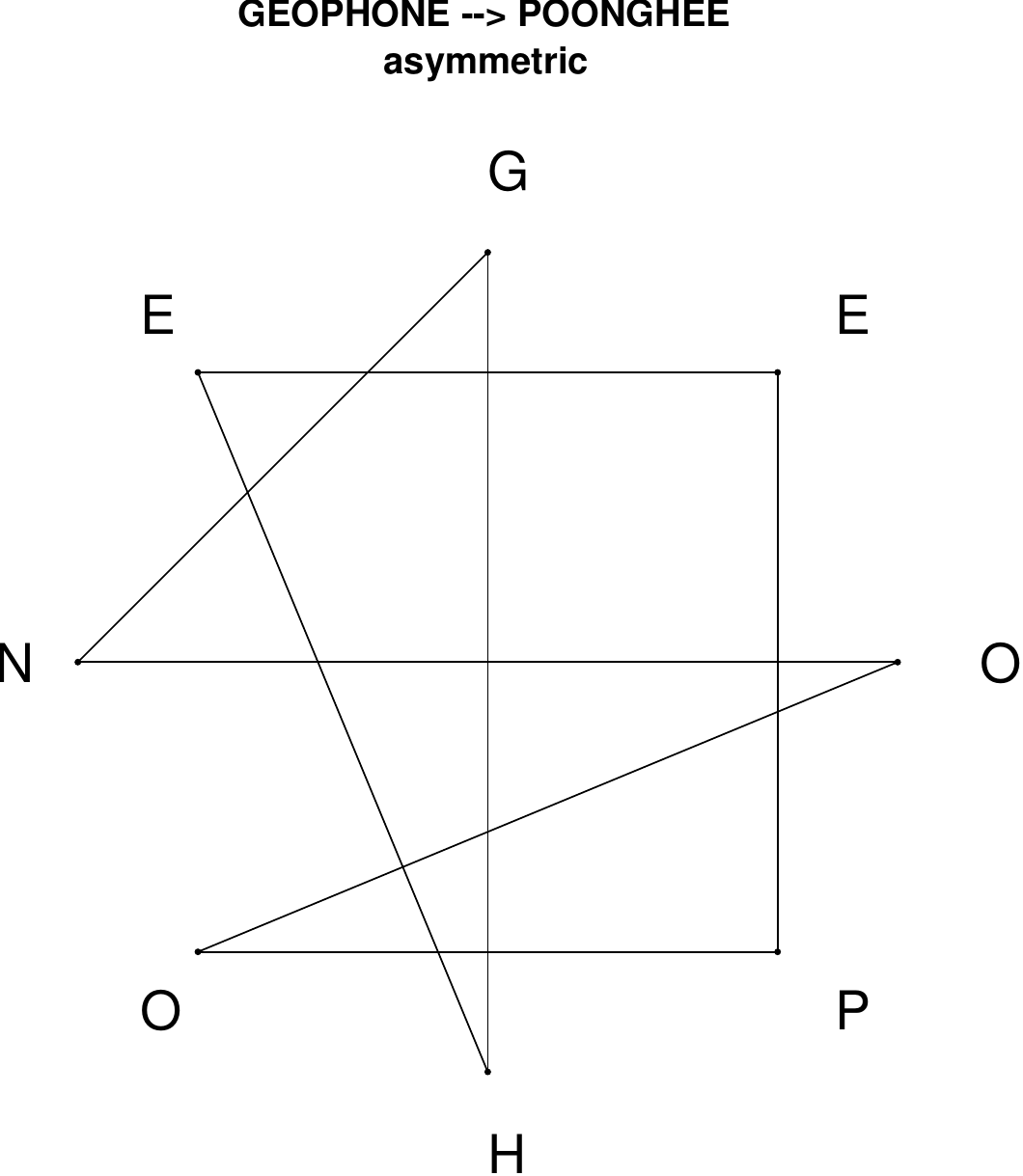}
\end{subfigure}
\hfill
\begin{subfigure}[T]{0.19\textwidth}
\centering
\includegraphics[width=\textwidth]{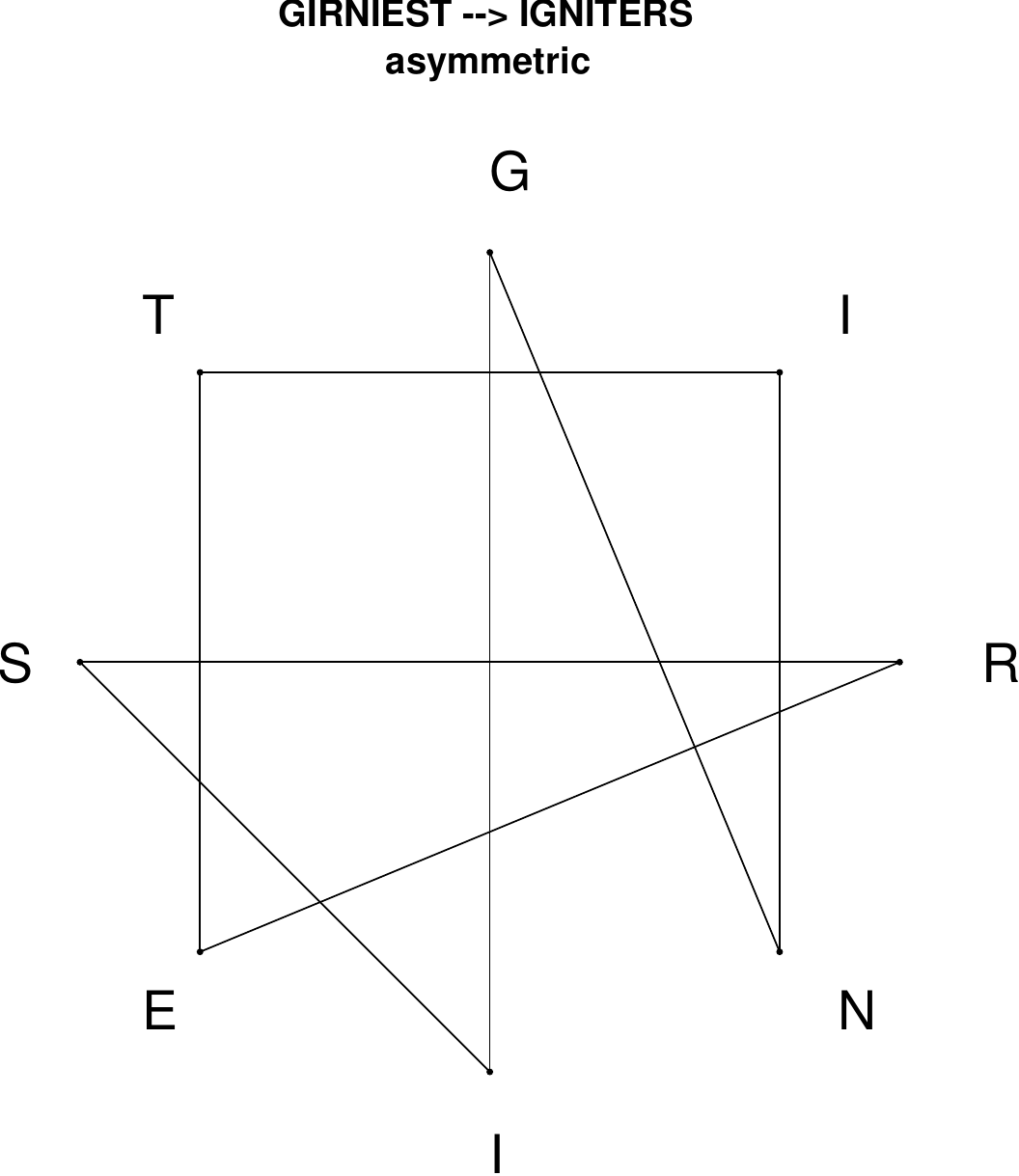}
\end{subfigure}
\hfill
\begin{subfigure}[T]{0.19\textwidth}
\centering
\includegraphics[width=\textwidth]{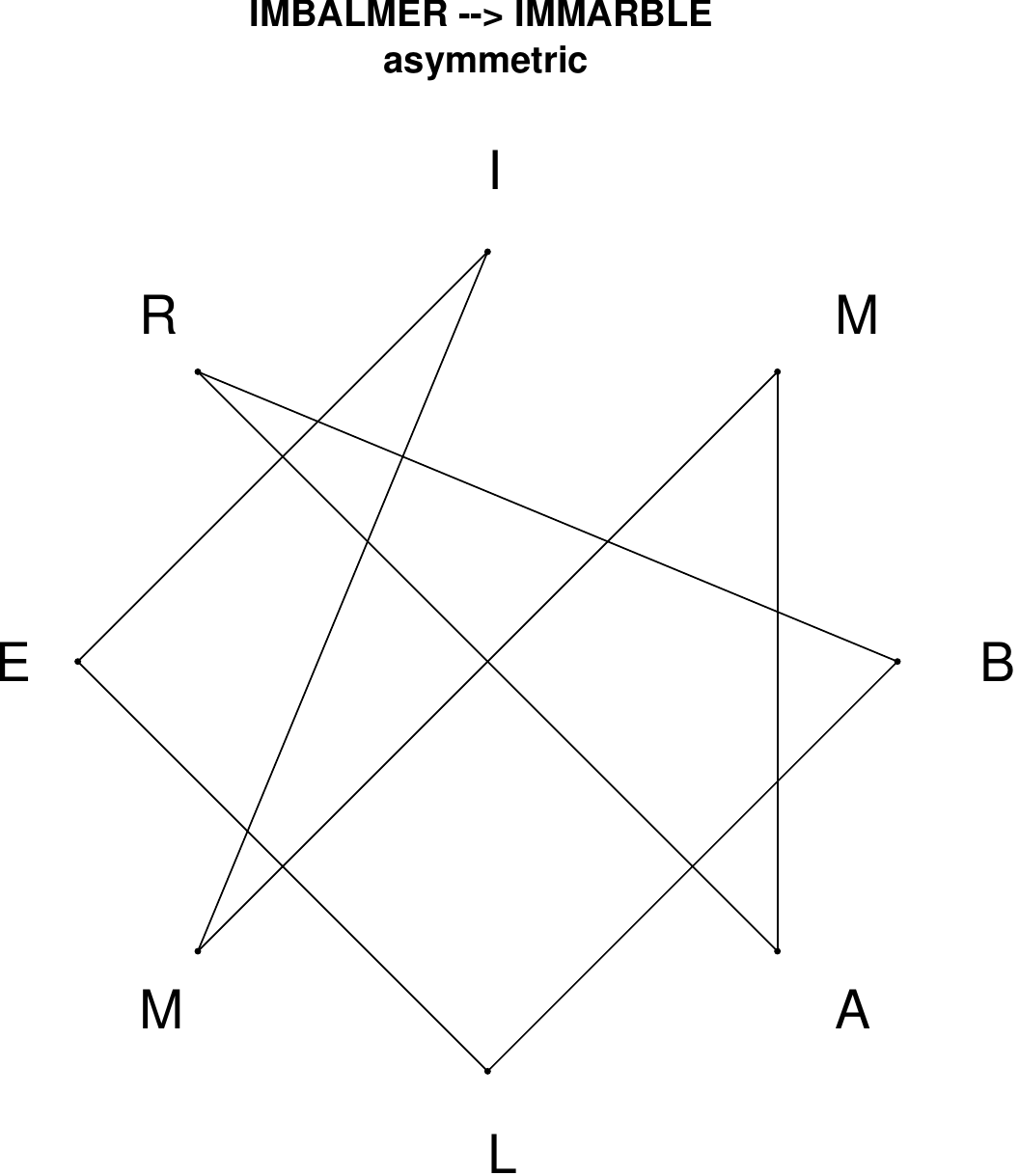}
\end{subfigure}
\hfill
\begin{subfigure}[T]{0.19\textwidth}
\centering
\includegraphics[width=\textwidth]{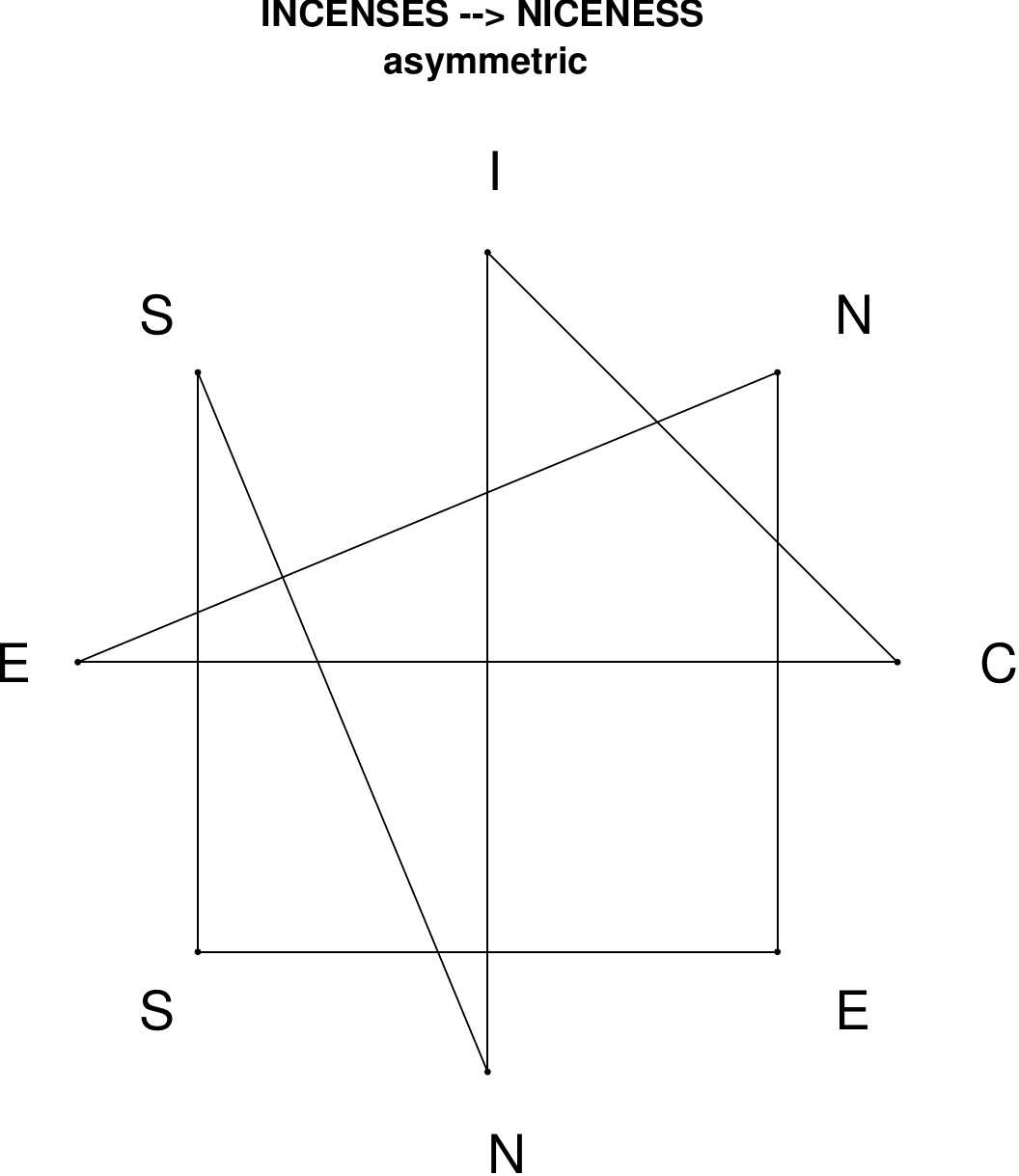}
\end{subfigure}
\hfill
\begin{subfigure}[T]{0.19\textwidth}
\centering
\includegraphics[width=\textwidth]{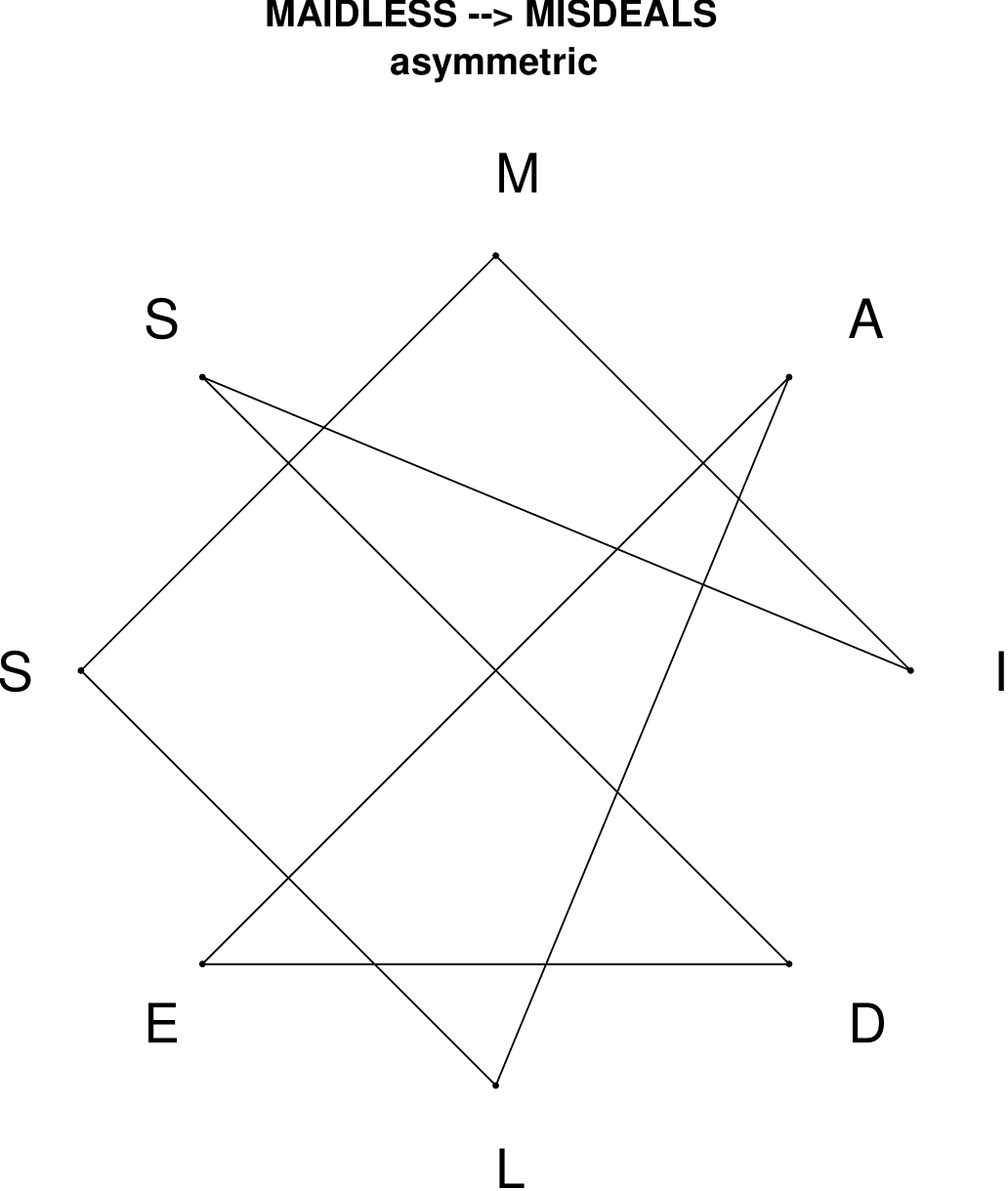}
\end{subfigure}
\end{figure}

\begin{figure}[H]
\centering
\begin{subfigure}[T]{0.19\textwidth}
\centering
\includegraphics[width=\textwidth]{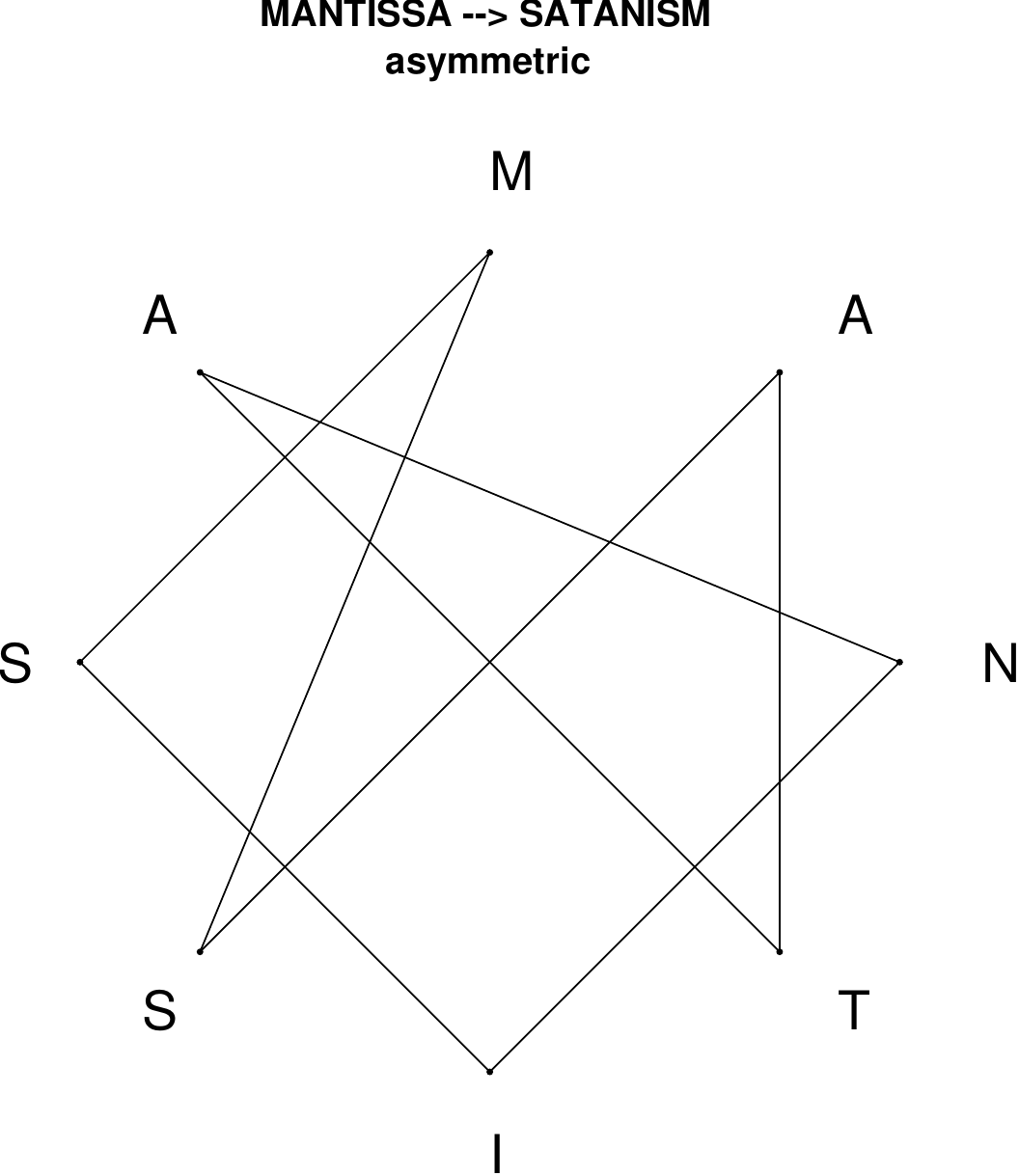}
\end{subfigure}
\hfill
\begin{subfigure}[T]{0.19\textwidth}
\centering
\includegraphics[width=\textwidth]{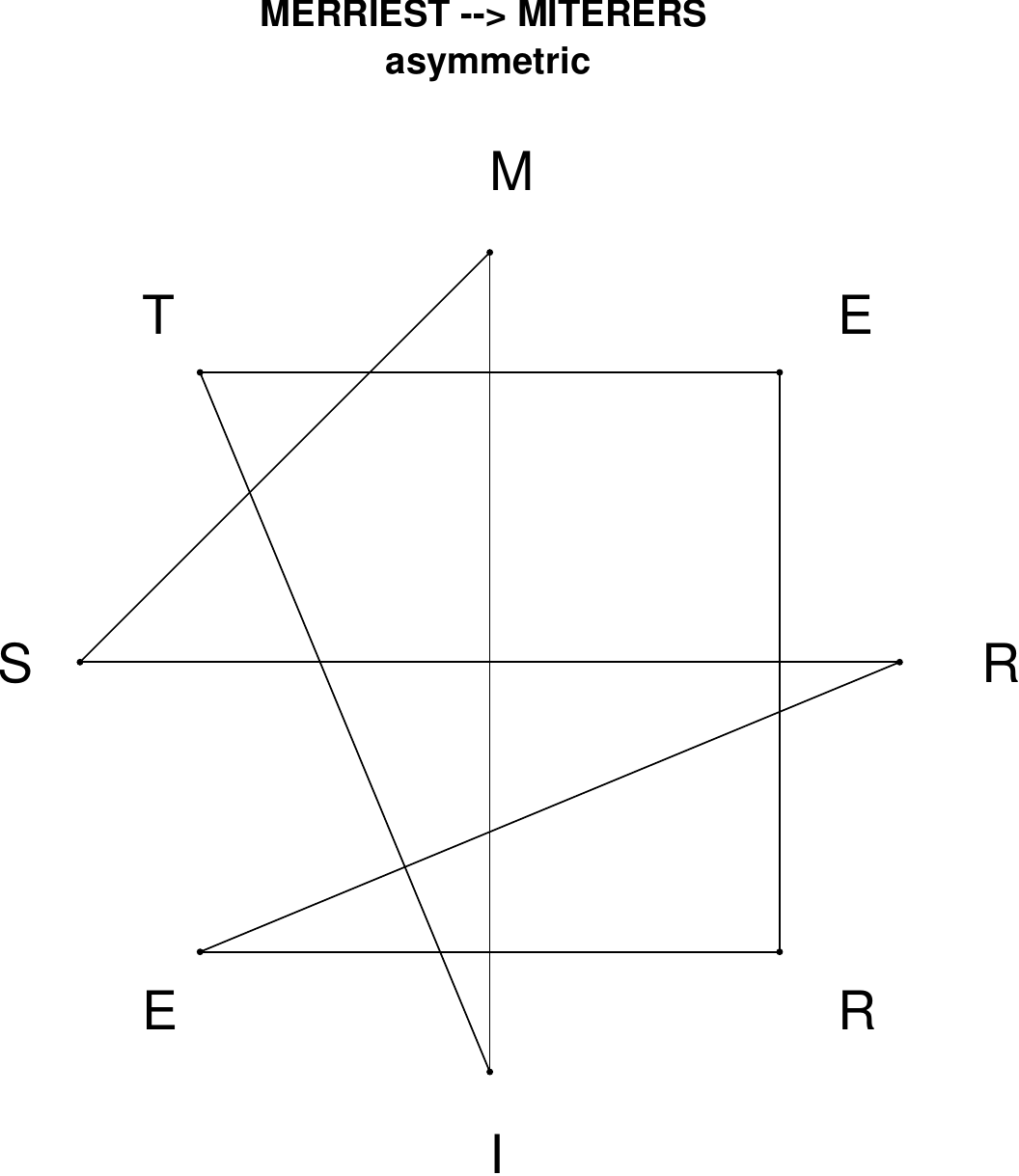}
\end{subfigure}
\hfill
\begin{subfigure}[T]{0.19\textwidth}
\centering
\includegraphics[width=\textwidth]{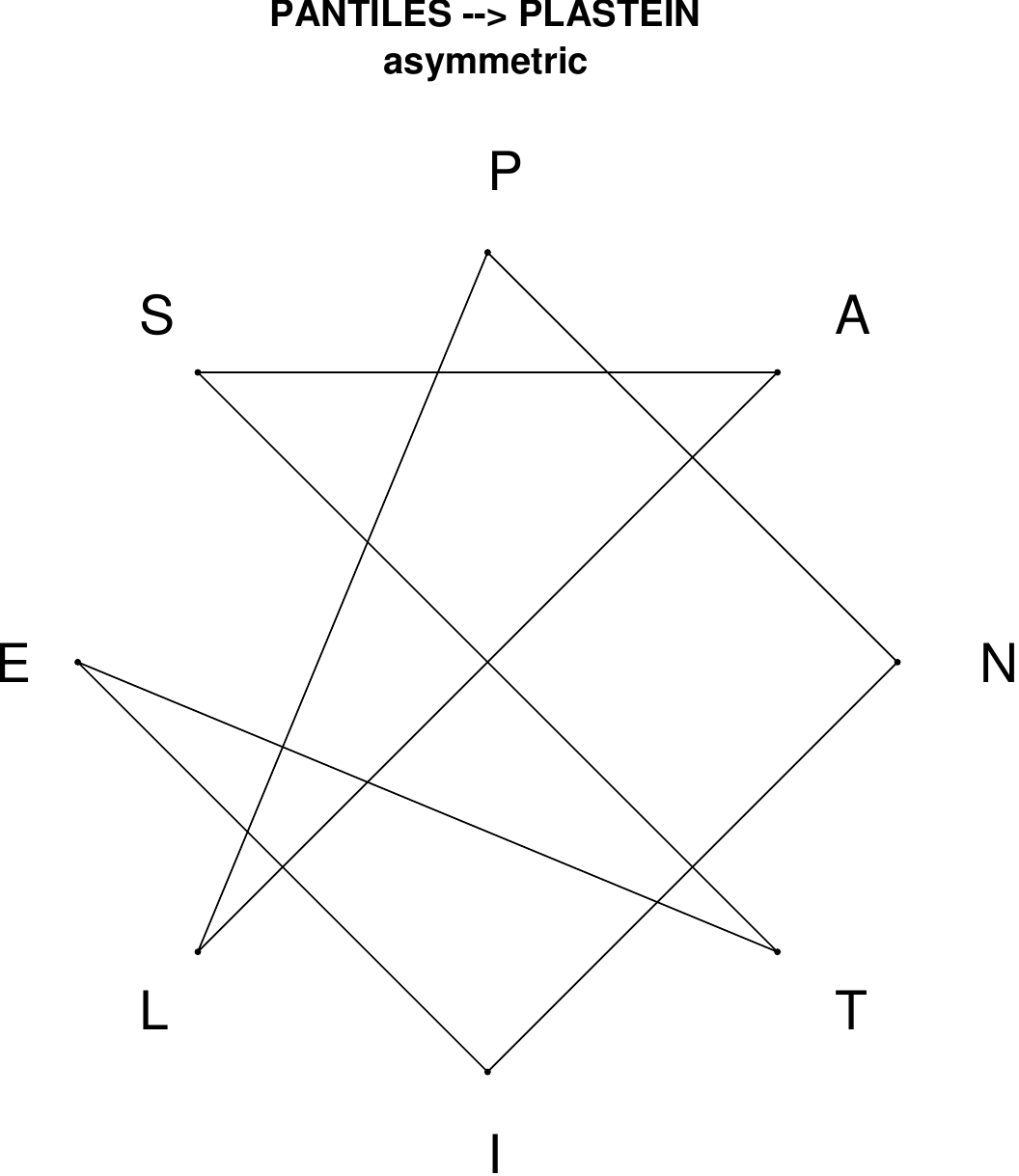}
\end{subfigure}
\hfill
\begin{subfigure}[T]{0.19\textwidth}
\centering
\includegraphics[width=\textwidth]{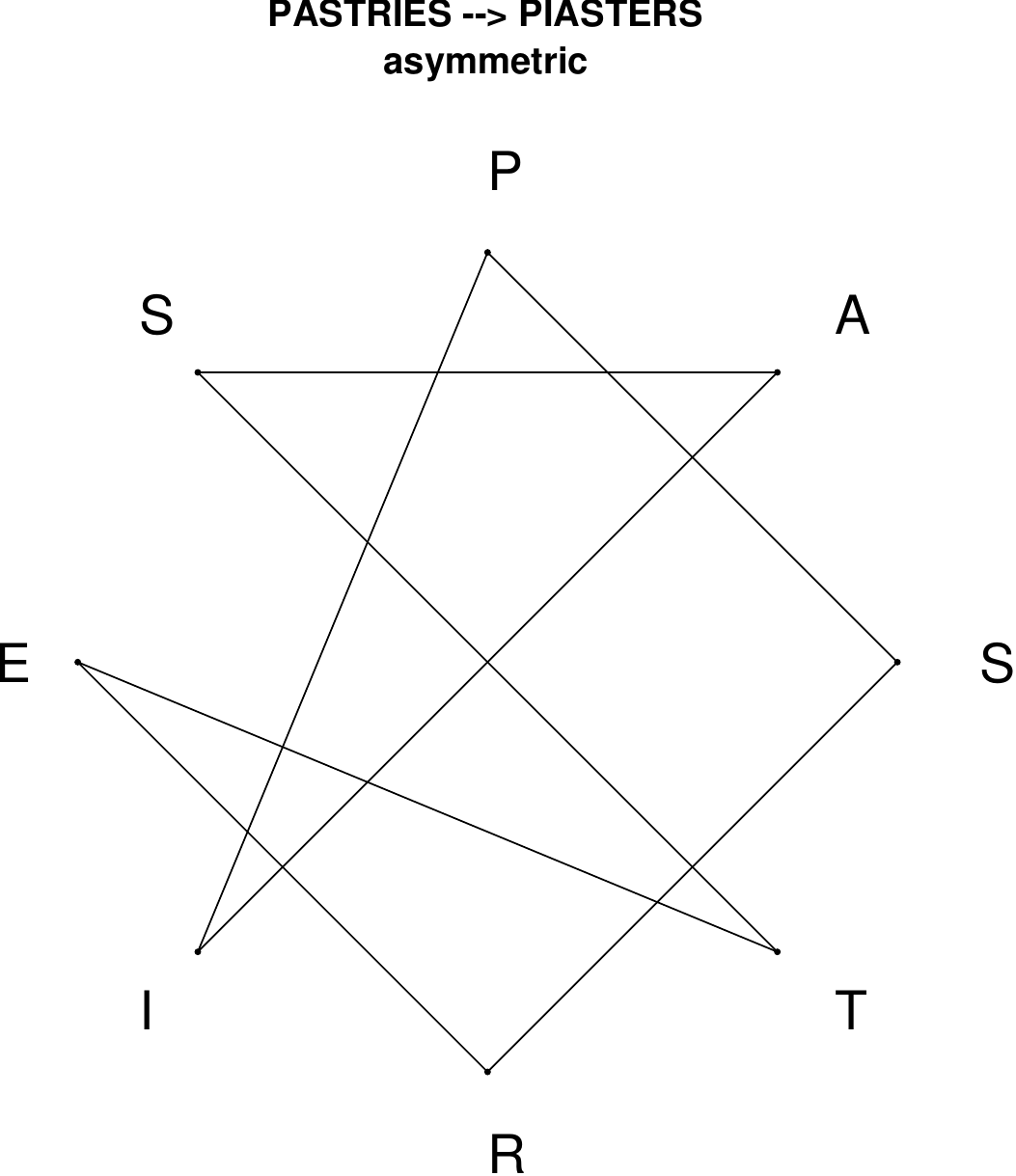}
\end{subfigure}
\hfill
\begin{subfigure}[T]{0.19\textwidth}
\centering
\includegraphics[width=\textwidth]{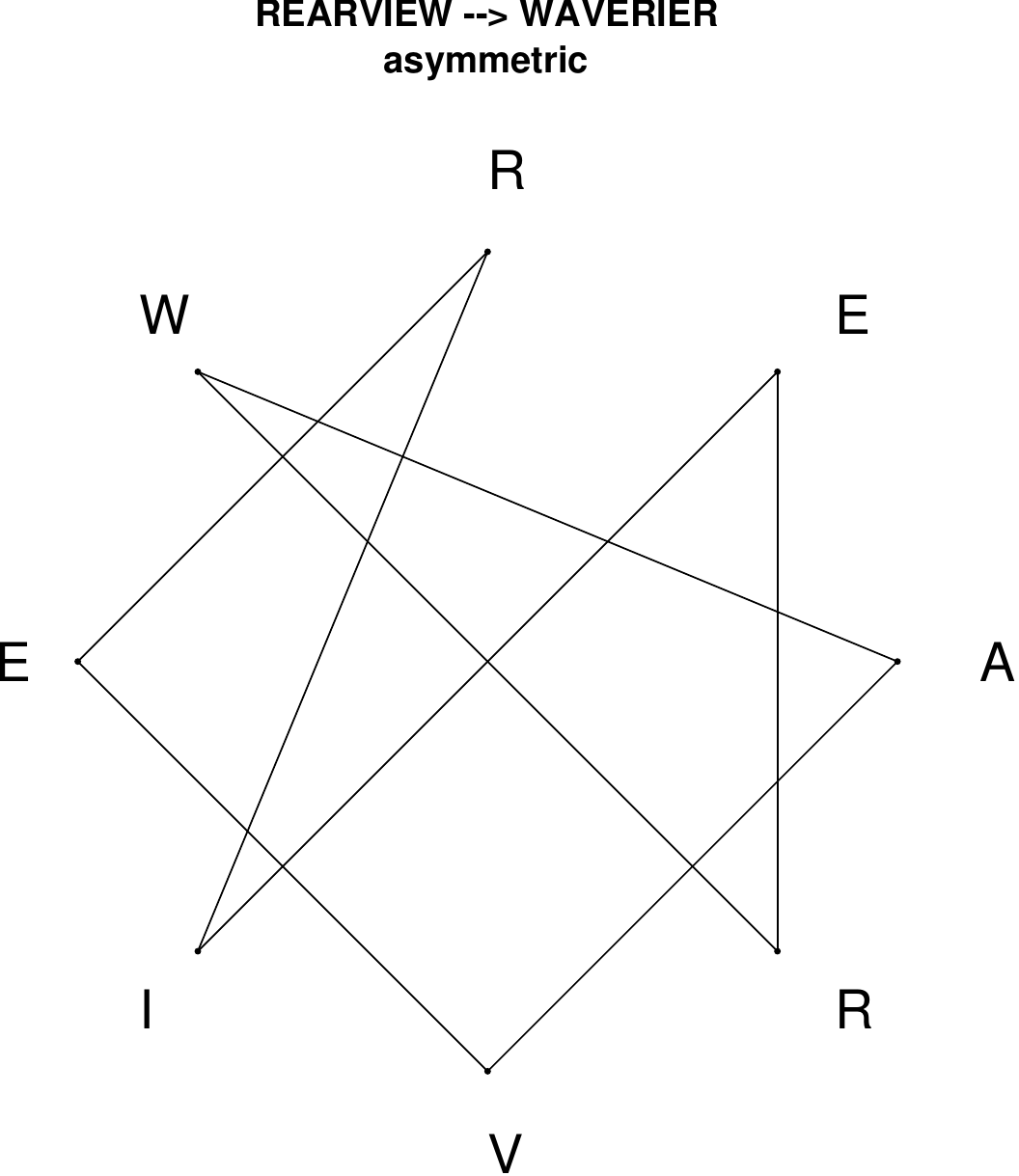}
\end{subfigure}
\end{figure}

\begin{figure}[H]
\centering
\begin{subfigure}[T]{0.19\textwidth}
\centering
\includegraphics[width=\textwidth]{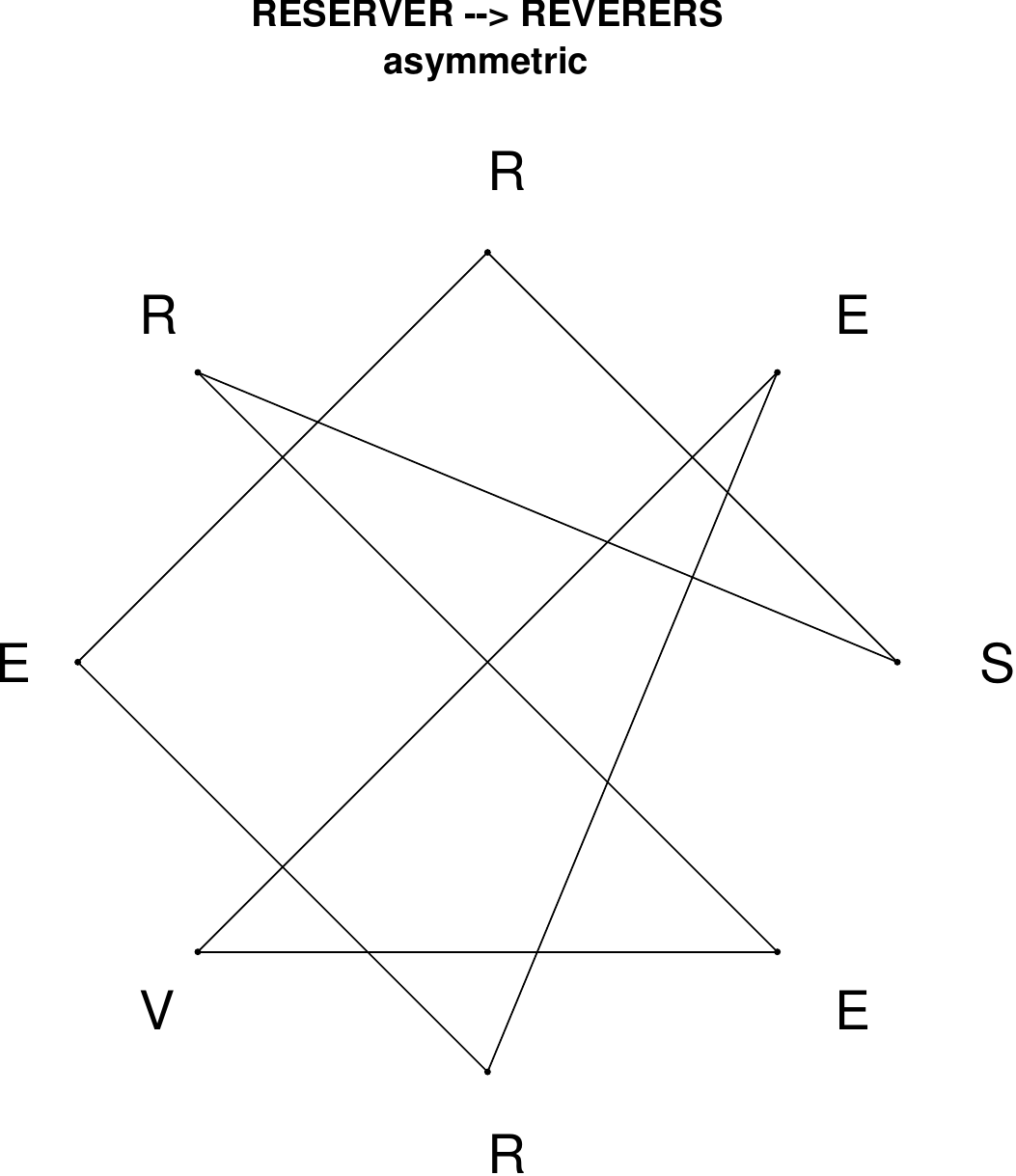}
\end{subfigure}
\hfill
\begin{subfigure}[T]{0.19\textwidth}
\centering
\includegraphics[width=\textwidth]{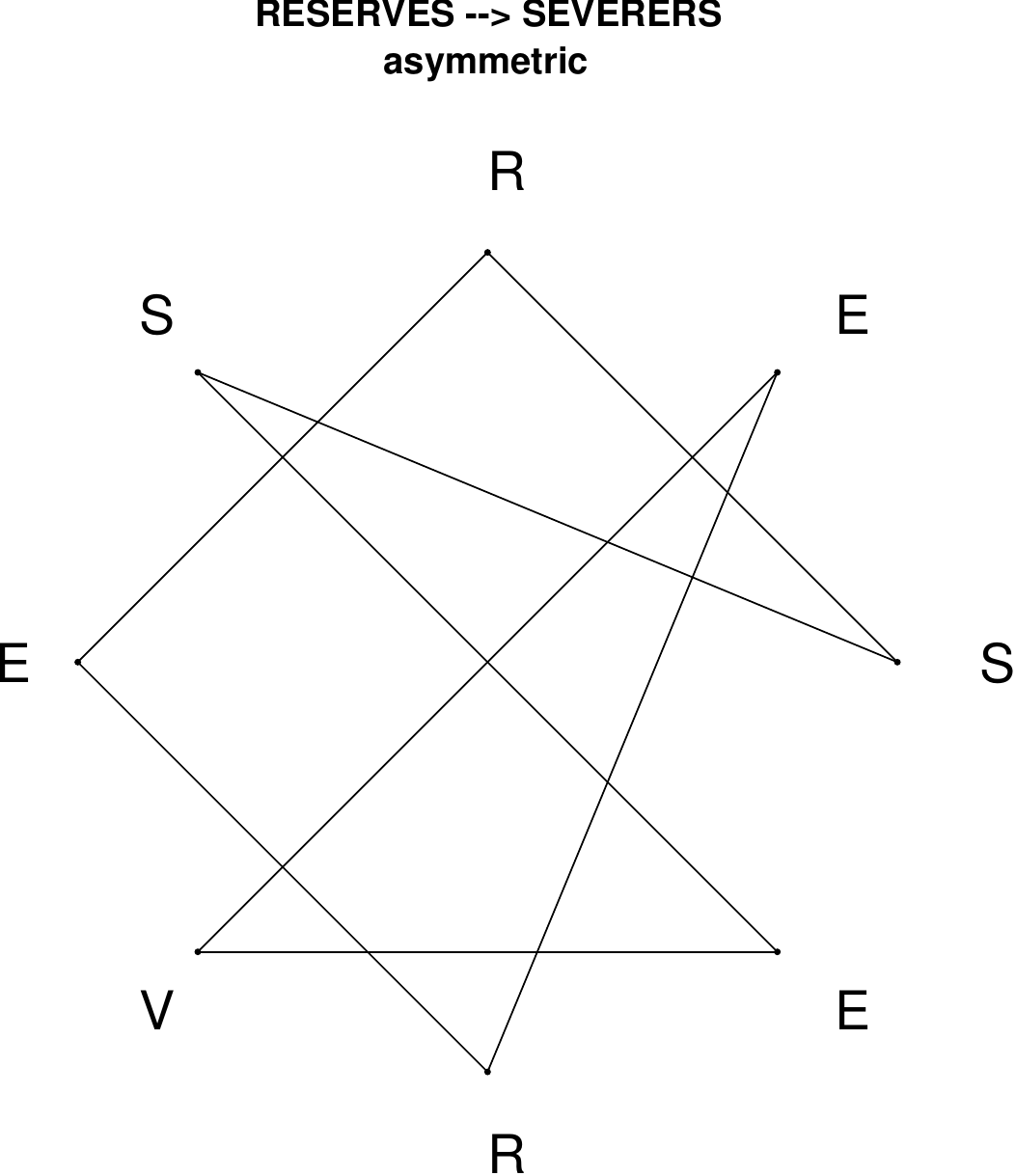}
\end{subfigure}
\hfill
\begin{subfigure}[T]{0.19\textwidth}
\centering
\includegraphics[width=\textwidth]{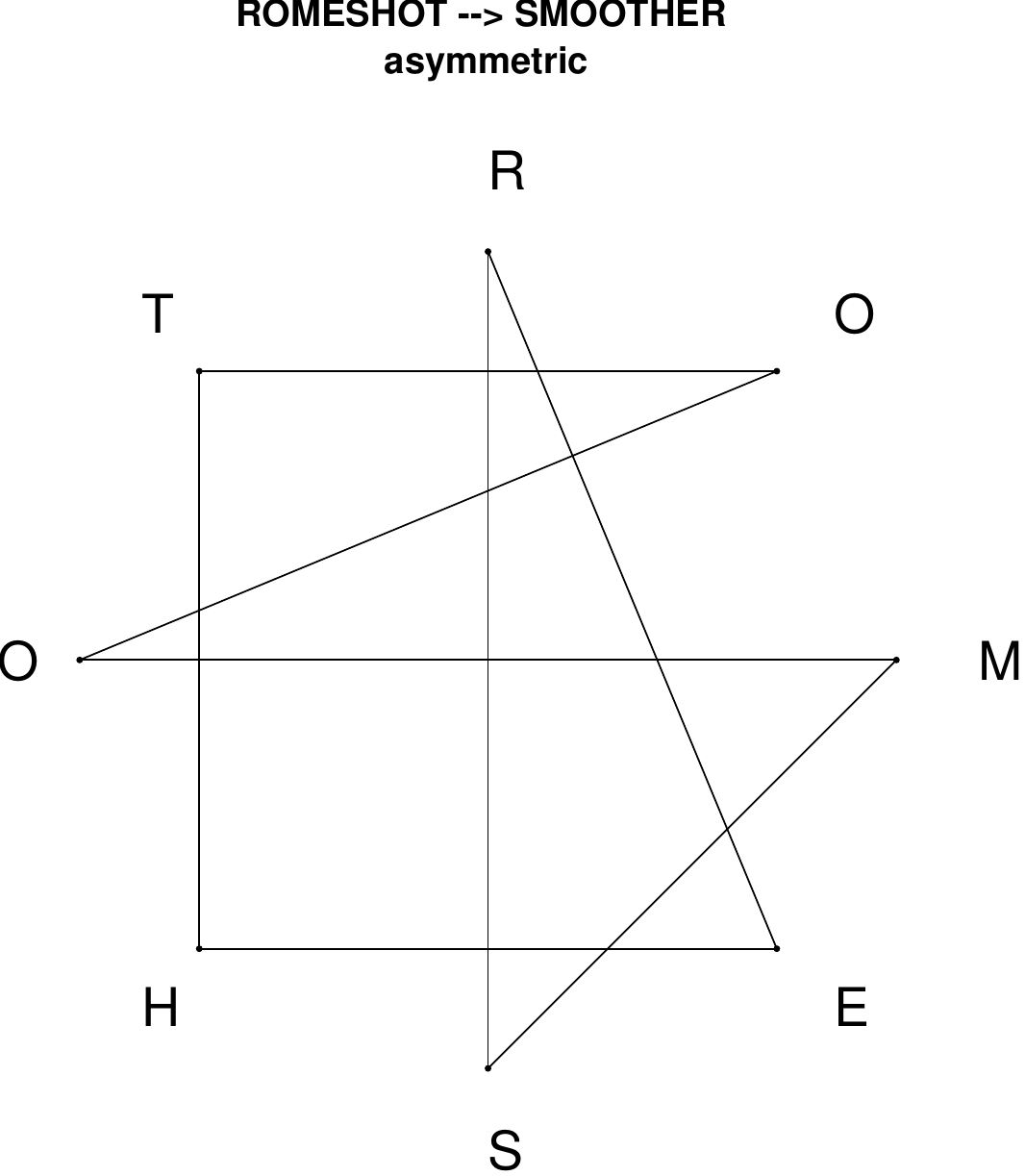}
\end{subfigure}
\hfill
\begin{subfigure}[T]{0.19\textwidth}
\centering
\includegraphics[width=\textwidth]{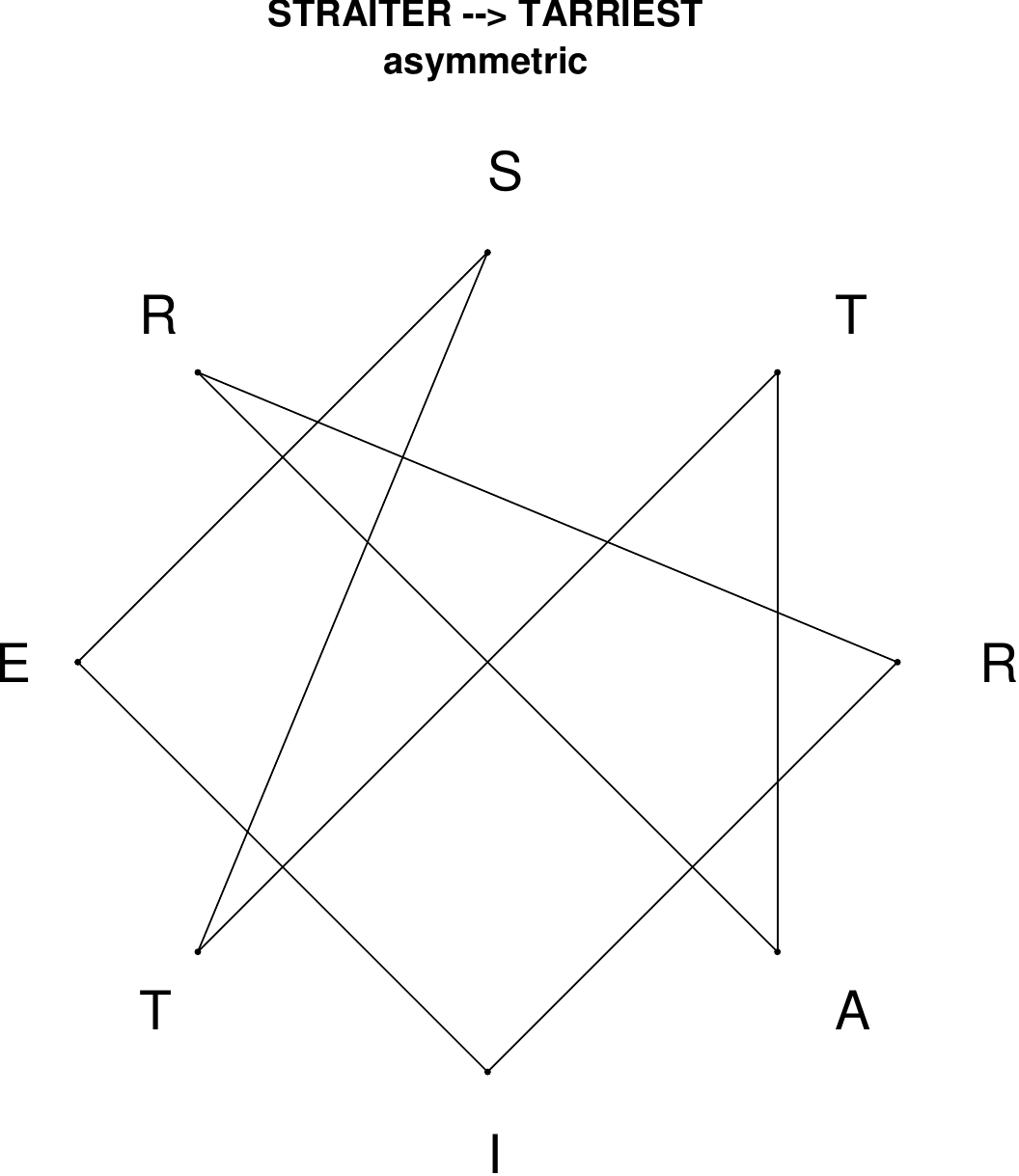}
\end{subfigure}
\hfill
\begin{subfigure}[T]{0.19\textwidth}
\centering
\includegraphics[width=\textwidth]{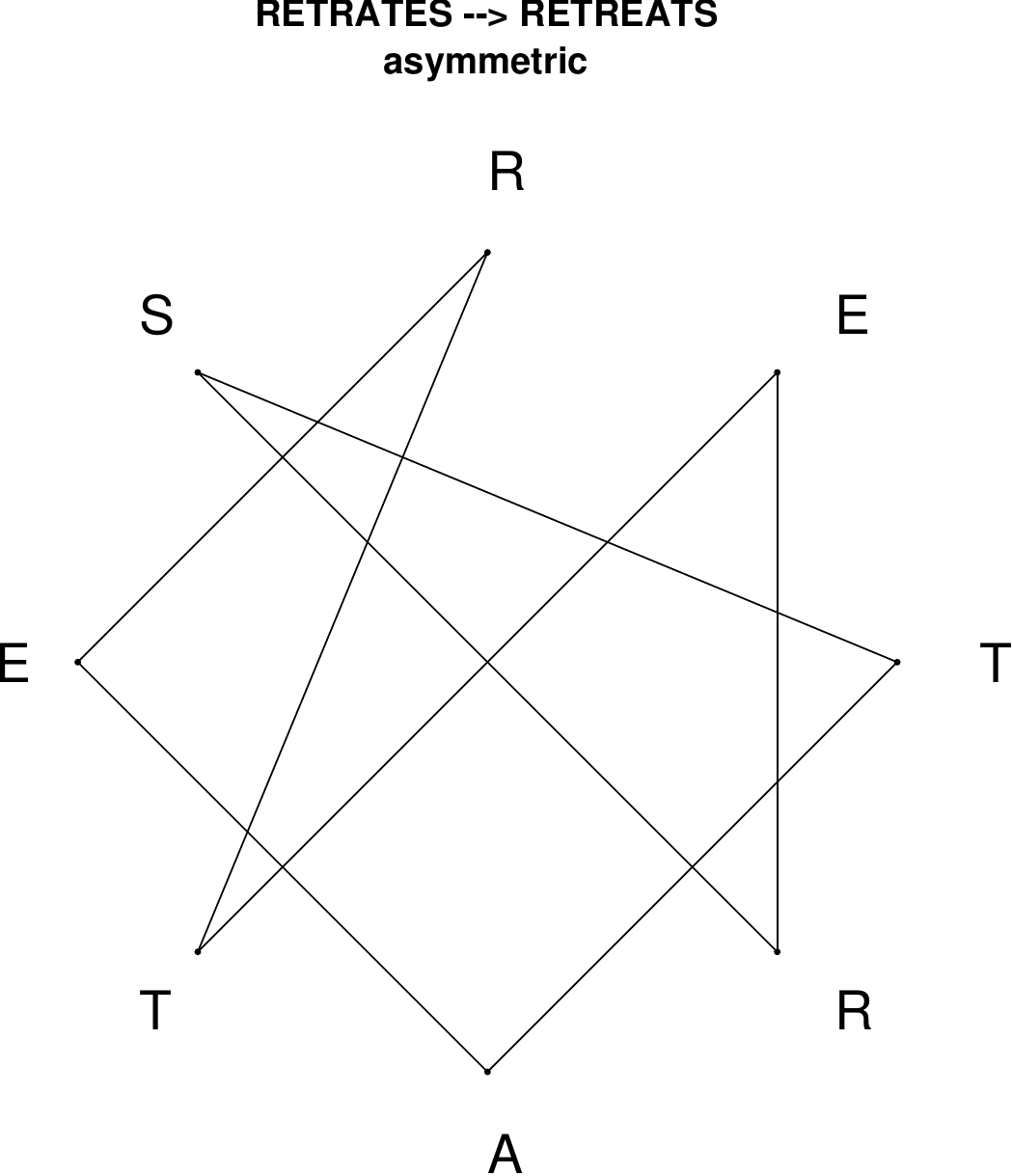}
\end{subfigure}
\end{figure}

\begin{figure}[H]
\centering
\begin{subfigure}[T]{0.19\textwidth}
\centering
\includegraphics[width=\textwidth]{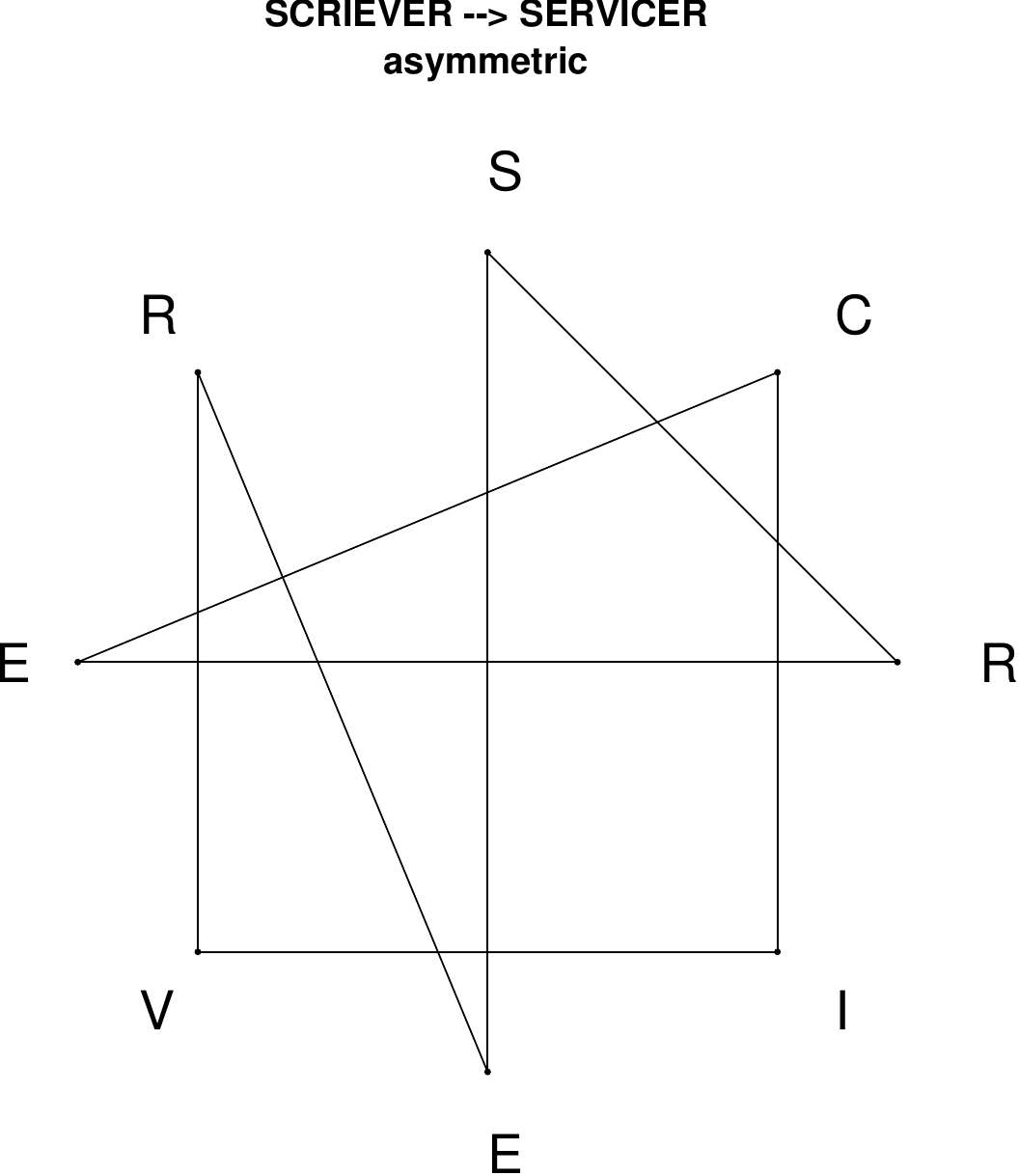}
\end{subfigure}
\hfill
\begin{subfigure}[T]{0.19\textwidth}
\centering
\includegraphics[width=\textwidth]{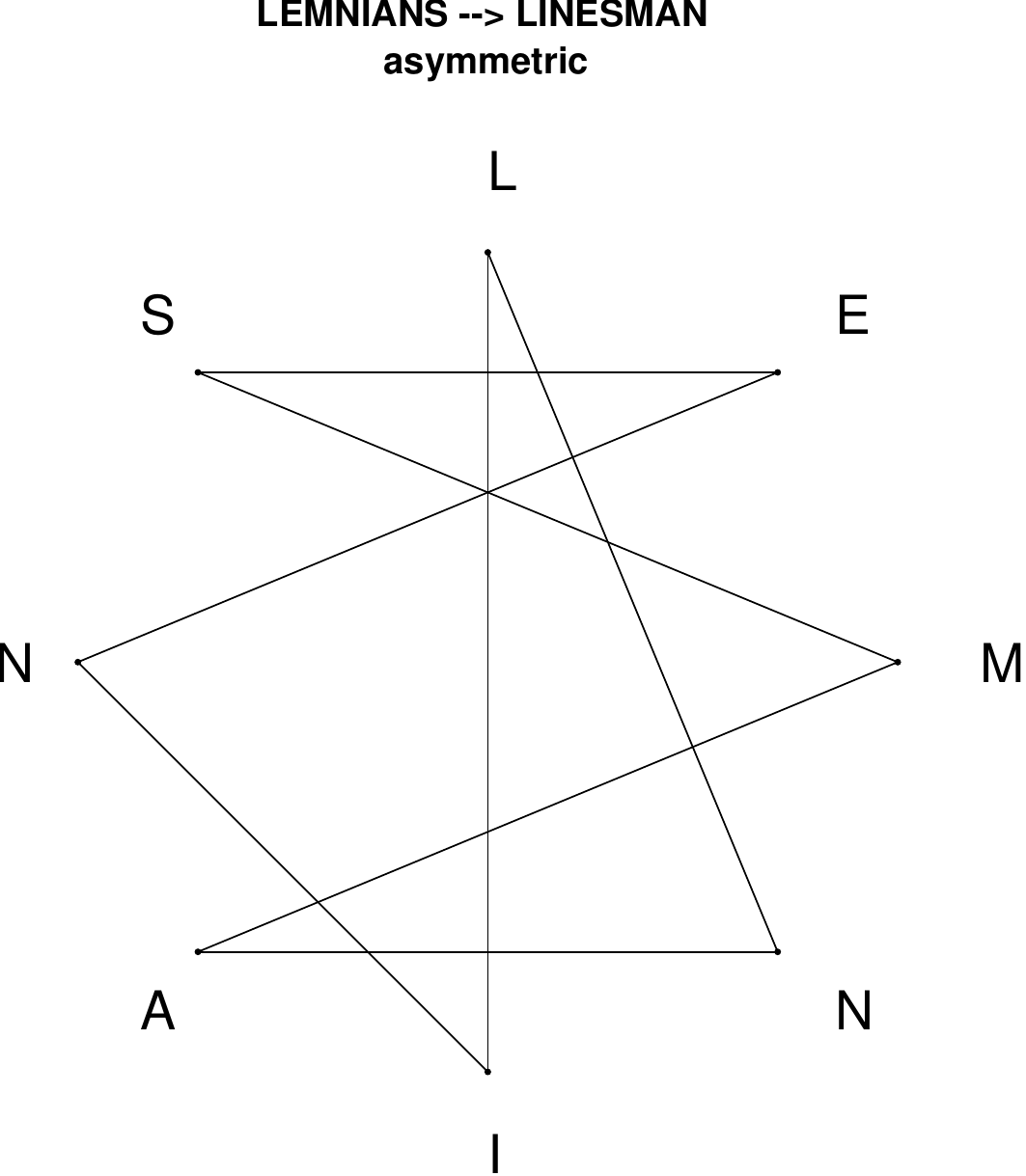}
\end{subfigure}
\hfill
\begin{subfigure}[T]{0.19\textwidth}
\centering
\includegraphics[width=\textwidth]{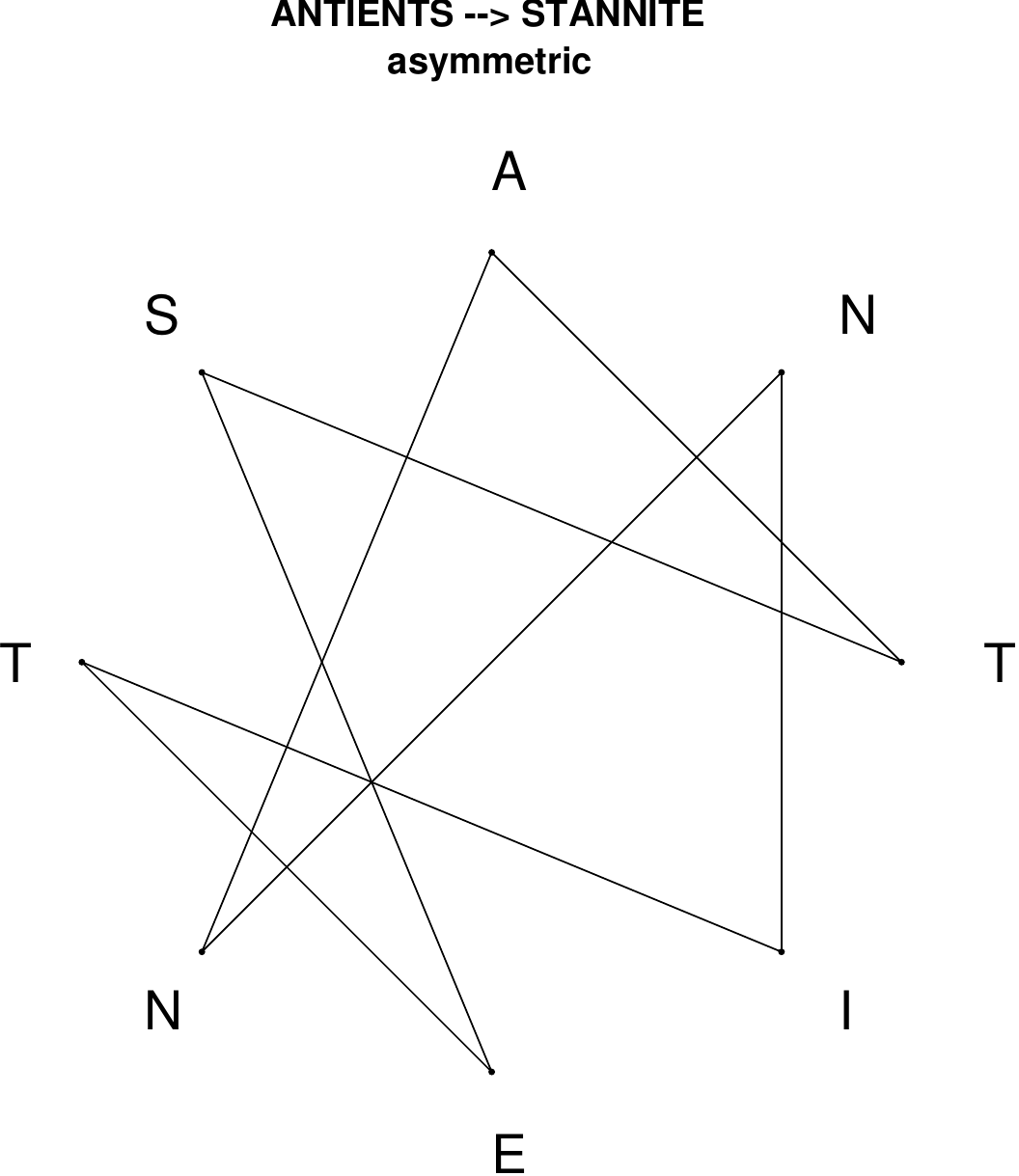}
\end{subfigure}
\hfill
\begin{subfigure}[T]{0.19\textwidth}
\centering
\includegraphics[width=\textwidth]{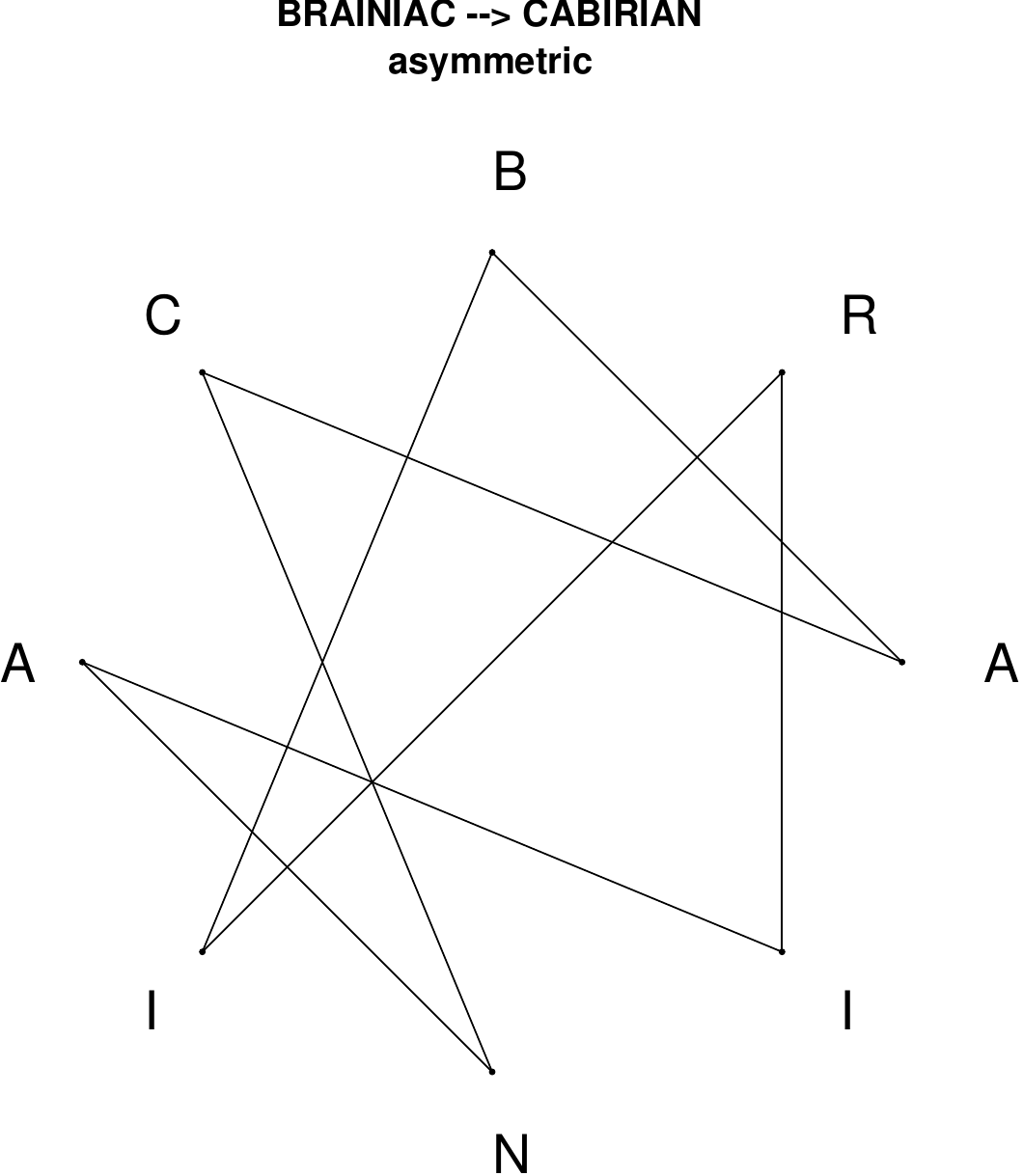}
\end{subfigure}
\hfill
\begin{subfigure}[T]{0.19\textwidth}
\centering
\includegraphics[width=\textwidth]{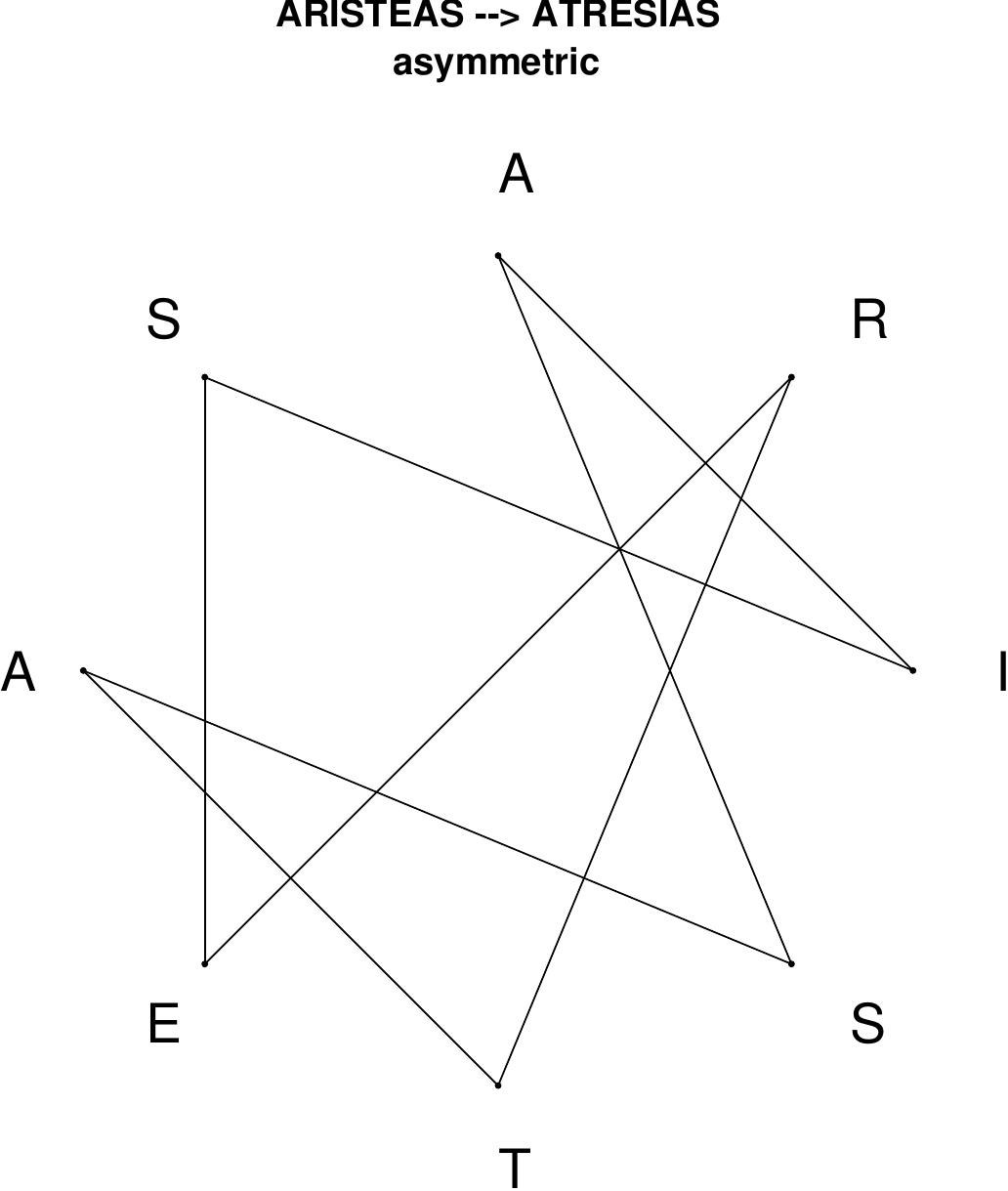}
\end{subfigure}
\end{figure}

\begin{figure}[H]
\centering
\begin{subfigure}[T]{0.19\textwidth}
\centering
\includegraphics[width=\textwidth]{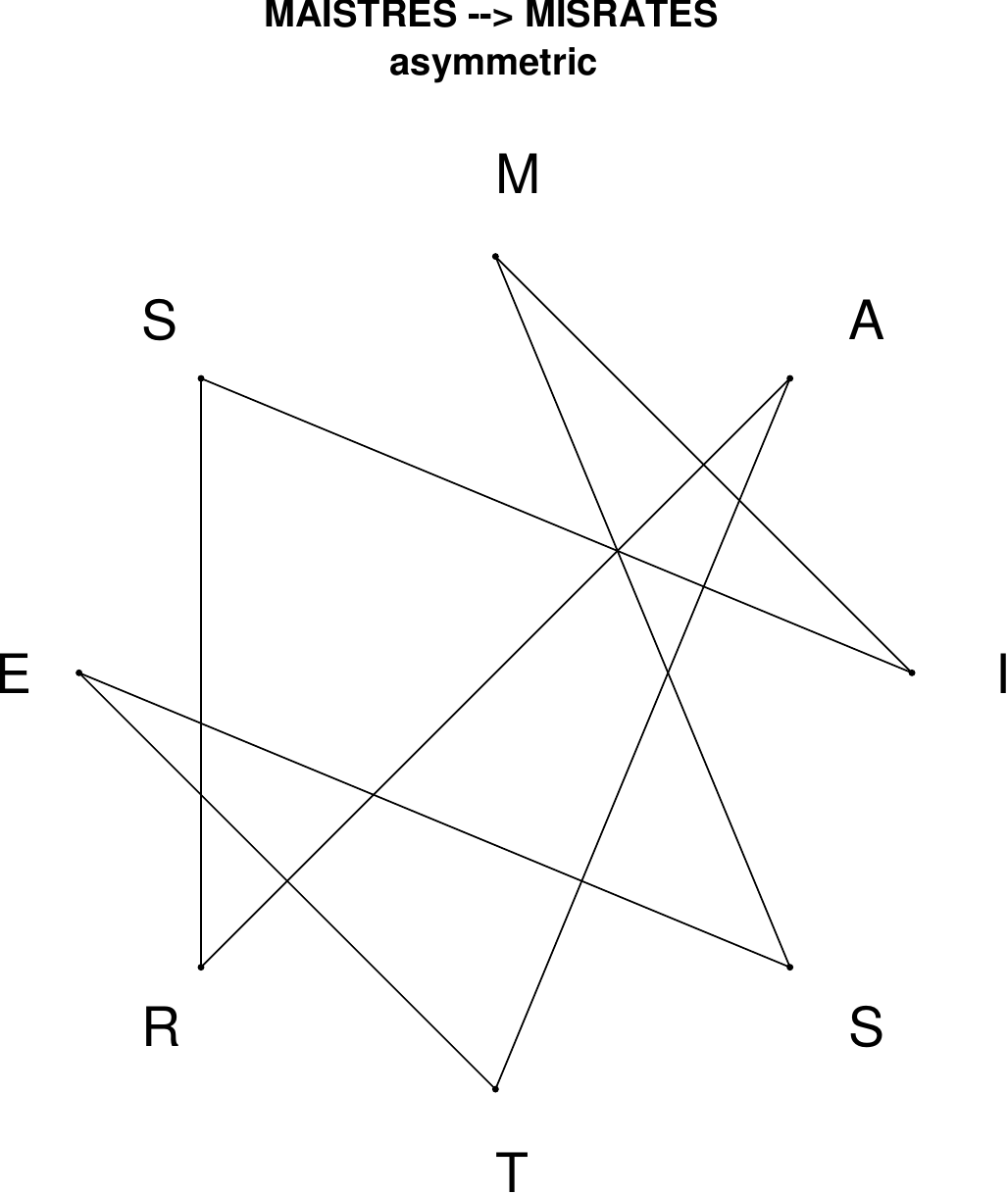}
\end{subfigure}
\hfill
\begin{subfigure}[T]{0.19\textwidth}
\centering
\includegraphics[width=\textwidth]{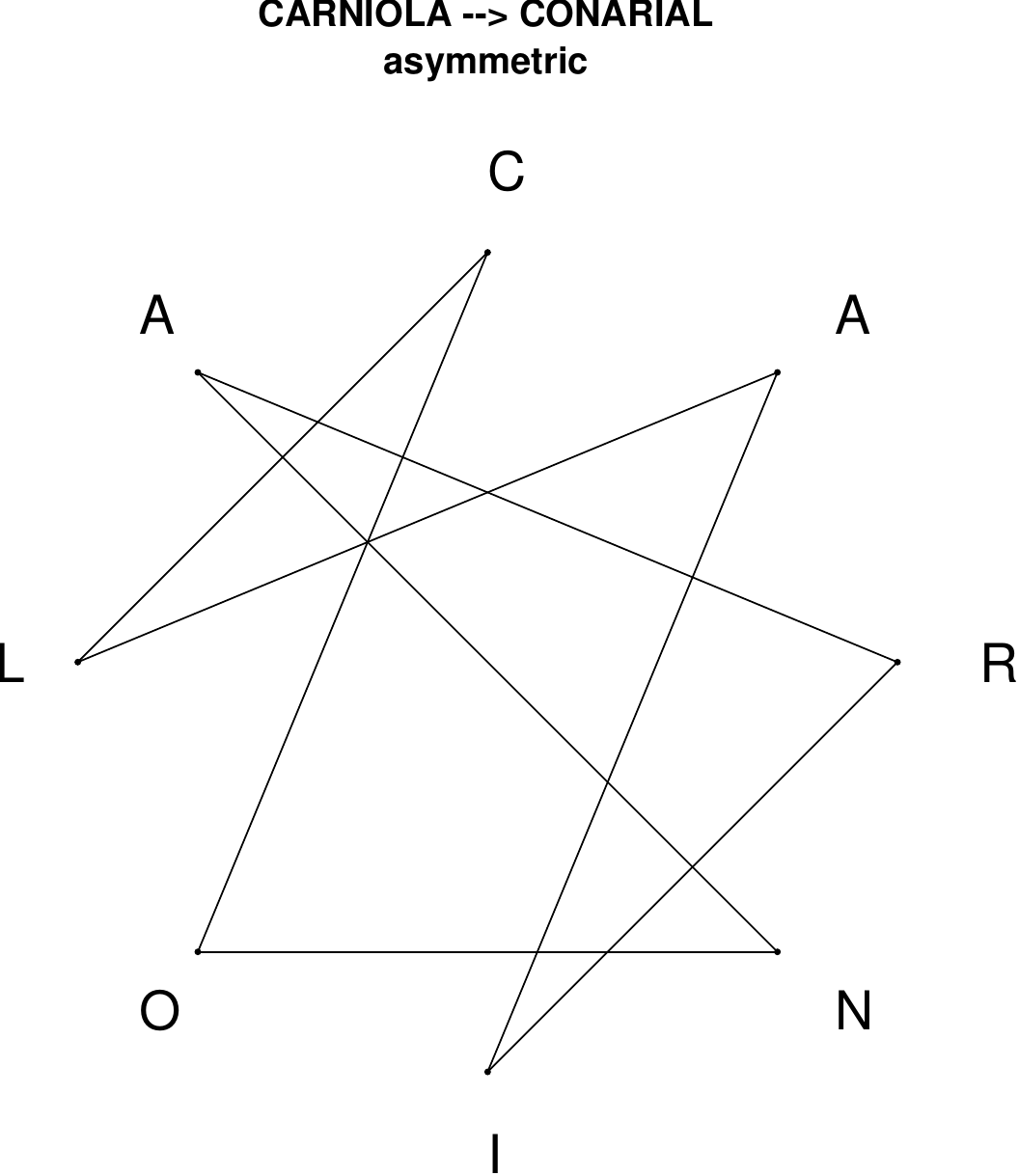}
\end{subfigure}
\hfill
\begin{subfigure}[T]{0.19\textwidth}
\centering
\includegraphics[width=\textwidth]{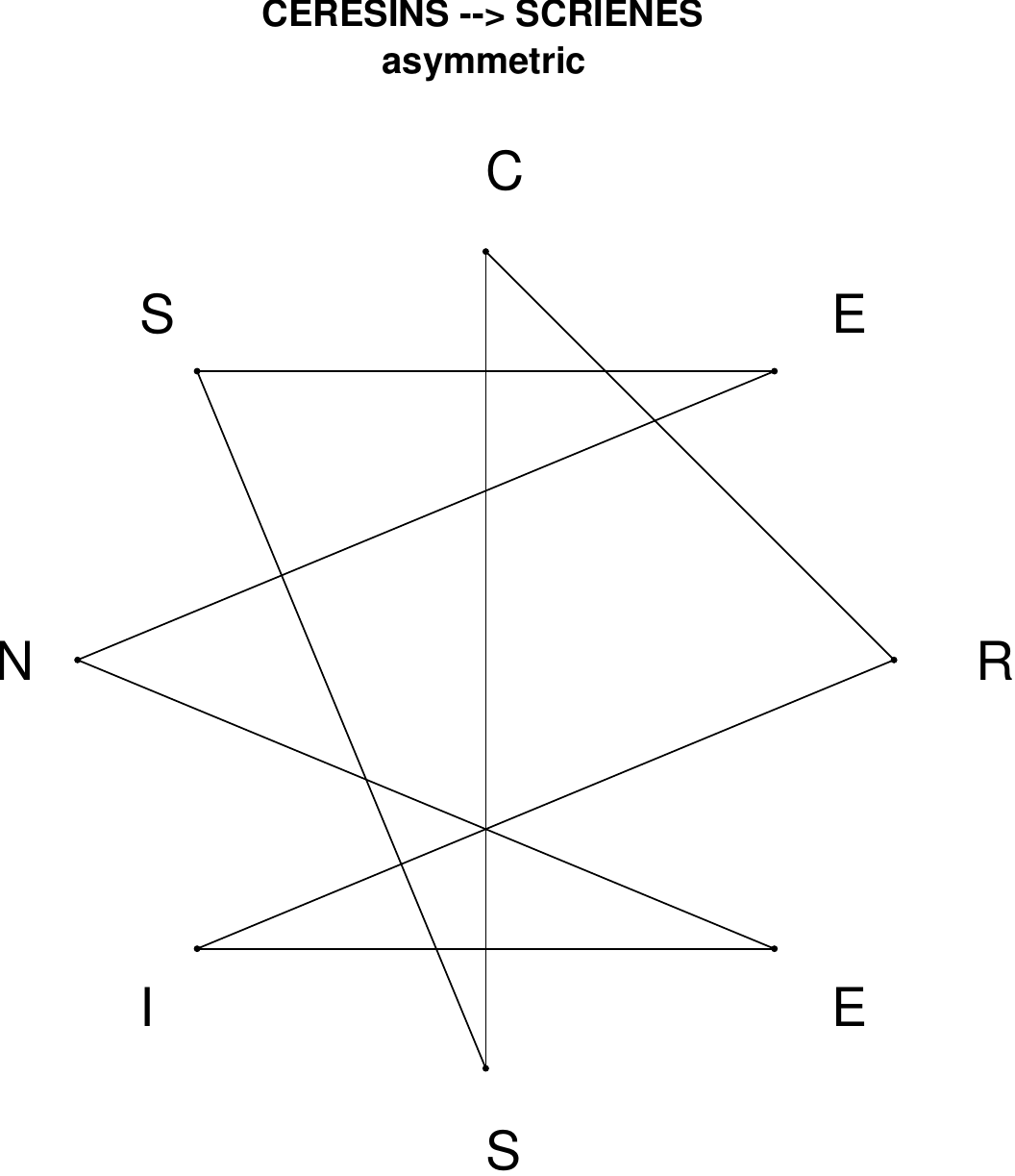}
\end{subfigure}
\hfill
\begin{subfigure}[T]{0.19\textwidth}
\centering
\includegraphics[width=\textwidth]{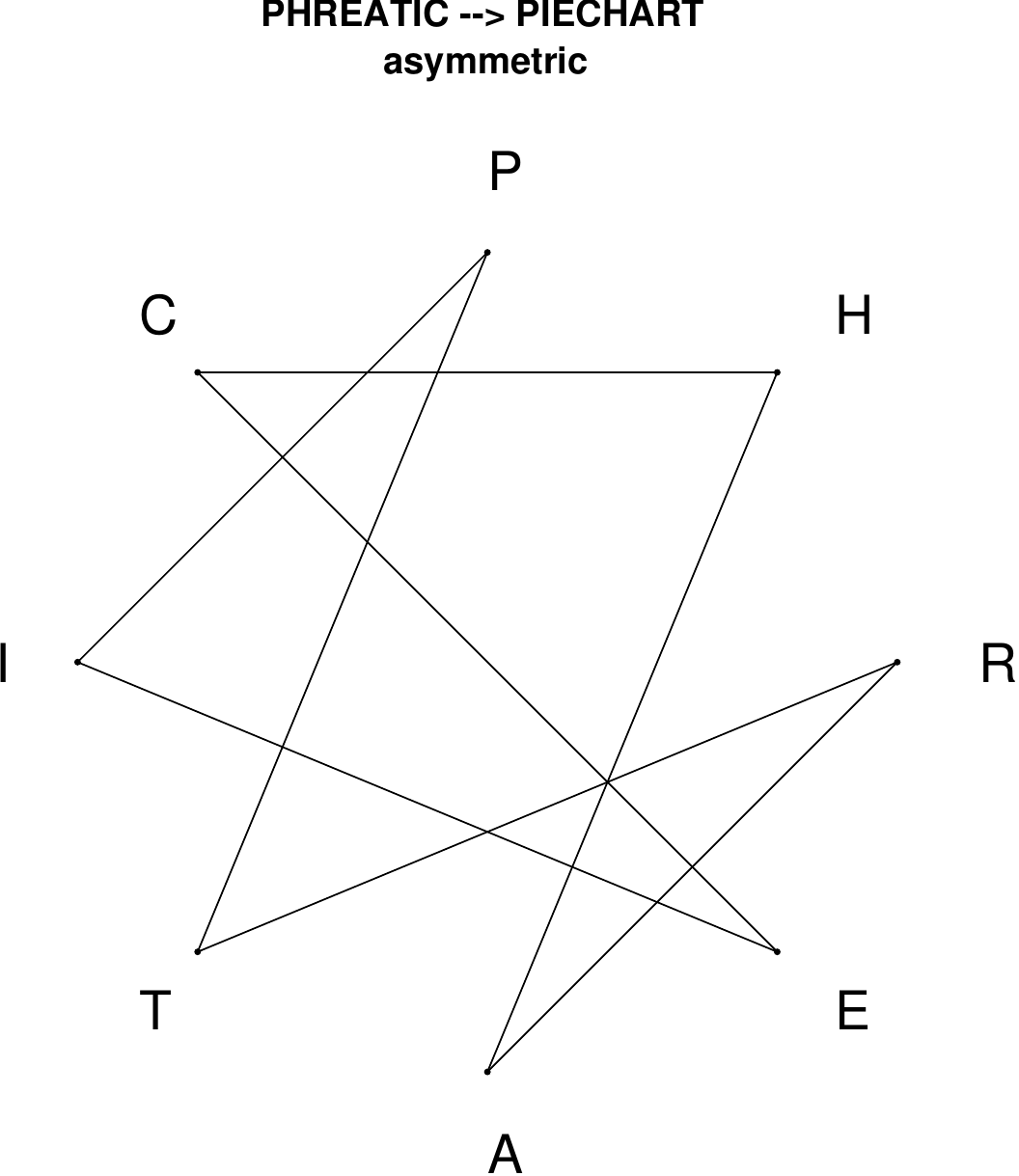}
\end{subfigure}
\hfill
\begin{subfigure}[T]{0.19\textwidth}
\centering
\includegraphics[width=\textwidth]{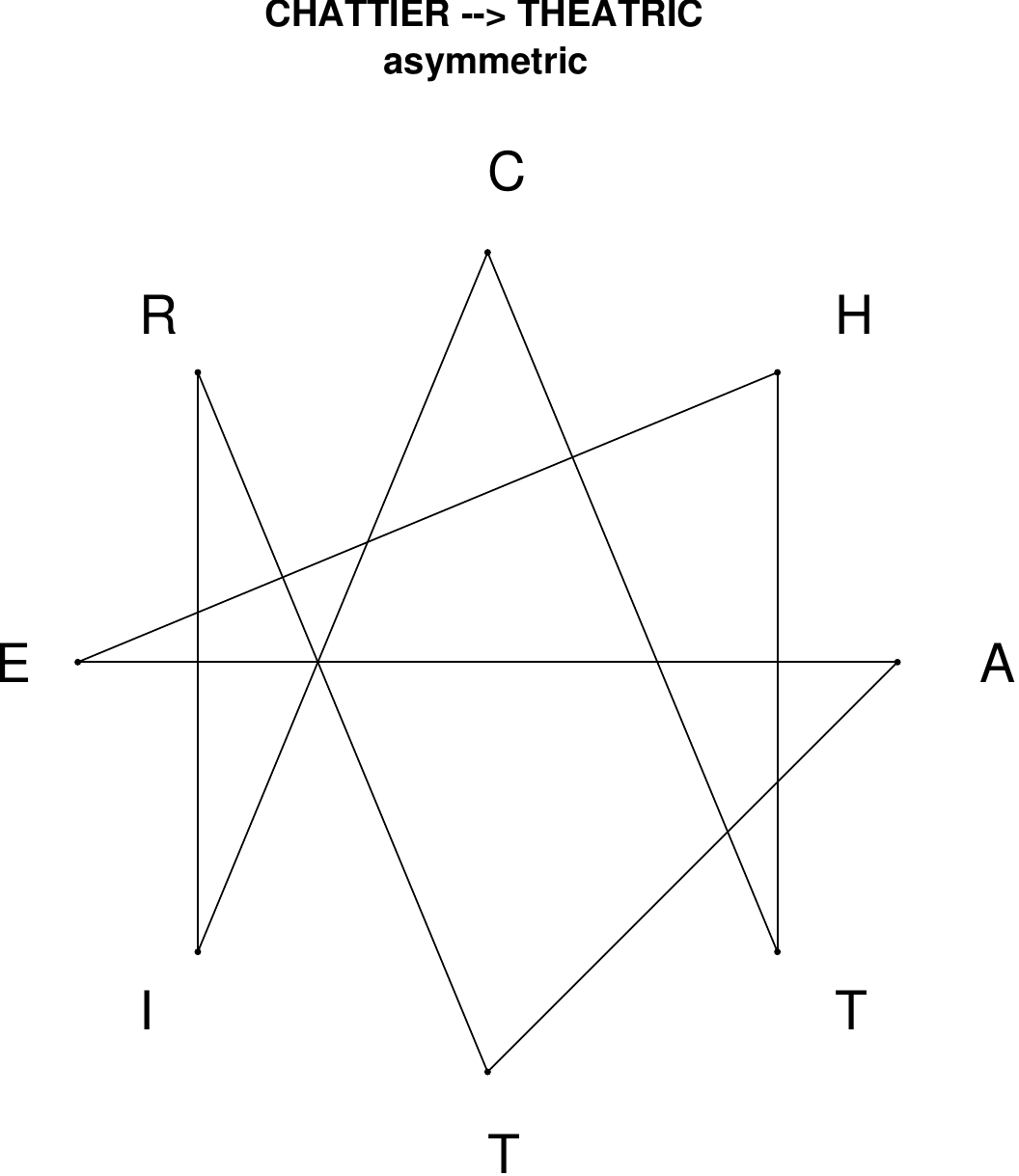}
\end{subfigure}
\end{figure}

\begin{figure}[H]
\centering
\begin{subfigure}[T]{0.19\textwidth}
\centering
\includegraphics[width=\textwidth]{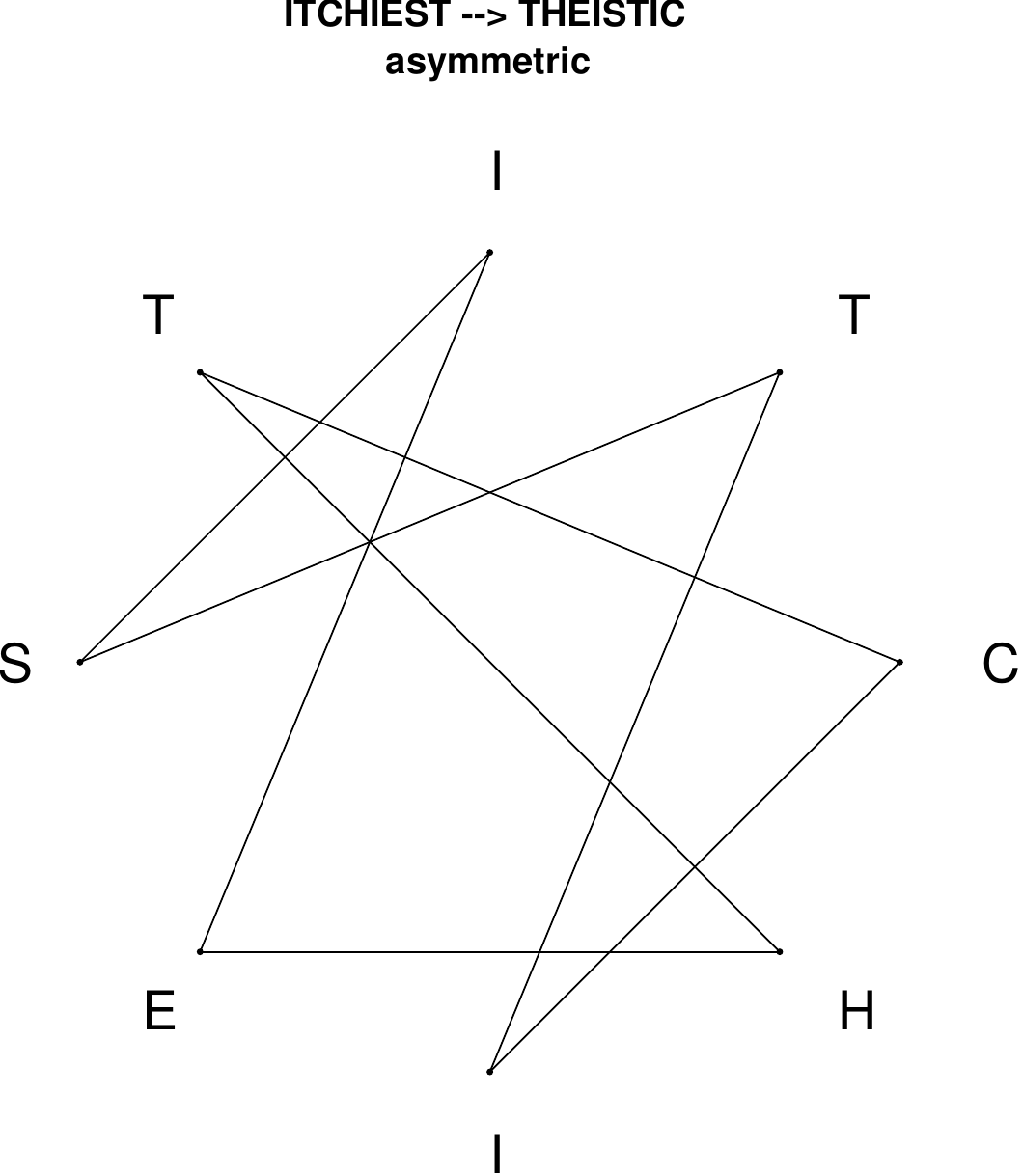}
\end{subfigure}
\hfill
\begin{subfigure}[T]{0.19\textwidth}
\centering
\includegraphics[width=\textwidth]{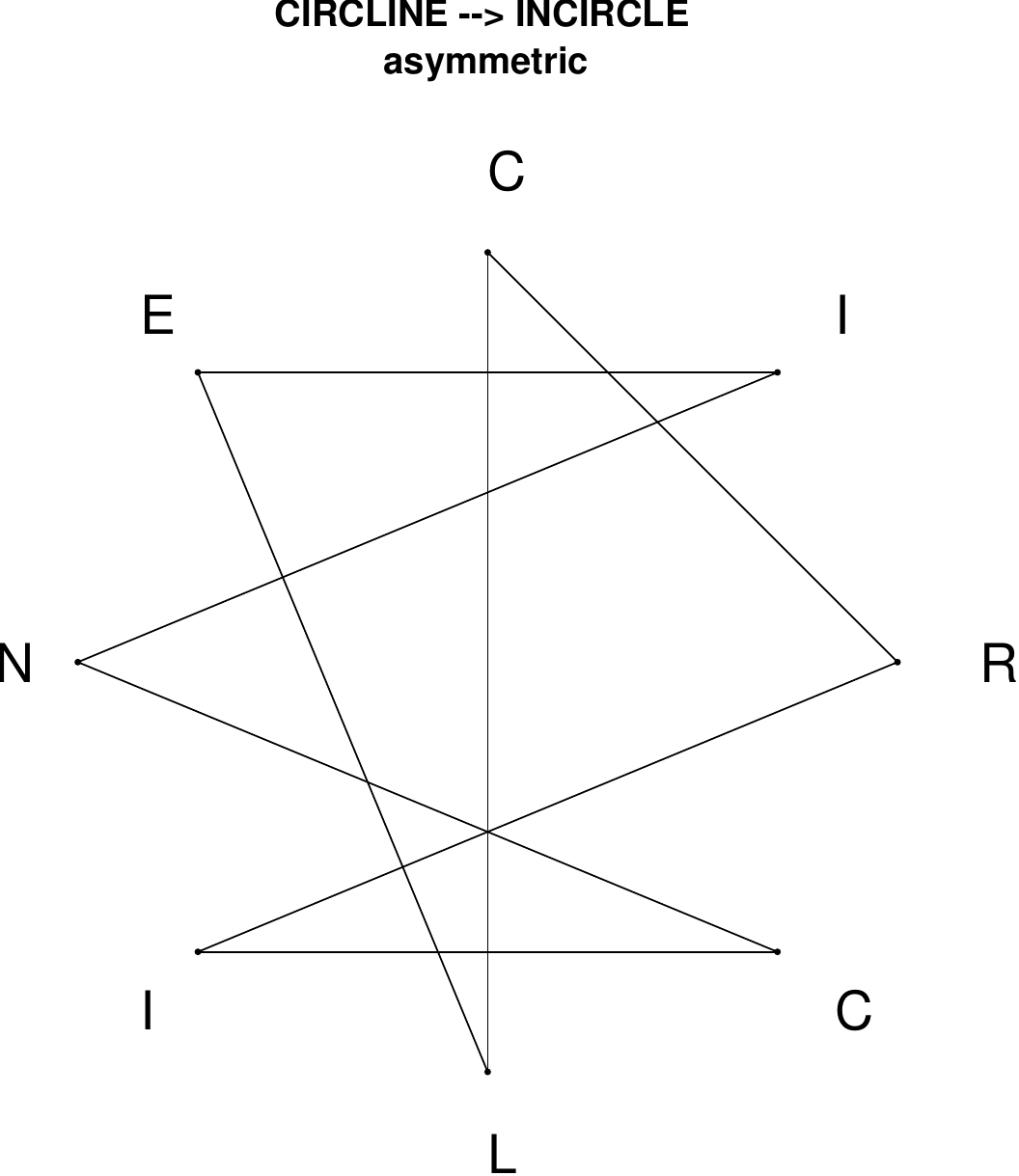}
\end{subfigure}
\hfill
\begin{subfigure}[T]{0.19\textwidth}
\centering
\includegraphics[width=\textwidth]{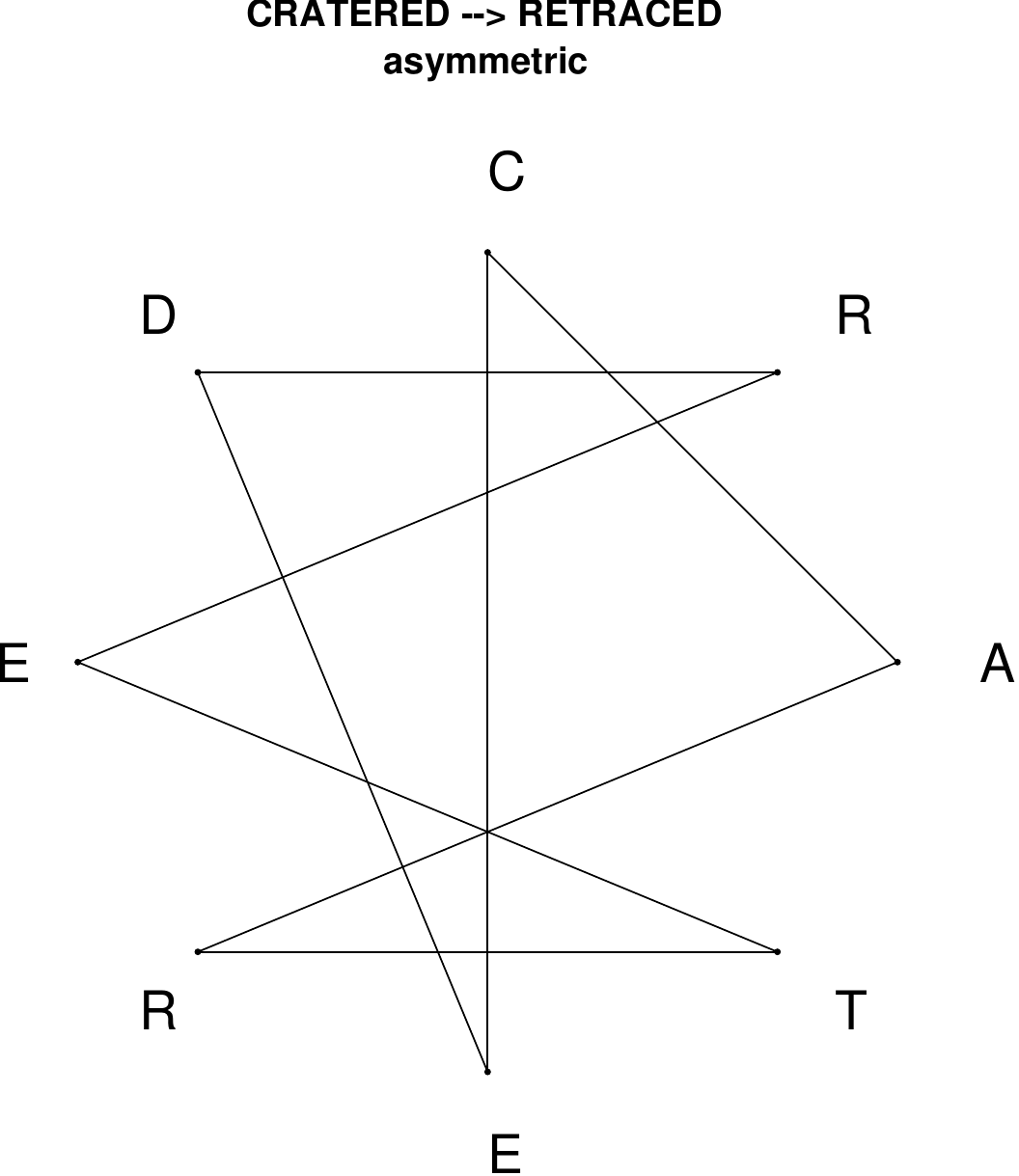}
\end{subfigure}
\hfill
\begin{subfigure}[T]{0.19\textwidth}
\centering
\includegraphics[width=\textwidth]{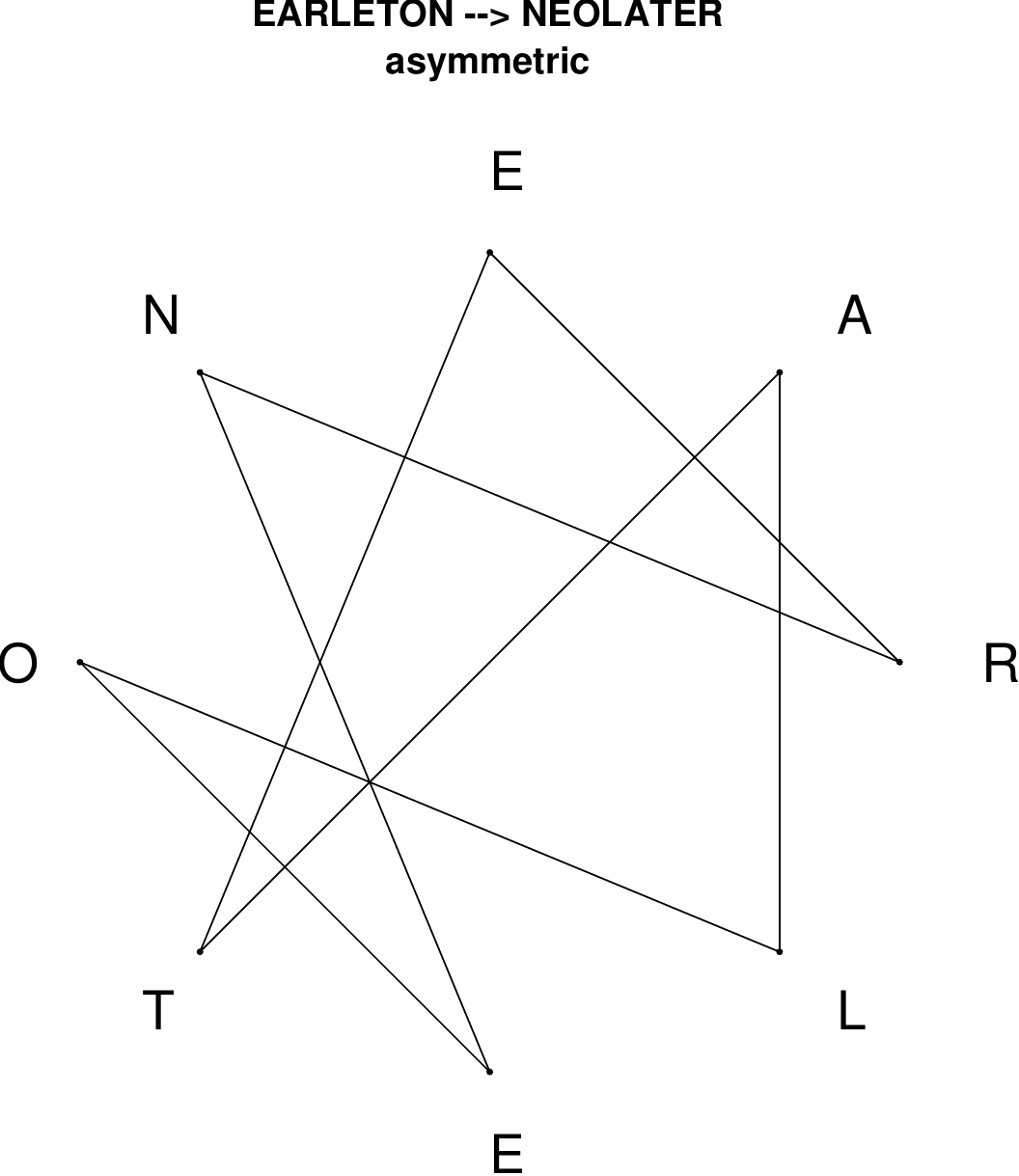}
\end{subfigure}
\hfill
\begin{subfigure}[T]{0.19\textwidth}
\centering
\includegraphics[width=\textwidth]{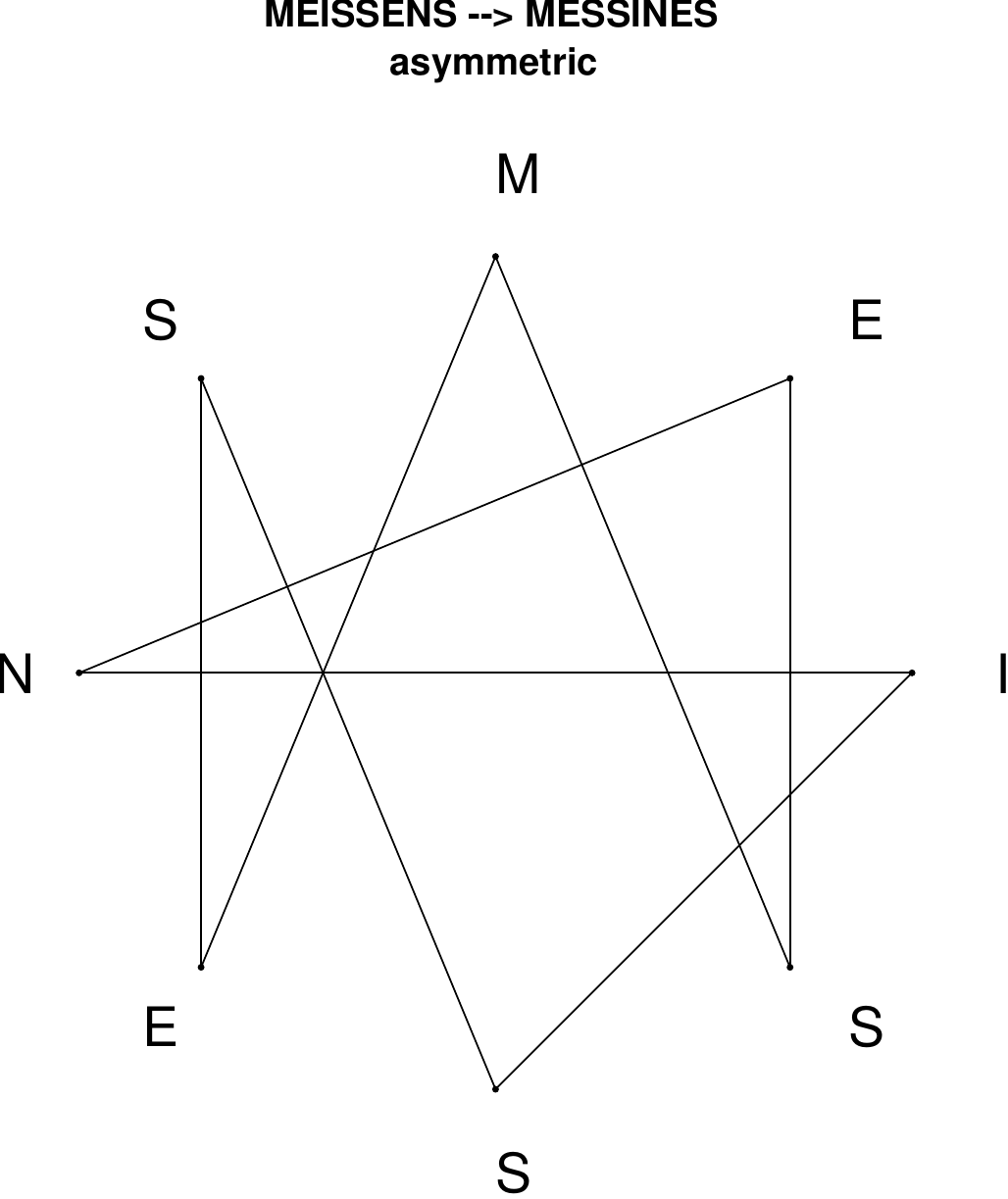}
\end{subfigure}
\end{figure}

\begin{figure}[H]
\centering
\begin{subfigure}[T]{0.19\textwidth}
\centering
\includegraphics[width=\textwidth]{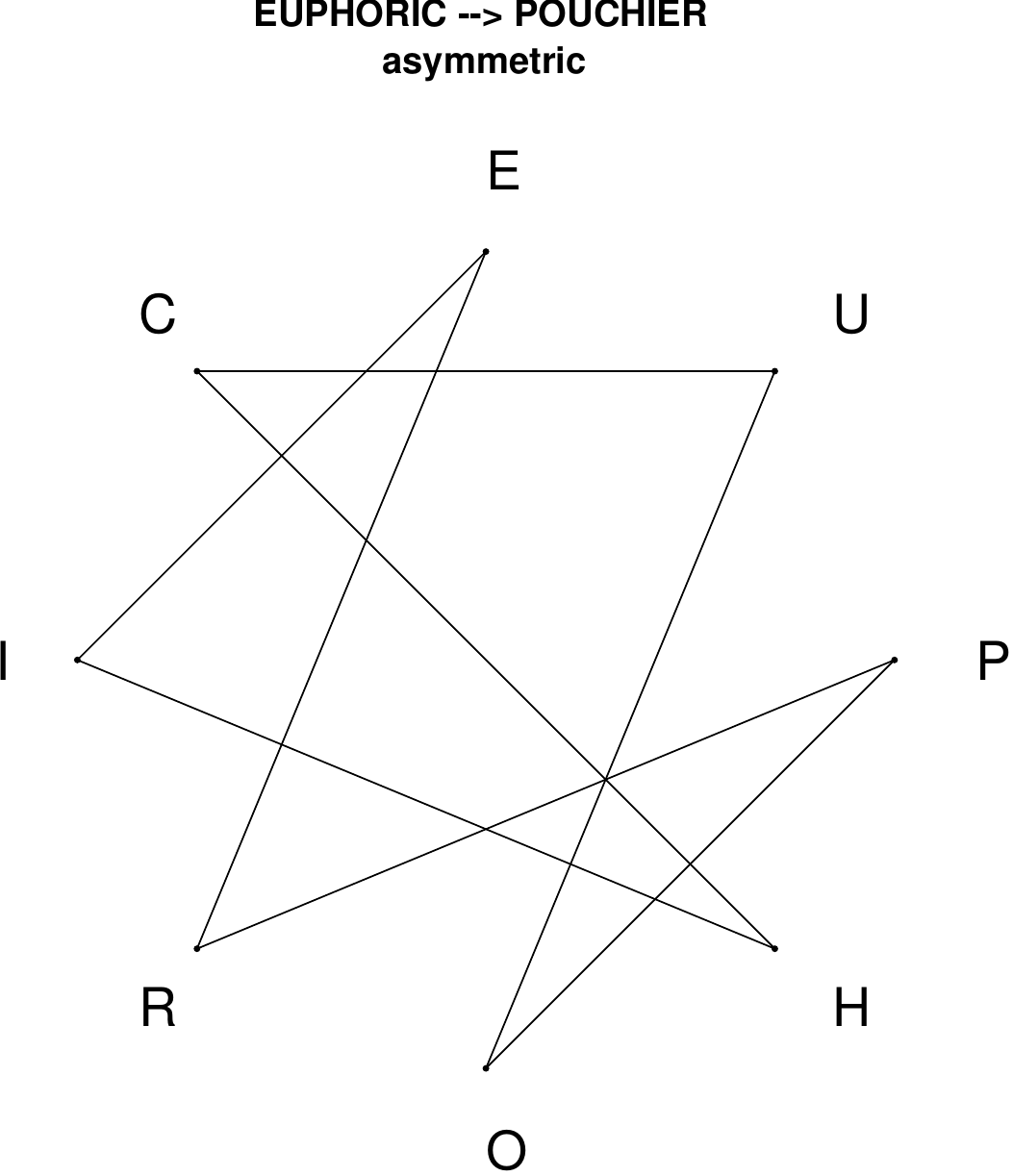}
\end{subfigure}
\hfill
\begin{subfigure}[T]{0.19\textwidth}
\centering
\includegraphics[width=\textwidth]{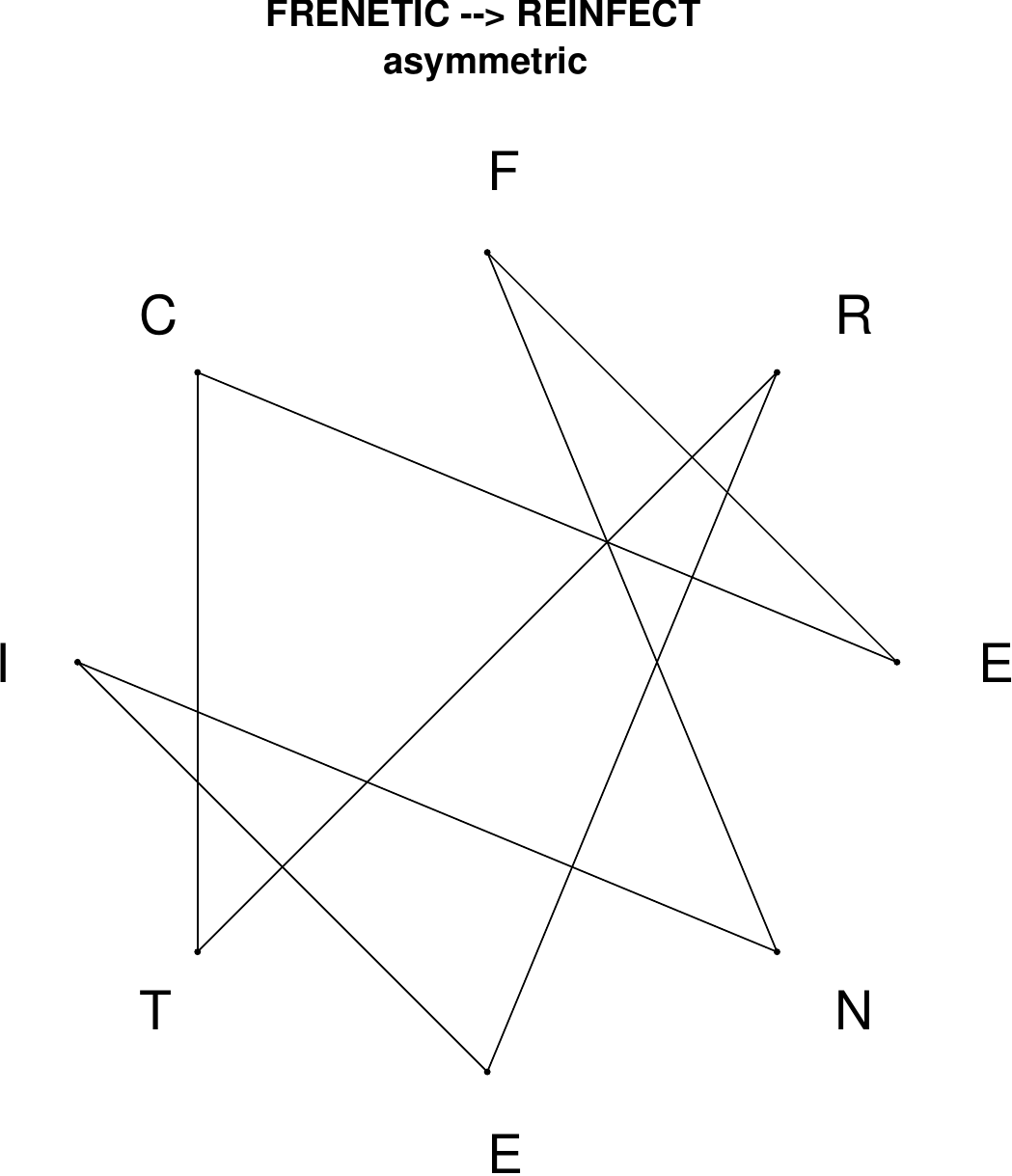}
\end{subfigure}
\hfill
\begin{subfigure}[T]{0.19\textwidth}
\centering
\includegraphics[width=\textwidth]{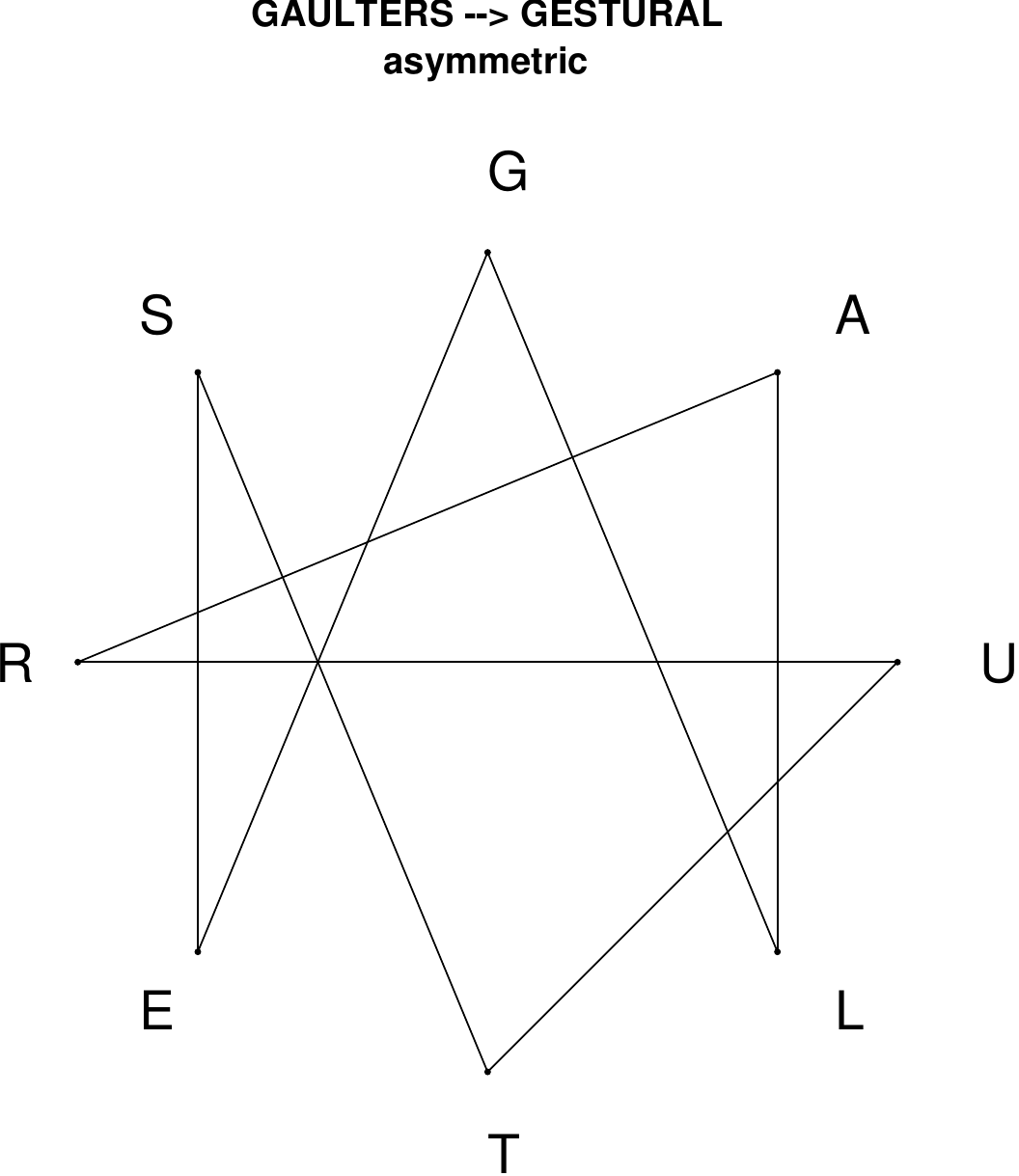}
\end{subfigure}
\hfill
\begin{subfigure}[T]{0.19\textwidth}
\centering
\includegraphics[width=\textwidth]{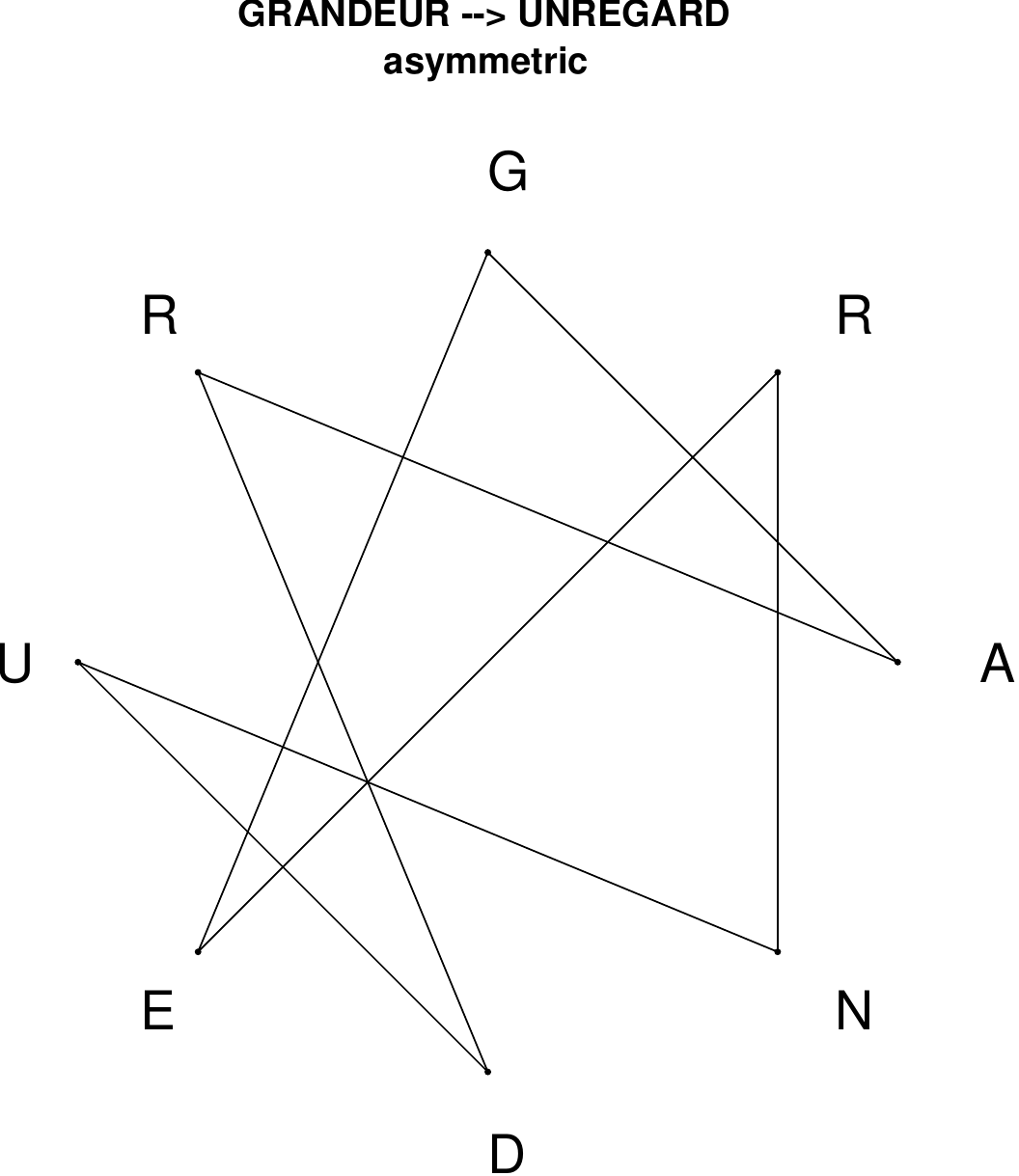}
\end{subfigure}
\hfill
\begin{subfigure}[T]{0.19\textwidth}
\centering
\includegraphics[width=\textwidth]{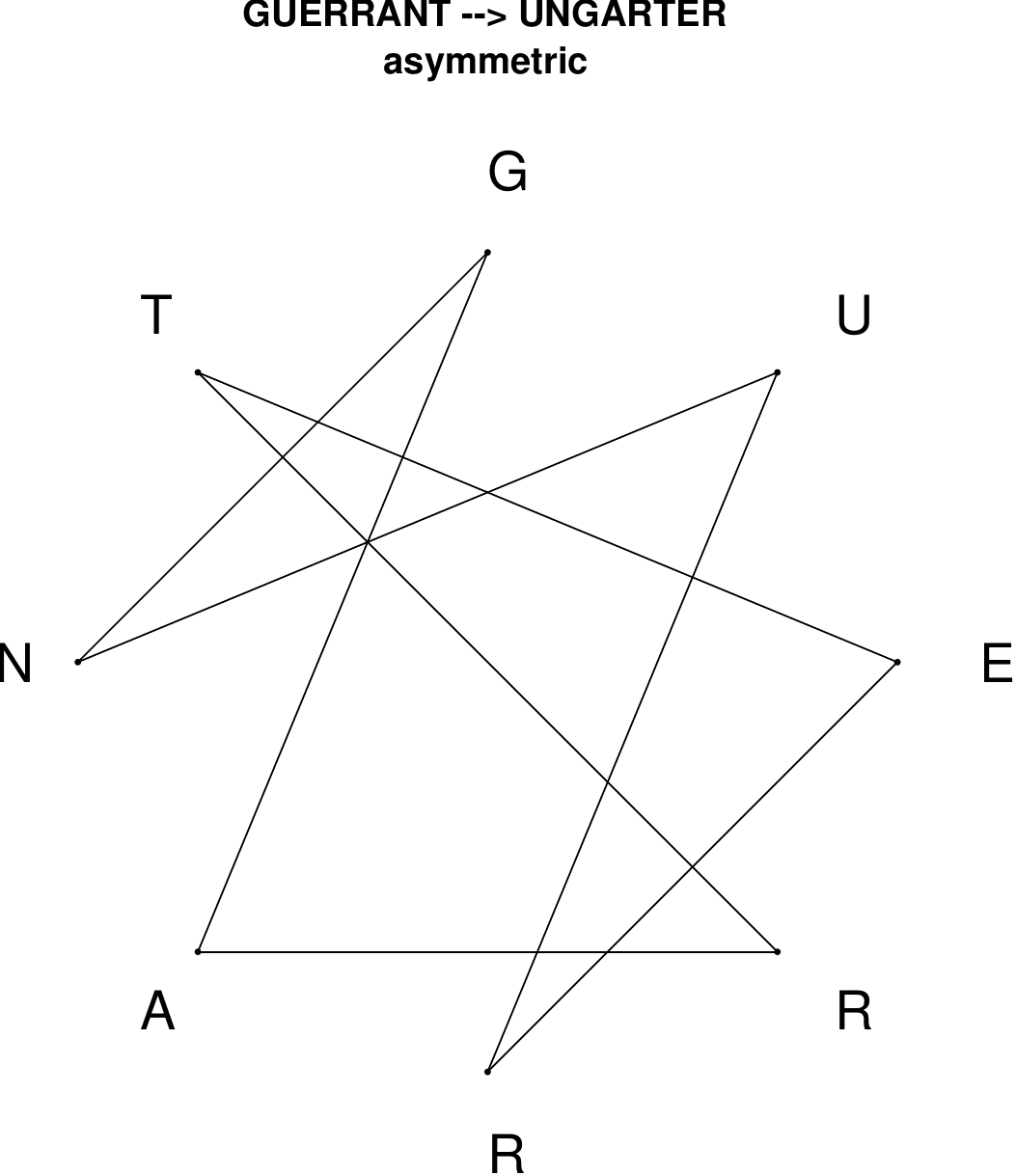}
\end{subfigure}
\end{figure}

\begin{figure}[H]
\centering
\begin{subfigure}[T]{0.19\textwidth}
\centering
\includegraphics[width=\textwidth]{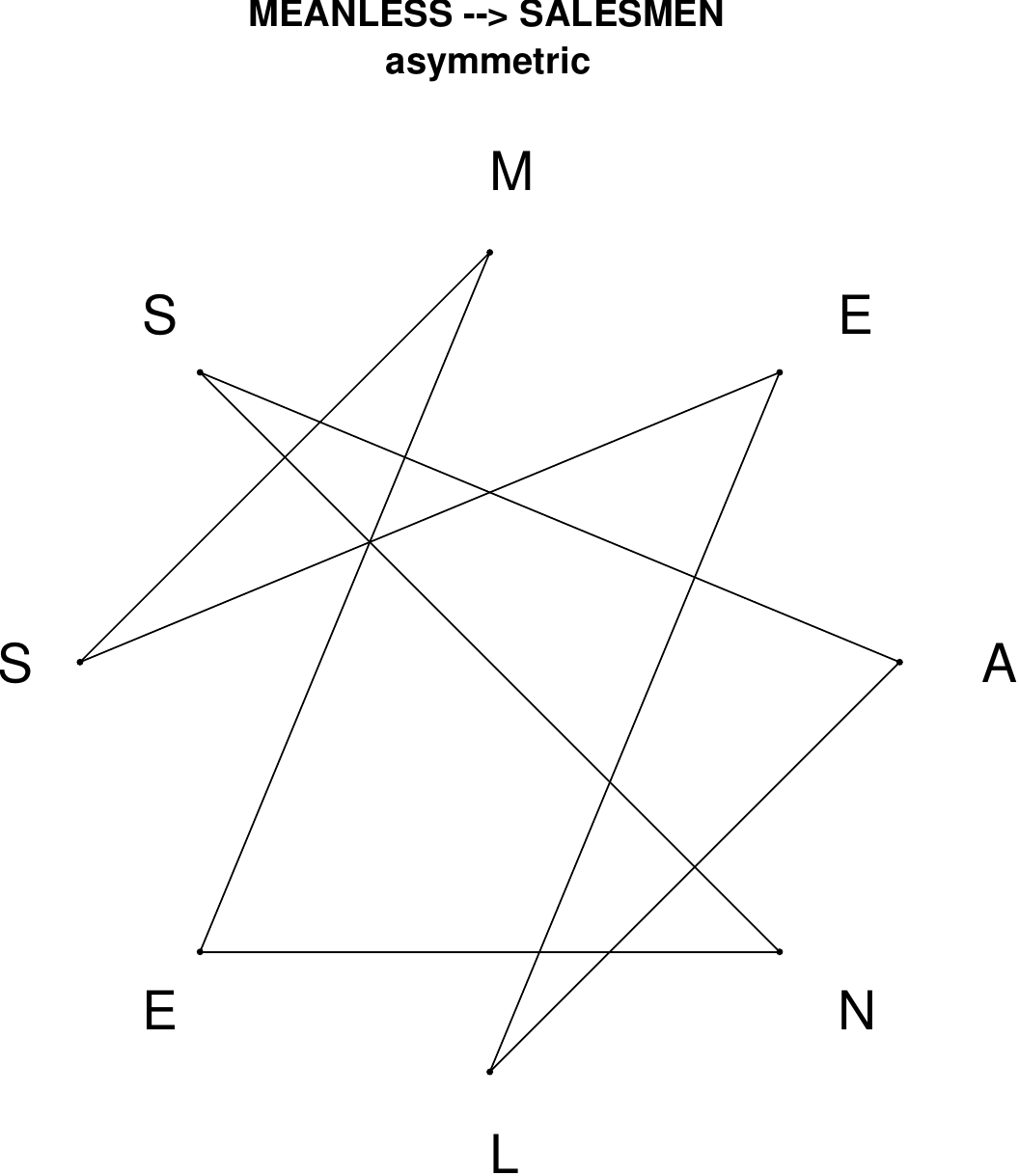}
\end{subfigure}
\hfill
\begin{subfigure}[T]{0.19\textwidth}
\centering
\includegraphics[width=\textwidth]{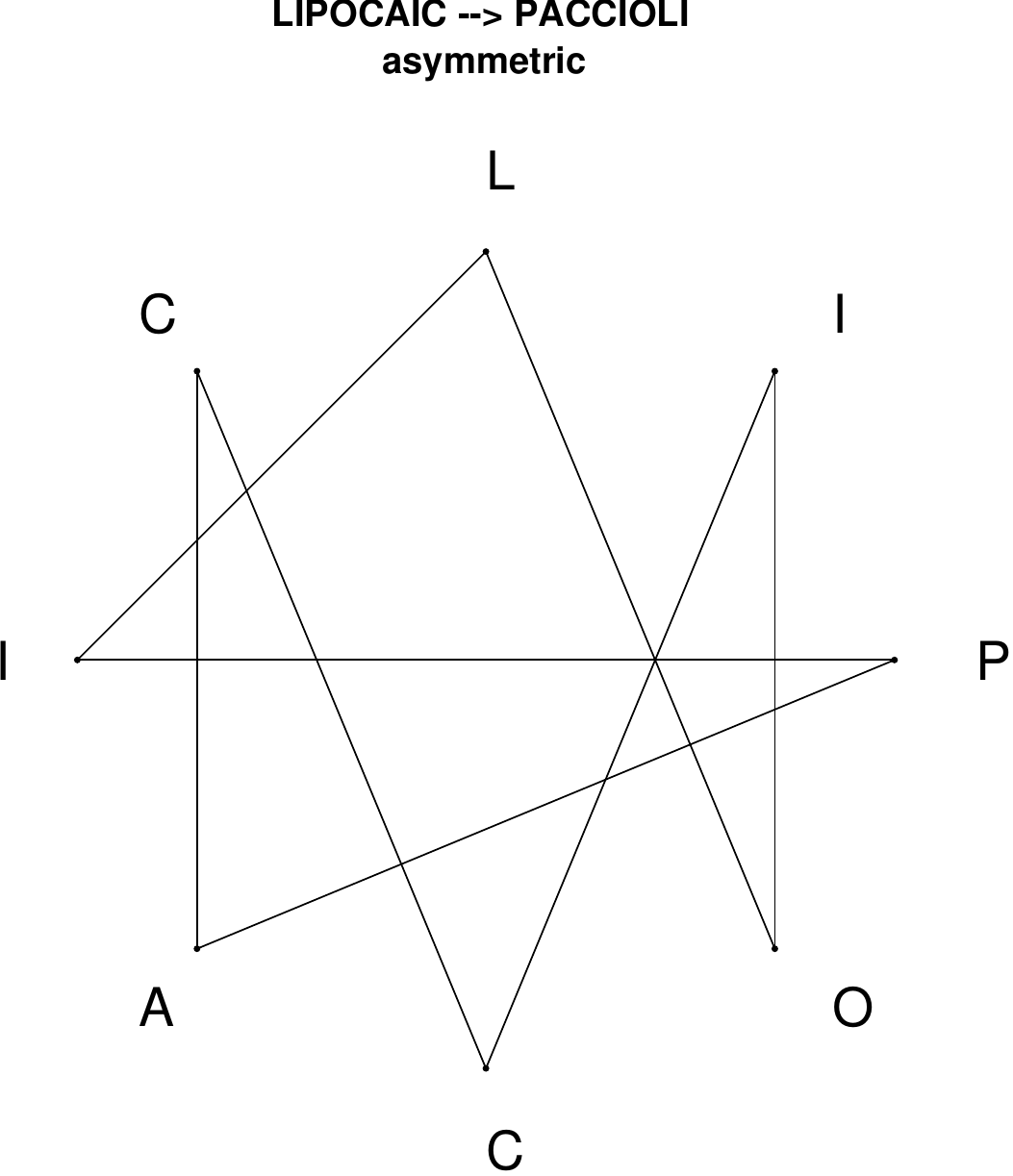}
\end{subfigure}
\hfill
\begin{subfigure}[T]{0.19\textwidth}
\centering
\includegraphics[width=\textwidth]{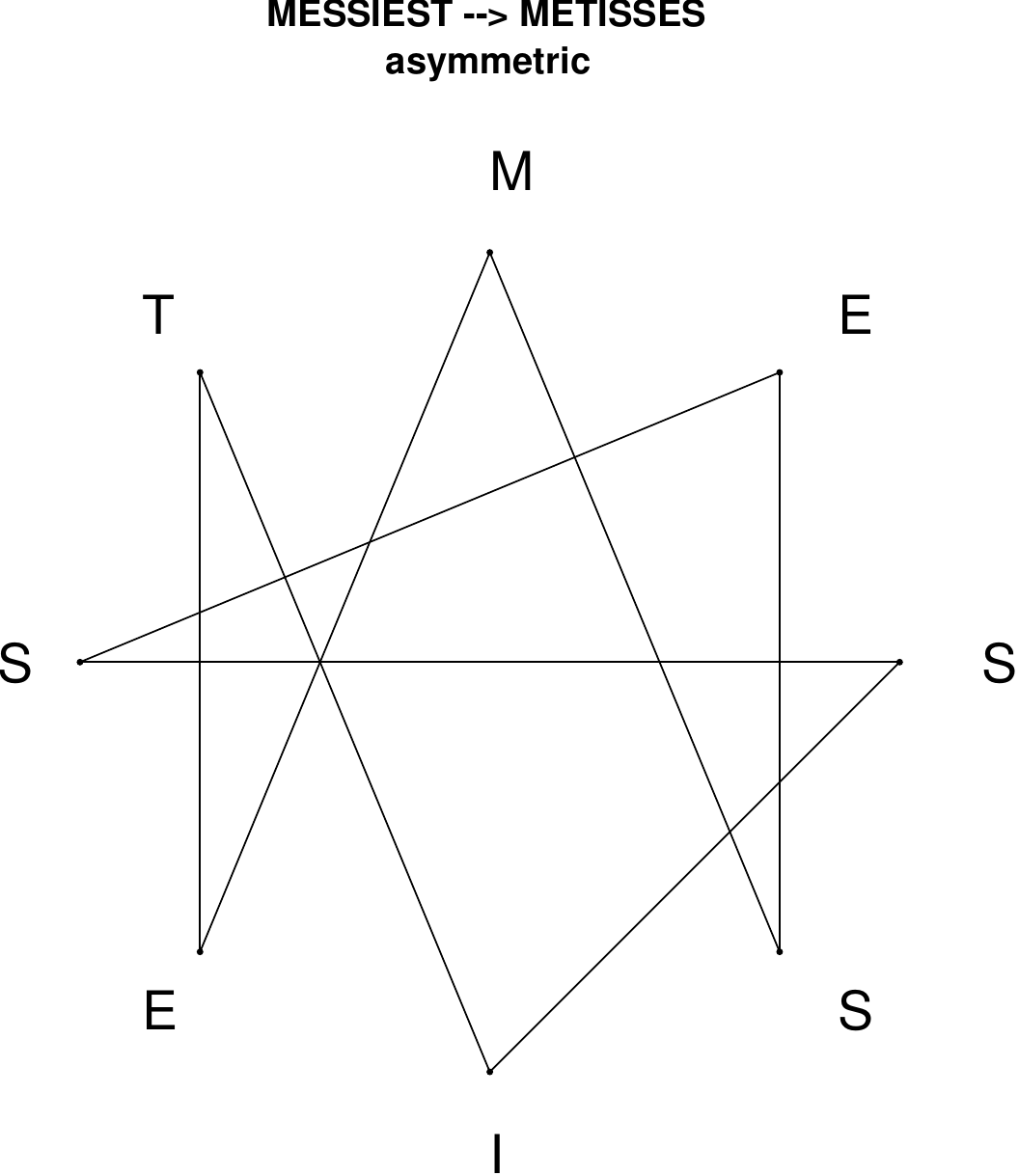}
\end{subfigure}
\hfill
\begin{subfigure}[T]{0.19\textwidth}
\centering
\includegraphics[width=\textwidth]{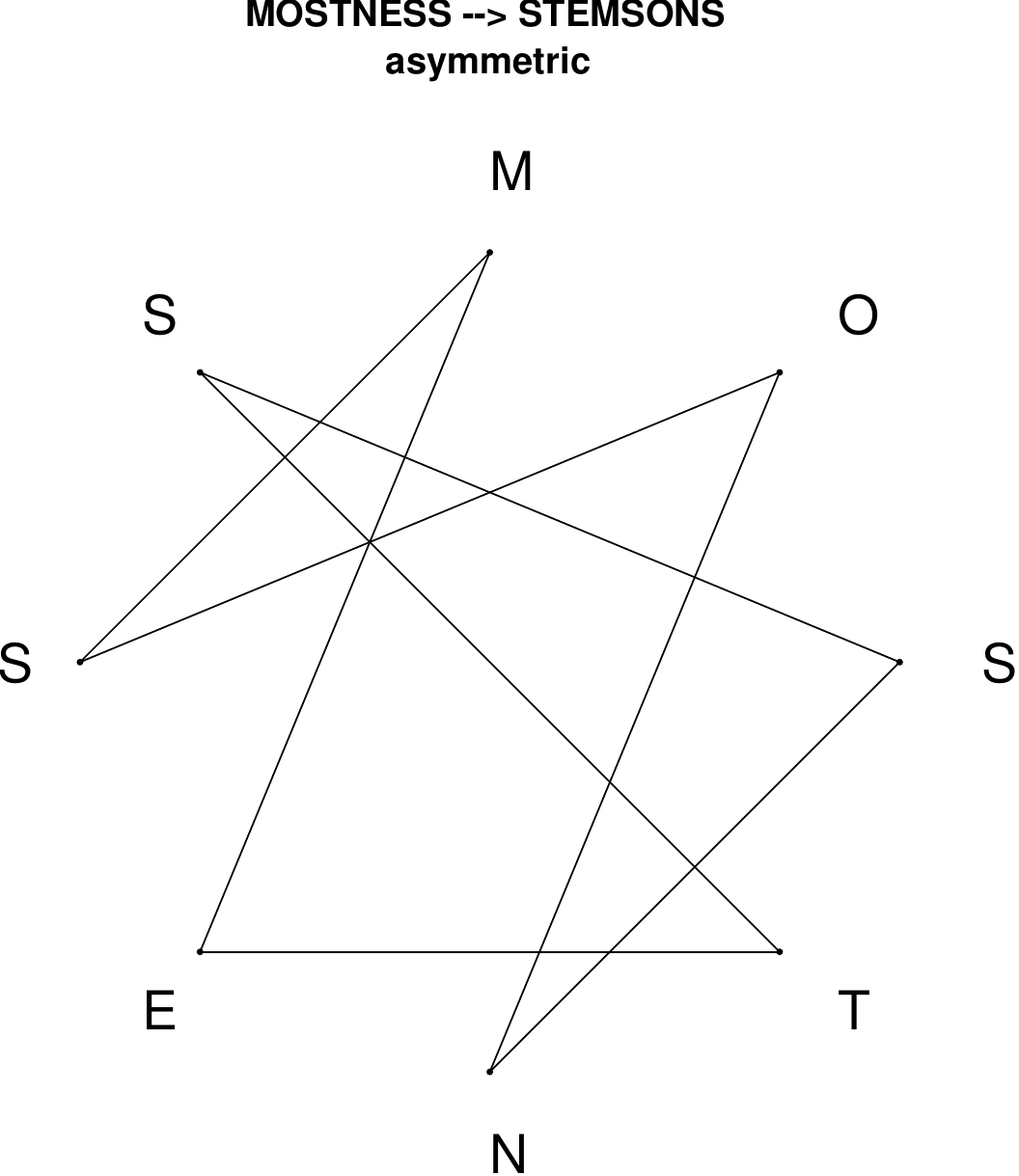}
\end{subfigure}
\hfill
\begin{subfigure}[T]{0.19\textwidth}
\centering
\includegraphics[width=\textwidth]{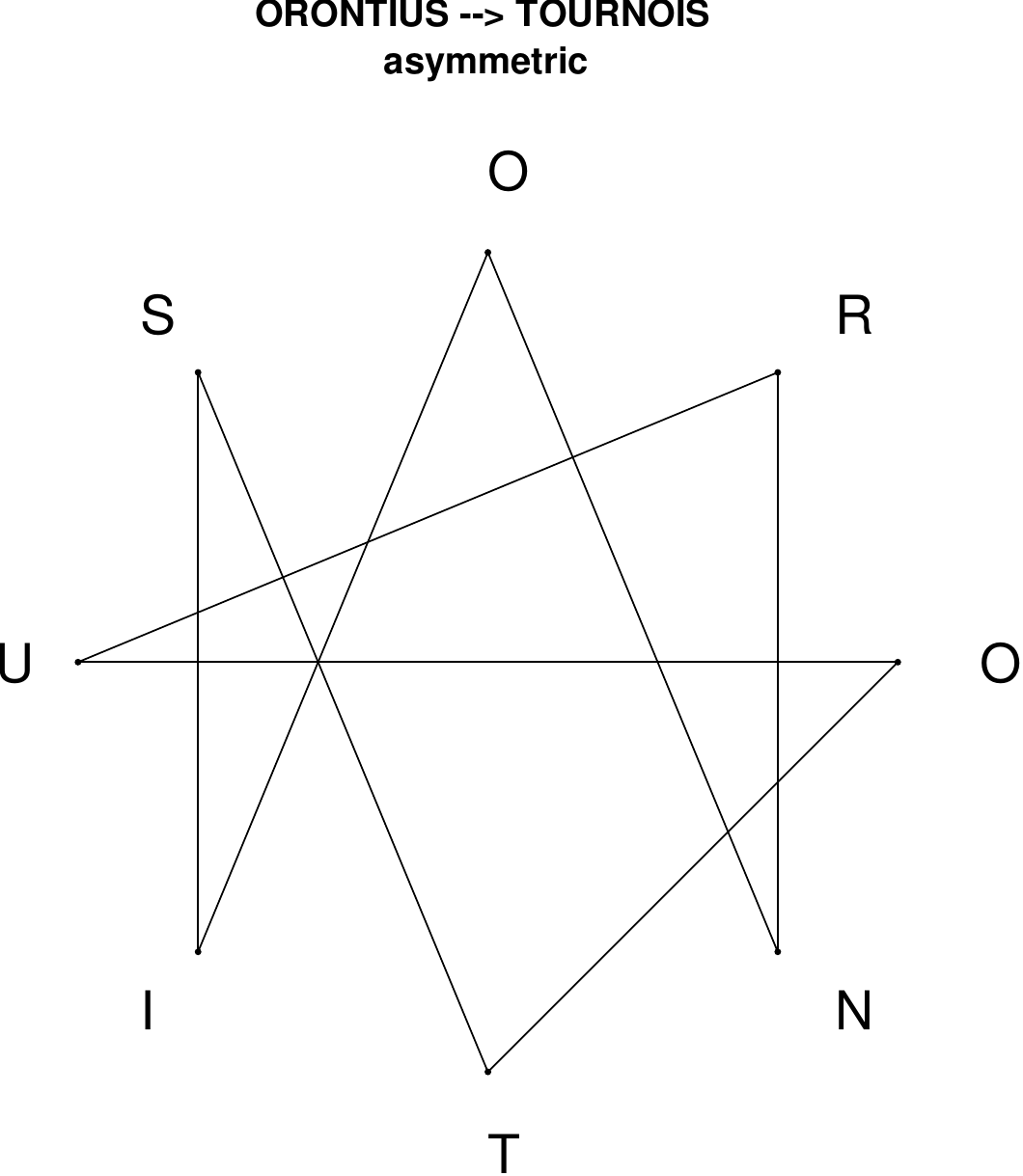}
\end{subfigure}
\end{figure}

\begin{figure}[H]
\centering
\begin{subfigure}[T]{0.19\textwidth}
\centering
\includegraphics[width=\textwidth]{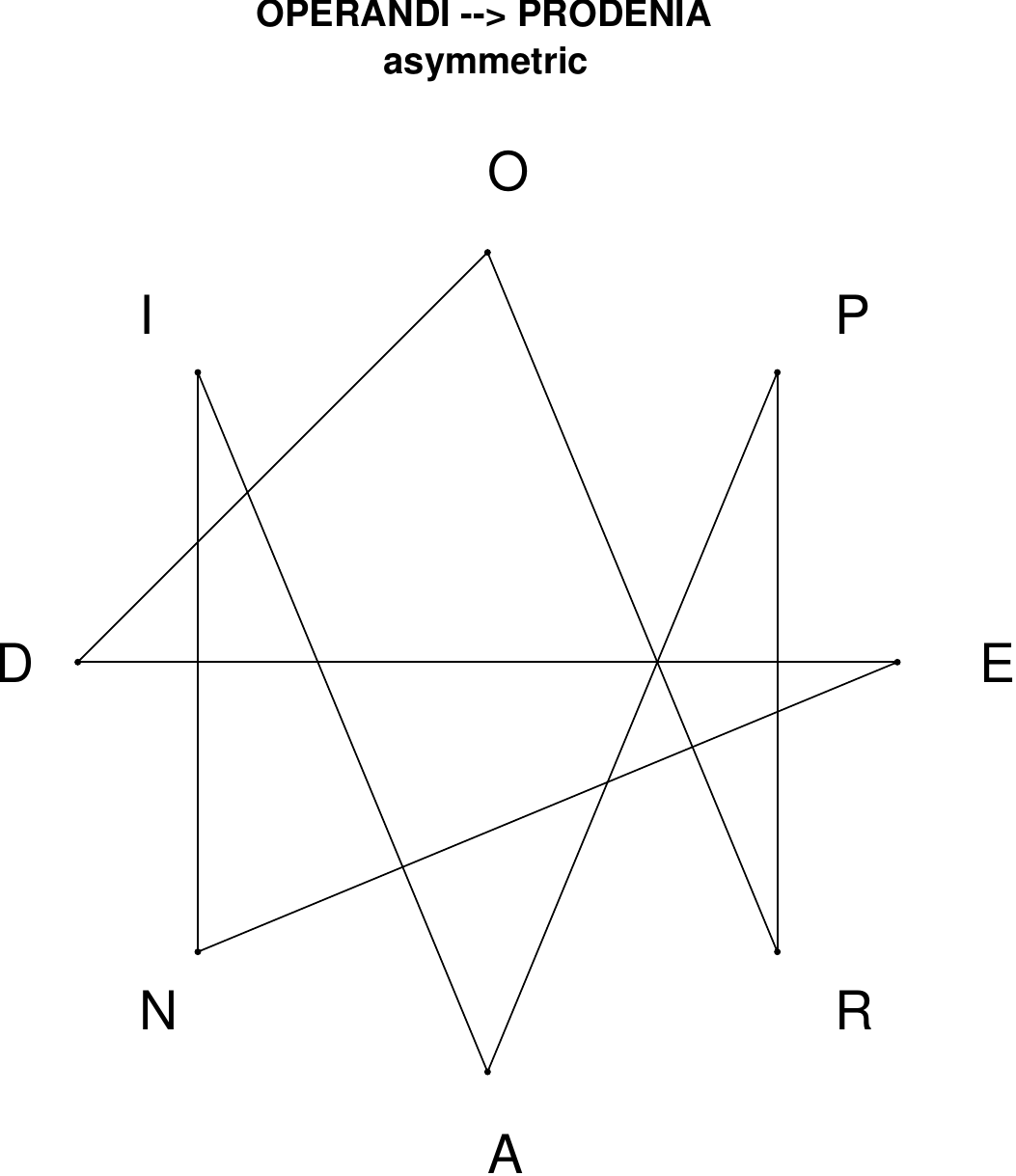}
\end{subfigure}
\hfill
\begin{subfigure}[T]{0.19\textwidth}
\centering
\includegraphics[width=\textwidth]{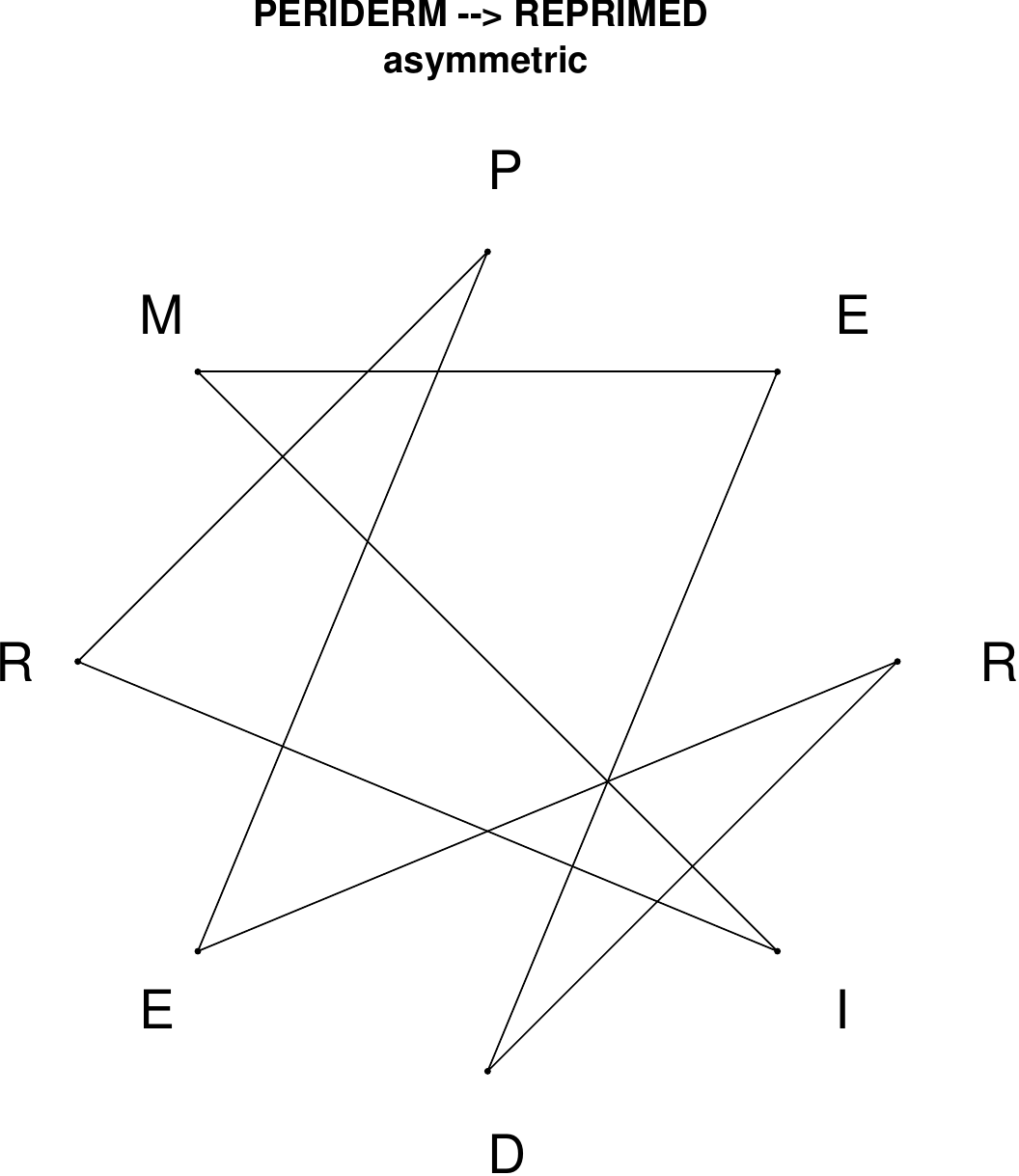}
\end{subfigure}
\hfill
\begin{subfigure}[T]{0.19\textwidth}
\centering
\includegraphics[width=\textwidth]{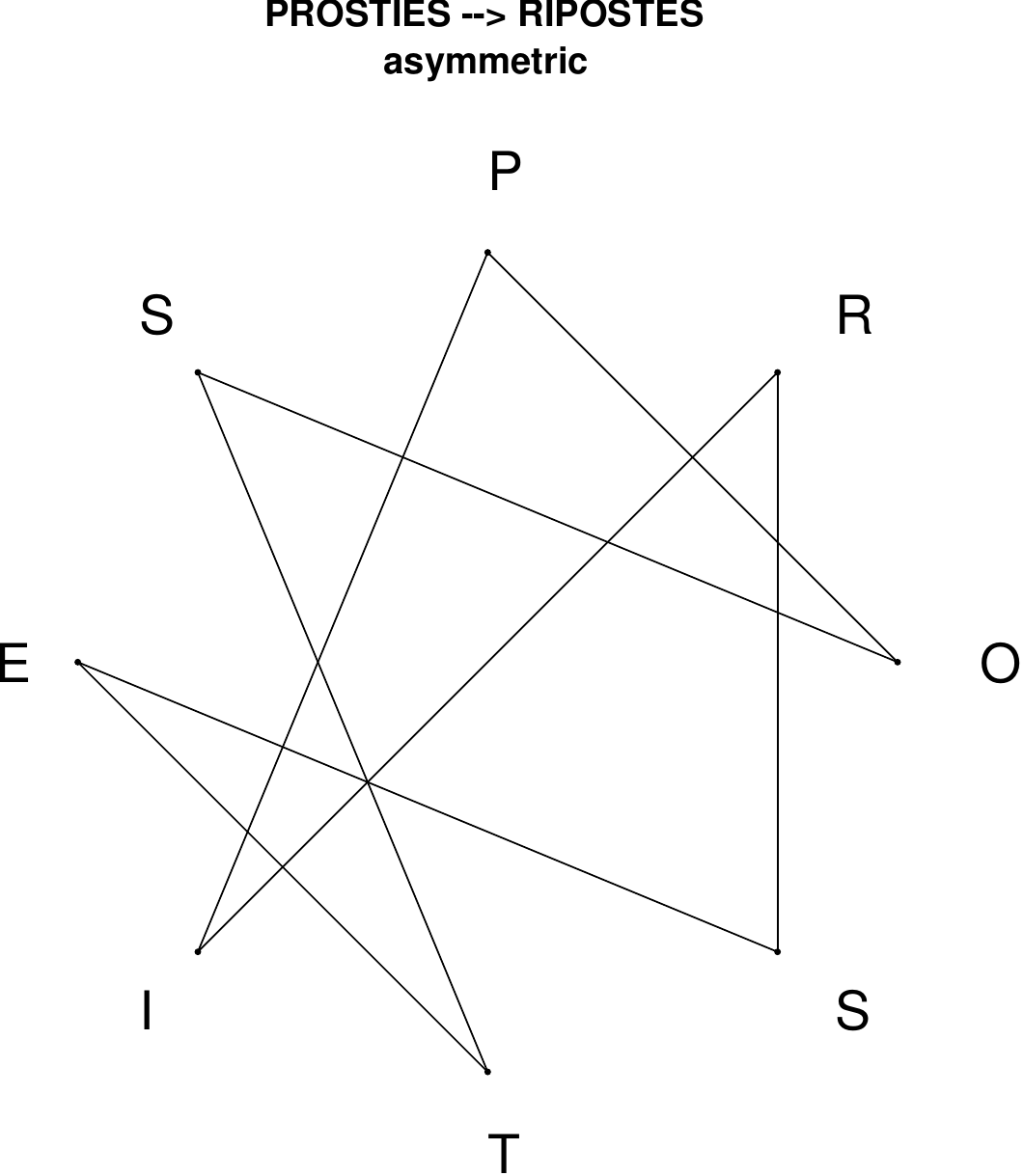}
\end{subfigure}
\hfill
\begin{subfigure}[T]{0.19\textwidth}
\centering
\includegraphics[width=\textwidth]{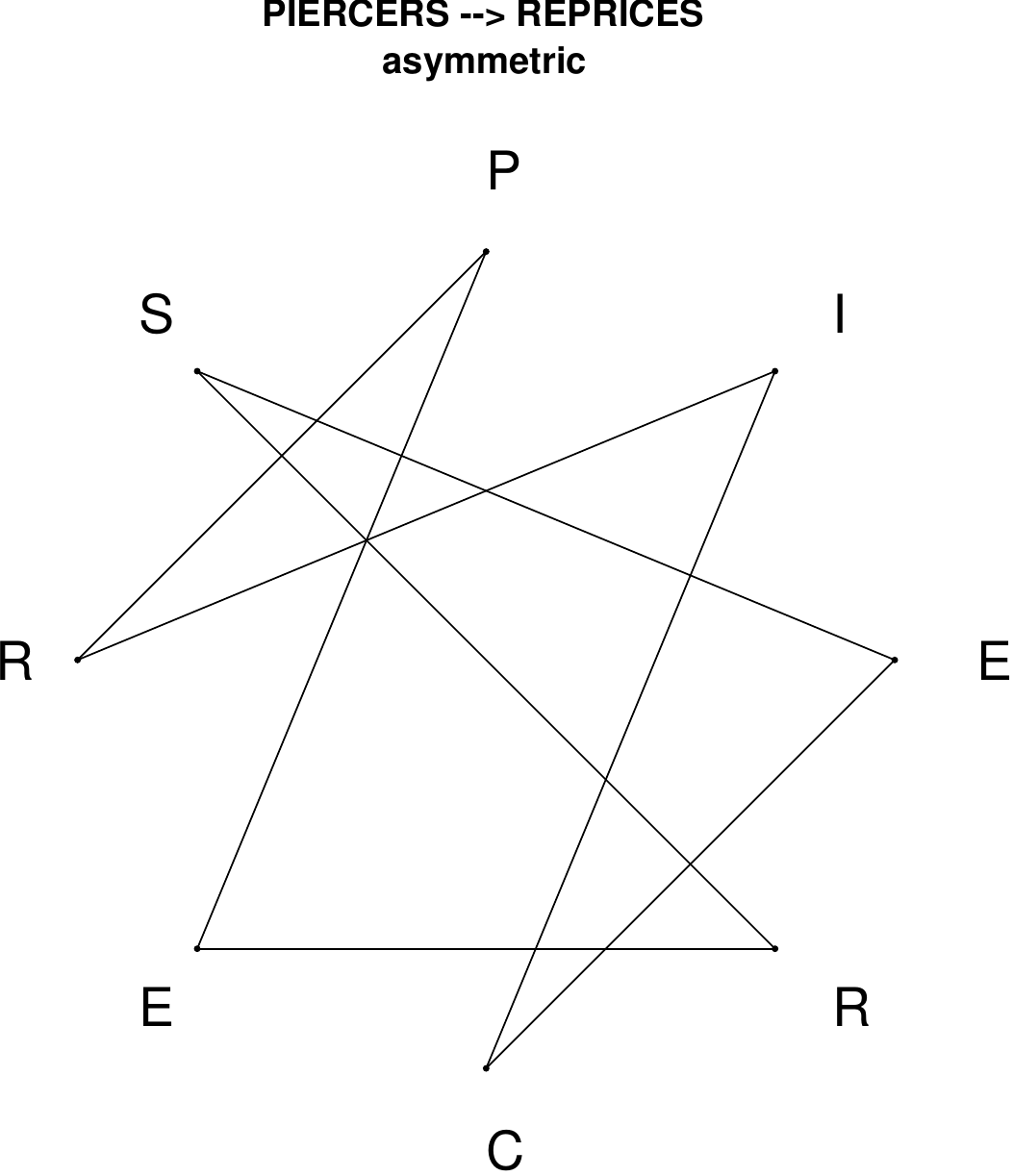}
\end{subfigure}
\hfill
\begin{subfigure}[T]{0.19\textwidth}
\centering
\includegraphics[width=\textwidth]{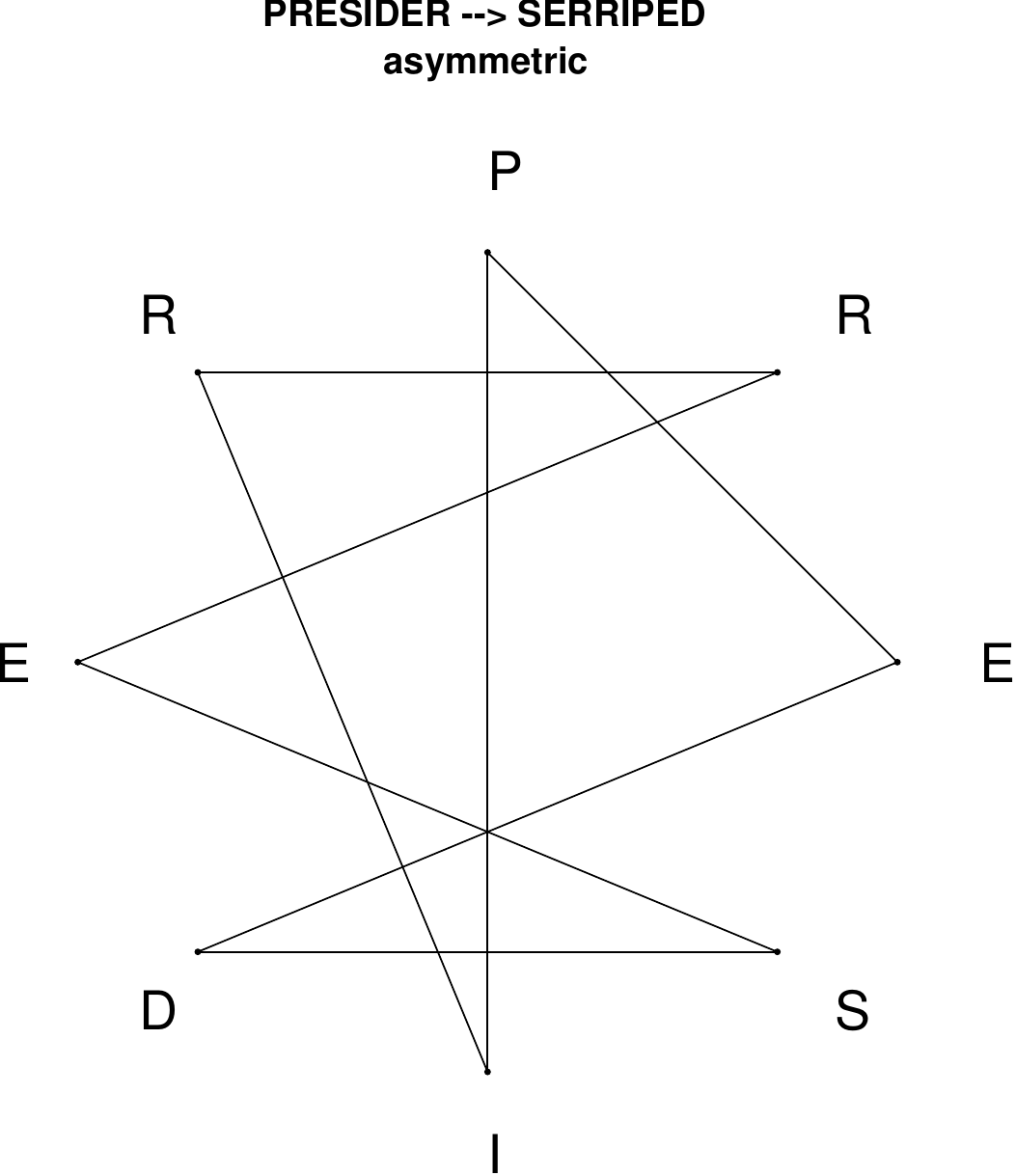}
\end{subfigure}
\end{figure}

\begin{figure}[H]
\centering
\begin{subfigure}[T]{0.19\textwidth}
\centering
\includegraphics[width=\textwidth]{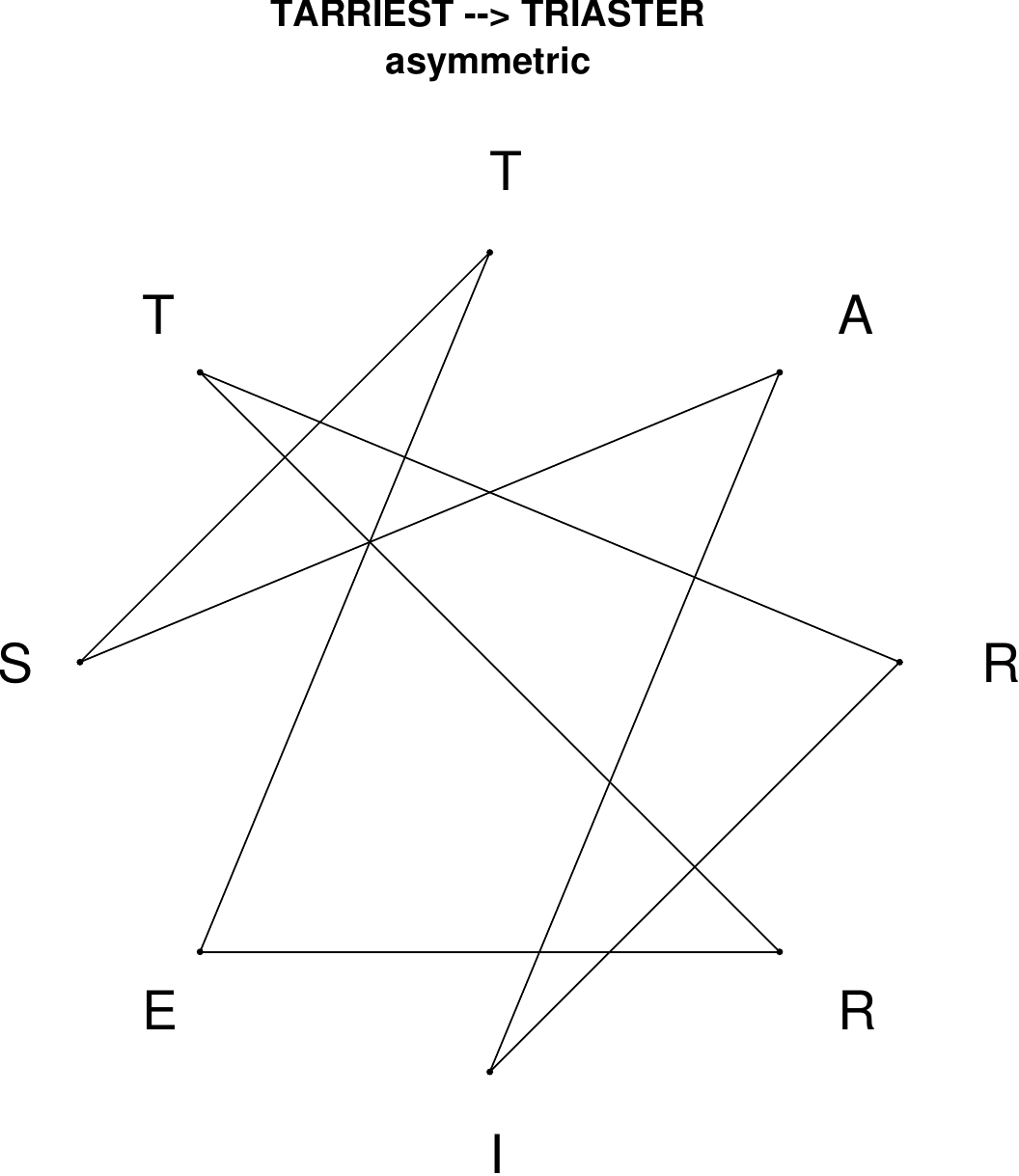}
\end{subfigure}
\hfill
\begin{subfigure}[T]{0.19\textwidth}
\centering
\includegraphics[width=\textwidth]{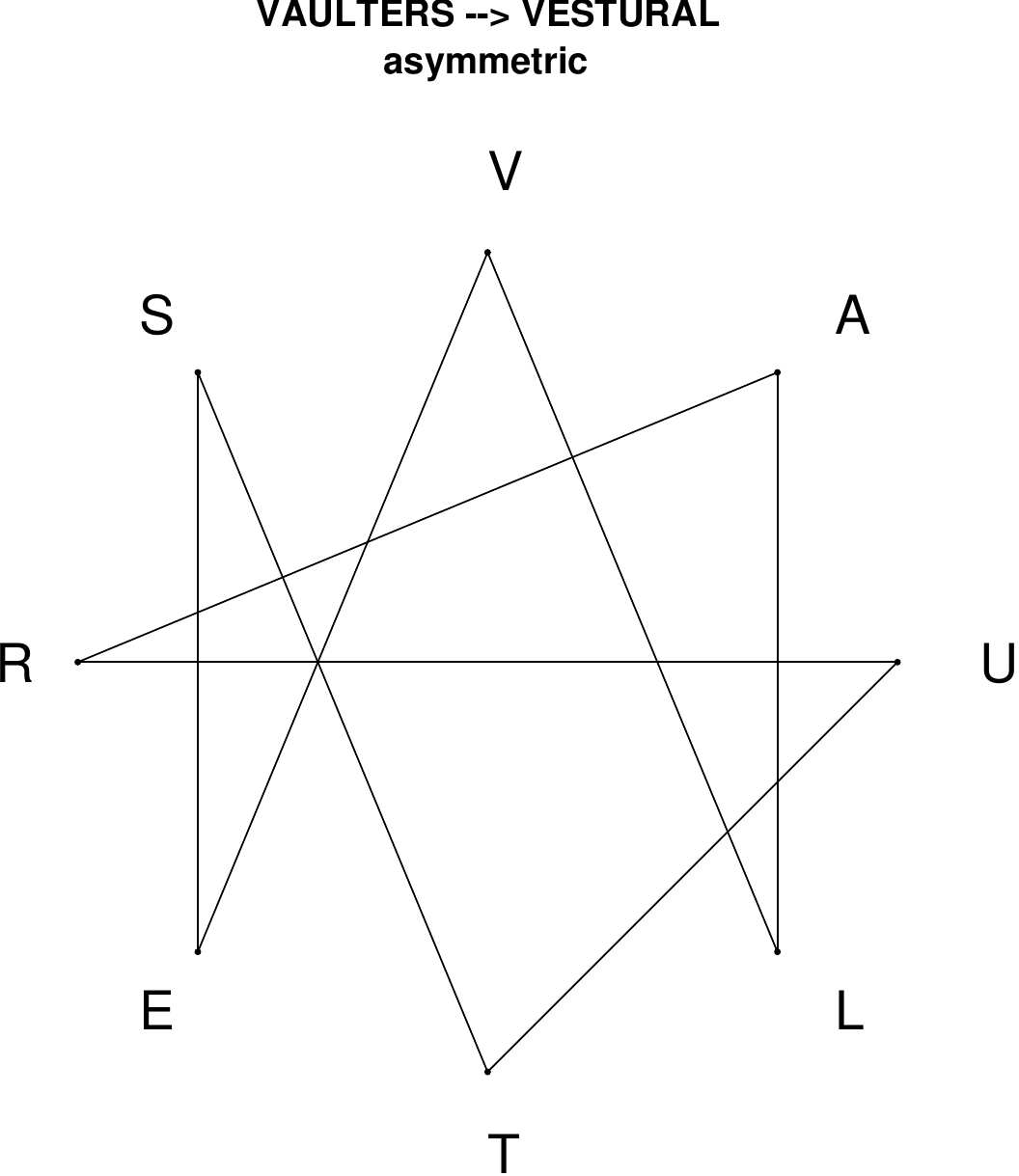}
\end{subfigure}
\hfill
\end{figure}

%%%%%%%%%%%%%%%%%%
\clearpage
\subsection{Star Anagrams $N = 7$}
All of the numerous stars with length $N=7$ are either symmetric or perfect. 

\subsubsection{Perfect Stars $N=7$}

\begin{figure}[H]
\centering
\begin{subfigure}[T]{0.19\textwidth}
\centering
\includegraphics[width=\textwidth]{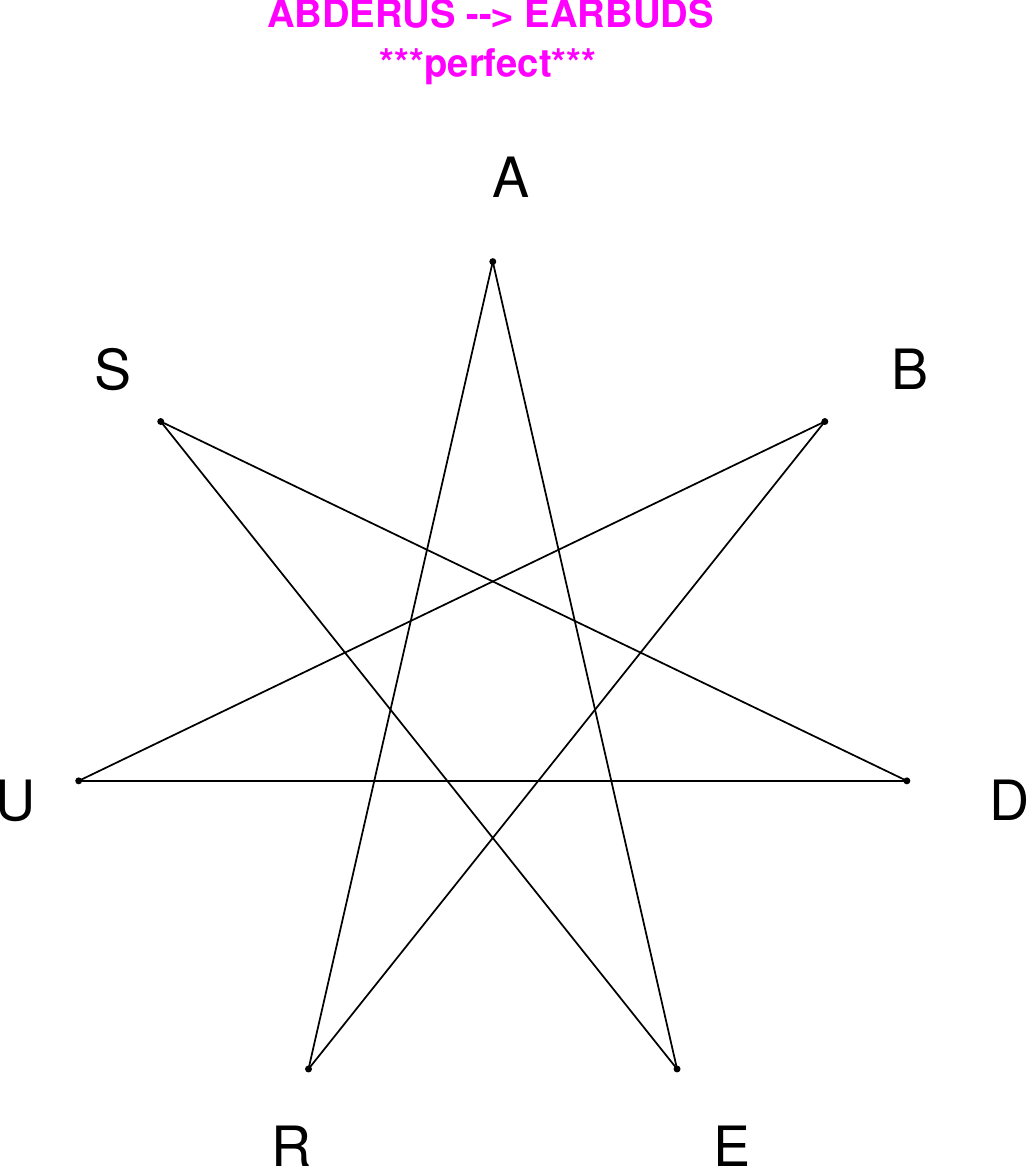}
\end{subfigure}
\hfill
\begin{subfigure}[T]{0.19\textwidth}
\centering
\includegraphics[width=\textwidth]{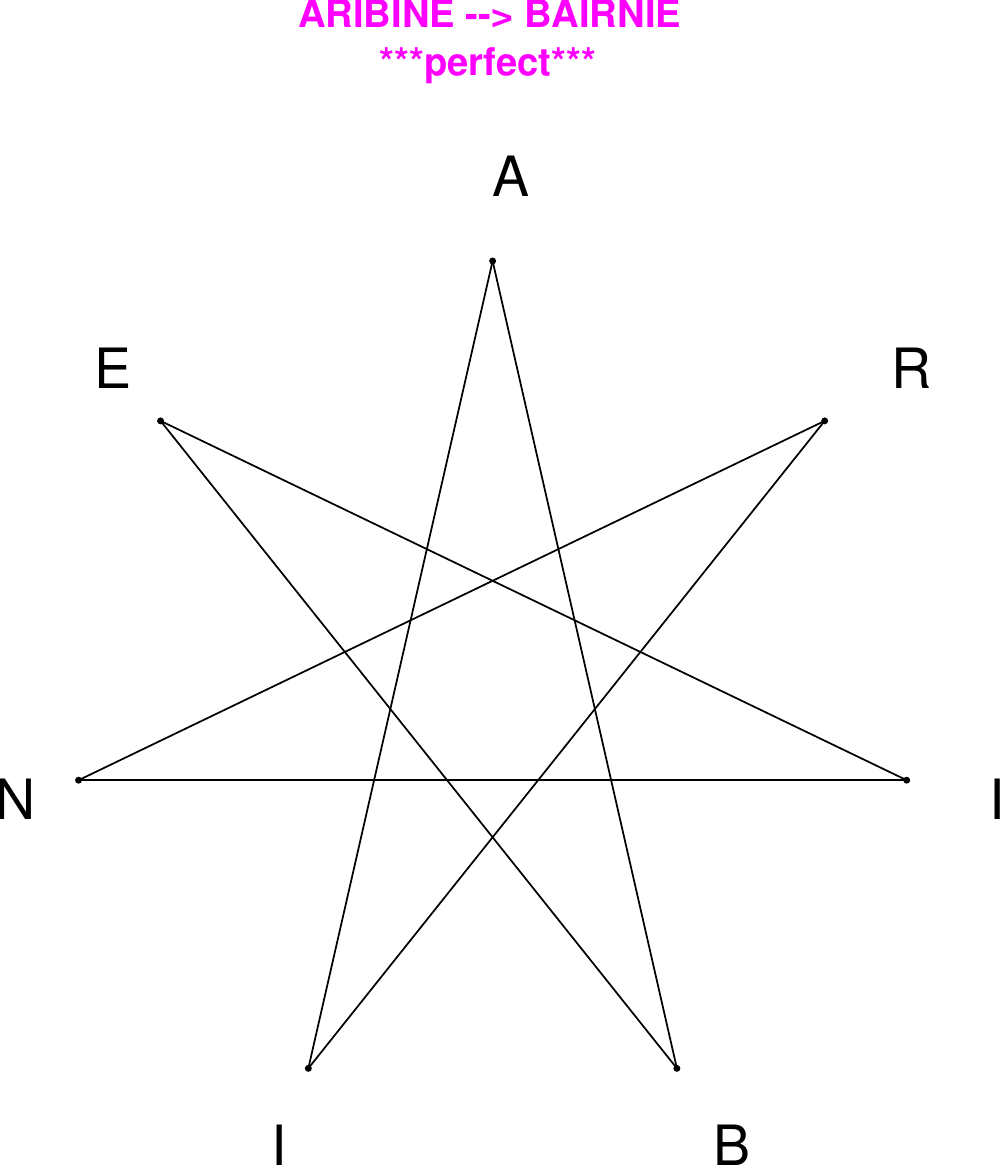}
\end{subfigure}
\hfill
\begin{subfigure}[T]{0.19\textwidth}
\centering
\includegraphics[width=\textwidth]{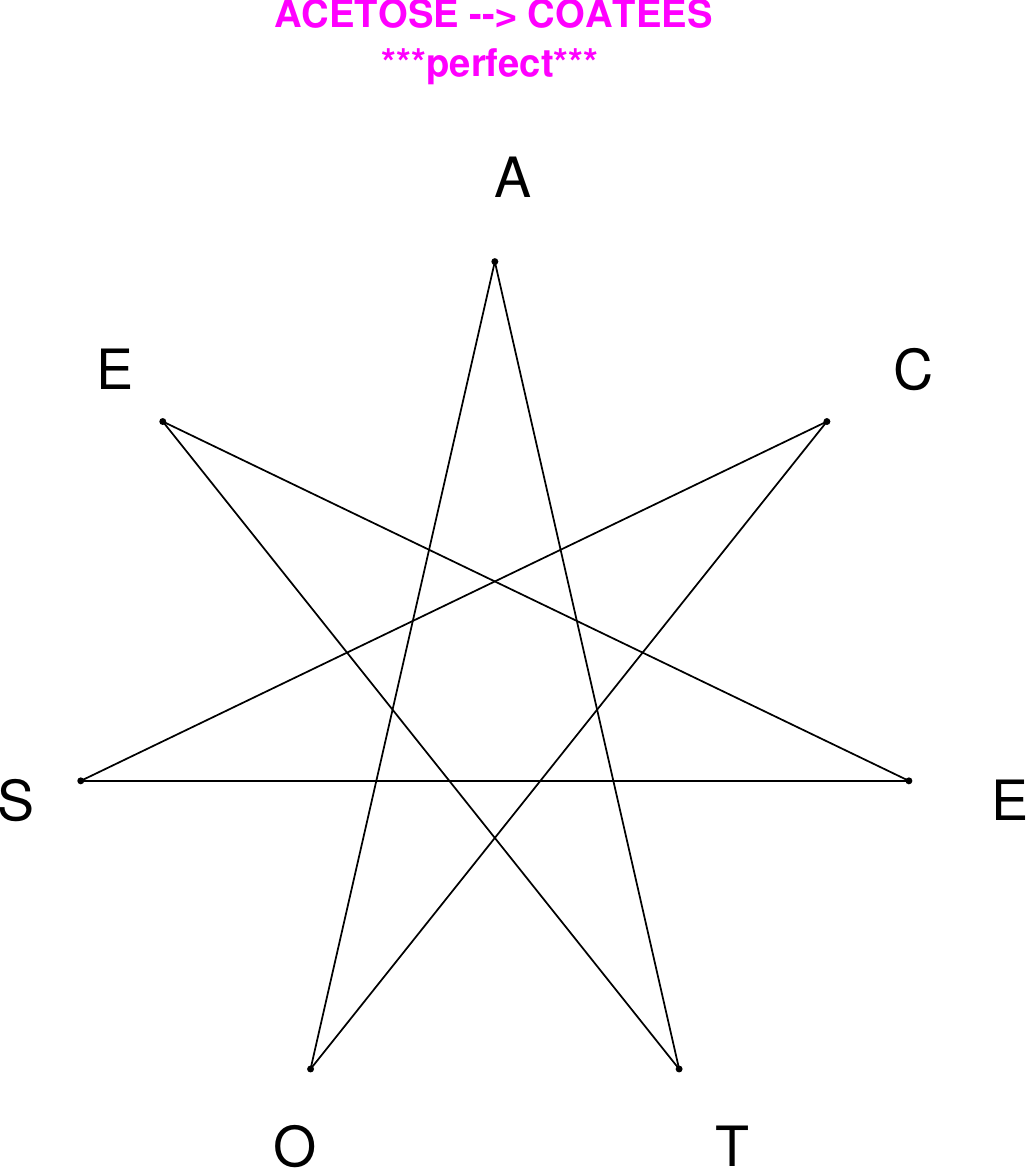}
\end{subfigure}
\hfill
\begin{subfigure}[T]{0.19\textwidth}
\centering
\includegraphics[width=\textwidth]{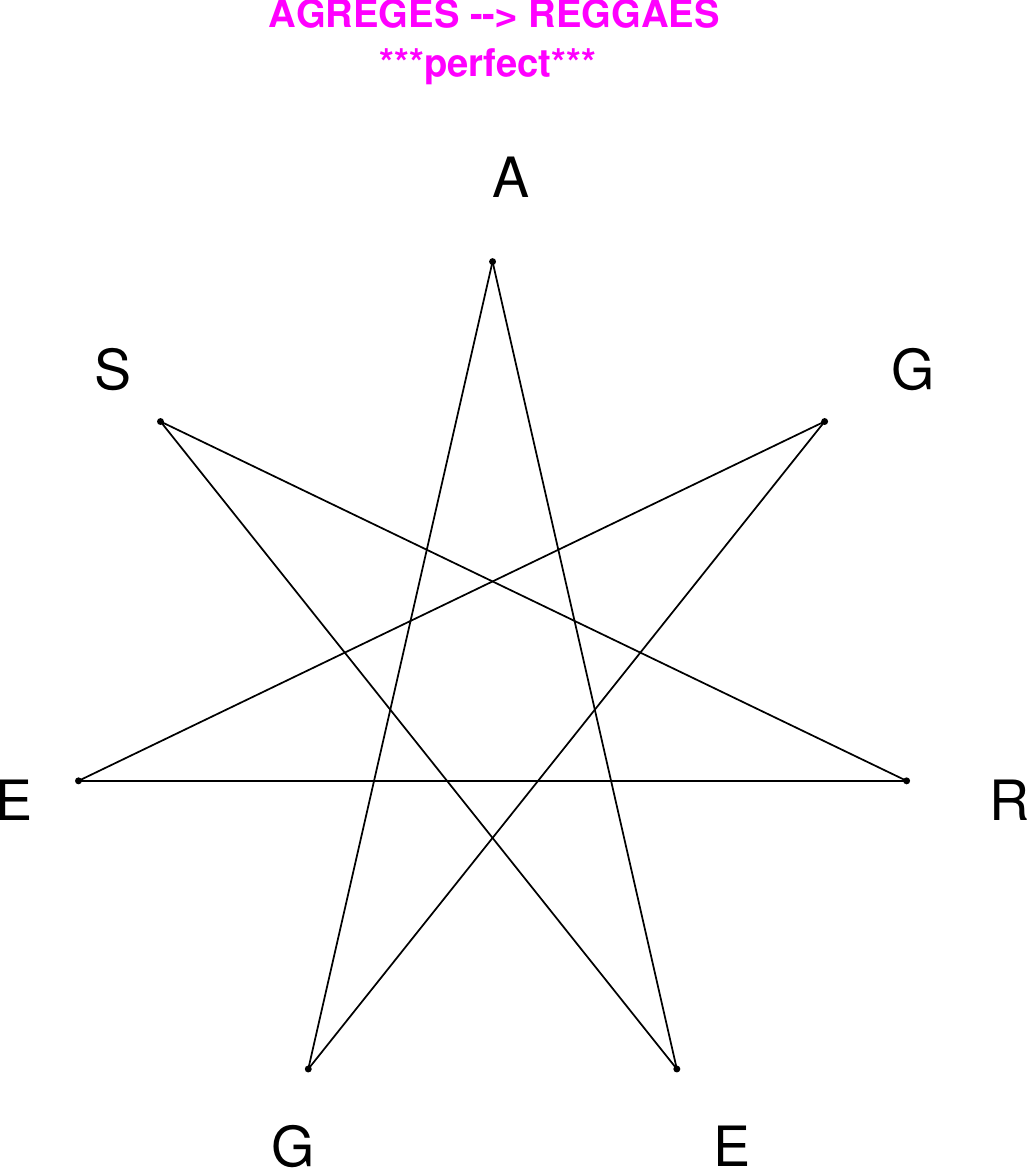}
\end{subfigure}
\hfill
\begin{subfigure}[T]{0.19\textwidth}
\centering
\includegraphics[width=\textwidth]{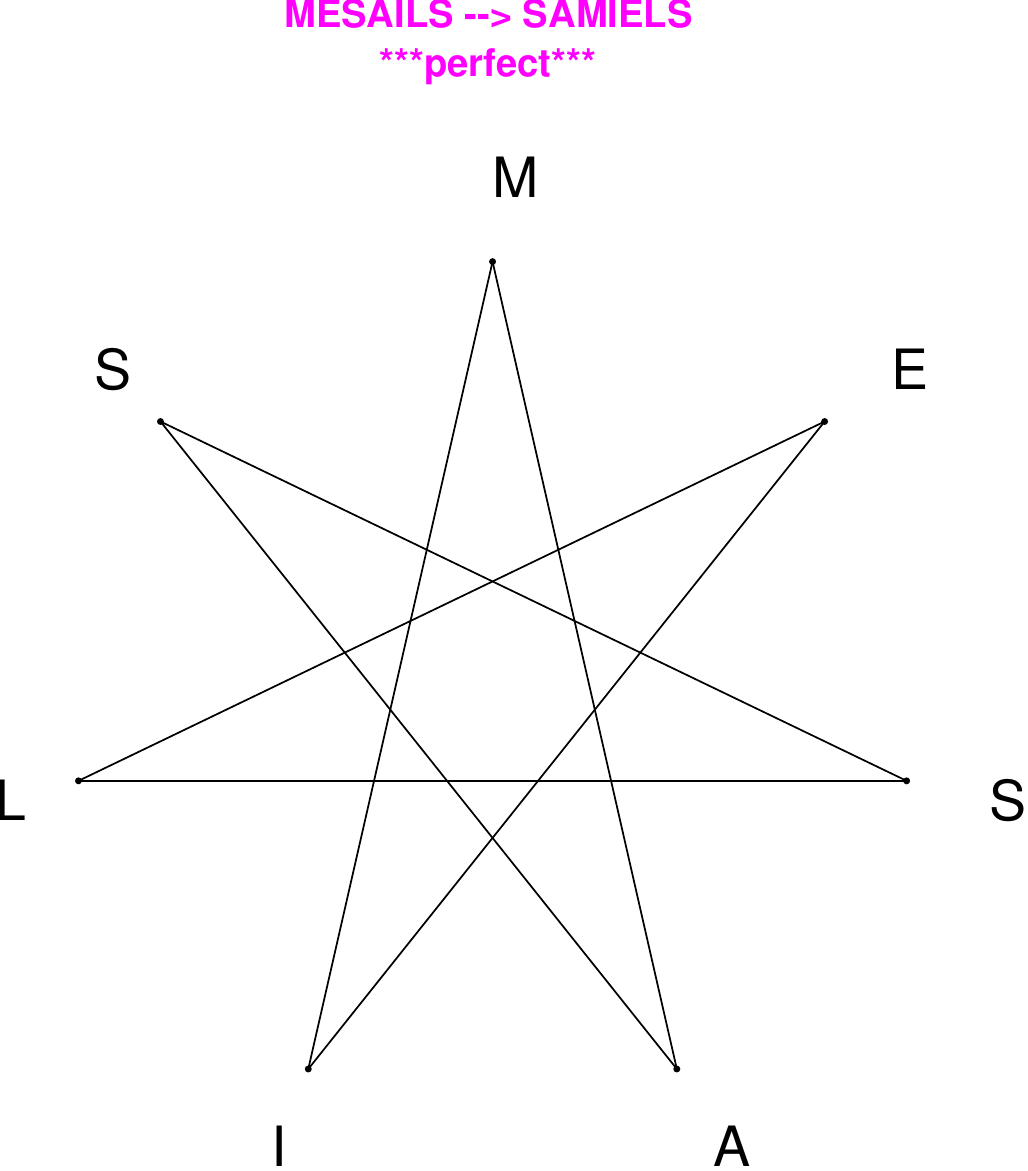}
\end{subfigure}
\end{figure}

\begin{figure}[H]
\centering
\begin{subfigure}[T]{0.19\textwidth}
\centering
\includegraphics[width=\textwidth]{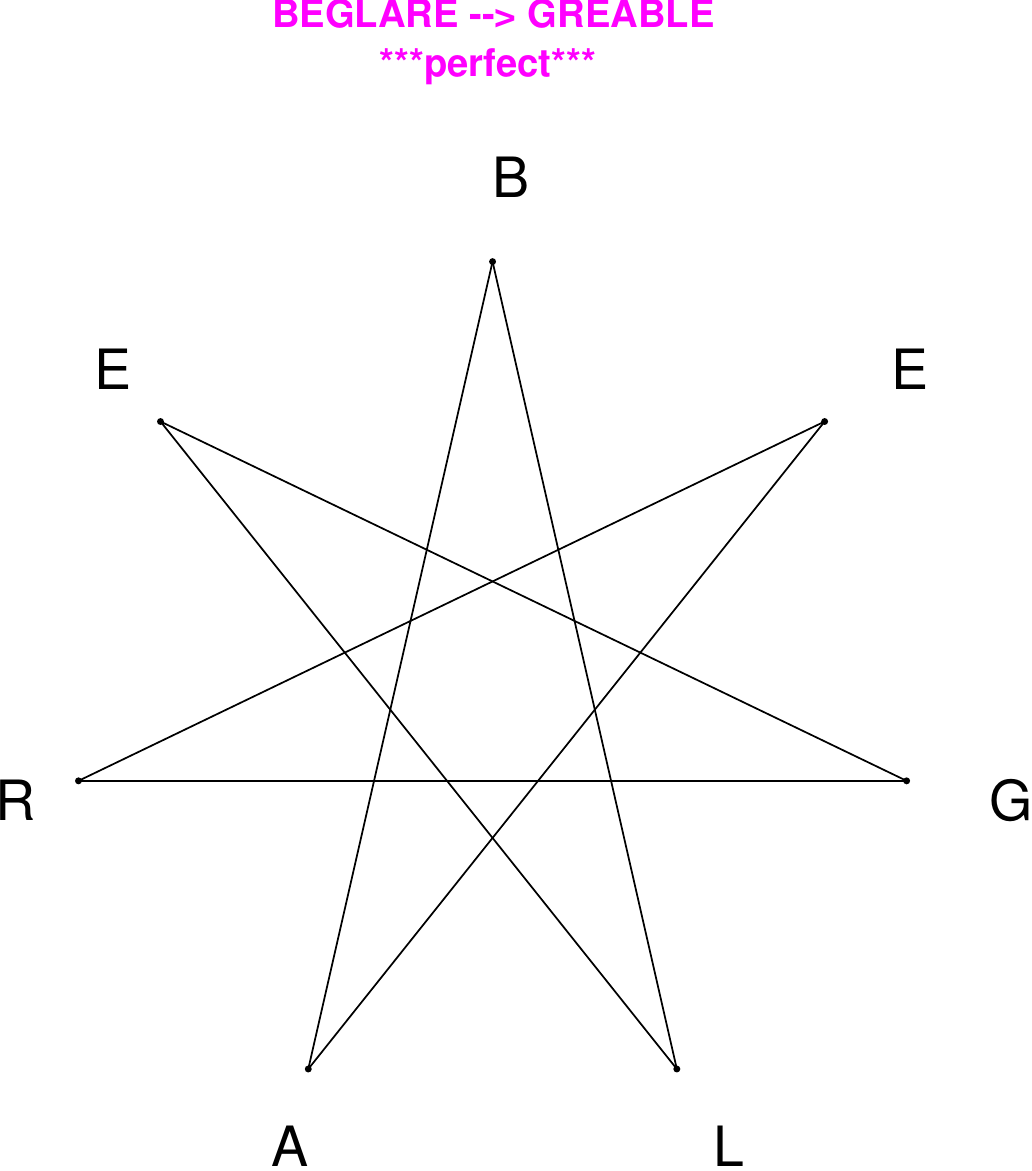}
\end{subfigure}
\hfill
\begin{subfigure}[T]{0.19\textwidth}
\centering
\includegraphics[width=\textwidth]{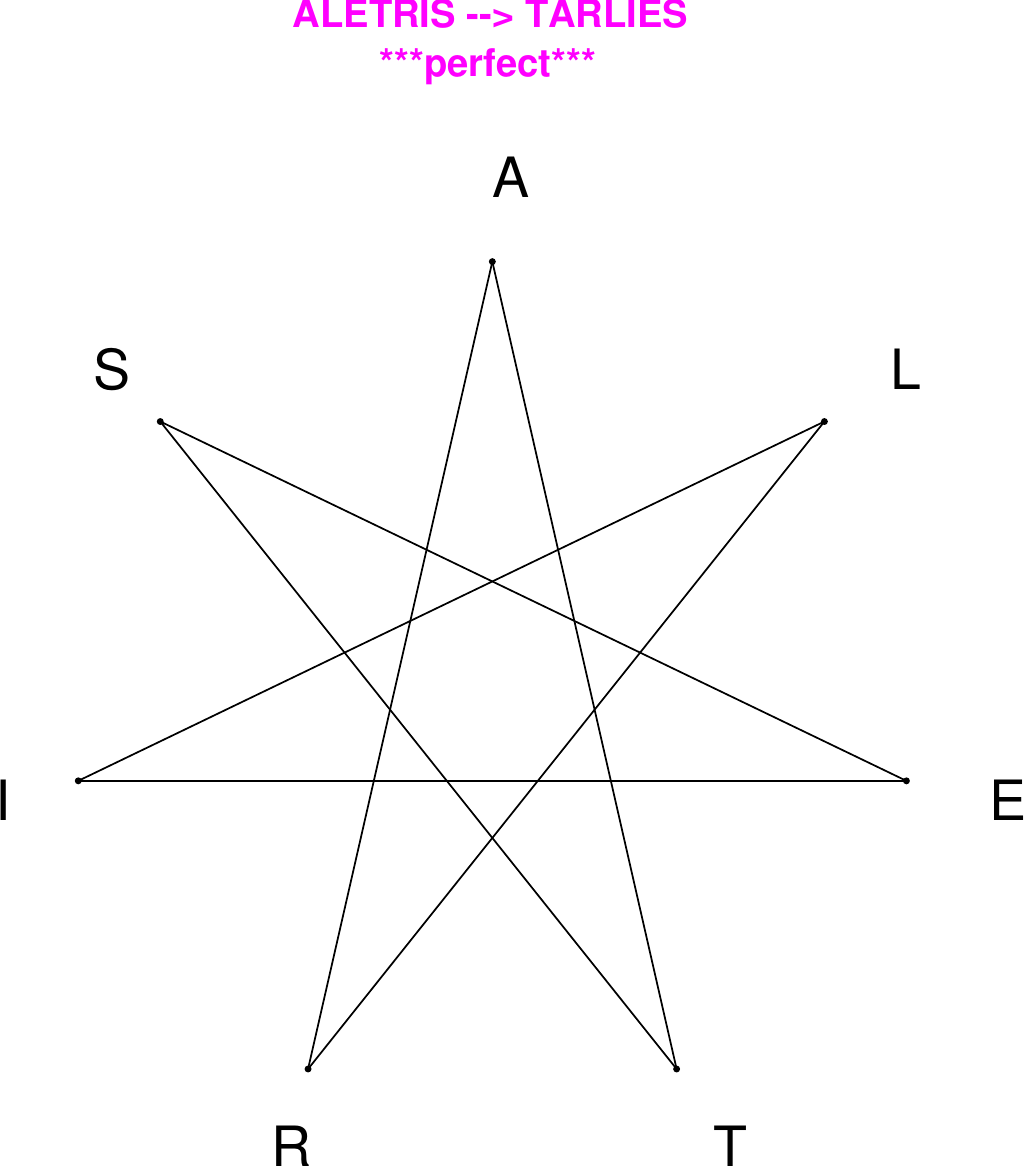}
\end{subfigure}
\hfill
\begin{subfigure}[T]{0.19\textwidth}
\centering
\includegraphics[width=\textwidth]{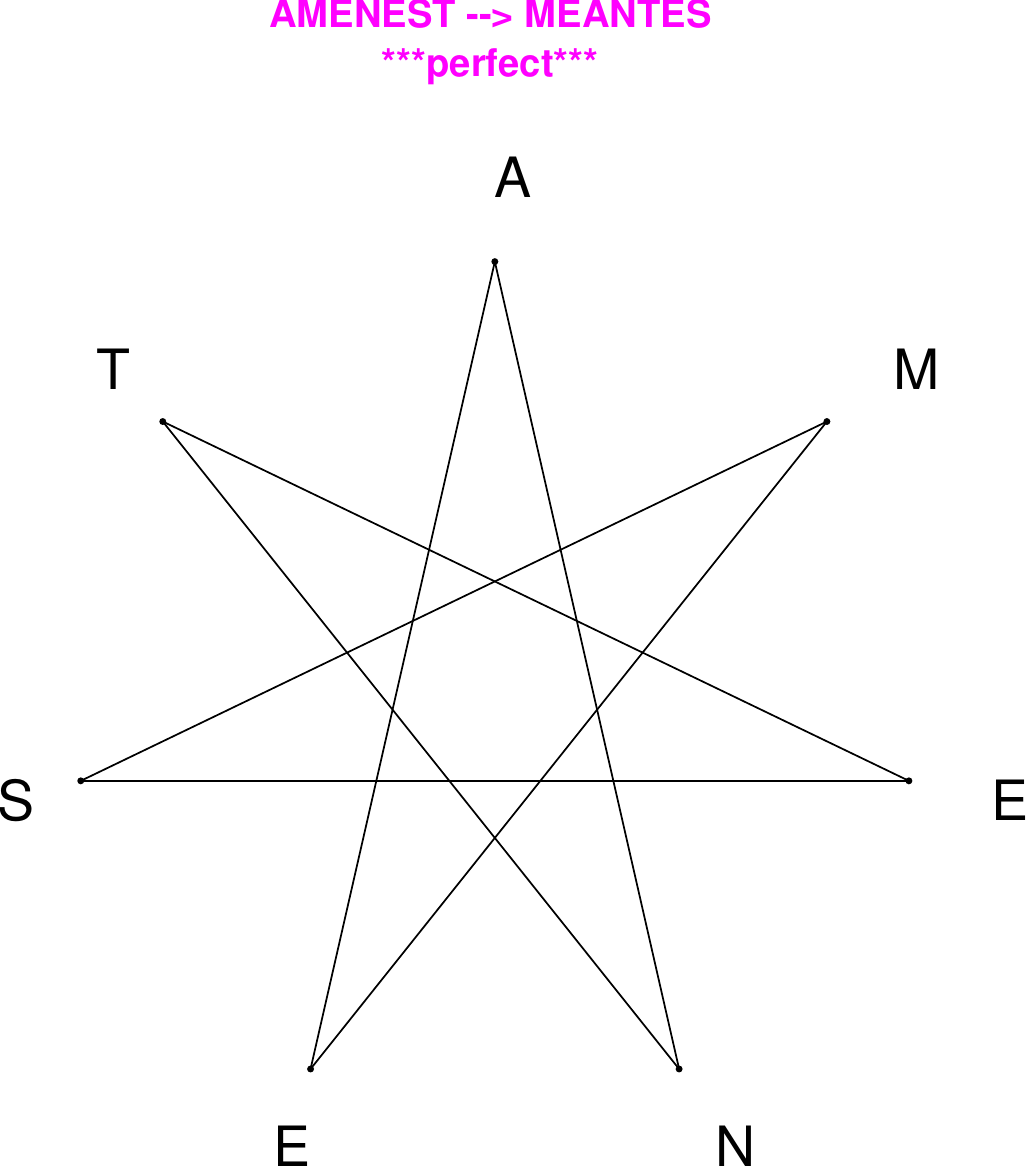}
\end{subfigure}
\hfill
\begin{subfigure}[T]{0.19\textwidth}
\centering
\includegraphics[width=\textwidth]{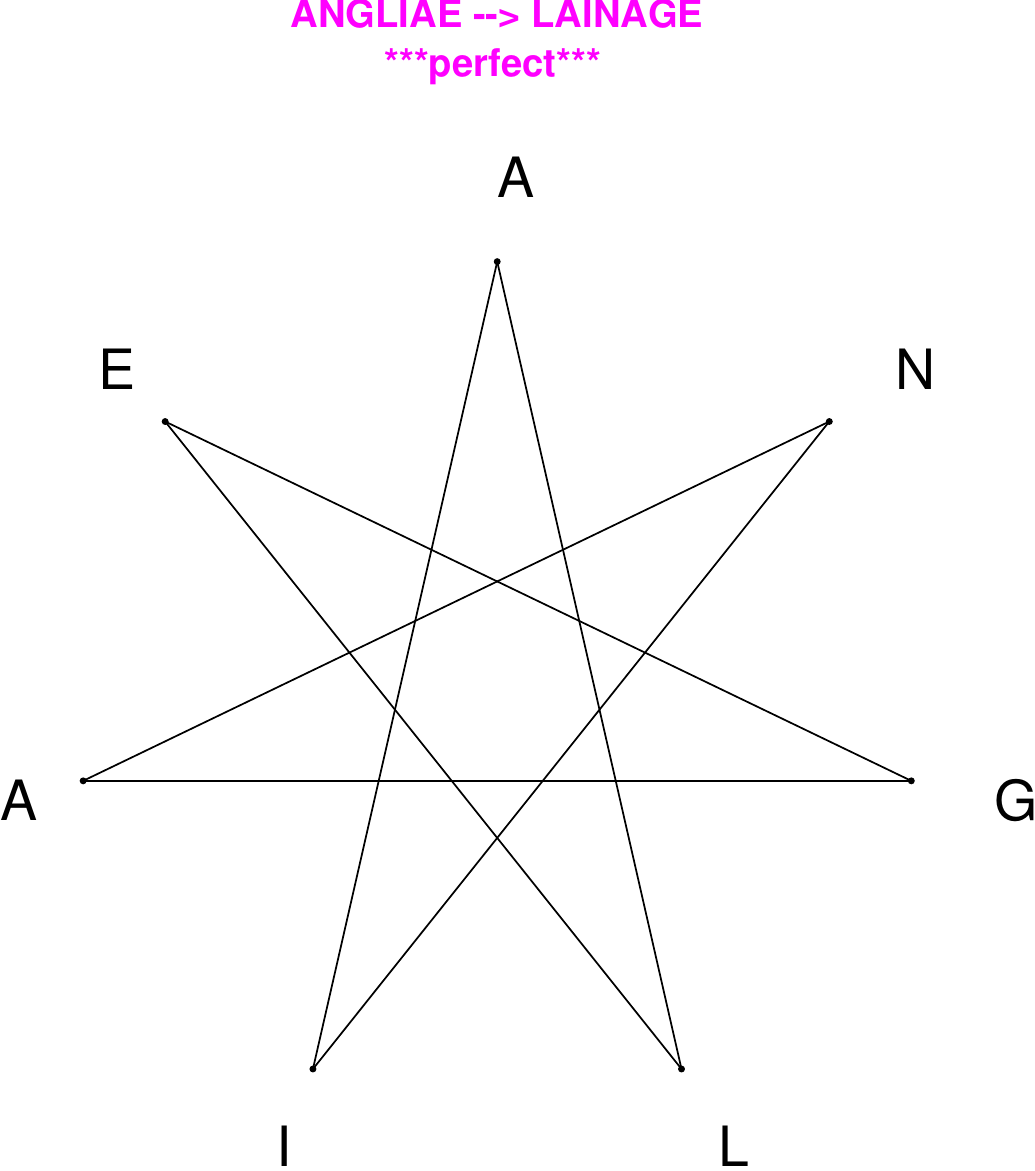}
\end{subfigure}
\hfill
\begin{subfigure}[T]{0.19\textwidth}
\centering
\includegraphics[width=\textwidth]{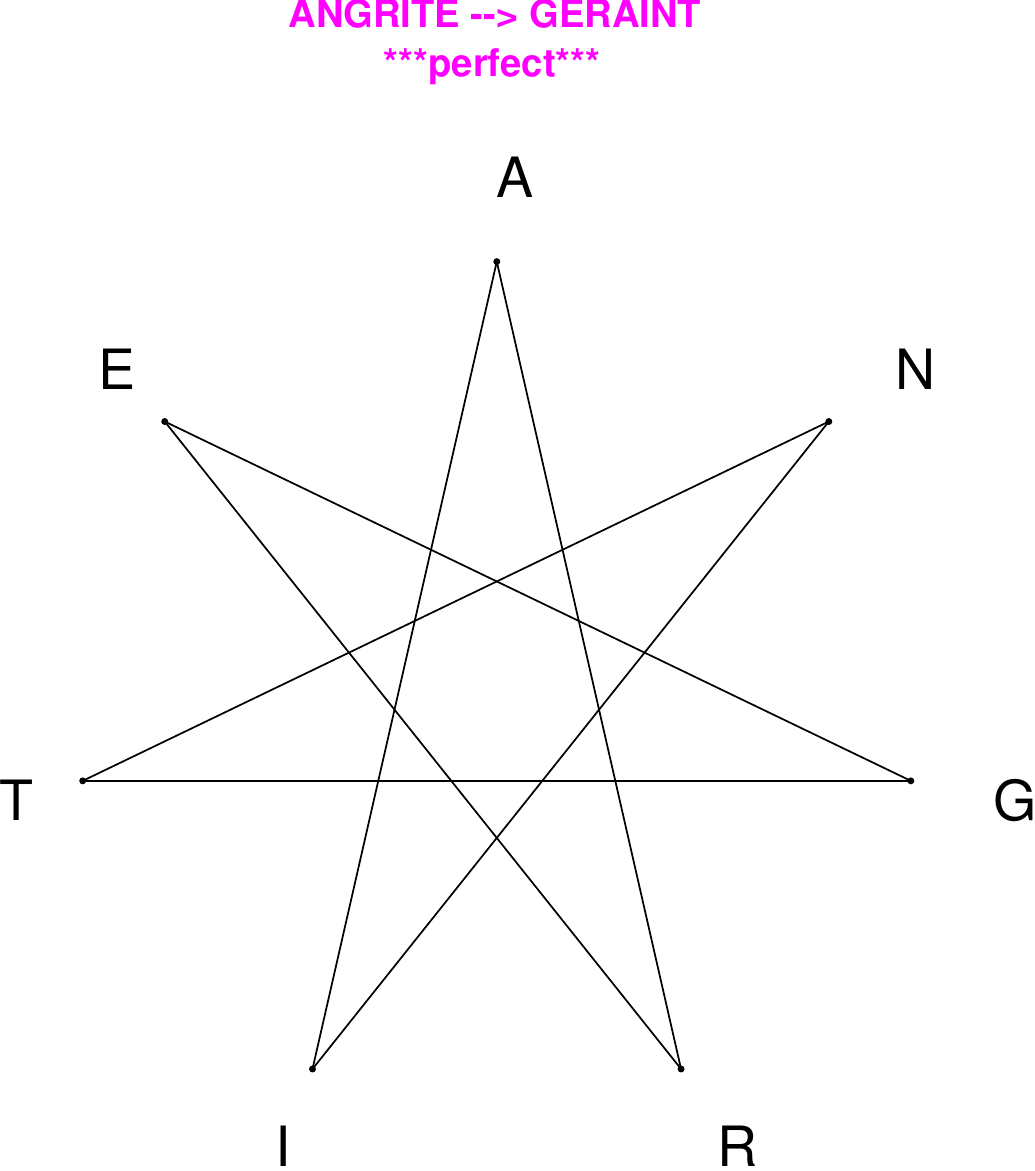}
\end{subfigure}
\end{figure}

\begin{figure}[H]
\centering
\begin{subfigure}[T]{0.19\textwidth}
\centering
\includegraphics[width=\textwidth]{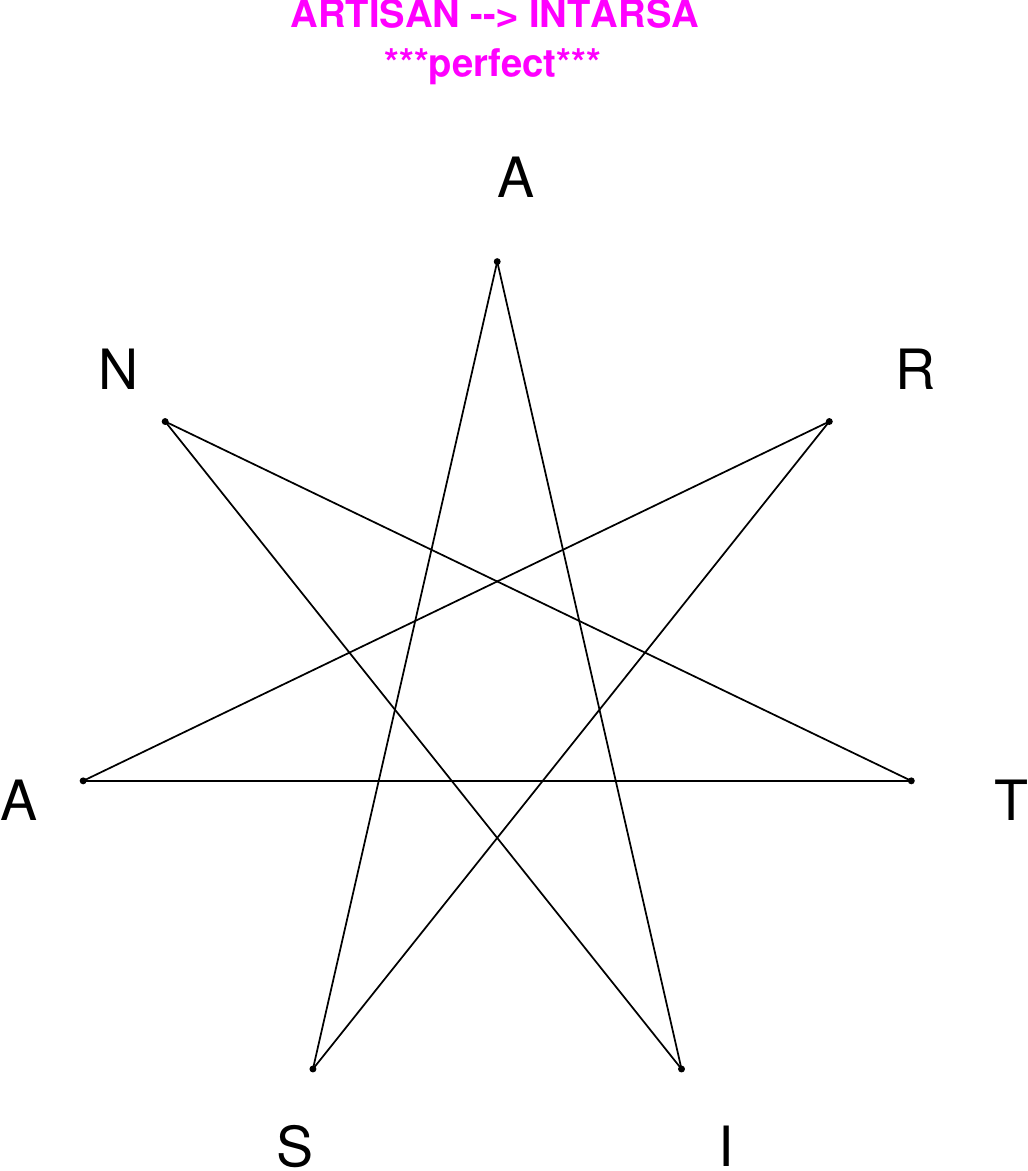}
\end{subfigure}
\hfill
\begin{subfigure}[T]{0.19\textwidth}
\centering
\includegraphics[width=\textwidth]{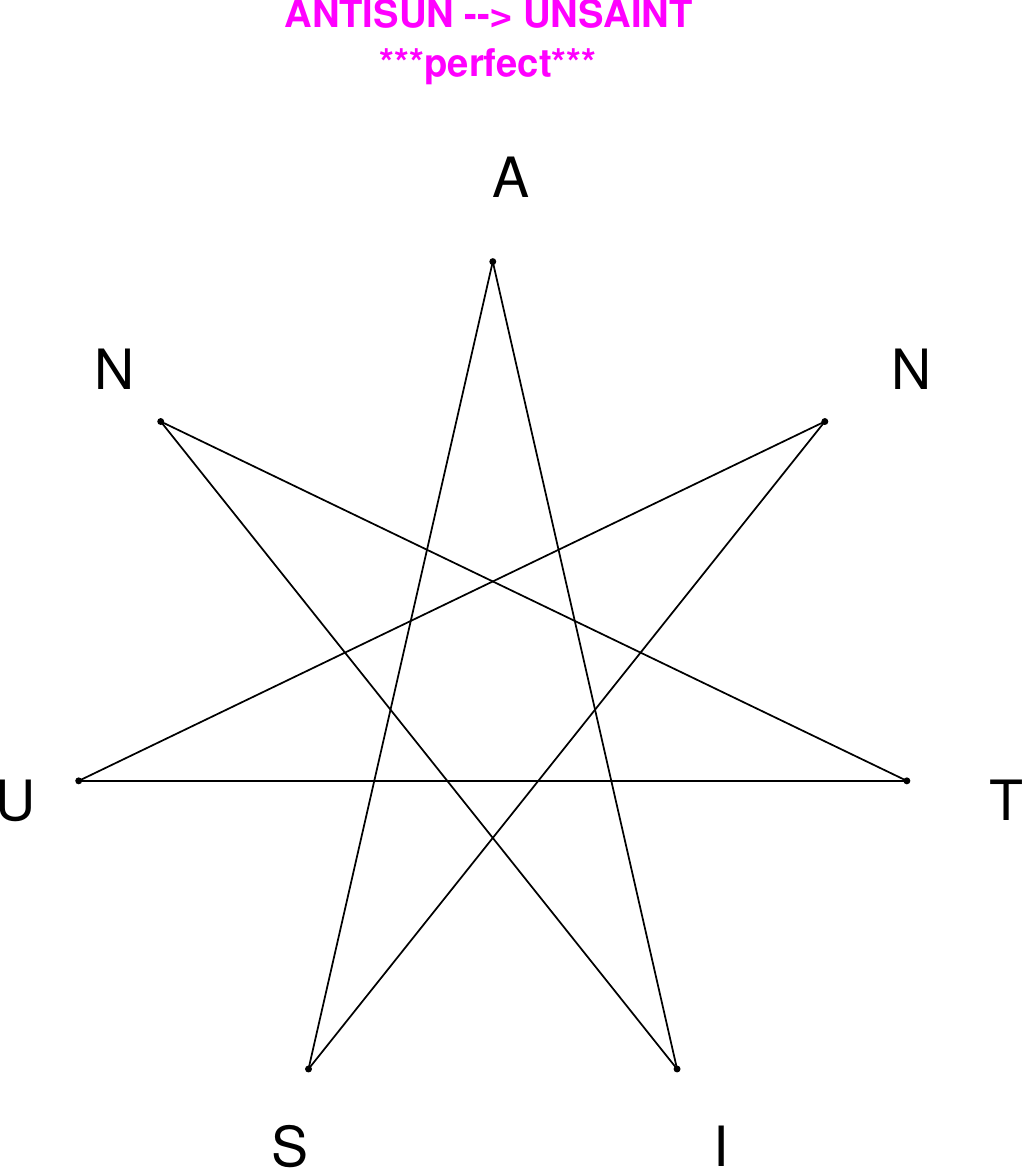}
\end{subfigure}
\hfill
\begin{subfigure}[T]{0.19\textwidth}
\centering
\includegraphics[width=\textwidth]{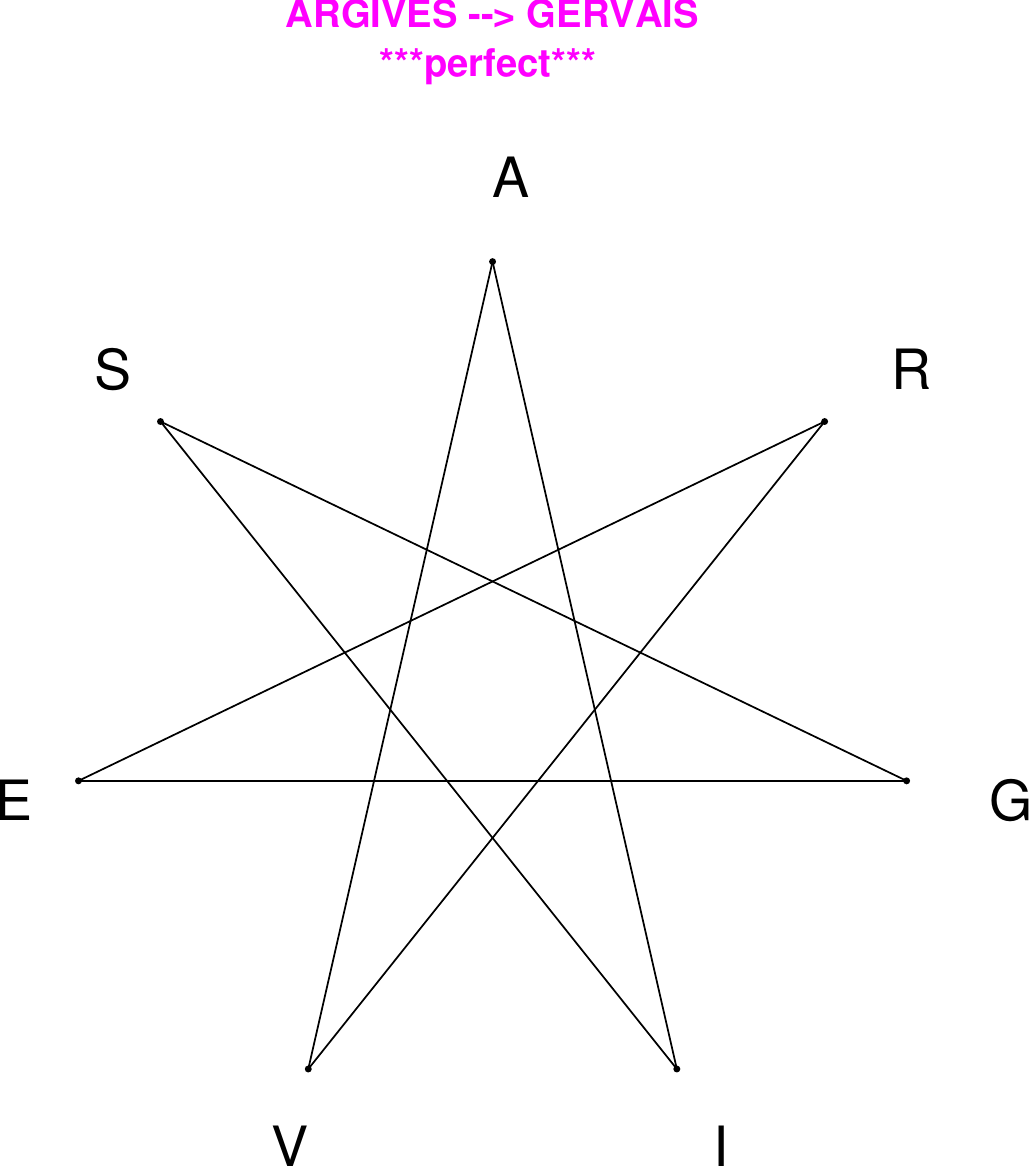}
\end{subfigure}
\hfill
\begin{subfigure}[T]{0.19\textwidth}
\centering
\includegraphics[width=\textwidth]{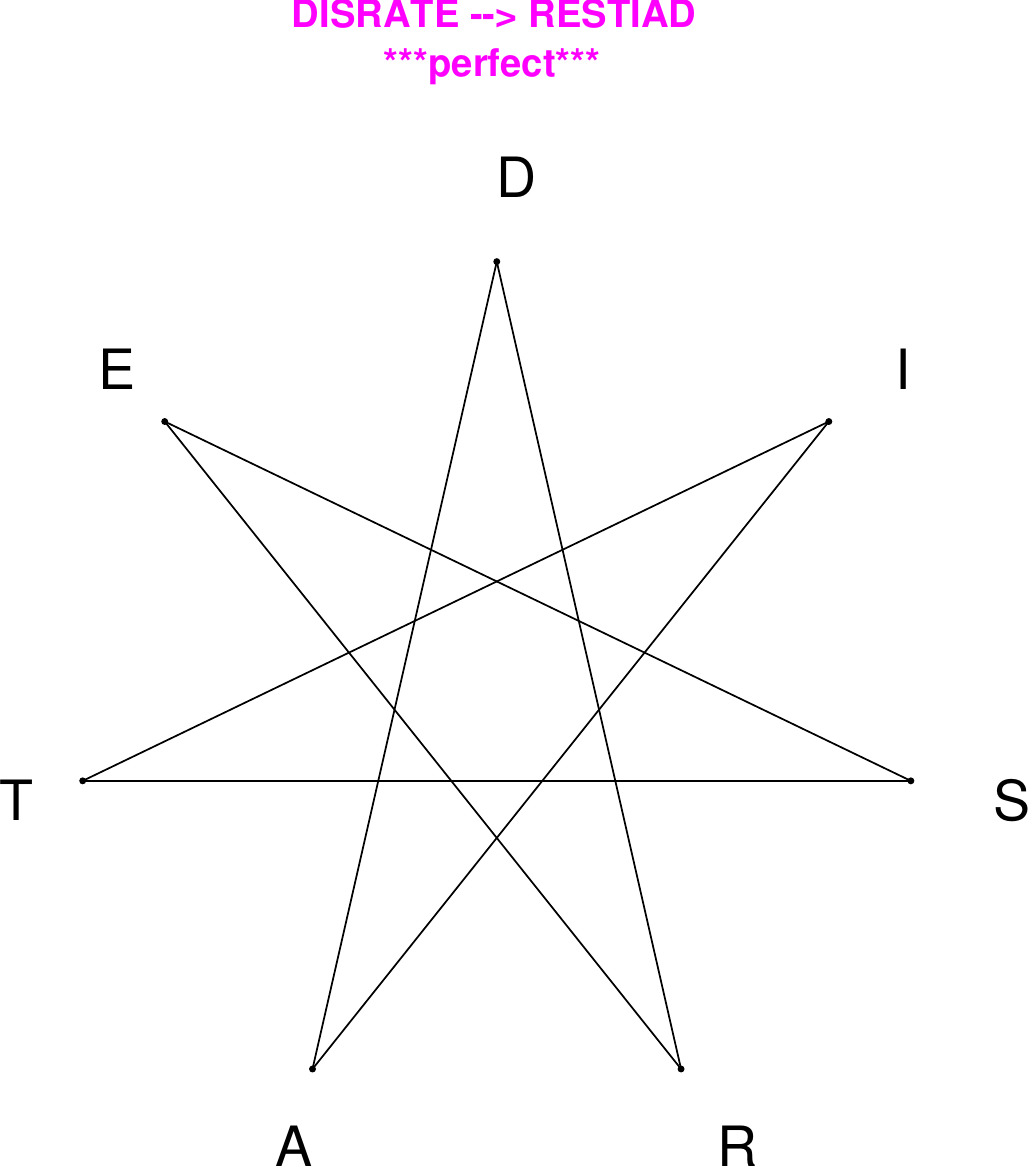}
\end{subfigure}
\hfill
\begin{subfigure}[T]{0.19\textwidth}
\centering
\includegraphics[width=\textwidth]{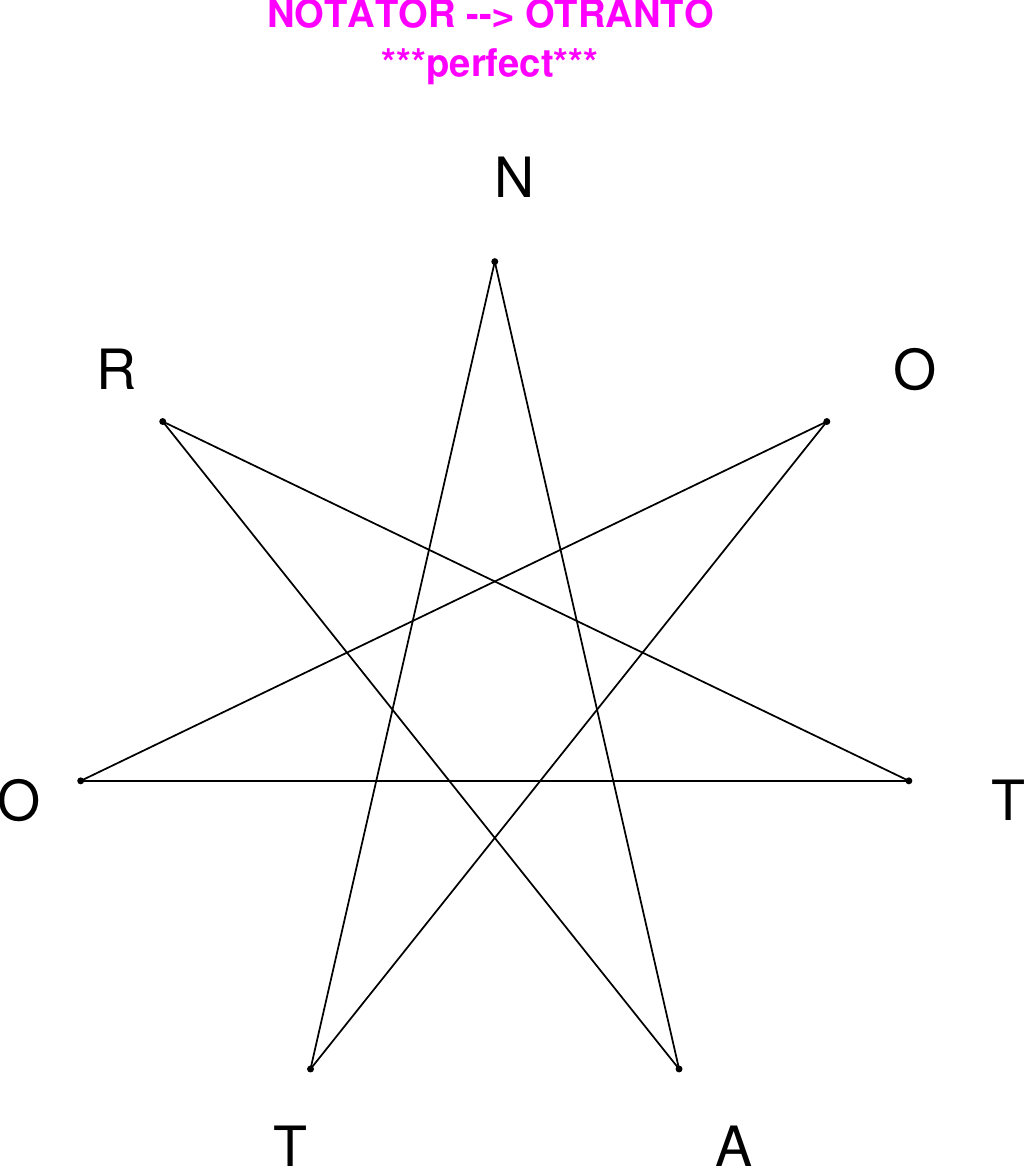}
\end{subfigure}
\end{figure}

\begin{figure}[H]
\centering
\begin{subfigure}[T]{0.19\textwidth}
\centering
\includegraphics[width=\textwidth]{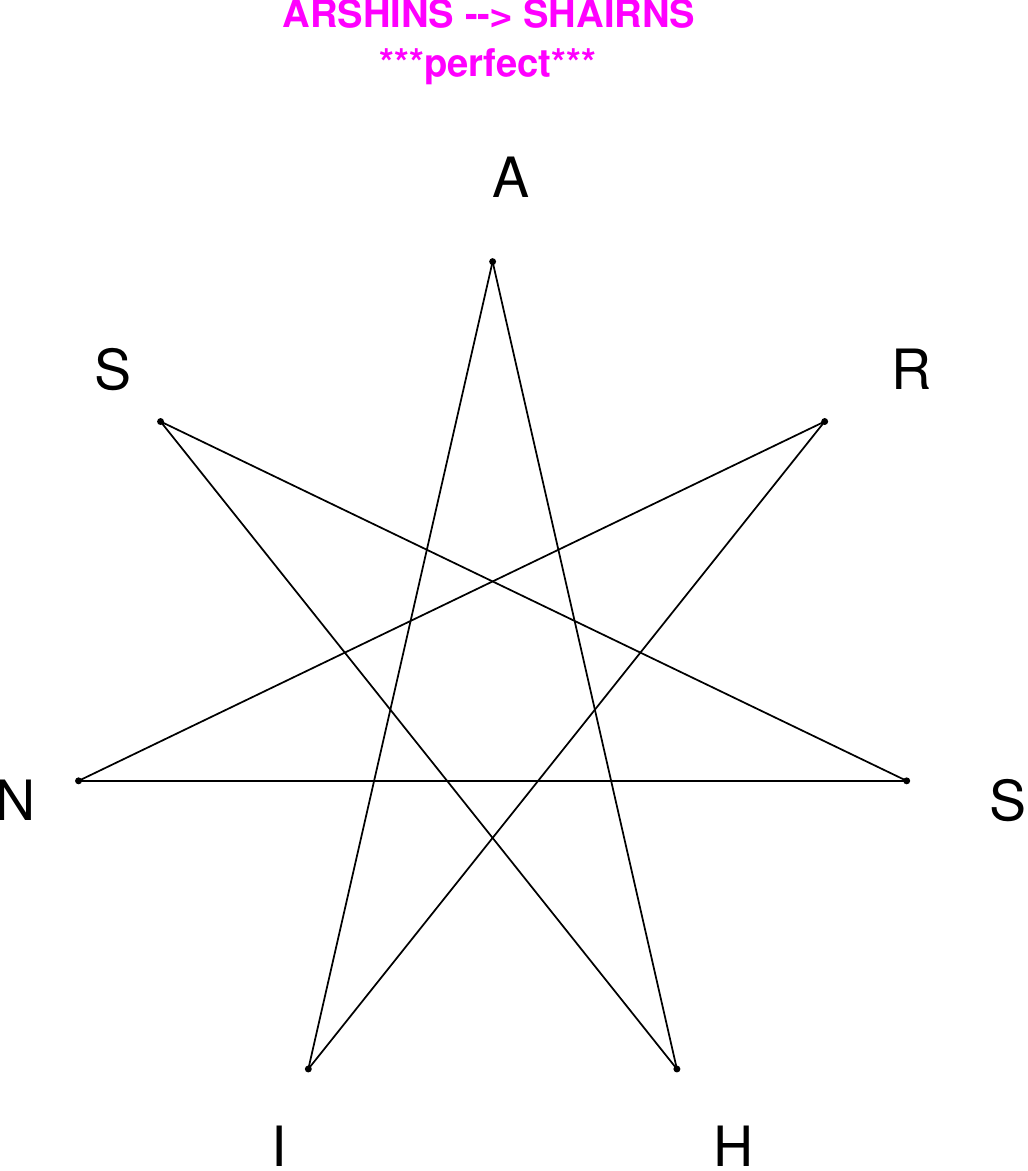}
\end{subfigure}
\hfill
\begin{subfigure}[T]{0.19\textwidth}
\centering
\includegraphics[width=\textwidth]{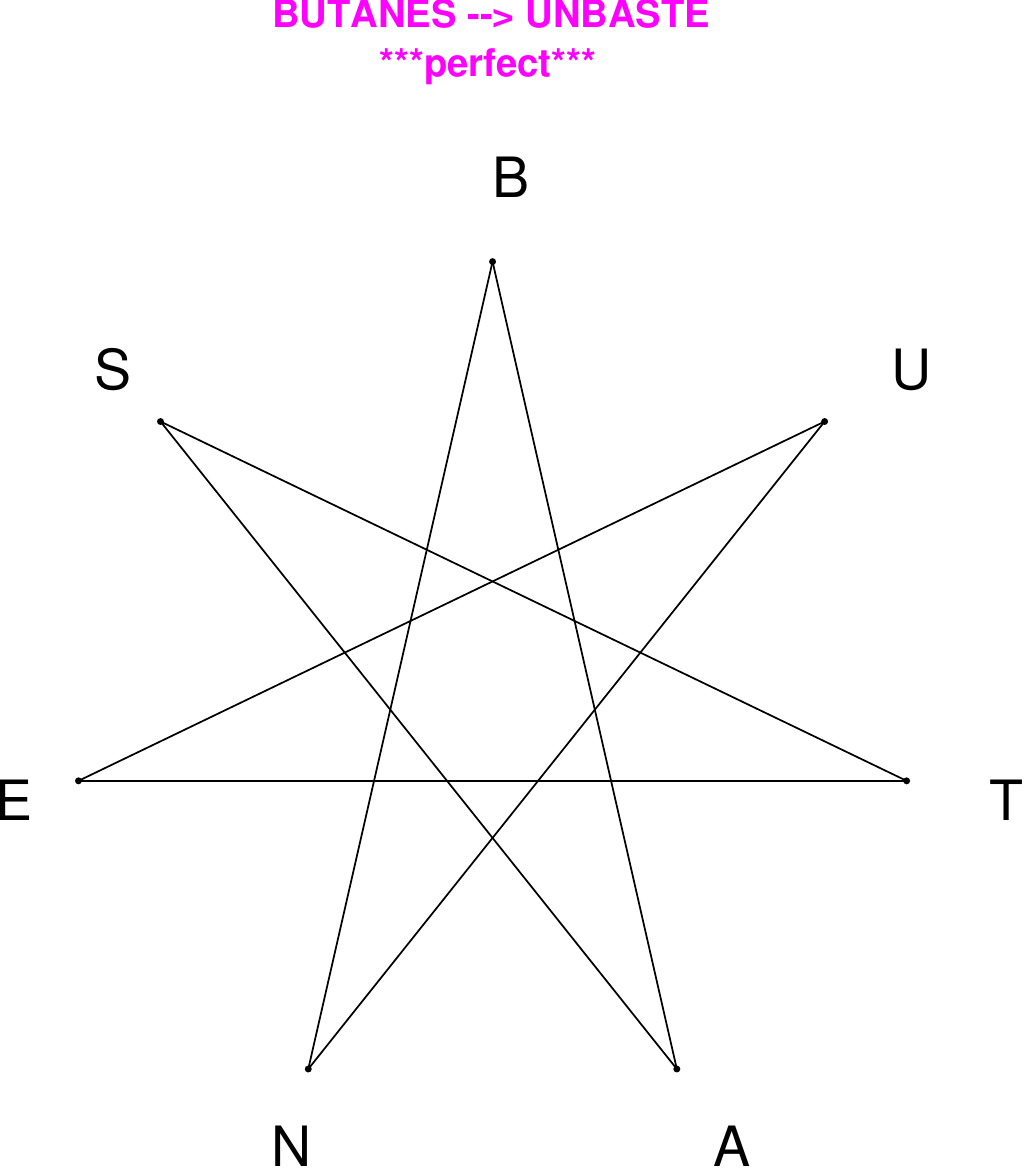}
\end{subfigure}
\hfill
\begin{subfigure}[T]{0.19\textwidth}
\centering
\includegraphics[width=\textwidth]{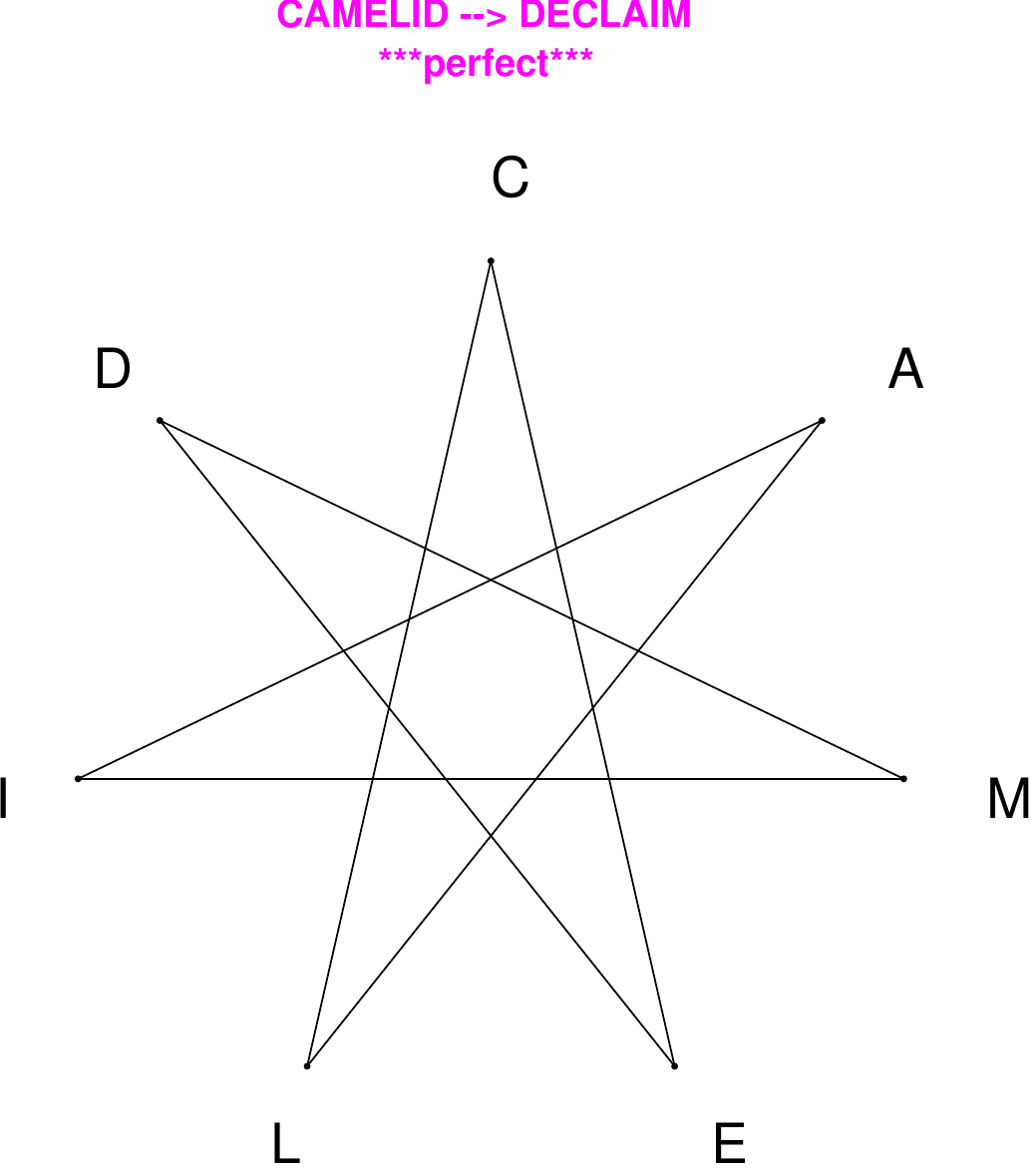}
\end{subfigure}
\hfill
\begin{subfigure}[T]{0.19\textwidth}
\centering
\includegraphics[width=\textwidth]{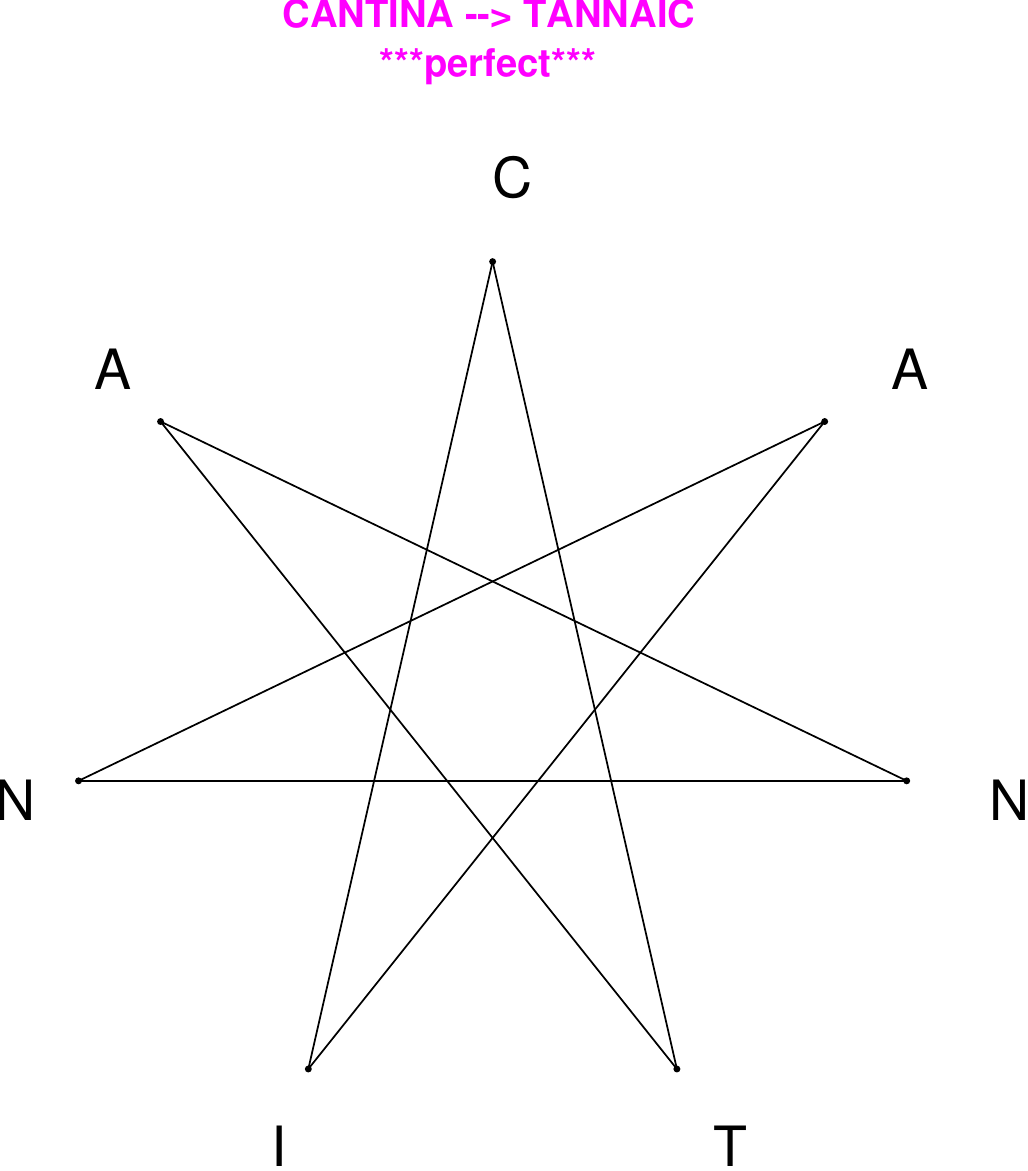}
\end{subfigure}
\hfill
\begin{subfigure}[T]{0.19\textwidth}
\centering
\includegraphics[width=\textwidth]{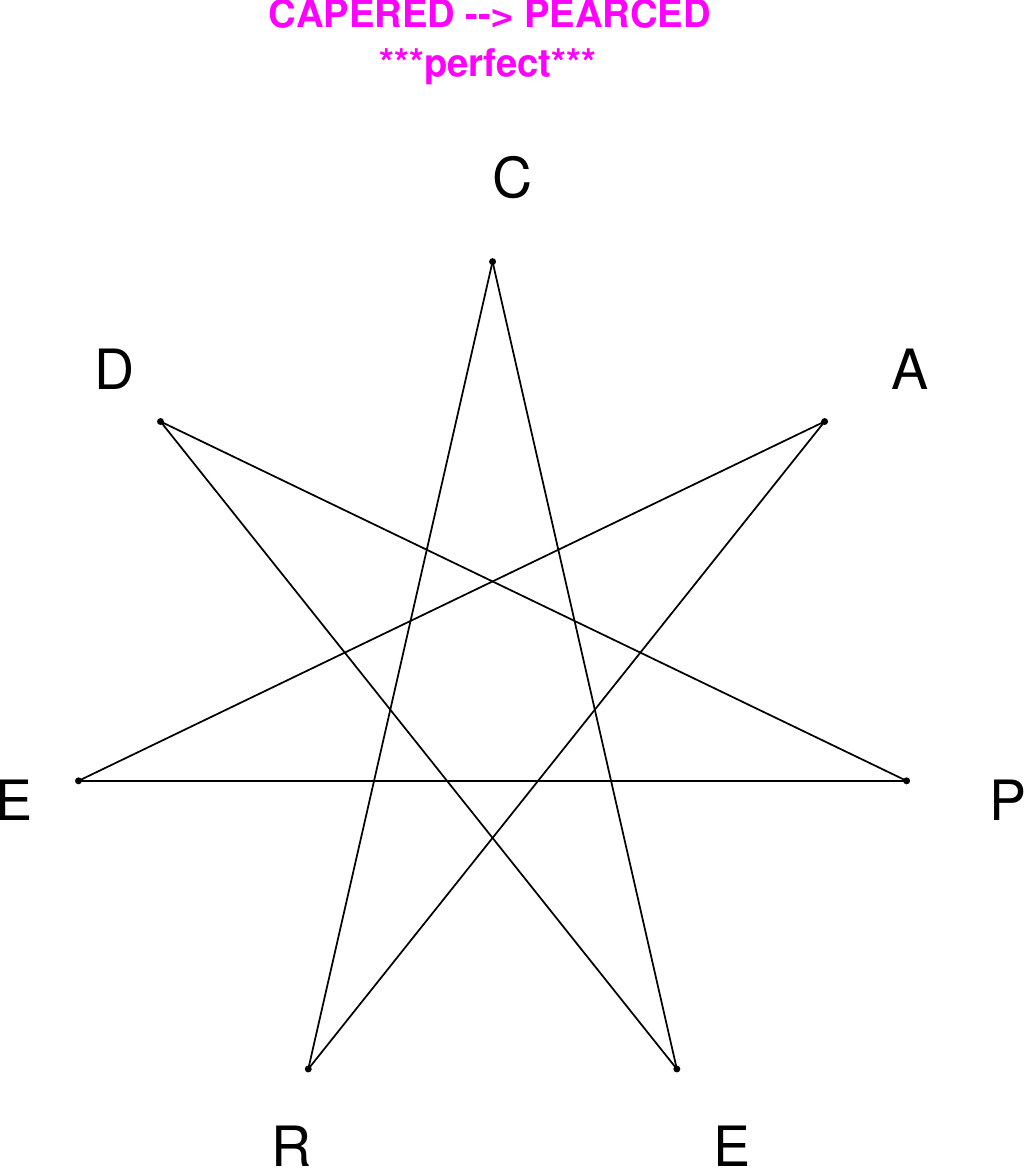}
\end{subfigure}
\end{figure}

\begin{figure}[H]
\centering
\begin{subfigure}[T]{0.19\textwidth}
\centering
\includegraphics[width=\textwidth]{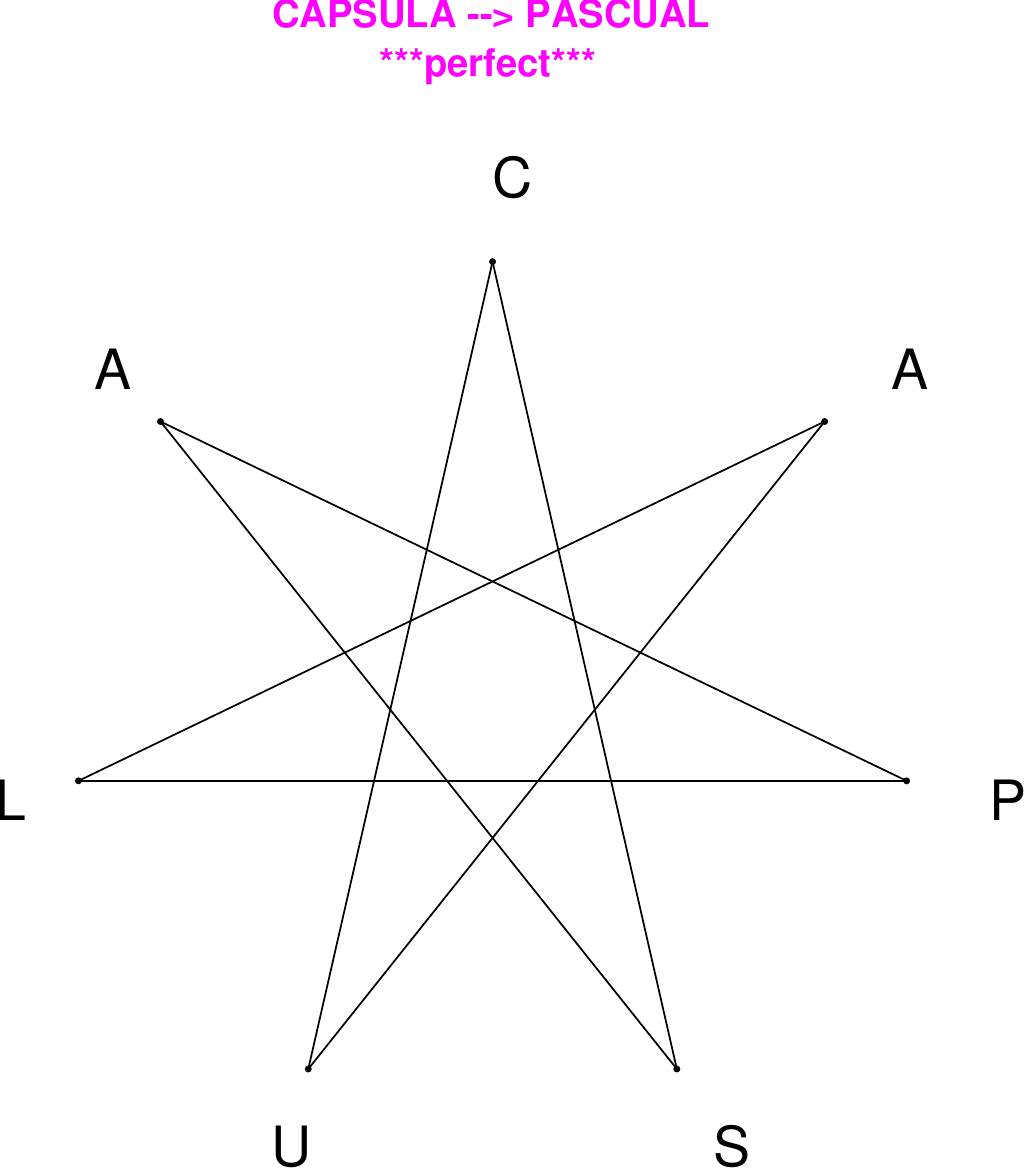}
\end{subfigure}
\hfill
\begin{subfigure}[T]{0.19\textwidth}
\centering
\includegraphics[width=\textwidth]{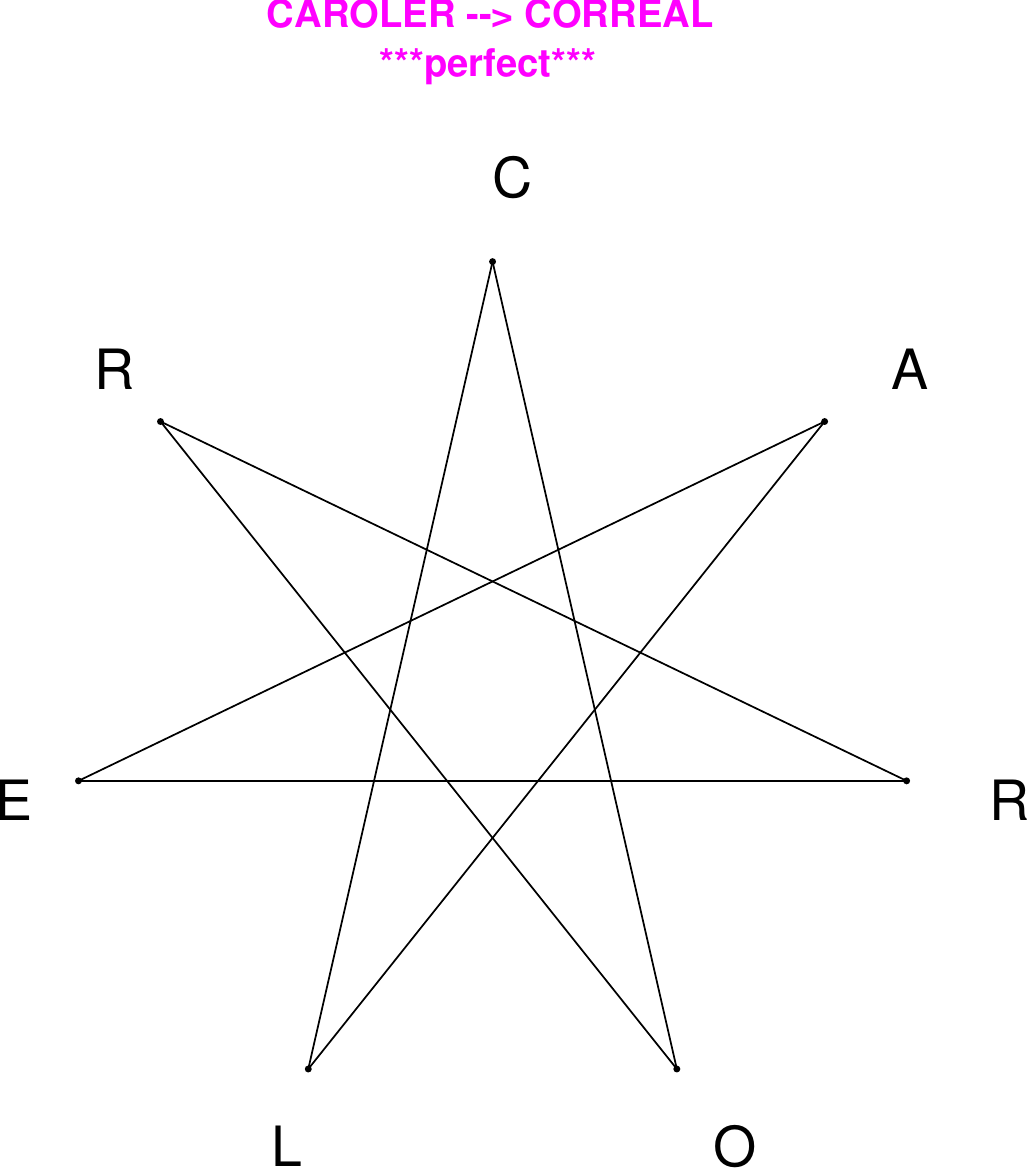}
\end{subfigure}
\hfill
\begin{subfigure}[T]{0.19\textwidth}
\centering
\includegraphics[width=\textwidth]{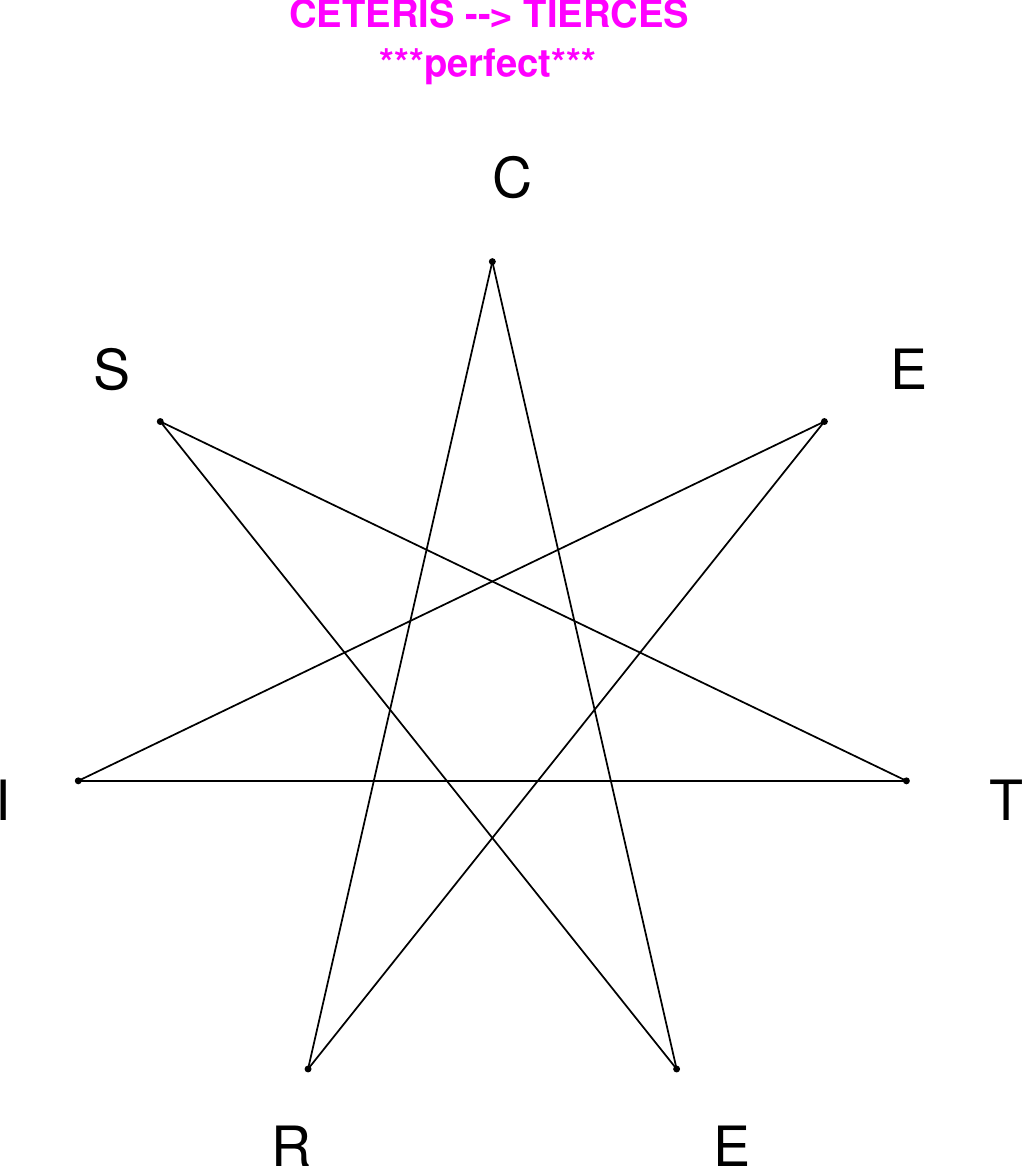}
\end{subfigure}
\hfill
\begin{subfigure}[T]{0.19\textwidth}
\centering
\includegraphics[width=\textwidth]{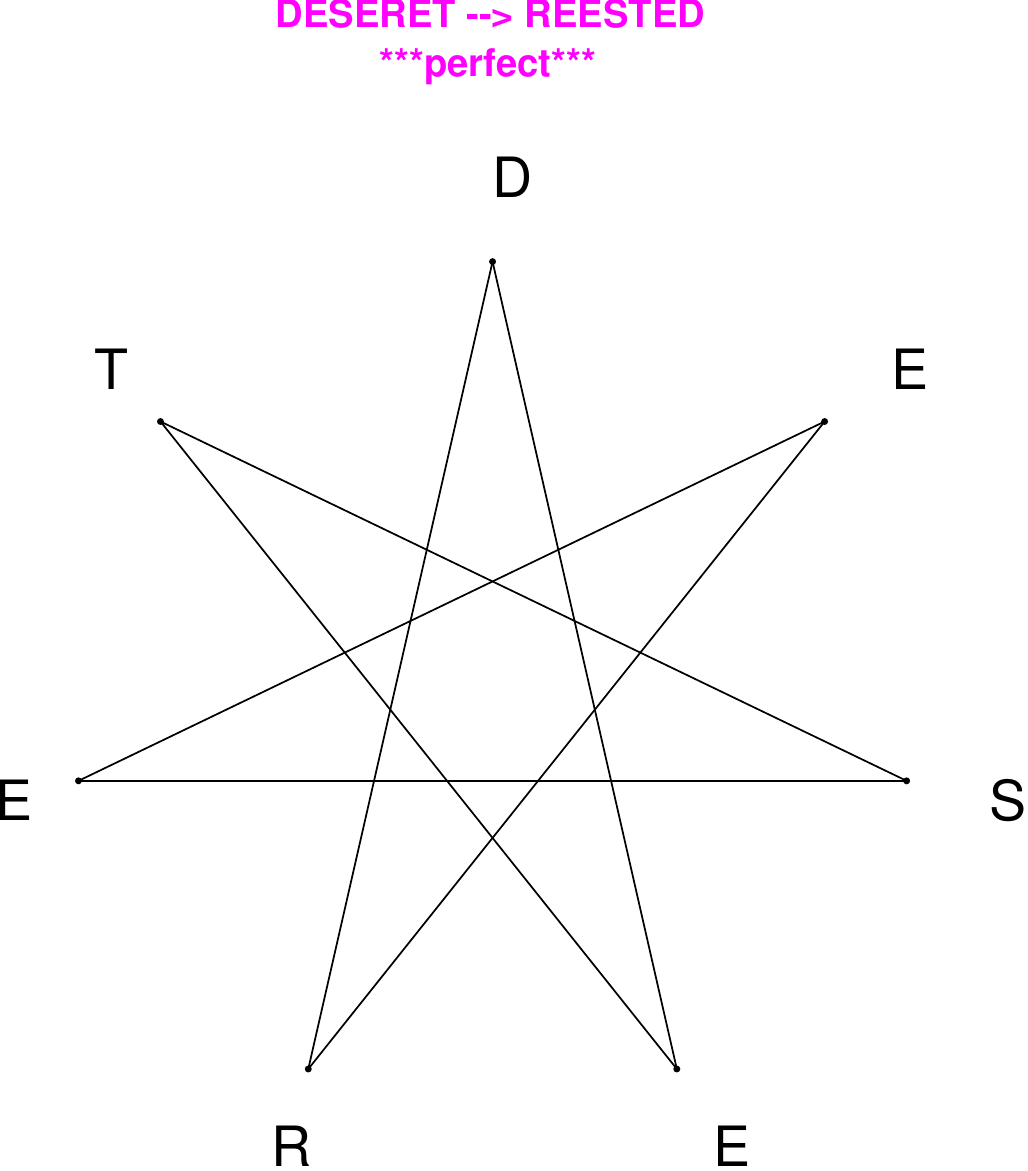}
\end{subfigure}
\hfill
\begin{subfigure}[T]{0.19\textwidth}
\centering
\includegraphics[width=\textwidth]{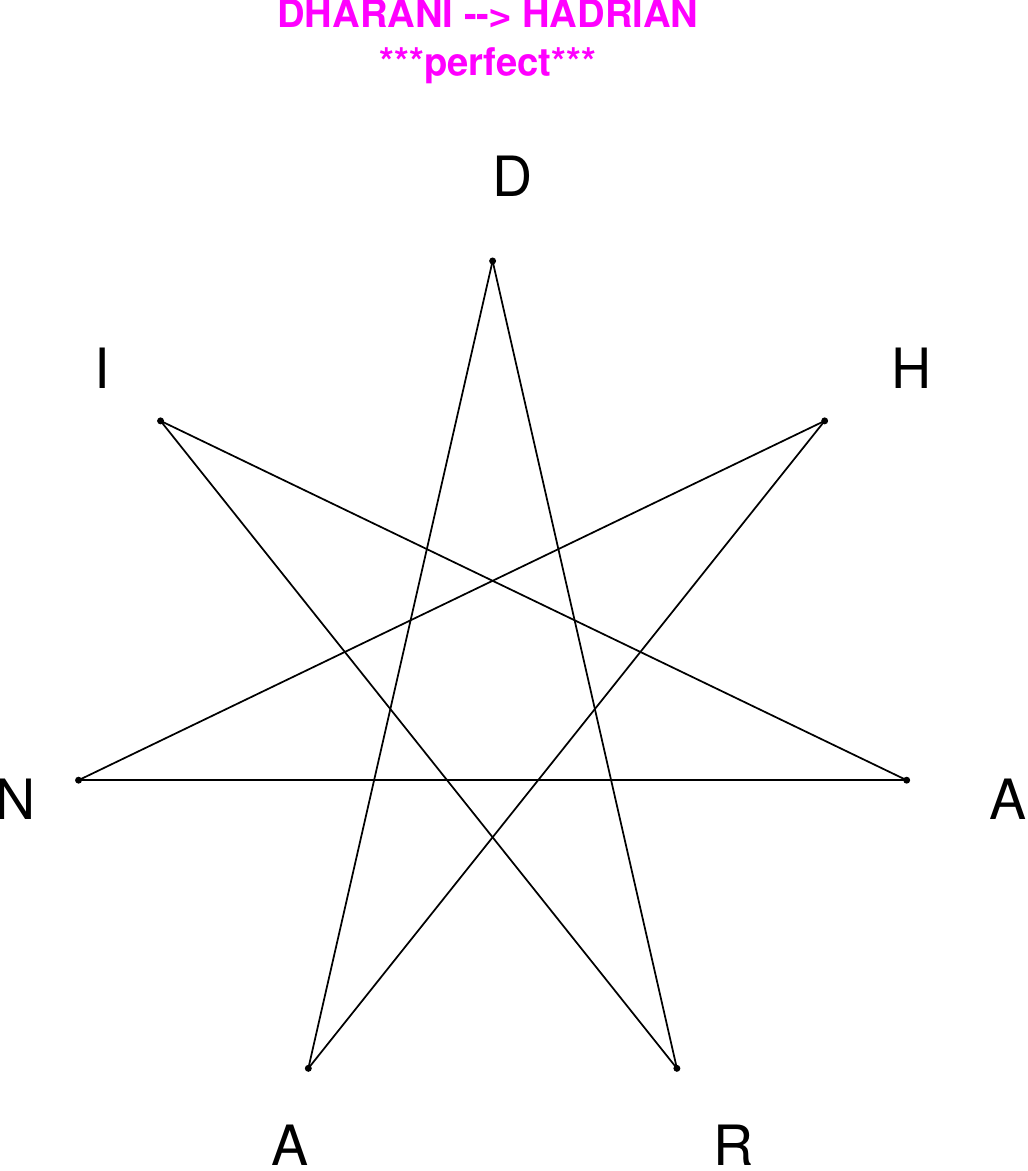}
\end{subfigure}
\end{figure}

\begin{figure}[H]
\centering
\begin{subfigure}[T]{0.19\textwidth}
\centering
\includegraphics[width=\textwidth]{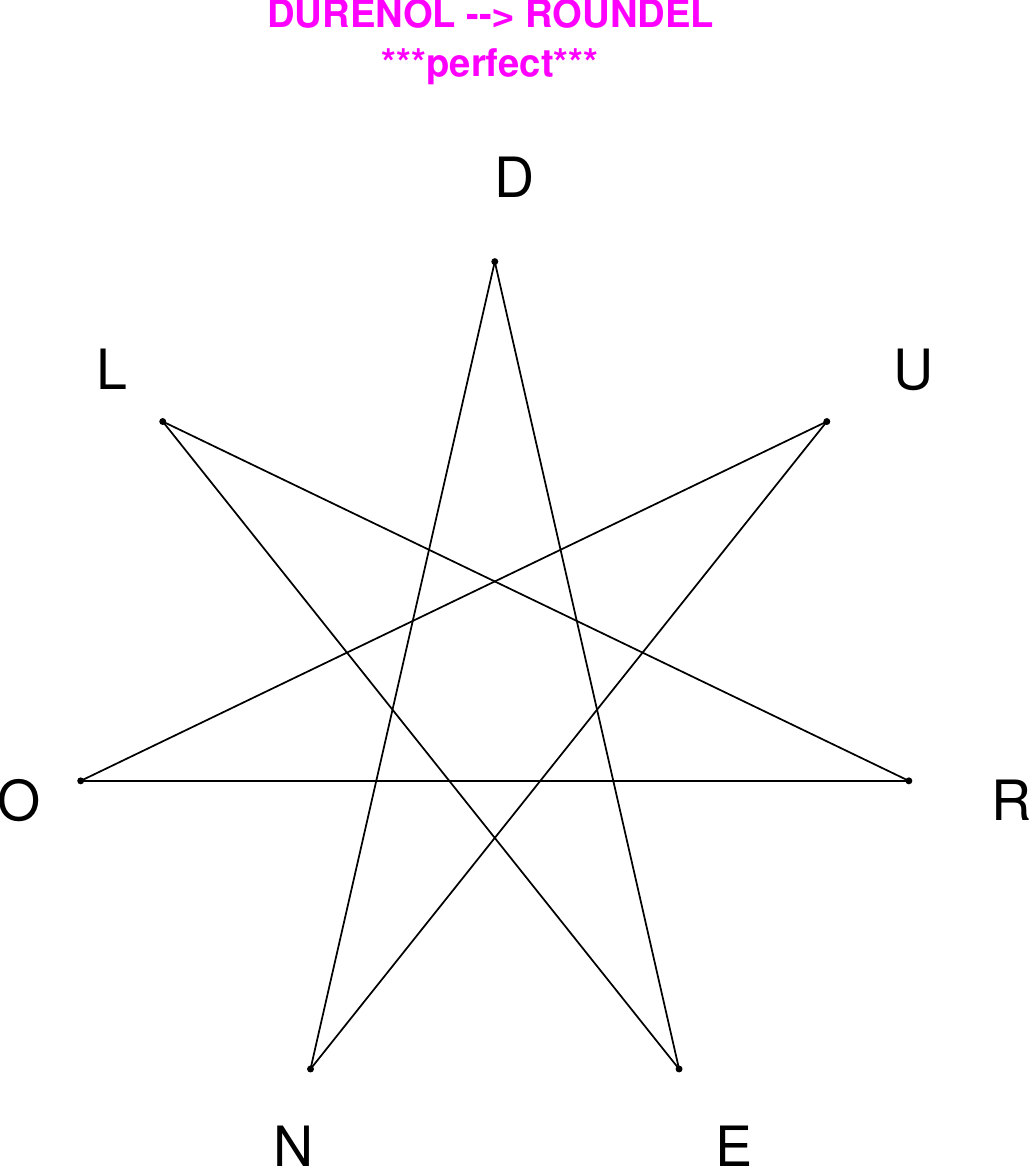}
\end{subfigure}
\hfill
\begin{subfigure}[T]{0.19\textwidth}
\centering
\includegraphics[width=\textwidth]{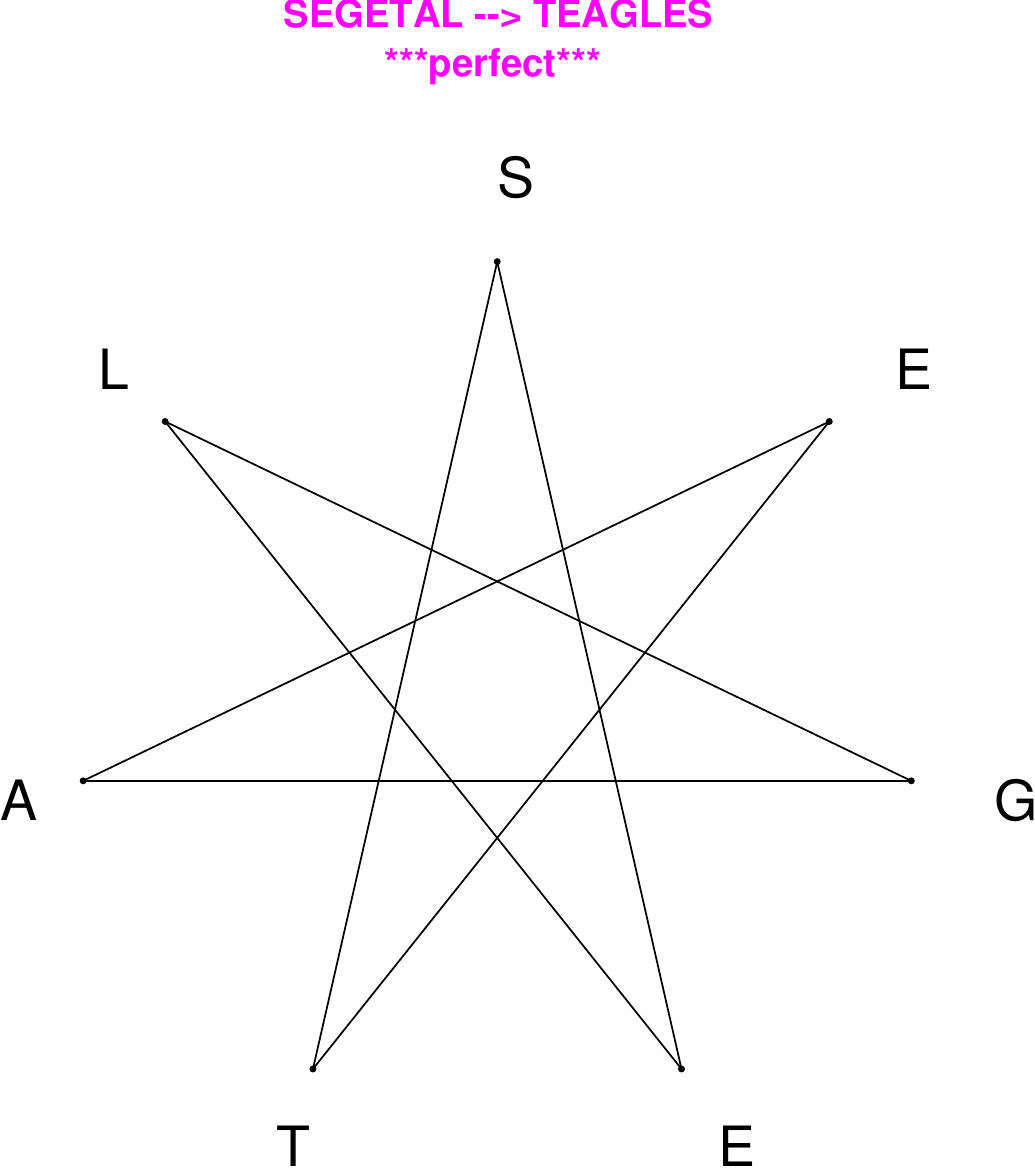}
\end{subfigure}
\hfill
\begin{subfigure}[T]{0.19\textwidth}
\centering
\includegraphics[width=\textwidth]{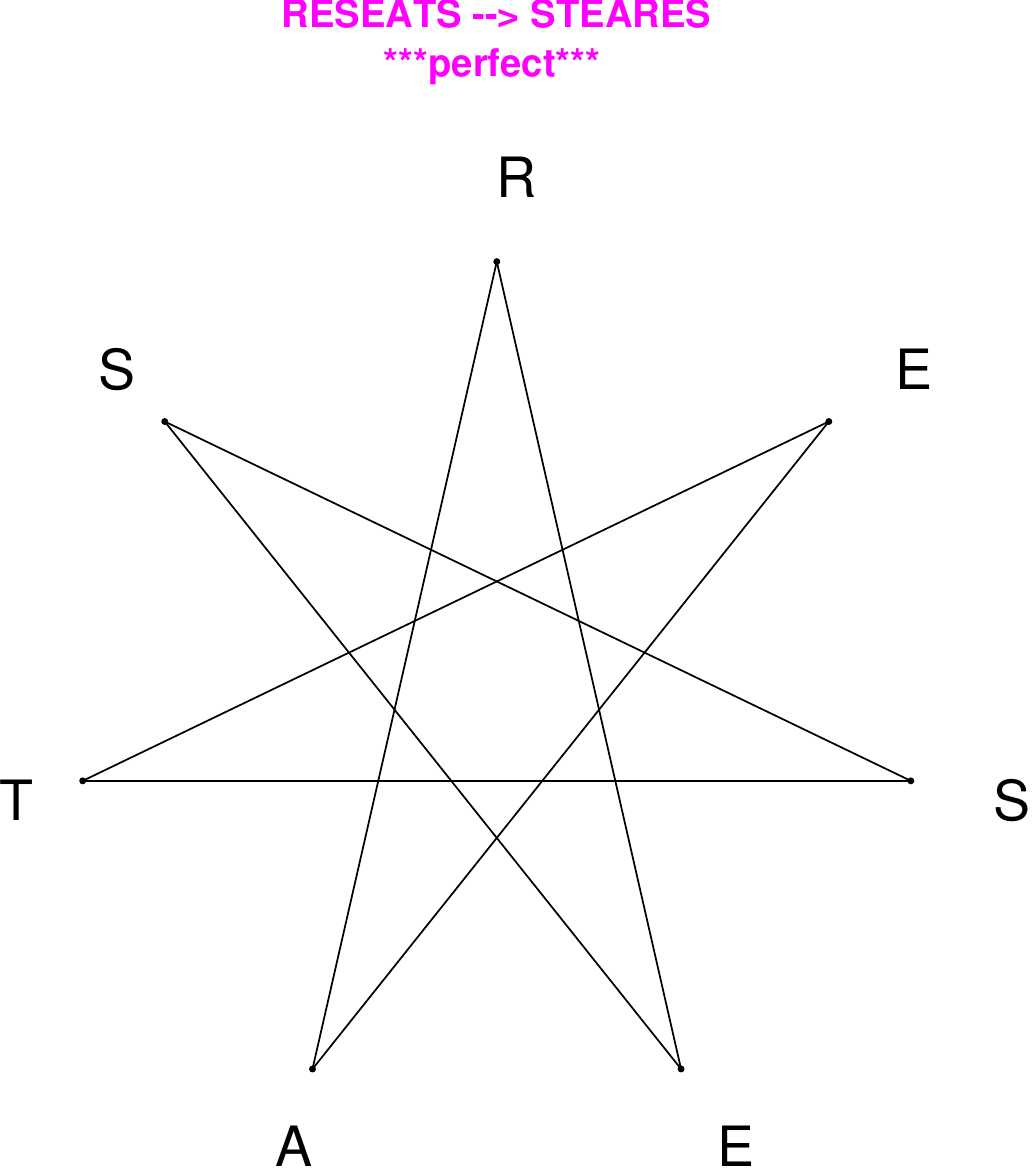}
\end{subfigure}
\hfill
\begin{subfigure}[T]{0.19\textwidth}
\centering
\includegraphics[width=\textwidth]{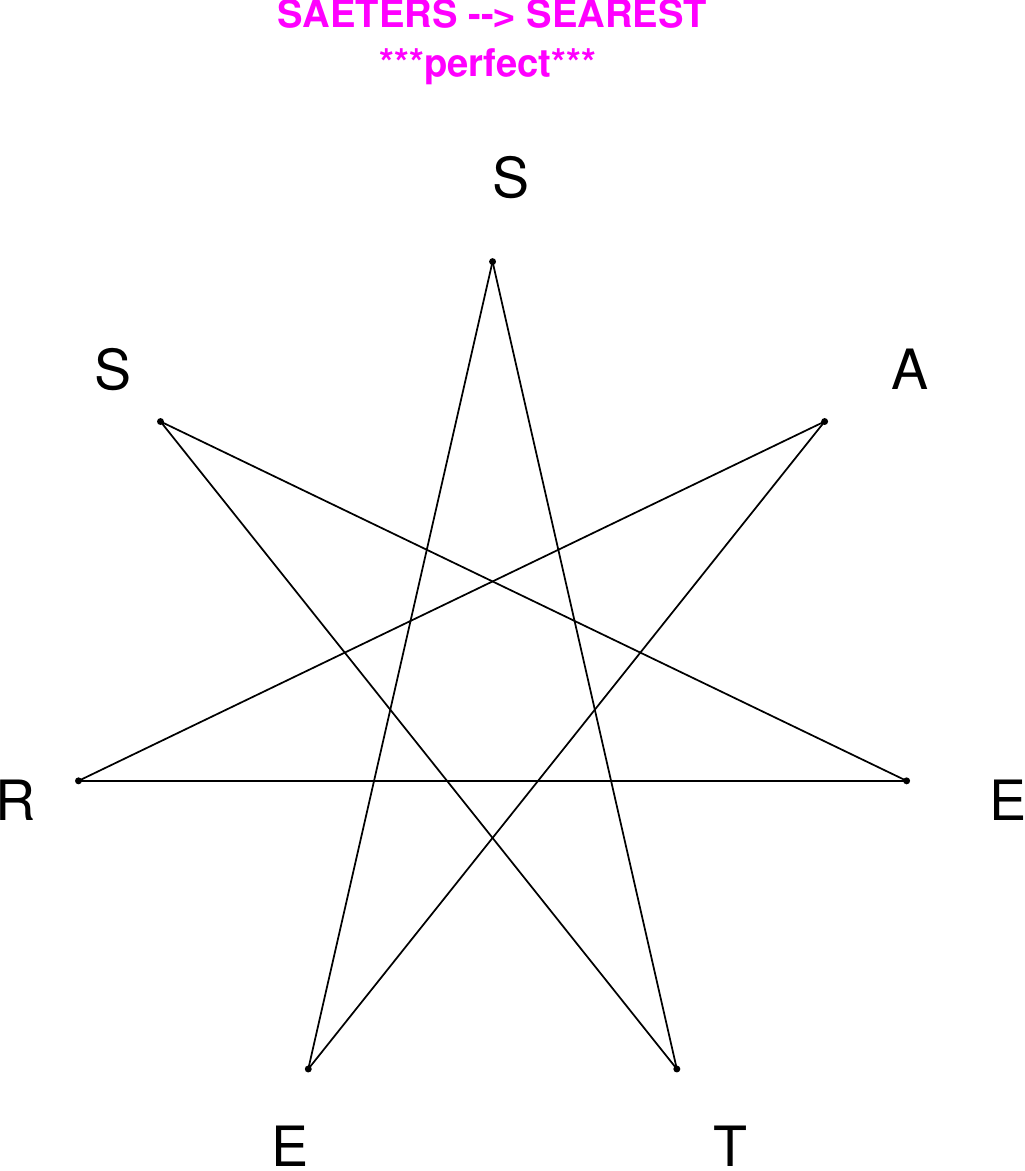}
\end{subfigure}
\hfill
\begin{subfigure}[T]{0.19\textwidth}
\centering
\includegraphics[width=\textwidth]{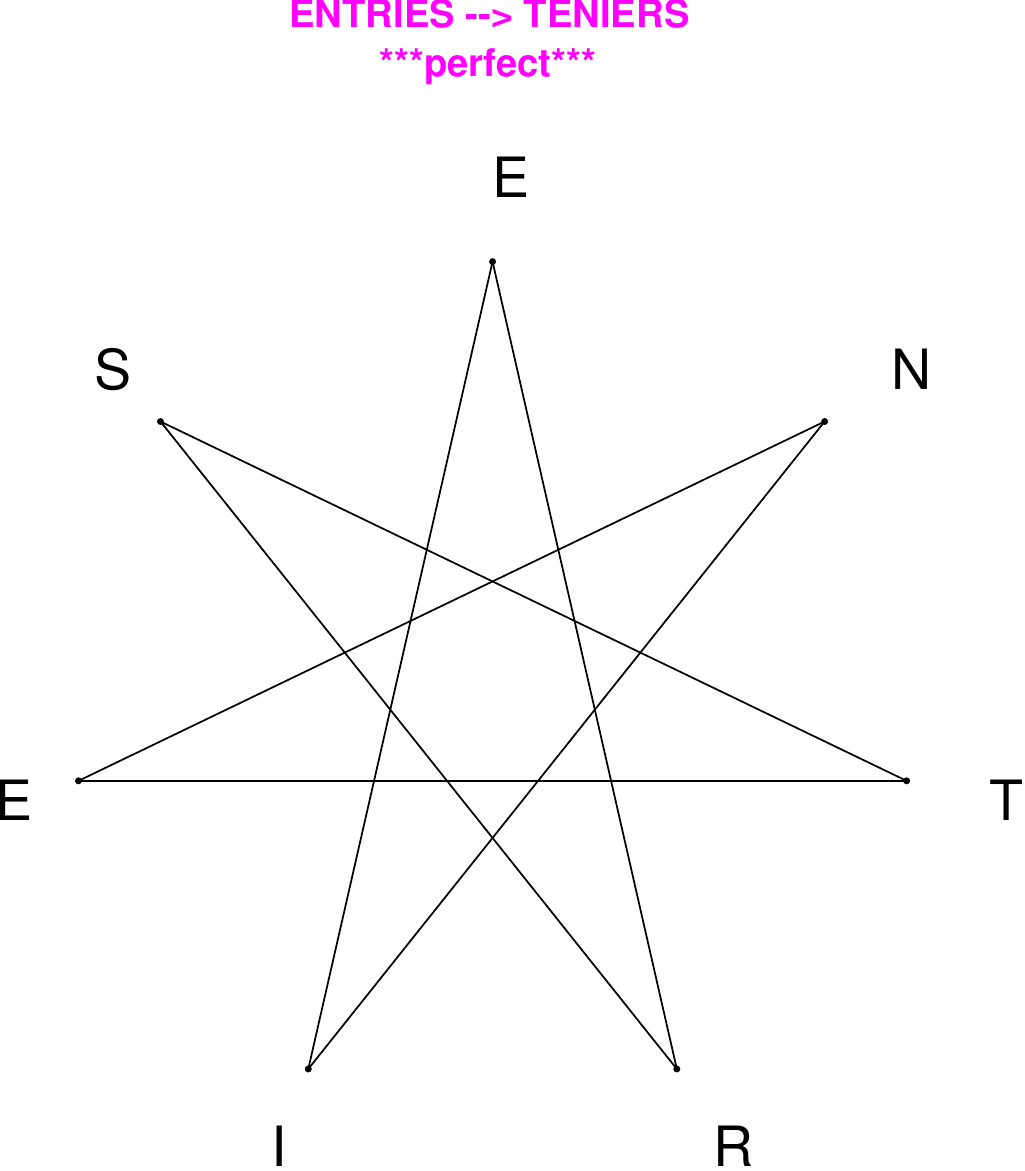}
\end{subfigure}
\end{figure}

\begin{figure}[H]
\centering
\begin{subfigure}[T]{0.19\textwidth}
\centering
\includegraphics[width=\textwidth]{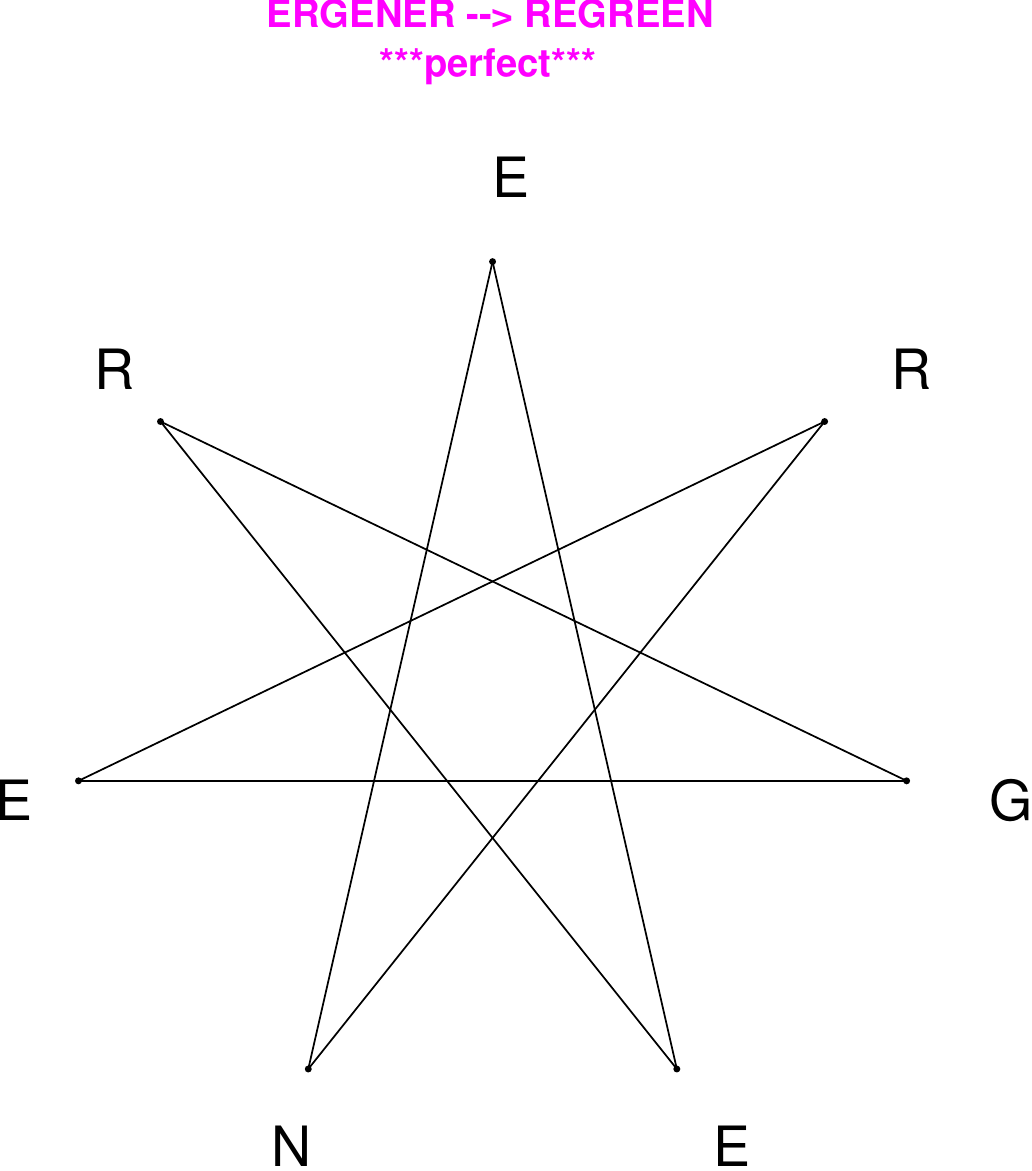}
\end{subfigure}
\hfill
\begin{subfigure}[T]{0.19\textwidth}
\centering
\includegraphics[width=\textwidth]{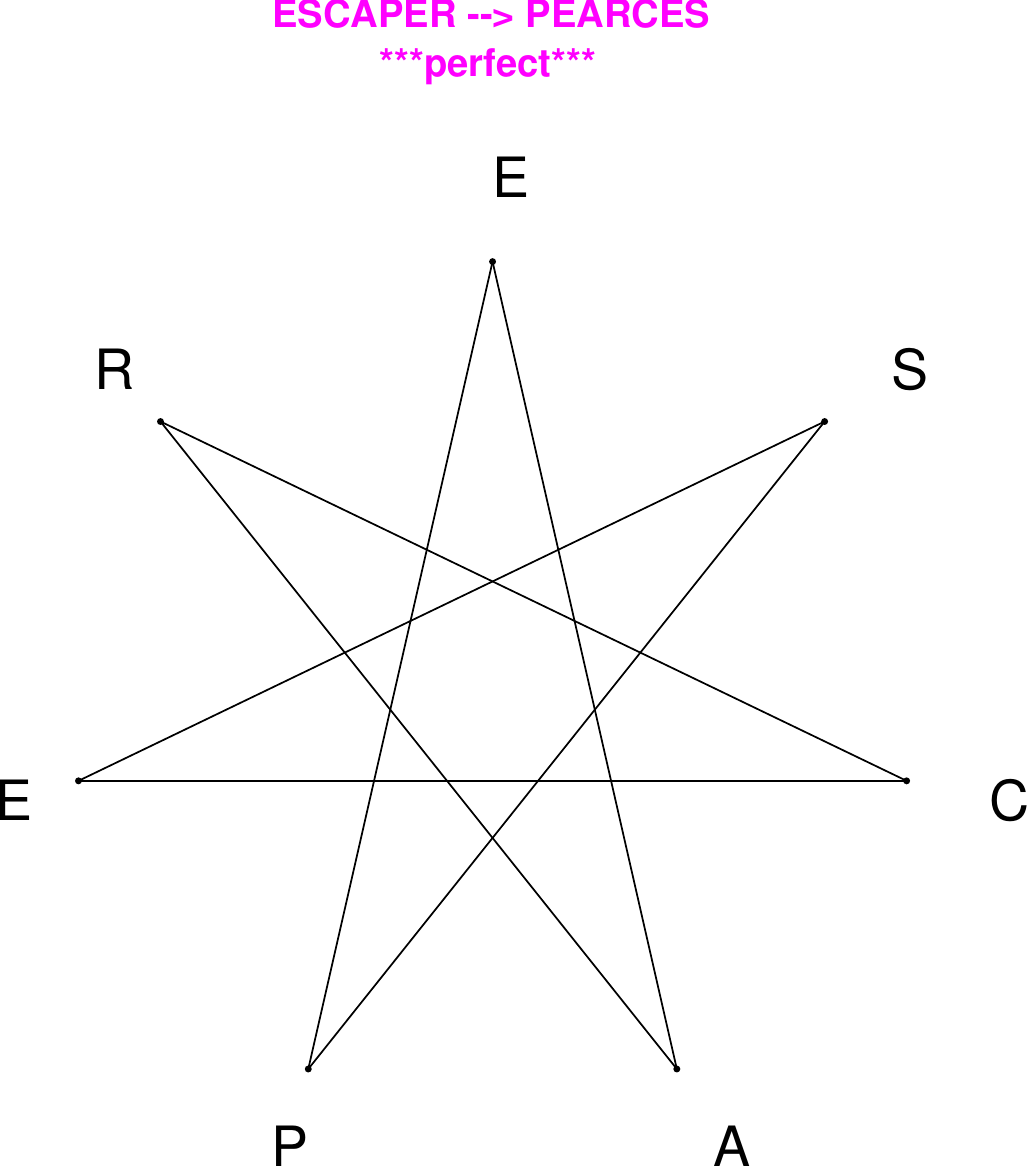}
\end{subfigure}
\hfill
\begin{subfigure}[T]{0.19\textwidth}
\centering
\includegraphics[width=\textwidth]{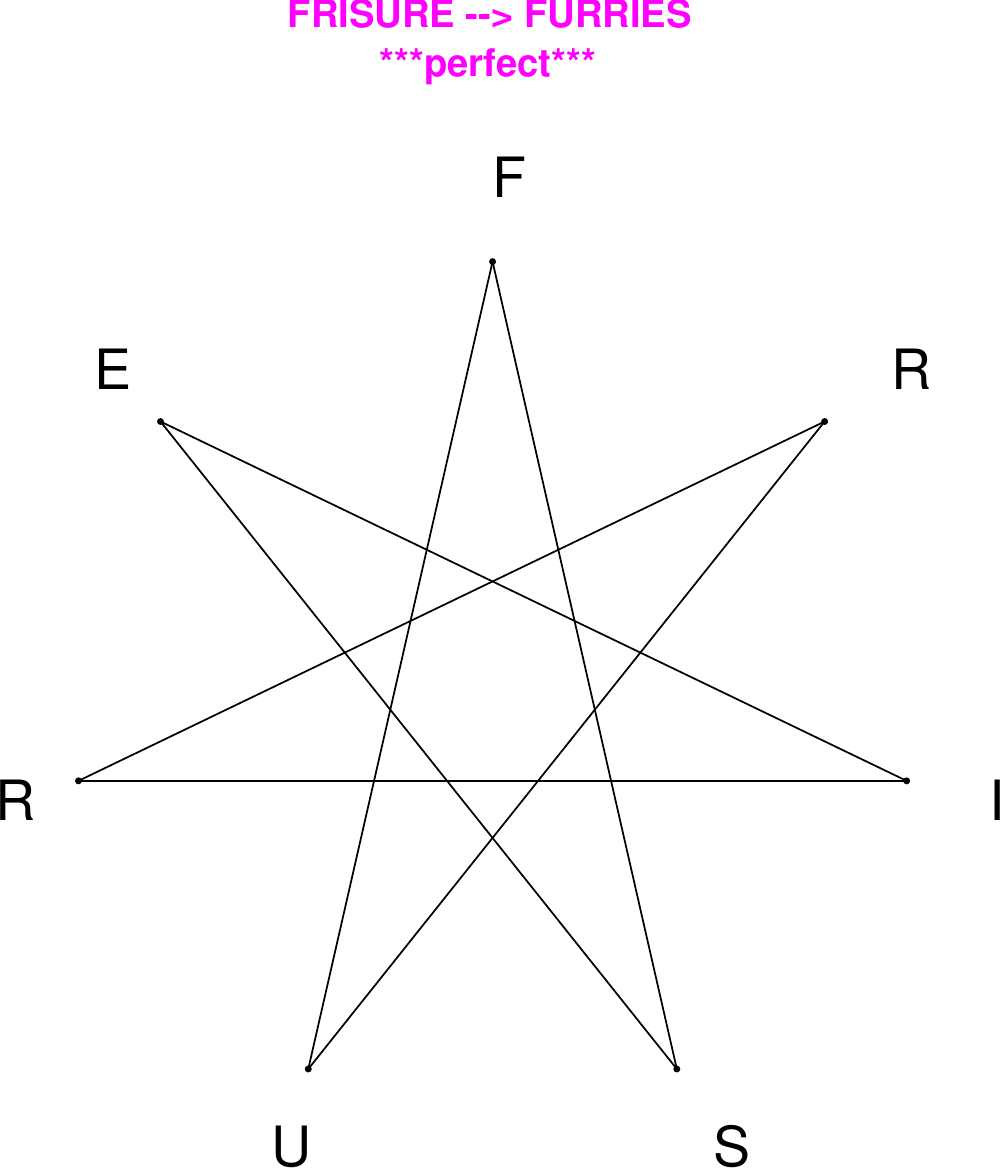}
\end{subfigure}
\hfill
\begin{subfigure}[T]{0.19\textwidth}
\centering
\includegraphics[width=\textwidth]{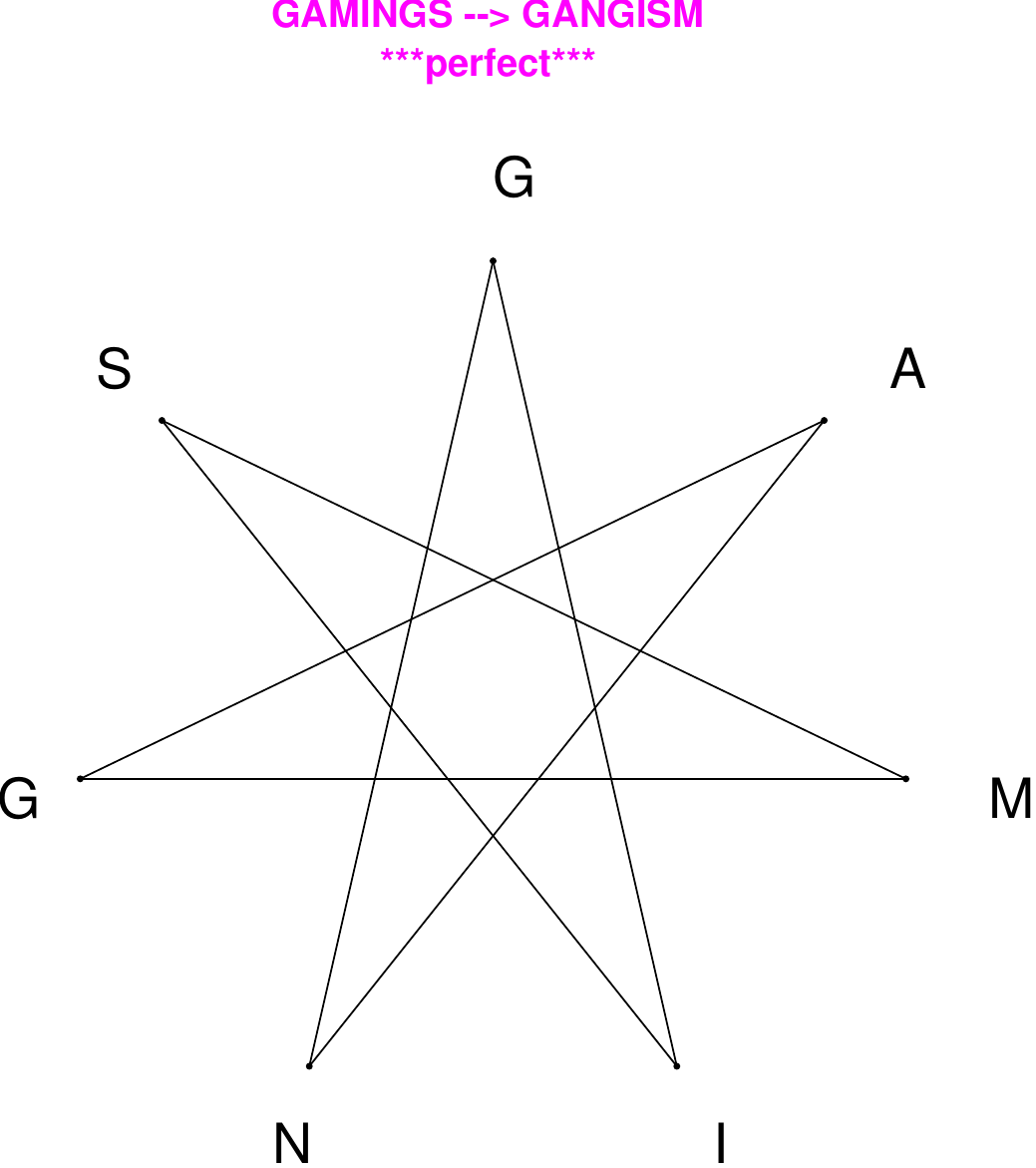}
\end{subfigure}
\hfill
\begin{subfigure}[T]{0.19\textwidth}
\centering
\includegraphics[width=\textwidth]{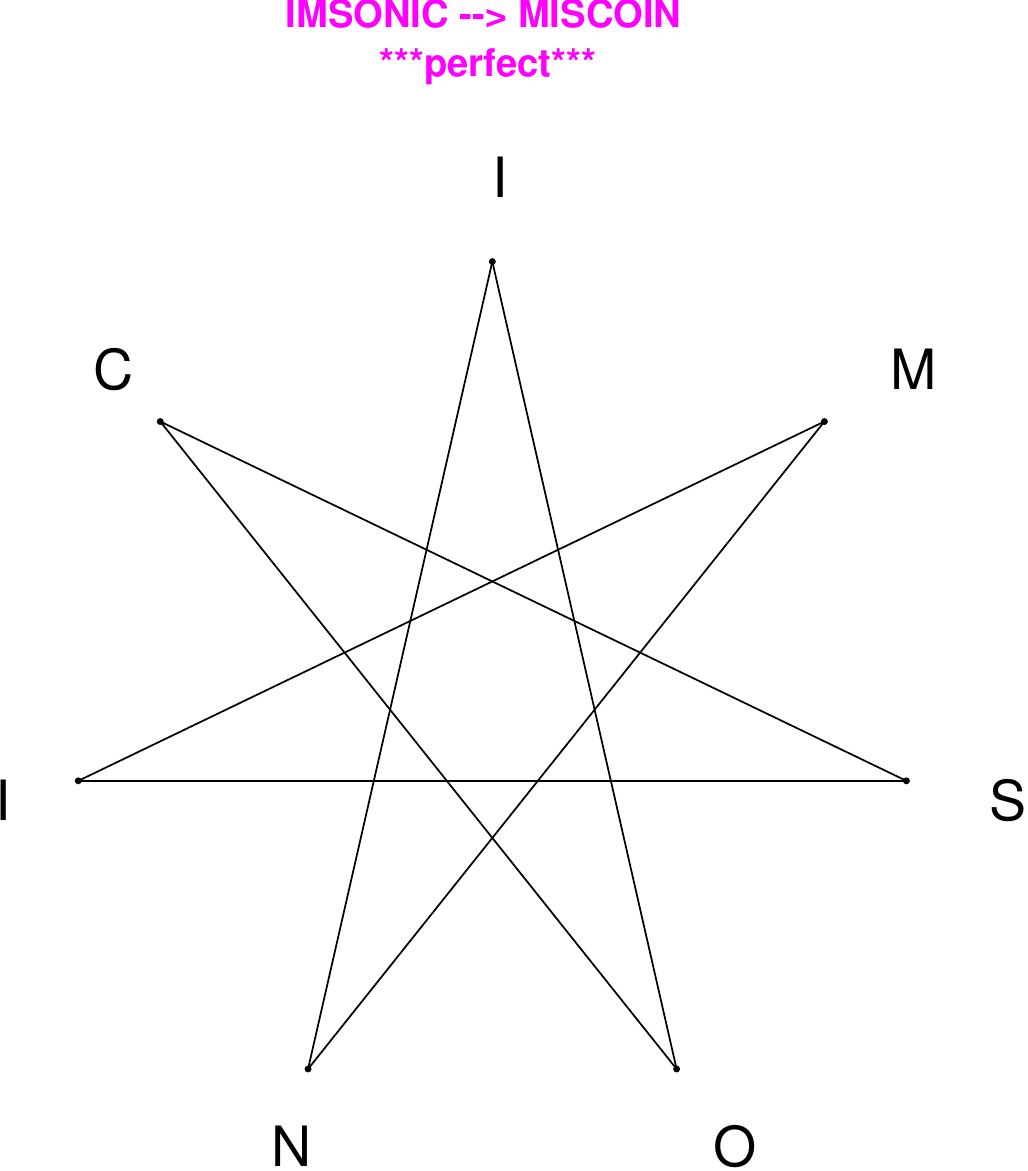}
\end{subfigure}
\end{figure}

\begin{figure}[H]
\centering
\begin{subfigure}[T]{0.19\textwidth}
\centering
\includegraphics[width=\textwidth]{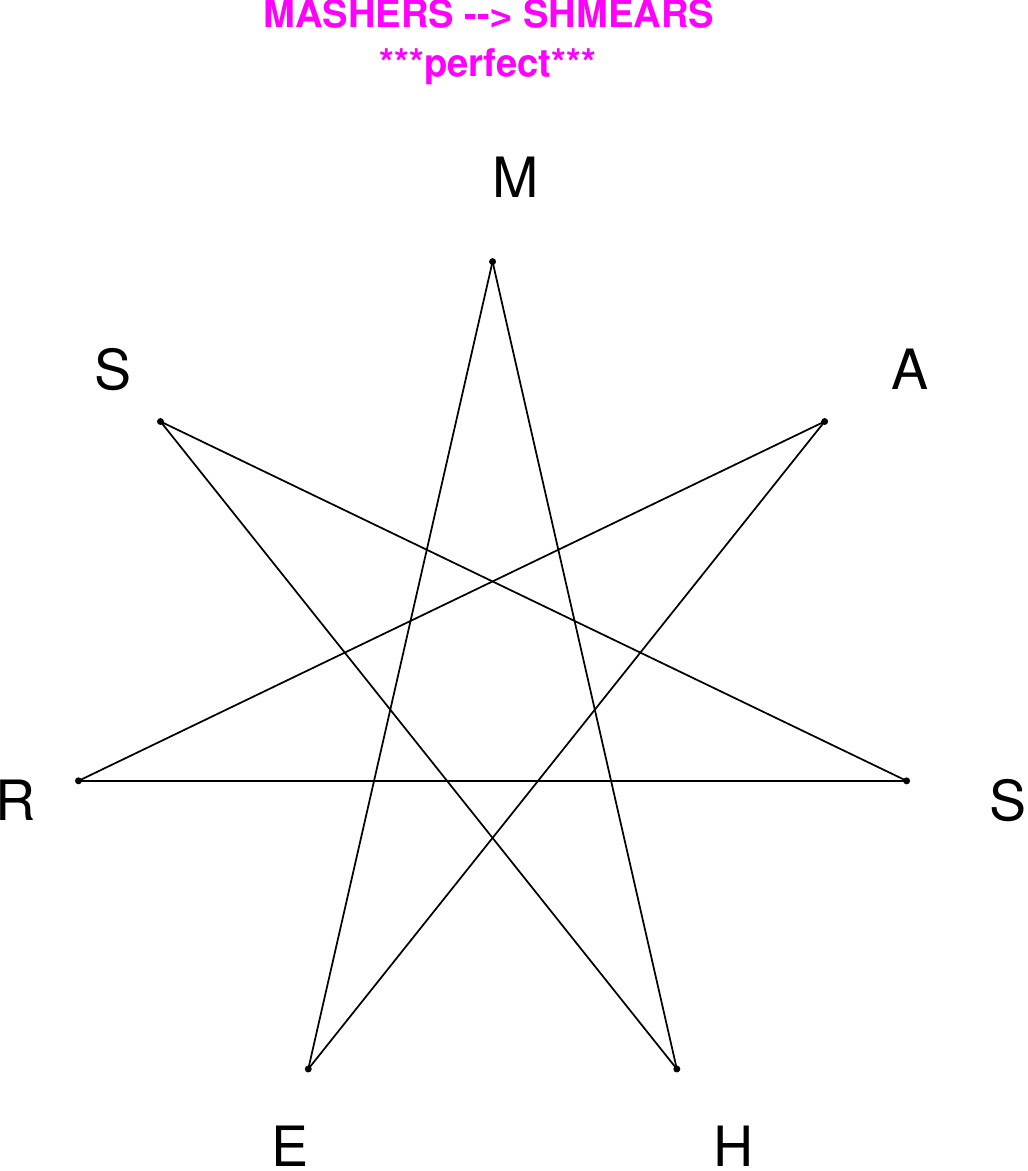}
\end{subfigure}
\hfill
\begin{subfigure}[T]{0.19\textwidth}
\centering
\includegraphics[width=\textwidth]{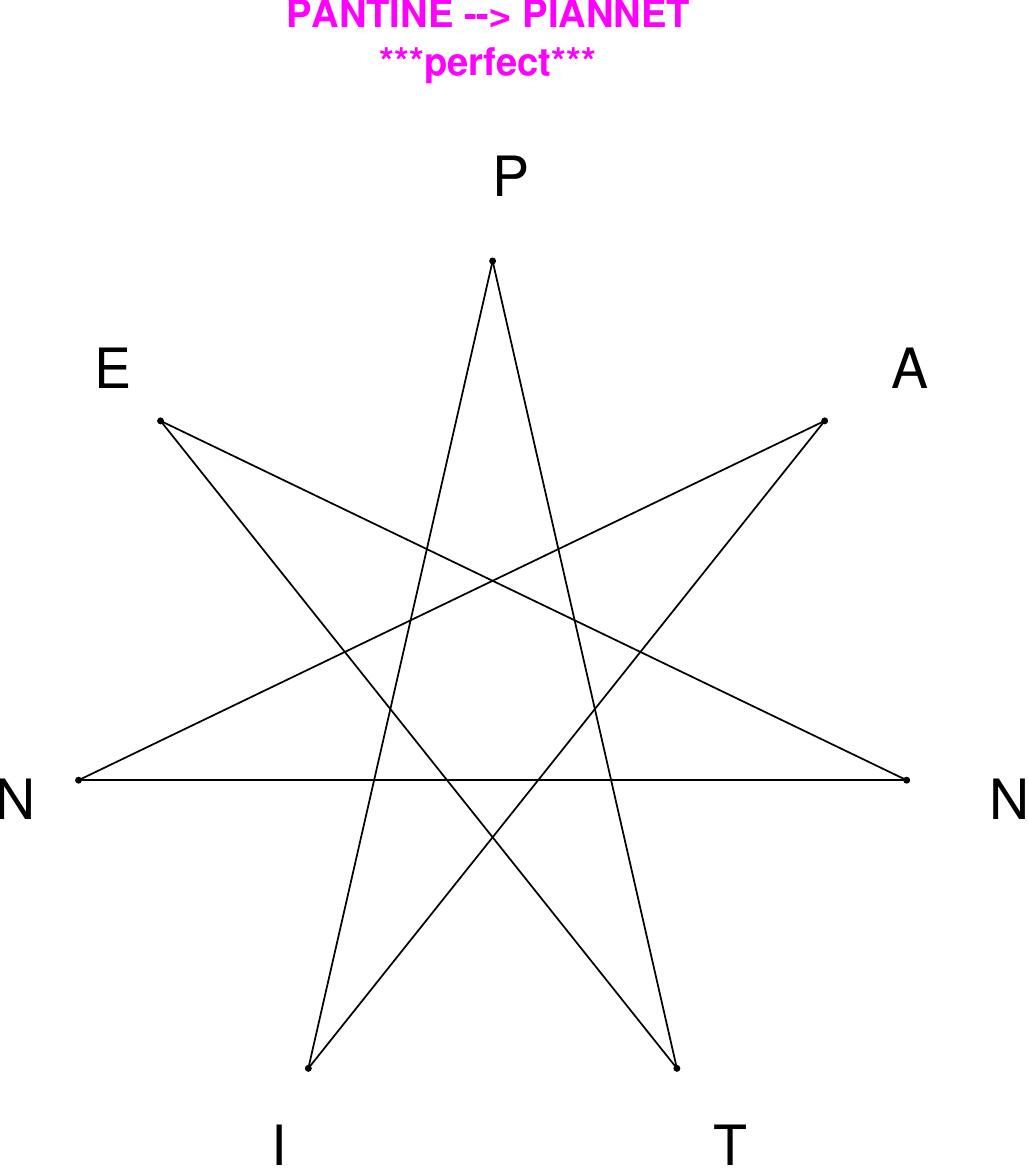}
\end{subfigure}
\hfill
\begin{subfigure}[T]{0.19\textwidth}
\centering
\includegraphics[width=\textwidth]{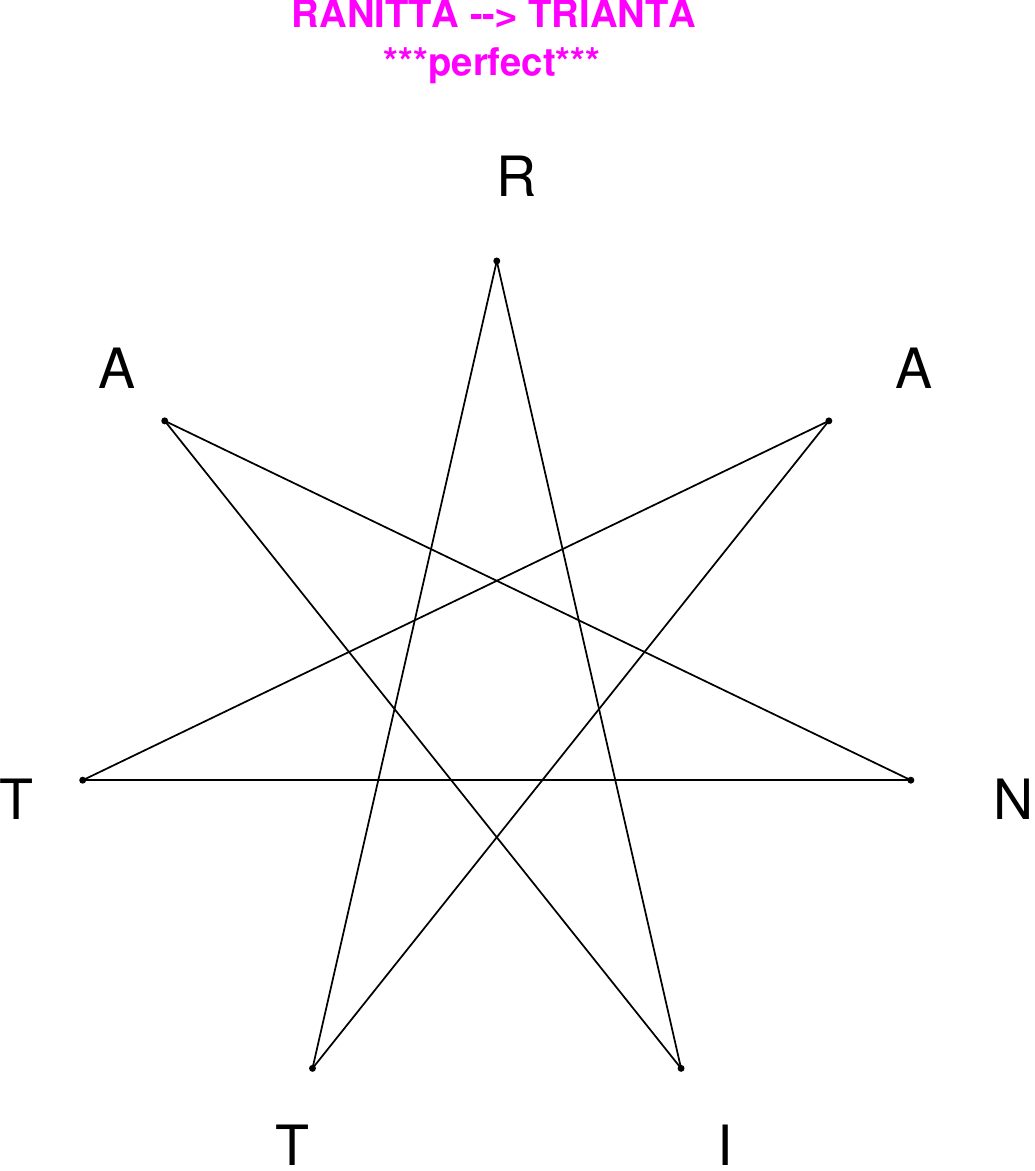}
\end{subfigure}
\hfill
\begin{subfigure}[T]{0.19\textwidth}
\centering
\includegraphics[width=\textwidth]{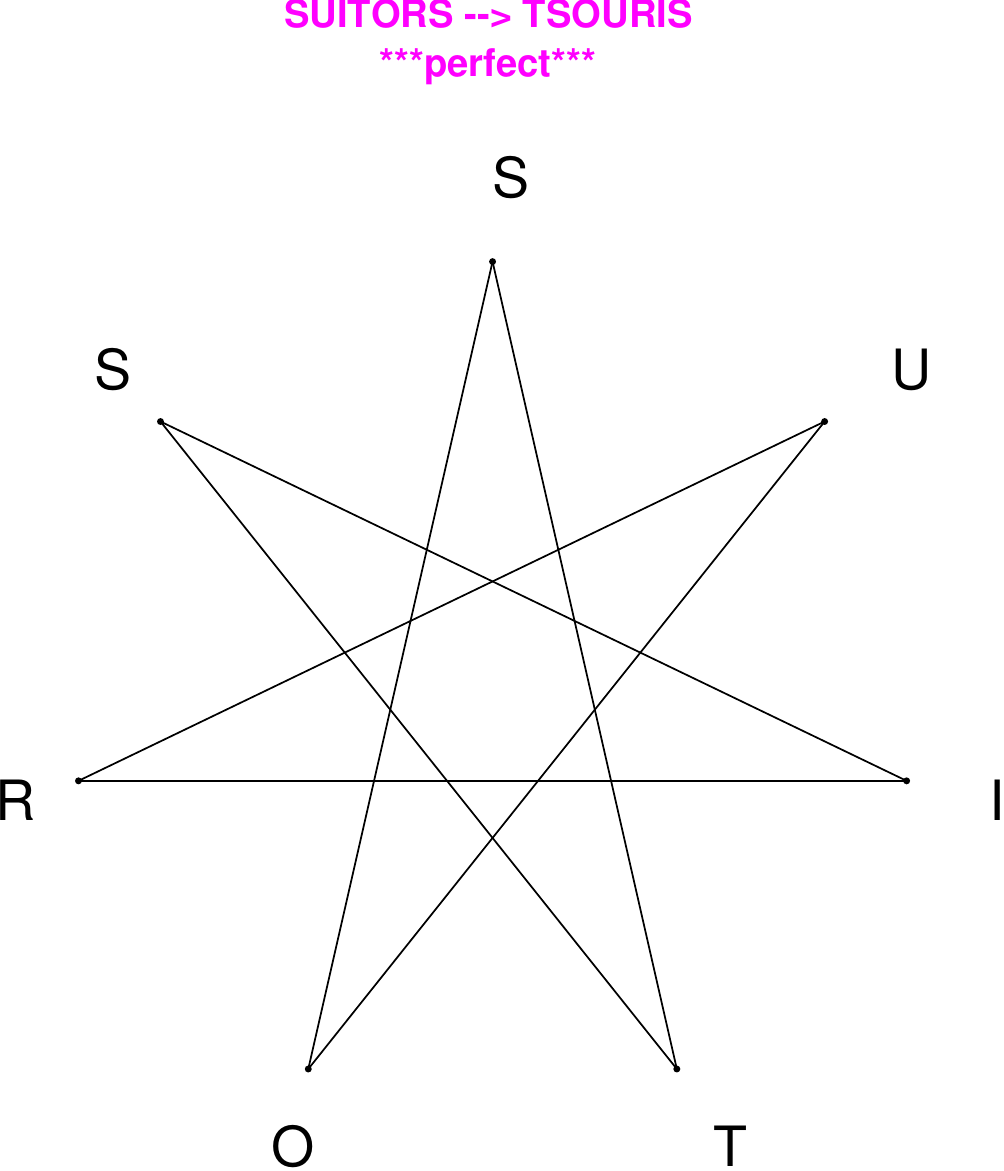}
\end{subfigure}
\hfill
\begin{subfigure}[T]{0.19\textwidth}
\centering
\includegraphics[width=\textwidth]{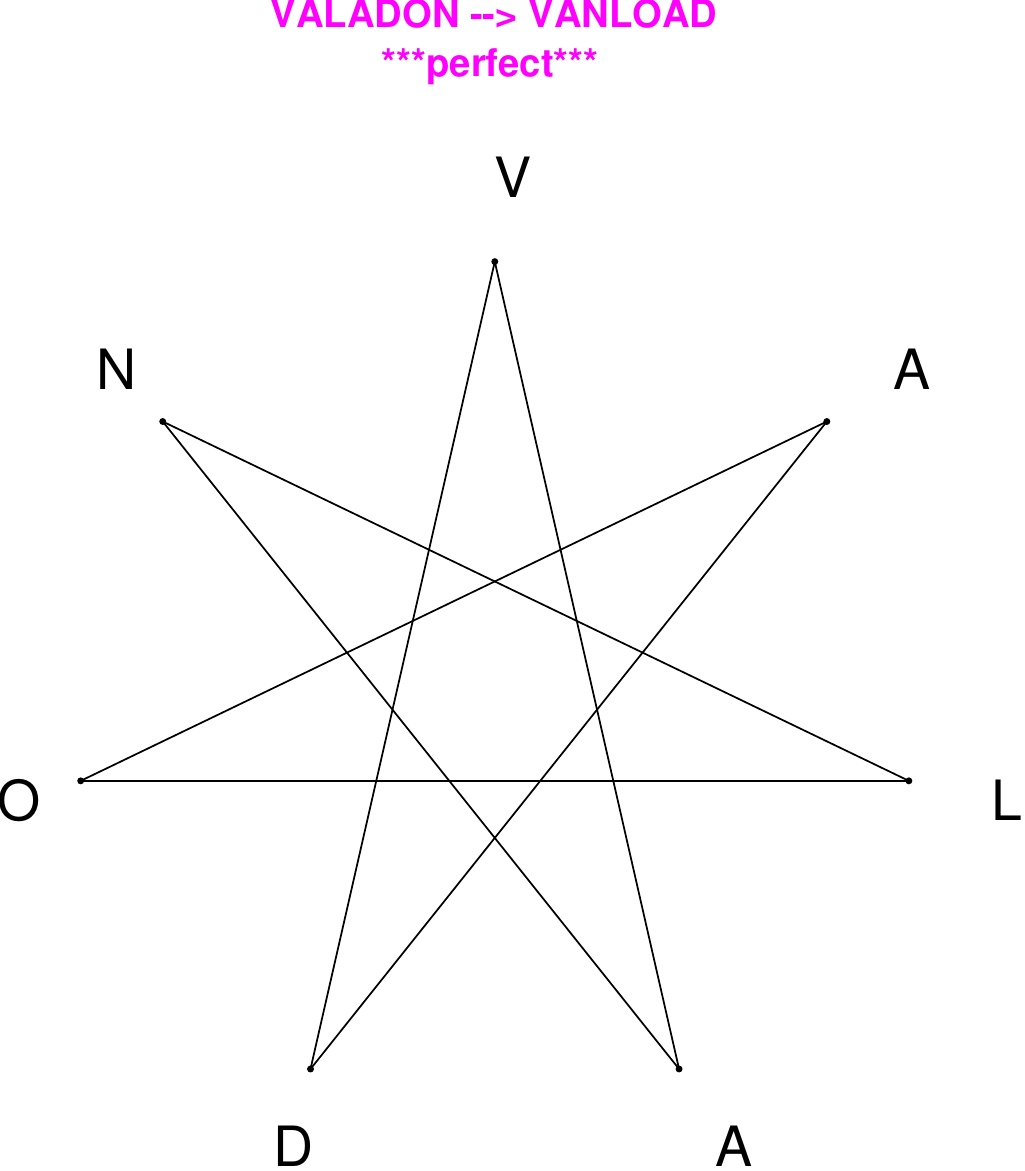}
\end{subfigure}
\end{figure}

\begin{figure}[H]
\centering
\begin{subfigure}[T]{0.19\textwidth}
\centering
\includegraphics[width=\textwidth]{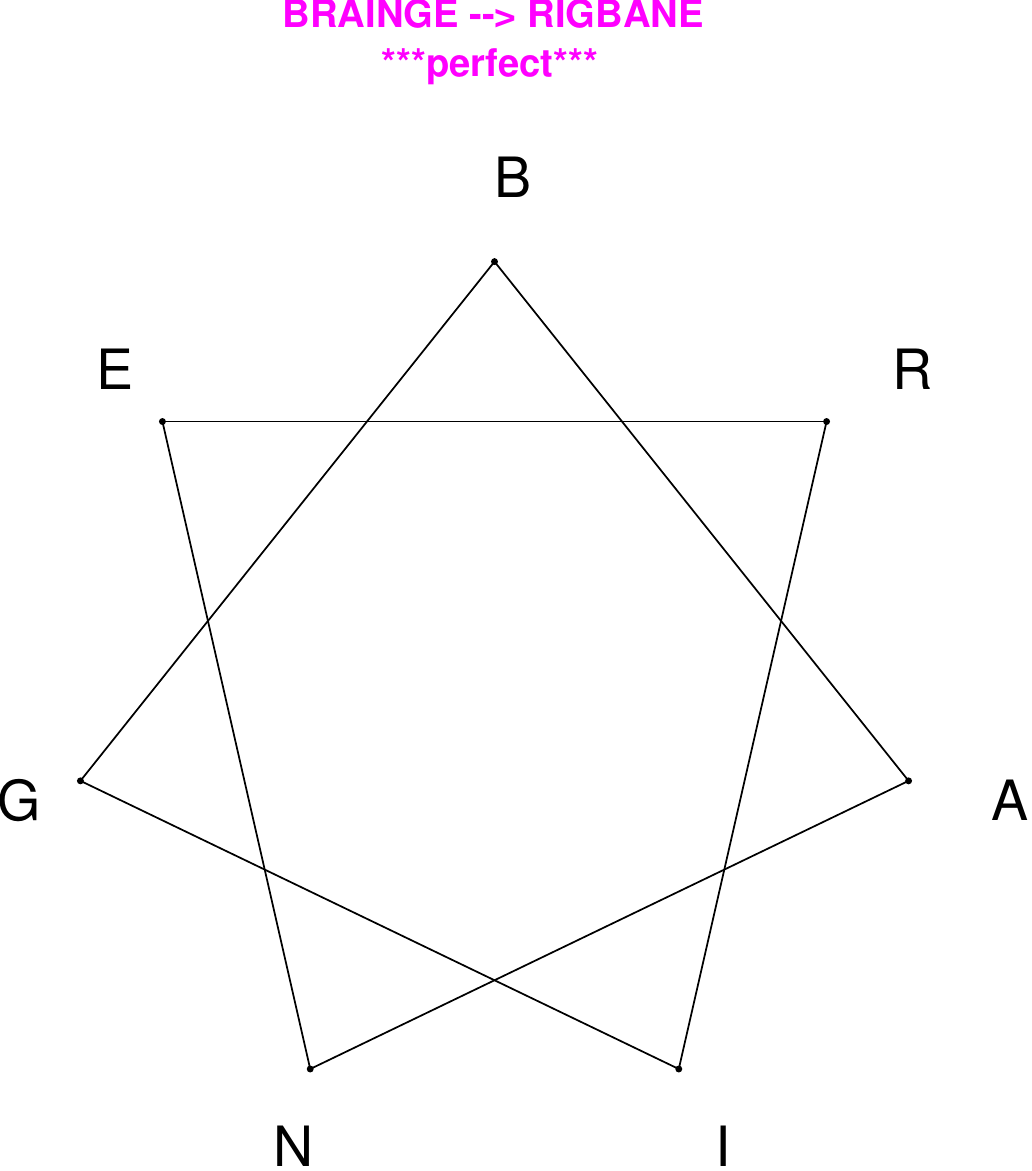}
\end{subfigure}
\hfill
\begin{subfigure}[T]{0.19\textwidth}
\centering
\includegraphics[width=\textwidth]{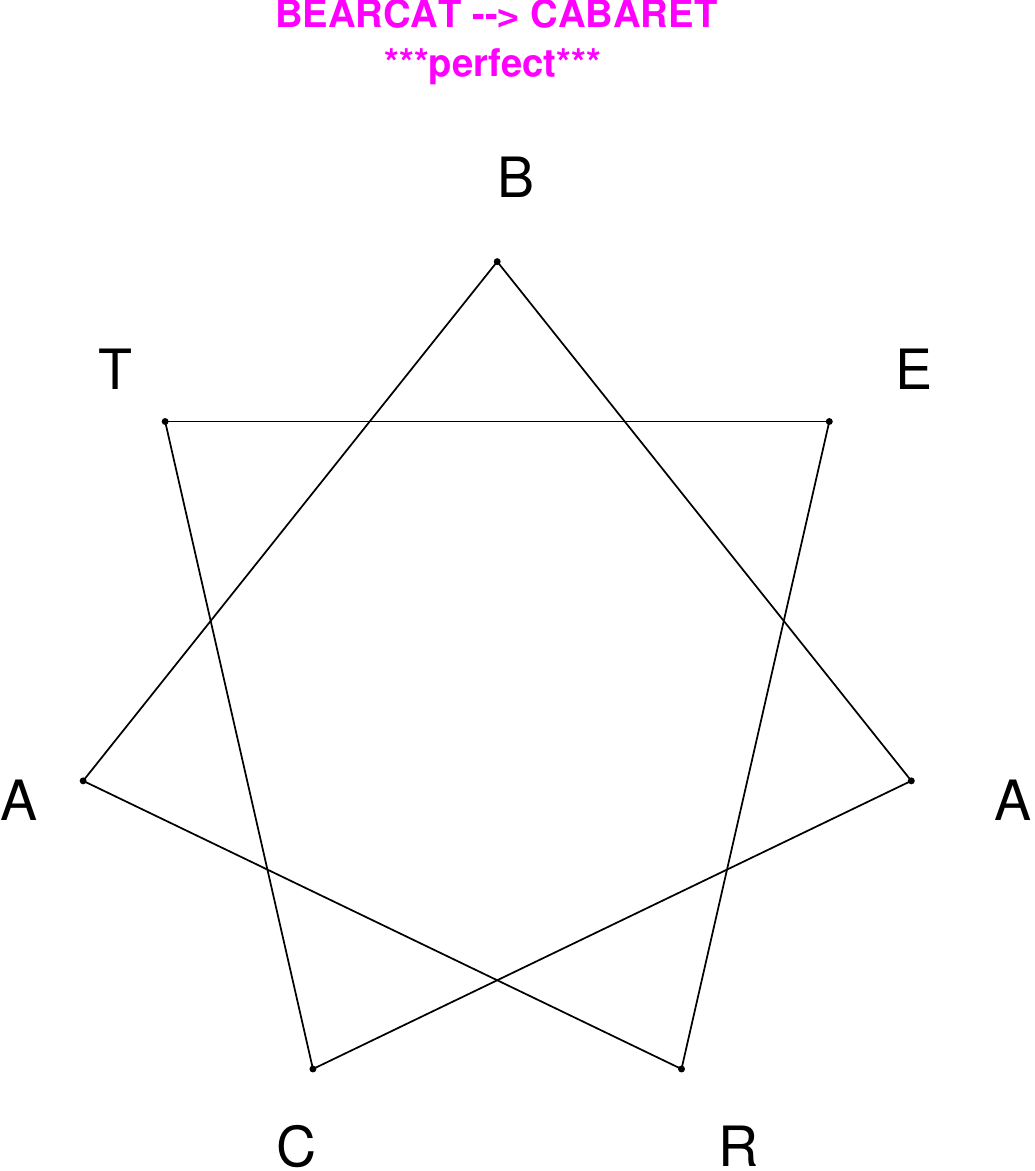}
\end{subfigure}
\hfill
\begin{subfigure}[T]{0.19\textwidth}
\centering
\includegraphics[width=\textwidth]{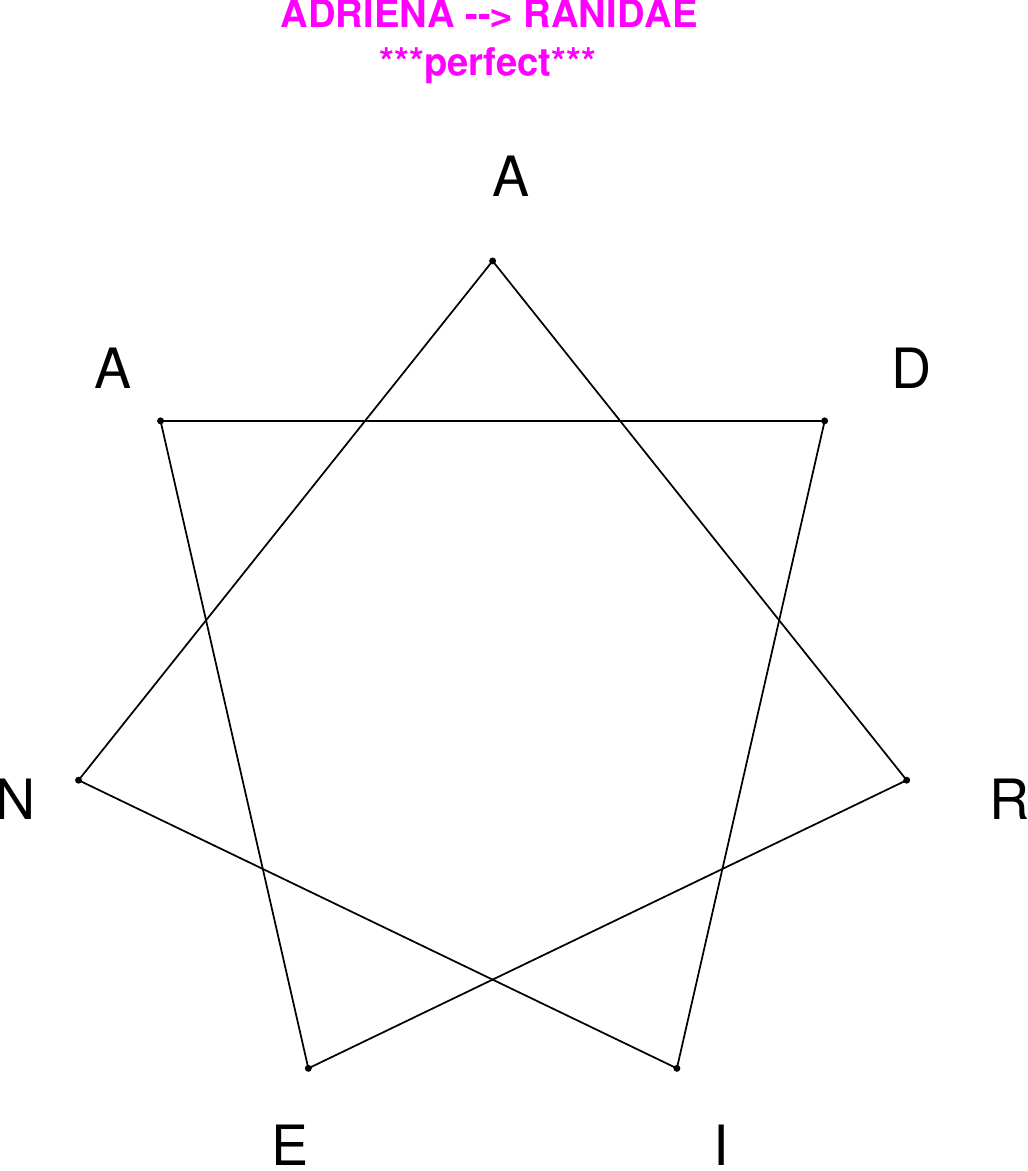}
\end{subfigure}
\hfill
\begin{subfigure}[T]{0.19\textwidth}
\centering
\includegraphics[width=\textwidth]{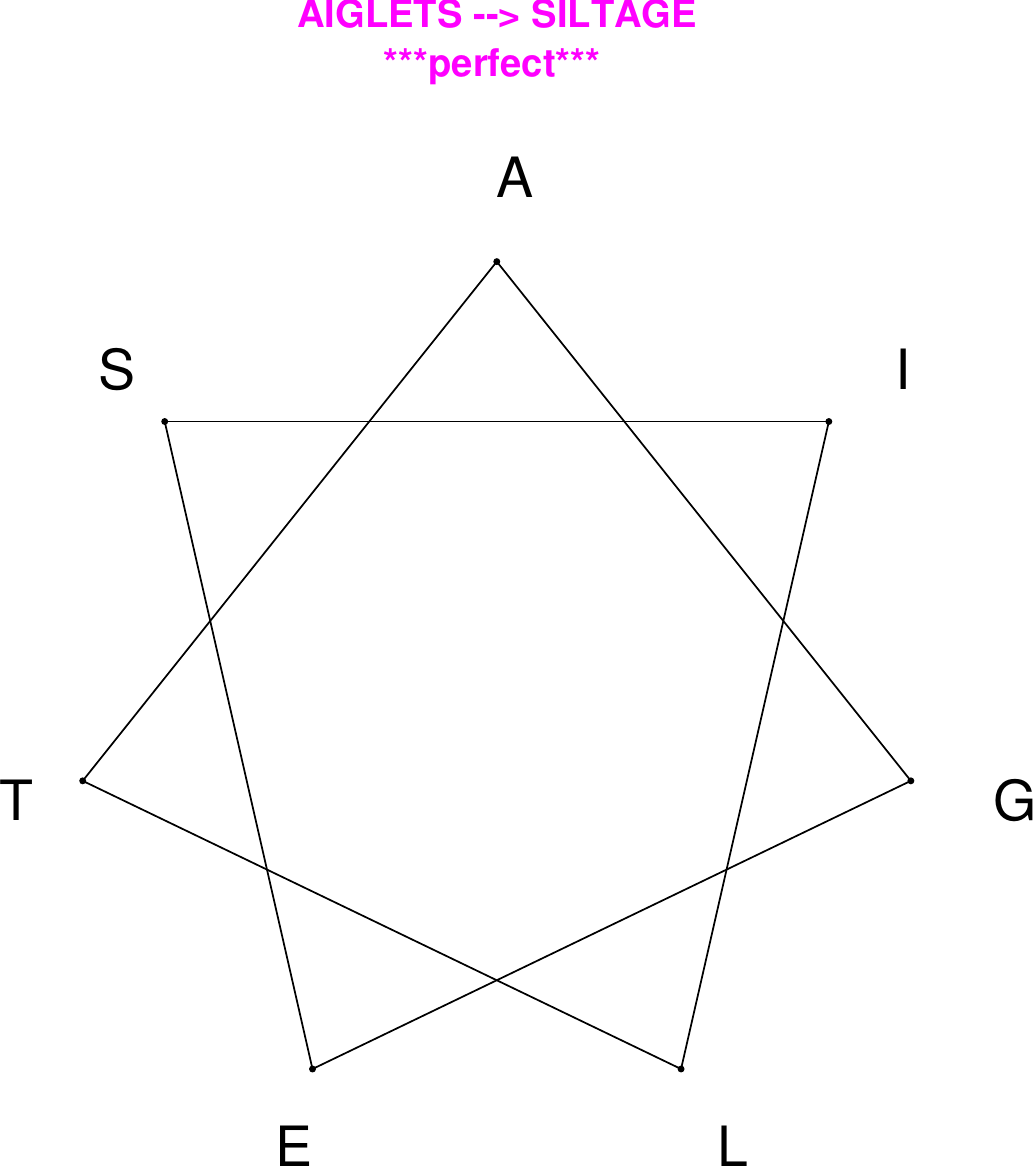}
\end{subfigure}
\hfill
\begin{subfigure}[T]{0.19\textwidth}
\centering
\includegraphics[width=\textwidth]{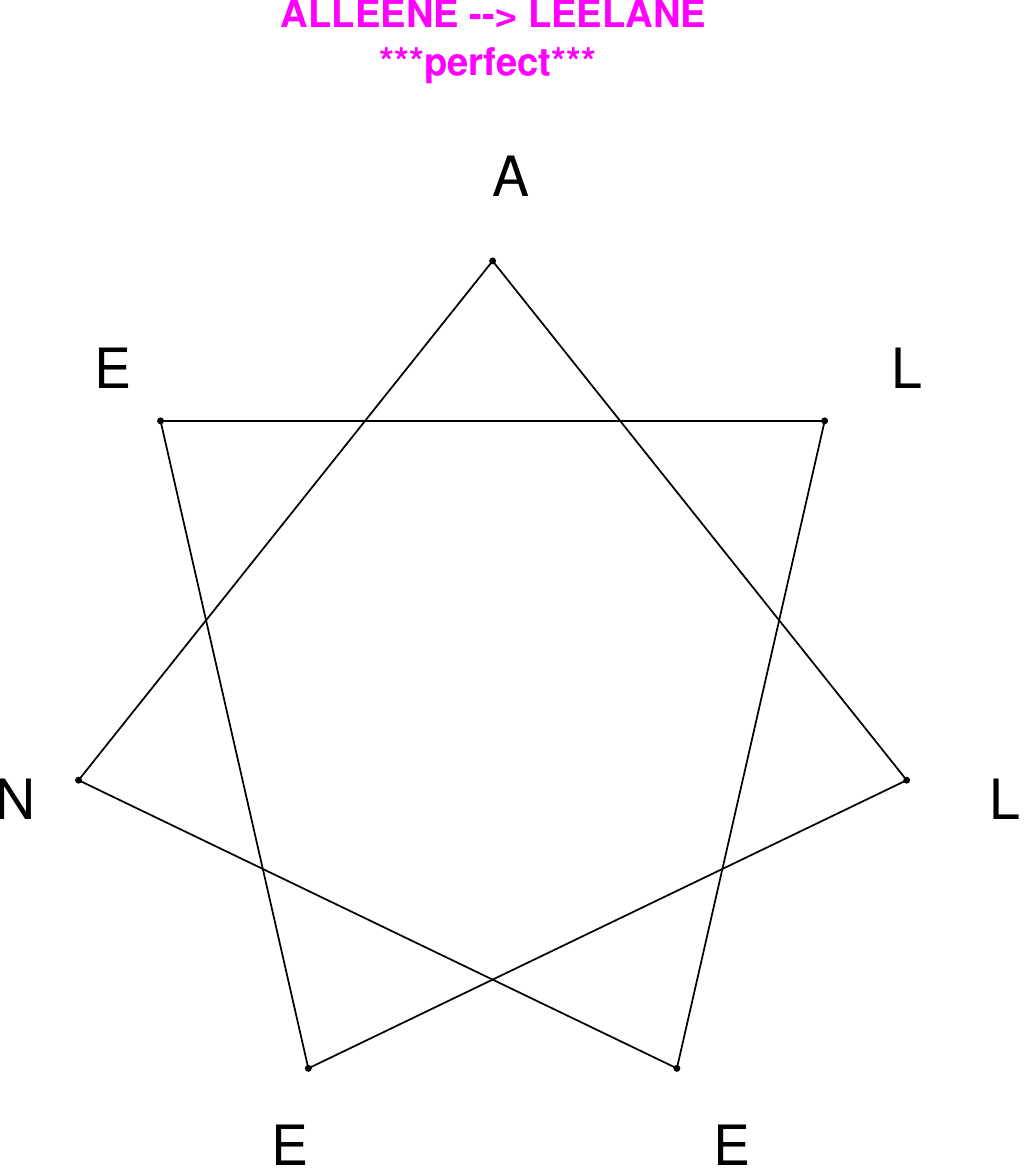}
\end{subfigure}
\end{figure}

\begin{figure}[H]
\centering
\begin{subfigure}[T]{0.19\textwidth}
\centering
\includegraphics[width=\textwidth]{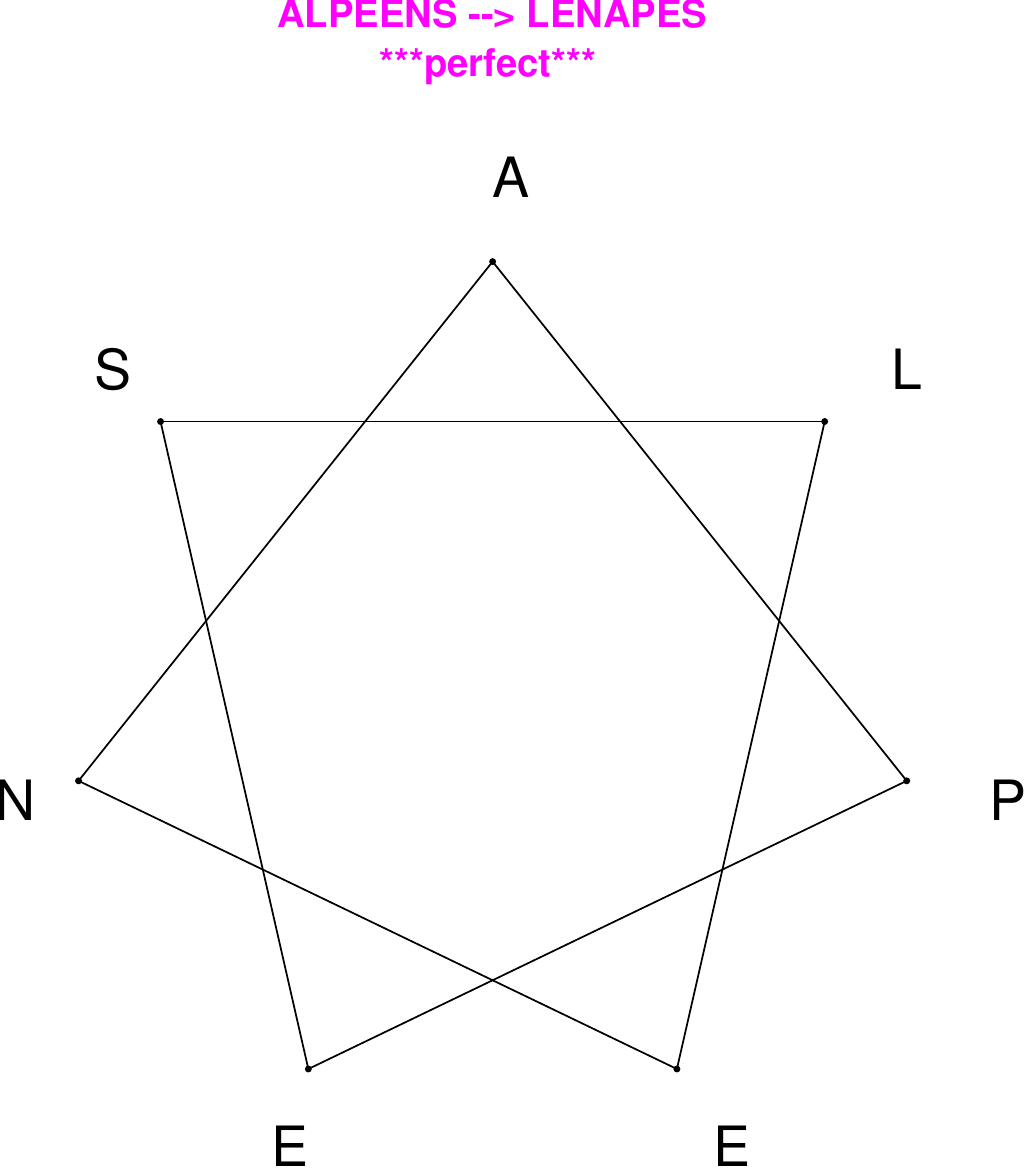}
\end{subfigure}
\hfill
\begin{subfigure}[T]{0.19\textwidth}
\centering
\includegraphics[width=\textwidth]{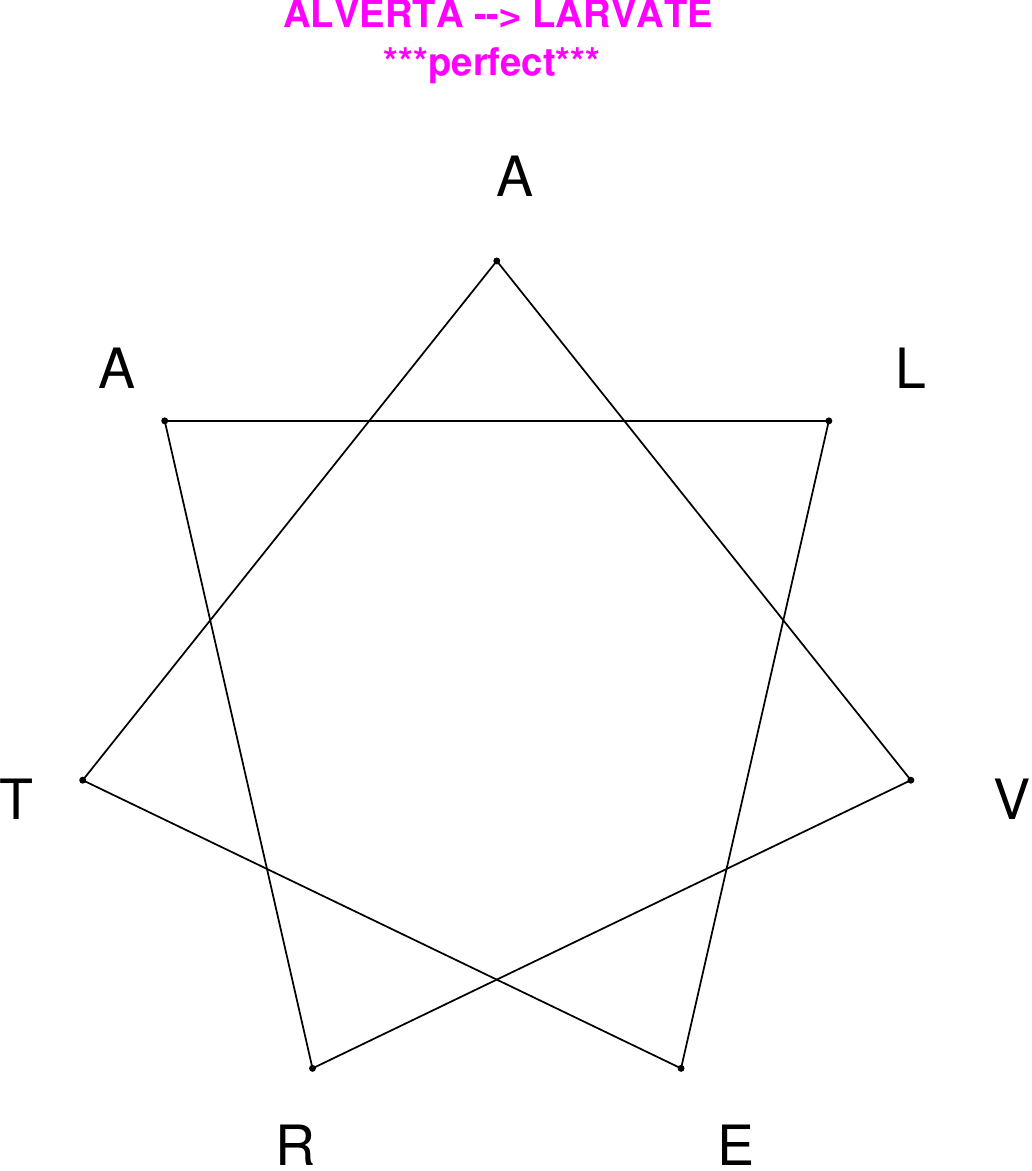}
\end{subfigure}
\hfill
\begin{subfigure}[T]{0.19\textwidth}
\centering
\includegraphics[width=\textwidth]{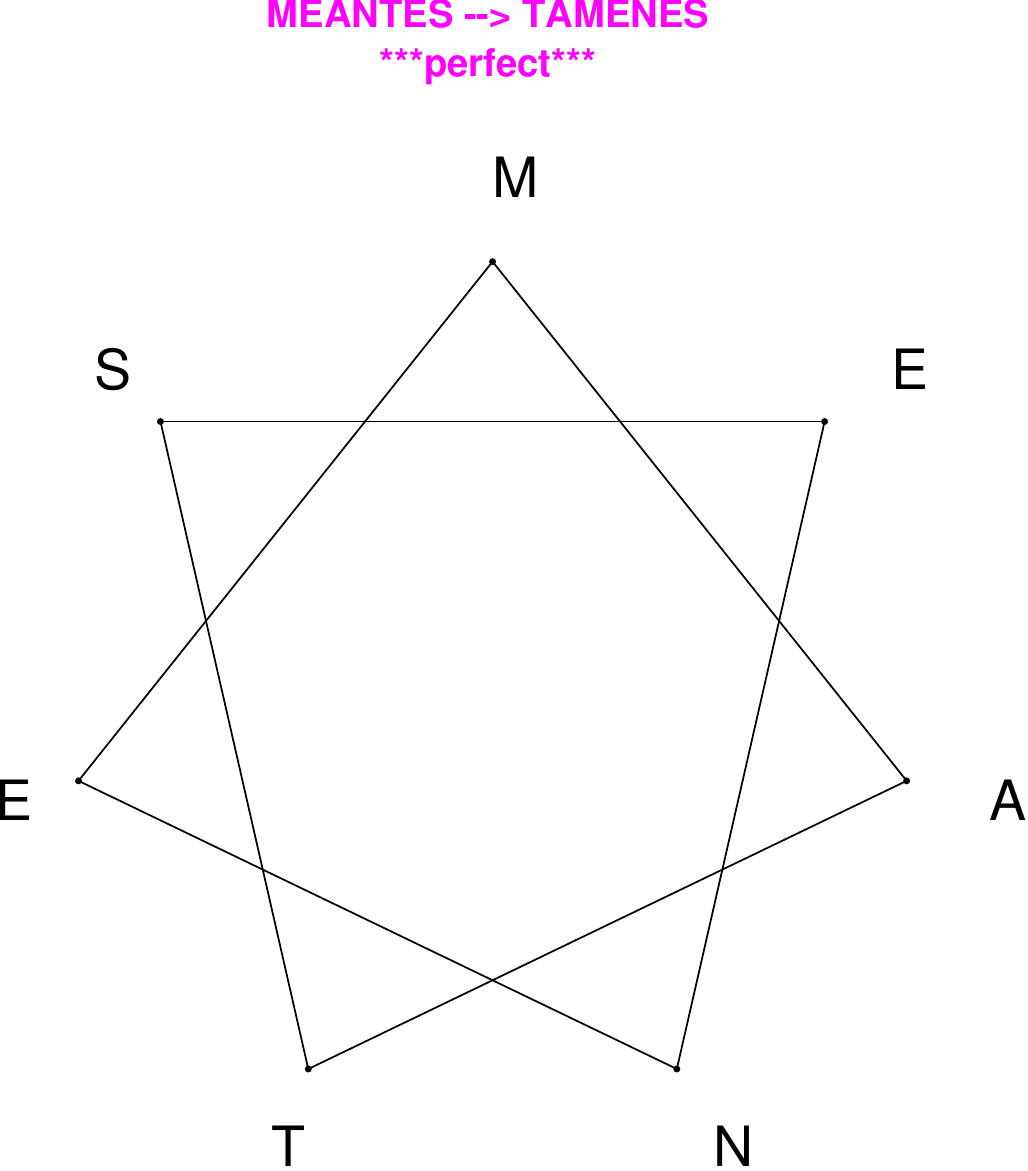}
\end{subfigure}
\hfill
\begin{subfigure}[T]{0.19\textwidth}
\centering
\includegraphics[width=\textwidth]{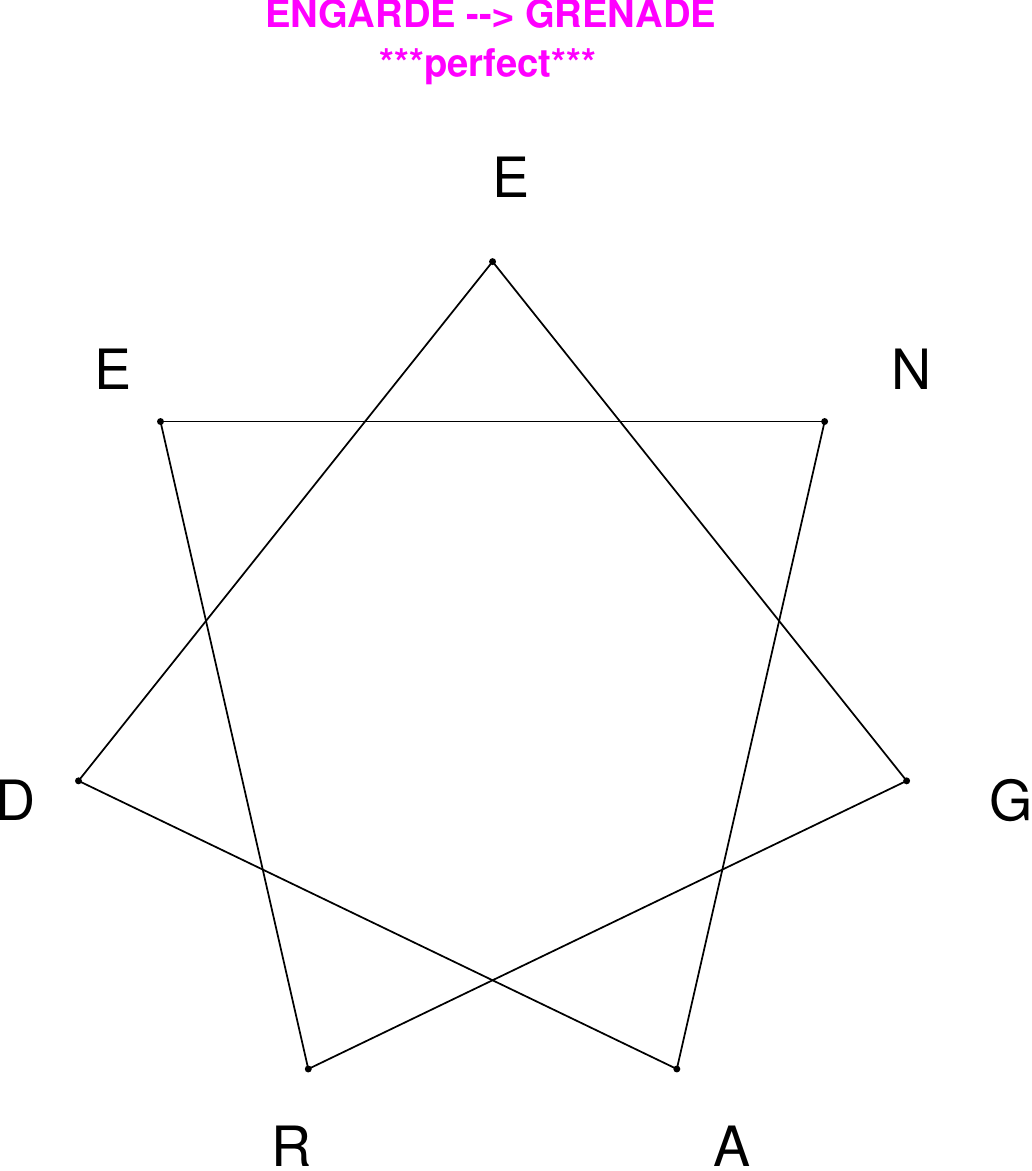}
\end{subfigure}
\hfill
\begin{subfigure}[T]{0.19\textwidth}
\centering
\includegraphics[width=\textwidth]{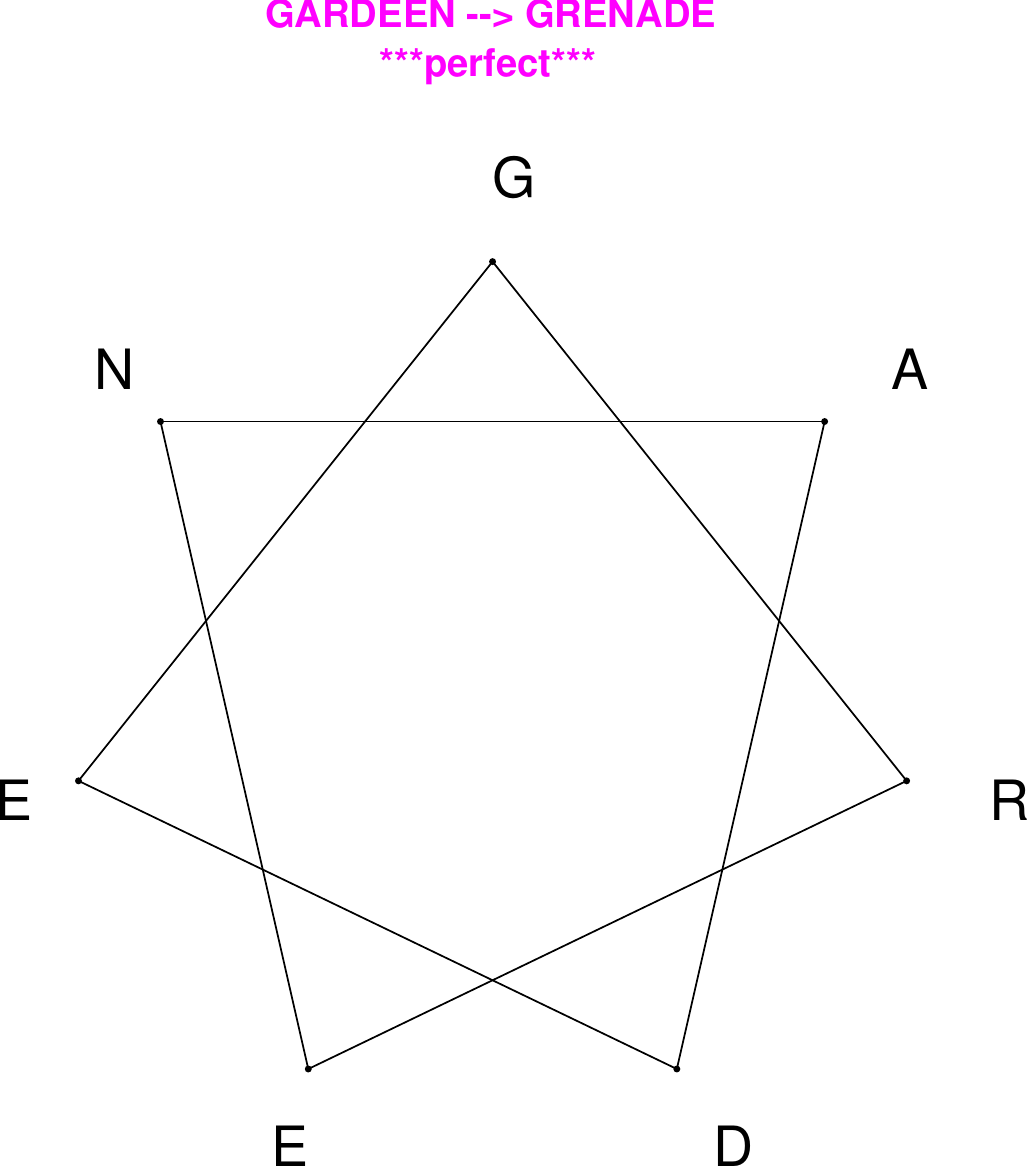}
\end{subfigure}
\end{figure}

\begin{figure}[H]
\centering
\begin{subfigure}[T]{0.19\textwidth}
\centering
\includegraphics[width=\textwidth]{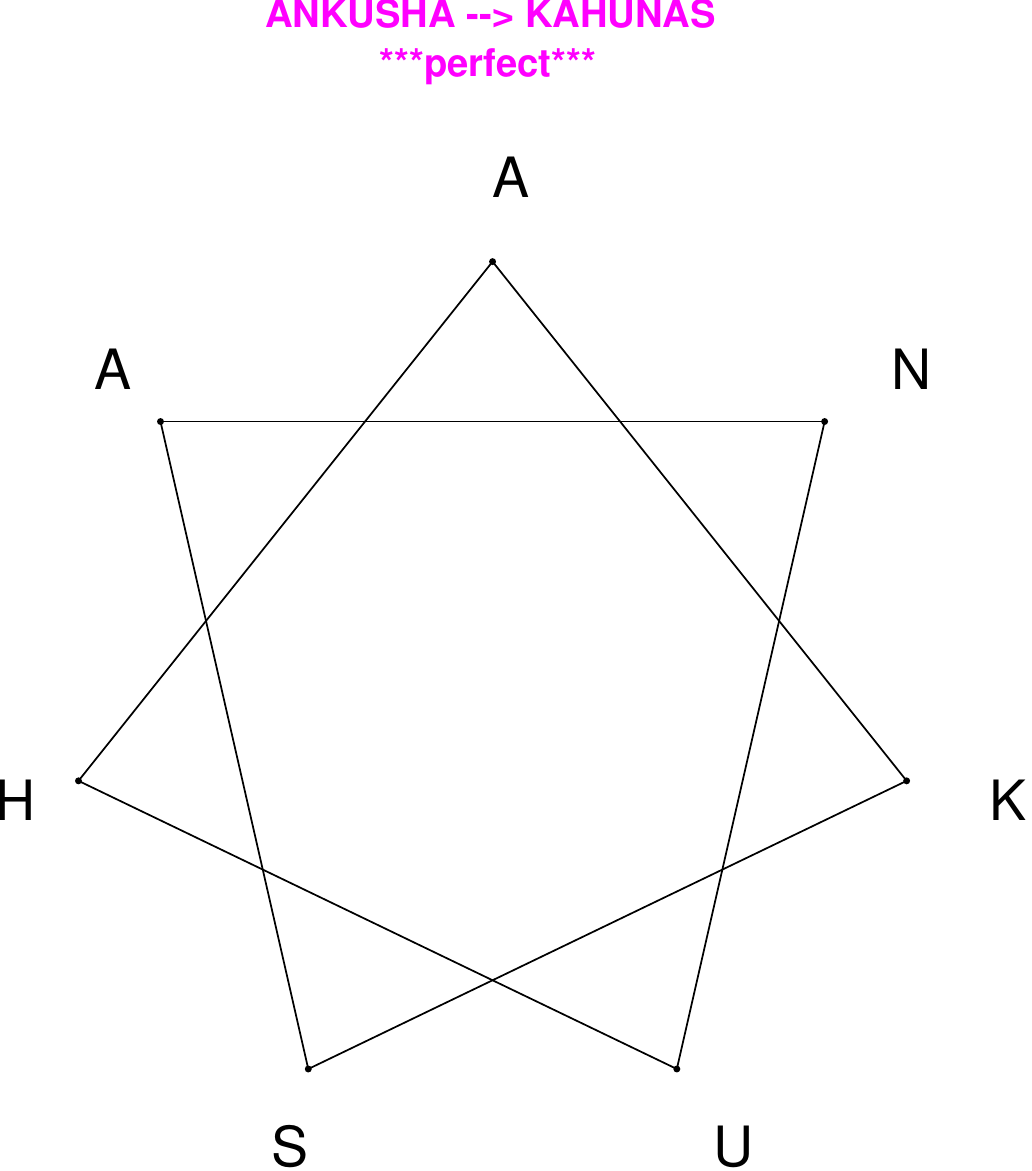}
\end{subfigure}
\hfill
\begin{subfigure}[T]{0.19\textwidth}
\centering
\includegraphics[width=\textwidth]{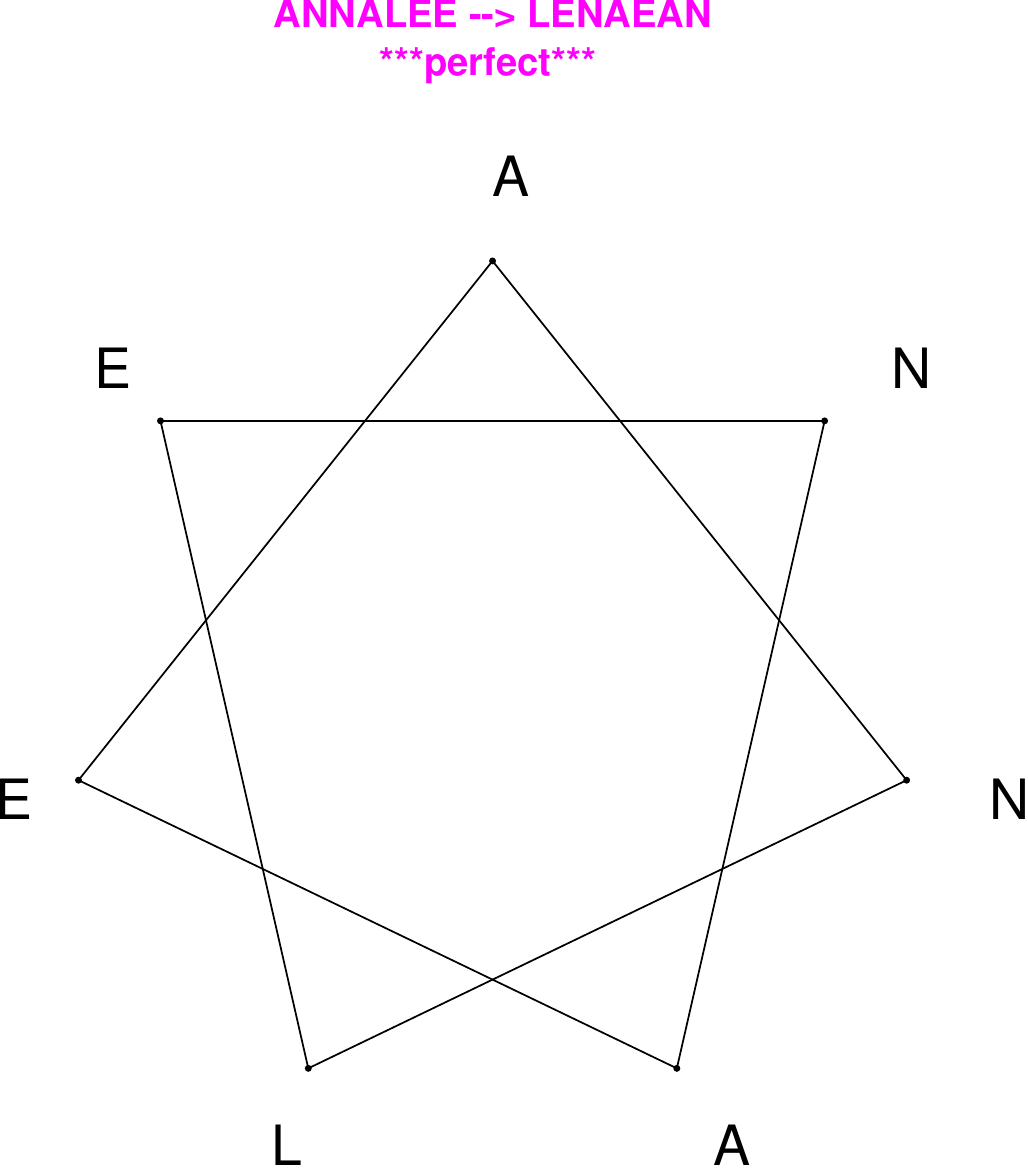}
\end{subfigure}
\hfill
\begin{subfigure}[T]{0.19\textwidth}
\centering
\includegraphics[width=\textwidth]{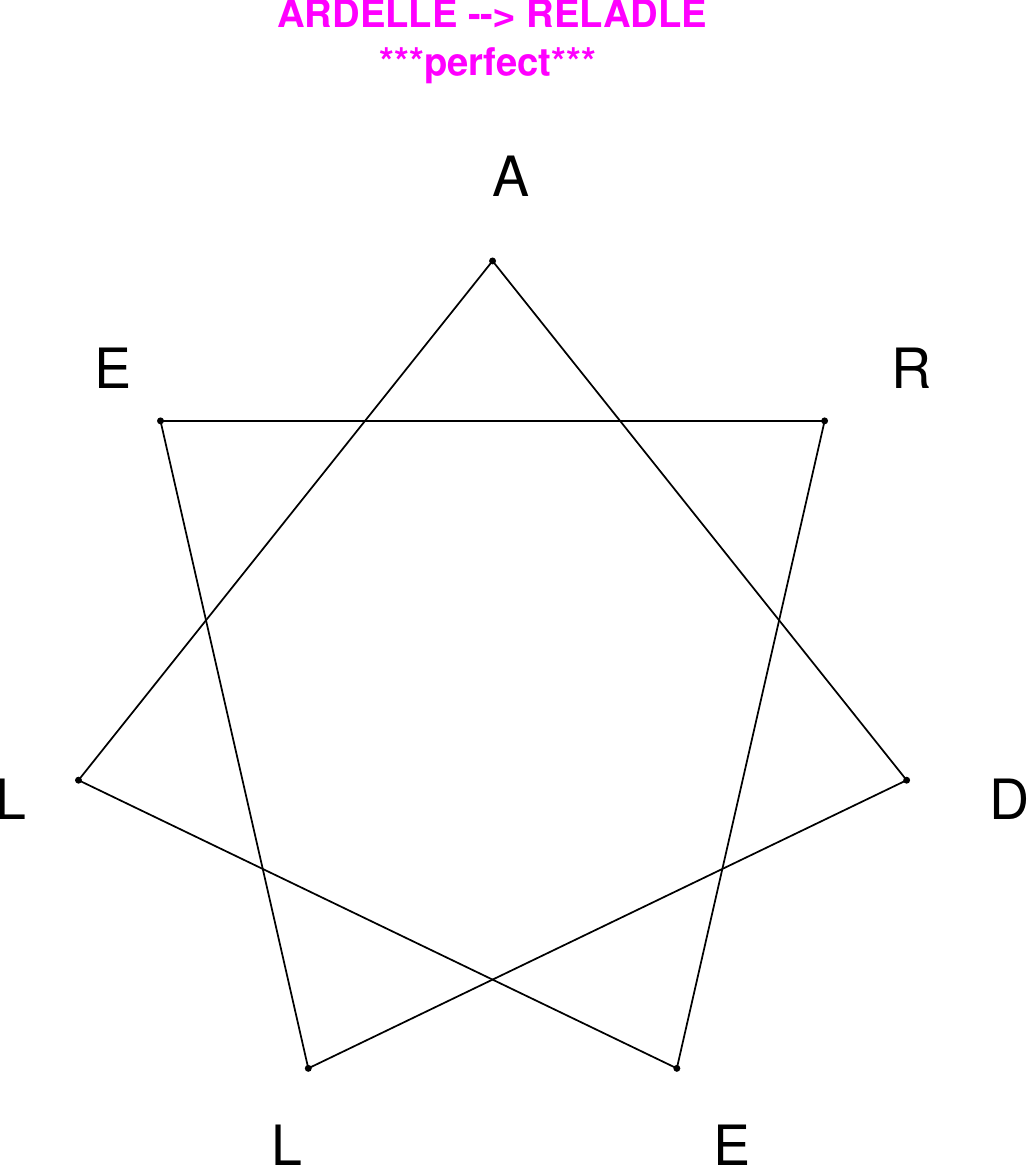}
\end{subfigure}
\hfill
\begin{subfigure}[T]{0.19\textwidth}
\centering
\includegraphics[width=\textwidth]{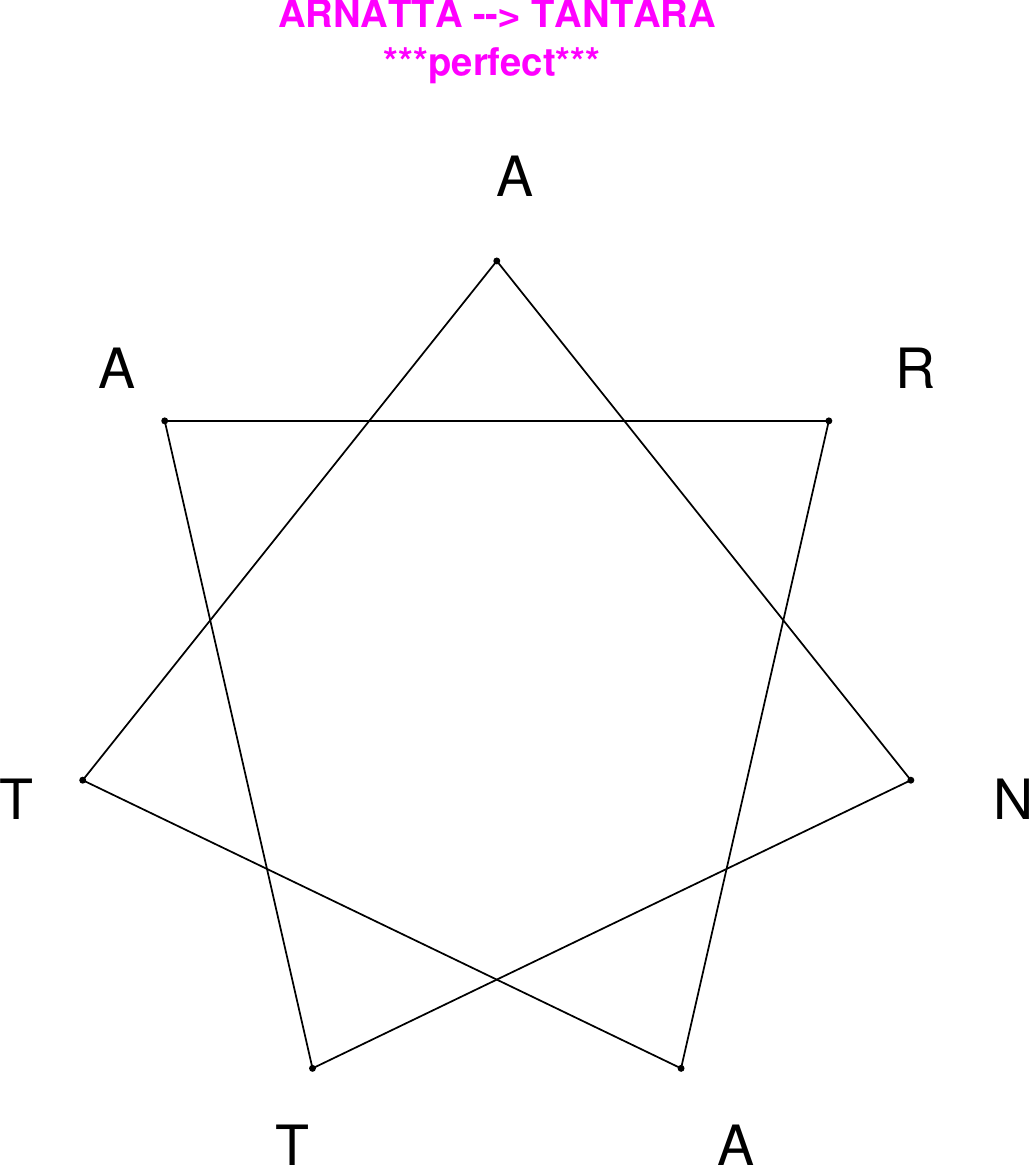}
\end{subfigure}
\hfill
\begin{subfigure}[T]{0.19\textwidth}
\centering
\includegraphics[width=\textwidth]{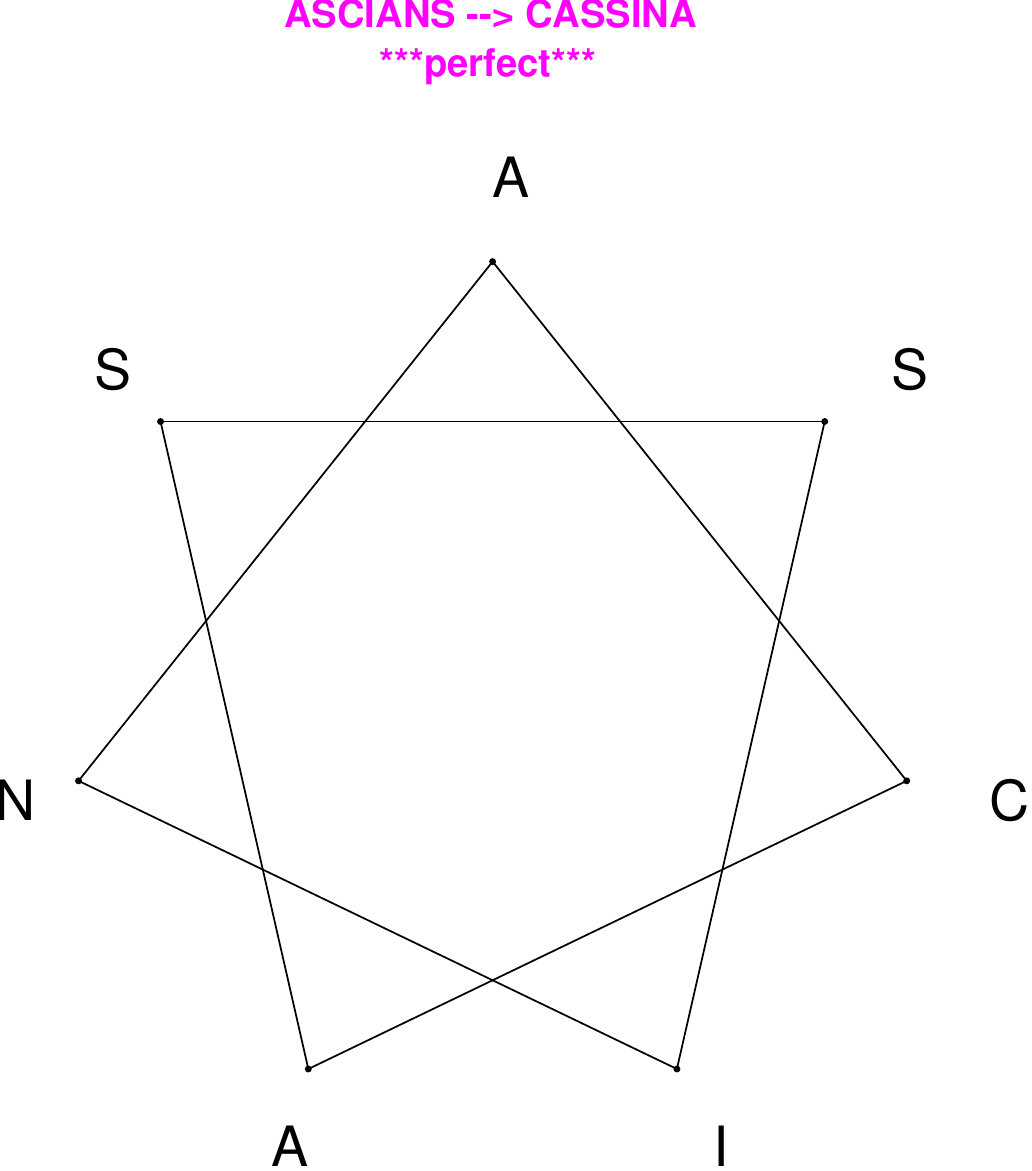}
\end{subfigure}
\end{figure}

\begin{figure}[H]
\centering
\begin{subfigure}[T]{0.19\textwidth}
\centering
\includegraphics[width=\textwidth]{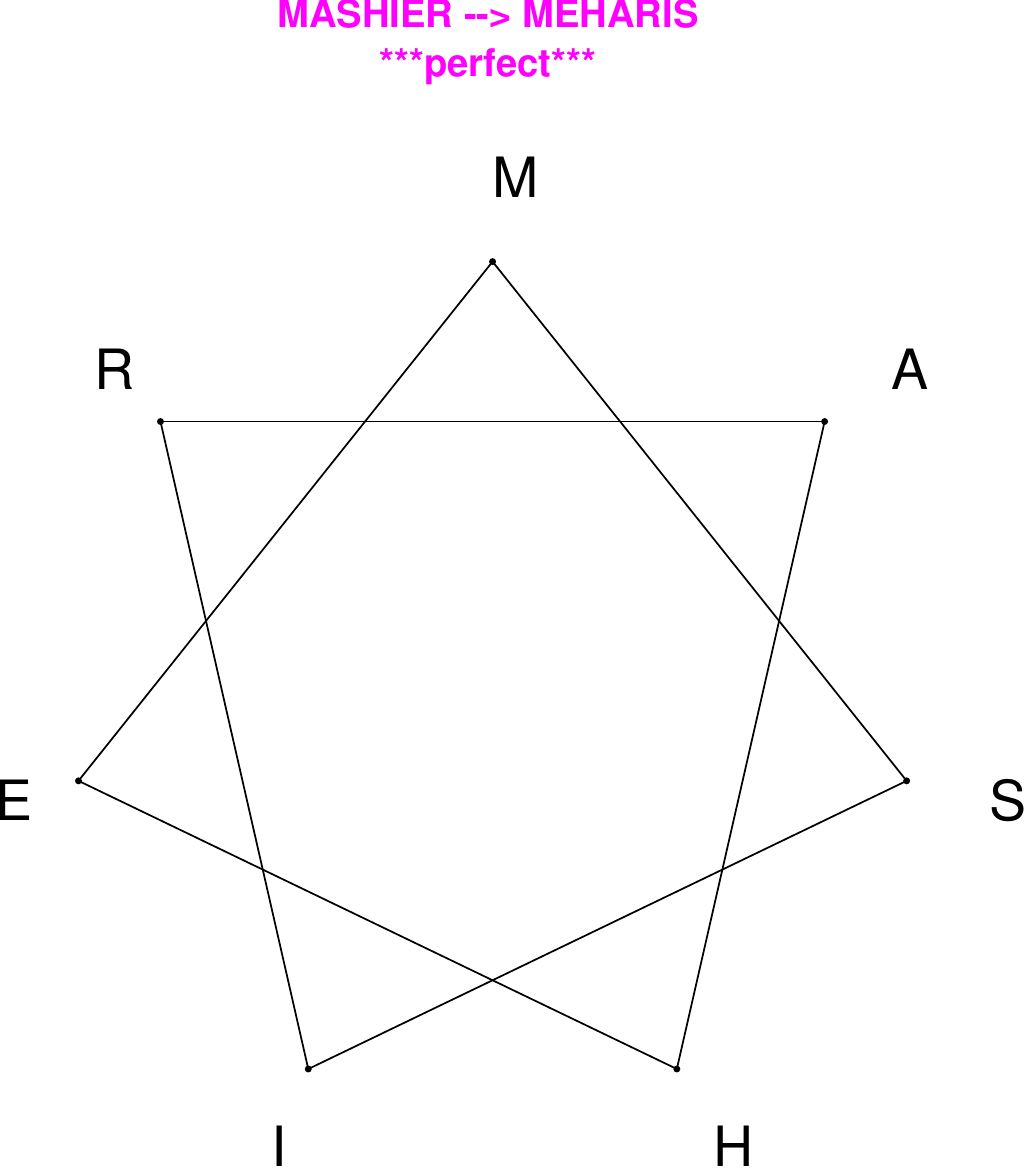}
\end{subfigure}
\hfill
\begin{subfigure}[T]{0.19\textwidth}
\centering
\includegraphics[width=\textwidth]{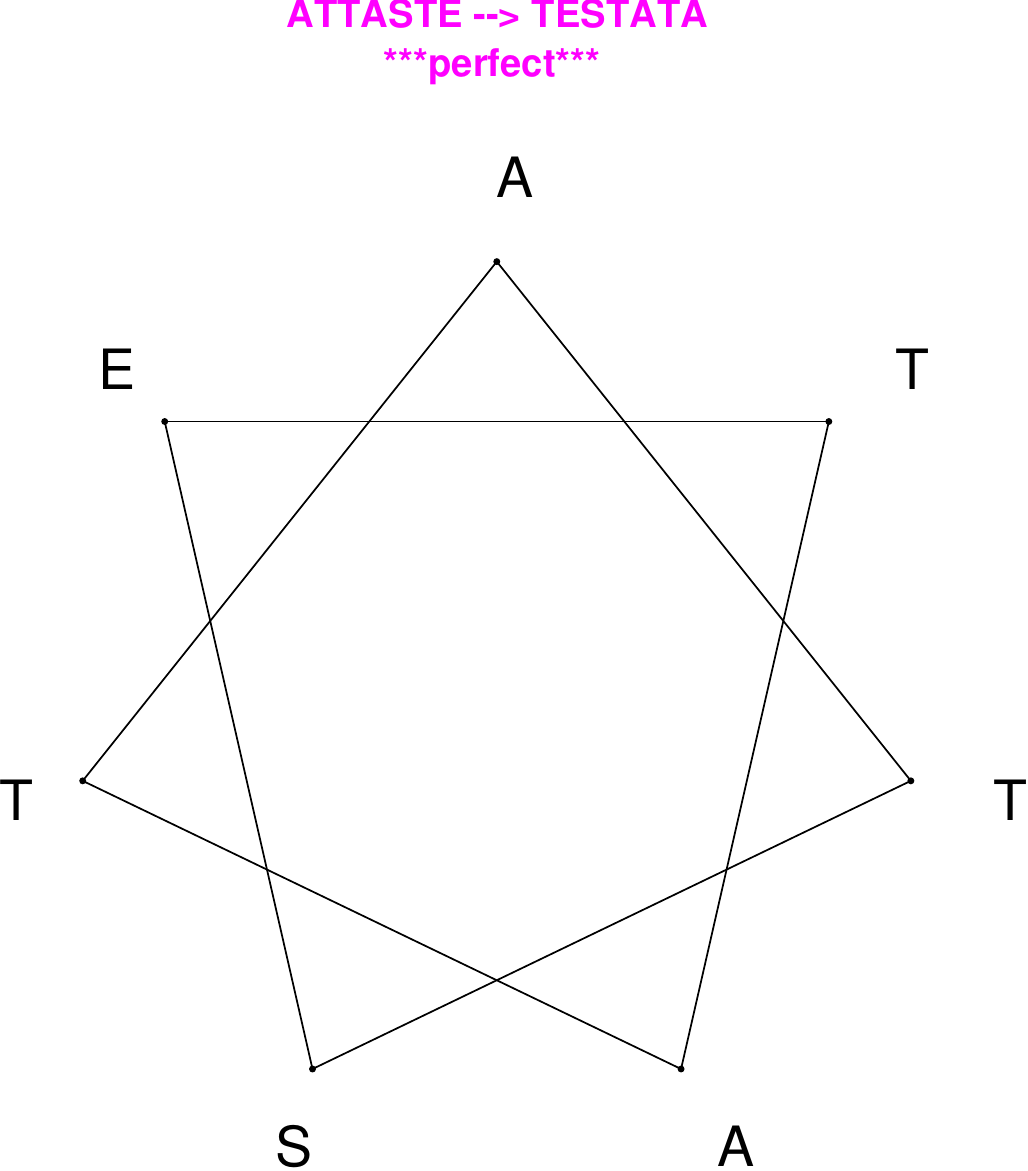}
\end{subfigure}
\hfill
\begin{subfigure}[T]{0.19\textwidth}
\centering
\includegraphics[width=\textwidth]{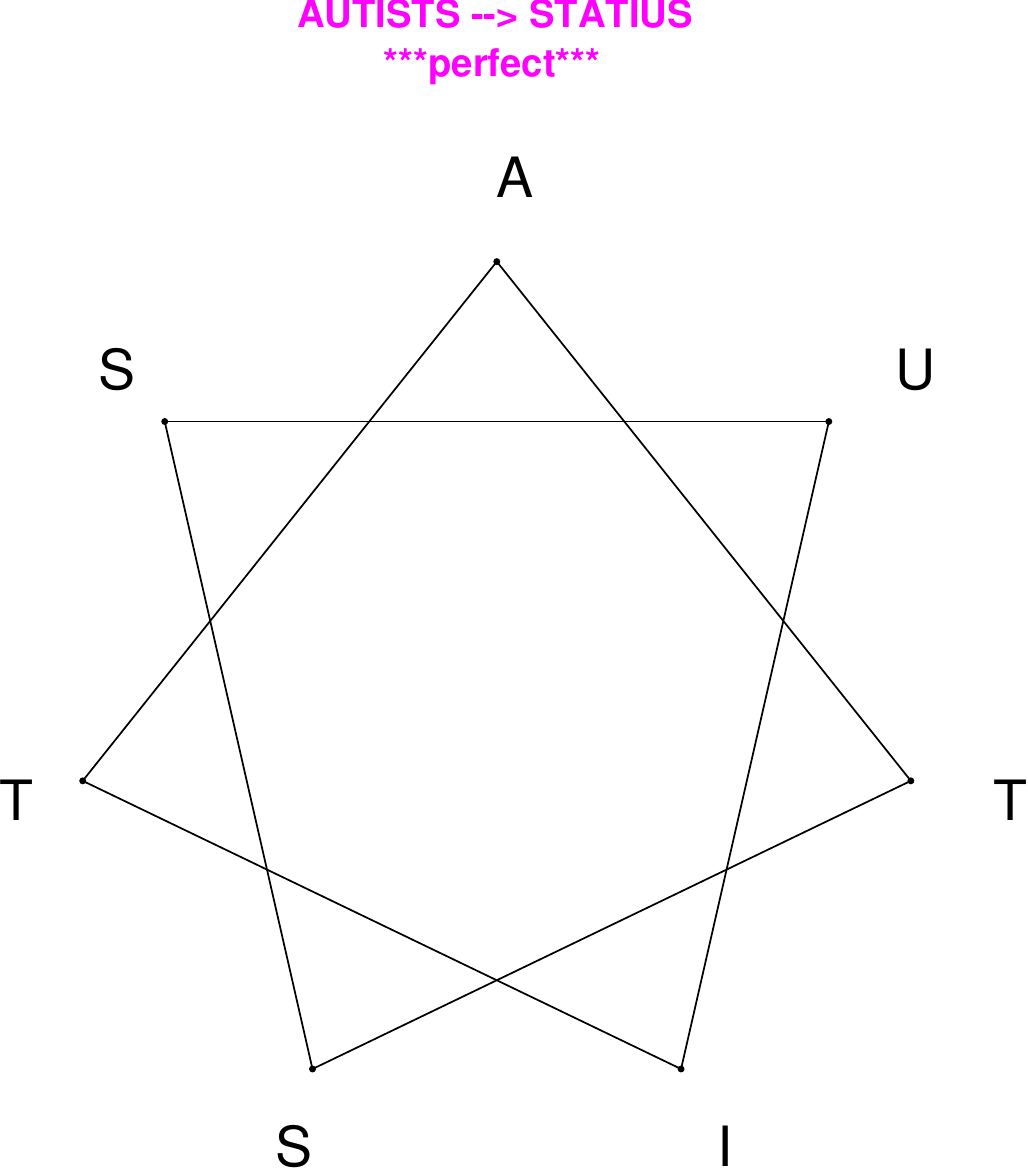}
\end{subfigure}
\hfill
\begin{subfigure}[T]{0.19\textwidth}
\centering
\includegraphics[width=\textwidth]{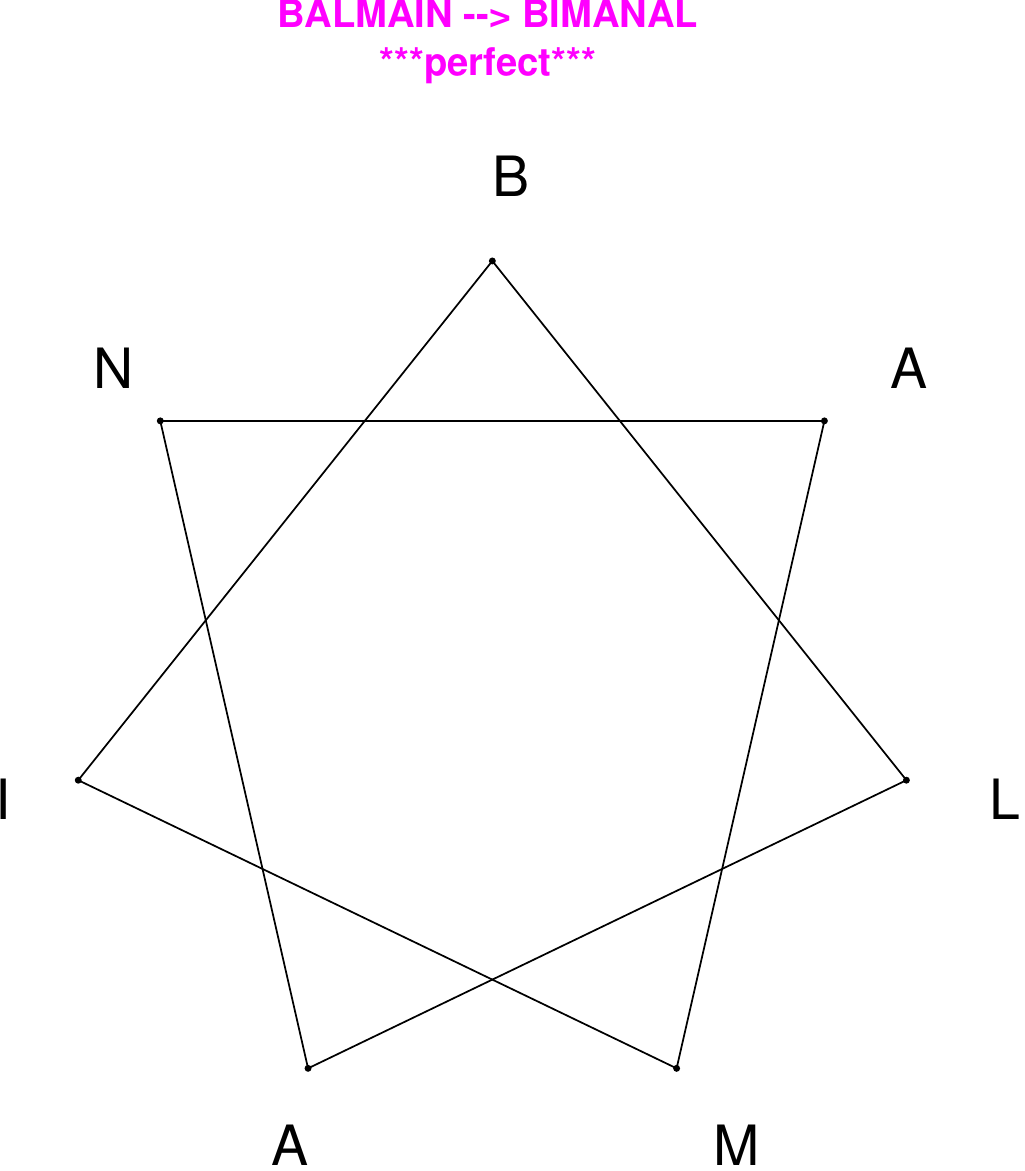}
\end{subfigure}
\hfill
\begin{subfigure}[T]{0.19\textwidth}
\centering
\includegraphics[width=\textwidth]{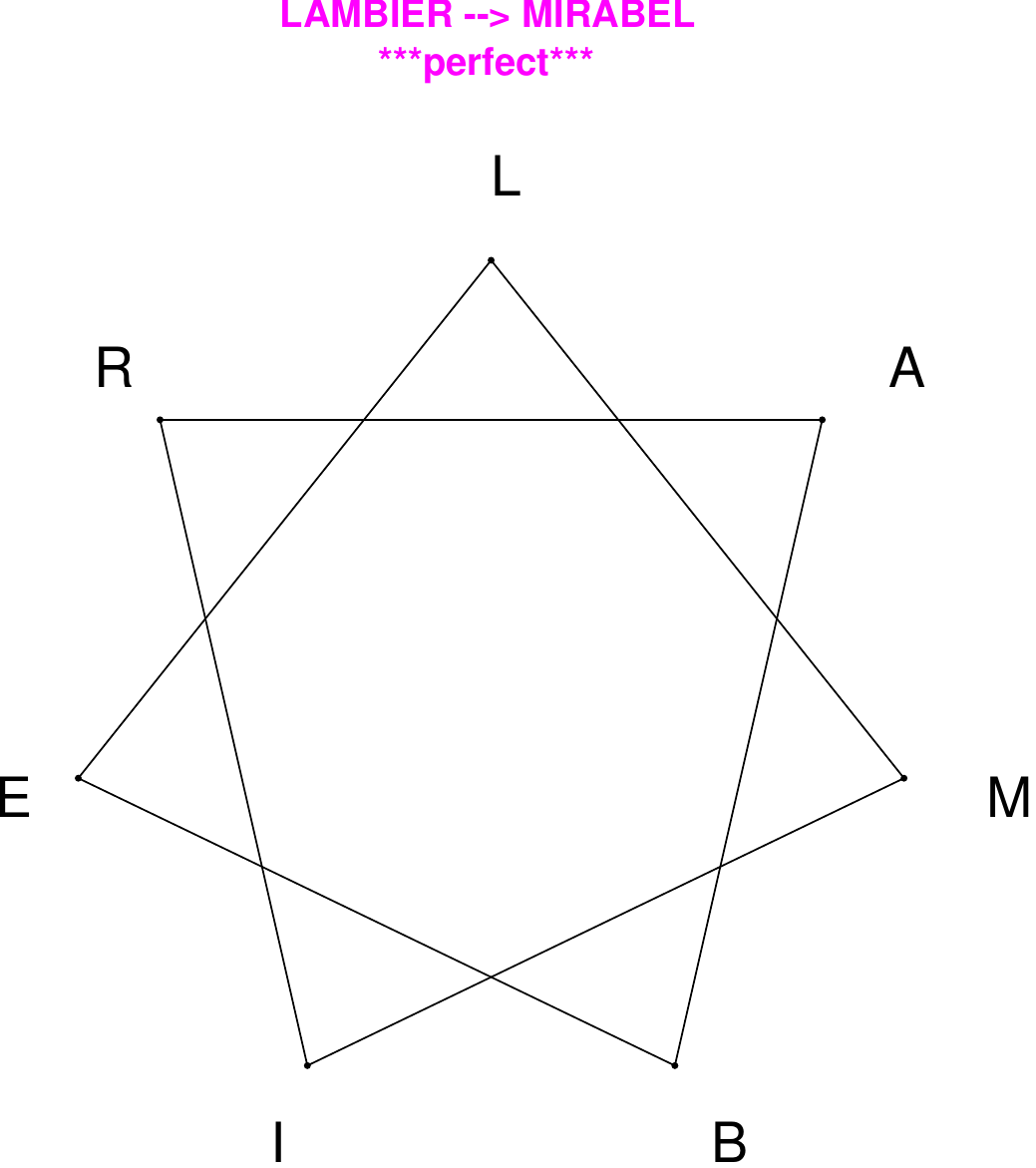}
\end{subfigure}
\end{figure}

\begin{figure}[H]
\centering
\begin{subfigure}[T]{0.19\textwidth}
\centering
\includegraphics[width=\textwidth]{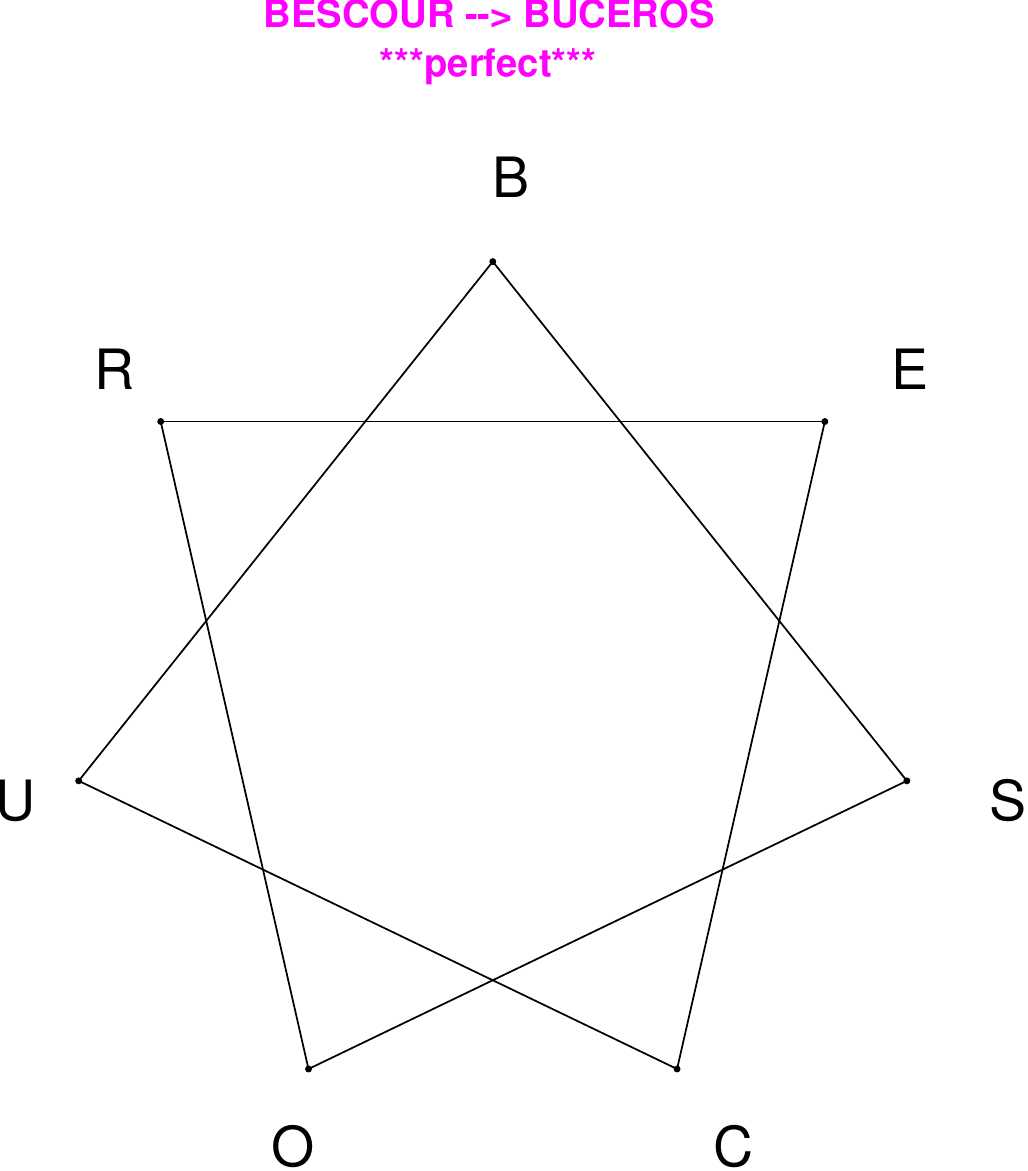}
\end{subfigure}
\hfill
\begin{subfigure}[T]{0.19\textwidth}
\centering
\includegraphics[width=\textwidth]{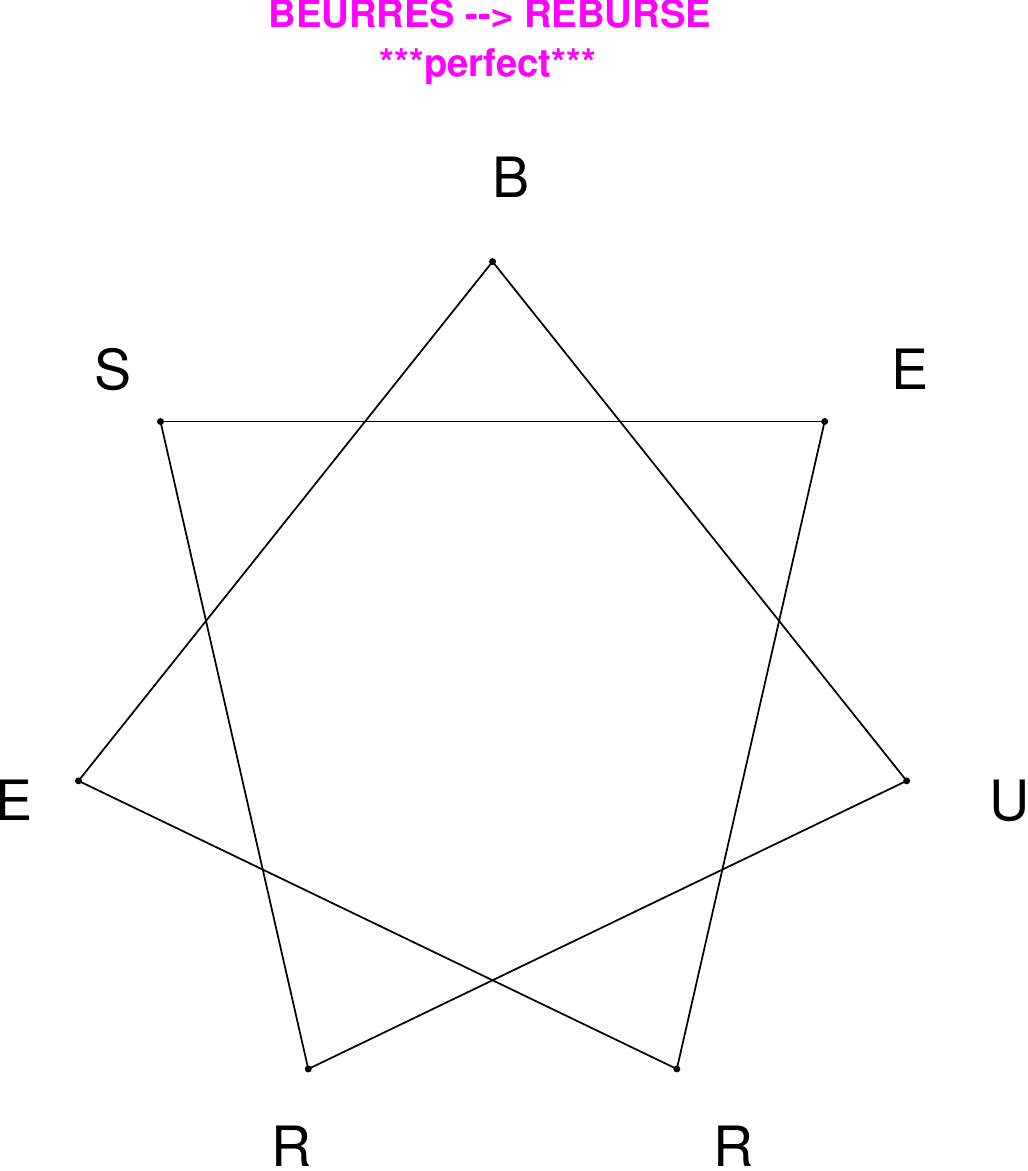}
\end{subfigure}
\hfill
\begin{subfigure}[T]{0.19\textwidth}
\centering
\includegraphics[width=\textwidth]{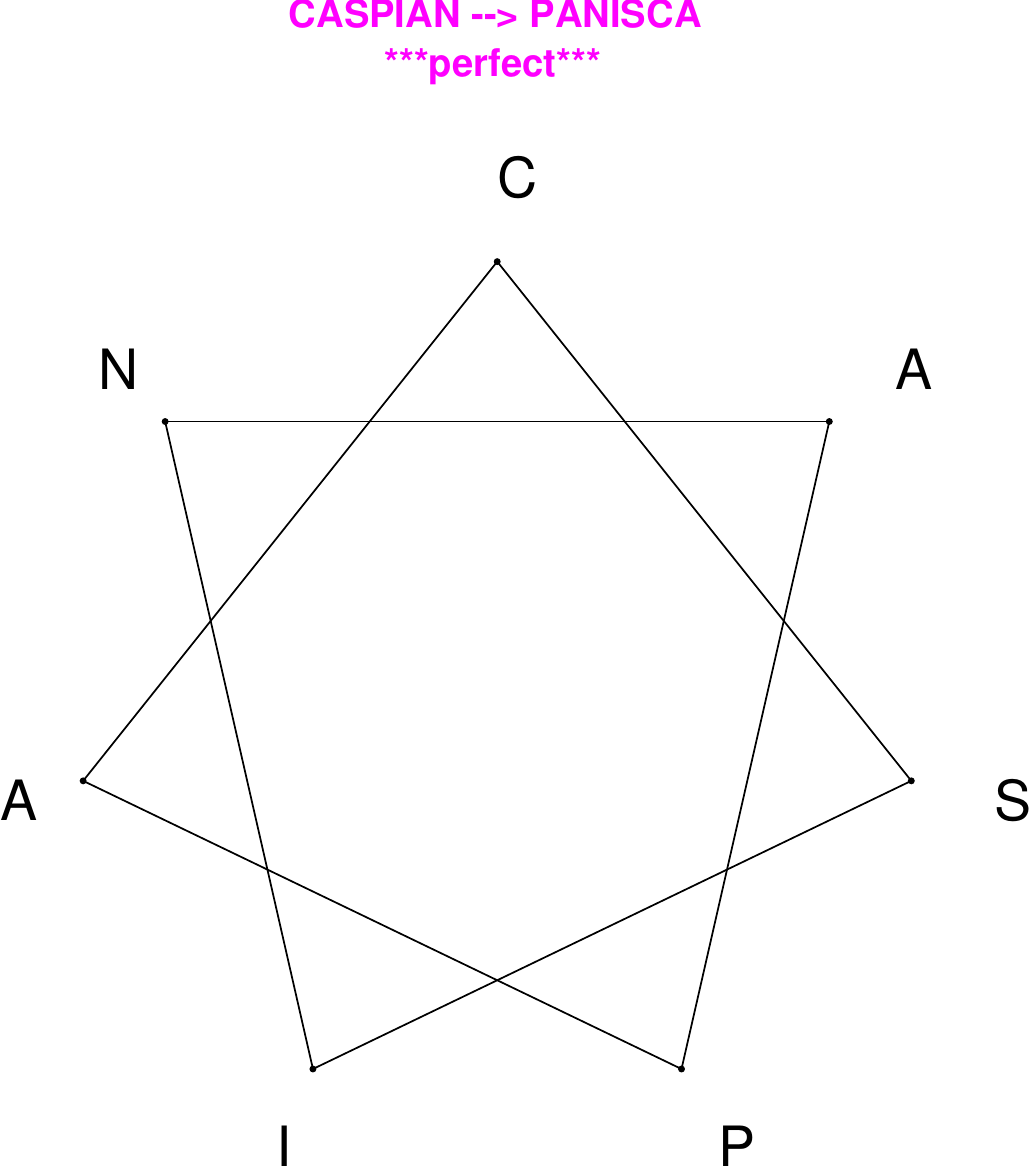}
\end{subfigure}
\hfill
\begin{subfigure}[T]{0.19\textwidth}
\centering
\includegraphics[width=\textwidth]{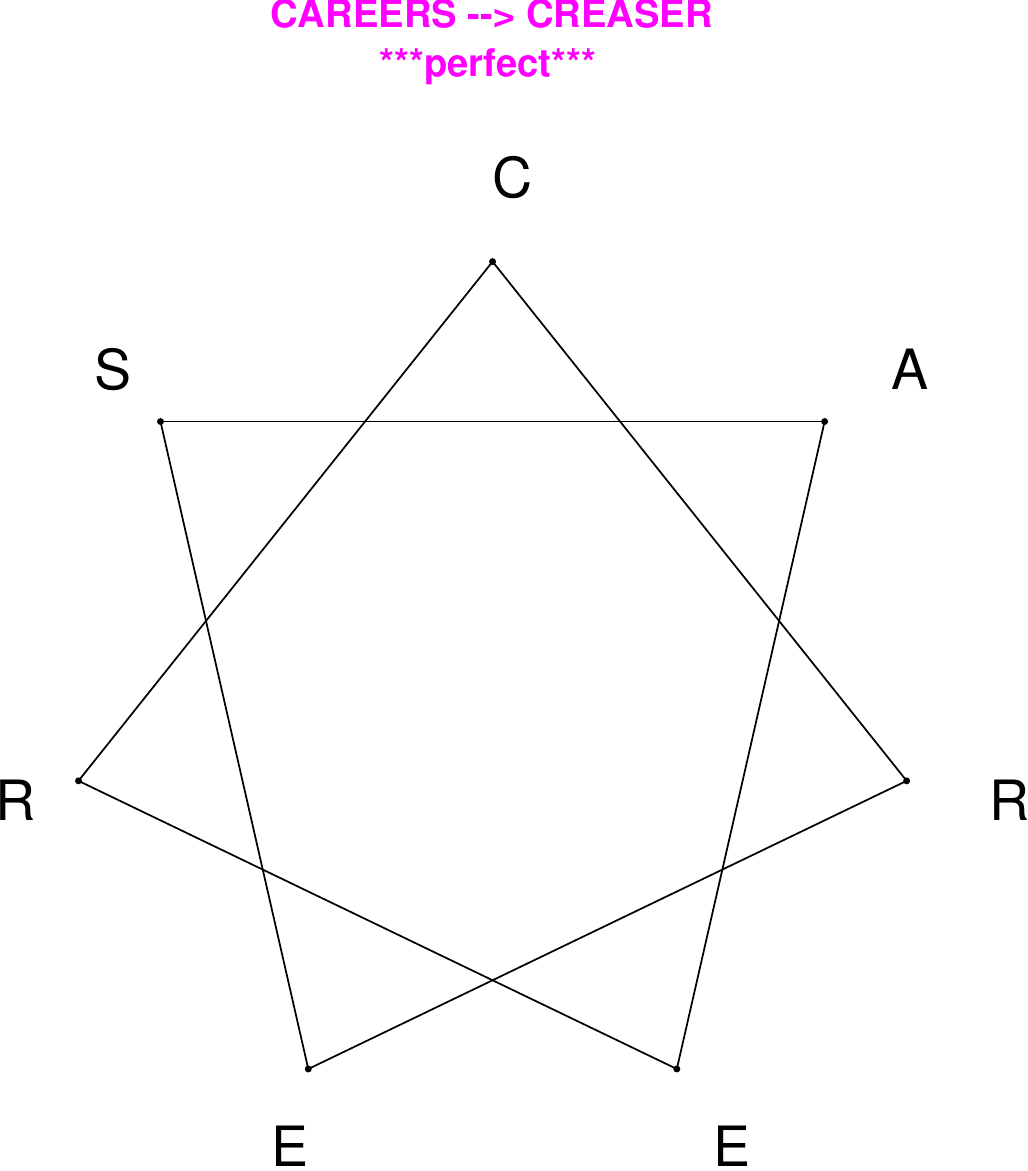}
\end{subfigure}
\hfill
\begin{subfigure}[T]{0.19\textwidth}
\centering
\includegraphics[width=\textwidth]{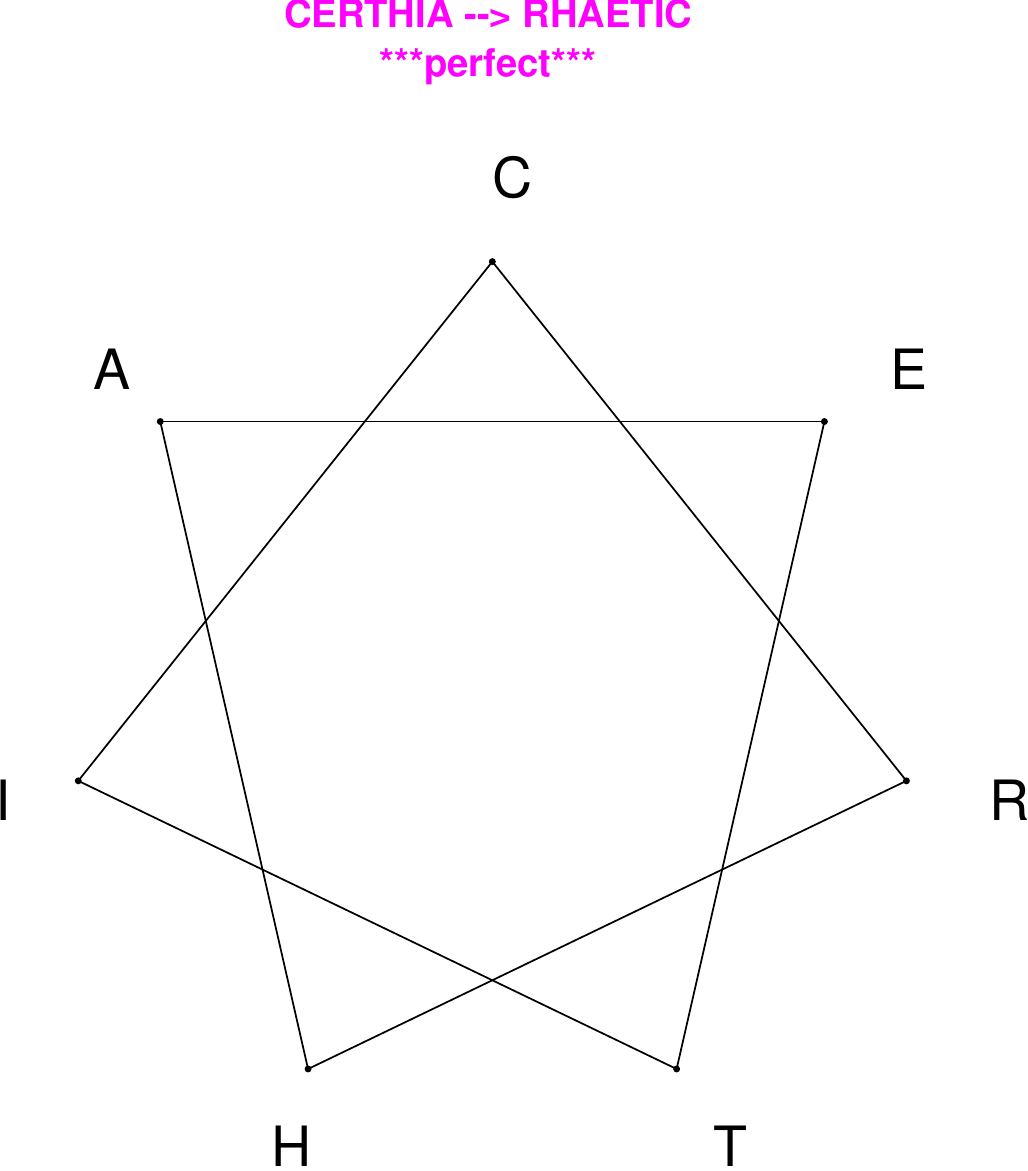}
\end{subfigure}
\end{figure}

\begin{figure}[H]
\centering
\begin{subfigure}[T]{0.19\textwidth}
\centering
\includegraphics[width=\textwidth]{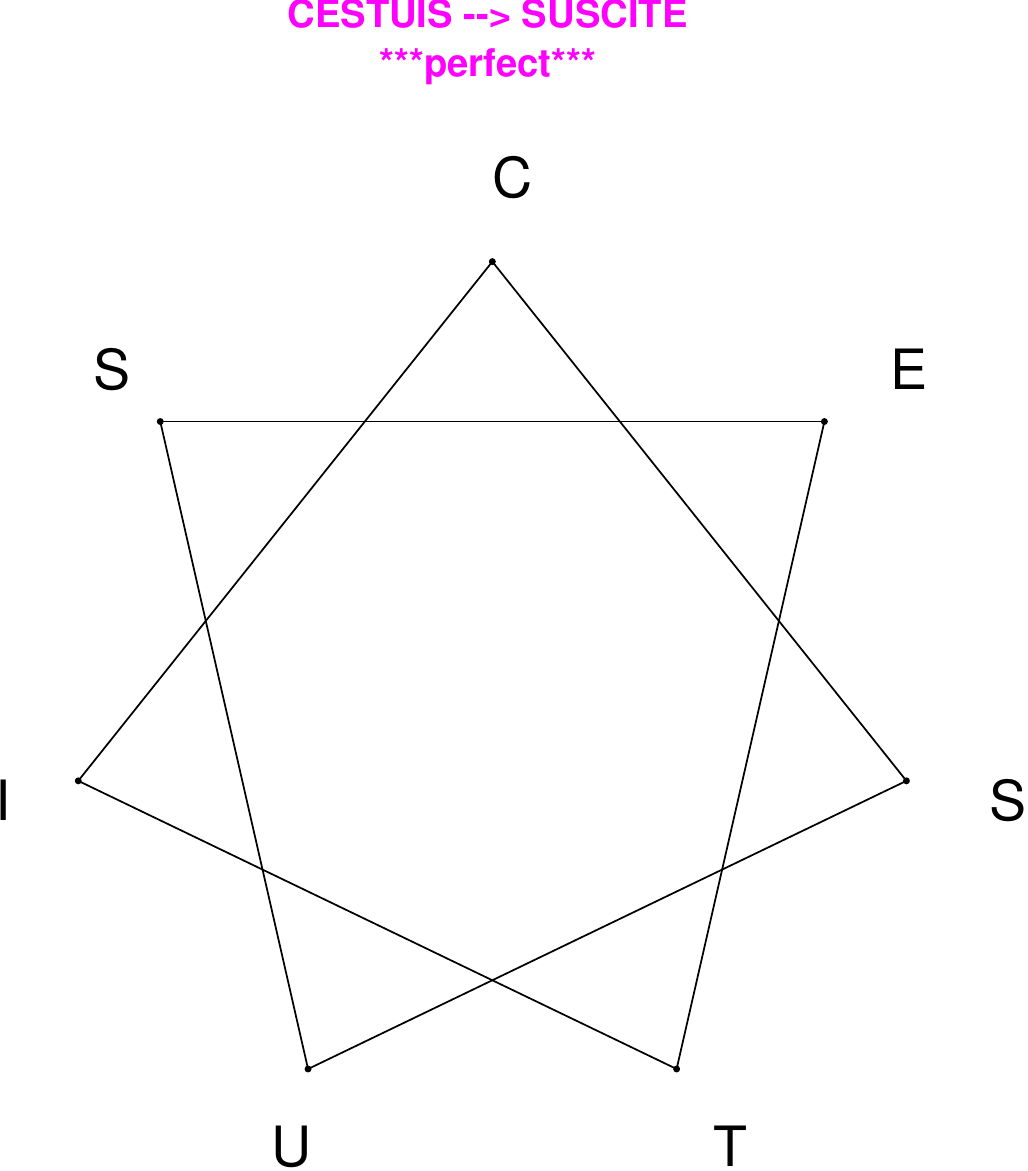}
\end{subfigure}
\hfill
\begin{subfigure}[T]{0.19\textwidth}
\centering
\includegraphics[width=\textwidth]{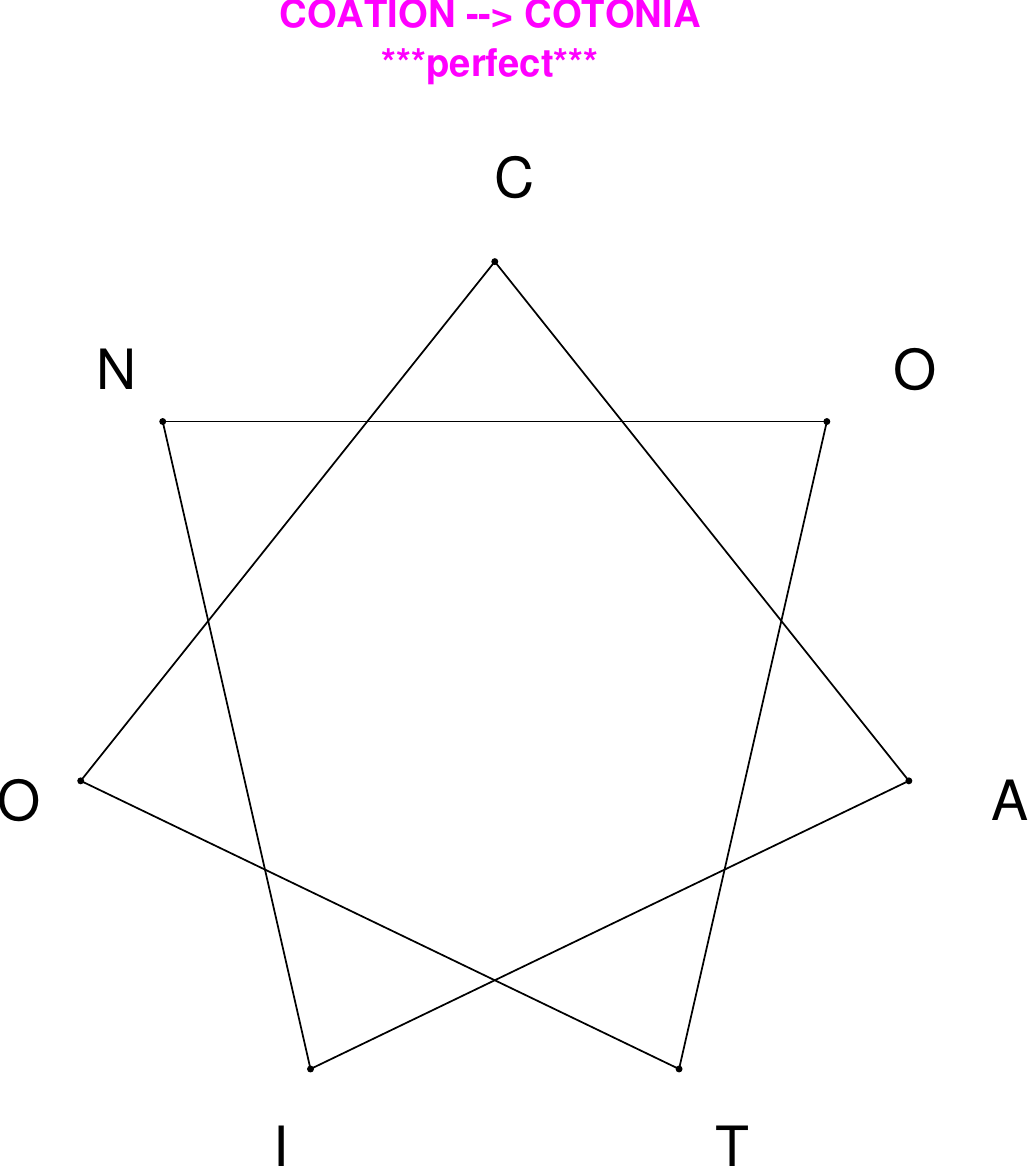}
\end{subfigure}
\hfill
\begin{subfigure}[T]{0.19\textwidth}
\centering
\includegraphics[width=\textwidth]{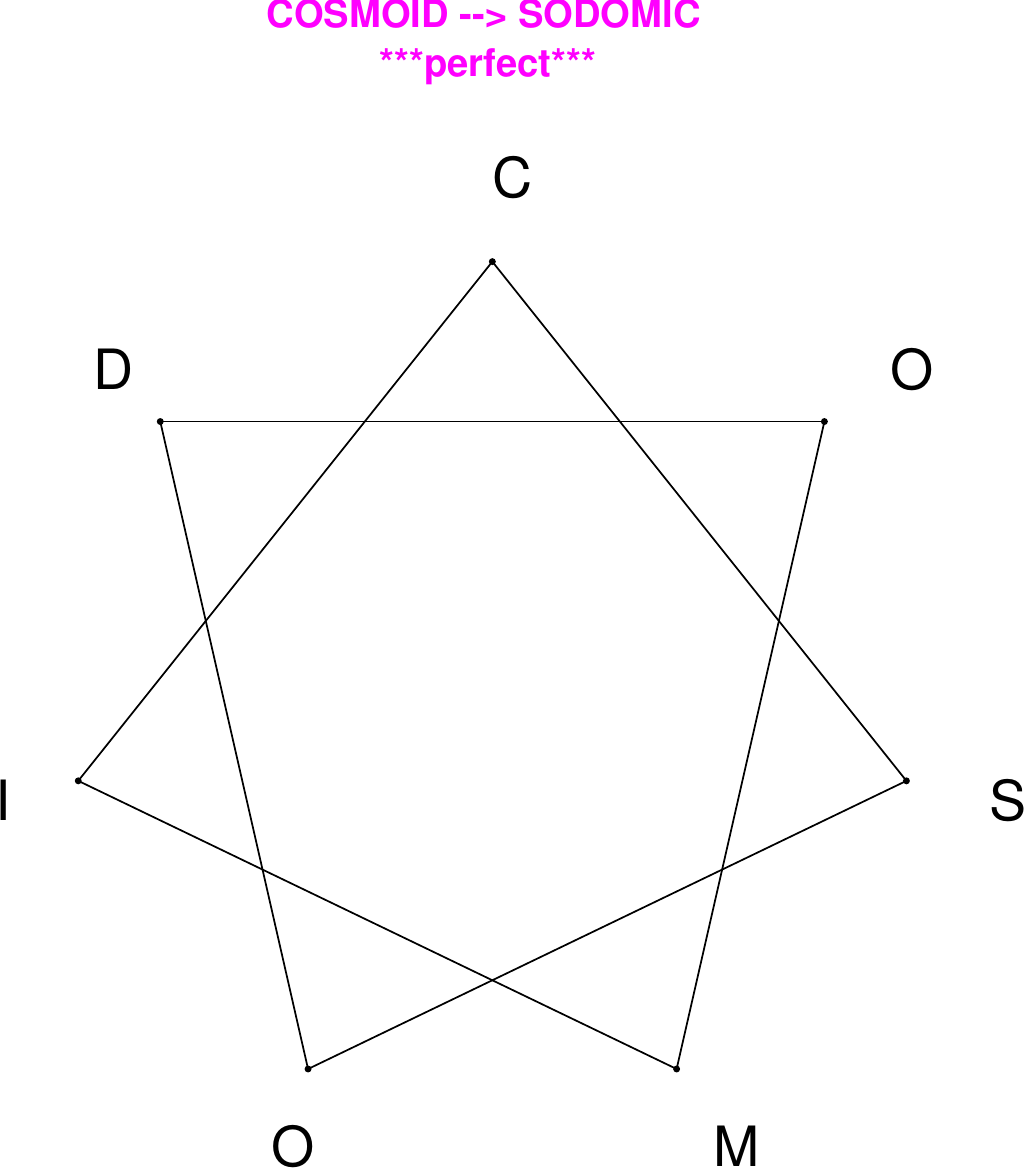}
\end{subfigure}
\hfill
\begin{subfigure}[T]{0.19\textwidth}
\centering
\includegraphics[width=\textwidth]{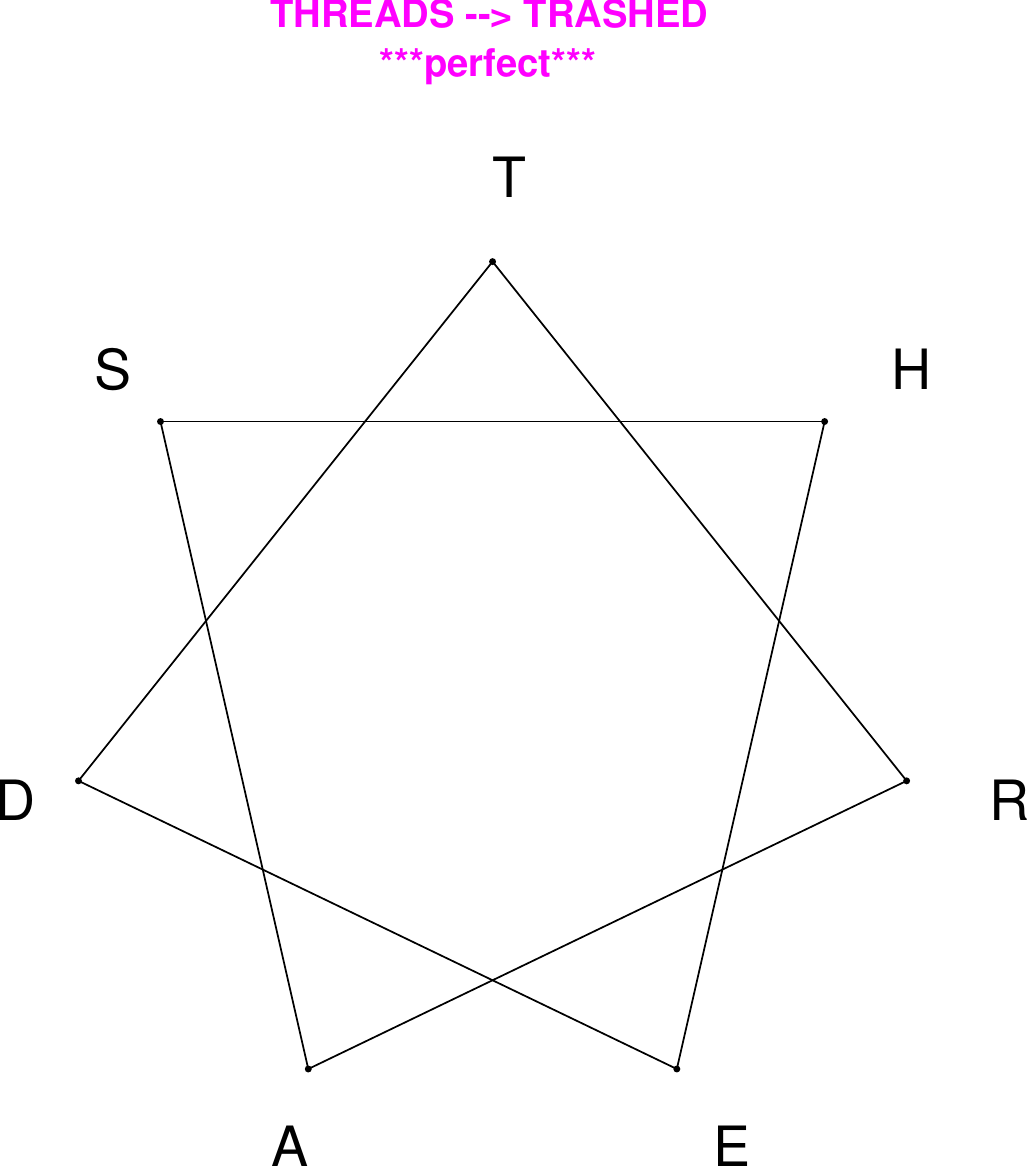}
\end{subfigure}
\hfill
\begin{subfigure}[T]{0.19\textwidth}
\centering
\includegraphics[width=\textwidth]{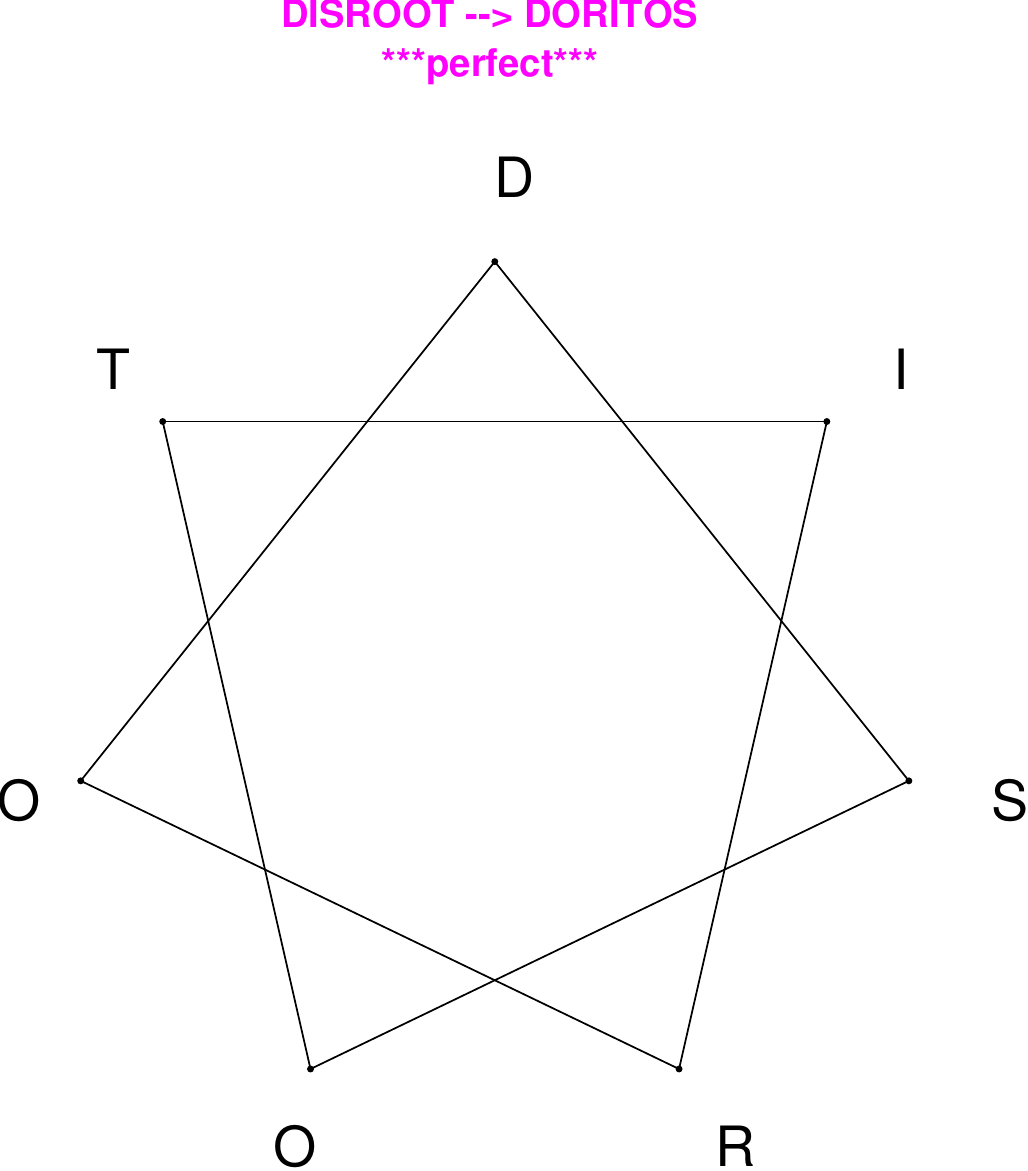}
\end{subfigure}
\end{figure}

\begin{figure}[H]
\centering
\begin{subfigure}[T]{0.19\textwidth}
\centering
\includegraphics[width=\textwidth]{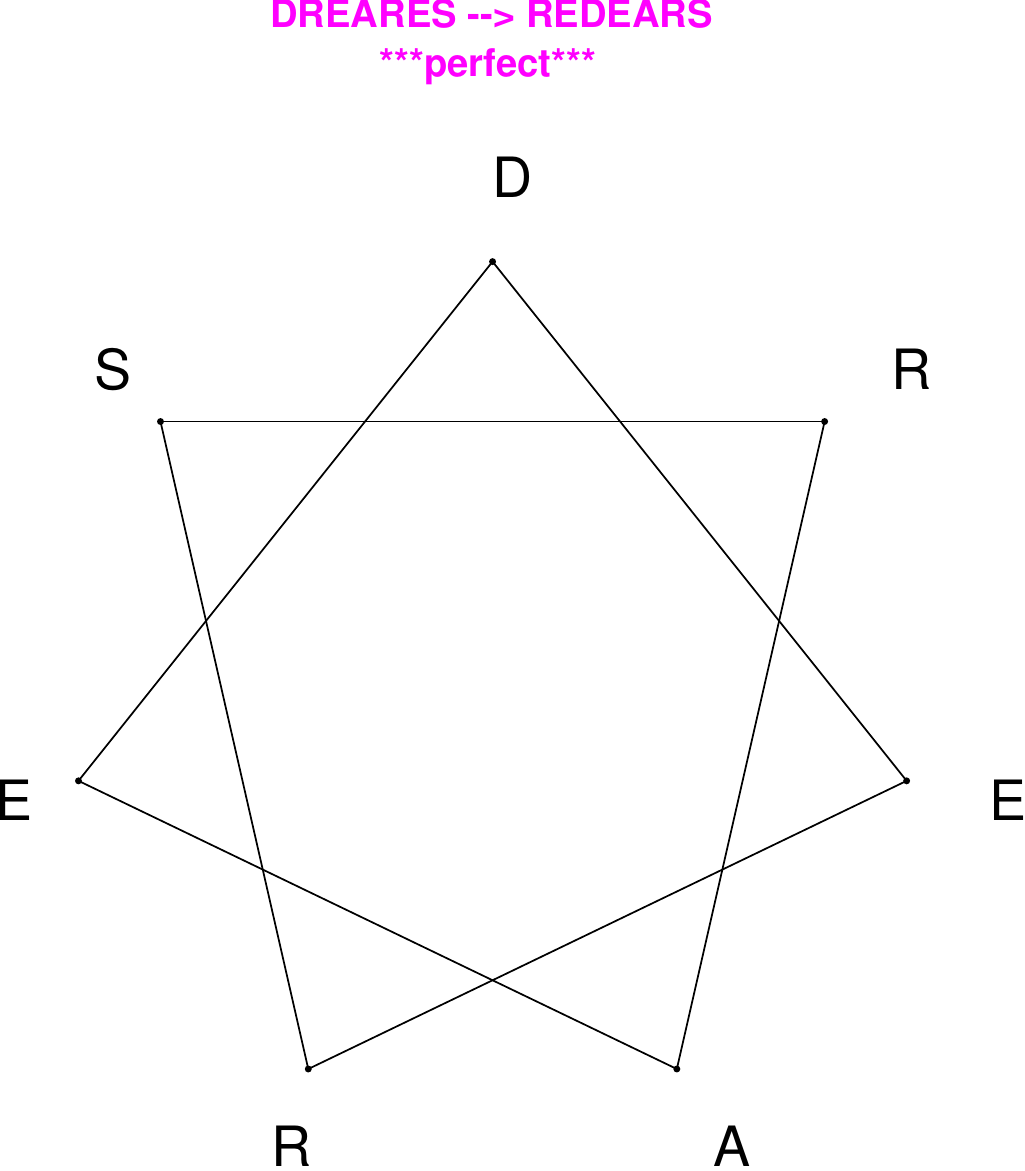}
\end{subfigure}
\hfill
\begin{subfigure}[T]{0.19\textwidth}
\centering
\includegraphics[width=\textwidth]{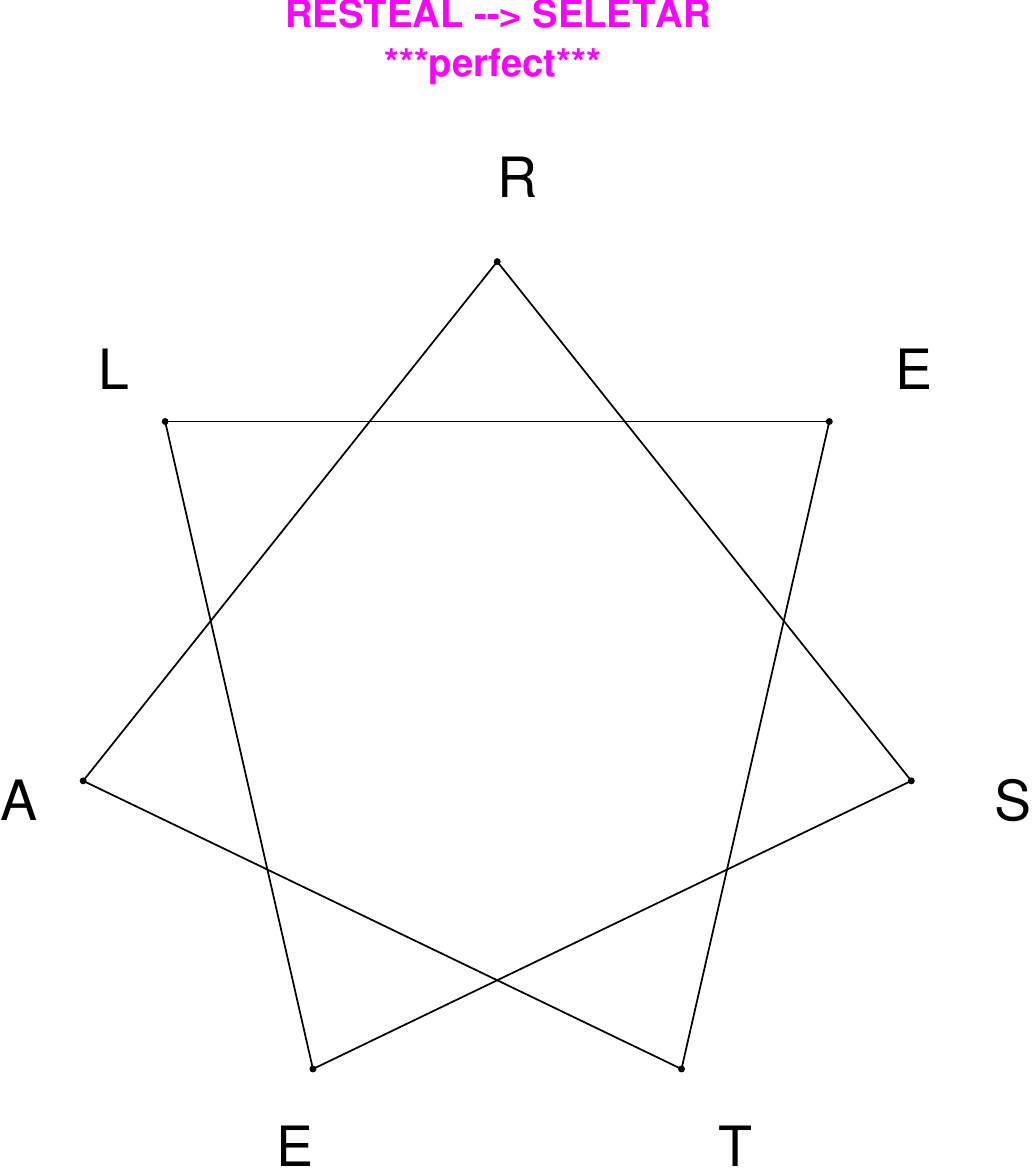}
\end{subfigure}
\hfill
\begin{subfigure}[T]{0.19\textwidth}
\centering
\includegraphics[width=\textwidth]{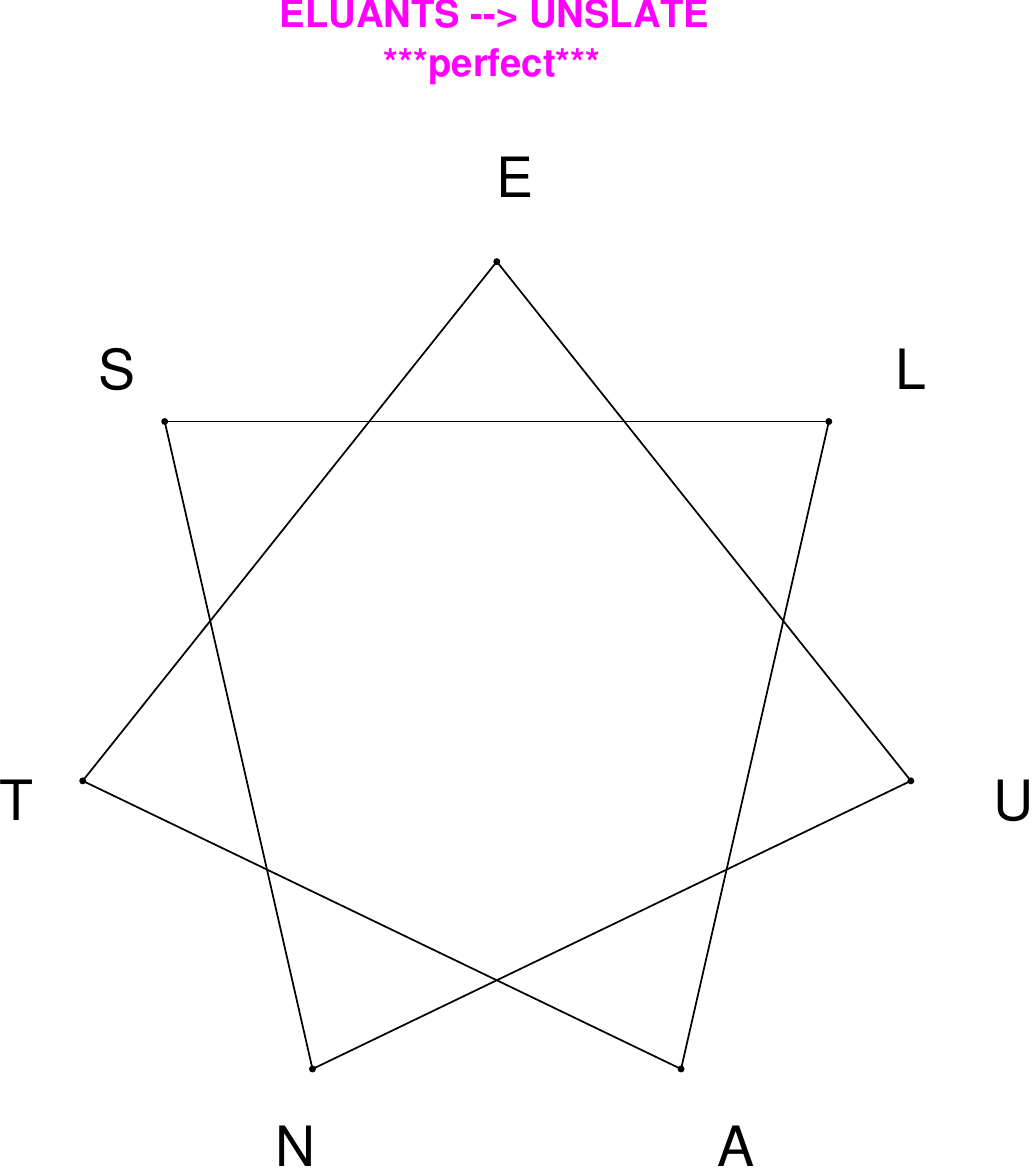}
\end{subfigure}
\hfill
\begin{subfigure}[T]{0.19\textwidth}
\centering
\includegraphics[width=\textwidth]{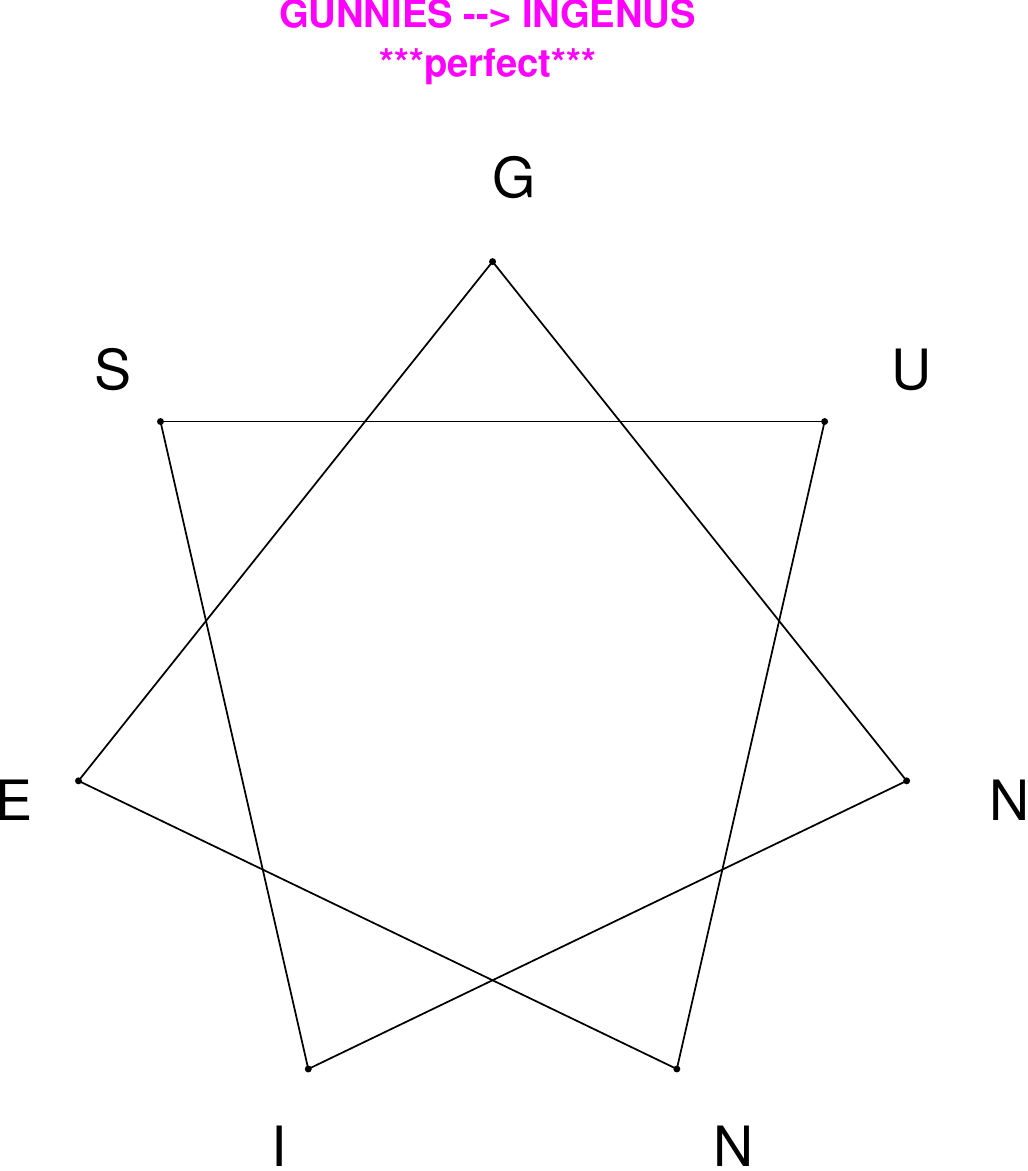}
\end{subfigure}
\hfill
\begin{subfigure}[T]{0.19\textwidth}
\centering
\includegraphics[width=\textwidth]{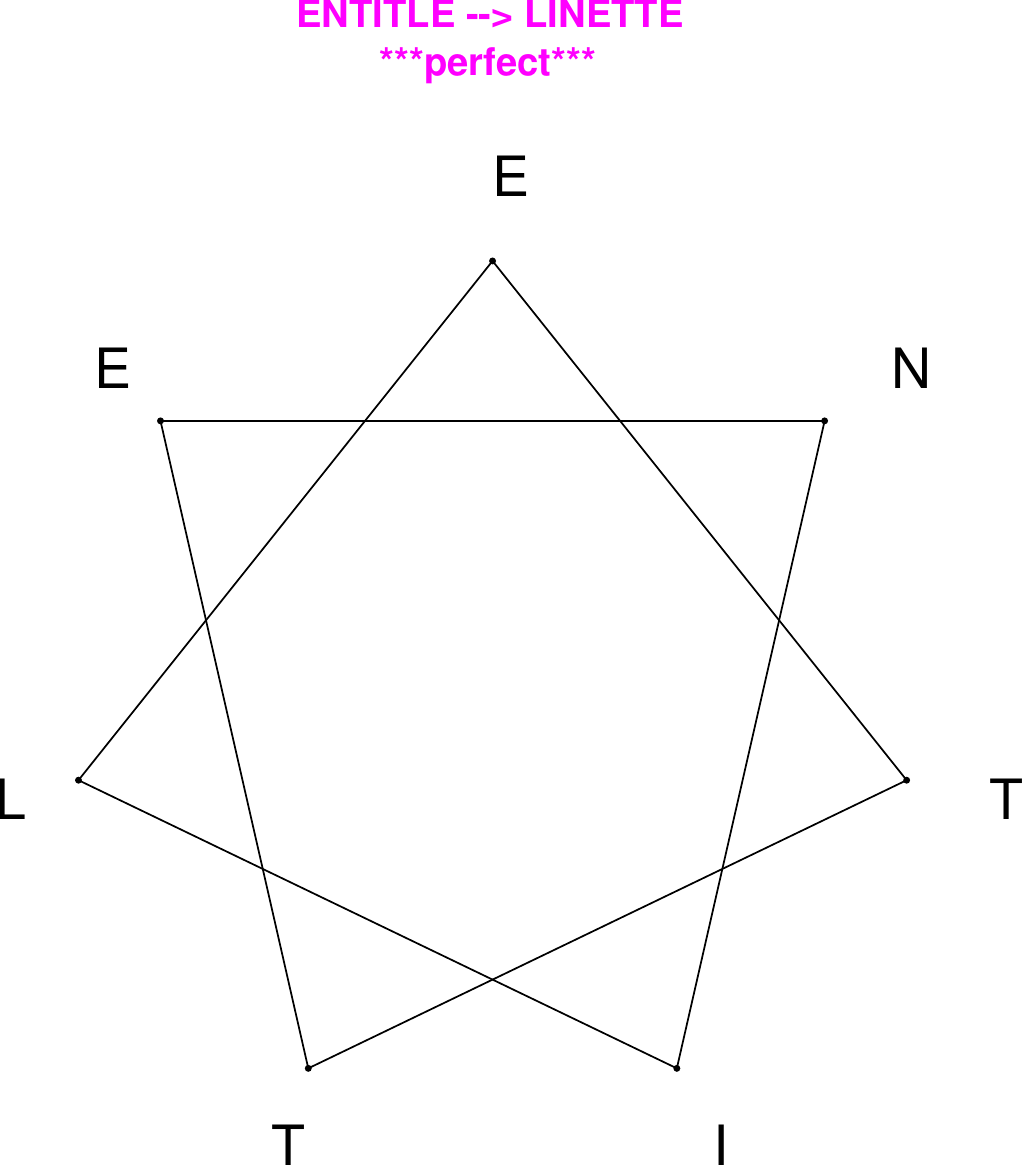}
\end{subfigure}
\end{figure}

\begin{figure}[H]
\centering
\begin{subfigure}[T]{0.19\textwidth}
\centering
\includegraphics[width=\textwidth]{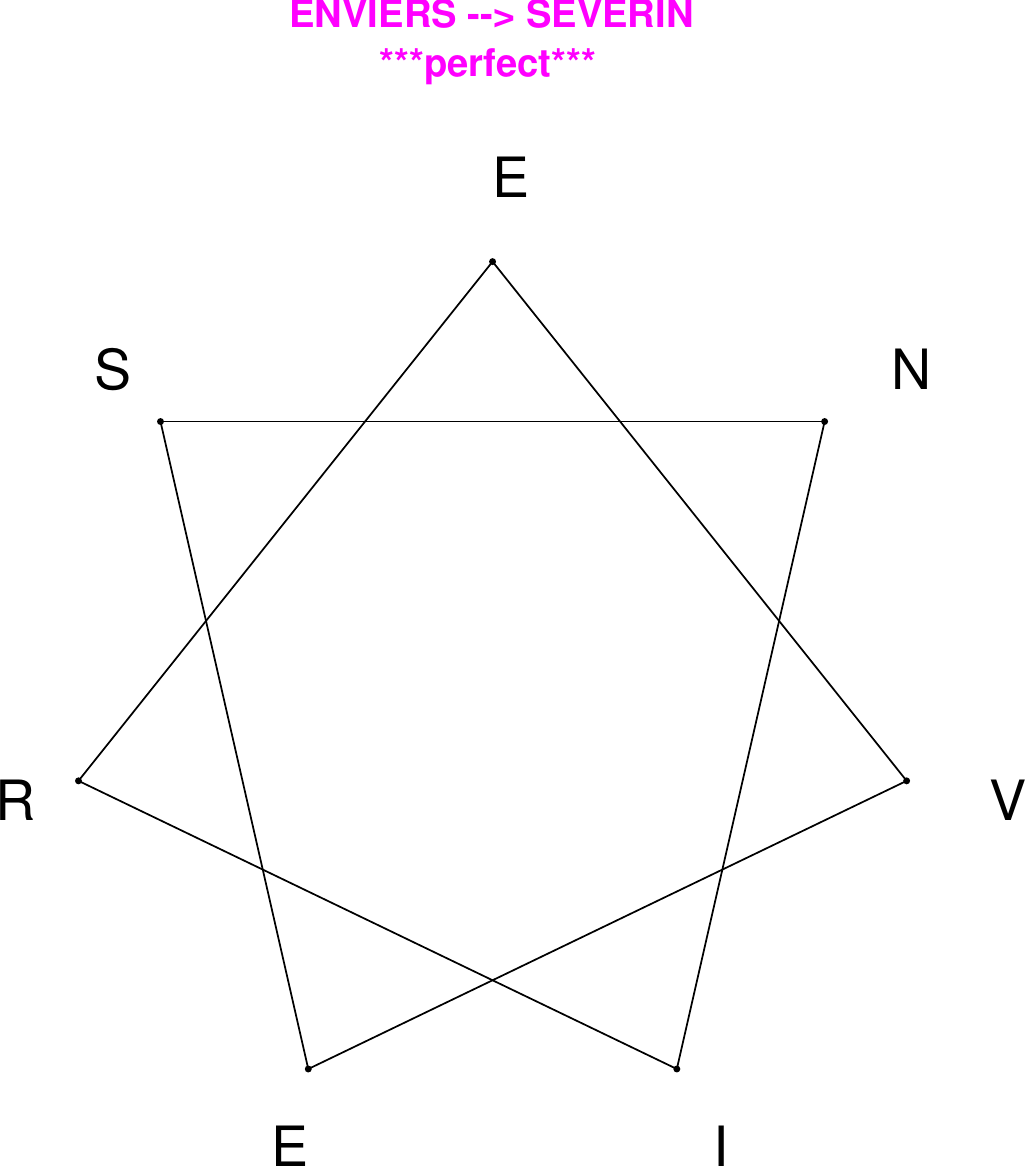}
\end{subfigure}
\hfill
\begin{subfigure}[T]{0.19\textwidth}
\centering
\includegraphics[width=\textwidth]{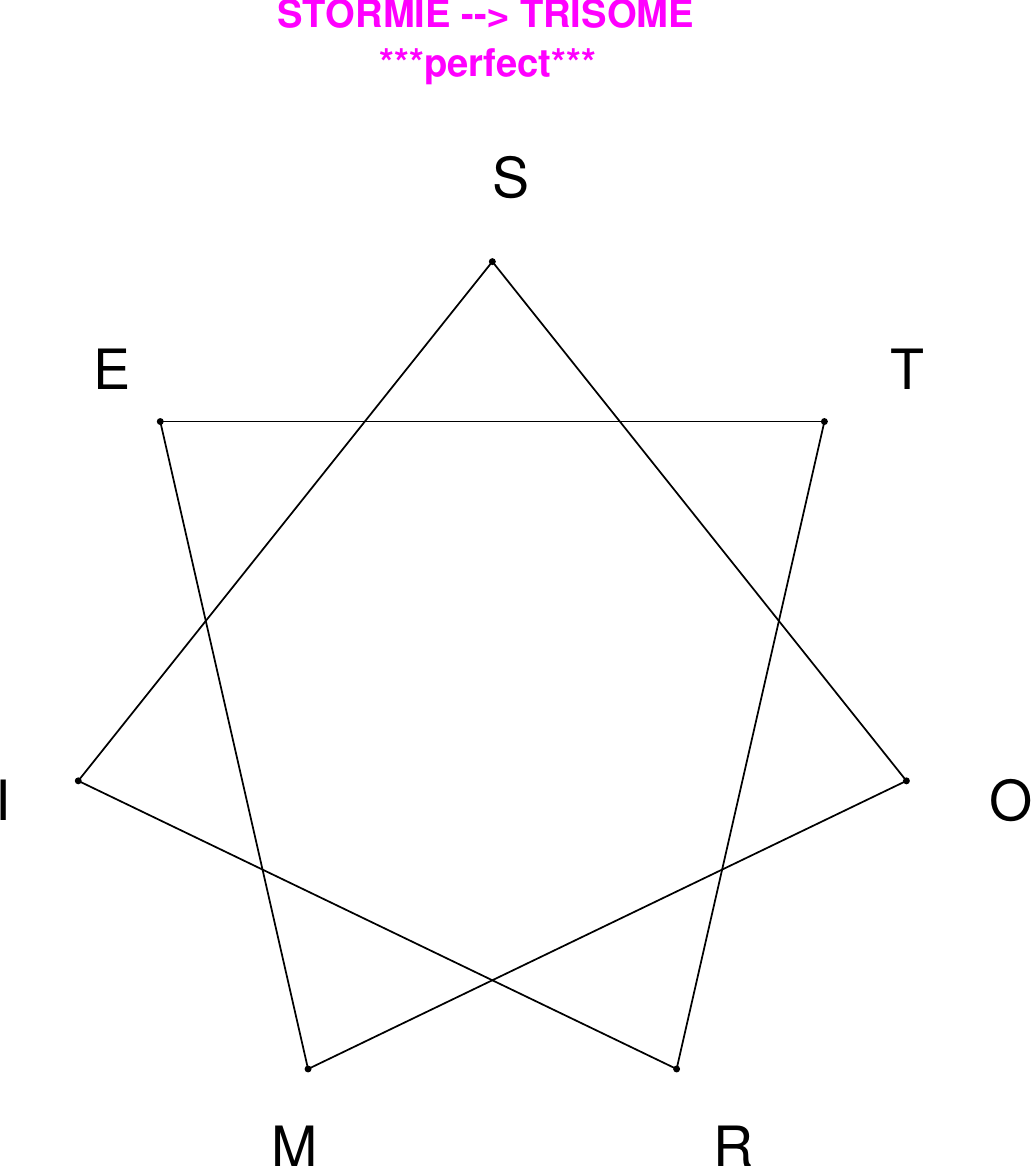}
\end{subfigure}
\hfill
\begin{subfigure}[T]{0.19\textwidth}
\centering
\includegraphics[width=\textwidth]{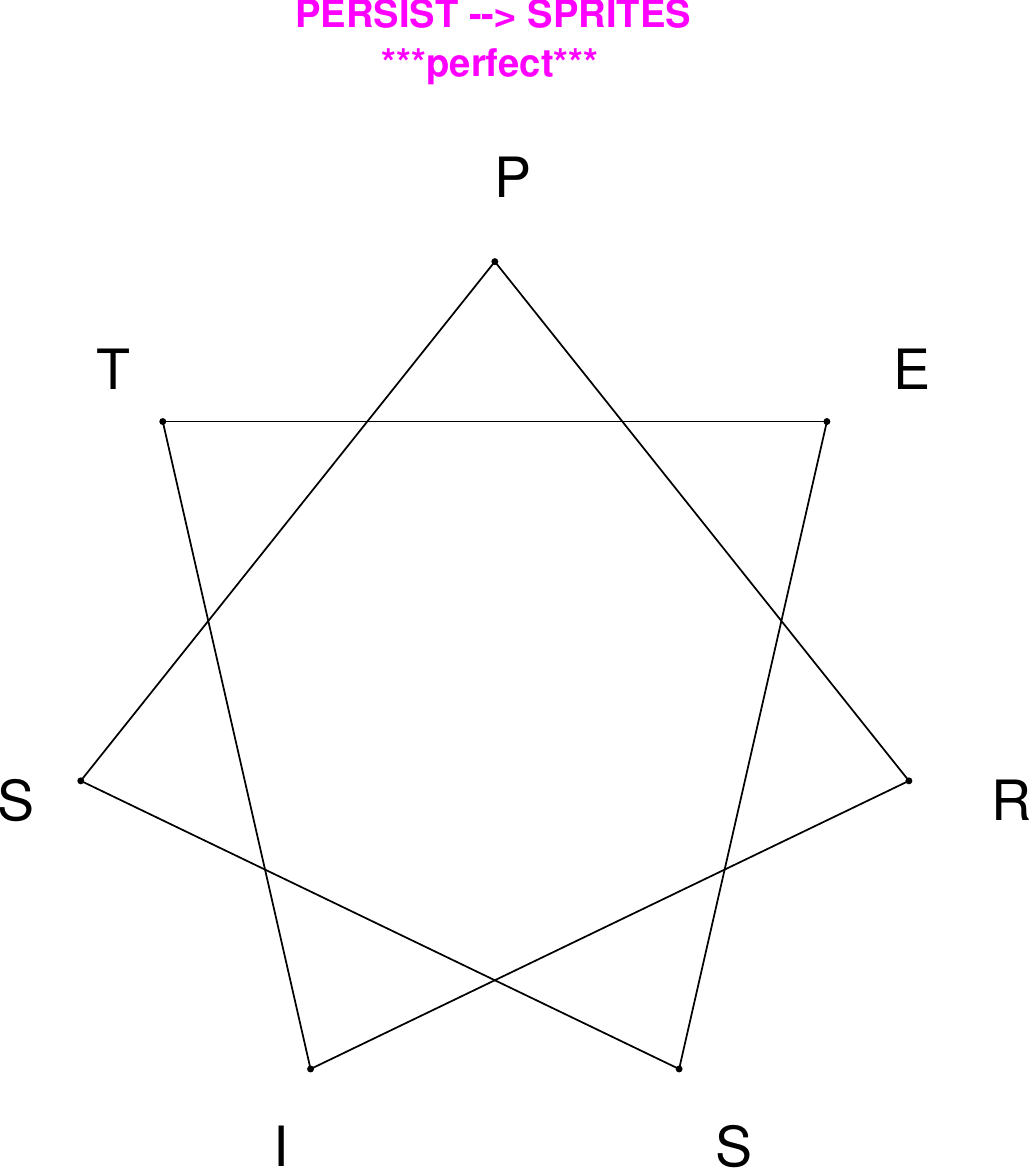}
\end{subfigure}
\hfill
\begin{subfigure}[T]{0.19\textwidth}
\centering
\includegraphics[width=\textwidth]{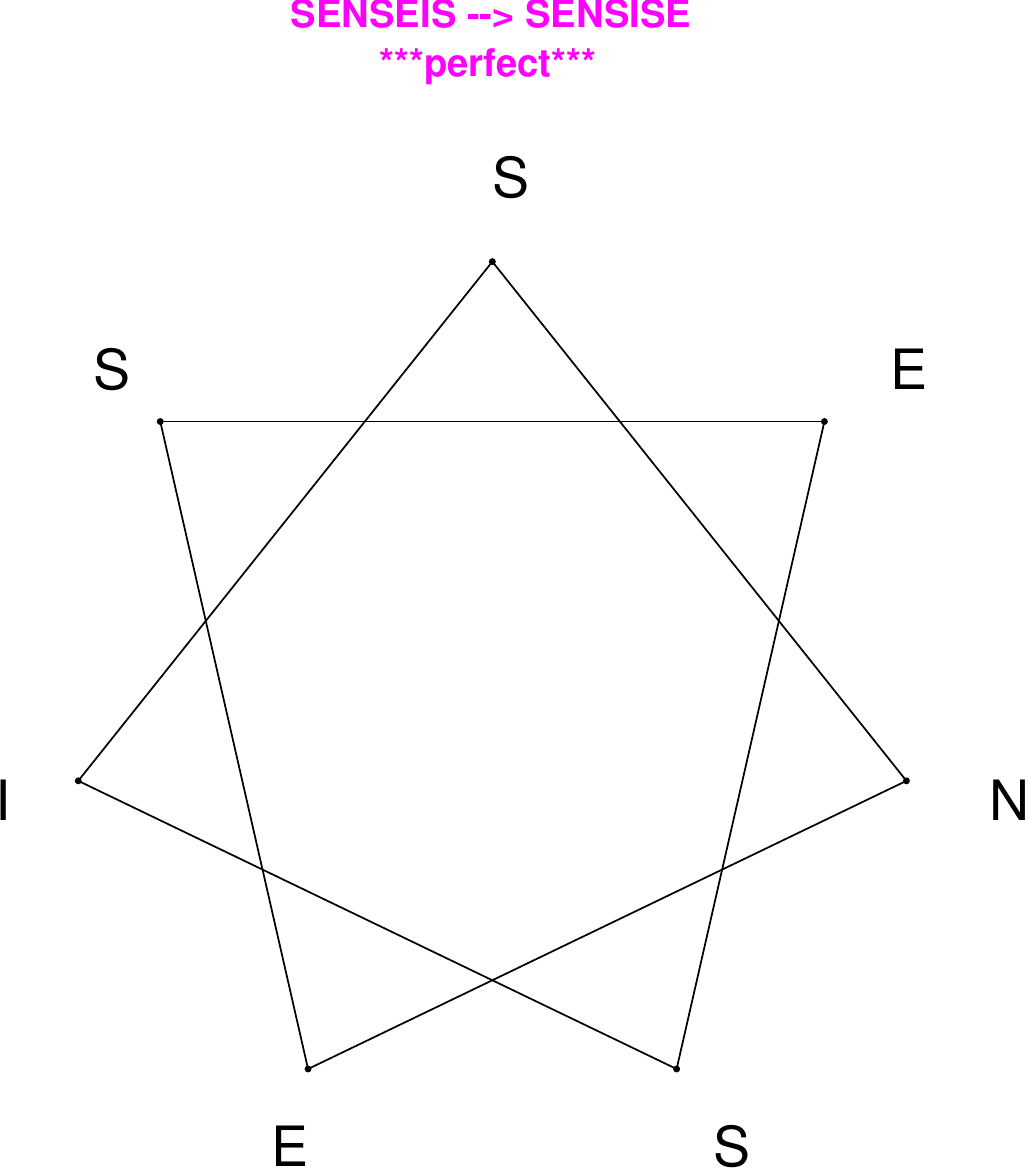}
\end{subfigure}
\hfill
\begin{subfigure}[T]{0.19\textwidth}
\centering
\includegraphics[width=\textwidth]{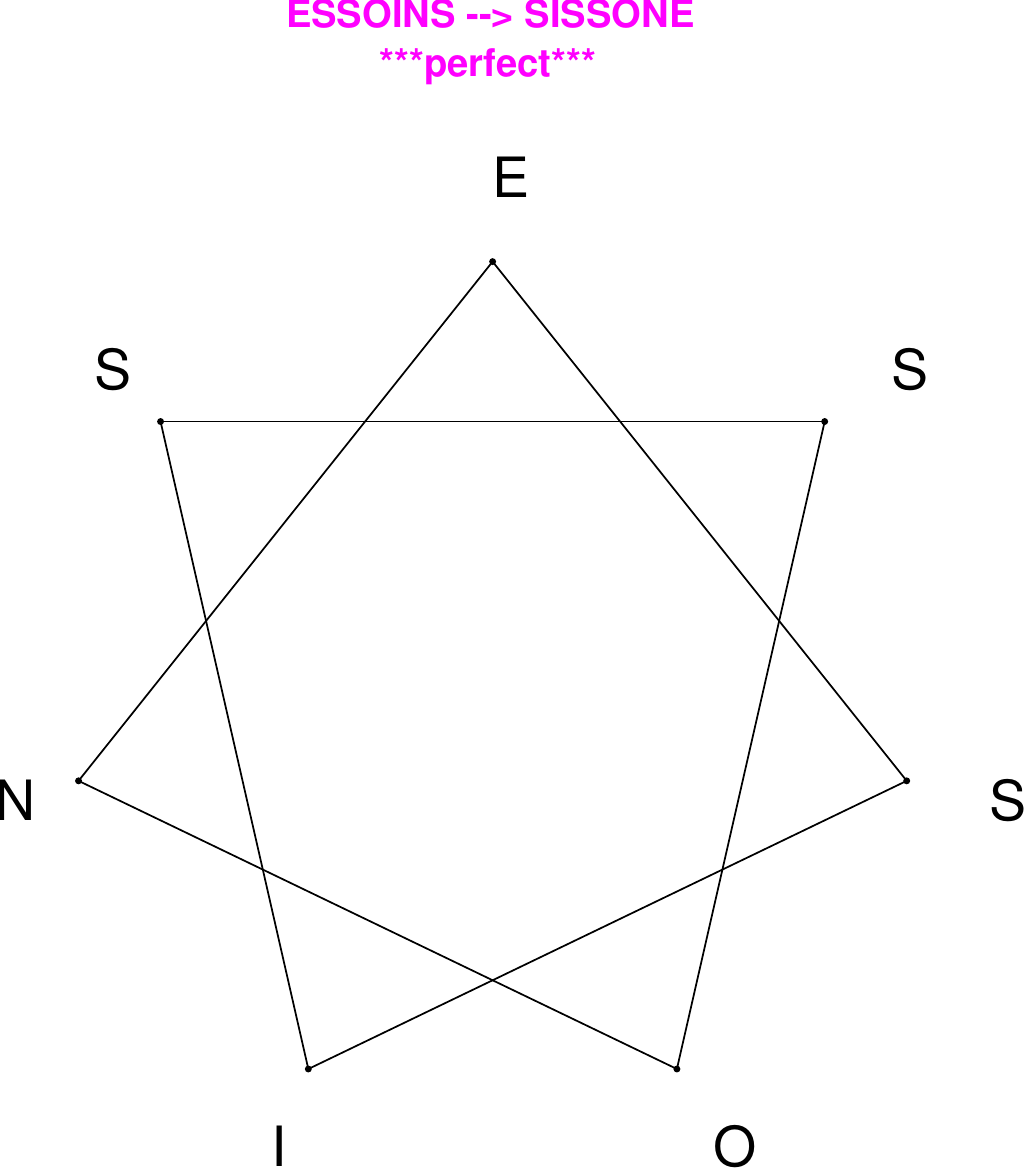}
\end{subfigure}
\end{figure}

\begin{figure}[H]
\centering
\begin{subfigure}[T]{0.19\textwidth}
\centering
\includegraphics[width=\textwidth]{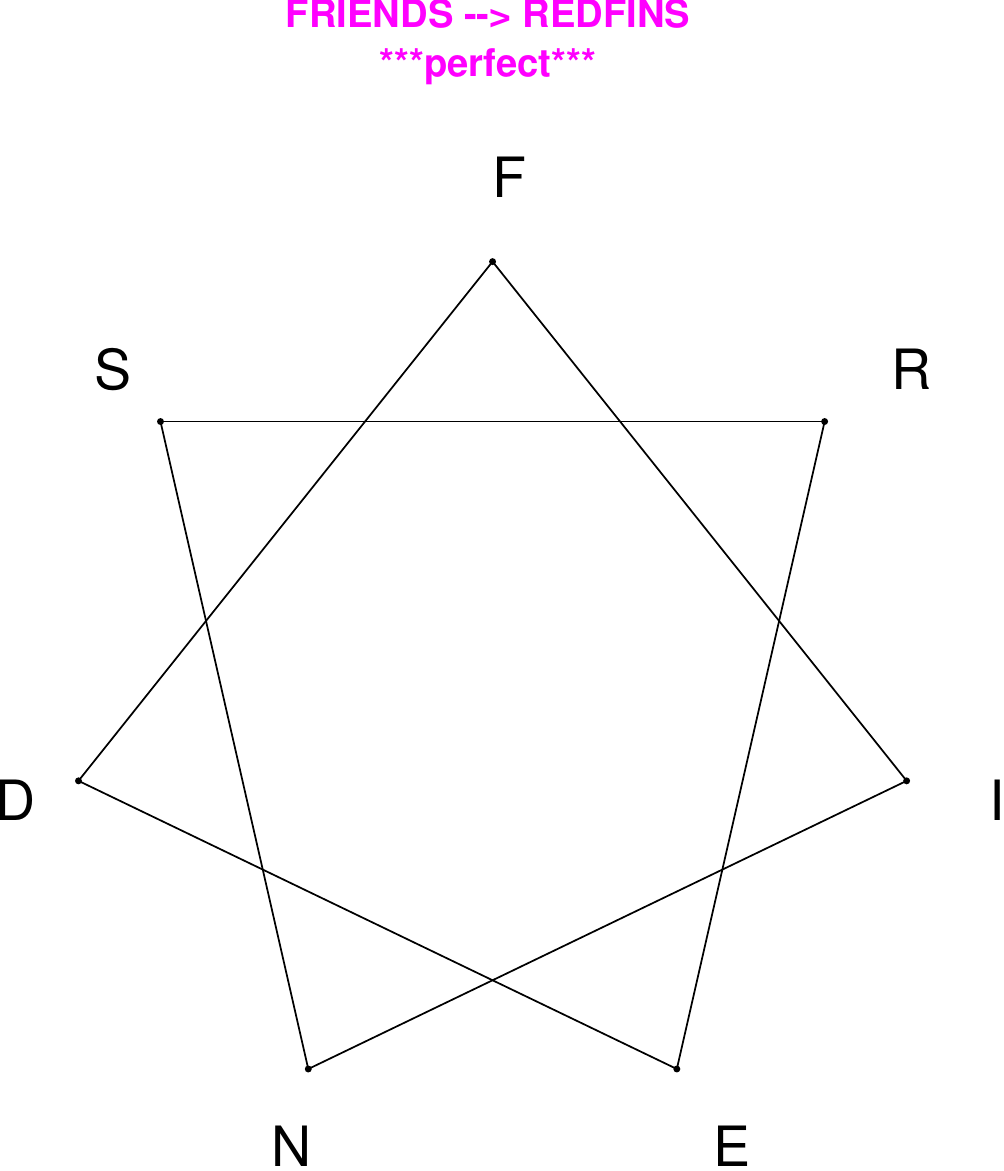}
\end{subfigure}
\hfill
\begin{subfigure}[T]{0.19\textwidth}
\centering
\includegraphics[width=\textwidth]{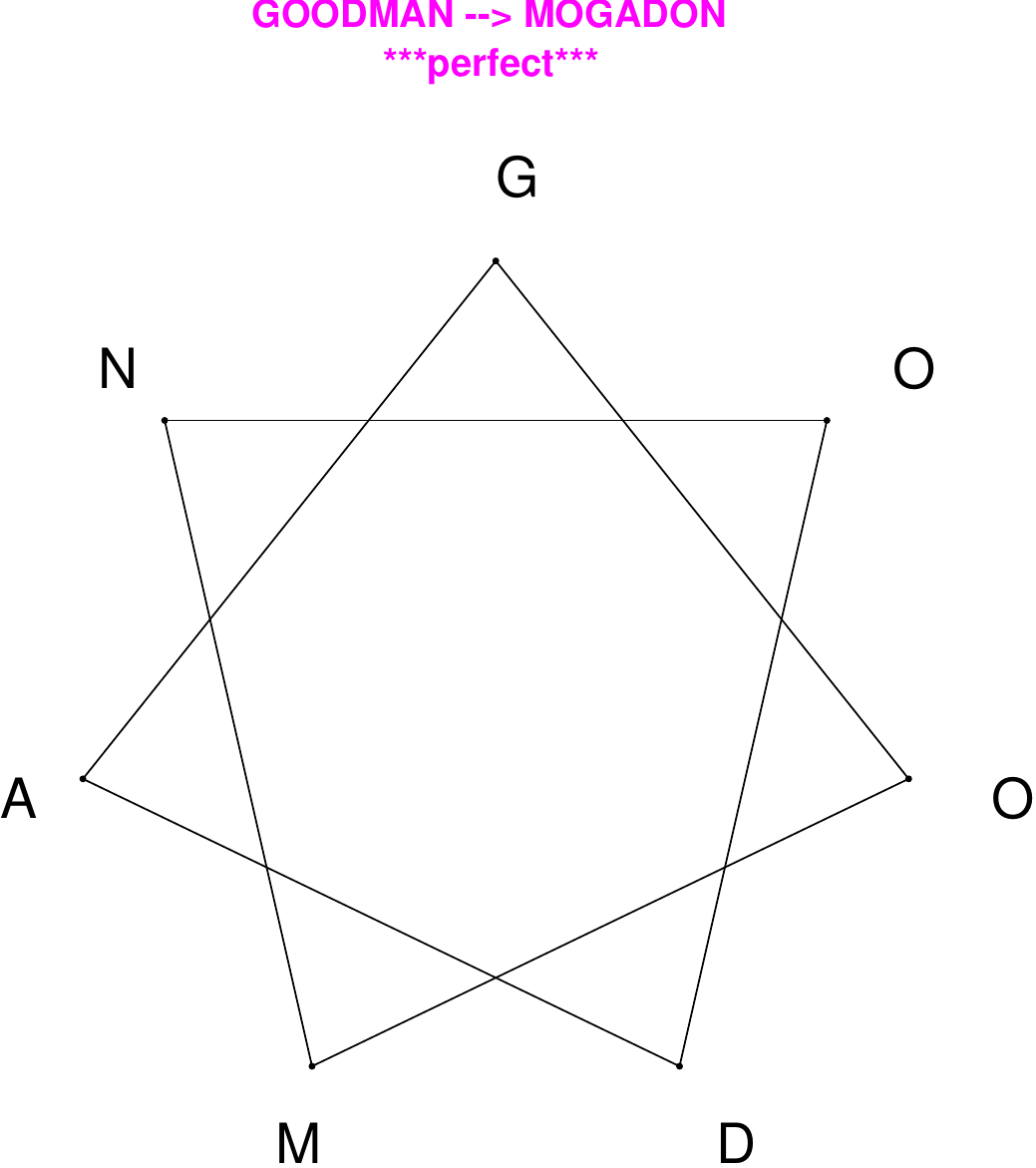}
\end{subfigure}
\hfill
\begin{subfigure}[T]{0.19\textwidth}
\centering
\includegraphics[width=\textwidth]{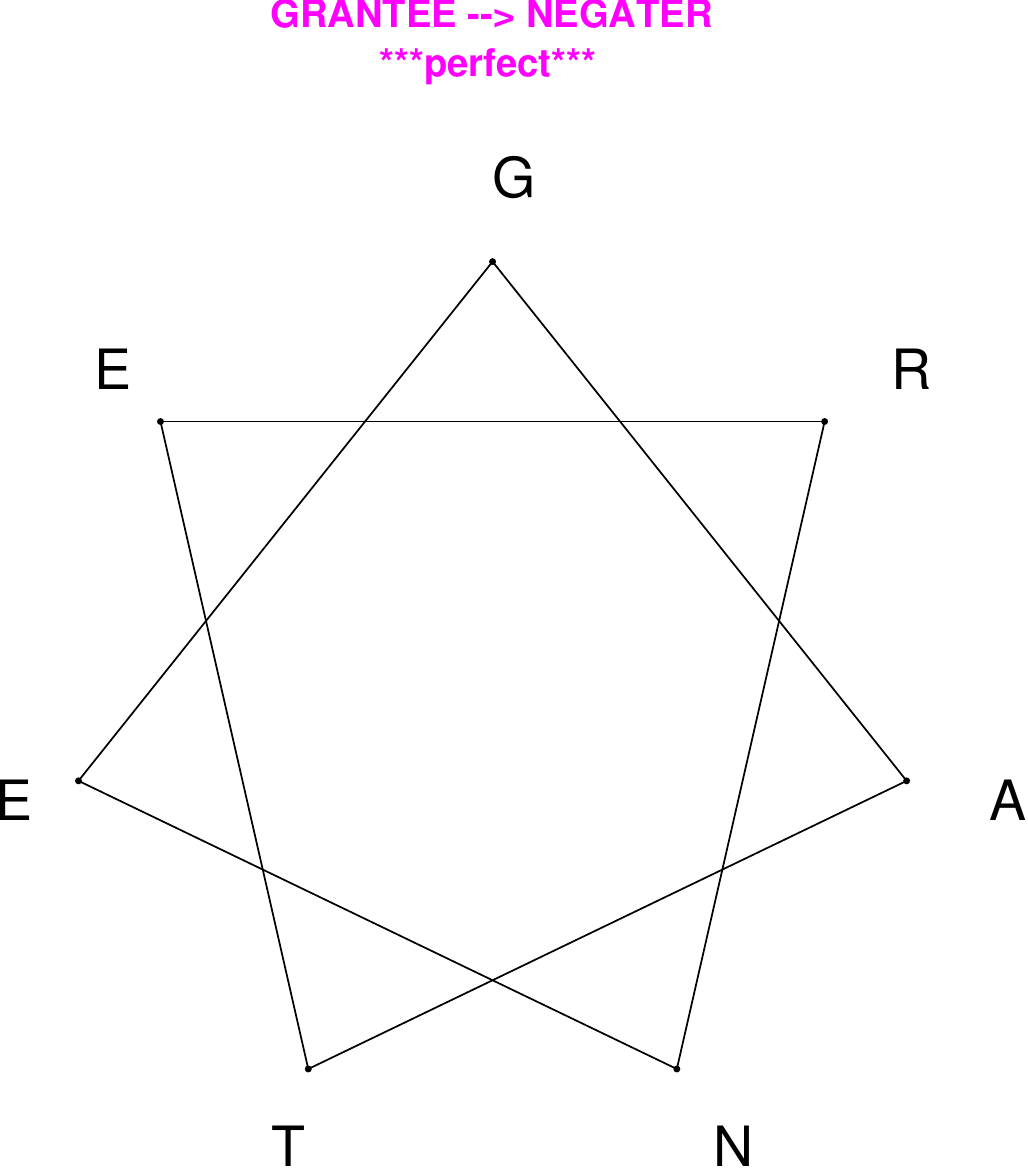}
\end{subfigure}
\hfill
\begin{subfigure}[T]{0.19\textwidth}
\centering
\includegraphics[width=\textwidth]{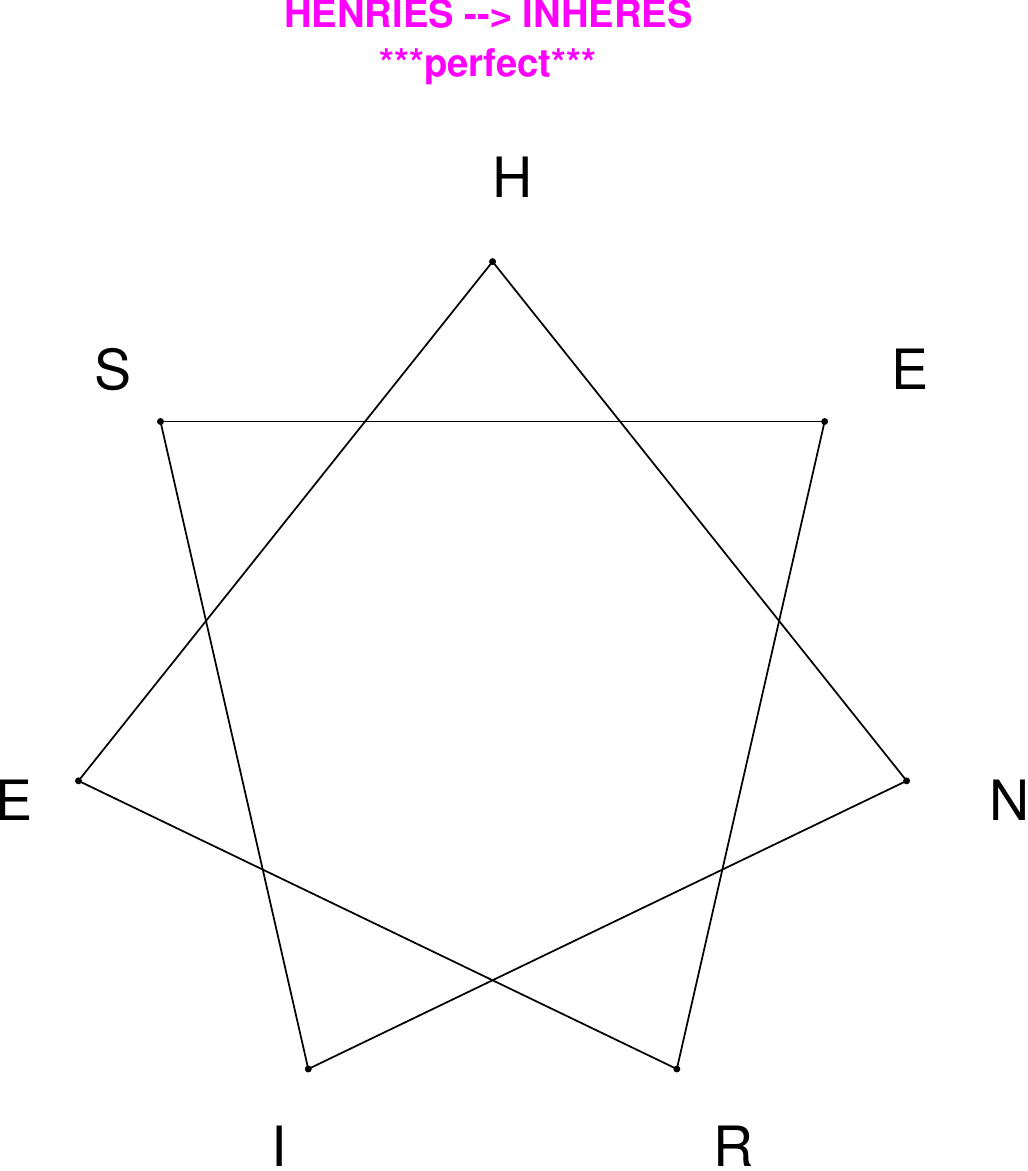}
\end{subfigure}
\hfill
\begin{subfigure}[T]{0.19\textwidth}
\centering
\includegraphics[width=\textwidth]{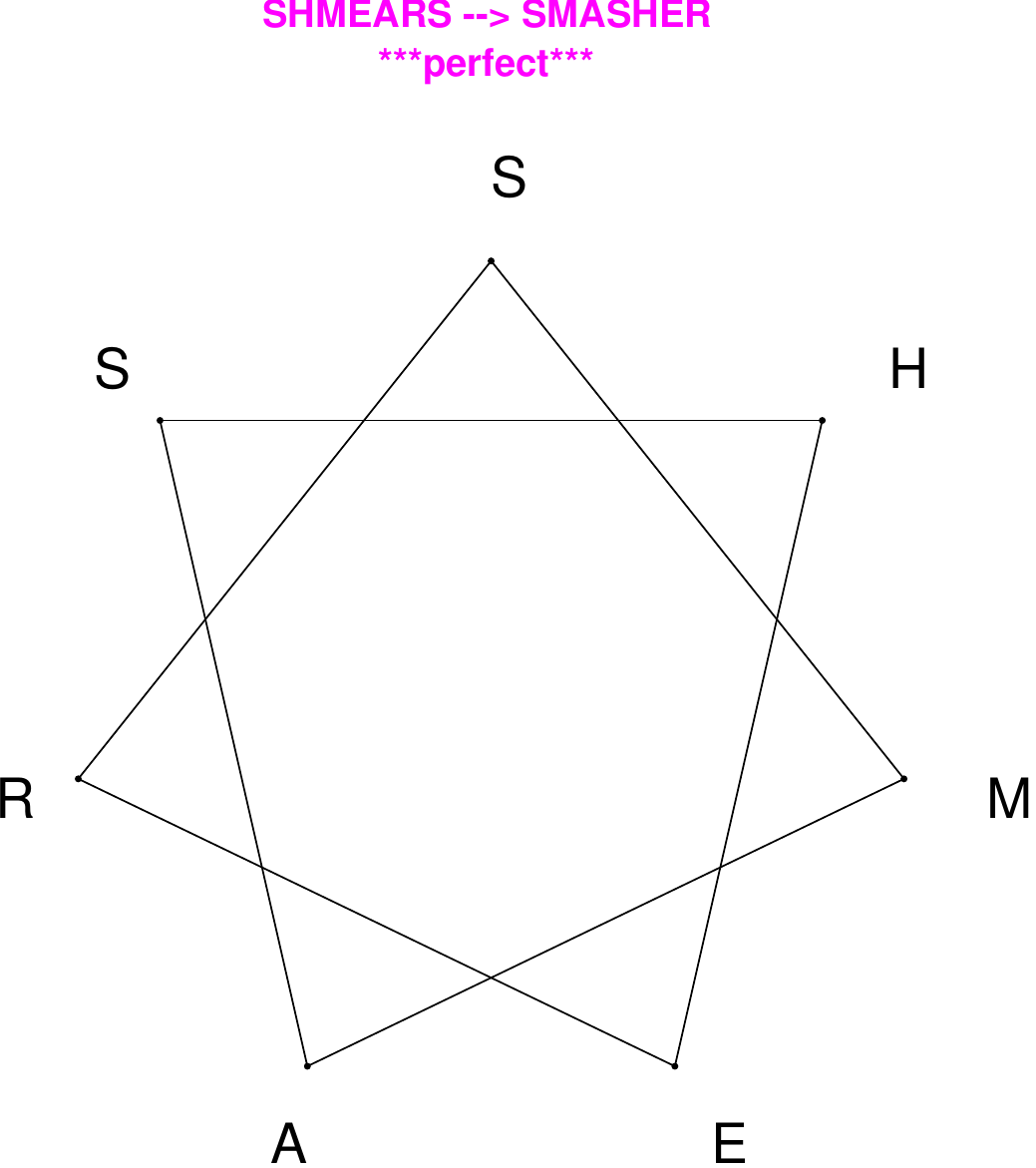}
\end{subfigure}
\end{figure}

\begin{figure}[H]
\centering
\begin{subfigure}[T]{0.19\textwidth}
\centering
\includegraphics[width=\textwidth]{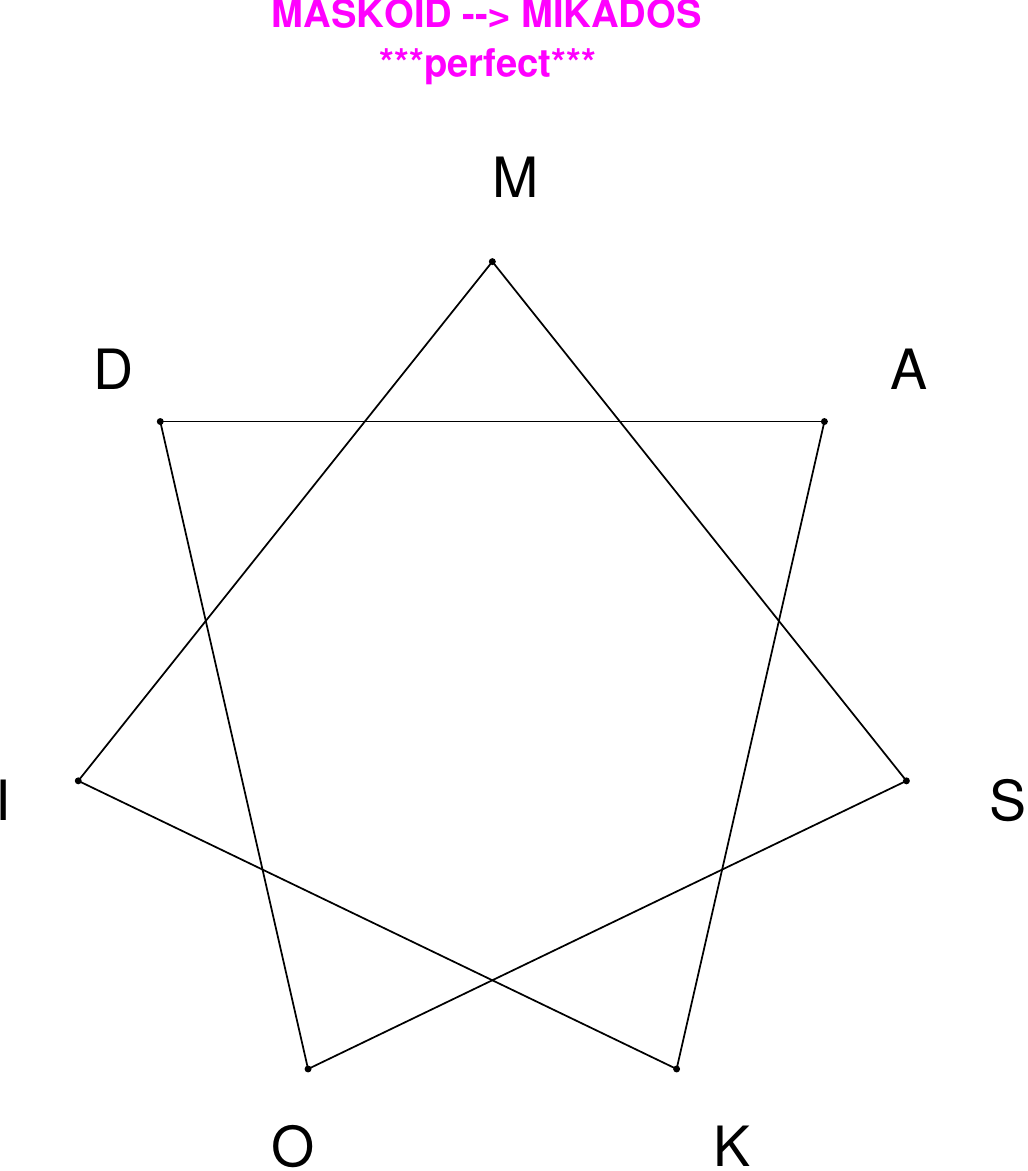}
\end{subfigure}
\hfill
\begin{subfigure}[T]{0.19\textwidth}
\centering
\includegraphics[width=\textwidth]{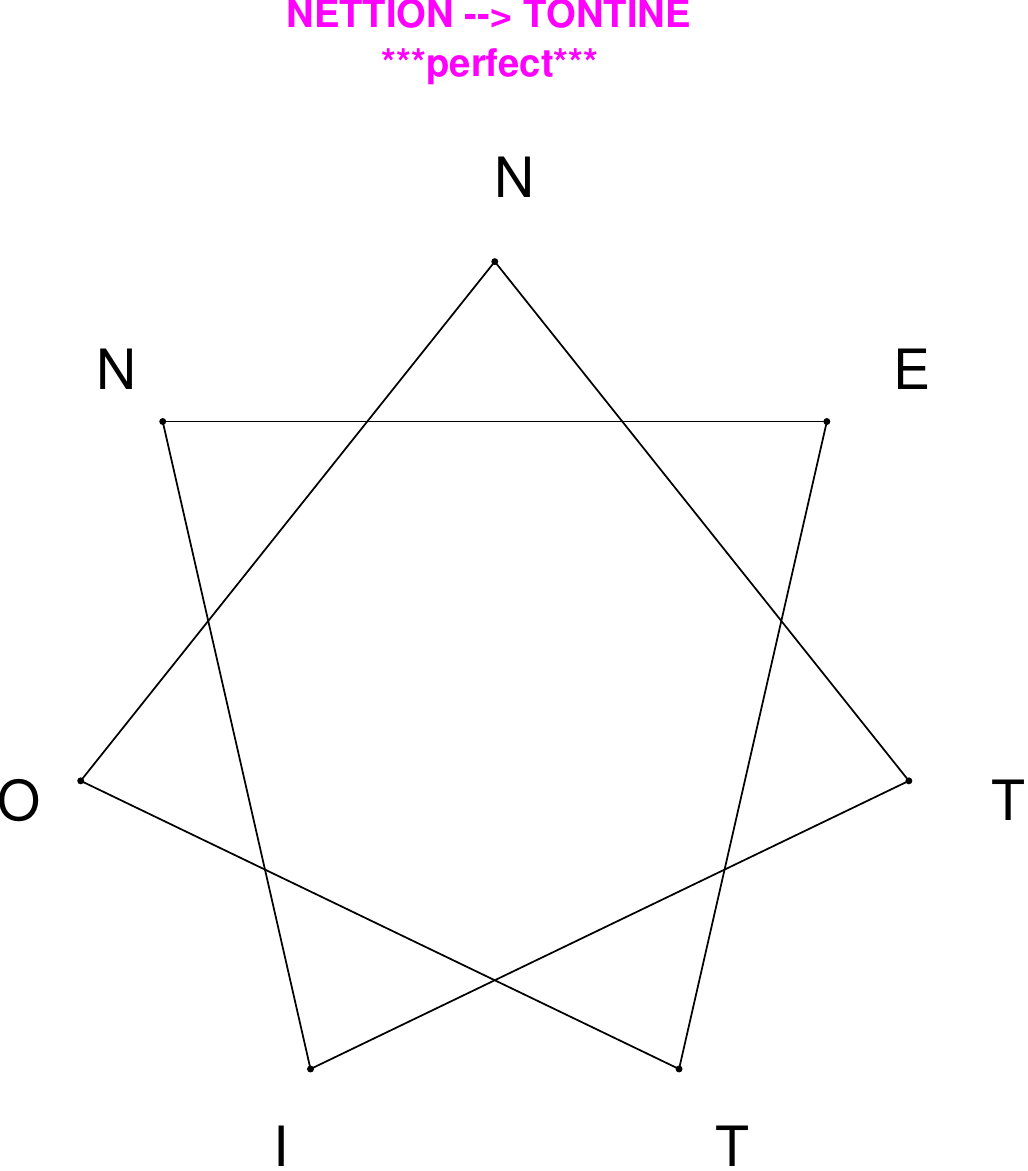}
\end{subfigure}
\hfill
\begin{subfigure}[T]{0.19\textwidth}
\centering
\includegraphics[width=\textwidth]{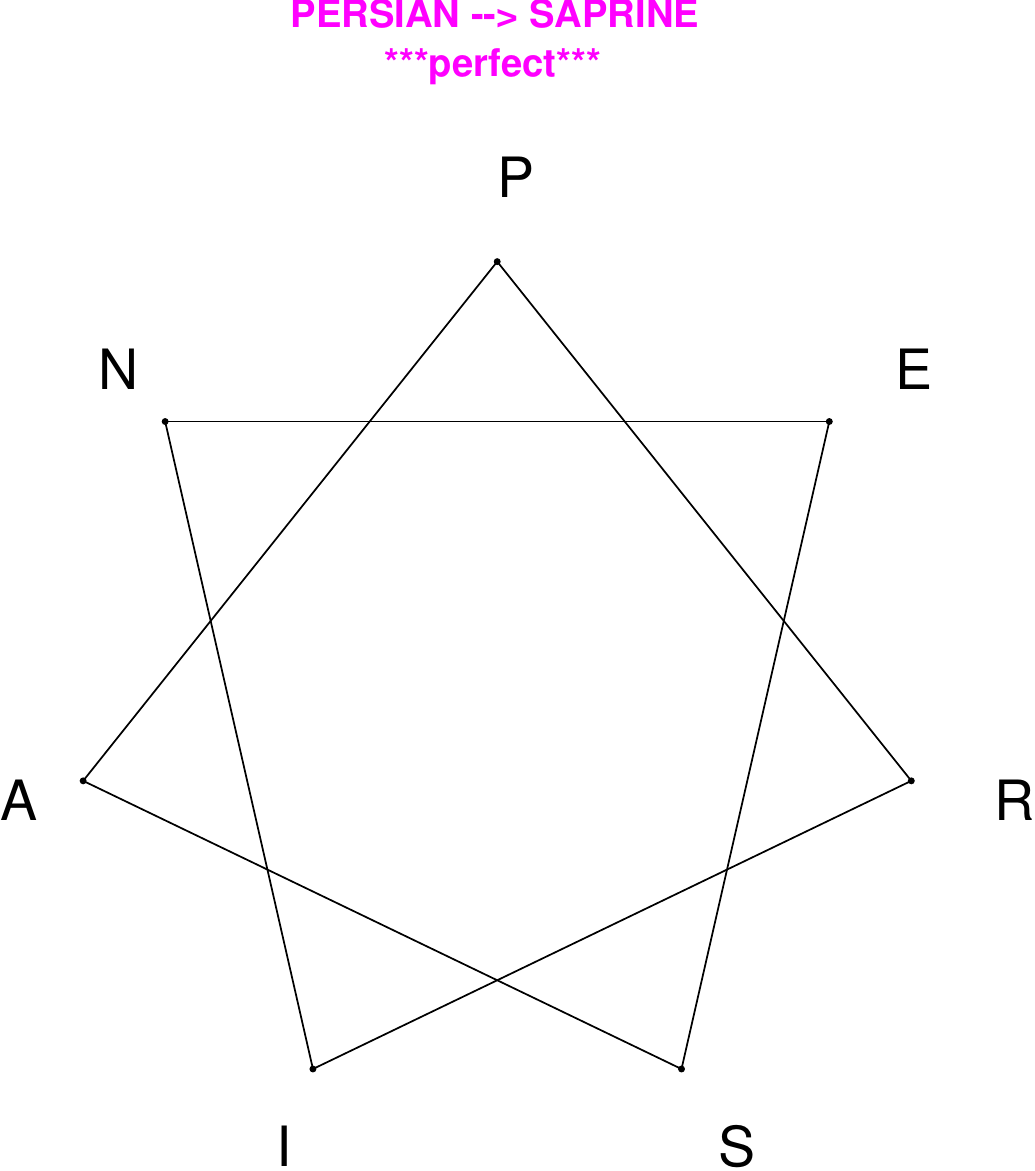}
\end{subfigure}
\hfill
\begin{subfigure}[T]{0.19\textwidth}
\centering
\includegraphics[width=\textwidth]{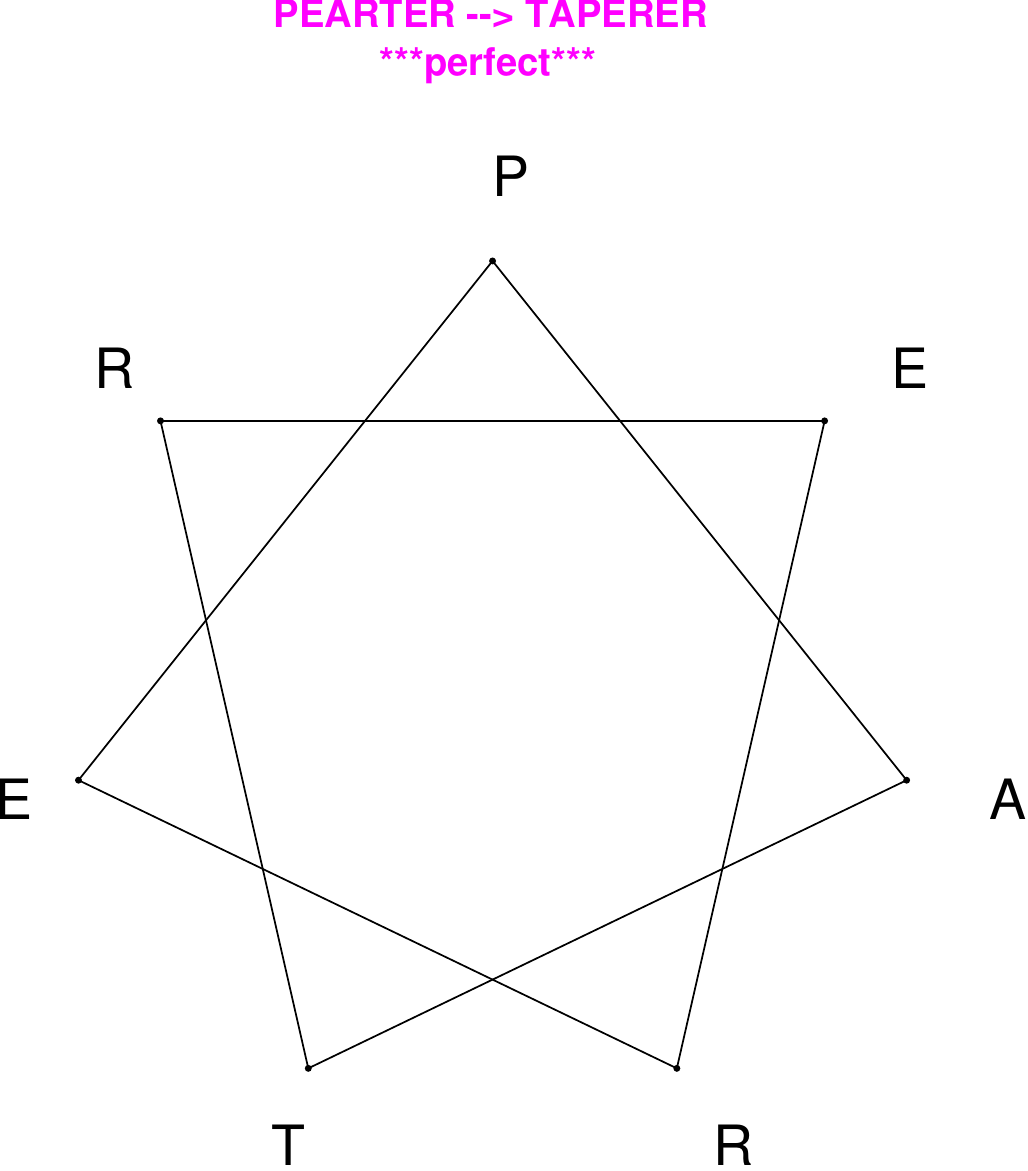}
\end{subfigure}
\hfill
\begin{subfigure}[T]{0.19\textwidth}
\centering
\includegraphics[width=\textwidth]{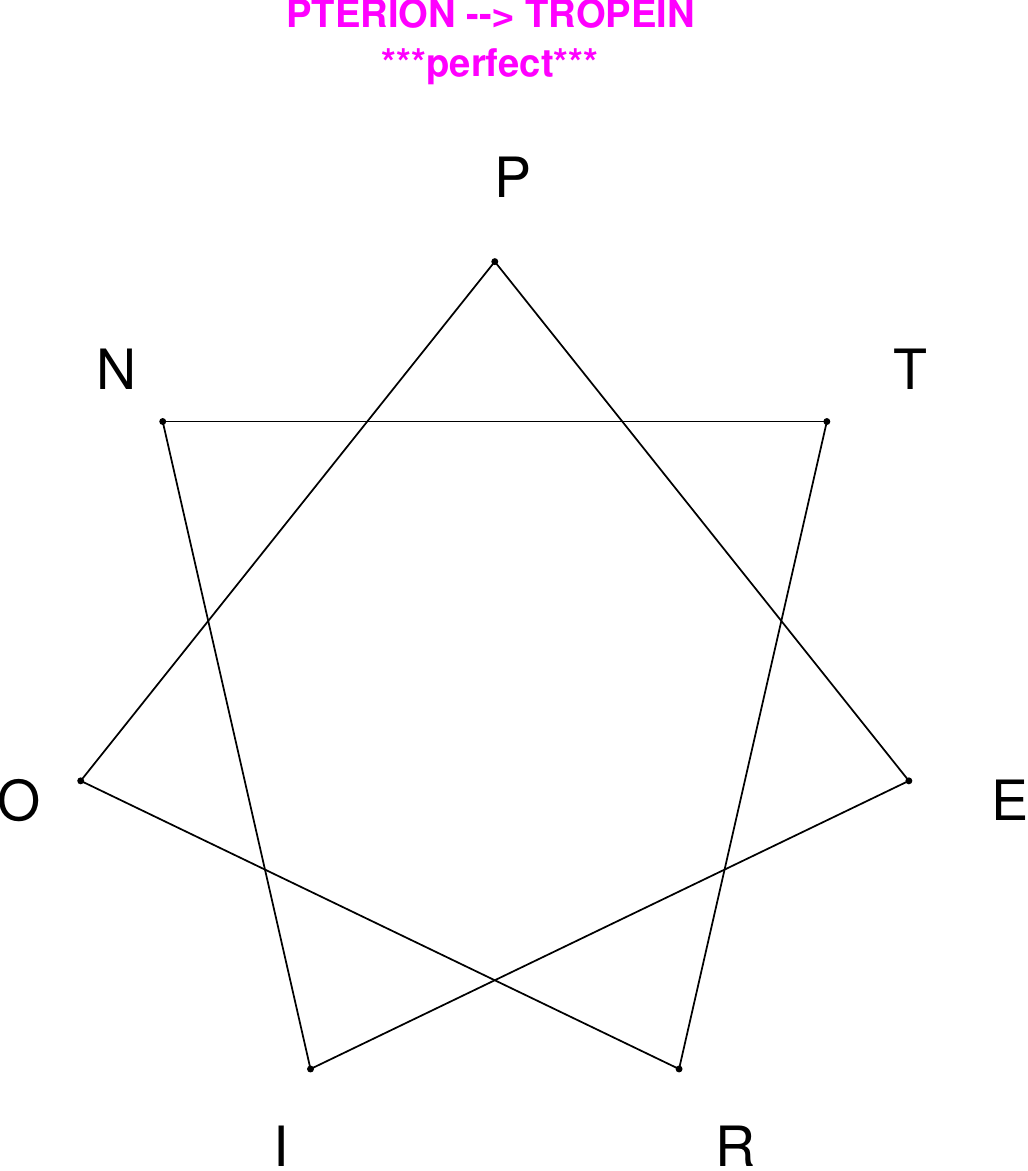}
\end{subfigure}
\end{figure}

\begin{figure}[H]
\centering
\begin{subfigure}[T]{0.19\textwidth}
\centering
\includegraphics[width=\textwidth]{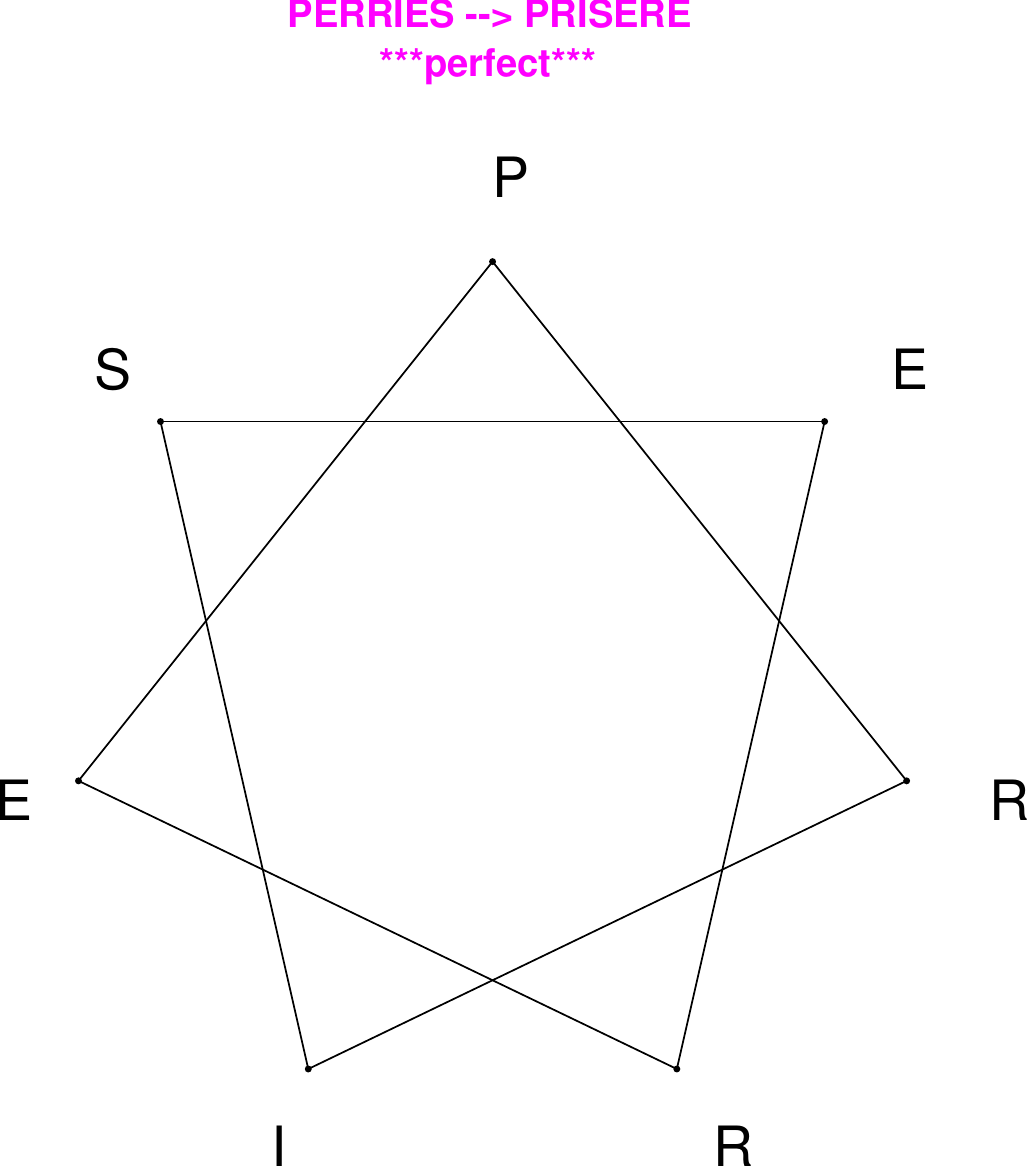}
\end{subfigure}
\hfill
\begin{subfigure}[T]{0.19\textwidth}
\centering
\includegraphics[width=\textwidth]{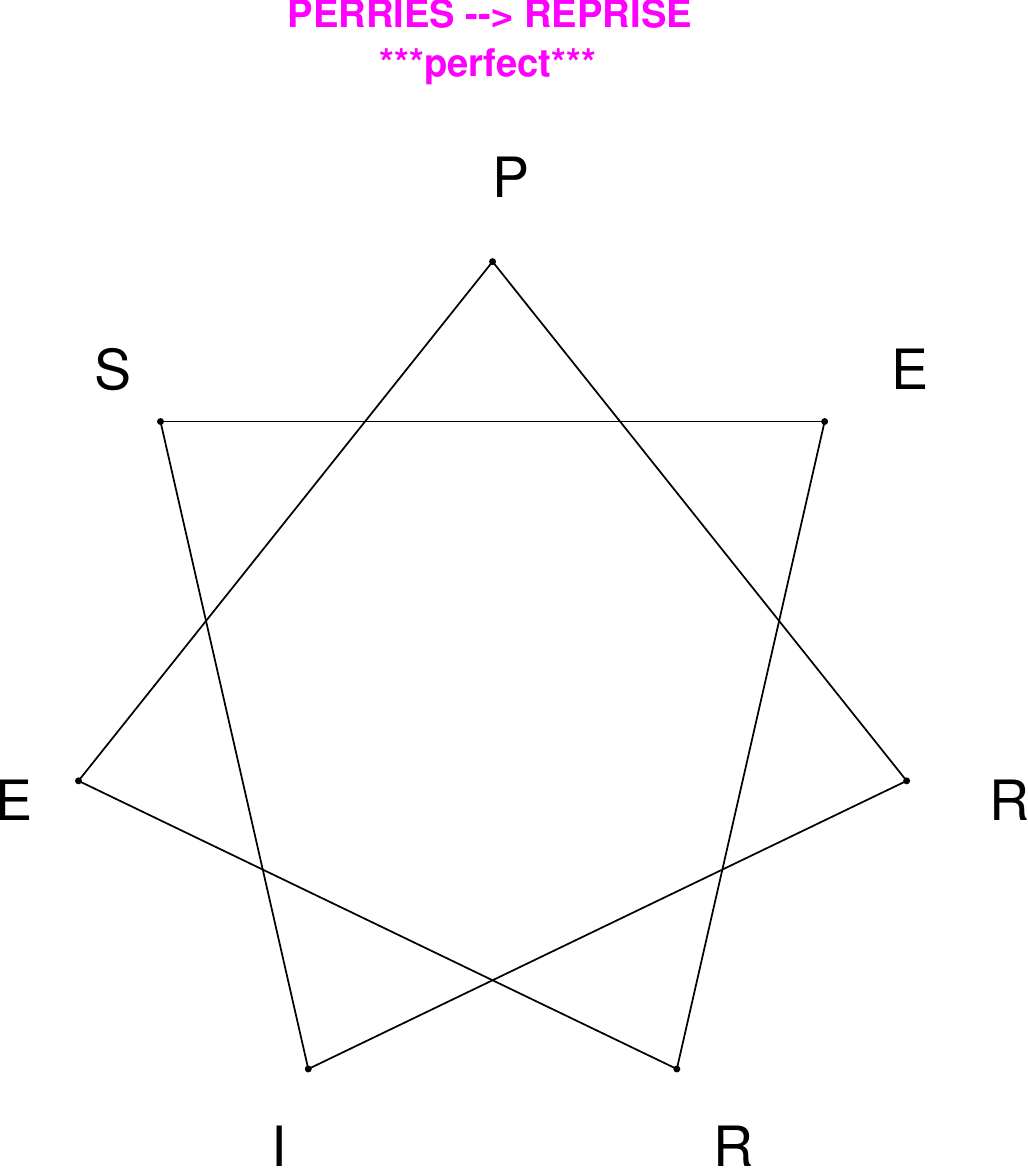}
\end{subfigure}
\hfill
\begin{subfigure}[T]{0.19\textwidth}
\centering
\includegraphics[width=\textwidth]{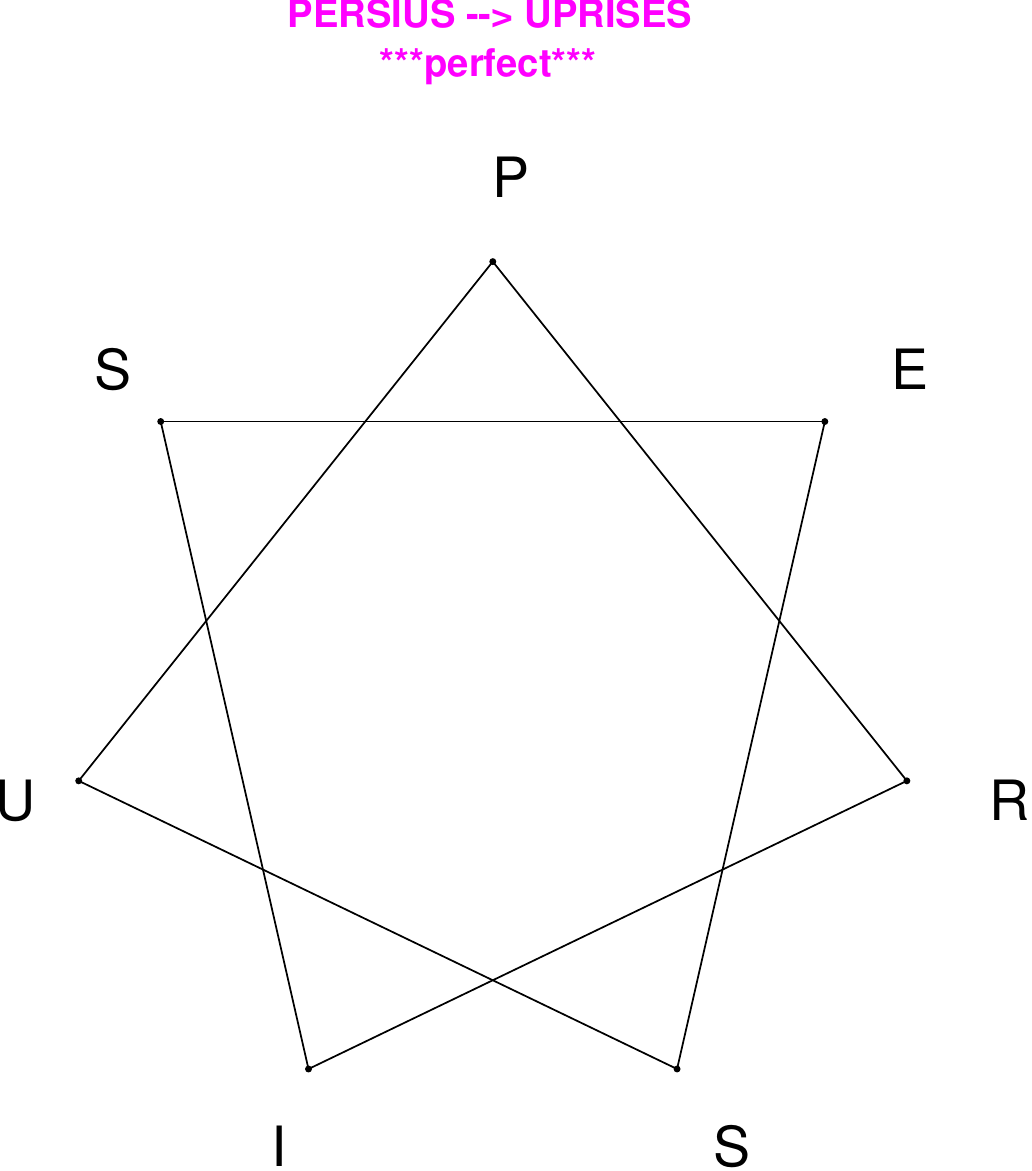}
\end{subfigure}
\hfill
\begin{subfigure}[T]{0.19\textwidth}
\centering
\includegraphics[width=\textwidth]{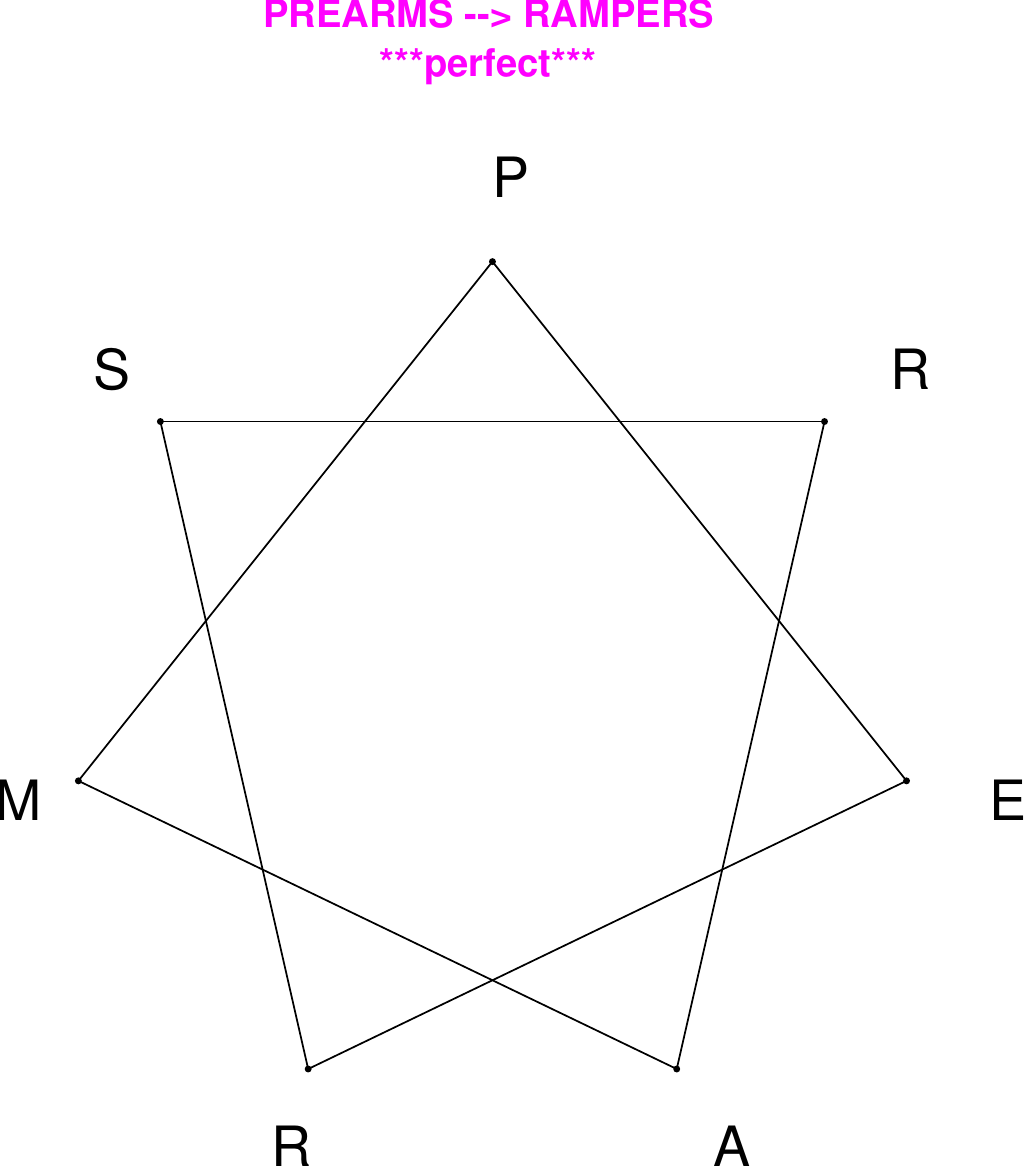}
\end{subfigure}
\hfill
\begin{subfigure}[T]{0.19\textwidth}
\centering
\includegraphics[width=\textwidth]{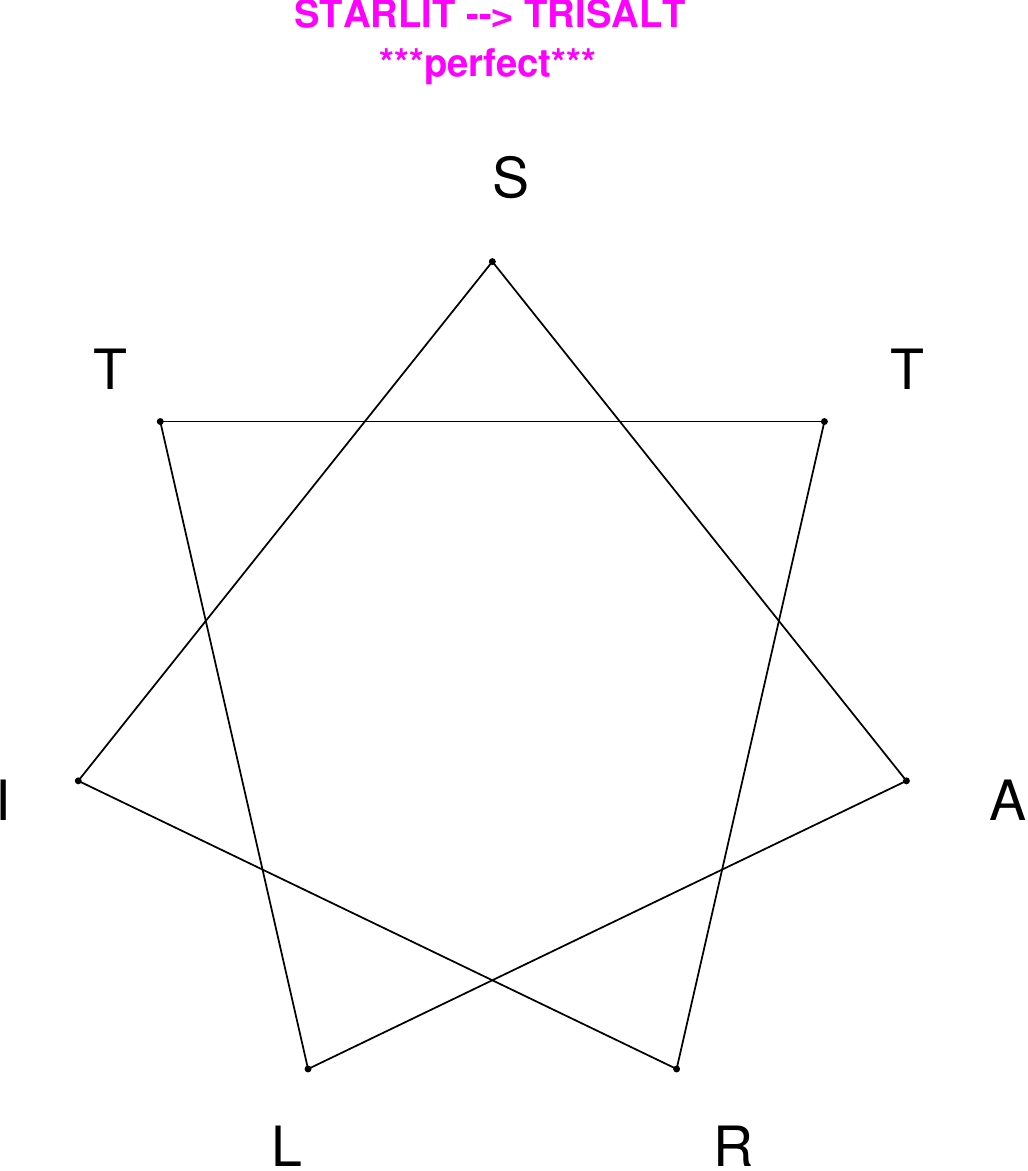}
\end{subfigure}
\end{figure}

\subsubsection{Symmetric Stars $N=7$}

\begin{figure}[H]
\centering
\begin{subfigure}[T]{0.19\textwidth}
\centering
\includegraphics[width=\textwidth]{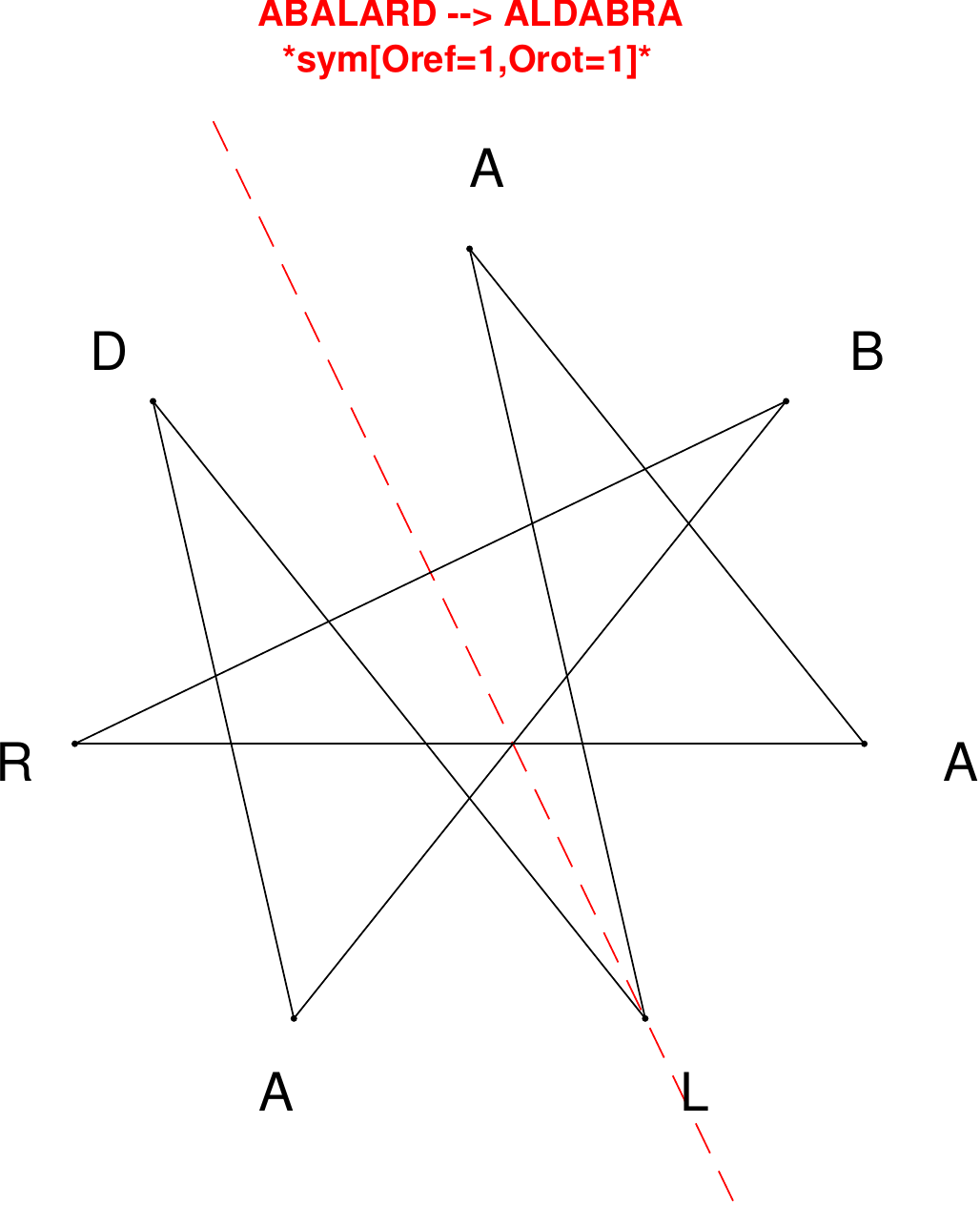}
\end{subfigure}
\hfill
\begin{subfigure}[T]{0.19\textwidth}
\centering
\includegraphics[width=\textwidth]{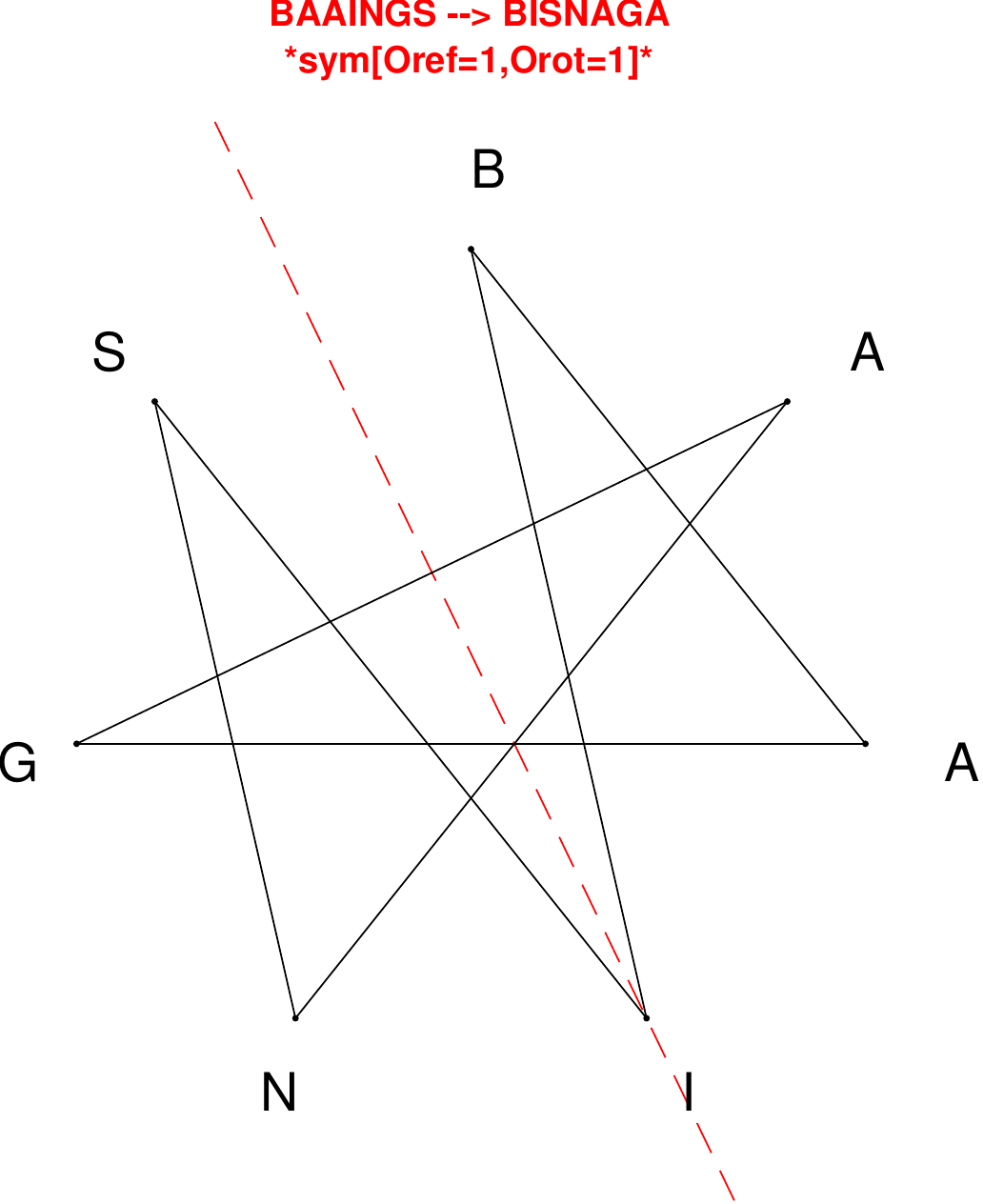}
\end{subfigure}
\hfill
\begin{subfigure}[T]{0.19\textwidth}
\centering
\includegraphics[width=\textwidth]{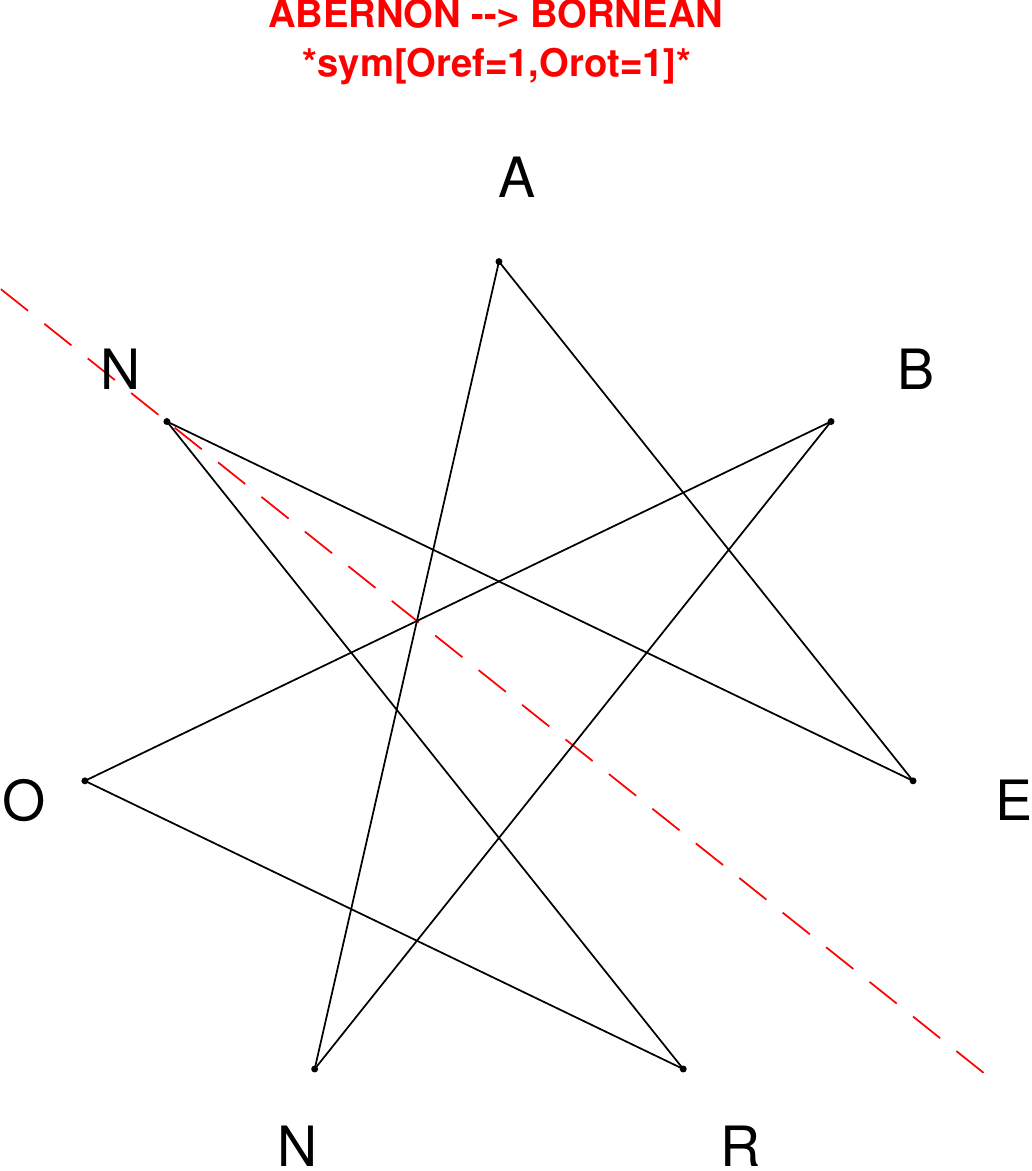}
\end{subfigure}
\hfill
\begin{subfigure}[T]{0.19\textwidth}
\centering
\includegraphics[width=\textwidth]{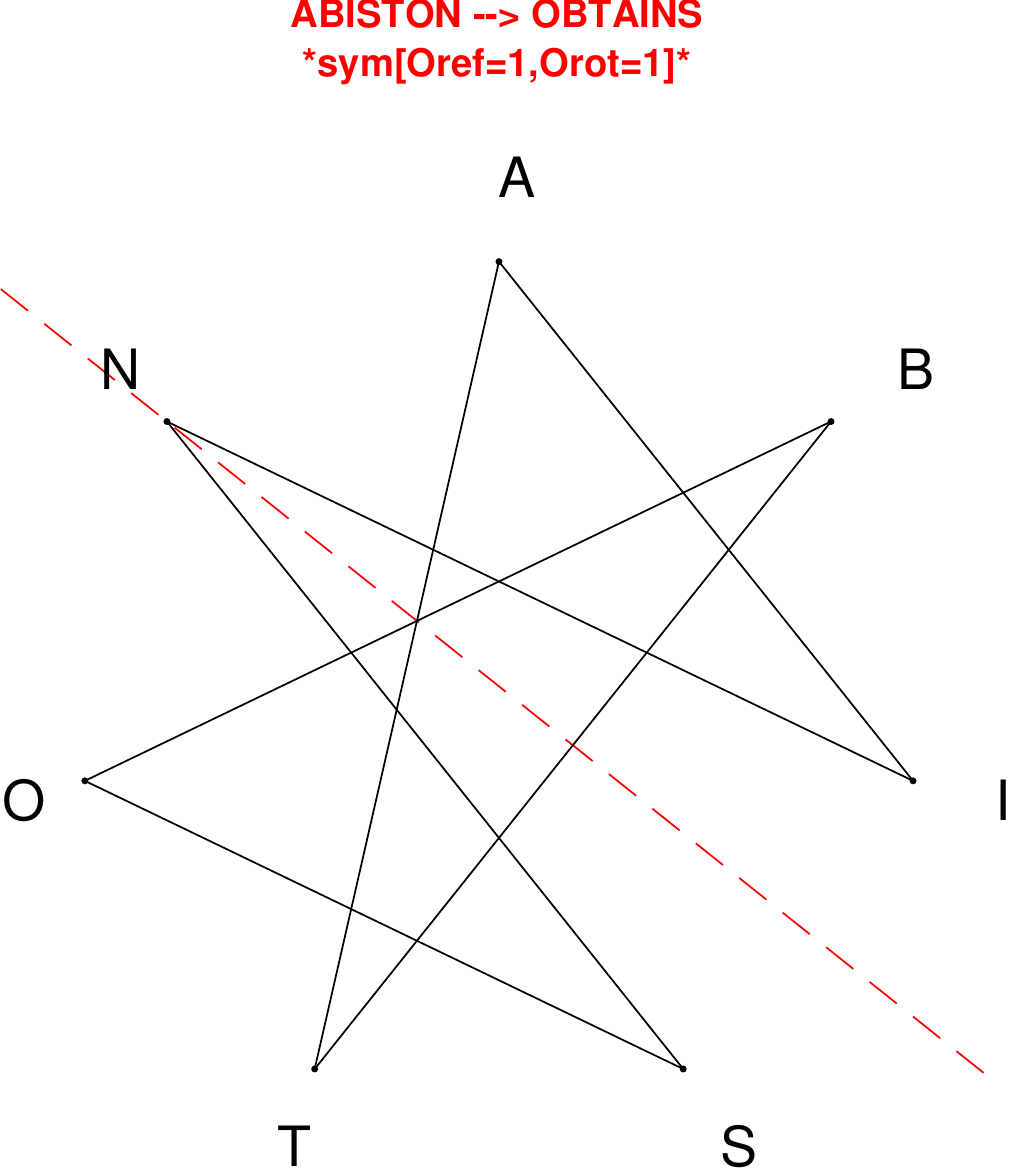}
\end{subfigure}
\hfill
\begin{subfigure}[T]{0.19\textwidth}
\centering
\includegraphics[width=\textwidth]{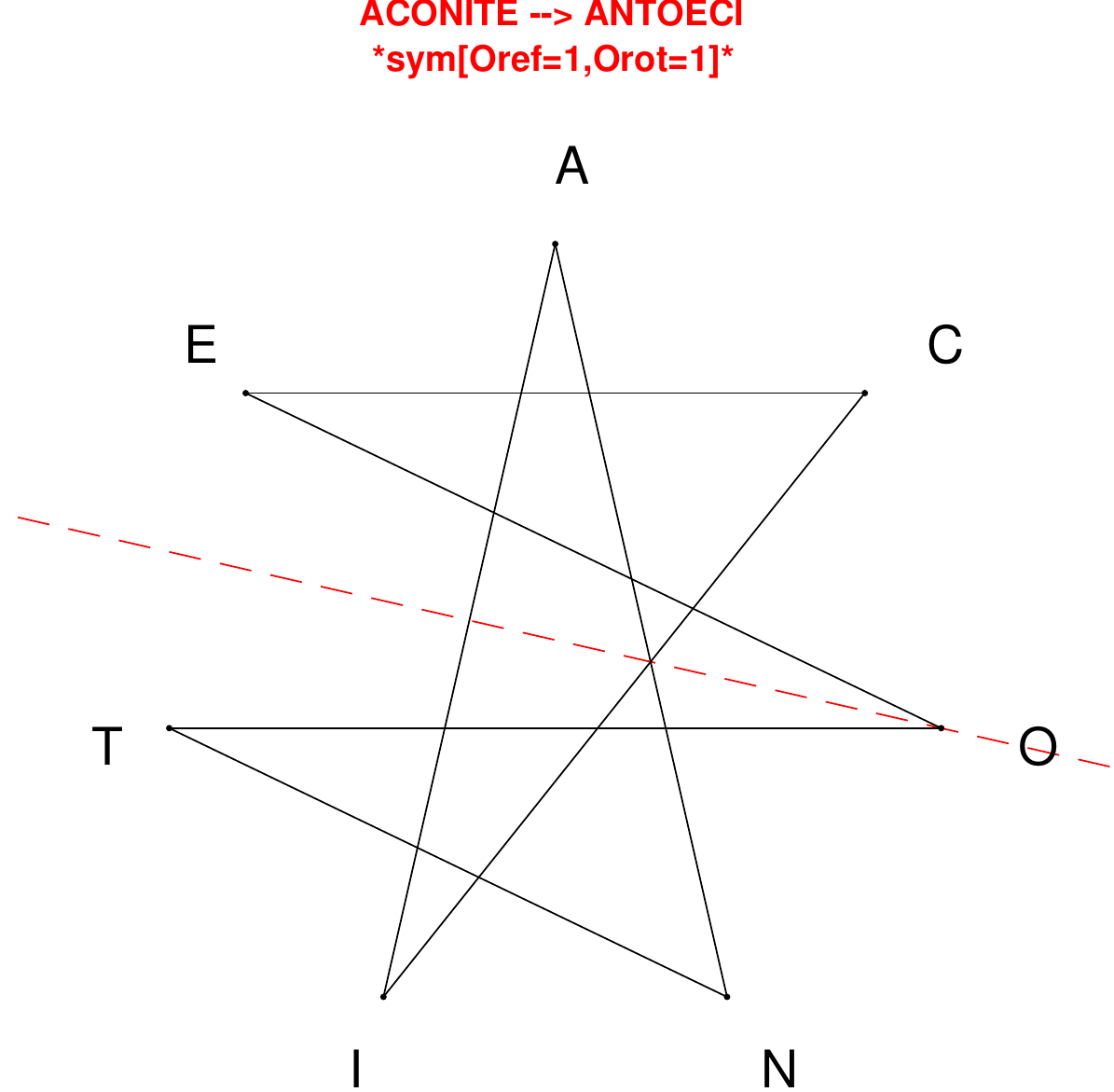}
\end{subfigure}
\end{figure}

\begin{figure}[H]
\centering
\begin{subfigure}[T]{0.19\textwidth}
\centering
\includegraphics[width=\textwidth]{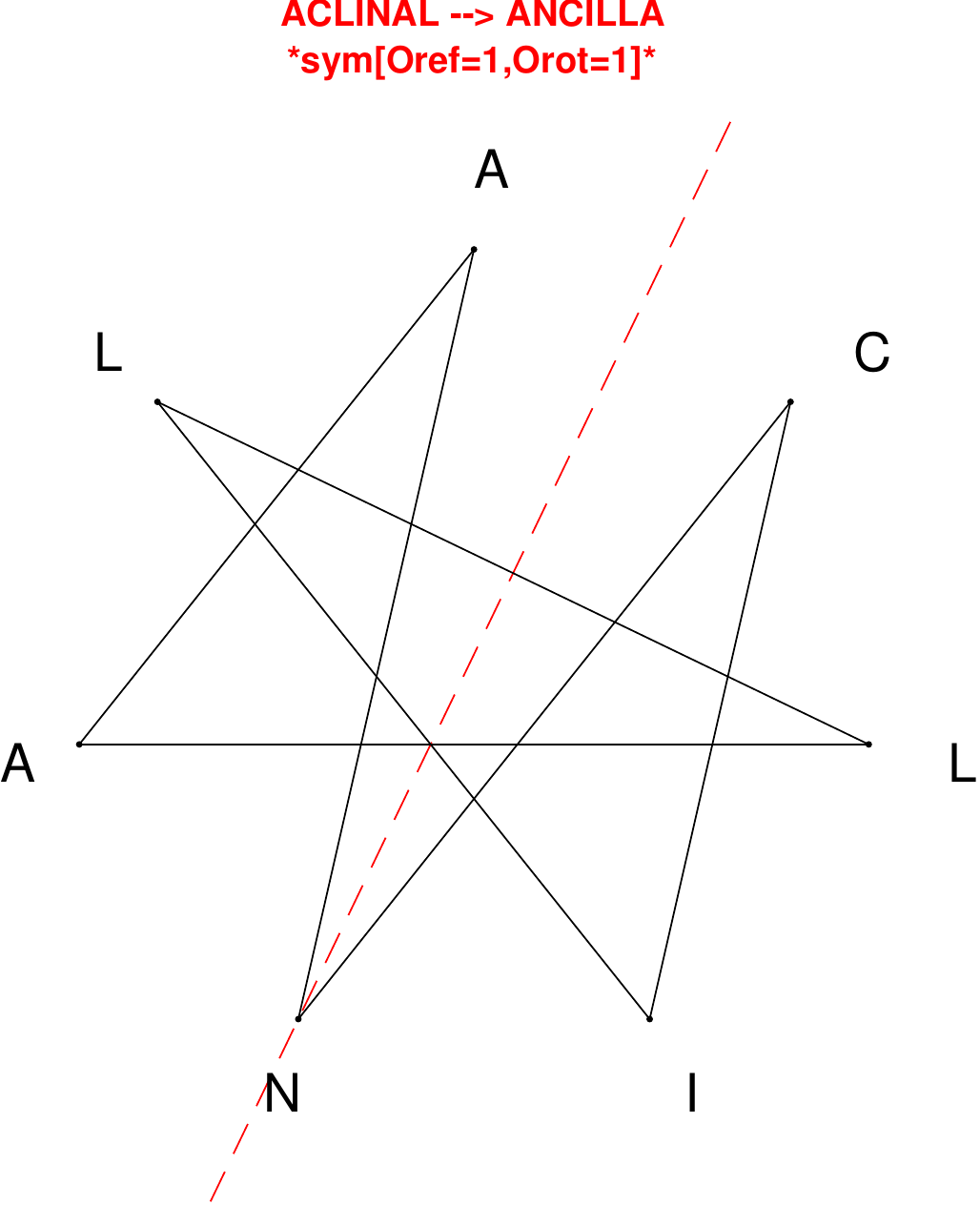}
\end{subfigure}
\hfill
\begin{subfigure}[T]{0.19\textwidth}
\centering
\includegraphics[width=\textwidth]{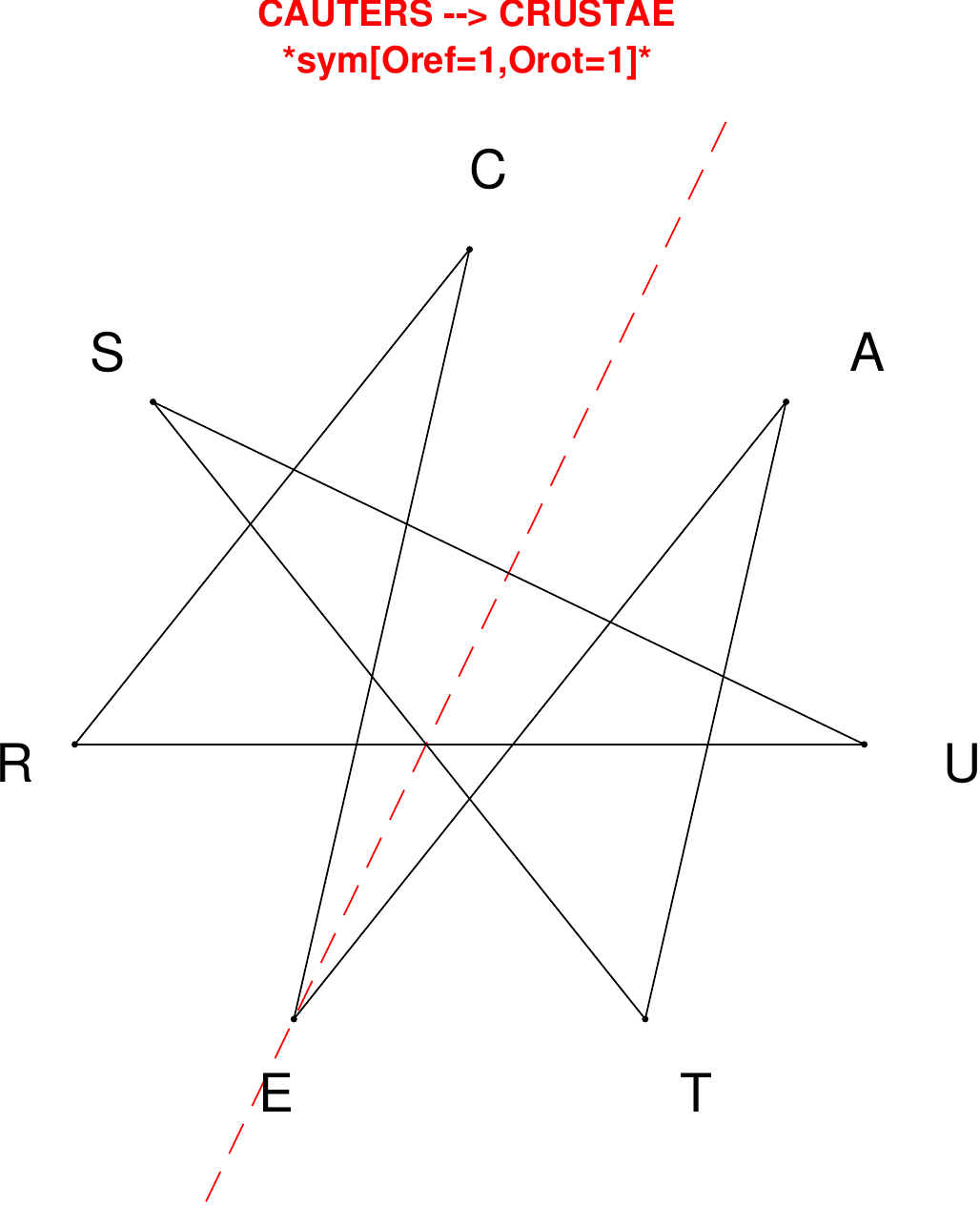}
\end{subfigure}
\hfill
\begin{subfigure}[T]{0.19\textwidth}
\centering
\includegraphics[width=\textwidth]{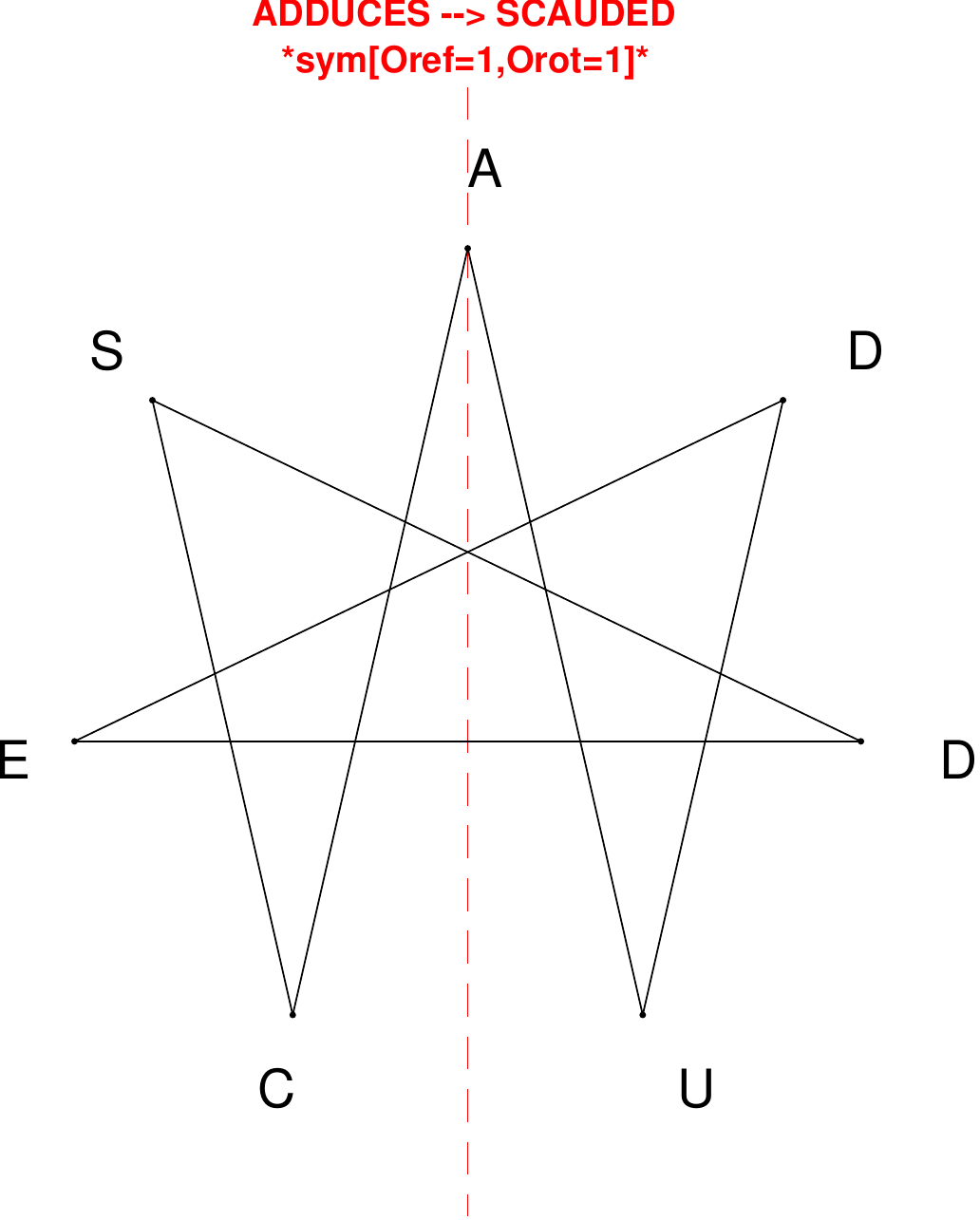}
\end{subfigure}
\hfill
\begin{subfigure}[T]{0.19\textwidth}
\centering
\includegraphics[width=\textwidth]{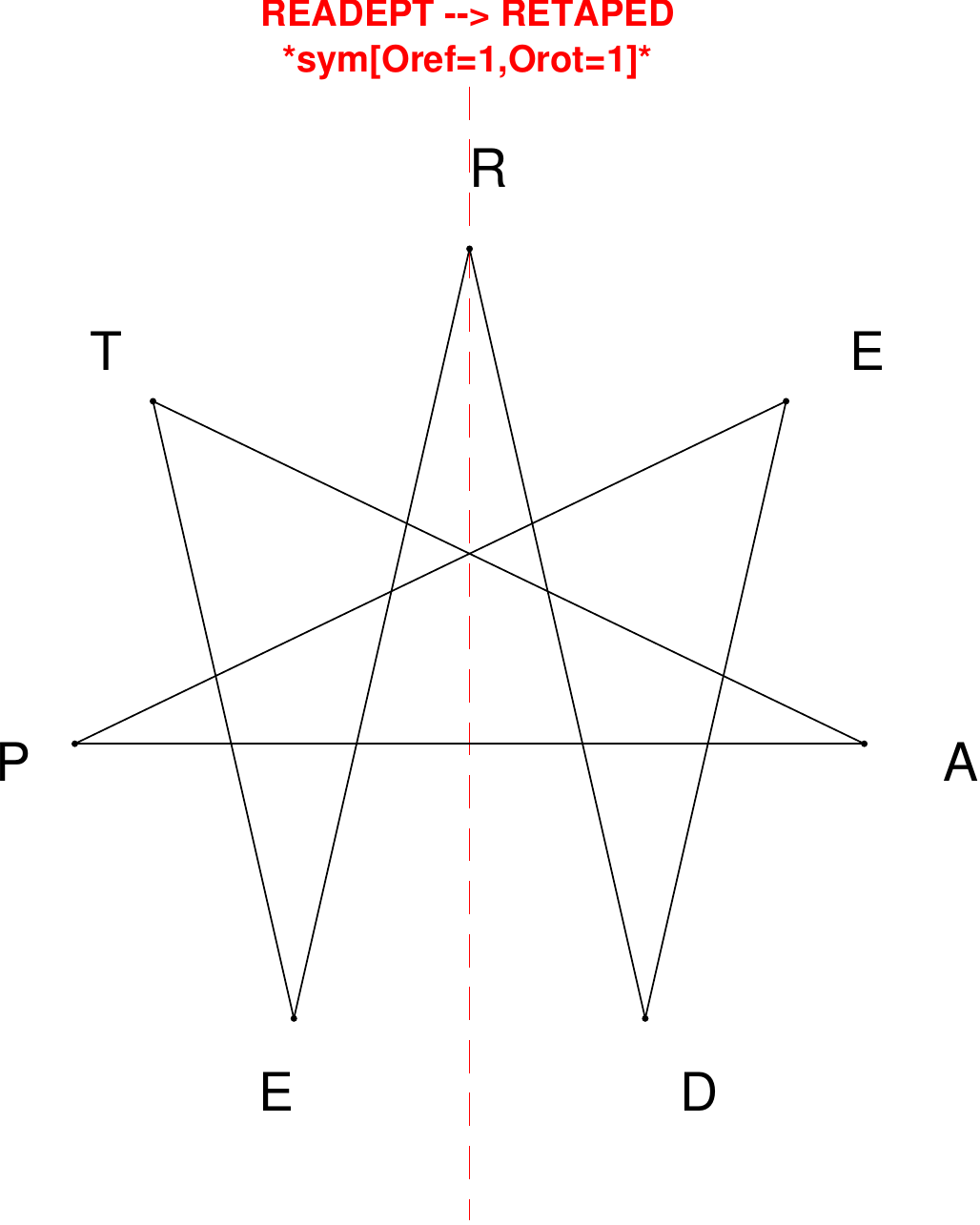}
\end{subfigure}
\hfill
\begin{subfigure}[T]{0.19\textwidth}
\centering
\includegraphics[width=\textwidth]{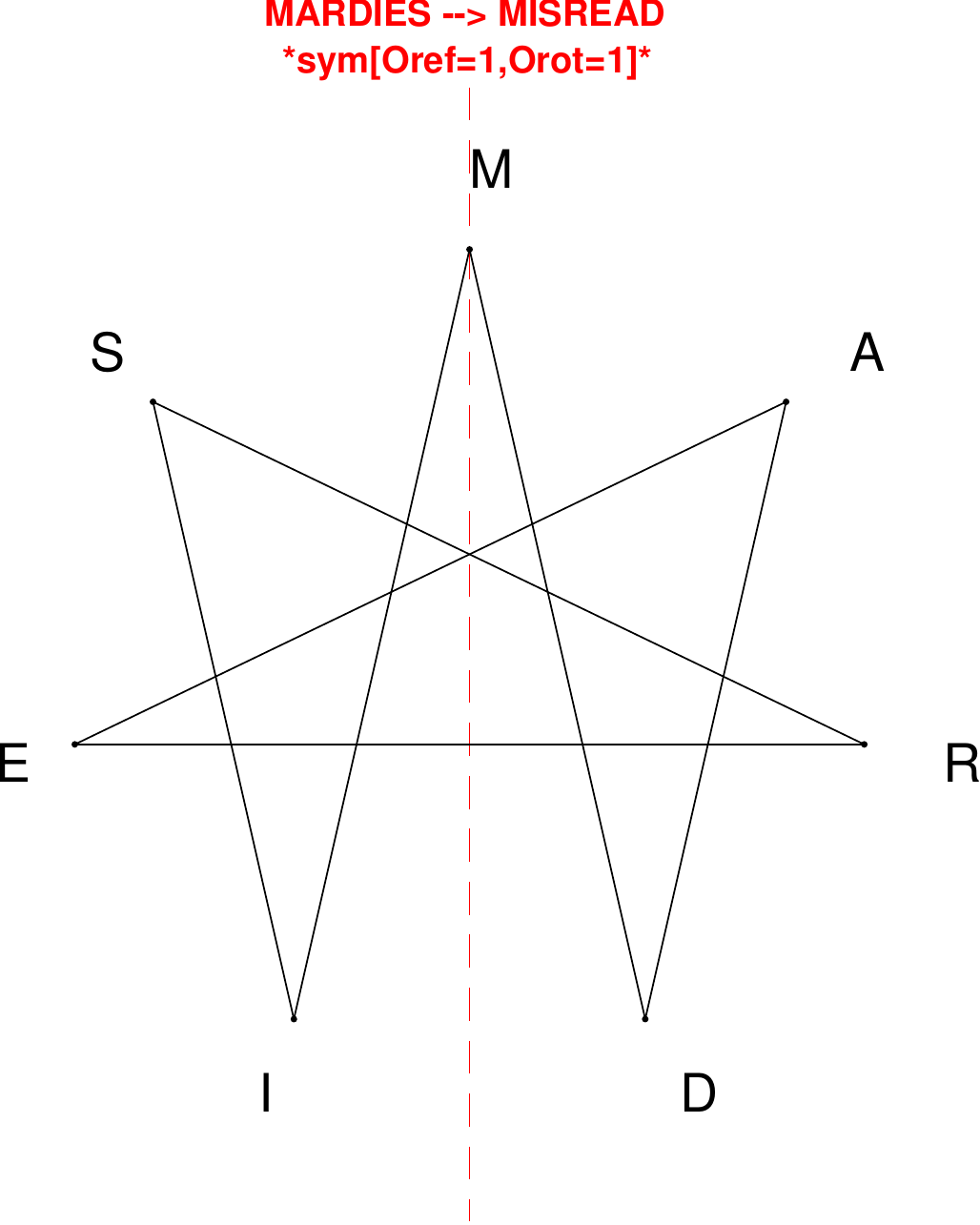}
\end{subfigure}
\end{figure}

\begin{figure}[H]
\centering
\begin{subfigure}[T]{0.19\textwidth}
\centering
\includegraphics[width=\textwidth]{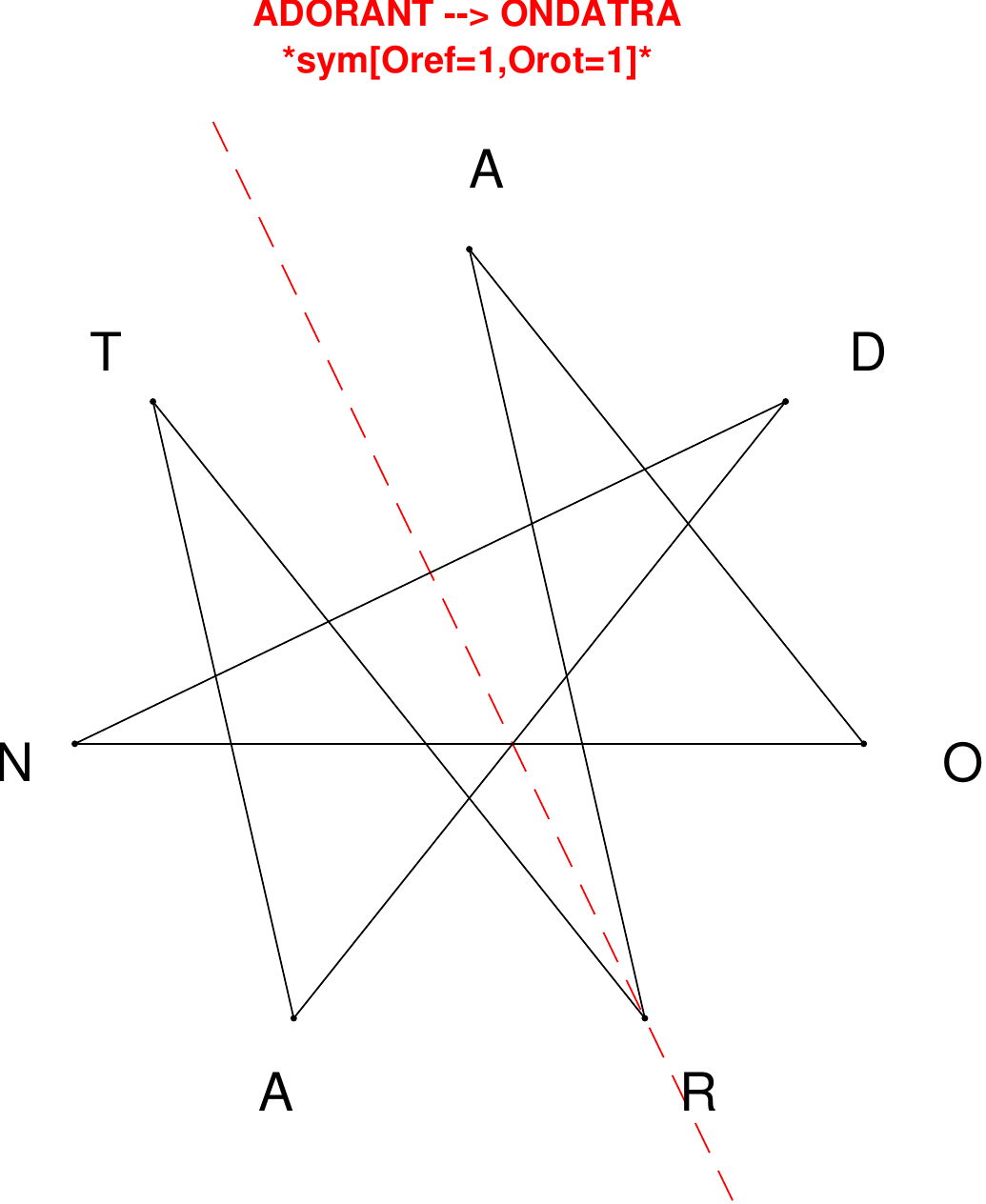}
\end{subfigure}
\hfill
\begin{subfigure}[T]{0.19\textwidth}
\centering
\includegraphics[width=\textwidth]{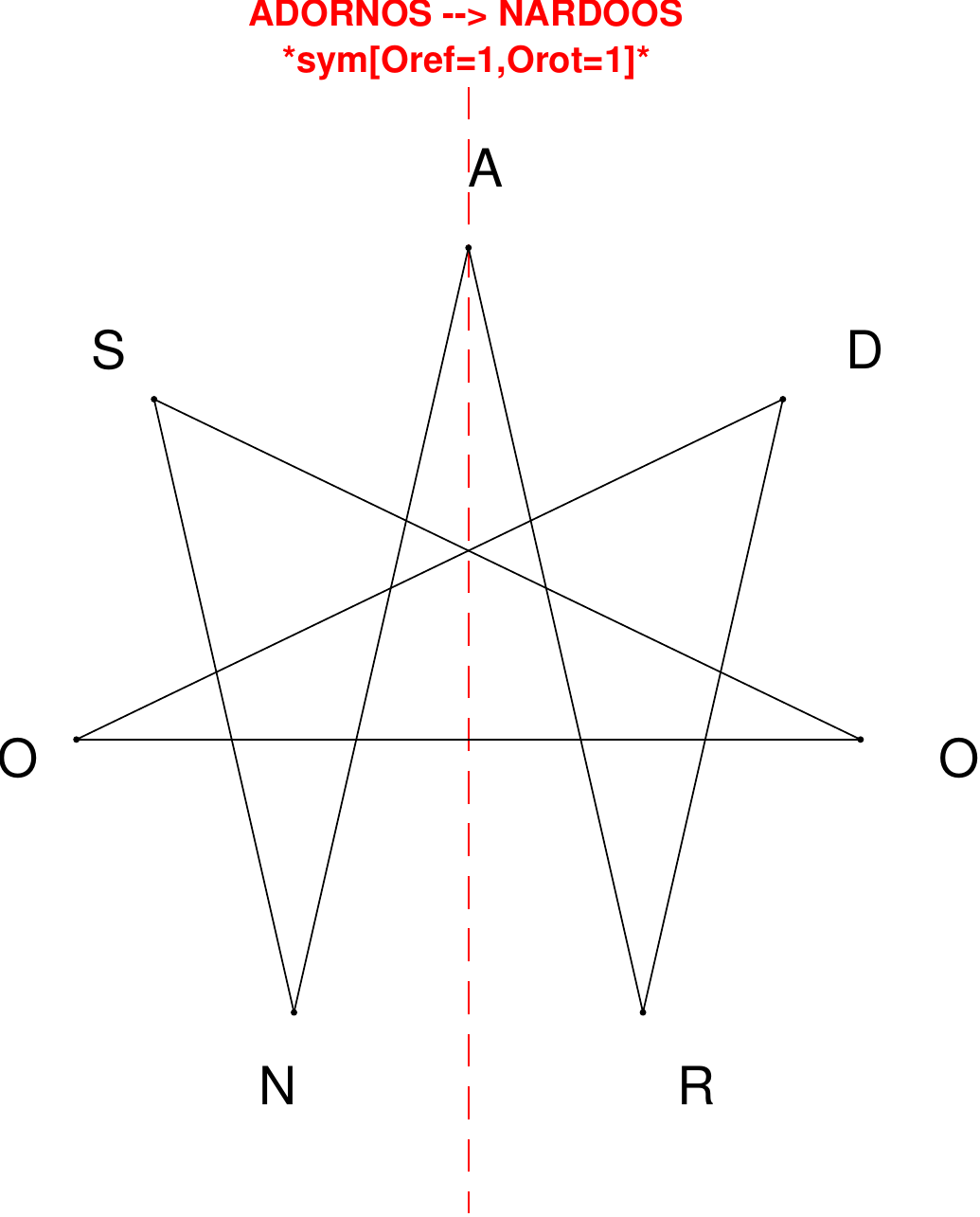}
\end{subfigure}
\hfill
\begin{subfigure}[T]{0.19\textwidth}
\centering
\includegraphics[width=\textwidth]{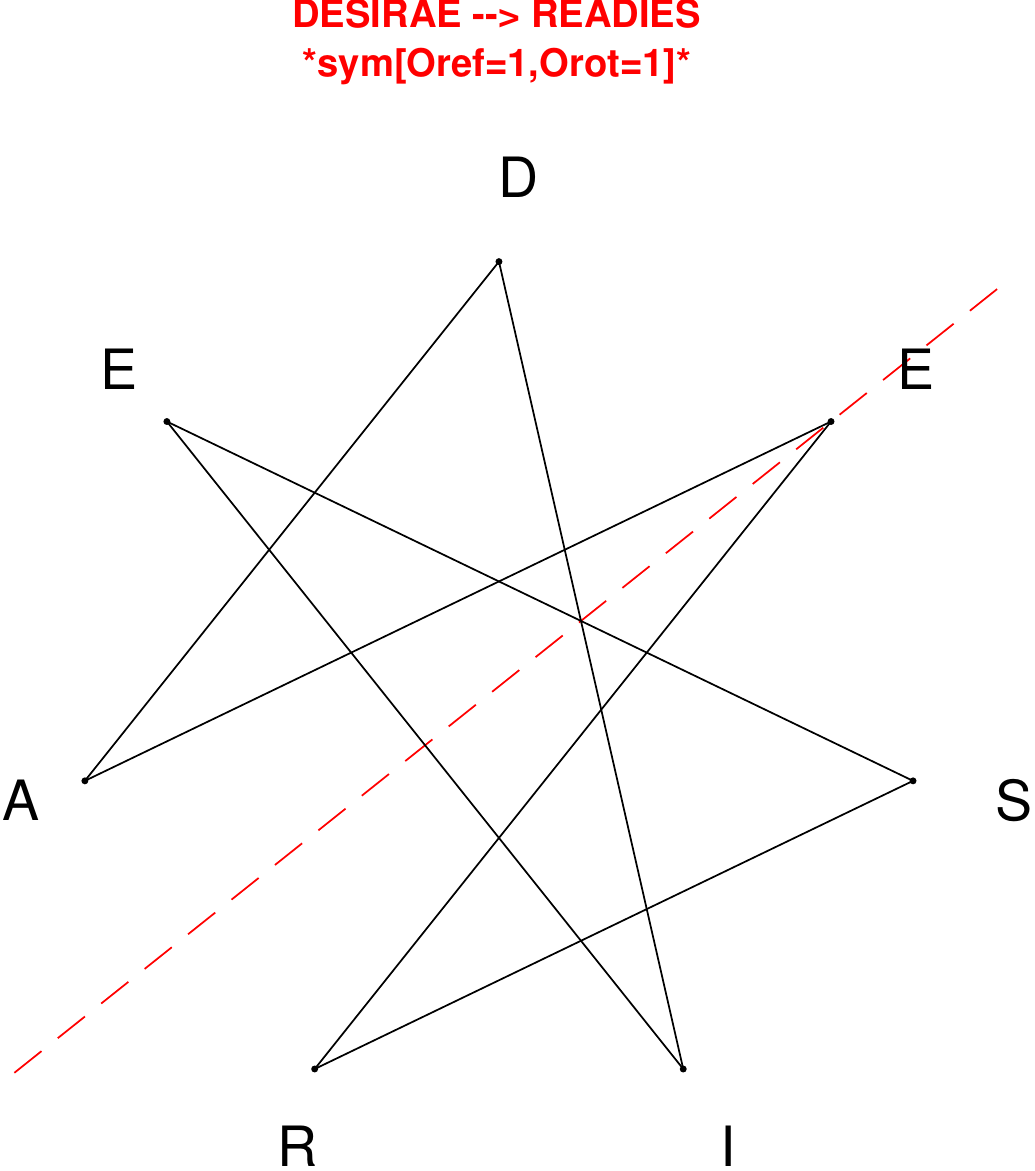}
\end{subfigure}
\hfill
\begin{subfigure}[T]{0.19\textwidth}
\centering
\includegraphics[width=\textwidth]{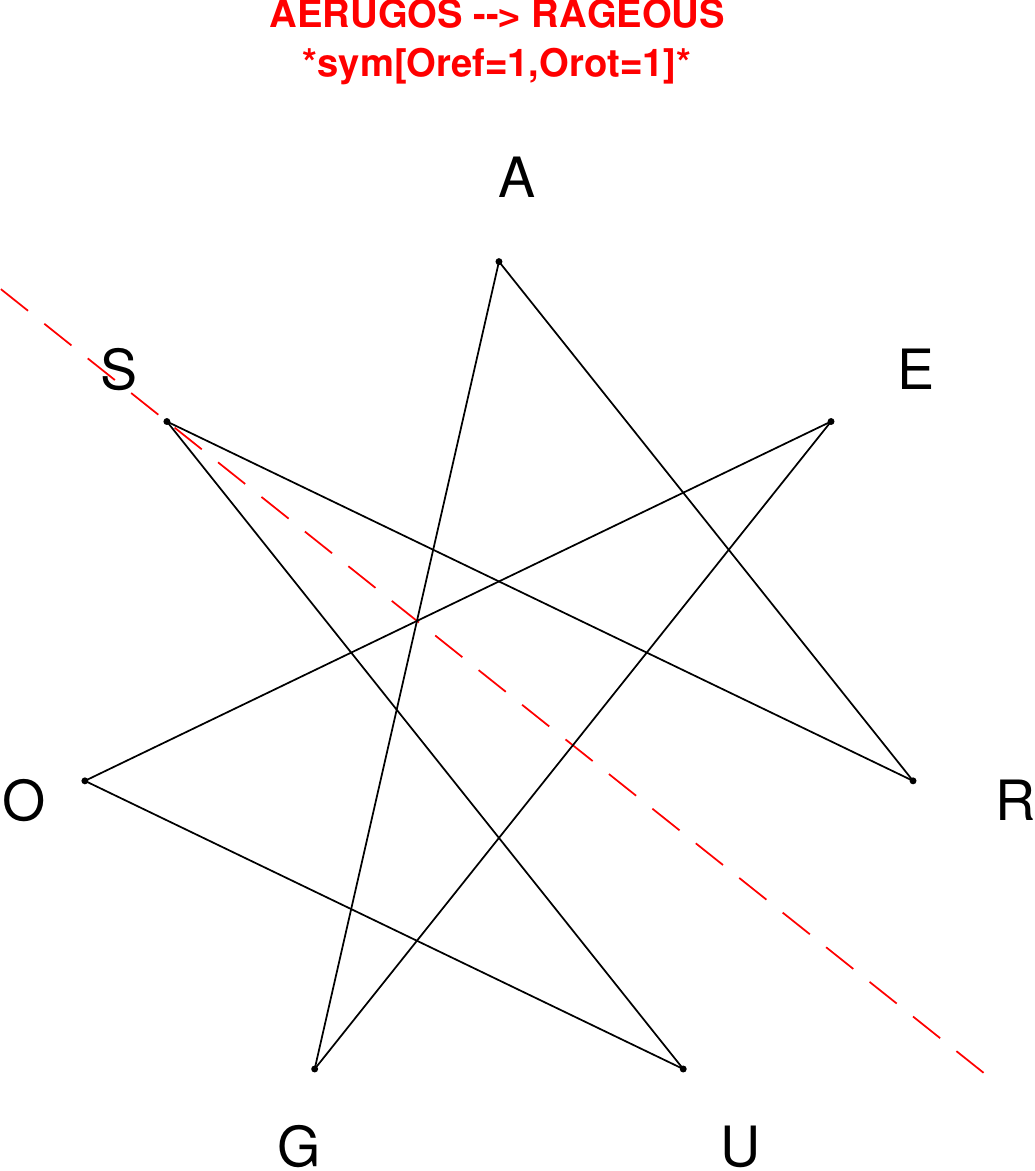}
\end{subfigure}
\hfill
\begin{subfigure}[T]{0.19\textwidth}
\centering
\includegraphics[width=\textwidth]{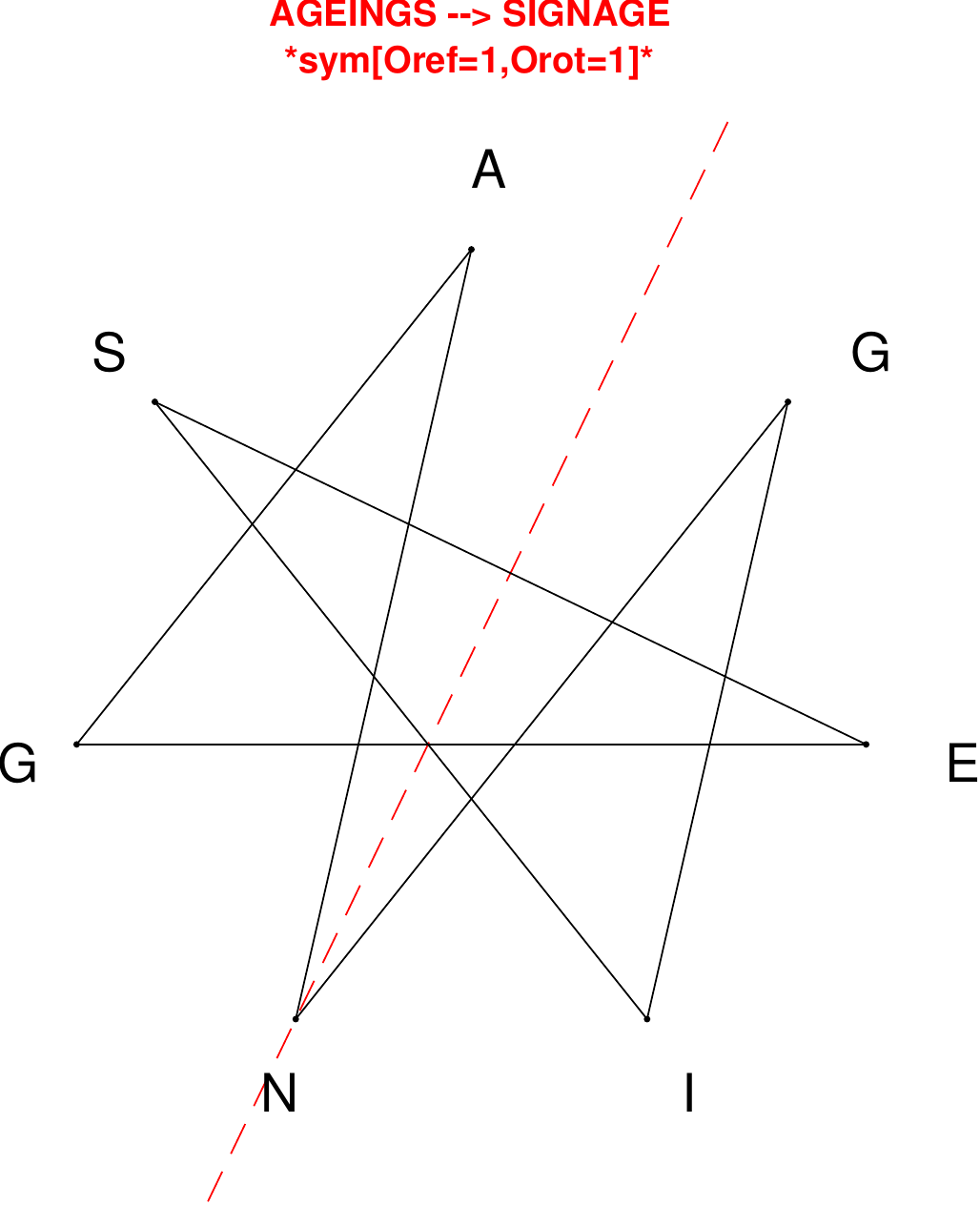}
\end{subfigure}
\end{figure}

\begin{figure}[H]
\centering
\begin{subfigure}[T]{0.19\textwidth}
\centering
\includegraphics[width=\textwidth]{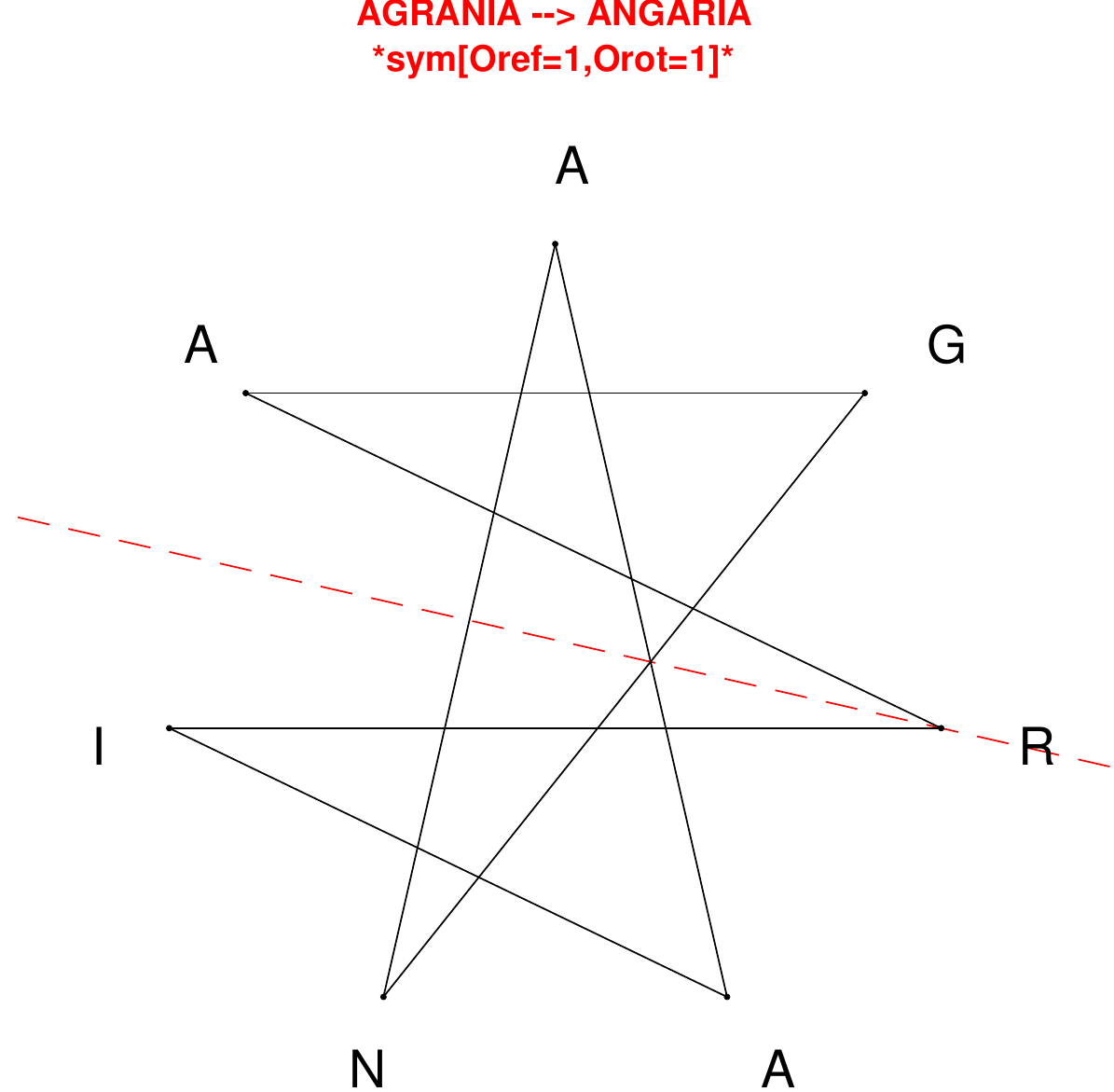}
\end{subfigure}
\hfill
\begin{subfigure}[T]{0.19\textwidth}
\centering
\includegraphics[width=\textwidth]{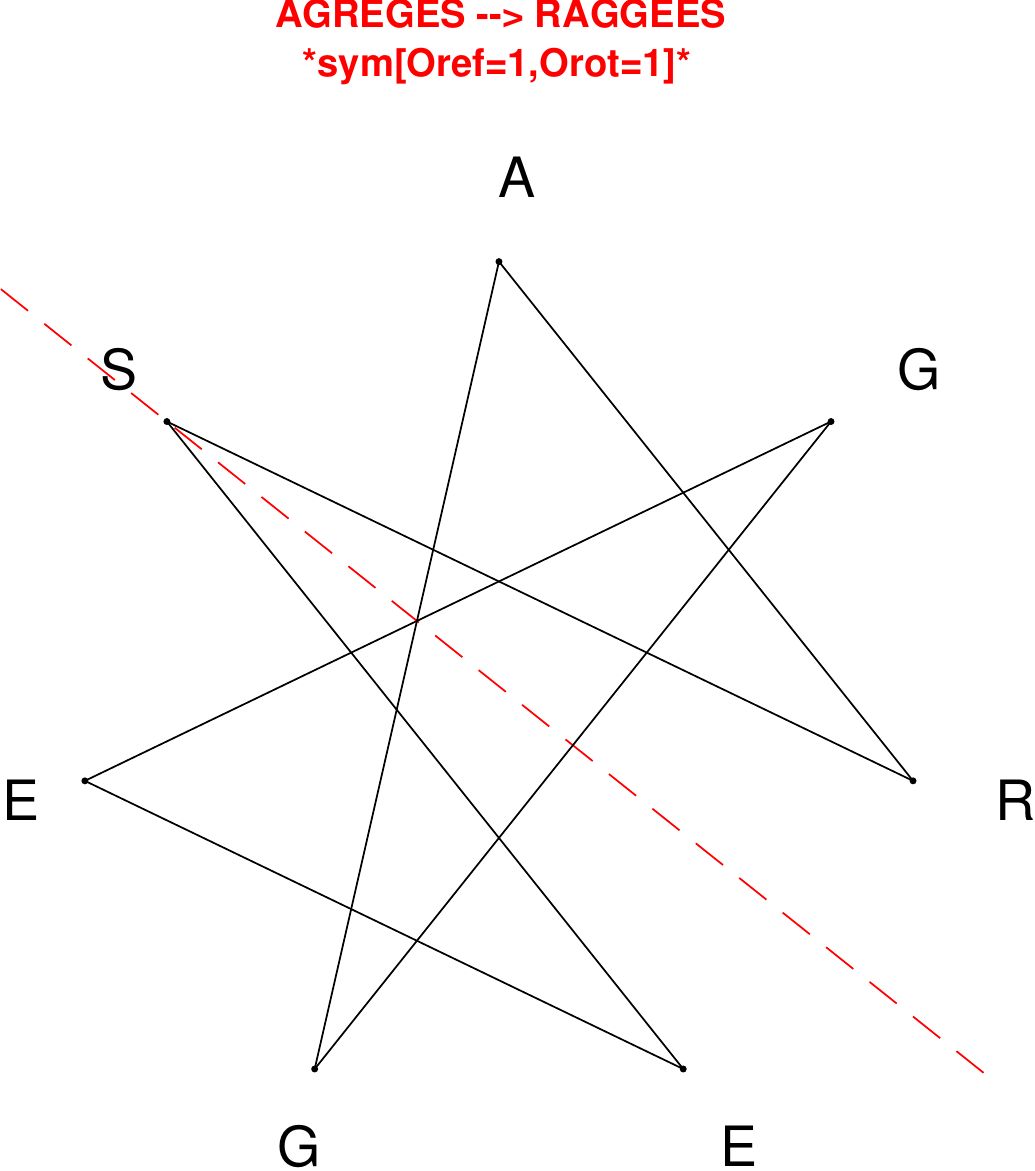}
\end{subfigure}
\hfill
\begin{subfigure}[T]{0.19\textwidth}
\centering
\includegraphics[width=\textwidth]{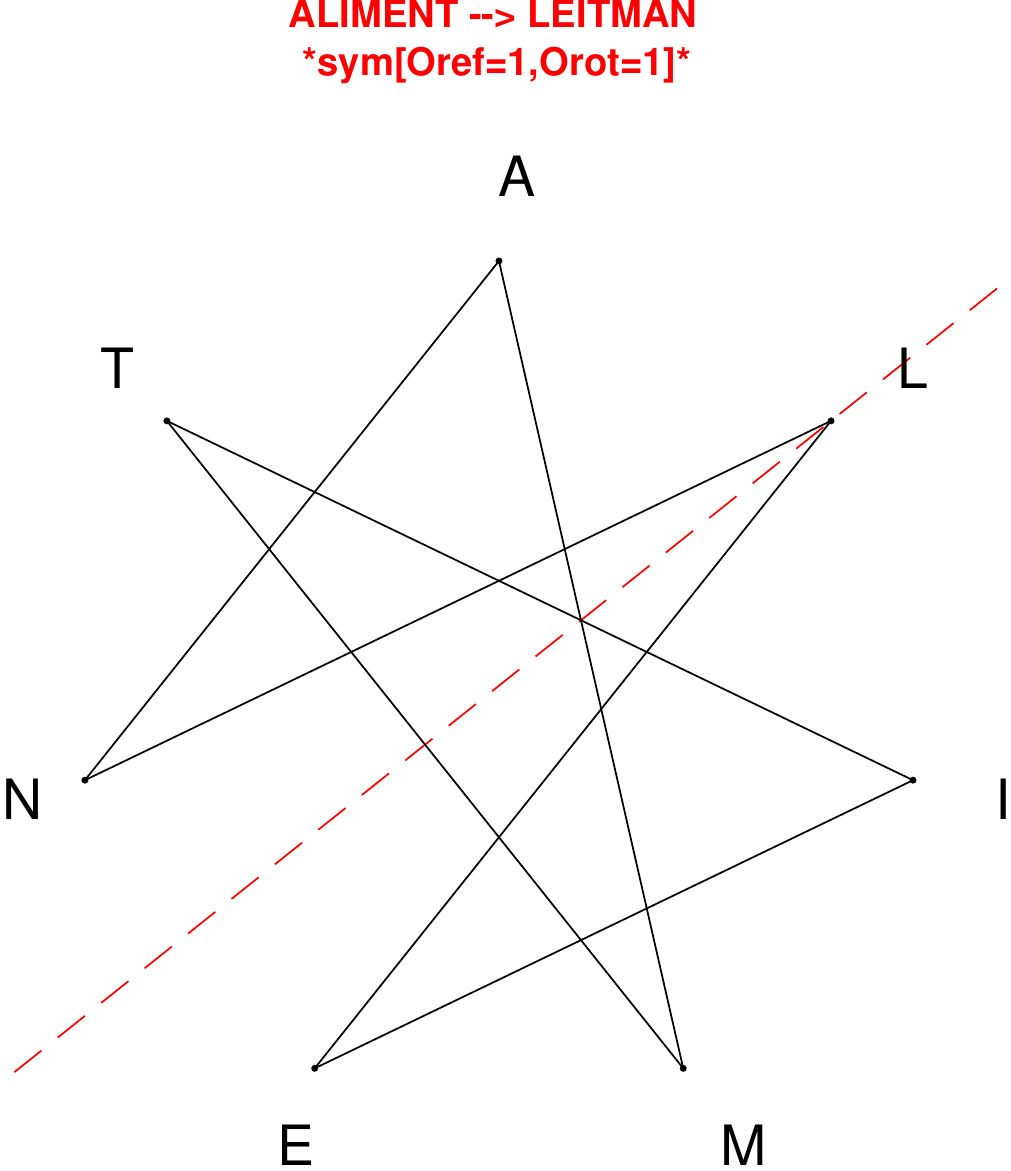}
\end{subfigure}
\hfill
\begin{subfigure}[T]{0.19\textwidth}
\centering
\includegraphics[width=\textwidth]{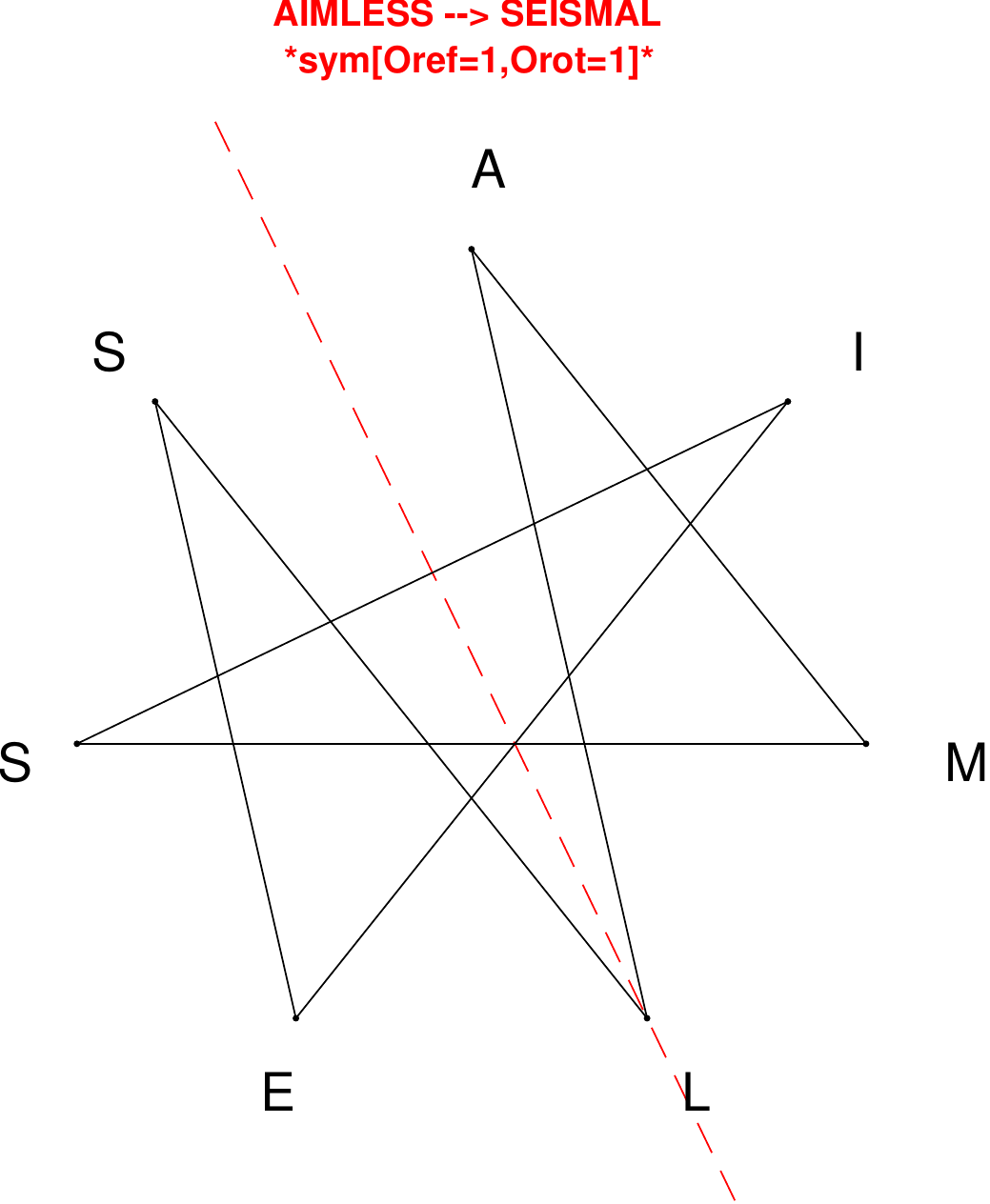}
\end{subfigure}
\hfill
\begin{subfigure}[T]{0.19\textwidth}
\centering
\includegraphics[width=\textwidth]{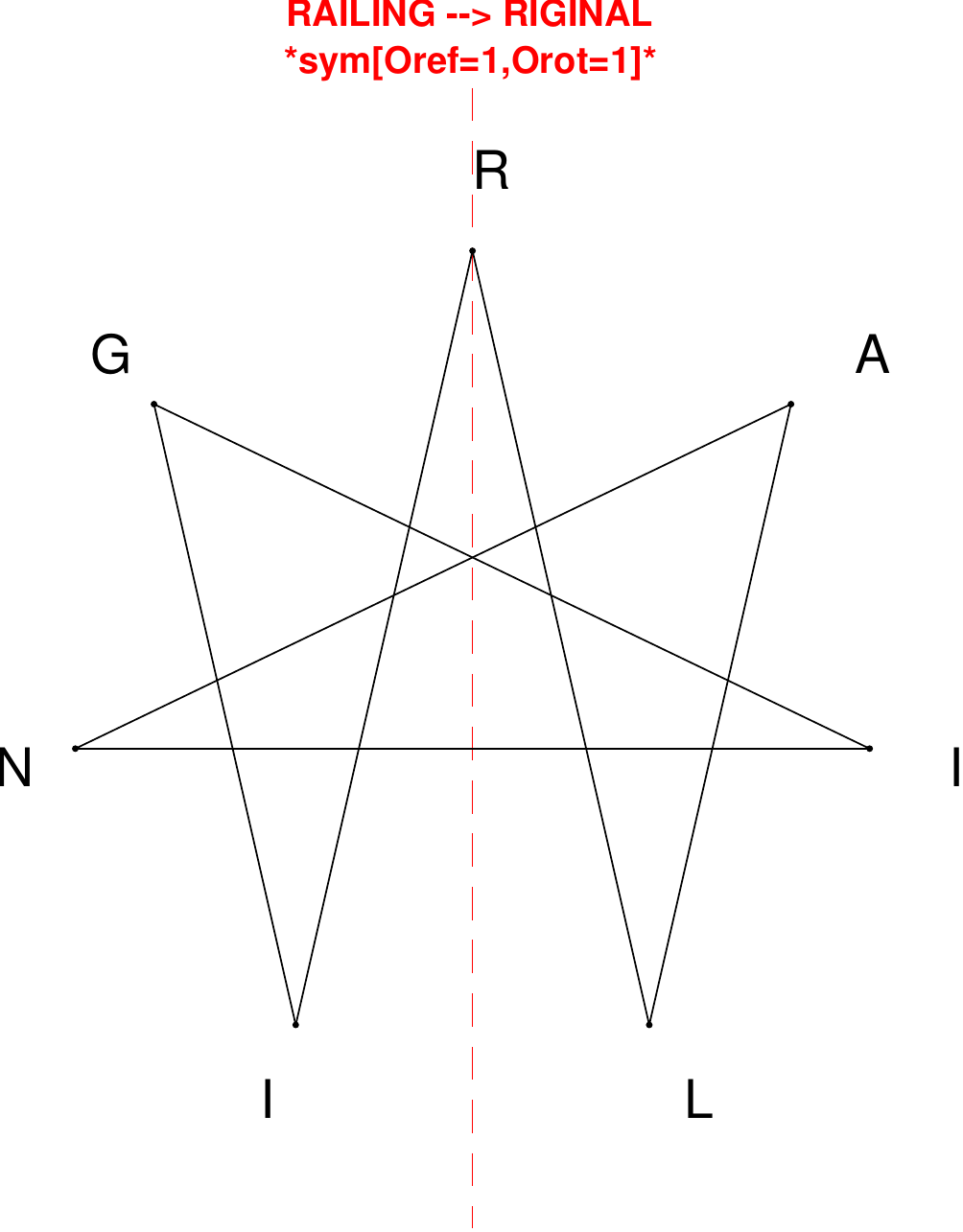}
\end{subfigure}
\end{figure}

\begin{figure}[H]
\centering
\begin{subfigure}[T]{0.19\textwidth}
\centering
\includegraphics[width=\textwidth]{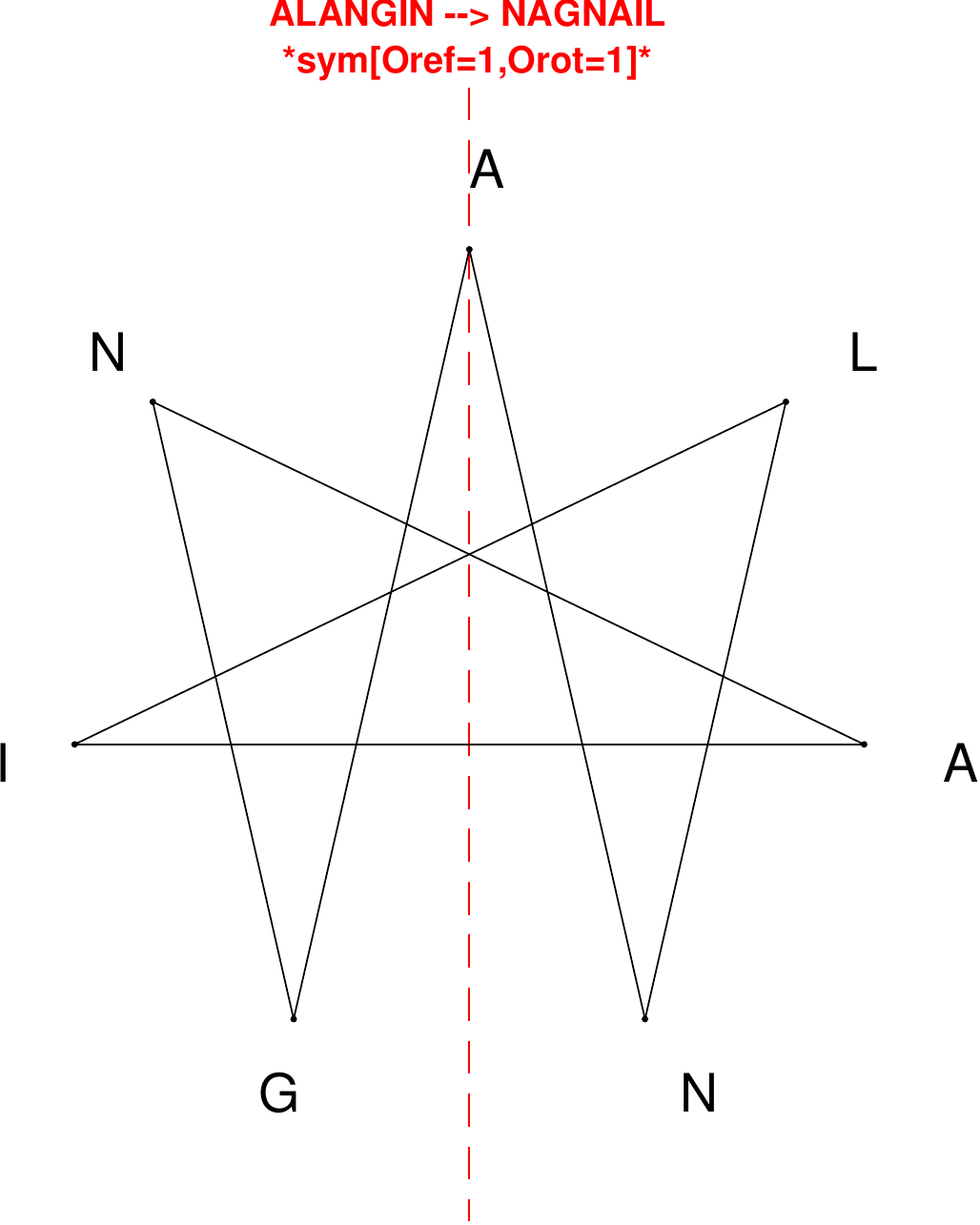}
\end{subfigure}
\hfill
\begin{subfigure}[T]{0.19\textwidth}
\centering
\includegraphics[width=\textwidth]{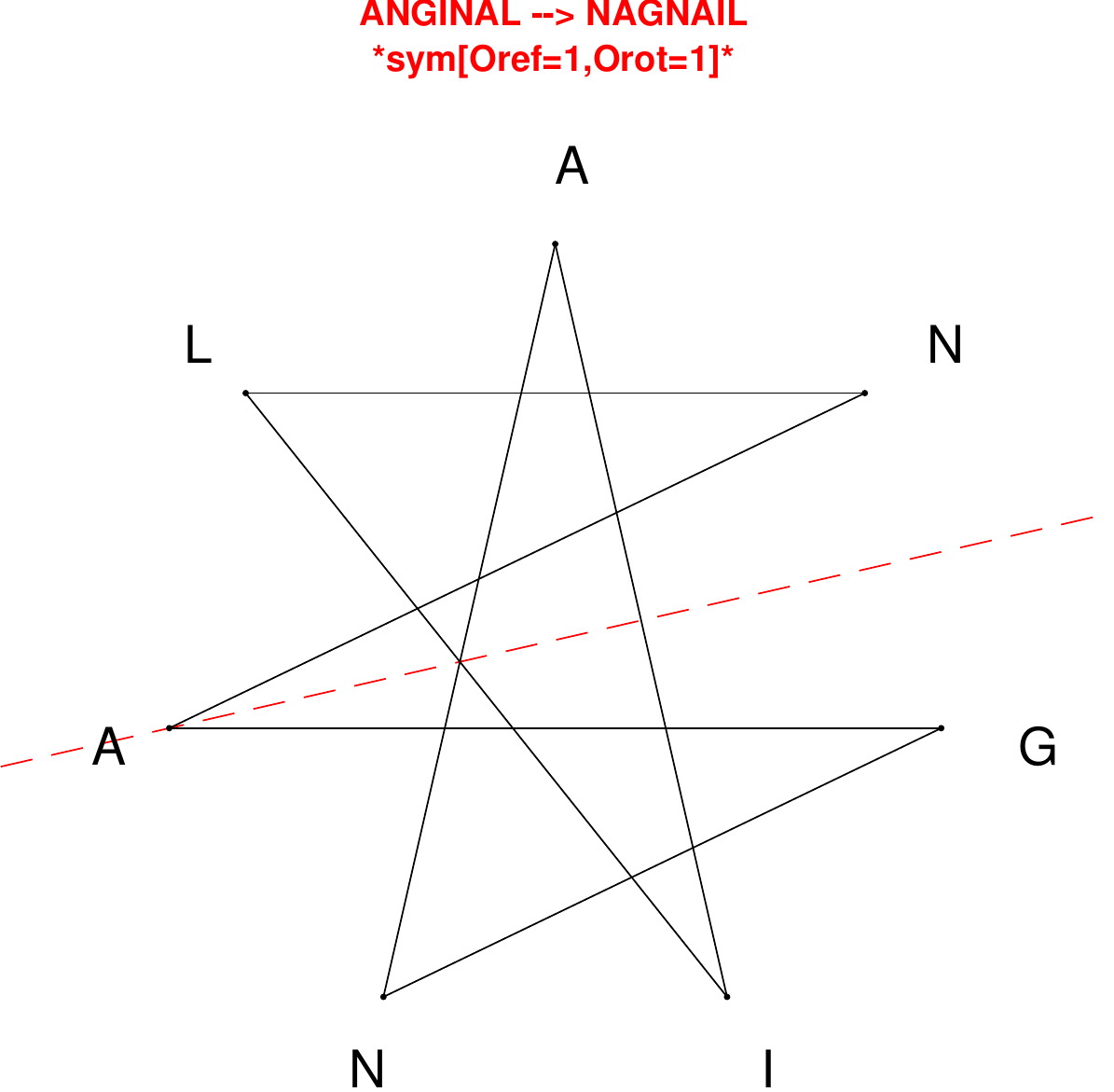}
\end{subfigure}
\hfill
\begin{subfigure}[T]{0.19\textwidth}
\centering
\includegraphics[width=\textwidth]{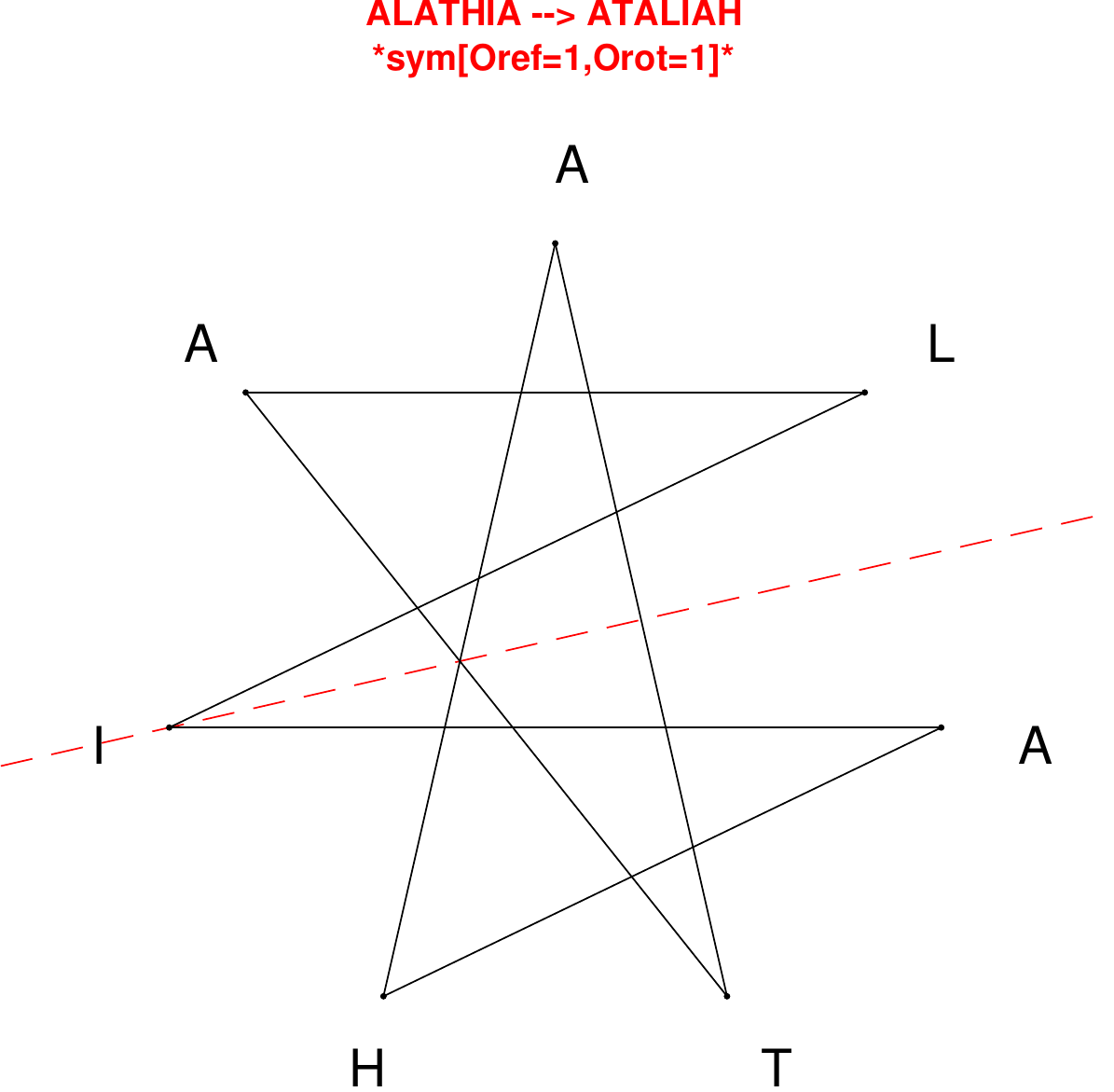}
\end{subfigure}
\hfill
\begin{subfigure}[T]{0.19\textwidth}
\centering
\includegraphics[width=\textwidth]{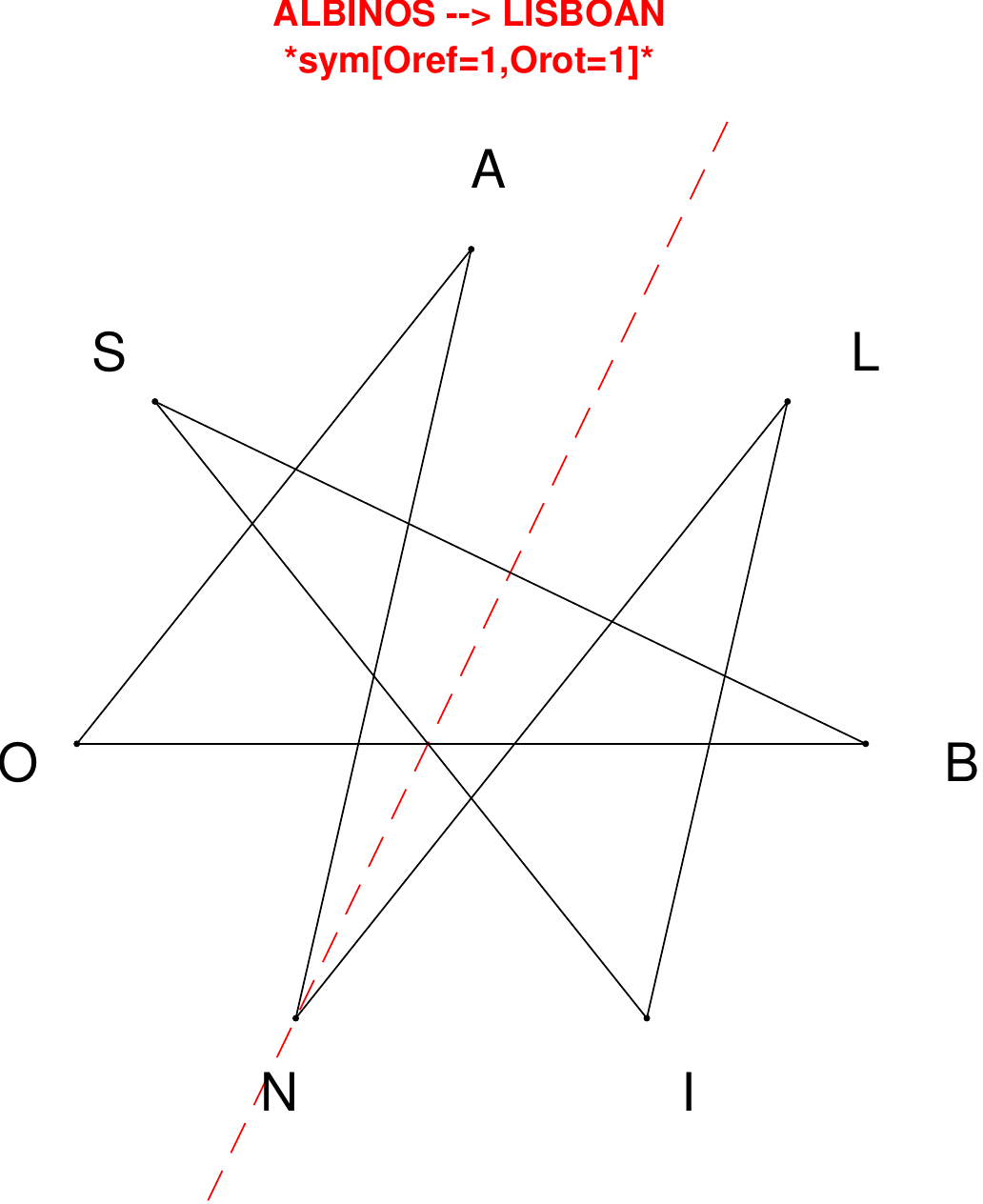}
\end{subfigure}
\hfill
\begin{subfigure}[T]{0.19\textwidth}
\centering
\includegraphics[width=\textwidth]{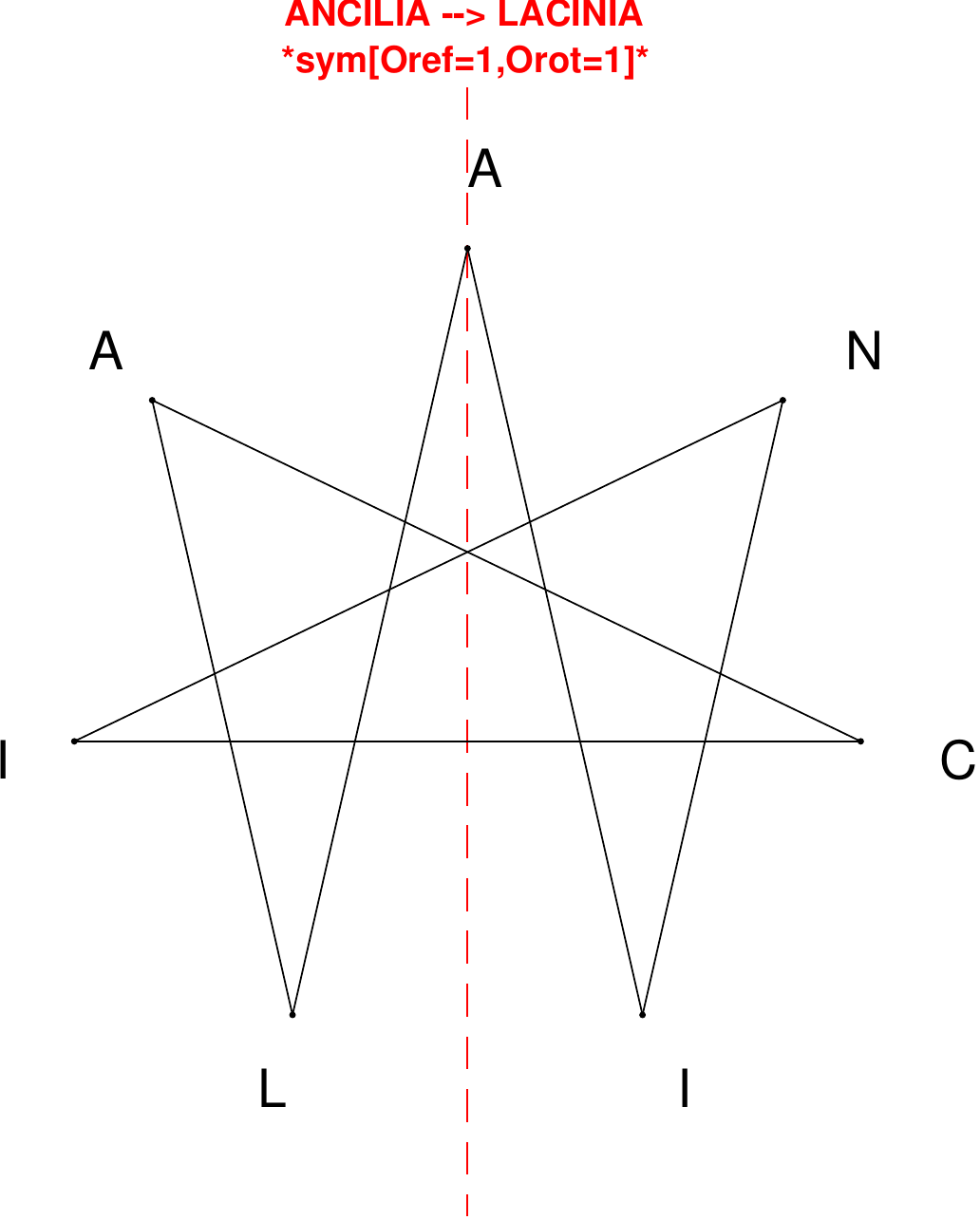}
\end{subfigure}
\end{figure}

\begin{figure}[H]
\centering
\begin{subfigure}[T]{0.19\textwidth}
\centering
\includegraphics[width=\textwidth]{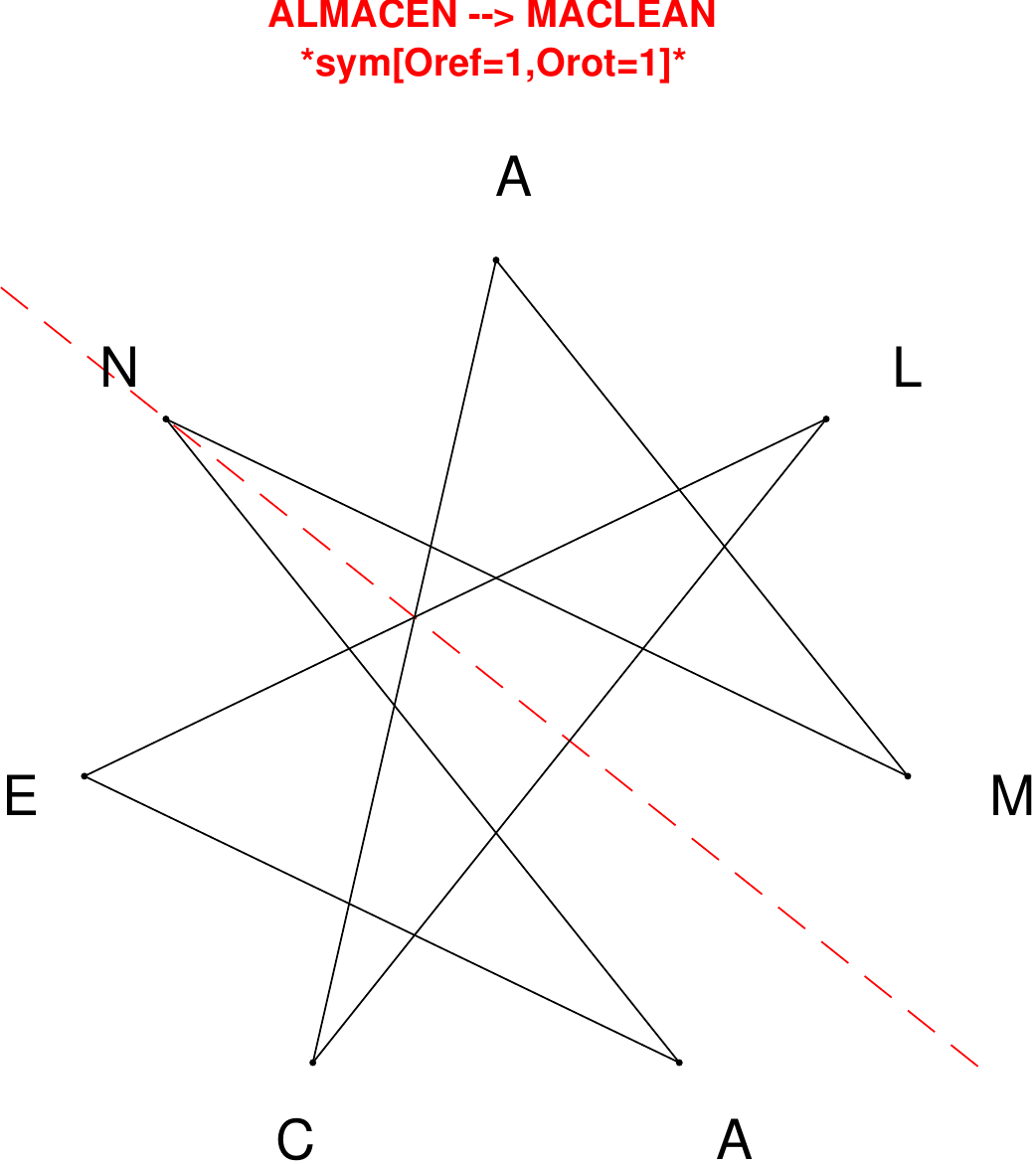}
\end{subfigure}
\hfill
\begin{subfigure}[T]{0.19\textwidth}
\centering
\includegraphics[width=\textwidth]{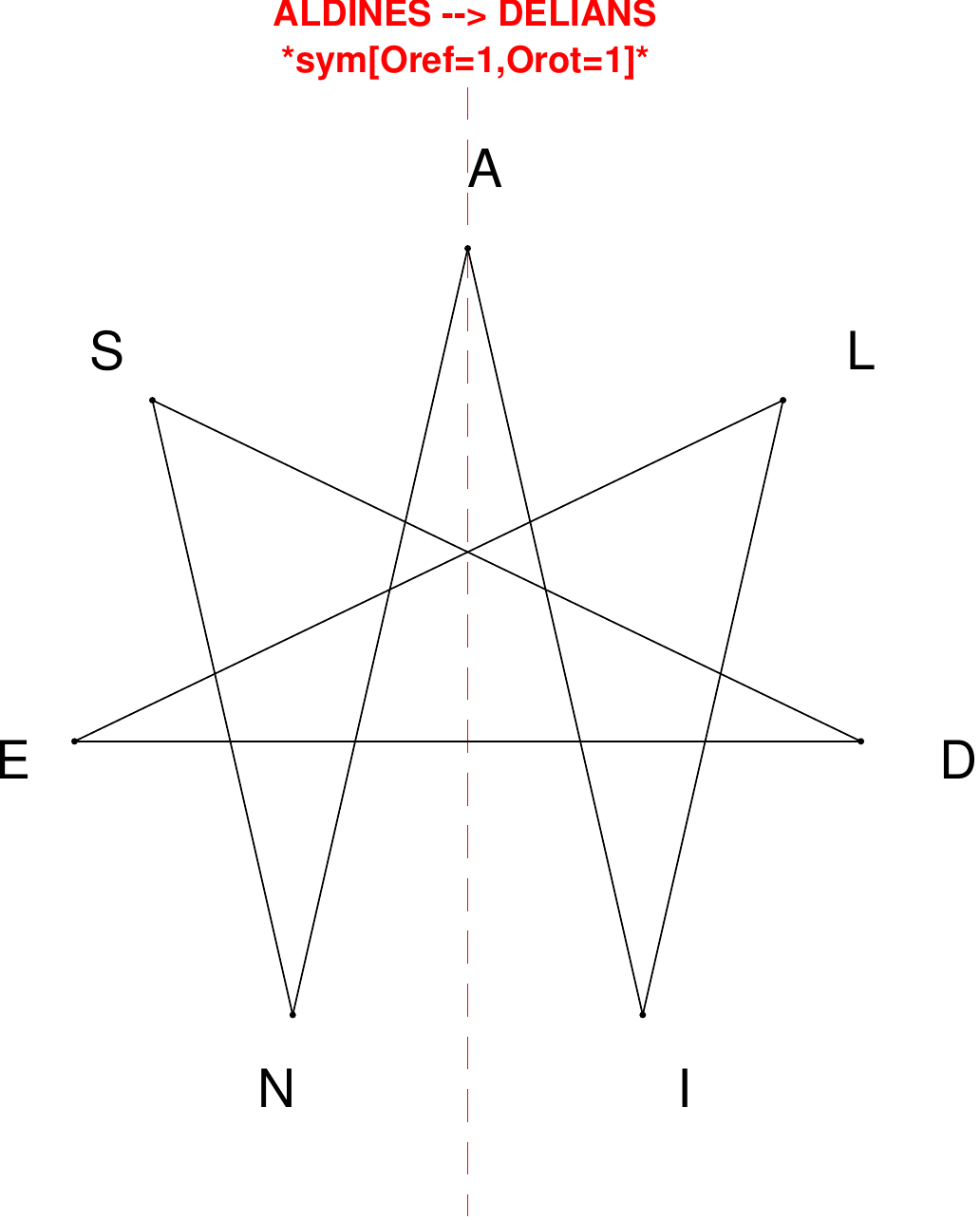}
\end{subfigure}
\hfill
\begin{subfigure}[T]{0.19\textwidth}
\centering
\includegraphics[width=\textwidth]{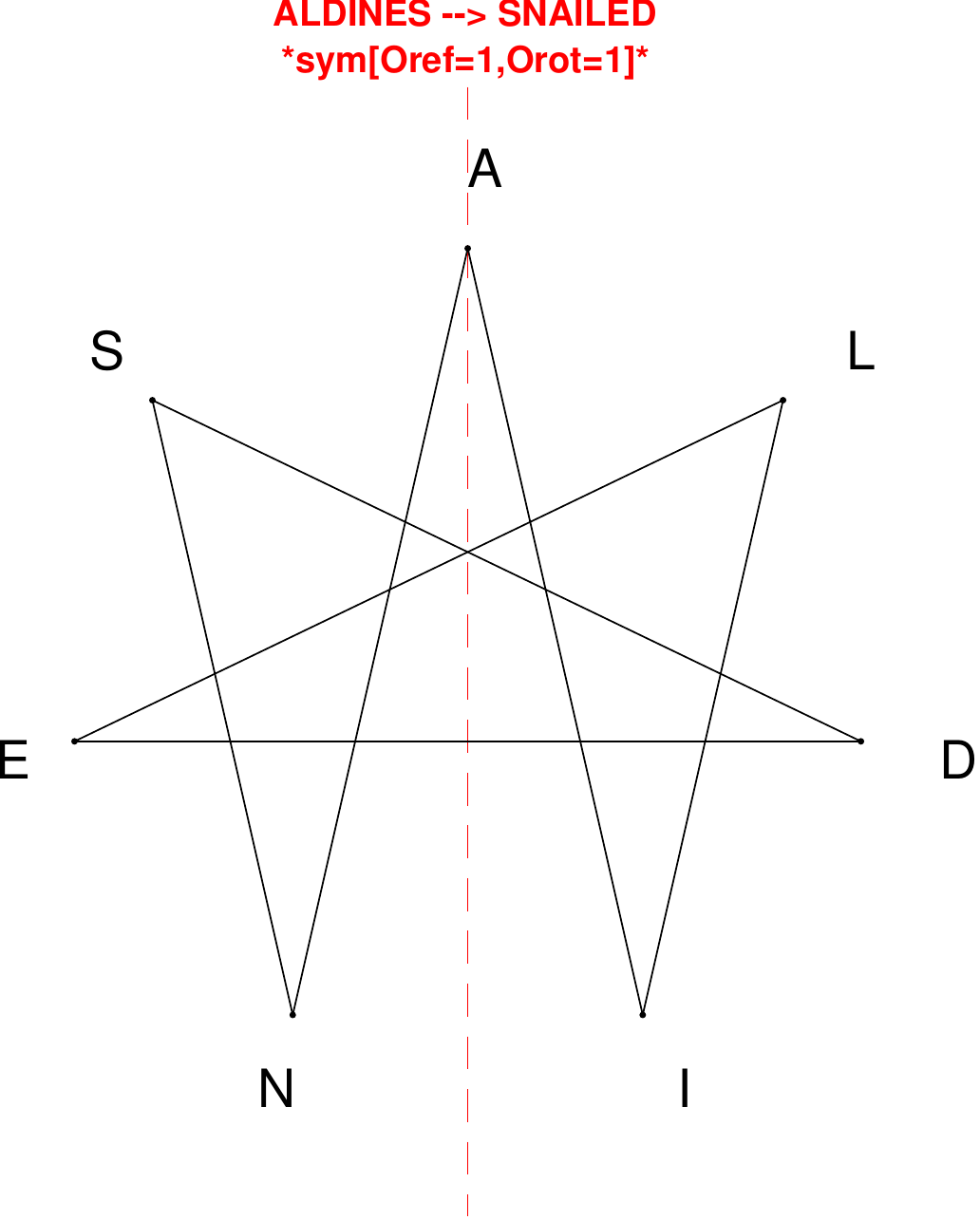}
\end{subfigure}
\hfill
\begin{subfigure}[T]{0.19\textwidth}
\centering
\includegraphics[width=\textwidth]{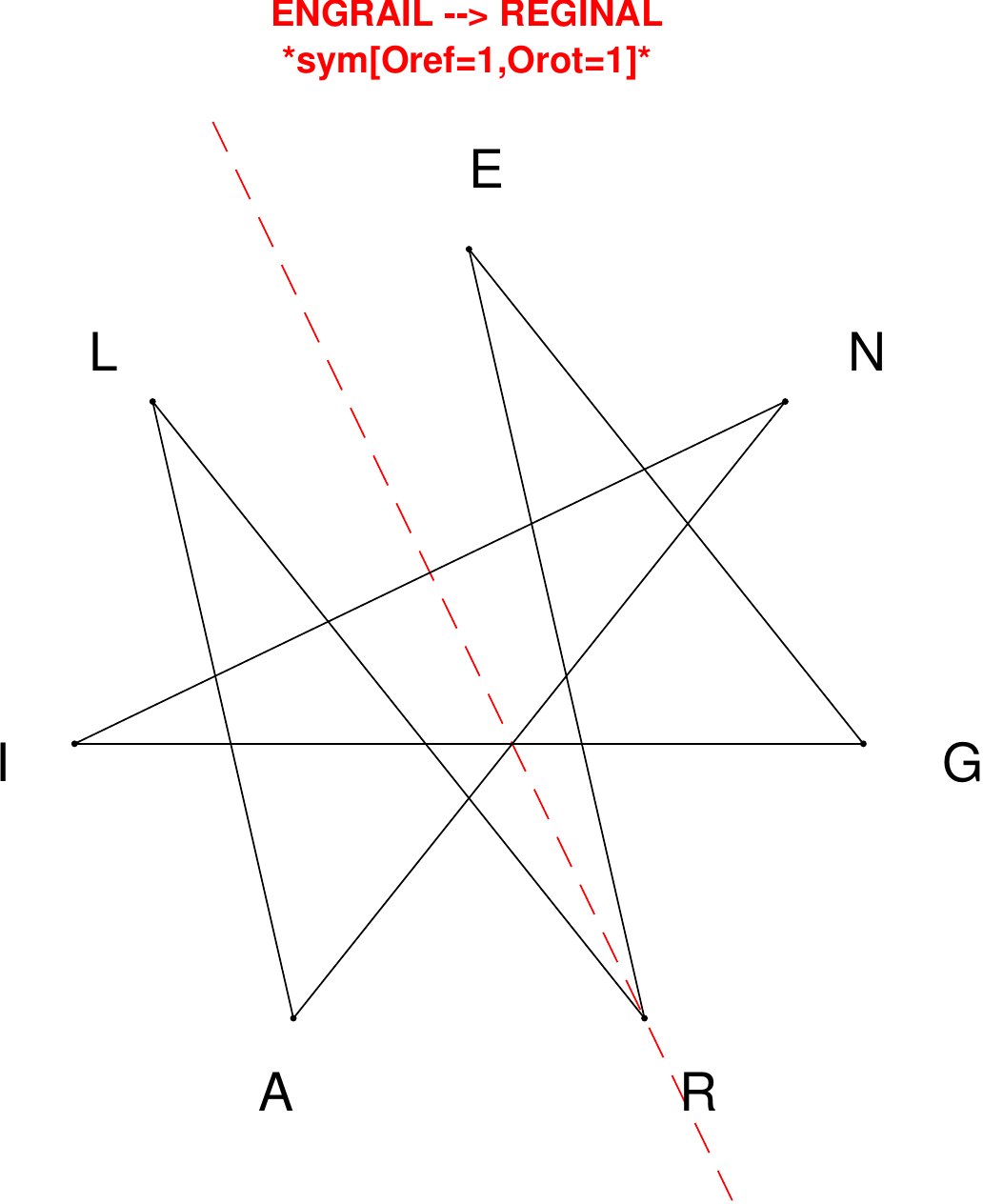}
\end{subfigure}
\hfill
\begin{subfigure}[T]{0.19\textwidth}
\centering
\includegraphics[width=\textwidth]{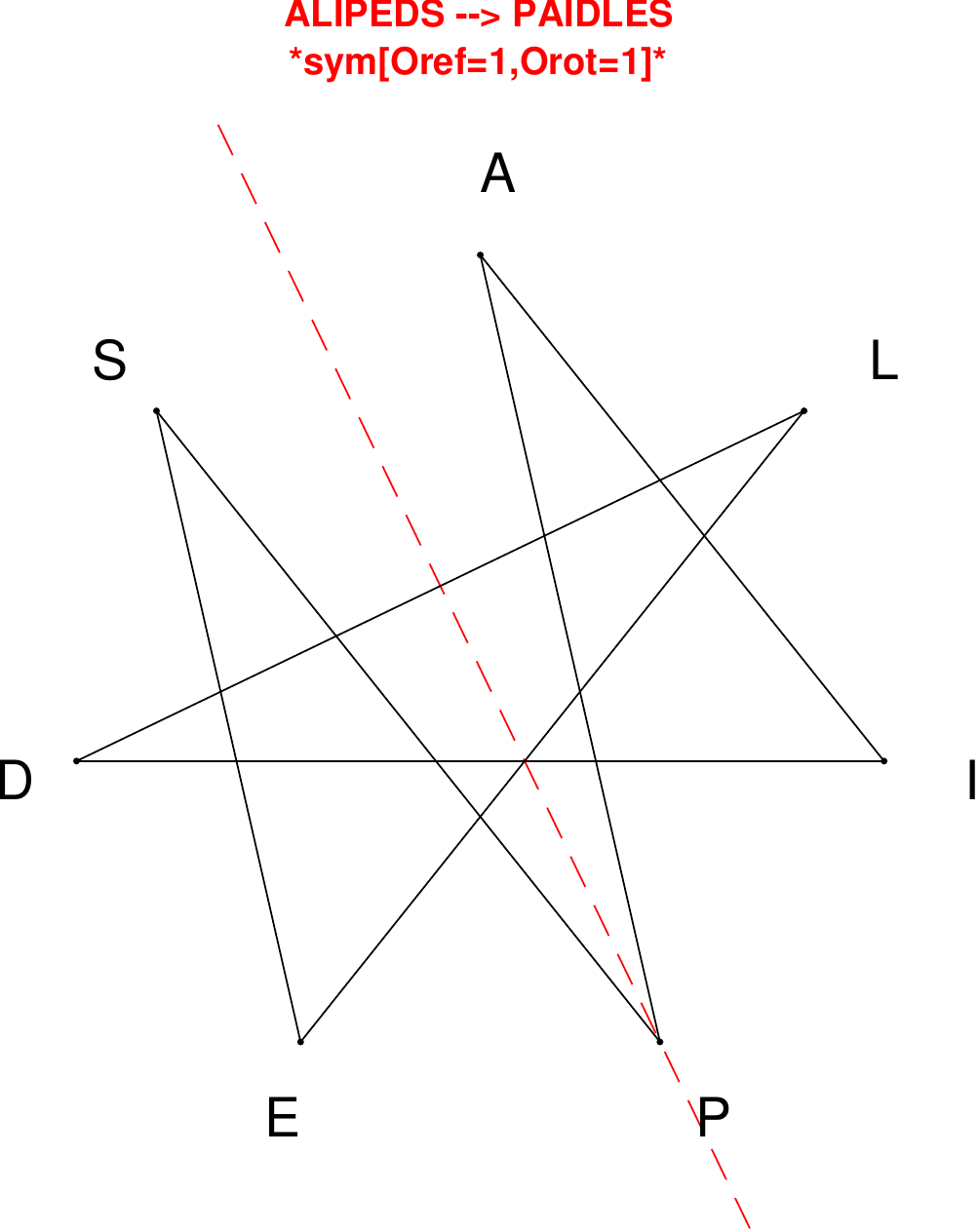}
\end{subfigure}
\end{figure}

\begin{figure}[H]
\centering
\begin{subfigure}[T]{0.19\textwidth}
\centering
\includegraphics[width=\textwidth]{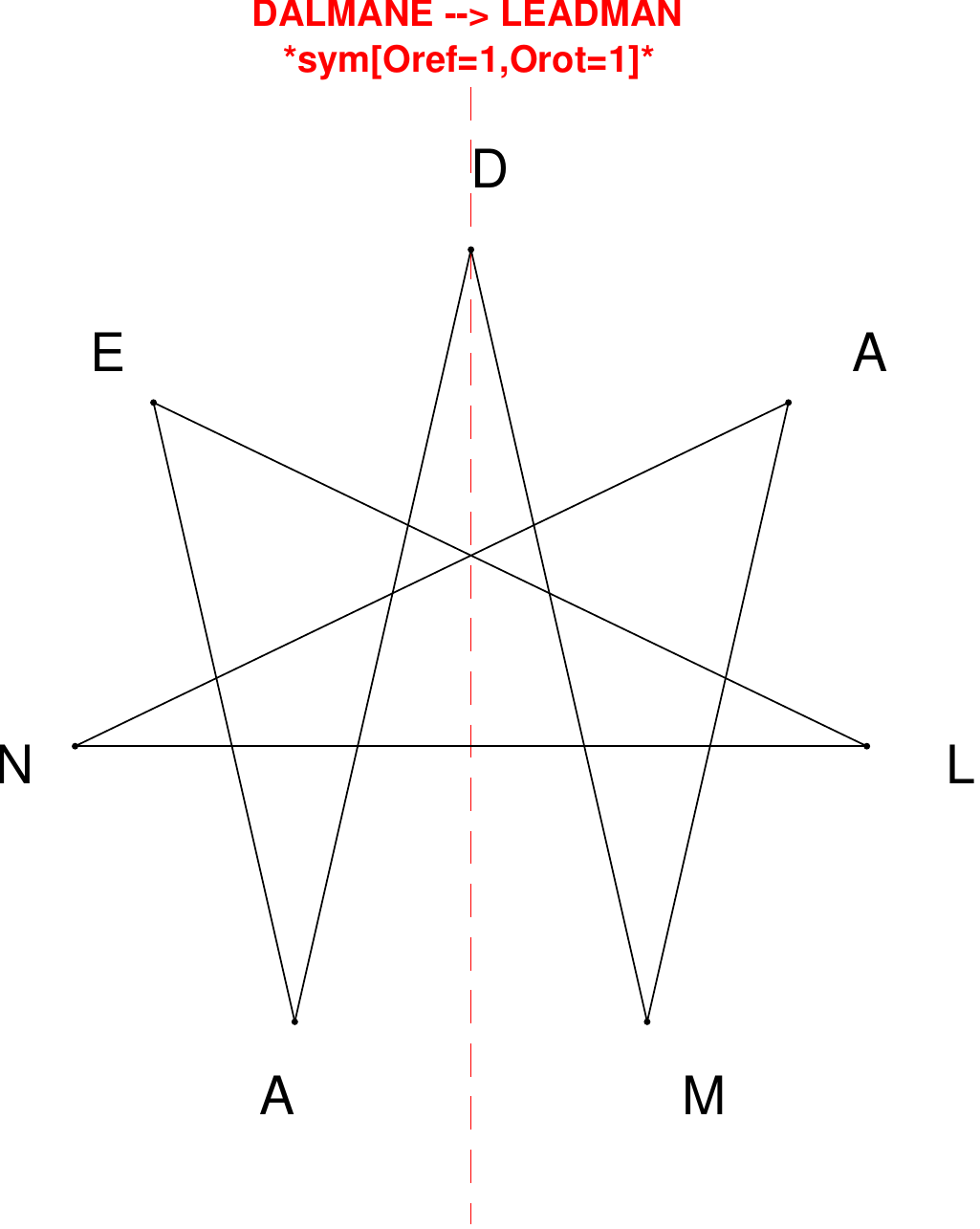}
\end{subfigure}
\hfill
\begin{subfigure}[T]{0.19\textwidth}
\centering
\includegraphics[width=\textwidth]{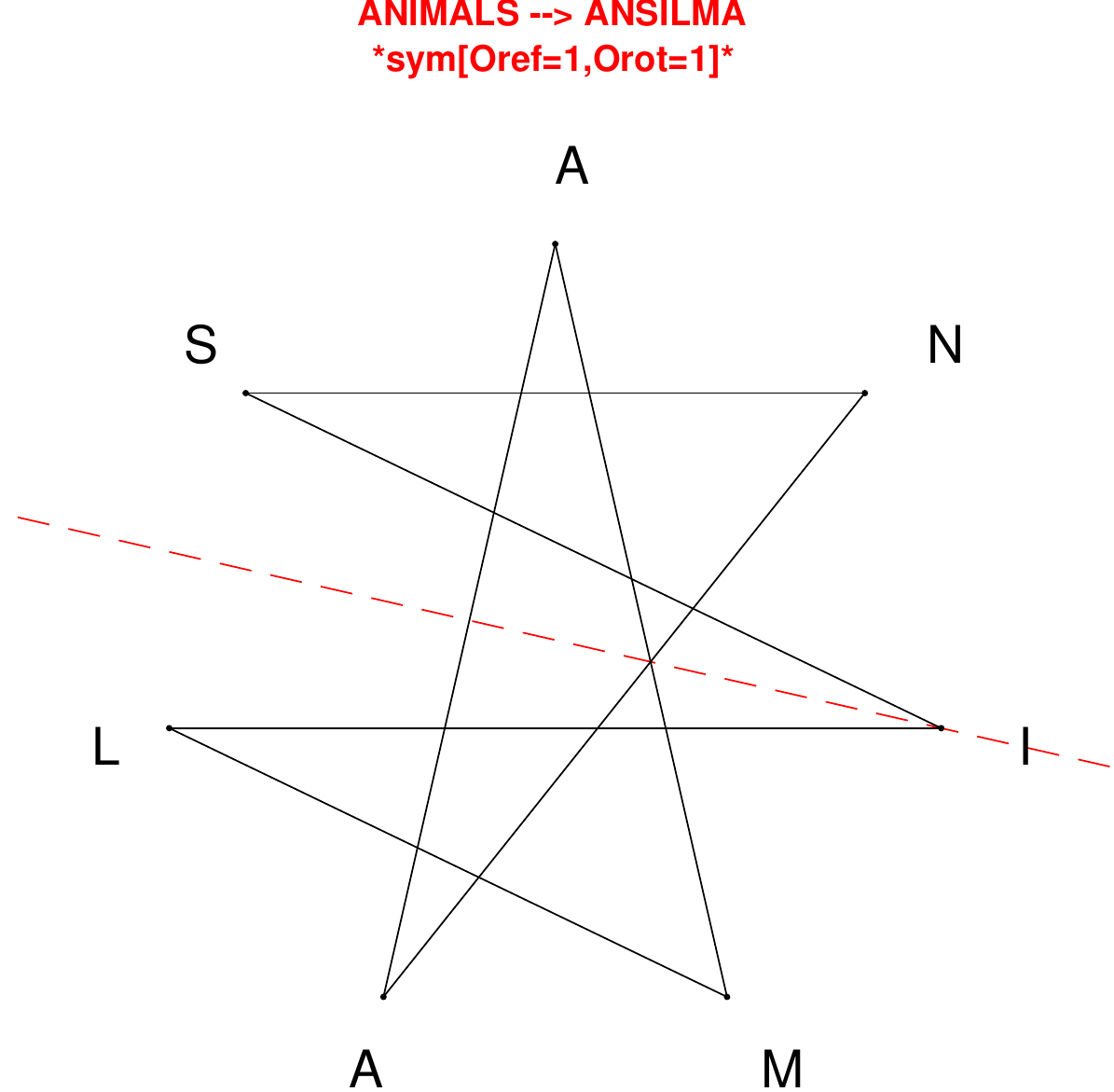}
\end{subfigure}
\hfill
\begin{subfigure}[T]{0.19\textwidth}
\centering
\includegraphics[width=\textwidth]{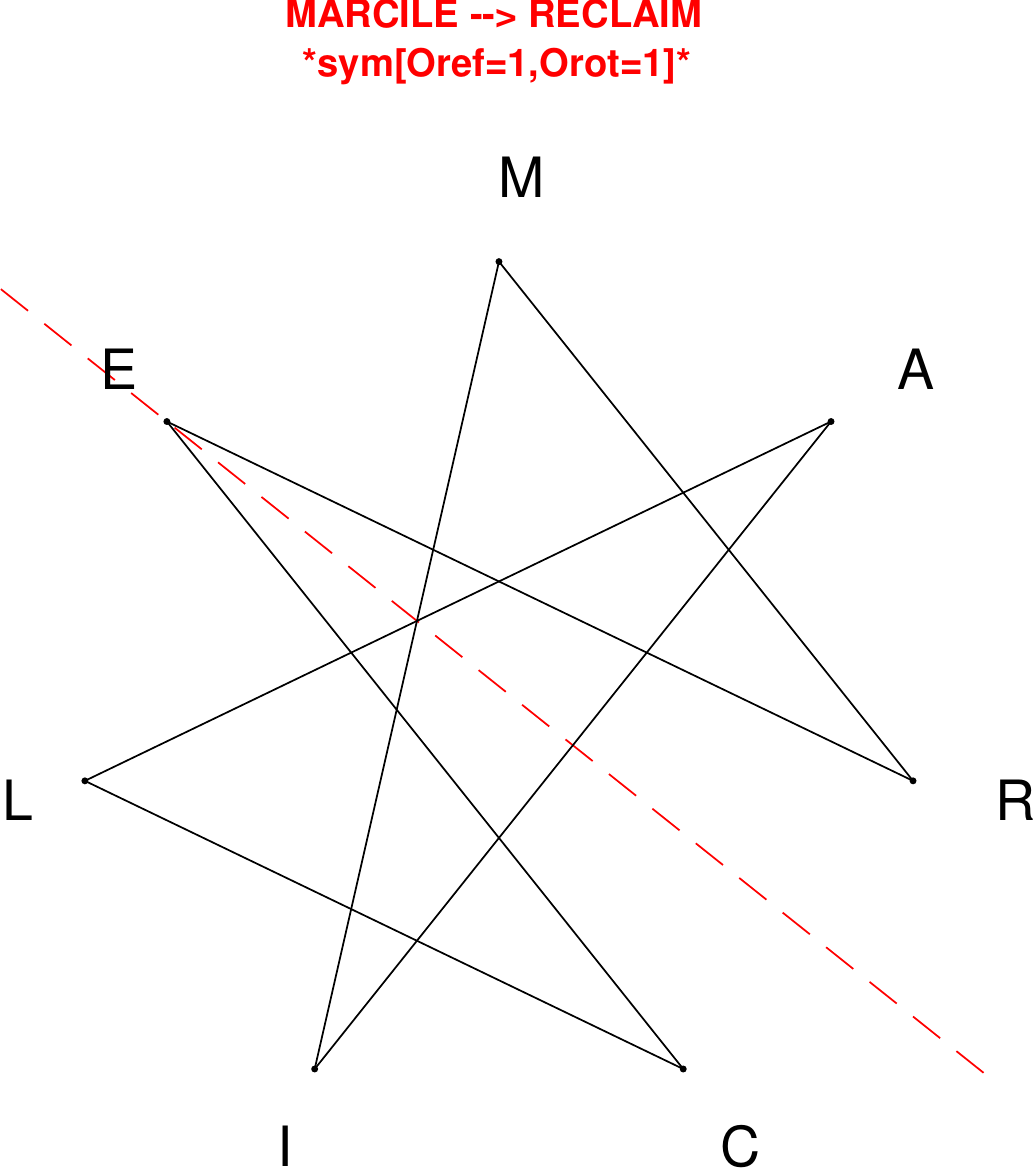}
\end{subfigure}
\hfill
\begin{subfigure}[T]{0.19\textwidth}
\centering
\includegraphics[width=\textwidth]{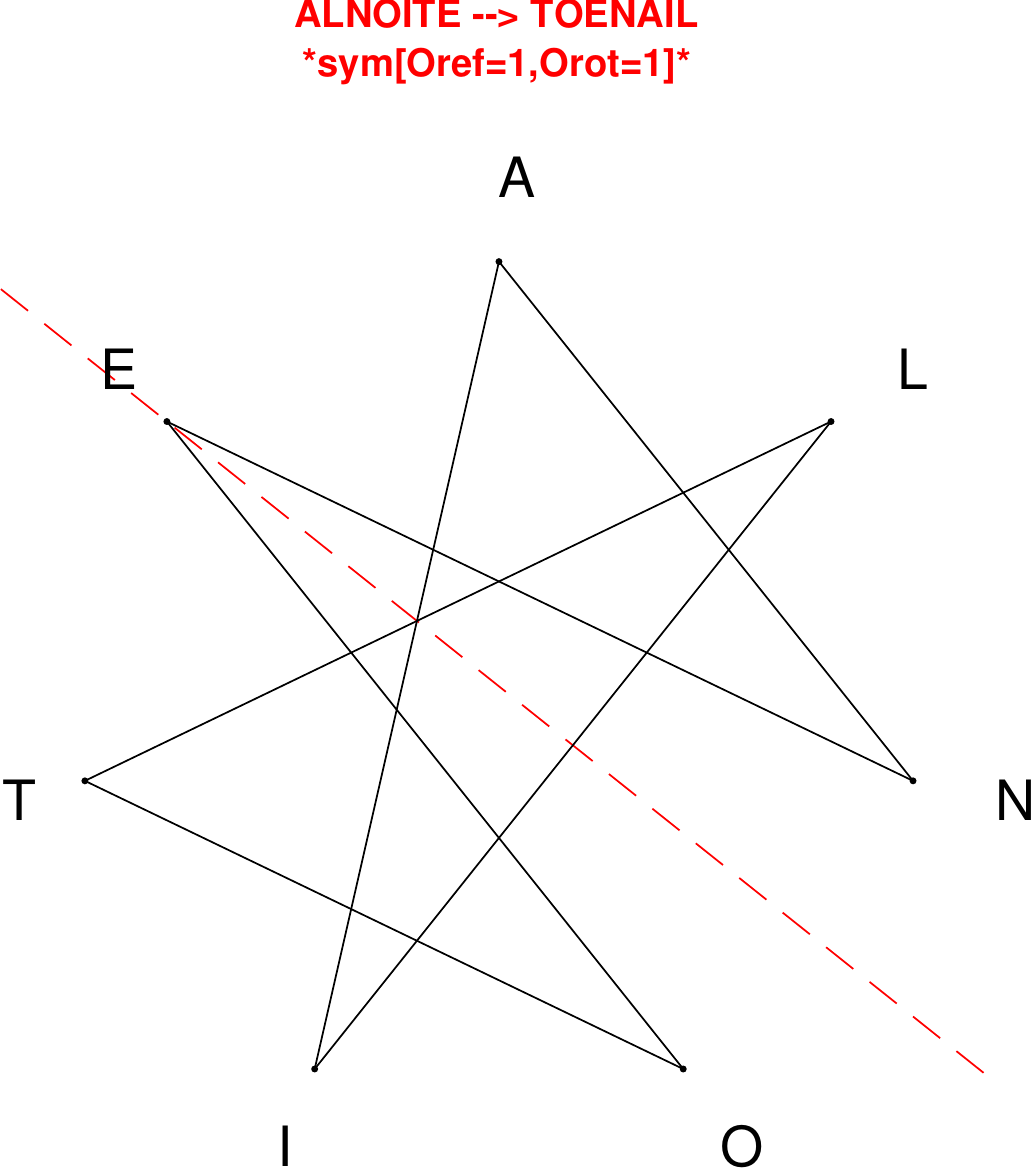}
\end{subfigure}
\hfill
\begin{subfigure}[T]{0.19\textwidth}
\centering
\includegraphics[width=\textwidth]{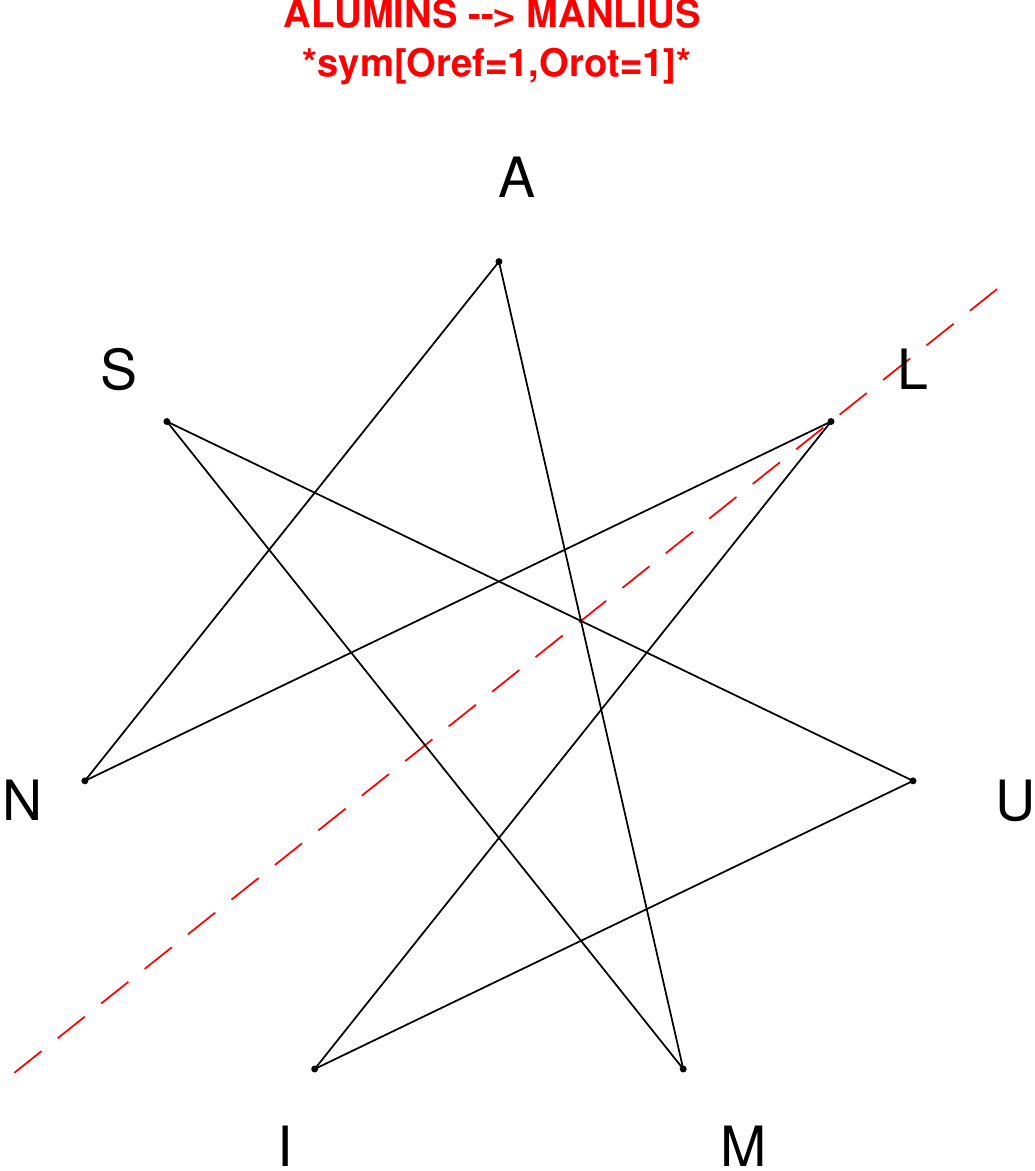}
\end{subfigure}
\end{figure}

\begin{figure}[H]
\centering
\begin{subfigure}[T]{0.19\textwidth}
\centering
\includegraphics[width=\textwidth]{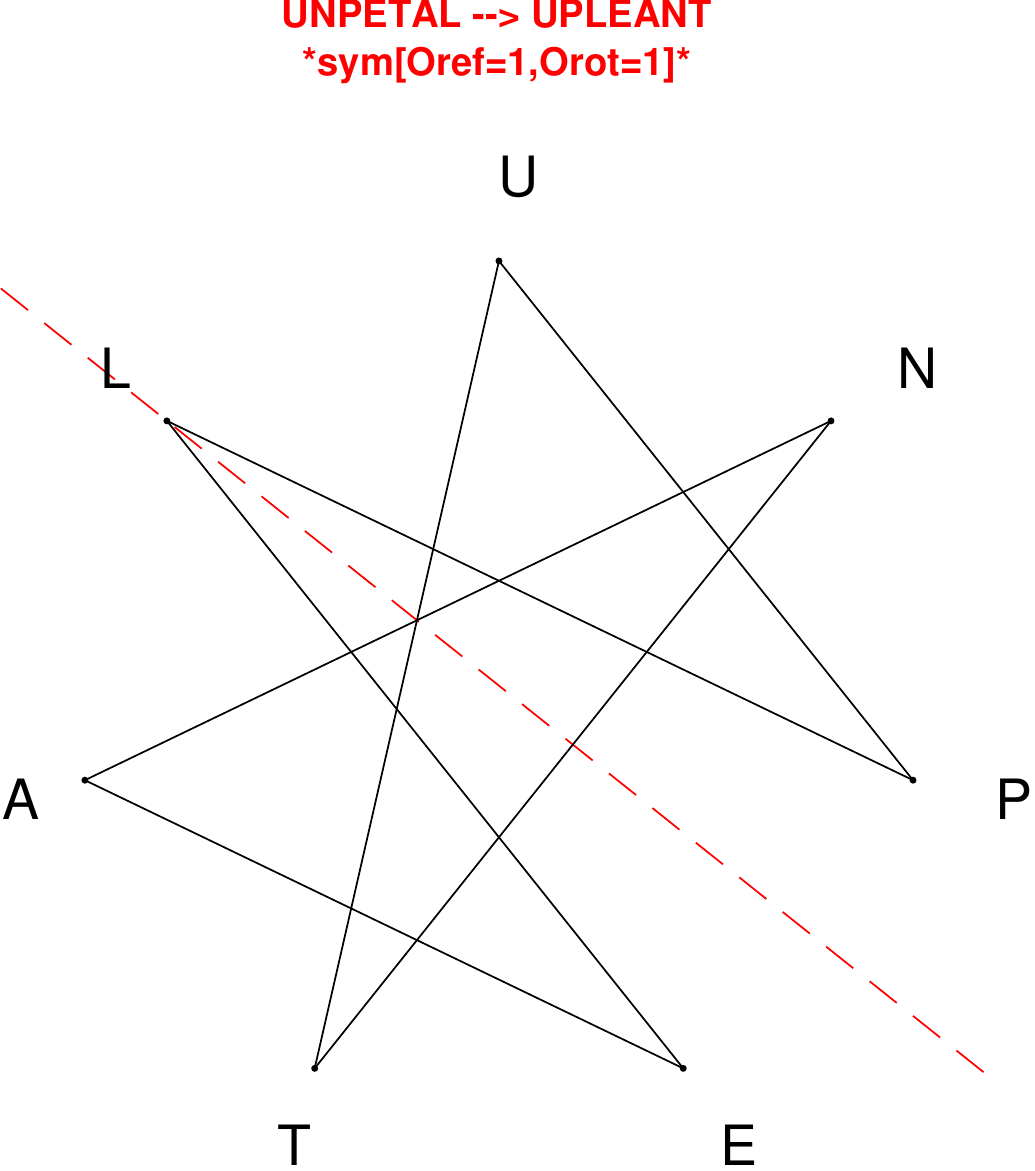}
\end{subfigure}
\hfill
\begin{subfigure}[T]{0.19\textwidth}
\centering
\includegraphics[width=\textwidth]{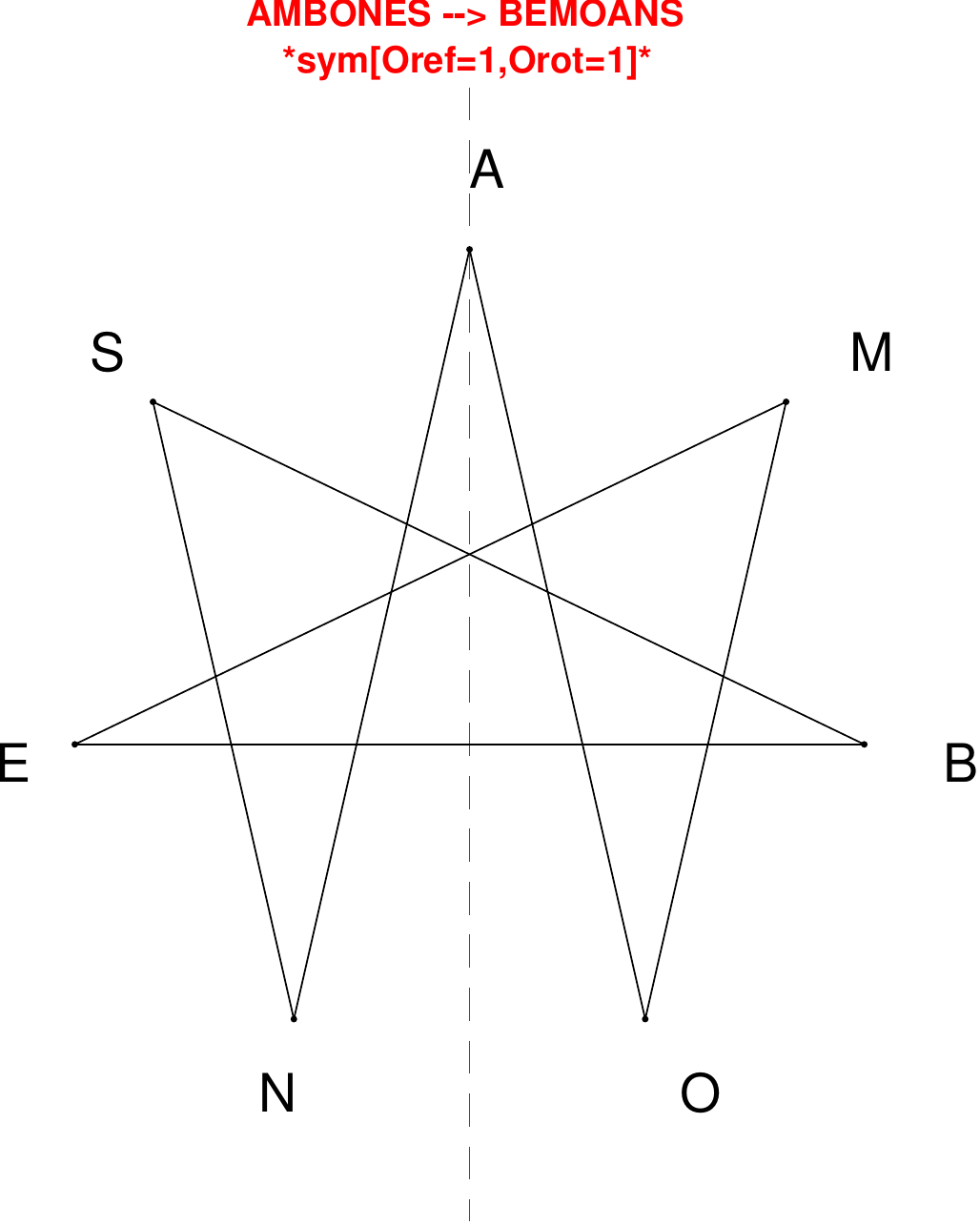}
\end{subfigure}
\hfill
\begin{subfigure}[T]{0.19\textwidth}
\centering
\includegraphics[width=\textwidth]{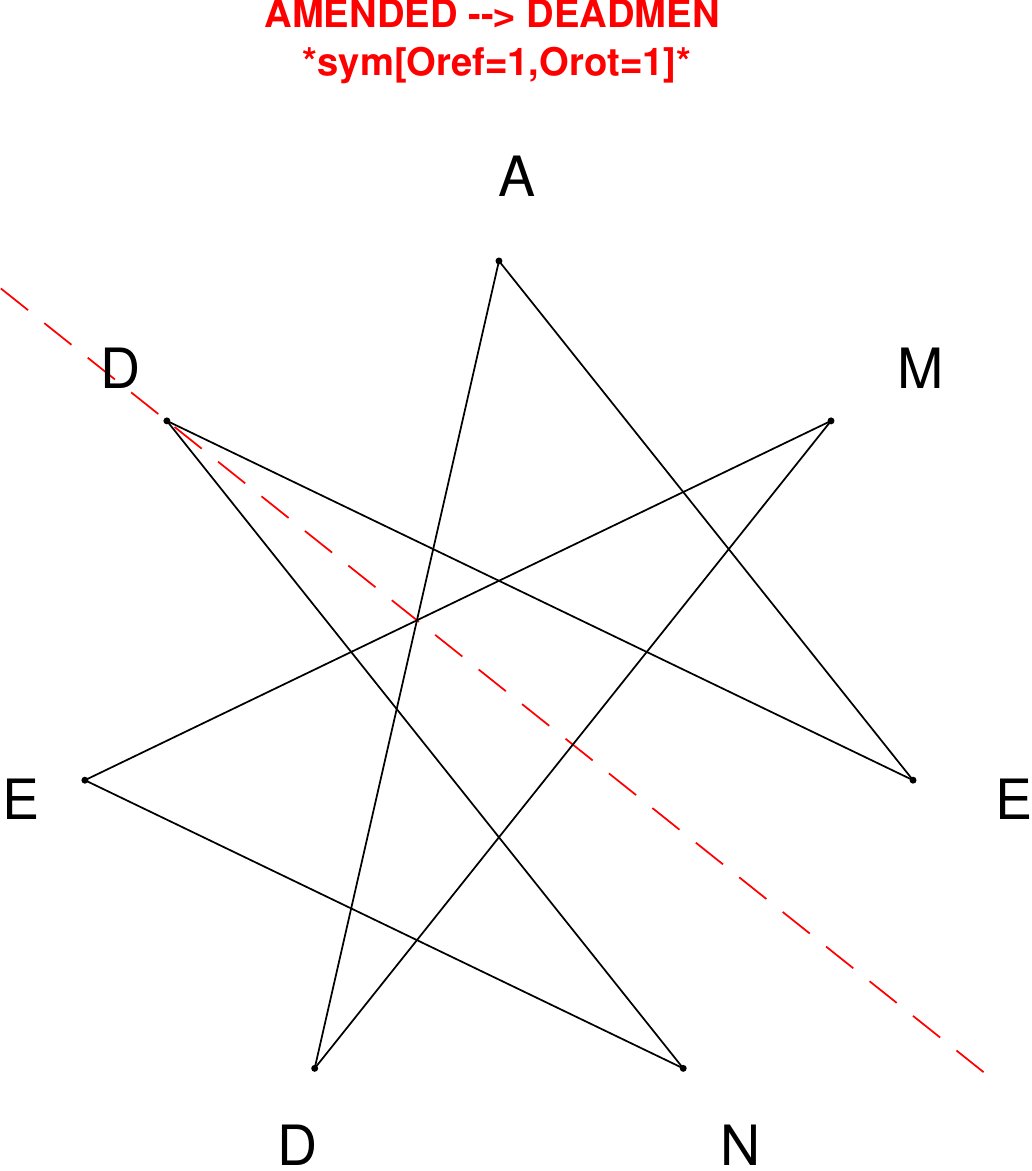}
\end{subfigure}
\hfill
\begin{subfigure}[T]{0.19\textwidth}
\centering
\includegraphics[width=\textwidth]{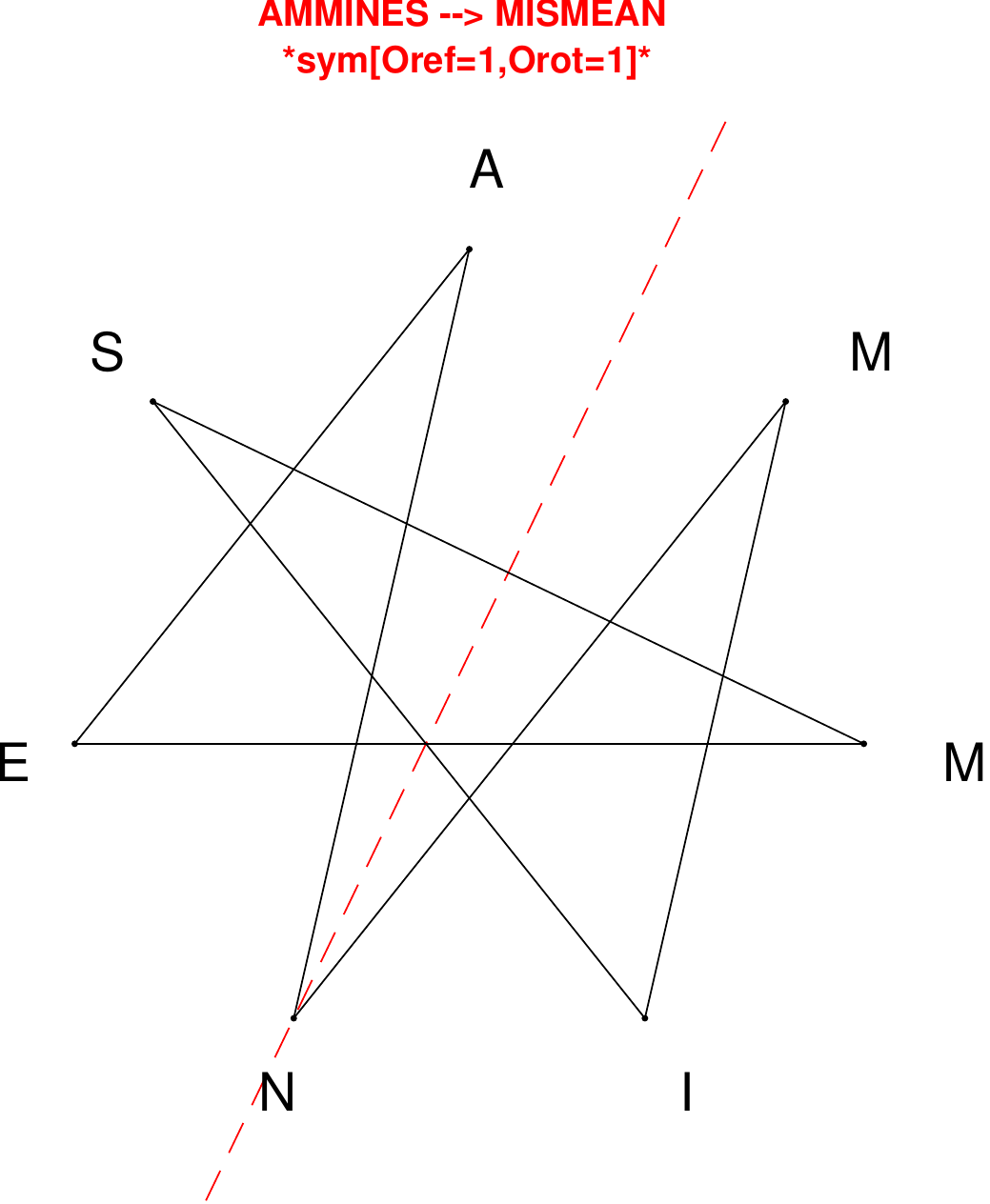}
\end{subfigure}
\hfill
\begin{subfigure}[T]{0.19\textwidth}
\centering
\includegraphics[width=\textwidth]{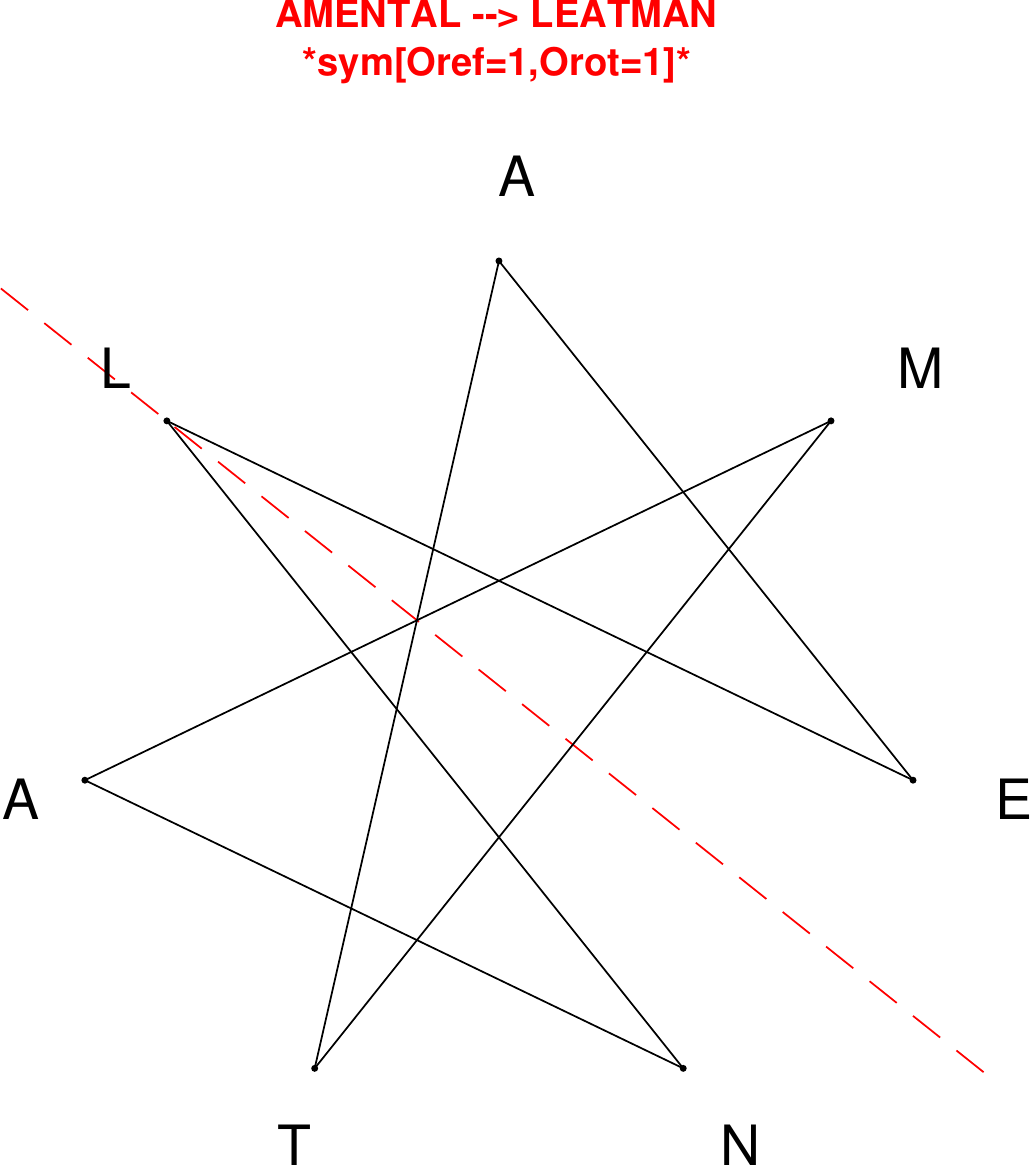}
\end{subfigure}
\end{figure}

\begin{figure}[H]
\centering
\begin{subfigure}[T]{0.19\textwidth}
\centering
\includegraphics[width=\textwidth]{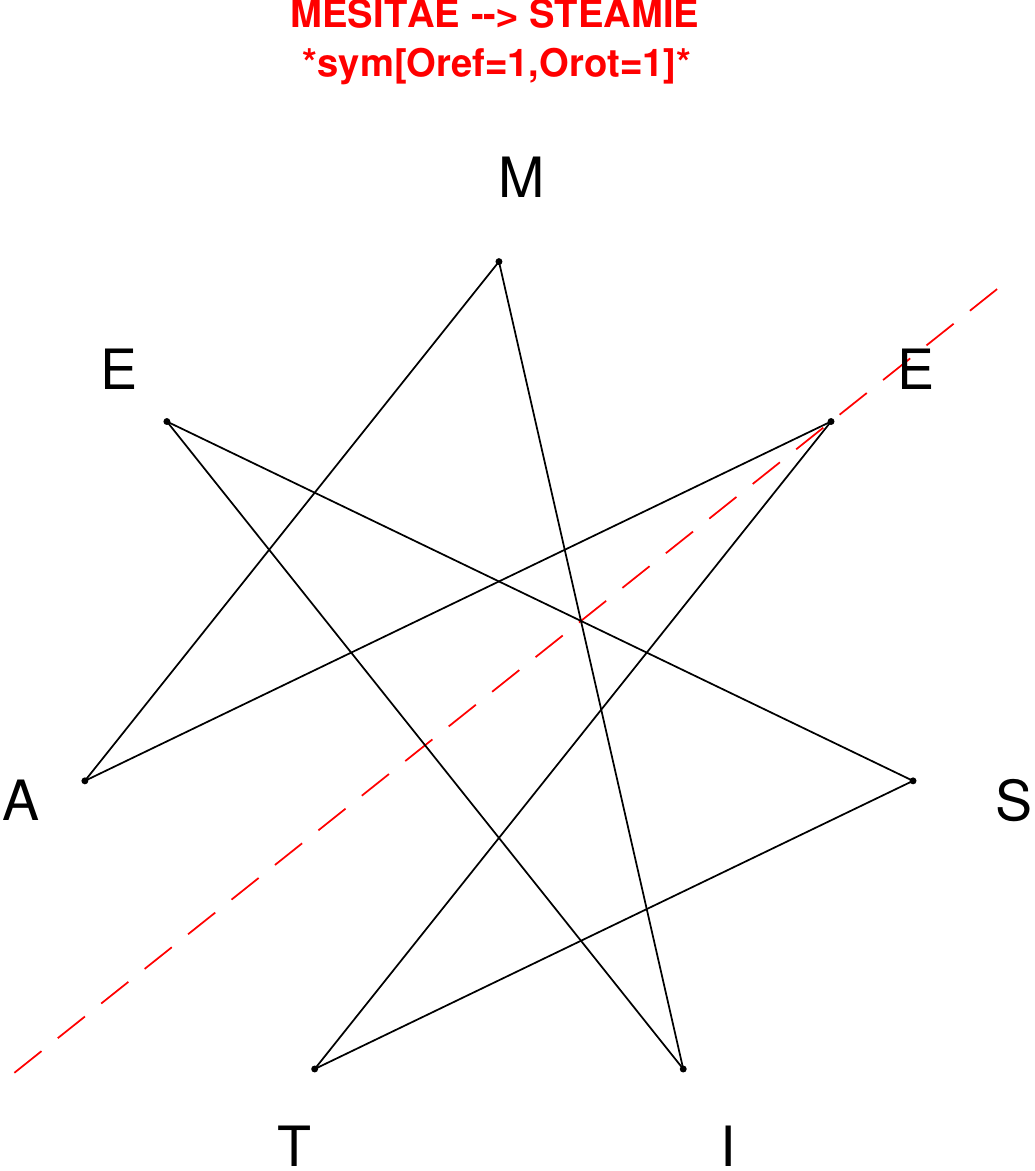}
\end{subfigure}
\hfill
\begin{subfigure}[T]{0.19\textwidth}
\centering
\includegraphics[width=\textwidth]{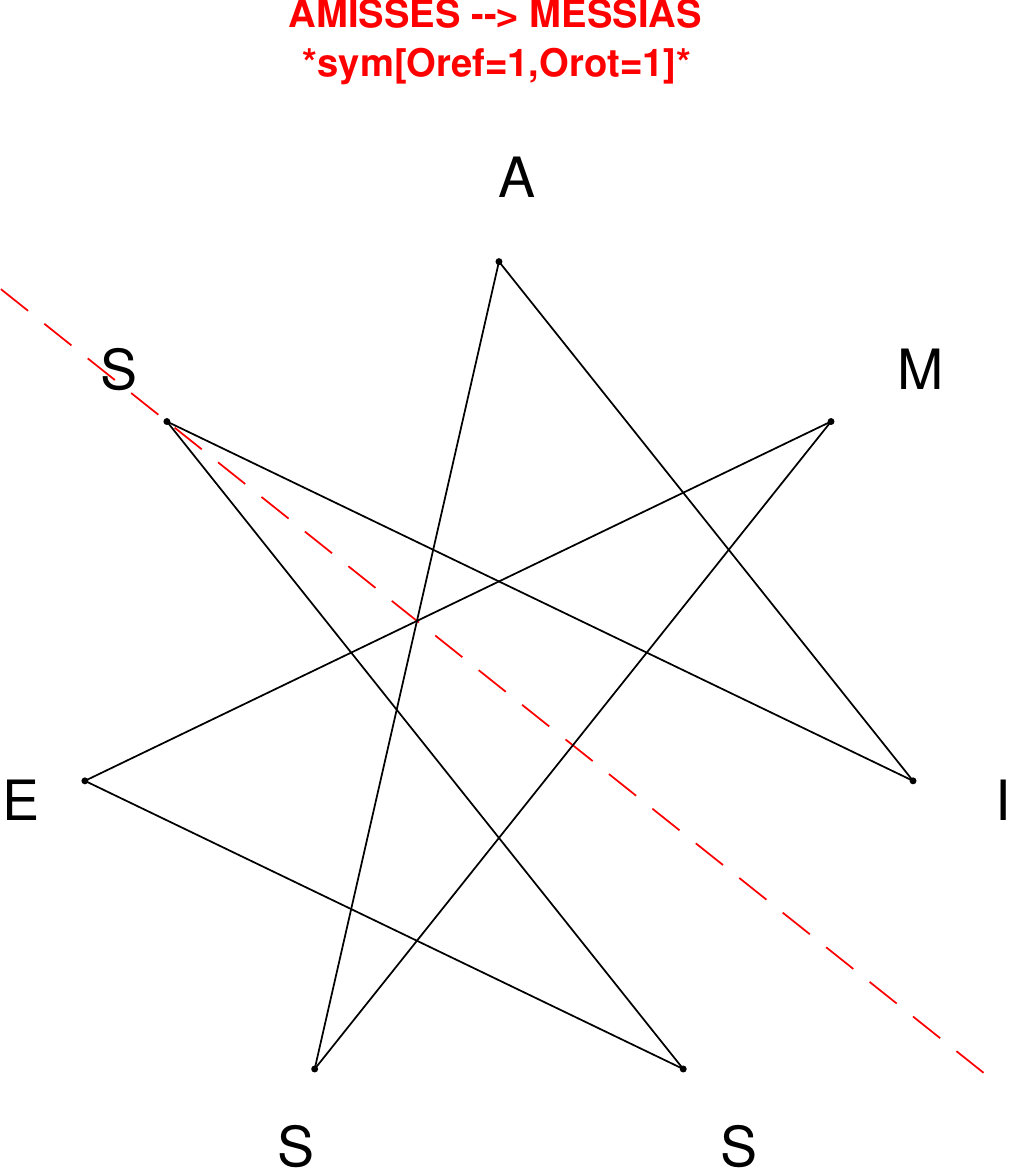}
\end{subfigure}
\hfill
\begin{subfigure}[T]{0.19\textwidth}
\centering
\includegraphics[width=\textwidth]{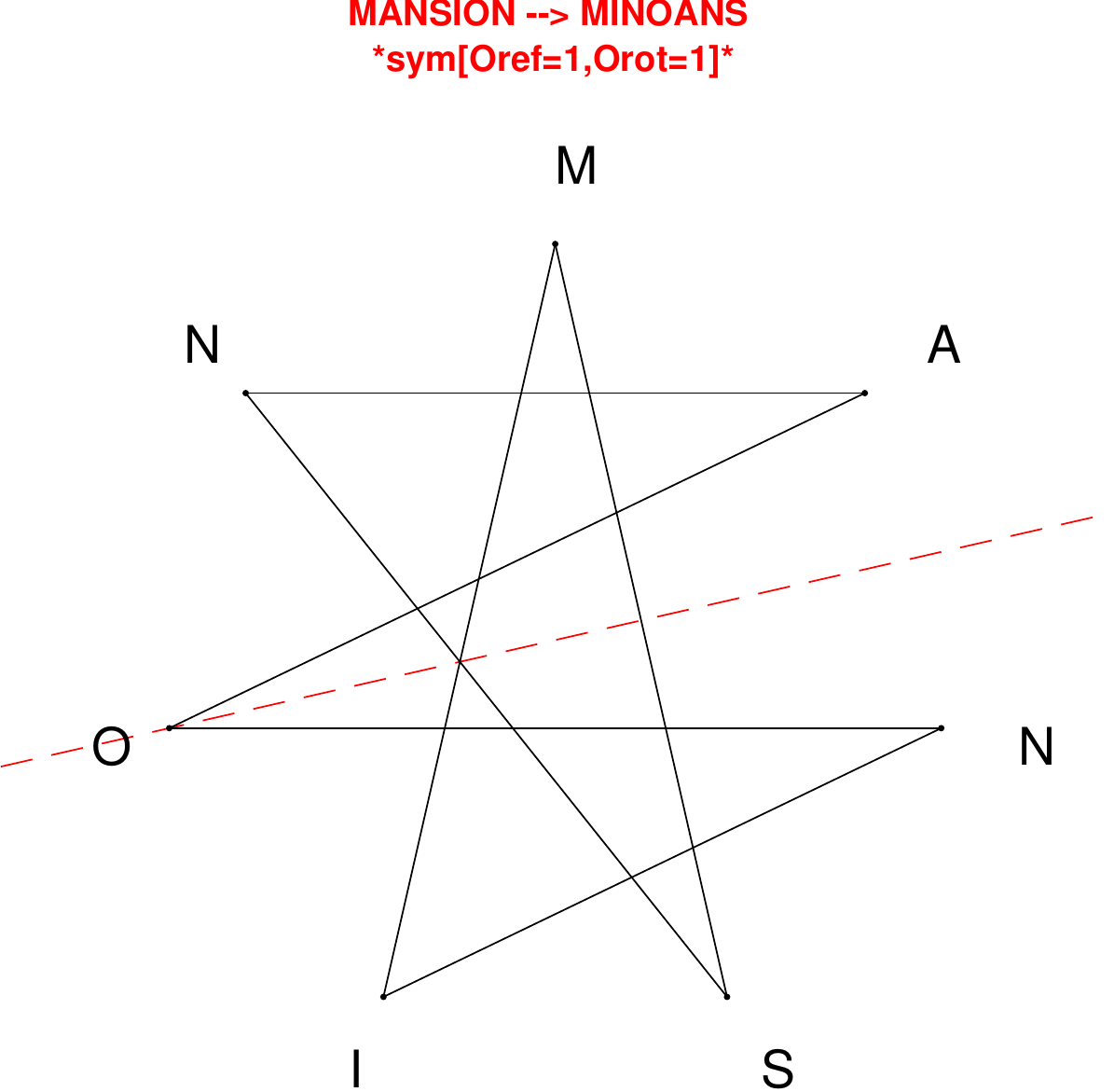}
\end{subfigure}
\hfill
\begin{subfigure}[T]{0.19\textwidth}
\centering
\includegraphics[width=\textwidth]{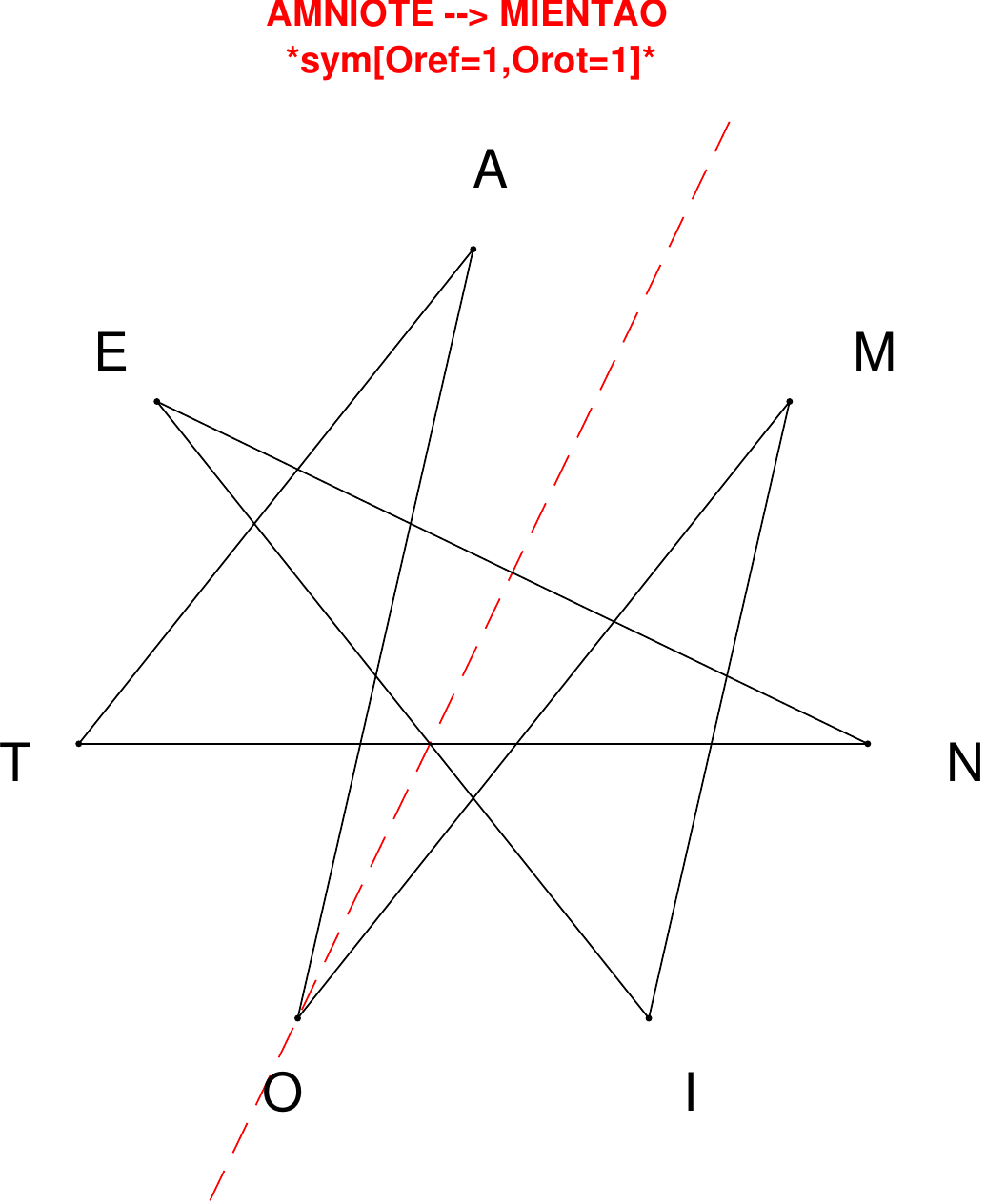}
\end{subfigure}
\hfill
\begin{subfigure}[T]{0.19\textwidth}
\centering
\includegraphics[width=\textwidth]{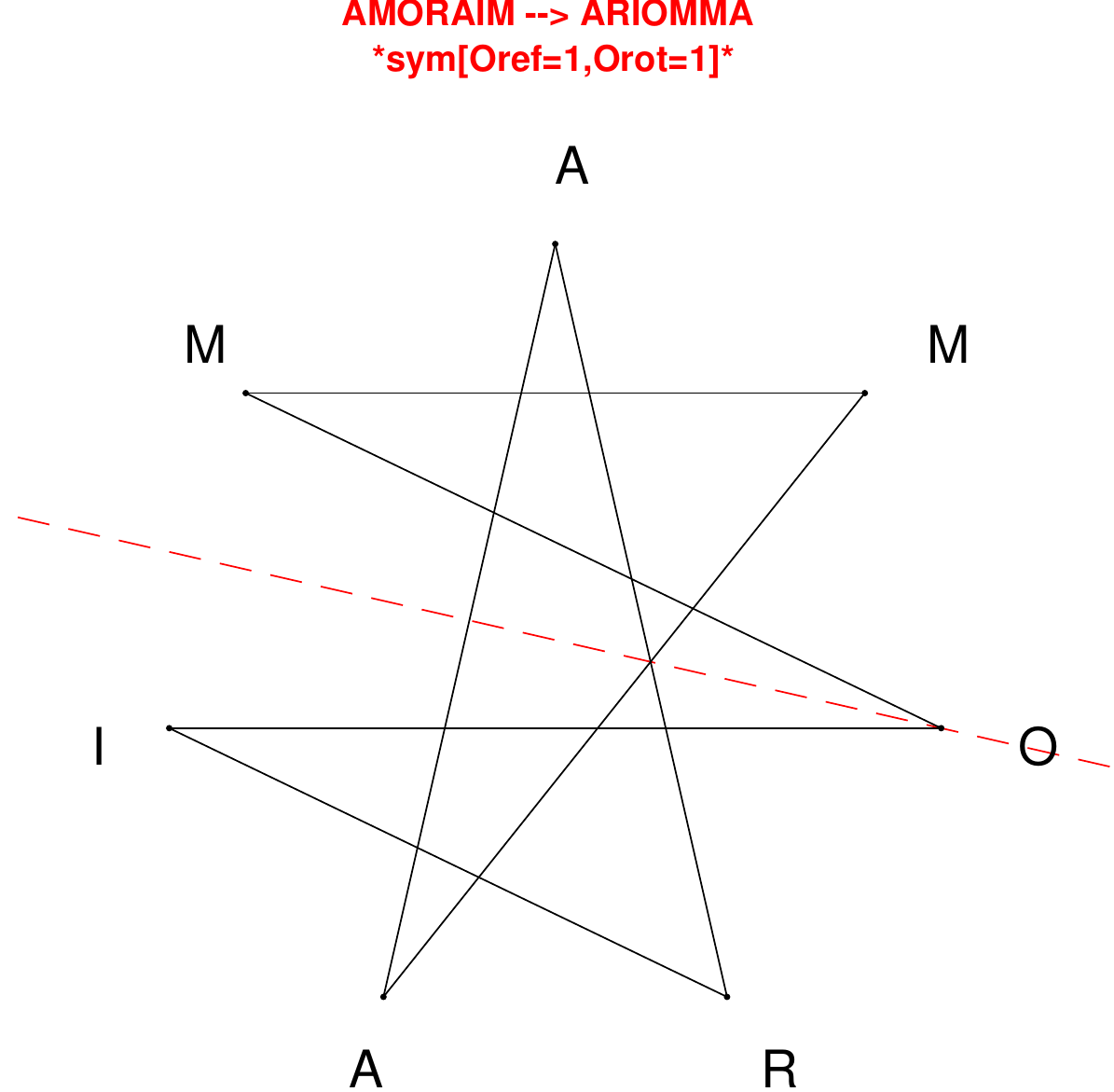}
\end{subfigure}
\end{figure}

\begin{figure}[H]
\centering
\begin{subfigure}[T]{0.19\textwidth}
\centering
\includegraphics[width=\textwidth]{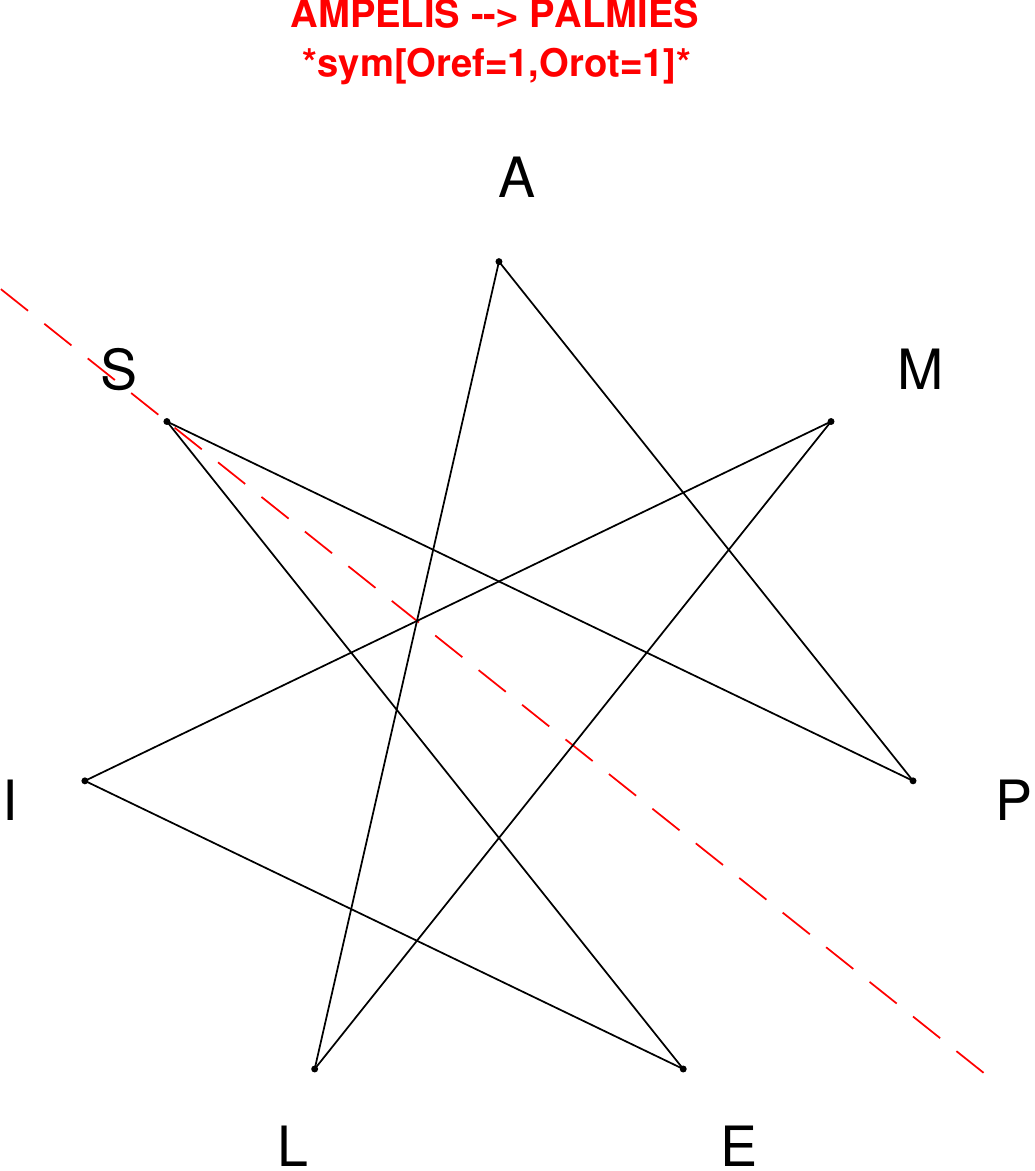}
\end{subfigure}
\hfill
\begin{subfigure}[T]{0.19\textwidth}
\centering
\includegraphics[width=\textwidth]{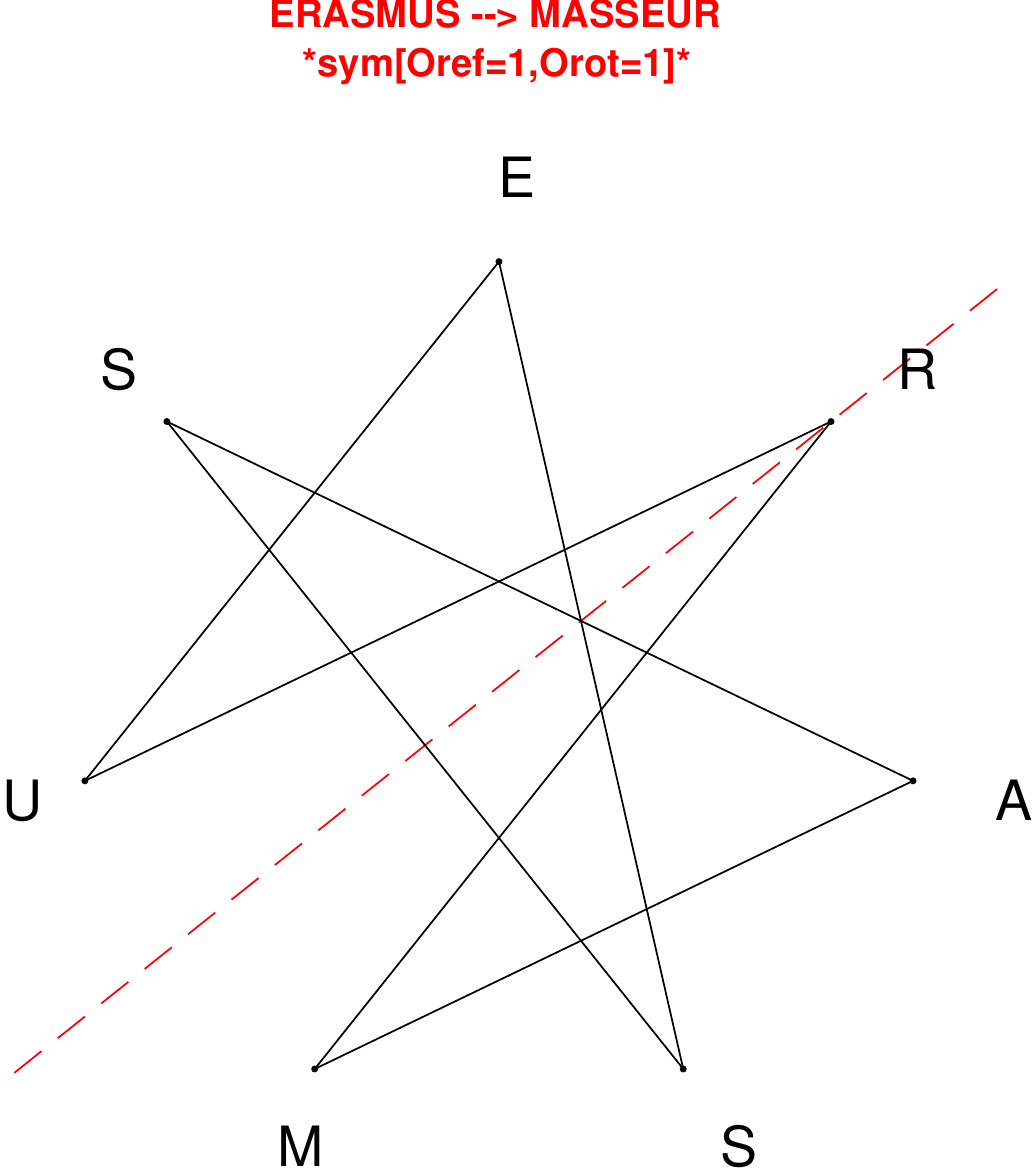}
\end{subfigure}
\hfill
\begin{subfigure}[T]{0.19\textwidth}
\centering
\includegraphics[width=\textwidth]{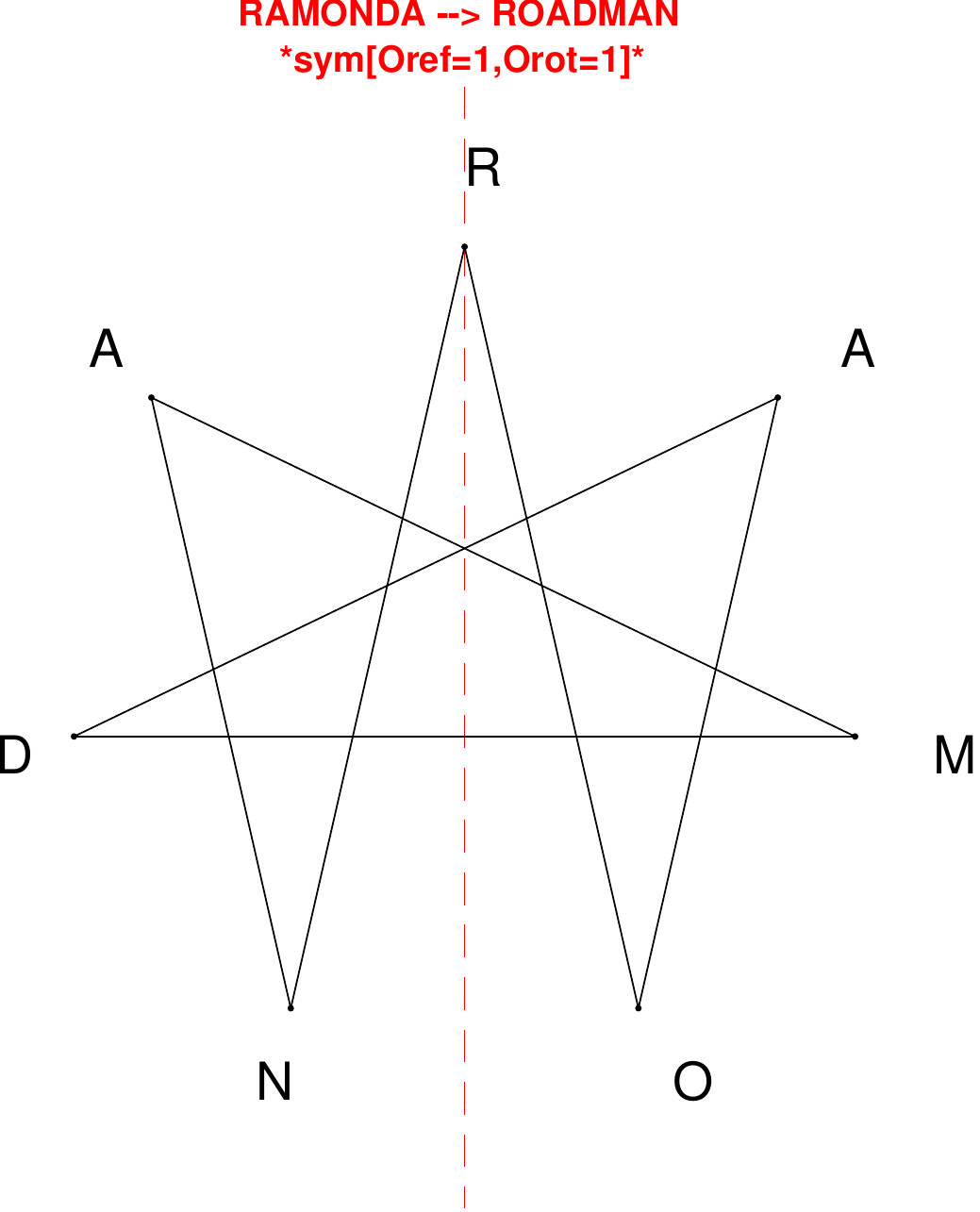}
\end{subfigure}
\hfill
\begin{subfigure}[T]{0.19\textwidth}
\centering
\includegraphics[width=\textwidth]{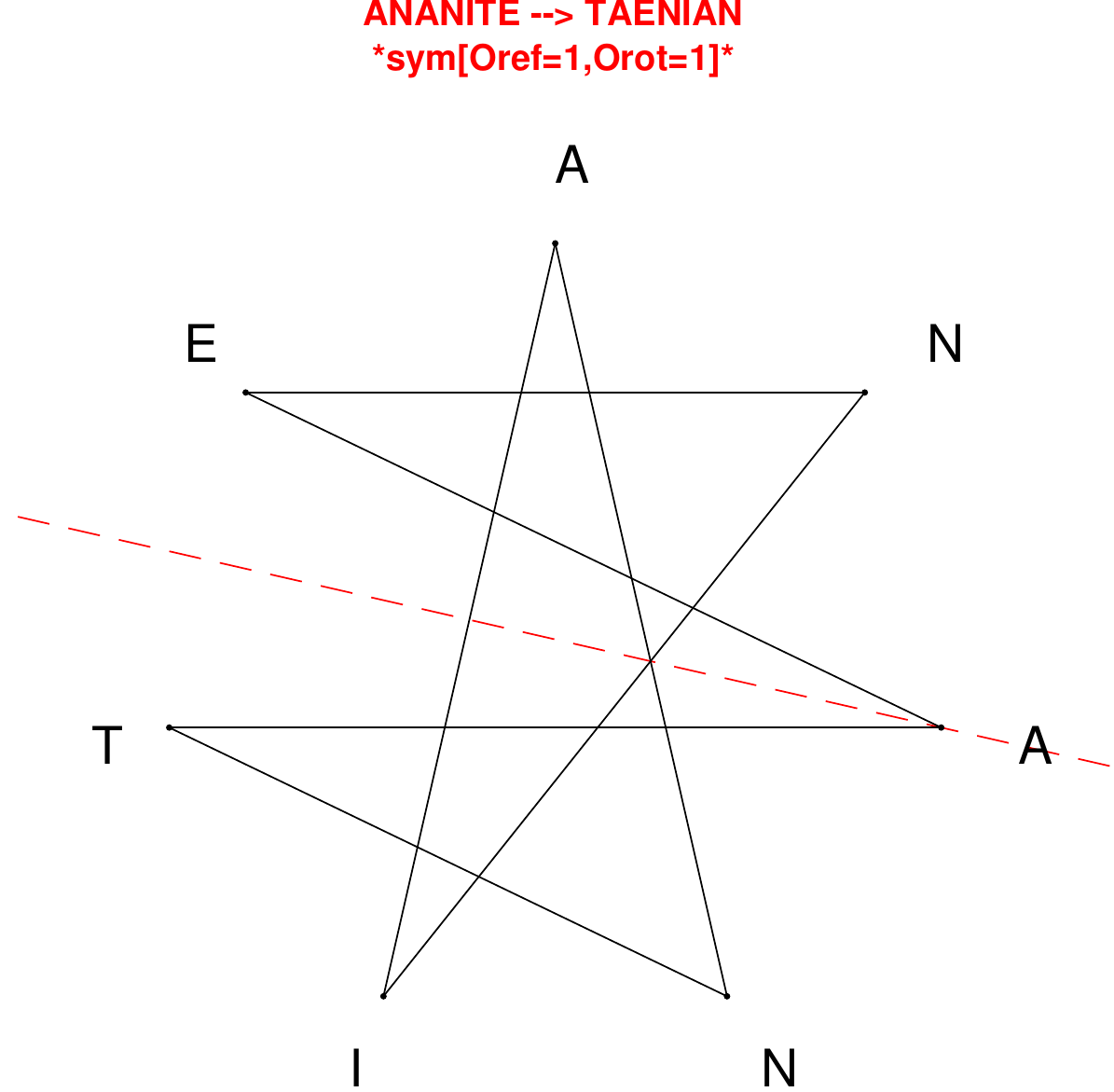}
\end{subfigure}
\hfill
\begin{subfigure}[T]{0.19\textwidth}
\centering
\includegraphics[width=\textwidth]{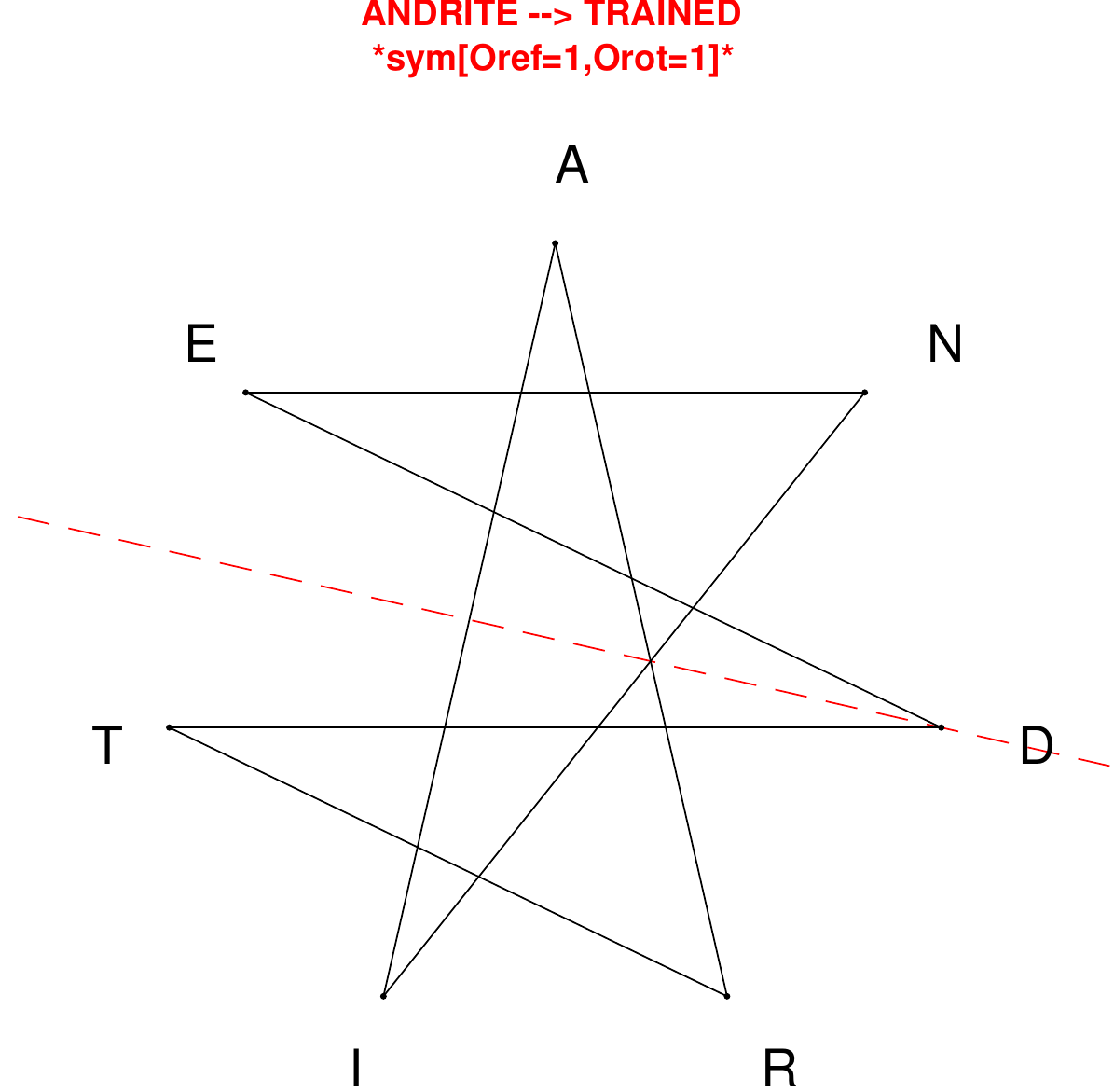}
\end{subfigure}
\end{figure}

\begin{figure}[H]
\centering
\begin{subfigure}[T]{0.19\textwidth}
\centering
\includegraphics[width=\textwidth]{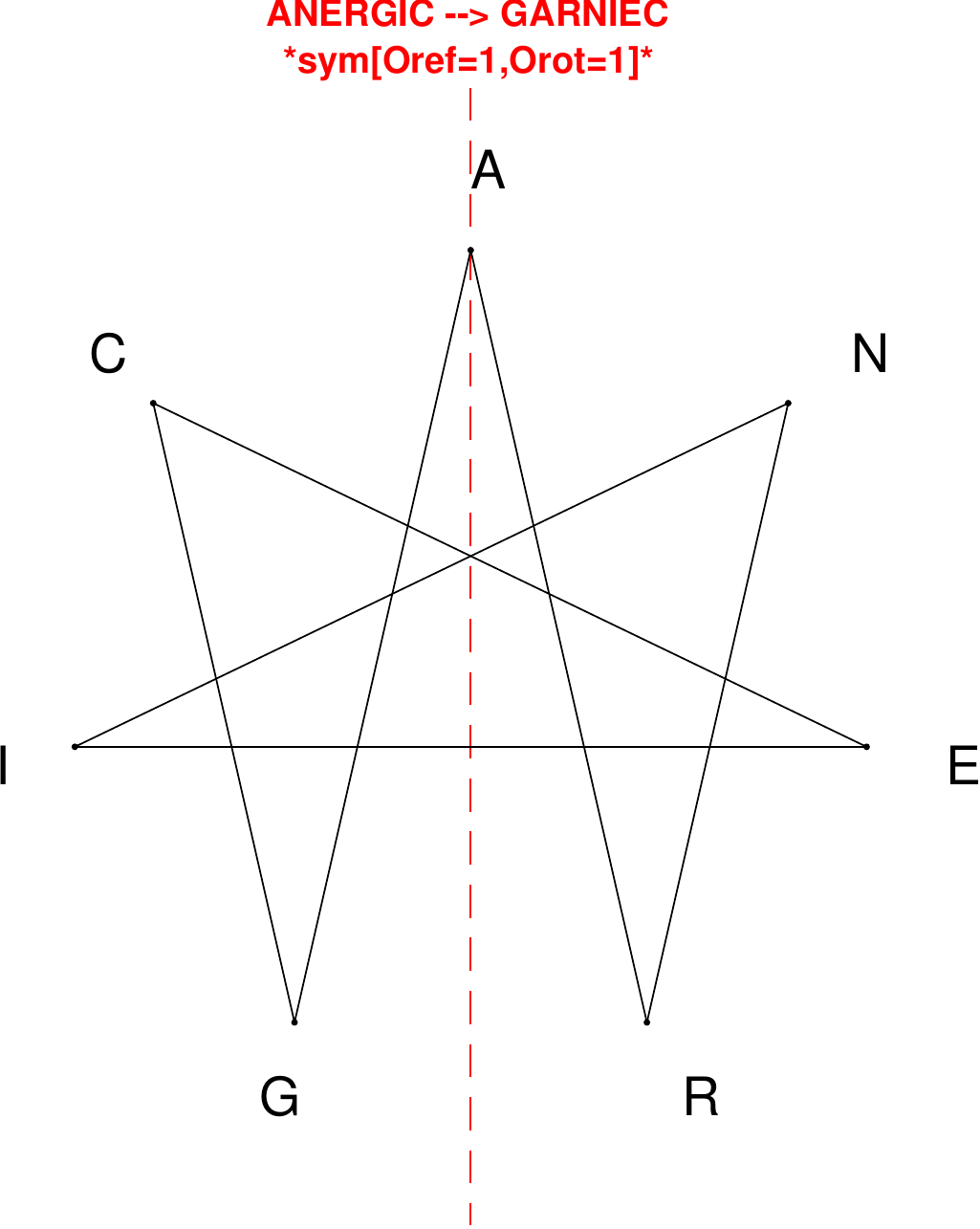}
\end{subfigure}
\hfill
\begin{subfigure}[T]{0.19\textwidth}
\centering
\includegraphics[width=\textwidth]{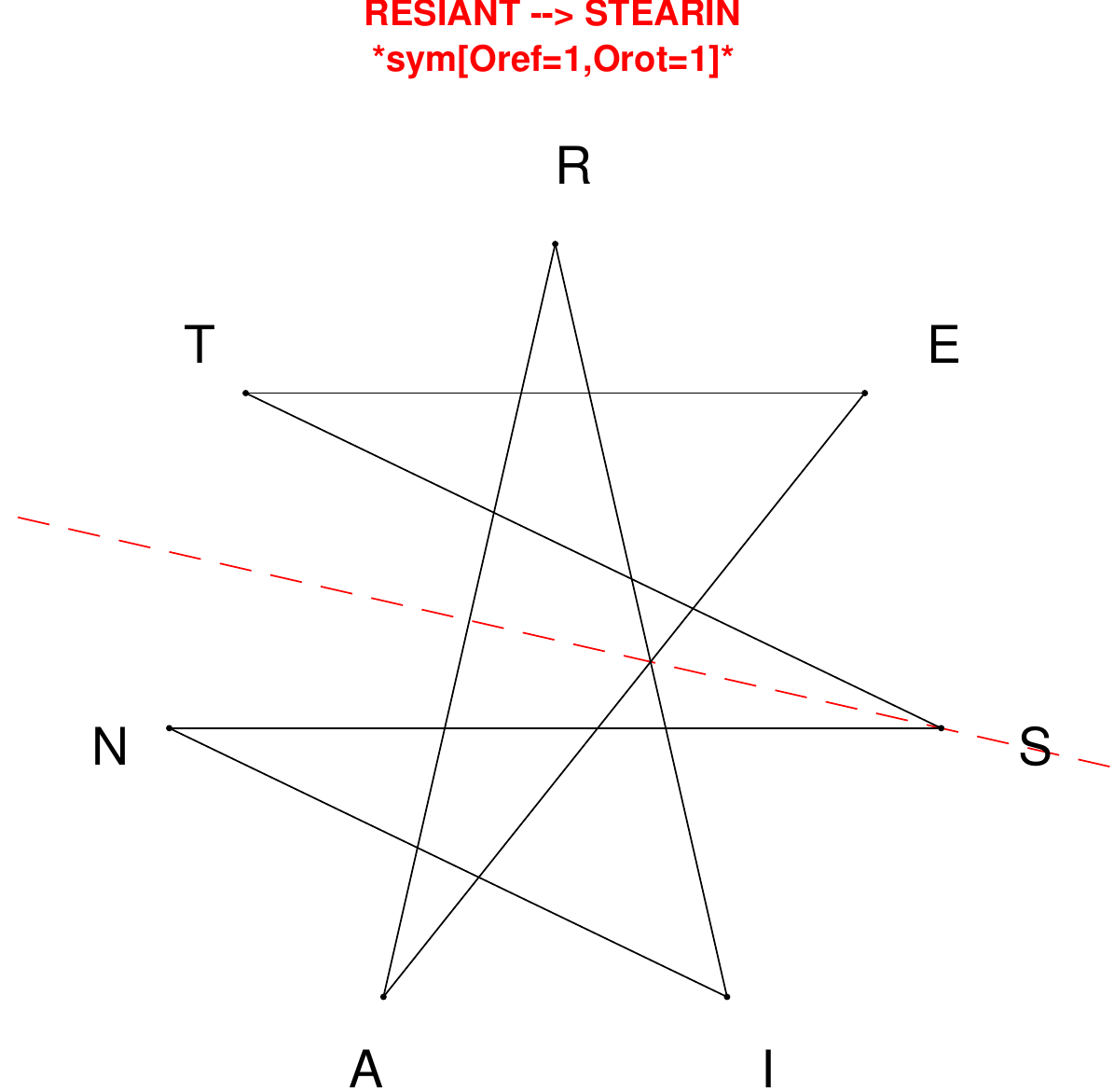}
\end{subfigure}
\hfill
\begin{subfigure}[T]{0.19\textwidth}
\centering
\includegraphics[width=\textwidth]{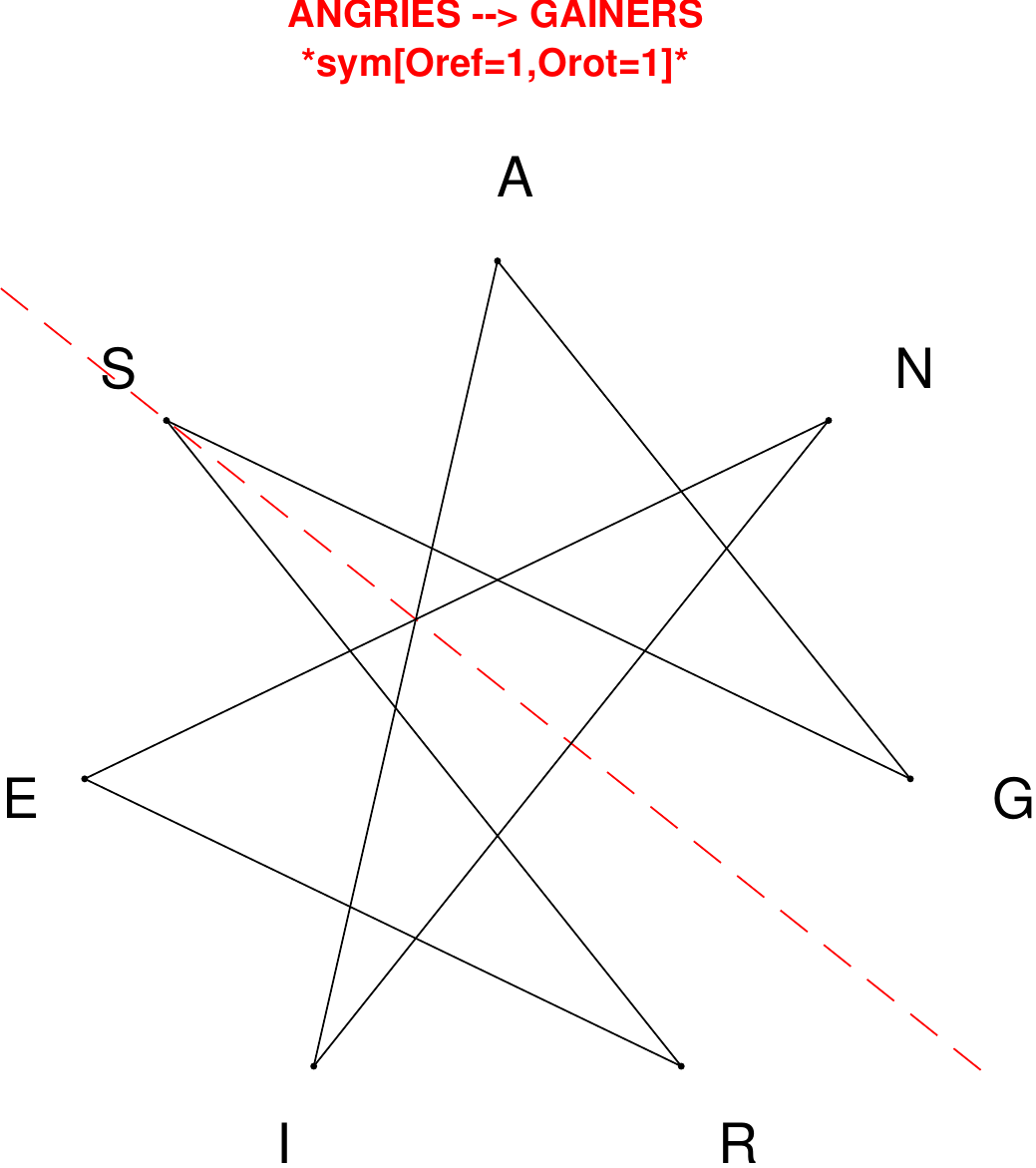}
\end{subfigure}
\hfill
\begin{subfigure}[T]{0.19\textwidth}
\centering
\includegraphics[width=\textwidth]{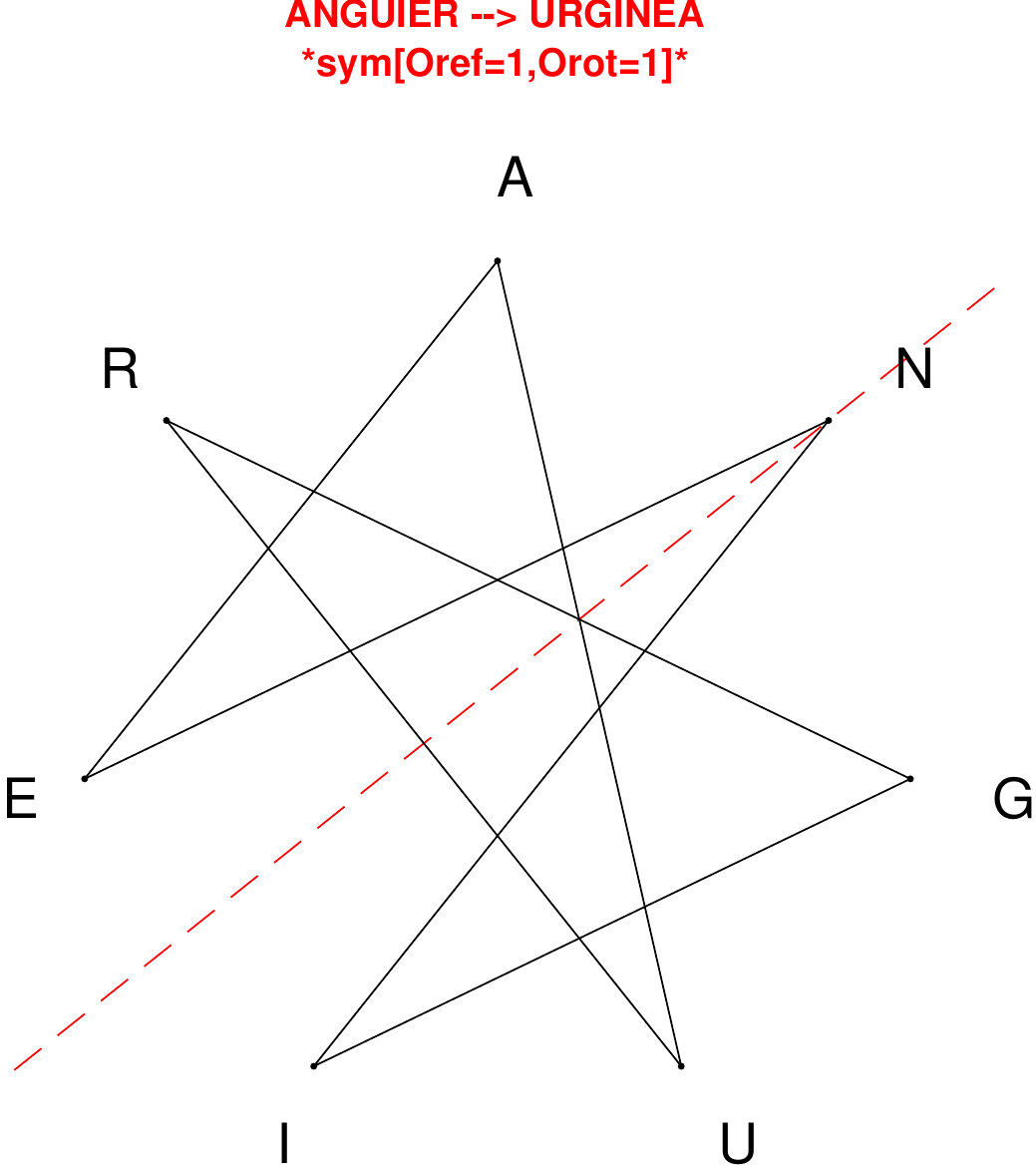}
\end{subfigure}
\hfill
\begin{subfigure}[T]{0.19\textwidth}
\centering
\includegraphics[width=\textwidth]{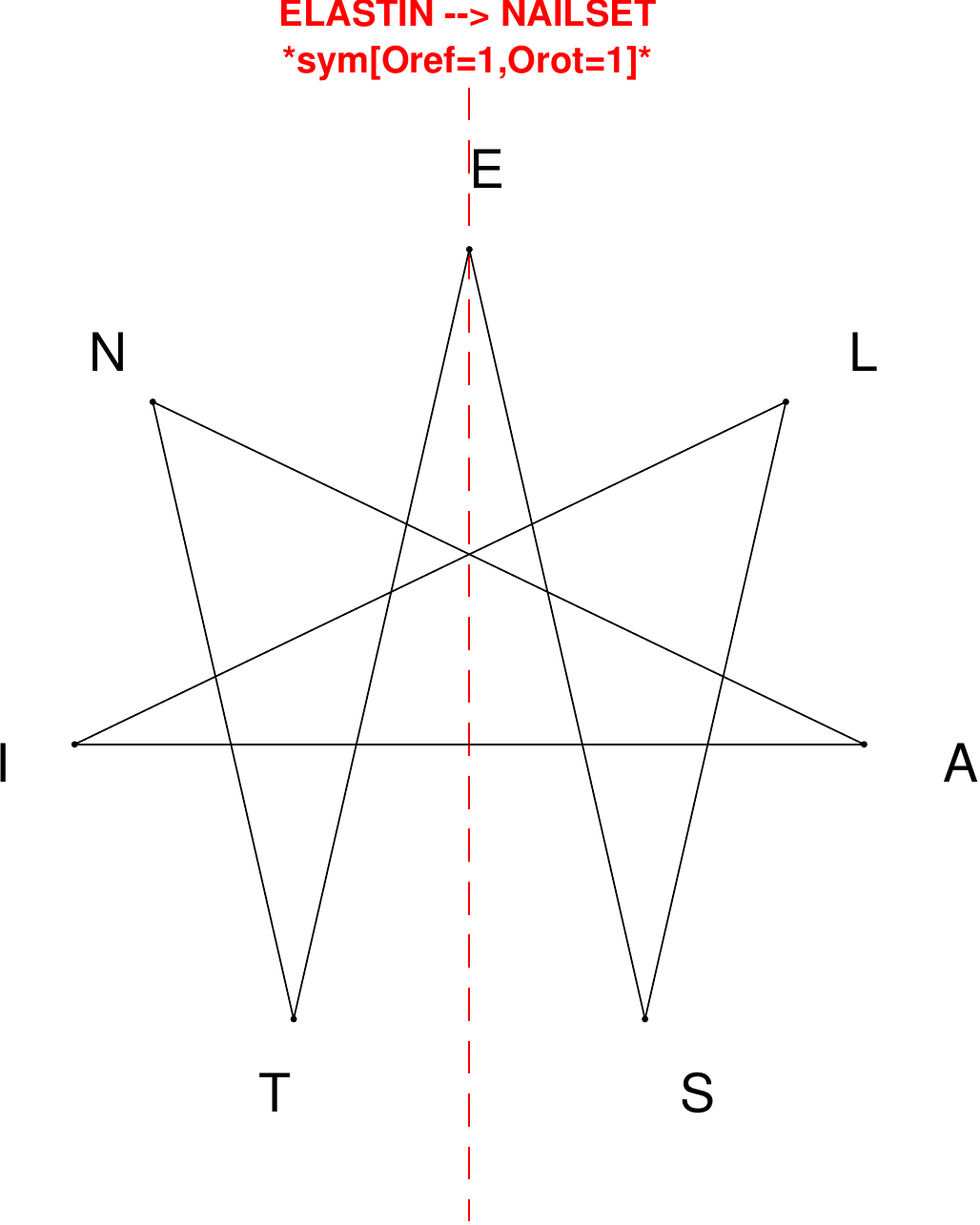}
\end{subfigure}
\end{figure}

\begin{figure}[H]
\centering
\begin{subfigure}[T]{0.19\textwidth}
\centering
\includegraphics[width=\textwidth]{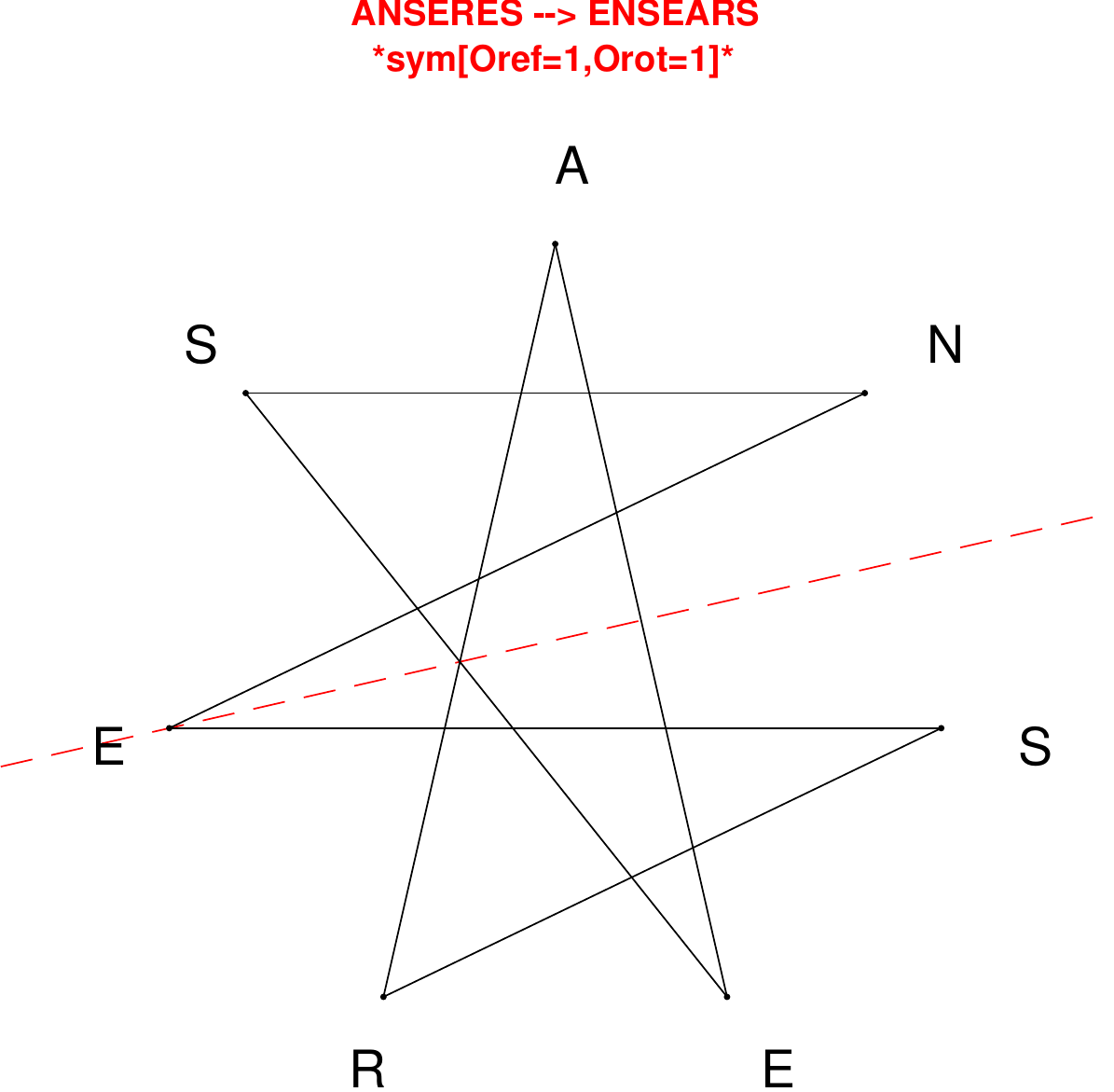}
\end{subfigure}
\hfill
\begin{subfigure}[T]{0.19\textwidth}
\centering
\includegraphics[width=\textwidth]{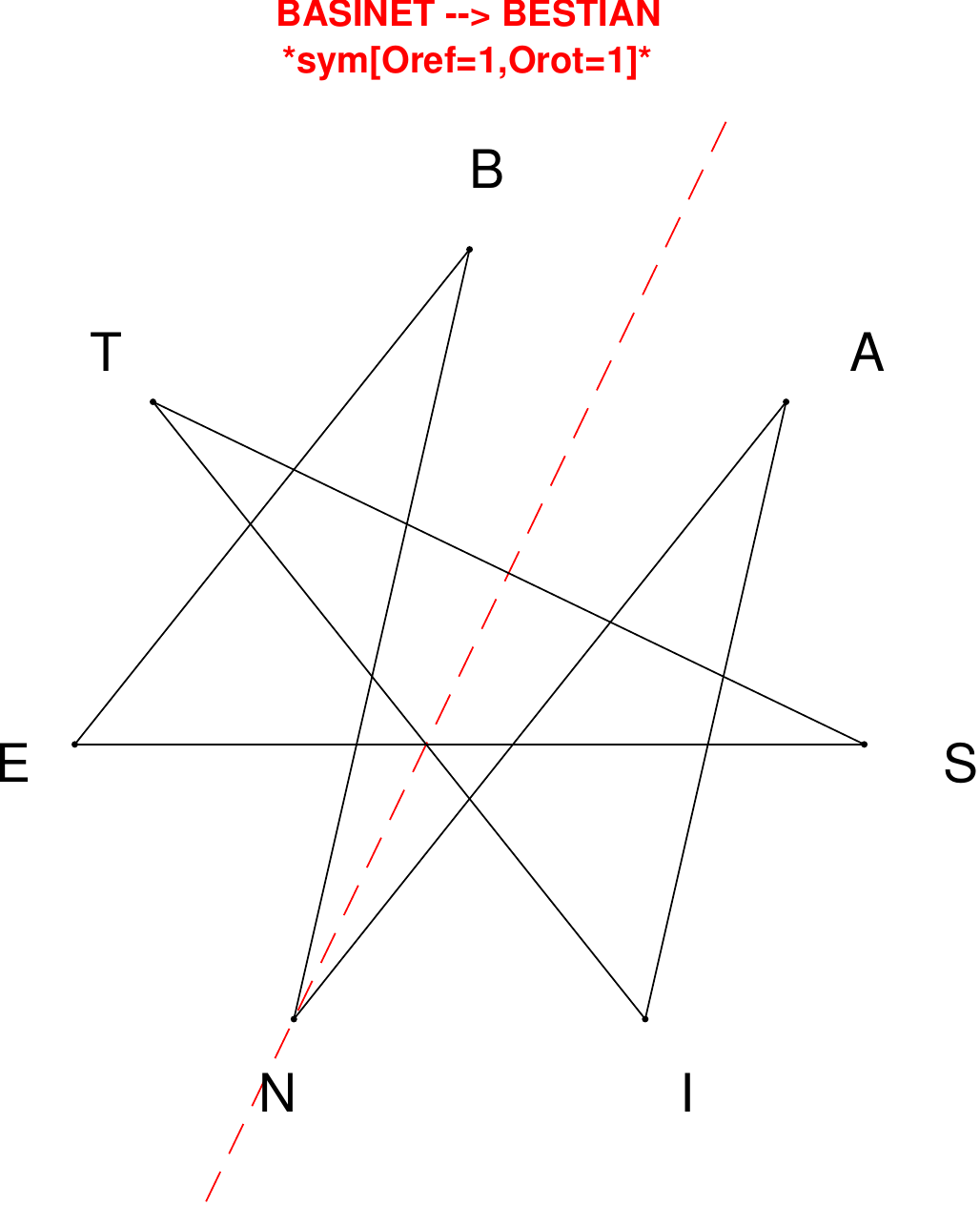}
\end{subfigure}
\hfill
\begin{subfigure}[T]{0.19\textwidth}
\centering
\includegraphics[width=\textwidth]{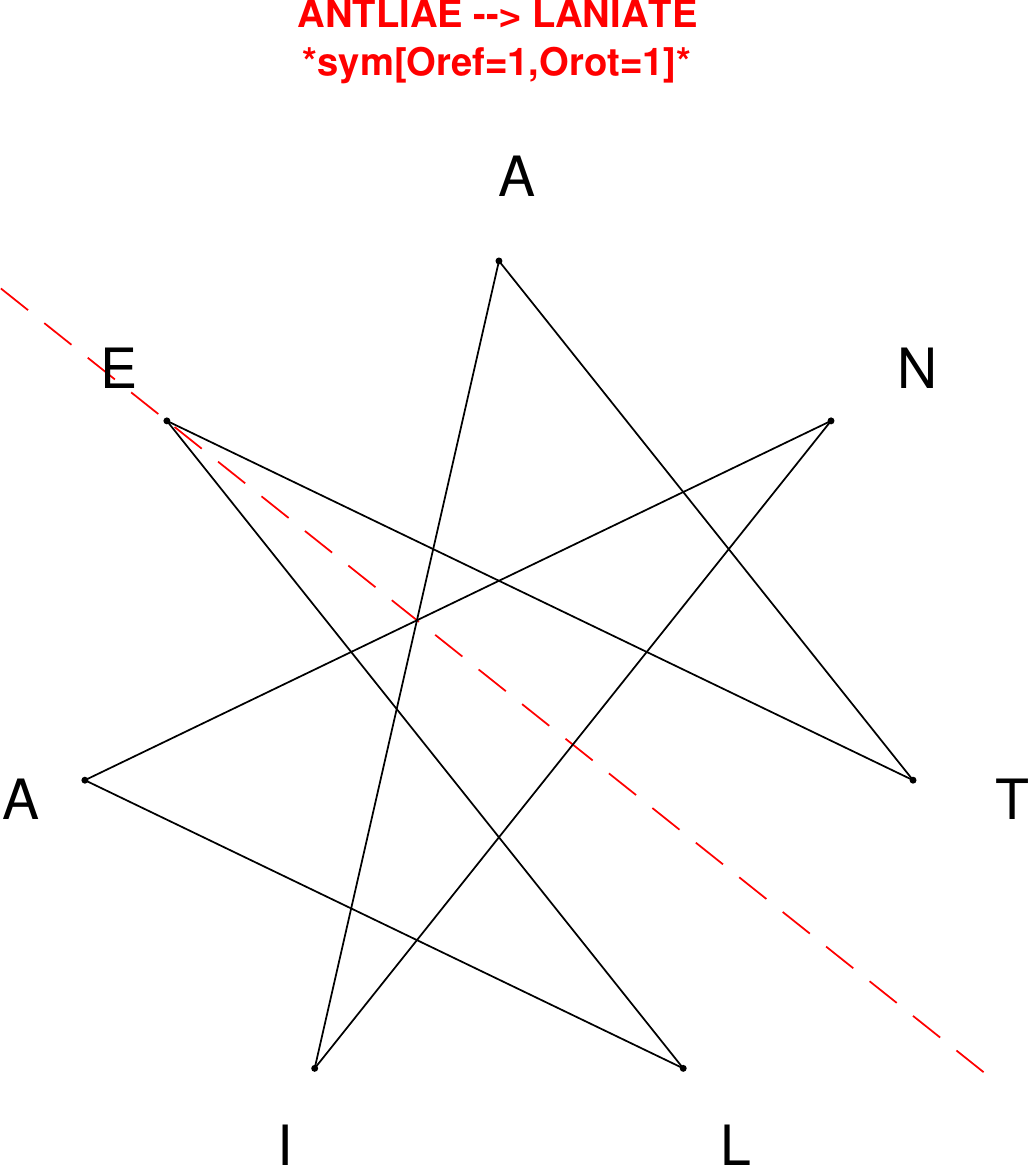}
\end{subfigure}
\hfill
\begin{subfigure}[T]{0.19\textwidth}
\centering
\includegraphics[width=\textwidth]{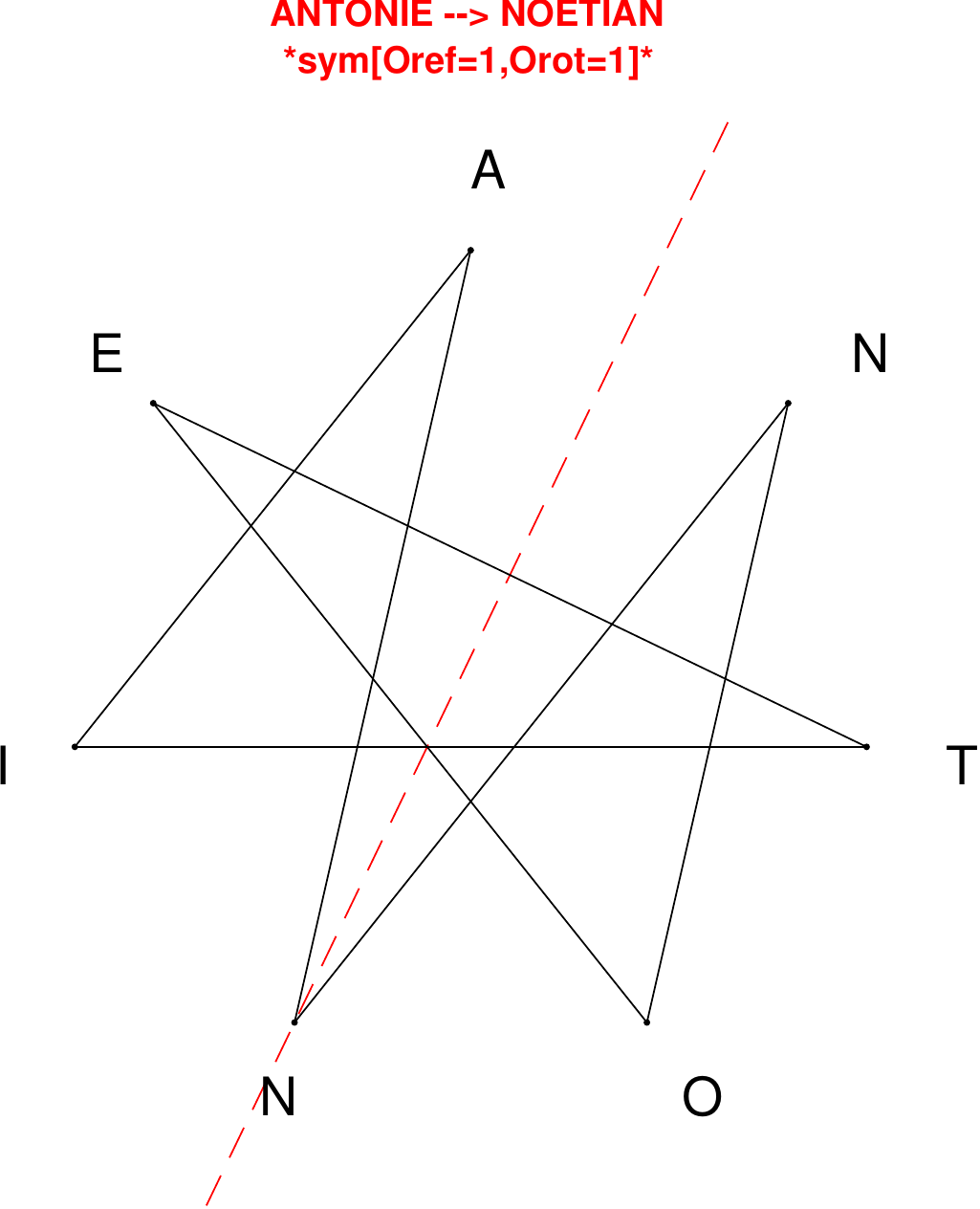}
\end{subfigure}
\hfill
\begin{subfigure}[T]{0.19\textwidth}
\centering
\includegraphics[width=\textwidth]{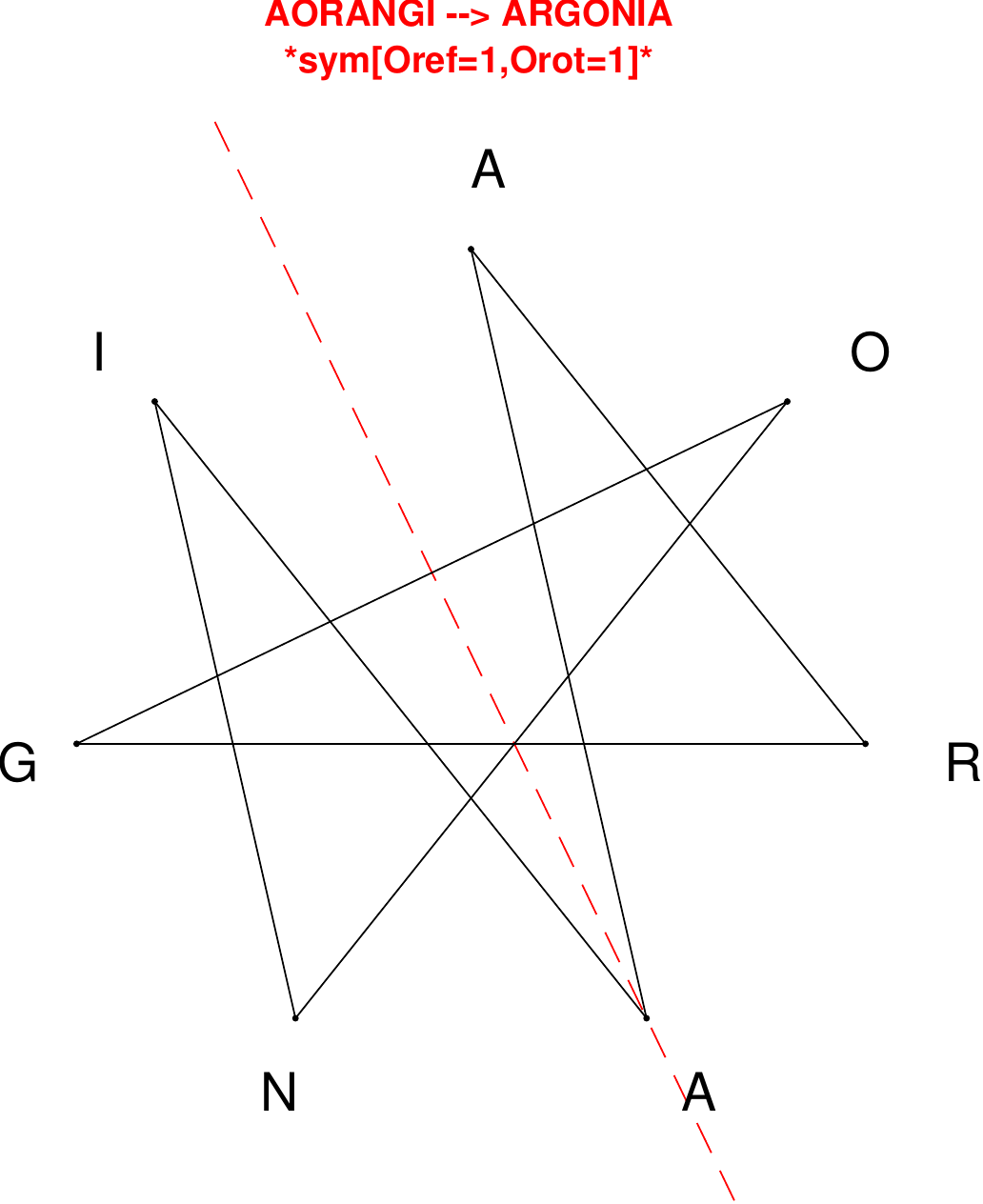}
\end{subfigure}
\end{figure}

\begin{figure}[H]
\centering
\begin{subfigure}[T]{0.19\textwidth}
\centering
\includegraphics[width=\textwidth]{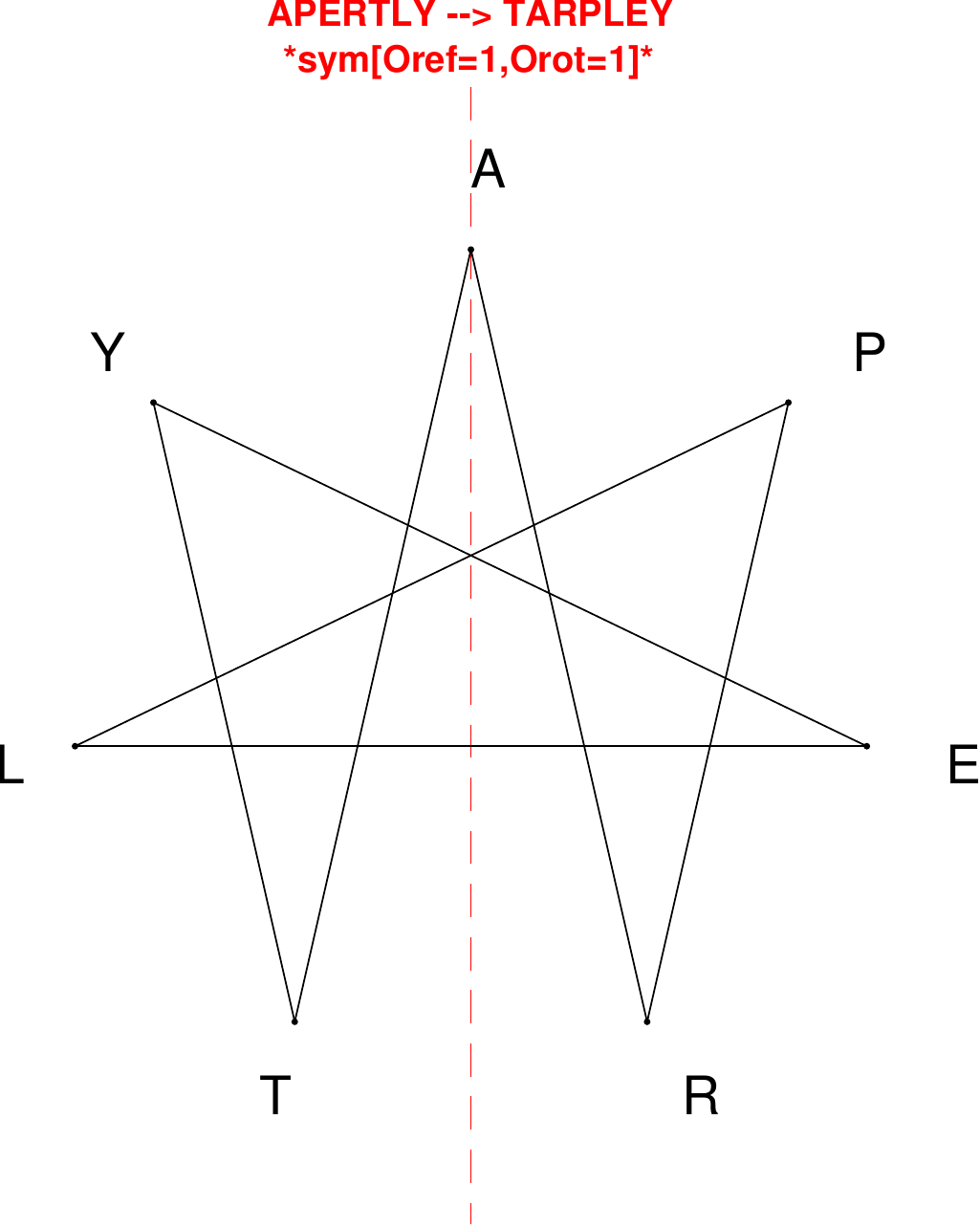}
\end{subfigure}
\hfill
\begin{subfigure}[T]{0.19\textwidth}
\centering
\includegraphics[width=\textwidth]{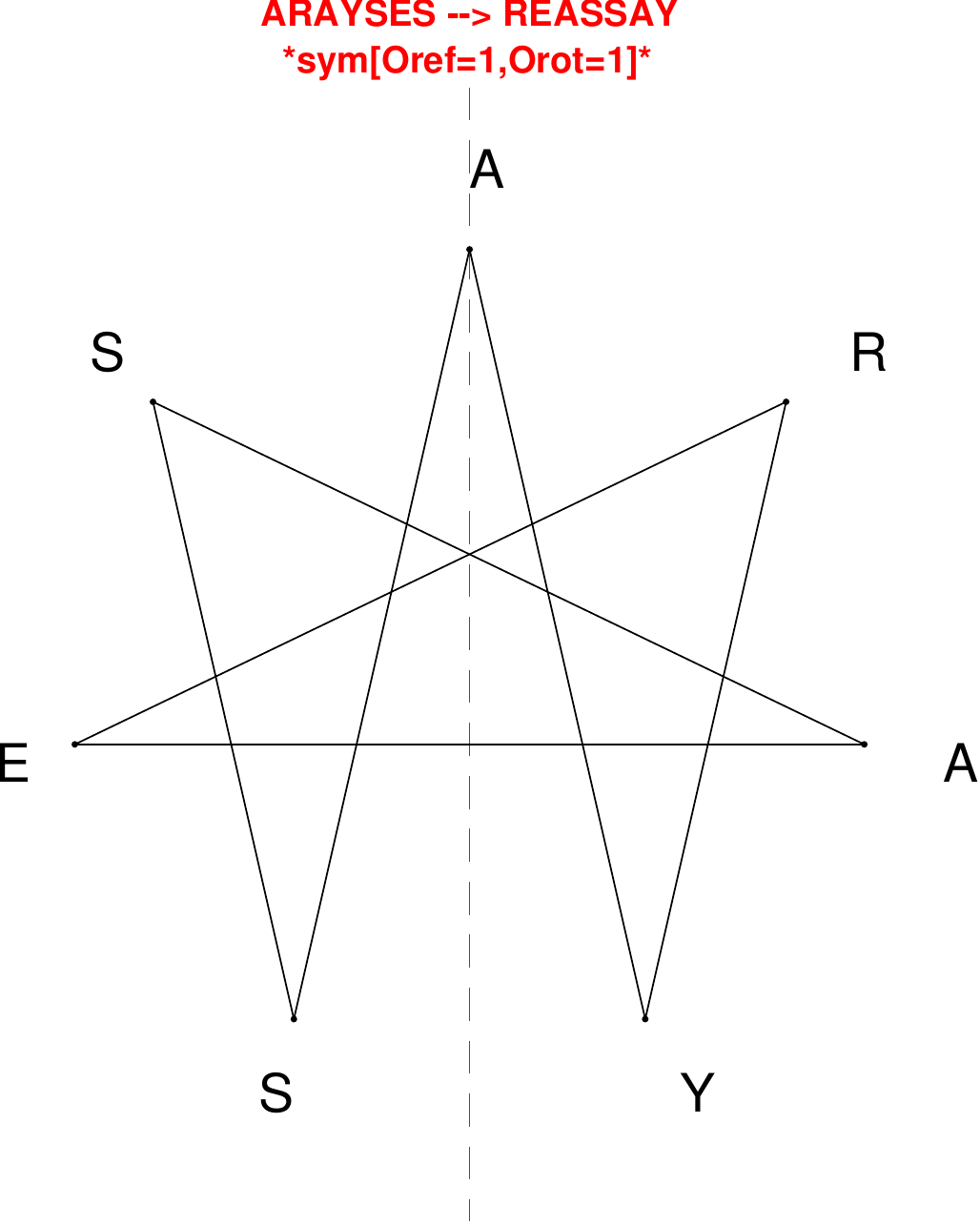}
\end{subfigure}
\hfill
\begin{subfigure}[T]{0.19\textwidth}
\centering
\includegraphics[width=\textwidth]{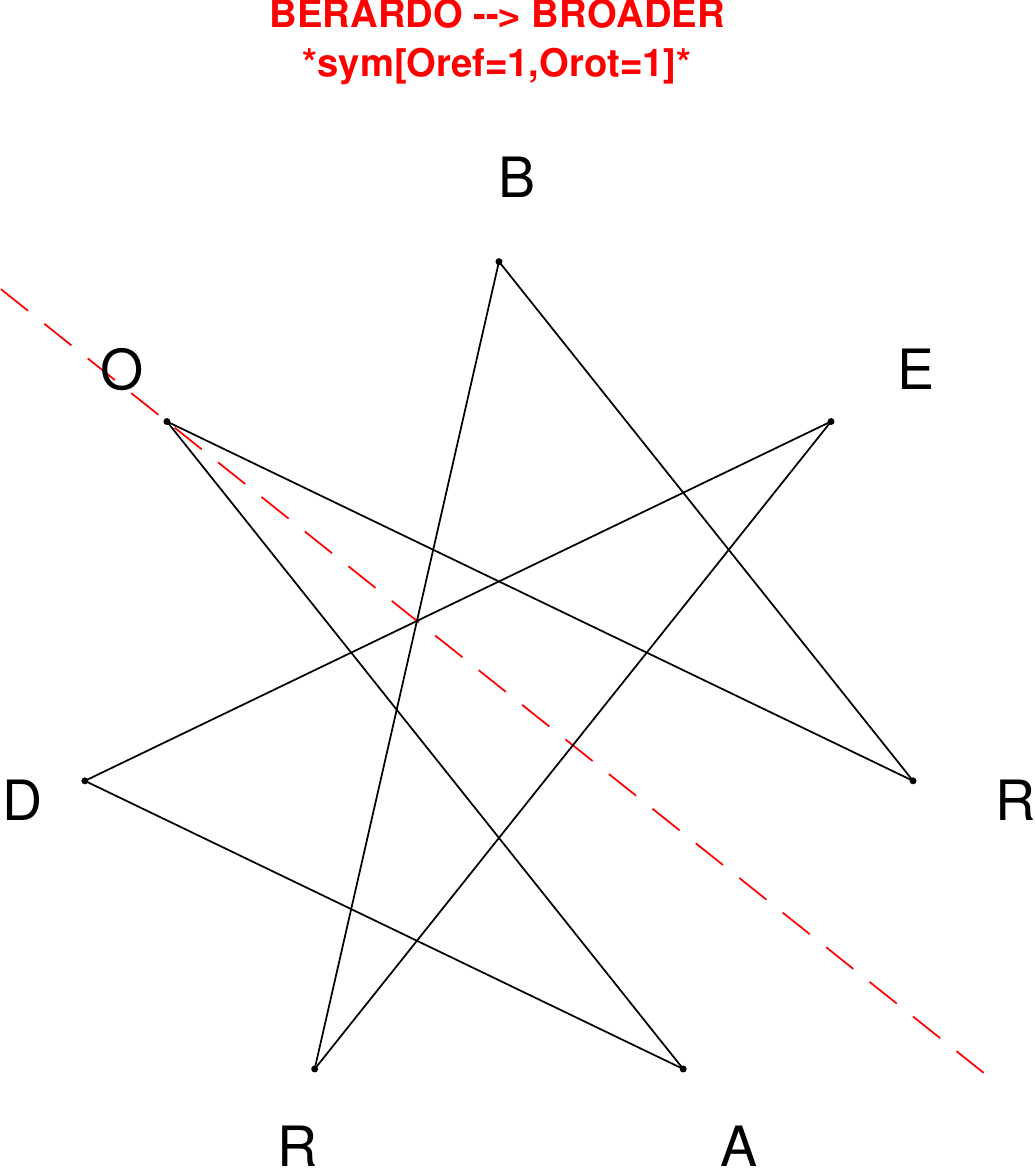}
\end{subfigure}
\hfill
\begin{subfigure}[T]{0.19\textwidth}
\centering
\includegraphics[width=\textwidth]{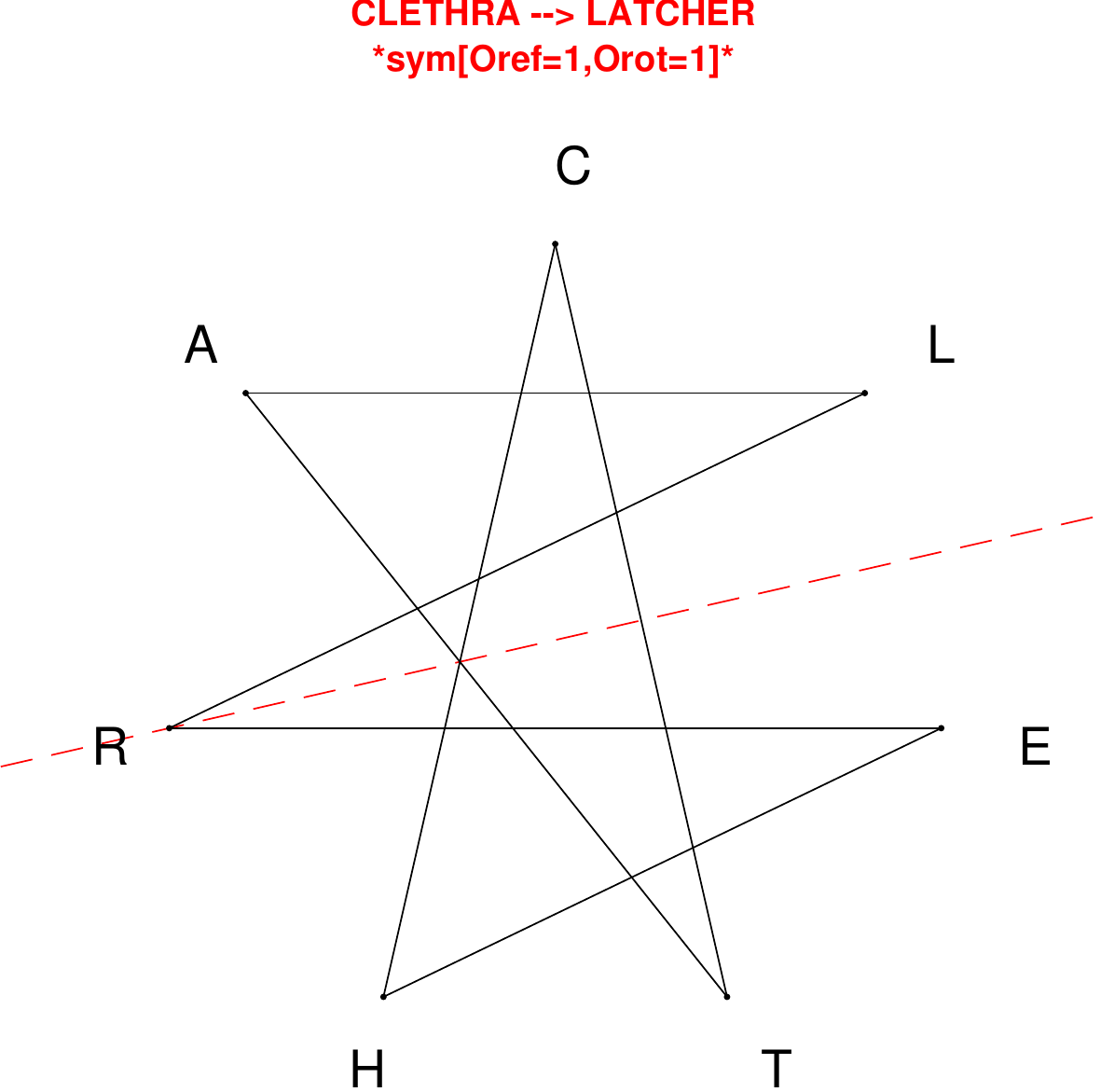}
\end{subfigure}
\hfill
\begin{subfigure}[T]{0.19\textwidth}
\centering
\includegraphics[width=\textwidth]{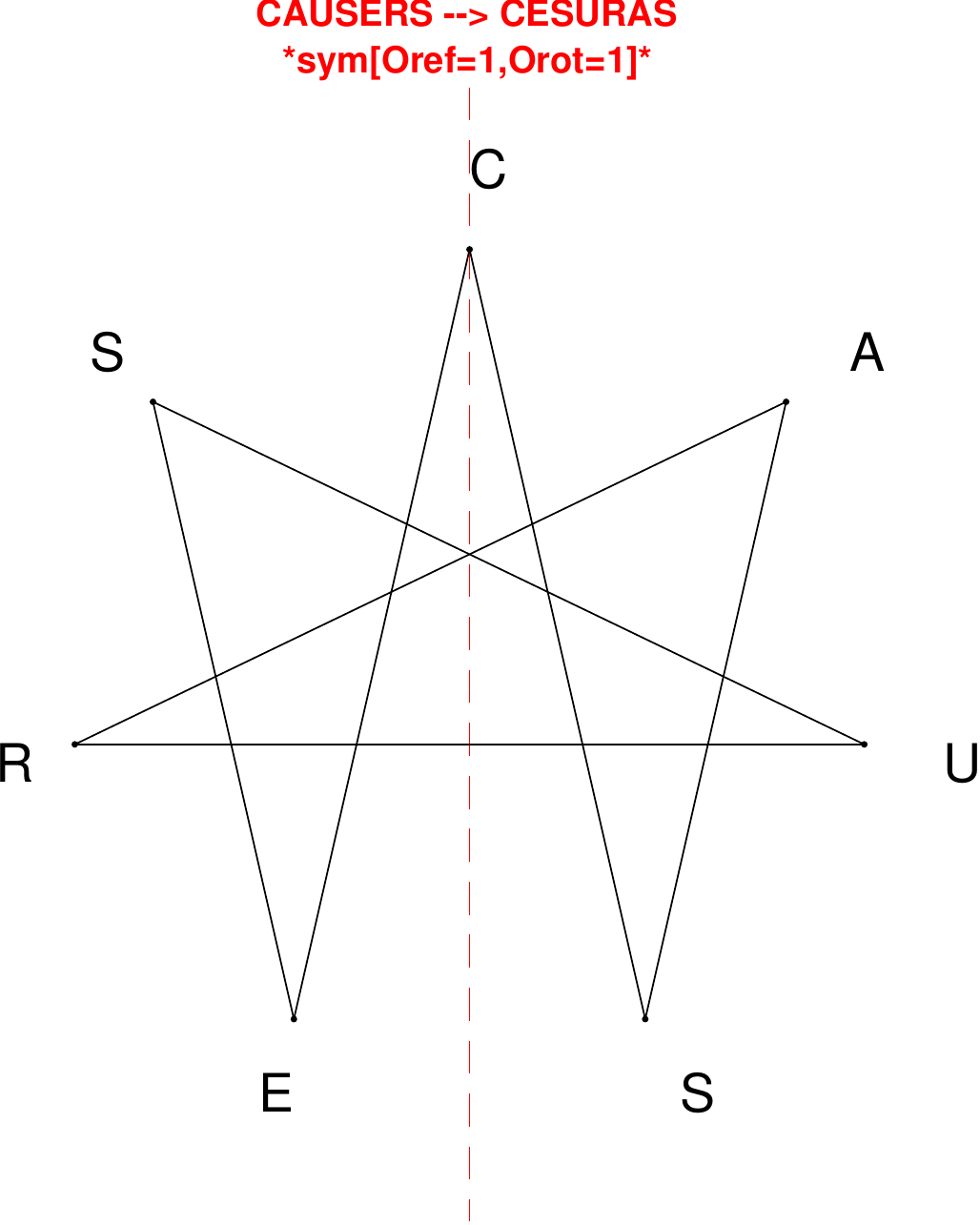}
\end{subfigure}
\end{figure}

\begin{figure}[H]
\centering
\begin{subfigure}[T]{0.19\textwidth}
\centering
\includegraphics[width=\textwidth]{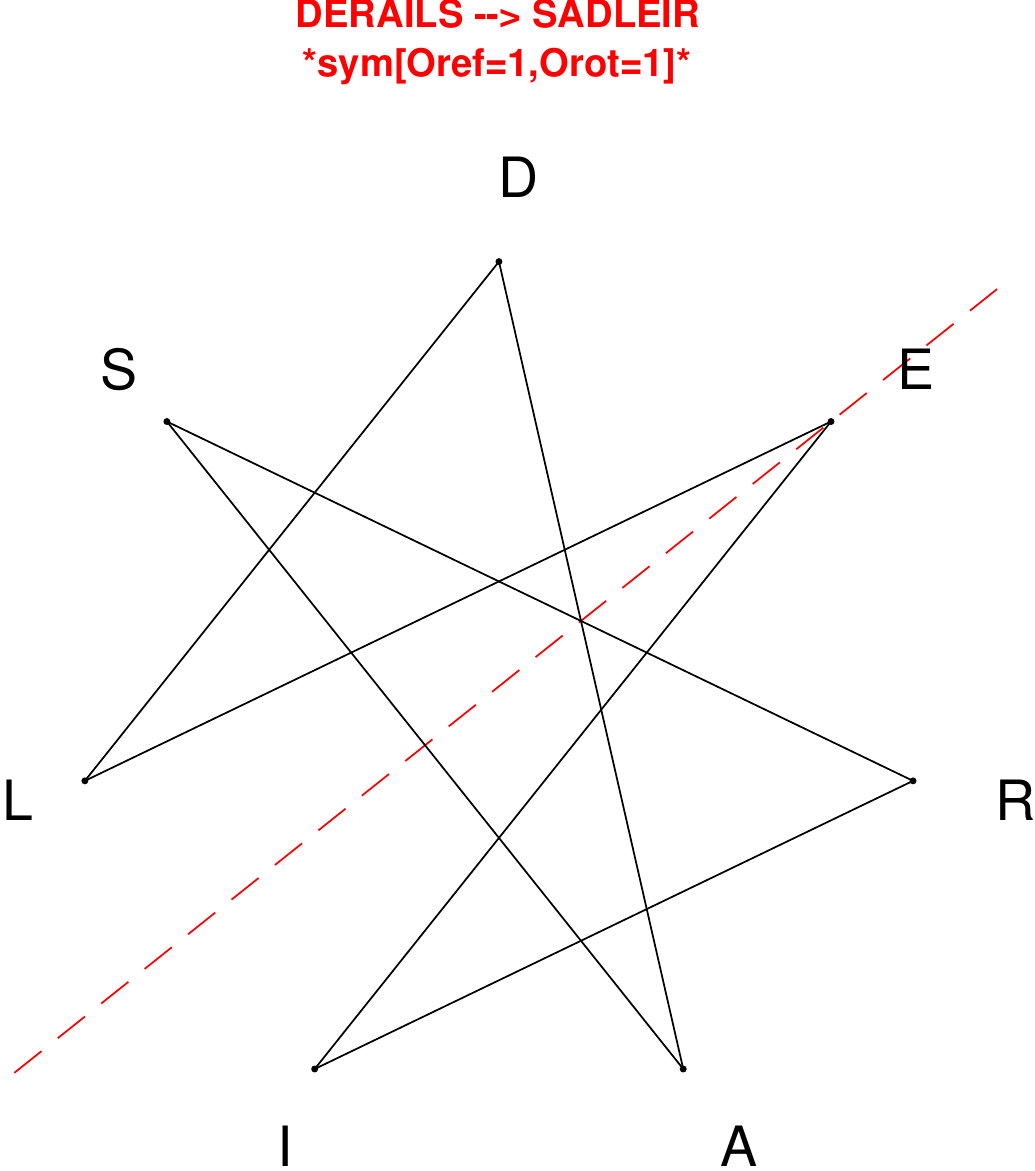}
\end{subfigure}
\hfill
\begin{subfigure}[T]{0.19\textwidth}
\centering
\includegraphics[width=\textwidth]{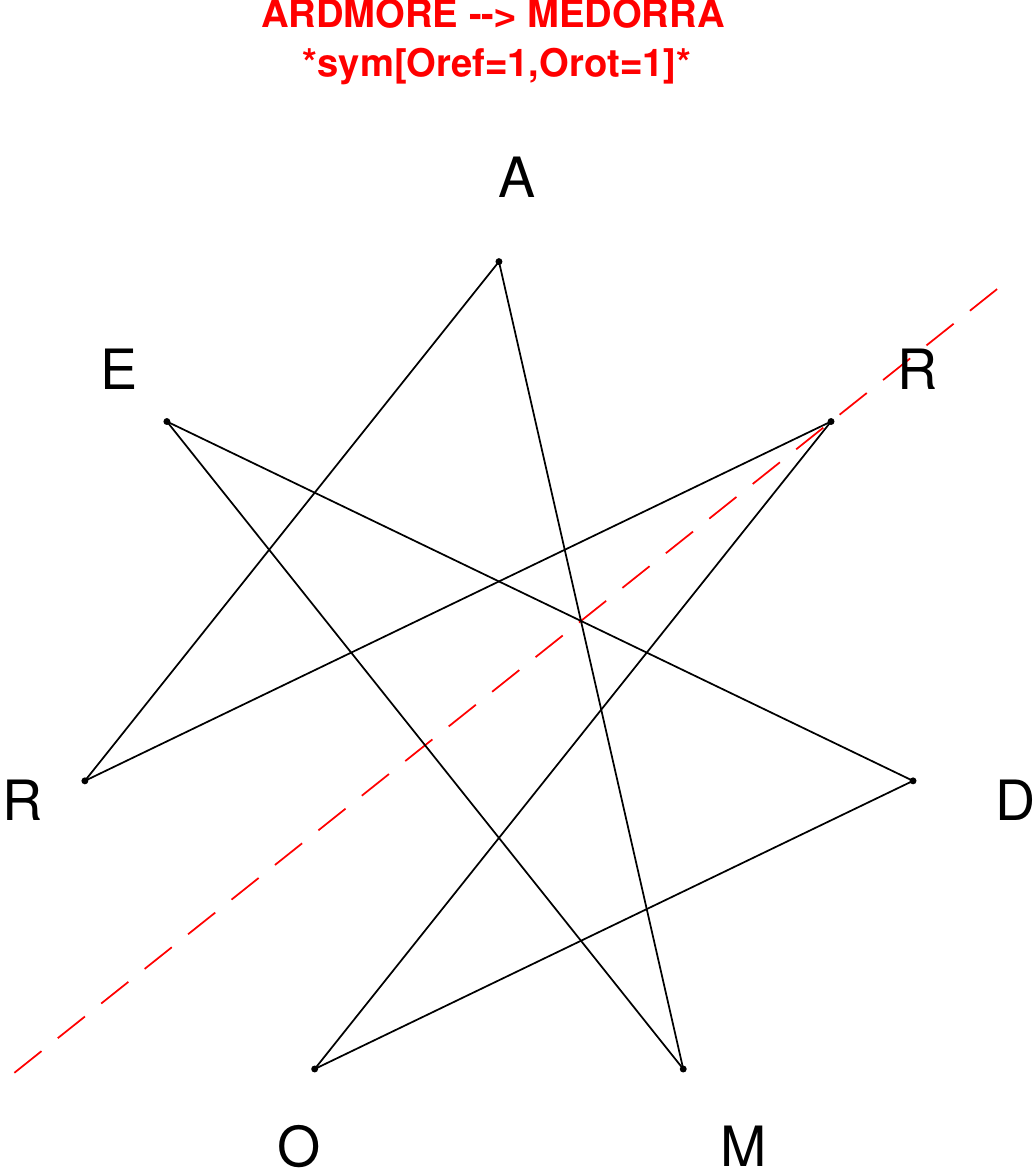}
\end{subfigure}
\hfill
\begin{subfigure}[T]{0.19\textwidth}
\centering
\includegraphics[width=\textwidth]{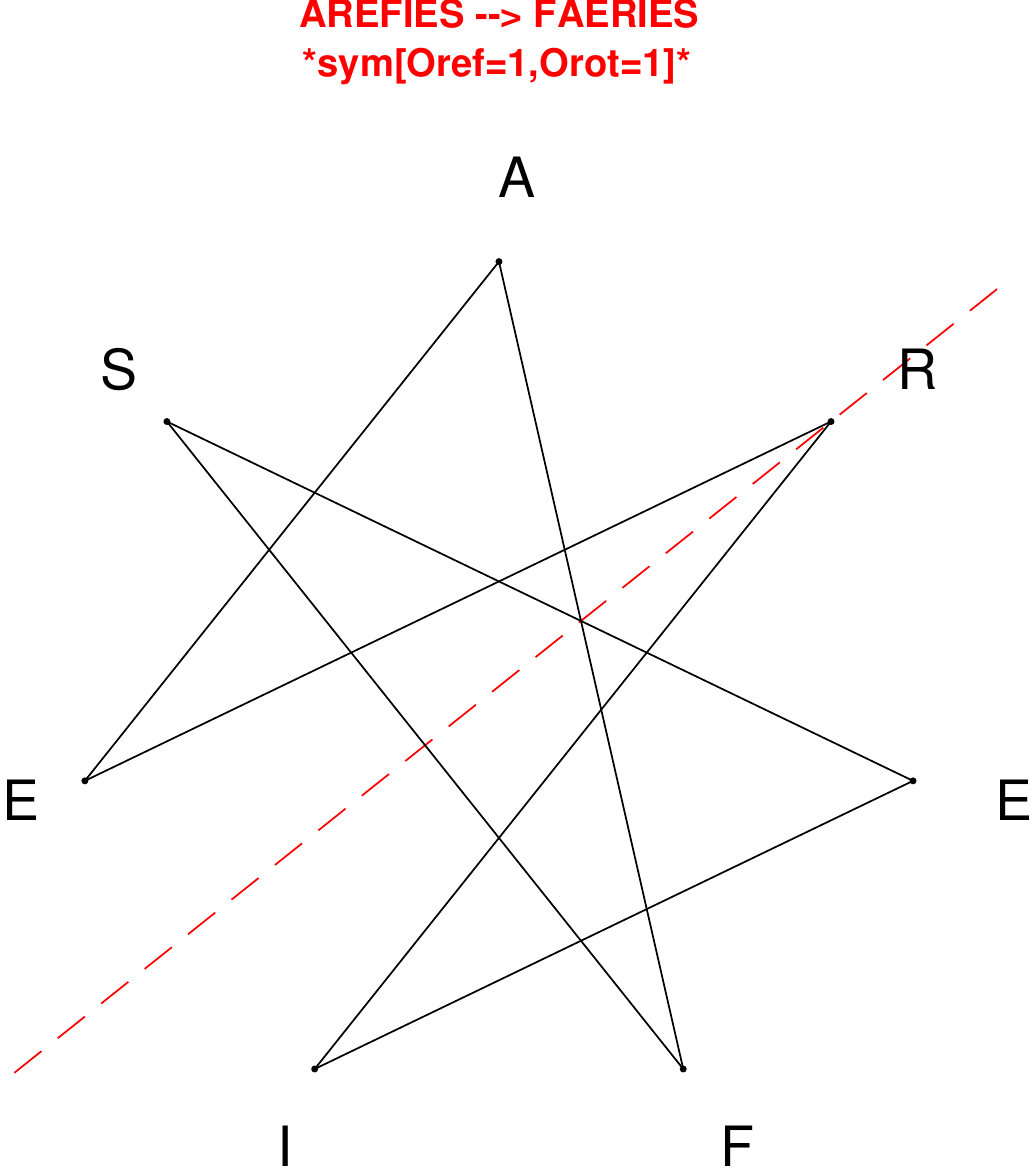}
\end{subfigure}
\hfill
\begin{subfigure}[T]{0.19\textwidth}
\centering
\includegraphics[width=\textwidth]{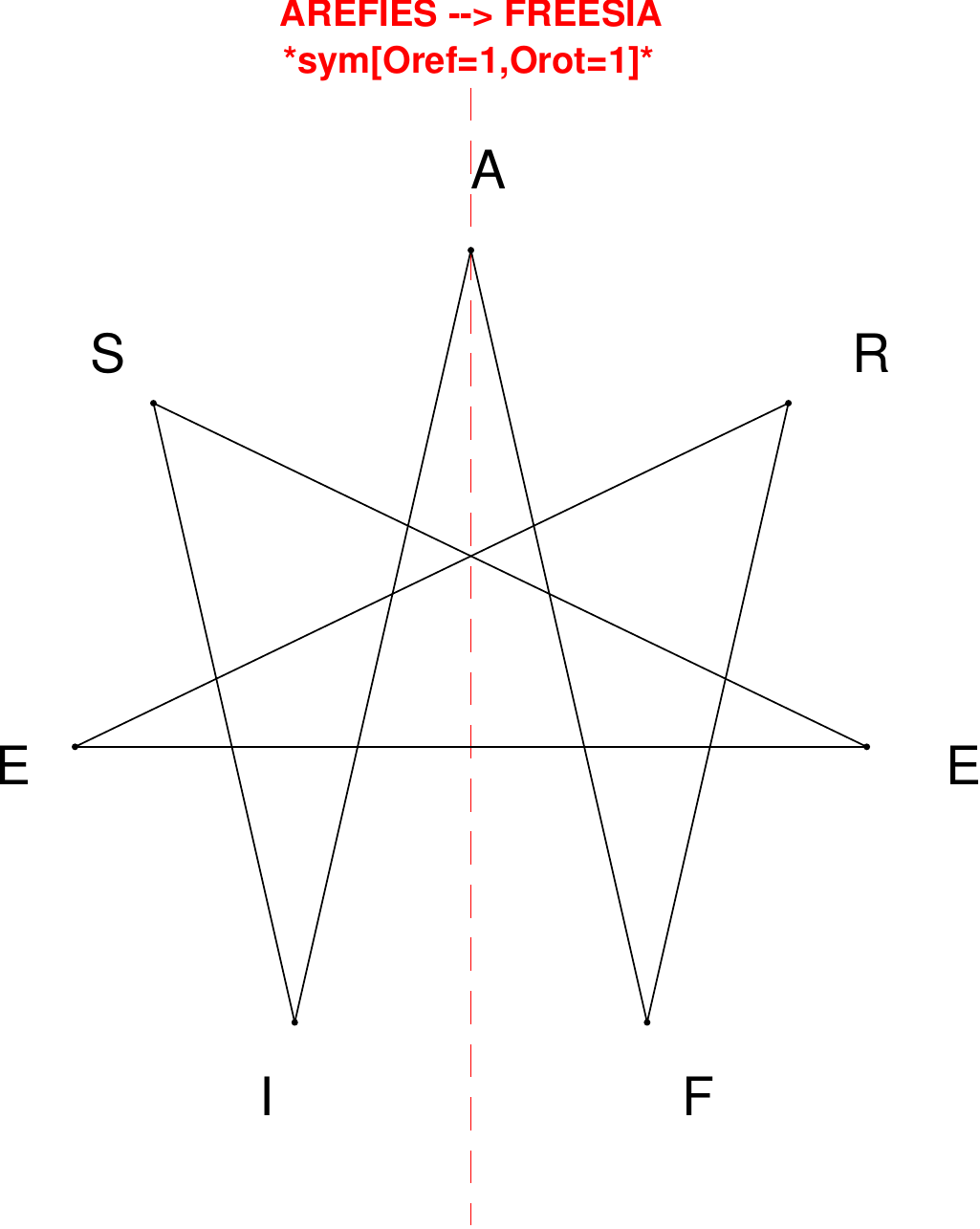}
\end{subfigure}
\hfill
\begin{subfigure}[T]{0.19\textwidth}
\centering
\includegraphics[width=\textwidth]{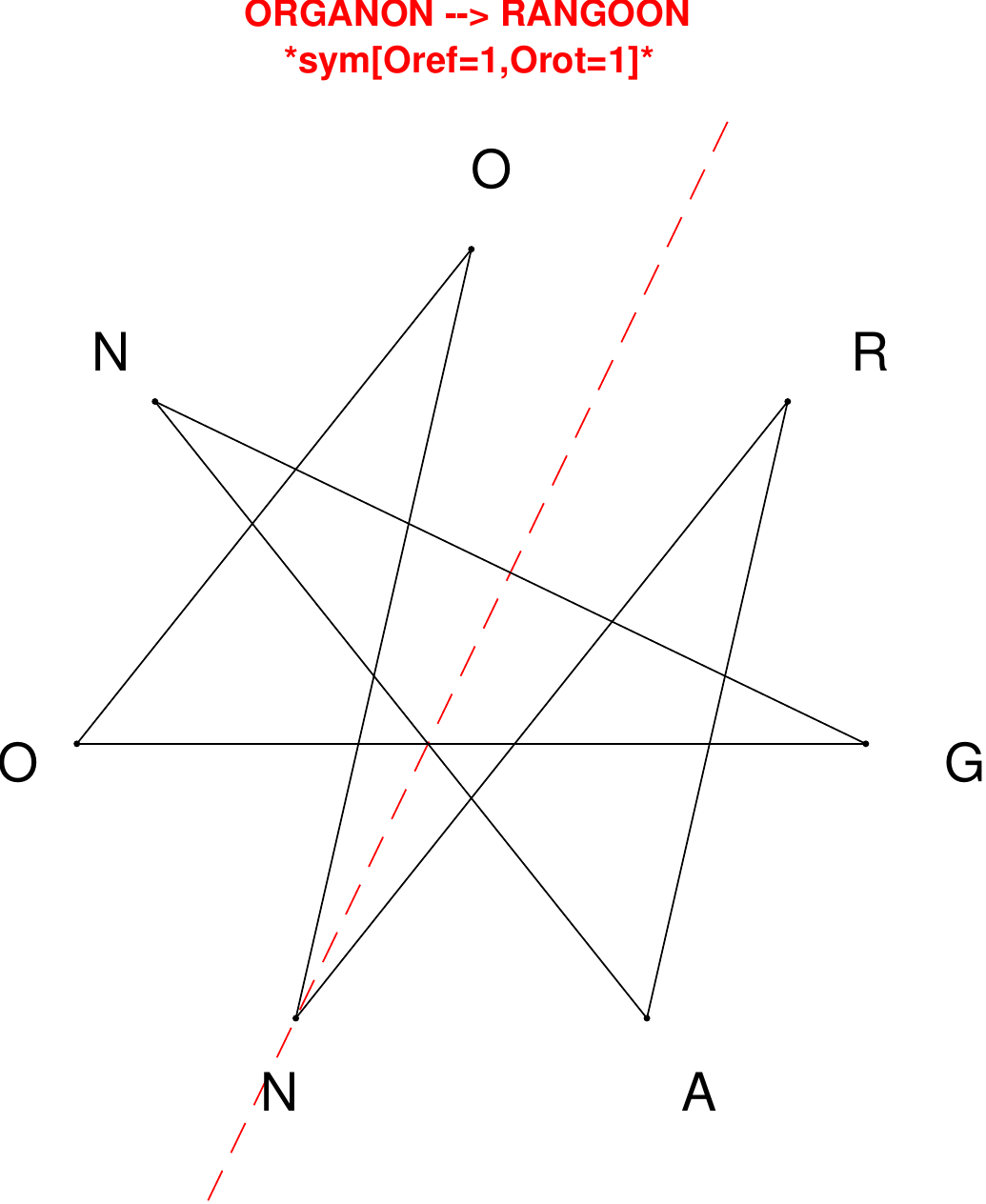}
\end{subfigure}
\end{figure}

\begin{figure}[H]
\centering
\begin{subfigure}[T]{0.19\textwidth}
\centering
\includegraphics[width=\textwidth]{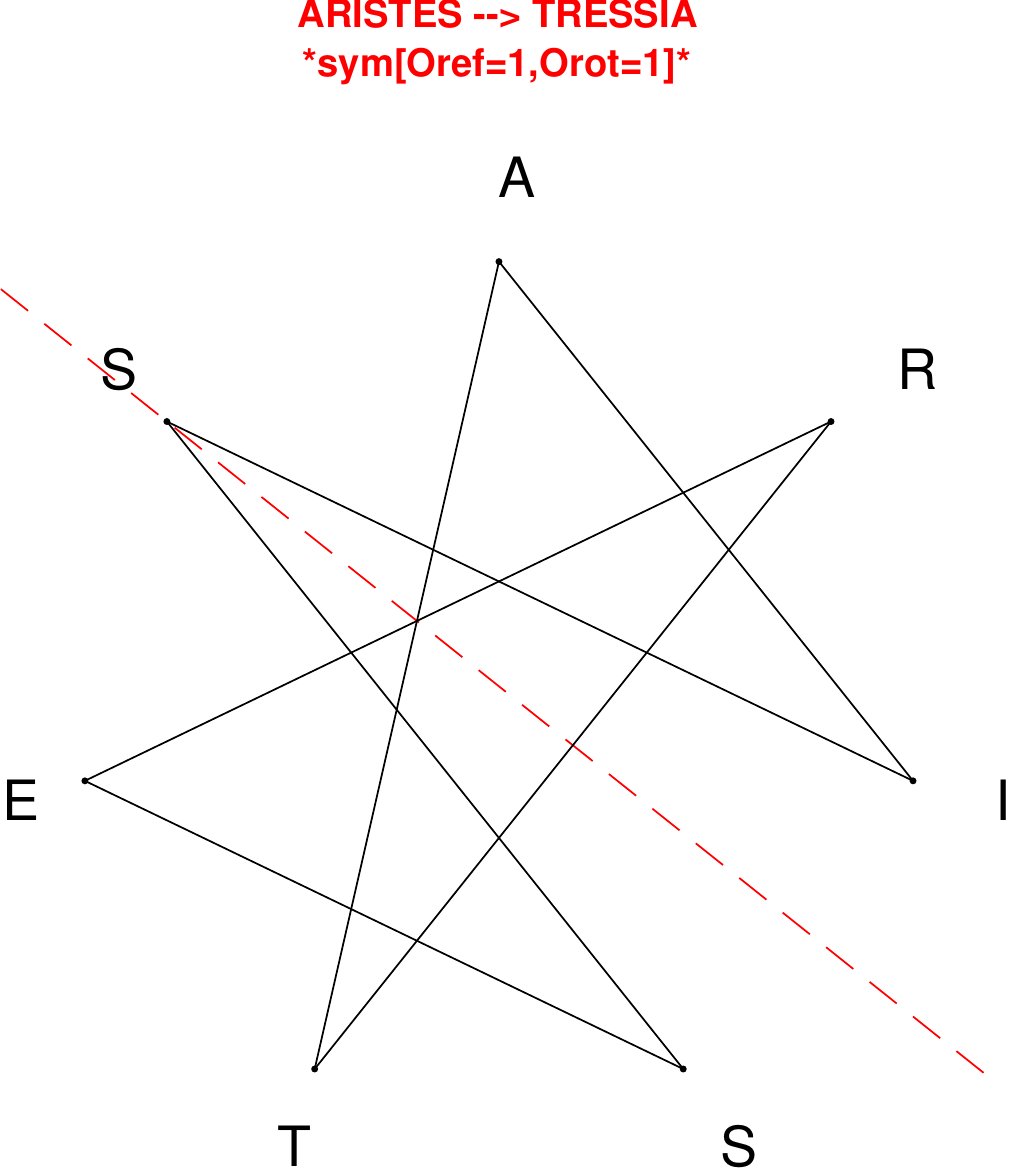}
\end{subfigure}
\hfill
\begin{subfigure}[T]{0.19\textwidth}
\centering
\includegraphics[width=\textwidth]{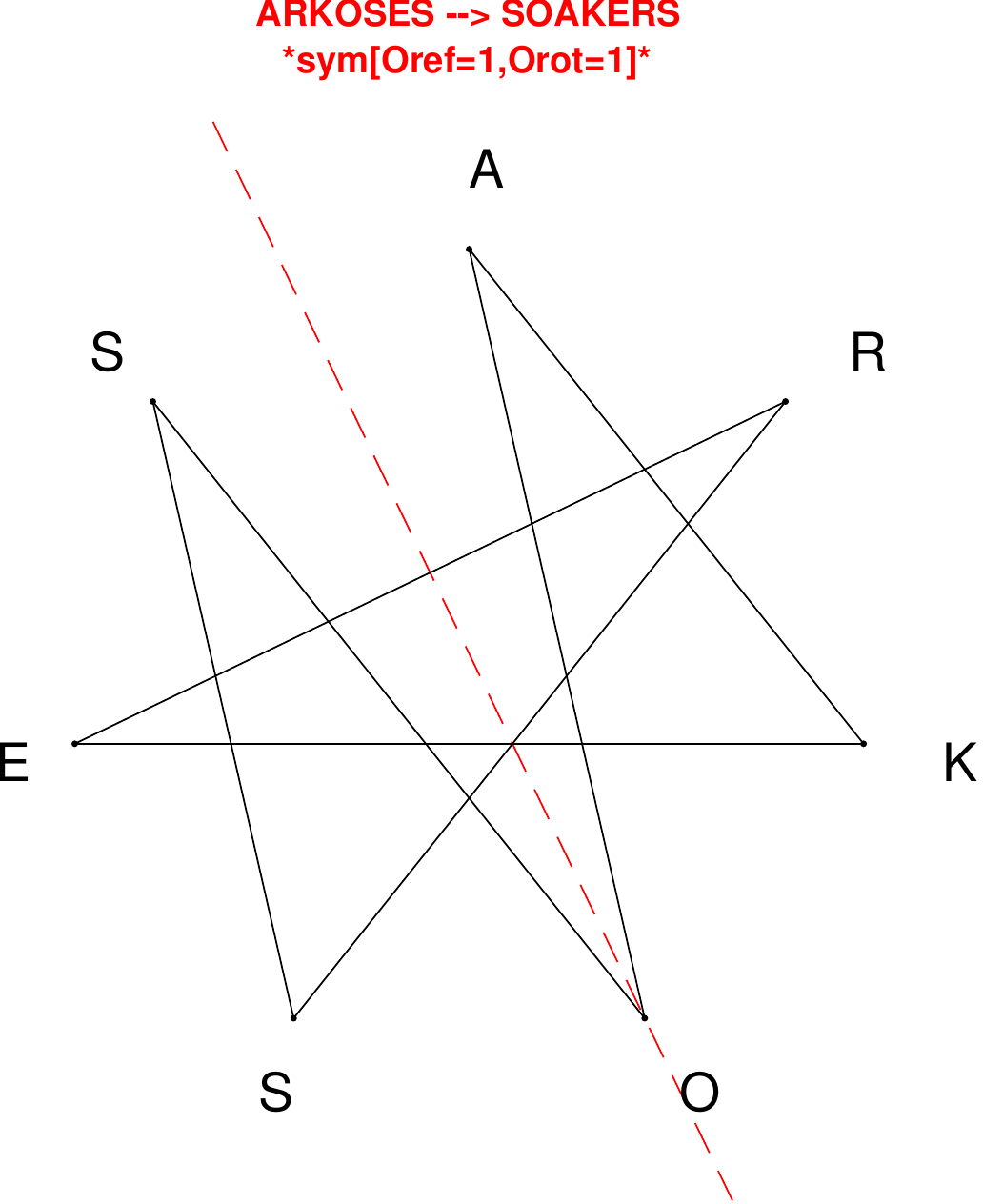}
\end{subfigure}
\hfill
\begin{subfigure}[T]{0.19\textwidth}
\centering
\includegraphics[width=\textwidth]{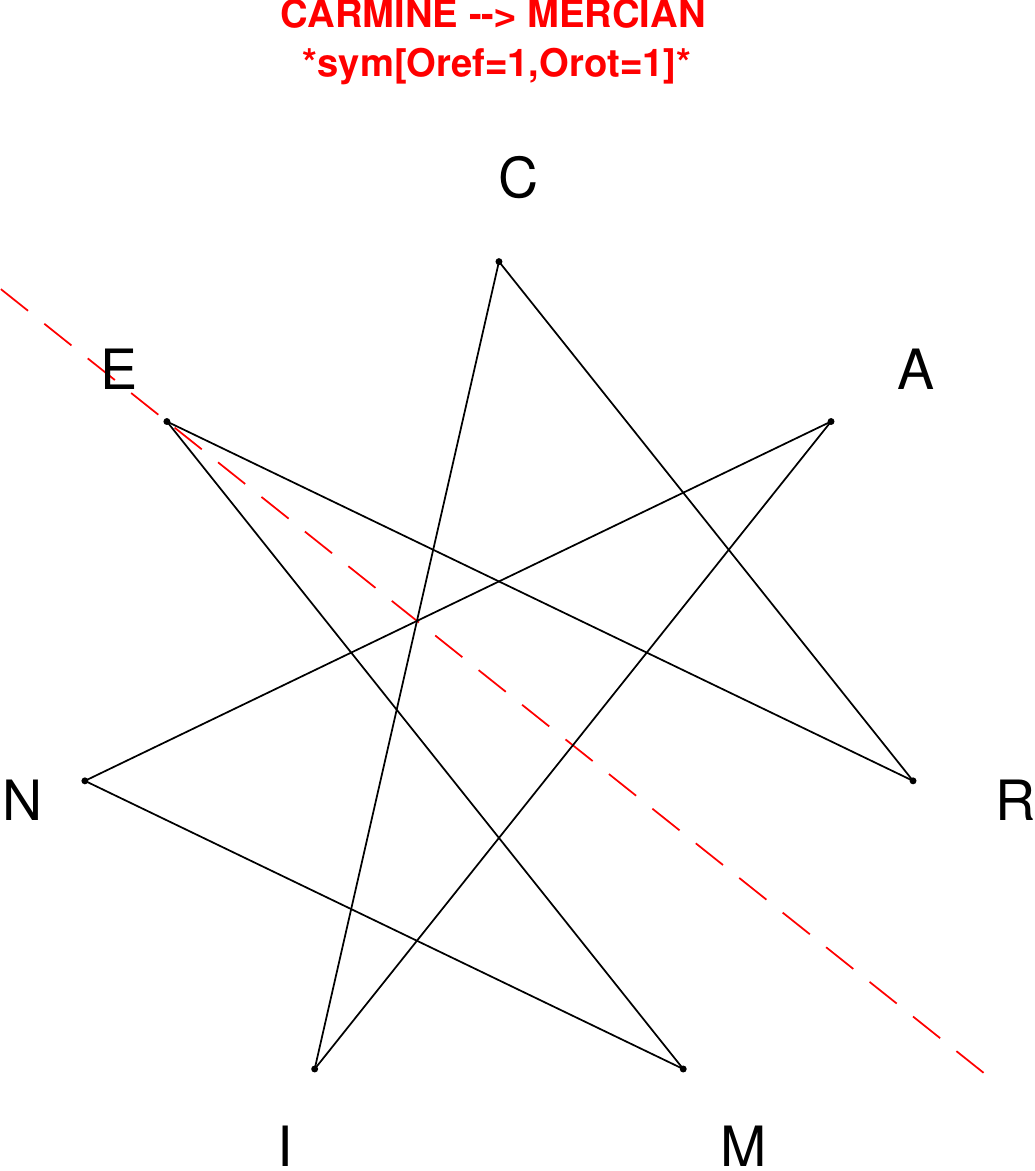}
\end{subfigure}
\hfill
\begin{subfigure}[T]{0.19\textwidth}
\centering
\includegraphics[width=\textwidth]{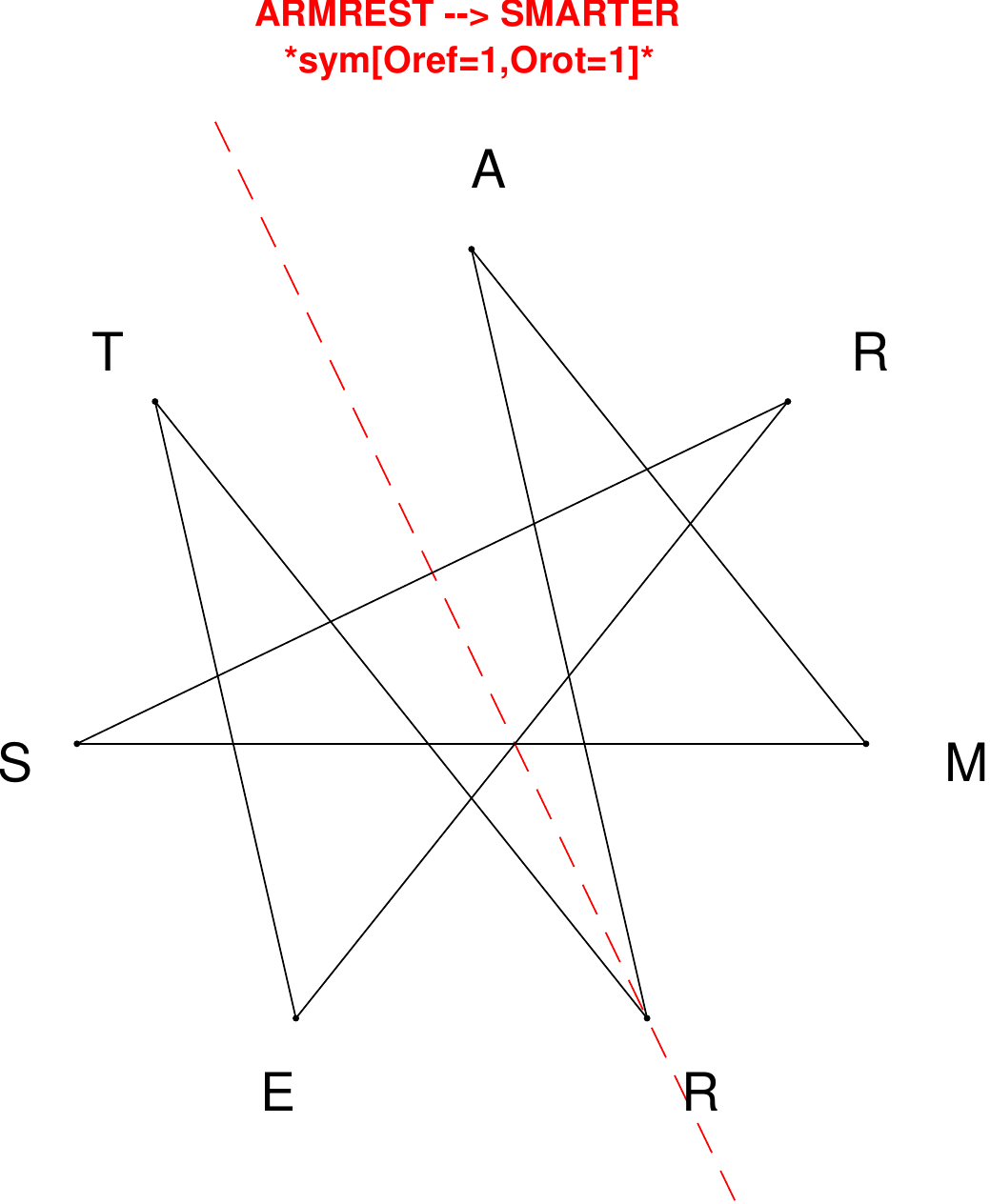}
\end{subfigure}
\hfill
\begin{subfigure}[T]{0.19\textwidth}
\centering
\includegraphics[width=\textwidth]{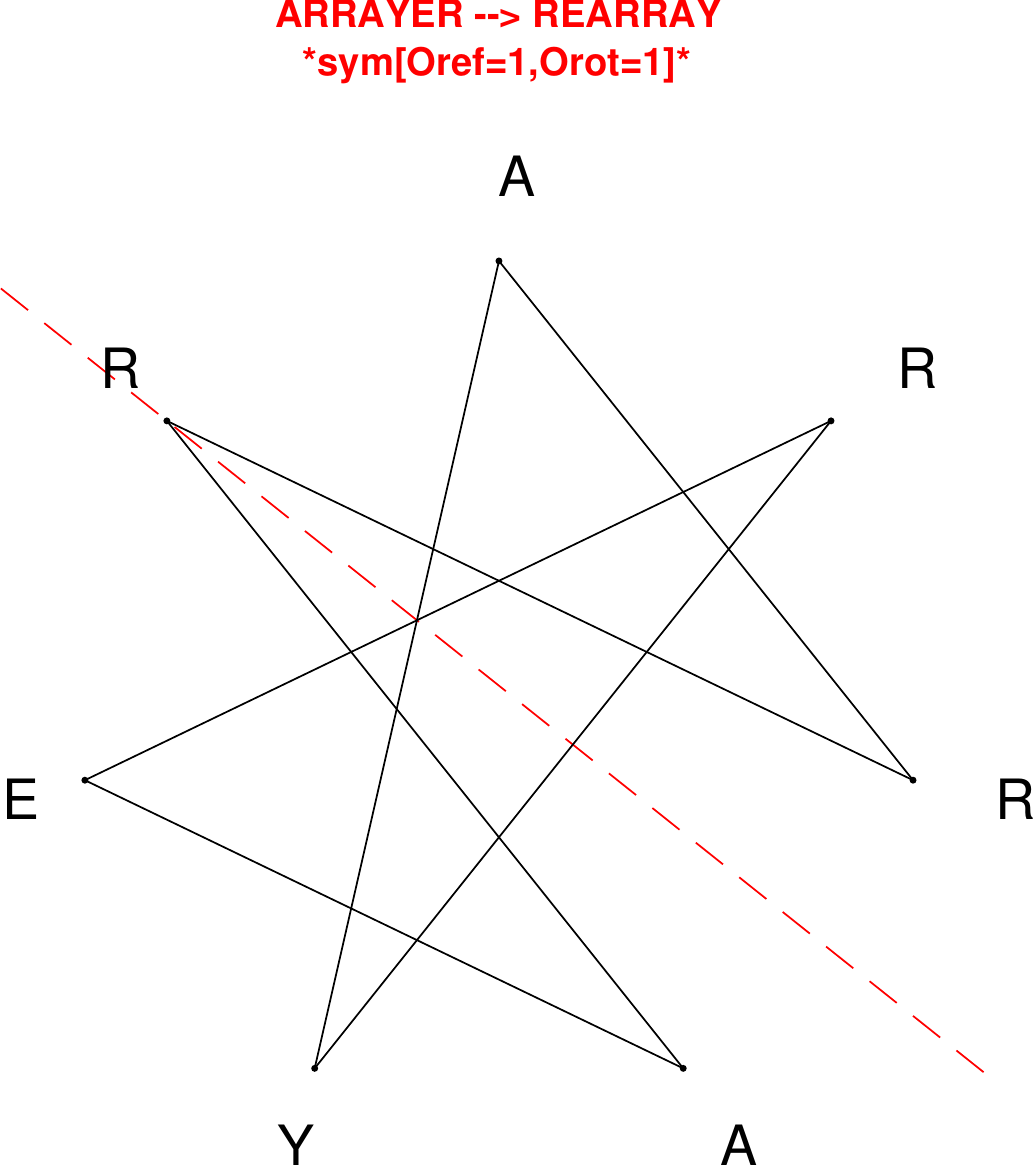}
\end{subfigure}
\end{figure}

\begin{figure}[H]
\centering
\begin{subfigure}[T]{0.19\textwidth}
\centering
\includegraphics[width=\textwidth]{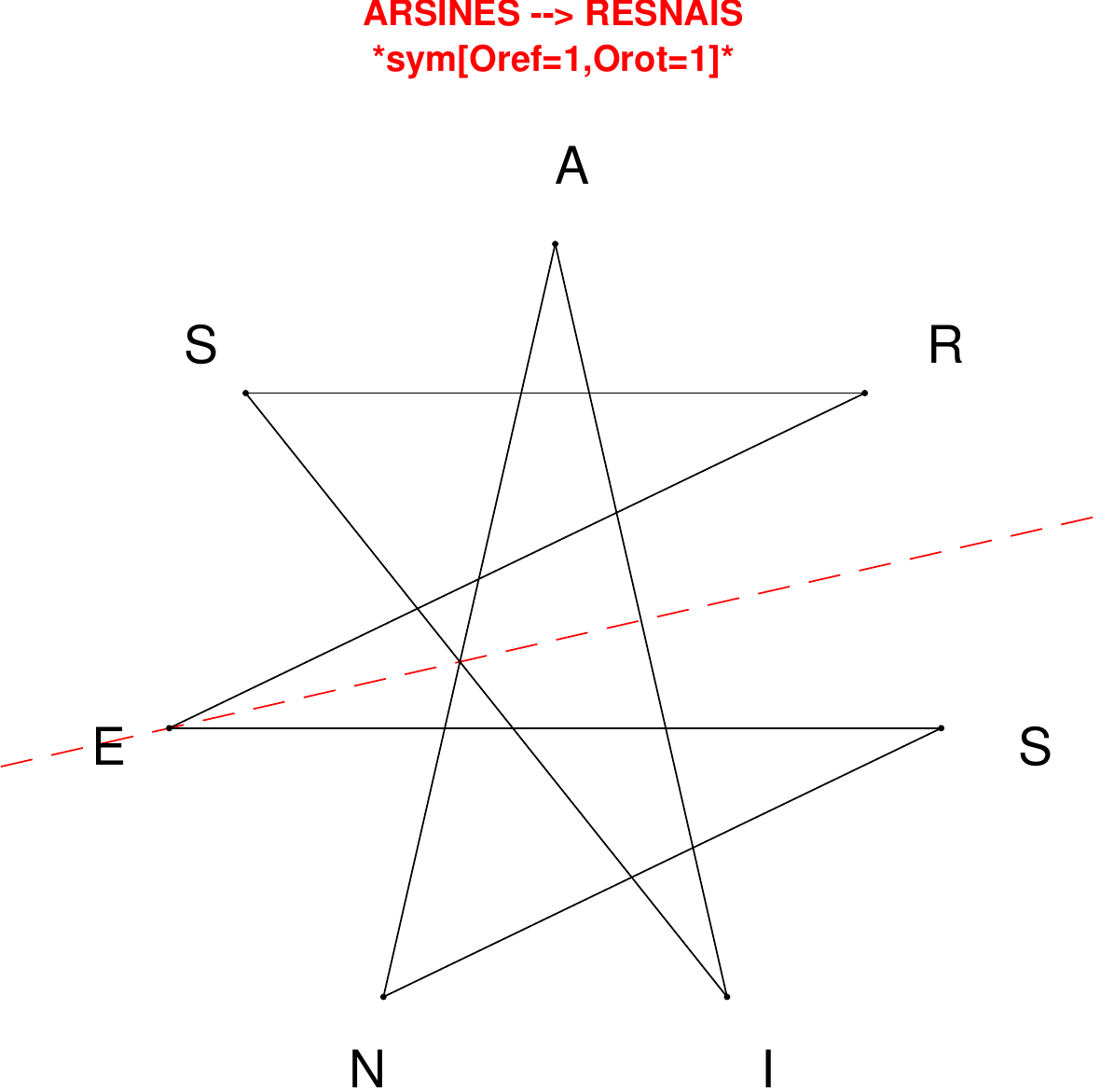}
\end{subfigure}
\hfill
\begin{subfigure}[T]{0.19\textwidth}
\centering
\includegraphics[width=\textwidth]{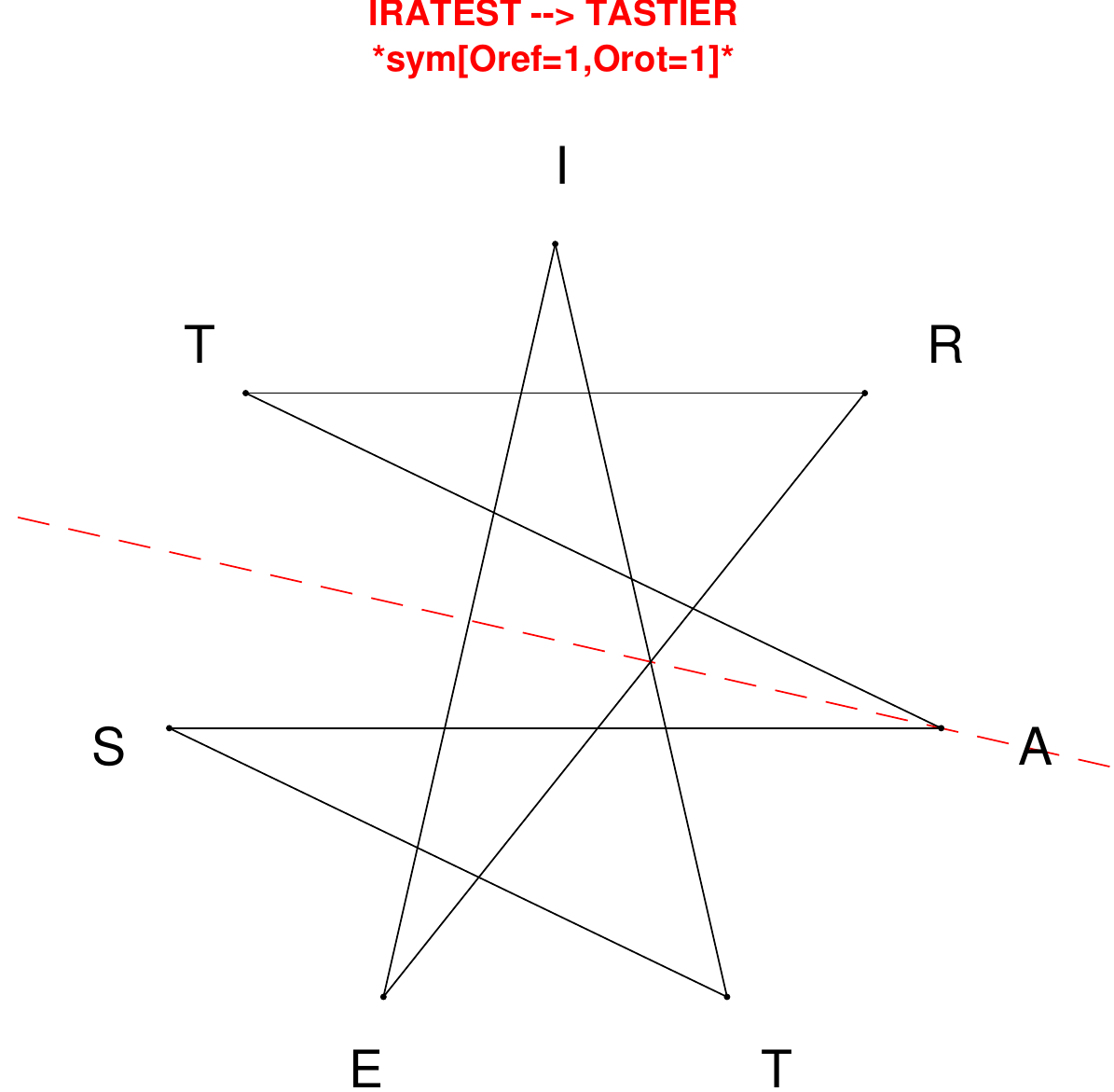}
\end{subfigure}
\hfill
\begin{subfigure}[T]{0.19\textwidth}
\centering
\includegraphics[width=\textwidth]{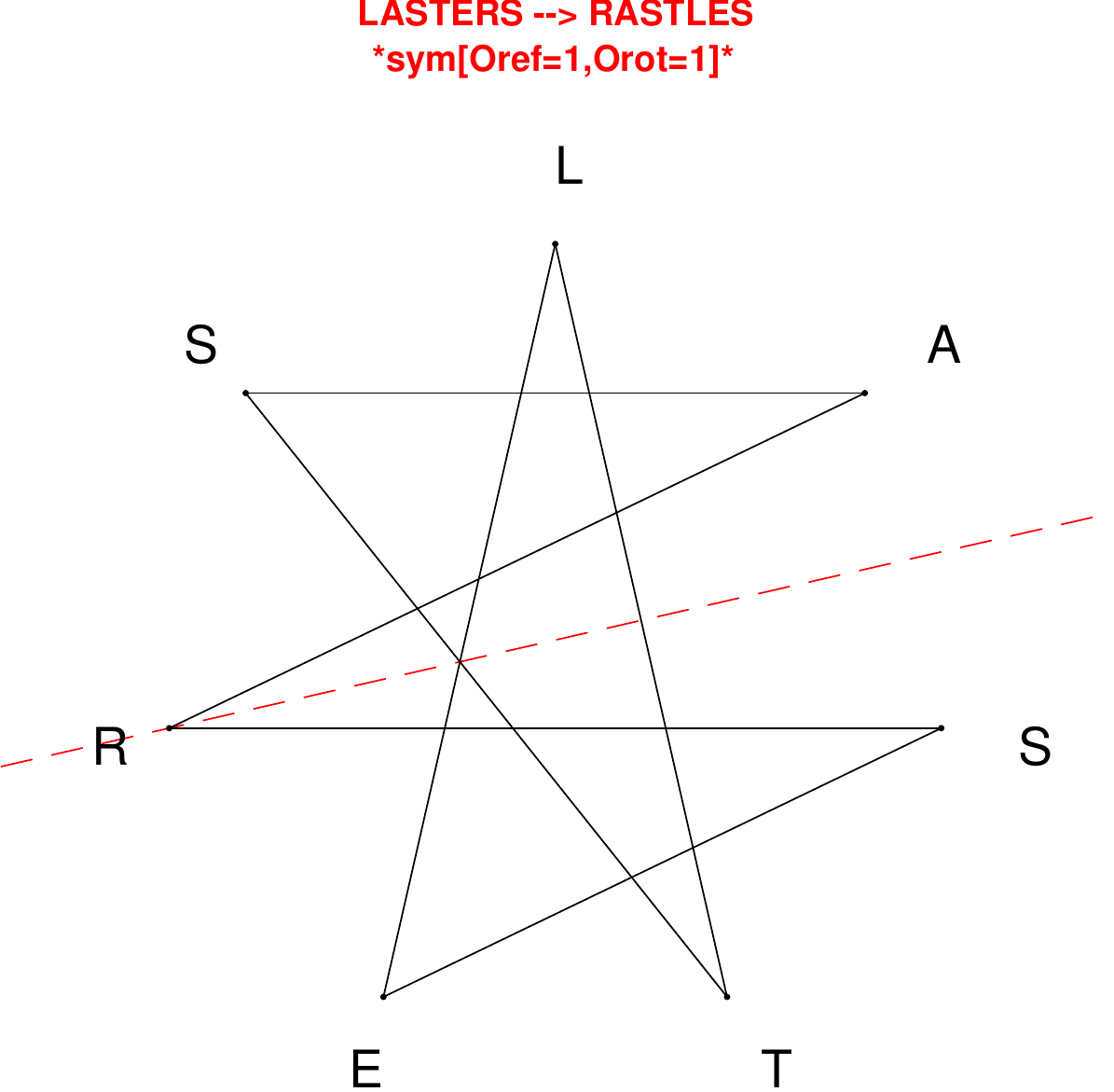}
\end{subfigure}
\hfill
\begin{subfigure}[T]{0.19\textwidth}
\centering
\includegraphics[width=\textwidth]{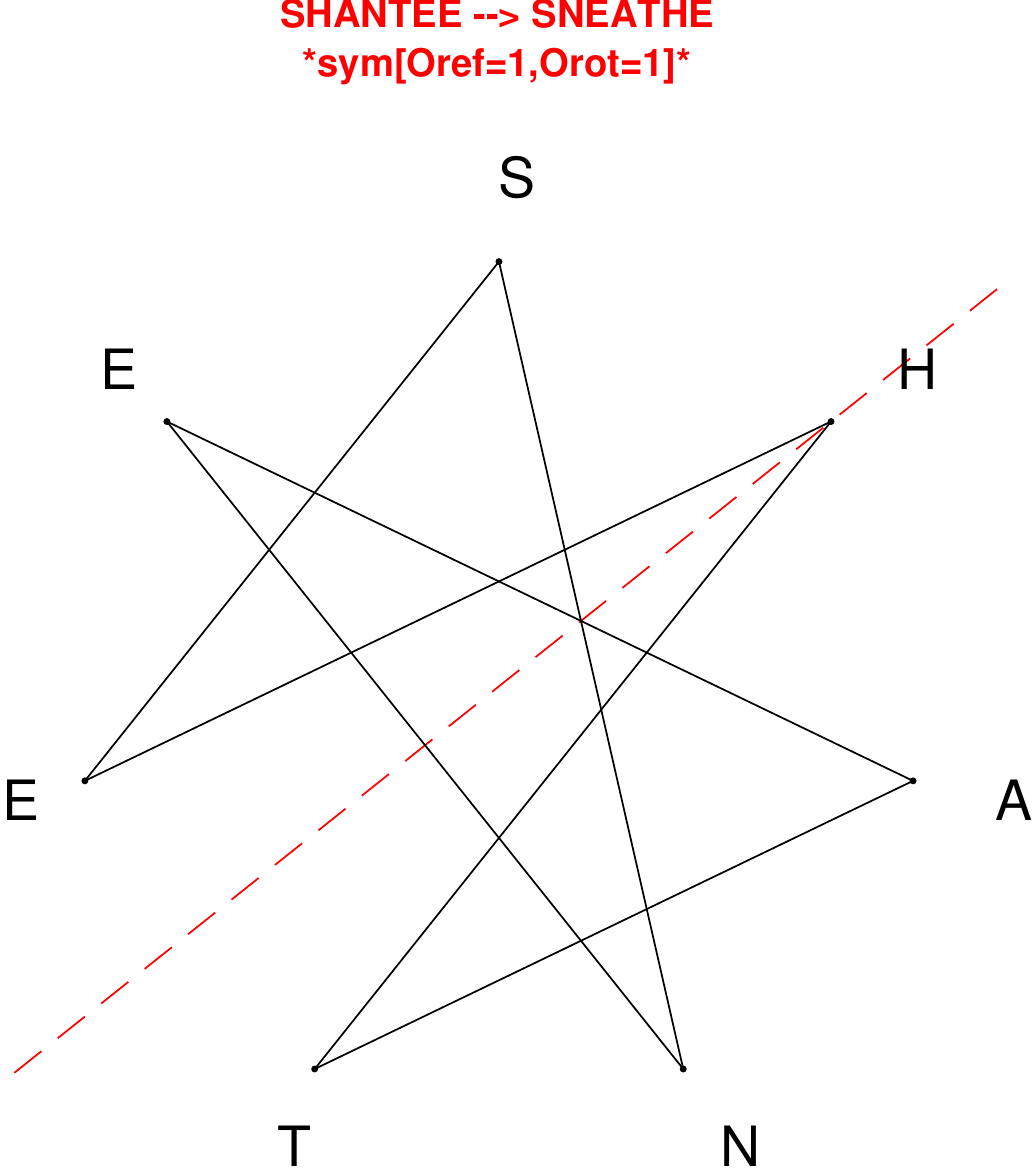}
\end{subfigure}
\hfill
\begin{subfigure}[T]{0.19\textwidth}
\centering
\includegraphics[width=\textwidth]{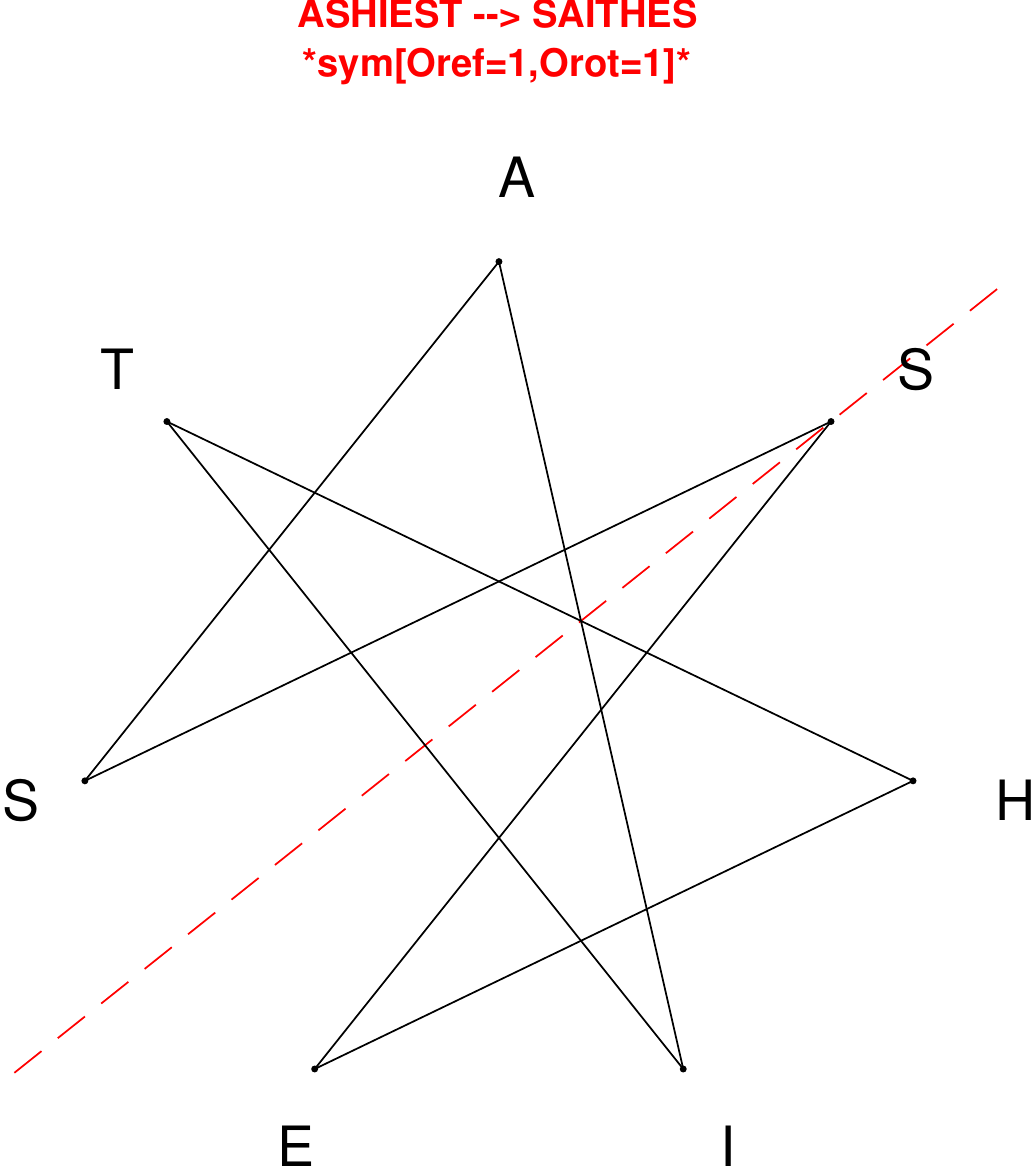}
\end{subfigure}
\end{figure}

\begin{figure}[H]
\centering
\begin{subfigure}[T]{0.19\textwidth}
\centering
\includegraphics[width=\textwidth]{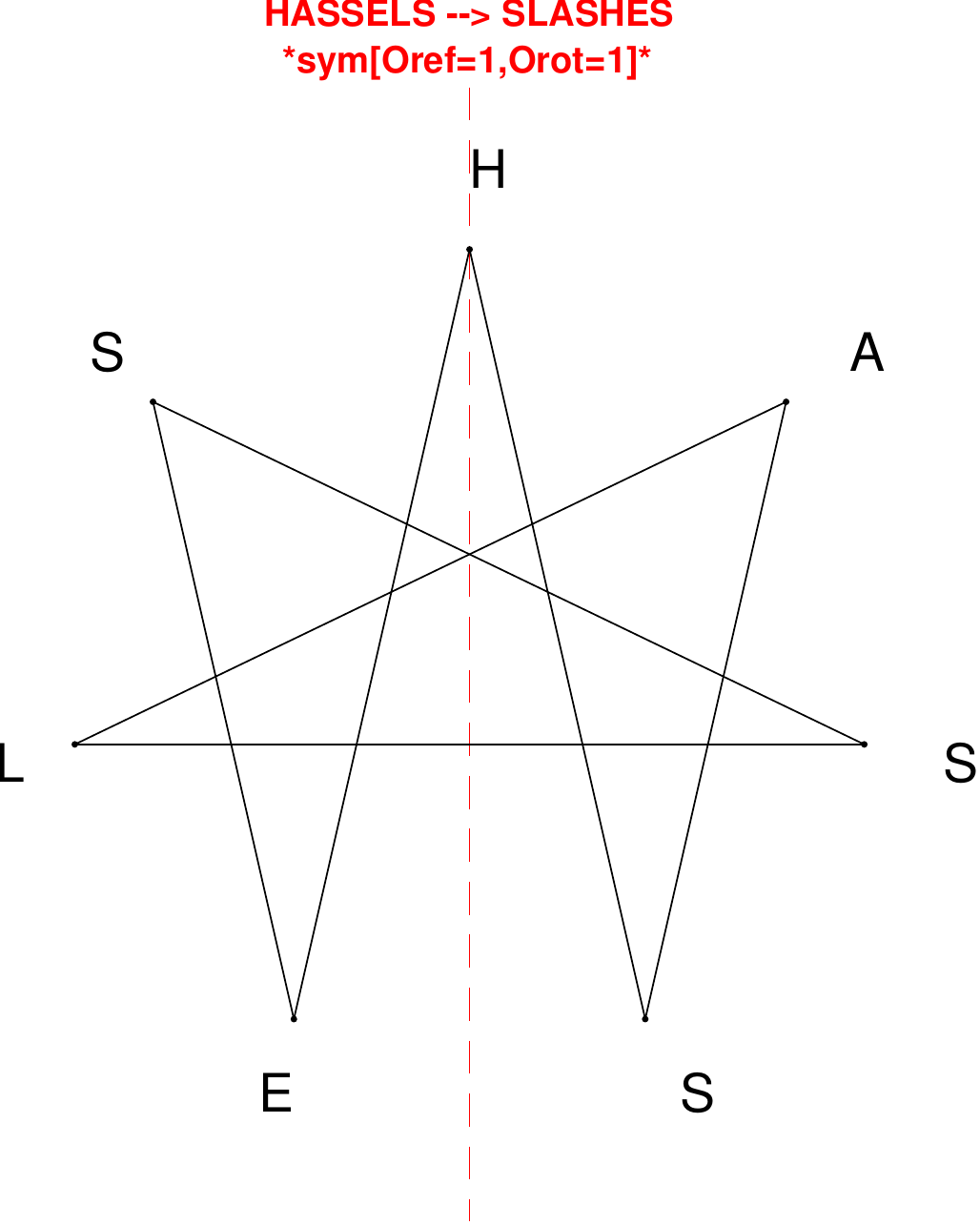}
\end{subfigure}
\hfill
\begin{subfigure}[T]{0.19\textwidth}
\centering
\includegraphics[width=\textwidth]{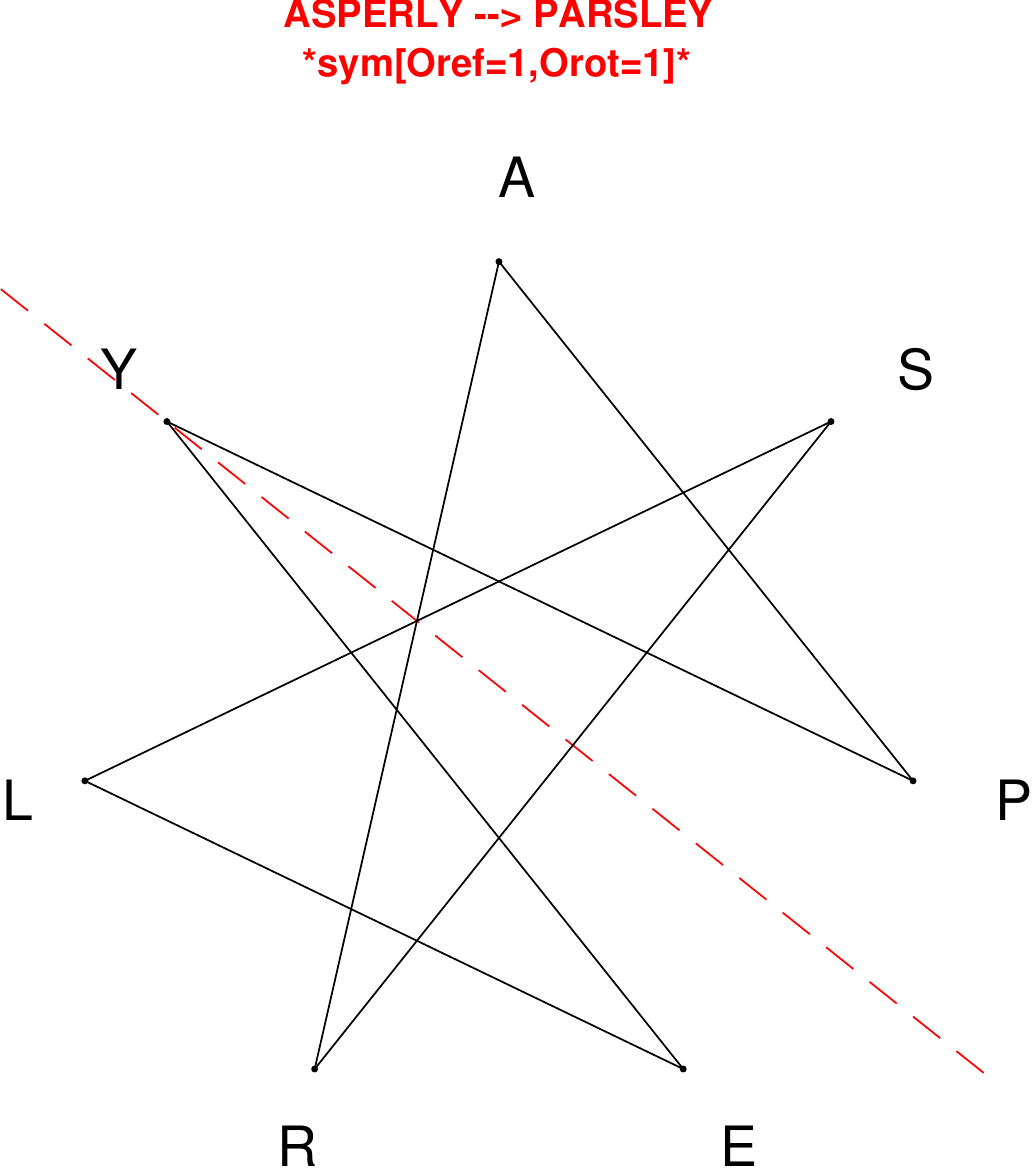}
\end{subfigure}
\hfill
\begin{subfigure}[T]{0.19\textwidth}
\centering
\includegraphics[width=\textwidth]{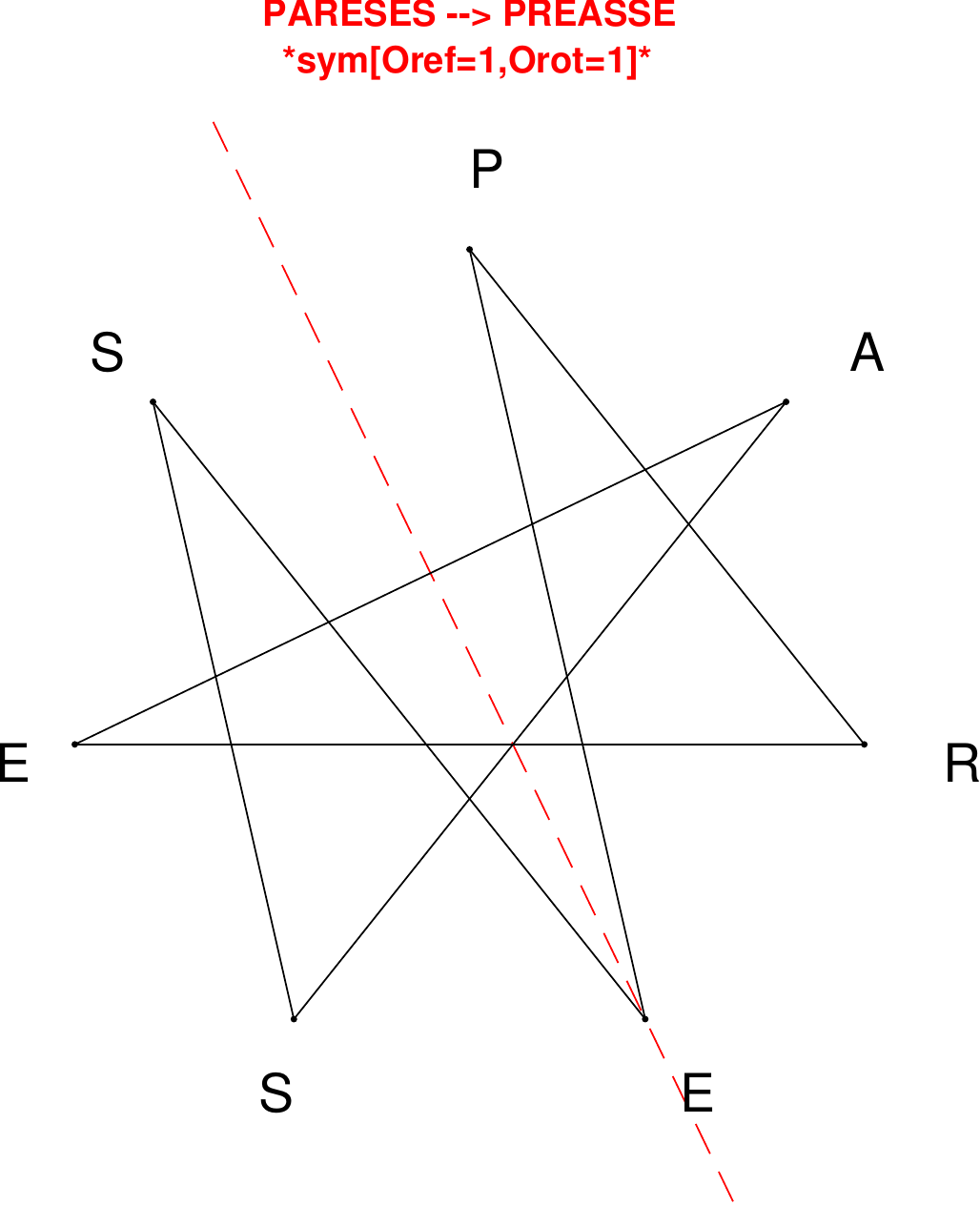}
\end{subfigure}
\hfill
\begin{subfigure}[T]{0.19\textwidth}
\centering
\includegraphics[width=\textwidth]{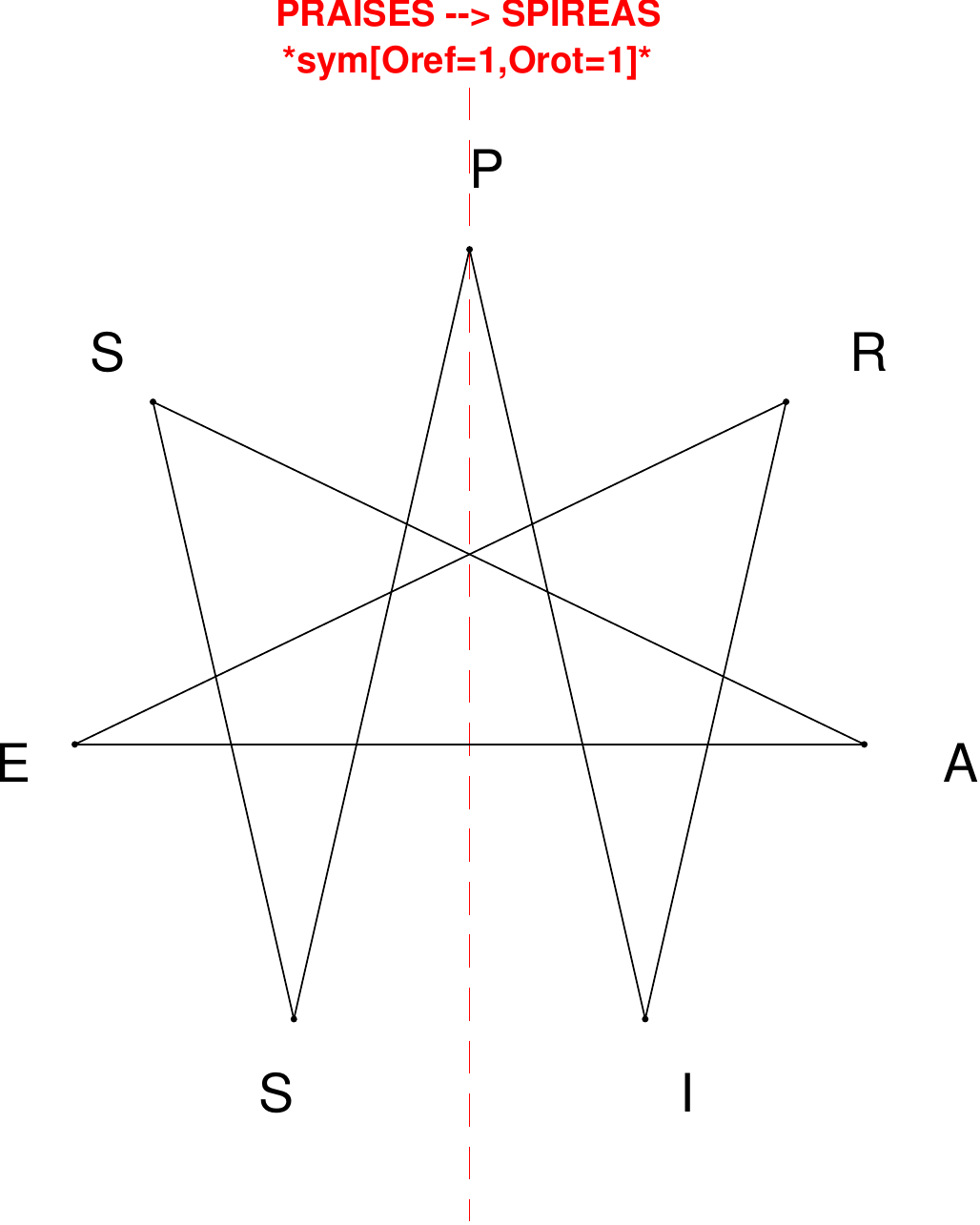}
\end{subfigure}
\hfill
\begin{subfigure}[T]{0.19\textwidth}
\centering
\includegraphics[width=\textwidth]{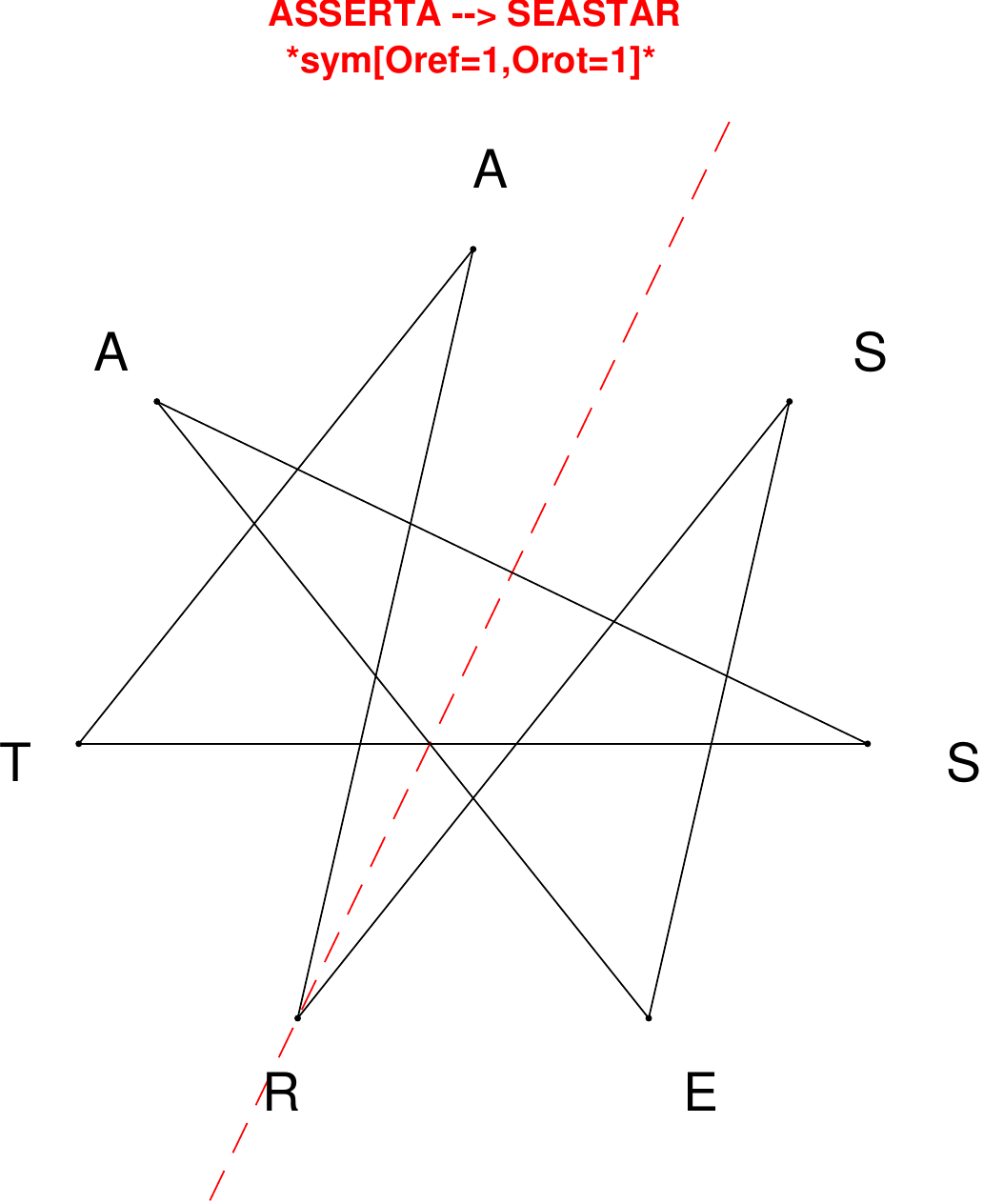}
\end{subfigure}
\end{figure}

\begin{figure}[H]
\centering
\begin{subfigure}[T]{0.19\textwidth}
\centering
\includegraphics[width=\textwidth]{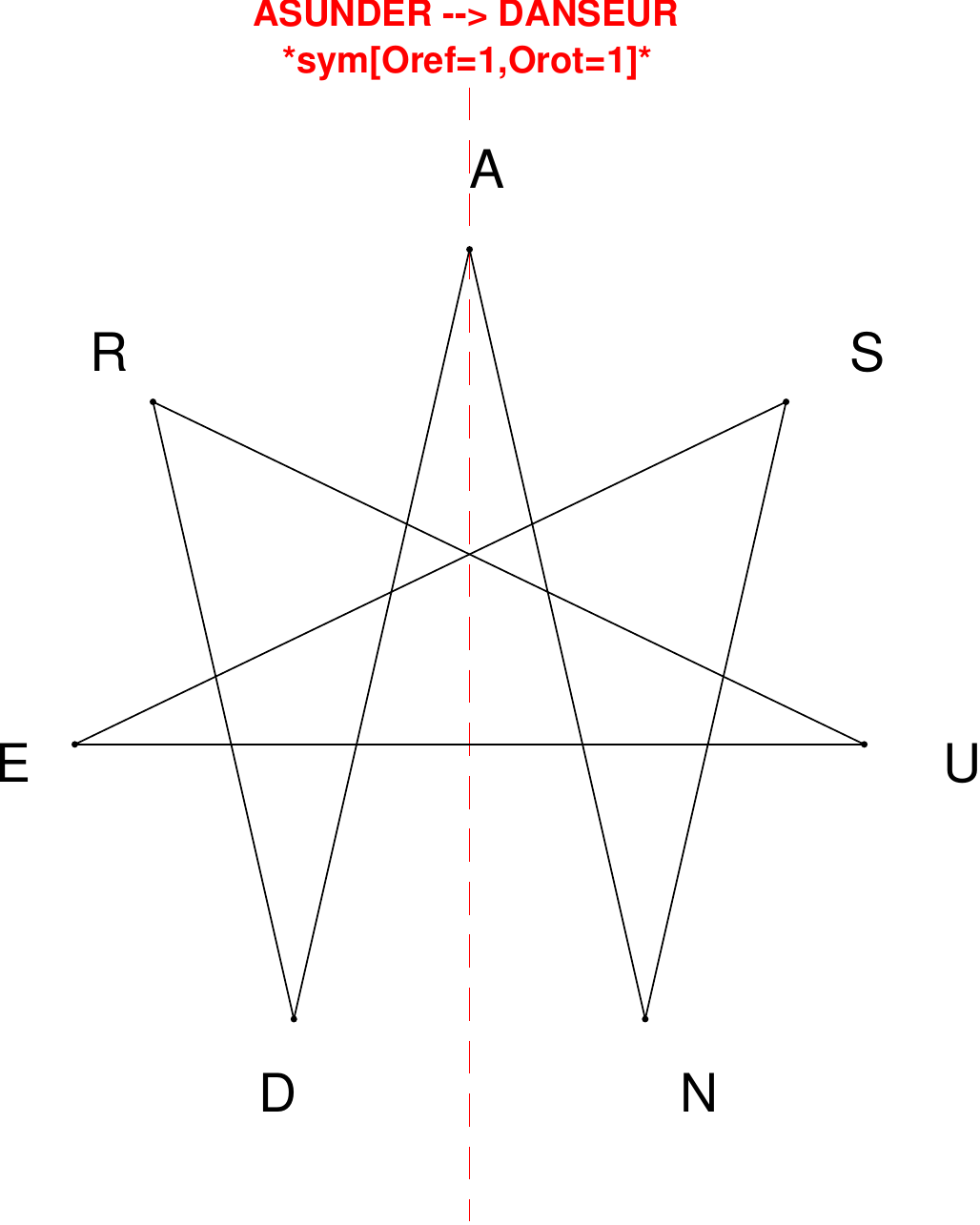}
\end{subfigure}
\hfill
\begin{subfigure}[T]{0.19\textwidth}
\centering
\includegraphics[width=\textwidth]{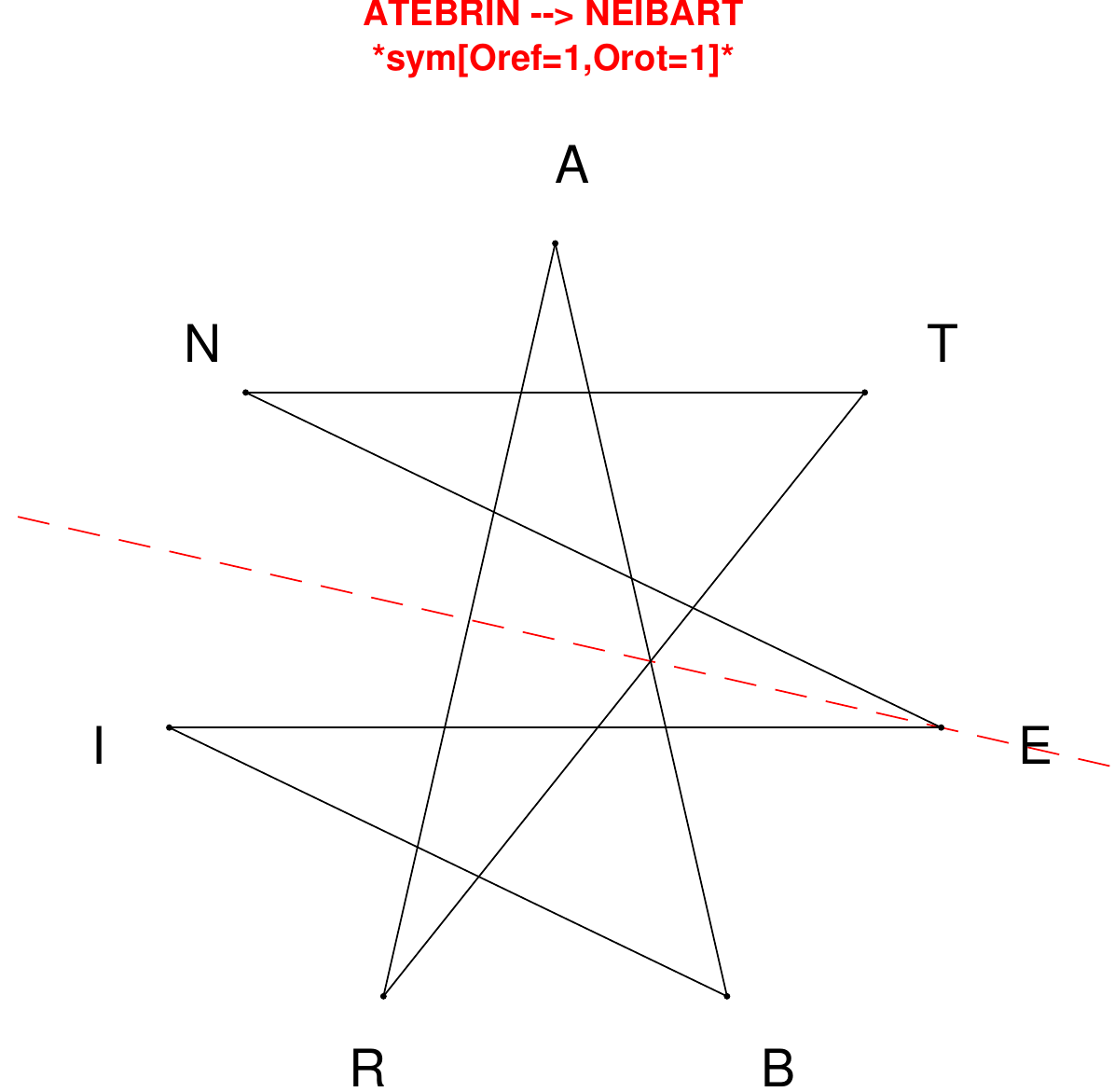}
\end{subfigure}
\hfill
\begin{subfigure}[T]{0.19\textwidth}
\centering
\includegraphics[width=\textwidth]{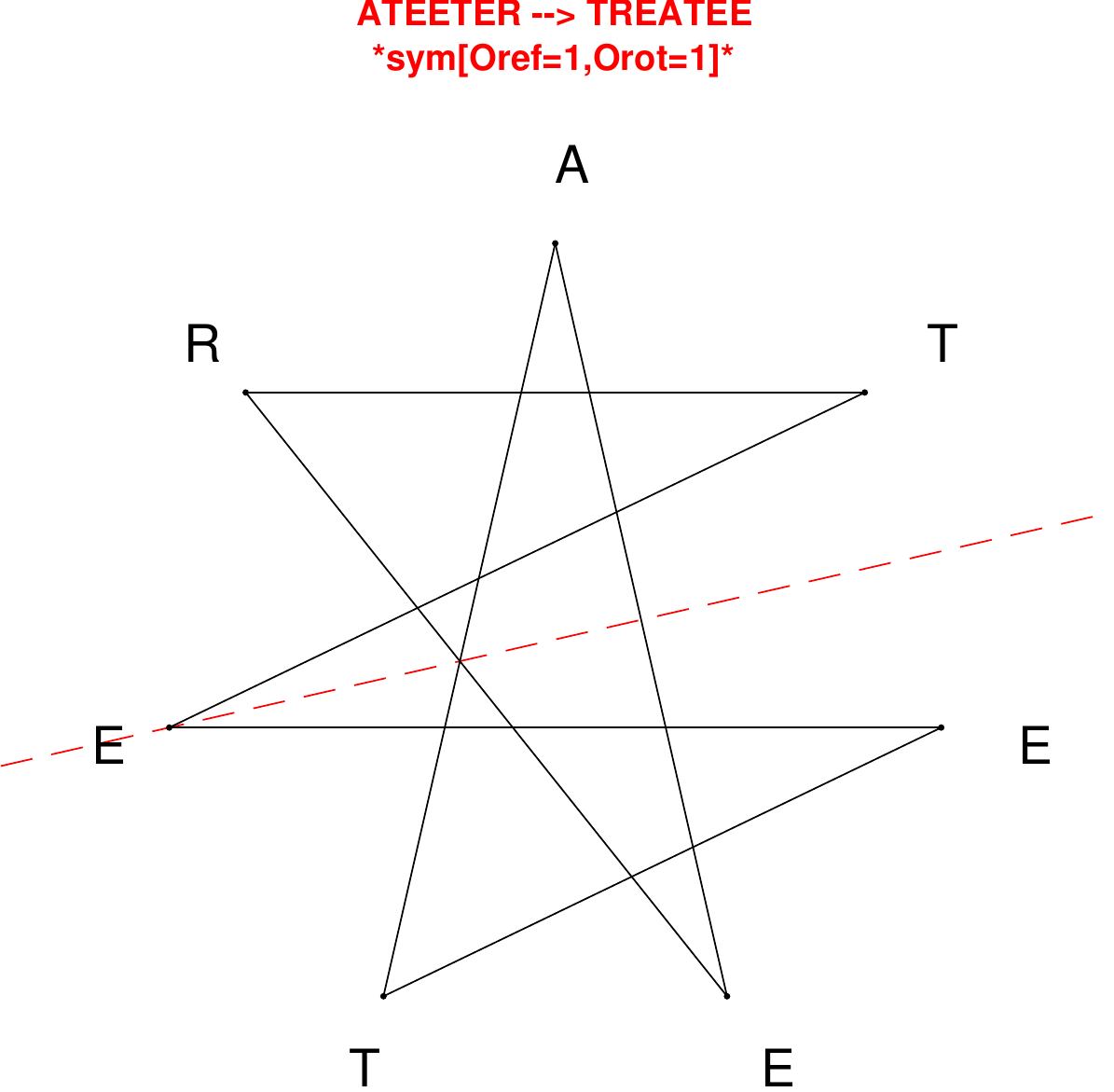}
\end{subfigure}
\hfill
\begin{subfigure}[T]{0.19\textwidth}
\centering
\includegraphics[width=\textwidth]{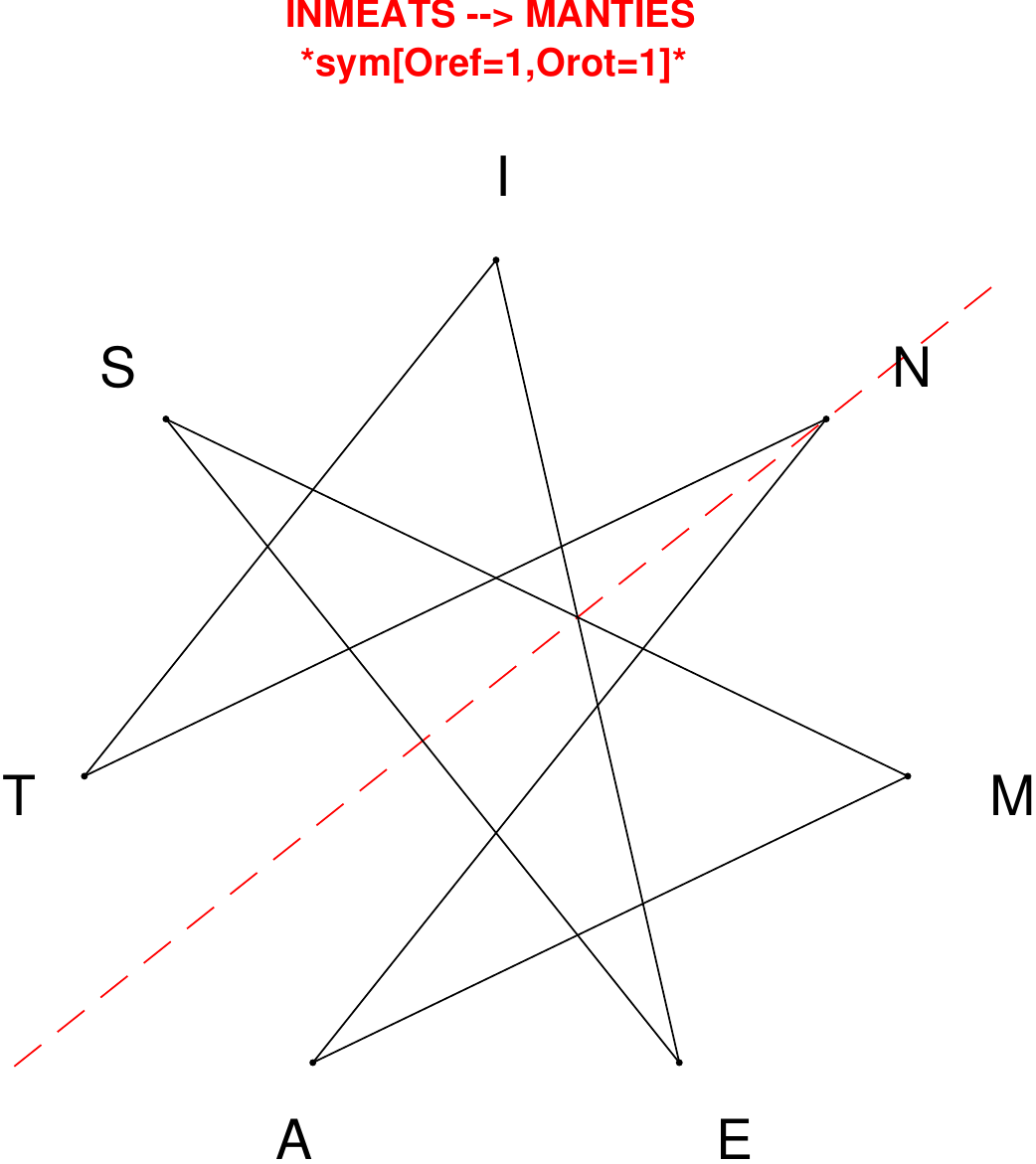}
\end{subfigure}
\hfill
\begin{subfigure}[T]{0.19\textwidth}
\centering
\includegraphics[width=\textwidth]{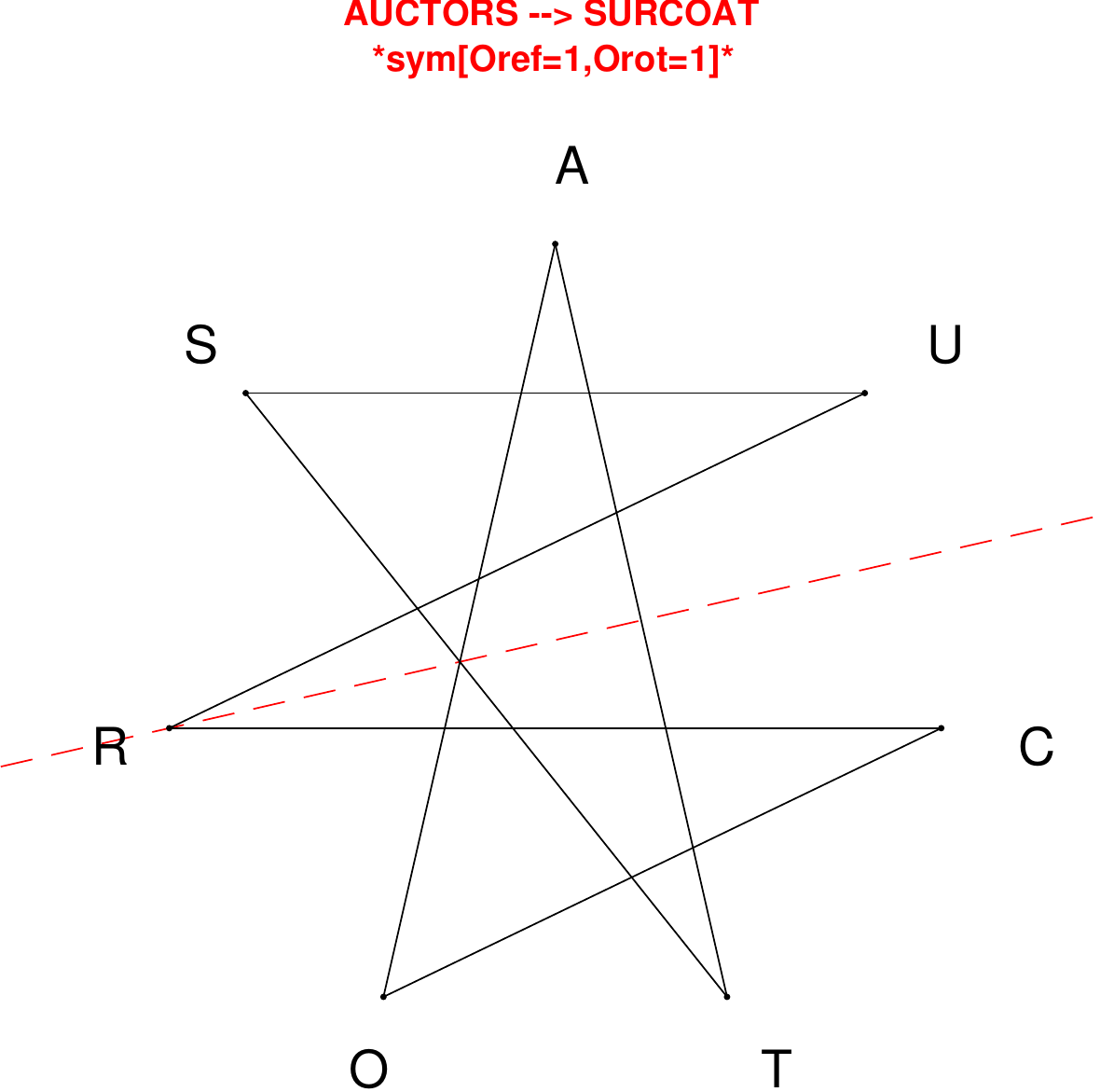}
\end{subfigure}
\end{figure}

\begin{figure}[H]
\centering
\begin{subfigure}[T]{0.19\textwidth}
\centering
\includegraphics[width=\textwidth]{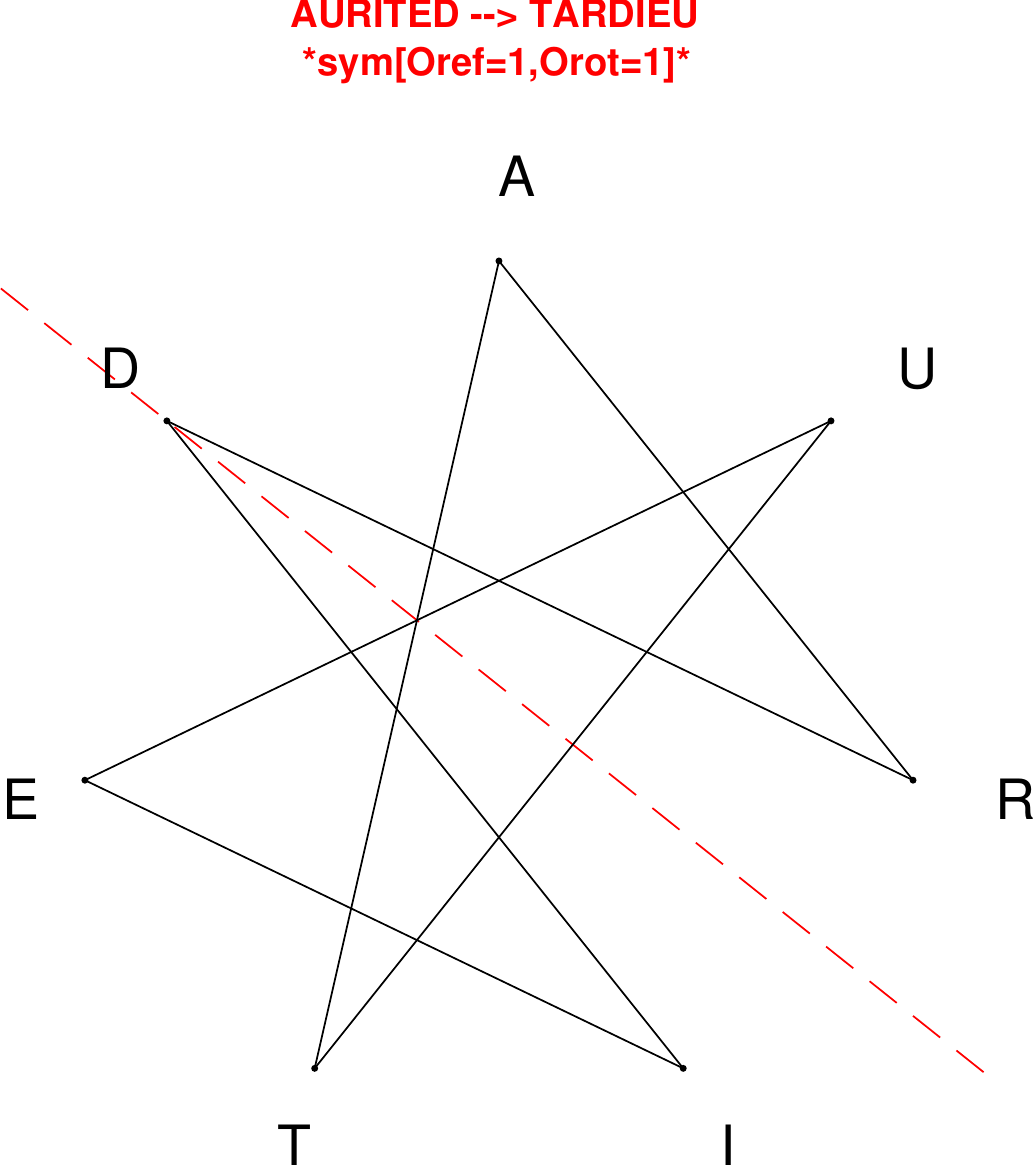}
\end{subfigure}
\hfill
\begin{subfigure}[T]{0.19\textwidth}
\centering
\includegraphics[width=\textwidth]{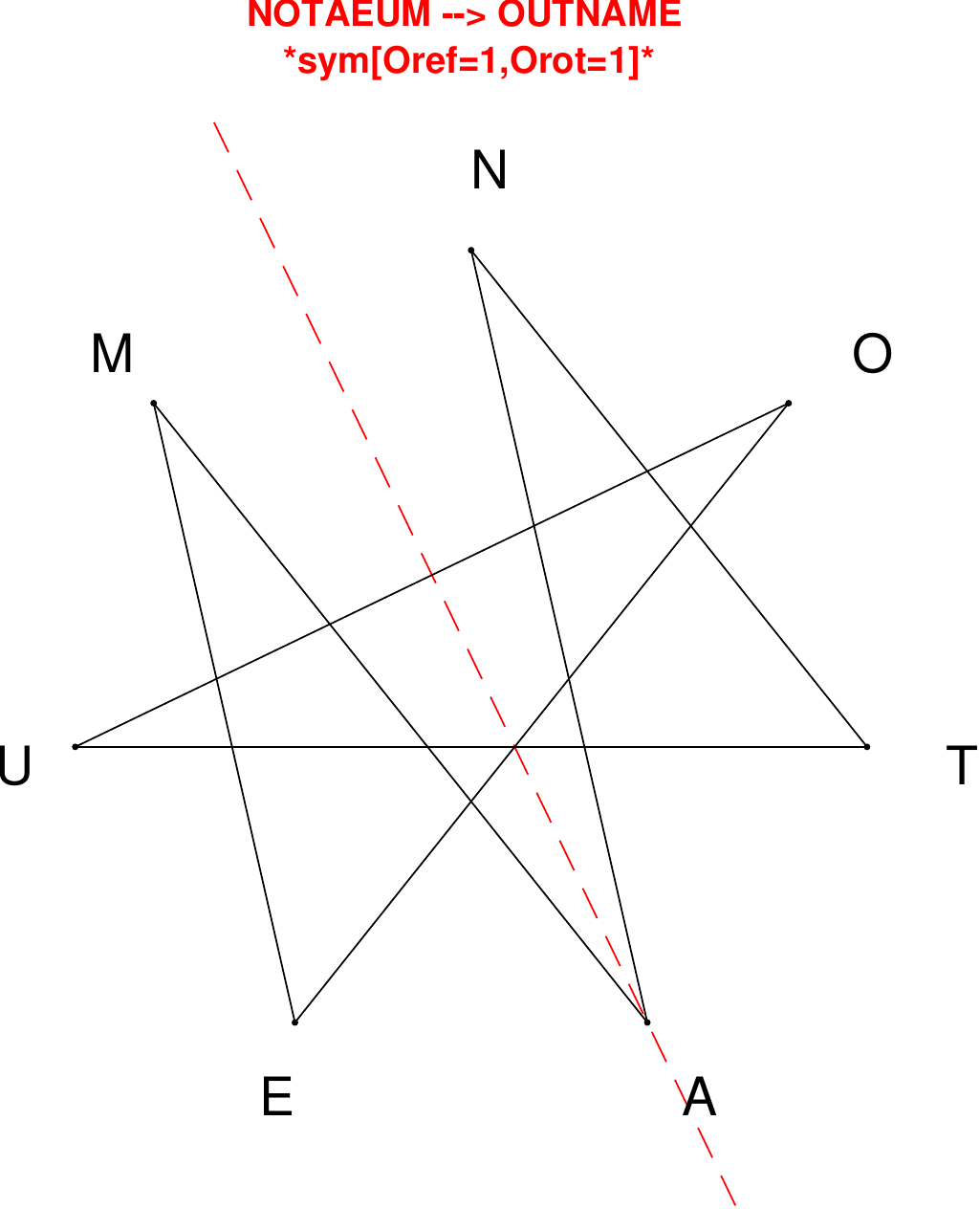}
\end{subfigure}
\hfill
\begin{subfigure}[T]{0.19\textwidth}
\centering
\includegraphics[width=\textwidth]{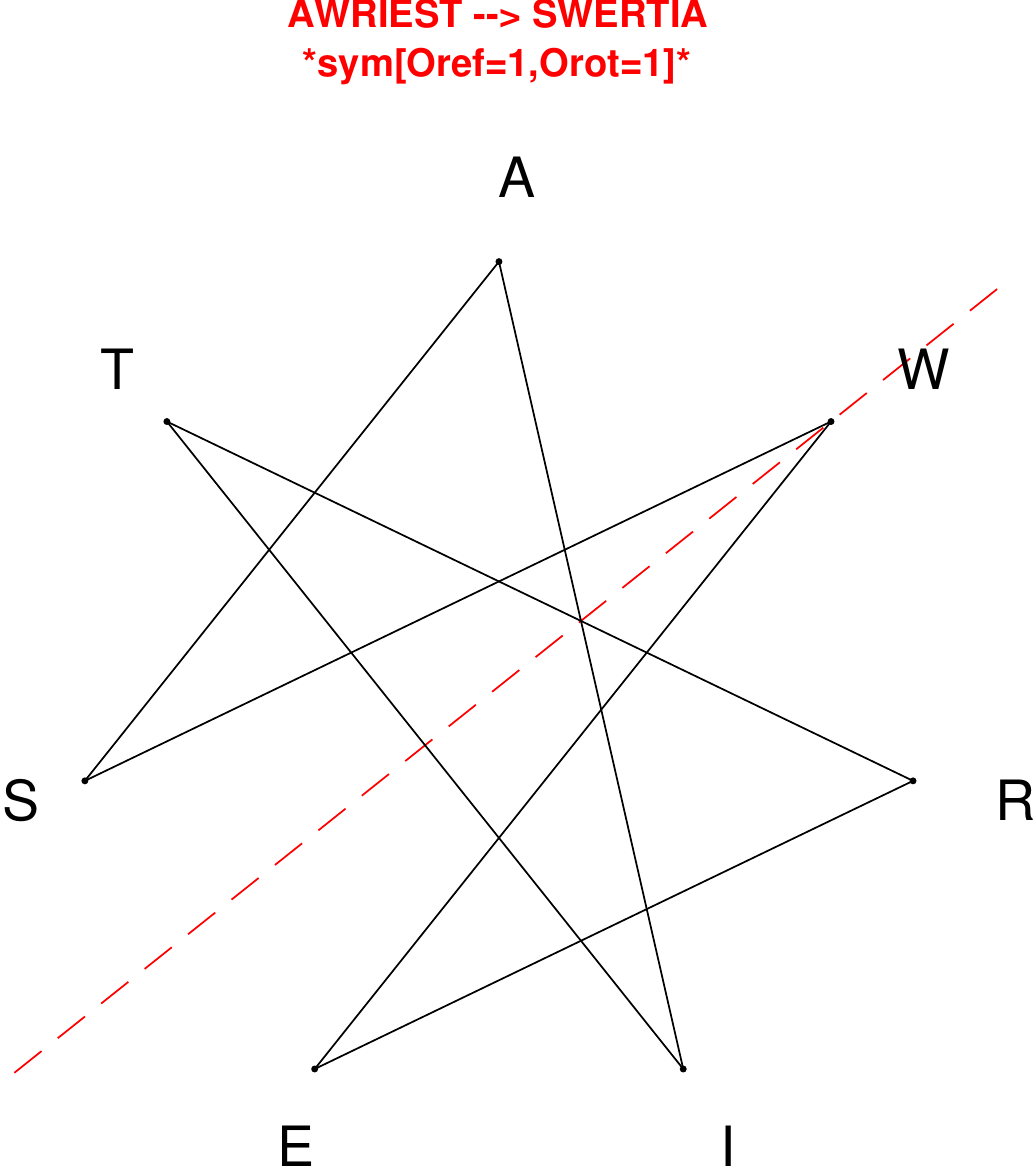}
\end{subfigure}
\hfill
\begin{subfigure}[T]{0.19\textwidth}
\centering
\includegraphics[width=\textwidth]{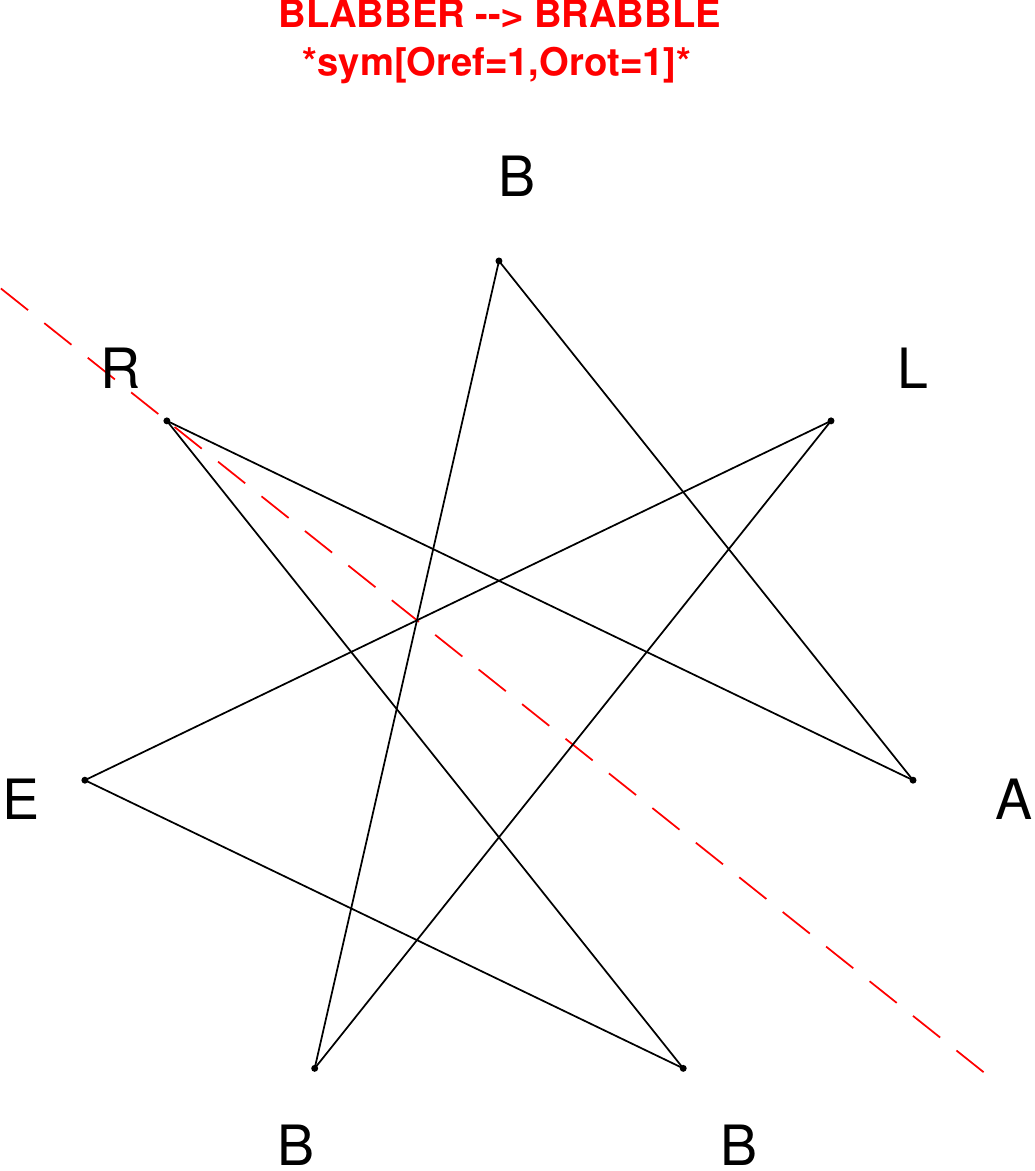}
\end{subfigure}
\hfill
\begin{subfigure}[T]{0.19\textwidth}
\centering
\includegraphics[width=\textwidth]{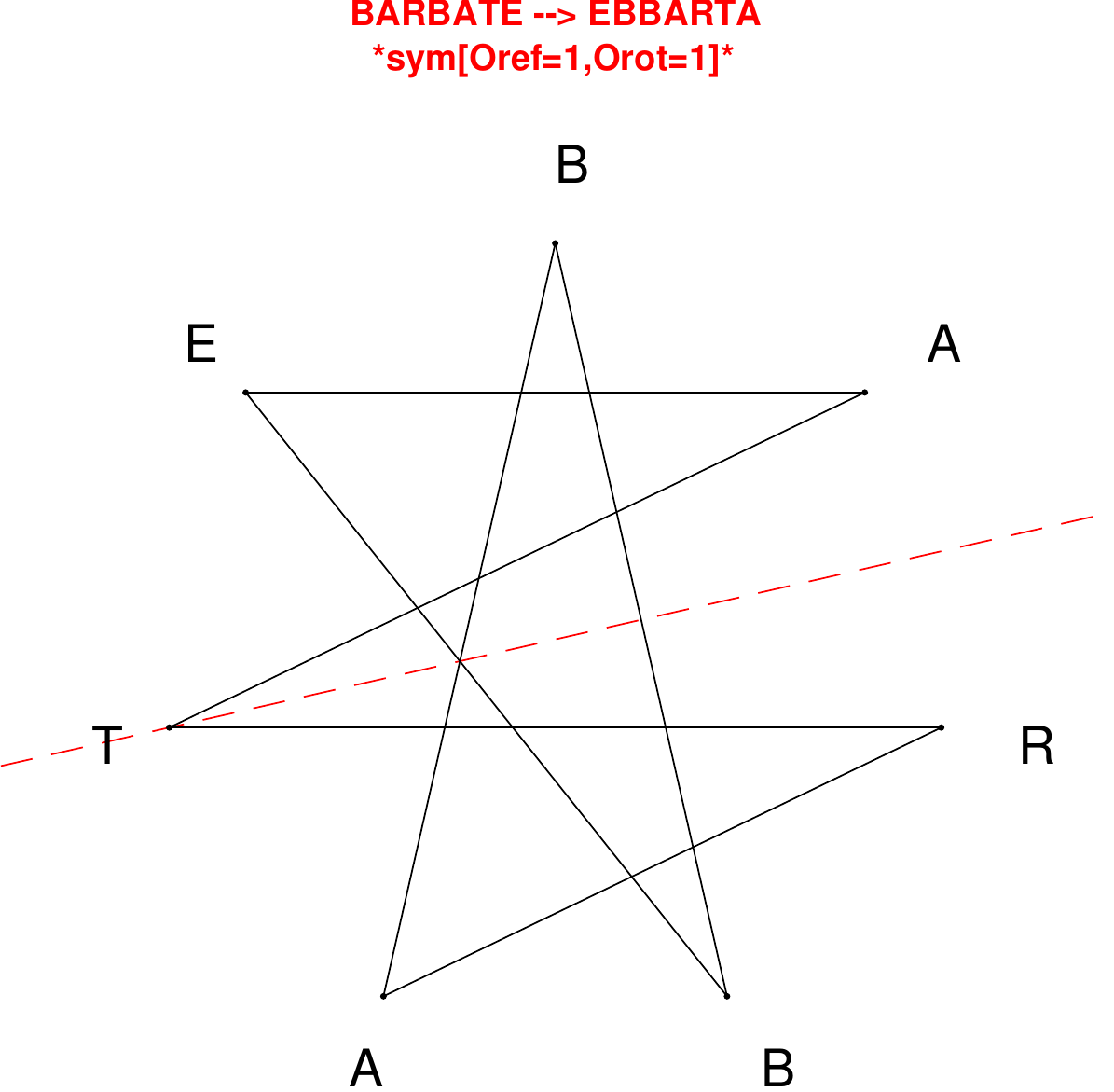}
\end{subfigure}
\end{figure}

\begin{figure}[H]
\centering
\begin{subfigure}[T]{0.19\textwidth}
\centering
\includegraphics[width=\textwidth]{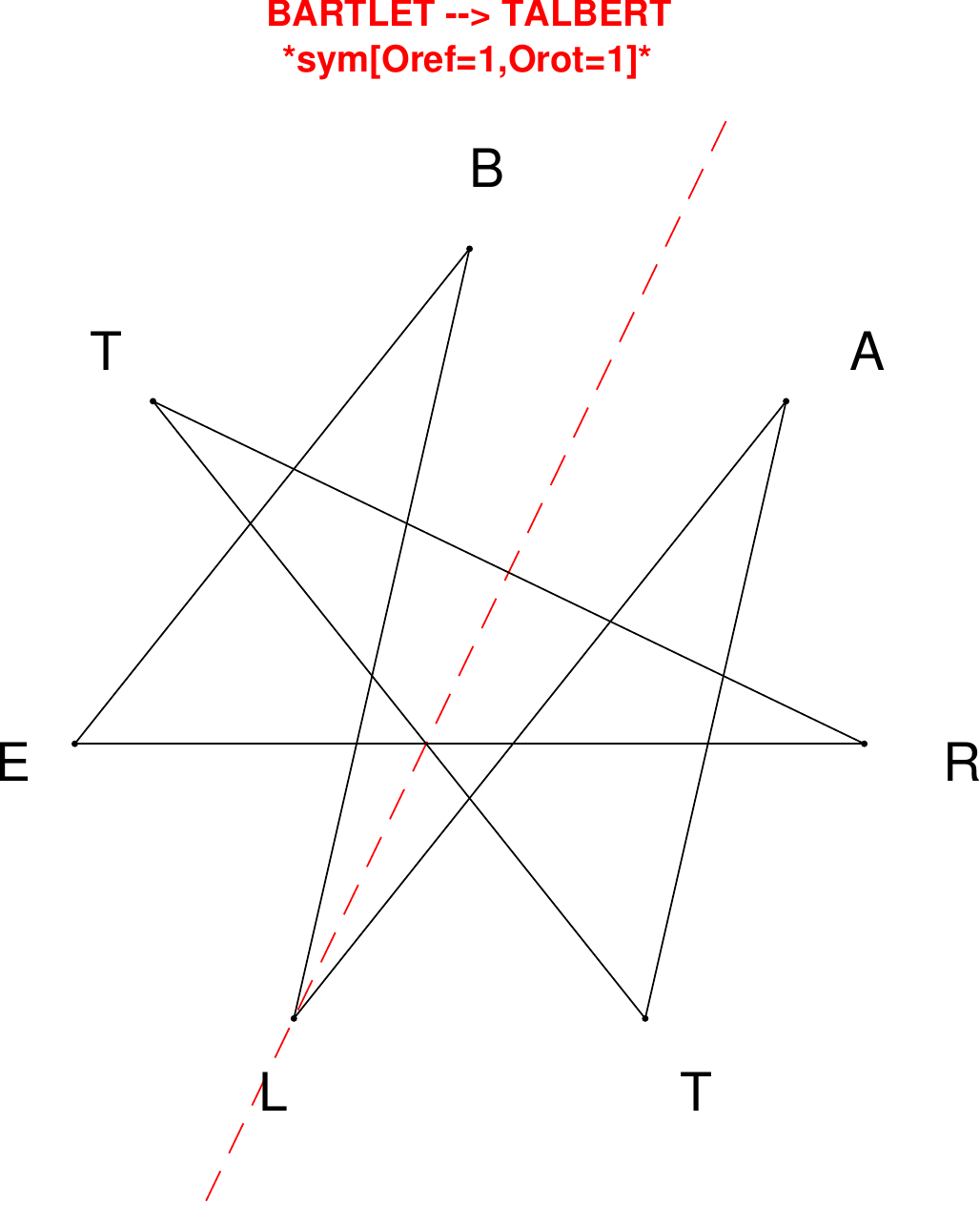}
\end{subfigure}
\hfill
\begin{subfigure}[T]{0.19\textwidth}
\centering
\includegraphics[width=\textwidth]{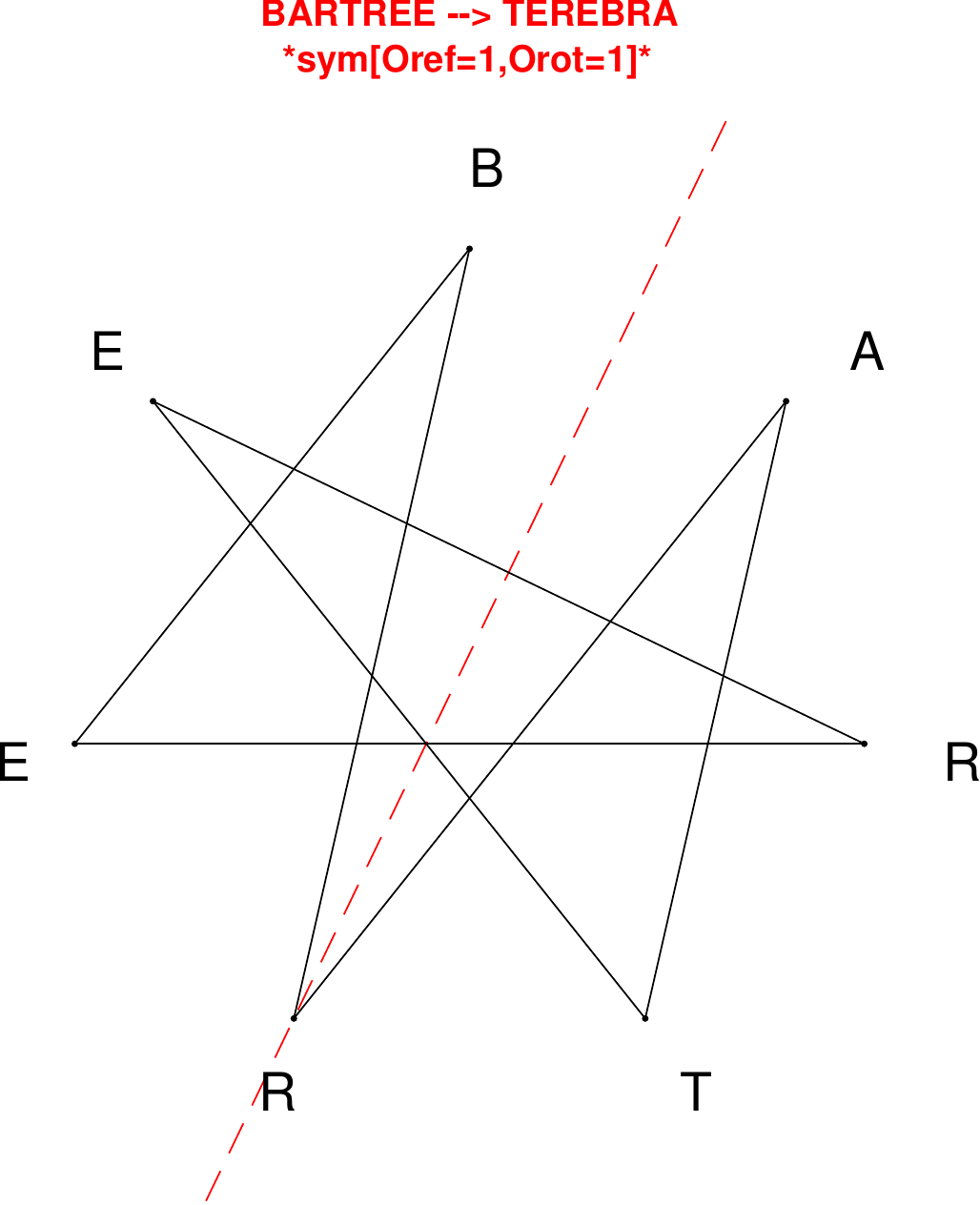}
\end{subfigure}
\hfill
\begin{subfigure}[T]{0.19\textwidth}
\centering
\includegraphics[width=\textwidth]{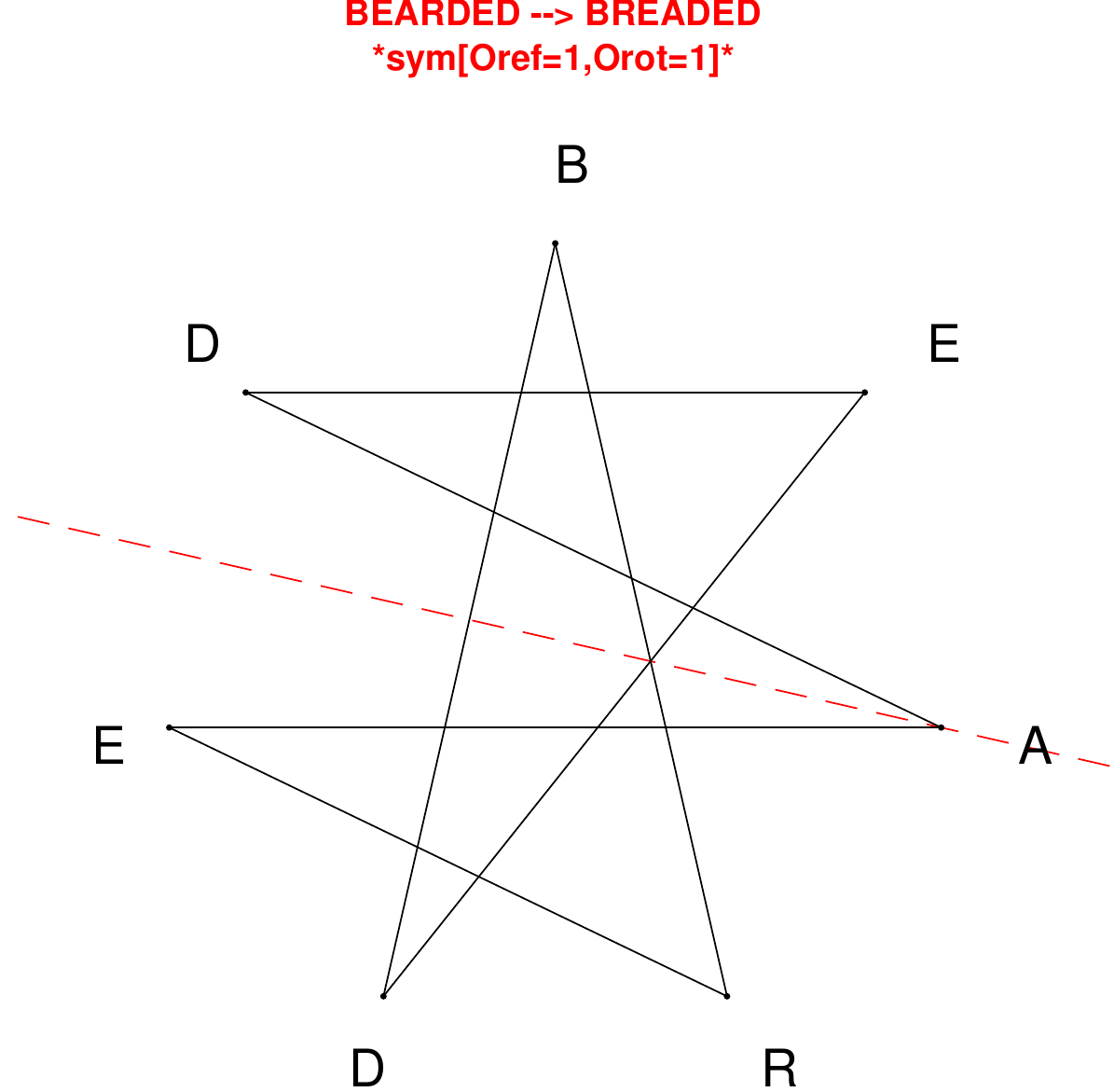}
\end{subfigure}
\hfill
\begin{subfigure}[T]{0.19\textwidth}
\centering
\includegraphics[width=\textwidth]{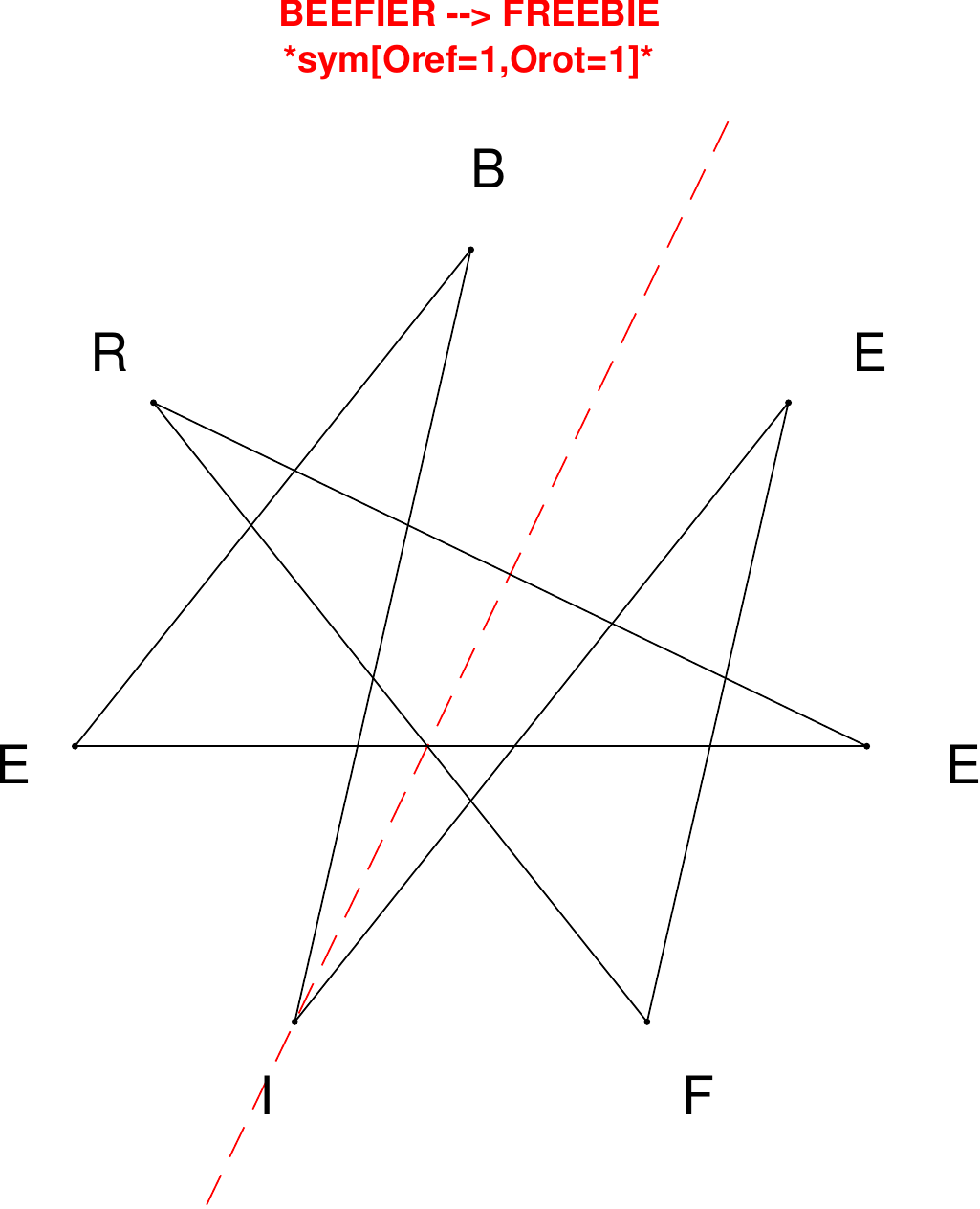}
\end{subfigure}
\hfill
\begin{subfigure}[T]{0.19\textwidth}
\centering
\includegraphics[width=\textwidth]{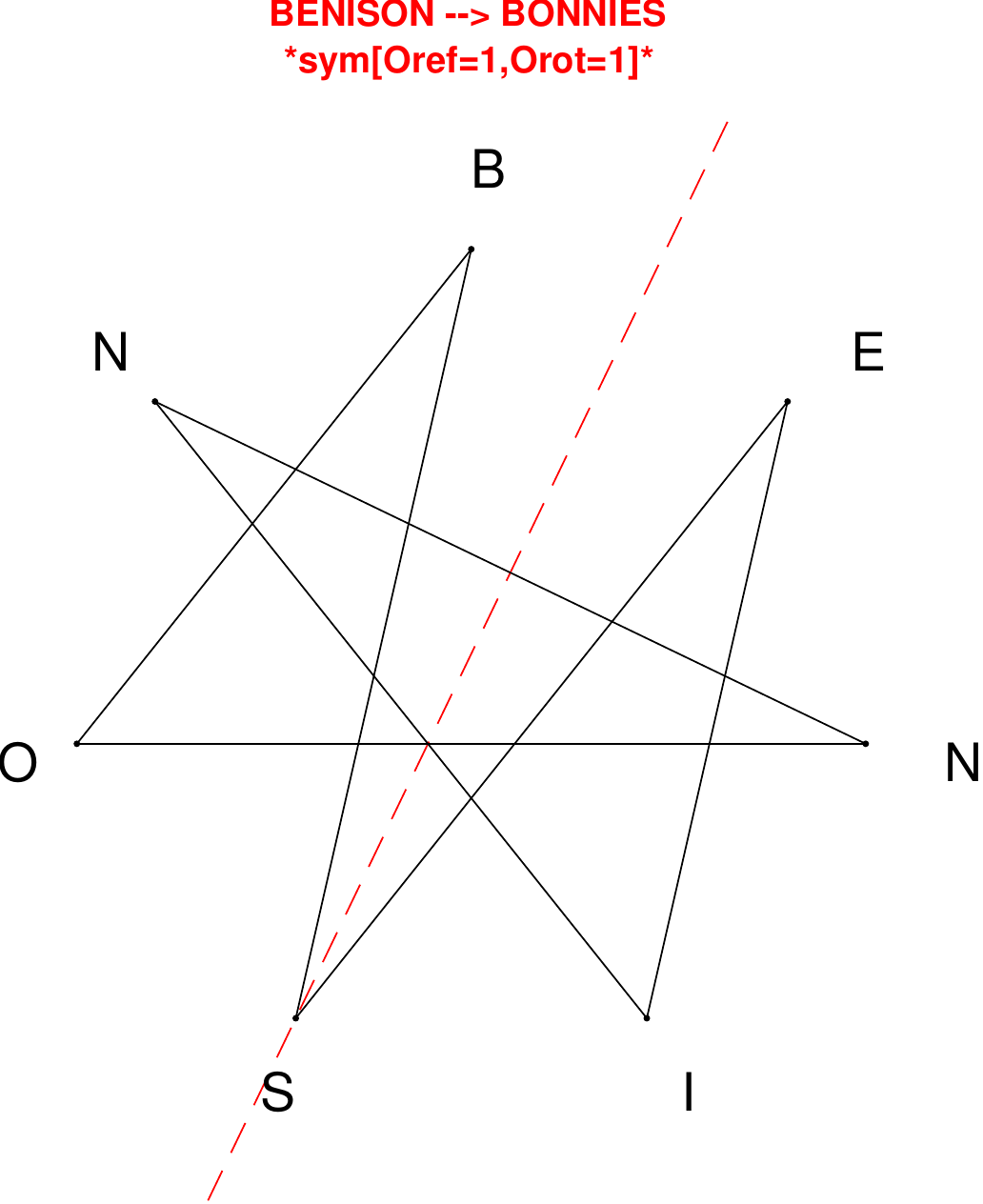}
\end{subfigure}
\end{figure}

\begin{figure}[H]
\centering
\begin{subfigure}[T]{0.19\textwidth}
\centering
\includegraphics[width=\textwidth]{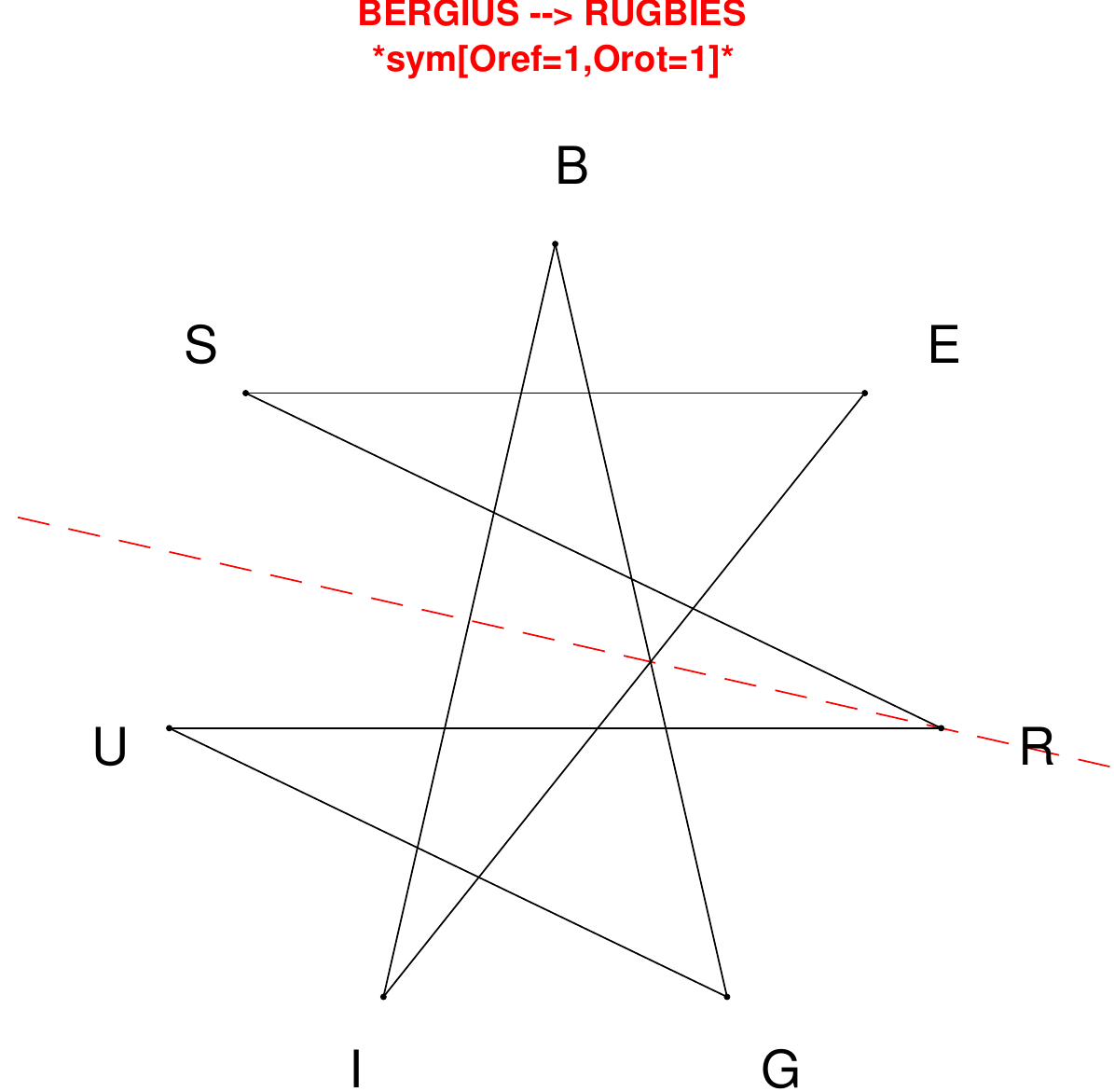}
\end{subfigure}
\hfill
\begin{subfigure}[T]{0.19\textwidth}
\centering
\includegraphics[width=\textwidth]{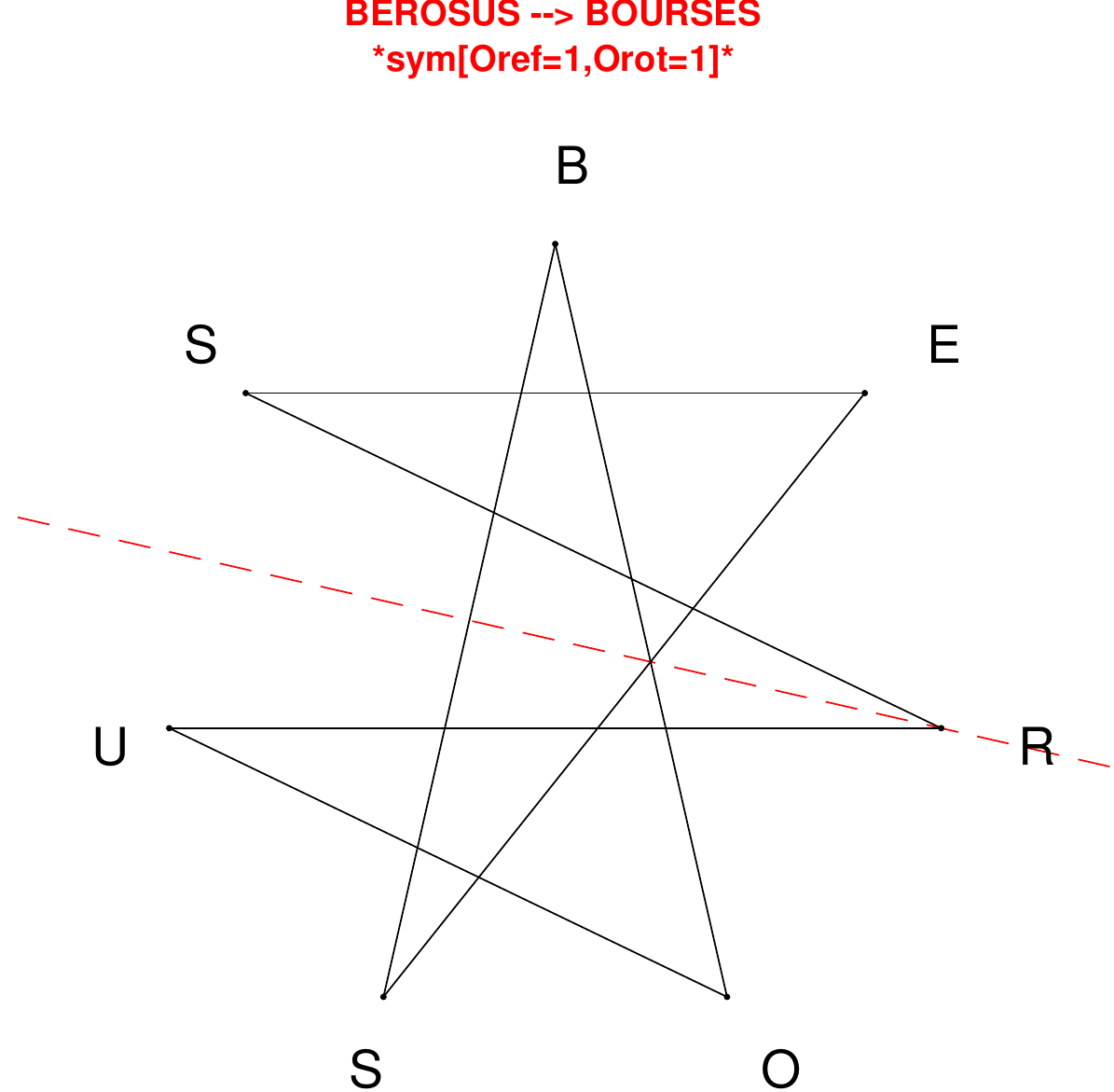}
\end{subfigure}
\hfill
\begin{subfigure}[T]{0.19\textwidth}
\centering
\includegraphics[width=\textwidth]{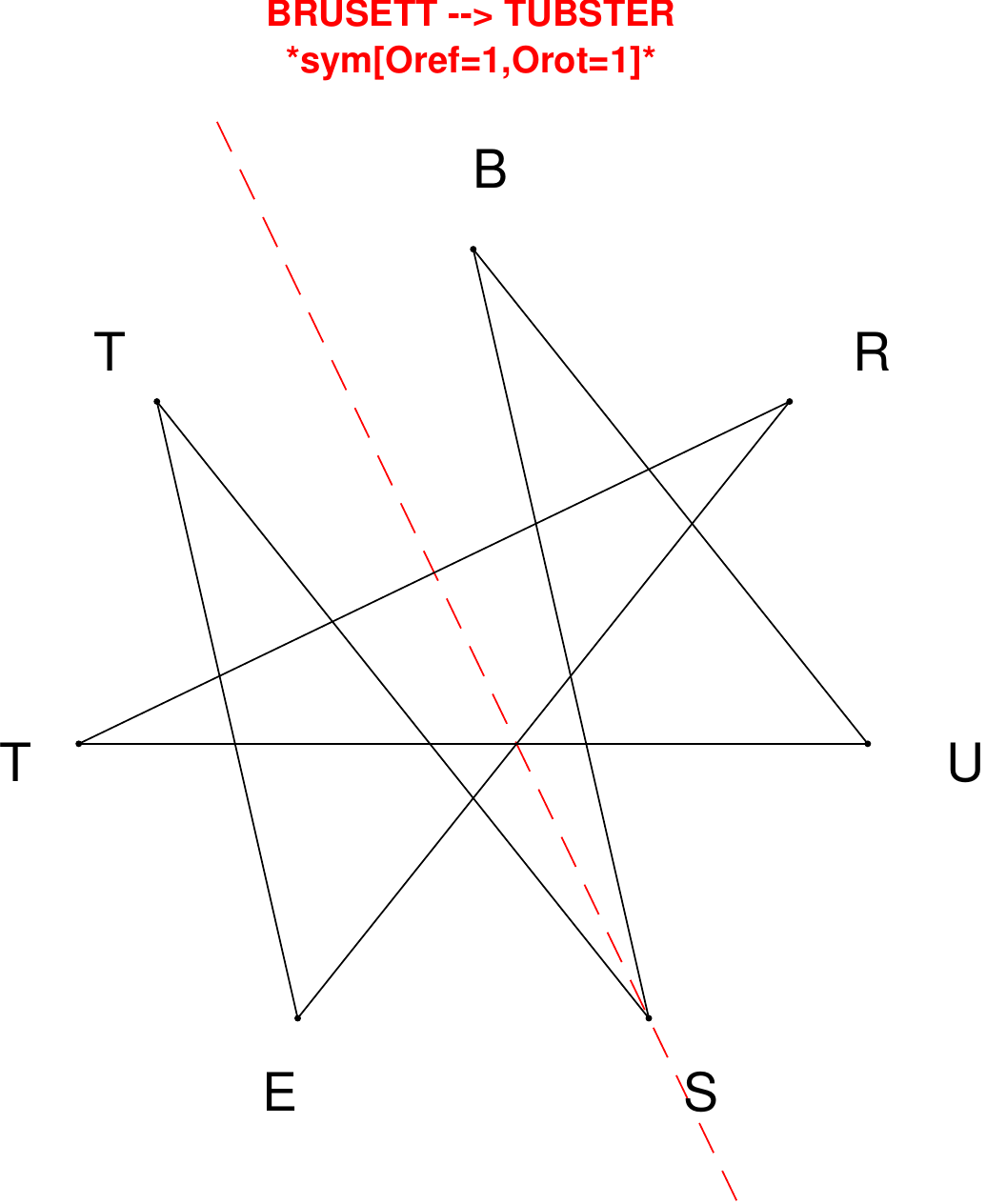}
\end{subfigure}
\hfill
\begin{subfigure}[T]{0.19\textwidth}
\centering
\includegraphics[width=\textwidth]{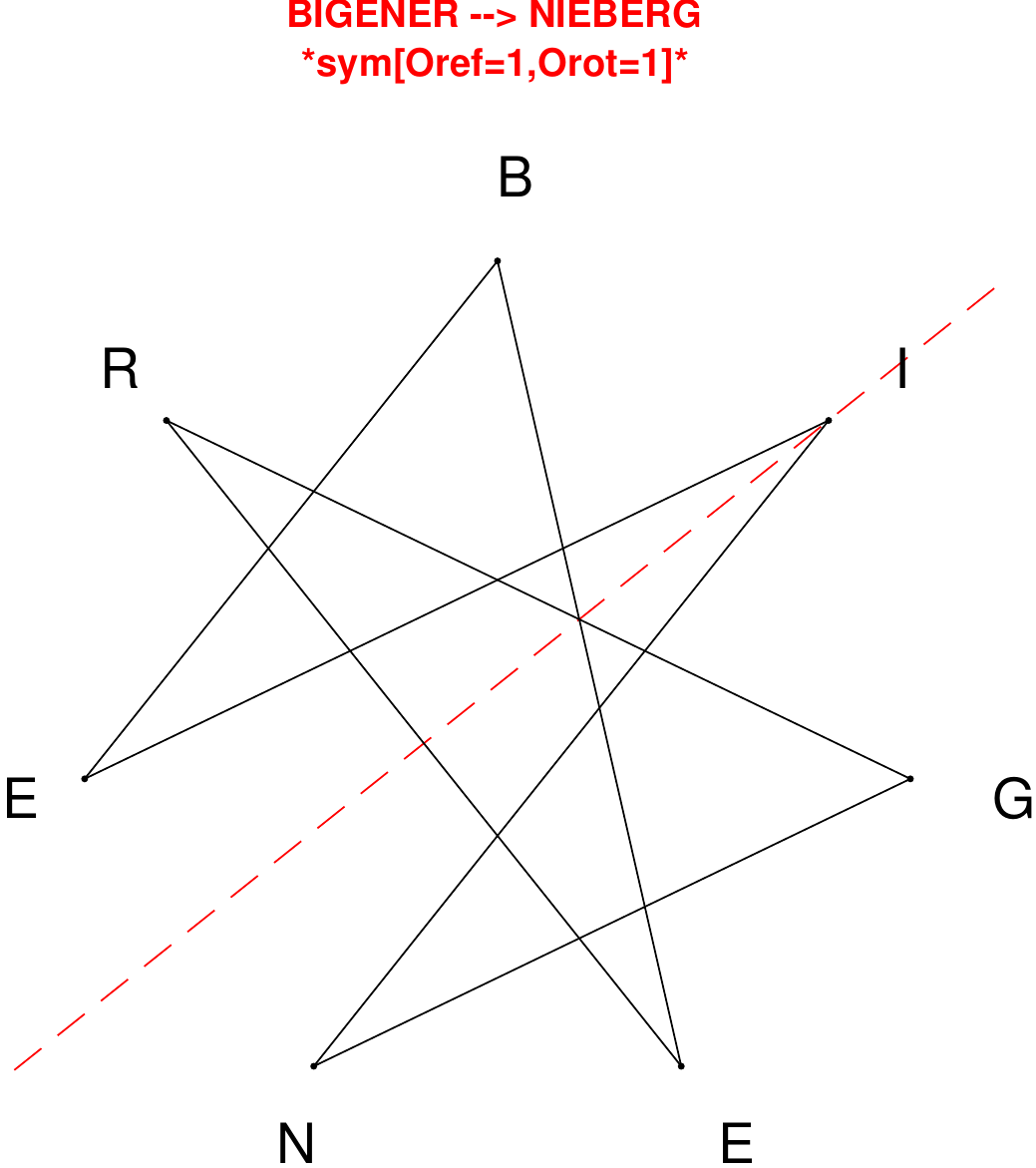}
\end{subfigure}
\hfill
\begin{subfigure}[T]{0.19\textwidth}
\centering
\includegraphics[width=\textwidth]{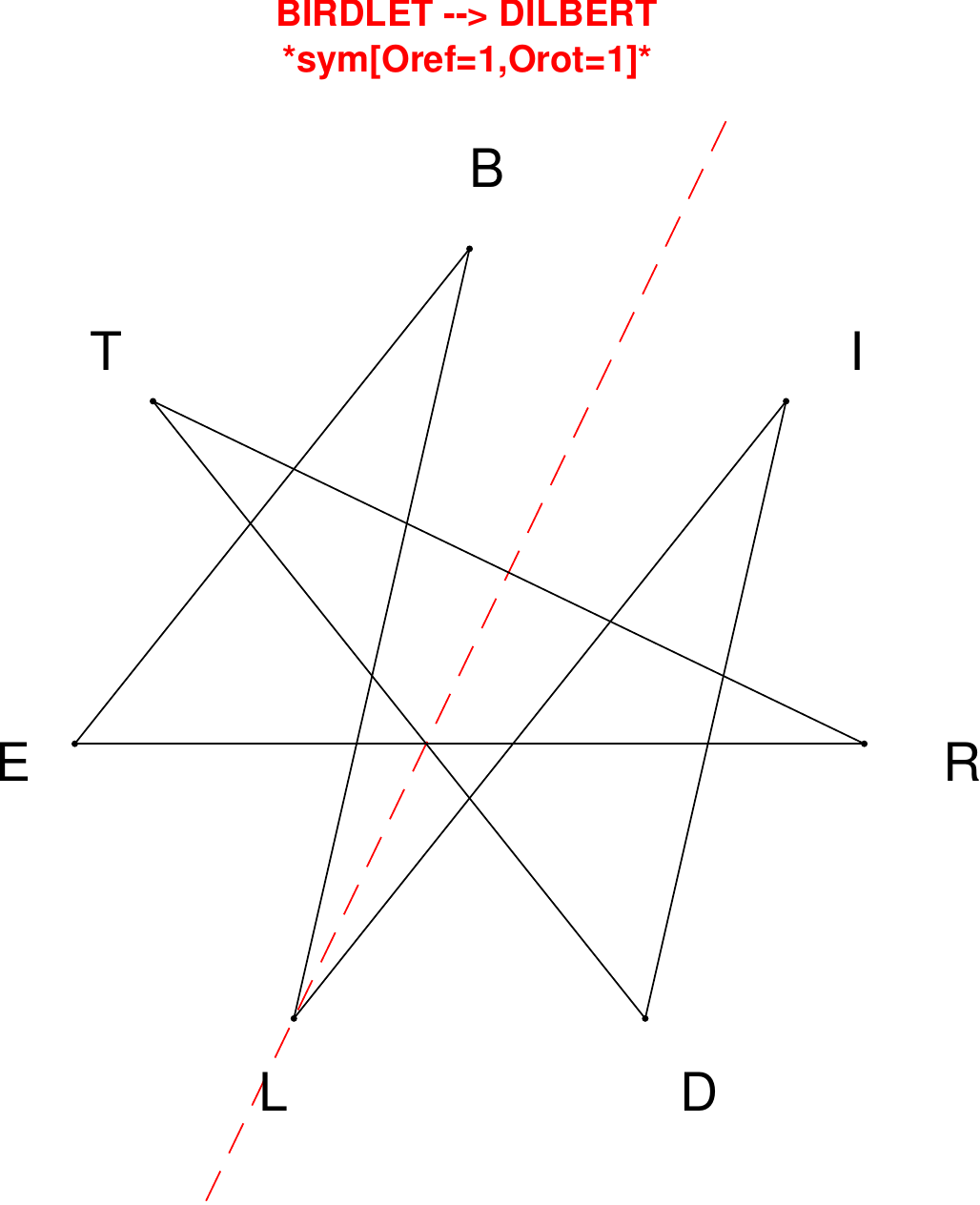}
\end{subfigure}
\end{figure}

\begin{figure}[H]
\centering
\begin{subfigure}[T]{0.19\textwidth}
\centering
\includegraphics[width=\textwidth]{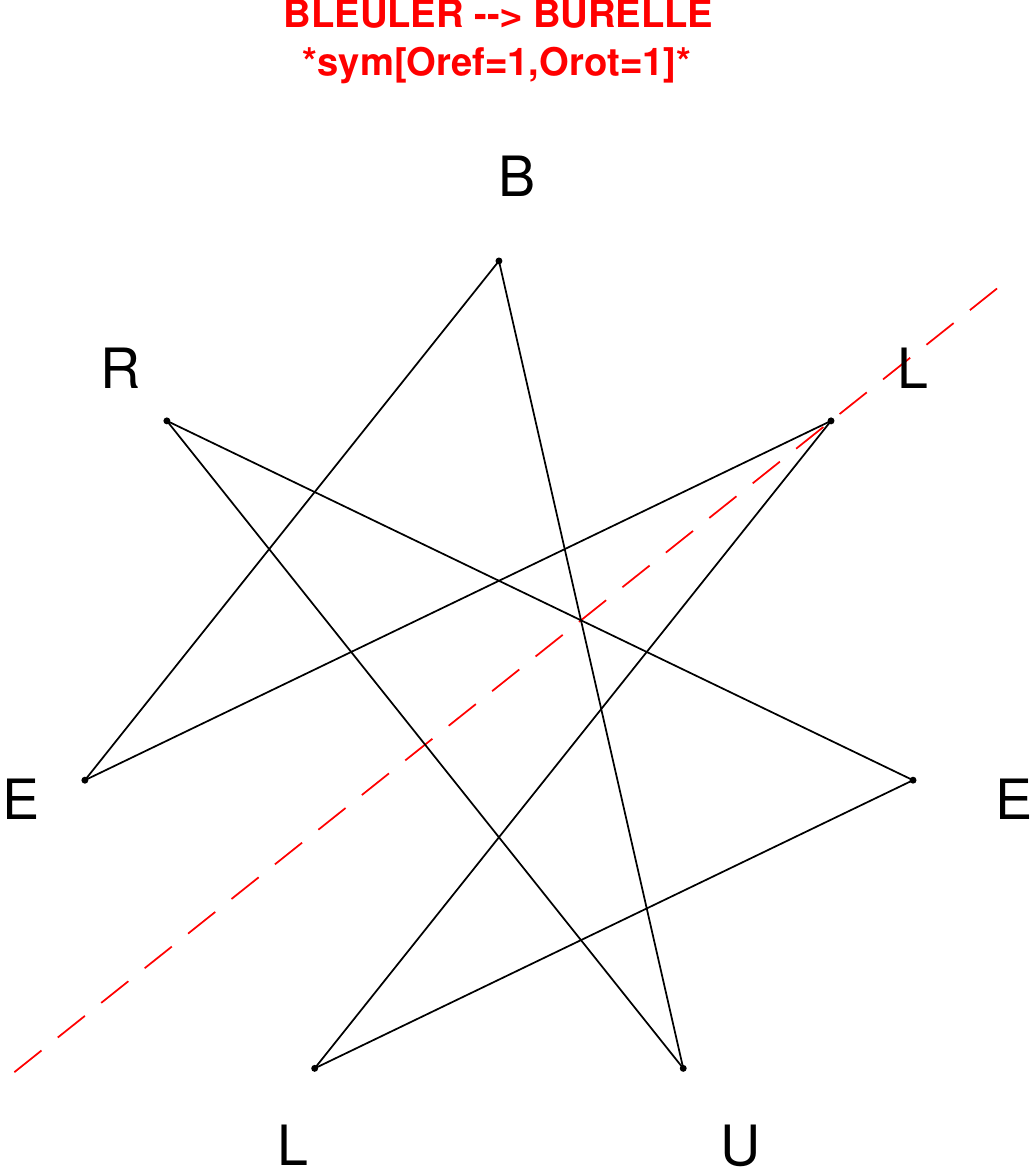}
\end{subfigure}
\hfill
\begin{subfigure}[T]{0.19\textwidth}
\centering
\includegraphics[width=\textwidth]{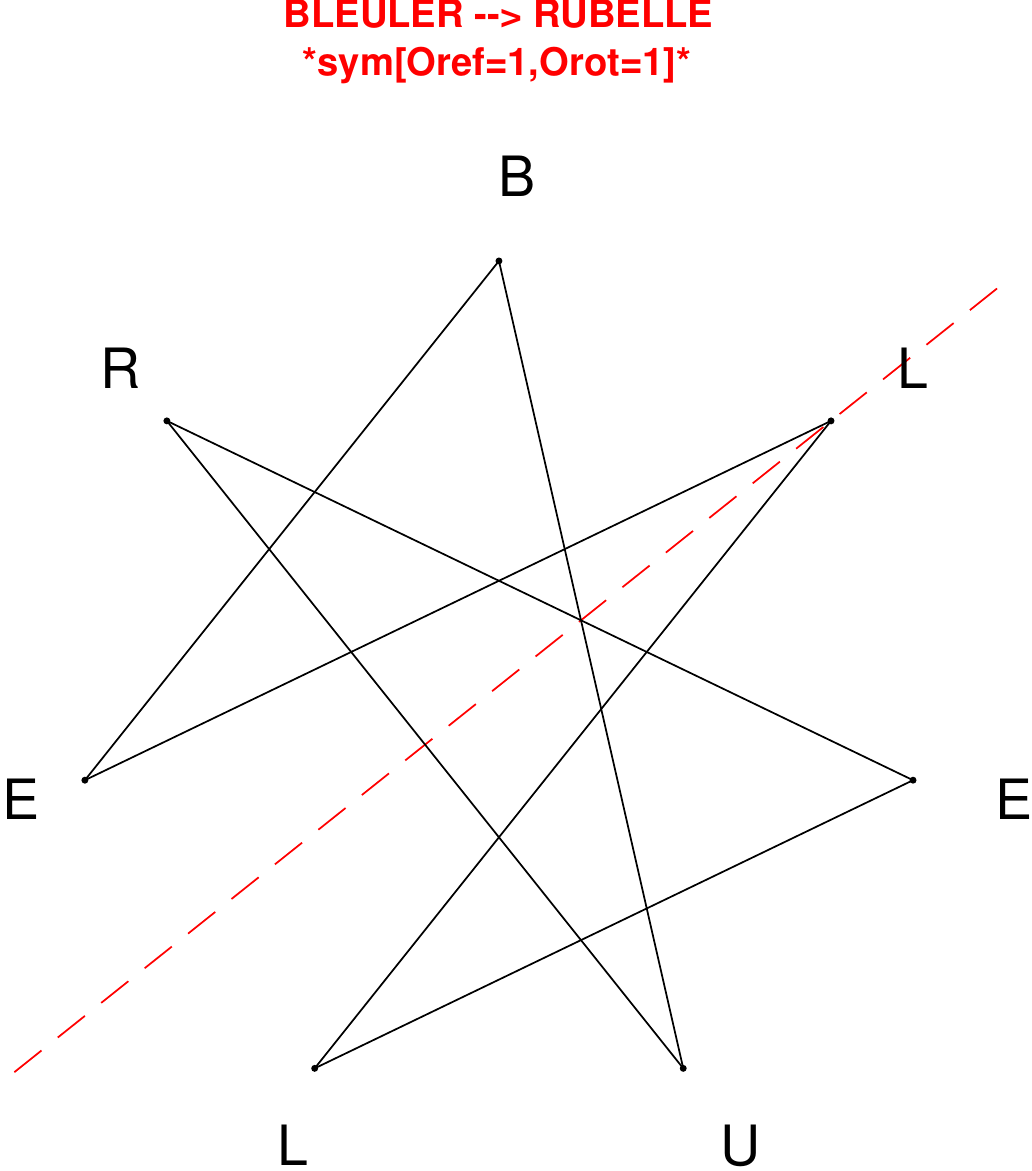}
\end{subfigure}
\hfill
\begin{subfigure}[T]{0.19\textwidth}
\centering
\includegraphics[width=\textwidth]{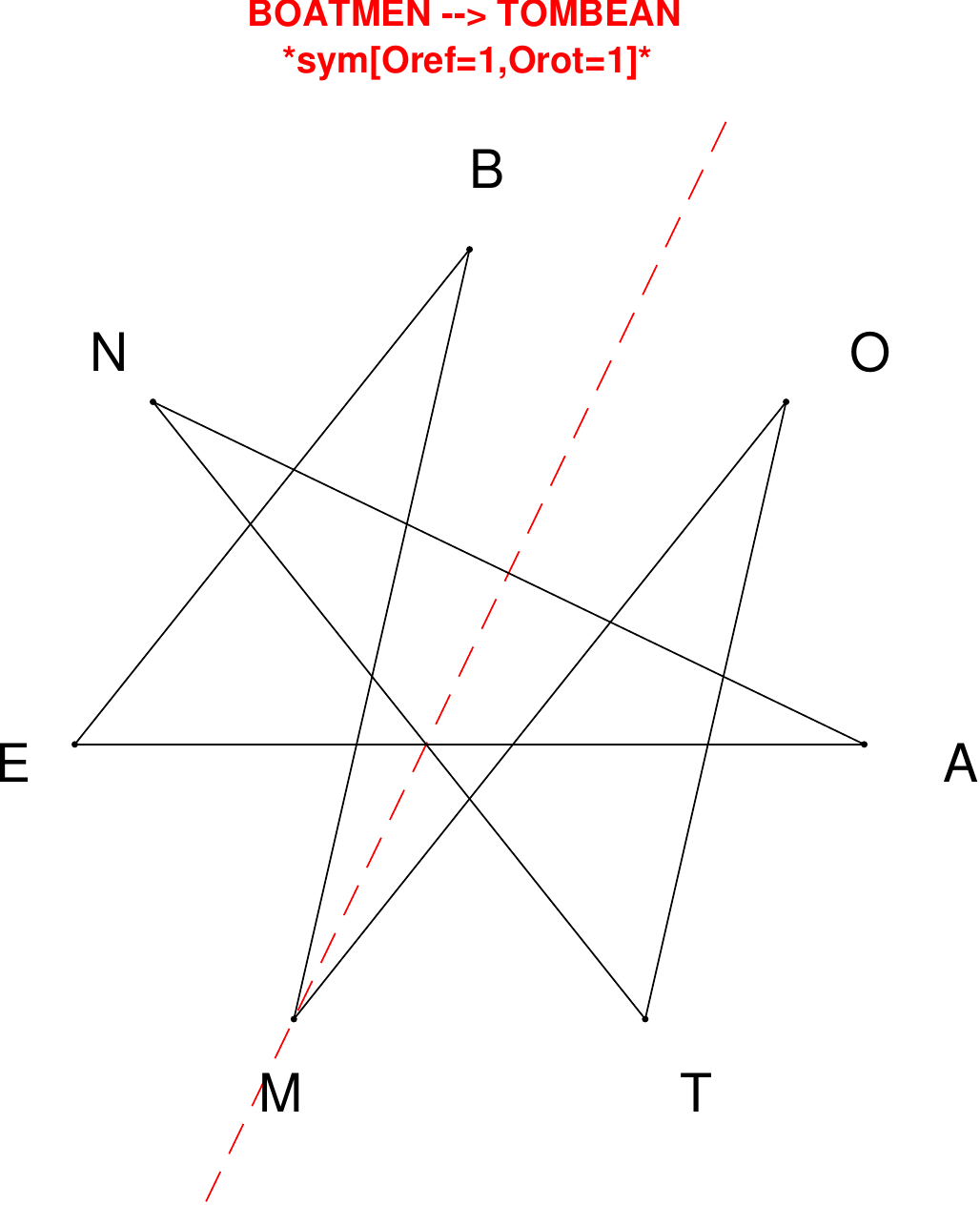}
\end{subfigure}
\hfill
\begin{subfigure}[T]{0.19\textwidth}
\centering
\includegraphics[width=\textwidth]{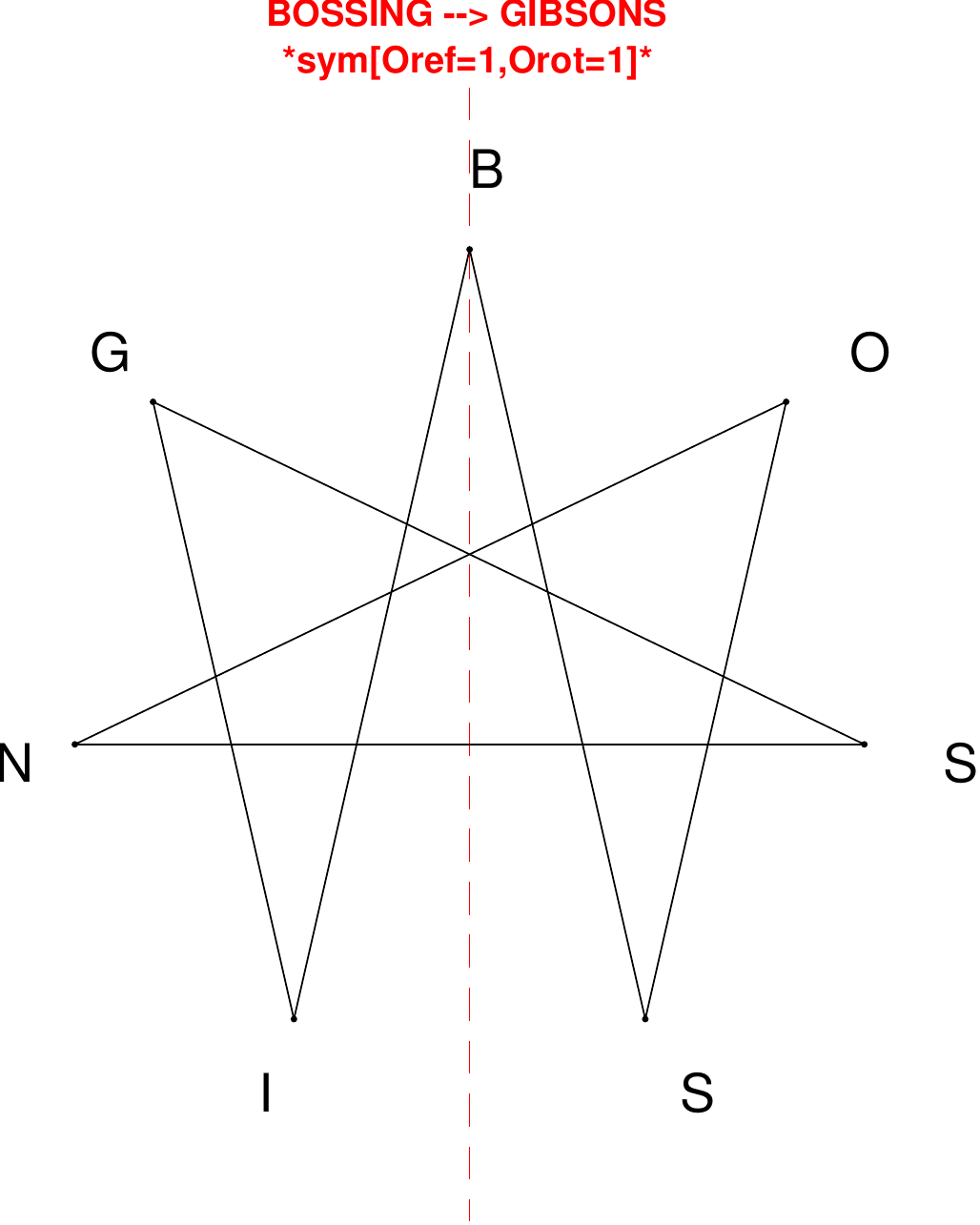}
\end{subfigure}
\hfill
\begin{subfigure}[T]{0.19\textwidth}
\centering
\includegraphics[width=\textwidth]{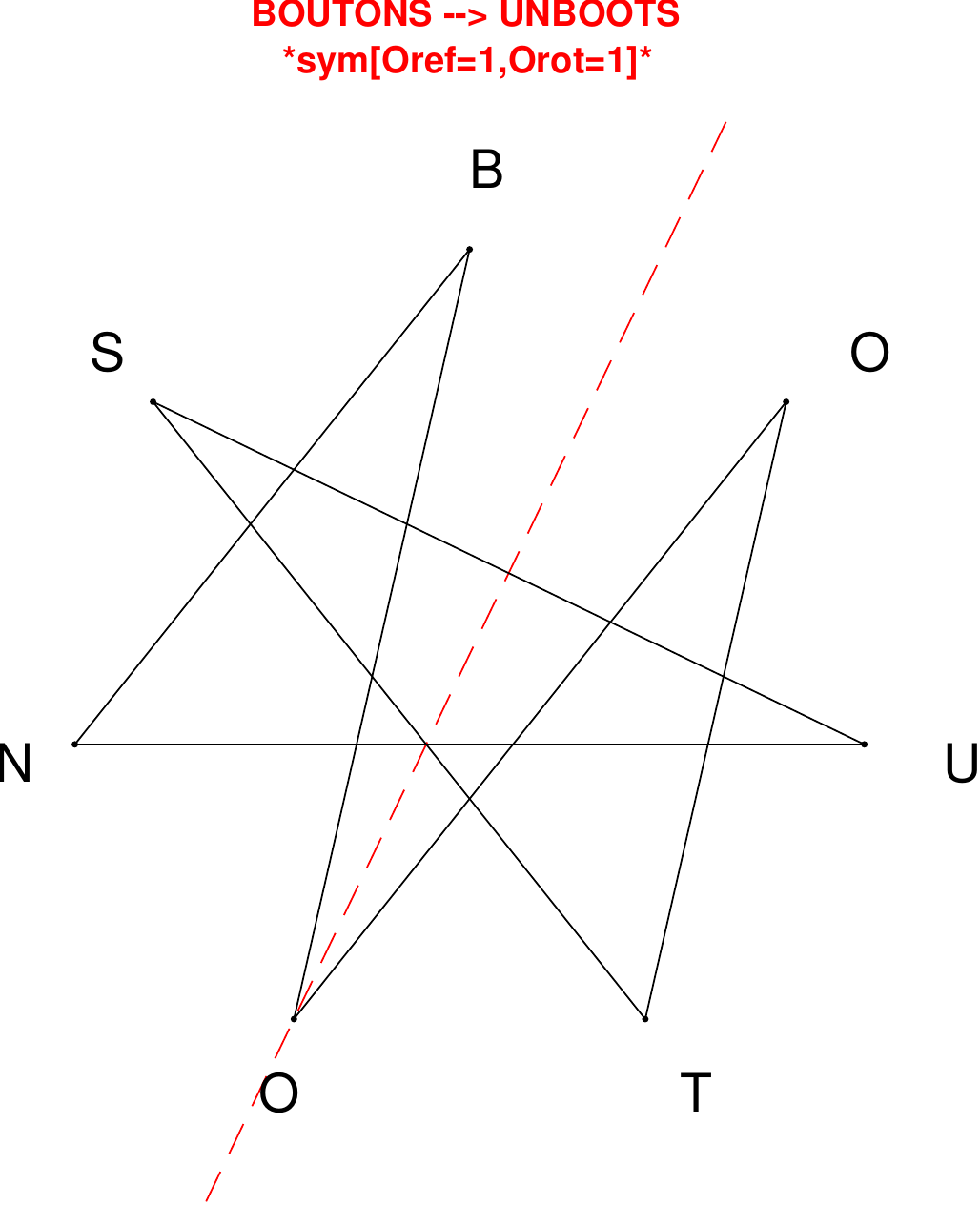}
\end{subfigure}
\end{figure}

\begin{figure}[H]
\centering
\begin{subfigure}[T]{0.19\textwidth}
\centering
\includegraphics[width=\textwidth]{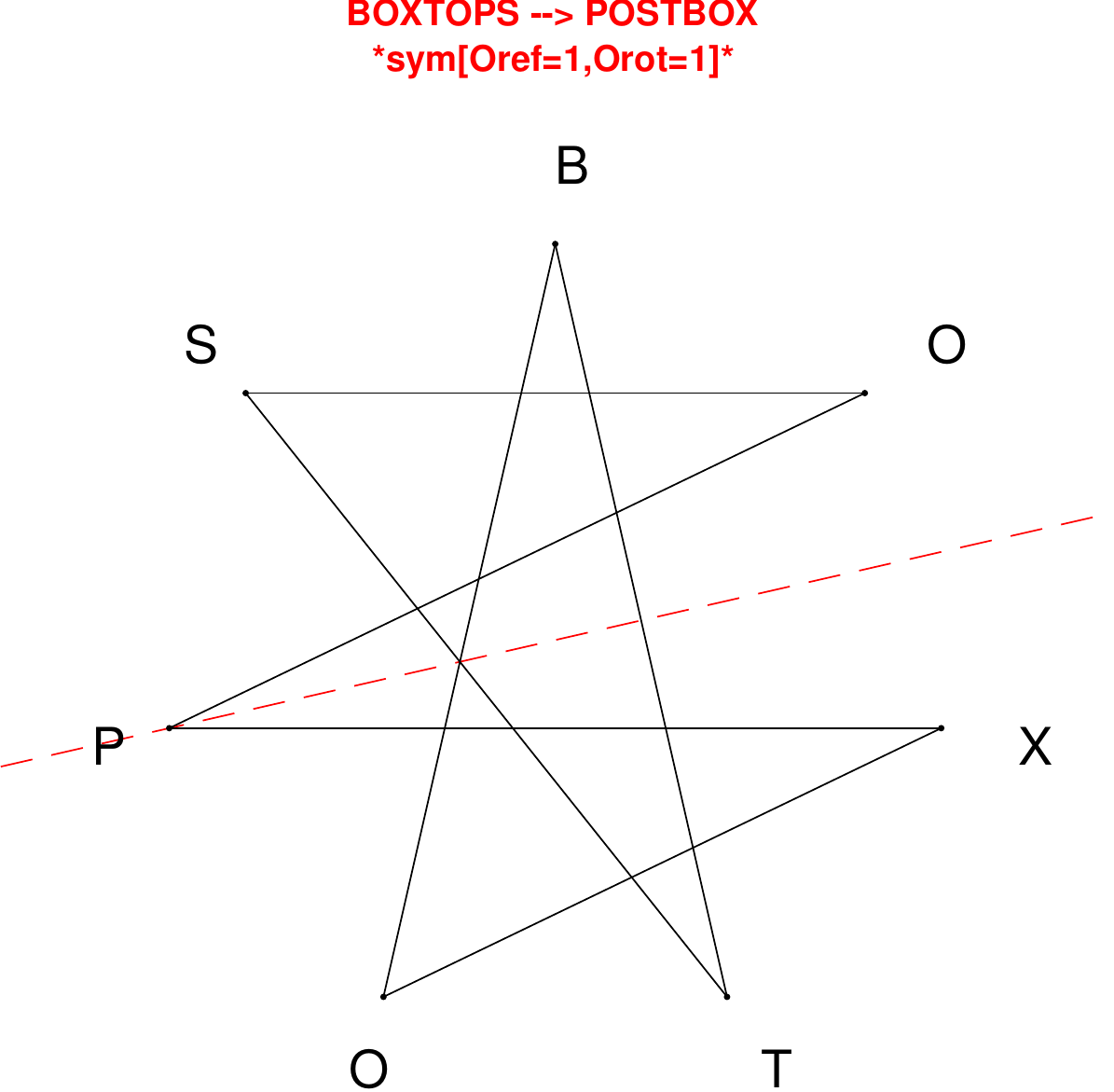}
\end{subfigure}
\hfill
\begin{subfigure}[T]{0.19\textwidth}
\centering
\includegraphics[width=\textwidth]{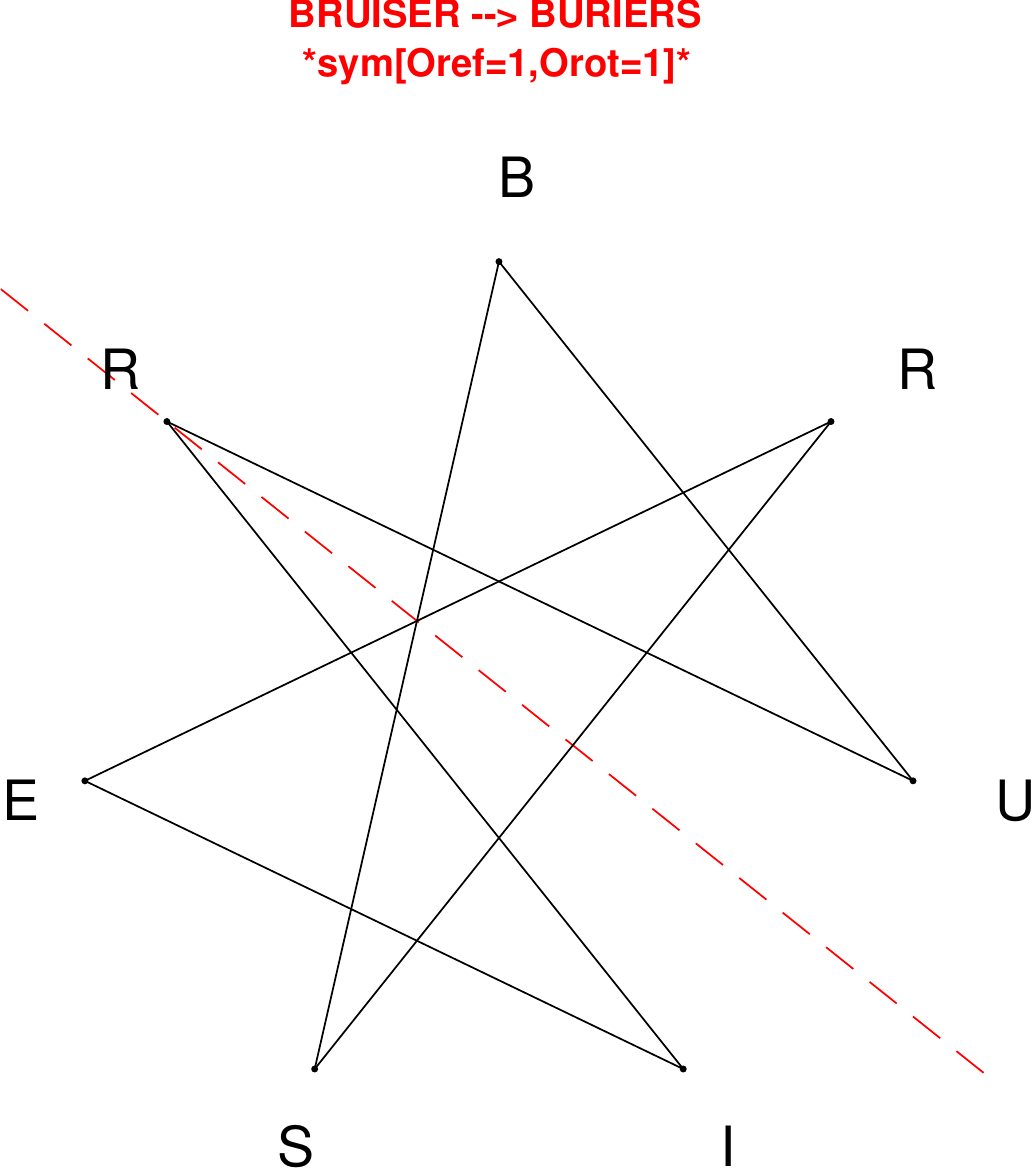}
\end{subfigure}
\hfill
\begin{subfigure}[T]{0.19\textwidth}
\centering
\includegraphics[width=\textwidth]{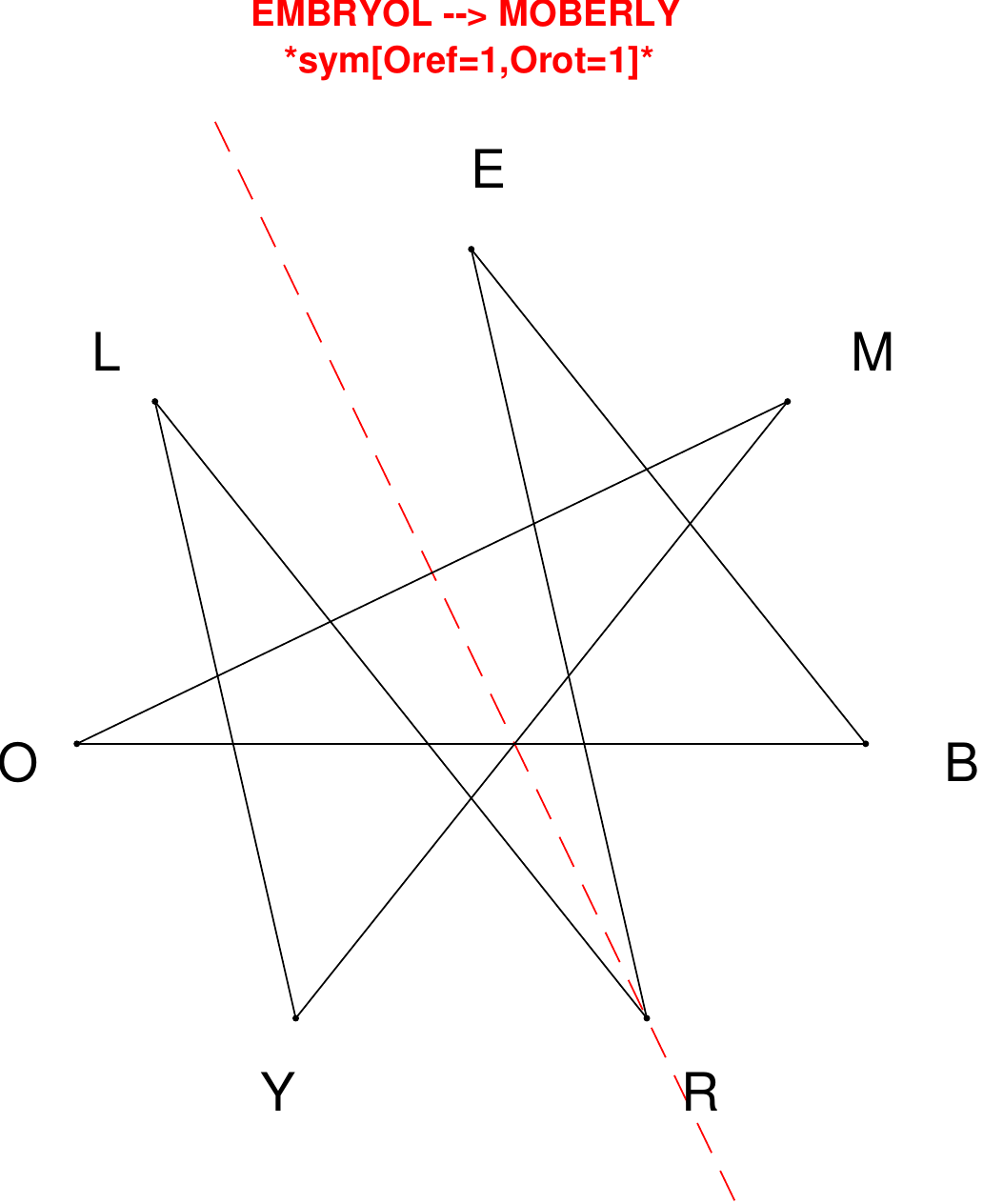}
\end{subfigure}
\hfill
\begin{subfigure}[T]{0.19\textwidth}
\centering
\includegraphics[width=\textwidth]{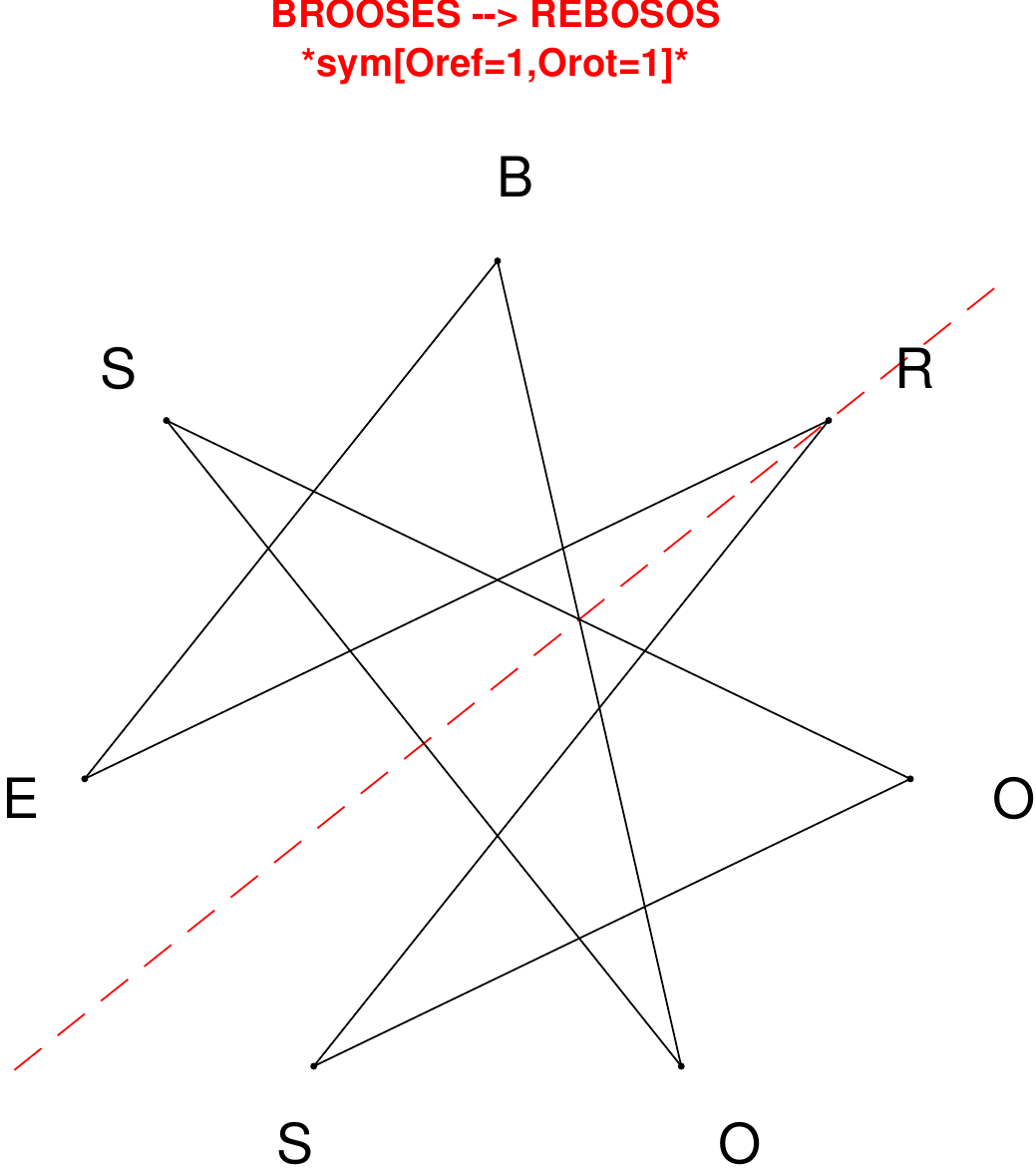}
\end{subfigure}
\hfill
\begin{subfigure}[T]{0.19\textwidth}
\centering
\includegraphics[width=\textwidth]{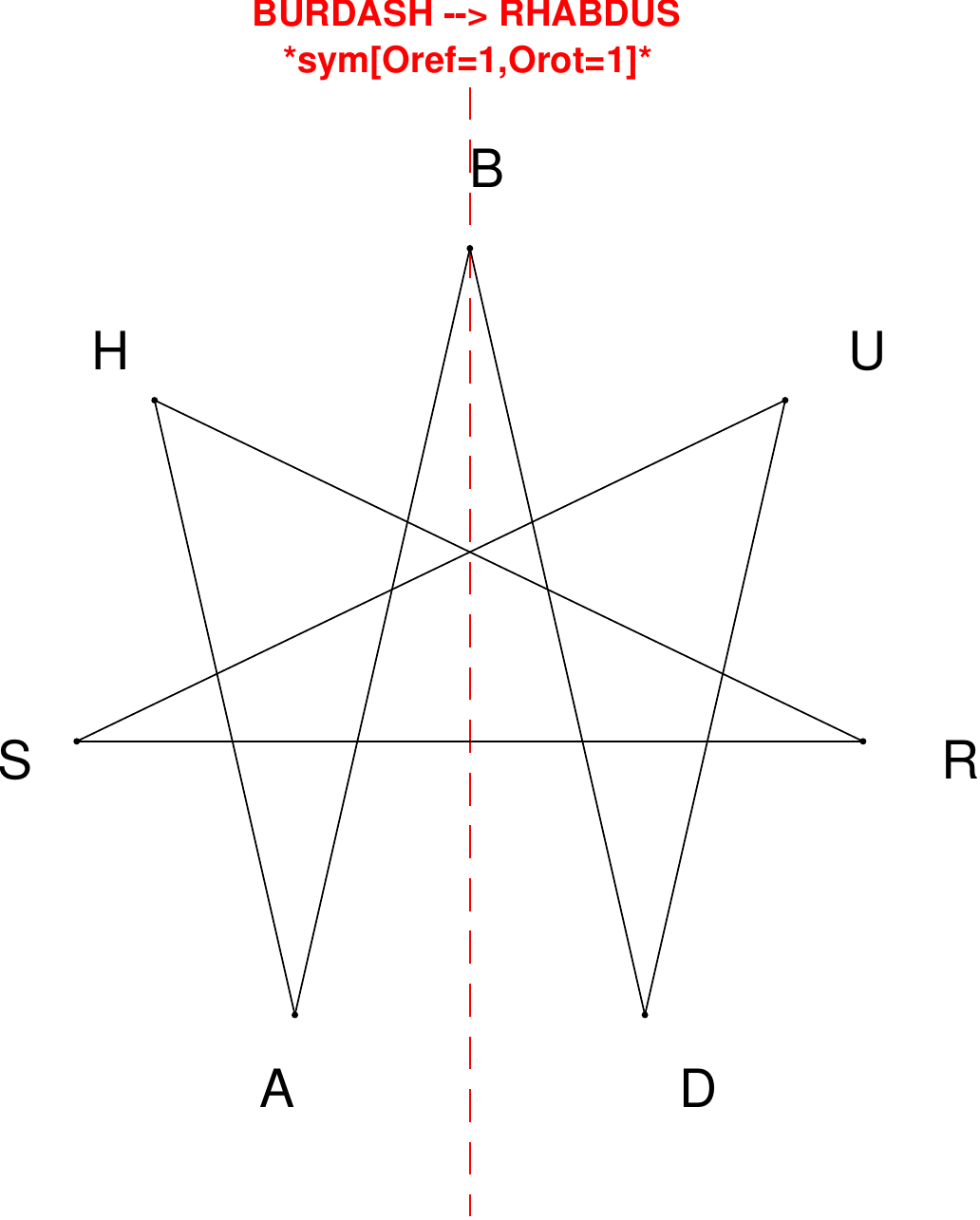}
\end{subfigure}
\end{figure}

\begin{figure}[H]
\centering
\begin{subfigure}[T]{0.19\textwidth}
\centering
\includegraphics[width=\textwidth]{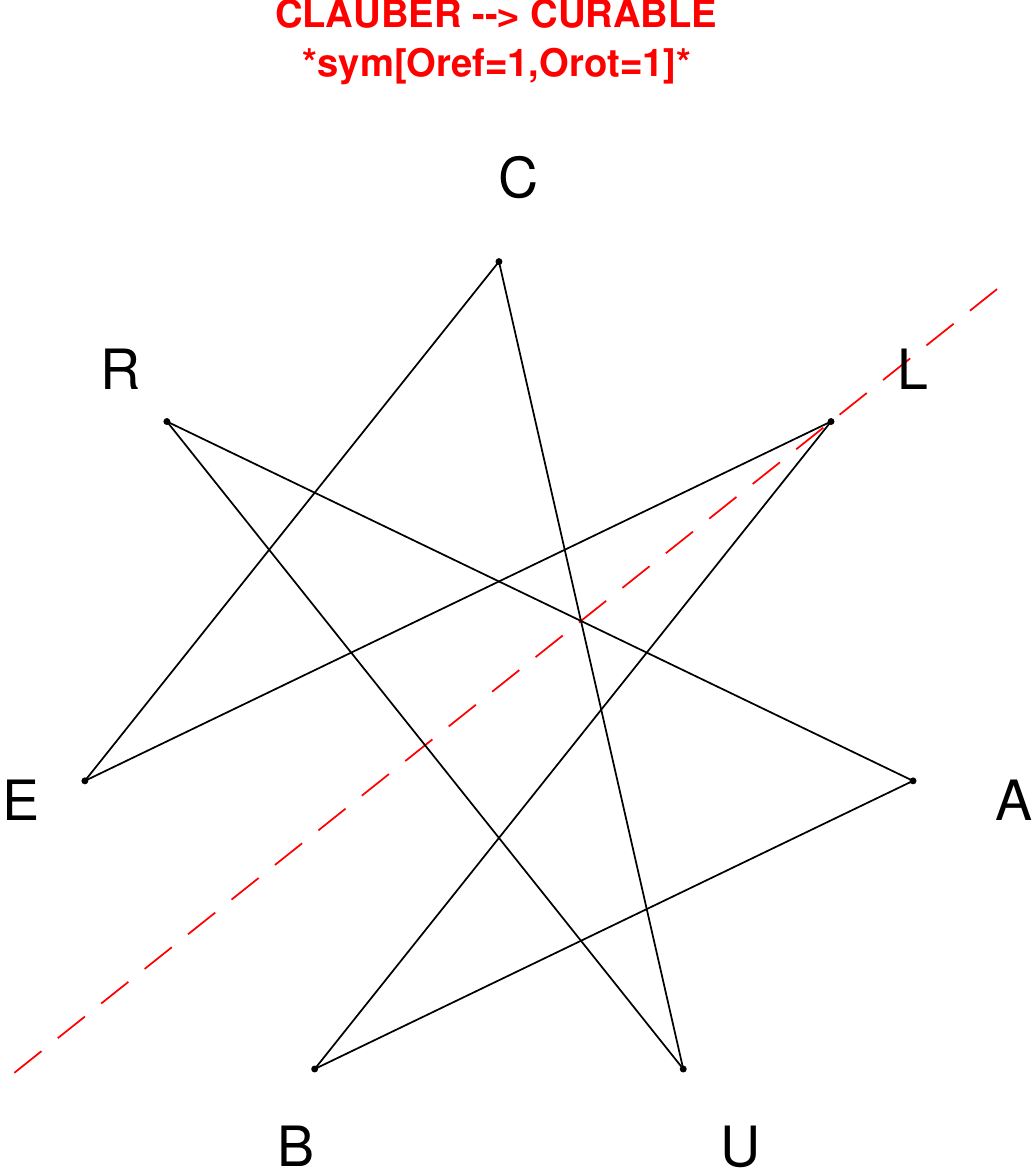}
\end{subfigure}
\hfill
\begin{subfigure}[T]{0.19\textwidth}
\centering
\includegraphics[width=\textwidth]{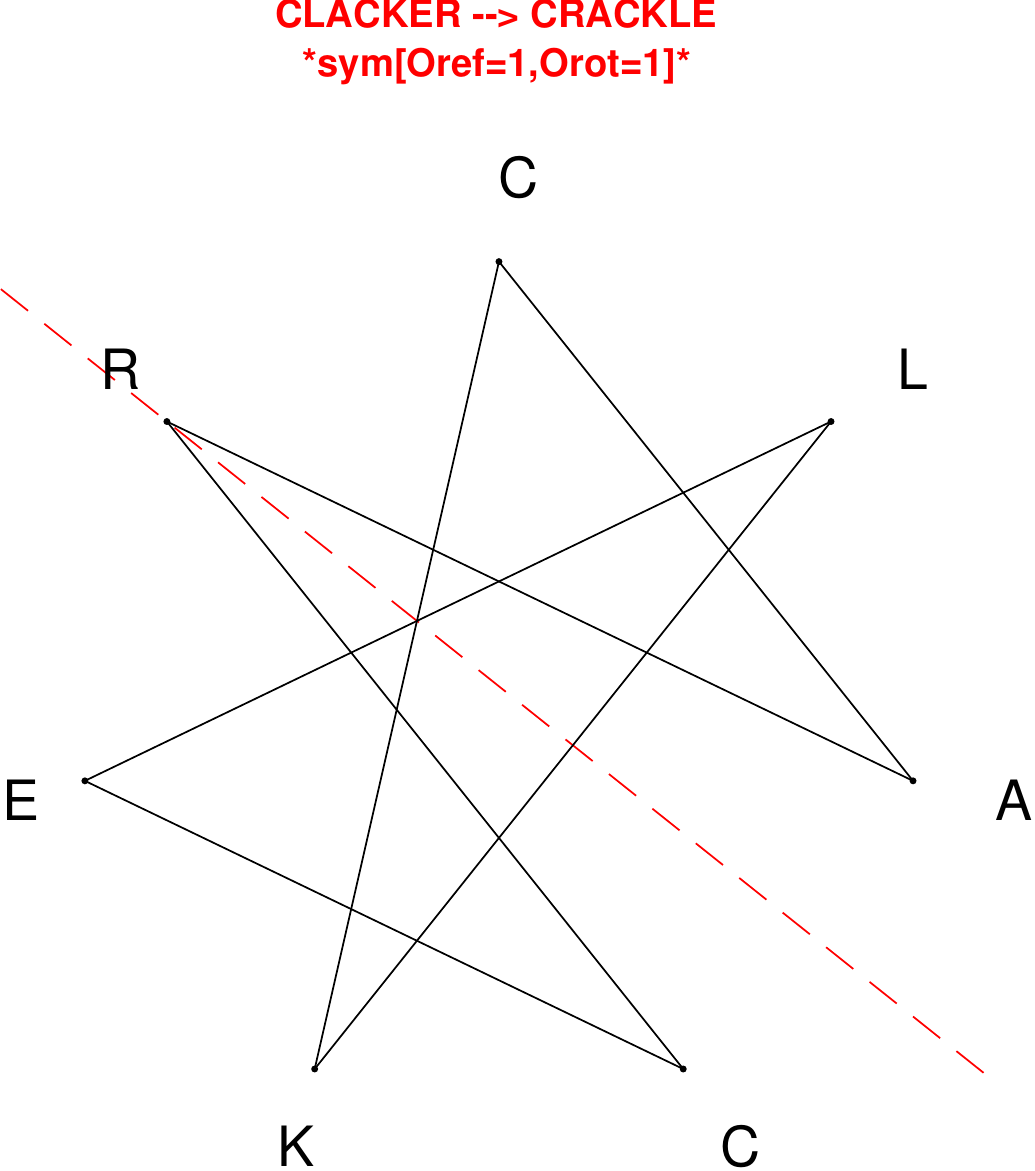}
\end{subfigure}
\hfill
\begin{subfigure}[T]{0.19\textwidth}
\centering
\includegraphics[width=\textwidth]{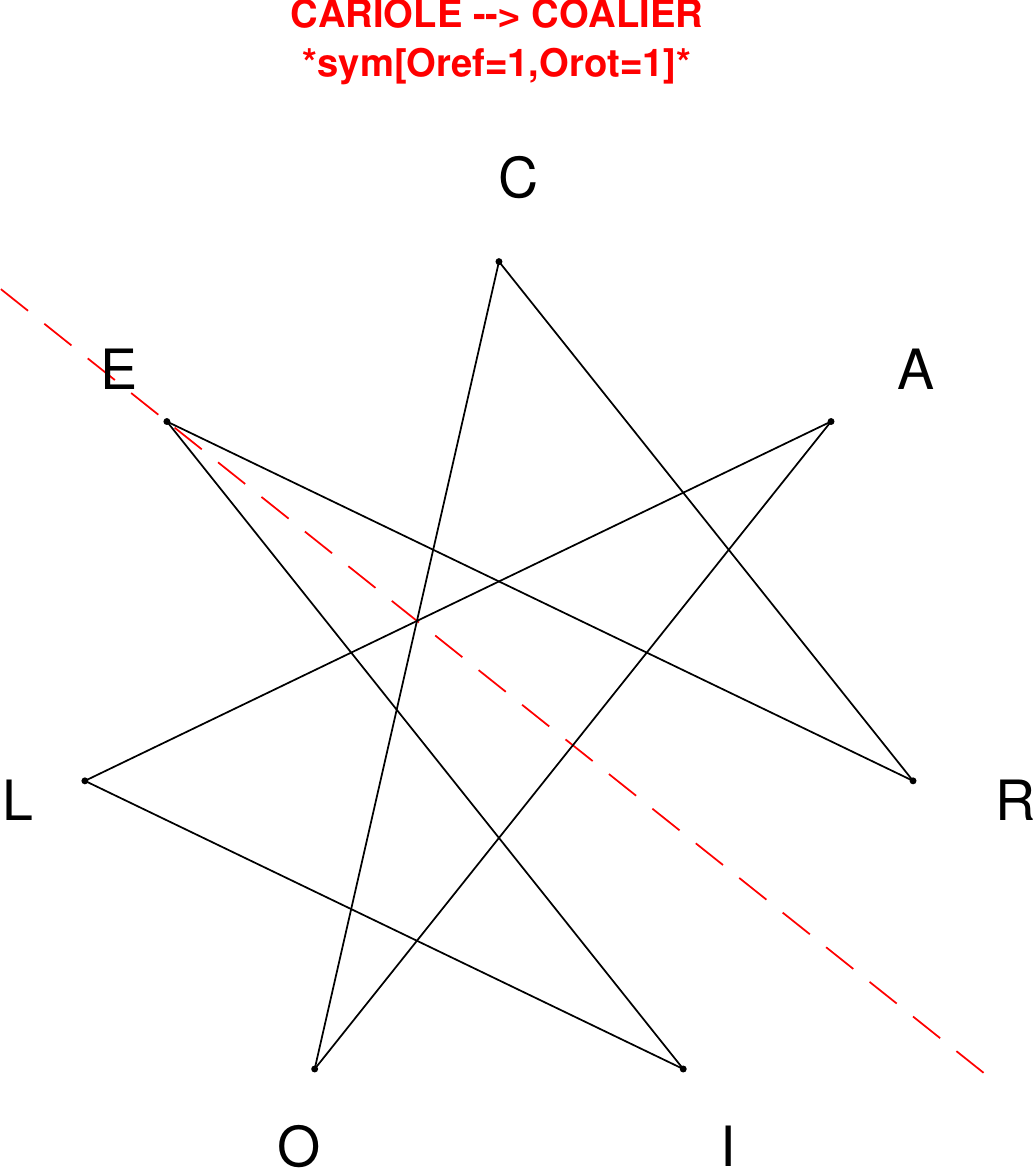}
\end{subfigure}
\hfill
\begin{subfigure}[T]{0.19\textwidth}
\centering
\includegraphics[width=\textwidth]{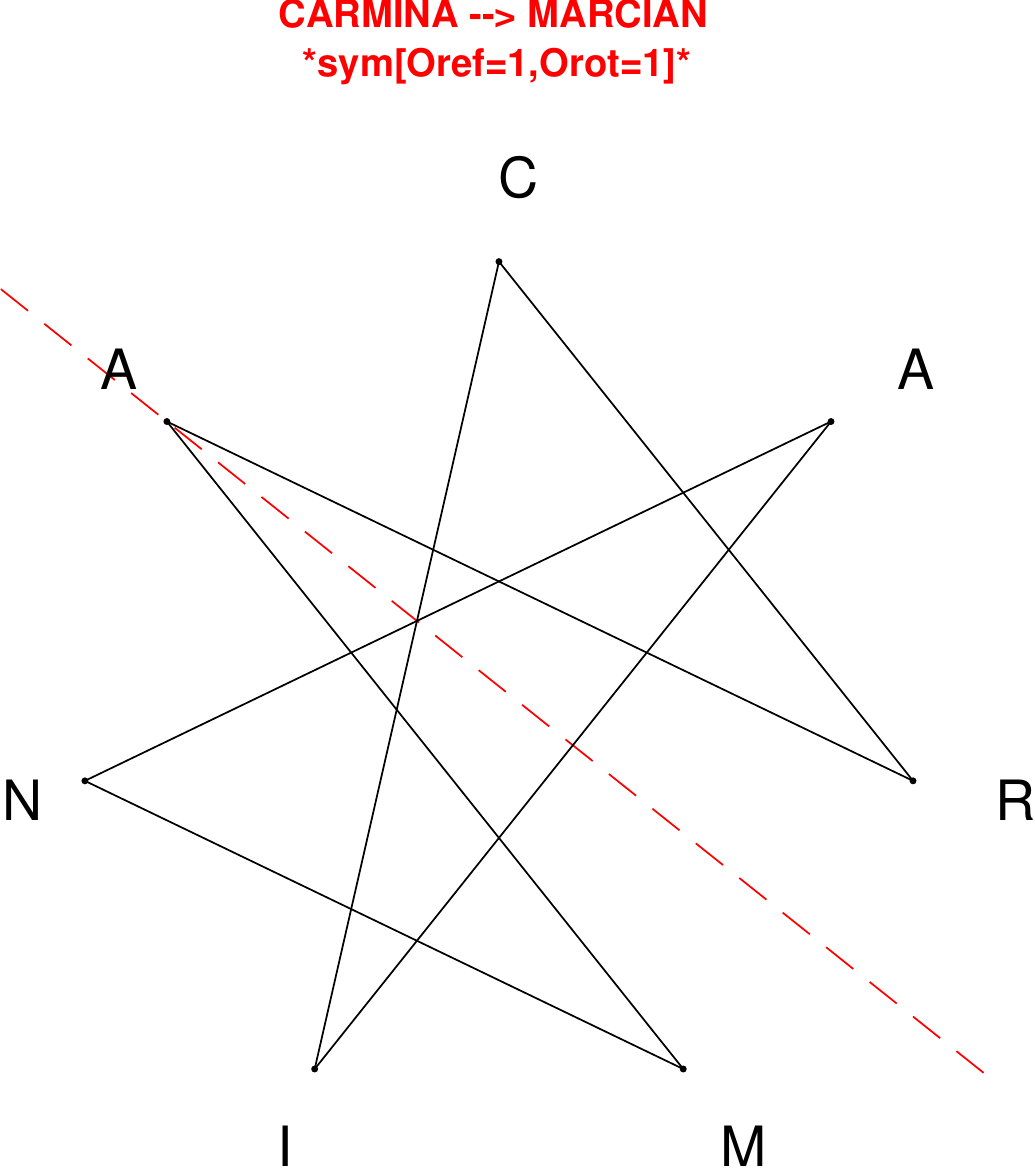}
\end{subfigure}
\hfill
\begin{subfigure}[T]{0.19\textwidth}
\centering
\includegraphics[width=\textwidth]{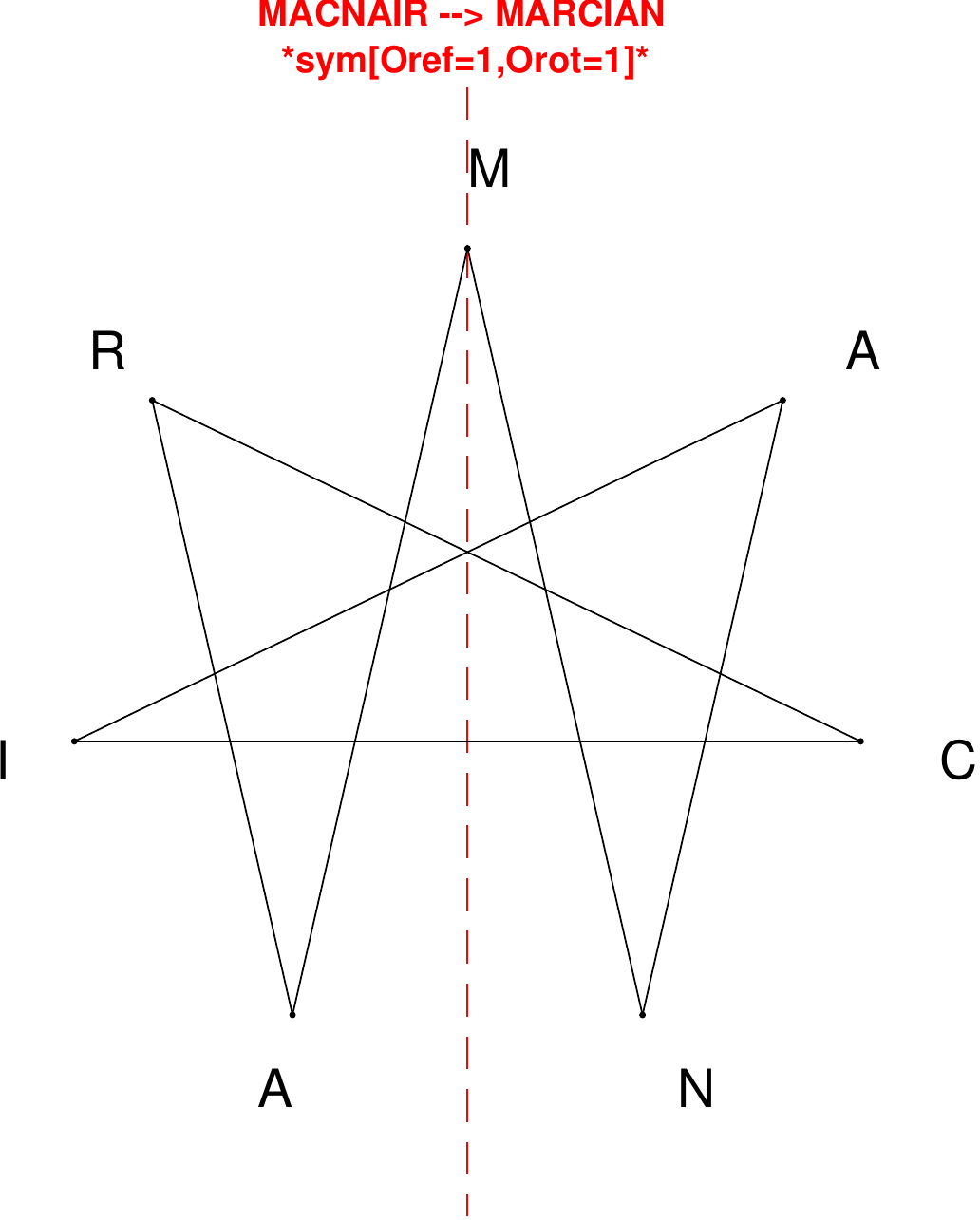}
\end{subfigure}
\end{figure}

\begin{figure}[H]
\centering
\begin{subfigure}[T]{0.19\textwidth}
\centering
\includegraphics[width=\textwidth]{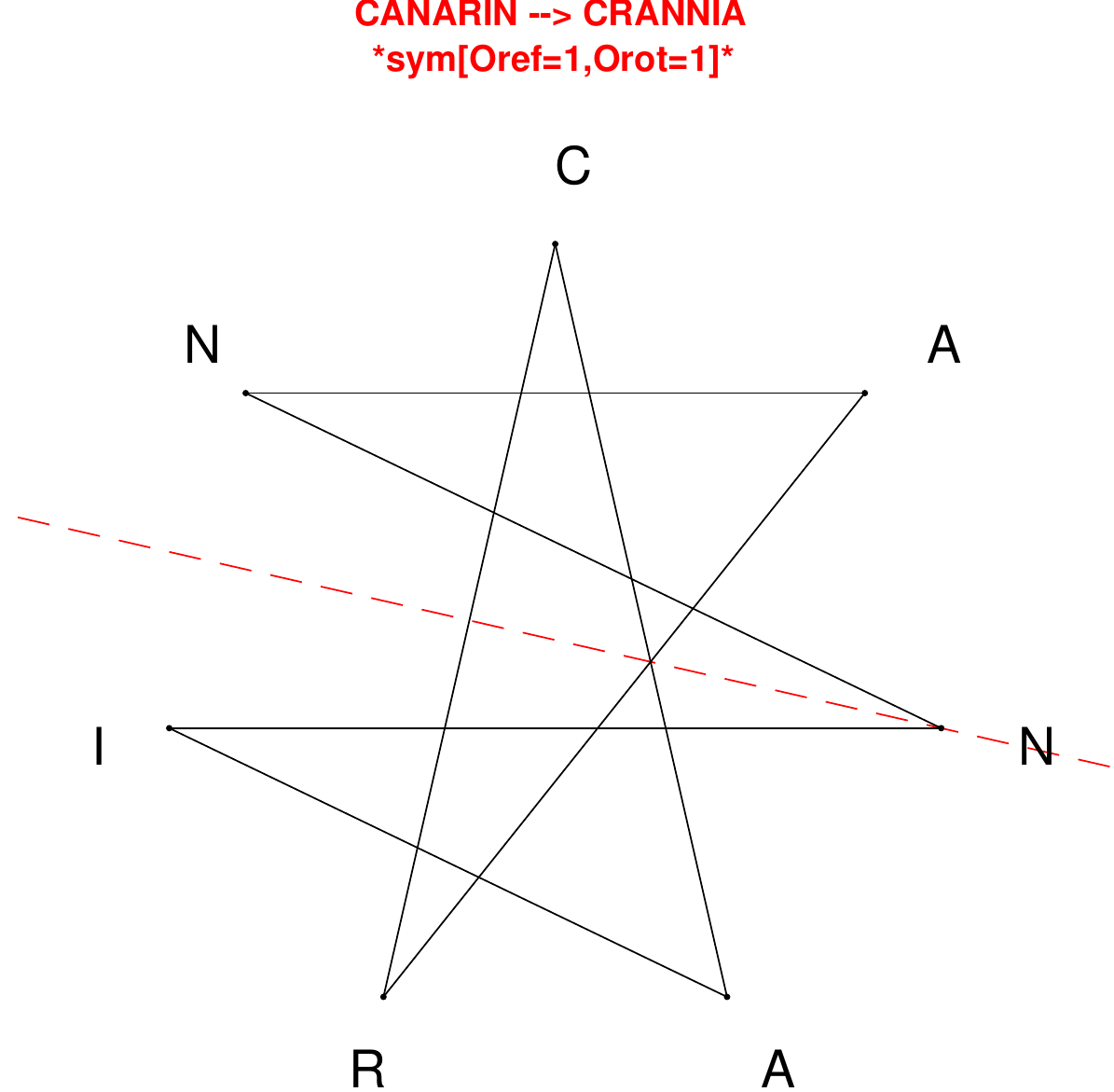}
\end{subfigure}
\hfill
\begin{subfigure}[T]{0.19\textwidth}
\centering
\includegraphics[width=\textwidth]{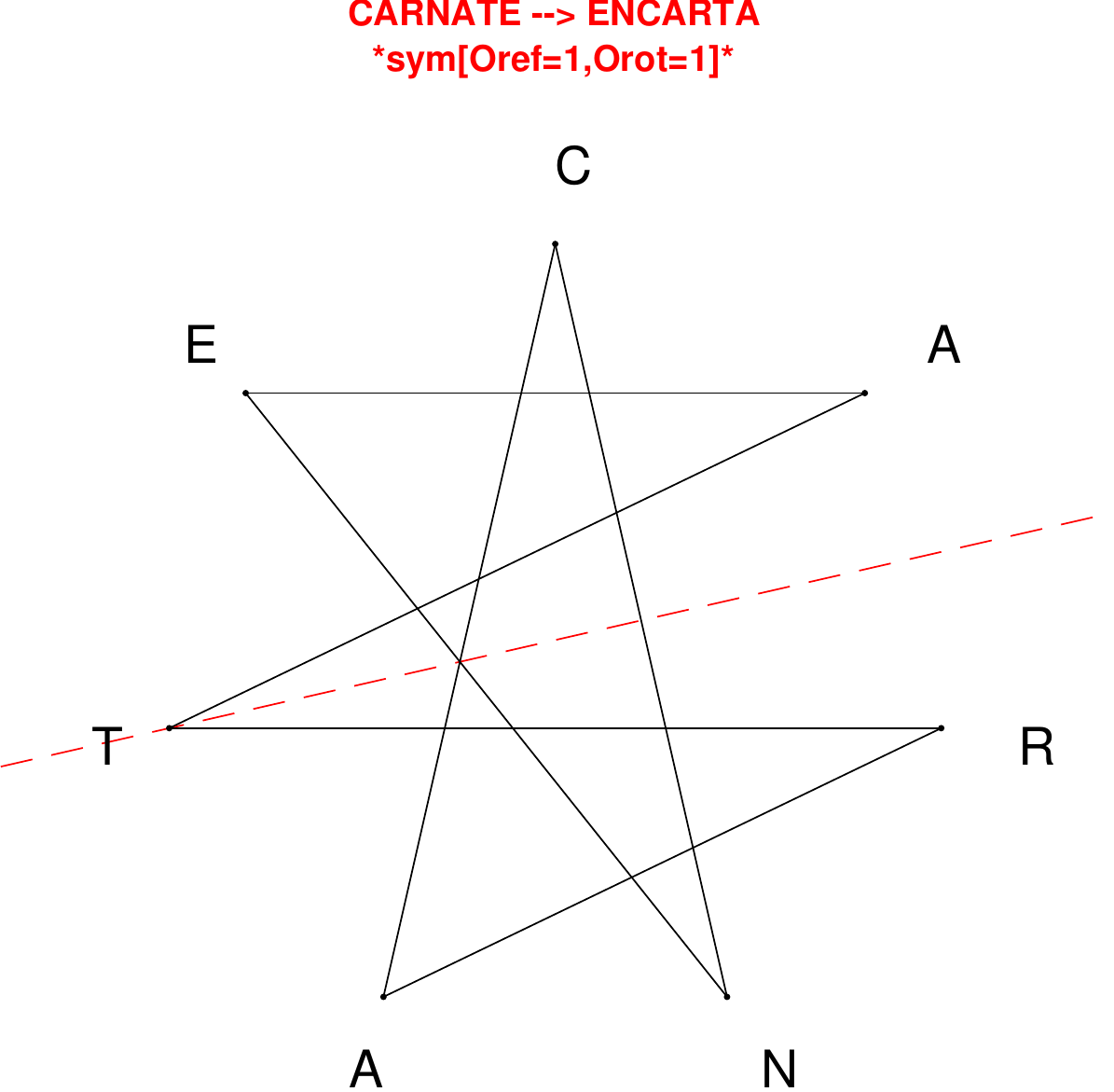}
\end{subfigure}
\hfill
\begin{subfigure}[T]{0.19\textwidth}
\centering
\includegraphics[width=\textwidth]{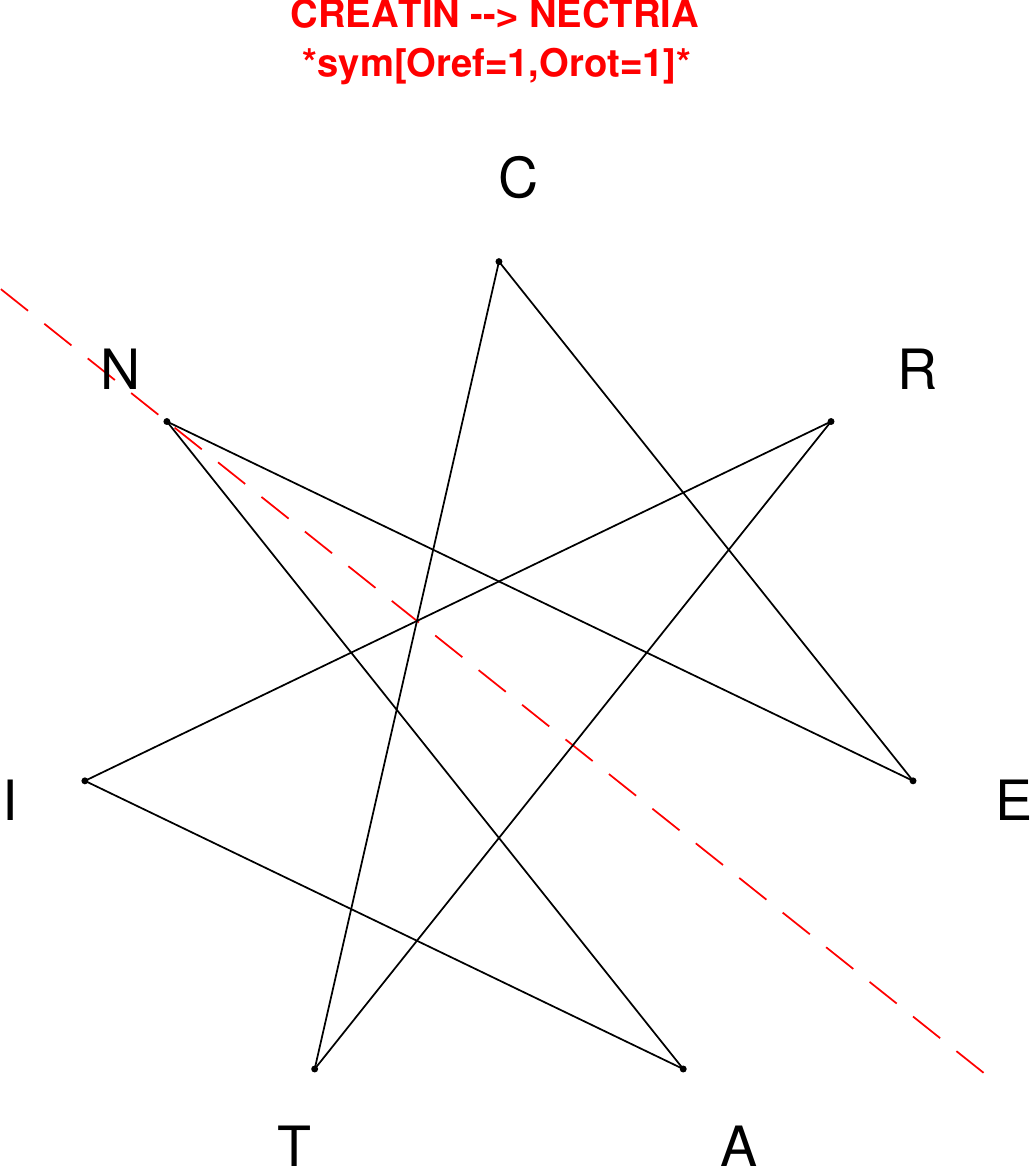}
\end{subfigure}
\hfill
\begin{subfigure}[T]{0.19\textwidth}
\centering
\includegraphics[width=\textwidth]{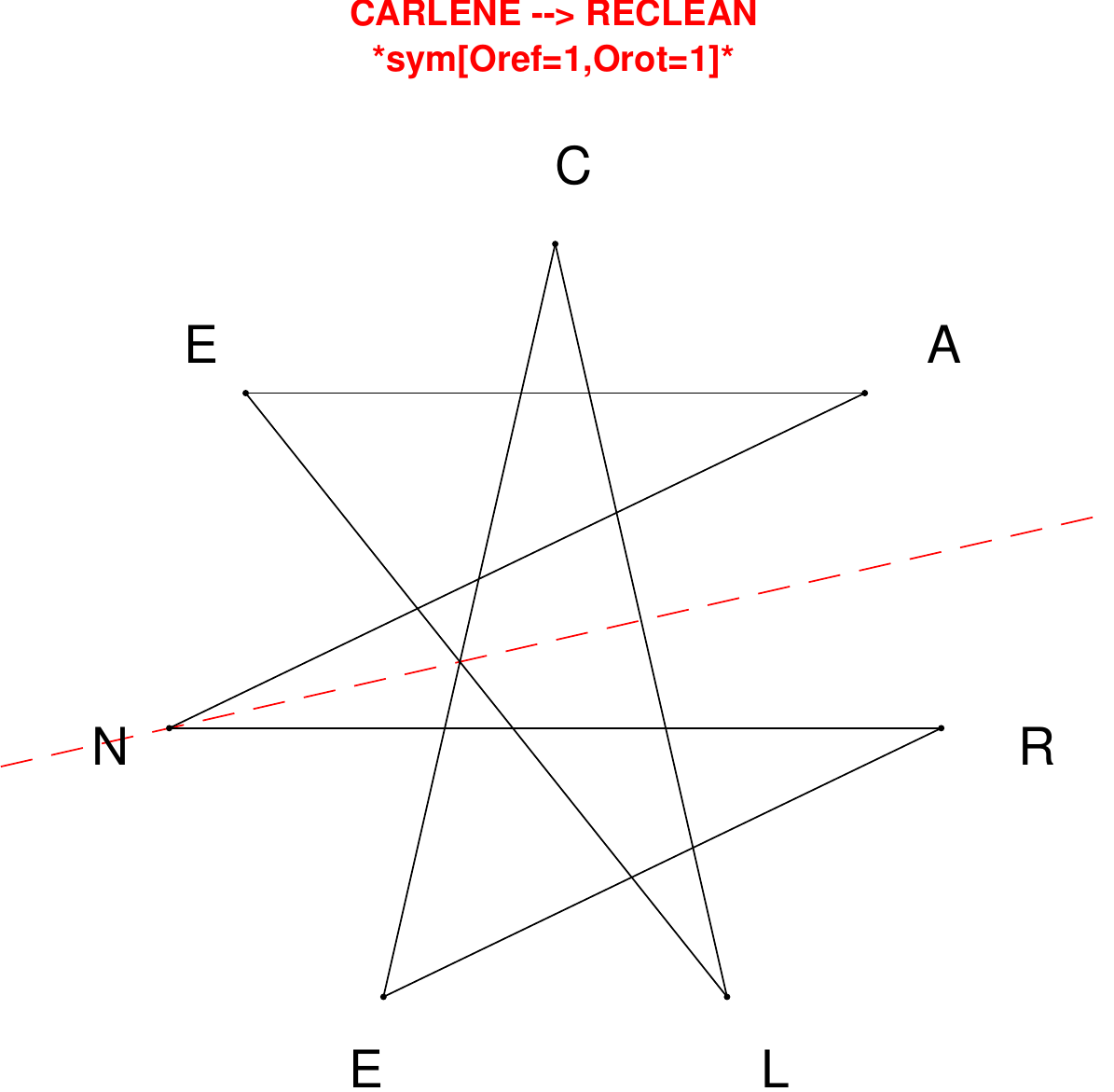}
\end{subfigure}
\hfill
\begin{subfigure}[T]{0.19\textwidth}
\centering
\includegraphics[width=\textwidth]{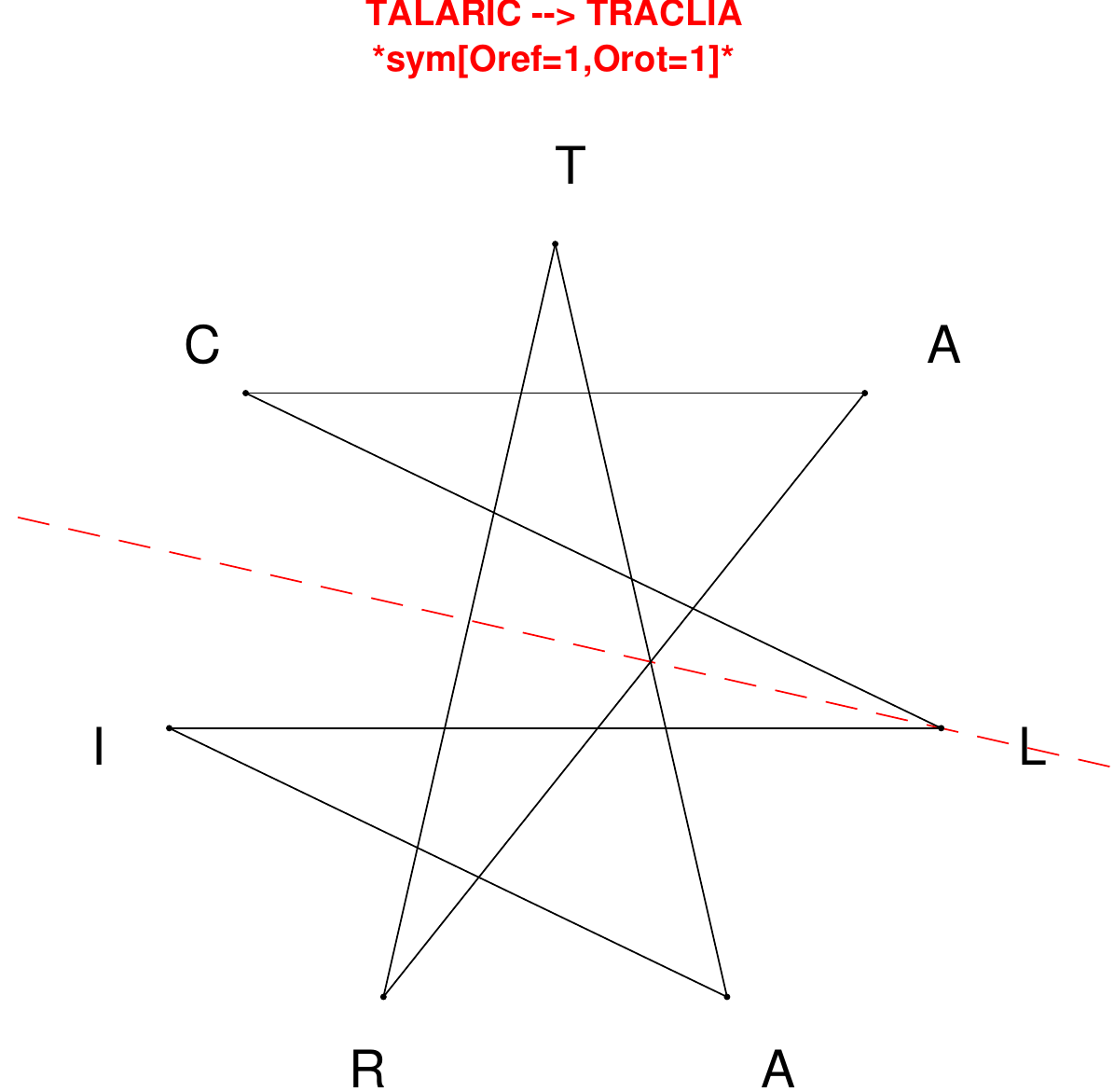}
\end{subfigure}
\end{figure}

\begin{figure}[H]
\centering
\begin{subfigure}[T]{0.19\textwidth}
\centering
\includegraphics[width=\textwidth]{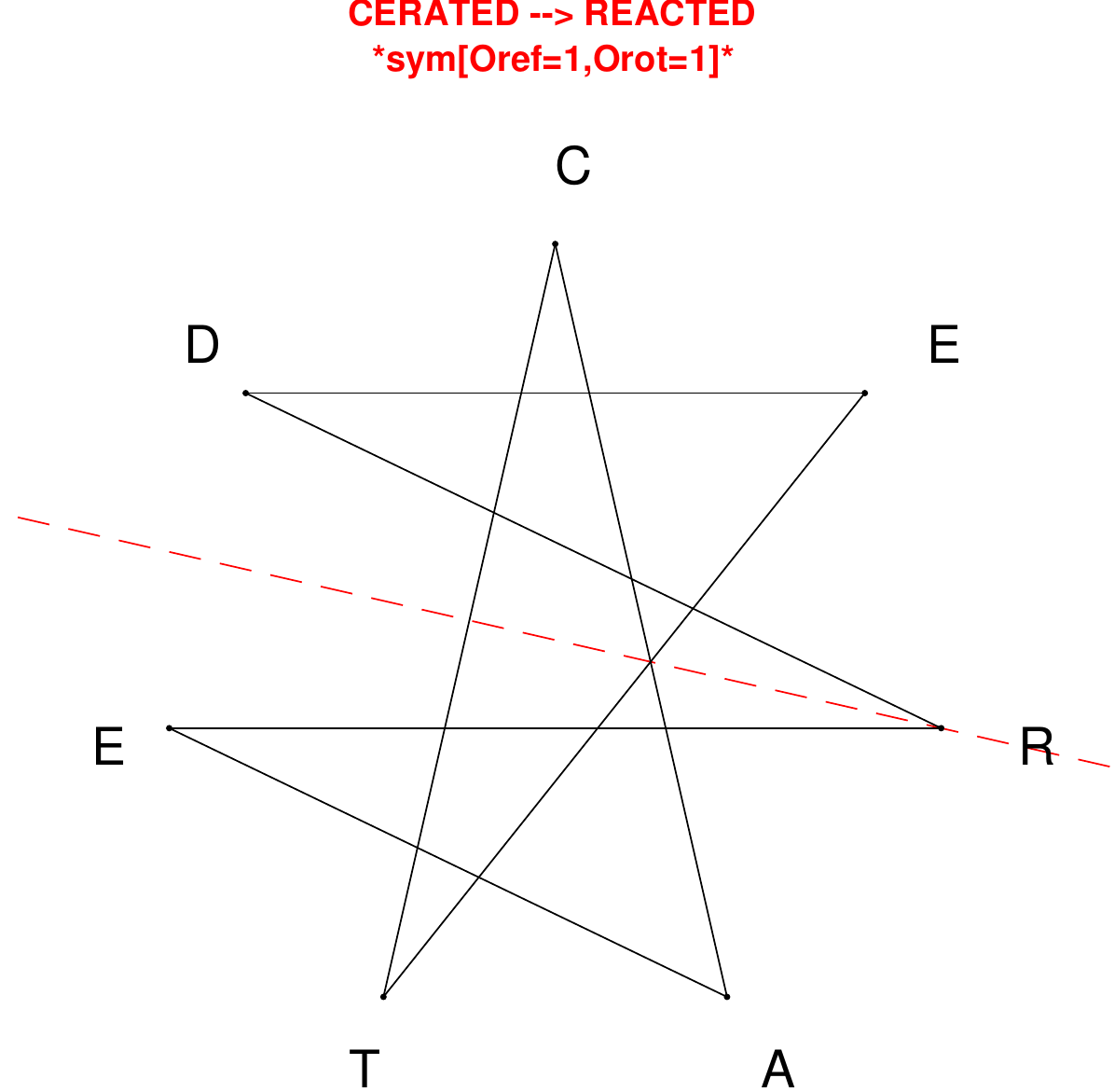}
\end{subfigure}
\hfill
\begin{subfigure}[T]{0.19\textwidth}
\centering
\includegraphics[width=\textwidth]{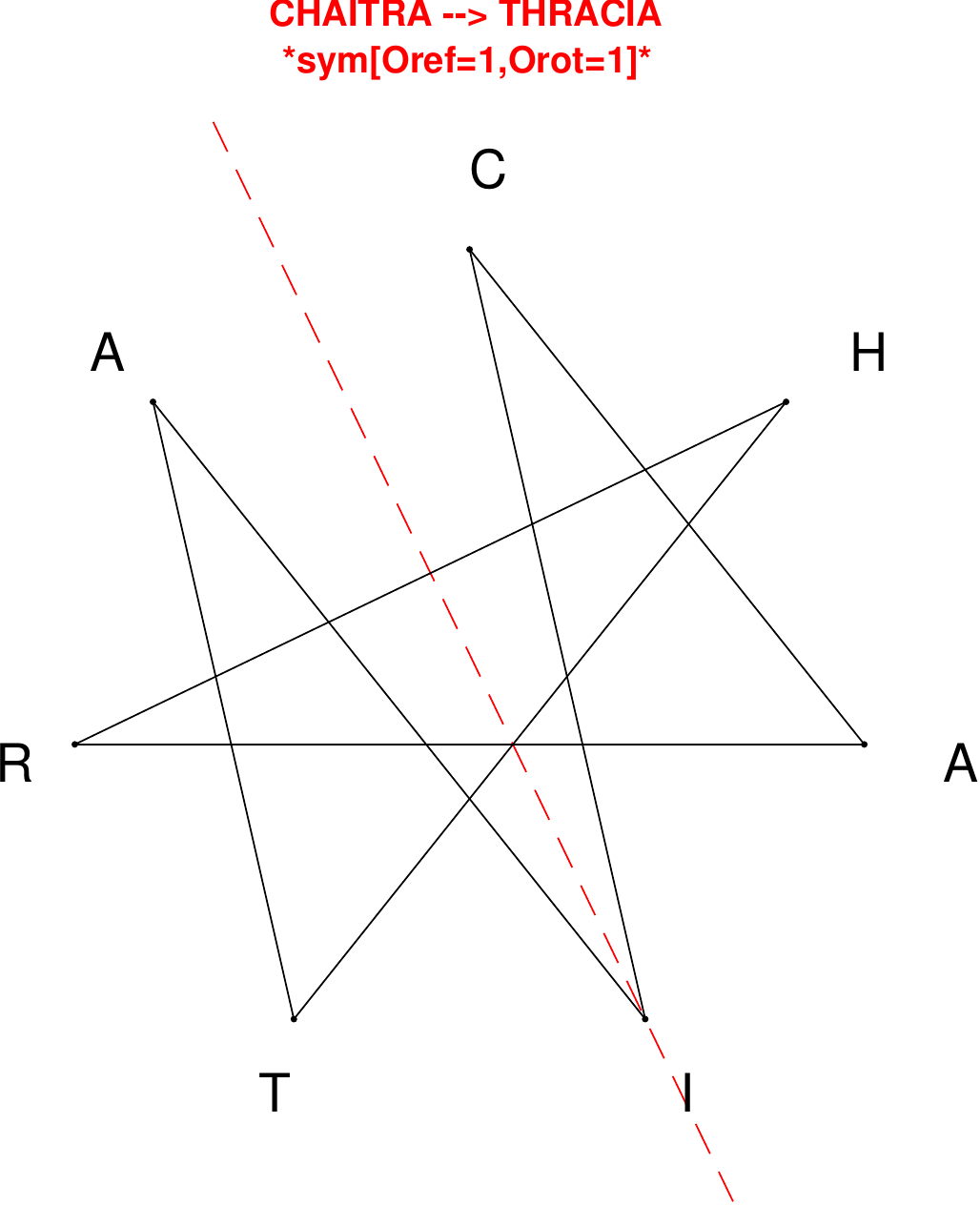}
\end{subfigure}
\hfill
\begin{subfigure}[T]{0.19\textwidth}
\centering
\includegraphics[width=\textwidth]{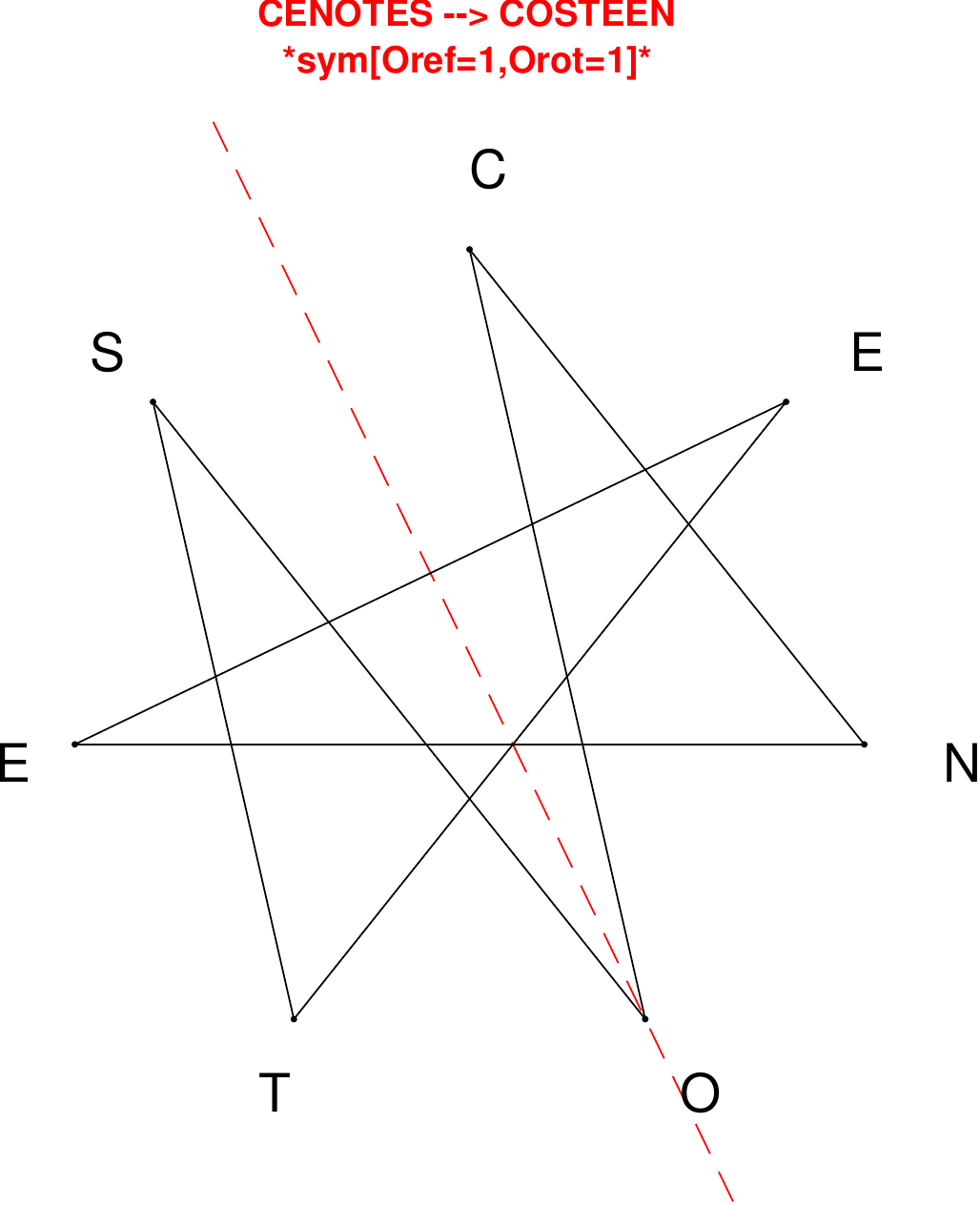}
\end{subfigure}
\hfill
\begin{subfigure}[T]{0.19\textwidth}
\centering
\includegraphics[width=\textwidth]{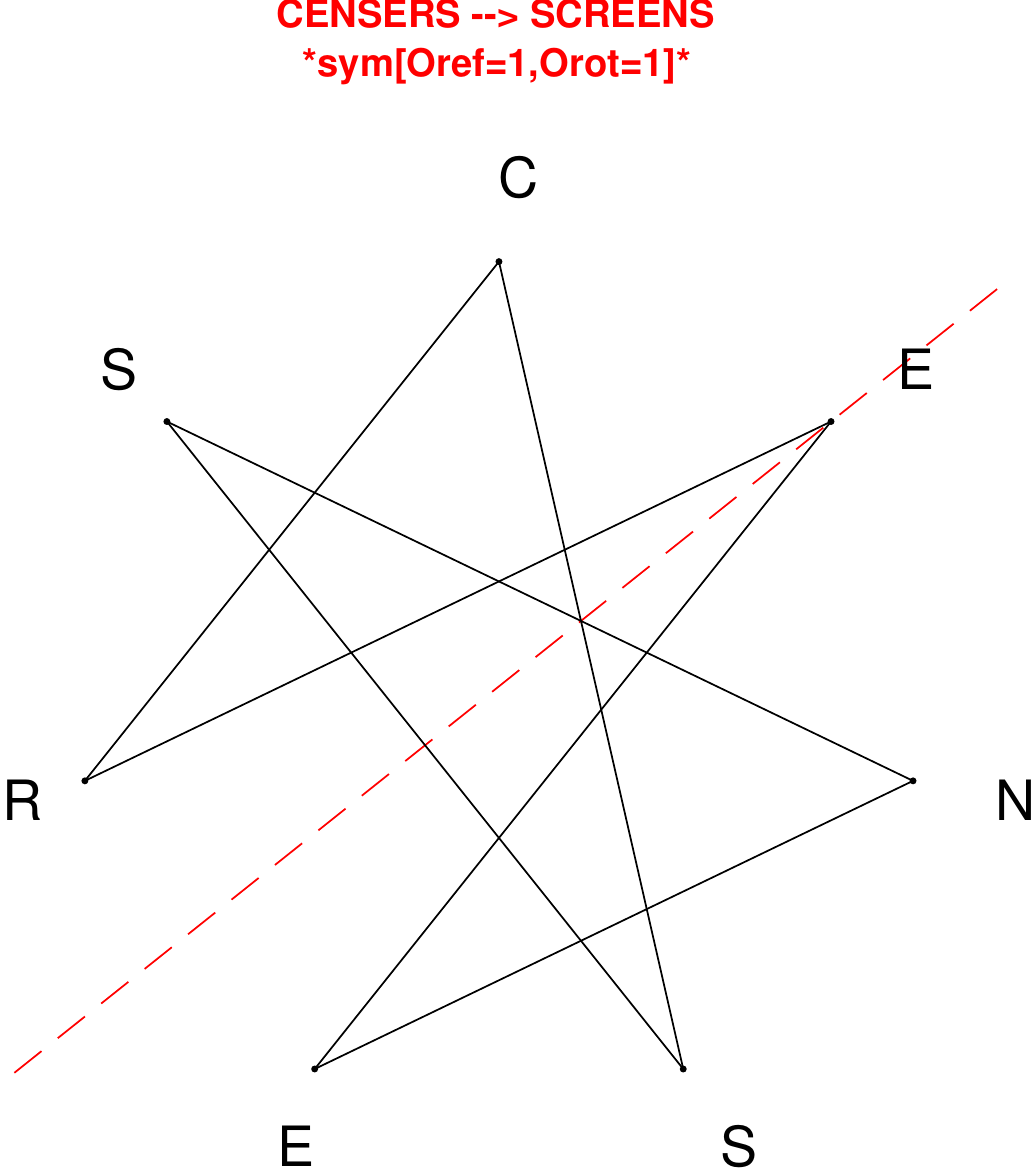}
\end{subfigure}
\hfill
\begin{subfigure}[T]{0.19\textwidth}
\centering
\includegraphics[width=\textwidth]{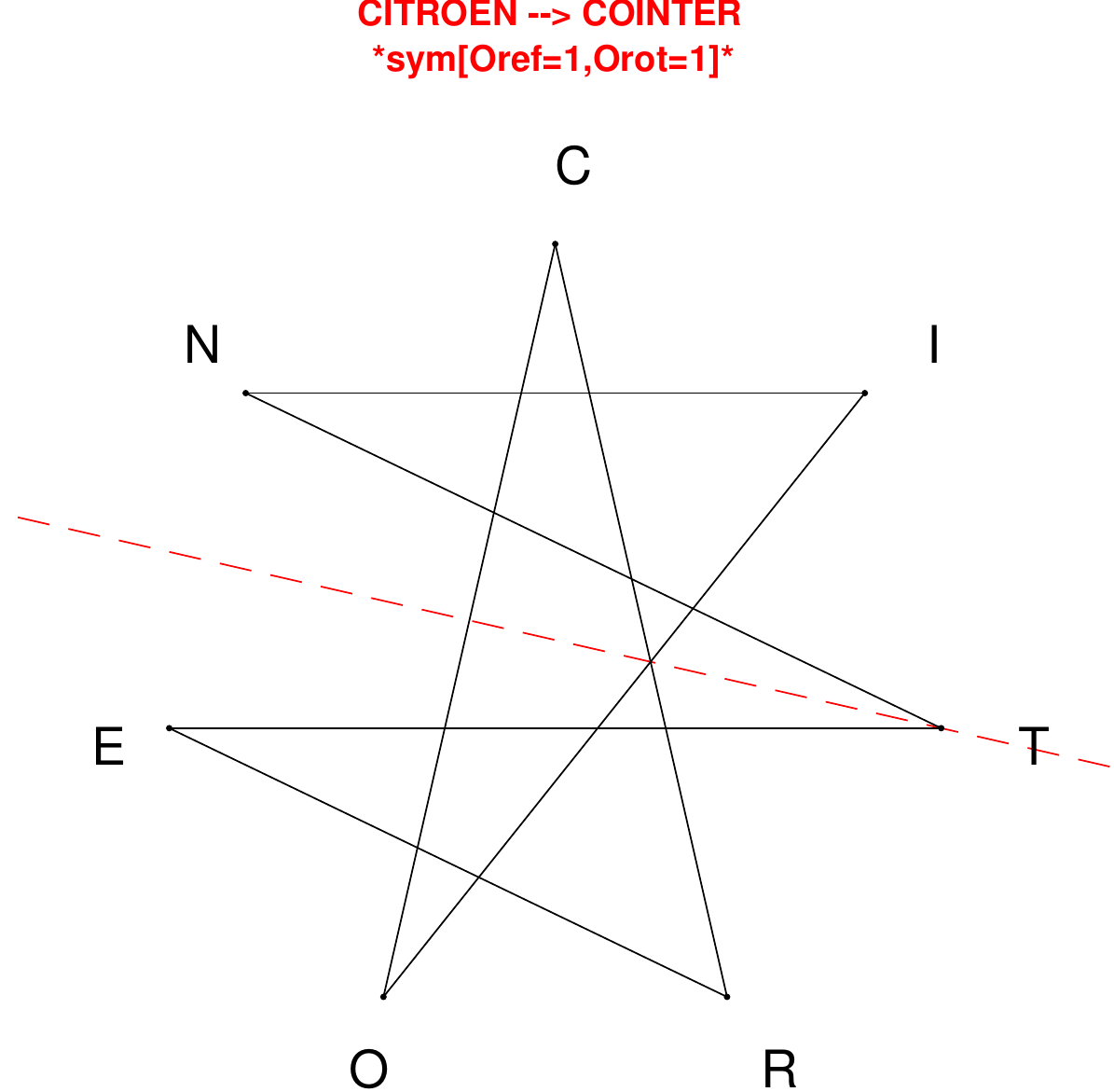}
\end{subfigure}
\end{figure}

\begin{figure}[H]
\centering
\begin{subfigure}[T]{0.19\textwidth}
\centering
\includegraphics[width=\textwidth]{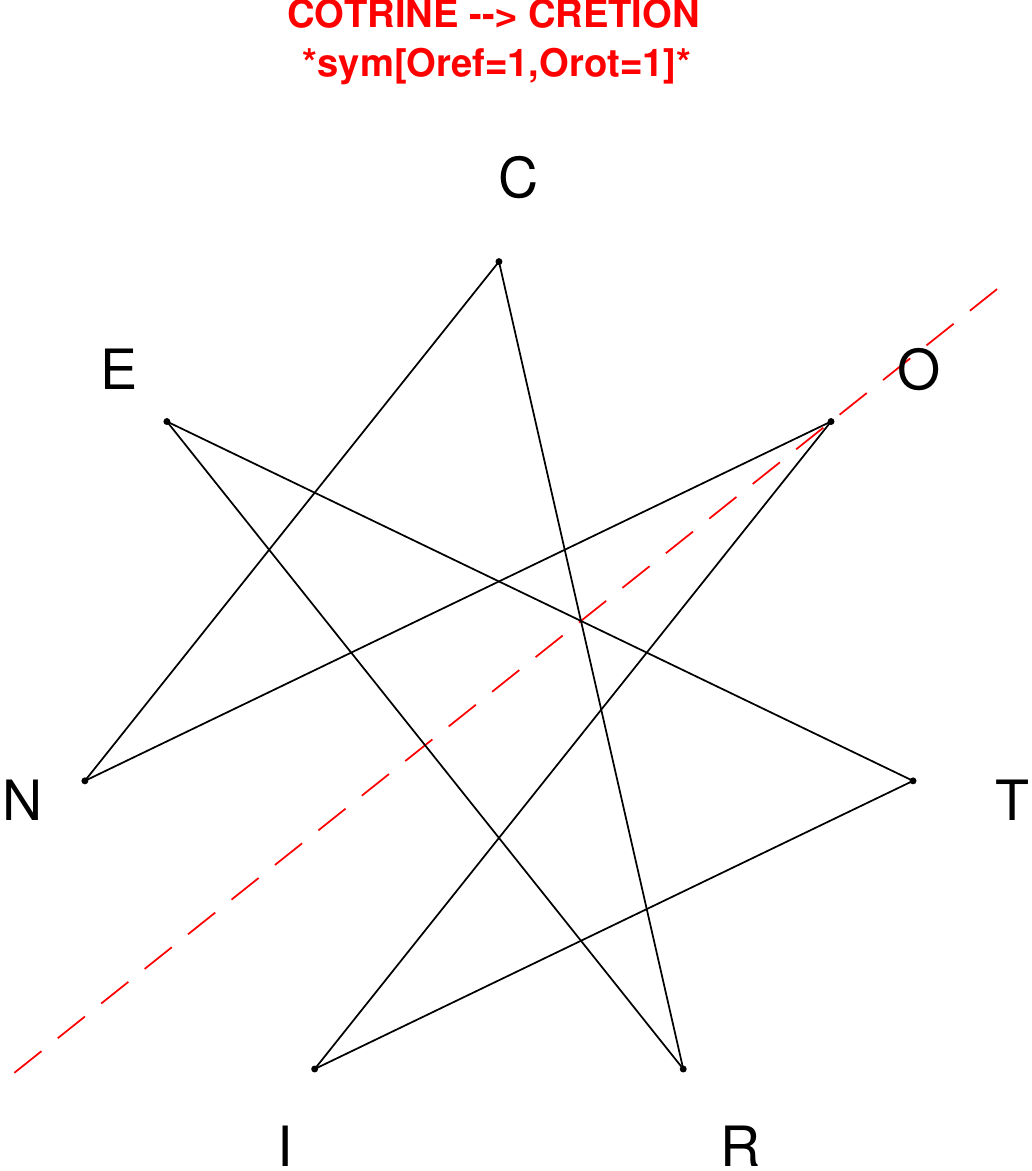}
\end{subfigure}
\hfill
\begin{subfigure}[T]{0.19\textwidth}
\centering
\includegraphics[width=\textwidth]{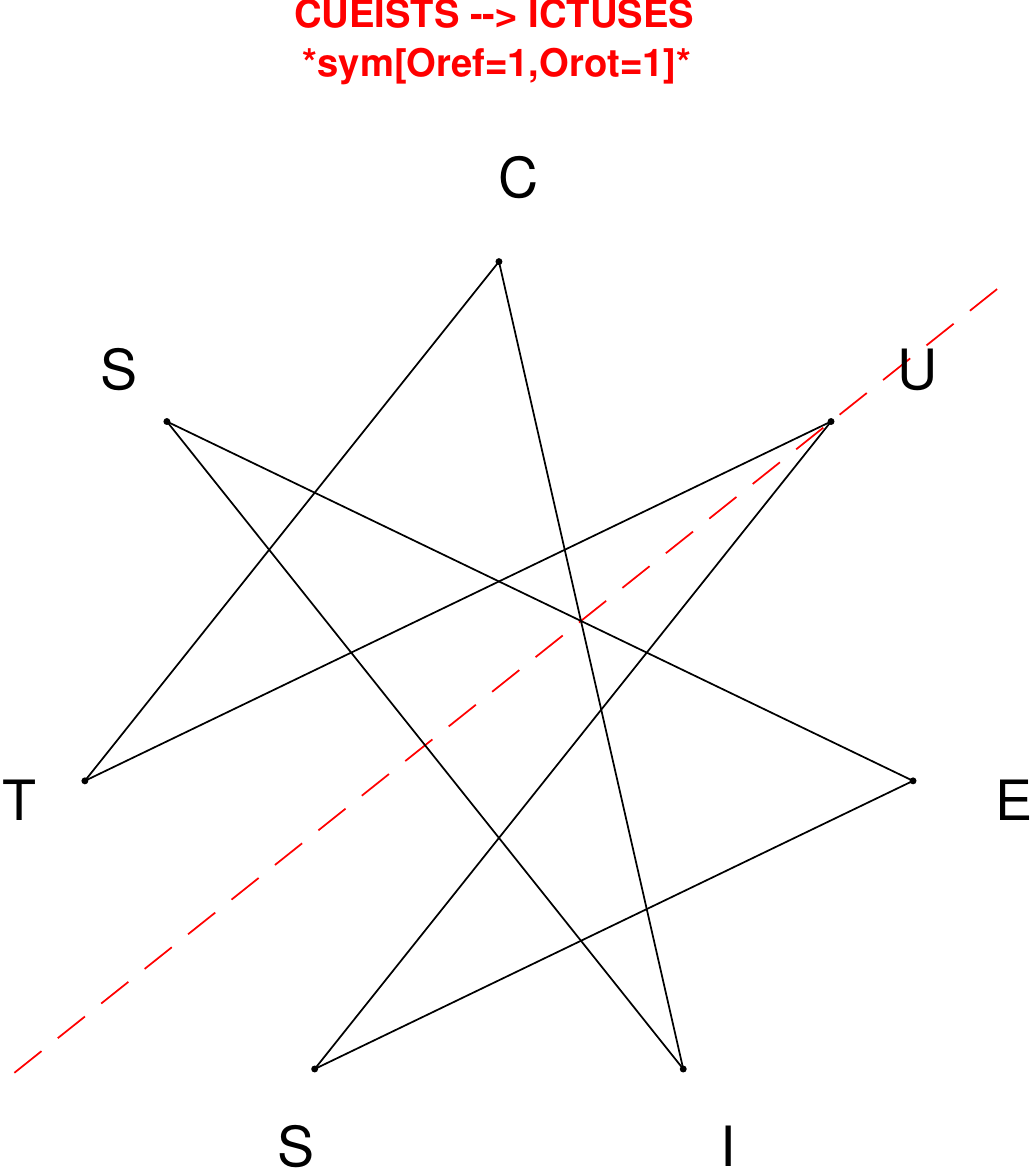}
\end{subfigure}
\hfill
\begin{subfigure}[T]{0.19\textwidth}
\centering
\includegraphics[width=\textwidth]{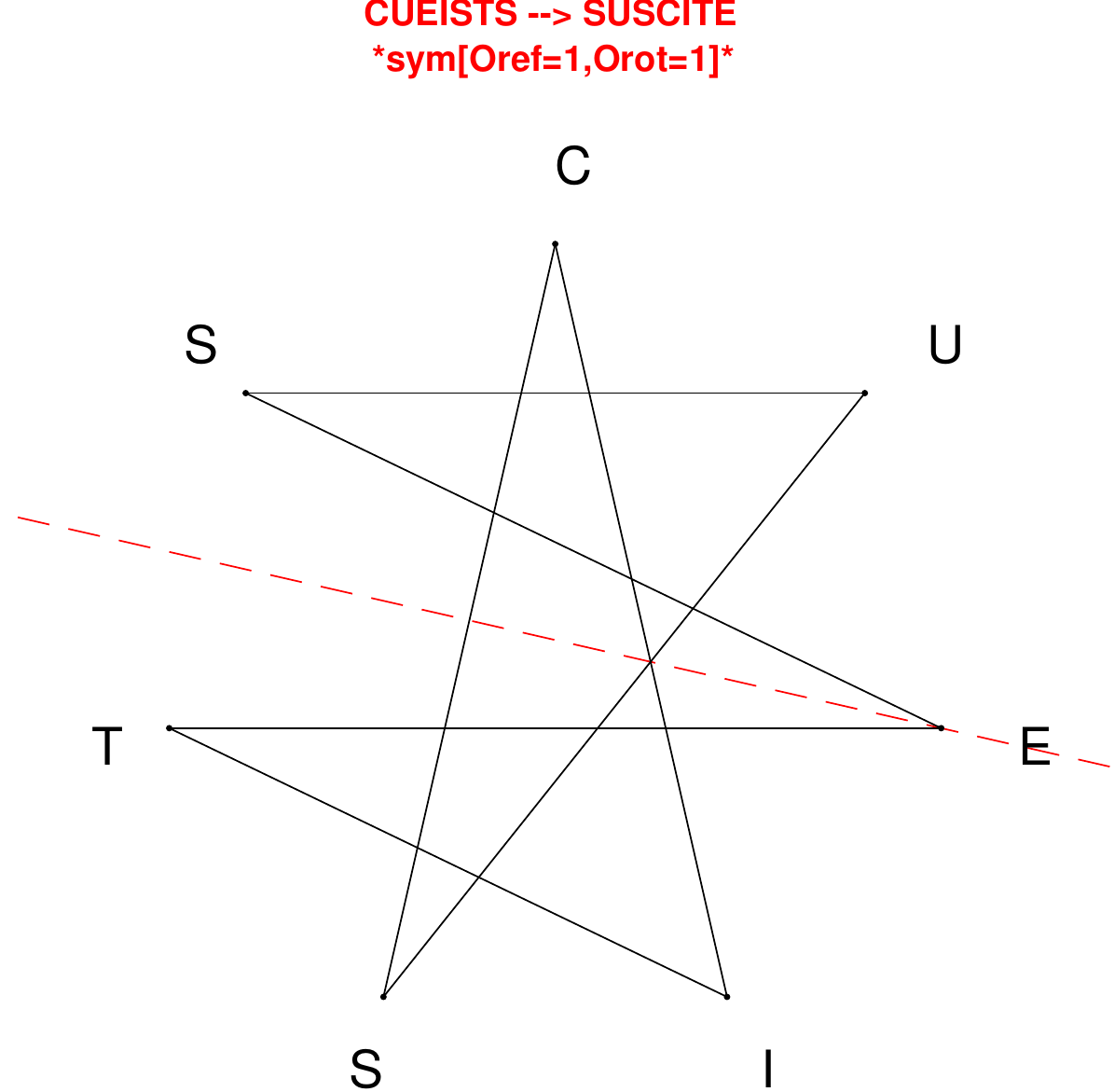}
\end{subfigure}
\hfill
\begin{subfigure}[T]{0.19\textwidth}
\centering
\includegraphics[width=\textwidth]{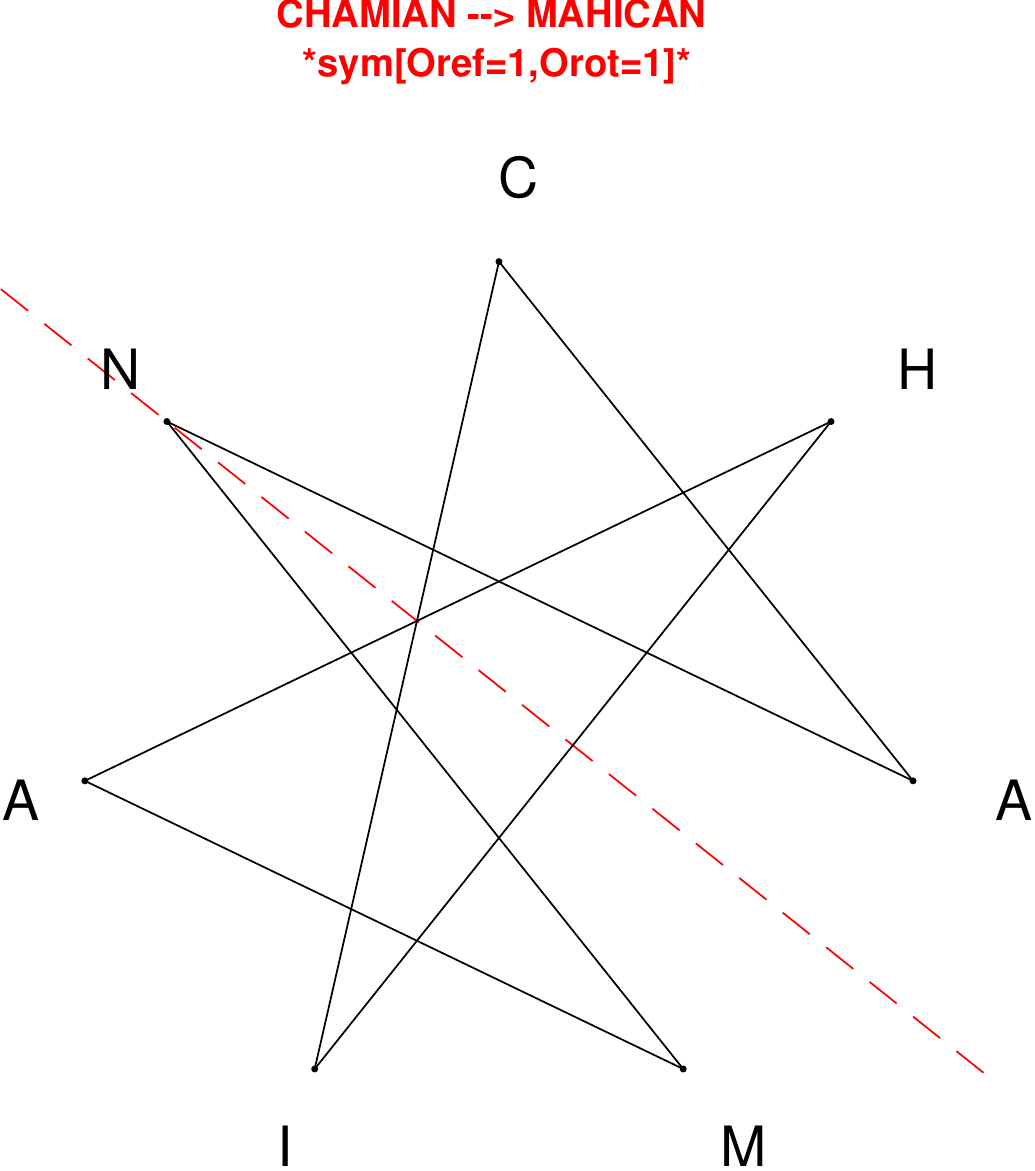}
\end{subfigure}
\hfill
\begin{subfigure}[T]{0.19\textwidth}
\centering
\includegraphics[width=\textwidth]{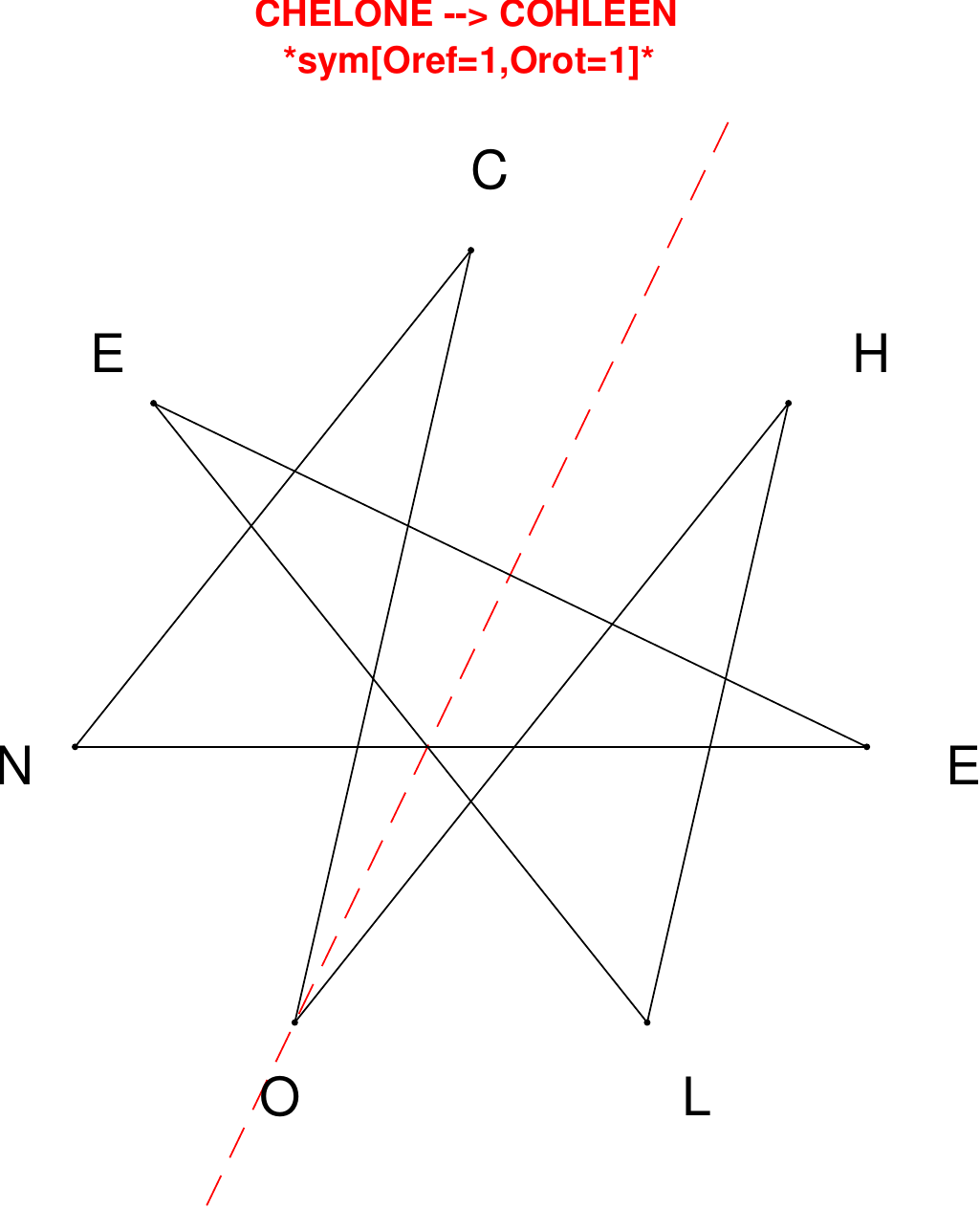}
\end{subfigure}
\end{figure}

\begin{figure}[H]
\centering
\begin{subfigure}[T]{0.19\textwidth}
\centering
\includegraphics[width=\textwidth]{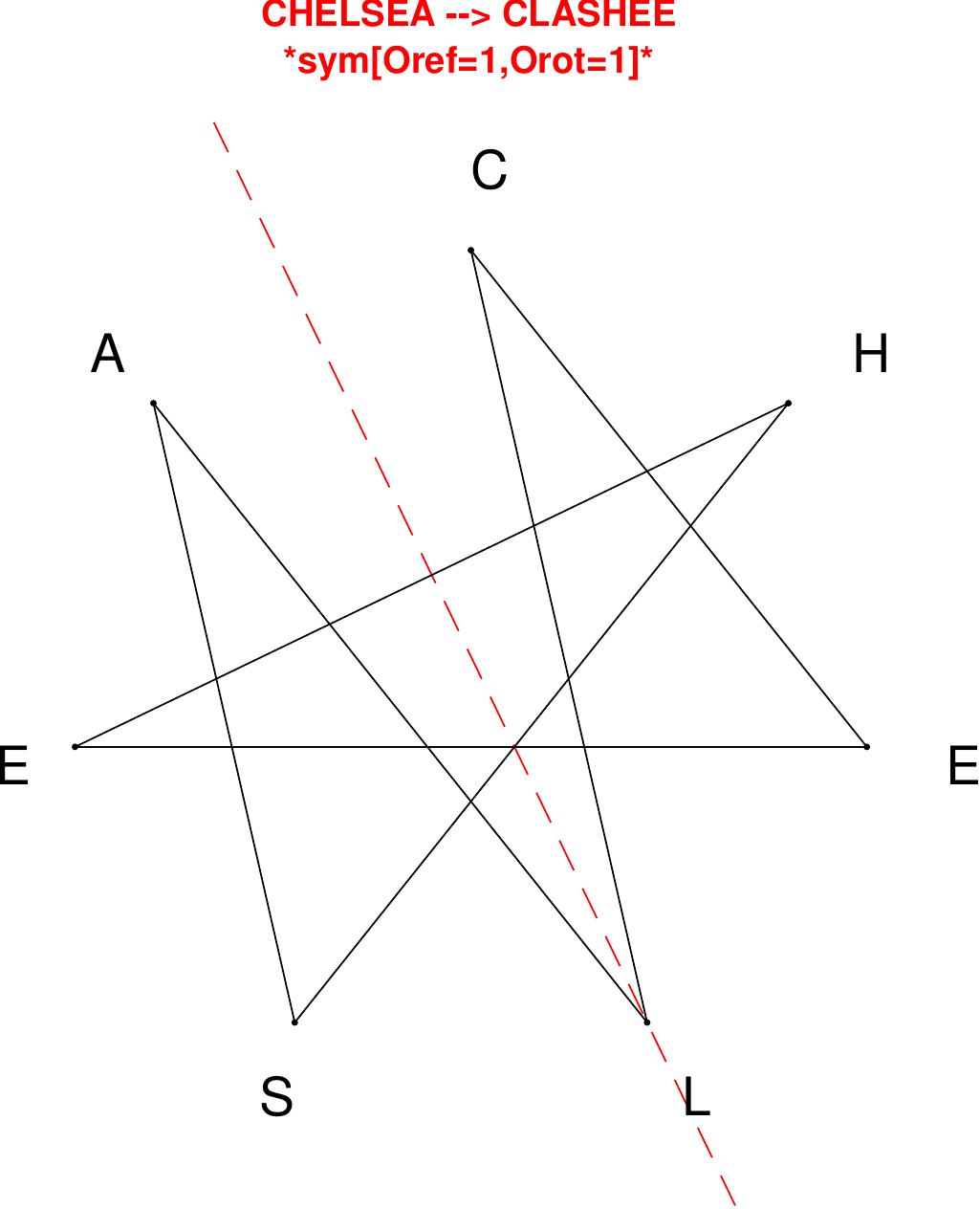}
\end{subfigure}
\hfill
\begin{subfigure}[T]{0.19\textwidth}
\centering
\includegraphics[width=\textwidth]{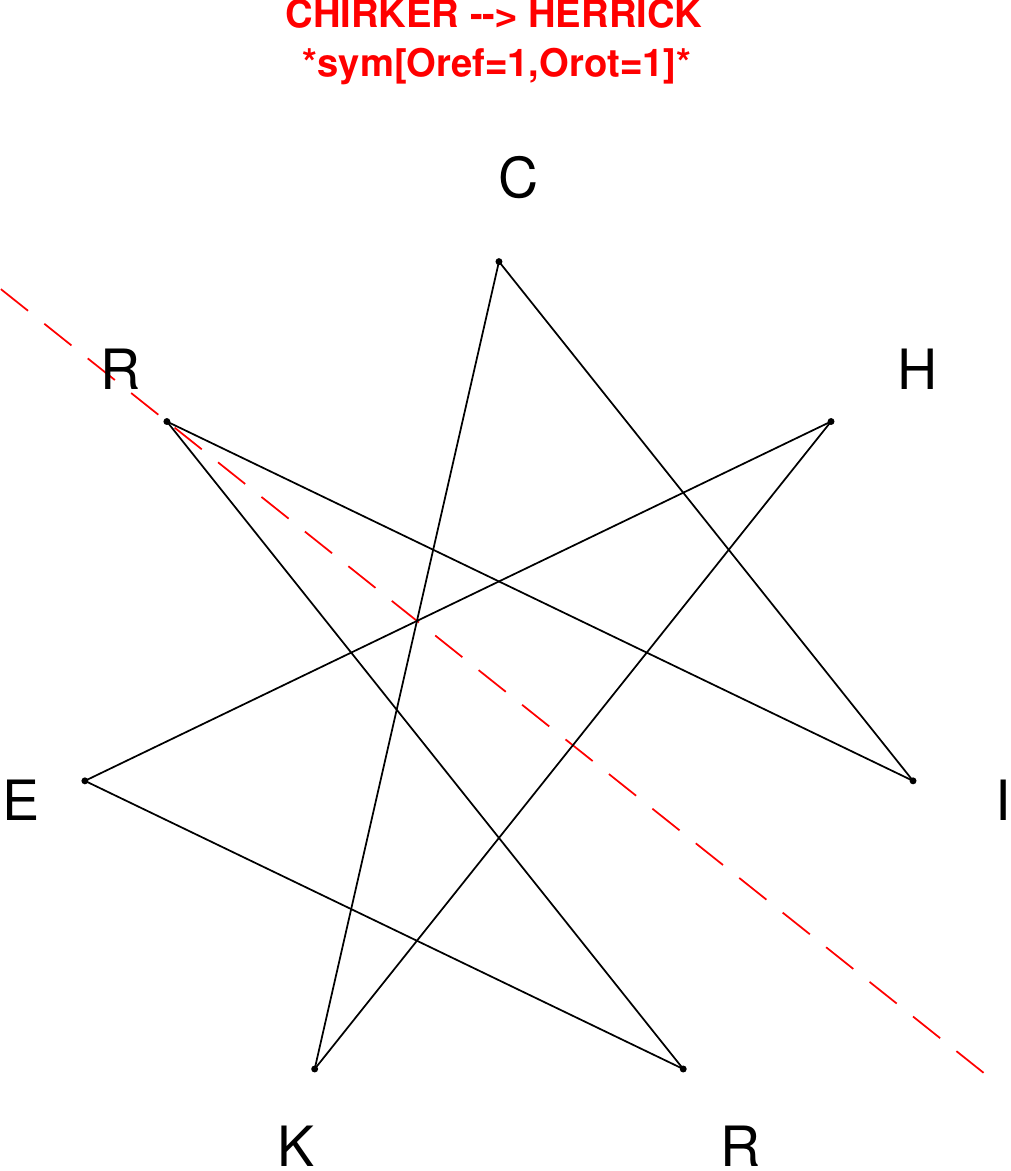}
\end{subfigure}
\hfill
\begin{subfigure}[T]{0.19\textwidth}
\centering
\includegraphics[width=\textwidth]{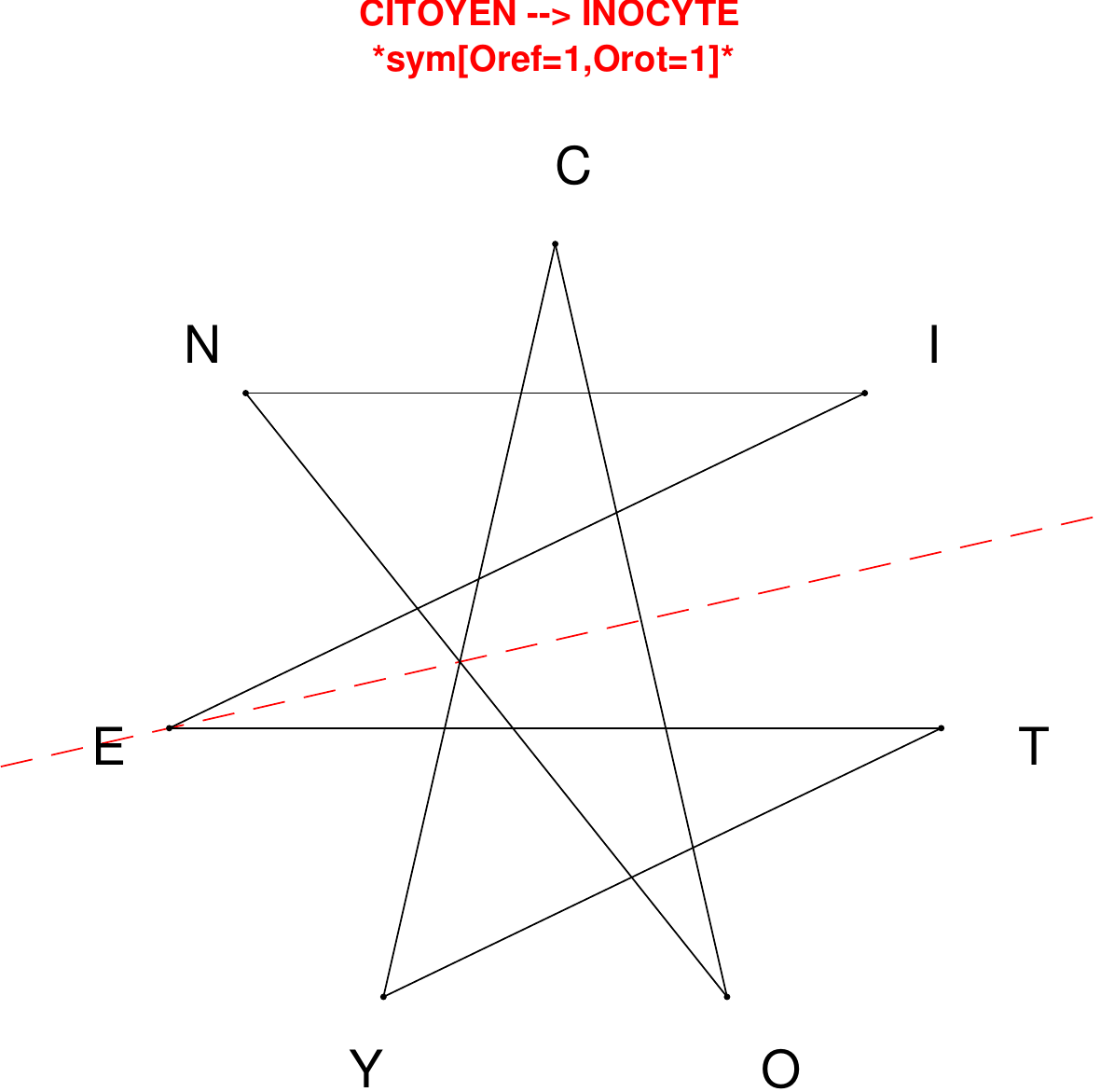}
\end{subfigure}
\hfill
\begin{subfigure}[T]{0.19\textwidth}
\centering
\includegraphics[width=\textwidth]{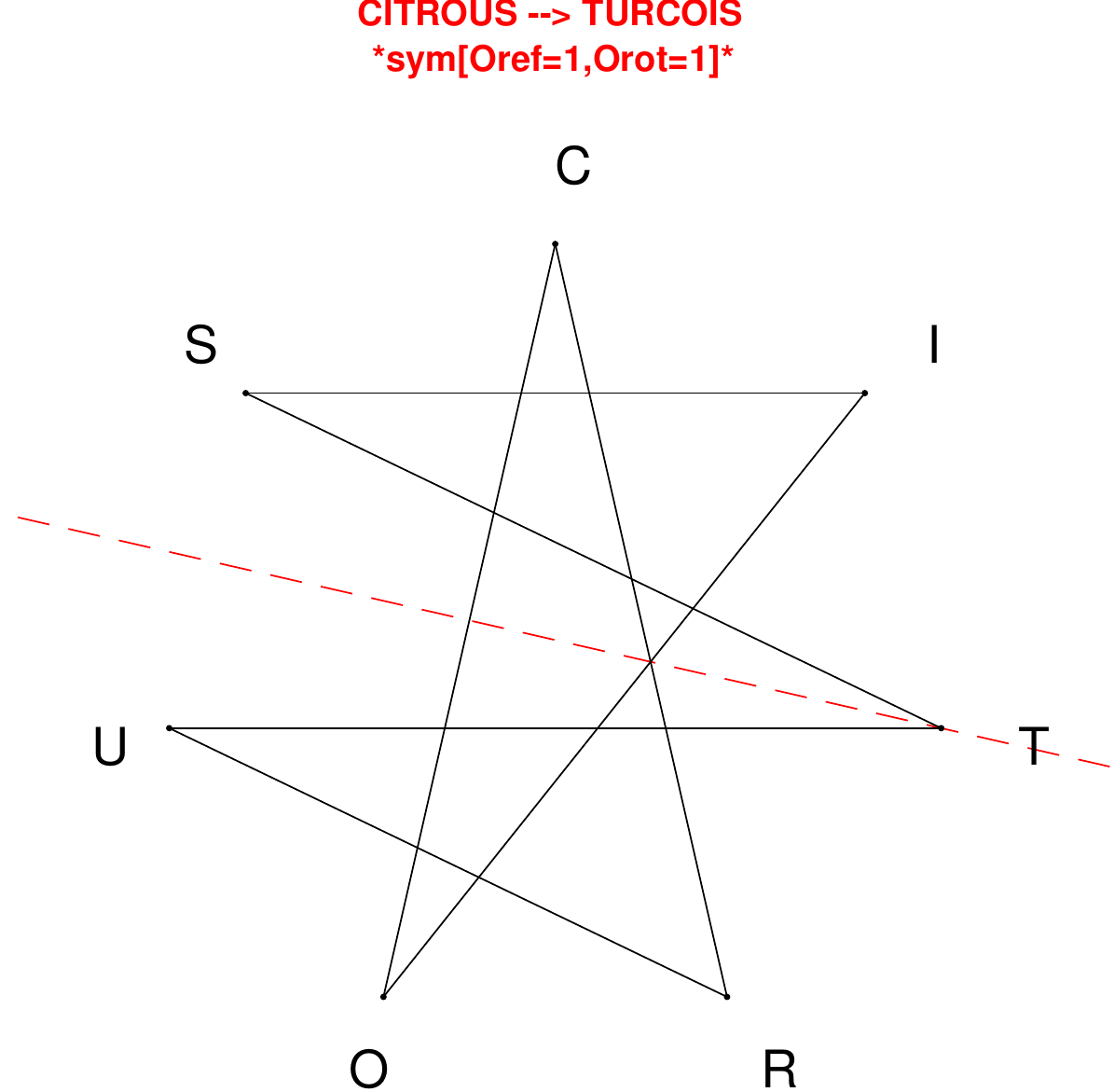}
\end{subfigure}
\hfill
\begin{subfigure}[T]{0.19\textwidth}
\centering
\includegraphics[width=\textwidth]{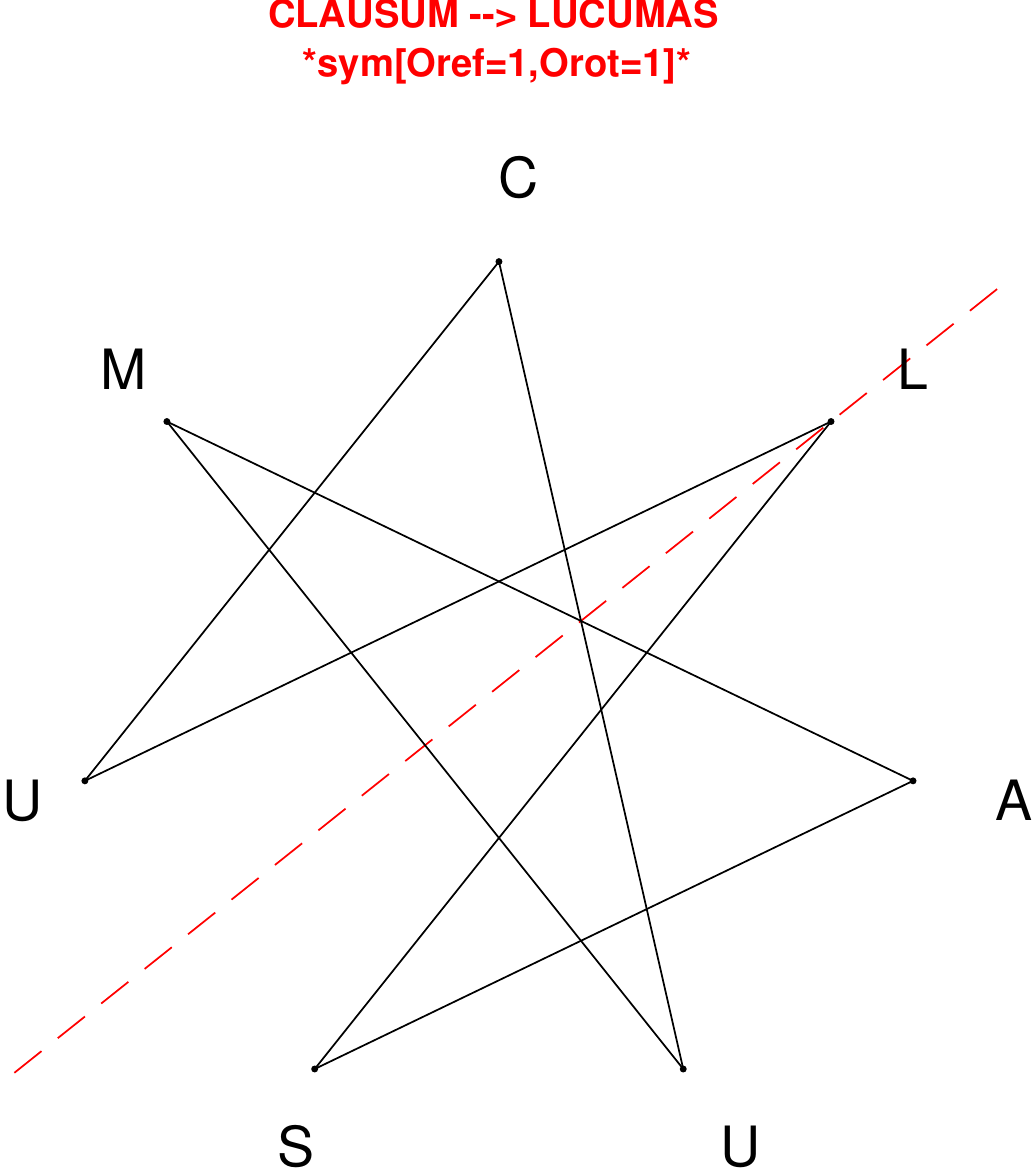}
\end{subfigure}
\end{figure}

\begin{figure}[H]
\centering
\begin{subfigure}[T]{0.19\textwidth}
\centering
\includegraphics[width=\textwidth]{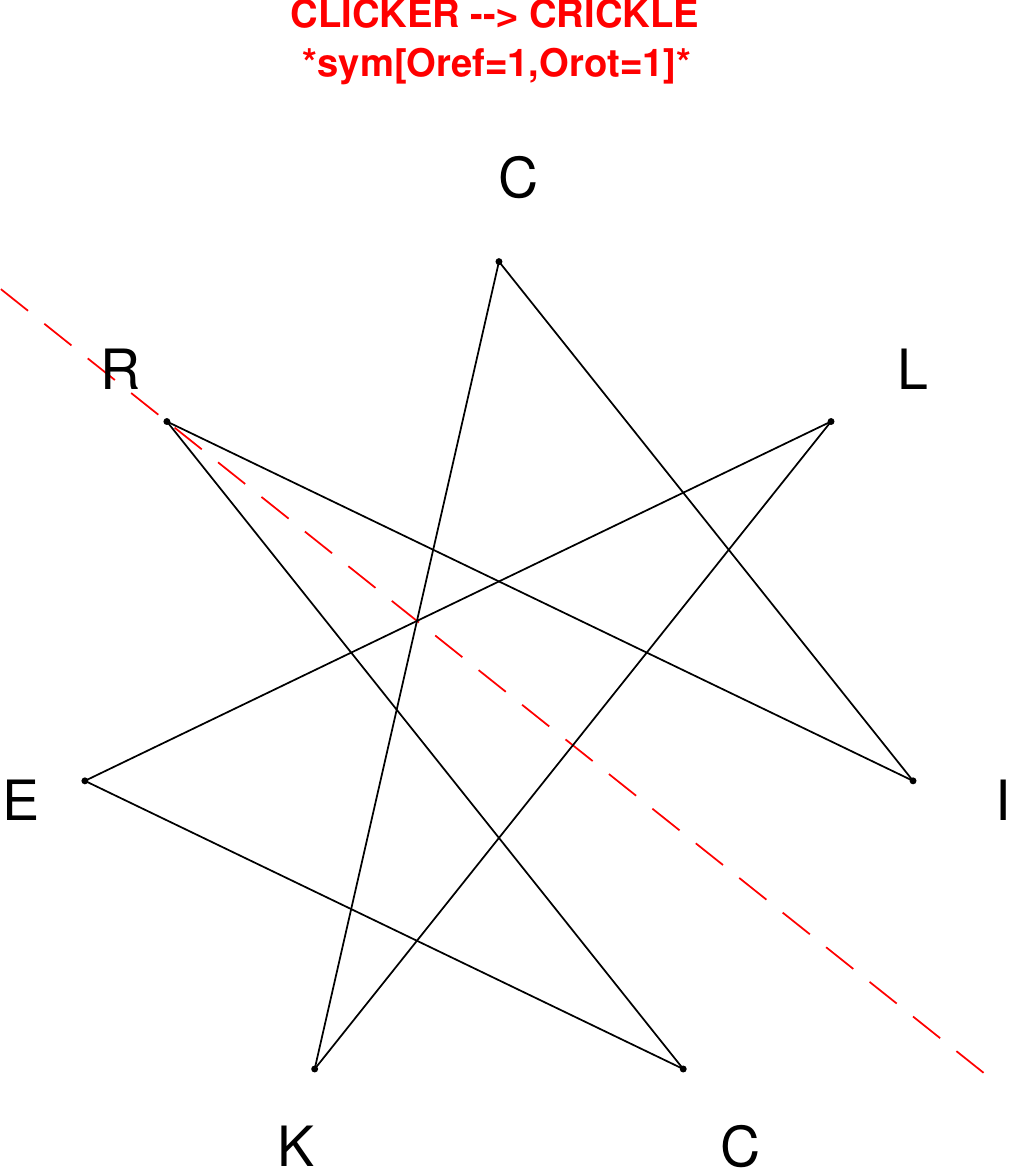}
\end{subfigure}
\hfill
\begin{subfigure}[T]{0.19\textwidth}
\centering
\includegraphics[width=\textwidth]{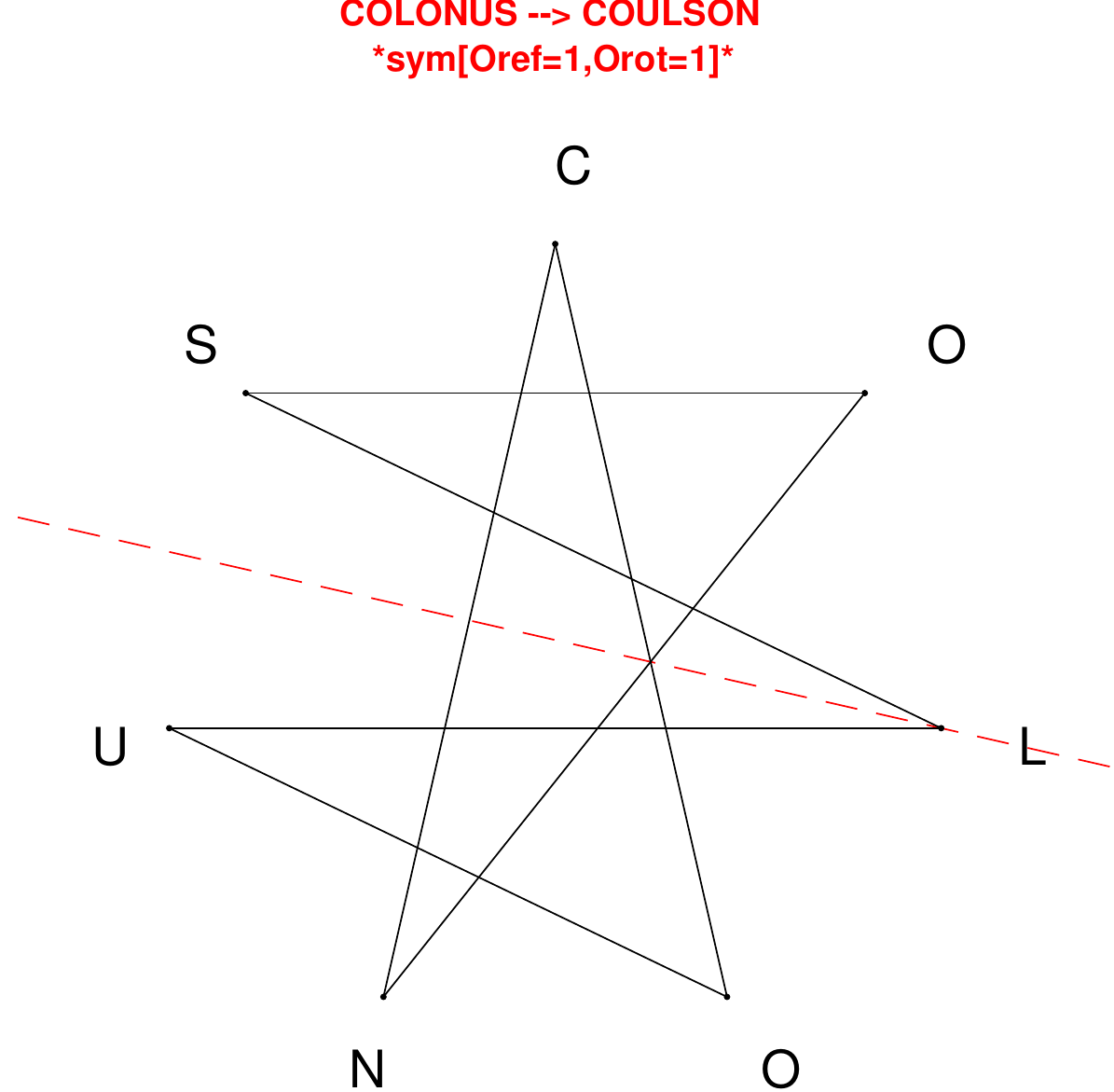}
\end{subfigure}
\hfill
\begin{subfigure}[T]{0.19\textwidth}
\centering
\includegraphics[width=\textwidth]{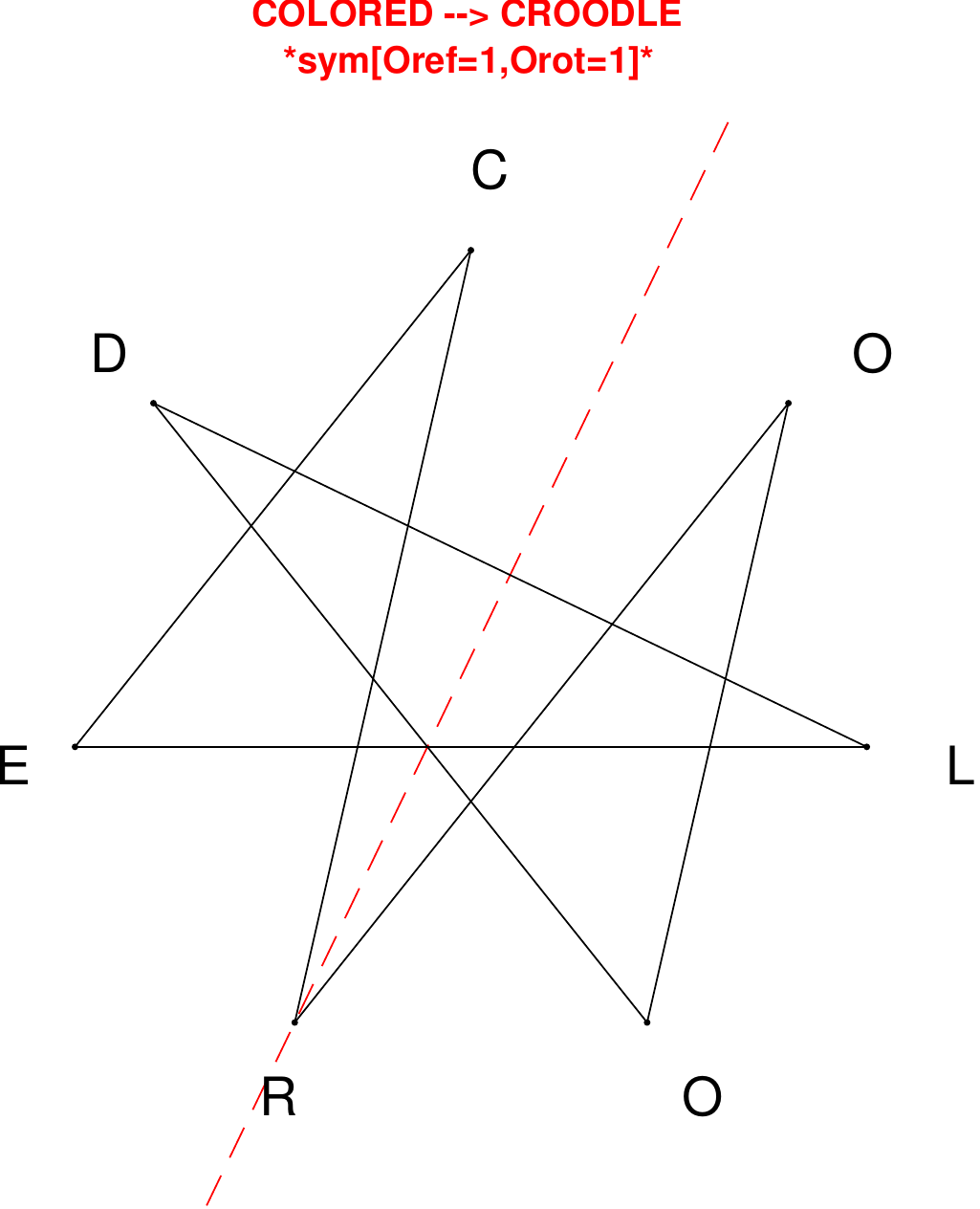}
\end{subfigure}
\hfill
\begin{subfigure}[T]{0.19\textwidth}
\centering
\includegraphics[width=\textwidth]{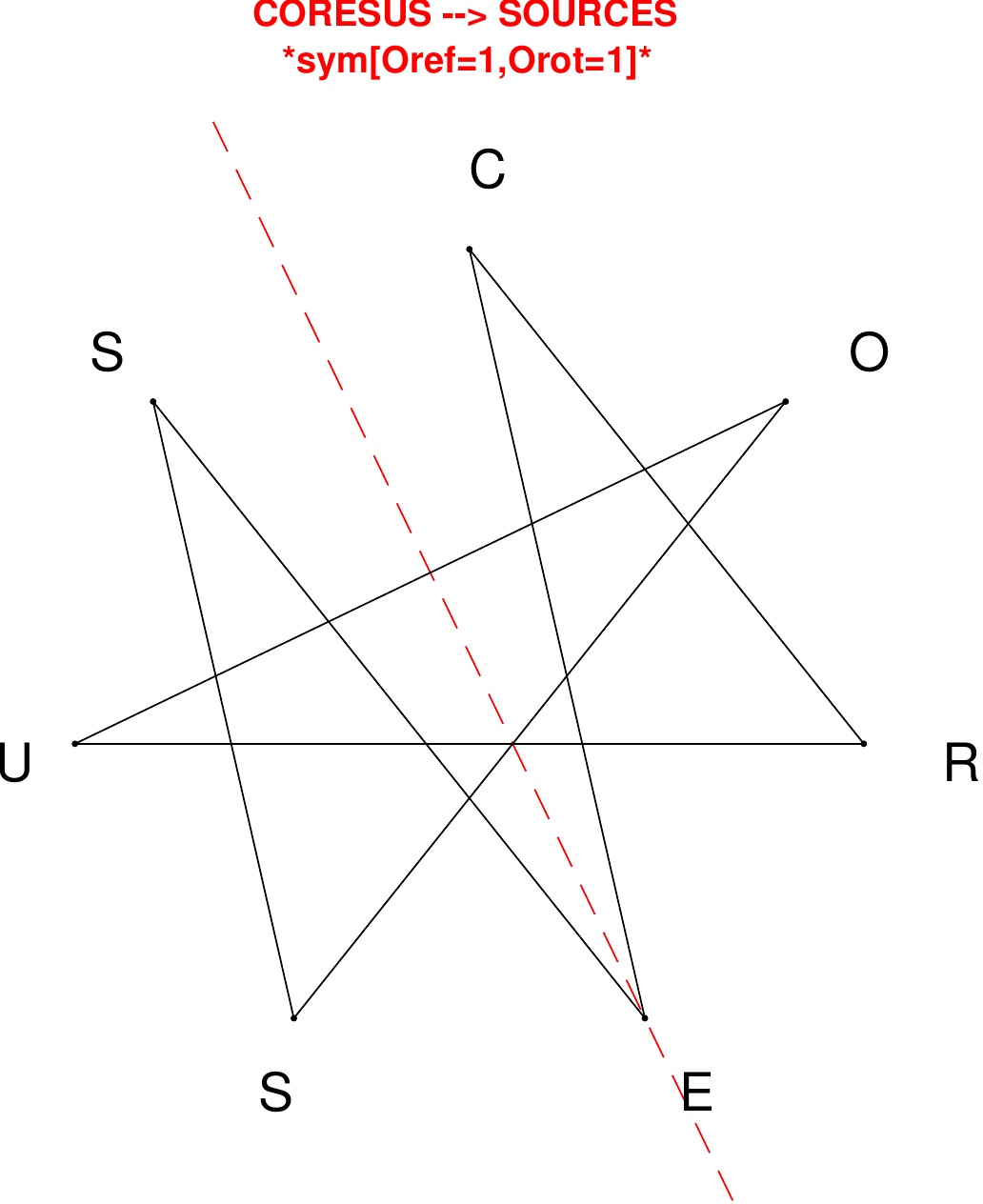}
\end{subfigure}
\hfill
\begin{subfigure}[T]{0.19\textwidth}
\centering
\includegraphics[width=\textwidth]{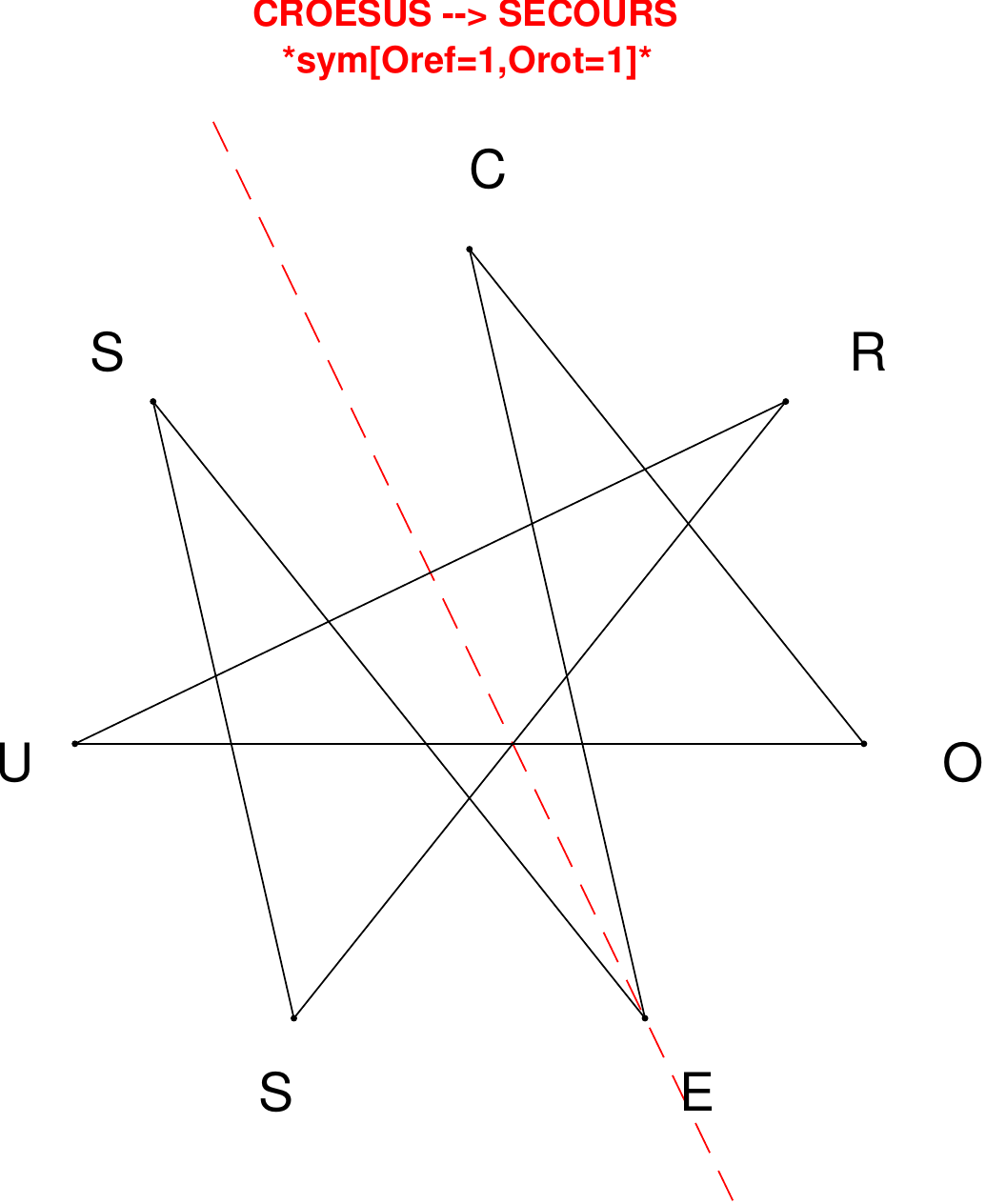}
\end{subfigure}
\end{figure}

\begin{figure}[H]
\centering
\begin{subfigure}[T]{0.19\textwidth}
\centering
\includegraphics[width=\textwidth]{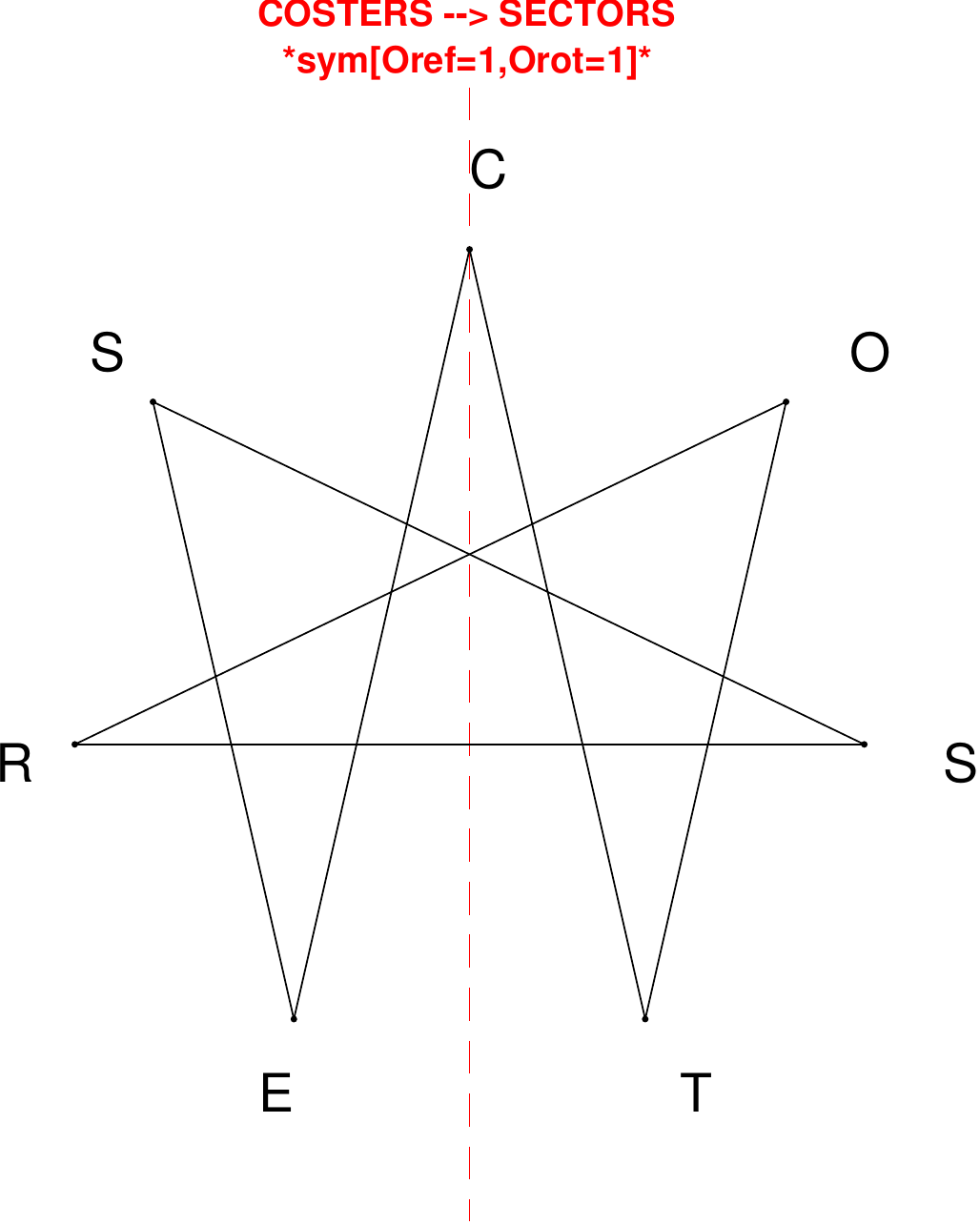}
\end{subfigure}
\hfill
\begin{subfigure}[T]{0.19\textwidth}
\centering
\includegraphics[width=\textwidth]{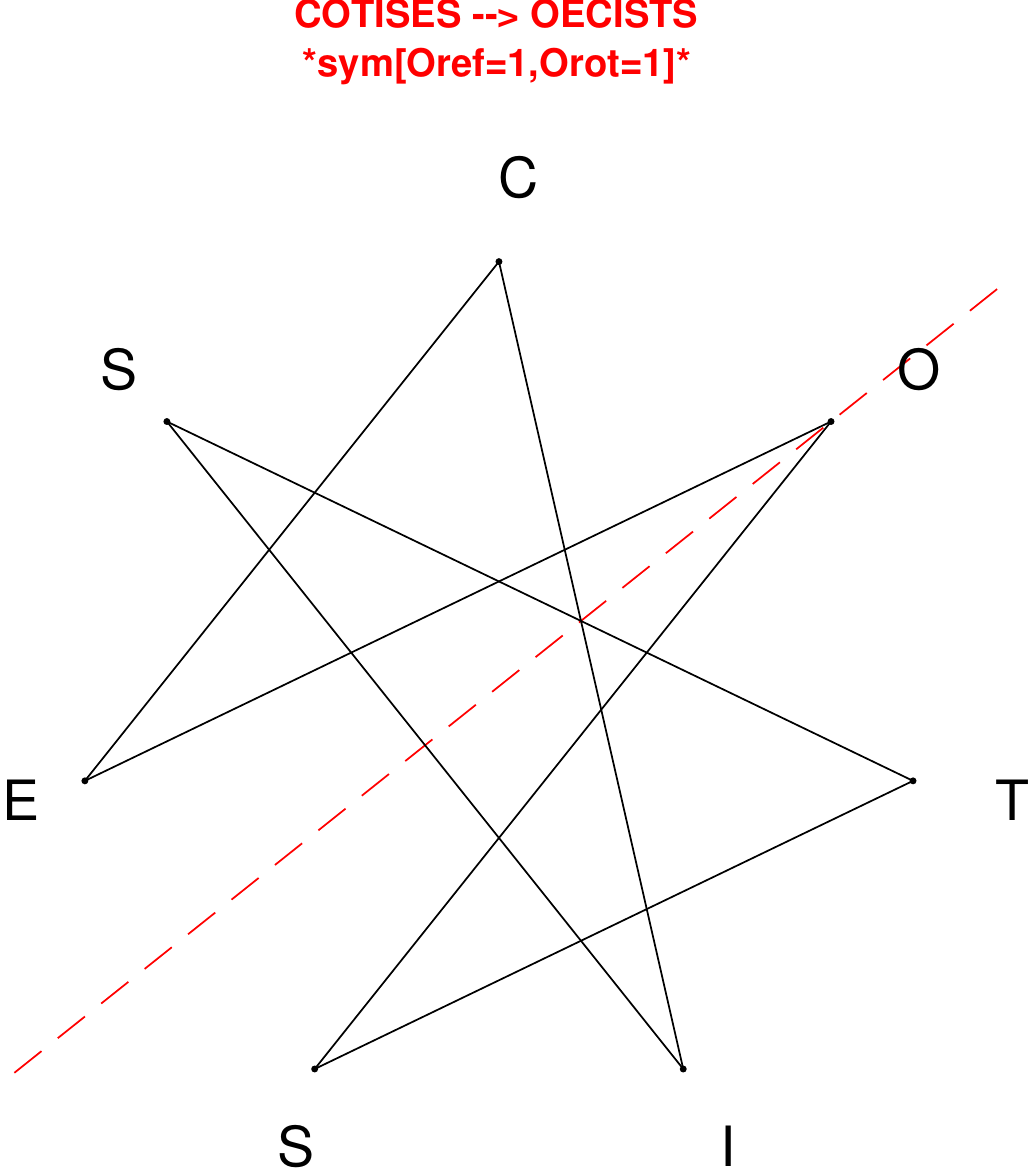}
\end{subfigure}
\hfill
\begin{subfigure}[T]{0.19\textwidth}
\centering
\includegraphics[width=\textwidth]{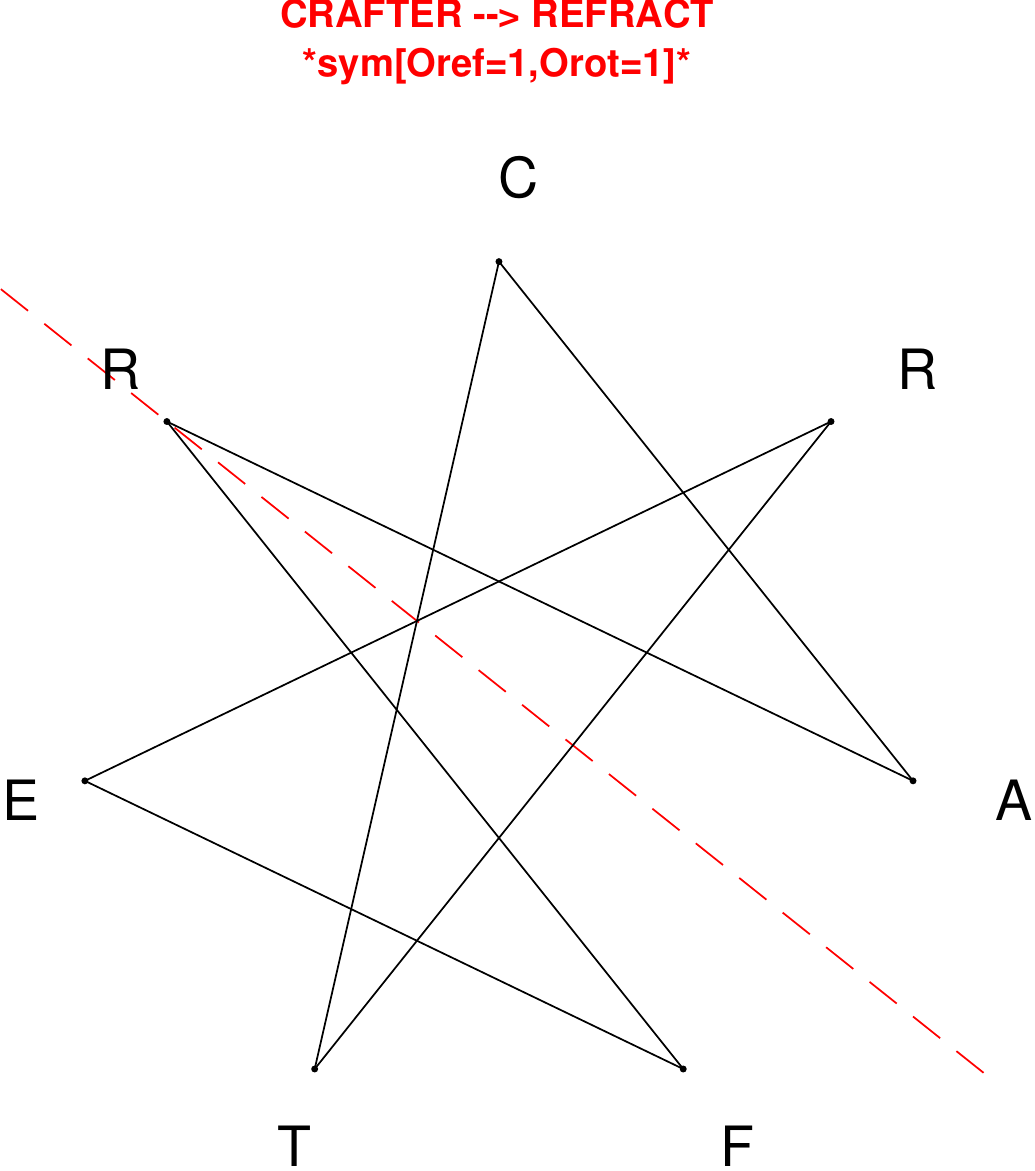}
\end{subfigure}
\hfill
\begin{subfigure}[T]{0.19\textwidth}
\centering
\includegraphics[width=\textwidth]{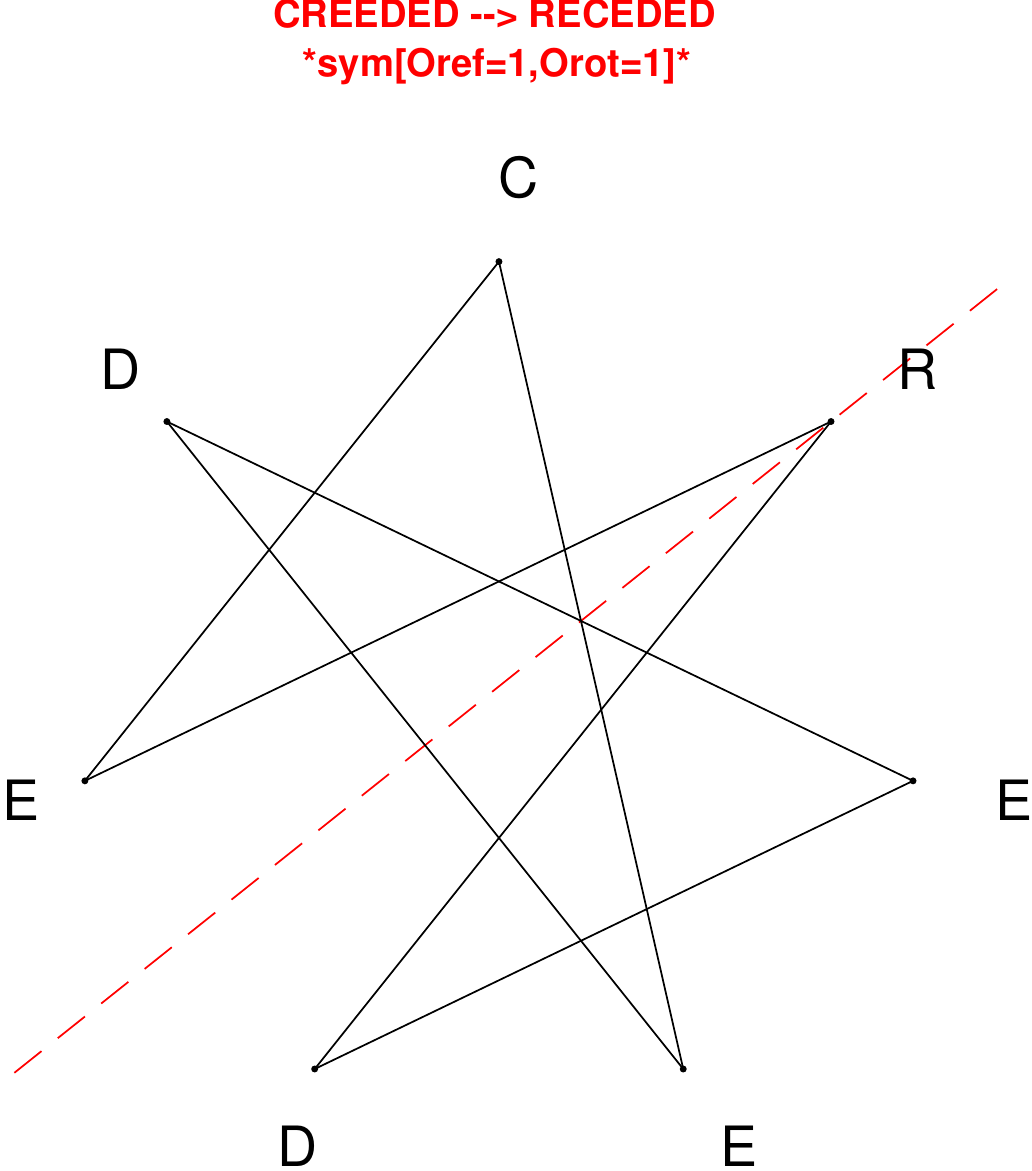}
\end{subfigure}
\hfill
\begin{subfigure}[T]{0.19\textwidth}
\centering
\includegraphics[width=\textwidth]{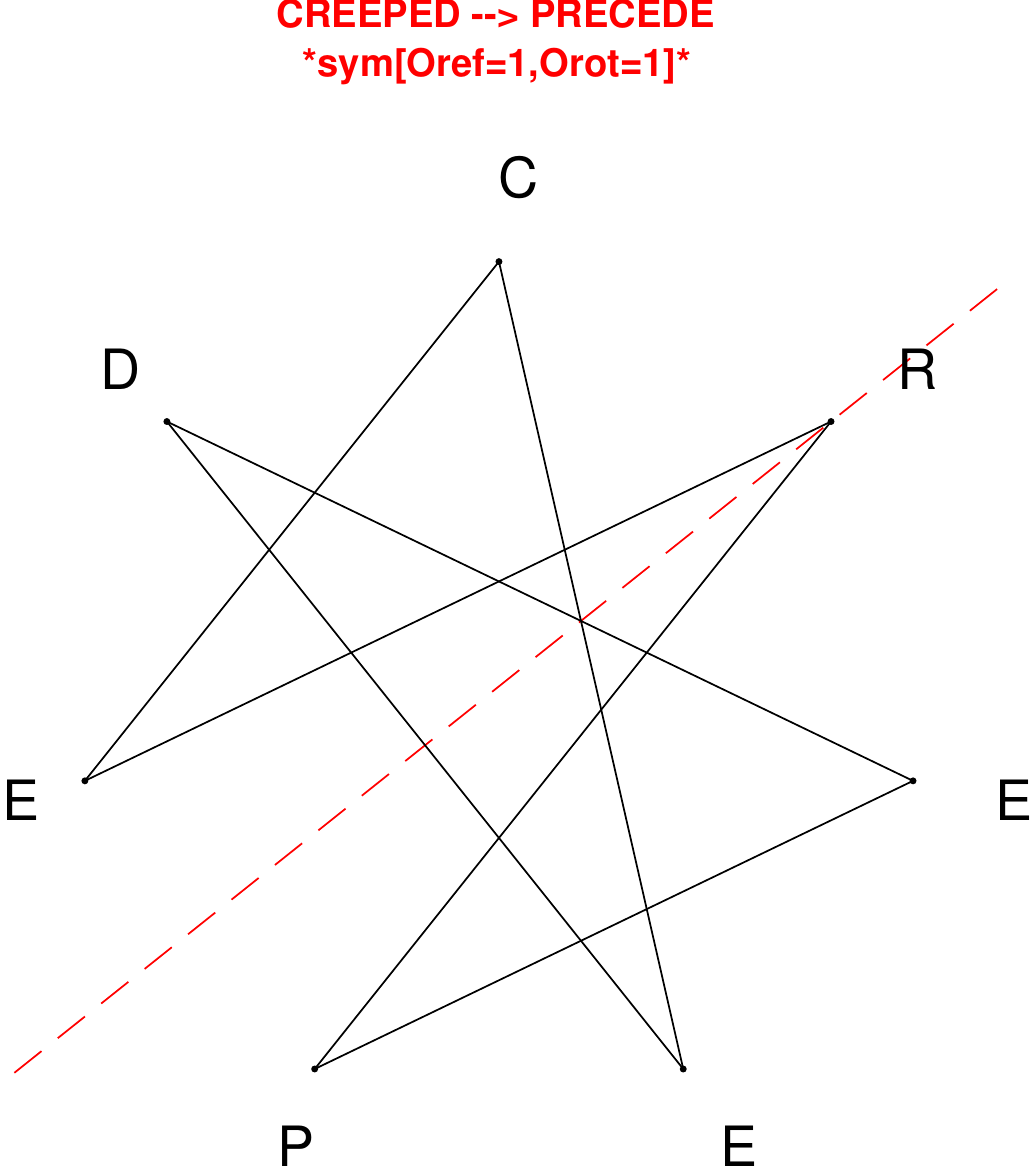}
\end{subfigure}
\end{figure}

\begin{figure}[H]
\centering
\begin{subfigure}[T]{0.19\textwidth}
\centering
\includegraphics[width=\textwidth]{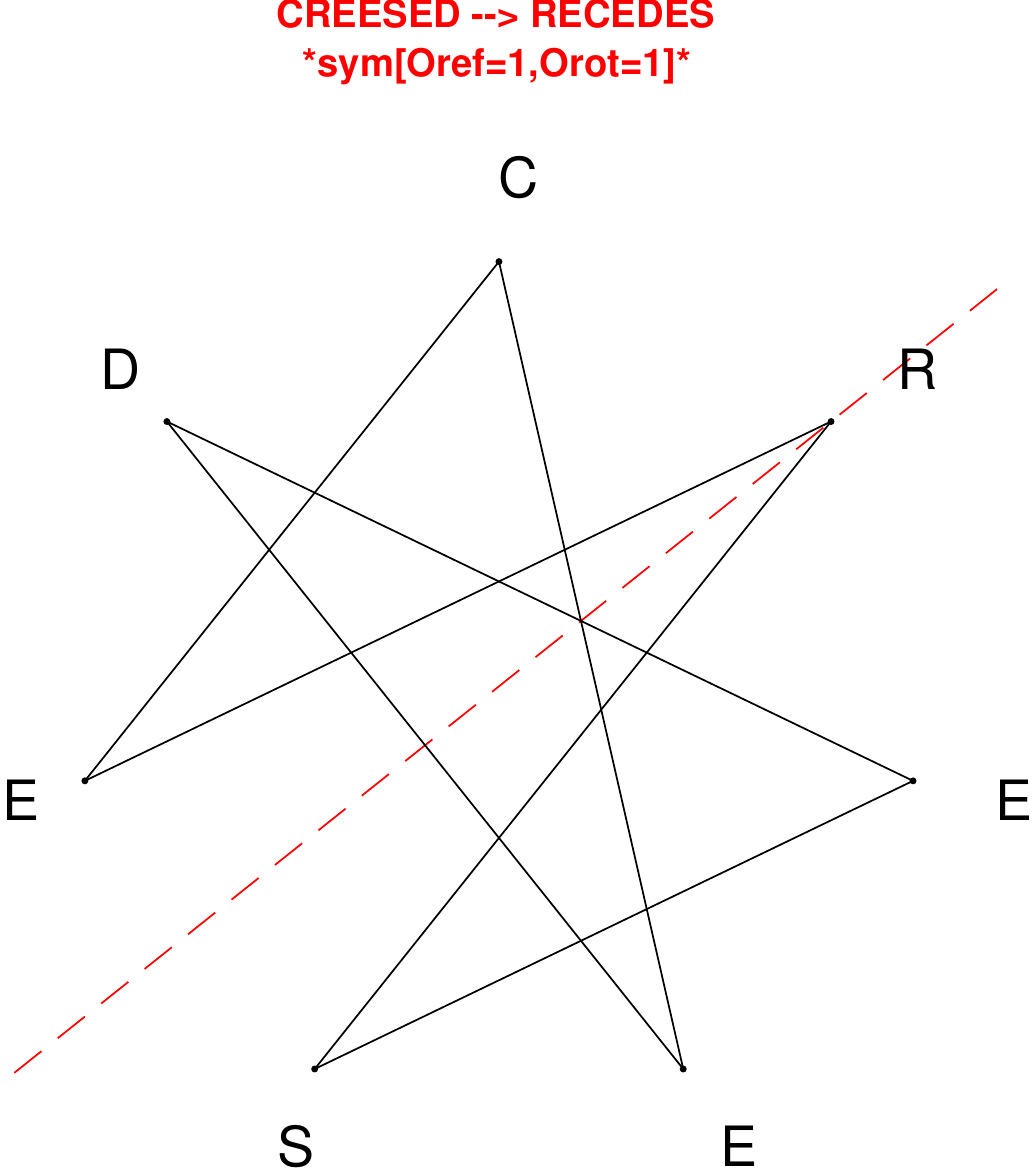}
\end{subfigure}
\hfill
\begin{subfigure}[T]{0.19\textwidth}
\centering
\includegraphics[width=\textwidth]{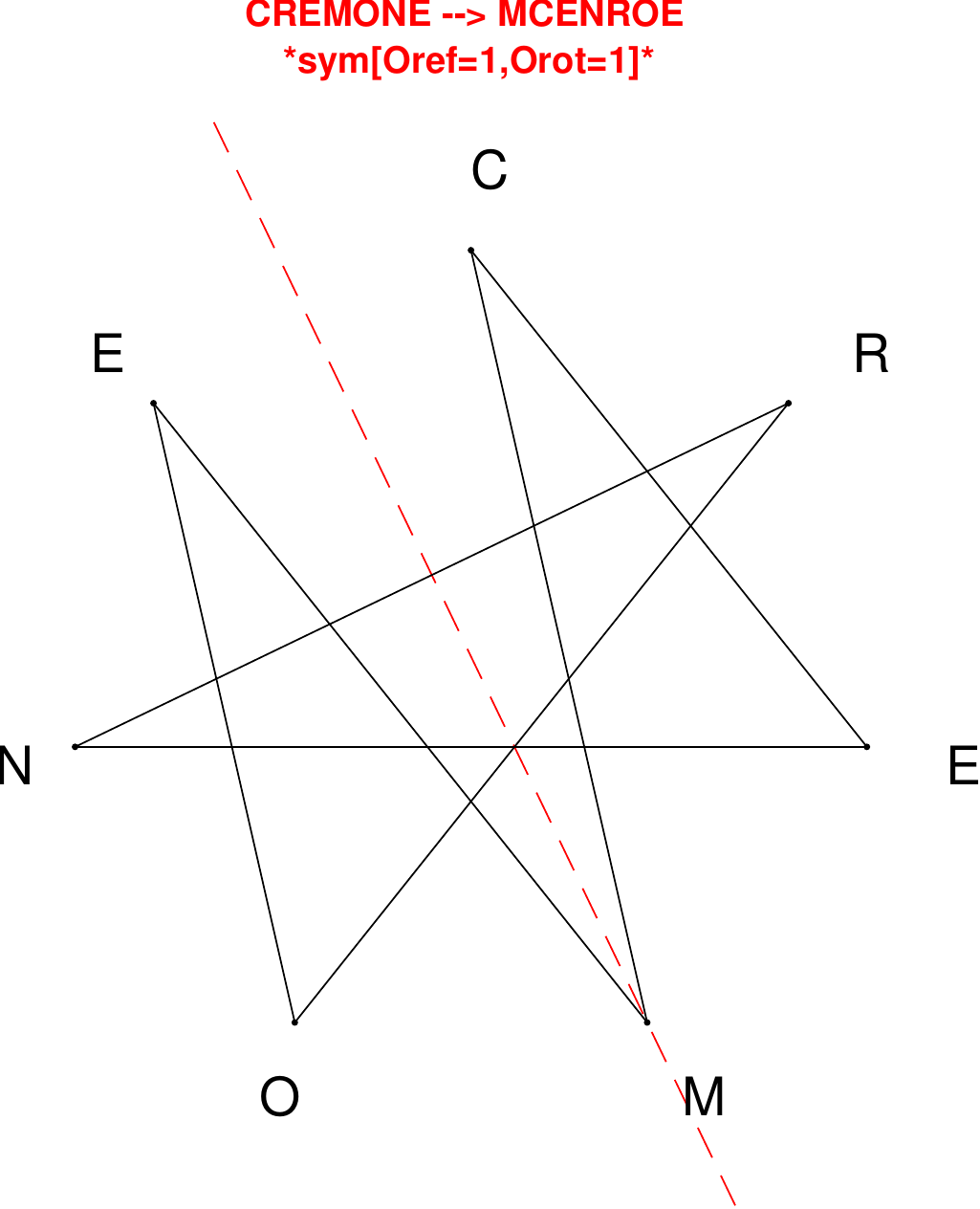}
\end{subfigure}
\hfill
\begin{subfigure}[T]{0.19\textwidth}
\centering
\includegraphics[width=\textwidth]{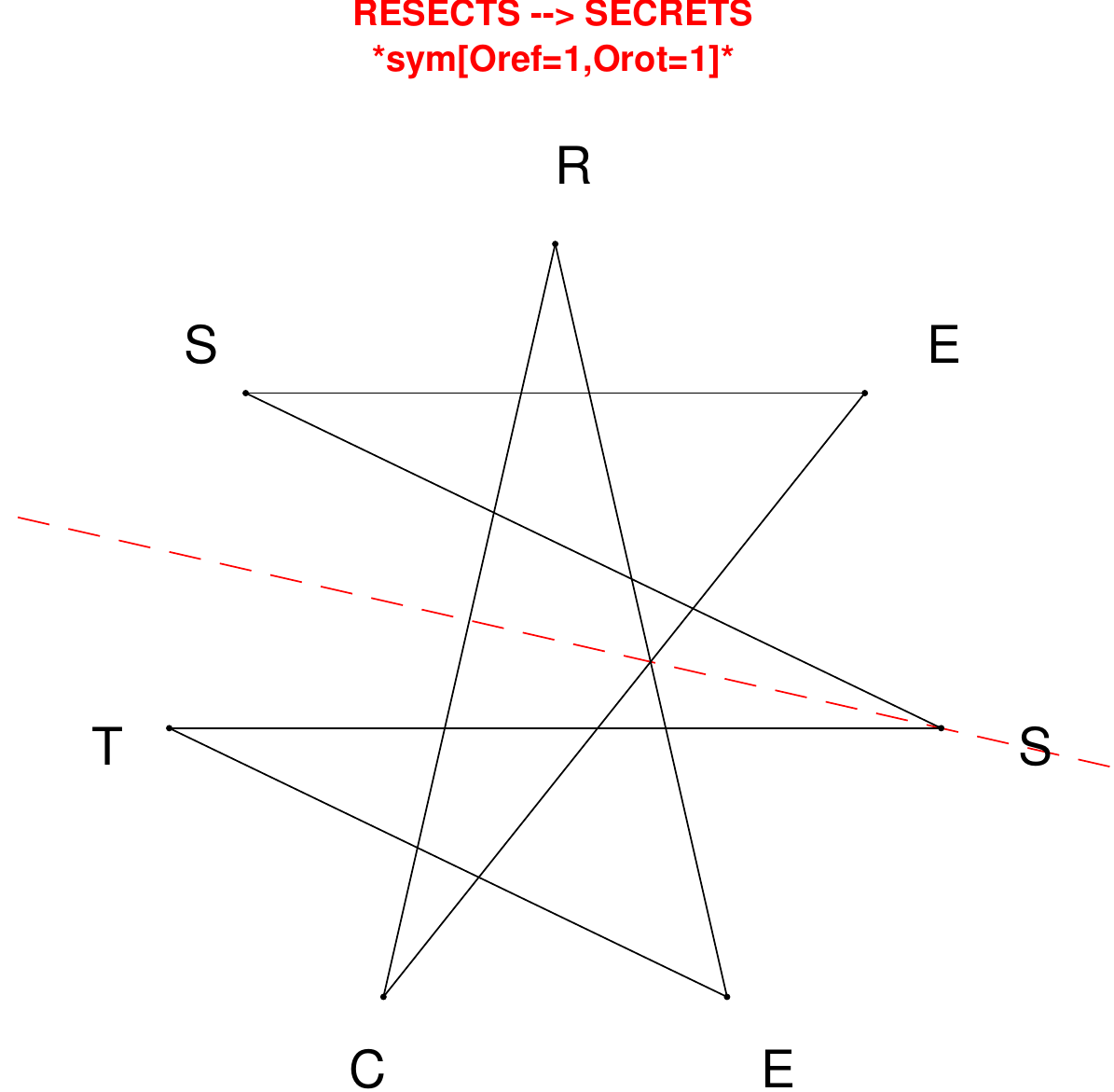}
\end{subfigure}
\hfill
\begin{subfigure}[T]{0.19\textwidth}
\centering
\includegraphics[width=\textwidth]{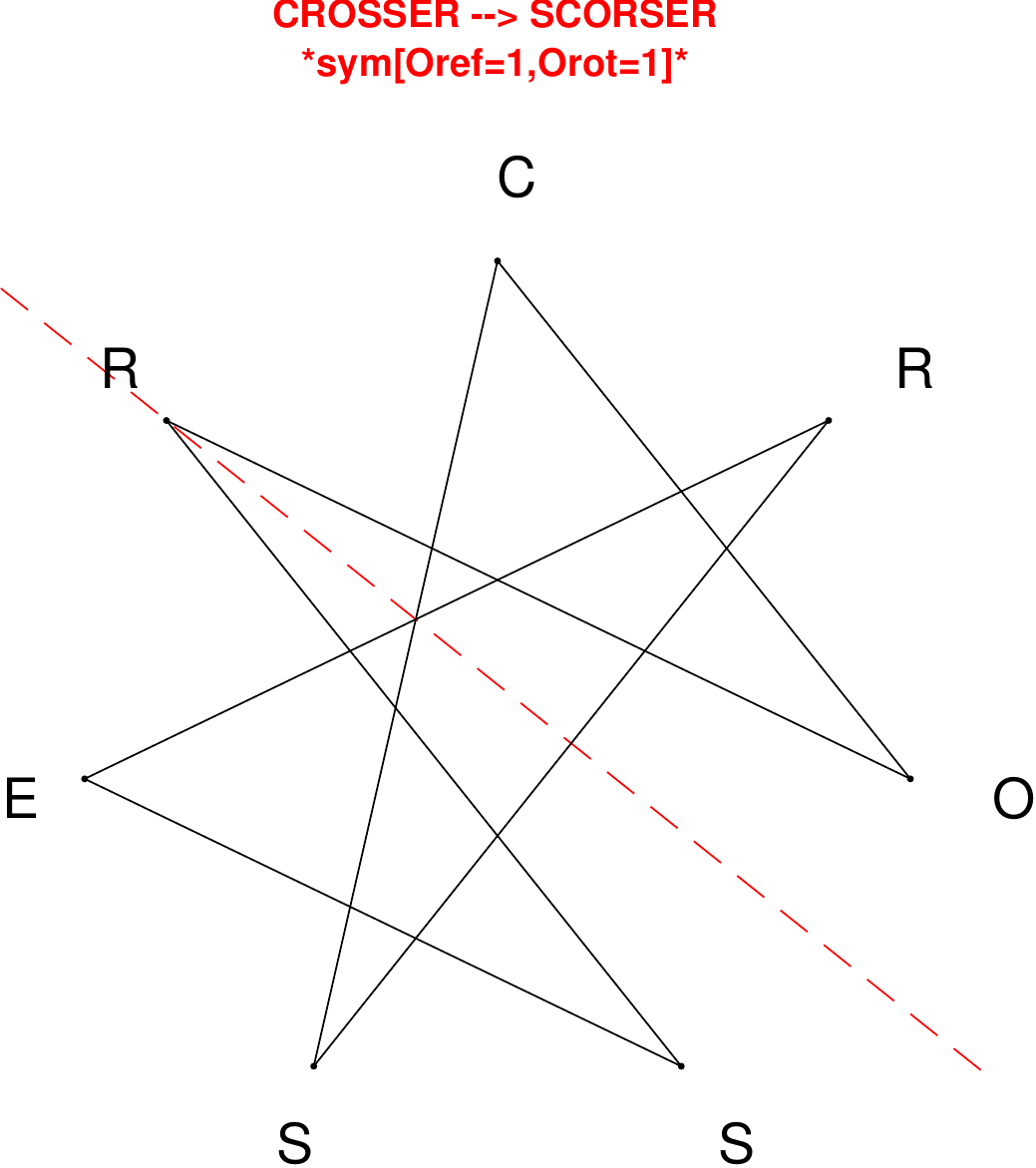}
\end{subfigure}
\hfill
\begin{subfigure}[T]{0.19\textwidth}
\centering
\includegraphics[width=\textwidth]{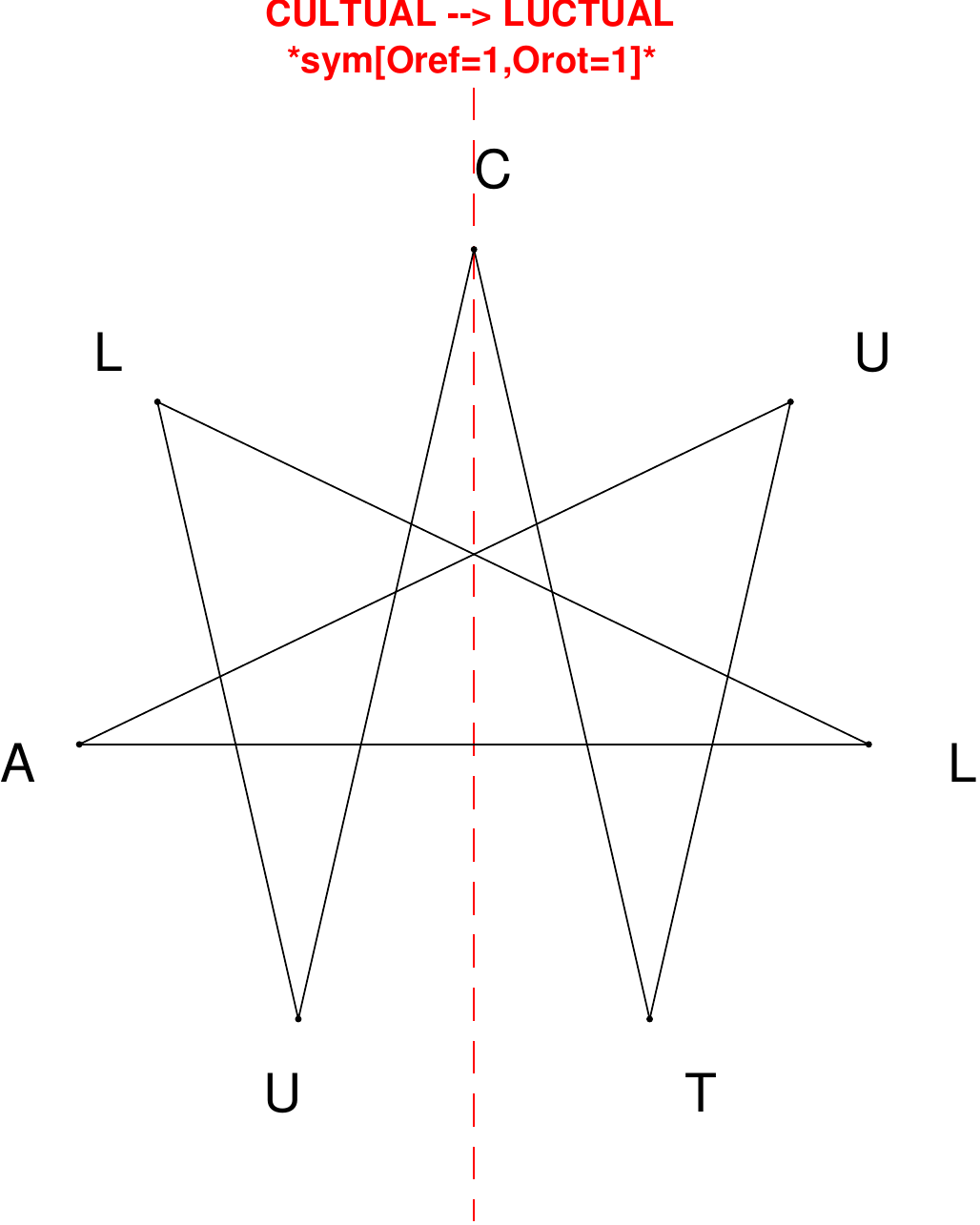}
\end{subfigure}
\end{figure}

\begin{figure}[H]
\centering
\begin{subfigure}[T]{0.19\textwidth}
\centering
\includegraphics[width=\textwidth]{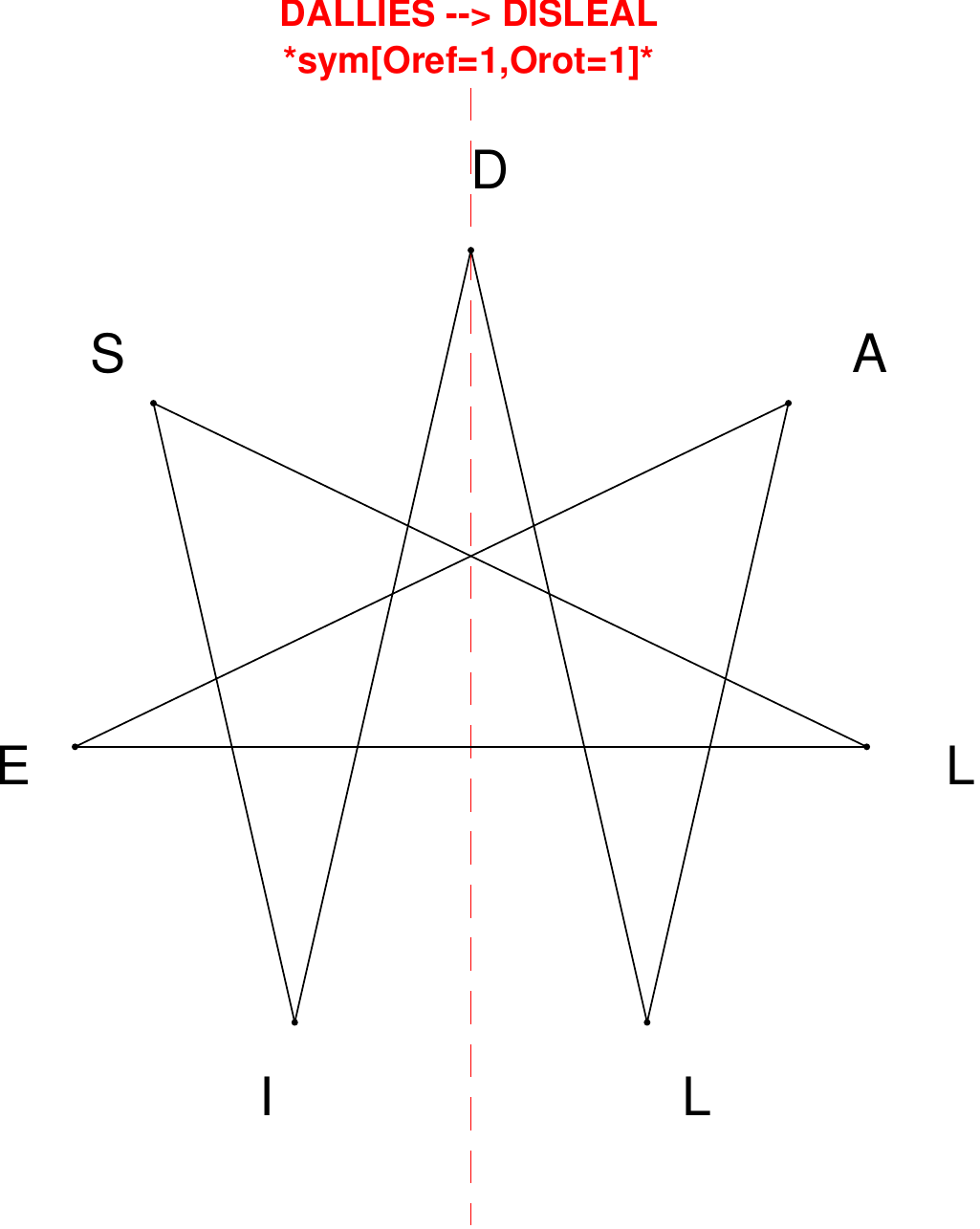}
\end{subfigure}
\hfill
\begin{subfigure}[T]{0.19\textwidth}
\centering
\includegraphics[width=\textwidth]{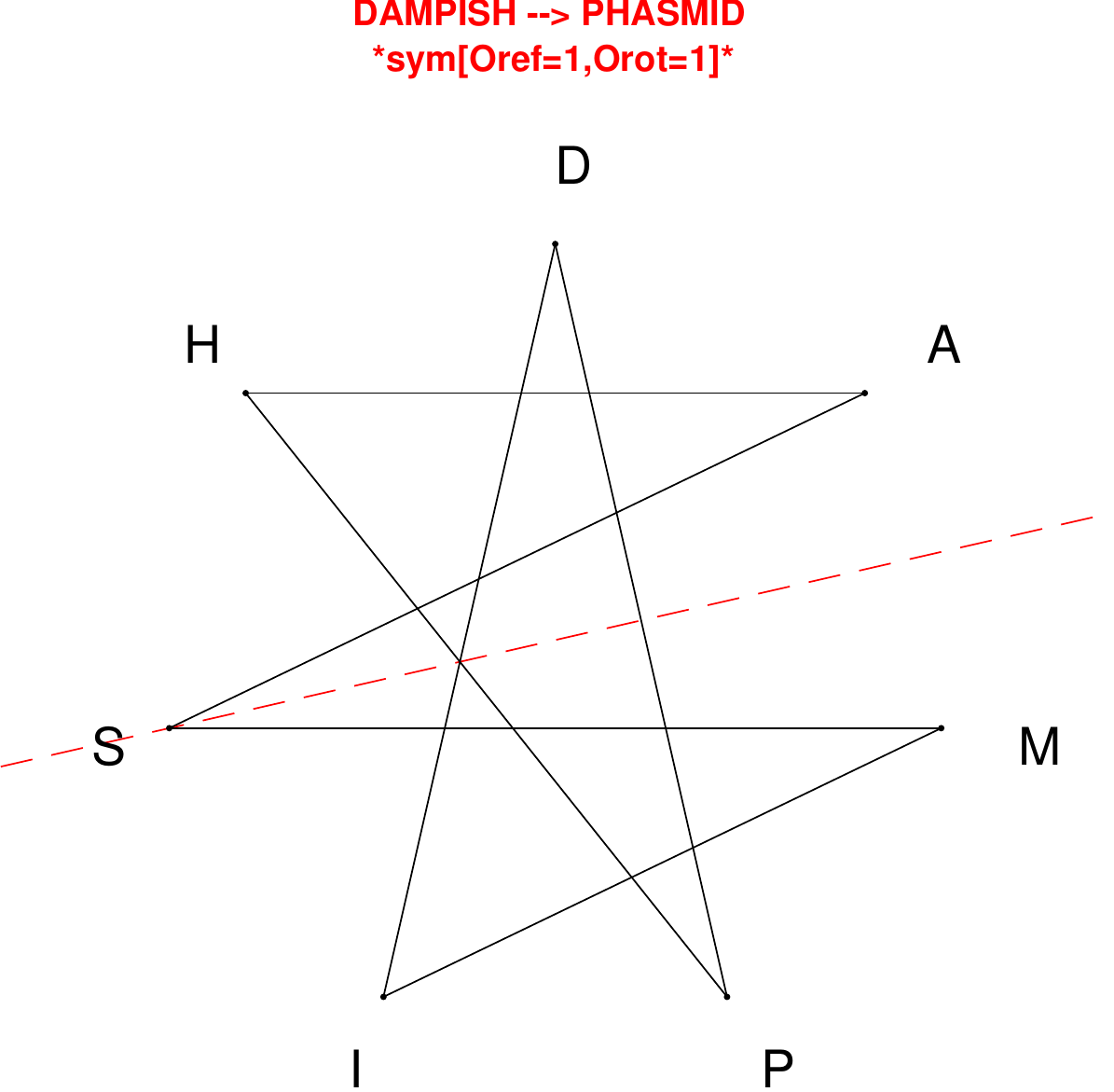}
\end{subfigure}
\hfill
\begin{subfigure}[T]{0.19\textwidth}
\centering
\includegraphics[width=\textwidth]{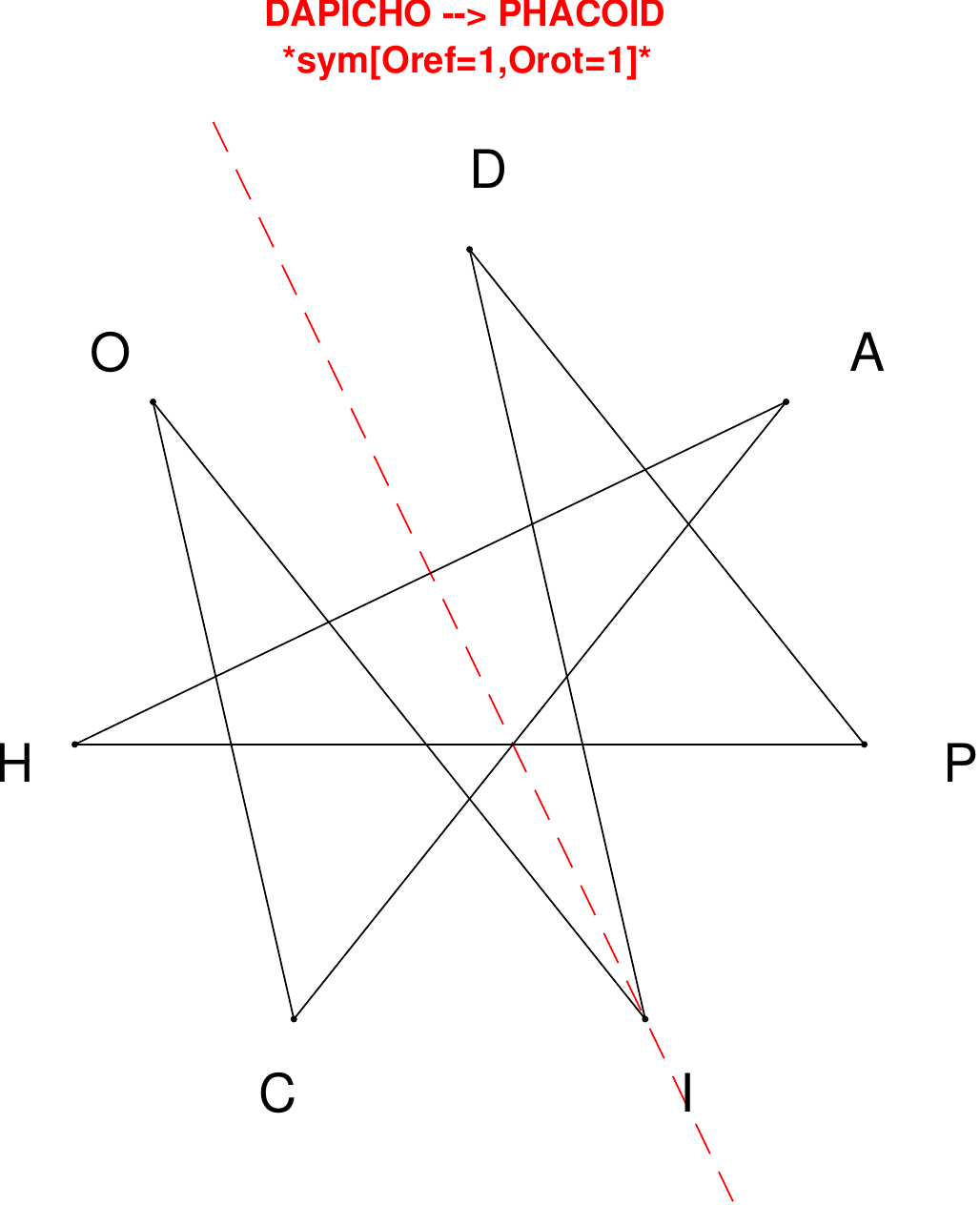}
\end{subfigure}
\hfill
\begin{subfigure}[T]{0.19\textwidth}
\centering
\includegraphics[width=\textwidth]{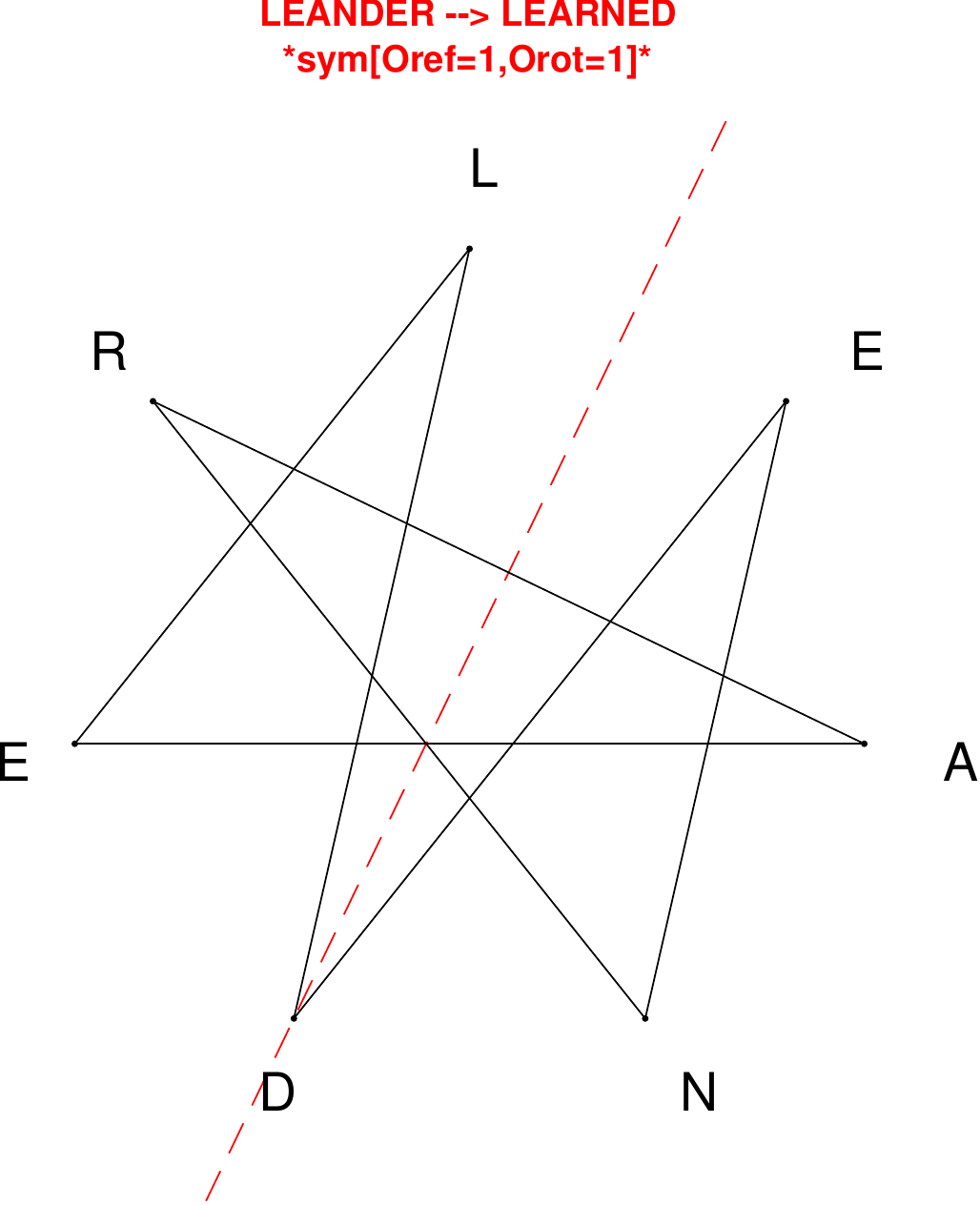}
\end{subfigure}
\hfill
\begin{subfigure}[T]{0.19\textwidth}
\centering
\includegraphics[width=\textwidth]{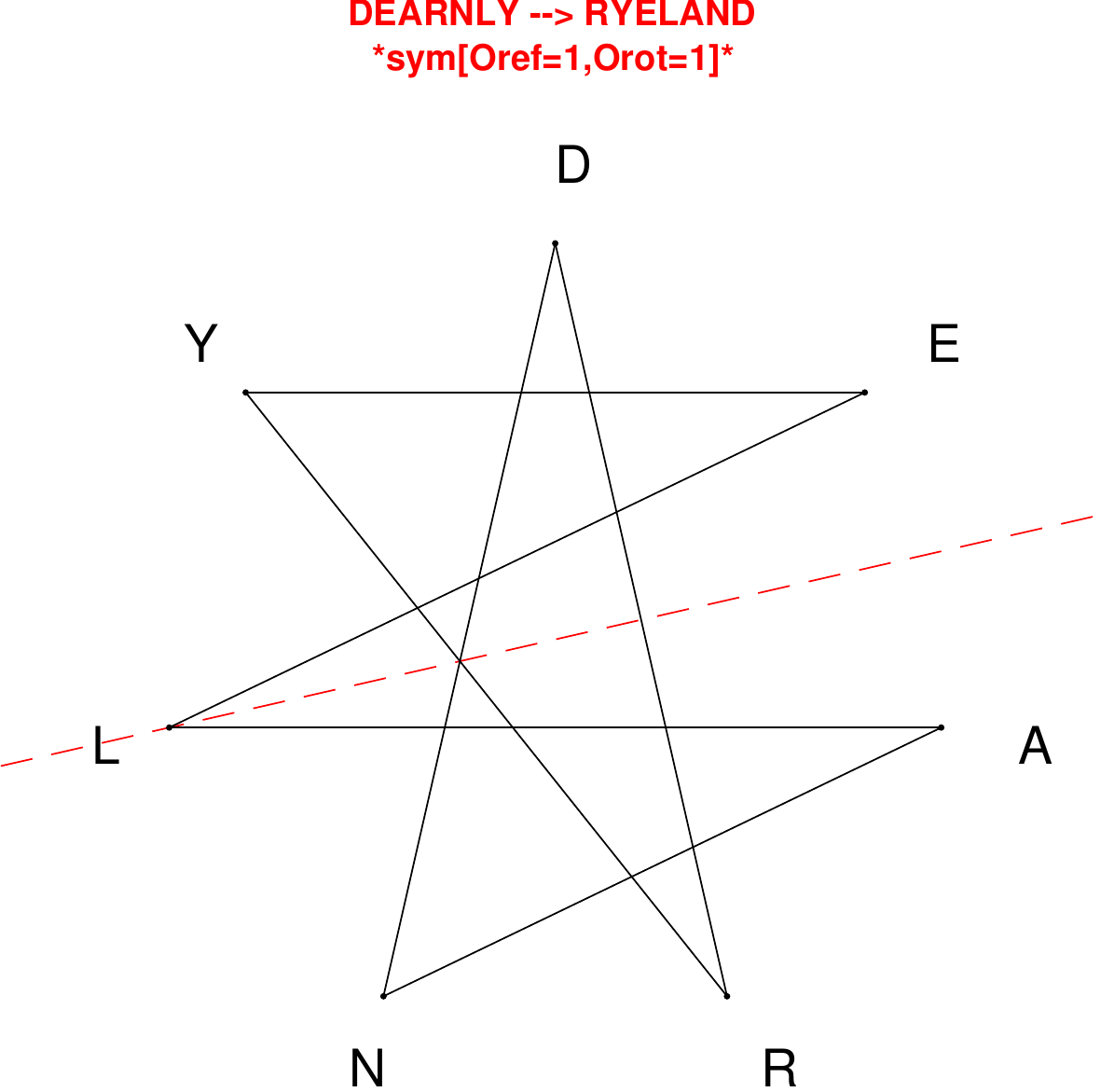}
\end{subfigure}
\end{figure}

\begin{figure}[H]
\centering
\begin{subfigure}[T]{0.19\textwidth}
\centering
\includegraphics[width=\textwidth]{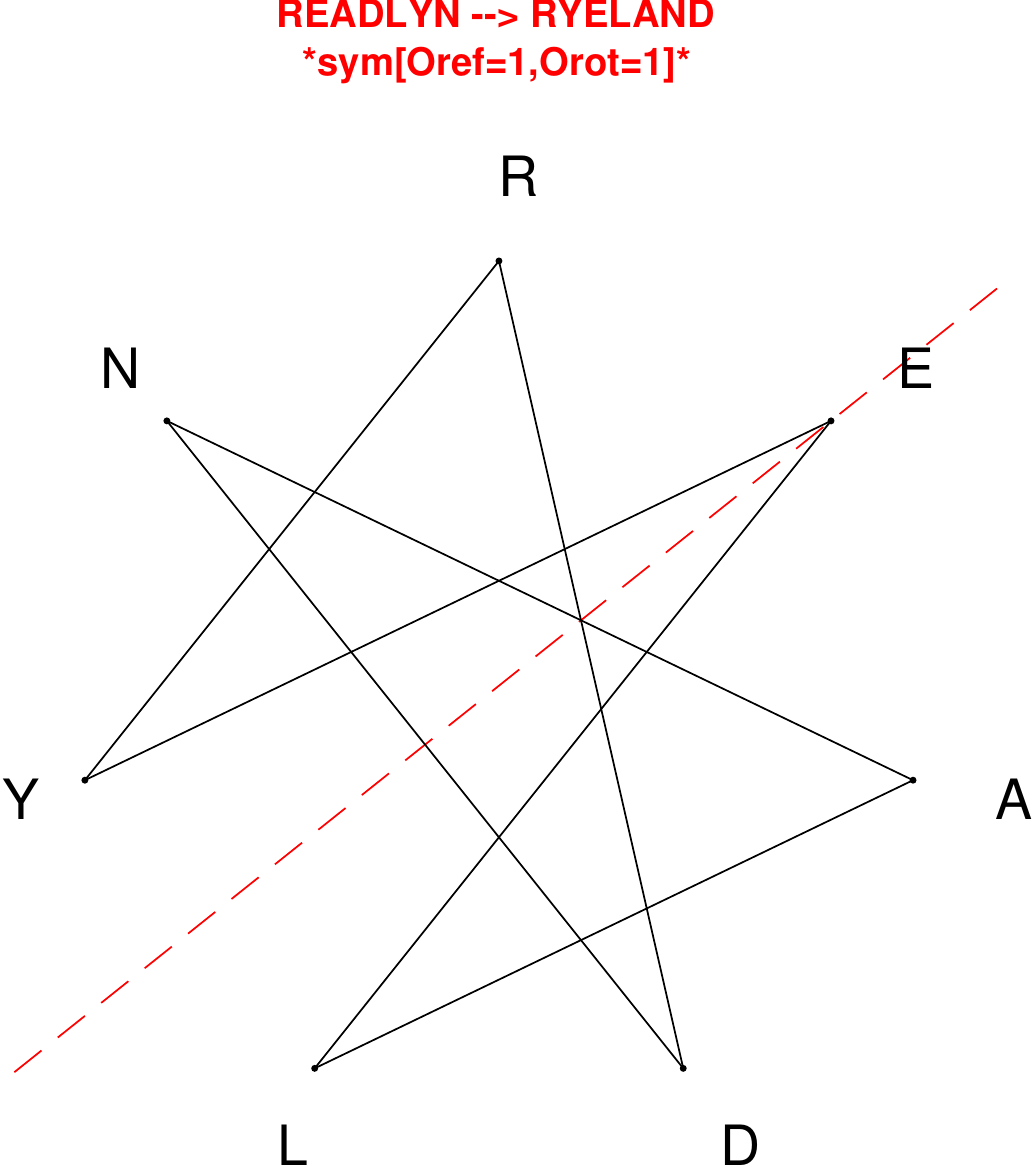}
\end{subfigure}
\hfill
\begin{subfigure}[T]{0.19\textwidth}
\centering
\includegraphics[width=\textwidth]{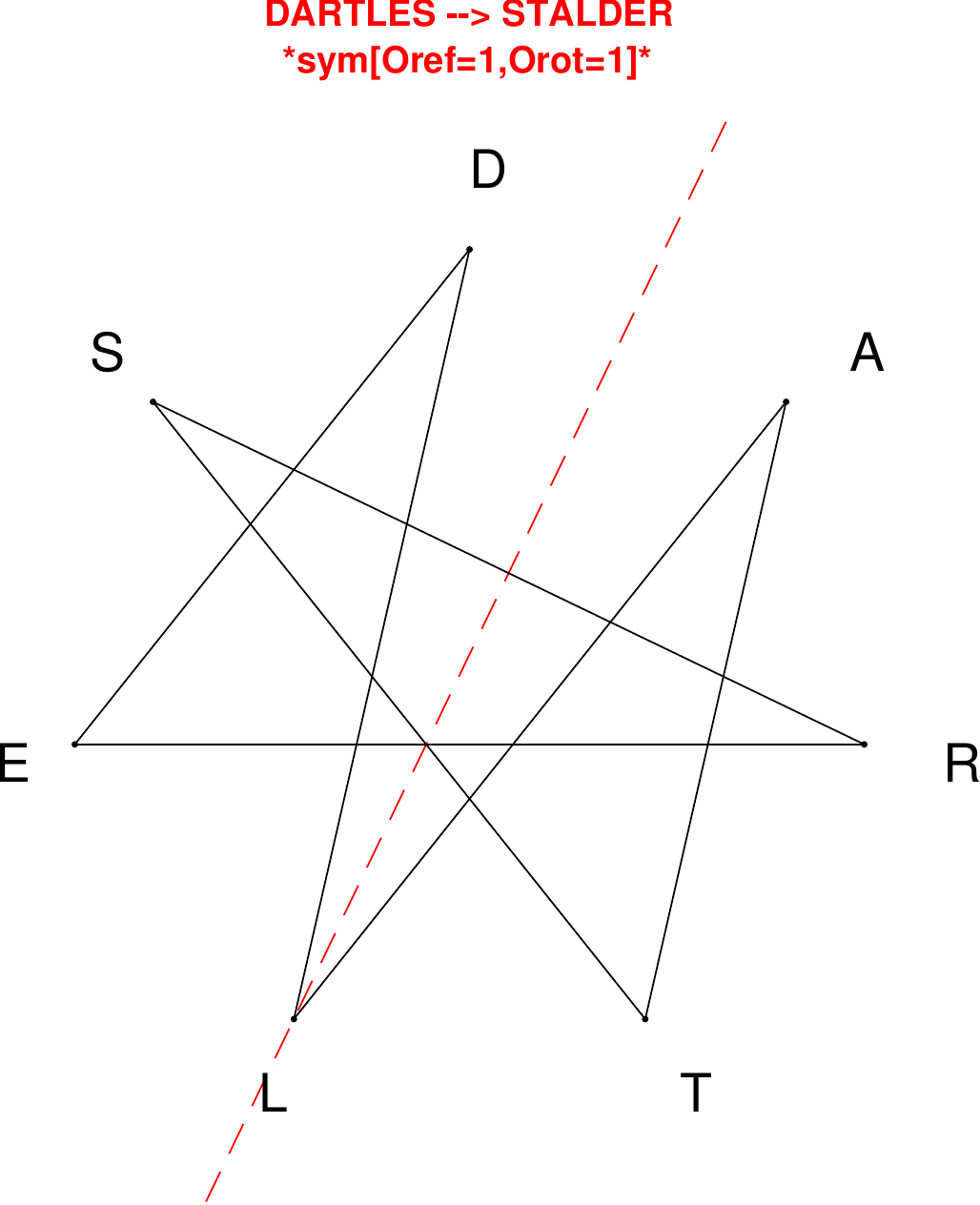}
\end{subfigure}
\hfill
\begin{subfigure}[T]{0.19\textwidth}
\centering
\includegraphics[width=\textwidth]{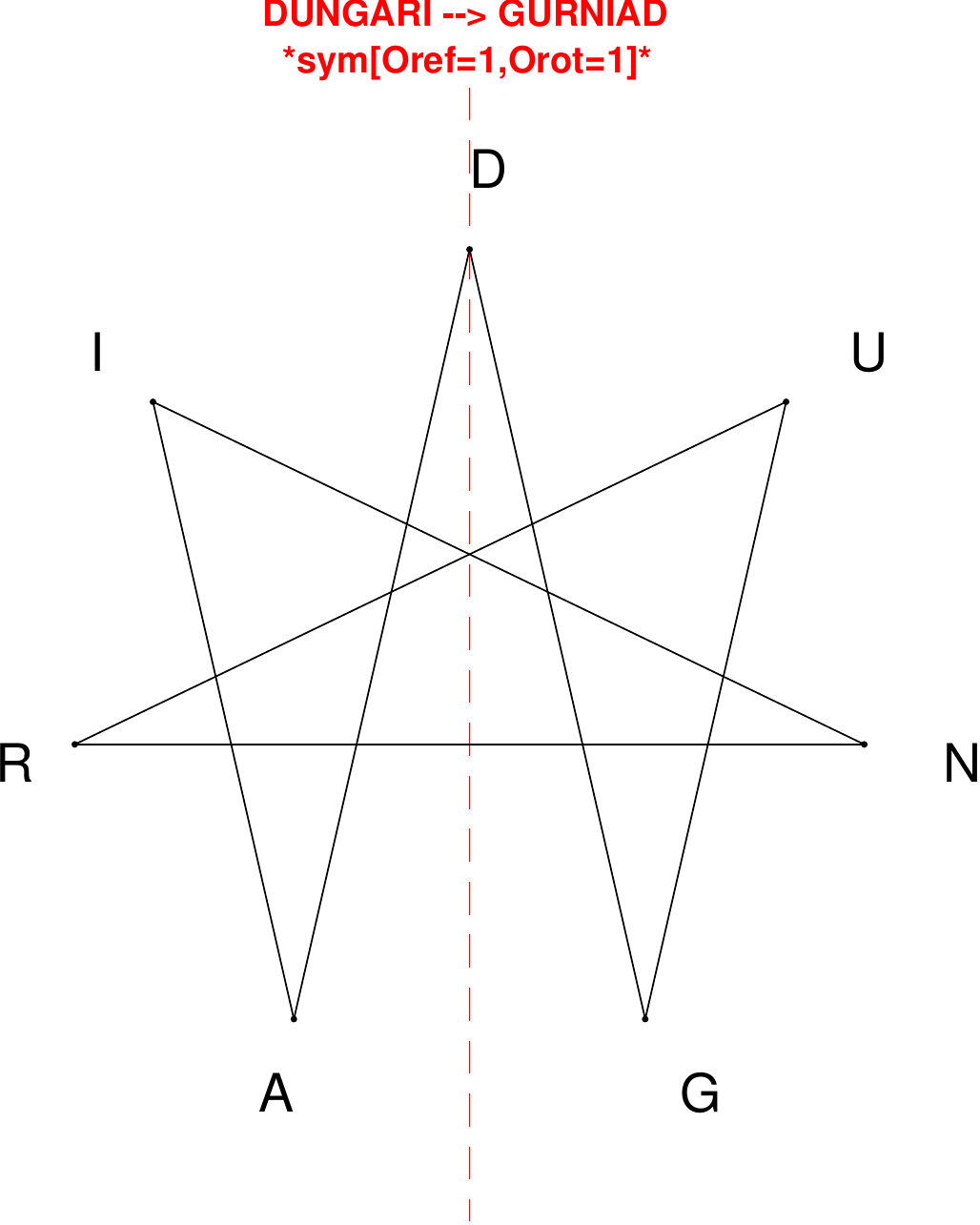}
\end{subfigure}
\hfill
\begin{subfigure}[T]{0.19\textwidth}
\centering
\includegraphics[width=\textwidth]{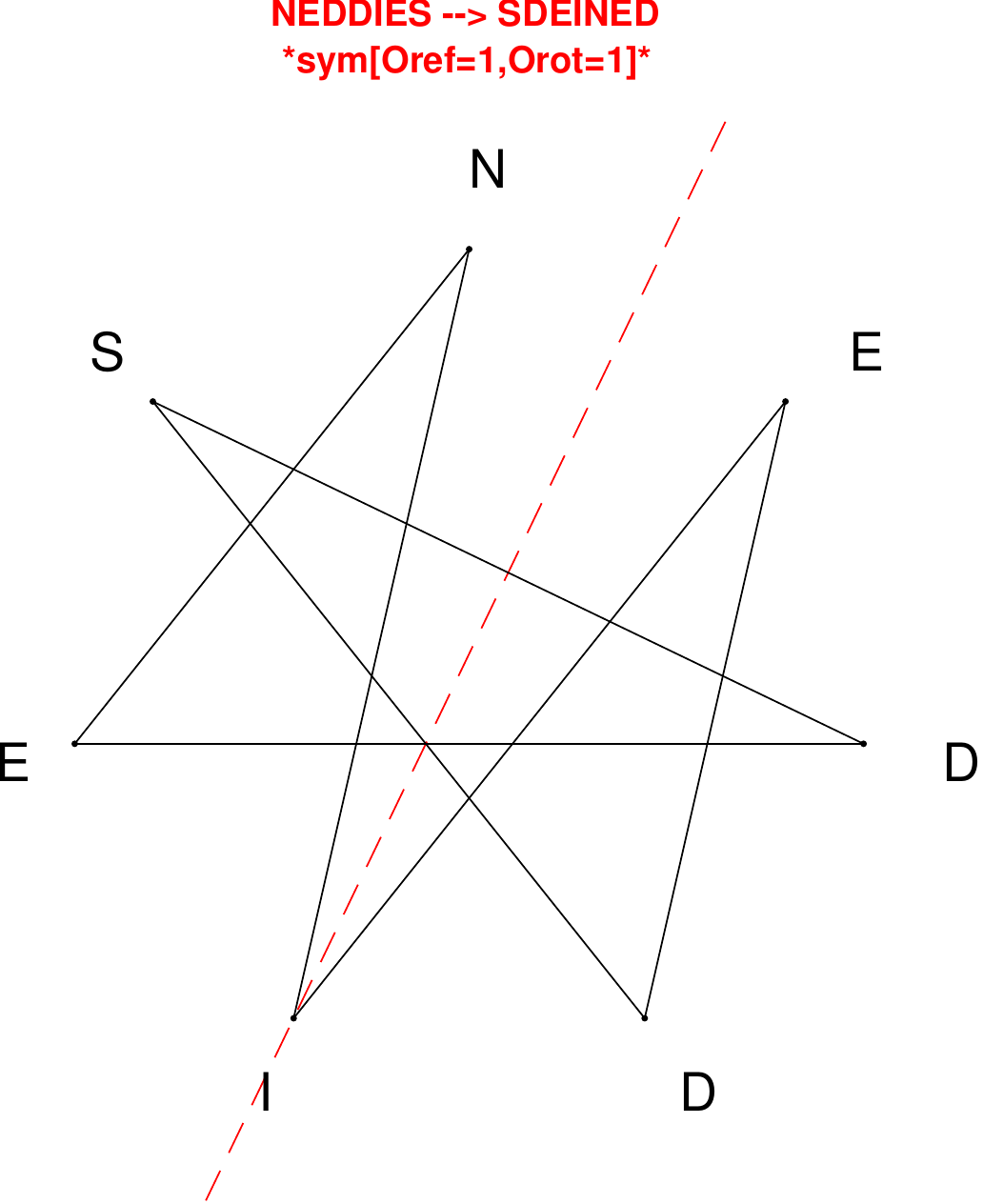}
\end{subfigure}
\hfill
\begin{subfigure}[T]{0.19\textwidth}
\centering
\includegraphics[width=\textwidth]{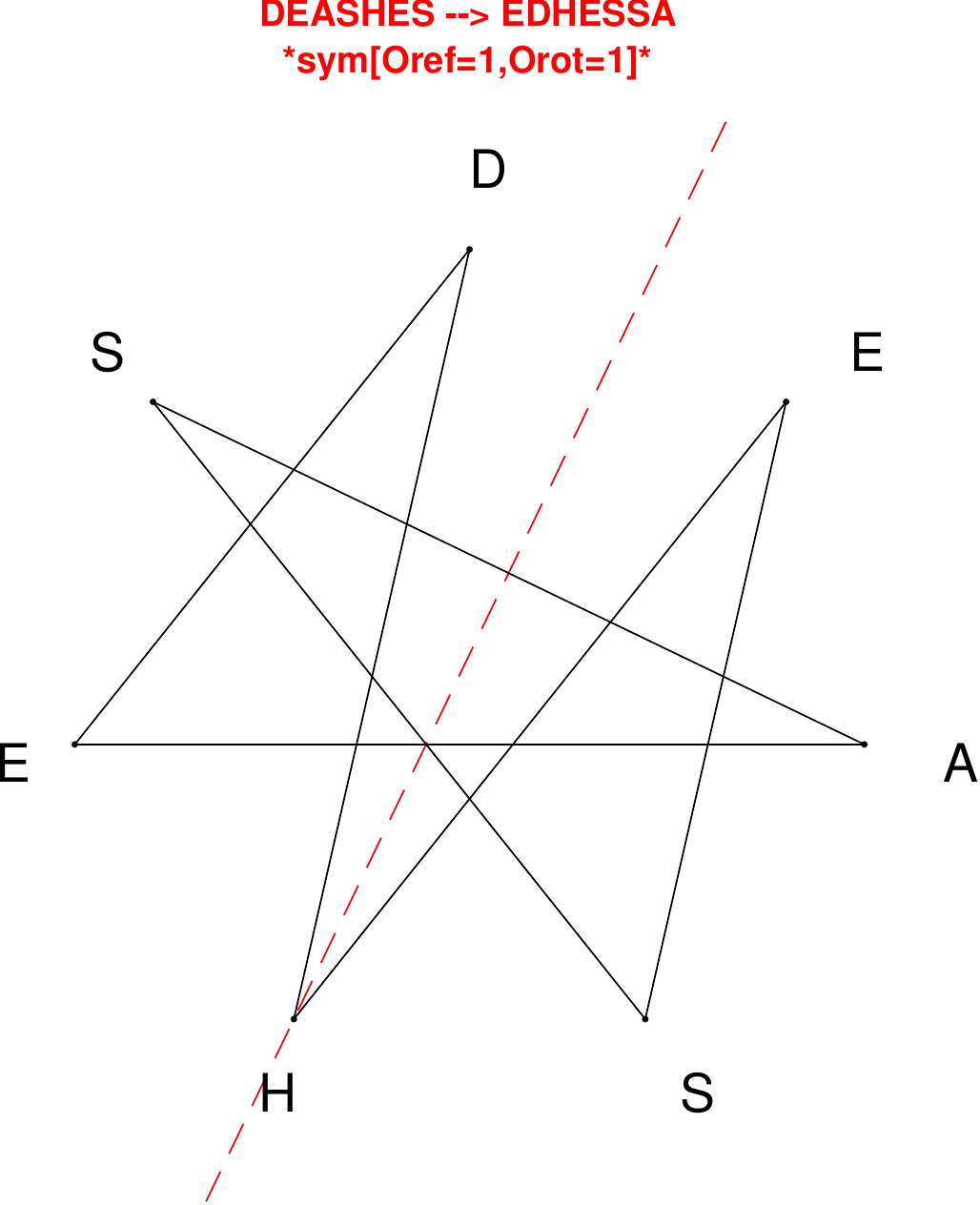}
\end{subfigure}
\end{figure}

\begin{figure}[H]
\centering
\begin{subfigure}[T]{0.19\textwidth}
\centering
\includegraphics[width=\textwidth]{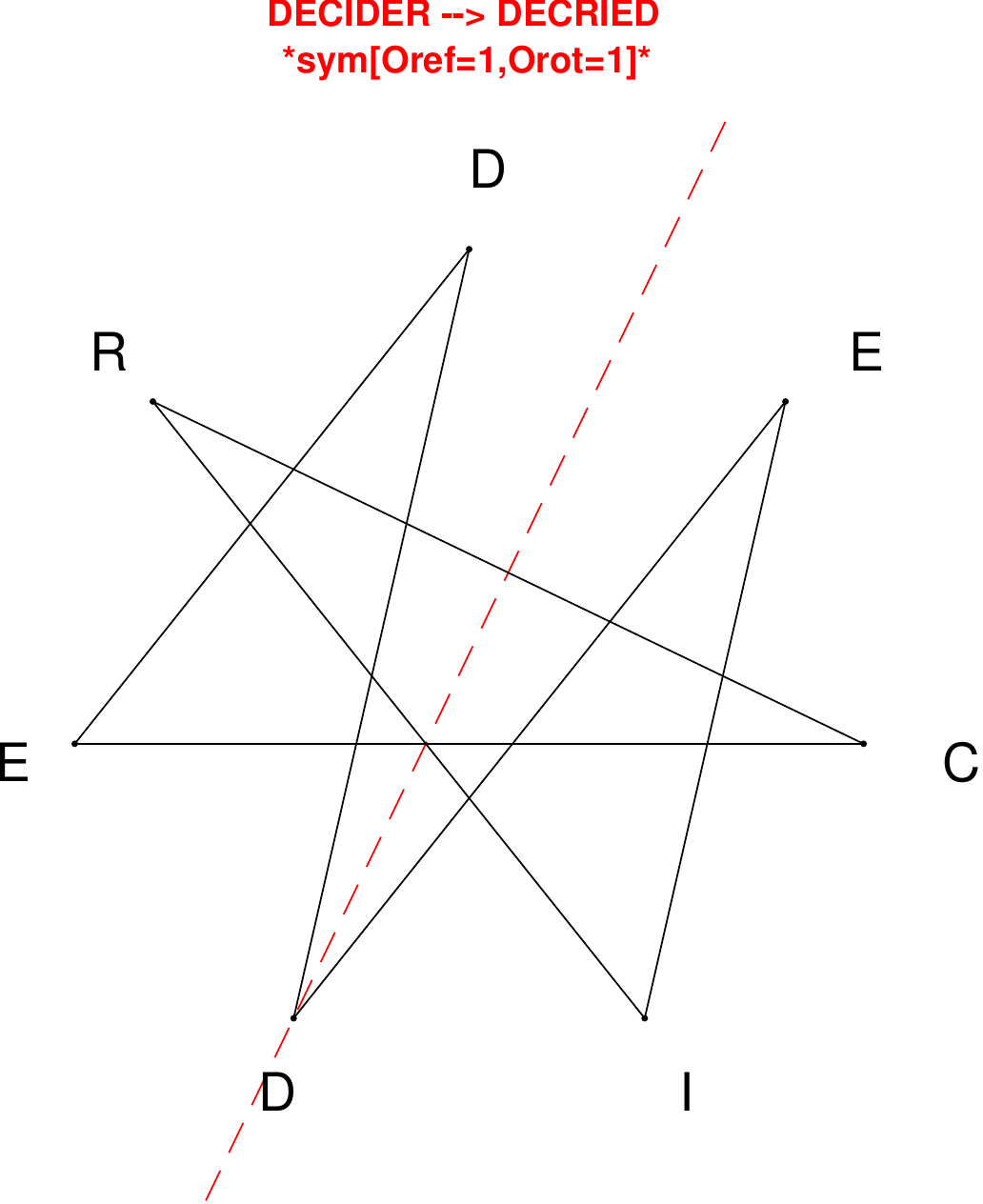}
\end{subfigure}
\hfill
\begin{subfigure}[T]{0.19\textwidth}
\centering
\includegraphics[width=\textwidth]{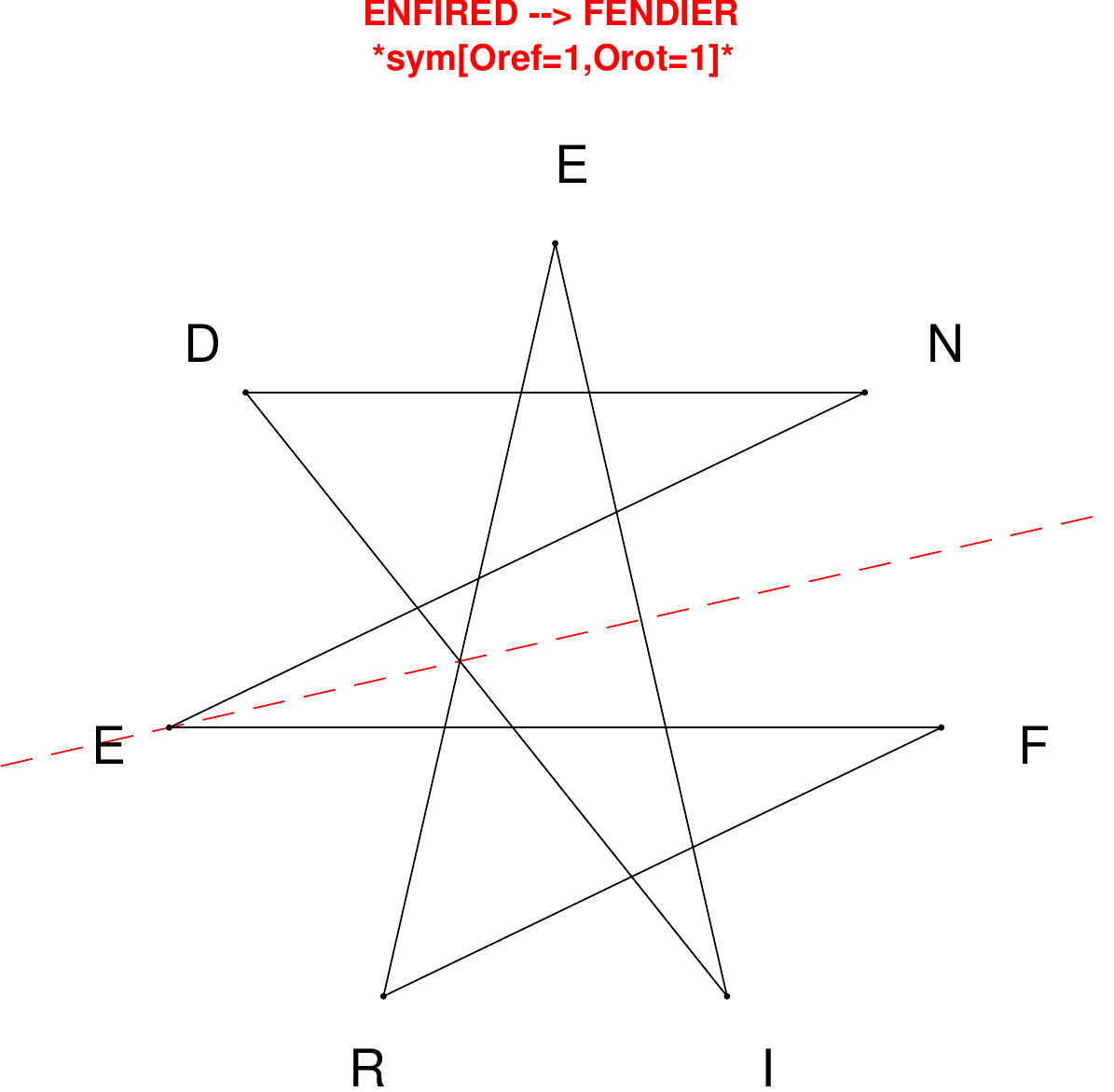}
\end{subfigure}
\hfill
\begin{subfigure}[T]{0.19\textwidth}
\centering
\includegraphics[width=\textwidth]{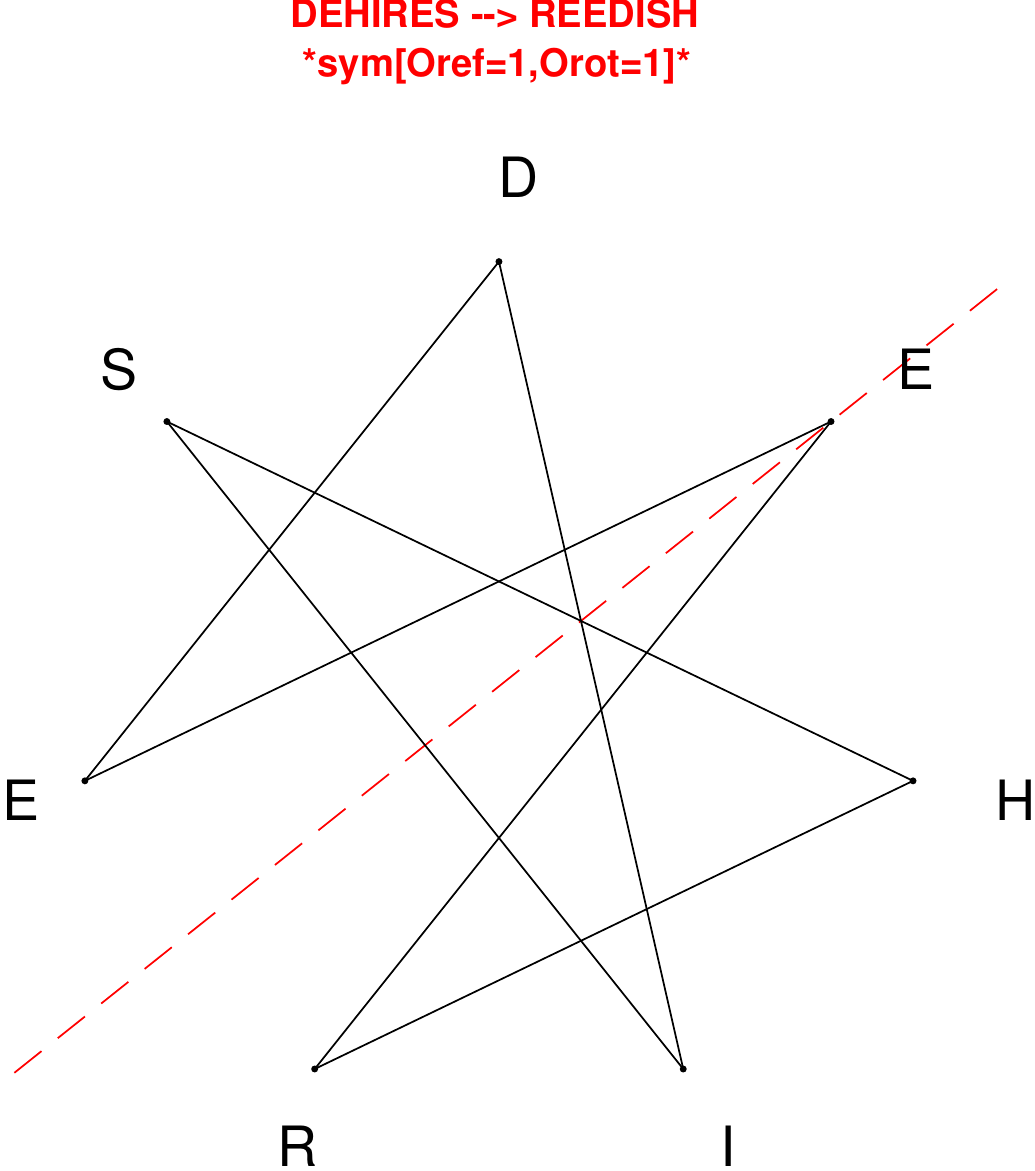}
\end{subfigure}
\hfill
\begin{subfigure}[T]{0.19\textwidth}
\centering
\includegraphics[width=\textwidth]{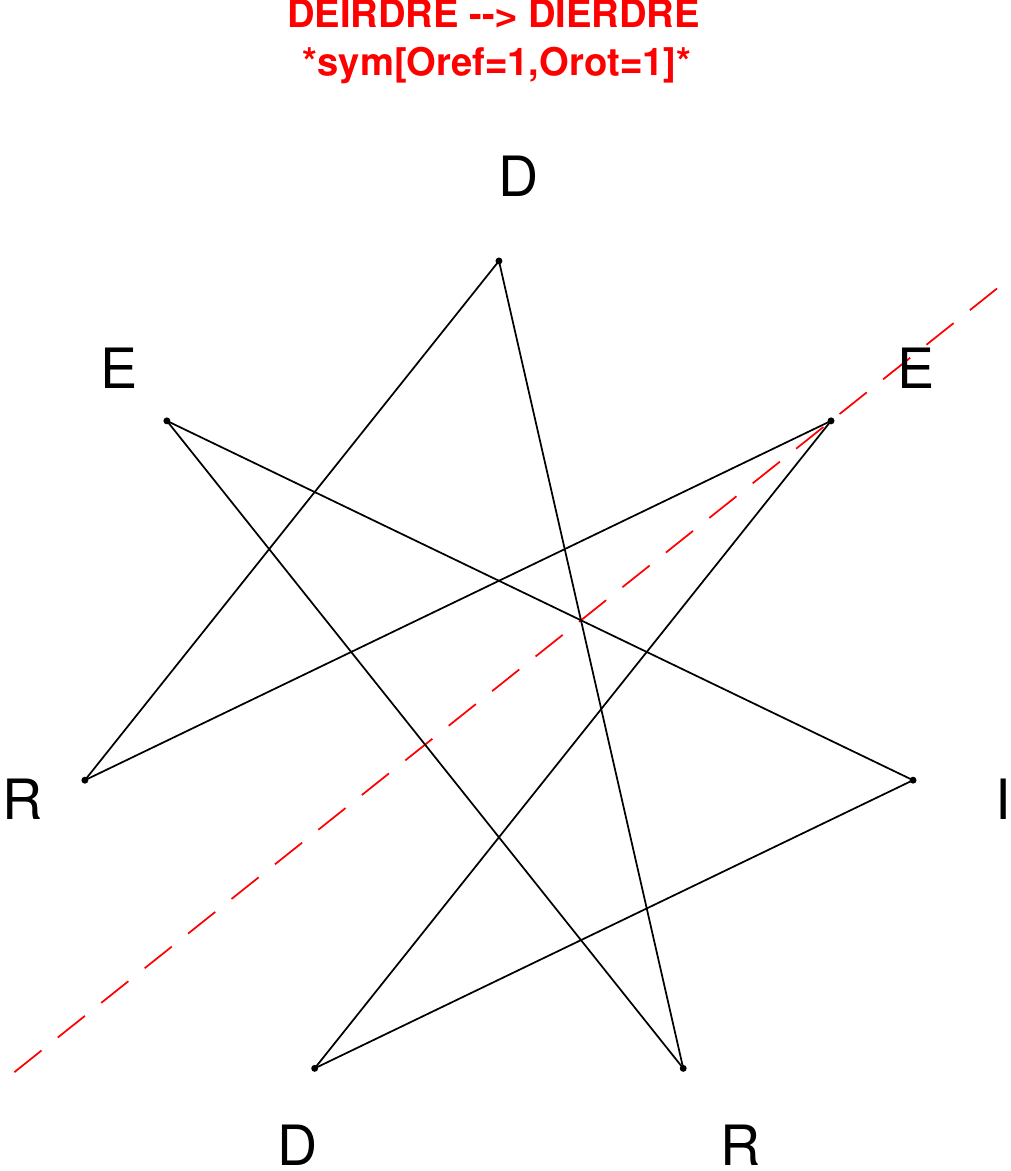}
\end{subfigure}
\hfill
\begin{subfigure}[T]{0.19\textwidth}
\centering
\includegraphics[width=\textwidth]{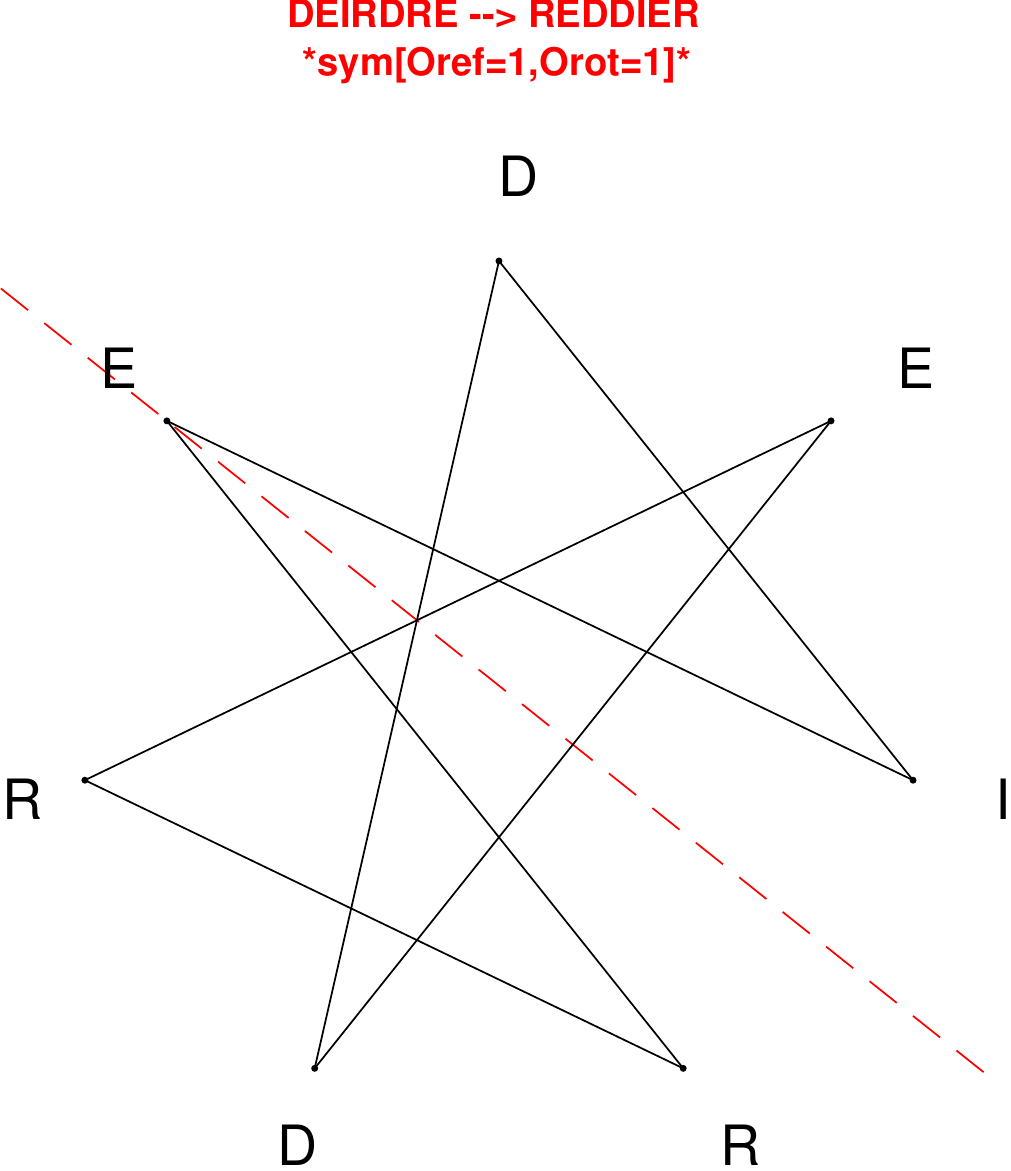}
\end{subfigure}
\end{figure}

\begin{figure}[H]
\centering
\begin{subfigure}[T]{0.19\textwidth}
\centering
\includegraphics[width=\textwidth]{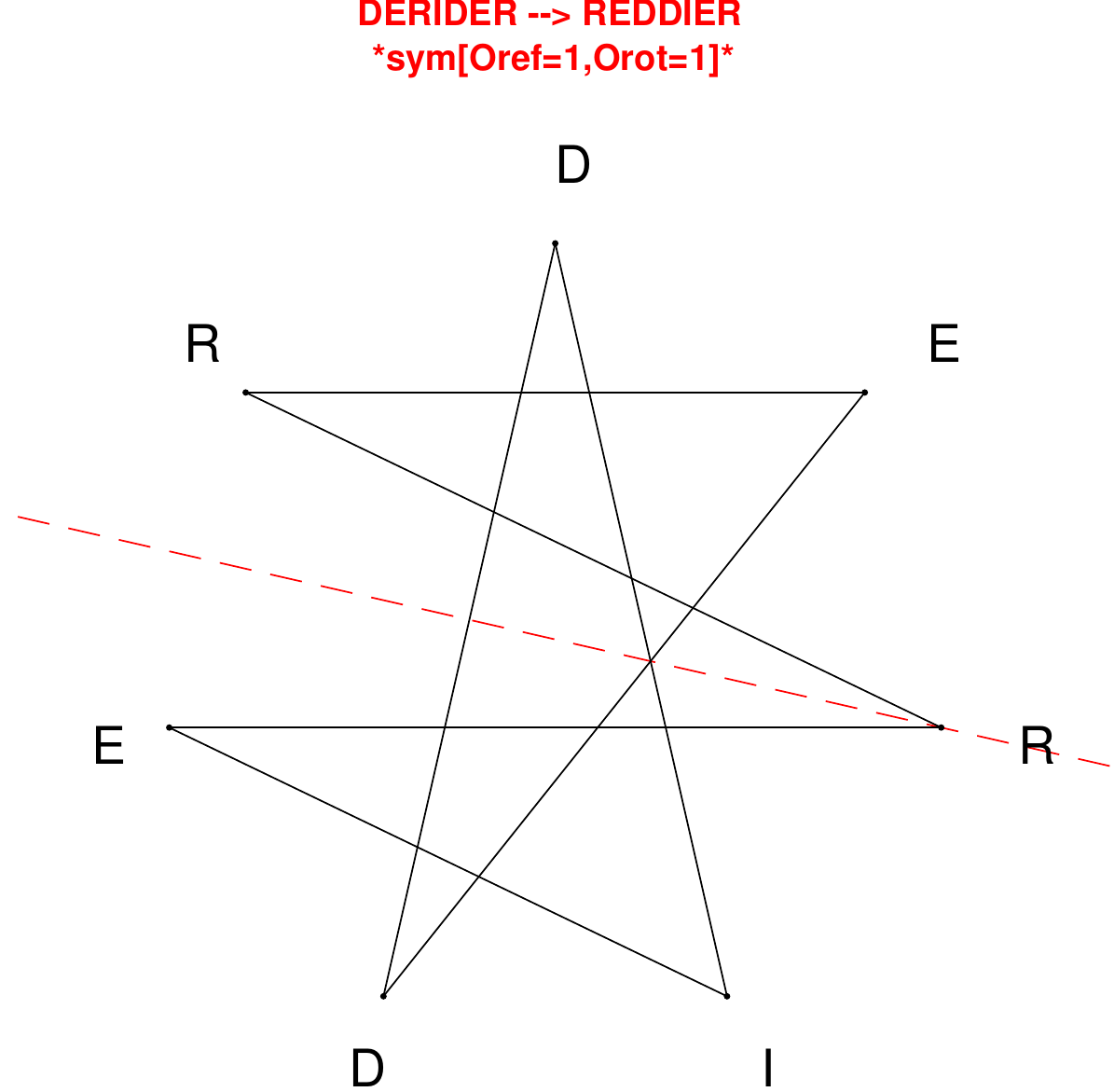}
\end{subfigure}
\hfill
\begin{subfigure}[T]{0.19\textwidth}
\centering
\includegraphics[width=\textwidth]{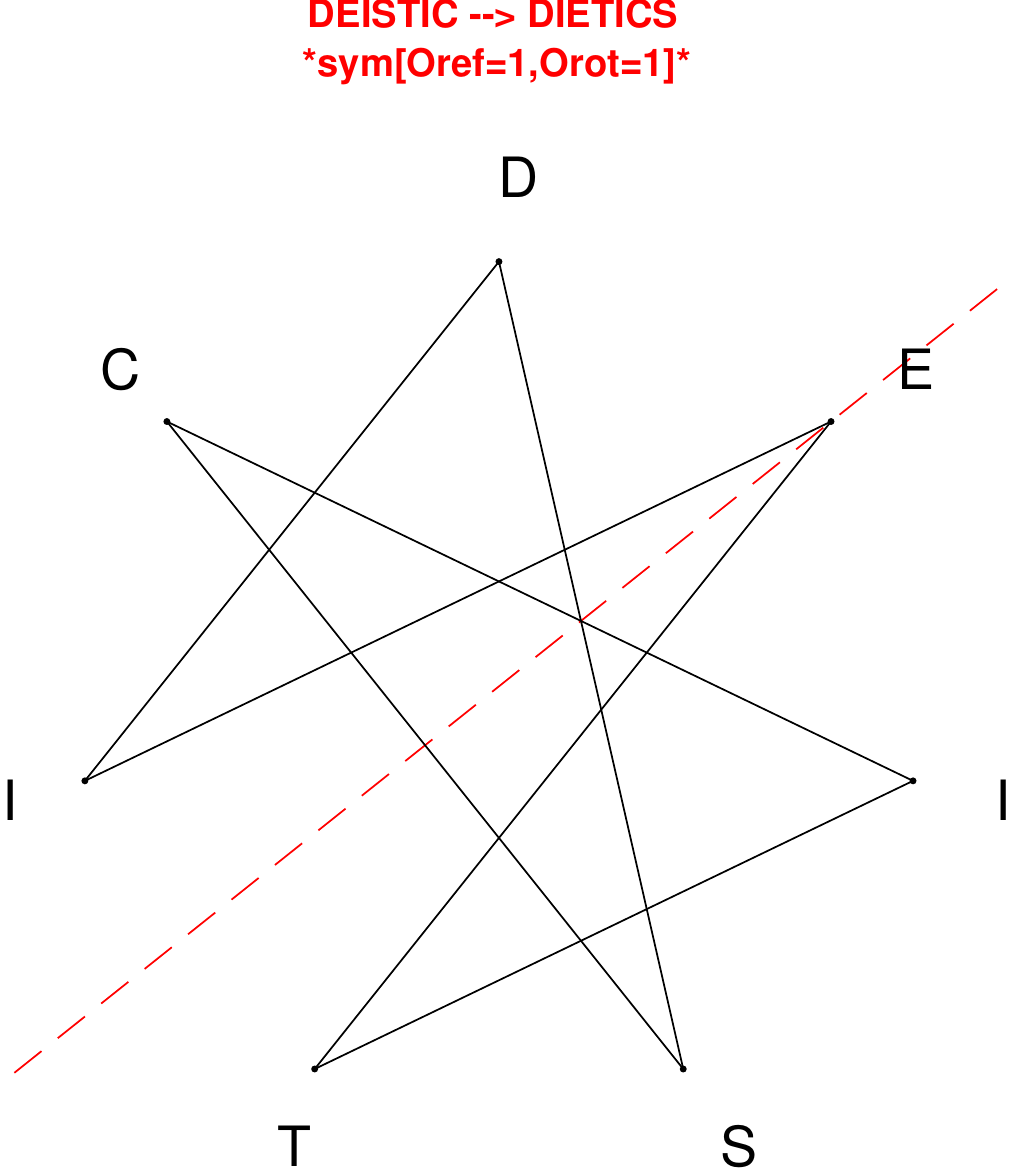}
\end{subfigure}
\hfill
\begin{subfigure}[T]{0.19\textwidth}
\centering
\includegraphics[width=\textwidth]{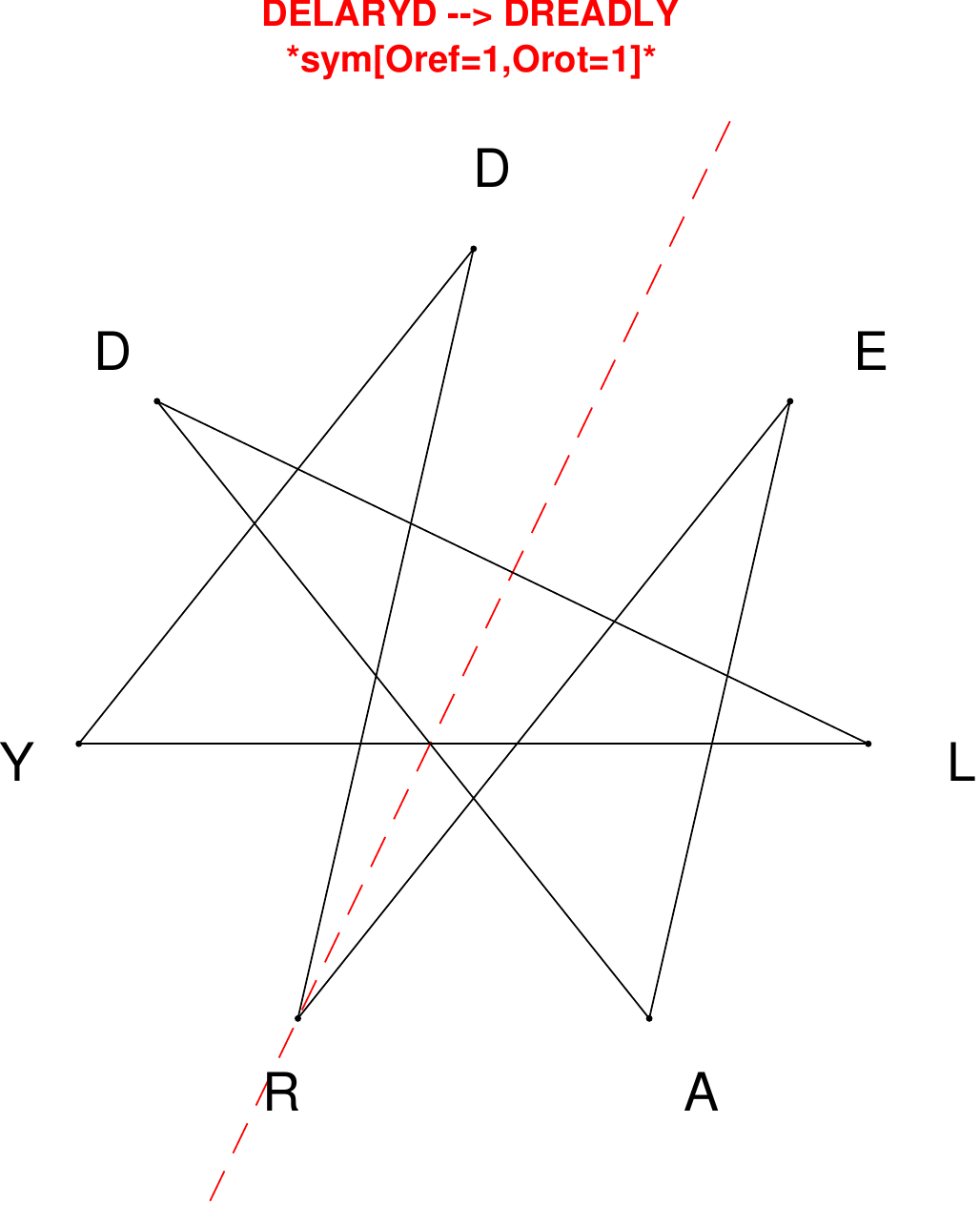}
\end{subfigure}
\hfill
\begin{subfigure}[T]{0.19\textwidth}
\centering
\includegraphics[width=\textwidth]{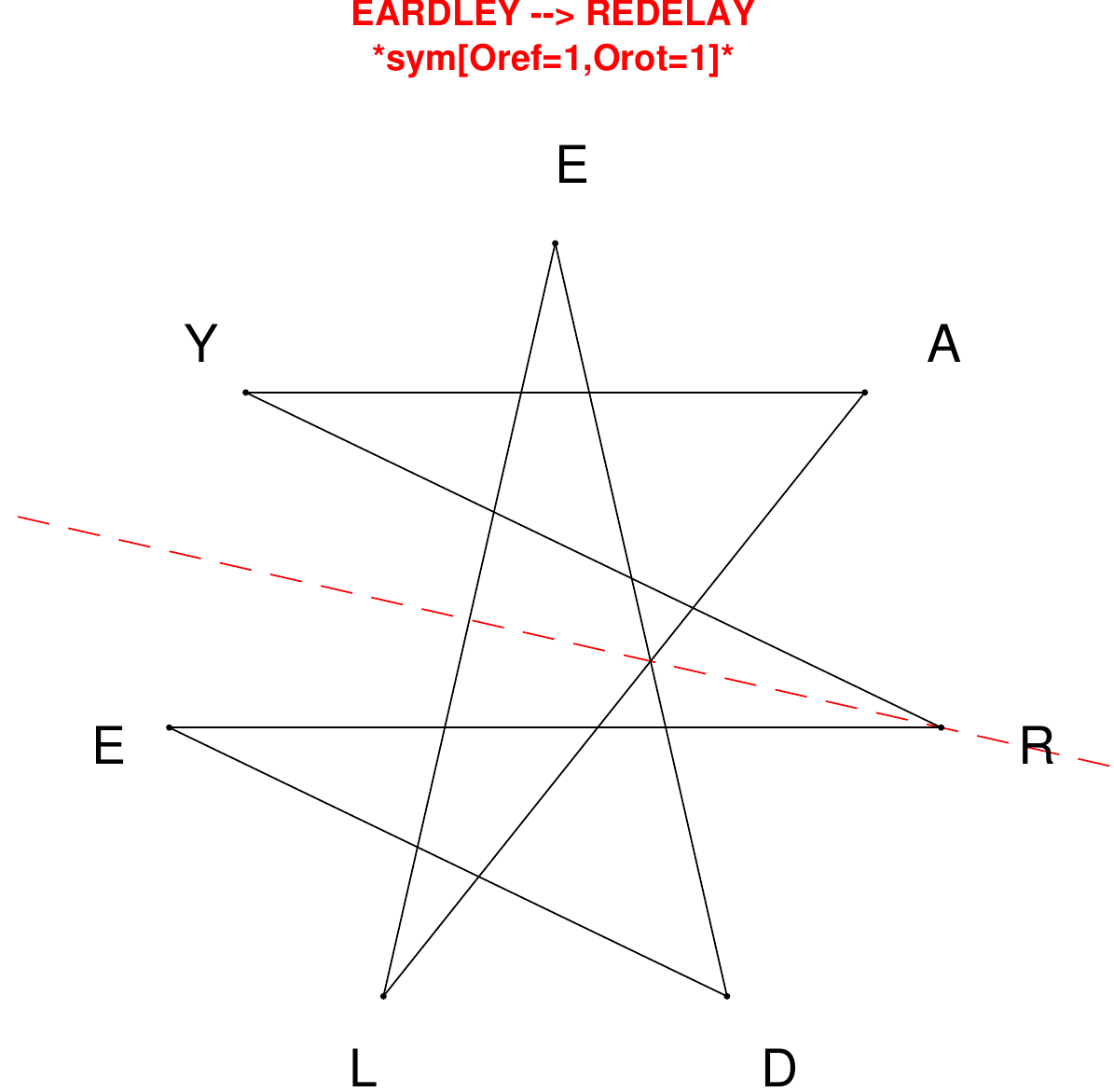}
\end{subfigure}
\hfill
\begin{subfigure}[T]{0.19\textwidth}
\centering
\includegraphics[width=\textwidth]{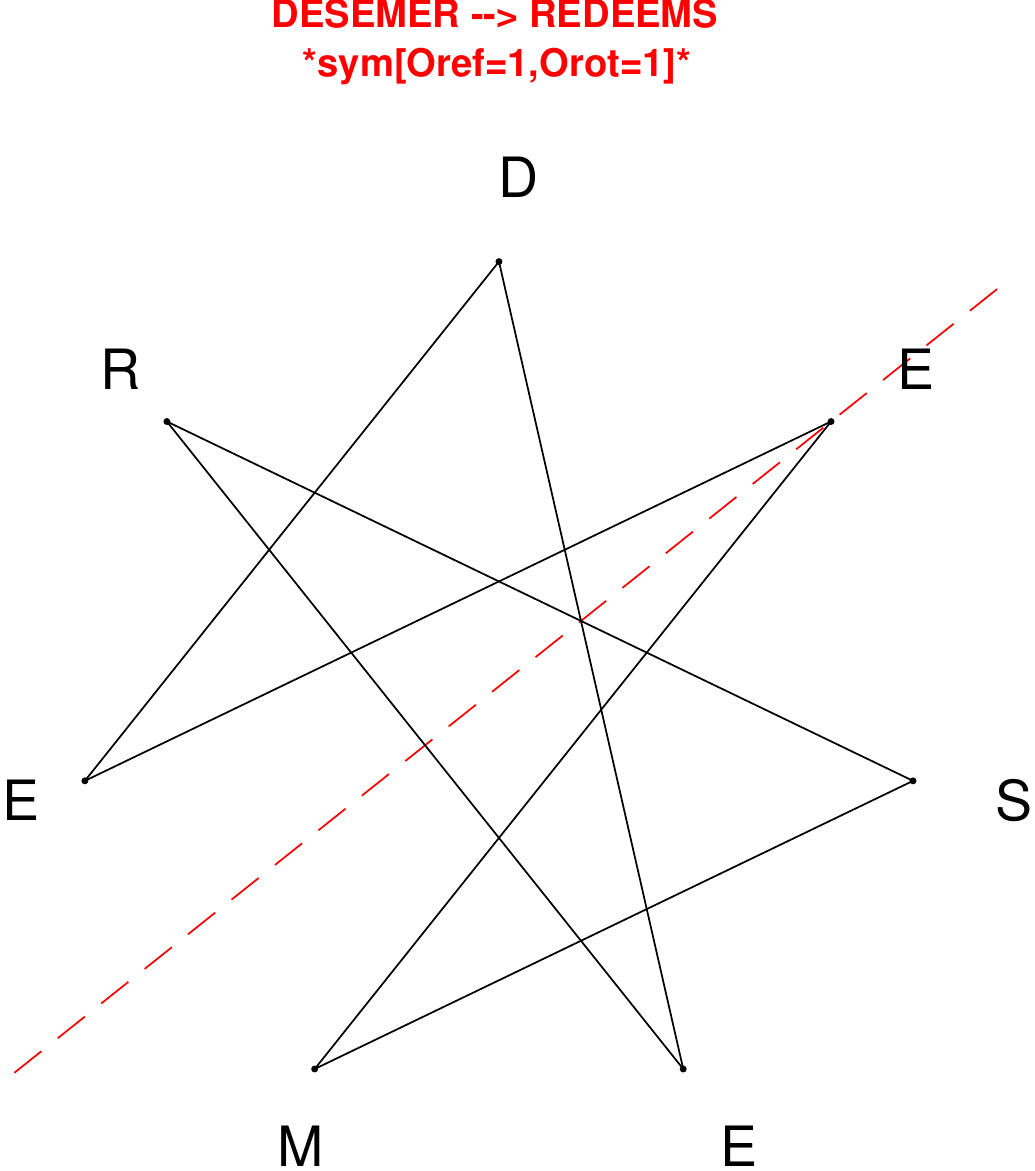}
\end{subfigure}
\end{figure}

\begin{figure}[H]
\centering
\begin{subfigure}[T]{0.19\textwidth}
\centering
\includegraphics[width=\textwidth]{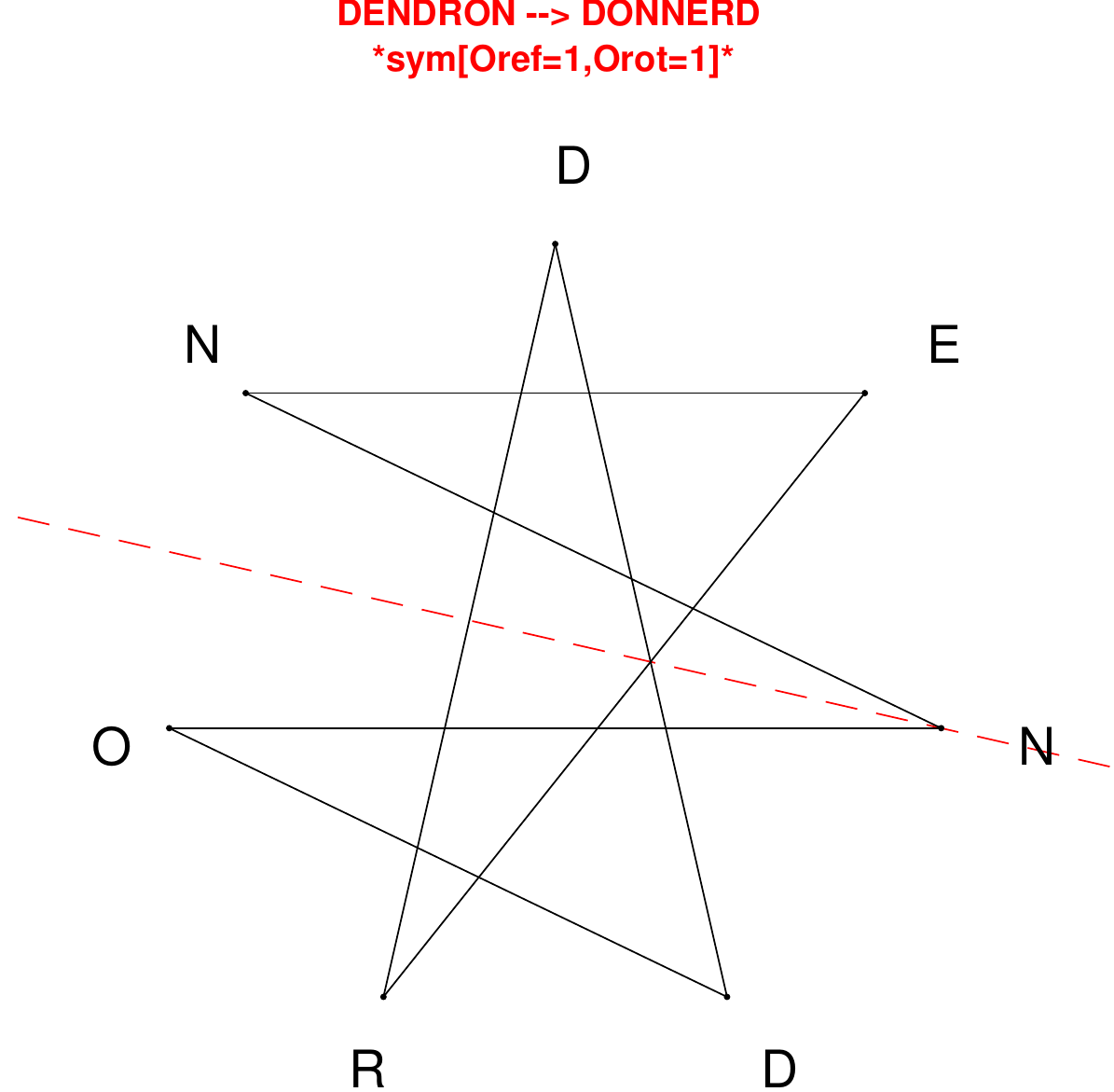}
\end{subfigure}
\hfill
\begin{subfigure}[T]{0.19\textwidth}
\centering
\includegraphics[width=\textwidth]{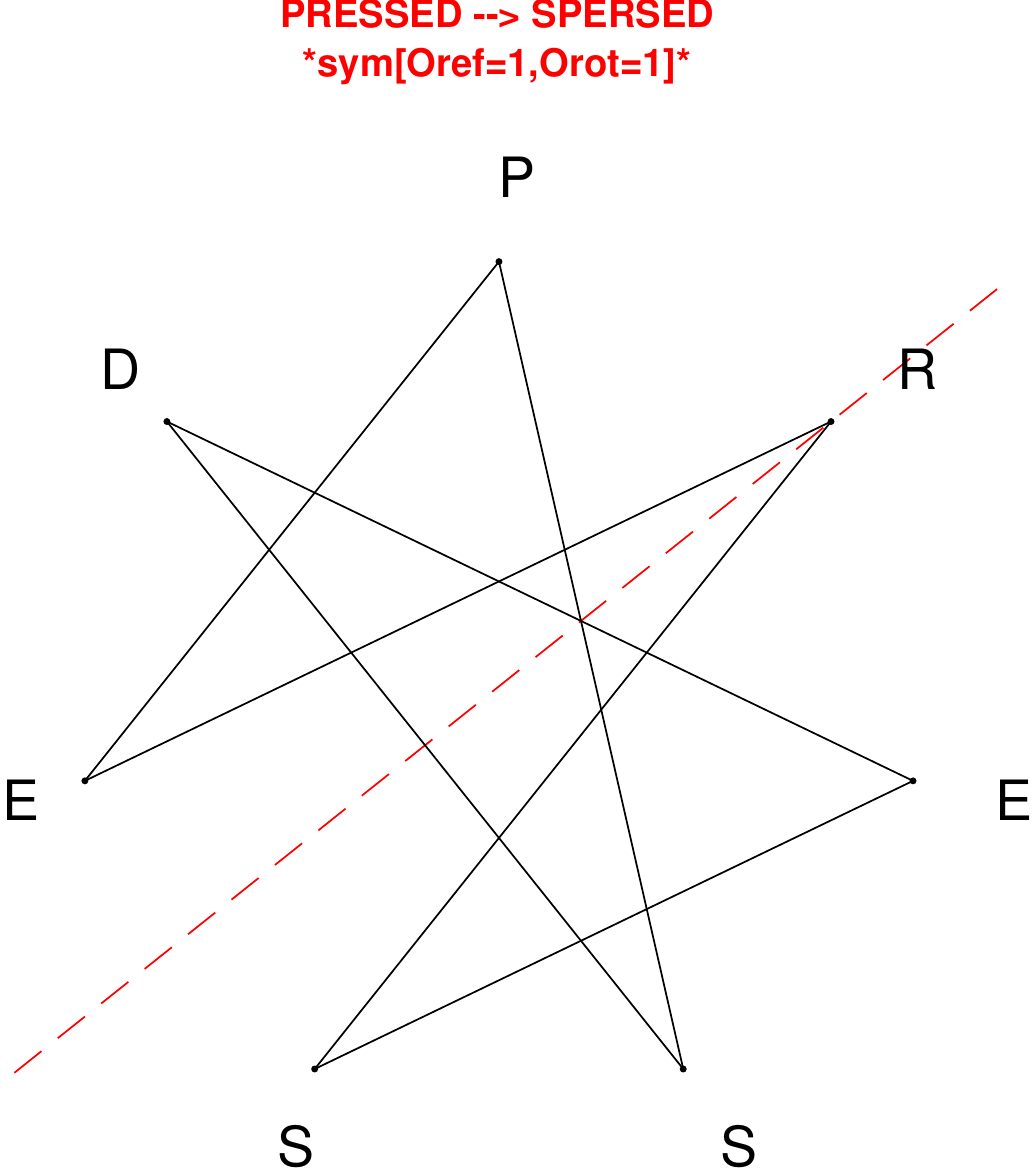}
\end{subfigure}
\hfill
\begin{subfigure}[T]{0.19\textwidth}
\centering
\includegraphics[width=\textwidth]{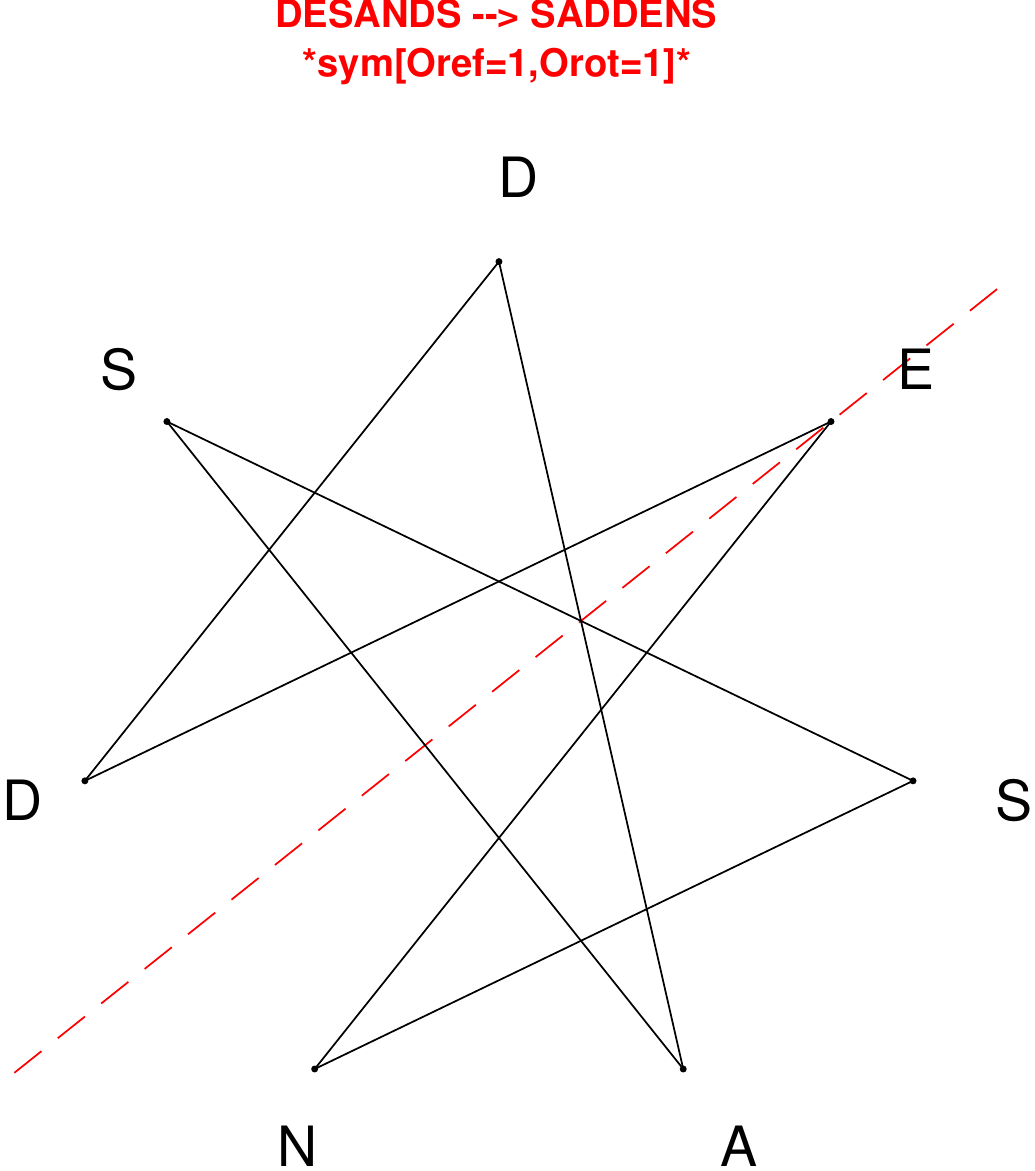}
\end{subfigure}
\hfill
\begin{subfigure}[T]{0.19\textwidth}
\centering
\includegraphics[width=\textwidth]{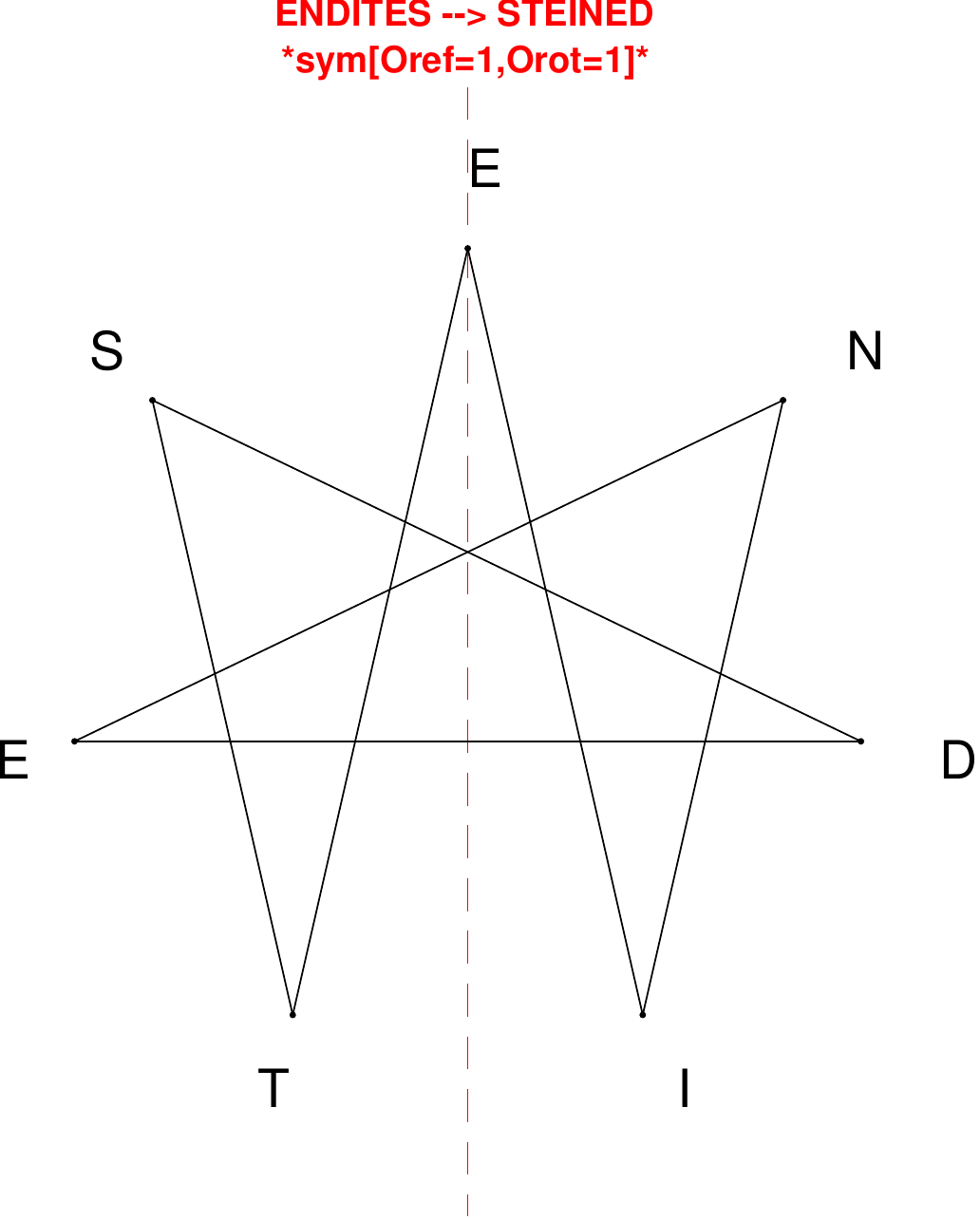}
\end{subfigure}
\hfill
\begin{subfigure}[T]{0.19\textwidth}
\centering
\includegraphics[width=\textwidth]{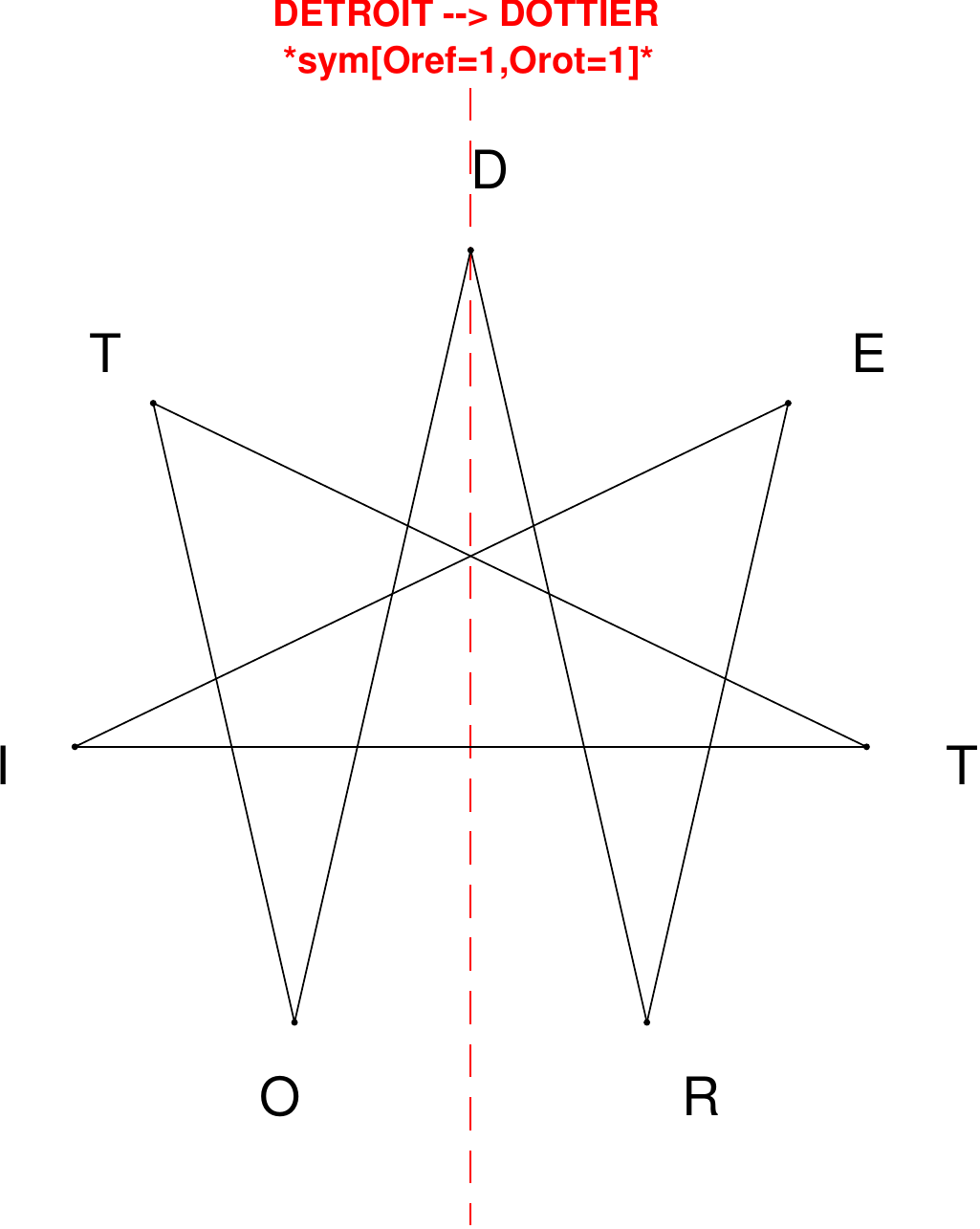}
\end{subfigure}
\end{figure}

\begin{figure}[H]
\centering
\begin{subfigure}[T]{0.19\textwidth}
\centering
\includegraphics[width=\textwidth]{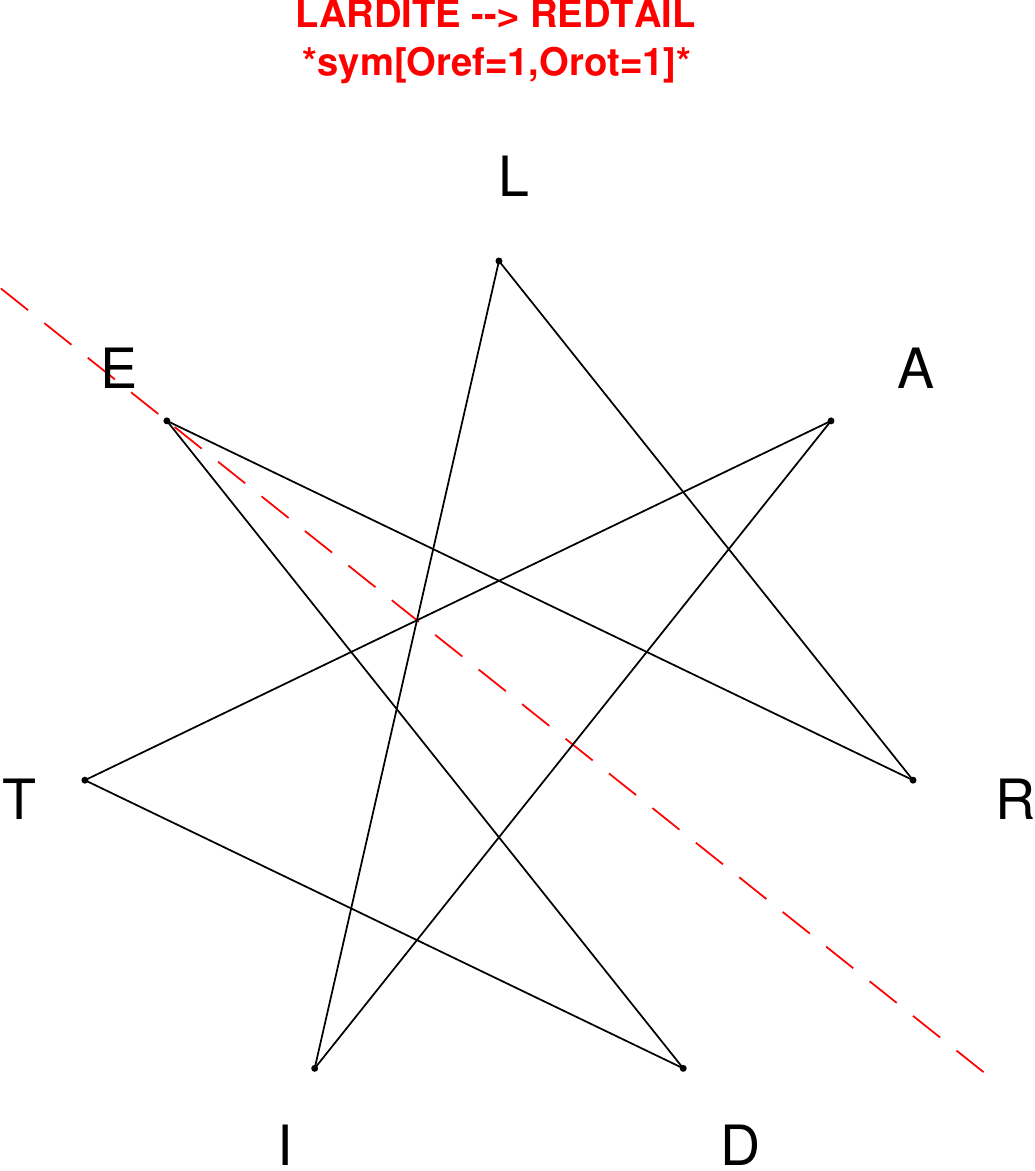}
\end{subfigure}
\hfill
\begin{subfigure}[T]{0.19\textwidth}
\centering
\includegraphics[width=\textwidth]{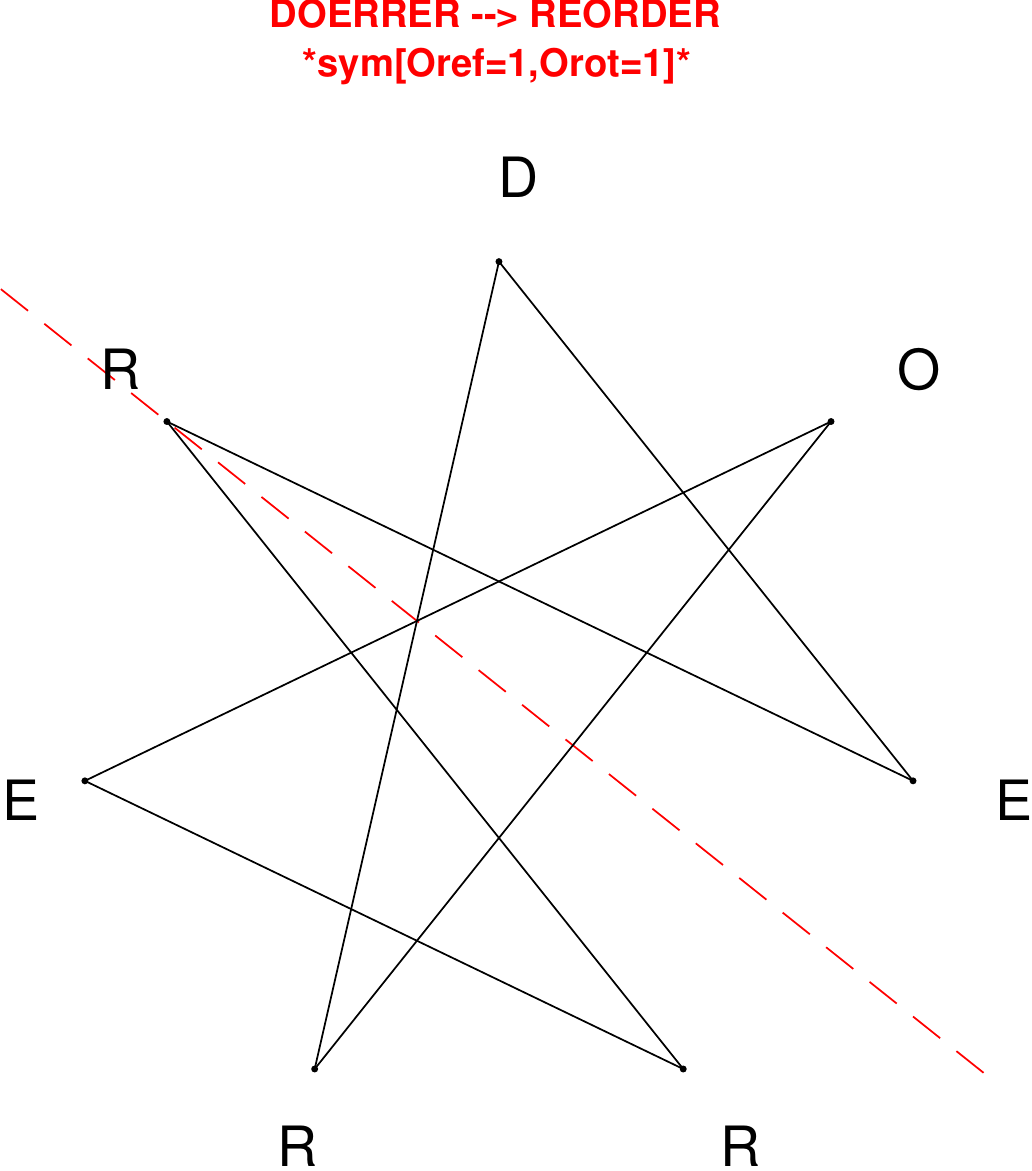}
\end{subfigure}
\hfill
\begin{subfigure}[T]{0.19\textwidth}
\centering
\includegraphics[width=\textwidth]{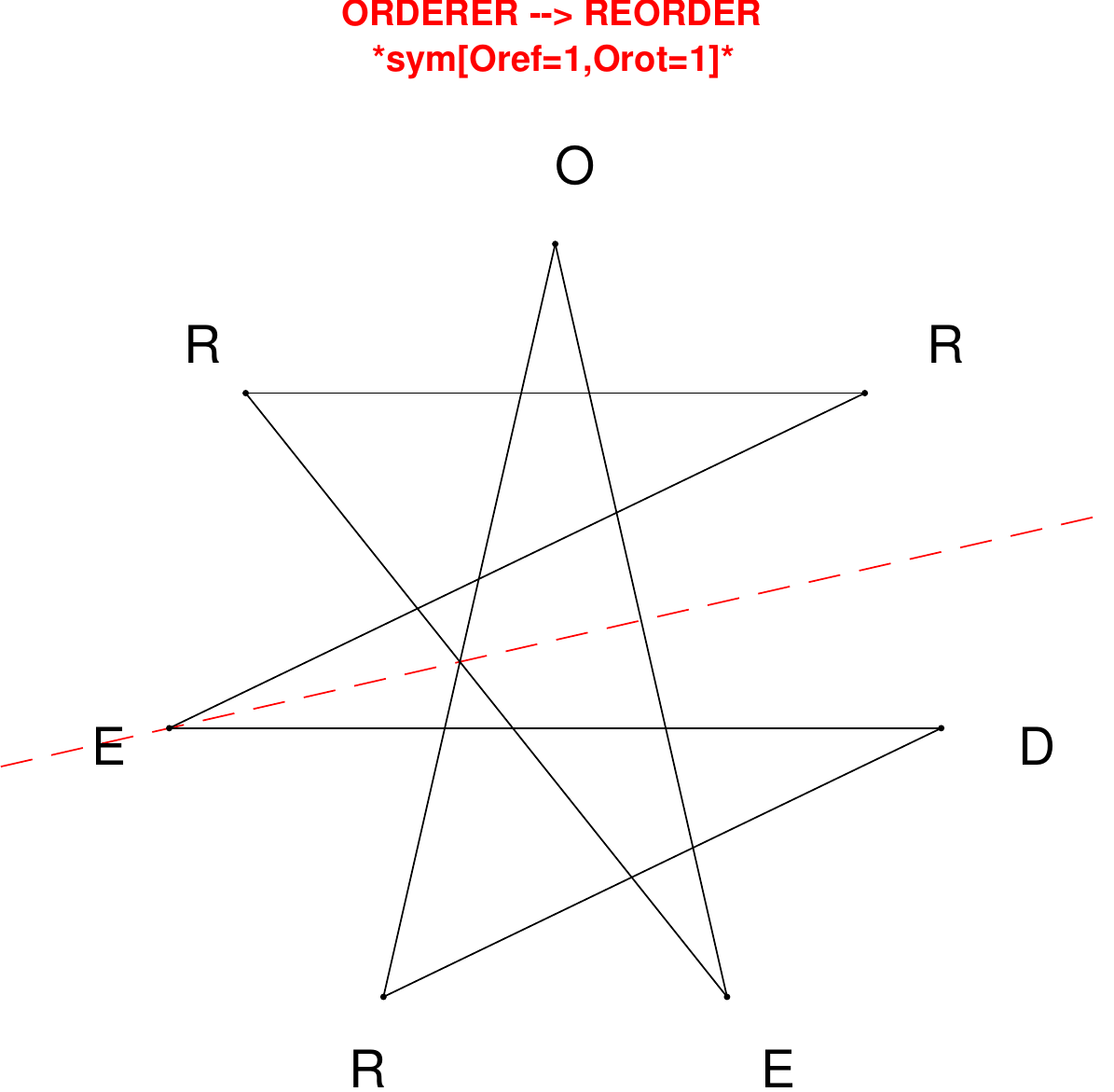}
\end{subfigure}
\hfill
\begin{subfigure}[T]{0.19\textwidth}
\centering
\includegraphics[width=\textwidth]{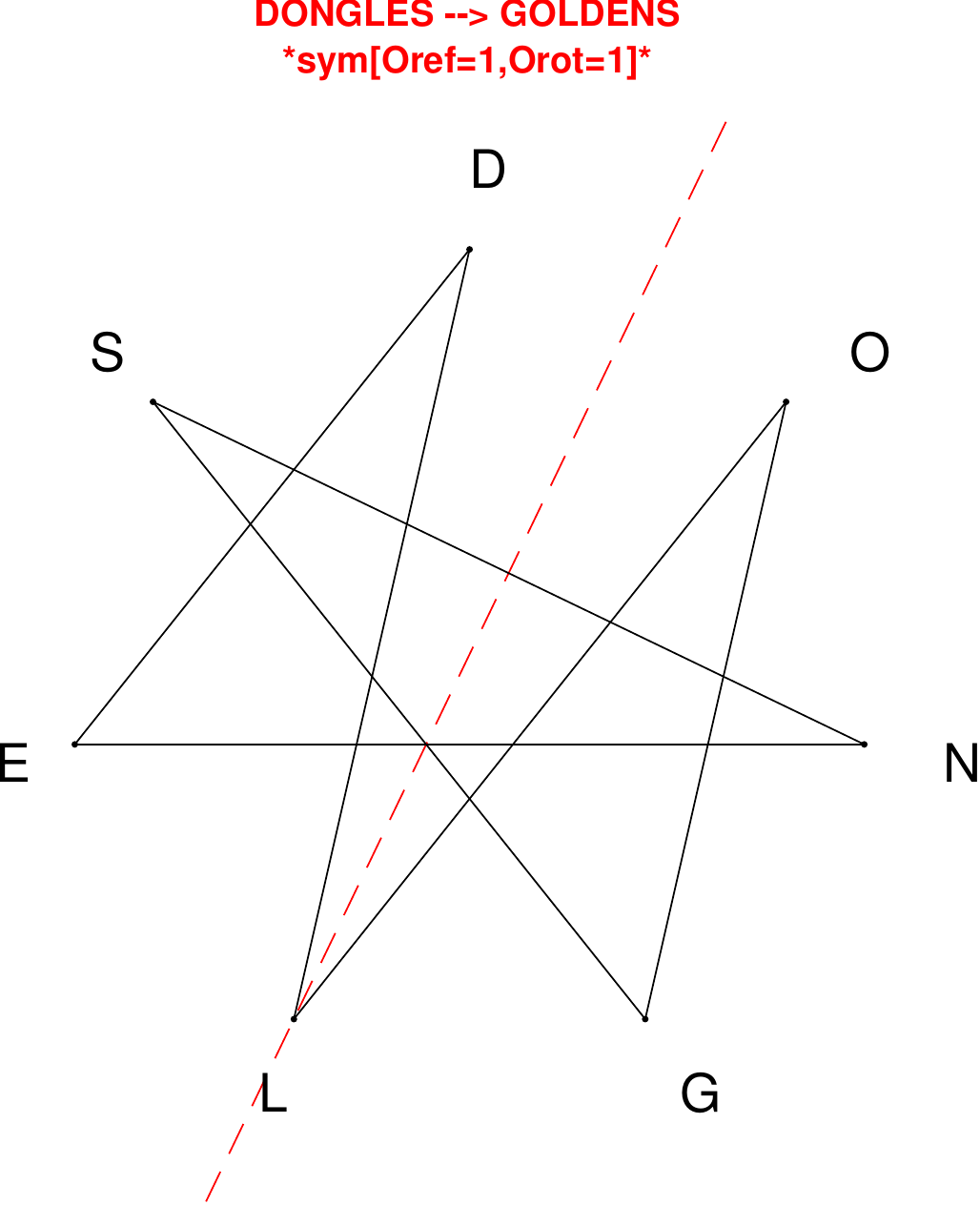}
\end{subfigure}
\hfill
\begin{subfigure}[T]{0.19\textwidth}
\centering
\includegraphics[width=\textwidth]{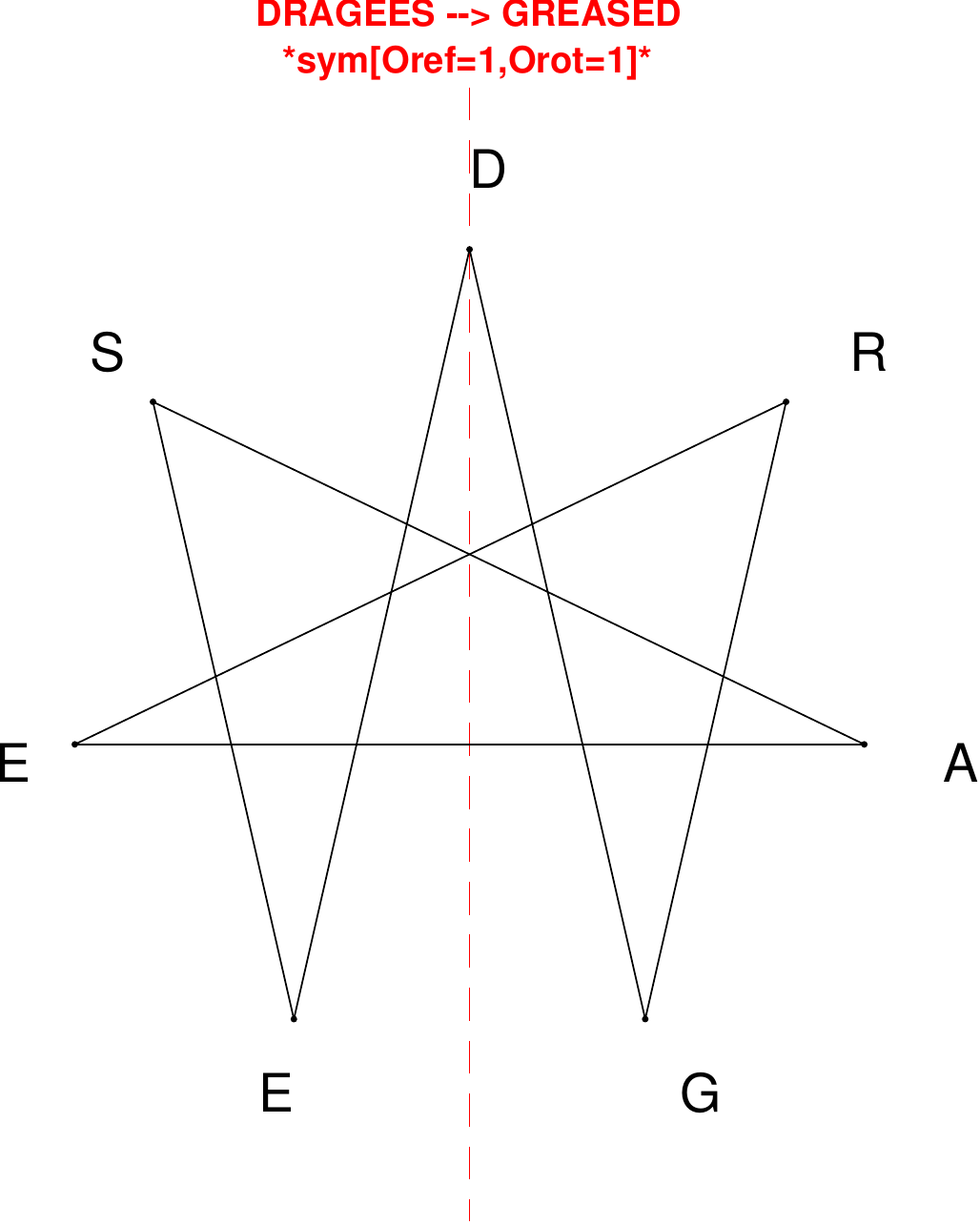}
\end{subfigure}
\end{figure}

\begin{figure}[H]
\centering
\begin{subfigure}[T]{0.19\textwidth}
\centering
\includegraphics[width=\textwidth]{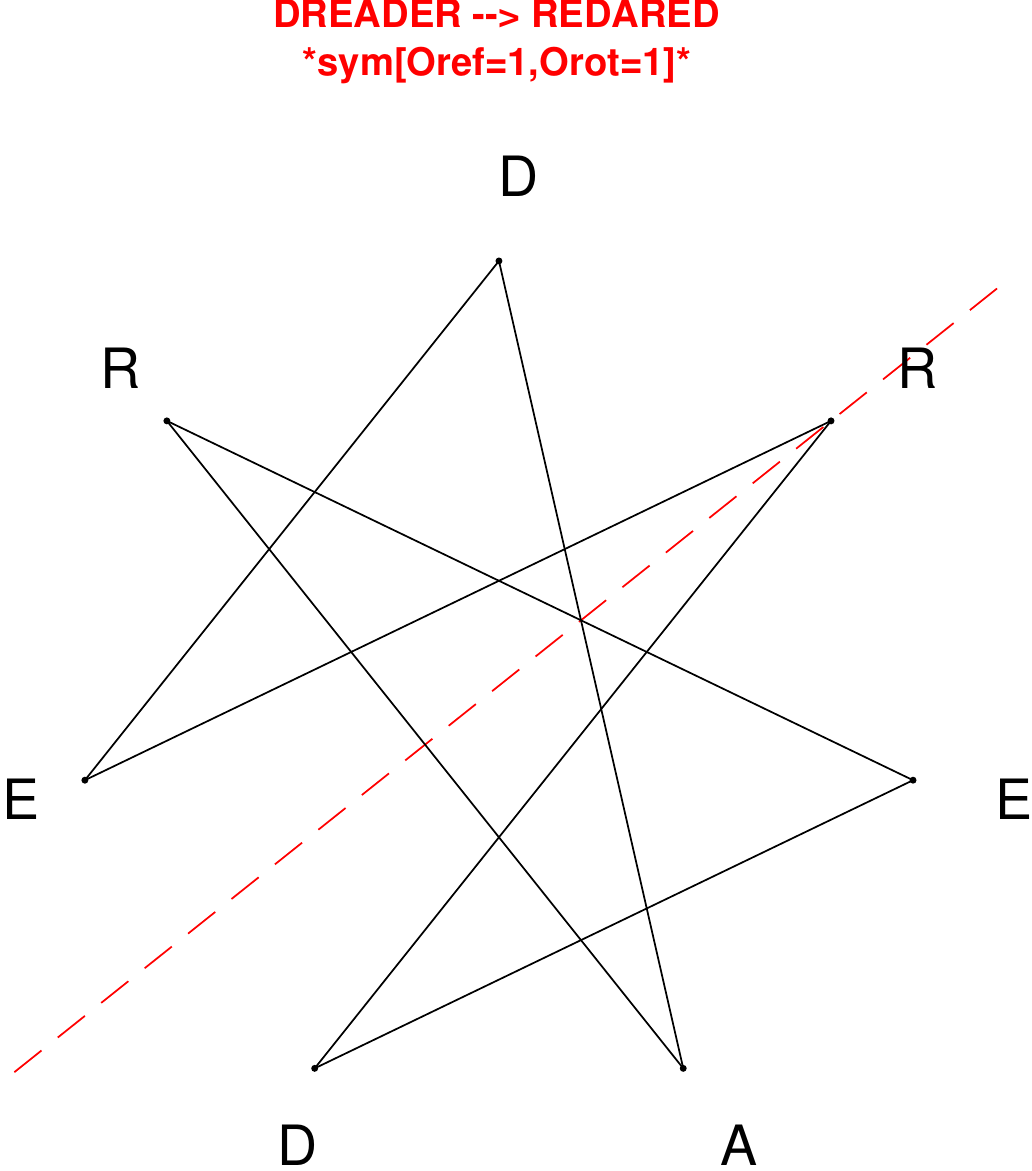}
\end{subfigure}
\hfill
\begin{subfigure}[T]{0.19\textwidth}
\centering
\includegraphics[width=\textwidth]{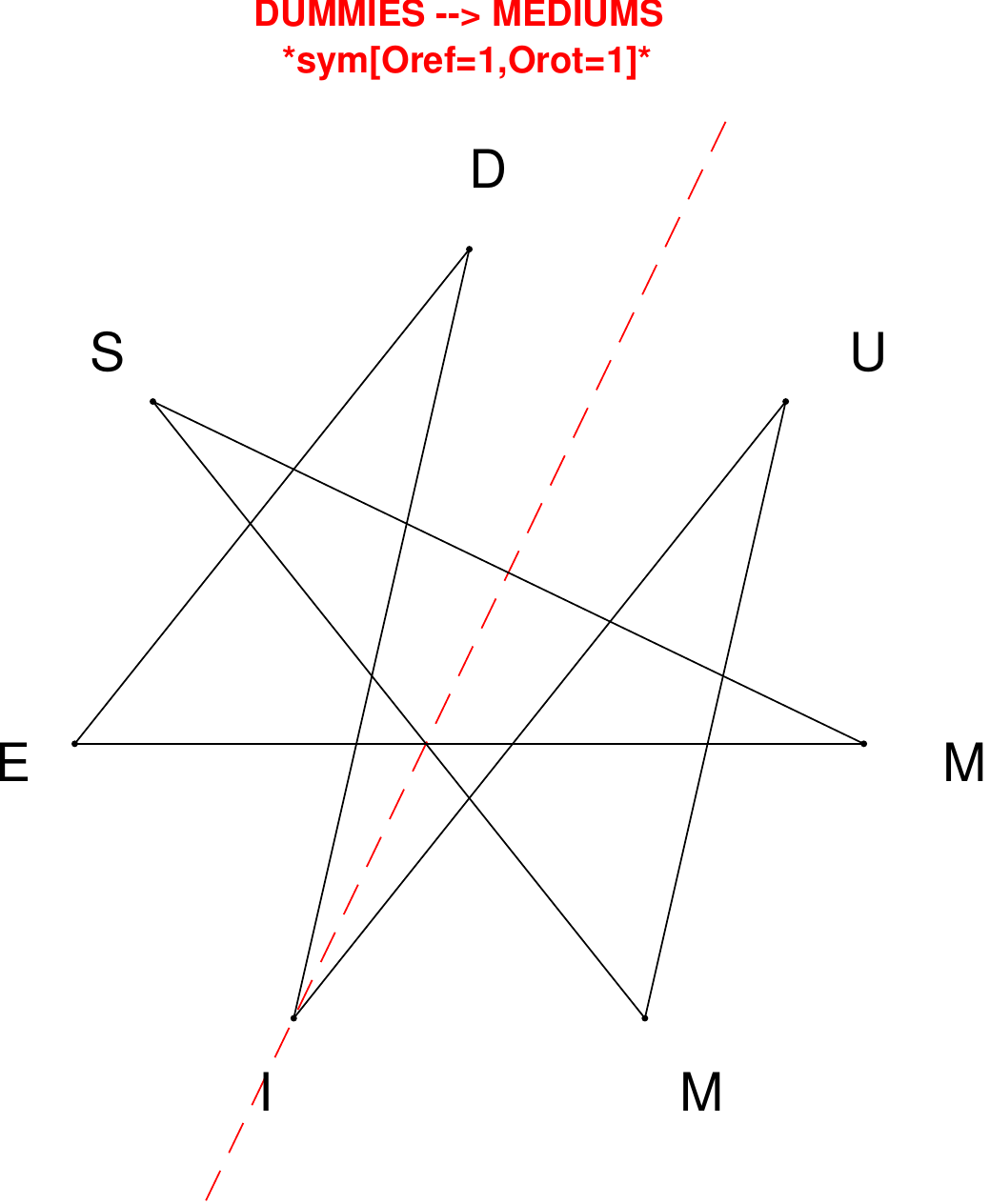}
\end{subfigure}
\hfill
\begin{subfigure}[T]{0.19\textwidth}
\centering
\includegraphics[width=\textwidth]{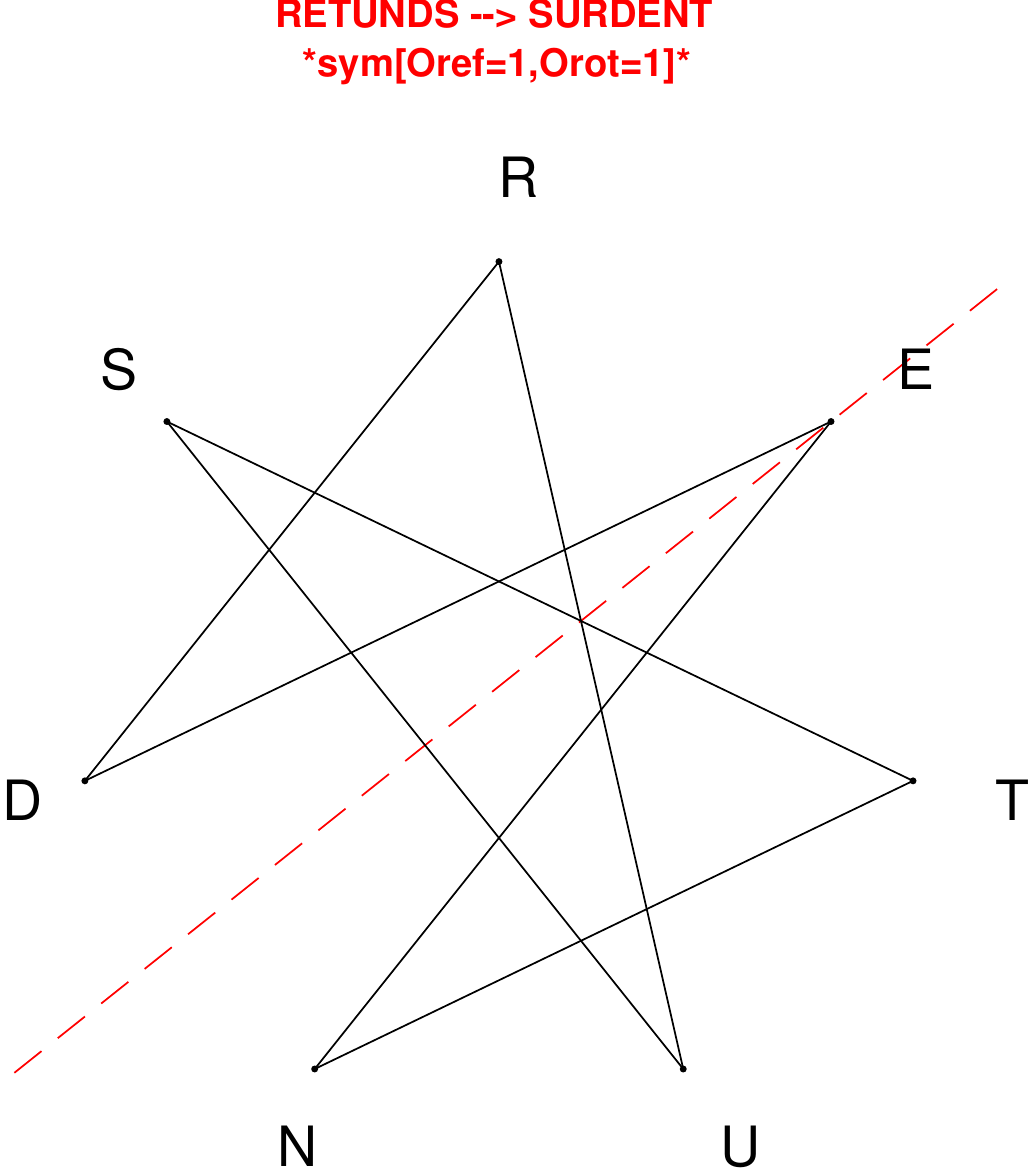}
\end{subfigure}
\hfill
\begin{subfigure}[T]{0.19\textwidth}
\centering
\includegraphics[width=\textwidth]{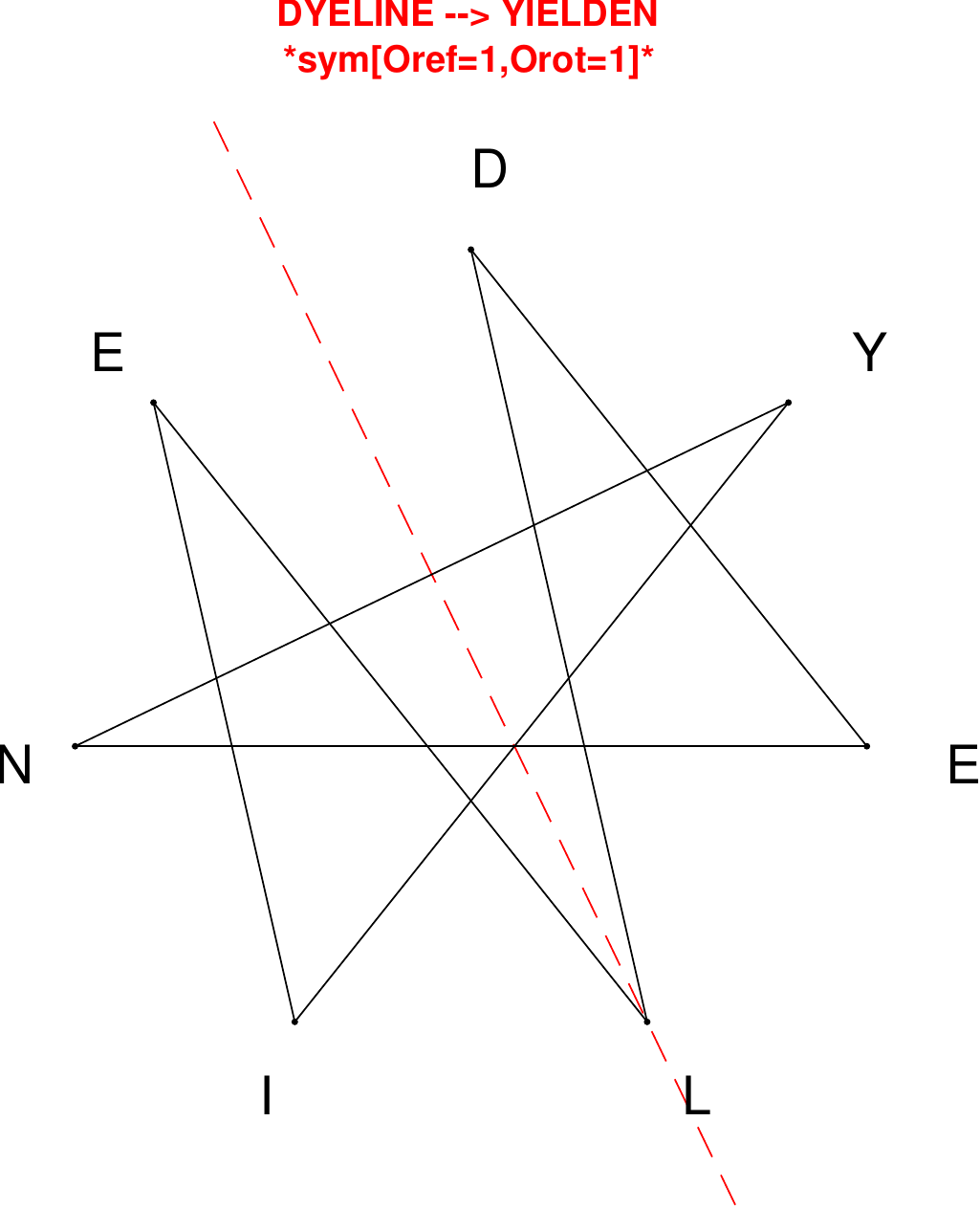}
\end{subfigure}
\hfill
\begin{subfigure}[T]{0.19\textwidth}
\centering
\includegraphics[width=\textwidth]{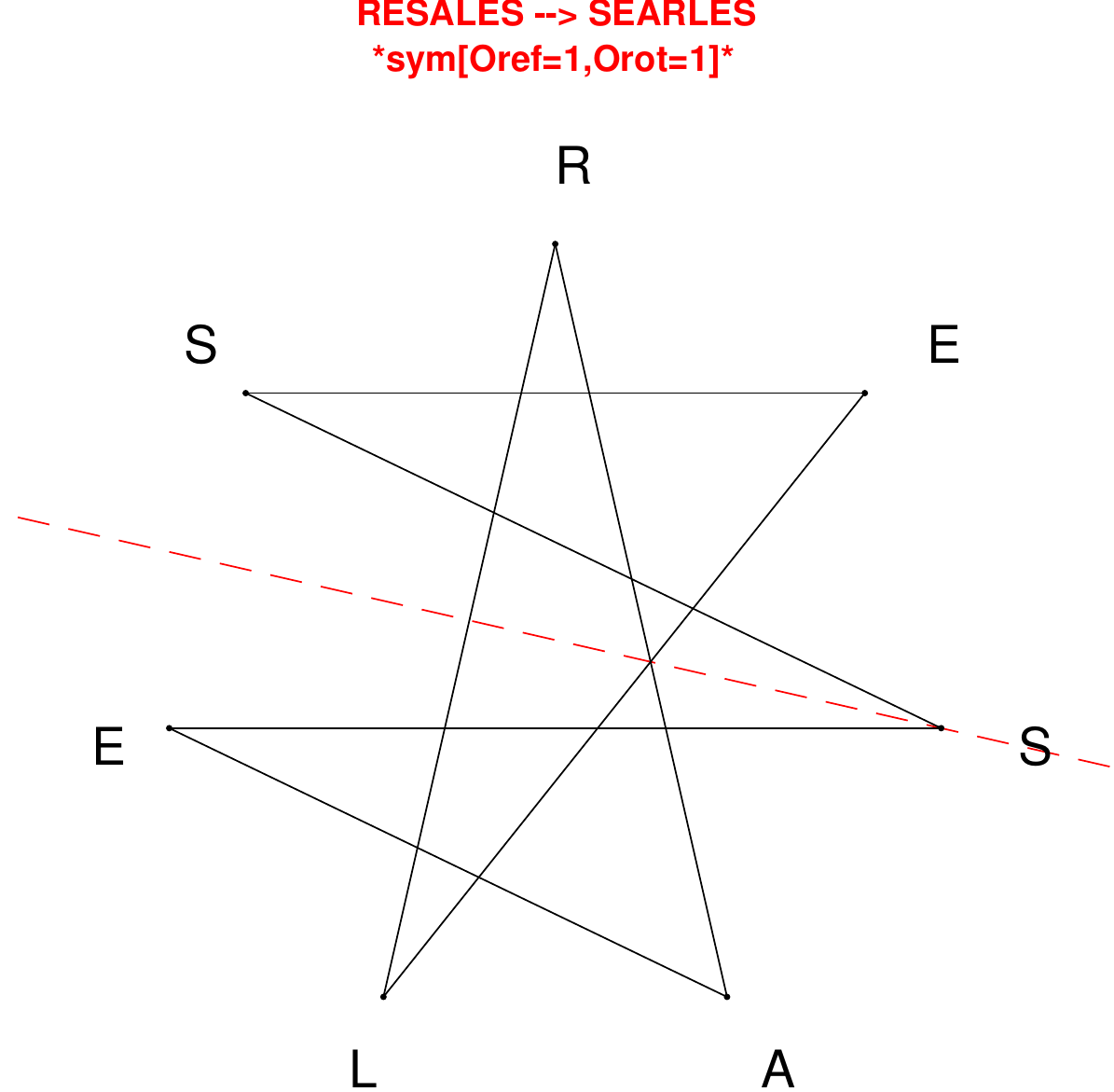}
\end{subfigure}
\end{figure}

\begin{figure}[H]
\centering
\begin{subfigure}[T]{0.19\textwidth}
\centering
\includegraphics[width=\textwidth]{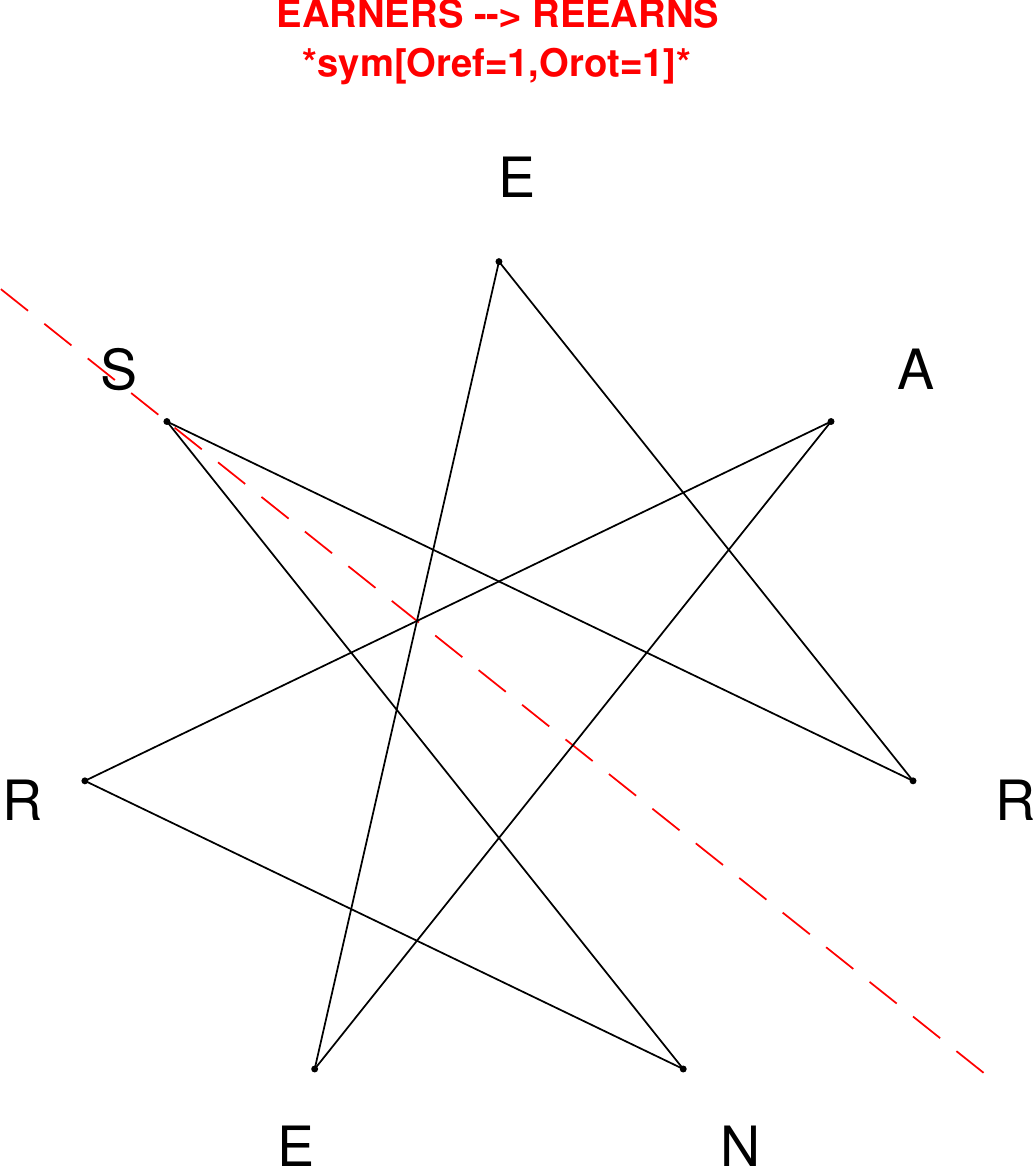}
\end{subfigure}
\hfill
\begin{subfigure}[T]{0.19\textwidth}
\centering
\includegraphics[width=\textwidth]{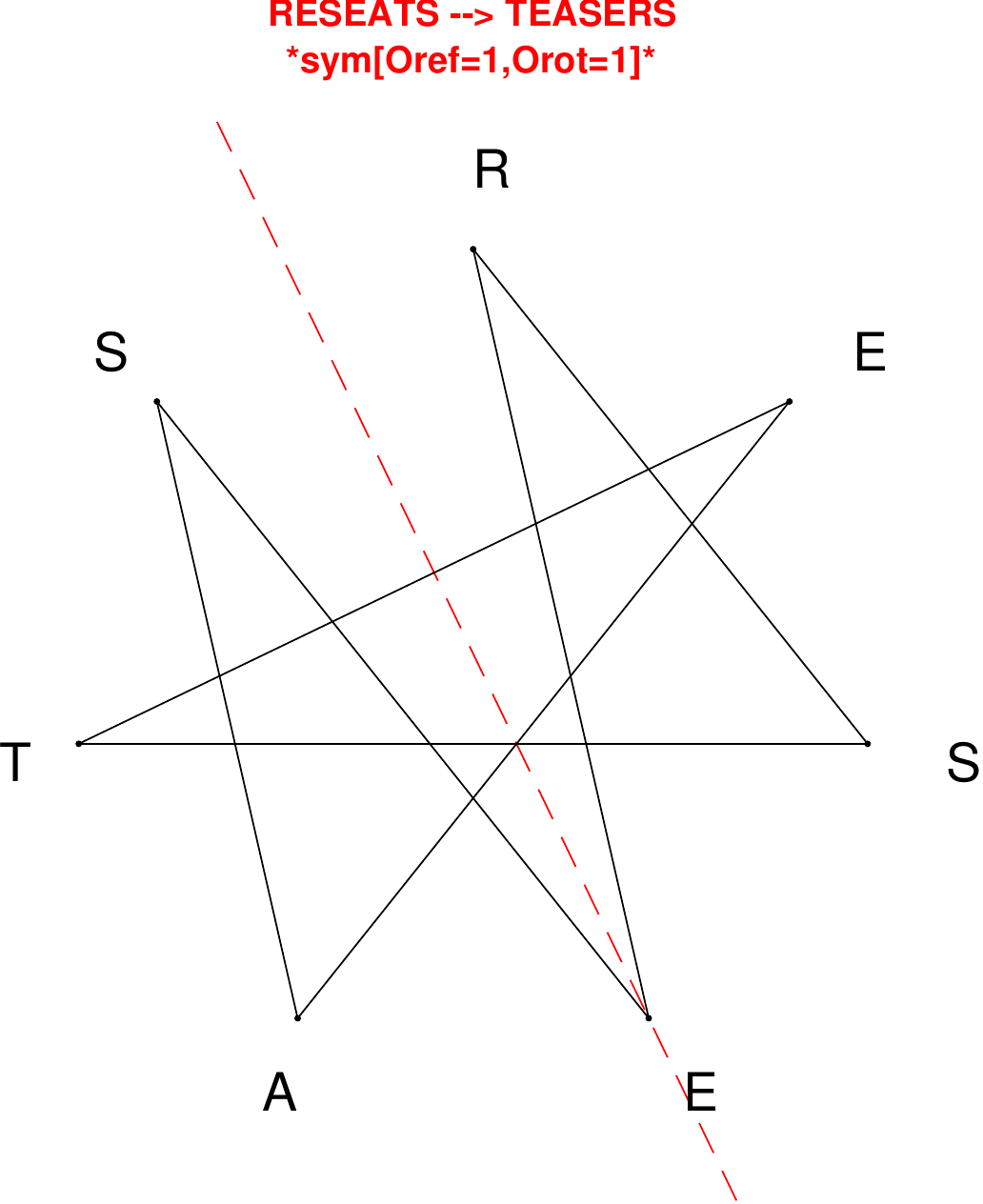}
\end{subfigure}
\hfill
\begin{subfigure}[T]{0.19\textwidth}
\centering
\includegraphics[width=\textwidth]{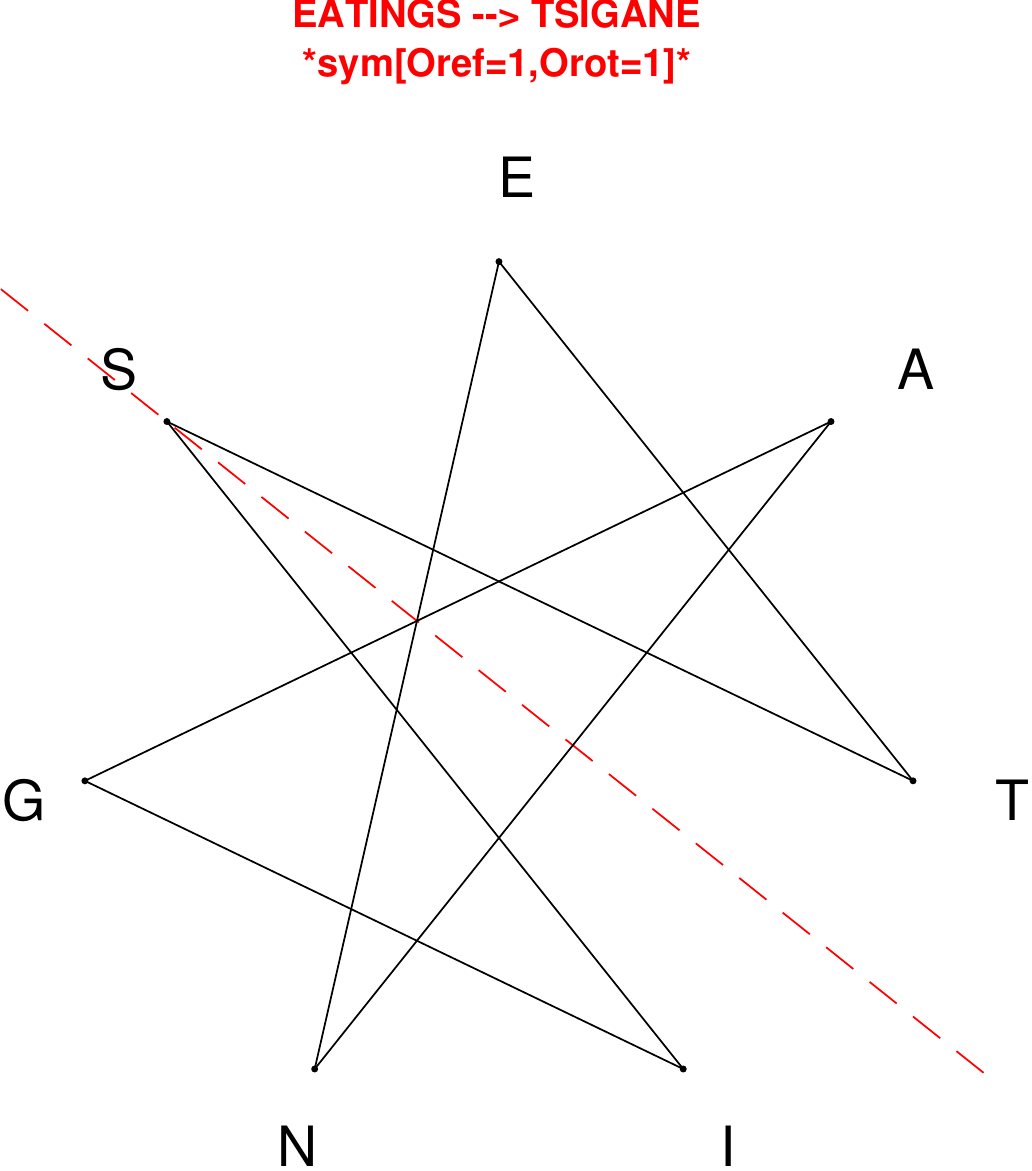}
\end{subfigure}
\hfill
\begin{subfigure}[T]{0.19\textwidth}
\centering
\includegraphics[width=\textwidth]{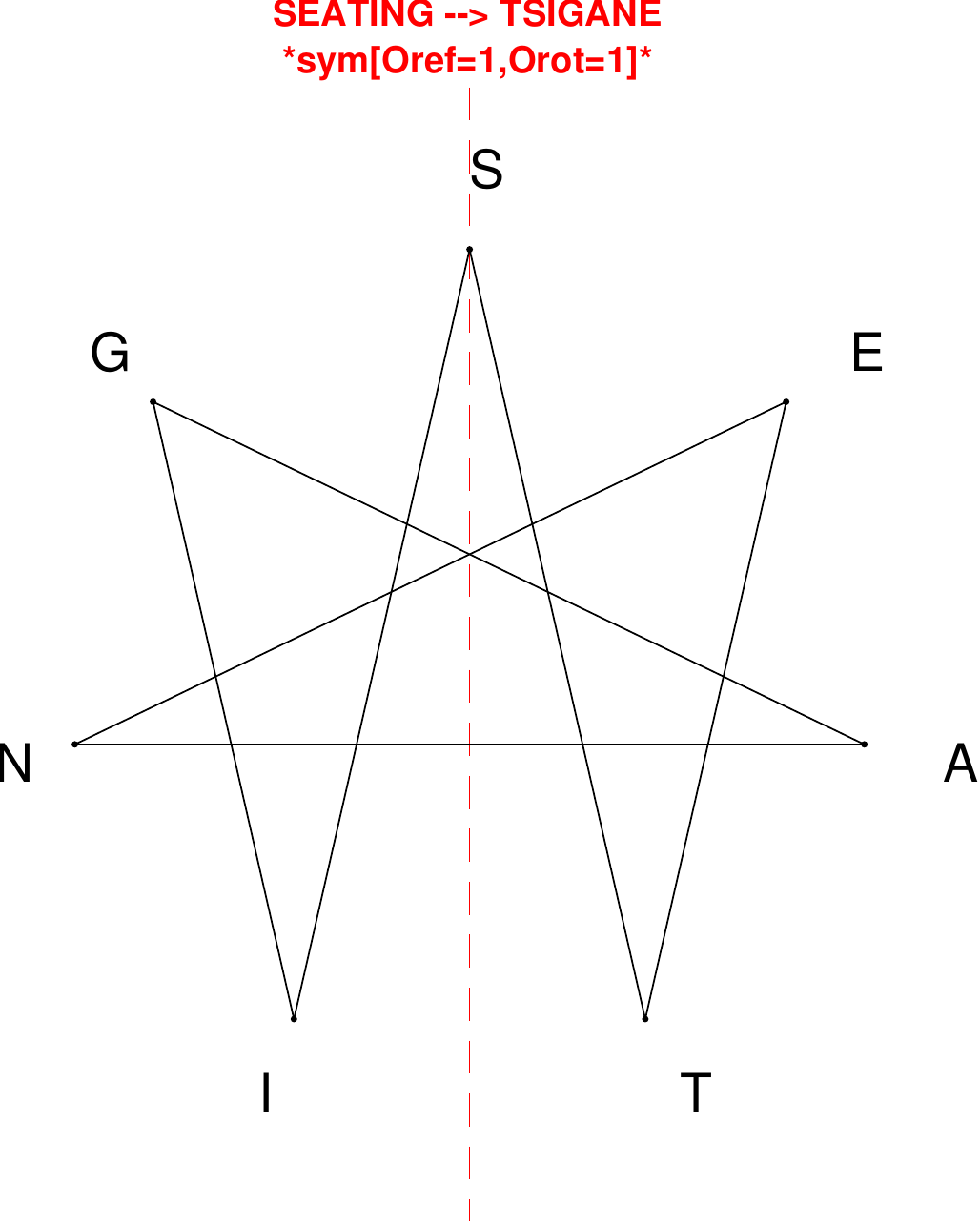}
\end{subfigure}
\hfill
\begin{subfigure}[T]{0.19\textwidth}
\centering
\includegraphics[width=\textwidth]{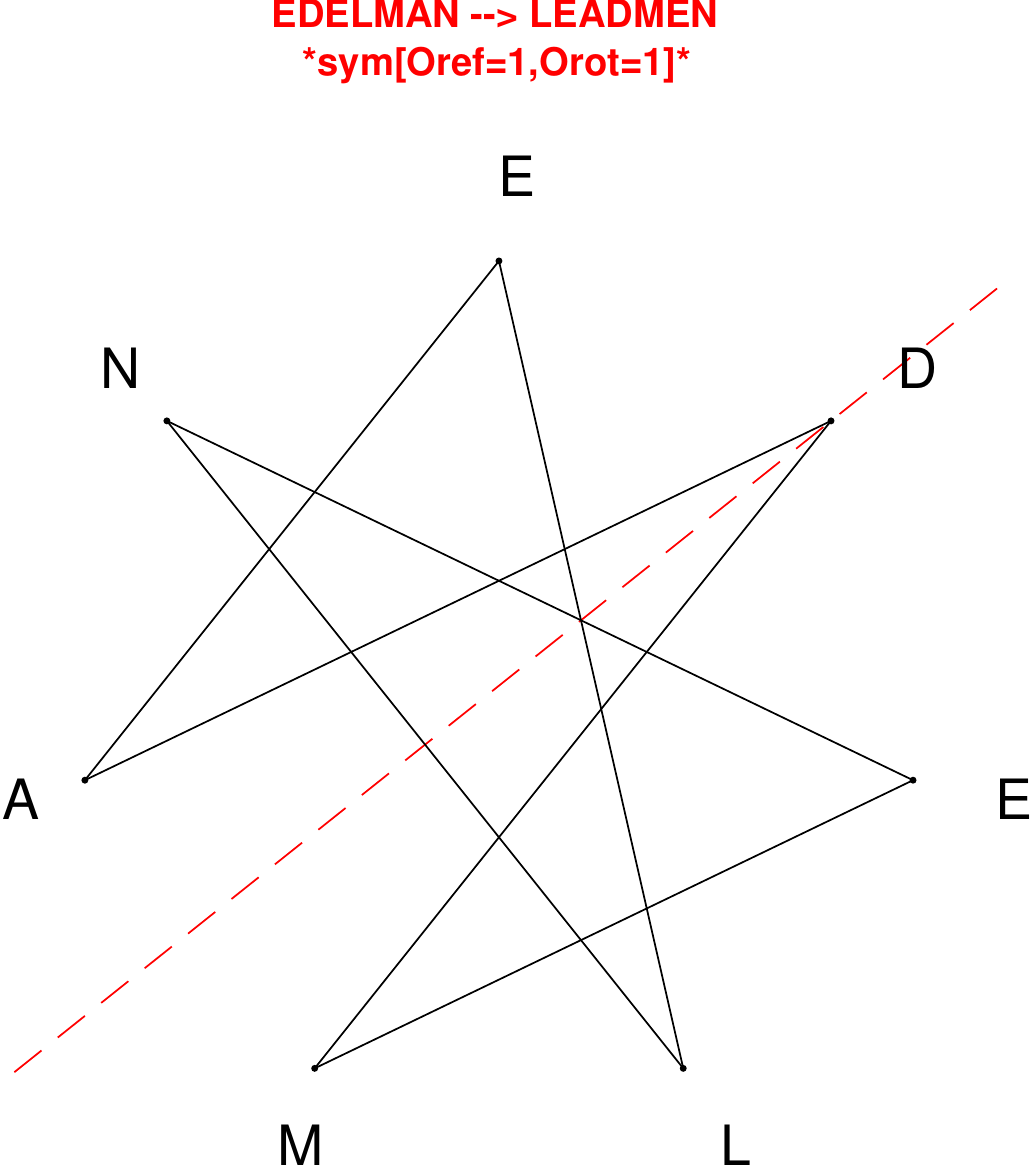}
\end{subfigure}
\end{figure}

\begin{figure}[H]
\centering
\begin{subfigure}[T]{0.19\textwidth}
\centering
\includegraphics[width=\textwidth]{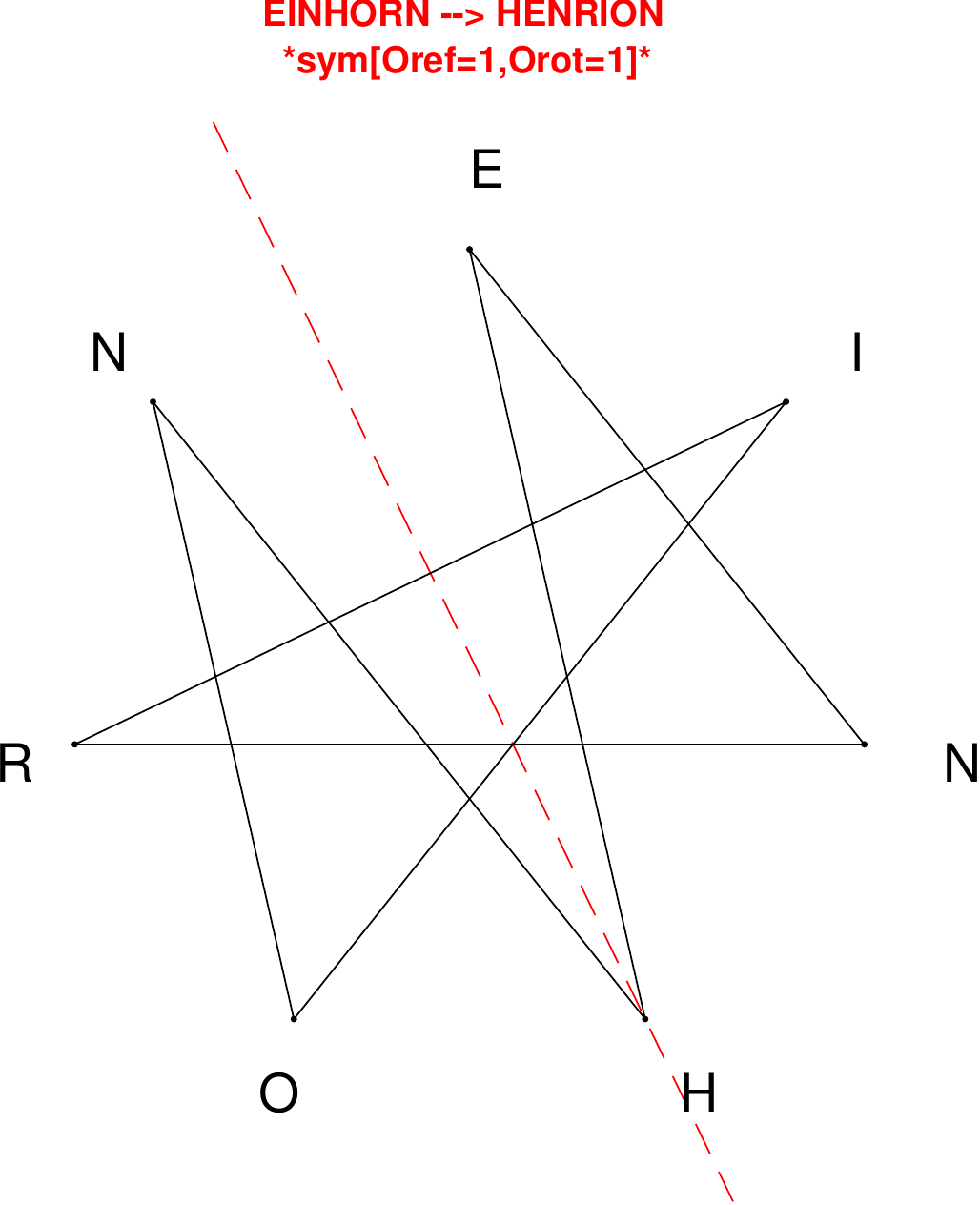}
\end{subfigure}
\hfill
\begin{subfigure}[T]{0.19\textwidth}
\centering
\includegraphics[width=\textwidth]{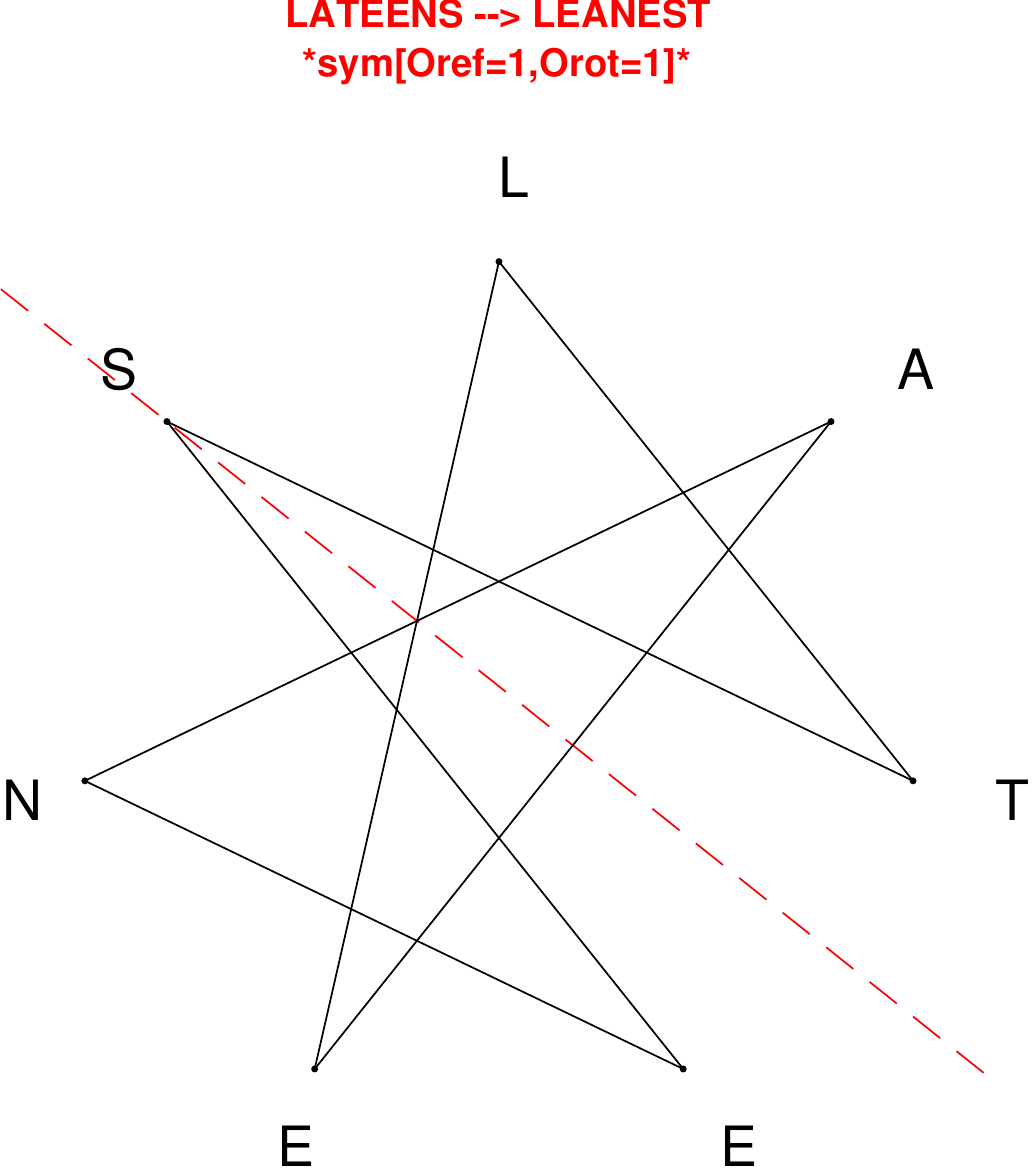}
\end{subfigure}
\hfill
\begin{subfigure}[T]{0.19\textwidth}
\centering
\includegraphics[width=\textwidth]{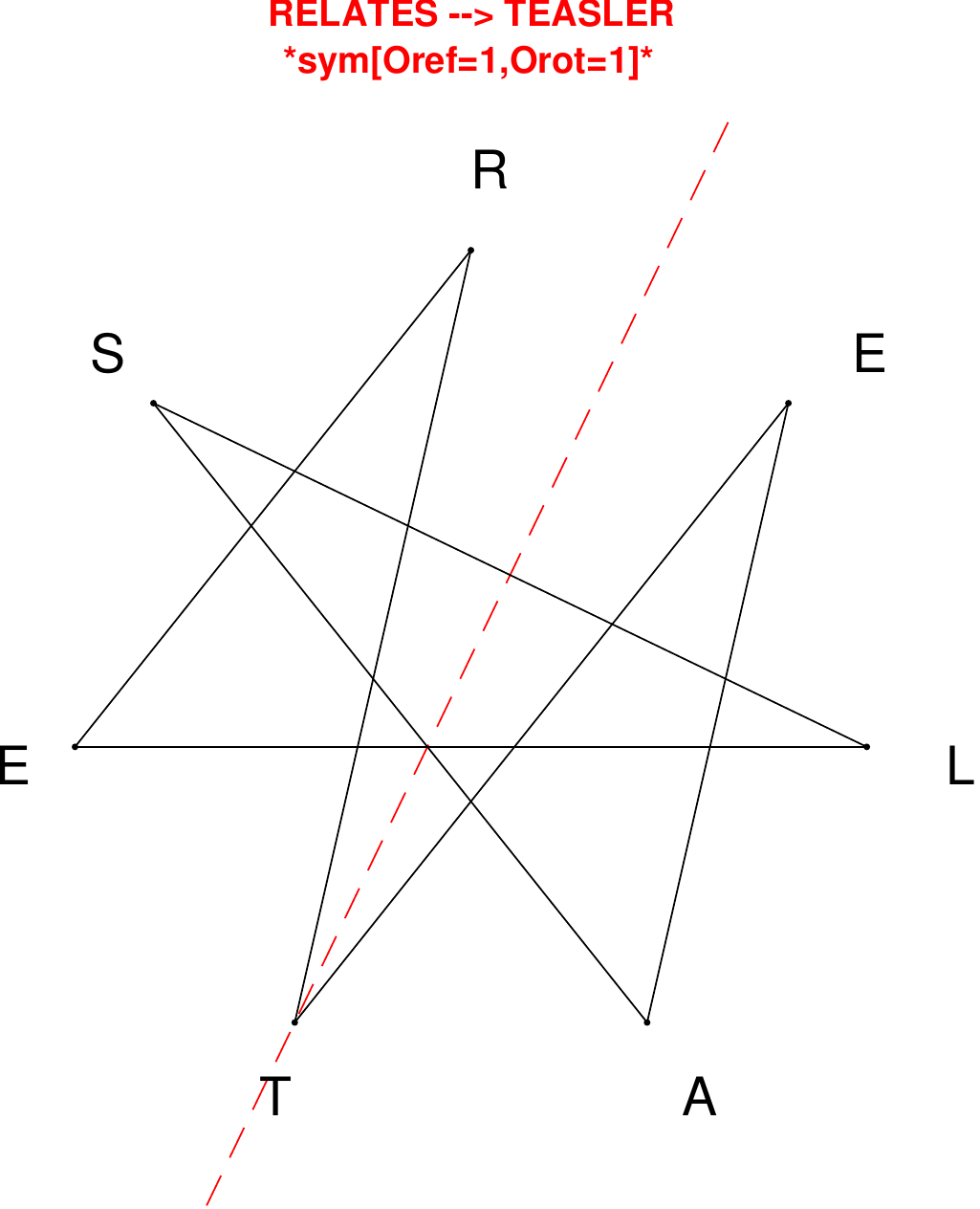}
\end{subfigure}
\hfill
\begin{subfigure}[T]{0.19\textwidth}
\centering
\includegraphics[width=\textwidth]{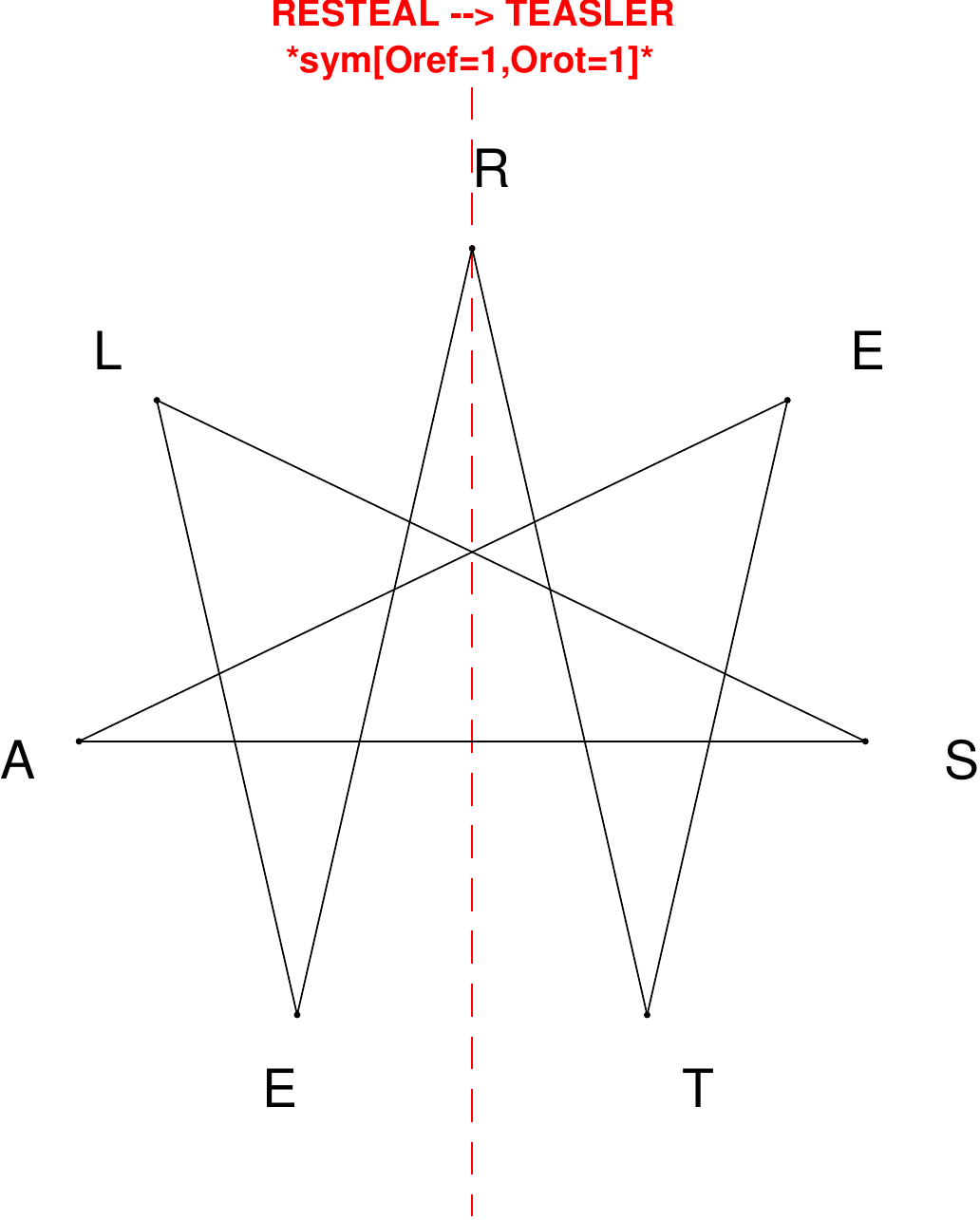}
\end{subfigure}
\hfill
\begin{subfigure}[T]{0.19\textwidth}
\centering
\includegraphics[width=\textwidth]{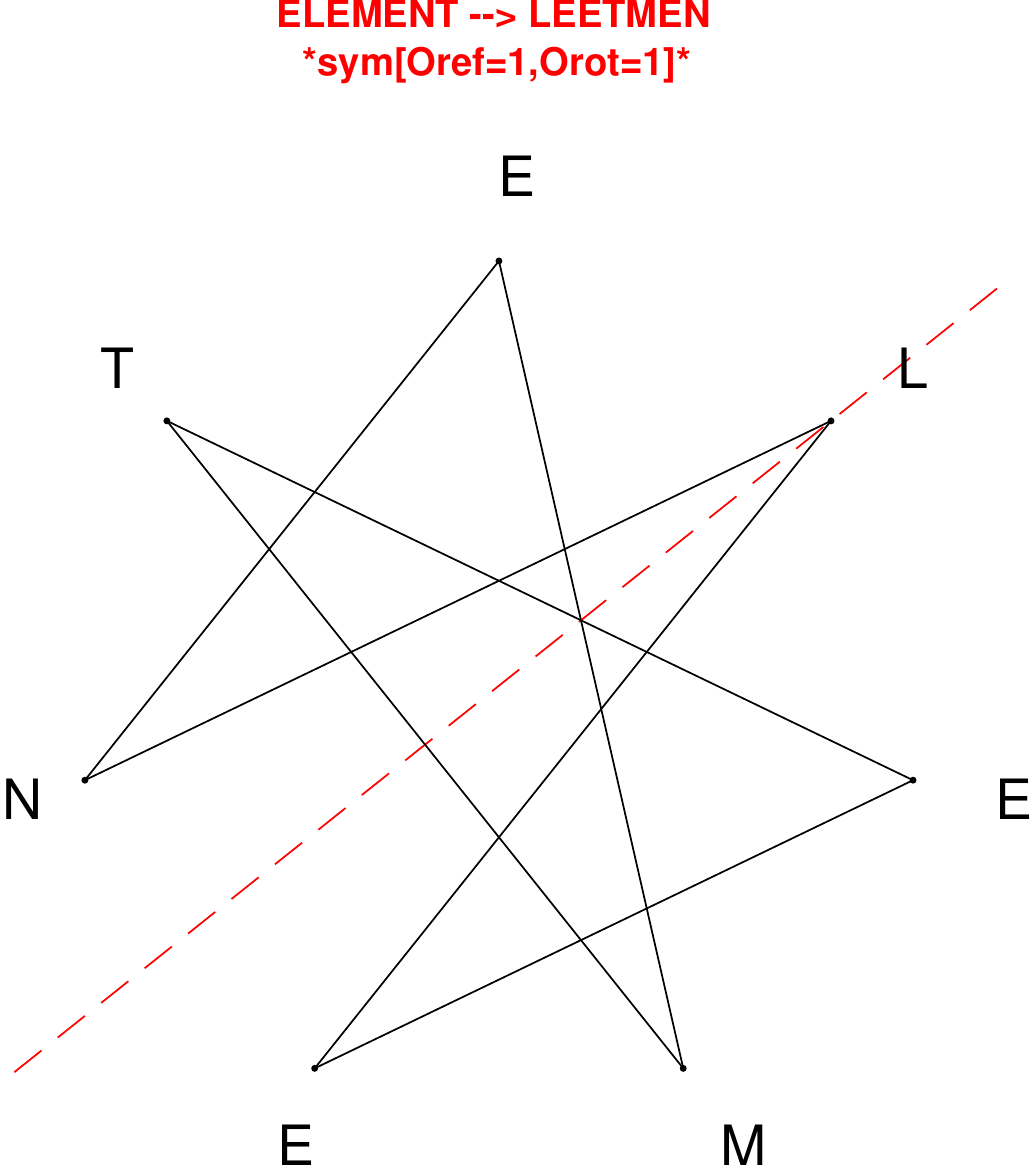}
\end{subfigure}
\end{figure}

\begin{figure}[H]
\centering
\begin{subfigure}[T]{0.19\textwidth}
\centering
\includegraphics[width=\textwidth]{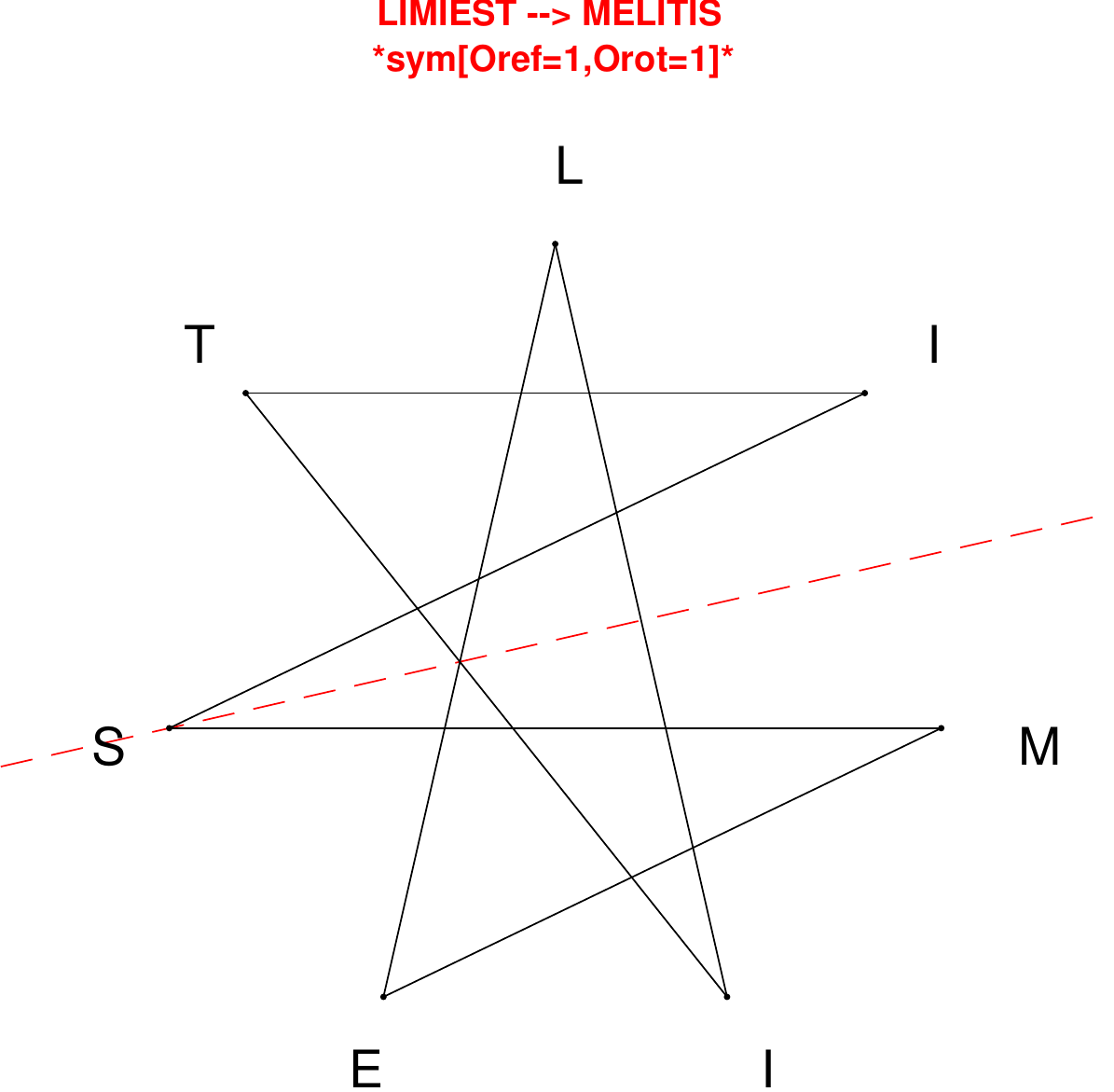}
\end{subfigure}
\hfill
\begin{subfigure}[T]{0.19\textwidth}
\centering
\includegraphics[width=\textwidth]{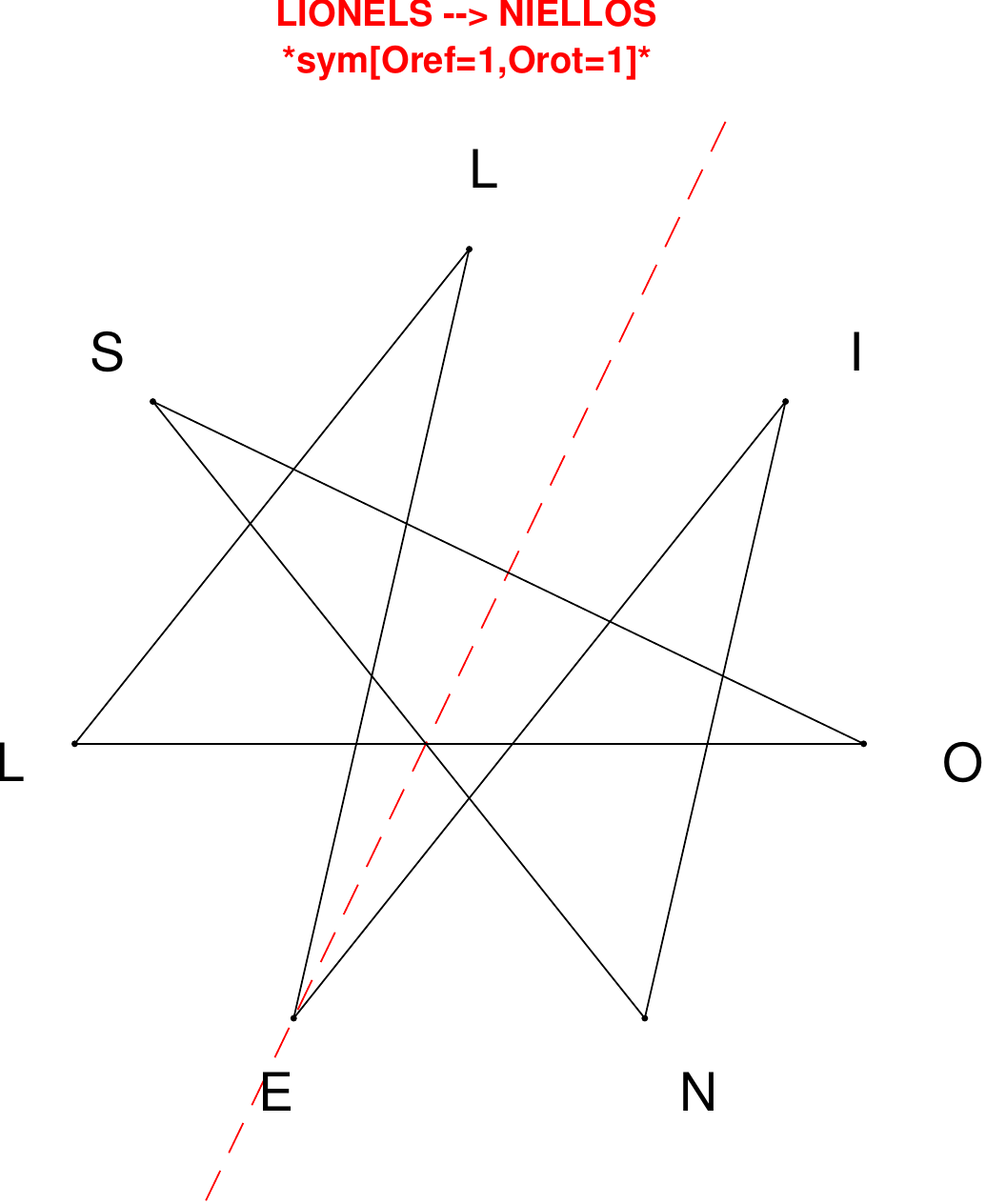}
\end{subfigure}
\hfill
\begin{subfigure}[T]{0.19\textwidth}
\centering
\includegraphics[width=\textwidth]{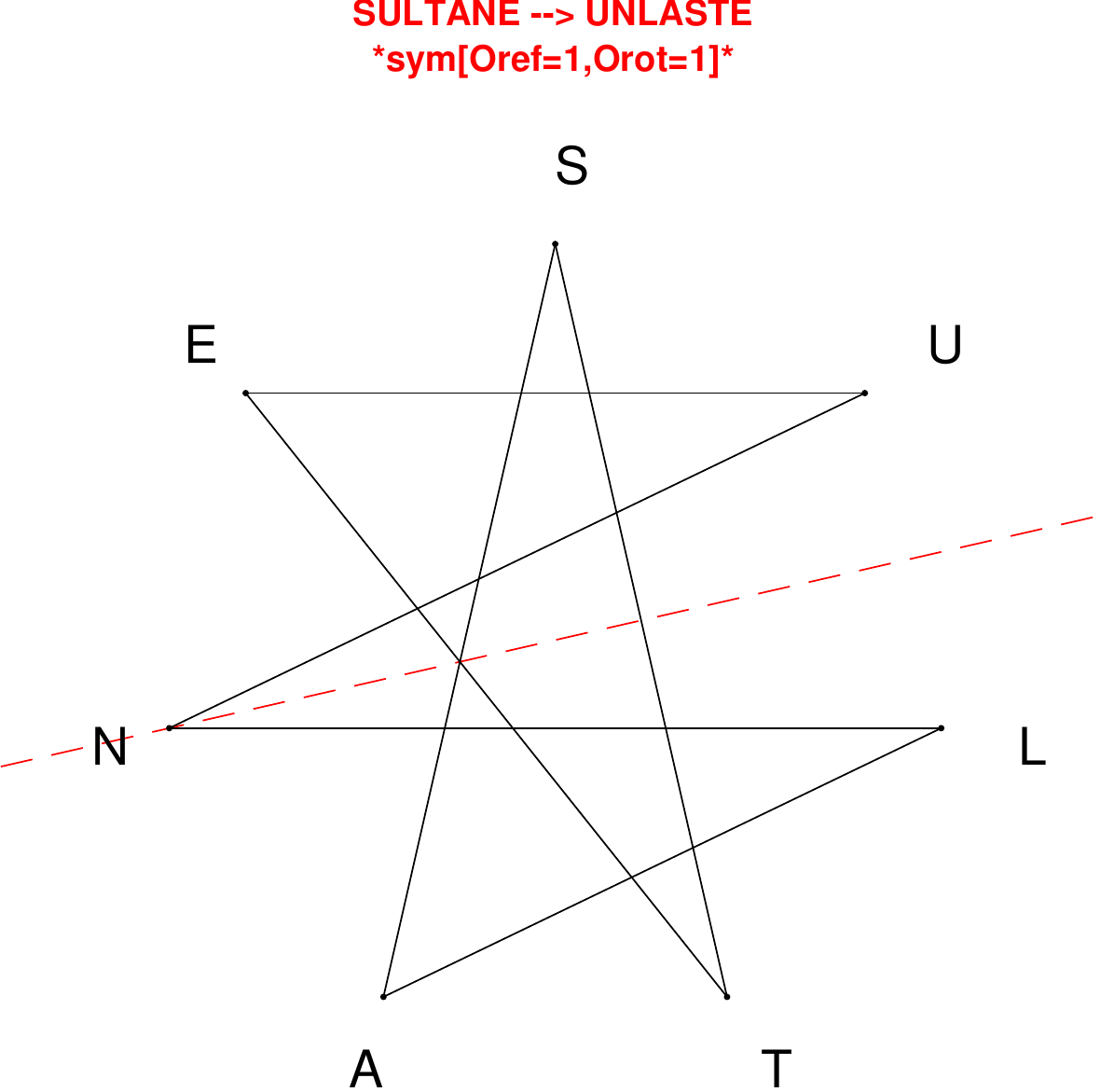}
\end{subfigure}
\hfill
\begin{subfigure}[T]{0.19\textwidth}
\centering
\includegraphics[width=\textwidth]{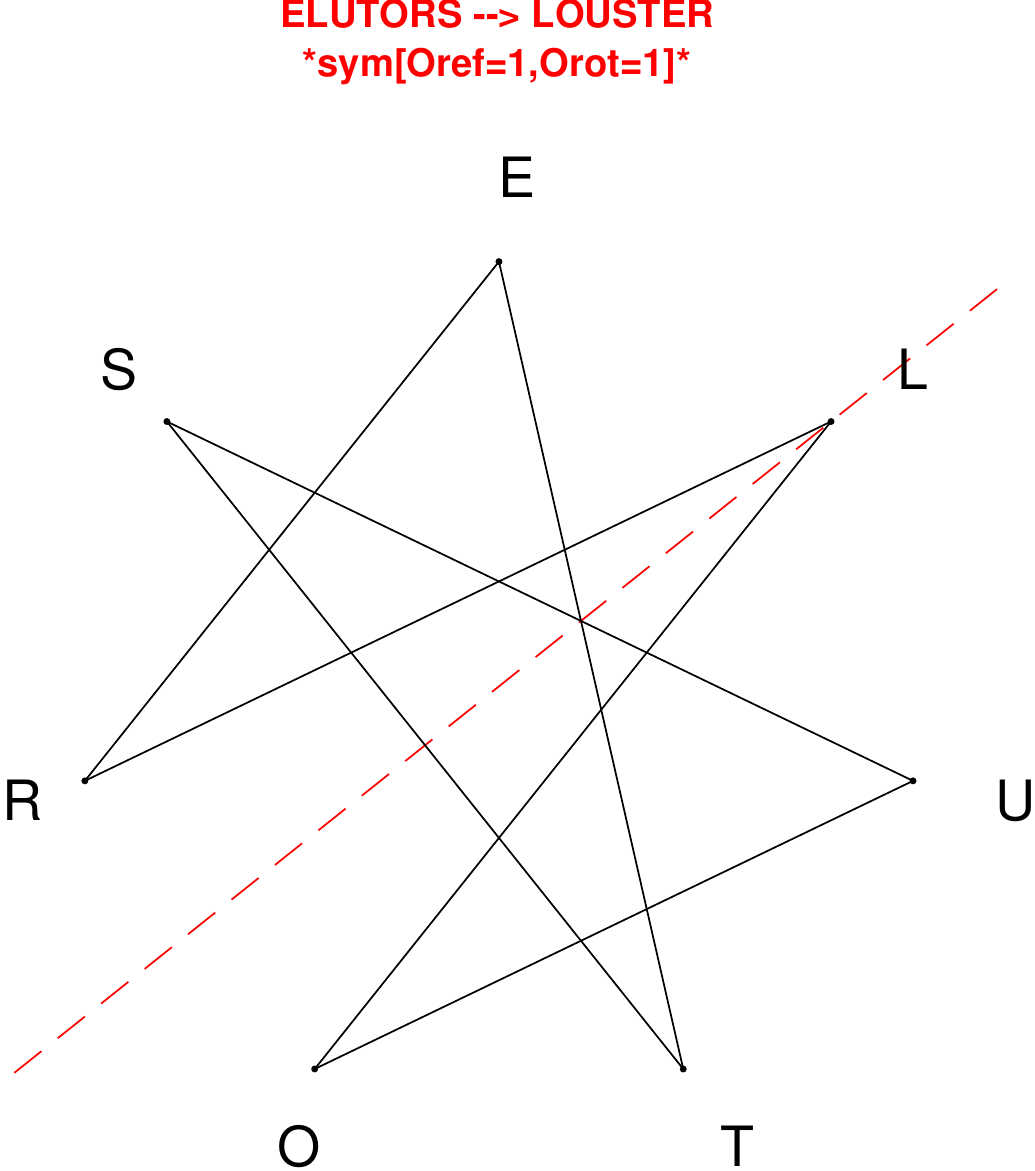}
\end{subfigure}
\hfill
\begin{subfigure}[T]{0.19\textwidth}
\centering
\includegraphics[width=\textwidth]{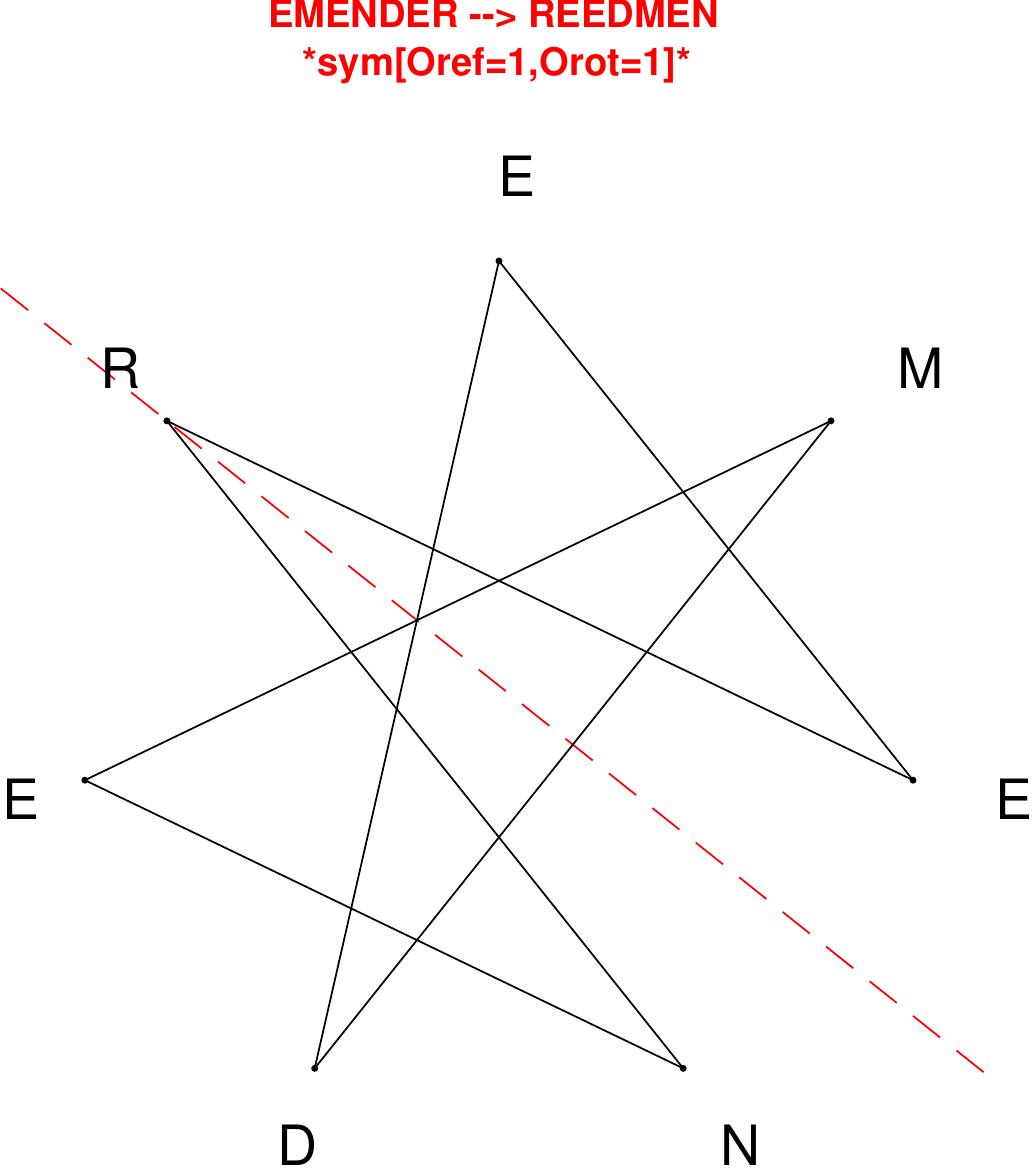}
\end{subfigure}
\end{figure}

\begin{figure}[H]
\centering
\begin{subfigure}[T]{0.19\textwidth}
\centering
\includegraphics[width=\textwidth]{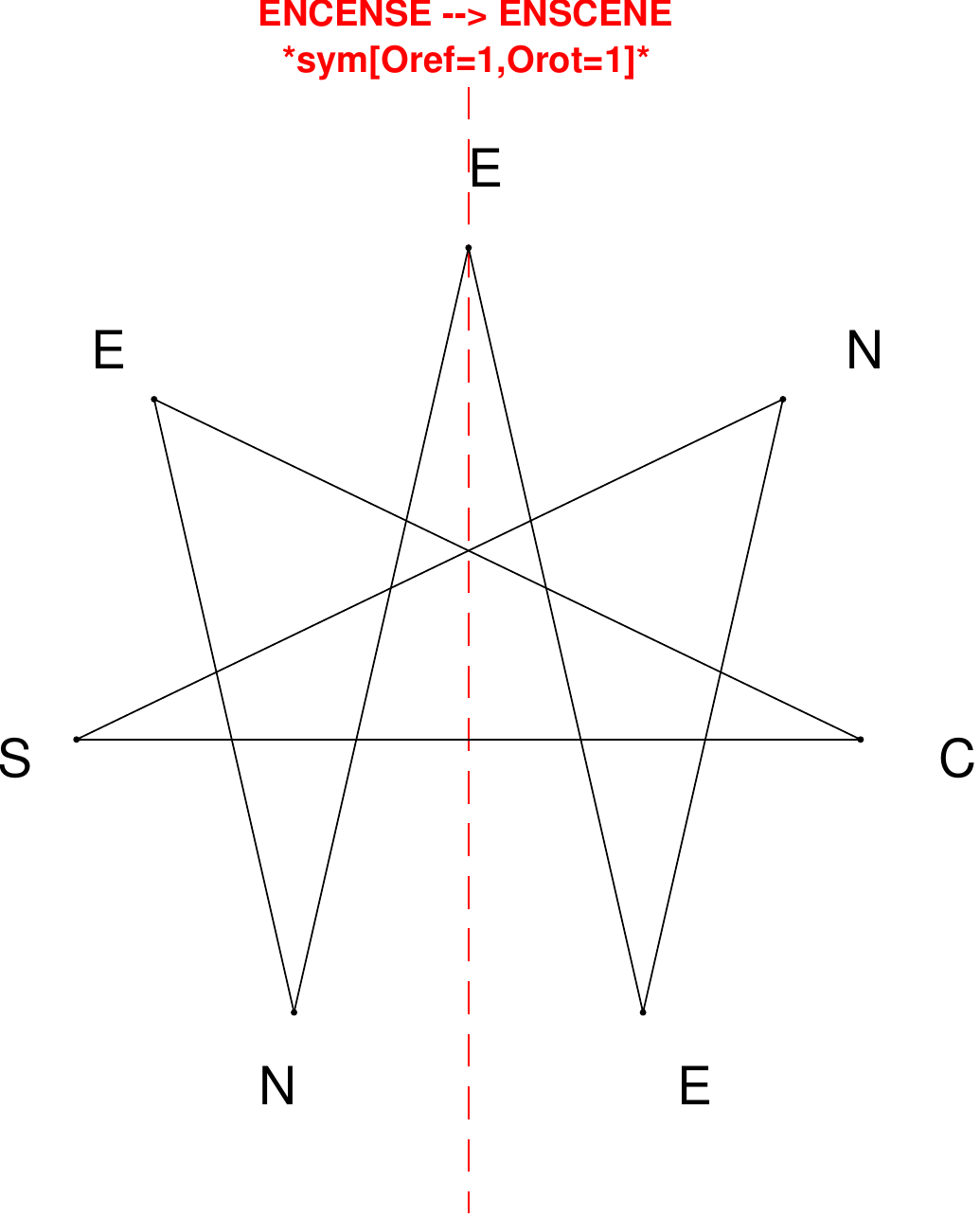}
\end{subfigure}
\hfill
\begin{subfigure}[T]{0.19\textwidth}
\centering
\includegraphics[width=\textwidth]{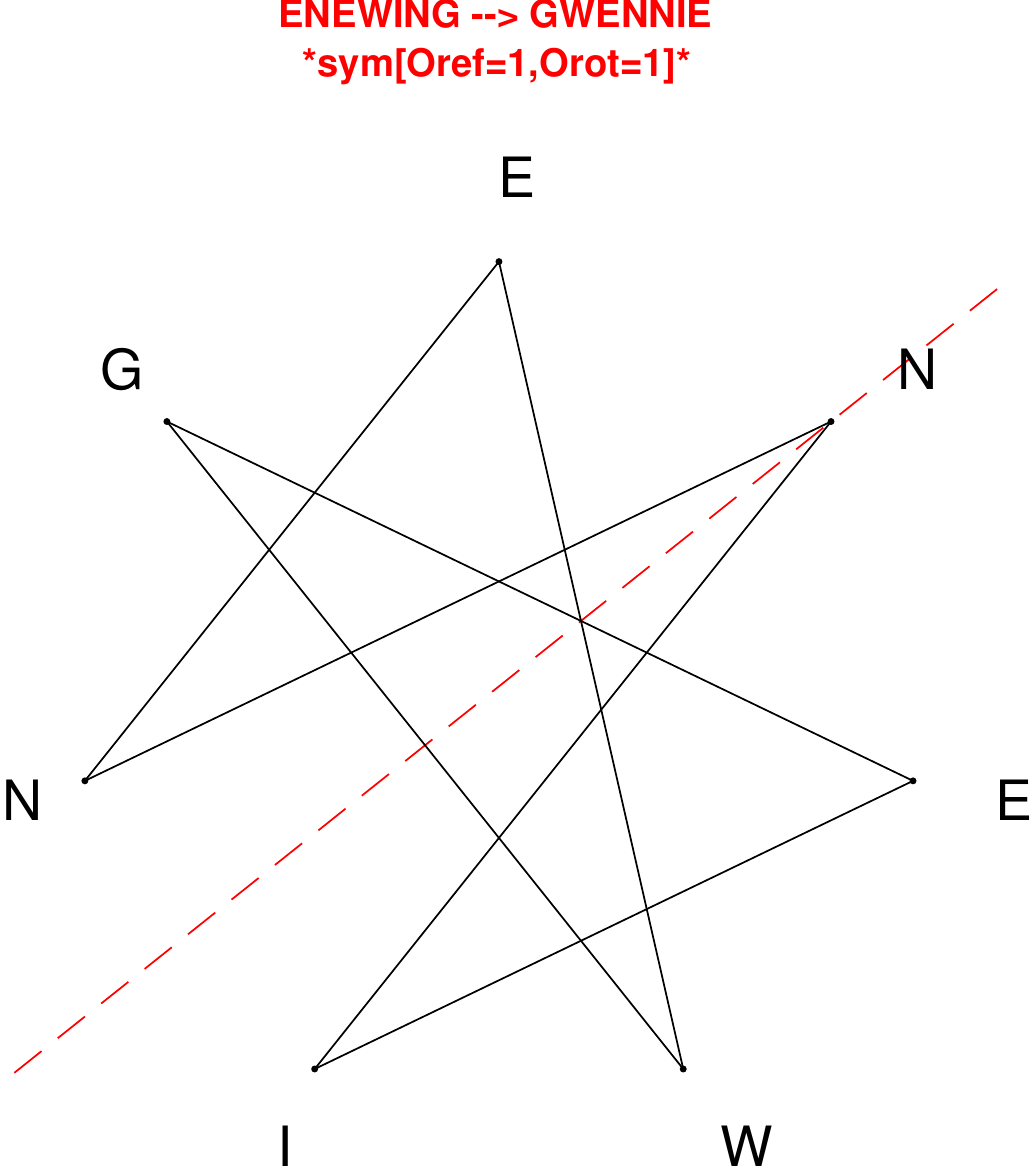}
\end{subfigure}
\hfill
\begin{subfigure}[T]{0.19\textwidth}
\centering
\includegraphics[width=\textwidth]{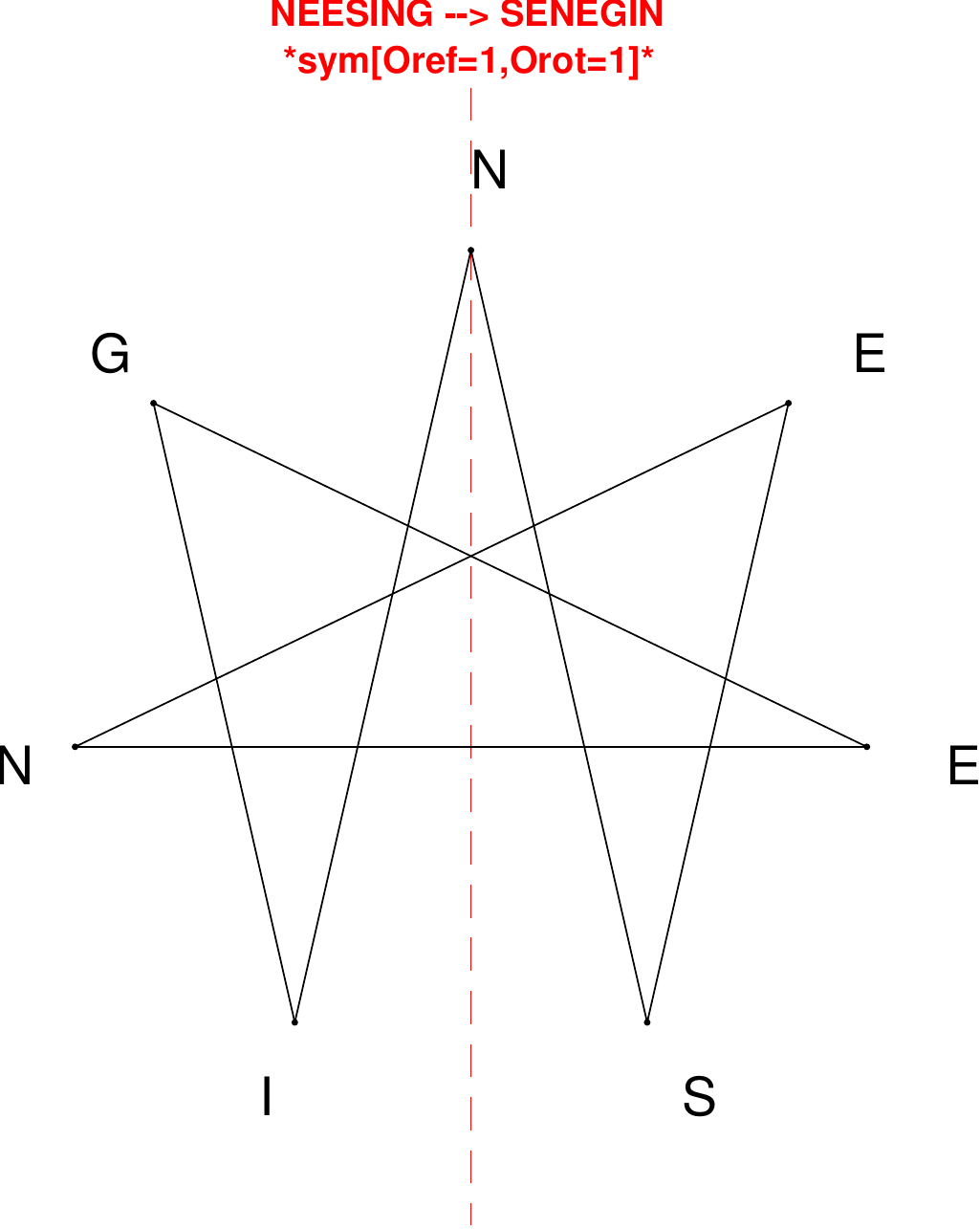}
\end{subfigure}
\hfill
\begin{subfigure}[T]{0.19\textwidth}
\centering
\includegraphics[width=\textwidth]{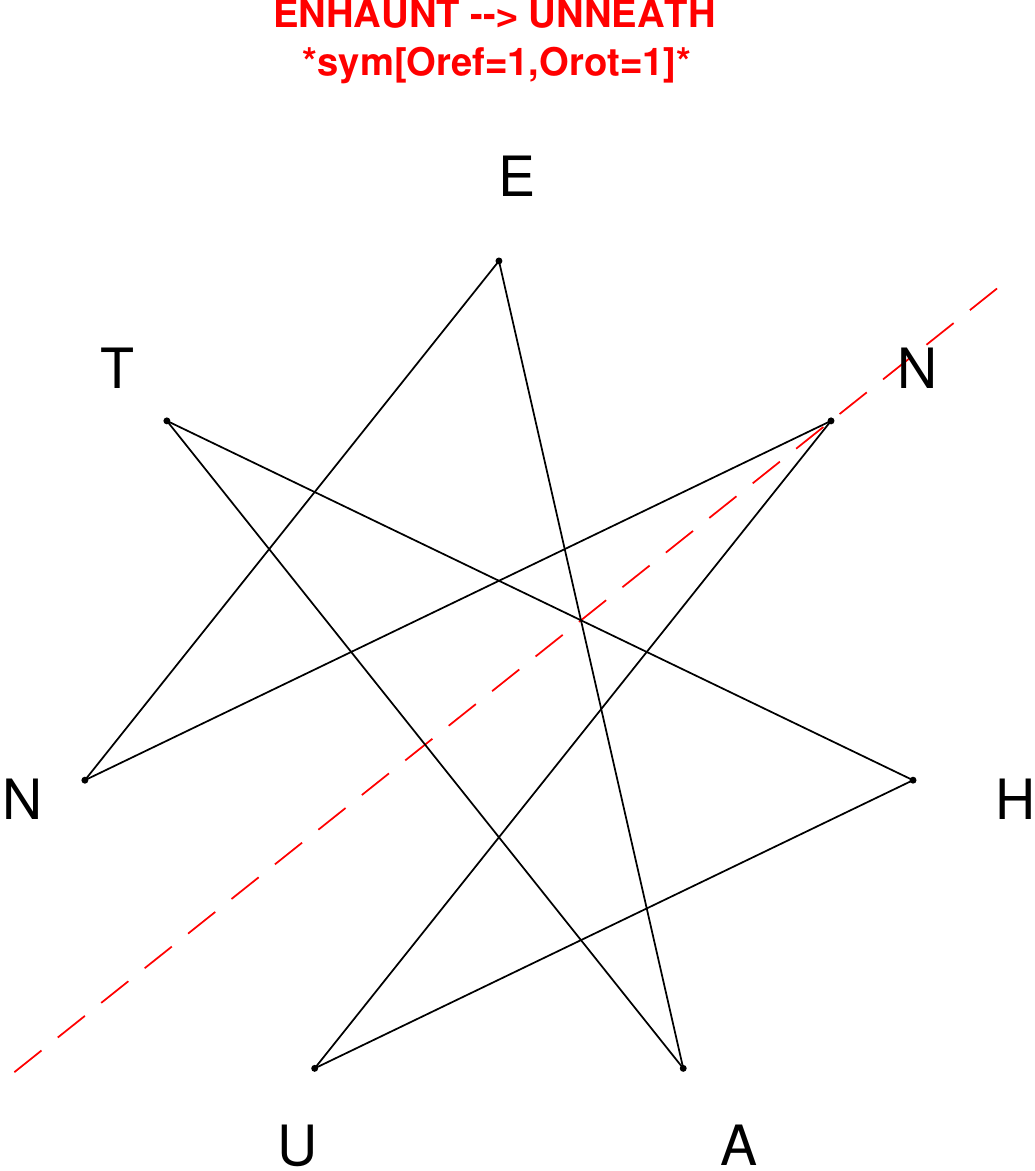}
\end{subfigure}
\hfill
\begin{subfigure}[T]{0.19\textwidth}
\centering
\includegraphics[width=\textwidth]{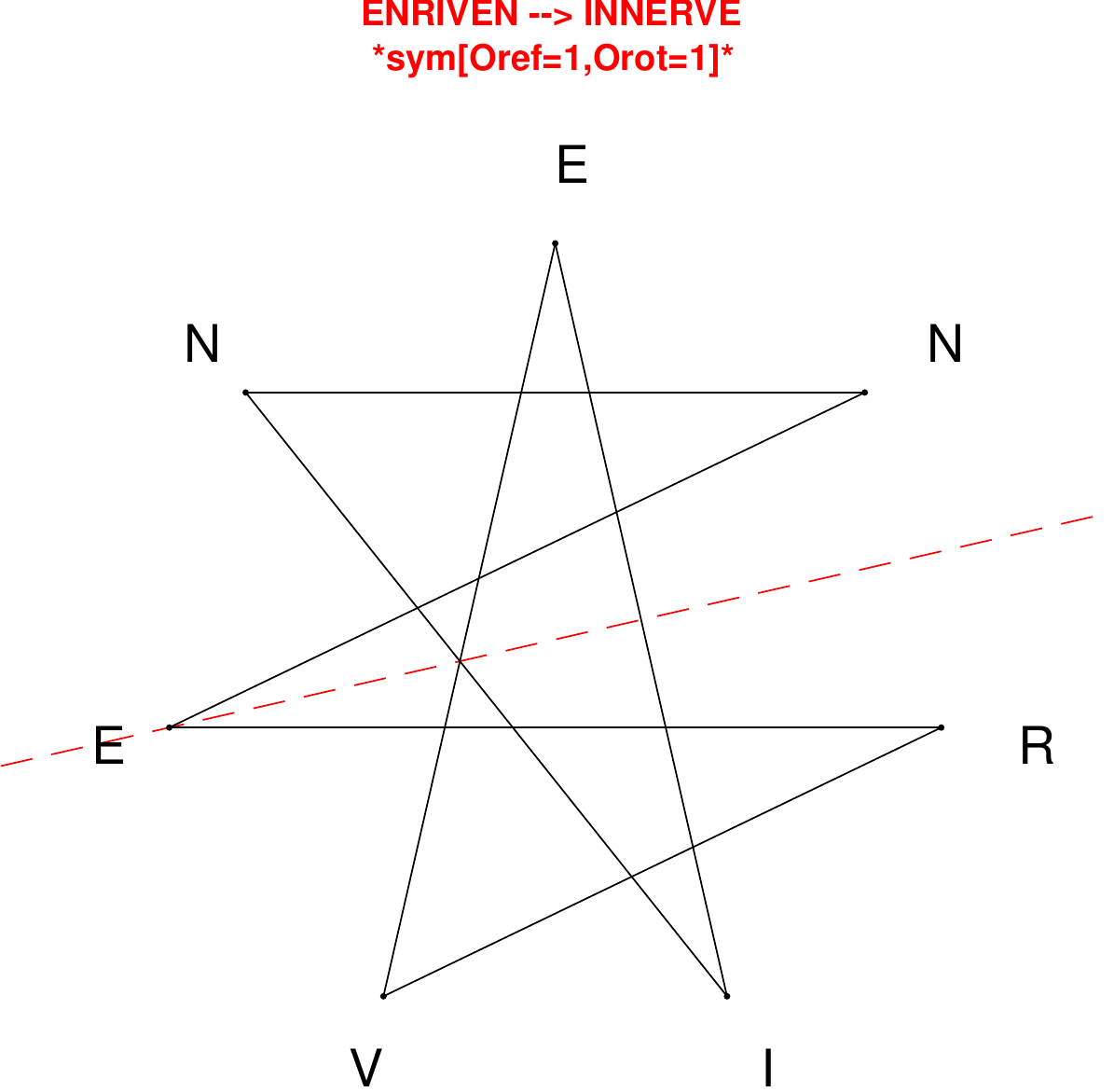}
\end{subfigure}
\end{figure}

\begin{figure}[H]
\centering
\begin{subfigure}[T]{0.19\textwidth}
\centering
\includegraphics[width=\textwidth]{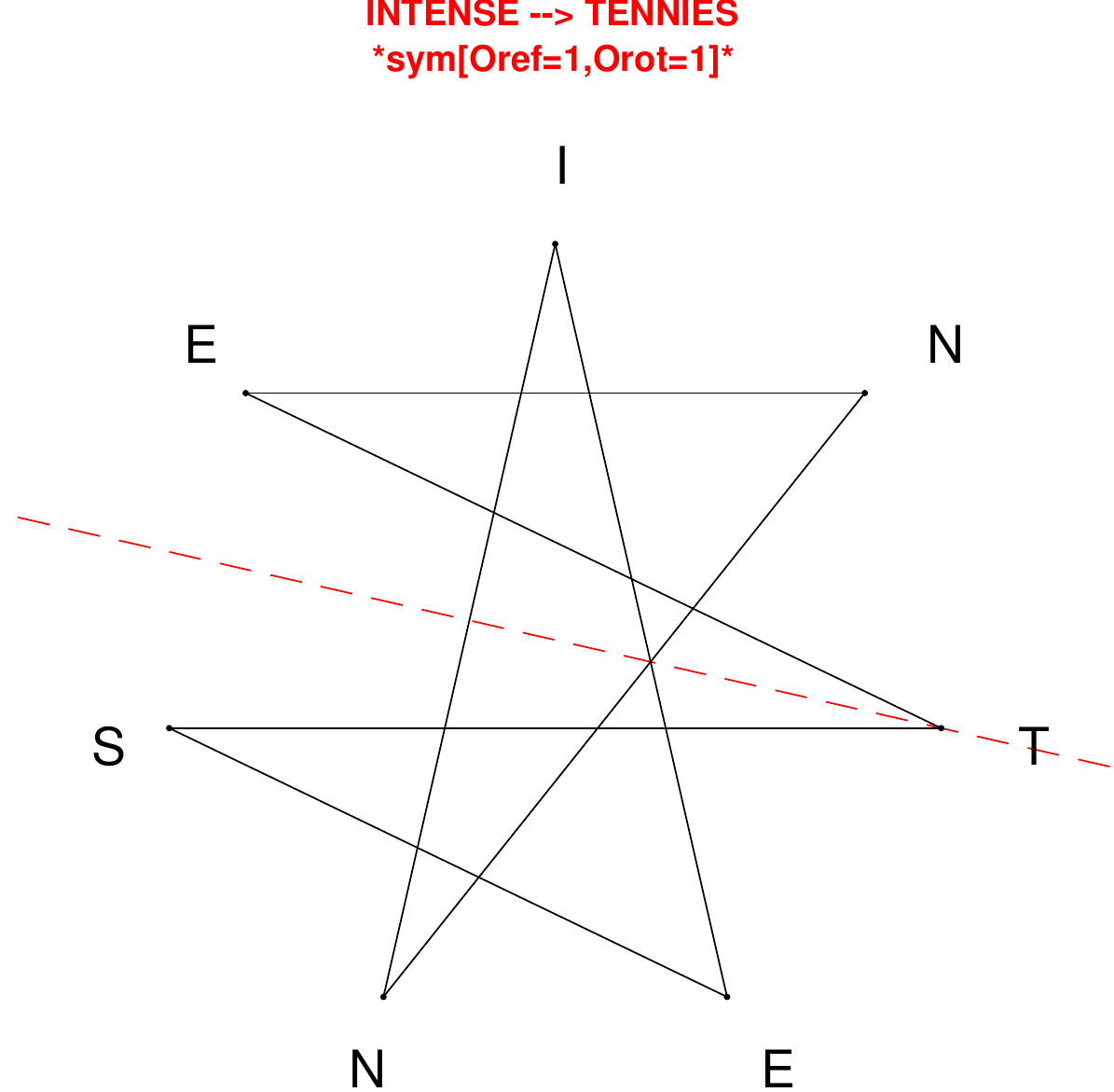}
\end{subfigure}
\hfill
\begin{subfigure}[T]{0.19\textwidth}
\centering
\includegraphics[width=\textwidth]{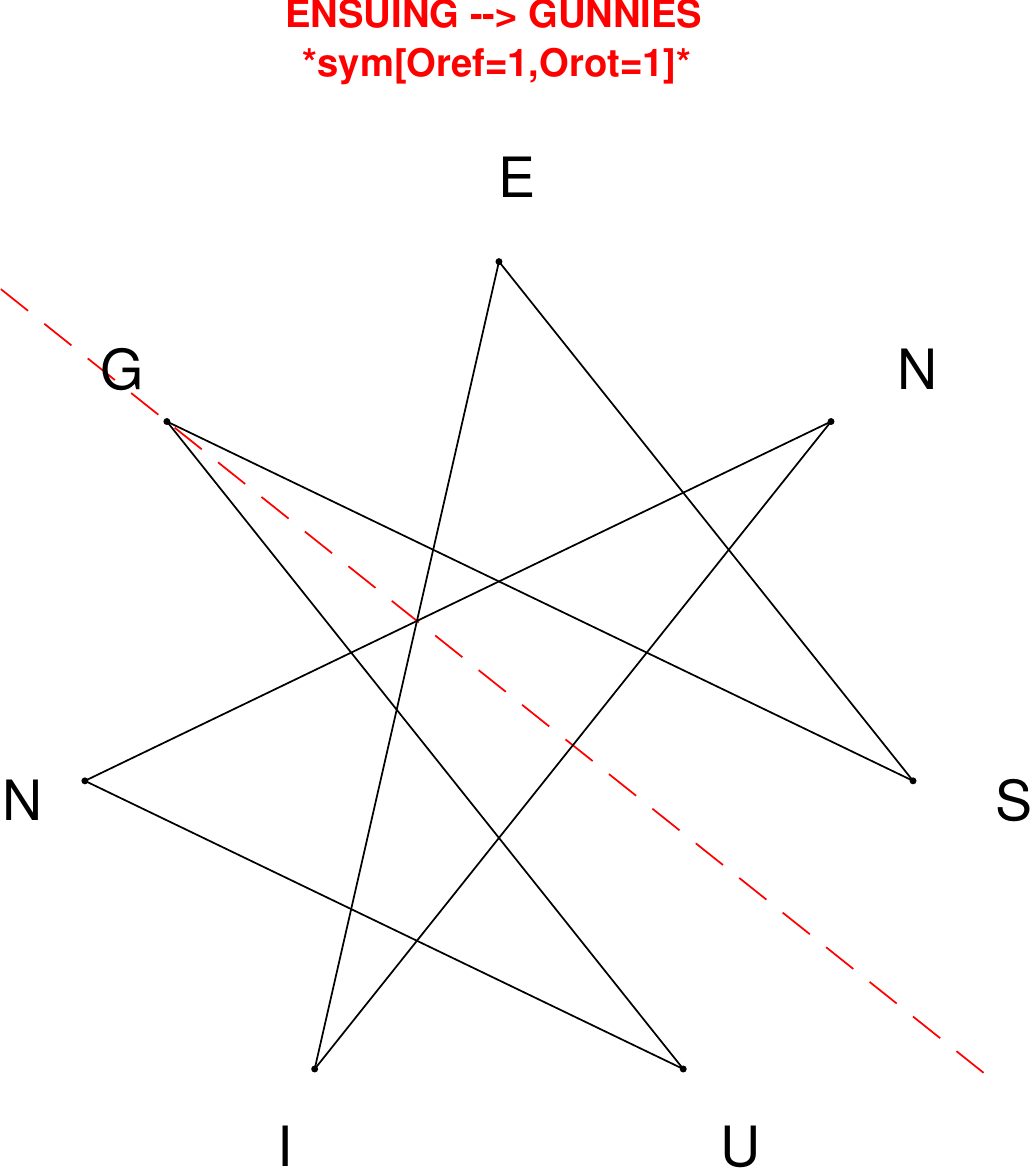}
\end{subfigure}
\hfill
\begin{subfigure}[T]{0.19\textwidth}
\centering
\includegraphics[width=\textwidth]{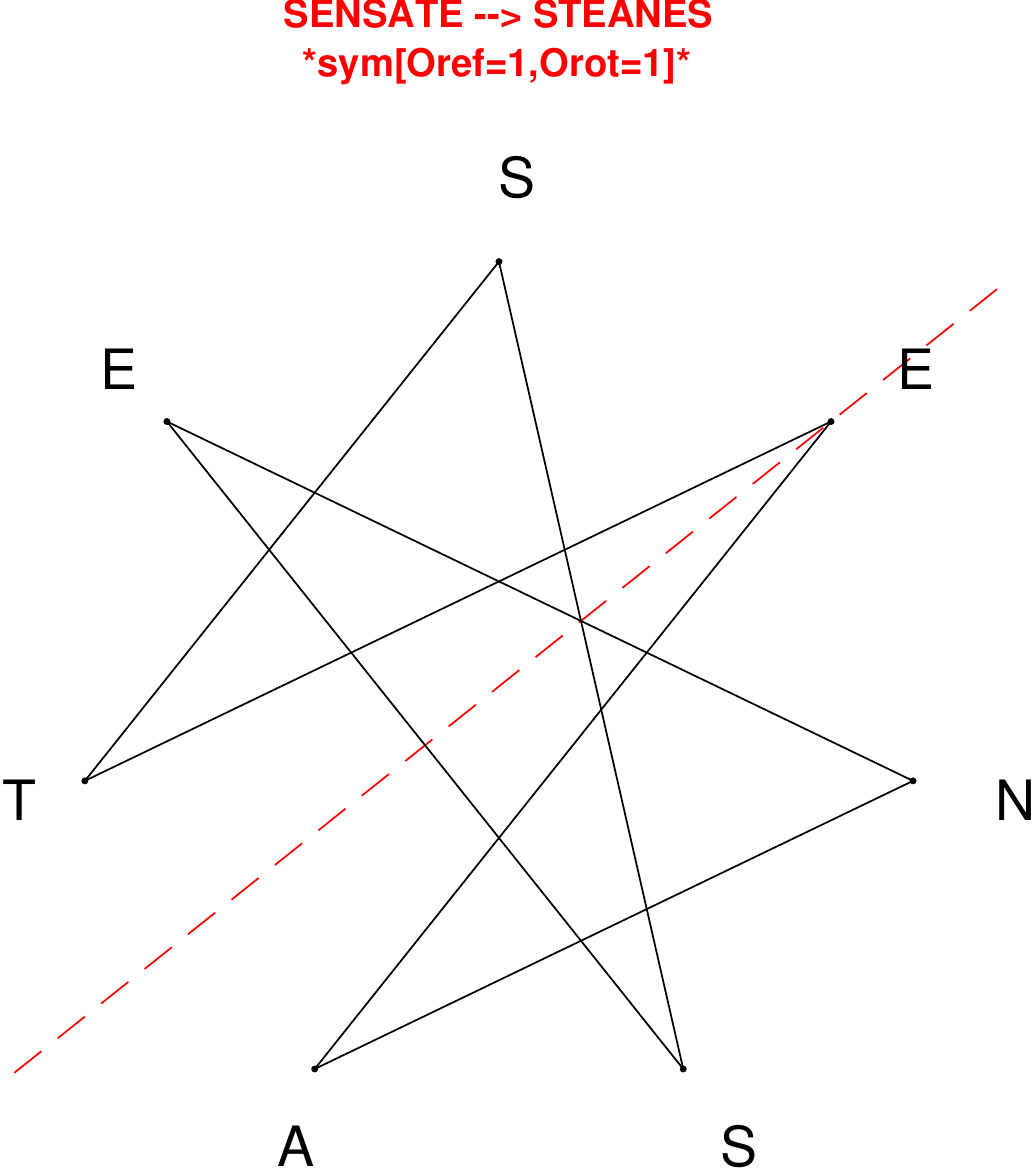}
\end{subfigure}
\hfill
\begin{subfigure}[T]{0.19\textwidth}
\centering
\includegraphics[width=\textwidth]{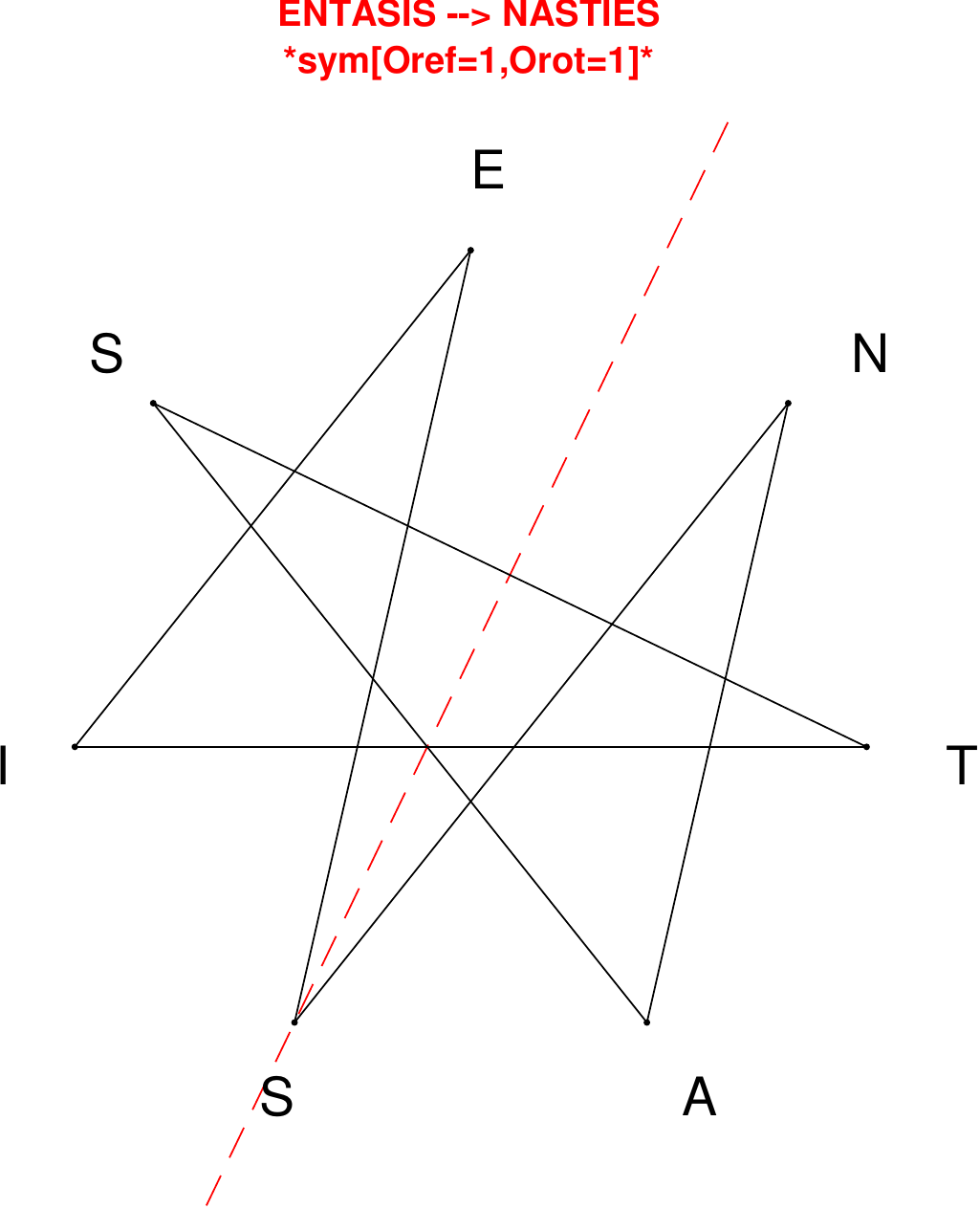}
\end{subfigure}
\hfill
\begin{subfigure}[T]{0.19\textwidth}
\centering
\includegraphics[width=\textwidth]{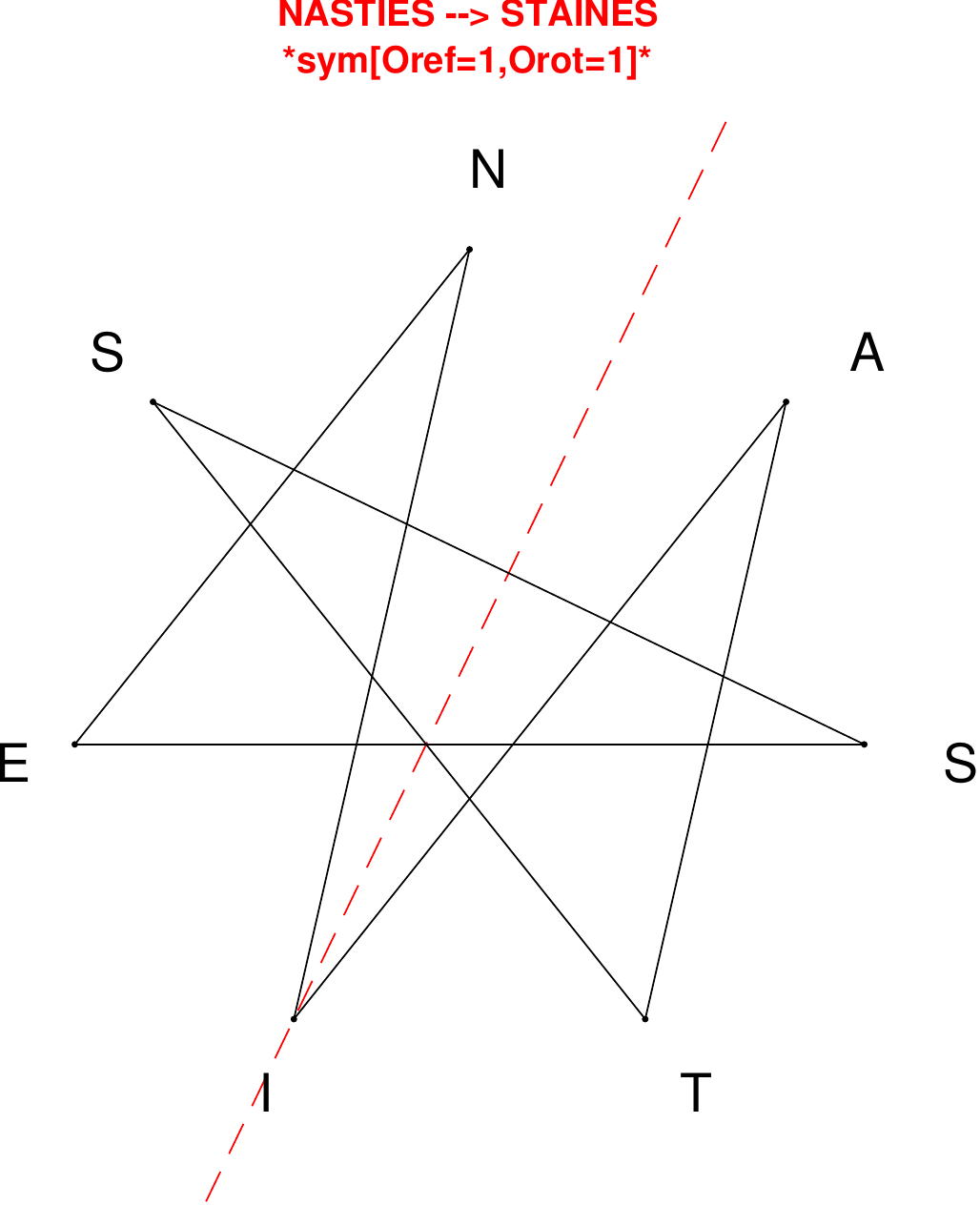}
\end{subfigure}
\end{figure}

\begin{figure}[H]
\centering
\begin{subfigure}[T]{0.19\textwidth}
\centering
\includegraphics[width=\textwidth]{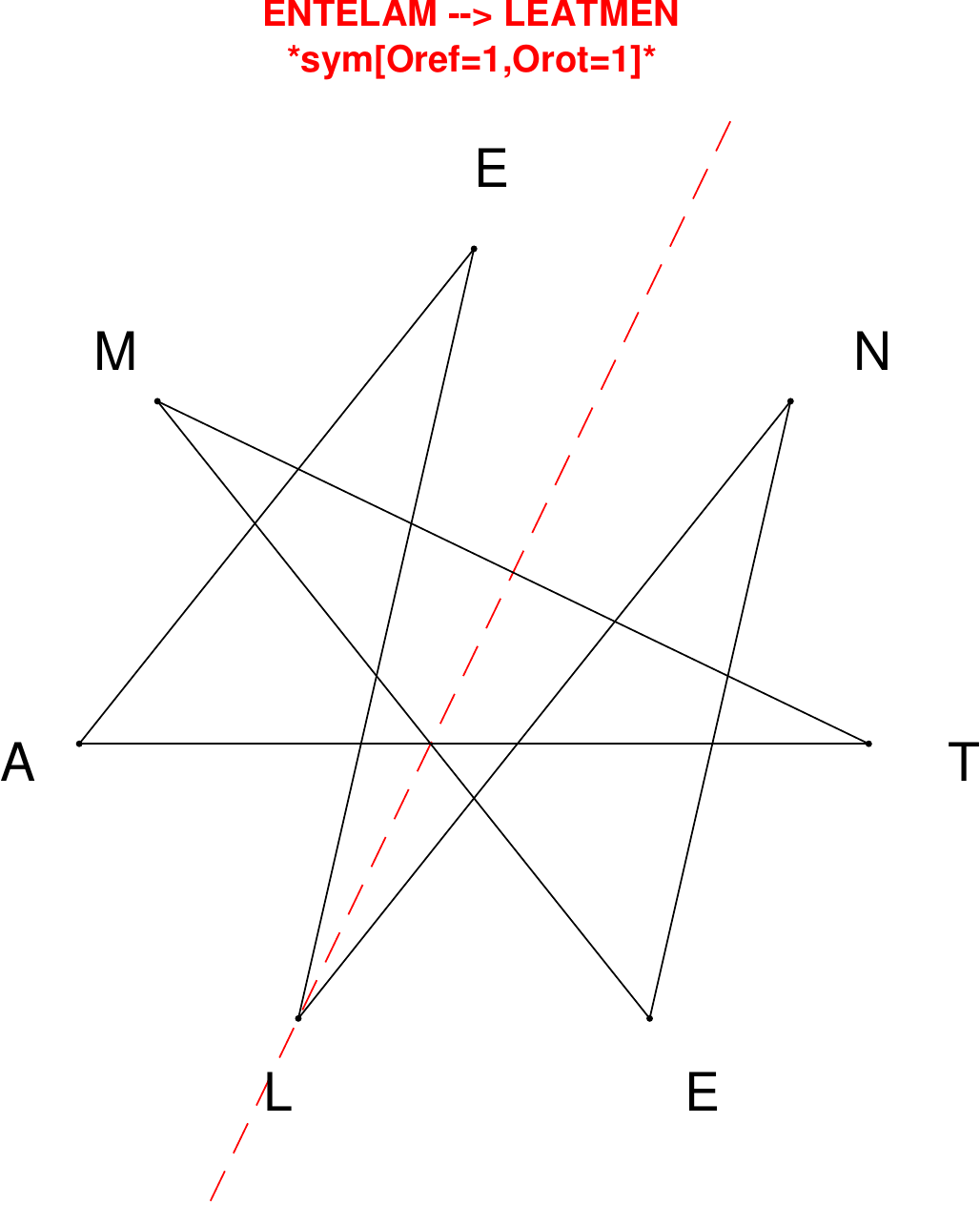}
\end{subfigure}
\hfill
\begin{subfigure}[T]{0.19\textwidth}
\centering
\includegraphics[width=\textwidth]{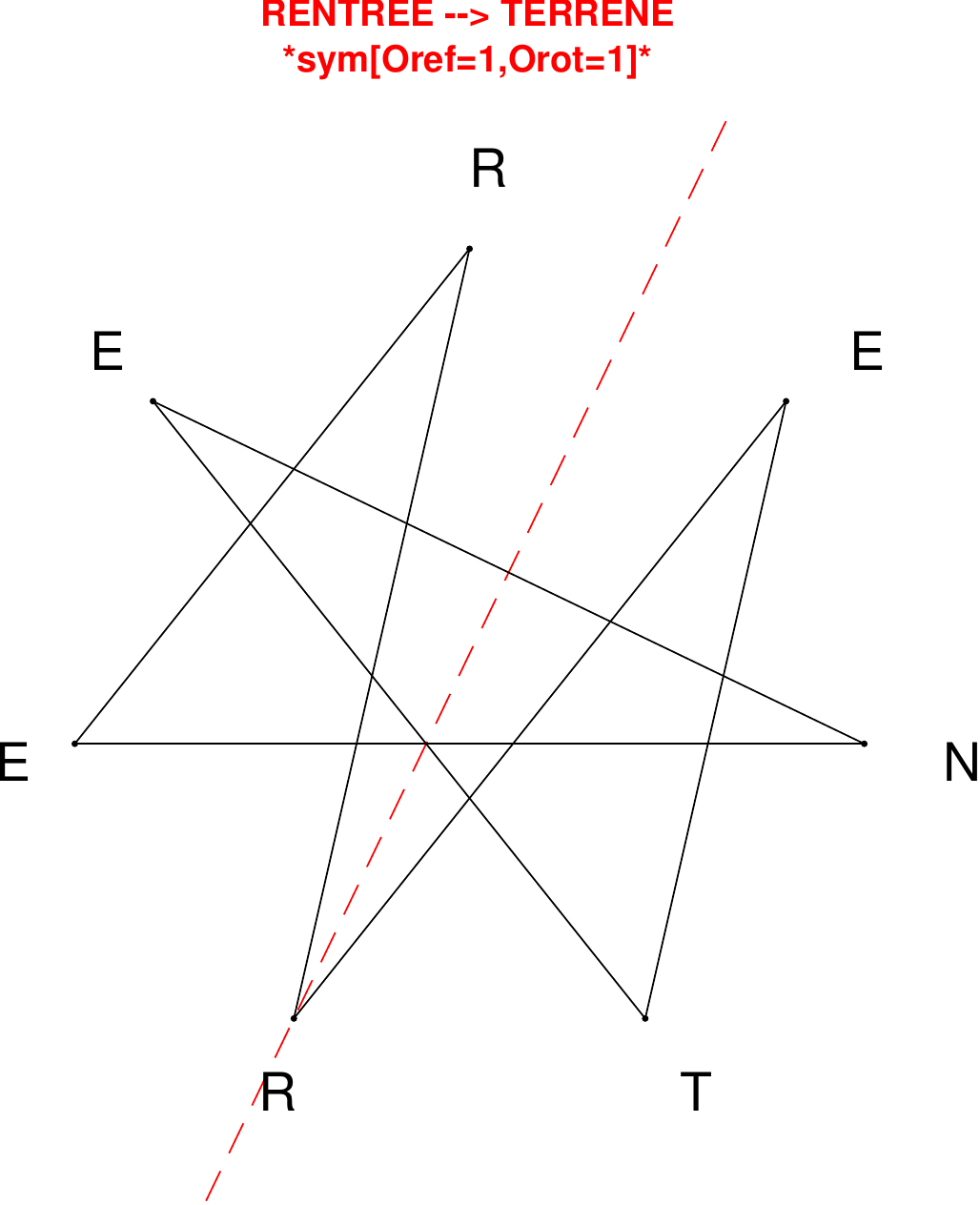}
\end{subfigure}
\hfill
\begin{subfigure}[T]{0.19\textwidth}
\centering
\includegraphics[width=\textwidth]{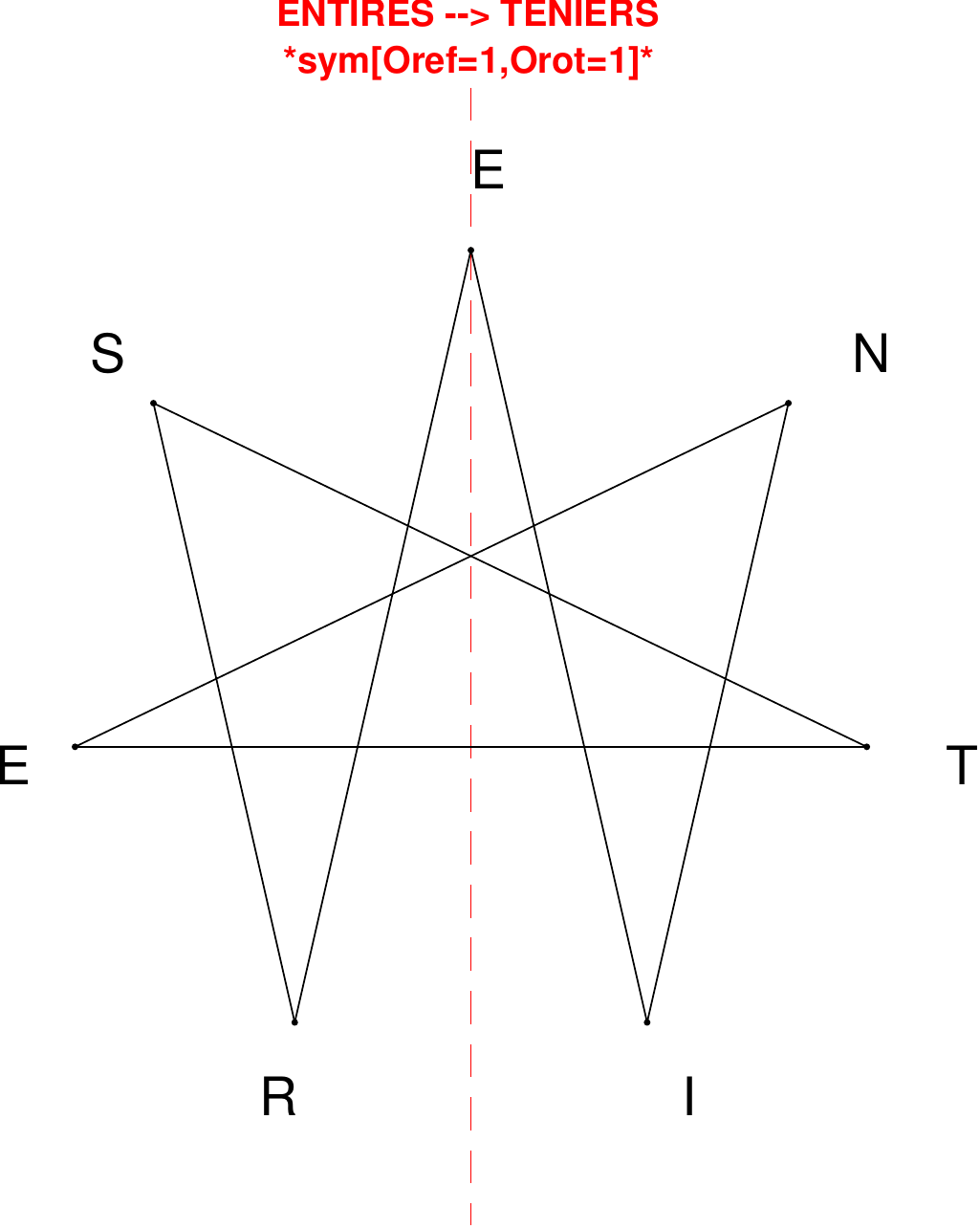}
\end{subfigure}
\hfill
\begin{subfigure}[T]{0.19\textwidth}
\centering
\includegraphics[width=\textwidth]{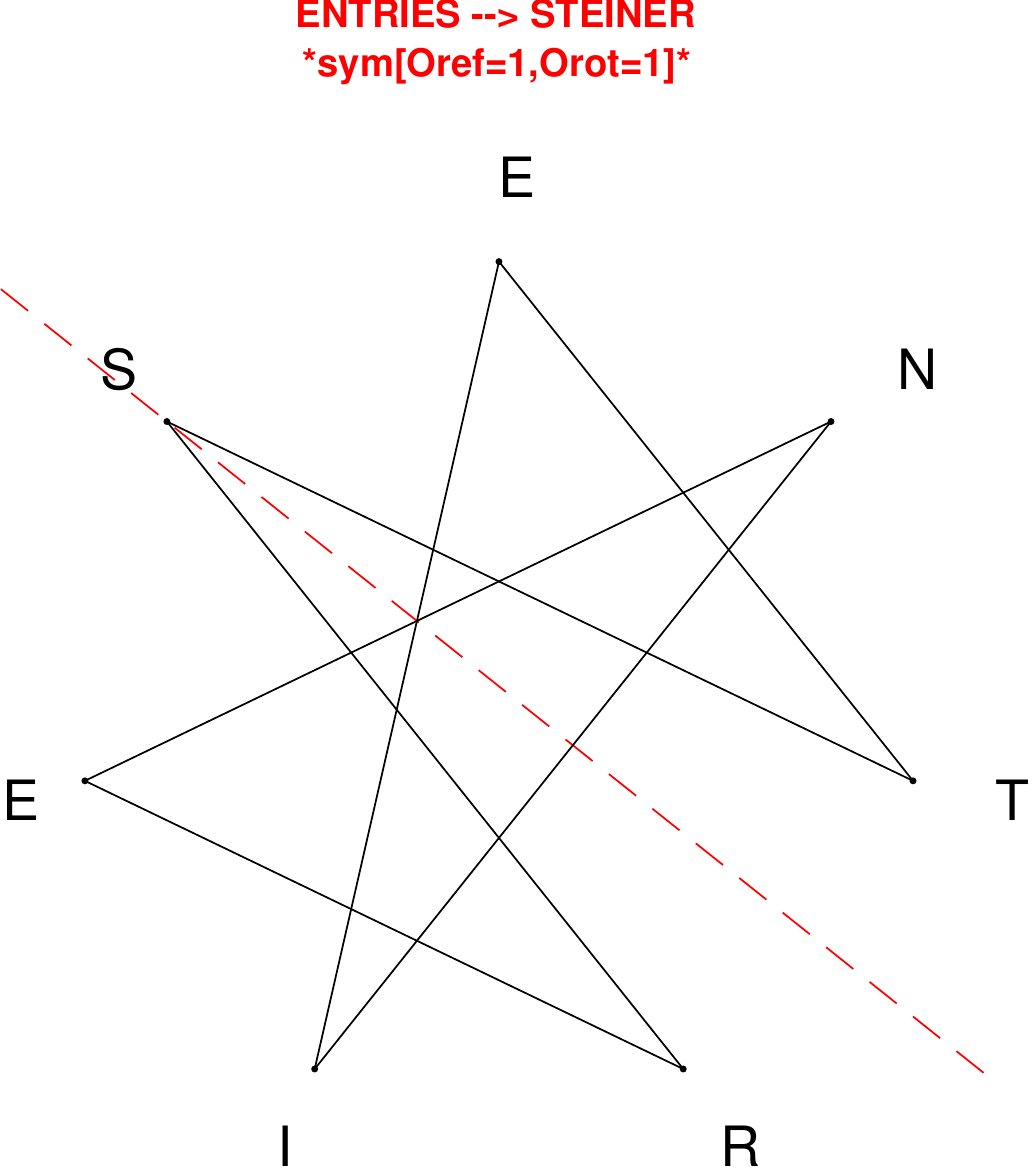}
\end{subfigure}
\hfill
\begin{subfigure}[T]{0.19\textwidth}
\centering
\includegraphics[width=\textwidth]{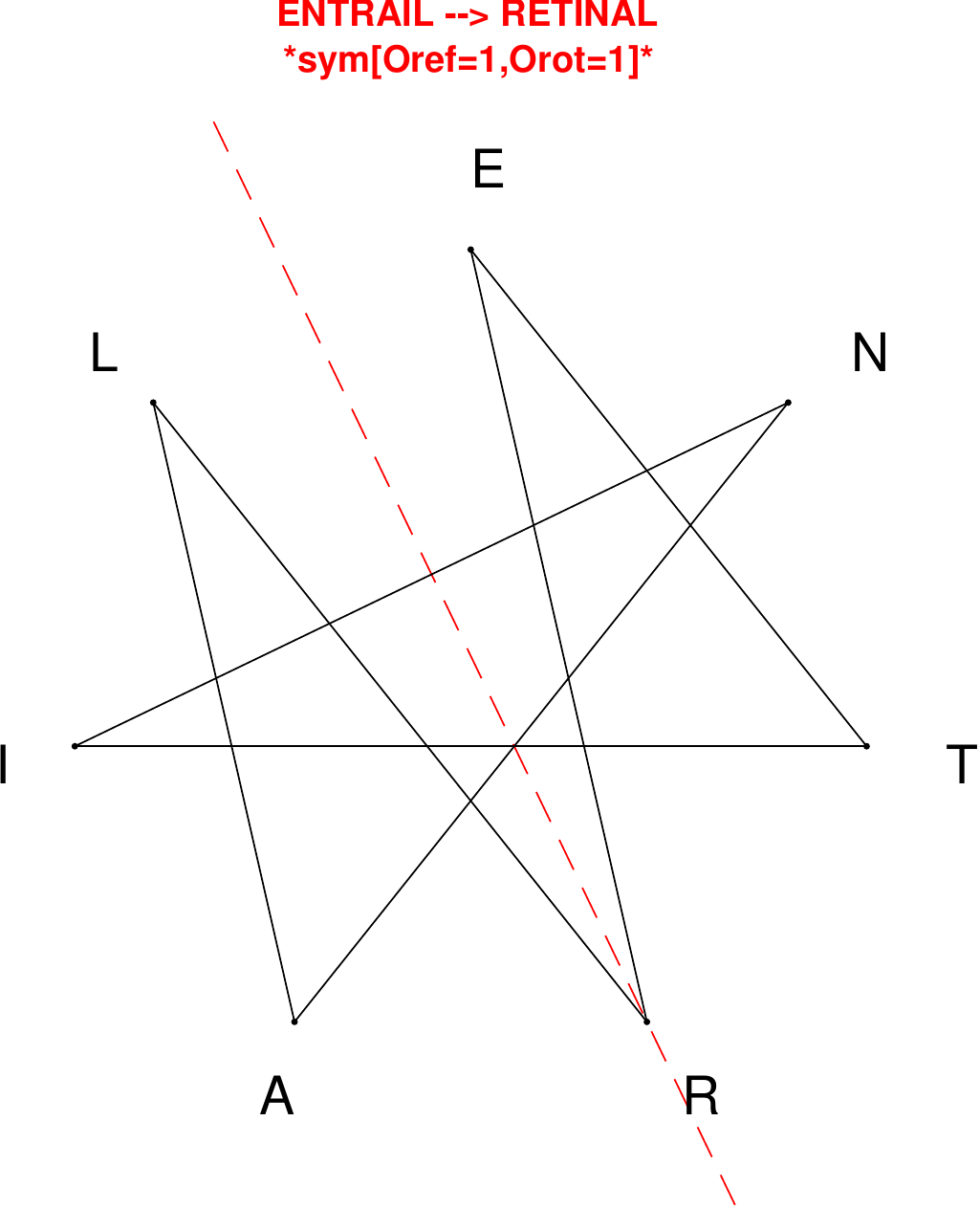}
\end{subfigure}
\end{figure}

\begin{figure}[H]
\centering
\begin{subfigure}[T]{0.19\textwidth}
\centering
\includegraphics[width=\textwidth]{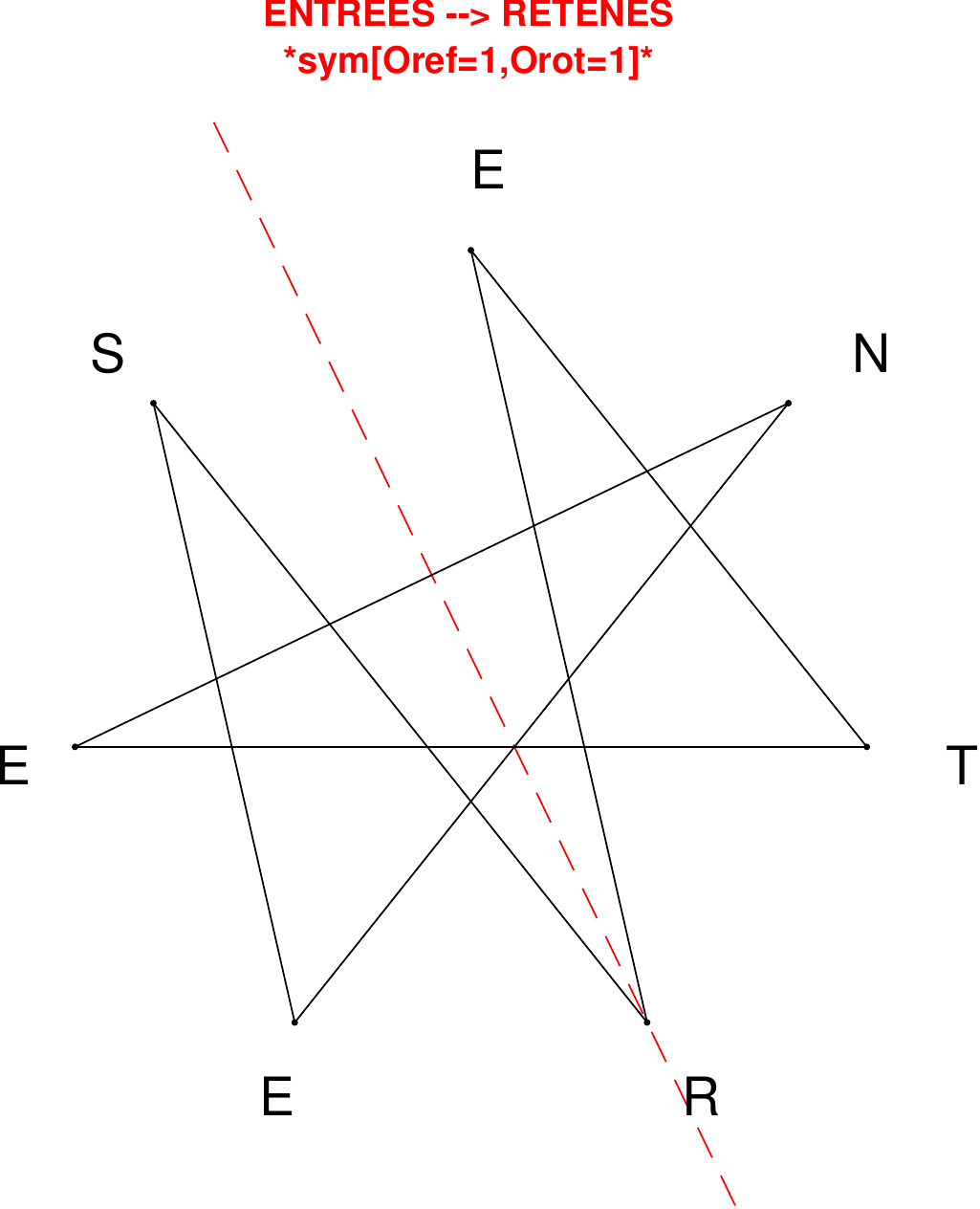}
\end{subfigure}
\hfill
\begin{subfigure}[T]{0.19\textwidth}
\centering
\includegraphics[width=\textwidth]{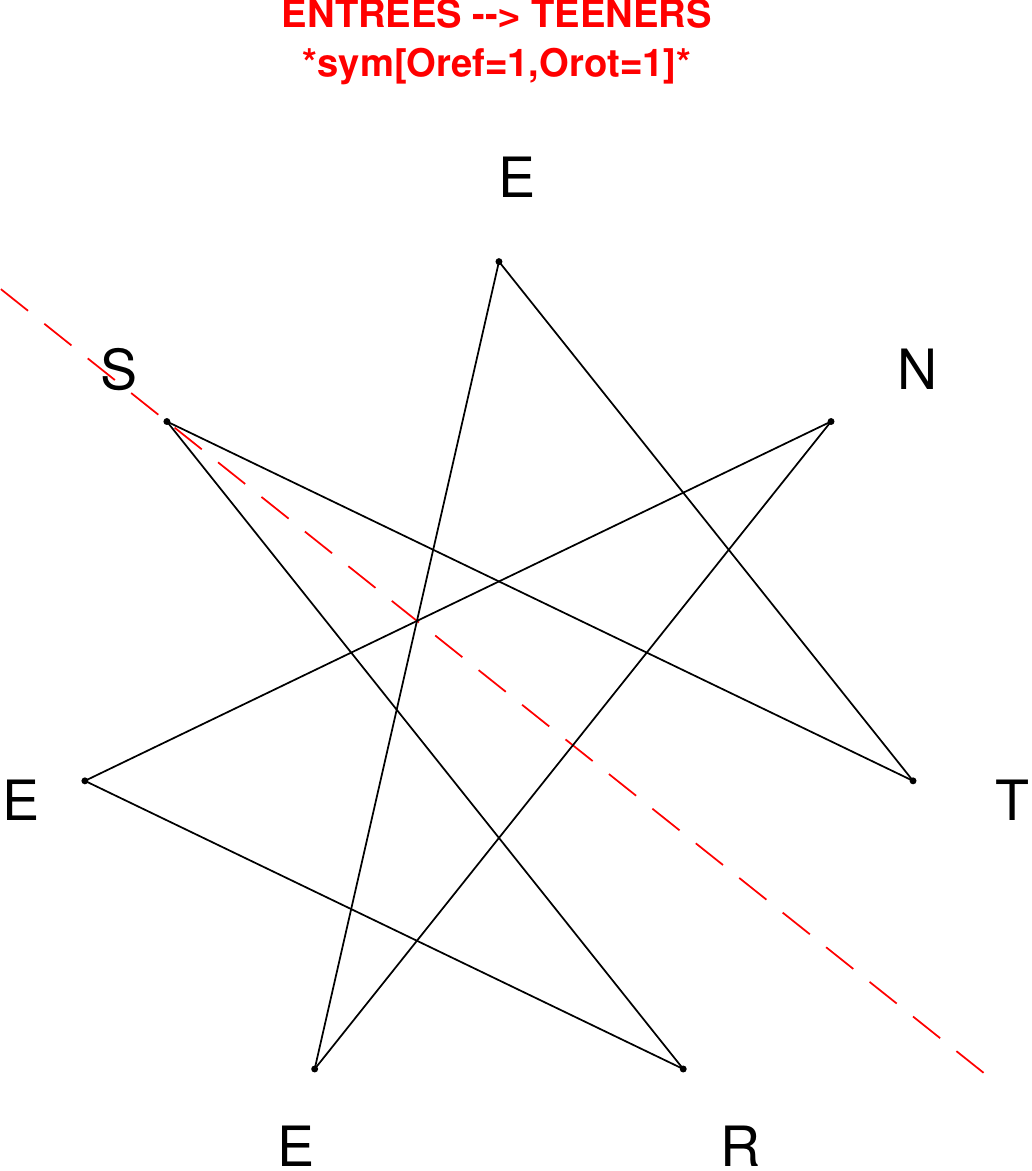}
\end{subfigure}
\hfill
\begin{subfigure}[T]{0.19\textwidth}
\centering
\includegraphics[width=\textwidth]{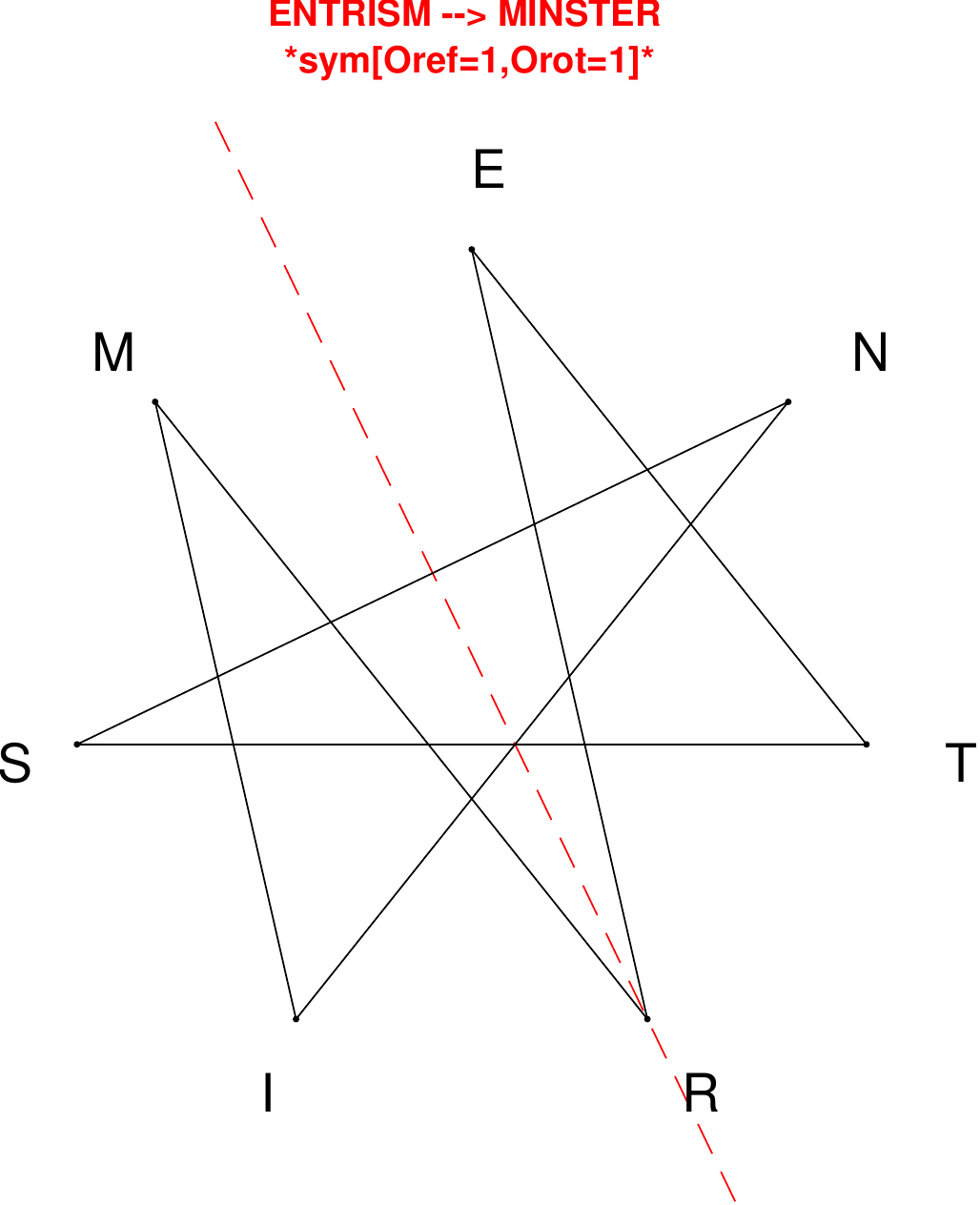}
\end{subfigure}
\hfill
\begin{subfigure}[T]{0.19\textwidth}
\centering
\includegraphics[width=\textwidth]{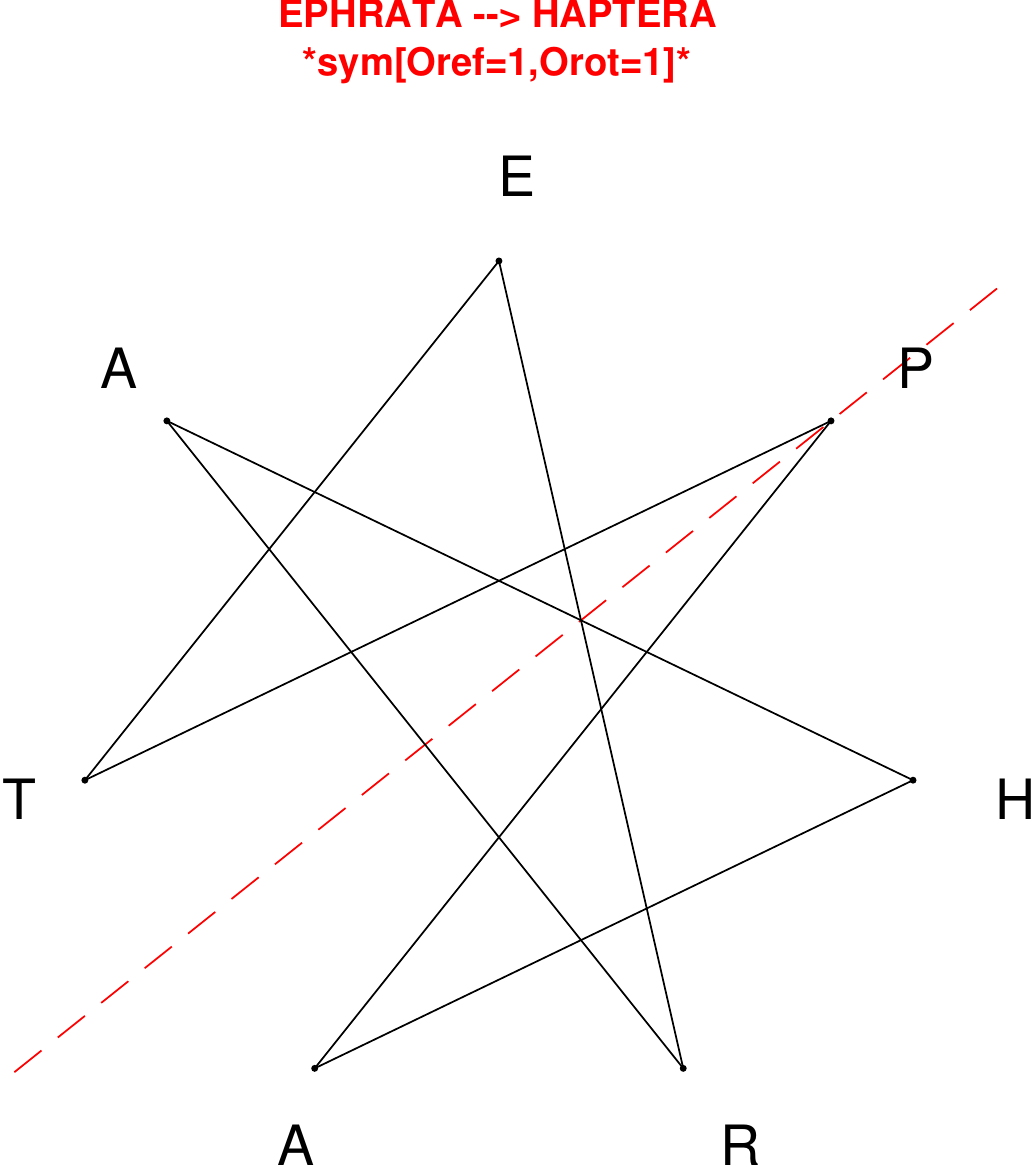}
\end{subfigure}
\hfill
\begin{subfigure}[T]{0.19\textwidth}
\centering
\includegraphics[width=\textwidth]{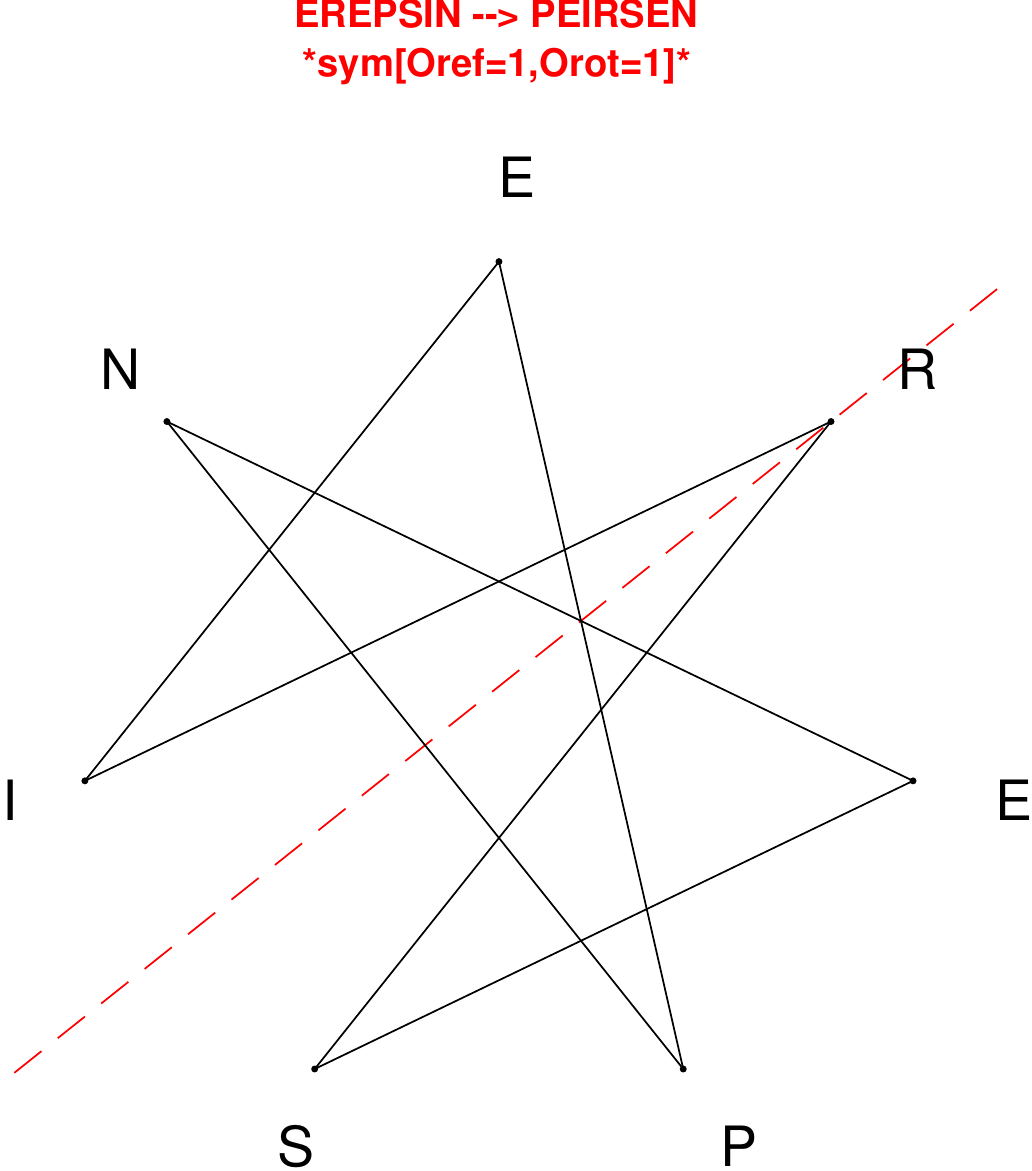}
\end{subfigure}
\end{figure}

\begin{figure}[H]
\centering
\begin{subfigure}[T]{0.19\textwidth}
\centering
\includegraphics[width=\textwidth]{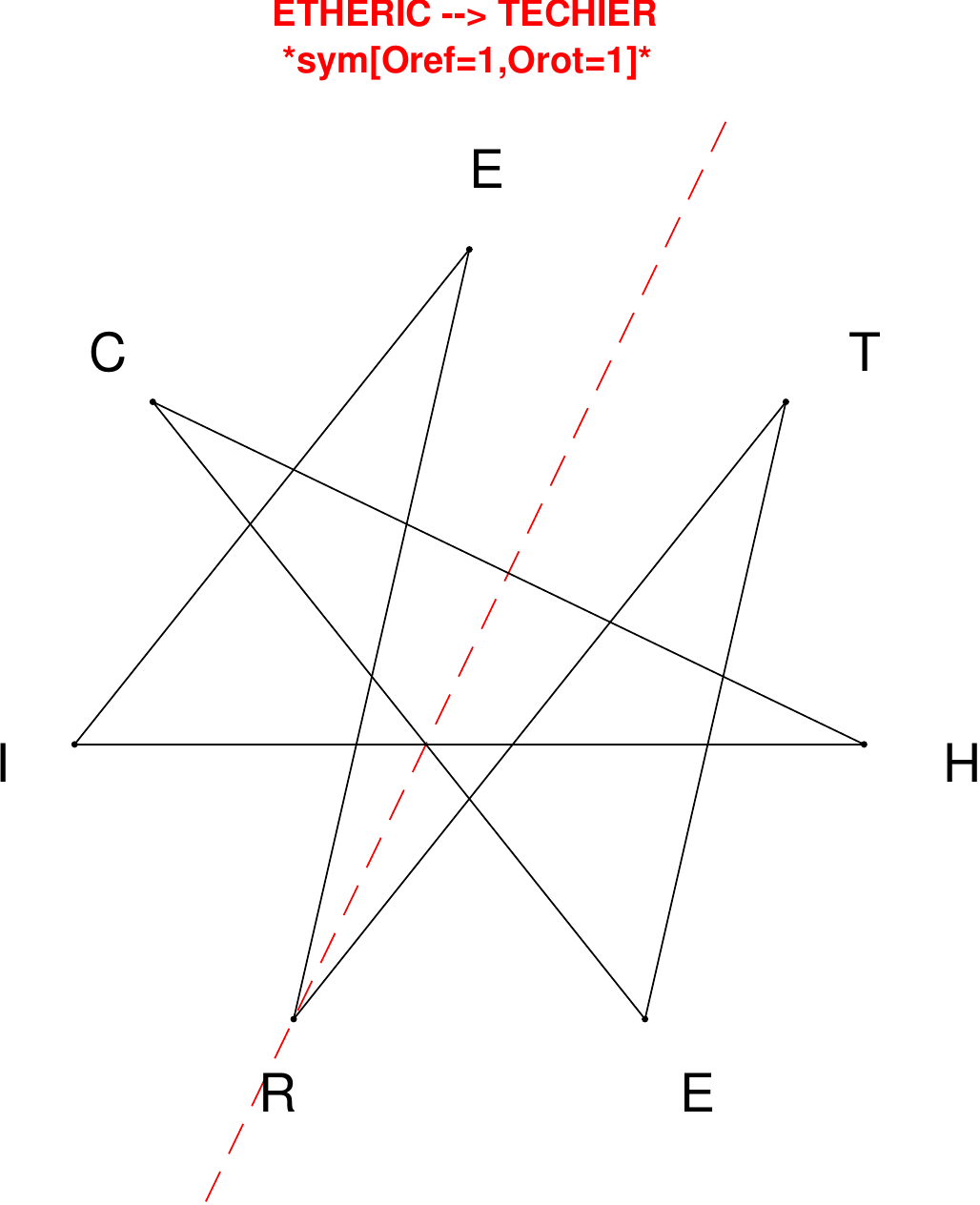}
\end{subfigure}
\hfill
\begin{subfigure}[T]{0.19\textwidth}
\centering
\includegraphics[width=\textwidth]{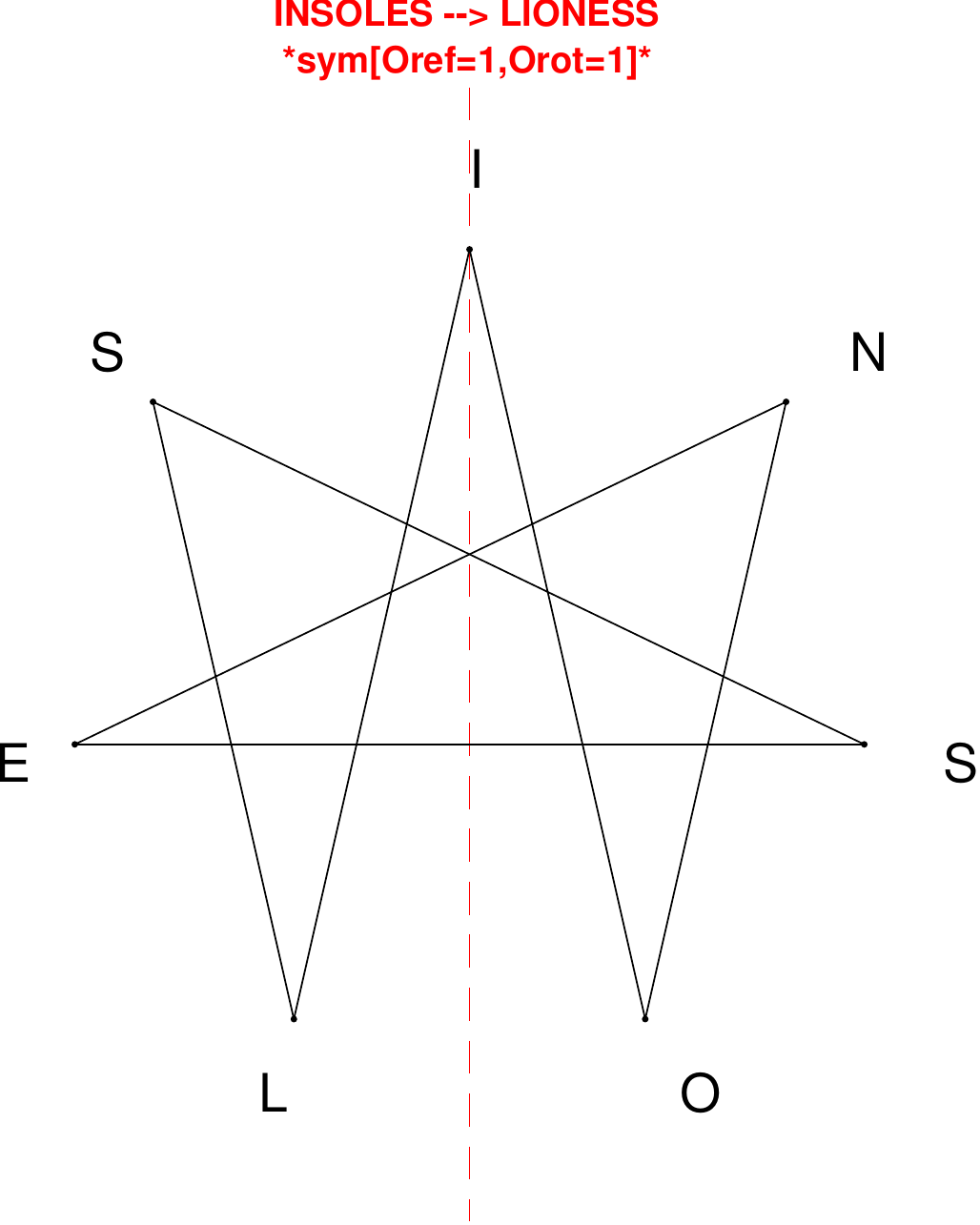}
\end{subfigure}
\hfill
\begin{subfigure}[T]{0.19\textwidth}
\centering
\includegraphics[width=\textwidth]{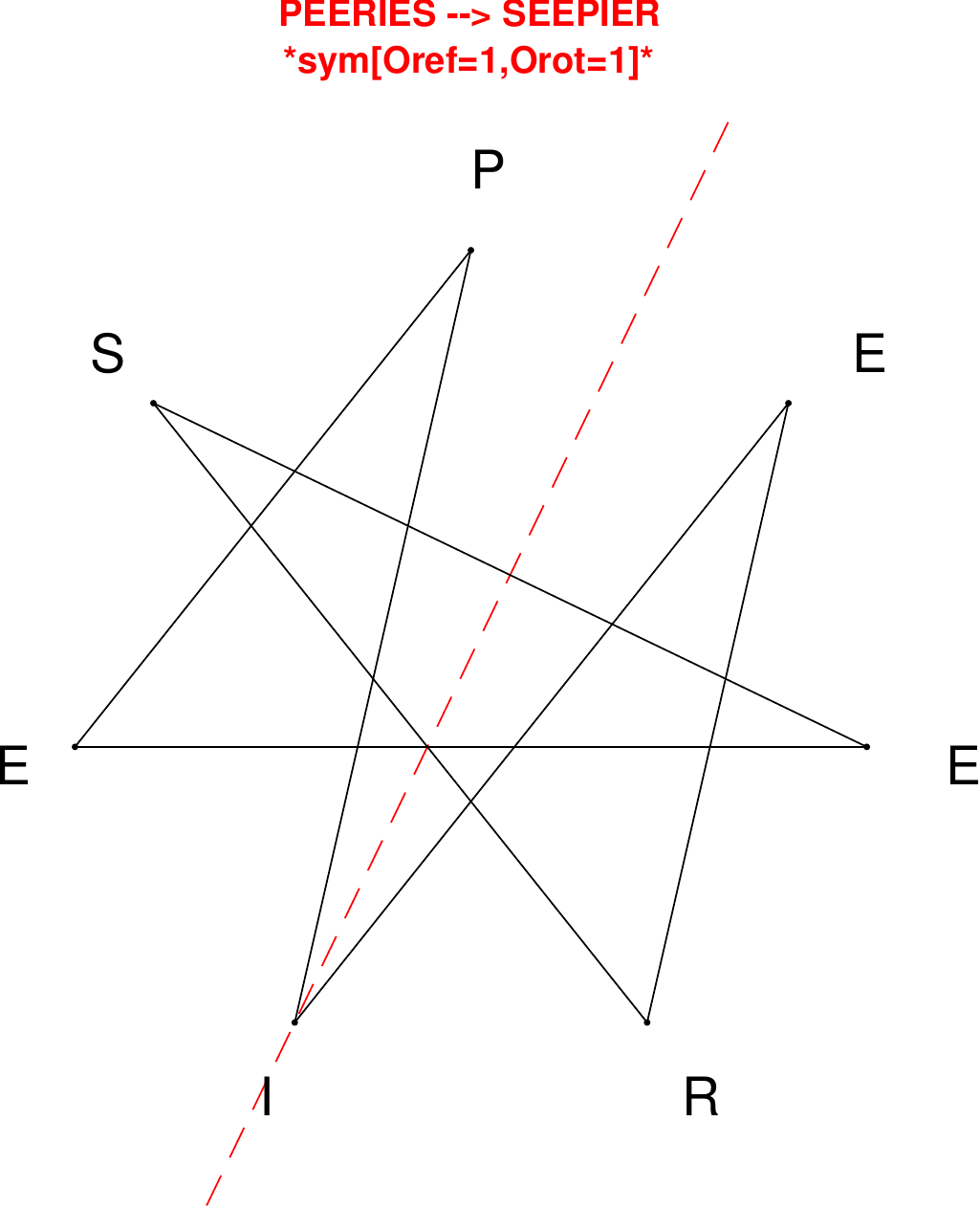}
\end{subfigure}
\hfill
\begin{subfigure}[T]{0.19\textwidth}
\centering
\includegraphics[width=\textwidth]{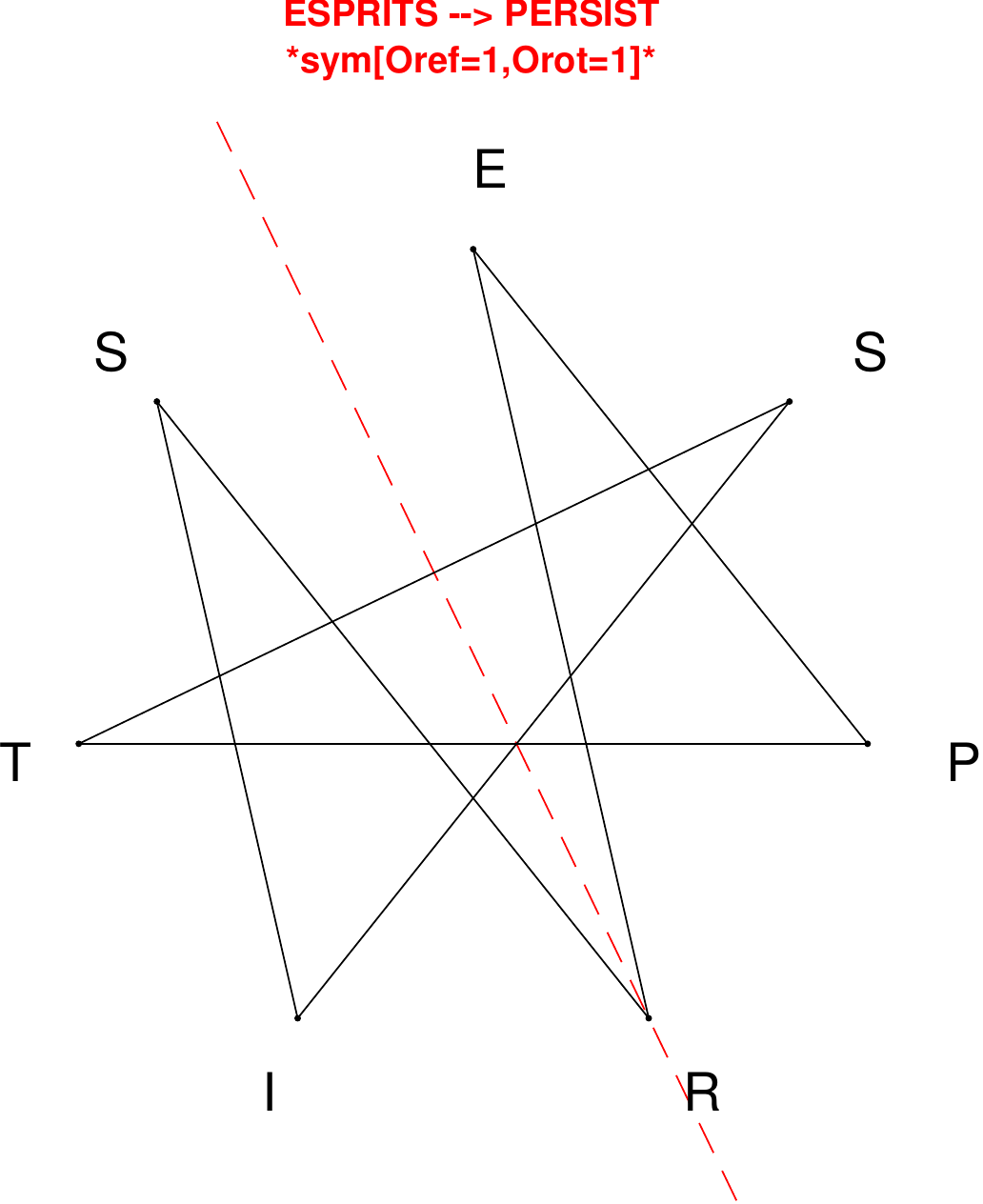}
\end{subfigure}
\hfill
\begin{subfigure}[T]{0.19\textwidth}
\centering
\includegraphics[width=\textwidth]{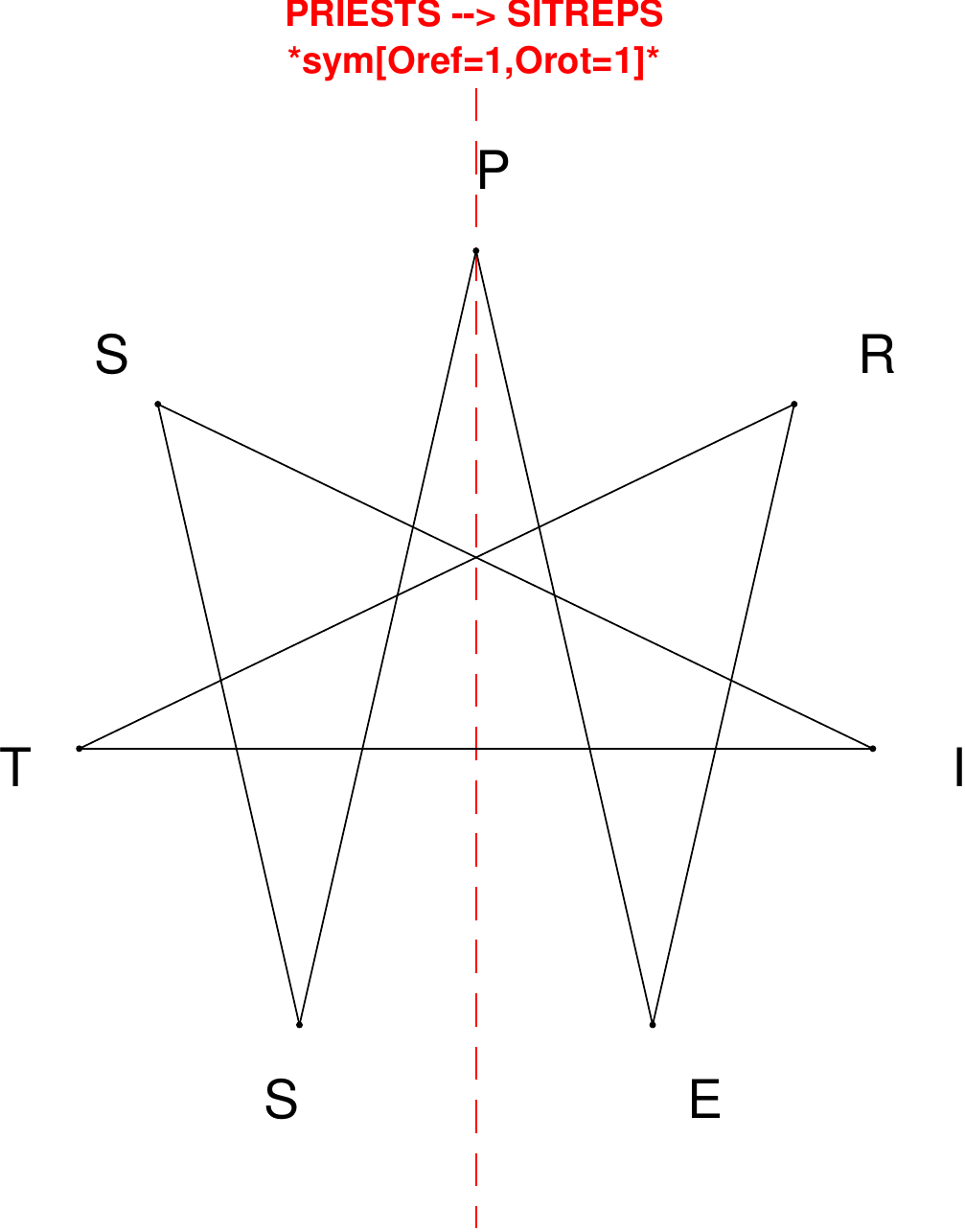}
\end{subfigure}
\end{figure}

\begin{figure}[H]
\centering
\begin{subfigure}[T]{0.19\textwidth}
\centering
\includegraphics[width=\textwidth]{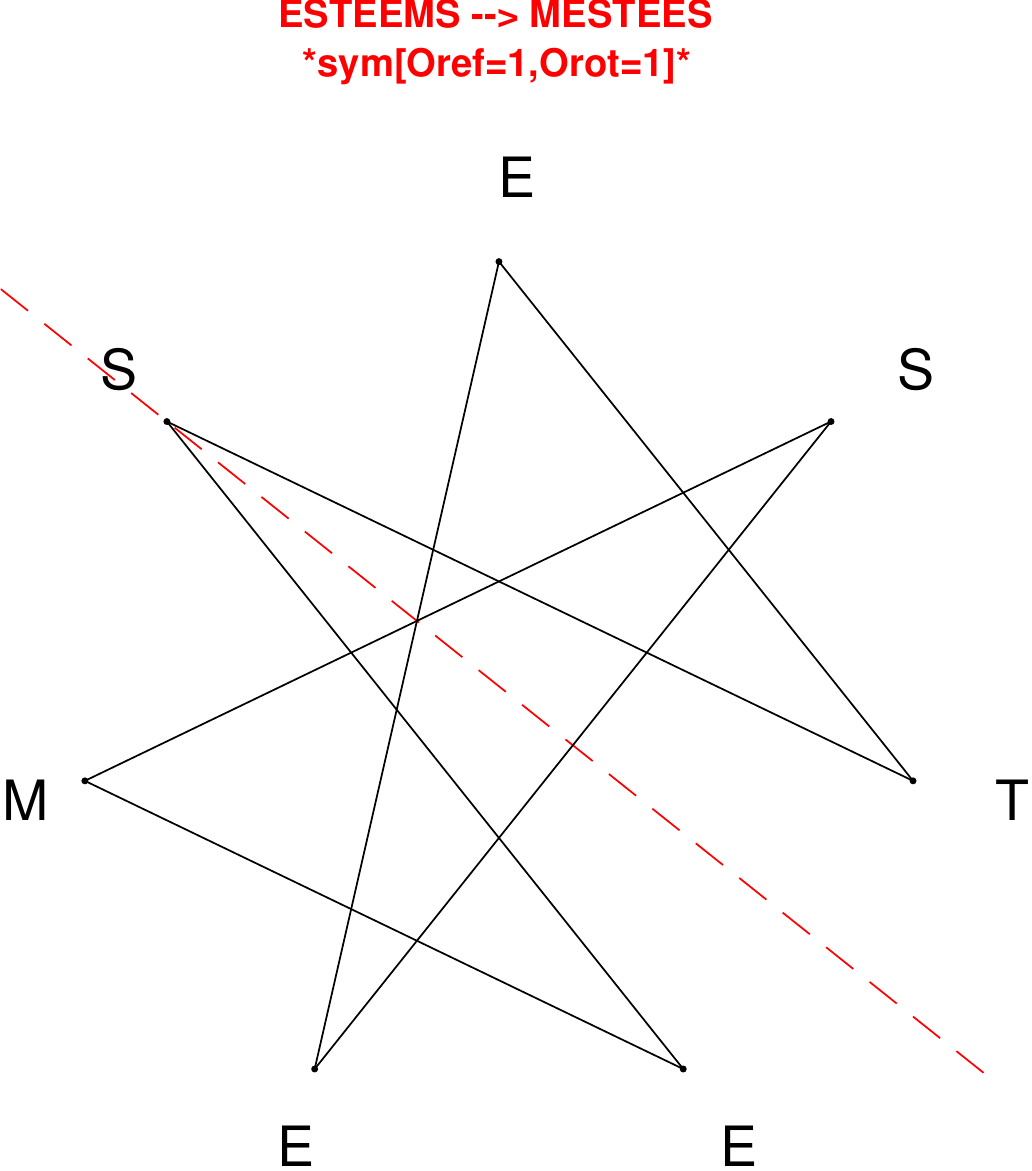}
\end{subfigure}
\hfill
\begin{subfigure}[T]{0.19\textwidth}
\centering
\includegraphics[width=\textwidth]{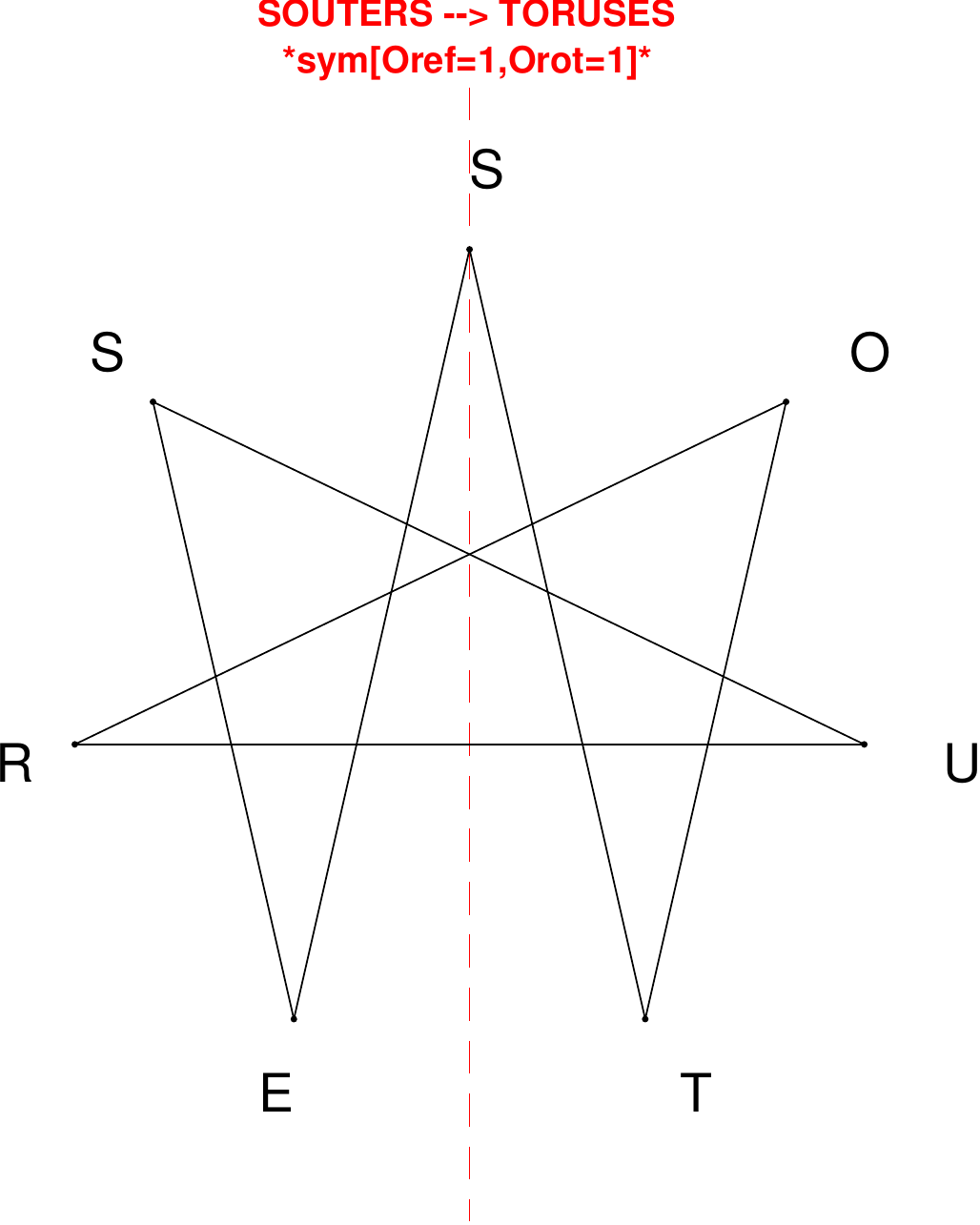}
\end{subfigure}
\hfill
\begin{subfigure}[T]{0.19\textwidth}
\centering
\includegraphics[width=\textwidth]{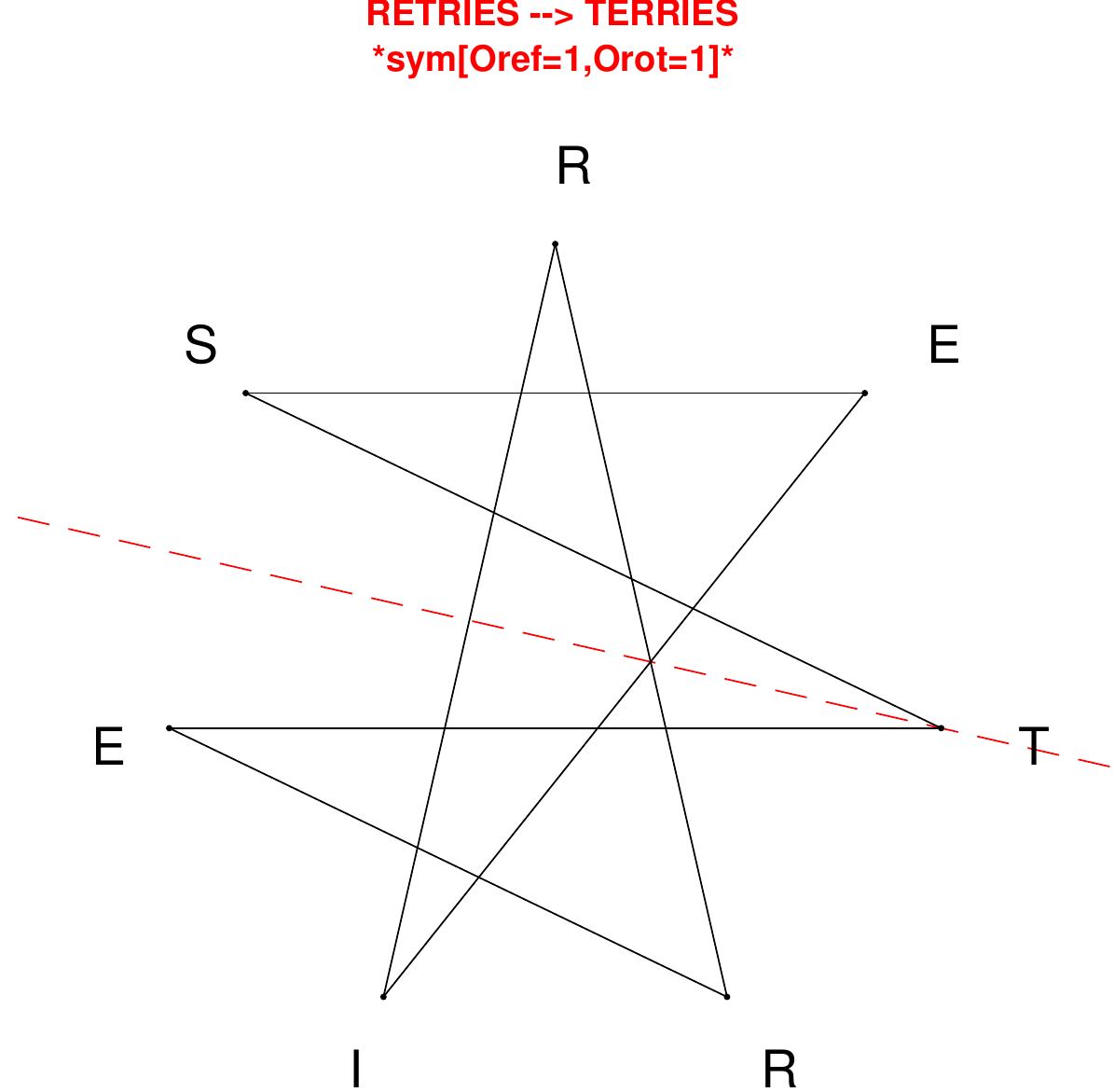}
\end{subfigure}
\hfill
\begin{subfigure}[T]{0.19\textwidth}
\centering
\includegraphics[width=\textwidth]{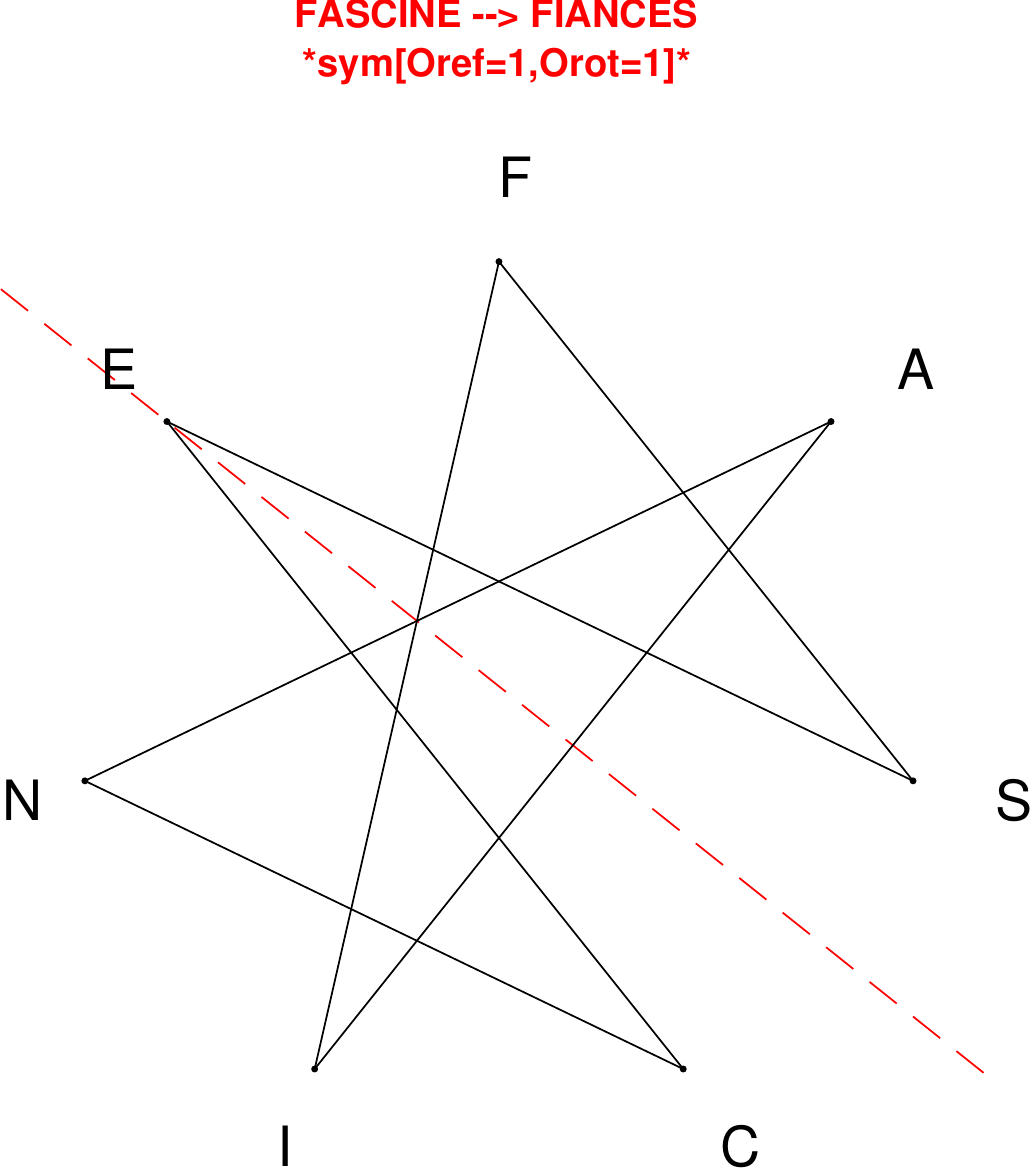}
\end{subfigure}
\hfill
\begin{subfigure}[T]{0.19\textwidth}
\centering
\includegraphics[width=\textwidth]{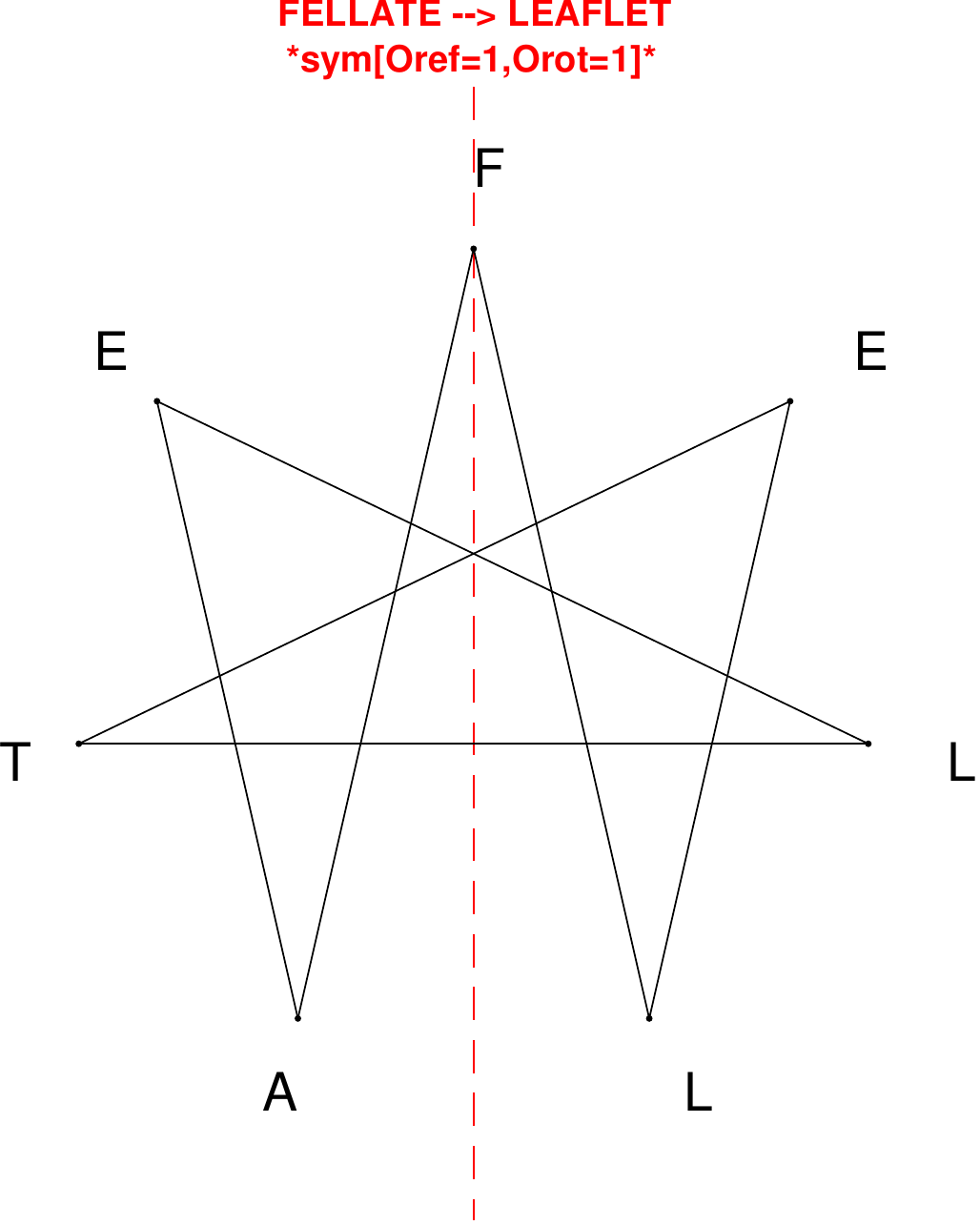}
\end{subfigure}
\end{figure}

\begin{figure}[H]
\centering
\begin{subfigure}[T]{0.19\textwidth}
\centering
\includegraphics[width=\textwidth]{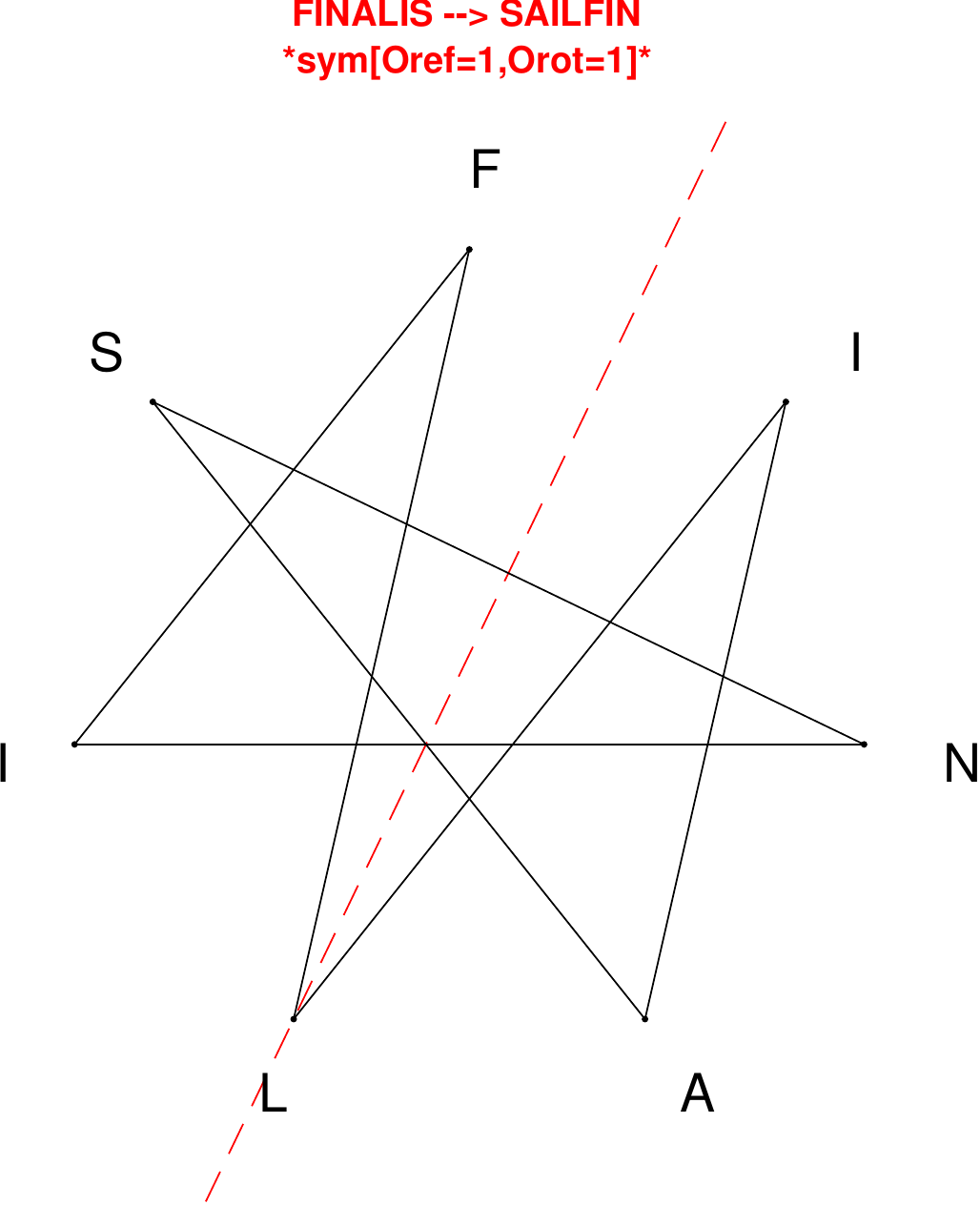}
\end{subfigure}
\hfill
\begin{subfigure}[T]{0.19\textwidth}
\centering
\includegraphics[width=\textwidth]{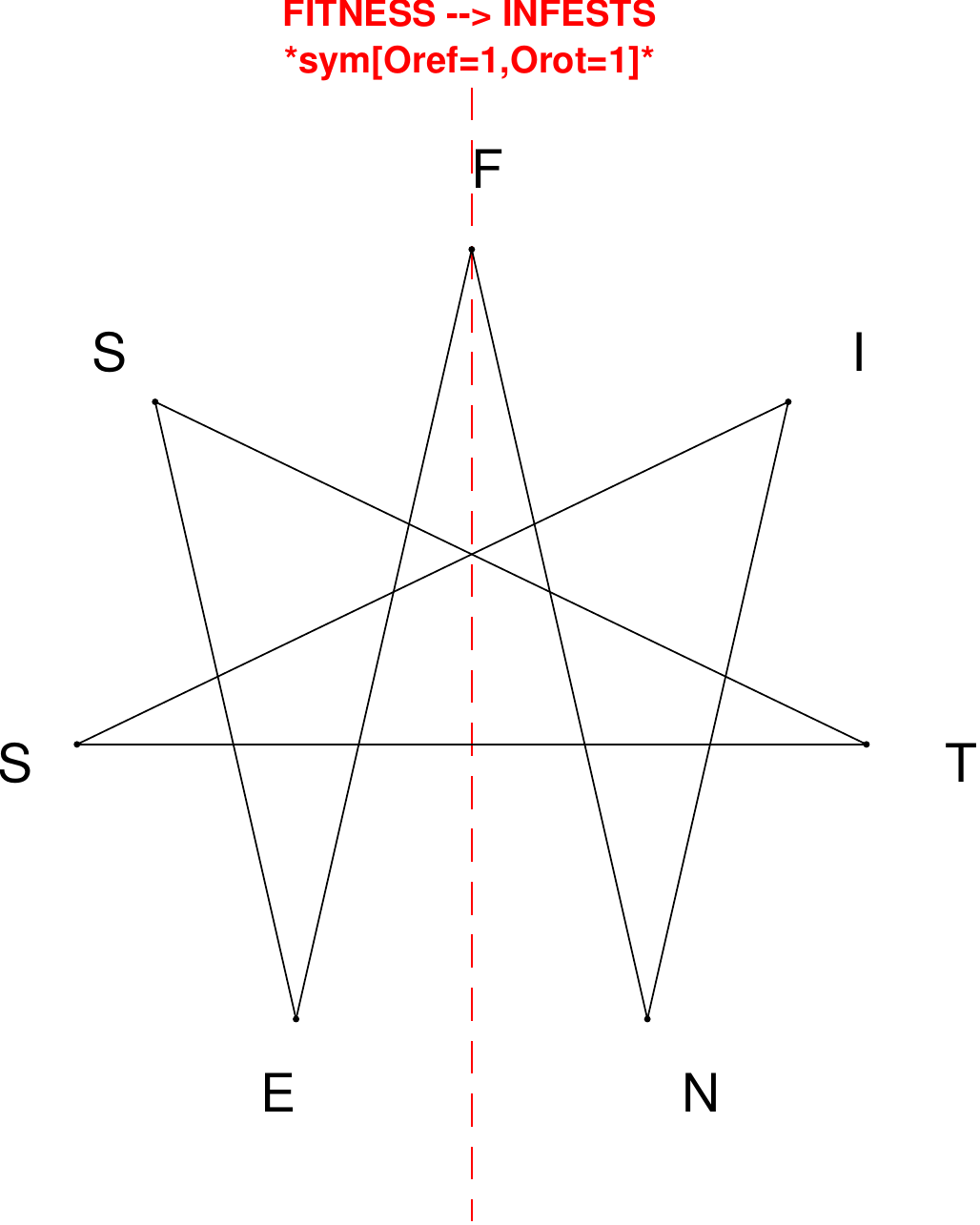}
\end{subfigure}
\hfill
\begin{subfigure}[T]{0.19\textwidth}
\centering
\includegraphics[width=\textwidth]{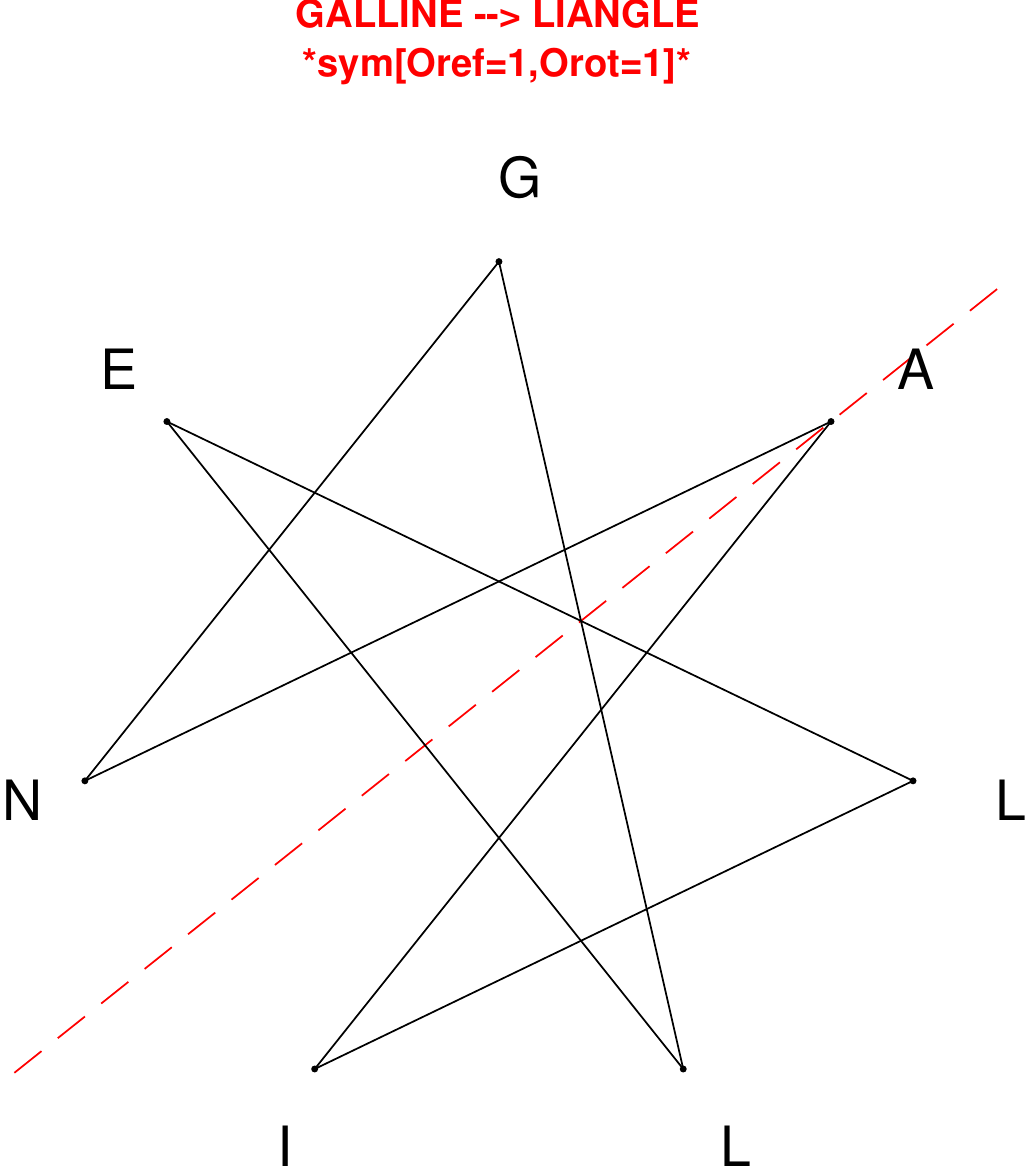}
\end{subfigure}
\hfill
\begin{subfigure}[T]{0.19\textwidth}
\centering
\includegraphics[width=\textwidth]{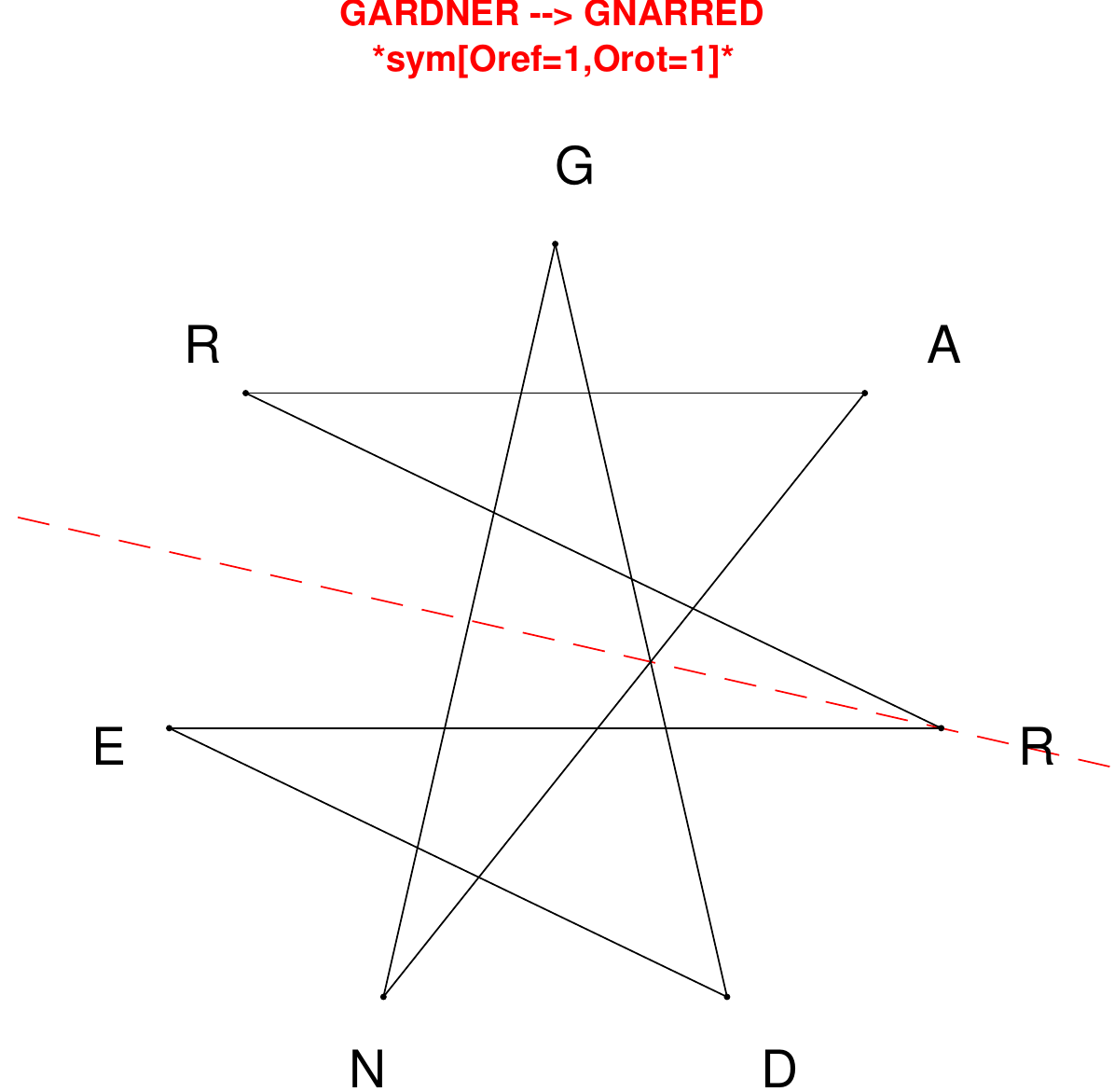}
\end{subfigure}
\hfill
\begin{subfigure}[T]{0.19\textwidth}
\centering
\includegraphics[width=\textwidth]{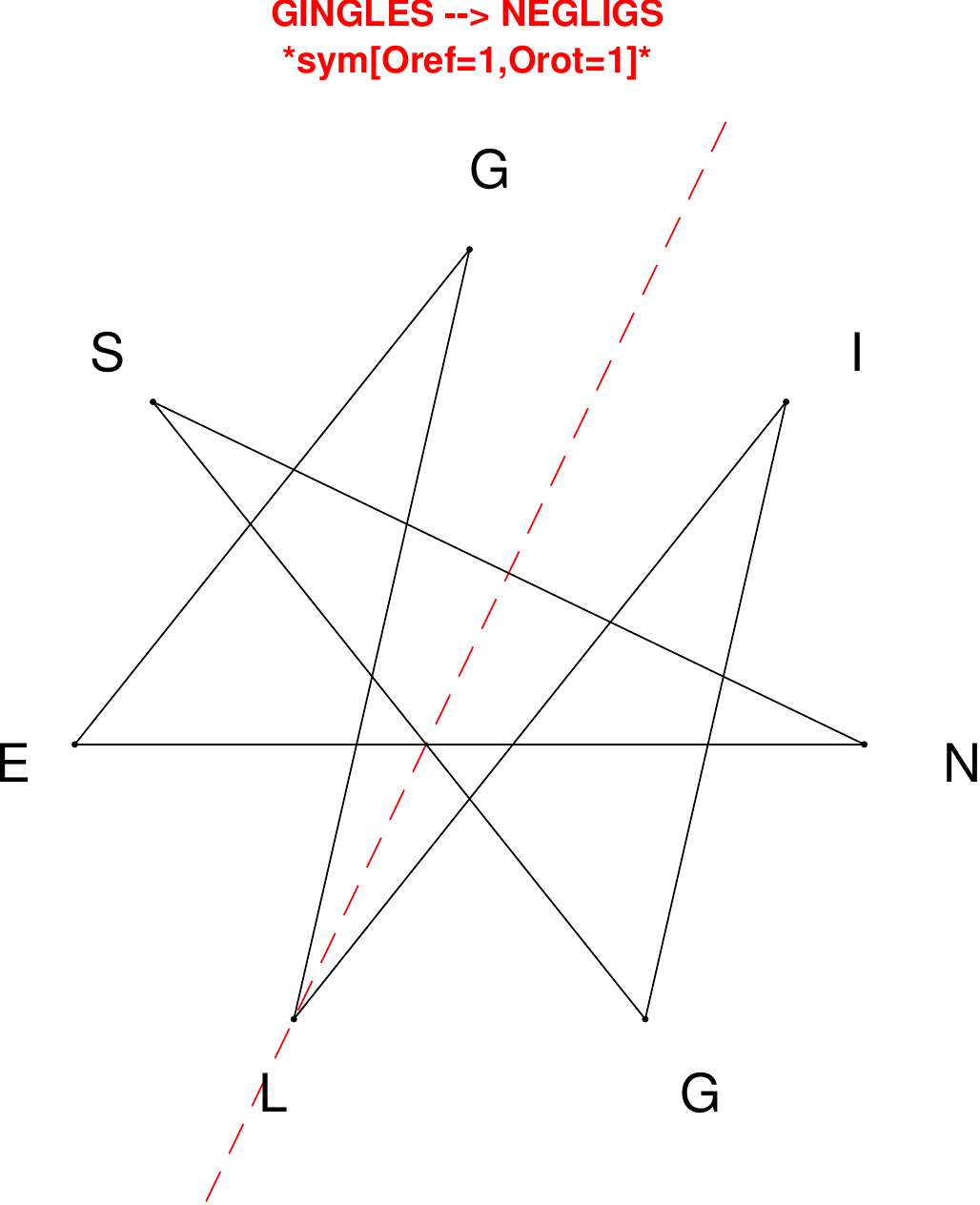}
\end{subfigure}
\end{figure}

\begin{figure}[H]
\centering
\begin{subfigure}[T]{0.19\textwidth}
\centering
\includegraphics[width=\textwidth]{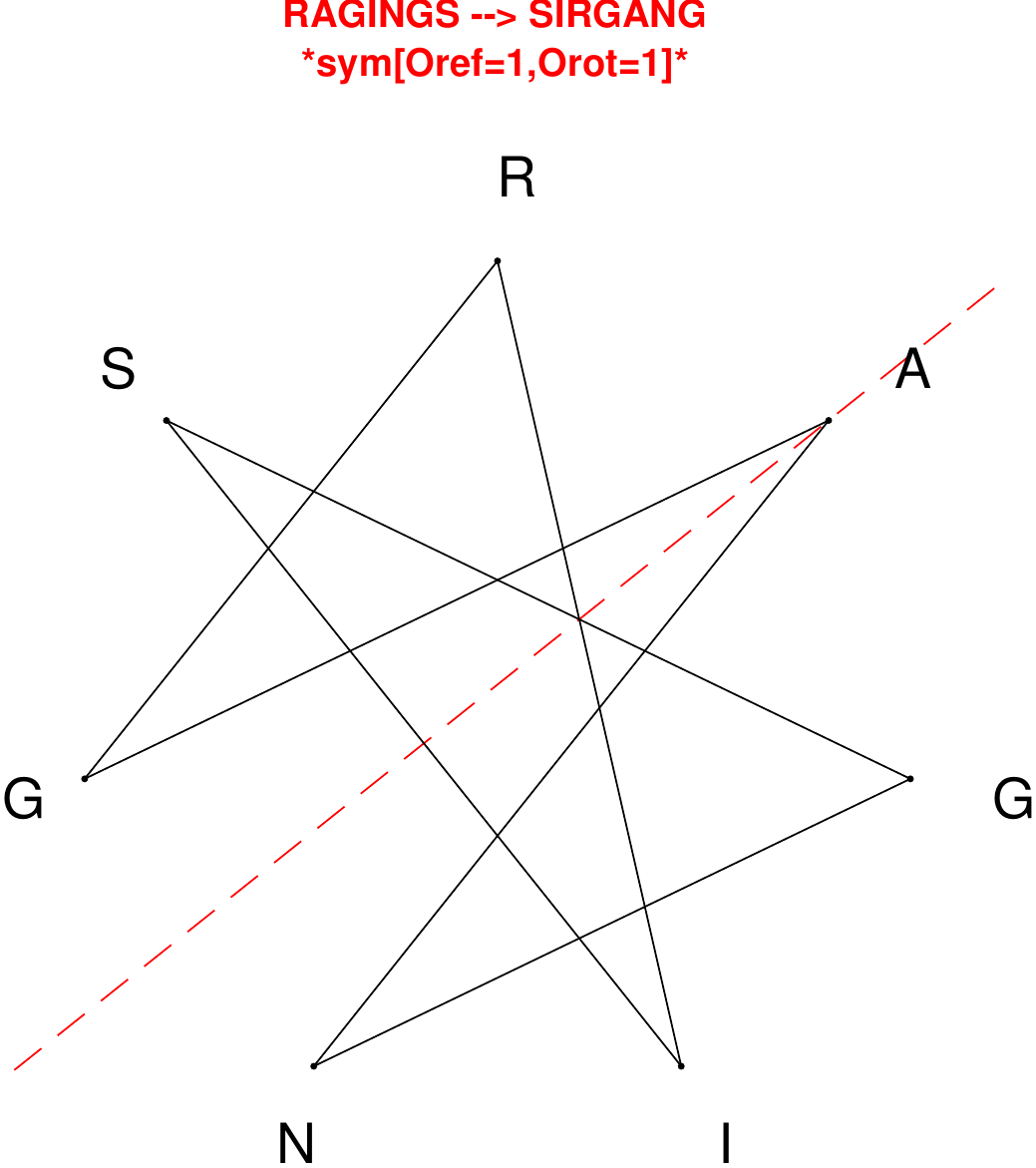}
\end{subfigure}
\hfill
\begin{subfigure}[T]{0.19\textwidth}
\centering
\includegraphics[width=\textwidth]{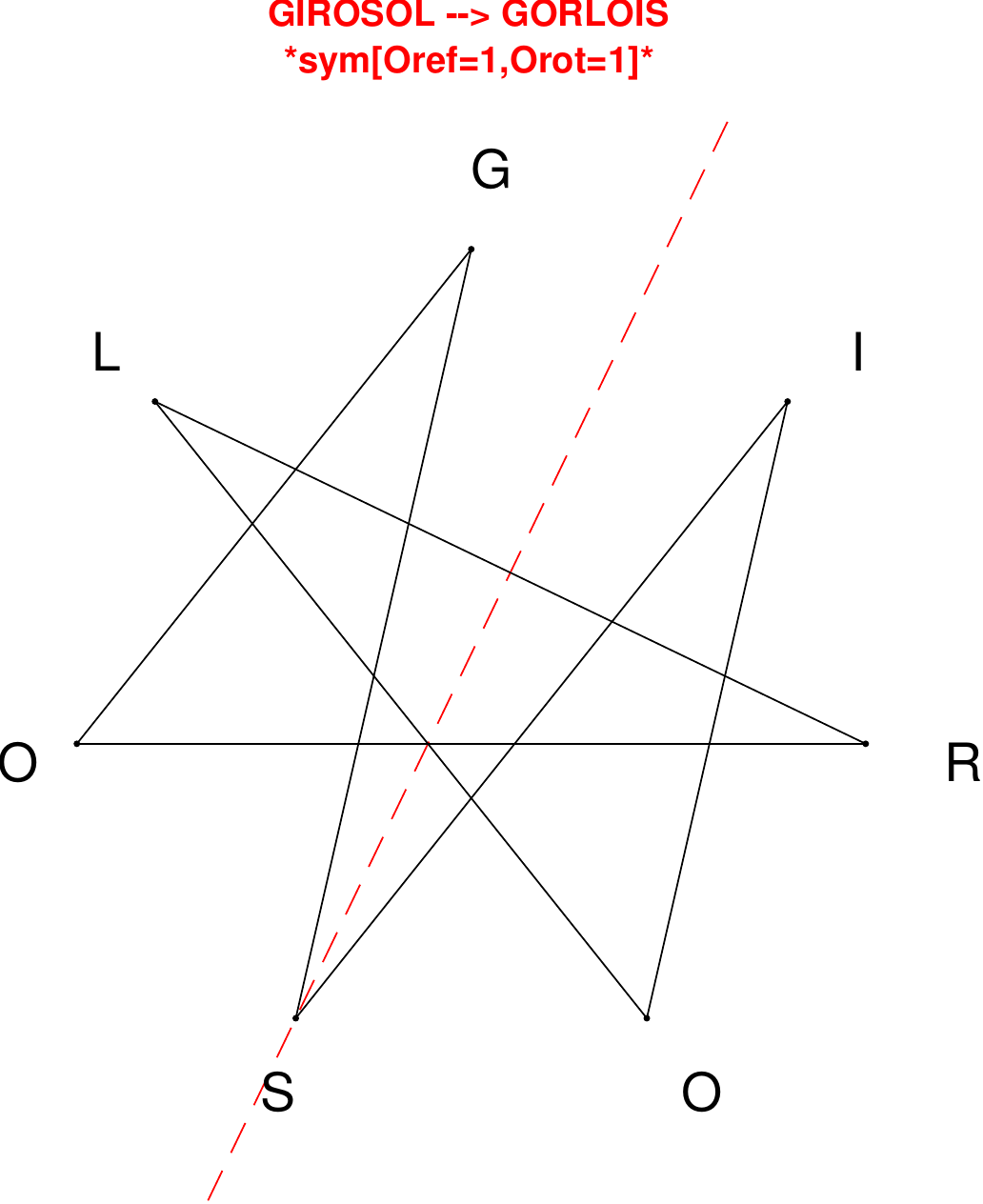}
\end{subfigure}
\hfill
\begin{subfigure}[T]{0.19\textwidth}
\centering
\includegraphics[width=\textwidth]{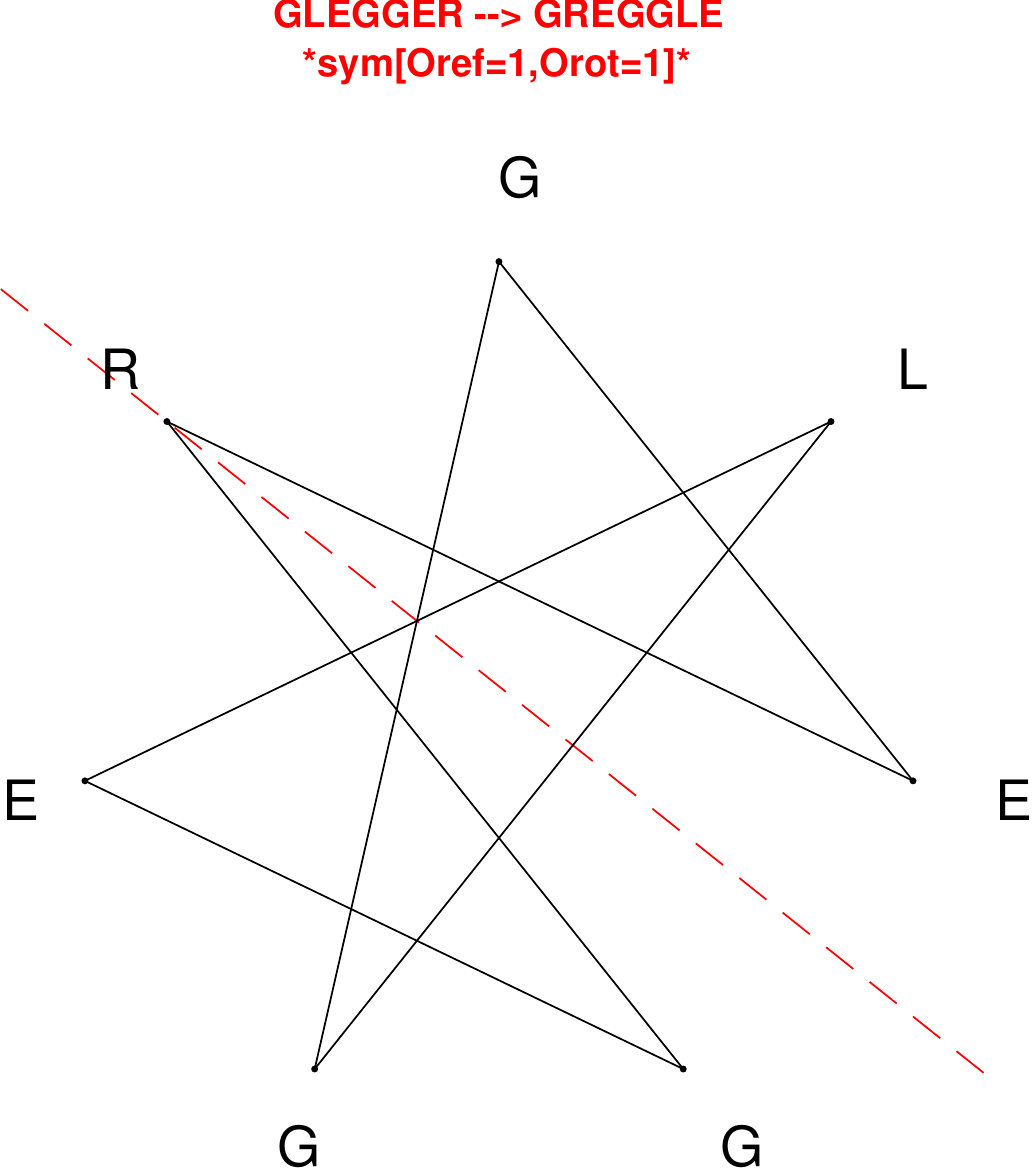}
\end{subfigure}
\hfill
\begin{subfigure}[T]{0.19\textwidth}
\centering
\includegraphics[width=\textwidth]{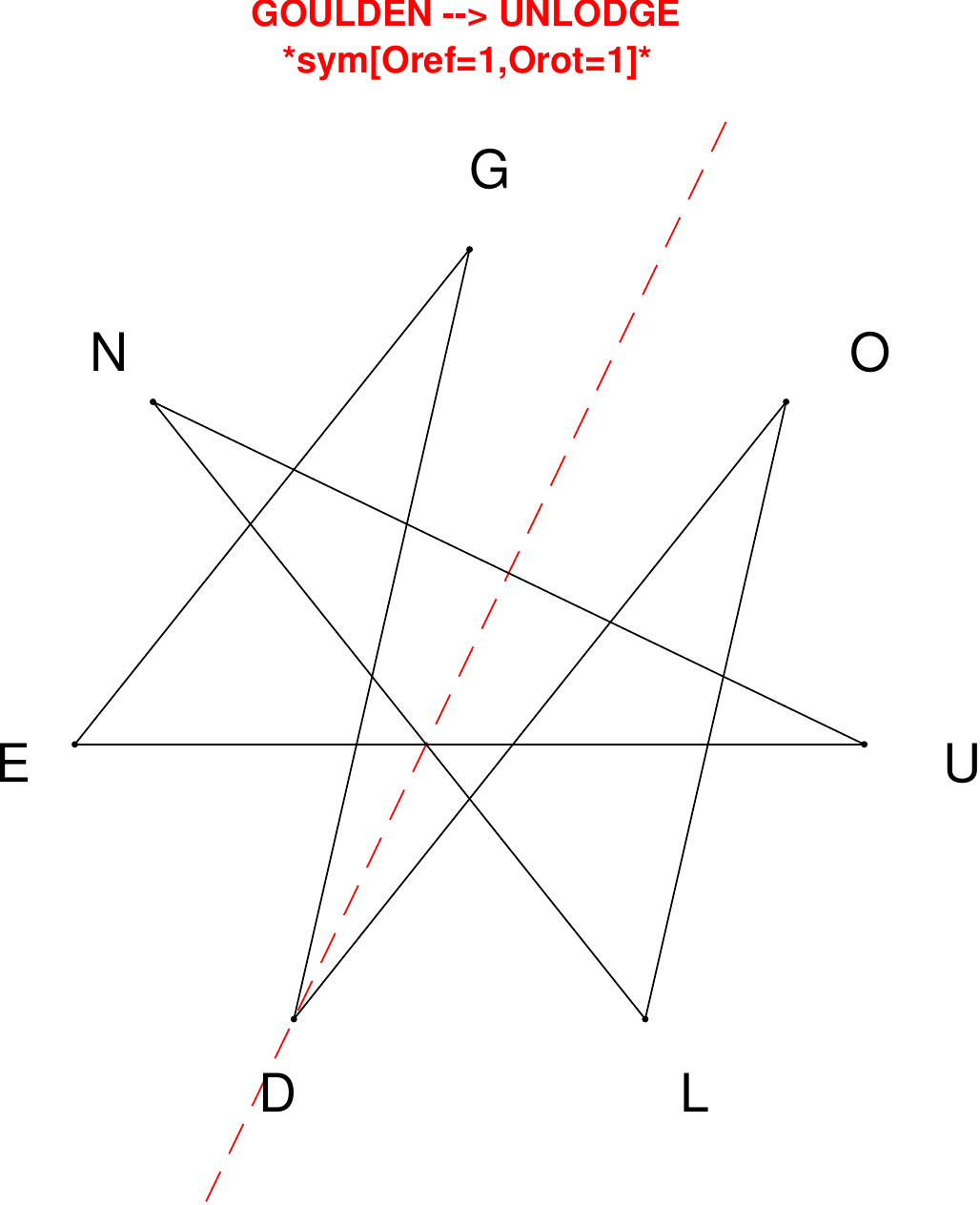}
\end{subfigure}
\hfill
\begin{subfigure}[T]{0.19\textwidth}
\centering
\includegraphics[width=\textwidth]{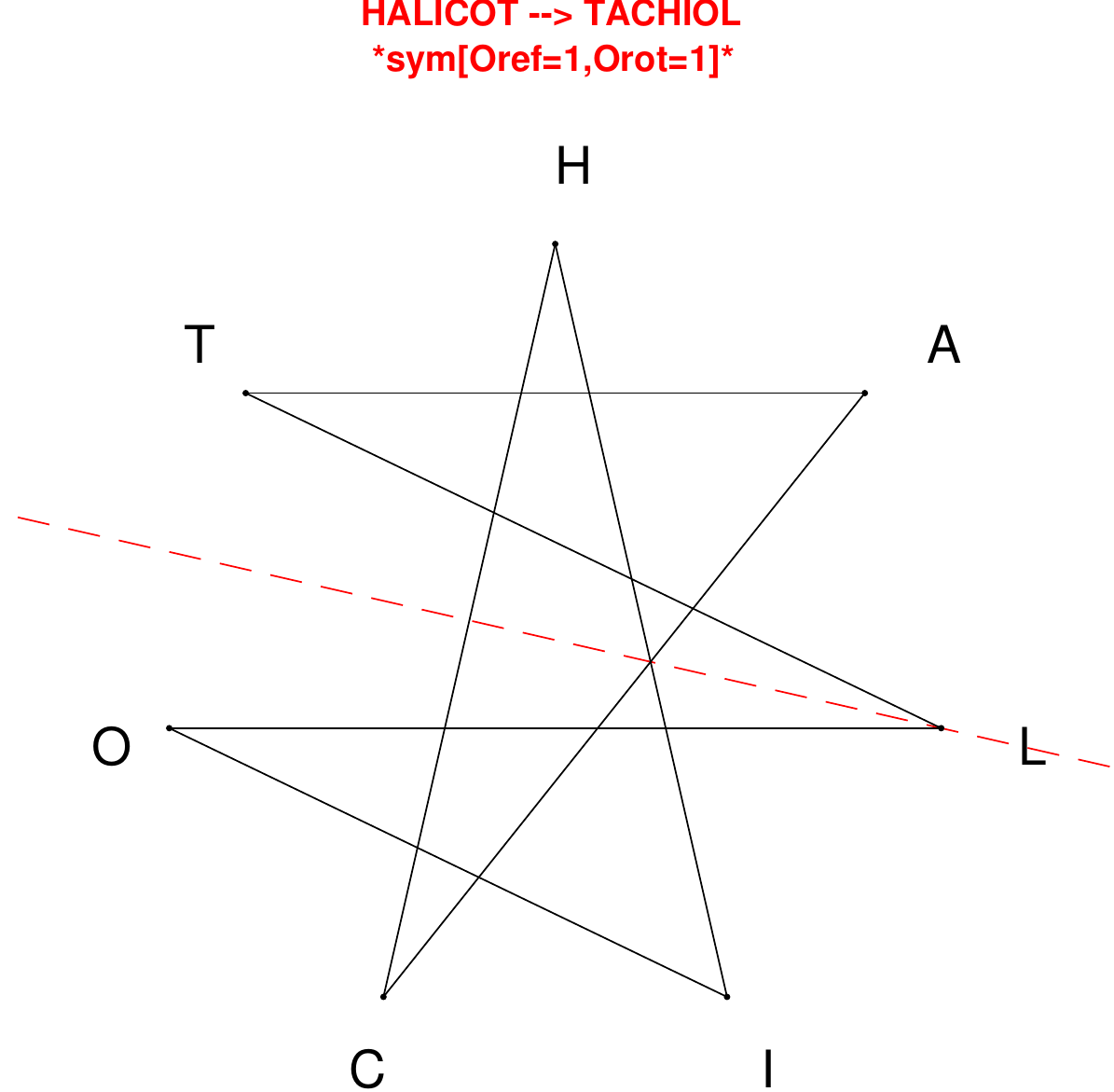}
\end{subfigure}
\end{figure}

\begin{figure}[H]
\centering
\begin{subfigure}[T]{0.19\textwidth}
\centering
\includegraphics[width=\textwidth]{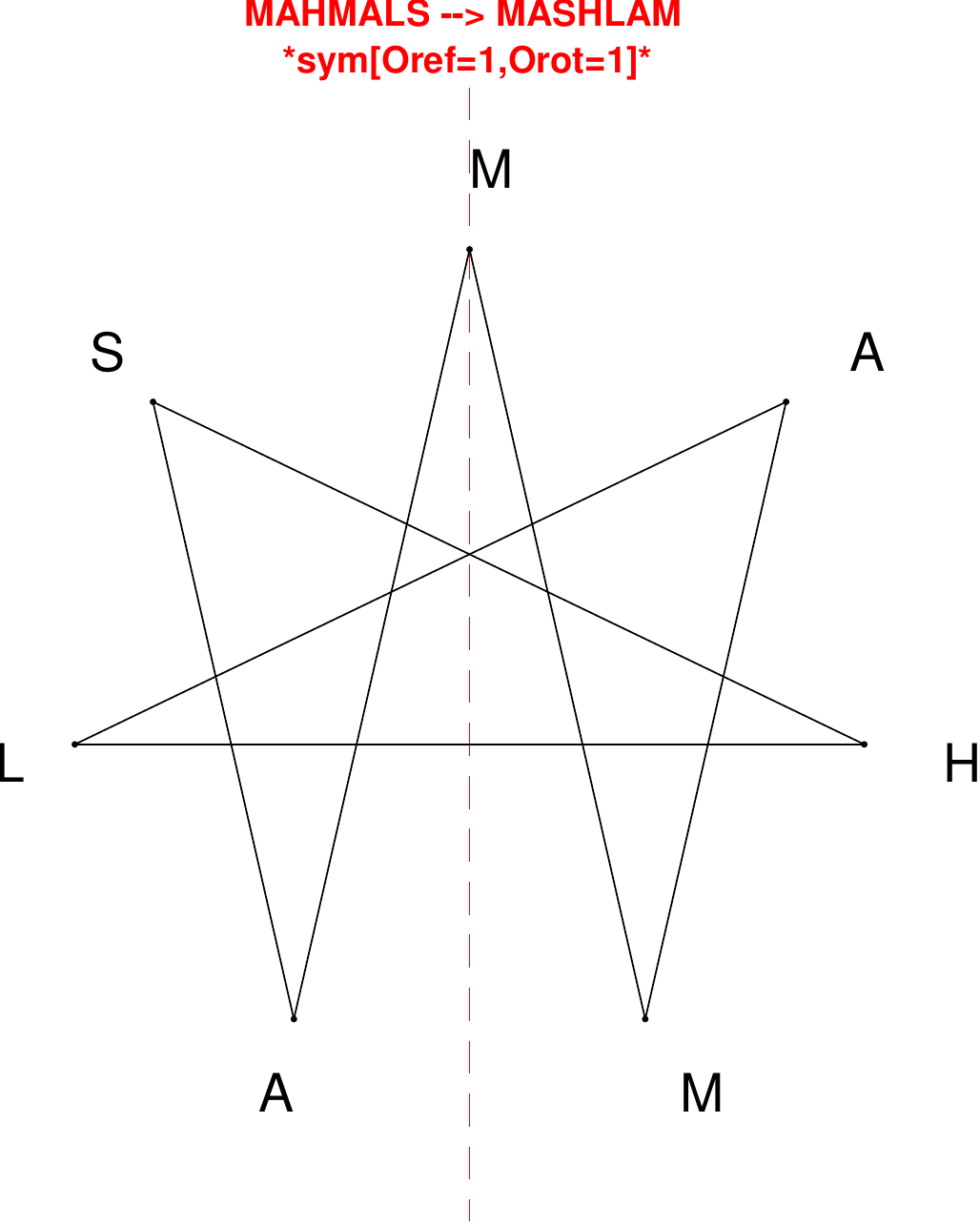}
\end{subfigure}
\hfill
\begin{subfigure}[T]{0.19\textwidth}
\centering
\includegraphics[width=\textwidth]{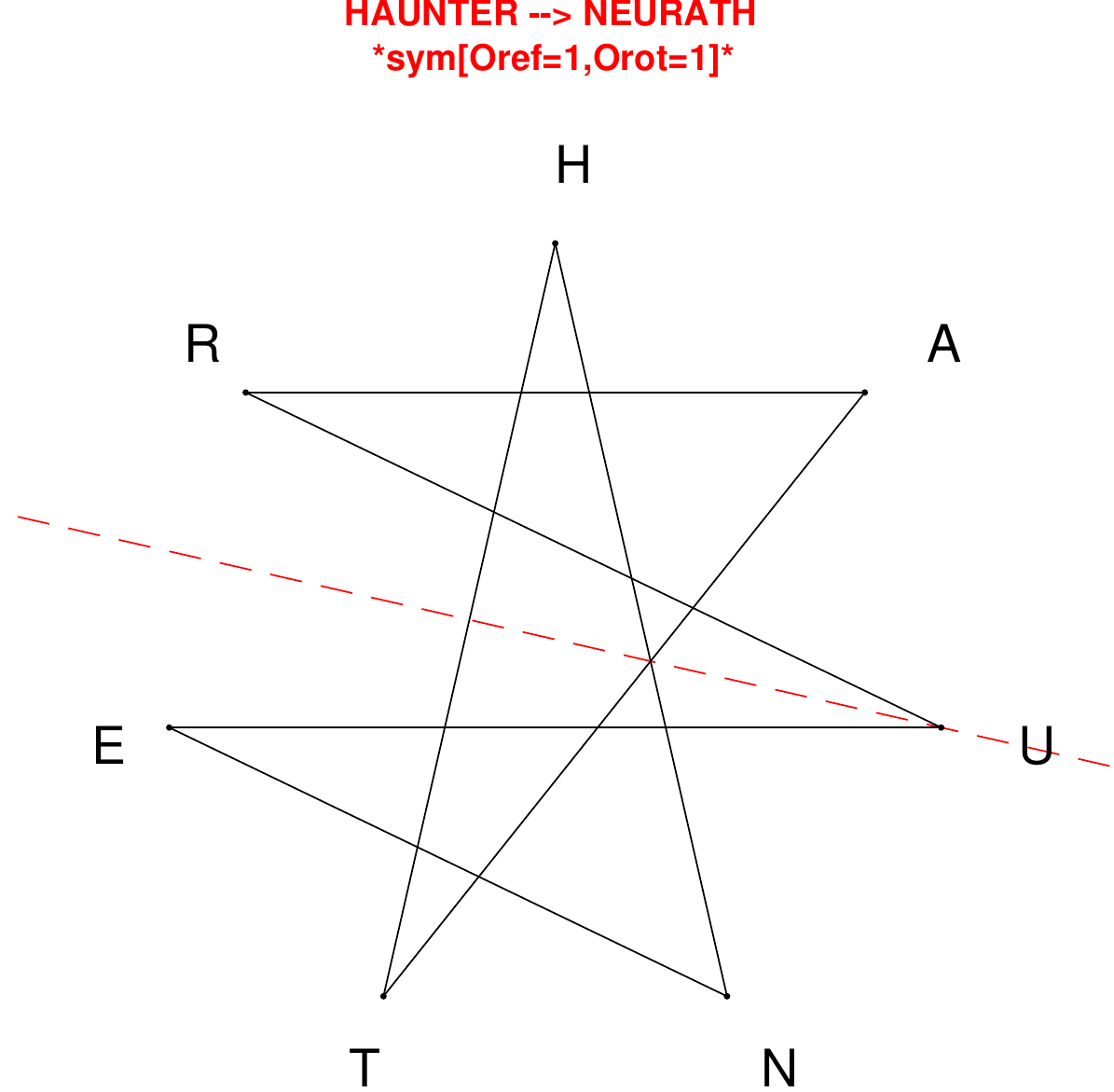}
\end{subfigure}
\hfill
\begin{subfigure}[T]{0.19\textwidth}
\centering
\includegraphics[width=\textwidth]{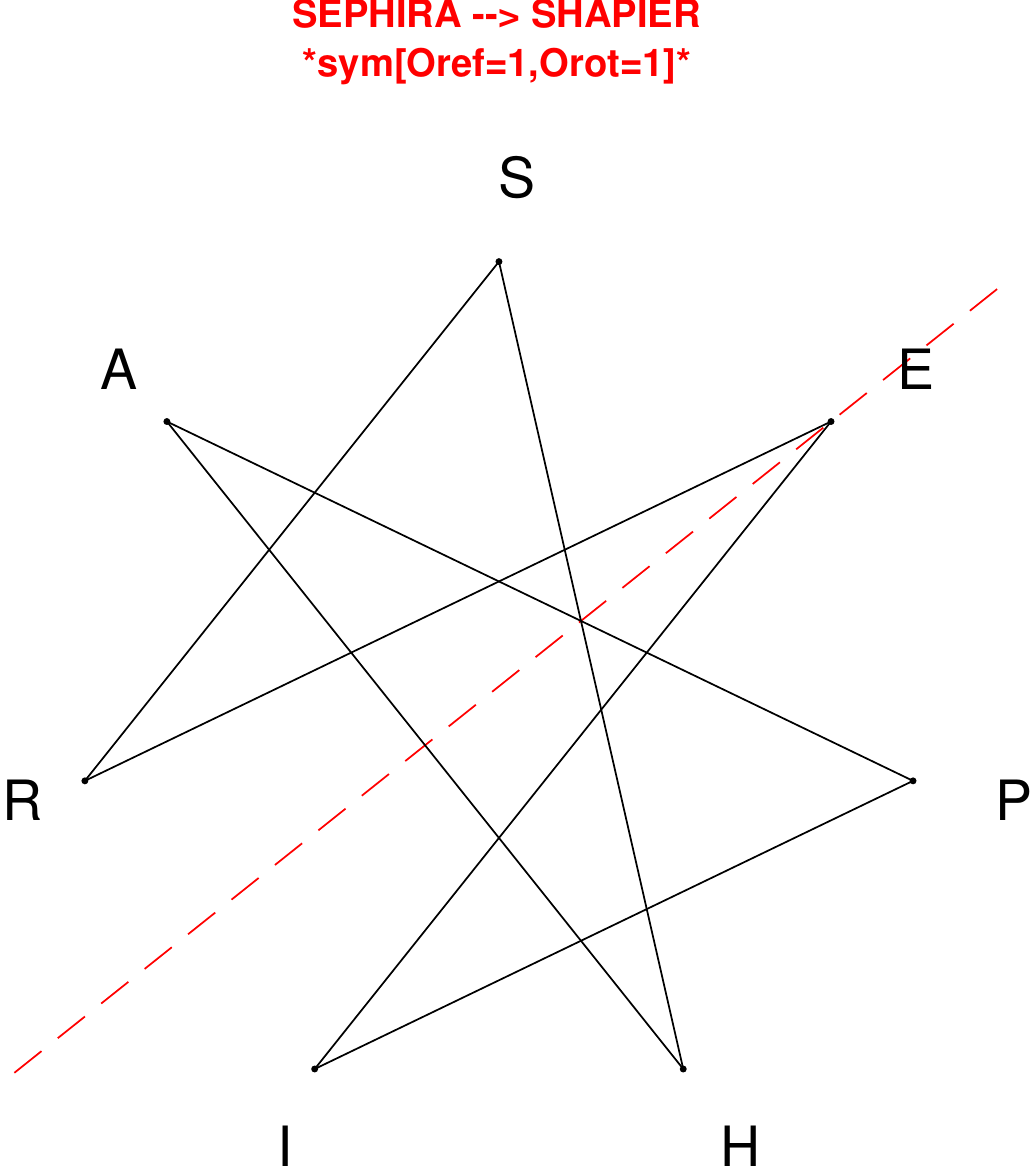}
\end{subfigure}
\hfill
\begin{subfigure}[T]{0.19\textwidth}
\centering
\includegraphics[width=\textwidth]{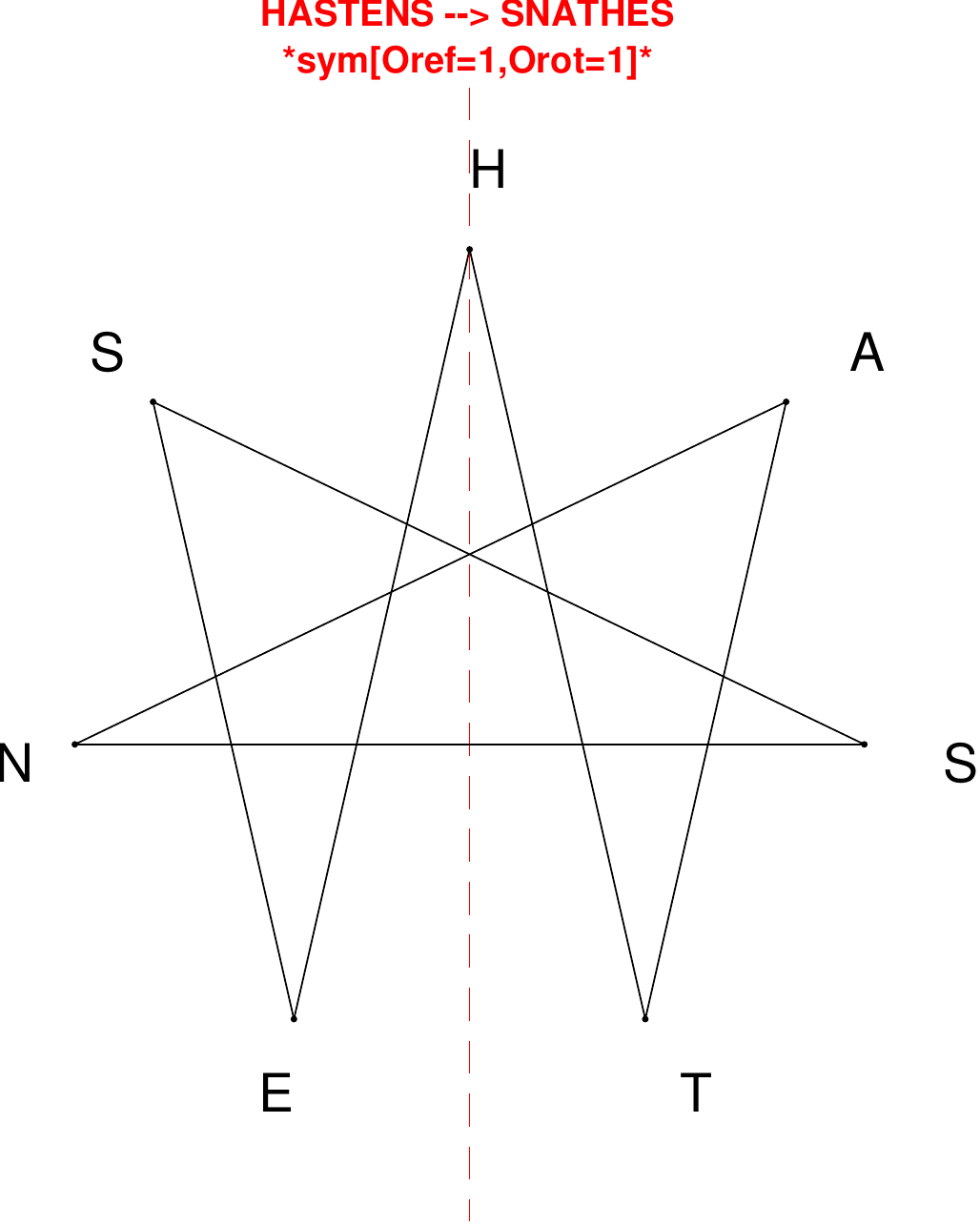}
\end{subfigure}
\hfill
\begin{subfigure}[T]{0.19\textwidth}
\centering
\includegraphics[width=\textwidth]{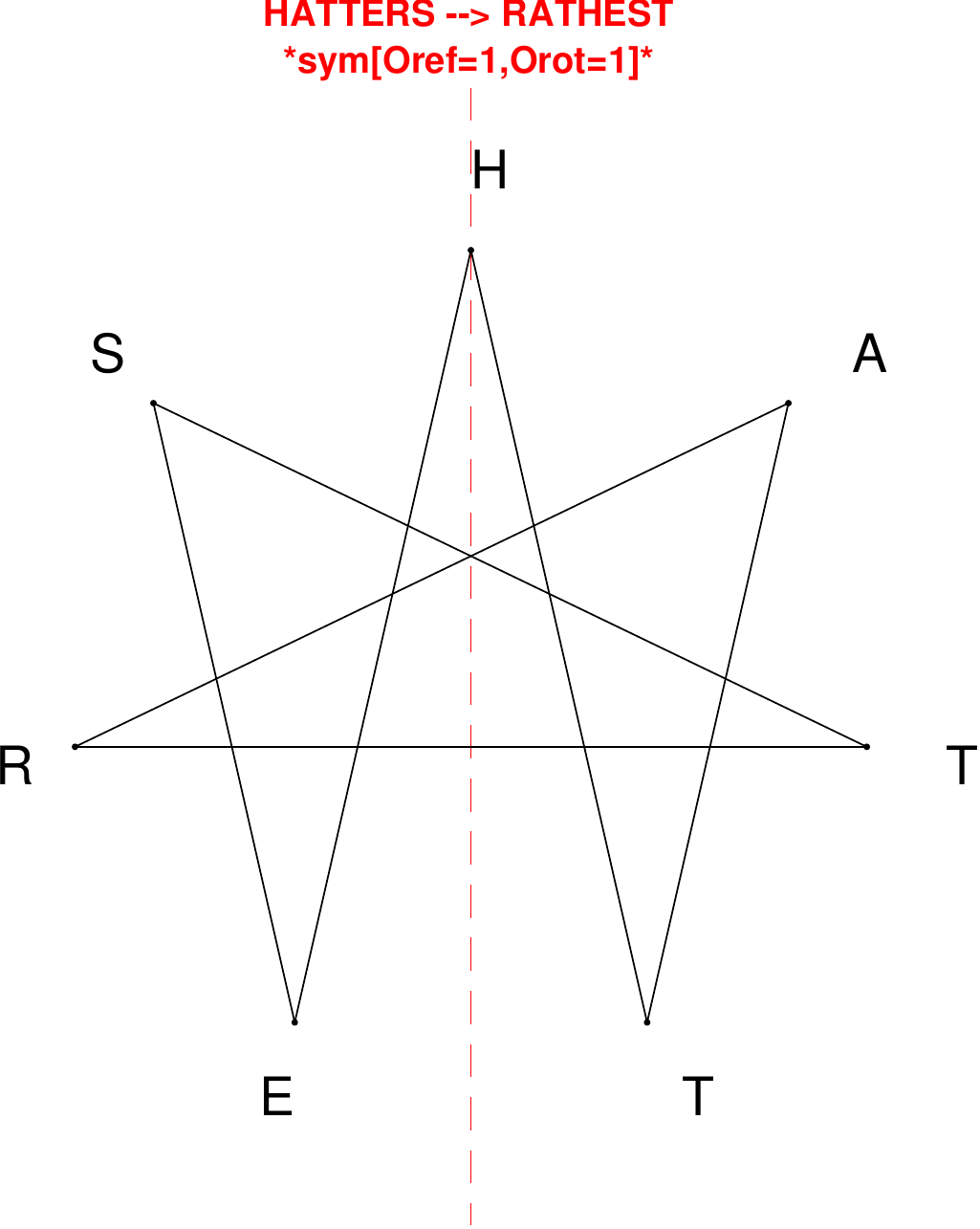}
\end{subfigure}
\end{figure}

\begin{figure}[H]
\centering
\begin{subfigure}[T]{0.19\textwidth}
\centering
\includegraphics[width=\textwidth]{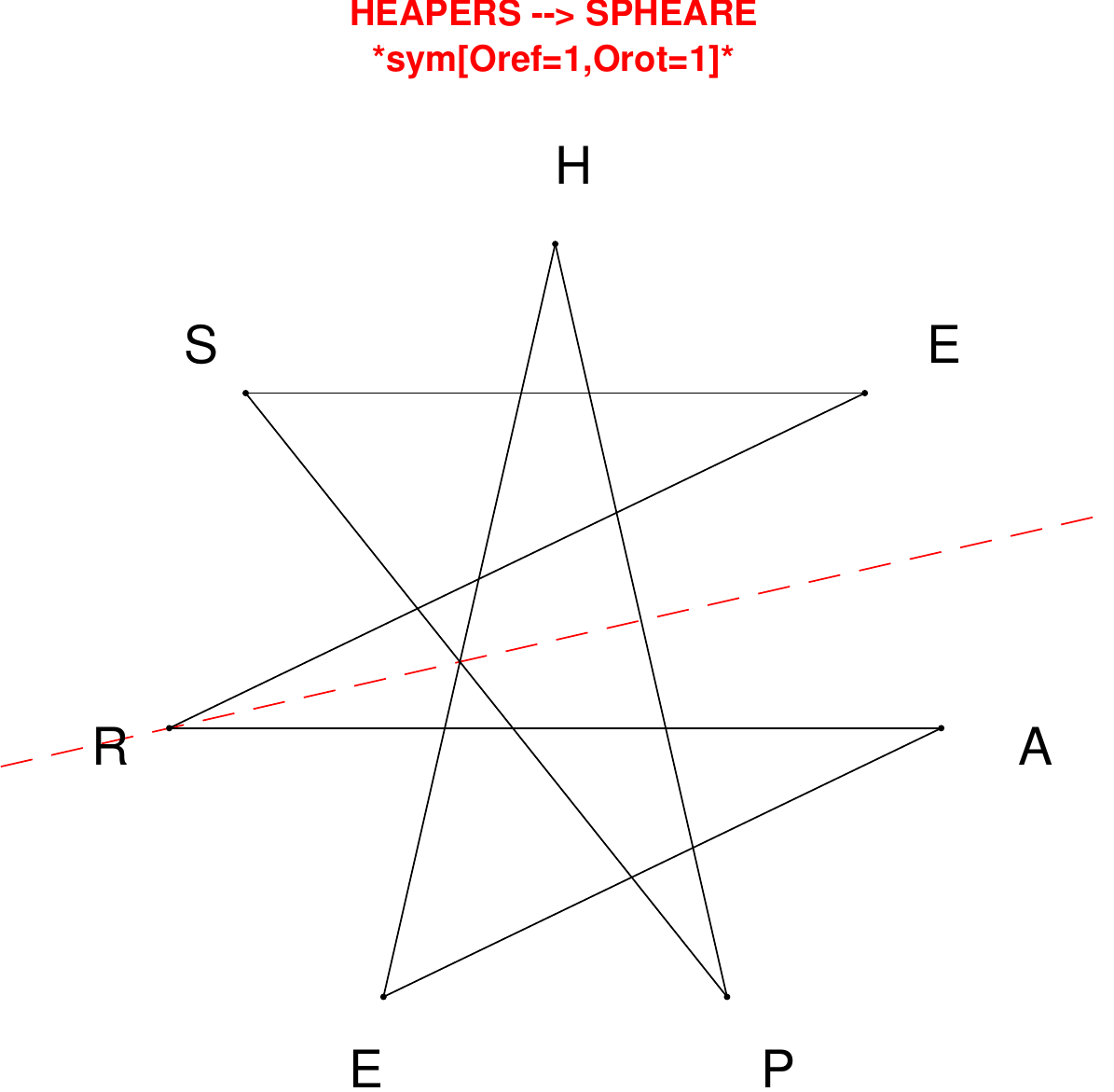}
\end{subfigure}
\hfill
\begin{subfigure}[T]{0.19\textwidth}
\centering
\includegraphics[width=\textwidth]{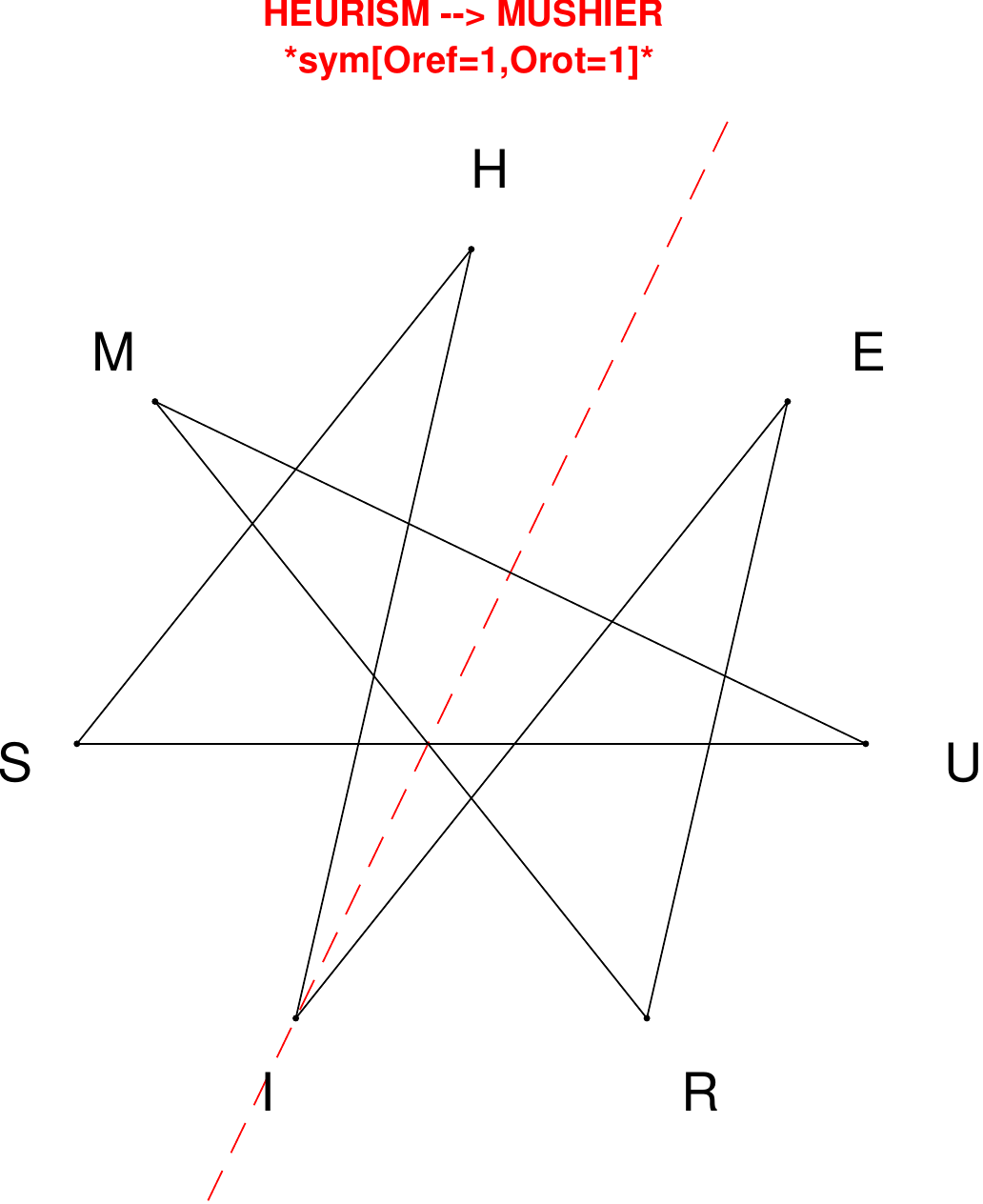}
\end{subfigure}
\hfill
\begin{subfigure}[T]{0.19\textwidth}
\centering
\includegraphics[width=\textwidth]{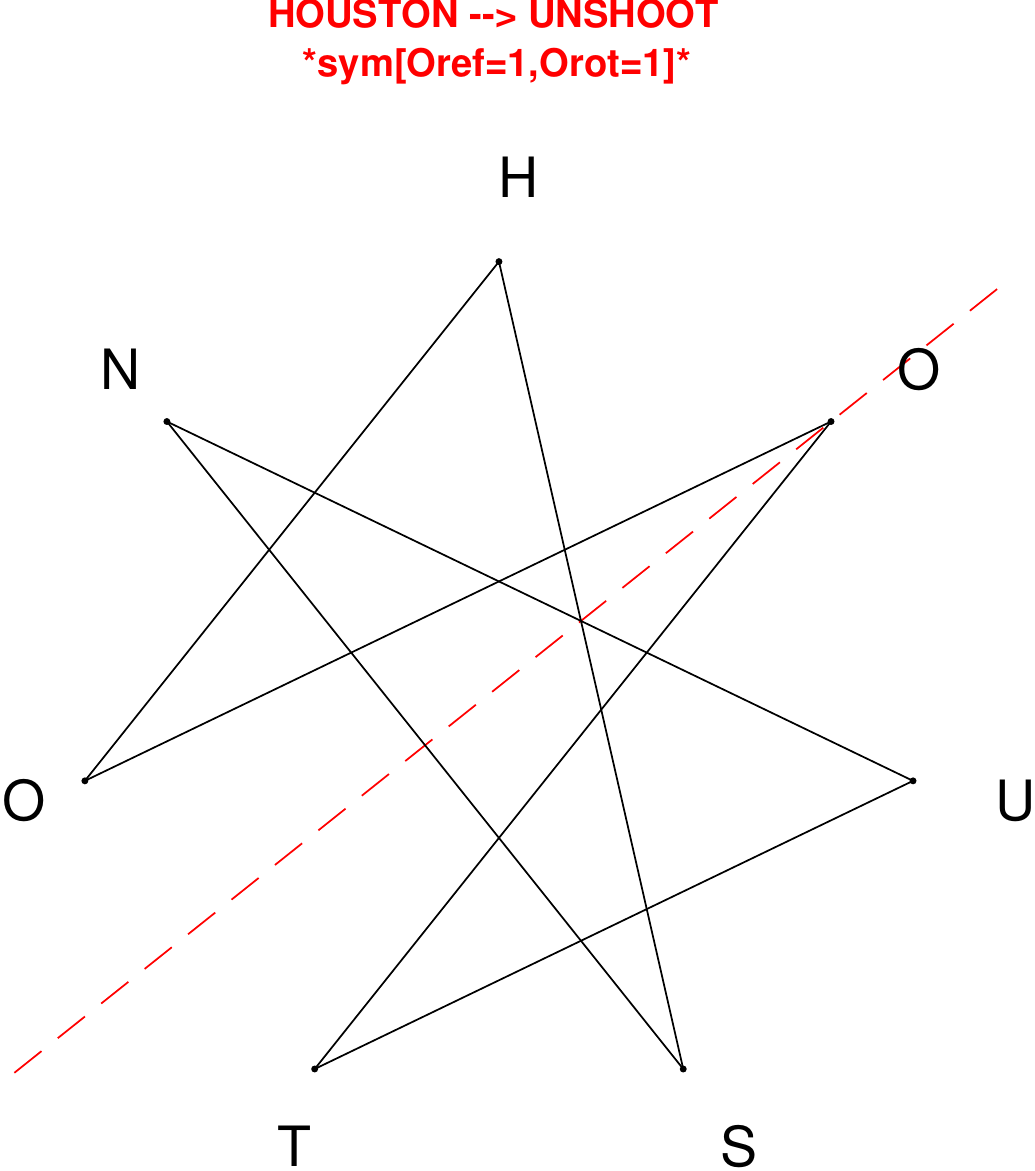}
\end{subfigure}
\hfill
\begin{subfigure}[T]{0.19\textwidth}
\centering
\includegraphics[width=\textwidth]{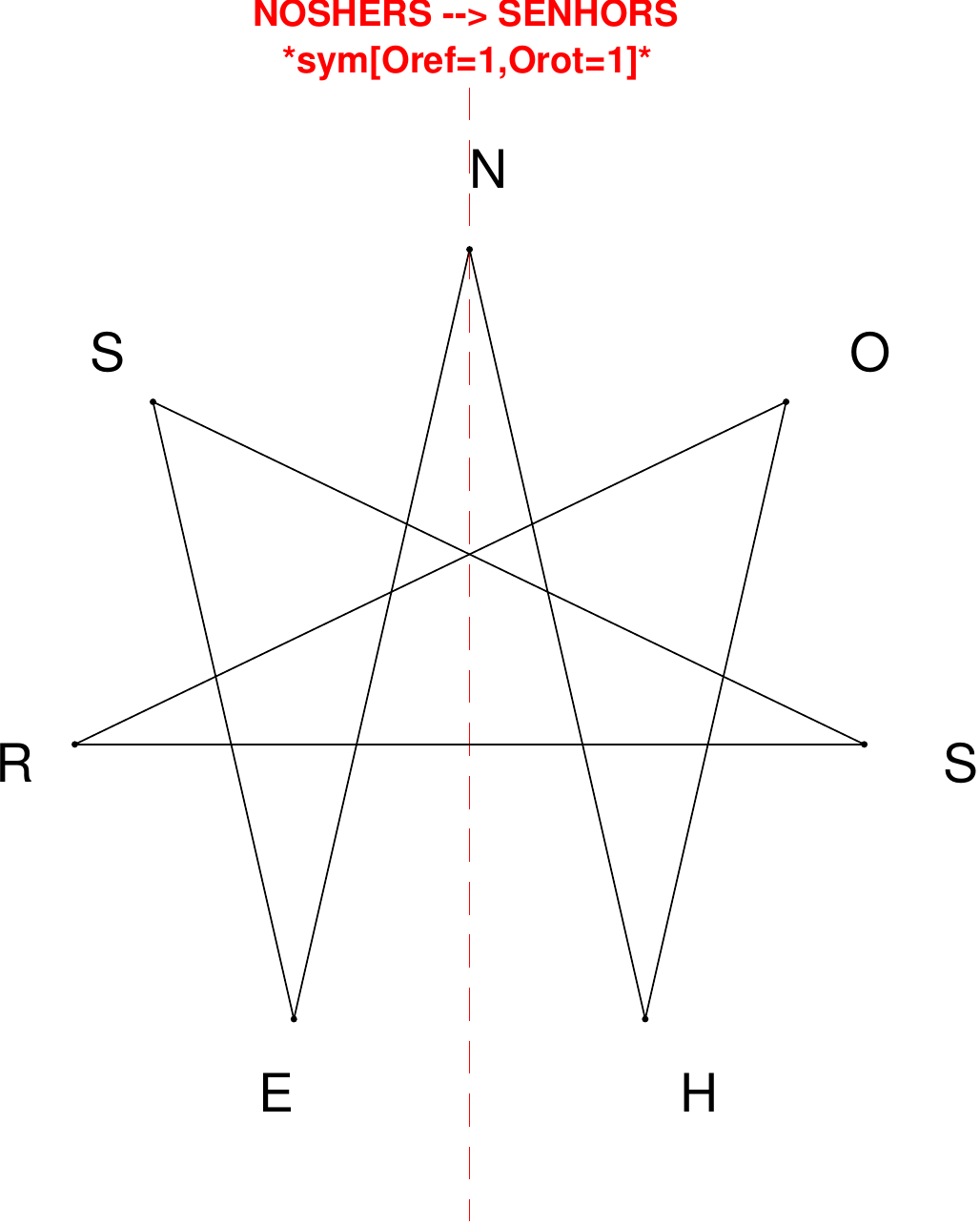}
\end{subfigure}
\hfill
\begin{subfigure}[T]{0.19\textwidth}
\centering
\includegraphics[width=\textwidth]{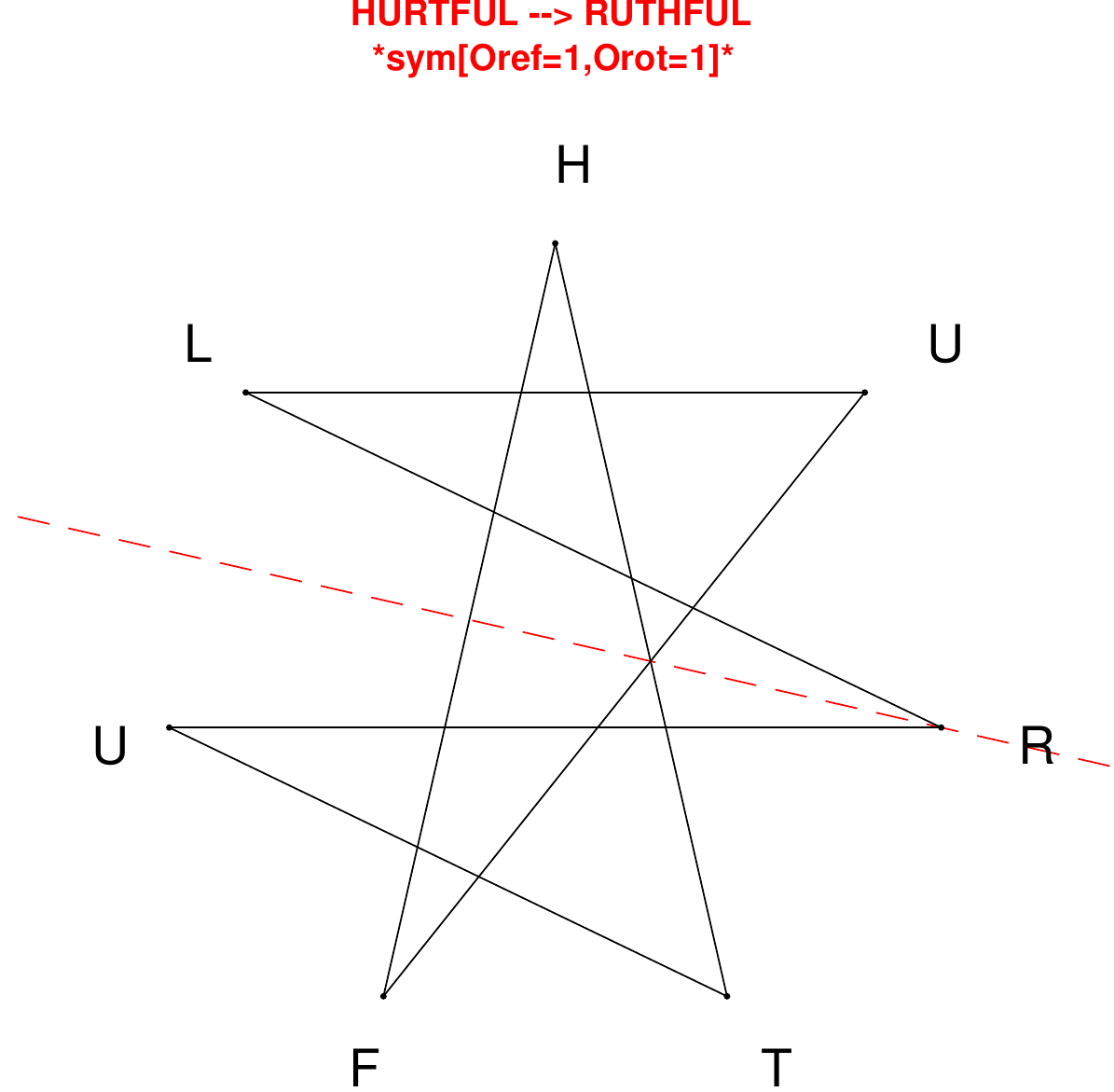}
\end{subfigure}
\end{figure}

\begin{figure}[H]
\centering
\begin{subfigure}[T]{0.19\textwidth}
\centering
\includegraphics[width=\textwidth]{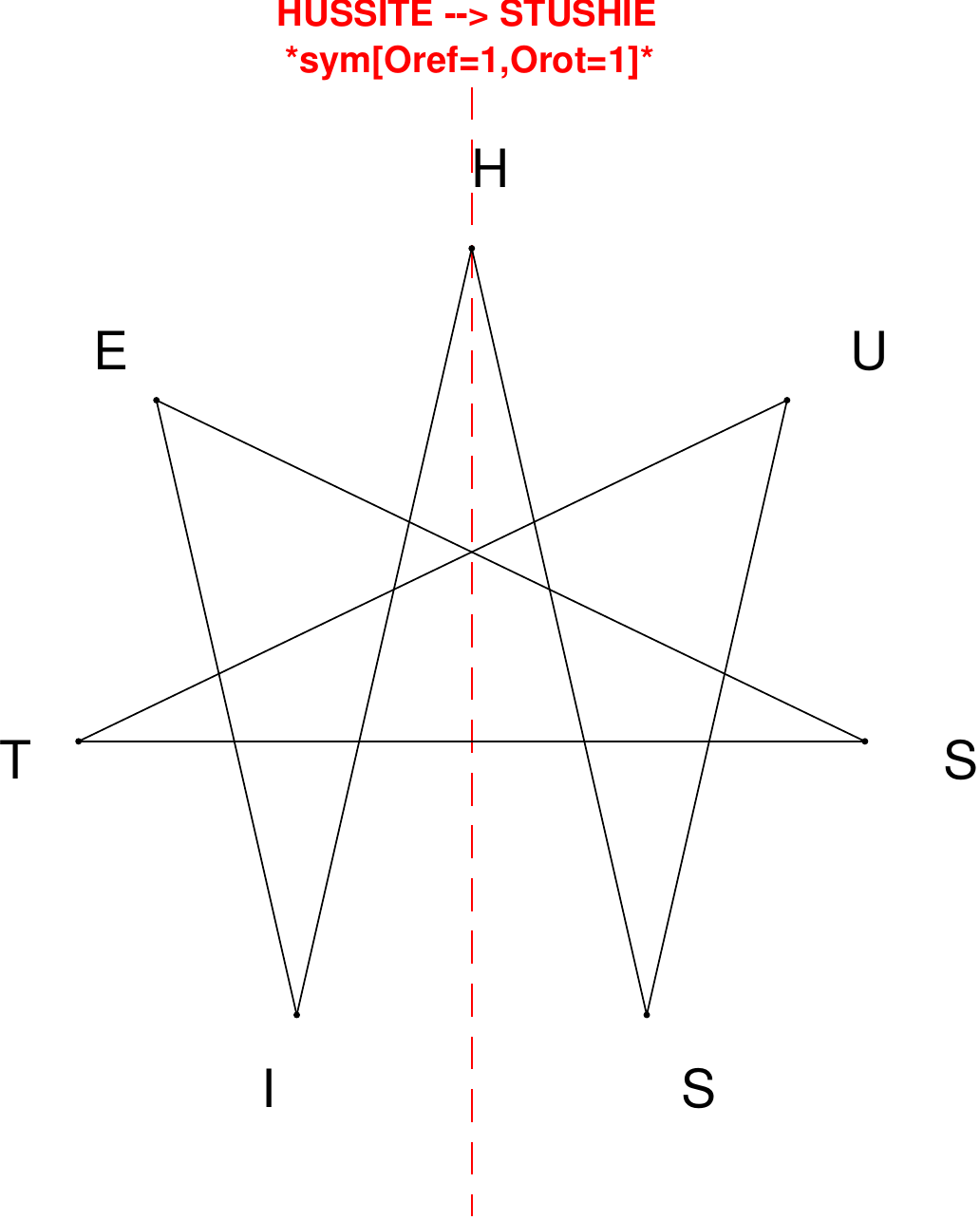}
\end{subfigure}
\hfill
\begin{subfigure}[T]{0.19\textwidth}
\centering
\includegraphics[width=\textwidth]{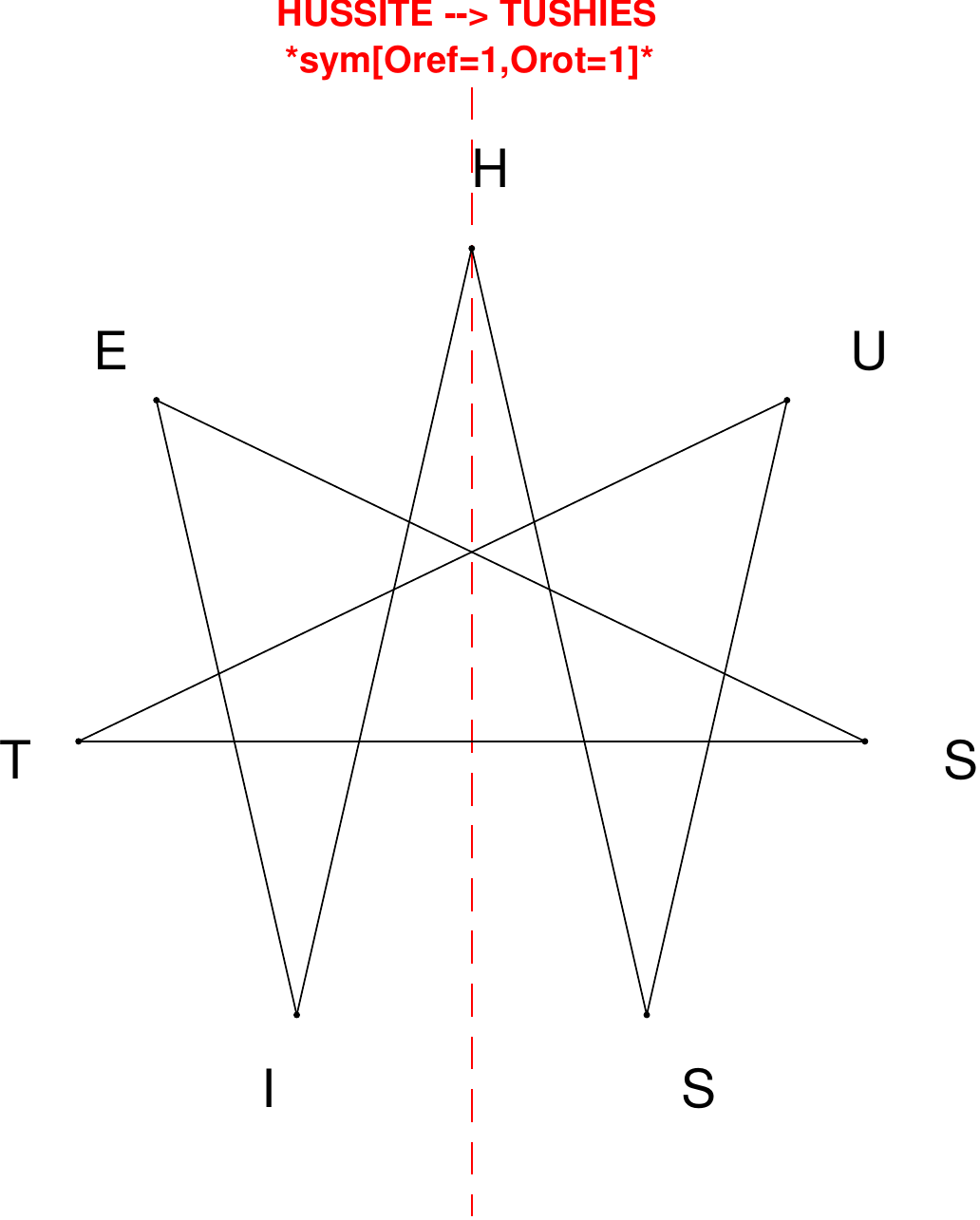}
\end{subfigure}
\hfill
\begin{subfigure}[T]{0.19\textwidth}
\centering
\includegraphics[width=\textwidth]{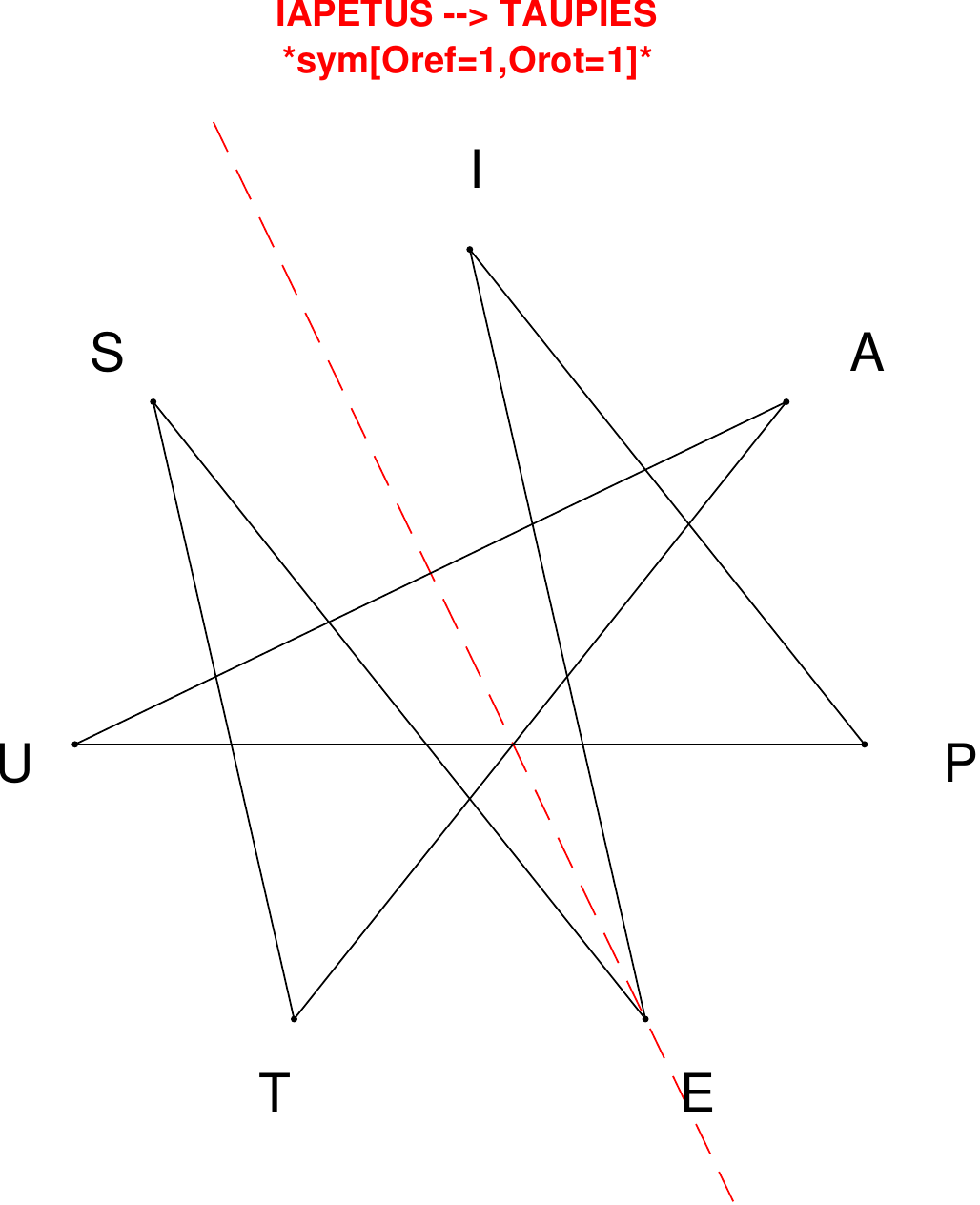}
\end{subfigure}
\hfill
\begin{subfigure}[T]{0.19\textwidth}
\centering
\includegraphics[width=\textwidth]{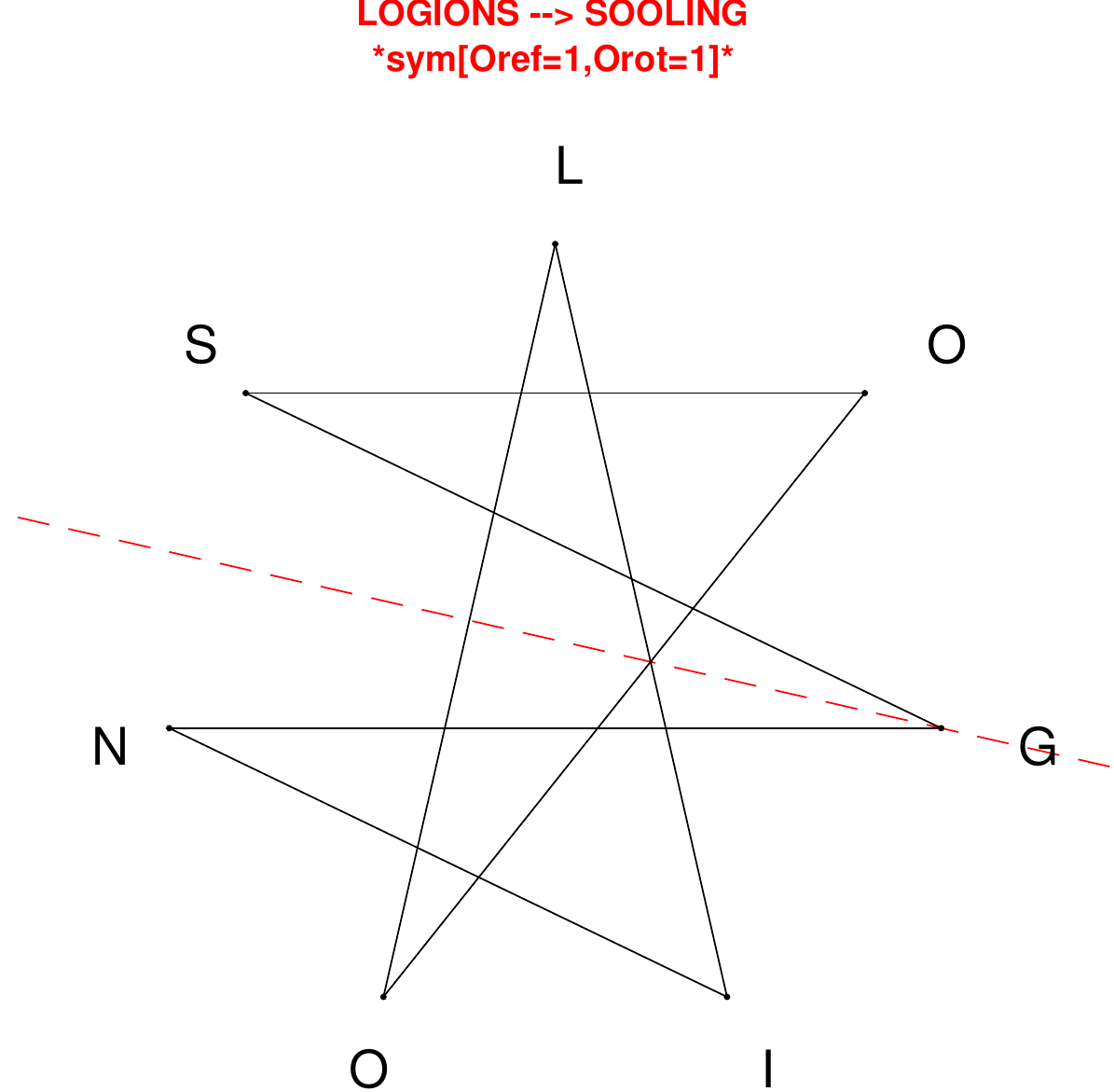}
\end{subfigure}
\hfill
\begin{subfigure}[T]{0.19\textwidth}
\centering
\includegraphics[width=\textwidth]{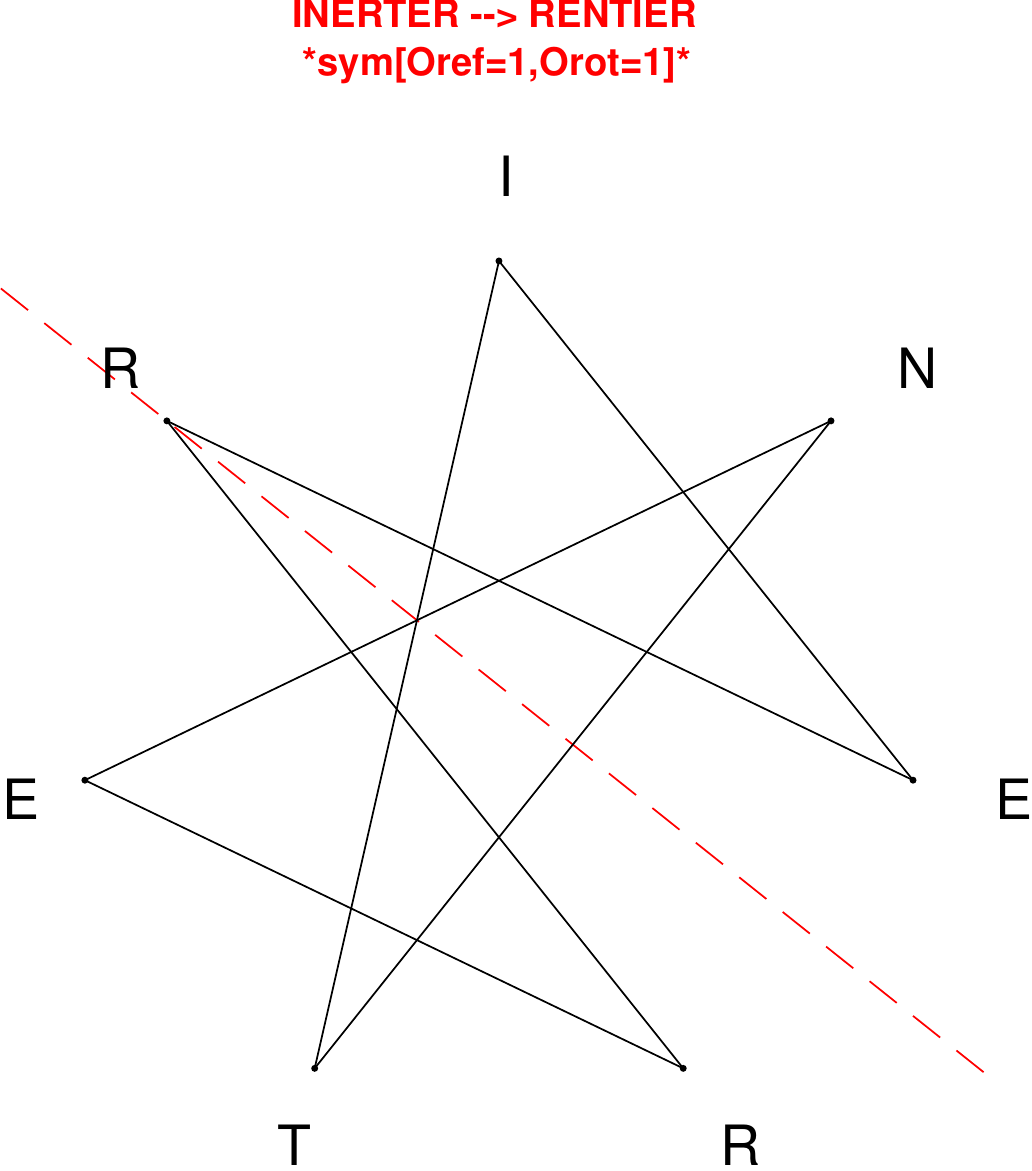}
\end{subfigure}
\end{figure}

\begin{figure}[H]
\centering
\begin{subfigure}[T]{0.19\textwidth}
\centering
\includegraphics[width=\textwidth]{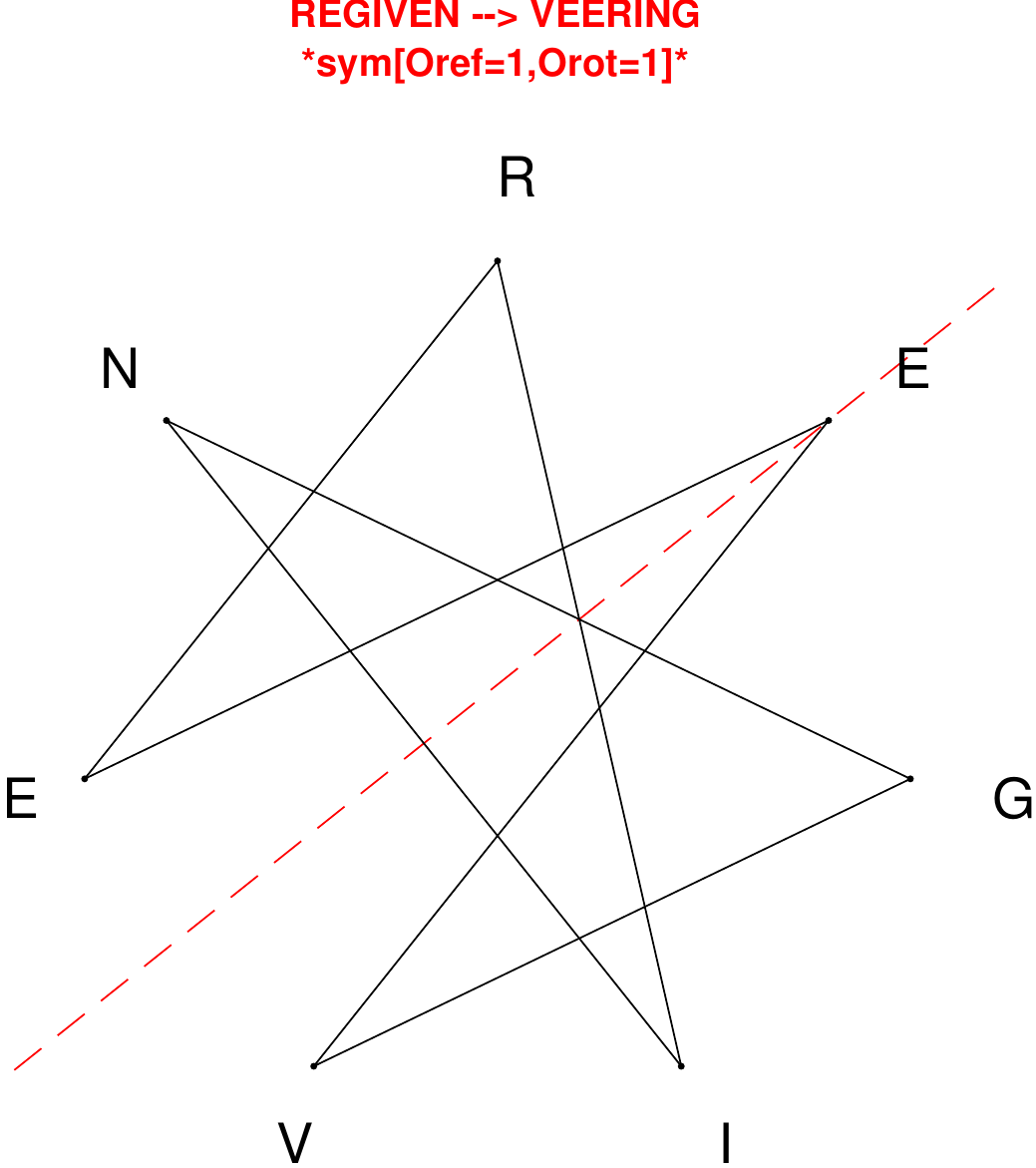}
\end{subfigure}
\hfill
\begin{subfigure}[T]{0.19\textwidth}
\centering
\includegraphics[width=\textwidth]{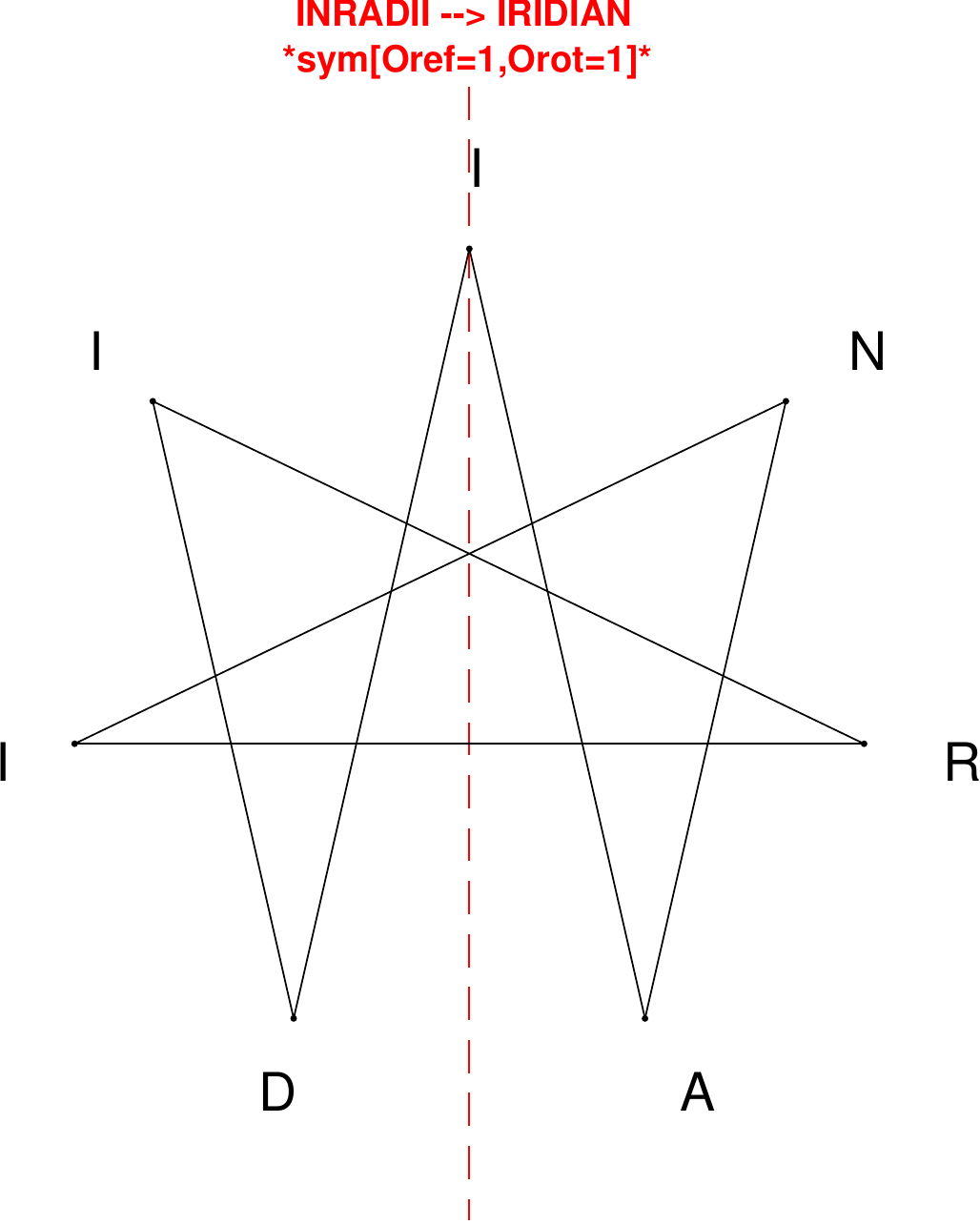}
\end{subfigure}
\hfill
\begin{subfigure}[T]{0.19\textwidth}
\centering
\includegraphics[width=\textwidth]{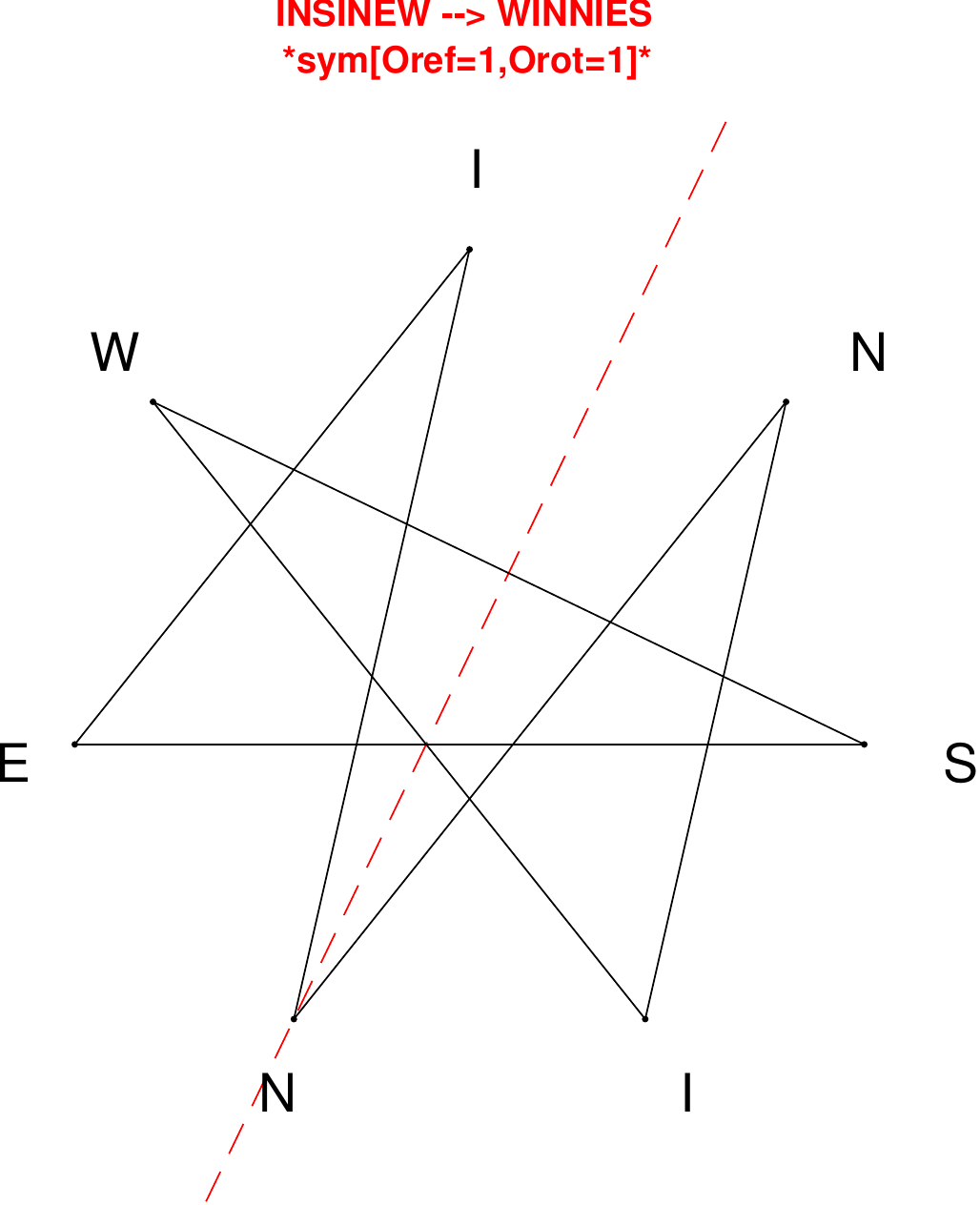}
\end{subfigure}
\hfill
\begin{subfigure}[T]{0.19\textwidth}
\centering
\includegraphics[width=\textwidth]{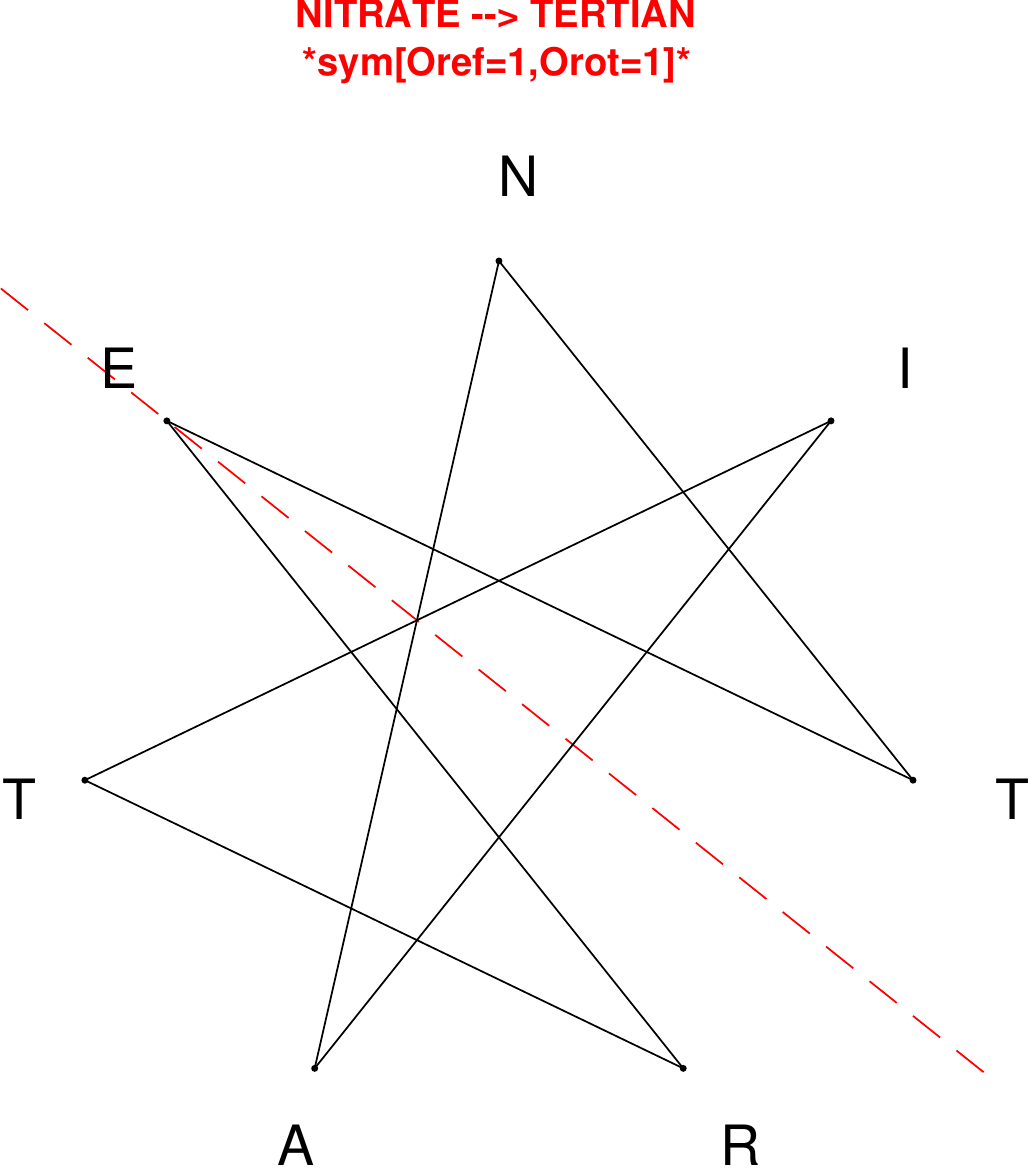}
\end{subfigure}
\hfill
\begin{subfigure}[T]{0.19\textwidth}
\centering
\includegraphics[width=\textwidth]{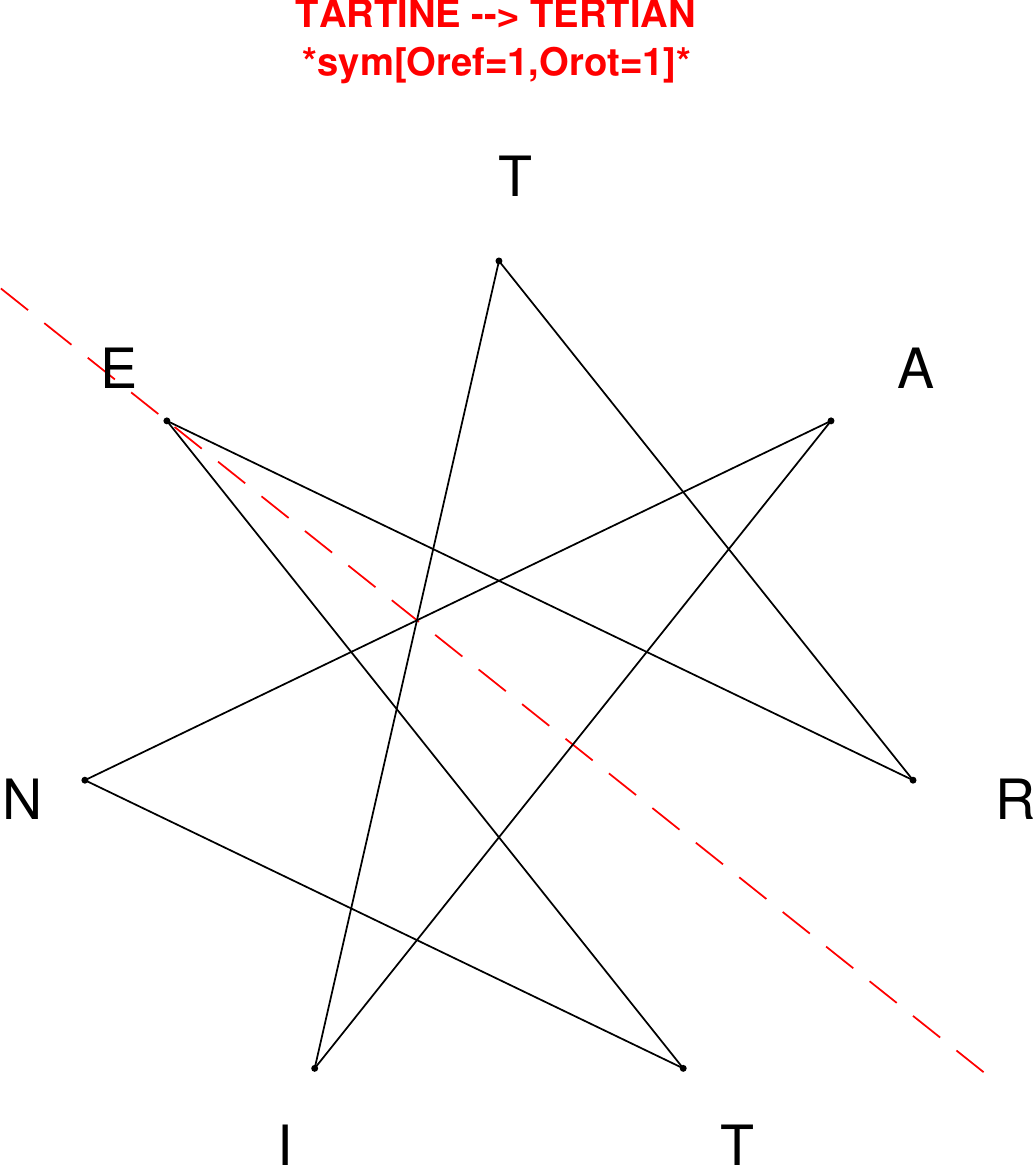}
\end{subfigure}
\end{figure}

\begin{figure}[H]
\centering
\begin{subfigure}[T]{0.19\textwidth}
\centering
\includegraphics[width=\textwidth]{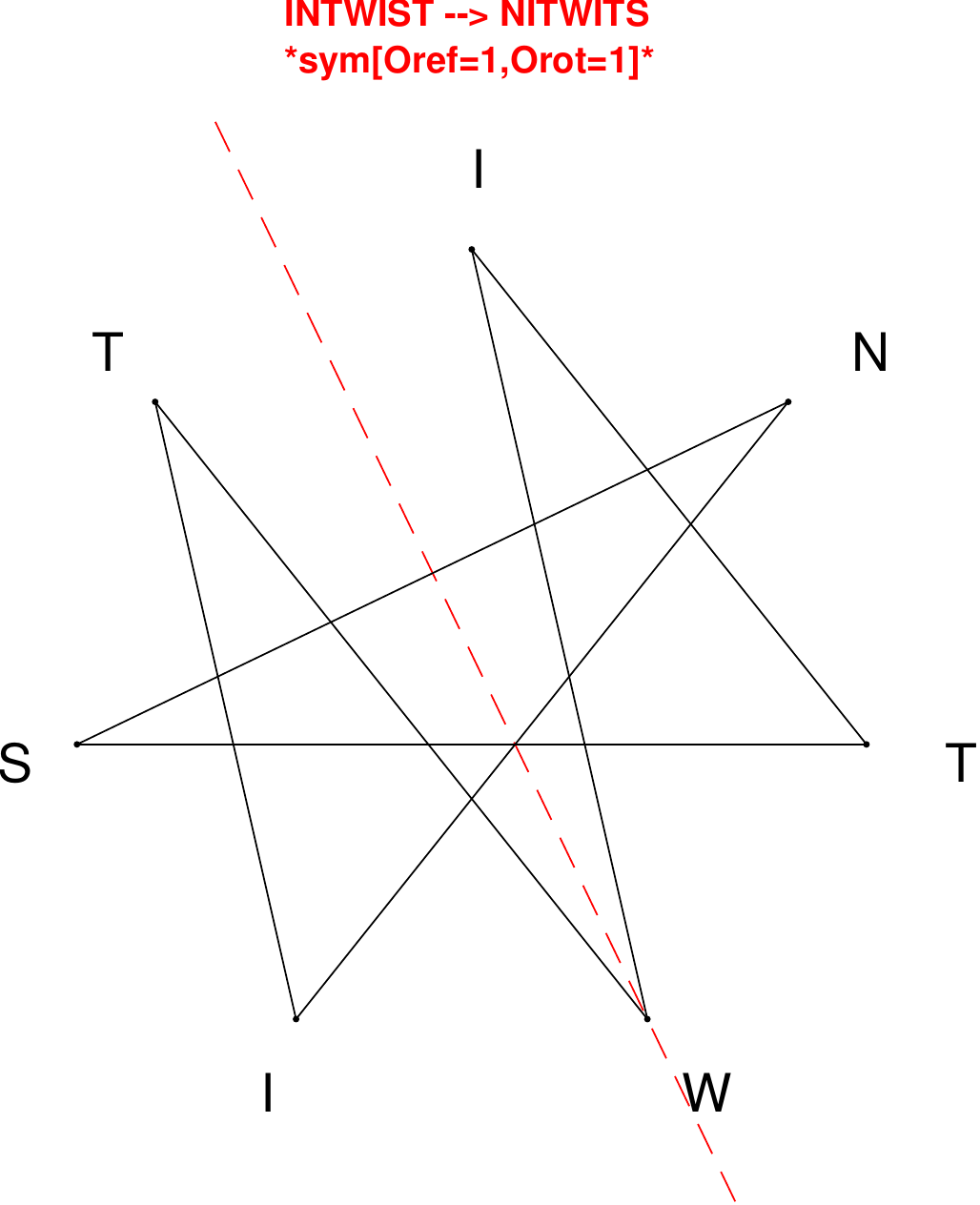}
\end{subfigure}
\hfill
\begin{subfigure}[T]{0.19\textwidth}
\centering
\includegraphics[width=\textwidth]{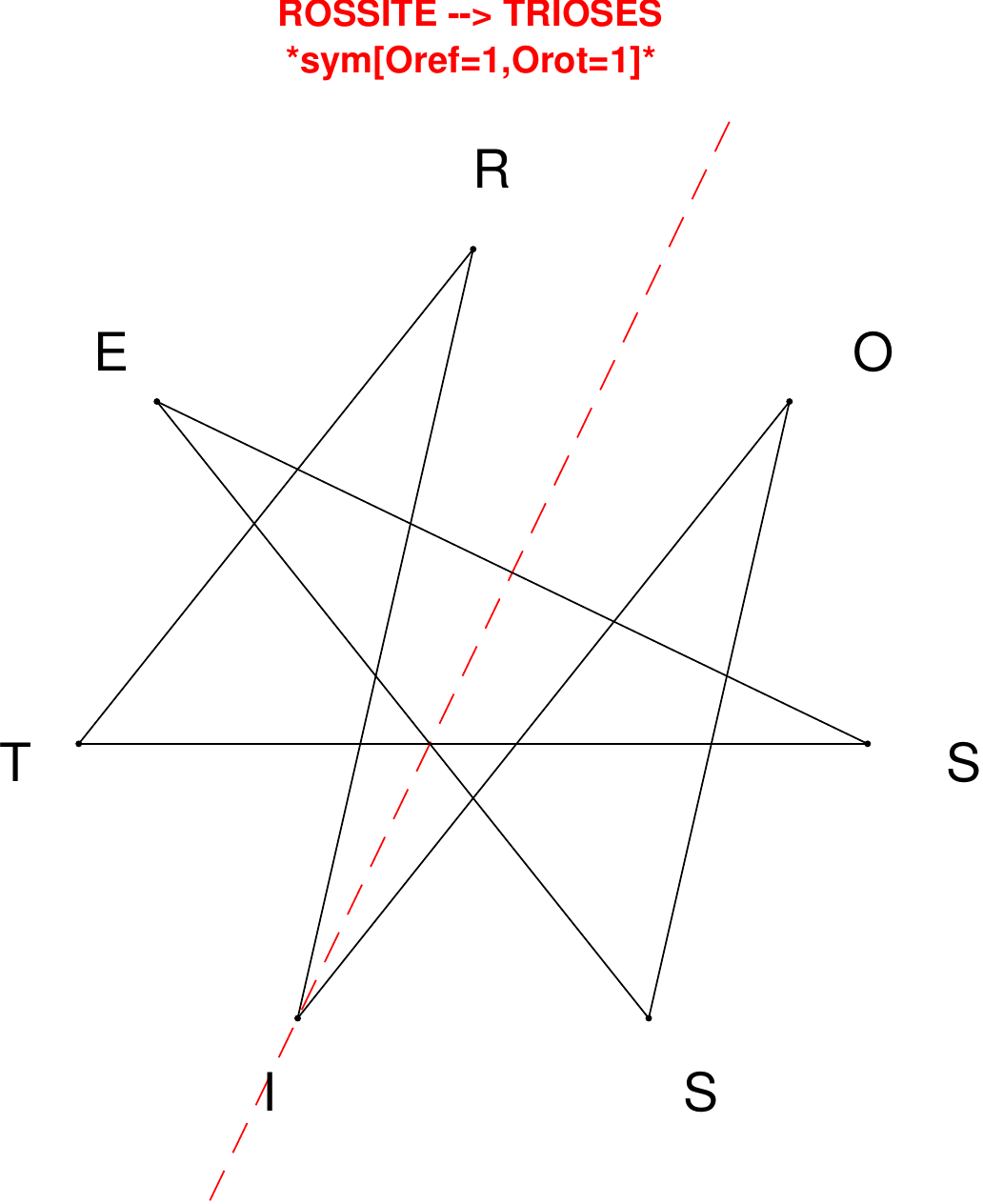}
\end{subfigure}
\hfill
\begin{subfigure}[T]{0.19\textwidth}
\centering
\includegraphics[width=\textwidth]{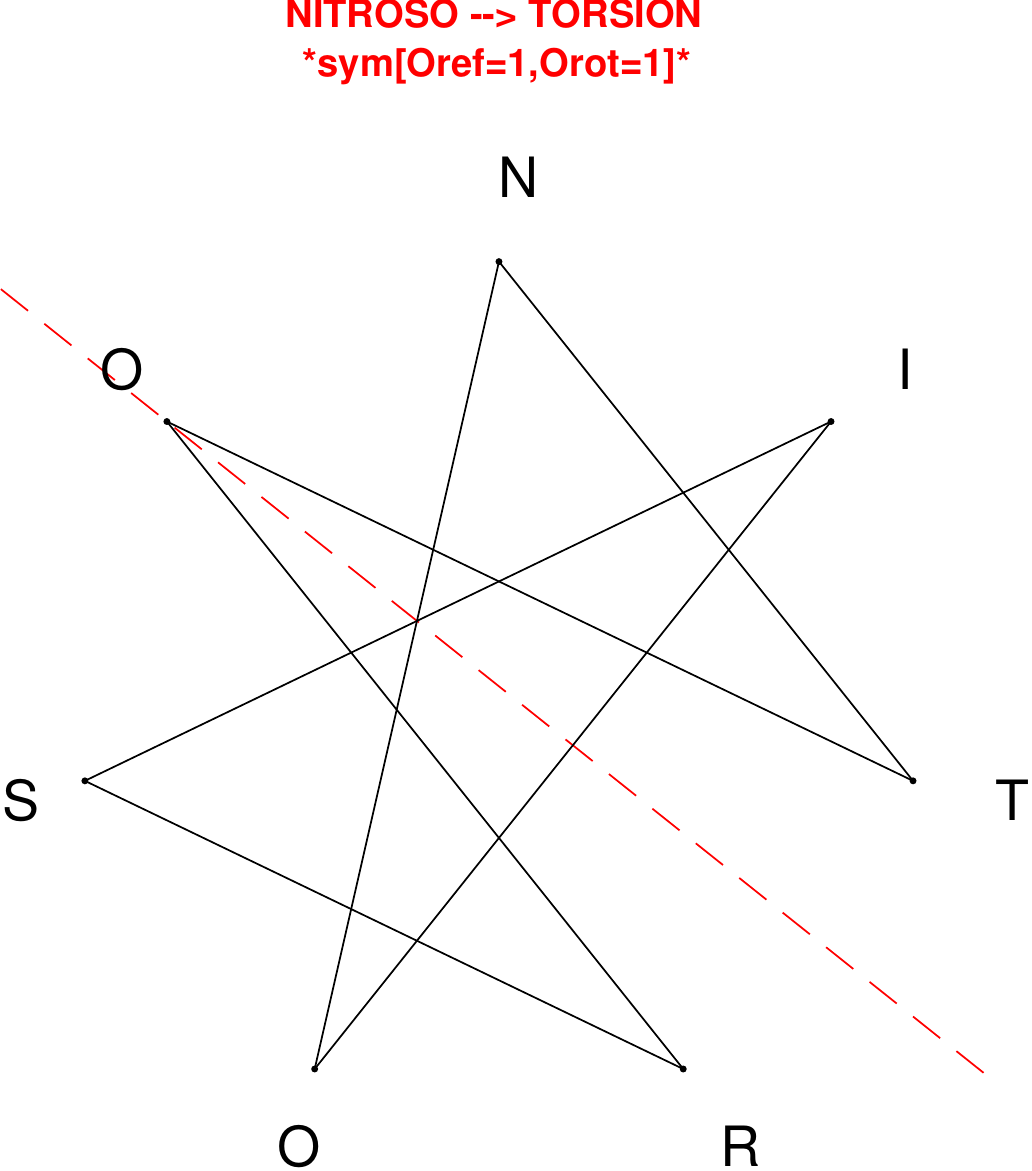}
\end{subfigure}
\hfill
\begin{subfigure}[T]{0.19\textwidth}
\centering
\includegraphics[width=\textwidth]{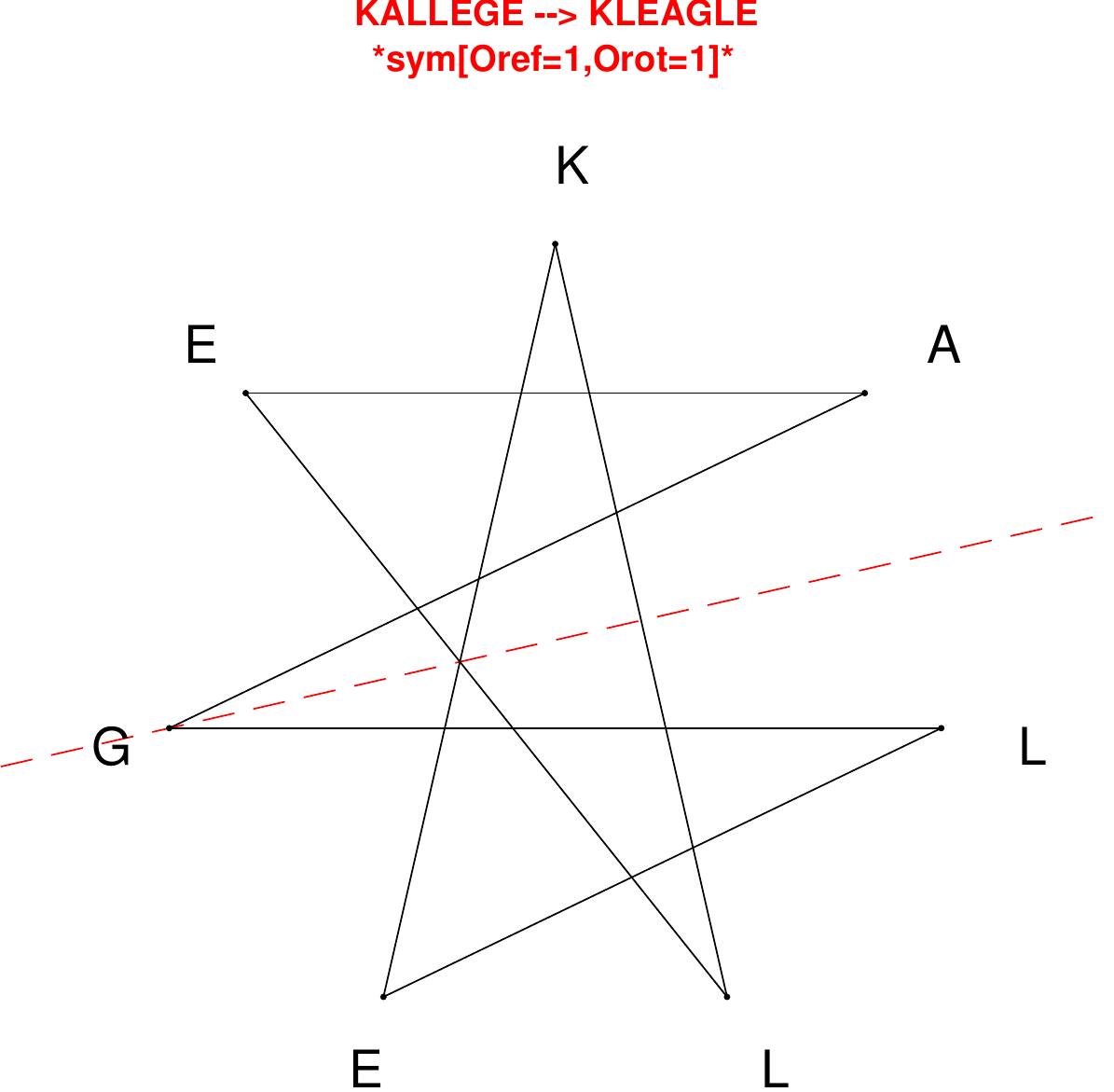}
\end{subfigure}
\hfill
\begin{subfigure}[T]{0.19\textwidth}
\centering
\includegraphics[width=\textwidth]{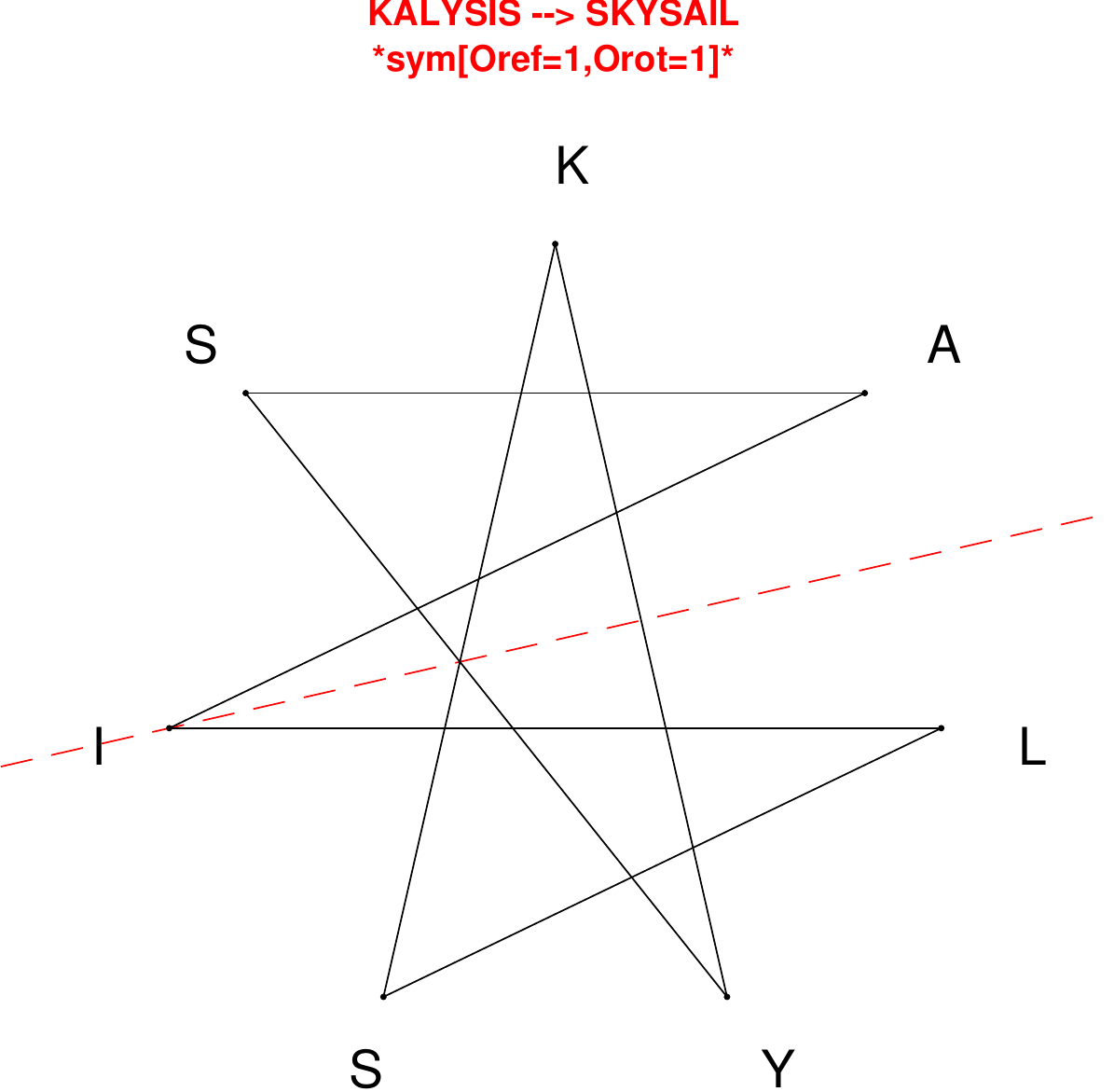}
\end{subfigure}
\end{figure}

\begin{figure}[H]
\centering
\begin{subfigure}[T]{0.19\textwidth}
\centering
\includegraphics[width=\textwidth]{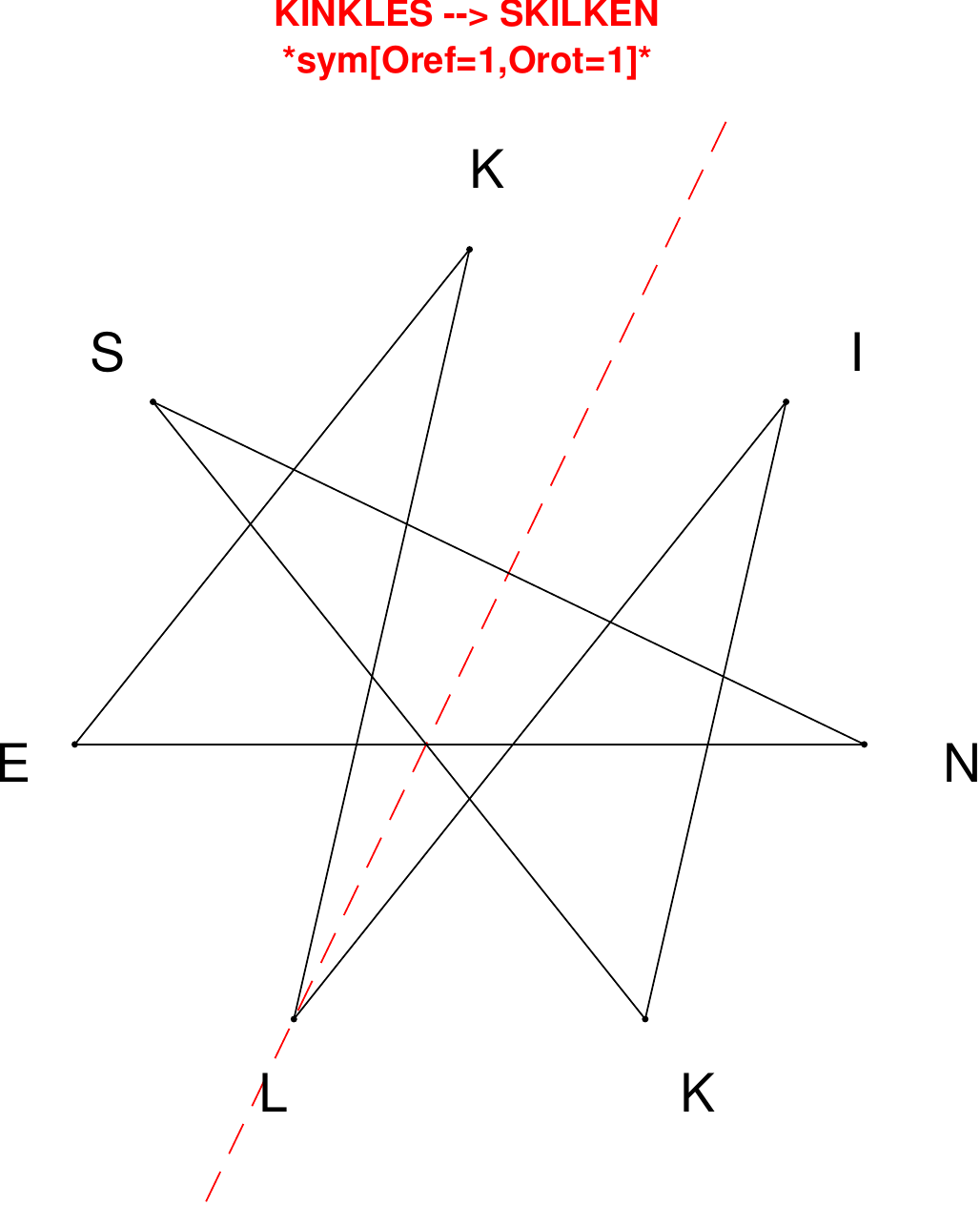}
\end{subfigure}
\hfill
\begin{subfigure}[T]{0.19\textwidth}
\centering
\includegraphics[width=\textwidth]{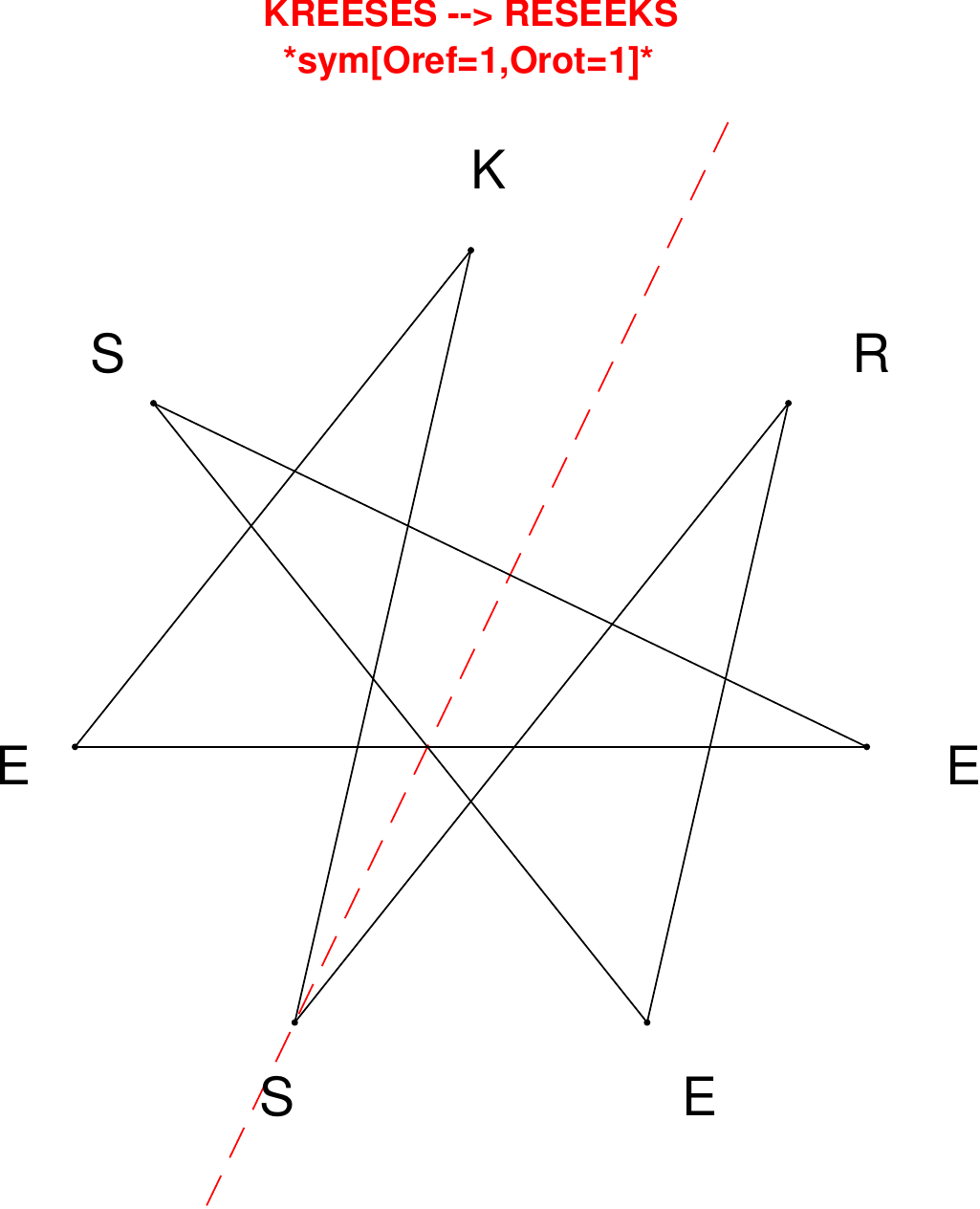}
\end{subfigure}
\hfill
\begin{subfigure}[T]{0.19\textwidth}
\centering
\includegraphics[width=\textwidth]{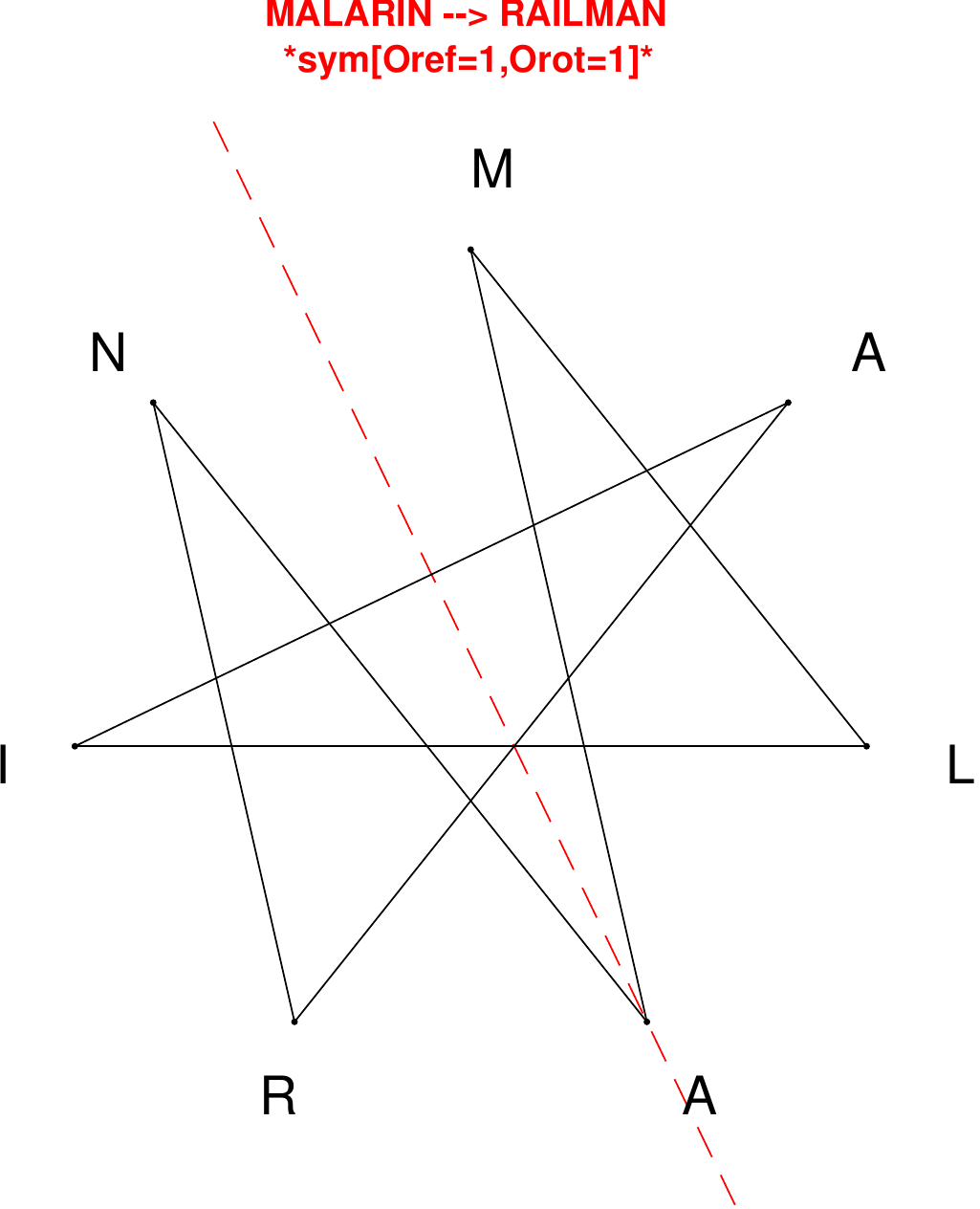}
\end{subfigure}
\hfill
\begin{subfigure}[T]{0.19\textwidth}
\centering
\includegraphics[width=\textwidth]{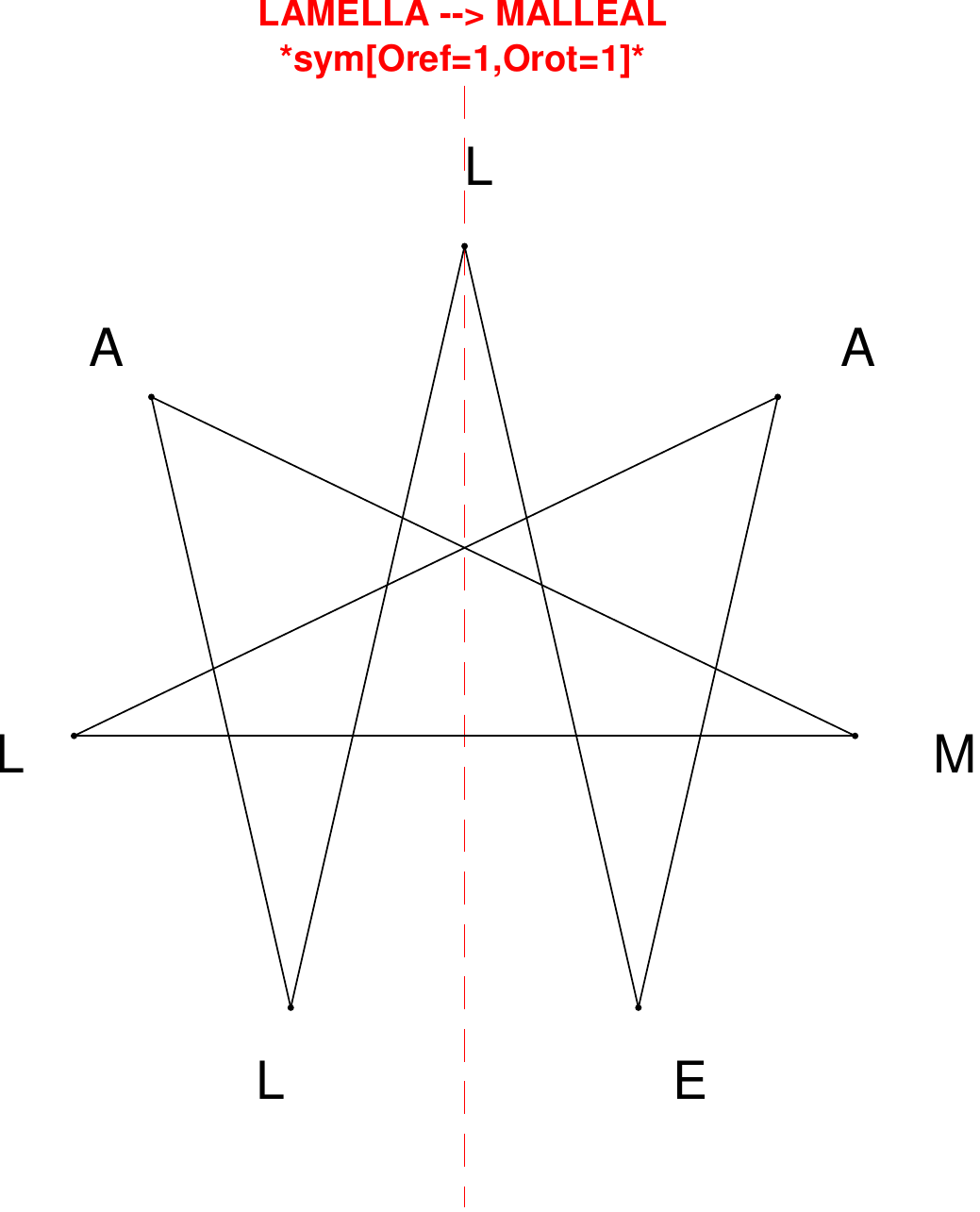}
\end{subfigure}
\hfill
\begin{subfigure}[T]{0.19\textwidth}
\centering
\includegraphics[width=\textwidth]{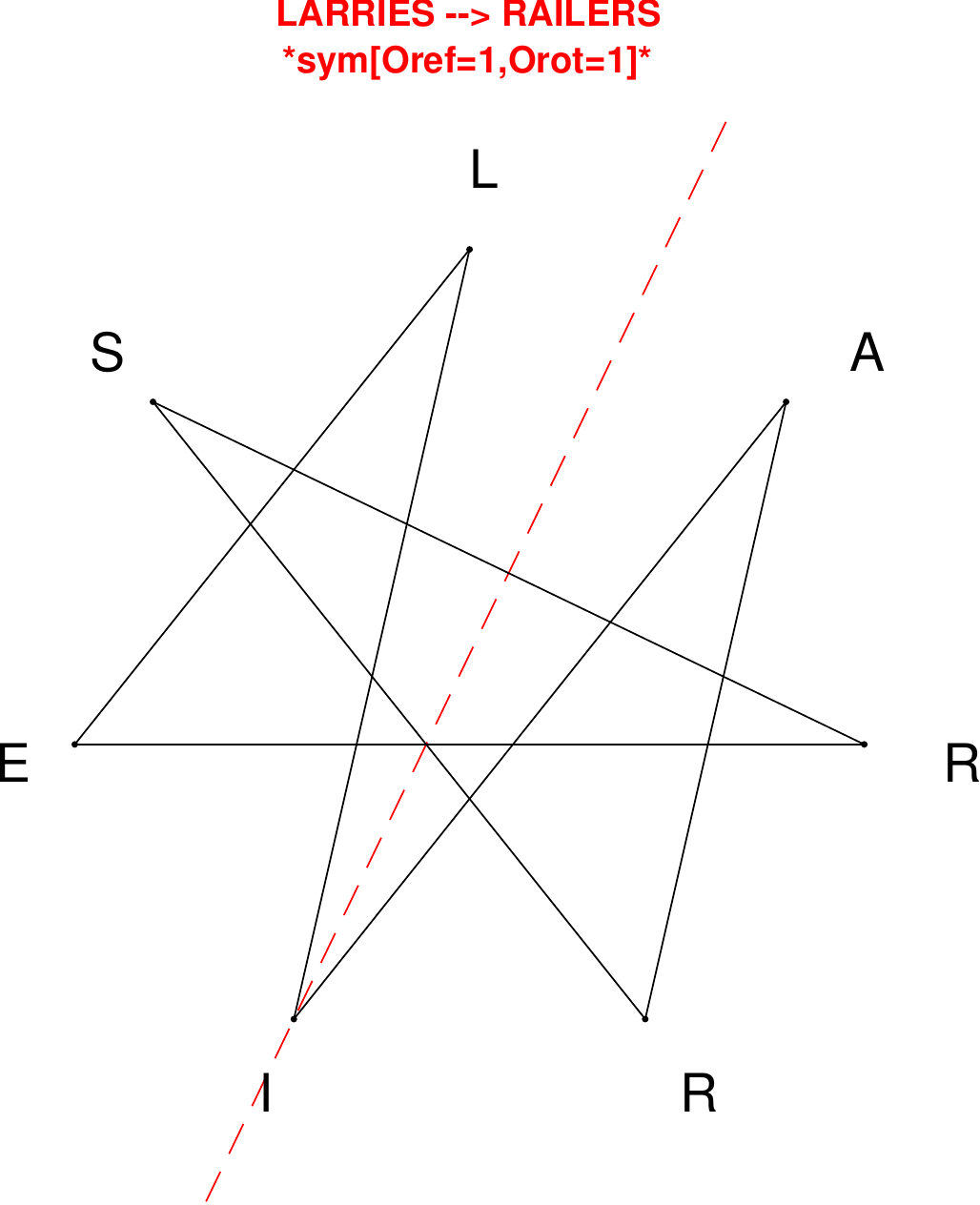}
\end{subfigure}
\end{figure}

\begin{figure}[H]
\centering
\begin{subfigure}[T]{0.19\textwidth}
\centering
\includegraphics[width=\textwidth]{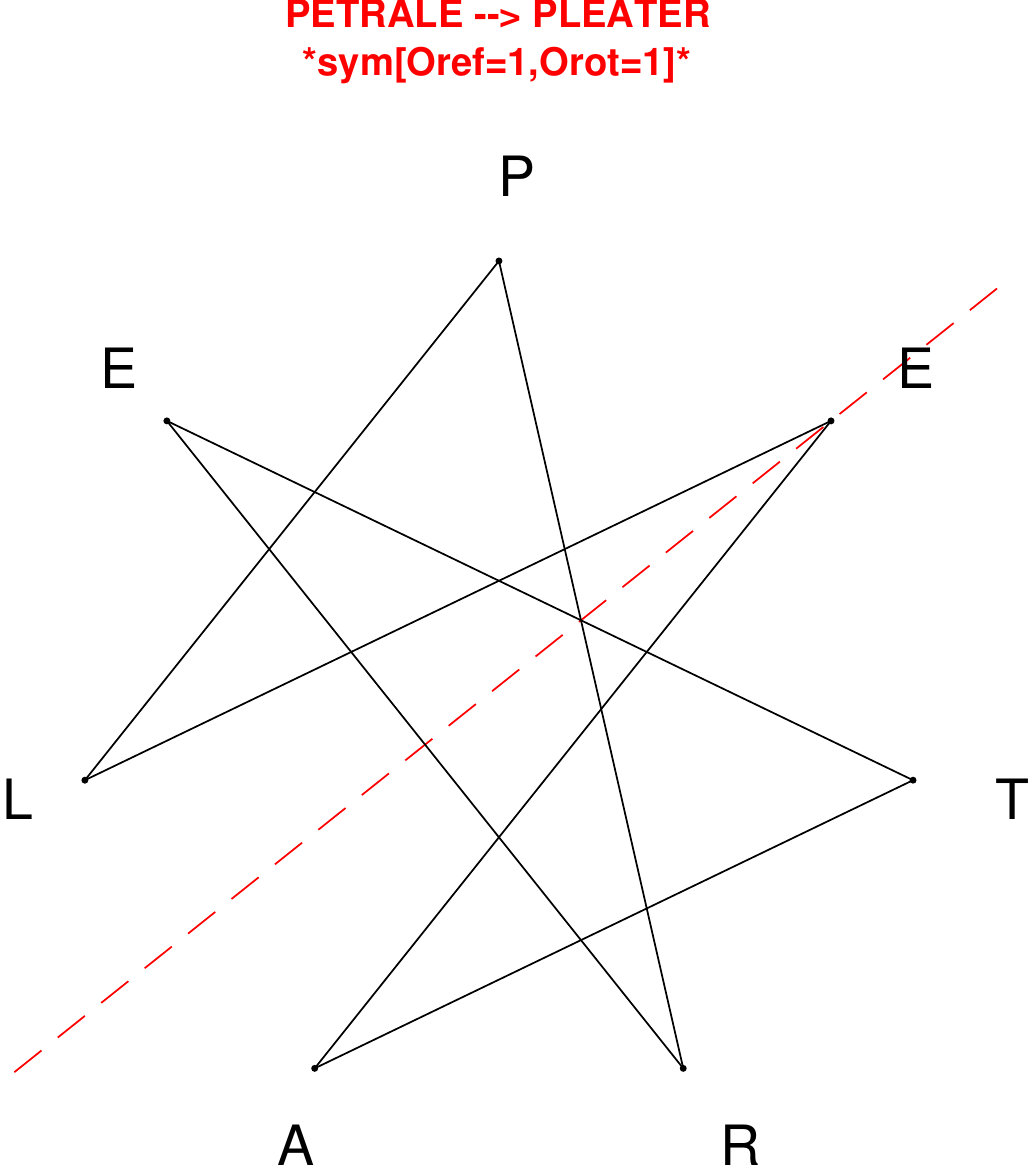}
\end{subfigure}
\hfill
\begin{subfigure}[T]{0.19\textwidth}
\centering
\includegraphics[width=\textwidth]{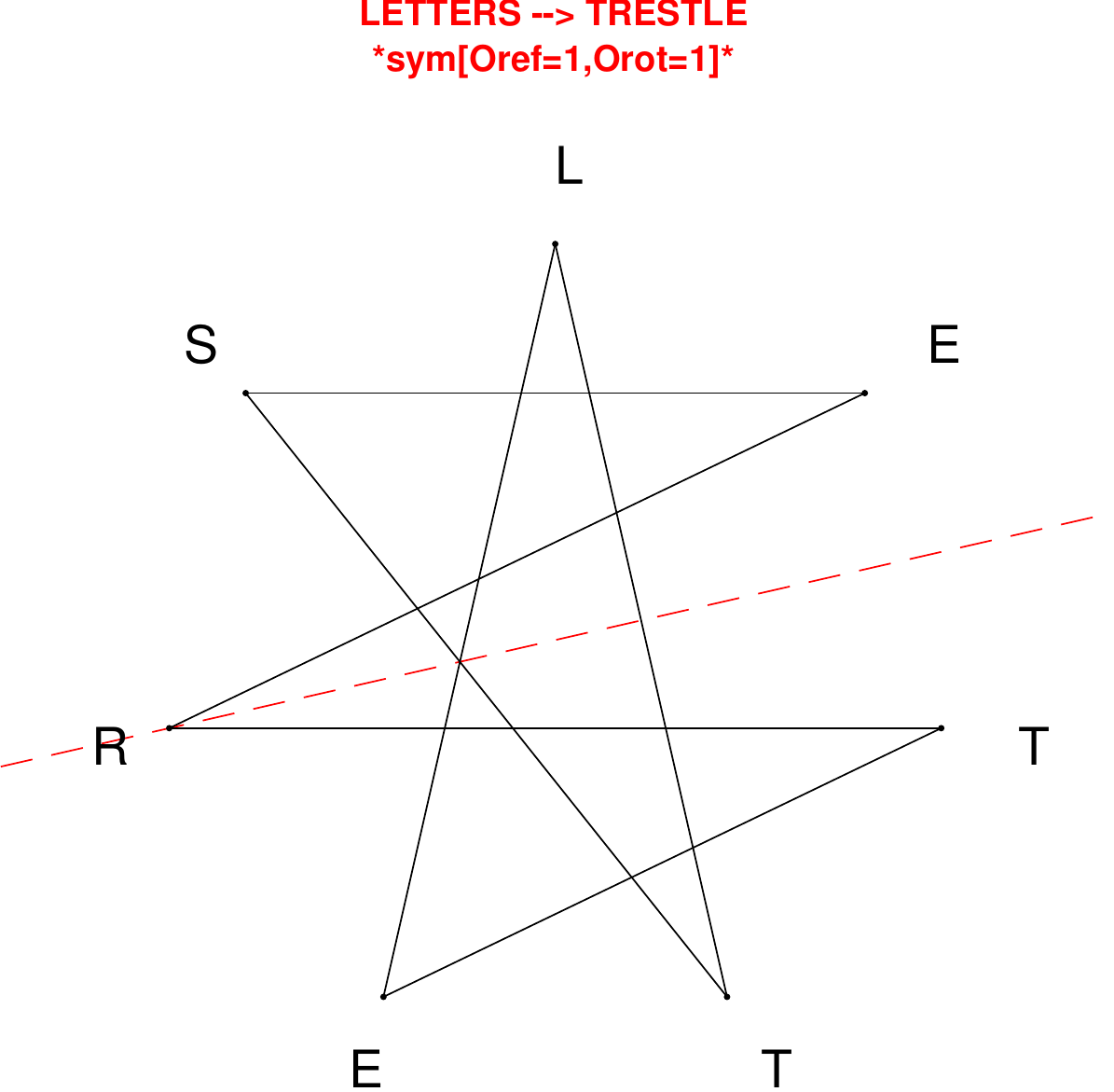}
\end{subfigure}
\hfill
\begin{subfigure}[T]{0.19\textwidth}
\centering
\includegraphics[width=\textwidth]{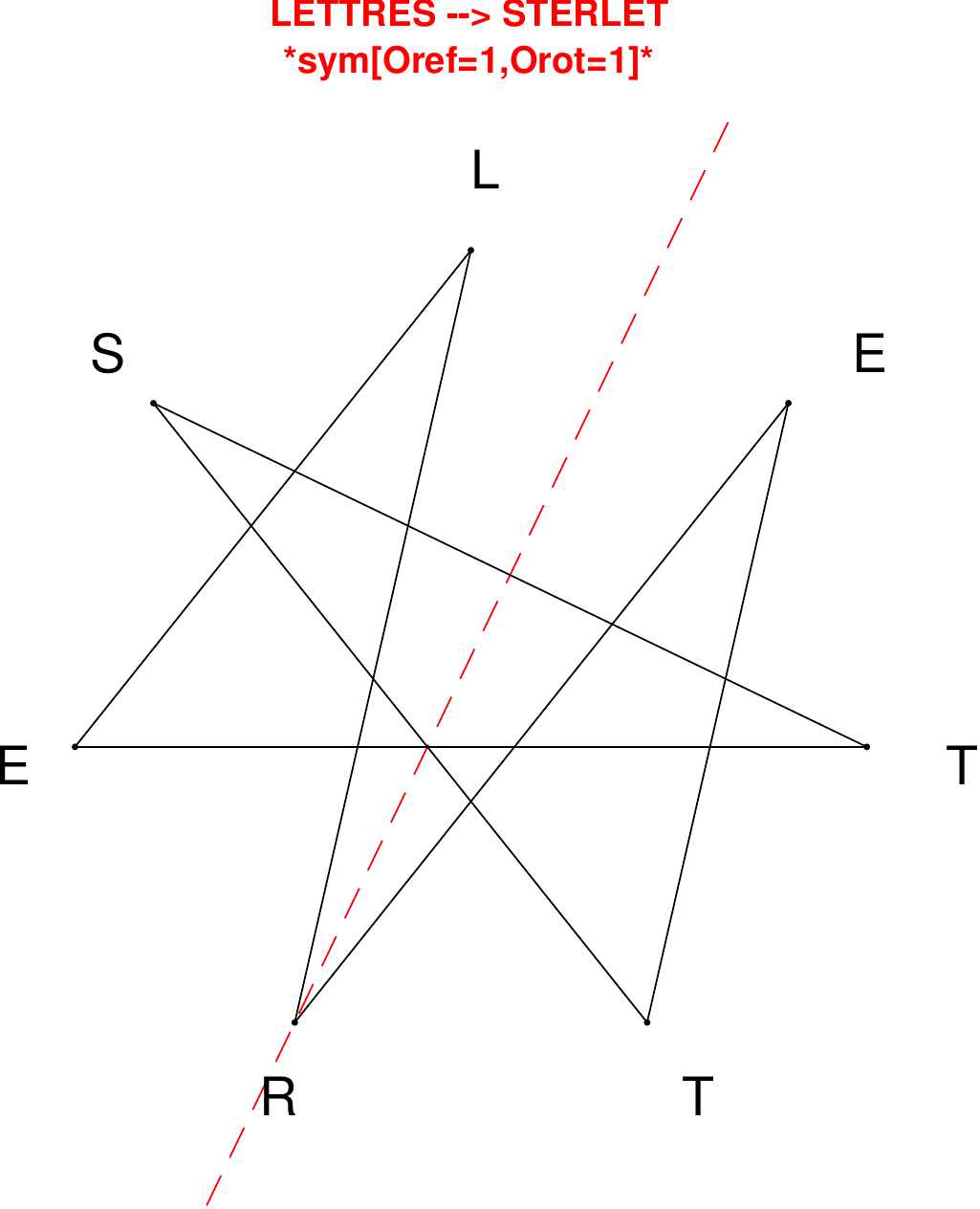}
\end{subfigure}
\hfill
\begin{subfigure}[T]{0.19\textwidth}
\centering
\includegraphics[width=\textwidth]{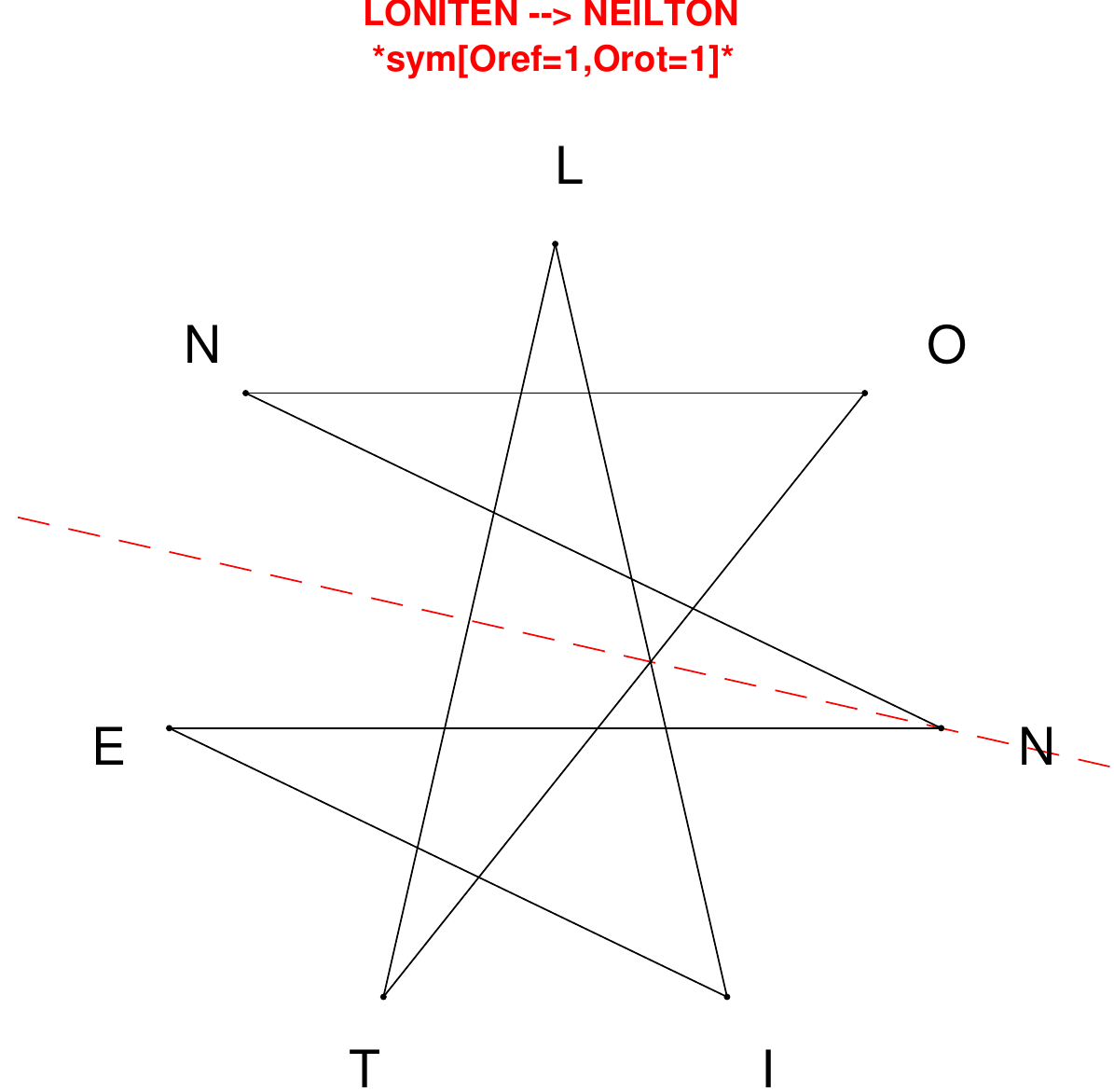}
\end{subfigure}
\hfill
\begin{subfigure}[T]{0.19\textwidth}
\centering
\includegraphics[width=\textwidth]{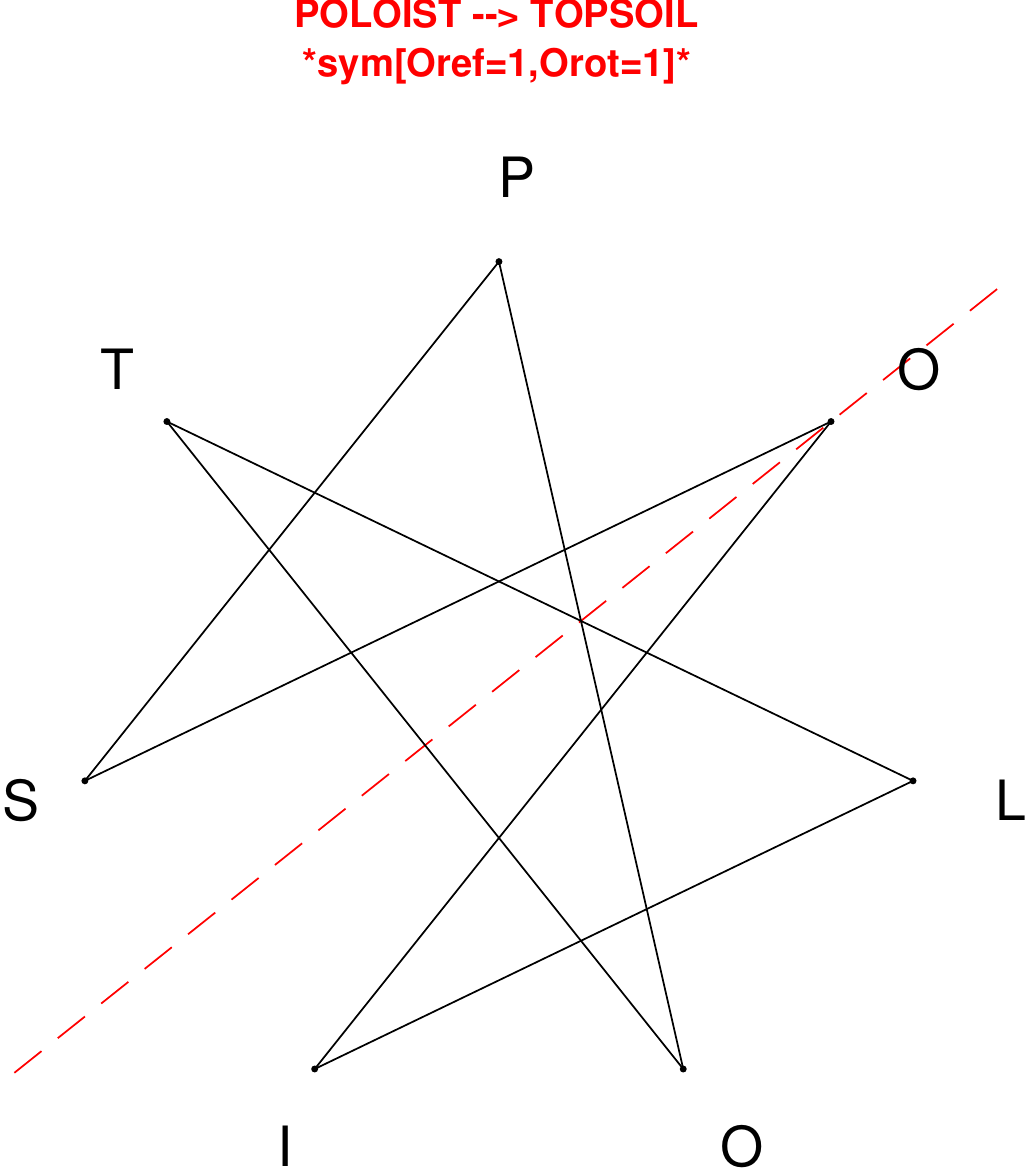}
\end{subfigure}
\end{figure}

\begin{figure}[H]
\centering
\begin{subfigure}[T]{0.19\textwidth}
\centering
\includegraphics[width=\textwidth]{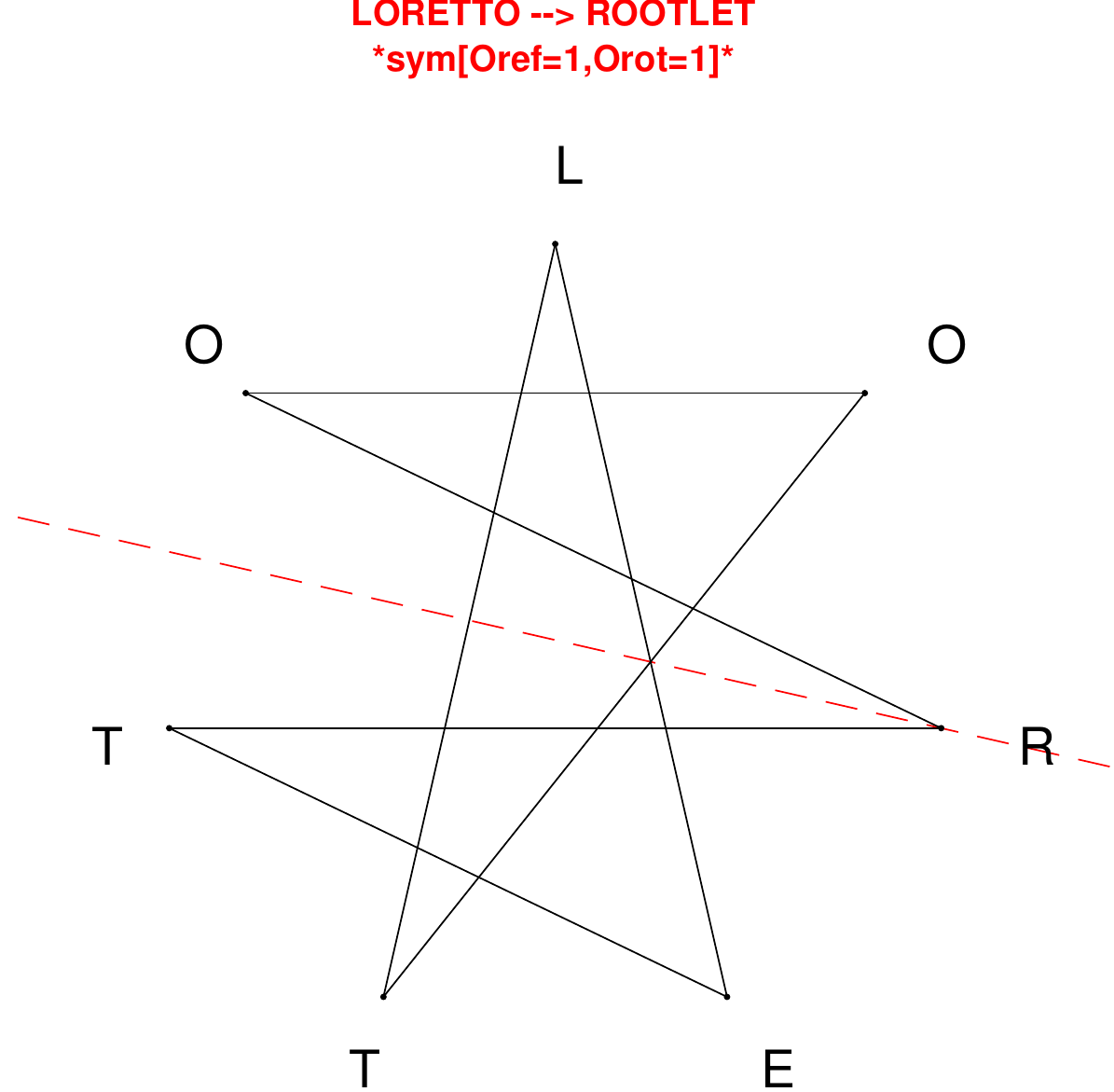}
\end{subfigure}
\hfill
\begin{subfigure}[T]{0.19\textwidth}
\centering
\includegraphics[width=\textwidth]{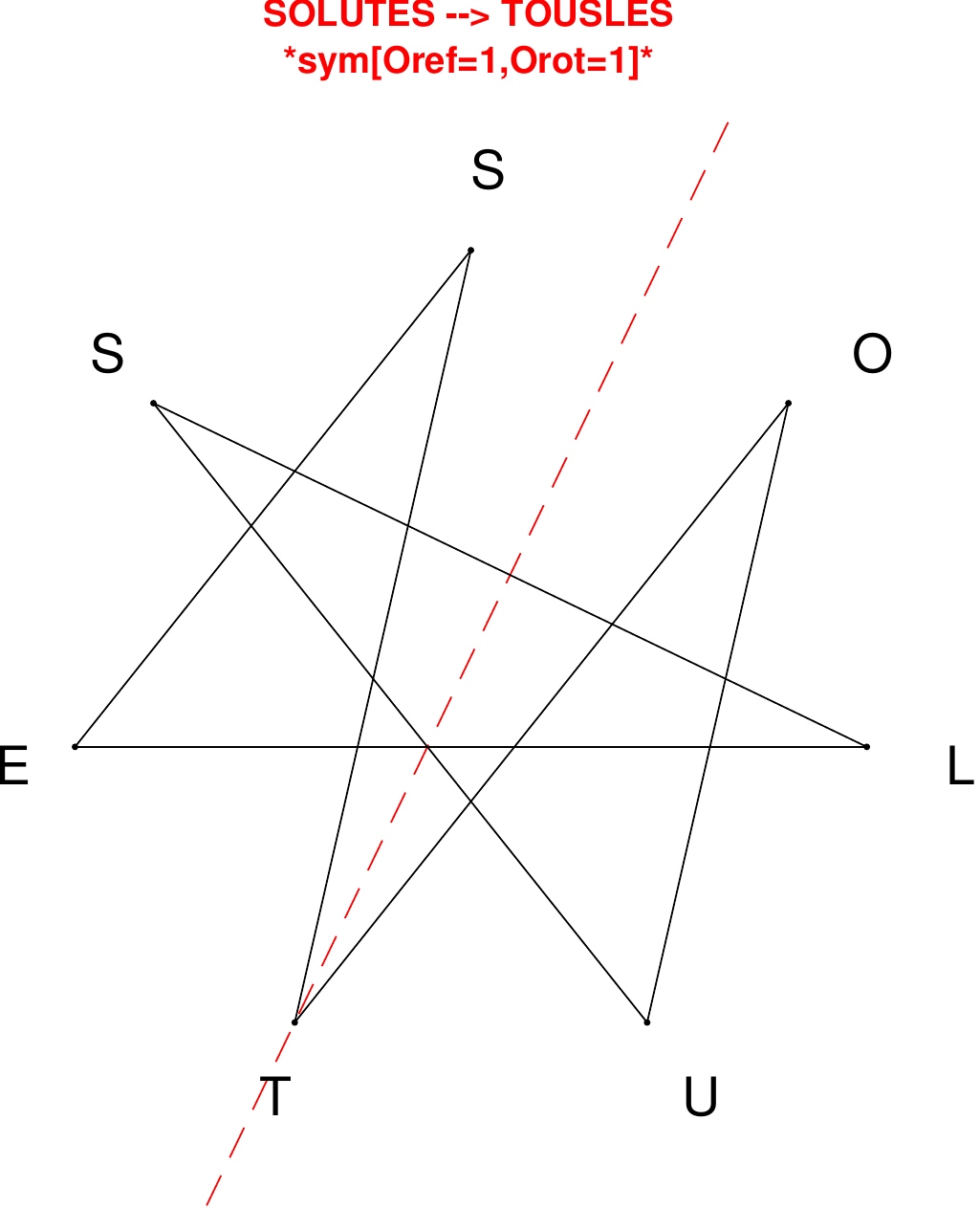}
\end{subfigure}
\hfill
\begin{subfigure}[T]{0.19\textwidth}
\centering
\includegraphics[width=\textwidth]{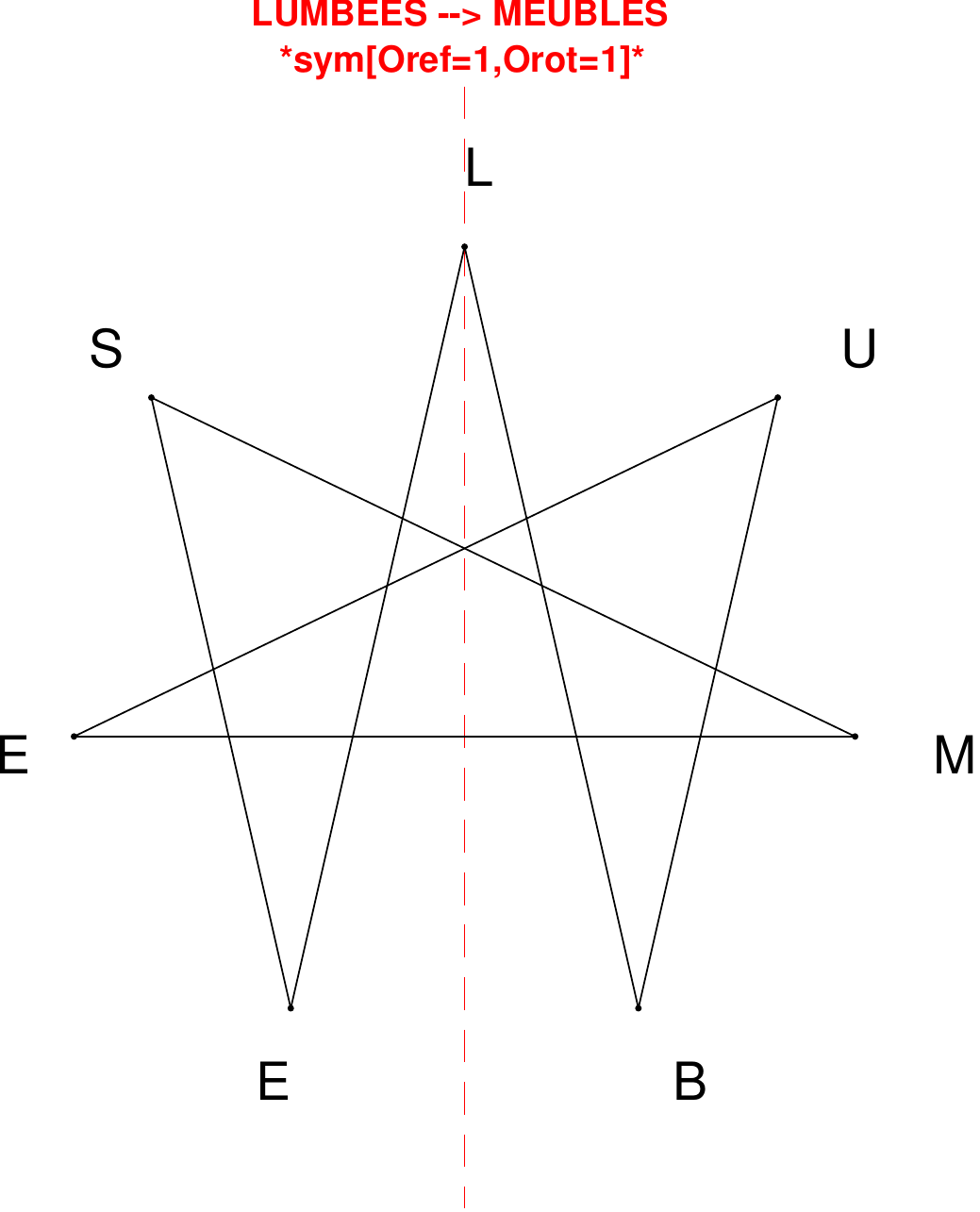}
\end{subfigure}
\hfill
\begin{subfigure}[T]{0.19\textwidth}
\centering
\includegraphics[width=\textwidth]{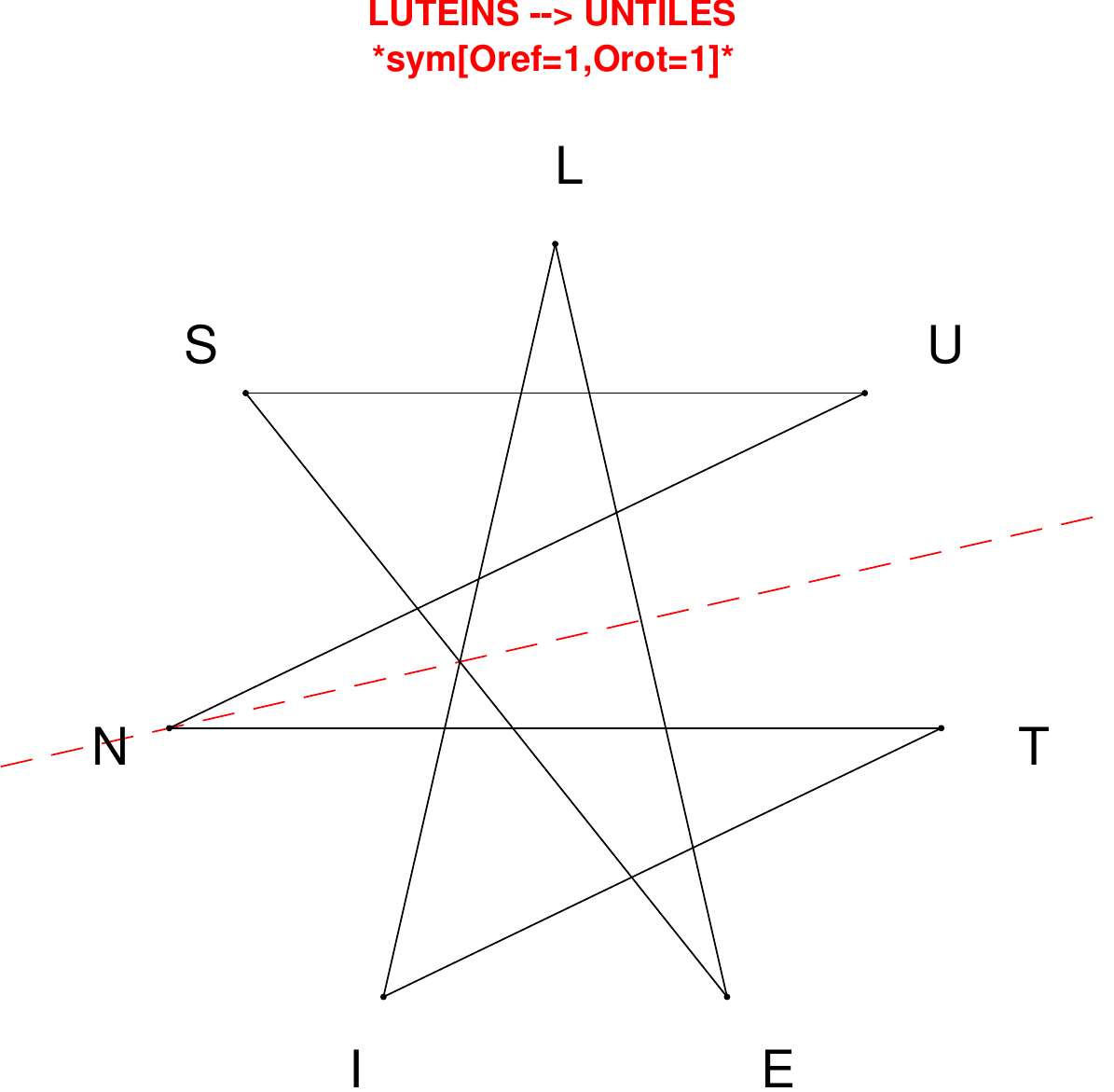}
\end{subfigure}
\hfill
\begin{subfigure}[T]{0.19\textwidth}
\centering
\includegraphics[width=\textwidth]{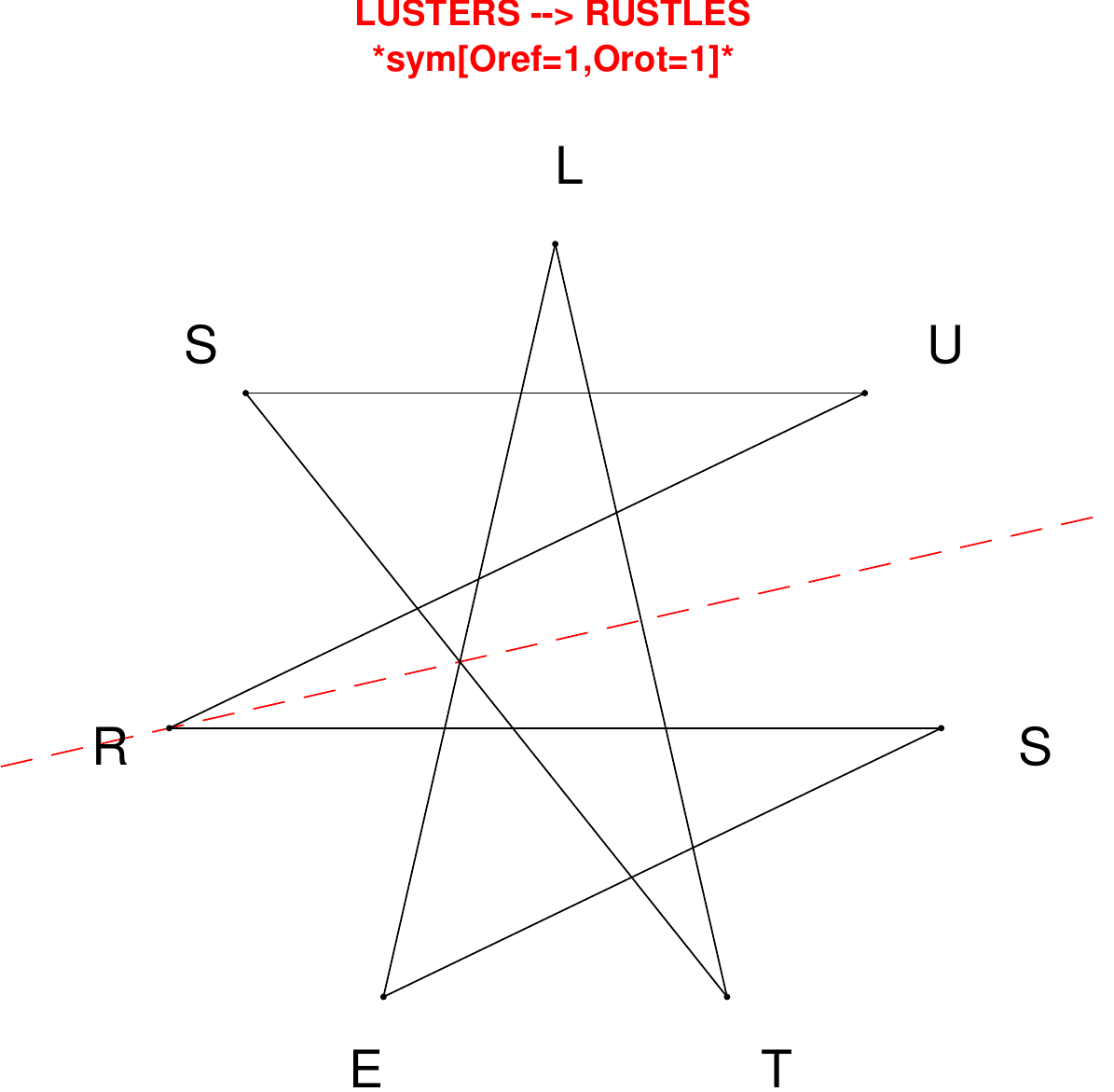}
\end{subfigure}
\end{figure}

\begin{figure}[H]
\centering
\begin{subfigure}[T]{0.19\textwidth}
\centering
\includegraphics[width=\textwidth]{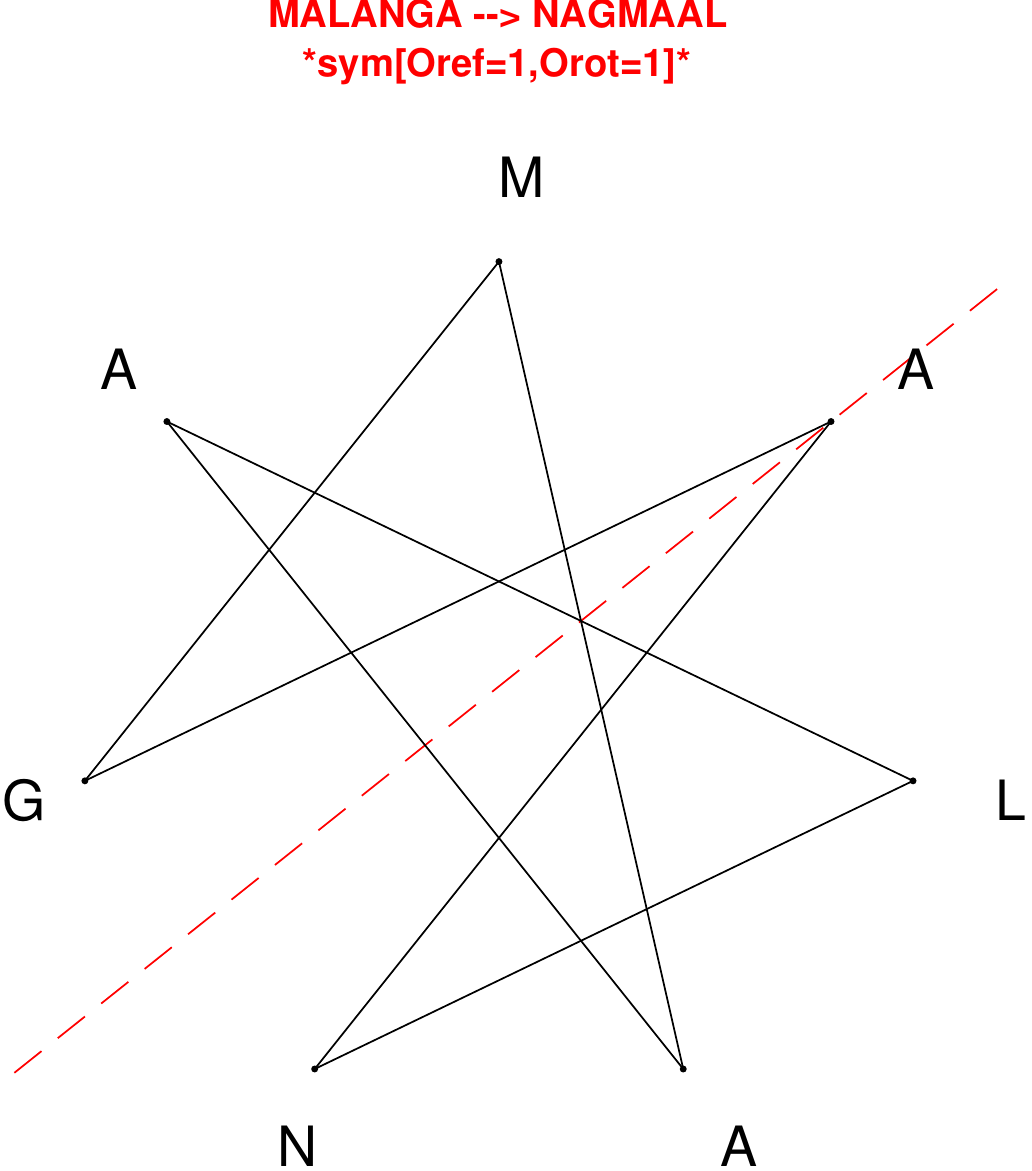}
\end{subfigure}
\hfill
\begin{subfigure}[T]{0.19\textwidth}
\centering
\includegraphics[width=\textwidth]{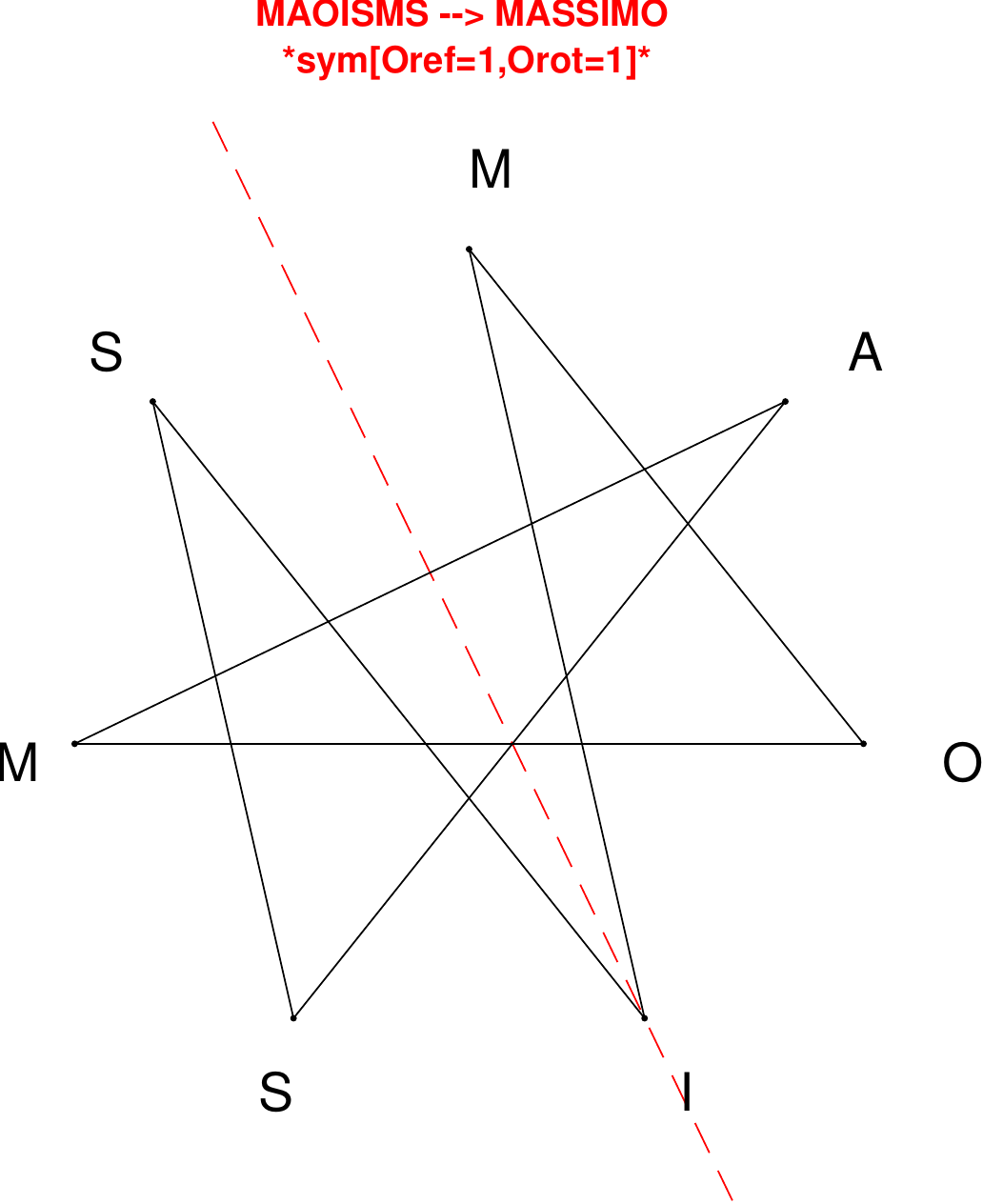}
\end{subfigure}
\hfill
\begin{subfigure}[T]{0.19\textwidth}
\centering
\includegraphics[width=\textwidth]{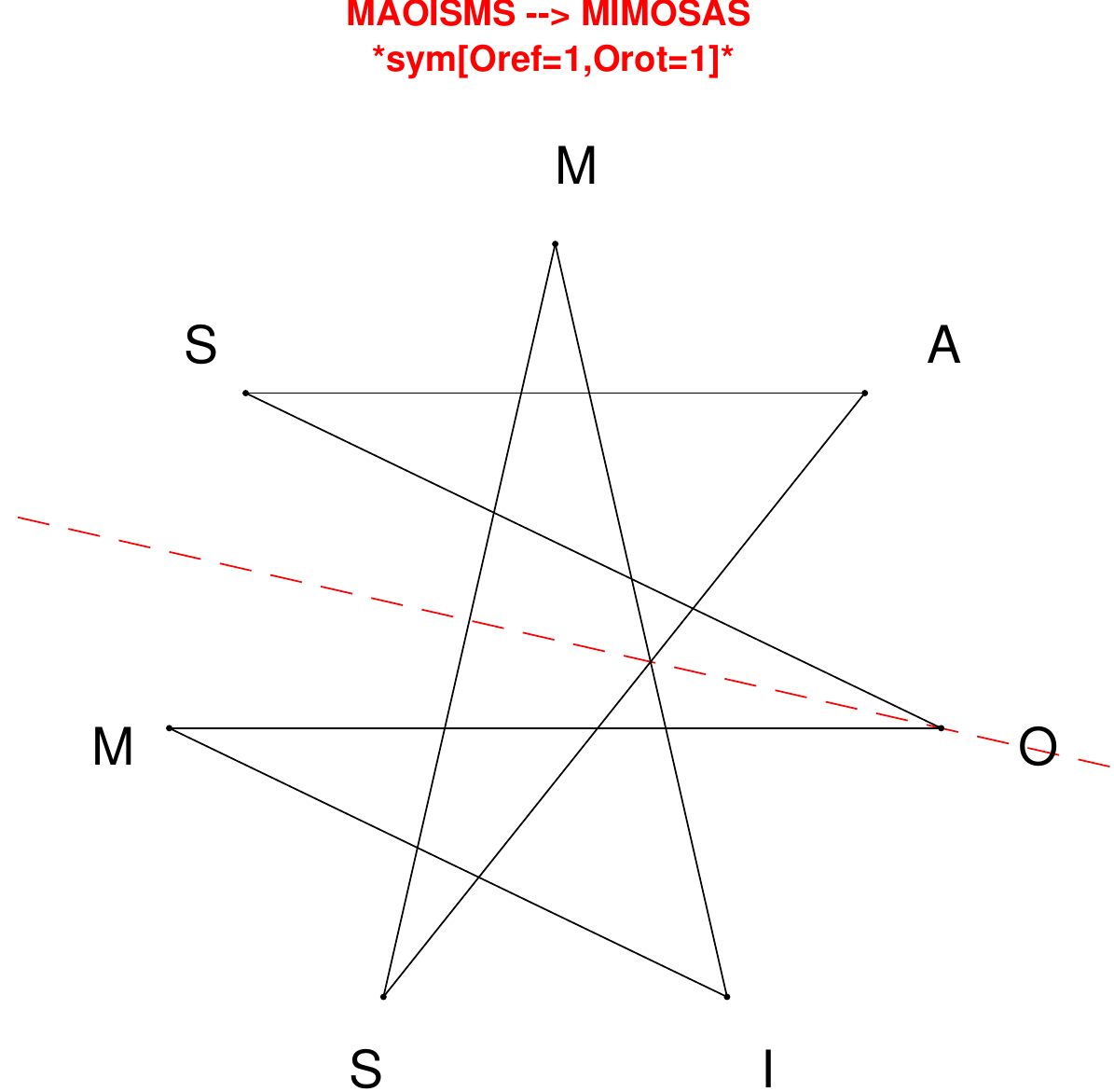}
\end{subfigure}
\hfill
\begin{subfigure}[T]{0.19\textwidth}
\centering
\includegraphics[width=\textwidth]{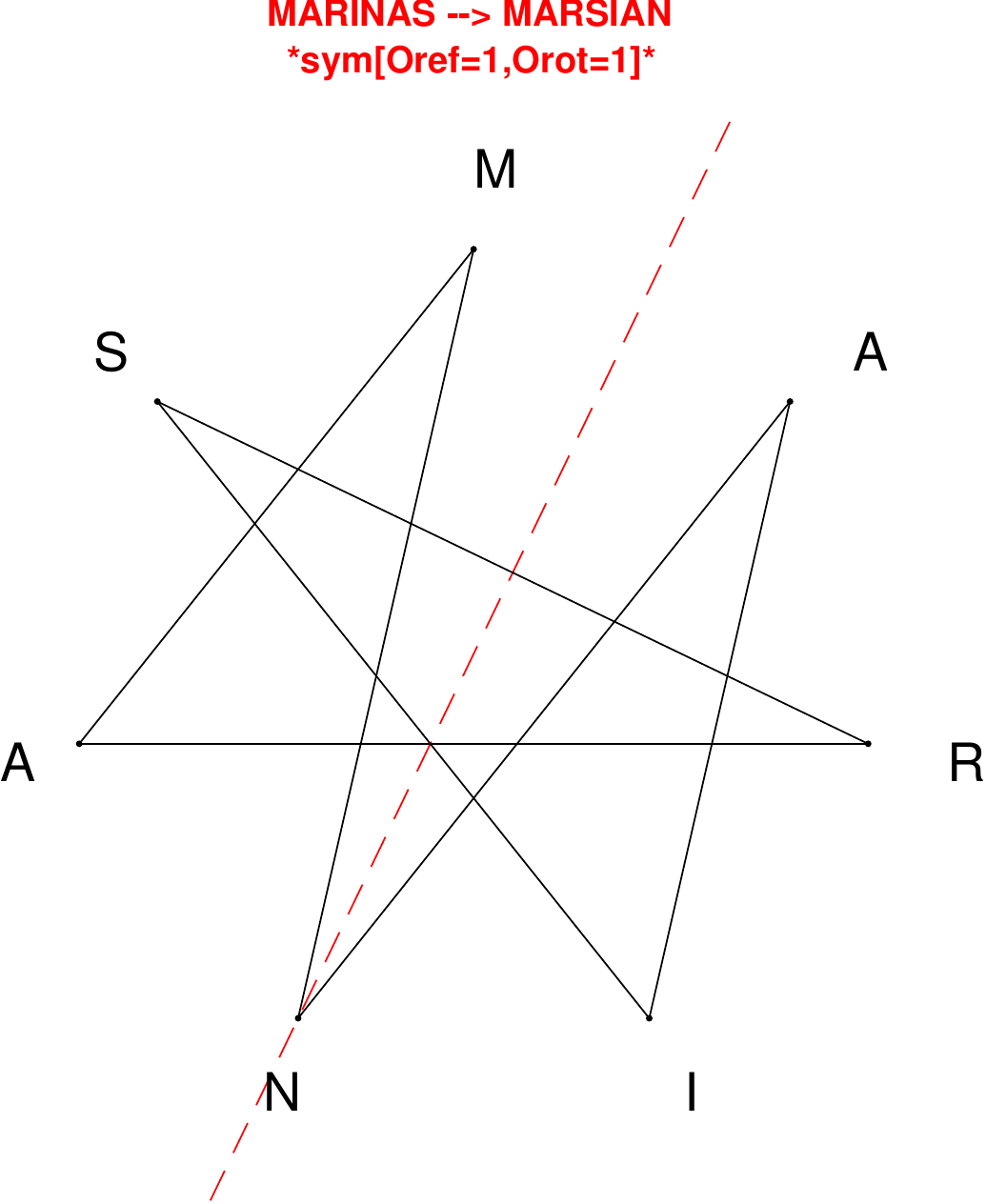}
\end{subfigure}
\hfill
\begin{subfigure}[T]{0.19\textwidth}
\centering
\includegraphics[width=\textwidth]{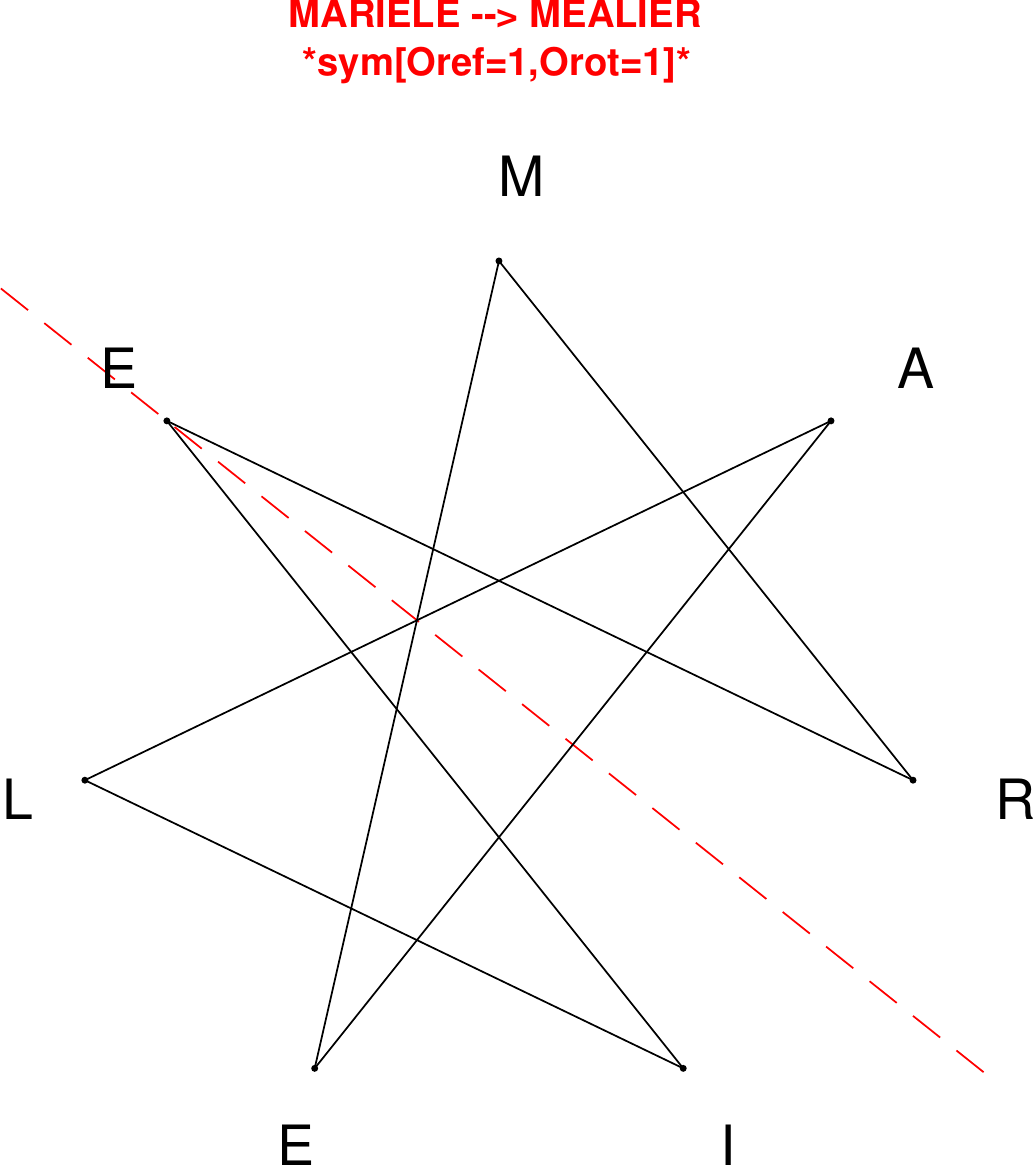}
\end{subfigure}
\end{figure}

\begin{figure}[H]
\centering
\begin{subfigure}[T]{0.19\textwidth}
\centering
\includegraphics[width=\textwidth]{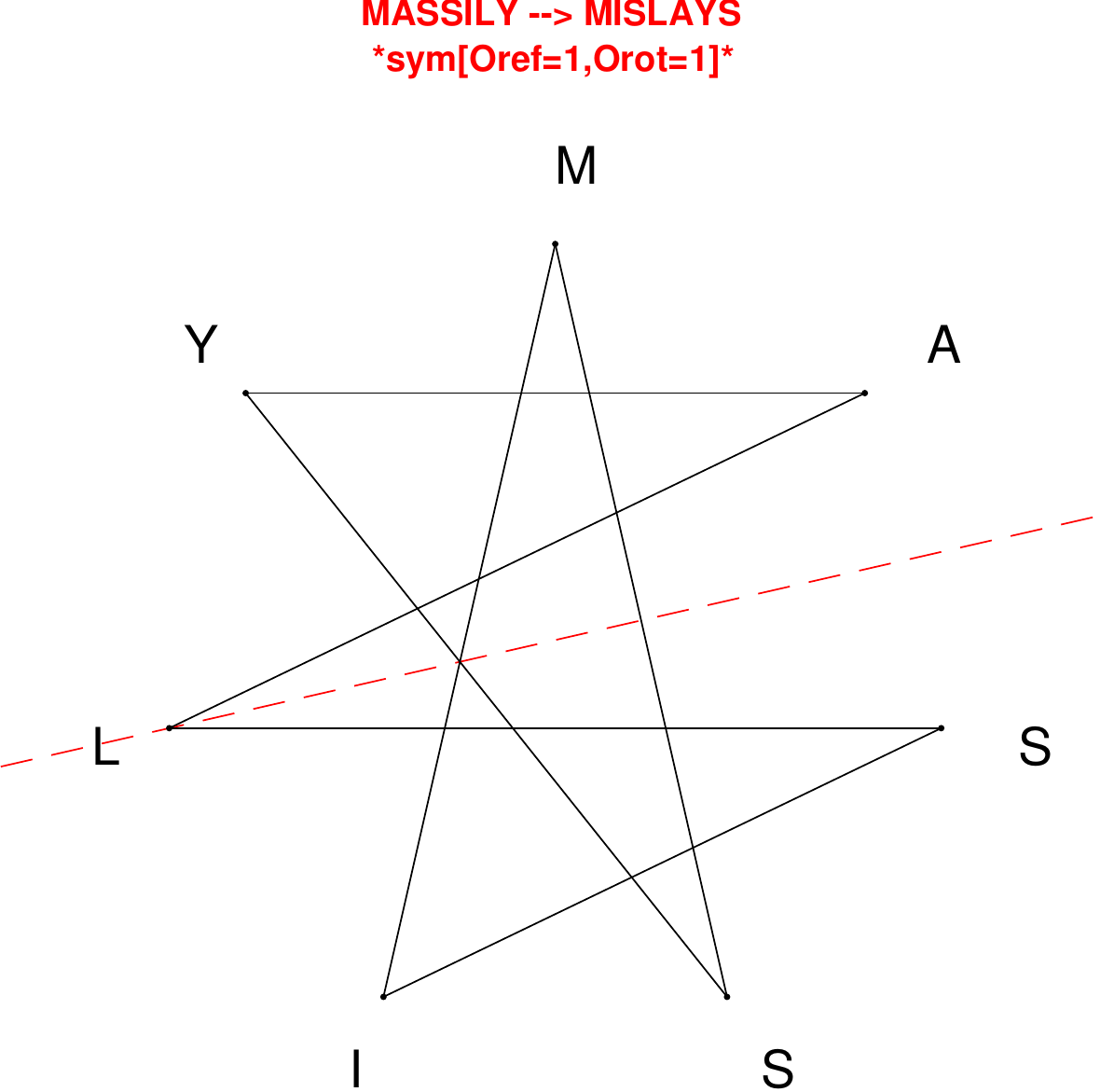}
\end{subfigure}
\hfill
\begin{subfigure}[T]{0.19\textwidth}
\centering
\includegraphics[width=\textwidth]{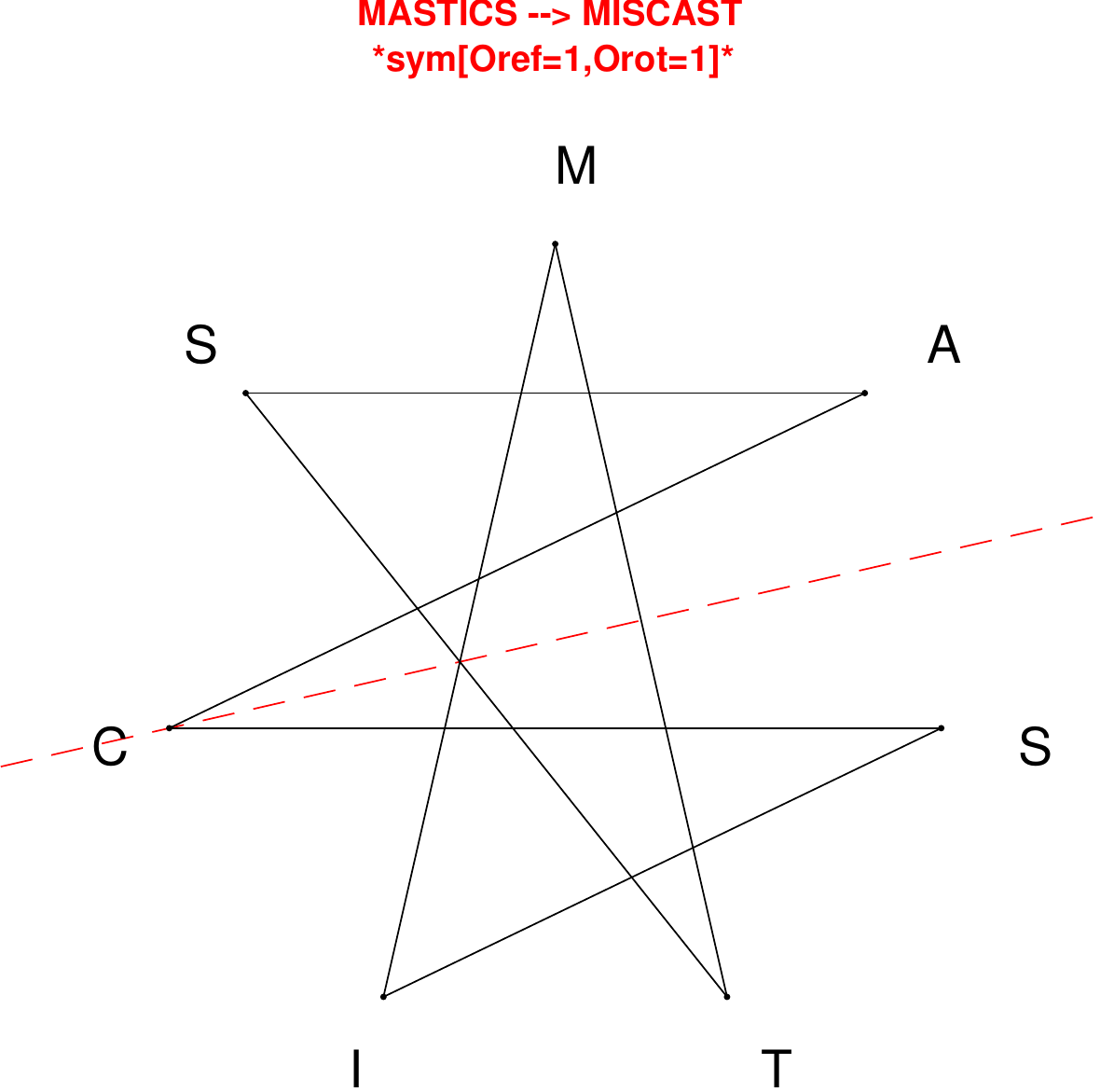}
\end{subfigure}
\hfill
\begin{subfigure}[T]{0.19\textwidth}
\centering
\includegraphics[width=\textwidth]{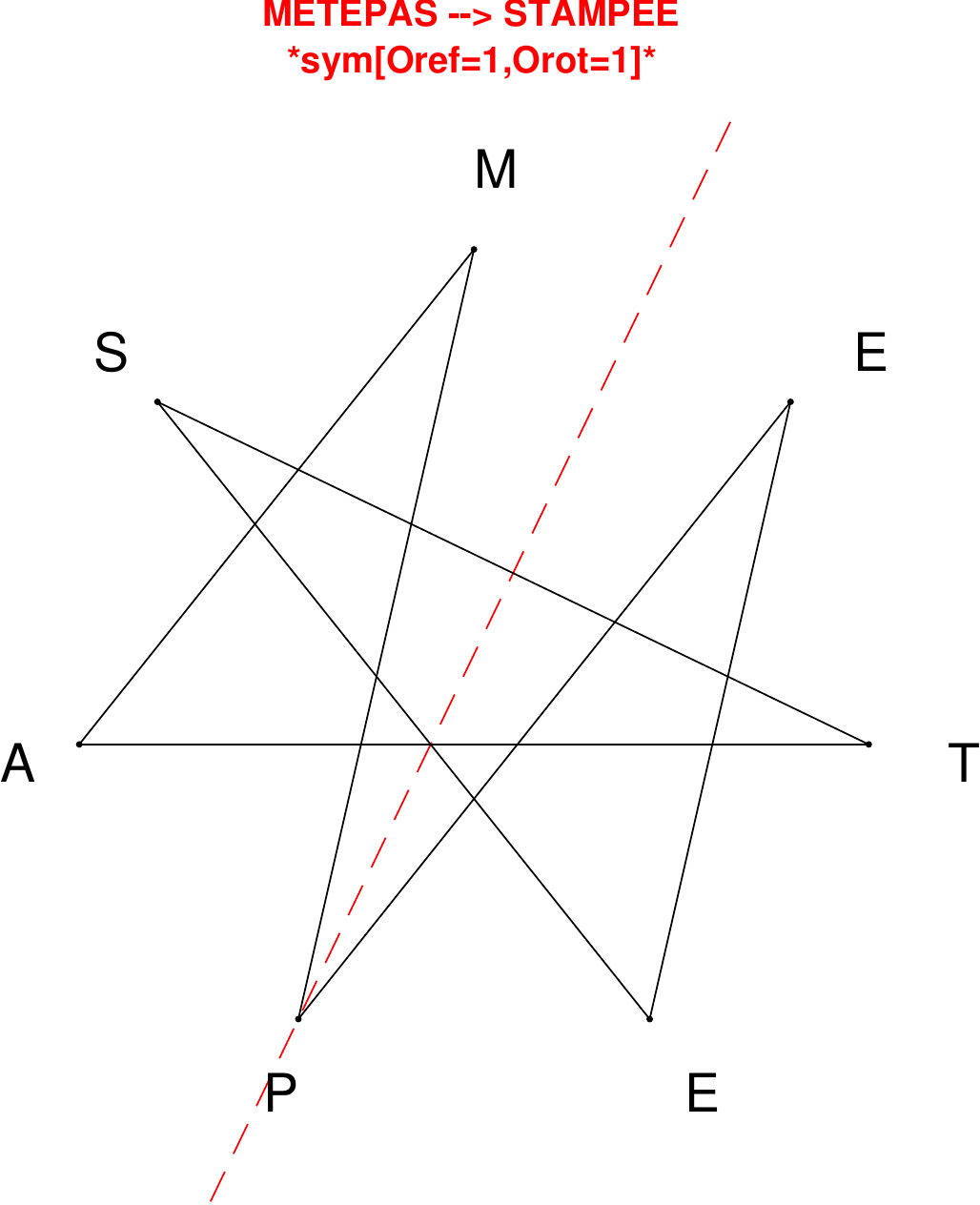}
\end{subfigure}
\hfill
\begin{subfigure}[T]{0.19\textwidth}
\centering
\includegraphics[width=\textwidth]{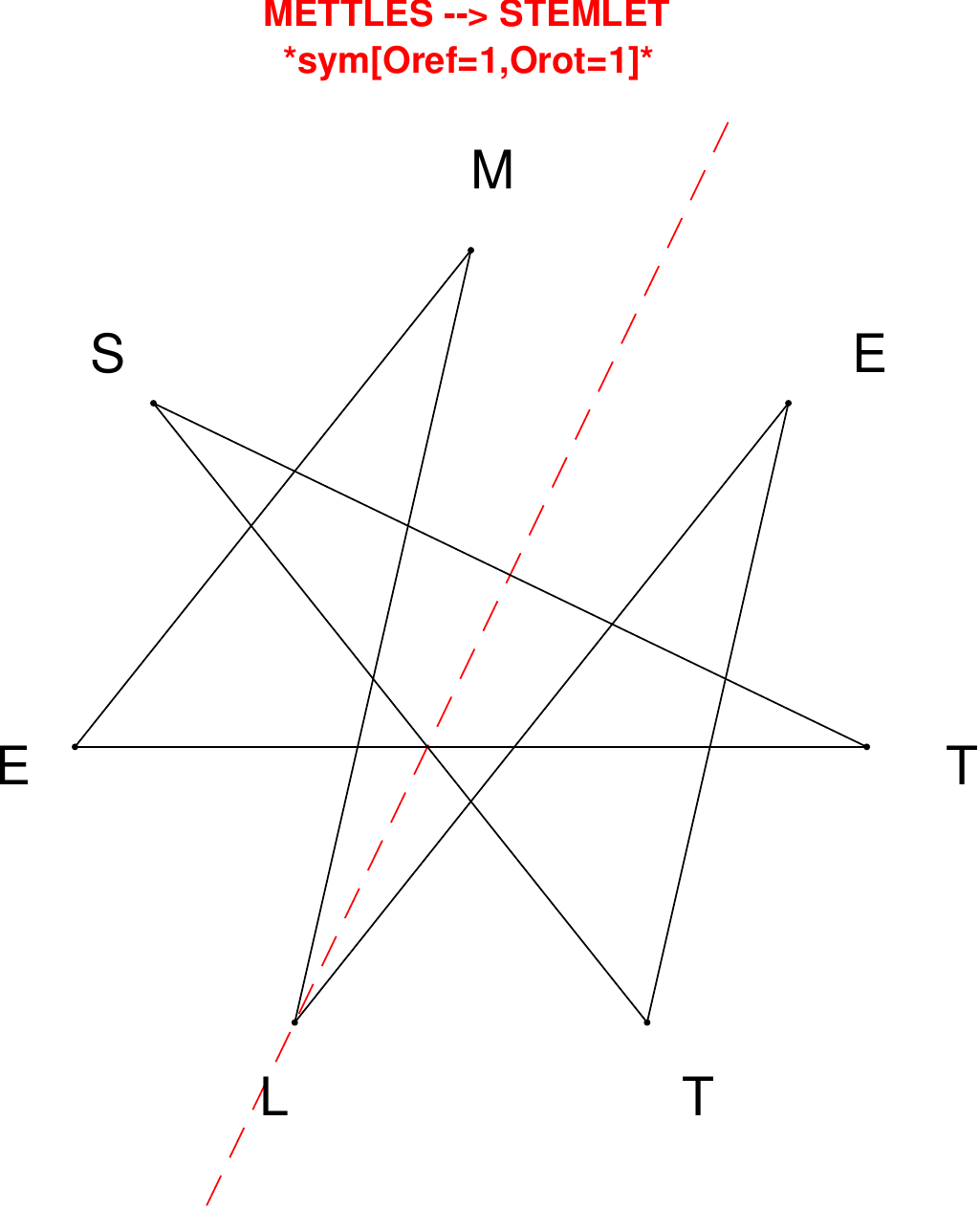}
\end{subfigure}
\hfill
\begin{subfigure}[T]{0.19\textwidth}
\centering
\includegraphics[width=\textwidth]{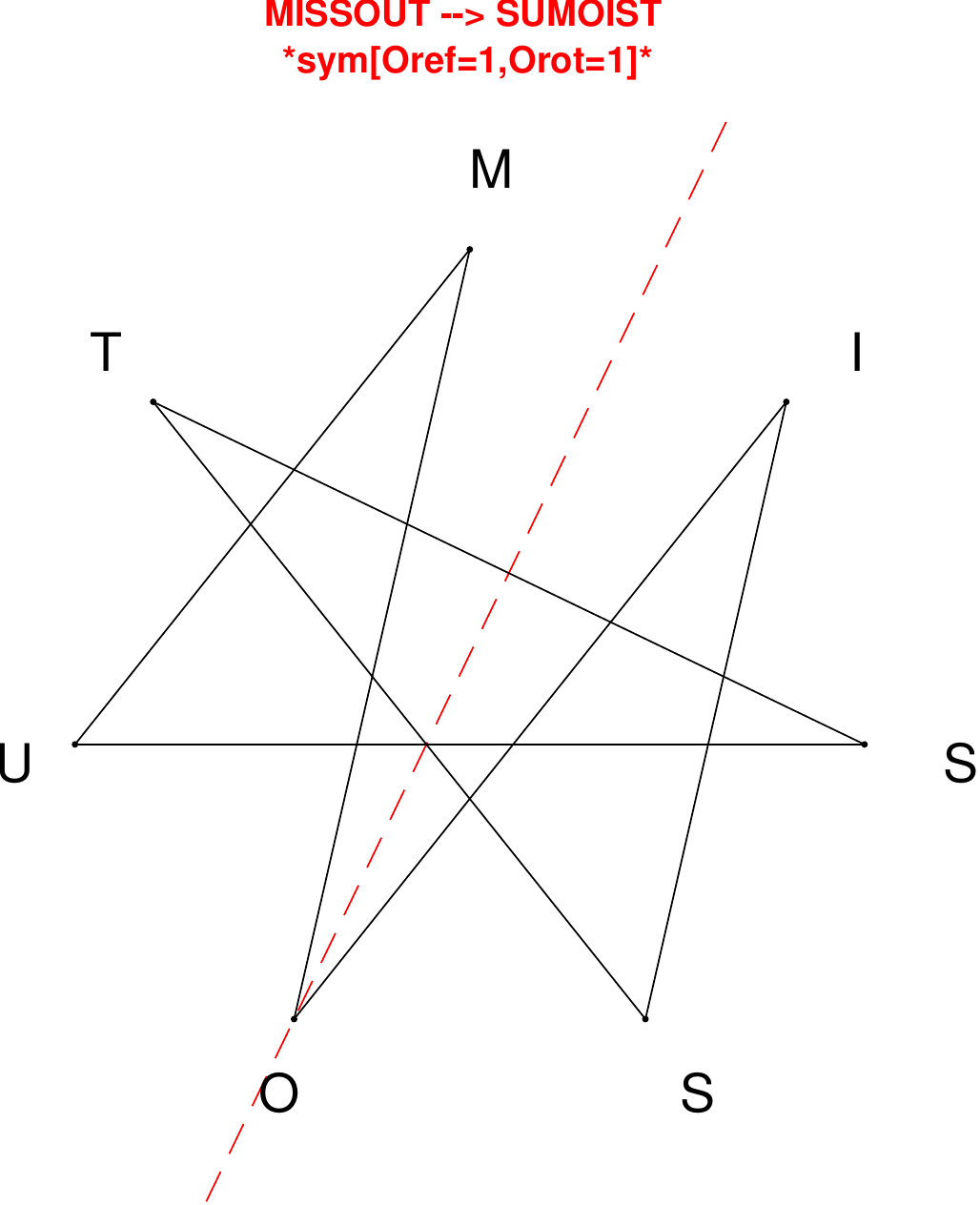}
\end{subfigure}
\end{figure}

\begin{figure}[H]
\centering
\begin{subfigure}[T]{0.19\textwidth}
\centering
\includegraphics[width=\textwidth]{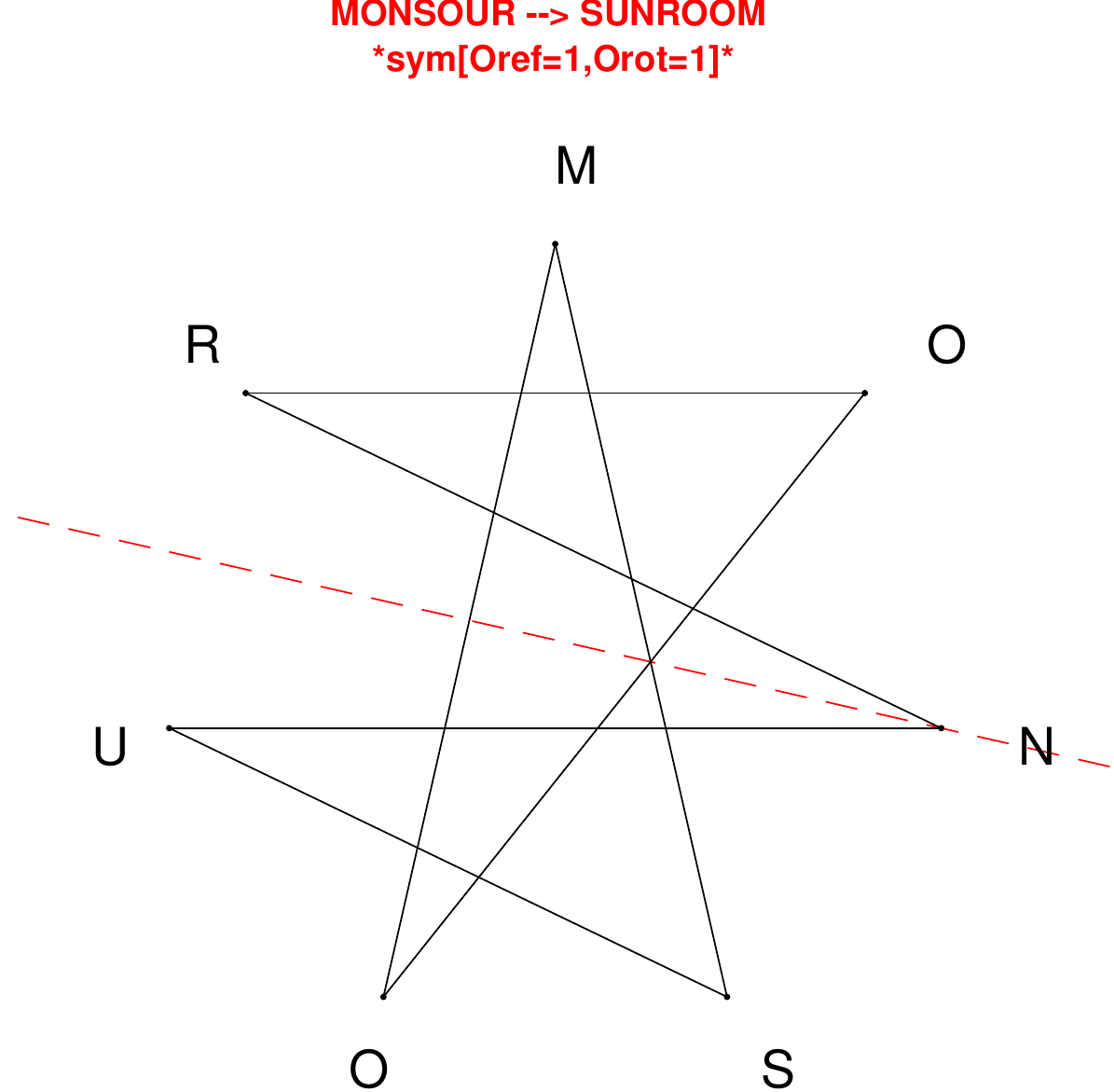}
\end{subfigure}
\hfill
\begin{subfigure}[T]{0.19\textwidth}
\centering
\includegraphics[width=\textwidth]{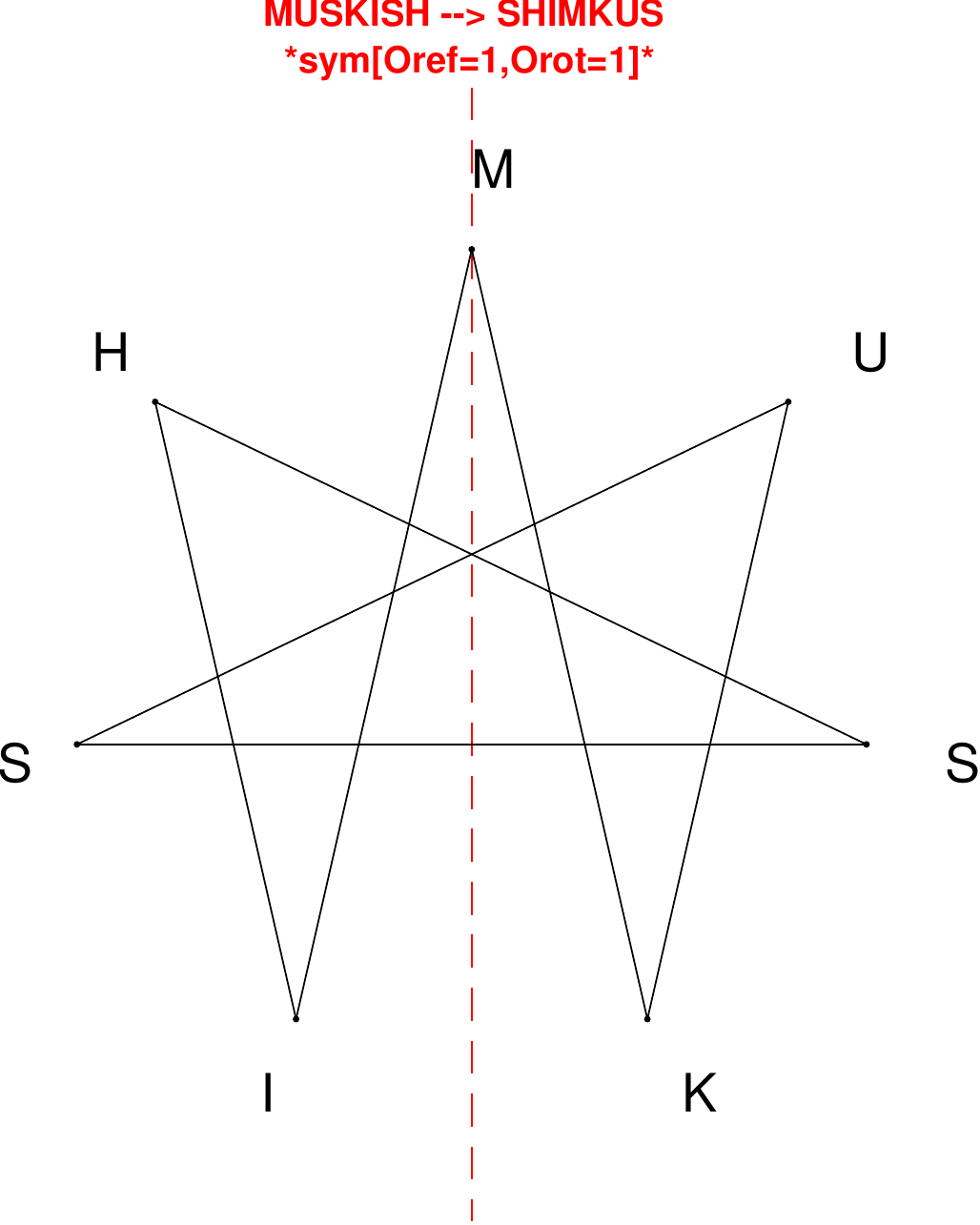}
\end{subfigure}
\hfill
\begin{subfigure}[T]{0.19\textwidth}
\centering
\includegraphics[width=\textwidth]{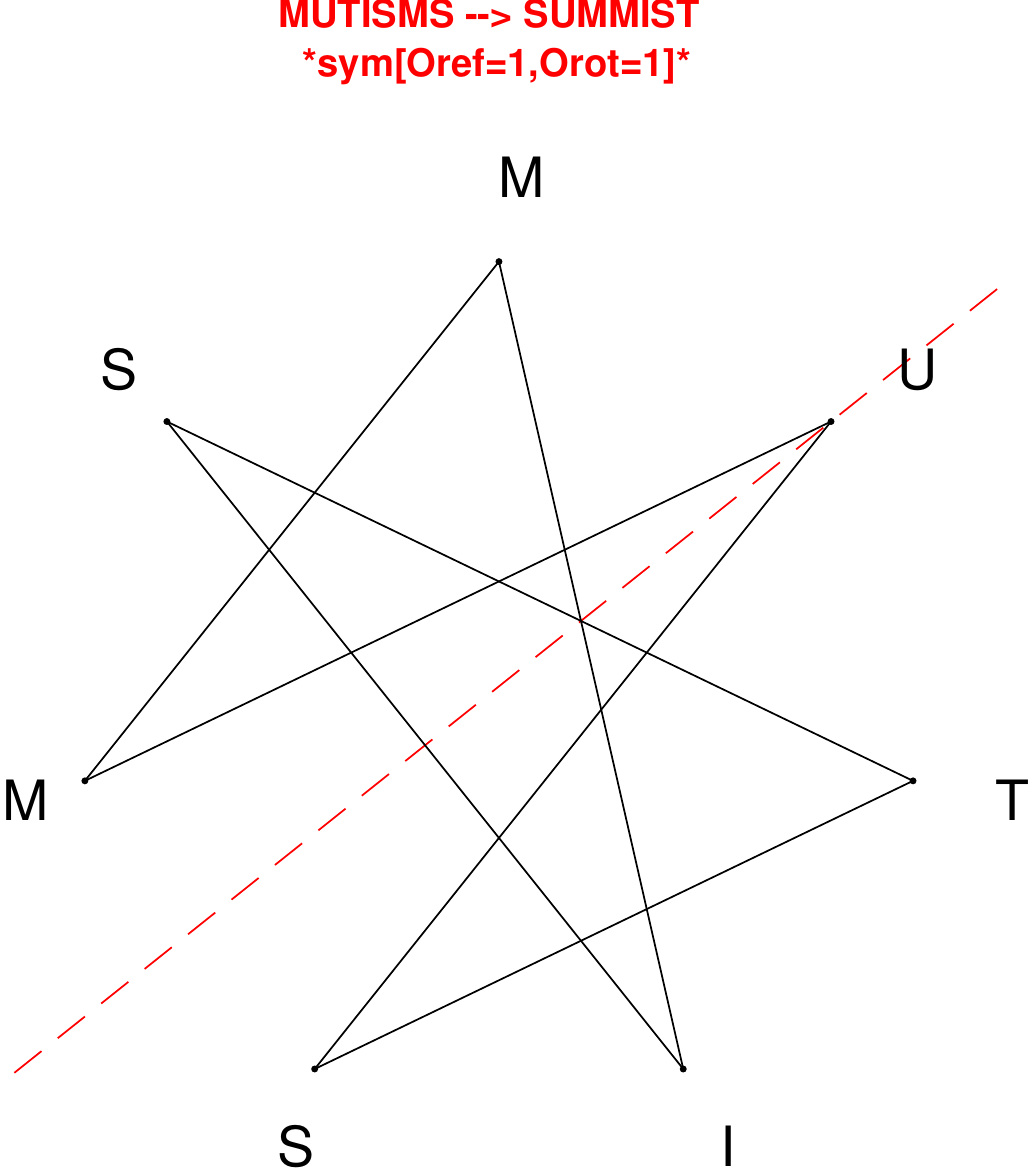}
\end{subfigure}
\hfill
\begin{subfigure}[T]{0.19\textwidth}
\centering
\includegraphics[width=\textwidth]{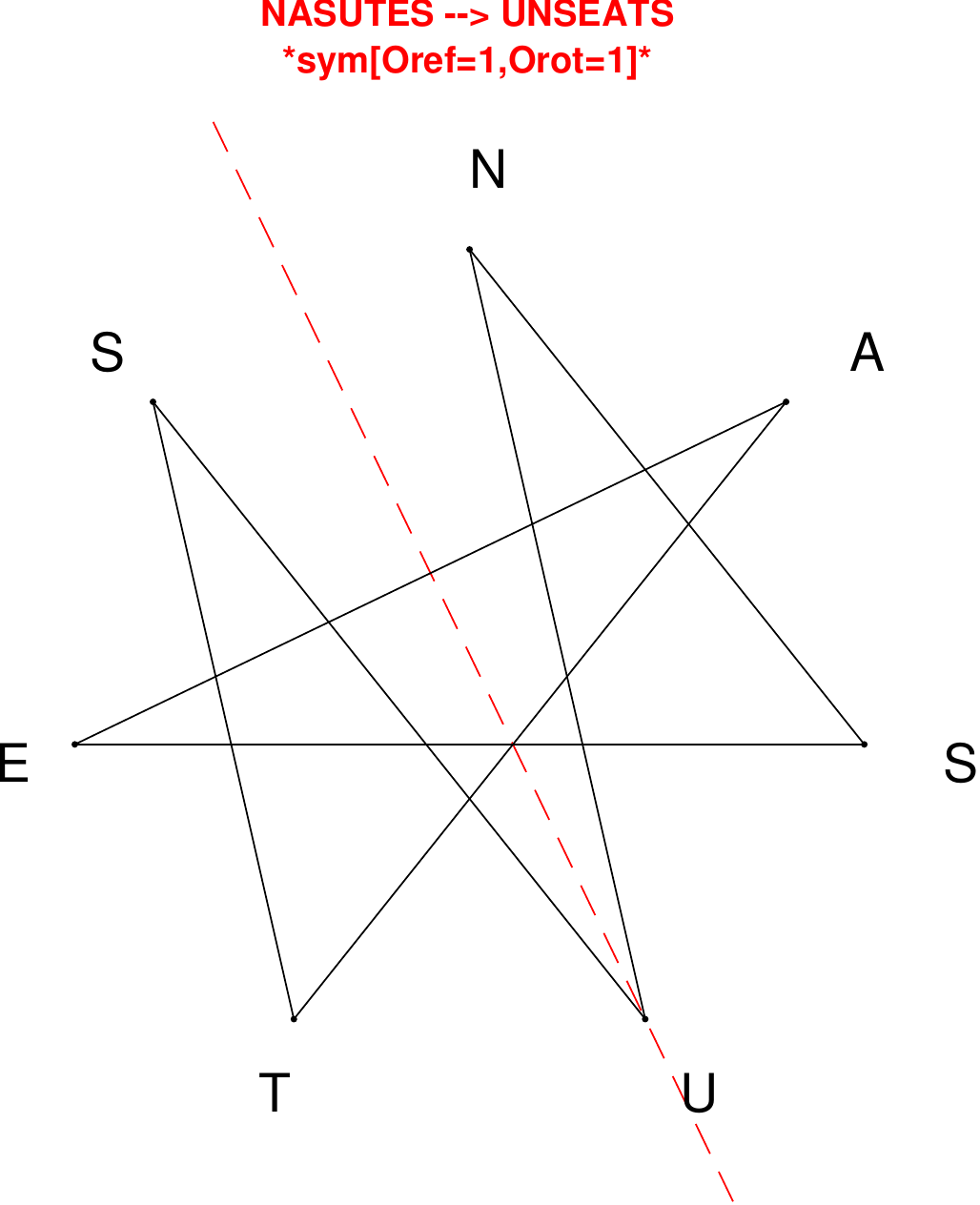}
\end{subfigure}
\hfill
\begin{subfigure}[T]{0.19\textwidth}
\centering
\includegraphics[width=\textwidth]{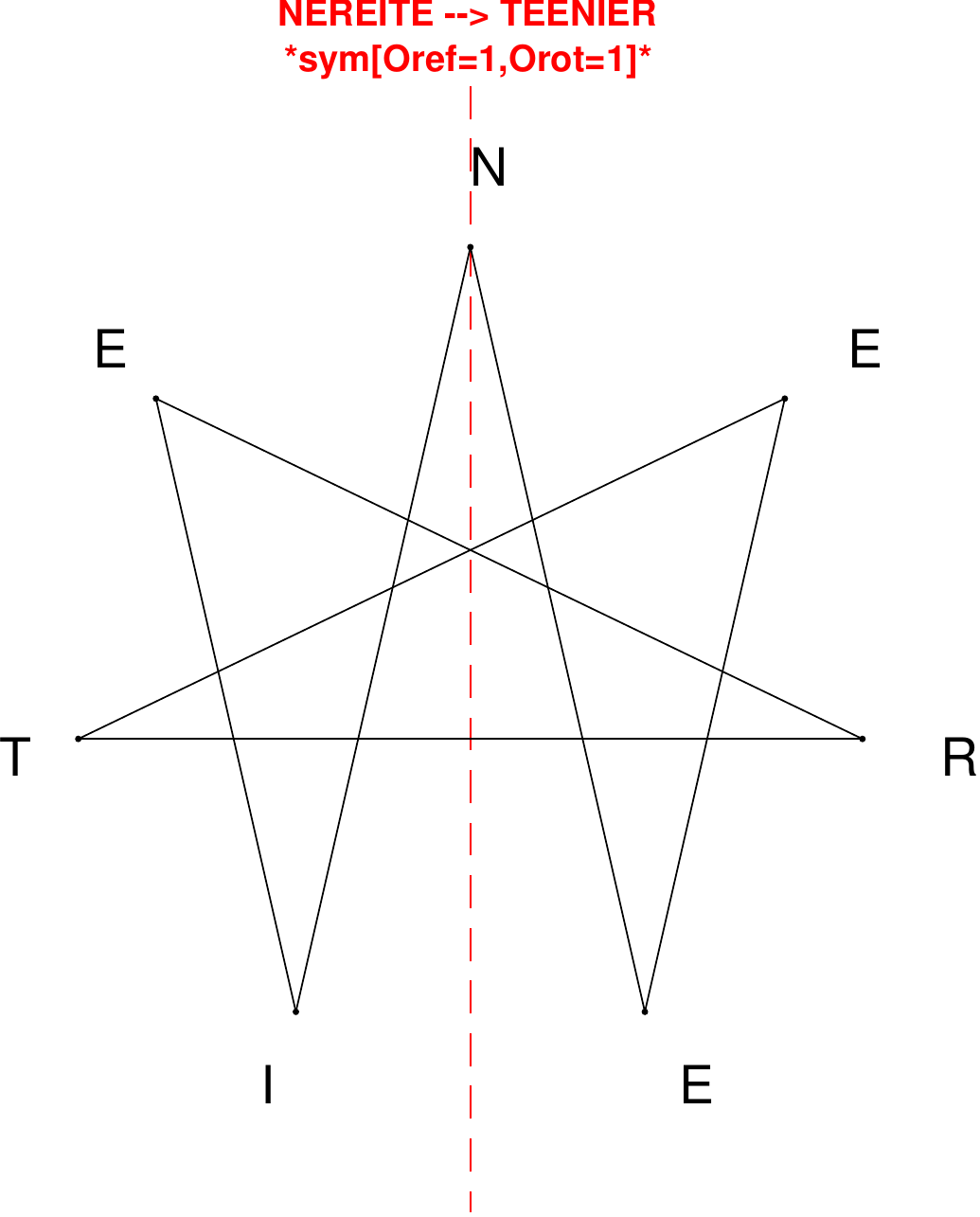}
\end{subfigure}
\end{figure}

\begin{figure}[H]
\centering
\begin{subfigure}[T]{0.19\textwidth}
\centering
\includegraphics[width=\textwidth]{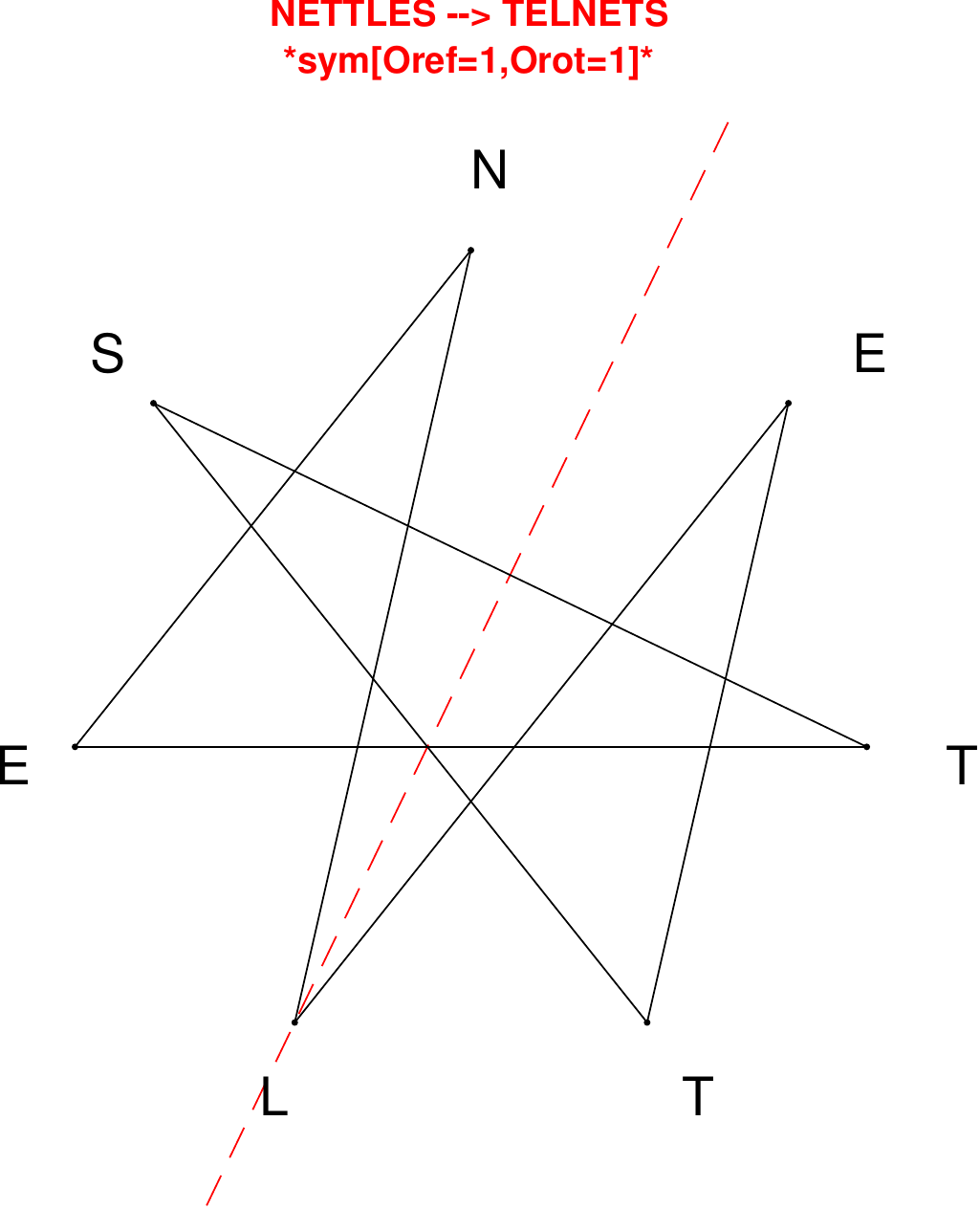}
\end{subfigure}
\hfill
\begin{subfigure}[T]{0.19\textwidth}
\centering
\includegraphics[width=\textwidth]{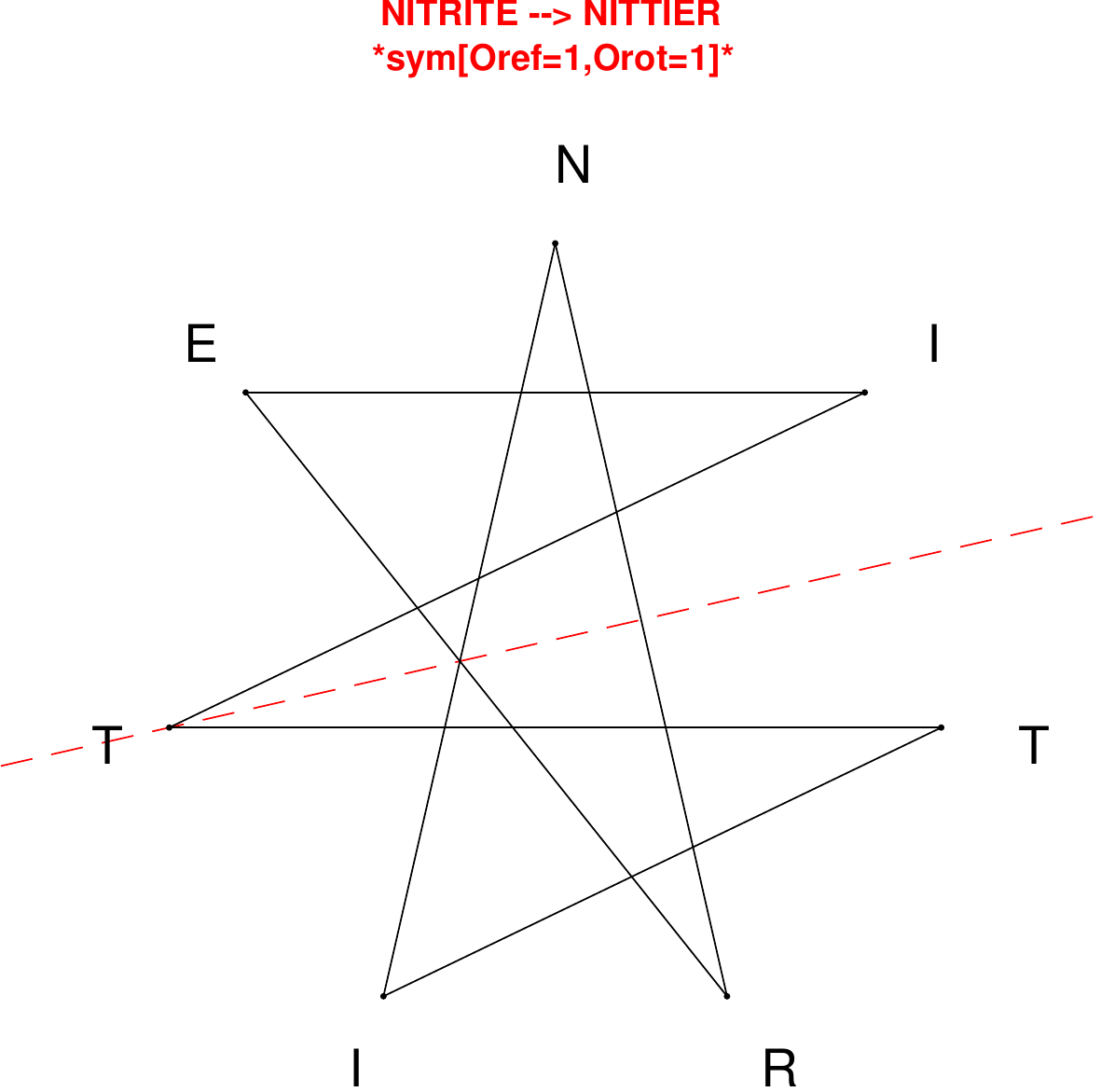}
\end{subfigure}
\hfill
\begin{subfigure}[T]{0.19\textwidth}
\centering
\includegraphics[width=\textwidth]{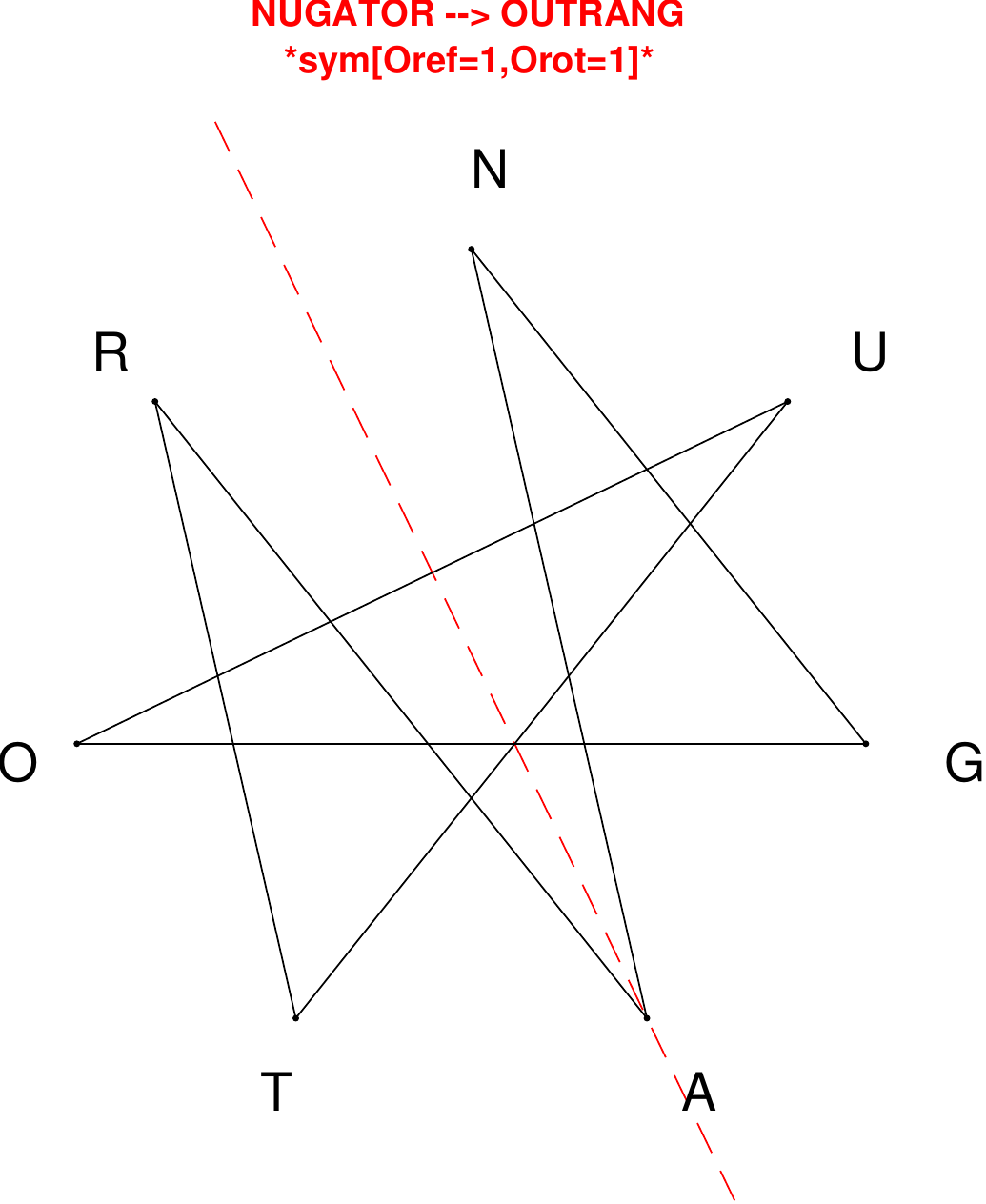}
\end{subfigure}
\hfill
\begin{subfigure}[T]{0.19\textwidth}
\centering
\includegraphics[width=\textwidth]{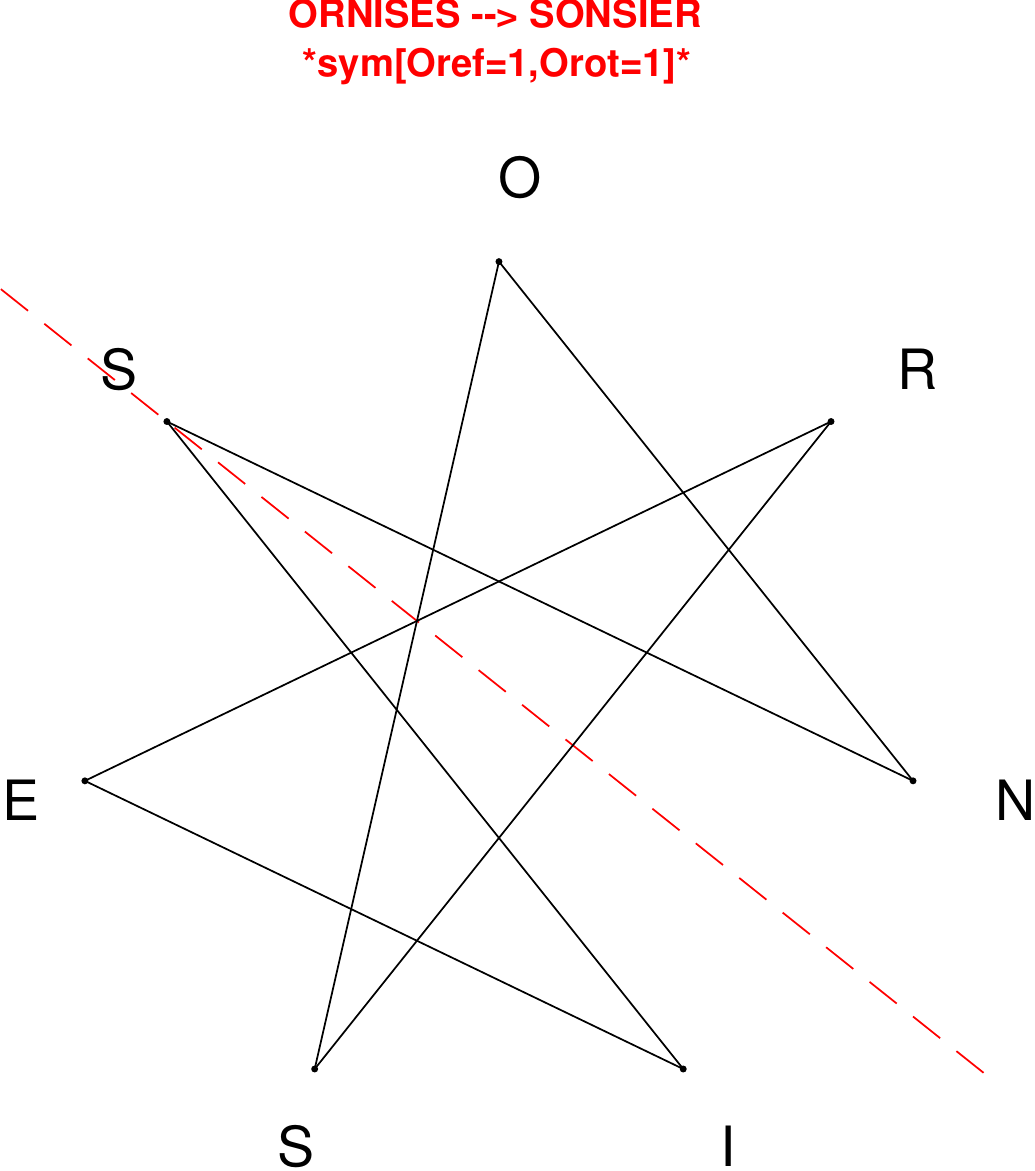}
\end{subfigure}
\hfill
\begin{subfigure}[T]{0.19\textwidth}
\centering
\includegraphics[width=\textwidth]{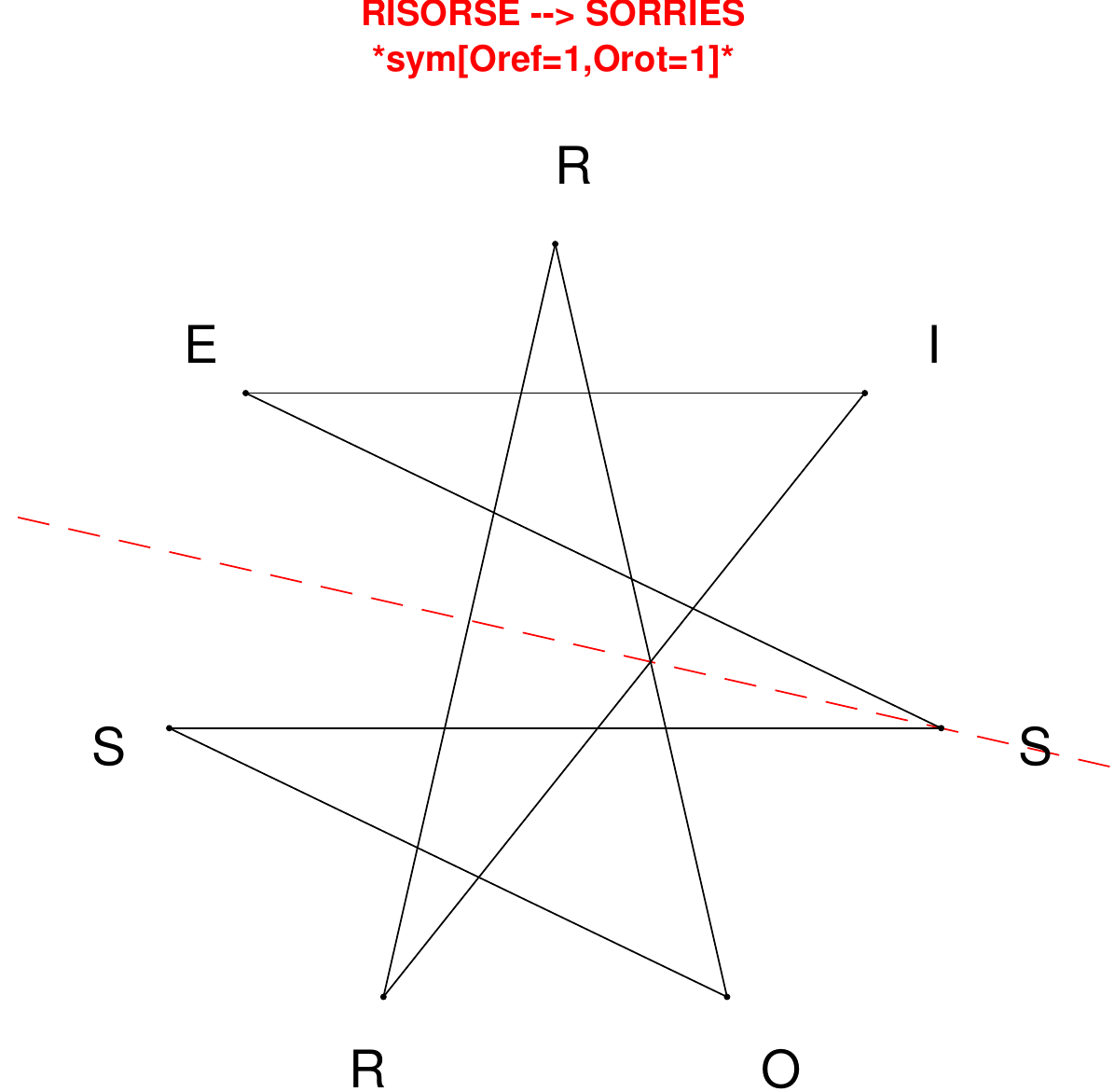}
\end{subfigure}
\end{figure}

\begin{figure}[H]
\centering
\begin{subfigure}[T]{0.19\textwidth}
\centering
\includegraphics[width=\textwidth]{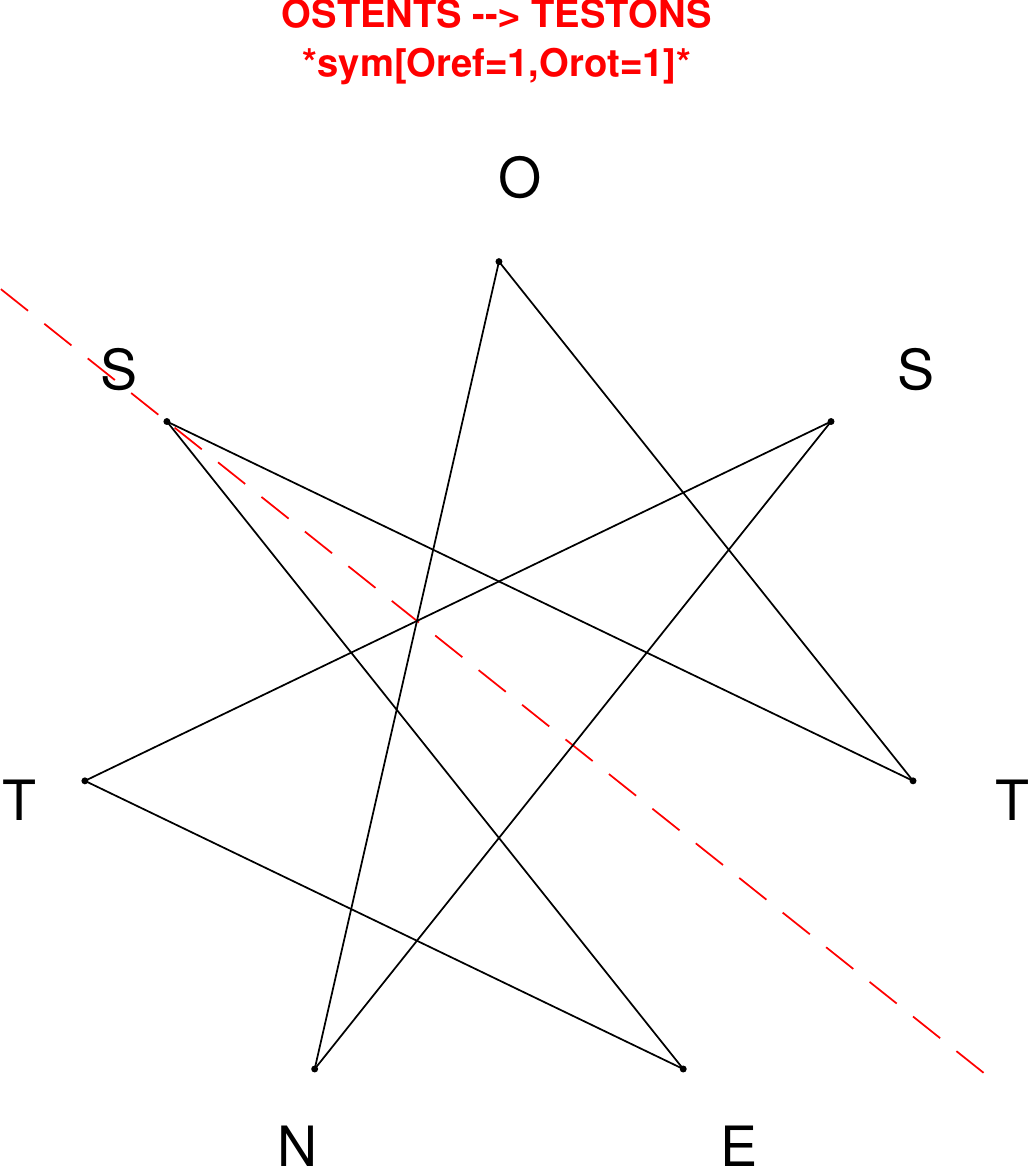}
\end{subfigure}
\hfill
\begin{subfigure}[T]{0.19\textwidth}
\centering
\includegraphics[width=\textwidth]{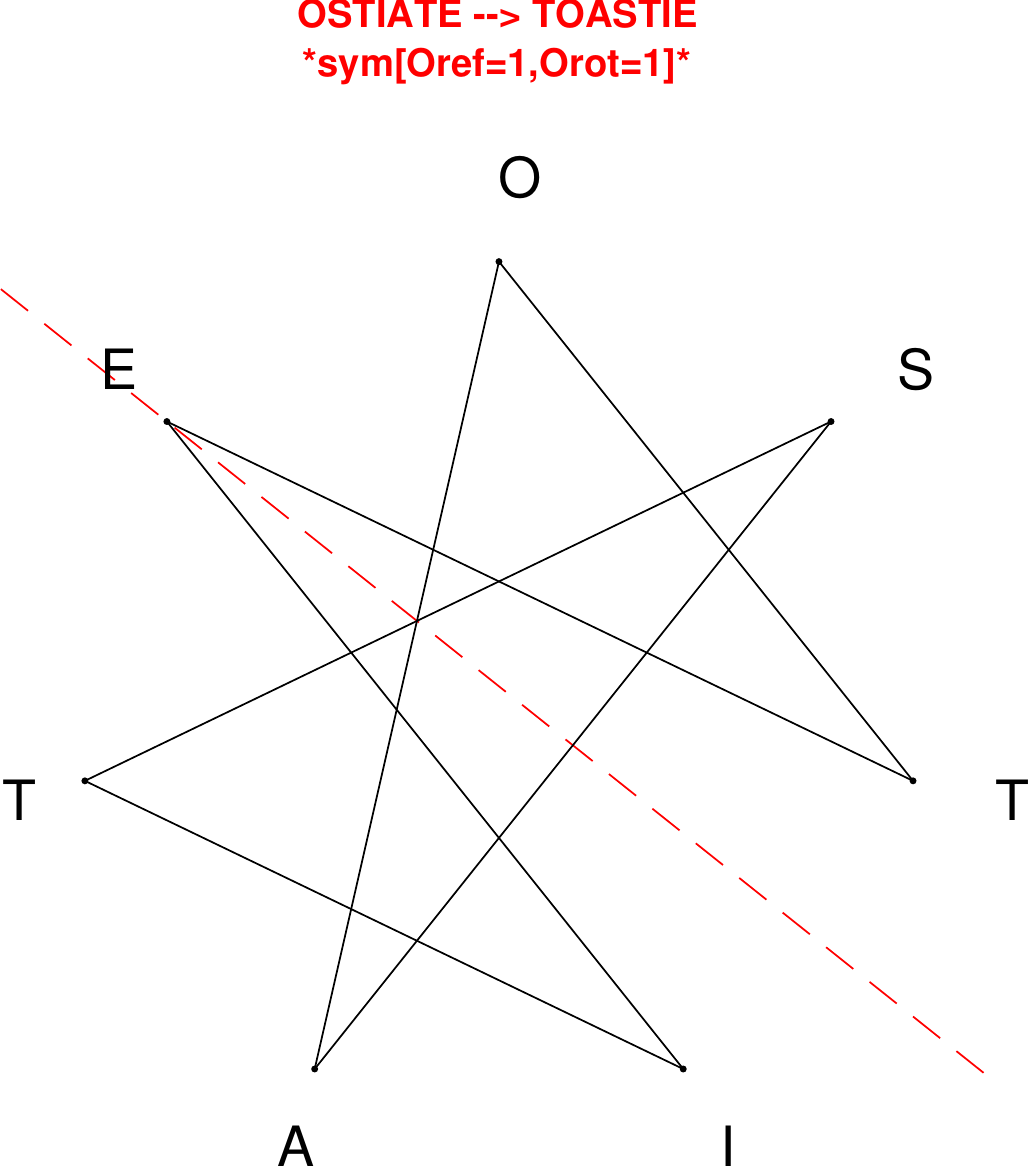}
\end{subfigure}
\hfill
\begin{subfigure}[T]{0.19\textwidth}
\centering
\includegraphics[width=\textwidth]{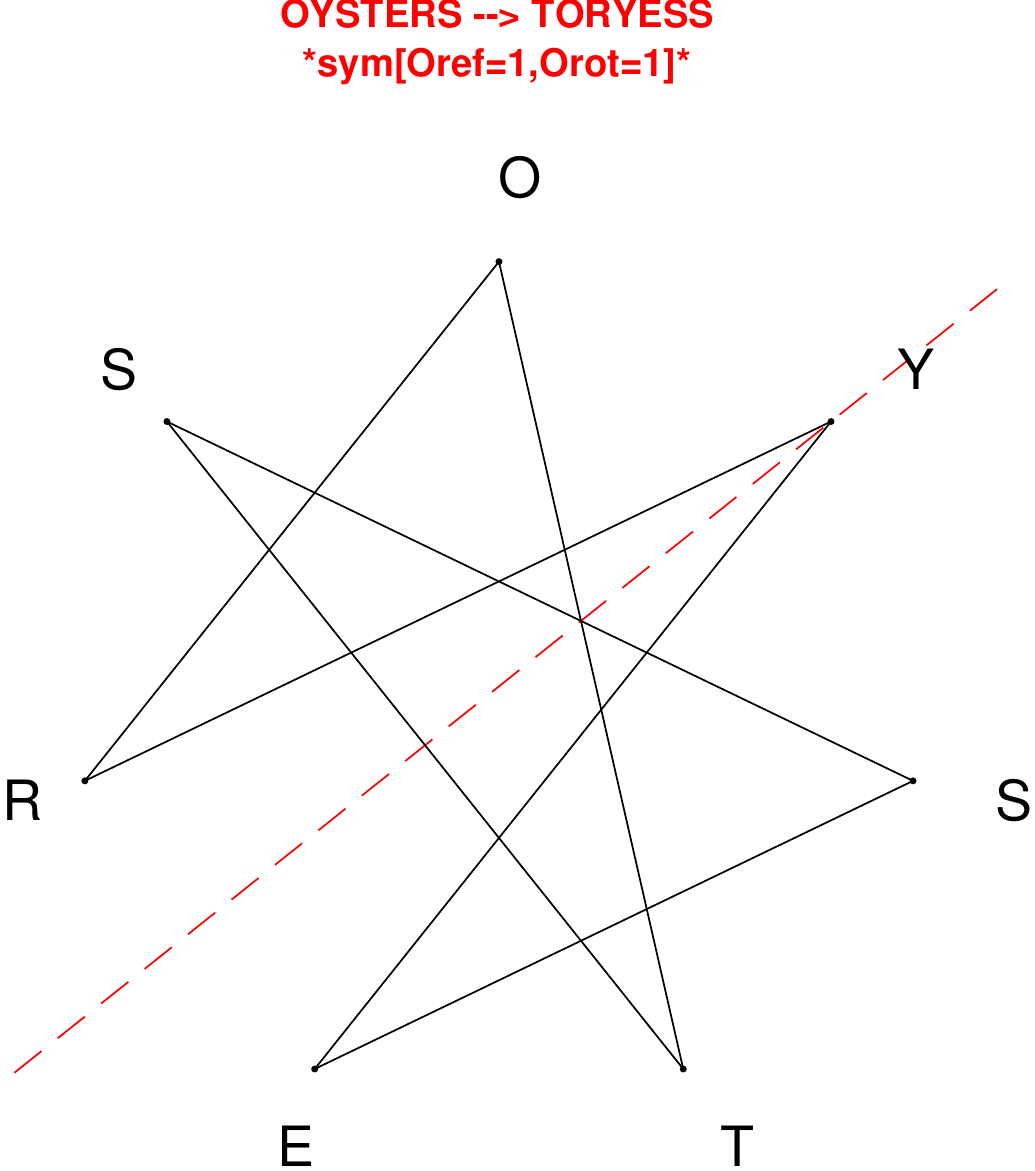}
\end{subfigure}
\hfill
\begin{subfigure}[T]{0.19\textwidth}
\centering
\includegraphics[width=\textwidth]{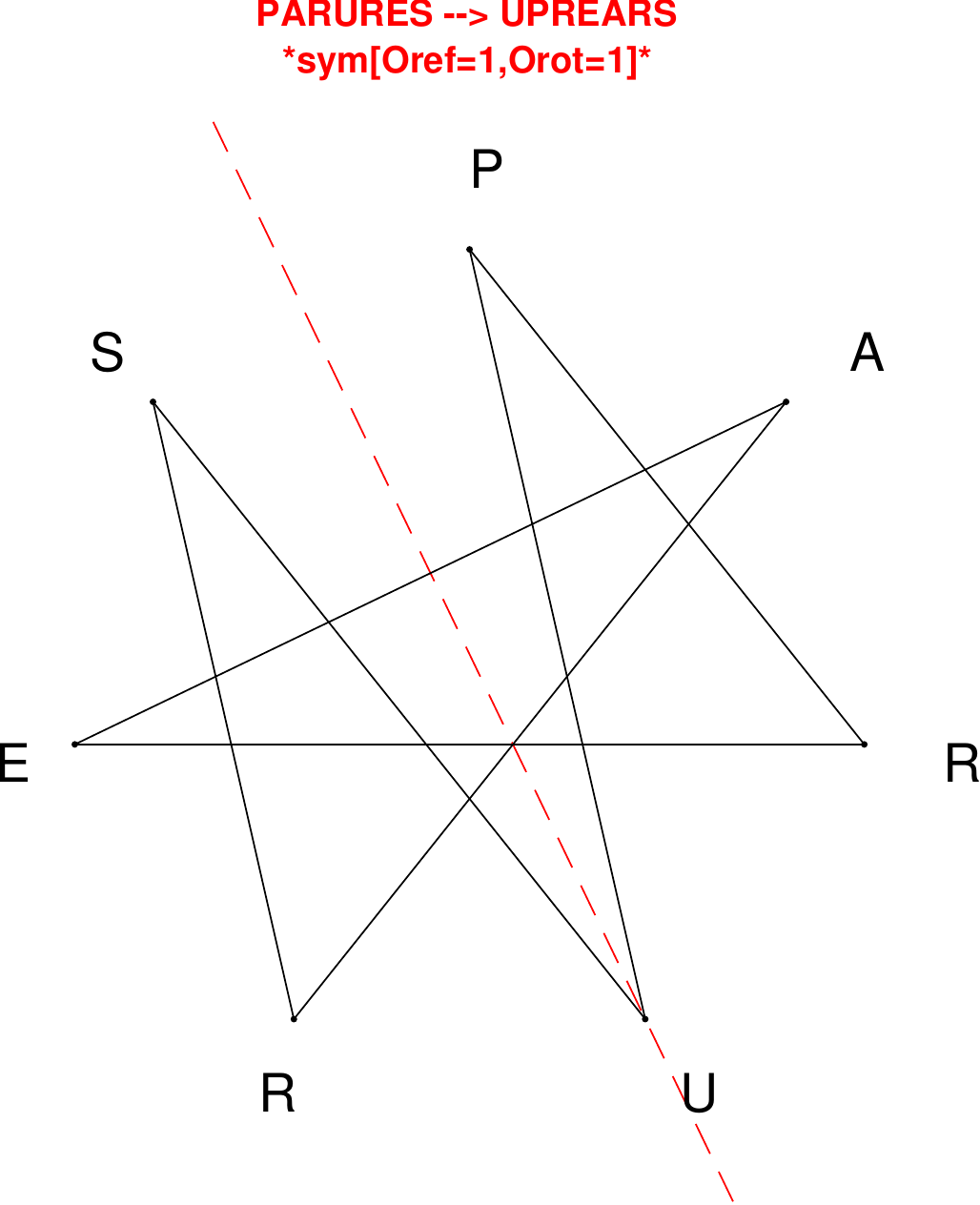}
\end{subfigure}
\hfill
\begin{subfigure}[T]{0.19\textwidth}
\centering
\includegraphics[width=\textwidth]{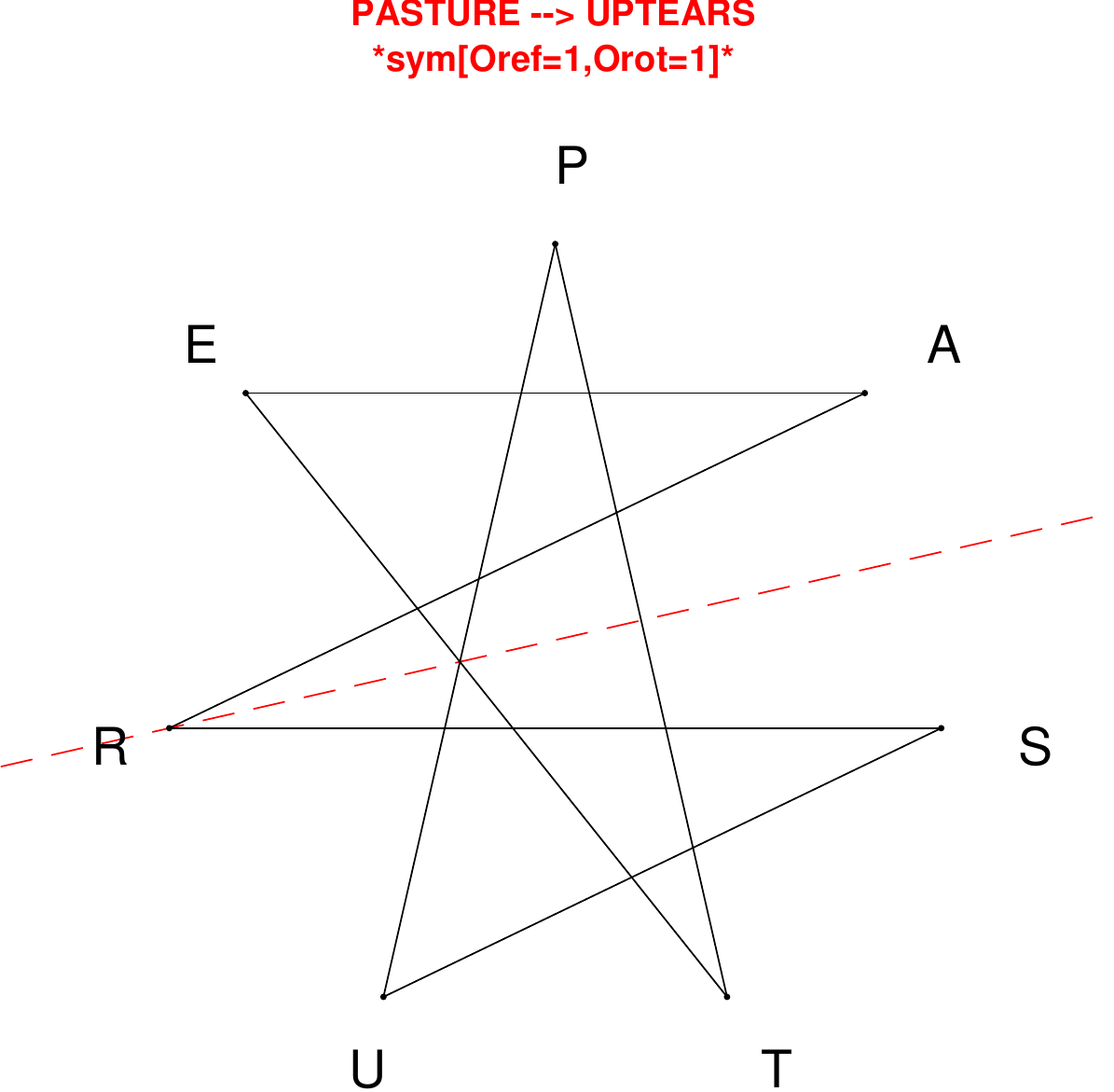}
\end{subfigure}
\end{figure}

\begin{figure}[H]
\centering
\begin{subfigure}[T]{0.19\textwidth}
\centering
\includegraphics[width=\textwidth]{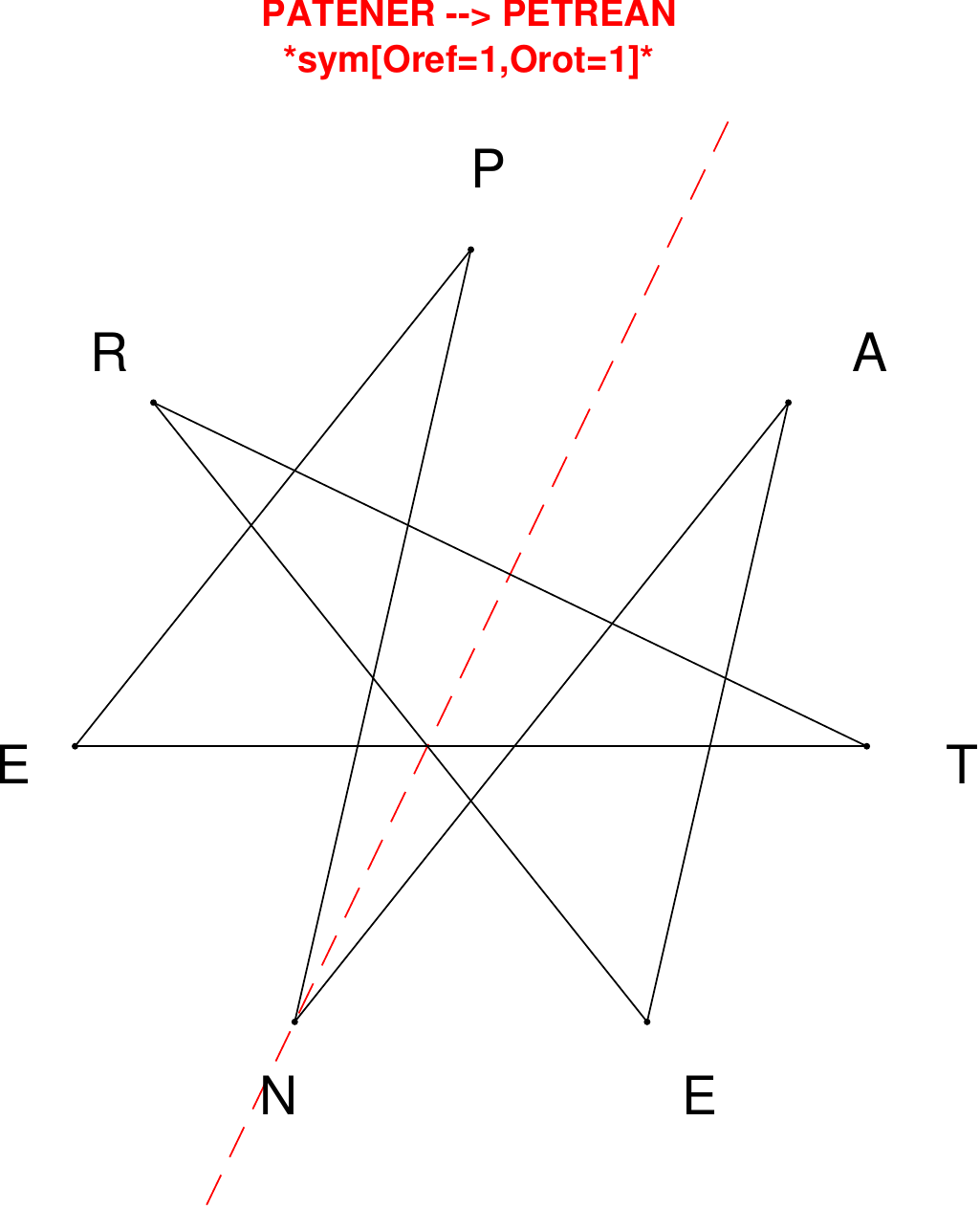}
\end{subfigure}
\hfill
\begin{subfigure}[T]{0.19\textwidth}
\centering
\includegraphics[width=\textwidth]{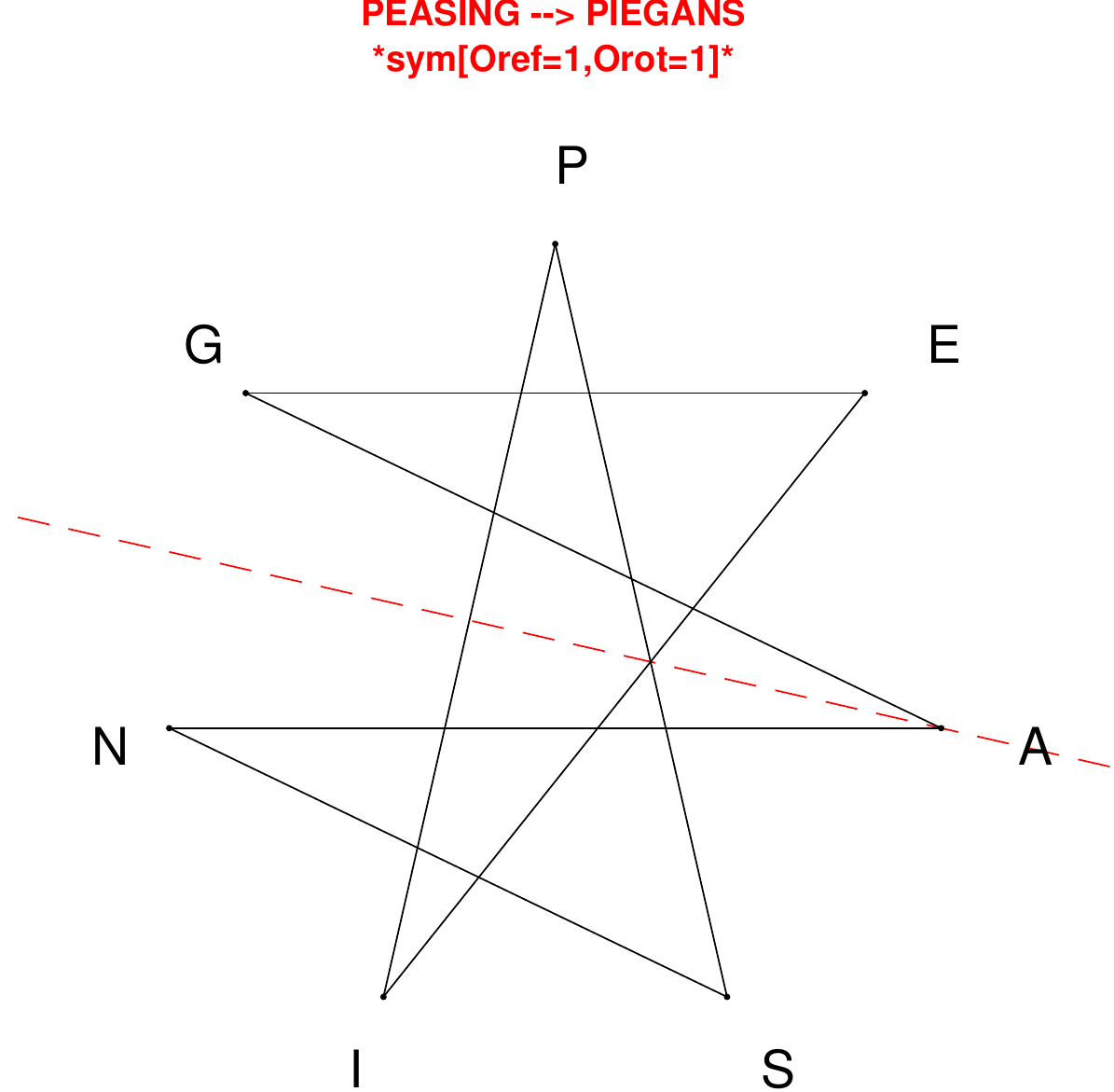}
\end{subfigure}
\hfill
\begin{subfigure}[T]{0.19\textwidth}
\centering
\includegraphics[width=\textwidth]{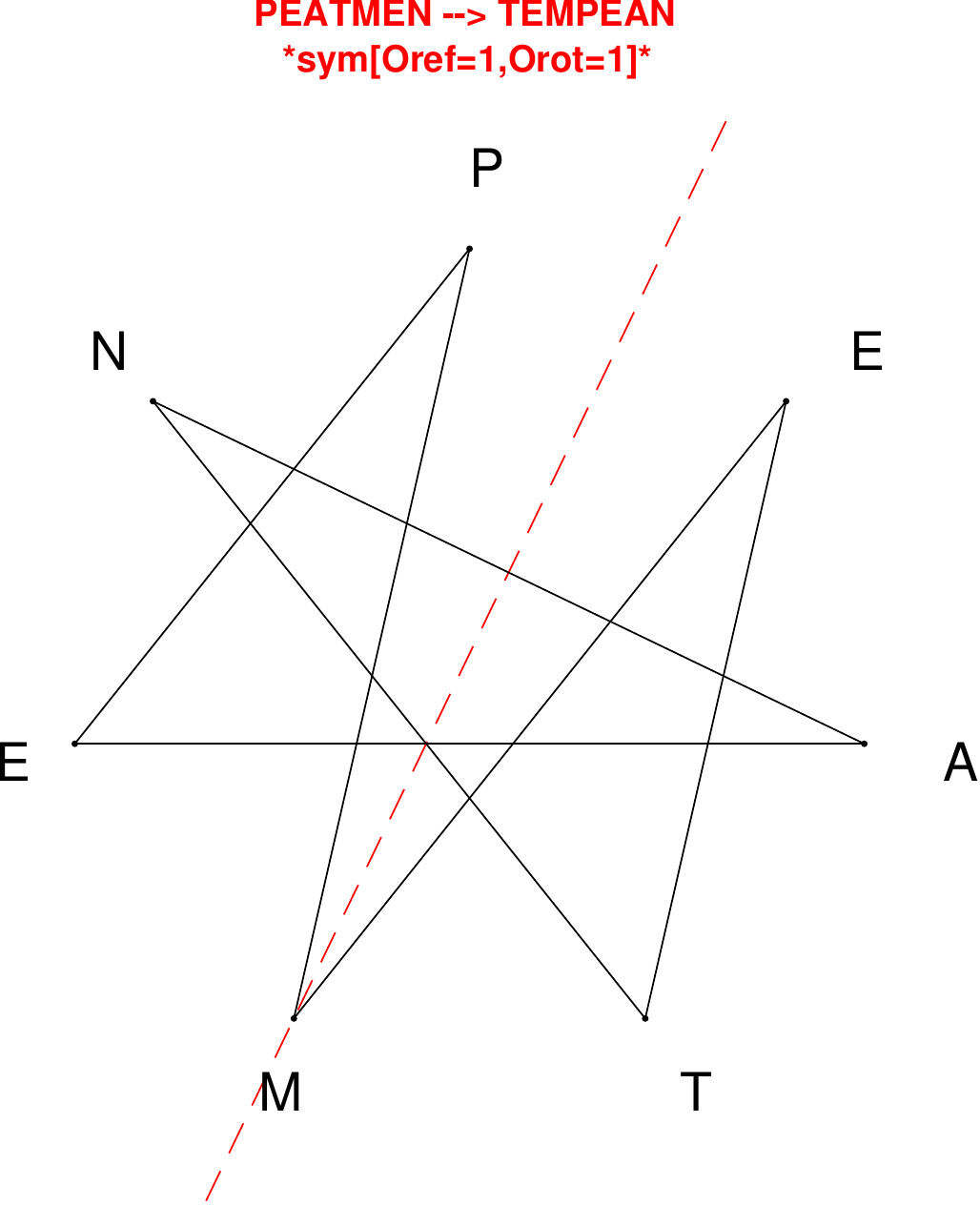}
\end{subfigure}
\hfill
\begin{subfigure}[T]{0.19\textwidth}
\centering
\includegraphics[width=\textwidth]{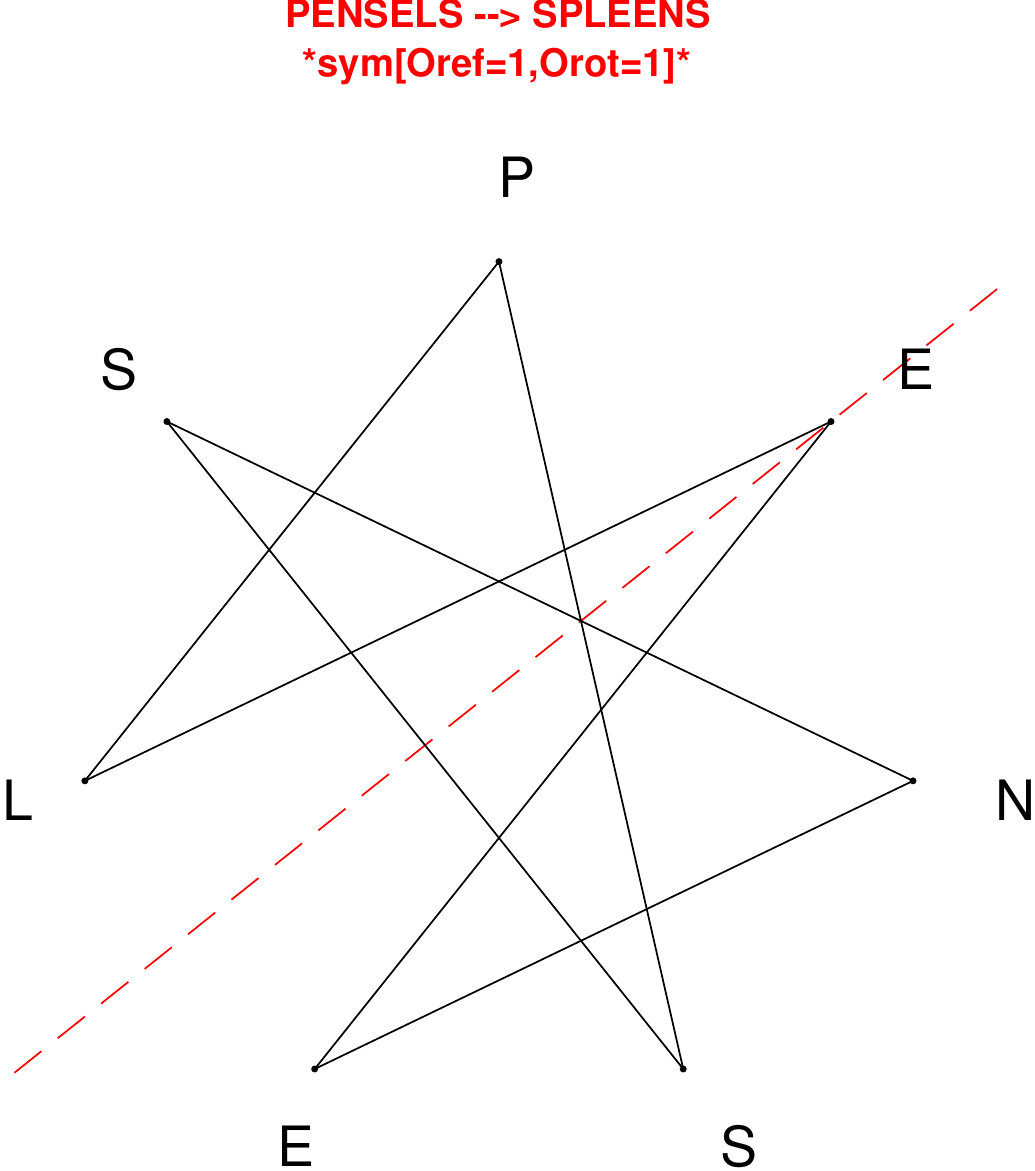}
\end{subfigure}
\hfill
\begin{subfigure}[T]{0.19\textwidth}
\centering
\includegraphics[width=\textwidth]{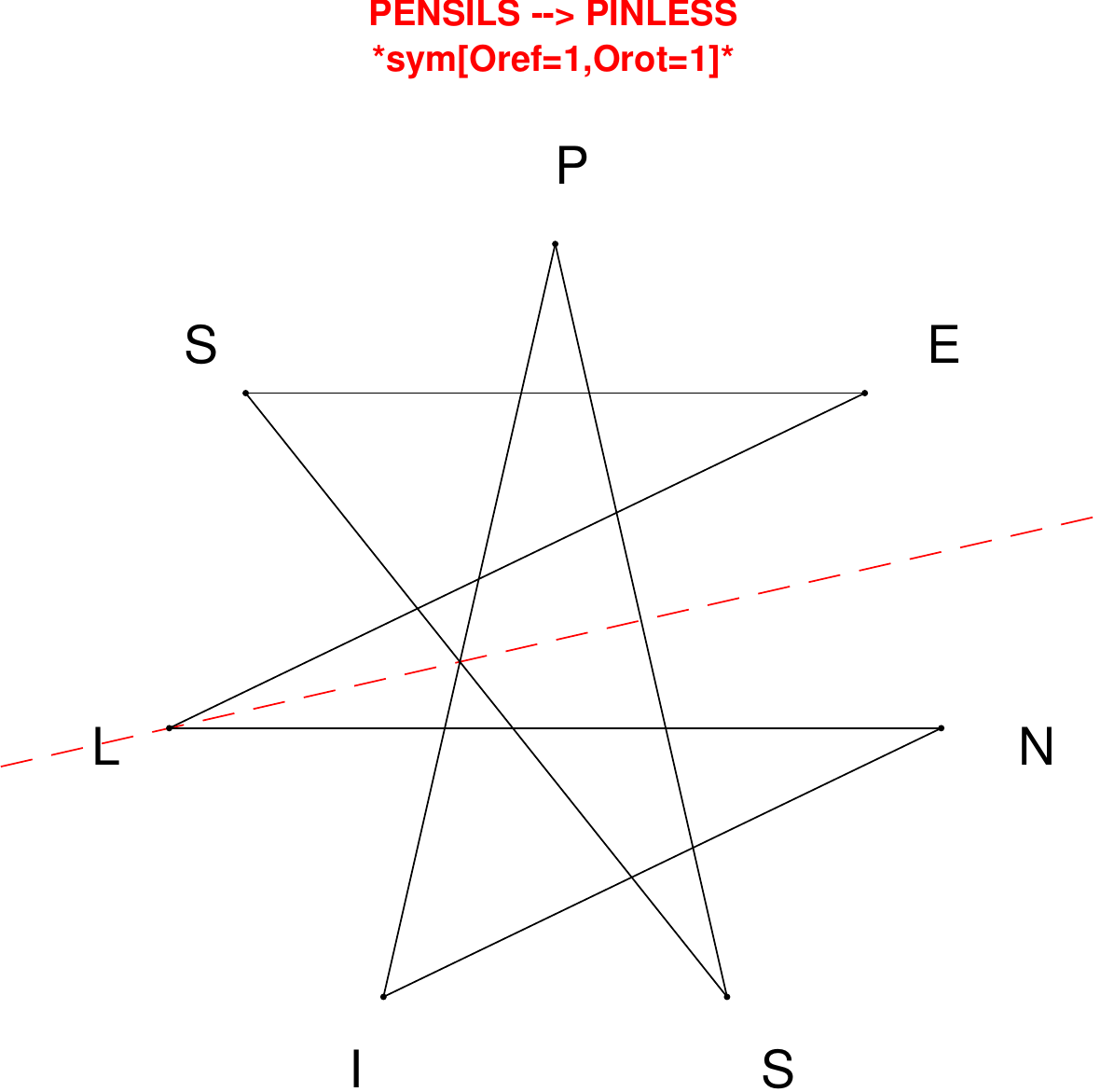}
\end{subfigure}
\end{figure}

\begin{figure}[H]
\centering
\begin{subfigure}[T]{0.19\textwidth}
\centering
\includegraphics[width=\textwidth]{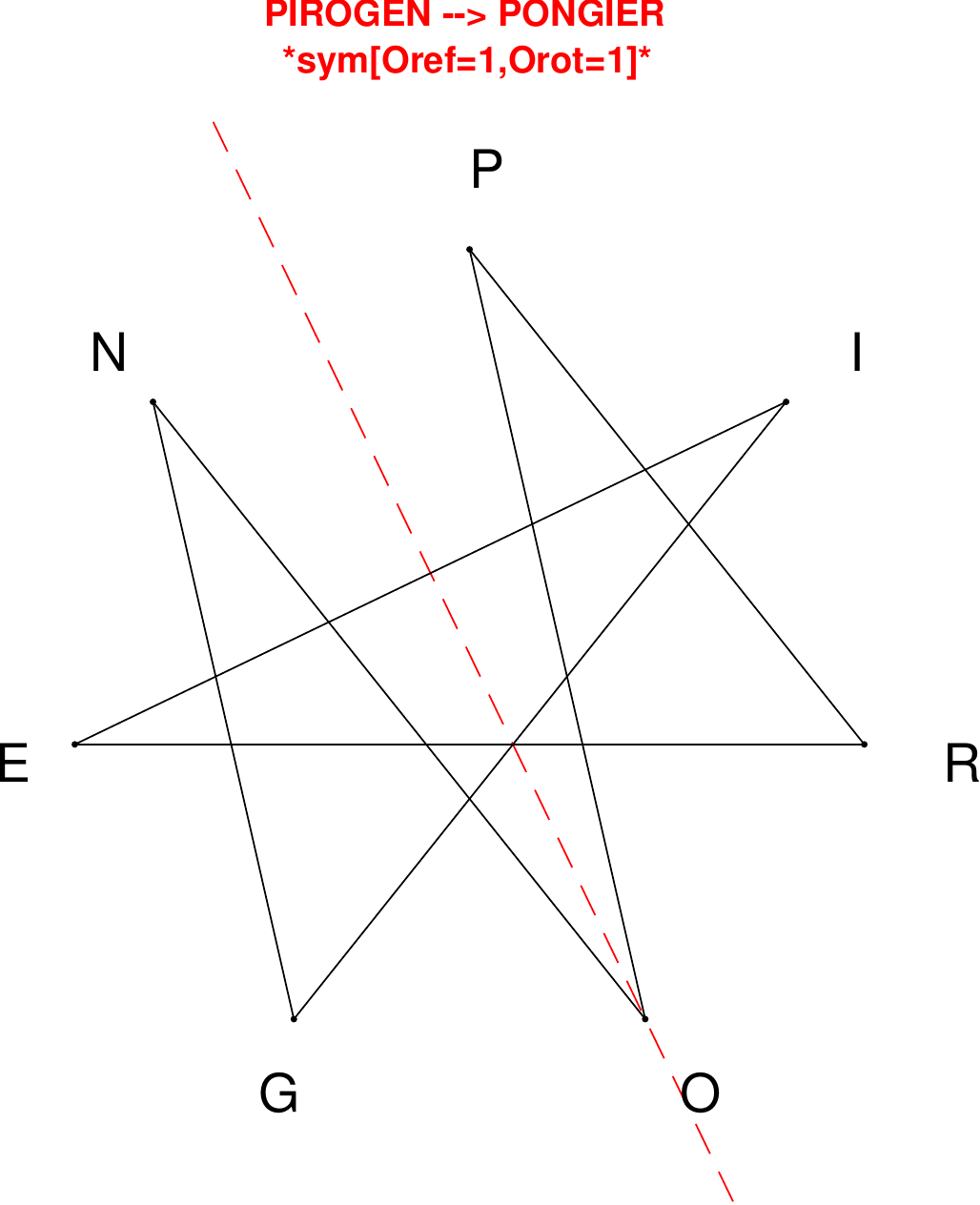}
\end{subfigure}
\hfill
\begin{subfigure}[T]{0.19\textwidth}
\centering
\includegraphics[width=\textwidth]{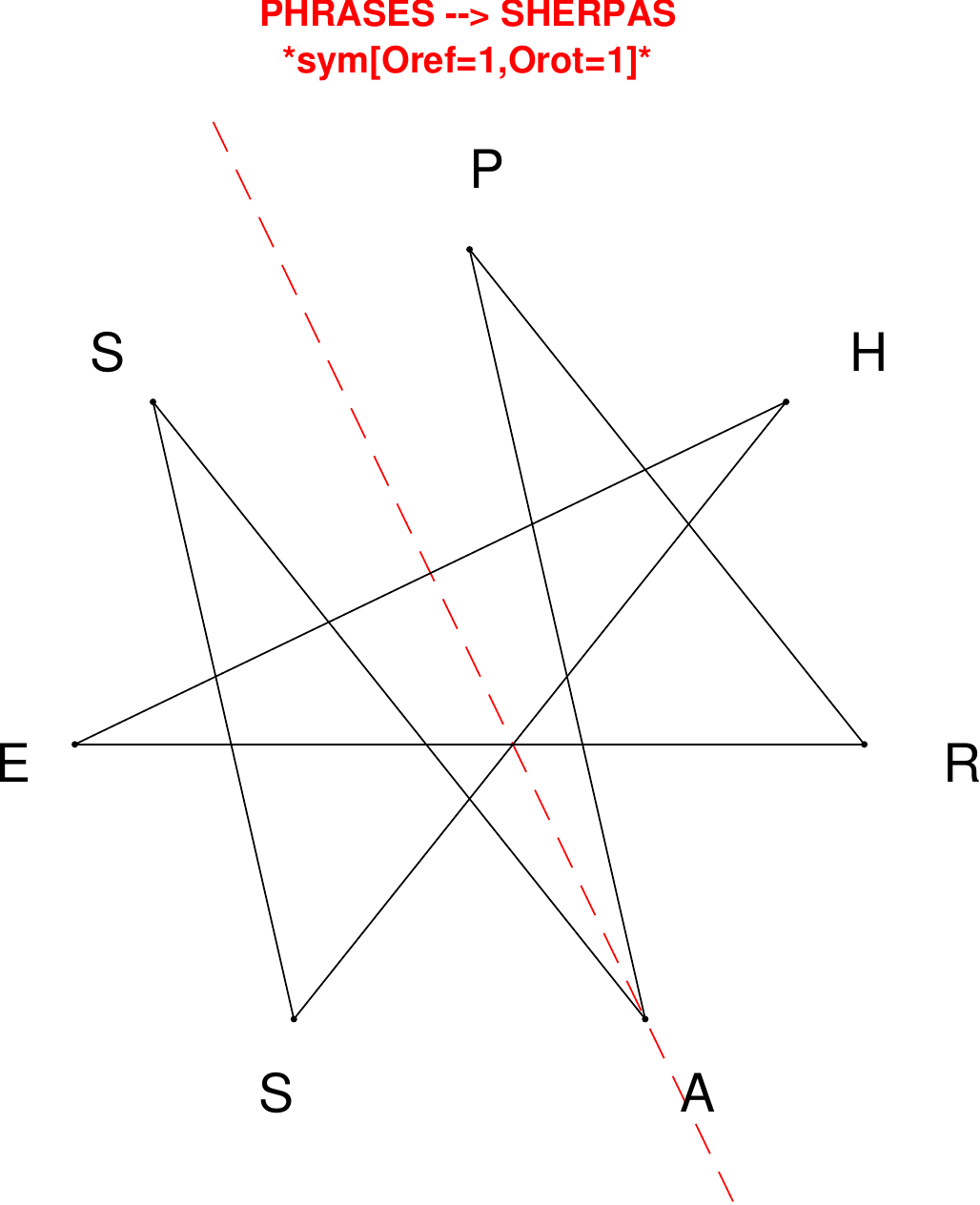}
\end{subfigure}
\hfill
\begin{subfigure}[T]{0.19\textwidth}
\centering
\includegraphics[width=\textwidth]{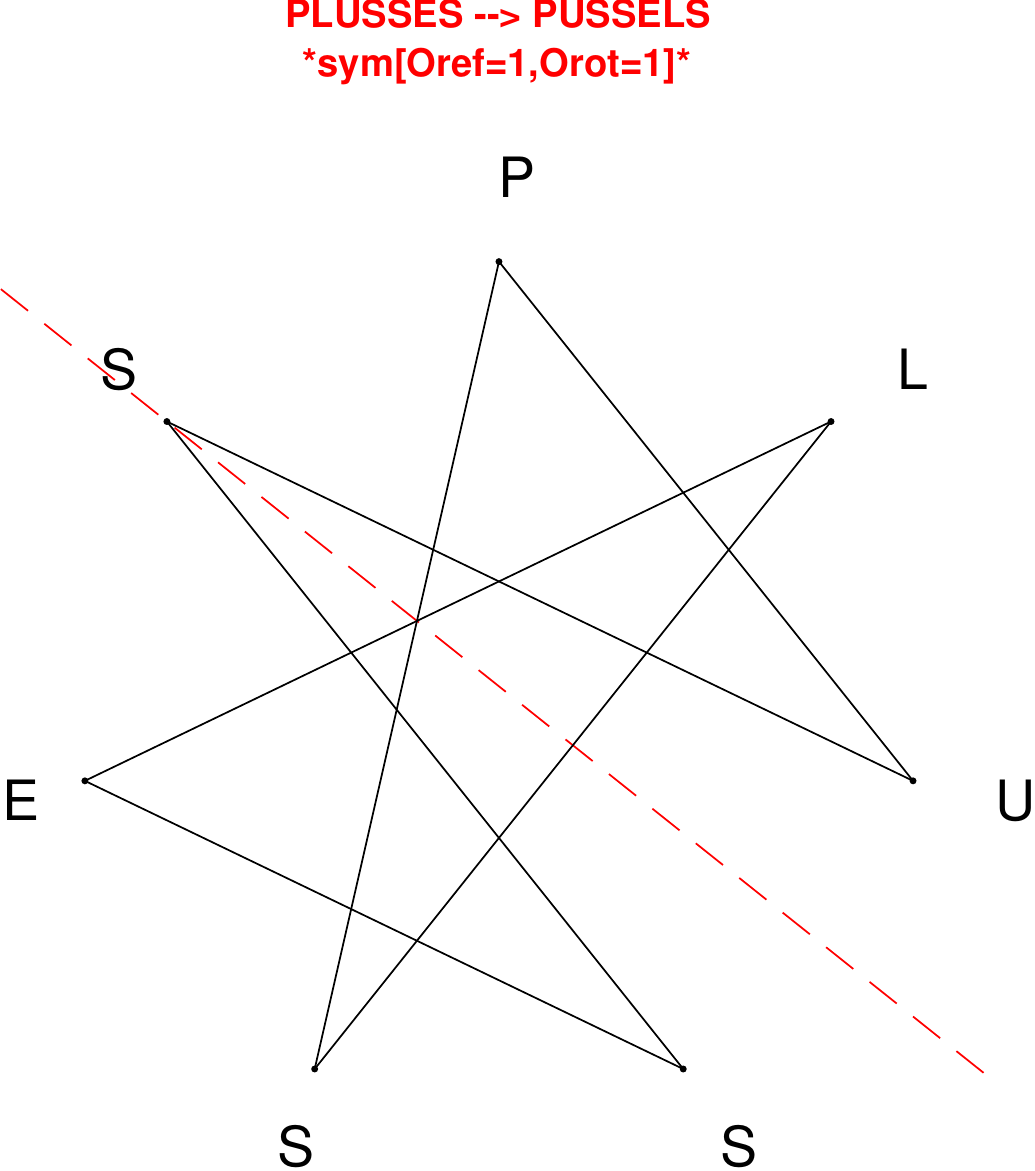}
\end{subfigure}
\hfill
\begin{subfigure}[T]{0.19\textwidth}
\centering
\includegraphics[width=\textwidth]{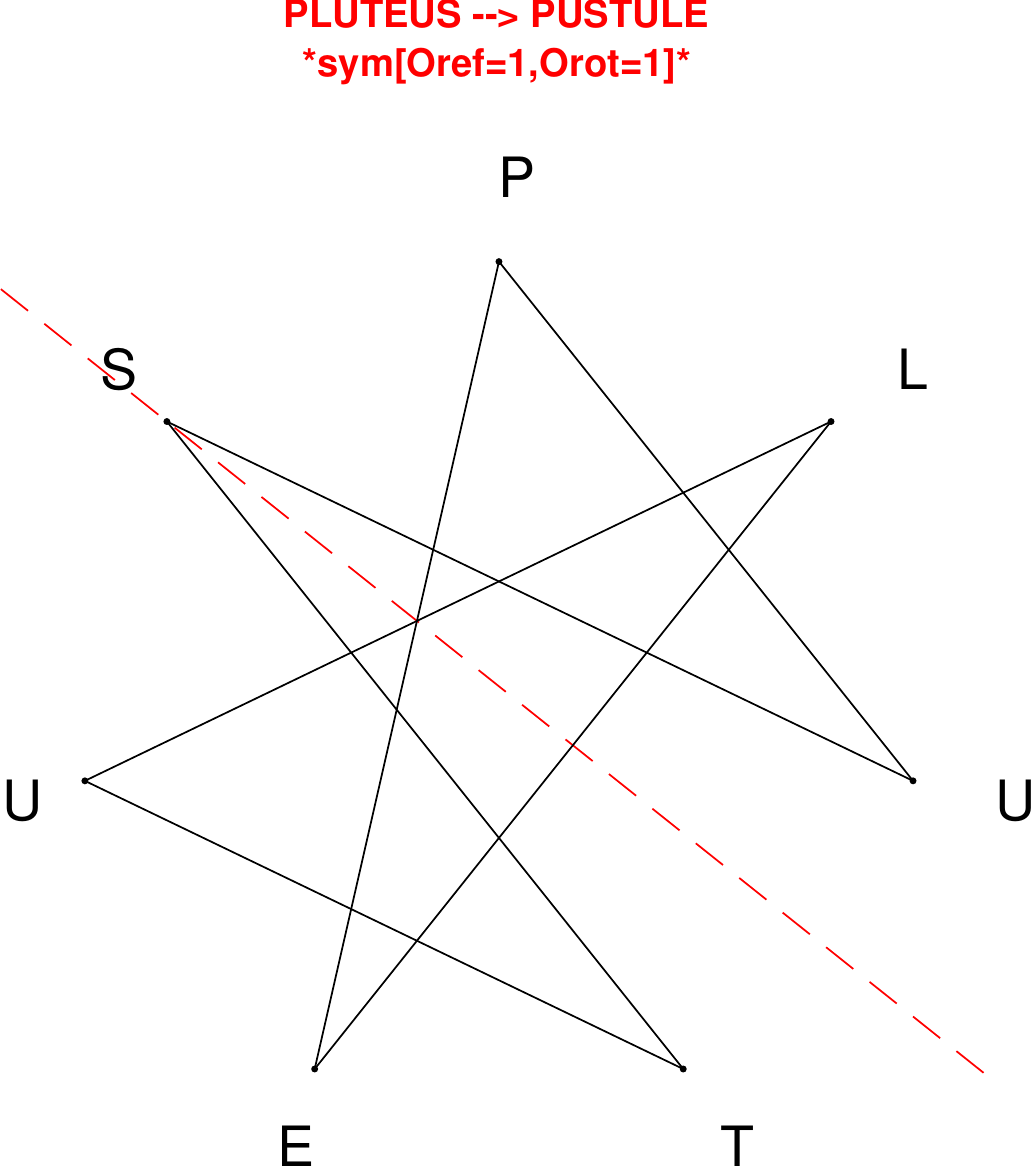}
\end{subfigure}
\hfill
\begin{subfigure}[T]{0.19\textwidth}
\centering
\includegraphics[width=\textwidth]{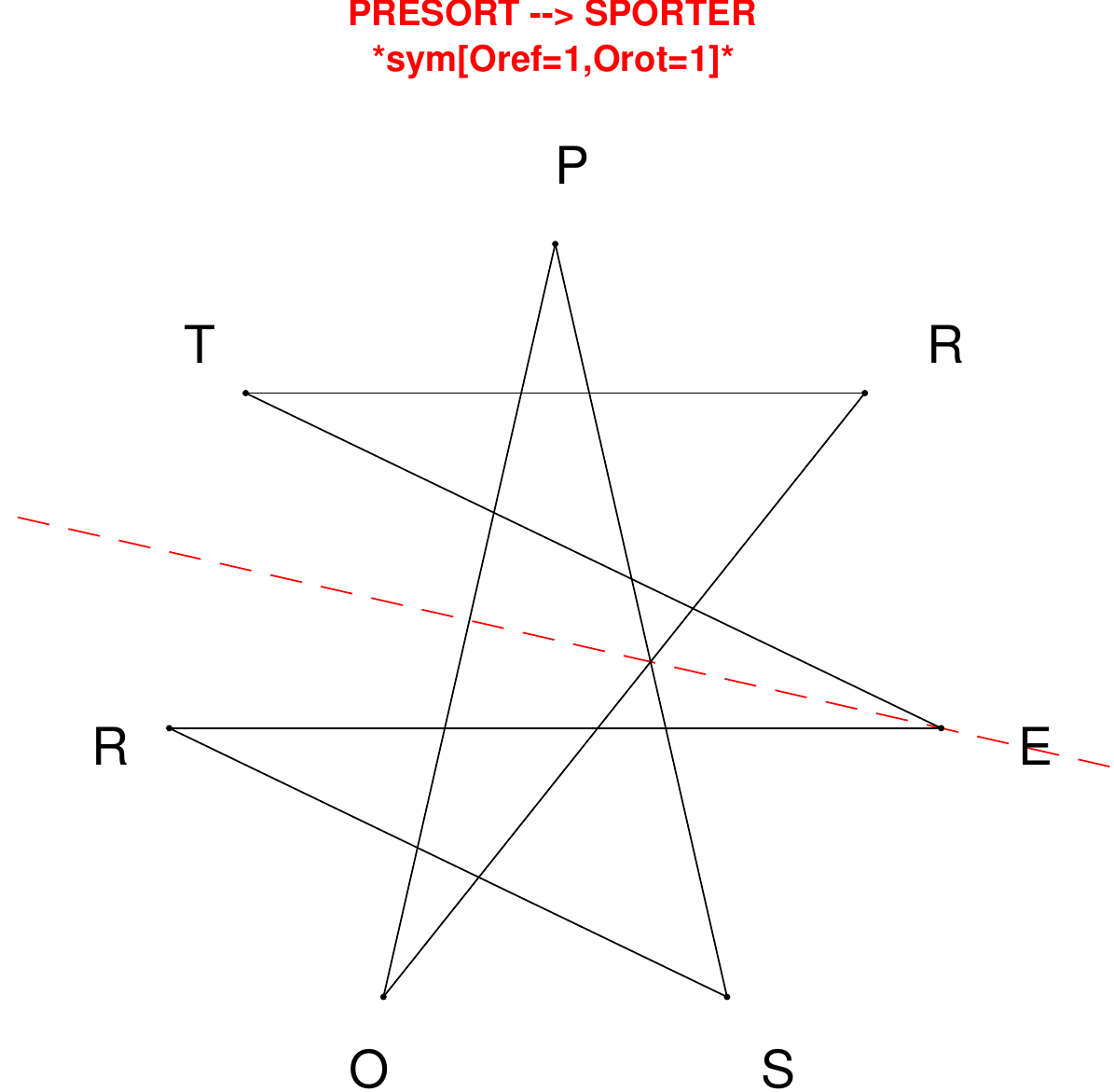}
\end{subfigure}
\end{figure}

\begin{figure}[H]
\centering
\begin{subfigure}[T]{0.19\textwidth}
\centering
\includegraphics[width=\textwidth]{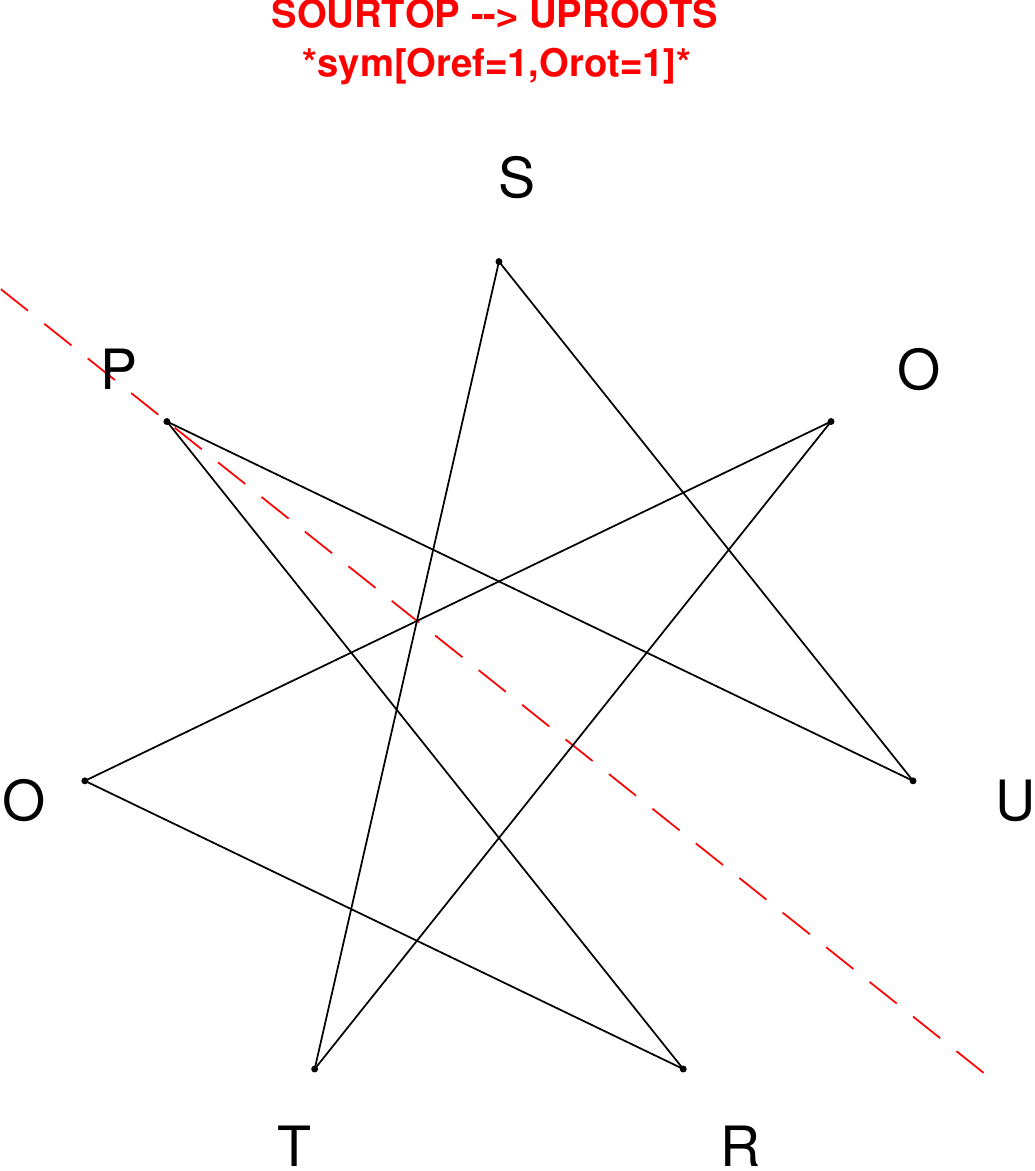}
\end{subfigure}
\hfill
\begin{subfigure}[T]{0.19\textwidth}
\centering
\includegraphics[width=\textwidth]{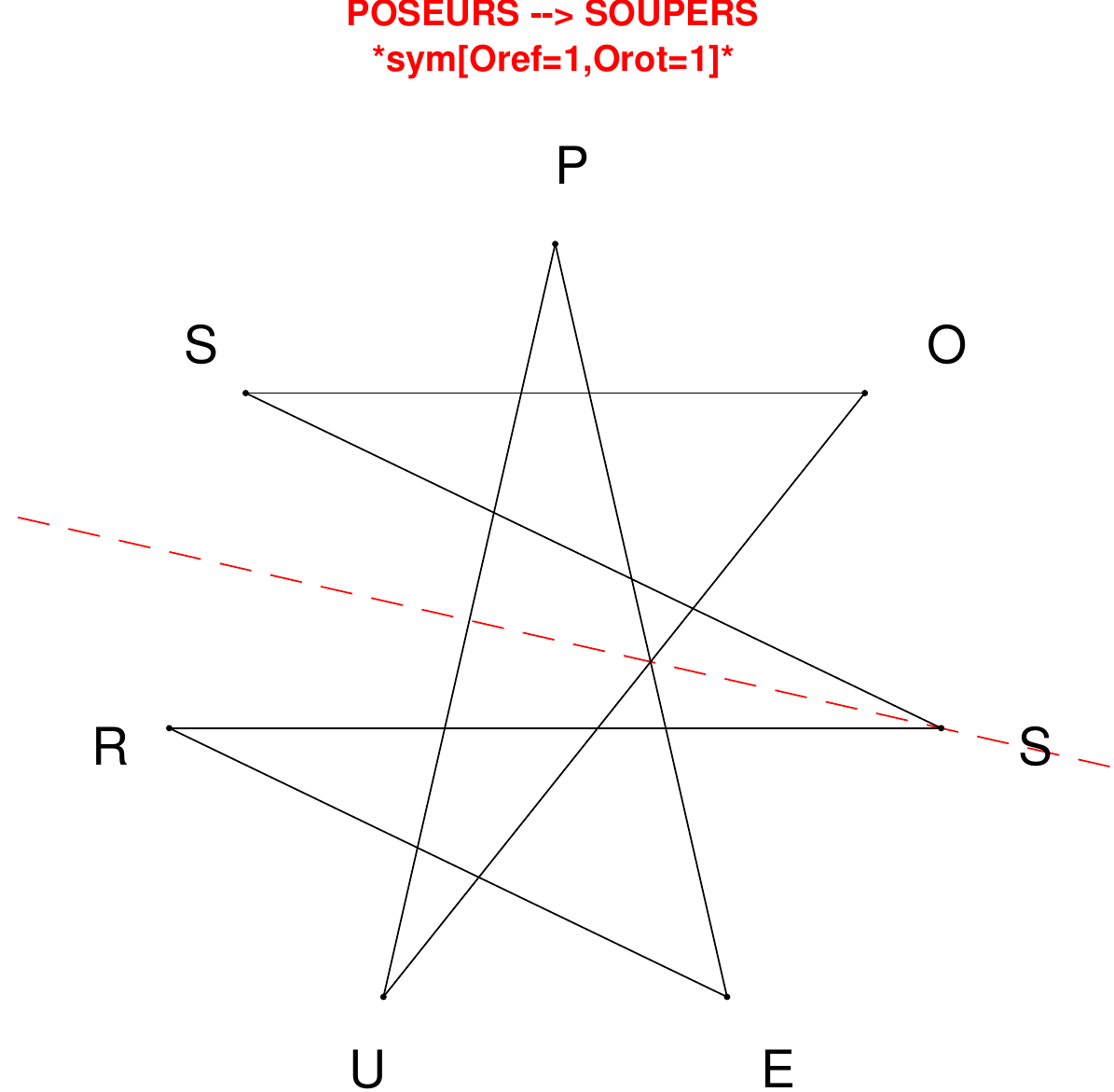}
\end{subfigure}
\hfill
\begin{subfigure}[T]{0.19\textwidth}
\centering
\includegraphics[width=\textwidth]{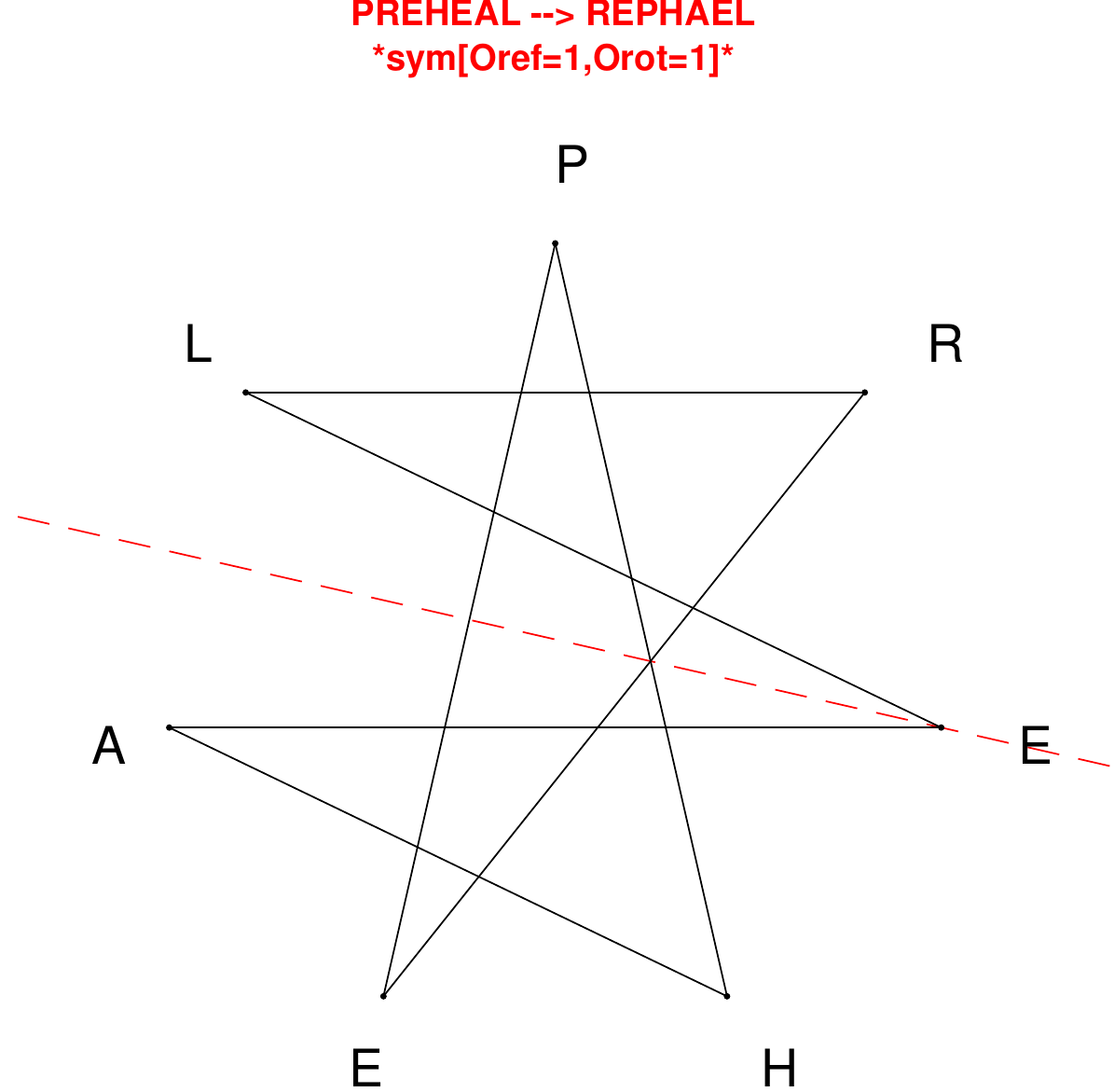}
\end{subfigure}
\hfill
\begin{subfigure}[T]{0.19\textwidth}
\centering
\includegraphics[width=\textwidth]{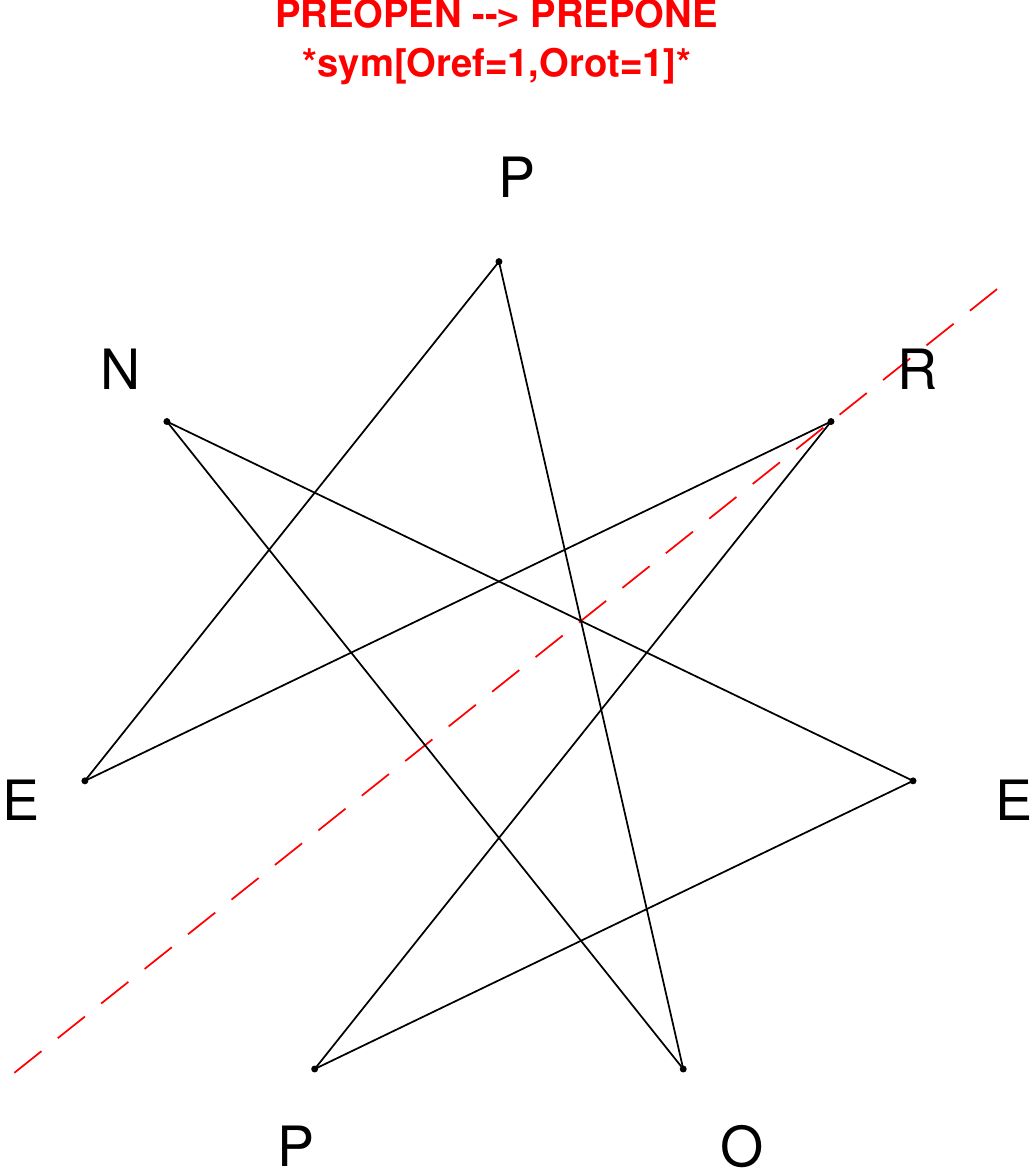}
\end{subfigure}
\hfill
\begin{subfigure}[T]{0.19\textwidth}
\centering
\includegraphics[width=\textwidth]{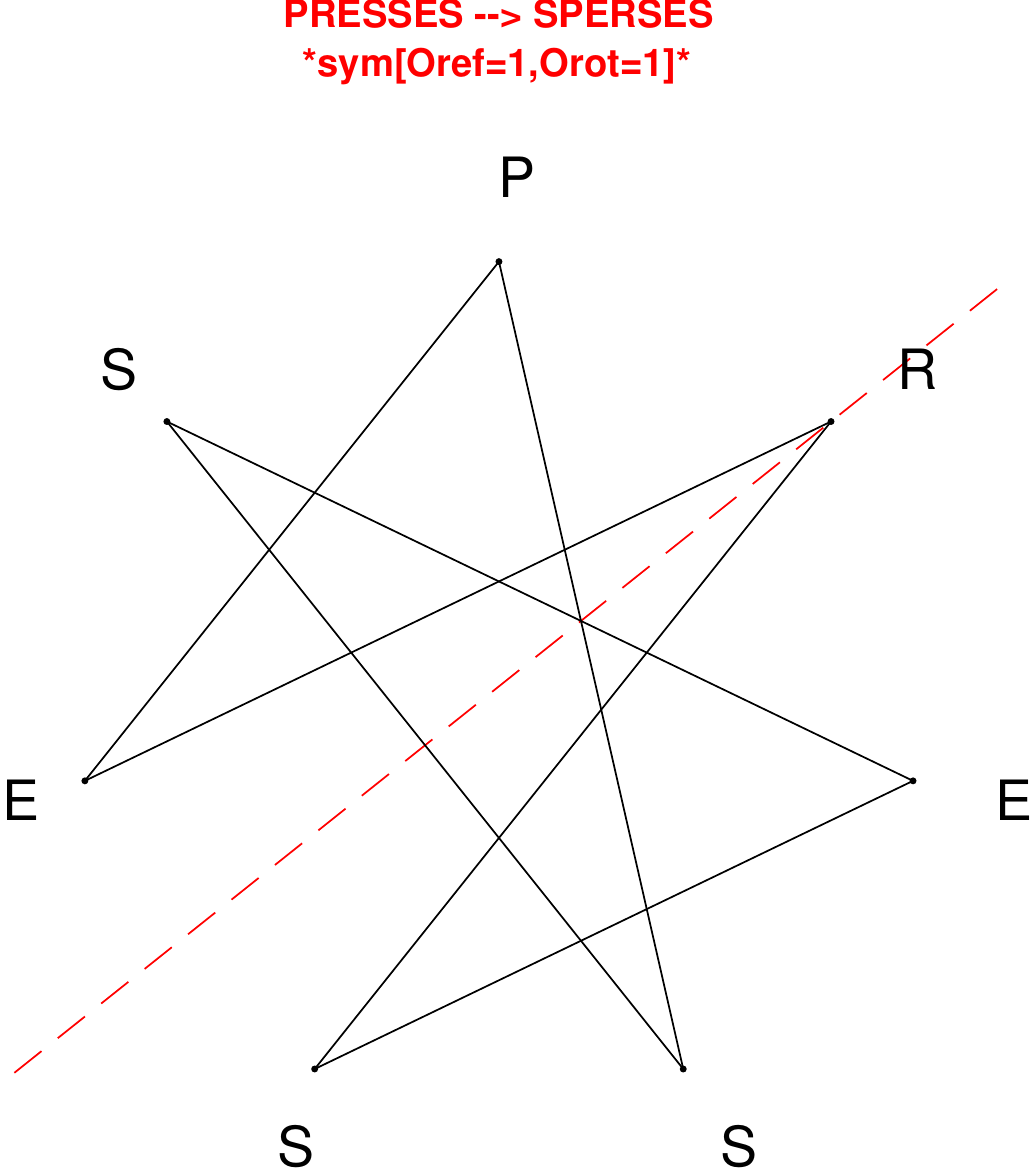}
\end{subfigure}
\end{figure}

\begin{figure}[H]
\centering
\begin{subfigure}[T]{0.19\textwidth}
\centering
\includegraphics[width=\textwidth]{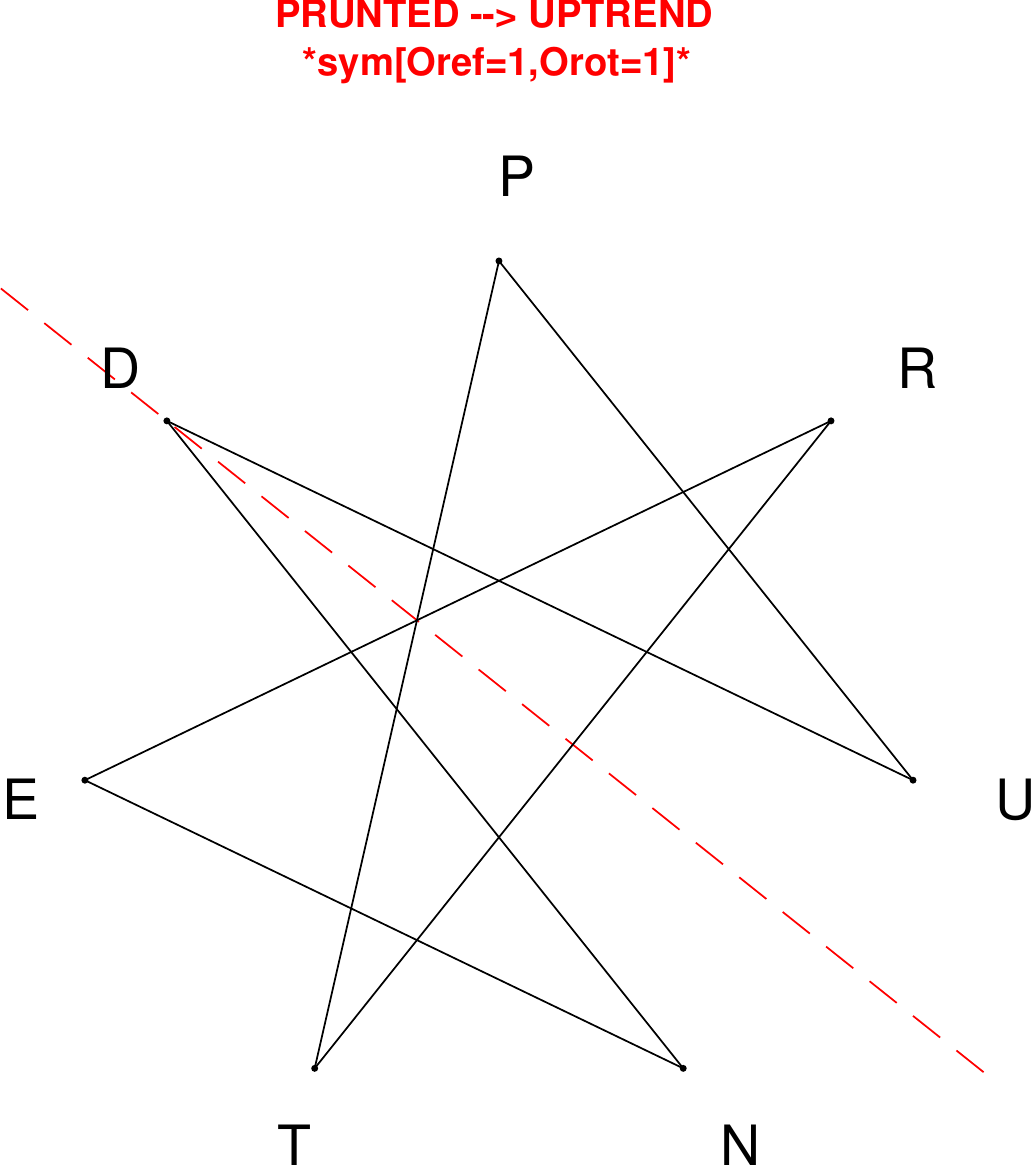}
\end{subfigure}
\hfill
\begin{subfigure}[T]{0.19\textwidth}
\centering
\includegraphics[width=\textwidth]{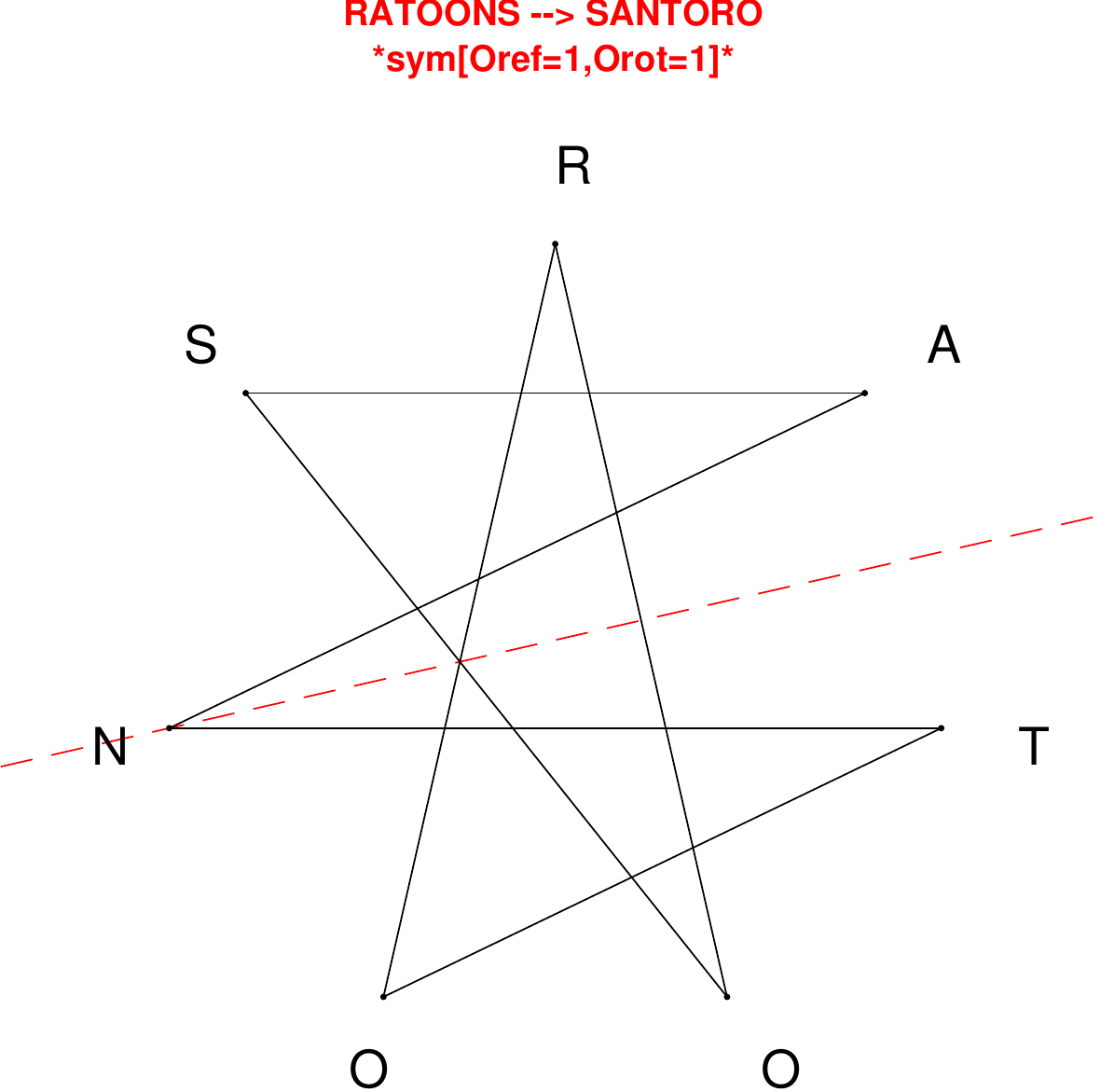}
\end{subfigure}
\hfill
\begin{subfigure}[T]{0.19\textwidth}
\centering
\includegraphics[width=\textwidth]{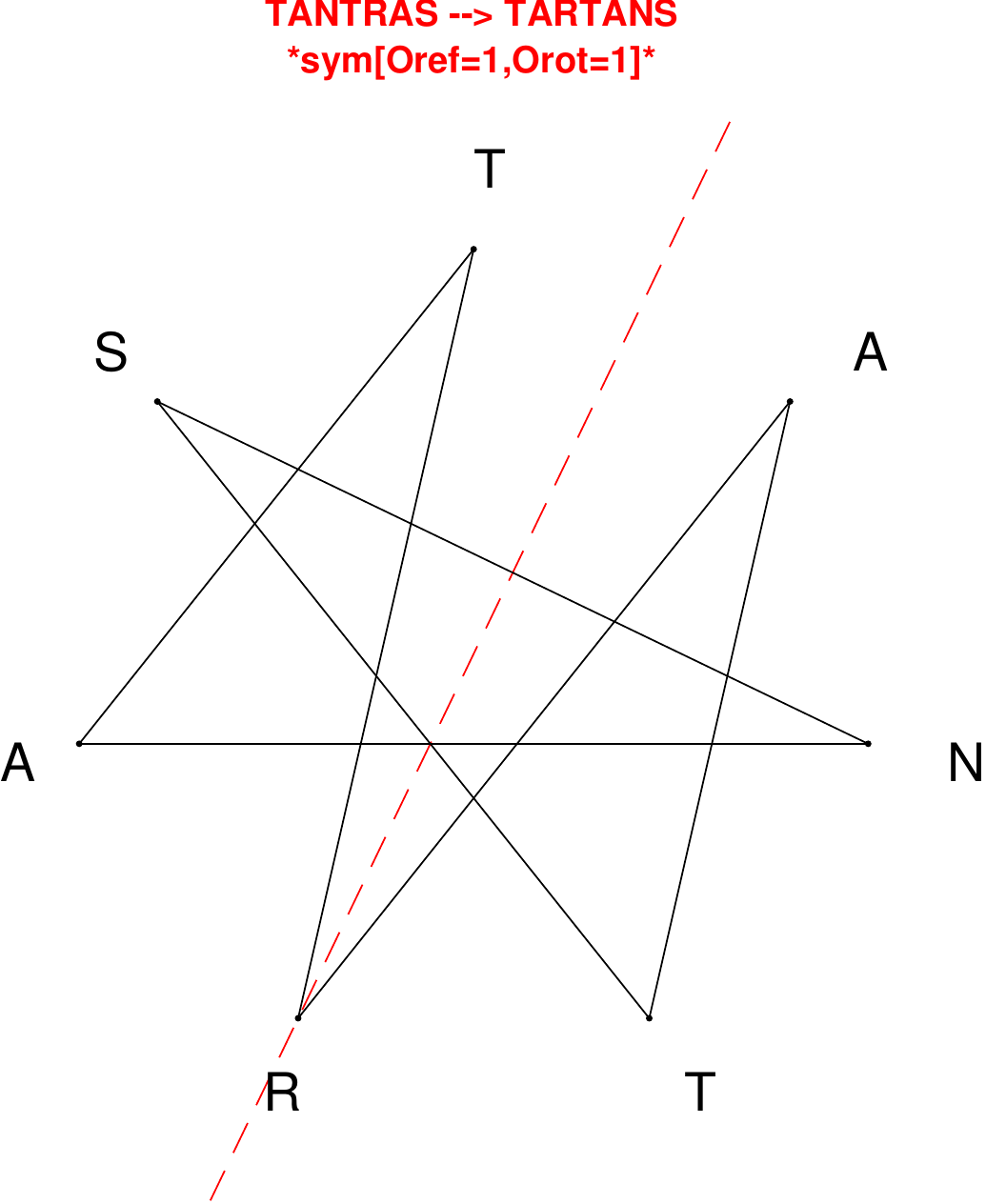}
\end{subfigure}
\hfill
\begin{subfigure}[T]{0.19\textwidth}
\centering
\includegraphics[width=\textwidth]{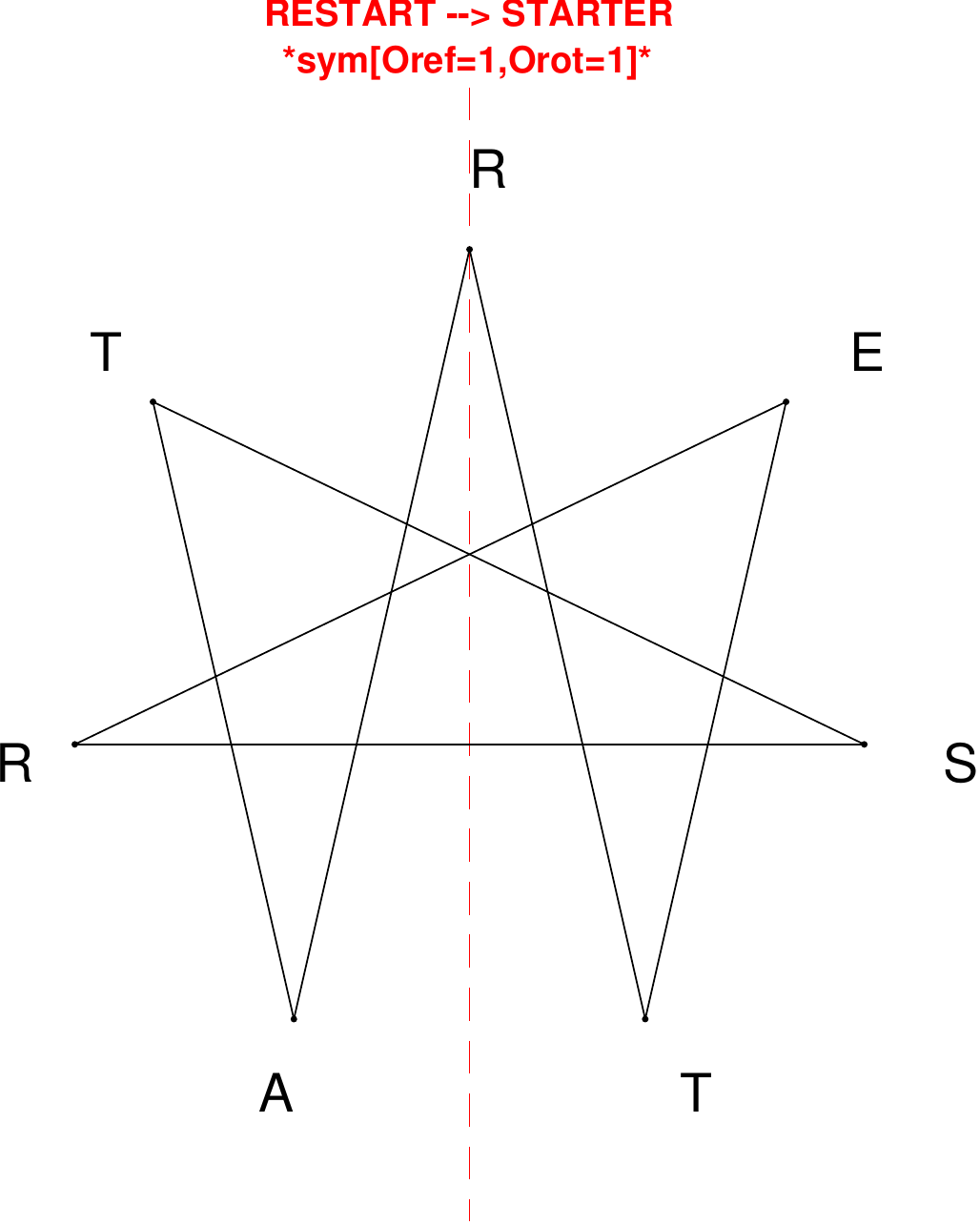}
\end{subfigure}
\hfill
\begin{subfigure}[T]{0.19\textwidth}
\centering
\includegraphics[width=\textwidth]{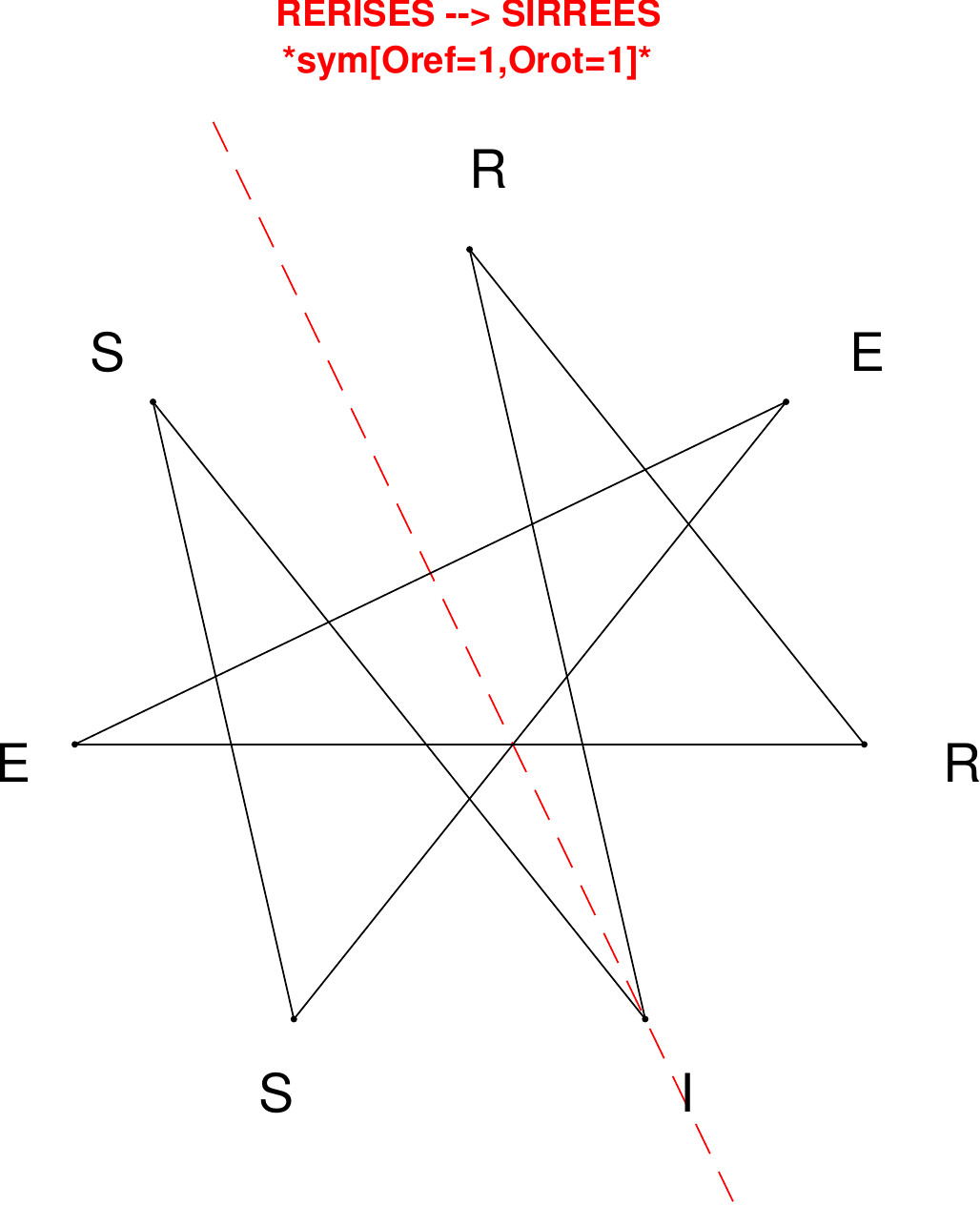}
\end{subfigure}
\end{figure}

\begin{figure}[H]
\centering
\begin{subfigure}[T]{0.19\textwidth}
\centering
\includegraphics[width=\textwidth]{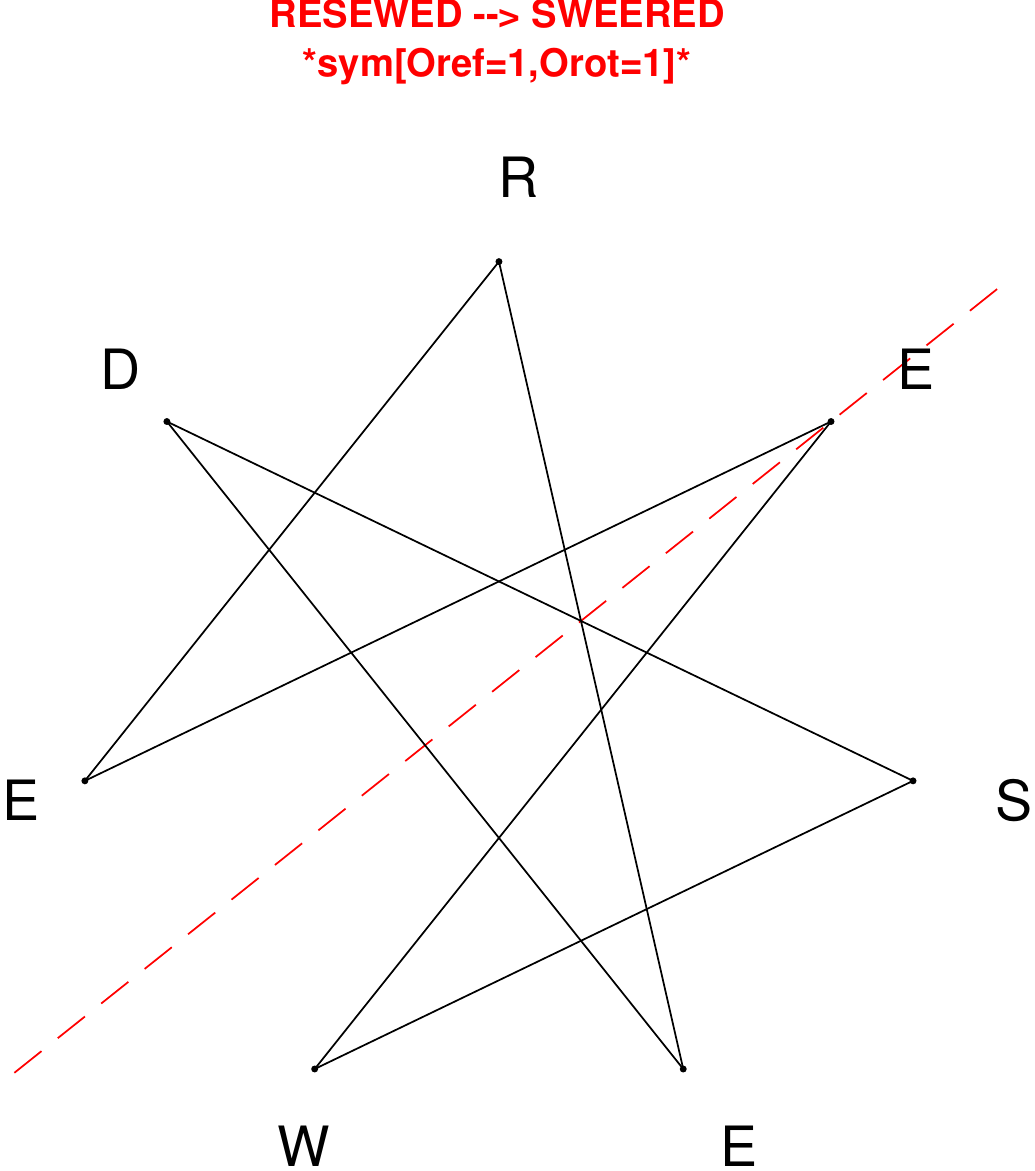}
\end{subfigure}
\hfill
\begin{subfigure}[T]{0.19\textwidth}
\centering
\includegraphics[width=\textwidth]{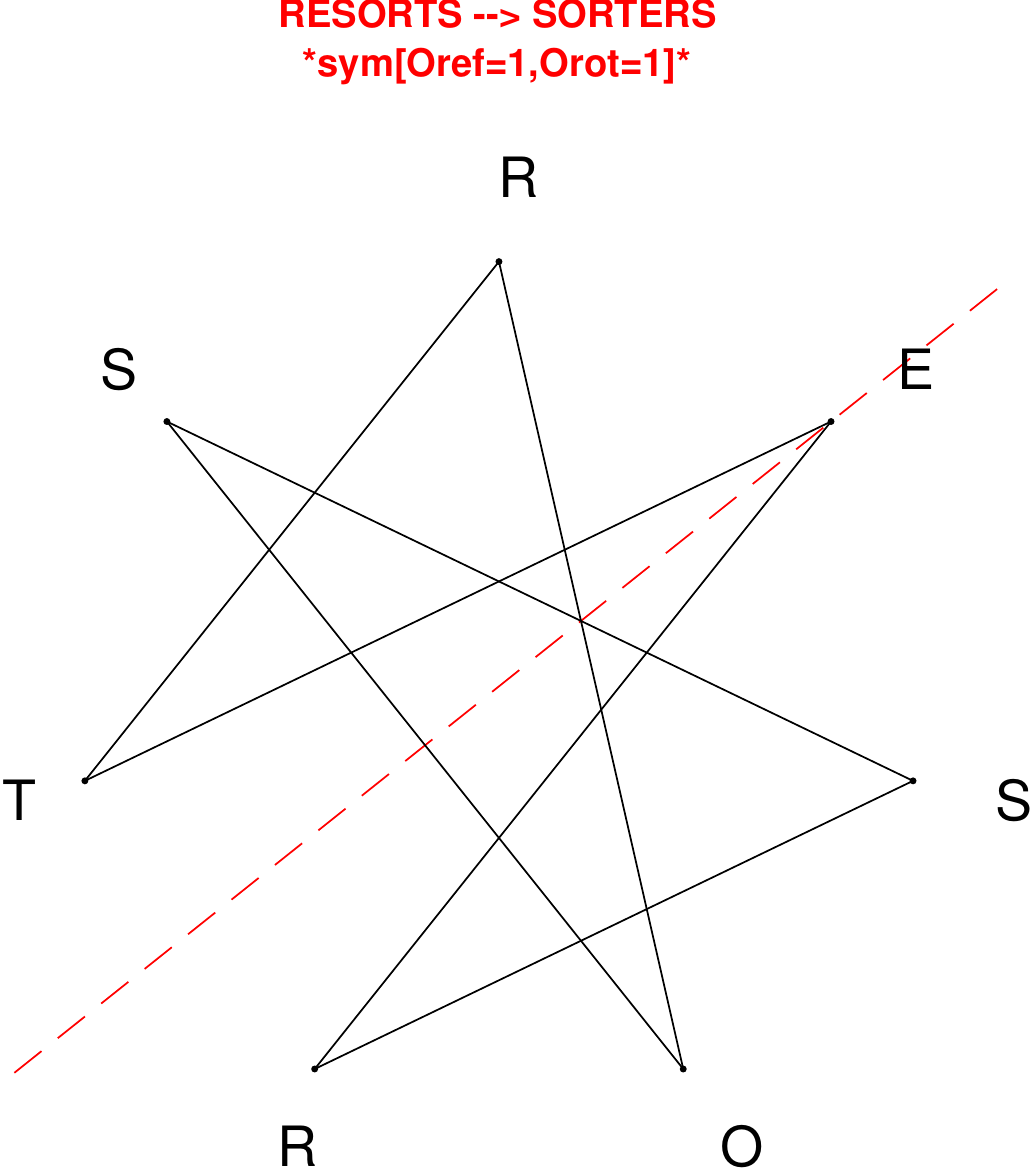}
\end{subfigure}
\hfill
\begin{subfigure}[T]{0.19\textwidth}
\centering
\includegraphics[width=\textwidth]{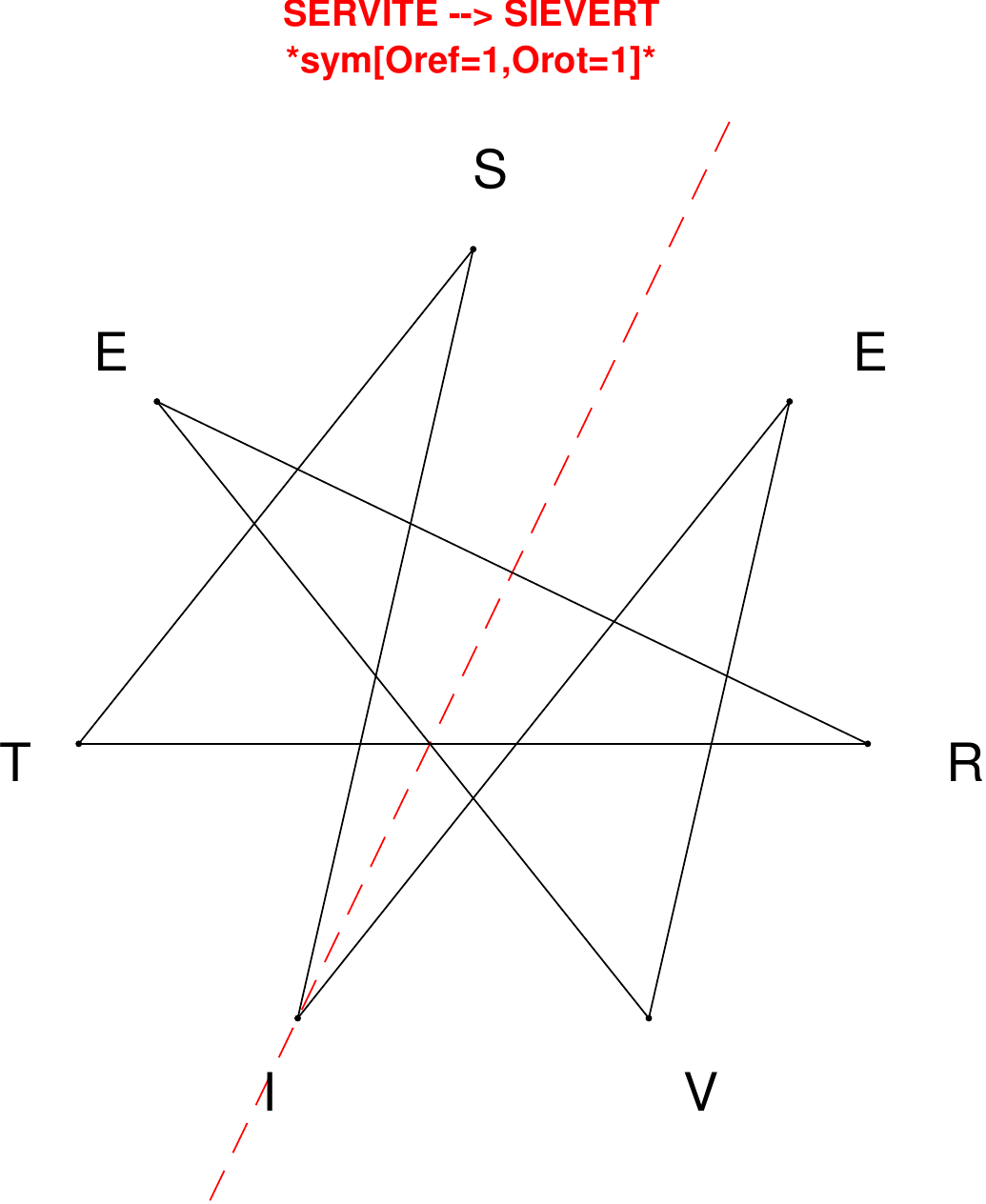}
\end{subfigure}
\hfill
\begin{subfigure}[T]{0.19\textwidth}
\centering
\includegraphics[width=\textwidth]{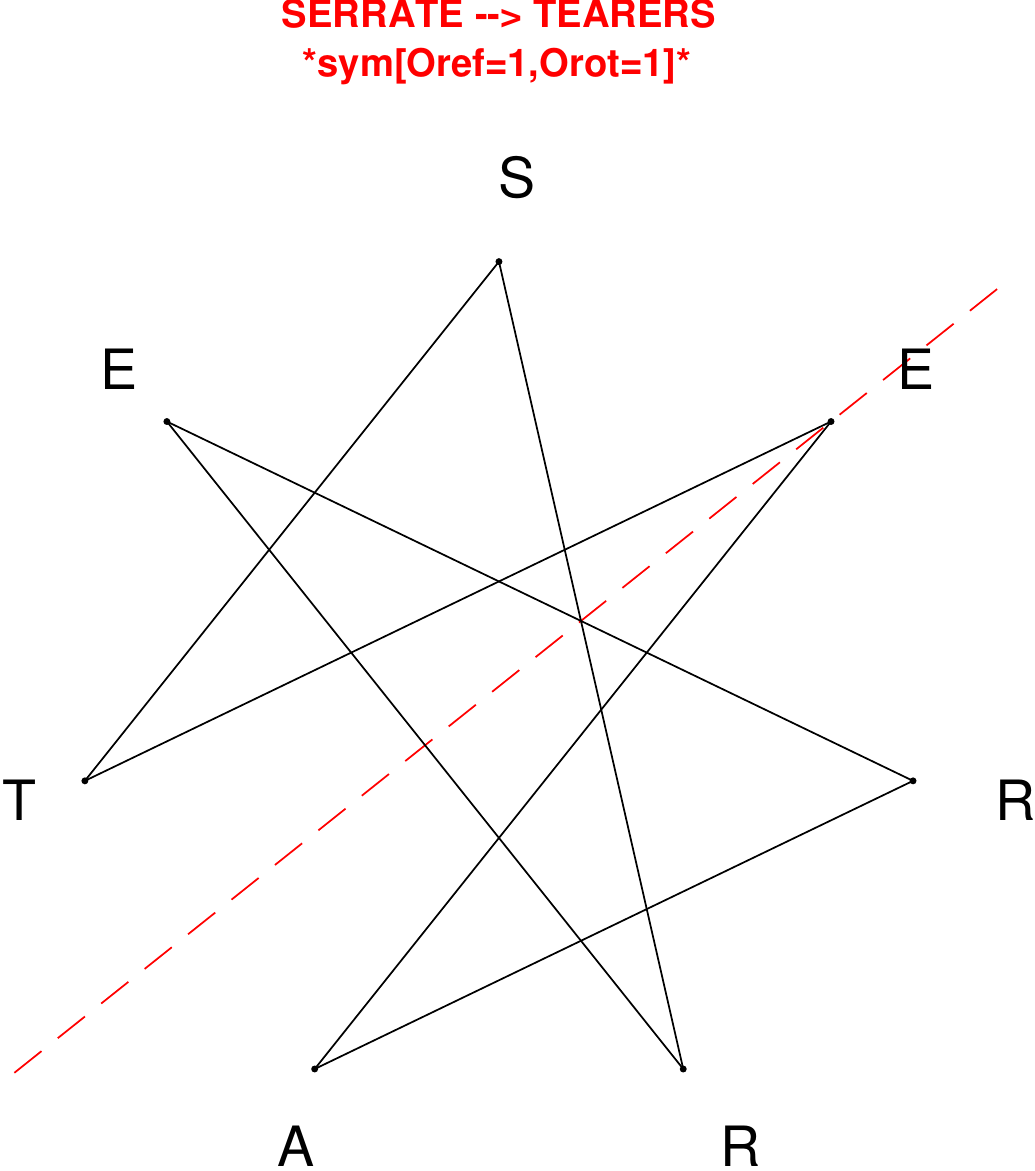}
\end{subfigure}
\hfill
\begin{subfigure}[T]{0.19\textwidth}
\centering
\includegraphics[width=\textwidth]{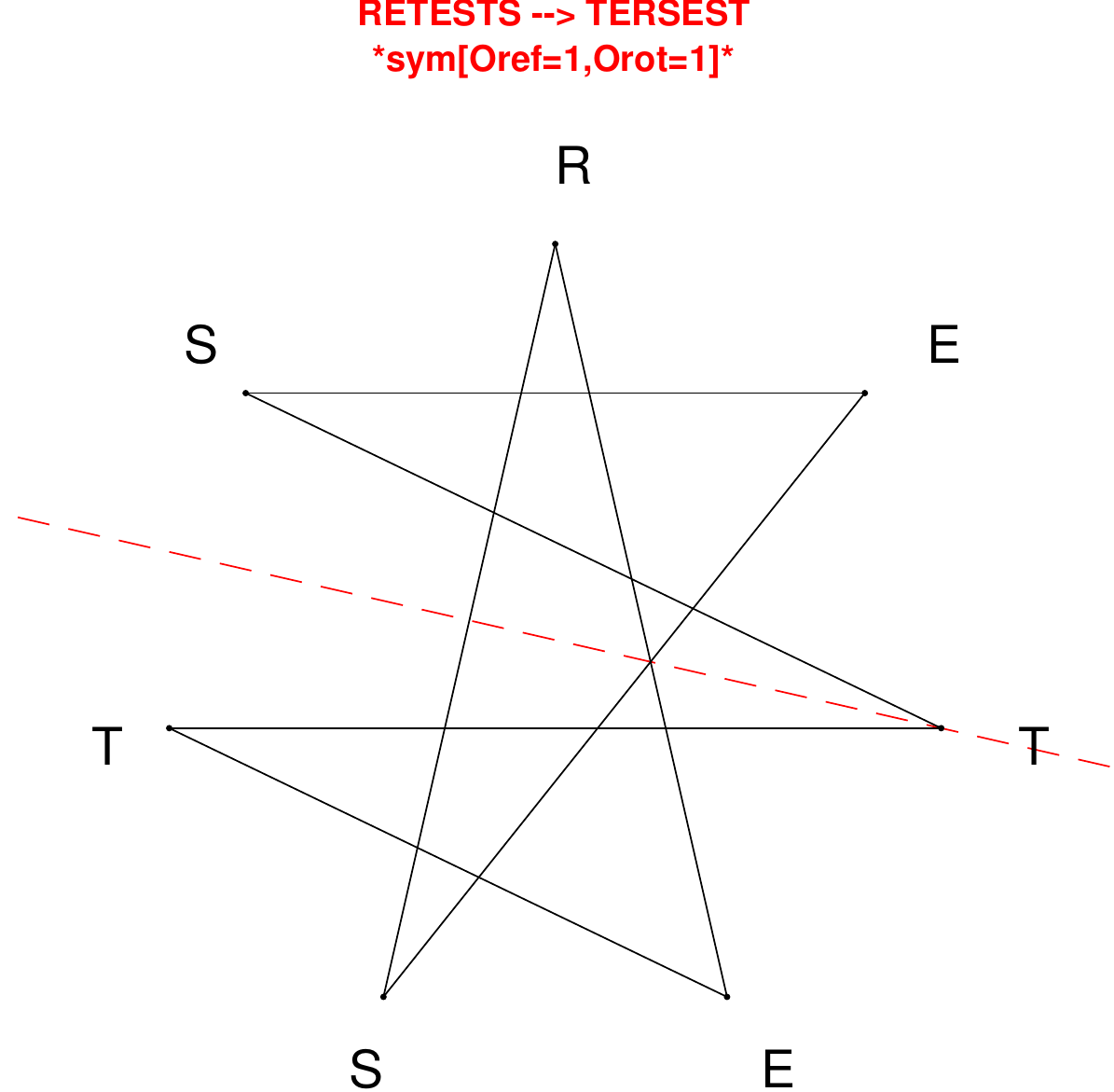}
\end{subfigure}
\end{figure}

\begin{figure}[H]
\centering
\begin{subfigure}[T]{0.19\textwidth}
\centering
\includegraphics[width=\textwidth]{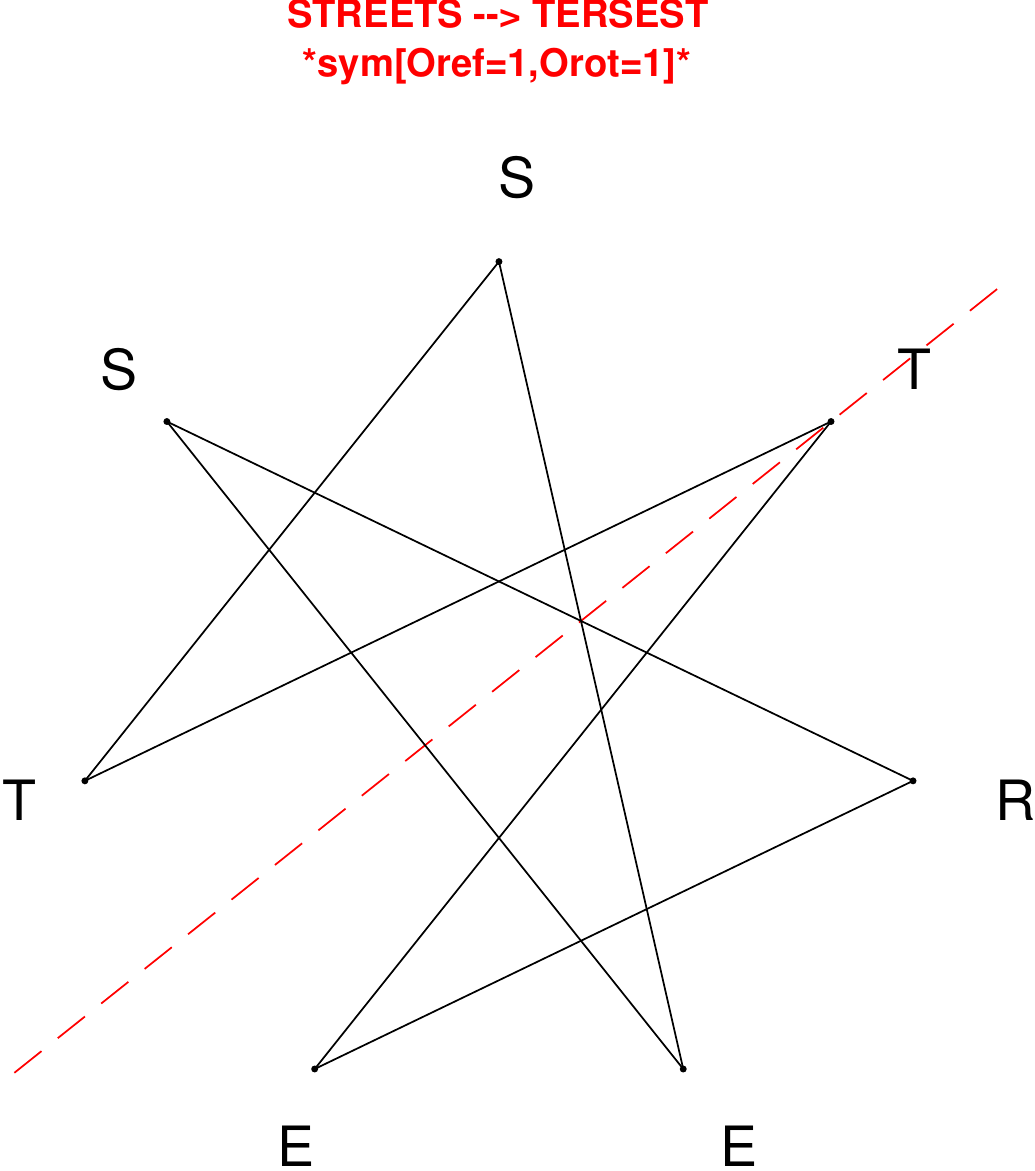}
\end{subfigure}
\hfill
\begin{subfigure}[T]{0.19\textwidth}
\centering
\includegraphics[width=\textwidth]{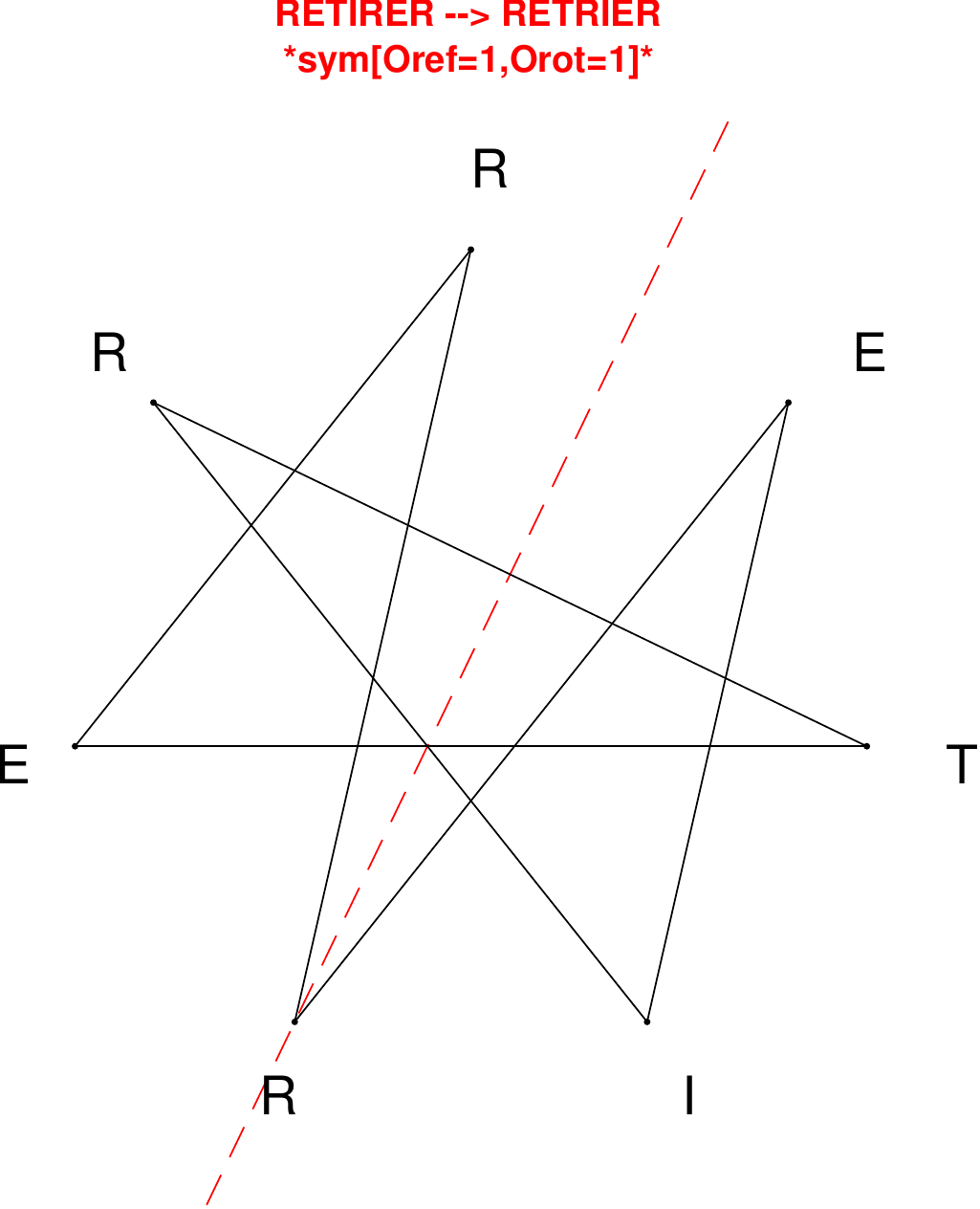}
\end{subfigure}
\hfill
\begin{subfigure}[T]{0.19\textwidth}
\centering
\includegraphics[width=\textwidth]{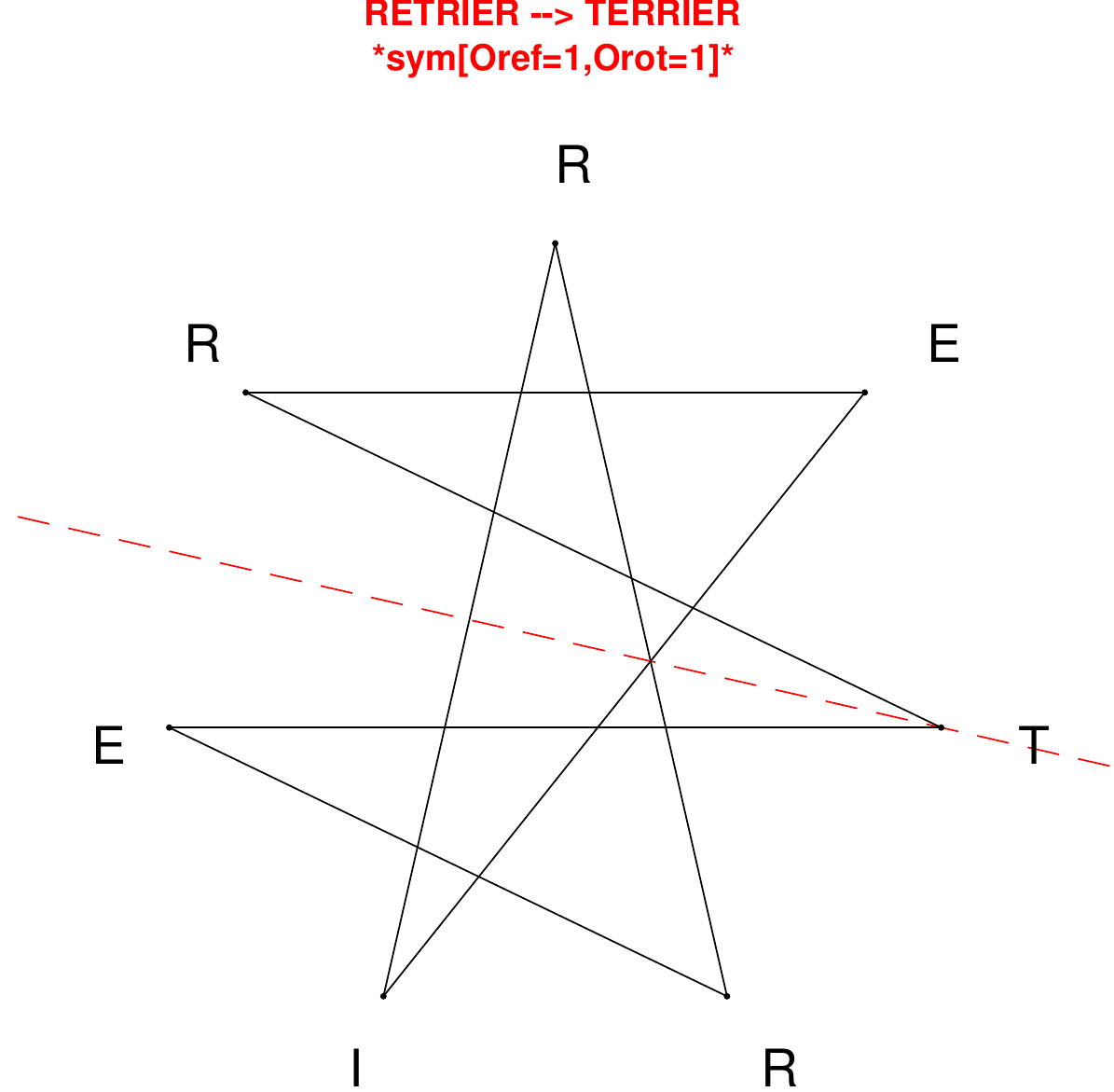}
\end{subfigure}
\hfill
\begin{subfigure}[T]{0.19\textwidth}
\centering
\includegraphics[width=\textwidth]{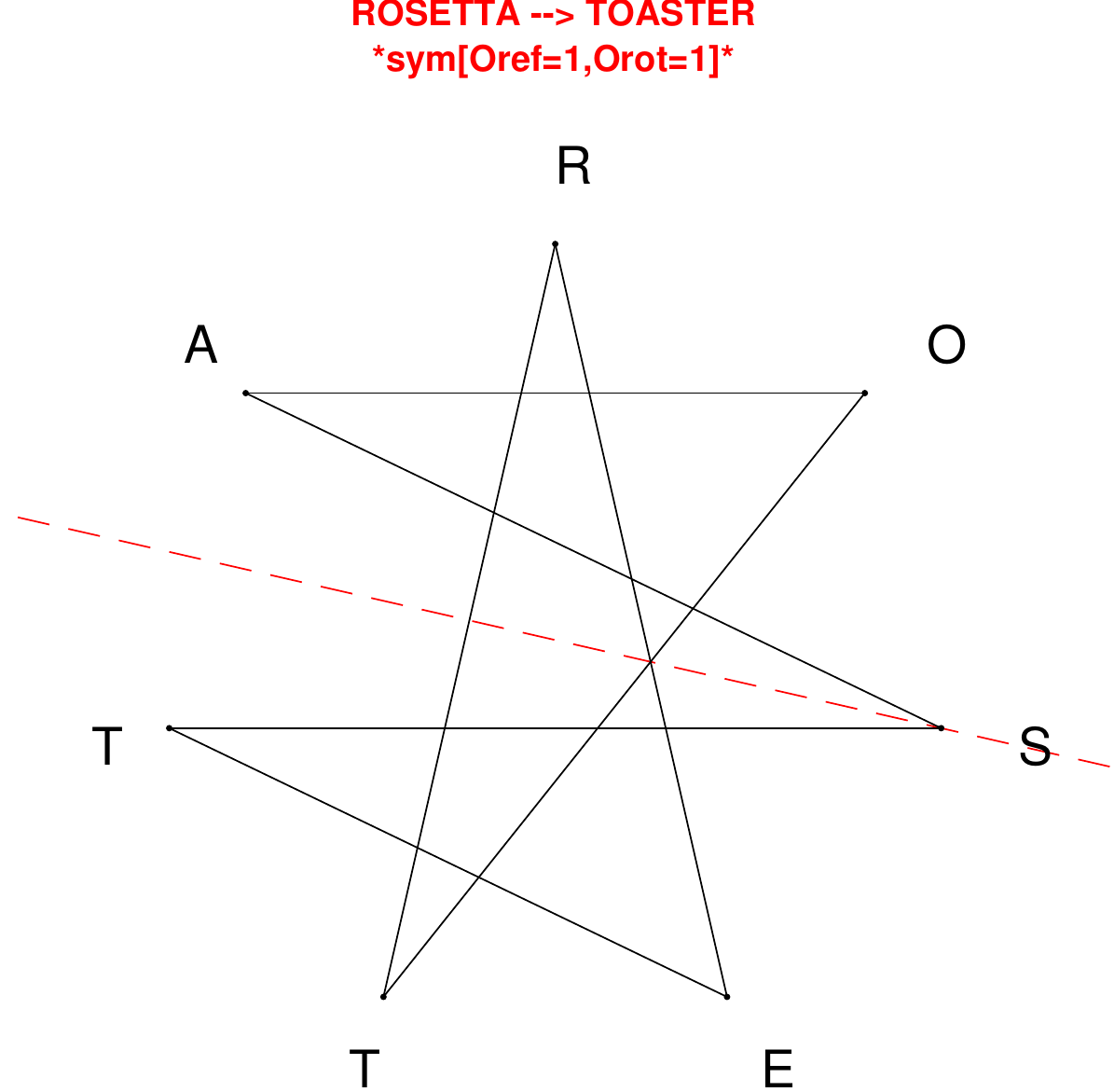}
\end{subfigure}
\hfill
\begin{subfigure}[T]{0.19\textwidth}
\centering
\includegraphics[width=\textwidth]{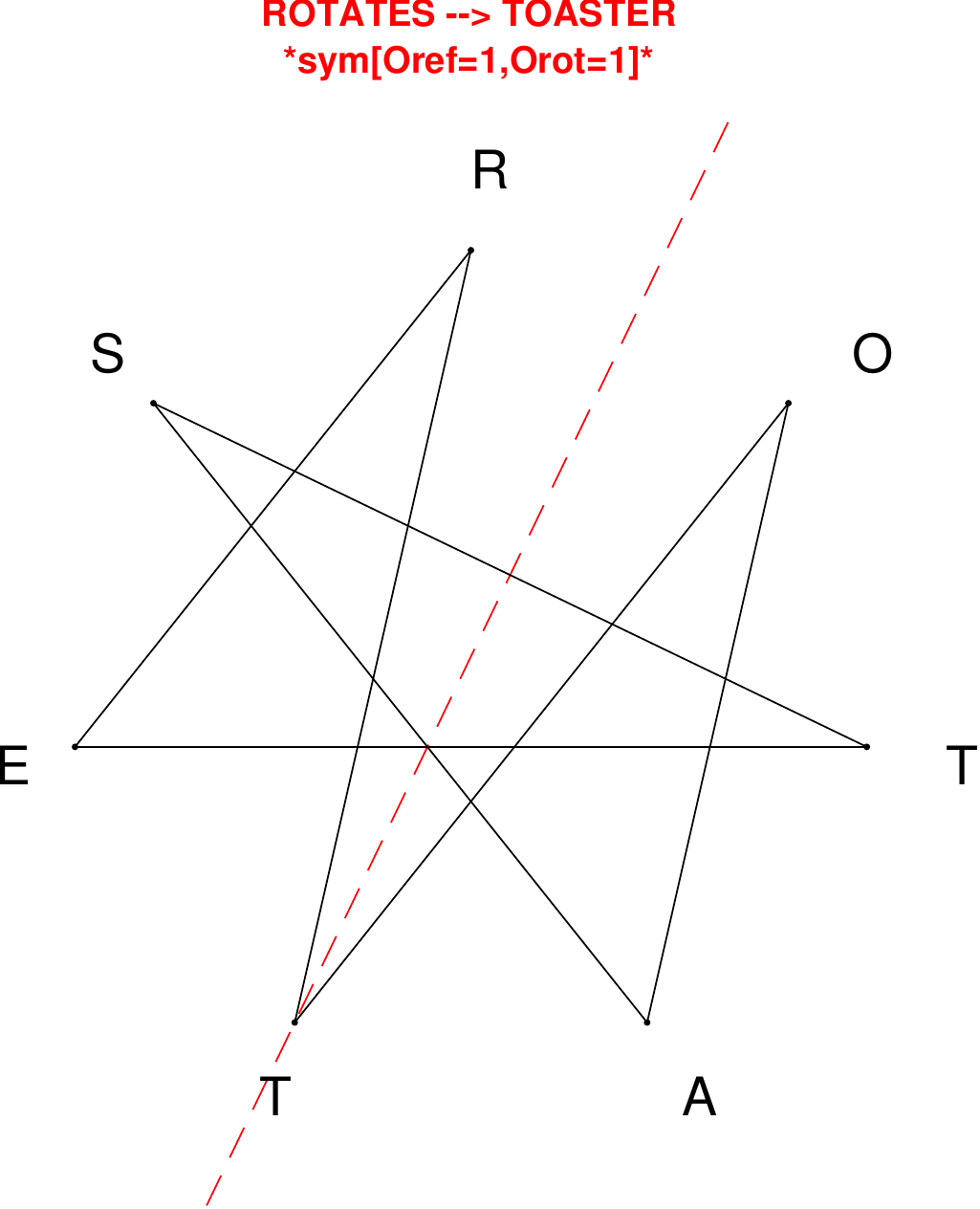}
\end{subfigure}
\end{figure}

\begin{figure}[H]
\centering
\begin{subfigure}[T]{0.19\textwidth}
\centering
\includegraphics[width=\textwidth]{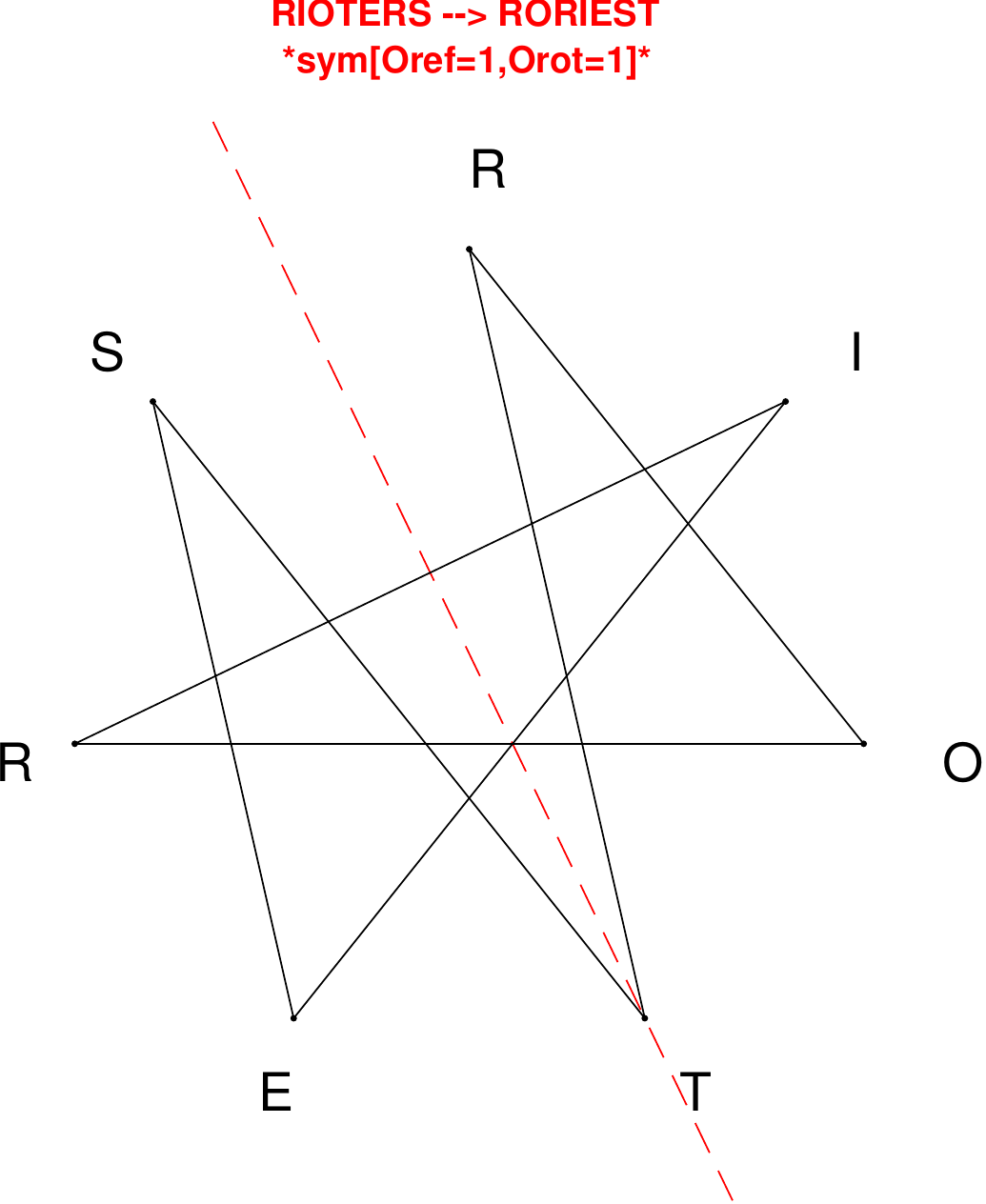}
\end{subfigure}
\hfill
\begin{subfigure}[T]{0.19\textwidth}
\centering
\includegraphics[width=\textwidth]{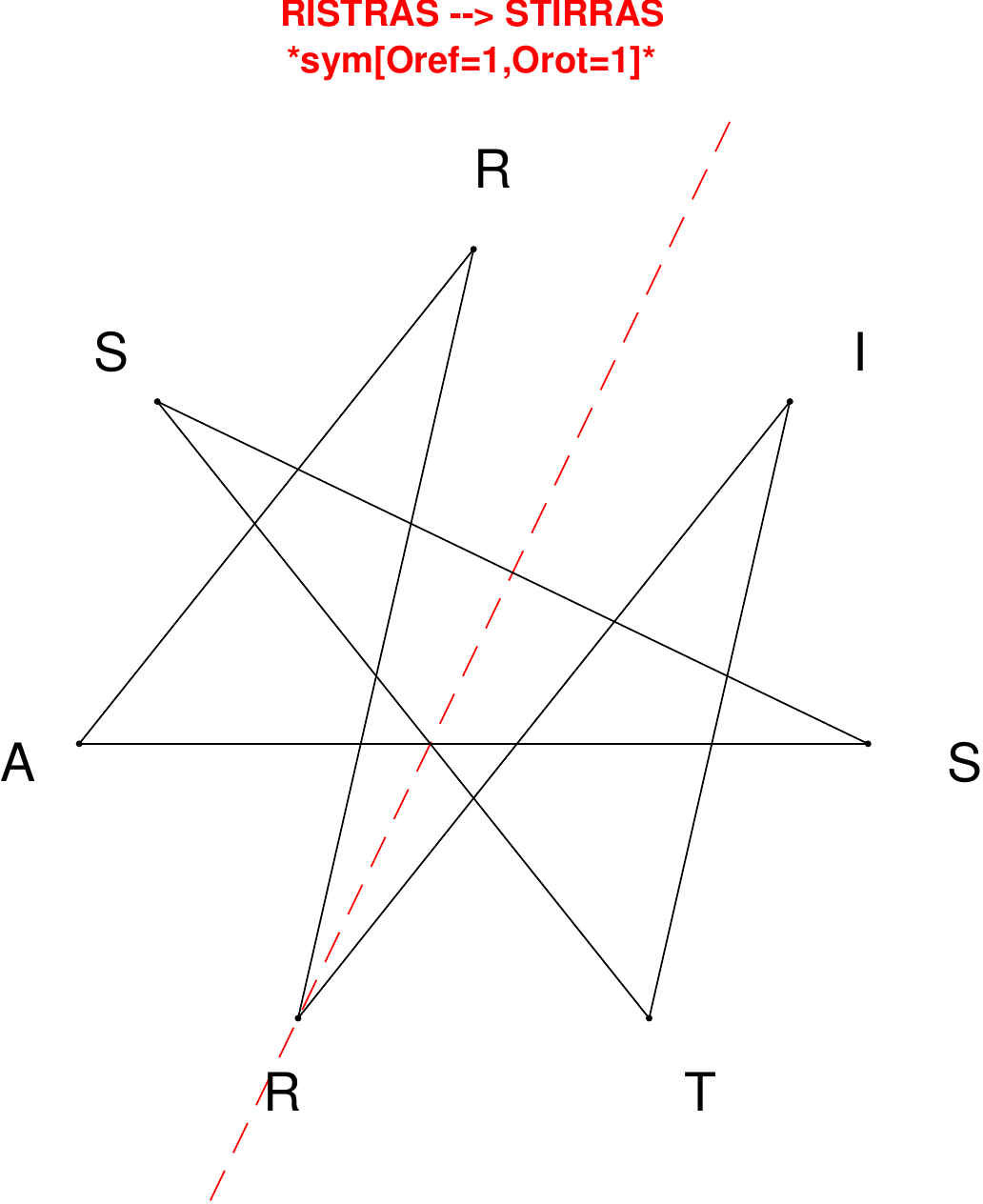}
\end{subfigure}
\hfill
\begin{subfigure}[T]{0.19\textwidth}
\centering
\includegraphics[width=\textwidth]{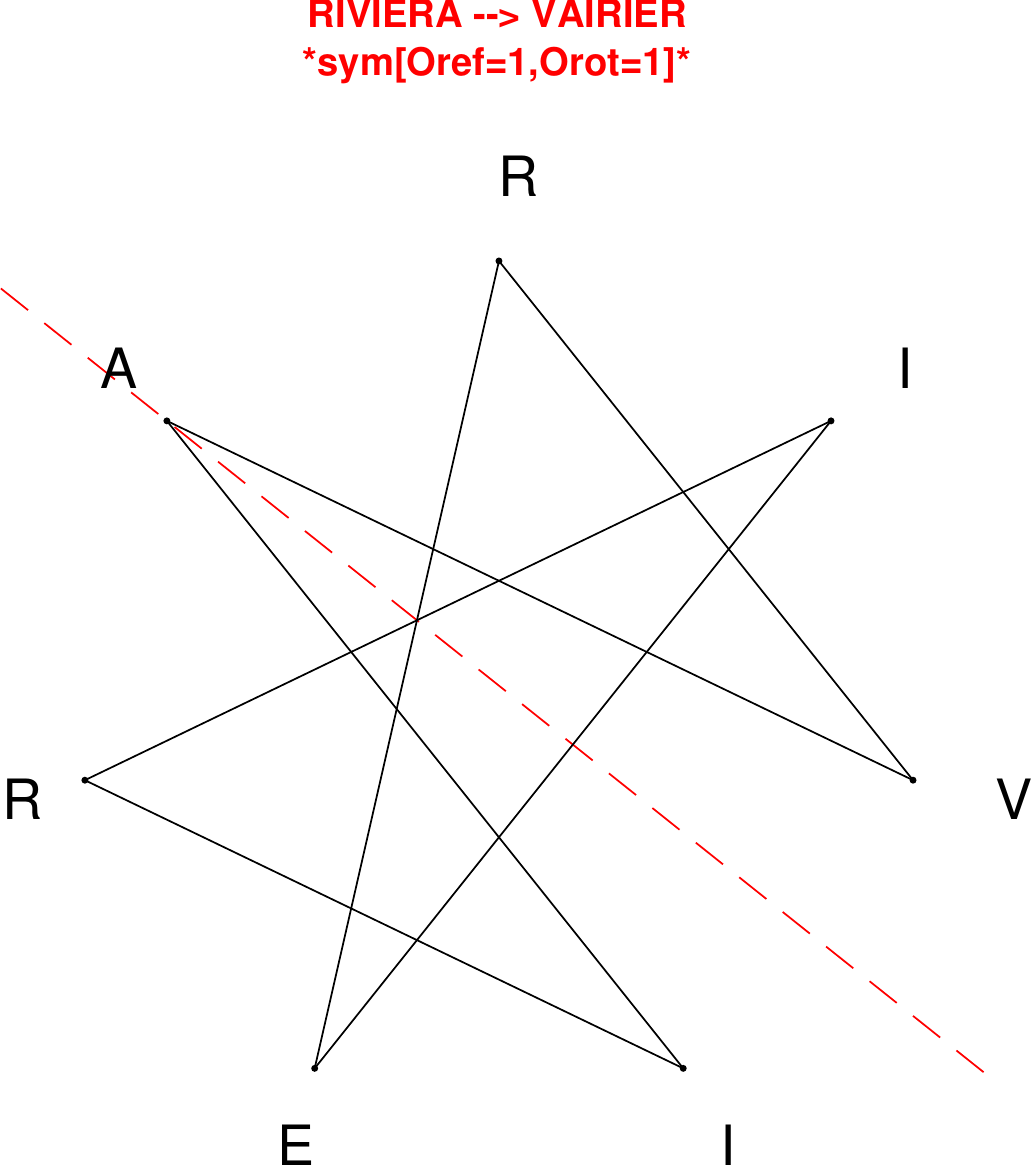}
\end{subfigure}
\hfill
\begin{subfigure}[T]{0.19\textwidth}
\centering
\includegraphics[width=\textwidth]{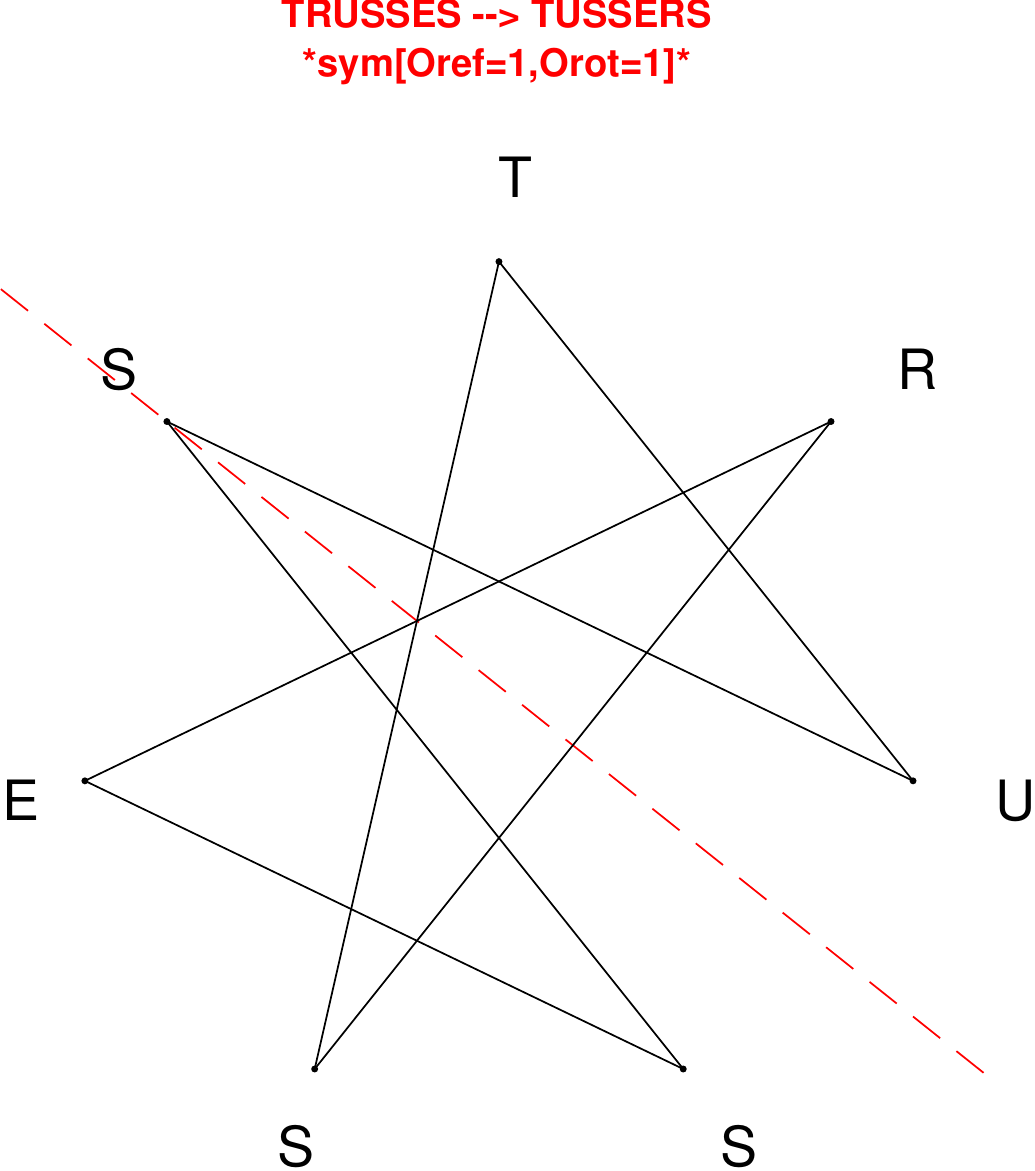}
\end{subfigure}
\hfill
\begin{subfigure}[T]{0.19\textwidth}
\centering
\includegraphics[width=\textwidth]{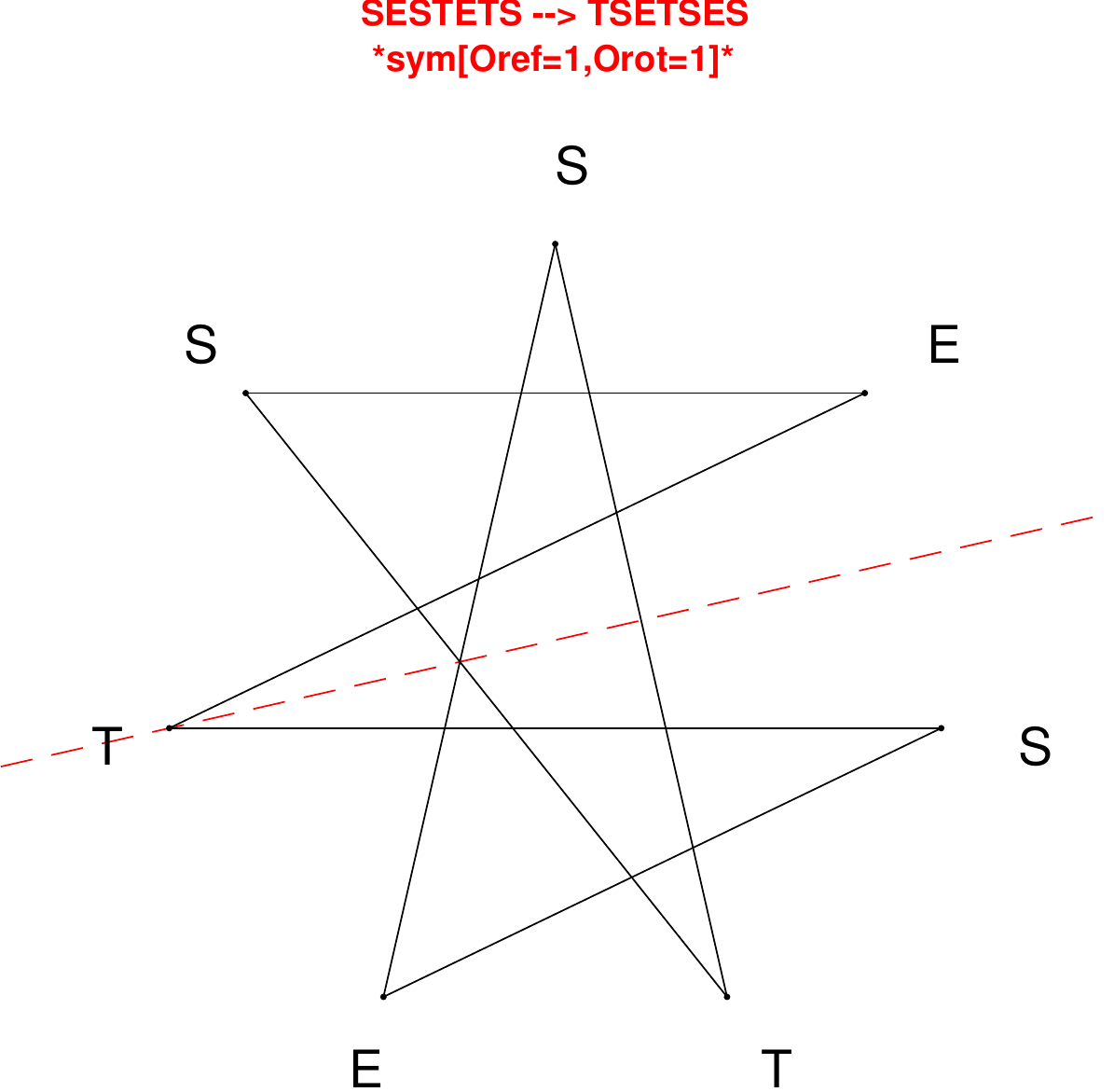}
\end{subfigure}
\end{figure}

\begin{figure}[H]
\centering
\begin{subfigure}[T]{0.19\textwidth}
\centering
\includegraphics[width=\textwidth]{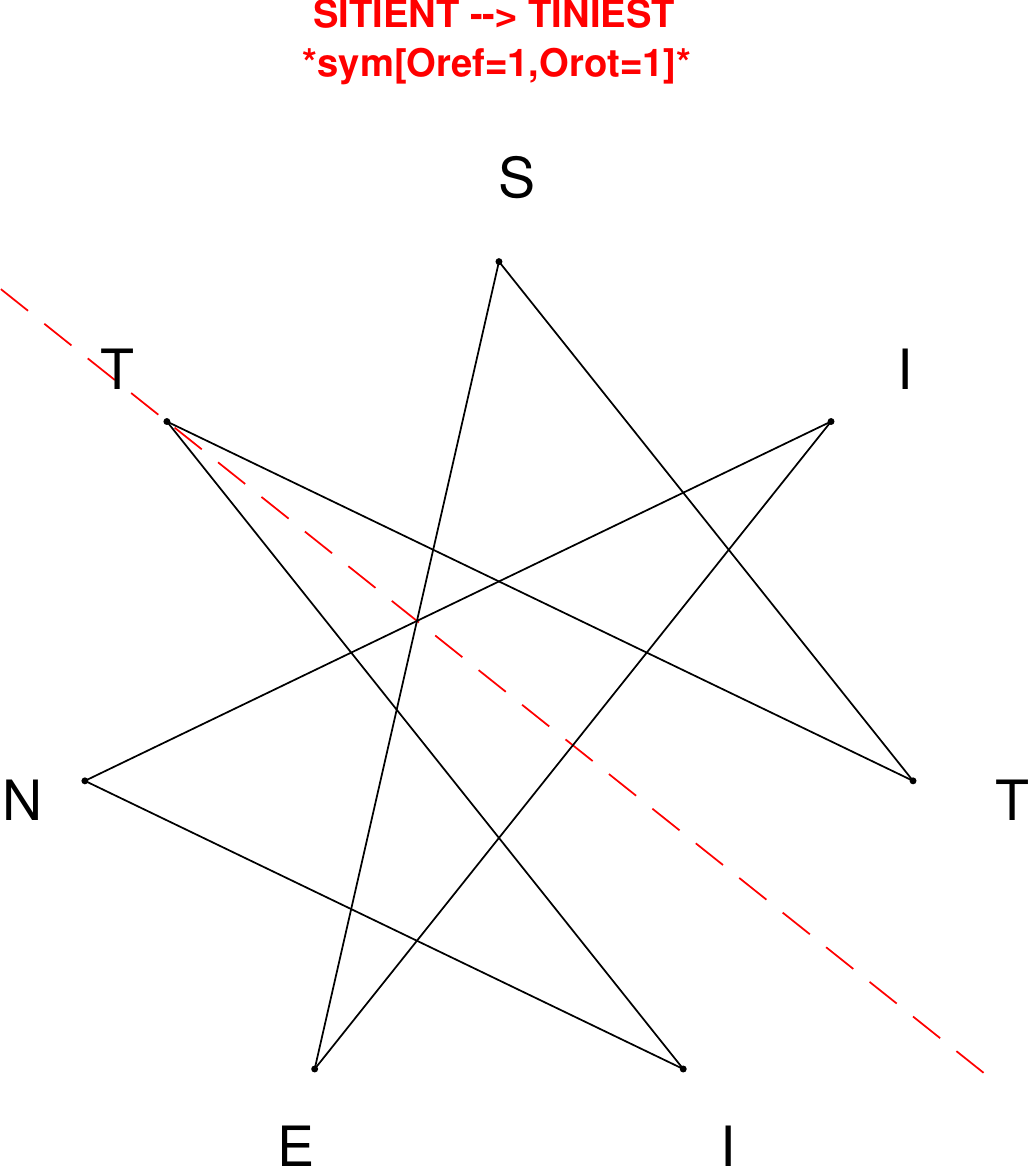}
\end{subfigure}
\hfill
\begin{subfigure}[T]{0.19\textwidth}
\centering
\includegraphics[width=\textwidth]{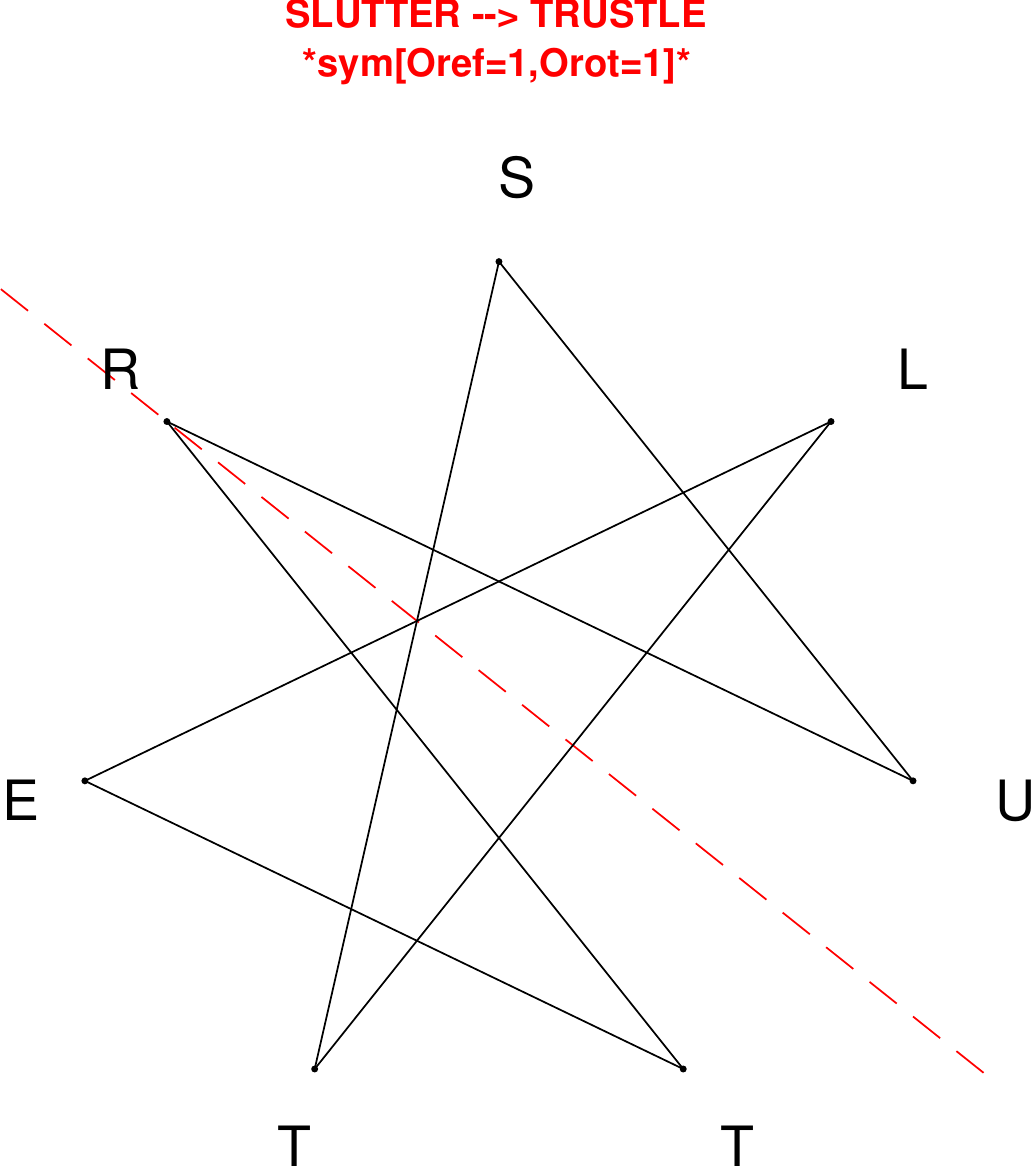}
\end{subfigure}
\hfill
\begin{subfigure}[T]{0.19\textwidth}
\centering
\includegraphics[width=\textwidth]{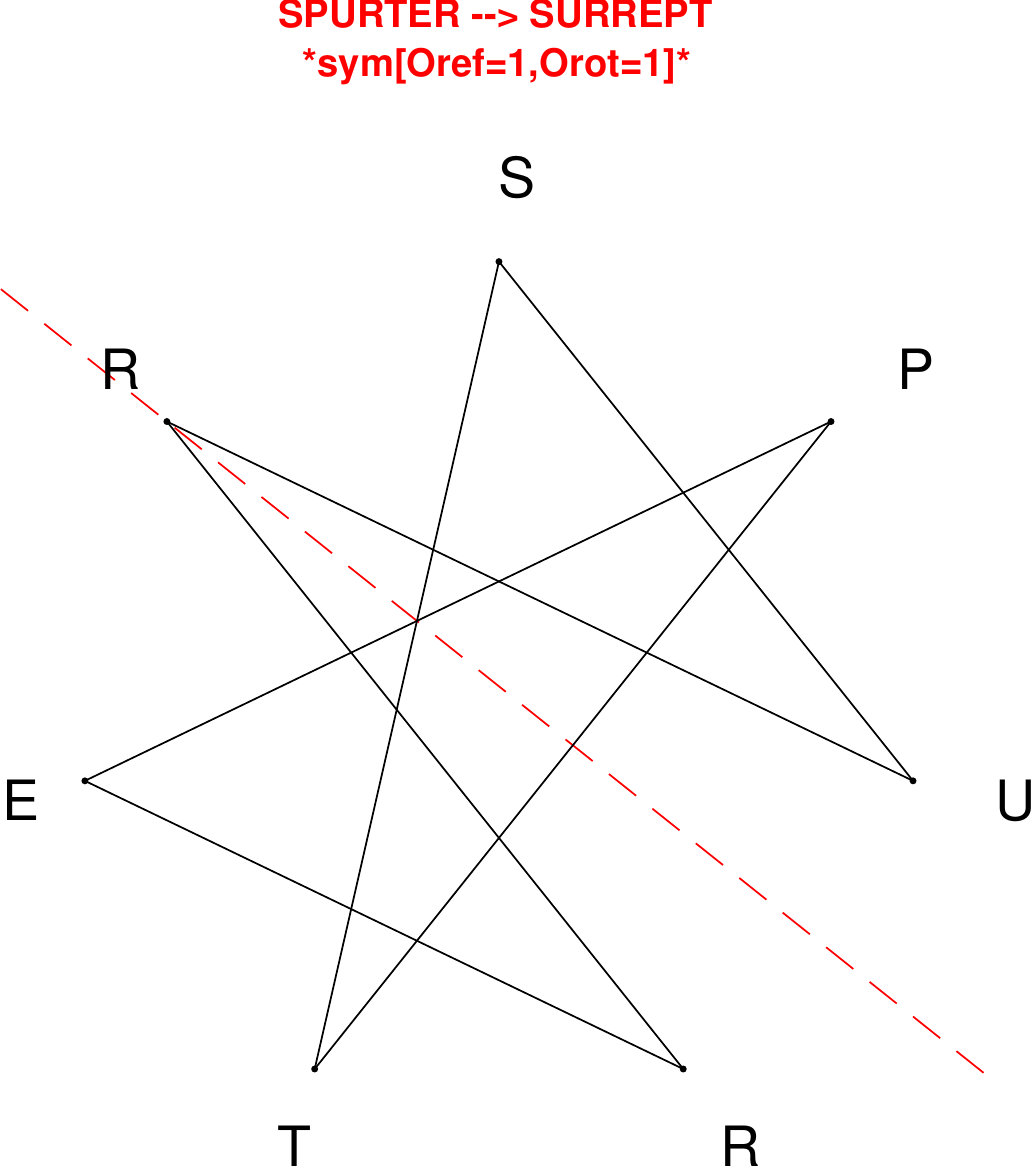}
\end{subfigure}
\hfill
\begin{subfigure}[T]{0.19\textwidth}
\centering
\includegraphics[width=\textwidth]{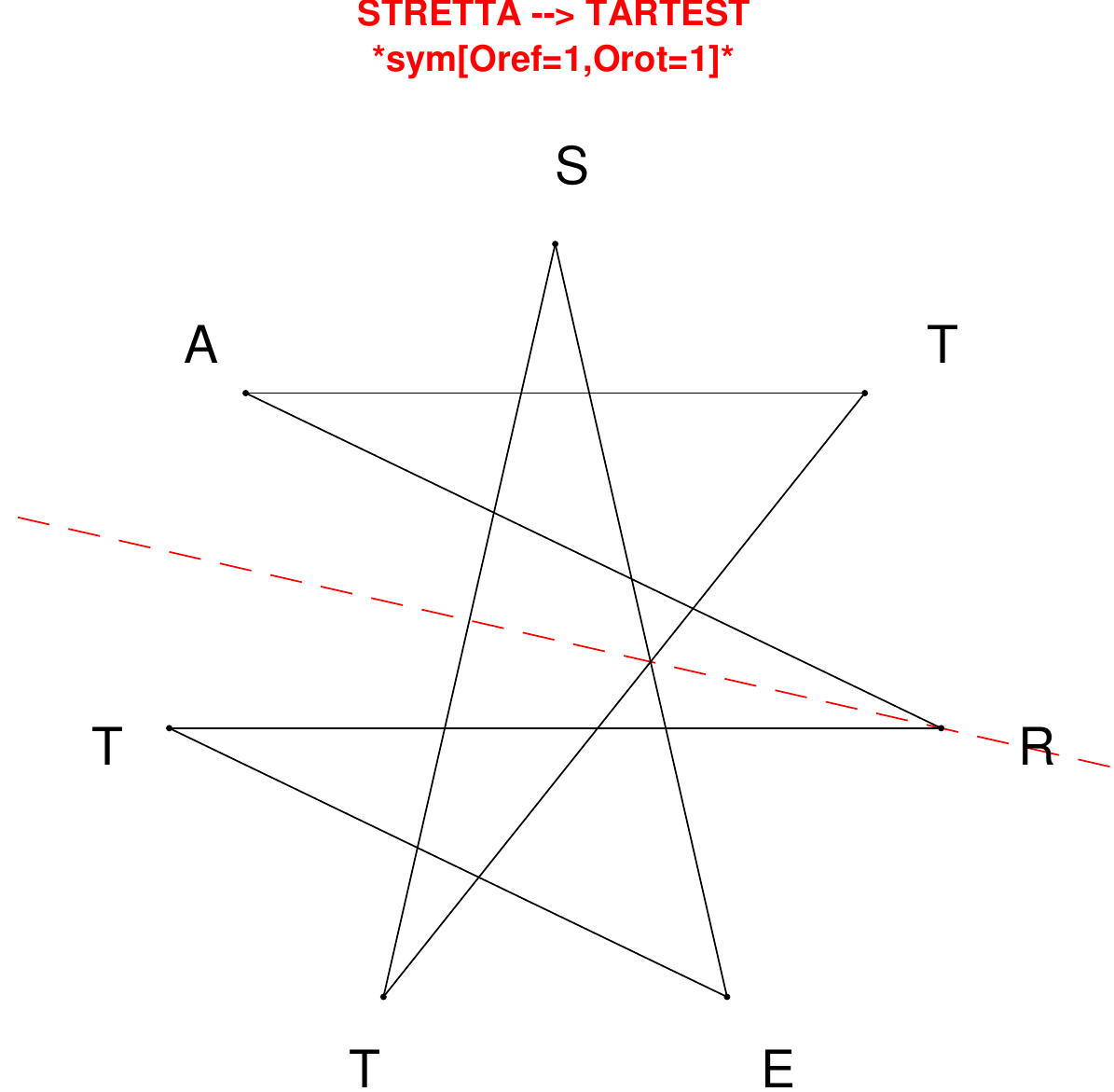}
\end{subfigure}
\hfill
\begin{subfigure}[T]{0.19\textwidth}
\centering
\includegraphics[width=\textwidth]{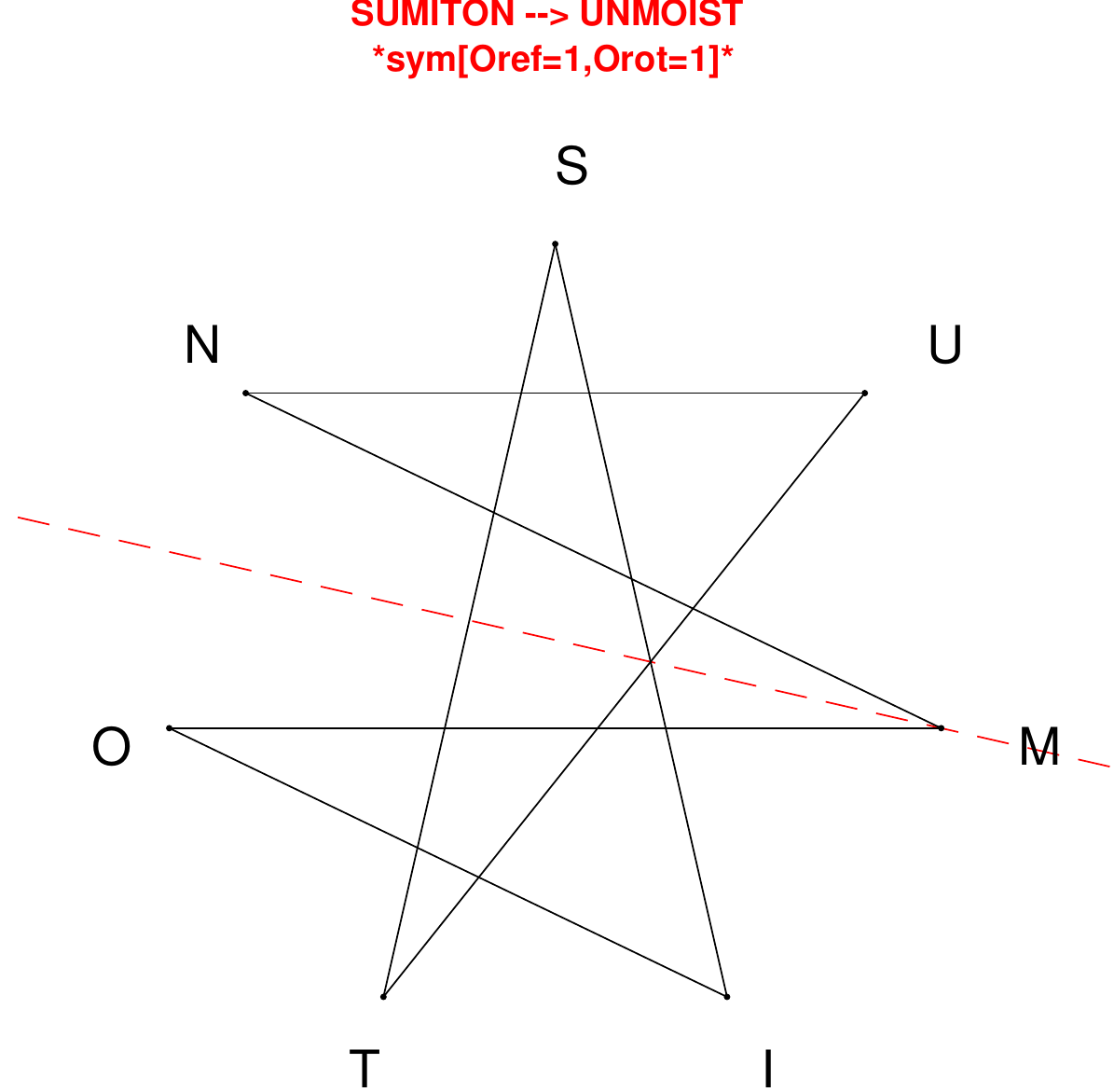}
\end{subfigure}
\end{figure}

\begin{figure}[H]
\centering
\begin{subfigure}[T]{0.19\textwidth}
\centering
\includegraphics[width=\textwidth]{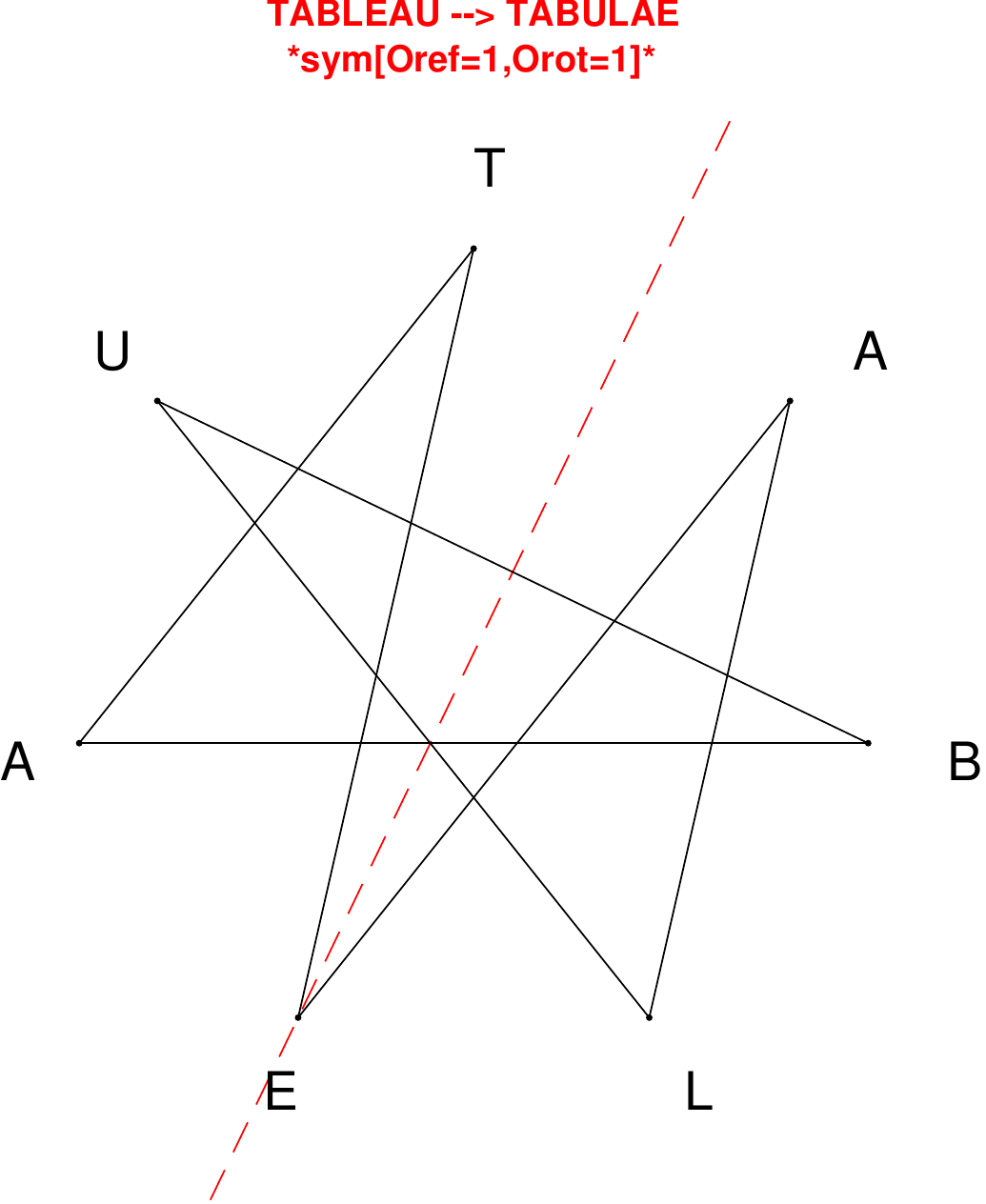}
\end{subfigure}
\hfill
\begin{subfigure}[T]{0.19\textwidth}
\centering
\includegraphics[width=\textwidth]{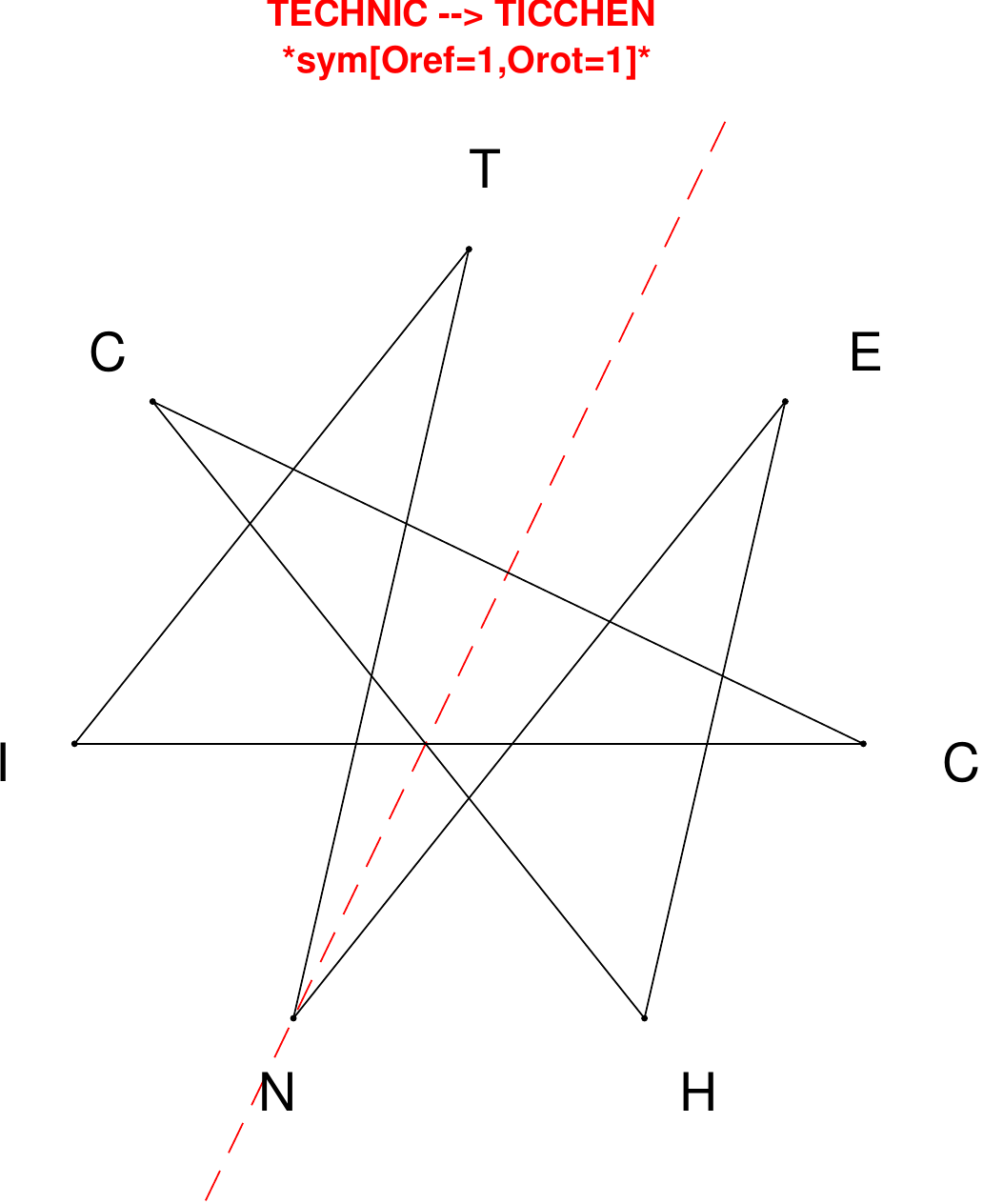}
\end{subfigure}
\hfill
\begin{subfigure}[T]{0.19\textwidth}
\centering
\includegraphics[width=\textwidth]{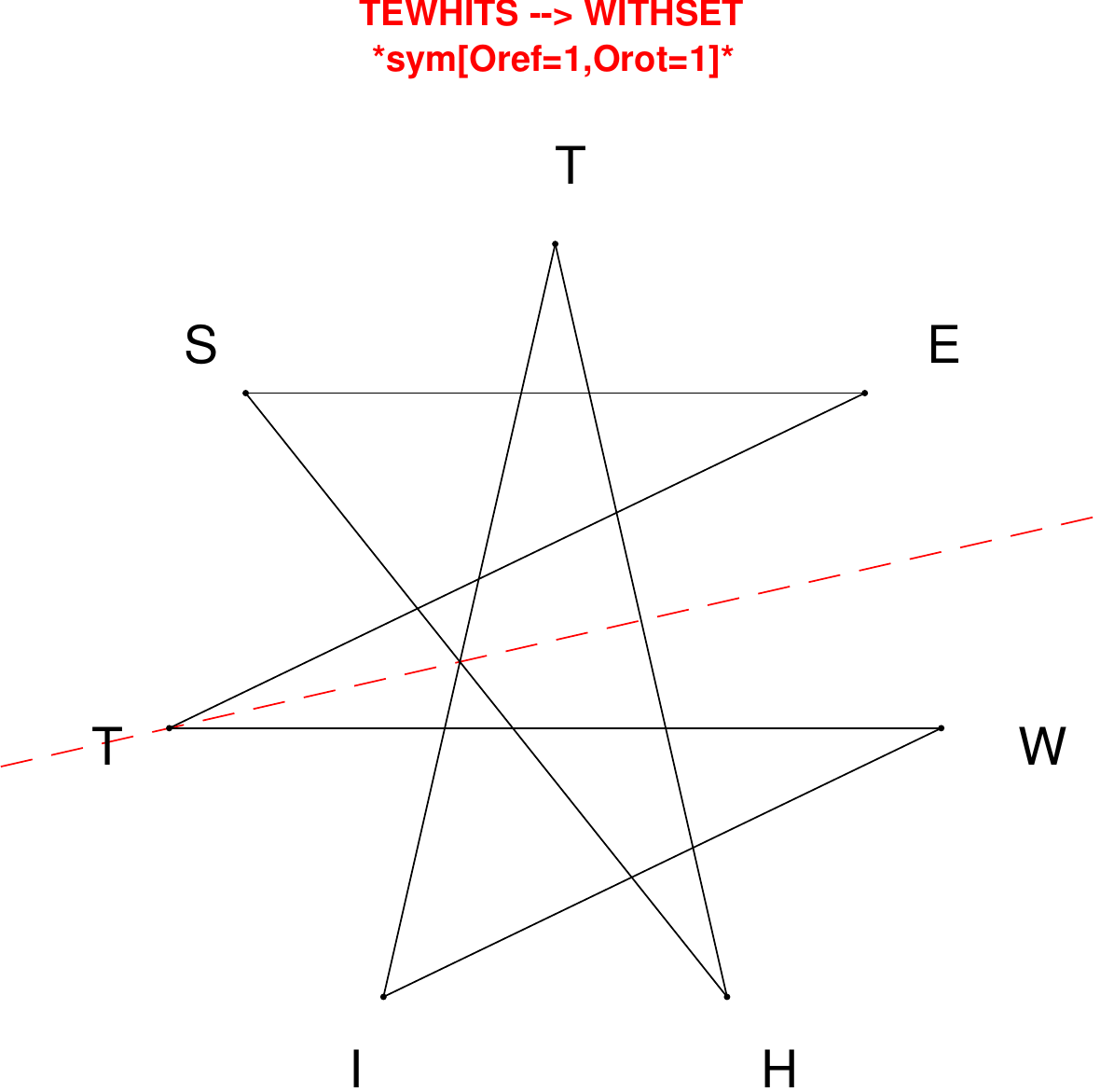}
\end{subfigure}
\hfill
\begin{subfigure}[T]{0.19\textwidth}
\centering
\includegraphics[width=\textwidth]{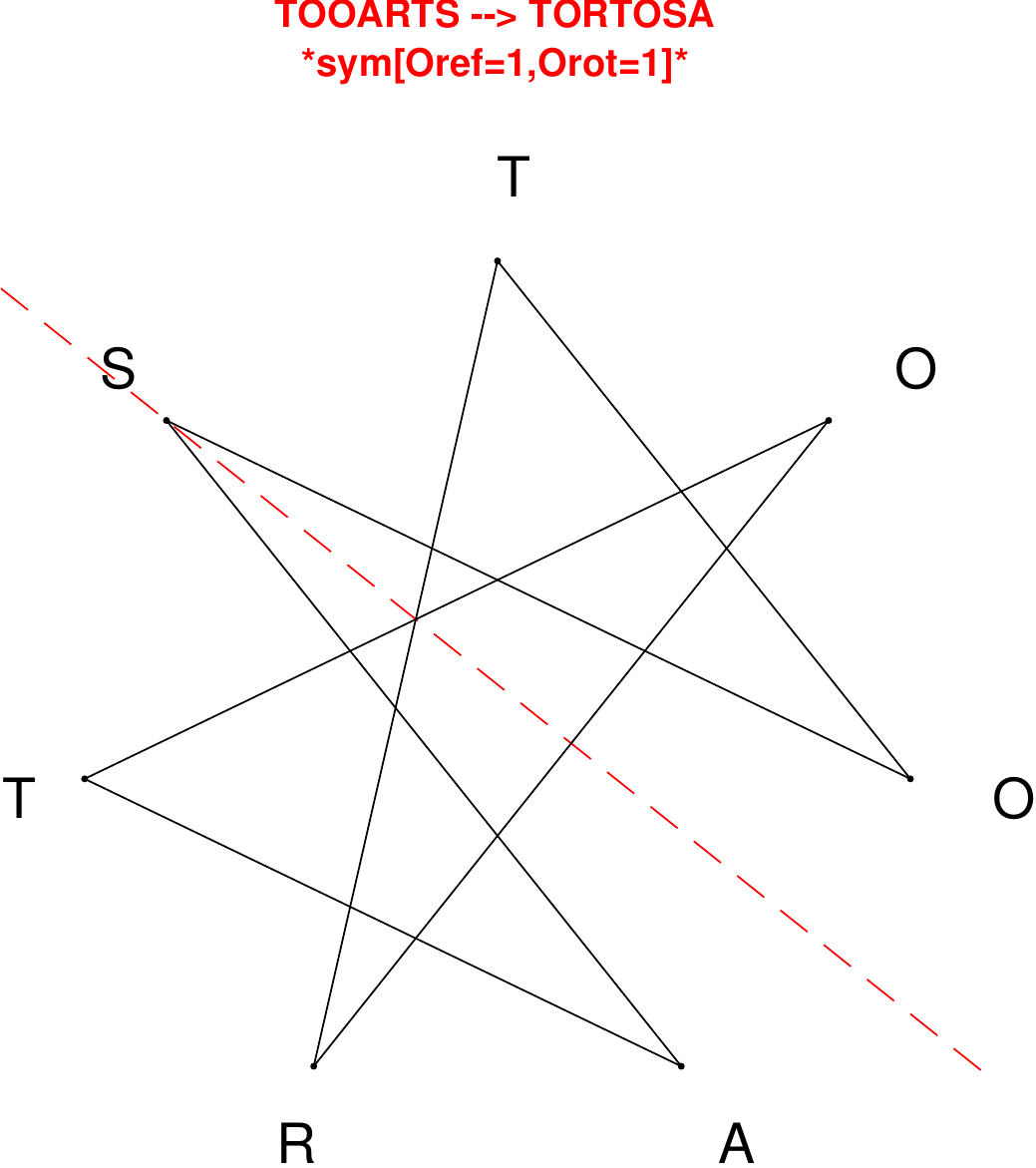}
\end{subfigure}
\hfill
\begin{subfigure}[T]{0.19\textwidth}
\centering
\includegraphics[width=\textwidth]{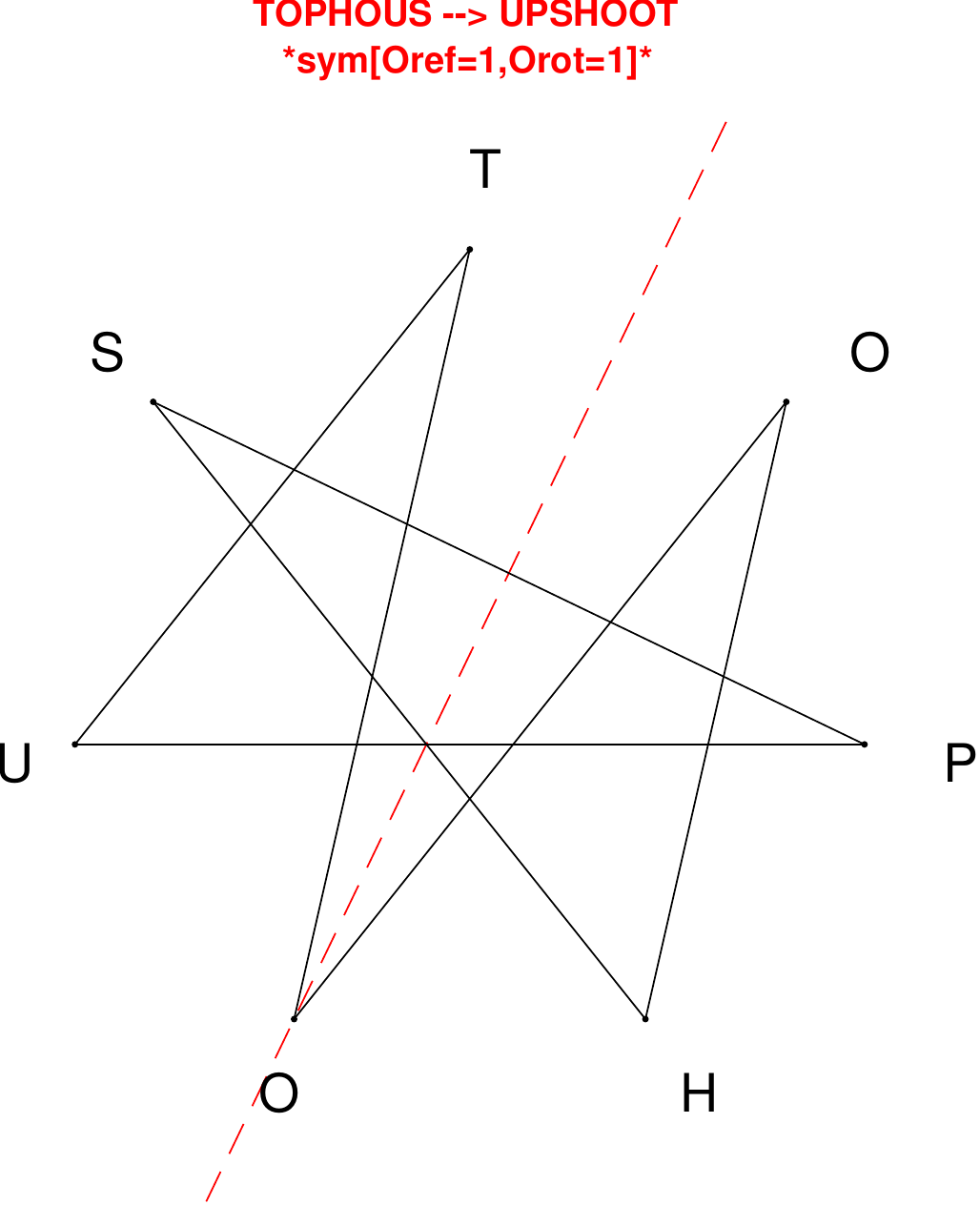}
\end{subfigure}
\end{figure}

\begin{figure}[H]
\centering
\begin{subfigure}[T]{0.19\textwidth}
\centering
\includegraphics[width=\textwidth]{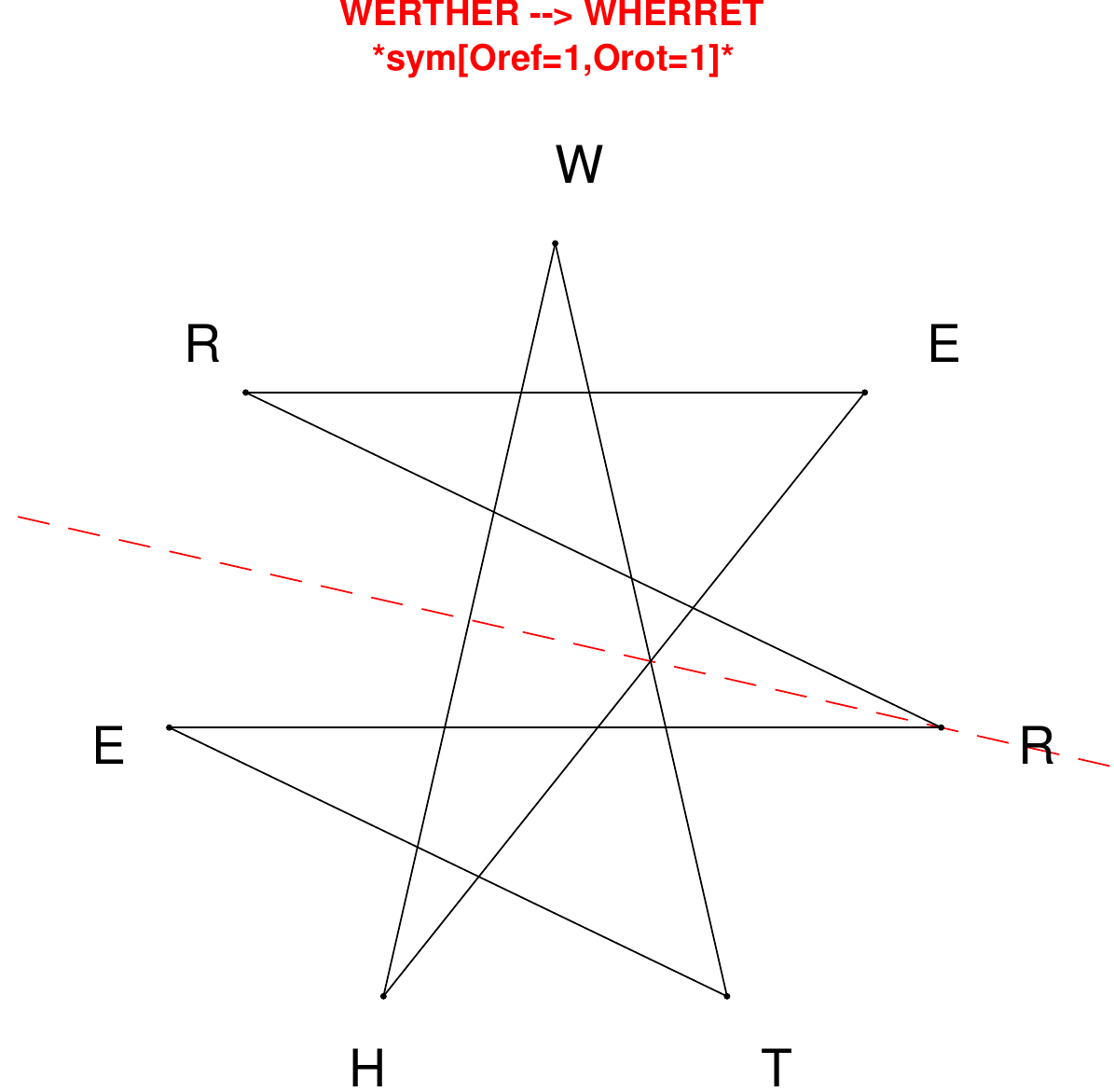}
\end{subfigure}
\hfill
\begin{subfigure}[T]{0.19\textwidth}
\centering
\includegraphics[width=\textwidth]{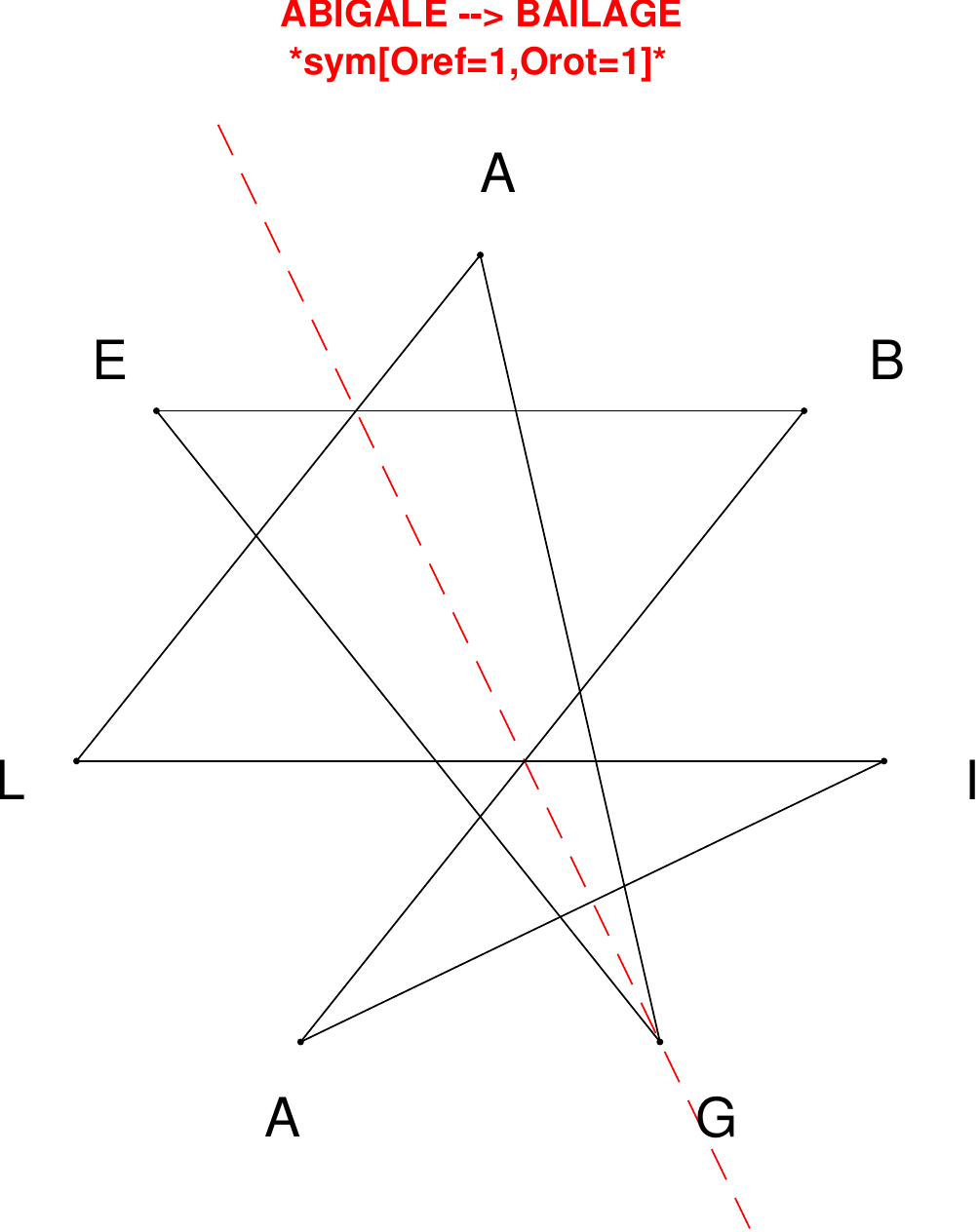}
\end{subfigure}
\hfill
\begin{subfigure}[T]{0.19\textwidth}
\centering
\includegraphics[width=\textwidth]{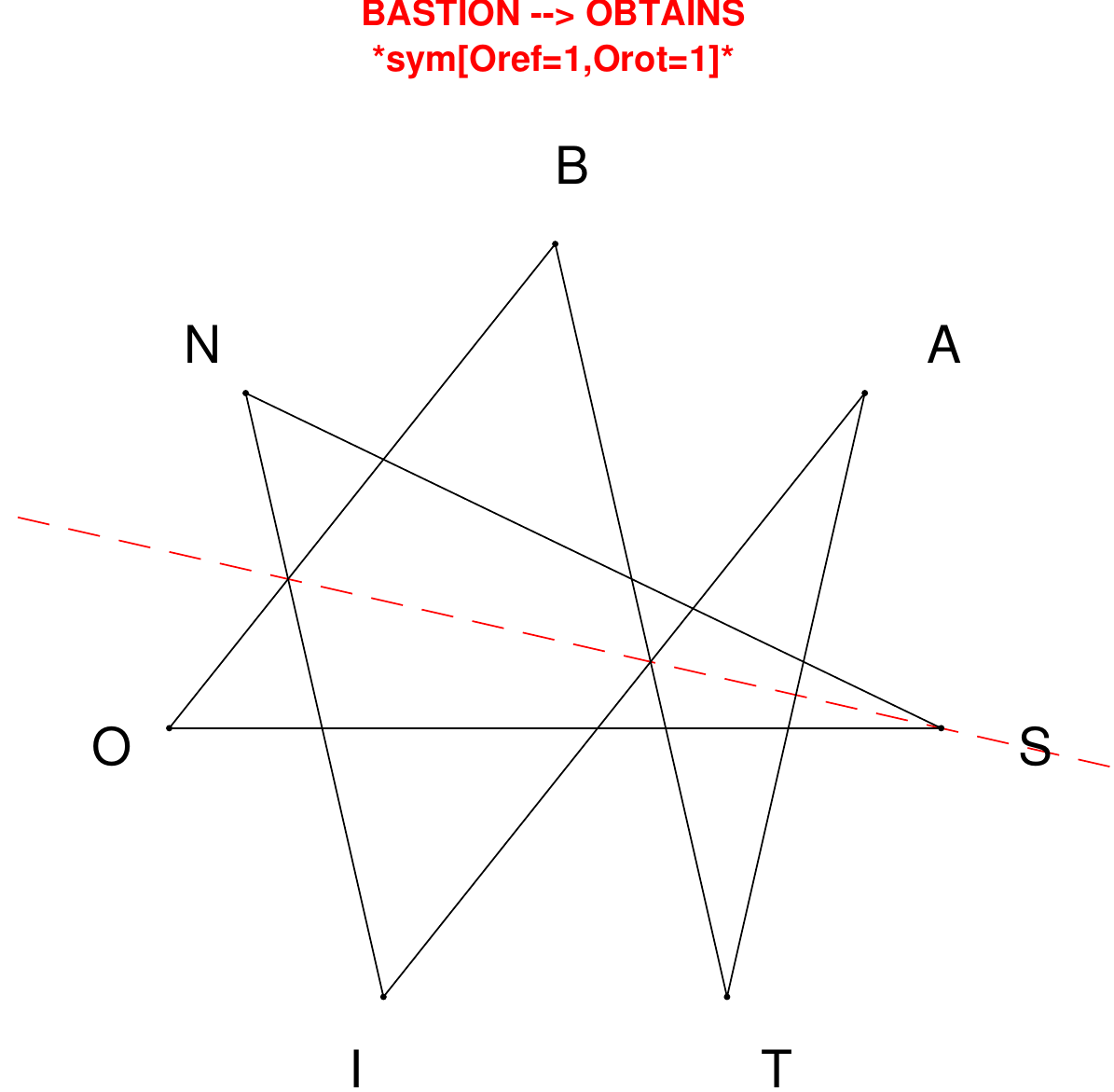}
\end{subfigure}
\hfill
\begin{subfigure}[T]{0.19\textwidth}
\centering
\includegraphics[width=\textwidth]{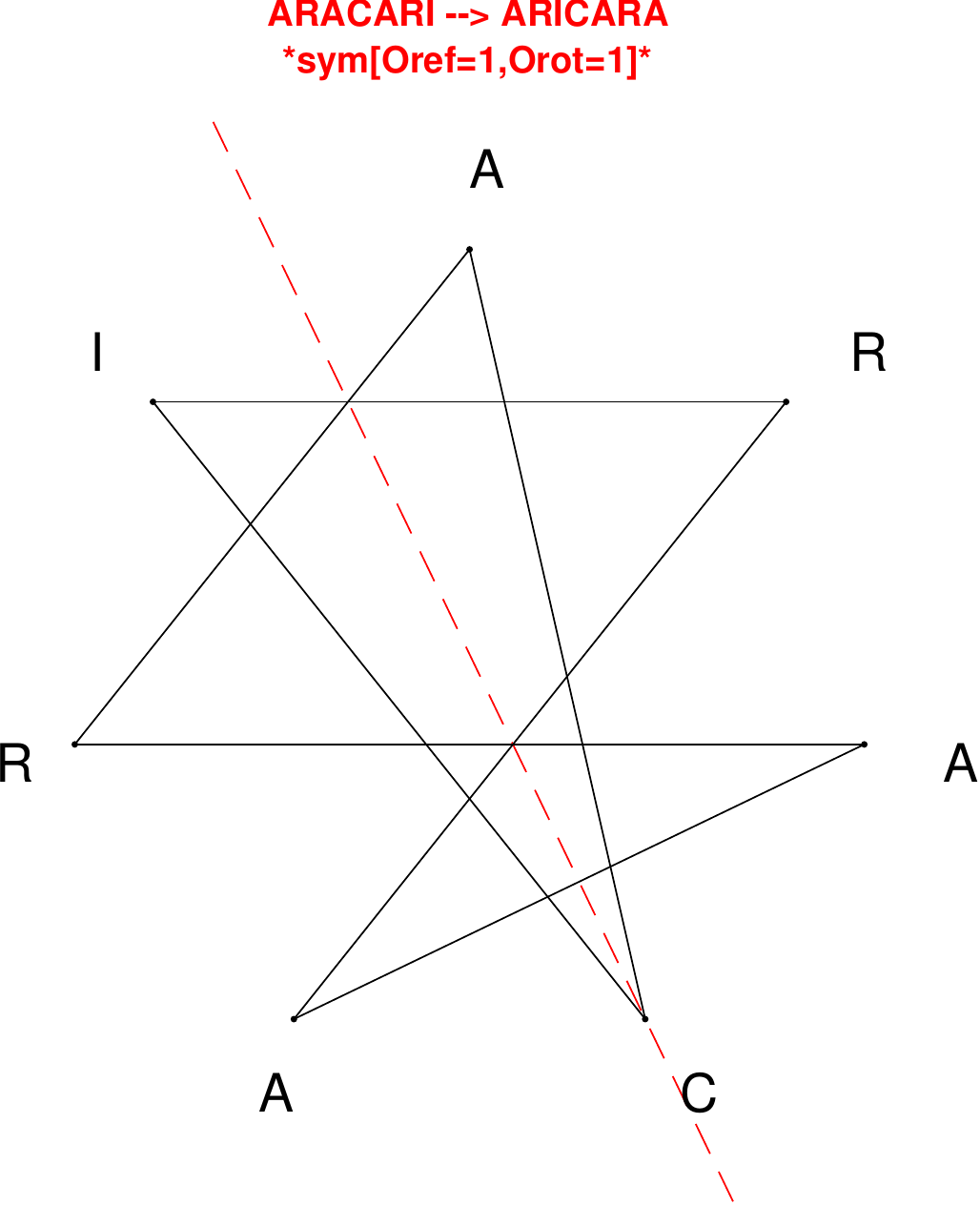}
\end{subfigure}
\hfill
\begin{subfigure}[T]{0.19\textwidth}
\centering
\includegraphics[width=\textwidth]{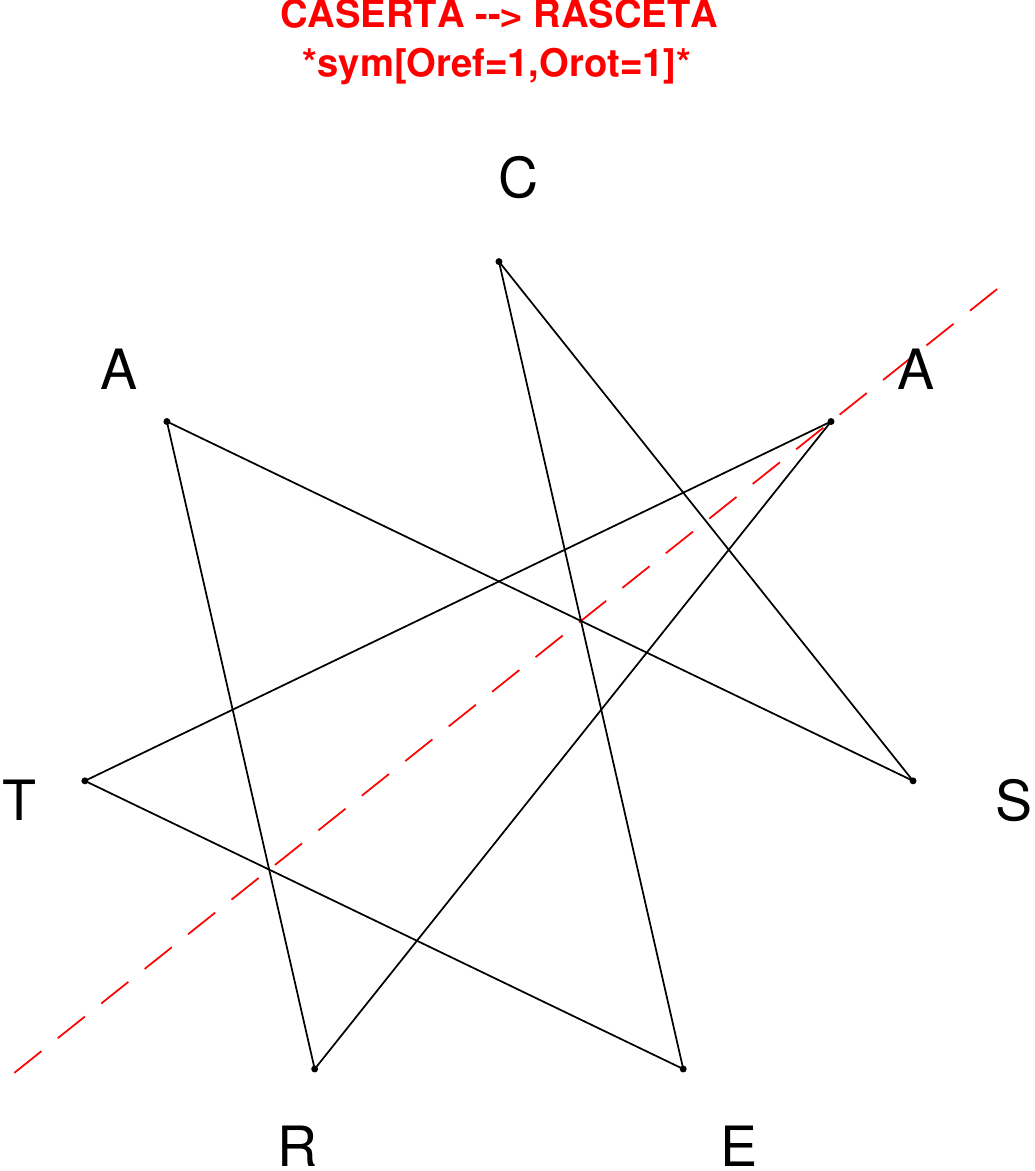}
\end{subfigure}
\end{figure}

\begin{figure}[H]
\centering
\begin{subfigure}[T]{0.19\textwidth}
\centering
\includegraphics[width=\textwidth]{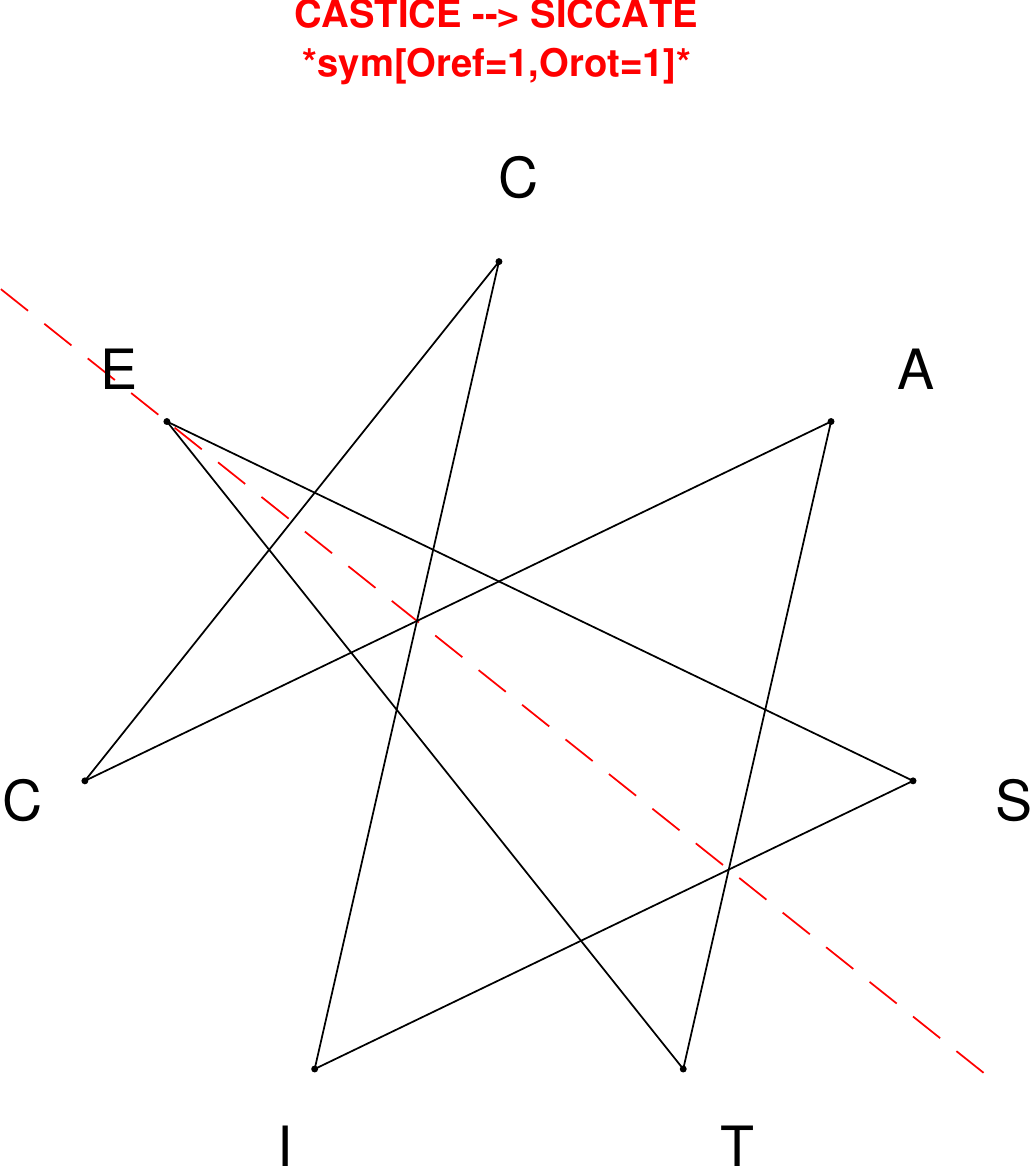}
\end{subfigure}
\hfill
\begin{subfigure}[T]{0.19\textwidth}
\centering
\includegraphics[width=\textwidth]{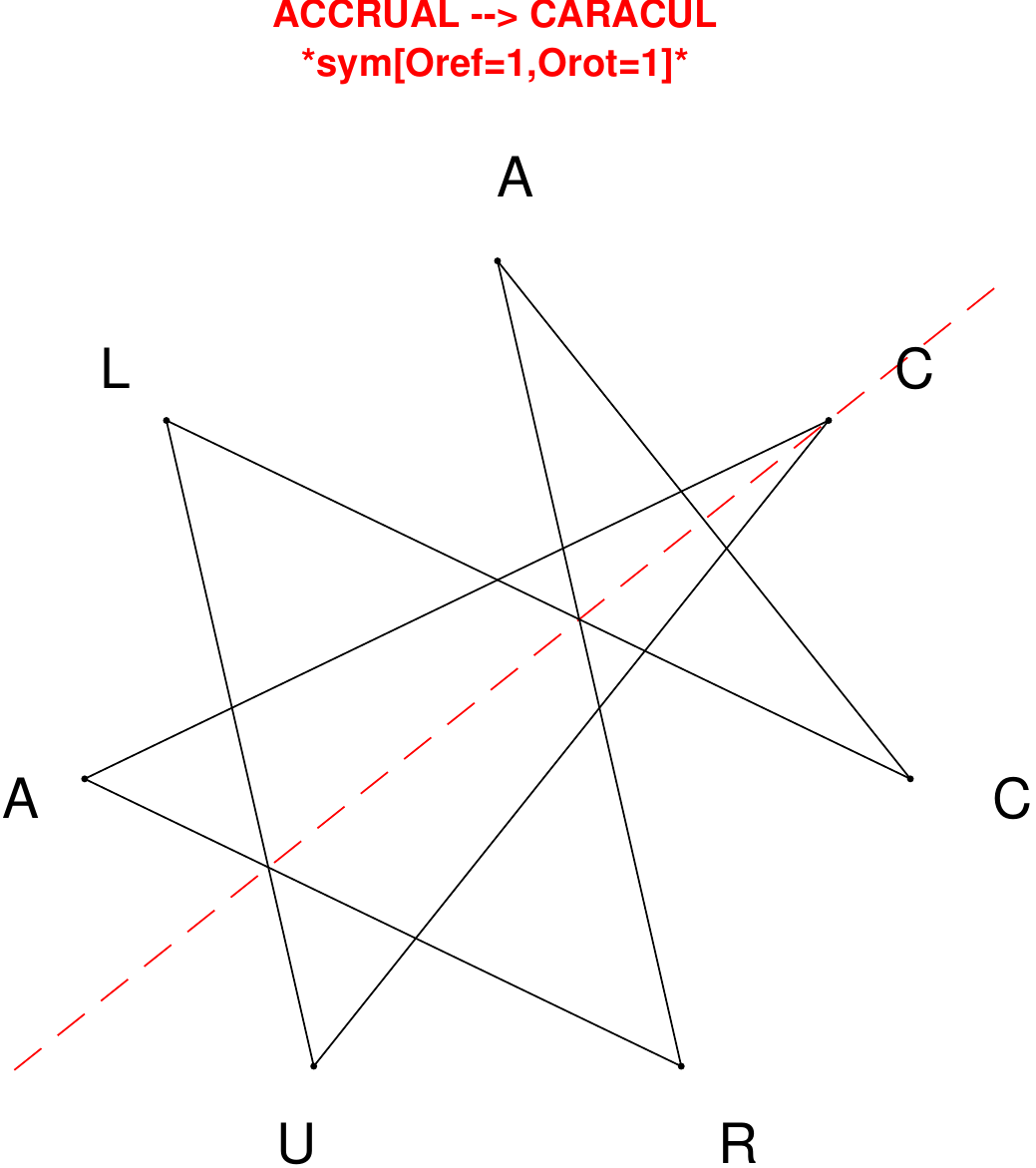}
\end{subfigure}
\hfill
\begin{subfigure}[T]{0.19\textwidth}
\centering
\includegraphics[width=\textwidth]{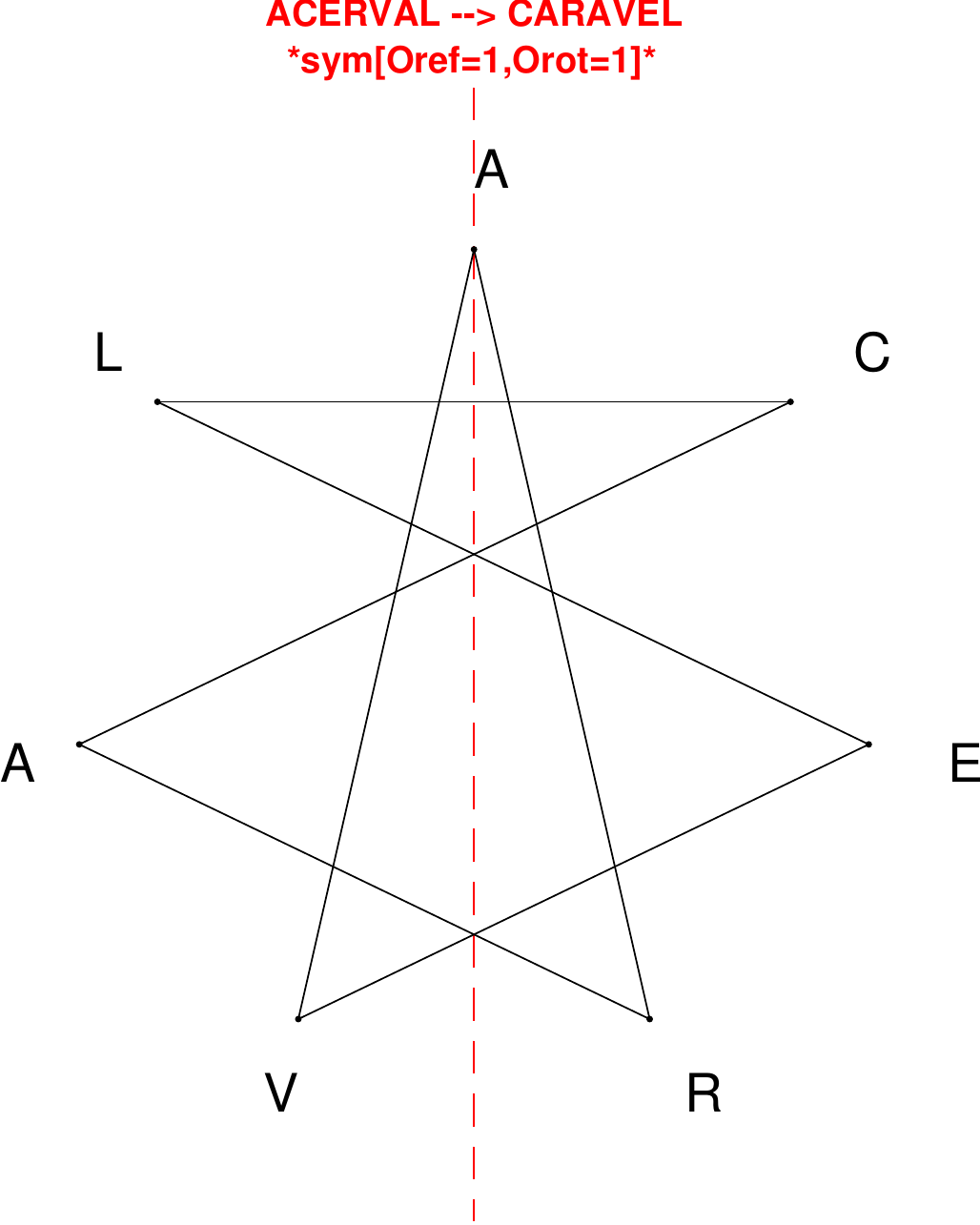}
\end{subfigure}
\hfill
\begin{subfigure}[T]{0.19\textwidth}
\centering
\includegraphics[width=\textwidth]{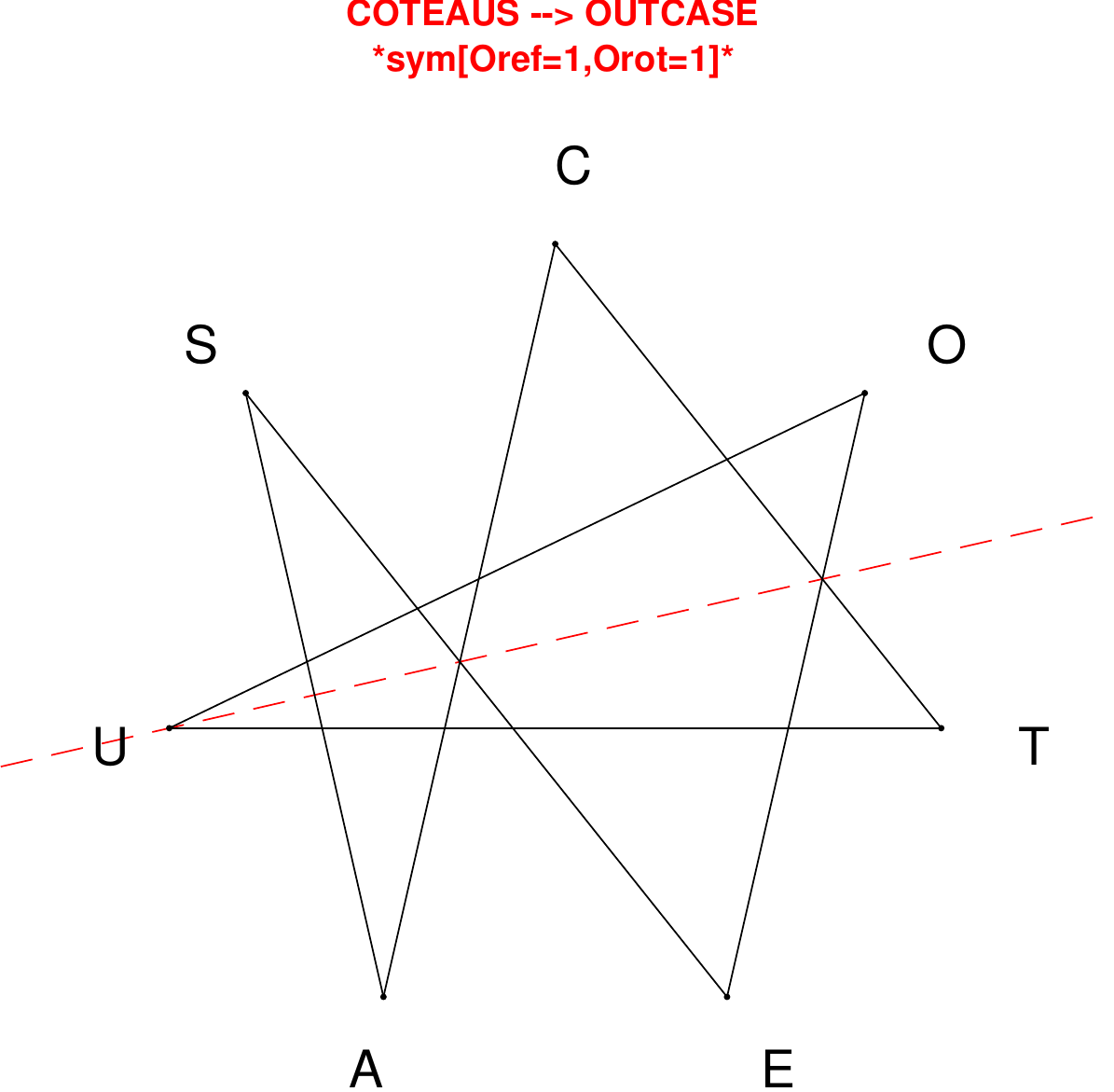}
\end{subfigure}
\hfill
\begin{subfigure}[T]{0.19\textwidth}
\centering
\includegraphics[width=\textwidth]{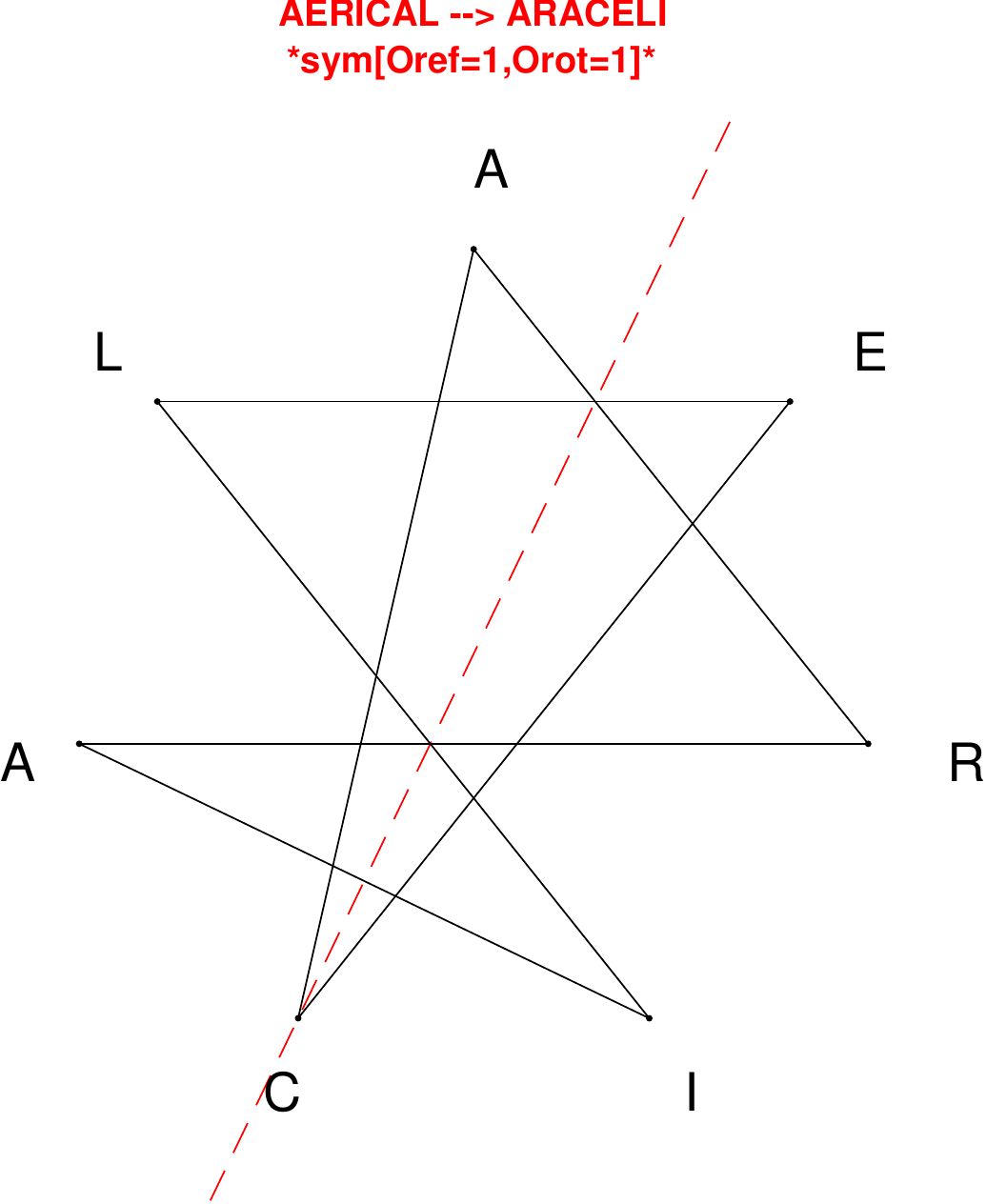}
\end{subfigure}
\end{figure}

\begin{figure}[H]
\centering
\begin{subfigure}[T]{0.19\textwidth}
\centering
\includegraphics[width=\textwidth]{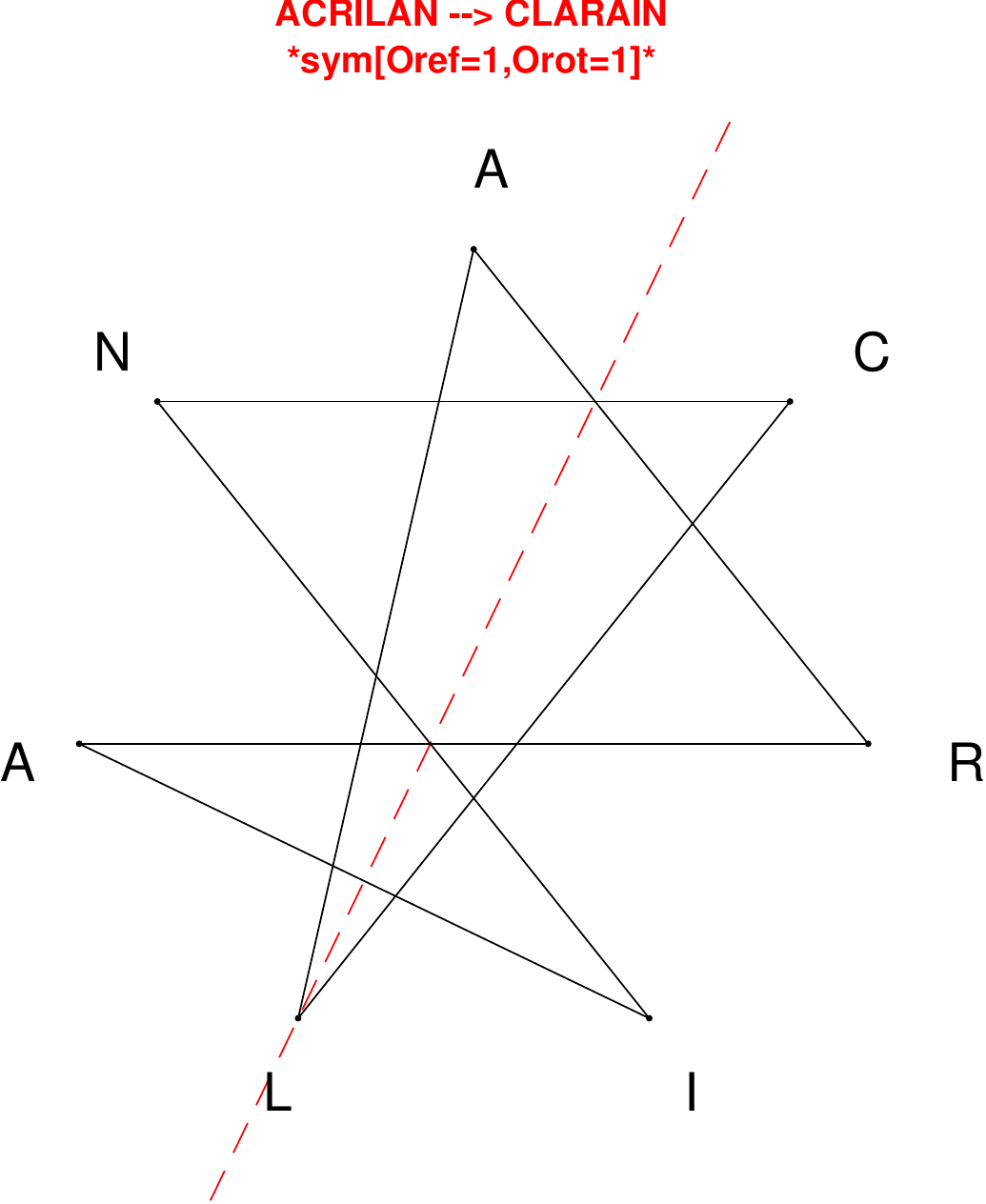}
\end{subfigure}
\hfill
\begin{subfigure}[T]{0.19\textwidth}
\centering
\includegraphics[width=\textwidth]{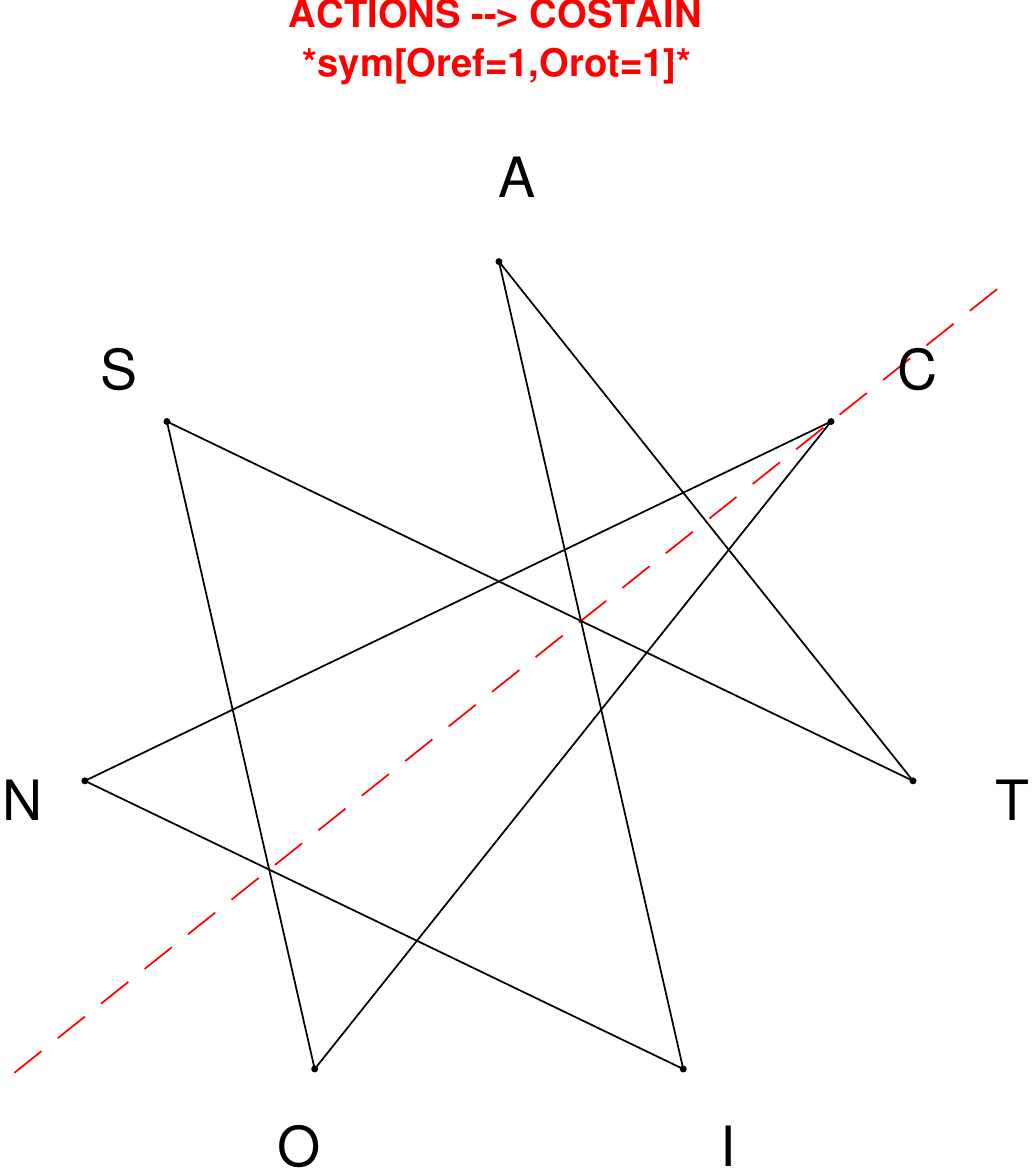}
\end{subfigure}
\hfill
\begin{subfigure}[T]{0.19\textwidth}
\centering
\includegraphics[width=\textwidth]{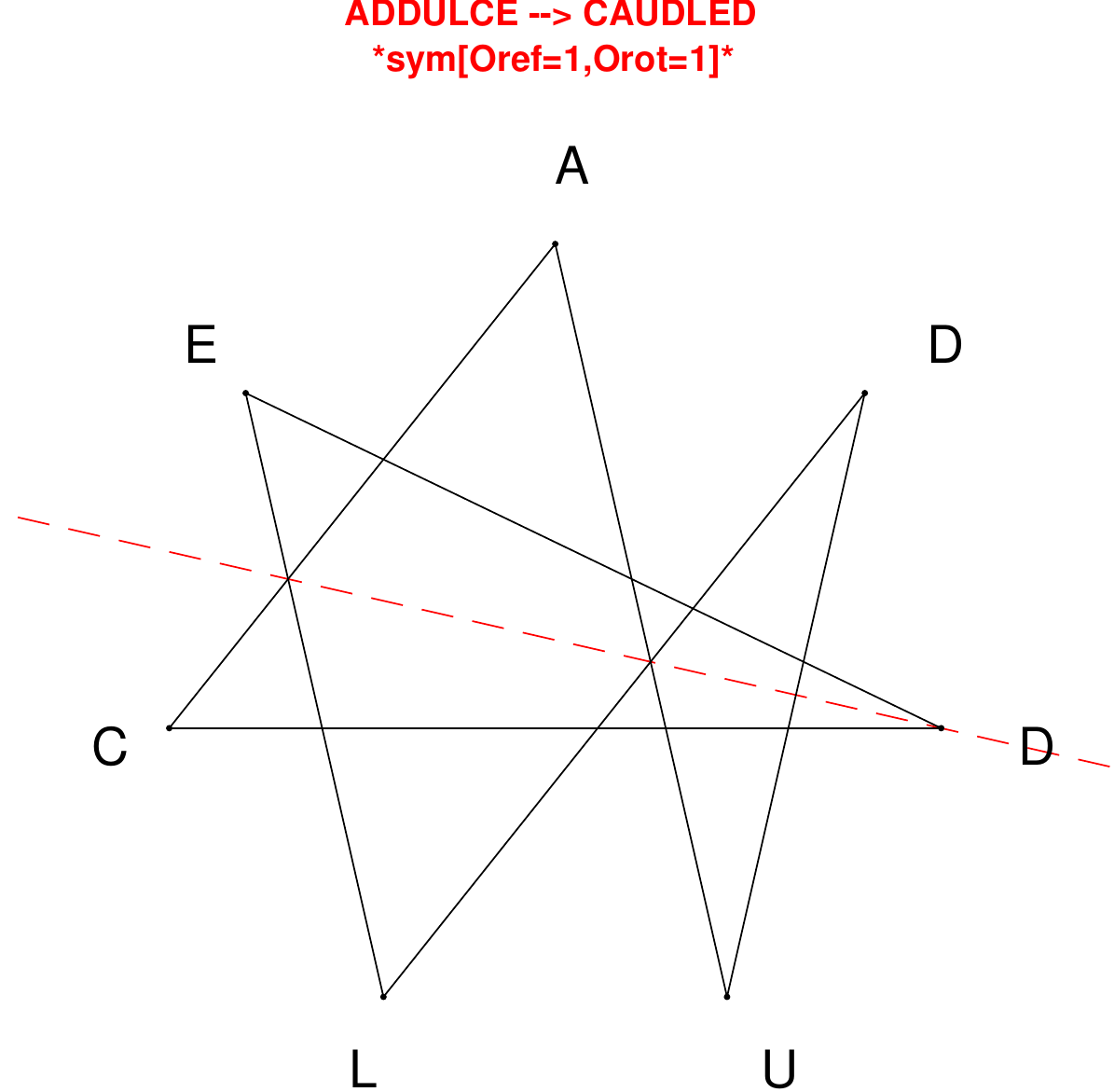}
\end{subfigure}
\hfill
\begin{subfigure}[T]{0.19\textwidth}
\centering
\includegraphics[width=\textwidth]{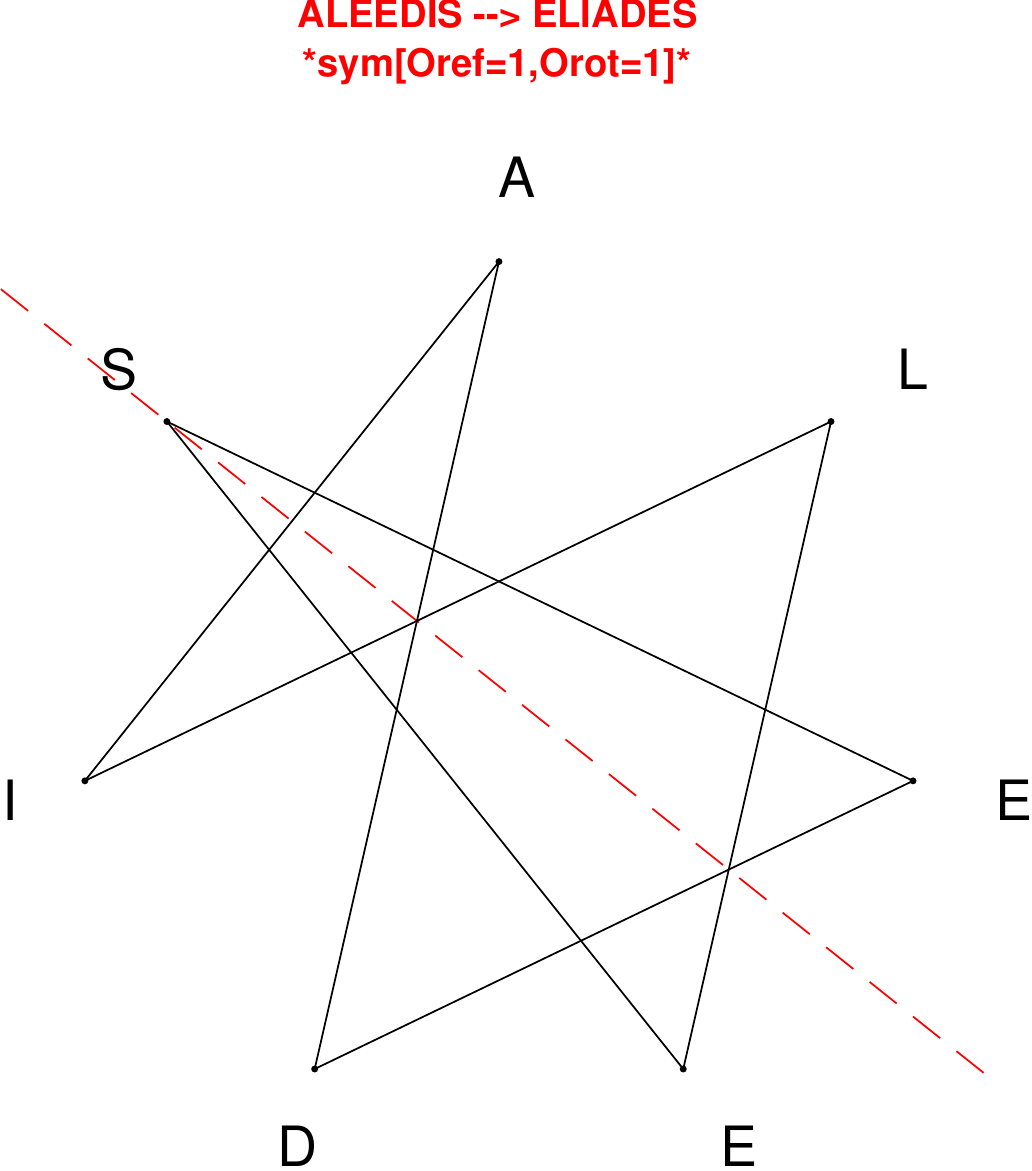}
\end{subfigure}
\hfill
\begin{subfigure}[T]{0.19\textwidth}
\centering
\includegraphics[width=\textwidth]{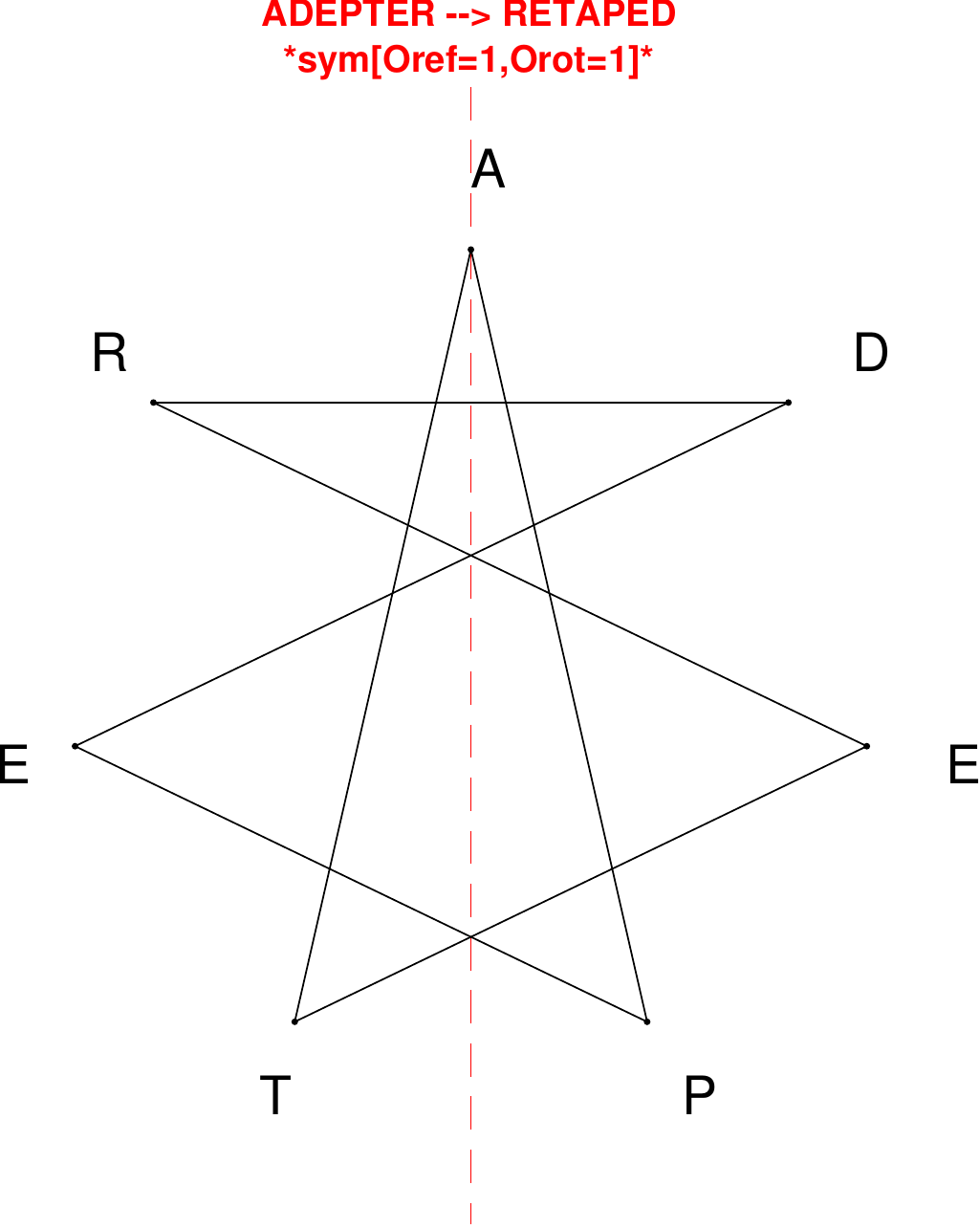}
\end{subfigure}
\end{figure}

\begin{figure}[H]
\centering
\begin{subfigure}[T]{0.19\textwidth}
\centering
\includegraphics[width=\textwidth]{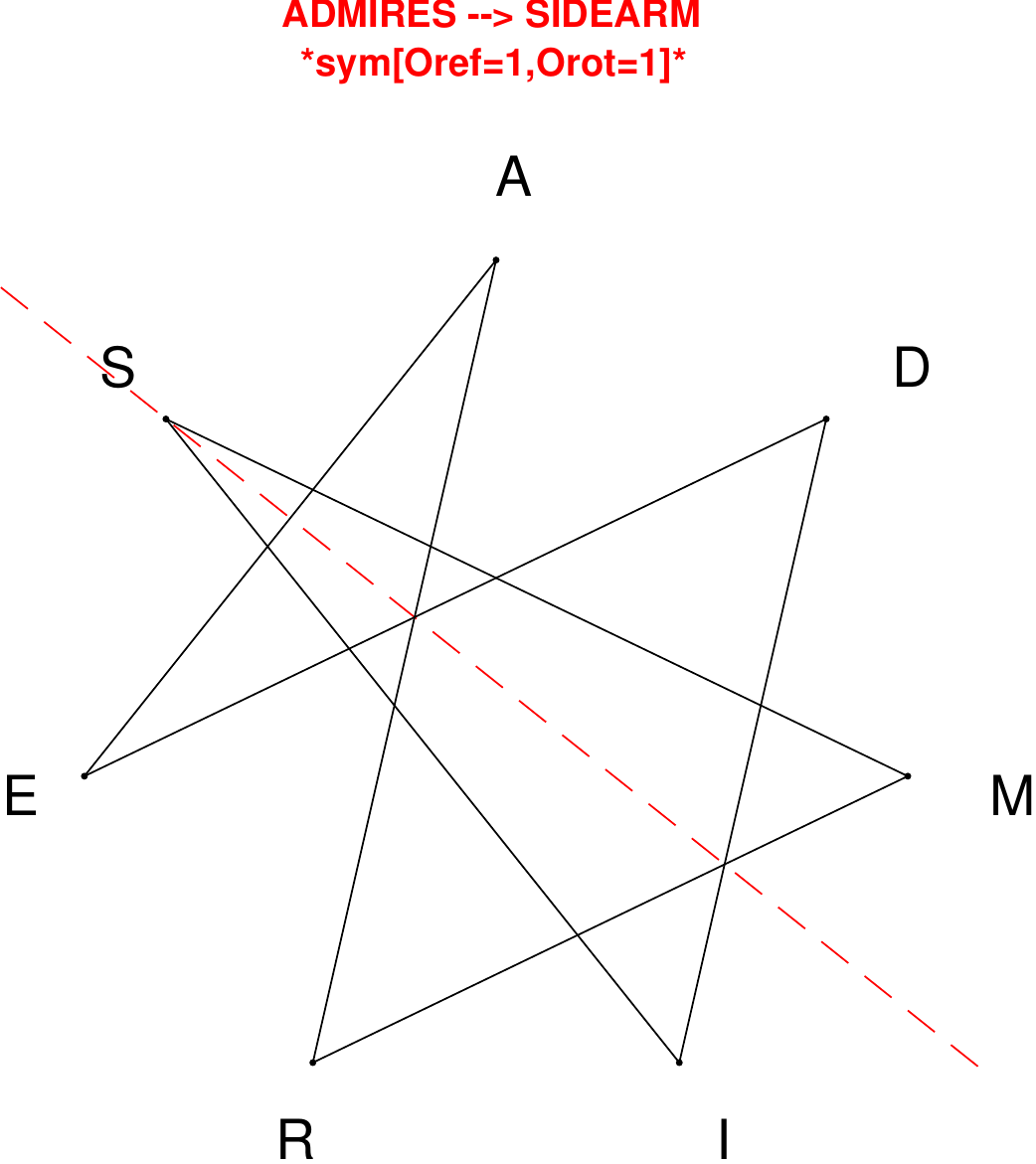}
\end{subfigure}
\hfill
\begin{subfigure}[T]{0.19\textwidth}
\centering
\includegraphics[width=\textwidth]{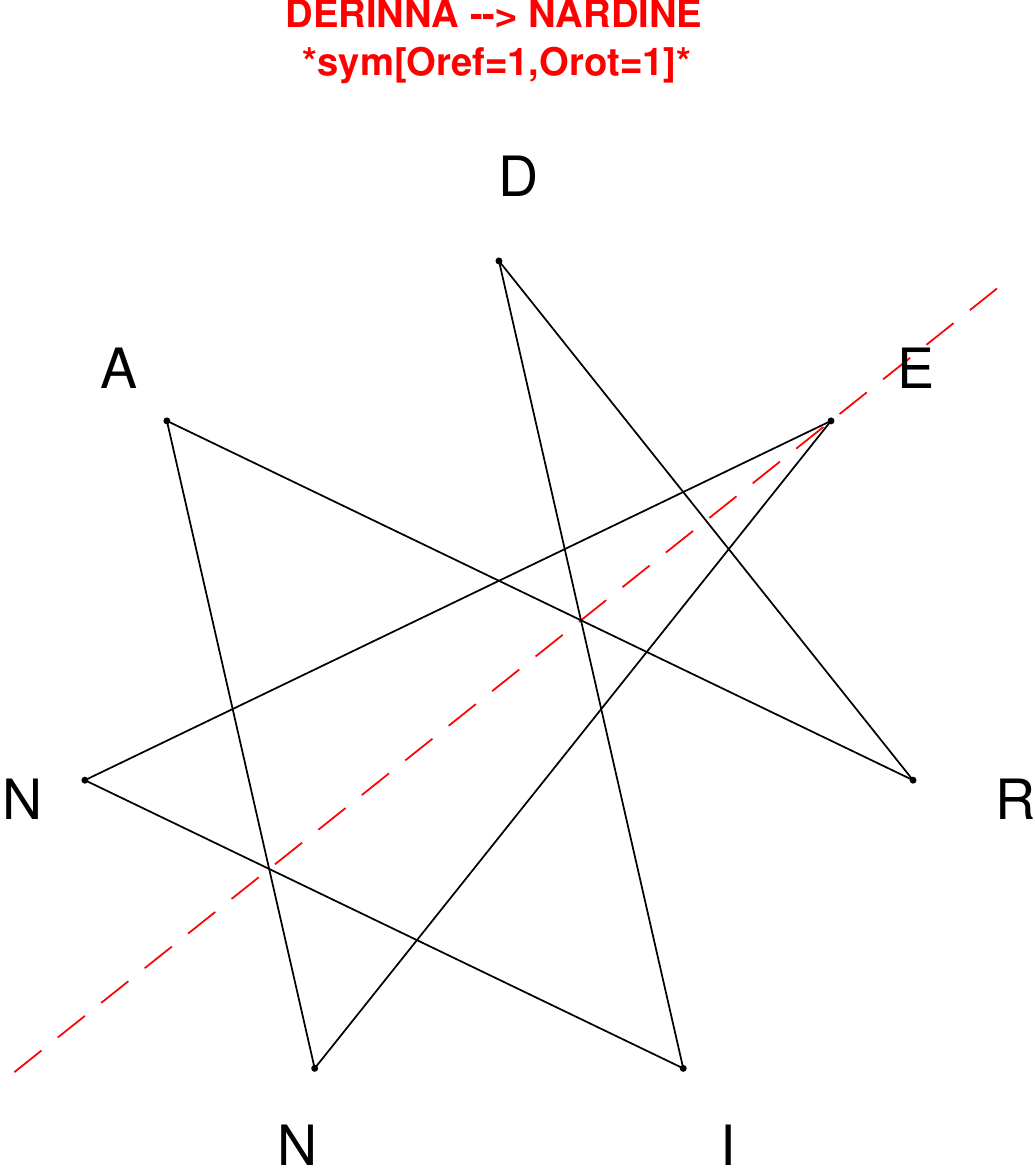}
\end{subfigure}
\hfill
\begin{subfigure}[T]{0.19\textwidth}
\centering
\includegraphics[width=\textwidth]{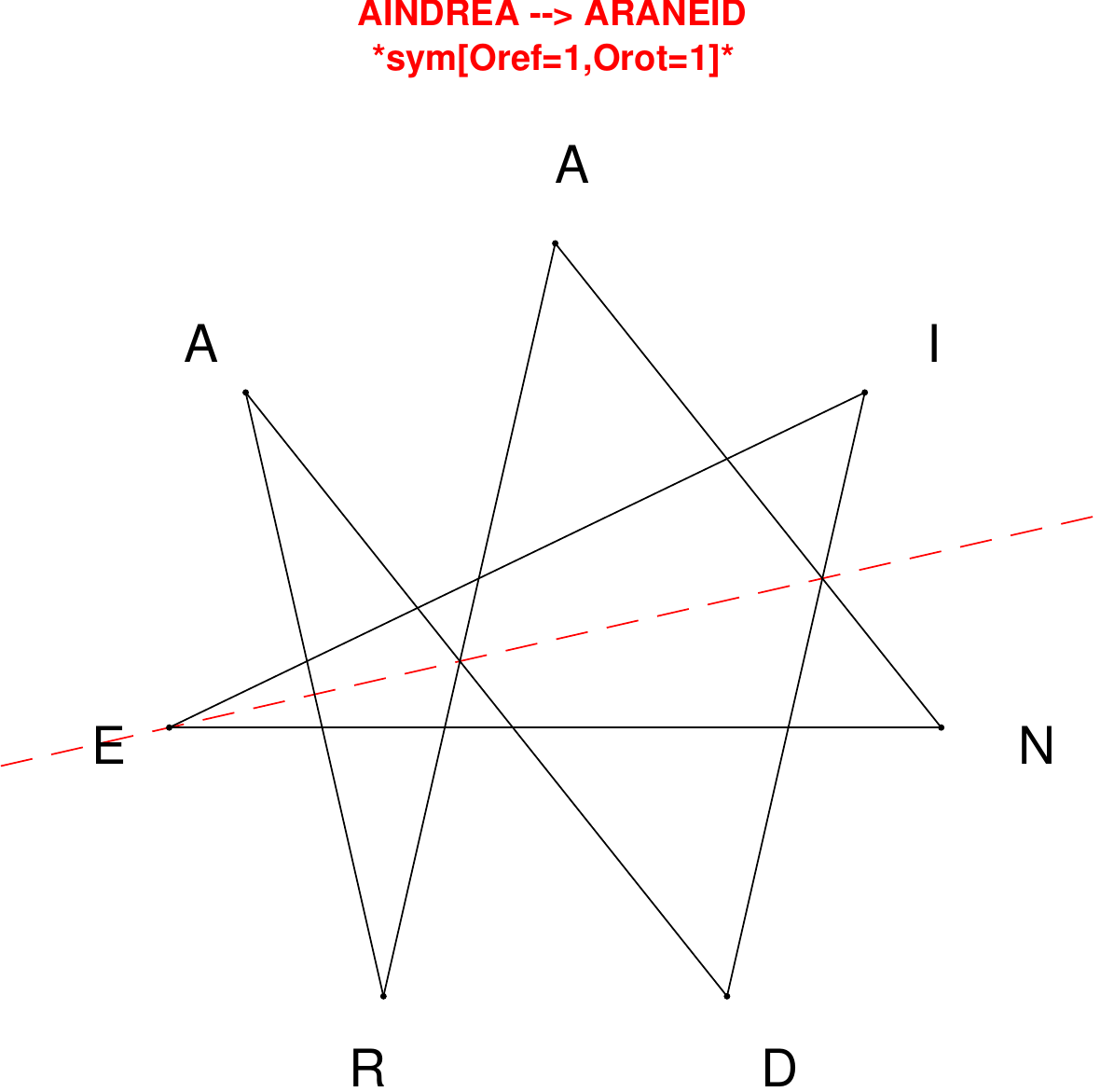}
\end{subfigure}
\hfill
\begin{subfigure}[T]{0.19\textwidth}
\centering
\includegraphics[width=\textwidth]{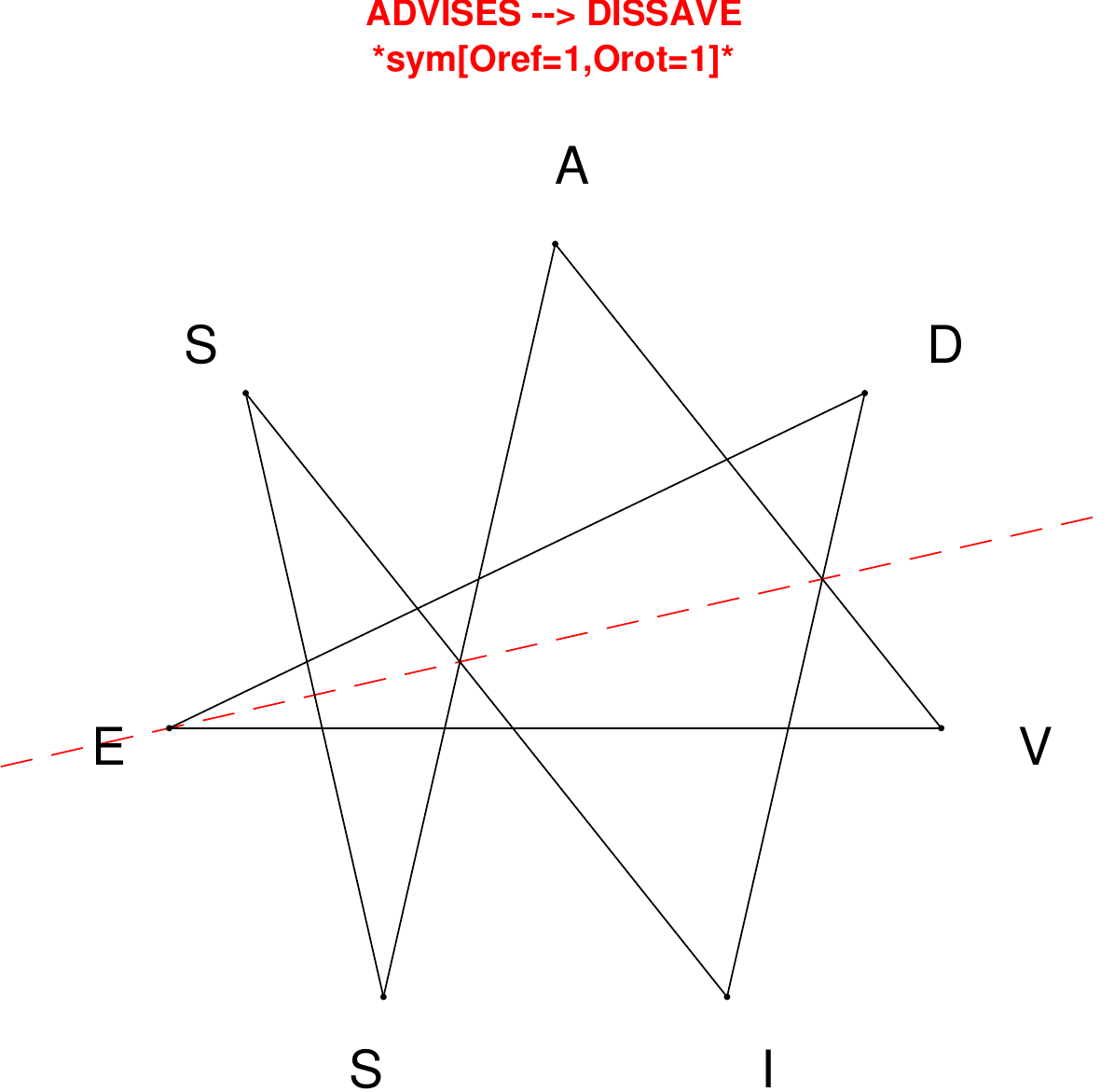}
\end{subfigure}
\hfill
\begin{subfigure}[T]{0.19\textwidth}
\centering
\includegraphics[width=\textwidth]{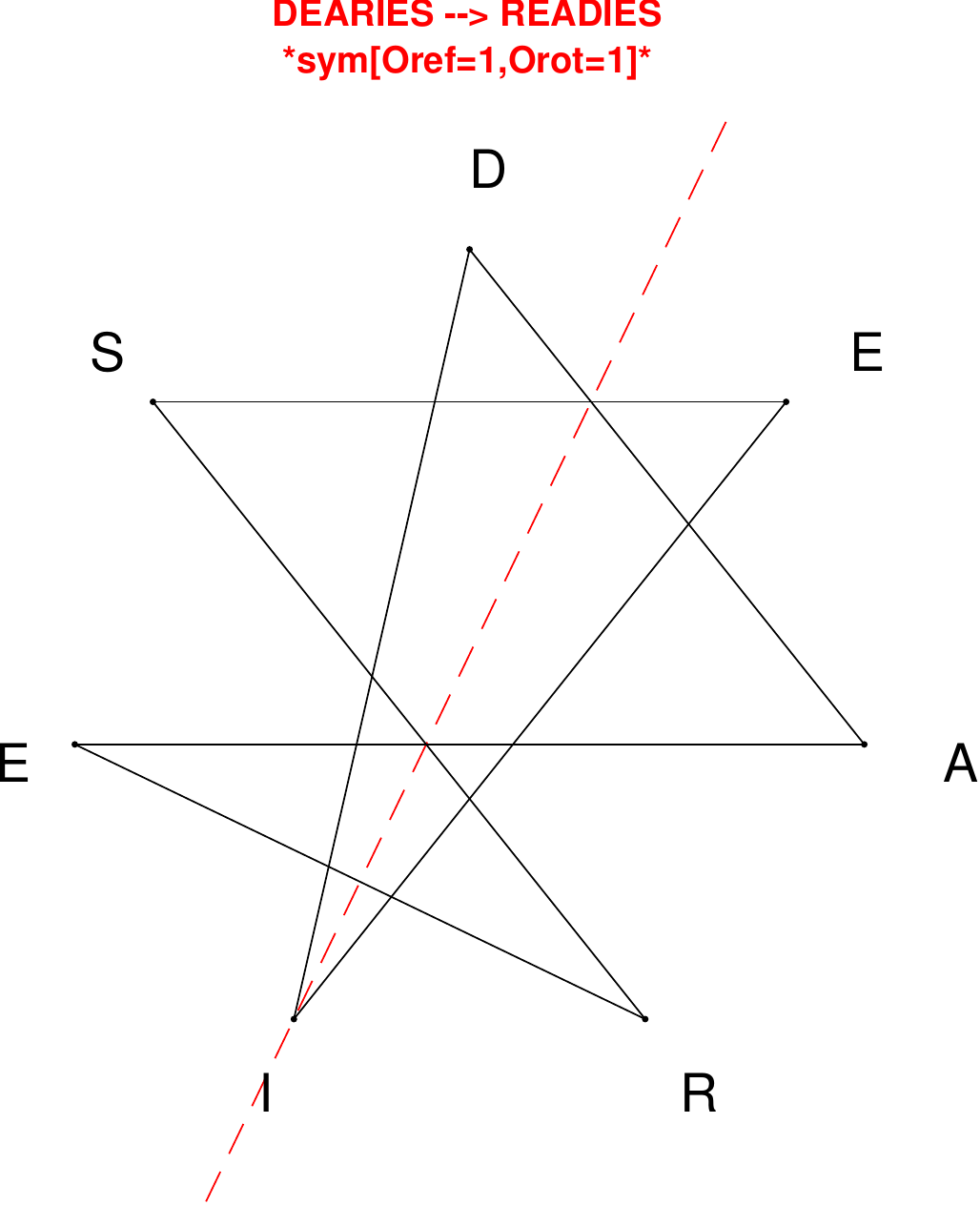}
\end{subfigure}
\end{figure}

\begin{figure}[H]
\centering
\begin{subfigure}[T]{0.19\textwidth}
\centering
\includegraphics[width=\textwidth]{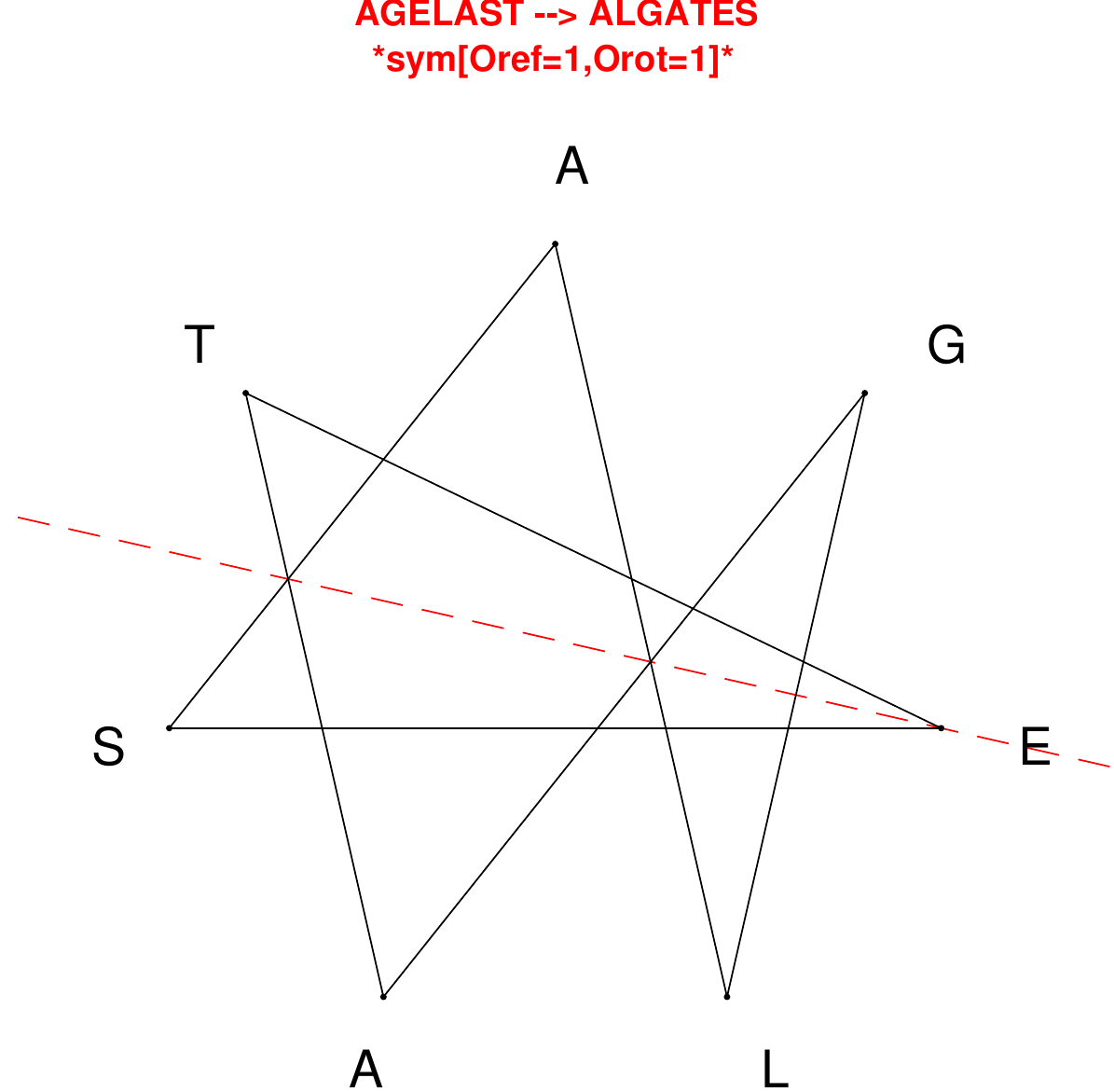}
\end{subfigure}
\hfill
\begin{subfigure}[T]{0.19\textwidth}
\centering
\includegraphics[width=\textwidth]{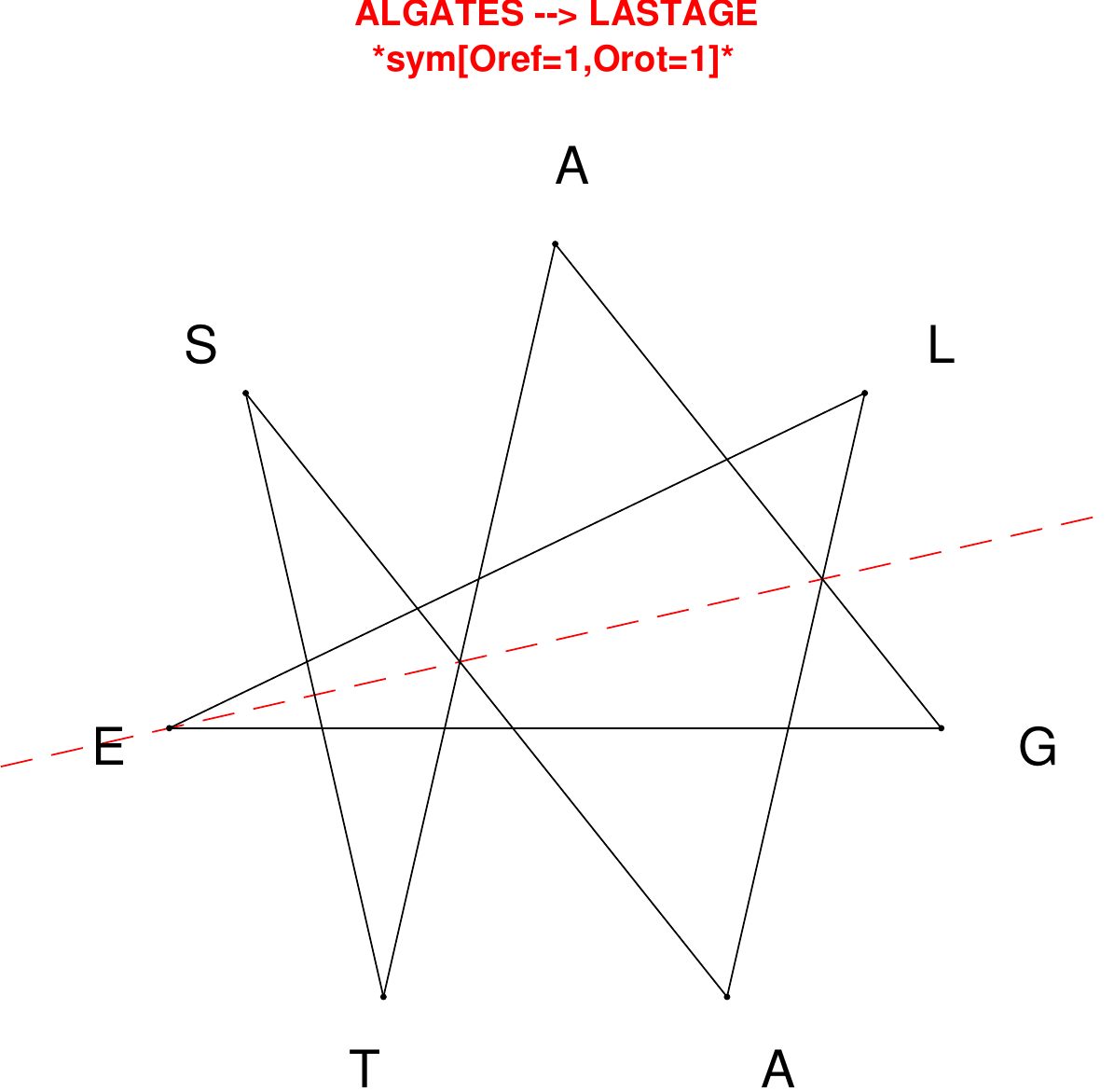}
\end{subfigure}
\hfill
\begin{subfigure}[T]{0.19\textwidth}
\centering
\includegraphics[width=\textwidth]{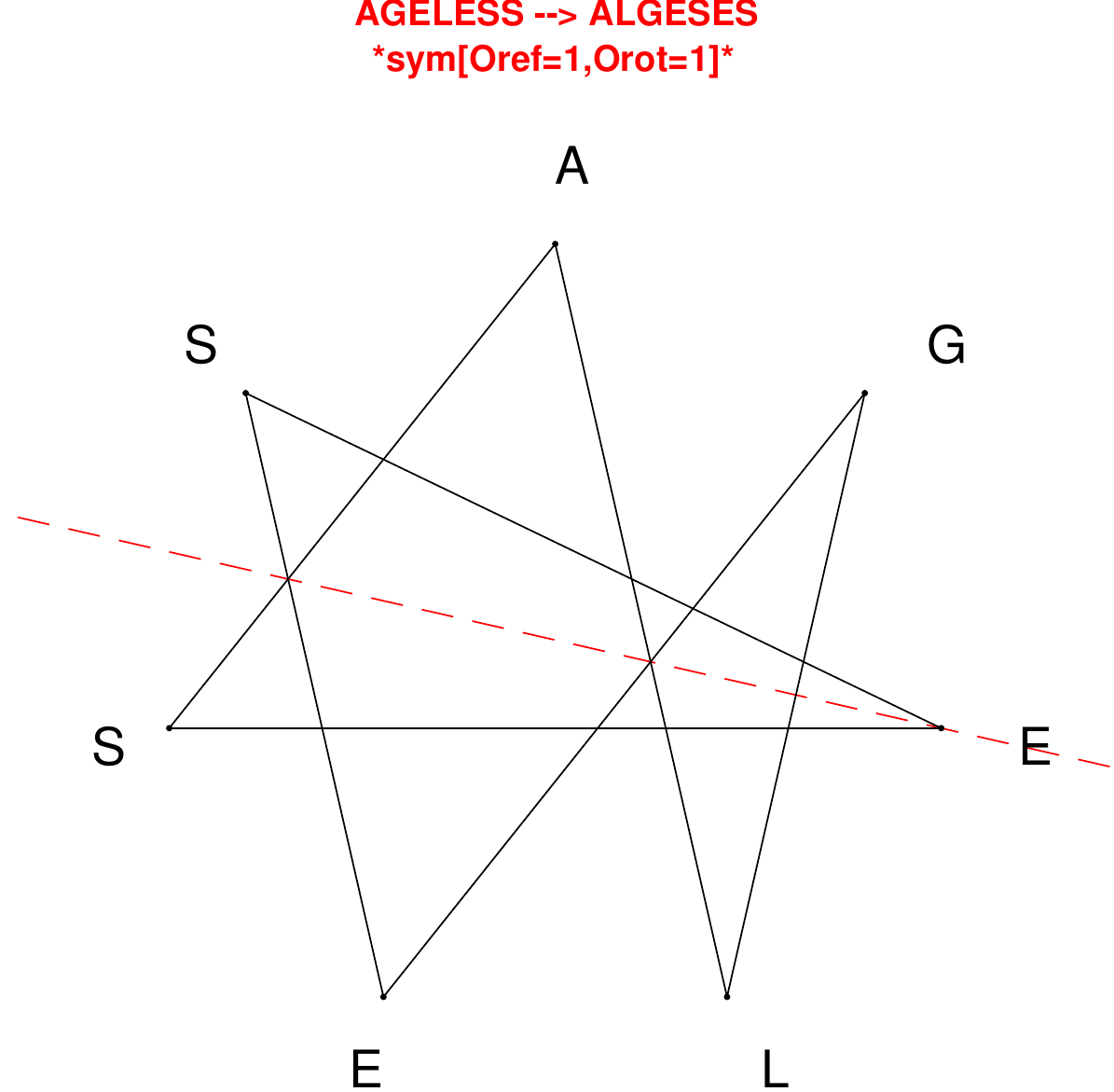}
\end{subfigure}
\hfill
\begin{subfigure}[T]{0.19\textwidth}
\centering
\includegraphics[width=\textwidth]{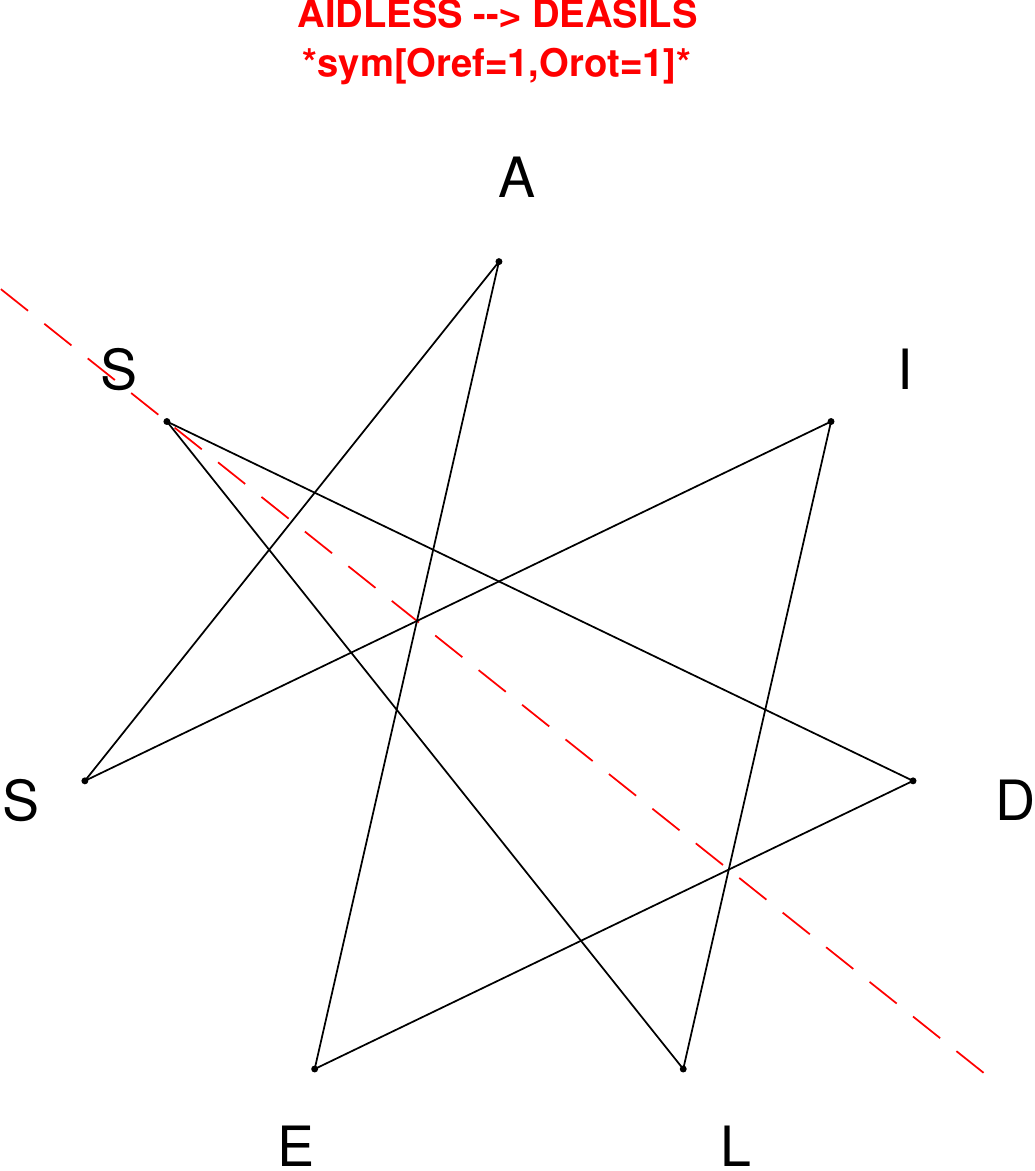}
\end{subfigure}
\hfill
\begin{subfigure}[T]{0.19\textwidth}
\centering
\includegraphics[width=\textwidth]{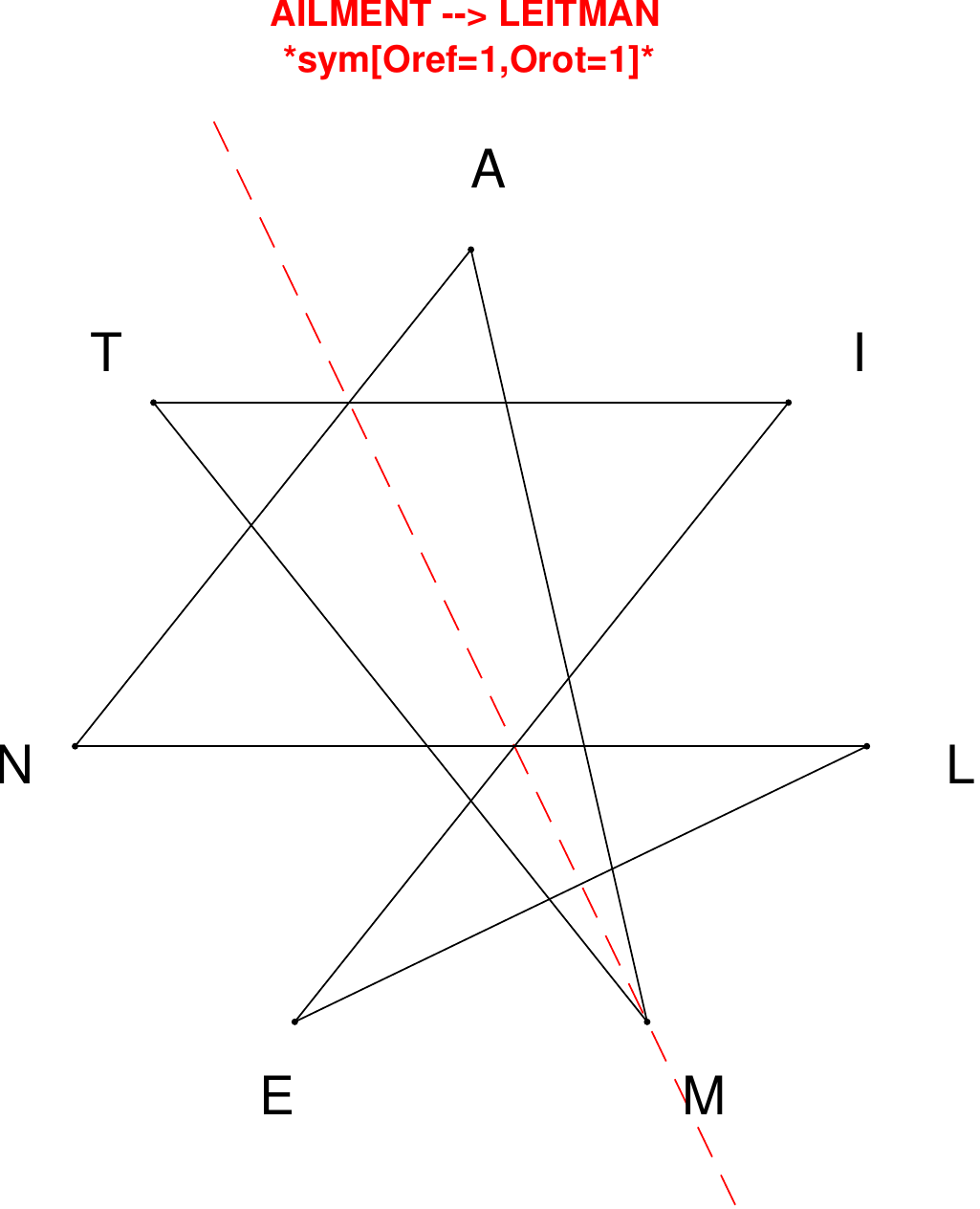}
\end{subfigure}
\end{figure}

\begin{figure}[H]
\centering
\begin{subfigure}[T]{0.19\textwidth}
\centering
\includegraphics[width=\textwidth]{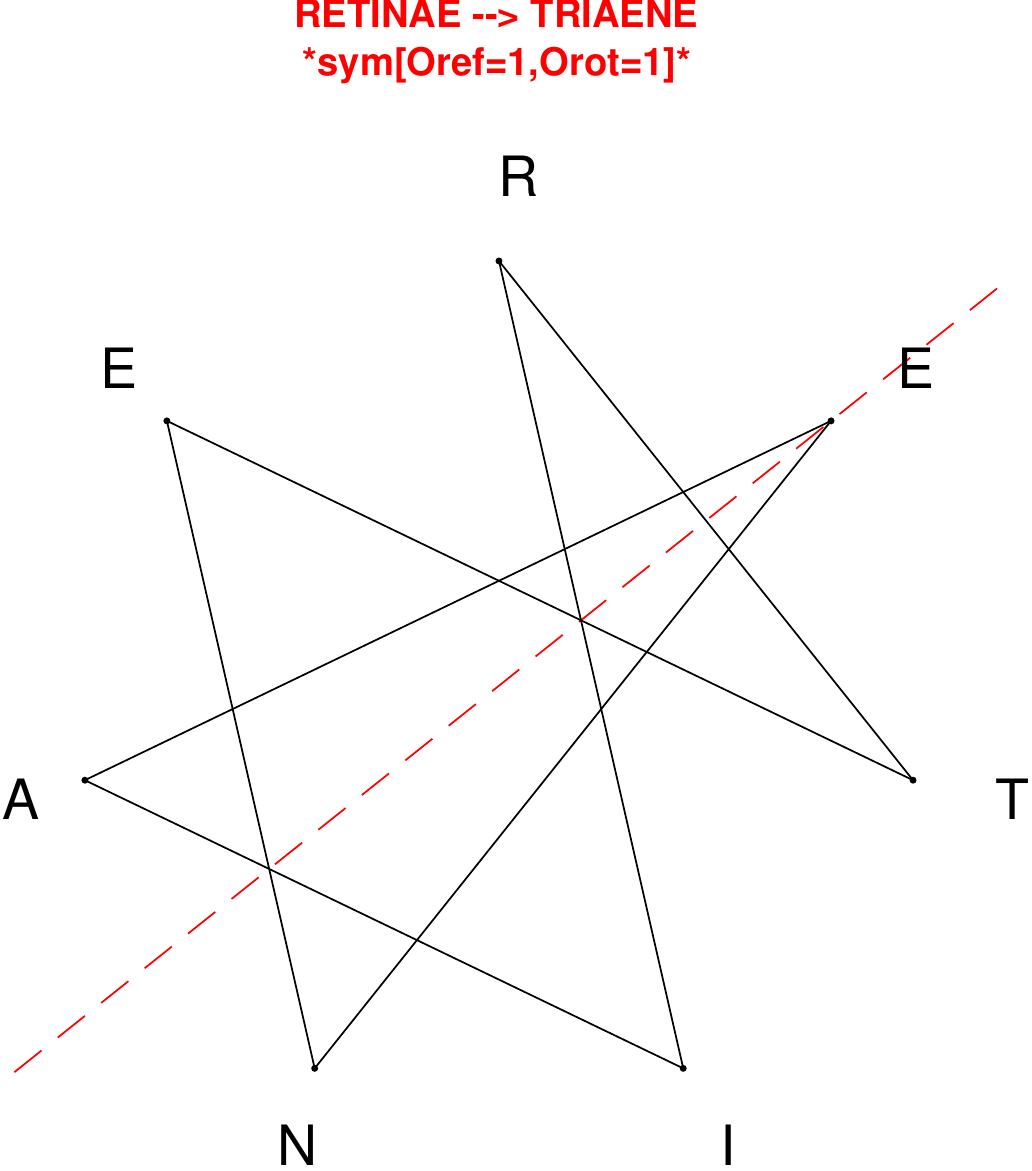}
\end{subfigure}
\hfill
\begin{subfigure}[T]{0.19\textwidth}
\centering
\includegraphics[width=\textwidth]{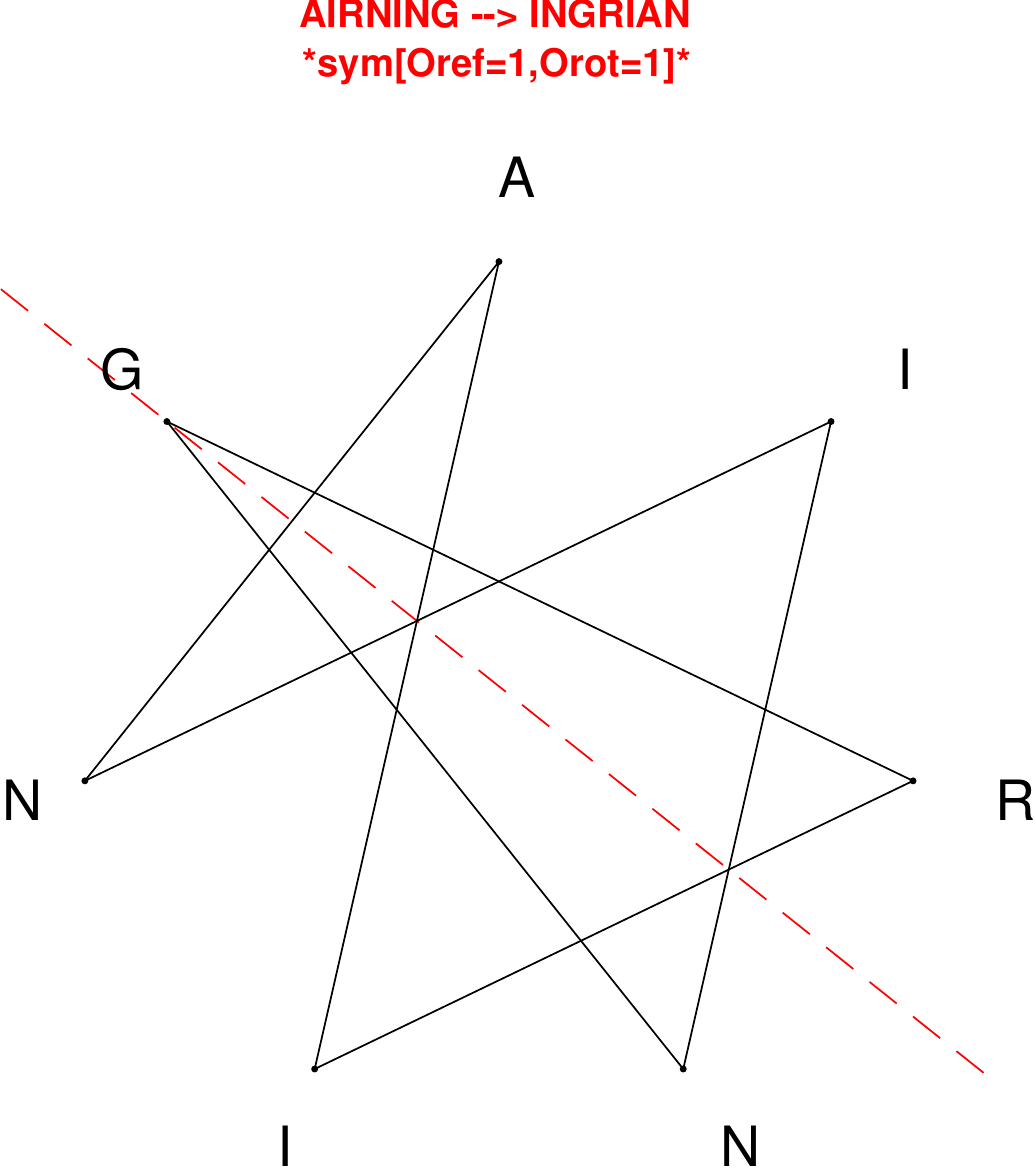}
\end{subfigure}
\hfill
\begin{subfigure}[T]{0.19\textwidth}
\centering
\includegraphics[width=\textwidth]{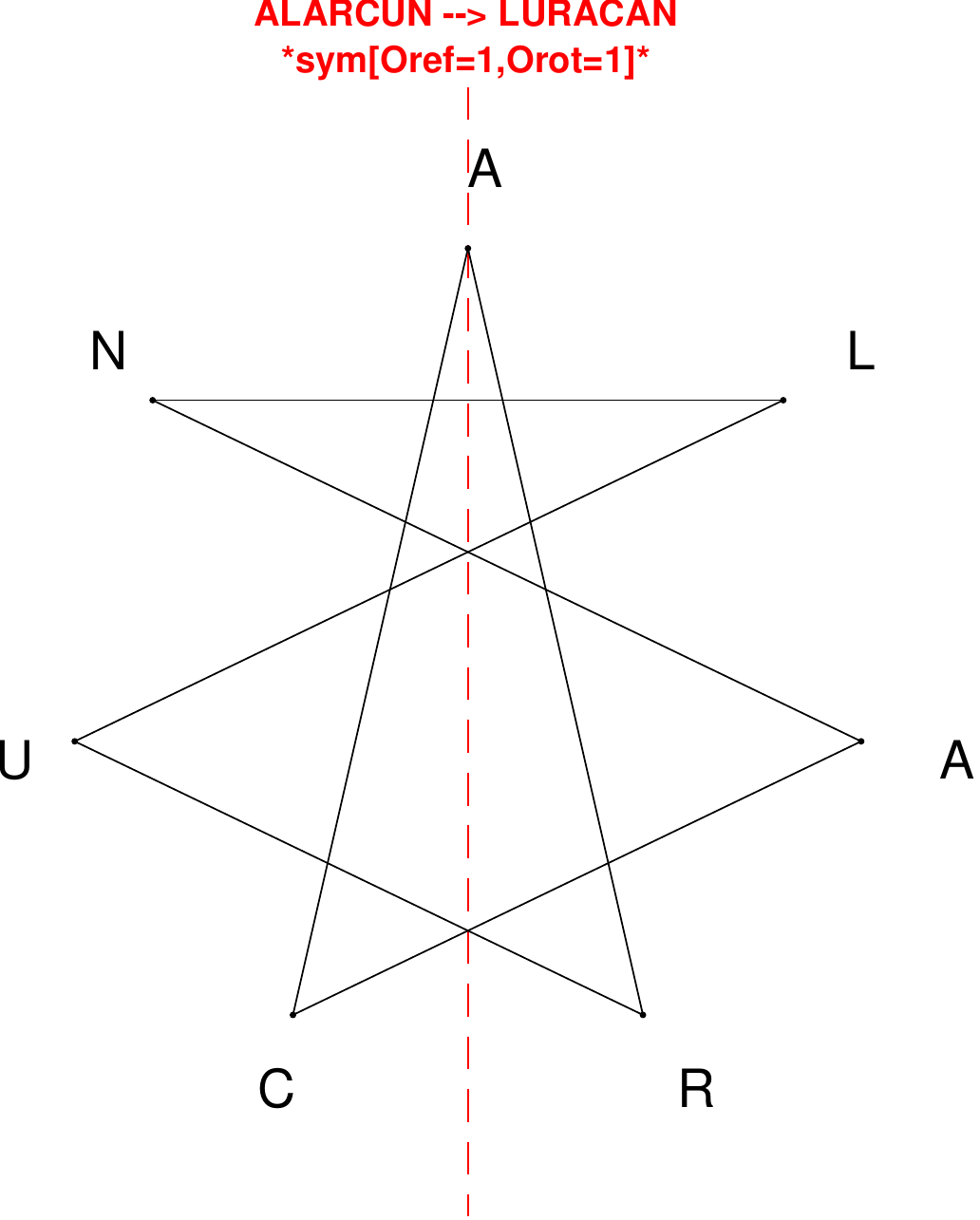}
\end{subfigure}
\hfill
\begin{subfigure}[T]{0.19\textwidth}
\centering
\includegraphics[width=\textwidth]{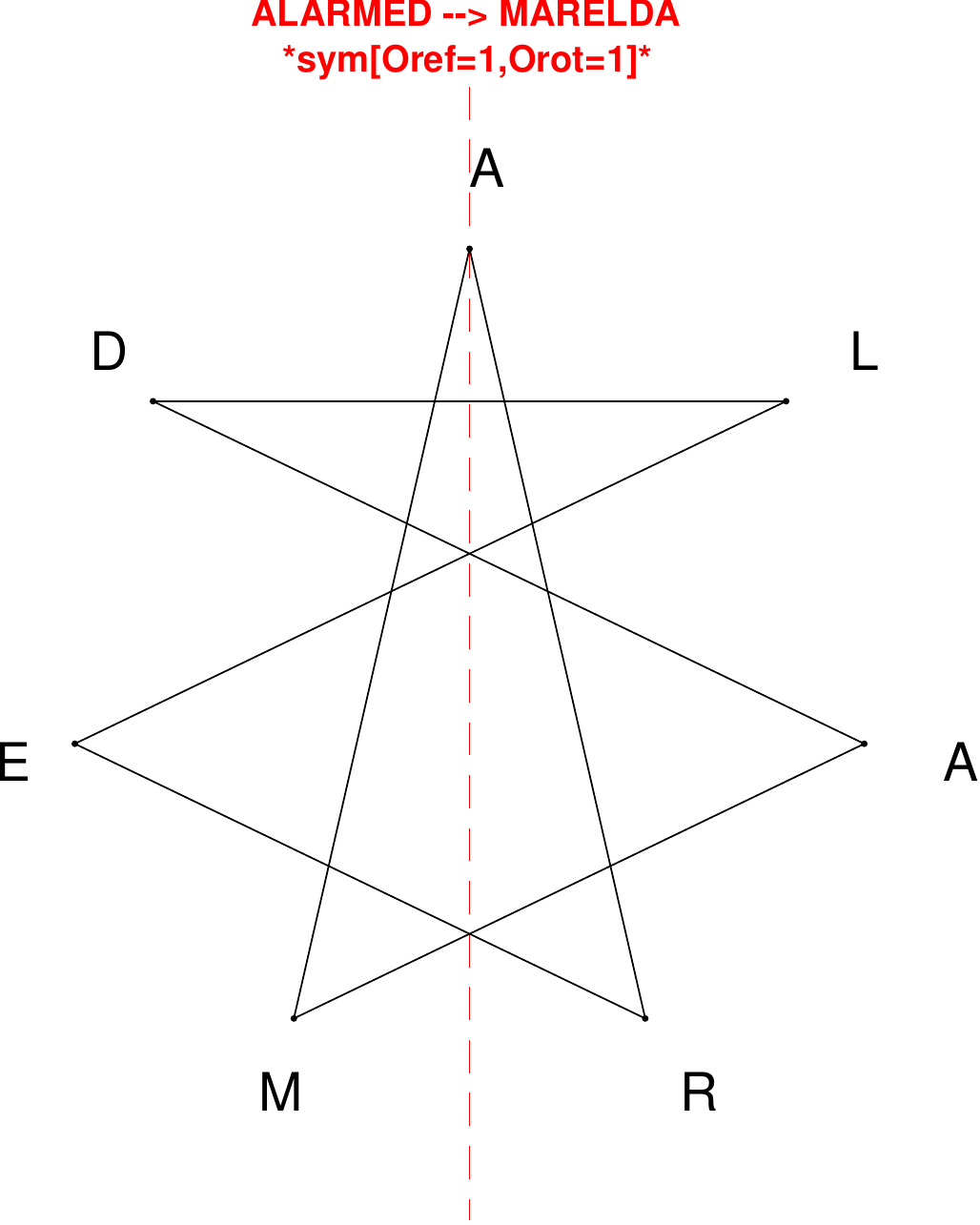}
\end{subfigure}
\hfill
\begin{subfigure}[T]{0.19\textwidth}
\centering
\includegraphics[width=\textwidth]{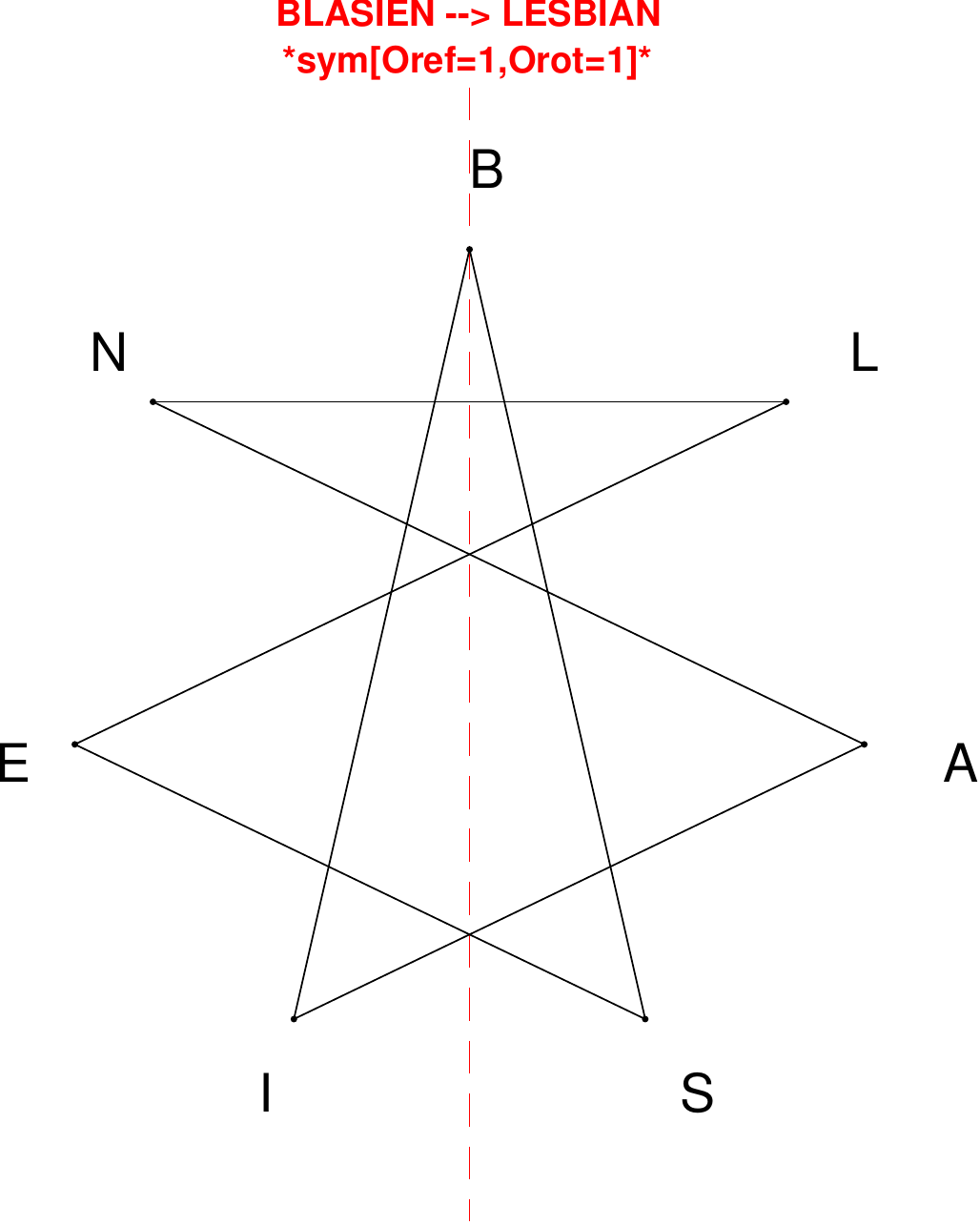}
\end{subfigure}
\end{figure}

\begin{figure}[H]
\centering
\begin{subfigure}[T]{0.19\textwidth}
\centering
\includegraphics[width=\textwidth]{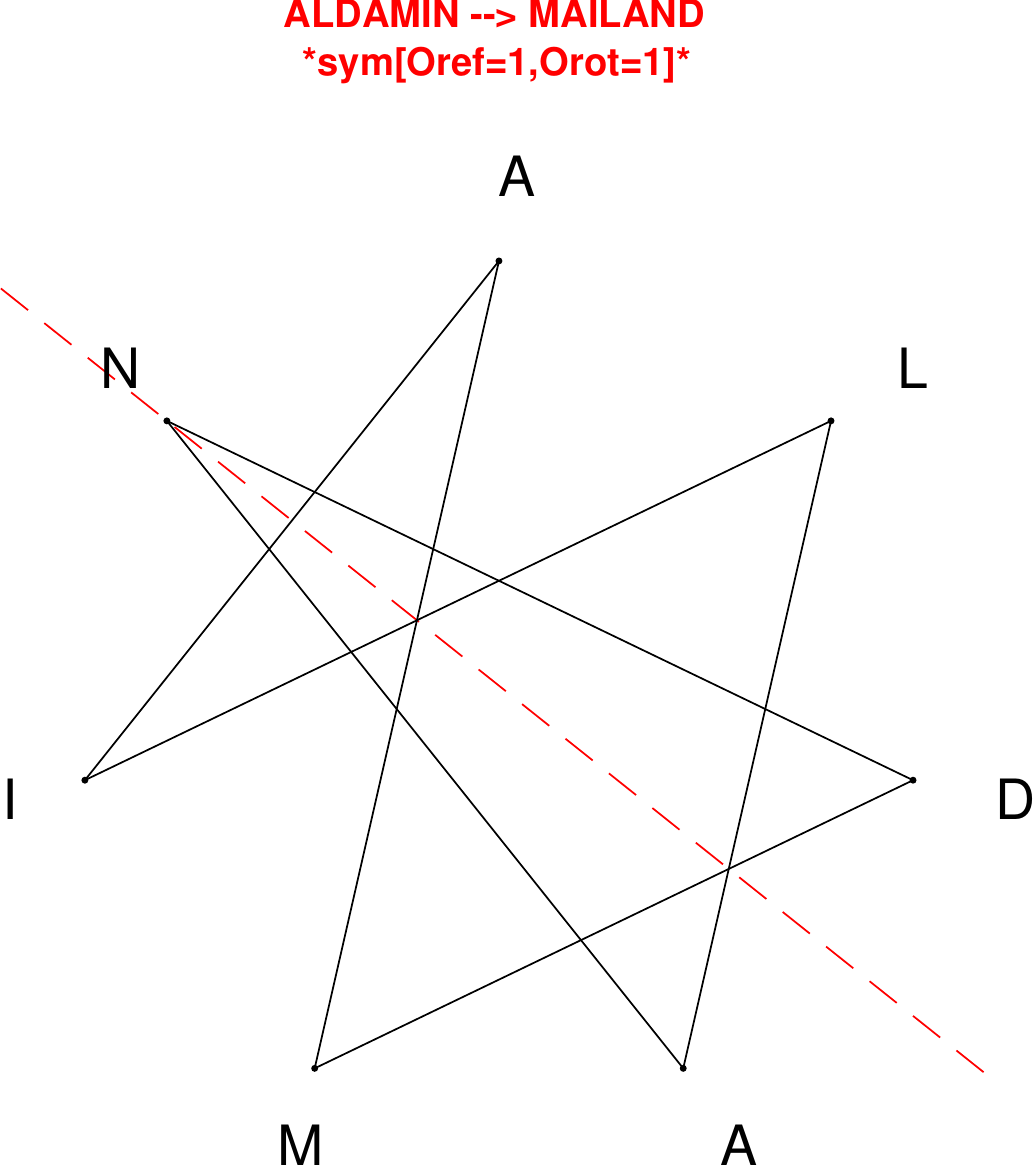}
\end{subfigure}
\hfill
\begin{subfigure}[T]{0.19\textwidth}
\centering
\includegraphics[width=\textwidth]{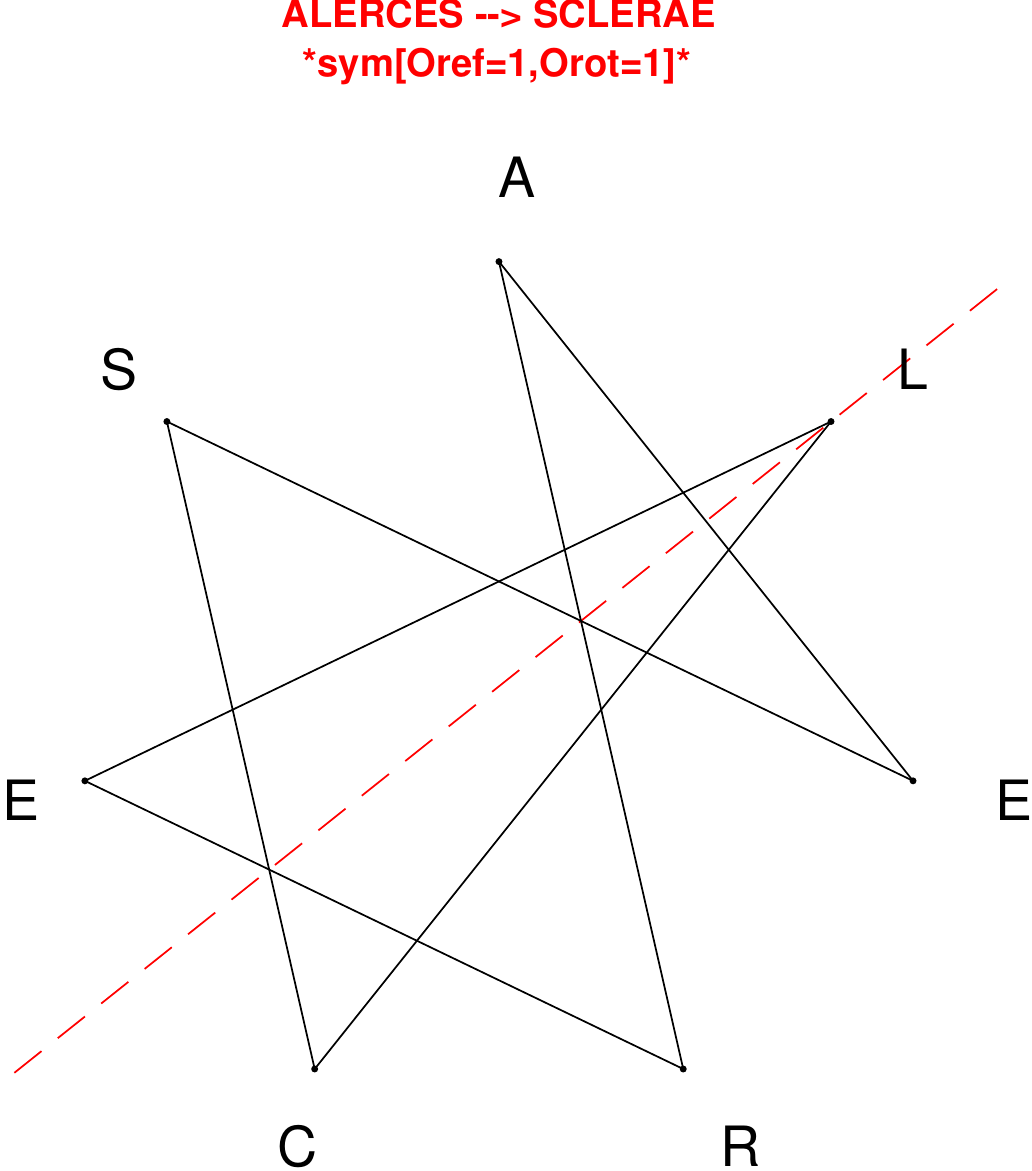}
\end{subfigure}
\hfill
\begin{subfigure}[T]{0.19\textwidth}
\centering
\includegraphics[width=\textwidth]{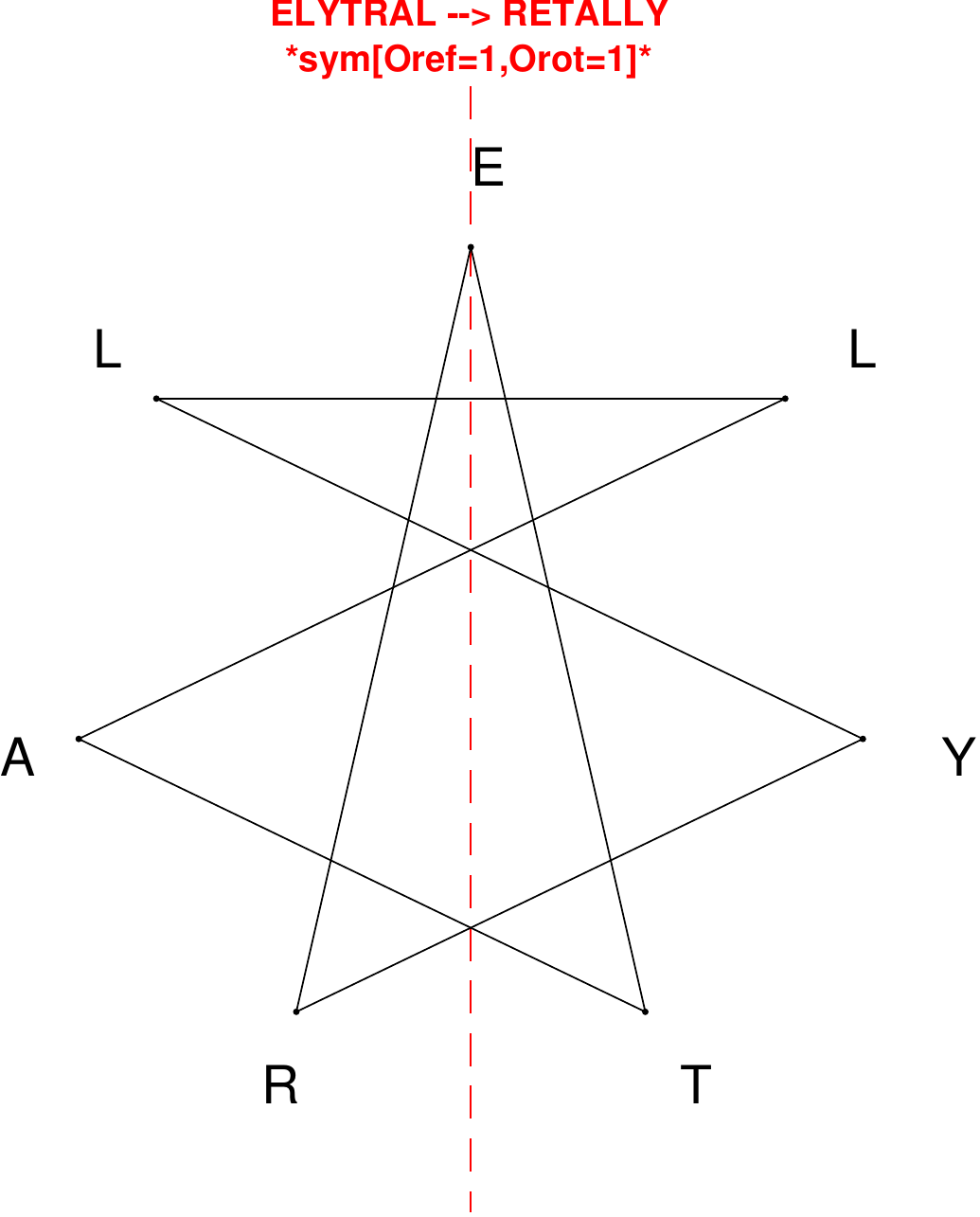}
\end{subfigure}
\hfill
\begin{subfigure}[T]{0.19\textwidth}
\centering
\includegraphics[width=\textwidth]{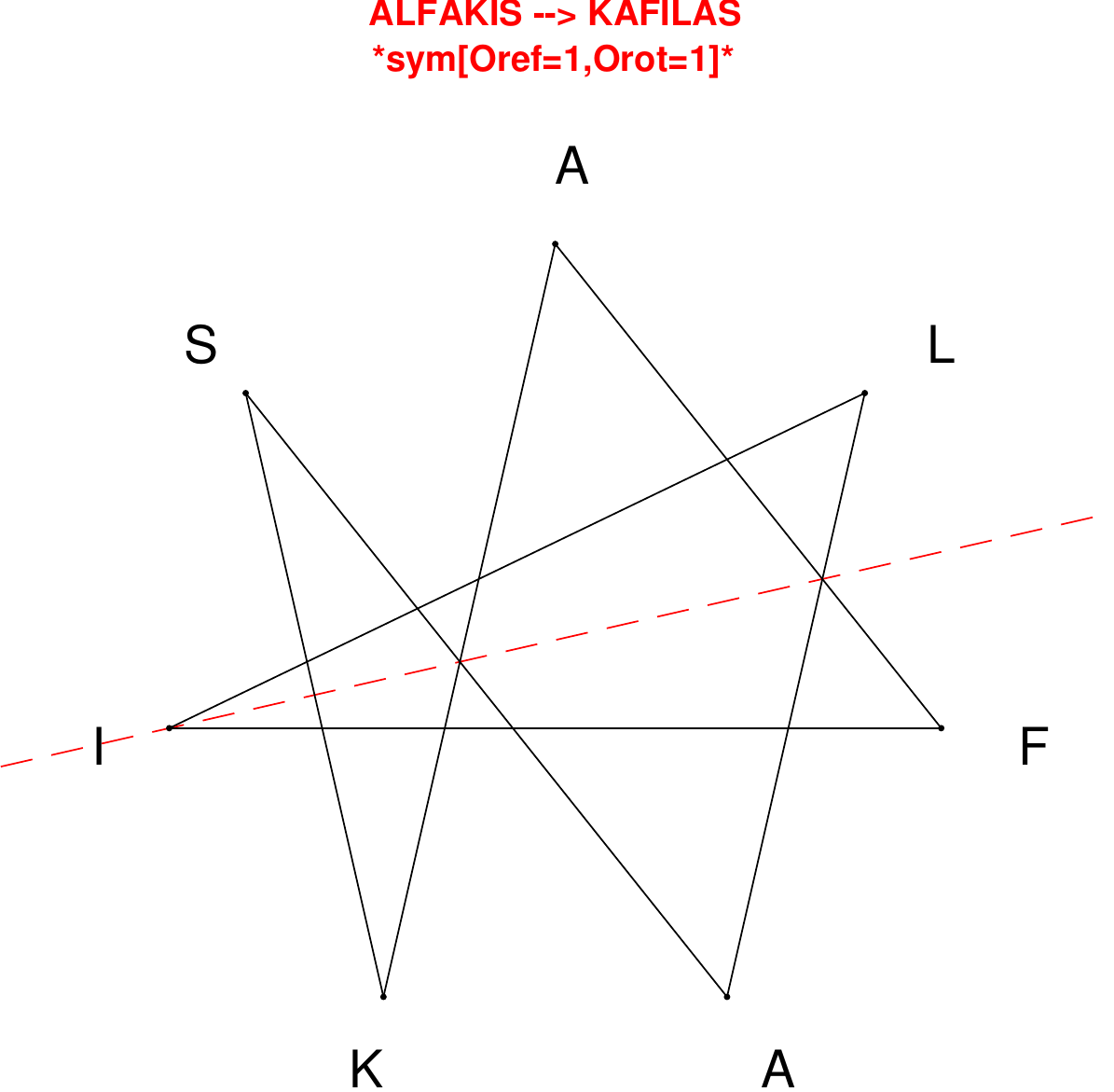}
\end{subfigure}
\hfill
\begin{subfigure}[T]{0.19\textwidth}
\centering
\includegraphics[width=\textwidth]{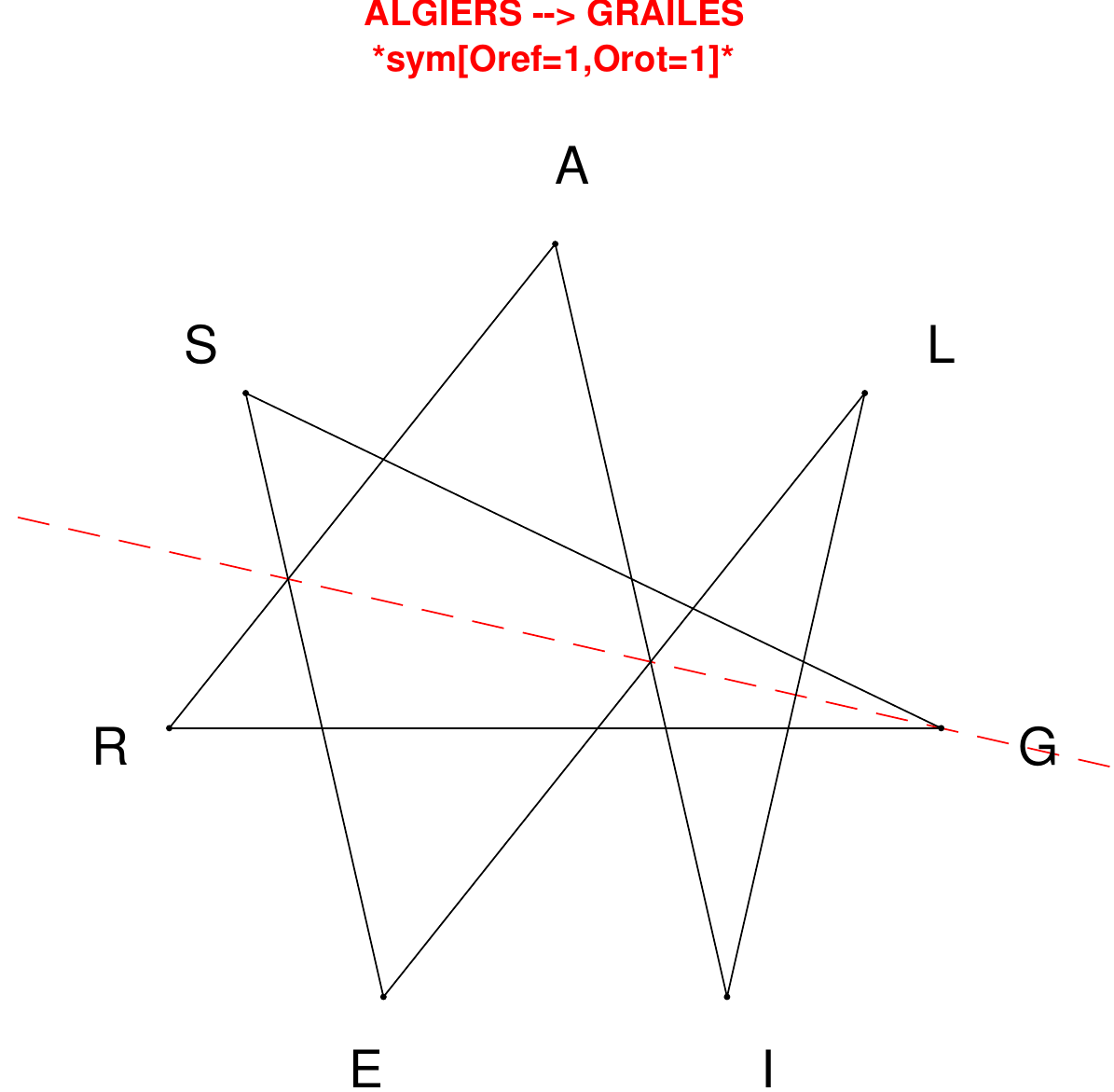}
\end{subfigure}
\end{figure}

\begin{figure}[H]
\centering
\begin{subfigure}[T]{0.19\textwidth}
\centering
\includegraphics[width=\textwidth]{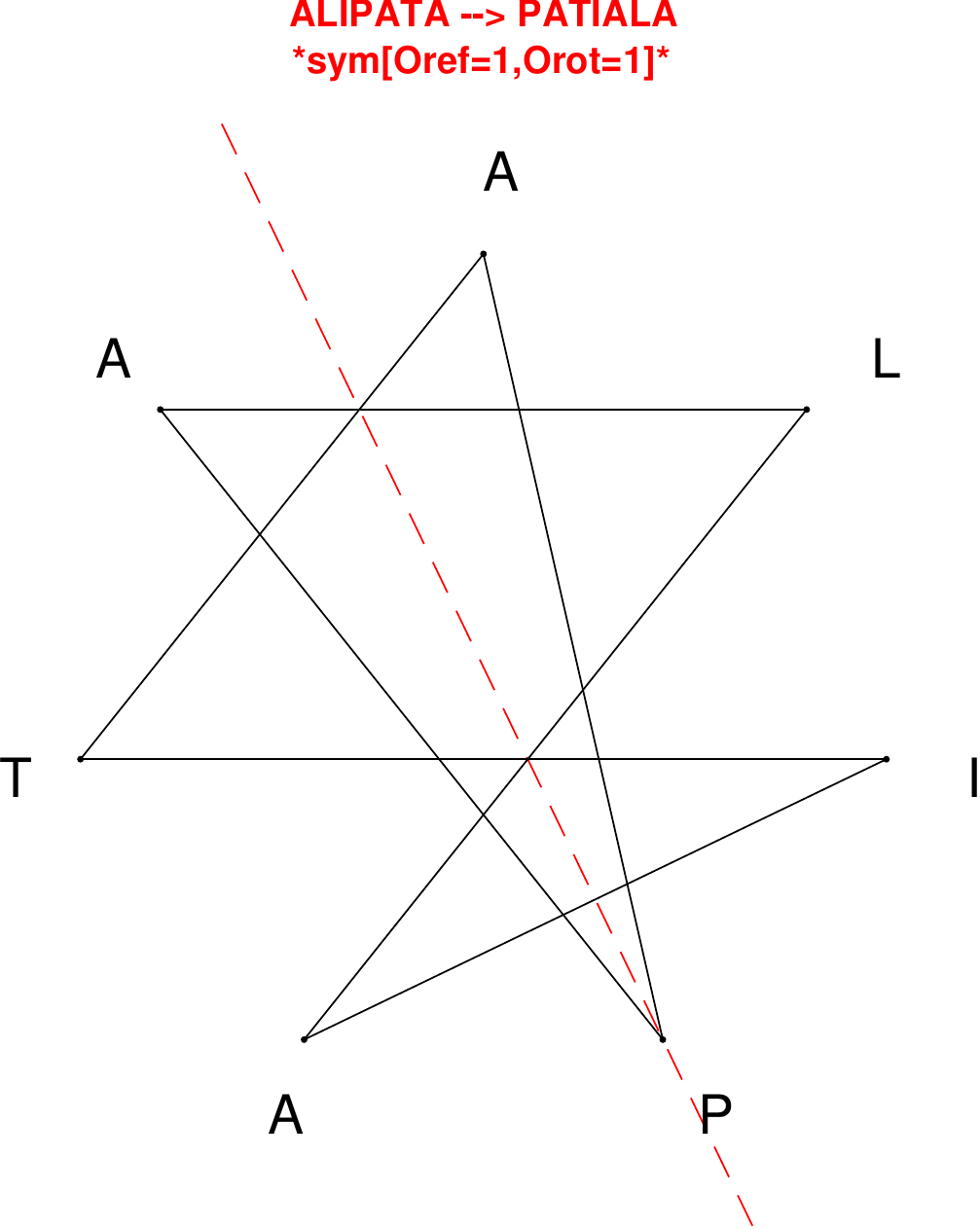}
\end{subfigure}
\hfill
\begin{subfigure}[T]{0.19\textwidth}
\centering
\includegraphics[width=\textwidth]{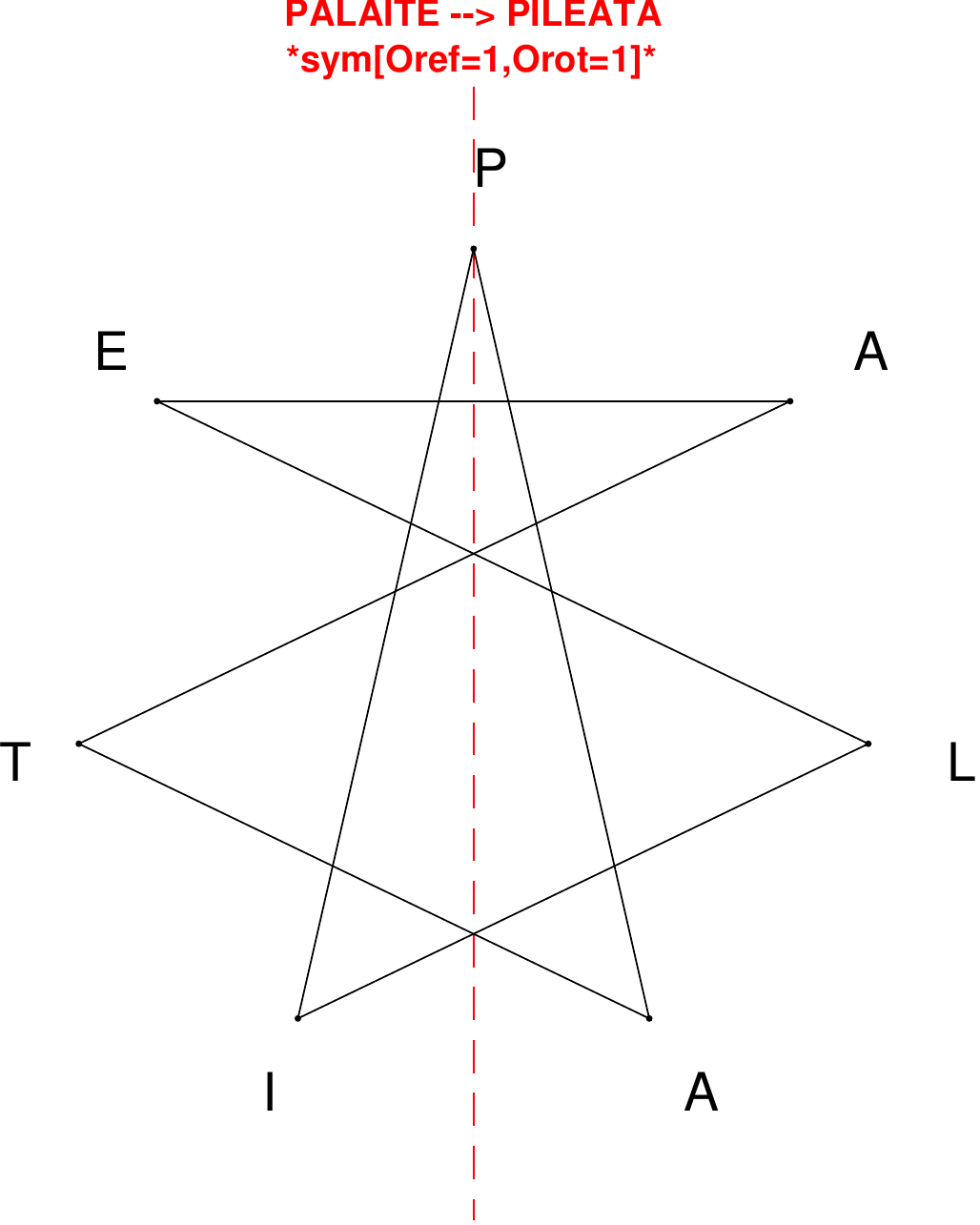}
\end{subfigure}
\hfill
\begin{subfigure}[T]{0.19\textwidth}
\centering
\includegraphics[width=\textwidth]{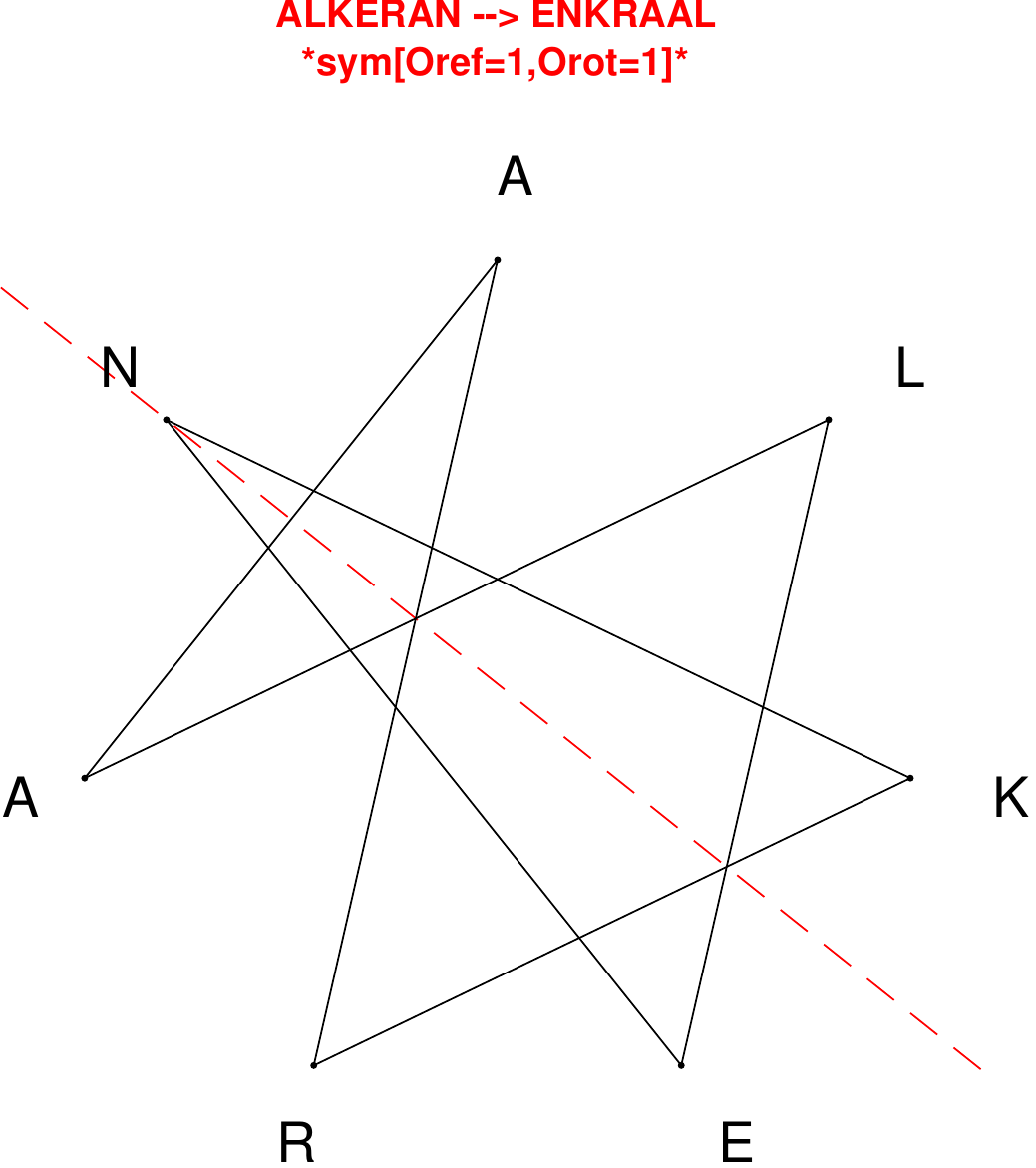}
\end{subfigure}
\hfill
\begin{subfigure}[T]{0.19\textwidth}
\centering
\includegraphics[width=\textwidth]{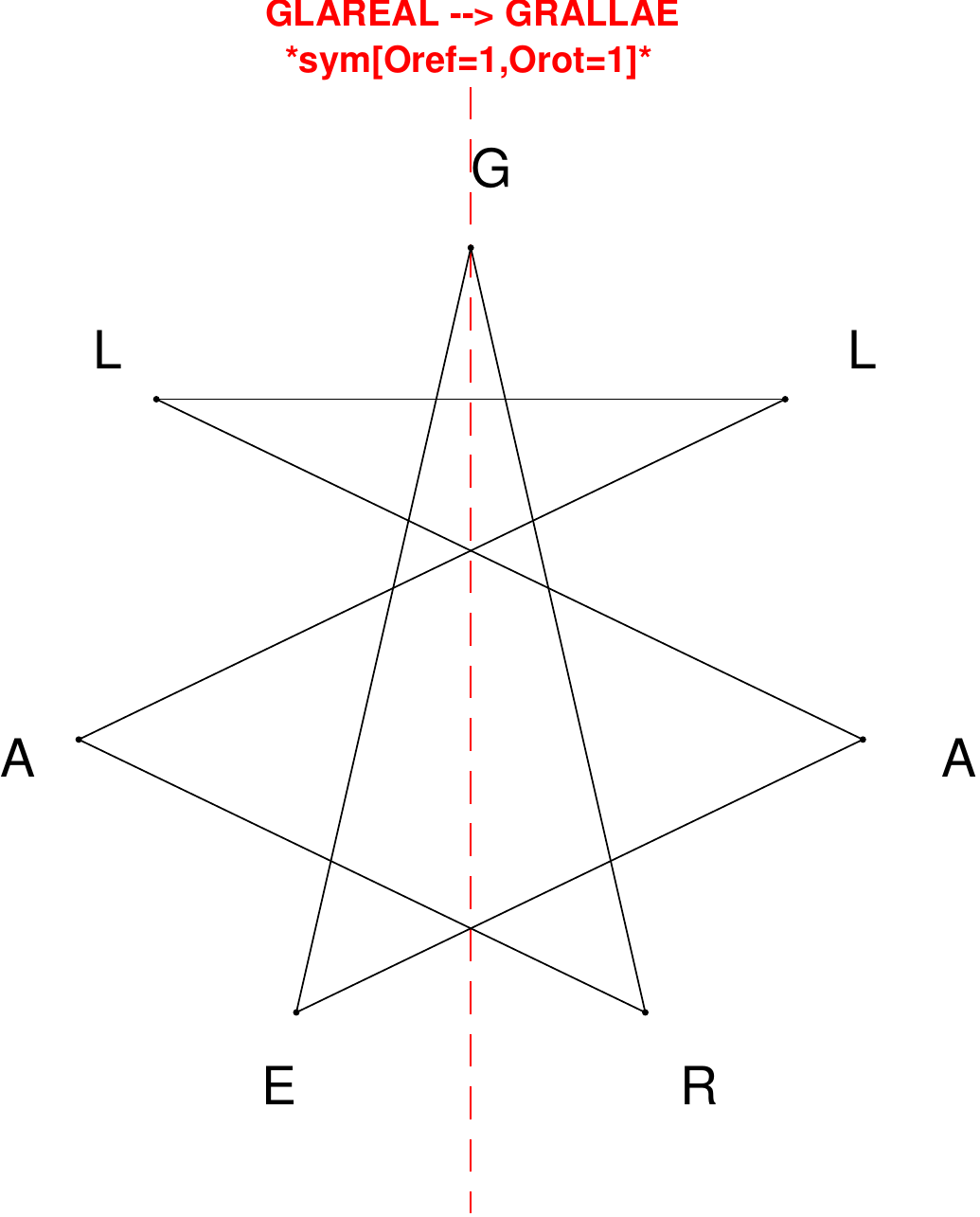}
\end{subfigure}
\hfill
\begin{subfigure}[T]{0.19\textwidth}
\centering
\includegraphics[width=\textwidth]{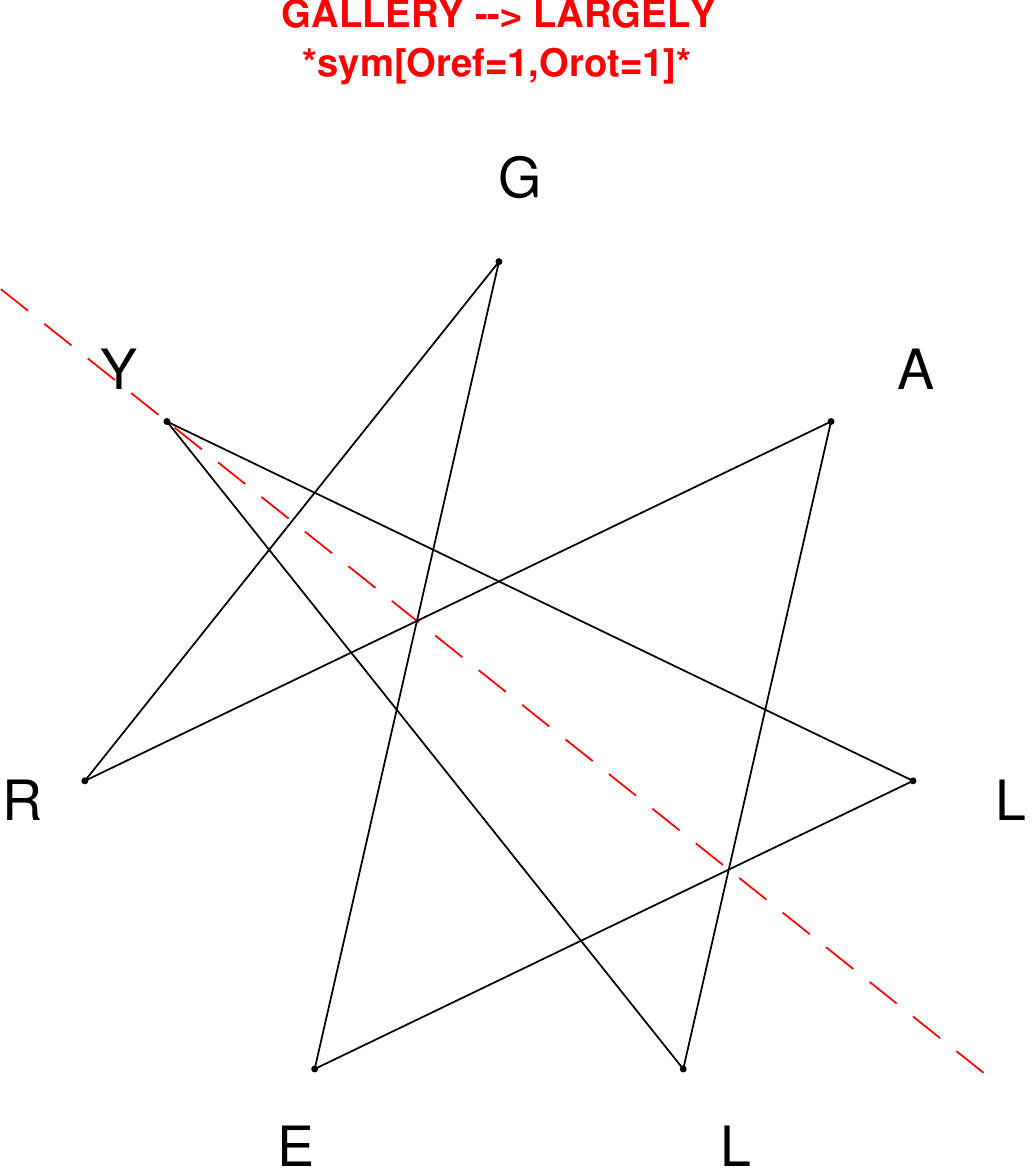}
\end{subfigure}
\end{figure}

\begin{figure}[H]
\centering
\begin{subfigure}[T]{0.19\textwidth}
\centering
\includegraphics[width=\textwidth]{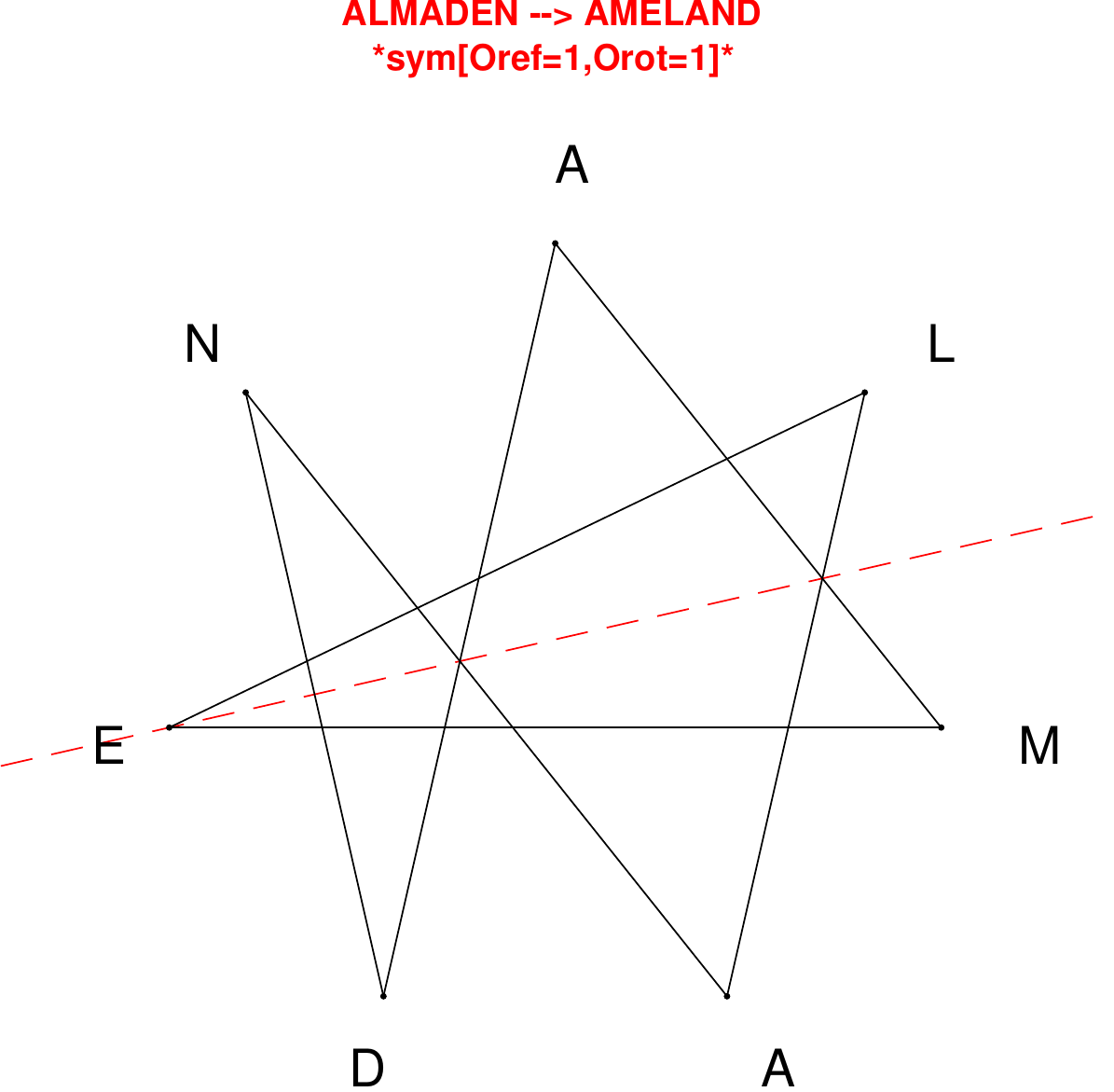}
\end{subfigure}
\hfill
\begin{subfigure}[T]{0.19\textwidth}
\centering
\includegraphics[width=\textwidth]{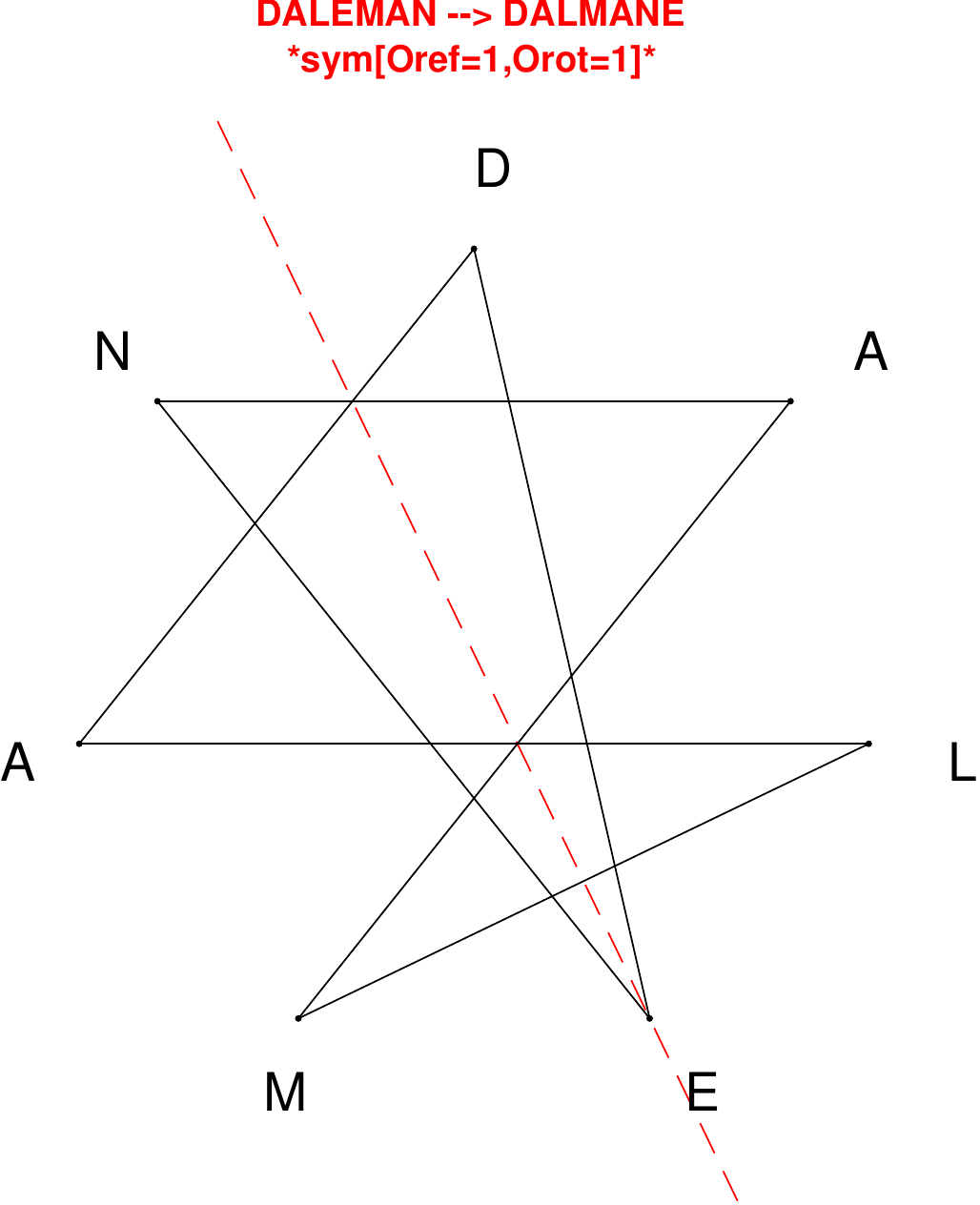}
\end{subfigure}
\hfill
\begin{subfigure}[T]{0.19\textwidth}
\centering
\includegraphics[width=\textwidth]{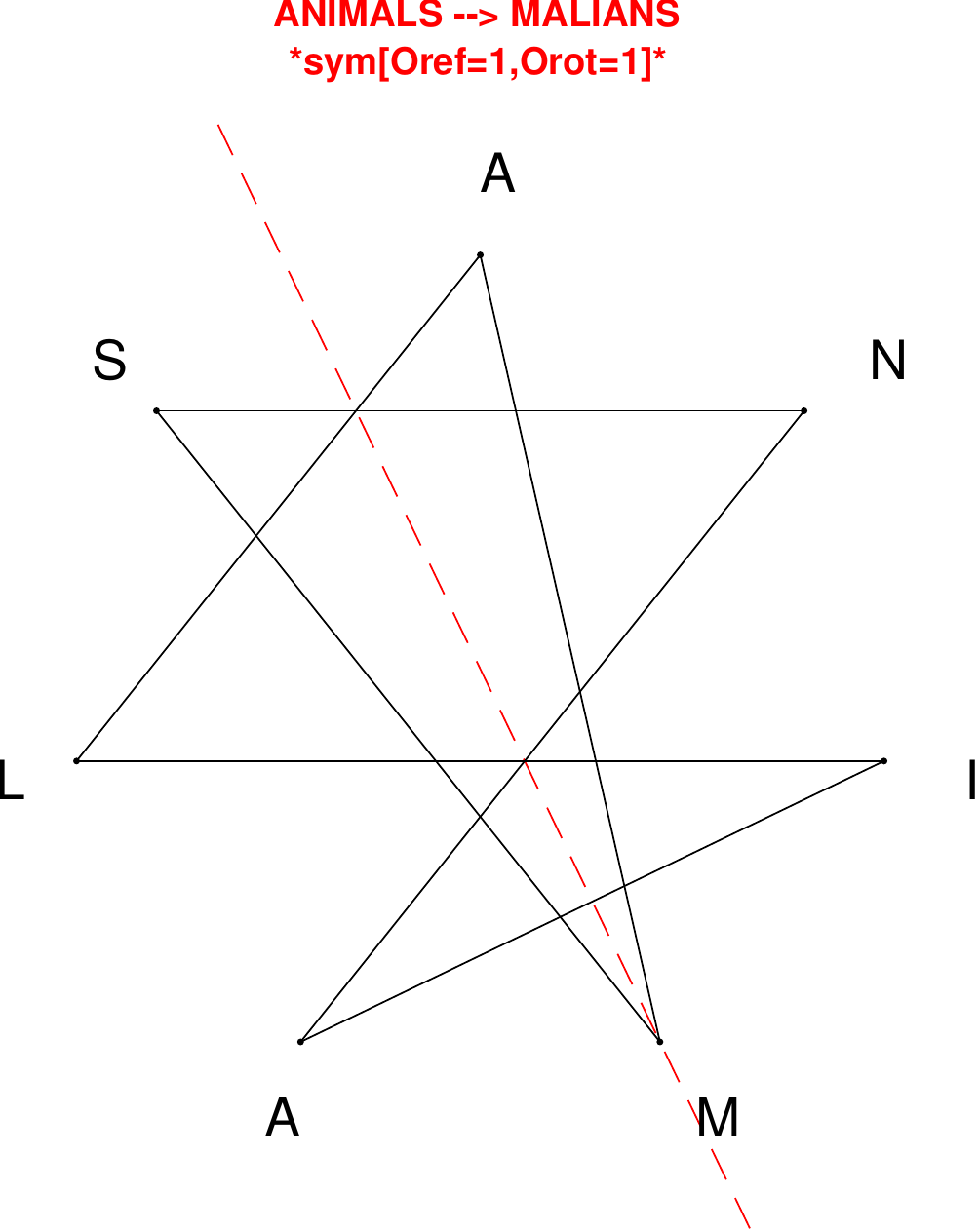}
\end{subfigure}
\hfill
\begin{subfigure}[T]{0.19\textwidth}
\centering
\includegraphics[width=\textwidth]{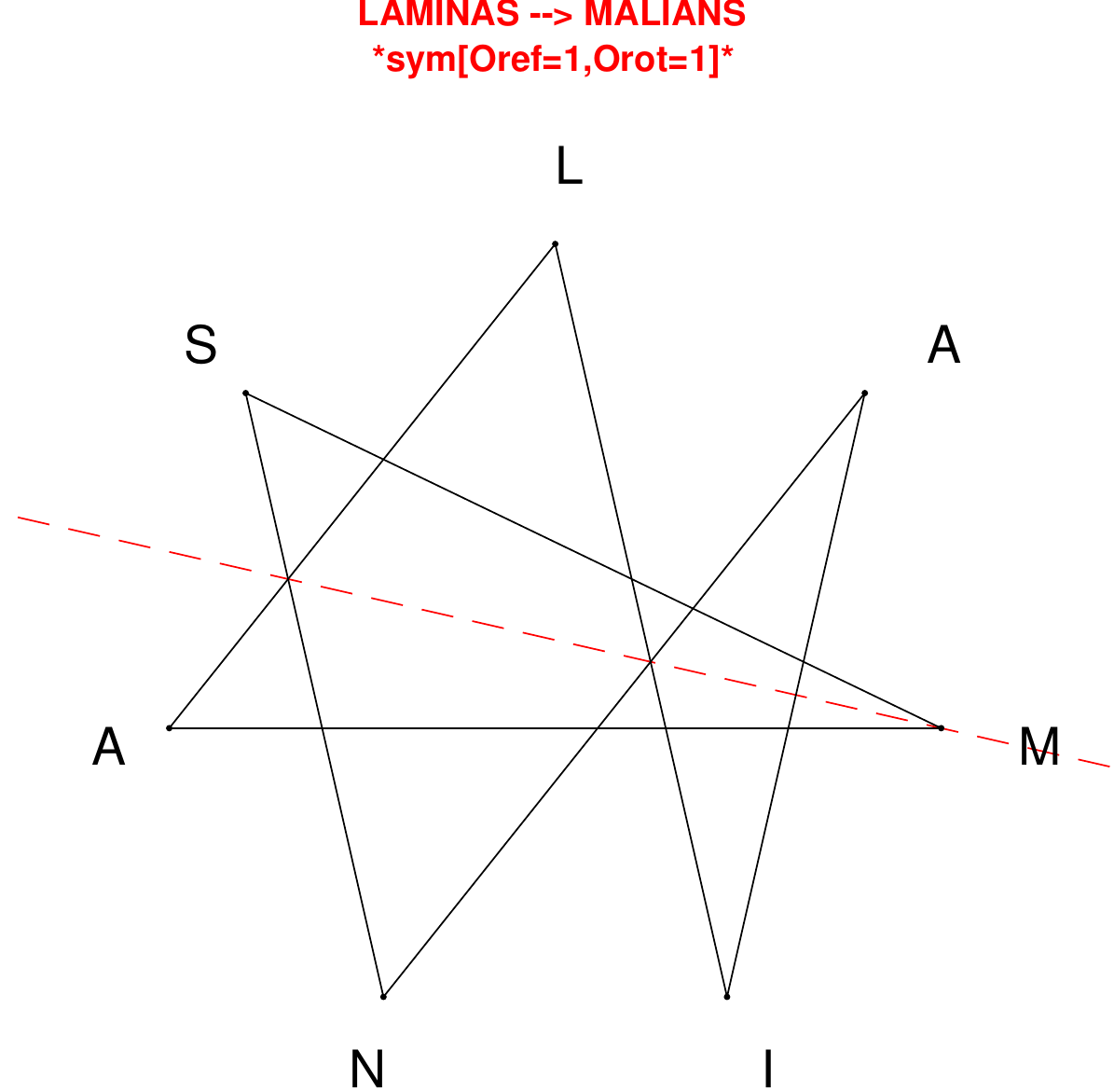}
\end{subfigure}
\hfill
\begin{subfigure}[T]{0.19\textwidth}
\centering
\includegraphics[width=\textwidth]{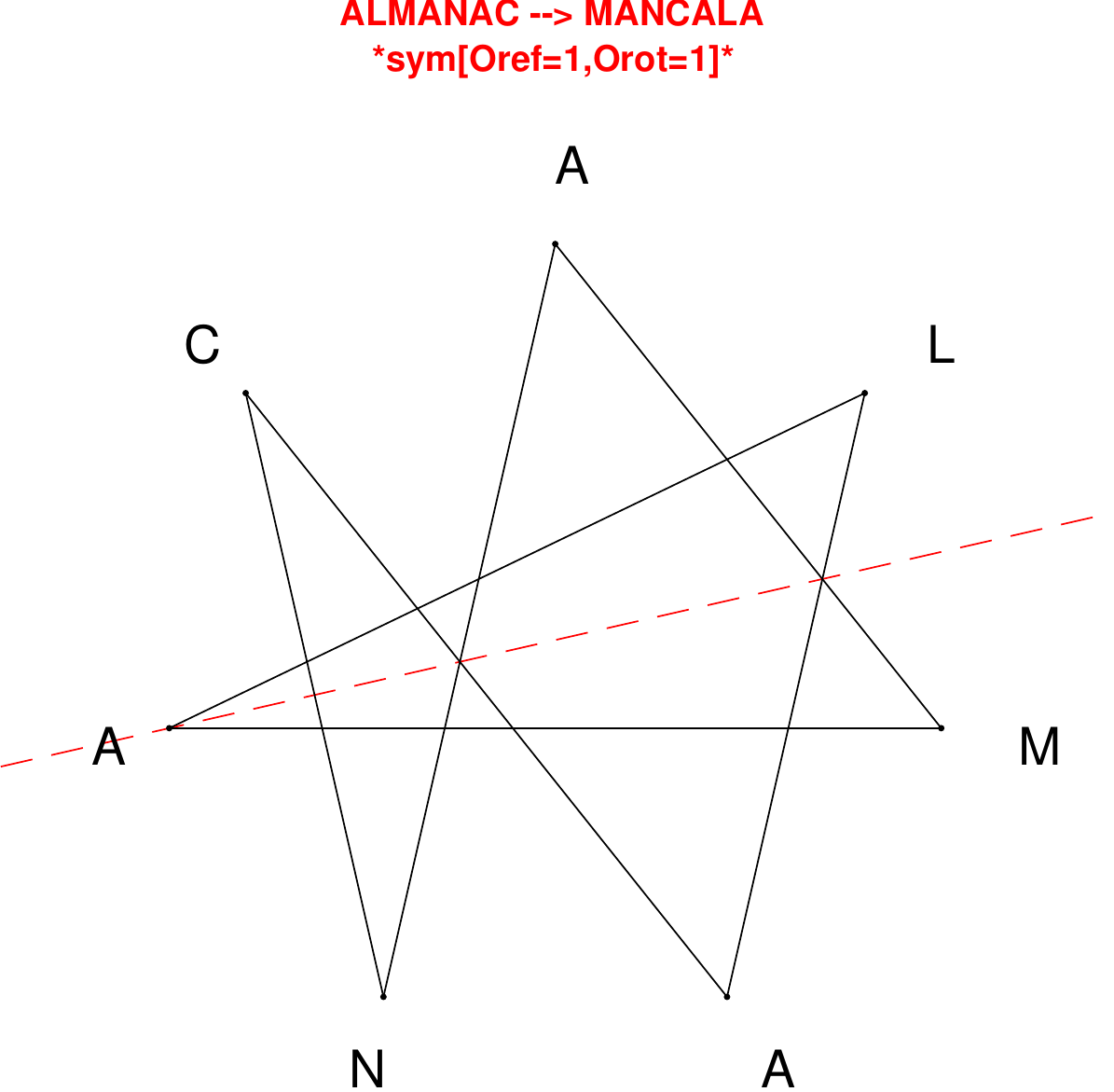}
\end{subfigure}
\end{figure}

\begin{figure}[H]
\centering
\begin{subfigure}[T]{0.19\textwidth}
\centering
\includegraphics[width=\textwidth]{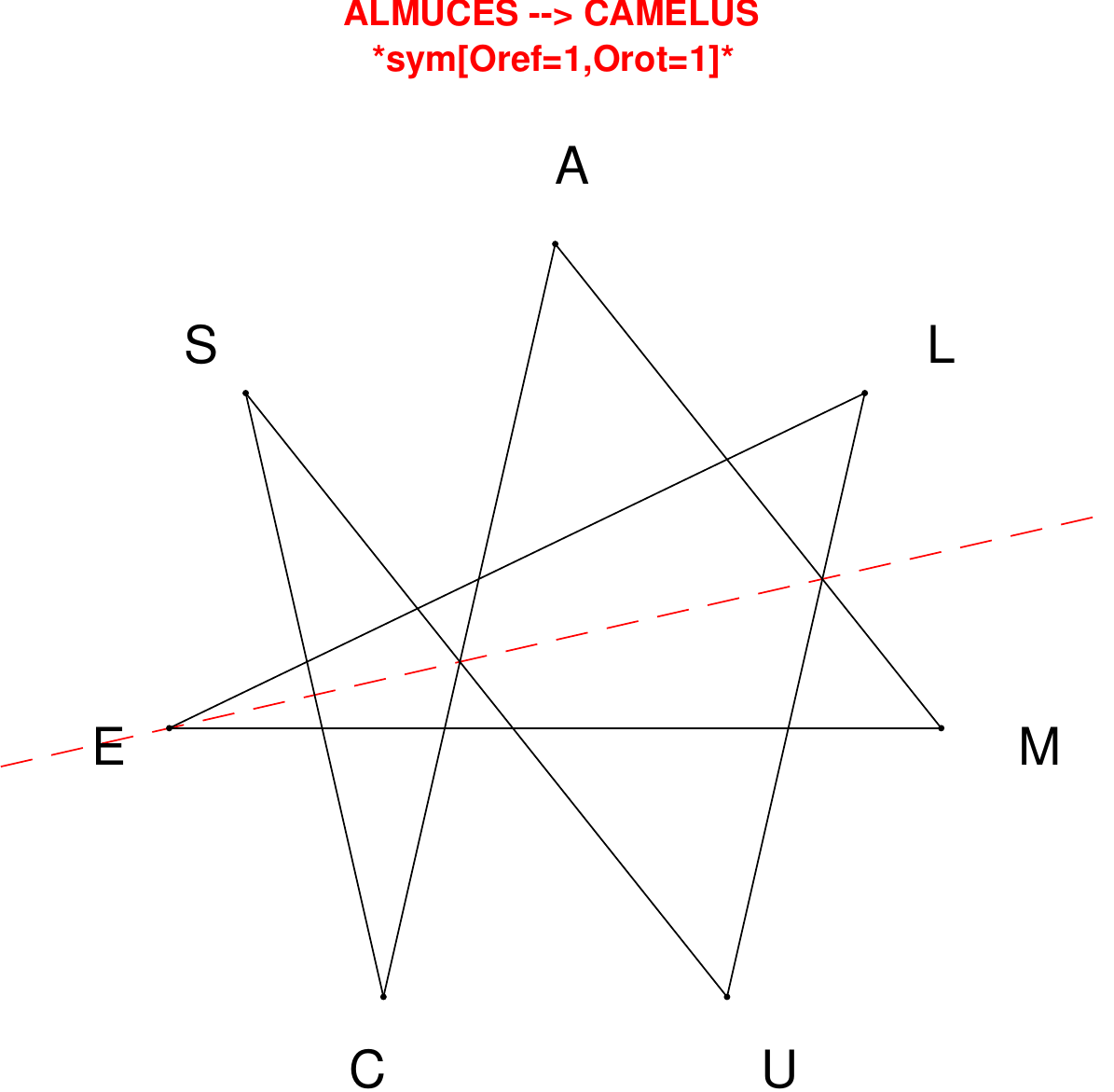}
\end{subfigure}
\hfill
\begin{subfigure}[T]{0.19\textwidth}
\centering
\includegraphics[width=\textwidth]{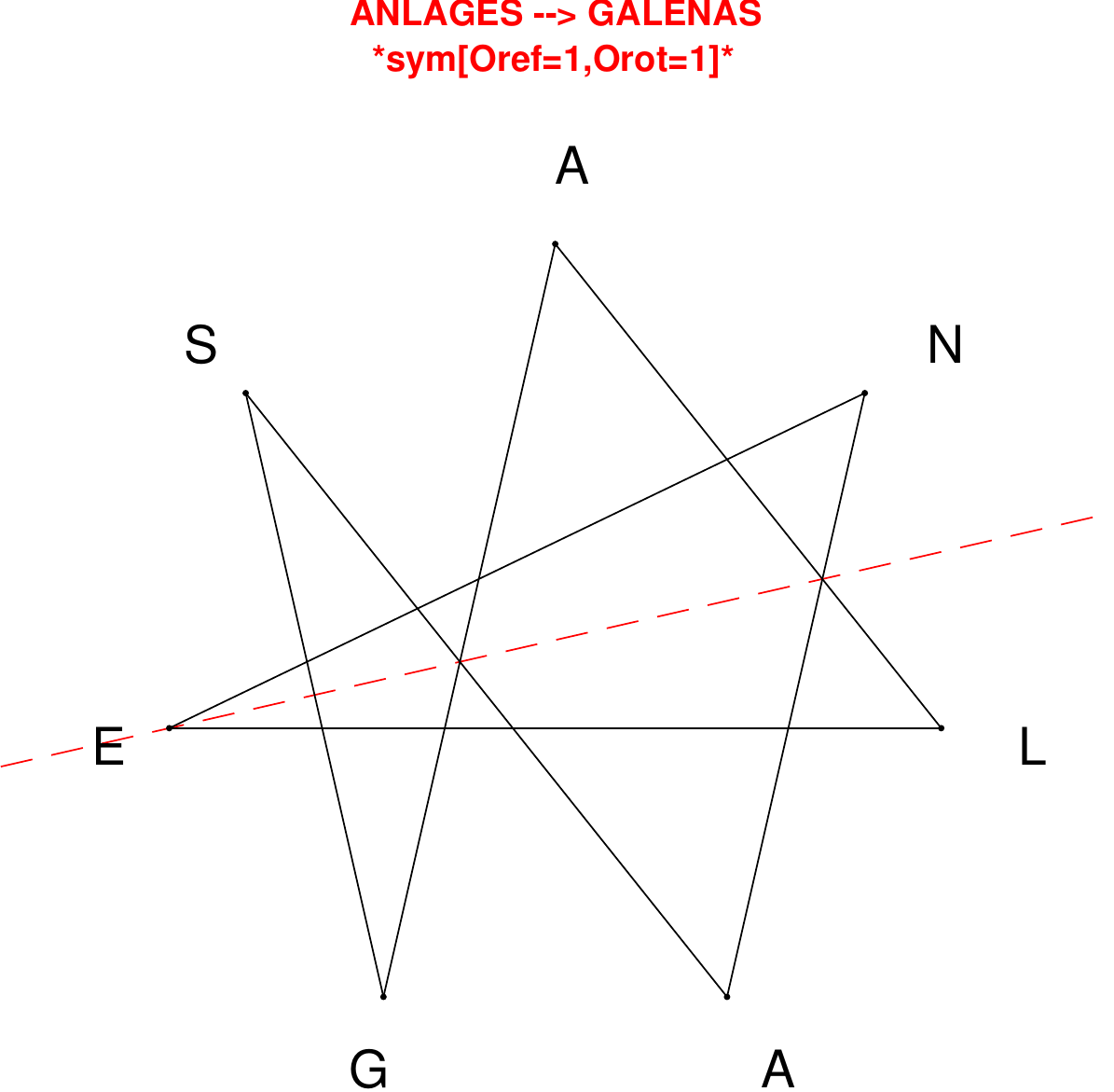}
\end{subfigure}
\hfill
\begin{subfigure}[T]{0.19\textwidth}
\centering
\includegraphics[width=\textwidth]{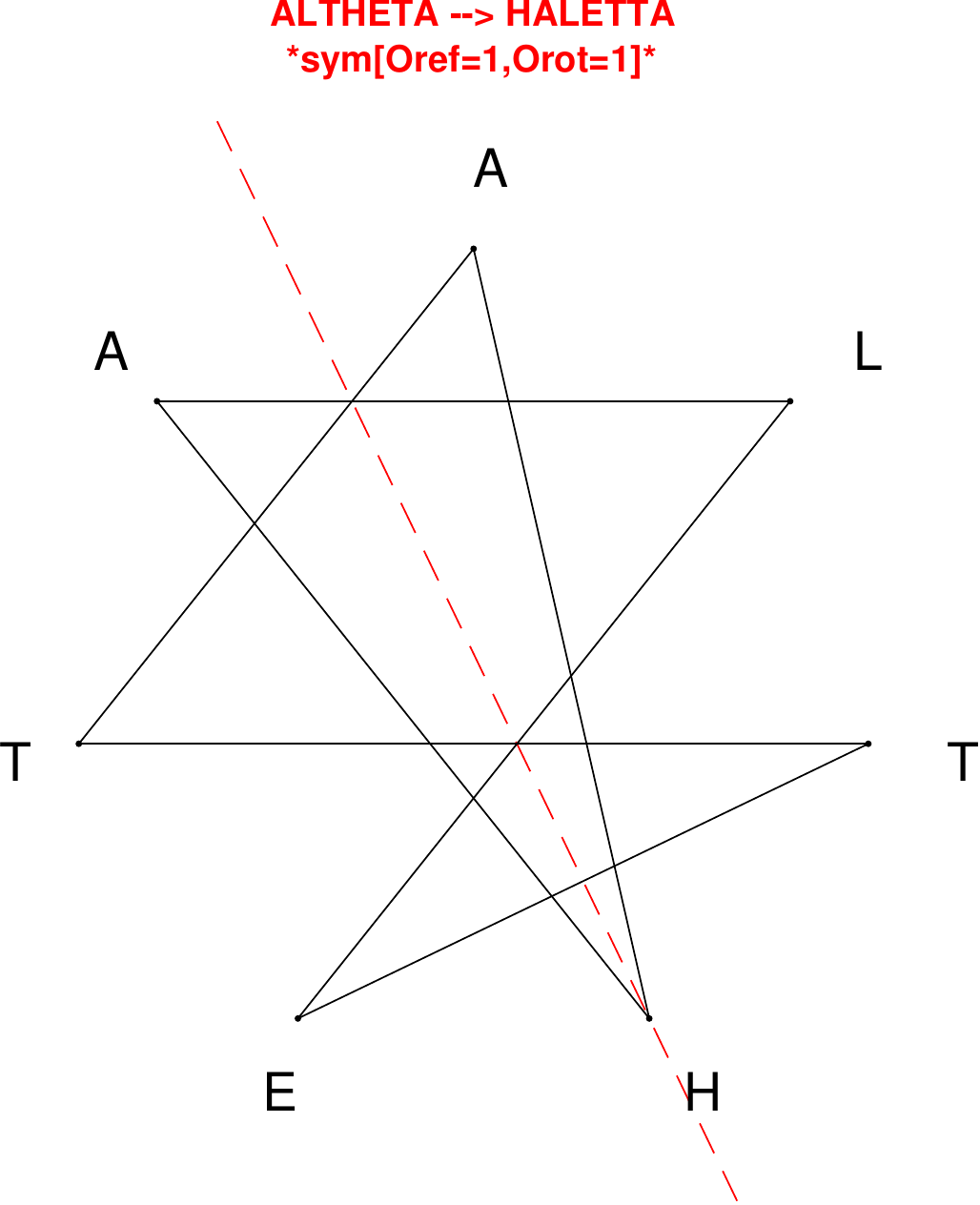}
\end{subfigure}
\hfill
\begin{subfigure}[T]{0.19\textwidth}
\centering
\includegraphics[width=\textwidth]{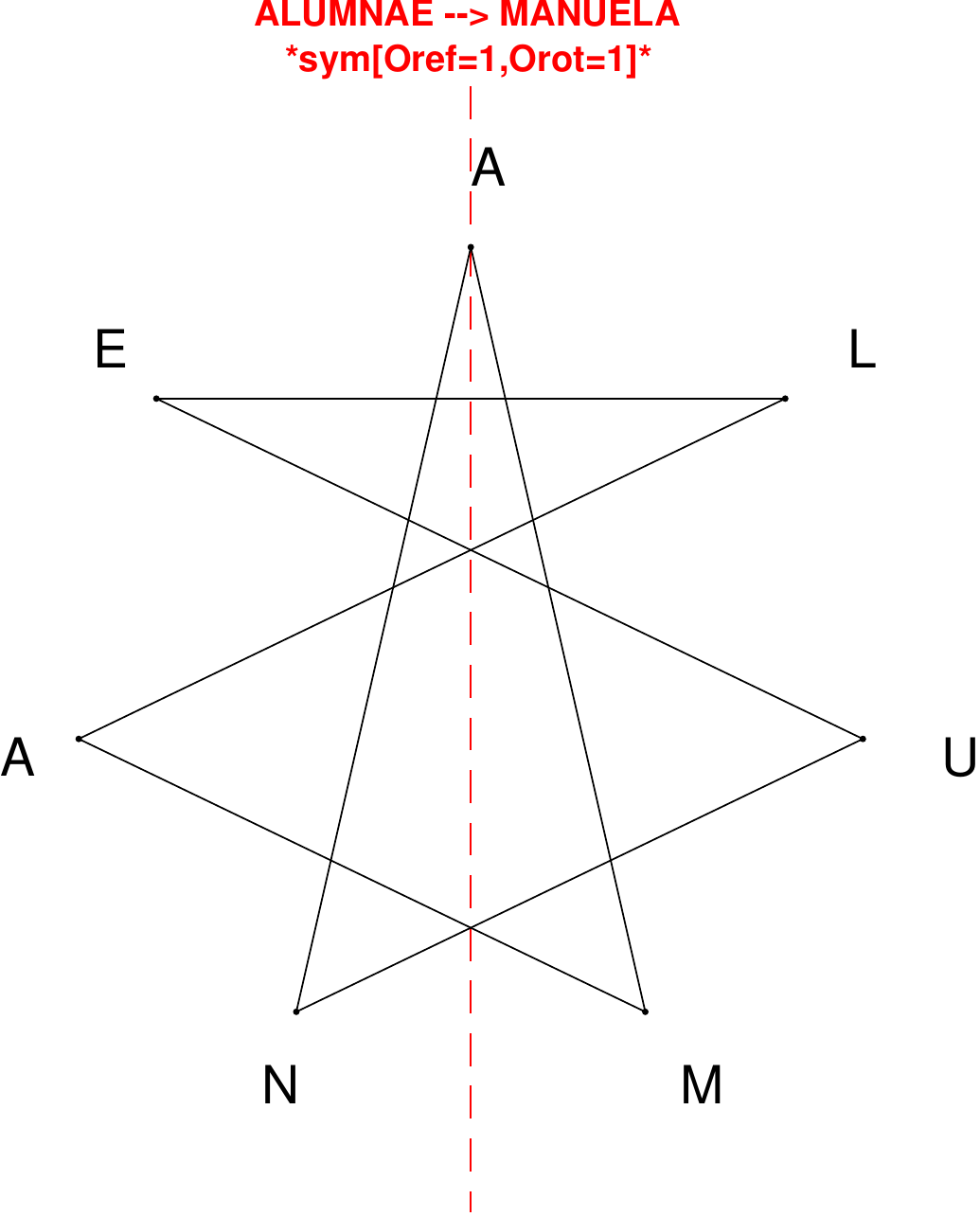}
\end{subfigure}
\hfill
\begin{subfigure}[T]{0.19\textwidth}
\centering
\includegraphics[width=\textwidth]{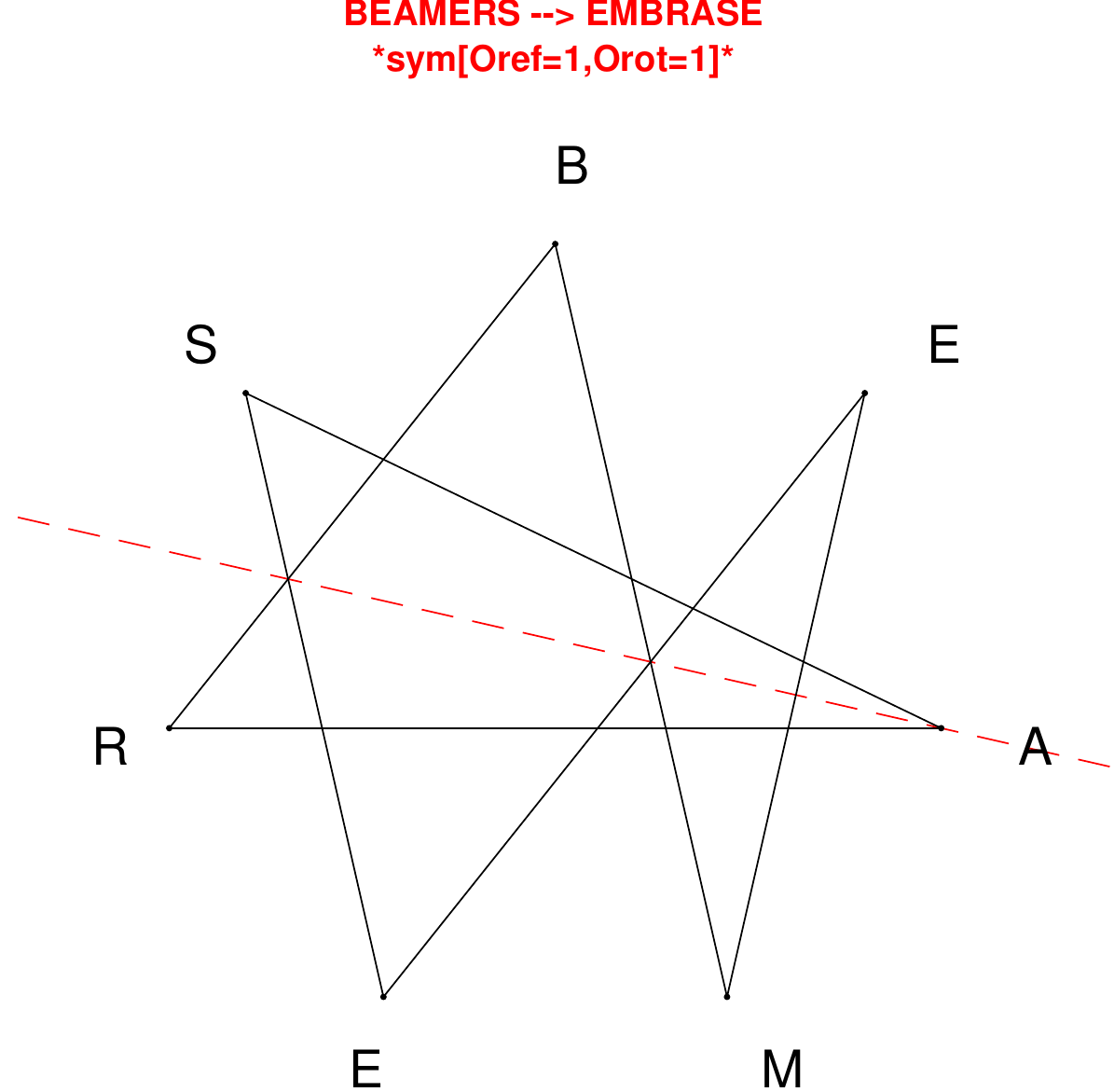}
\end{subfigure}
\end{figure}

\begin{figure}[H]
\centering
\begin{subfigure}[T]{0.19\textwidth}
\centering
\includegraphics[width=\textwidth]{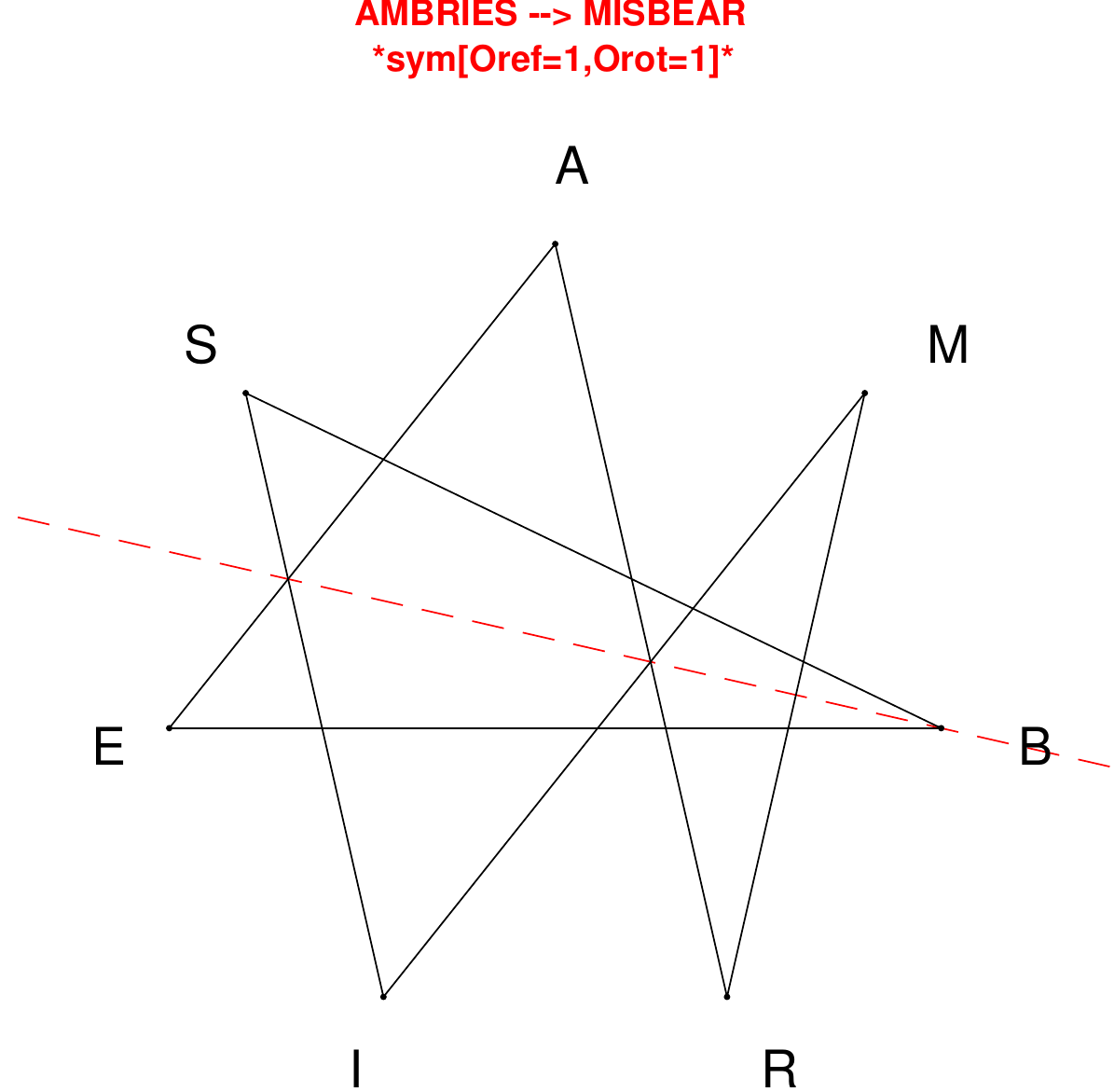}
\end{subfigure}
\hfill
\begin{subfigure}[T]{0.19\textwidth}
\centering
\includegraphics[width=\textwidth]{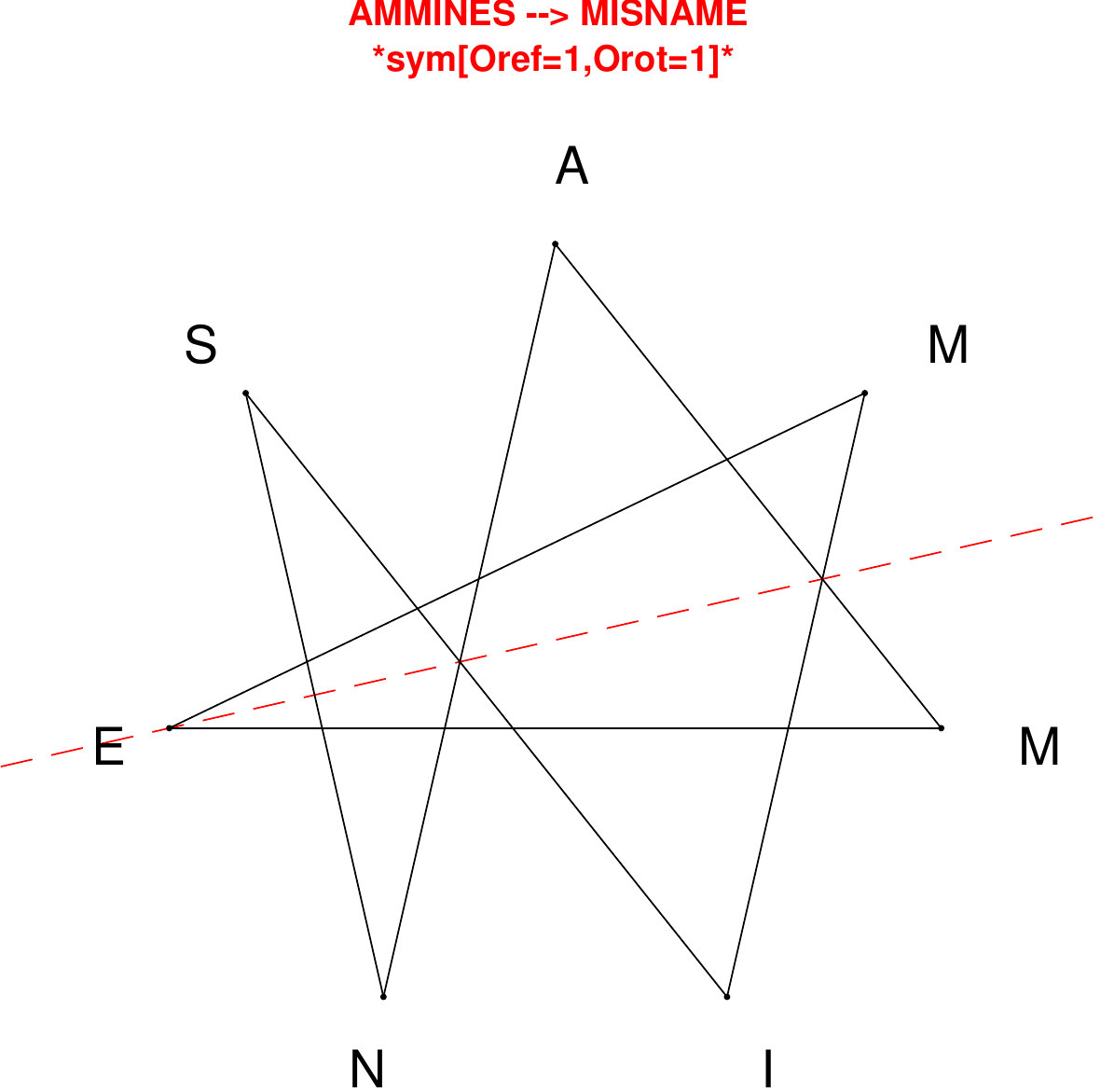}
\end{subfigure}
\hfill
\begin{subfigure}[T]{0.19\textwidth}
\centering
\includegraphics[width=\textwidth]{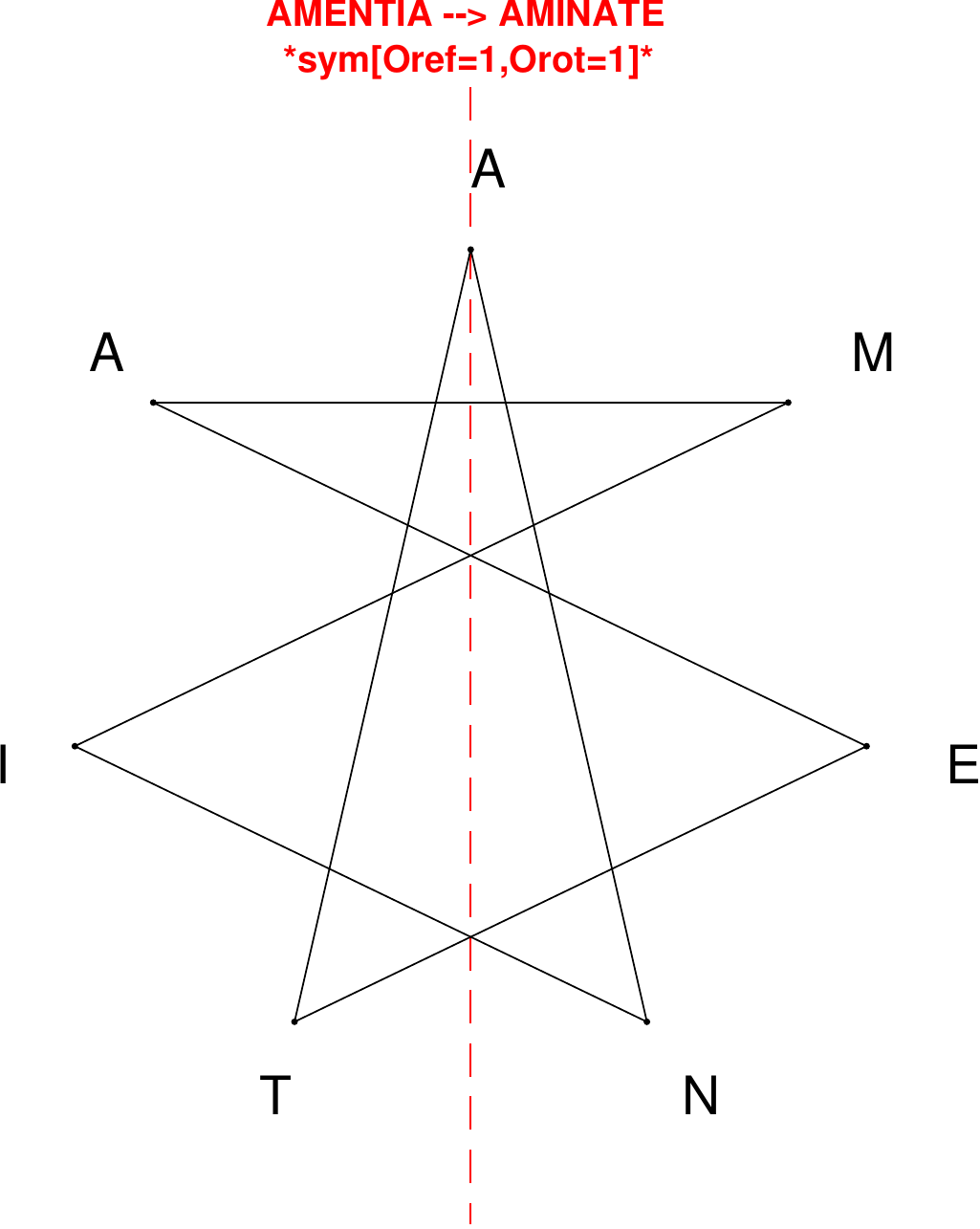}
\end{subfigure}
\hfill
\begin{subfigure}[T]{0.19\textwidth}
\centering
\includegraphics[width=\textwidth]{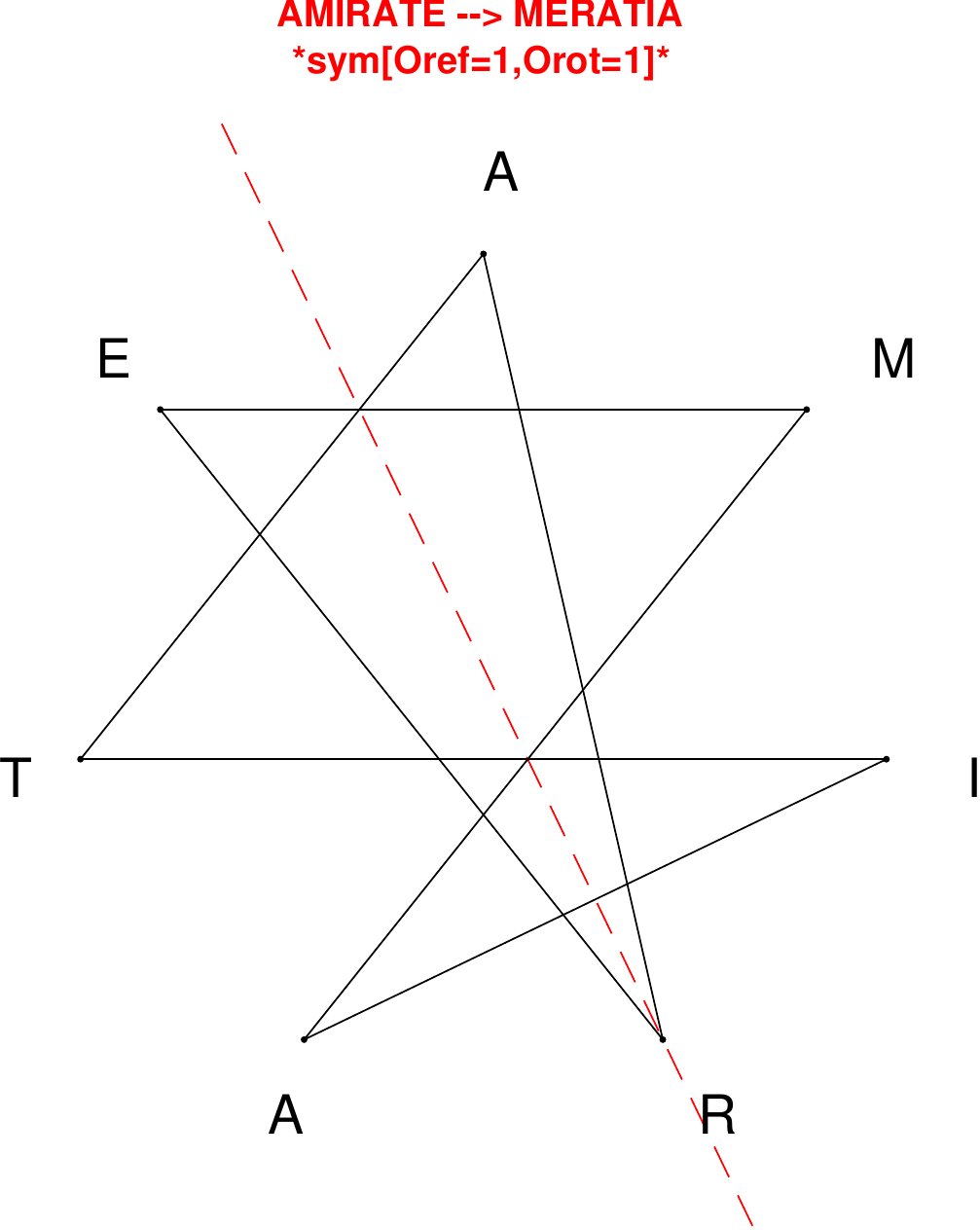}
\end{subfigure}
\hfill
\begin{subfigure}[T]{0.19\textwidth}
\centering
\includegraphics[width=\textwidth]{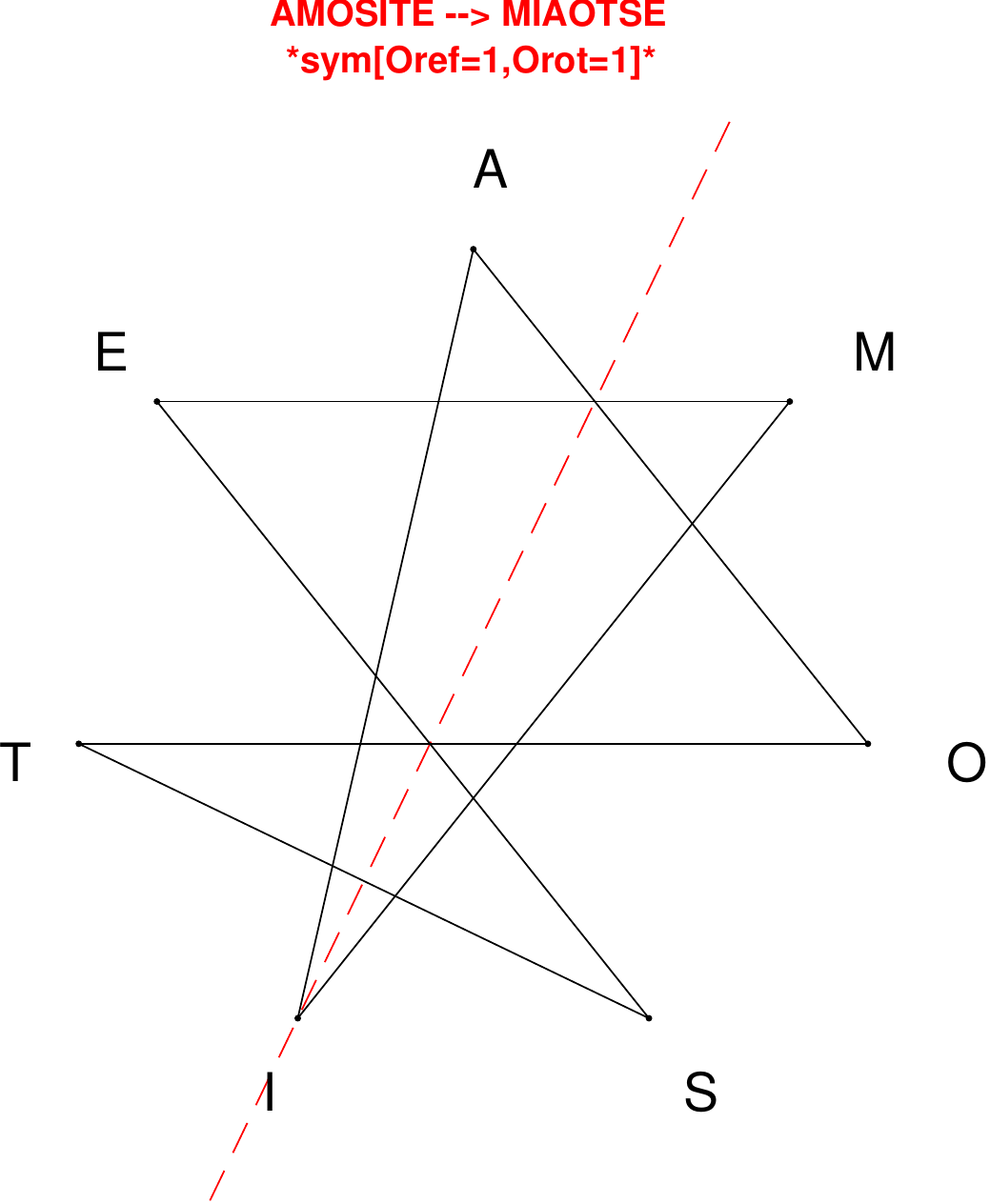}
\end{subfigure}
\end{figure}

\begin{figure}[H]
\centering
\begin{subfigure}[T]{0.19\textwidth}
\centering
\includegraphics[width=\textwidth]{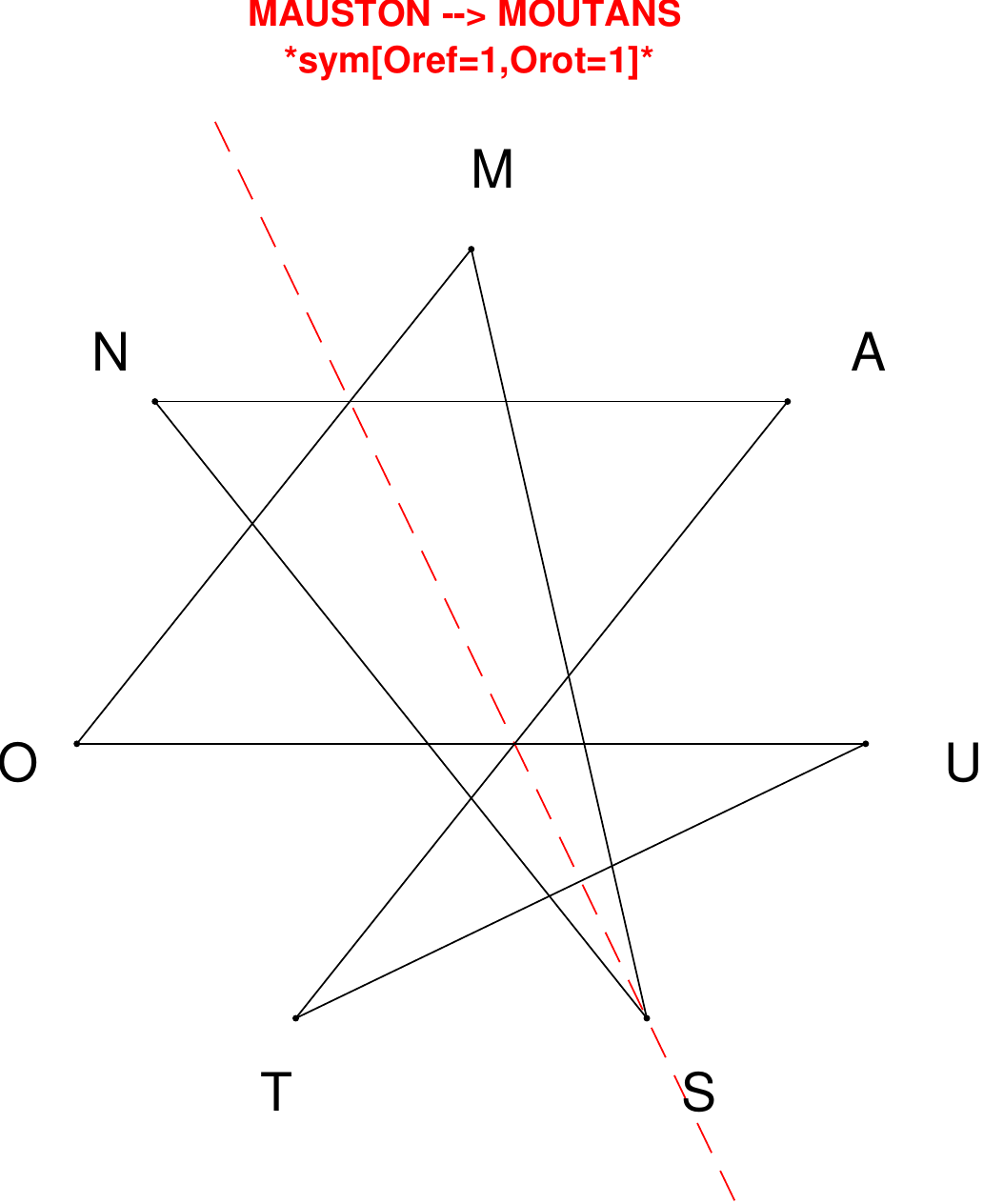}
\end{subfigure}
\hfill
\begin{subfigure}[T]{0.19\textwidth}
\centering
\includegraphics[width=\textwidth]{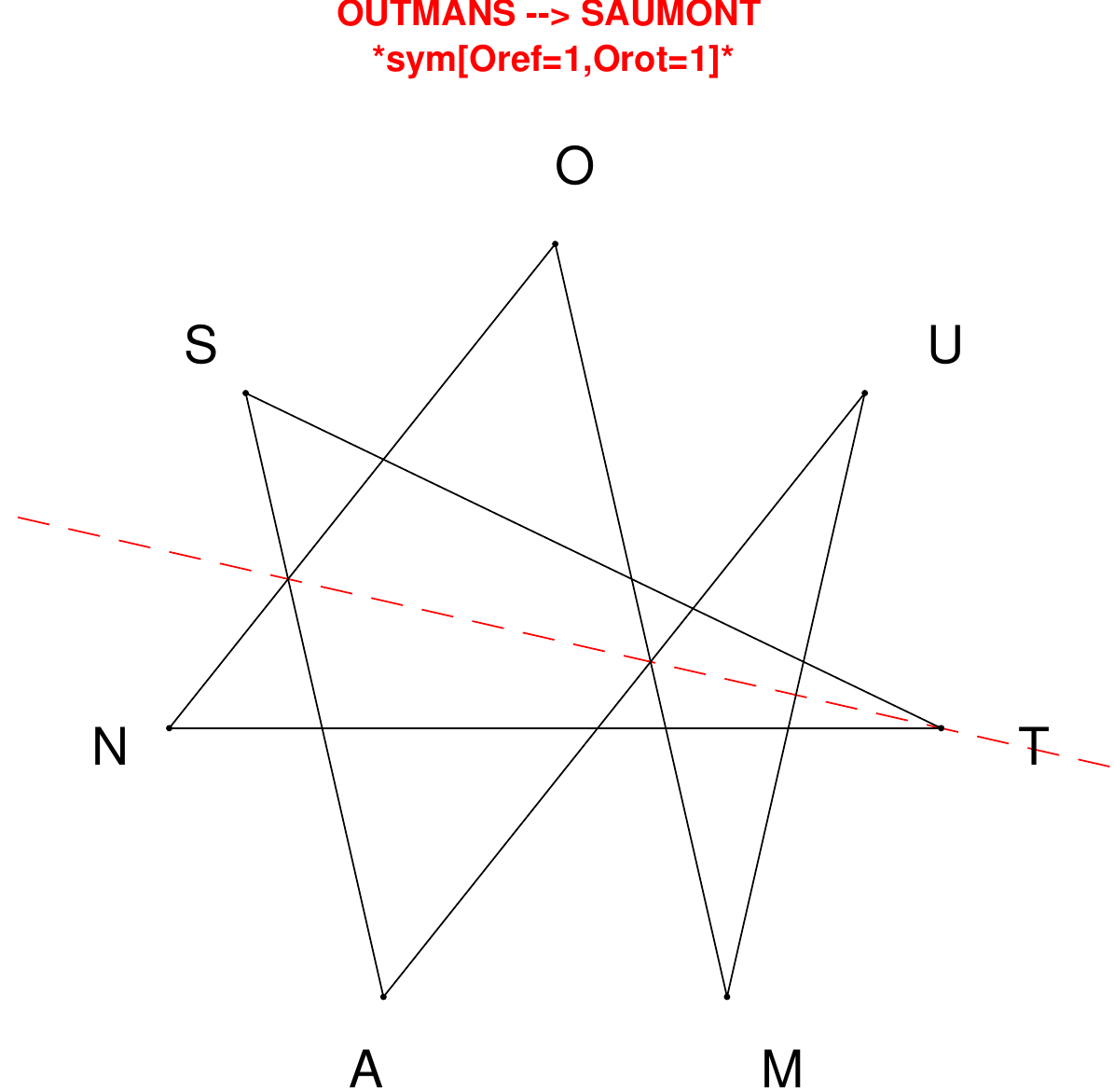}
\end{subfigure}
\hfill
\begin{subfigure}[T]{0.19\textwidth}
\centering
\includegraphics[width=\textwidth]{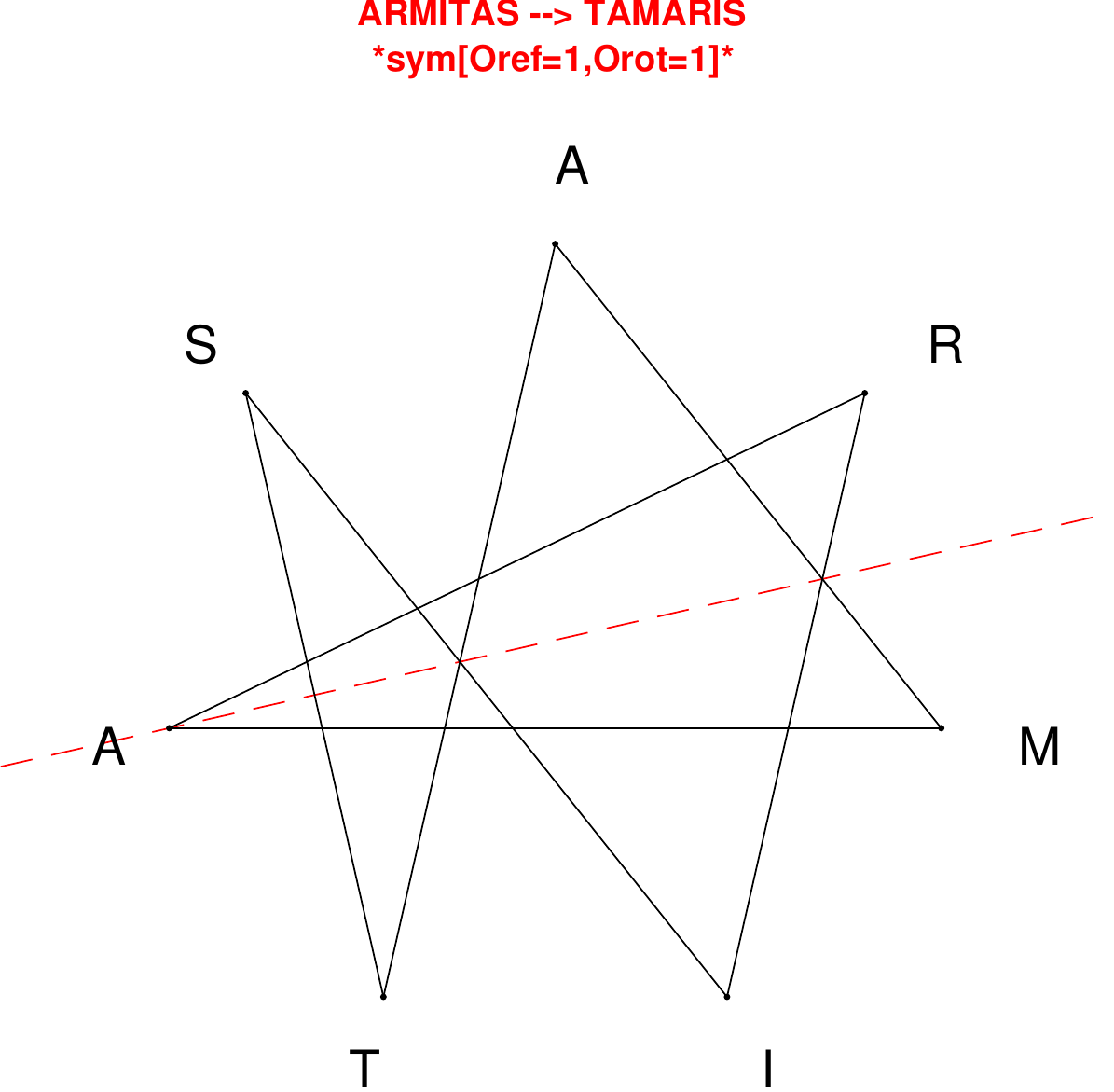}
\end{subfigure}
\hfill
\begin{subfigure}[T]{0.19\textwidth}
\centering
\includegraphics[width=\textwidth]{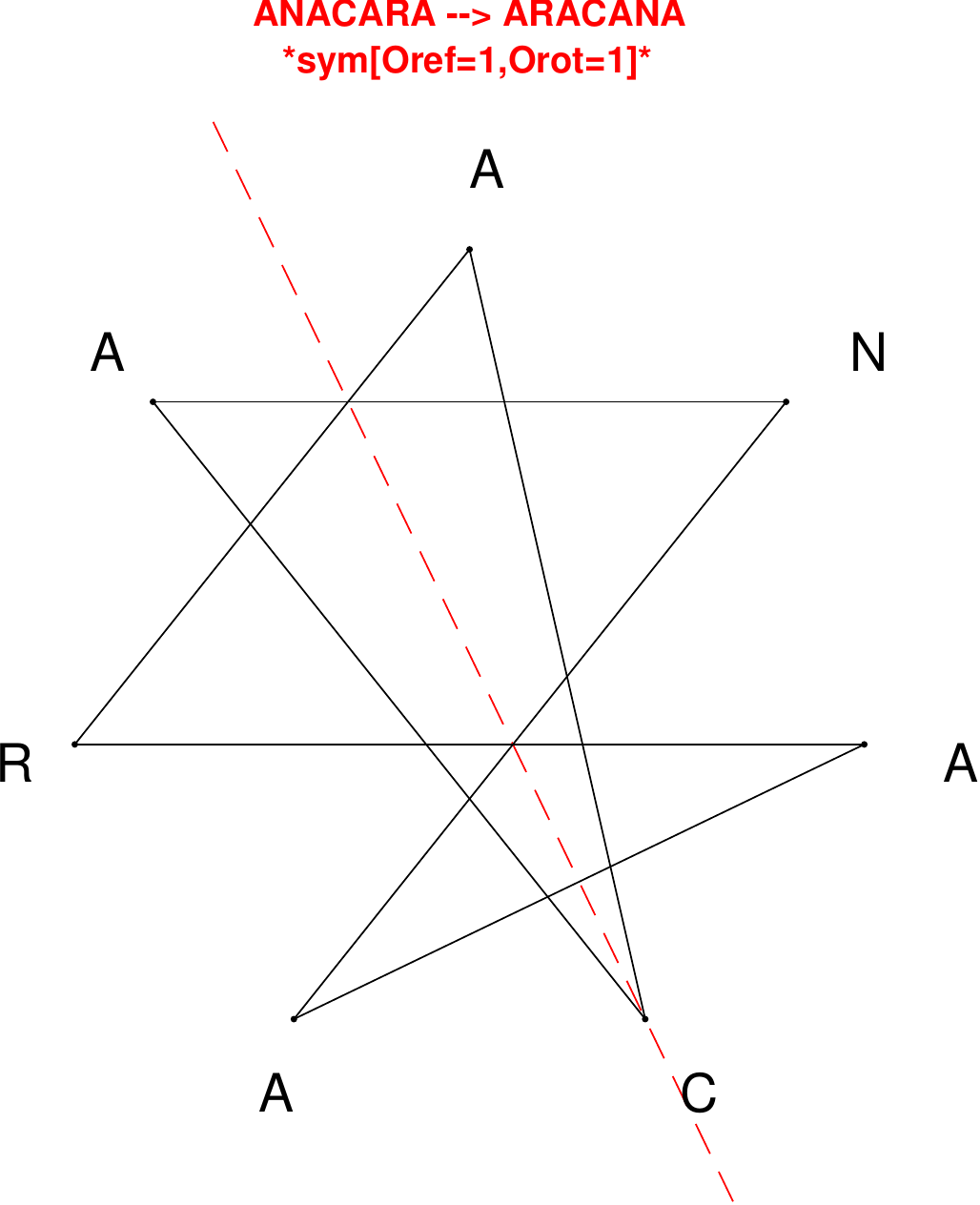}
\end{subfigure}
\hfill
\begin{subfigure}[T]{0.19\textwidth}
\centering
\includegraphics[width=\textwidth]{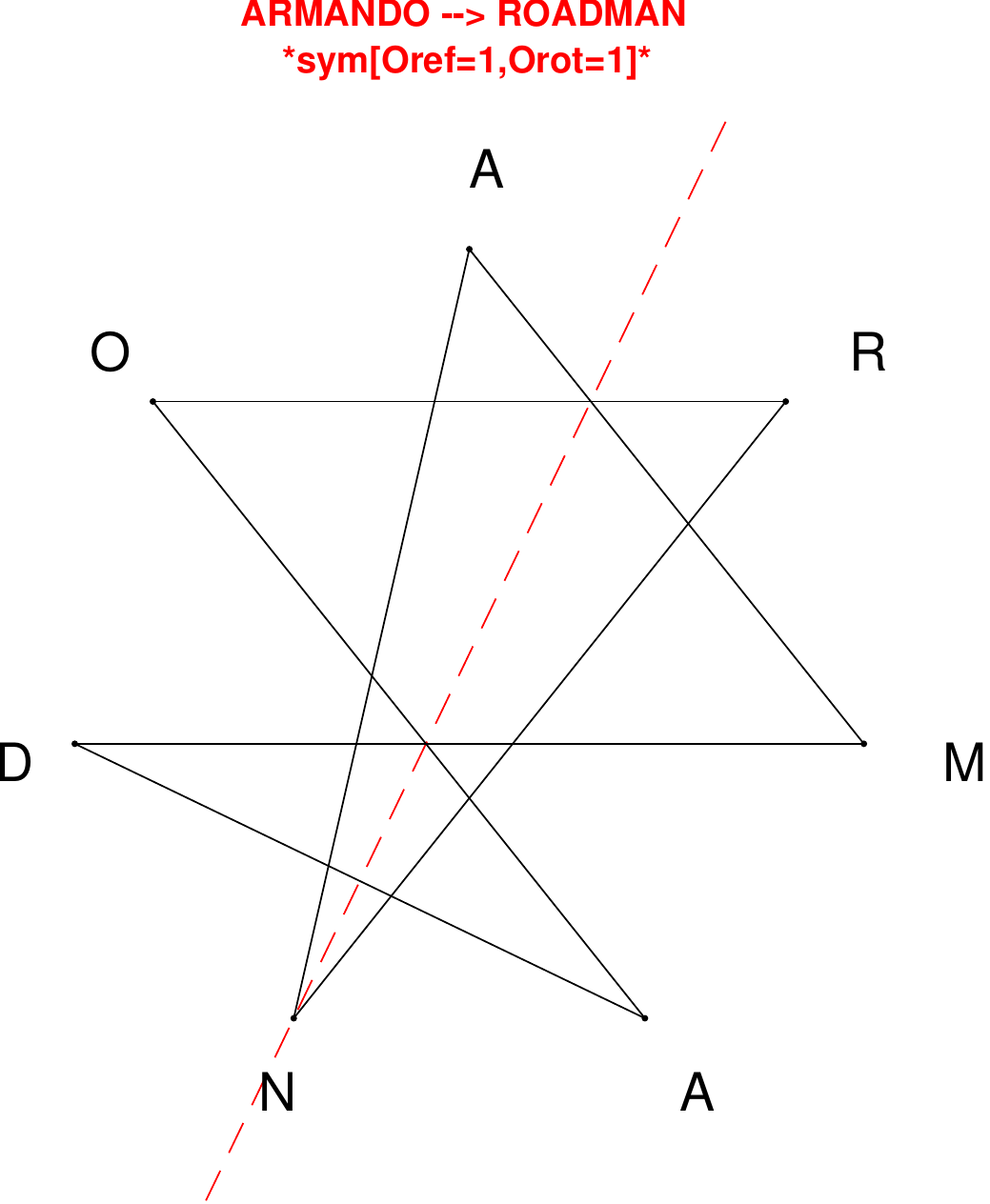}
\end{subfigure}
\end{figure}

\begin{figure}[H]
\centering
\begin{subfigure}[T]{0.19\textwidth}
\centering
\includegraphics[width=\textwidth]{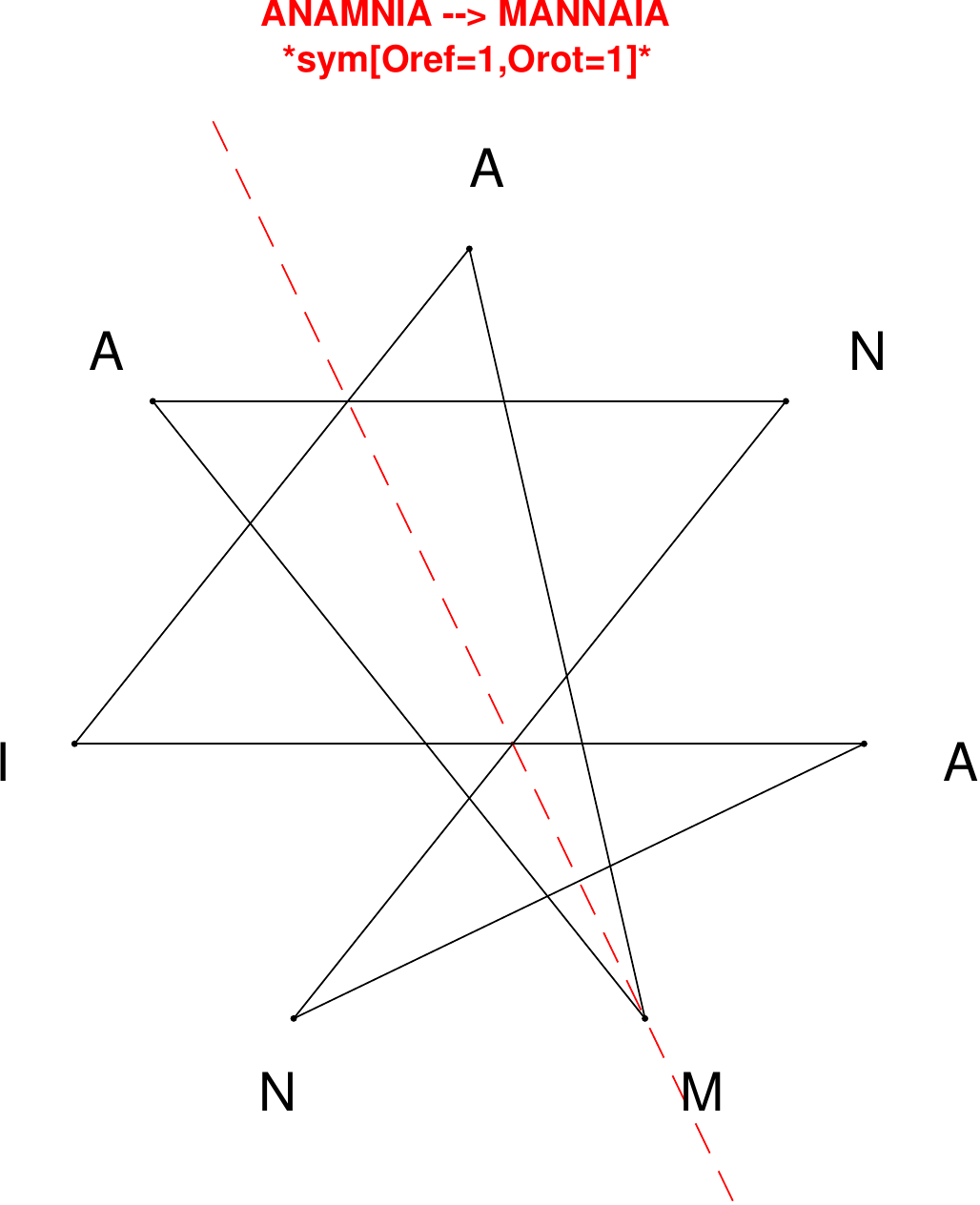}
\end{subfigure}
\hfill
\begin{subfigure}[T]{0.19\textwidth}
\centering
\includegraphics[width=\textwidth]{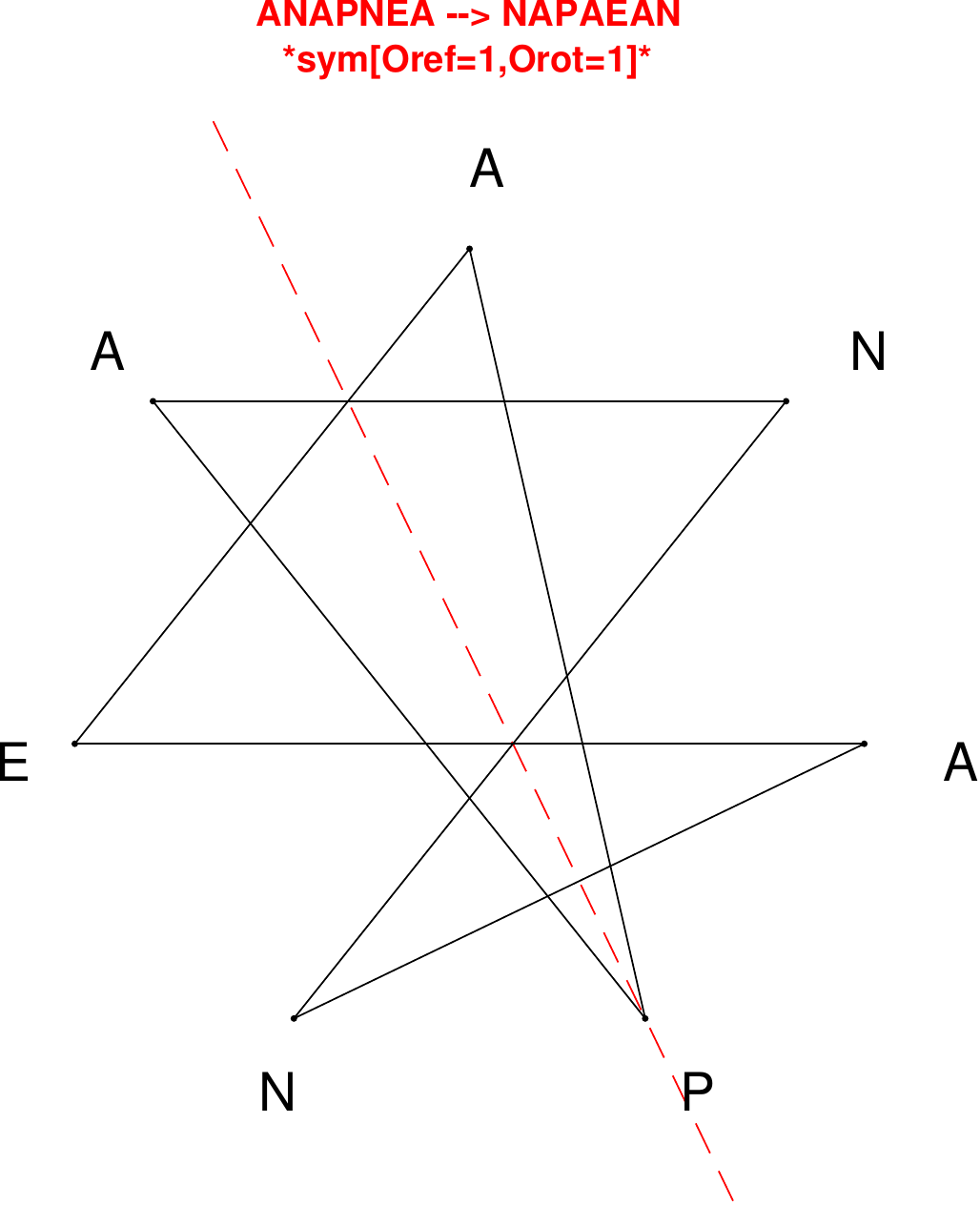}
\end{subfigure}
\hfill
\begin{subfigure}[T]{0.19\textwidth}
\centering
\includegraphics[width=\textwidth]{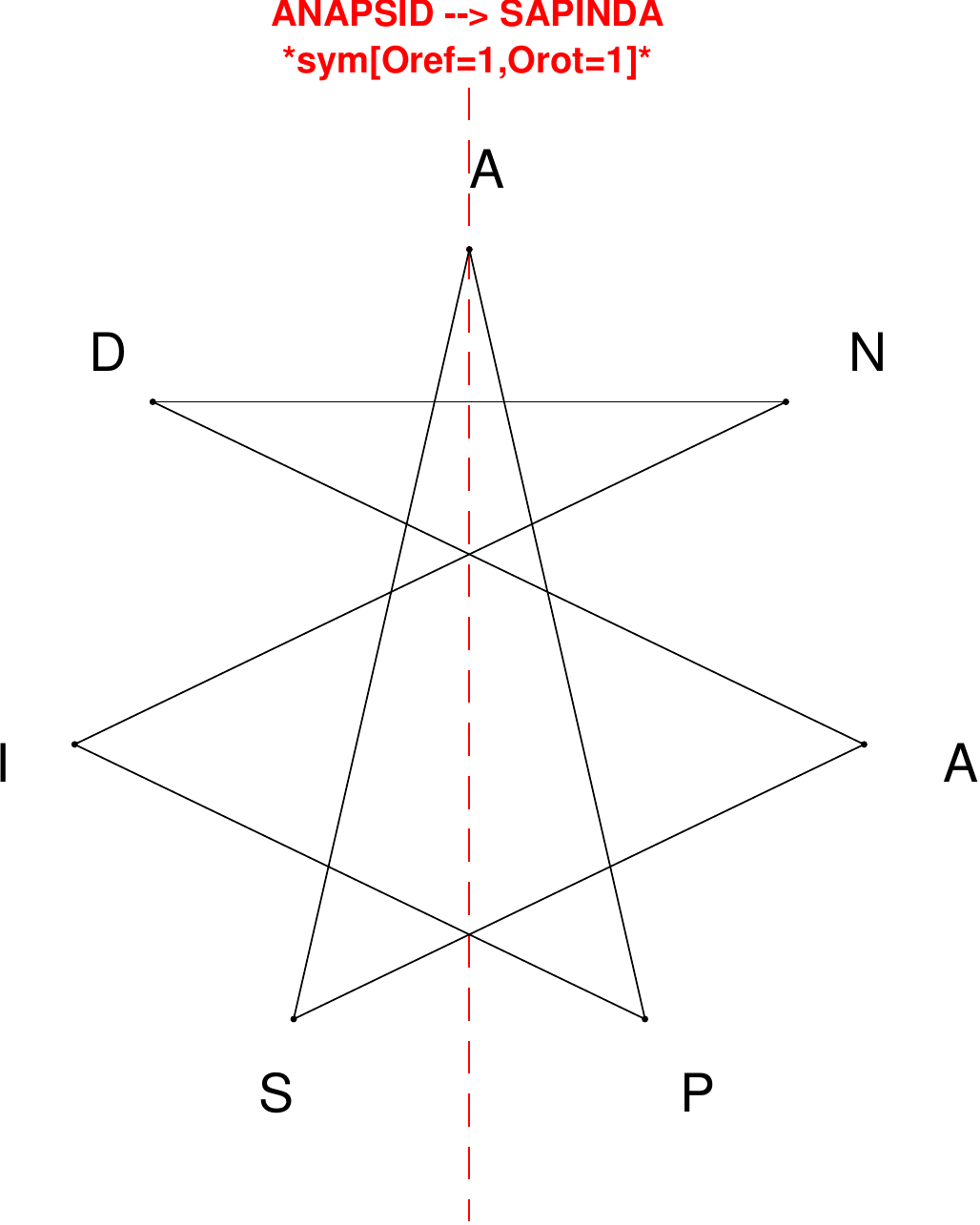}
\end{subfigure}
\hfill
\begin{subfigure}[T]{0.19\textwidth}
\centering
\includegraphics[width=\textwidth]{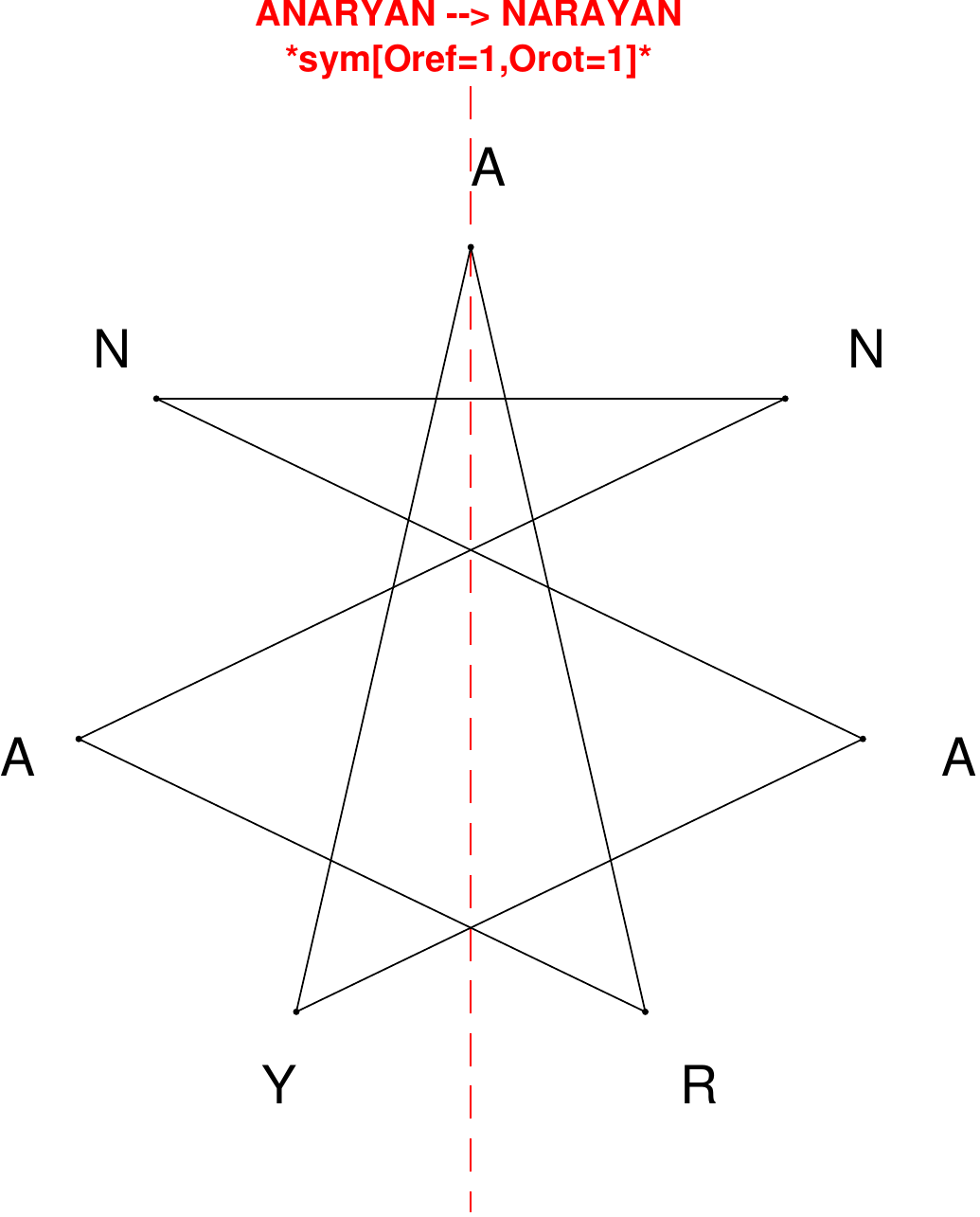}
\end{subfigure}
\hfill
\begin{subfigure}[T]{0.19\textwidth}
\centering
\includegraphics[width=\textwidth]{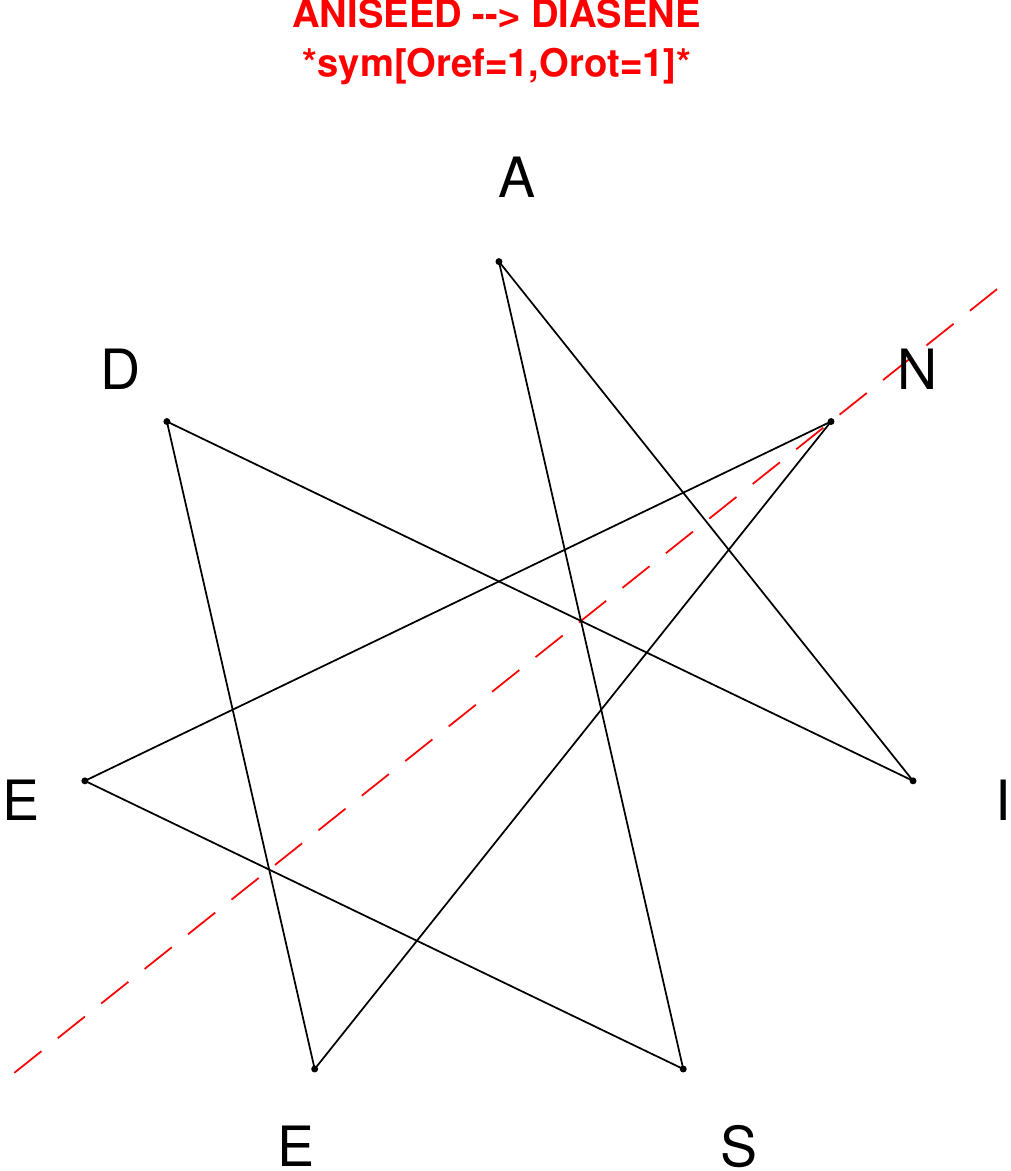}
\end{subfigure}
\end{figure}

\begin{figure}[H]
\centering
\begin{subfigure}[T]{0.19\textwidth}
\centering
\includegraphics[width=\textwidth]{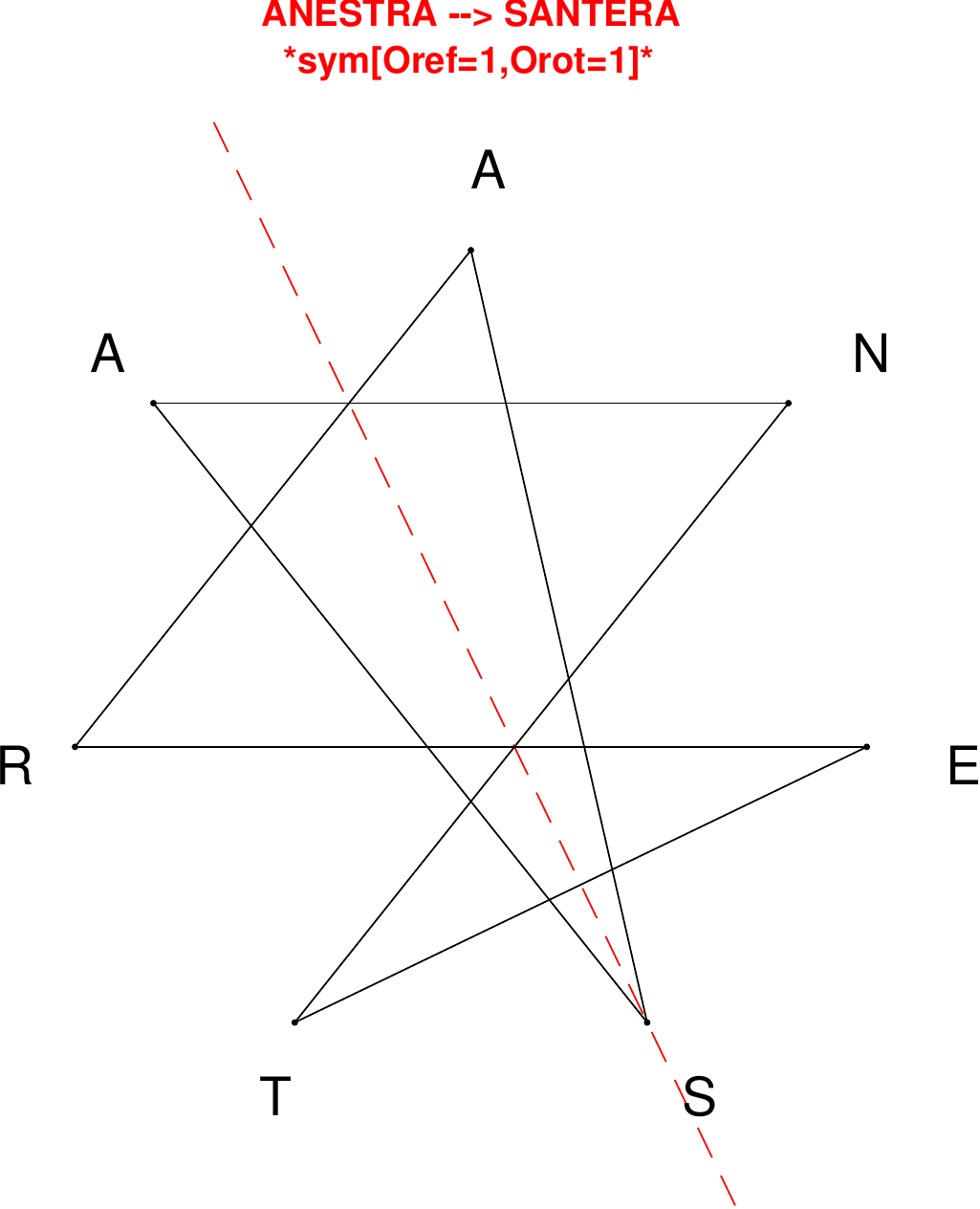}
\end{subfigure}
\hfill
\begin{subfigure}[T]{0.19\textwidth}
\centering
\includegraphics[width=\textwidth]{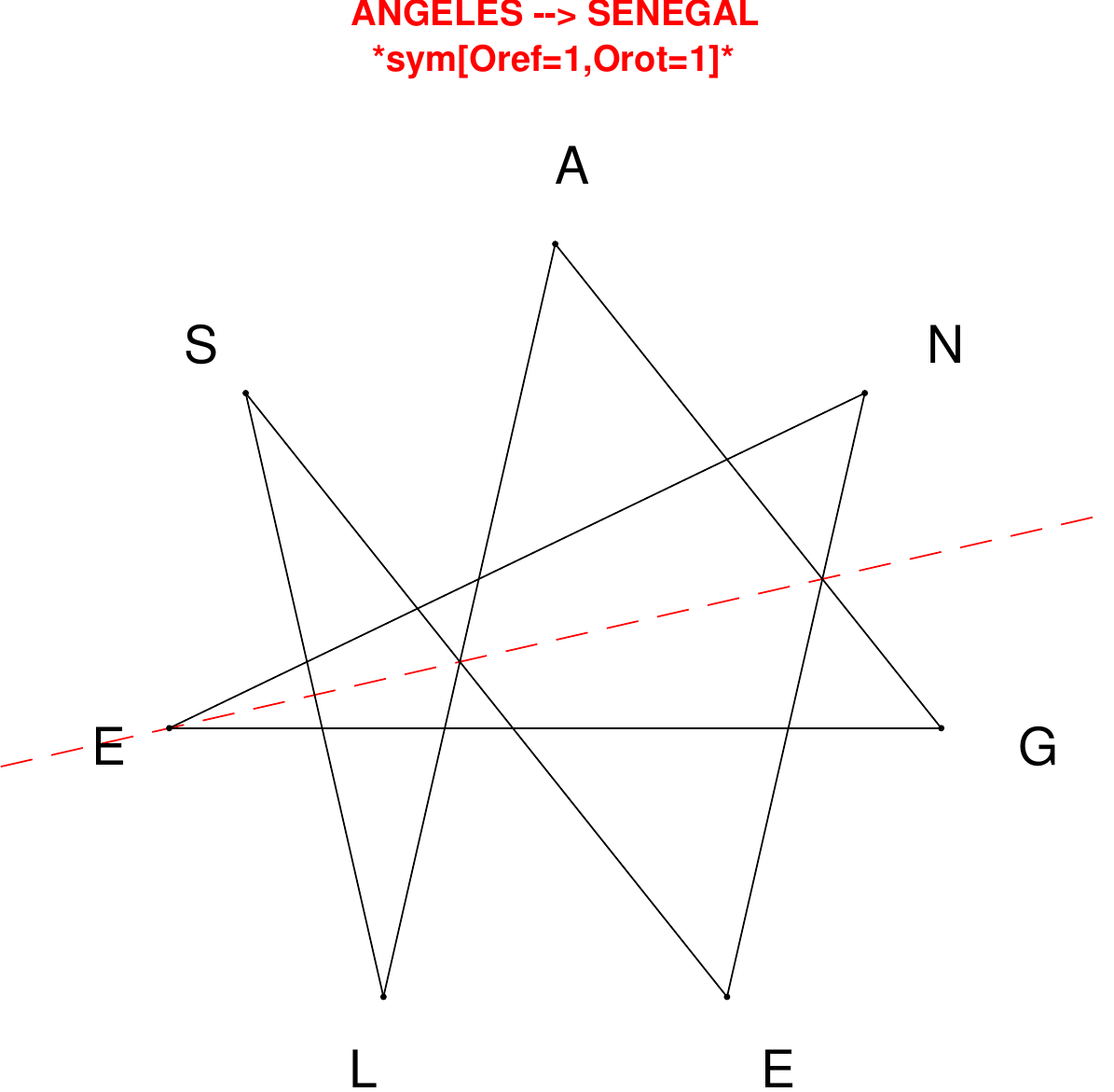}
\end{subfigure}
\hfill
\begin{subfigure}[T]{0.19\textwidth}
\centering
\includegraphics[width=\textwidth]{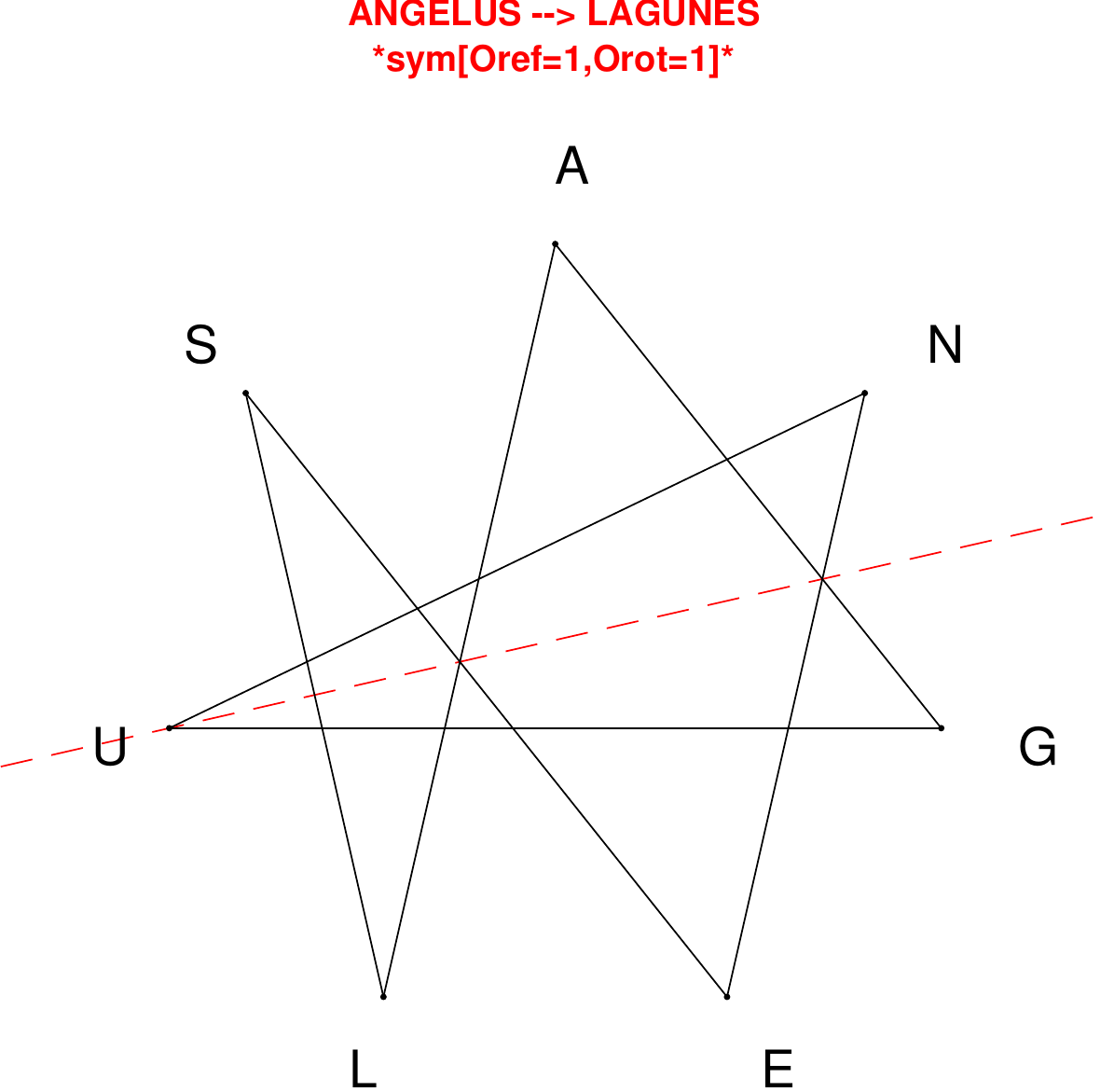}
\end{subfigure}
\hfill
\begin{subfigure}[T]{0.19\textwidth}
\centering
\includegraphics[width=\textwidth]{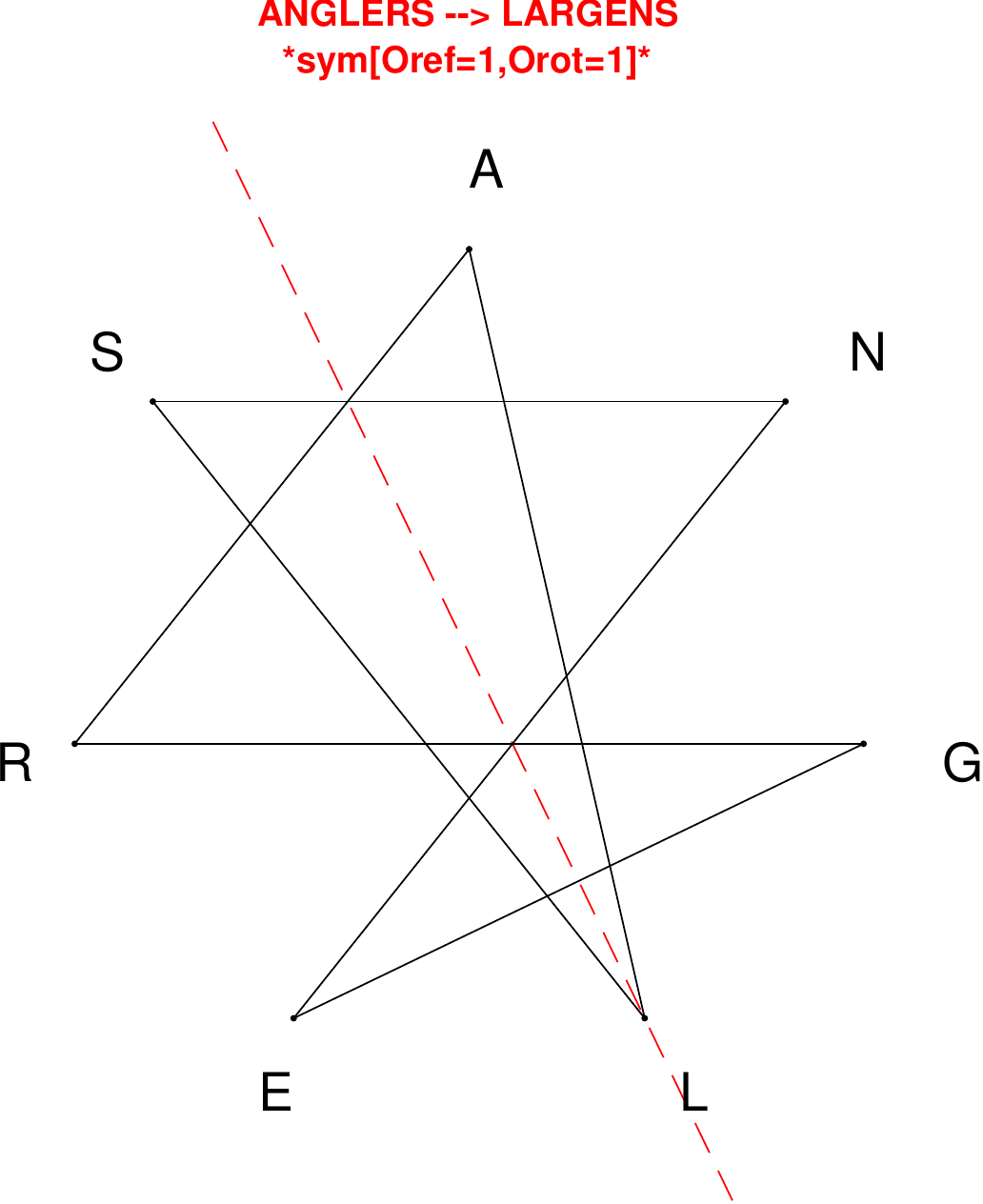}
\end{subfigure}
\hfill
\begin{subfigure}[T]{0.19\textwidth}
\centering
\includegraphics[width=\textwidth]{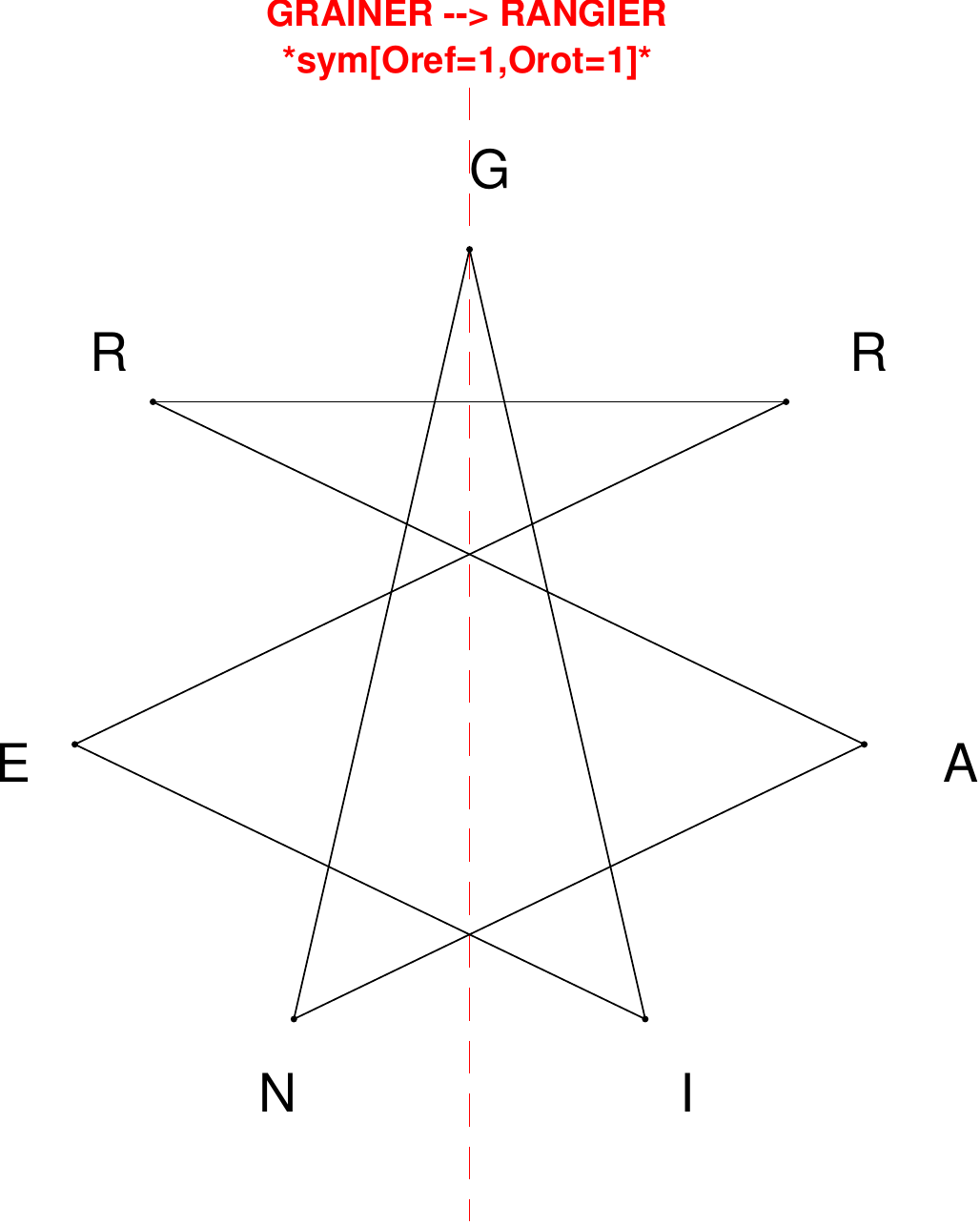}
\end{subfigure}
\end{figure}

\begin{figure}[H]
\centering
\begin{subfigure}[T]{0.19\textwidth}
\centering
\includegraphics[width=\textwidth]{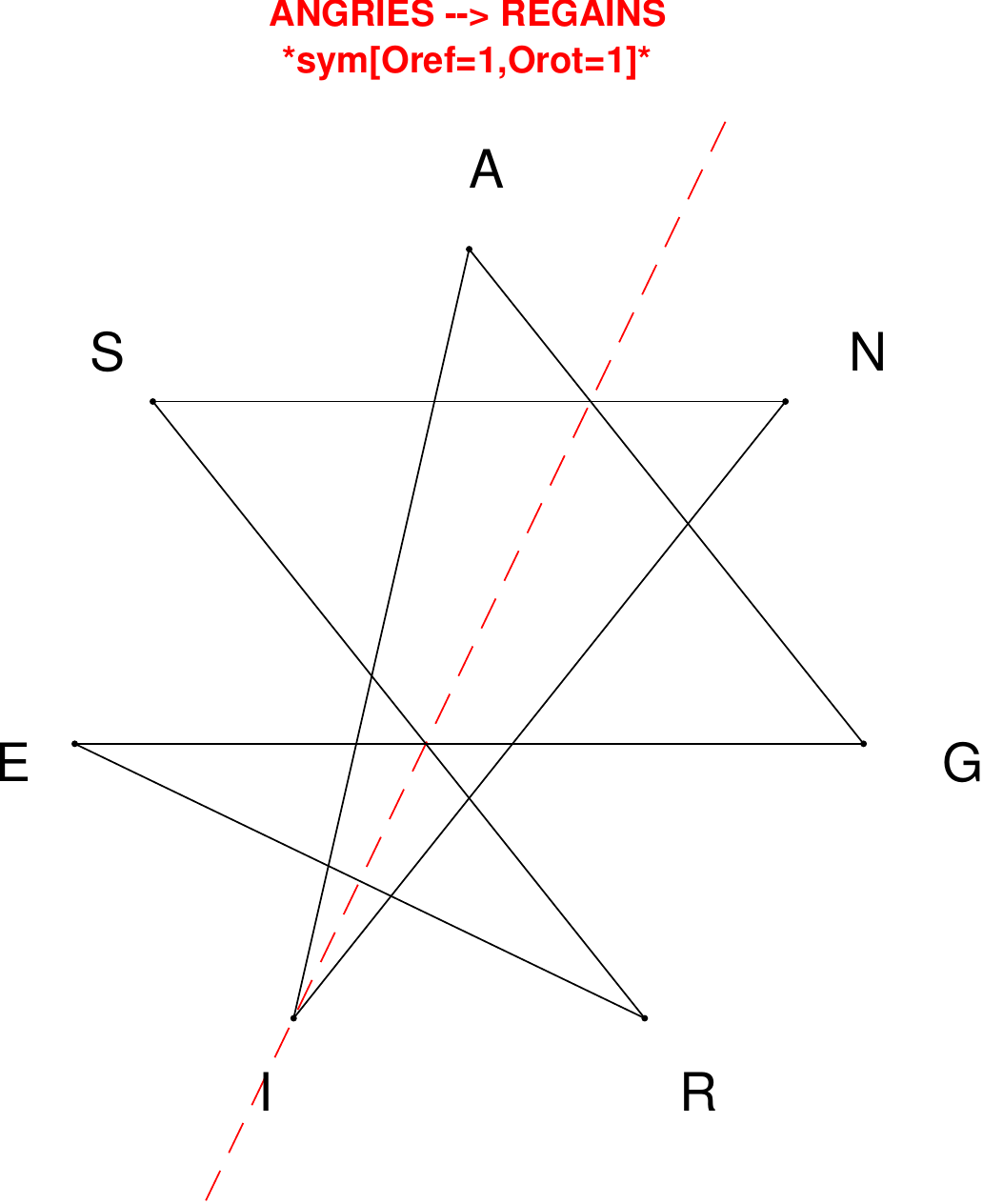}
\end{subfigure}
\hfill
\begin{subfigure}[T]{0.19\textwidth}
\centering
\includegraphics[width=\textwidth]{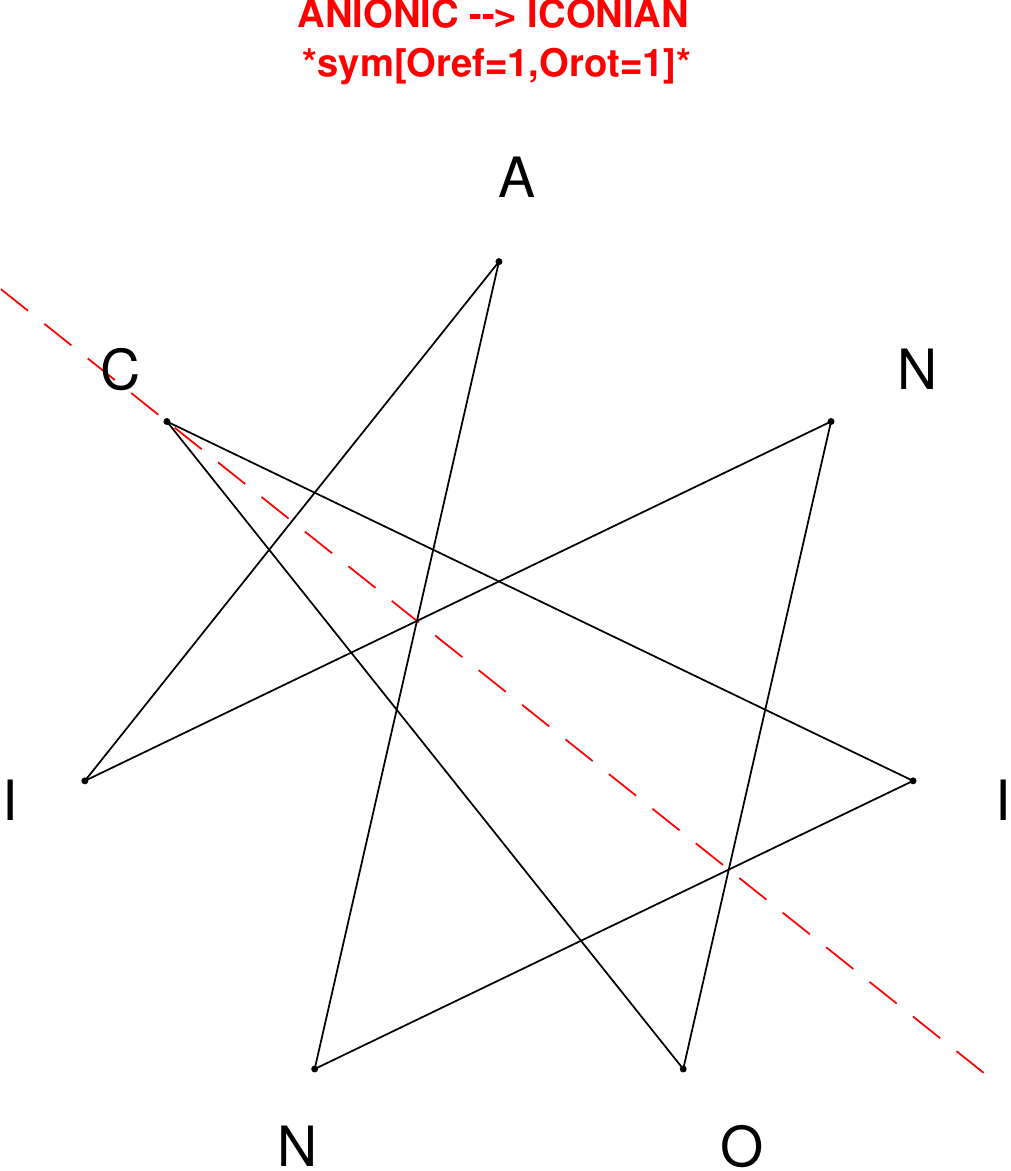}
\end{subfigure}
\hfill
\begin{subfigure}[T]{0.19\textwidth}
\centering
\includegraphics[width=\textwidth]{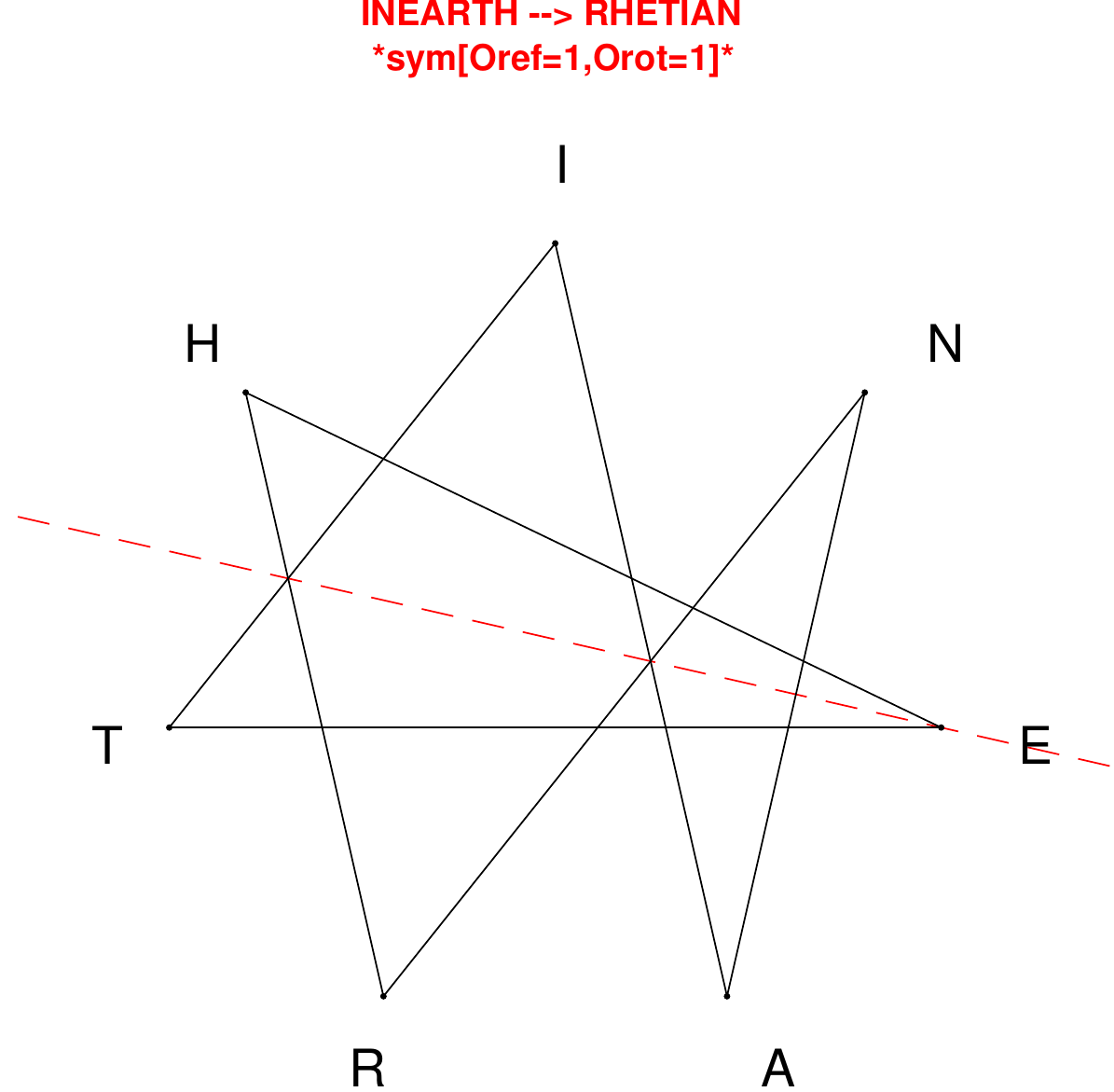}
\end{subfigure}
\hfill
\begin{subfigure}[T]{0.19\textwidth}
\centering
\includegraphics[width=\textwidth]{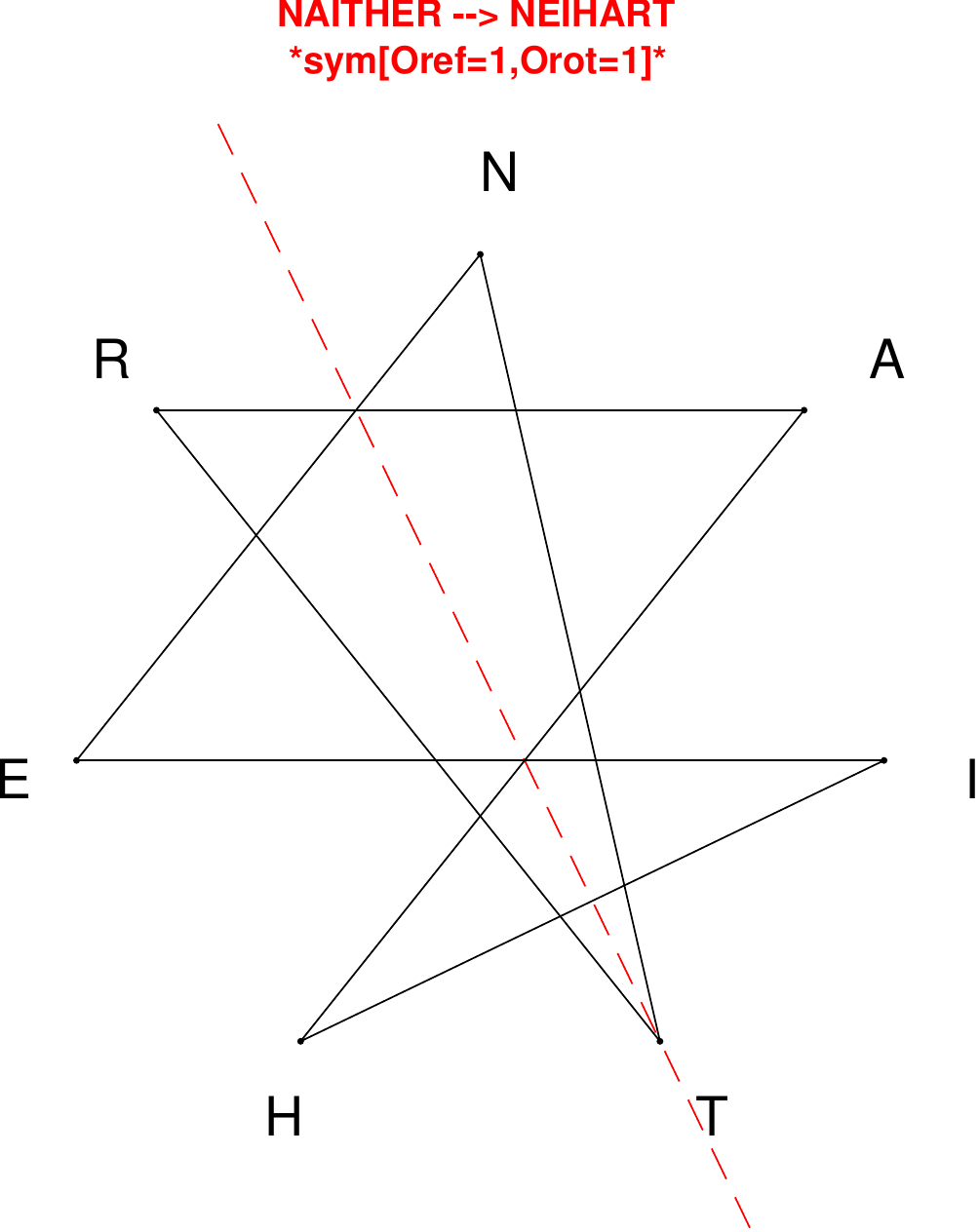}
\end{subfigure}
\hfill
\begin{subfigure}[T]{0.19\textwidth}
\centering
\includegraphics[width=\textwidth]{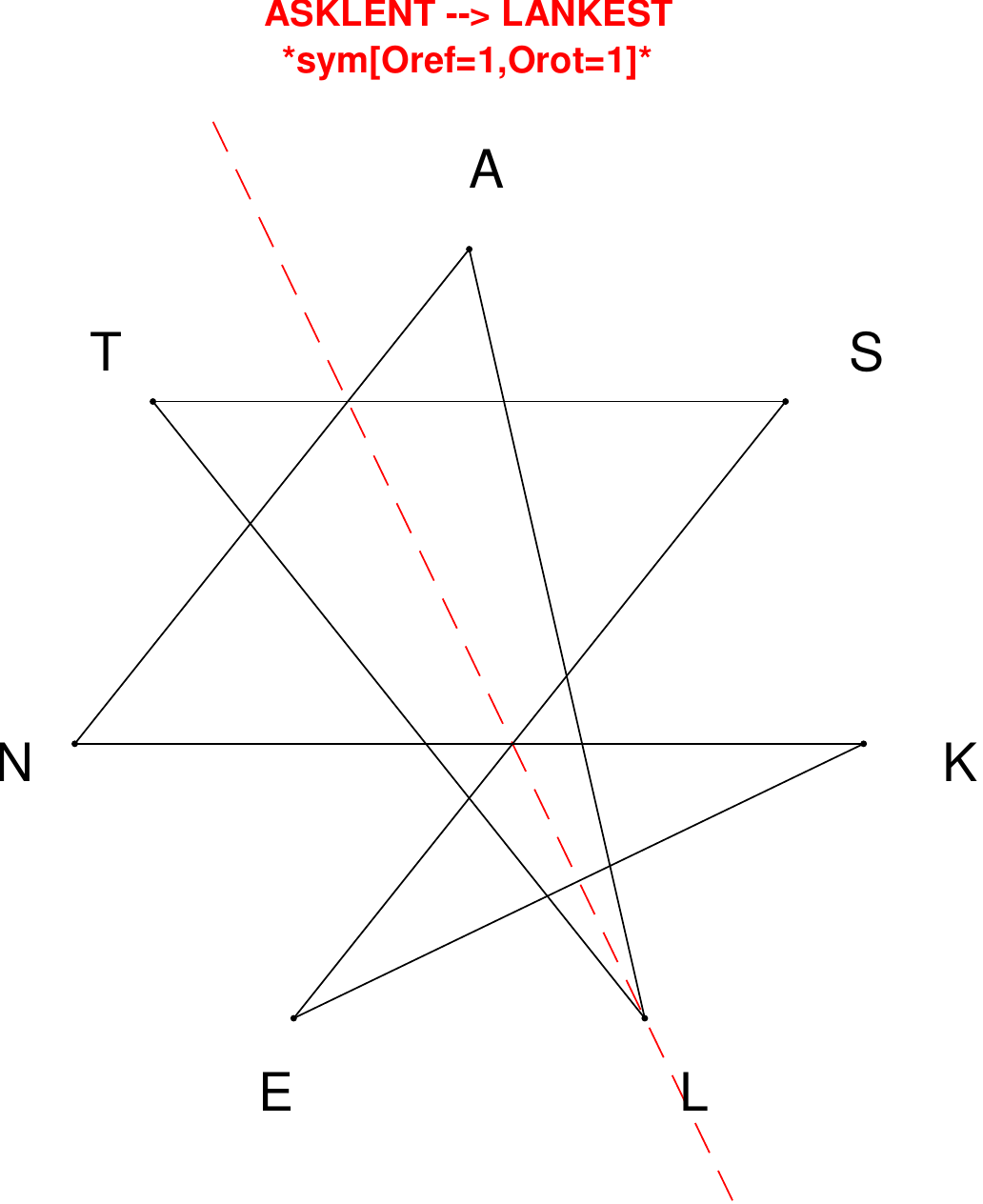}
\end{subfigure}
\end{figure}

\begin{figure}[H]
\centering
\begin{subfigure}[T]{0.19\textwidth}
\centering
\includegraphics[width=\textwidth]{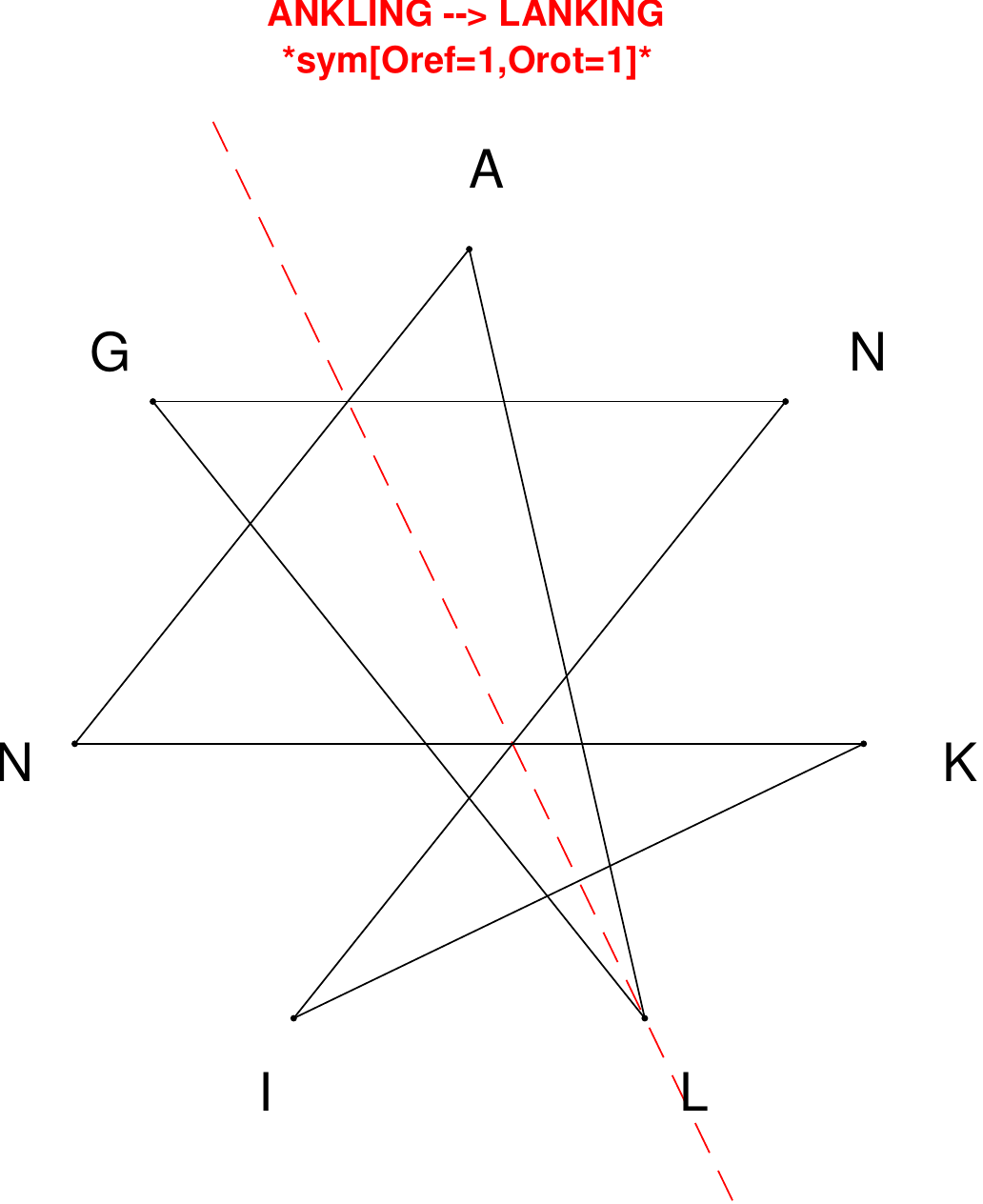}
\end{subfigure}
\hfill
\begin{subfigure}[T]{0.19\textwidth}
\centering
\includegraphics[width=\textwidth]{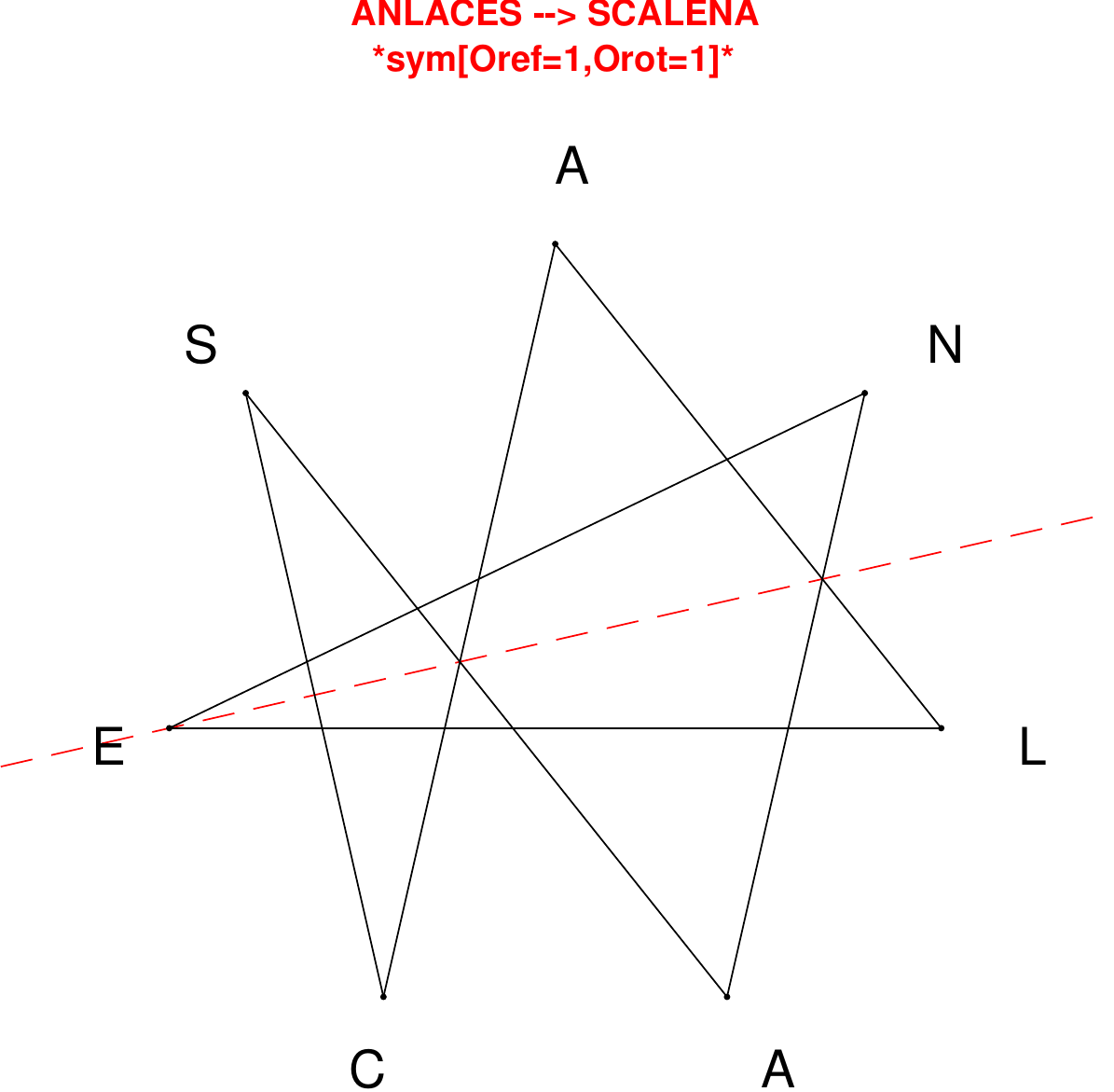}
\end{subfigure}
\hfill
\begin{subfigure}[T]{0.19\textwidth}
\centering
\includegraphics[width=\textwidth]{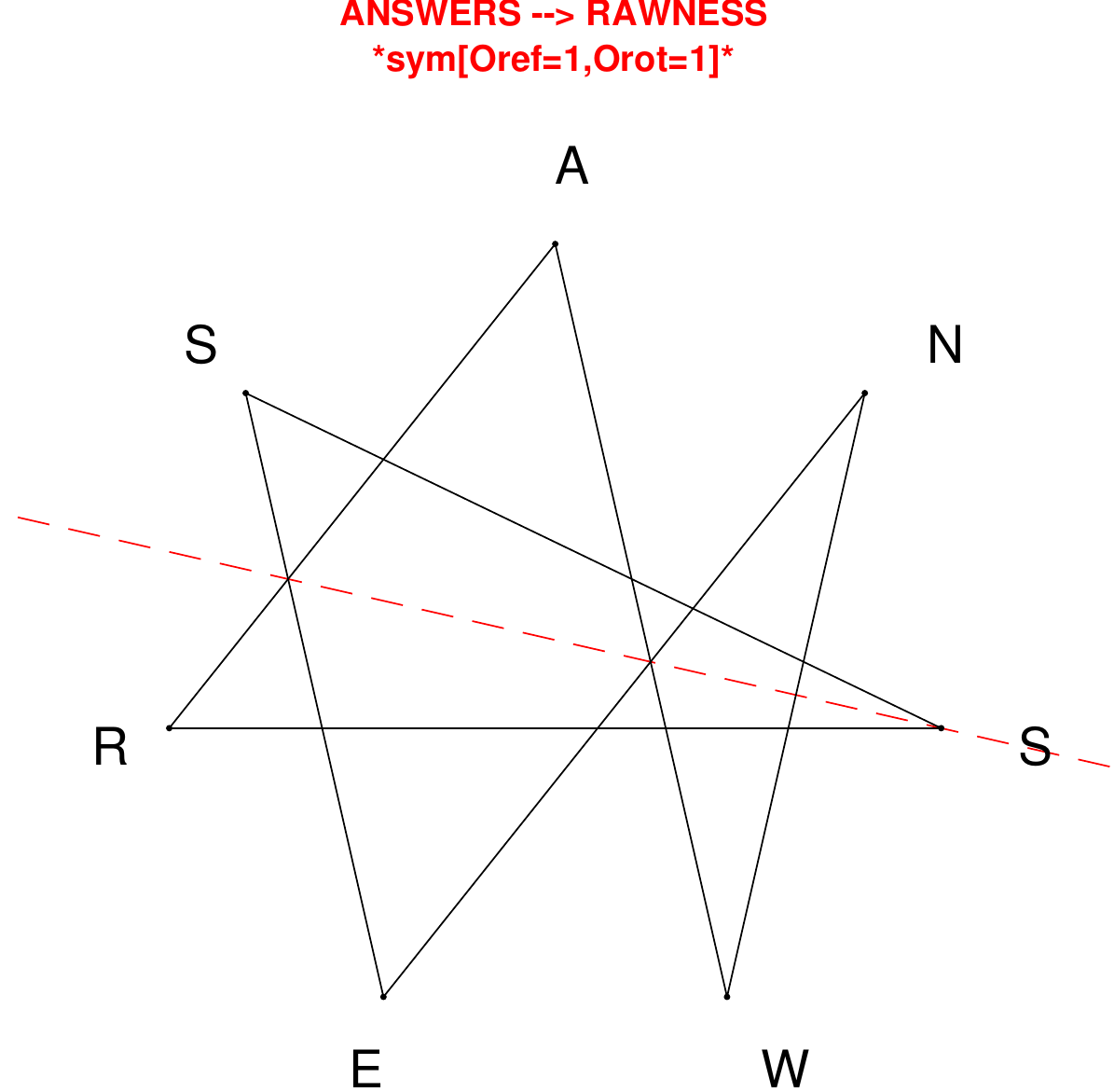}
\end{subfigure}
\hfill
\begin{subfigure}[T]{0.19\textwidth}
\centering
\includegraphics[width=\textwidth]{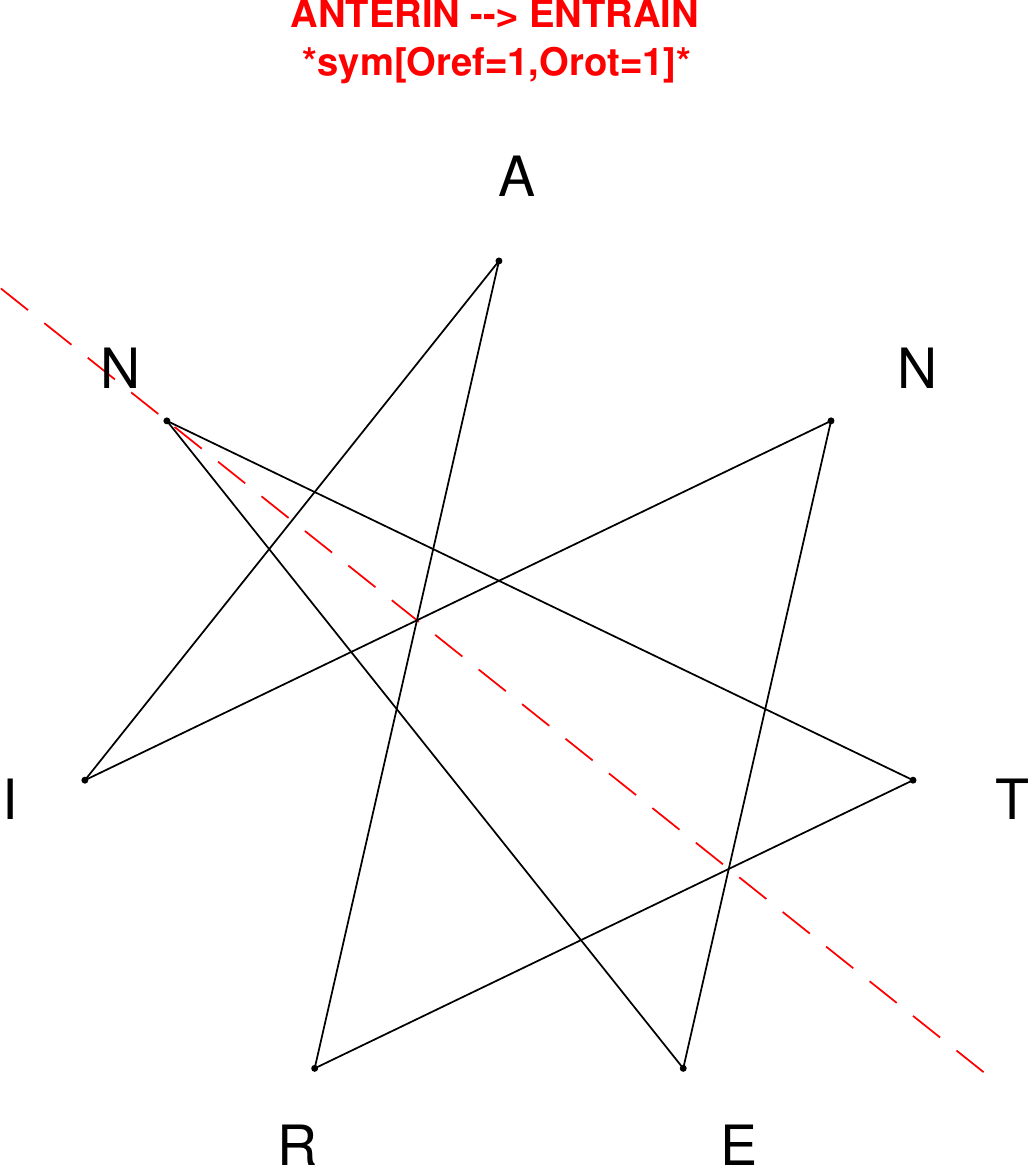}
\end{subfigure}
\hfill
\begin{subfigure}[T]{0.19\textwidth}
\centering
\includegraphics[width=\textwidth]{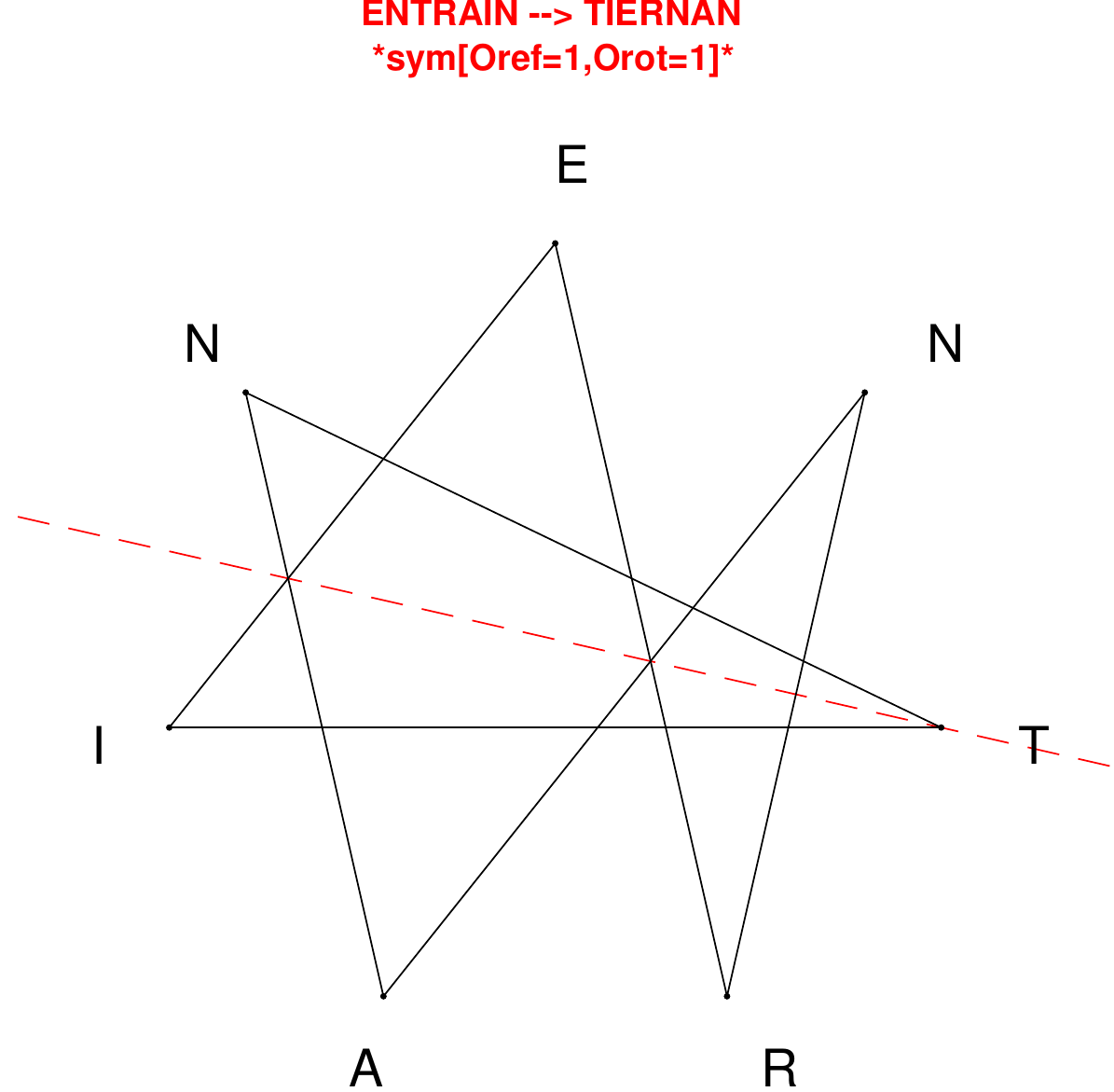}
\end{subfigure}
\end{figure}

\begin{figure}[H]
\centering
\begin{subfigure}[T]{0.19\textwidth}
\centering
\includegraphics[width=\textwidth]{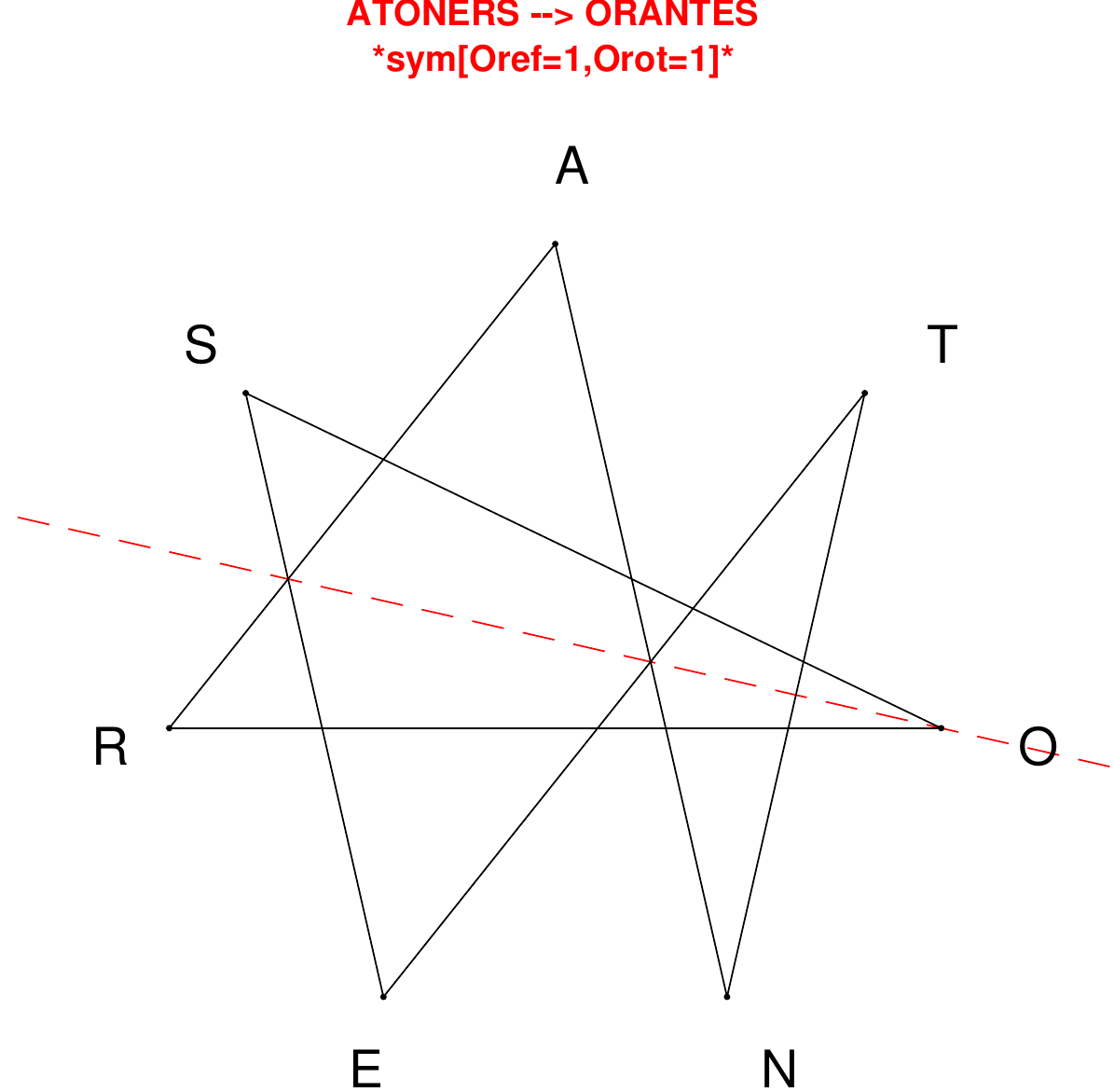}
\end{subfigure}
\hfill
\begin{subfigure}[T]{0.19\textwidth}
\centering
\includegraphics[width=\textwidth]{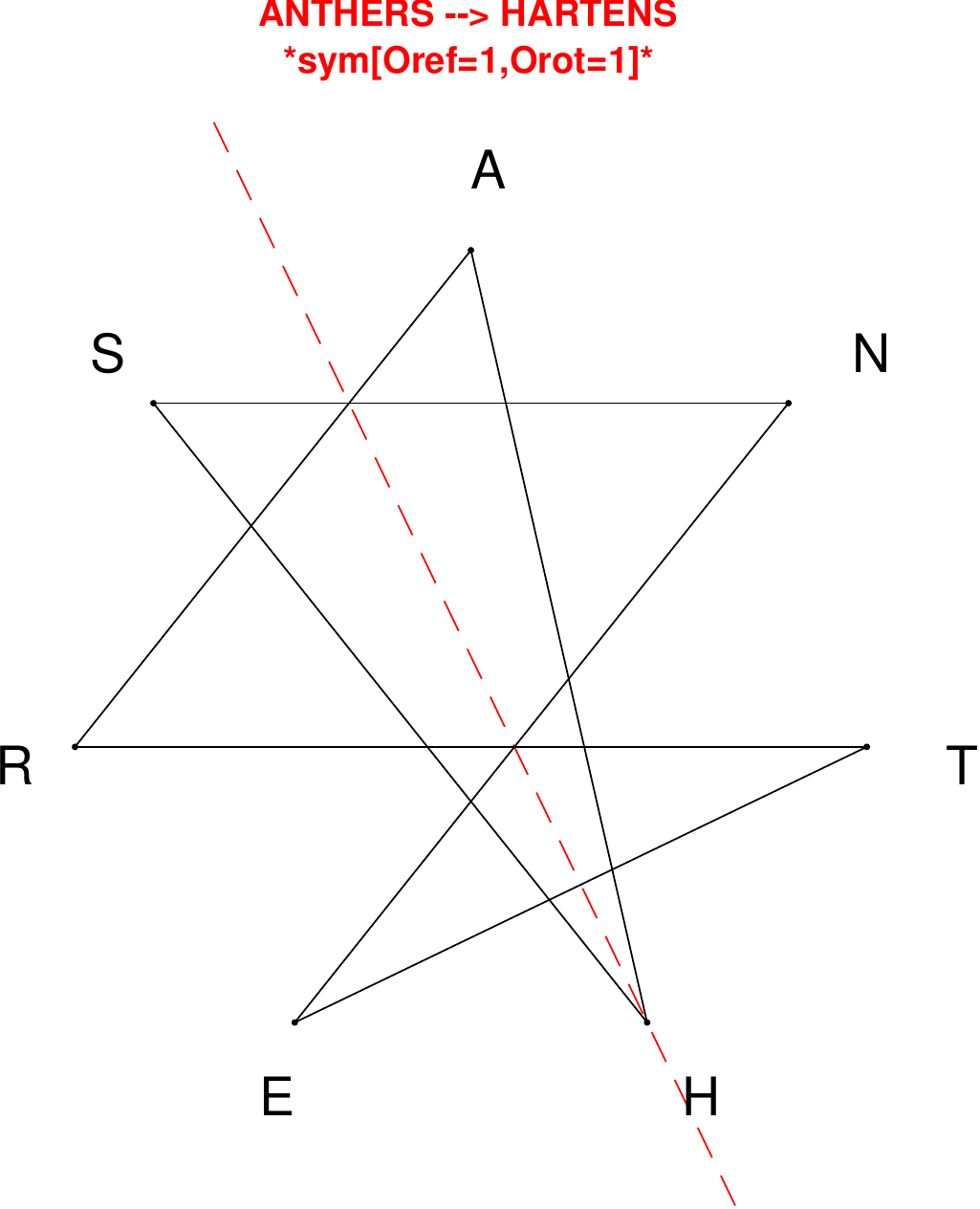}
\end{subfigure}
\hfill
\begin{subfigure}[T]{0.19\textwidth}
\centering
\includegraphics[width=\textwidth]{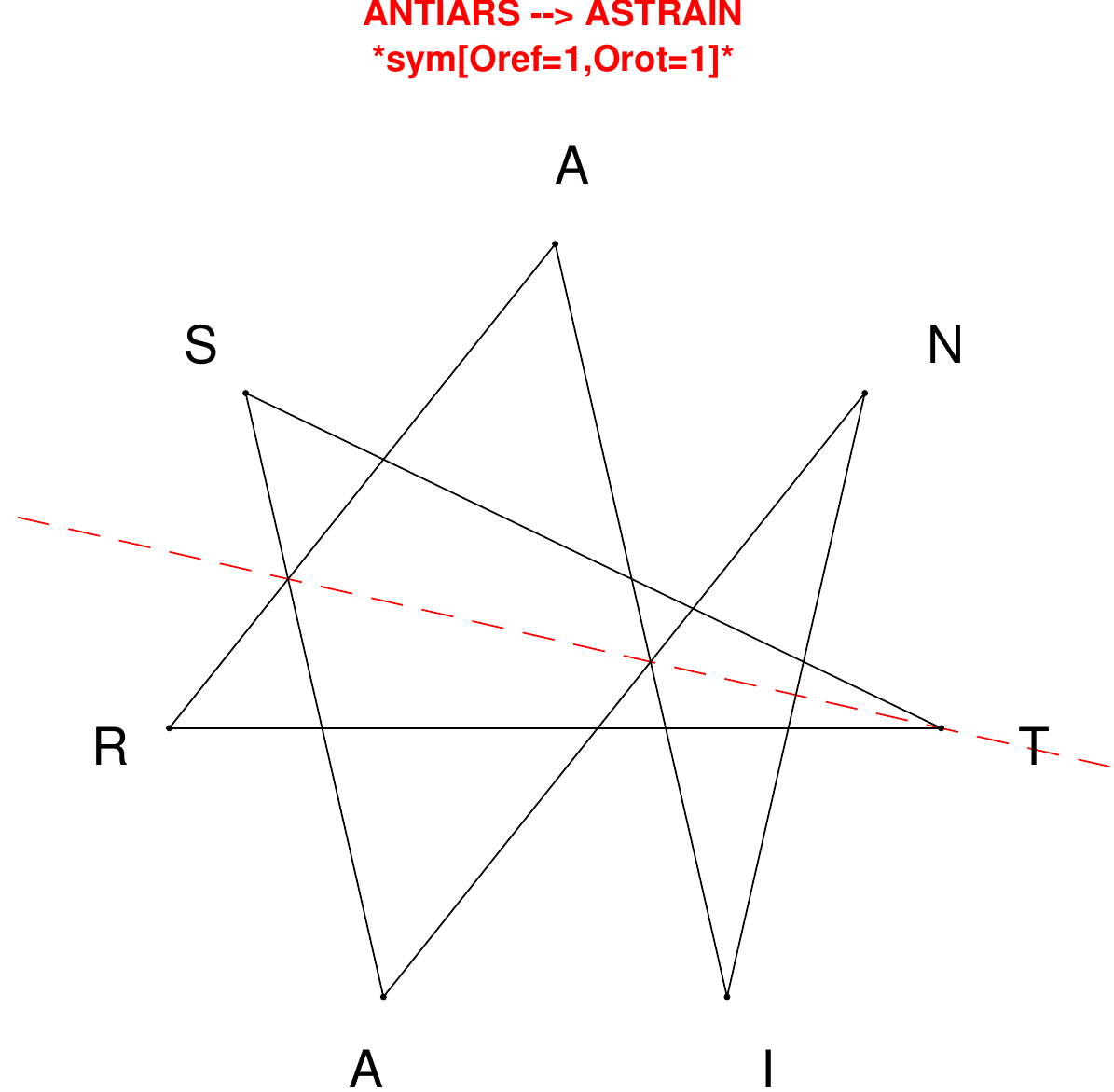}
\end{subfigure}
\hfill
\begin{subfigure}[T]{0.19\textwidth}
\centering
\includegraphics[width=\textwidth]{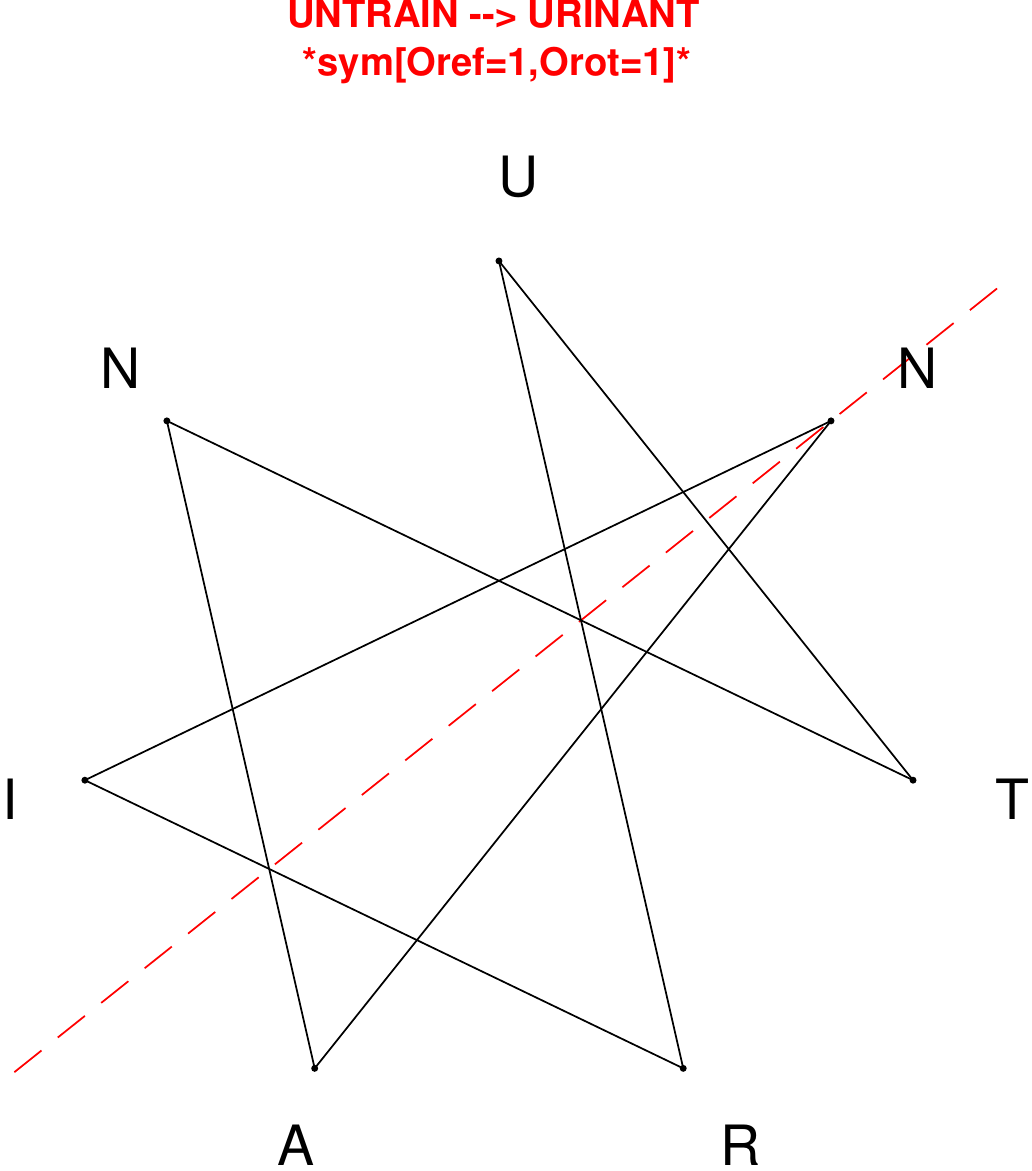}
\end{subfigure}
\hfill
\begin{subfigure}[T]{0.19\textwidth}
\centering
\includegraphics[width=\textwidth]{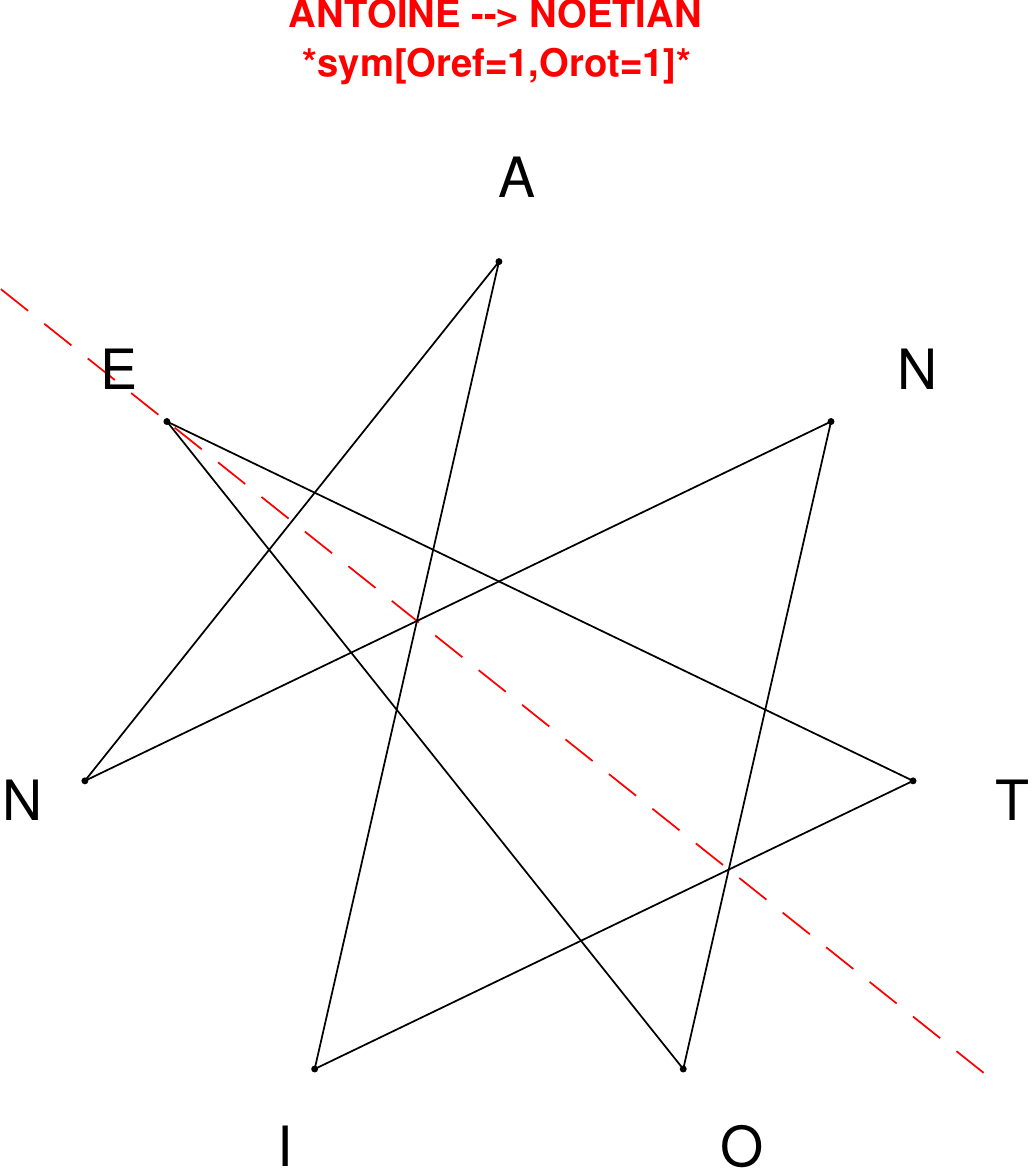}
\end{subfigure}
\end{figure}

\begin{figure}[H]
\centering
\begin{subfigure}[T]{0.19\textwidth}
\centering
\includegraphics[width=\textwidth]{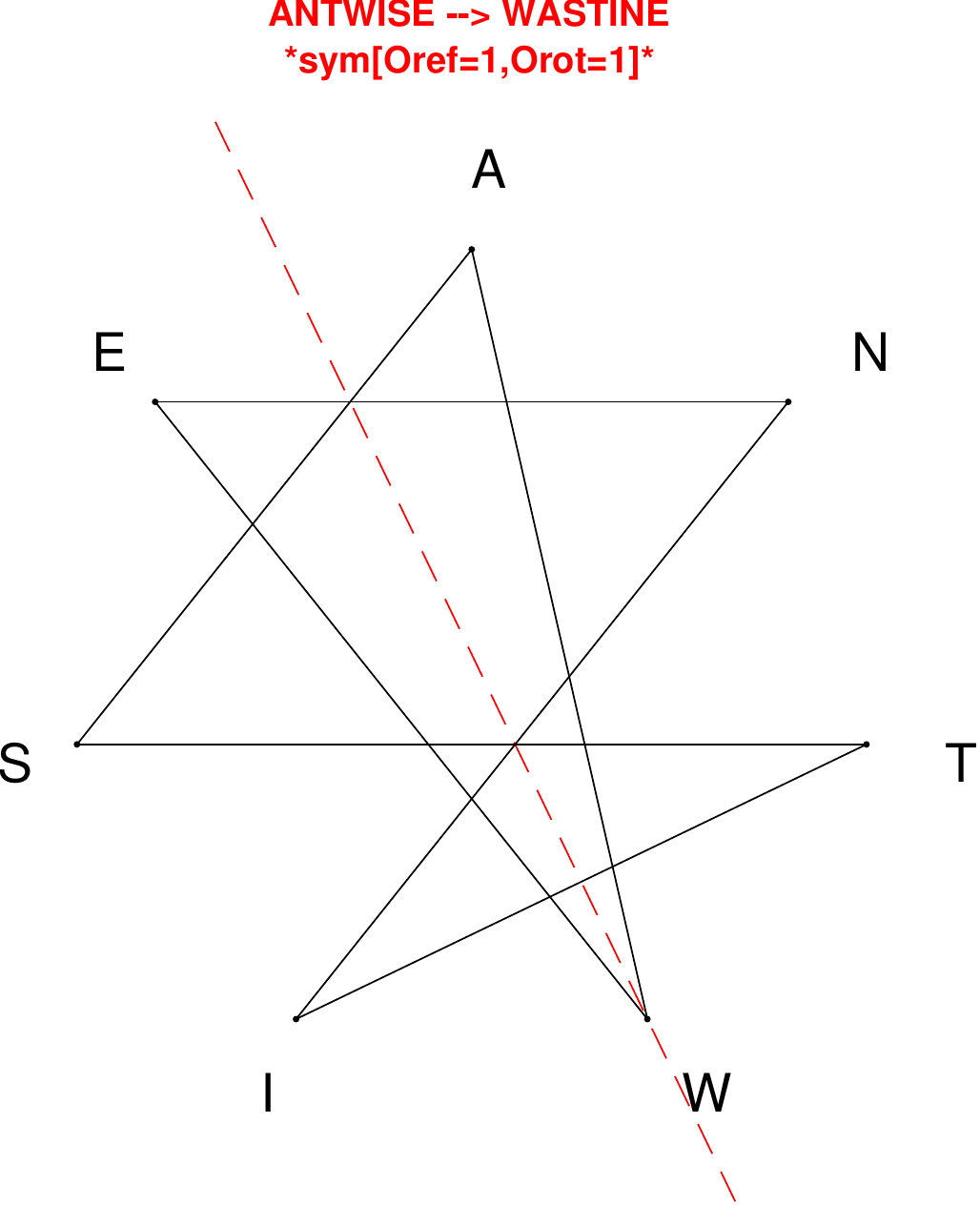}
\end{subfigure}
\hfill
\begin{subfigure}[T]{0.19\textwidth}
\centering
\includegraphics[width=\textwidth]{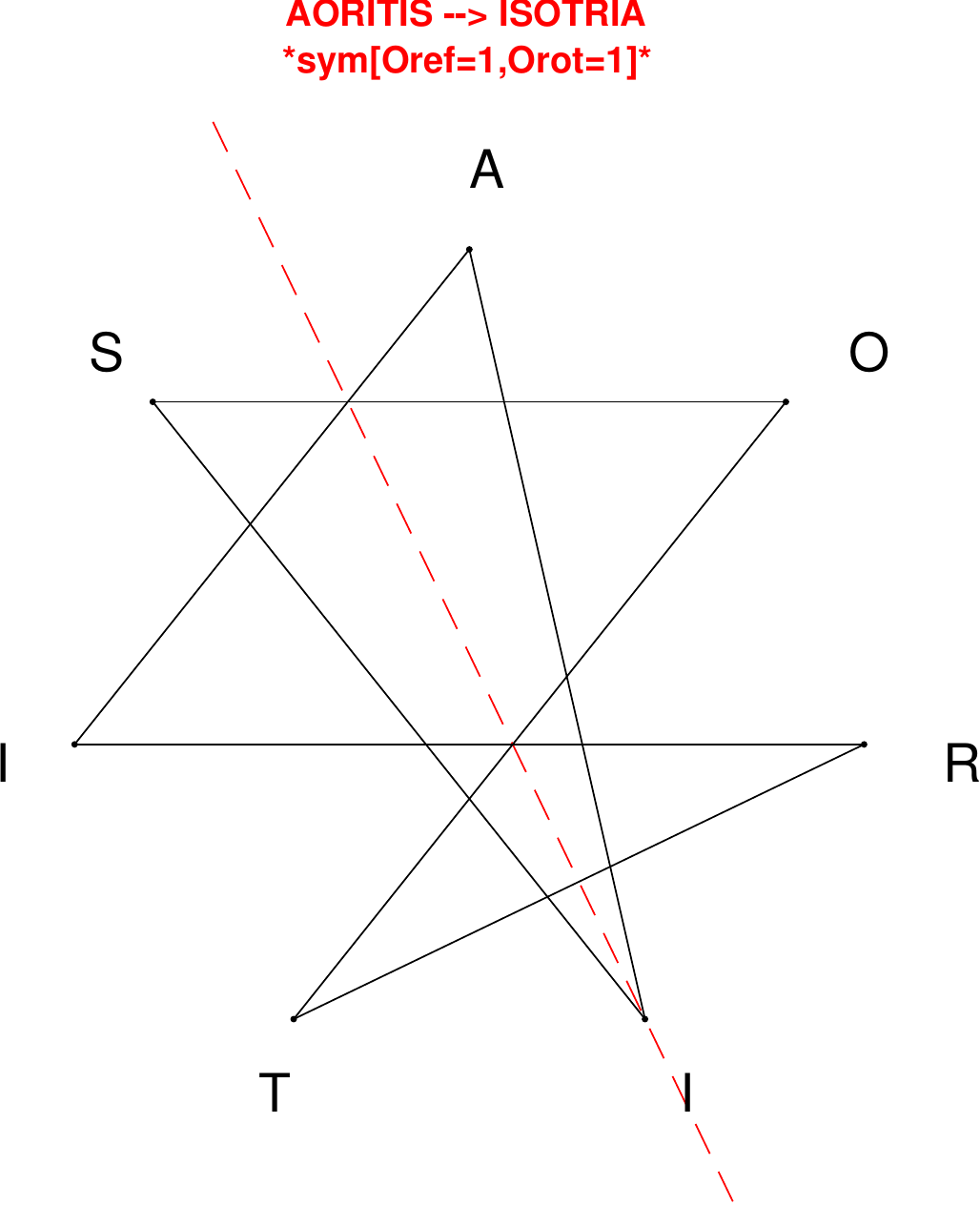}
\end{subfigure}
\hfill
\begin{subfigure}[T]{0.19\textwidth}
\centering
\includegraphics[width=\textwidth]{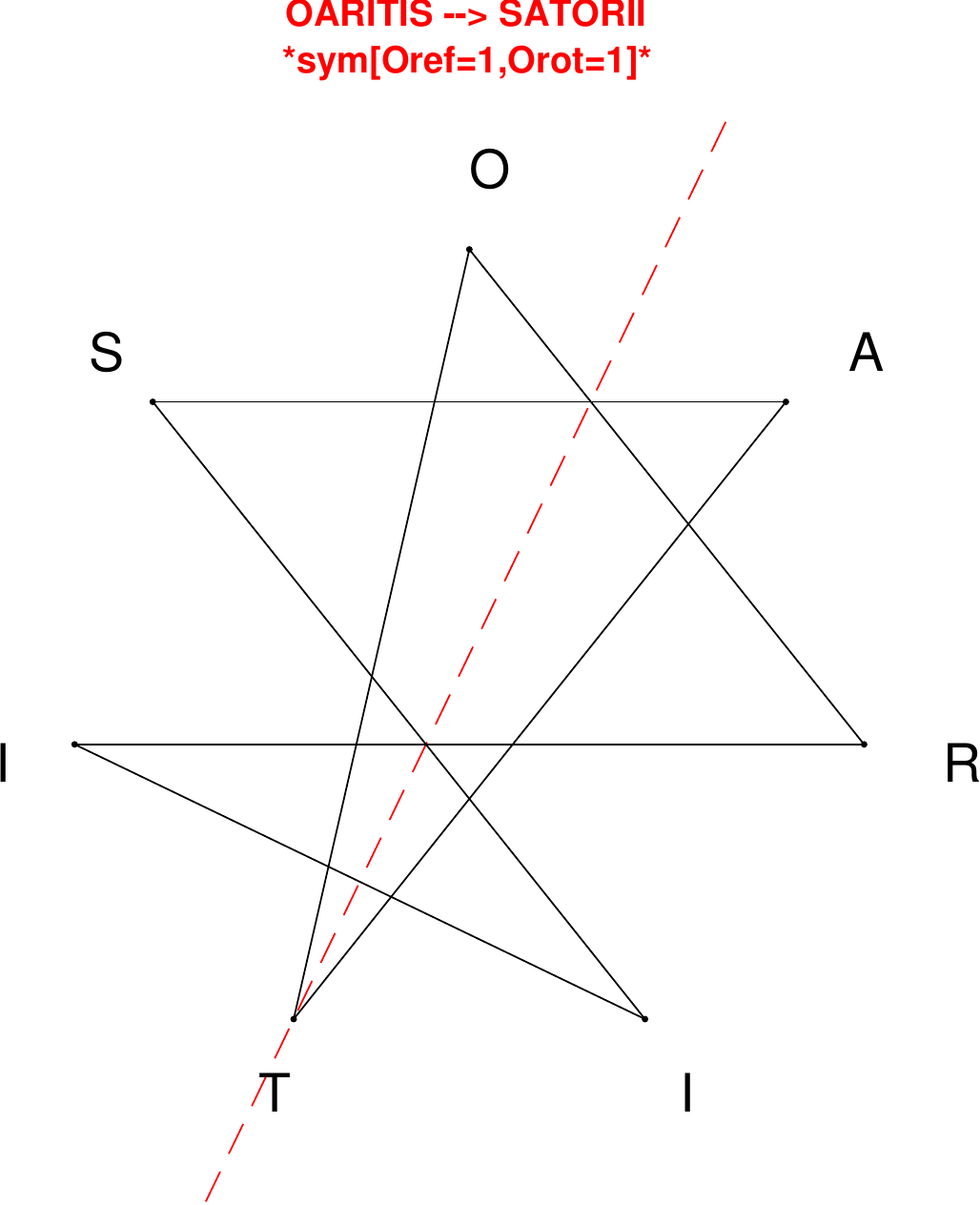}
\end{subfigure}
\hfill
\begin{subfigure}[T]{0.19\textwidth}
\centering
\includegraphics[width=\textwidth]{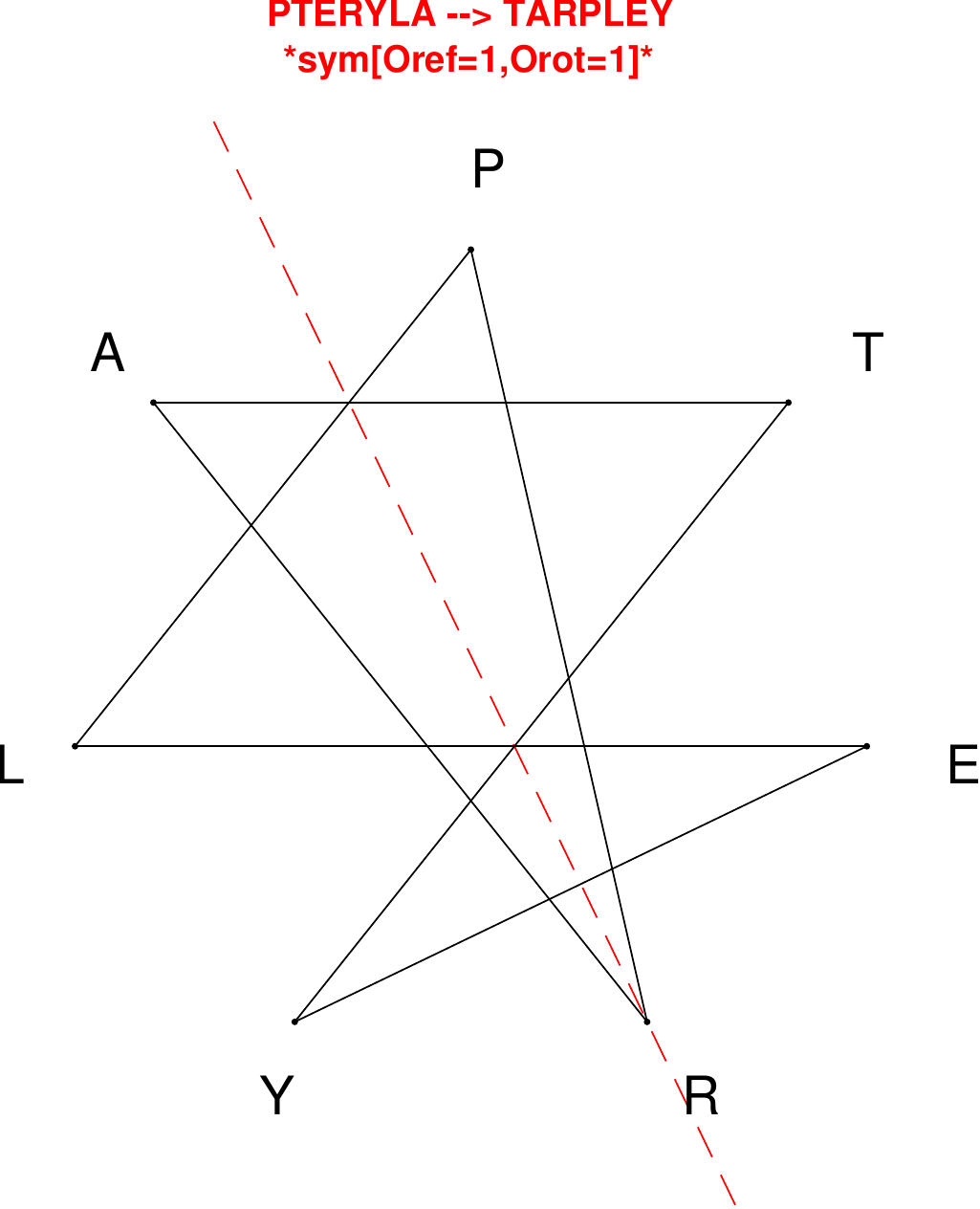}
\end{subfigure}
\hfill
\begin{subfigure}[T]{0.19\textwidth}
\centering
\includegraphics[width=\textwidth]{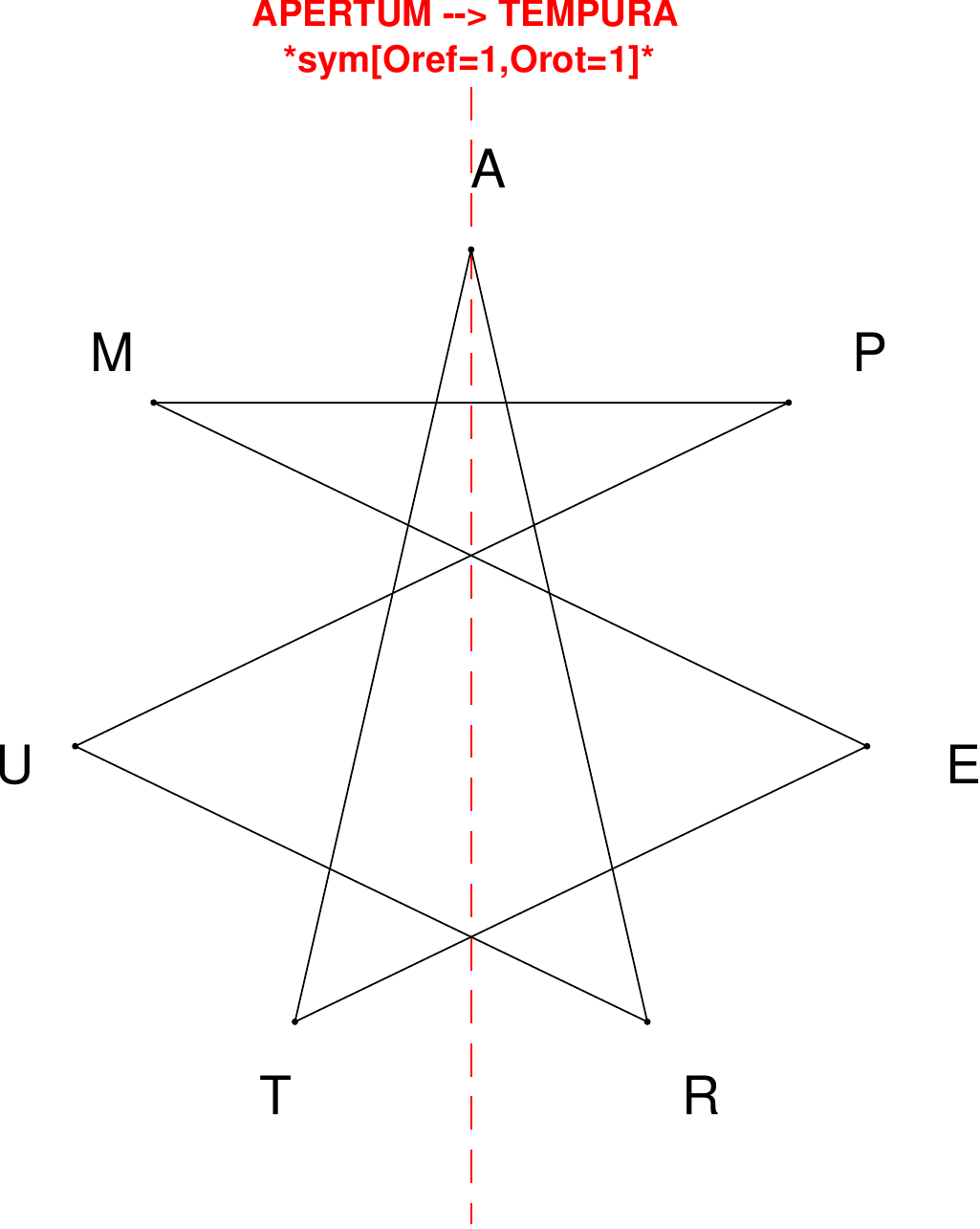}
\end{subfigure}
\end{figure}

\begin{figure}[H]
\centering
\begin{subfigure}[T]{0.19\textwidth}
\centering
\includegraphics[width=\textwidth]{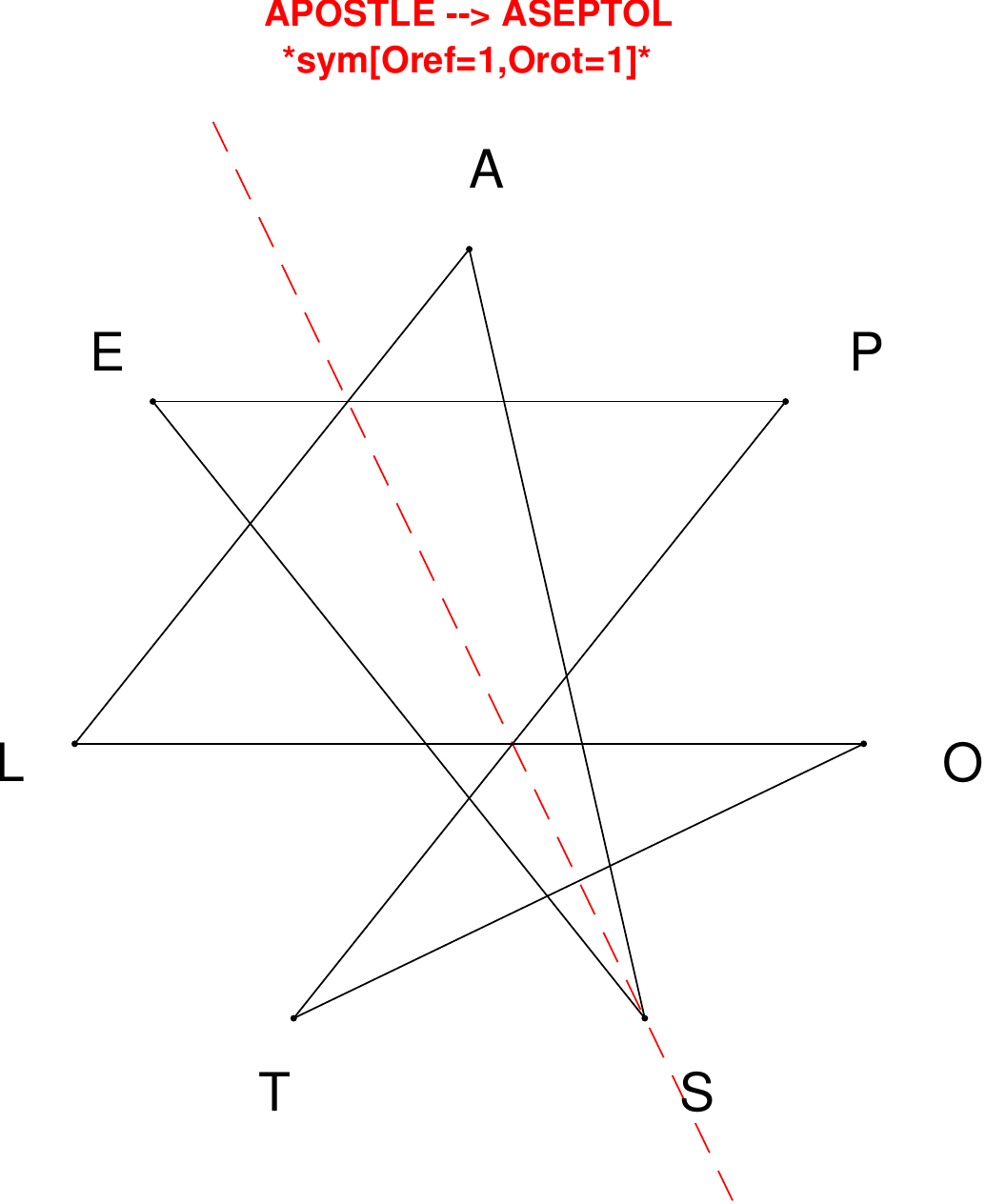}
\end{subfigure}
\hfill
\begin{subfigure}[T]{0.19\textwidth}
\centering
\includegraphics[width=\textwidth]{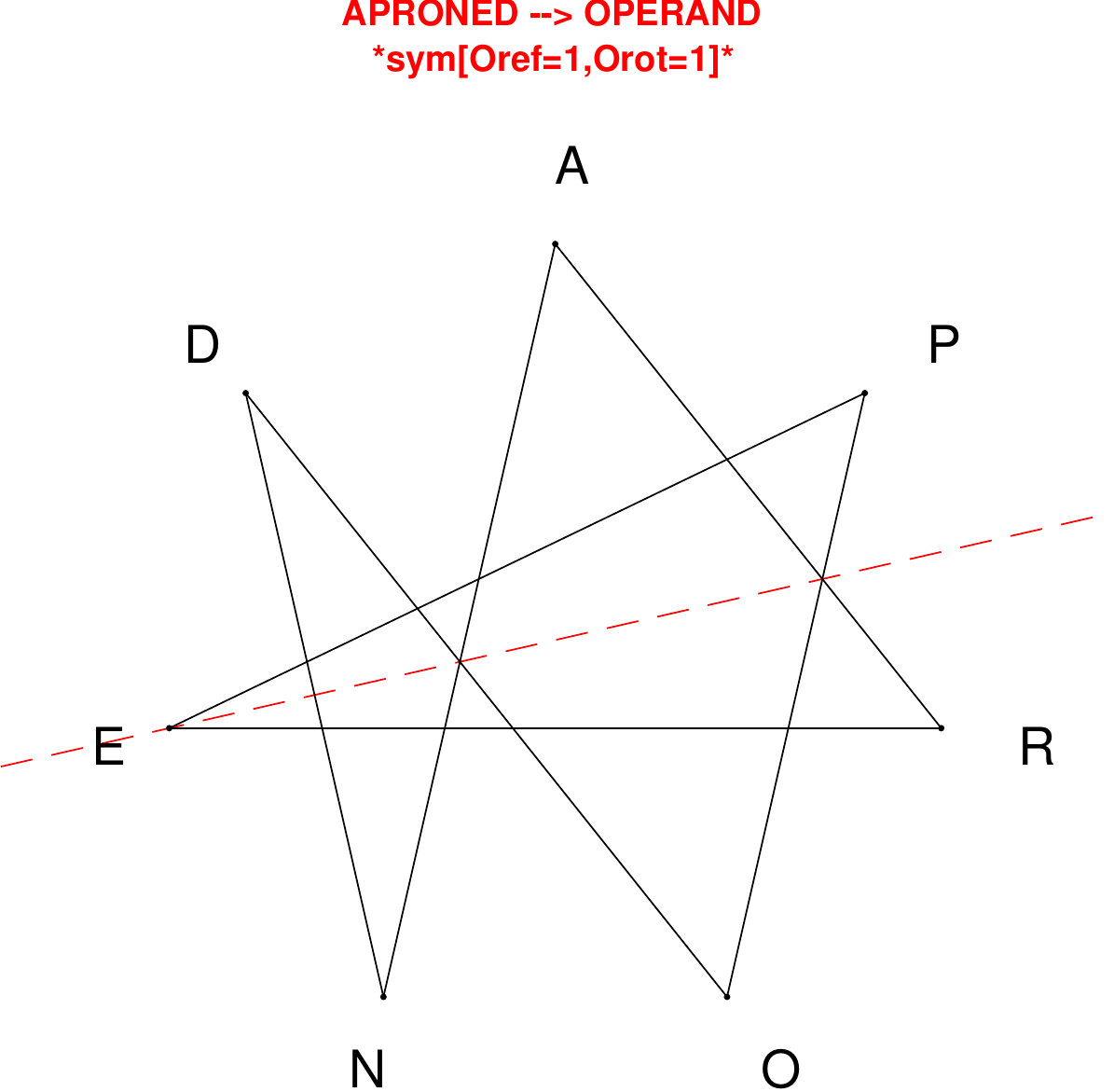}
\end{subfigure}
\hfill
\begin{subfigure}[T]{0.19\textwidth}
\centering
\includegraphics[width=\textwidth]{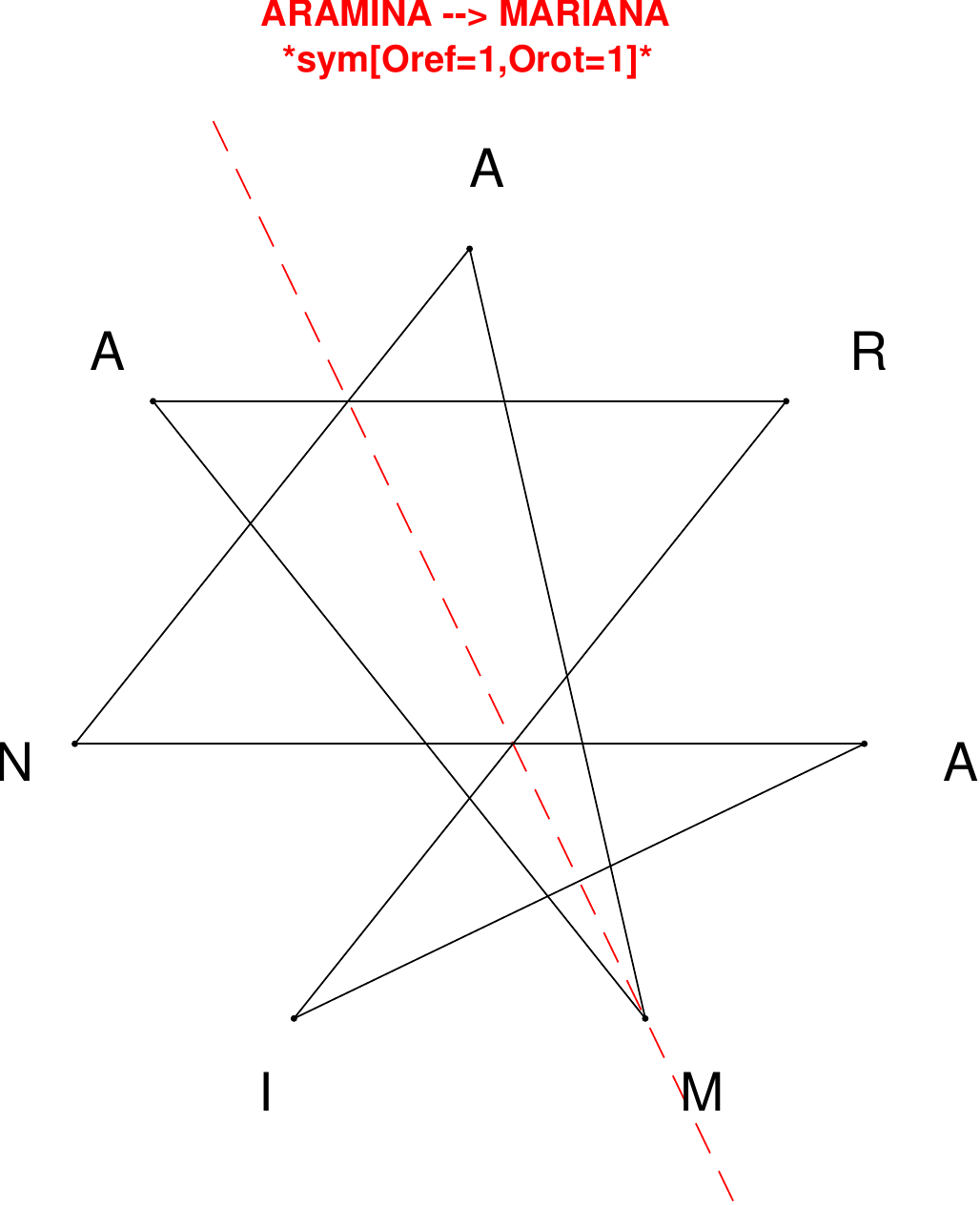}
\end{subfigure}
\hfill
\begin{subfigure}[T]{0.19\textwidth}
\centering
\includegraphics[width=\textwidth]{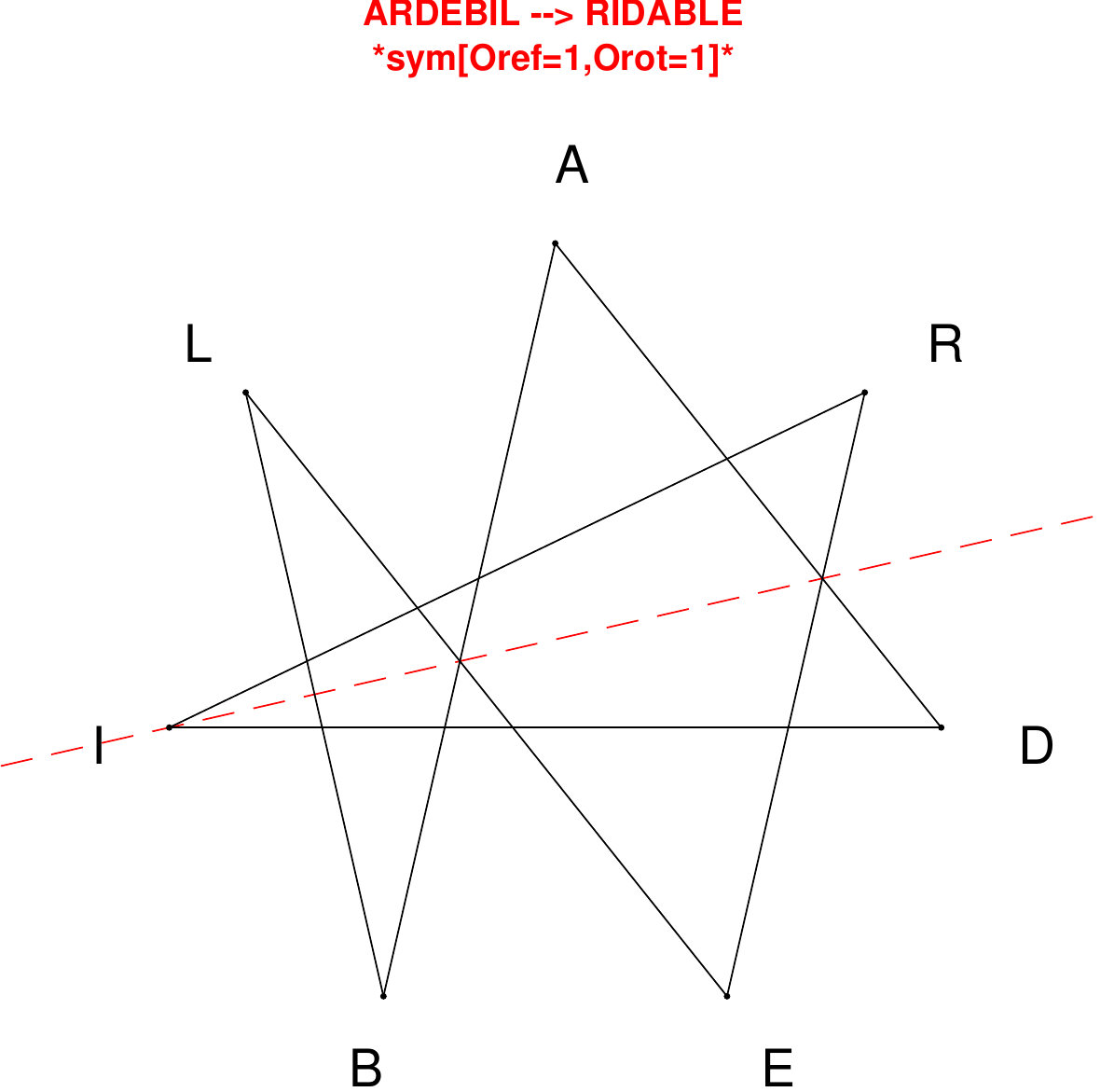}
\end{subfigure}
\hfill
\begin{subfigure}[T]{0.19\textwidth}
\centering
\includegraphics[width=\textwidth]{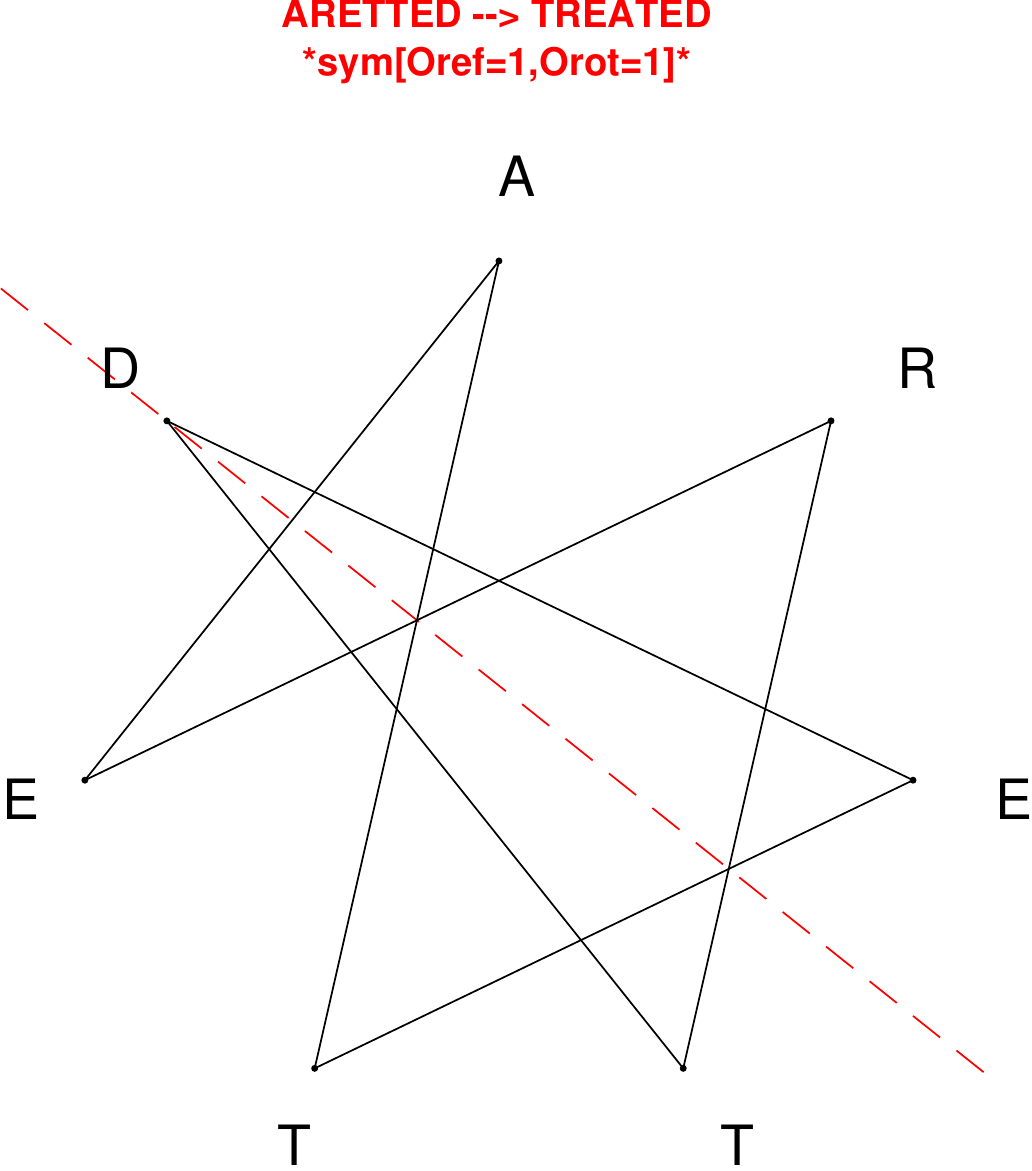}
\end{subfigure}
\end{figure}

\begin{figure}[H]
\centering
\begin{subfigure}[T]{0.19\textwidth}
\centering
\includegraphics[width=\textwidth]{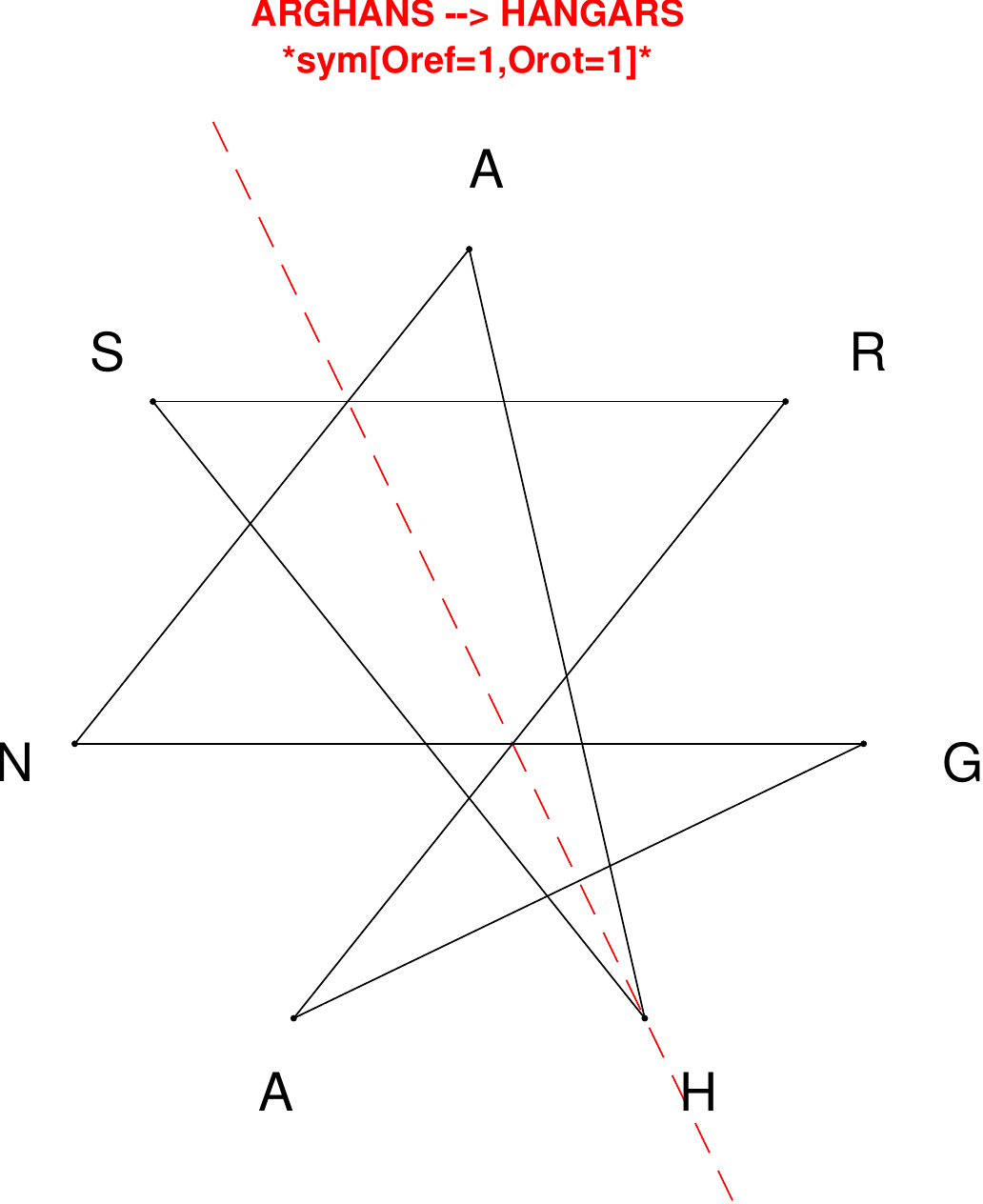}
\end{subfigure}
\hfill
\begin{subfigure}[T]{0.19\textwidth}
\centering
\includegraphics[width=\textwidth]{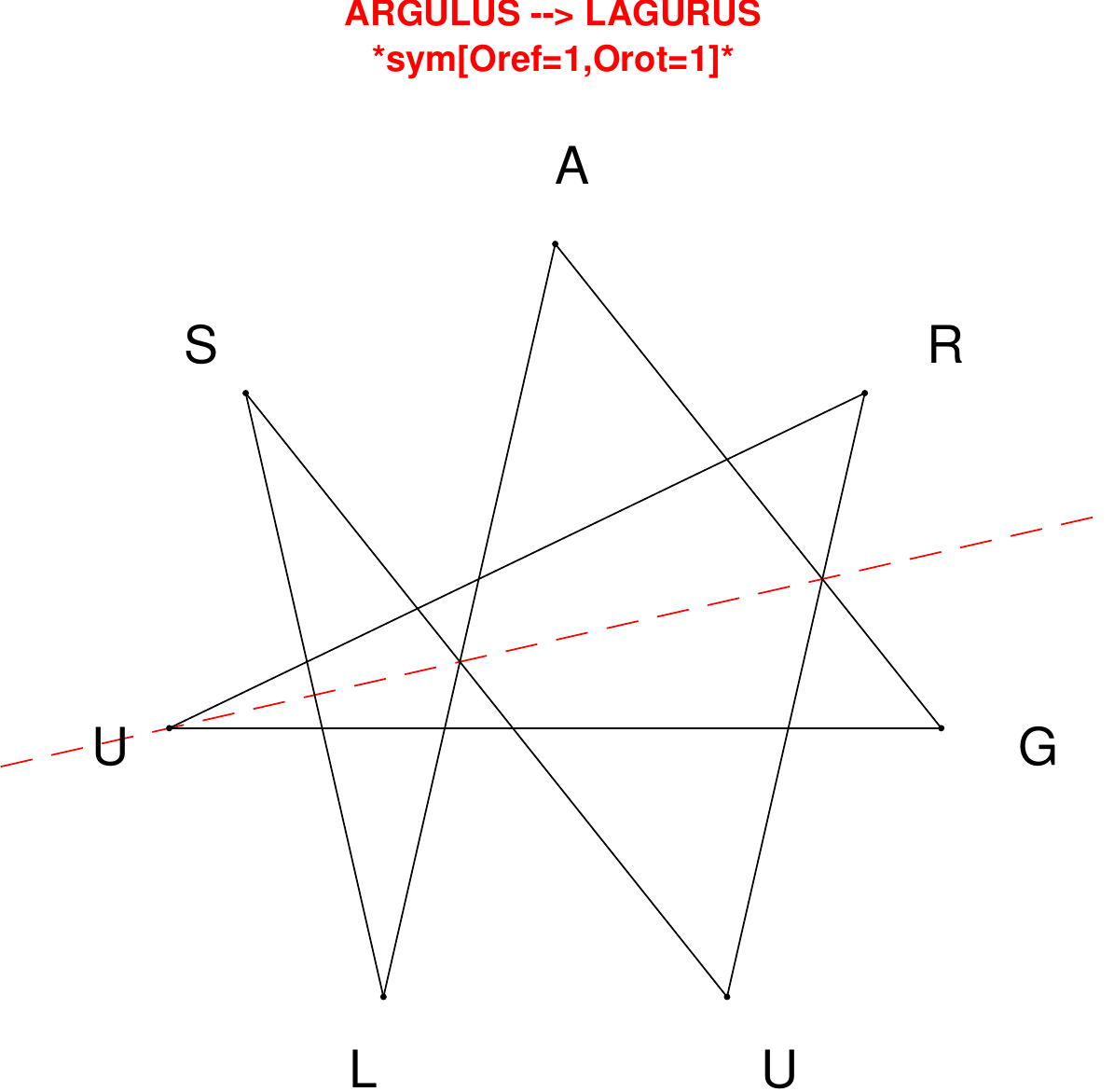}
\end{subfigure}
\hfill
\begin{subfigure}[T]{0.19\textwidth}
\centering
\includegraphics[width=\textwidth]{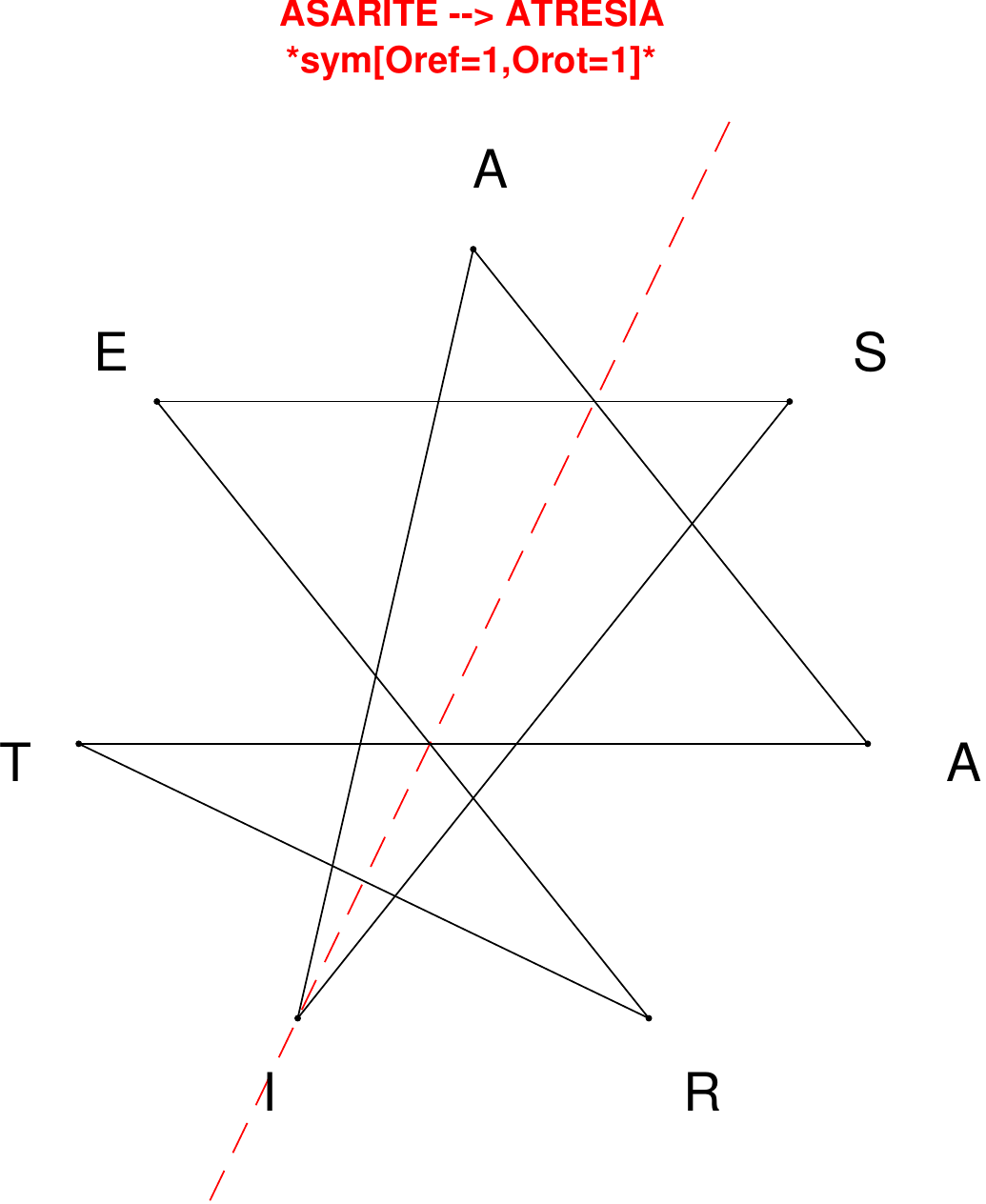}
\end{subfigure}
\hfill
\begin{subfigure}[T]{0.19\textwidth}
\centering
\includegraphics[width=\textwidth]{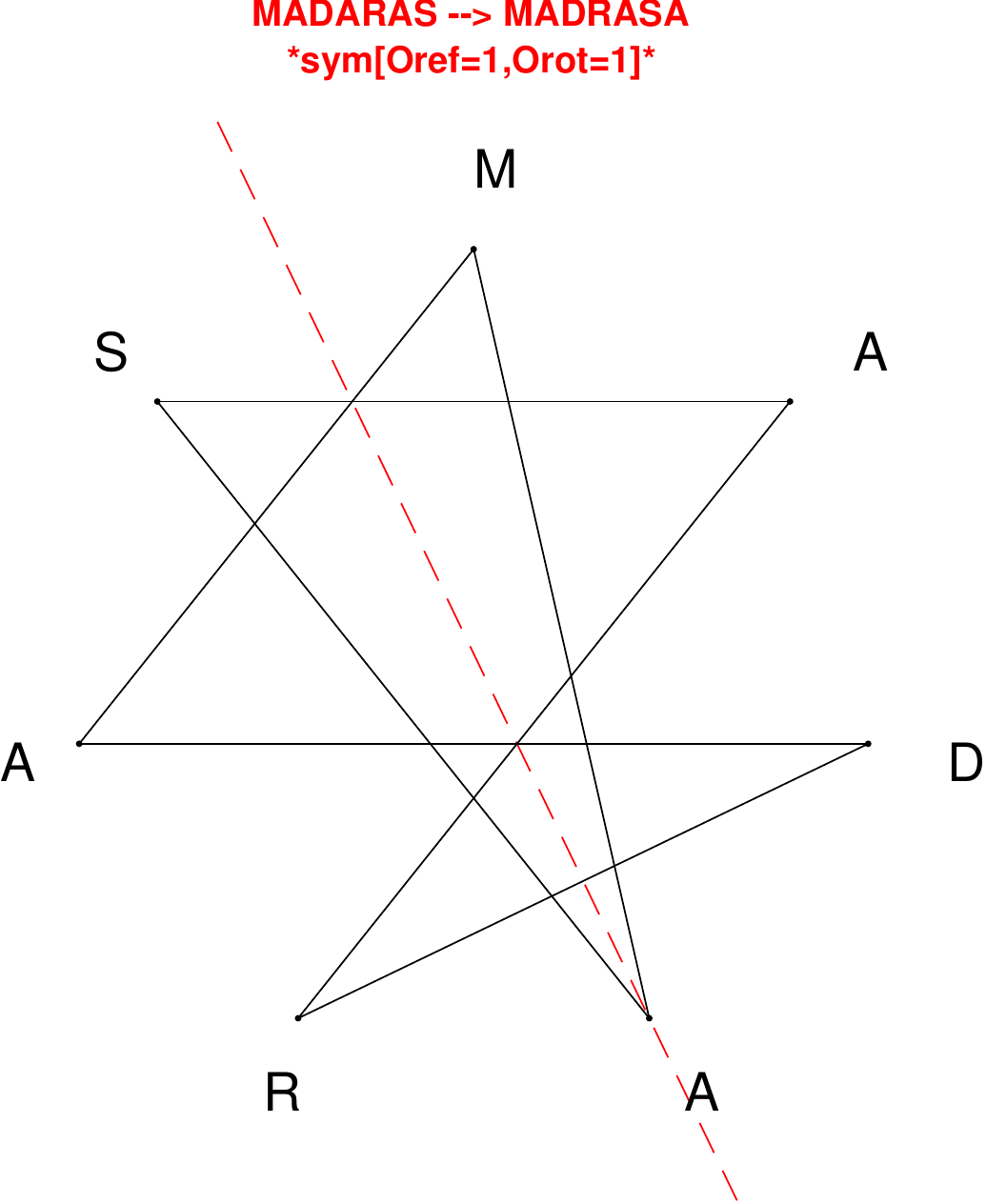}
\end{subfigure}
\hfill
\begin{subfigure}[T]{0.19\textwidth}
\centering
\includegraphics[width=\textwidth]{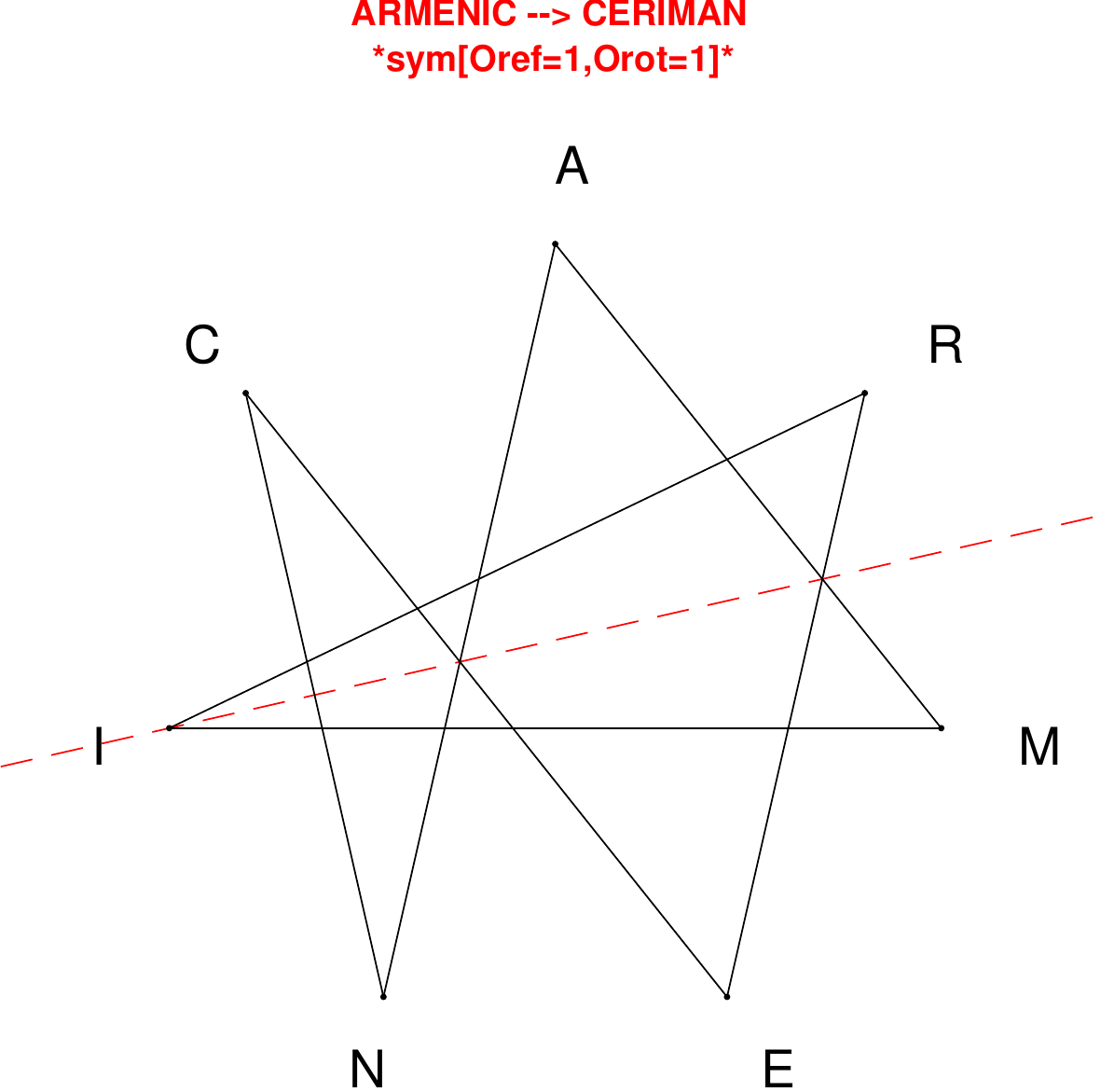}
\end{subfigure}
\end{figure}

\begin{figure}[H]
\centering
\begin{subfigure}[T]{0.19\textwidth}
\centering
\includegraphics[width=\textwidth]{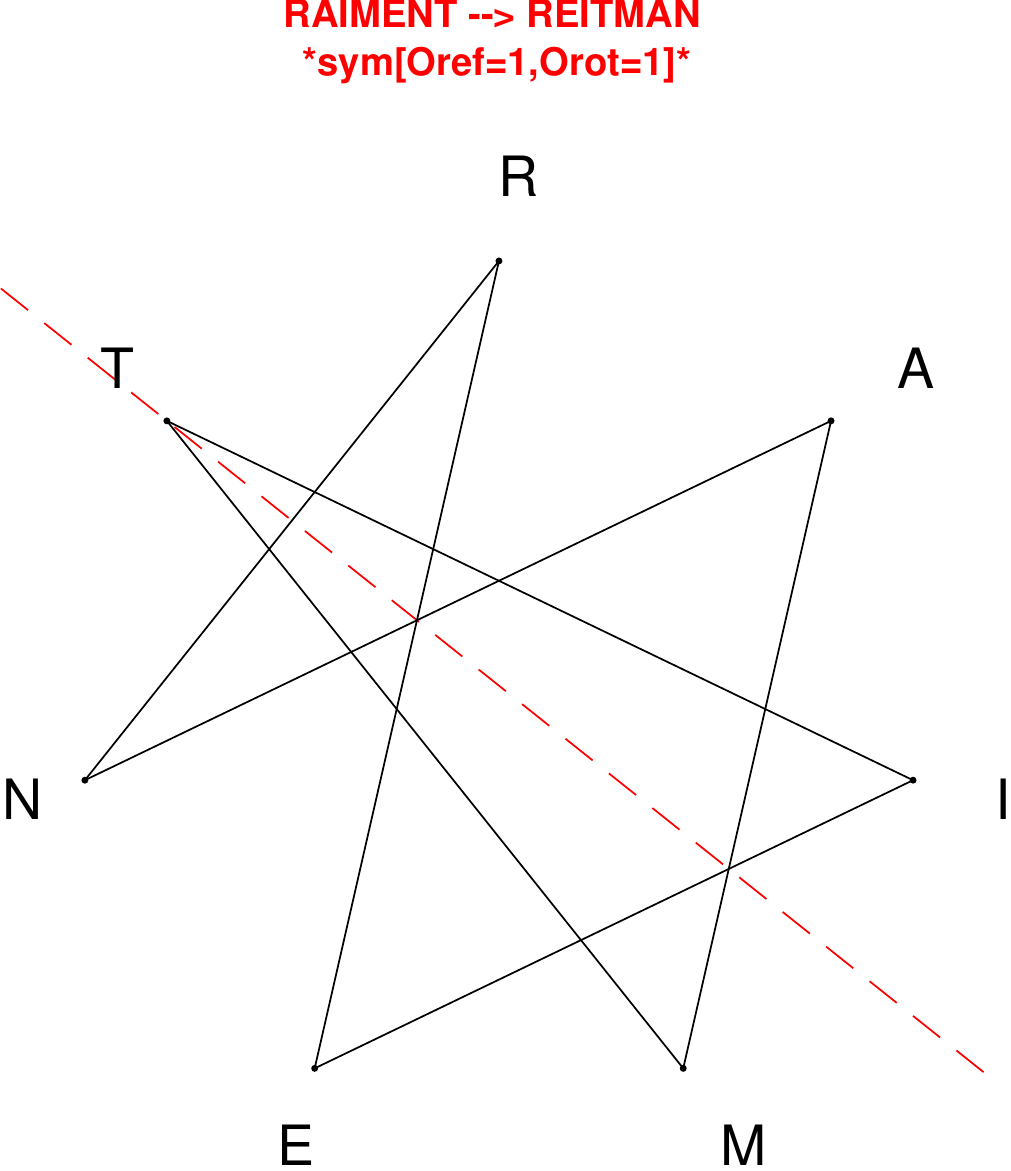}
\end{subfigure}
\hfill
\begin{subfigure}[T]{0.19\textwidth}
\centering
\includegraphics[width=\textwidth]{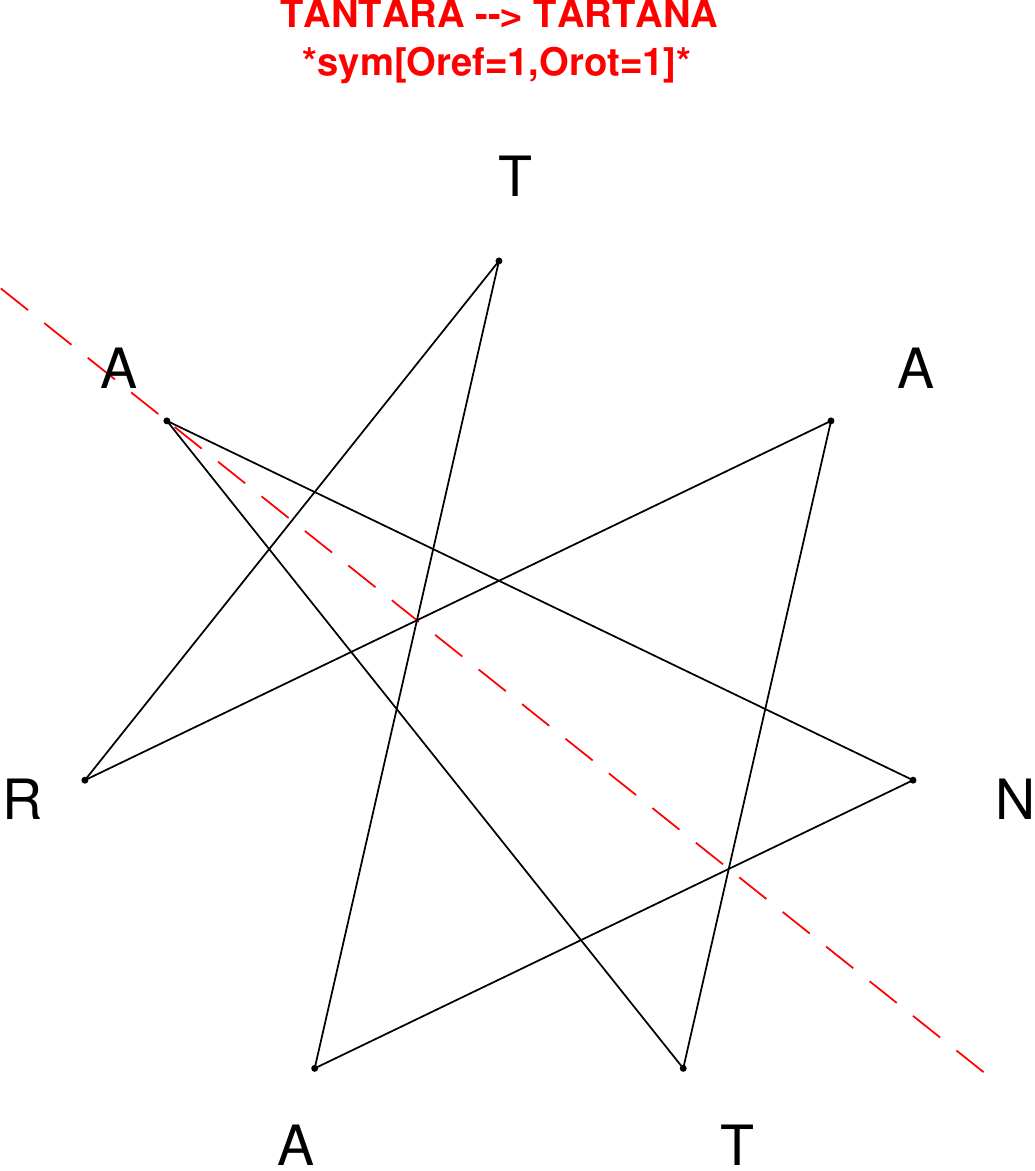}
\end{subfigure}
\hfill
\begin{subfigure}[T]{0.19\textwidth}
\centering
\includegraphics[width=\textwidth]{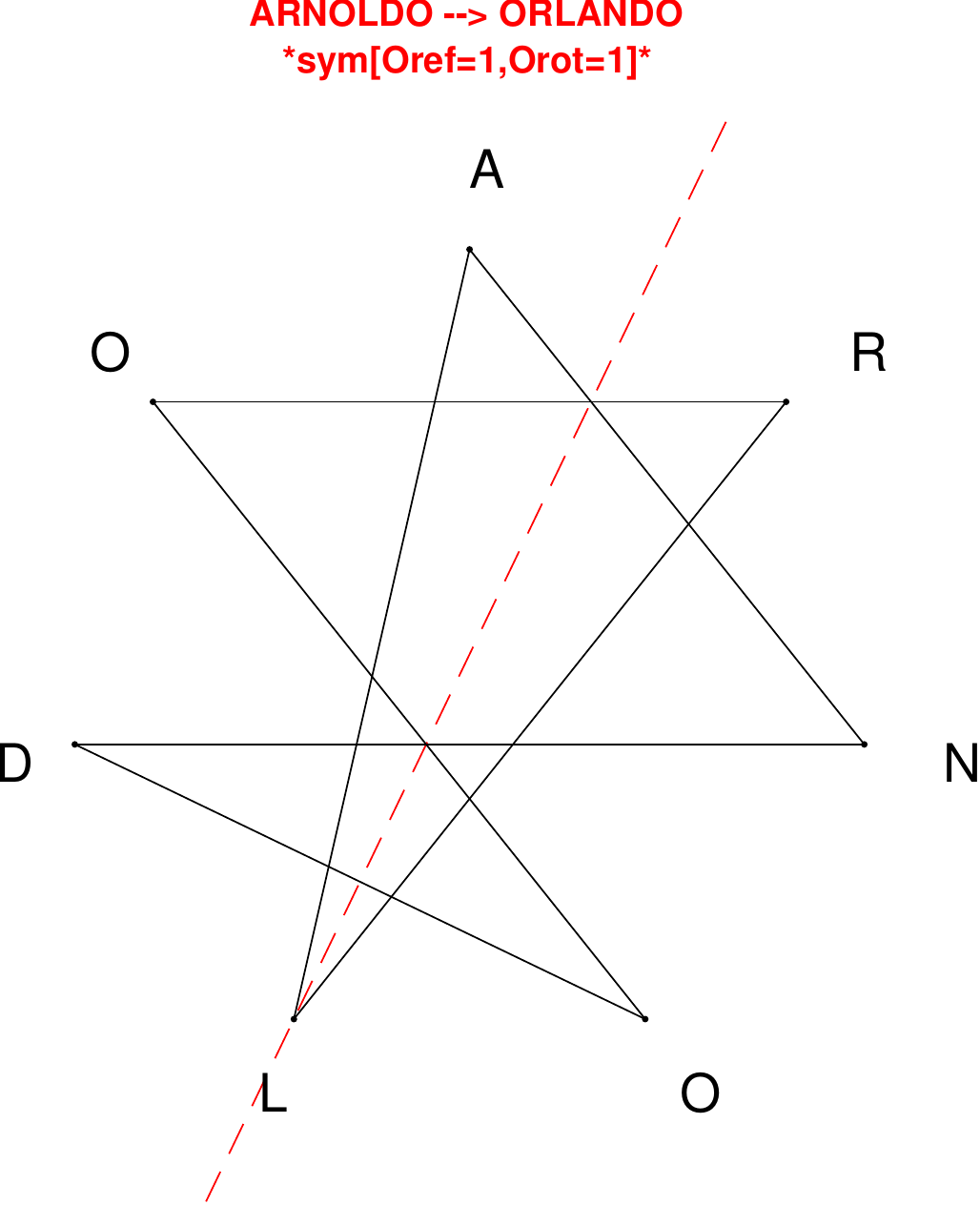}
\end{subfigure}
\hfill
\begin{subfigure}[T]{0.19\textwidth}
\centering
\includegraphics[width=\textwidth]{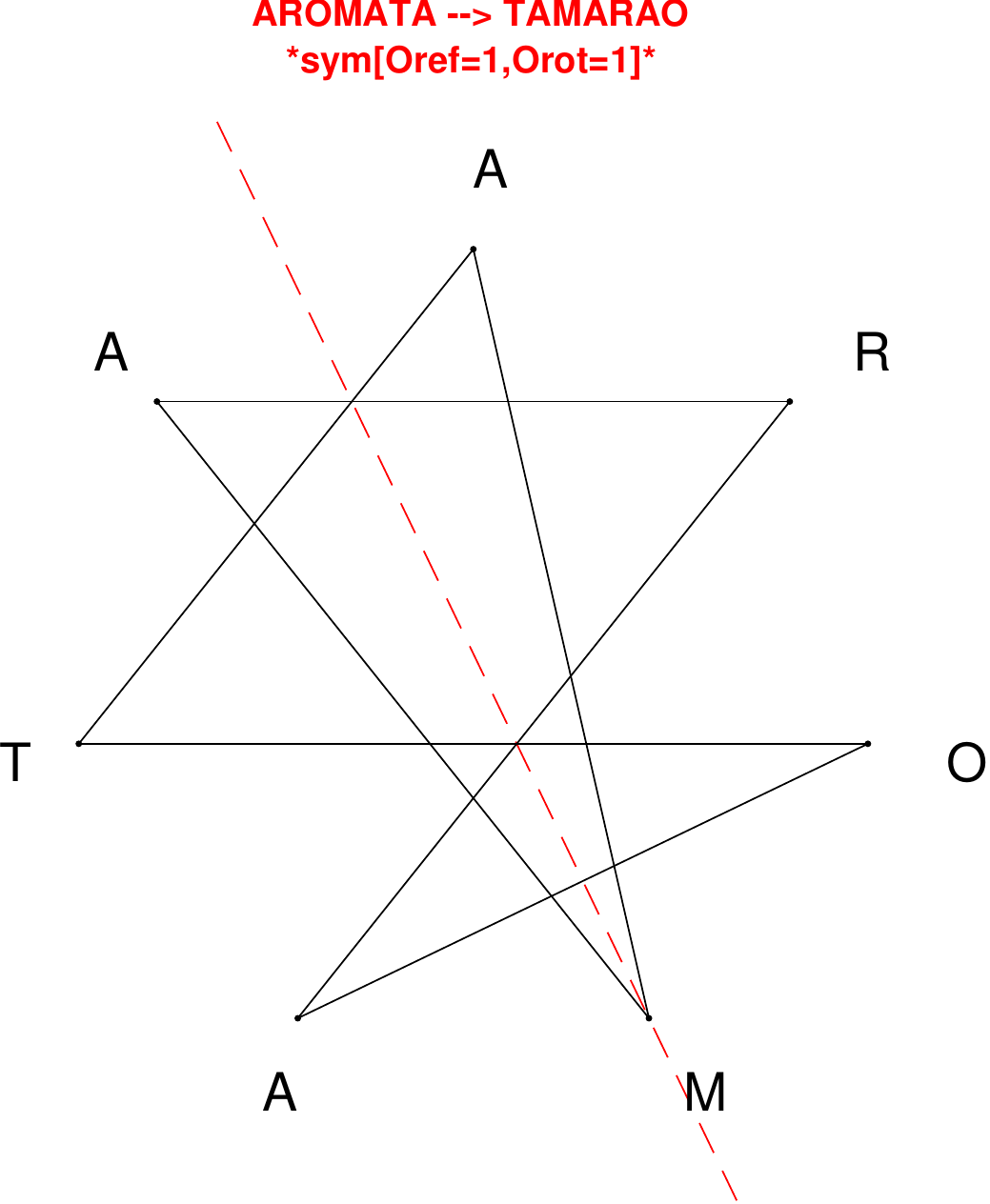}
\end{subfigure}
\hfill
\begin{subfigure}[T]{0.19\textwidth}
\centering
\includegraphics[width=\textwidth]{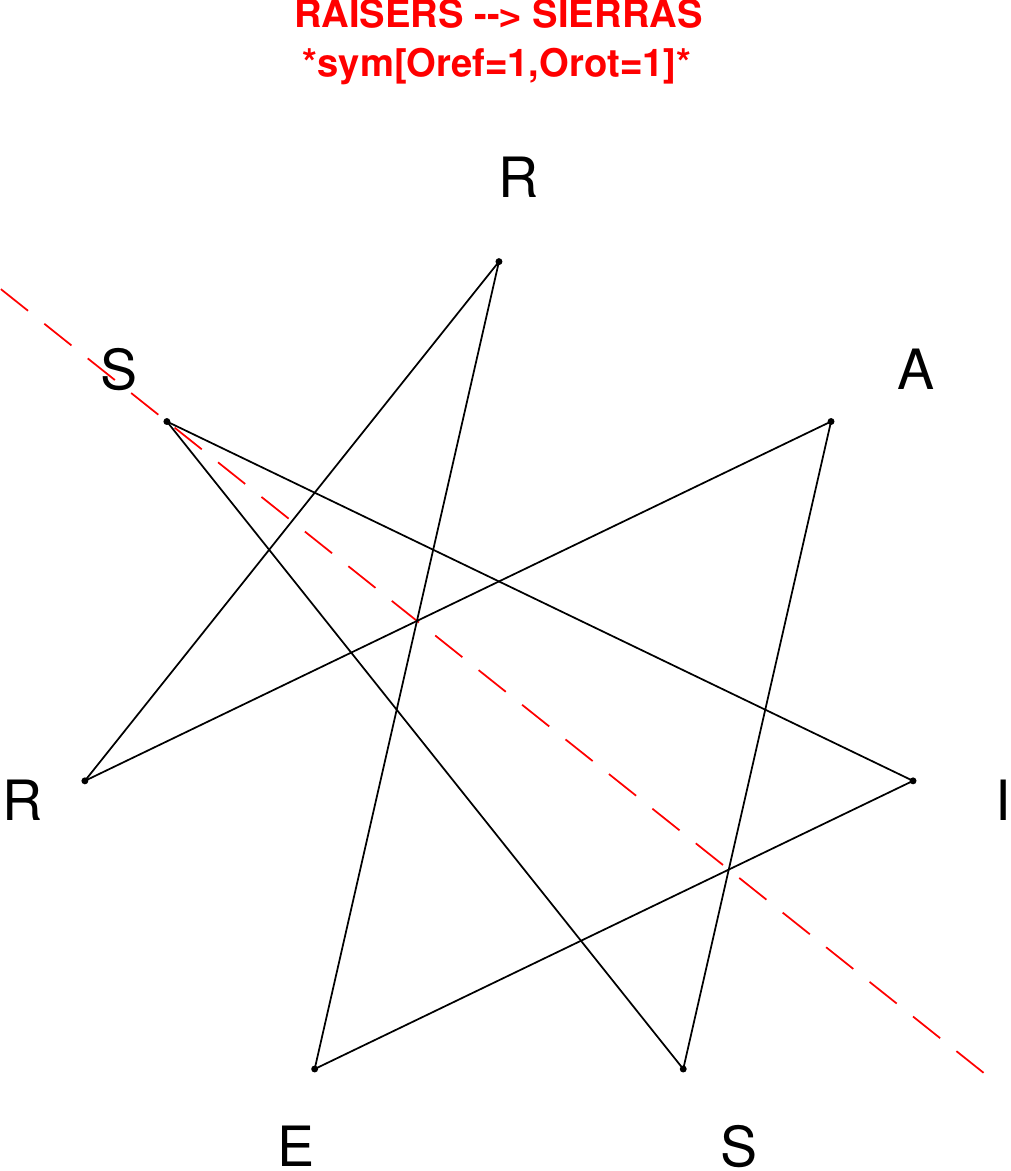}
\end{subfigure}
\end{figure}

\begin{figure}[H]
\centering
\begin{subfigure}[T]{0.19\textwidth}
\centering
\includegraphics[width=\textwidth]{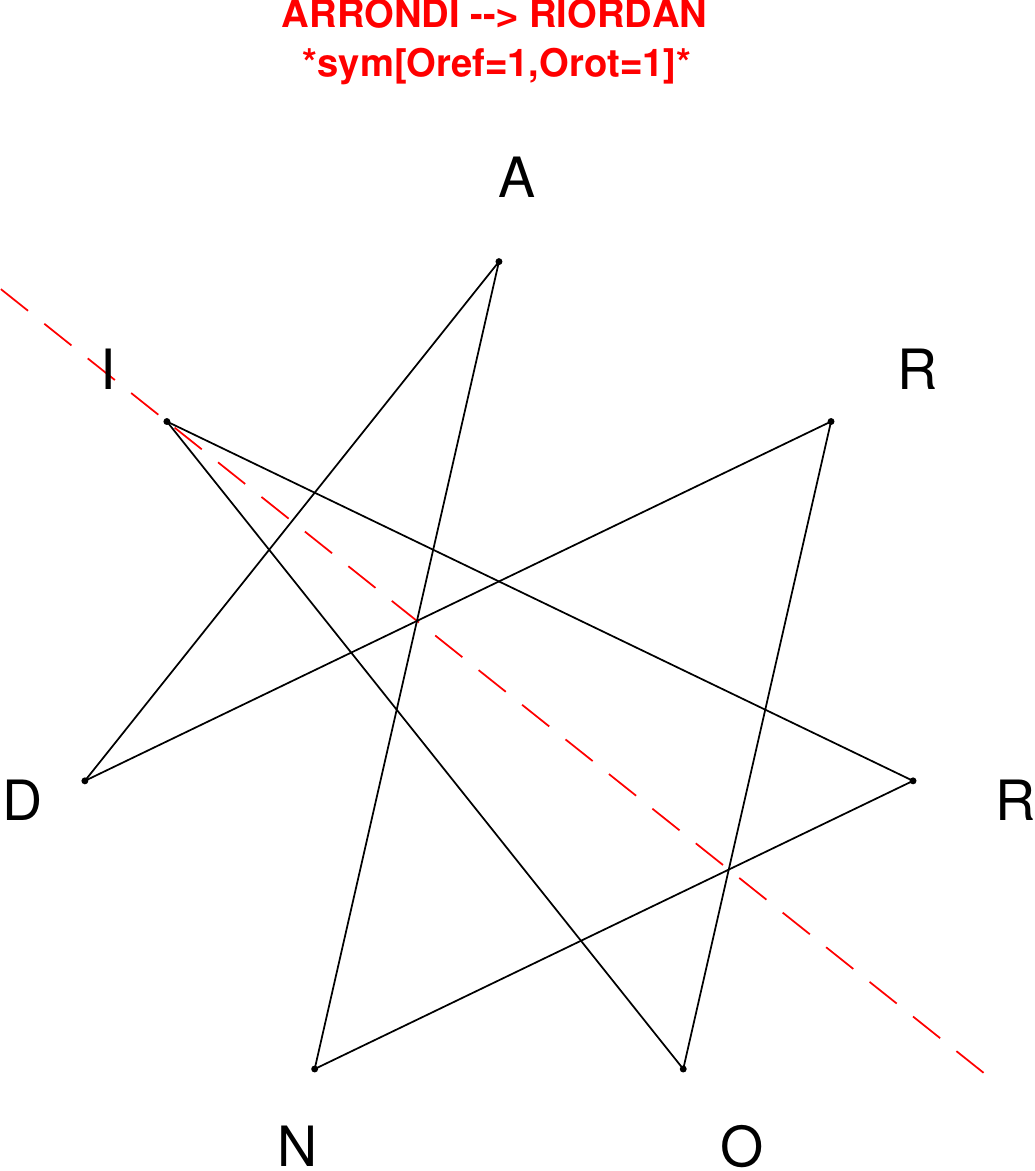}
\end{subfigure}
\hfill
\begin{subfigure}[T]{0.19\textwidth}
\centering
\includegraphics[width=\textwidth]{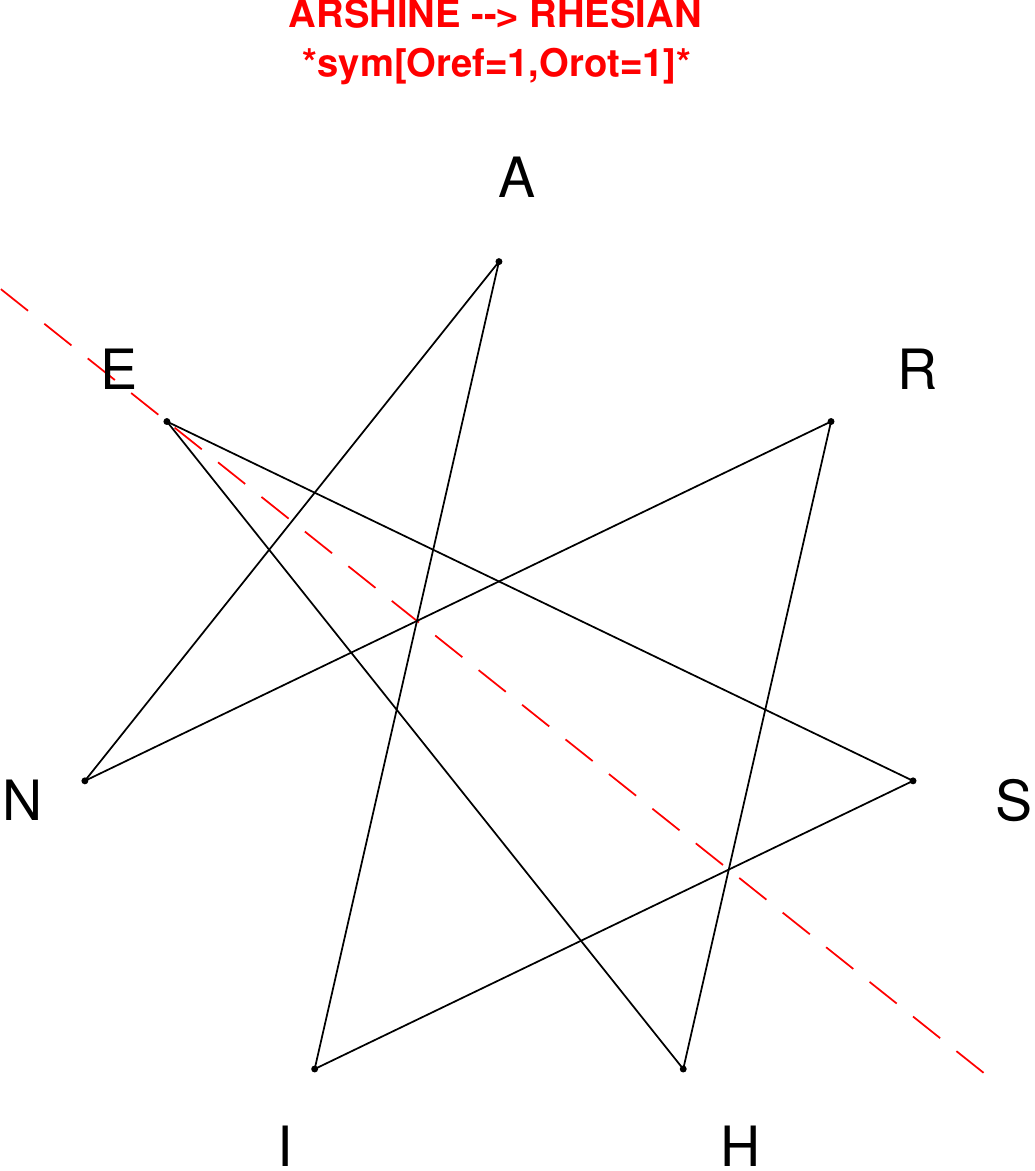}
\end{subfigure}
\hfill
\begin{subfigure}[T]{0.19\textwidth}
\centering
\includegraphics[width=\textwidth]{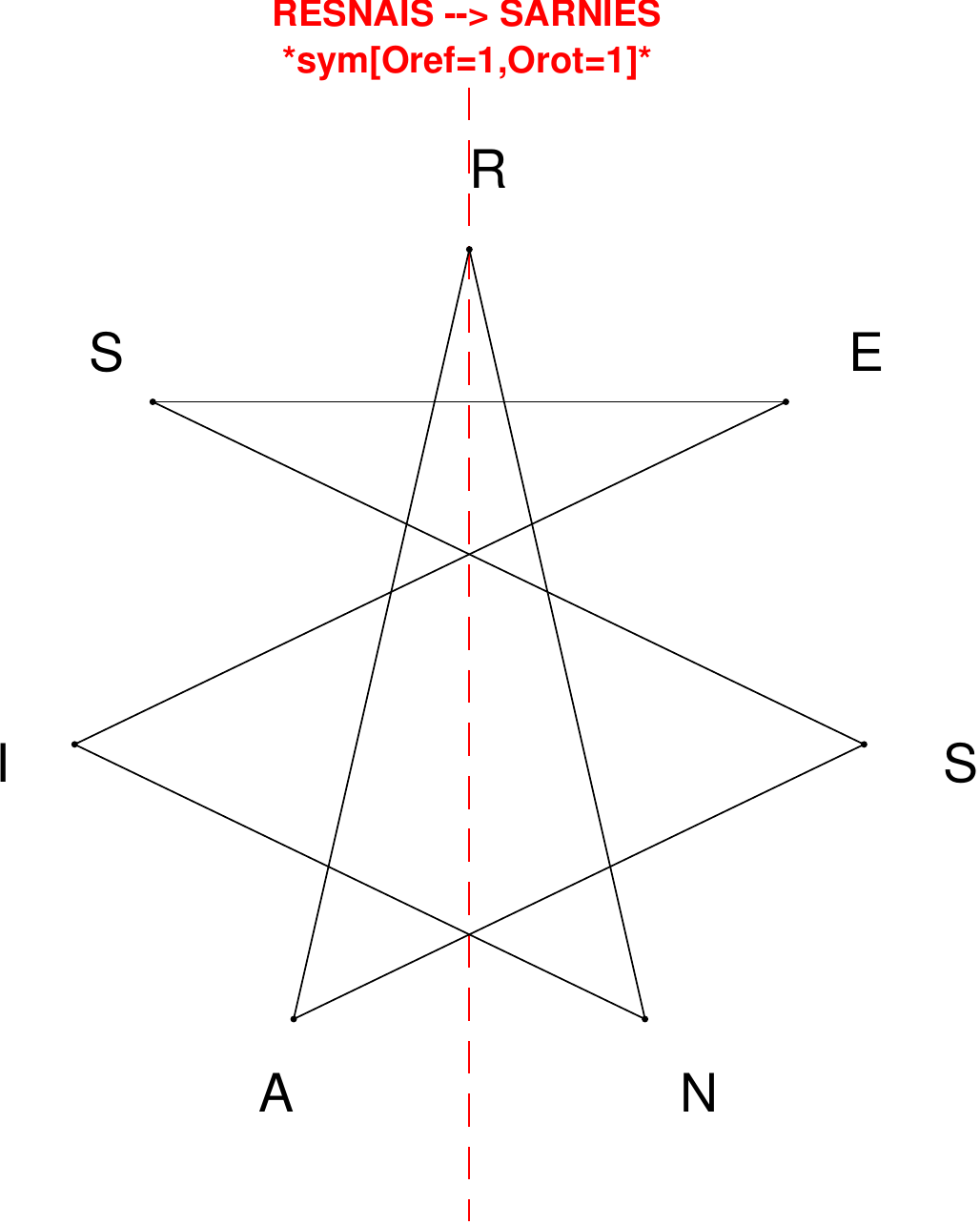}
\end{subfigure}
\hfill
\begin{subfigure}[T]{0.19\textwidth}
\centering
\includegraphics[width=\textwidth]{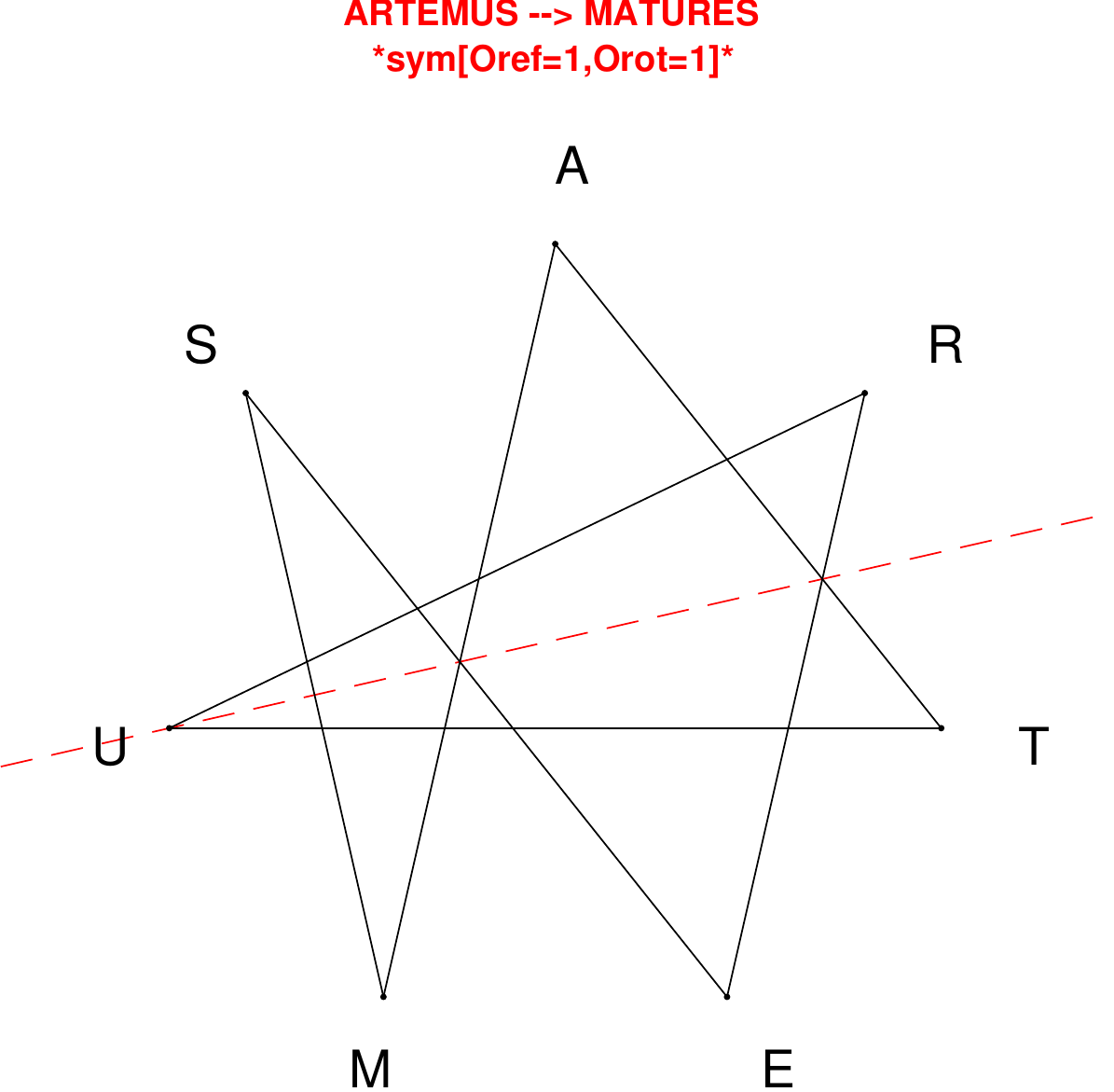}
\end{subfigure}
\hfill
\begin{subfigure}[T]{0.19\textwidth}
\centering
\includegraphics[width=\textwidth]{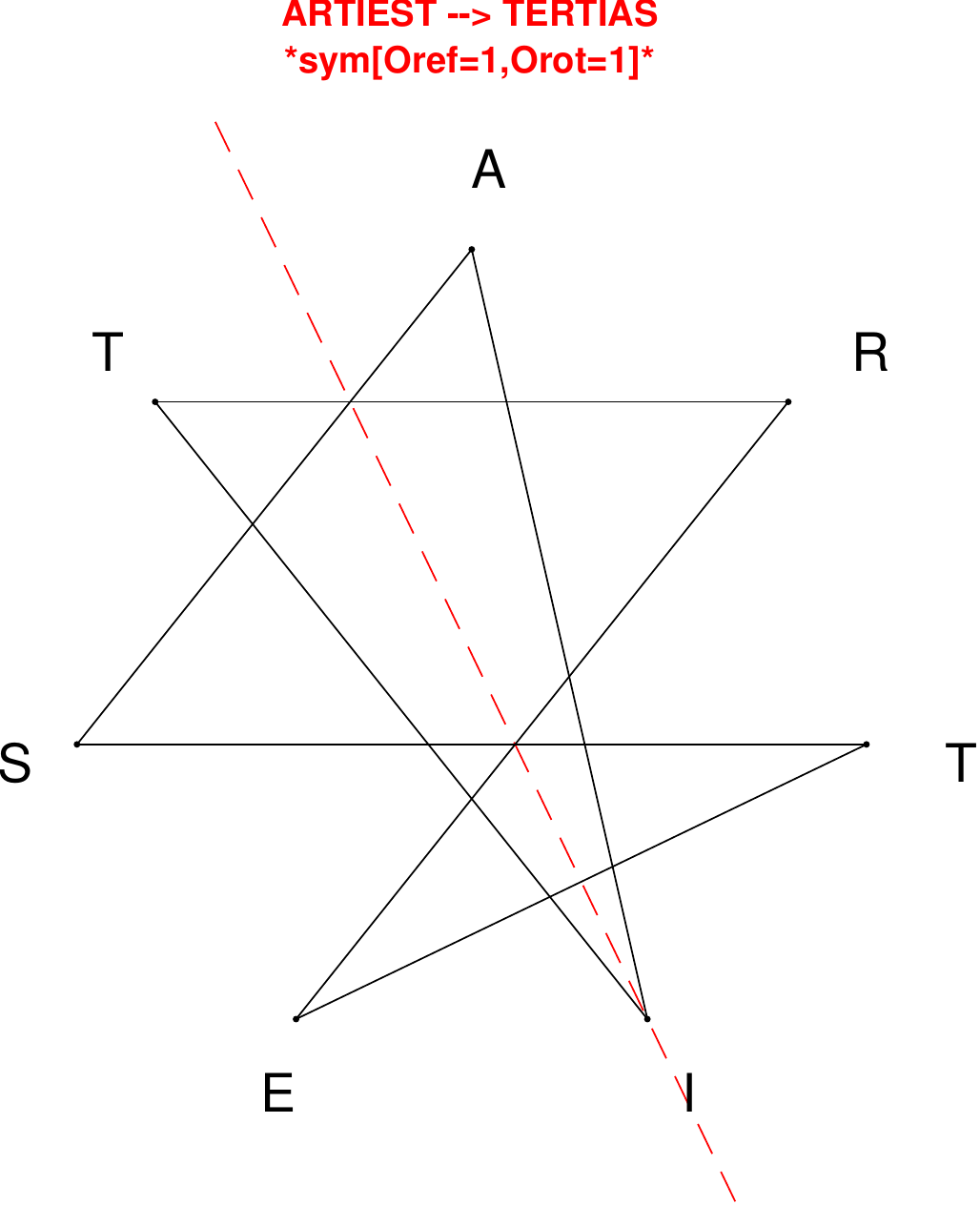}
\end{subfigure}
\end{figure}

\begin{figure}[H]
\centering
\begin{subfigure}[T]{0.19\textwidth}
\centering
\includegraphics[width=\textwidth]{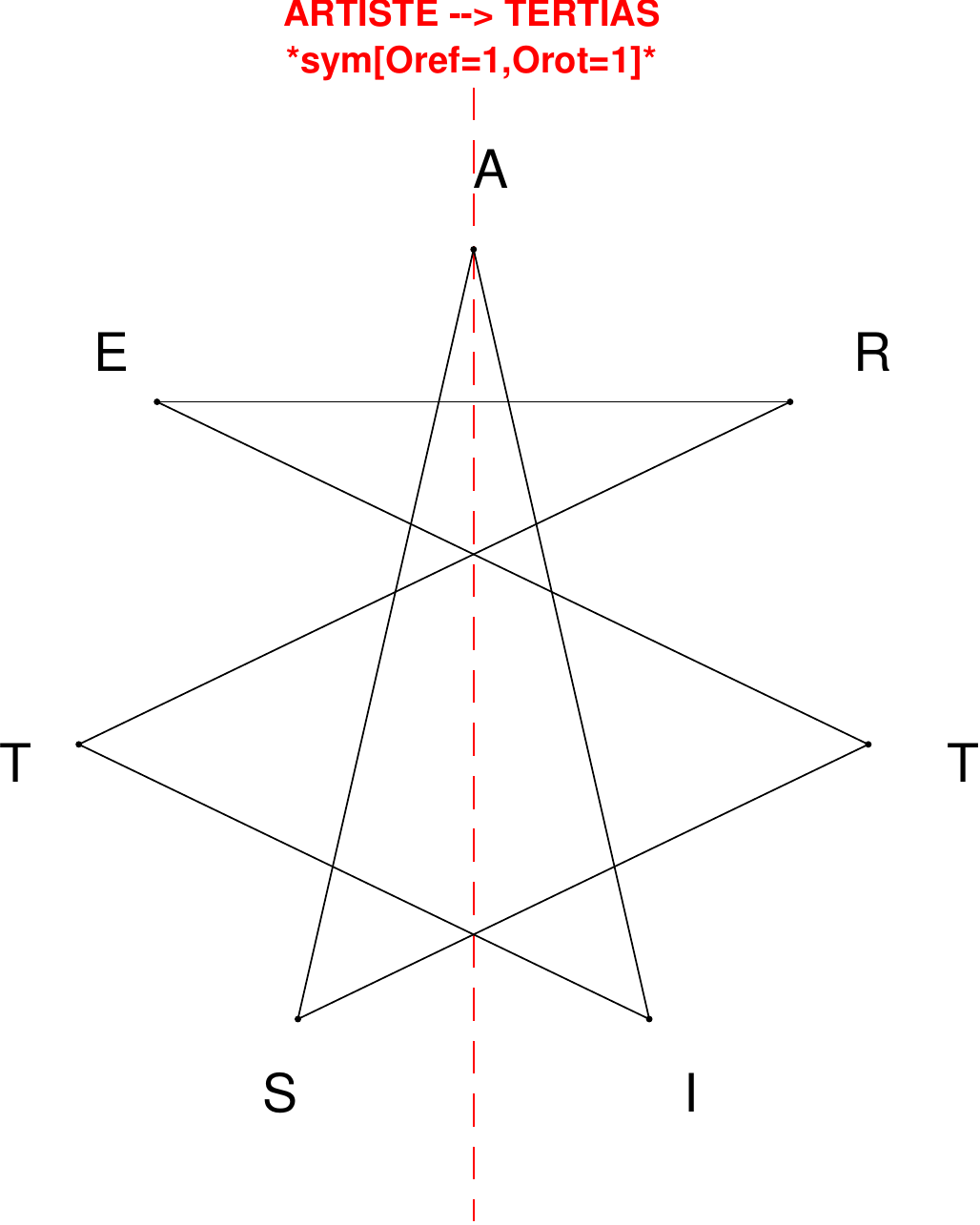}
\end{subfigure}
\hfill
\begin{subfigure}[T]{0.19\textwidth}
\centering
\includegraphics[width=\textwidth]{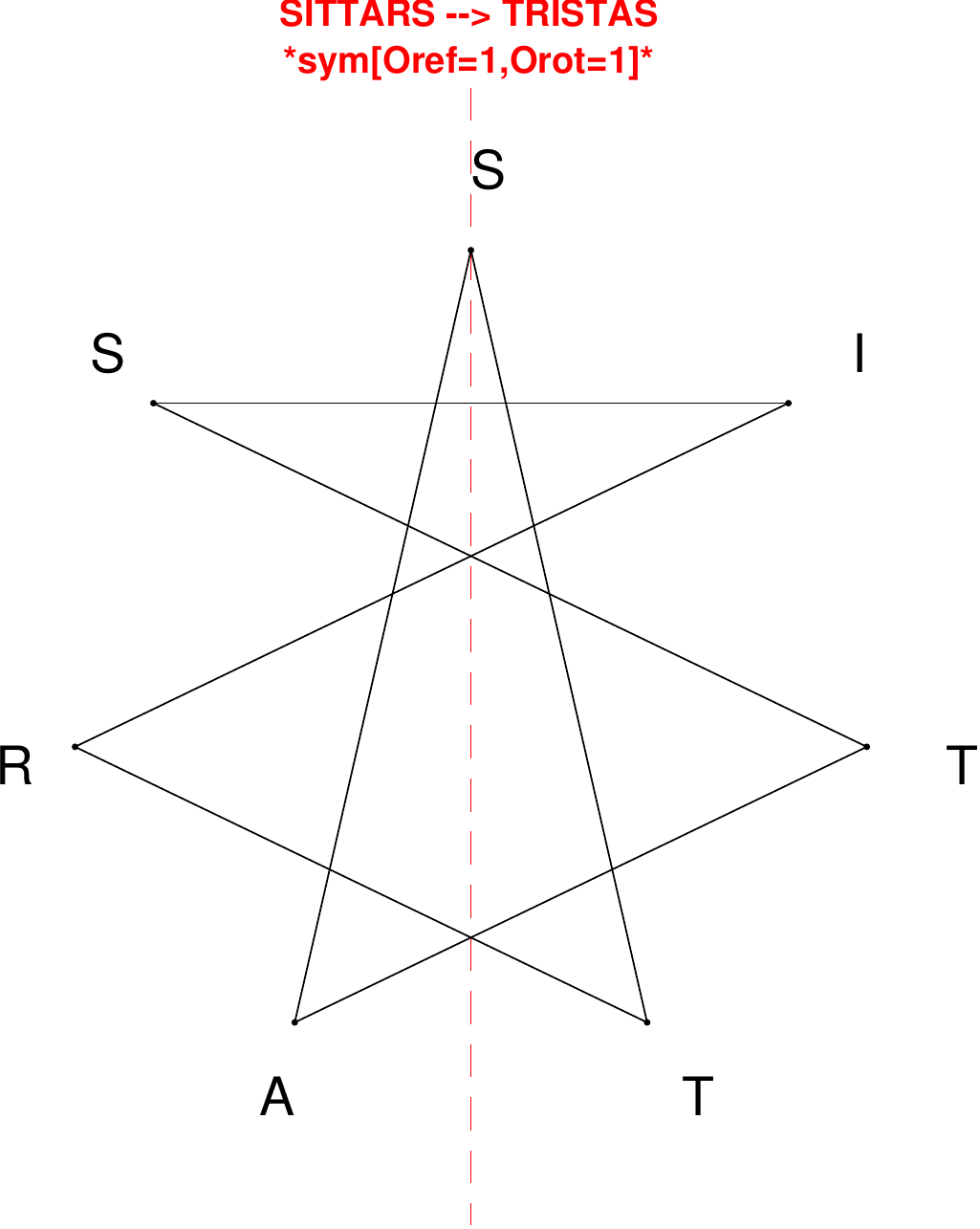}
\end{subfigure}
\hfill
\begin{subfigure}[T]{0.19\textwidth}
\centering
\includegraphics[width=\textwidth]{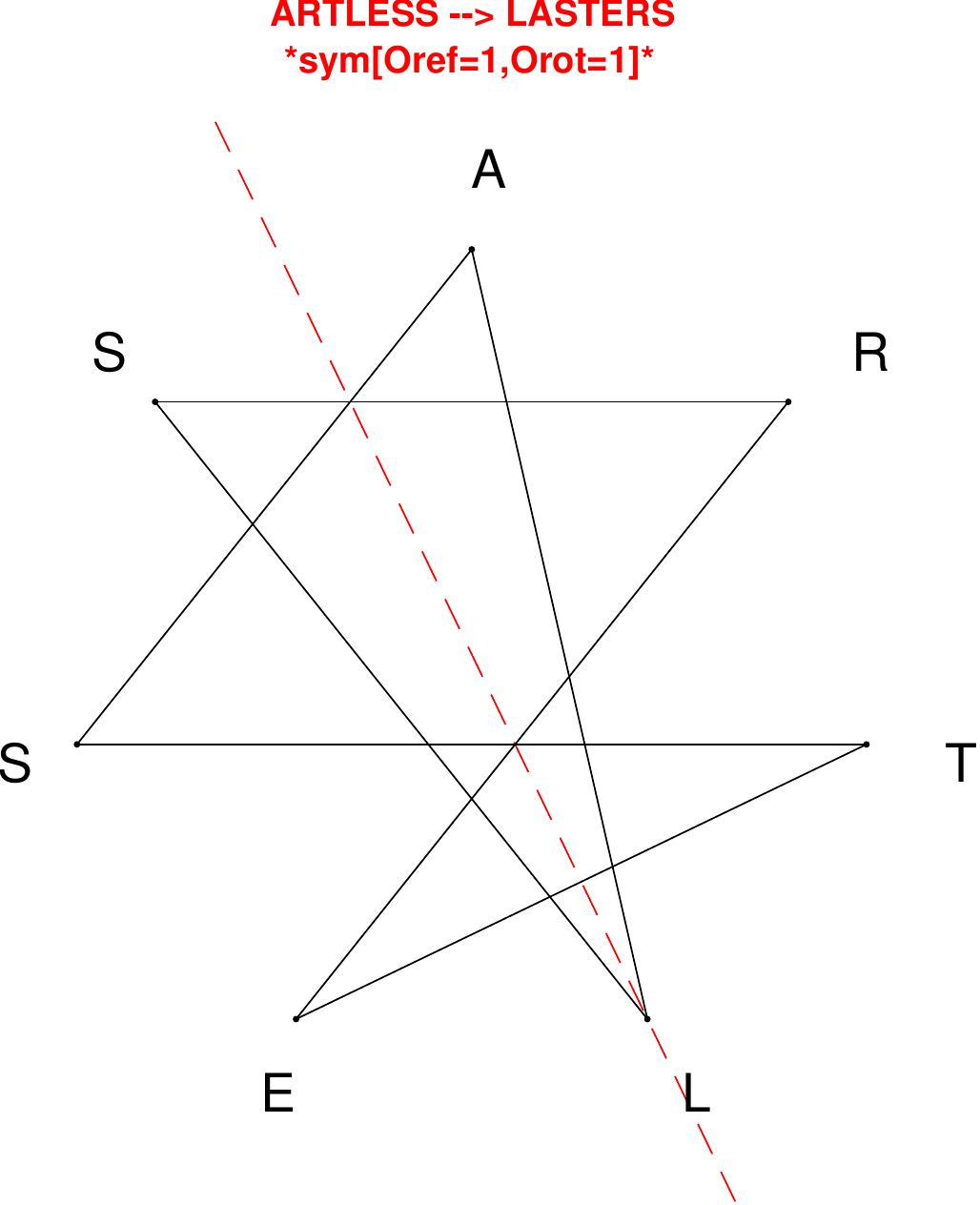}
\end{subfigure}
\hfill
\begin{subfigure}[T]{0.19\textwidth}
\centering
\includegraphics[width=\textwidth]{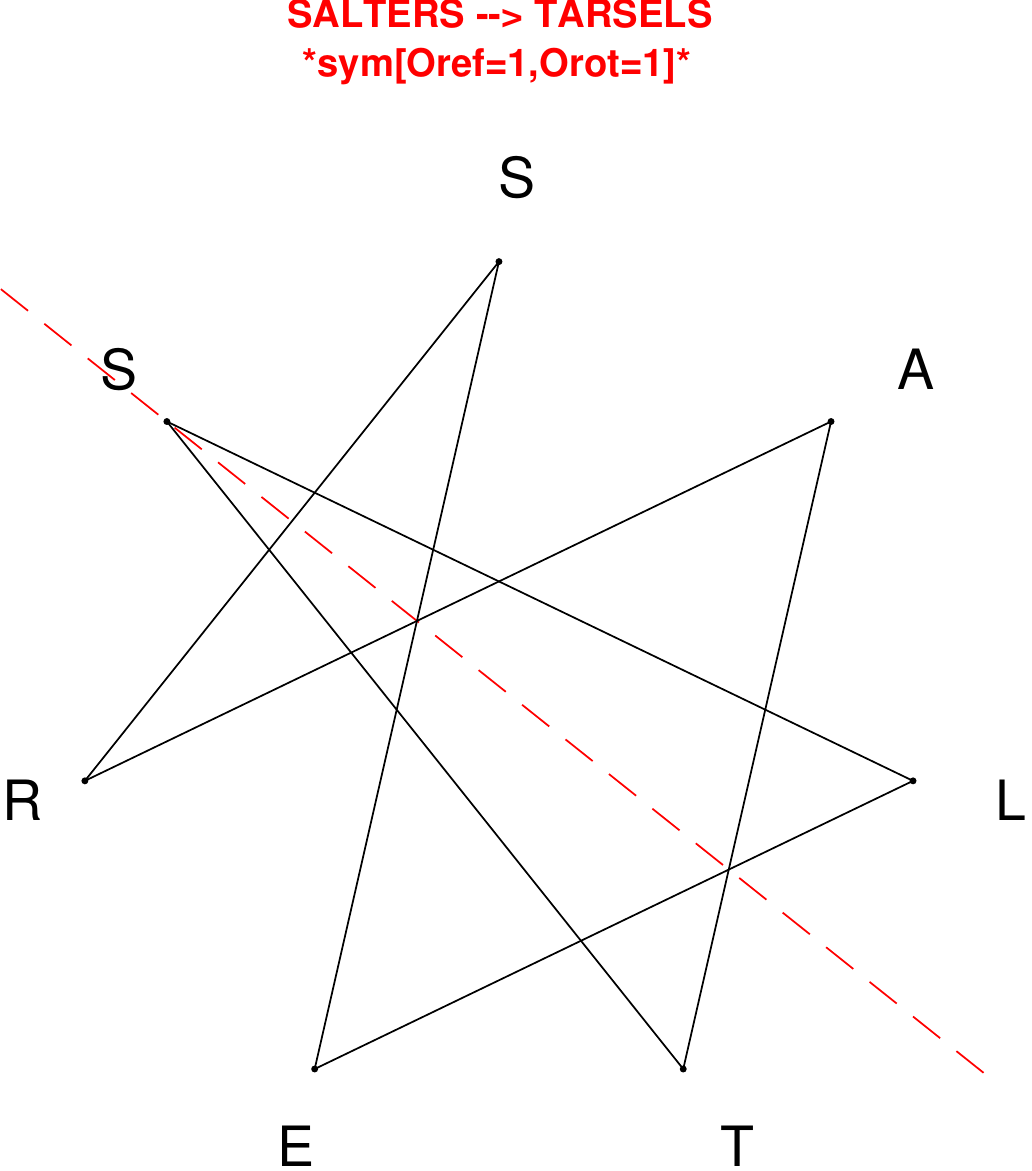}
\end{subfigure}
\hfill
\begin{subfigure}[T]{0.19\textwidth}
\centering
\includegraphics[width=\textwidth]{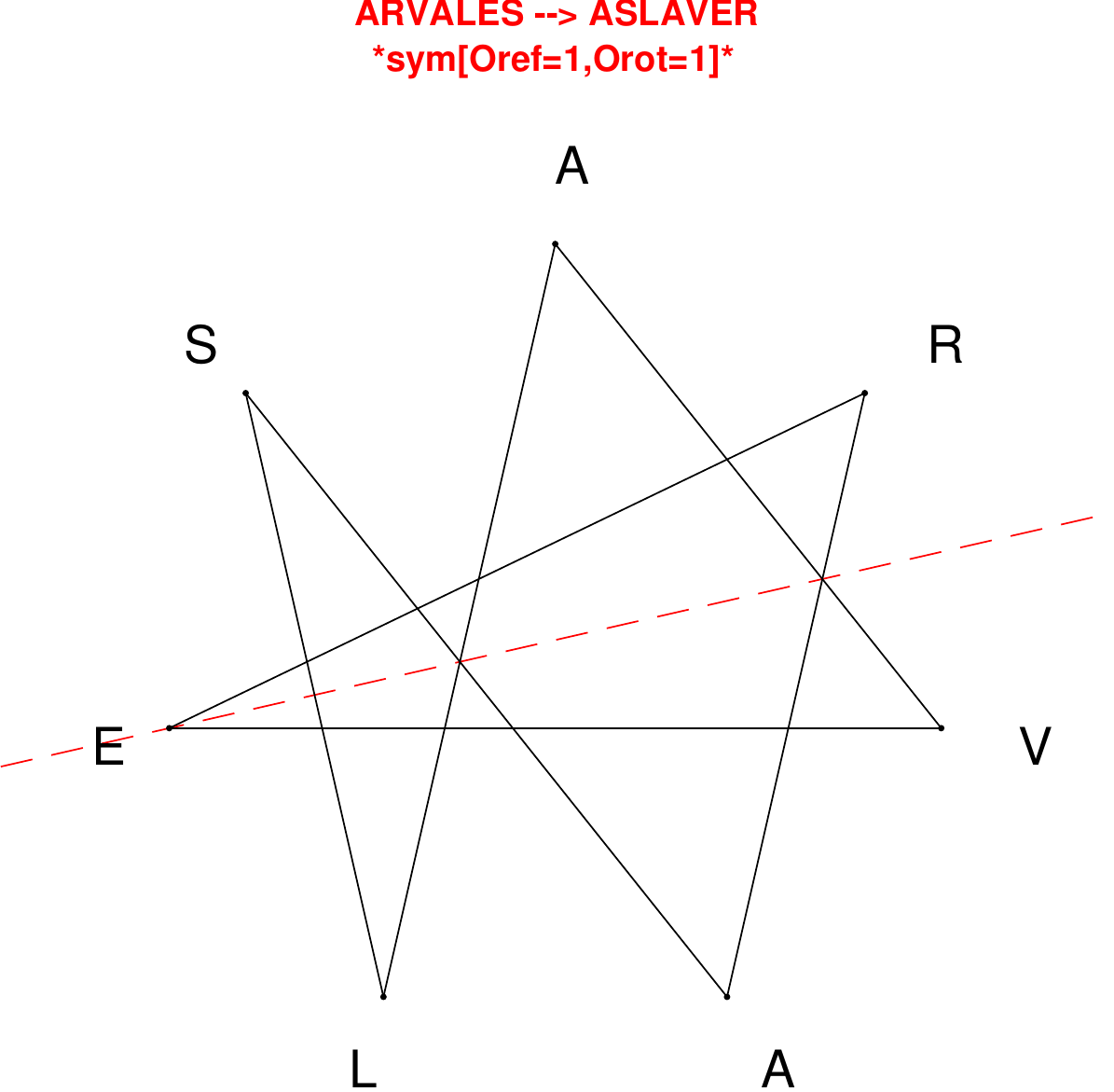}
\end{subfigure}
\end{figure}

\begin{figure}[H]
\centering
\begin{subfigure}[T]{0.19\textwidth}
\centering
\includegraphics[width=\textwidth]{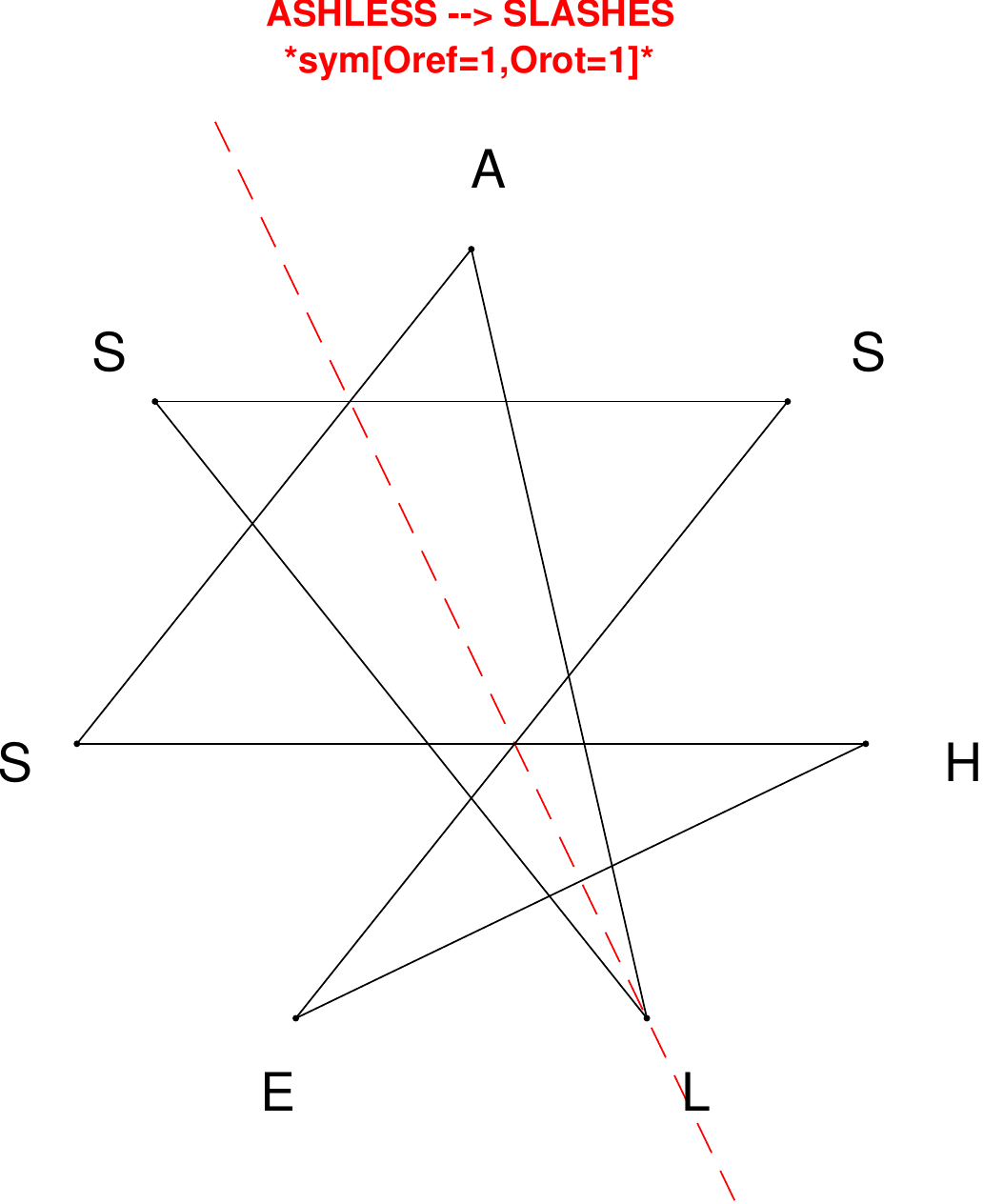}
\end{subfigure}
\hfill
\begin{subfigure}[T]{0.19\textwidth}
\centering
\includegraphics[width=\textwidth]{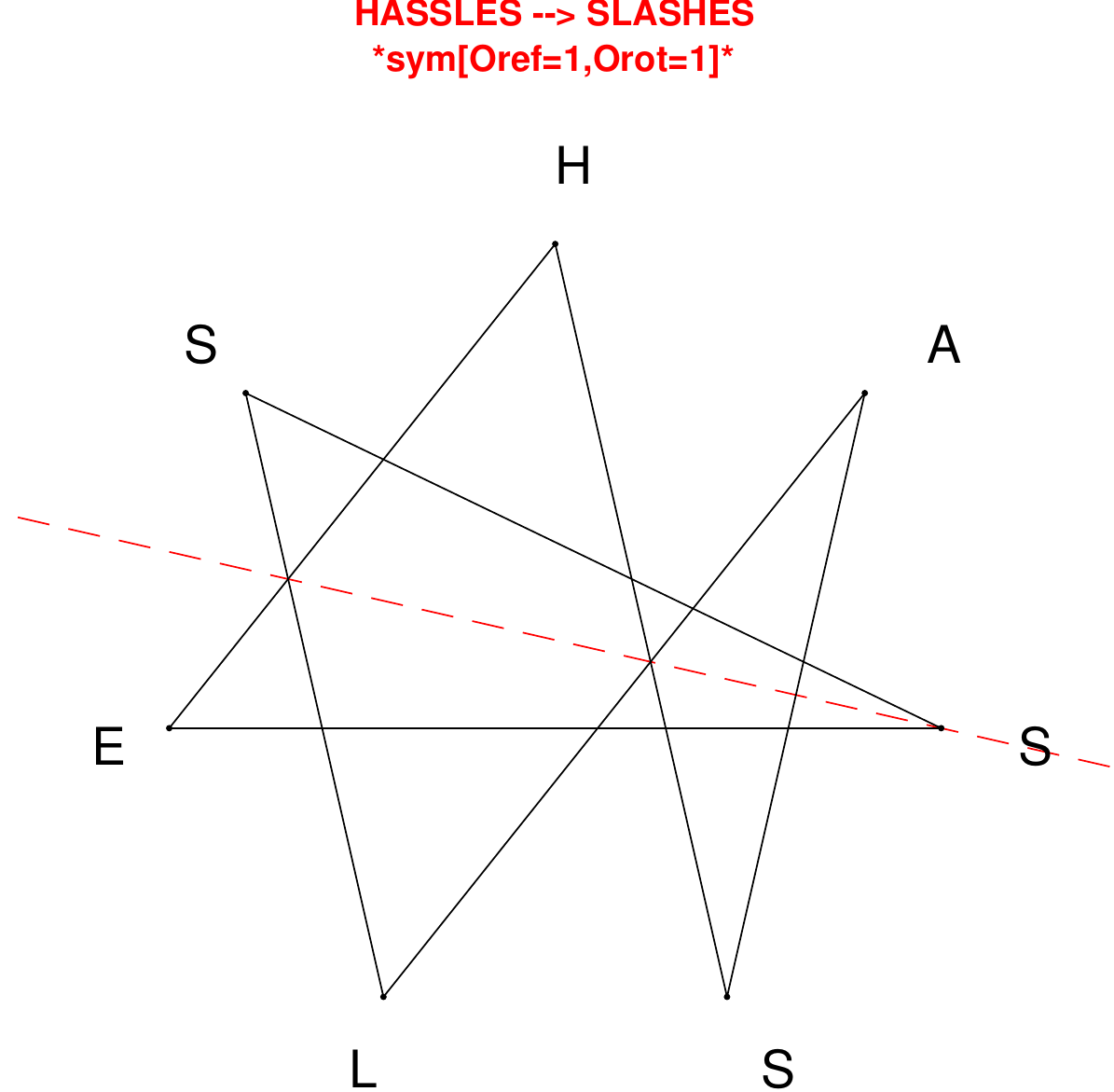}
\end{subfigure}
\hfill
\begin{subfigure}[T]{0.19\textwidth}
\centering
\includegraphics[width=\textwidth]{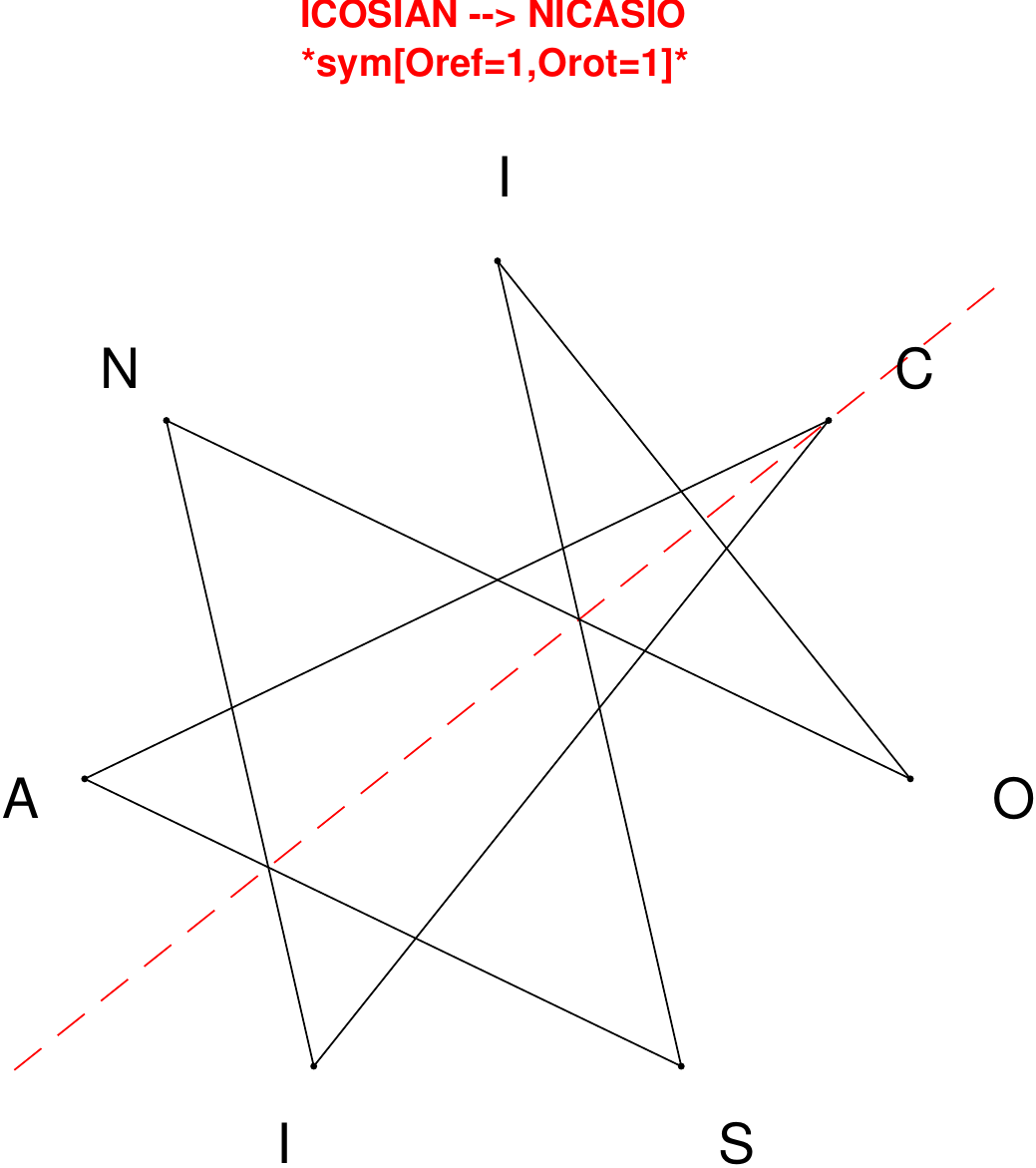}
\end{subfigure}
\hfill
\begin{subfigure}[T]{0.19\textwidth}
\centering
\includegraphics[width=\textwidth]{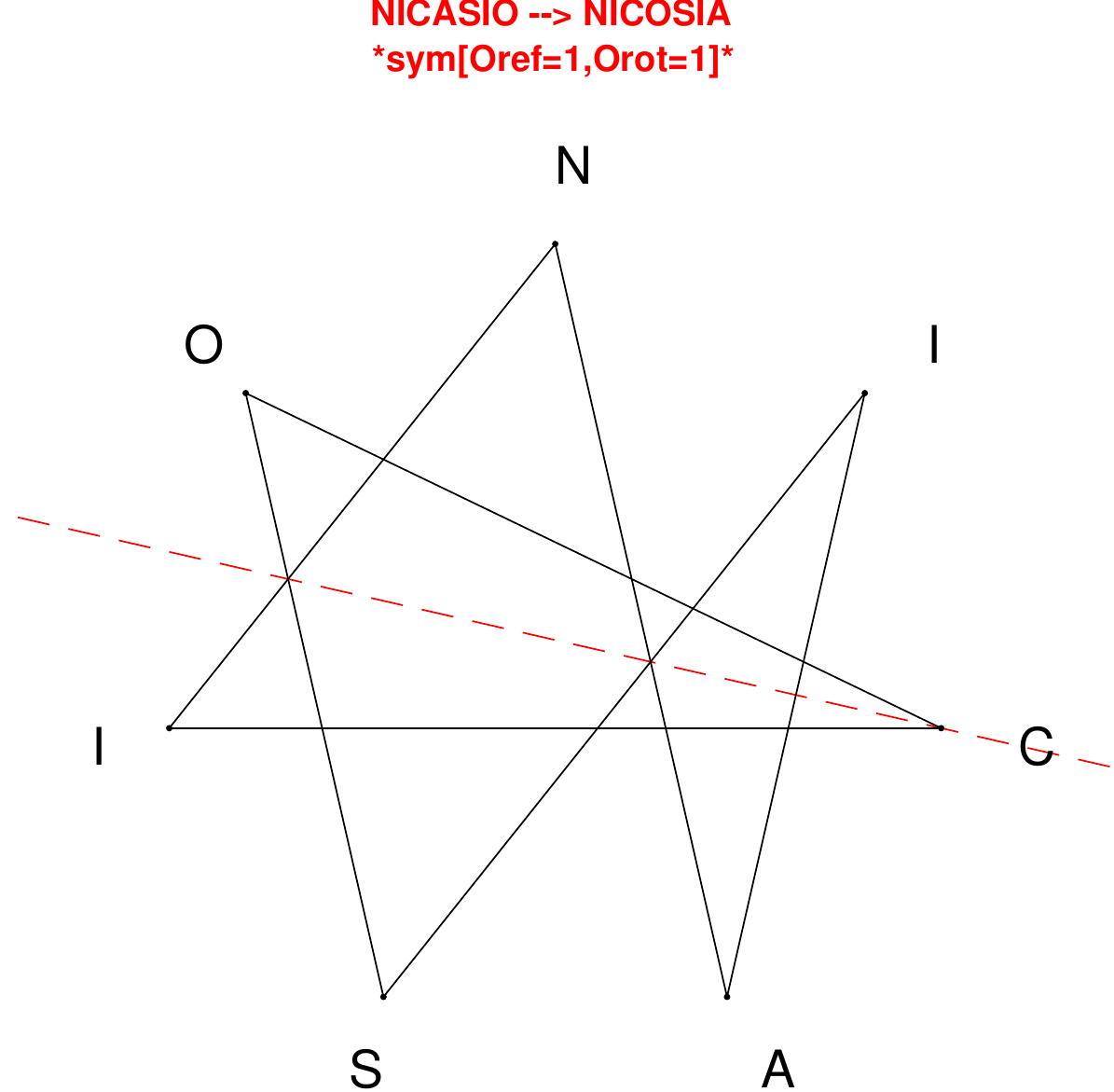}
\end{subfigure}
\hfill
\begin{subfigure}[T]{0.19\textwidth}
\centering
\includegraphics[width=\textwidth]{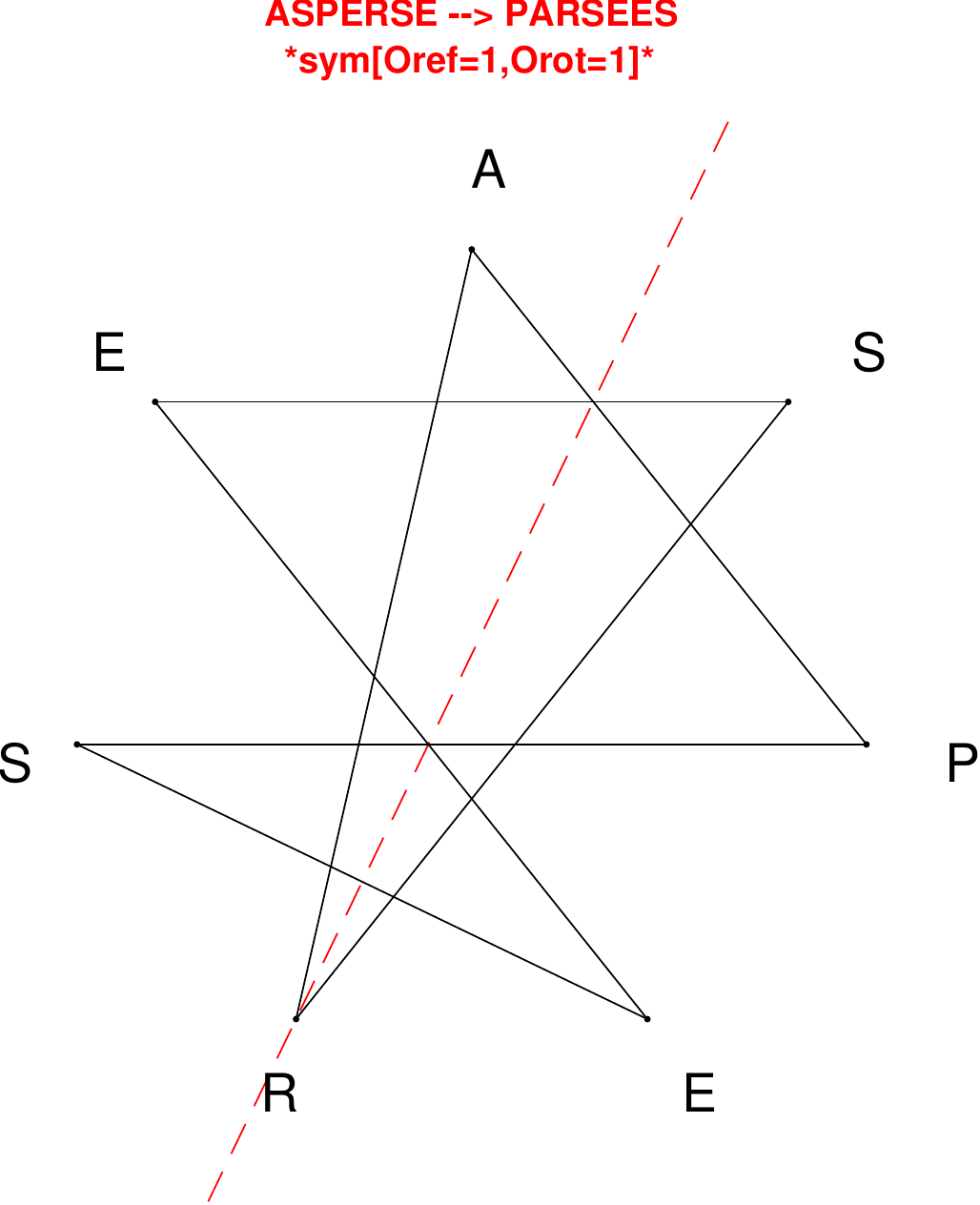}
\end{subfigure}
\end{figure}

\begin{figure}[H]
\centering
\begin{subfigure}[T]{0.19\textwidth}
\centering
\includegraphics[width=\textwidth]{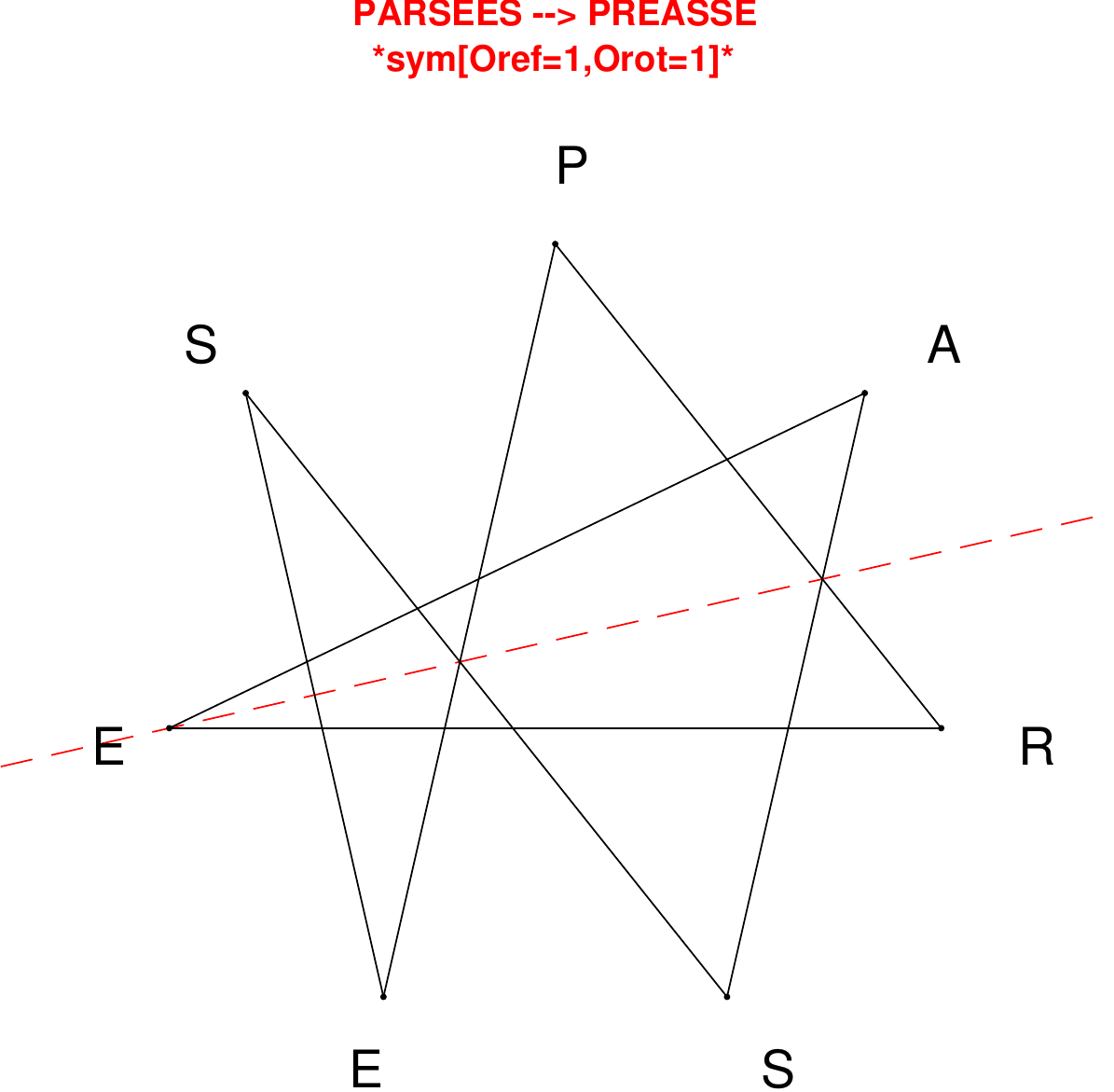}
\end{subfigure}
\hfill
\begin{subfigure}[T]{0.19\textwidth}
\centering
\includegraphics[width=\textwidth]{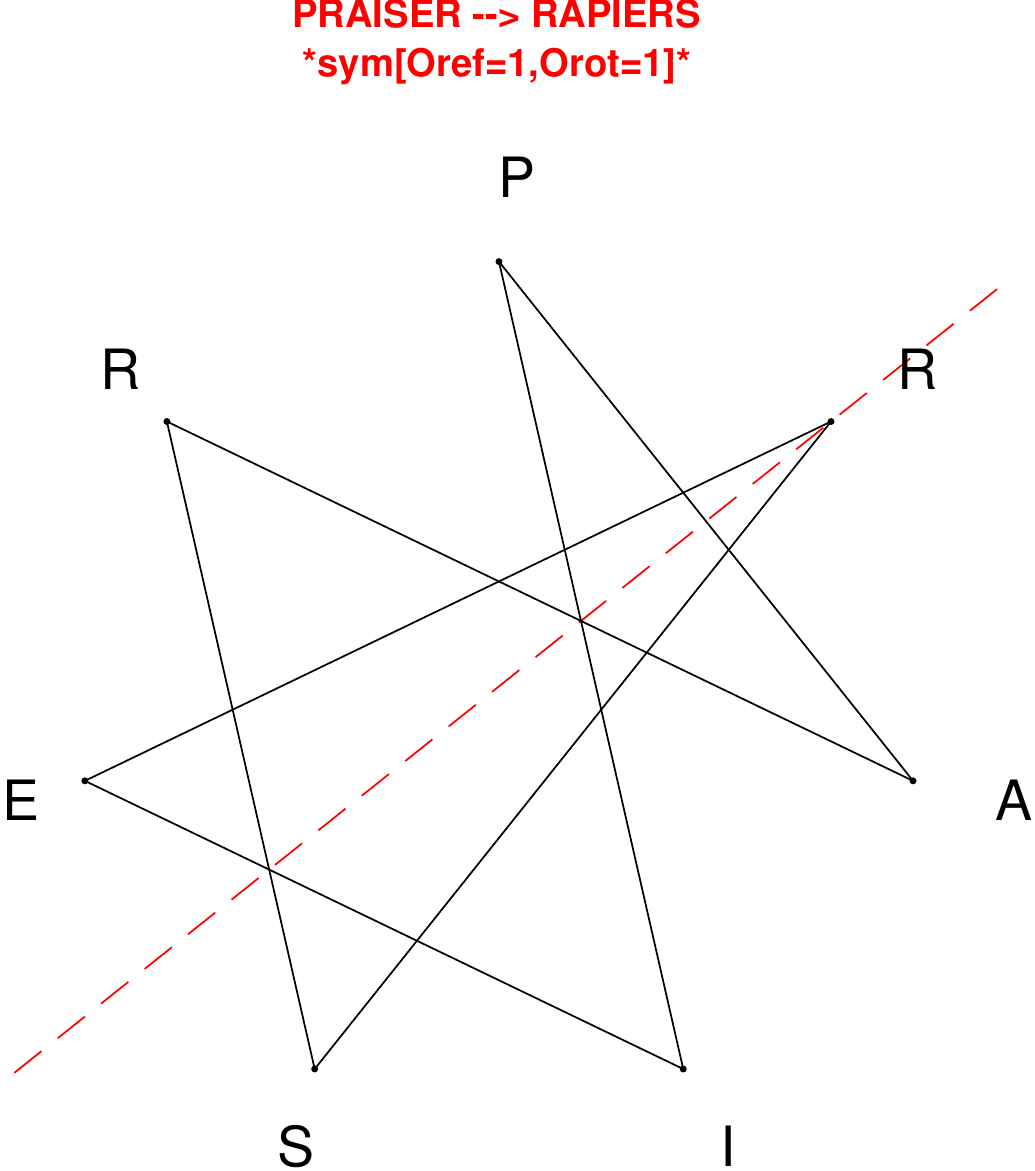}
\end{subfigure}
\hfill
\begin{subfigure}[T]{0.19\textwidth}
\centering
\includegraphics[width=\textwidth]{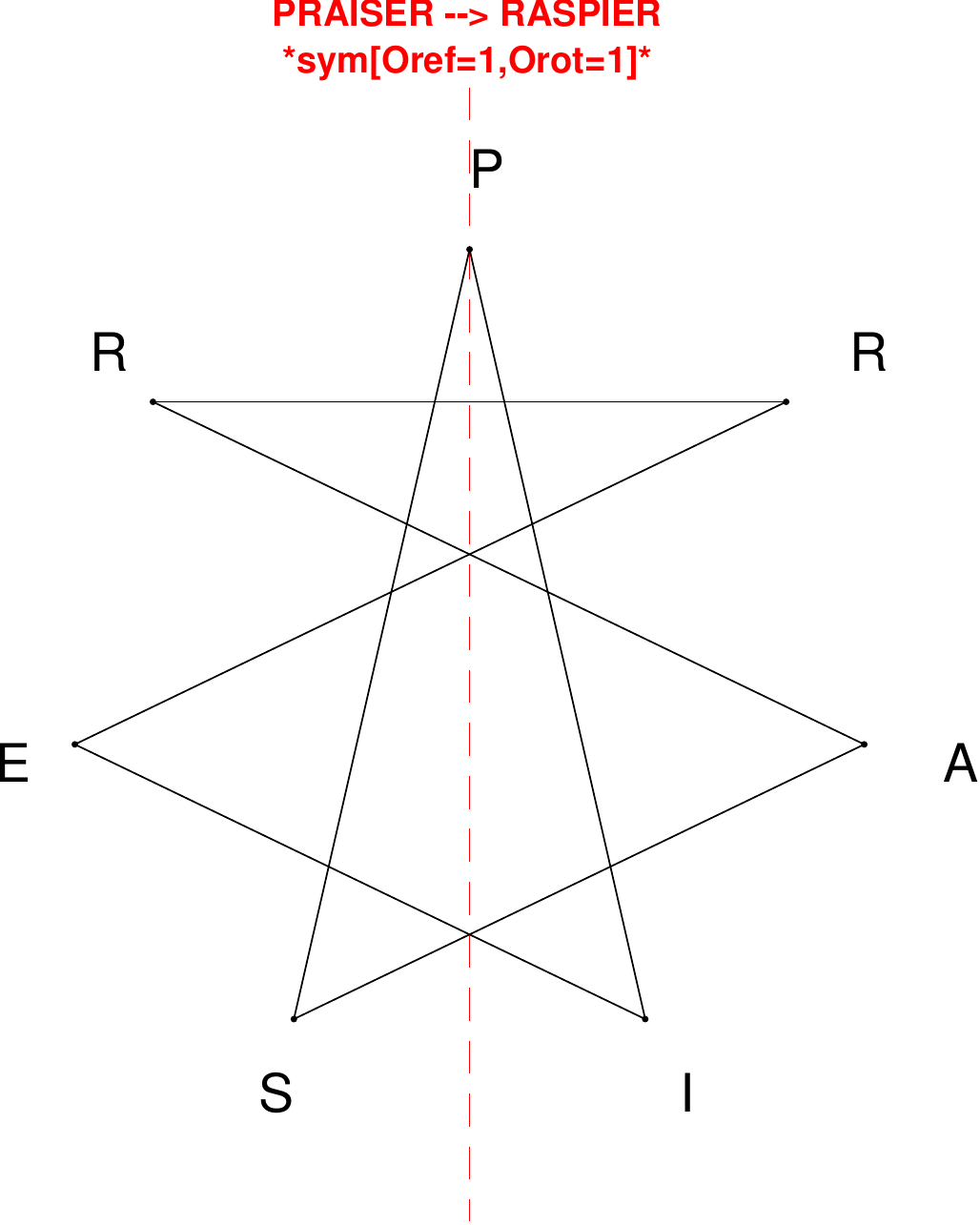}
\end{subfigure}
\hfill
\begin{subfigure}[T]{0.19\textwidth}
\centering
\includegraphics[width=\textwidth]{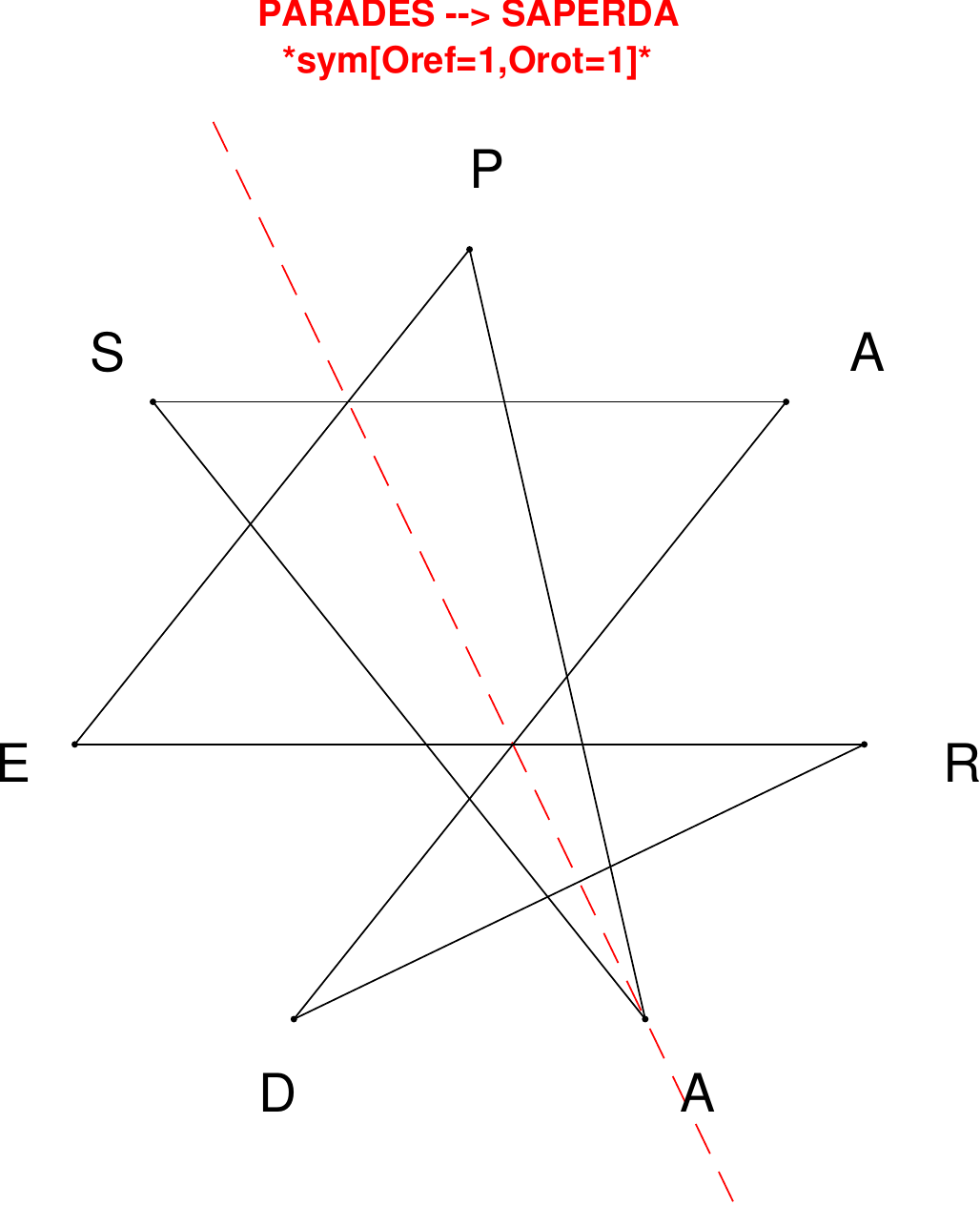}
\end{subfigure}
\hfill
\begin{subfigure}[T]{0.19\textwidth}
\centering
\includegraphics[width=\textwidth]{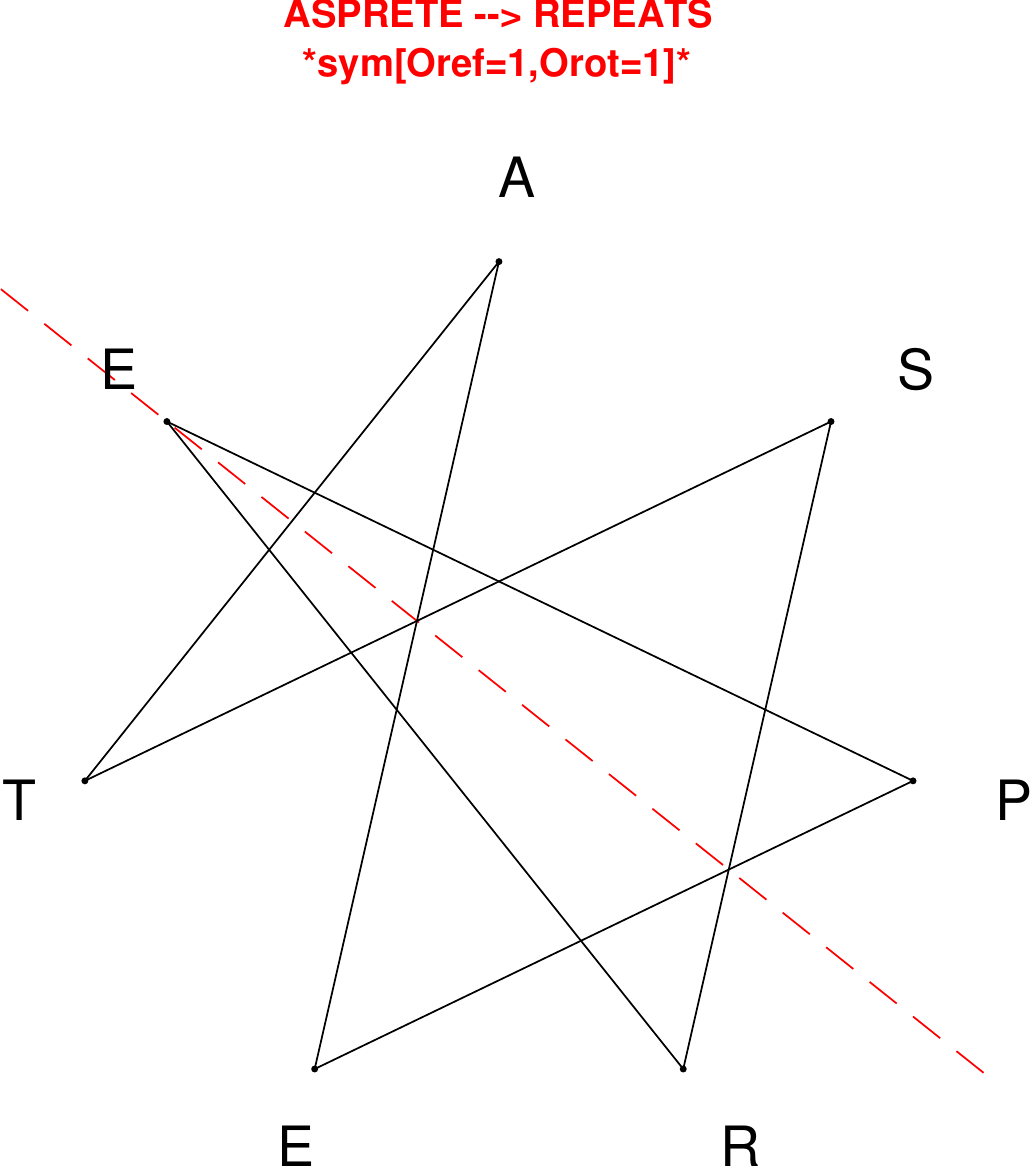}
\end{subfigure}
\end{figure}

\begin{figure}[H]
\centering
\begin{subfigure}[T]{0.19\textwidth}
\centering
\includegraphics[width=\textwidth]{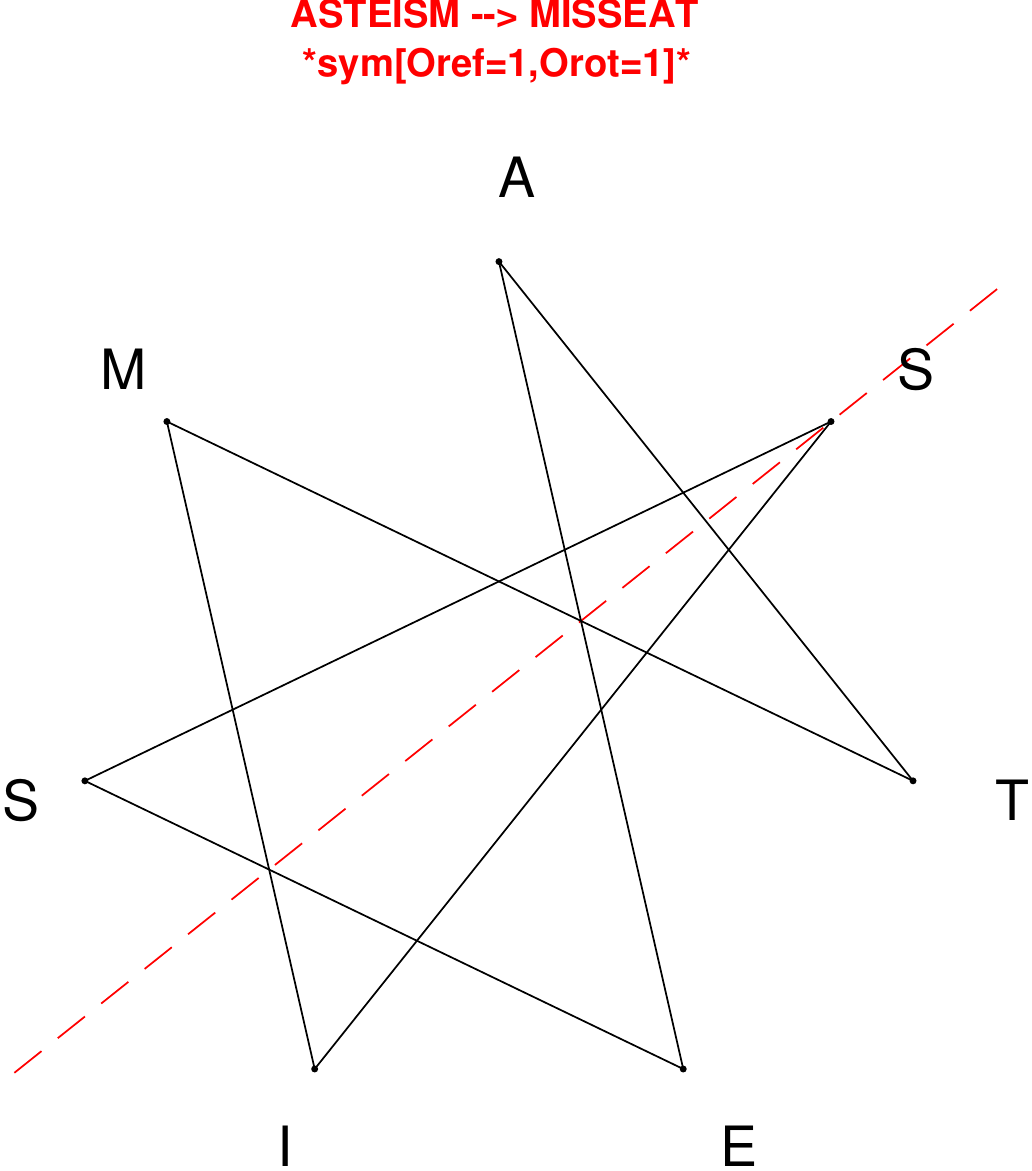}
\end{subfigure}
\hfill
\begin{subfigure}[T]{0.19\textwidth}
\centering
\includegraphics[width=\textwidth]{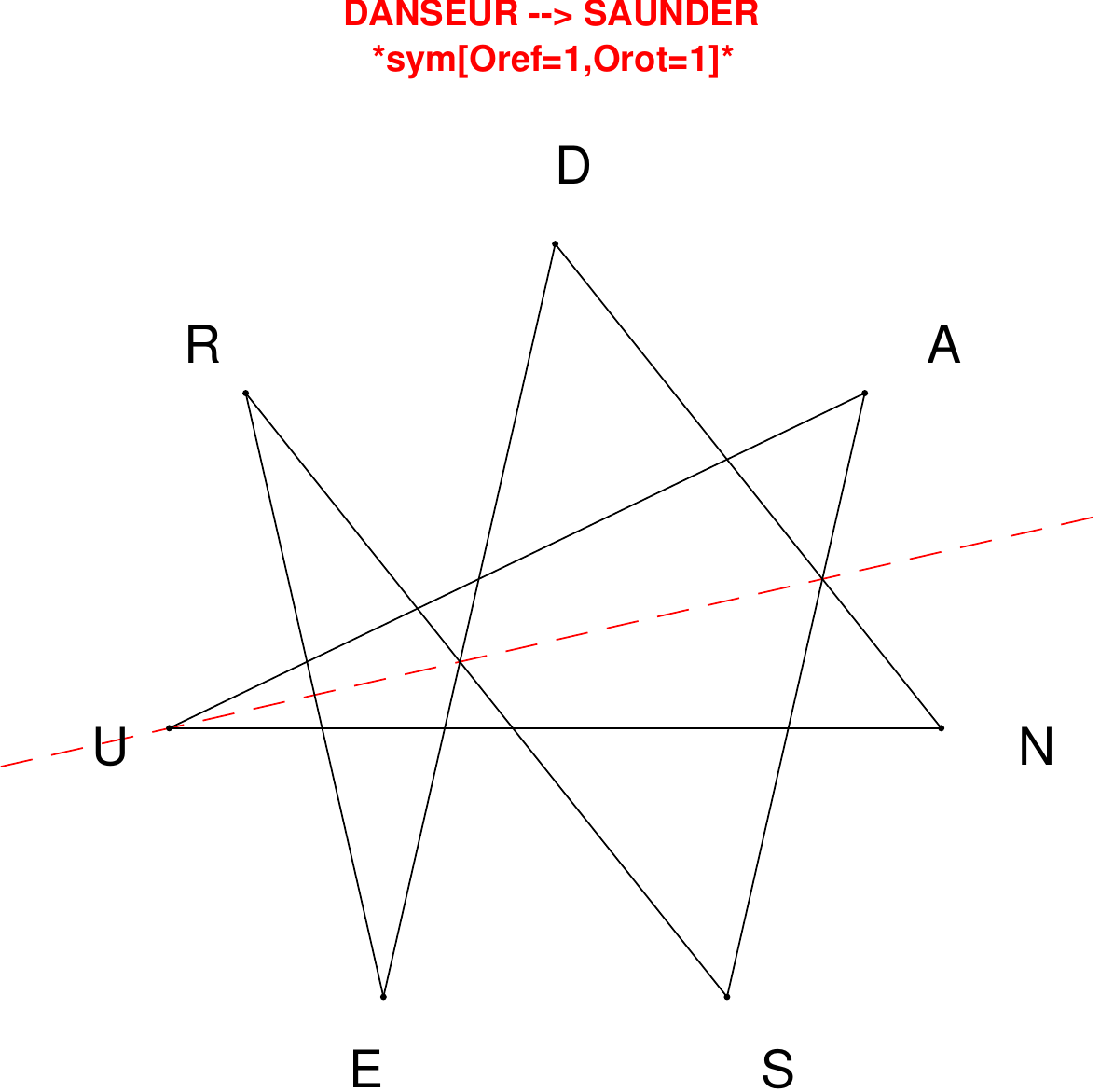}
\end{subfigure}
\hfill
\begin{subfigure}[T]{0.19\textwidth}
\centering
\includegraphics[width=\textwidth]{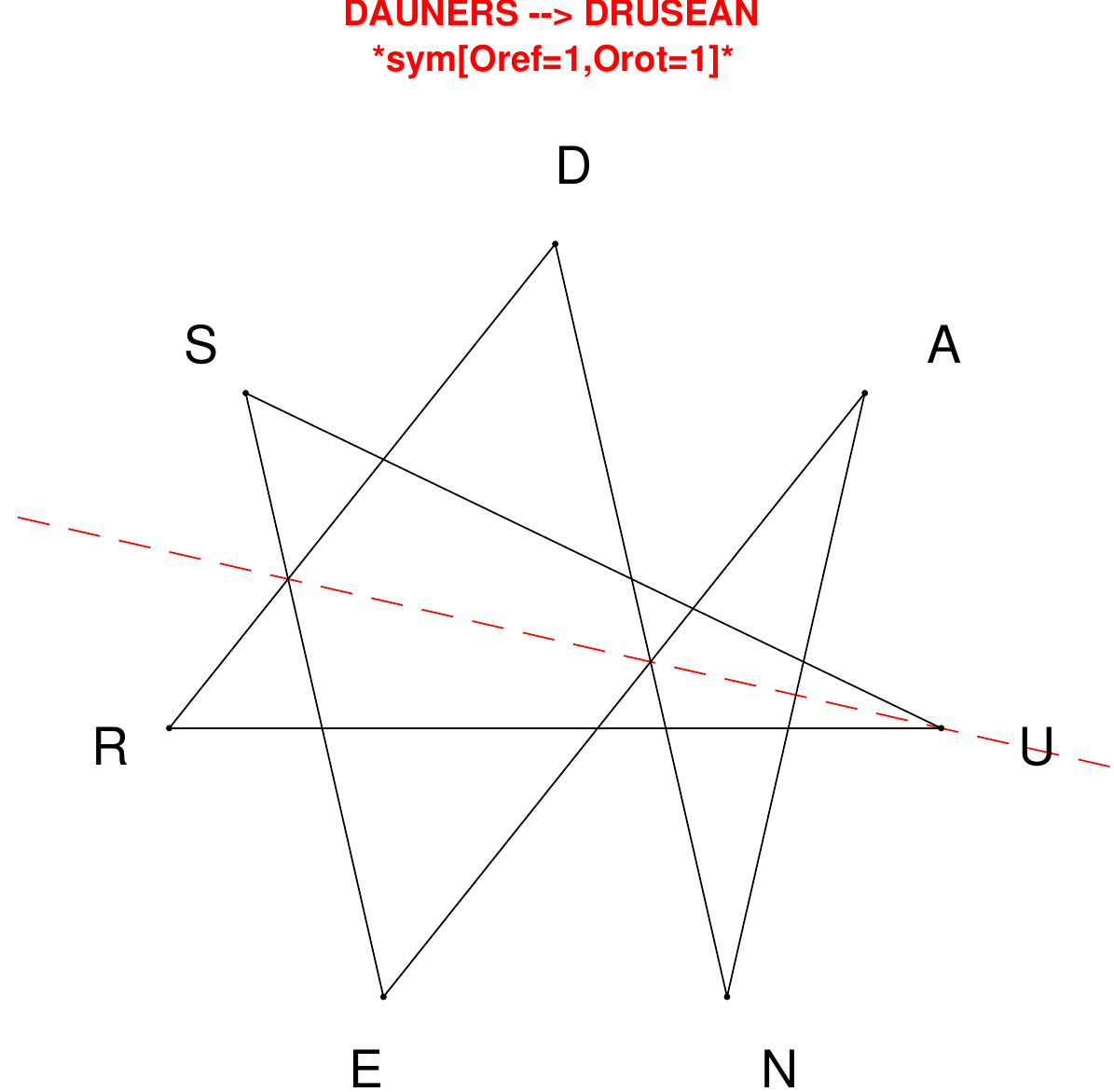}
\end{subfigure}
\hfill
\begin{subfigure}[T]{0.19\textwidth}
\centering
\includegraphics[width=\textwidth]{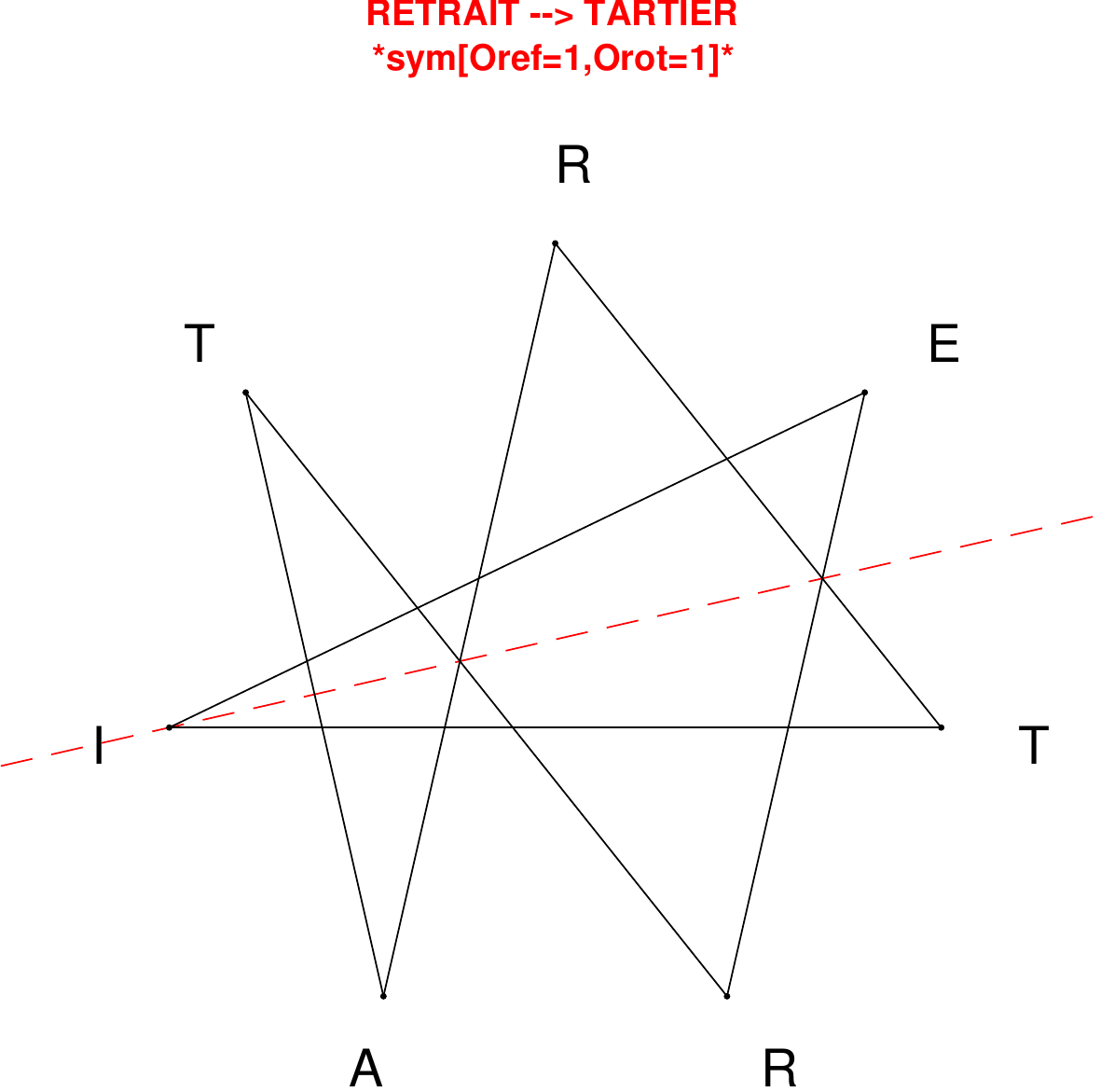}
\end{subfigure}
\hfill
\begin{subfigure}[T]{0.19\textwidth}
\centering
\includegraphics[width=\textwidth]{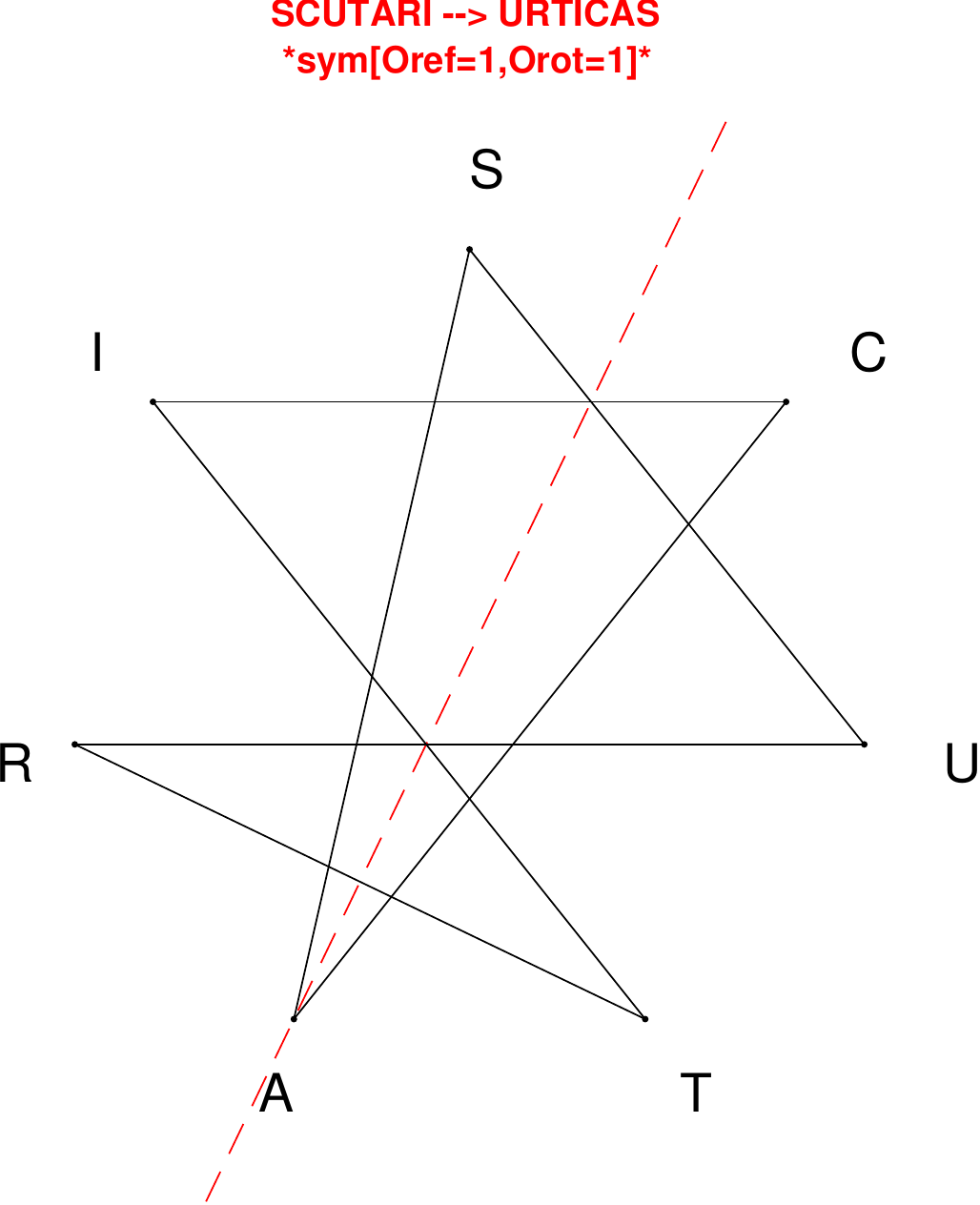}
\end{subfigure}
\end{figure}

\begin{figure}[H]
\centering
\begin{subfigure}[T]{0.19\textwidth}
\centering
\includegraphics[width=\textwidth]{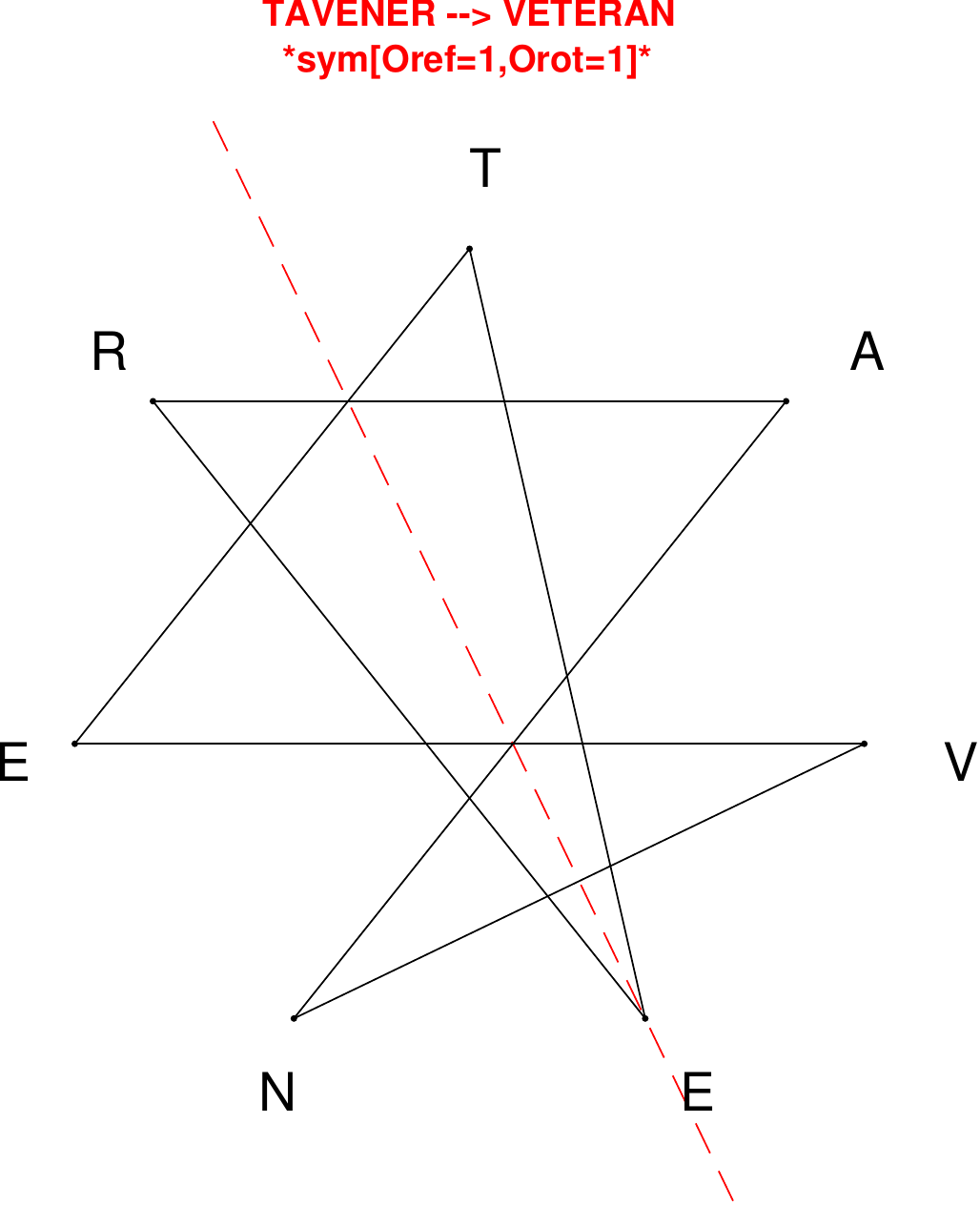}
\end{subfigure}
\hfill
\begin{subfigure}[T]{0.19\textwidth}
\centering
\includegraphics[width=\textwidth]{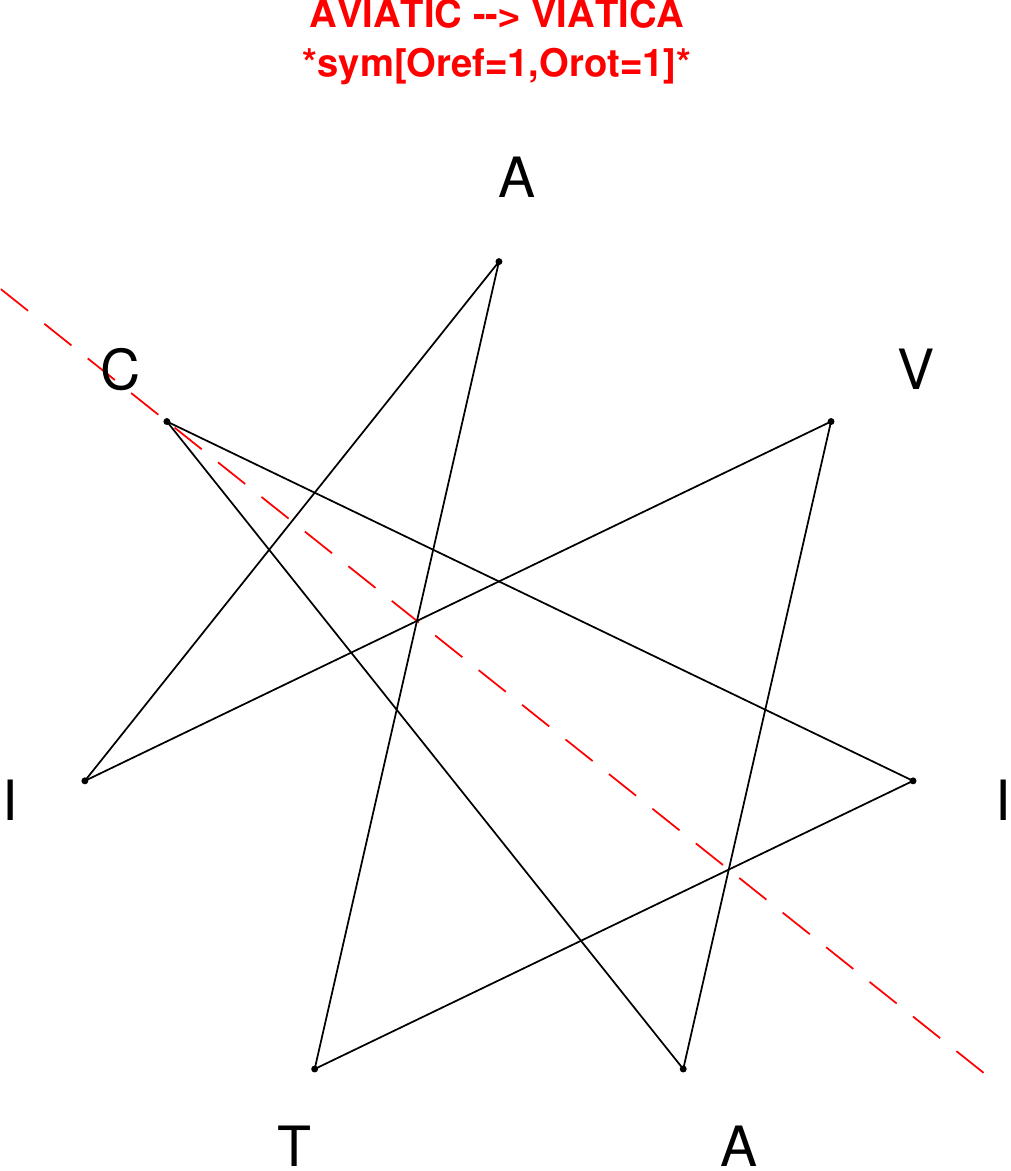}
\end{subfigure}
\hfill
\begin{subfigure}[T]{0.19\textwidth}
\centering
\includegraphics[width=\textwidth]{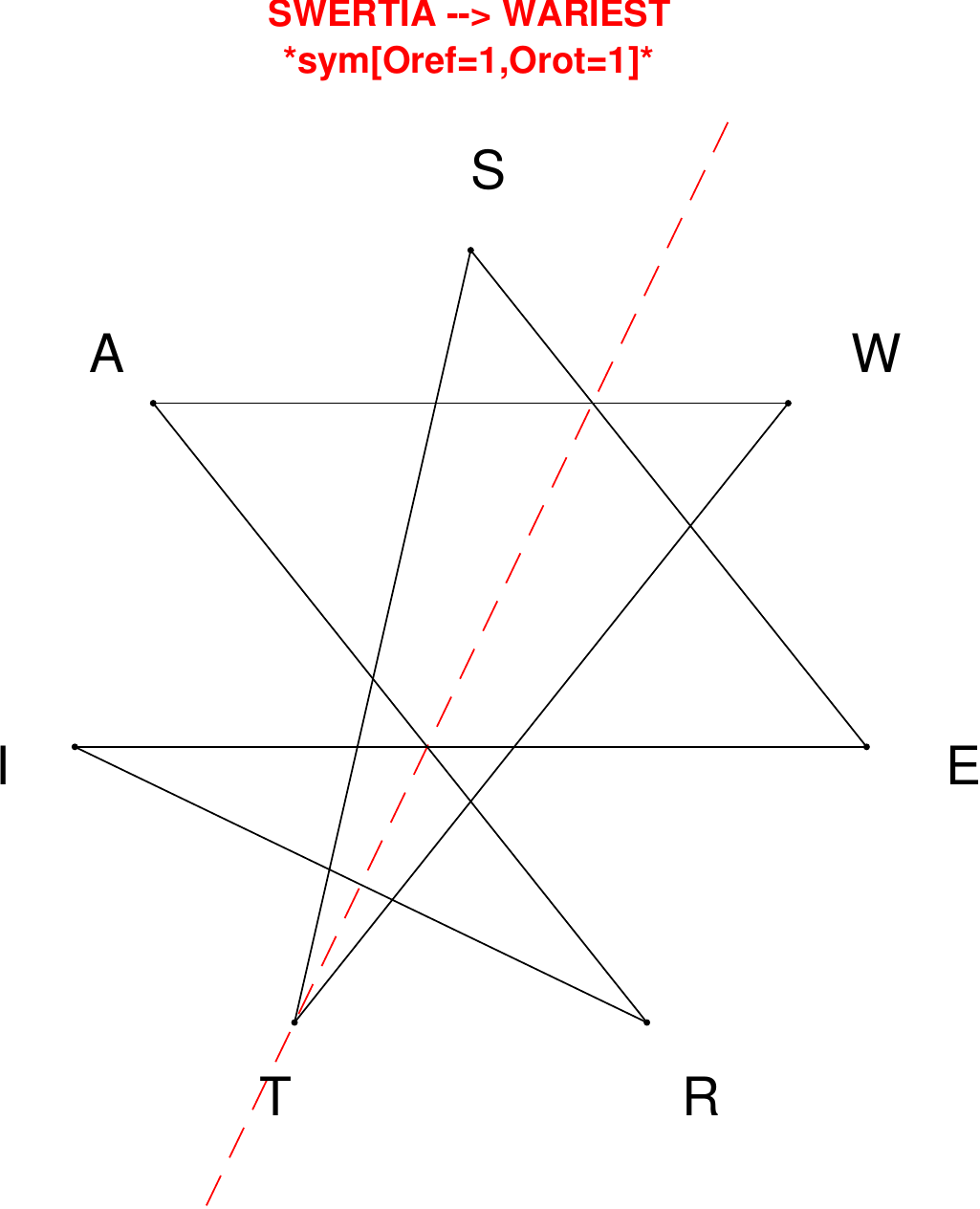}
\end{subfigure}
\hfill
\begin{subfigure}[T]{0.19\textwidth}
\centering
\includegraphics[width=\textwidth]{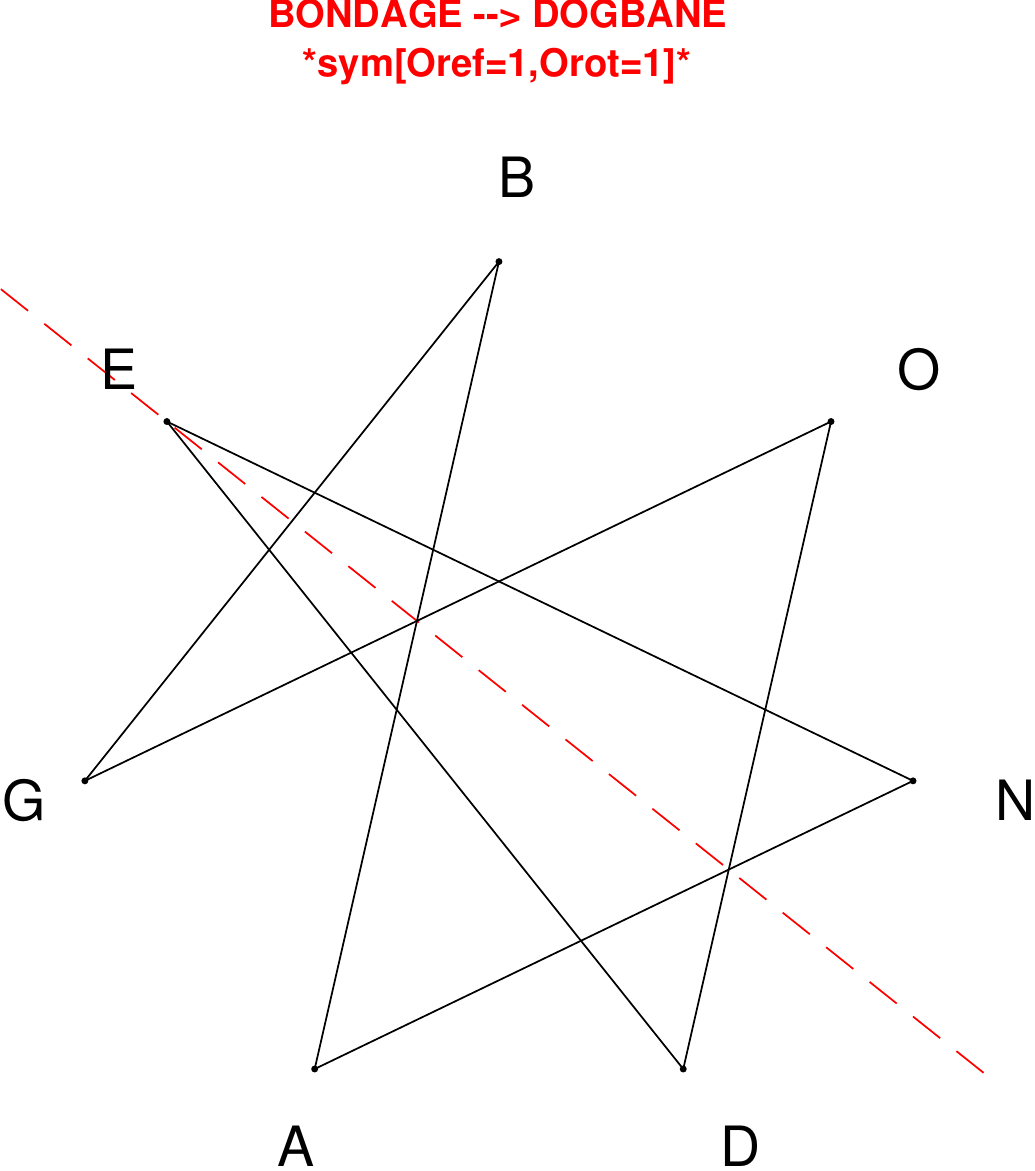}
\end{subfigure}
\hfill
\begin{subfigure}[T]{0.19\textwidth}
\centering
\includegraphics[width=\textwidth]{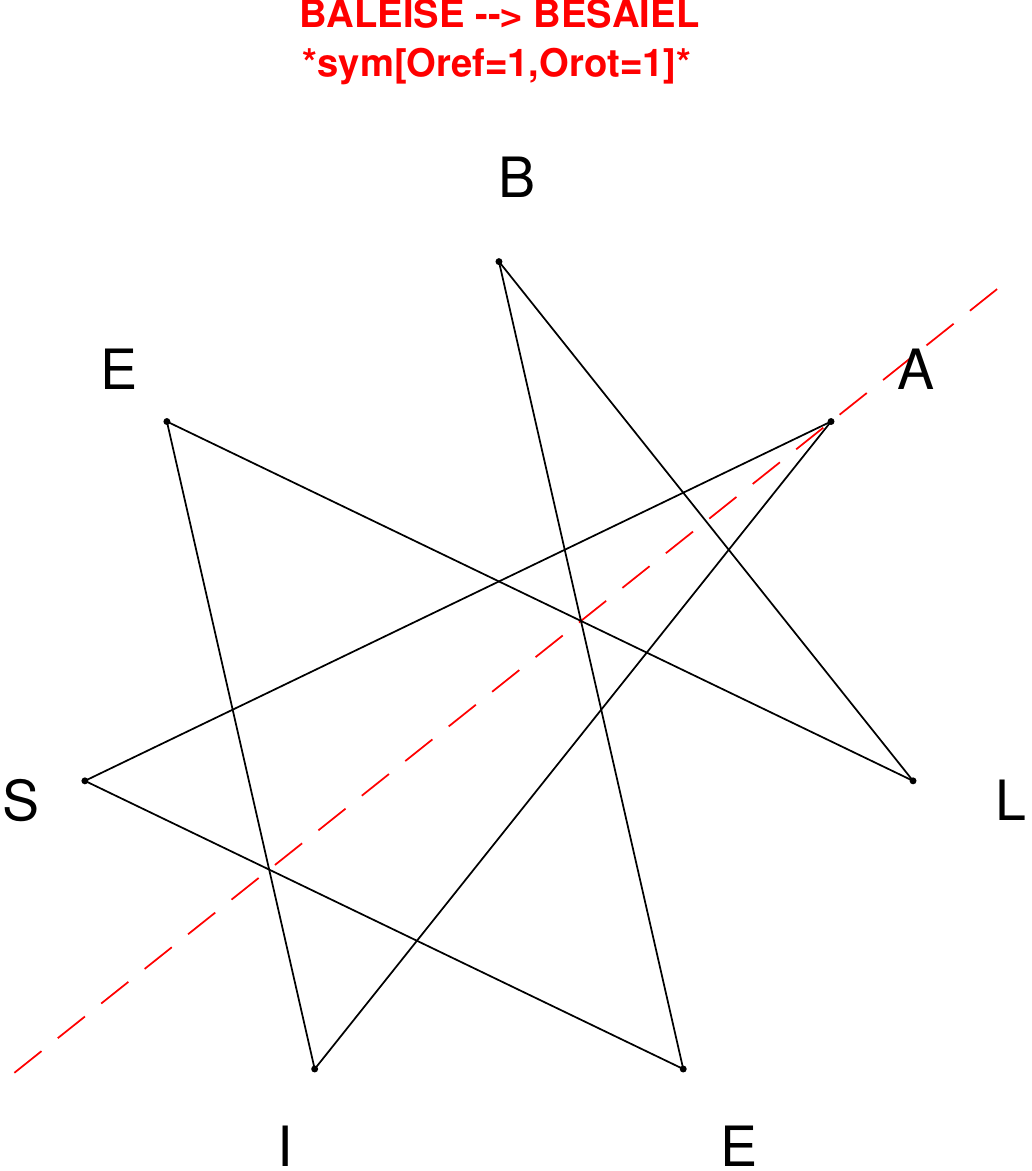}
\end{subfigure}
\end{figure}

\begin{figure}[H]
\centering
\begin{subfigure}[T]{0.19\textwidth}
\centering
\includegraphics[width=\textwidth]{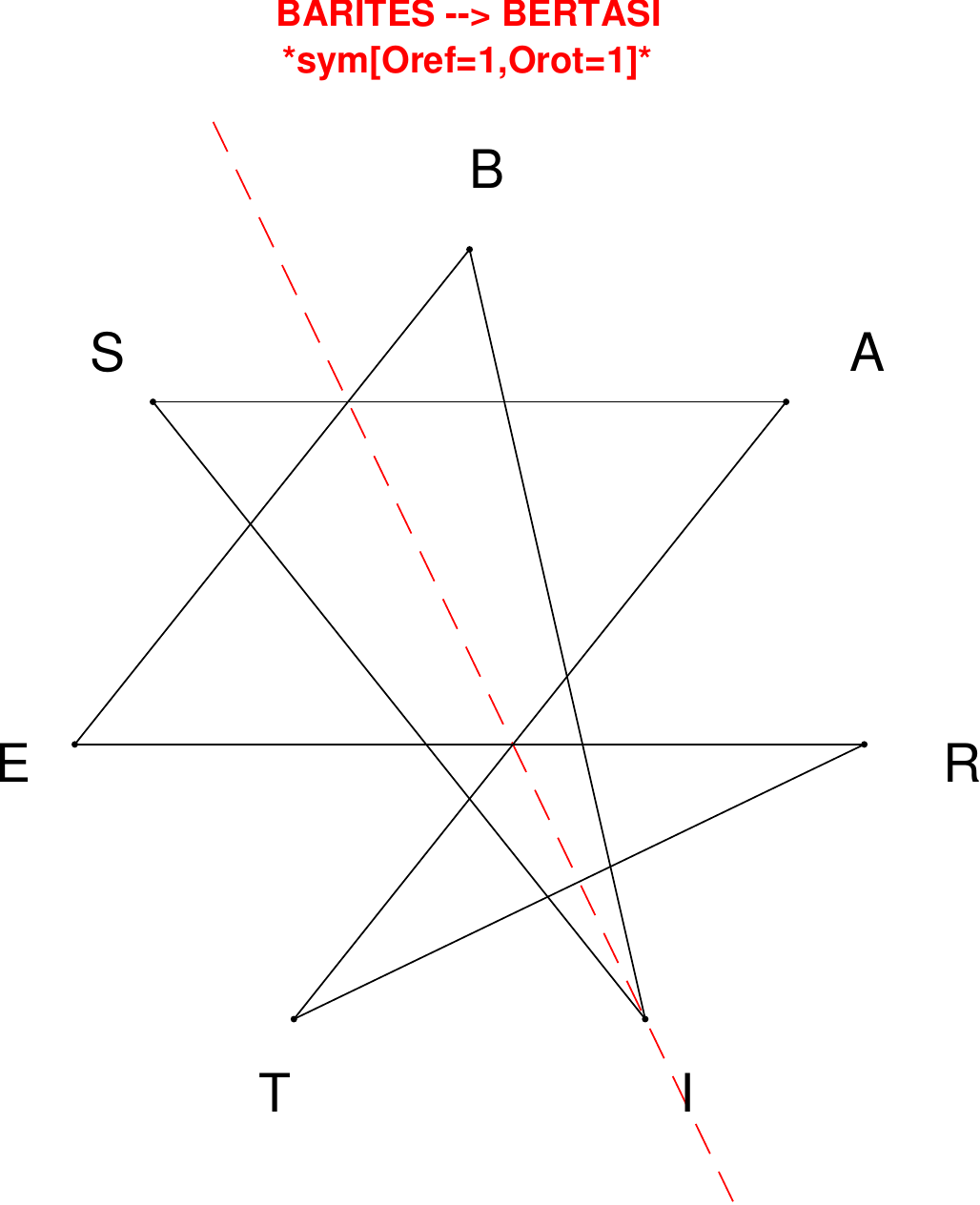}
\end{subfigure}
\hfill
\begin{subfigure}[T]{0.19\textwidth}
\centering
\includegraphics[width=\textwidth]{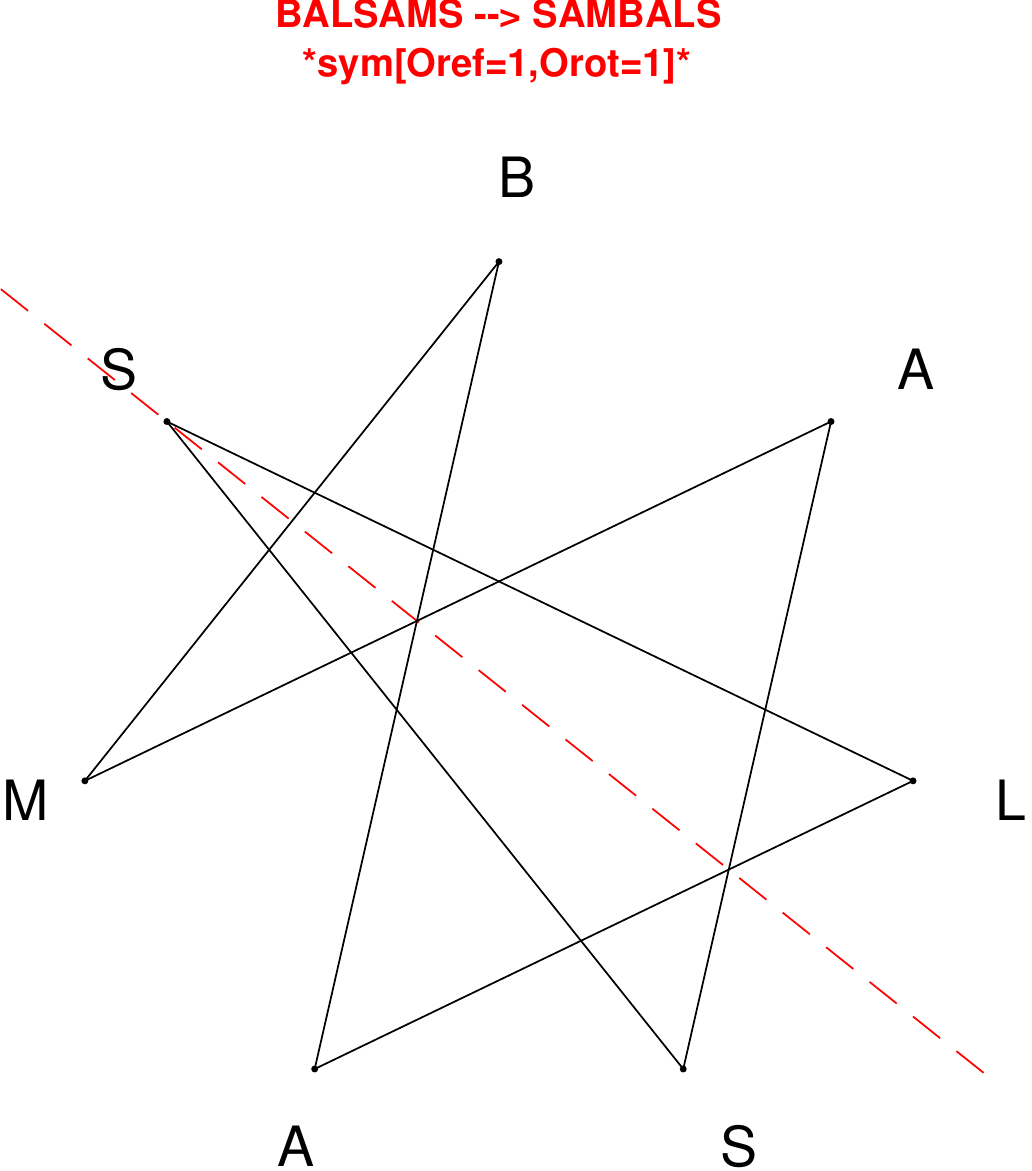}
\end{subfigure}
\hfill
\begin{subfigure}[T]{0.19\textwidth}
\centering
\includegraphics[width=\textwidth]{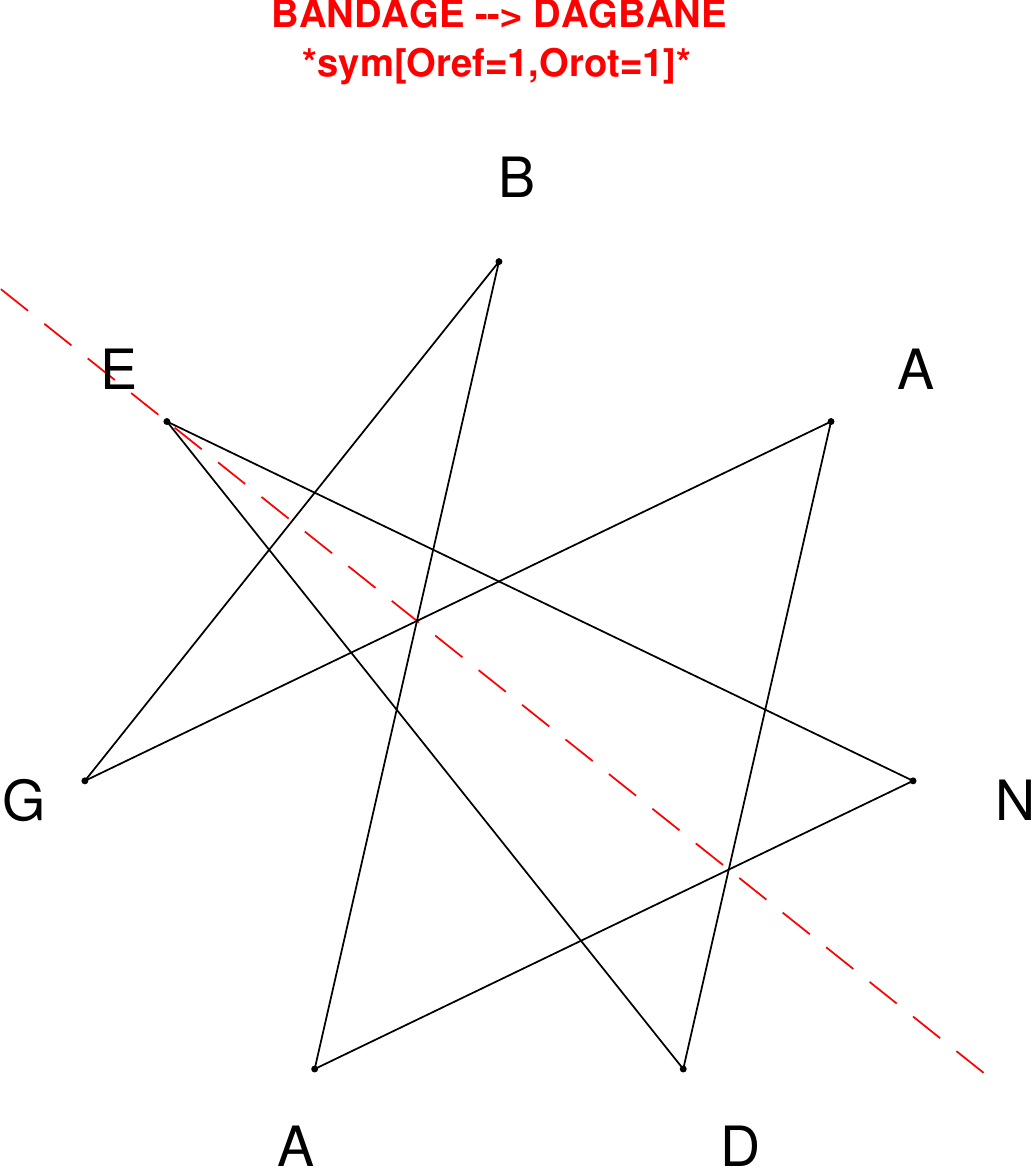}
\end{subfigure}
\hfill
\begin{subfigure}[T]{0.19\textwidth}
\centering
\includegraphics[width=\textwidth]{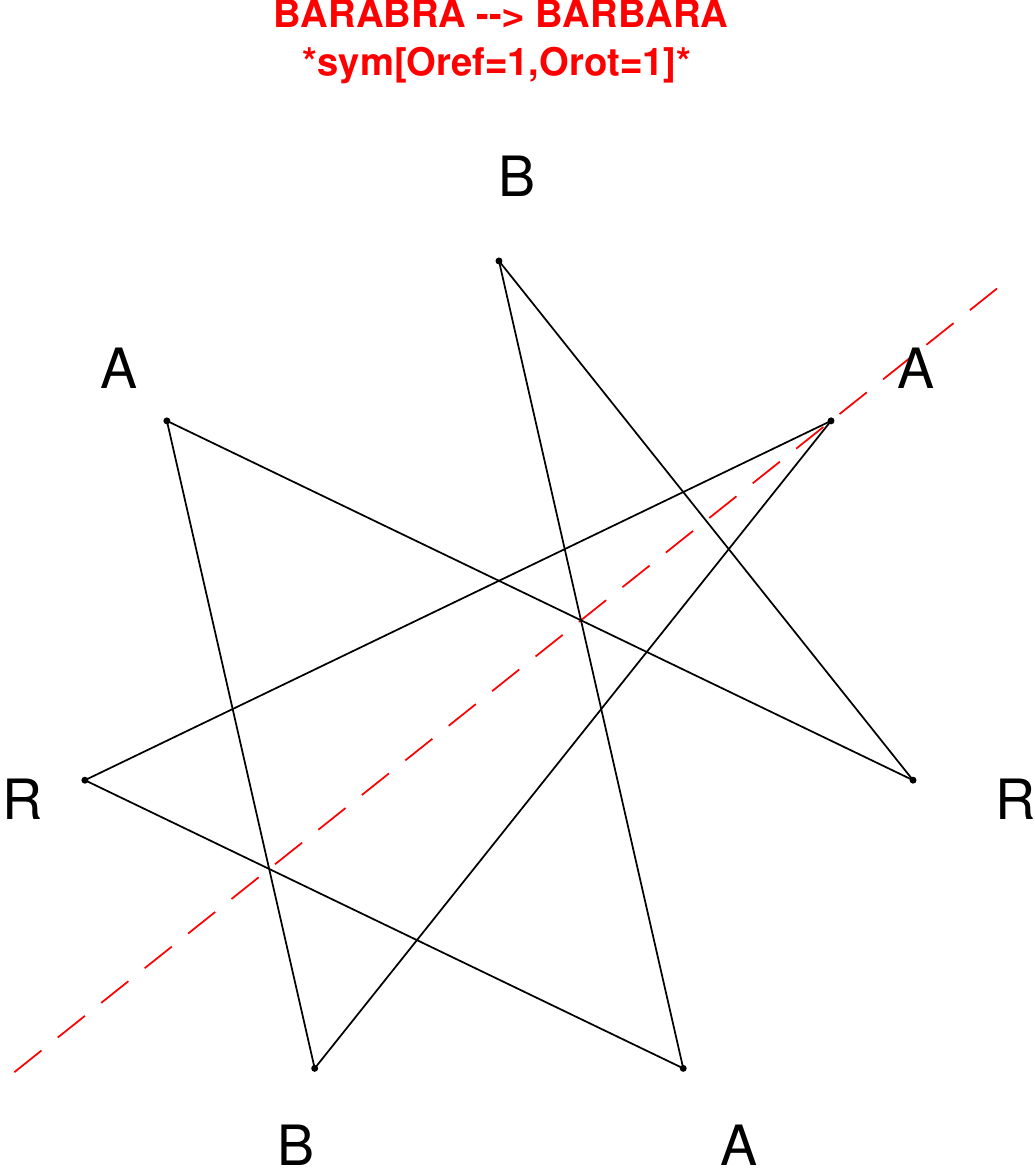}
\end{subfigure}
\hfill
\begin{subfigure}[T]{0.19\textwidth}
\centering
\includegraphics[width=\textwidth]{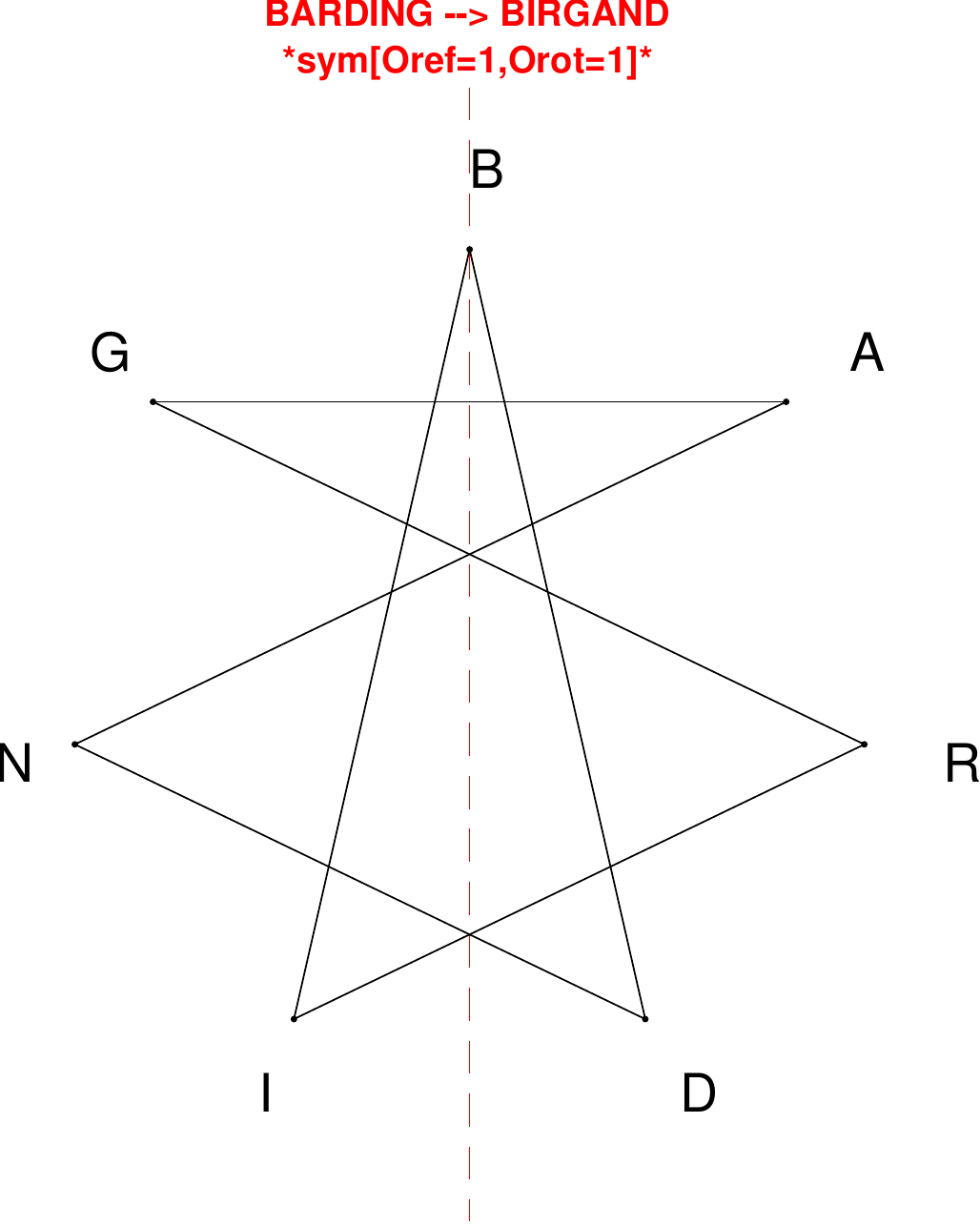}
\end{subfigure}
\end{figure}

\begin{figure}[H]
\centering
\begin{subfigure}[T]{0.19\textwidth}
\centering
\includegraphics[width=\textwidth]{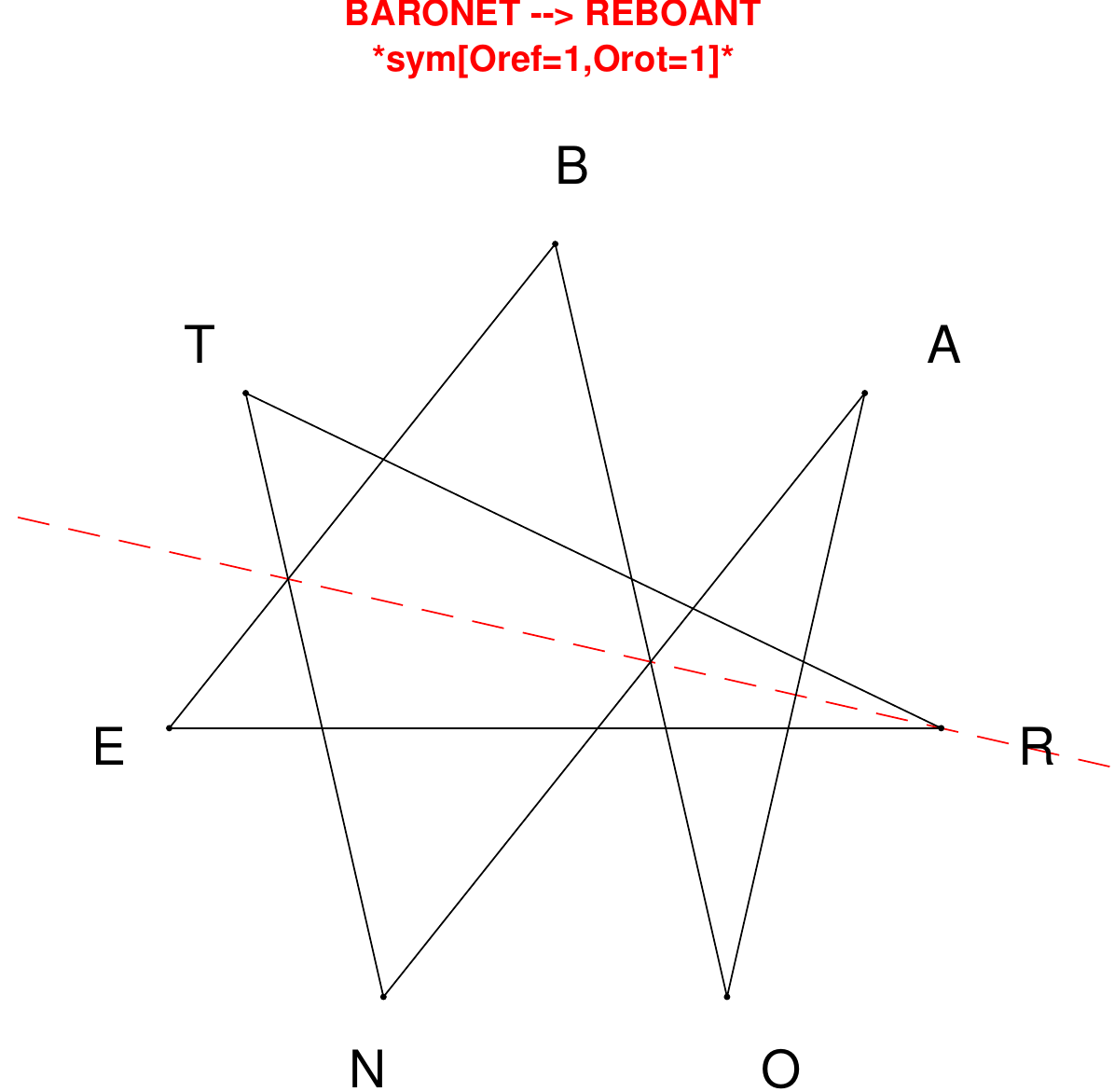}
\end{subfigure}
\hfill
\begin{subfigure}[T]{0.19\textwidth}
\centering
\includegraphics[width=\textwidth]{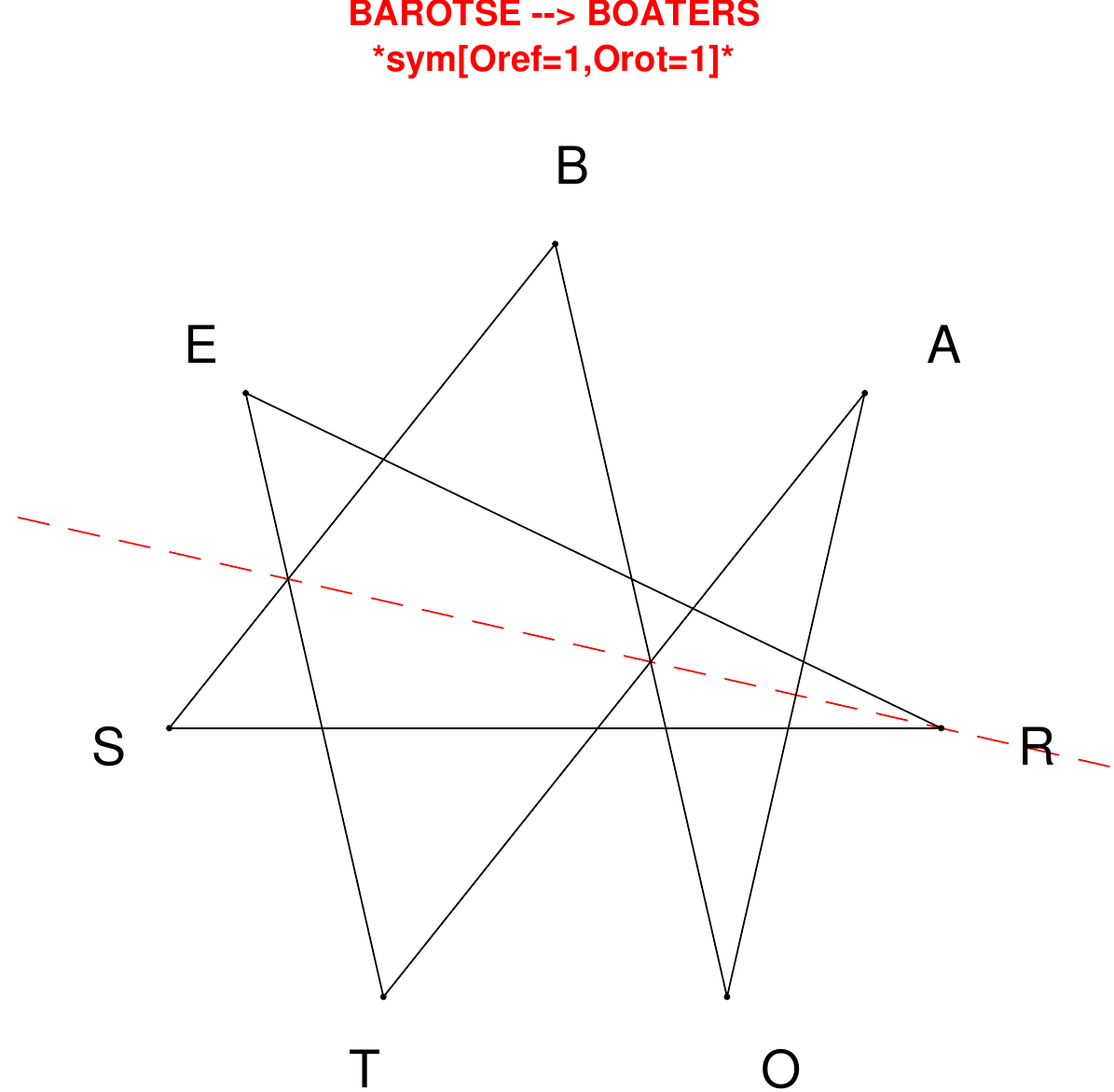}
\end{subfigure}
\hfill
\begin{subfigure}[T]{0.19\textwidth}
\centering
\includegraphics[width=\textwidth]{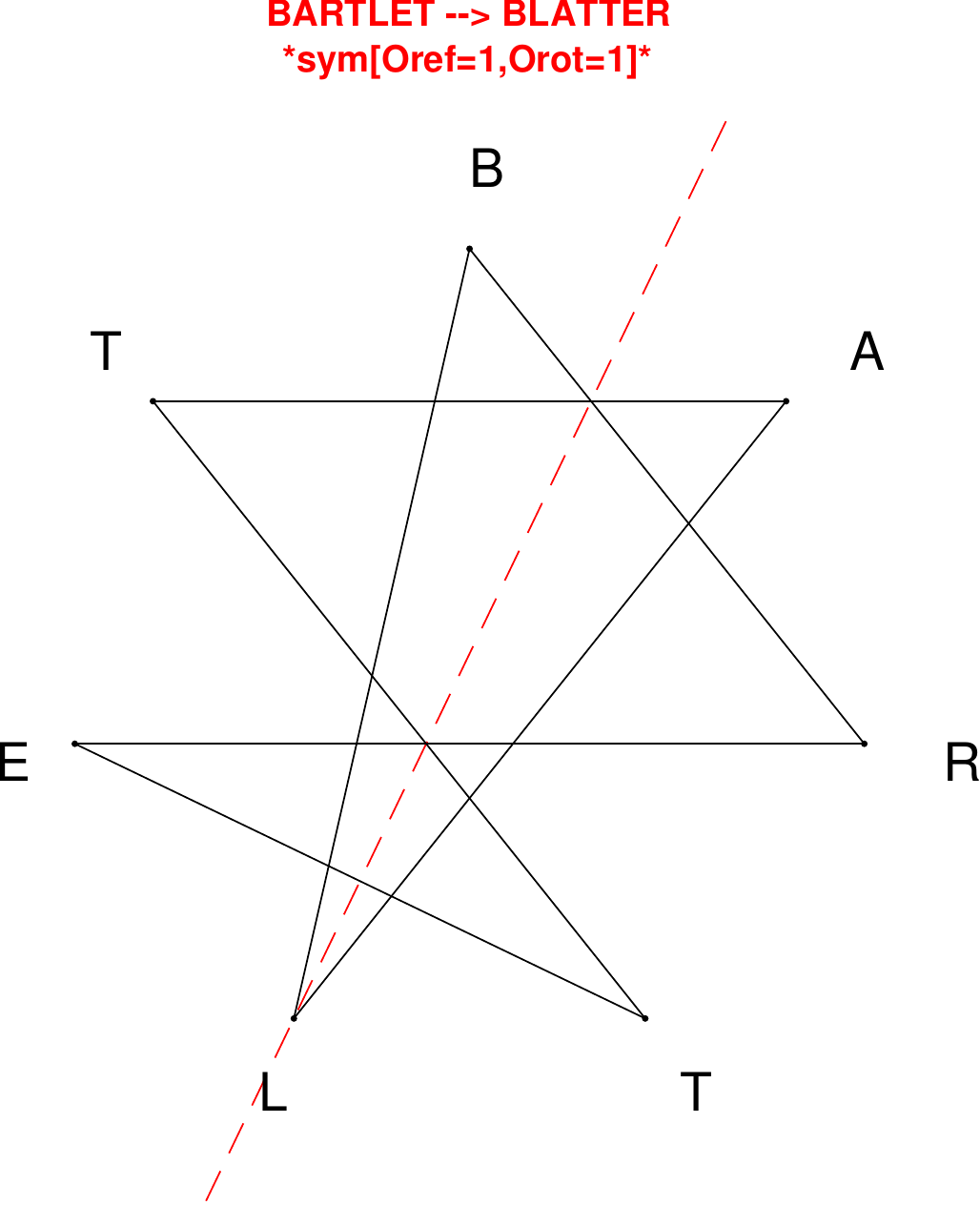}
\end{subfigure}
\hfill
\begin{subfigure}[T]{0.19\textwidth}
\centering
\includegraphics[width=\textwidth]{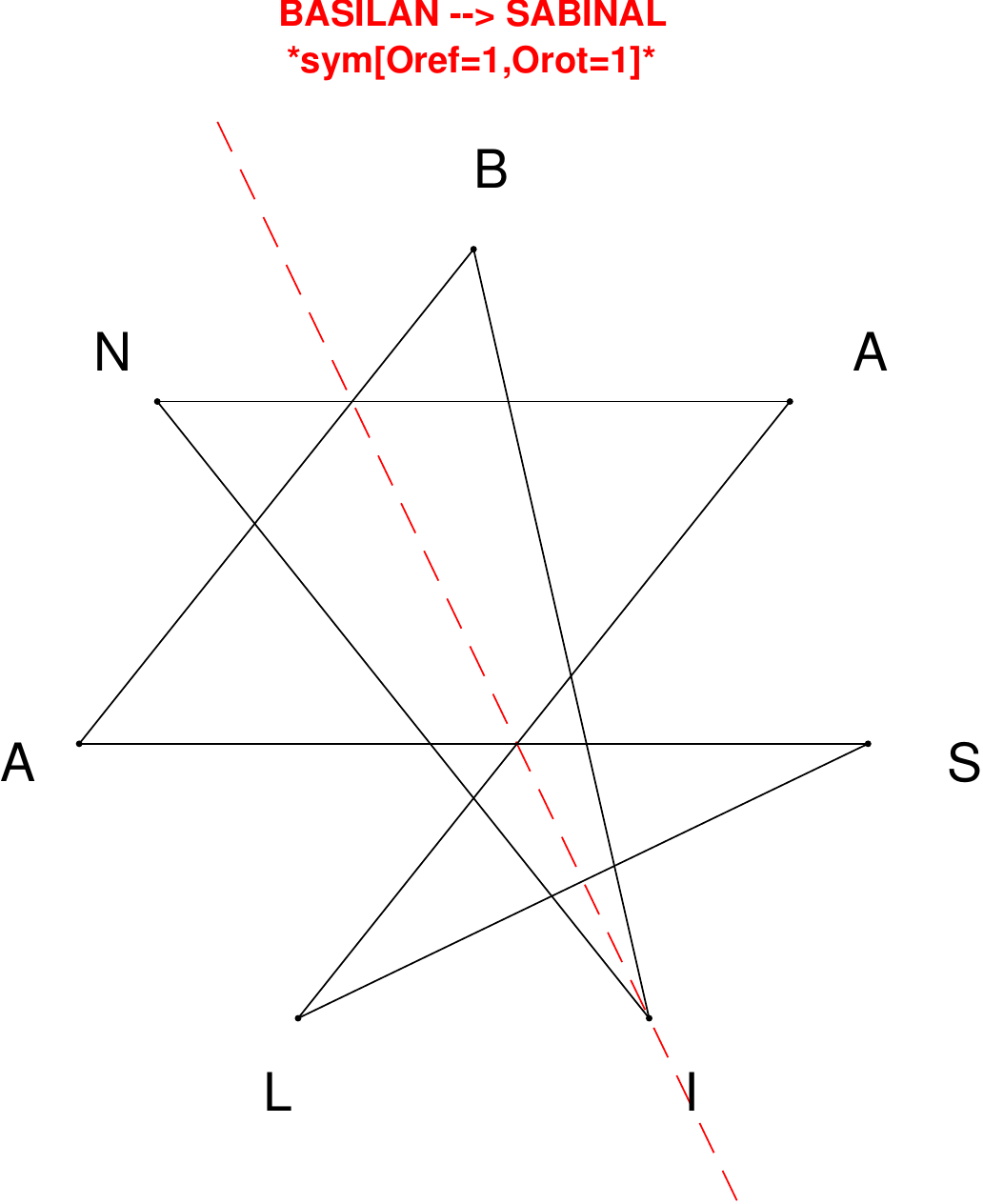}
\end{subfigure}
\hfill
\begin{subfigure}[T]{0.19\textwidth}
\centering
\includegraphics[width=\textwidth]{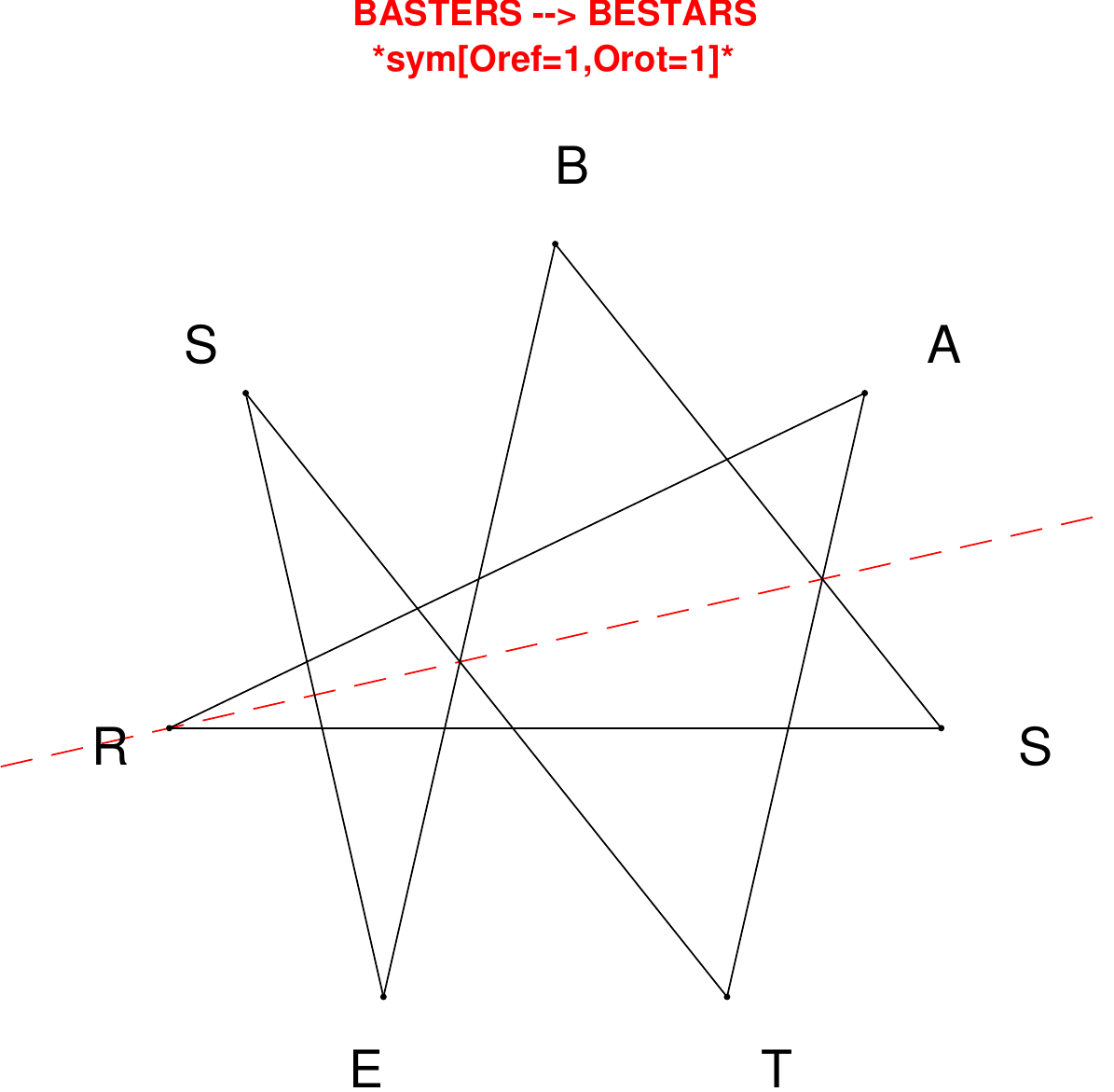}
\end{subfigure}
\end{figure}

\begin{figure}[H]
\centering
\begin{subfigure}[T]{0.19\textwidth}
\centering
\includegraphics[width=\textwidth]{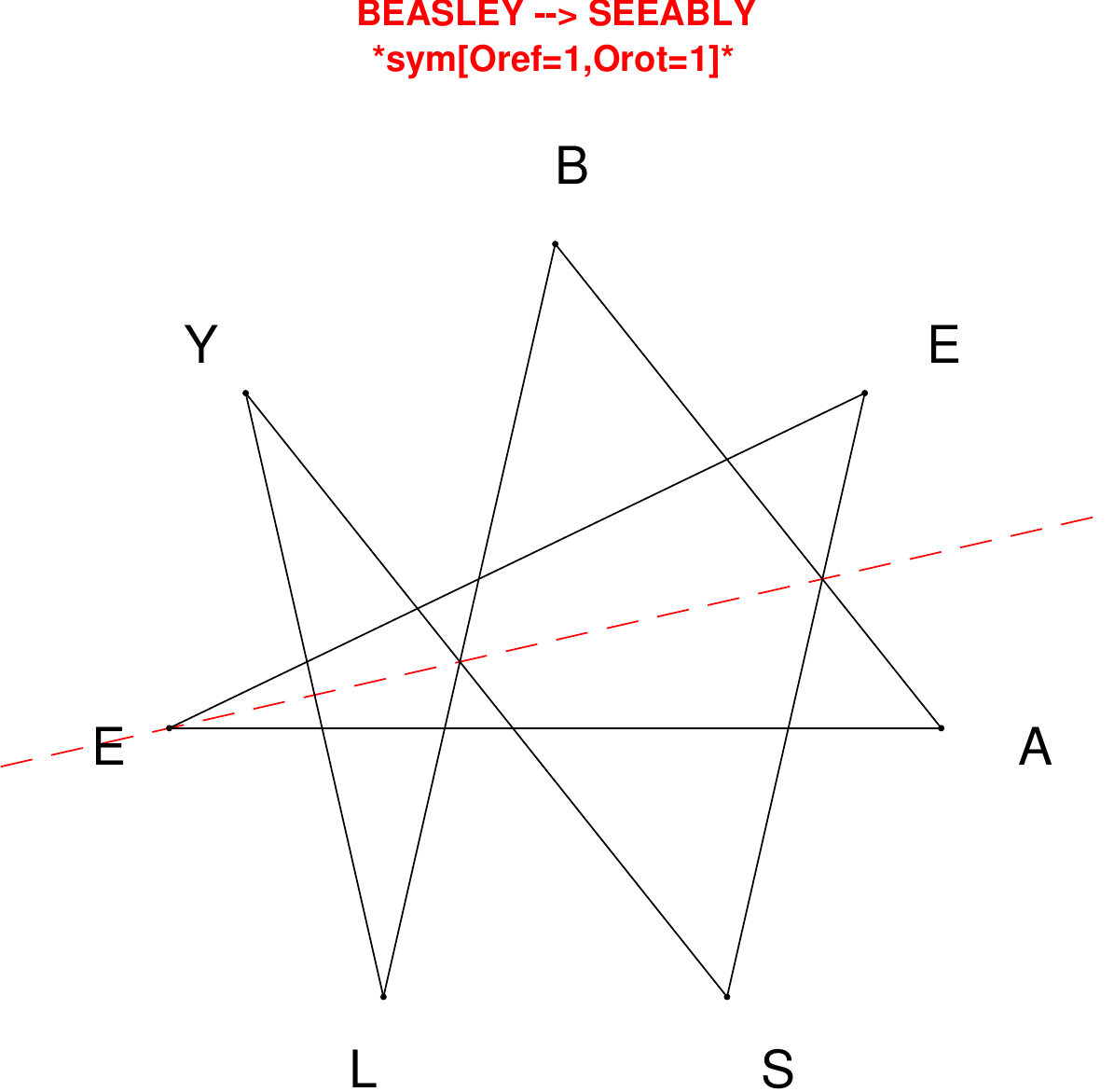}
\end{subfigure}
\hfill
\begin{subfigure}[T]{0.19\textwidth}
\centering
\includegraphics[width=\textwidth]{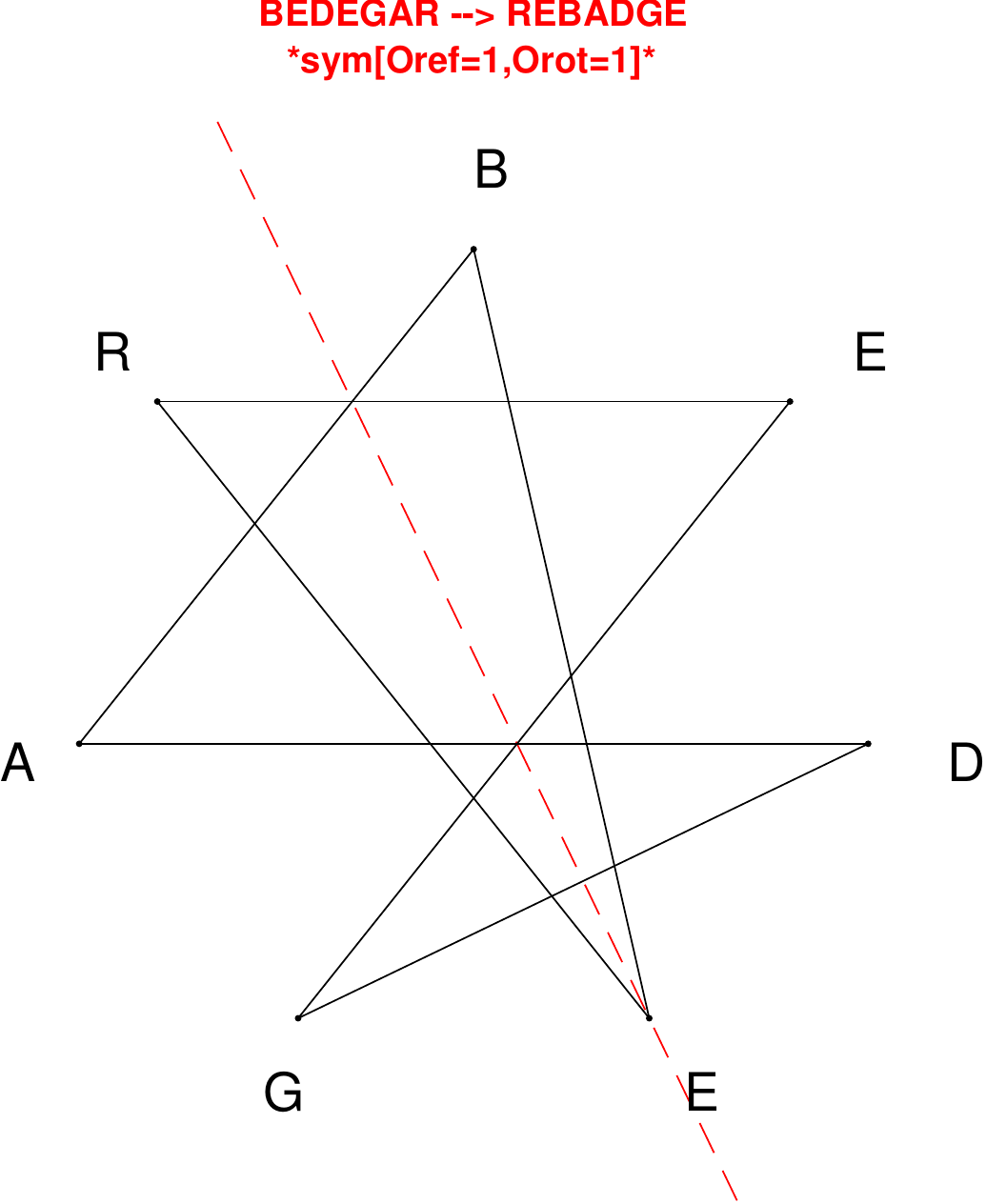}
\end{subfigure}
\hfill
\begin{subfigure}[T]{0.19\textwidth}
\centering
\includegraphics[width=\textwidth]{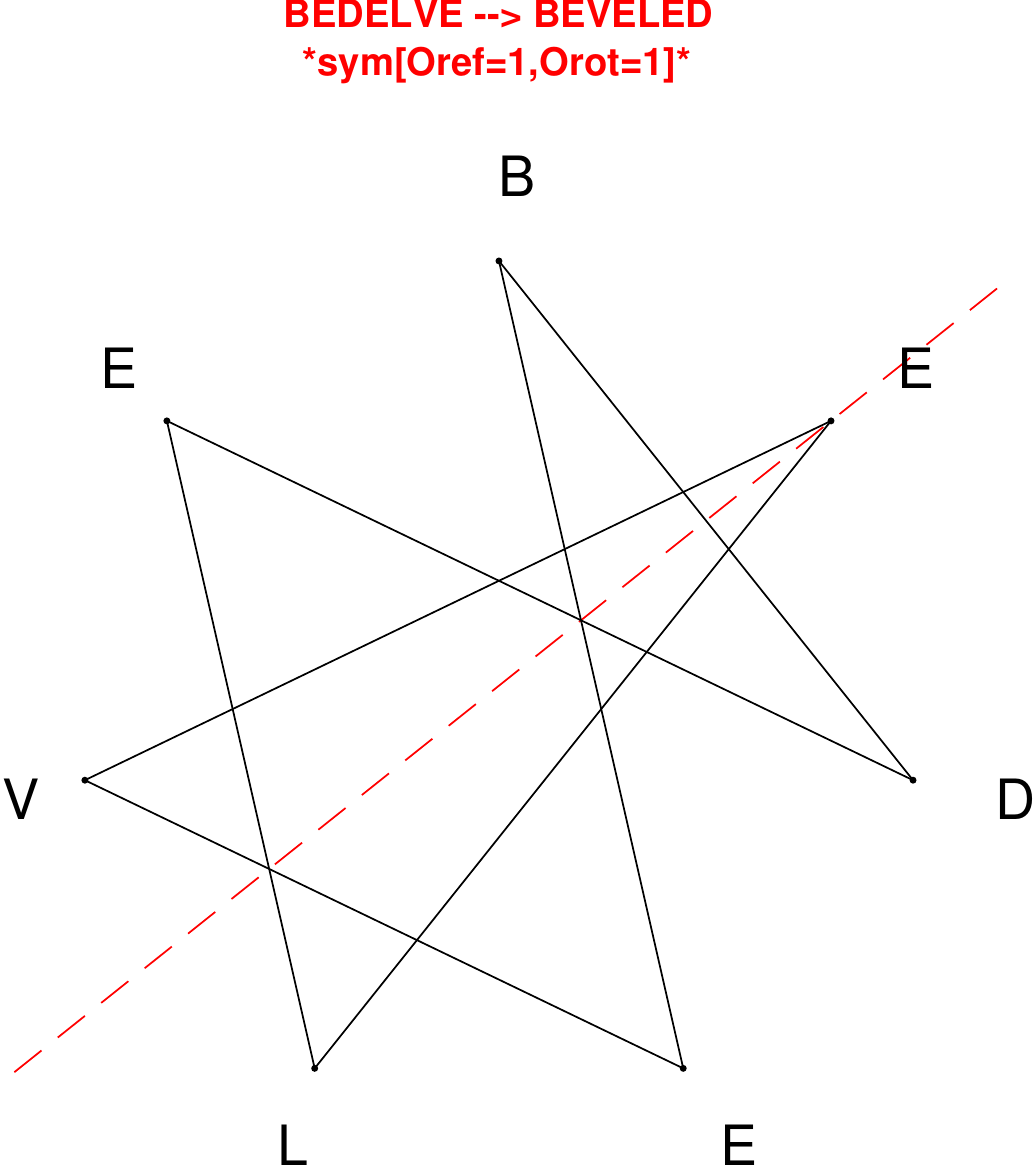}
\end{subfigure}
\hfill
\begin{subfigure}[T]{0.19\textwidth}
\centering
\includegraphics[width=\textwidth]{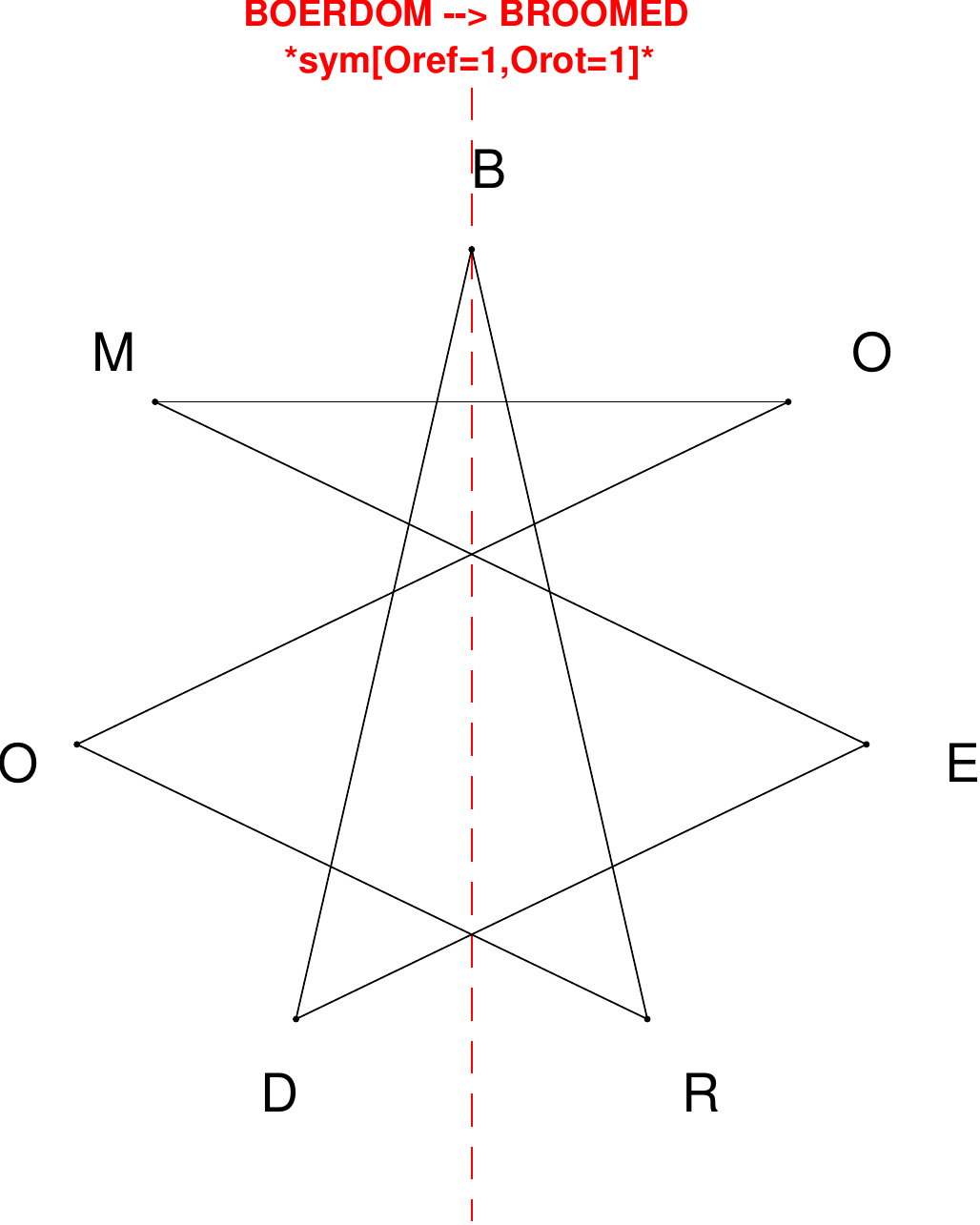}
\end{subfigure}
\hfill
\begin{subfigure}[T]{0.19\textwidth}
\centering
\includegraphics[width=\textwidth]{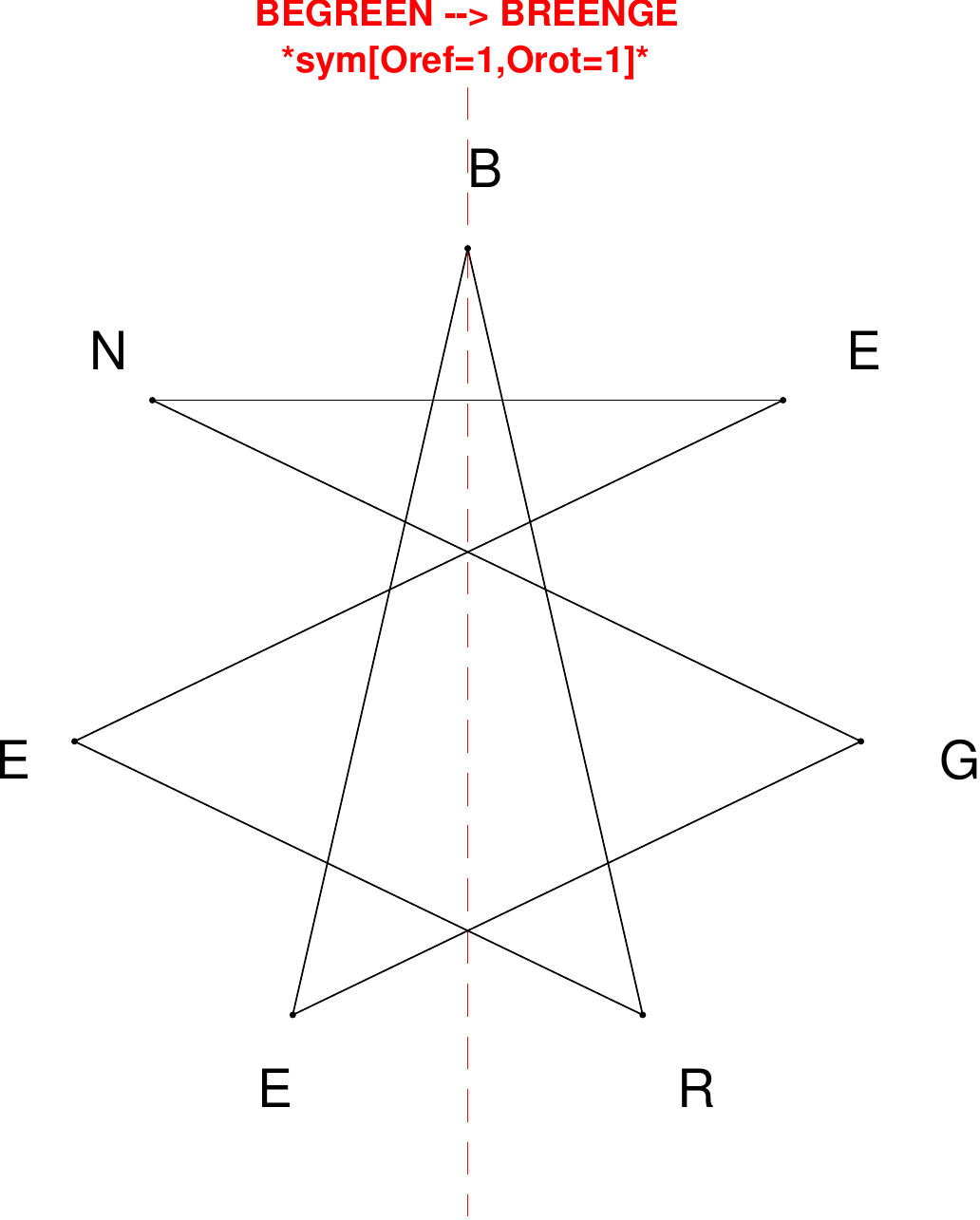}
\end{subfigure}
\end{figure}

\begin{figure}[H]
\centering
\begin{subfigure}[T]{0.19\textwidth}
\centering
\includegraphics[width=\textwidth]{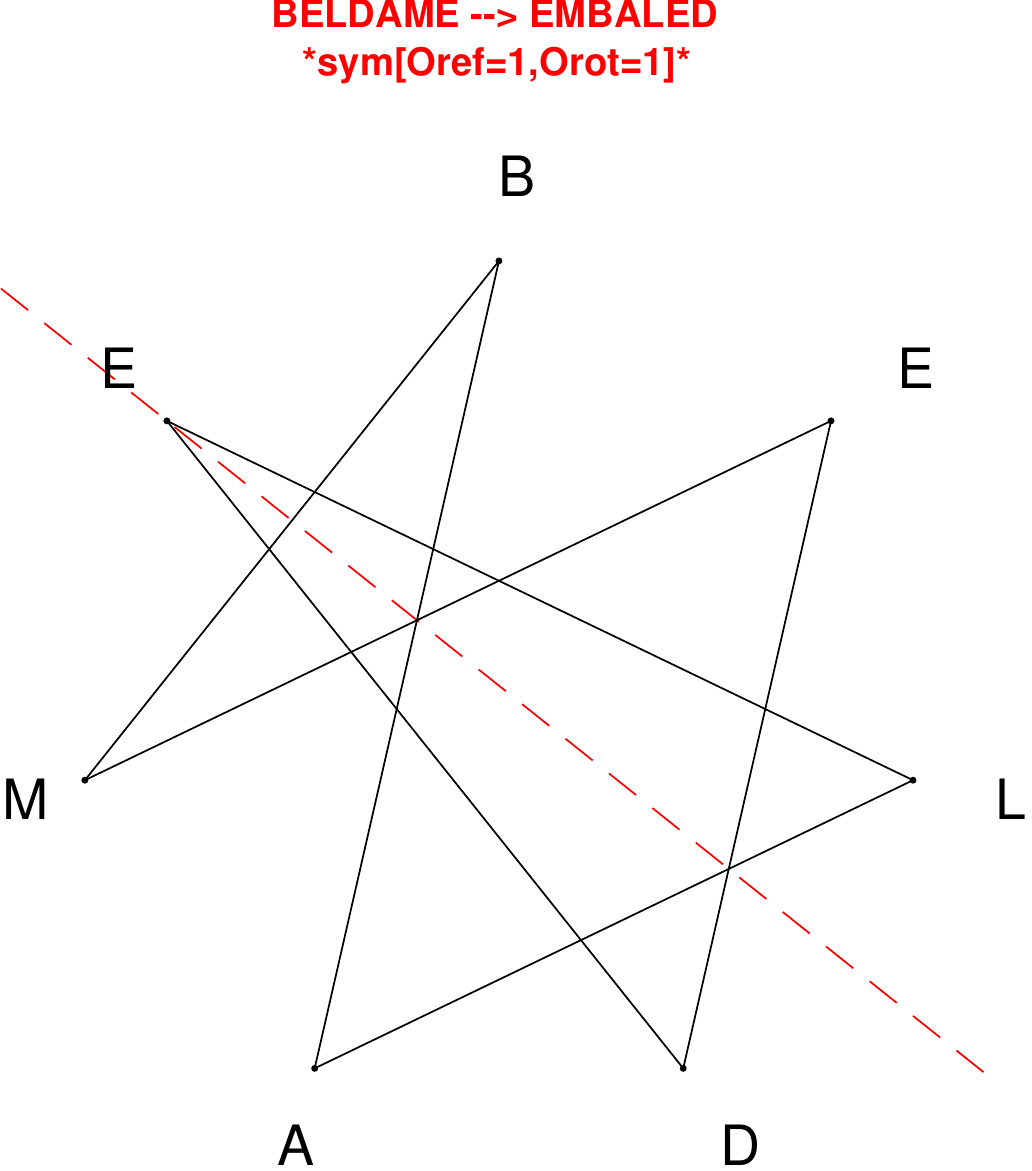}
\end{subfigure}
\hfill
\begin{subfigure}[T]{0.19\textwidth}
\centering
\includegraphics[width=\textwidth]{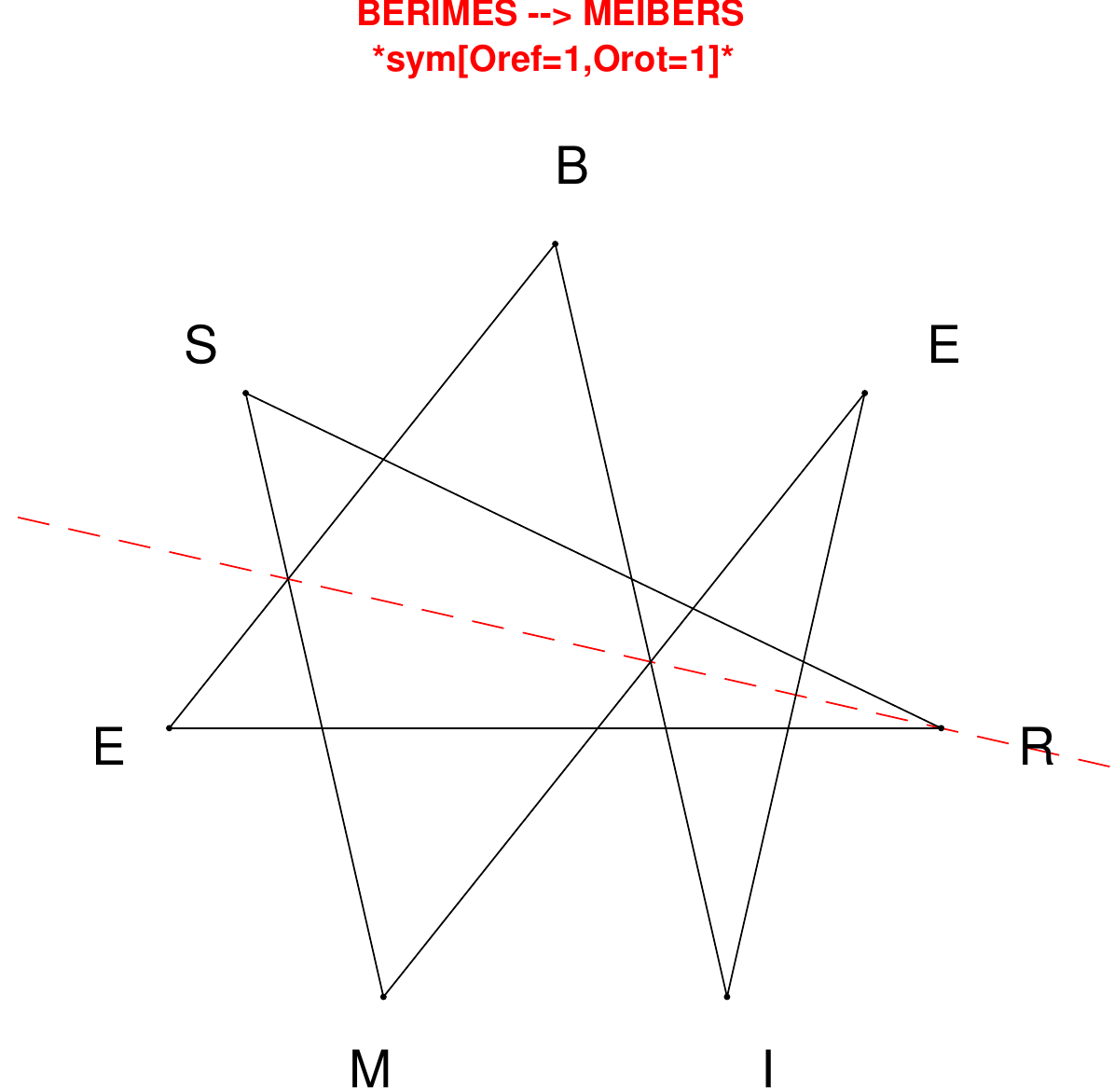}
\end{subfigure}
\hfill
\begin{subfigure}[T]{0.19\textwidth}
\centering
\includegraphics[width=\textwidth]{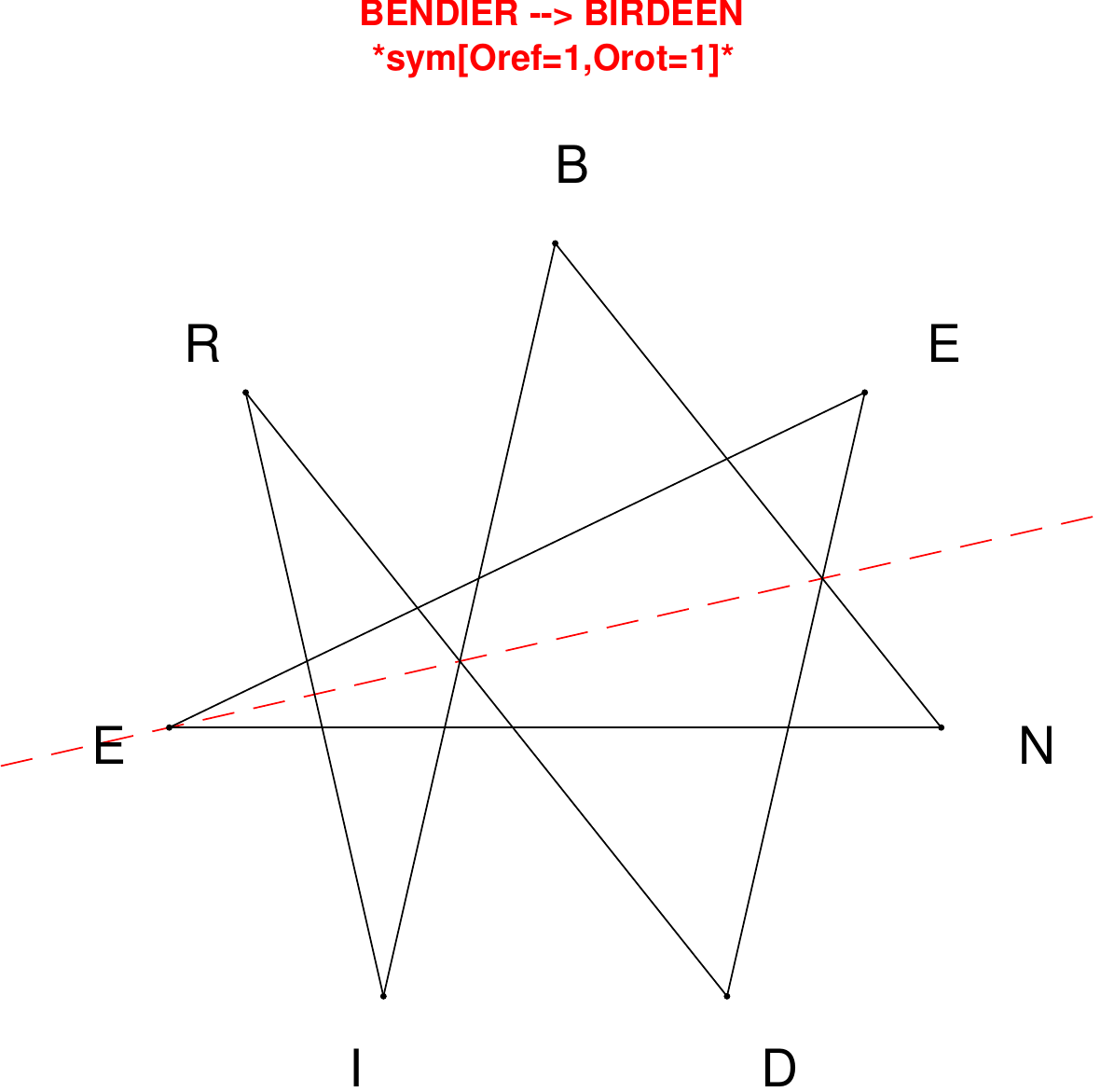}
\end{subfigure}
\hfill
\begin{subfigure}[T]{0.19\textwidth}
\centering
\includegraphics[width=\textwidth]{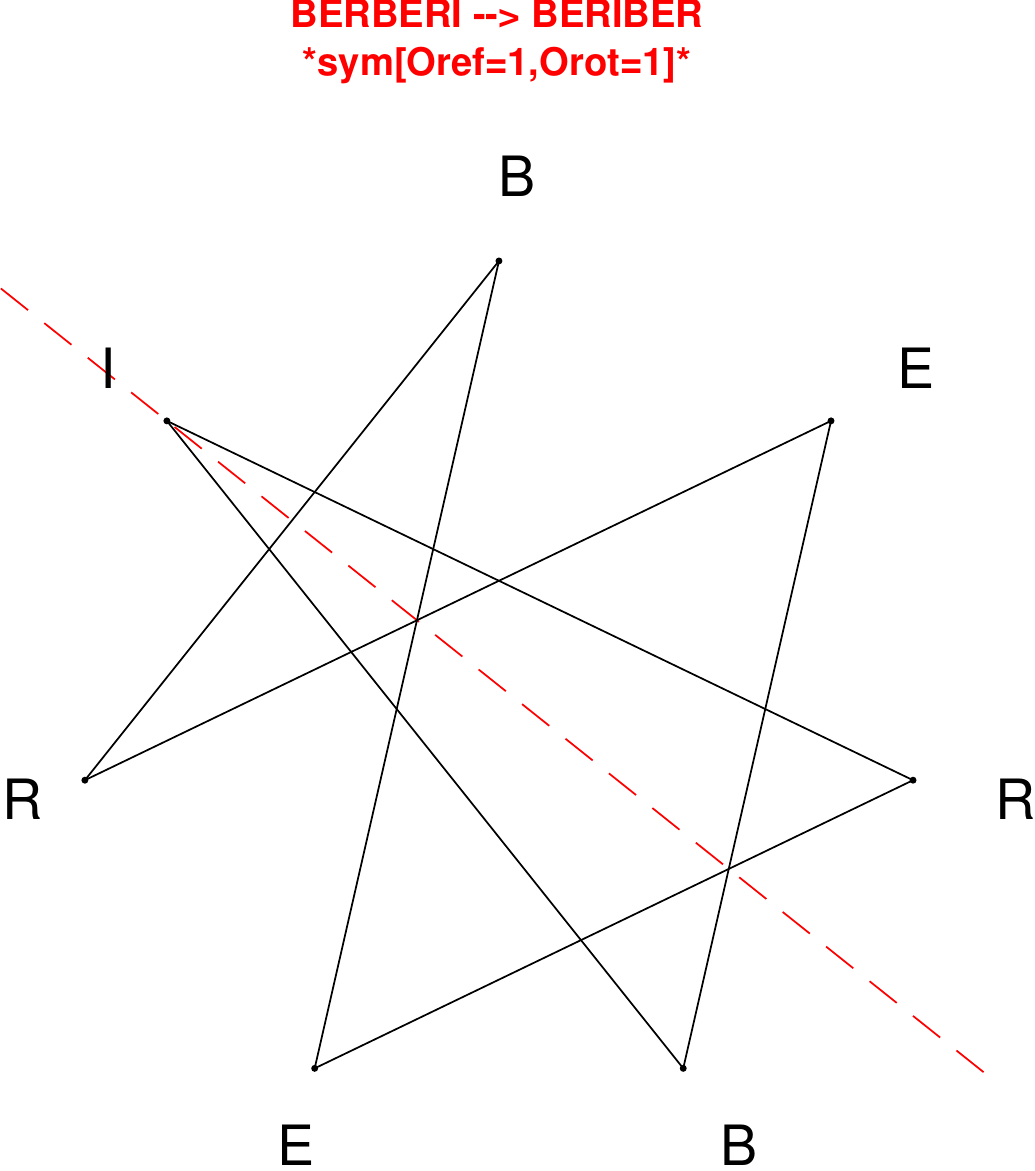}
\end{subfigure}
\hfill
\begin{subfigure}[T]{0.19\textwidth}
\centering
\includegraphics[width=\textwidth]{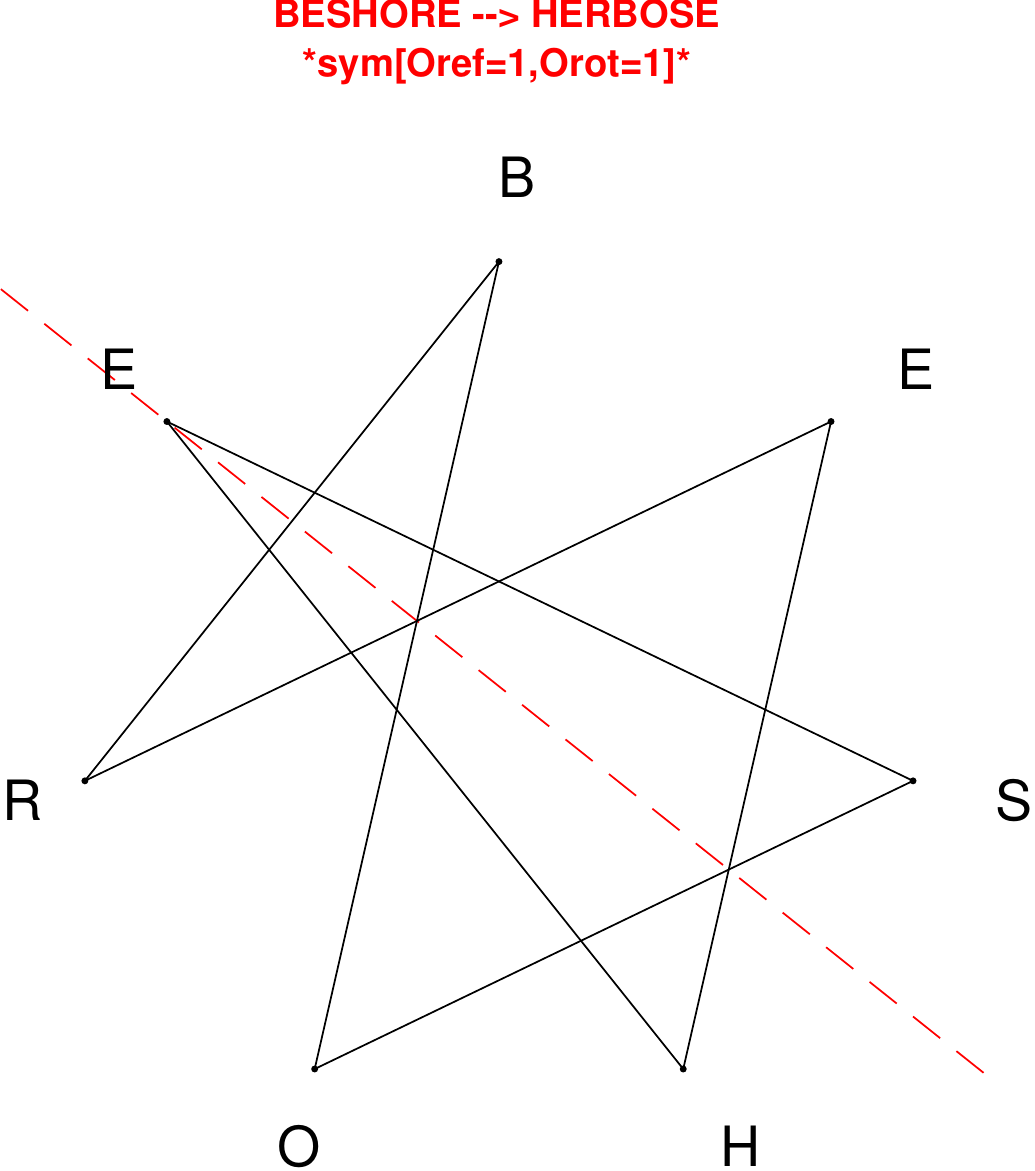}
\end{subfigure}
\end{figure}

\begin{figure}[H]
\centering
\begin{subfigure}[T]{0.19\textwidth}
\centering
\includegraphics[width=\textwidth]{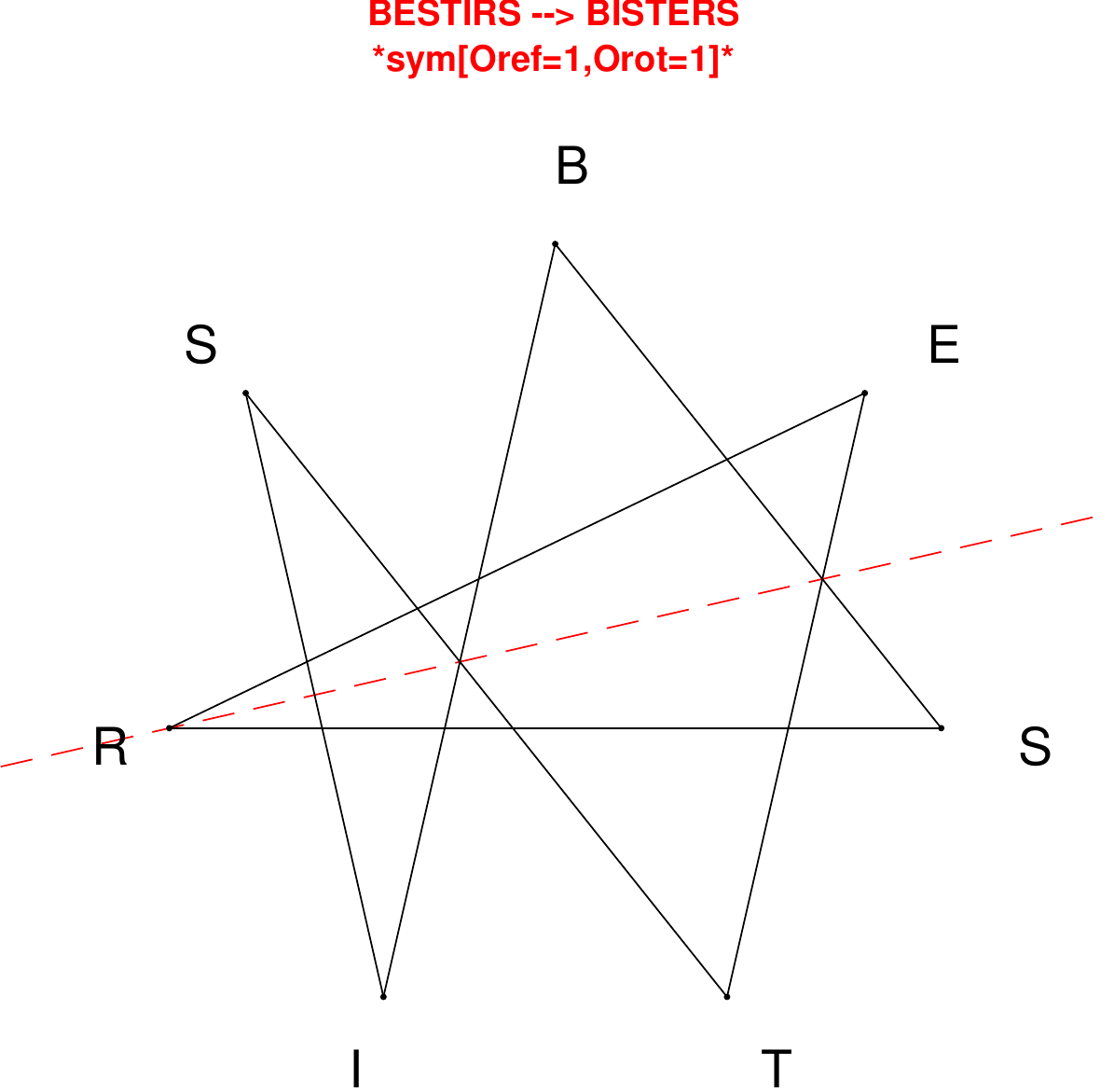}
\end{subfigure}
\hfill
\begin{subfigure}[T]{0.19\textwidth}
\centering
\includegraphics[width=\textwidth]{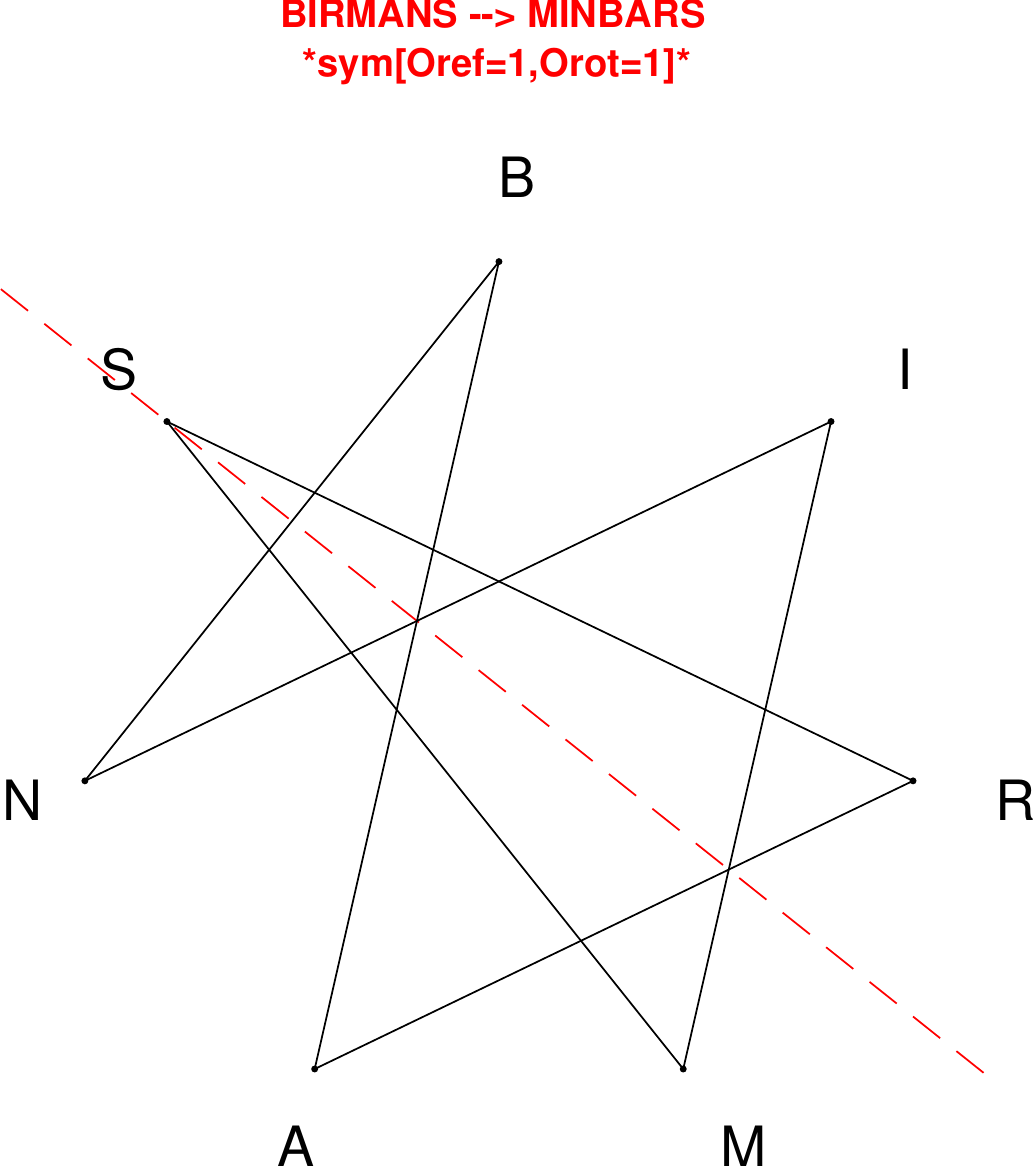}
\end{subfigure}
\hfill
\begin{subfigure}[T]{0.19\textwidth}
\centering
\includegraphics[width=\textwidth]{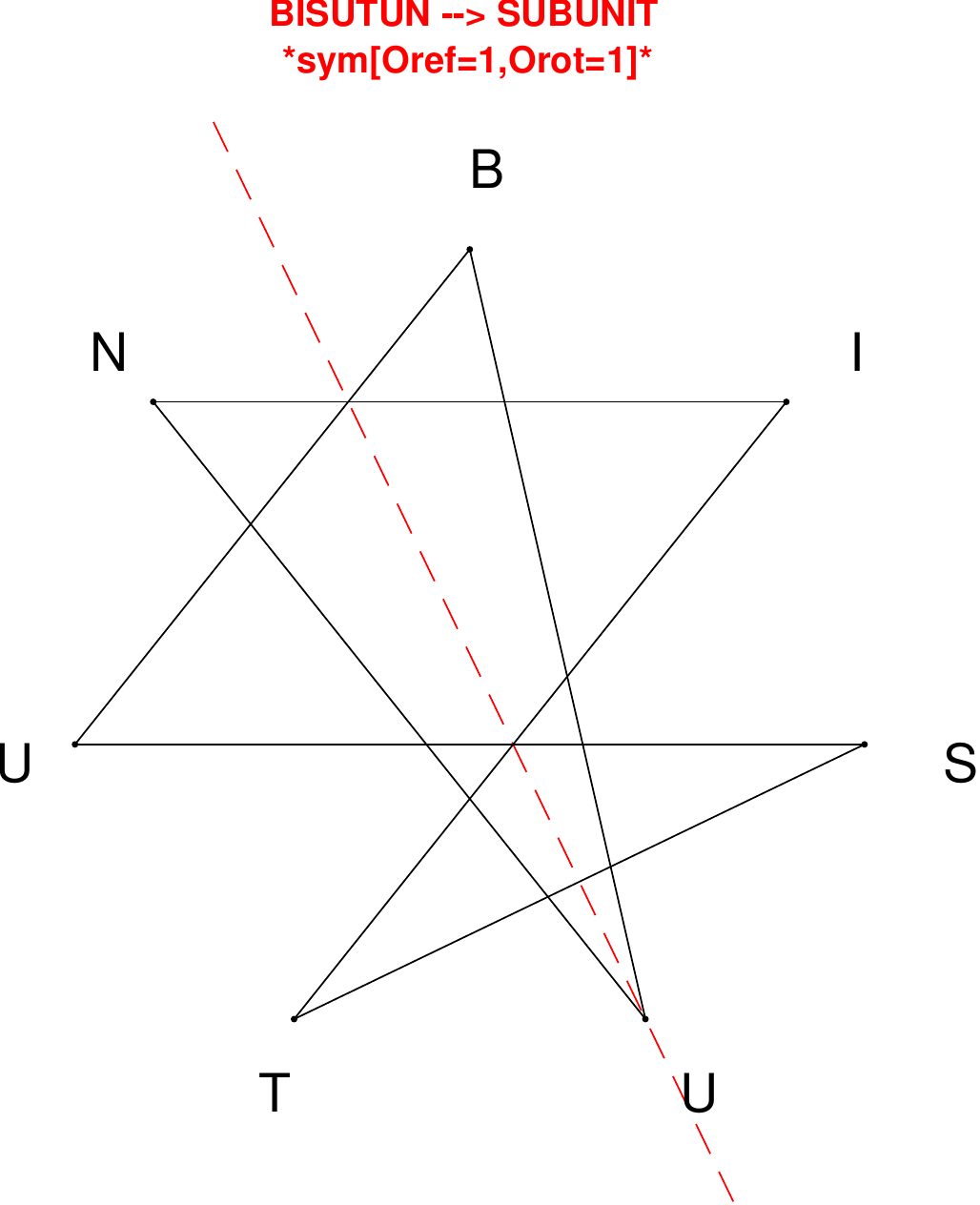}
\end{subfigure}
\hfill
\begin{subfigure}[T]{0.19\textwidth}
\centering
\includegraphics[width=\textwidth]{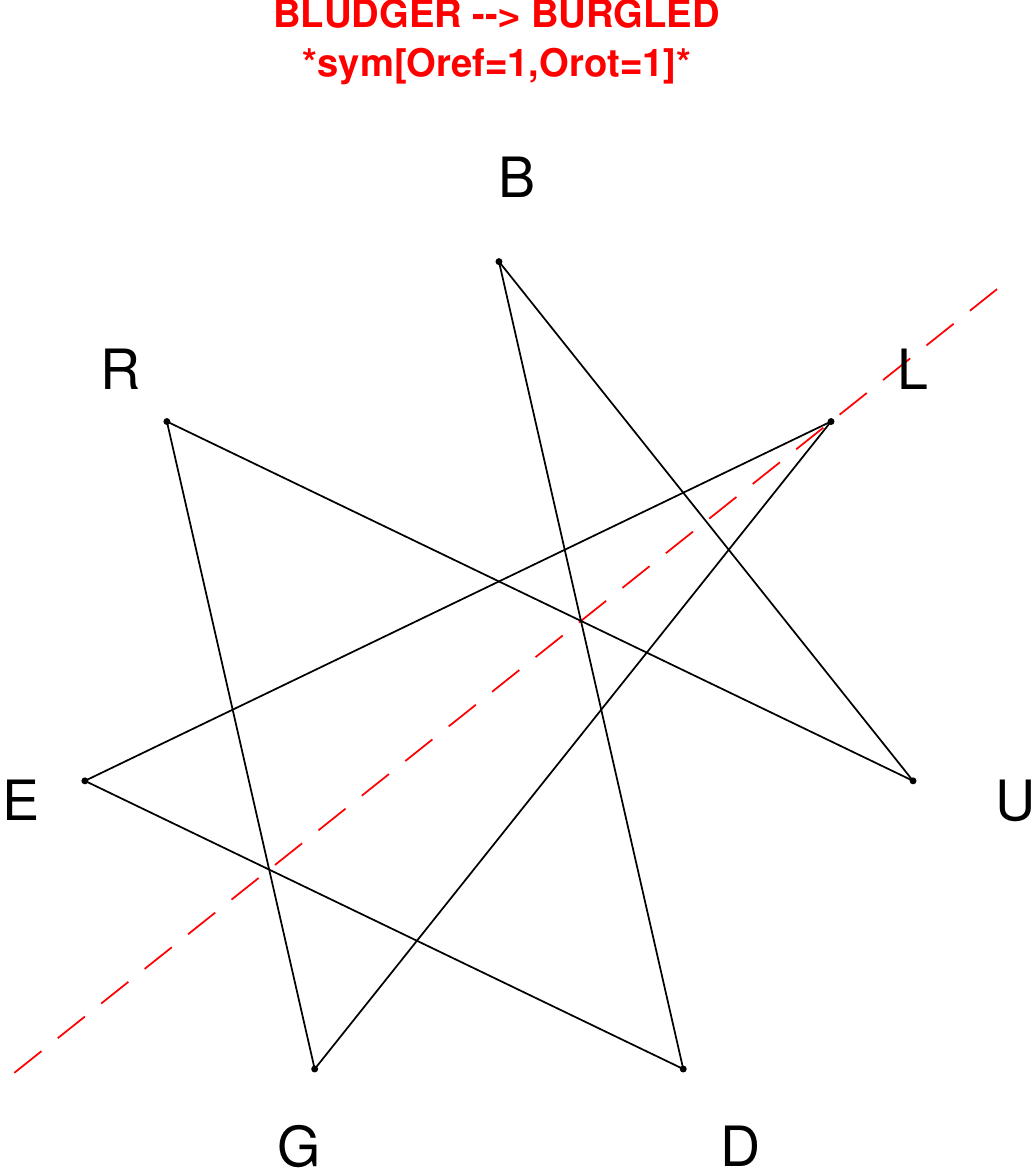}
\end{subfigure}
\hfill
\begin{subfigure}[T]{0.19\textwidth}
\centering
\includegraphics[width=\textwidth]{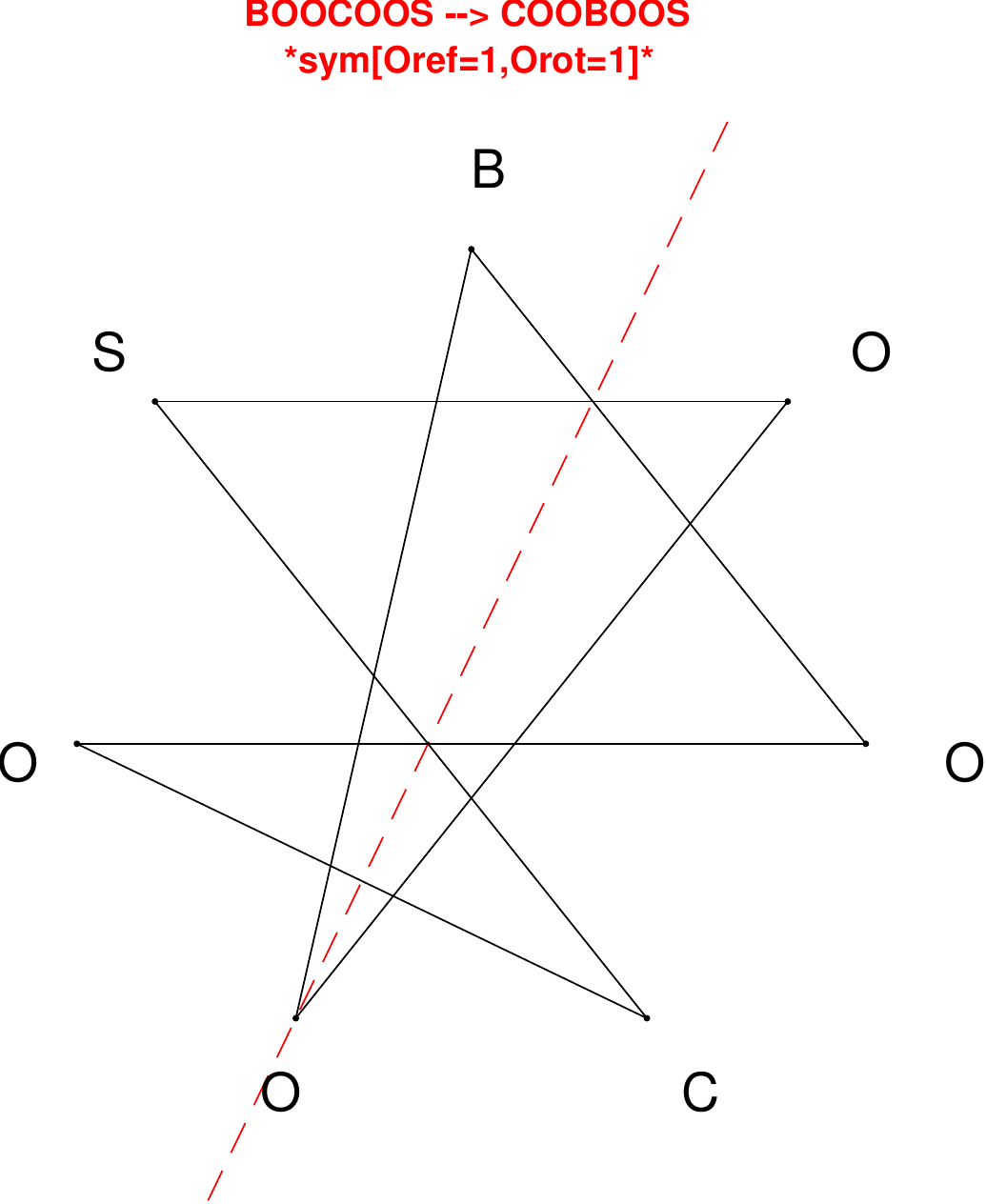}
\end{subfigure}
\end{figure}

\begin{figure}[H]
\centering
\begin{subfigure}[T]{0.19\textwidth}
\centering
\includegraphics[width=\textwidth]{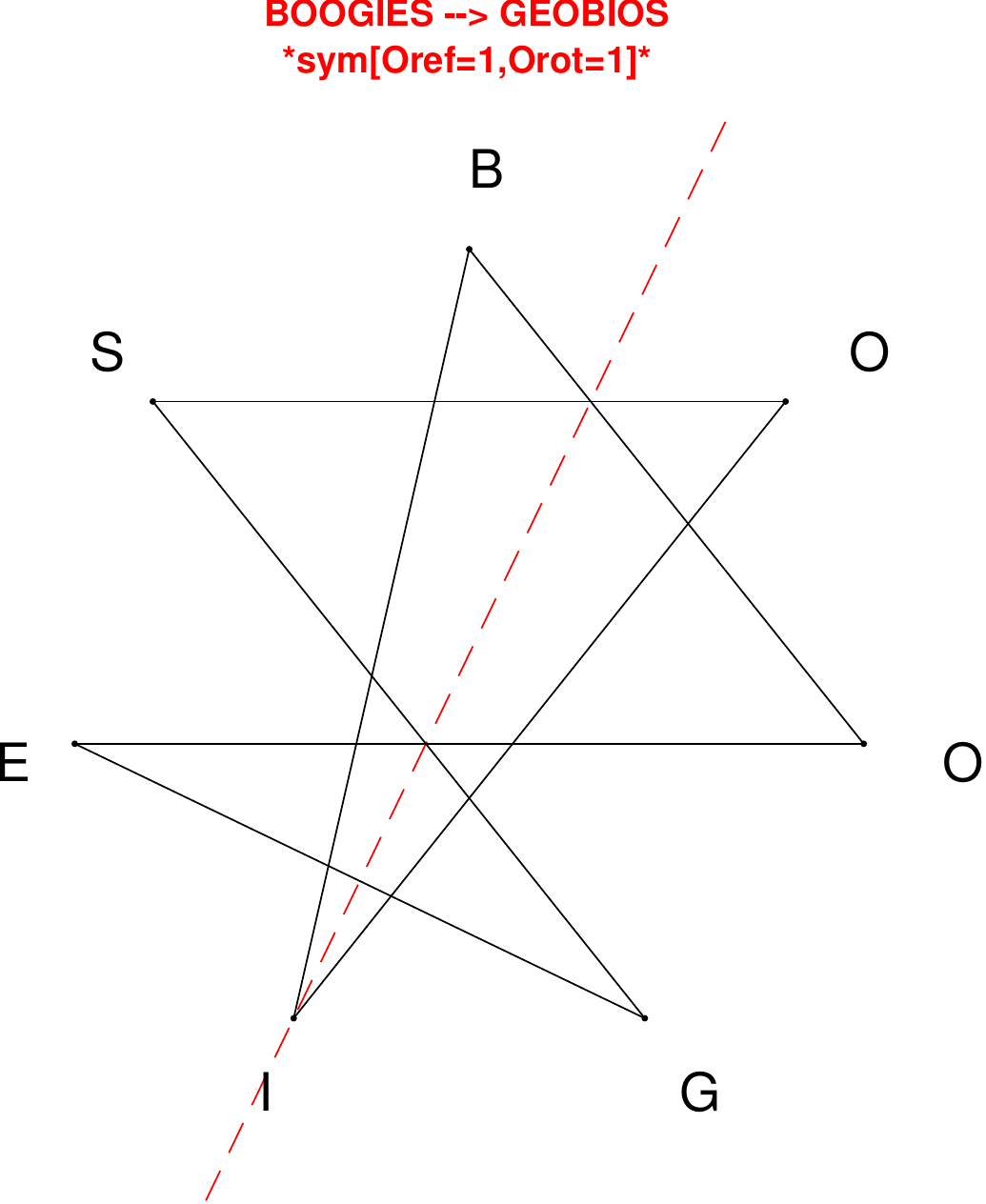}
\end{subfigure}
\hfill
\begin{subfigure}[T]{0.19\textwidth}
\centering
\includegraphics[width=\textwidth]{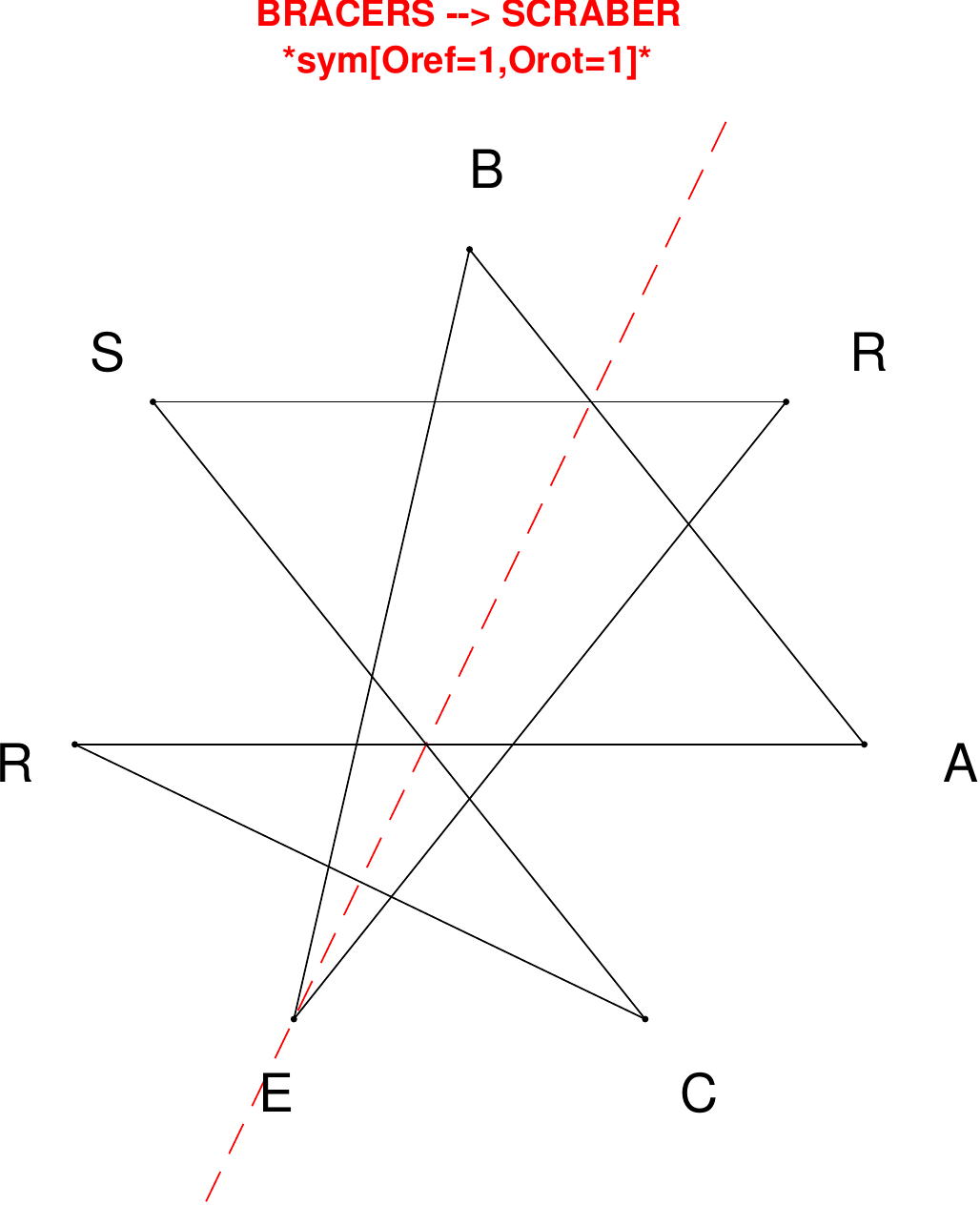}
\end{subfigure}
\hfill
\begin{subfigure}[T]{0.19\textwidth}
\centering
\includegraphics[width=\textwidth]{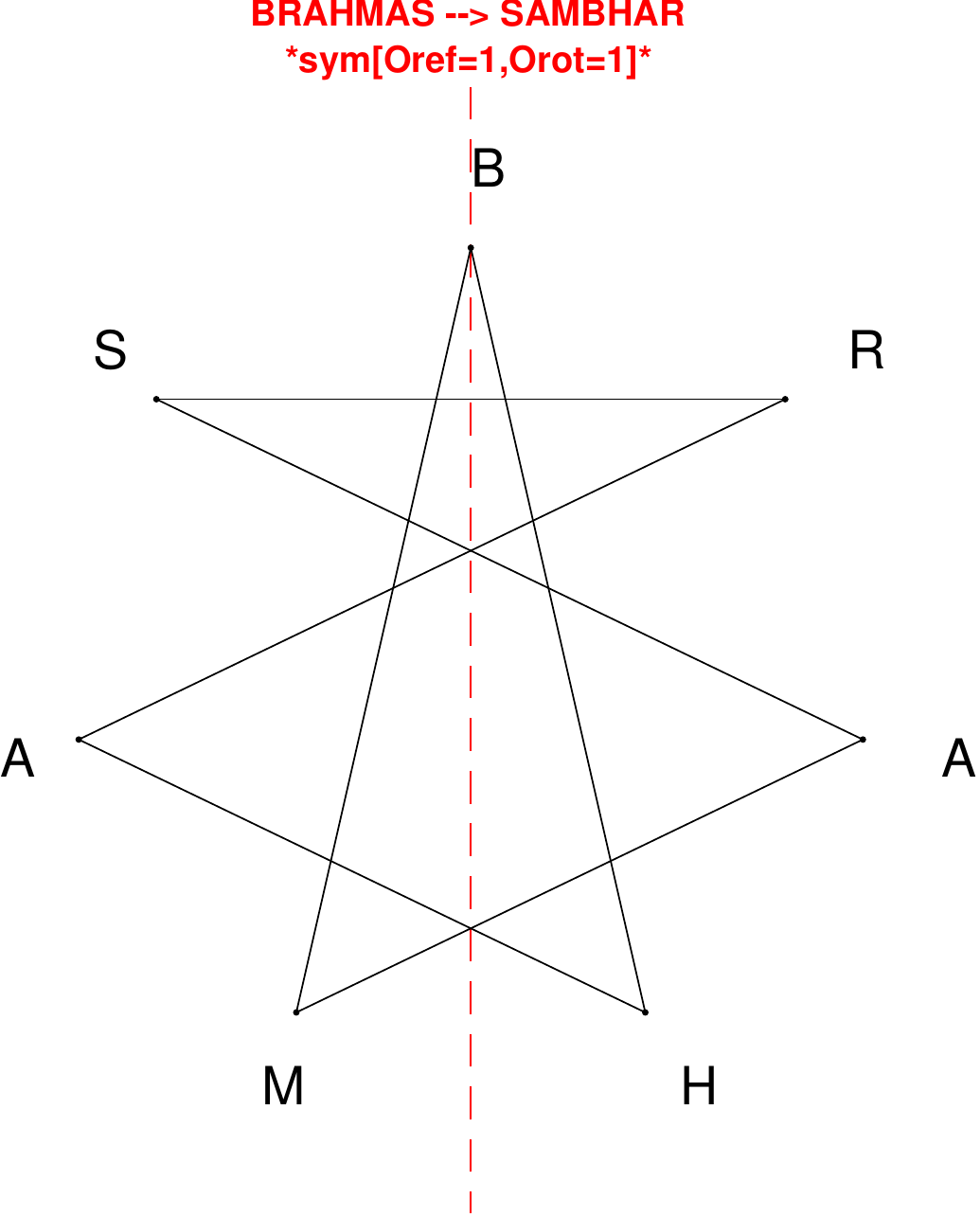}
\end{subfigure}
\hfill
\begin{subfigure}[T]{0.19\textwidth}
\centering
\includegraphics[width=\textwidth]{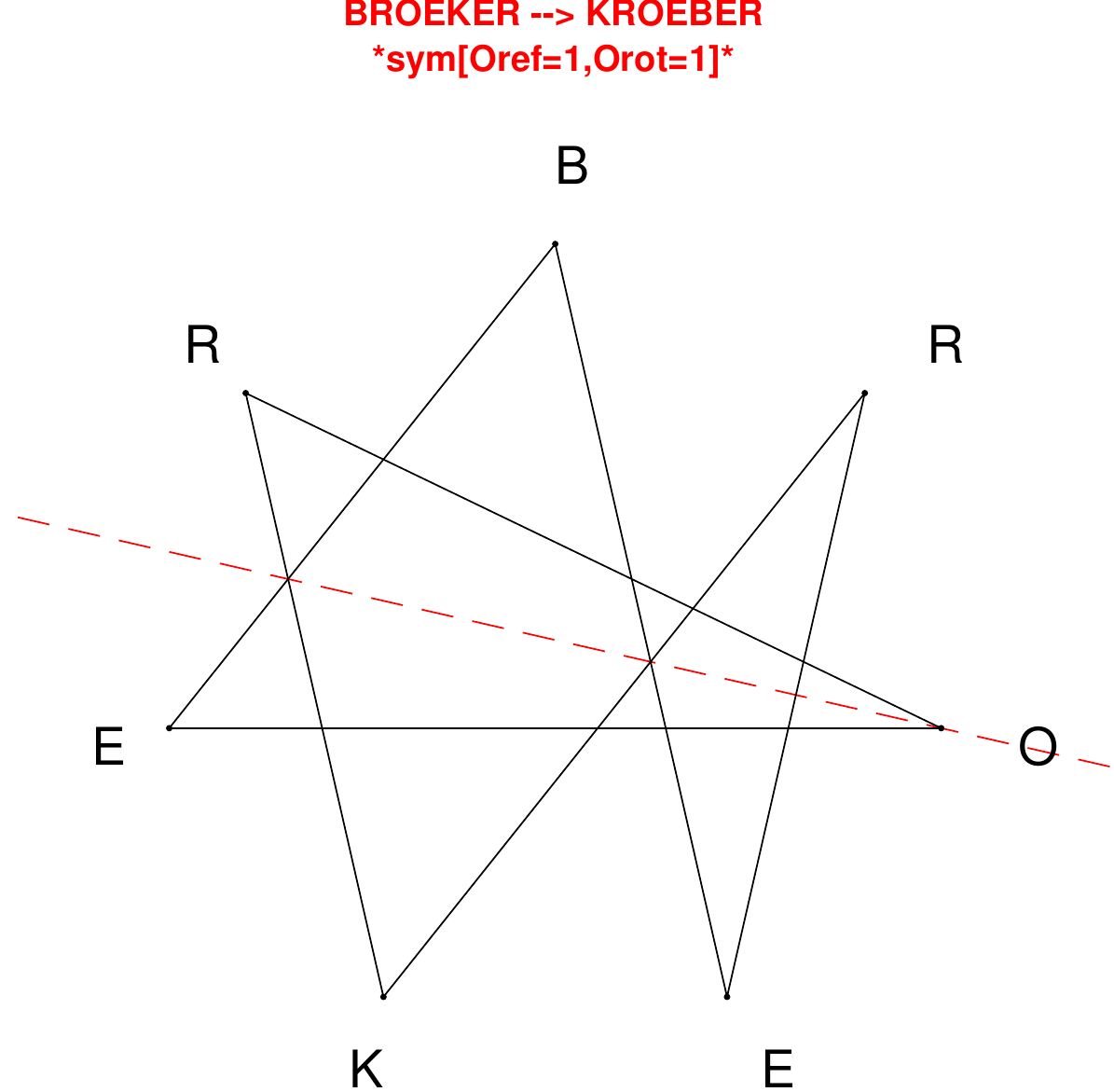}
\end{subfigure}
\hfill
\begin{subfigure}[T]{0.19\textwidth}
\centering
\includegraphics[width=\textwidth]{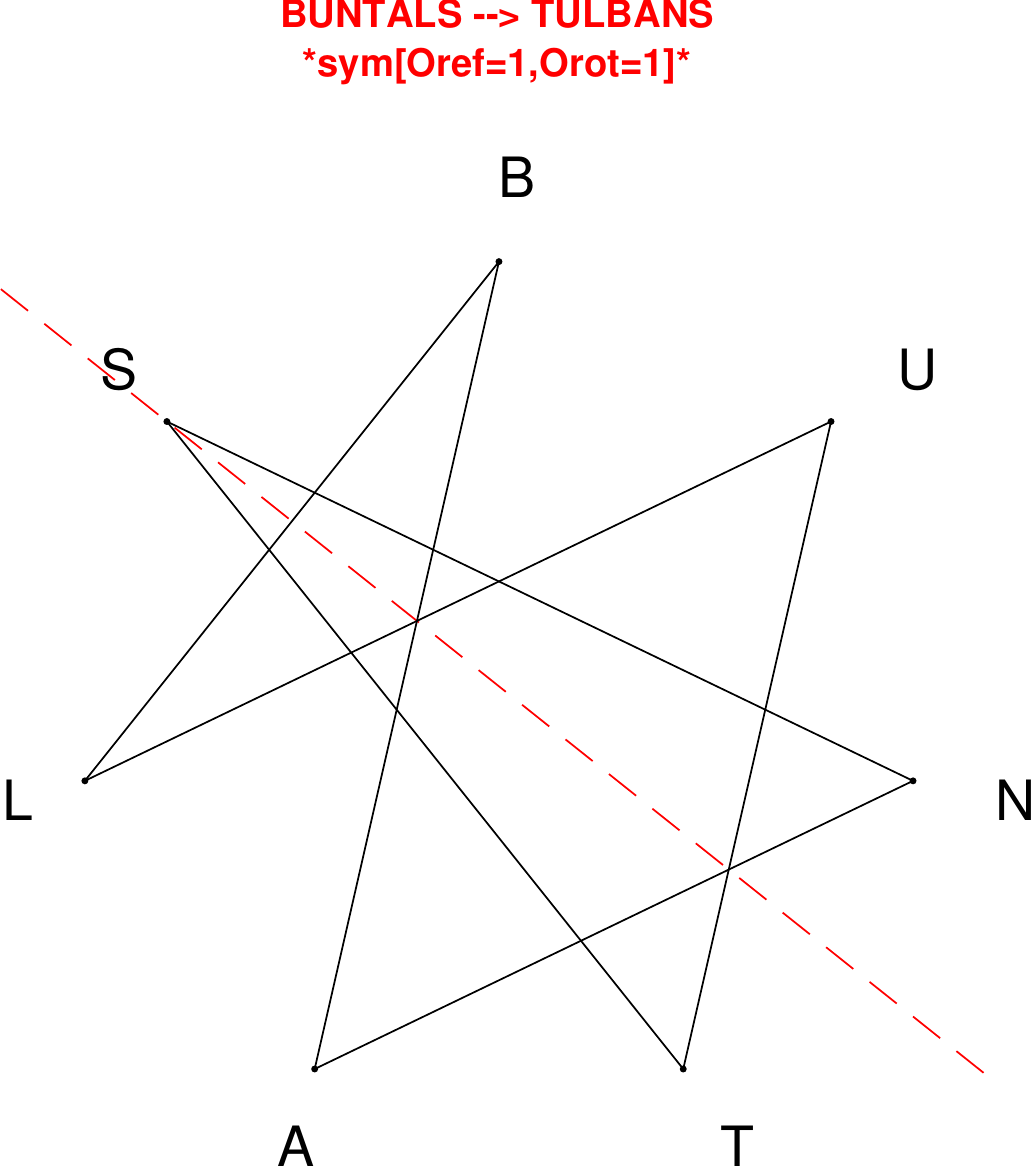}
\end{subfigure}
\end{figure}

\begin{figure}[H]
\centering
\begin{subfigure}[T]{0.19\textwidth}
\centering
\includegraphics[width=\textwidth]{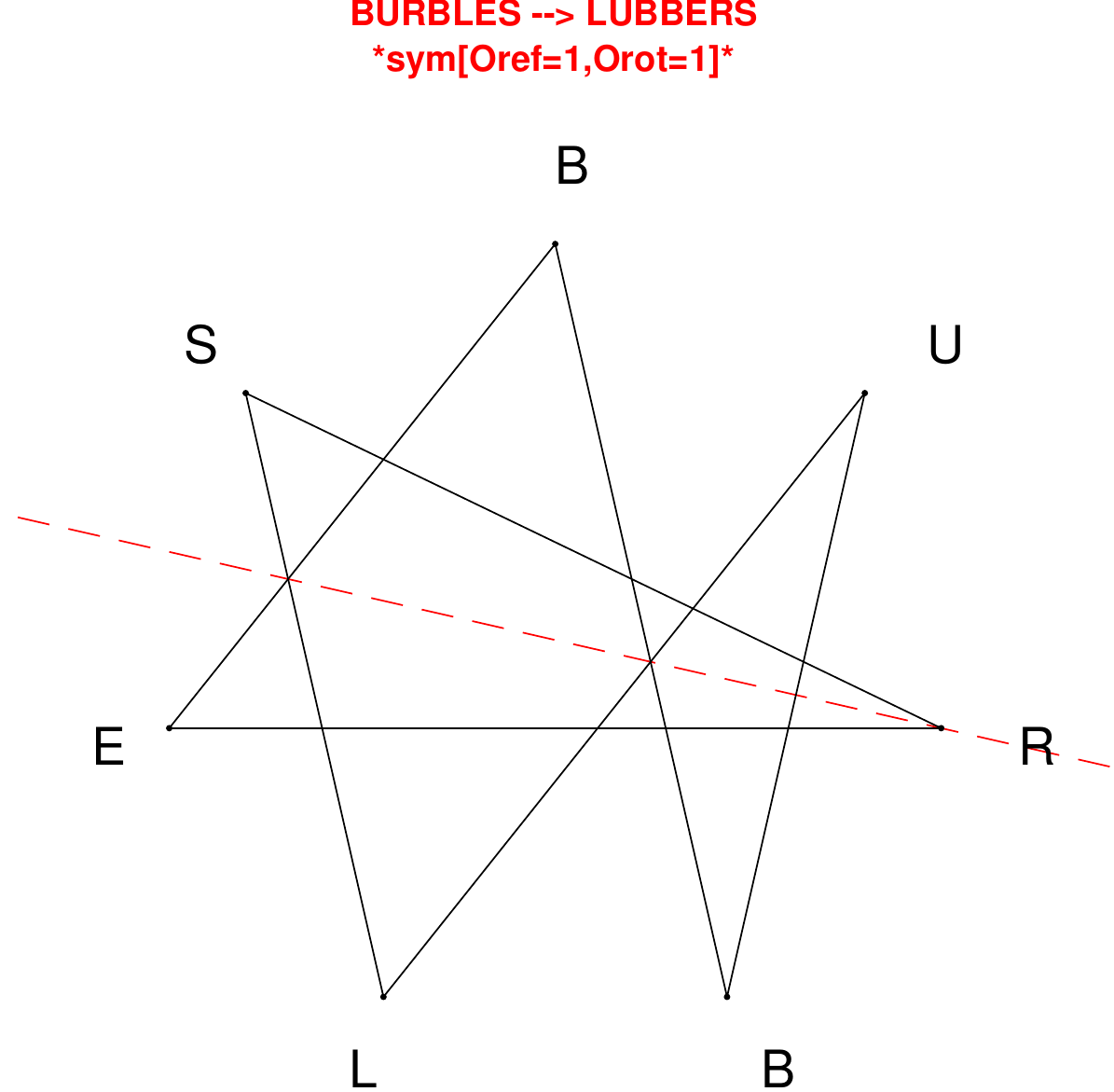}
\end{subfigure}
\hfill
\begin{subfigure}[T]{0.19\textwidth}
\centering
\includegraphics[width=\textwidth]{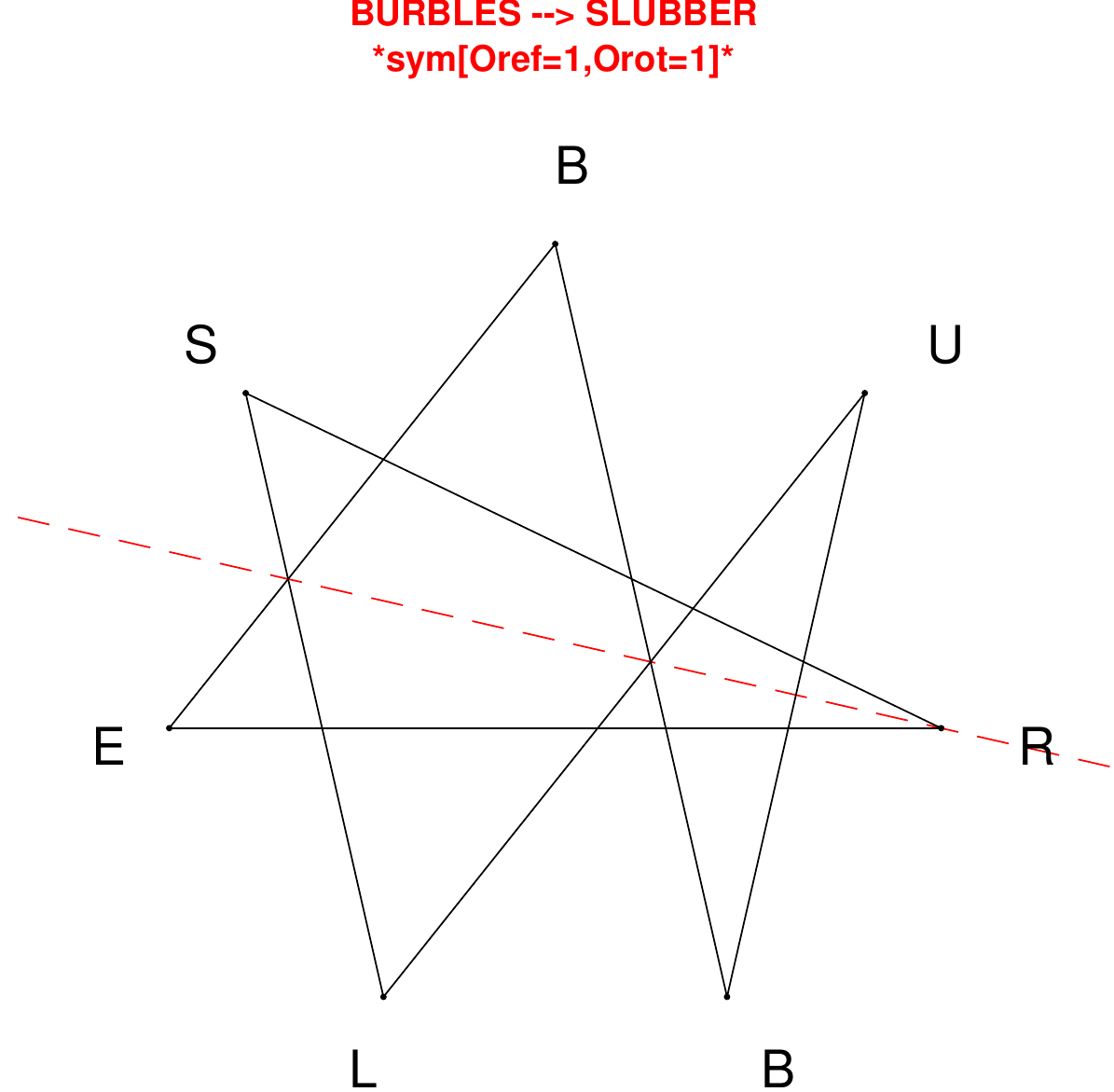}
\end{subfigure}
\hfill
\begin{subfigure}[T]{0.19\textwidth}
\centering
\includegraphics[width=\textwidth]{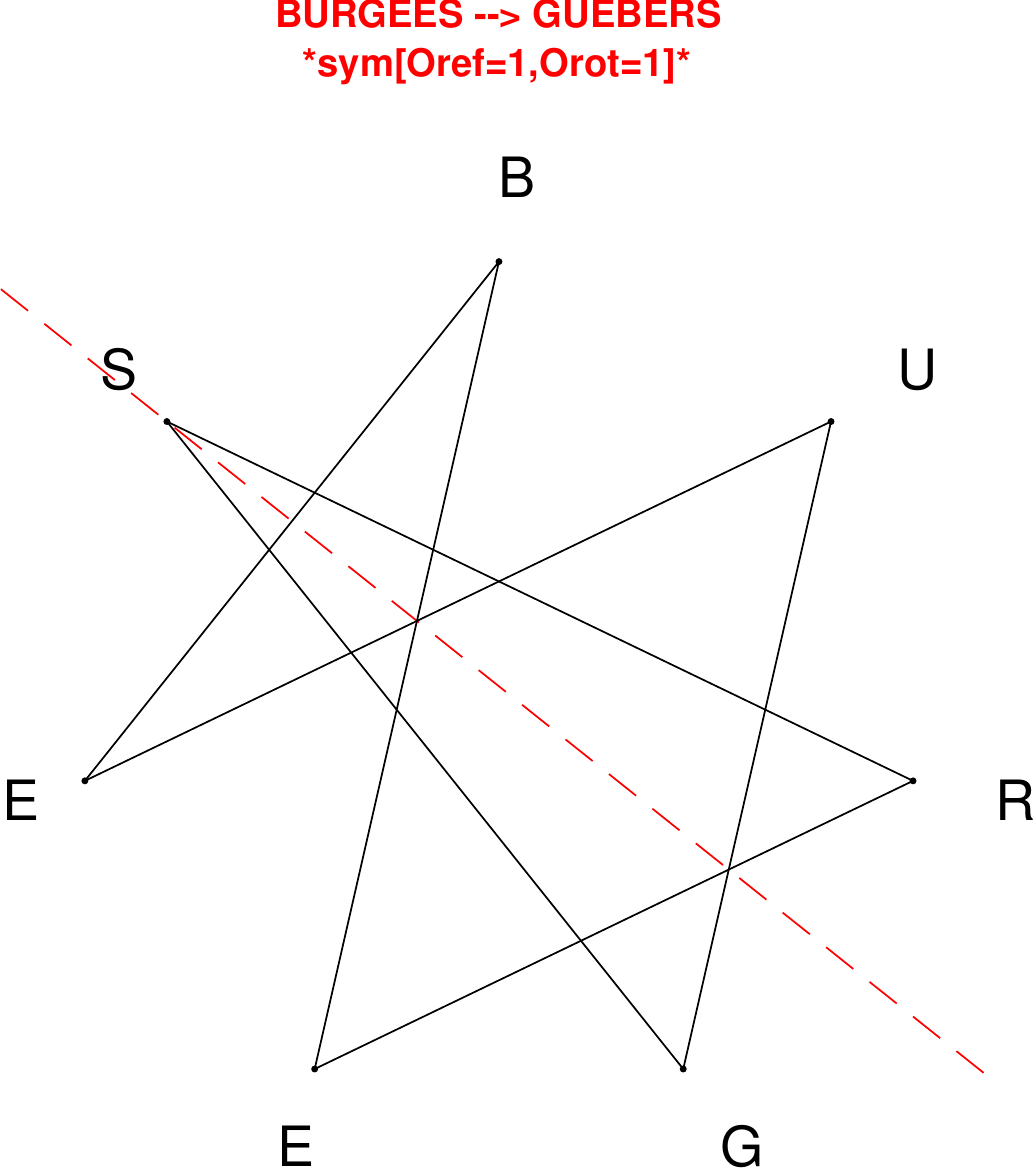}
\end{subfigure}
\hfill
\begin{subfigure}[T]{0.19\textwidth}
\centering
\includegraphics[width=\textwidth]{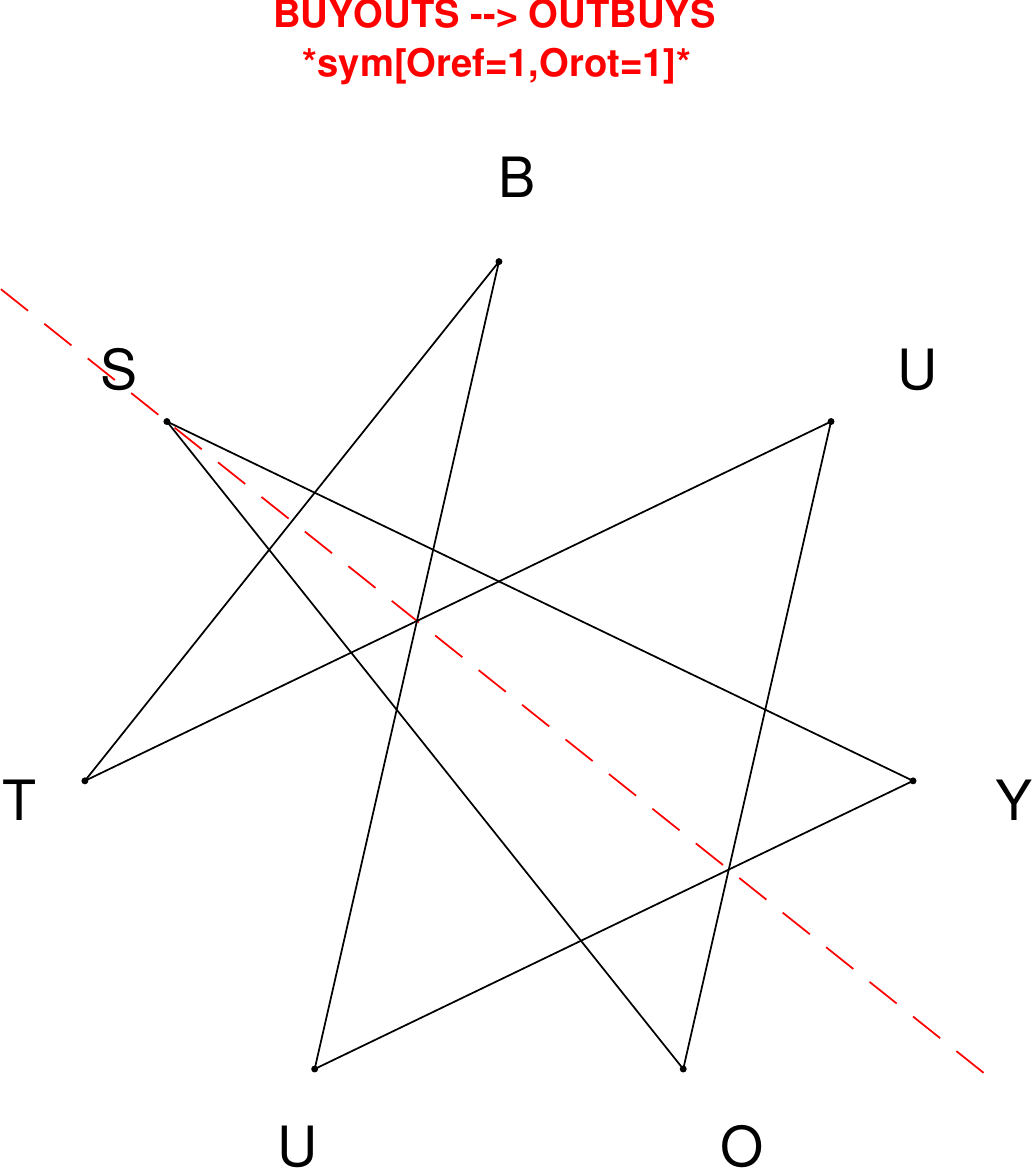}
\end{subfigure}
\hfill
\begin{subfigure}[T]{0.19\textwidth}
\centering
\includegraphics[width=\textwidth]{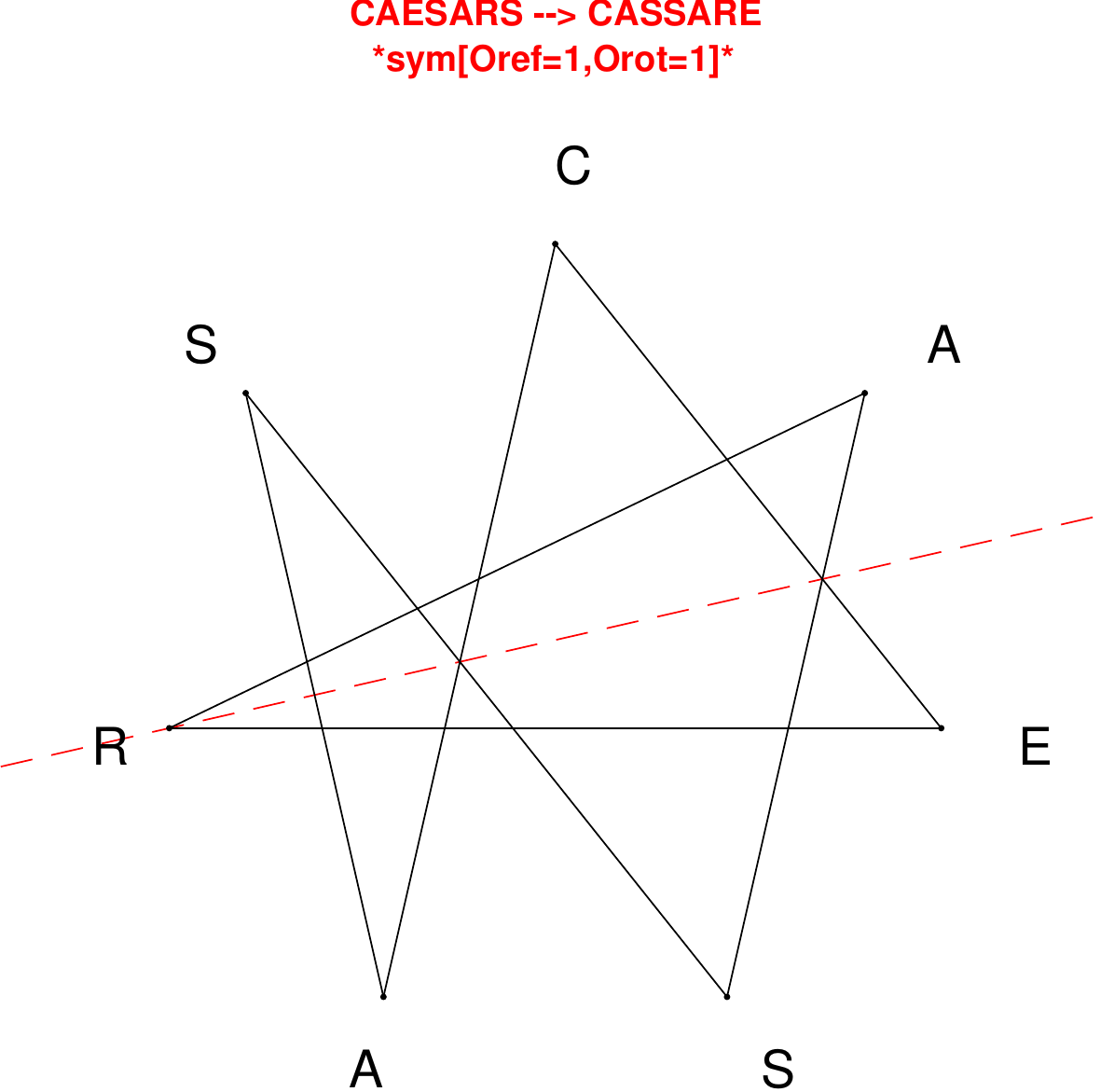}
\end{subfigure}
\end{figure}

\begin{figure}[H]
\centering
\begin{subfigure}[T]{0.19\textwidth}
\centering
\includegraphics[width=\textwidth]{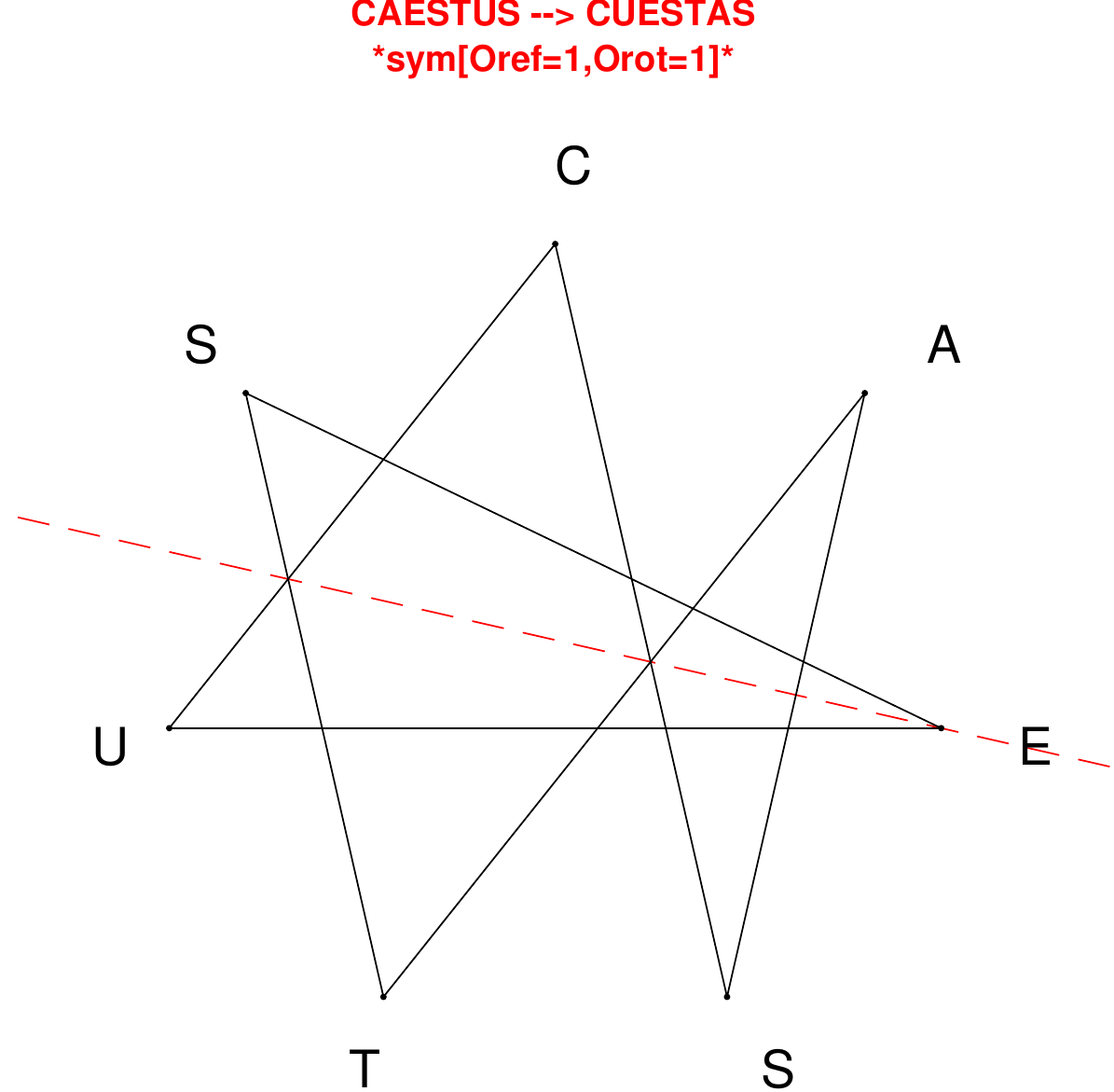}
\end{subfigure}
\hfill
\begin{subfigure}[T]{0.19\textwidth}
\centering
\includegraphics[width=\textwidth]{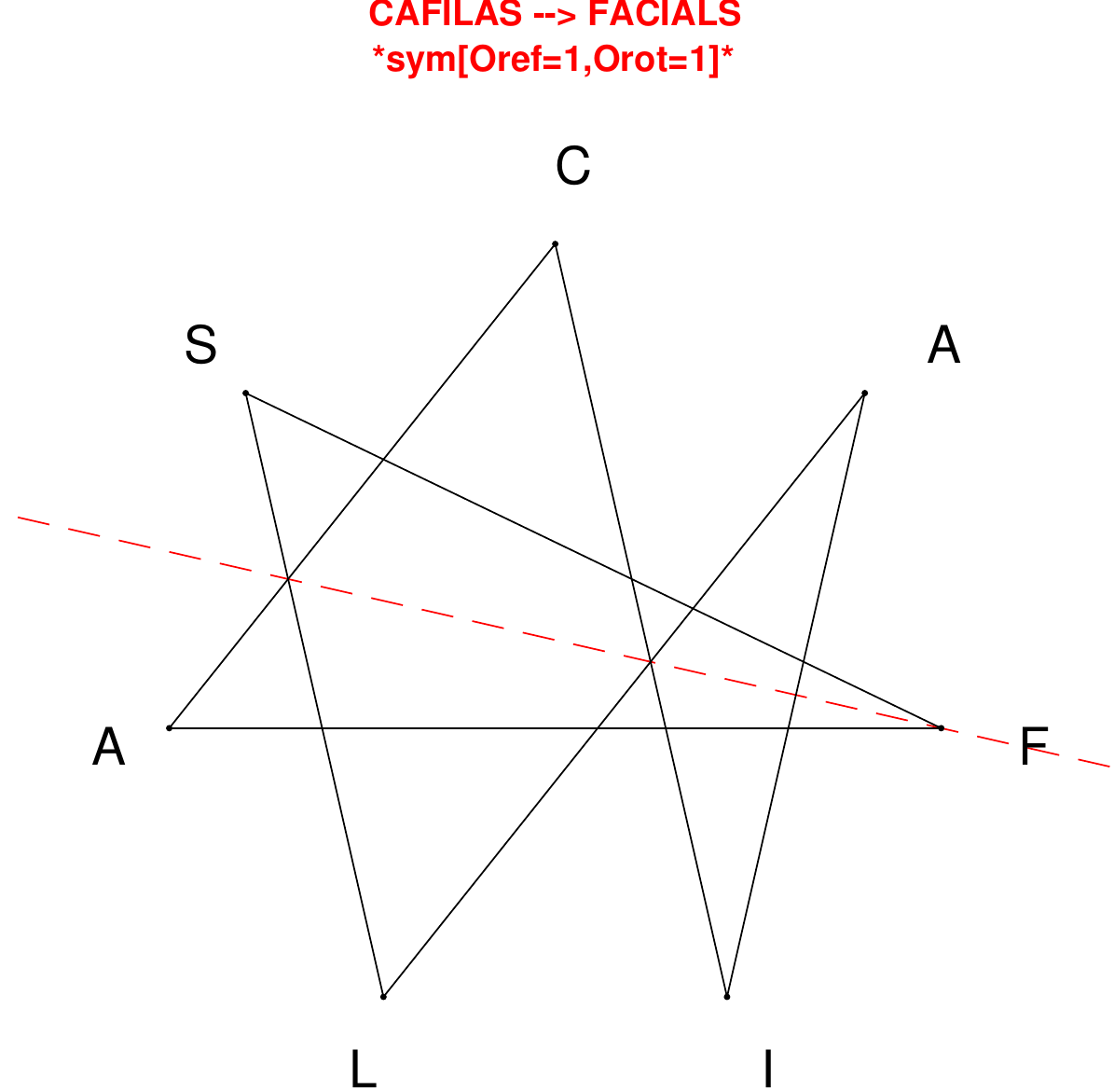}
\end{subfigure}
\hfill
\begin{subfigure}[T]{0.19\textwidth}
\centering
\includegraphics[width=\textwidth]{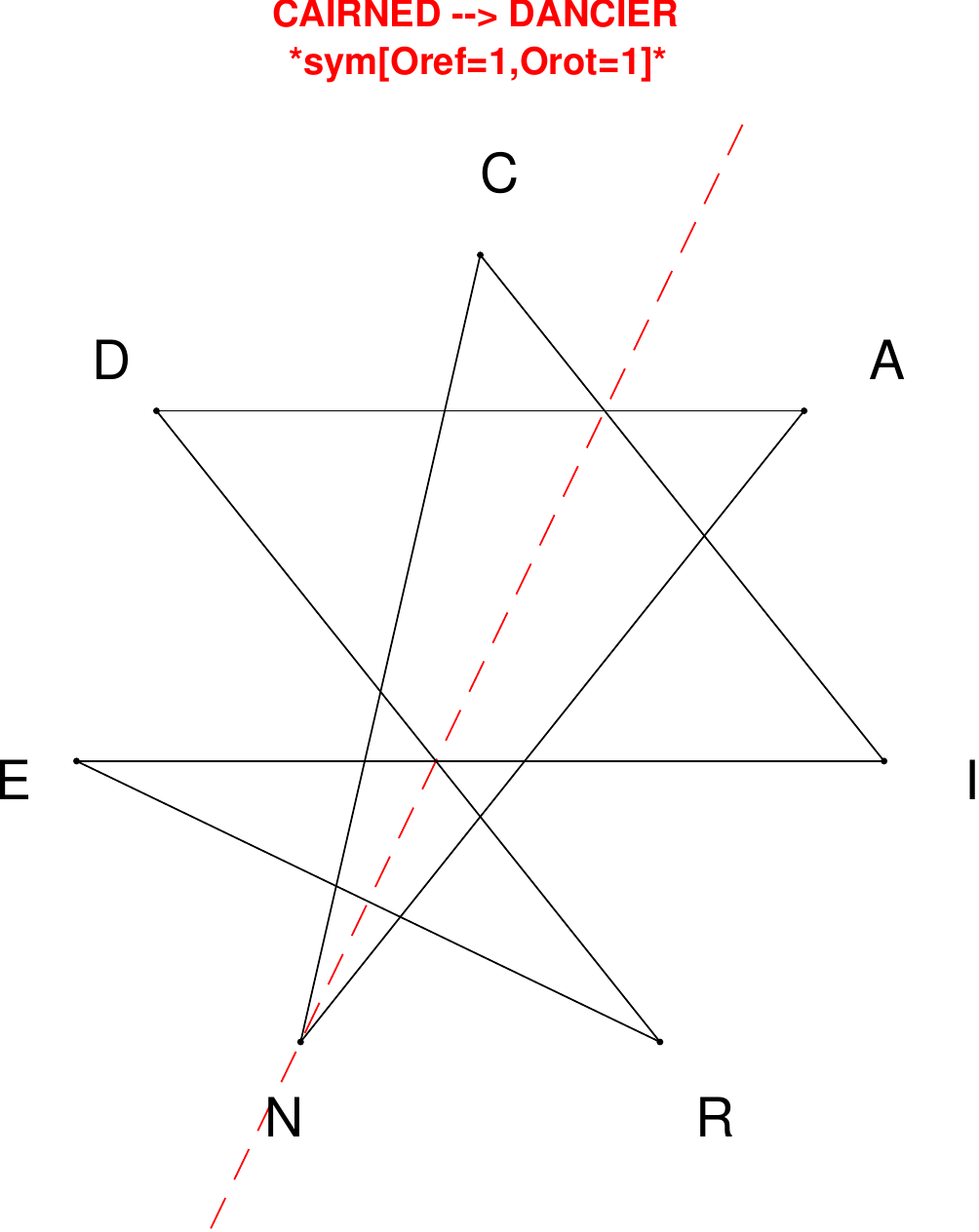}
\end{subfigure}
\hfill
\begin{subfigure}[T]{0.19\textwidth}
\centering
\includegraphics[width=\textwidth]{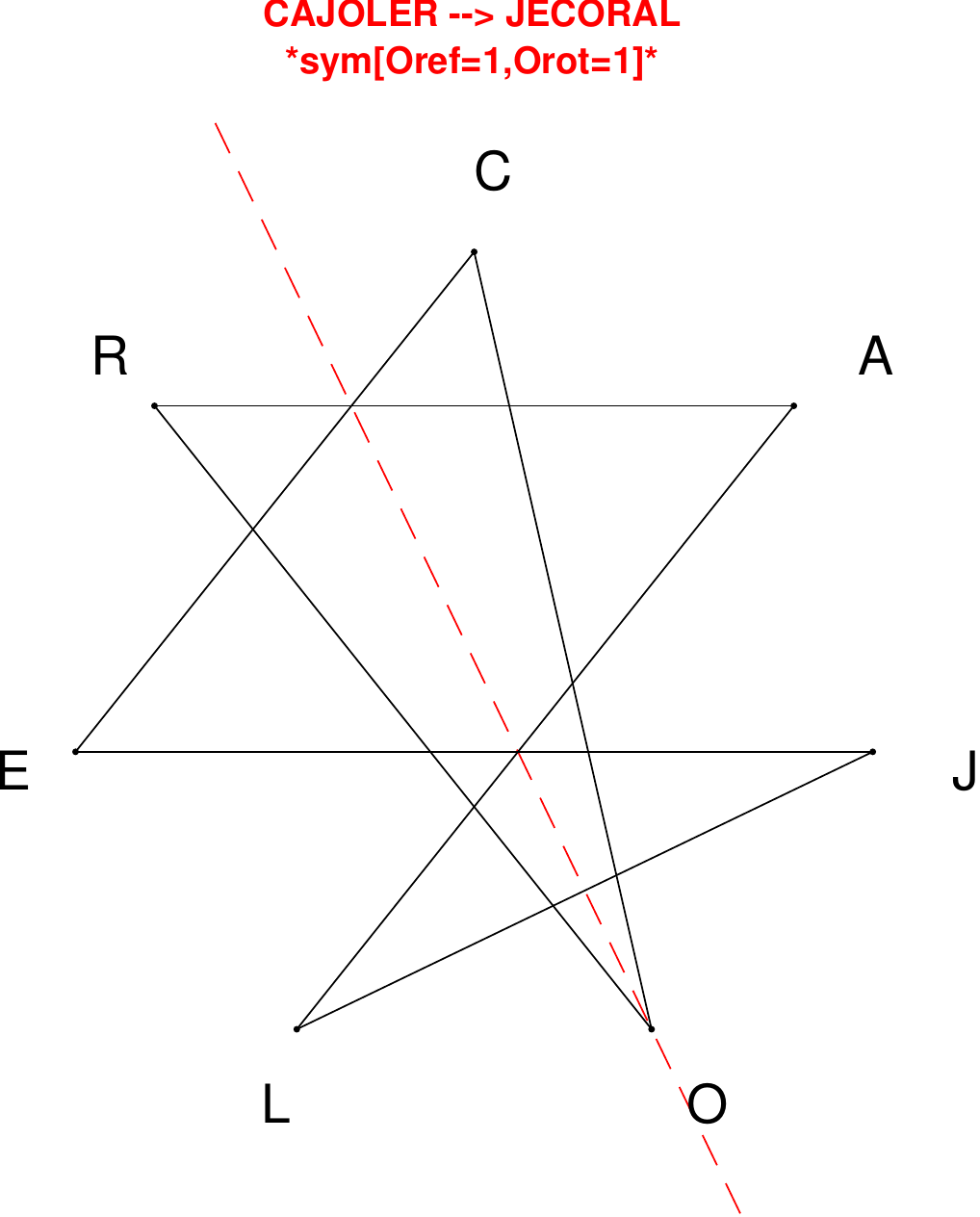}
\end{subfigure}
\hfill
\begin{subfigure}[T]{0.19\textwidth}
\centering
\includegraphics[width=\textwidth]{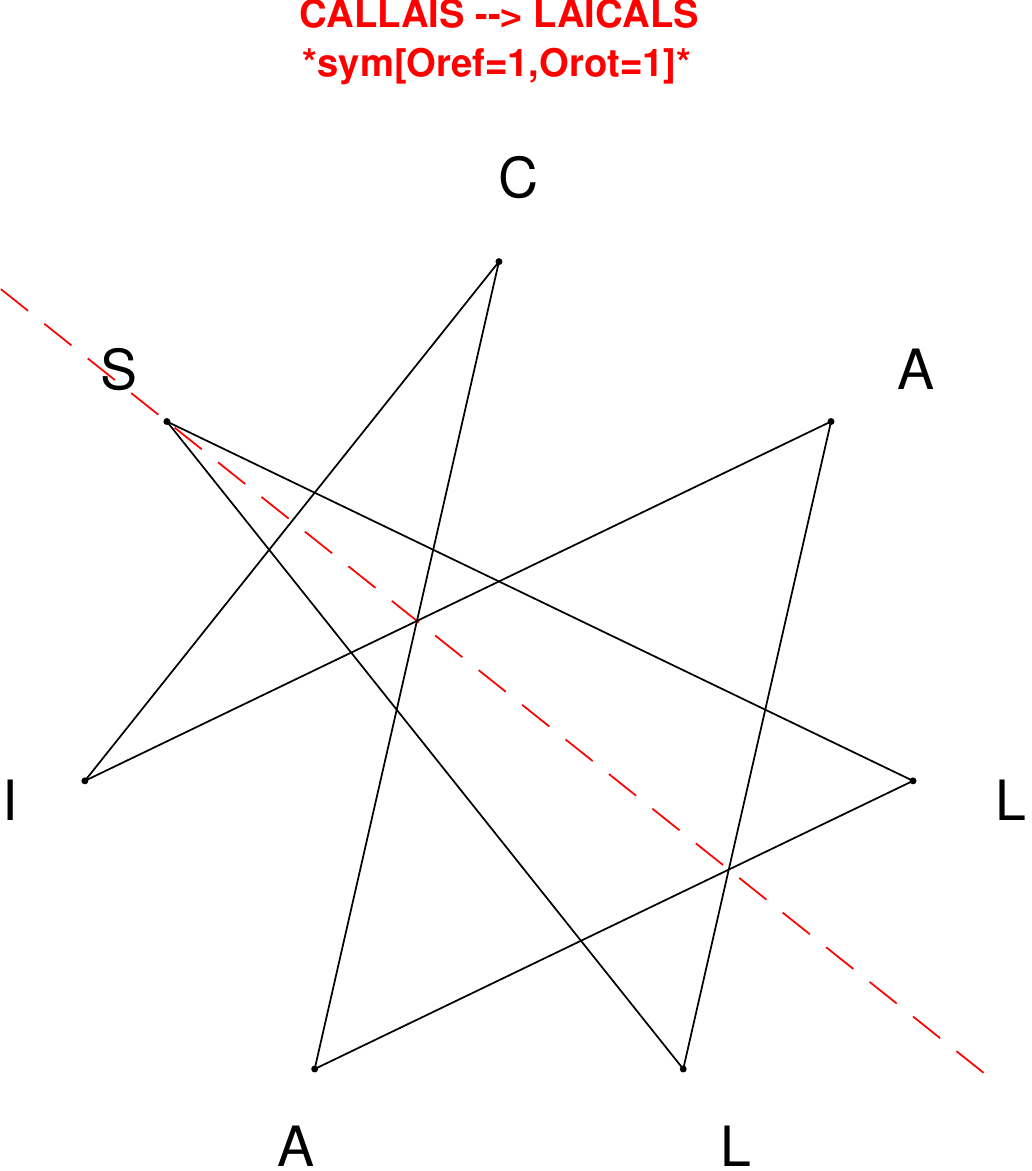}
\end{subfigure}
\end{figure}

\begin{figure}[H]
\centering
\begin{subfigure}[T]{0.19\textwidth}
\centering
\includegraphics[width=\textwidth]{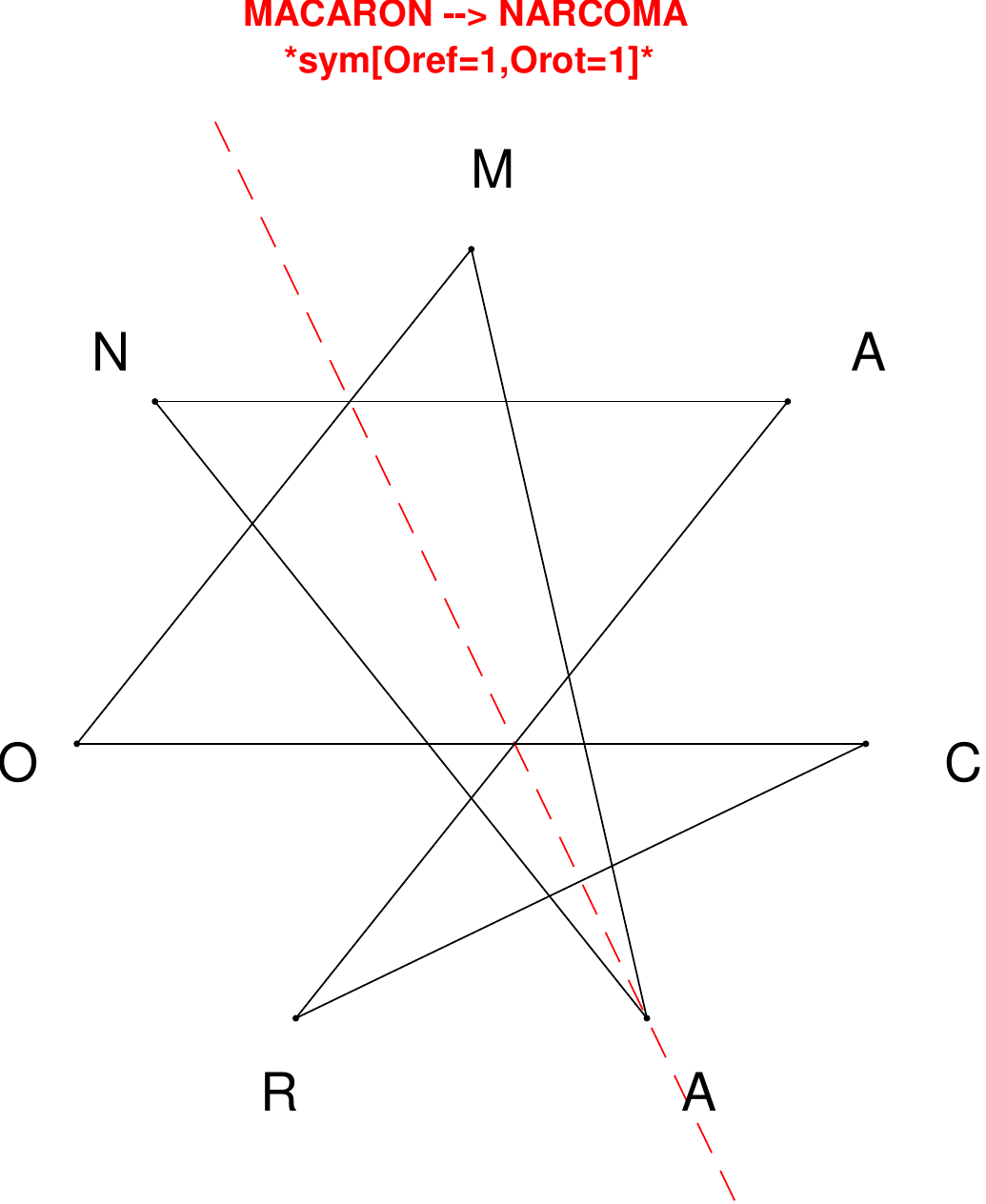}
\end{subfigure}
\hfill
\begin{subfigure}[T]{0.19\textwidth}
\centering
\includegraphics[width=\textwidth]{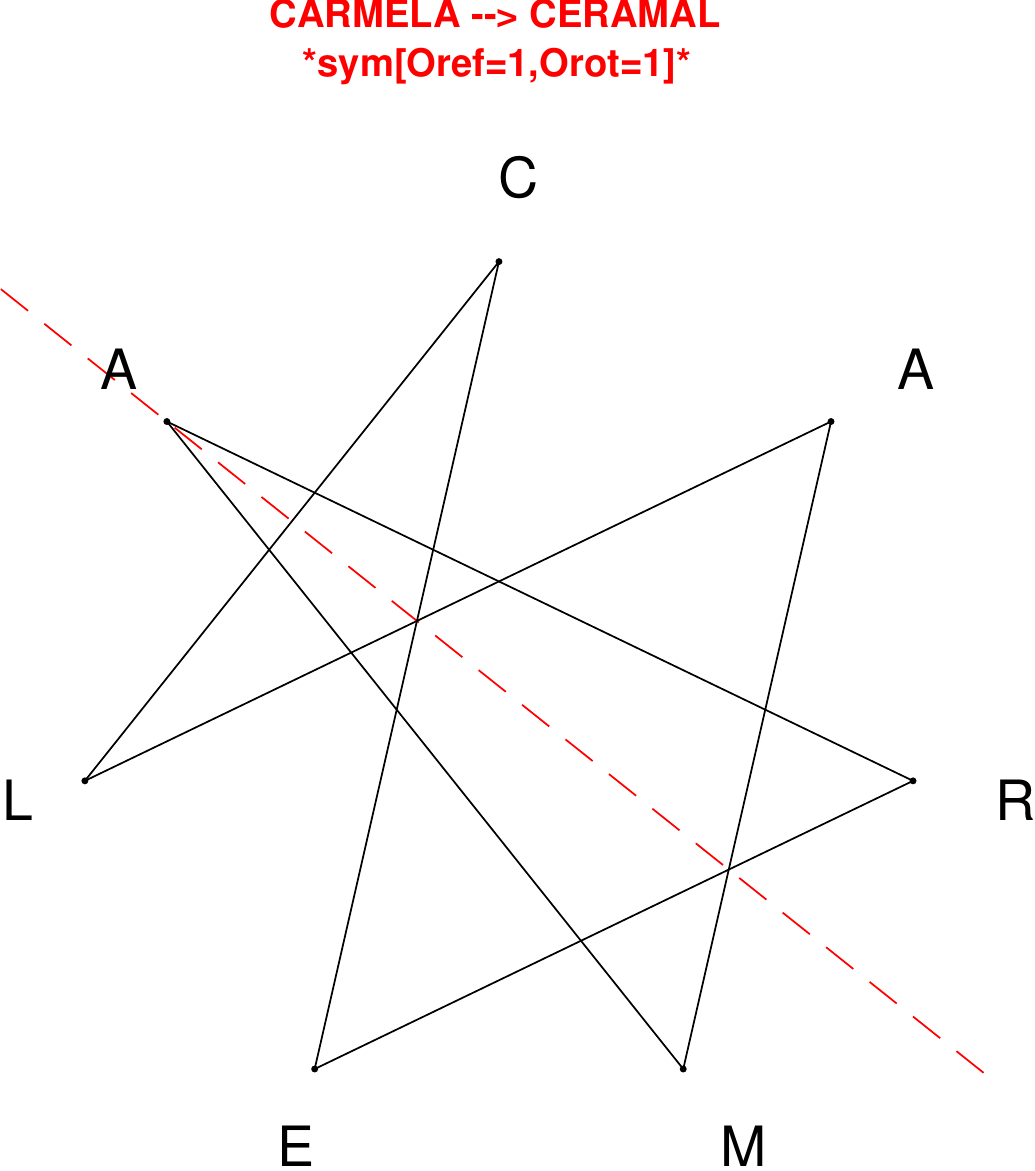}
\end{subfigure}
\hfill
\begin{subfigure}[T]{0.19\textwidth}
\centering
\includegraphics[width=\textwidth]{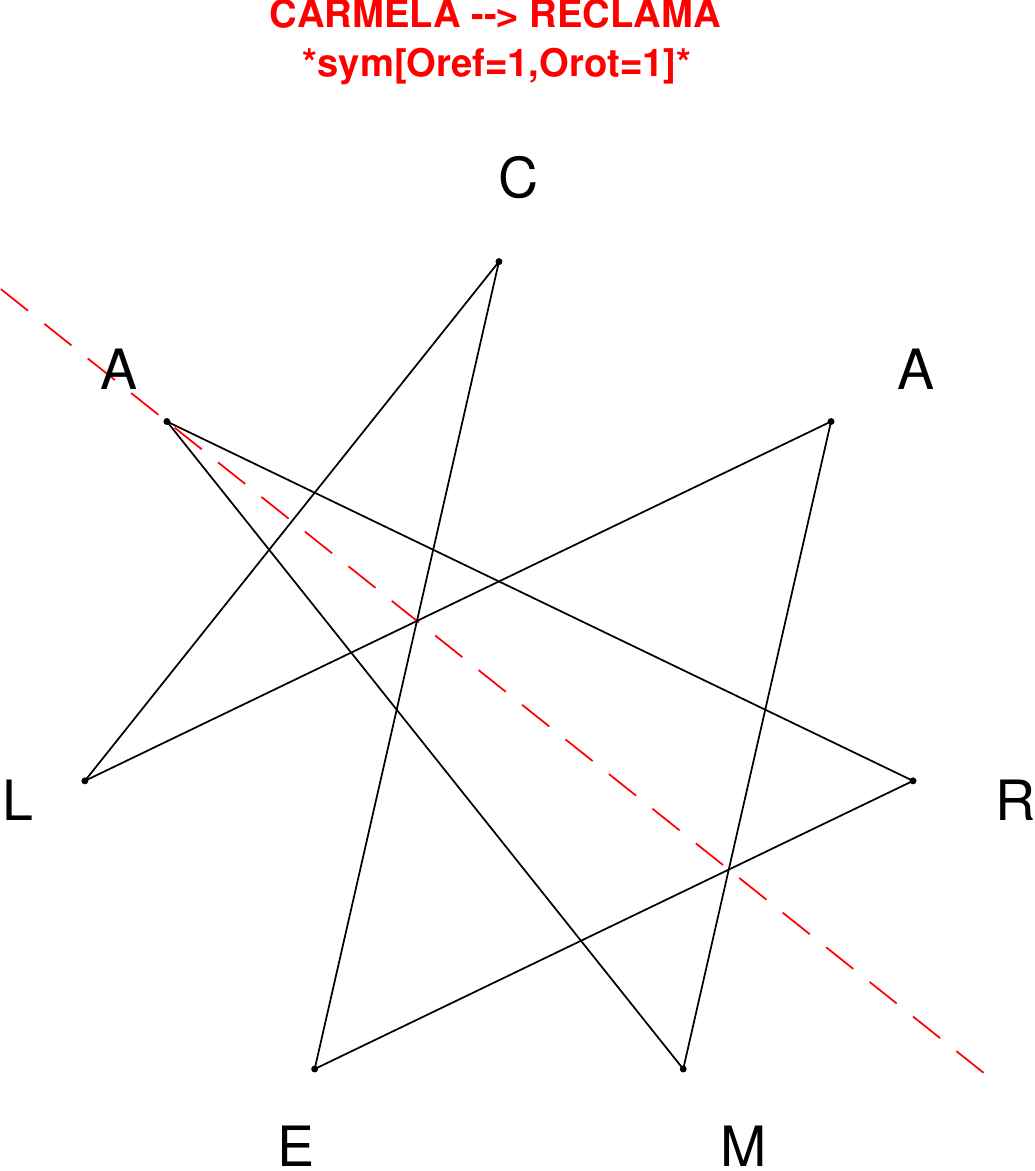}
\end{subfigure}
\hfill
\begin{subfigure}[T]{0.19\textwidth}
\centering
\includegraphics[width=\textwidth]{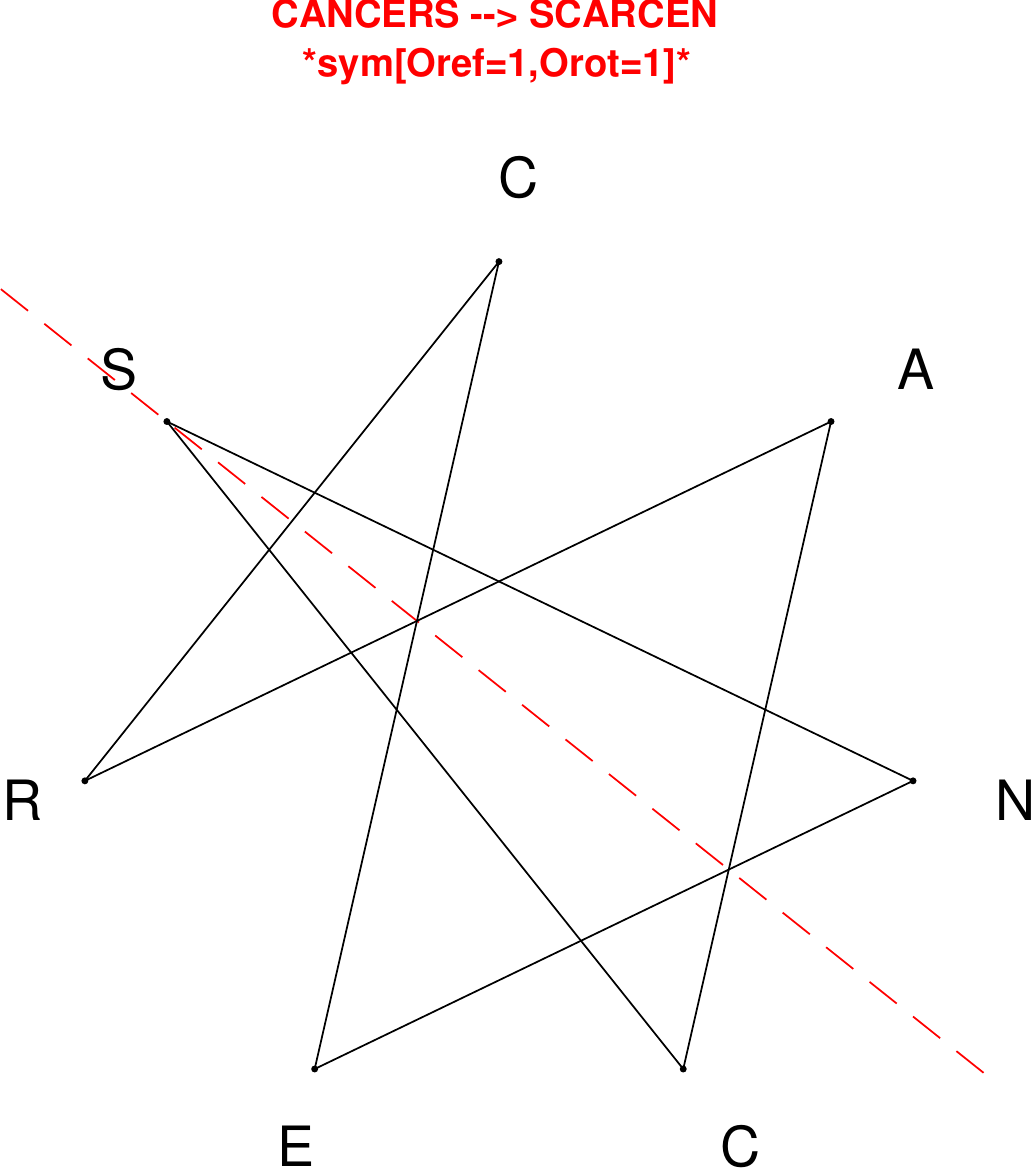}
\end{subfigure}
\hfill
\begin{subfigure}[T]{0.19\textwidth}
\centering
\includegraphics[width=\textwidth]{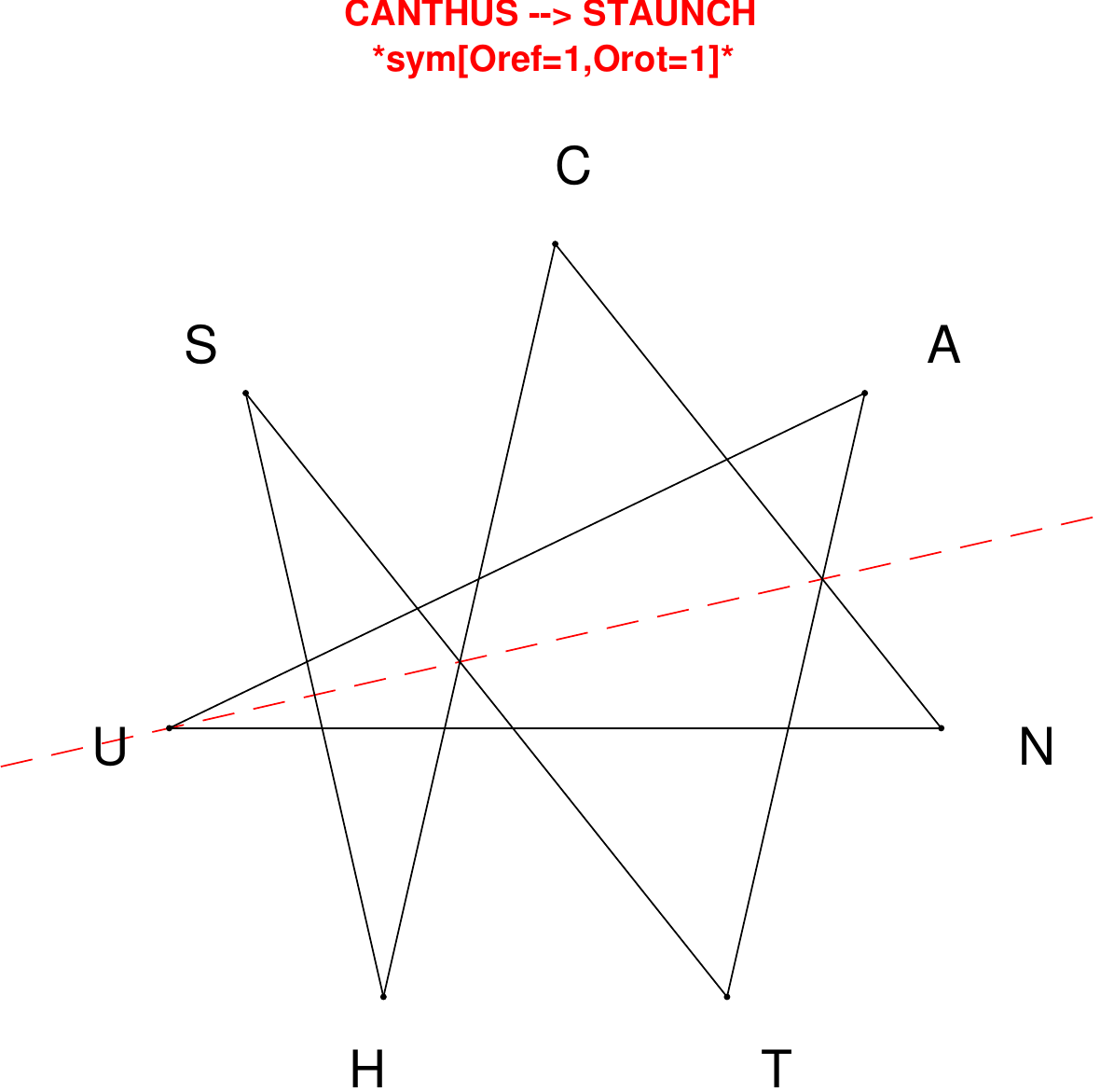}
\end{subfigure}
\end{figure}

\begin{figure}[H]
\centering
\begin{subfigure}[T]{0.19\textwidth}
\centering
\includegraphics[width=\textwidth]{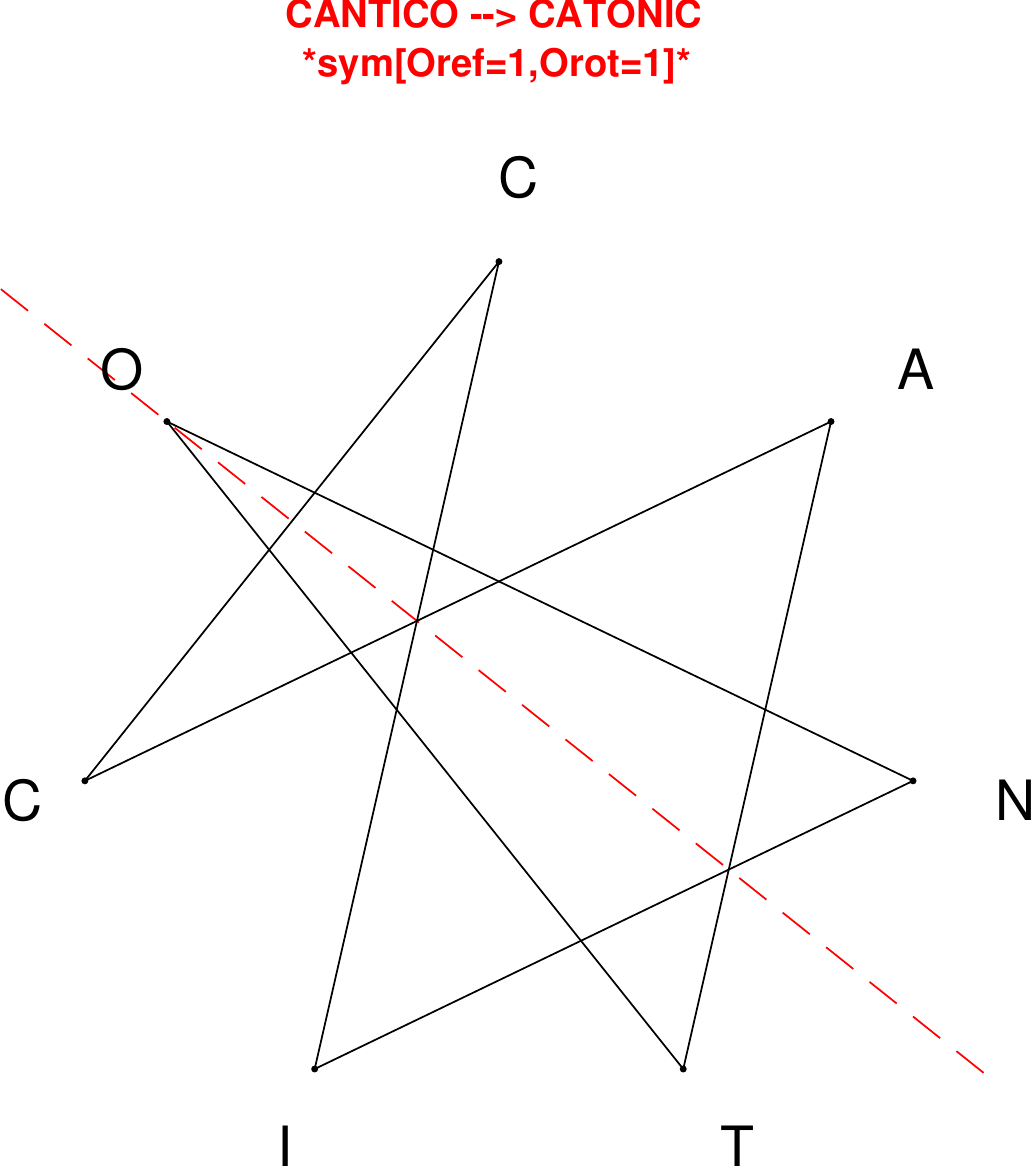}
\end{subfigure}
\hfill
\begin{subfigure}[T]{0.19\textwidth}
\centering
\includegraphics[width=\textwidth]{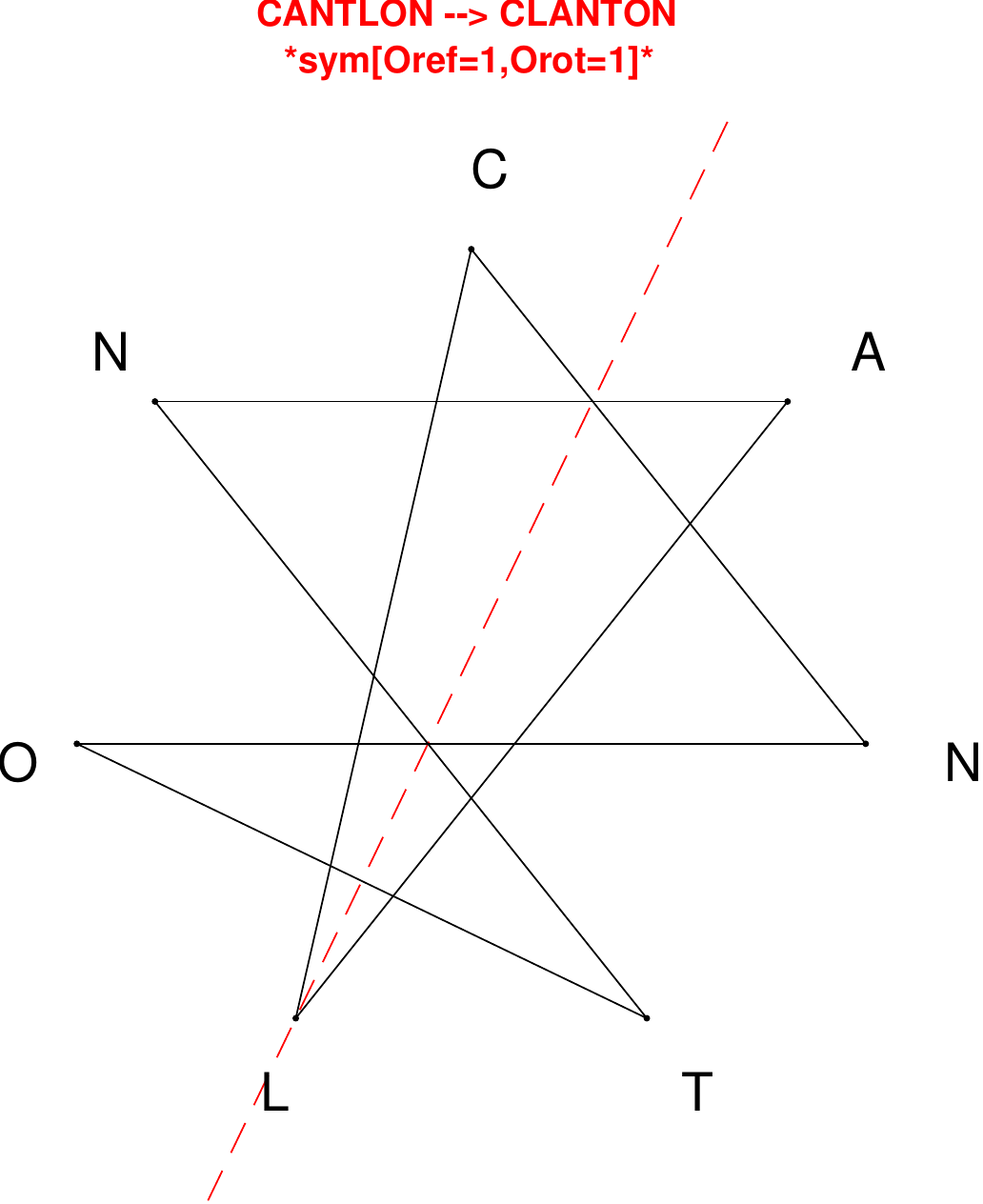}
\end{subfigure}
\hfill
\begin{subfigure}[T]{0.19\textwidth}
\centering
\includegraphics[width=\textwidth]{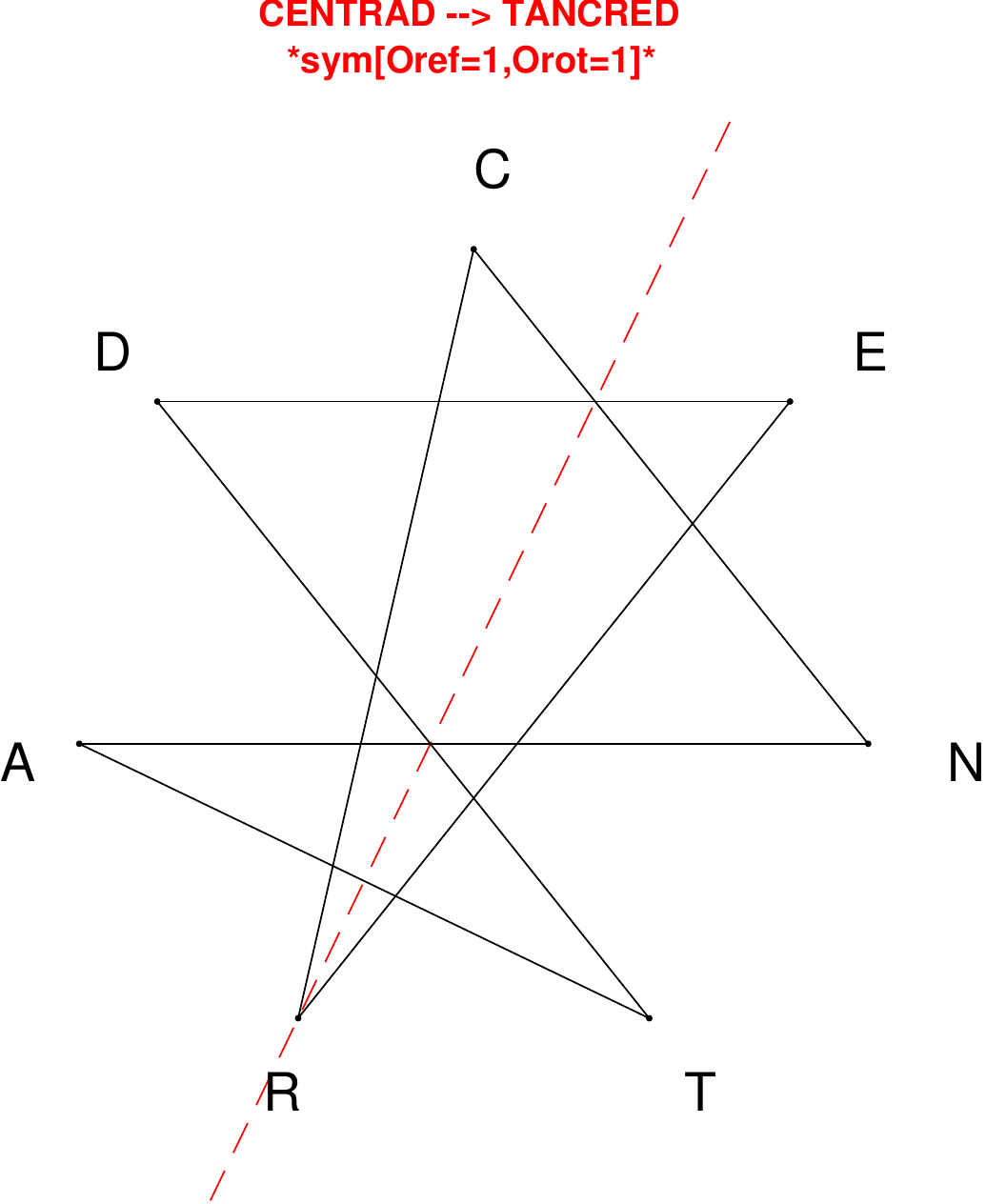}
\end{subfigure}
\hfill
\begin{subfigure}[T]{0.19\textwidth}
\centering
\includegraphics[width=\textwidth]{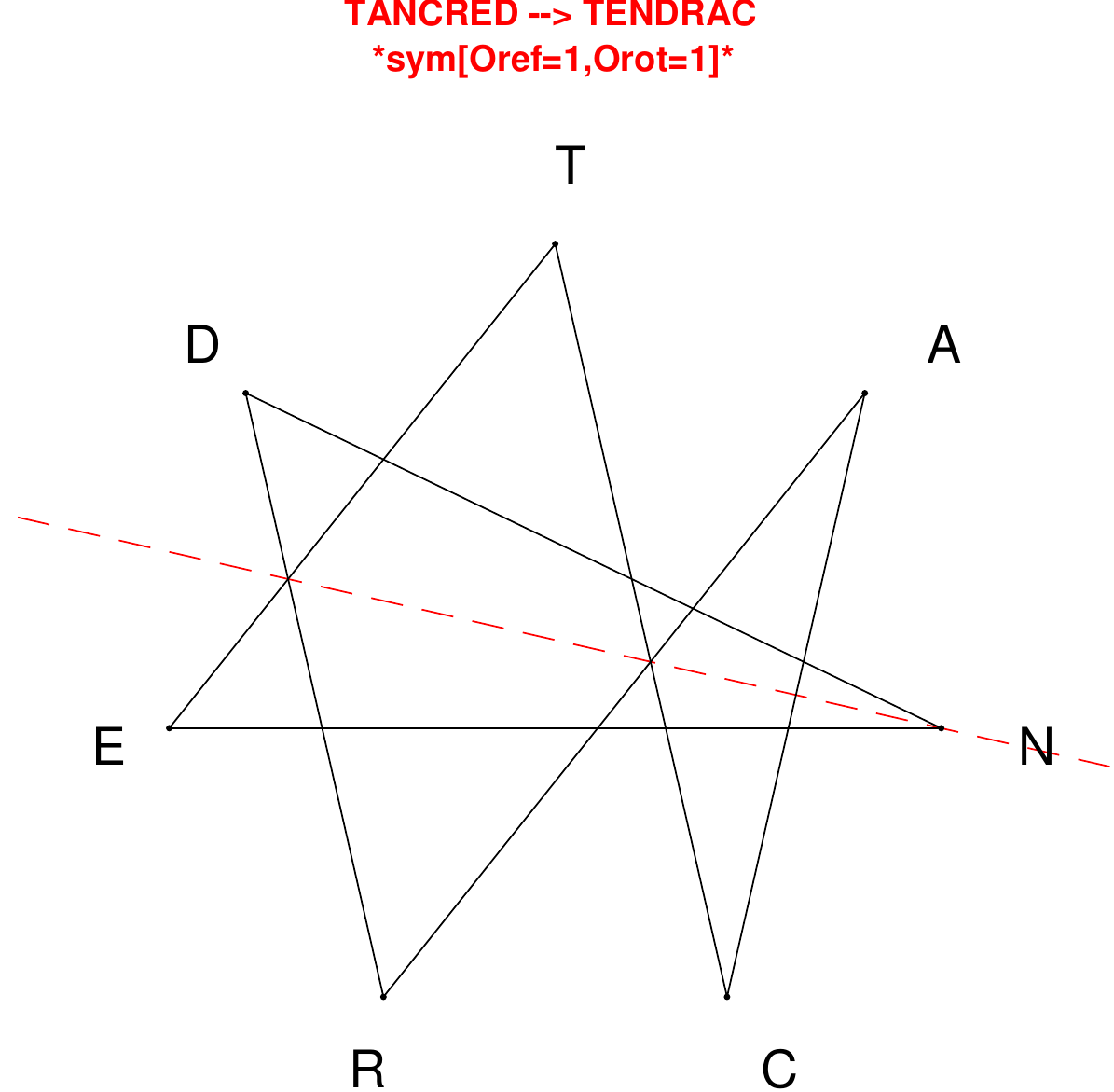}
\end{subfigure}
\hfill
\begin{subfigure}[T]{0.19\textwidth}
\centering
\includegraphics[width=\textwidth]{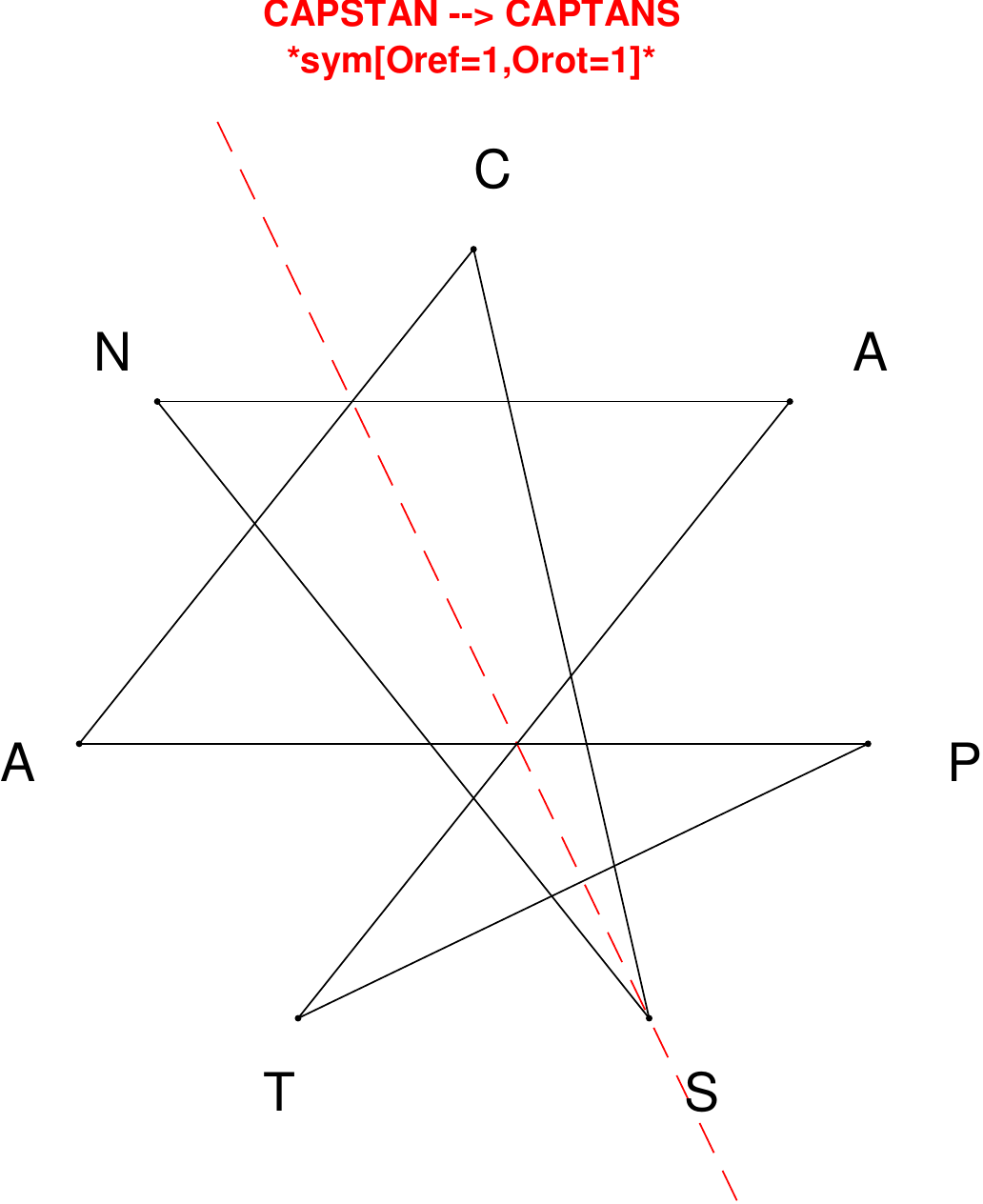}
\end{subfigure}
\end{figure}

\begin{figure}[H]
\centering
\begin{subfigure}[T]{0.19\textwidth}
\centering
\includegraphics[width=\textwidth]{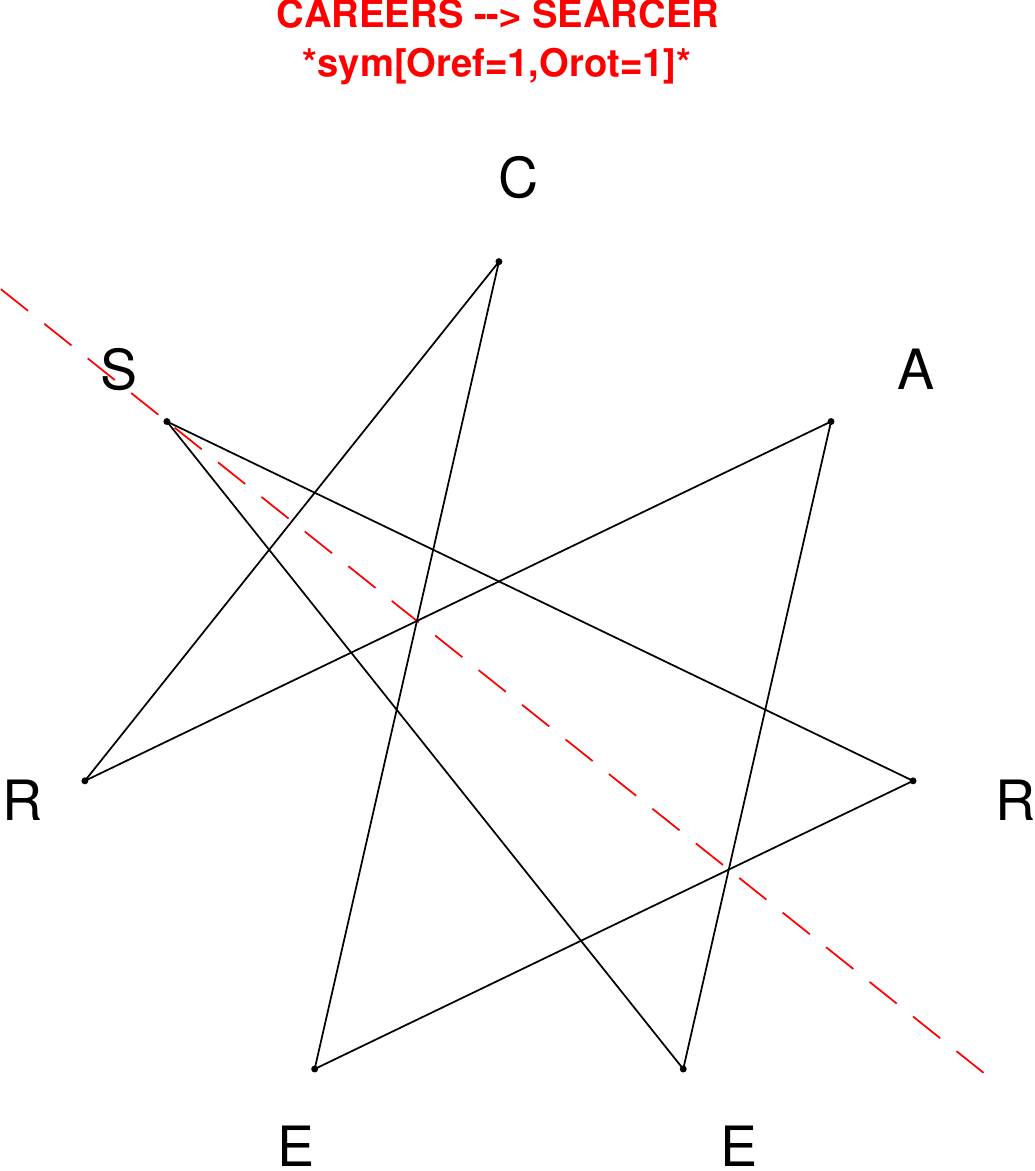}
\end{subfigure}
\hfill
\begin{subfigure}[T]{0.19\textwidth}
\centering
\includegraphics[width=\textwidth]{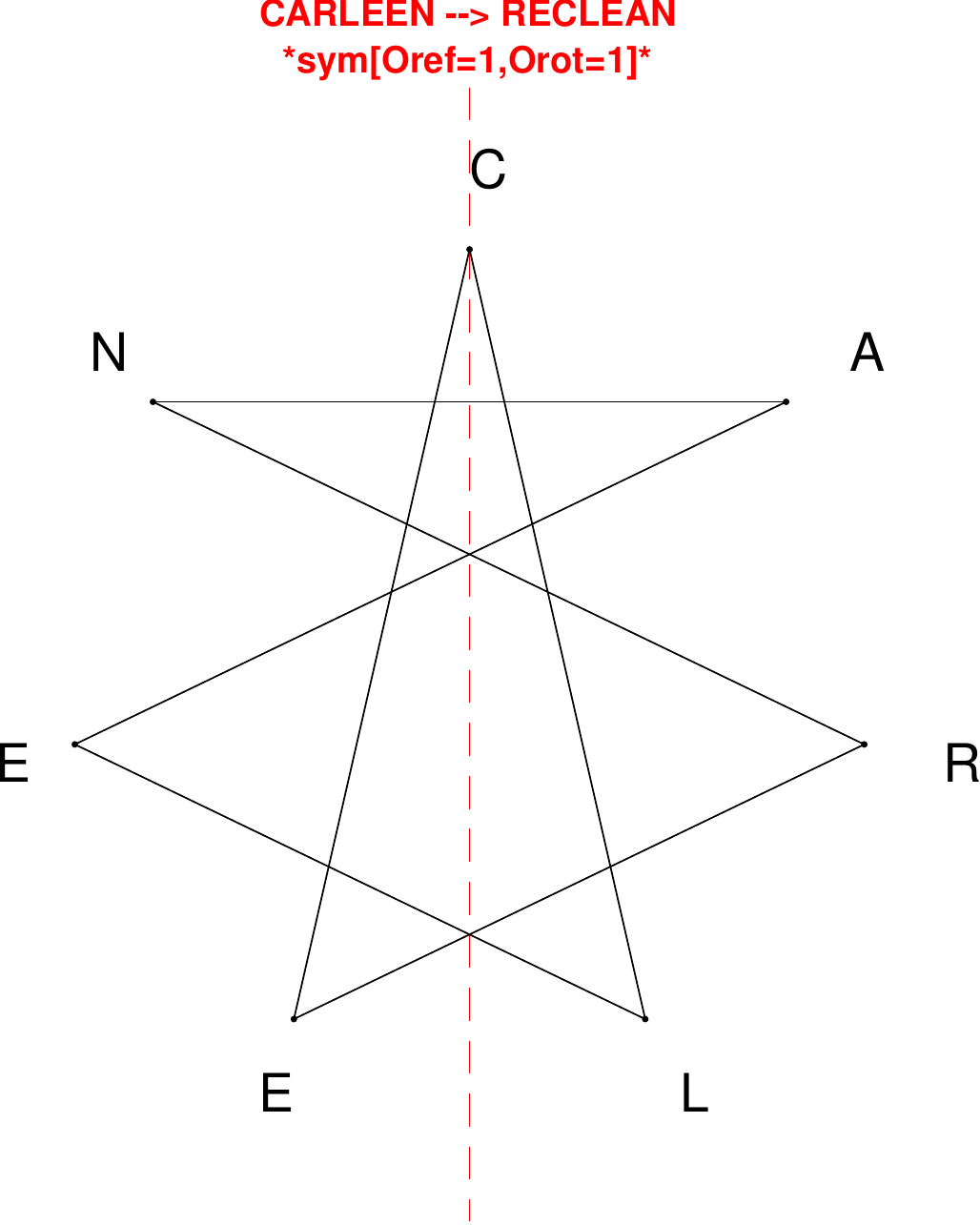}
\end{subfigure}
\hfill
\begin{subfigure}[T]{0.19\textwidth}
\centering
\includegraphics[width=\textwidth]{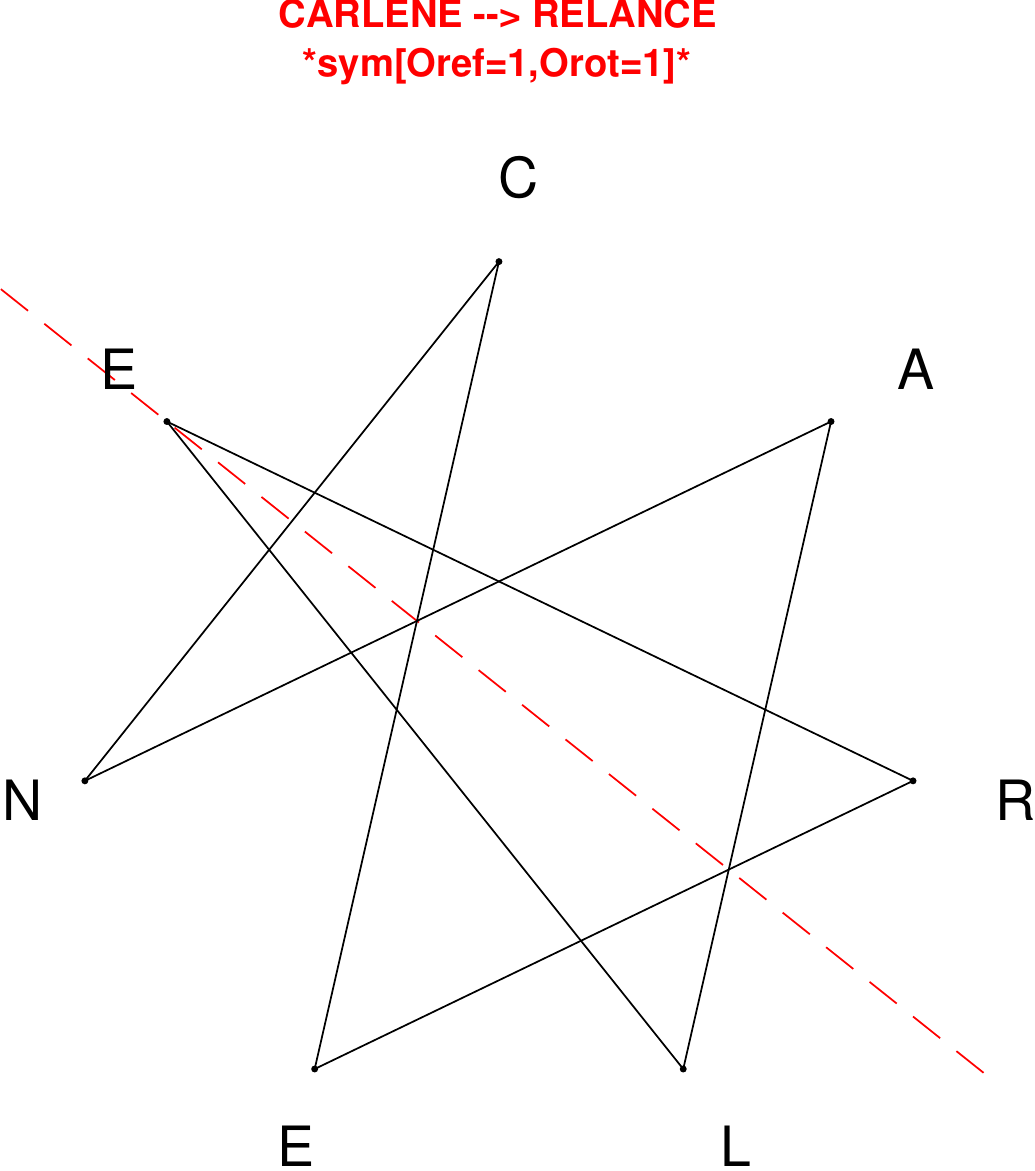}
\end{subfigure}
\hfill
\begin{subfigure}[T]{0.19\textwidth}
\centering
\includegraphics[width=\textwidth]{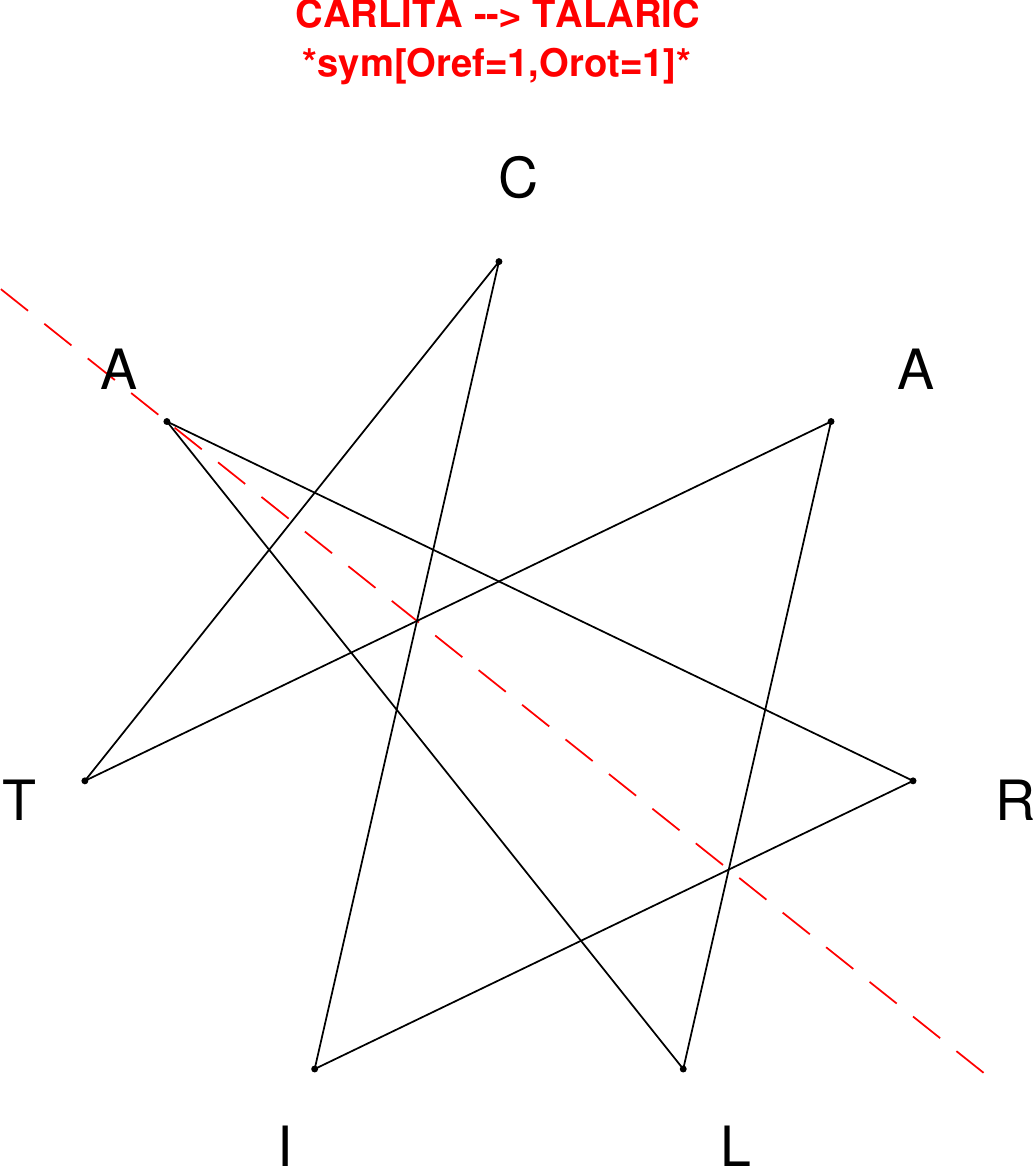}
\end{subfigure}
\hfill
\begin{subfigure}[T]{0.19\textwidth}
\centering
\includegraphics[width=\textwidth]{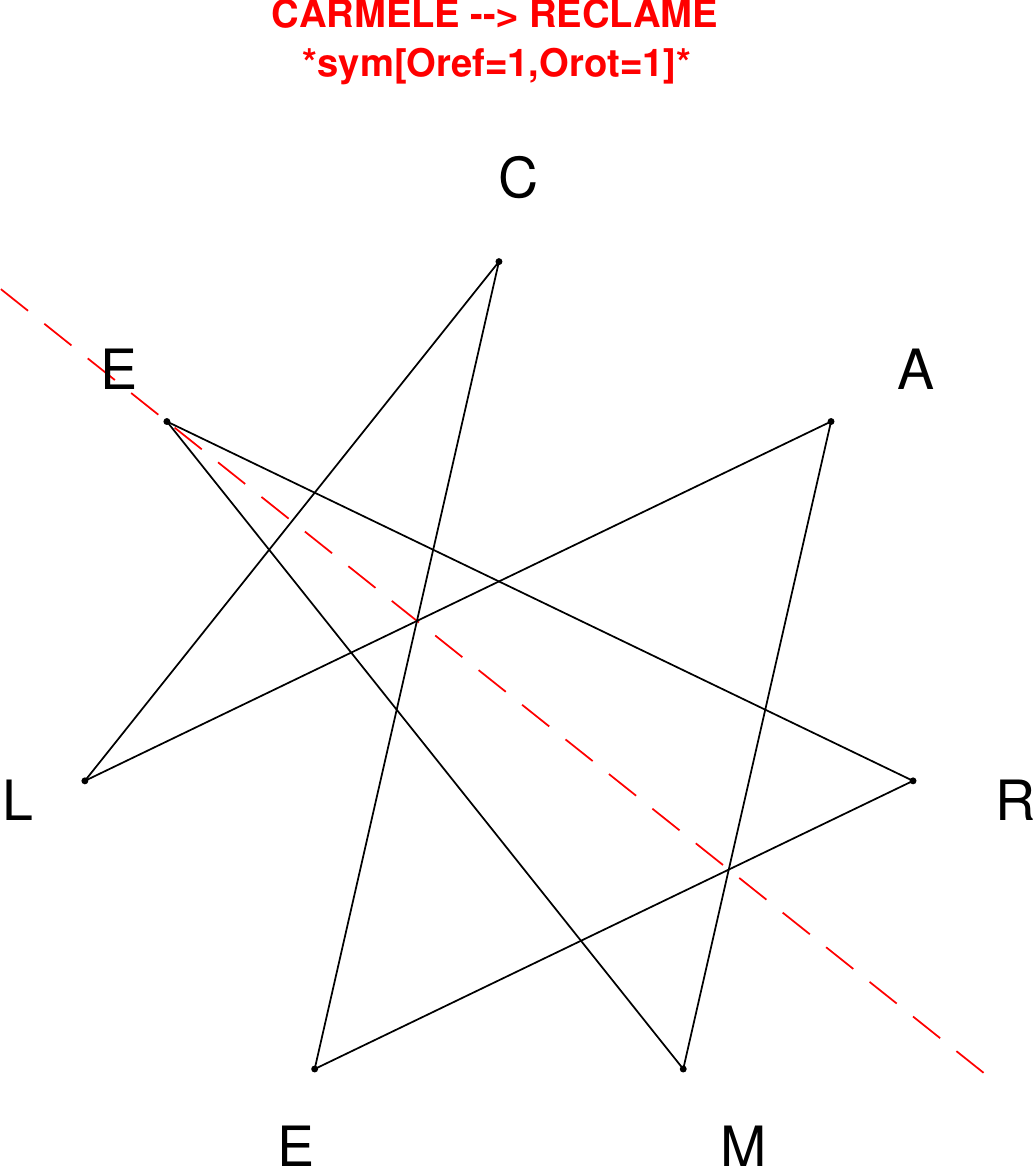}
\end{subfigure}
\end{figure}

\begin{figure}[H]
\centering
\begin{subfigure}[T]{0.19\textwidth}
\centering
\includegraphics[width=\textwidth]{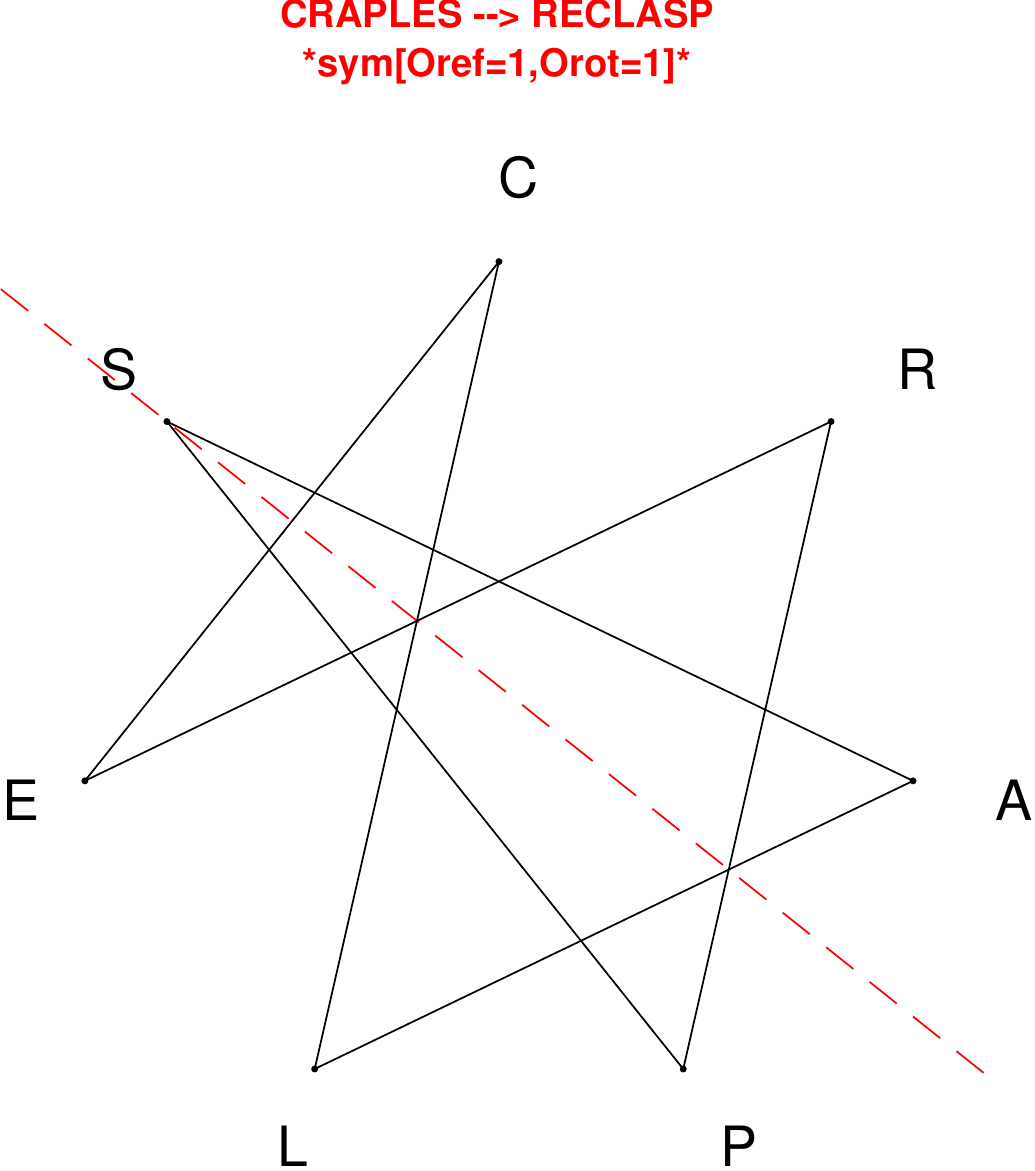}
\end{subfigure}
\hfill
\begin{subfigure}[T]{0.19\textwidth}
\centering
\includegraphics[width=\textwidth]{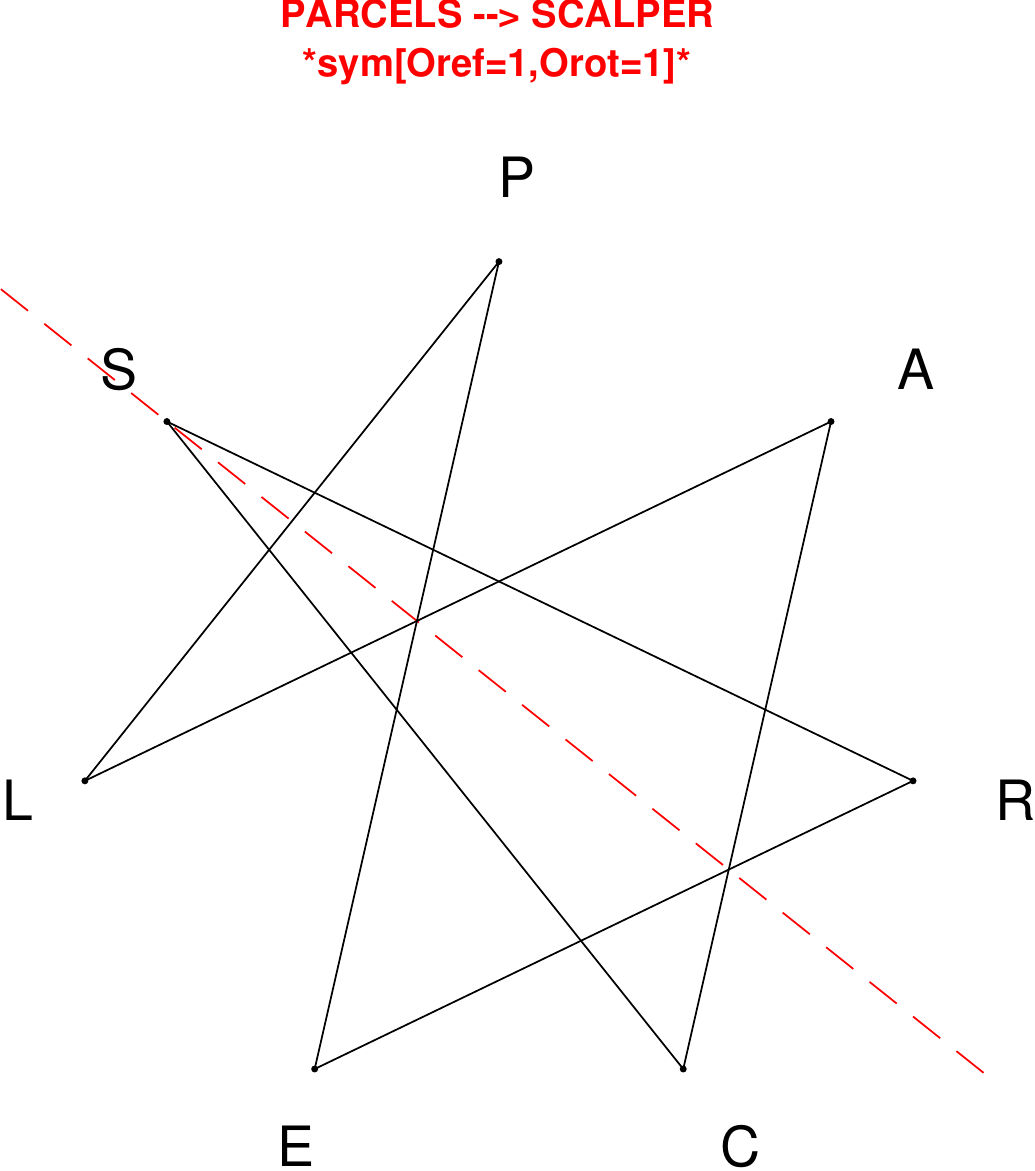}
\end{subfigure}
\hfill
\begin{subfigure}[T]{0.19\textwidth}
\centering
\includegraphics[width=\textwidth]{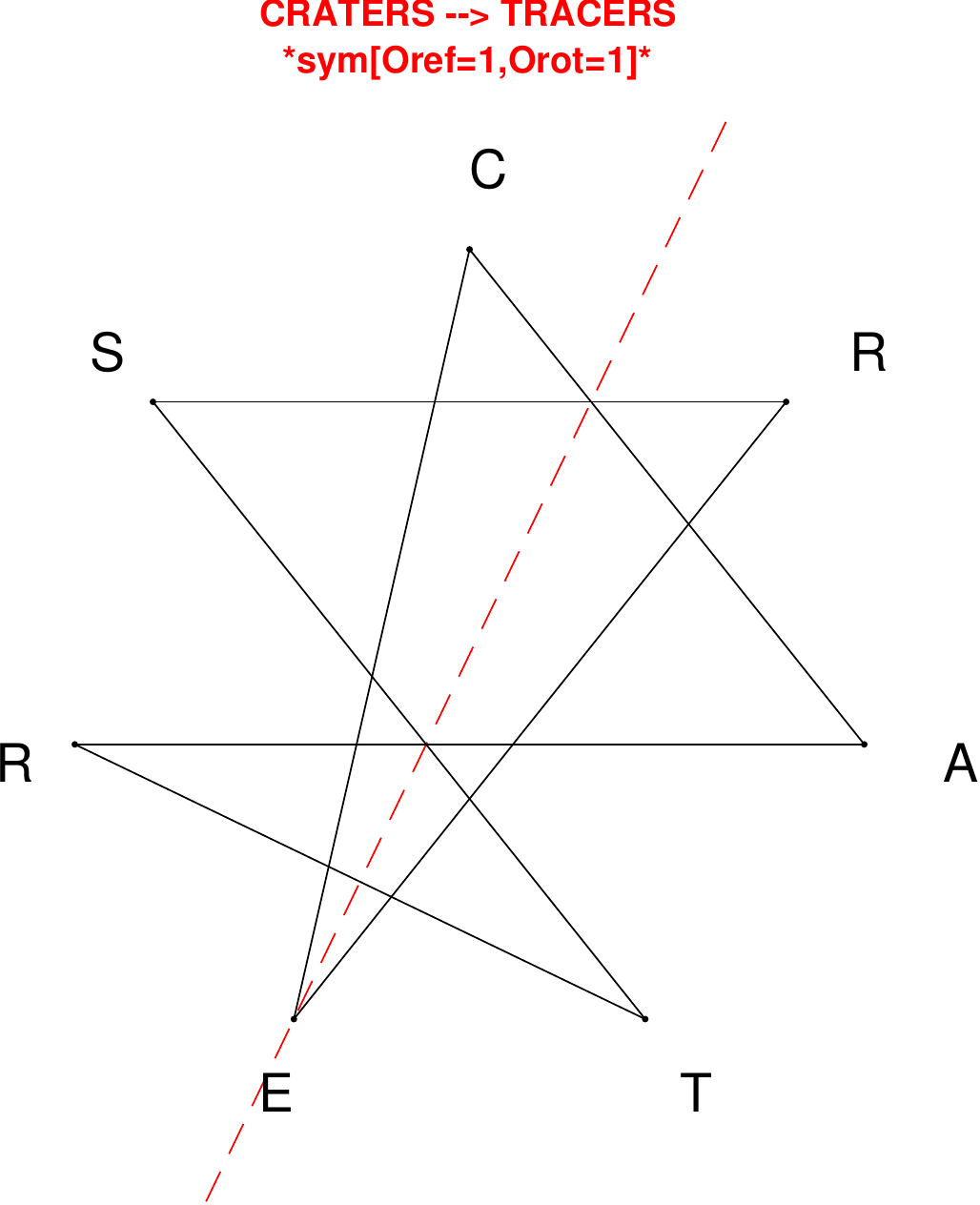}
\end{subfigure}
\hfill
\begin{subfigure}[T]{0.19\textwidth}
\centering
\includegraphics[width=\textwidth]{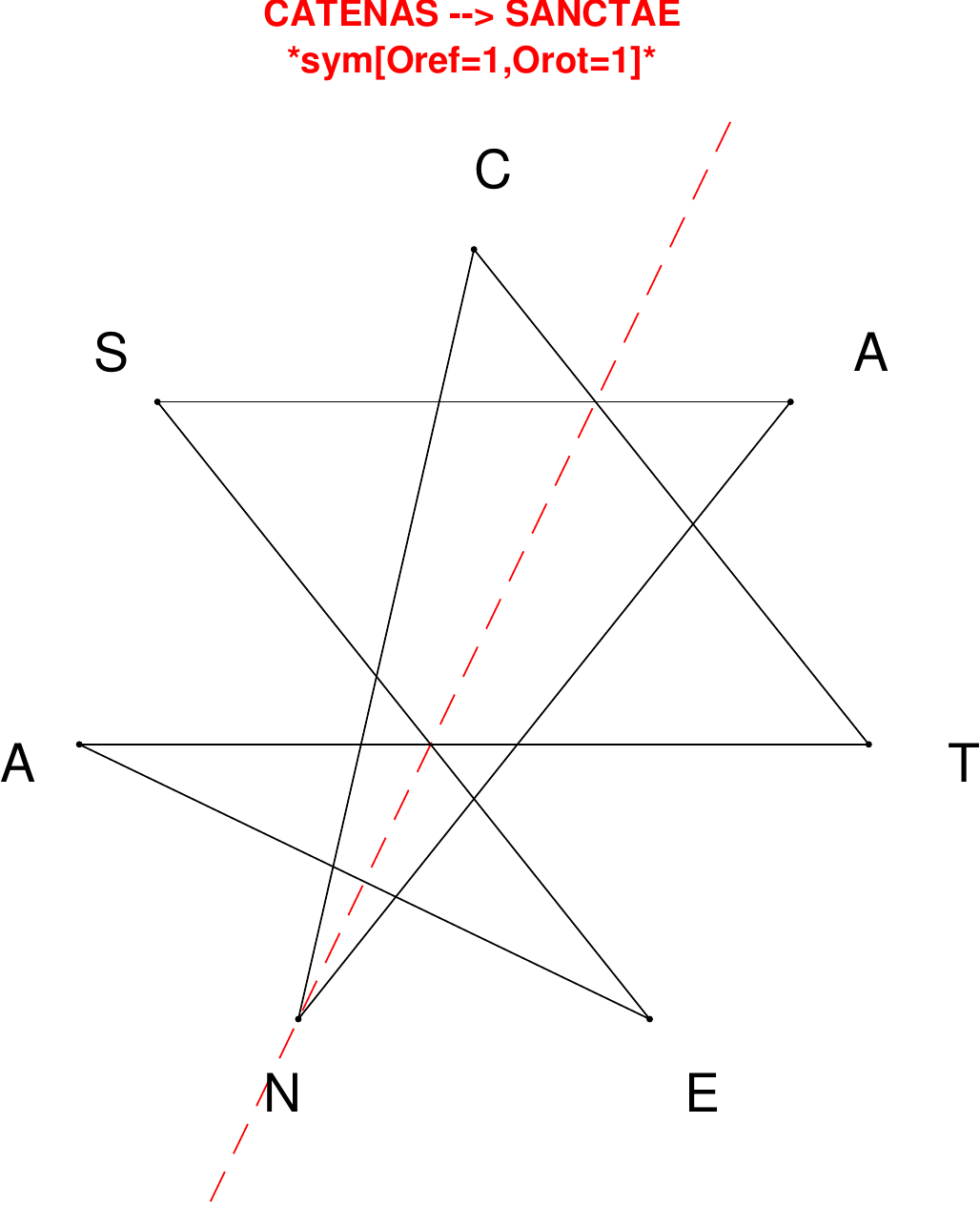}
\end{subfigure}
\hfill
\begin{subfigure}[T]{0.19\textwidth}
\centering
\includegraphics[width=\textwidth]{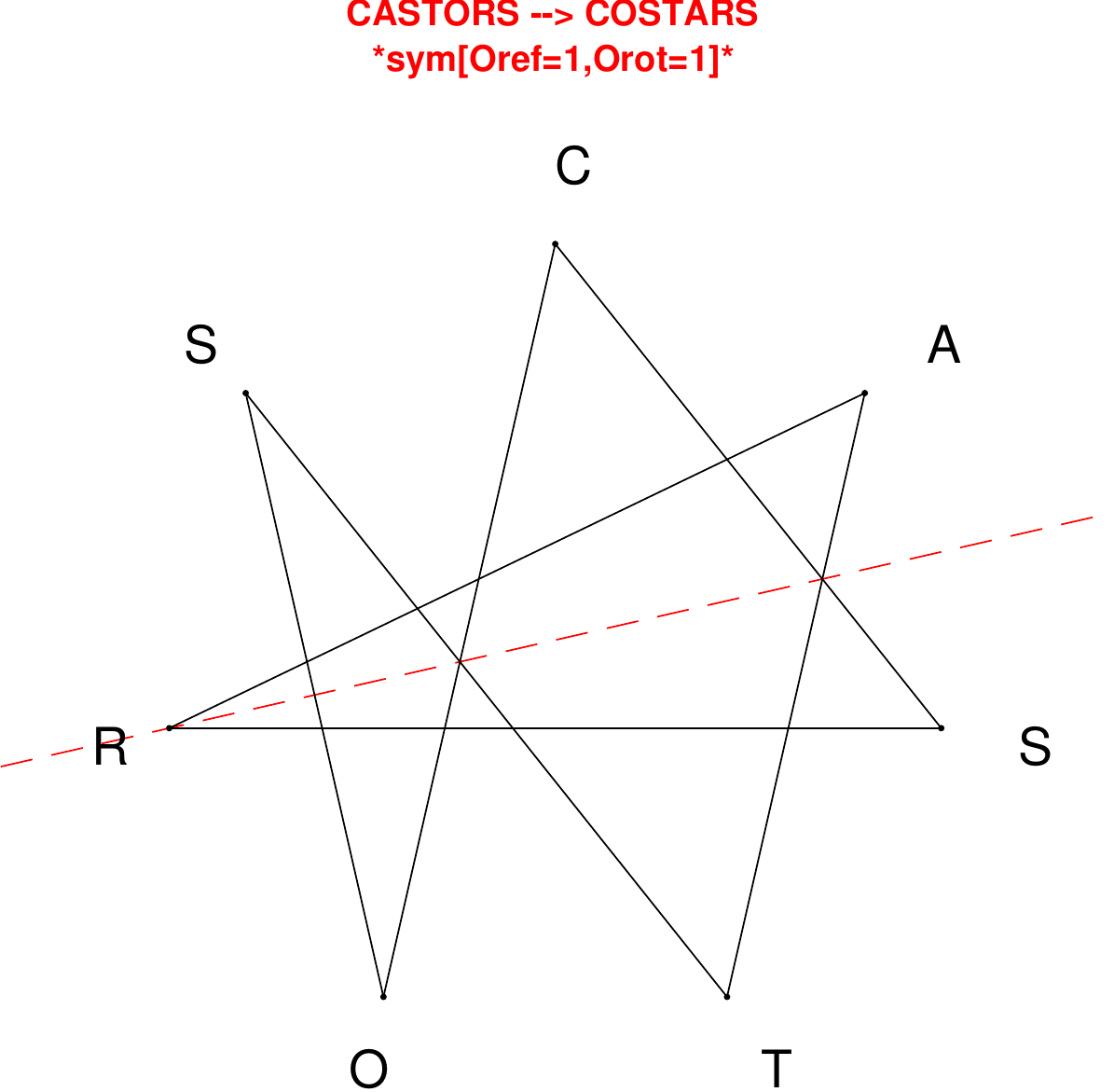}
\end{subfigure}
\end{figure}

\begin{figure}[H]
\centering
\begin{subfigure}[T]{0.19\textwidth}
\centering
\includegraphics[width=\textwidth]{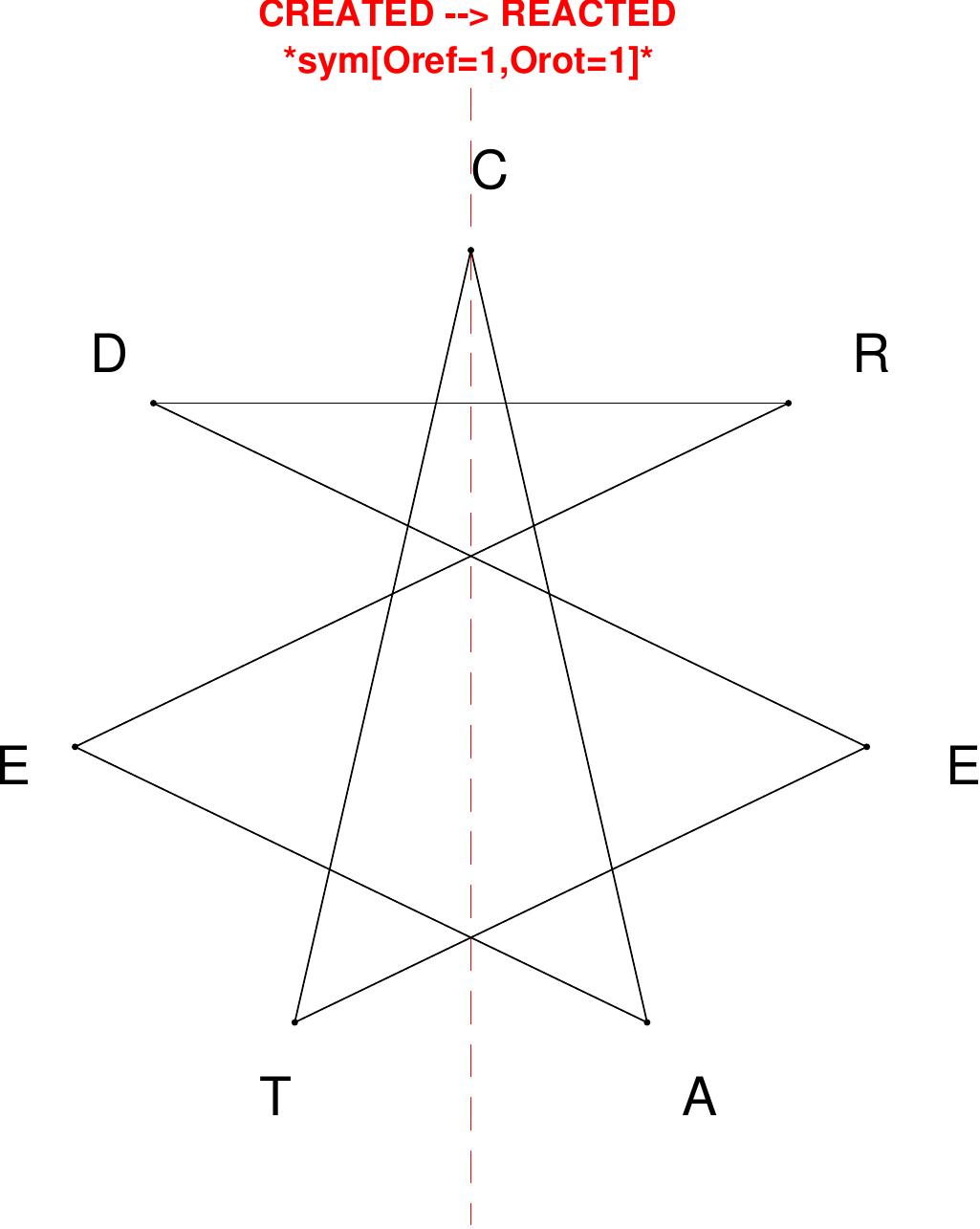}
\end{subfigure}
\hfill
\begin{subfigure}[T]{0.19\textwidth}
\centering
\includegraphics[width=\textwidth]{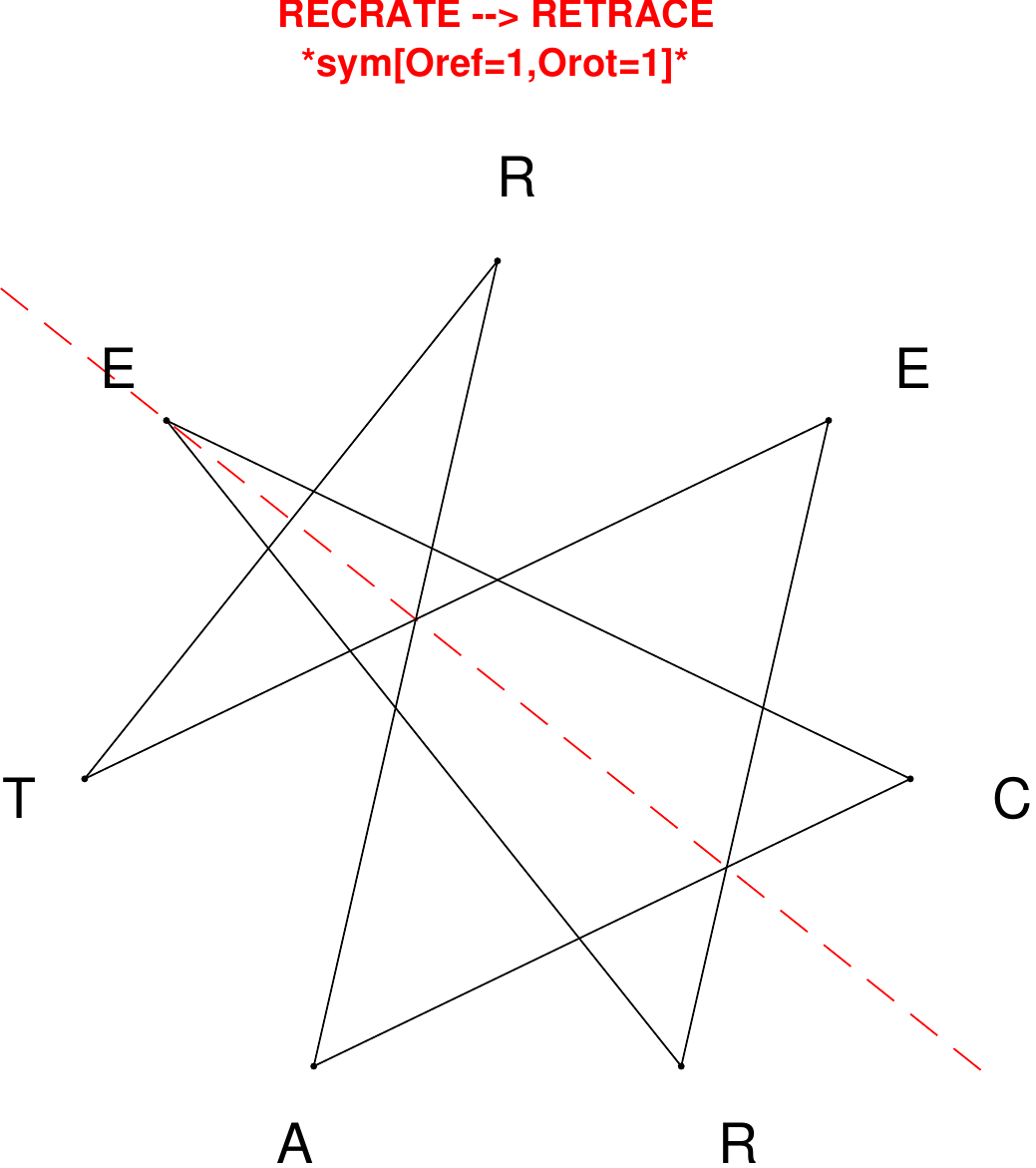}
\end{subfigure}
\hfill
\begin{subfigure}[T]{0.19\textwidth}
\centering
\includegraphics[width=\textwidth]{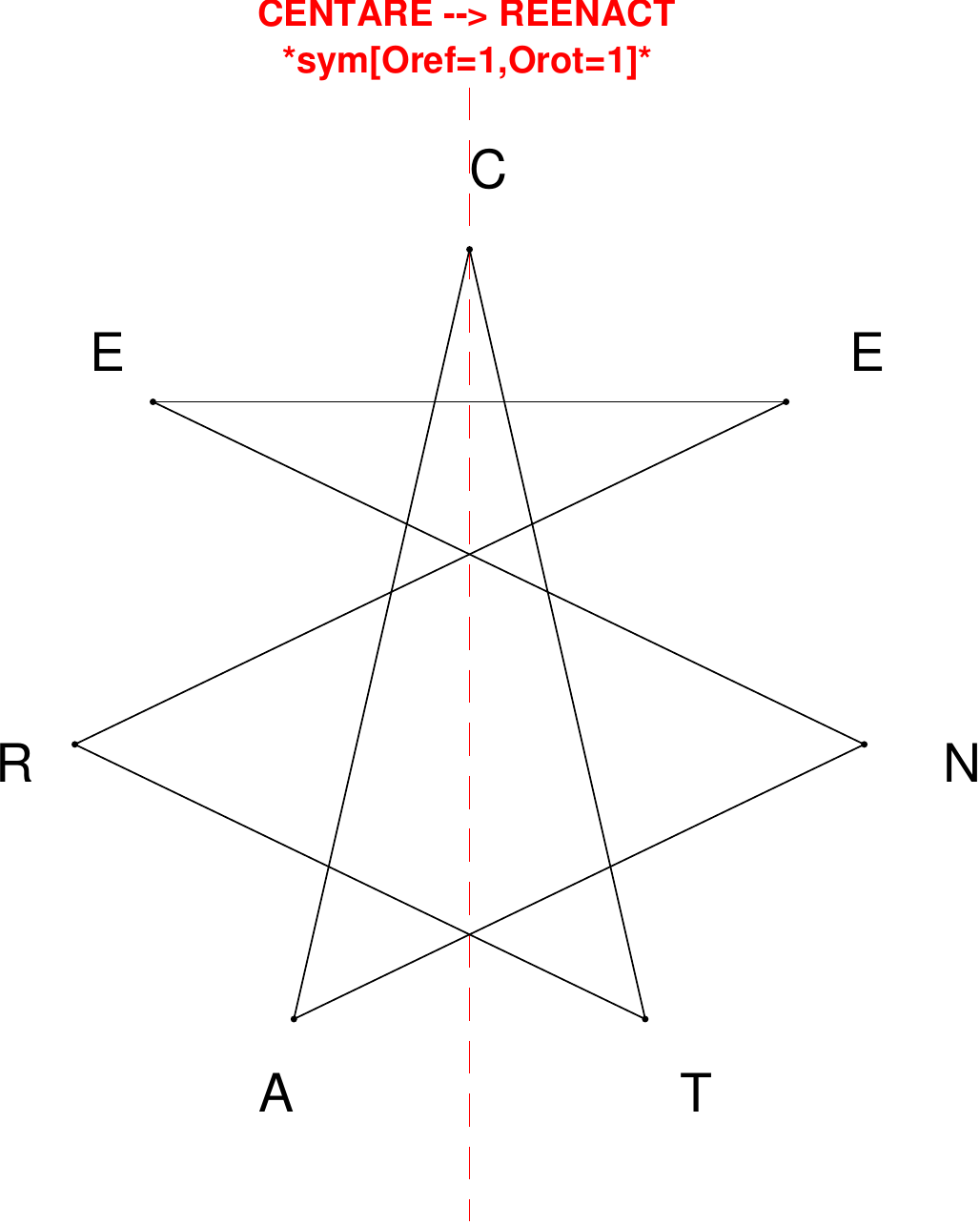}
\end{subfigure}
\hfill
\begin{subfigure}[T]{0.19\textwidth}
\centering
\includegraphics[width=\textwidth]{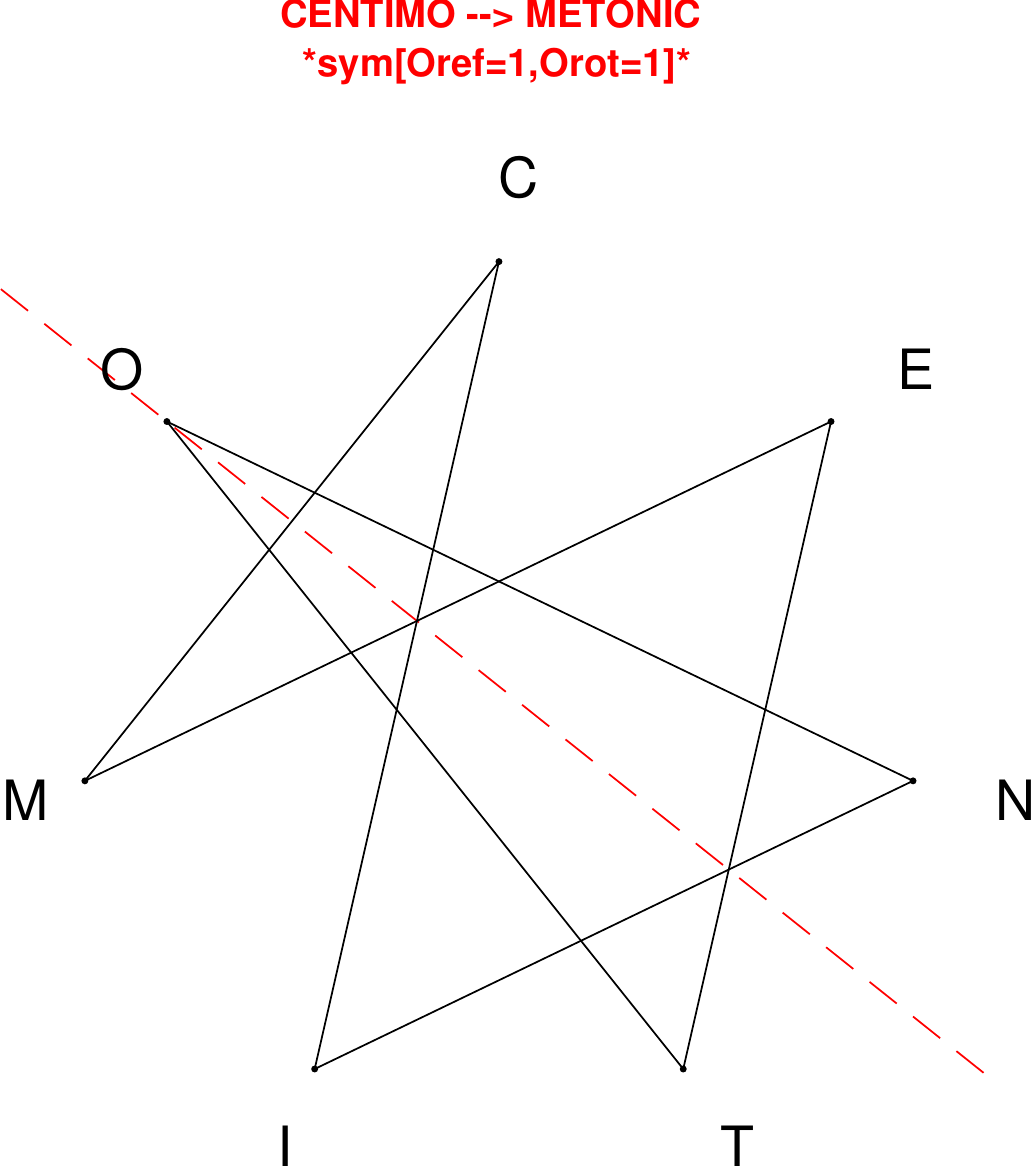}
\end{subfigure}
\hfill
\begin{subfigure}[T]{0.19\textwidth}
\centering
\includegraphics[width=\textwidth]{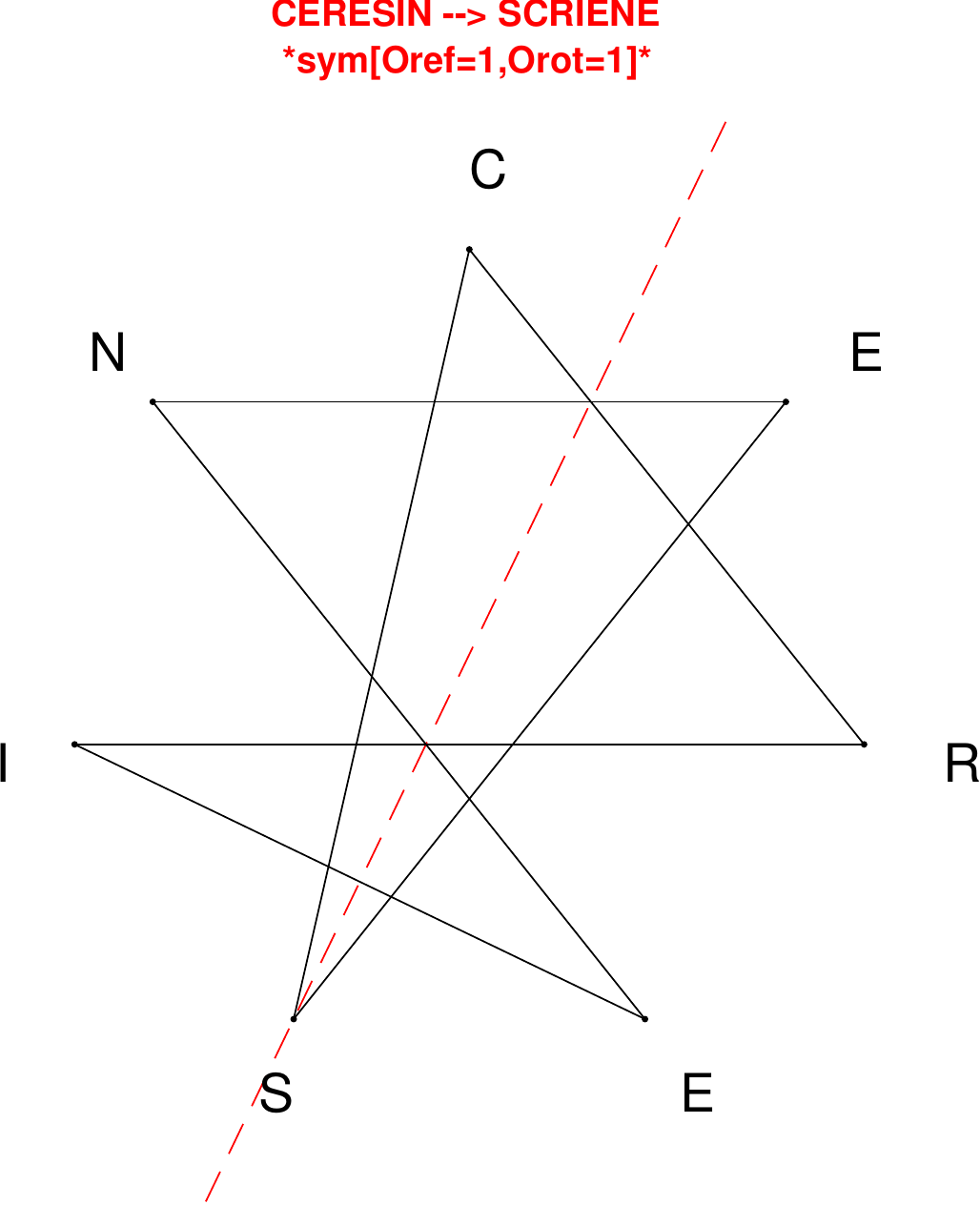}
\end{subfigure}
\end{figure}

\begin{figure}[H]
\centering
\begin{subfigure}[T]{0.19\textwidth}
\centering
\includegraphics[width=\textwidth]{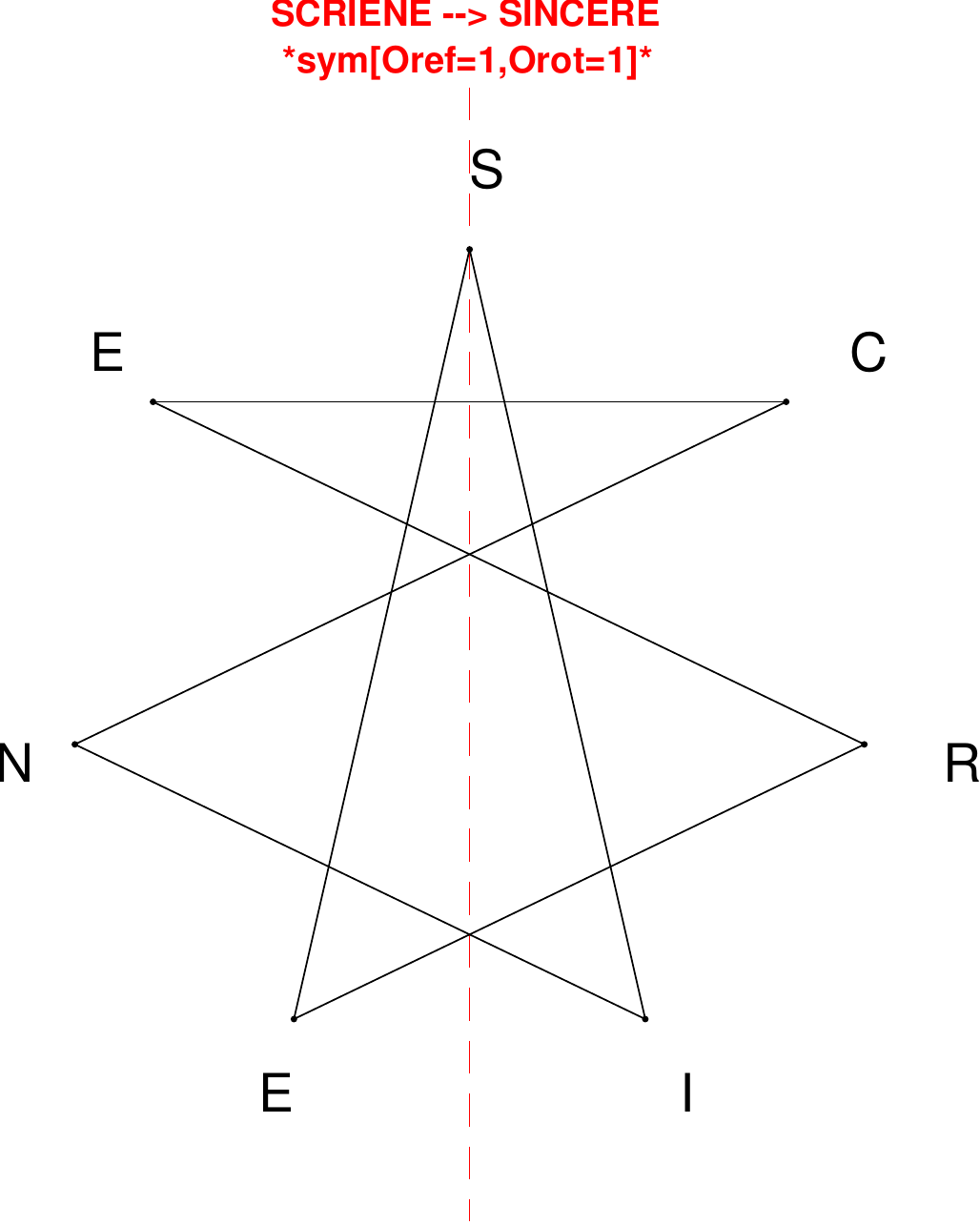}
\end{subfigure}
\hfill
\begin{subfigure}[T]{0.19\textwidth}
\centering
\includegraphics[width=\textwidth]{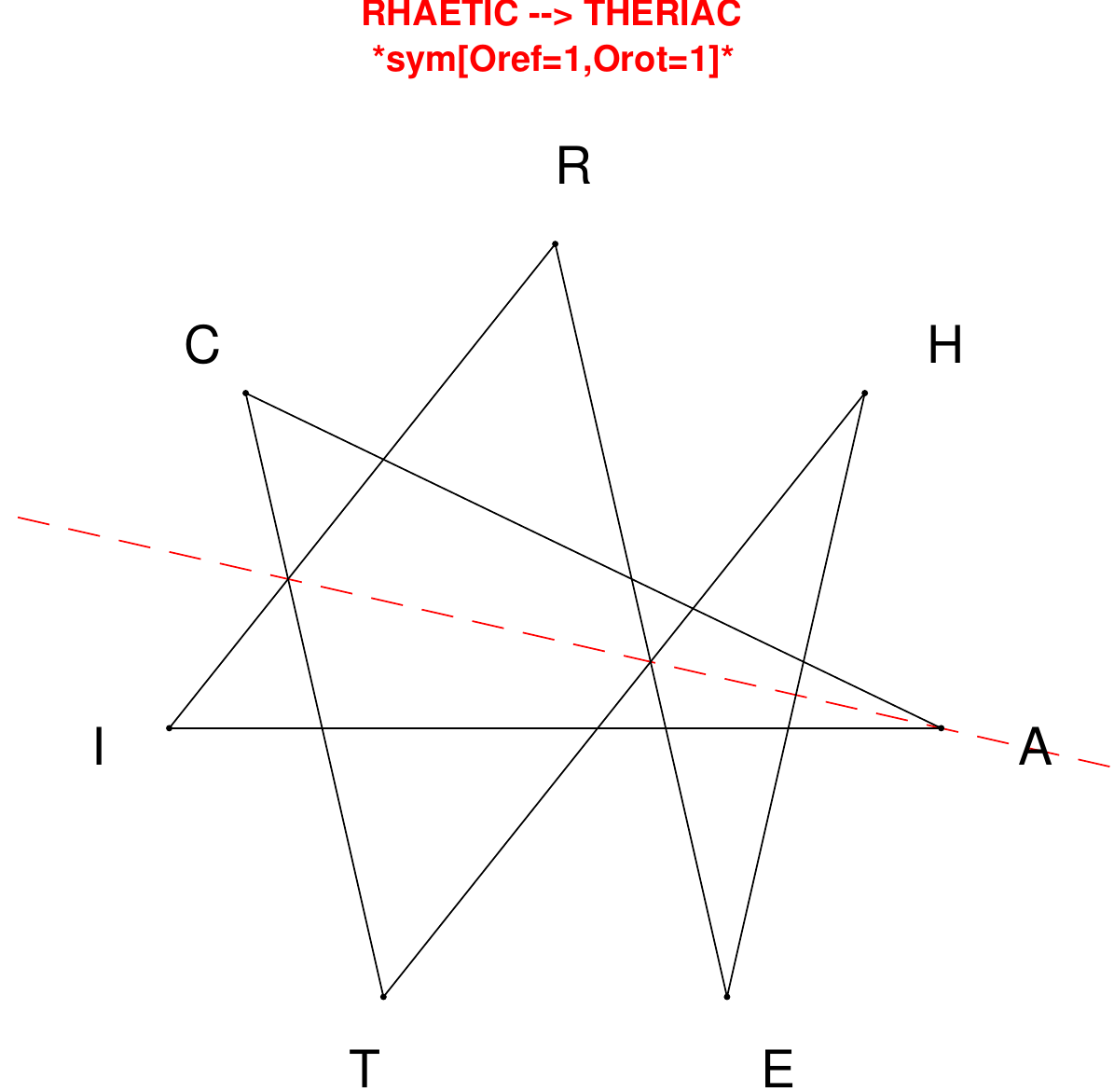}
\end{subfigure}
\hfill
\begin{subfigure}[T]{0.19\textwidth}
\centering
\includegraphics[width=\textwidth]{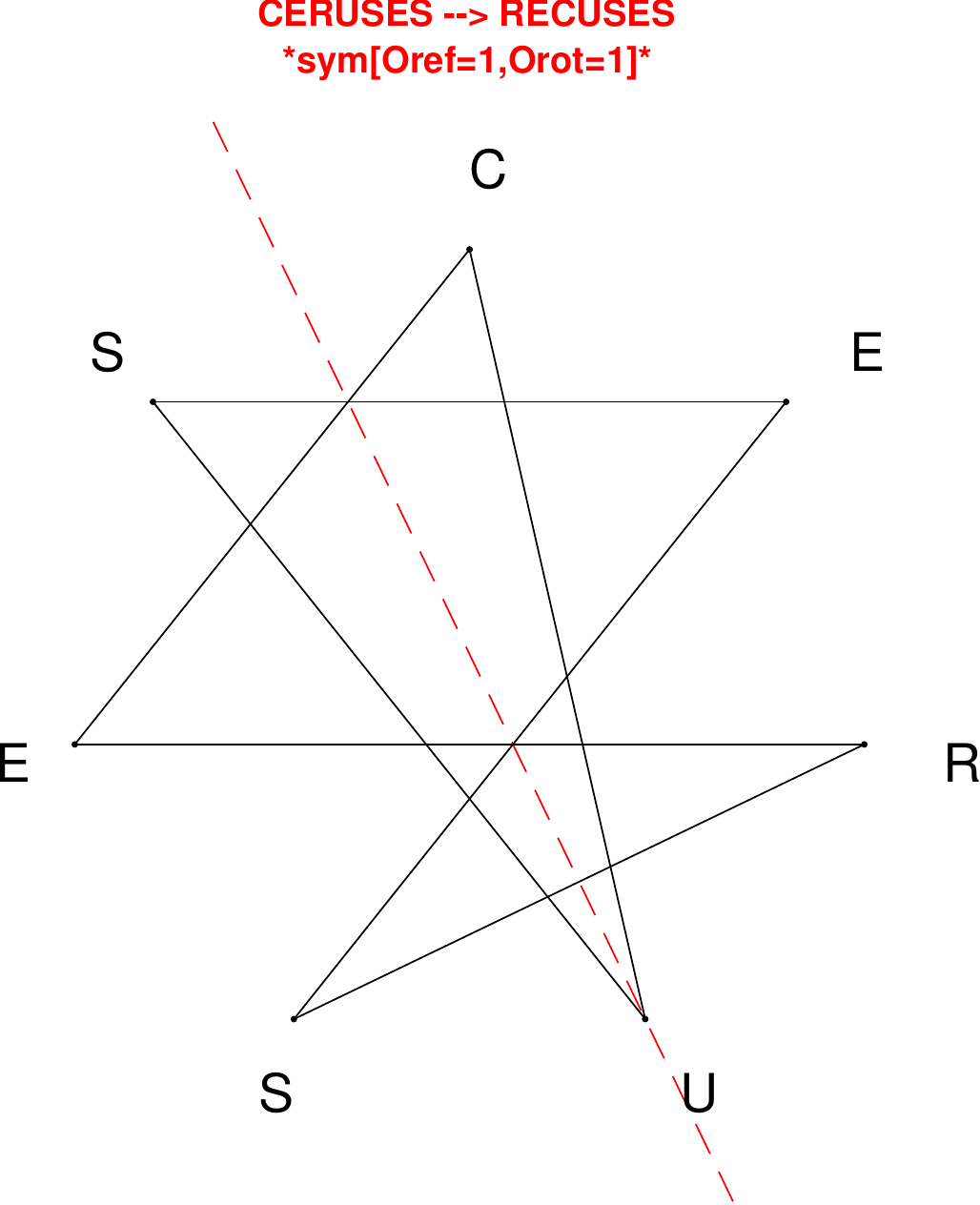}
\end{subfigure}
\hfill
\begin{subfigure}[T]{0.19\textwidth}
\centering
\includegraphics[width=\textwidth]{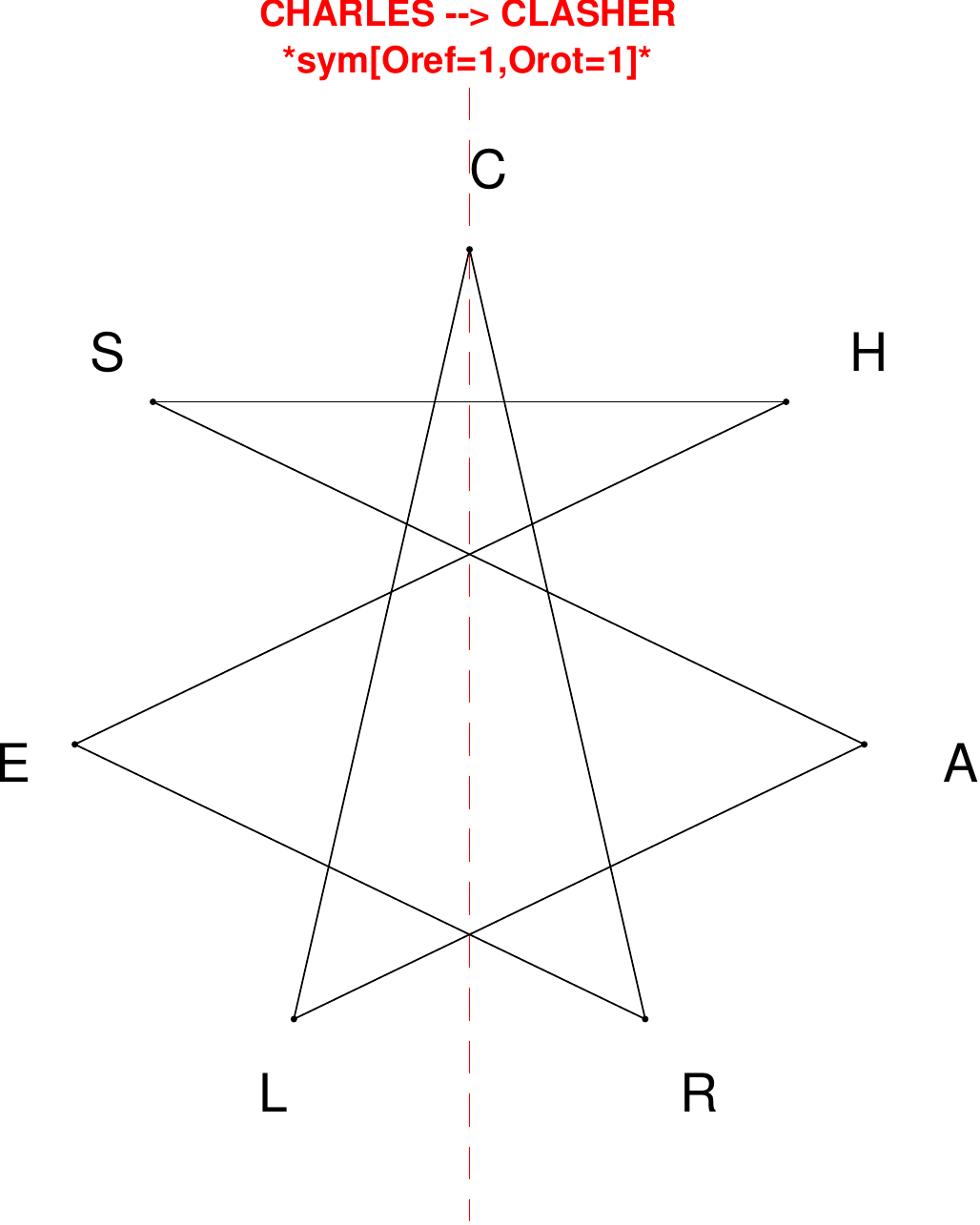}
\end{subfigure}
\hfill
\begin{subfigure}[T]{0.19\textwidth}
\centering
\includegraphics[width=\textwidth]{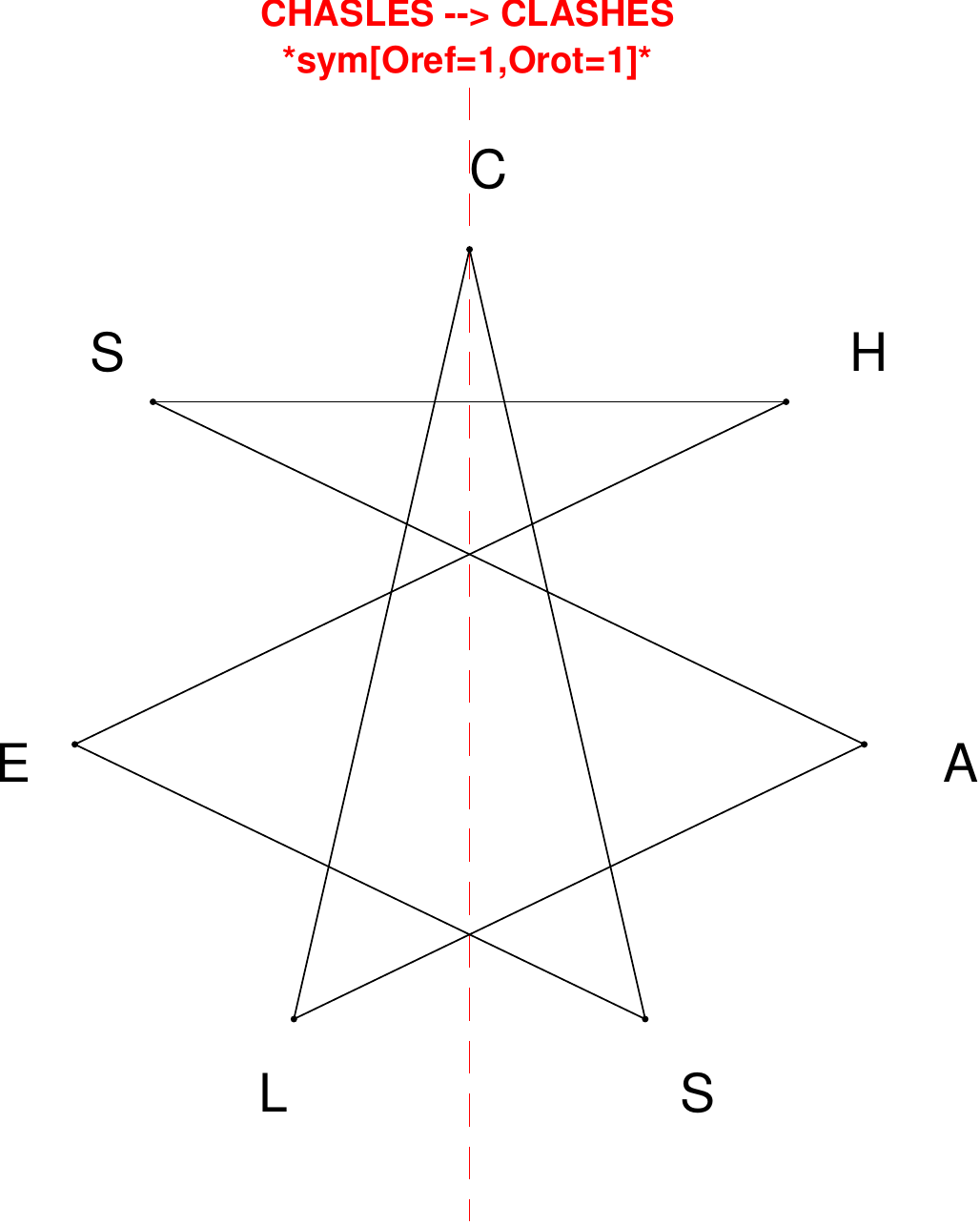}
\end{subfigure}
\end{figure}

\begin{figure}[H]
\centering
\begin{subfigure}[T]{0.19\textwidth}
\centering
\includegraphics[width=\textwidth]{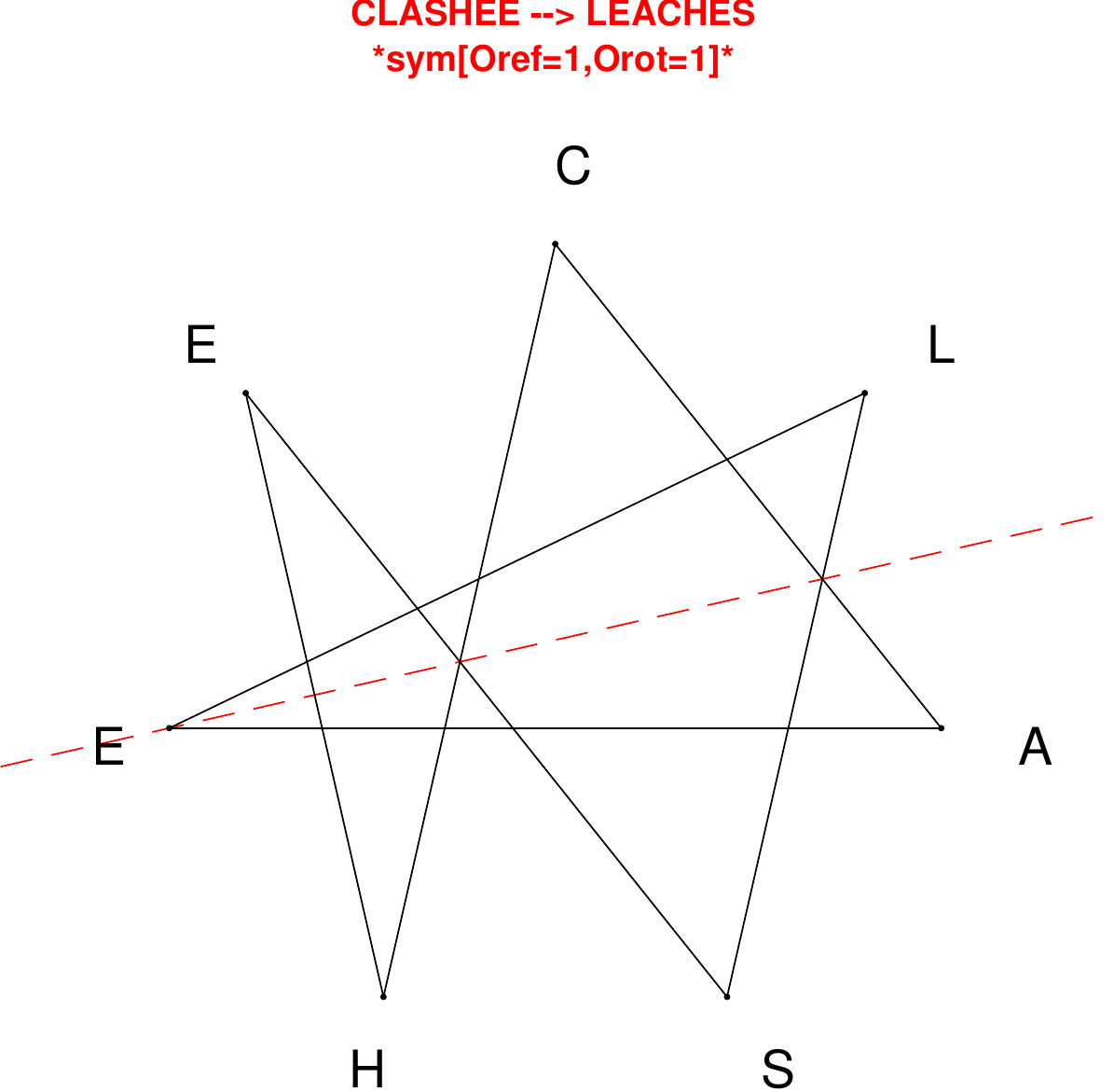}
\end{subfigure}
\hfill
\begin{subfigure}[T]{0.19\textwidth}
\centering
\includegraphics[width=\textwidth]{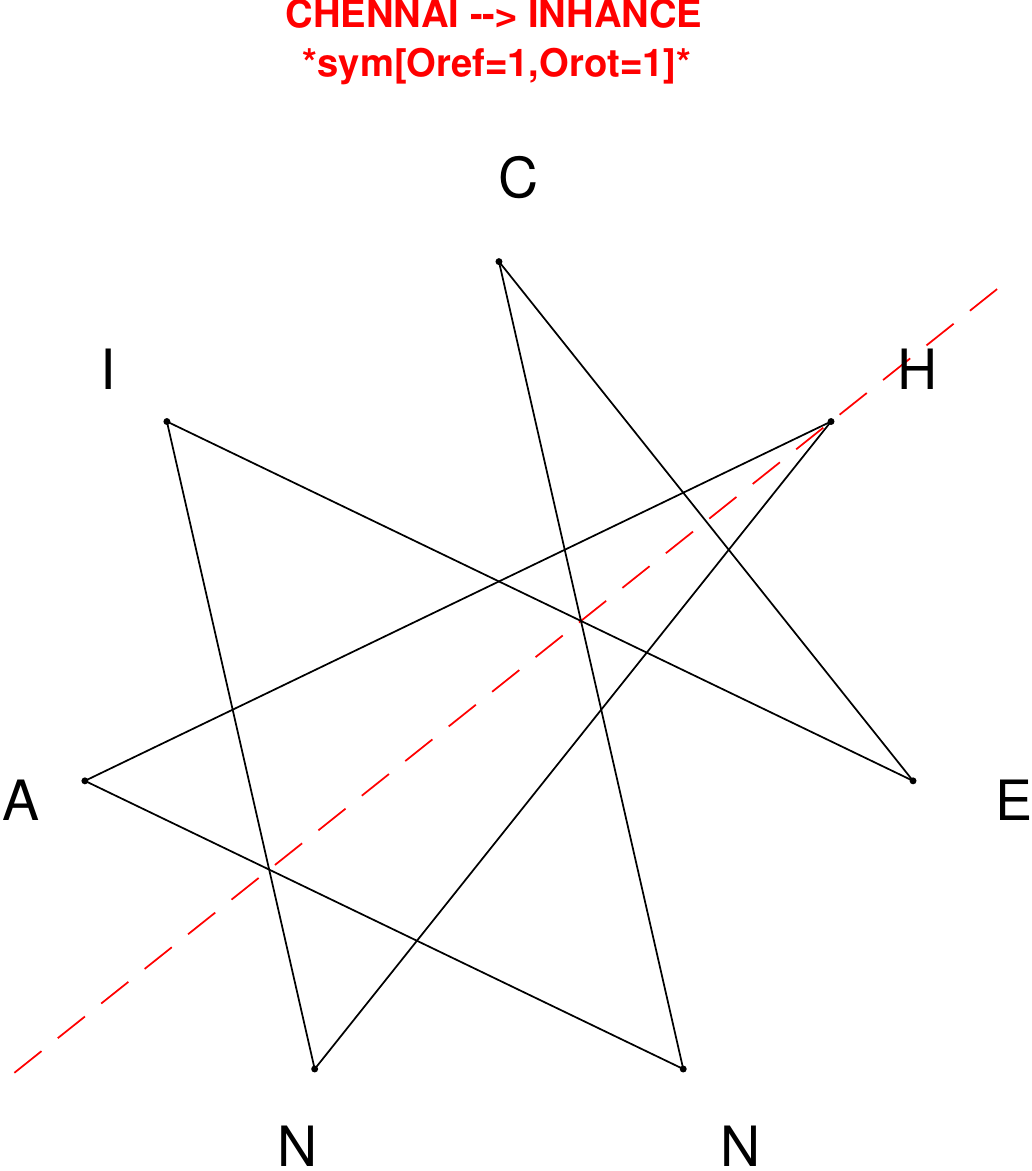}
\end{subfigure}
\hfill
\begin{subfigure}[T]{0.19\textwidth}
\centering
\includegraphics[width=\textwidth]{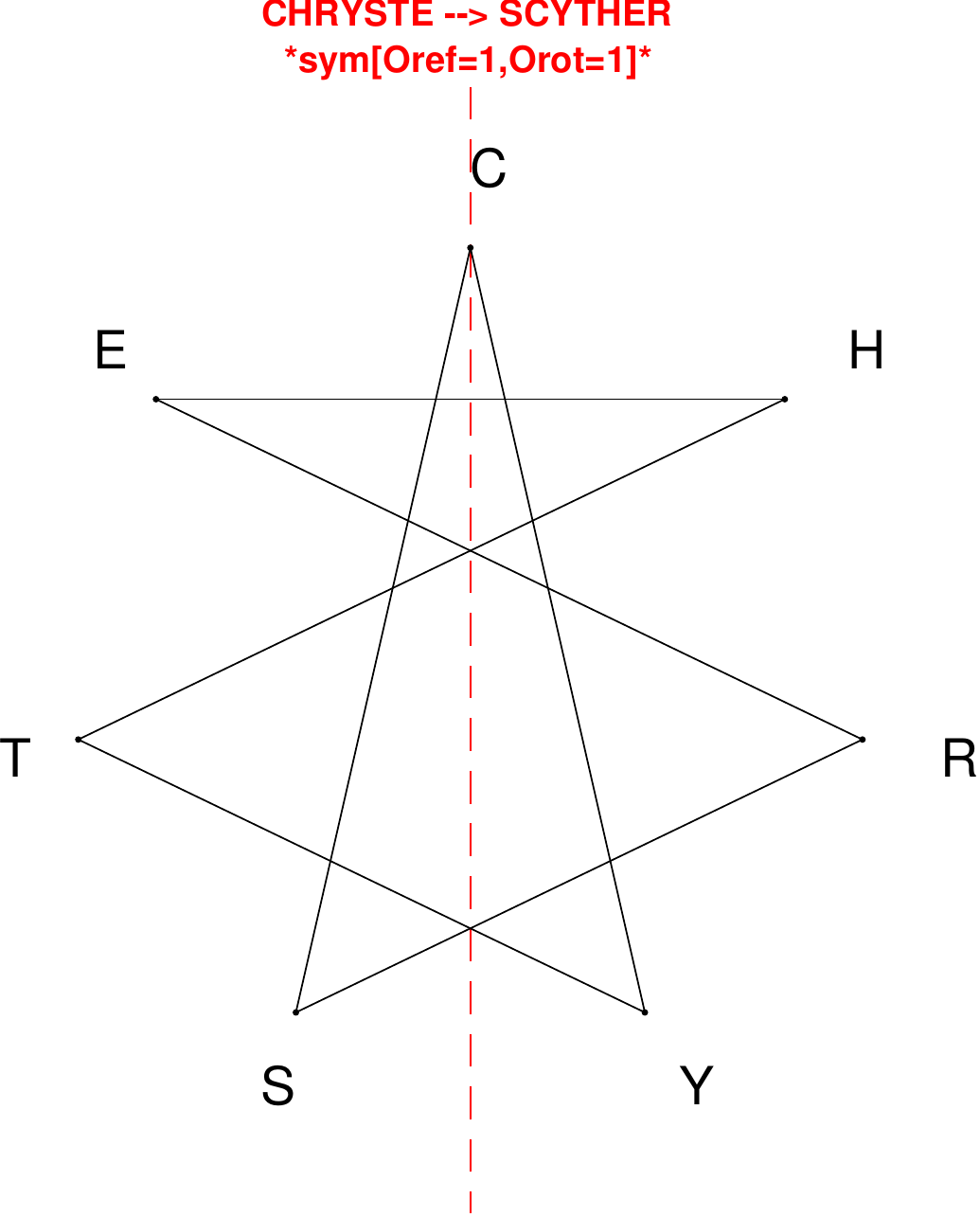}
\end{subfigure}
\hfill
\begin{subfigure}[T]{0.19\textwidth}
\centering
\includegraphics[width=\textwidth]{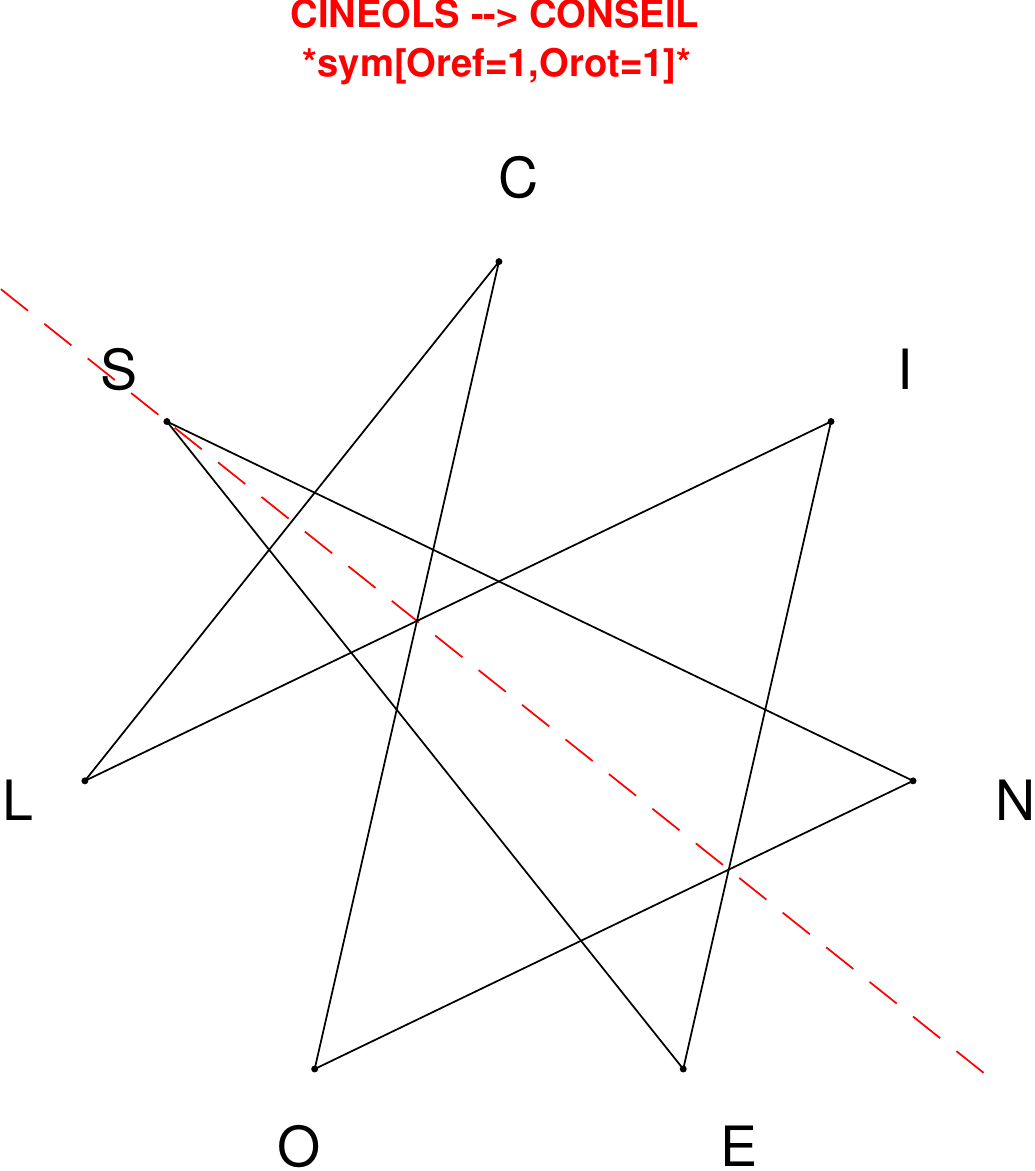}
\end{subfigure}
\hfill
\begin{subfigure}[T]{0.19\textwidth}
\centering
\includegraphics[width=\textwidth]{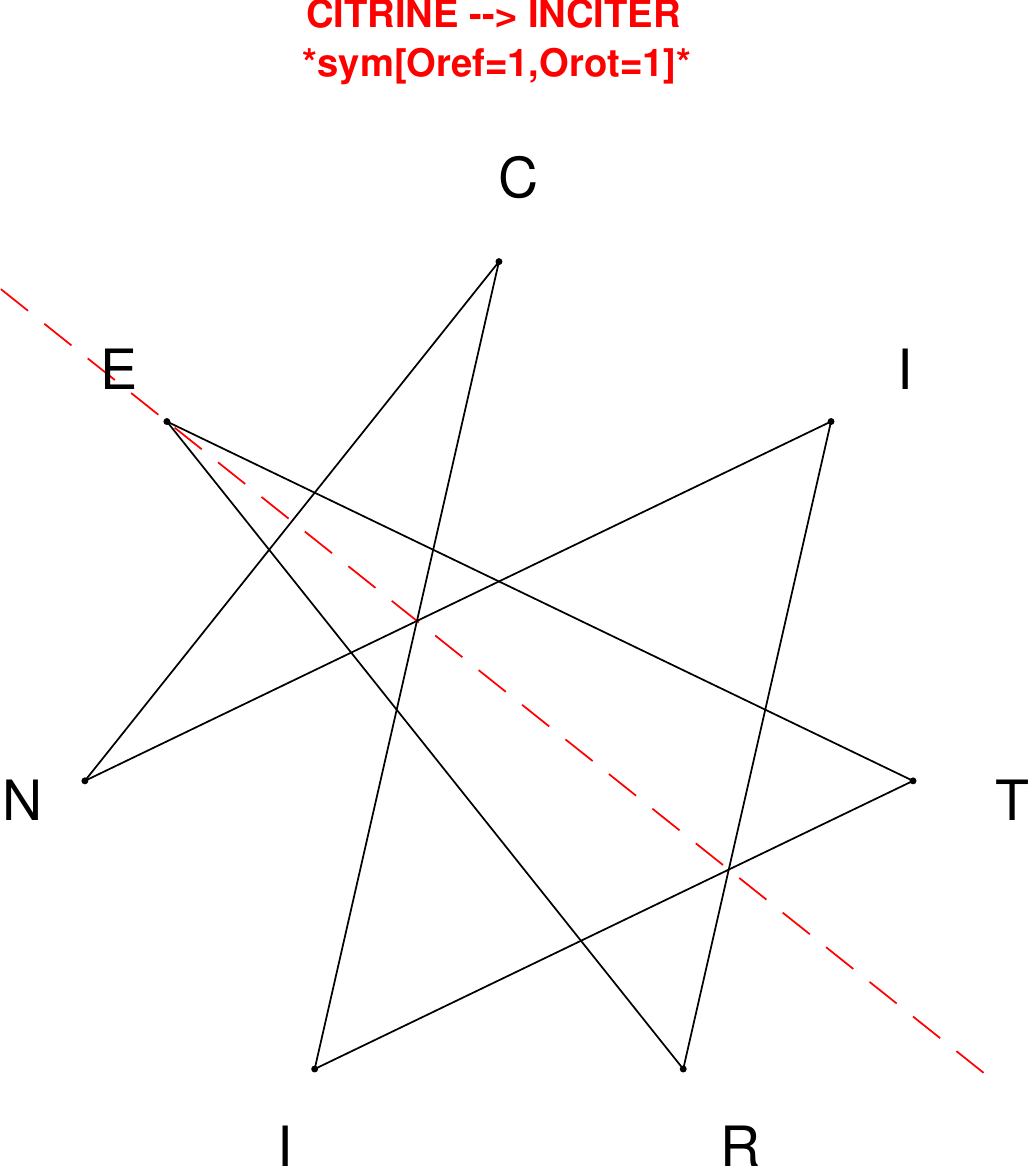}
\end{subfigure}
\end{figure}

\begin{figure}[H]
\centering
\begin{subfigure}[T]{0.19\textwidth}
\centering
\includegraphics[width=\textwidth]{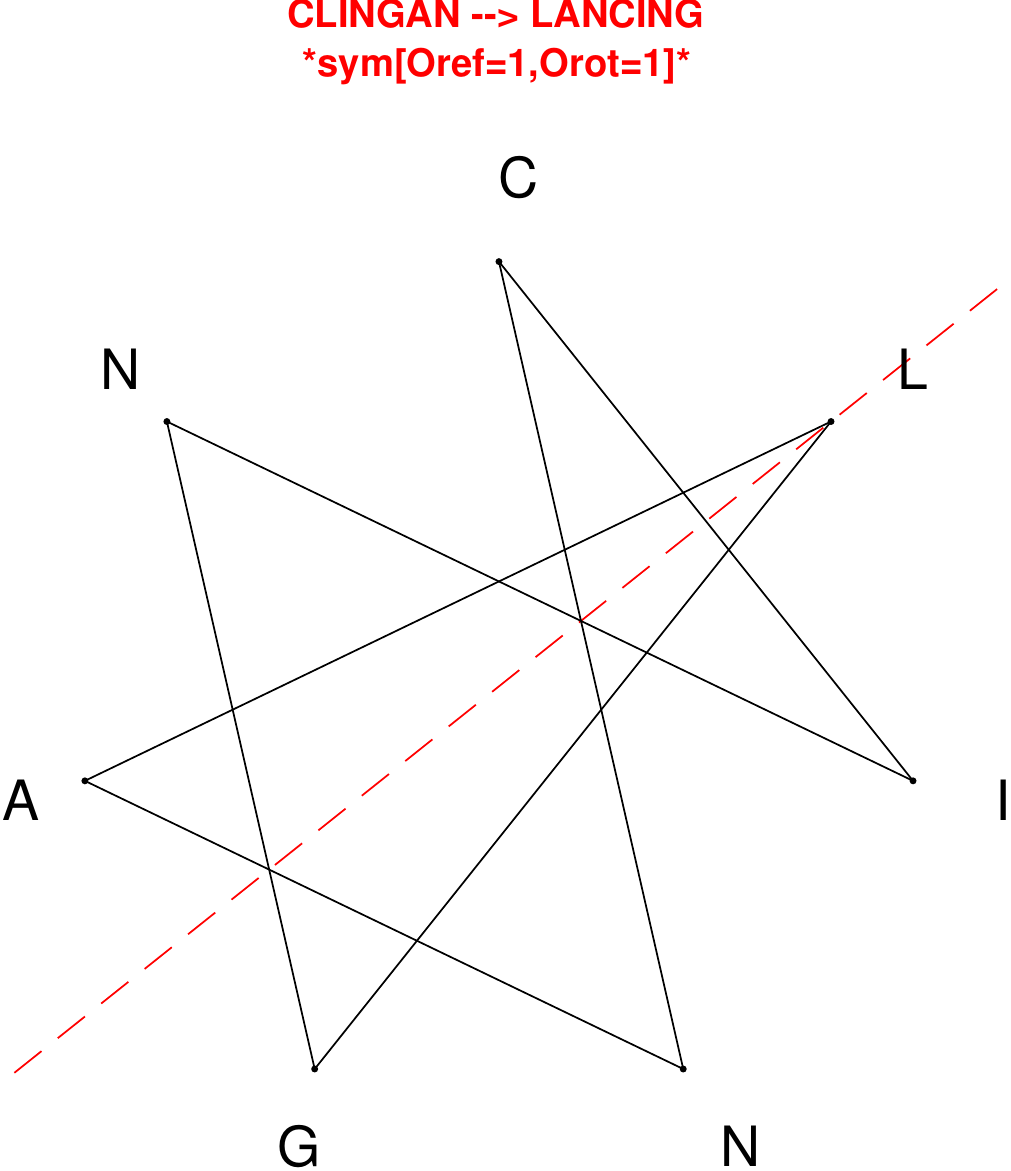}
\end{subfigure}
\hfill
\begin{subfigure}[T]{0.19\textwidth}
\centering
\includegraphics[width=\textwidth]{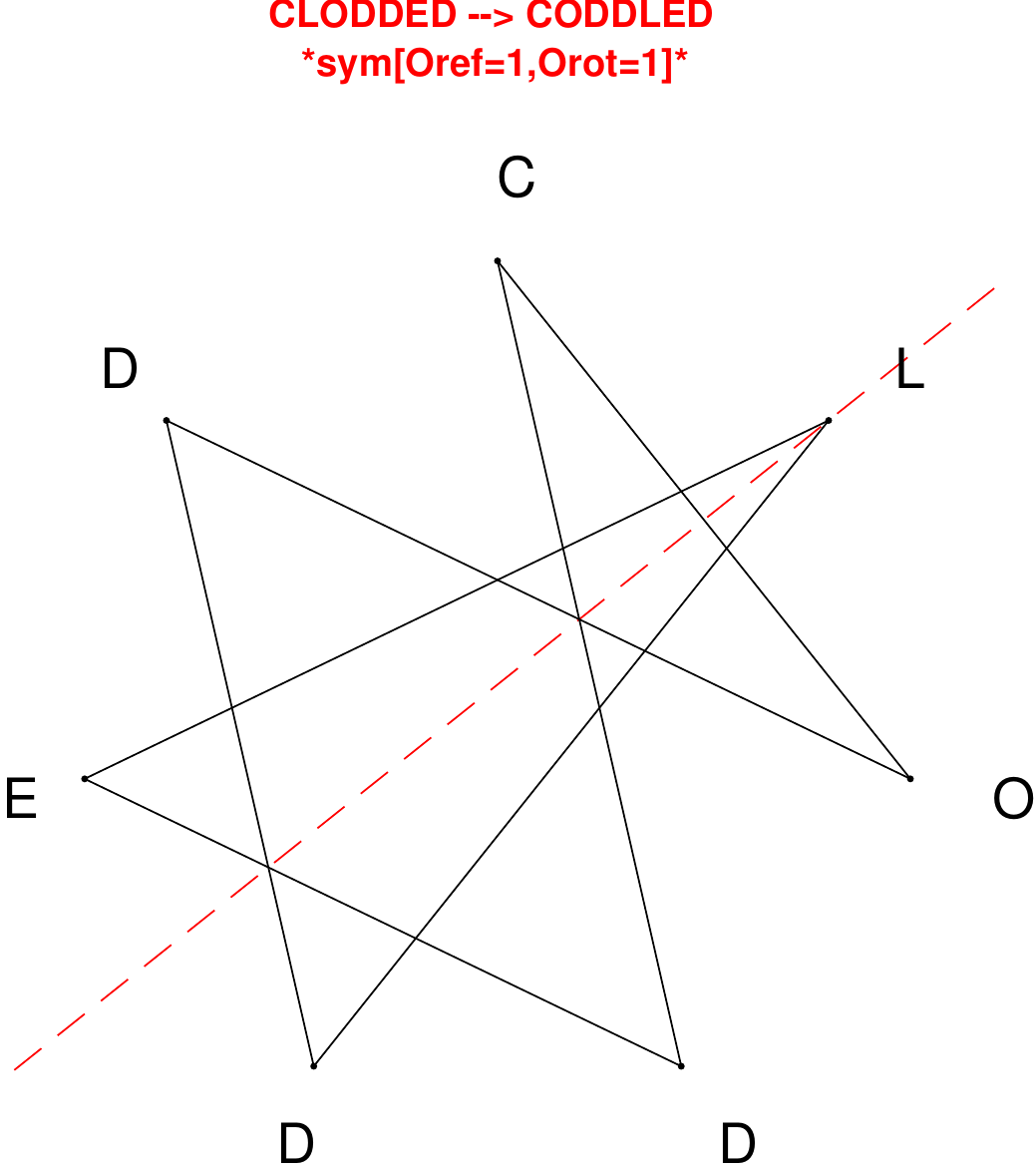}
\end{subfigure}
\hfill
\begin{subfigure}[T]{0.19\textwidth}
\centering
\includegraphics[width=\textwidth]{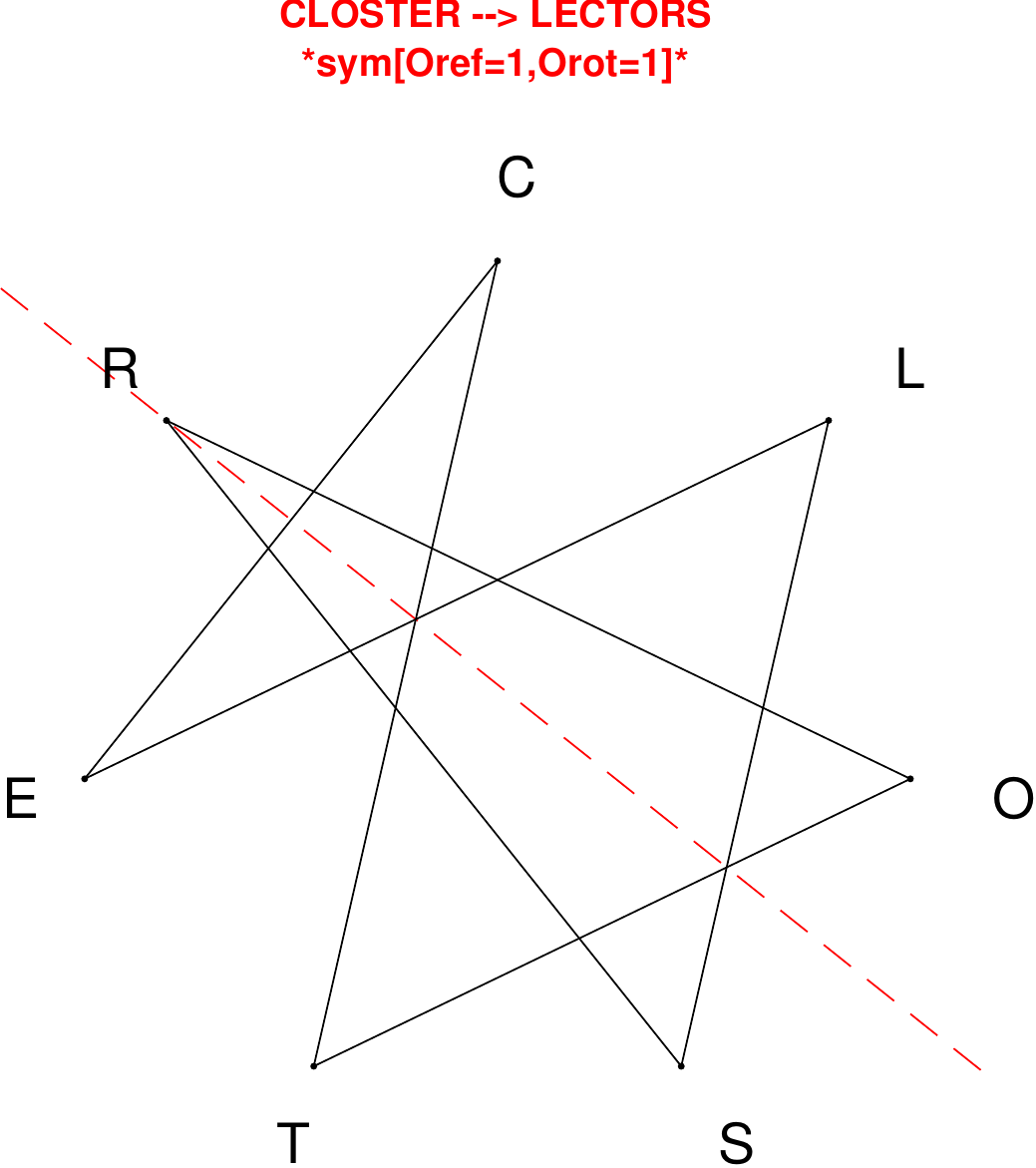}
\end{subfigure}
\hfill
\begin{subfigure}[T]{0.19\textwidth}
\centering
\includegraphics[width=\textwidth]{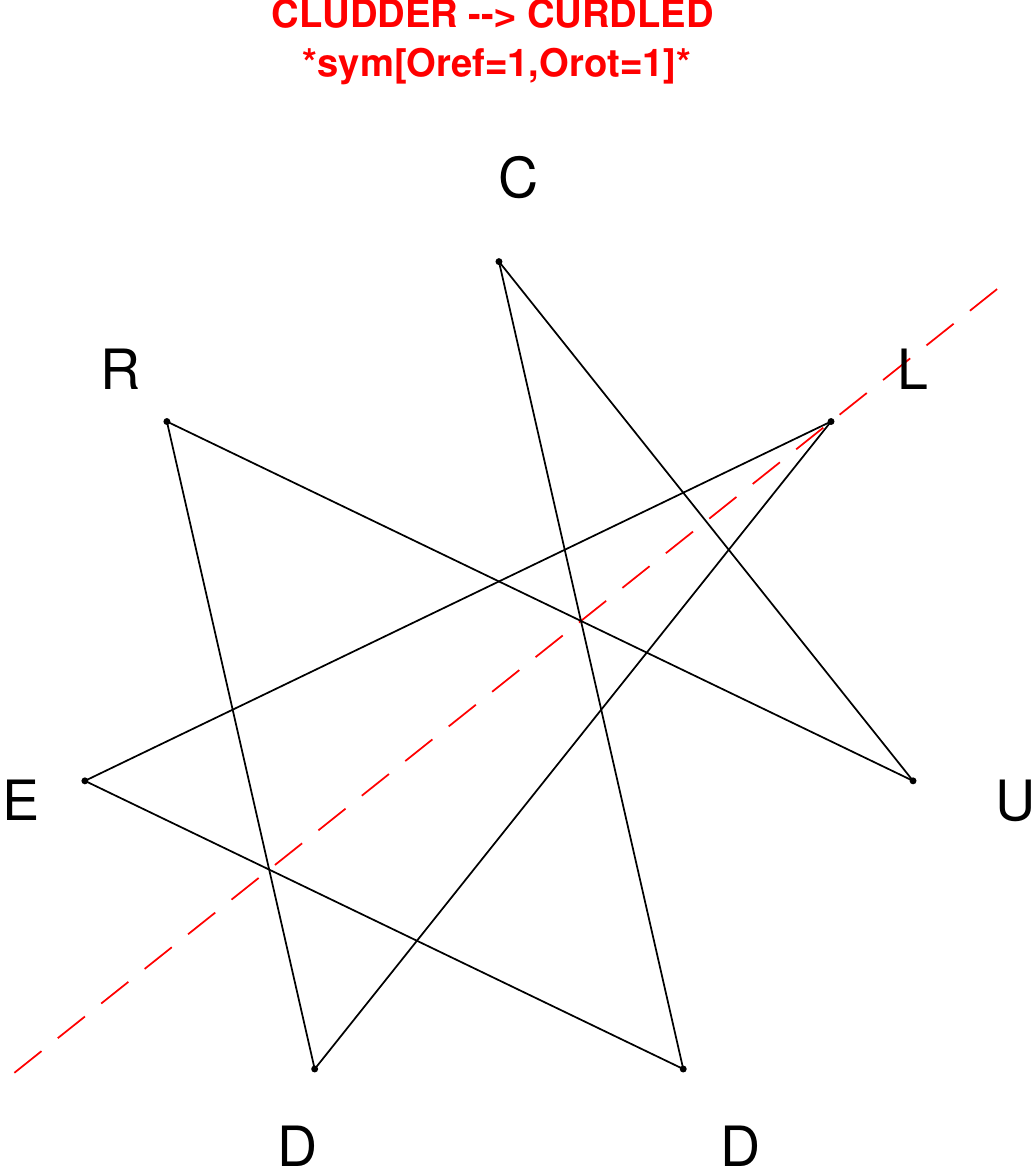}
\end{subfigure}
\hfill
\begin{subfigure}[T]{0.19\textwidth}
\centering
\includegraphics[width=\textwidth]{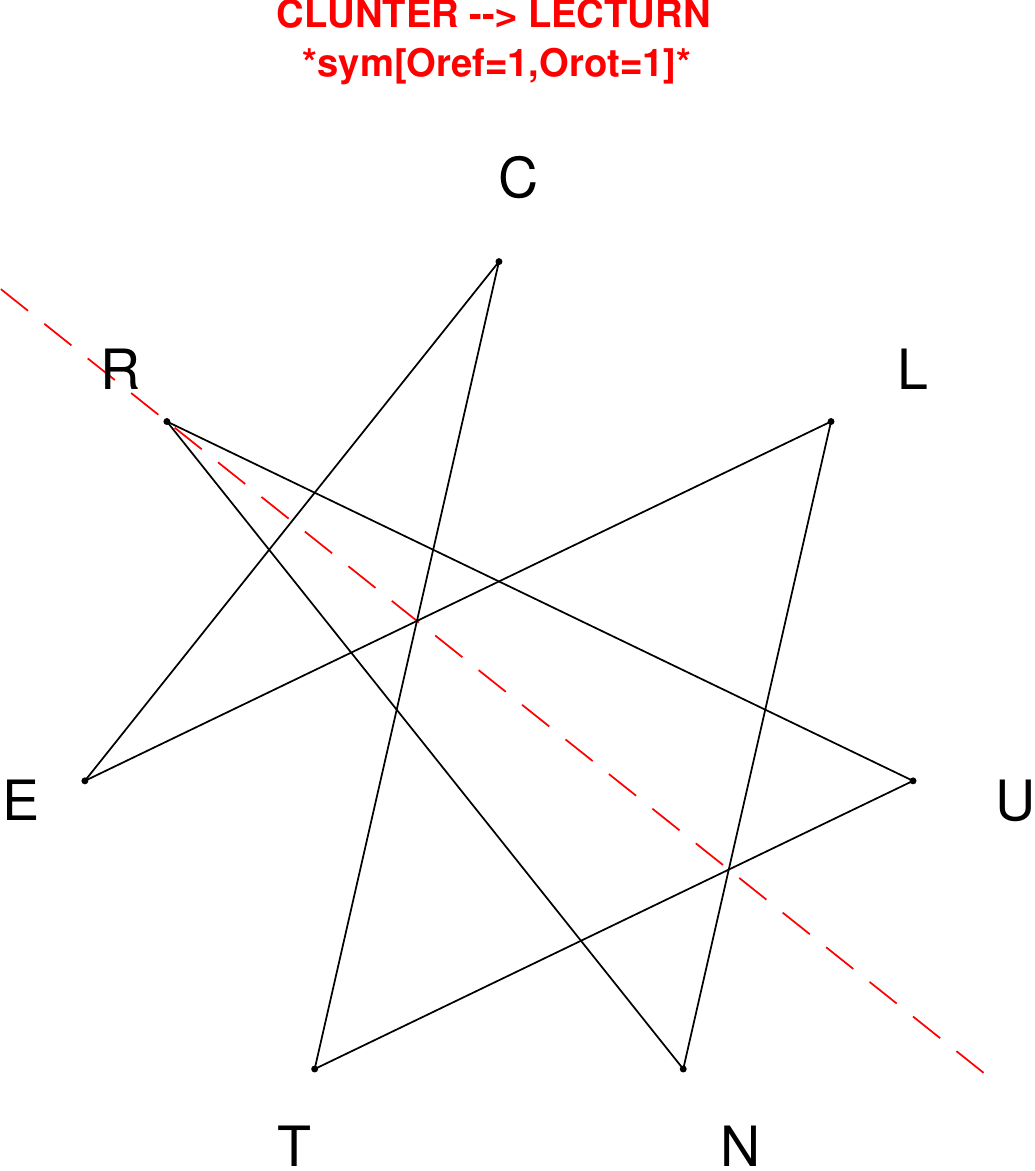}
\end{subfigure}
\end{figure}

\begin{figure}[H]
\centering
\begin{subfigure}[T]{0.19\textwidth}
\centering
\includegraphics[width=\textwidth]{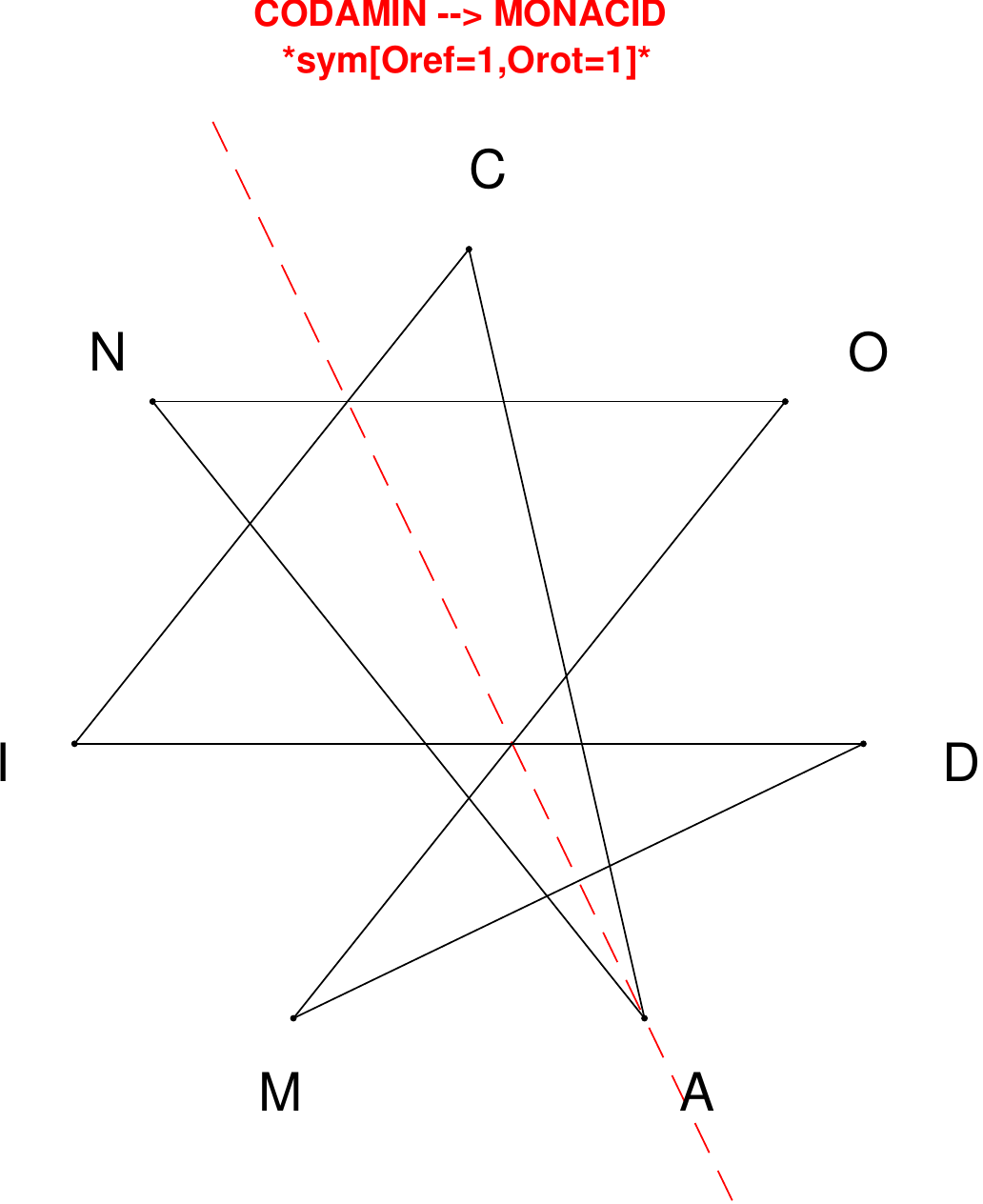}
\end{subfigure}
\hfill
\begin{subfigure}[T]{0.19\textwidth}
\centering
\includegraphics[width=\textwidth]{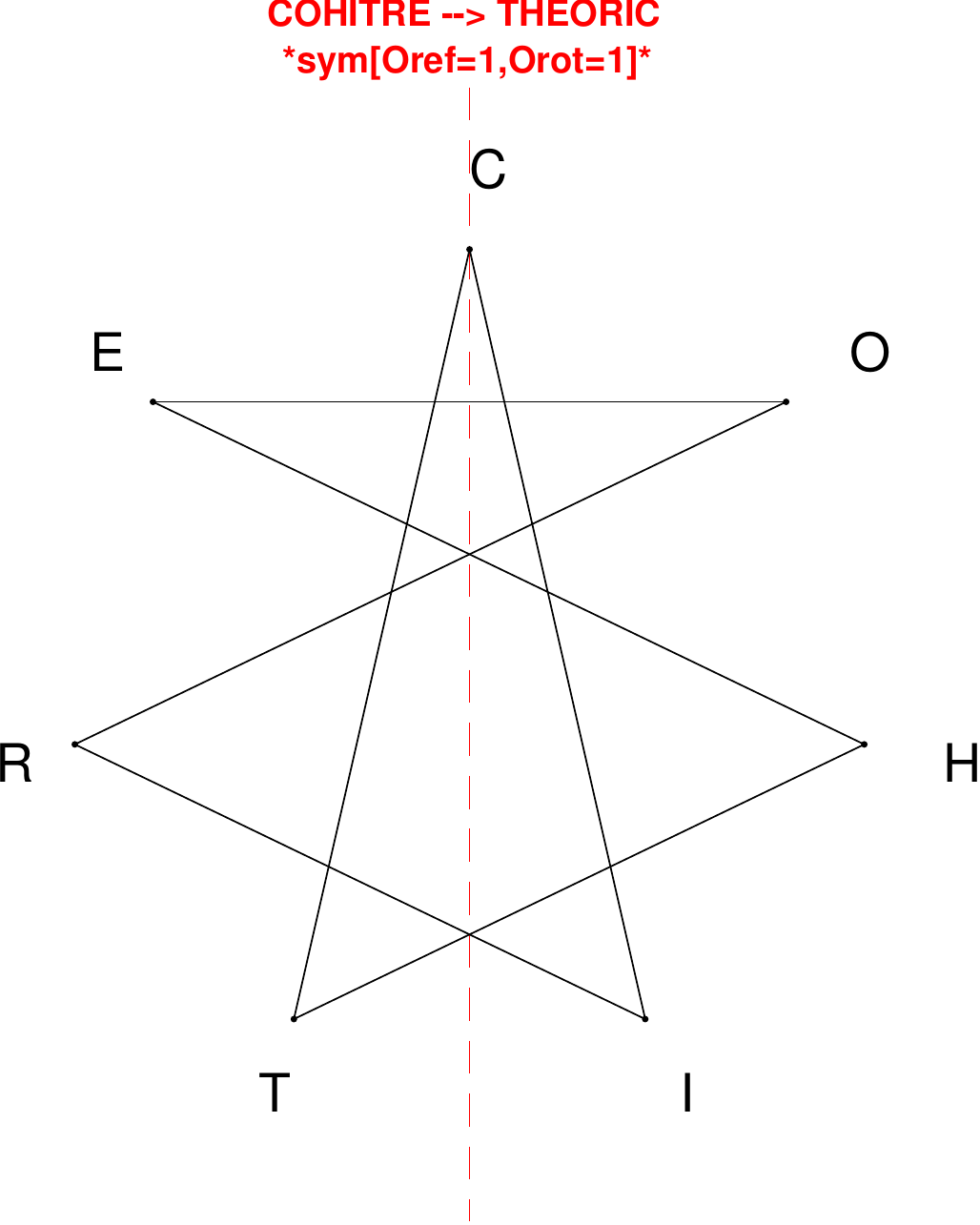}
\end{subfigure}
\hfill
\begin{subfigure}[T]{0.19\textwidth}
\centering
\includegraphics[width=\textwidth]{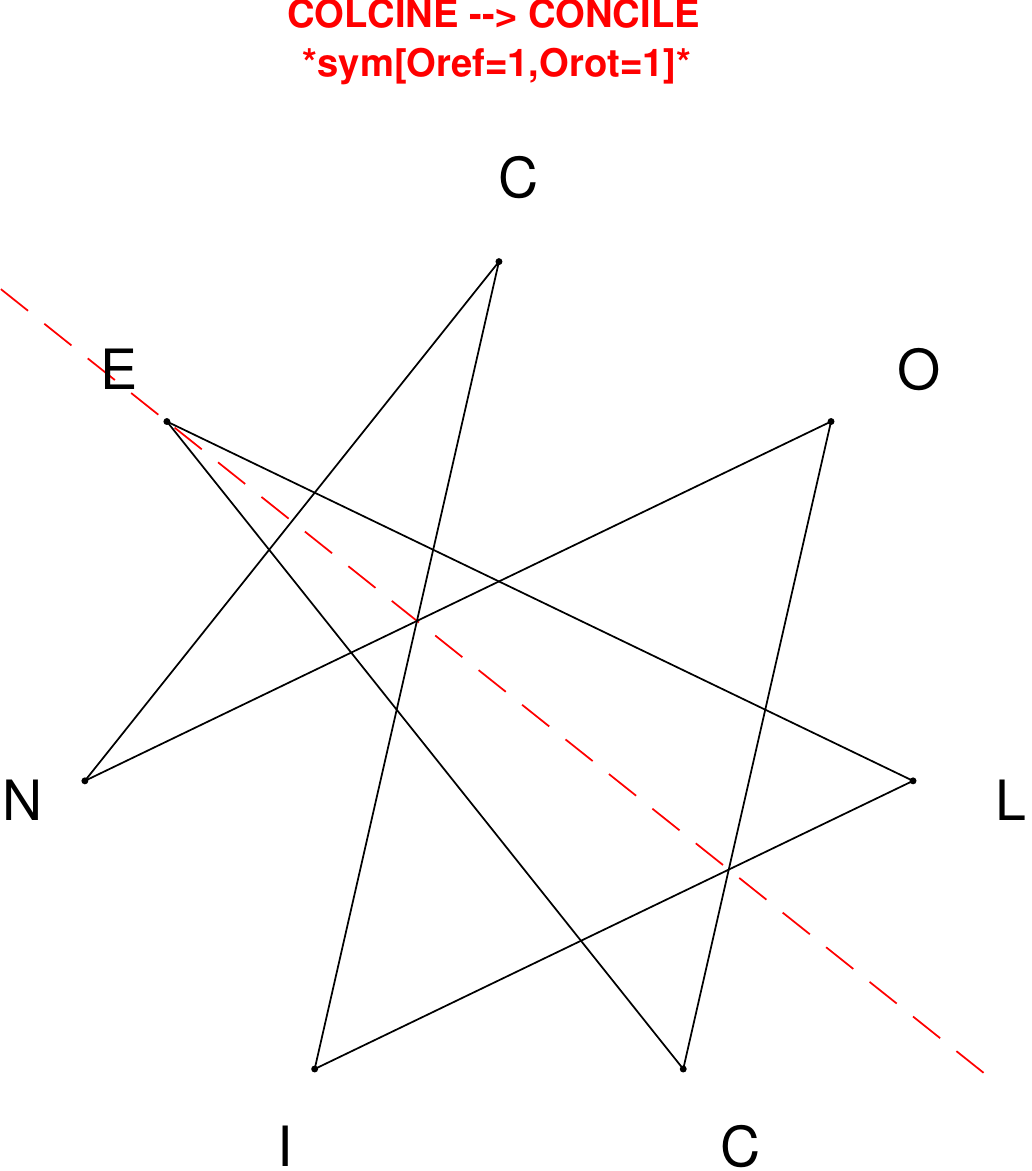}
\end{subfigure}
\hfill
\begin{subfigure}[T]{0.19\textwidth}
\centering
\includegraphics[width=\textwidth]{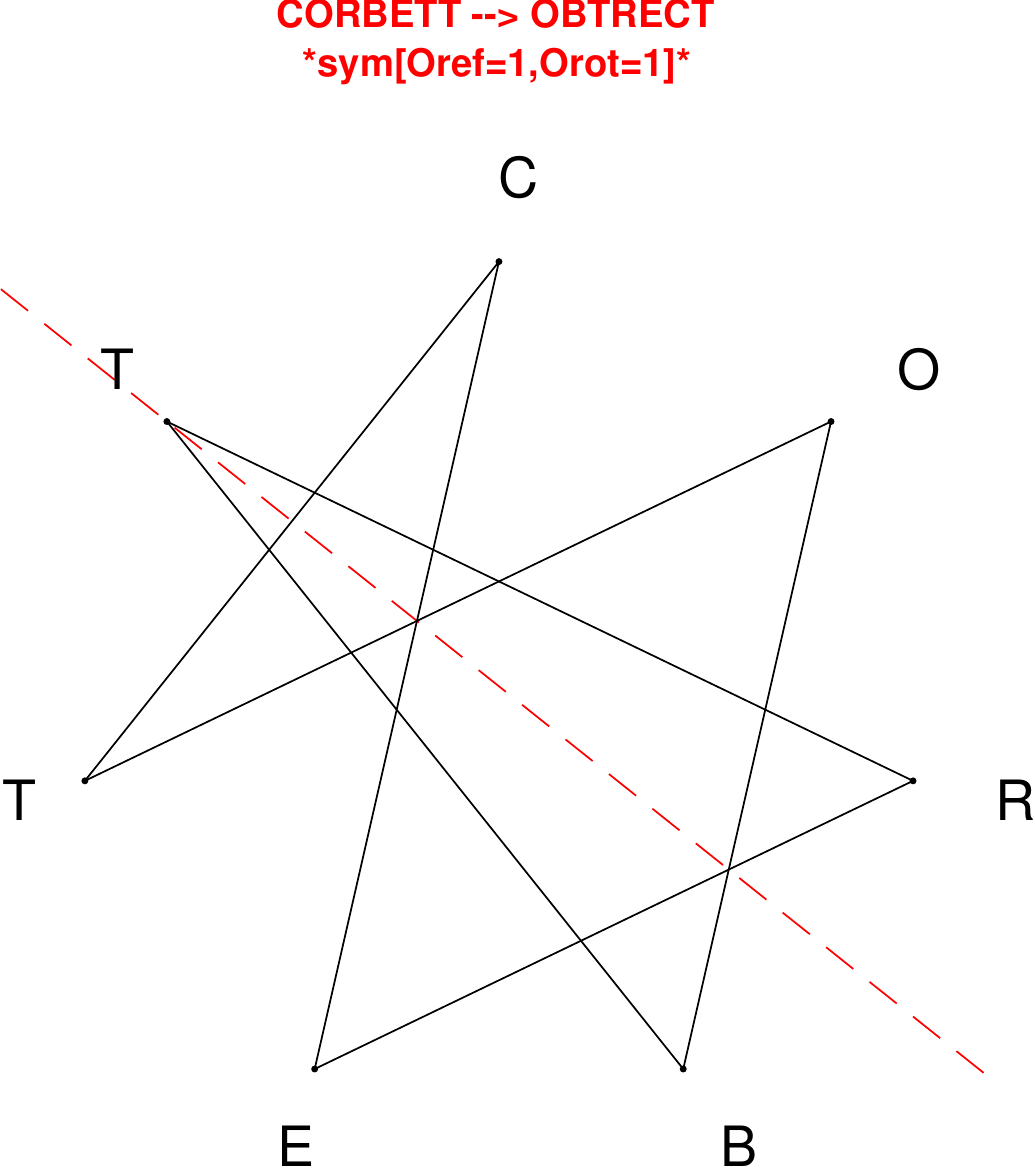}
\end{subfigure}
\hfill
\begin{subfigure}[T]{0.19\textwidth}
\centering
\includegraphics[width=\textwidth]{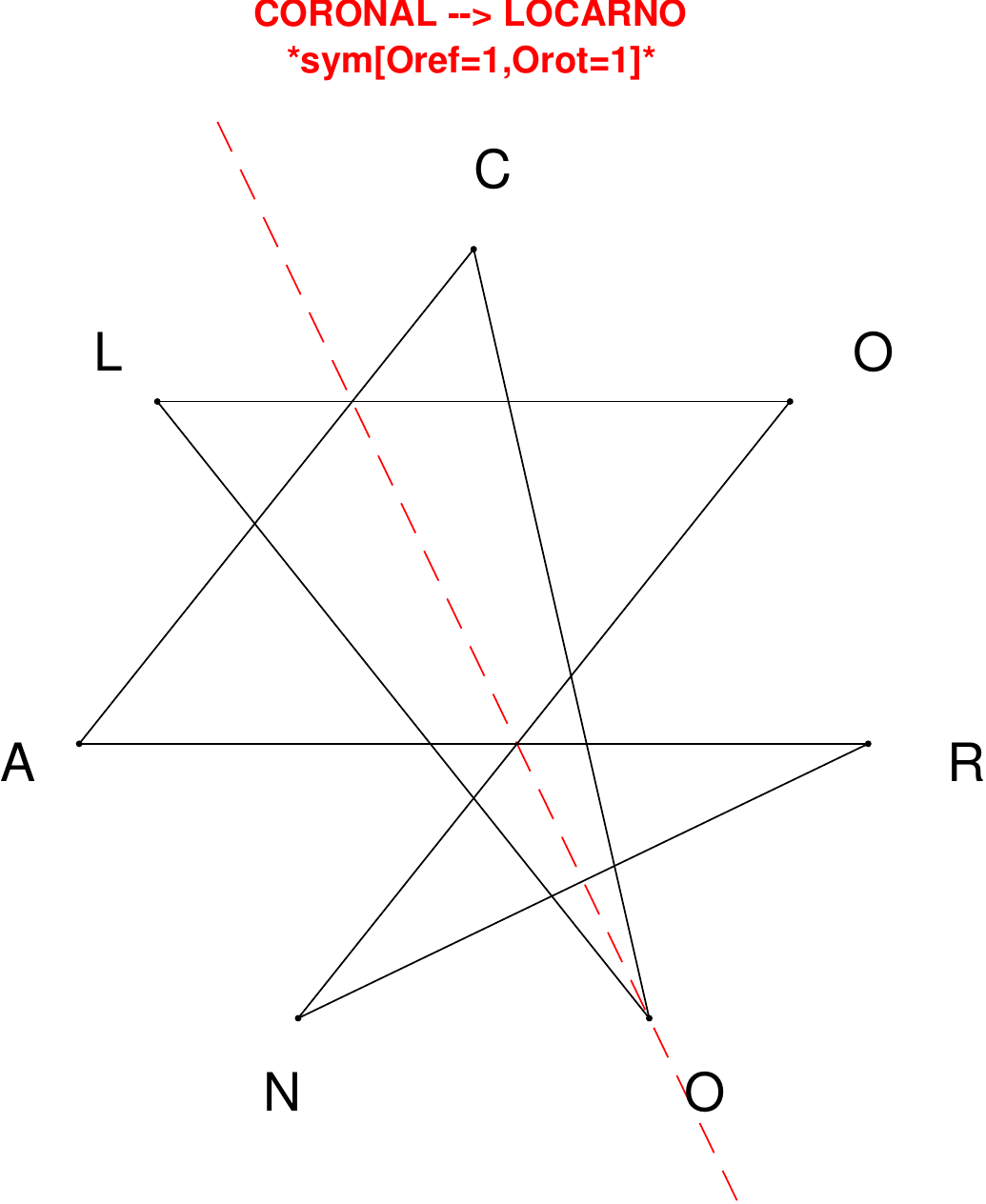}
\end{subfigure}
\end{figure}

\begin{figure}[H]
\centering
\begin{subfigure}[T]{0.19\textwidth}
\centering
\includegraphics[width=\textwidth]{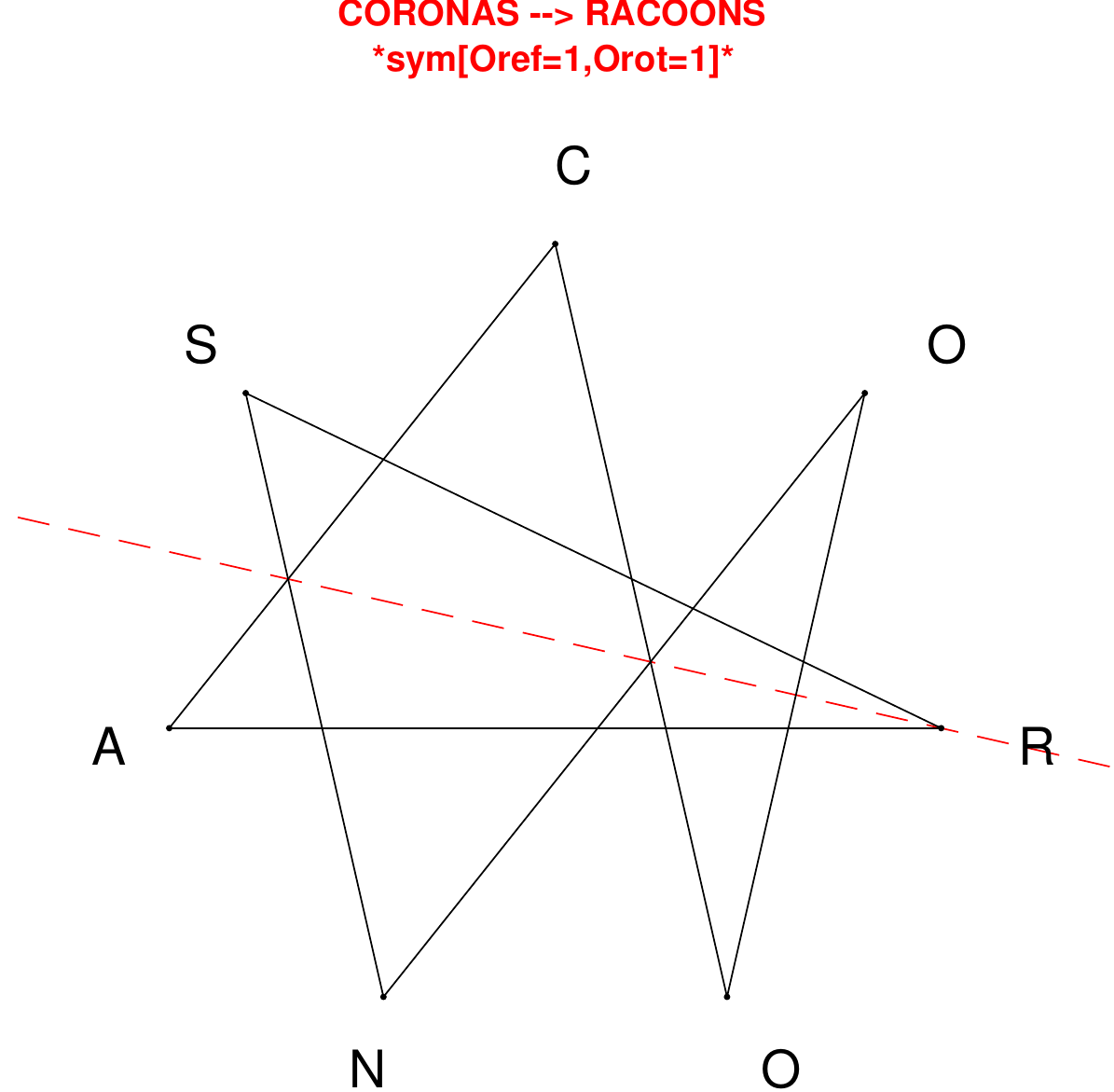}
\end{subfigure}
\hfill
\begin{subfigure}[T]{0.19\textwidth}
\centering
\includegraphics[width=\textwidth]{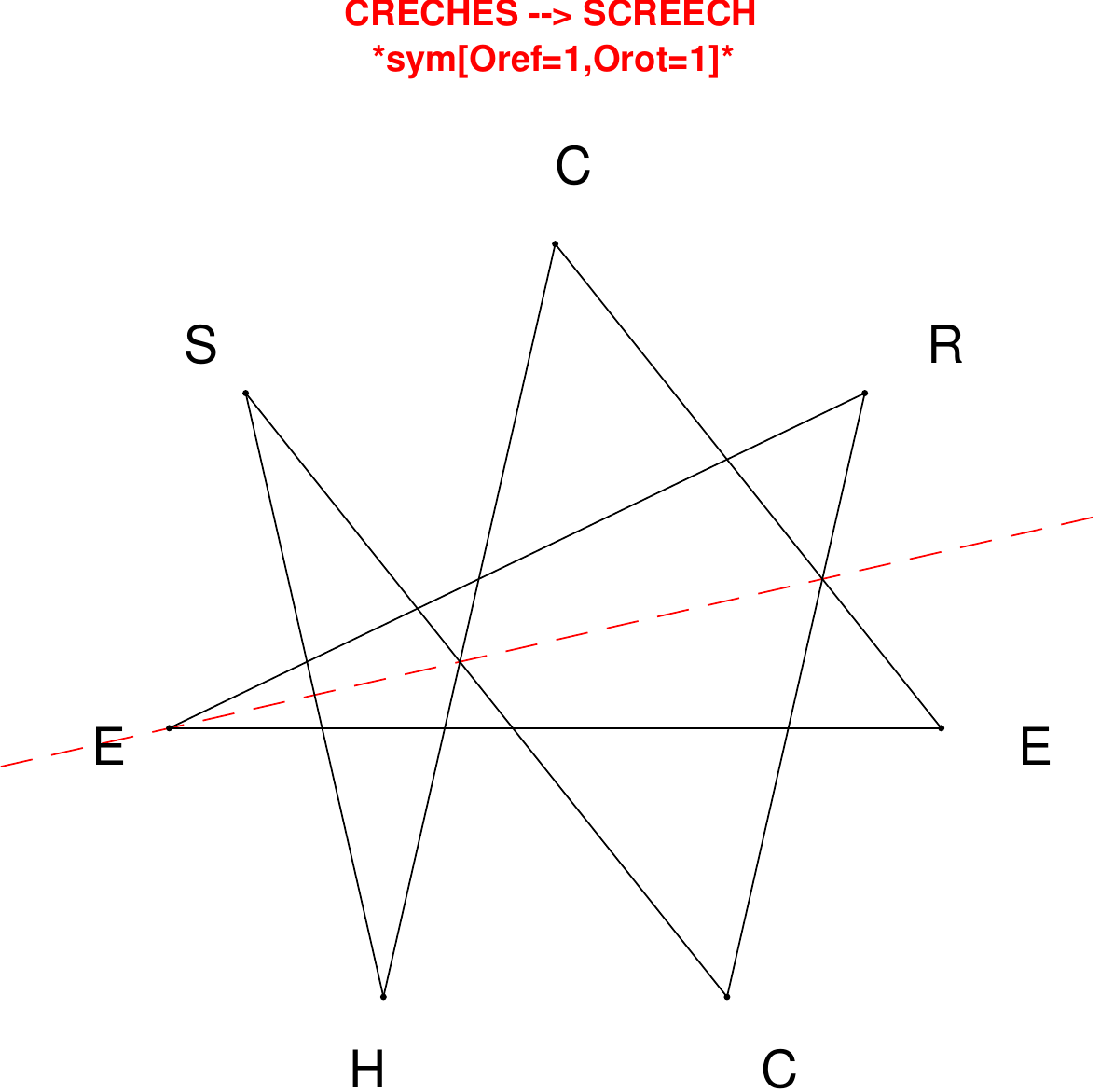}
\end{subfigure}
\hfill
\begin{subfigure}[T]{0.19\textwidth}
\centering
\includegraphics[width=\textwidth]{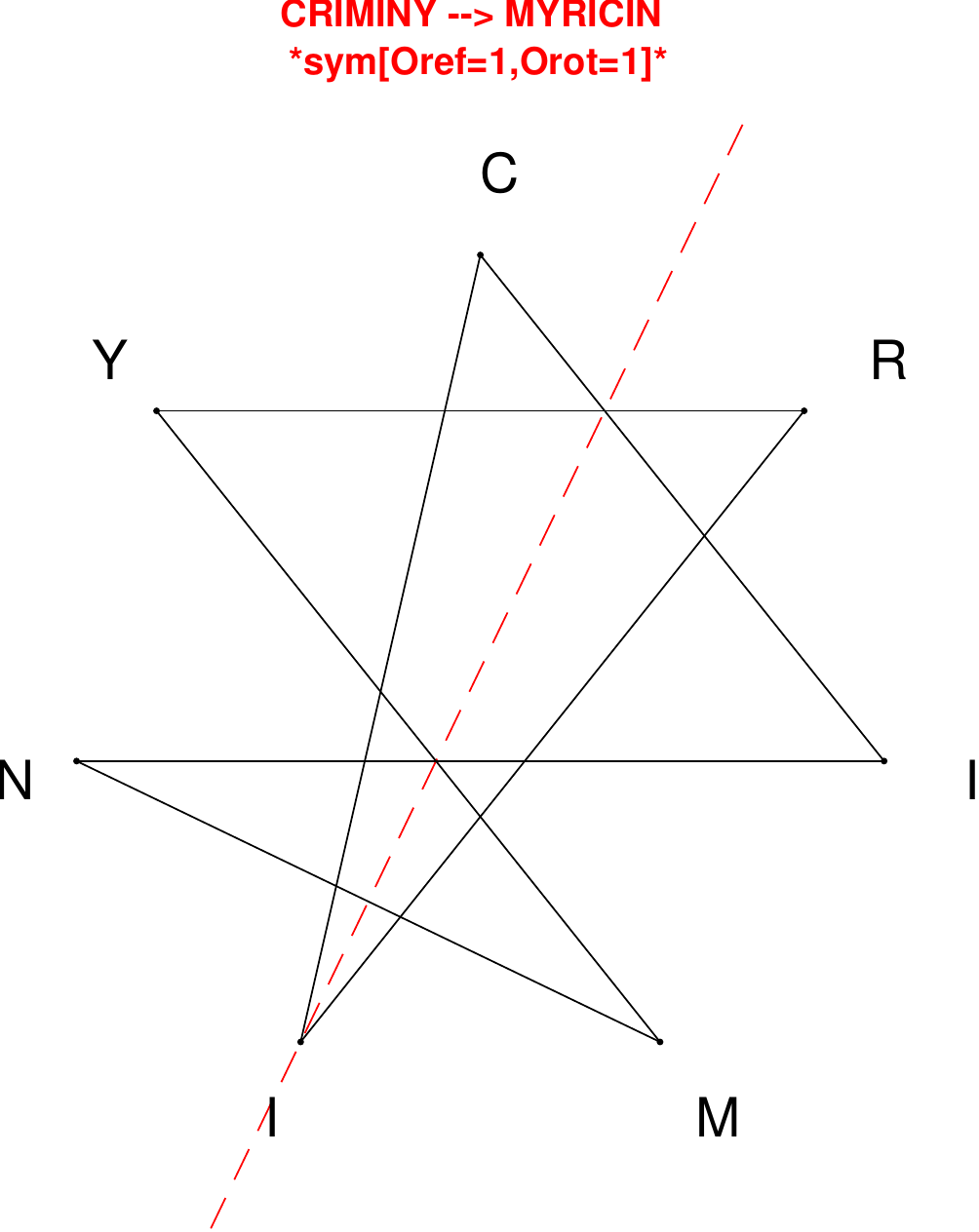}
\end{subfigure}
\hfill
\begin{subfigure}[T]{0.19\textwidth}
\centering
\includegraphics[width=\textwidth]{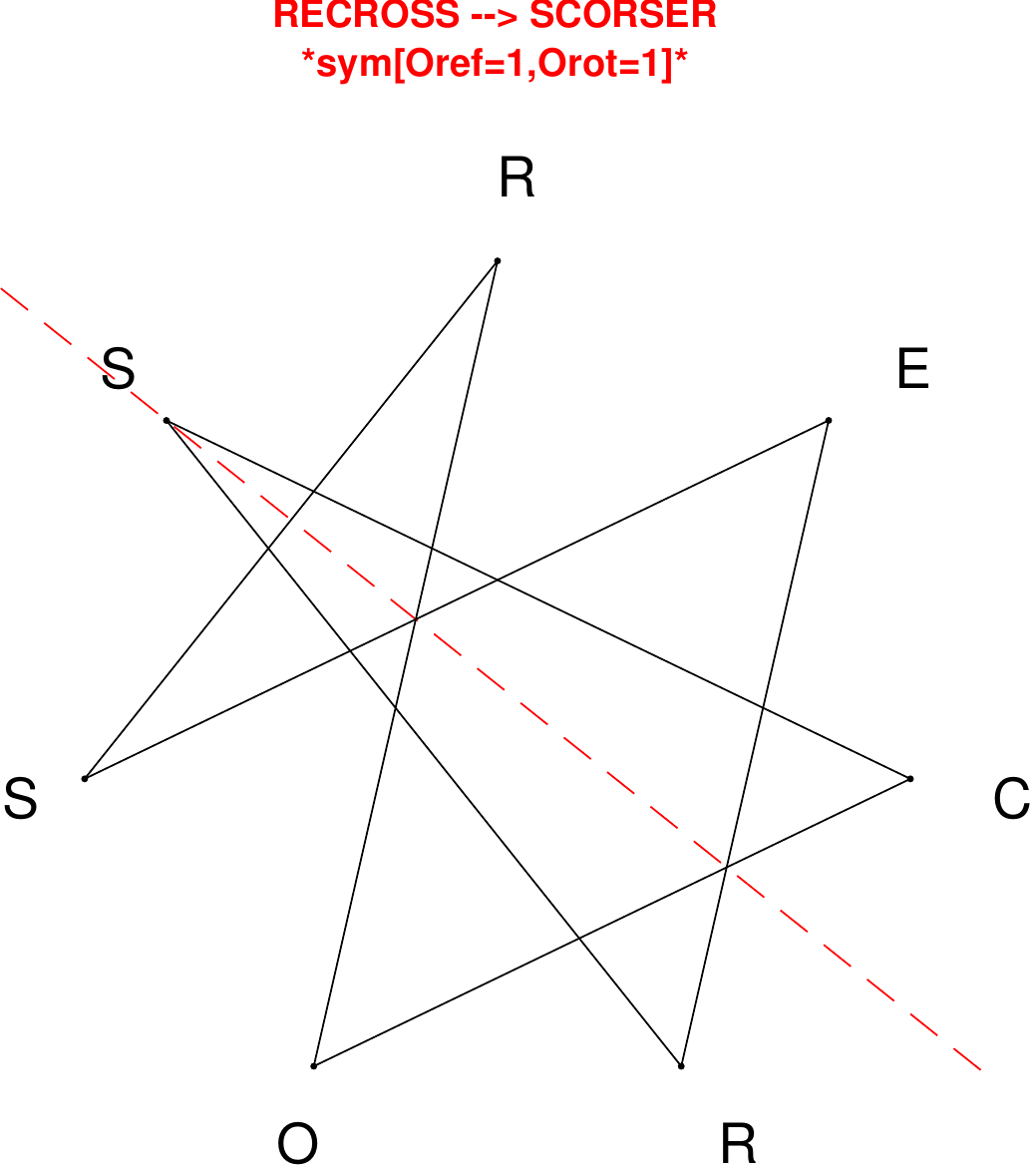}
\end{subfigure}
\hfill
\begin{subfigure}[T]{0.19\textwidth}
\centering
\includegraphics[width=\textwidth]{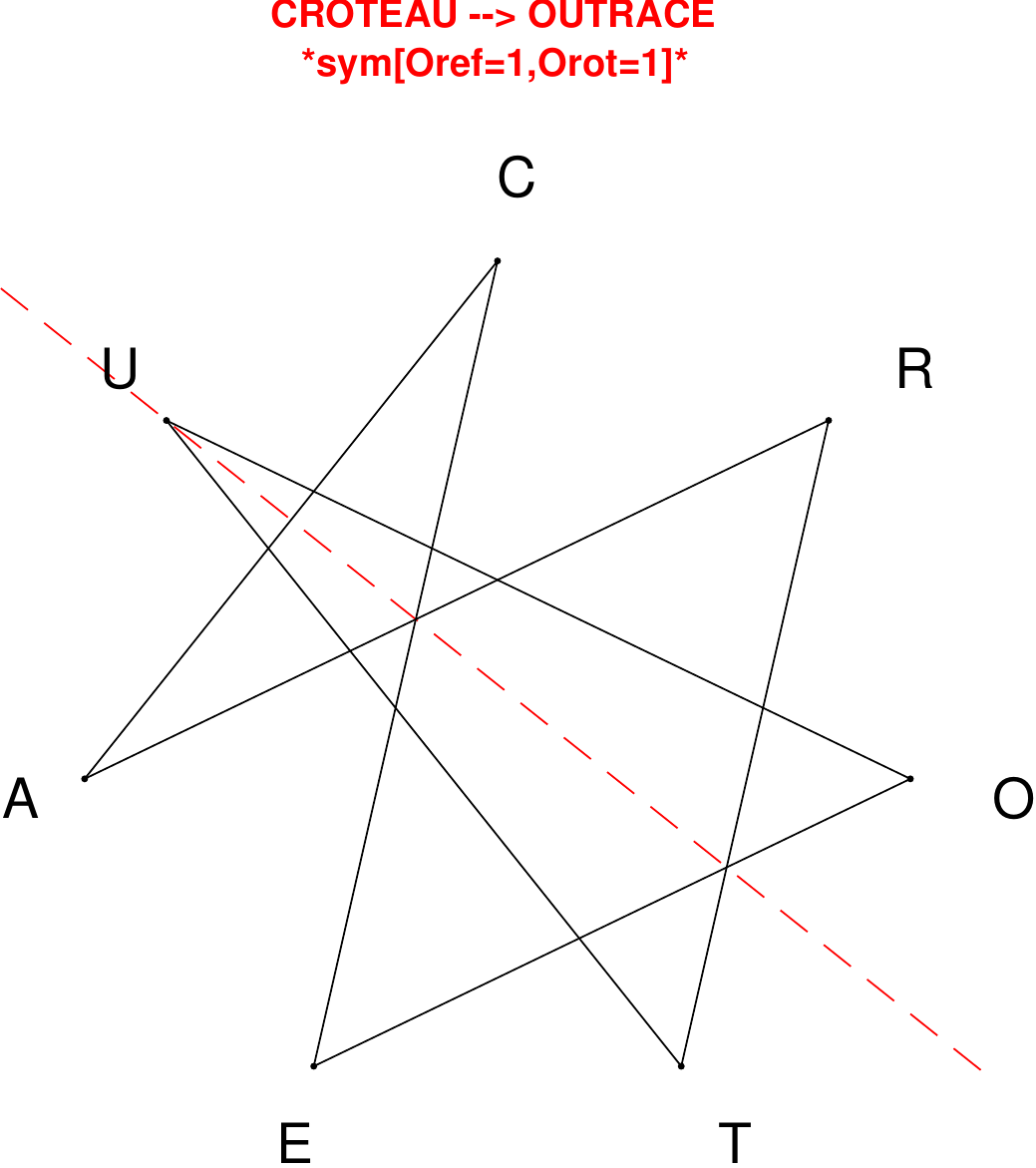}
\end{subfigure}
\end{figure}

\begin{figure}[H]
\centering
\begin{subfigure}[T]{0.19\textwidth}
\centering
\includegraphics[width=\textwidth]{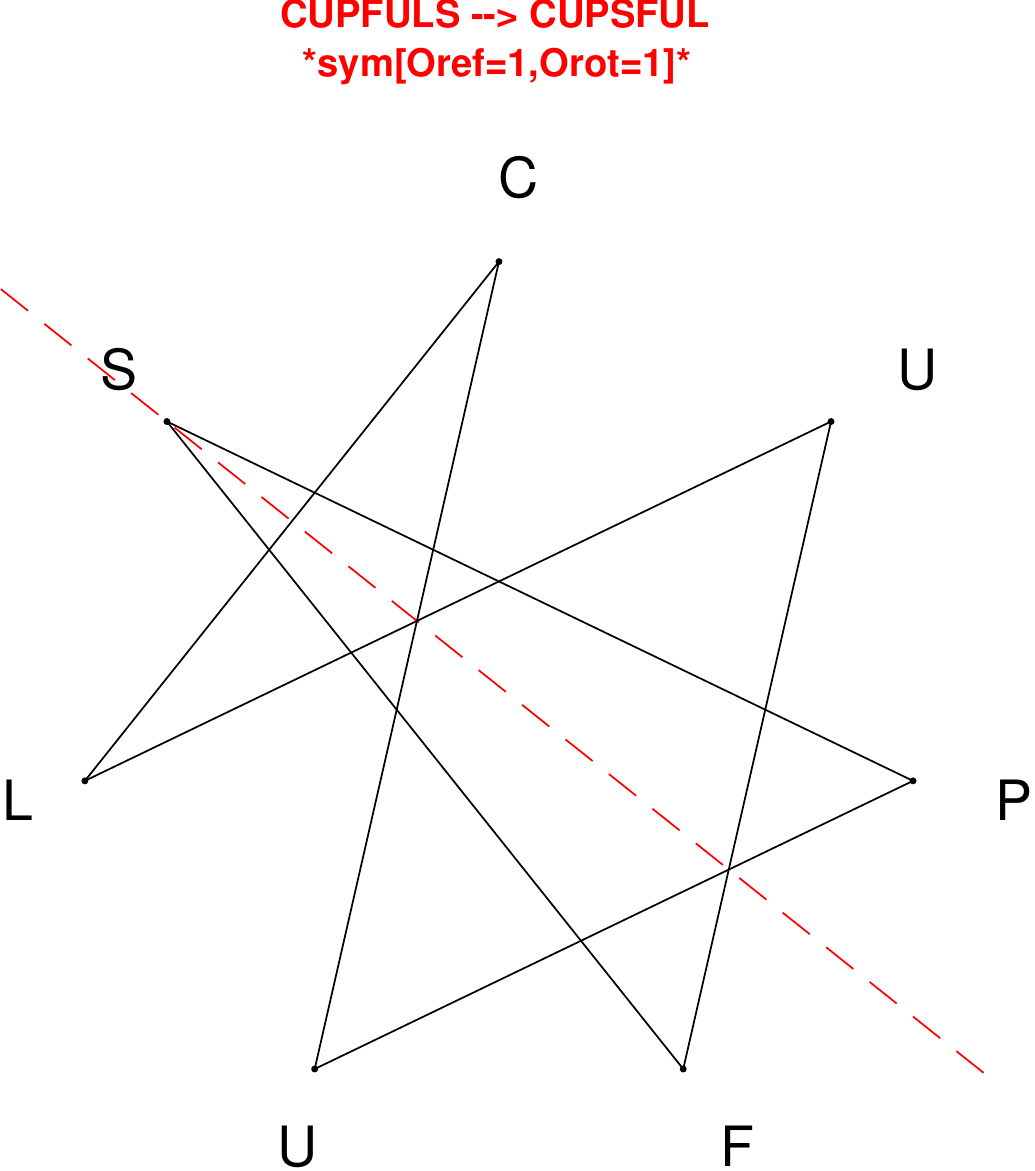}
\end{subfigure}
\hfill
\begin{subfigure}[T]{0.19\textwidth}
\centering
\includegraphics[width=\textwidth]{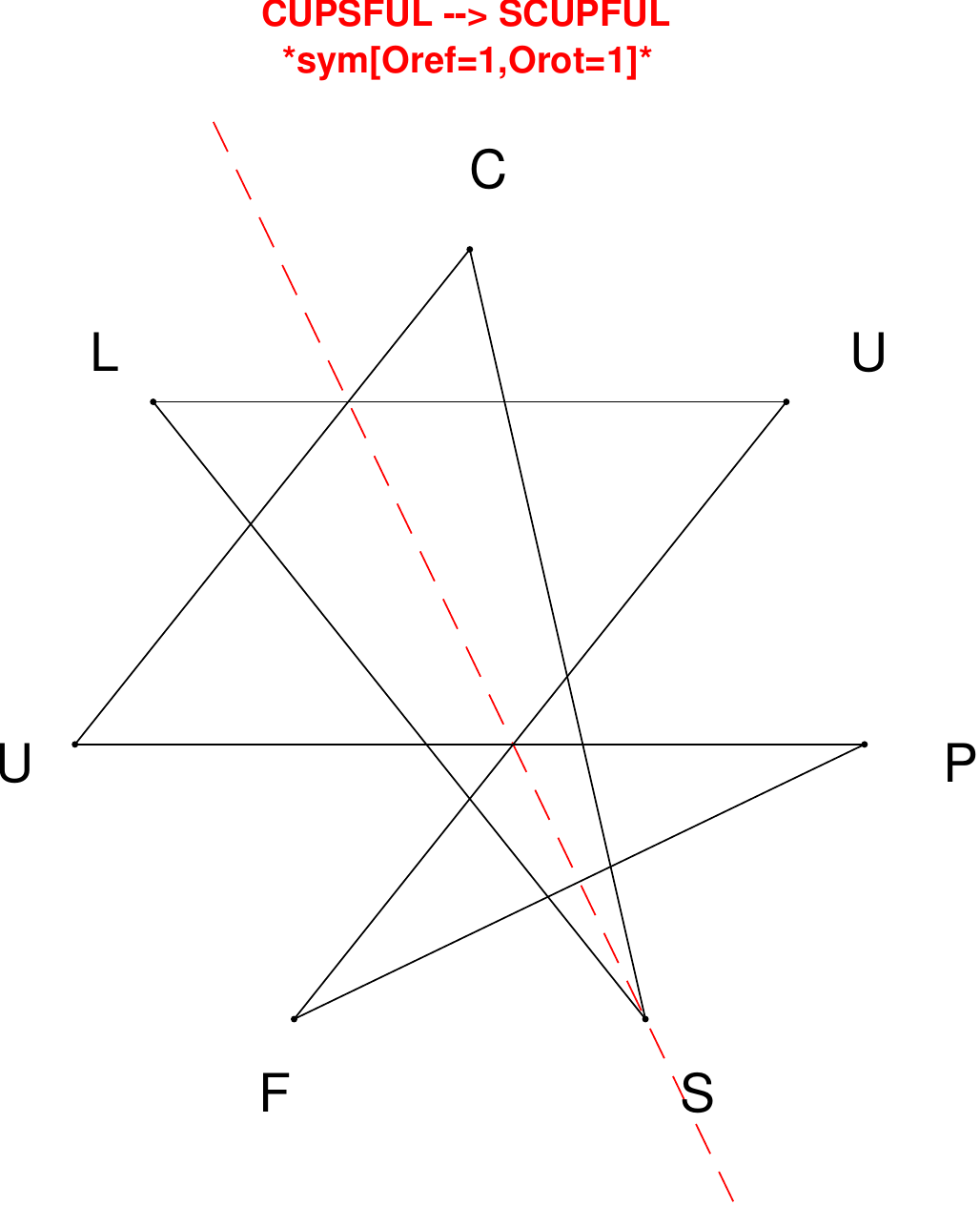}
\end{subfigure}
\hfill
\begin{subfigure}[T]{0.19\textwidth}
\centering
\includegraphics[width=\textwidth]{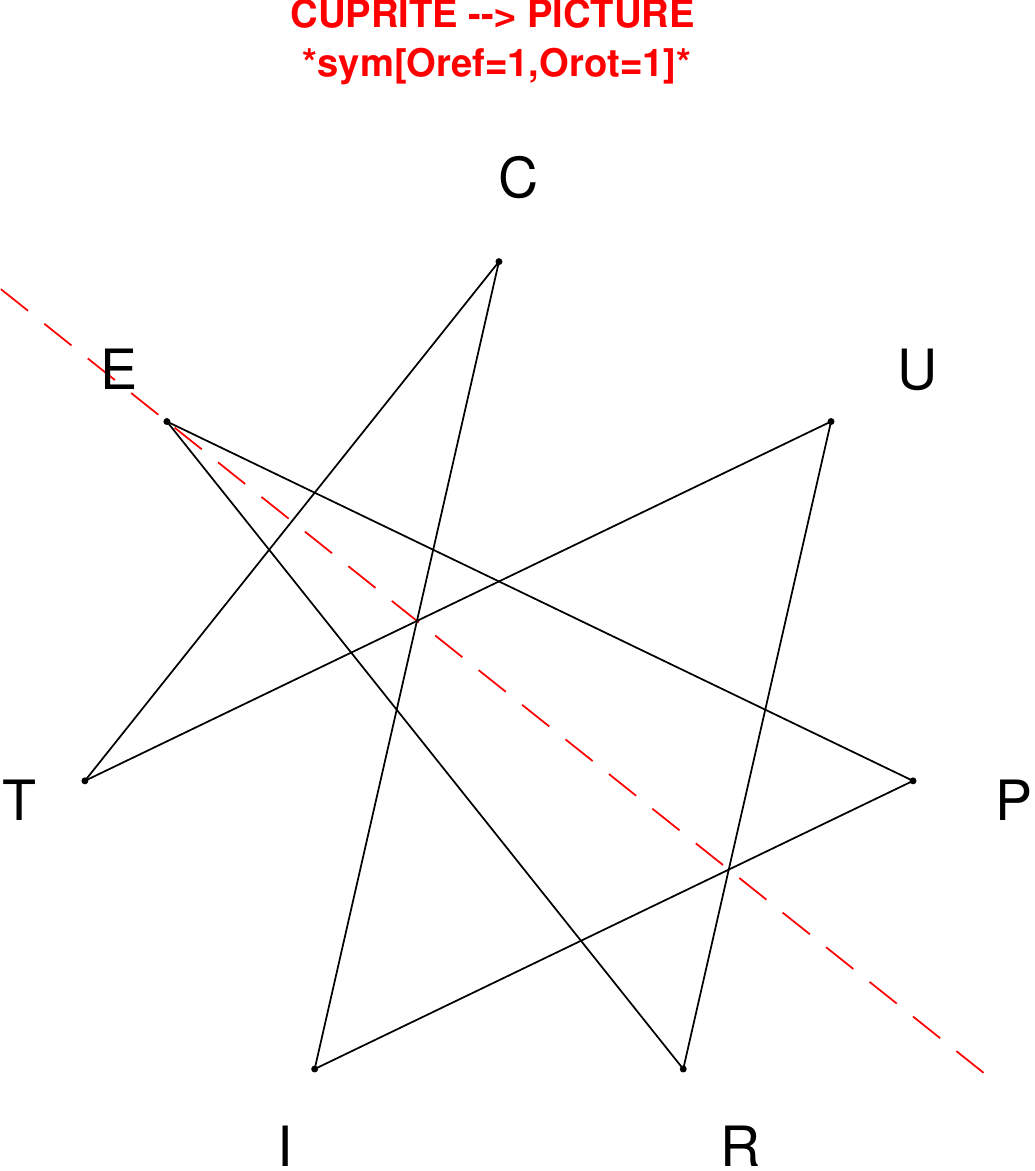}
\end{subfigure}
\hfill
\begin{subfigure}[T]{0.19\textwidth}
\centering
\includegraphics[width=\textwidth]{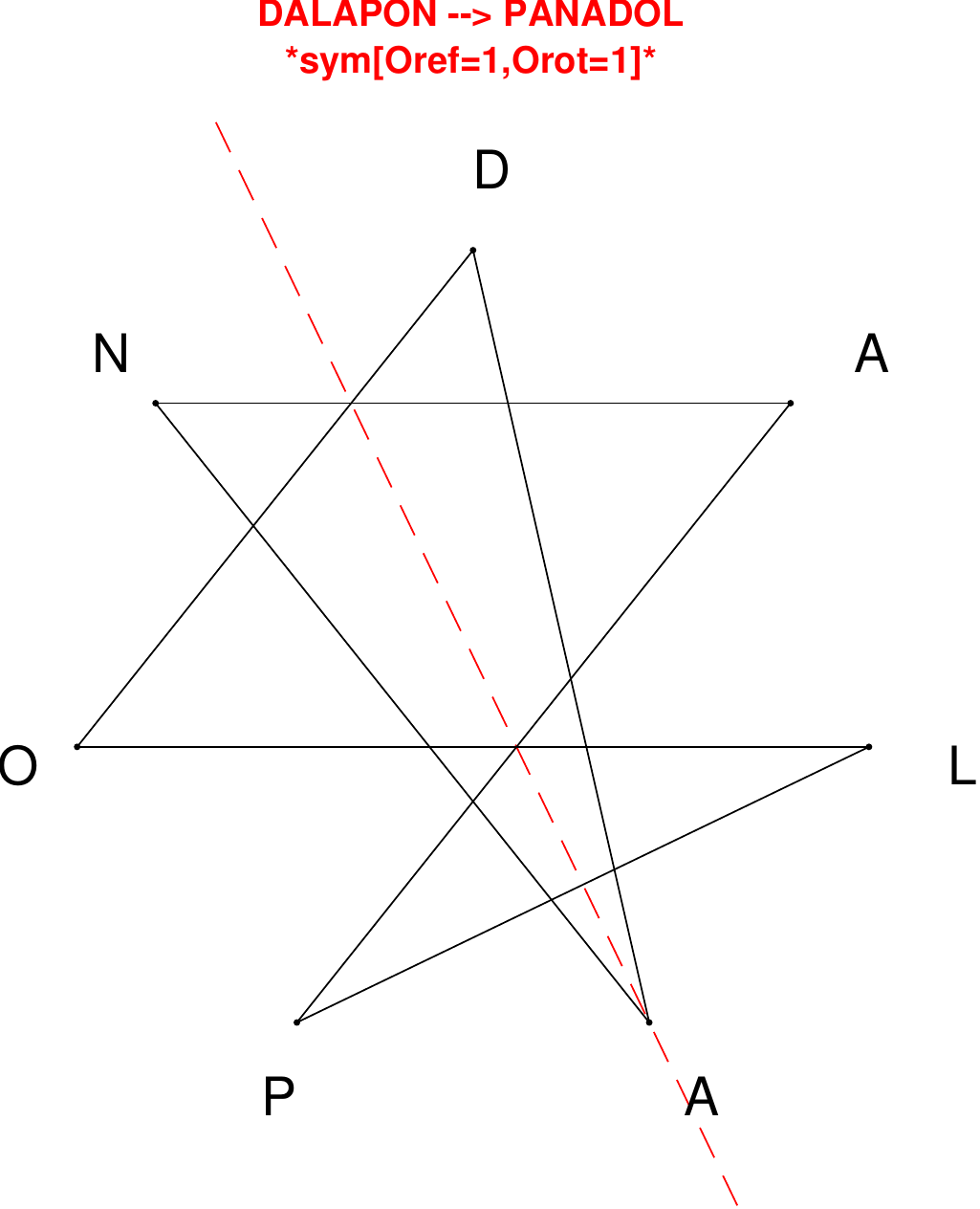}
\end{subfigure}
\hfill
\begin{subfigure}[T]{0.19\textwidth}
\centering
\includegraphics[width=\textwidth]{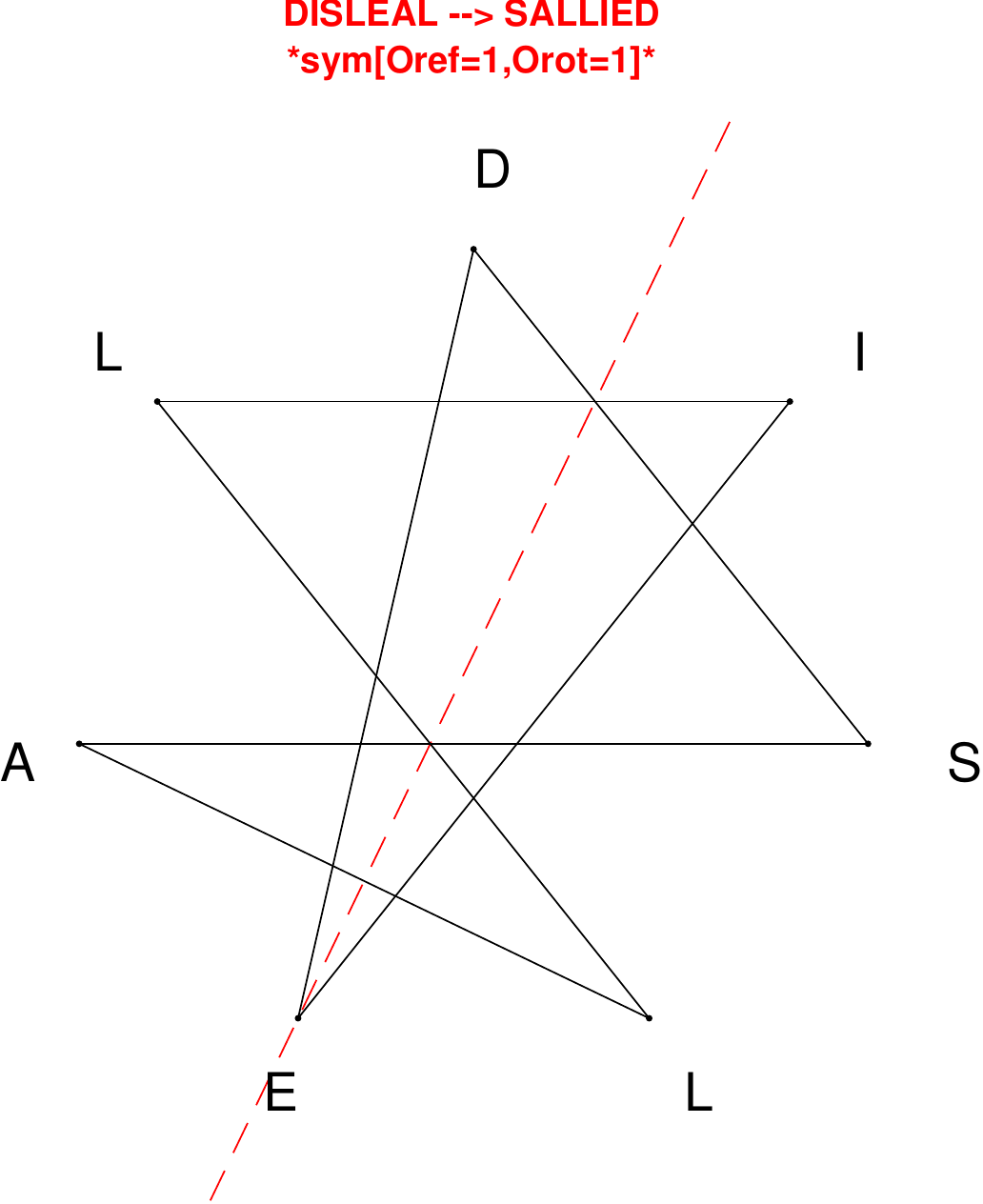}
\end{subfigure}
\end{figure}

\begin{figure}[H]
\centering
\begin{subfigure}[T]{0.19\textwidth}
\centering
\includegraphics[width=\textwidth]{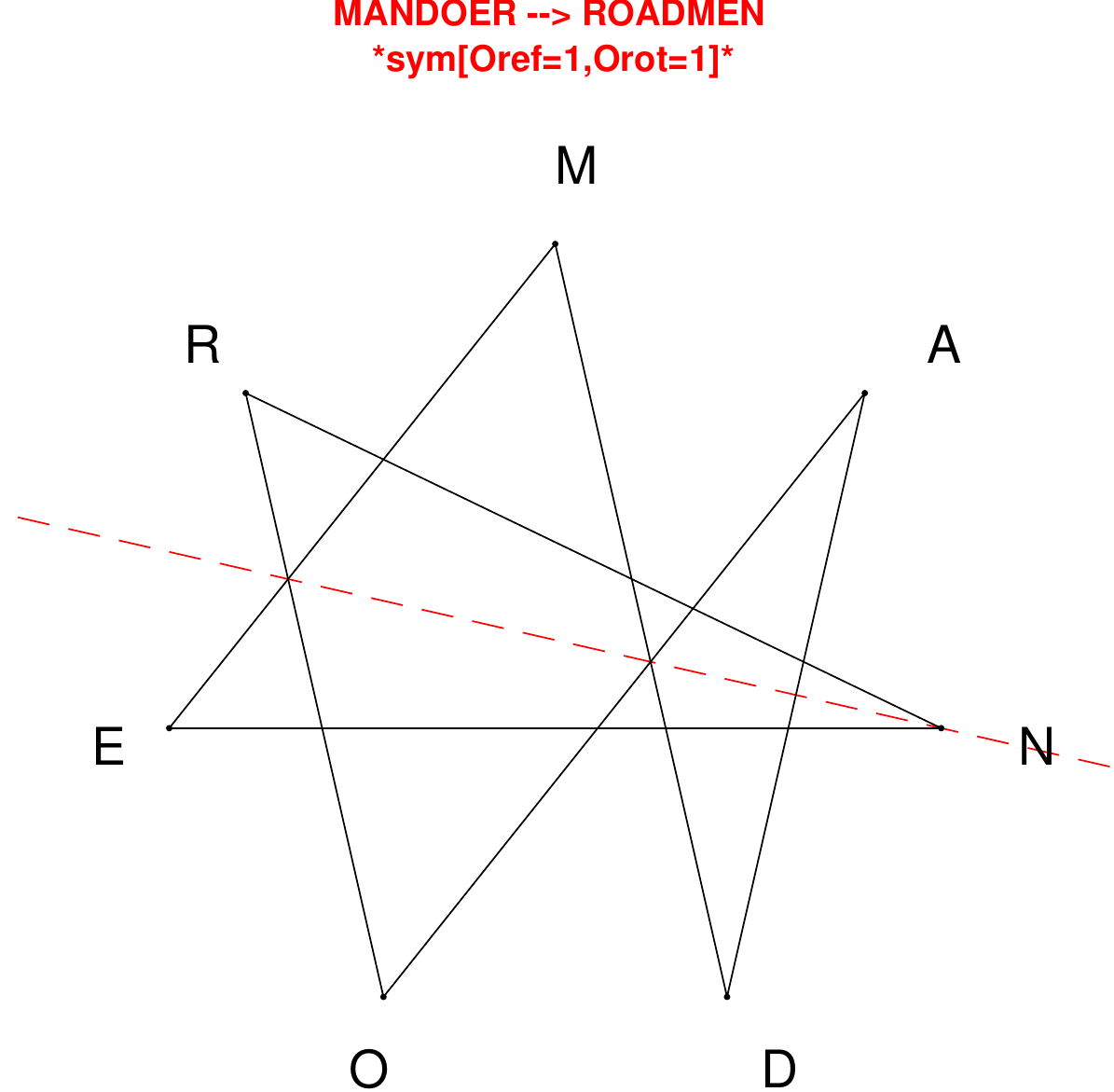}
\end{subfigure}
\hfill
\begin{subfigure}[T]{0.19\textwidth}
\centering
\includegraphics[width=\textwidth]{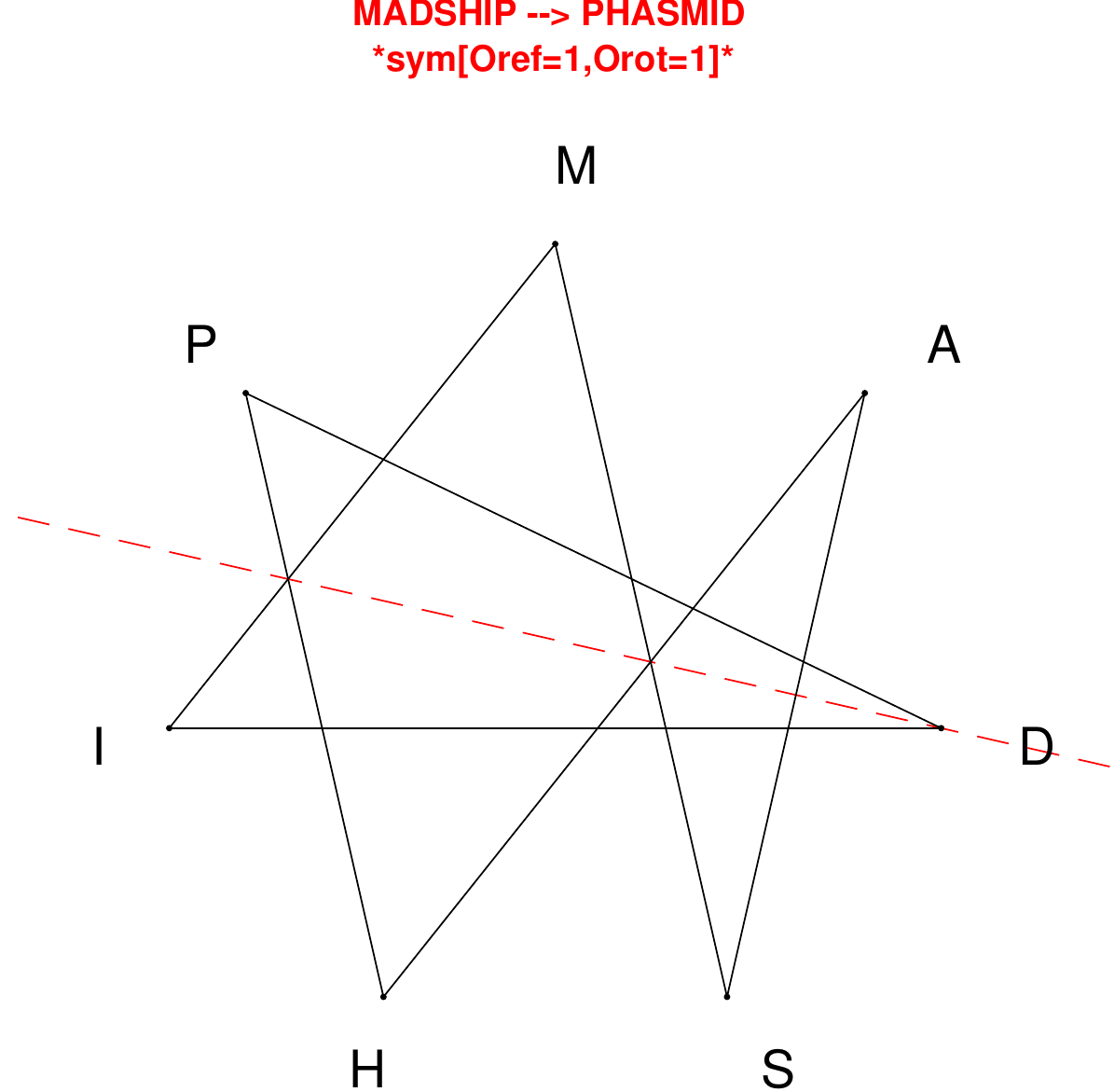}
\end{subfigure}
\hfill
\begin{subfigure}[T]{0.19\textwidth}
\centering
\includegraphics[width=\textwidth]{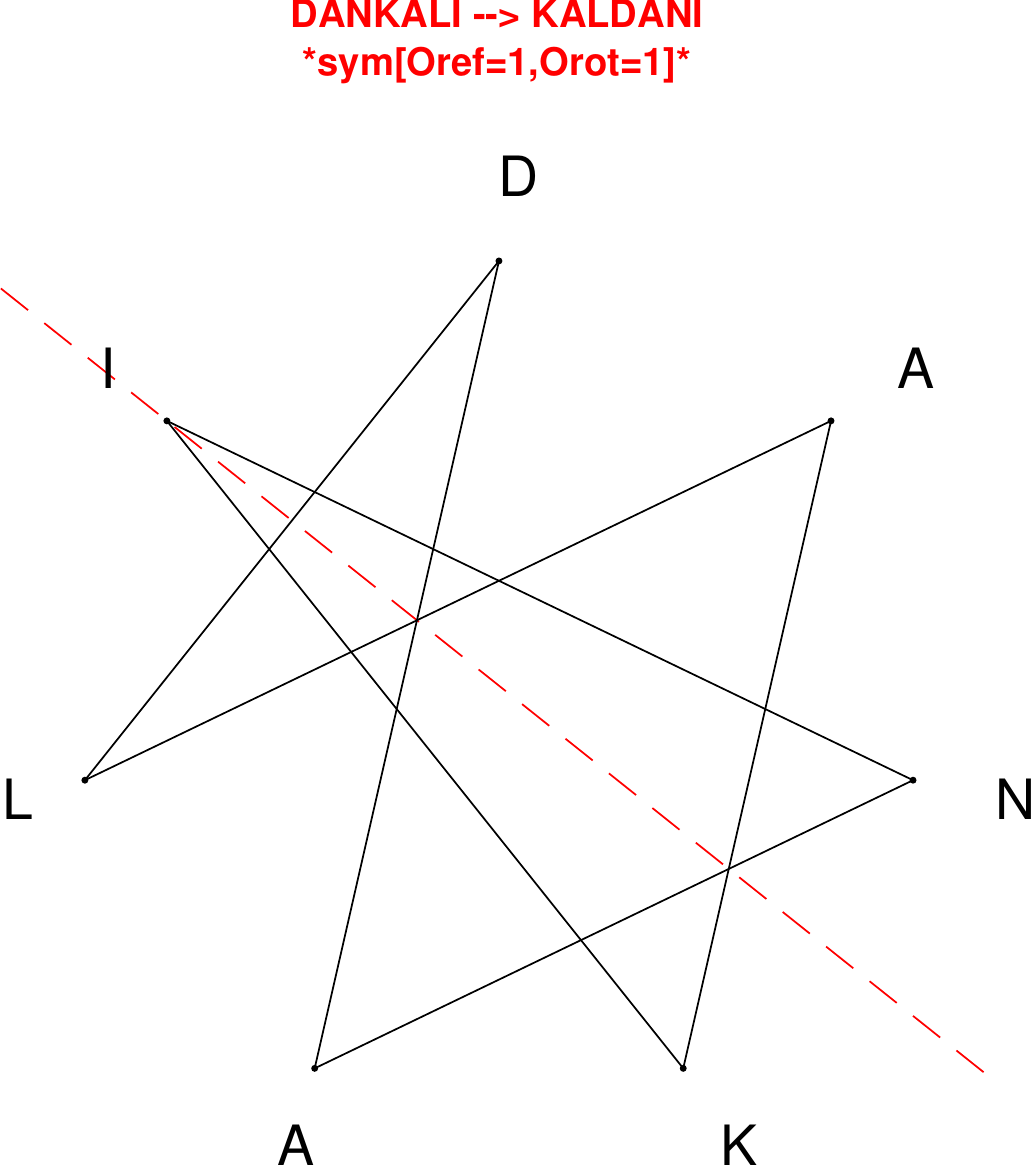}
\end{subfigure}
\hfill
\begin{subfigure}[T]{0.19\textwidth}
\centering
\includegraphics[width=\textwidth]{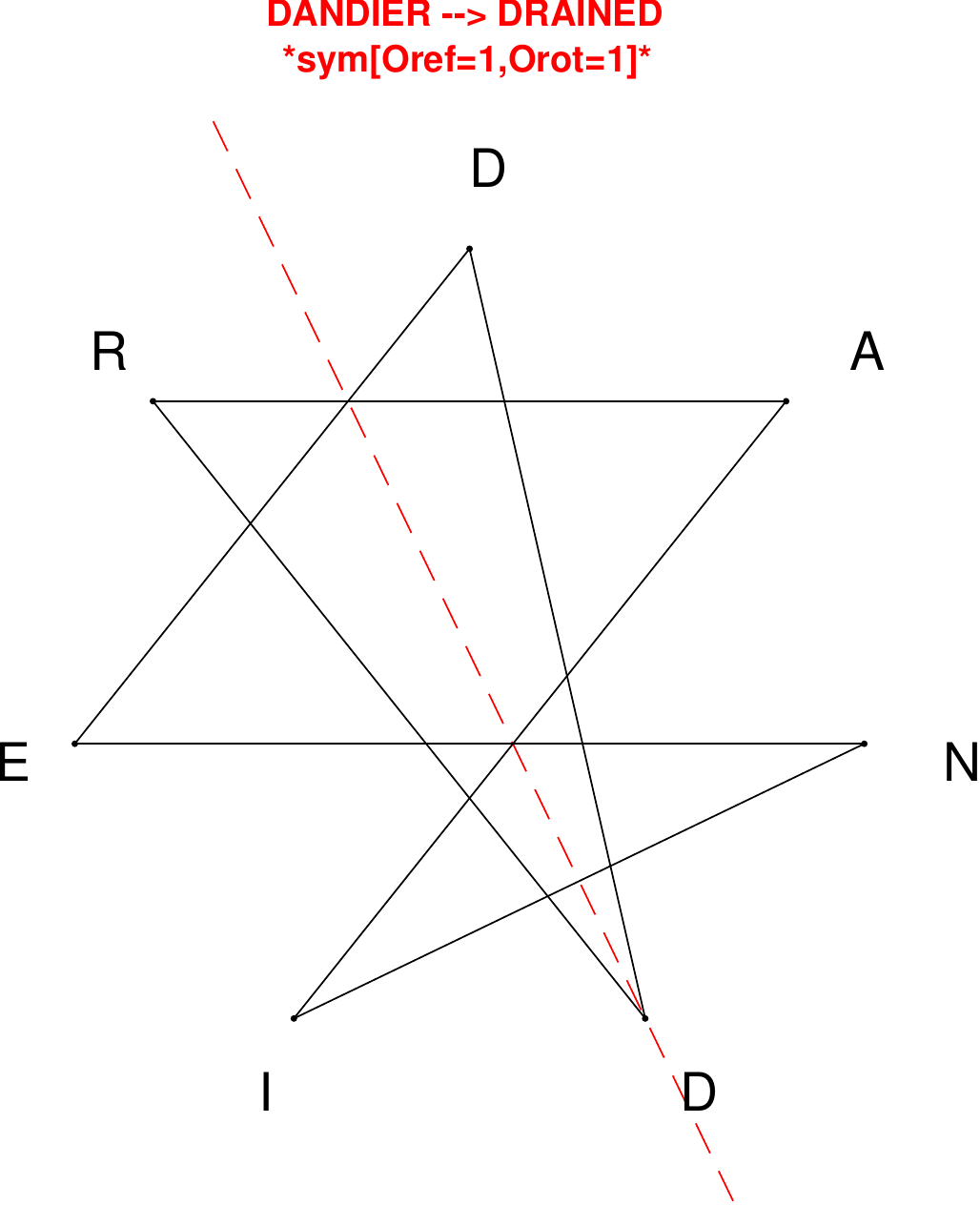}
\end{subfigure}
\hfill
\begin{subfigure}[T]{0.19\textwidth}
\centering
\includegraphics[width=\textwidth]{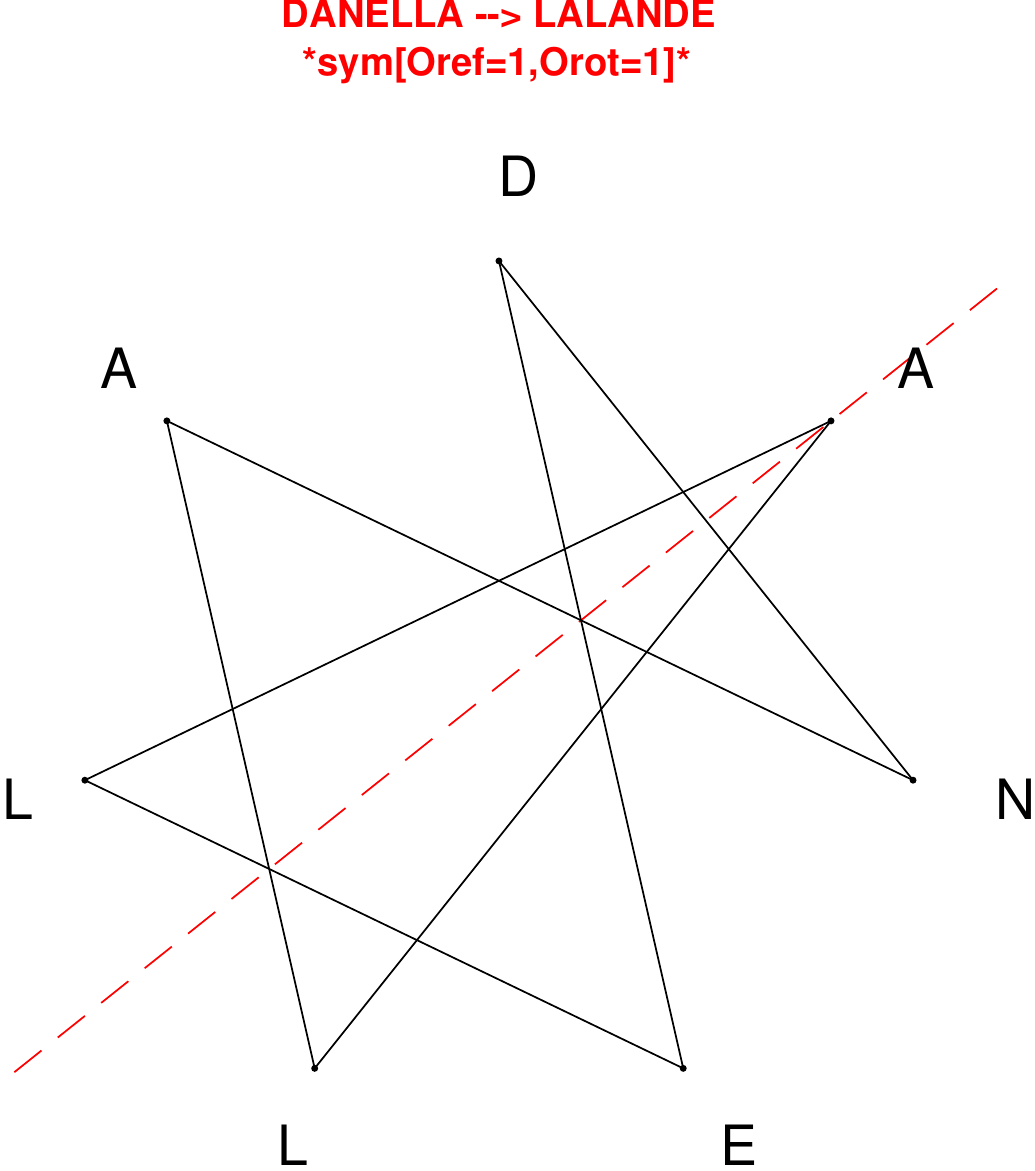}
\end{subfigure}
\end{figure}

\begin{figure}[H]
\centering
\begin{subfigure}[T]{0.19\textwidth}
\centering
\includegraphics[width=\textwidth]{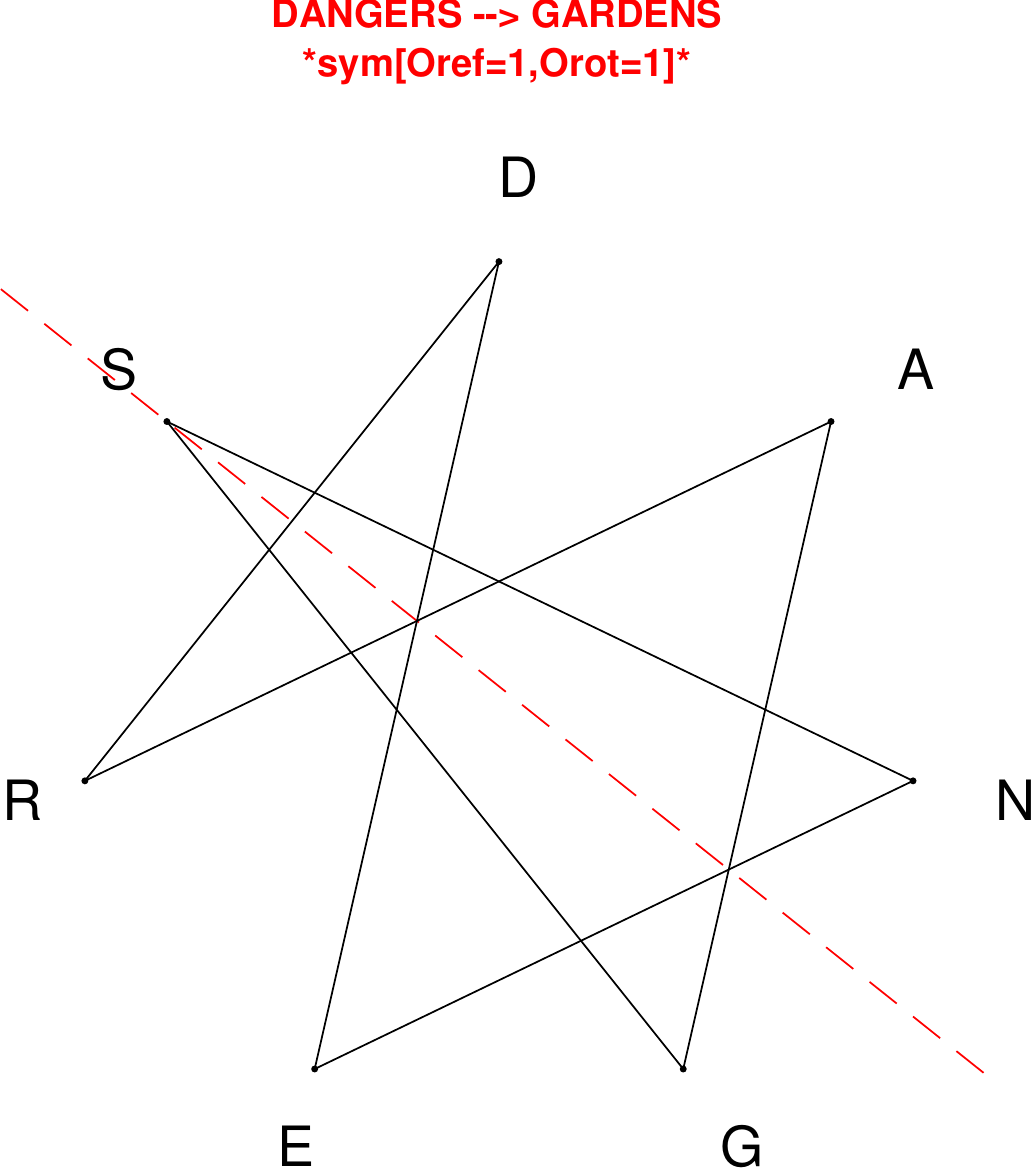}
\end{subfigure}
\hfill
\begin{subfigure}[T]{0.19\textwidth}
\centering
\includegraphics[width=\textwidth]{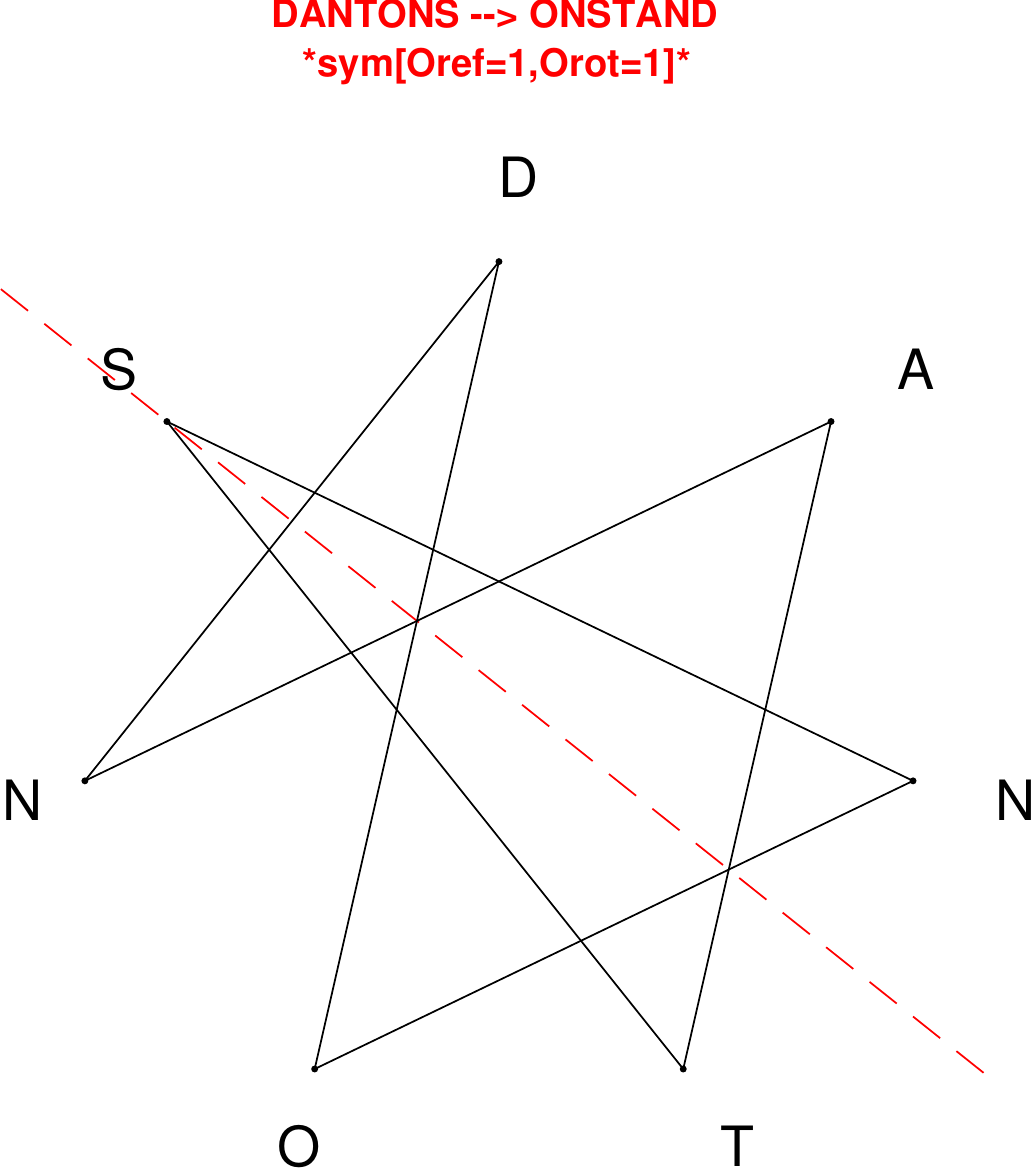}
\end{subfigure}
\hfill
\begin{subfigure}[T]{0.19\textwidth}
\centering
\includegraphics[width=\textwidth]{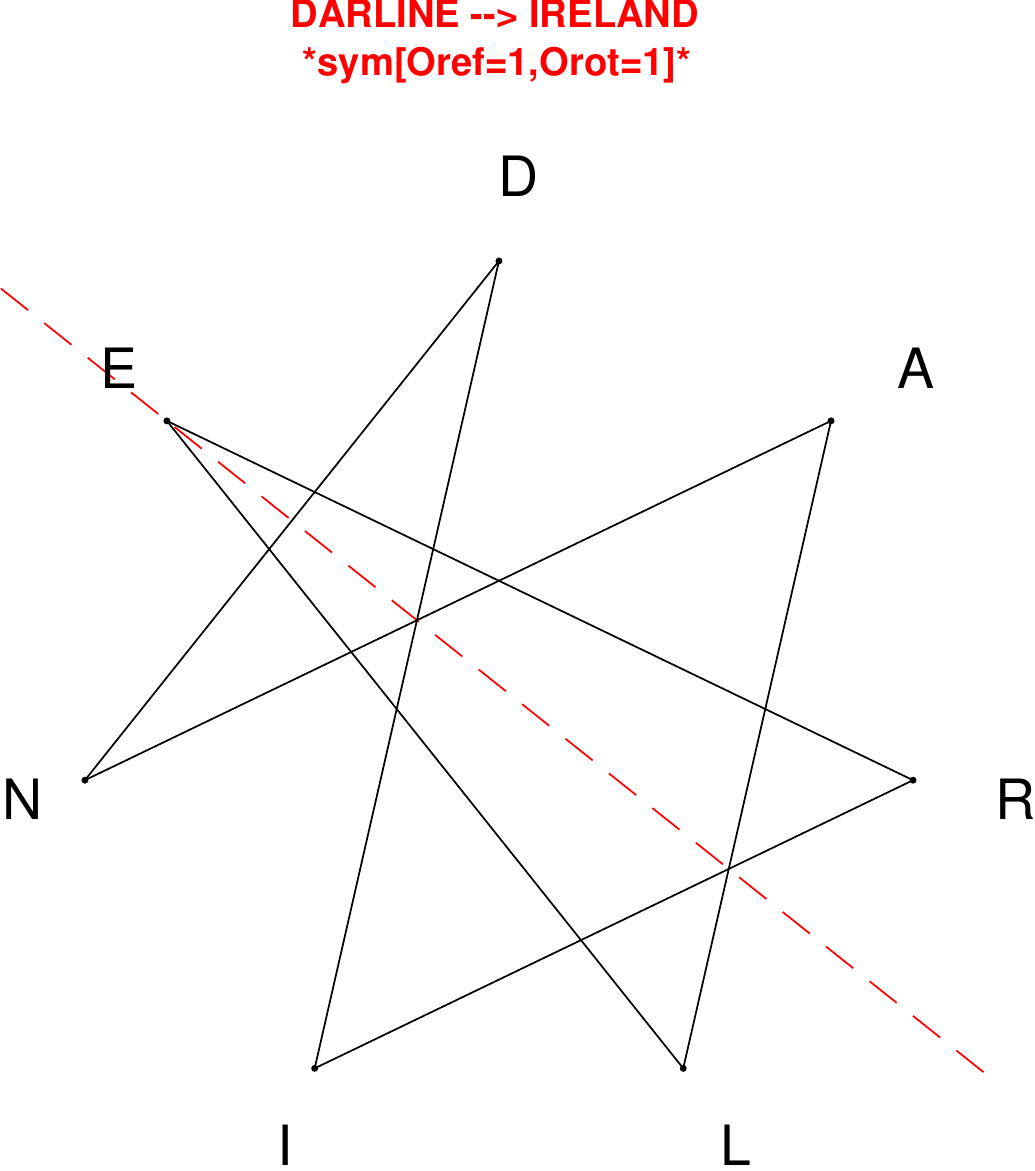}
\end{subfigure}
\hfill
\begin{subfigure}[T]{0.19\textwidth}
\centering
\includegraphics[width=\textwidth]{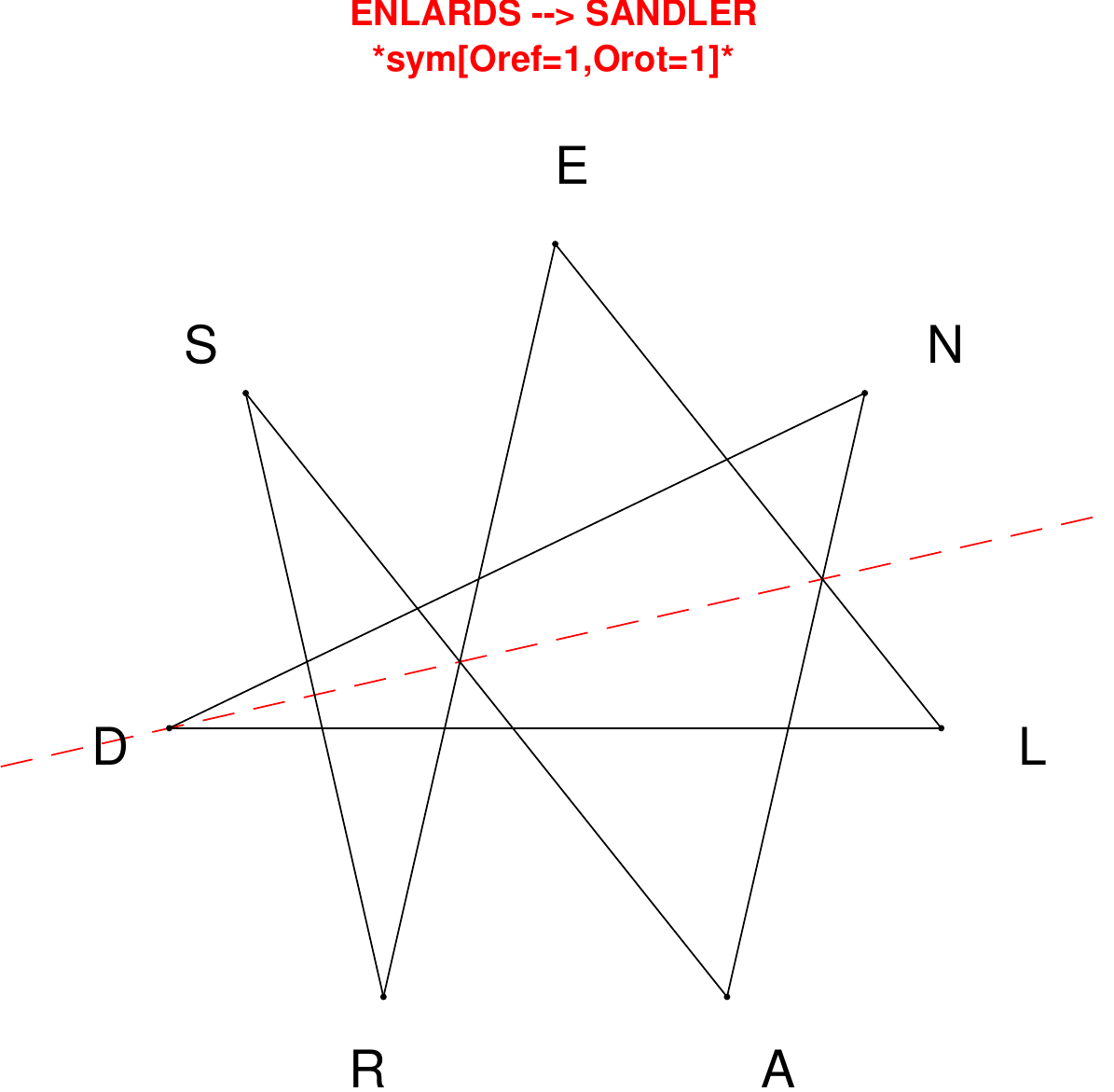}
\end{subfigure}
\hfill
\begin{subfigure}[T]{0.19\textwidth}
\centering
\includegraphics[width=\textwidth]{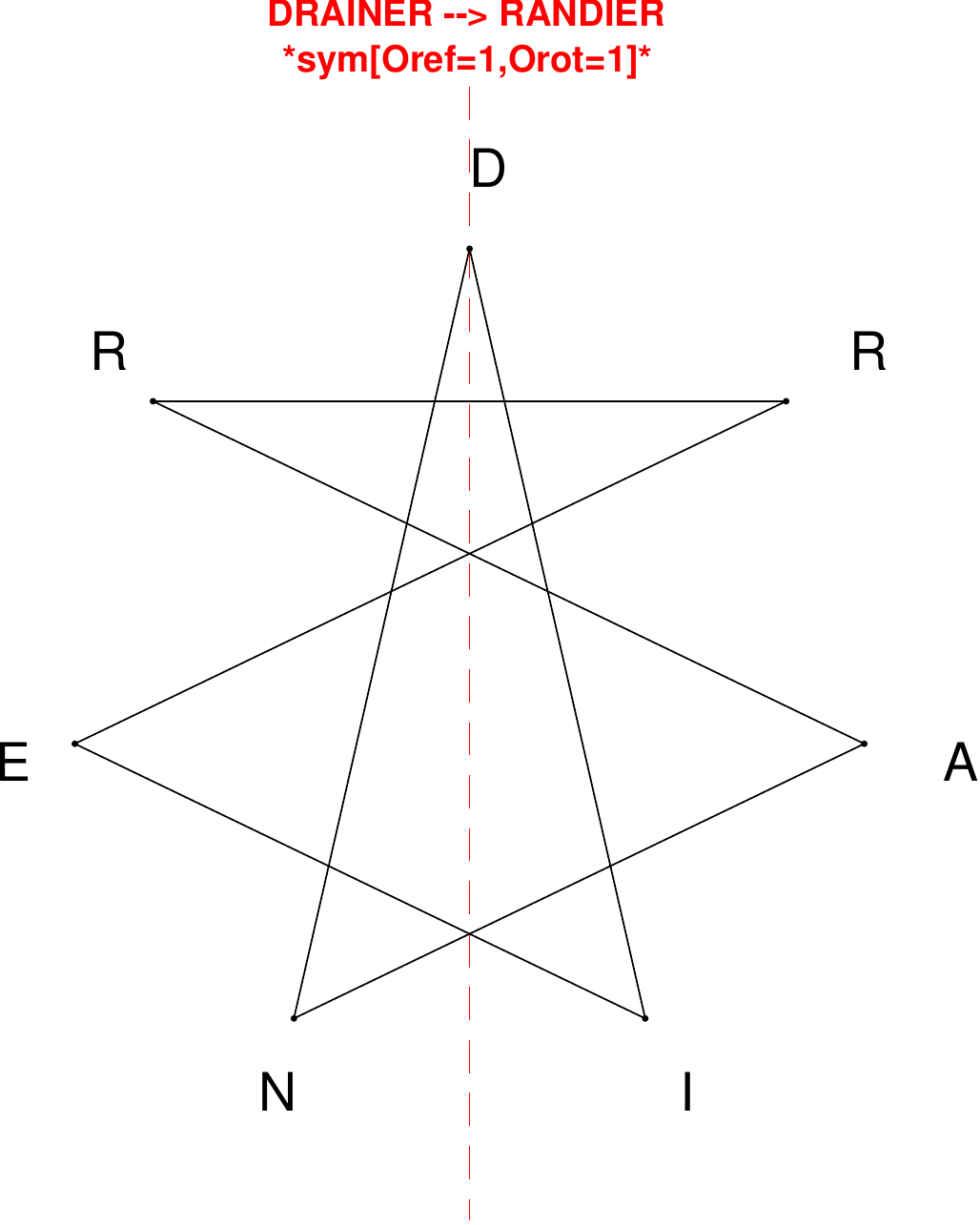}
\end{subfigure}
\end{figure}

\begin{figure}[H]
\centering
\begin{subfigure}[T]{0.19\textwidth}
\centering
\includegraphics[width=\textwidth]{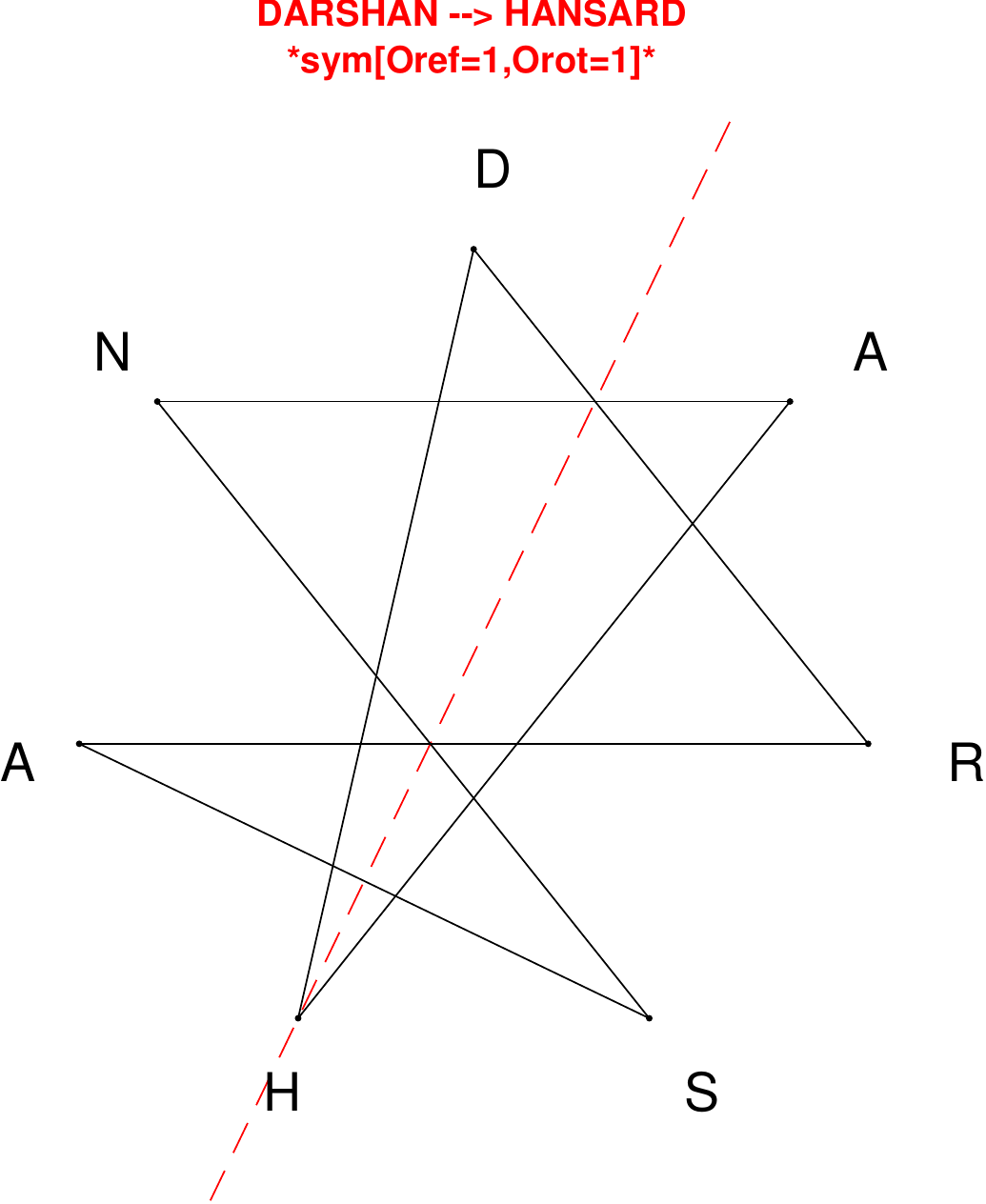}
\end{subfigure}
\hfill
\begin{subfigure}[T]{0.19\textwidth}
\centering
\includegraphics[width=\textwidth]{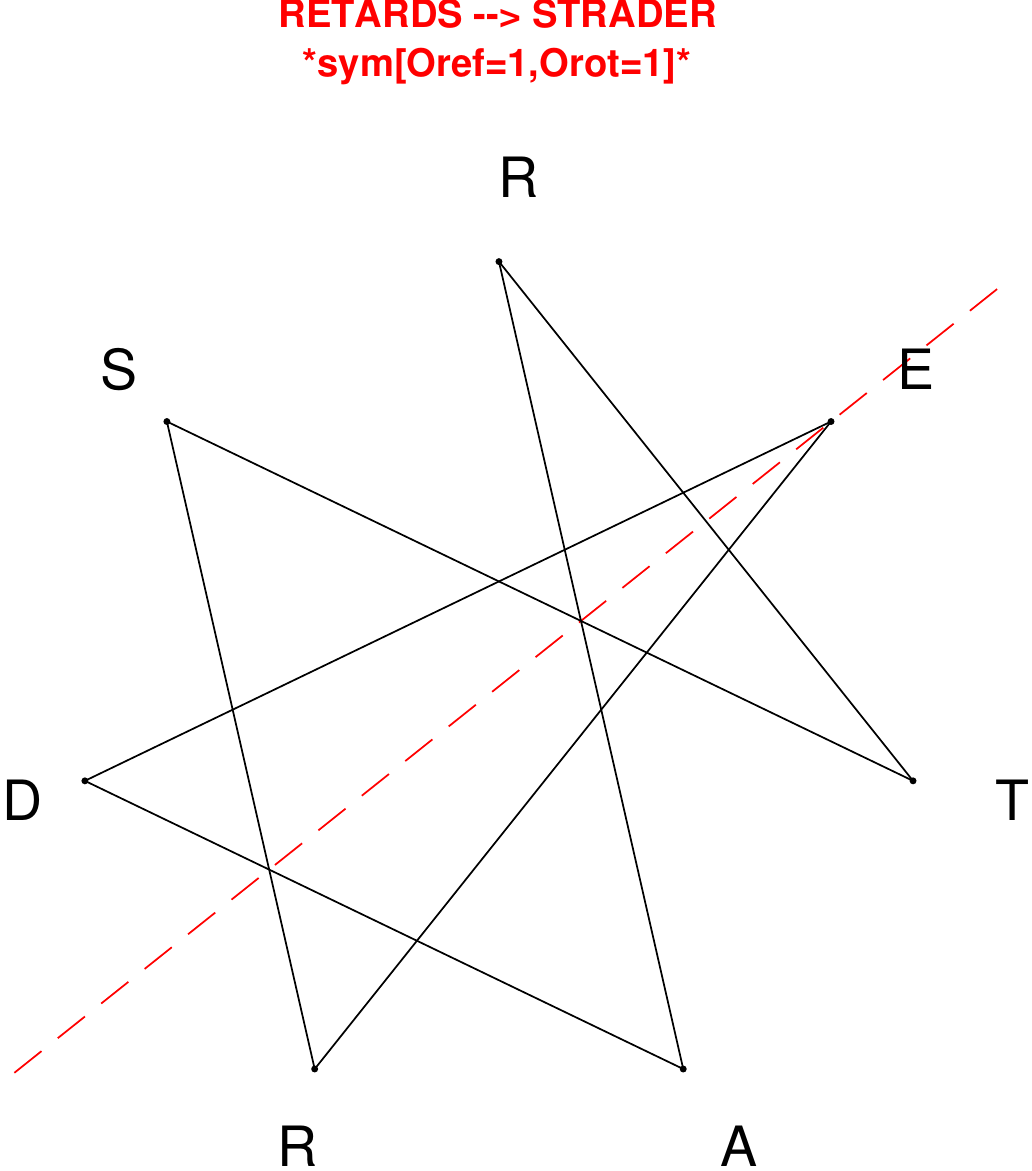}
\end{subfigure}
\hfill
\begin{subfigure}[T]{0.19\textwidth}
\centering
\includegraphics[width=\textwidth]{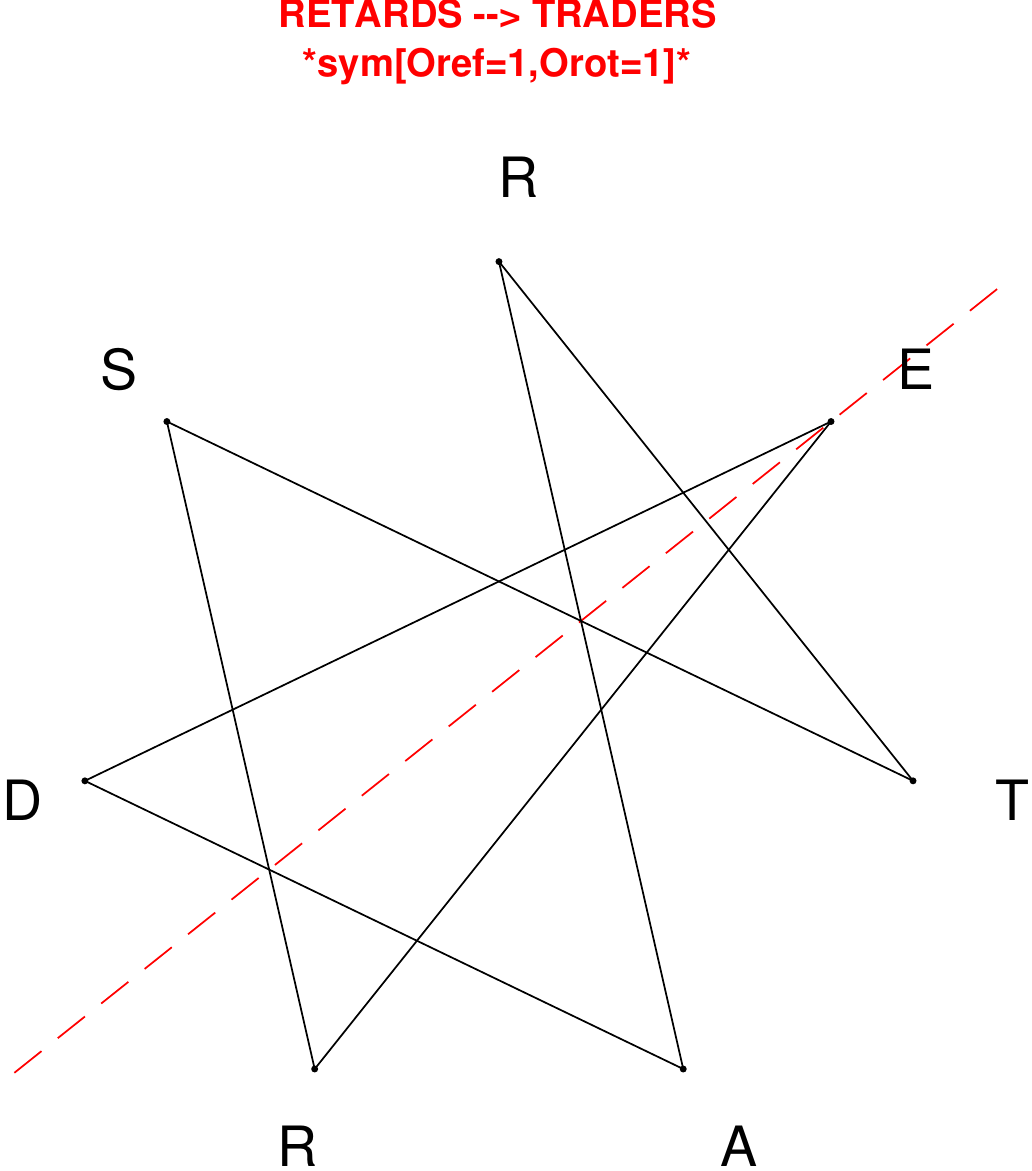}
\end{subfigure}
\hfill
\begin{subfigure}[T]{0.19\textwidth}
\centering
\includegraphics[width=\textwidth]{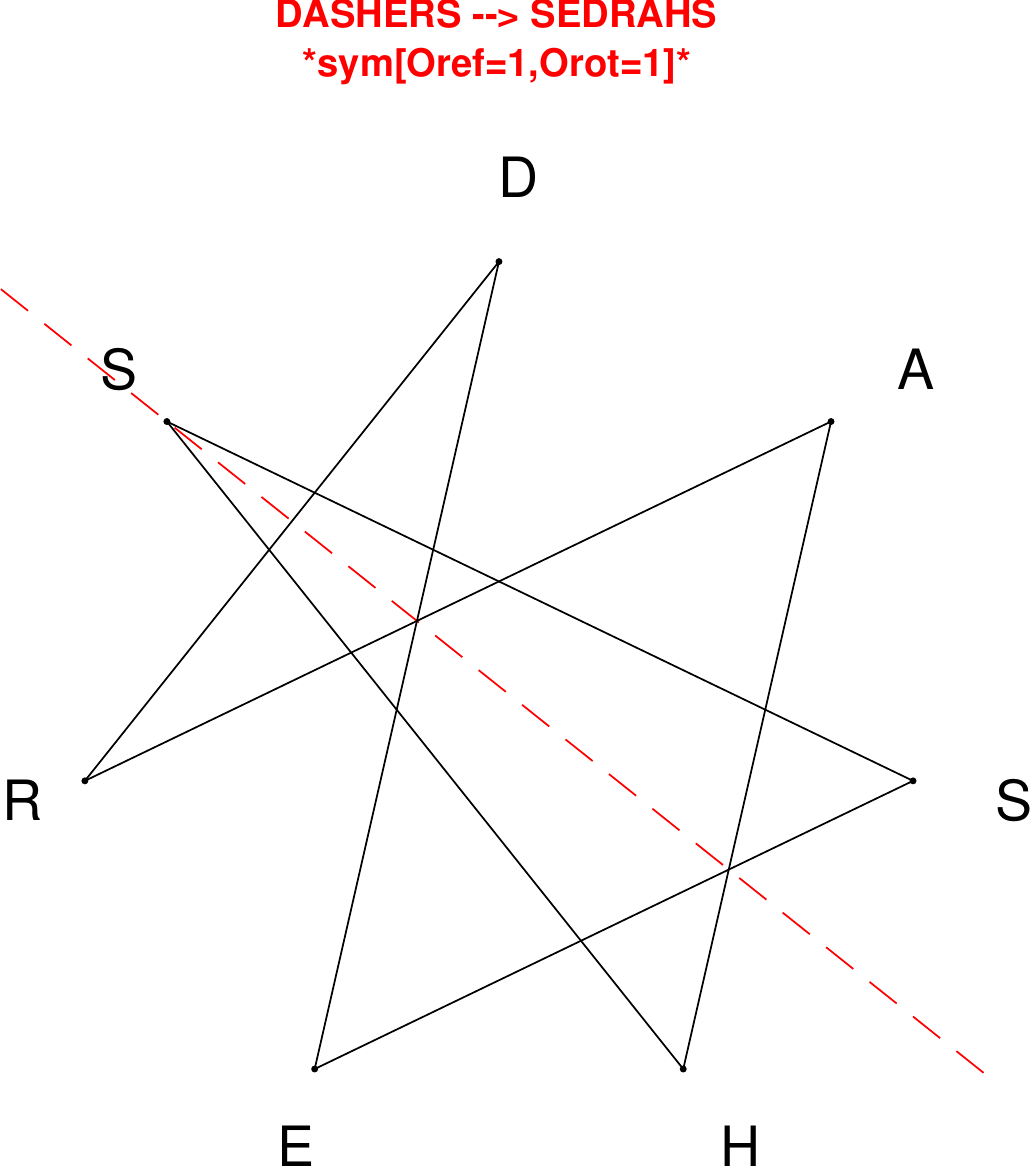}
\end{subfigure}
\hfill
\begin{subfigure}[T]{0.19\textwidth}
\centering
\includegraphics[width=\textwidth]{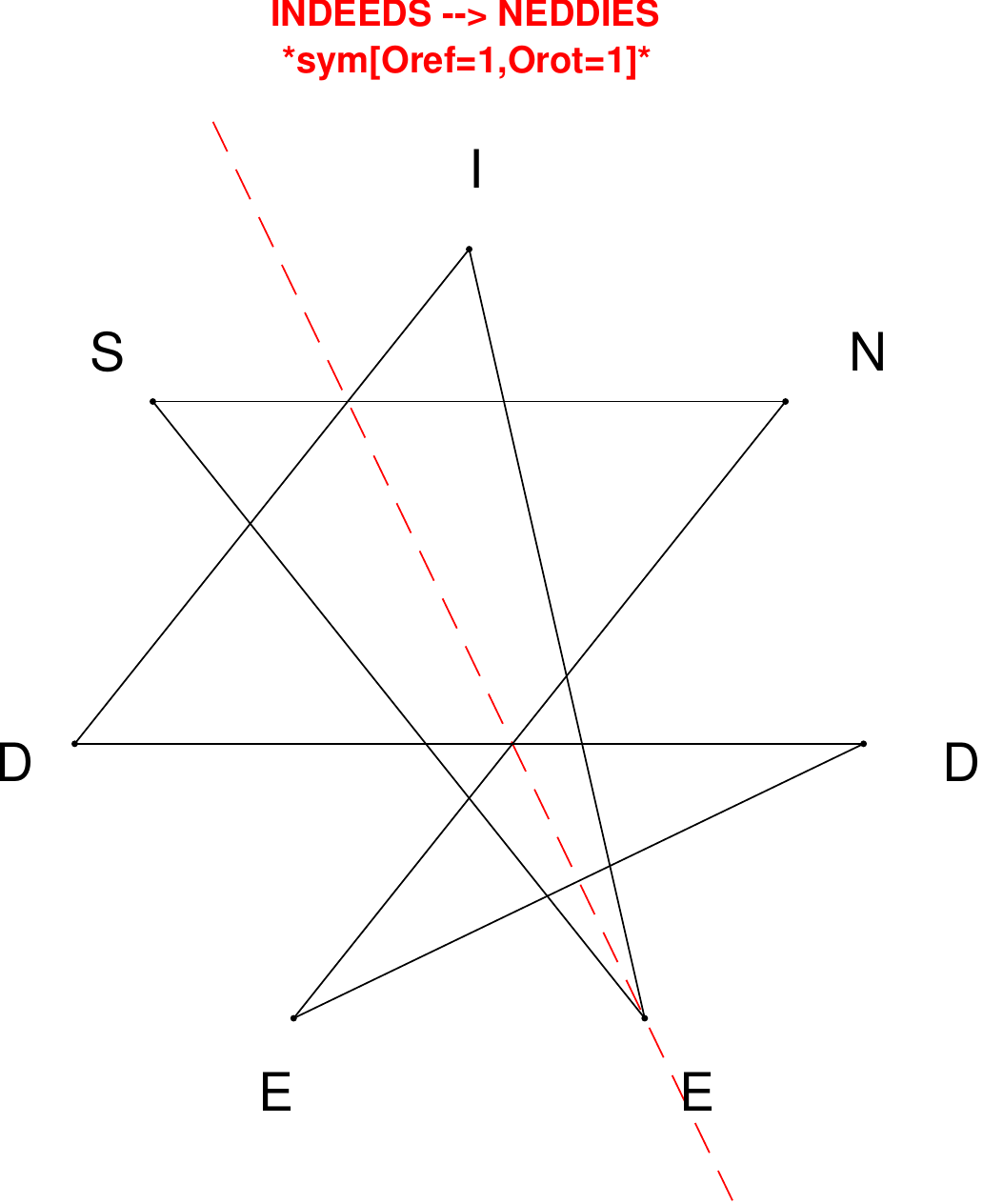}
\end{subfigure}
\end{figure}

\begin{figure}[H]
\centering
\begin{subfigure}[T]{0.19\textwidth}
\centering
\includegraphics[width=\textwidth]{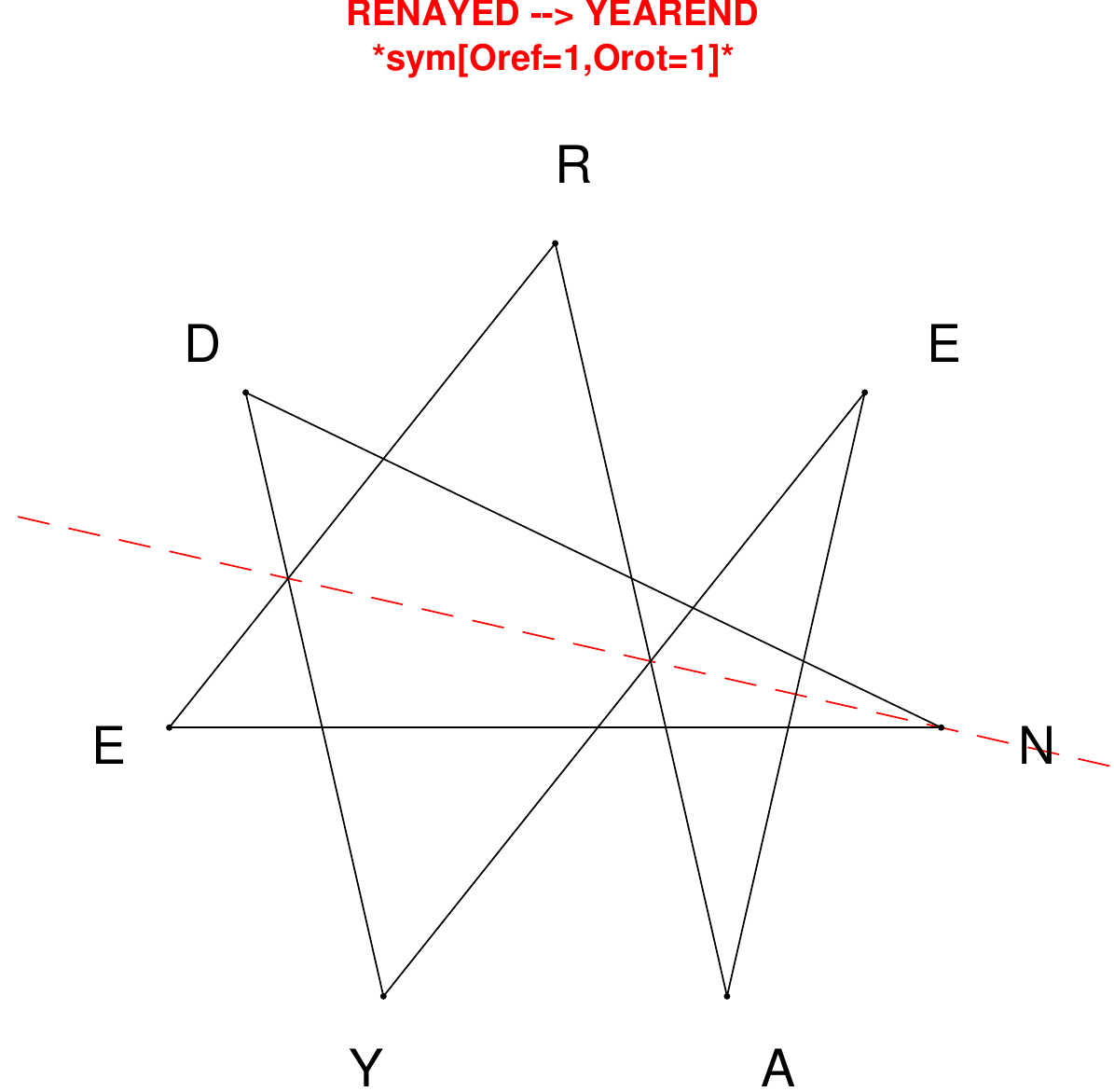}
\end{subfigure}
\hfill
\begin{subfigure}[T]{0.19\textwidth}
\centering
\includegraphics[width=\textwidth]{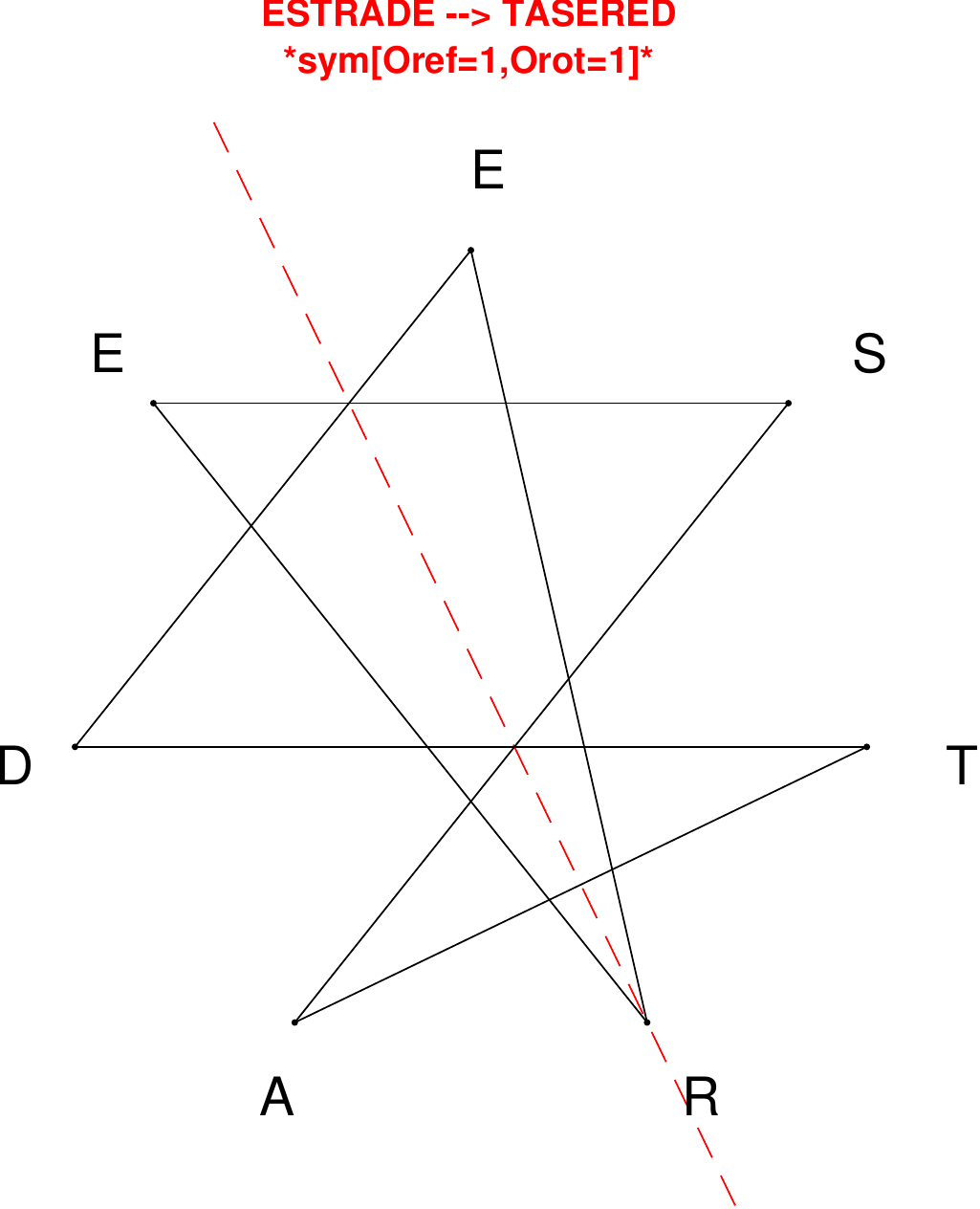}
\end{subfigure}
\hfill
\begin{subfigure}[T]{0.19\textwidth}
\centering
\includegraphics[width=\textwidth]{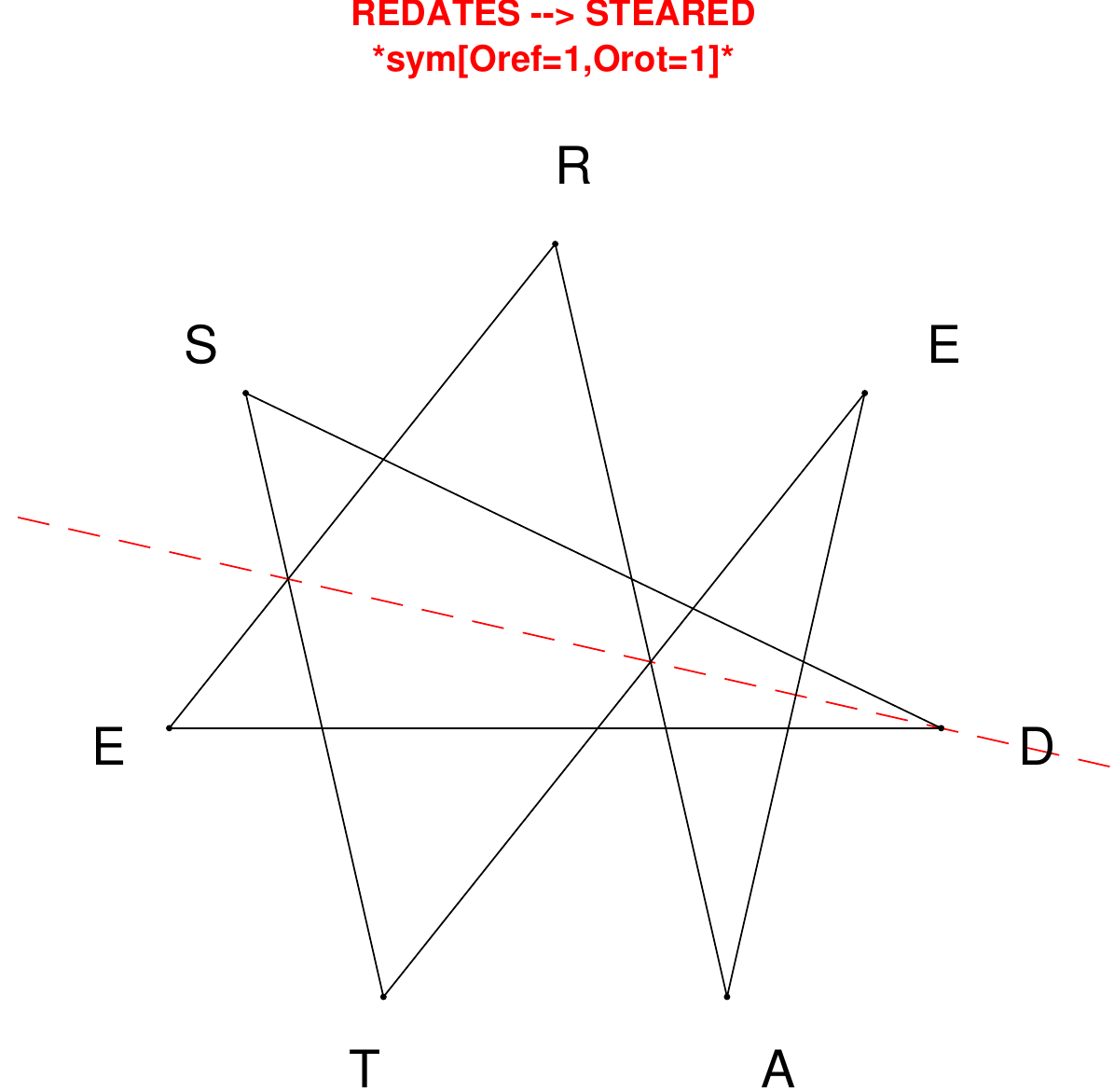}
\end{subfigure}
\hfill
\begin{subfigure}[T]{0.19\textwidth}
\centering
\includegraphics[width=\textwidth]{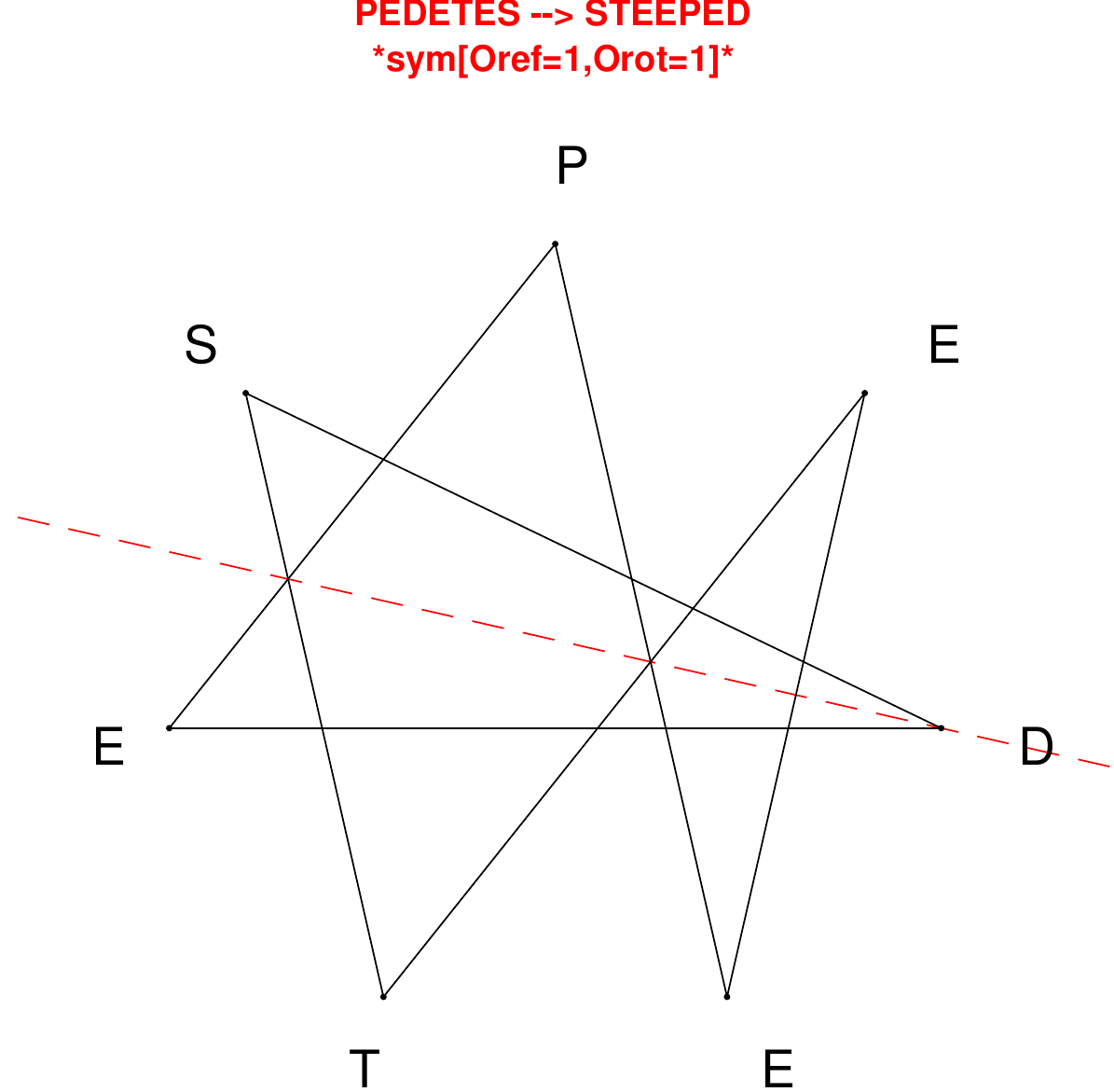}
\end{subfigure}
\hfill
\begin{subfigure}[T]{0.19\textwidth}
\centering
\includegraphics[width=\textwidth]{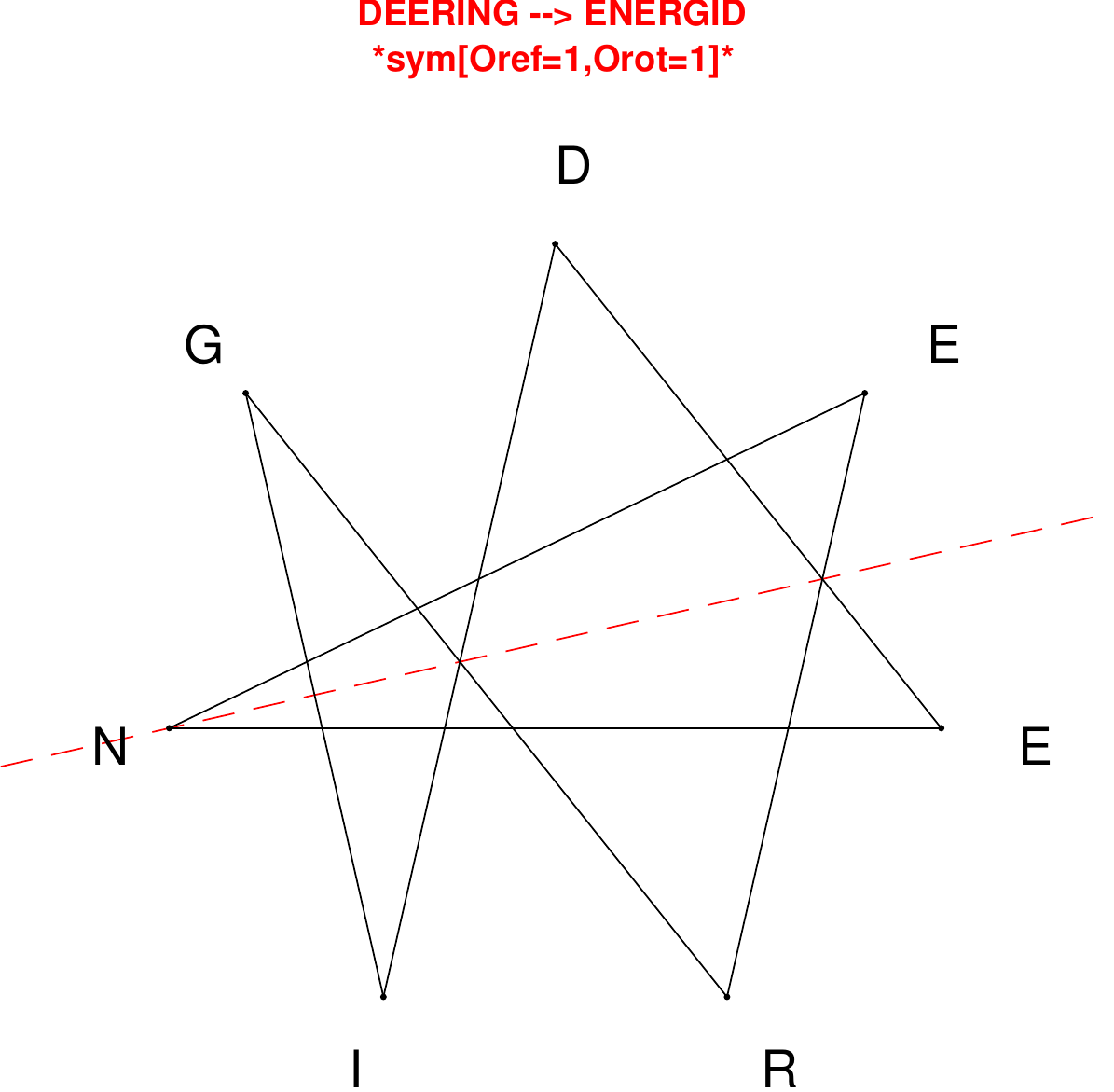}
\end{subfigure}
\end{figure}

\begin{figure}[H]
\centering
\begin{subfigure}[T]{0.19\textwidth}
\centering
\includegraphics[width=\textwidth]{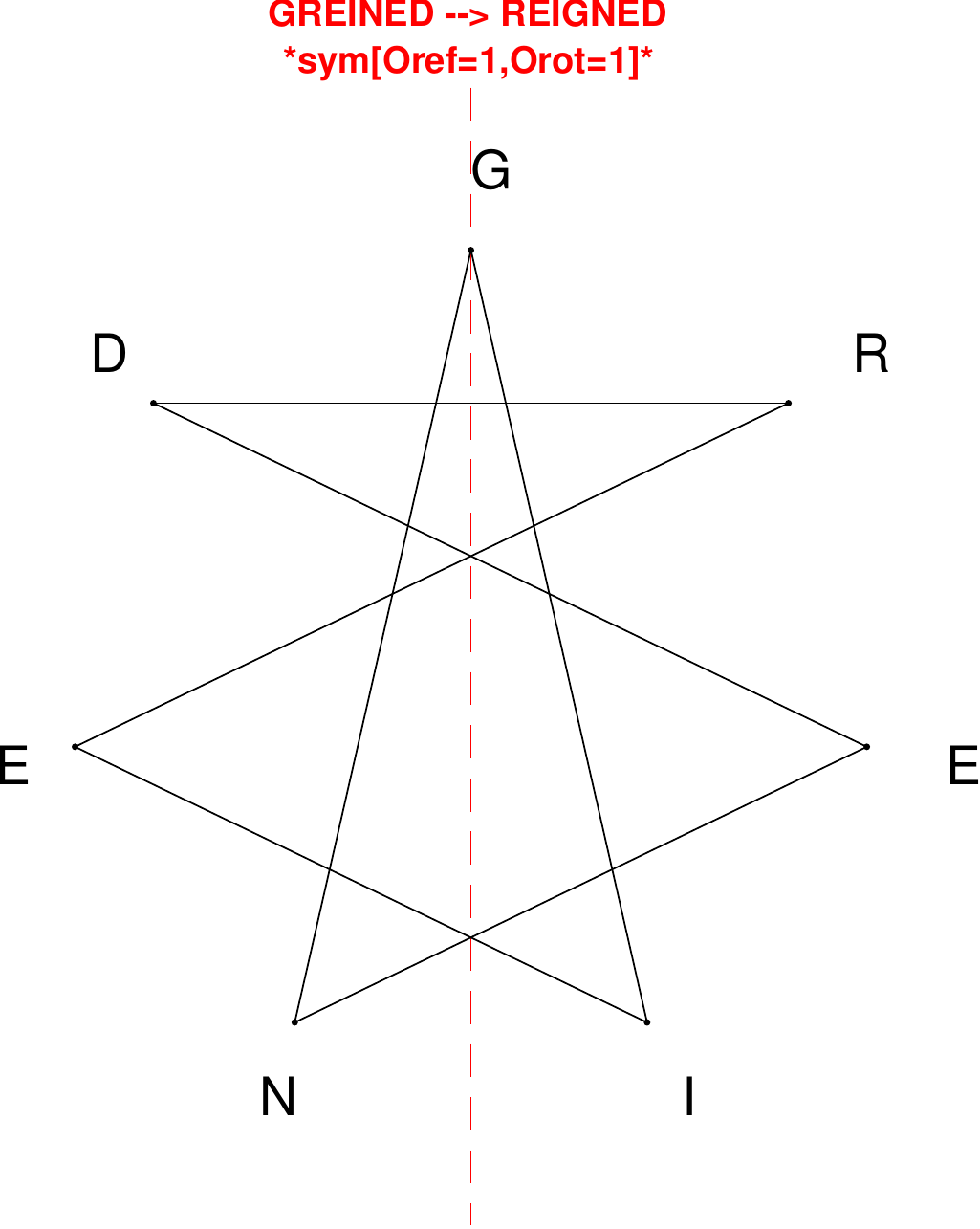}
\end{subfigure}
\hfill
\begin{subfigure}[T]{0.19\textwidth}
\centering
\includegraphics[width=\textwidth]{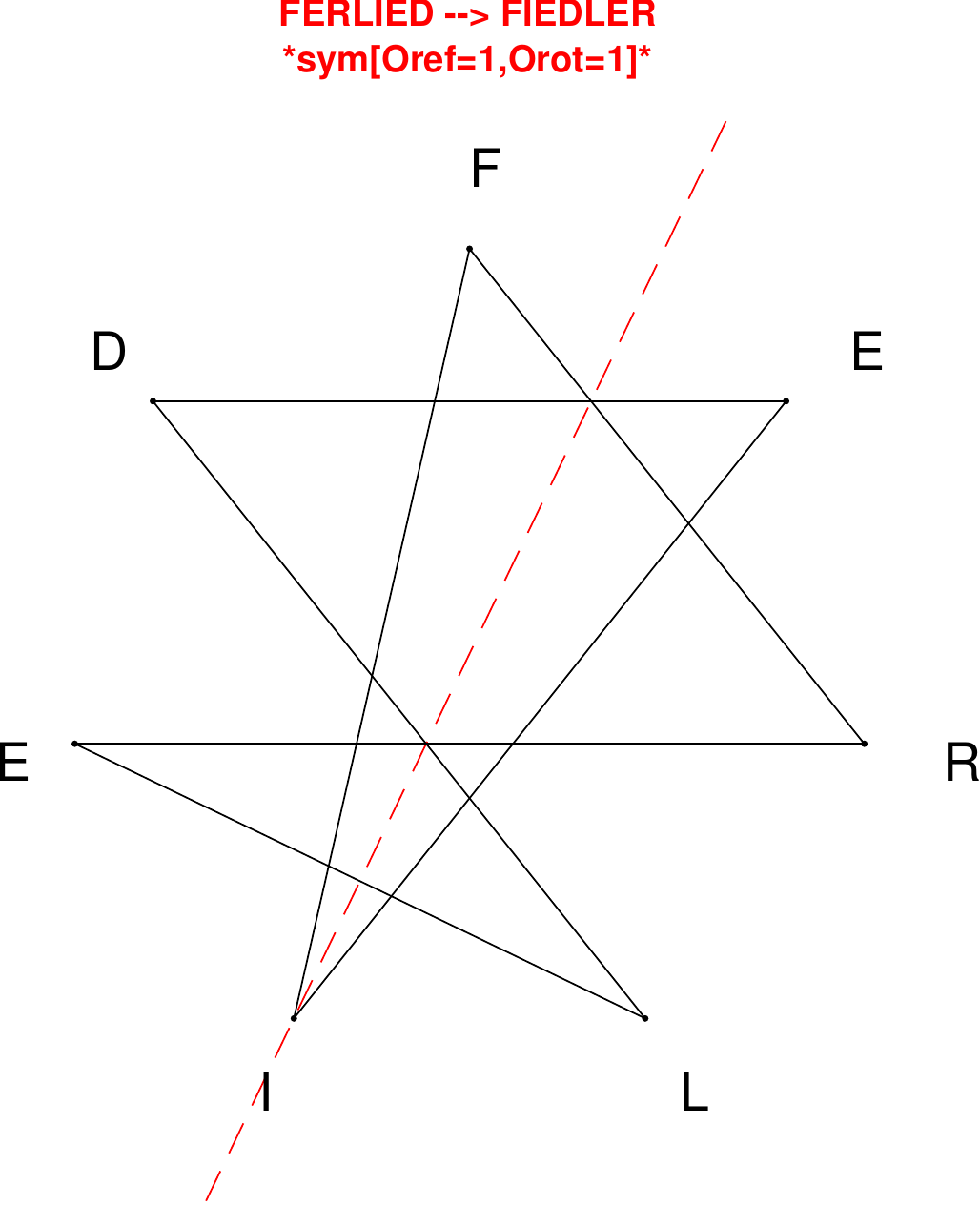}
\end{subfigure}
\hfill
\begin{subfigure}[T]{0.19\textwidth}
\centering
\includegraphics[width=\textwidth]{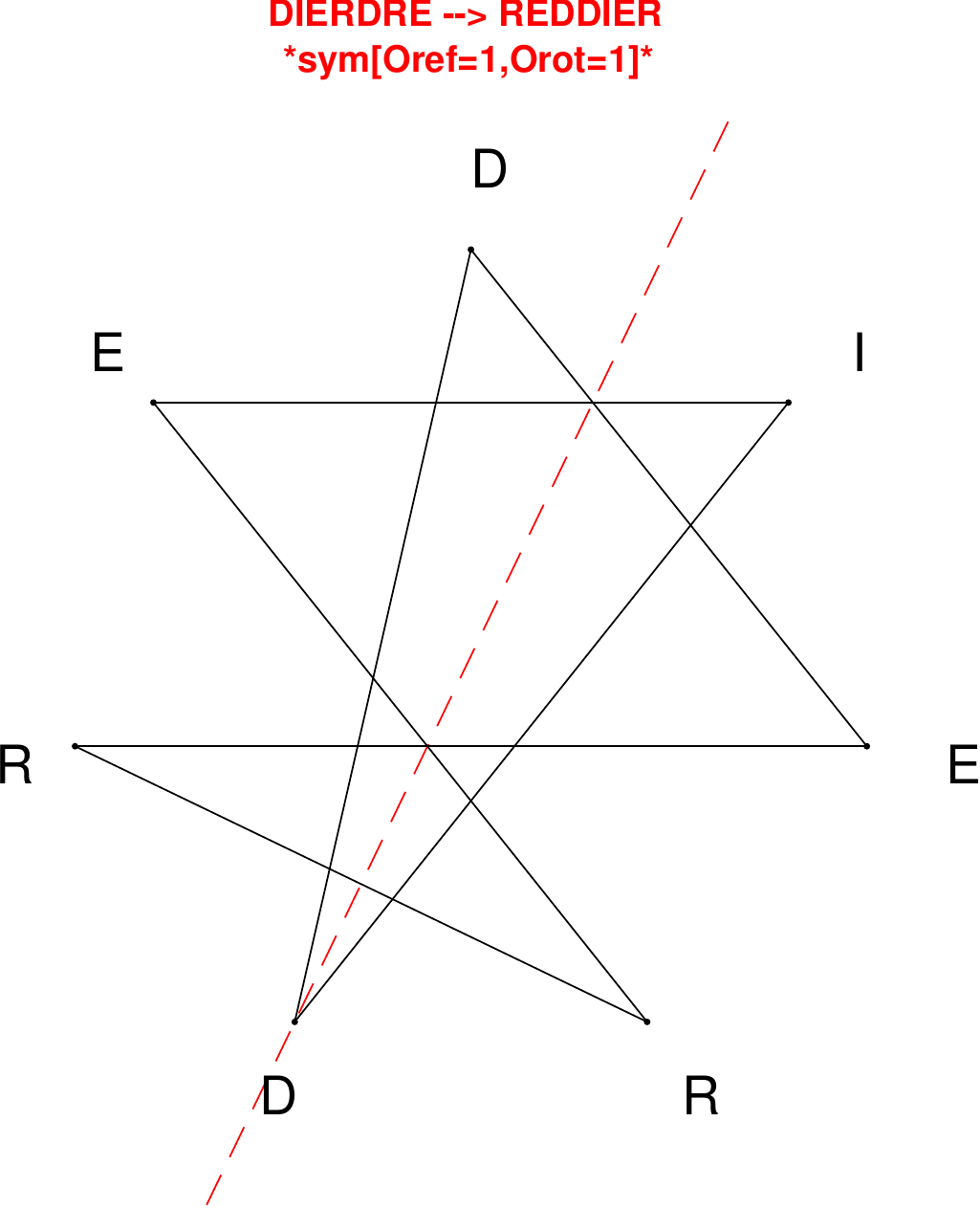}
\end{subfigure}
\hfill
\begin{subfigure}[T]{0.19\textwidth}
\centering
\includegraphics[width=\textwidth]{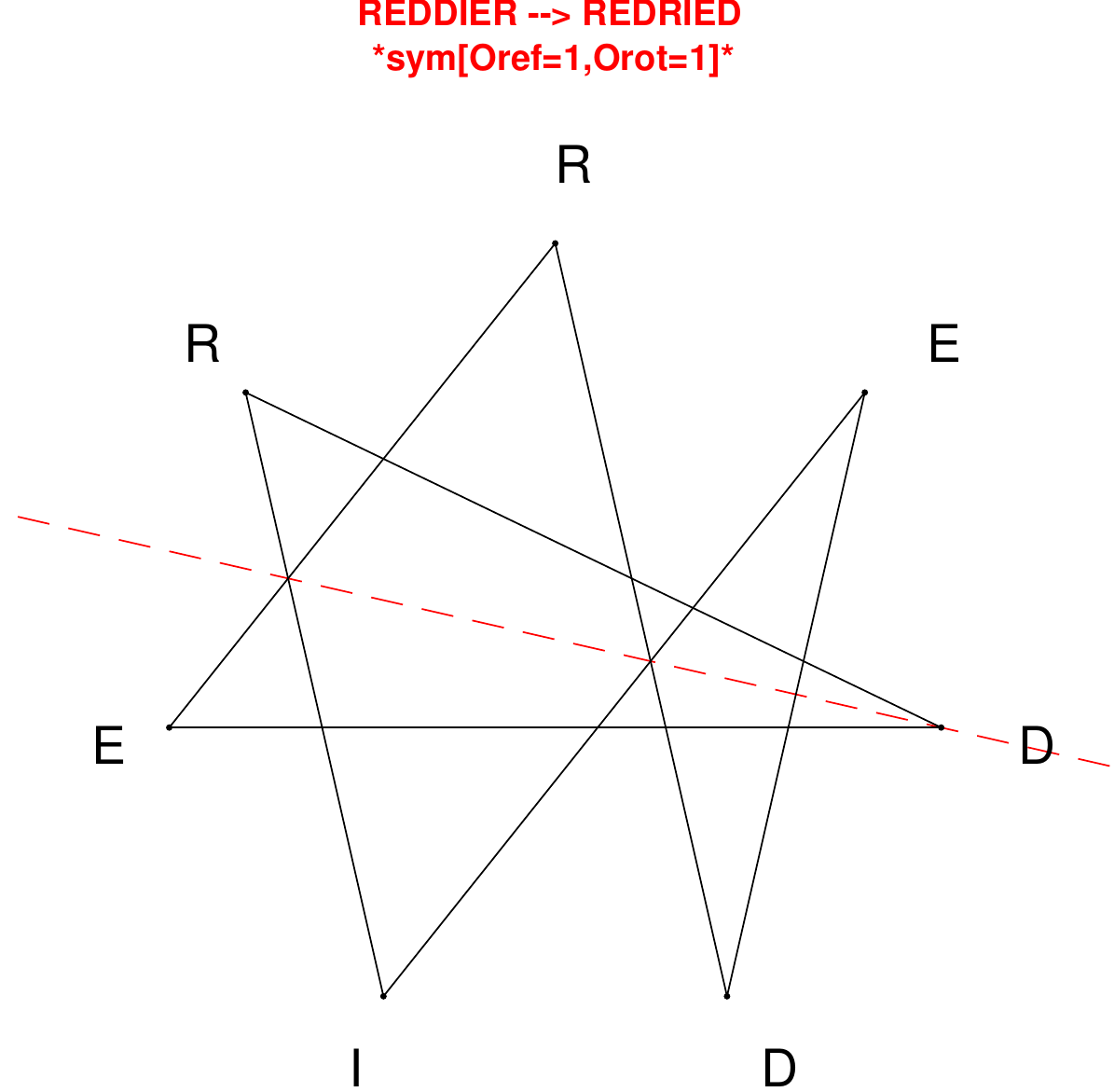}
\end{subfigure}
\hfill
\begin{subfigure}[T]{0.19\textwidth}
\centering
\includegraphics[width=\textwidth]{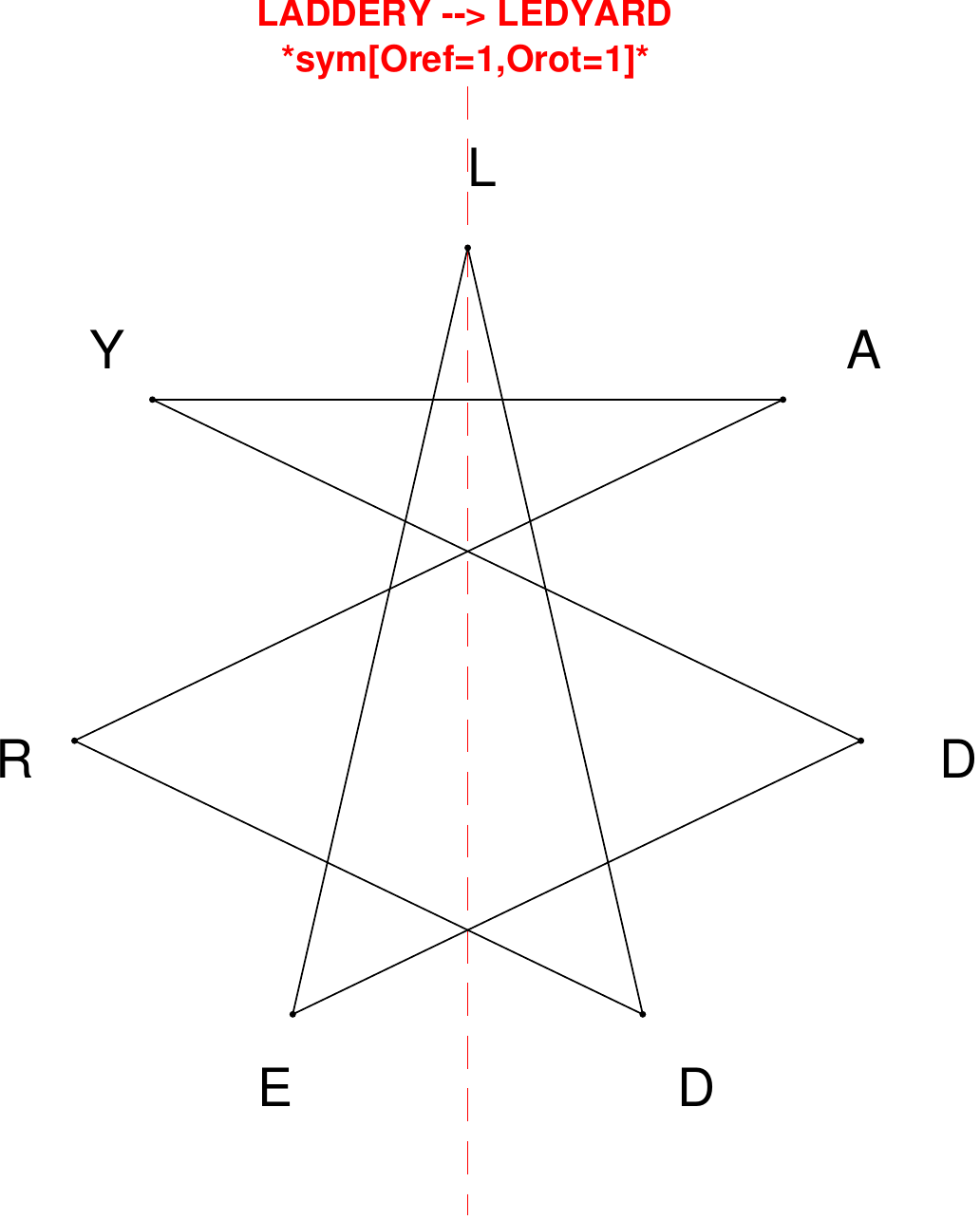}
\end{subfigure}
\end{figure}

\begin{figure}[H]
\centering
\begin{subfigure}[T]{0.19\textwidth}
\centering
\includegraphics[width=\textwidth]{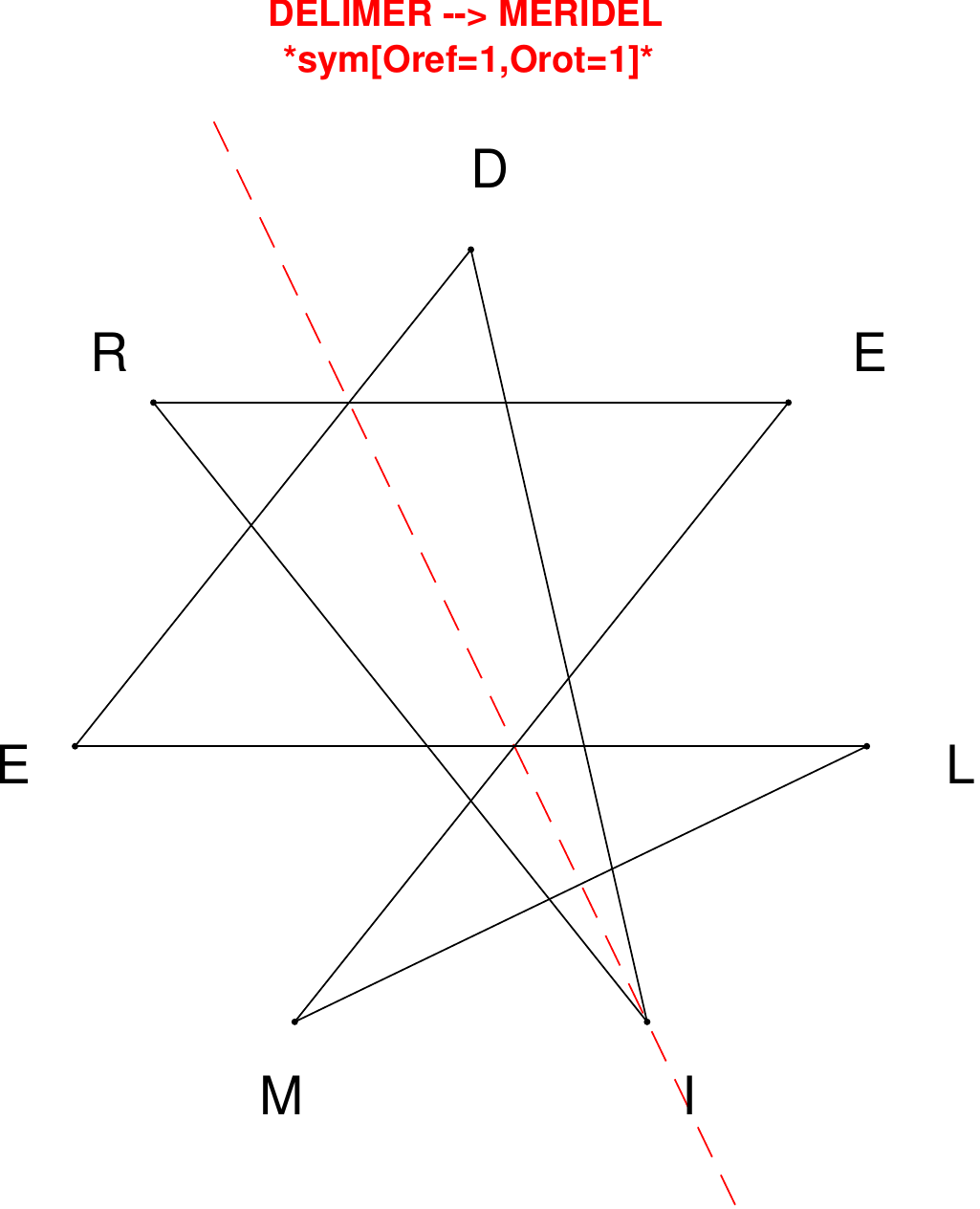}
\end{subfigure}
\hfill
\begin{subfigure}[T]{0.19\textwidth}
\centering
\includegraphics[width=\textwidth]{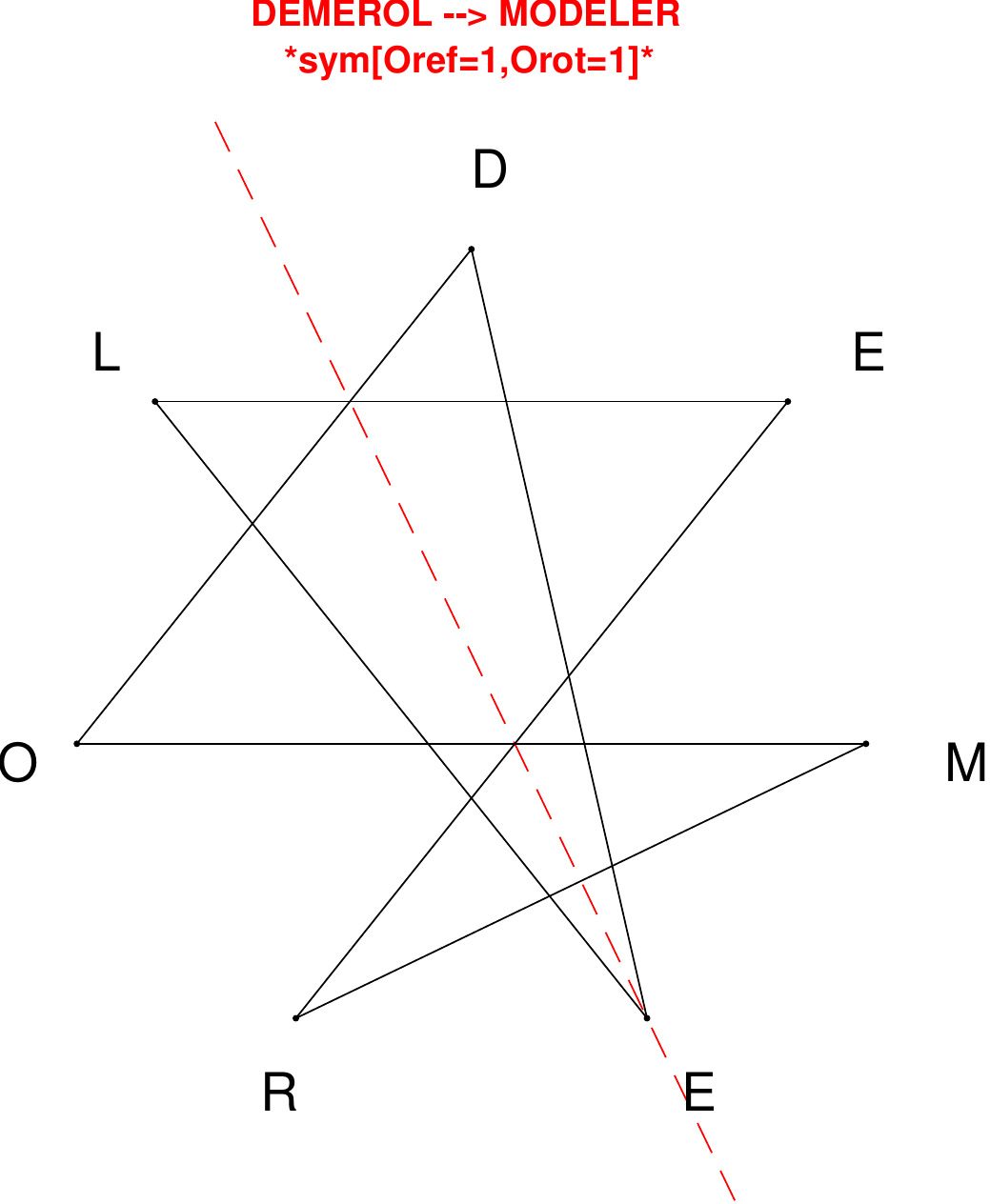}
\end{subfigure}
\hfill
\begin{subfigure}[T]{0.19\textwidth}
\centering
\includegraphics[width=\textwidth]{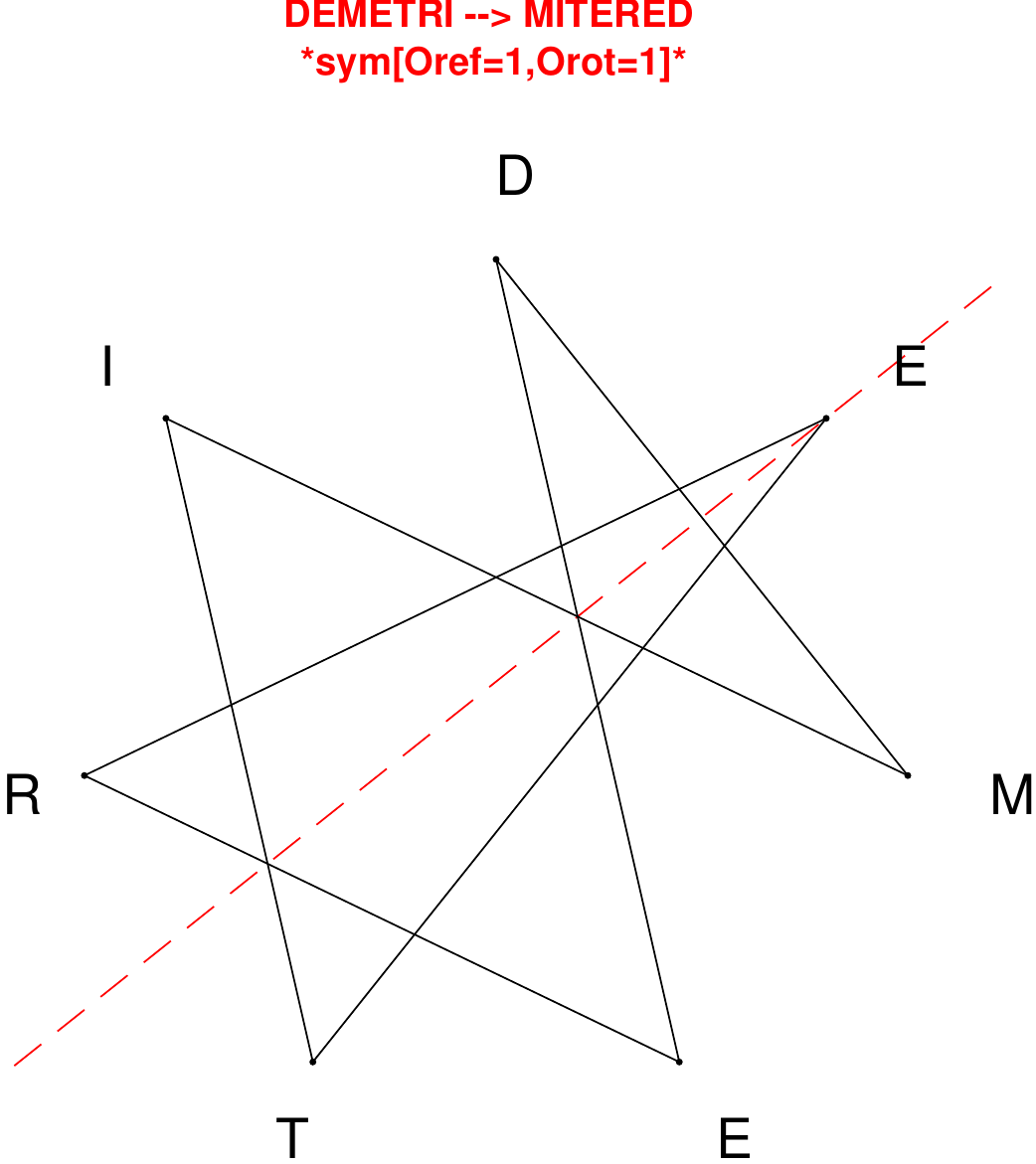}
\end{subfigure}
\hfill
\begin{subfigure}[T]{0.19\textwidth}
\centering
\includegraphics[width=\textwidth]{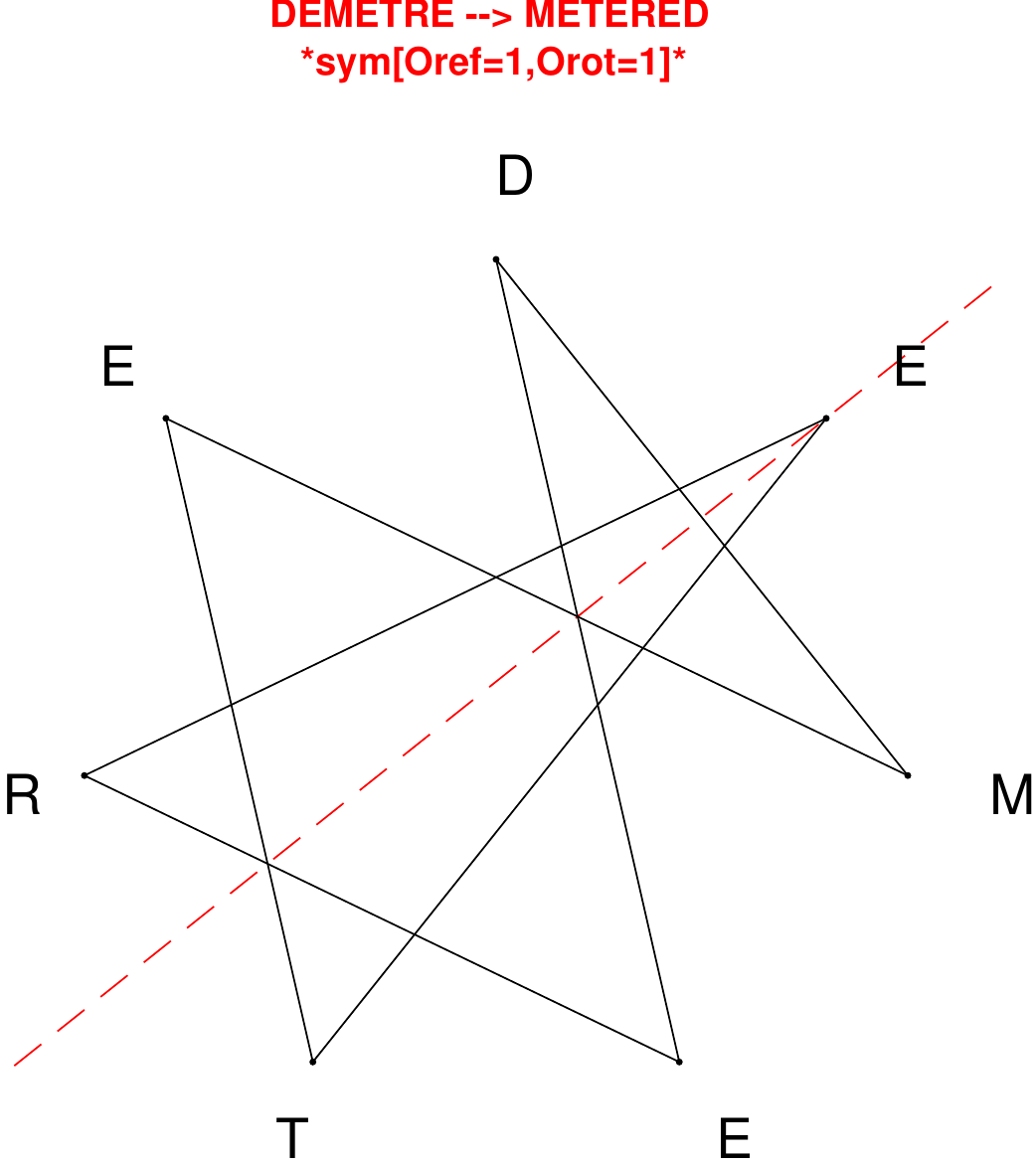}
\end{subfigure}
\hfill
\begin{subfigure}[T]{0.19\textwidth}
\centering
\includegraphics[width=\textwidth]{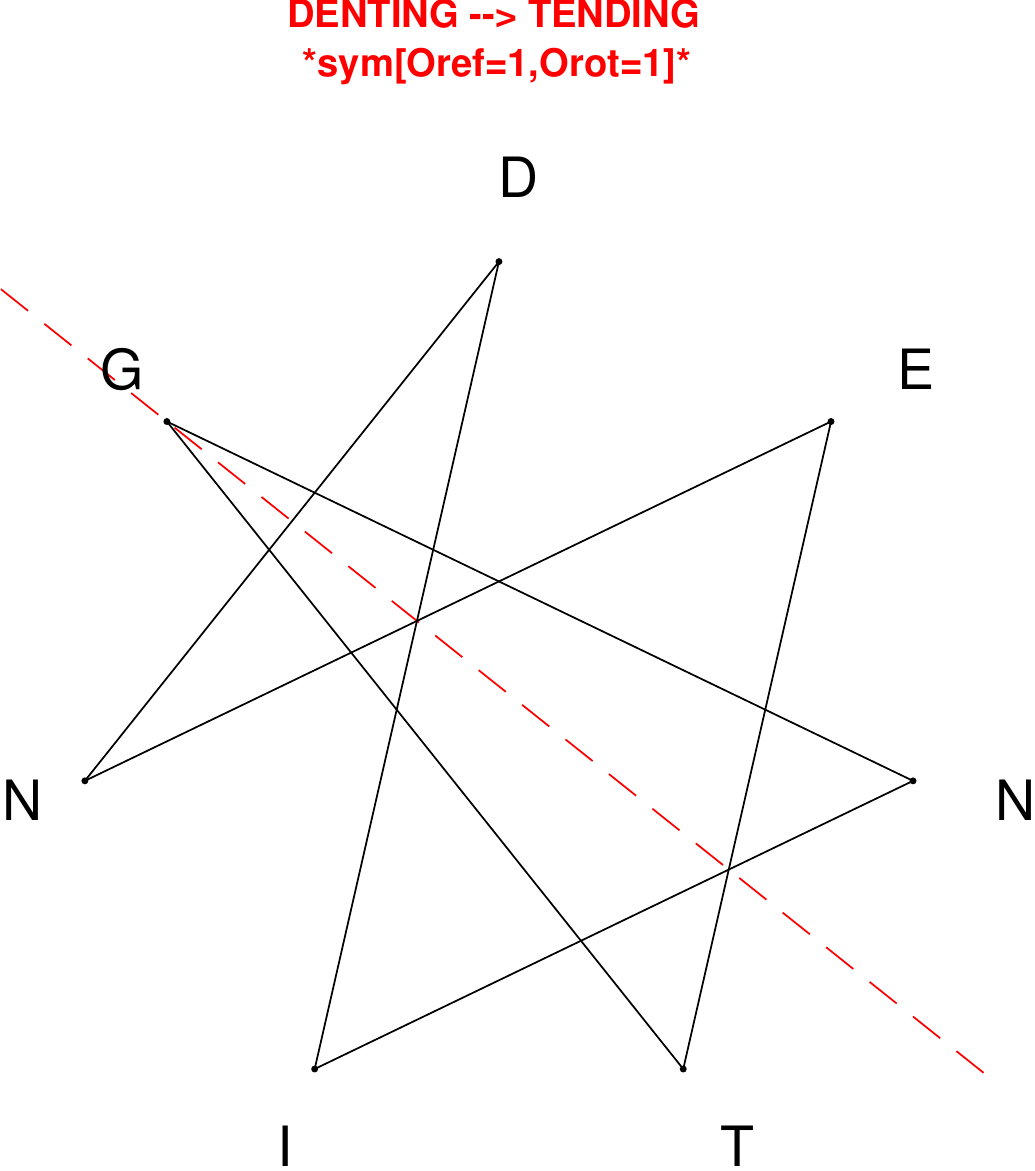}
\end{subfigure}
\end{figure}

\begin{figure}[H]
\centering
\begin{subfigure}[T]{0.19\textwidth}
\centering
\includegraphics[width=\textwidth]{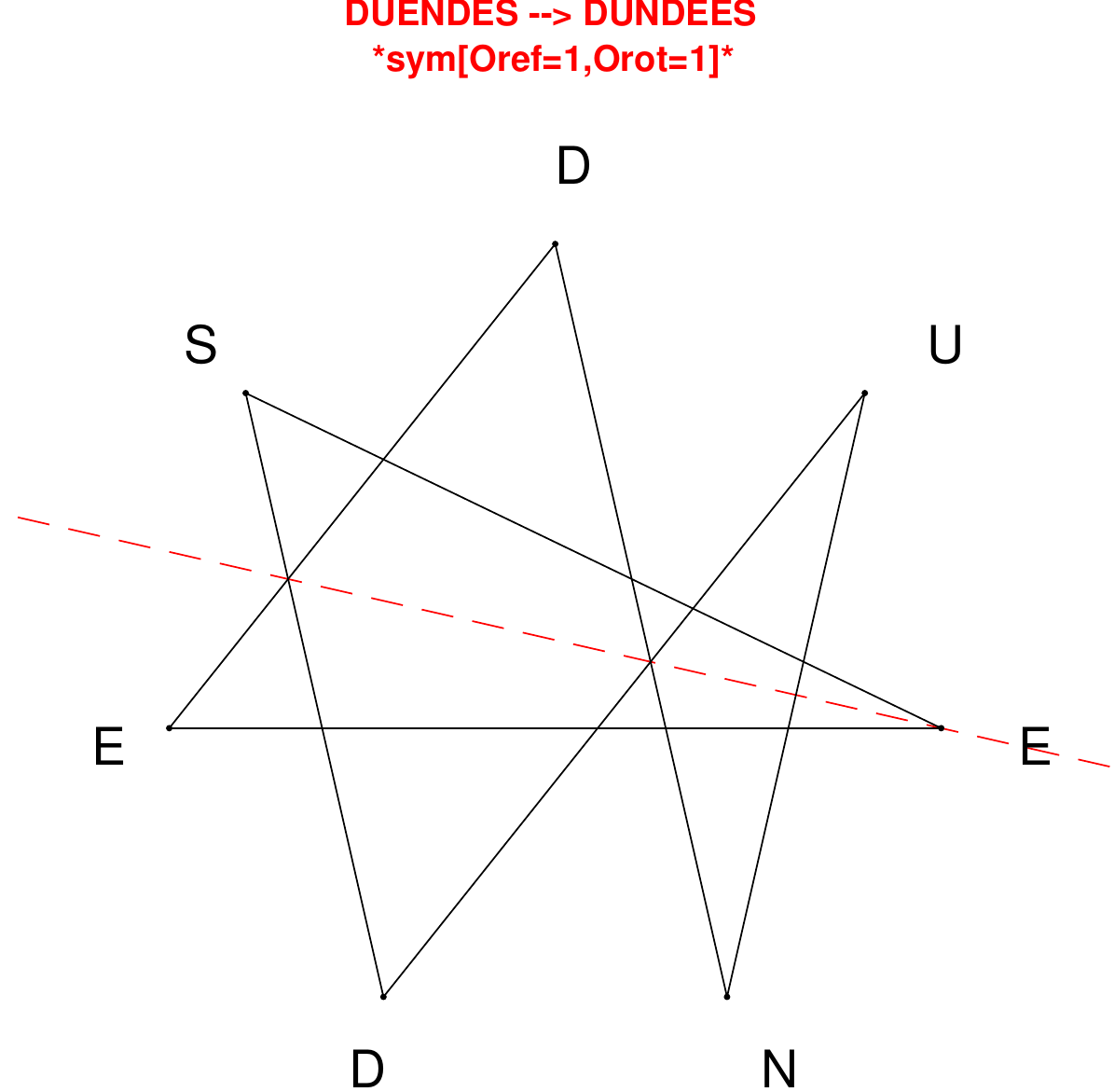}
\end{subfigure}
\hfill
\begin{subfigure}[T]{0.19\textwidth}
\centering
\includegraphics[width=\textwidth]{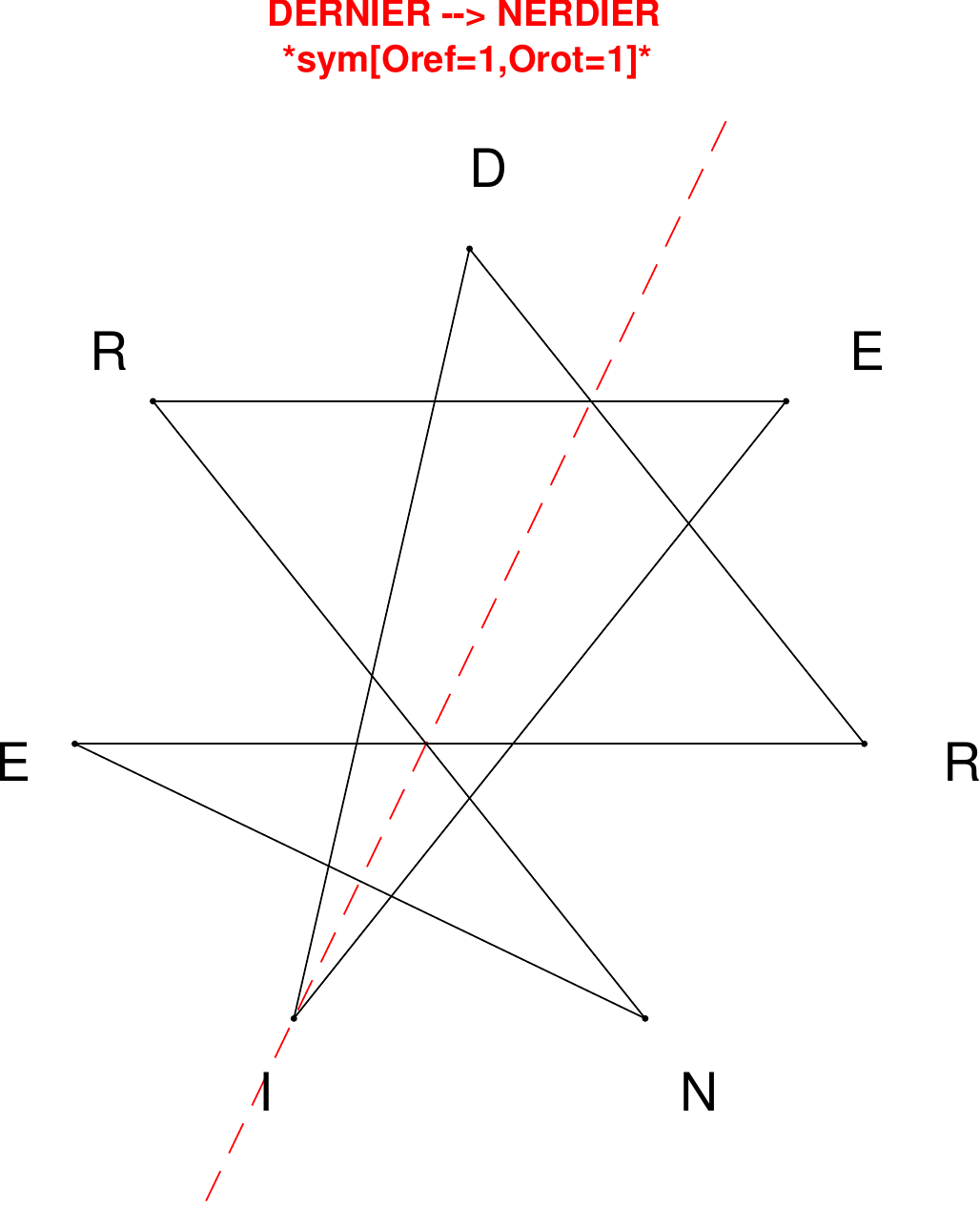}
\end{subfigure}
\hfill
\begin{subfigure}[T]{0.19\textwidth}
\centering
\includegraphics[width=\textwidth]{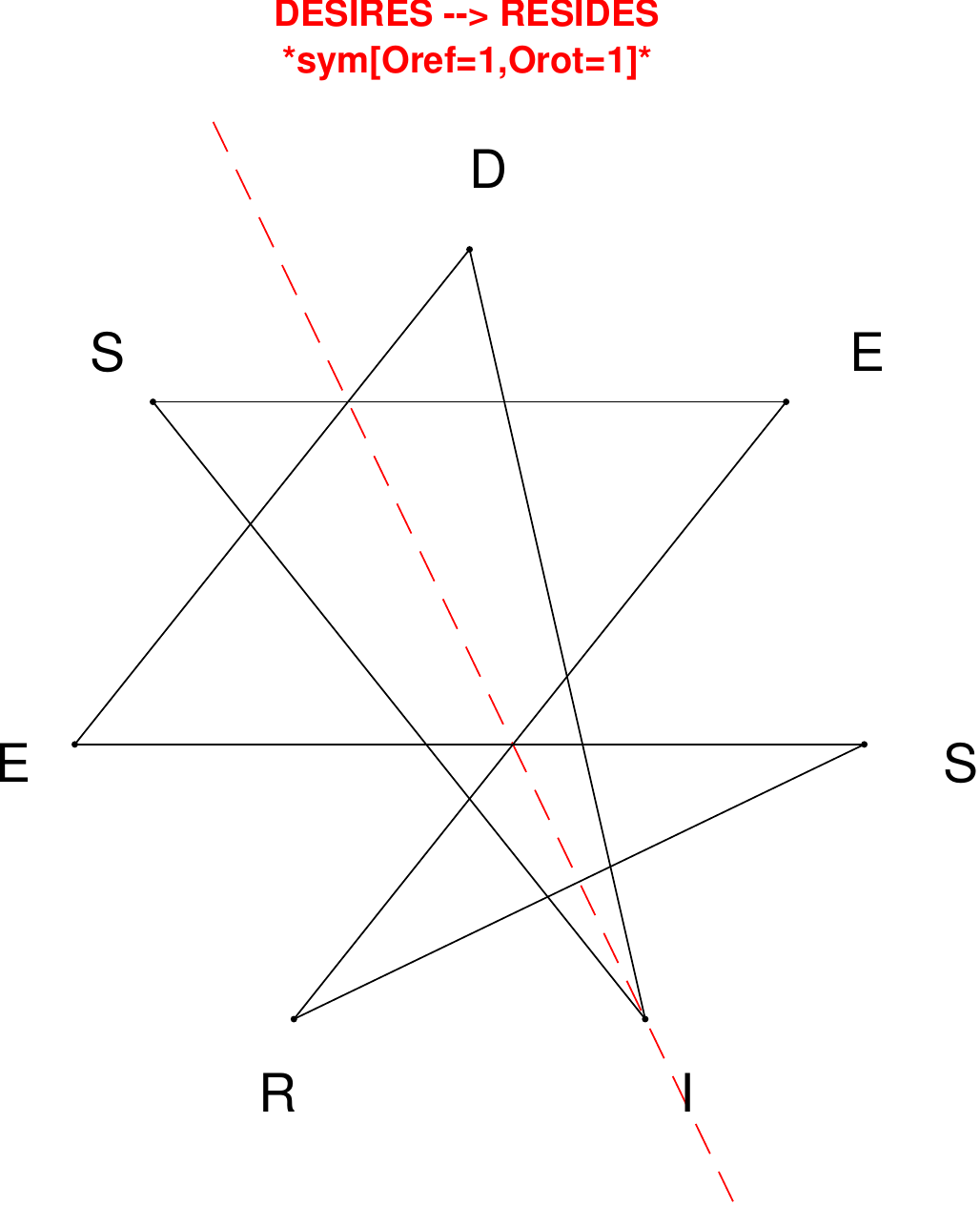}
\end{subfigure}
\hfill
\begin{subfigure}[T]{0.19\textwidth}
\centering
\includegraphics[width=\textwidth]{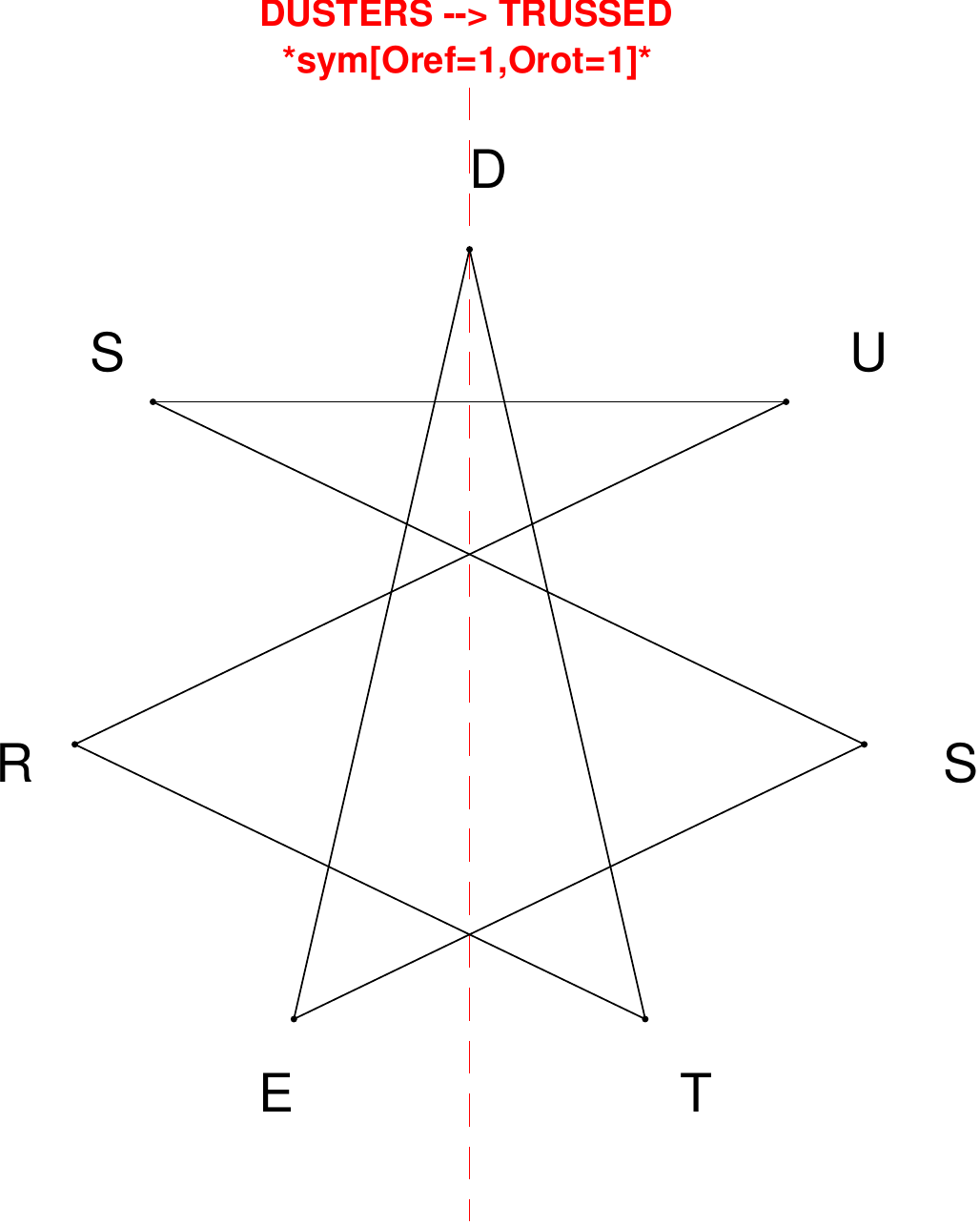}
\end{subfigure}
\hfill
\begin{subfigure}[T]{0.19\textwidth}
\centering
\includegraphics[width=\textwidth]{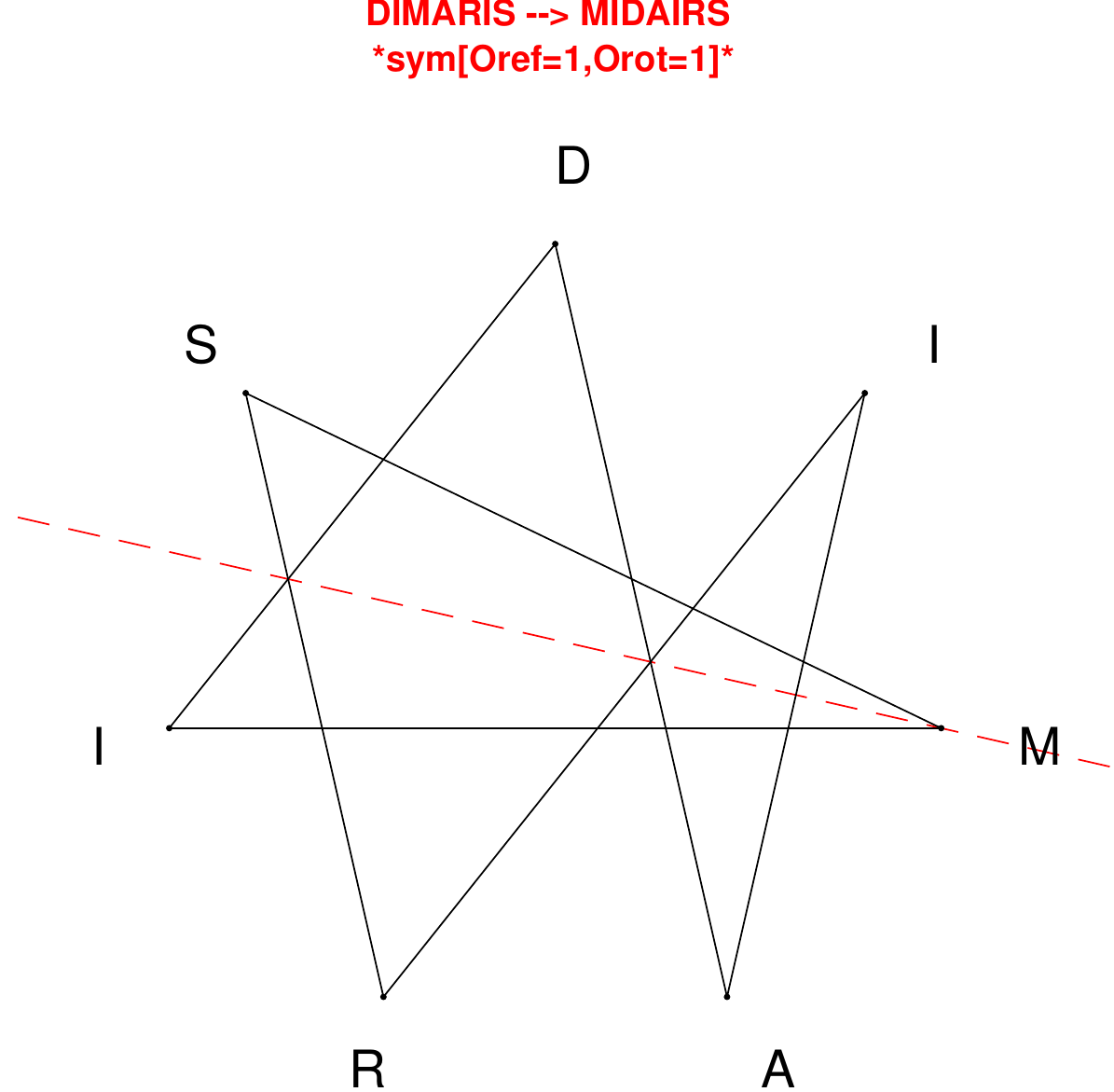}
\end{subfigure}
\end{figure}

\begin{figure}[H]
\centering
\begin{subfigure}[T]{0.19\textwidth}
\centering
\includegraphics[width=\textwidth]{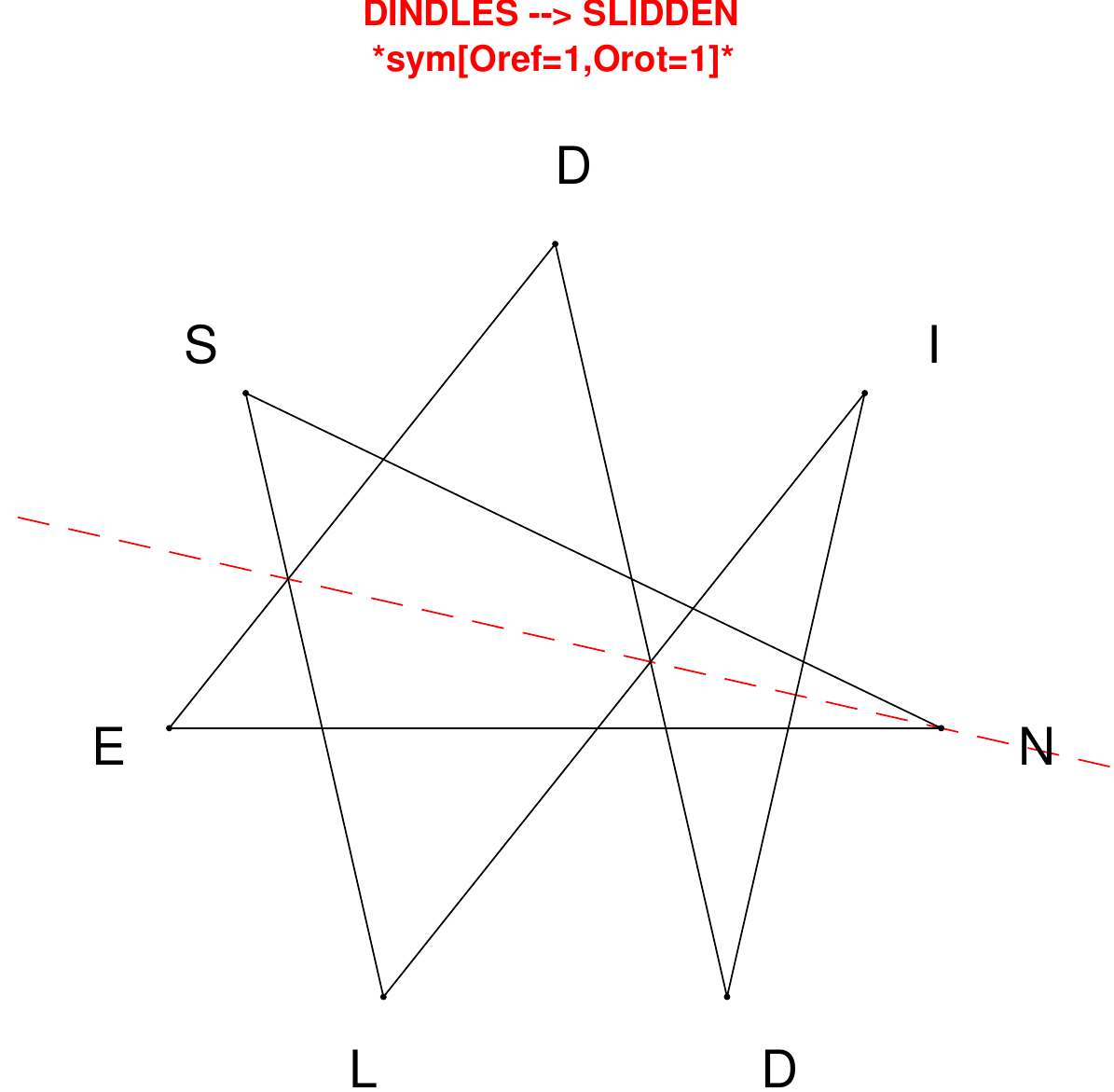}
\end{subfigure}
\hfill
\begin{subfigure}[T]{0.19\textwidth}
\centering
\includegraphics[width=\textwidth]{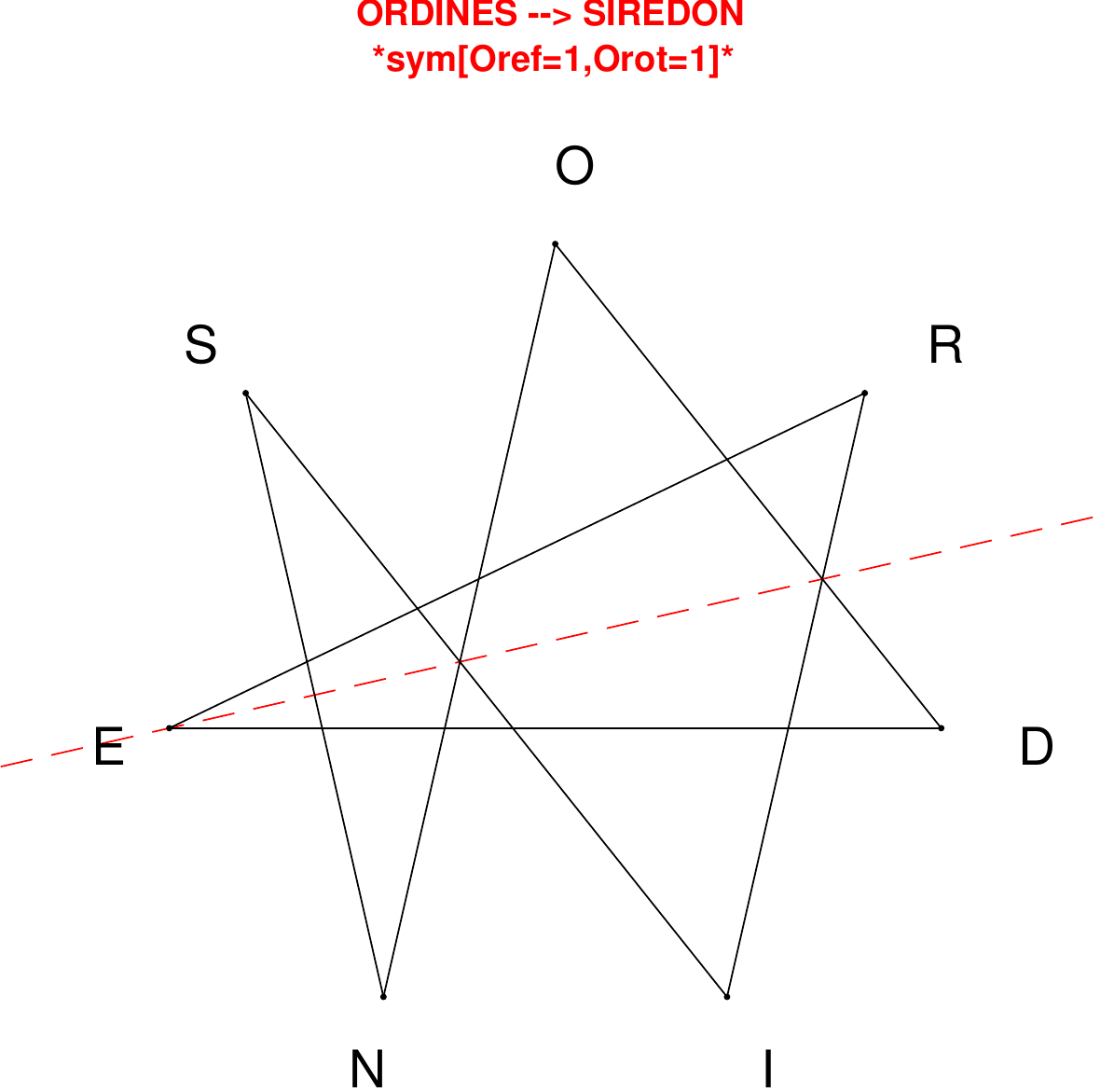}
\end{subfigure}
\hfill
\begin{subfigure}[T]{0.19\textwidth}
\centering
\includegraphics[width=\textwidth]{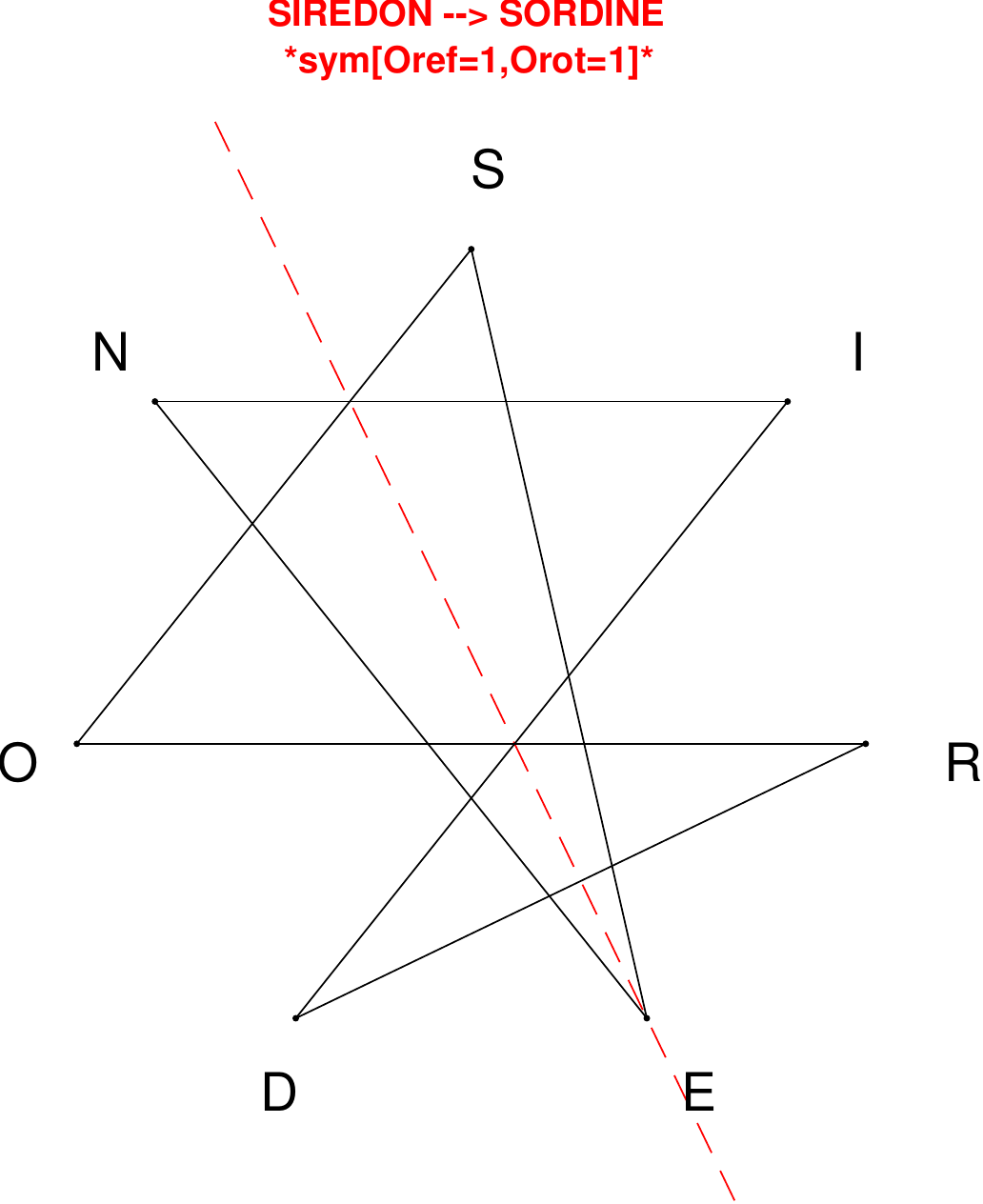}
\end{subfigure}
\hfill
\begin{subfigure}[T]{0.19\textwidth}
\centering
\includegraphics[width=\textwidth]{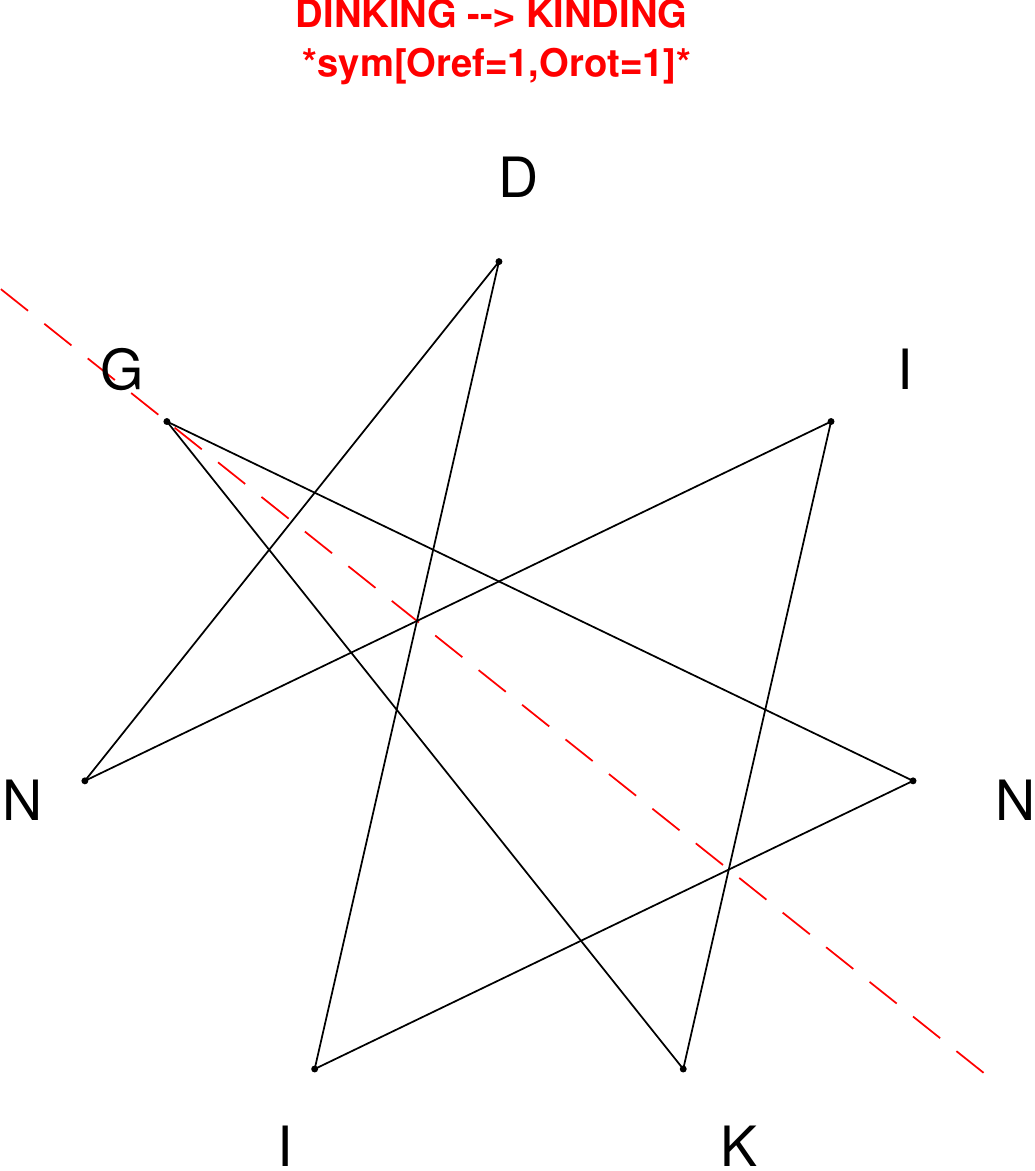}
\end{subfigure}
\hfill
\begin{subfigure}[T]{0.19\textwidth}
\centering
\includegraphics[width=\textwidth]{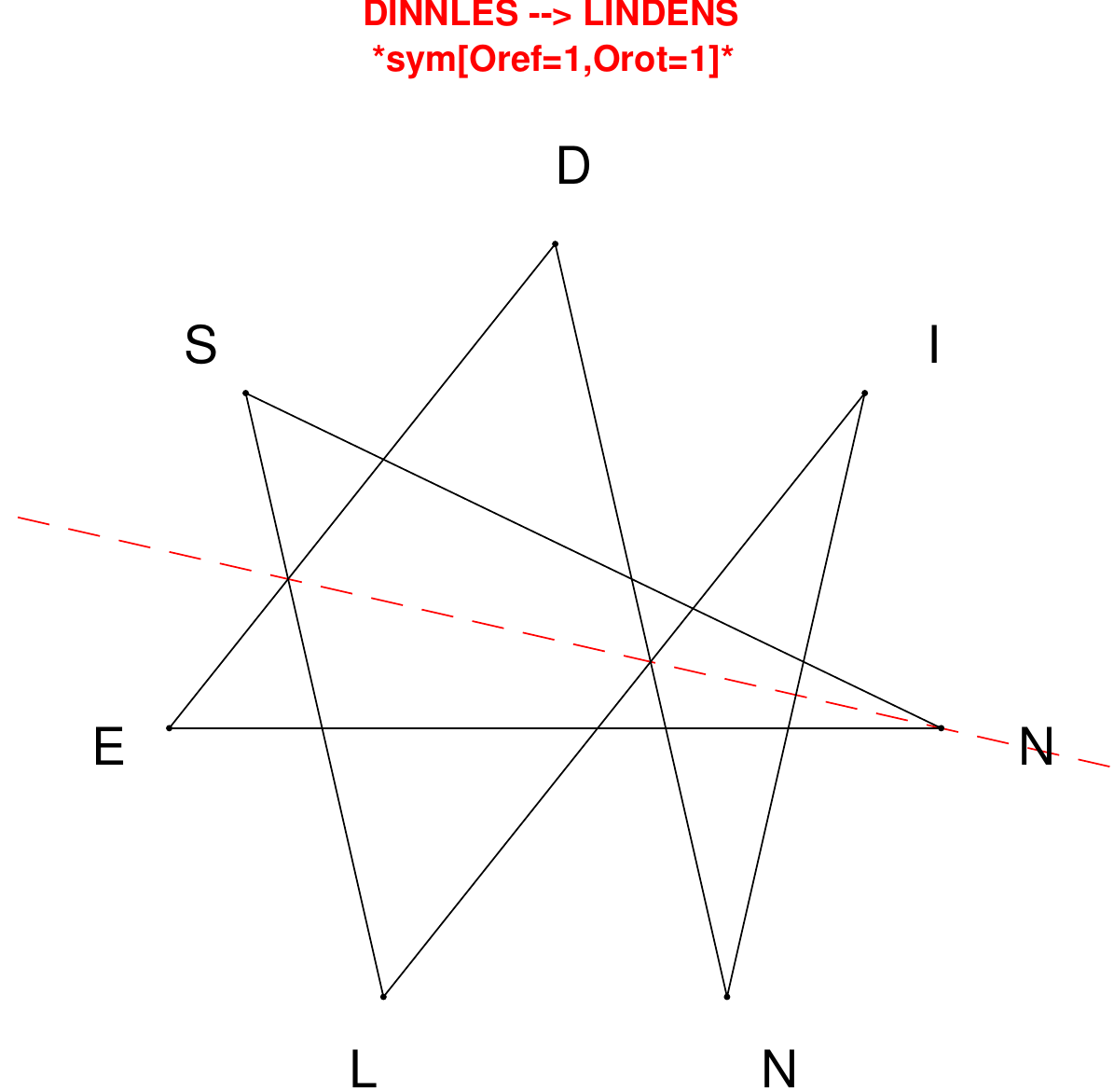}
\end{subfigure}
\end{figure}

\begin{figure}[H]
\centering
\begin{subfigure}[T]{0.19\textwidth}
\centering
\includegraphics[width=\textwidth]{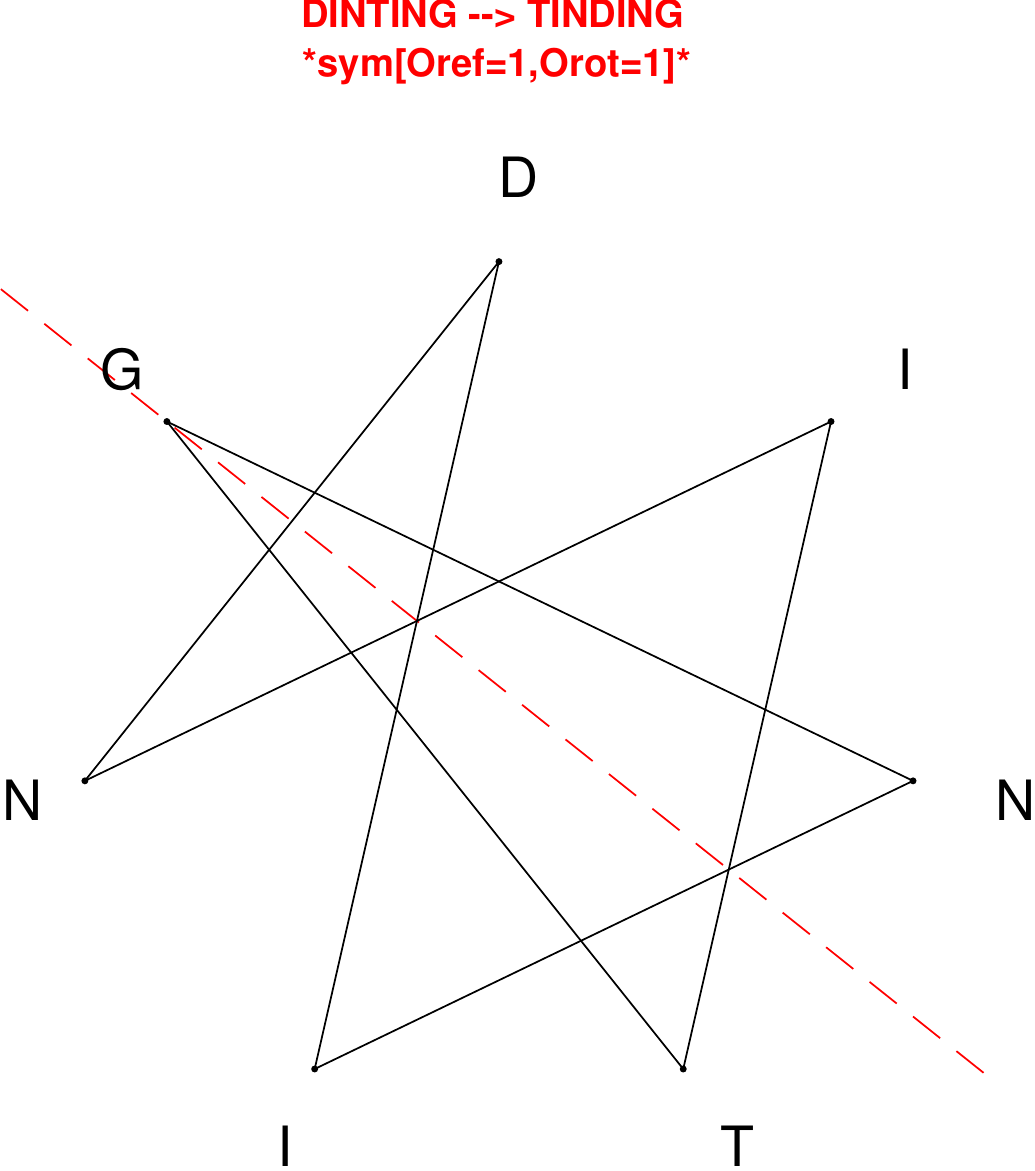}
\end{subfigure}
\hfill
\begin{subfigure}[T]{0.19\textwidth}
\centering
\includegraphics[width=\textwidth]{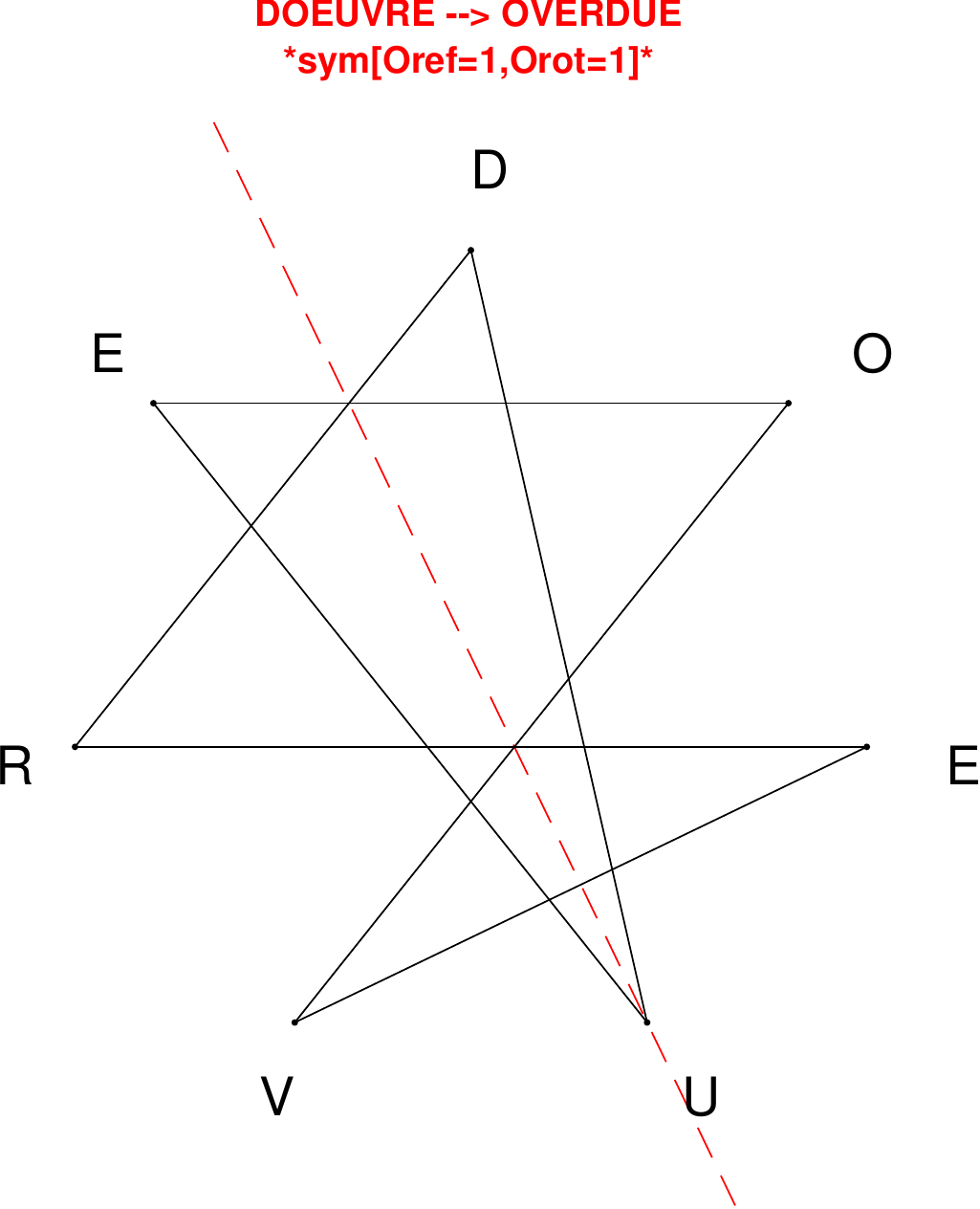}
\end{subfigure}
\hfill
\begin{subfigure}[T]{0.19\textwidth}
\centering
\includegraphics[width=\textwidth]{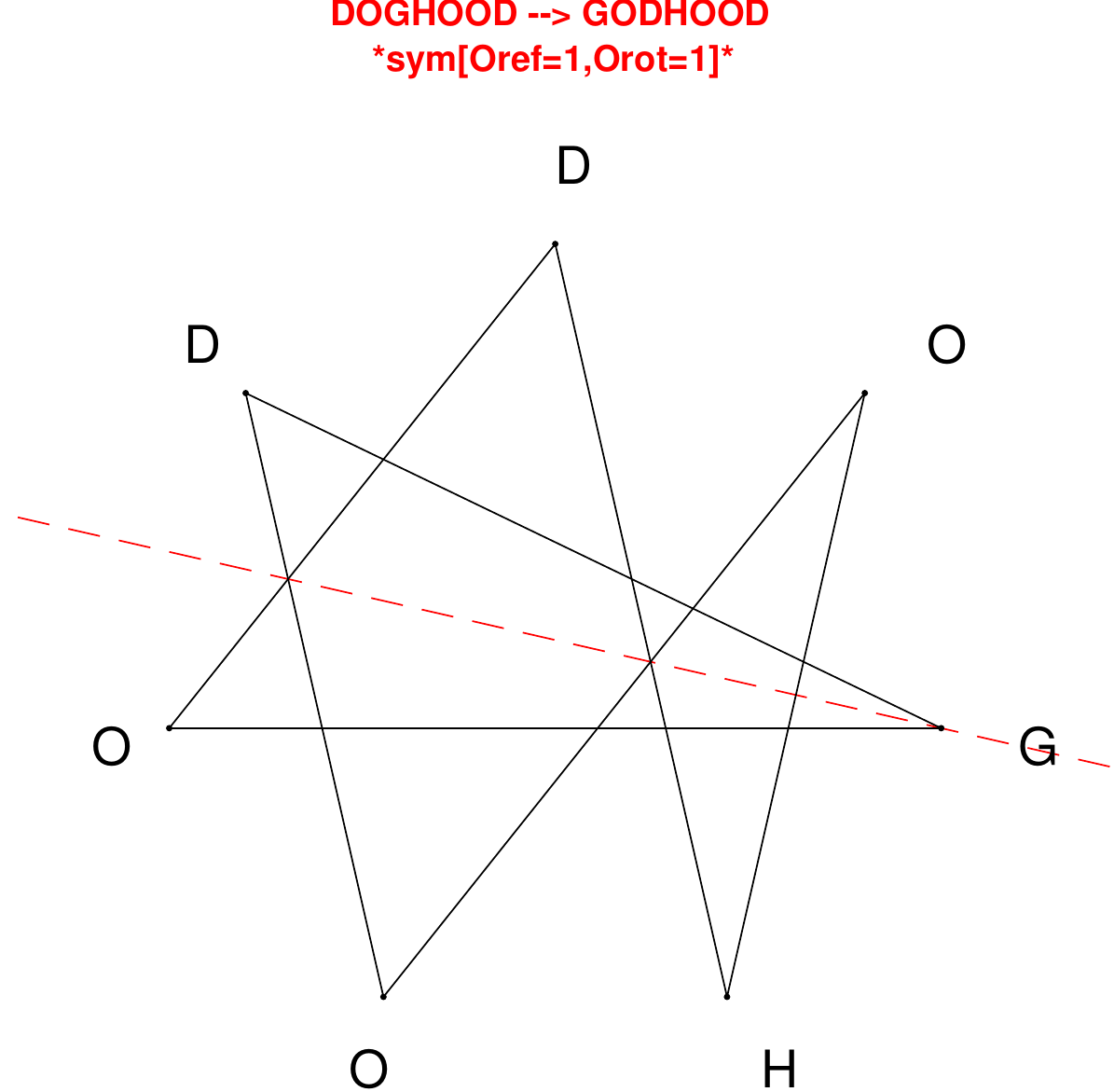}
\end{subfigure}
\hfill
\begin{subfigure}[T]{0.19\textwidth}
\centering
\includegraphics[width=\textwidth]{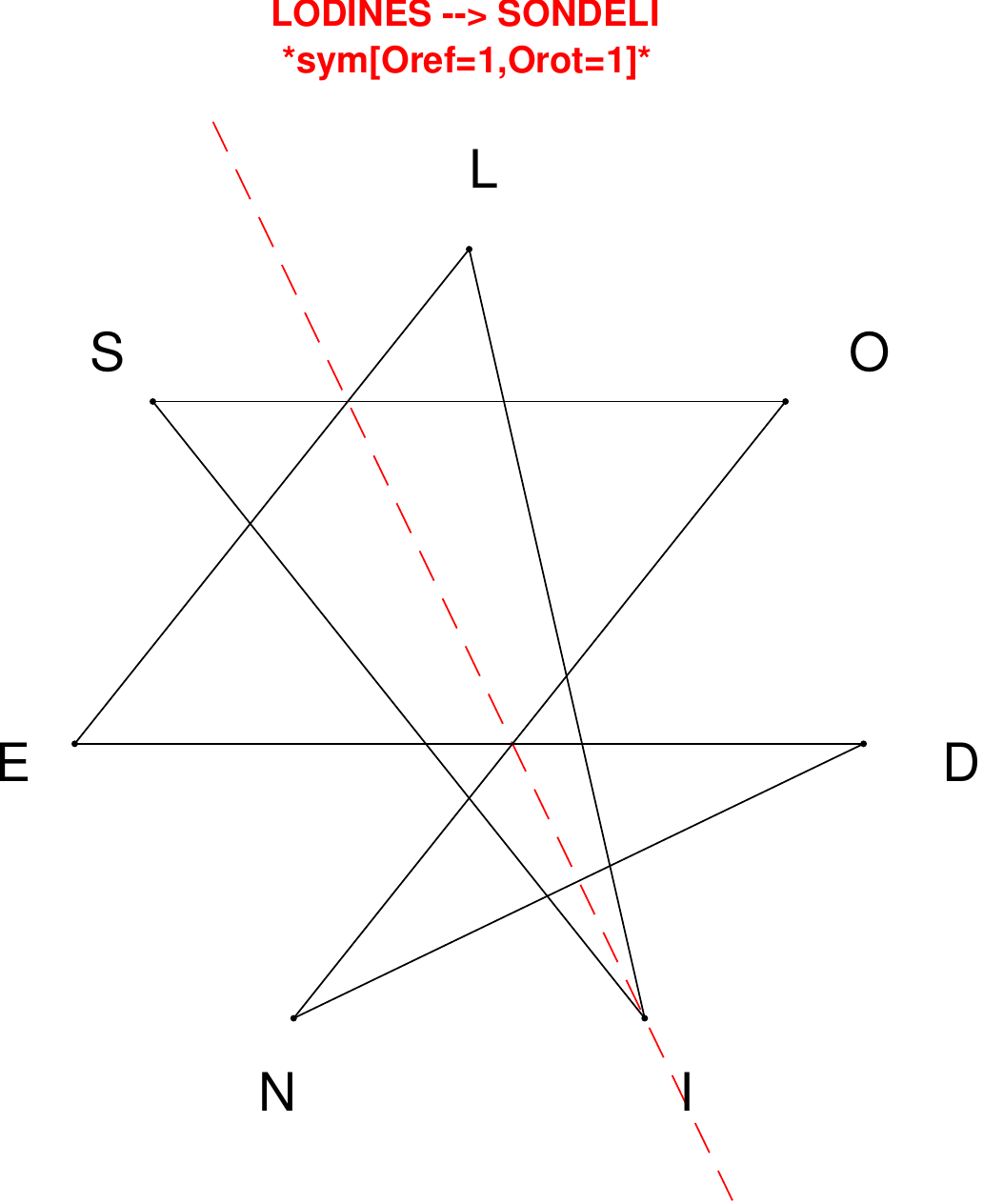}
\end{subfigure}
\hfill
\begin{subfigure}[T]{0.19\textwidth}
\centering
\includegraphics[width=\textwidth]{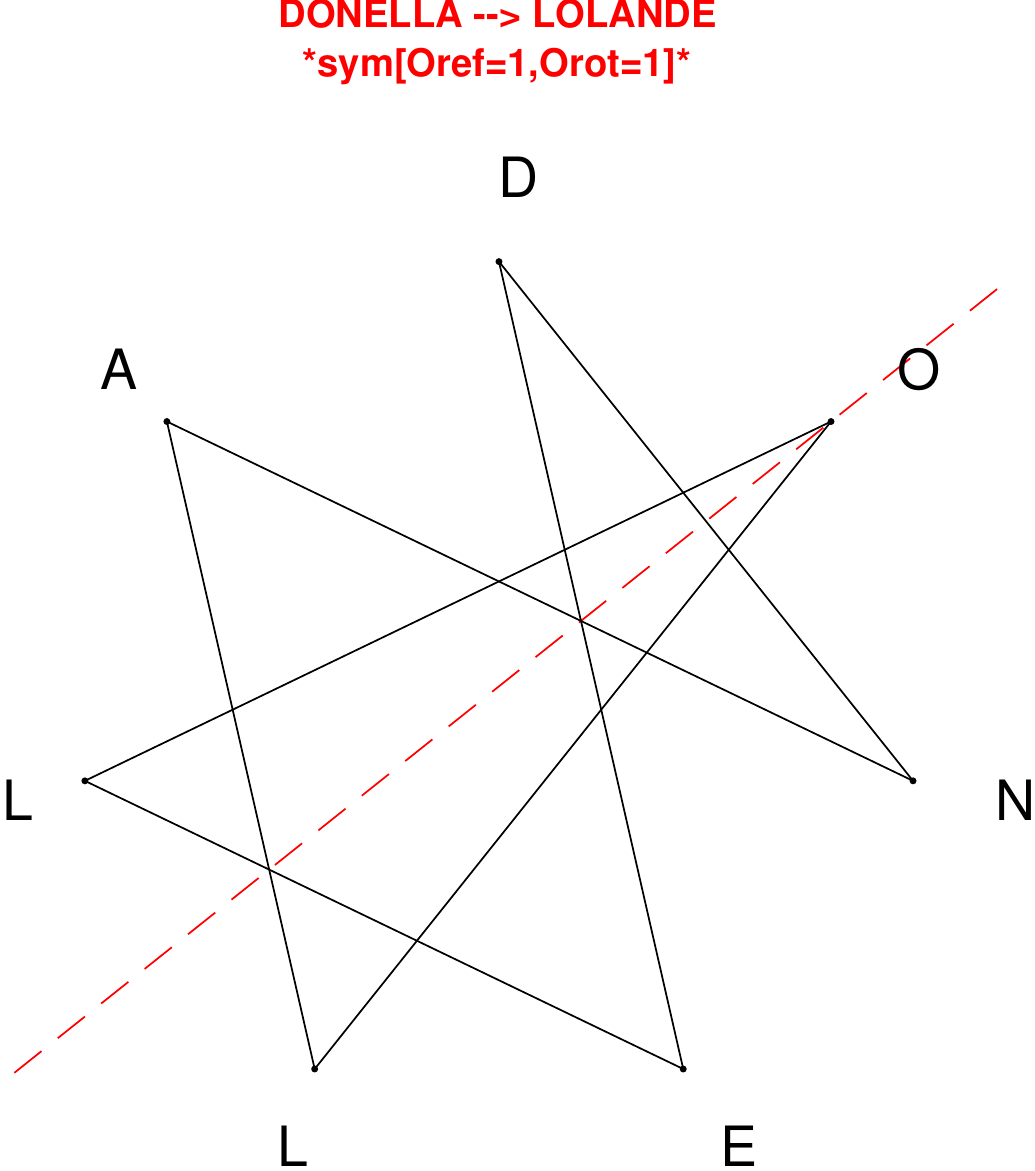}
\end{subfigure}
\end{figure}

\begin{figure}[H]
\centering
\begin{subfigure}[T]{0.19\textwidth}
\centering
\includegraphics[width=\textwidth]{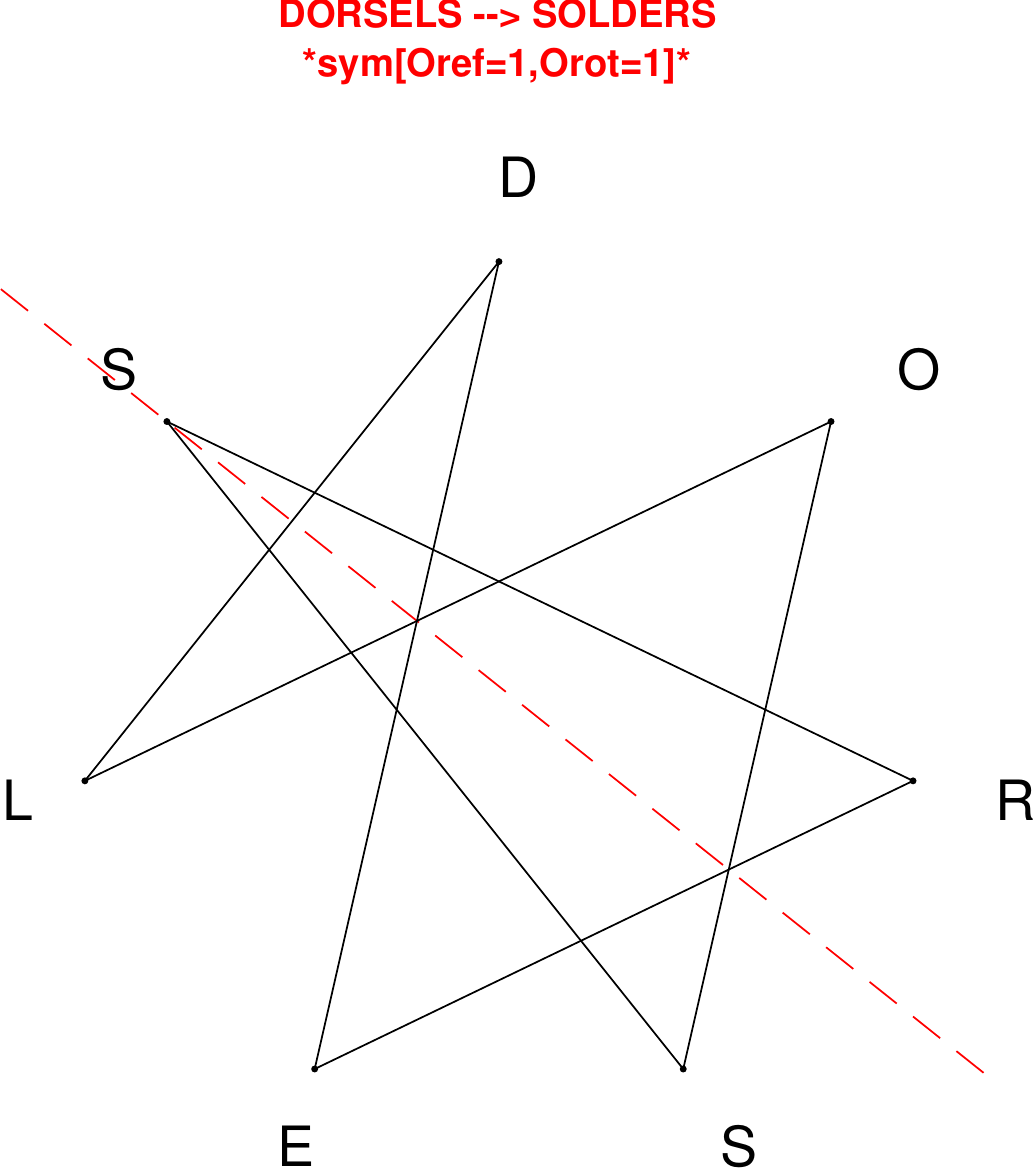}
\end{subfigure}
\hfill
\begin{subfigure}[T]{0.19\textwidth}
\centering
\includegraphics[width=\textwidth]{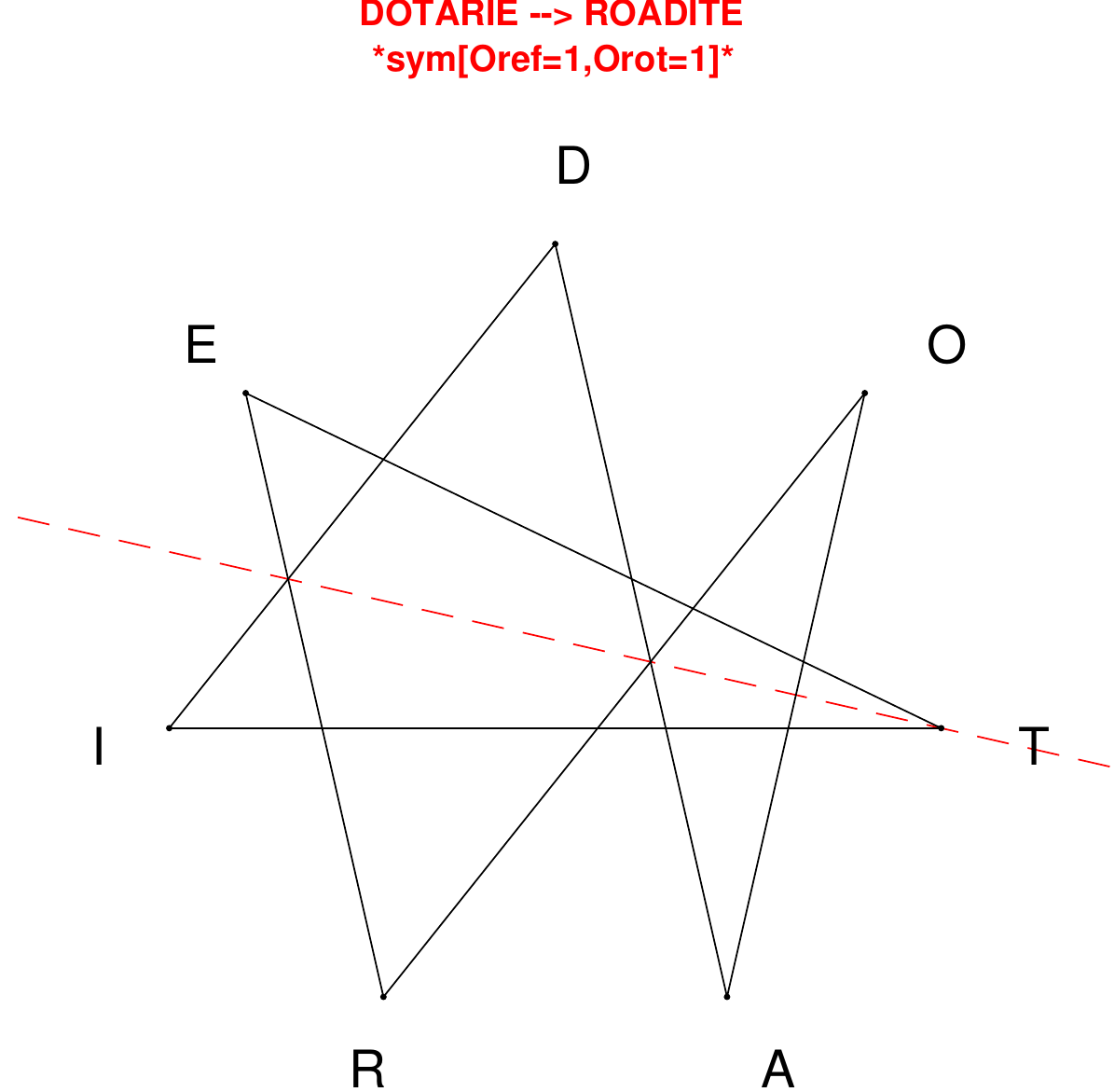}
\end{subfigure}
\hfill
\begin{subfigure}[T]{0.19\textwidth}
\centering
\includegraphics[width=\textwidth]{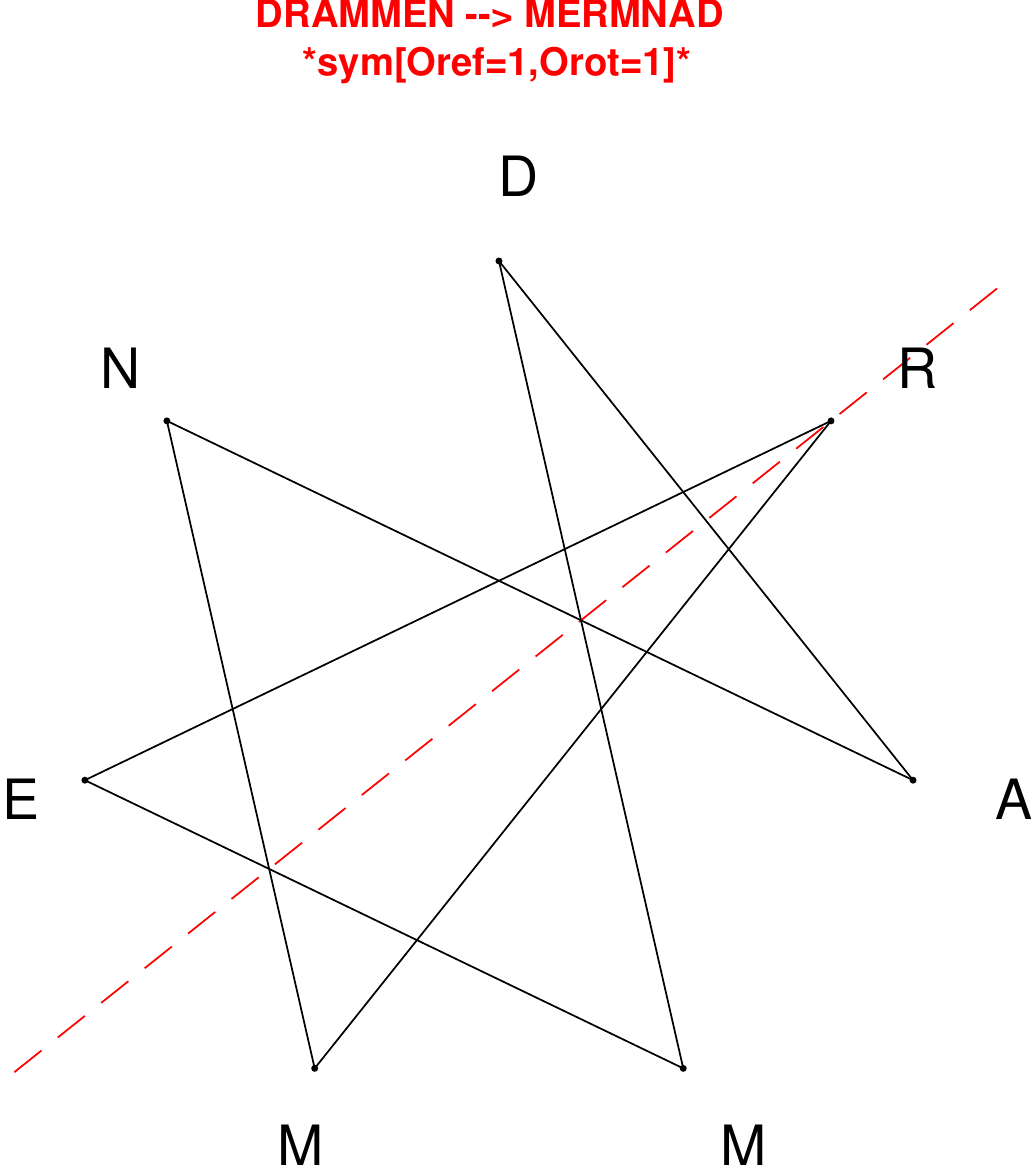}
\end{subfigure}
\hfill
\begin{subfigure}[T]{0.19\textwidth}
\centering
\includegraphics[width=\textwidth]{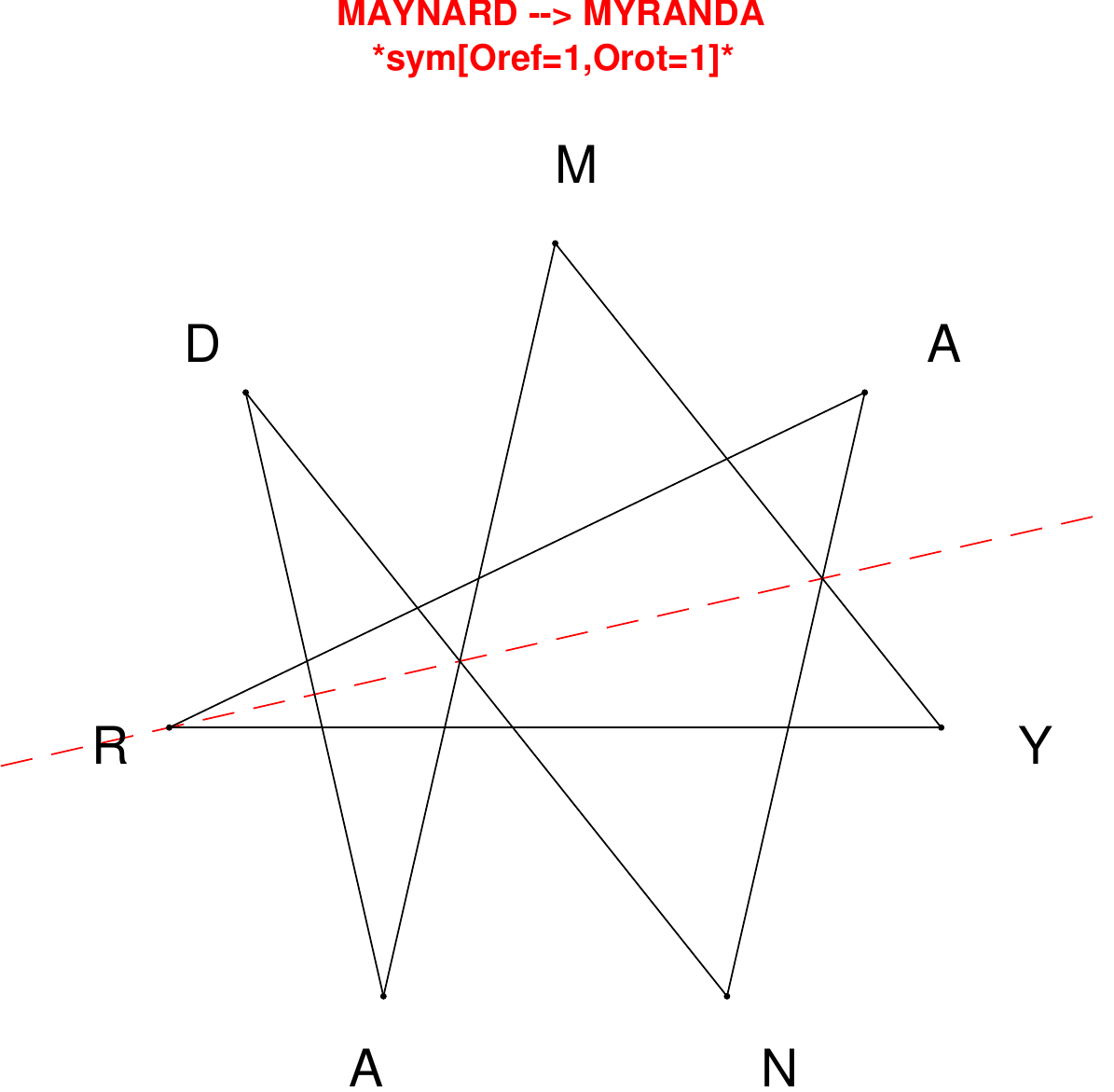}
\end{subfigure}
\hfill
\begin{subfigure}[T]{0.19\textwidth}
\centering
\includegraphics[width=\textwidth]{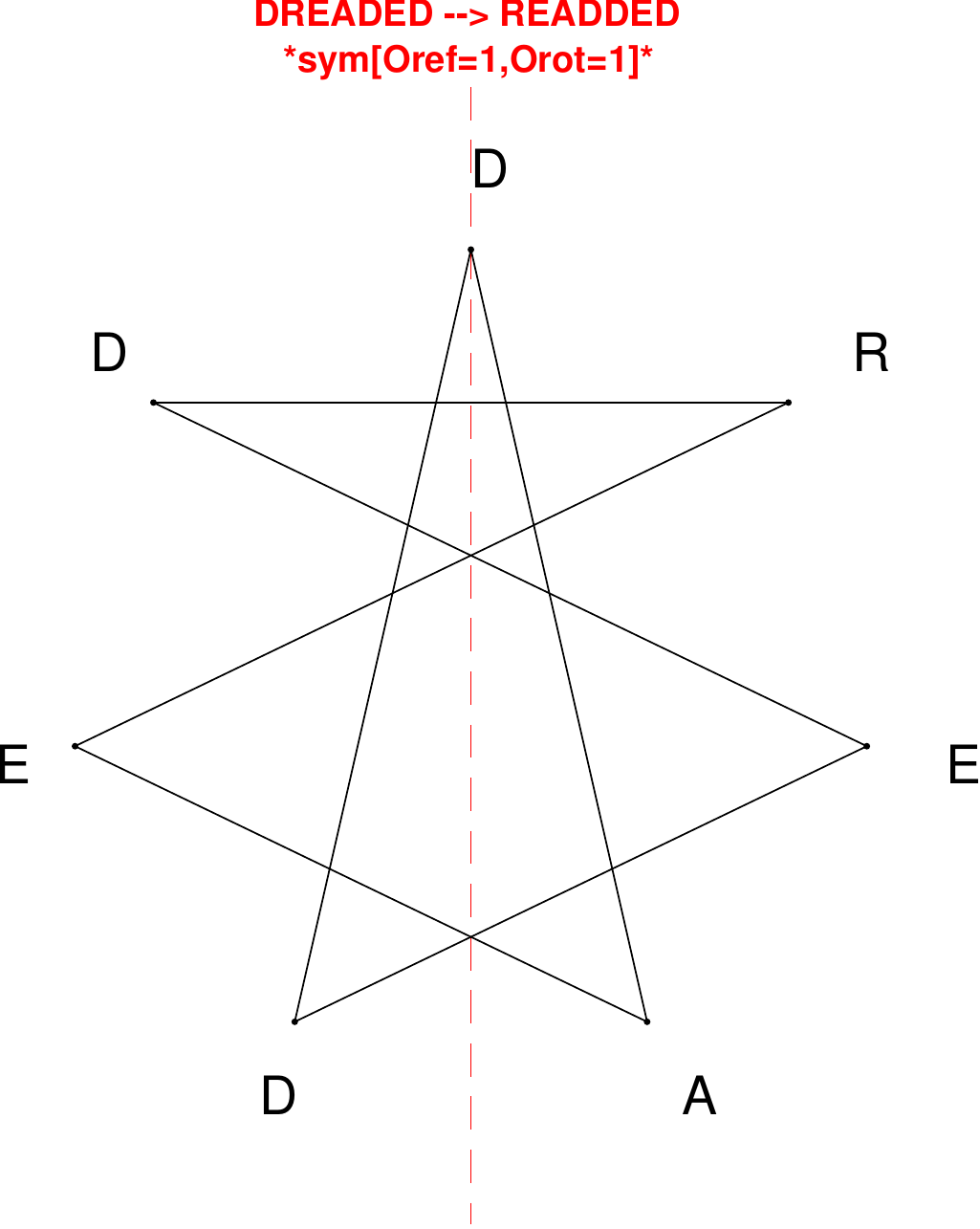}
\end{subfigure}
\end{figure}

\begin{figure}[H]
\centering
\begin{subfigure}[T]{0.19\textwidth}
\centering
\includegraphics[width=\textwidth]{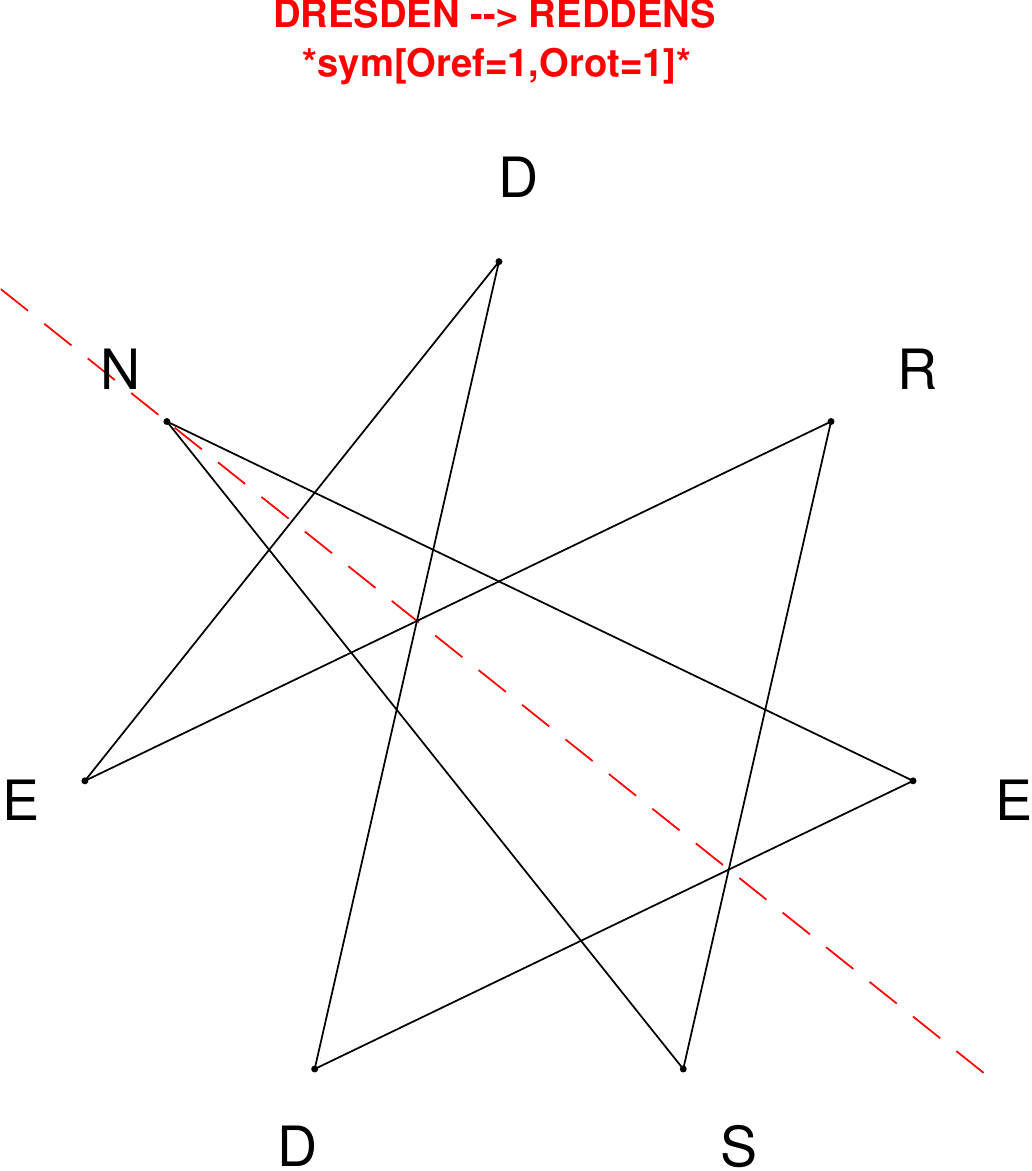}
\end{subfigure}
\hfill
\begin{subfigure}[T]{0.19\textwidth}
\centering
\includegraphics[width=\textwidth]{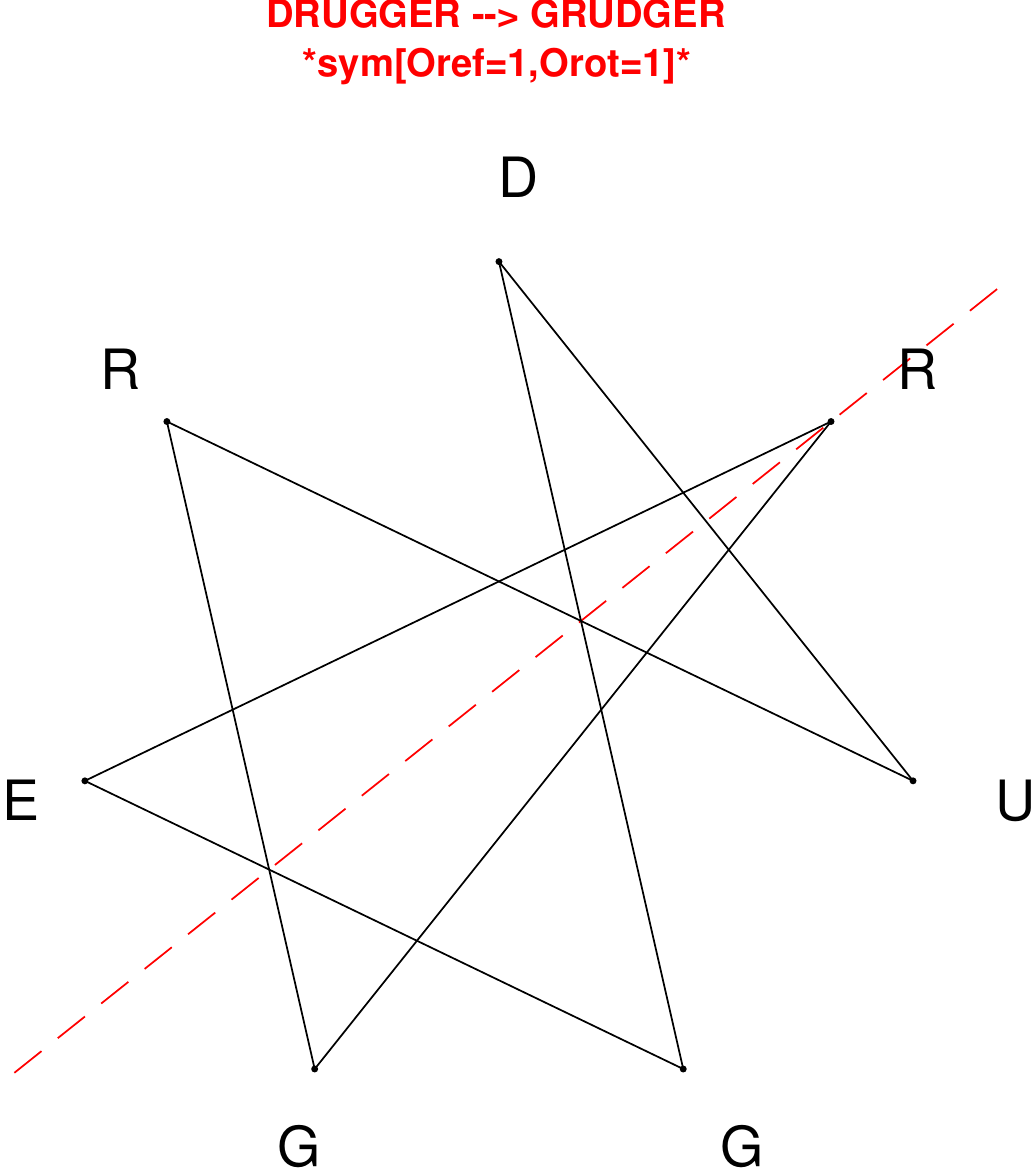}
\end{subfigure}
\hfill
\begin{subfigure}[T]{0.19\textwidth}
\centering
\includegraphics[width=\textwidth]{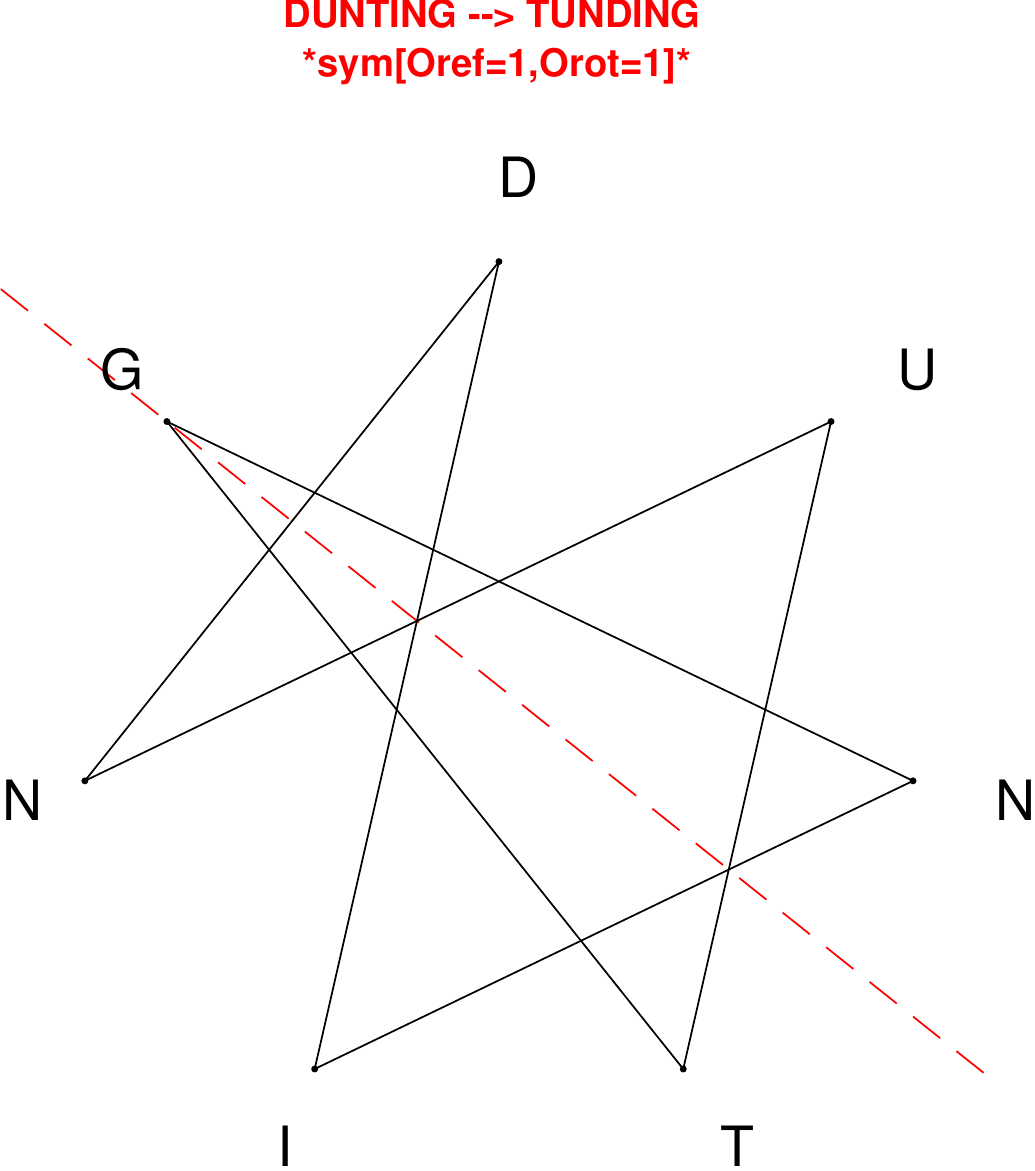}
\end{subfigure}
\hfill
\begin{subfigure}[T]{0.19\textwidth}
\centering
\includegraphics[width=\textwidth]{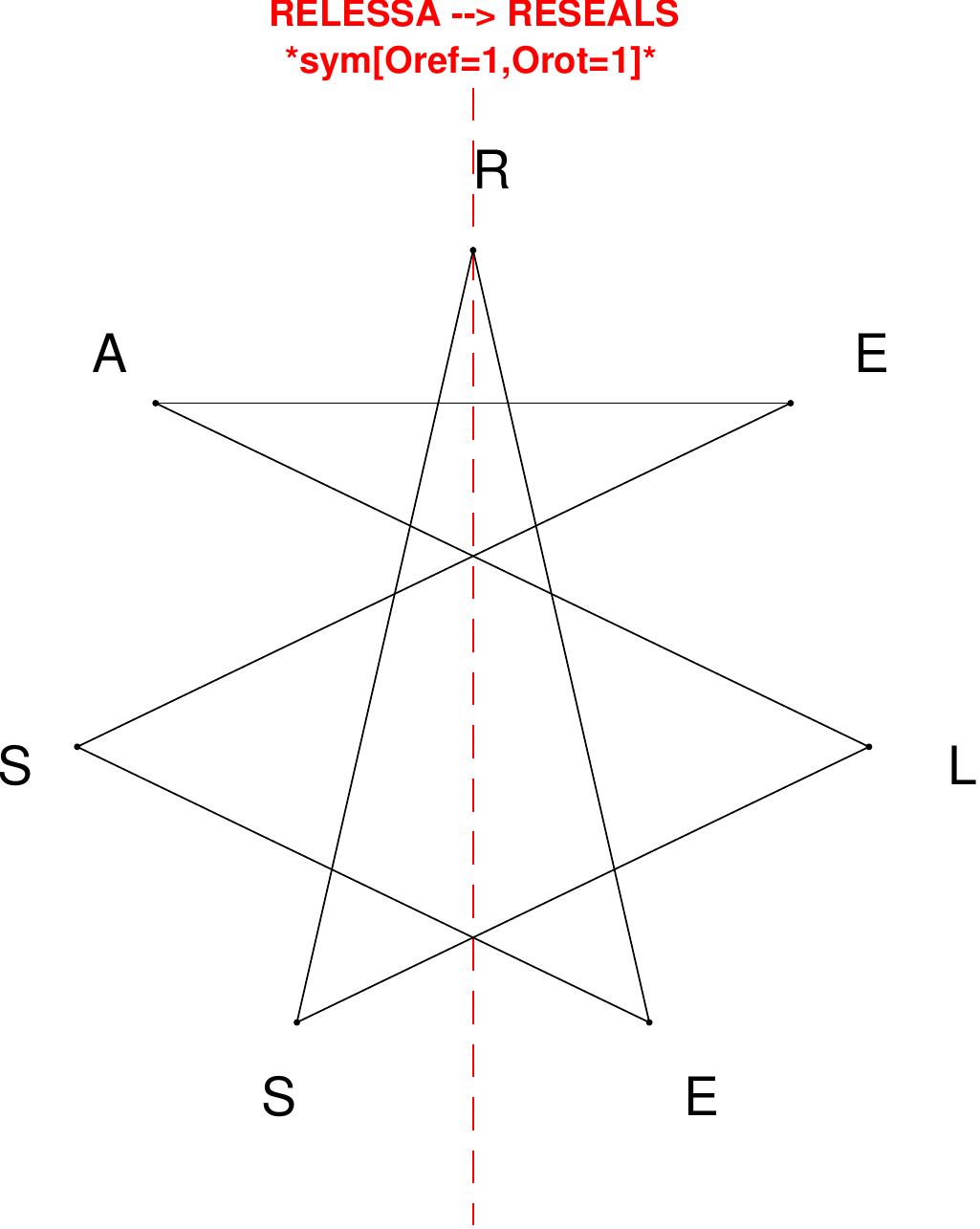}
\end{subfigure}
\hfill
\begin{subfigure}[T]{0.19\textwidth}
\centering
\includegraphics[width=\textwidth]{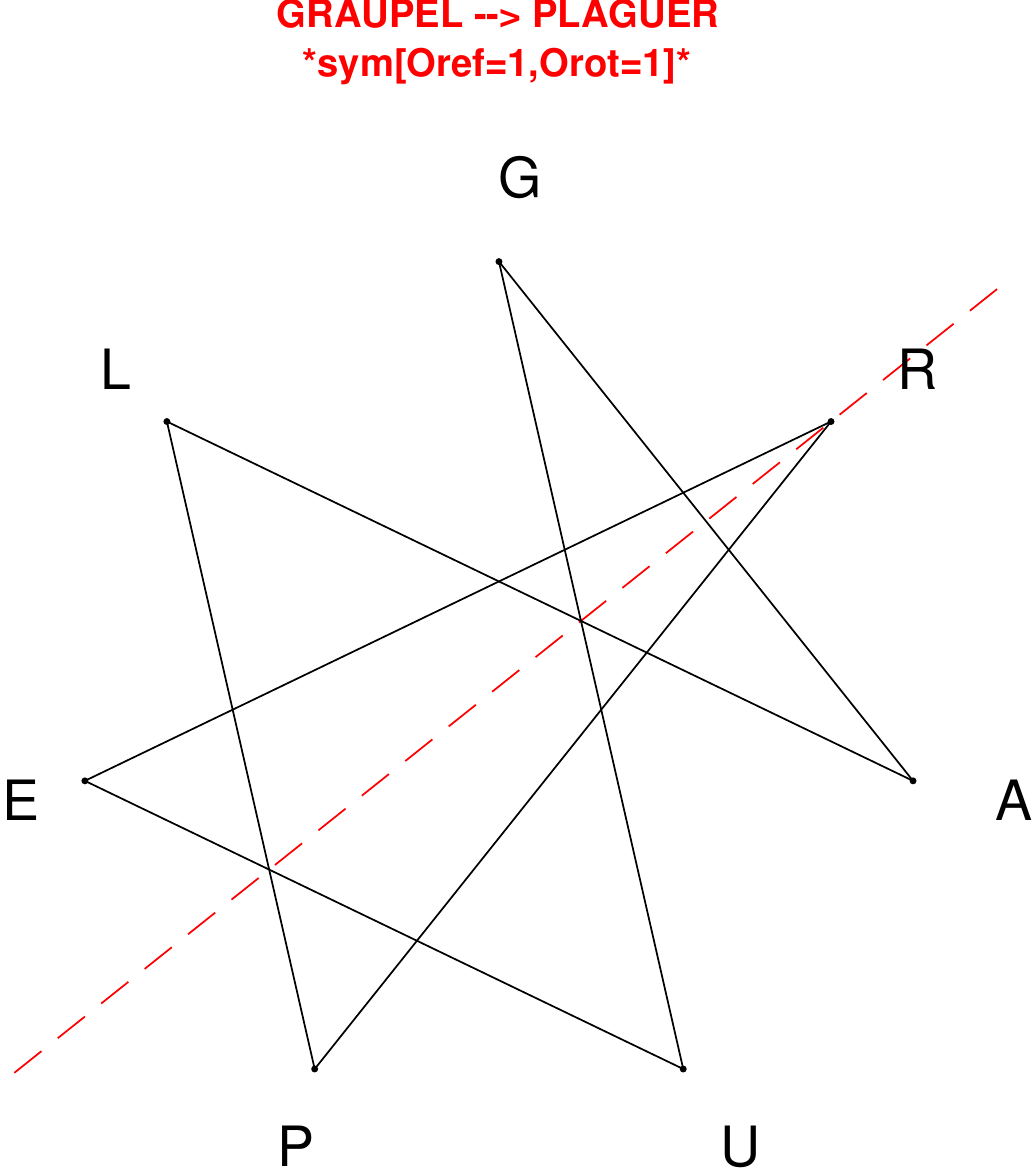}
\end{subfigure}
\end{figure}

\begin{figure}[H]
\centering
\begin{subfigure}[T]{0.19\textwidth}
\centering
\includegraphics[width=\textwidth]{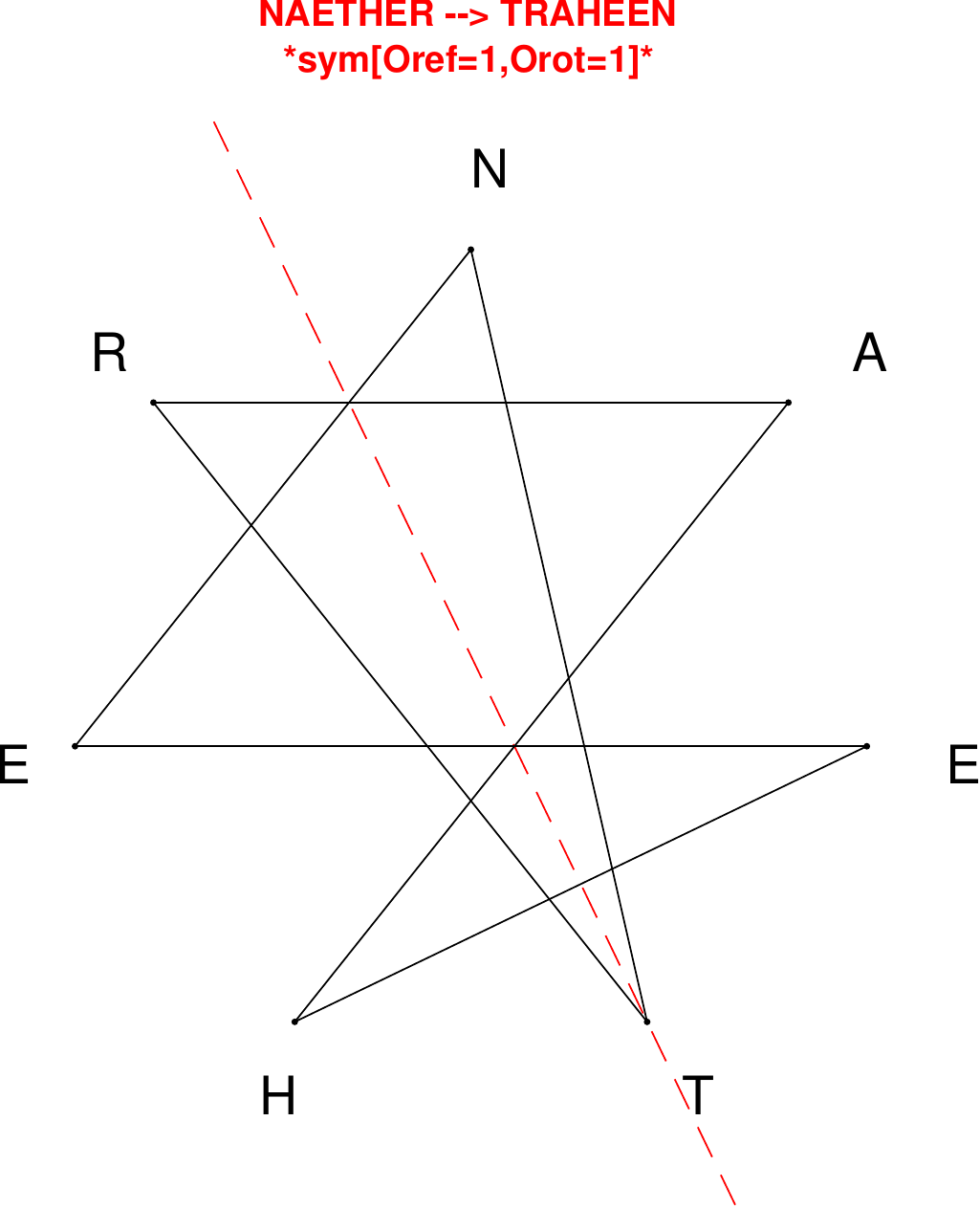}
\end{subfigure}
\hfill
\begin{subfigure}[T]{0.19\textwidth}
\centering
\includegraphics[width=\textwidth]{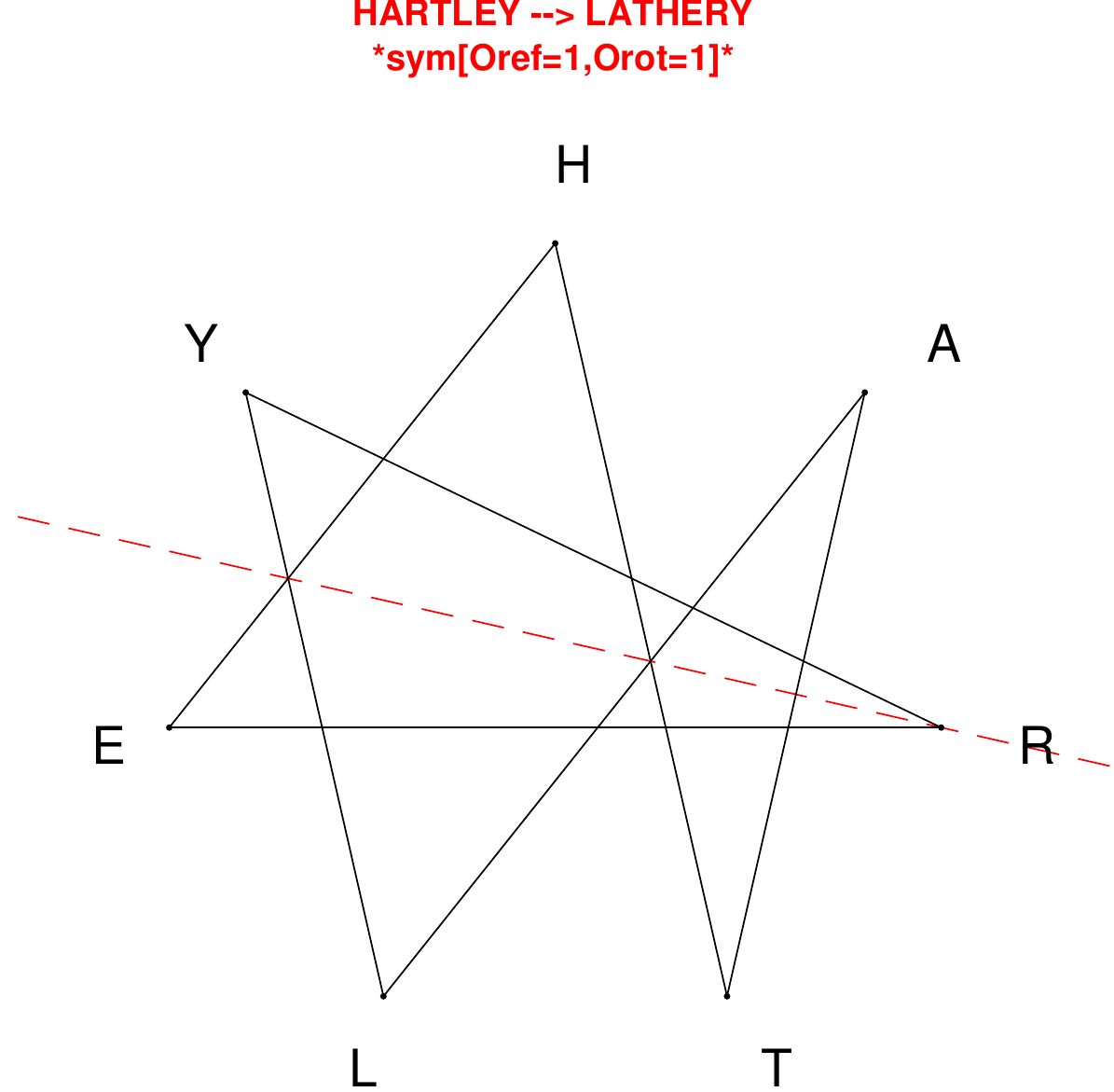}
\end{subfigure}
\hfill
\begin{subfigure}[T]{0.19\textwidth}
\centering
\includegraphics[width=\textwidth]{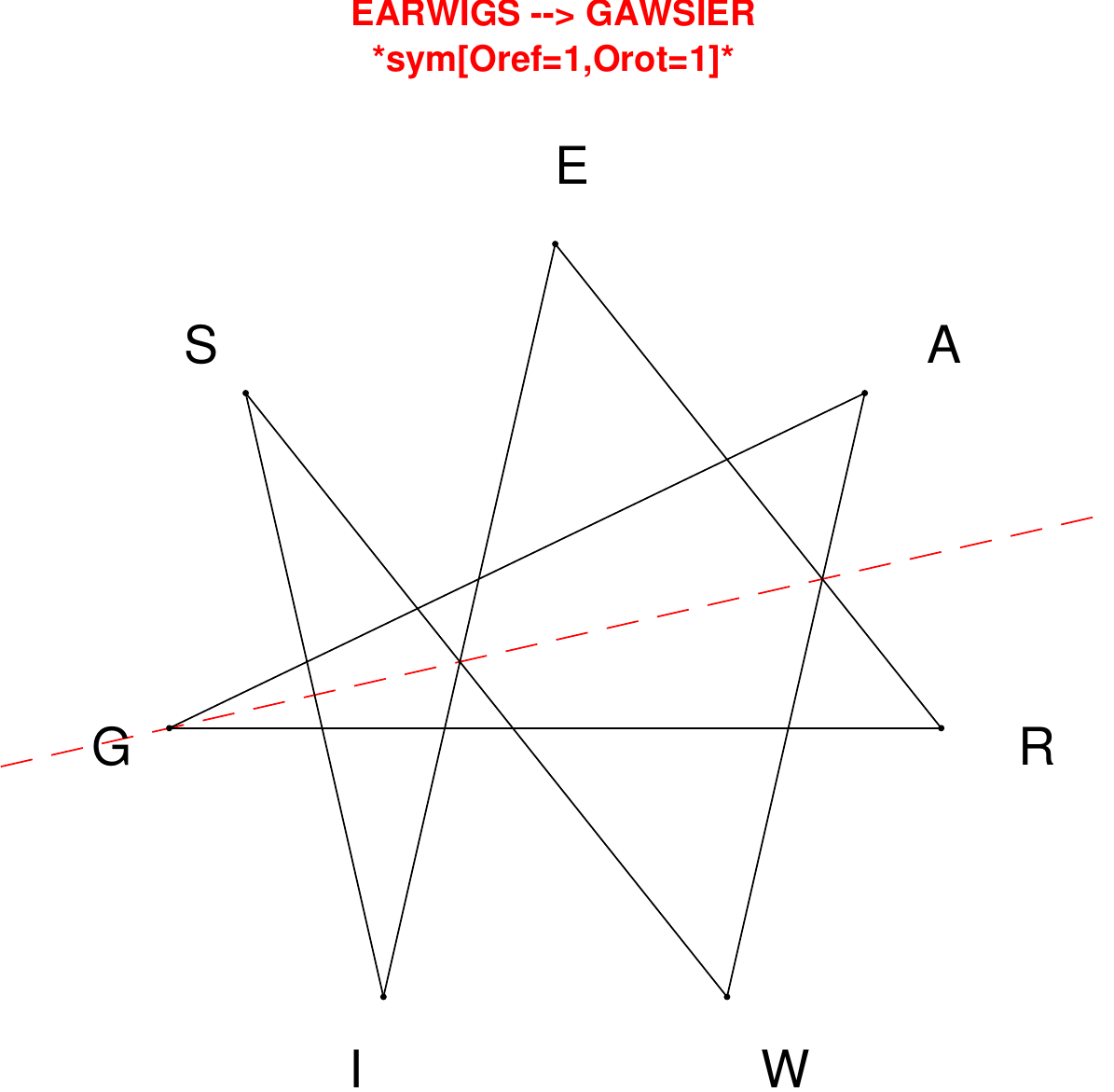}
\end{subfigure}
\hfill
\begin{subfigure}[T]{0.19\textwidth}
\centering
\includegraphics[width=\textwidth]{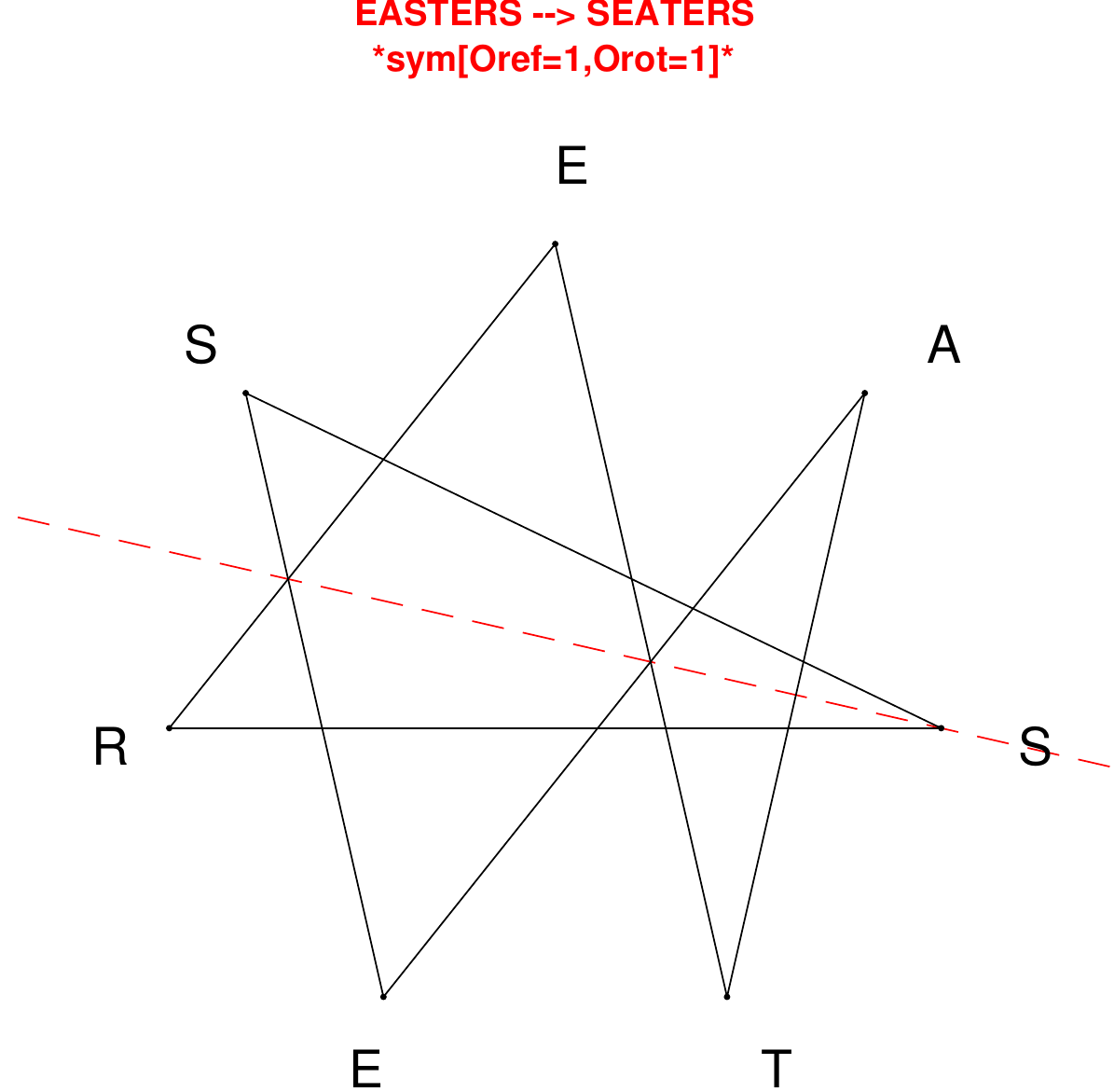}
\end{subfigure}
\hfill
\begin{subfigure}[T]{0.19\textwidth}
\centering
\includegraphics[width=\textwidth]{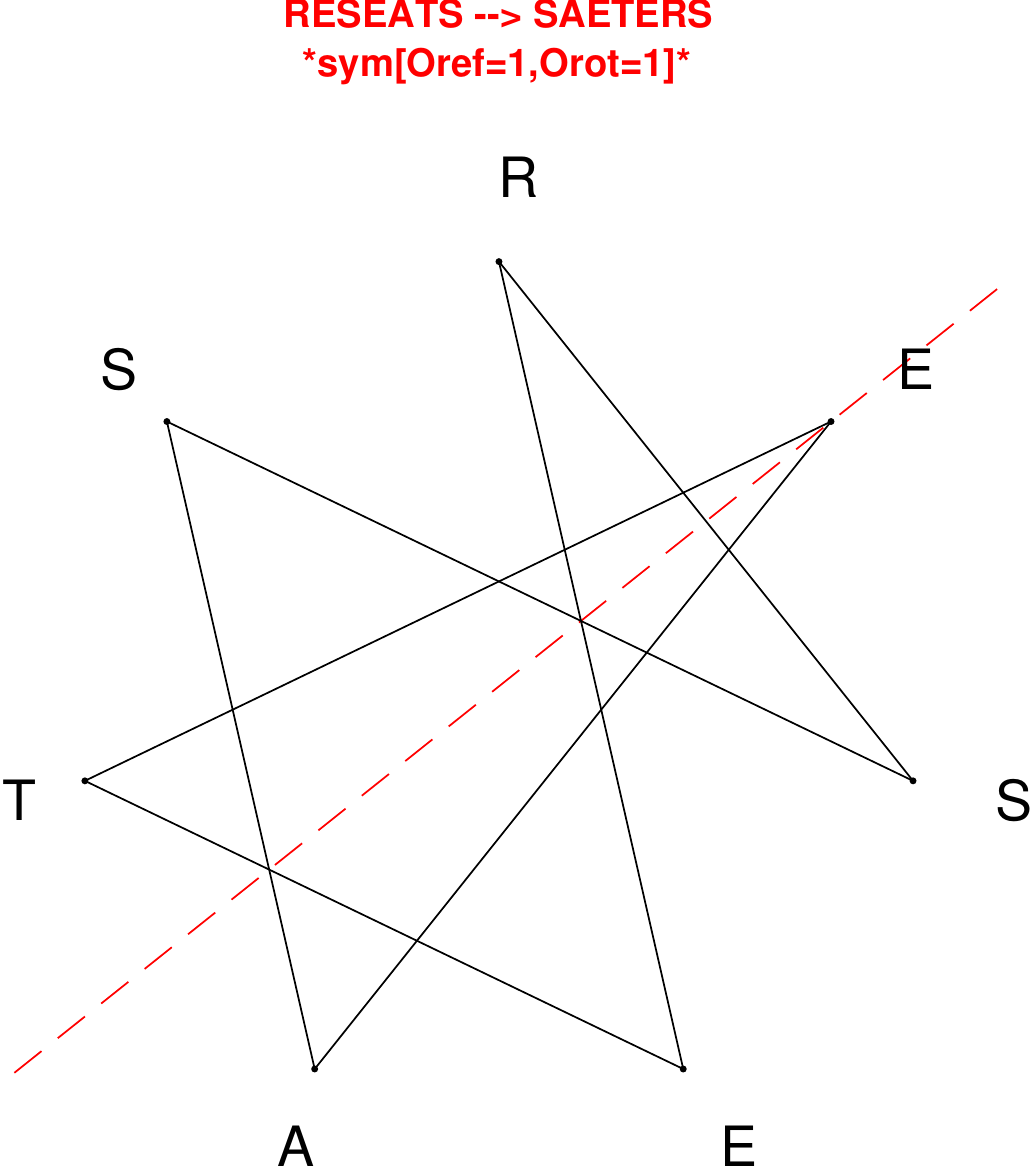}
\end{subfigure}
\end{figure}

\begin{figure}[H]
\centering
\begin{subfigure}[T]{0.19\textwidth}
\centering
\includegraphics[width=\textwidth]{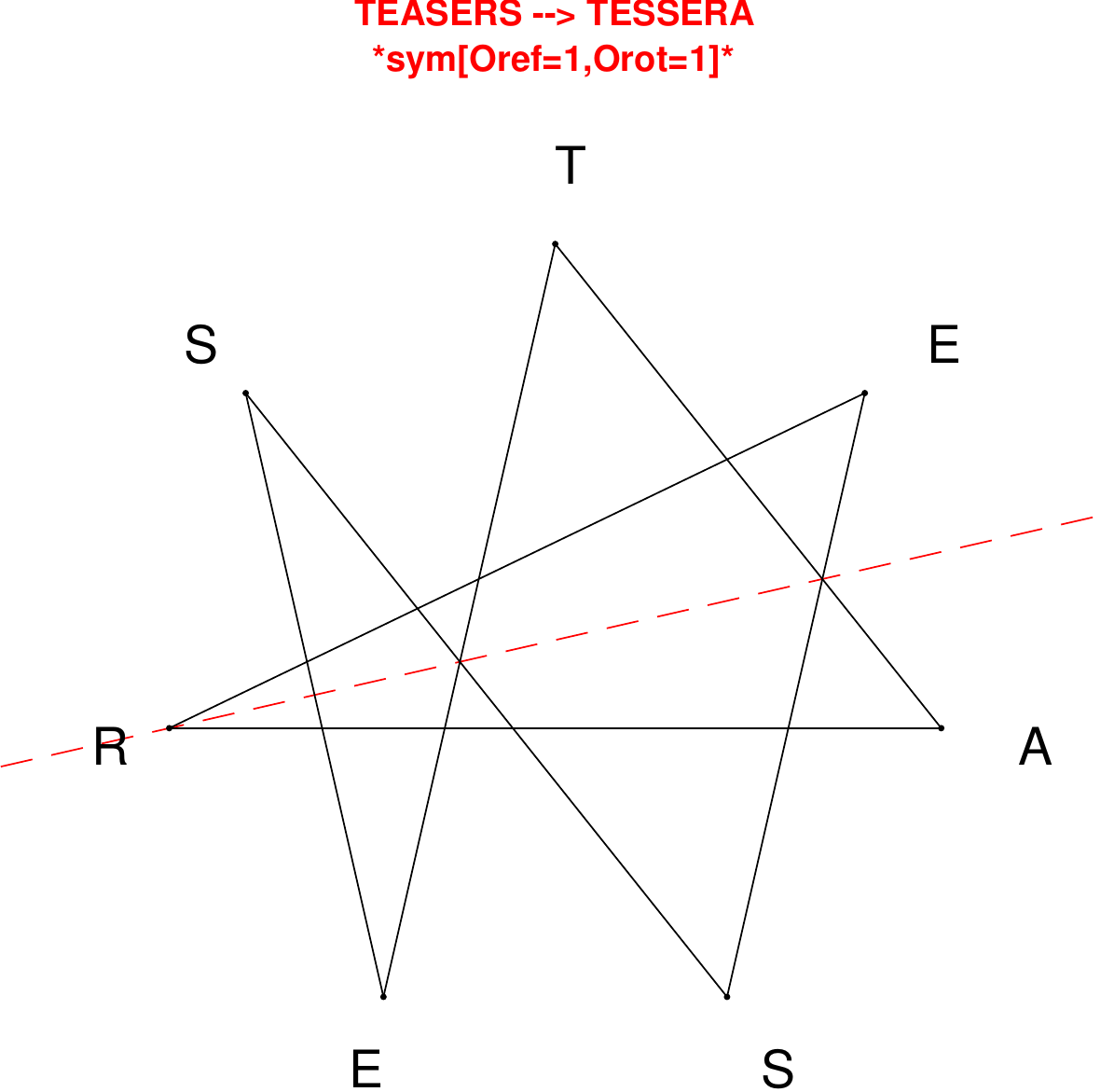}
\end{subfigure}
\hfill
\begin{subfigure}[T]{0.19\textwidth}
\centering
\includegraphics[width=\textwidth]{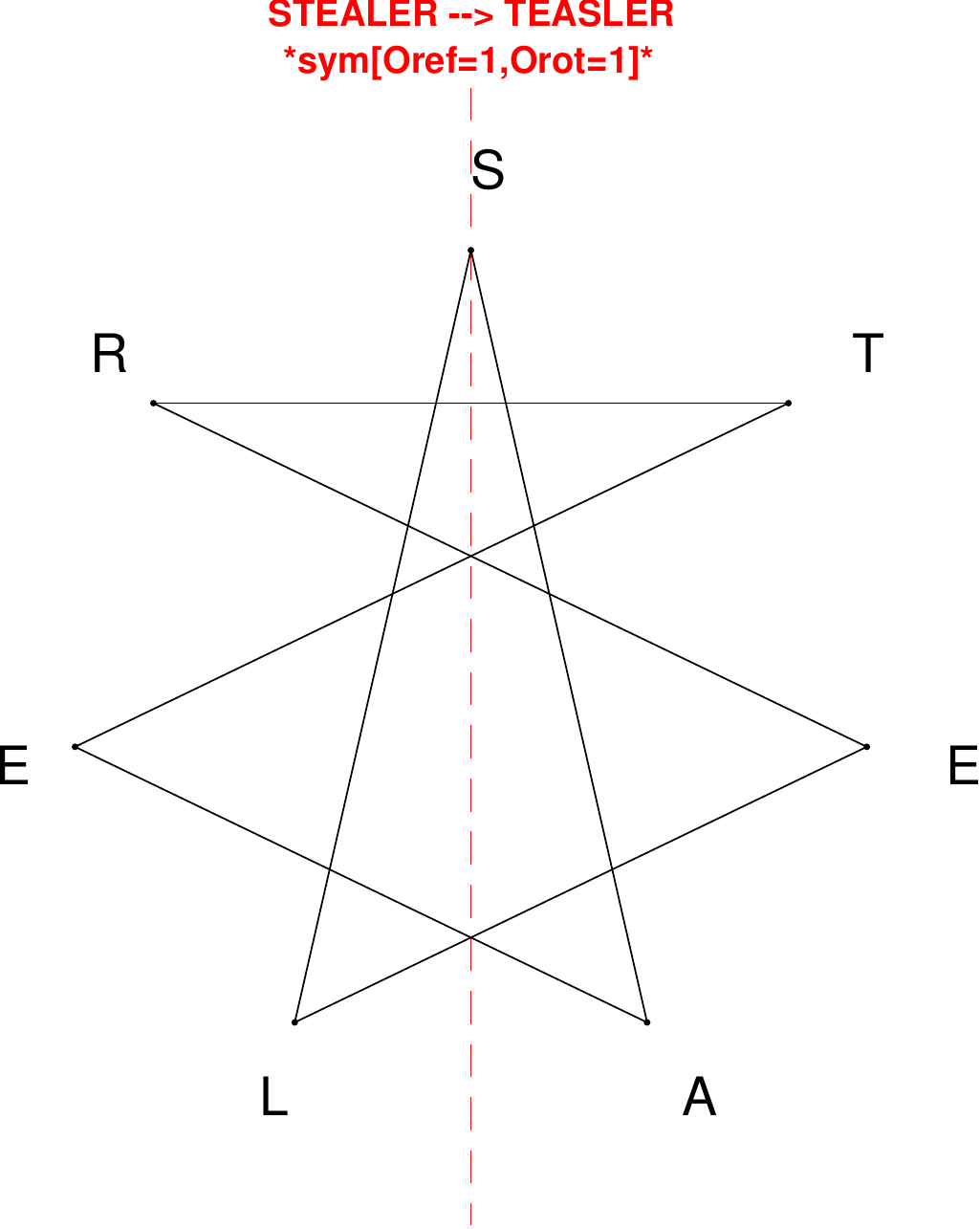}
\end{subfigure}
\hfill
\begin{subfigure}[T]{0.19\textwidth}
\centering
\includegraphics[width=\textwidth]{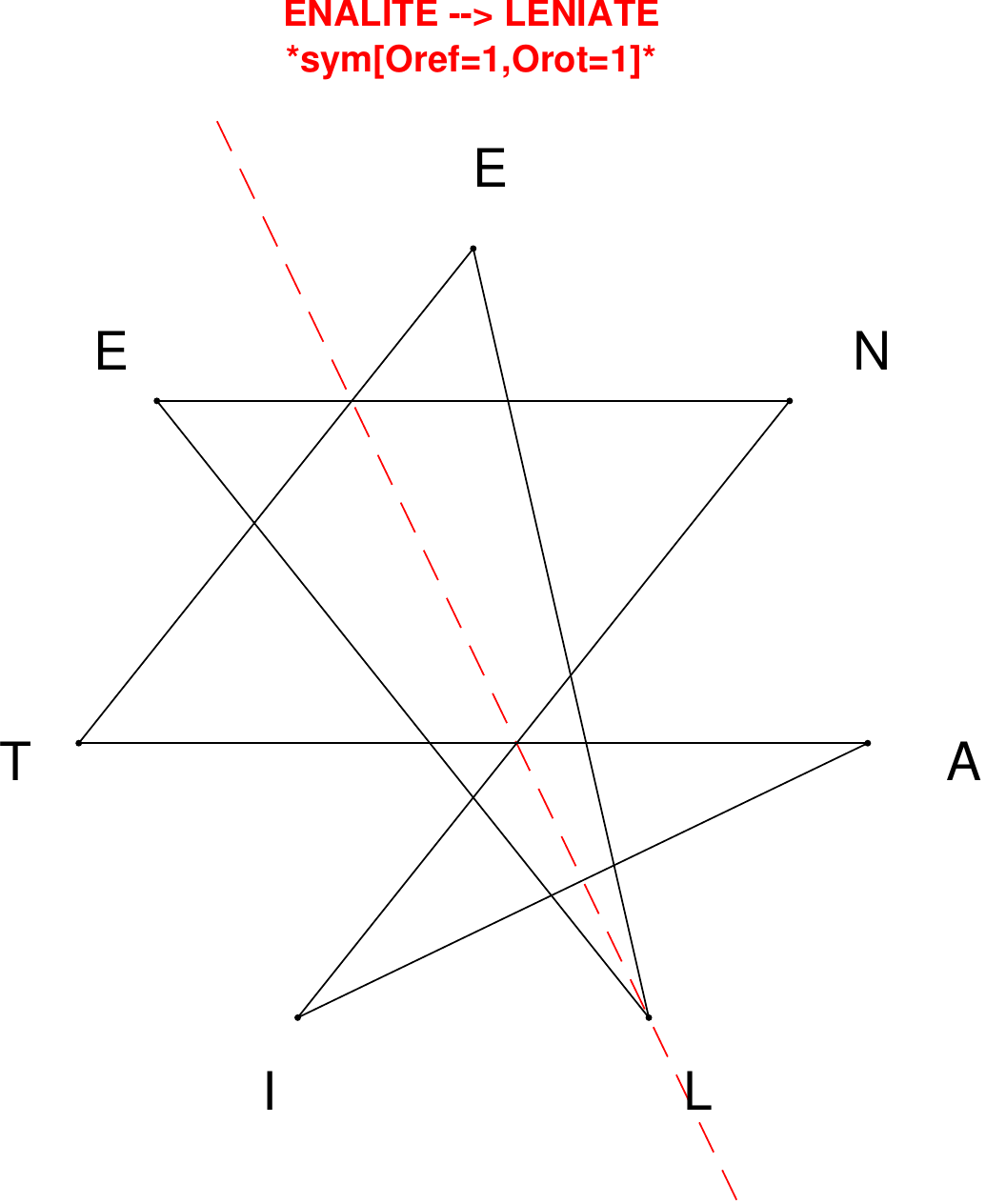}
\end{subfigure}
\hfill
\begin{subfigure}[T]{0.19\textwidth}
\centering
\includegraphics[width=\textwidth]{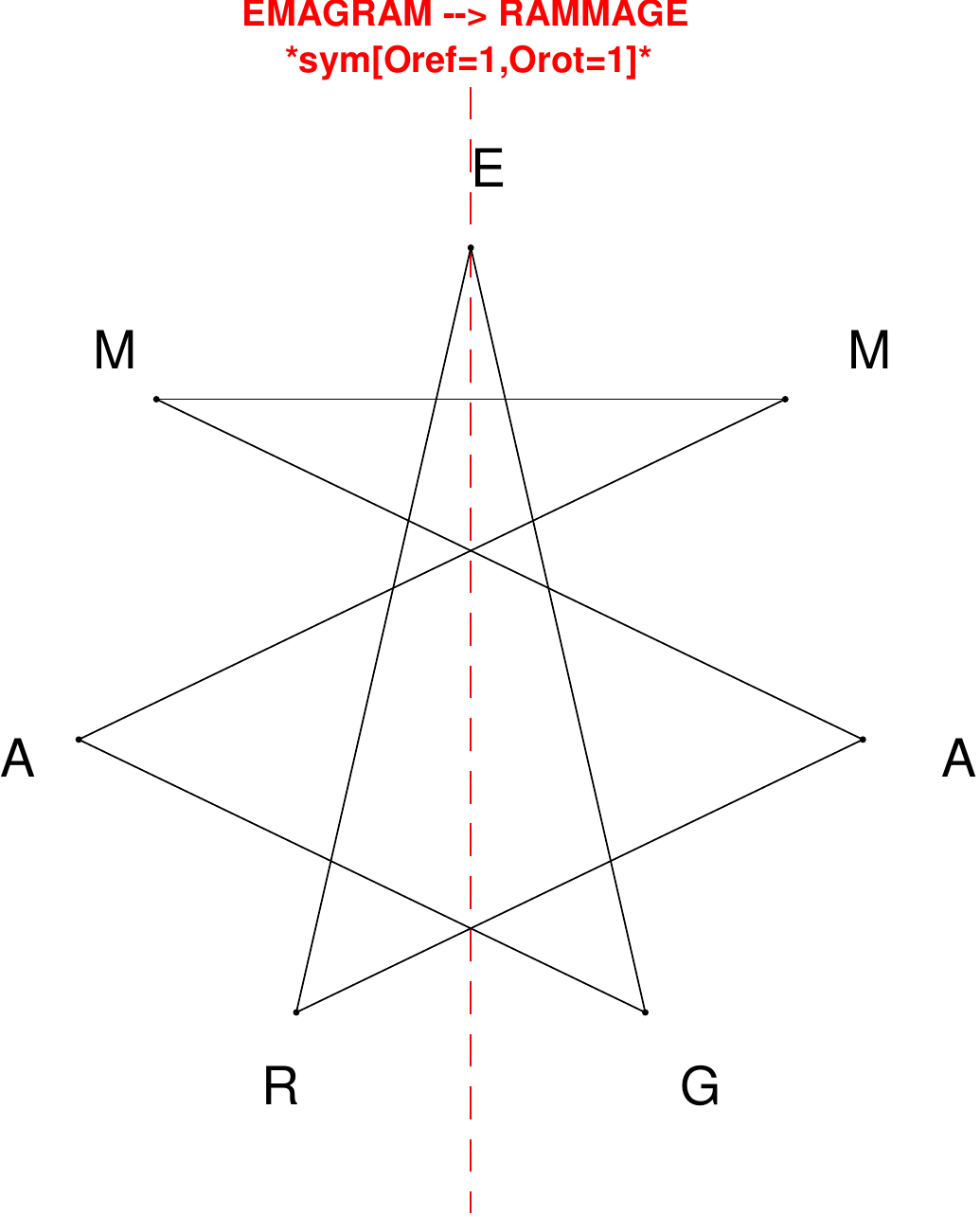}
\end{subfigure}
\hfill
\begin{subfigure}[T]{0.19\textwidth}
\centering
\includegraphics[width=\textwidth]{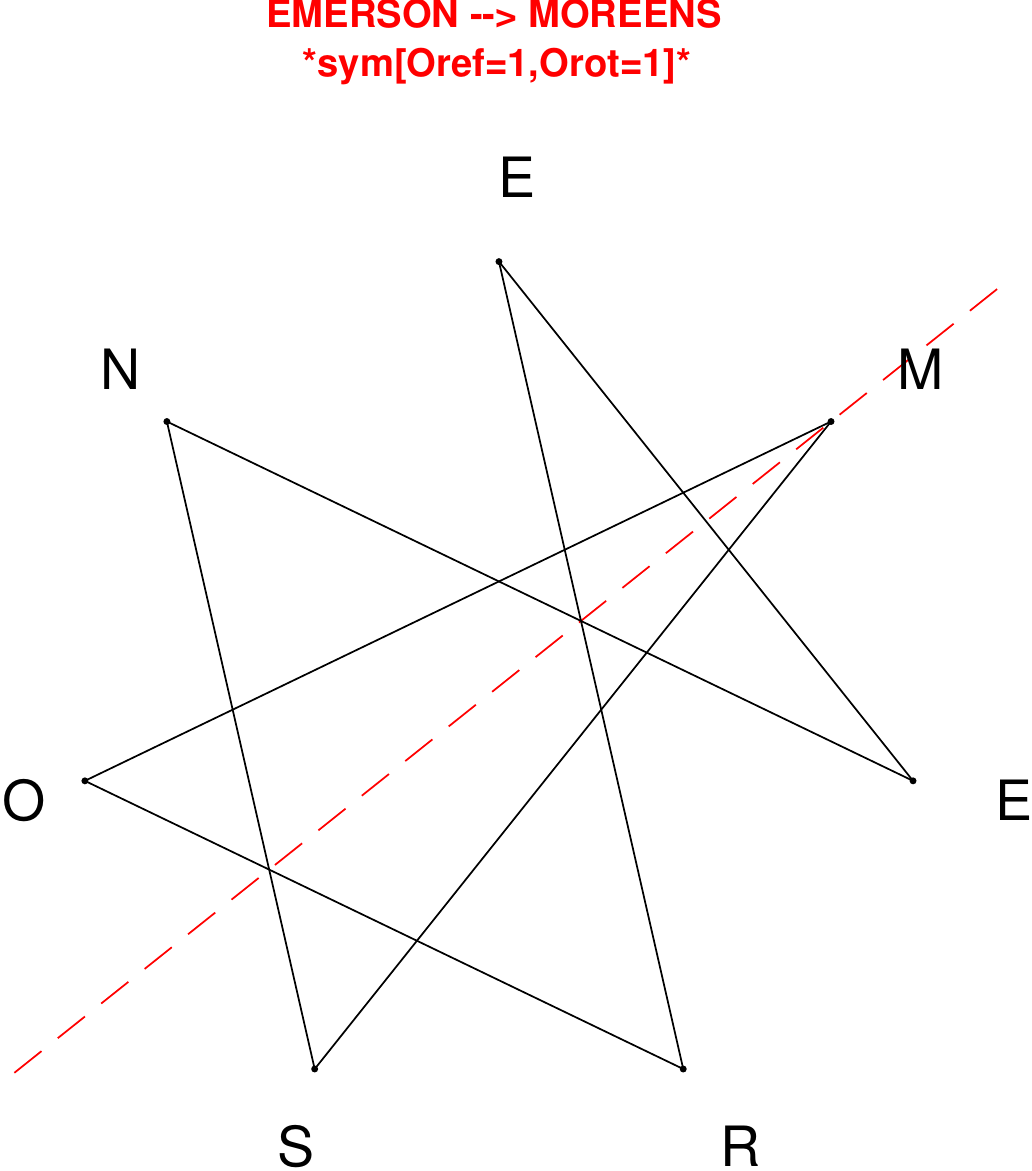}
\end{subfigure}
\end{figure}

\begin{figure}[H]
\centering
\begin{subfigure}[T]{0.19\textwidth}
\centering
\includegraphics[width=\textwidth]{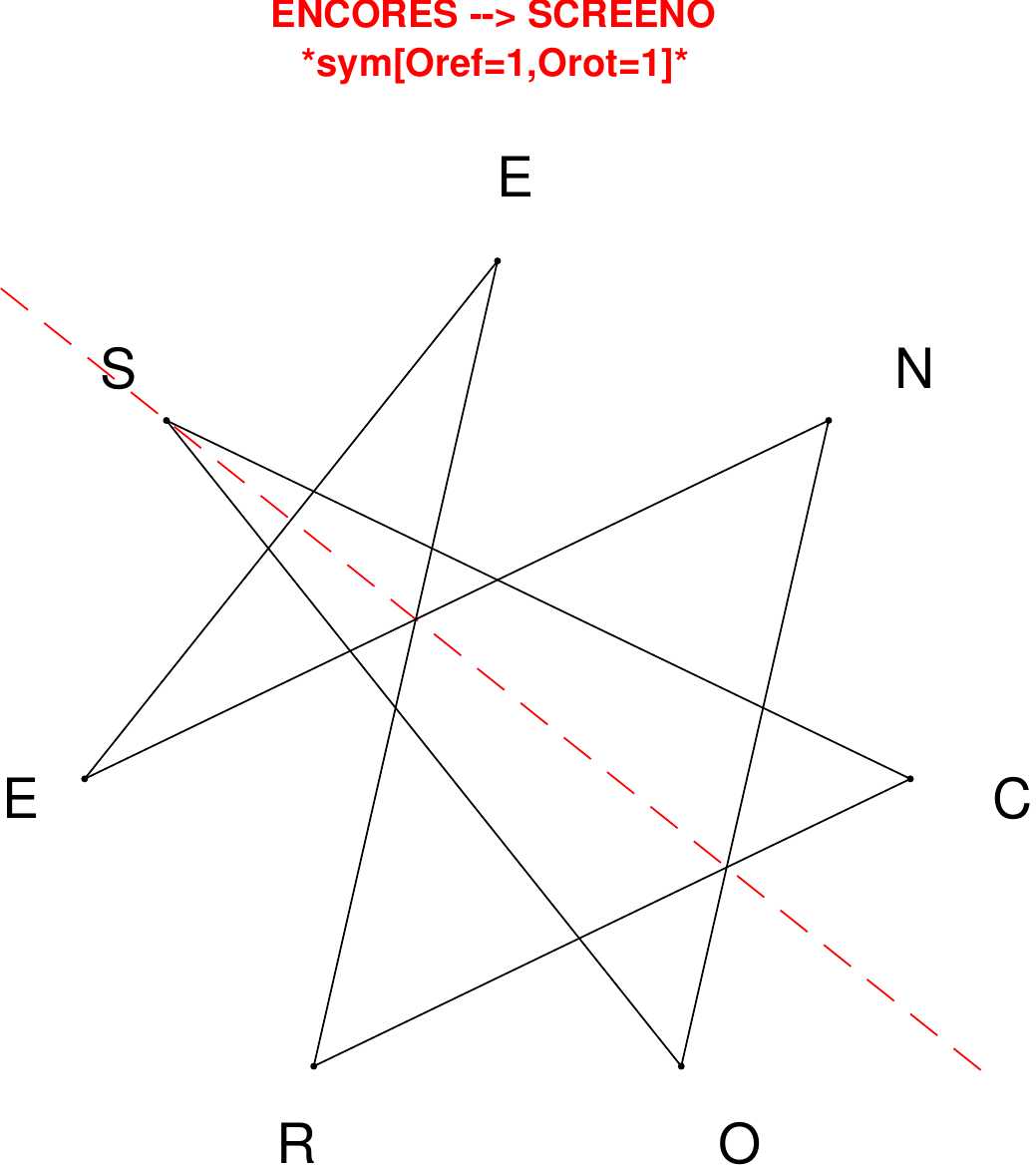}
\end{subfigure}
\hfill
\begin{subfigure}[T]{0.19\textwidth}
\centering
\includegraphics[width=\textwidth]{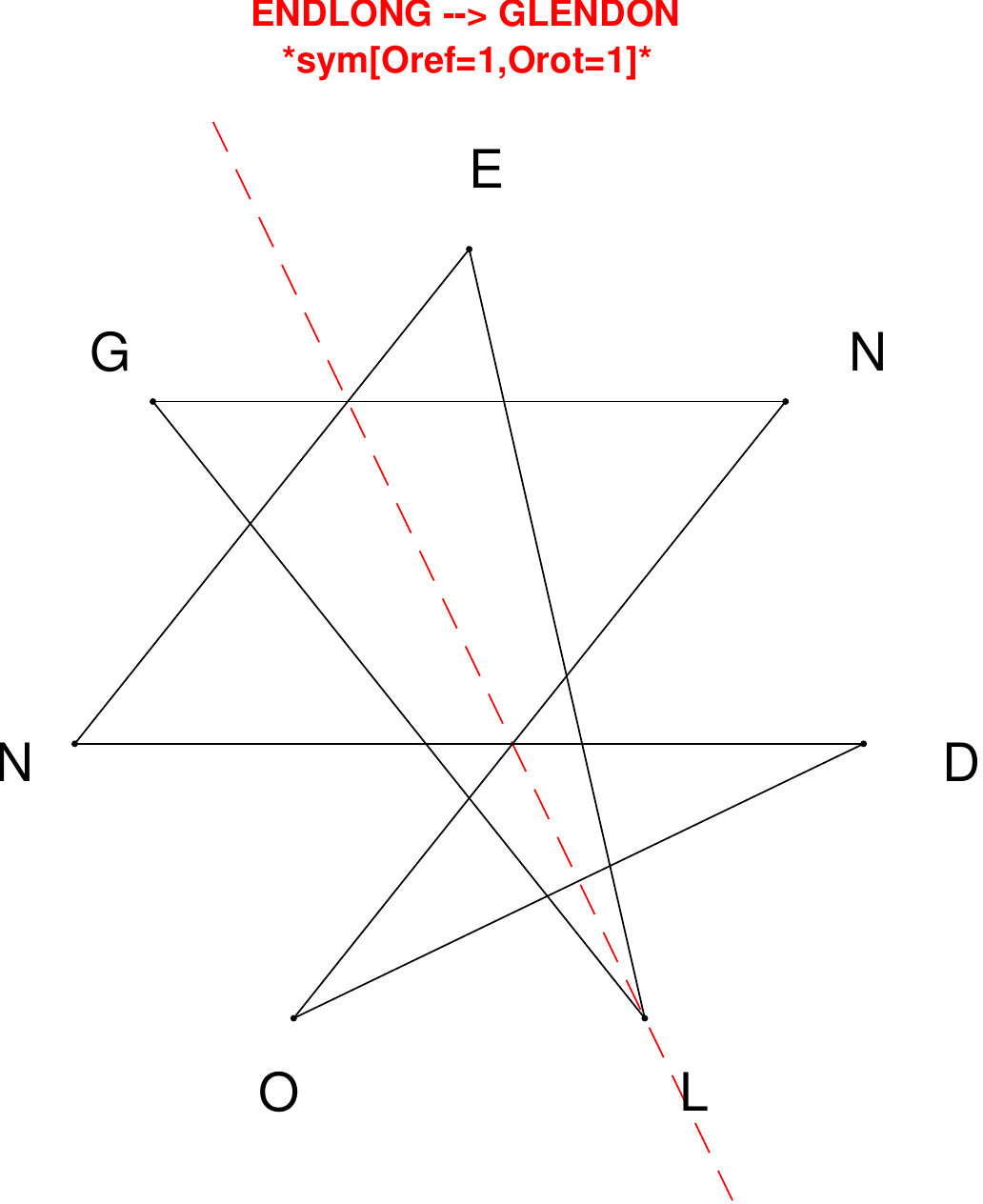}
\end{subfigure}
\hfill
\begin{subfigure}[T]{0.19\textwidth}
\centering
\includegraphics[width=\textwidth]{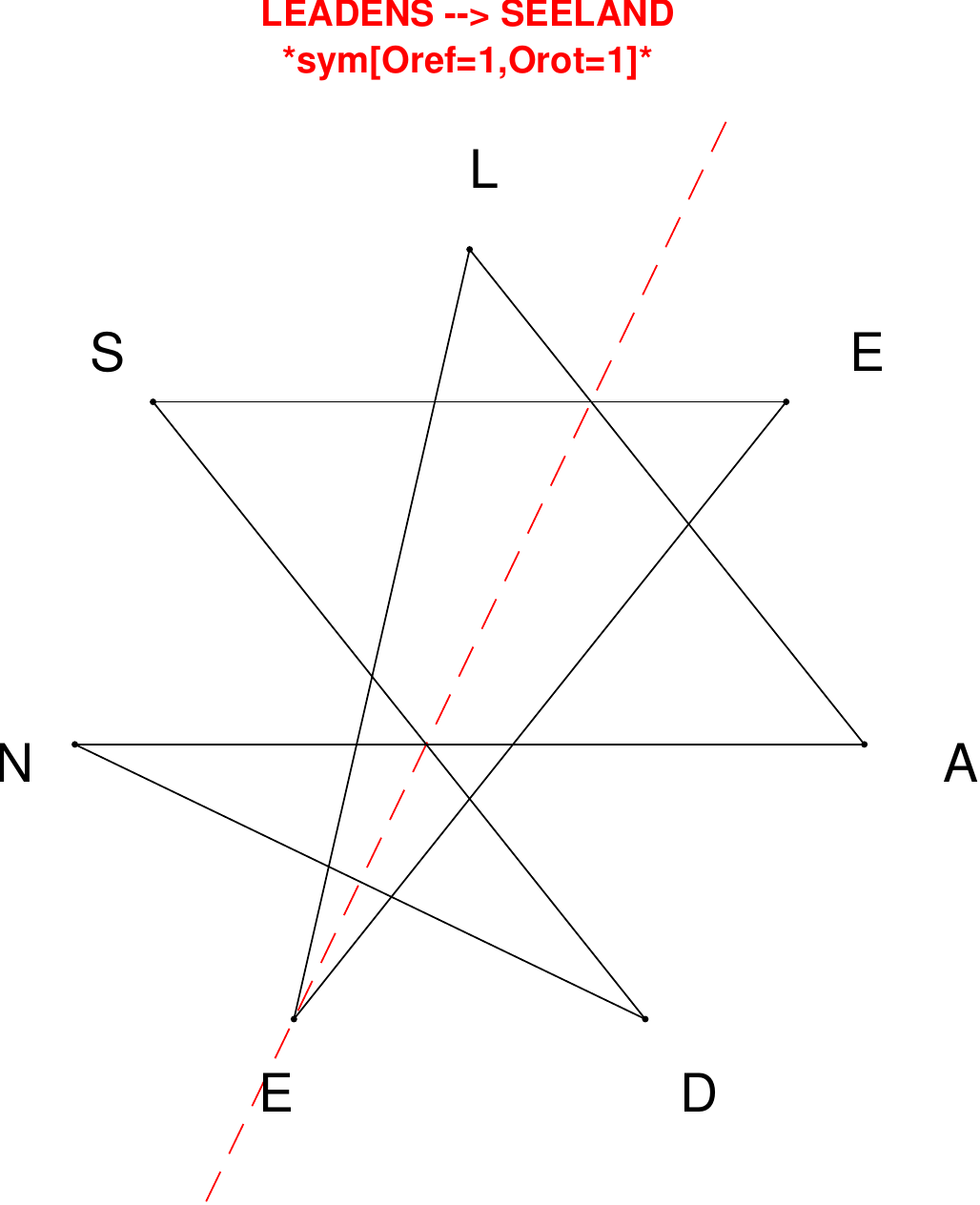}
\end{subfigure}
\hfill
\begin{subfigure}[T]{0.19\textwidth}
\centering
\includegraphics[width=\textwidth]{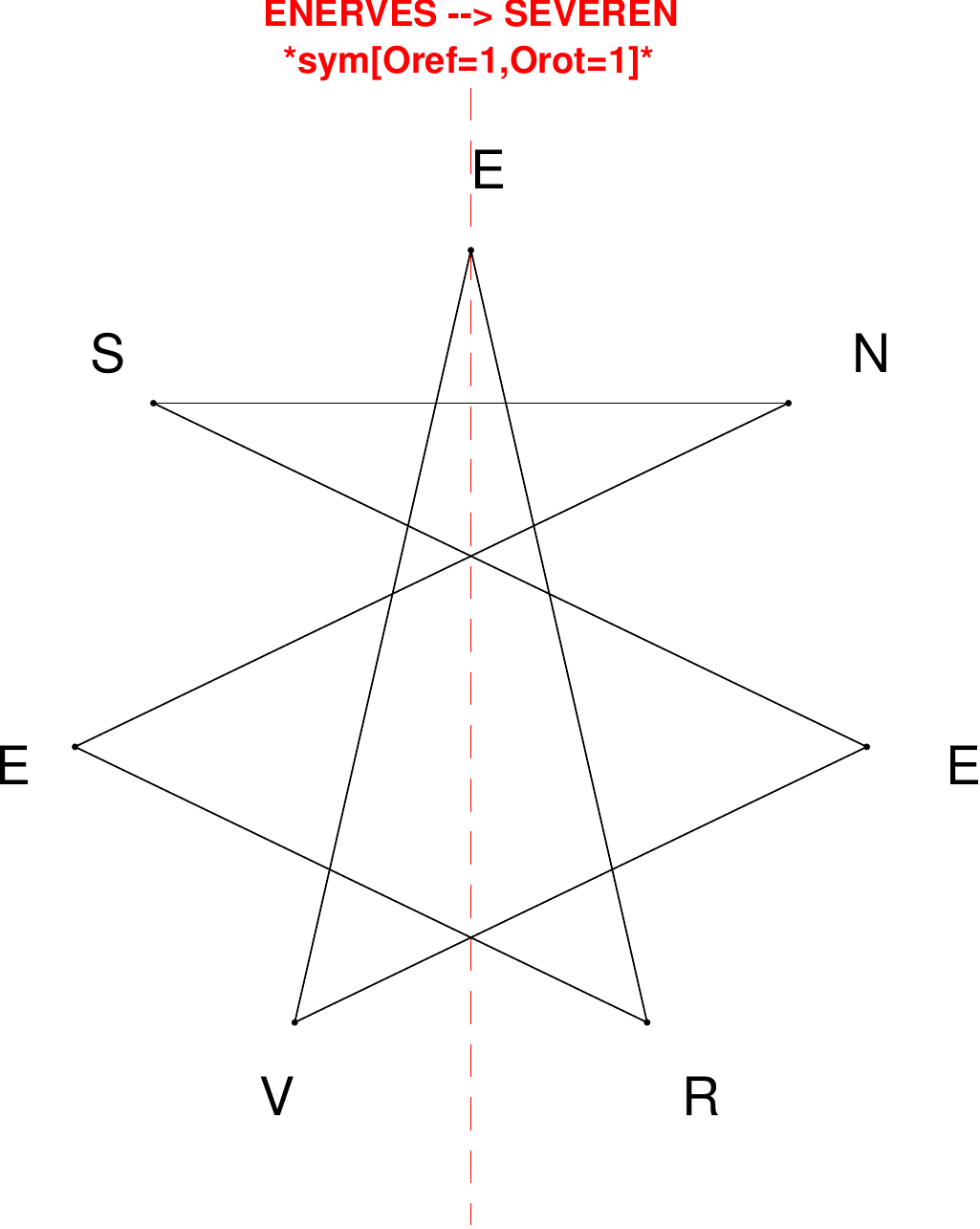}
\end{subfigure}
\hfill
\begin{subfigure}[T]{0.19\textwidth}
\centering
\includegraphics[width=\textwidth]{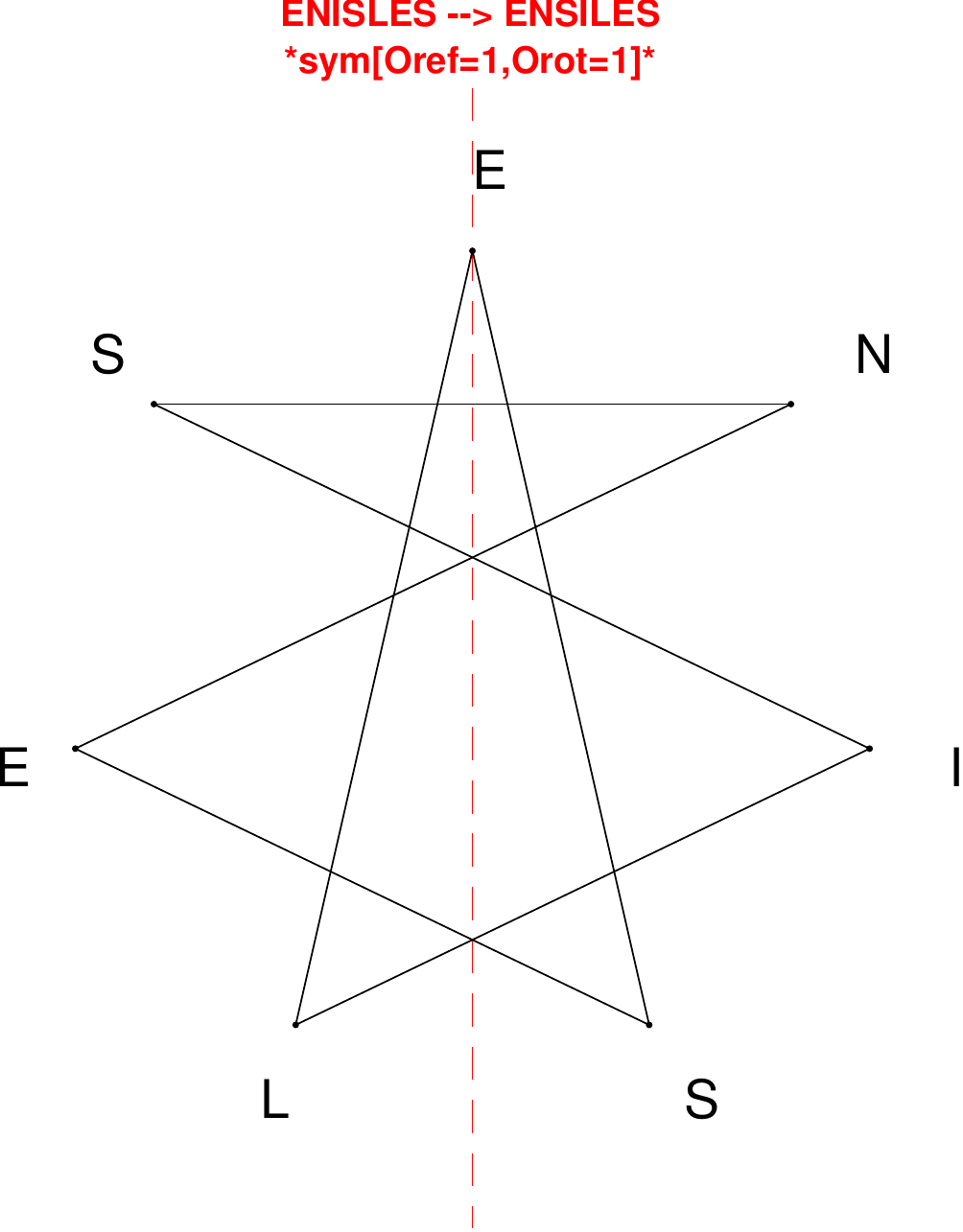}
\end{subfigure}
\end{figure}

\begin{figure}[H]
\centering
\begin{subfigure}[T]{0.19\textwidth}
\centering
\includegraphics[width=\textwidth]{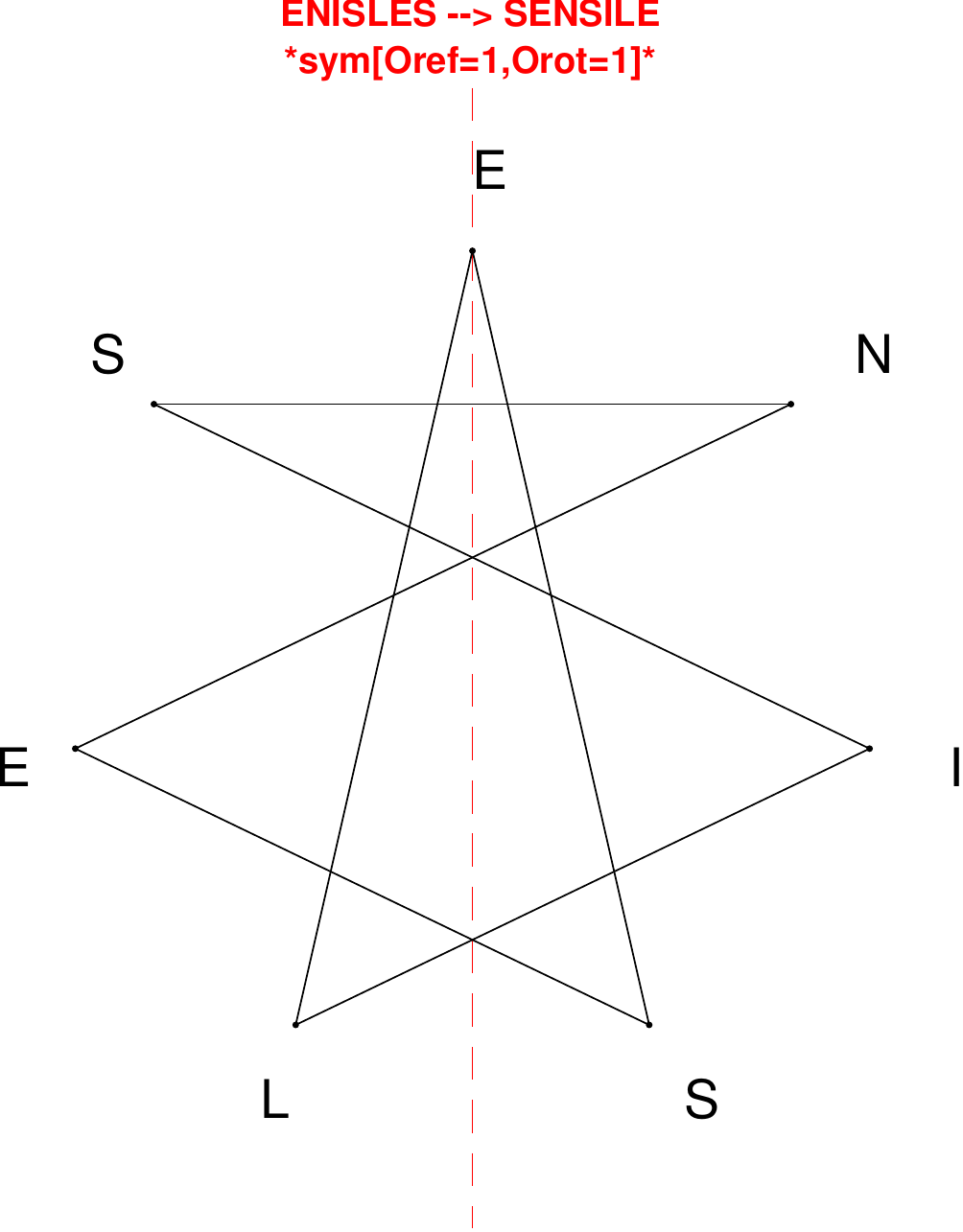}
\end{subfigure}
\hfill
\begin{subfigure}[T]{0.19\textwidth}
\centering
\includegraphics[width=\textwidth]{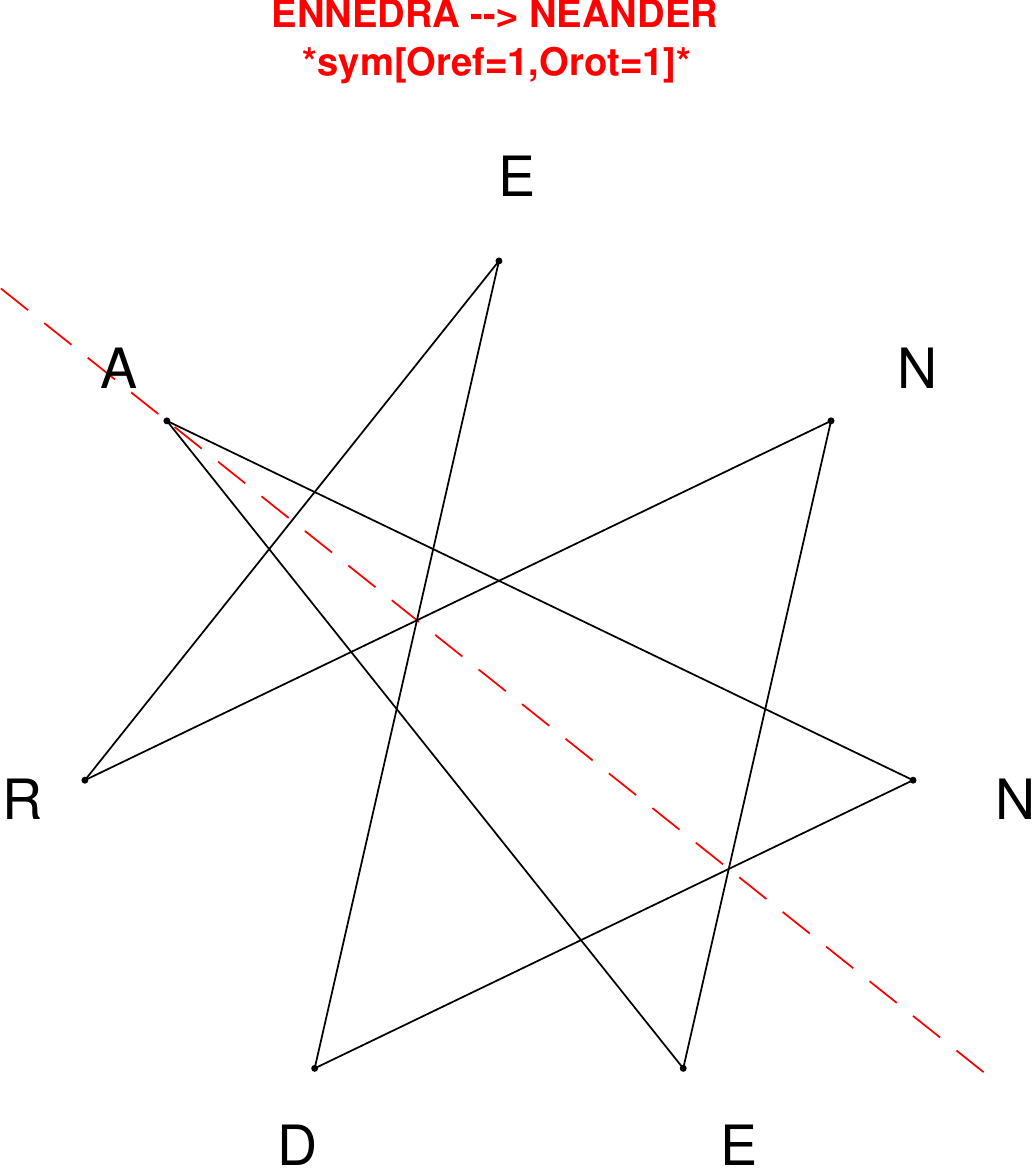}
\end{subfigure}
\hfill
\begin{subfigure}[T]{0.19\textwidth}
\centering
\includegraphics[width=\textwidth]{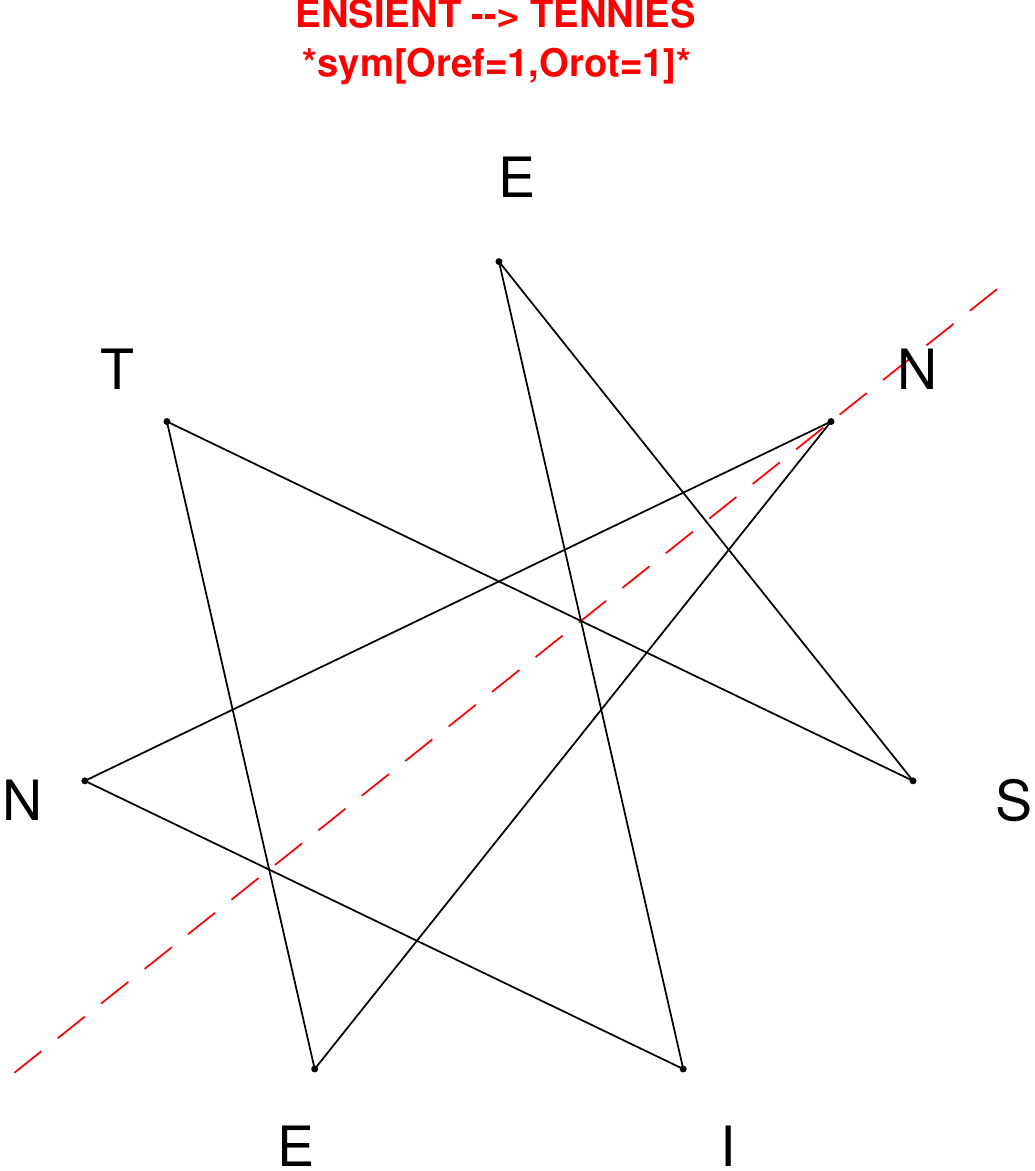}
\end{subfigure}
\hfill
\begin{subfigure}[T]{0.19\textwidth}
\centering
\includegraphics[width=\textwidth]{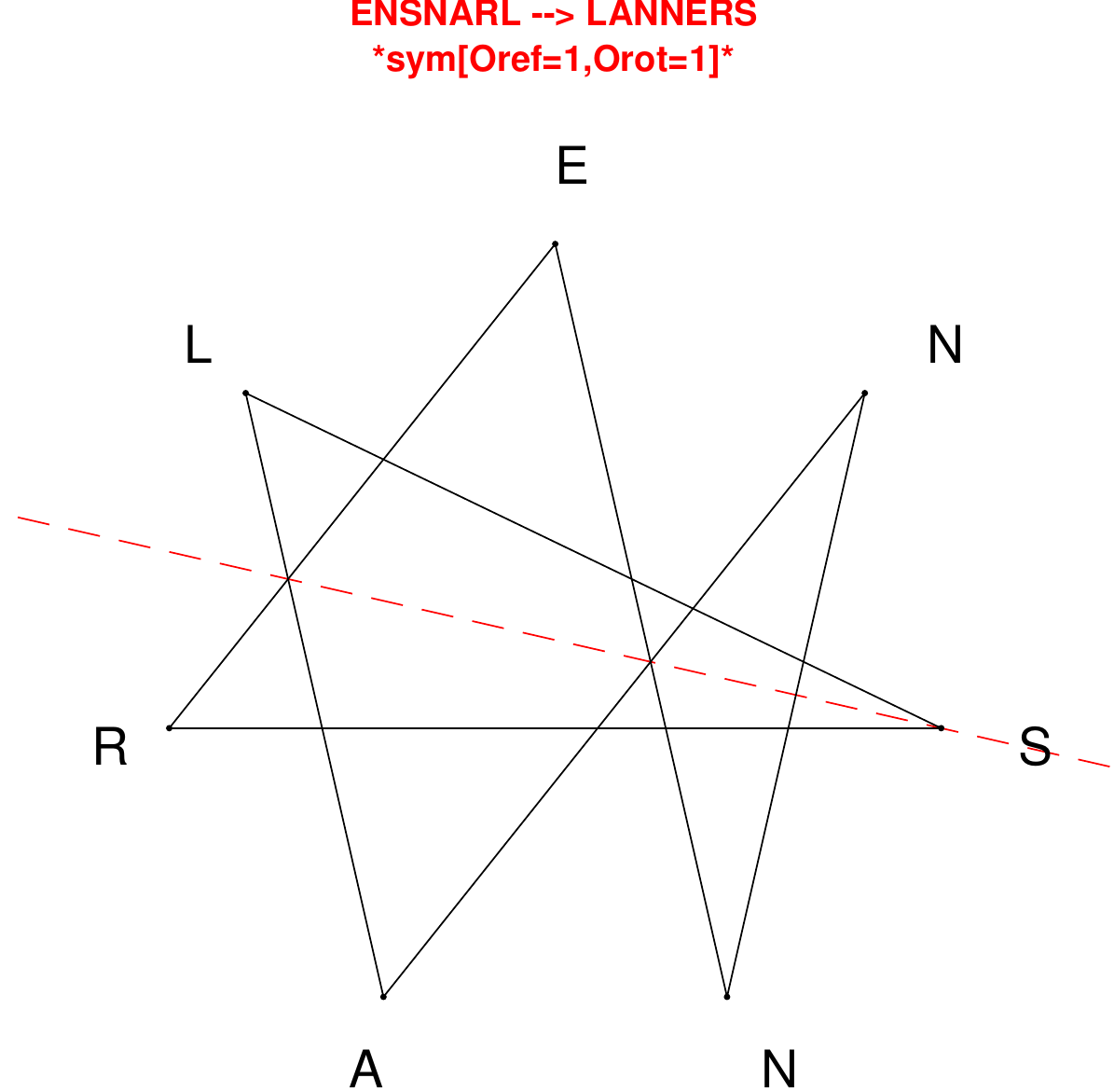}
\end{subfigure}
\hfill
\begin{subfigure}[T]{0.19\textwidth}
\centering
\includegraphics[width=\textwidth]{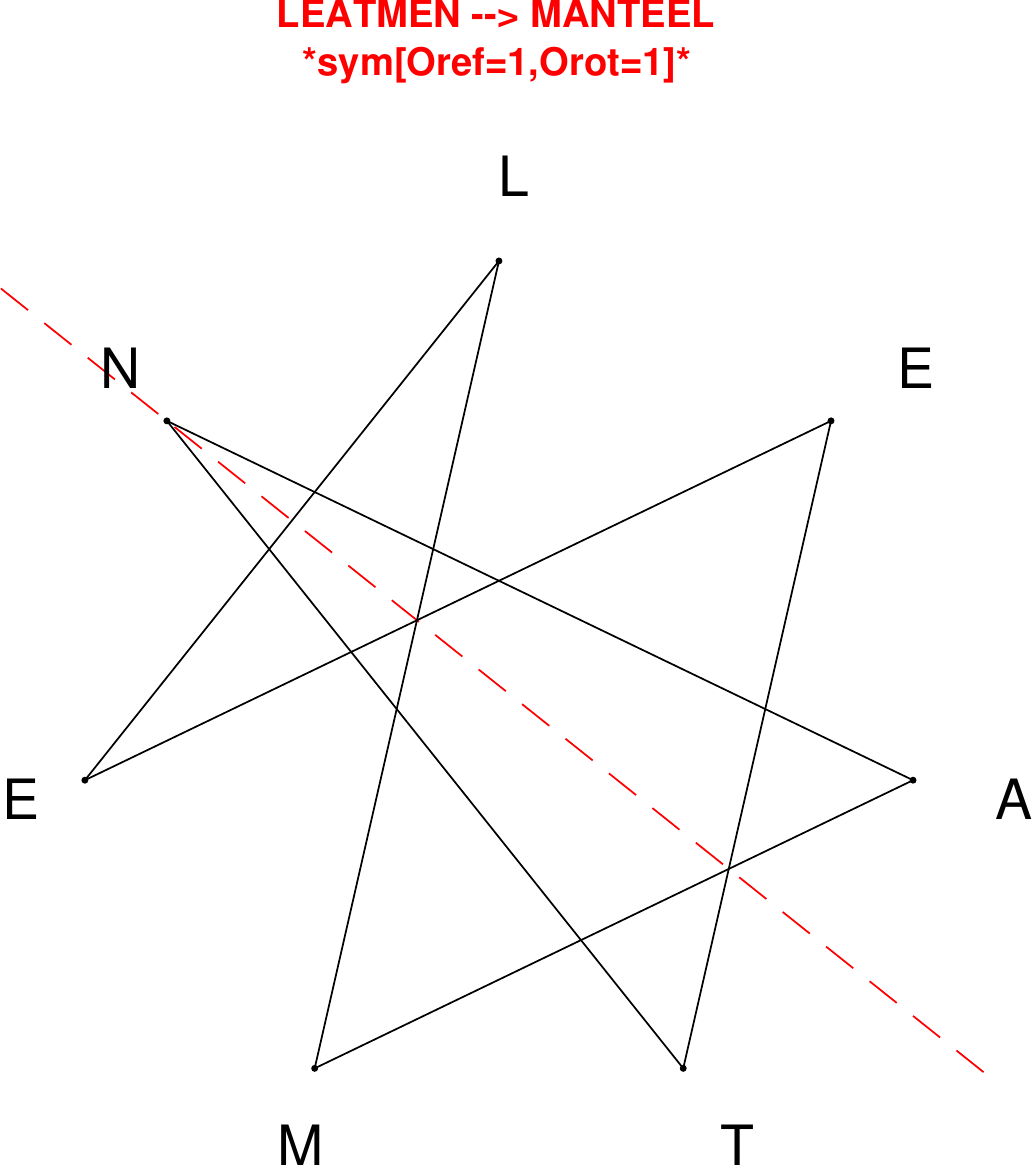}
\end{subfigure}
\end{figure}

\begin{figure}[H]
\centering
\begin{subfigure}[T]{0.19\textwidth}
\centering
\includegraphics[width=\textwidth]{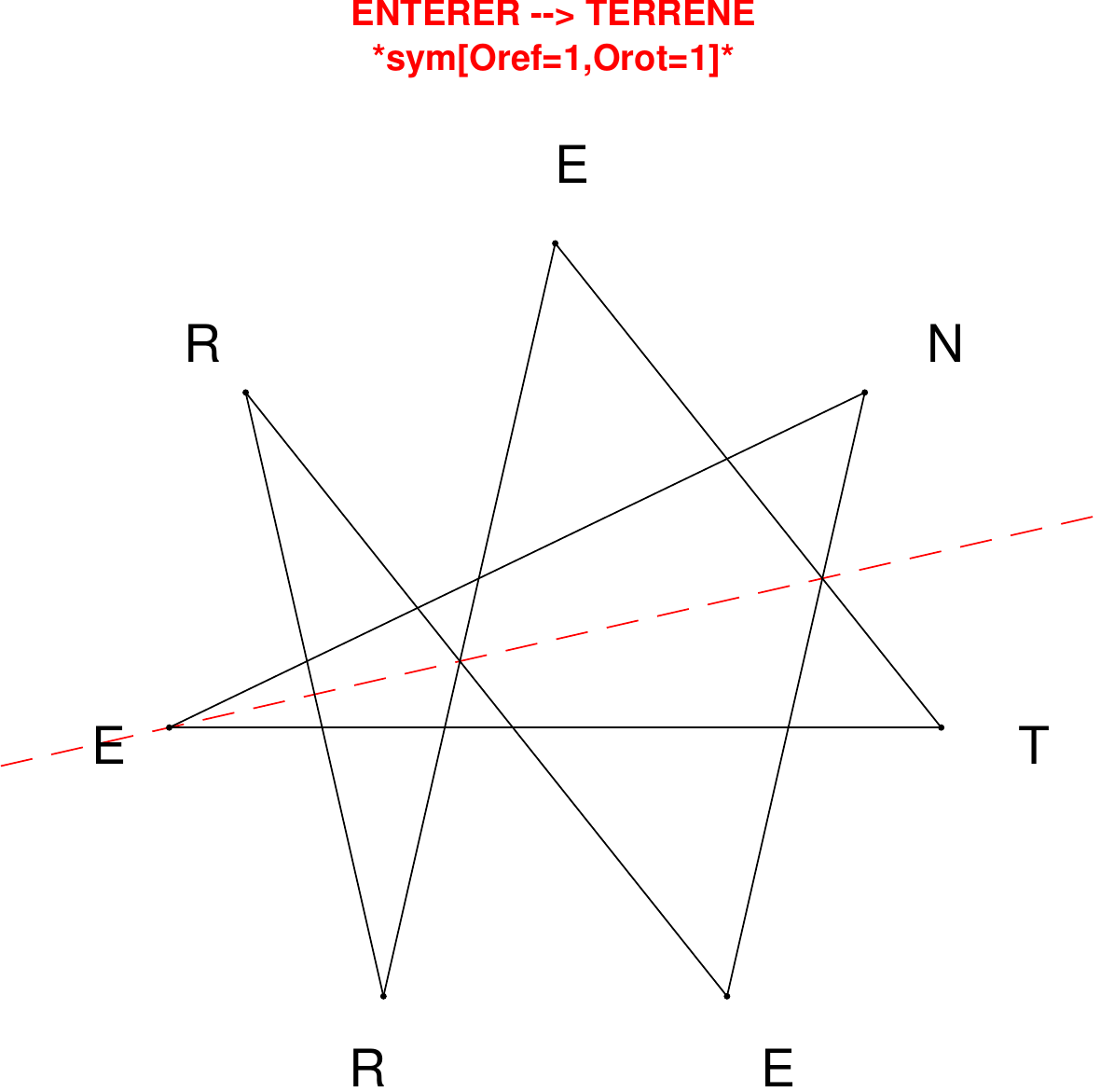}
\end{subfigure}
\hfill
\begin{subfigure}[T]{0.19\textwidth}
\centering
\includegraphics[width=\textwidth]{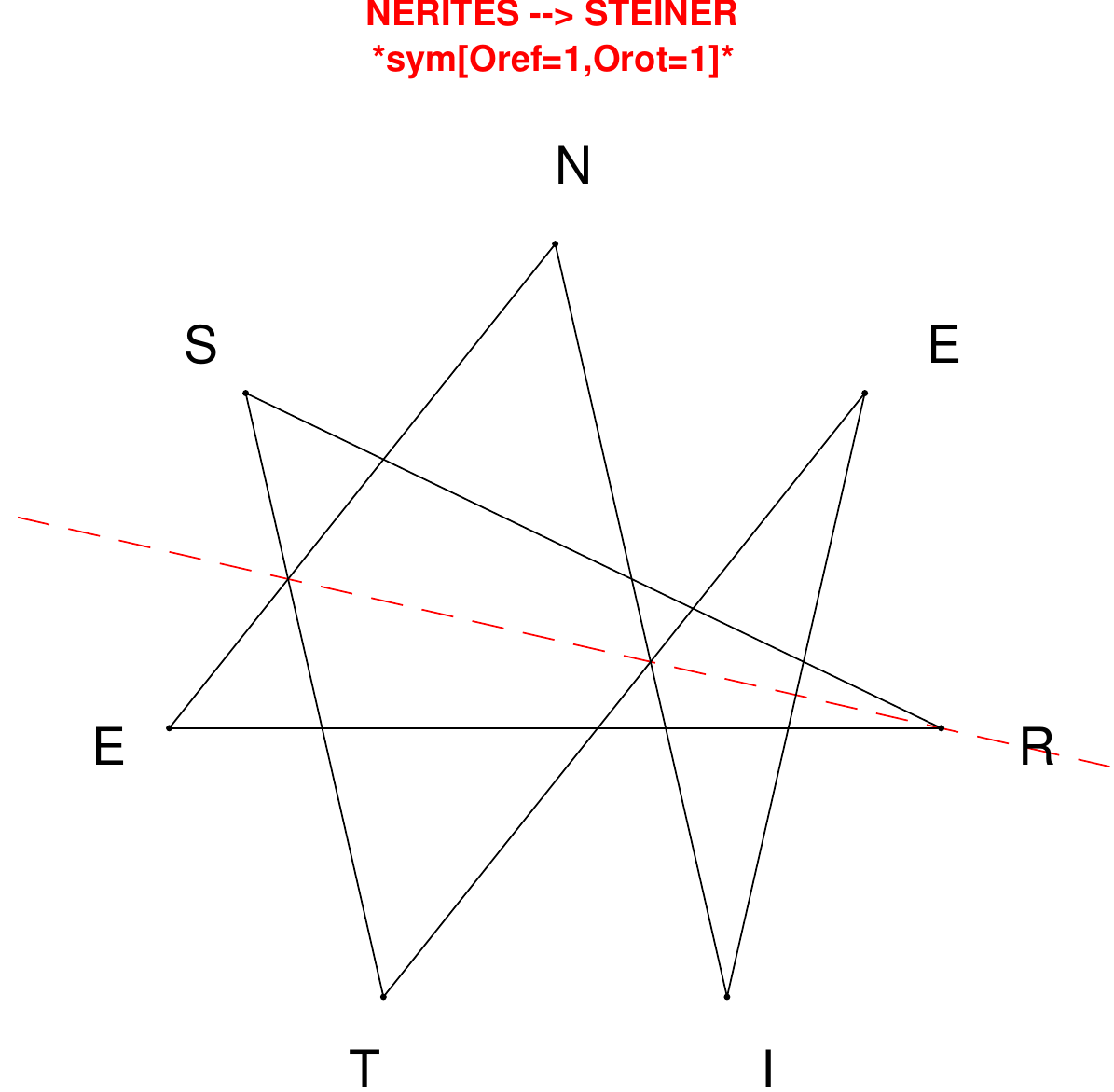}
\end{subfigure}
\hfill
\begin{subfigure}[T]{0.19\textwidth}
\centering
\includegraphics[width=\textwidth]{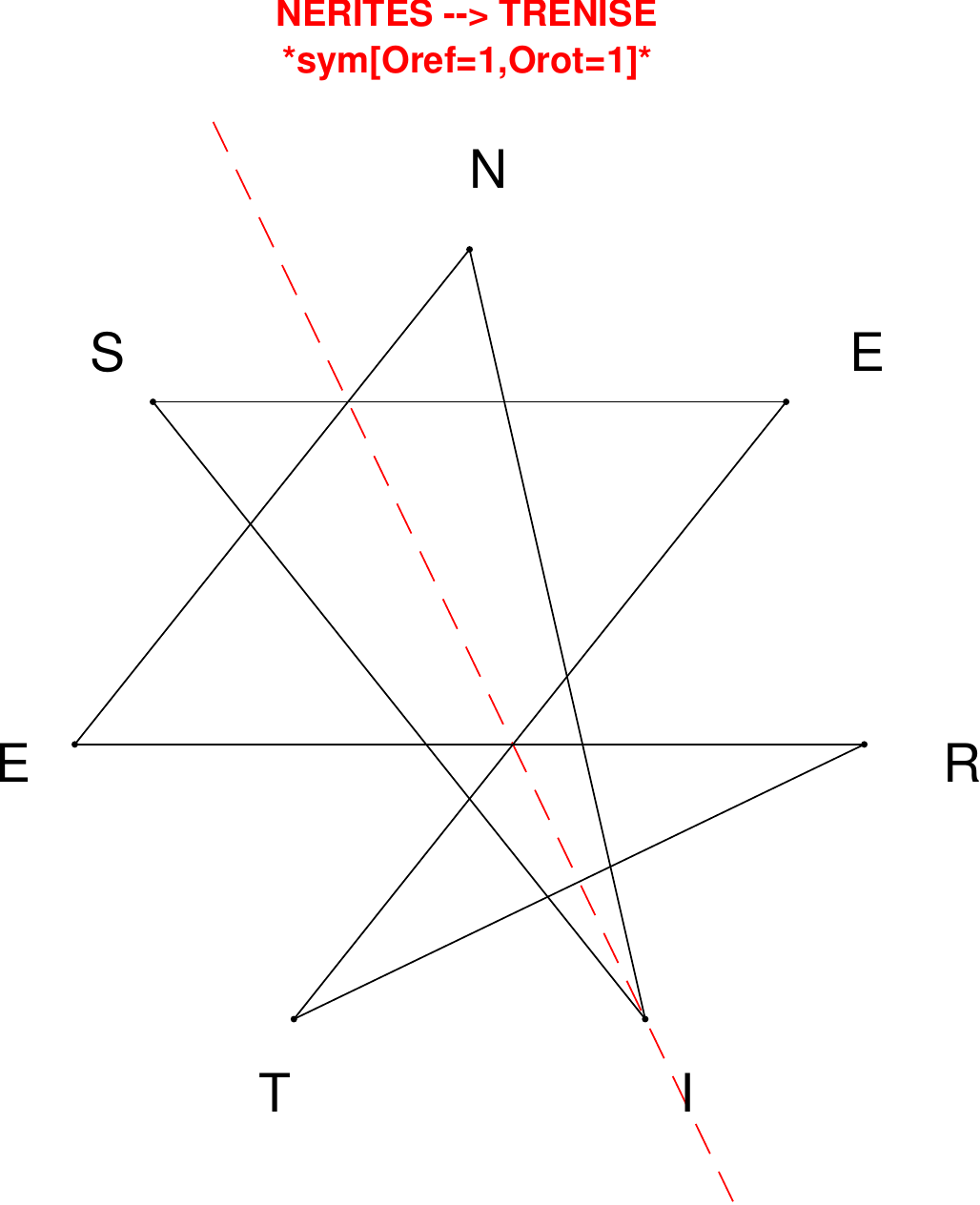}
\end{subfigure}
\hfill
\begin{subfigure}[T]{0.19\textwidth}
\centering
\includegraphics[width=\textwidth]{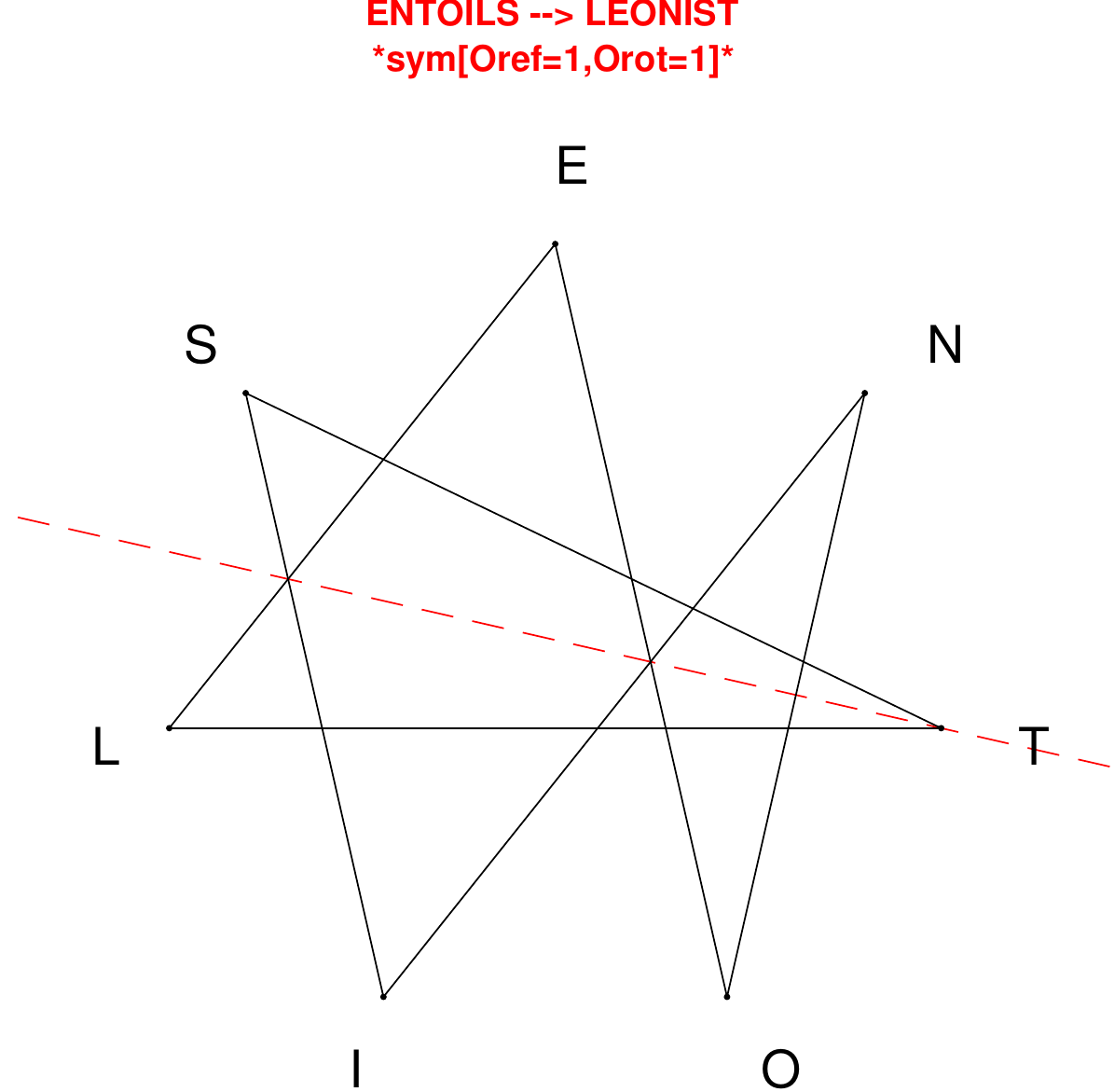}
\end{subfigure}
\hfill
\begin{subfigure}[T]{0.19\textwidth}
\centering
\includegraphics[width=\textwidth]{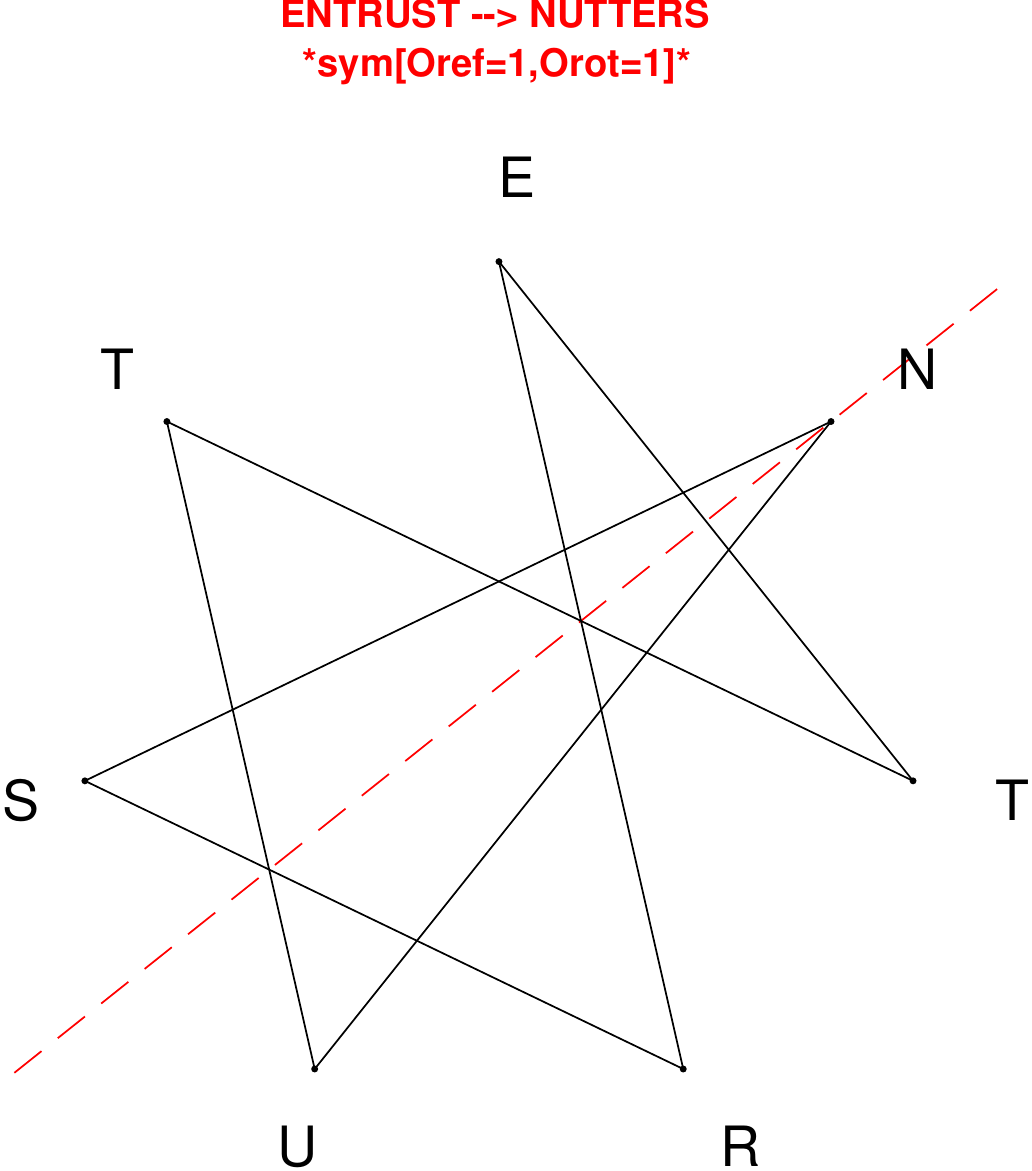}
\end{subfigure}
\end{figure}

\begin{figure}[H]
\centering
\begin{subfigure}[T]{0.19\textwidth}
\centering
\includegraphics[width=\textwidth]{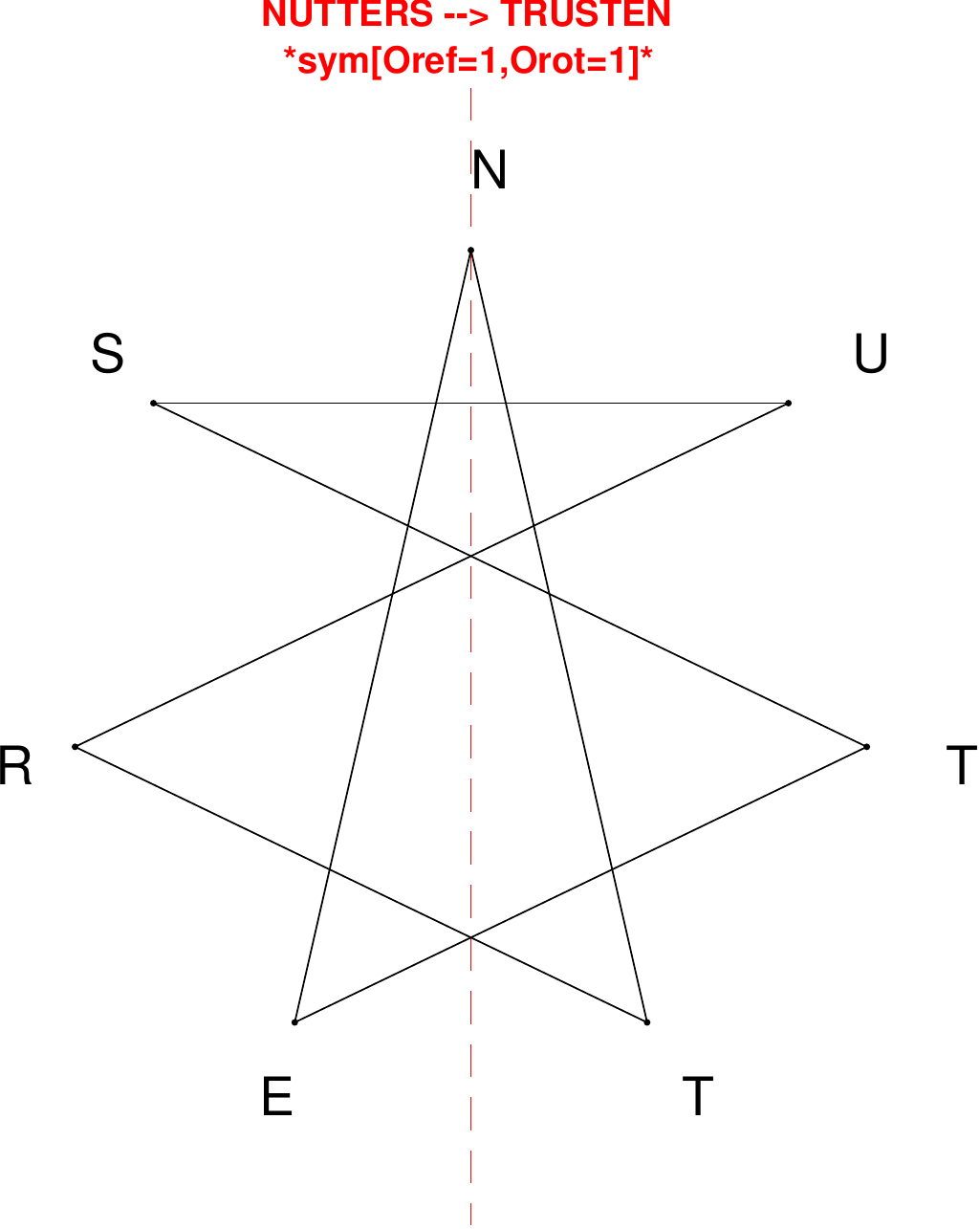}
\end{subfigure}
\hfill
\begin{subfigure}[T]{0.19\textwidth}
\centering
\includegraphics[width=\textwidth]{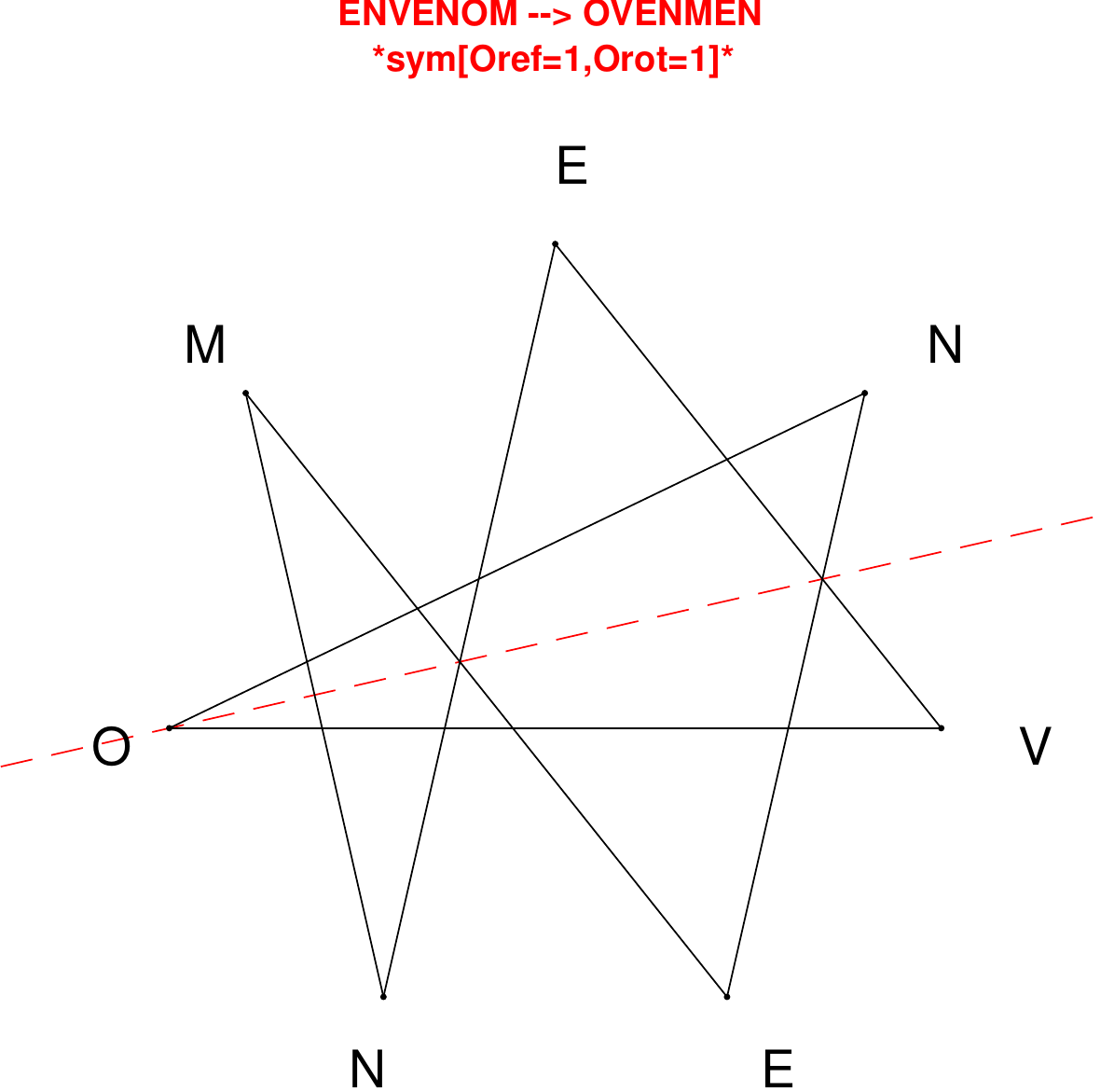}
\end{subfigure}
\hfill
\begin{subfigure}[T]{0.19\textwidth}
\centering
\includegraphics[width=\textwidth]{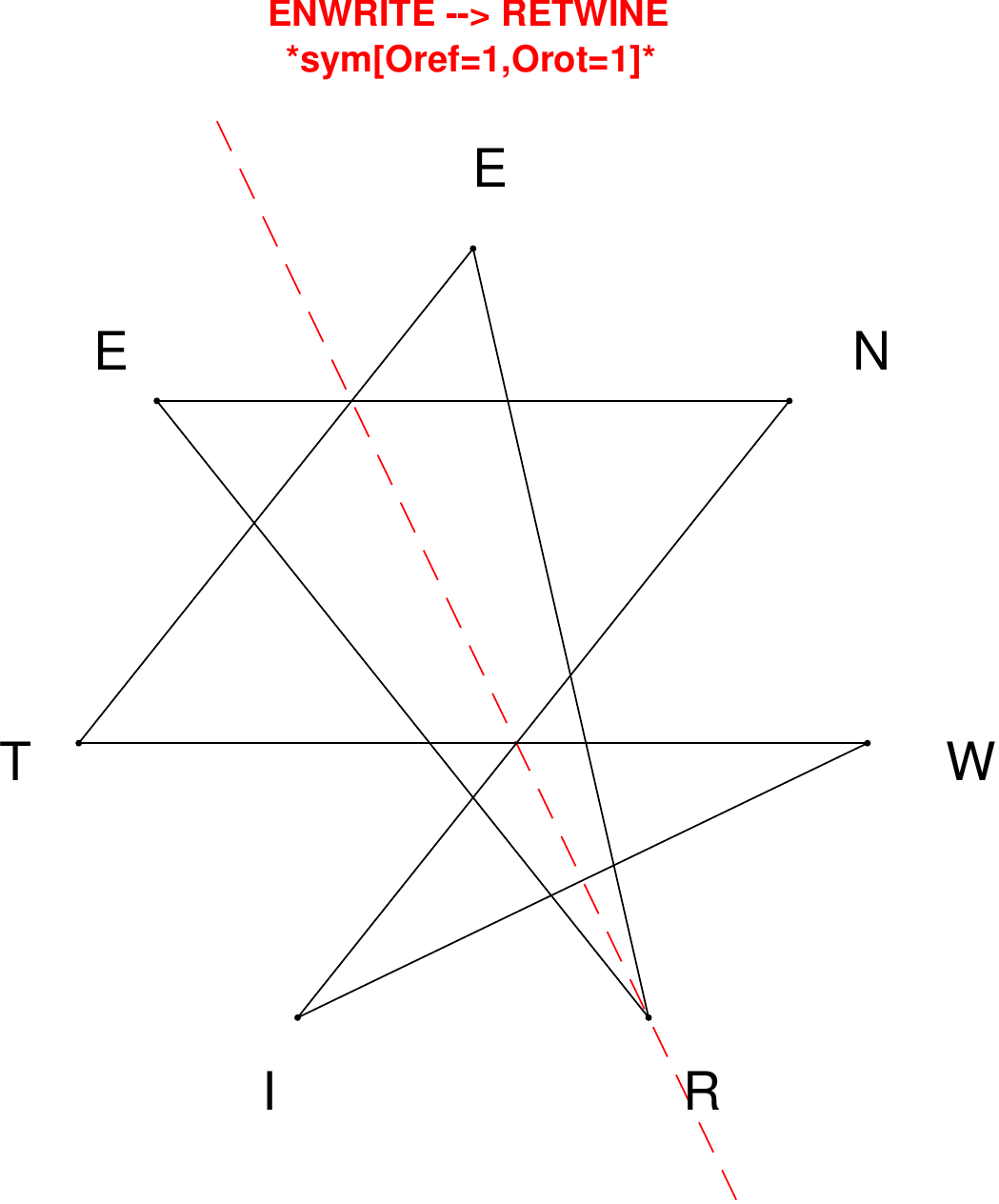}
\end{subfigure}
\hfill
\begin{subfigure}[T]{0.19\textwidth}
\centering
\includegraphics[width=\textwidth]{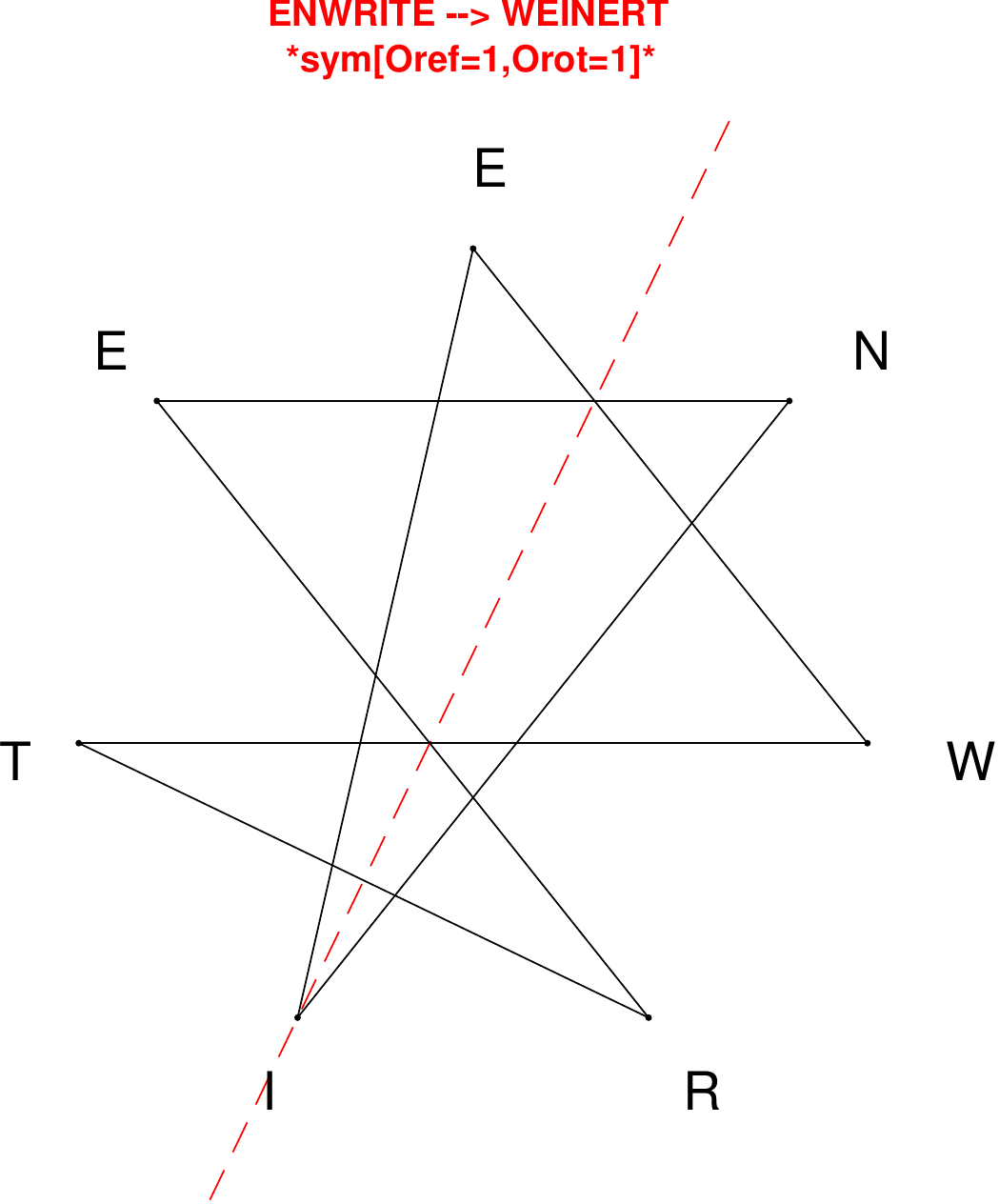}
\end{subfigure}
\hfill
\begin{subfigure}[T]{0.19\textwidth}
\centering
\includegraphics[width=\textwidth]{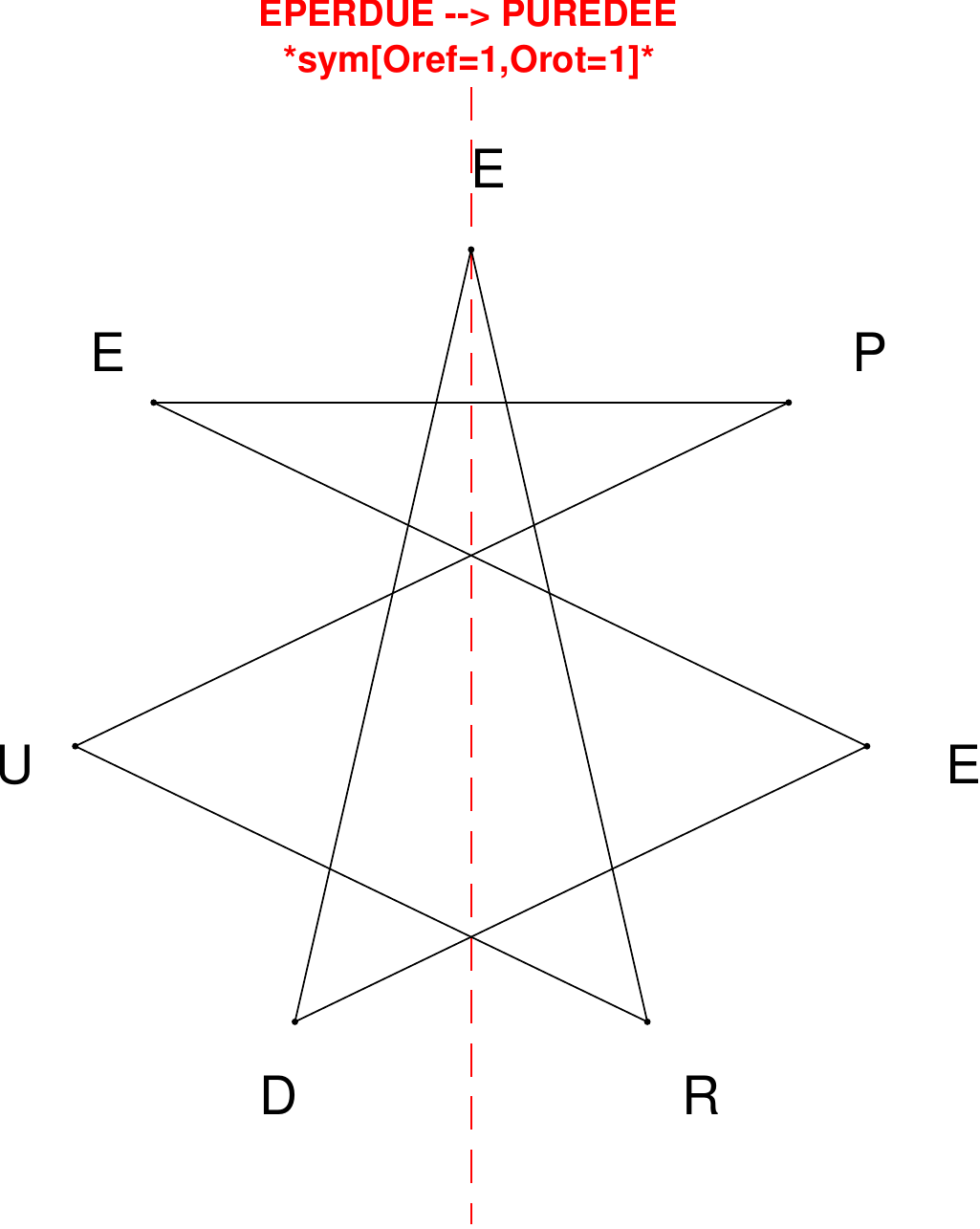}
\end{subfigure}
\end{figure}

\begin{figure}[H]
\centering
\begin{subfigure}[T]{0.19\textwidth}
\centering
\includegraphics[width=\textwidth]{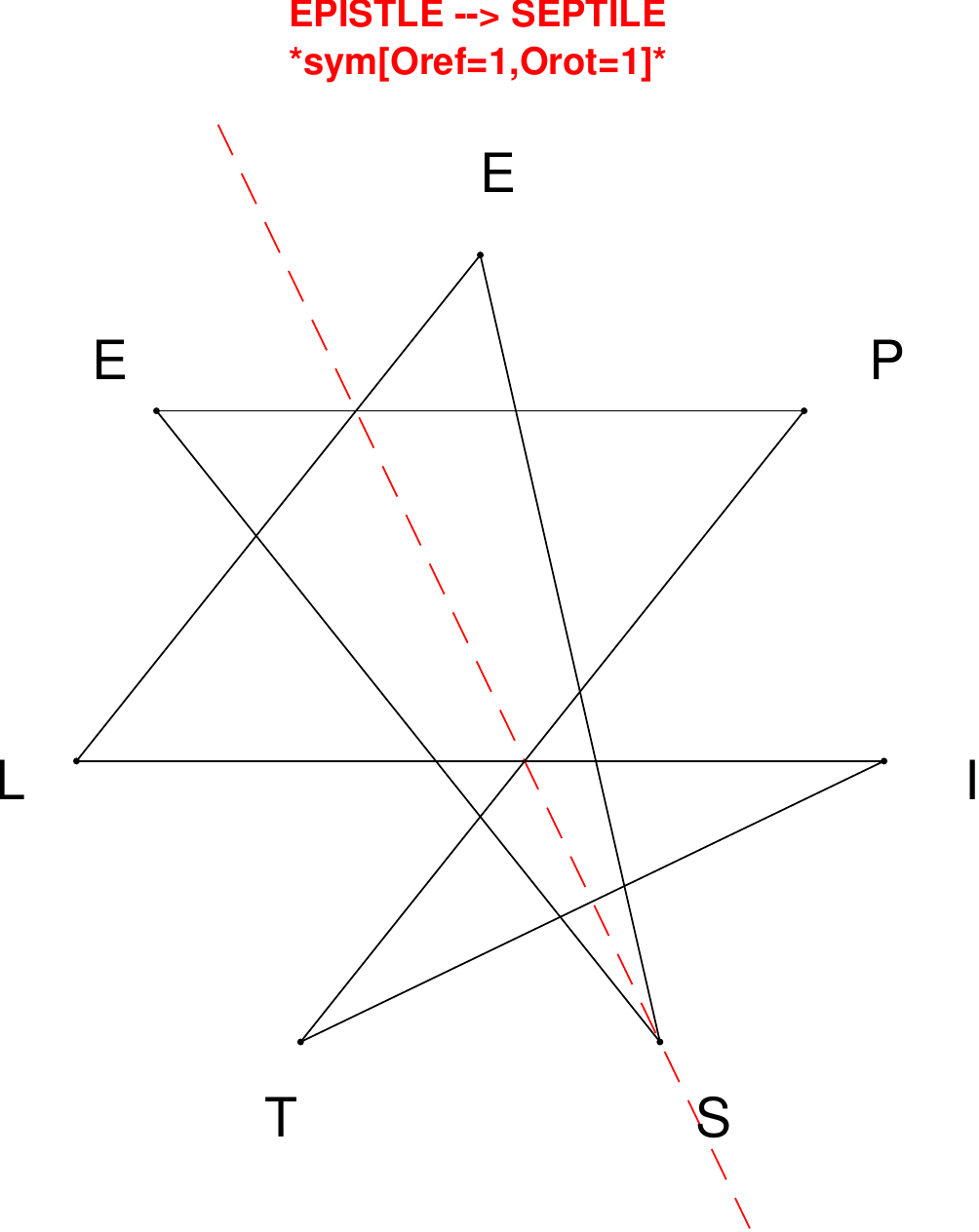}
\end{subfigure}
\hfill
\begin{subfigure}[T]{0.19\textwidth}
\centering
\includegraphics[width=\textwidth]{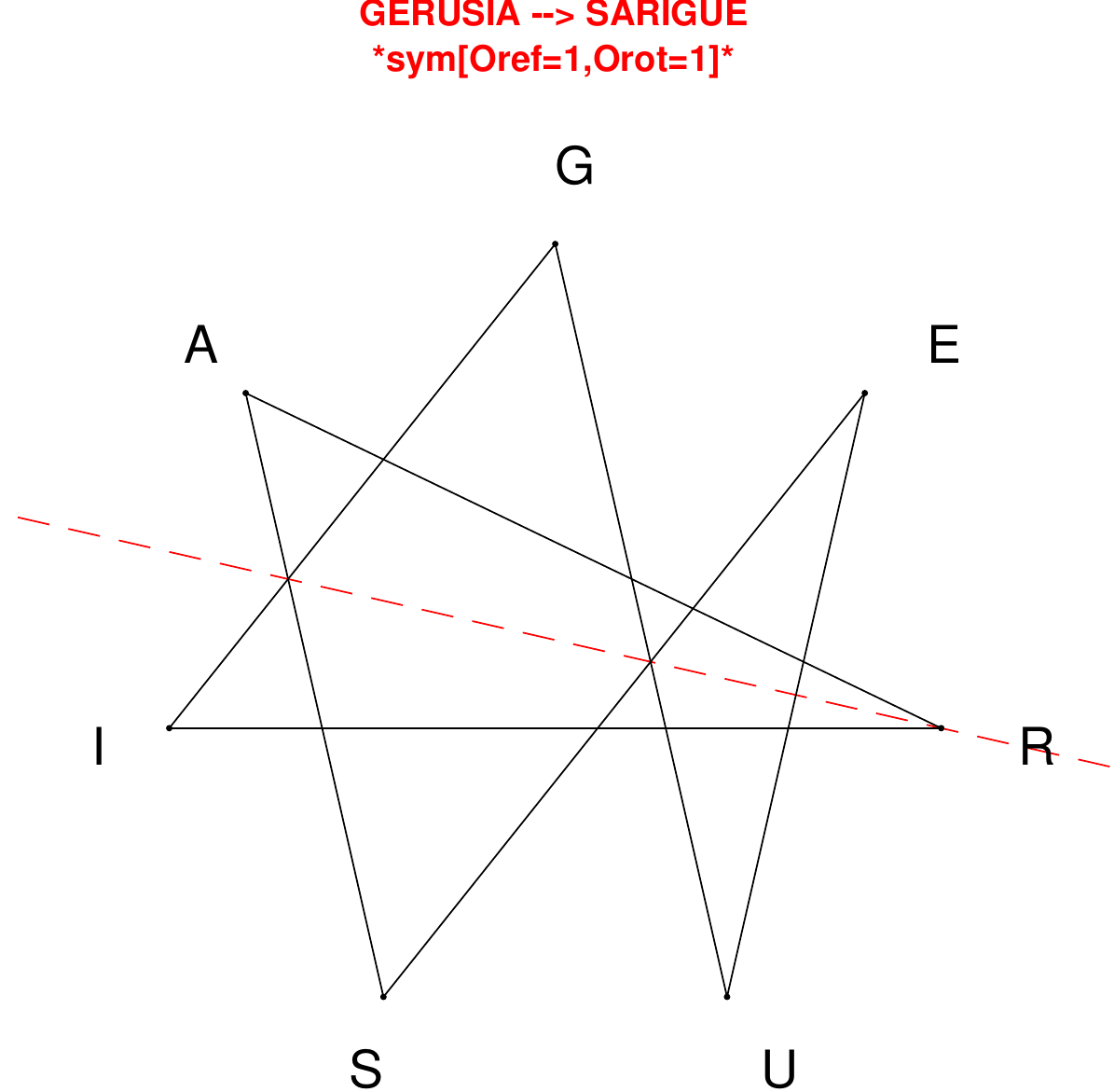}
\end{subfigure}
\hfill
\begin{subfigure}[T]{0.19\textwidth}
\centering
\includegraphics[width=\textwidth]{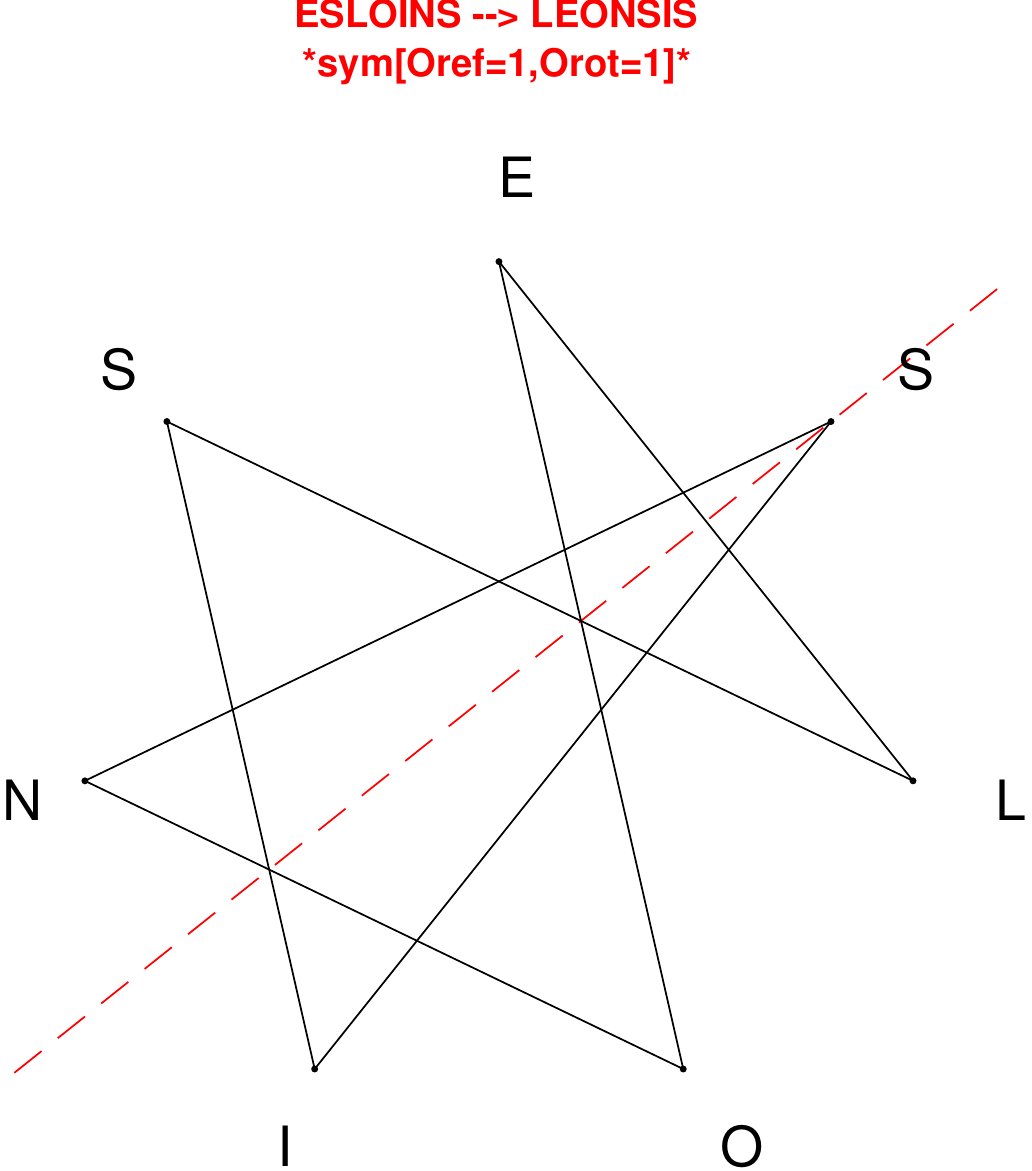}
\end{subfigure}
\hfill
\begin{subfigure}[T]{0.19\textwidth}
\centering
\includegraphics[width=\textwidth]{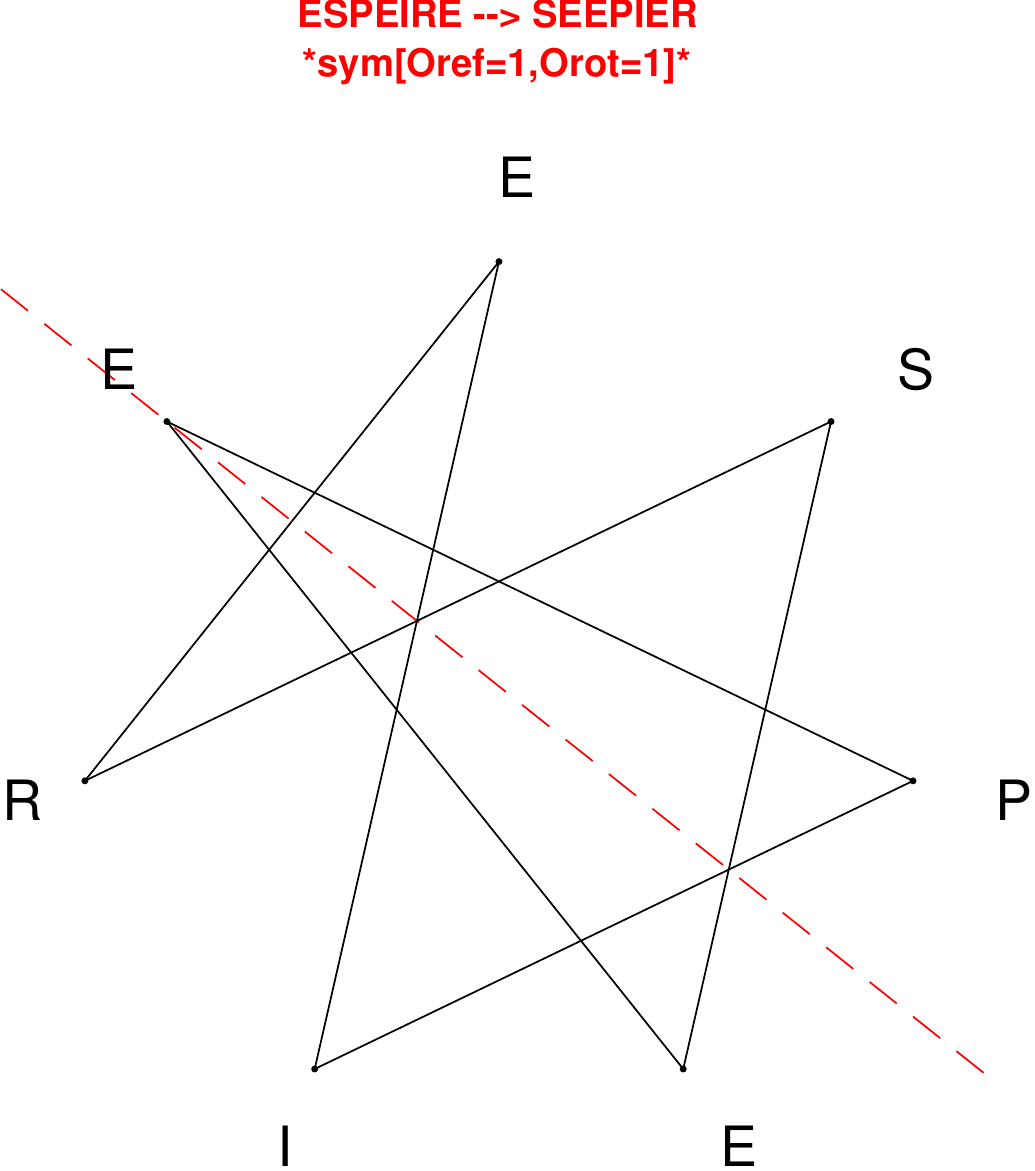}
\end{subfigure}
\hfill
\begin{subfigure}[T]{0.19\textwidth}
\centering
\includegraphics[width=\textwidth]{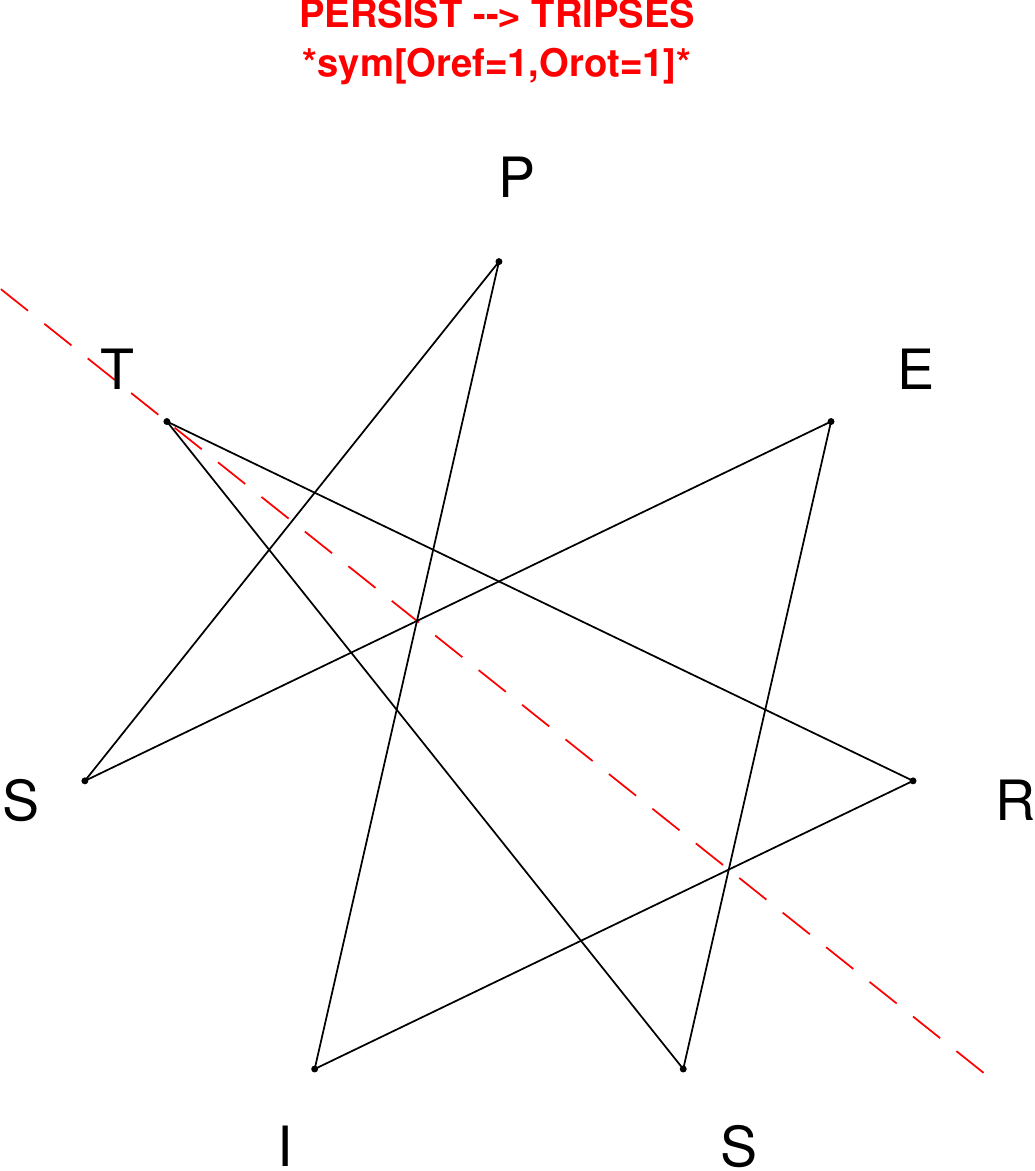}
\end{subfigure}
\end{figure}

\begin{figure}[H]
\centering
\begin{subfigure}[T]{0.19\textwidth}
\centering
\includegraphics[width=\textwidth]{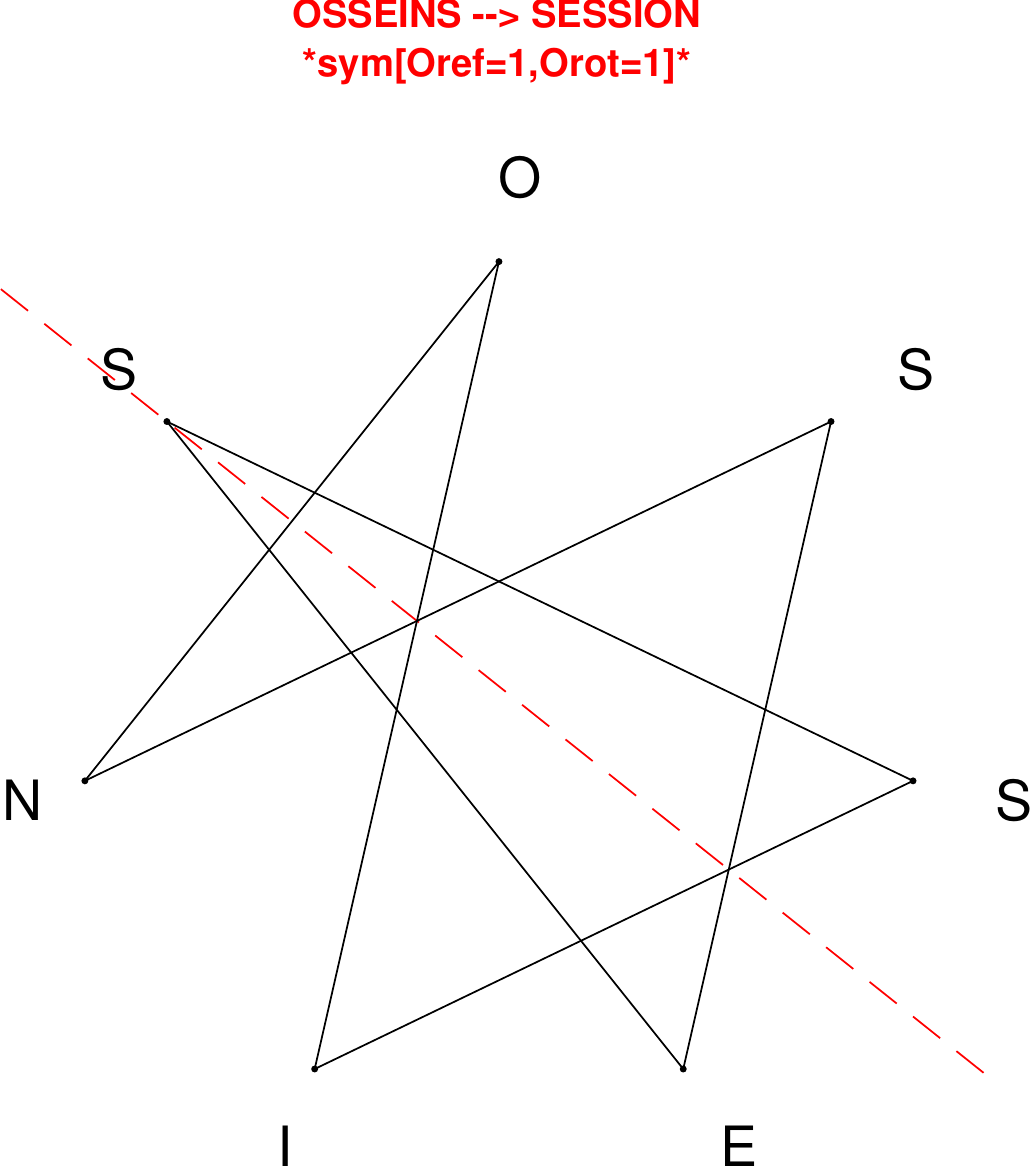}
\end{subfigure}
\hfill
\begin{subfigure}[T]{0.19\textwidth}
\centering
\includegraphics[width=\textwidth]{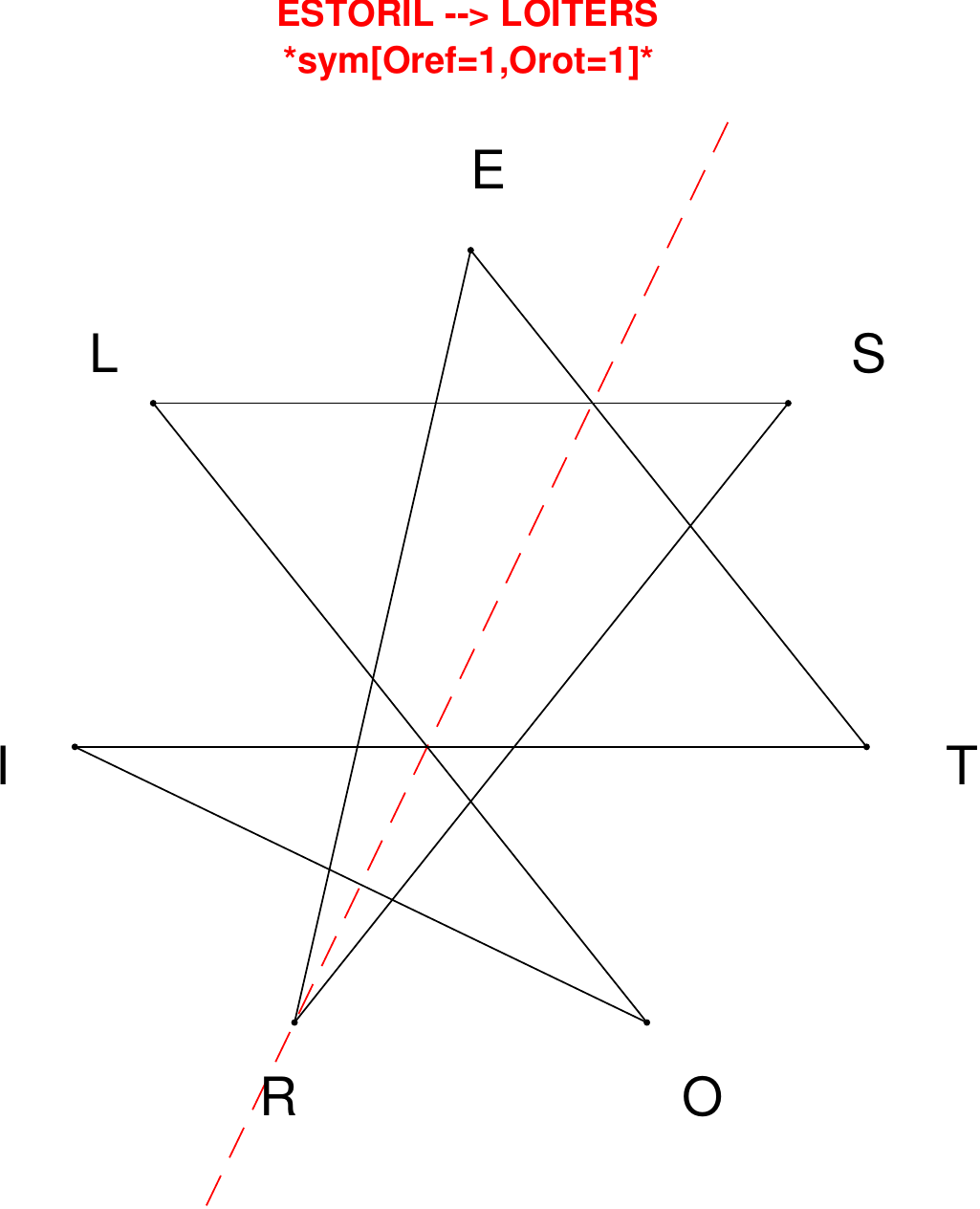}
\end{subfigure}
\hfill
\begin{subfigure}[T]{0.19\textwidth}
\centering
\includegraphics[width=\textwidth]{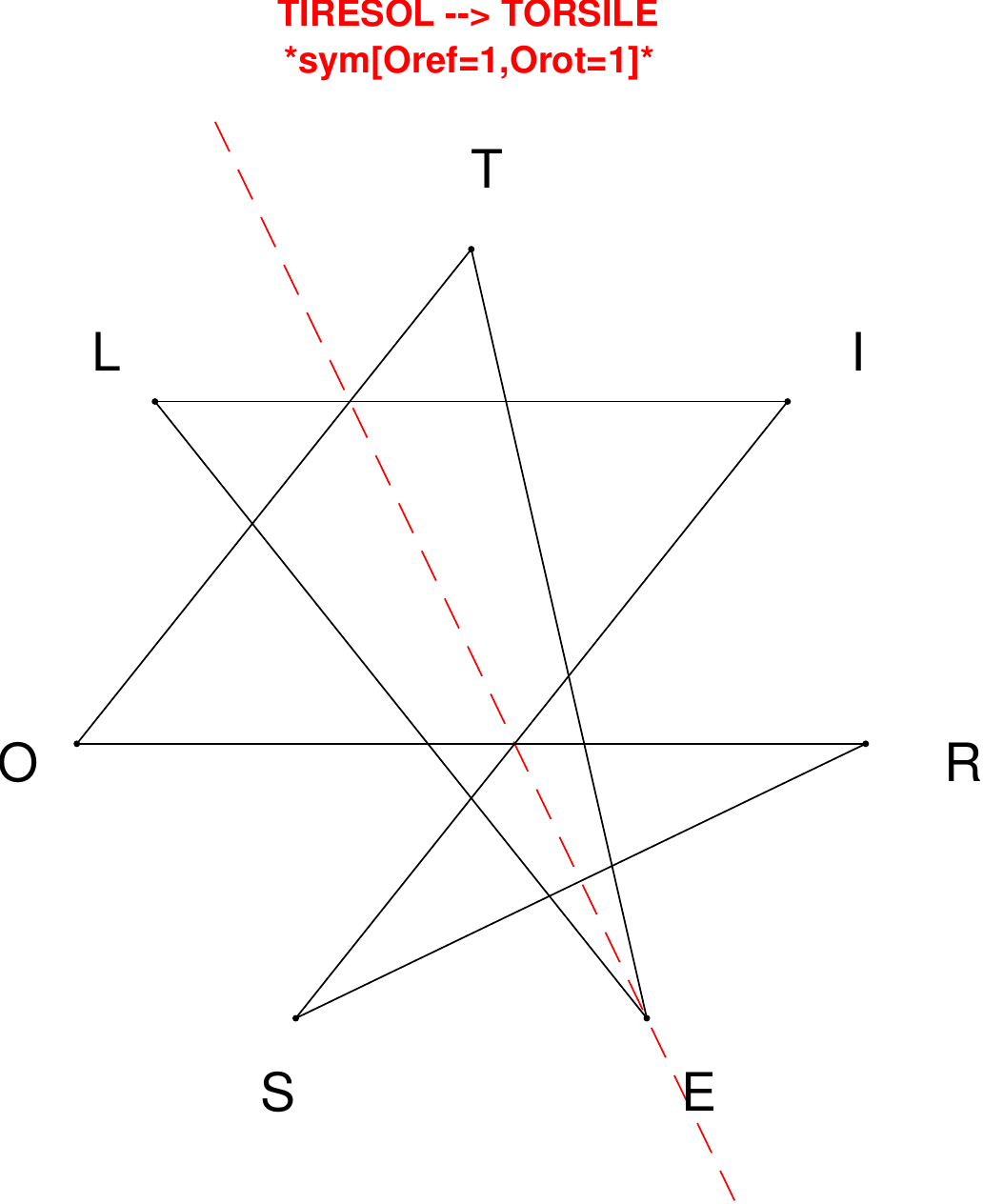}
\end{subfigure}
\hfill
\begin{subfigure}[T]{0.19\textwidth}
\centering
\includegraphics[width=\textwidth]{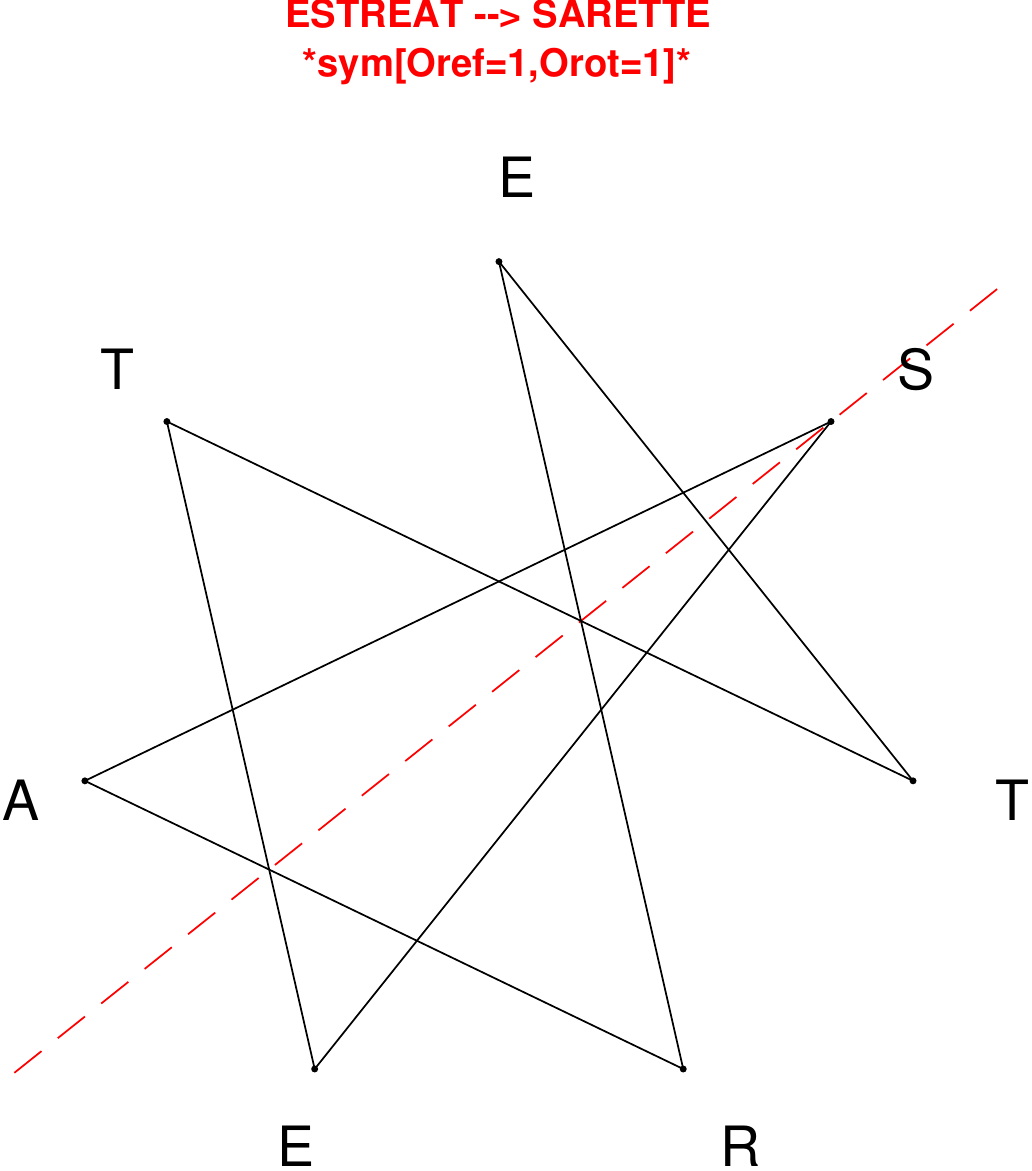}
\end{subfigure}
\hfill
\begin{subfigure}[T]{0.19\textwidth}
\centering
\includegraphics[width=\textwidth]{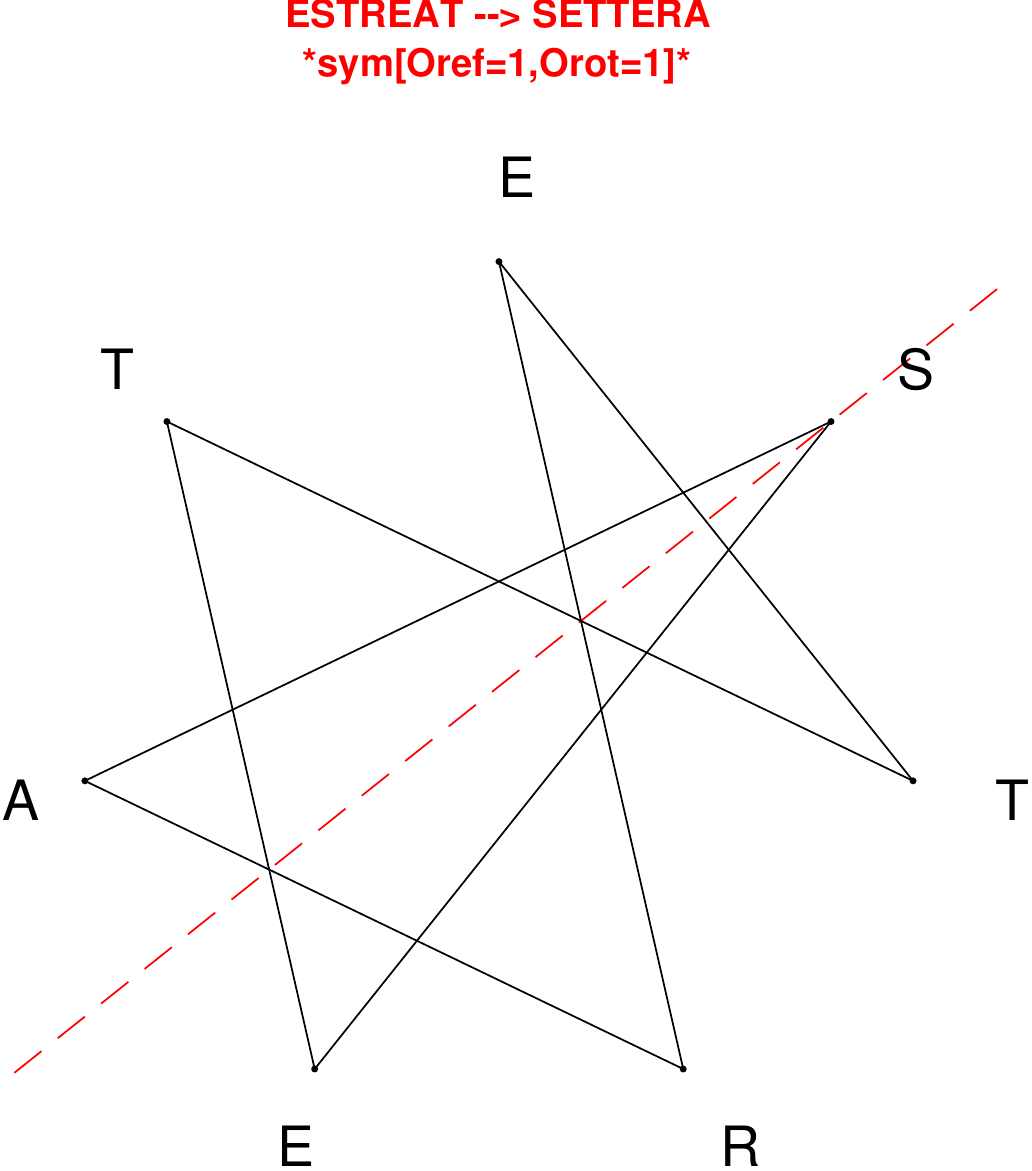}
\end{subfigure}
\end{figure}

\begin{figure}[H]
\centering
\begin{subfigure}[T]{0.19\textwidth}
\centering
\includegraphics[width=\textwidth]{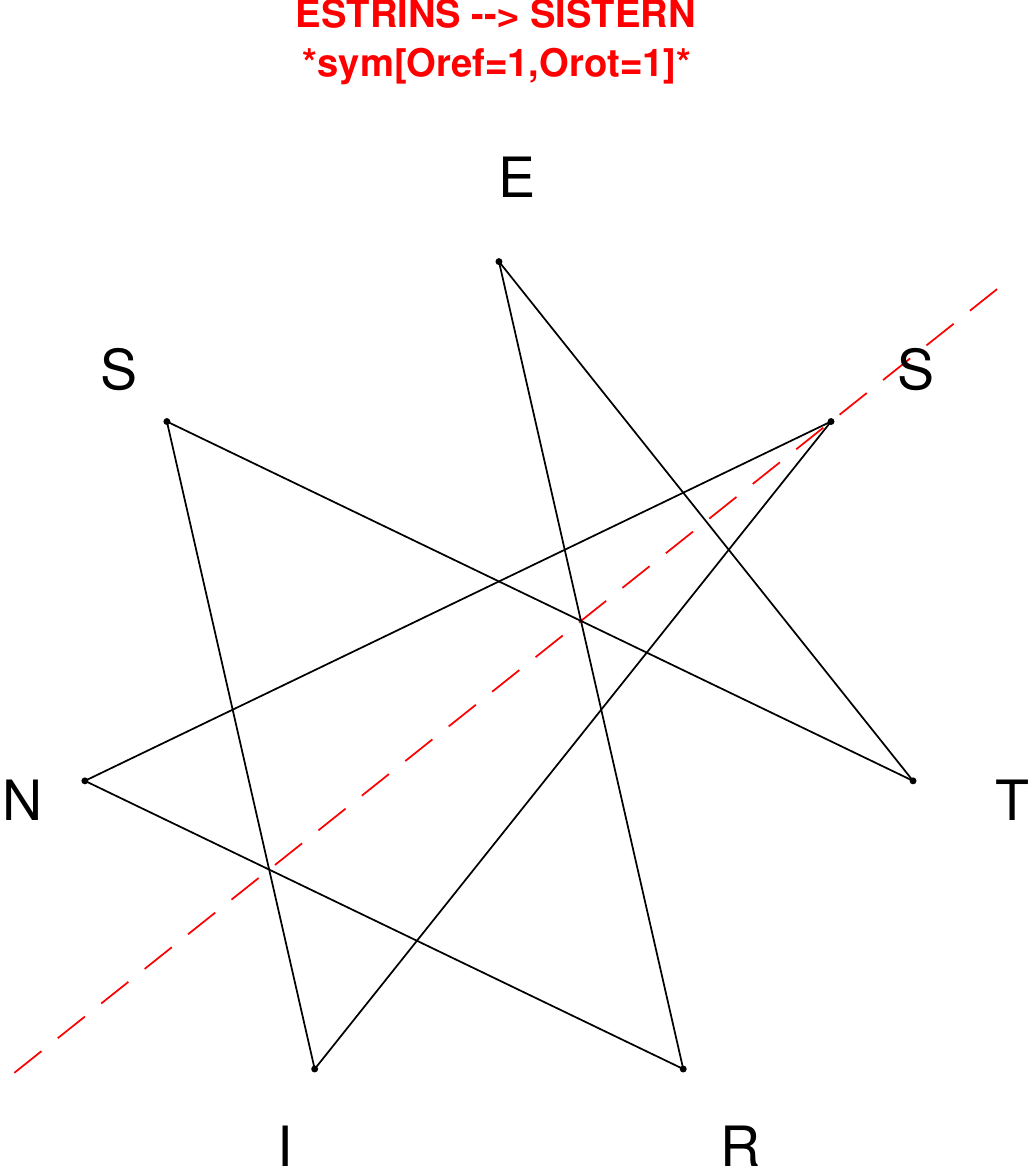}
\end{subfigure}
\hfill
\begin{subfigure}[T]{0.19\textwidth}
\centering
\includegraphics[width=\textwidth]{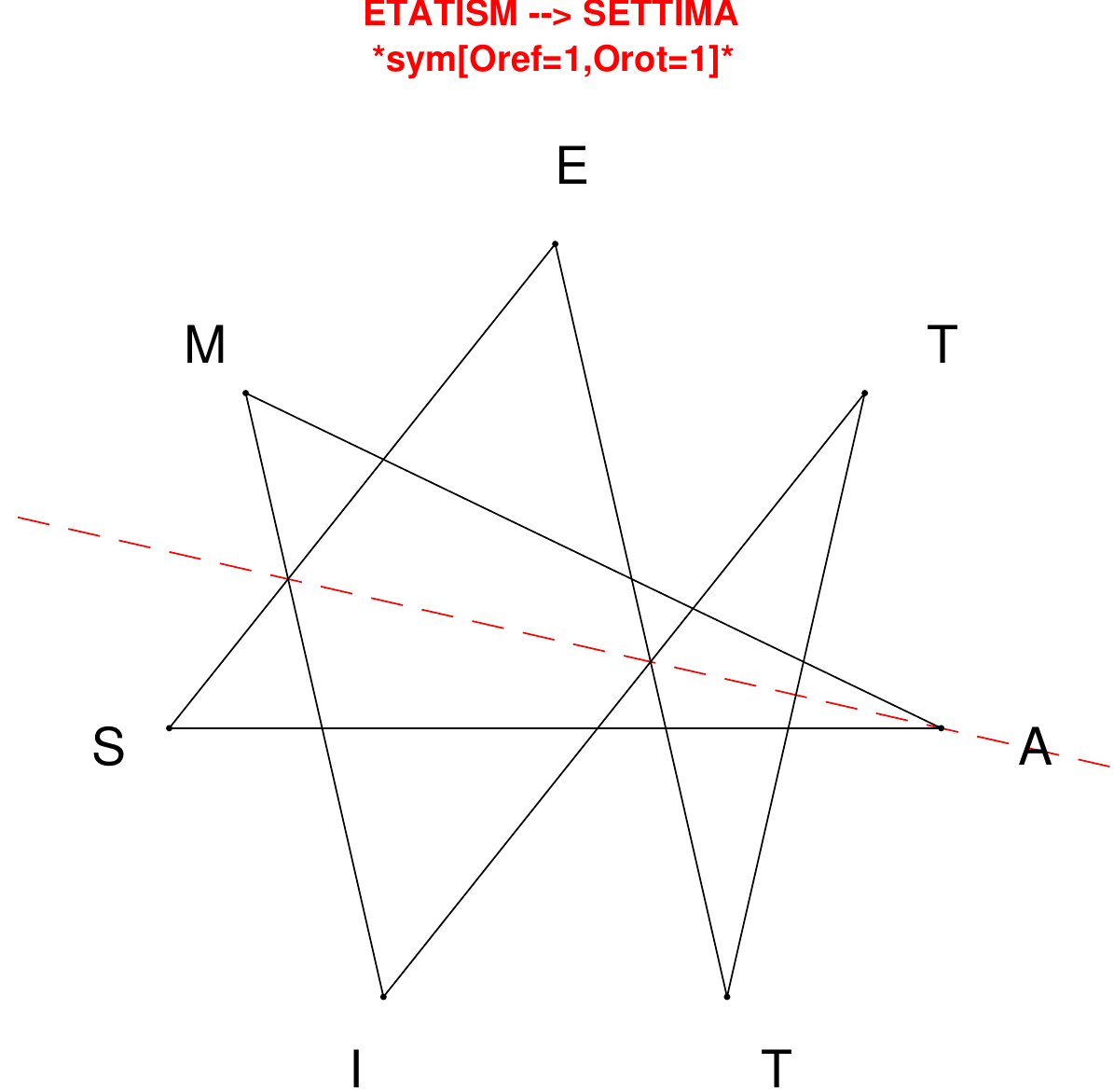}
\end{subfigure}
\hfill
\begin{subfigure}[T]{0.19\textwidth}
\centering
\includegraphics[width=\textwidth]{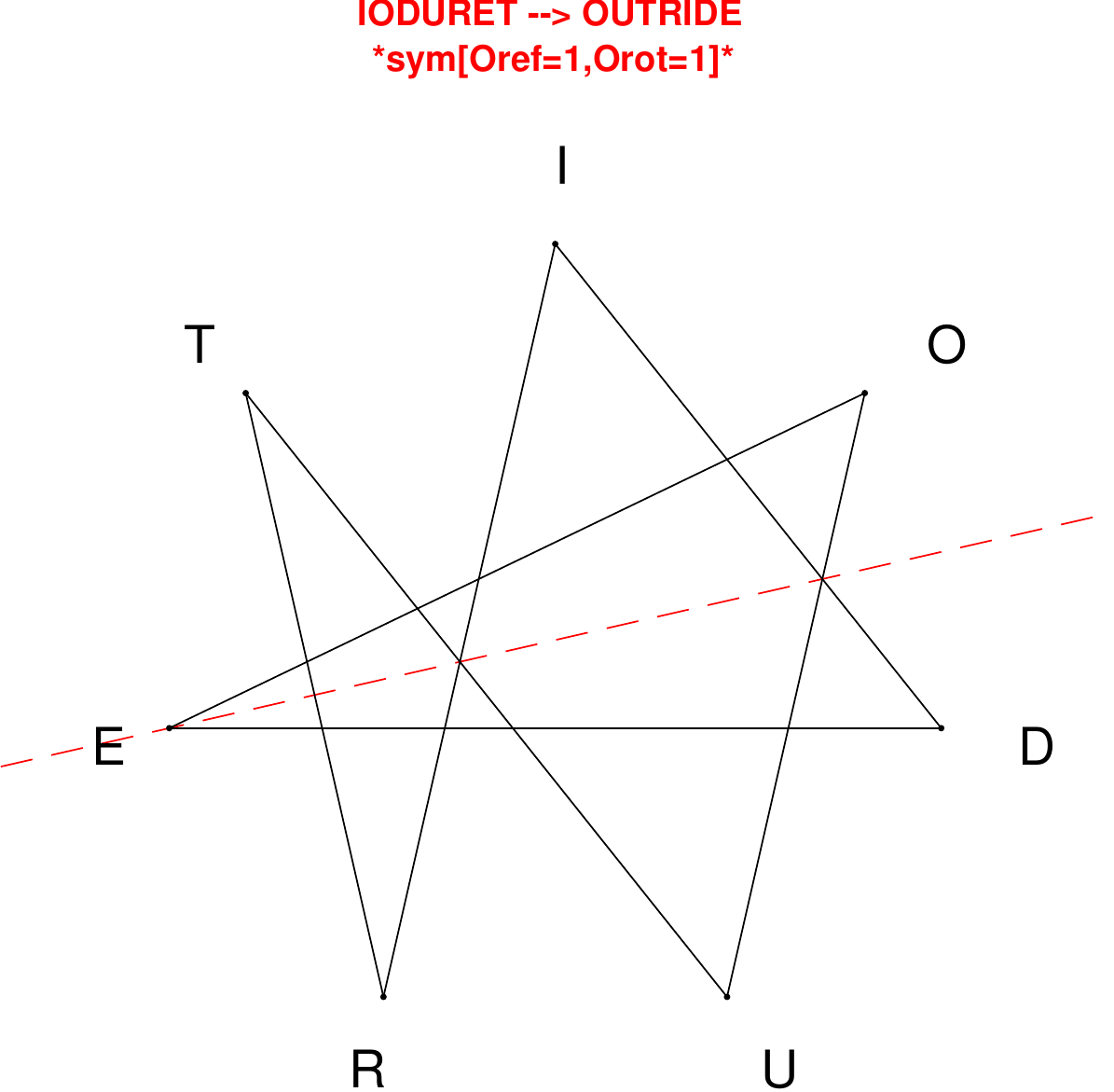}
\end{subfigure}
\hfill
\begin{subfigure}[T]{0.19\textwidth}
\centering
\includegraphics[width=\textwidth]{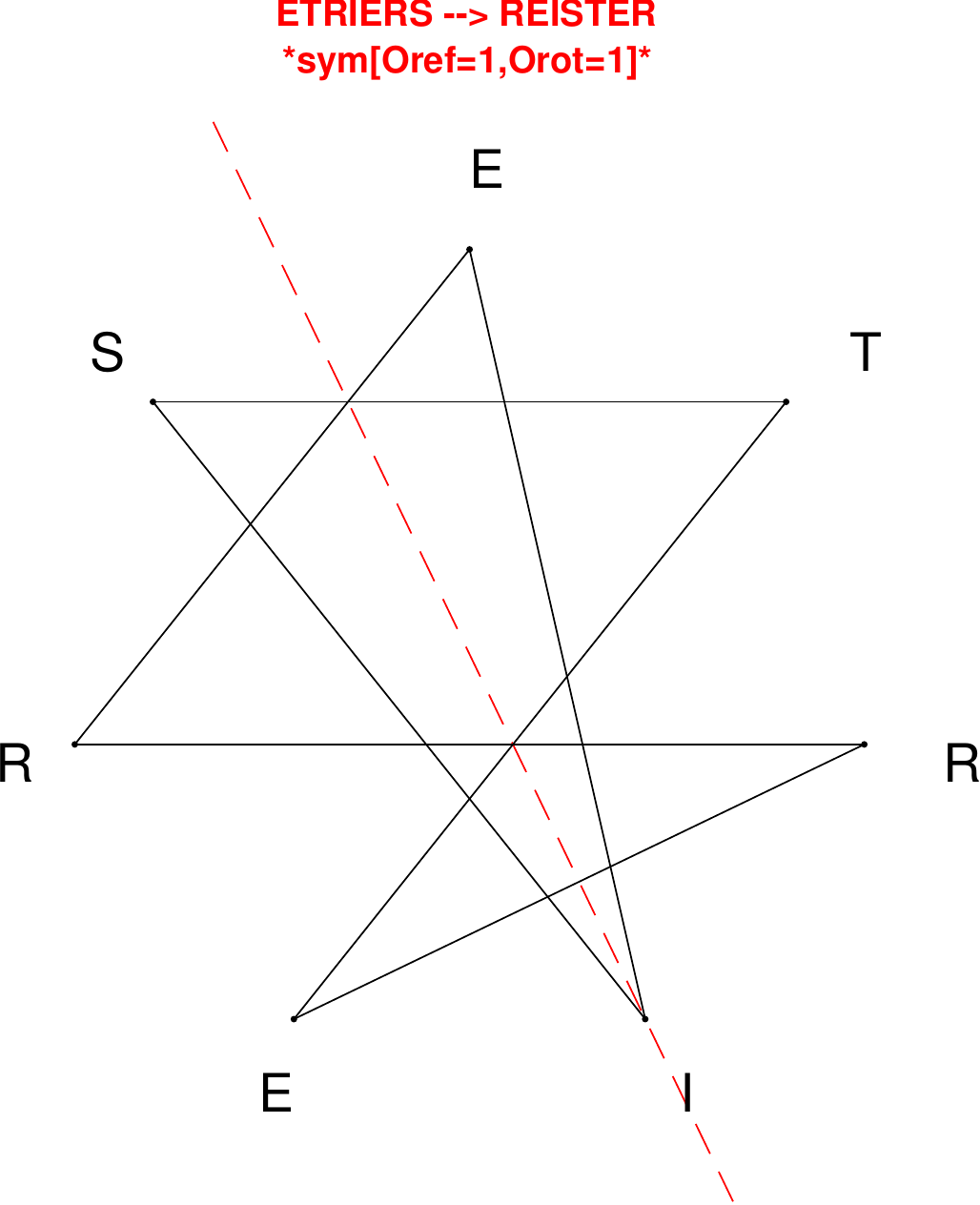}
\end{subfigure}
\hfill
\begin{subfigure}[T]{0.19\textwidth}
\centering
\includegraphics[width=\textwidth]{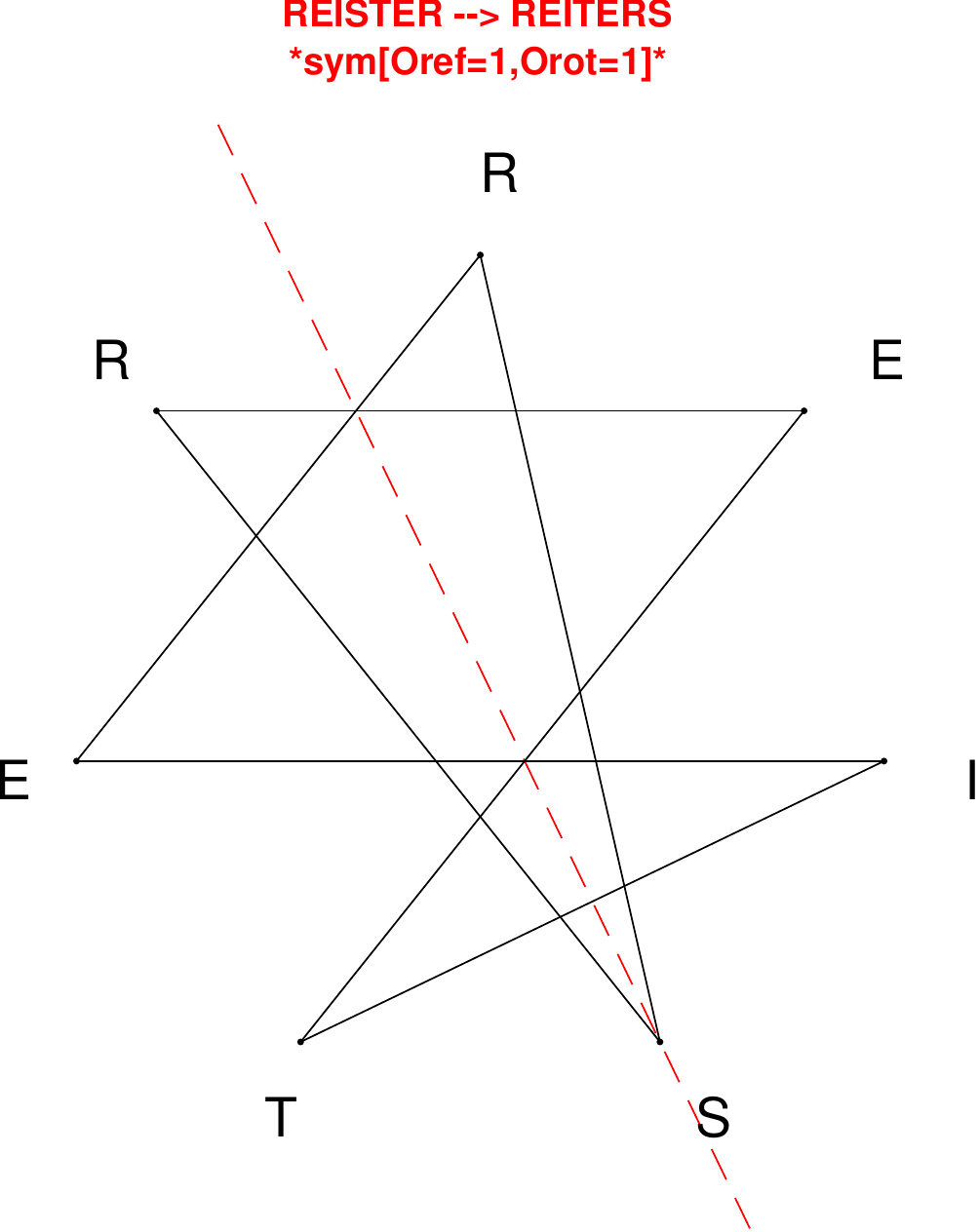}
\end{subfigure}
\end{figure}

\begin{figure}[H]
\centering
\begin{subfigure}[T]{0.19\textwidth}
\centering
\includegraphics[width=\textwidth]{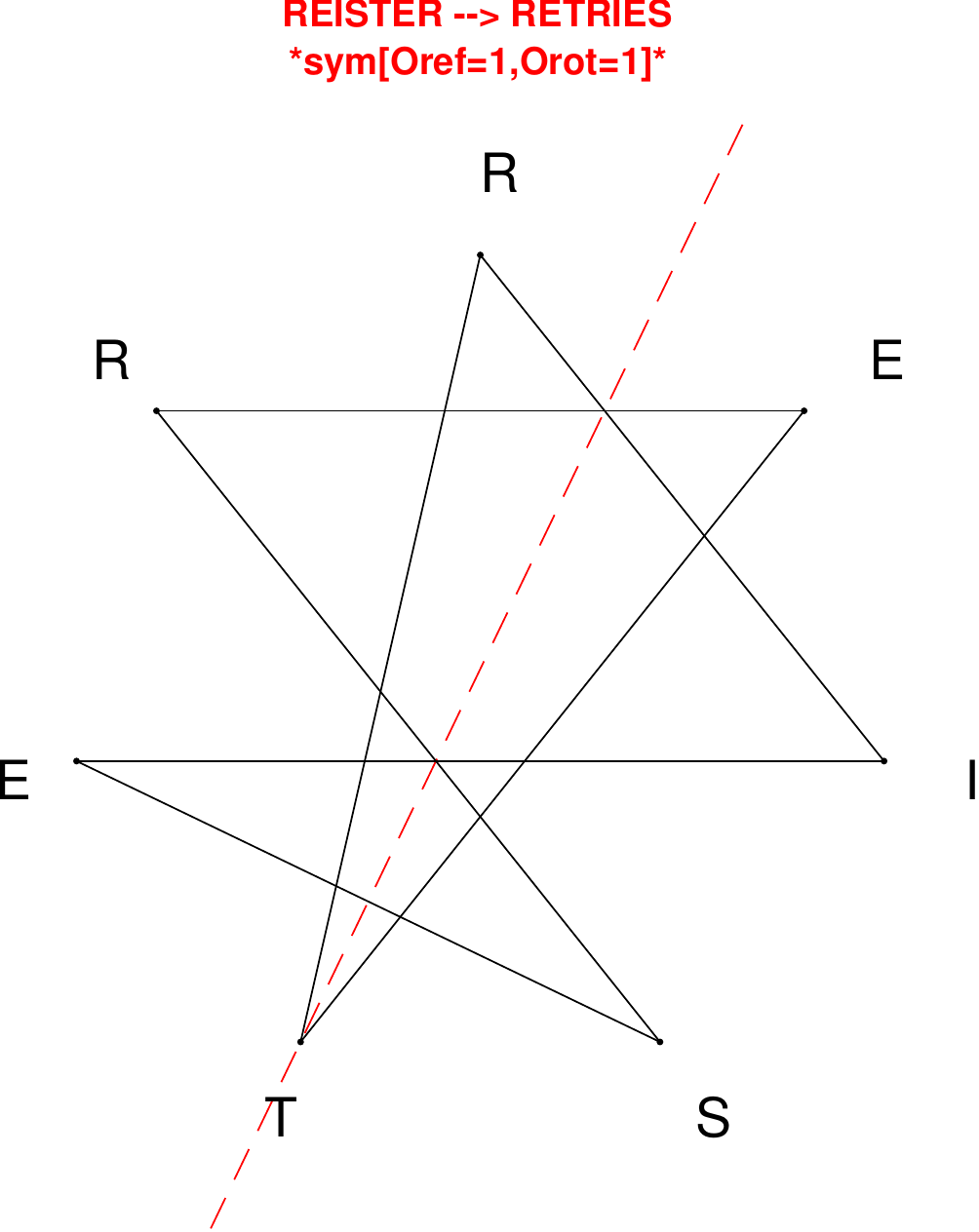}
\end{subfigure}
\hfill
\begin{subfigure}[T]{0.19\textwidth}
\centering
\includegraphics[width=\textwidth]{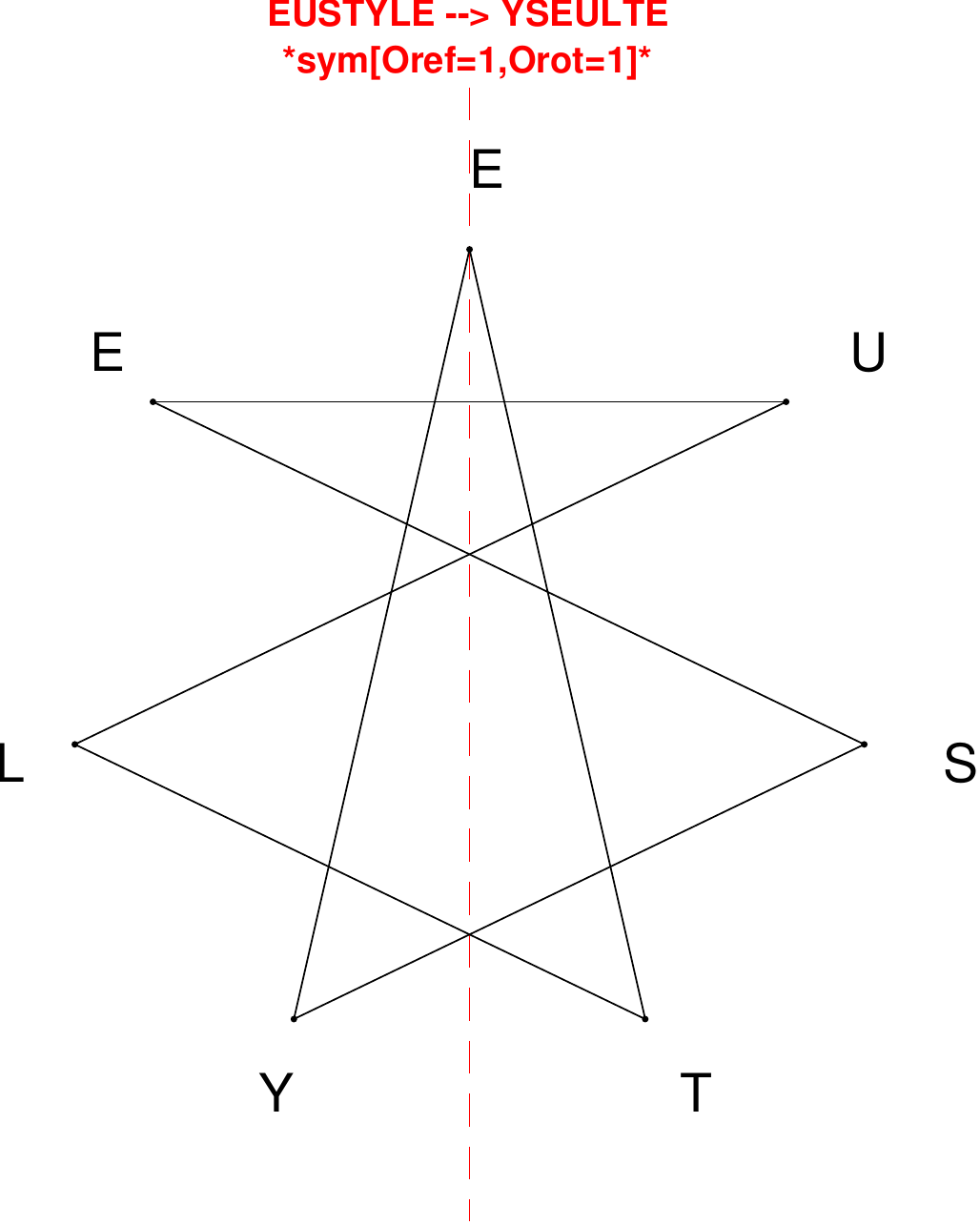}
\end{subfigure}
\hfill
\begin{subfigure}[T]{0.19\textwidth}
\centering
\includegraphics[width=\textwidth]{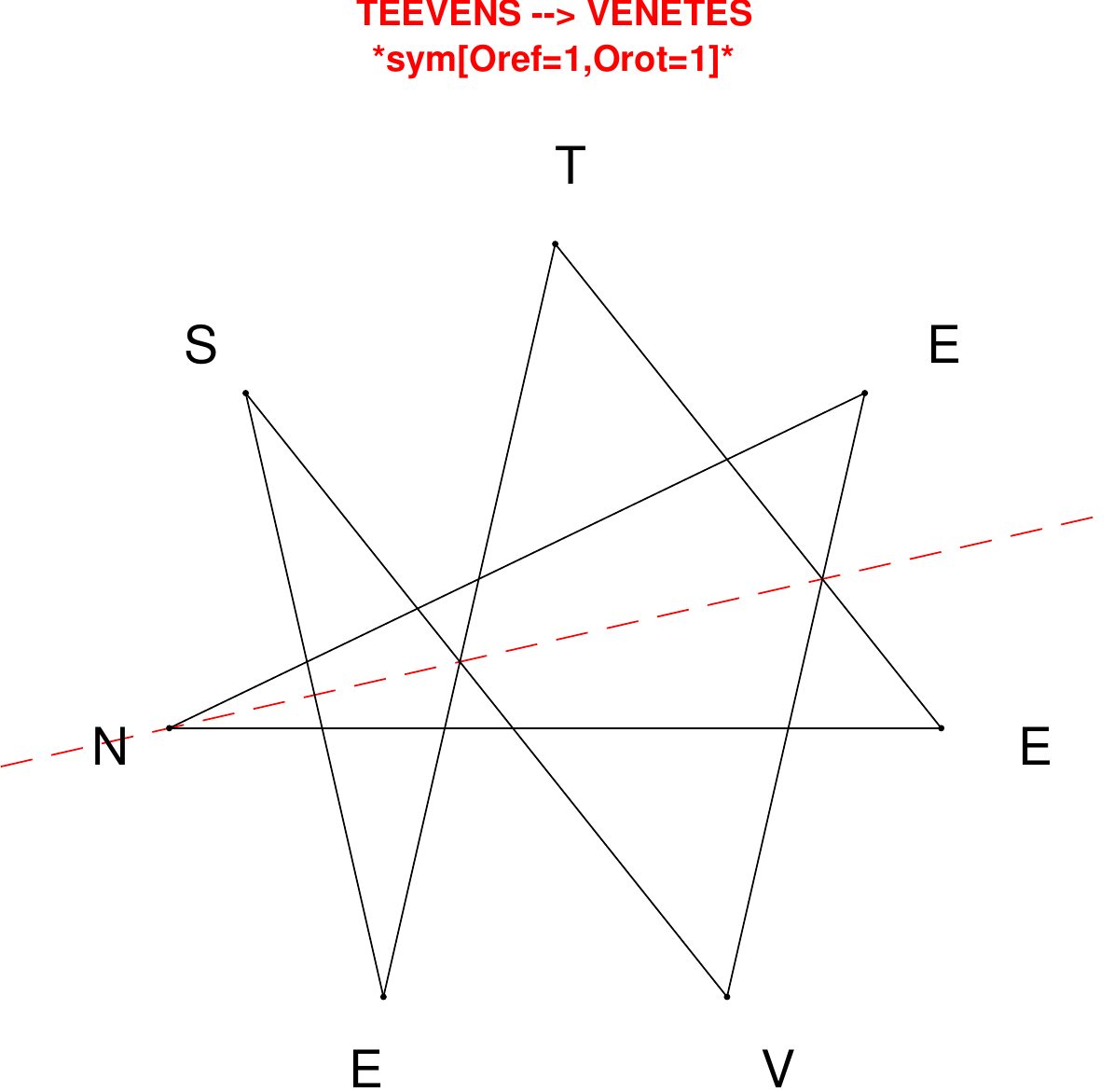}
\end{subfigure}
\hfill
\begin{subfigure}[T]{0.19\textwidth}
\centering
\includegraphics[width=\textwidth]{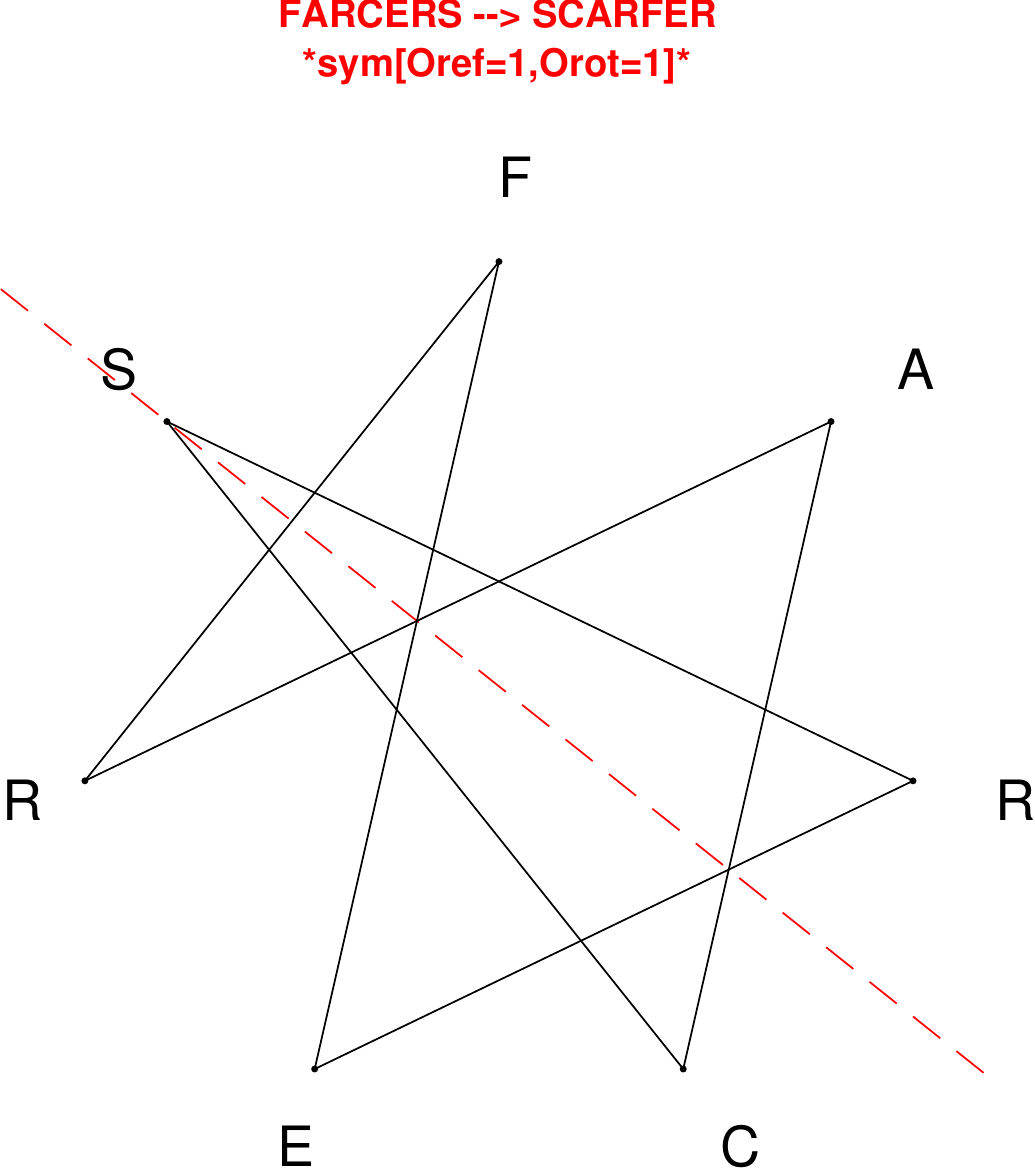}
\end{subfigure}
\hfill
\begin{subfigure}[T]{0.19\textwidth}
\centering
\includegraphics[width=\textwidth]{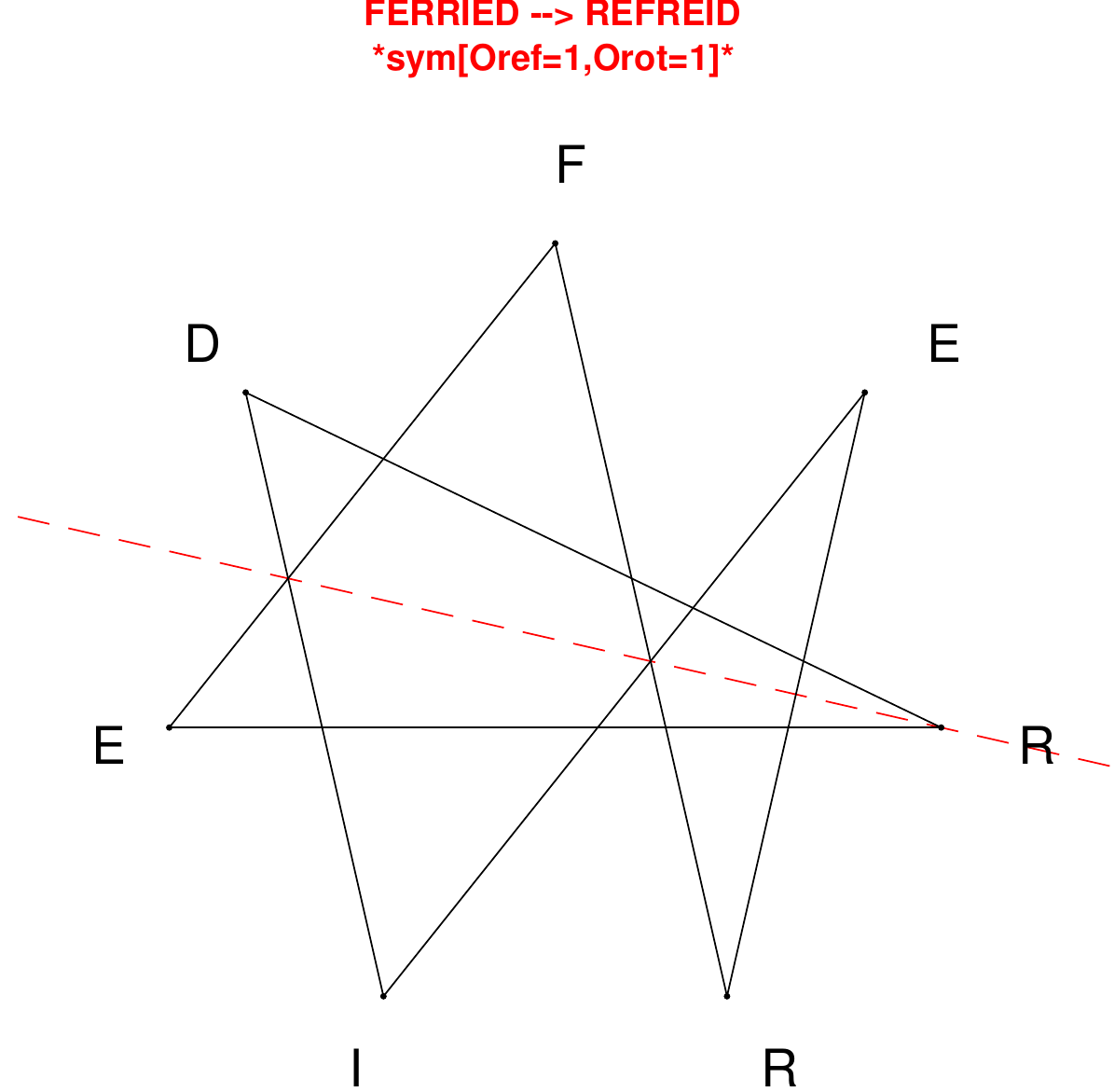}
\end{subfigure}
\end{figure}

\begin{figure}[H]
\centering
\begin{subfigure}[T]{0.19\textwidth}
\centering
\includegraphics[width=\textwidth]{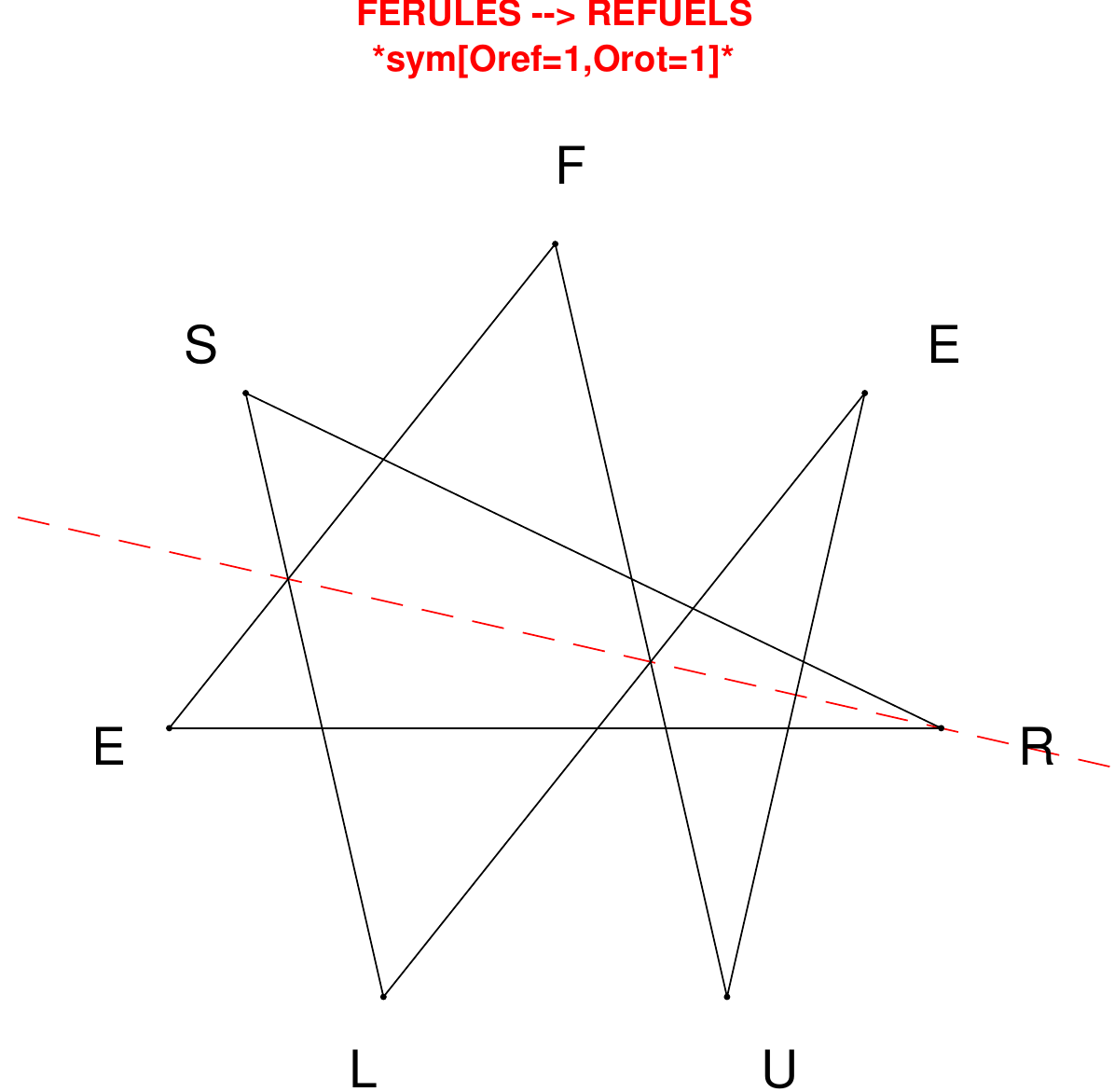}
\end{subfigure}
\hfill
\begin{subfigure}[T]{0.19\textwidth}
\centering
\includegraphics[width=\textwidth]{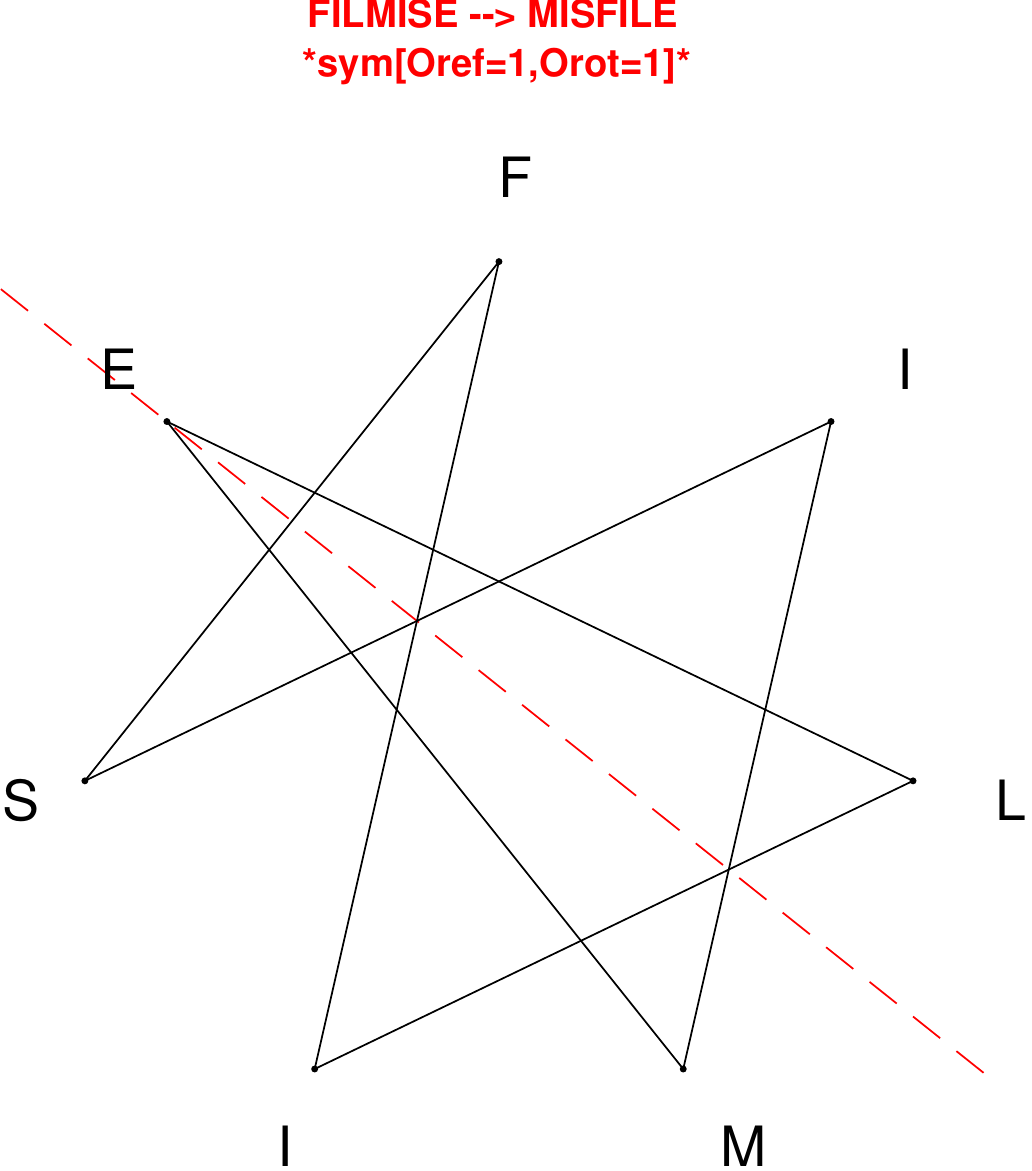}
\end{subfigure}
\hfill
\begin{subfigure}[T]{0.19\textwidth}
\centering
\includegraphics[width=\textwidth]{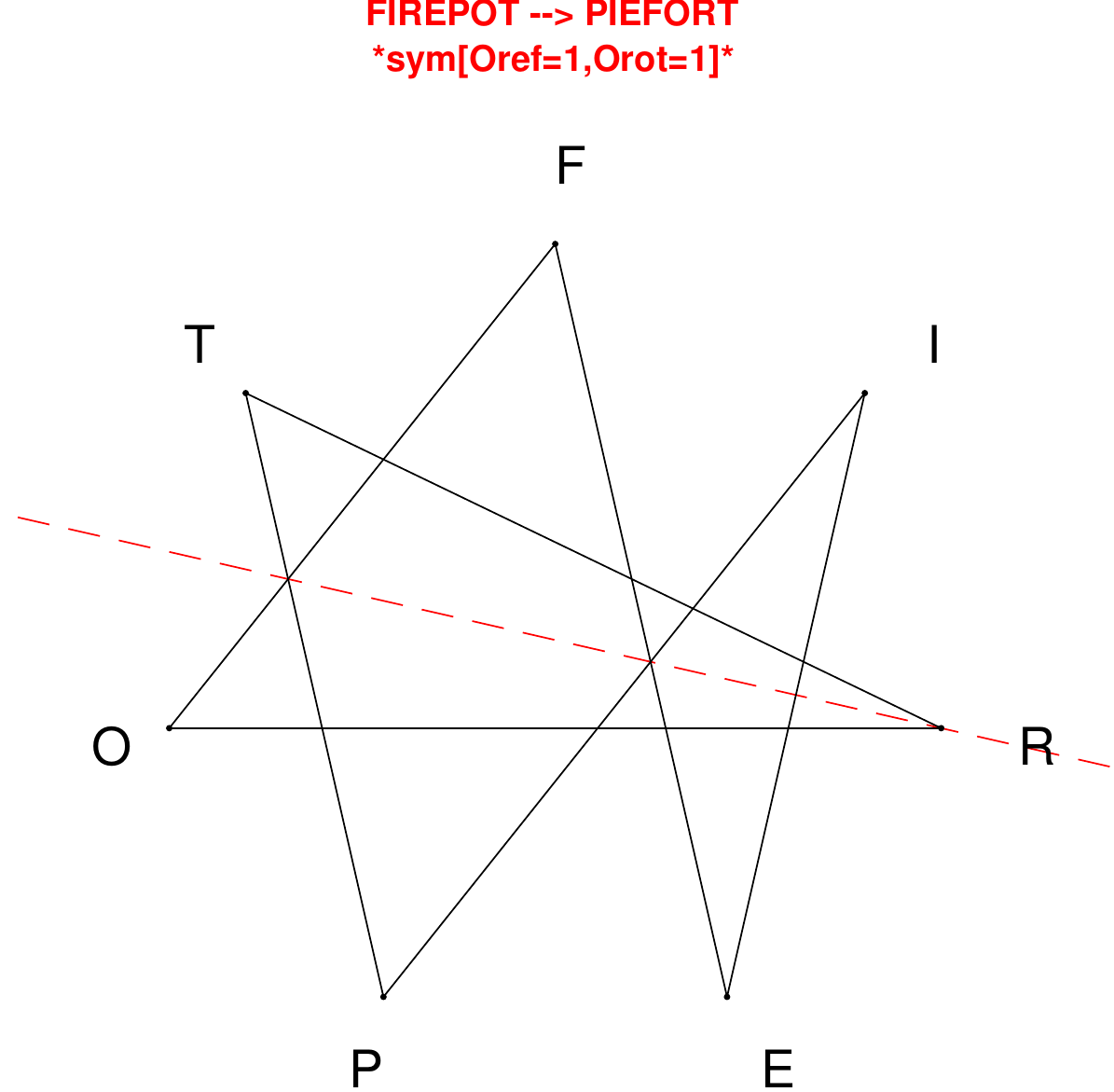}
\end{subfigure}
\hfill
\begin{subfigure}[T]{0.19\textwidth}
\centering
\includegraphics[width=\textwidth]{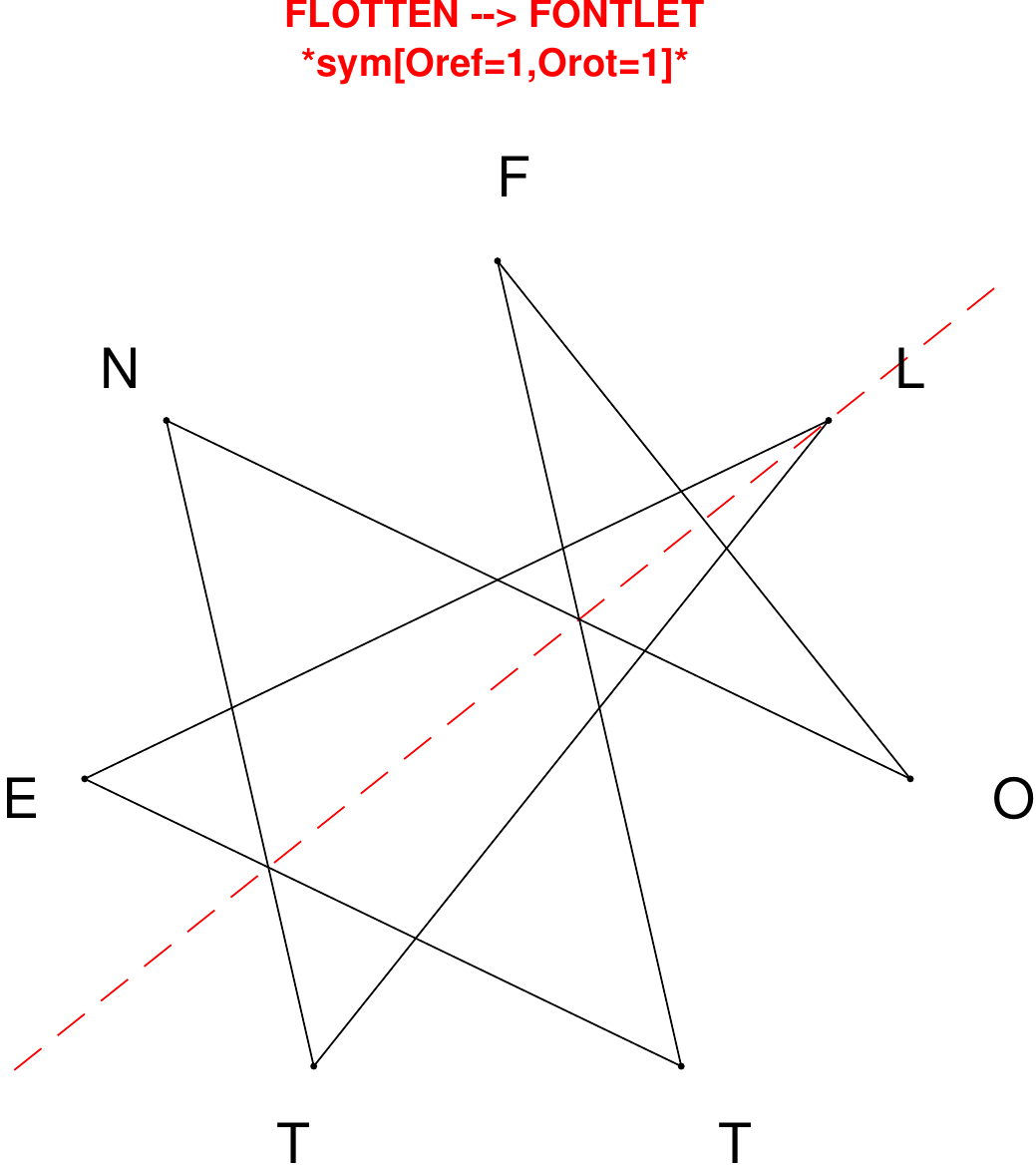}
\end{subfigure}
\hfill
\begin{subfigure}[T]{0.19\textwidth}
\centering
\includegraphics[width=\textwidth]{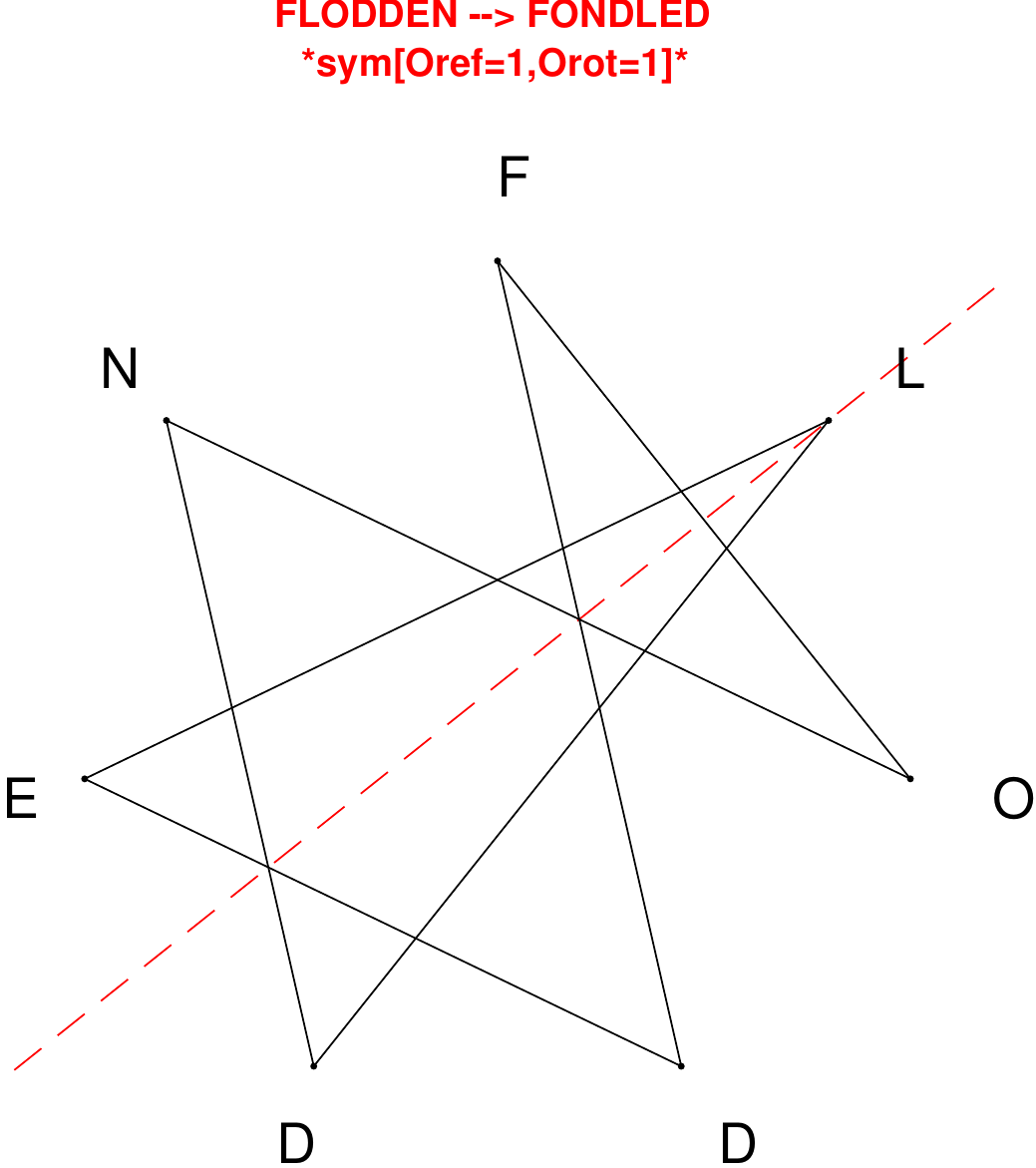}
\end{subfigure}
\end{figure}

\begin{figure}[H]
\centering
\begin{subfigure}[T]{0.19\textwidth}
\centering
\includegraphics[width=\textwidth]{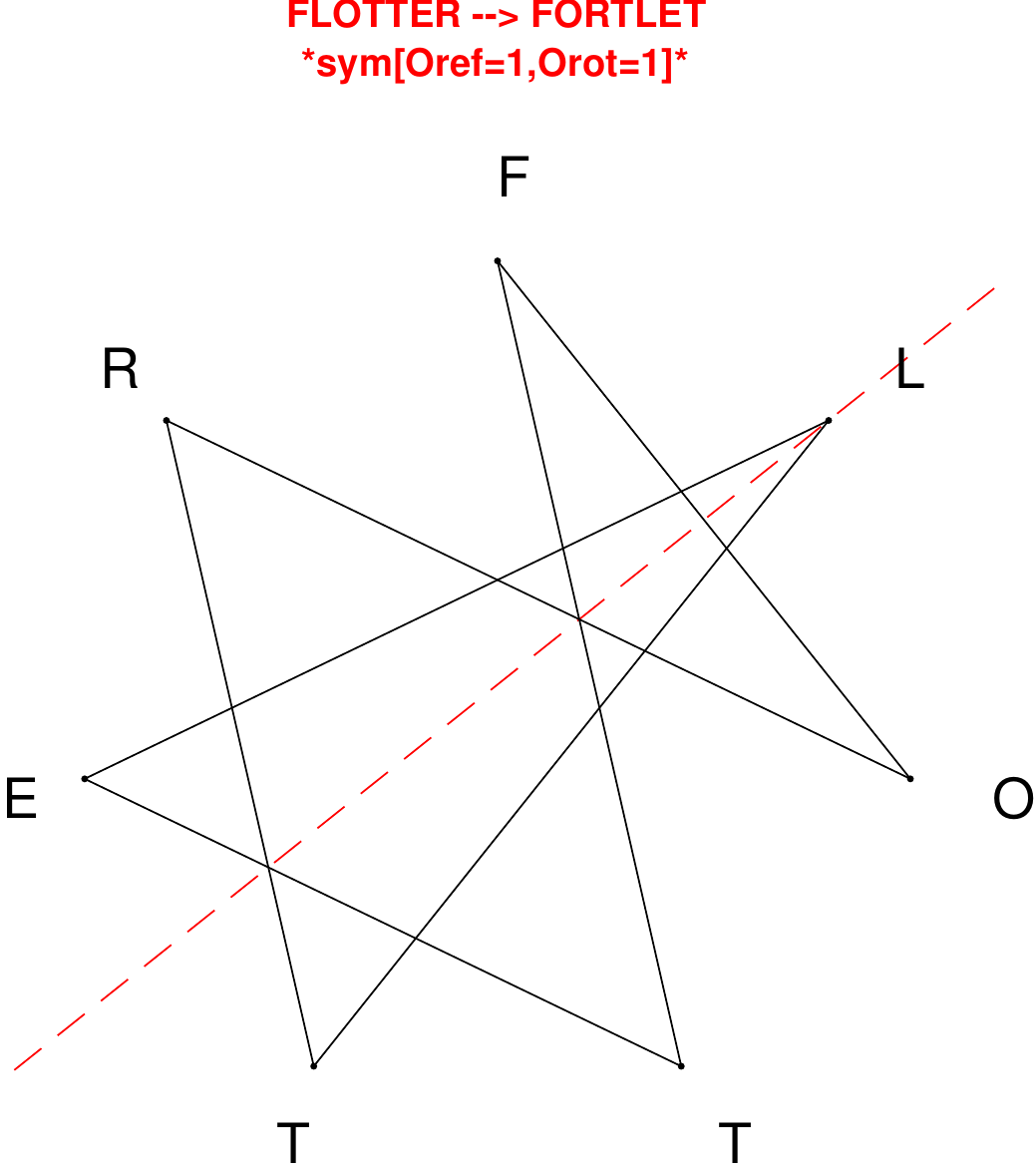}
\end{subfigure}
\hfill
\begin{subfigure}[T]{0.19\textwidth}
\centering
\includegraphics[width=\textwidth]{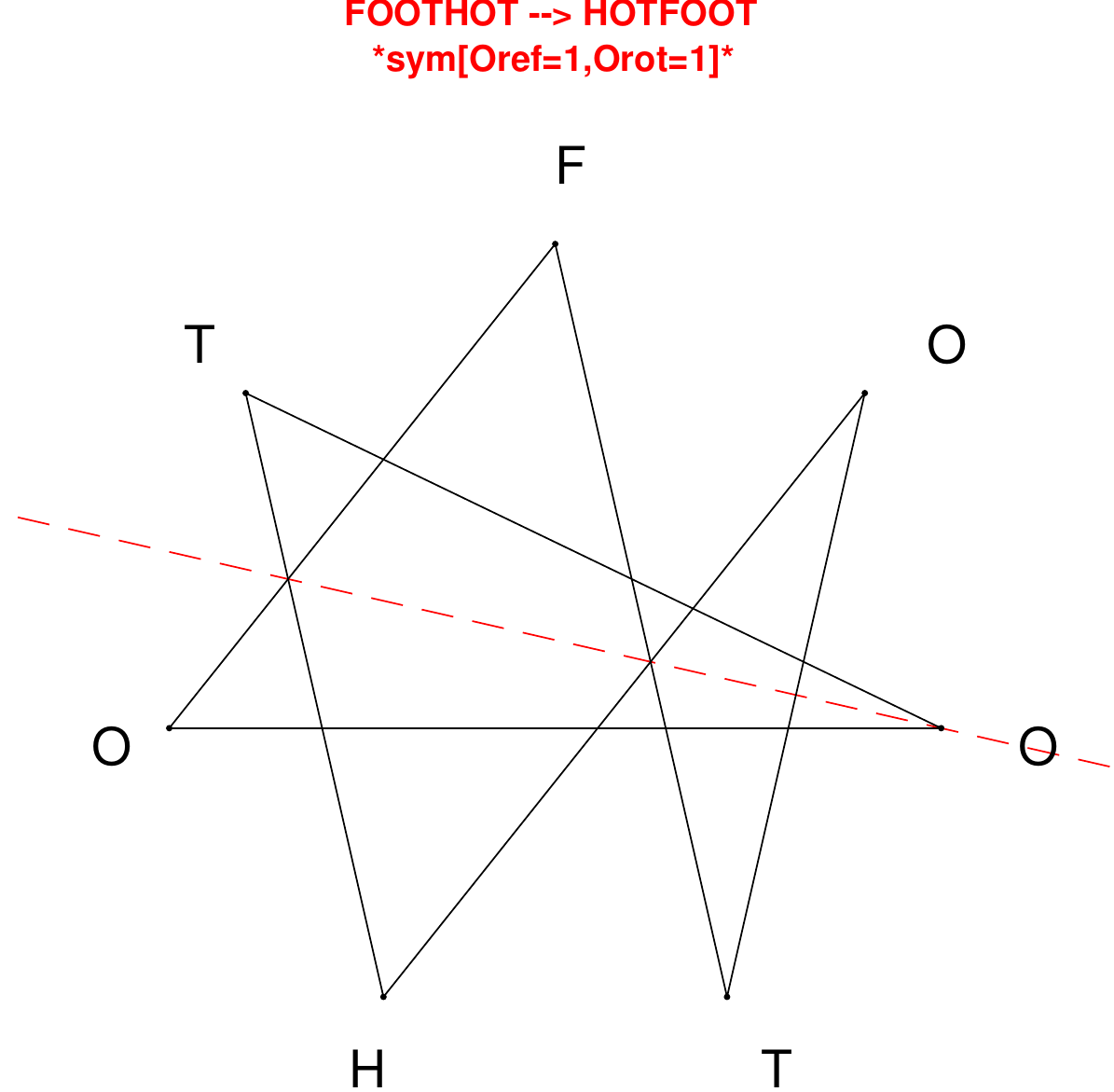}
\end{subfigure}
\hfill
\begin{subfigure}[T]{0.19\textwidth}
\centering
\includegraphics[width=\textwidth]{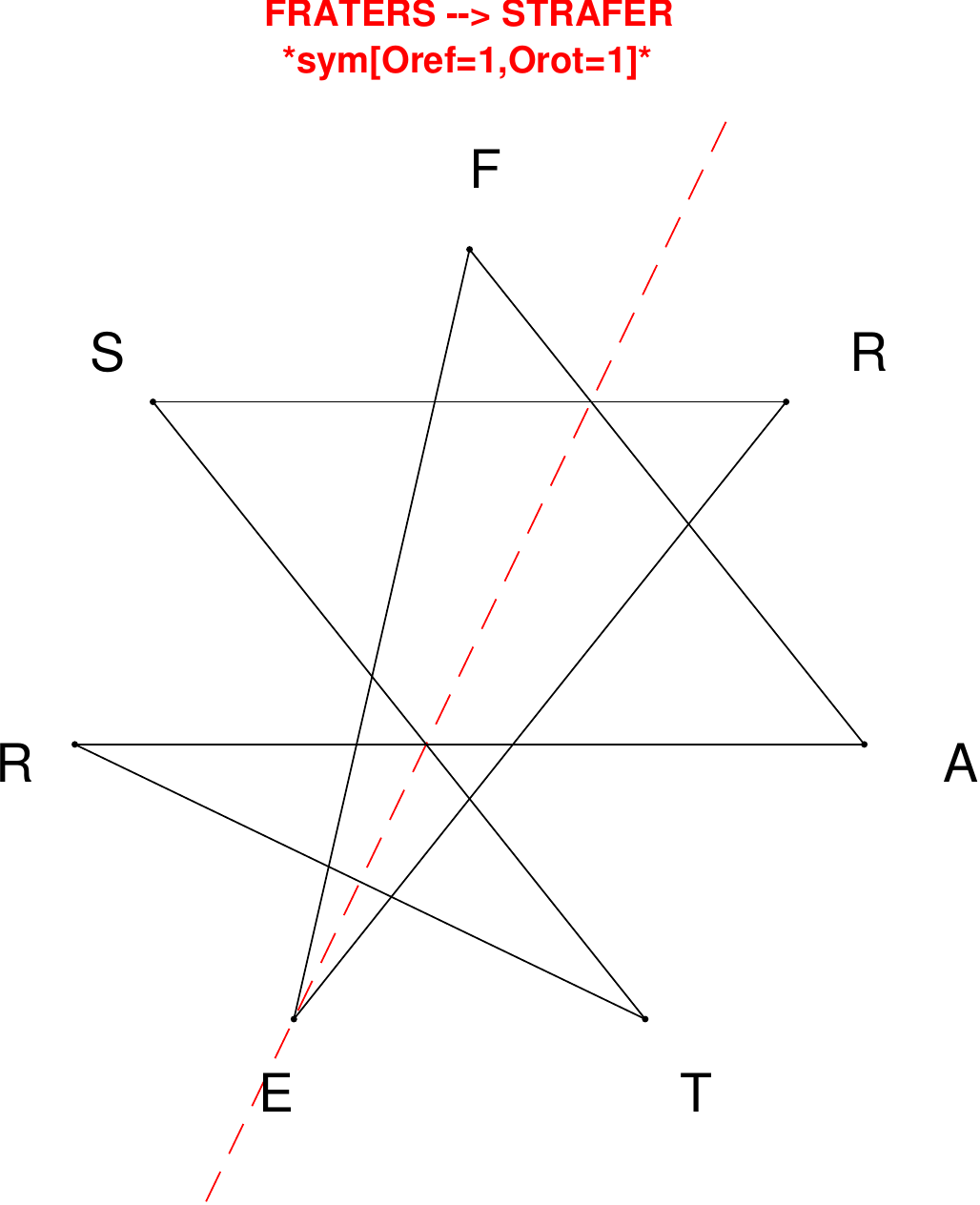}
\end{subfigure}
\hfill
\begin{subfigure}[T]{0.19\textwidth}
\centering
\includegraphics[width=\textwidth]{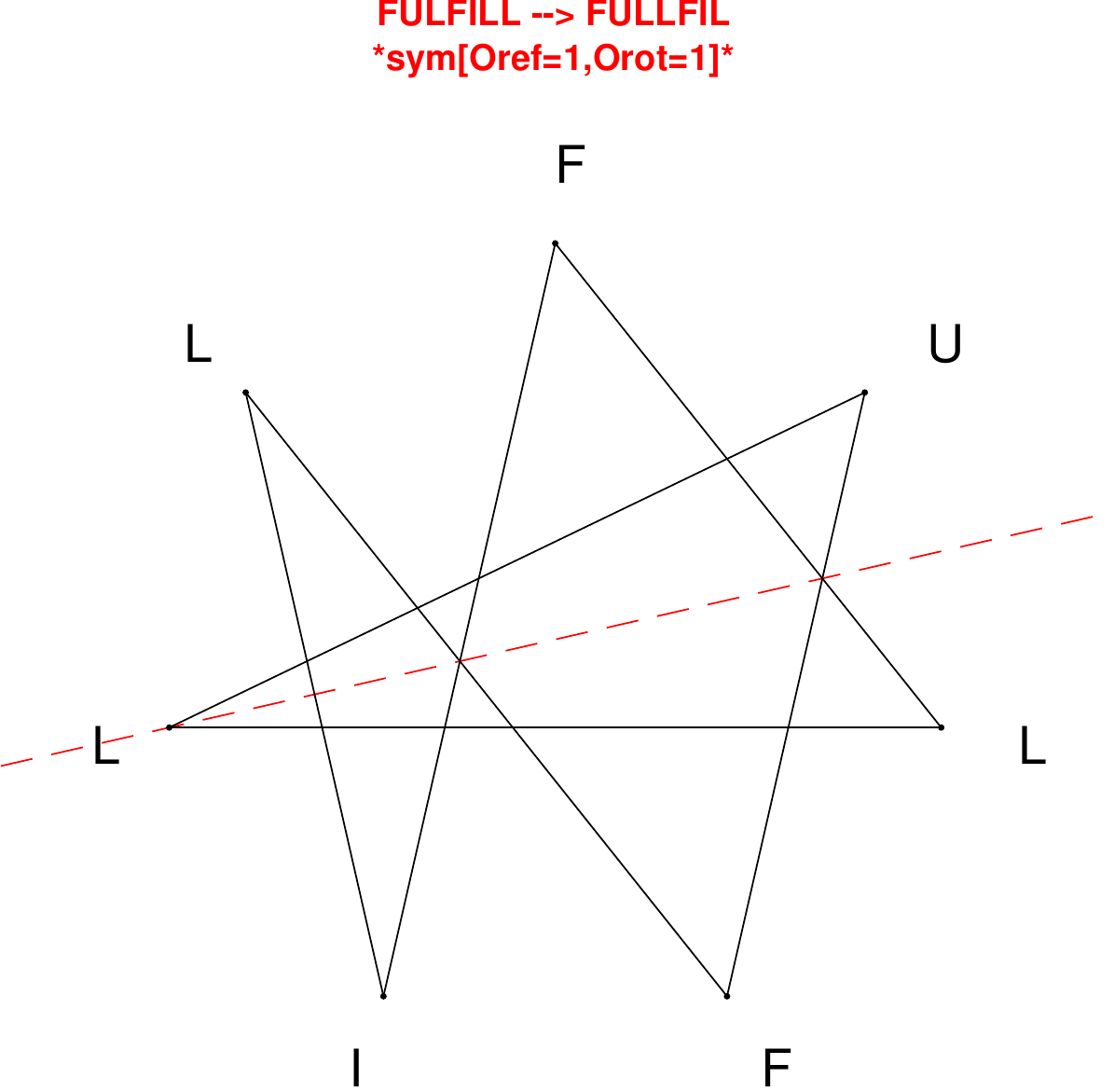}
\end{subfigure}
\hfill
\begin{subfigure}[T]{0.19\textwidth}
\centering
\includegraphics[width=\textwidth]{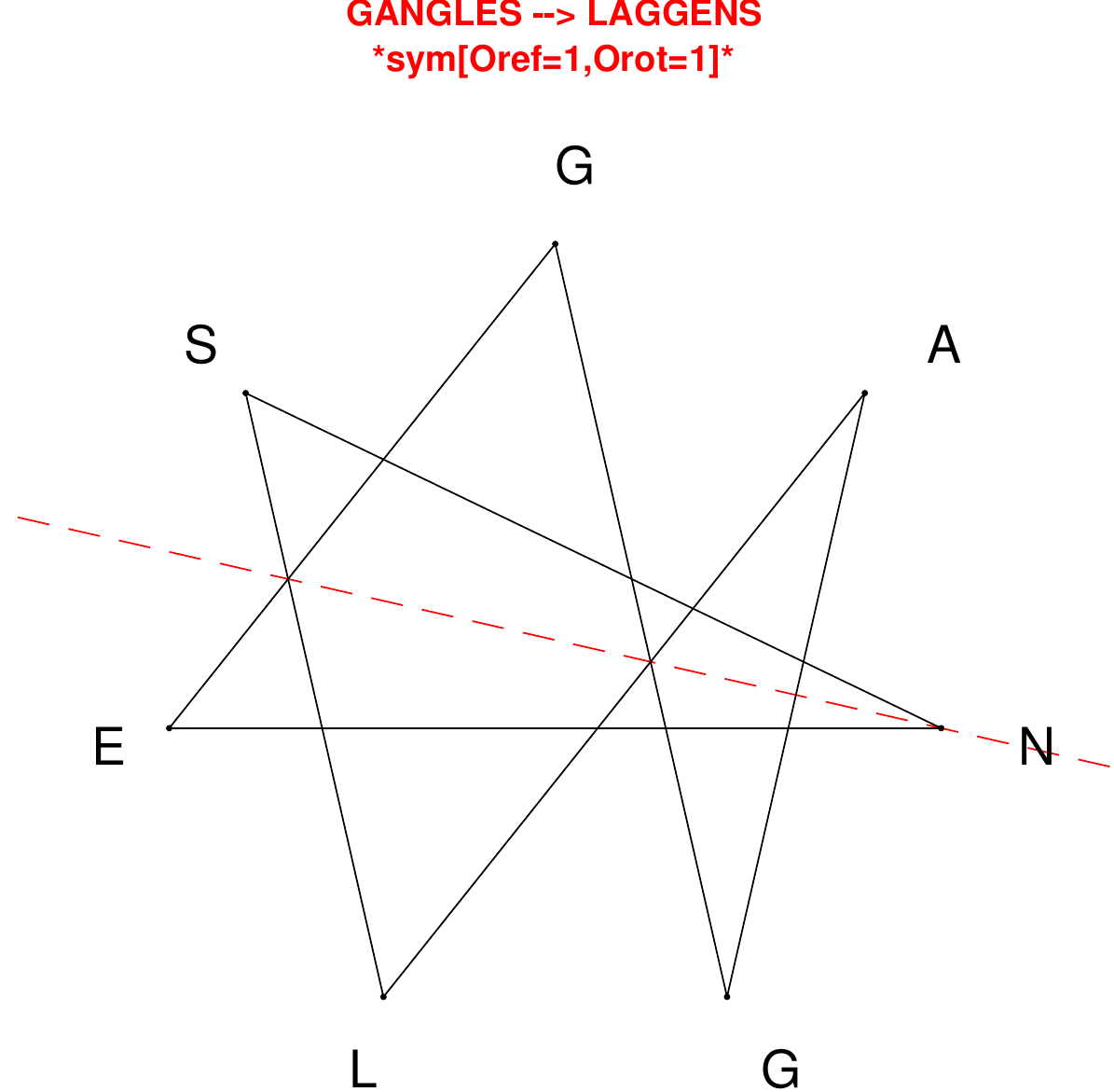}
\end{subfigure}
\end{figure}

\begin{figure}[H]
\centering
\begin{subfigure}[T]{0.19\textwidth}
\centering
\includegraphics[width=\textwidth]{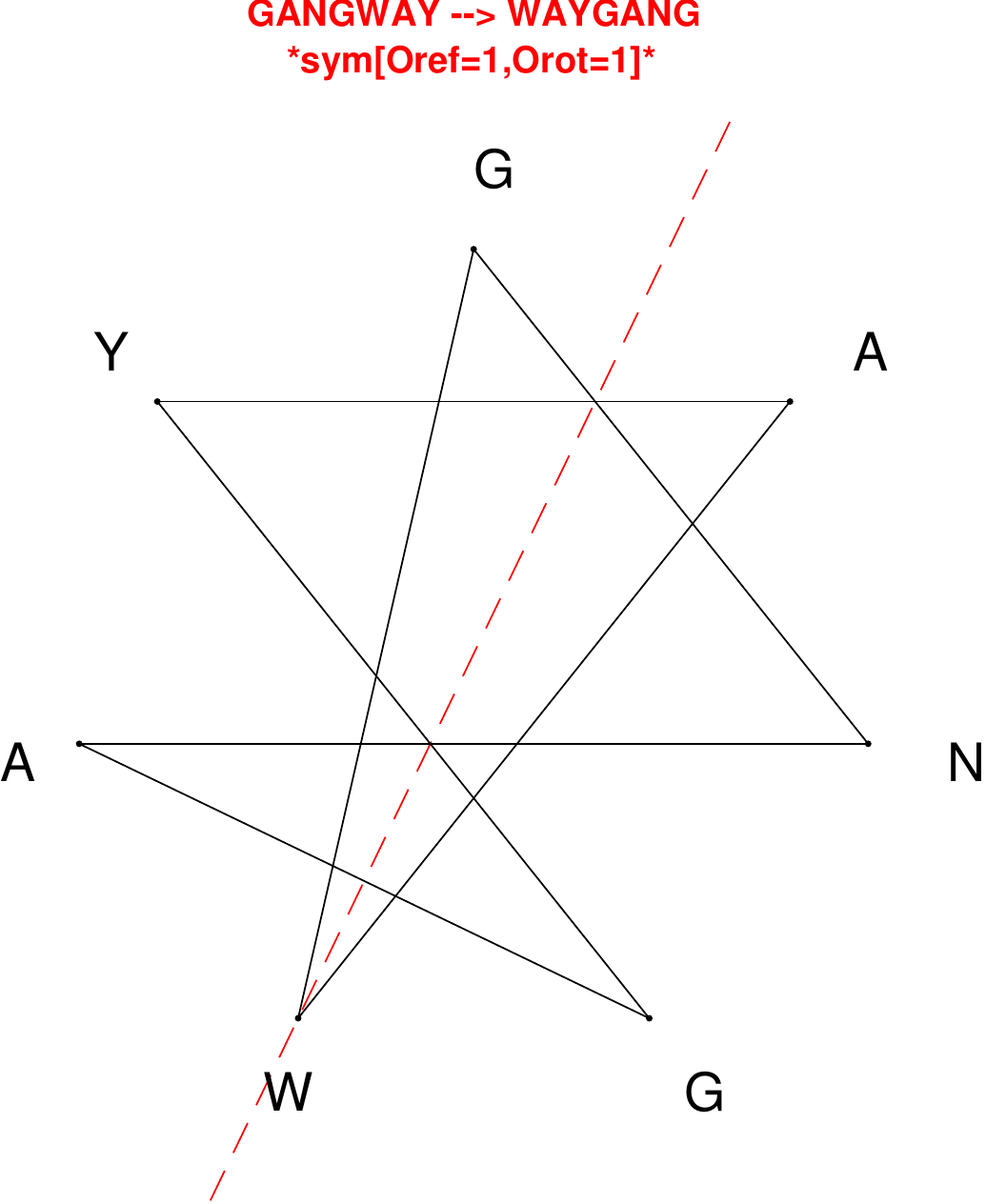}
\end{subfigure}
\hfill
\begin{subfigure}[T]{0.19\textwidth}
\centering
\includegraphics[width=\textwidth]{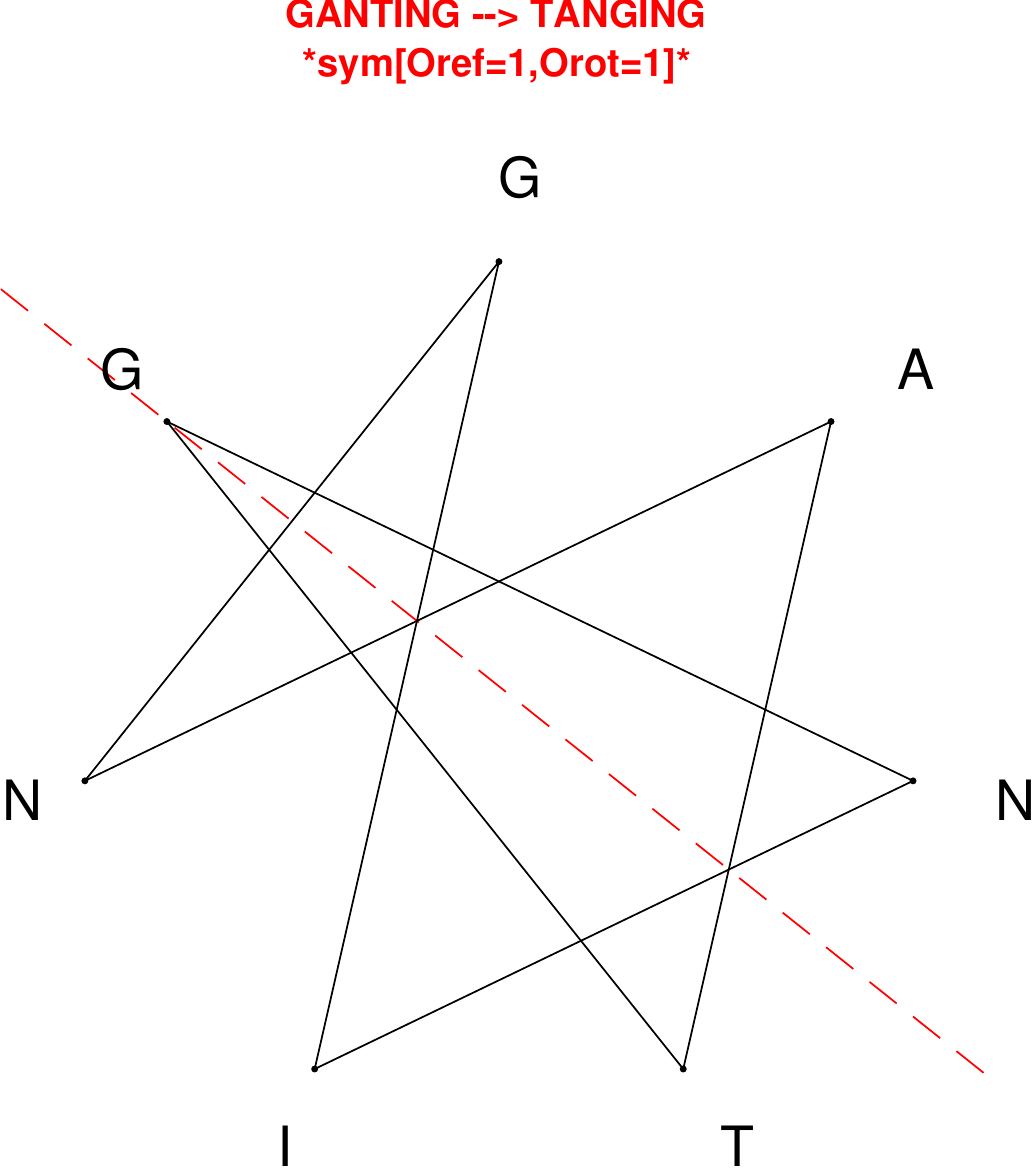}
\end{subfigure}
\hfill
\begin{subfigure}[T]{0.19\textwidth}
\centering
\includegraphics[width=\textwidth]{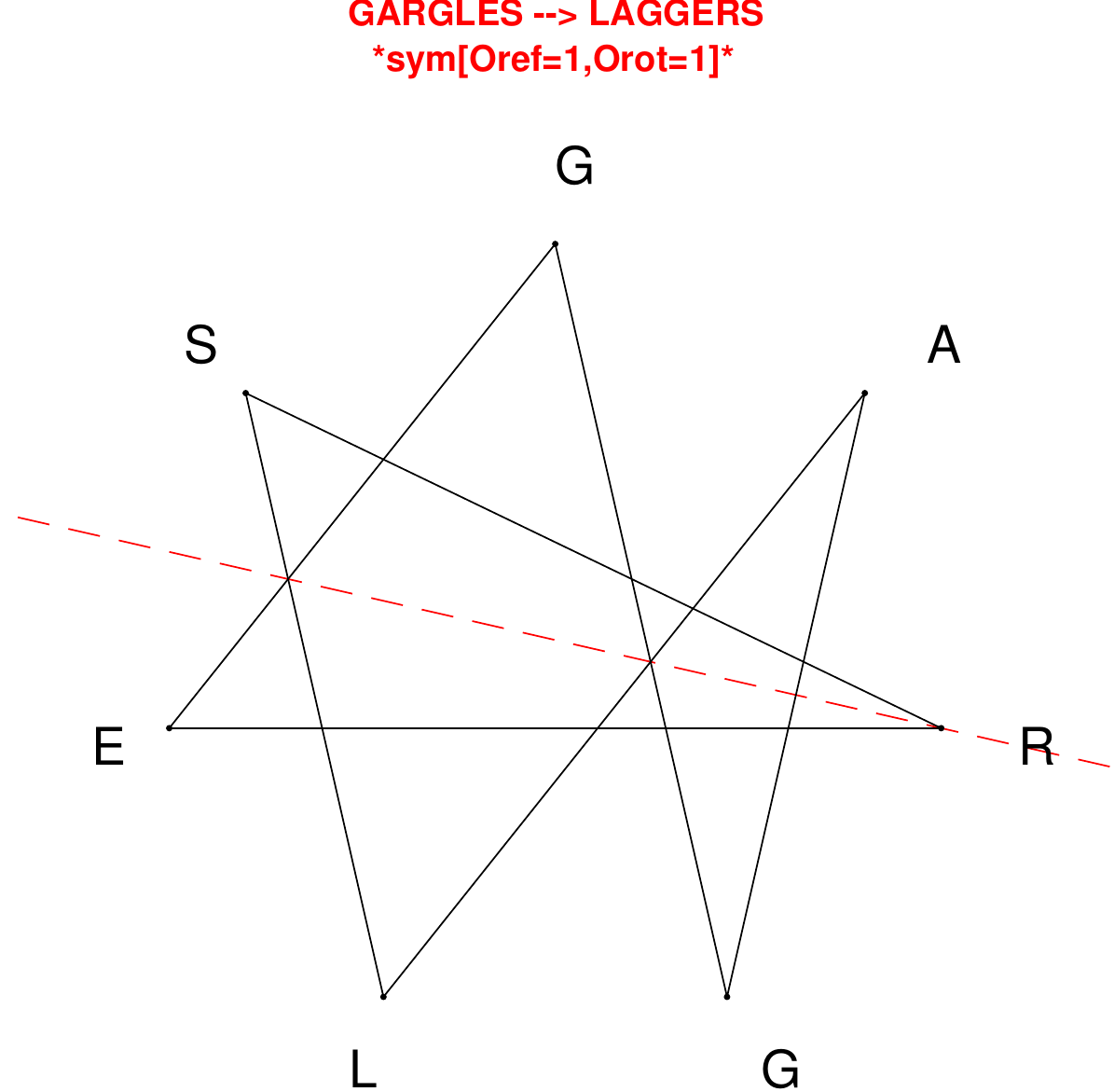}
\end{subfigure}
\hfill
\begin{subfigure}[T]{0.19\textwidth}
\centering
\includegraphics[width=\textwidth]{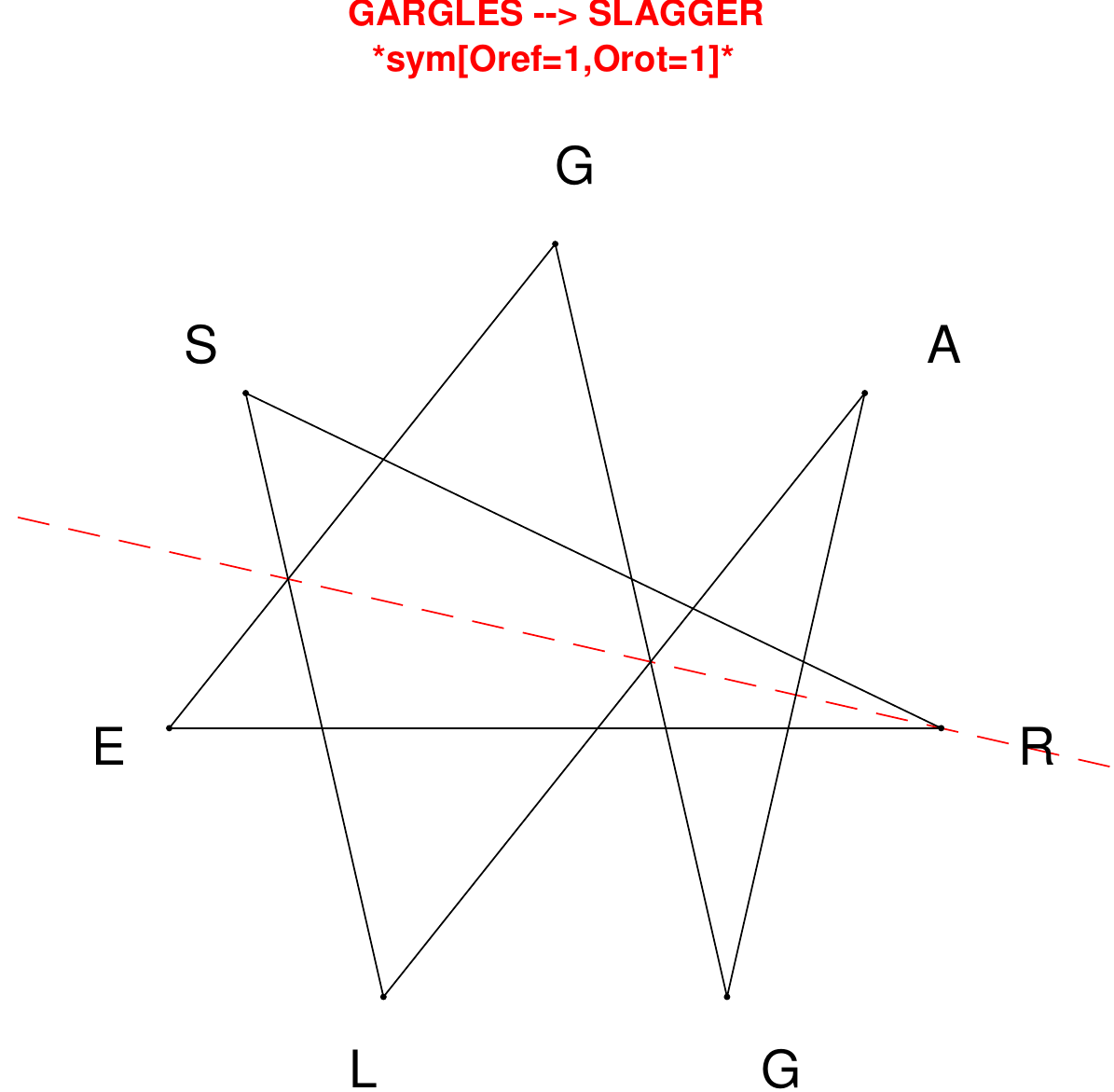}
\end{subfigure}
\hfill
\begin{subfigure}[T]{0.19\textwidth}
\centering
\includegraphics[width=\textwidth]{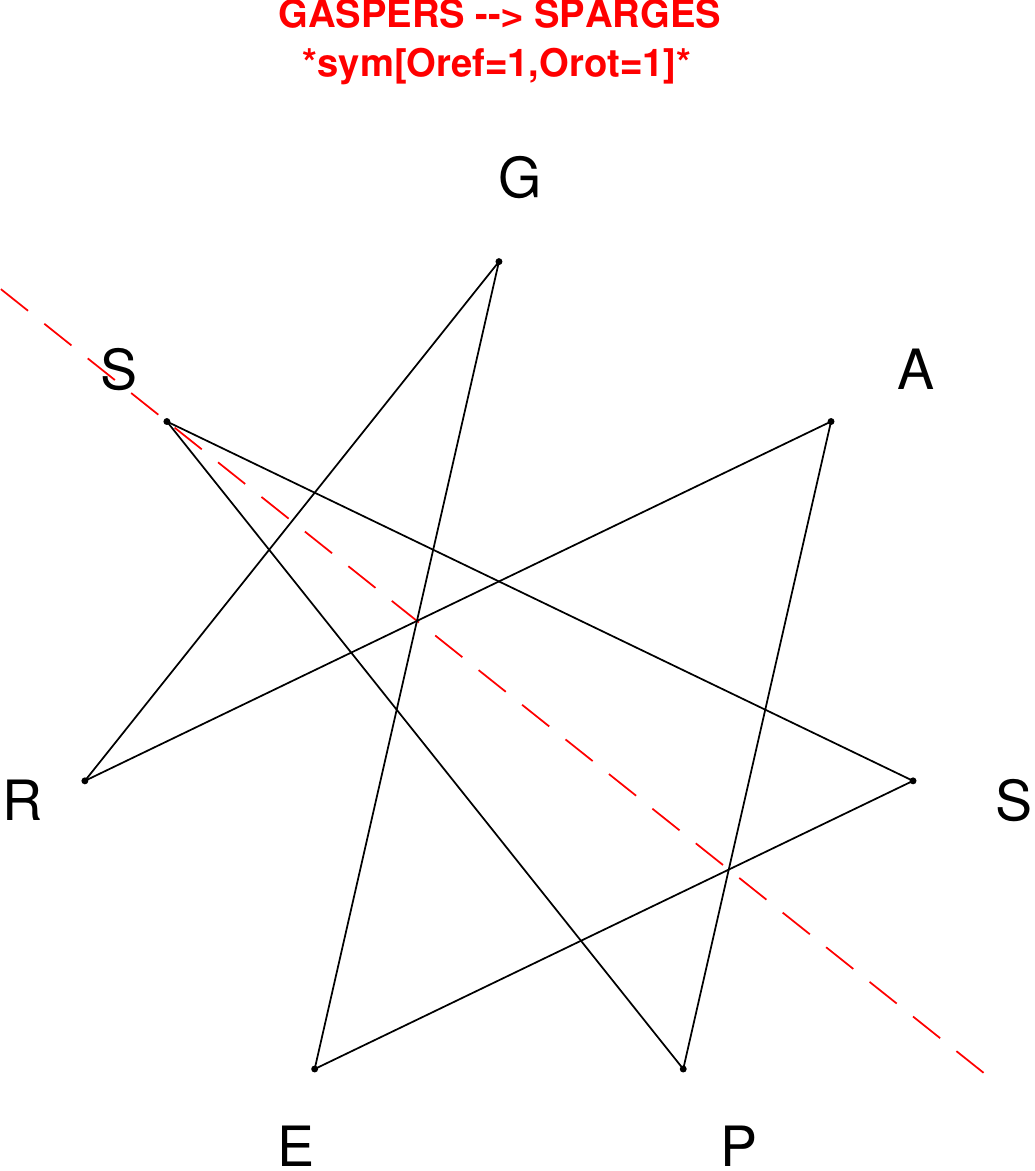}
\end{subfigure}
\end{figure}

\begin{figure}[H]
\centering
\begin{subfigure}[T]{0.19\textwidth}
\centering
\includegraphics[width=\textwidth]{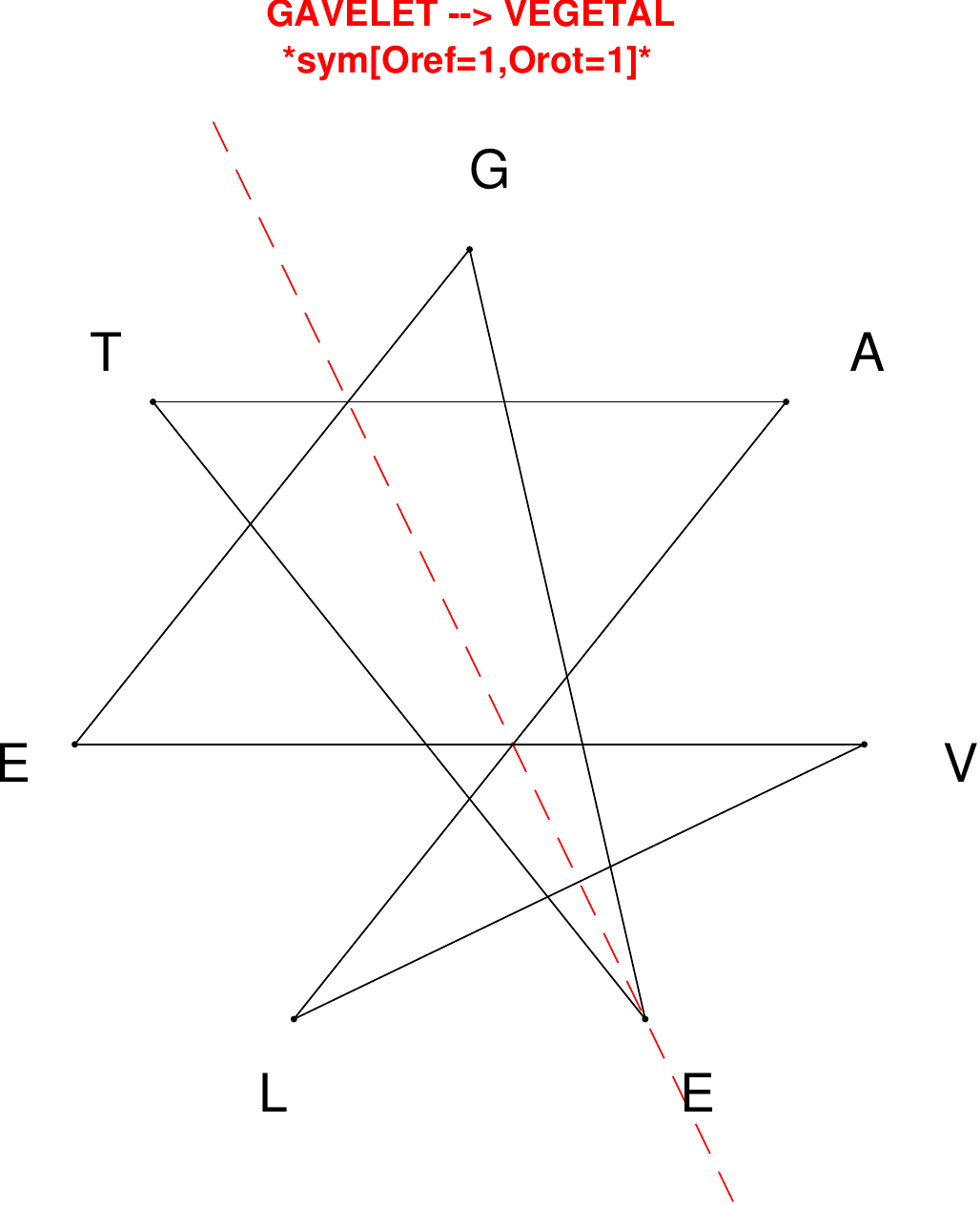}
\end{subfigure}
\hfill
\begin{subfigure}[T]{0.19\textwidth}
\centering
\includegraphics[width=\textwidth]{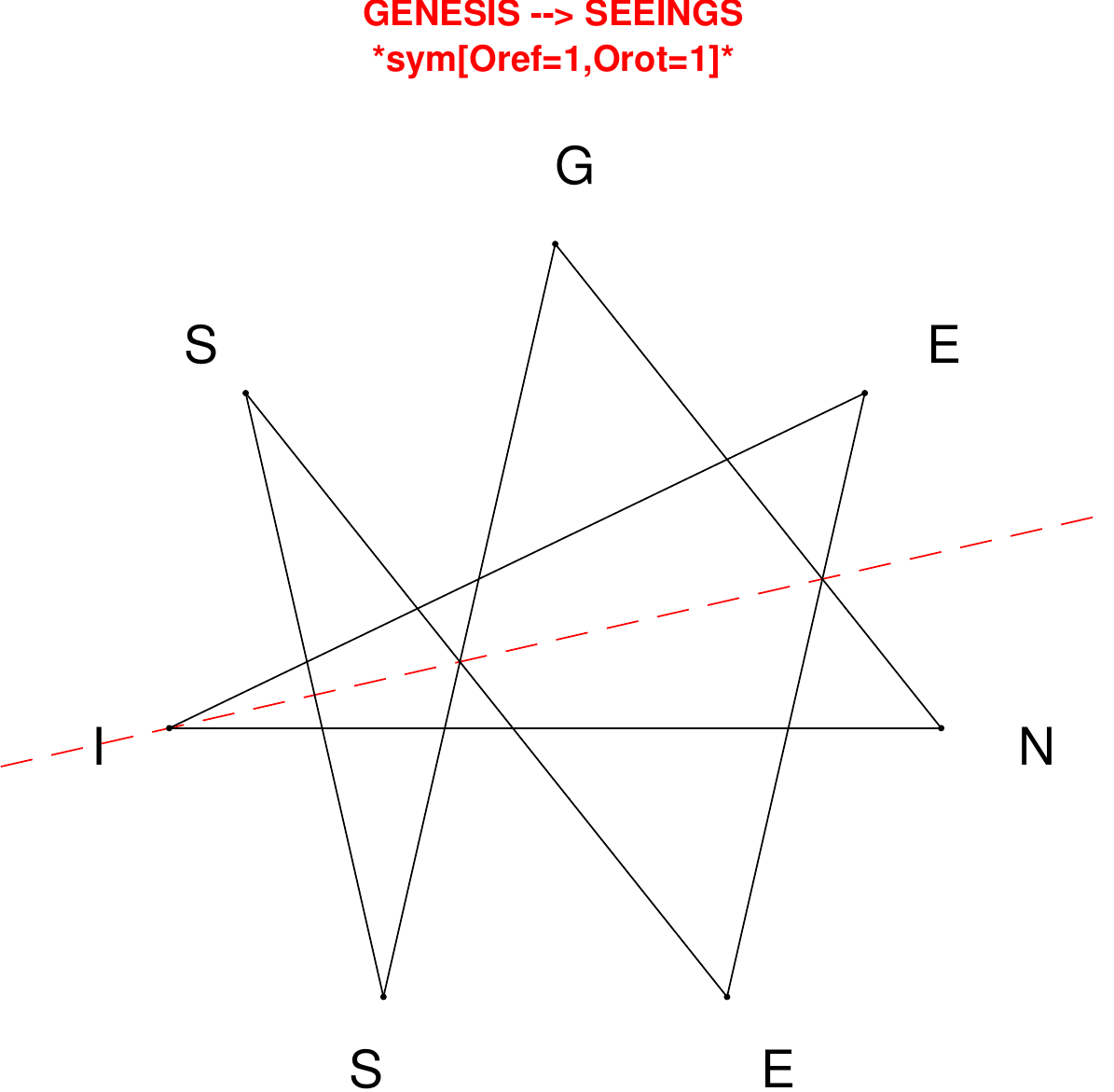}
\end{subfigure}
\hfill
\begin{subfigure}[T]{0.19\textwidth}
\centering
\includegraphics[width=\textwidth]{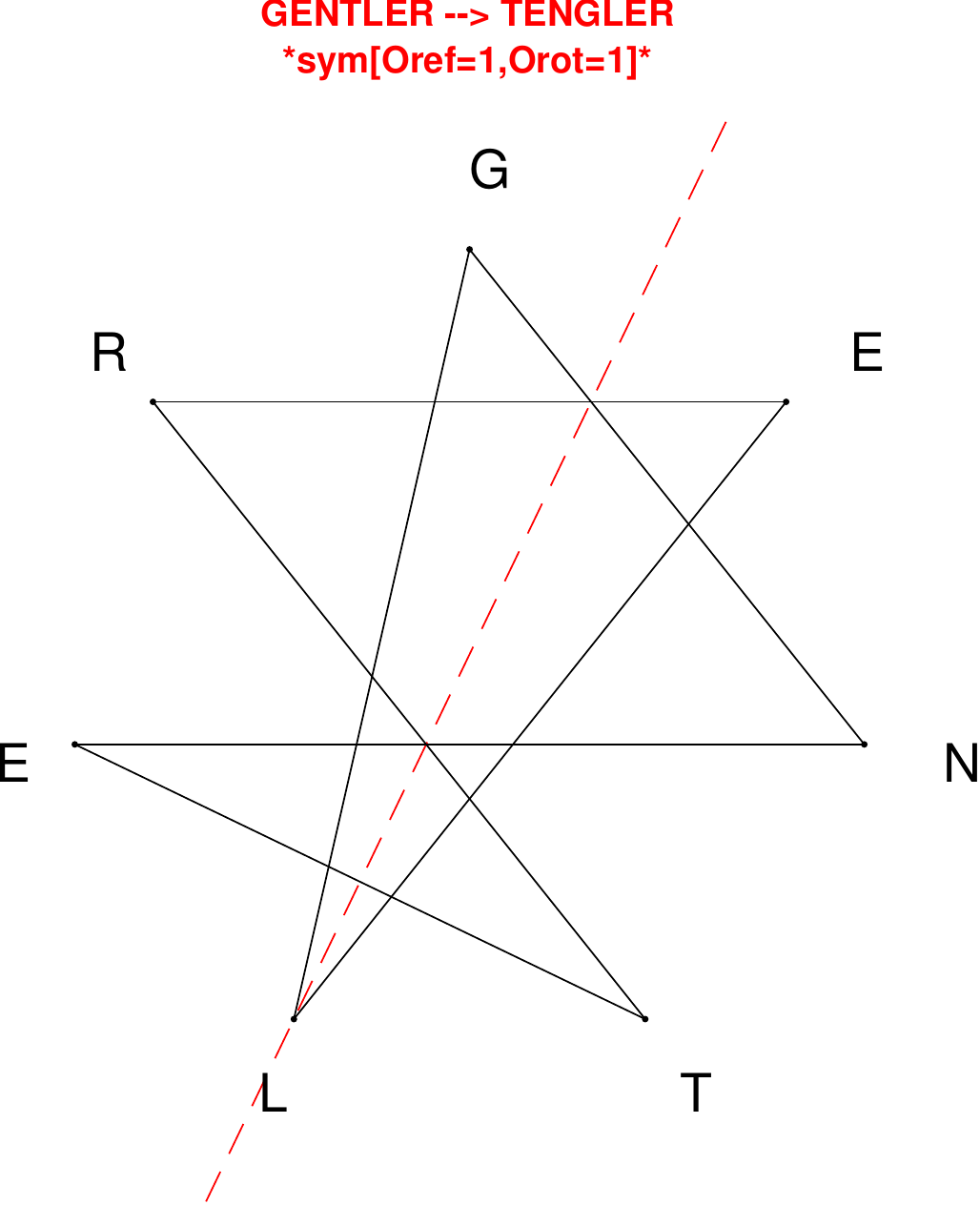}
\end{subfigure}
\hfill
\begin{subfigure}[T]{0.19\textwidth}
\centering
\includegraphics[width=\textwidth]{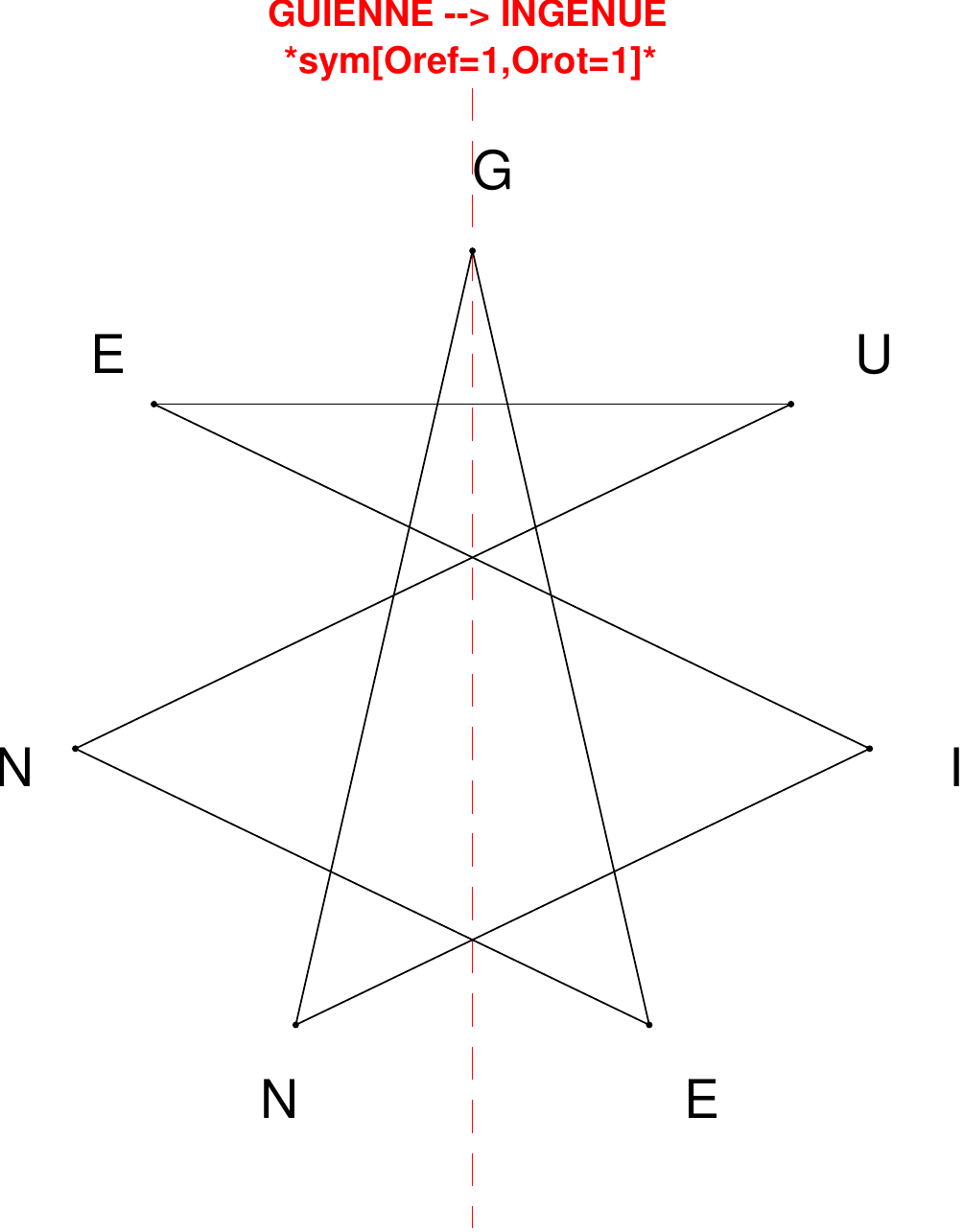}
\end{subfigure}
\hfill
\begin{subfigure}[T]{0.19\textwidth}
\centering
\includegraphics[width=\textwidth]{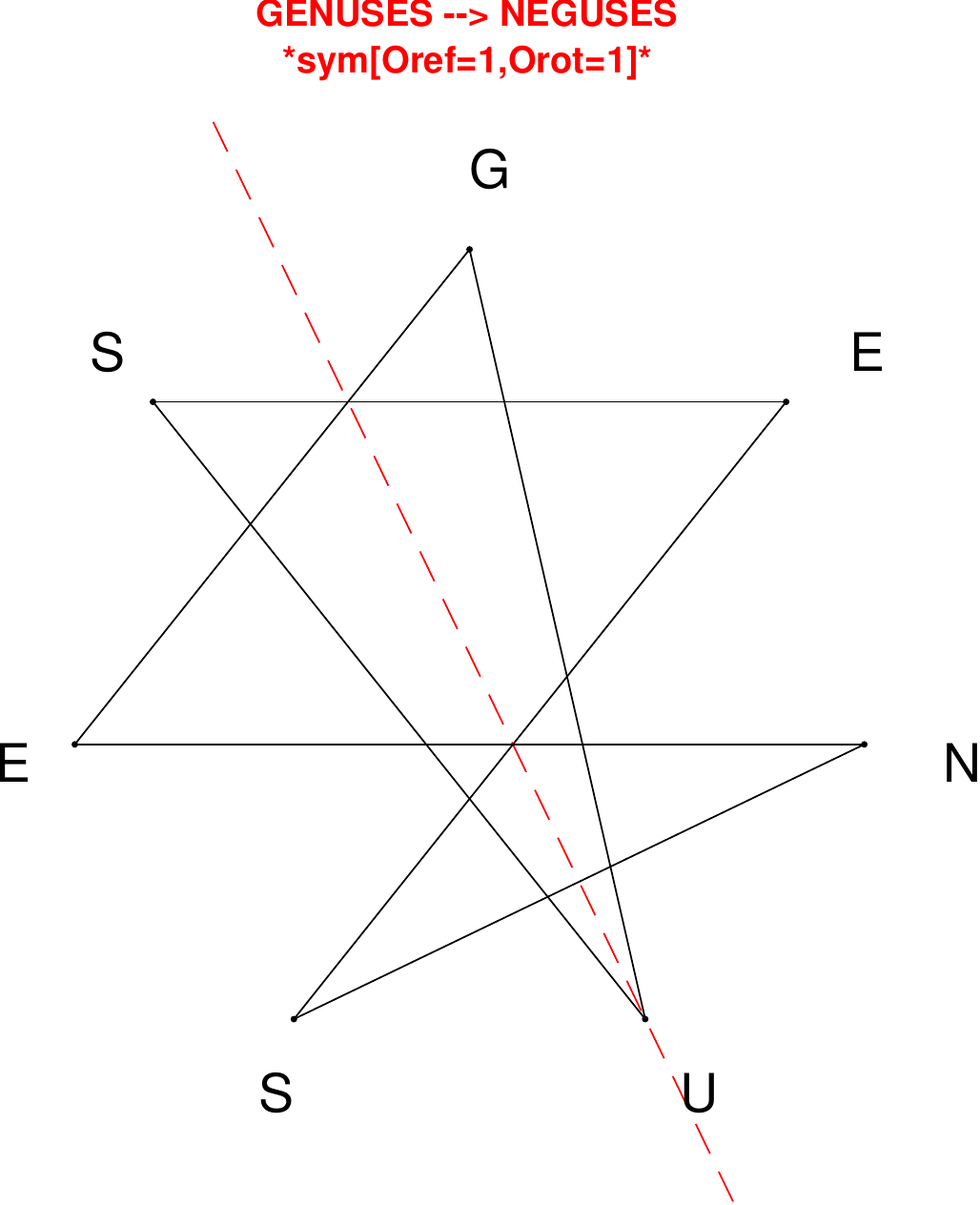}
\end{subfigure}
\end{figure}

\begin{figure}[H]
\centering
\begin{subfigure}[T]{0.19\textwidth}
\centering
\includegraphics[width=\textwidth]{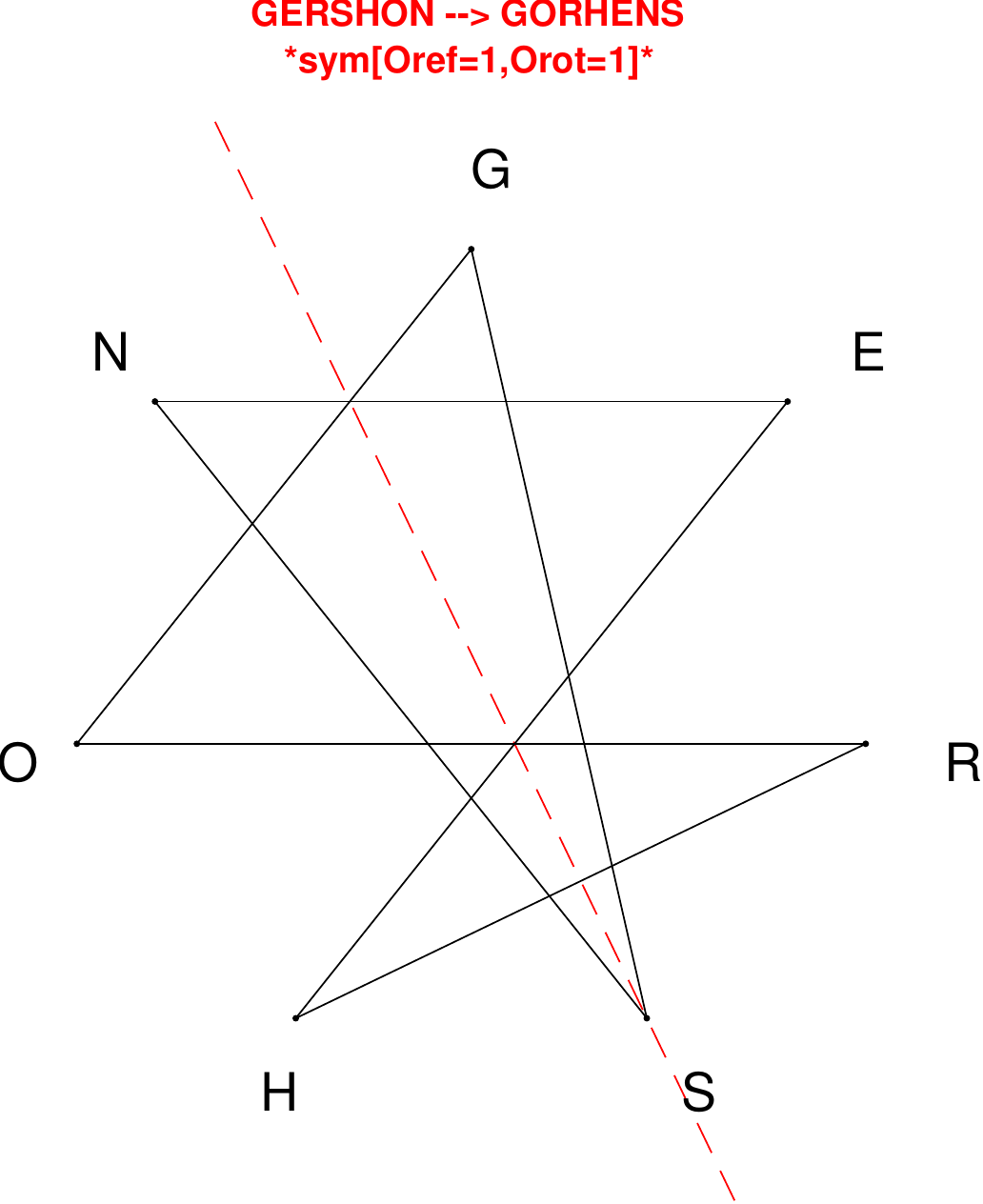}
\end{subfigure}
\hfill
\begin{subfigure}[T]{0.19\textwidth}
\centering
\includegraphics[width=\textwidth]{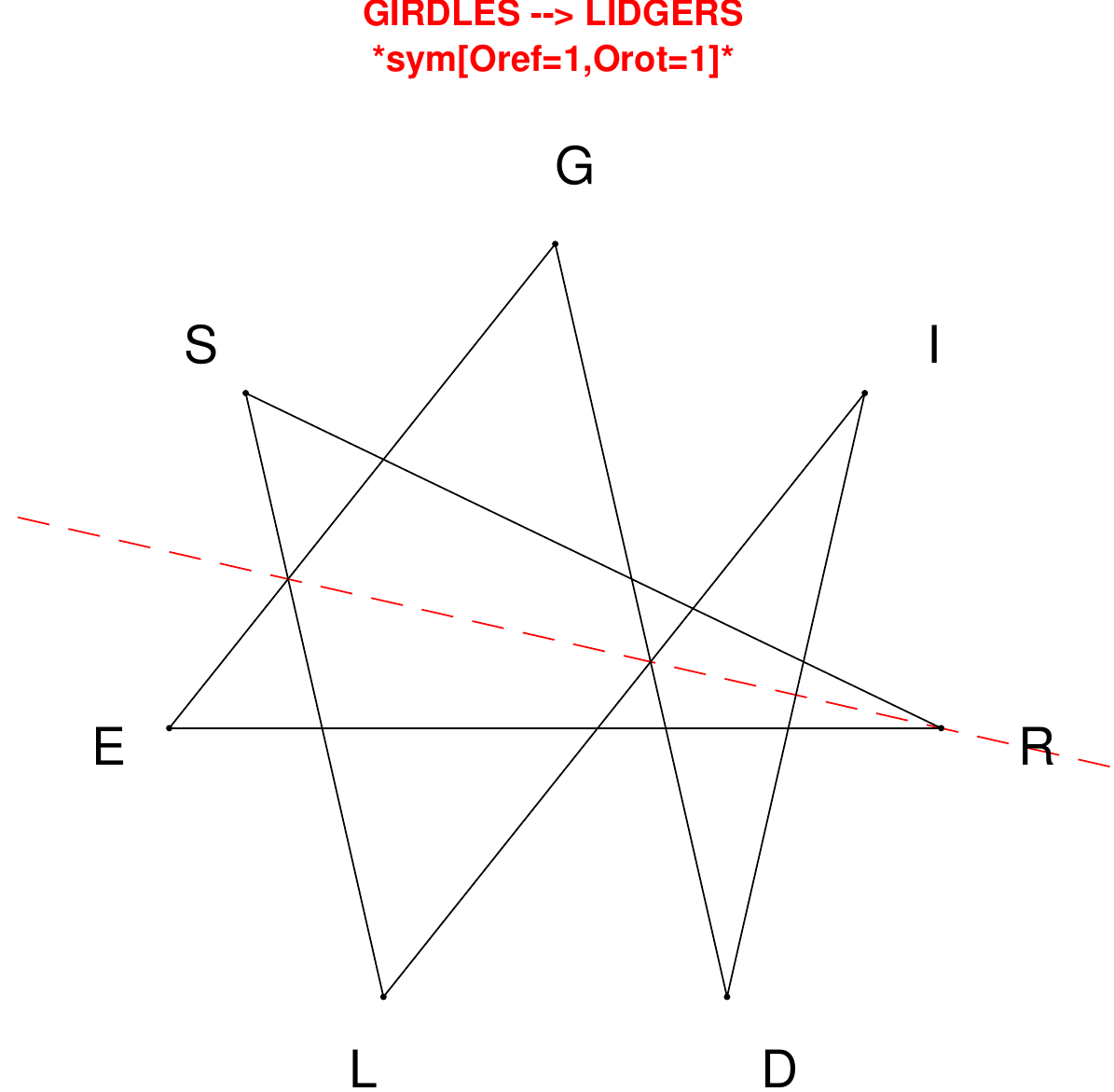}
\end{subfigure}
\hfill
\begin{subfigure}[T]{0.19\textwidth}
\centering
\includegraphics[width=\textwidth]{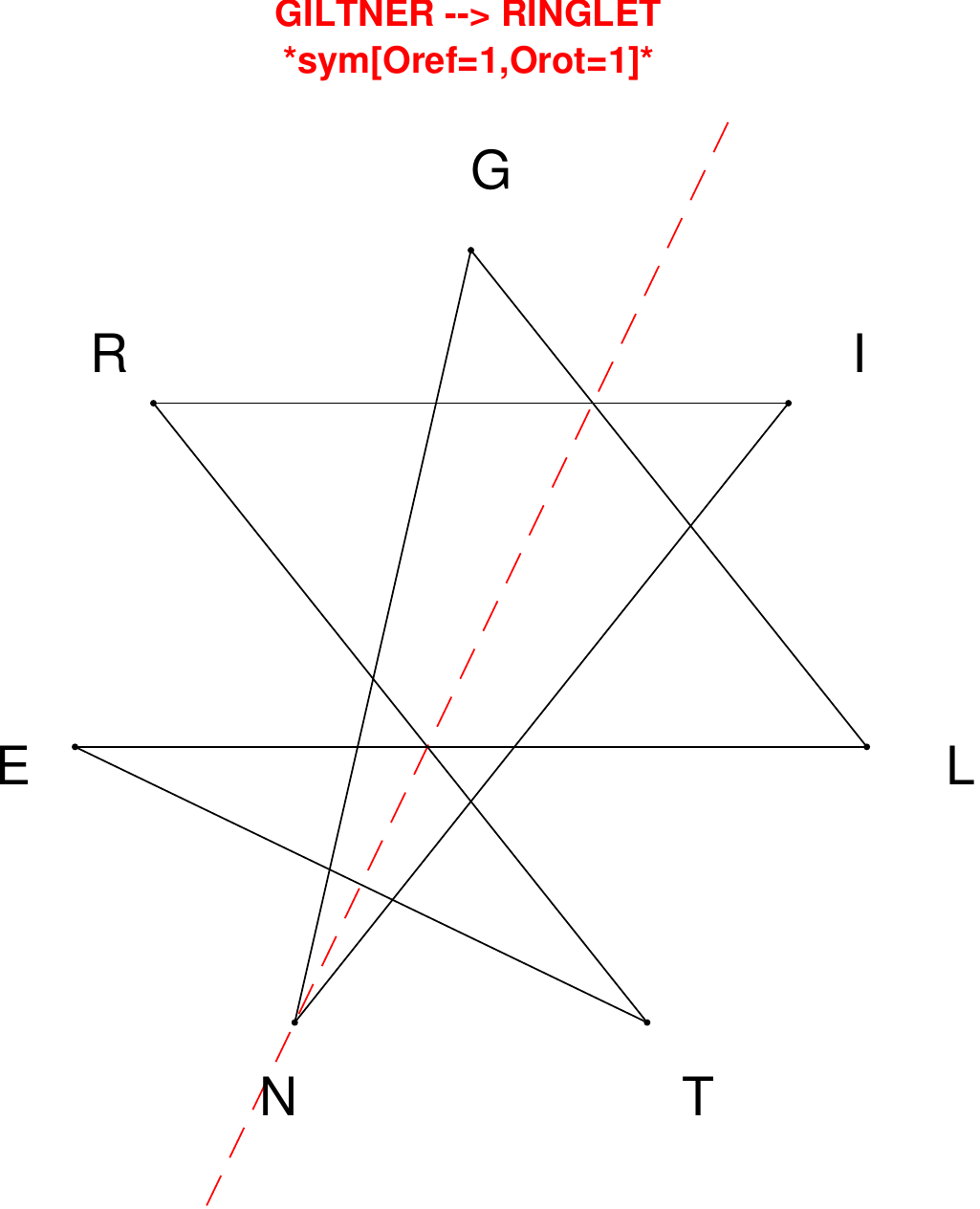}
\end{subfigure}
\hfill
\begin{subfigure}[T]{0.19\textwidth}
\centering
\includegraphics[width=\textwidth]{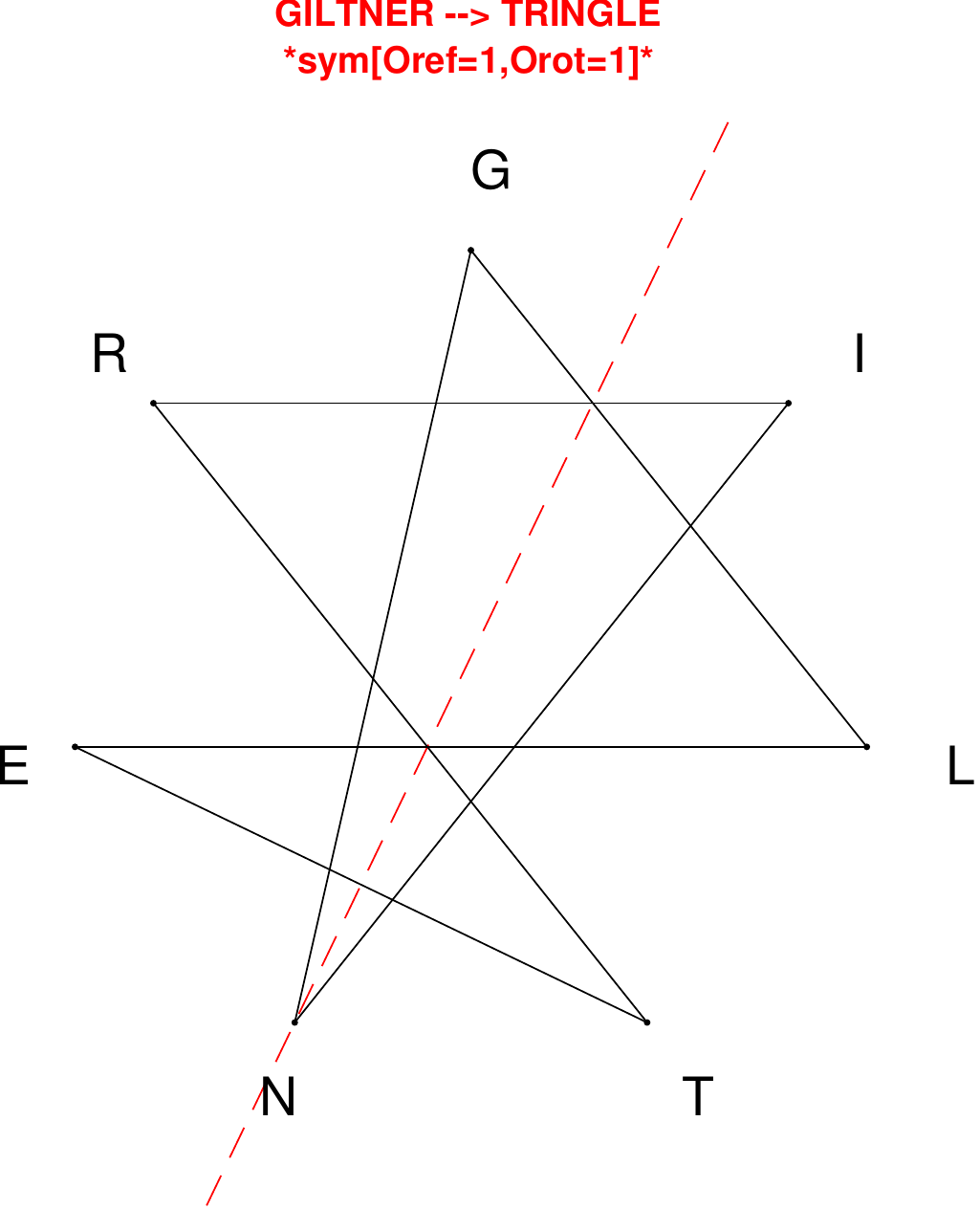}
\end{subfigure}
\hfill
\begin{subfigure}[T]{0.19\textwidth}
\centering
\includegraphics[width=\textwidth]{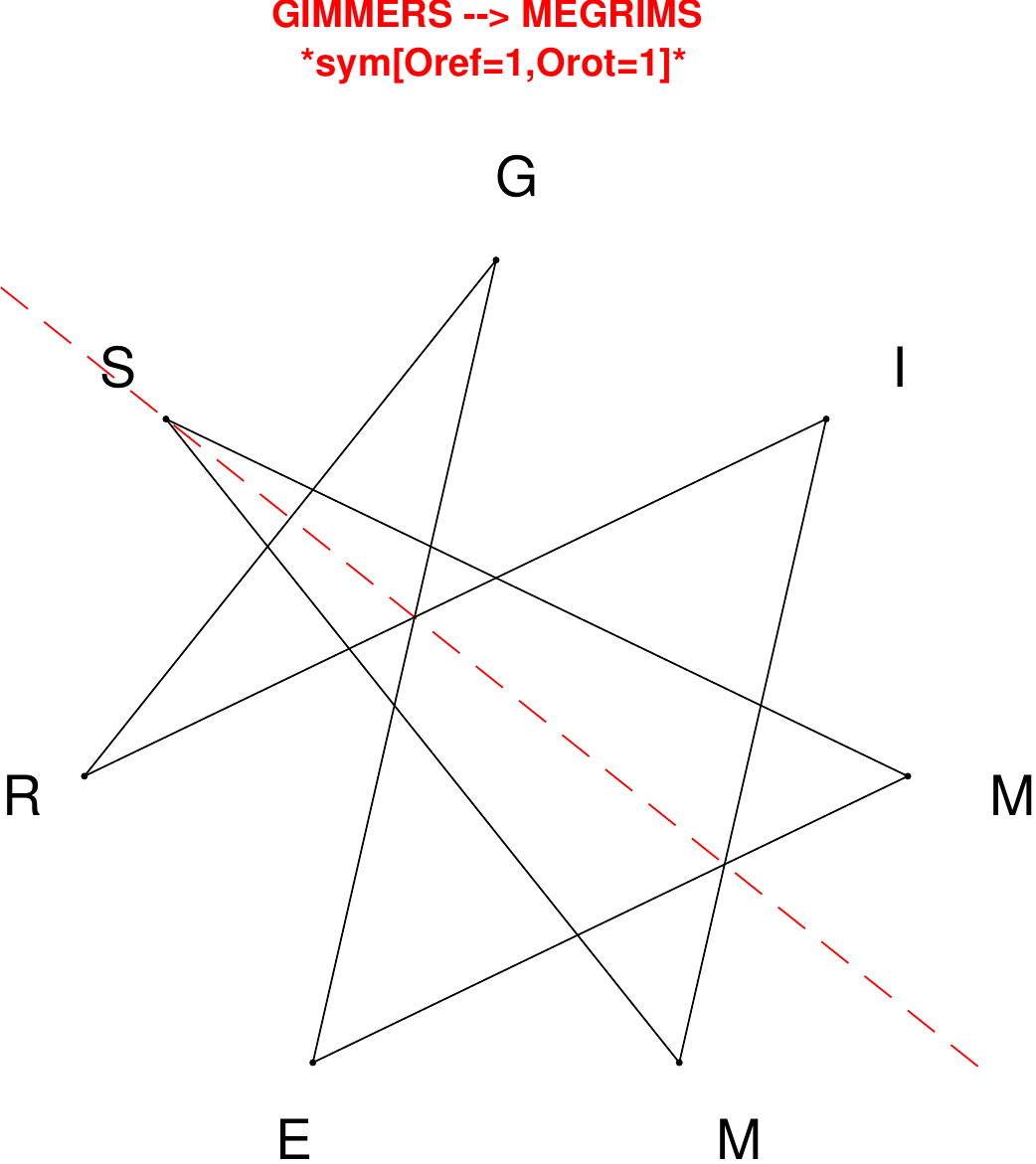}
\end{subfigure}
\end{figure}

\begin{figure}[H]
\centering
\begin{subfigure}[T]{0.19\textwidth}
\centering
\includegraphics[width=\textwidth]{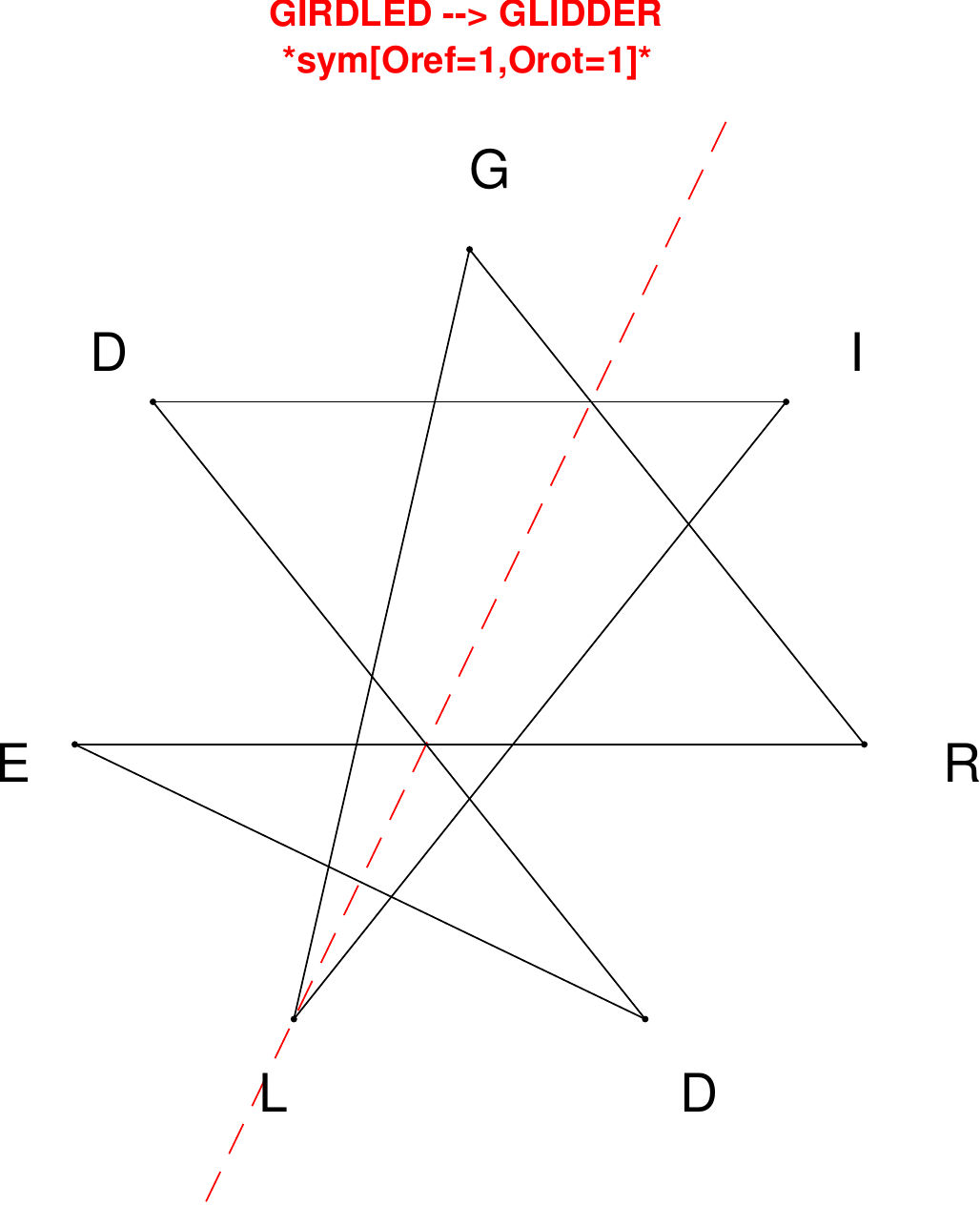}
\end{subfigure}
\hfill
\begin{subfigure}[T]{0.19\textwidth}
\centering
\includegraphics[width=\textwidth]{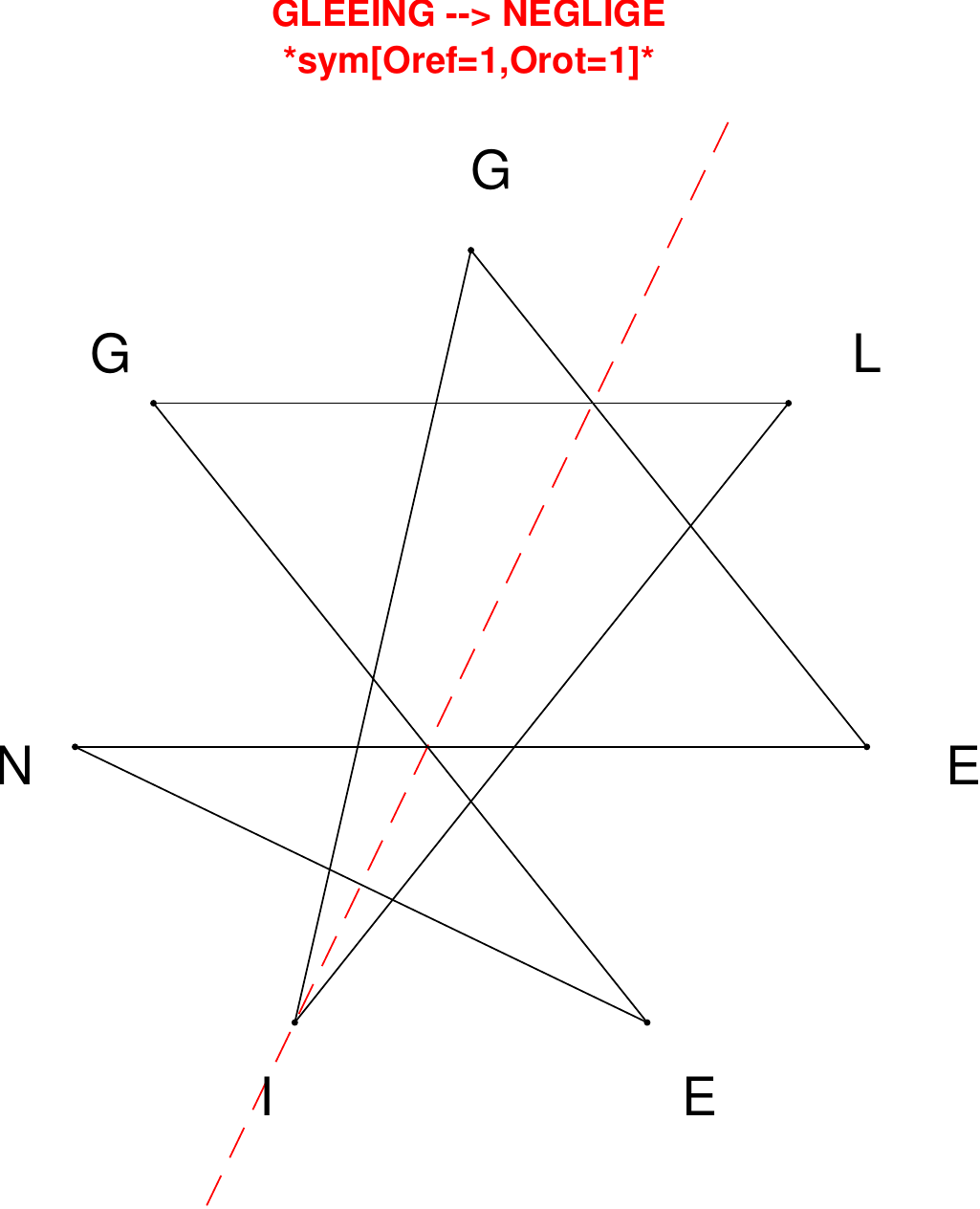}
\end{subfigure}
\hfill
\begin{subfigure}[T]{0.19\textwidth}
\centering
\includegraphics[width=\textwidth]{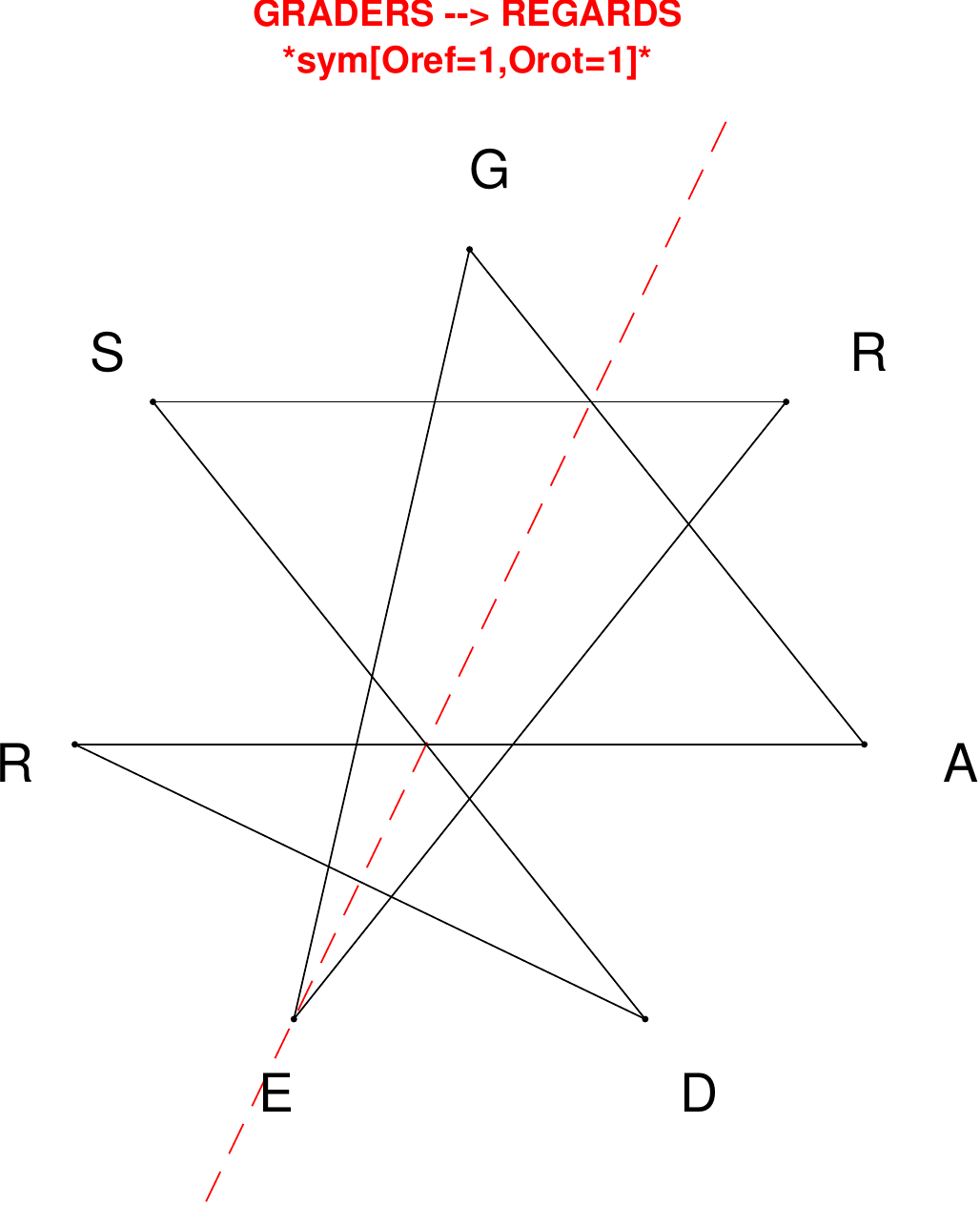}
\end{subfigure}
\hfill
\begin{subfigure}[T]{0.19\textwidth}
\centering
\includegraphics[width=\textwidth]{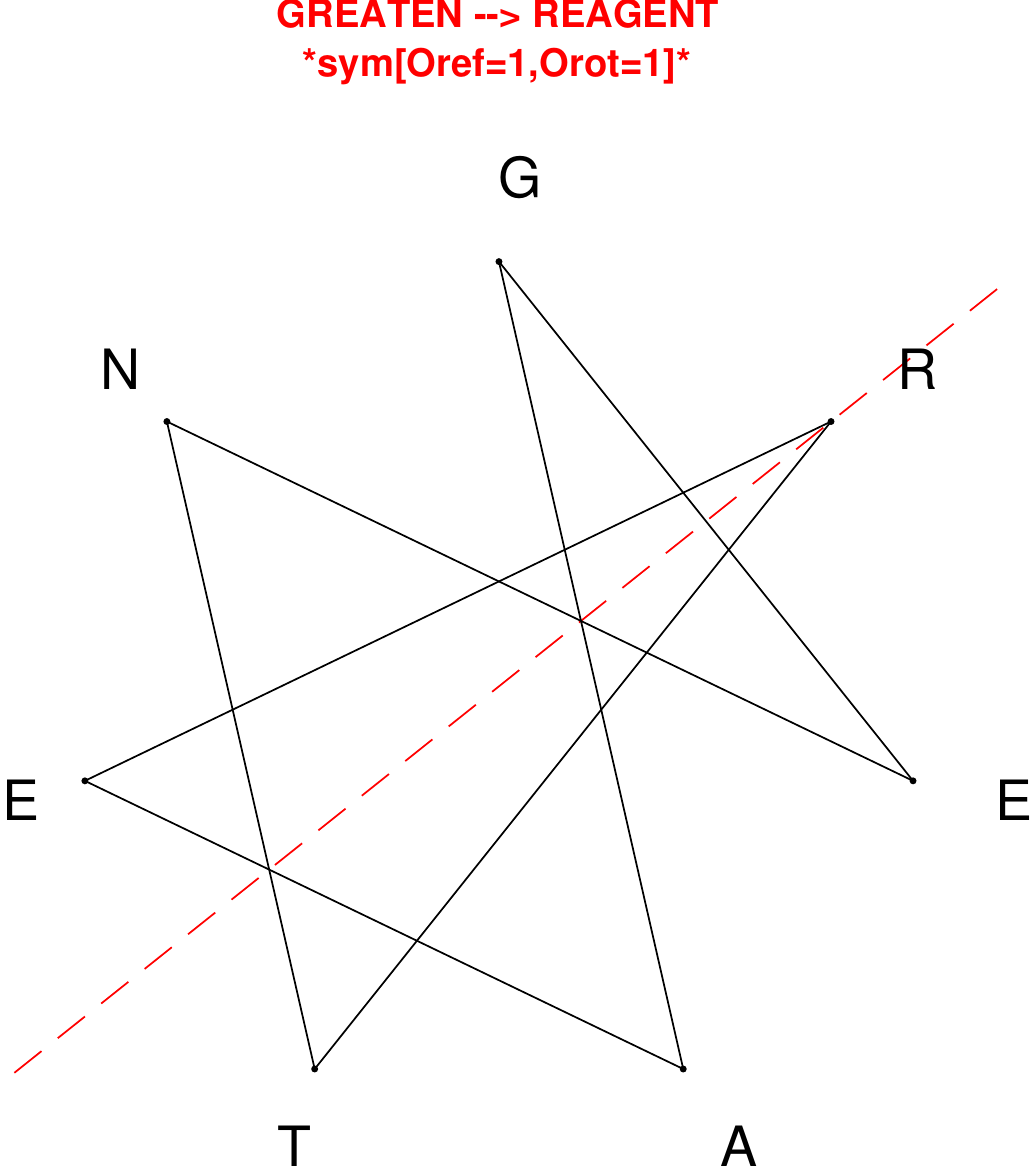}
\end{subfigure}
\hfill
\begin{subfigure}[T]{0.19\textwidth}
\centering
\includegraphics[width=\textwidth]{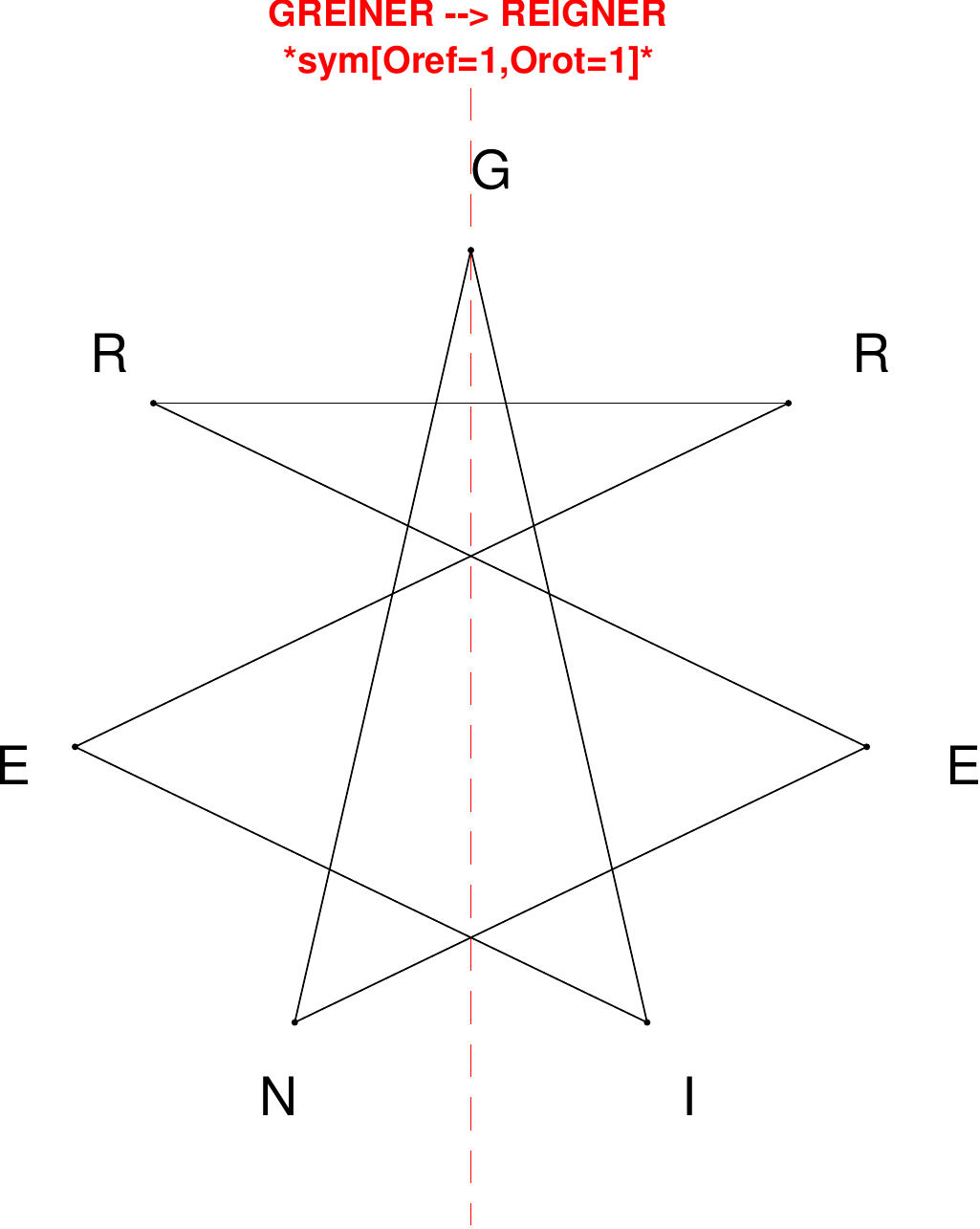}
\end{subfigure}
\end{figure}

\begin{figure}[H]
\centering
\begin{subfigure}[T]{0.19\textwidth}
\centering
\includegraphics[width=\textwidth]{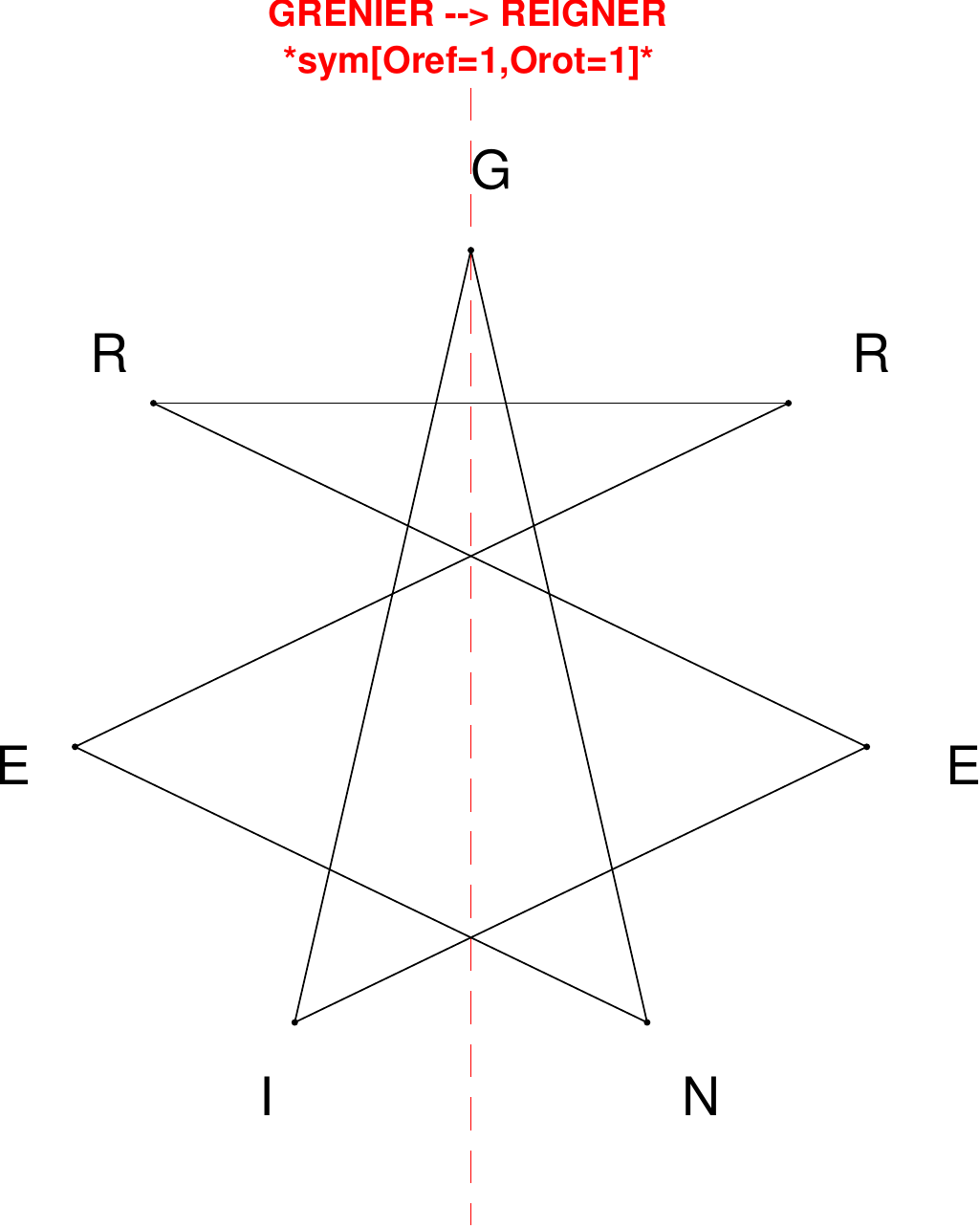}
\end{subfigure}
\hfill
\begin{subfigure}[T]{0.19\textwidth}
\centering
\includegraphics[width=\textwidth]{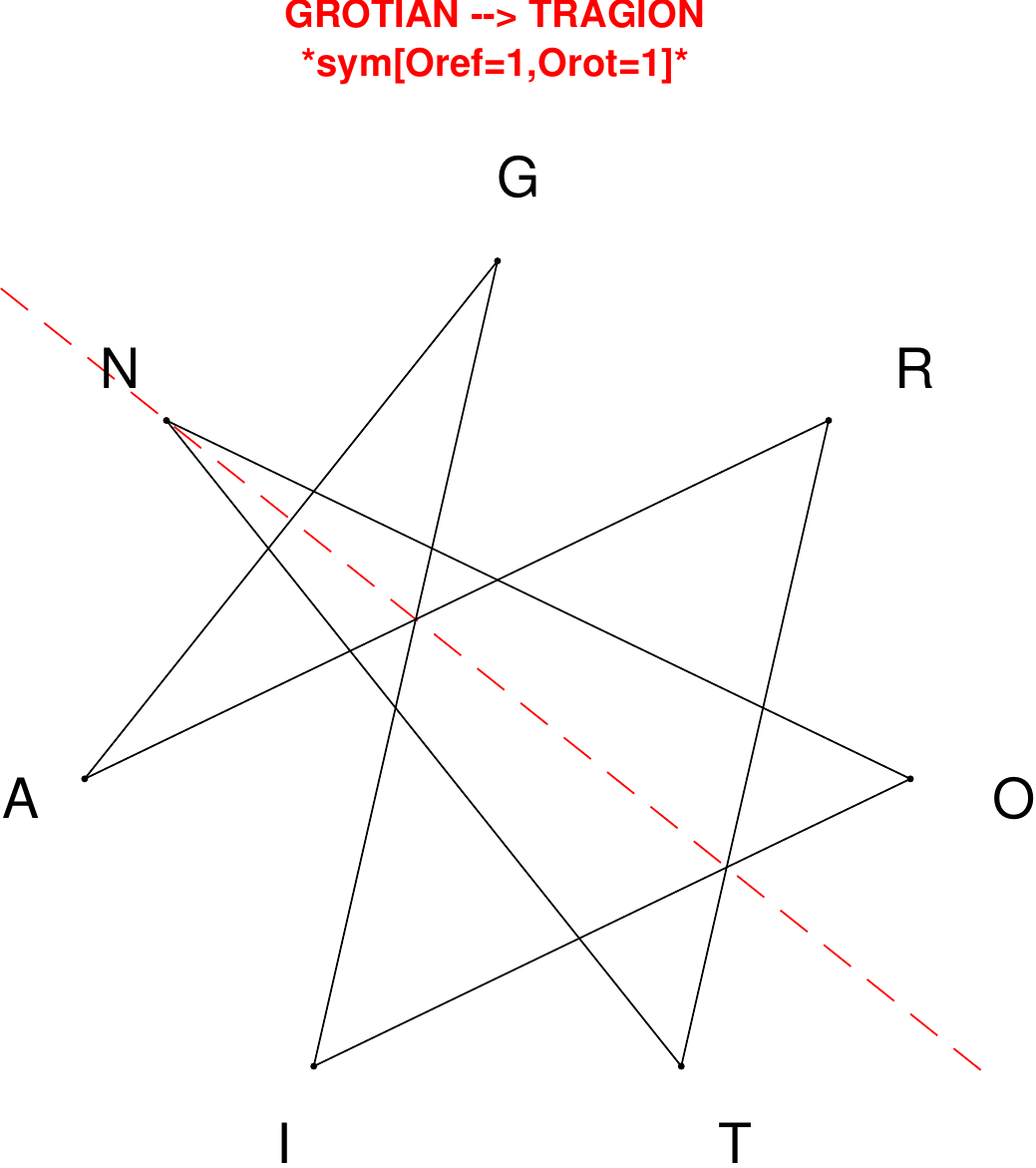}
\end{subfigure}
\hfill
\begin{subfigure}[T]{0.19\textwidth}
\centering
\includegraphics[width=\textwidth]{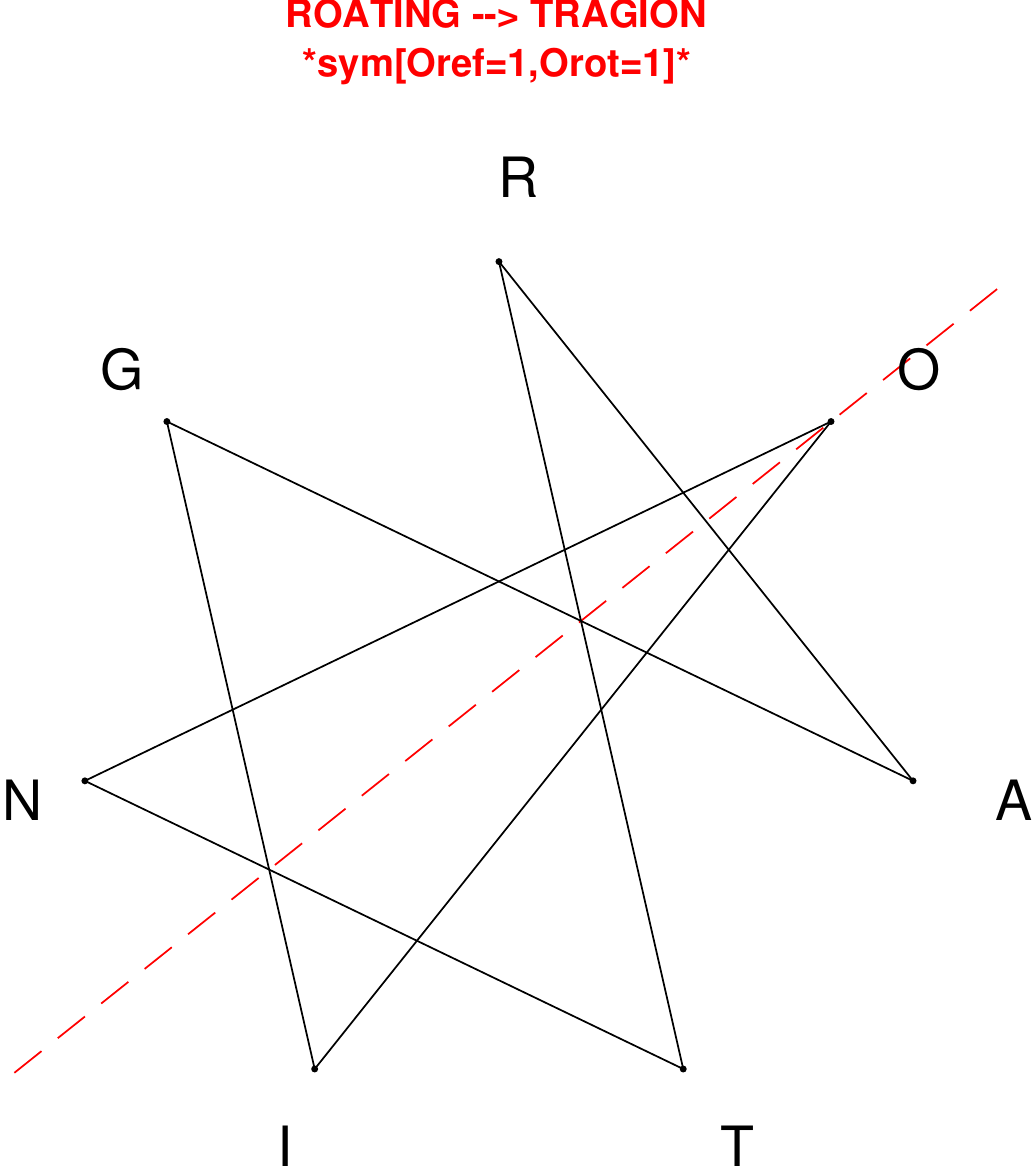}
\end{subfigure}
\hfill
\begin{subfigure}[T]{0.19\textwidth}
\centering
\includegraphics[width=\textwidth]{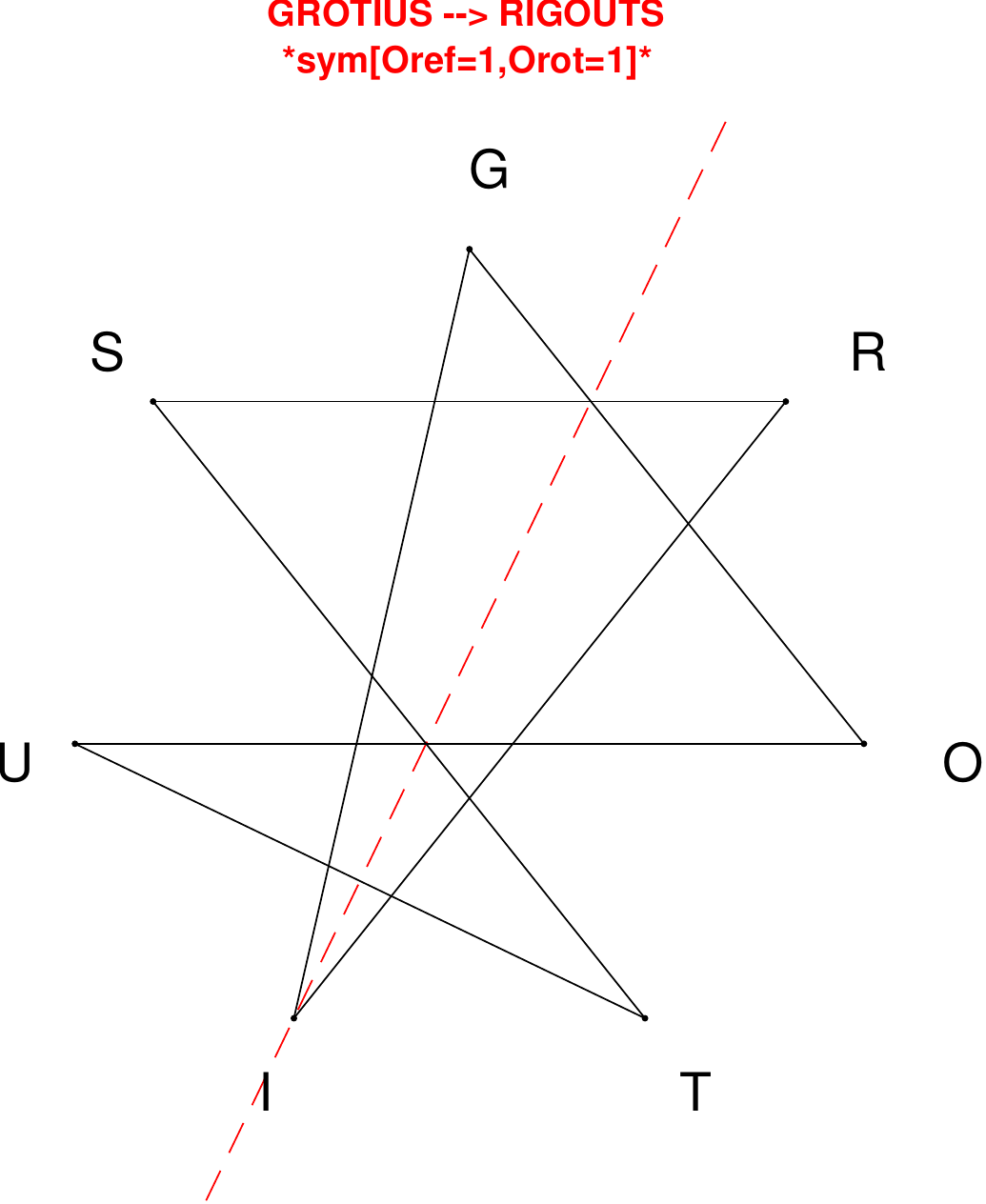}
\end{subfigure}
\hfill
\begin{subfigure}[T]{0.19\textwidth}
\centering
\includegraphics[width=\textwidth]{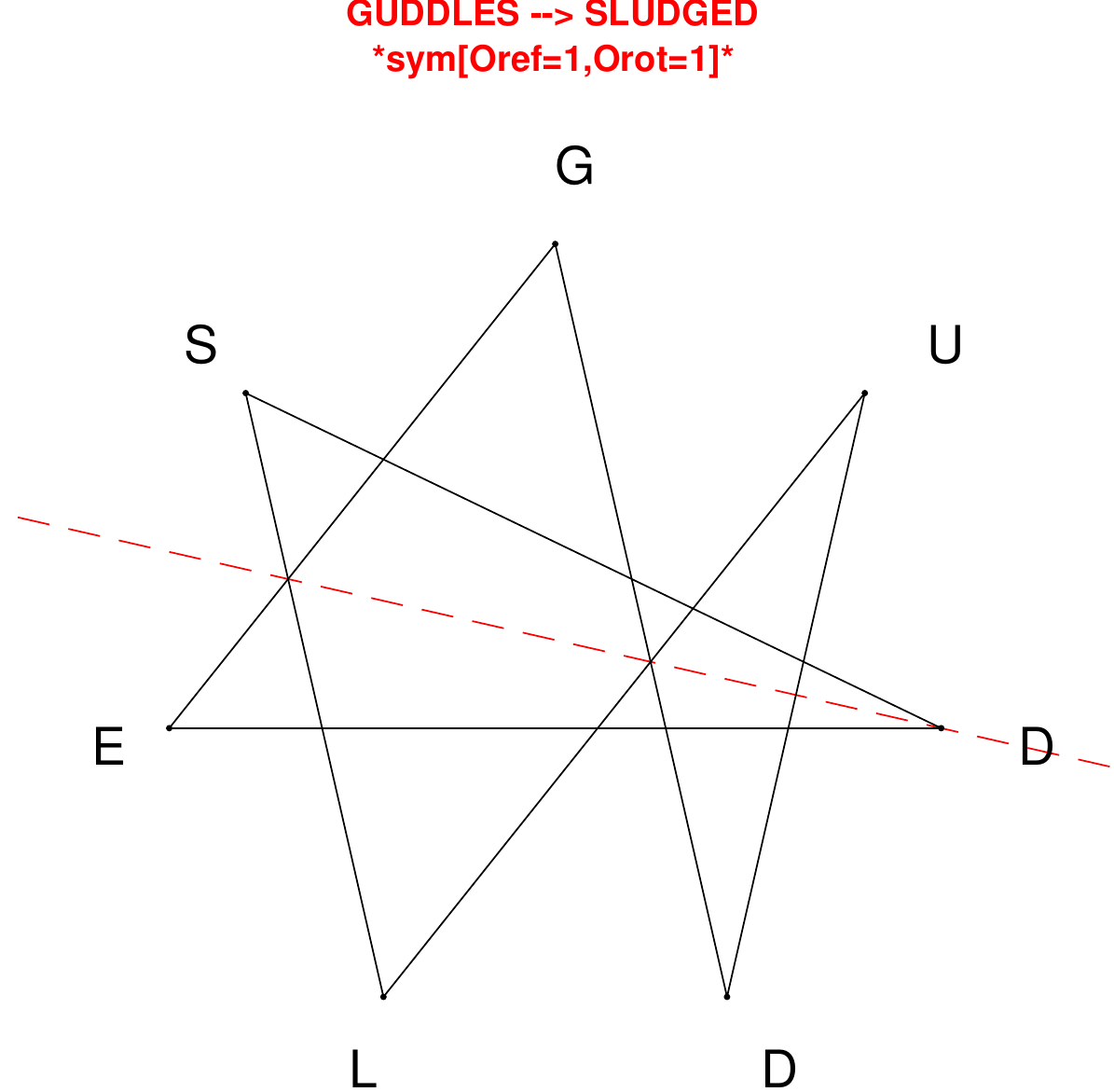}
\end{subfigure}
\end{figure}

\begin{figure}[H]
\centering
\begin{subfigure}[T]{0.19\textwidth}
\centering
\includegraphics[width=\textwidth]{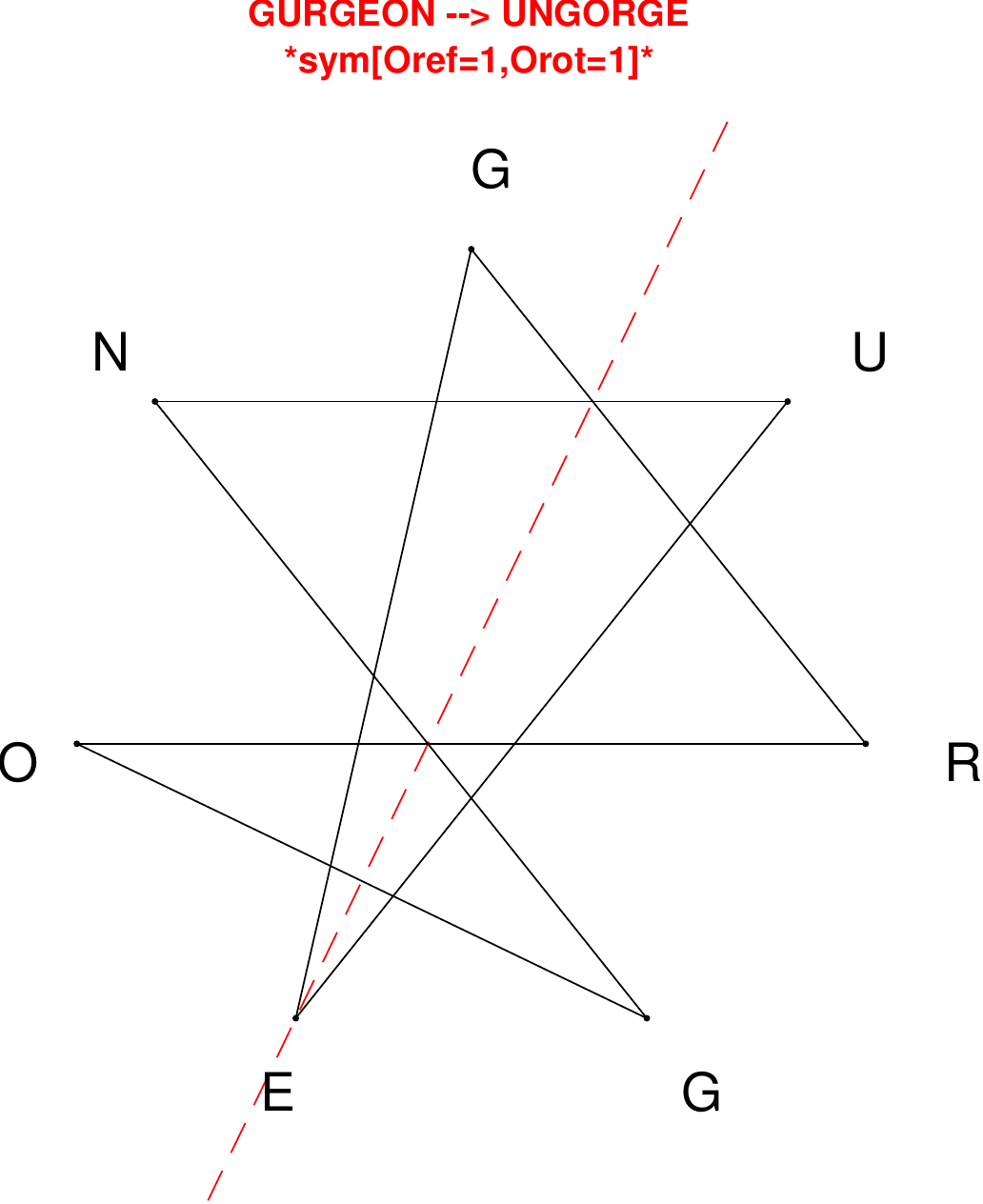}
\end{subfigure}
\hfill
\begin{subfigure}[T]{0.19\textwidth}
\centering
\includegraphics[width=\textwidth]{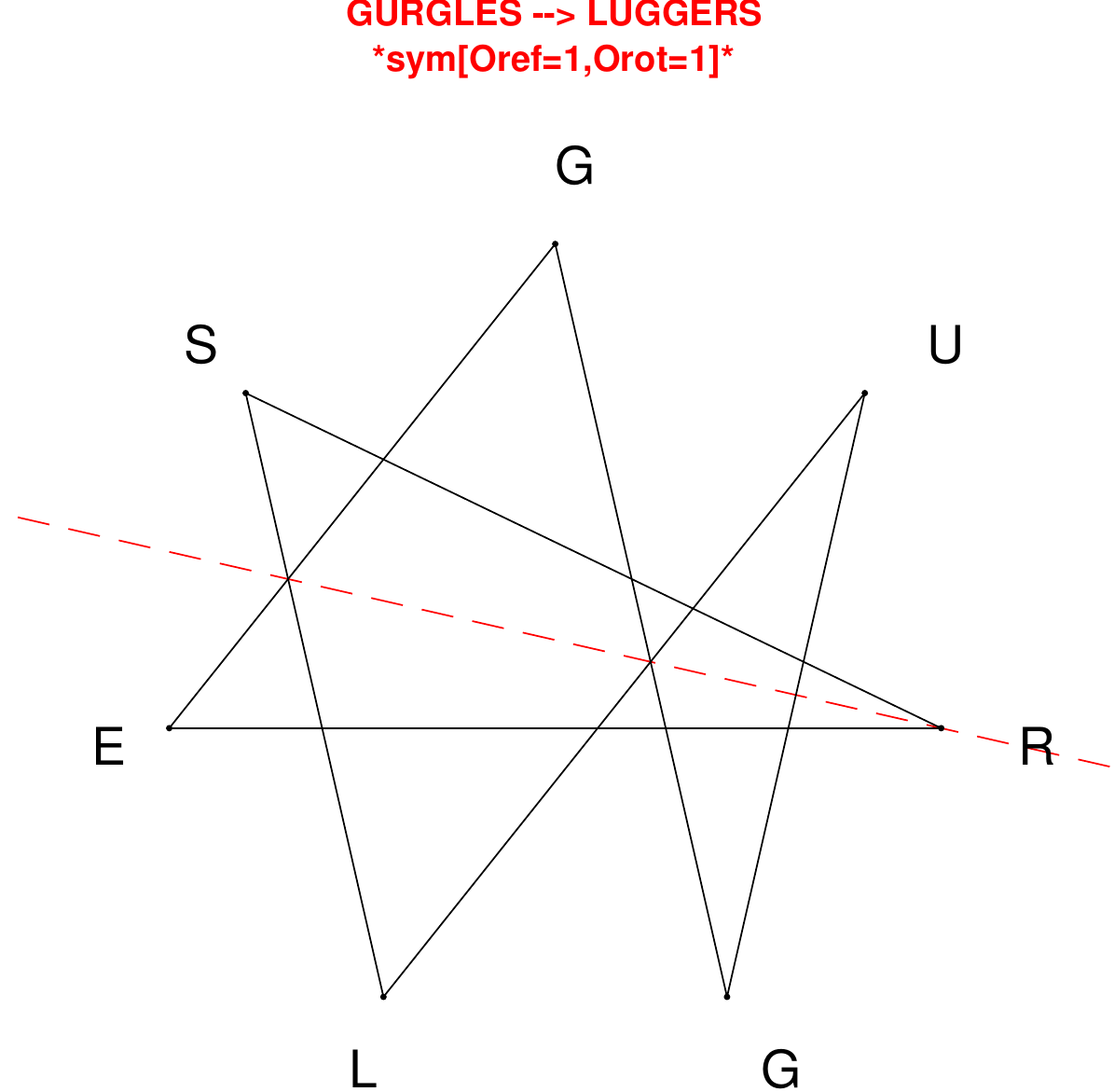}
\end{subfigure}
\hfill
\begin{subfigure}[T]{0.19\textwidth}
\centering
\includegraphics[width=\textwidth]{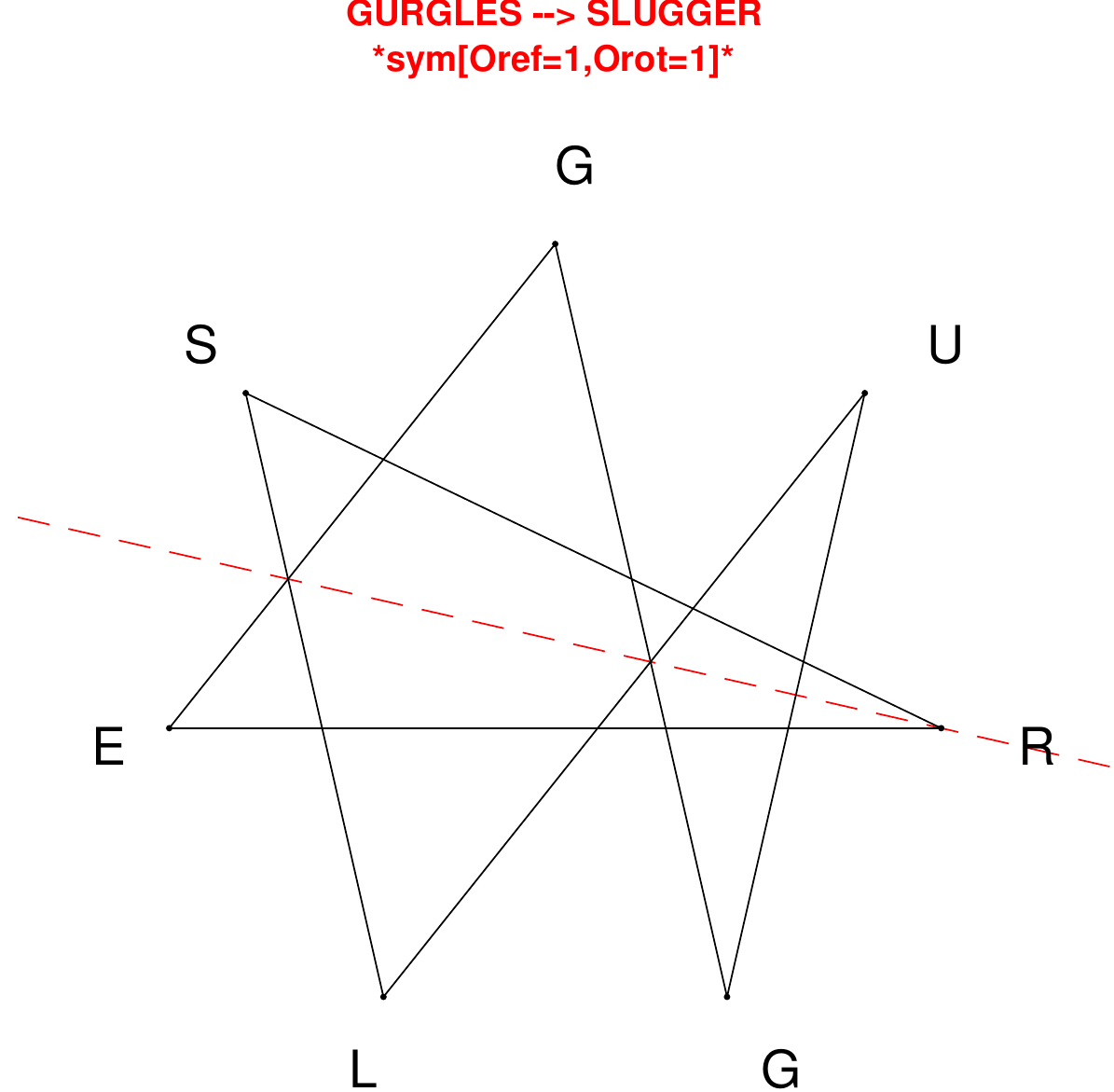}
\end{subfigure}
\hfill
\begin{subfigure}[T]{0.19\textwidth}
\centering
\includegraphics[width=\textwidth]{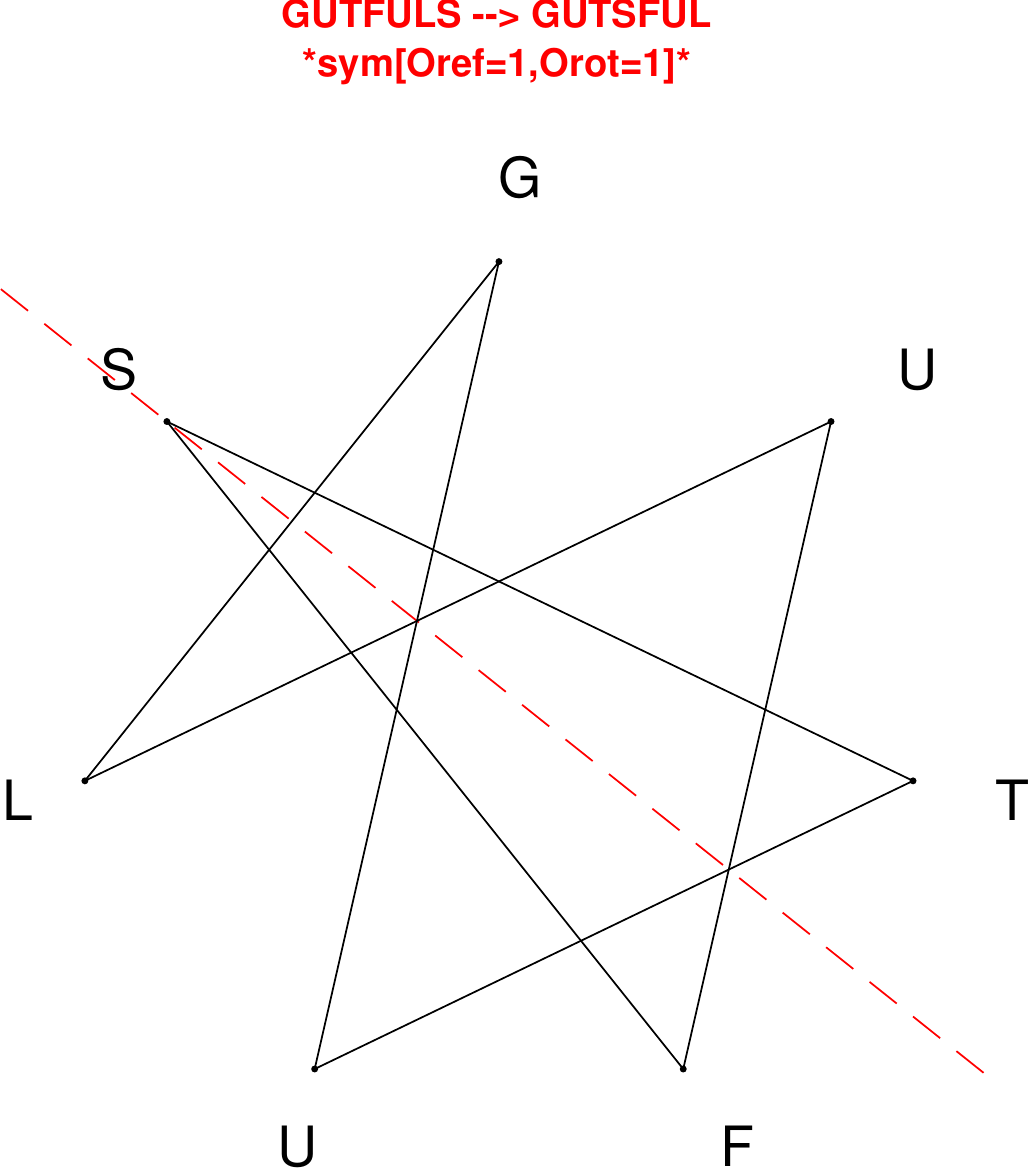}
\end{subfigure}
\hfill
\begin{subfigure}[T]{0.19\textwidth}
\centering
\includegraphics[width=\textwidth]{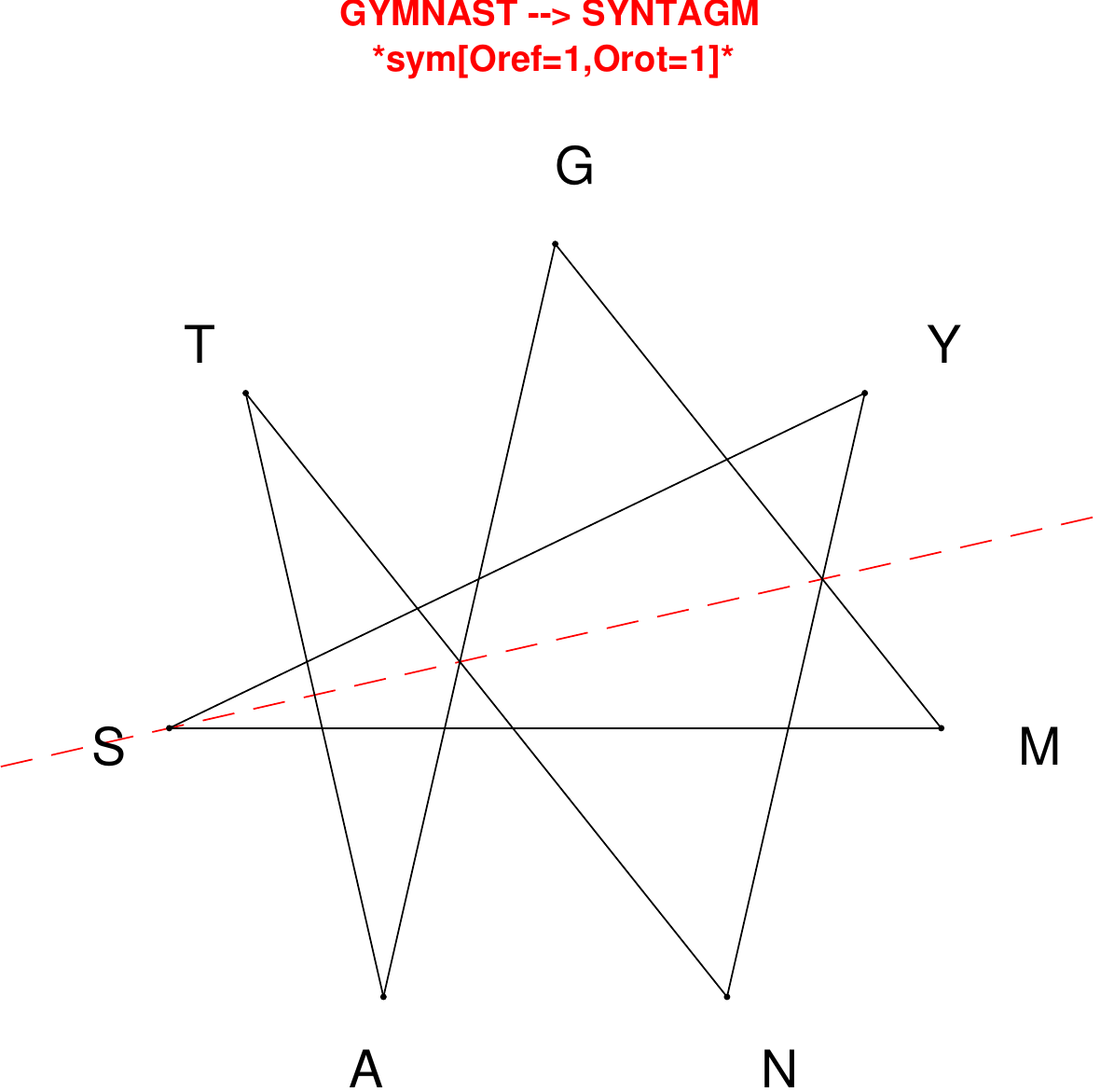}
\end{subfigure}
\end{figure}

\begin{figure}[H]
\centering
\begin{subfigure}[T]{0.19\textwidth}
\centering
\includegraphics[width=\textwidth]{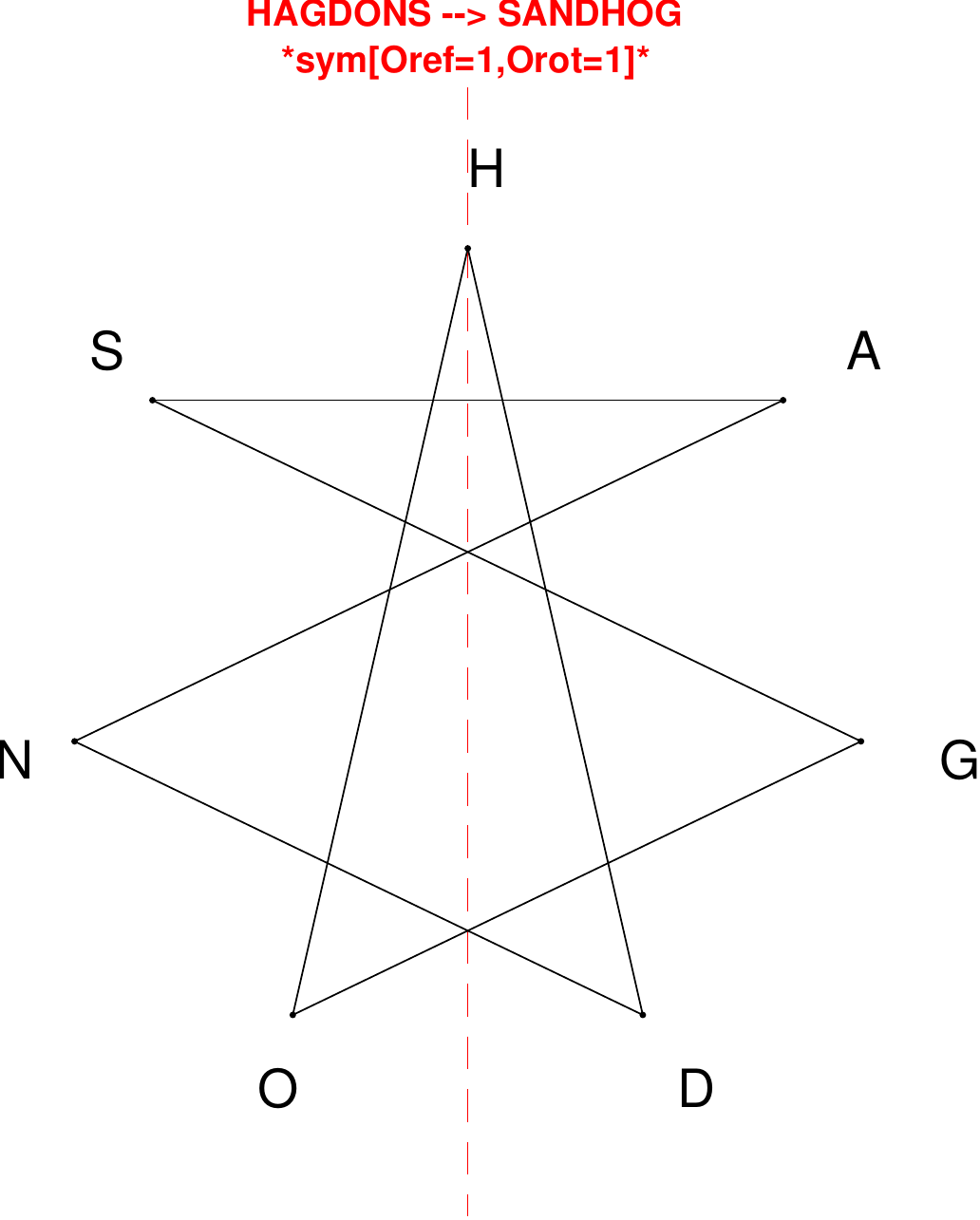}
\end{subfigure}
\hfill
\begin{subfigure}[T]{0.19\textwidth}
\centering
\includegraphics[width=\textwidth]{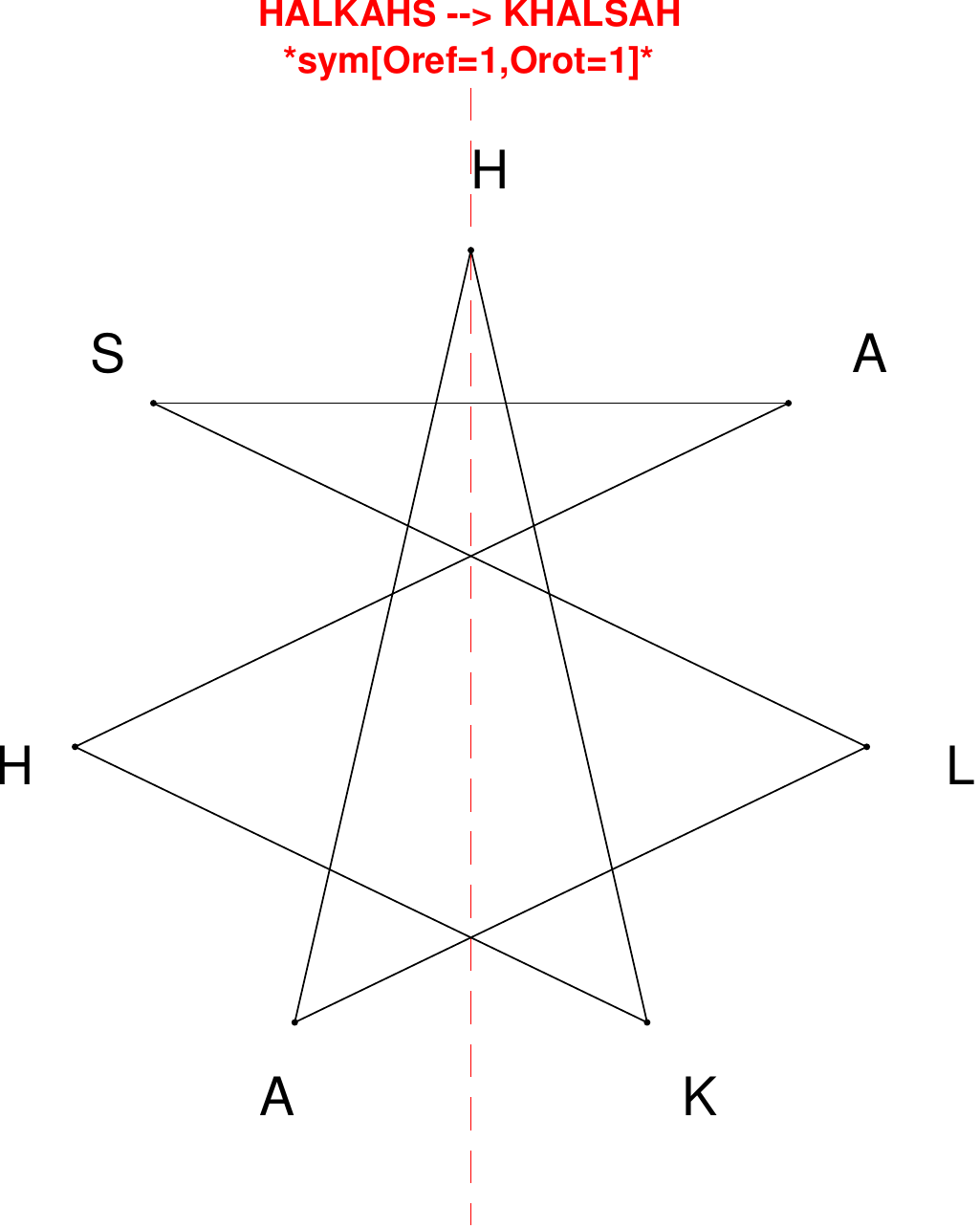}
\end{subfigure}
\hfill
\begin{subfigure}[T]{0.19\textwidth}
\centering
\includegraphics[width=\textwidth]{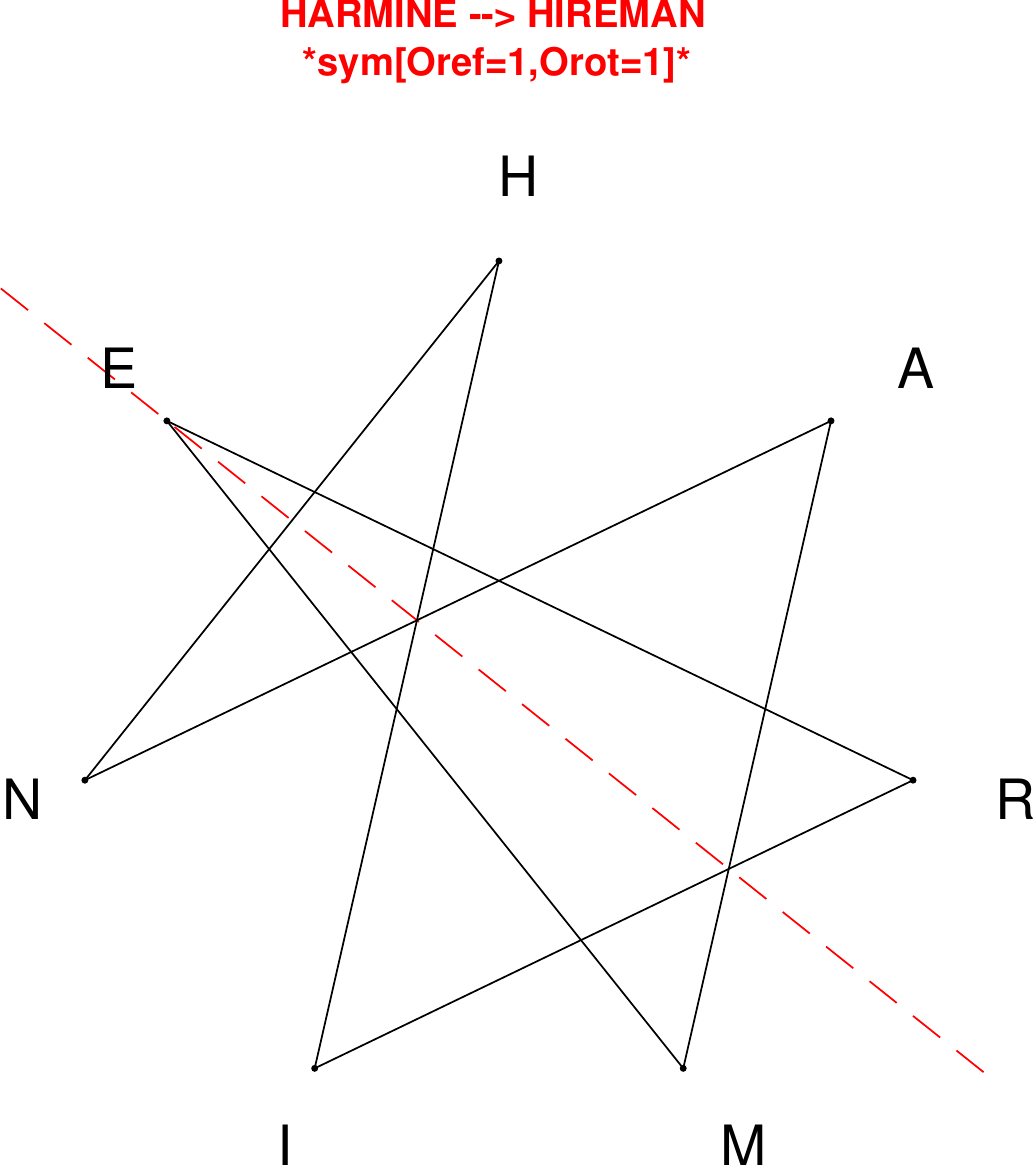}
\end{subfigure}
\hfill
\begin{subfigure}[T]{0.19\textwidth}
\centering
\includegraphics[width=\textwidth]{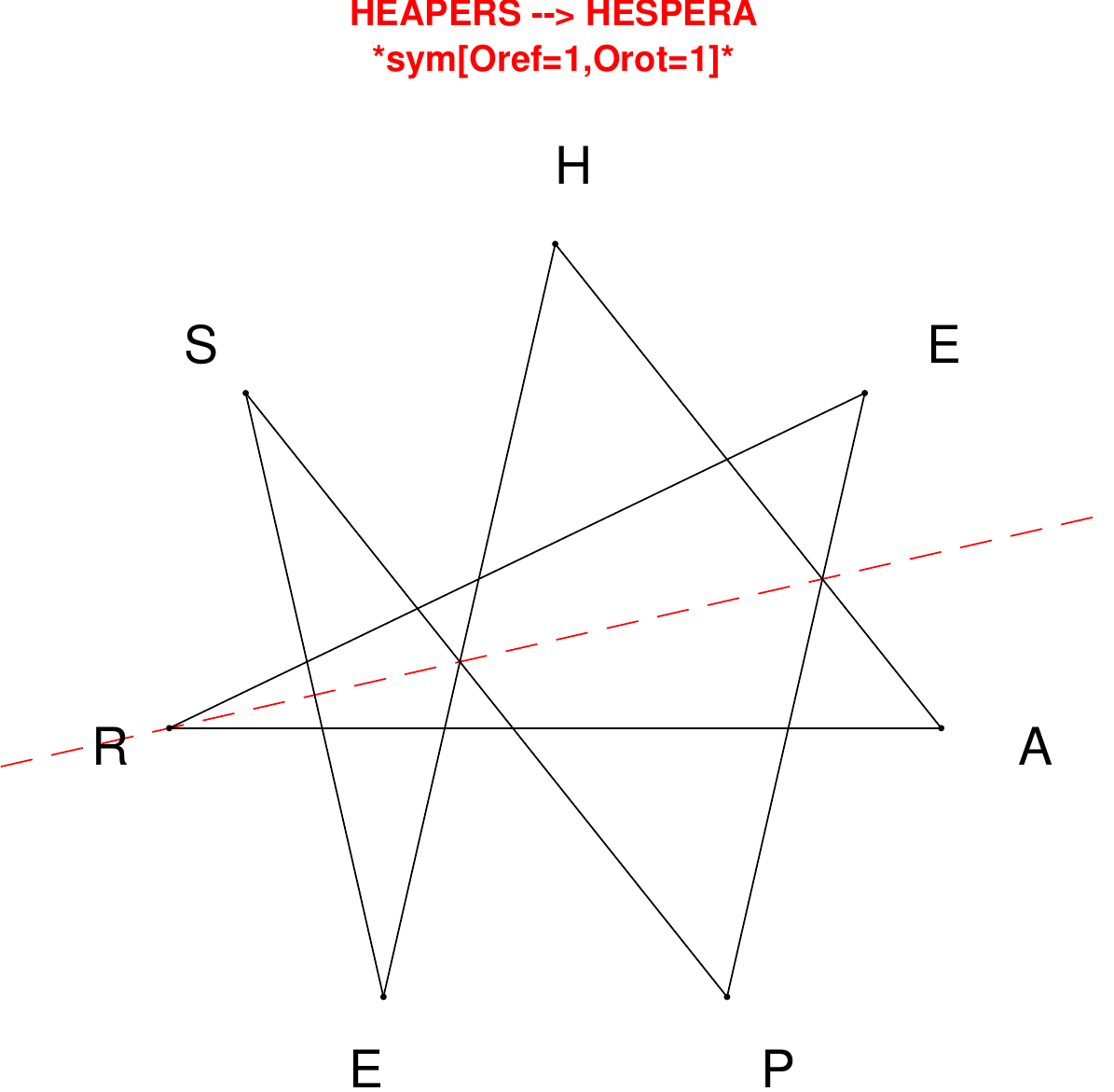}
\end{subfigure}
\hfill
\begin{subfigure}[T]{0.19\textwidth}
\centering
\includegraphics[width=\textwidth]{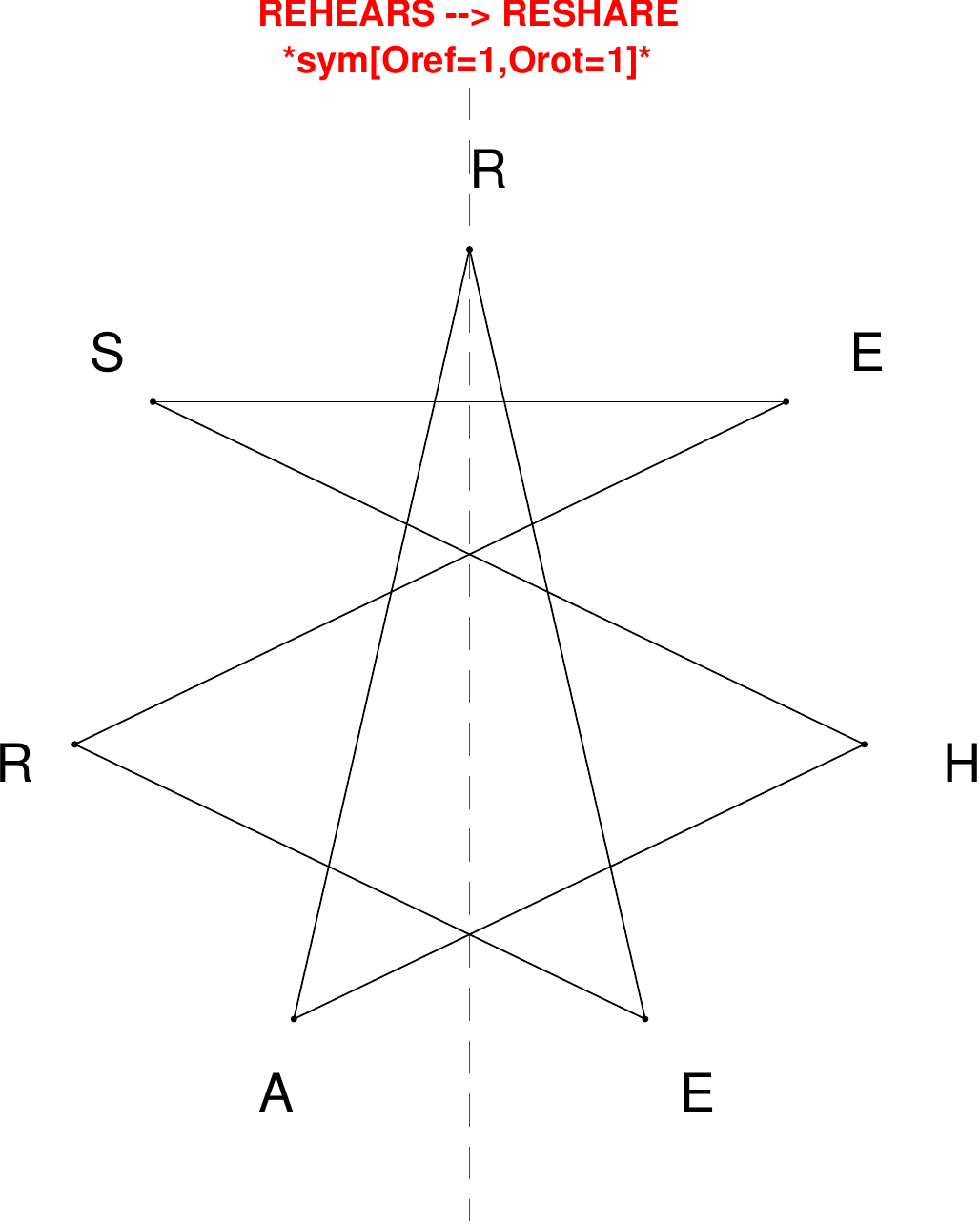}
\end{subfigure}
\end{figure}

\begin{figure}[H]
\centering
\begin{subfigure}[T]{0.19\textwidth}
\centering
\includegraphics[width=\textwidth]{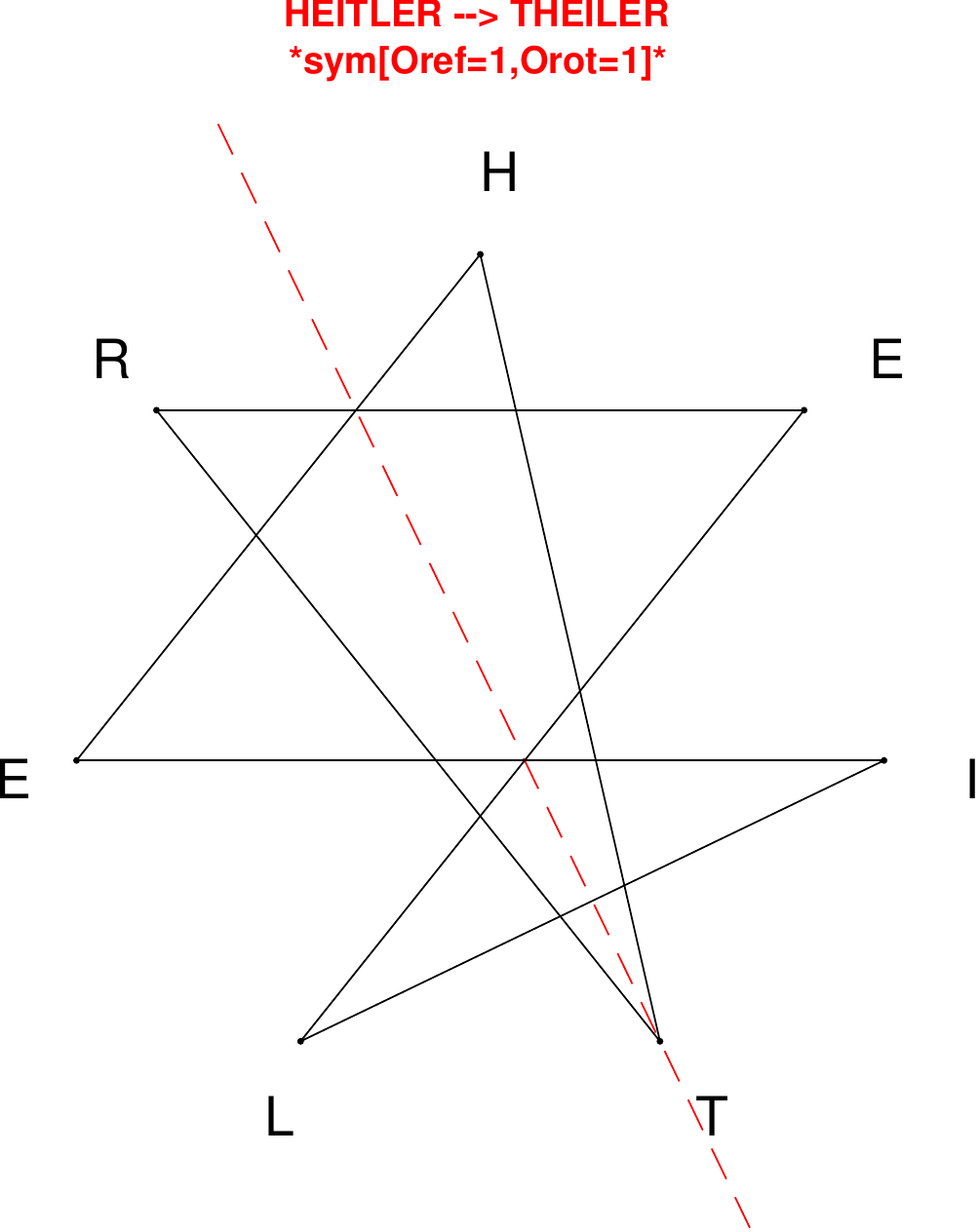}
\end{subfigure}
\hfill
\begin{subfigure}[T]{0.19\textwidth}
\centering
\includegraphics[width=\textwidth]{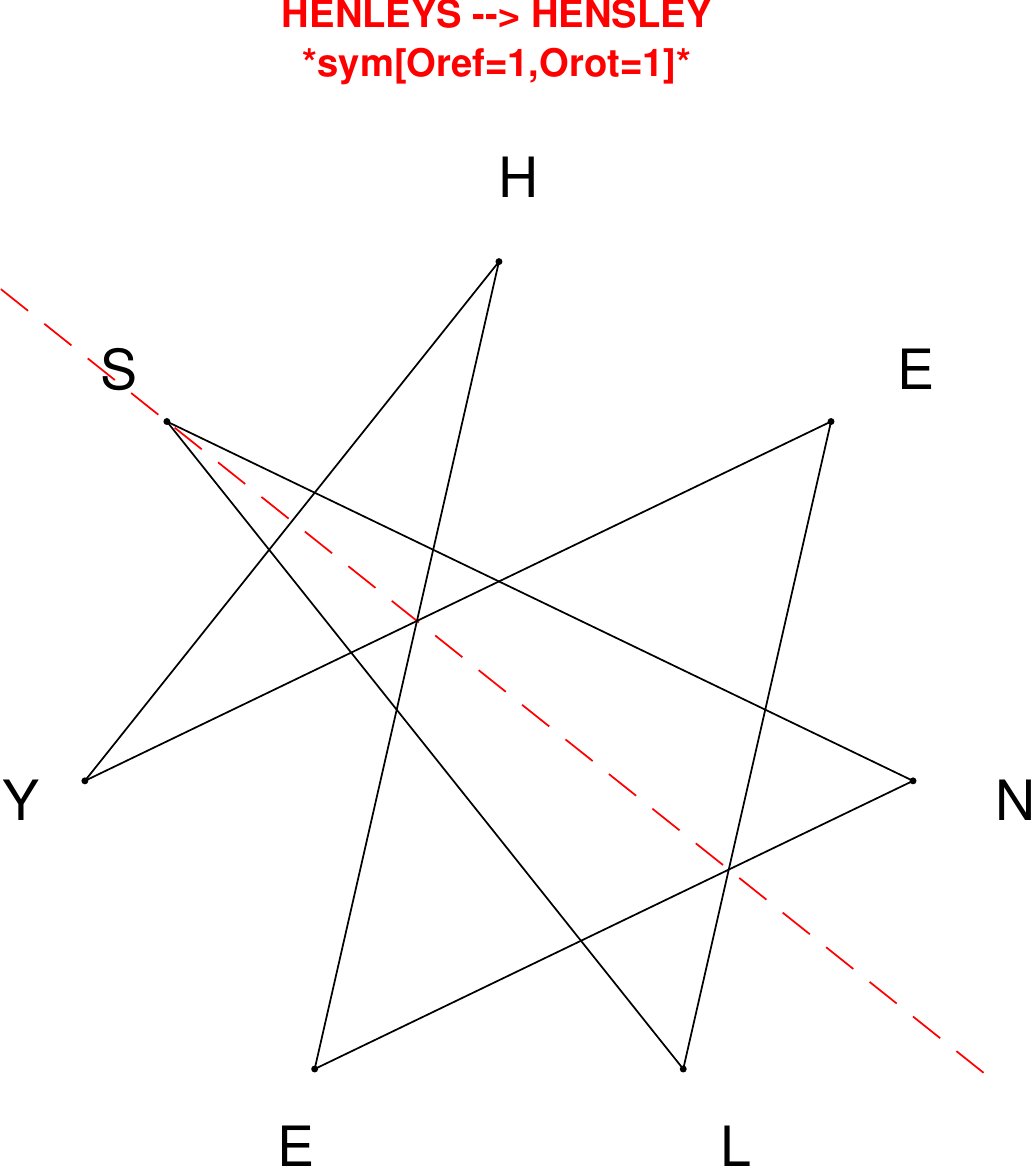}
\end{subfigure}
\hfill
\begin{subfigure}[T]{0.19\textwidth}
\centering
\includegraphics[width=\textwidth]{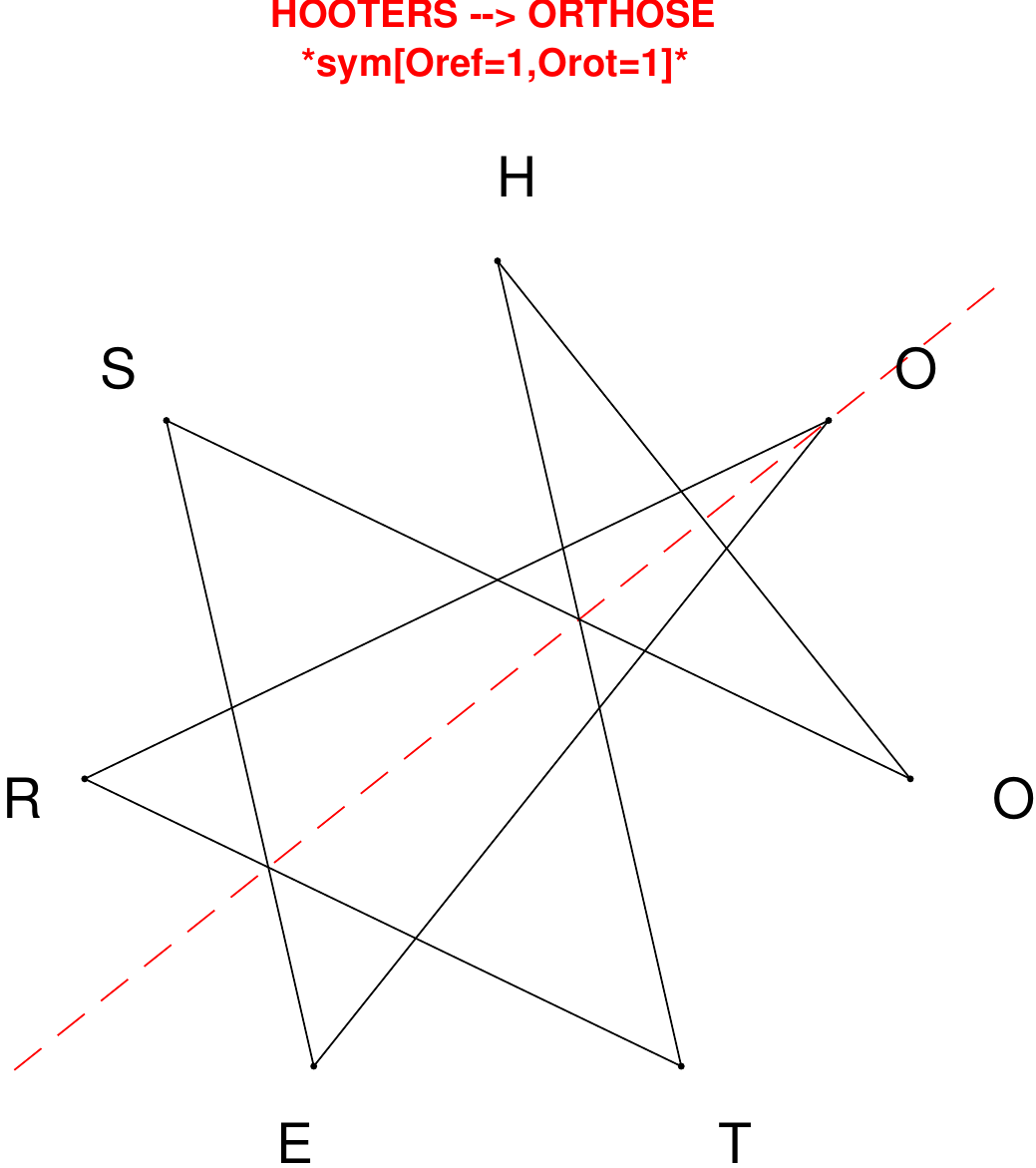}
\end{subfigure}
\hfill
\begin{subfigure}[T]{0.19\textwidth}
\centering
\includegraphics[width=\textwidth]{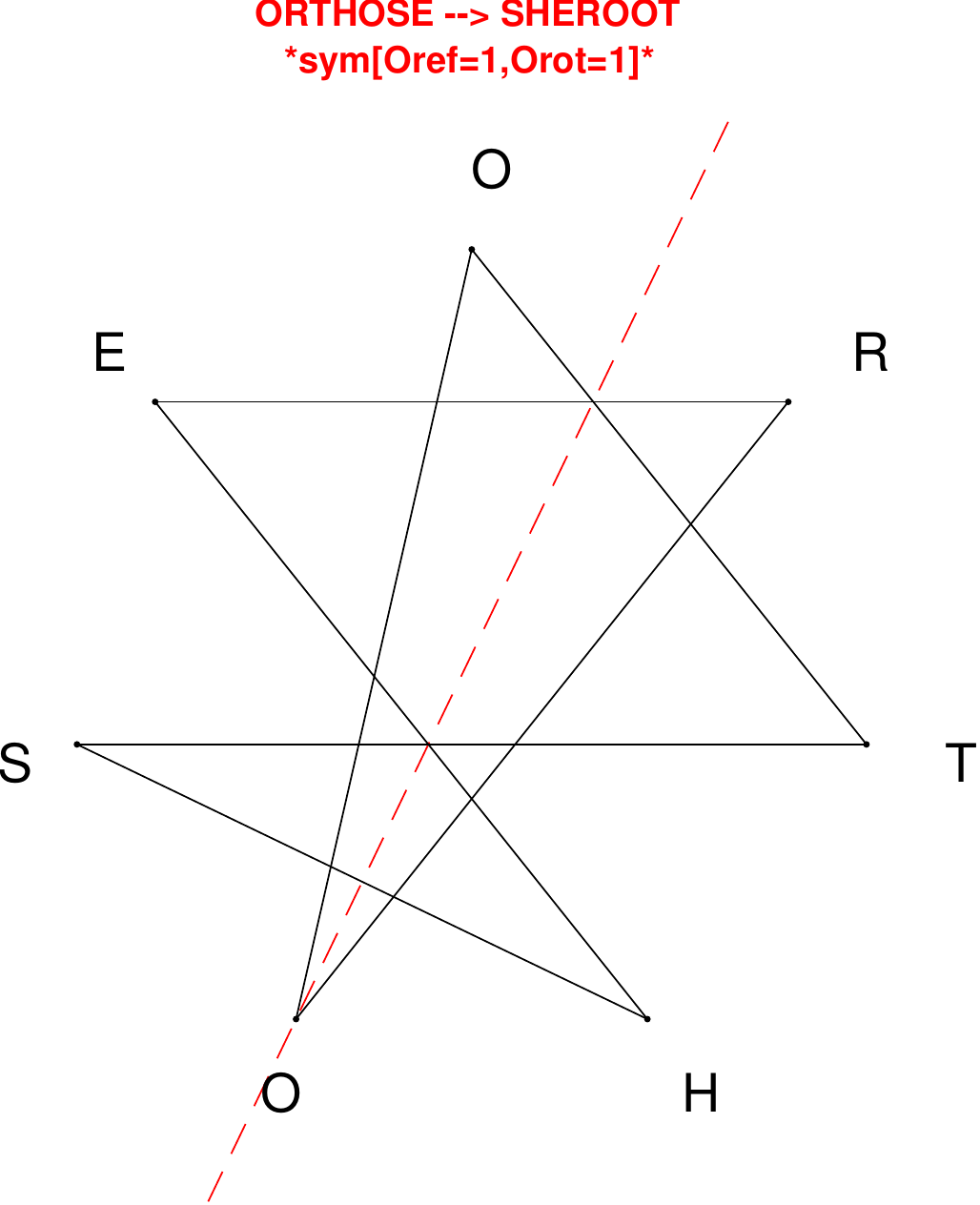}
\end{subfigure}
\hfill
\begin{subfigure}[T]{0.19\textwidth}
\centering
\includegraphics[width=\textwidth]{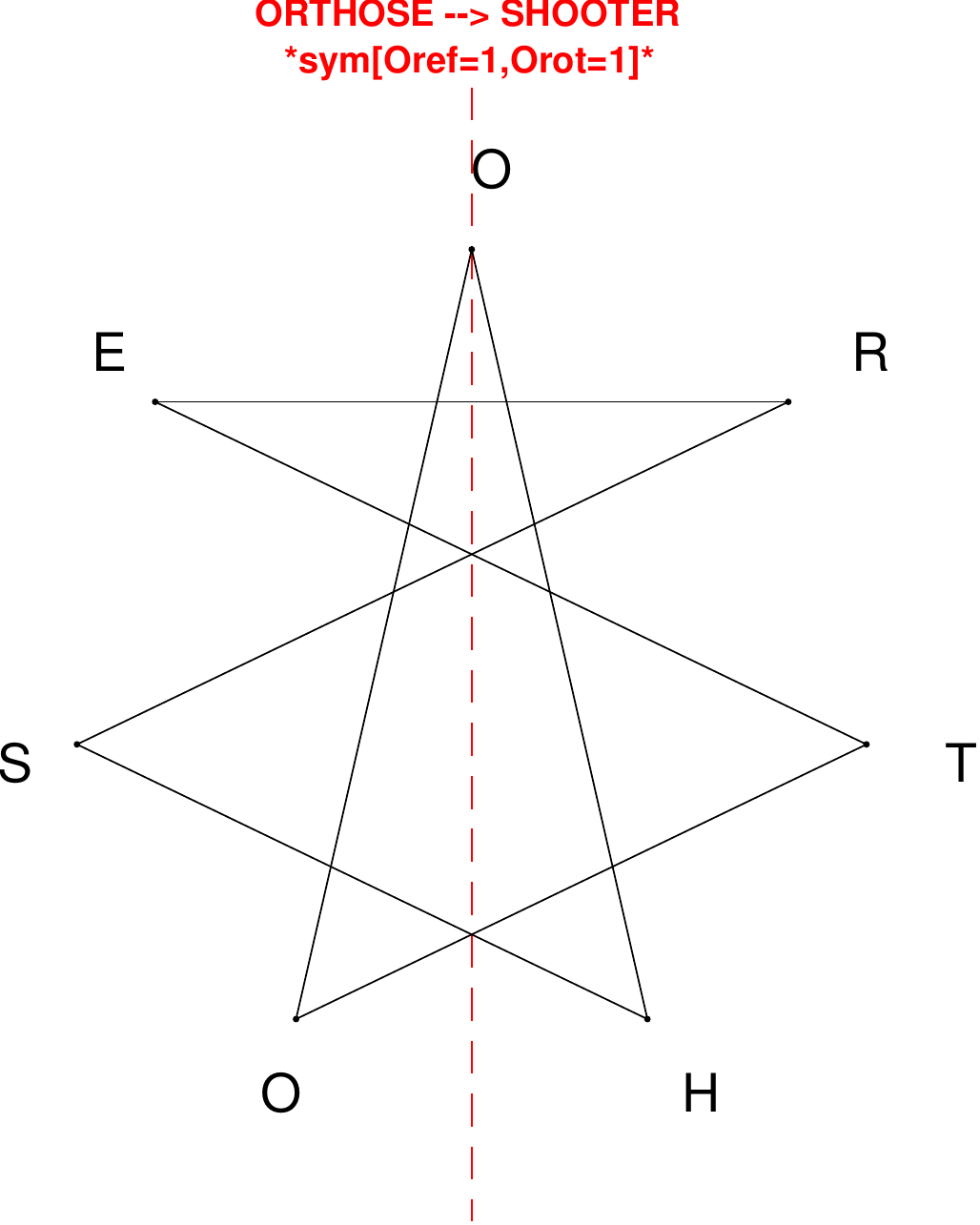}
\end{subfigure}
\end{figure}

\begin{figure}[H]
\centering
\begin{subfigure}[T]{0.19\textwidth}
\centering
\includegraphics[width=\textwidth]{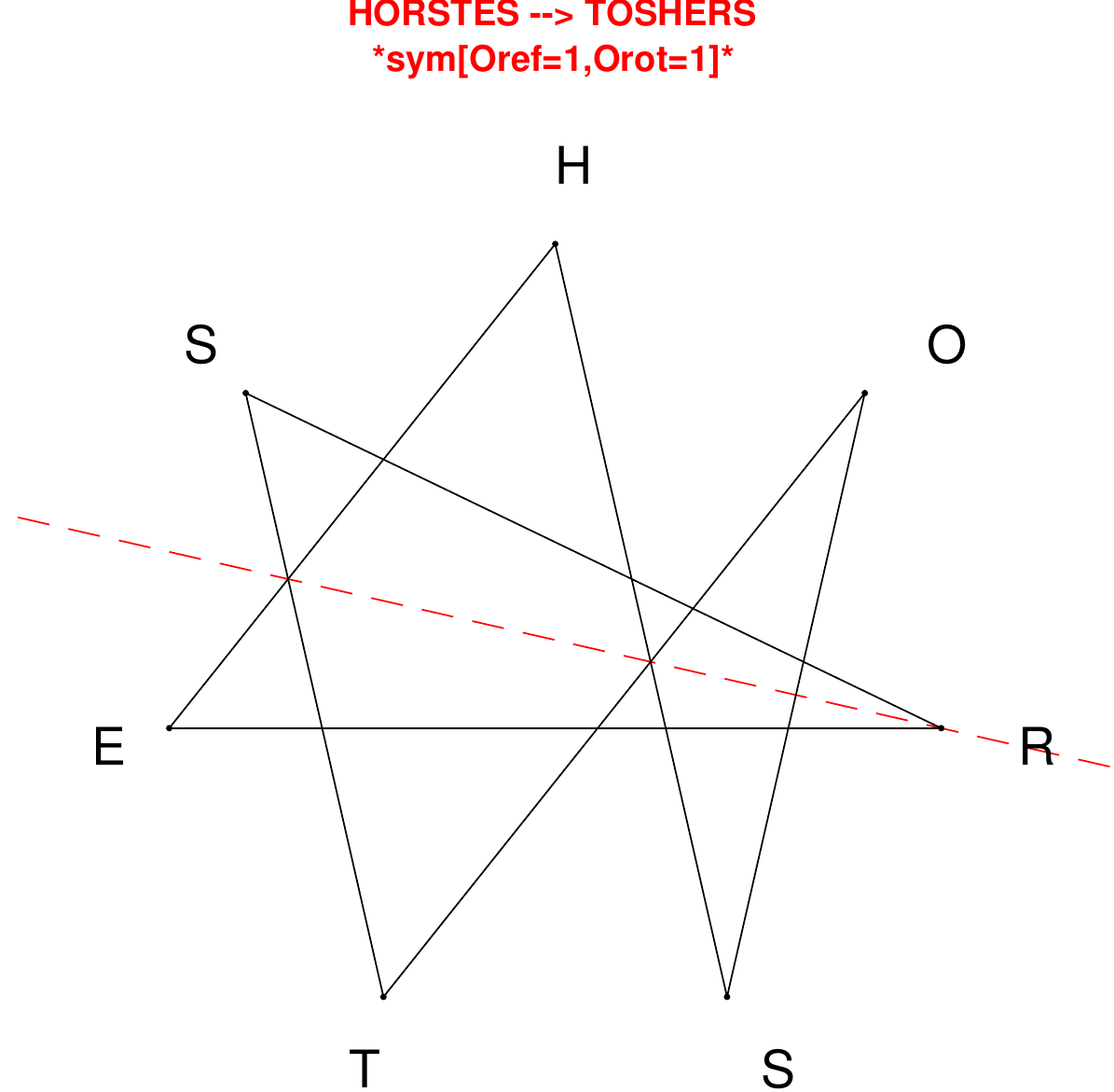}
\end{subfigure}
\hfill
\begin{subfigure}[T]{0.19\textwidth}
\centering
\includegraphics[width=\textwidth]{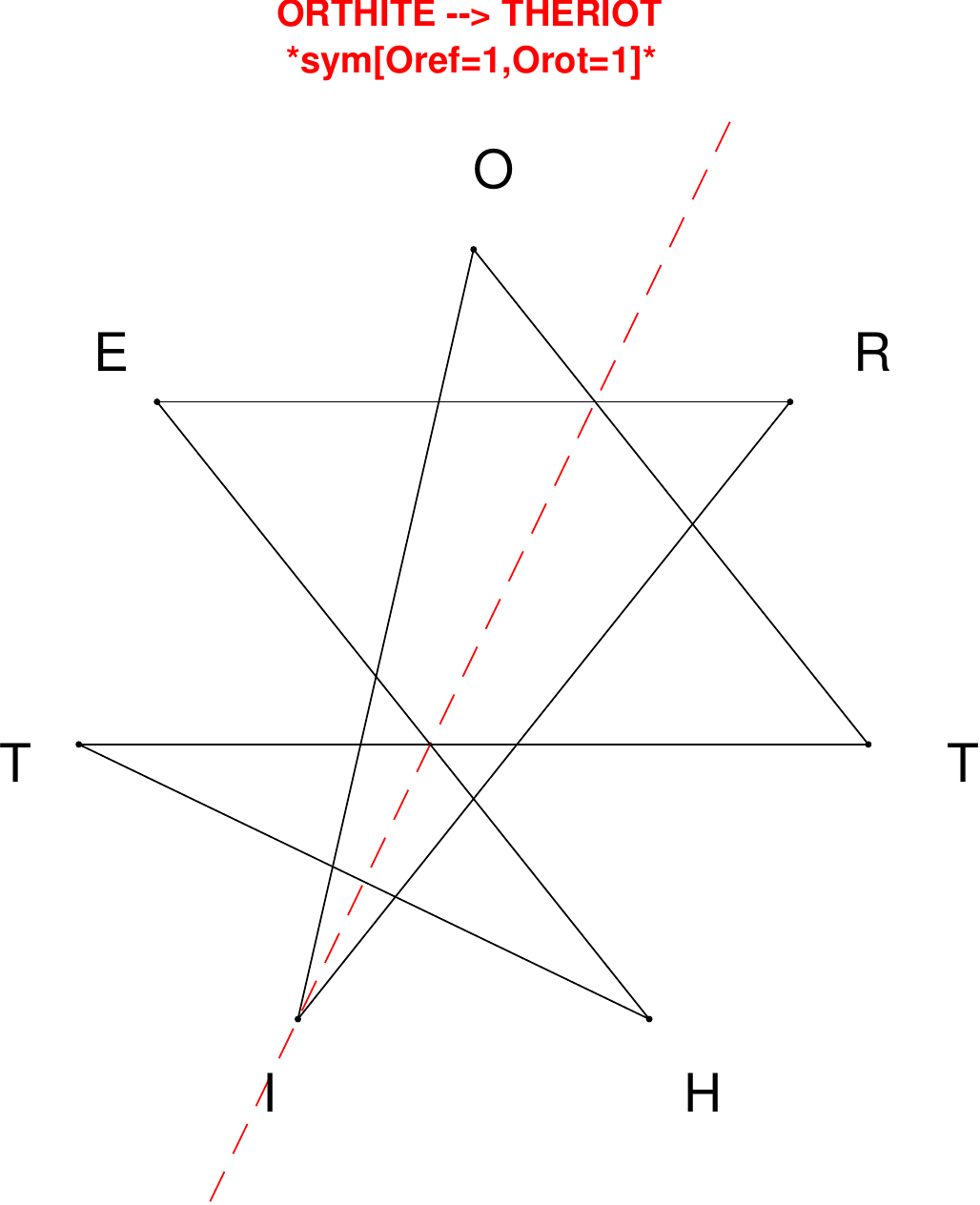}
\end{subfigure}
\hfill
\begin{subfigure}[T]{0.19\textwidth}
\centering
\includegraphics[width=\textwidth]{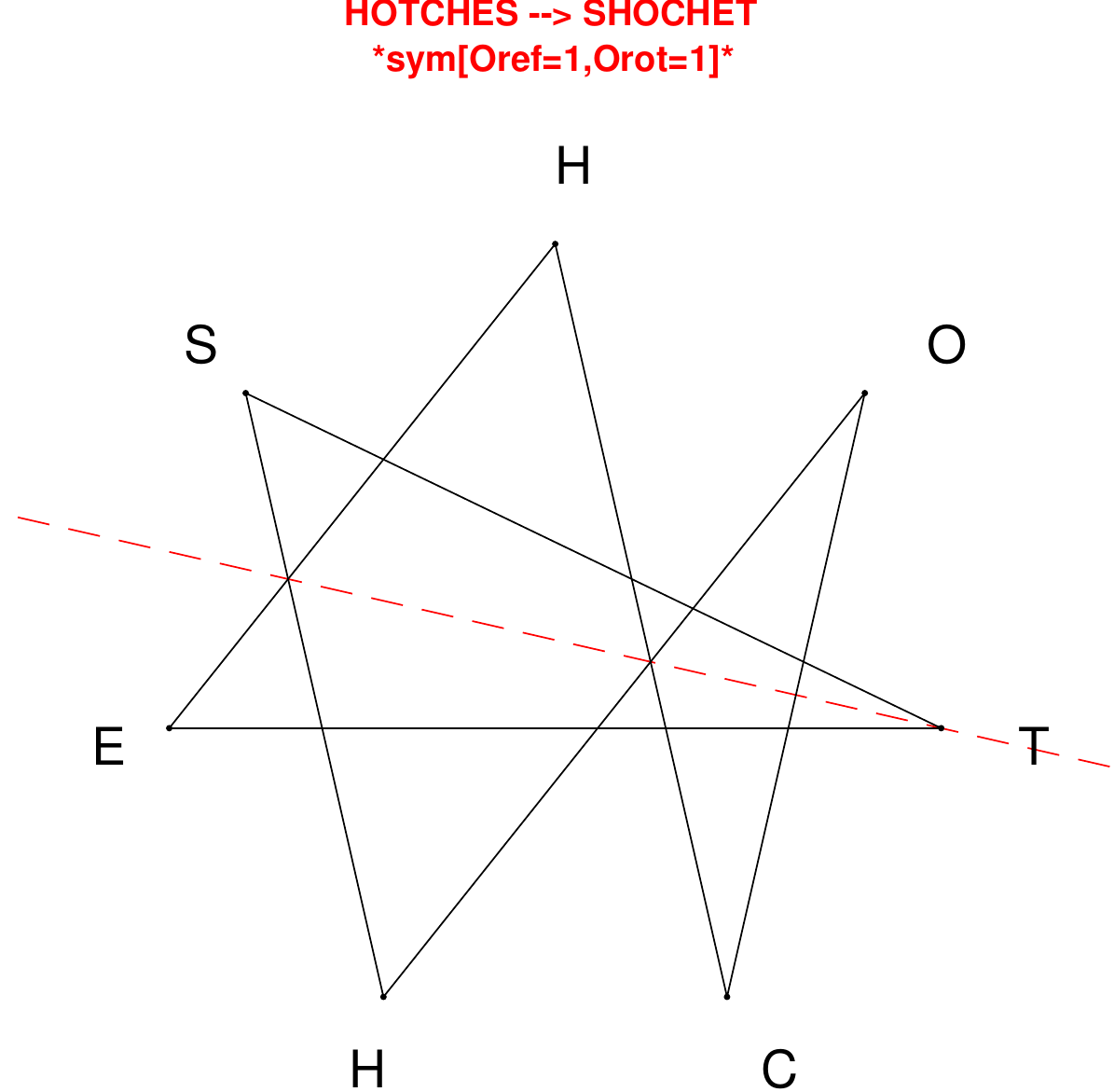}
\end{subfigure}
\hfill
\begin{subfigure}[T]{0.19\textwidth}
\centering
\includegraphics[width=\textwidth]{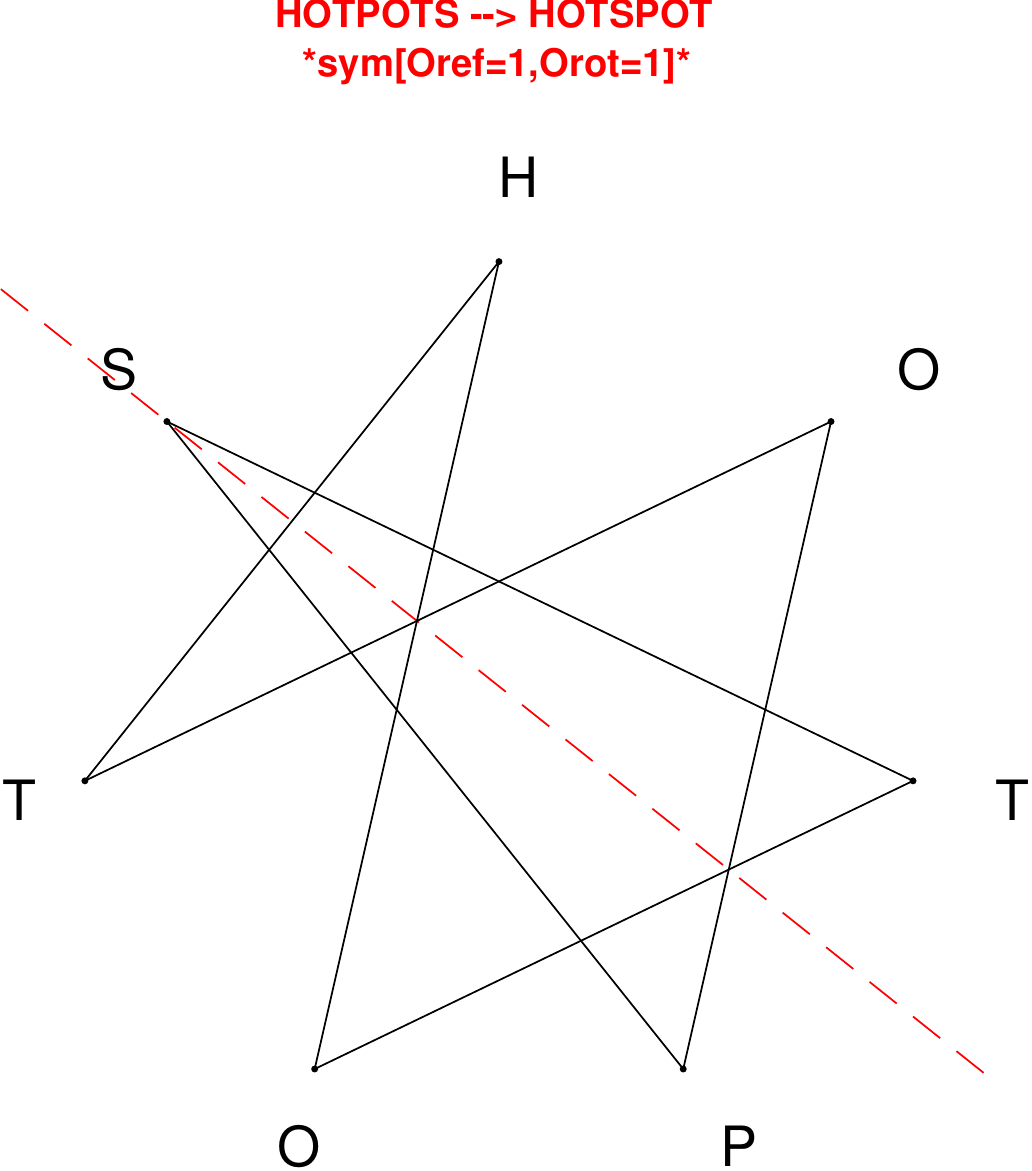}
\end{subfigure}
\hfill
\begin{subfigure}[T]{0.19\textwidth}
\centering
\includegraphics[width=\textwidth]{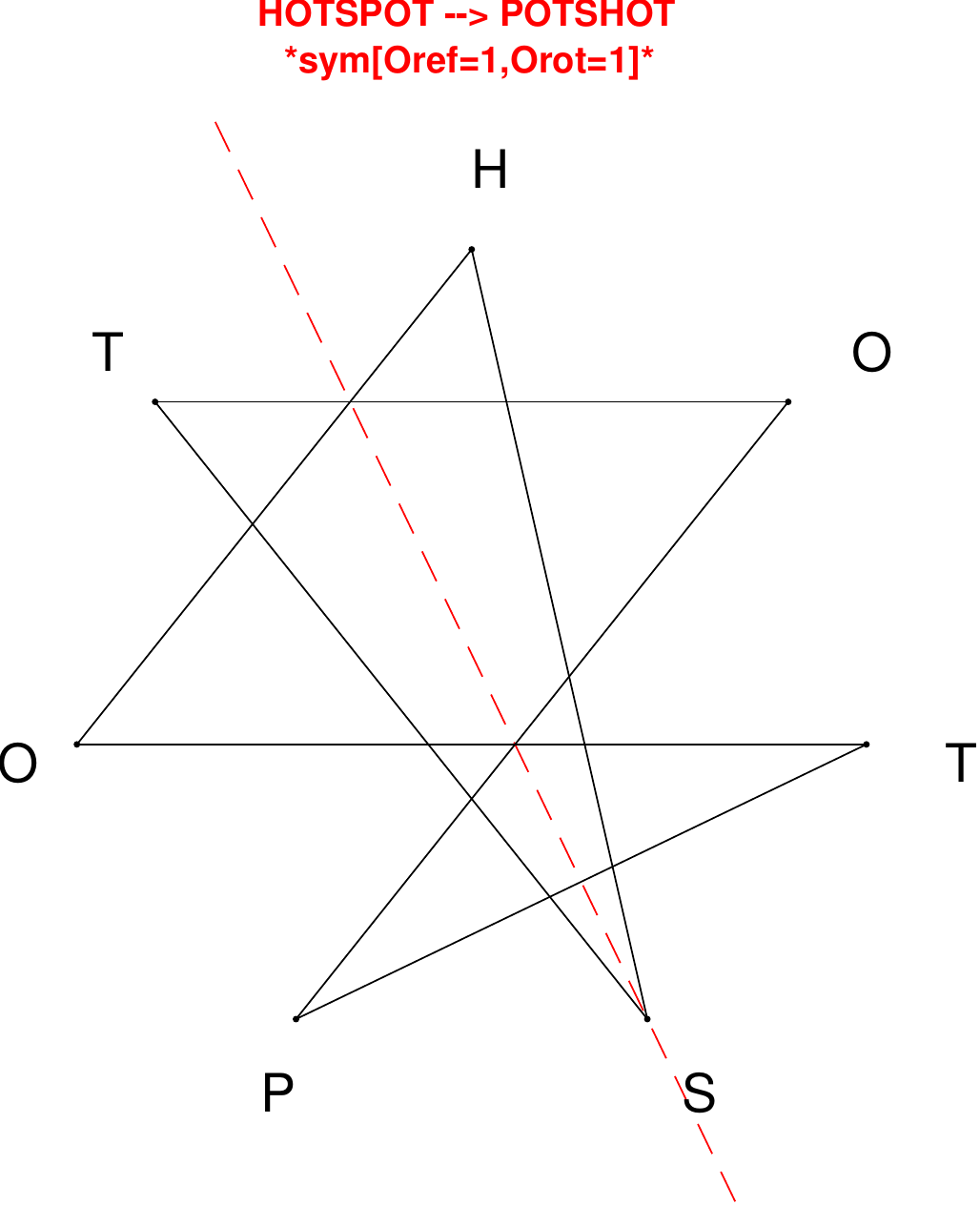}
\end{subfigure}
\end{figure}

\begin{figure}[H]
\centering
\begin{subfigure}[T]{0.19\textwidth}
\centering
\includegraphics[width=\textwidth]{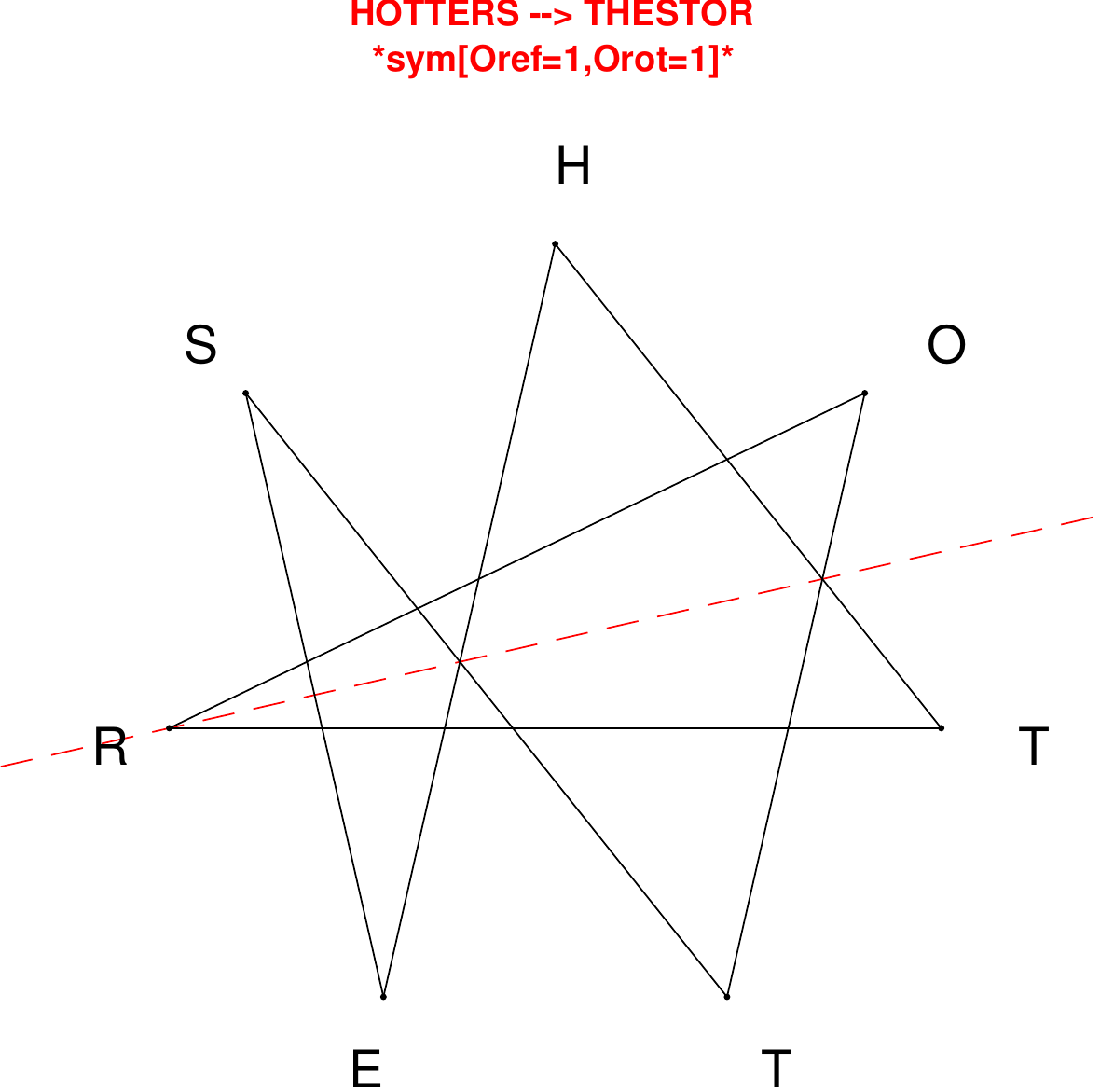}
\end{subfigure}
\hfill
\begin{subfigure}[T]{0.19\textwidth}
\centering
\includegraphics[width=\textwidth]{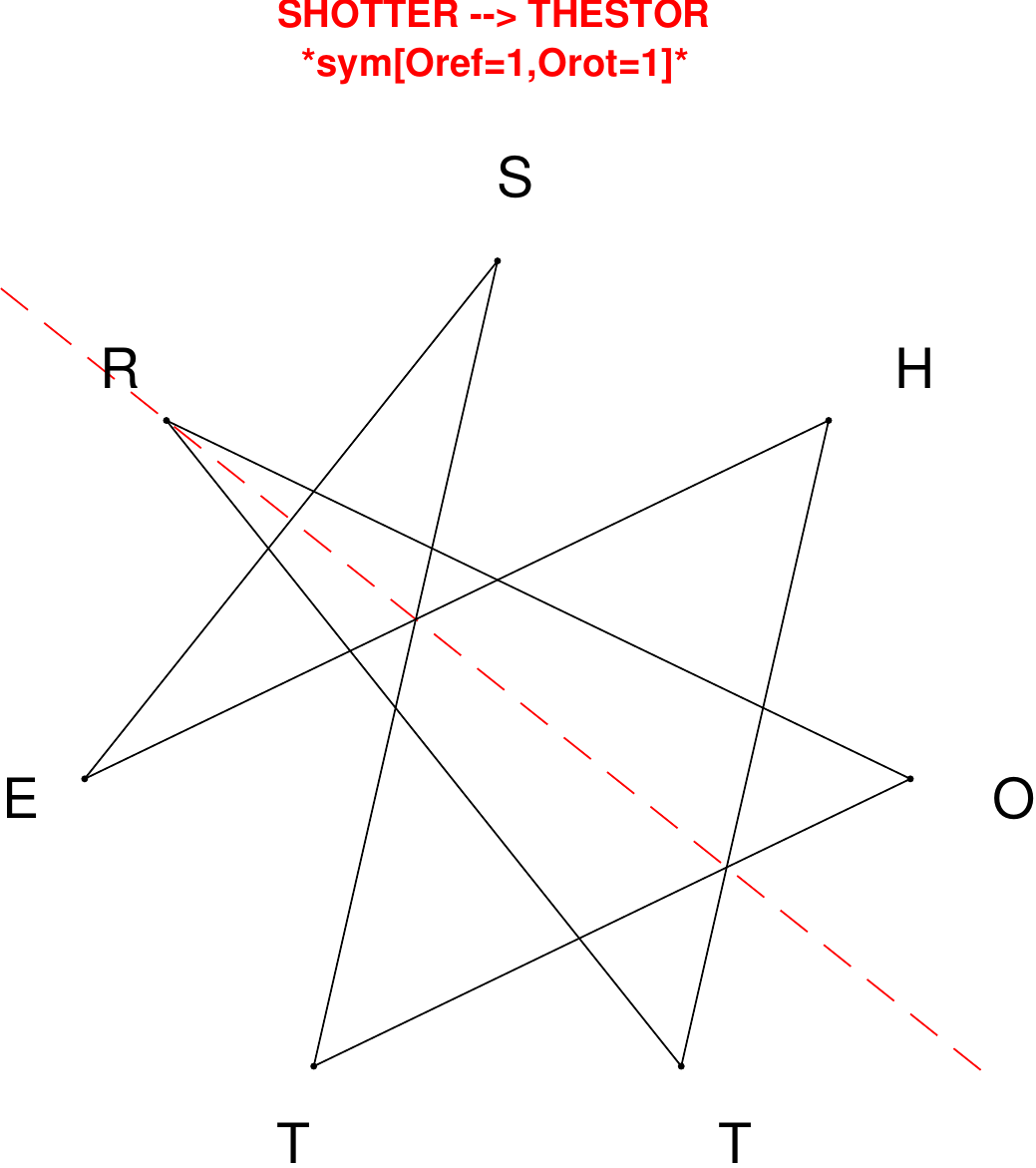}
\end{subfigure}
\hfill
\begin{subfigure}[T]{0.19\textwidth}
\centering
\includegraphics[width=\textwidth]{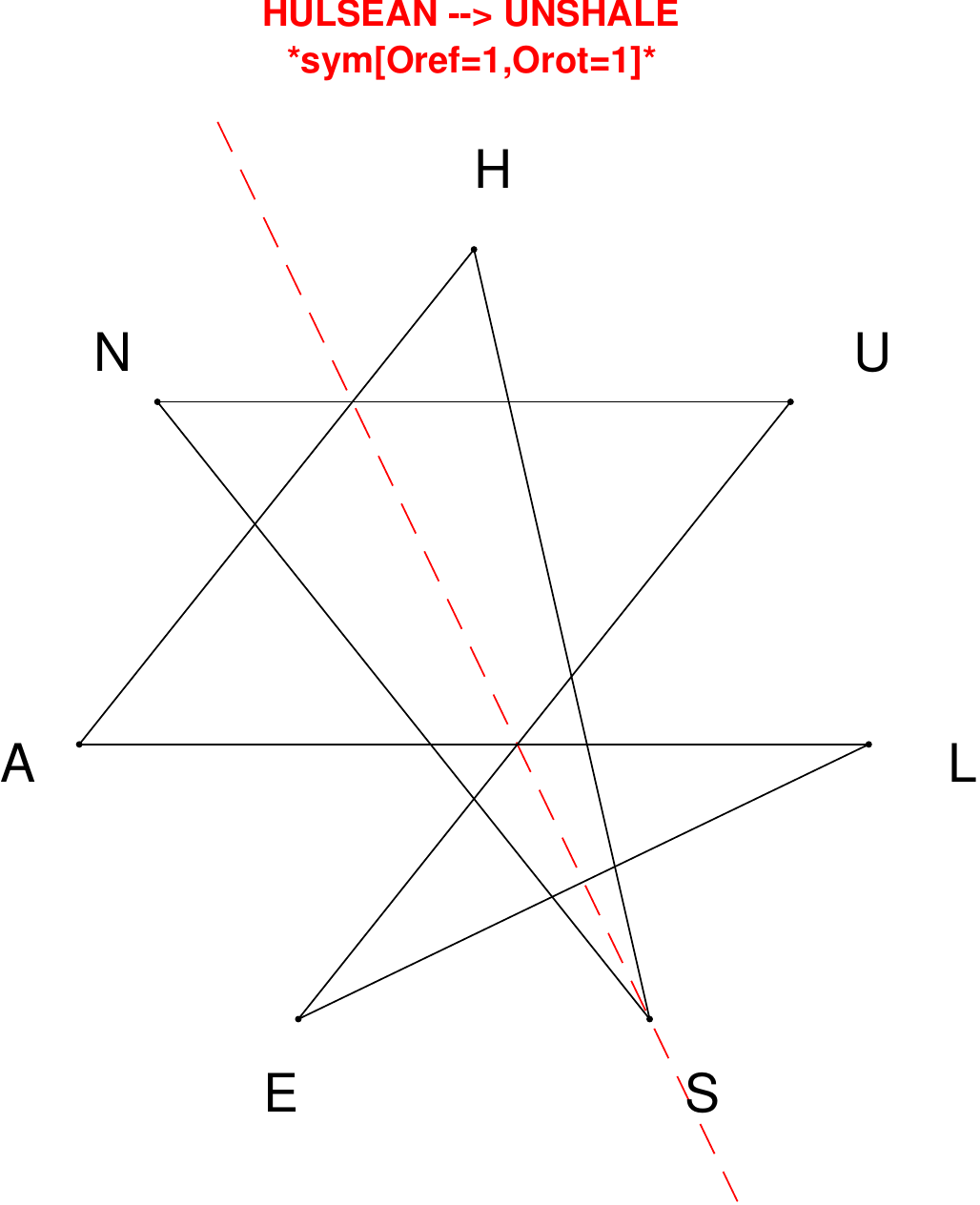}
\end{subfigure}
\hfill
\begin{subfigure}[T]{0.19\textwidth}
\centering
\includegraphics[width=\textwidth]{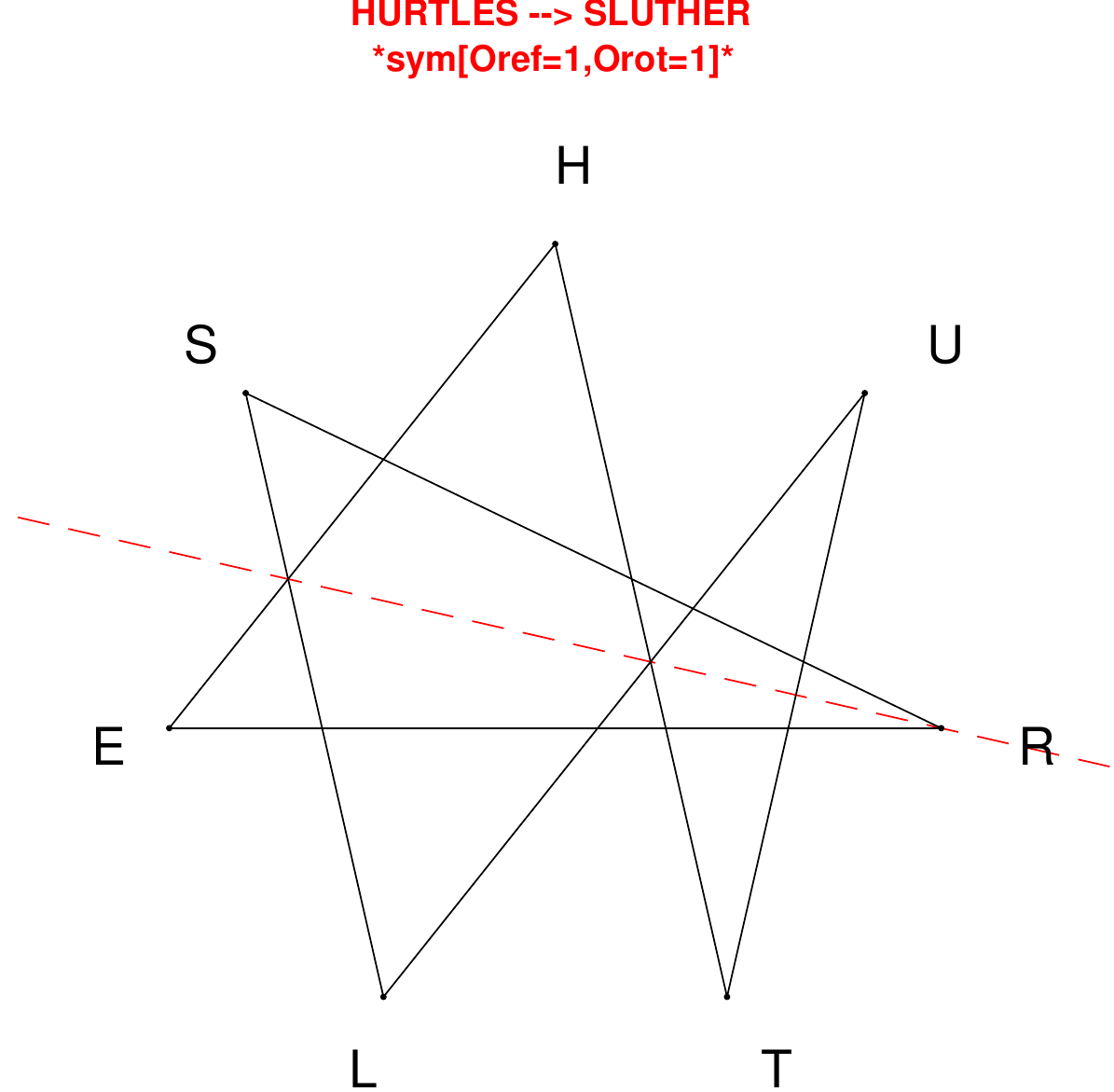}
\end{subfigure}
\hfill
\begin{subfigure}[T]{0.19\textwidth}
\centering
\includegraphics[width=\textwidth]{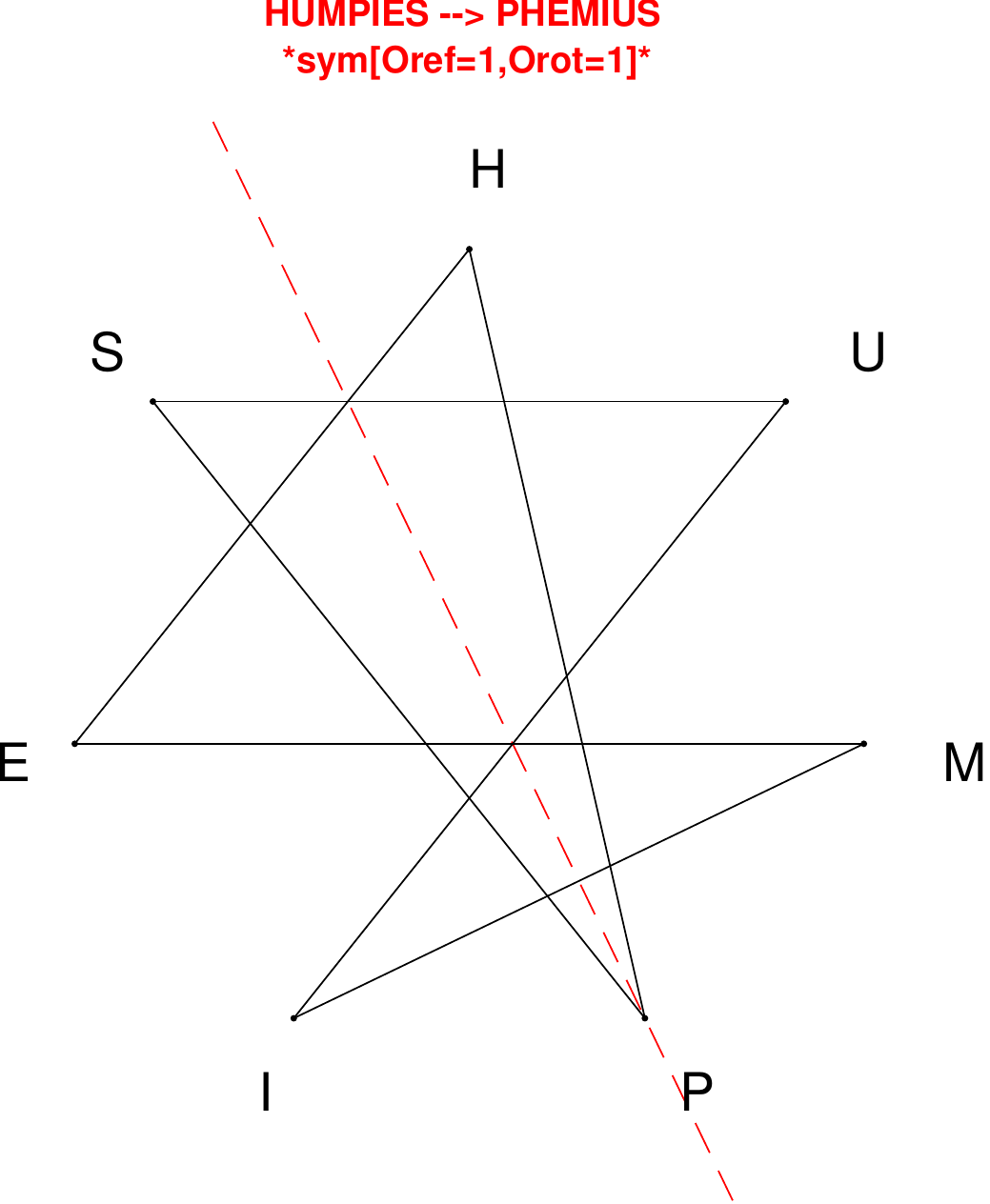}
\end{subfigure}
\end{figure}

\begin{figure}[H]
\centering
\begin{subfigure}[T]{0.19\textwidth}
\centering
\includegraphics[width=\textwidth]{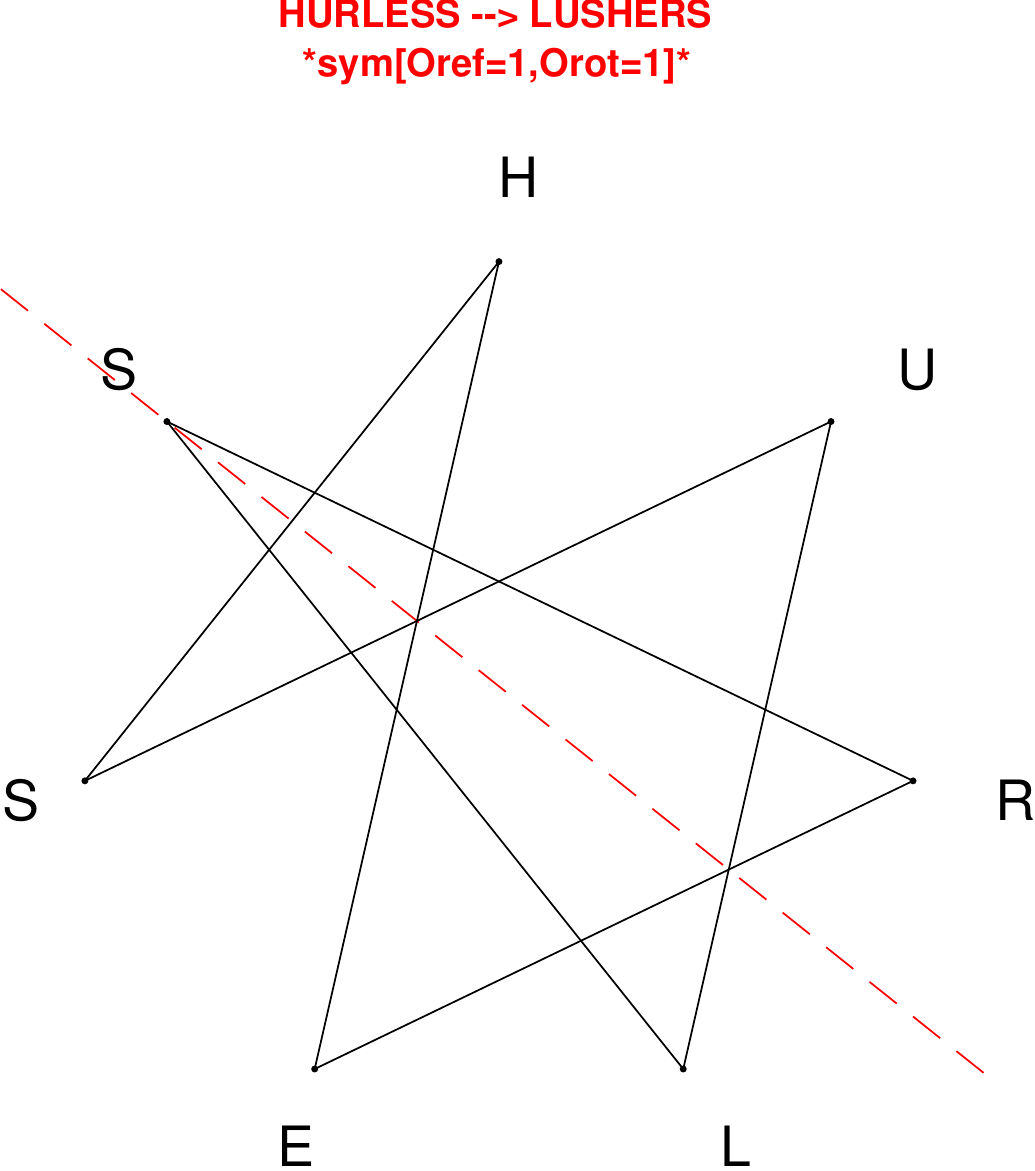}
\end{subfigure}
\hfill
\begin{subfigure}[T]{0.19\textwidth}
\centering
\includegraphics[width=\textwidth]{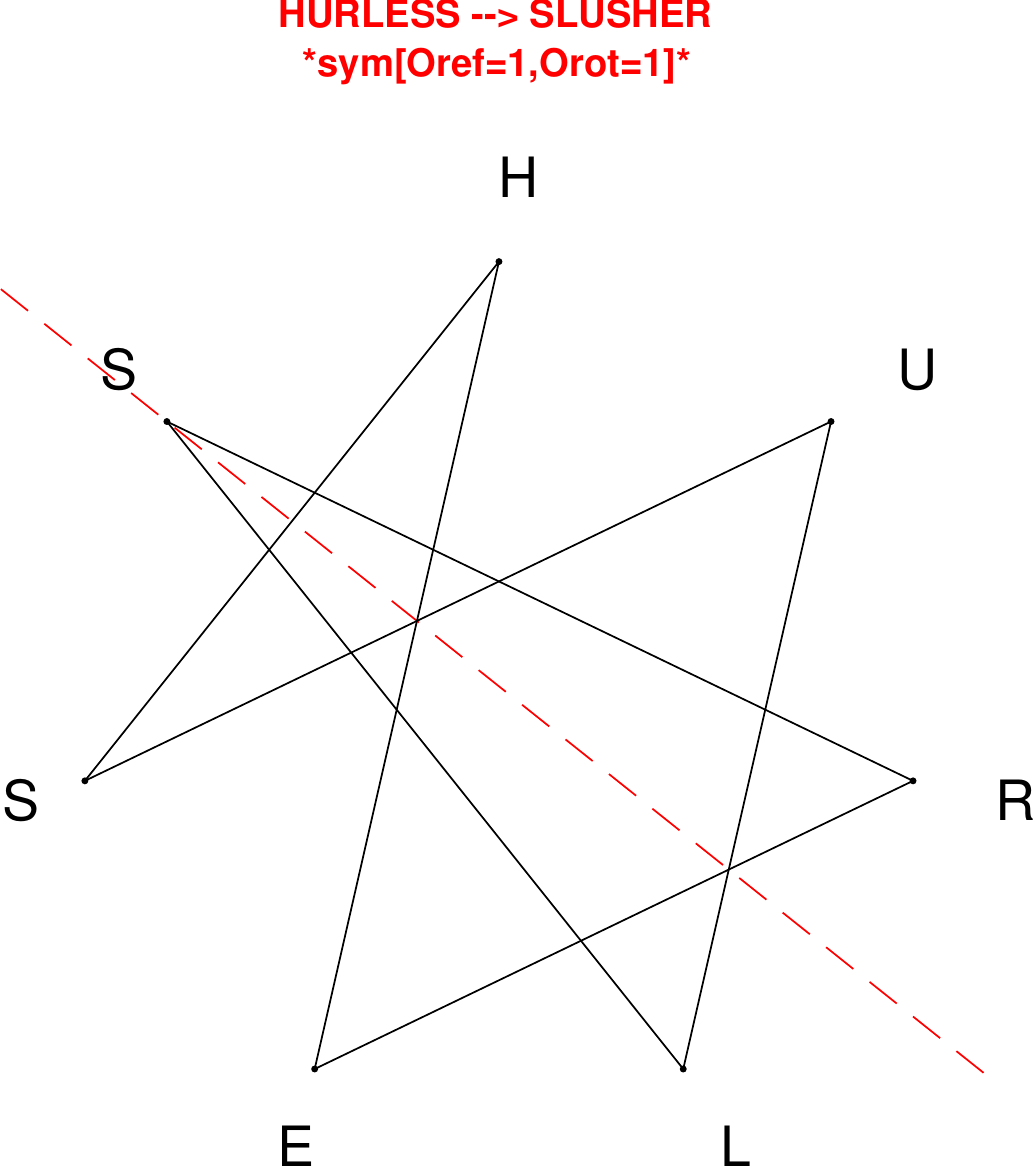}
\end{subfigure}
\hfill
\begin{subfigure}[T]{0.19\textwidth}
\centering
\includegraphics[width=\textwidth]{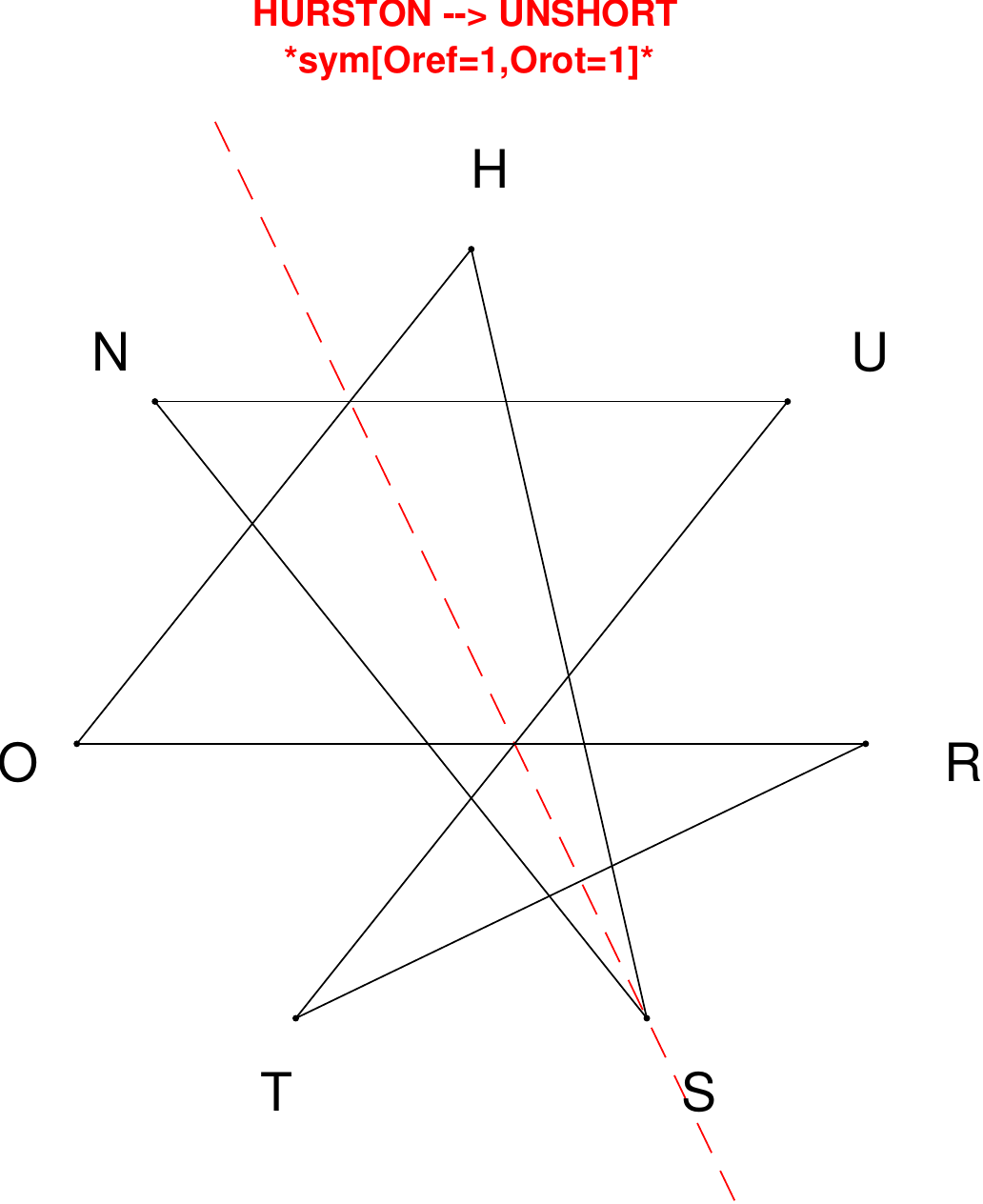}
\end{subfigure}
\hfill
\begin{subfigure}[T]{0.19\textwidth}
\centering
\includegraphics[width=\textwidth]{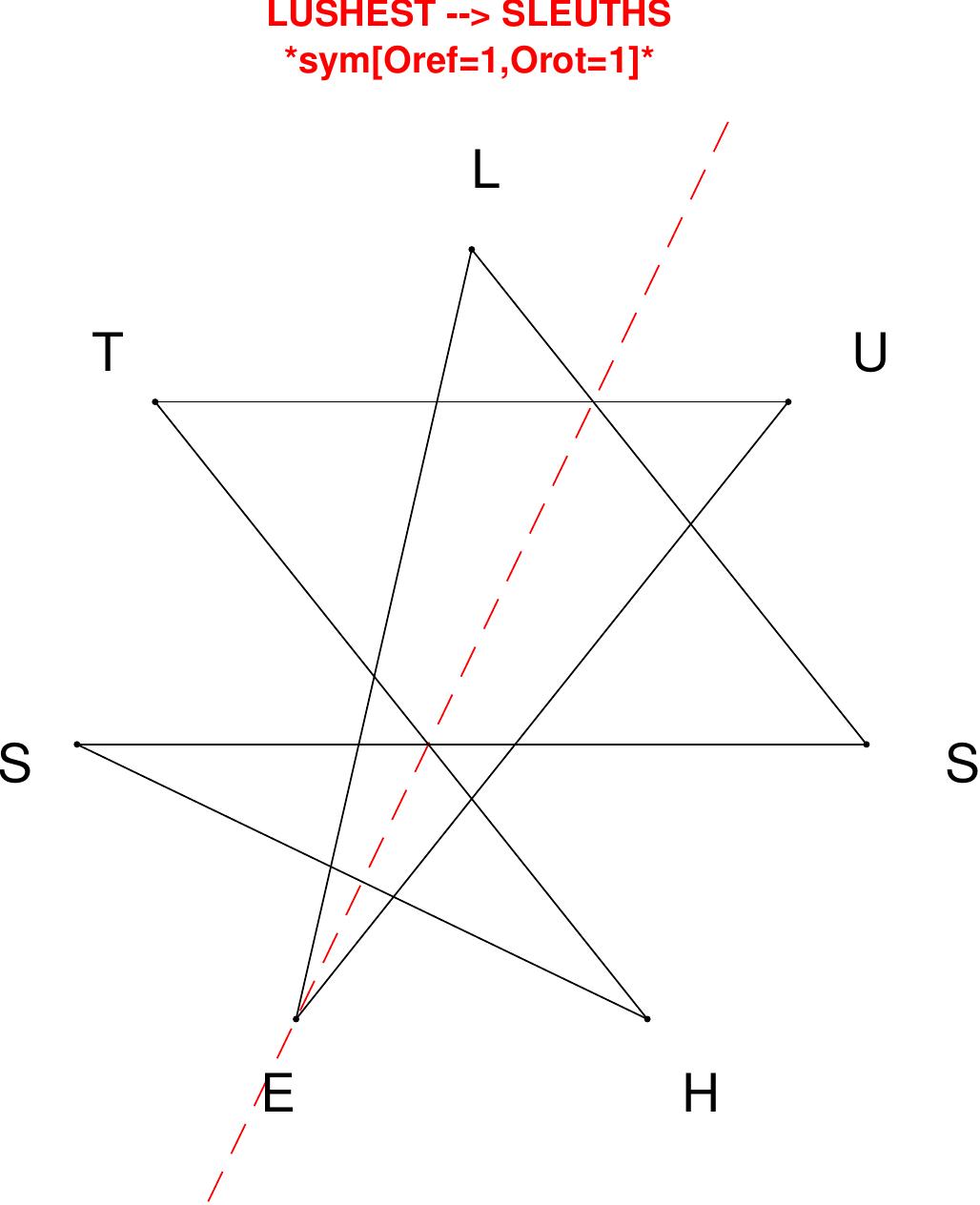}
\end{subfigure}
\hfill
\begin{subfigure}[T]{0.19\textwidth}
\centering
\includegraphics[width=\textwidth]{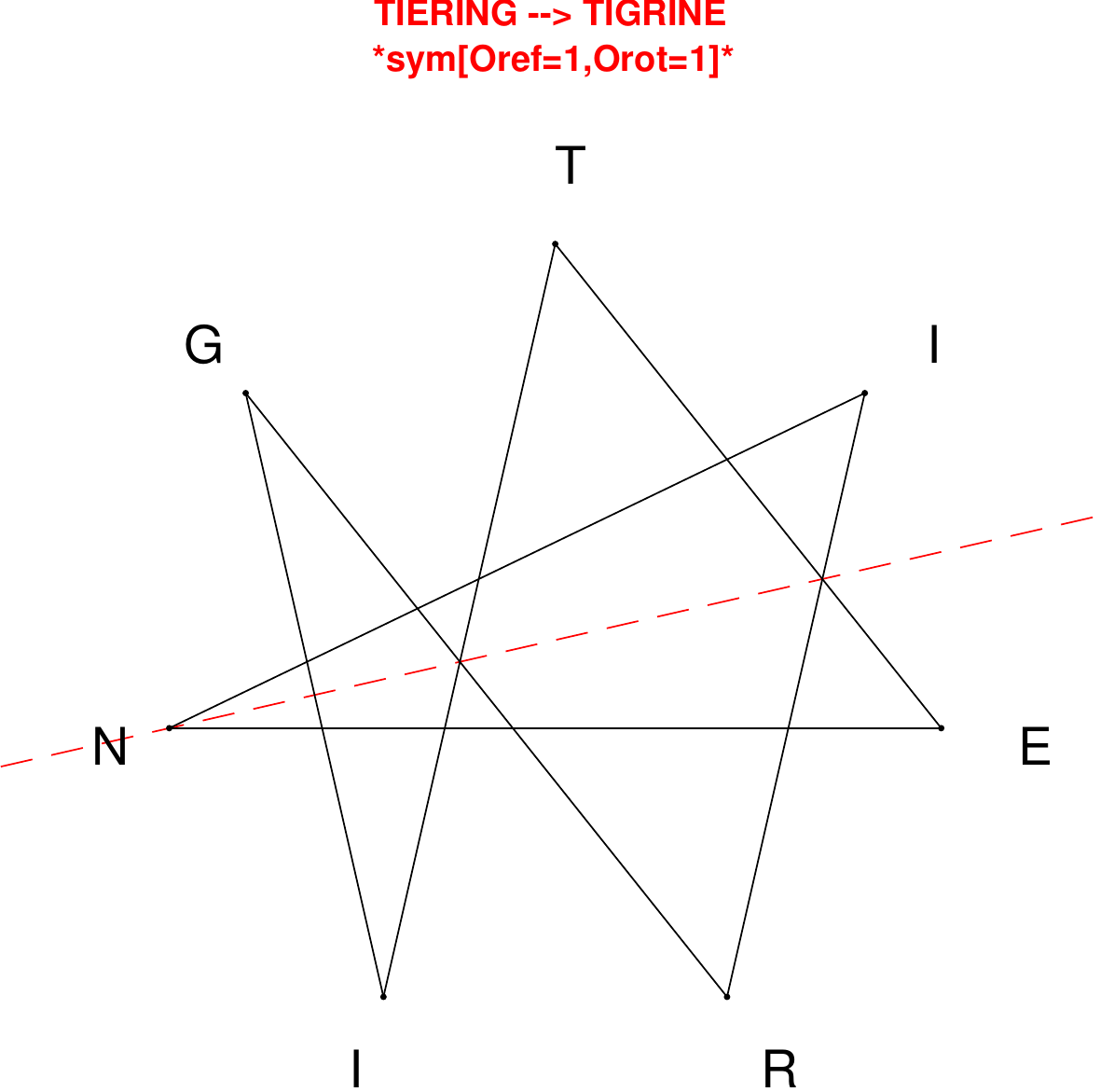}
\end{subfigure}
\end{figure}

\begin{figure}[H]
\centering
\begin{subfigure}[T]{0.19\textwidth}
\centering
\includegraphics[width=\textwidth]{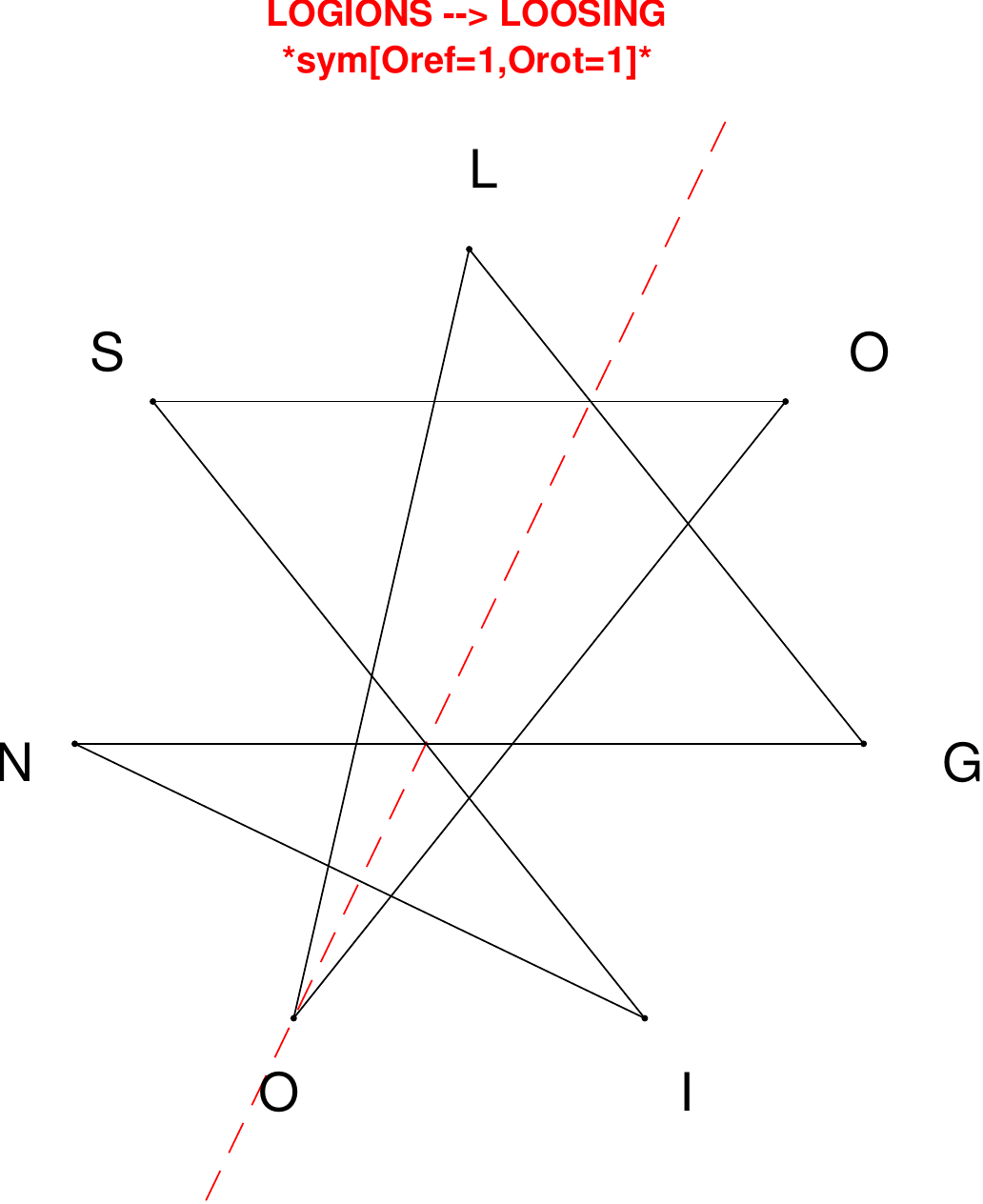}
\end{subfigure}
\hfill
\begin{subfigure}[T]{0.19\textwidth}
\centering
\includegraphics[width=\textwidth]{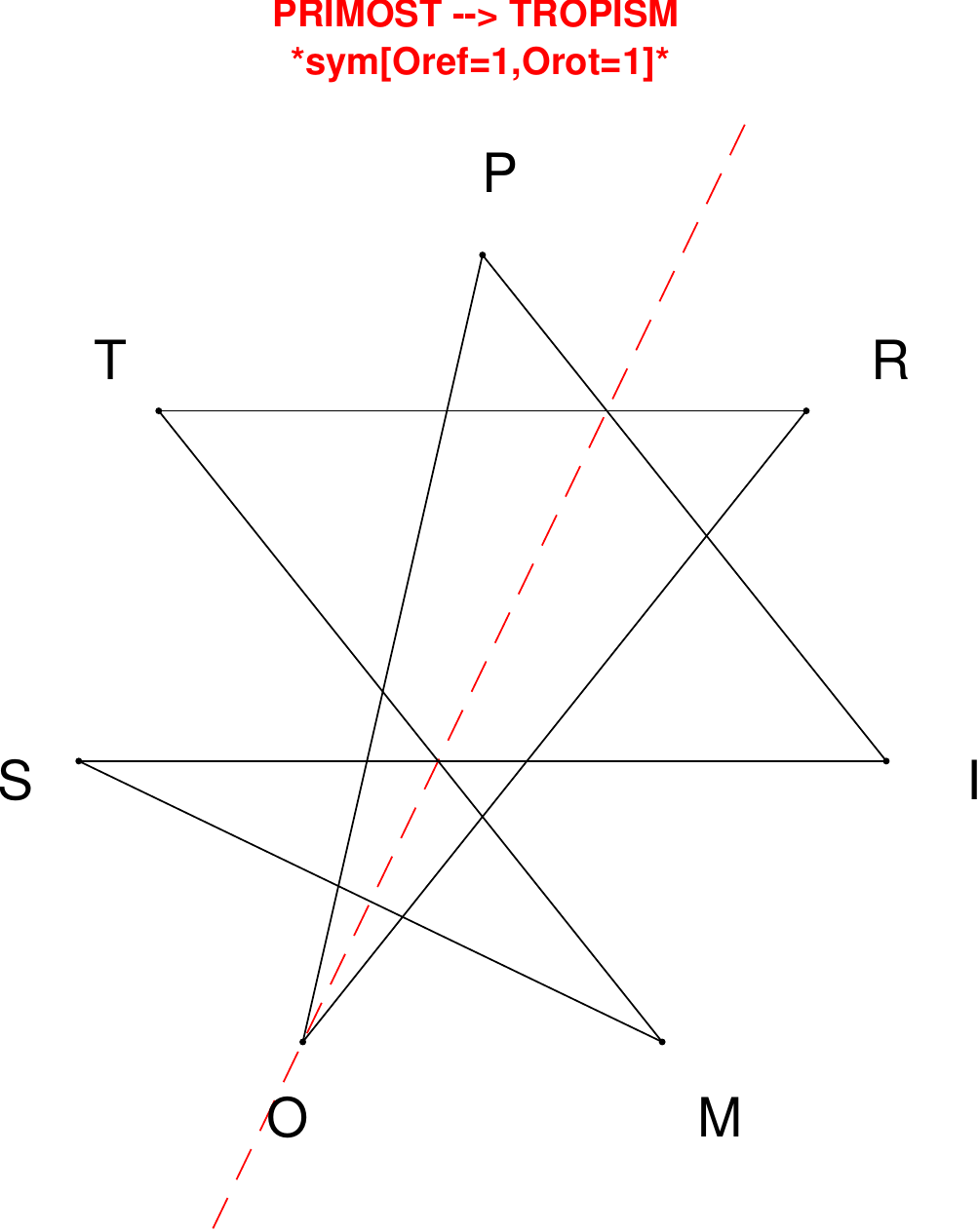}
\end{subfigure}
\hfill
\begin{subfigure}[T]{0.19\textwidth}
\centering
\includegraphics[width=\textwidth]{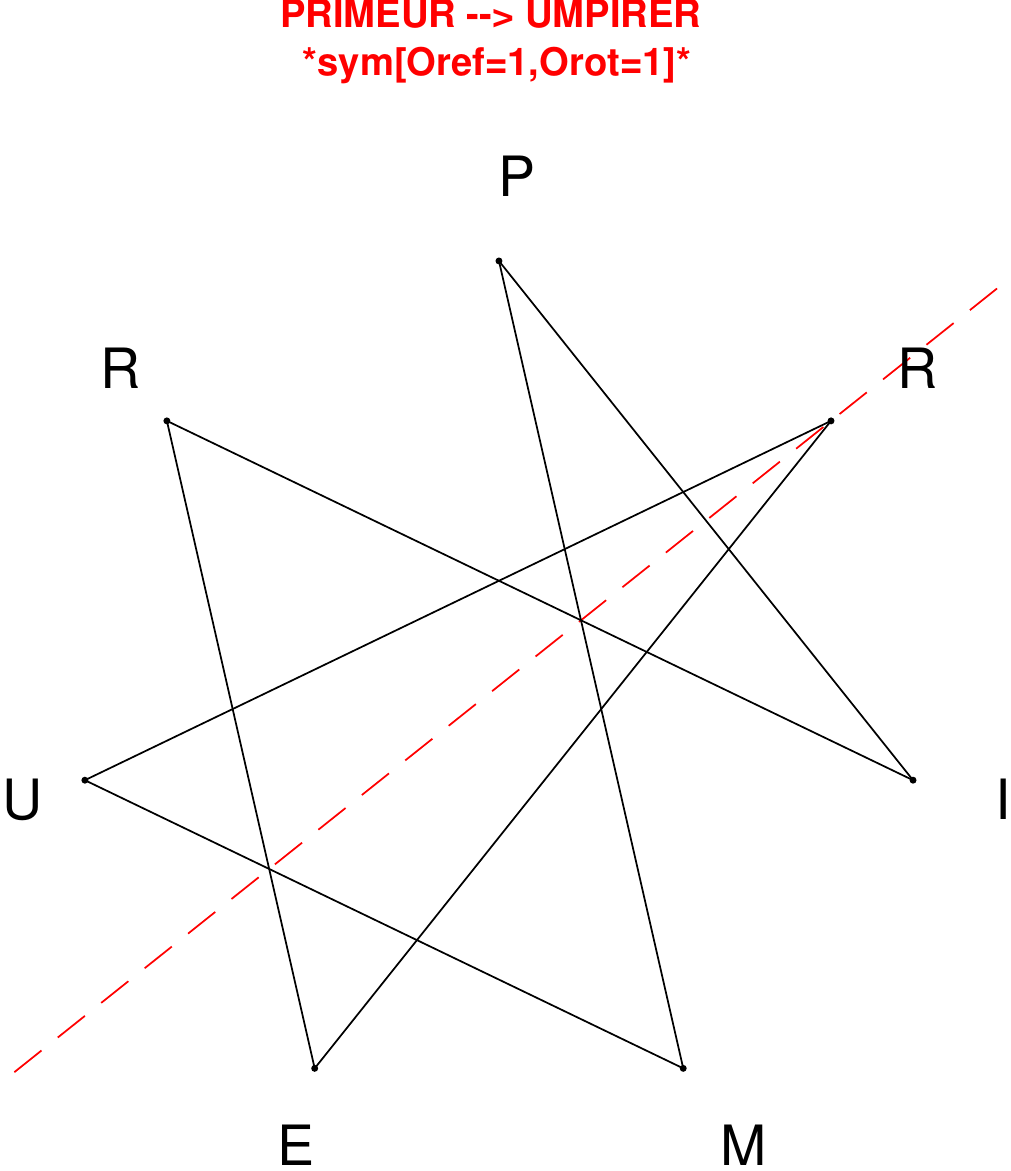}
\end{subfigure}
\hfill
\begin{subfigure}[T]{0.19\textwidth}
\centering
\includegraphics[width=\textwidth]{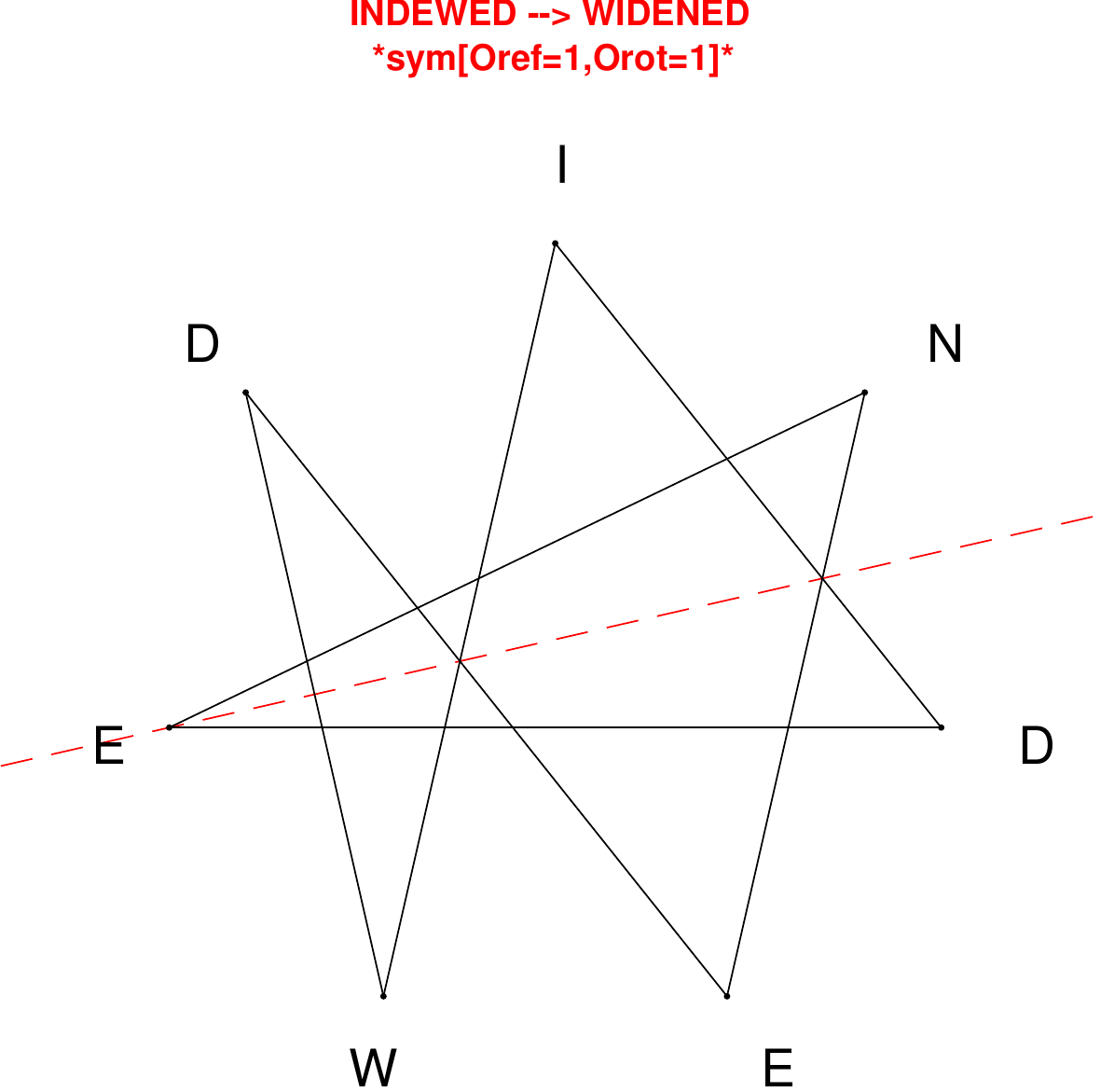}
\end{subfigure}
\hfill
\begin{subfigure}[T]{0.19\textwidth}
\centering
\includegraphics[width=\textwidth]{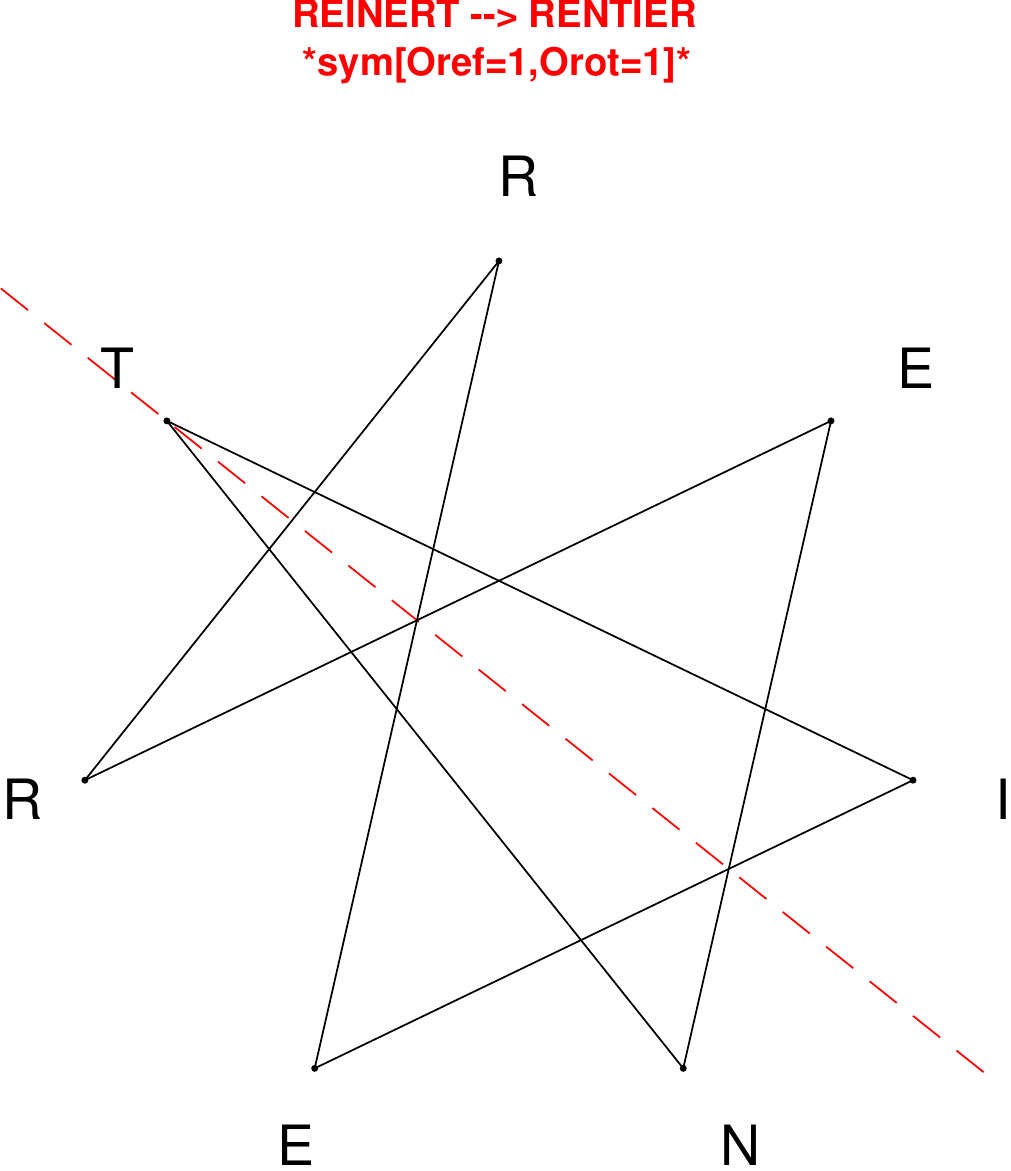}
\end{subfigure}
\end{figure}

\begin{figure}[H]
\centering
\begin{subfigure}[T]{0.19\textwidth}
\centering
\includegraphics[width=\textwidth]{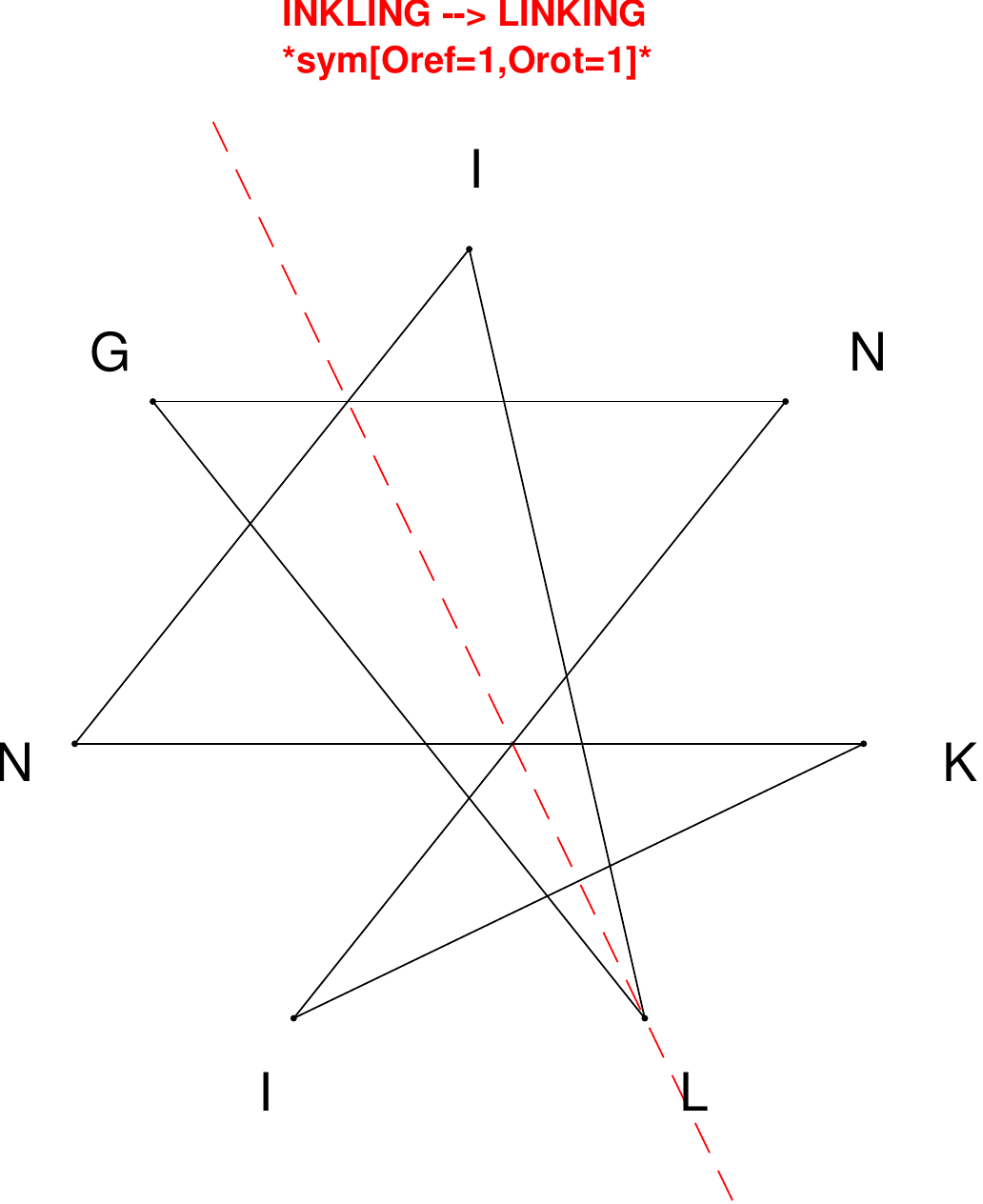}
\end{subfigure}
\hfill
\begin{subfigure}[T]{0.19\textwidth}
\centering
\includegraphics[width=\textwidth]{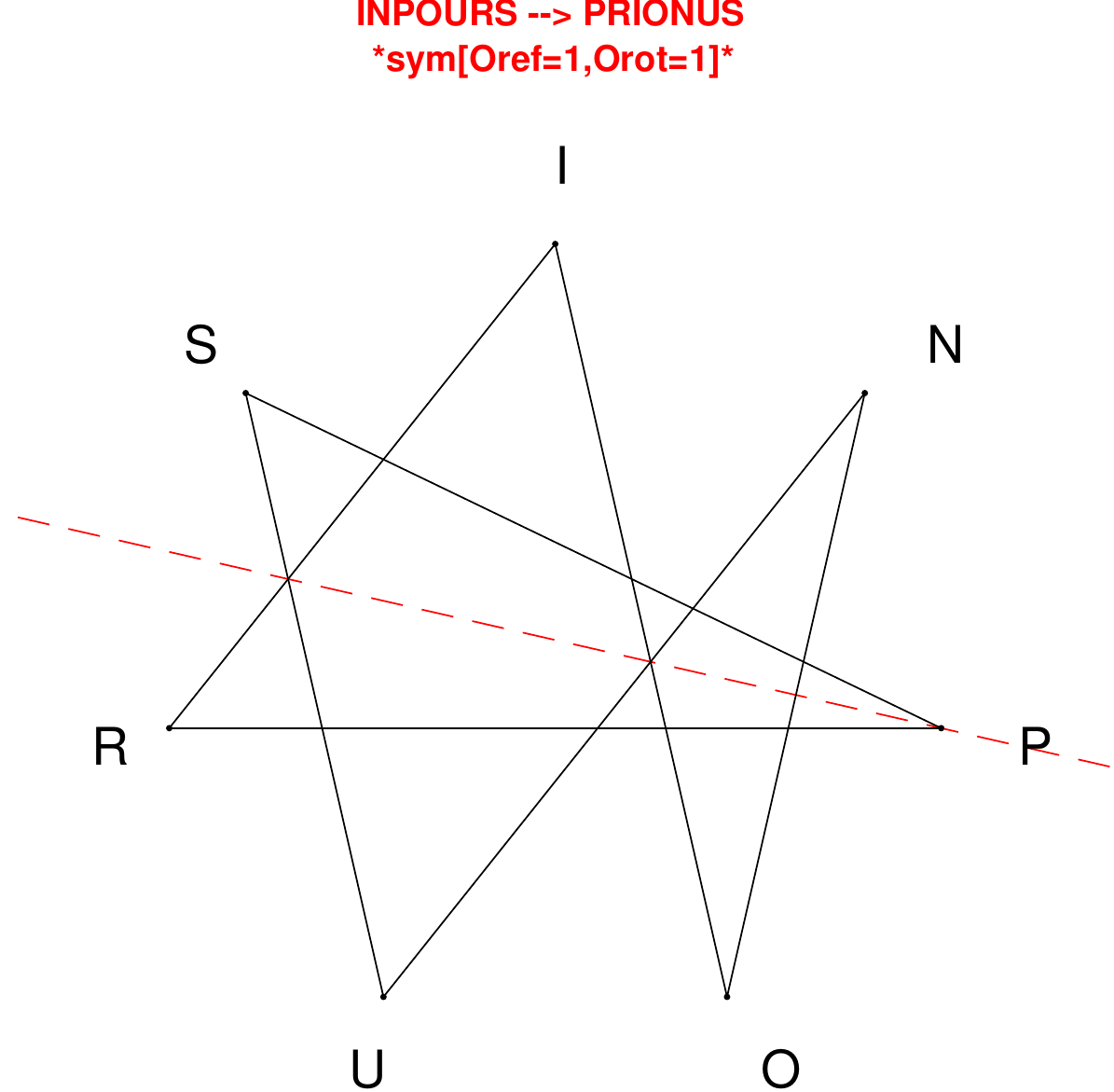}
\end{subfigure}
\hfill
\begin{subfigure}[T]{0.19\textwidth}
\centering
\includegraphics[width=\textwidth]{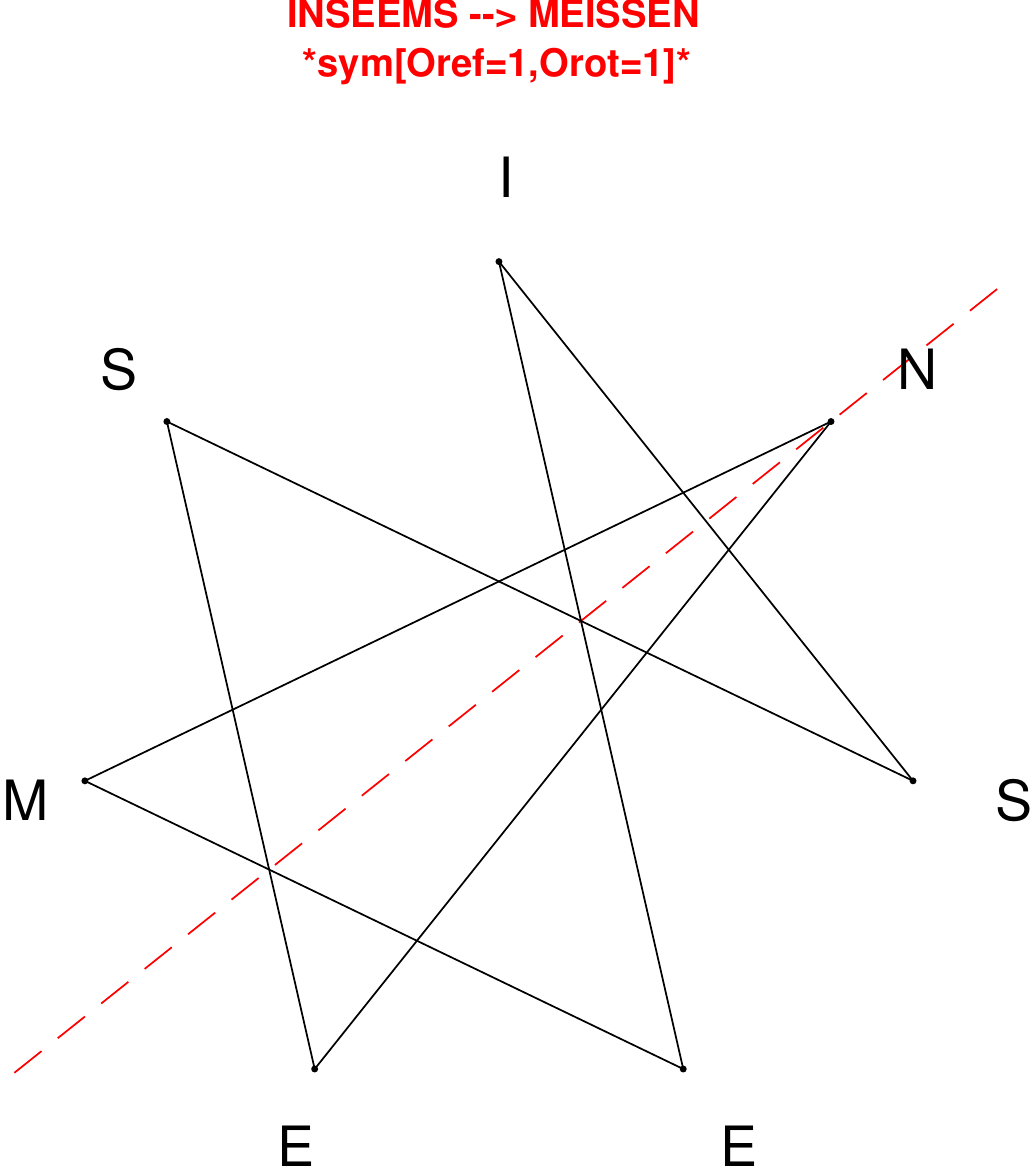}
\end{subfigure}
\hfill
\begin{subfigure}[T]{0.19\textwidth}
\centering
\includegraphics[width=\textwidth]{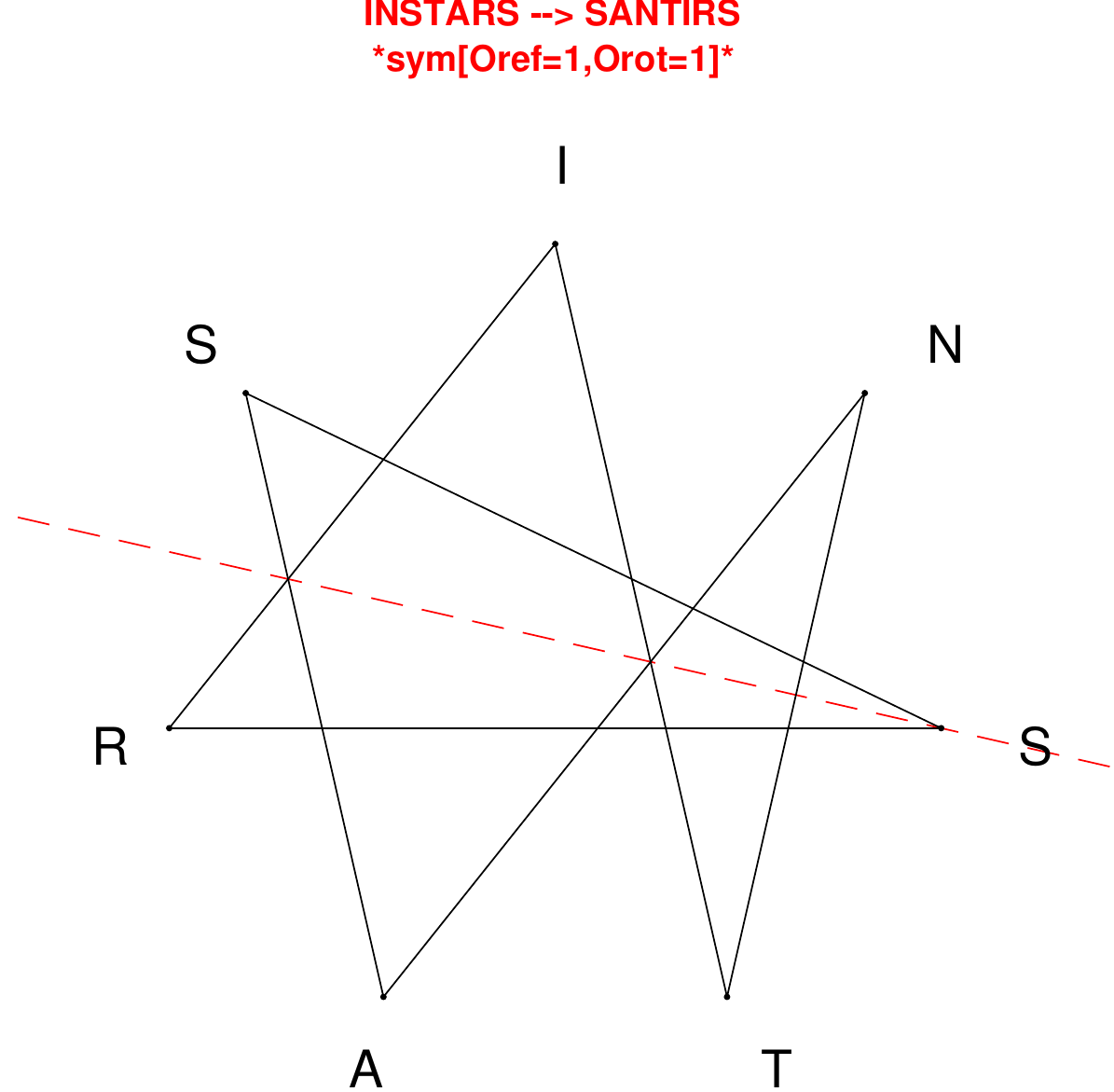}
\end{subfigure}
\hfill
\begin{subfigure}[T]{0.19\textwidth}
\centering
\includegraphics[width=\textwidth]{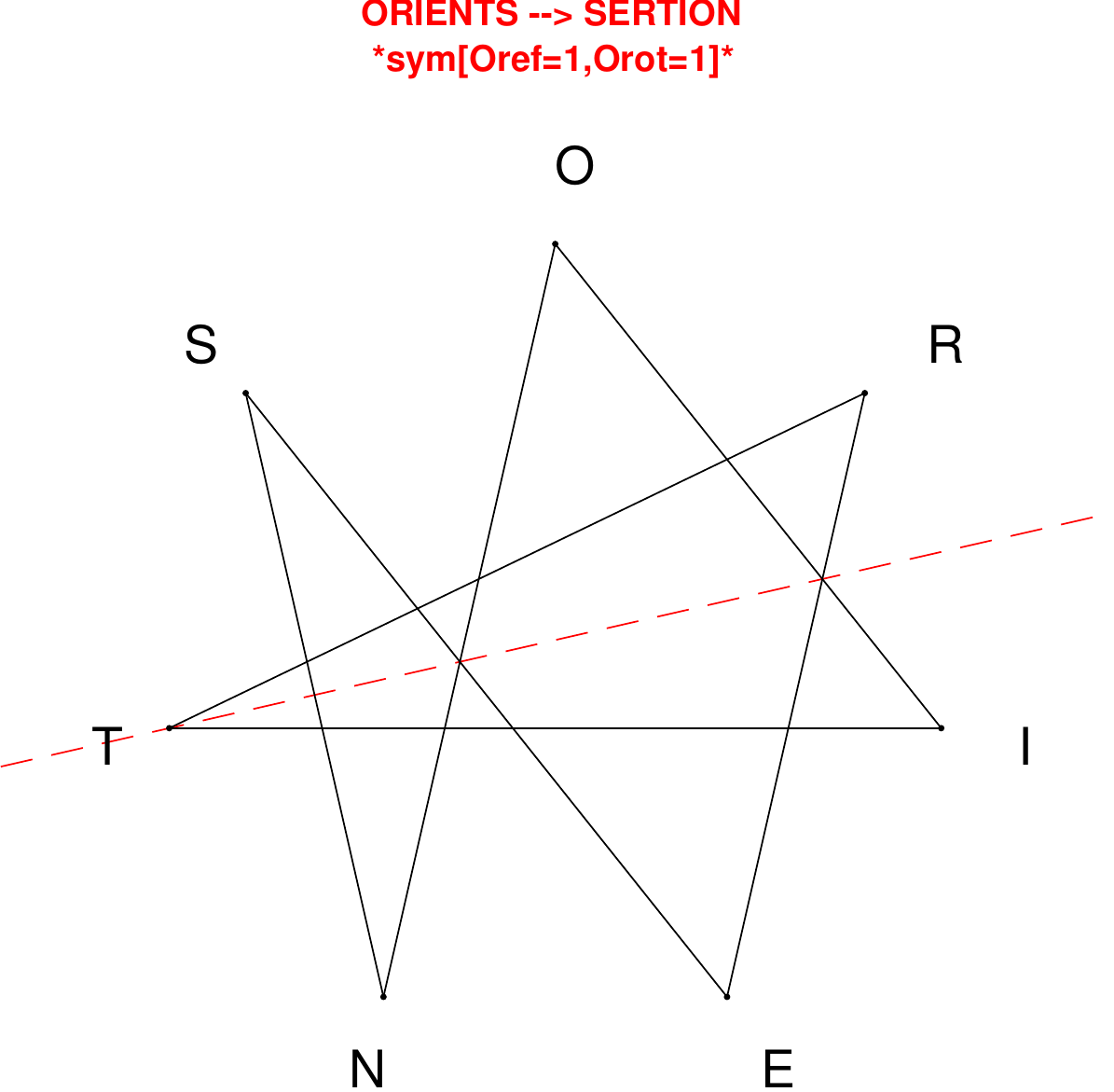}
\end{subfigure}
\end{figure}

\begin{figure}[H]
\centering
\begin{subfigure}[T]{0.19\textwidth}
\centering
\includegraphics[width=\textwidth]{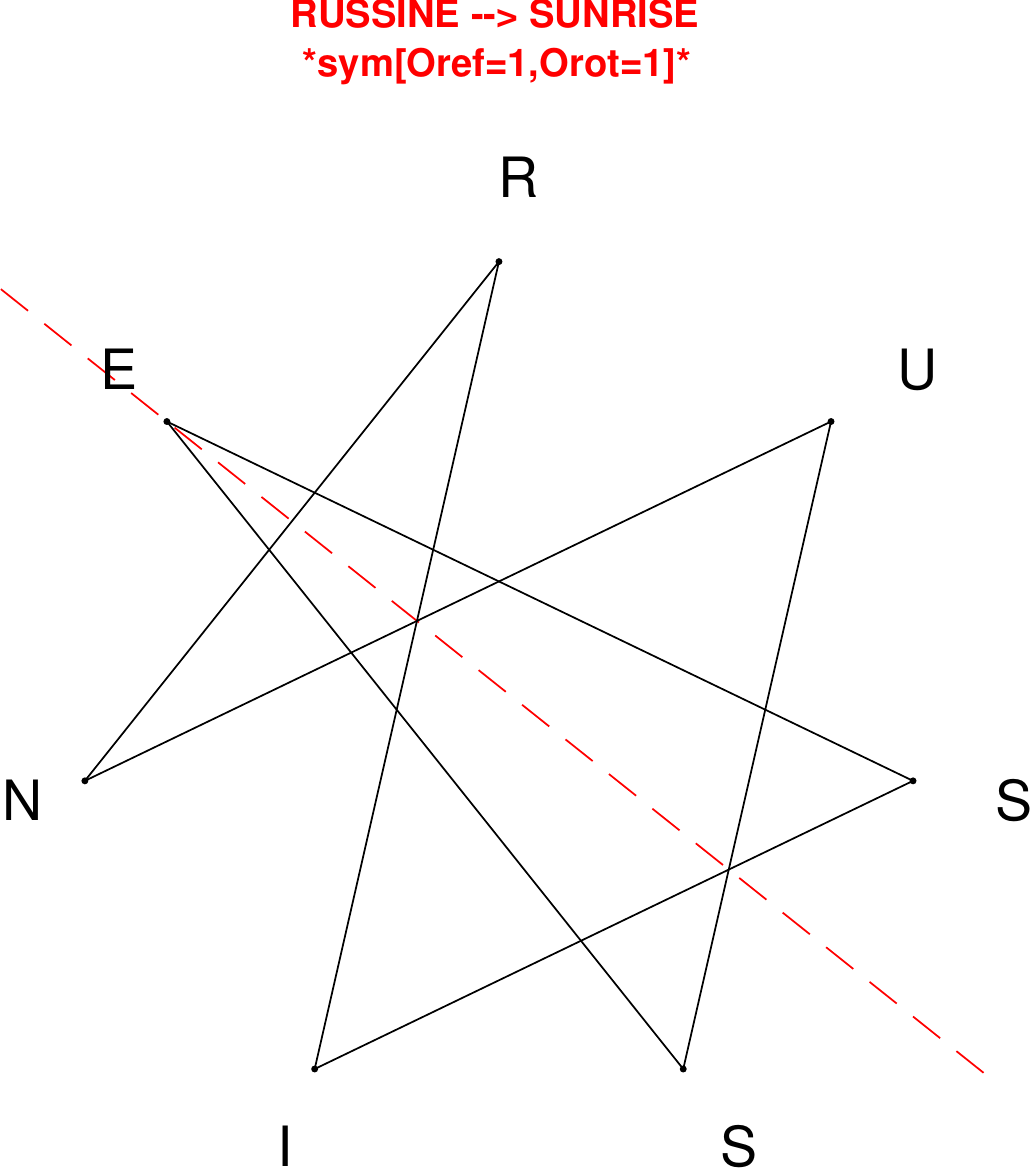}
\end{subfigure}
\hfill
\begin{subfigure}[T]{0.19\textwidth}
\centering
\includegraphics[width=\textwidth]{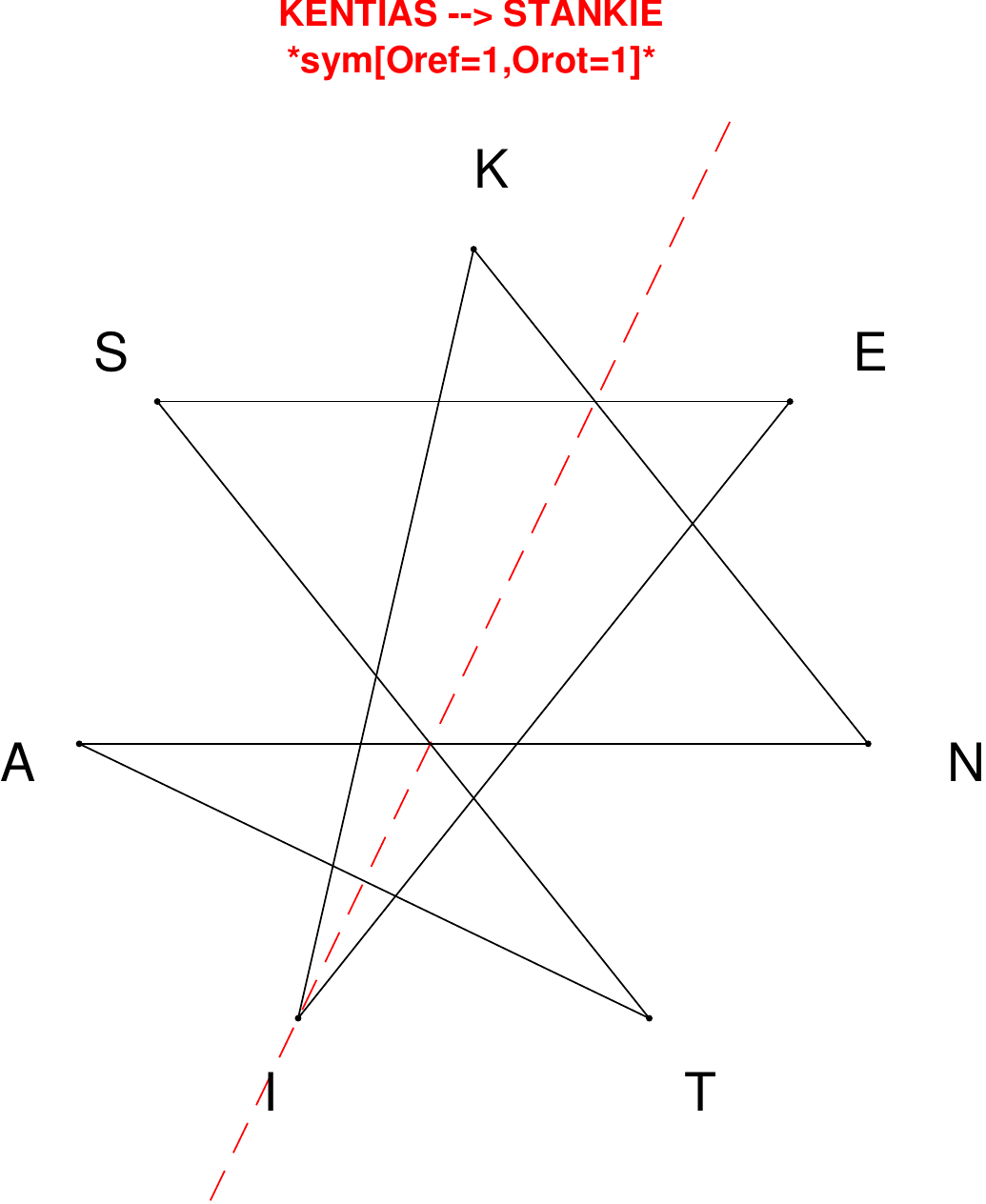}
\end{subfigure}
\hfill
\begin{subfigure}[T]{0.19\textwidth}
\centering
\includegraphics[width=\textwidth]{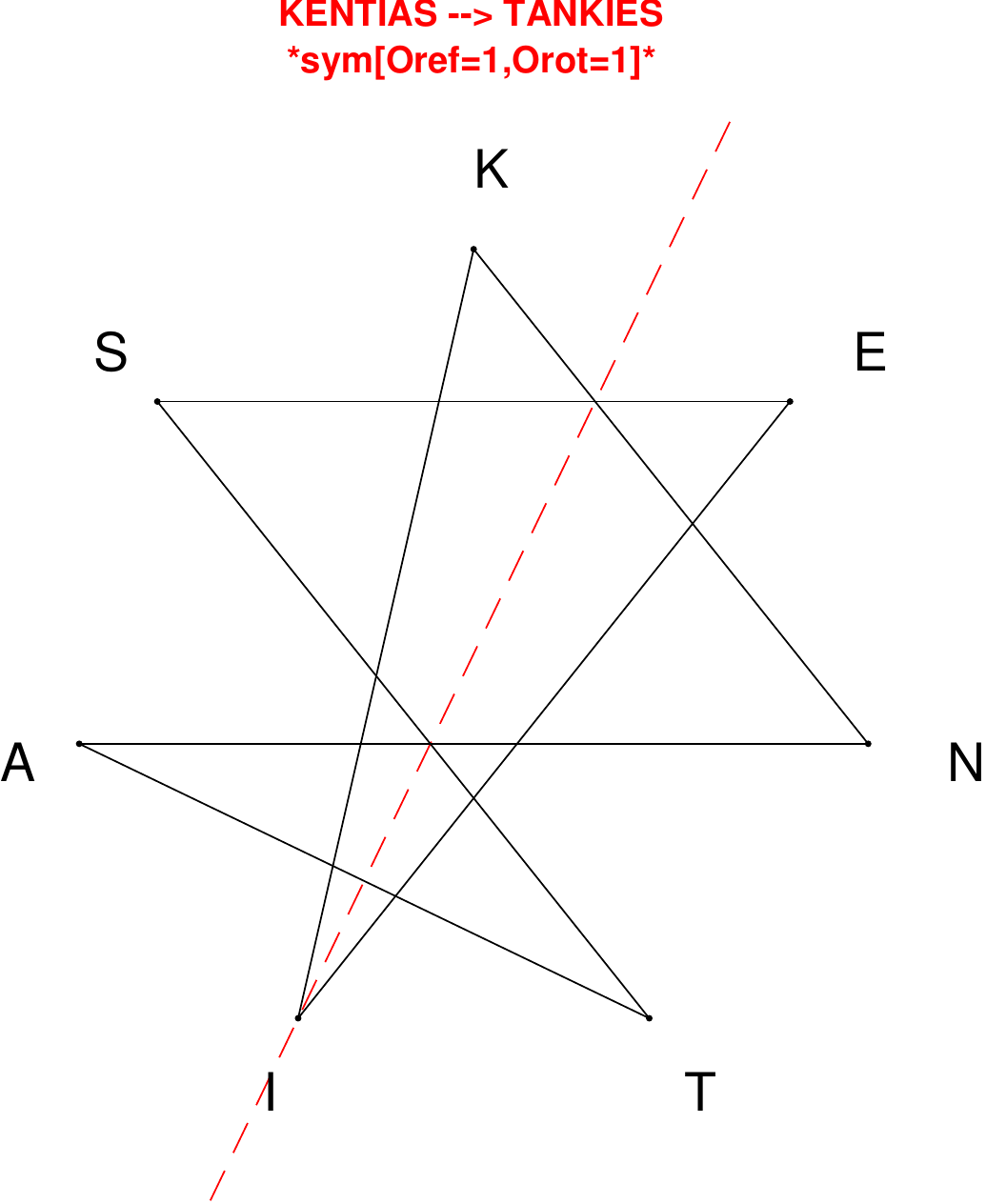}
\end{subfigure}
\hfill
\begin{subfigure}[T]{0.19\textwidth}
\centering
\includegraphics[width=\textwidth]{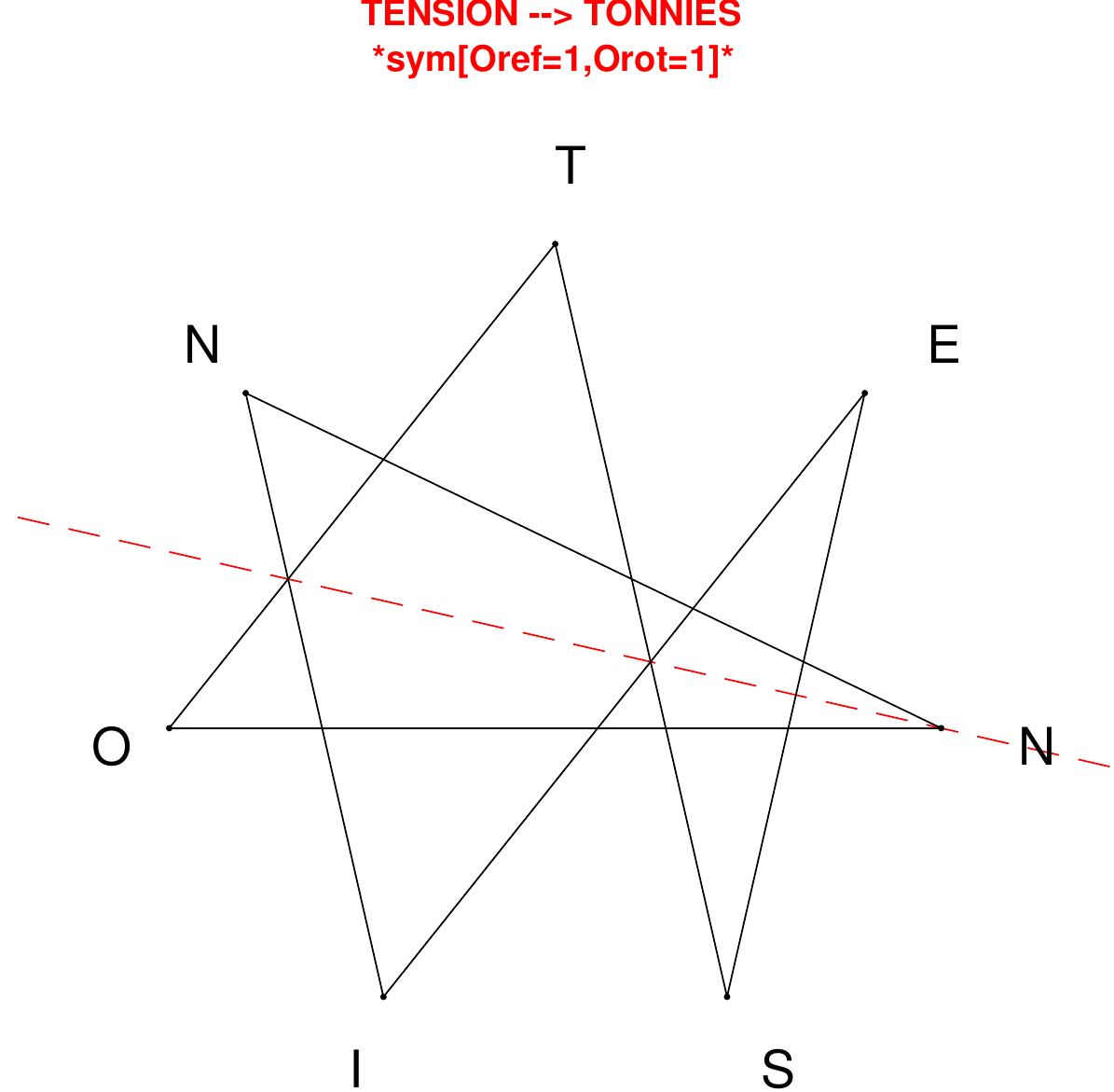}
\end{subfigure}
\hfill
\begin{subfigure}[T]{0.19\textwidth}
\centering
\includegraphics[width=\textwidth]{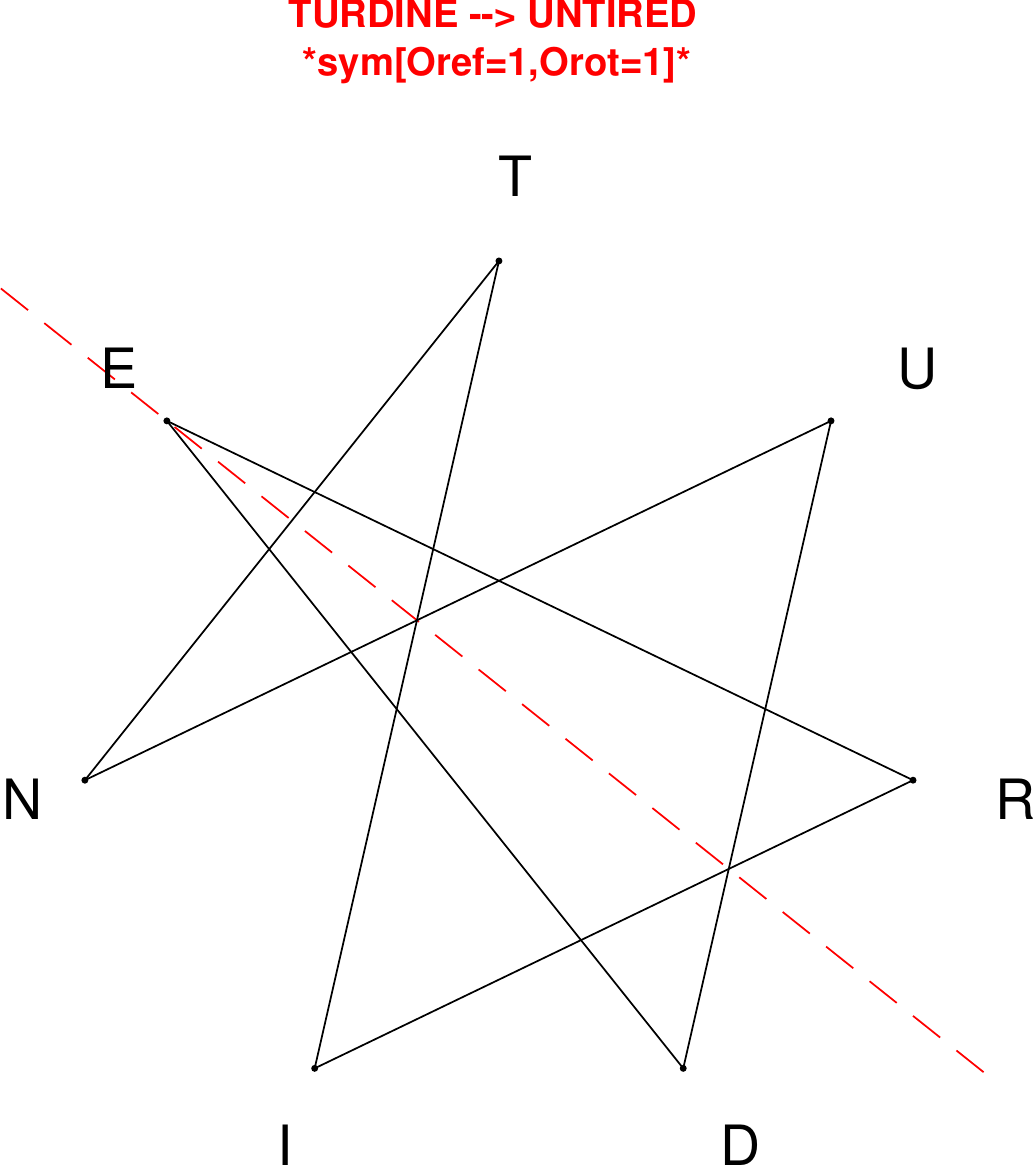}
\end{subfigure}
\end{figure}

\begin{figure}[H]
\centering
\begin{subfigure}[T]{0.19\textwidth}
\centering
\includegraphics[width=\textwidth]{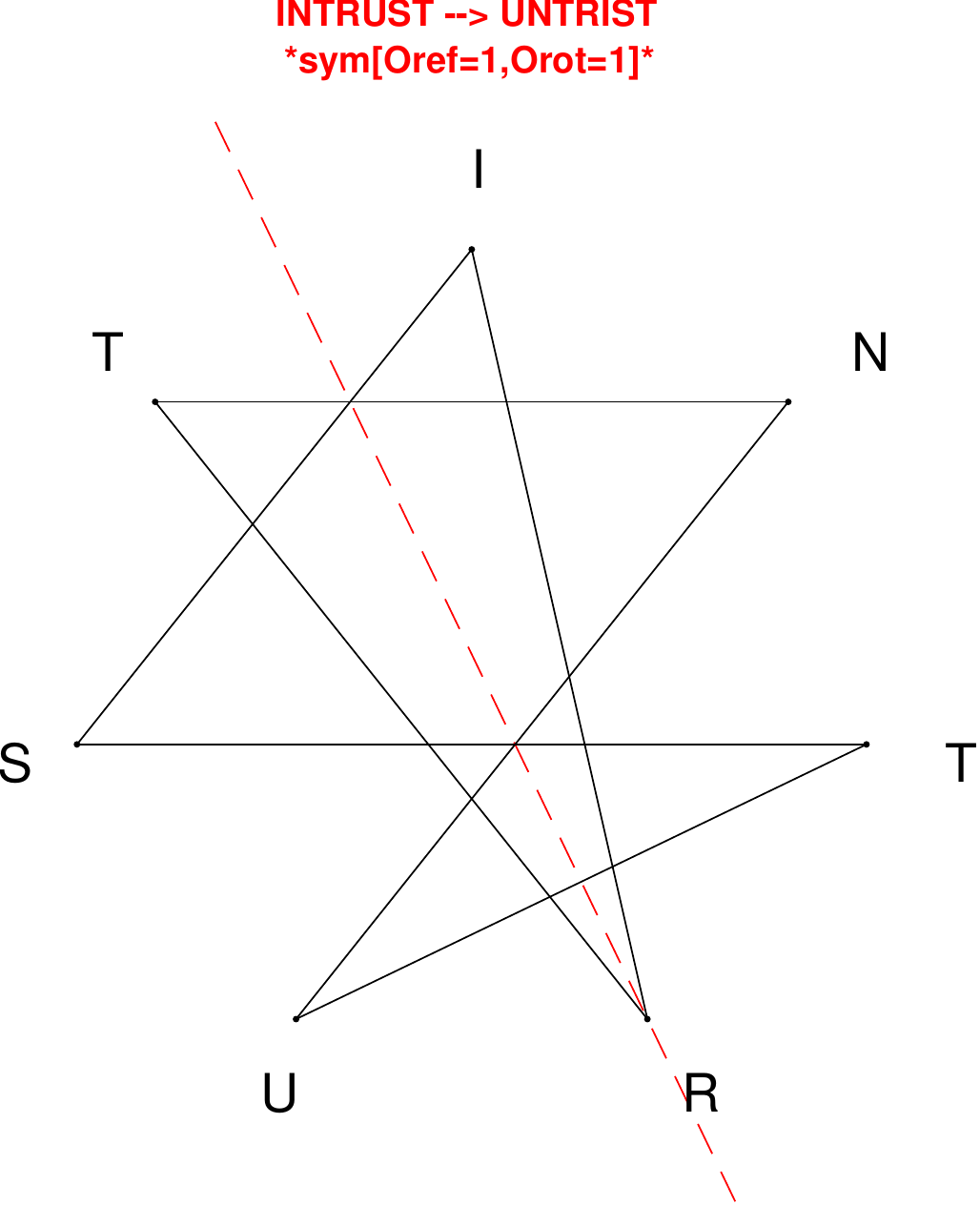}
\end{subfigure}
\hfill
\begin{subfigure}[T]{0.19\textwidth}
\centering
\includegraphics[width=\textwidth]{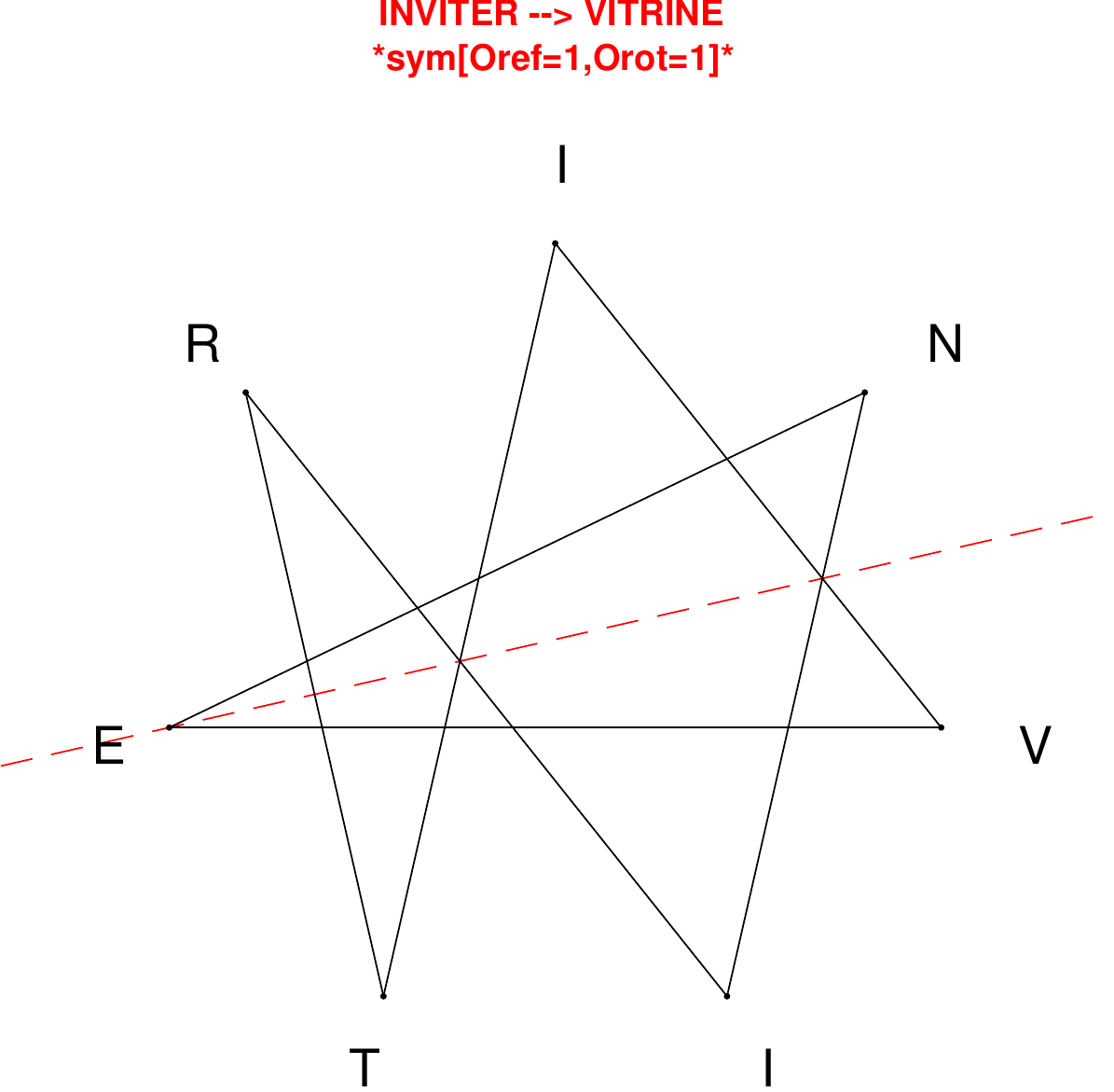}
\end{subfigure}
\hfill
\begin{subfigure}[T]{0.19\textwidth}
\centering
\includegraphics[width=\textwidth]{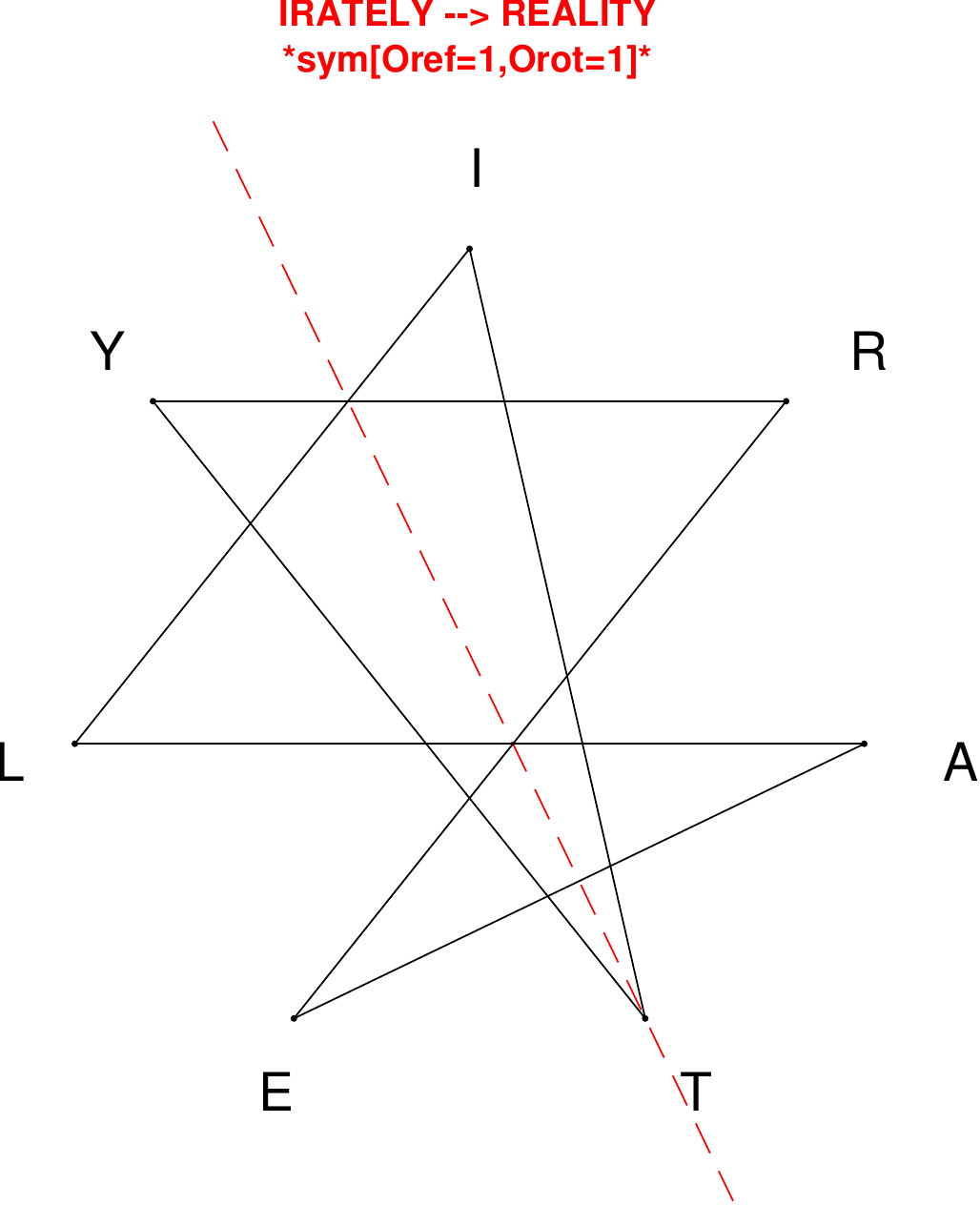}
\end{subfigure}
\hfill
\begin{subfigure}[T]{0.19\textwidth}
\centering
\includegraphics[width=\textwidth]{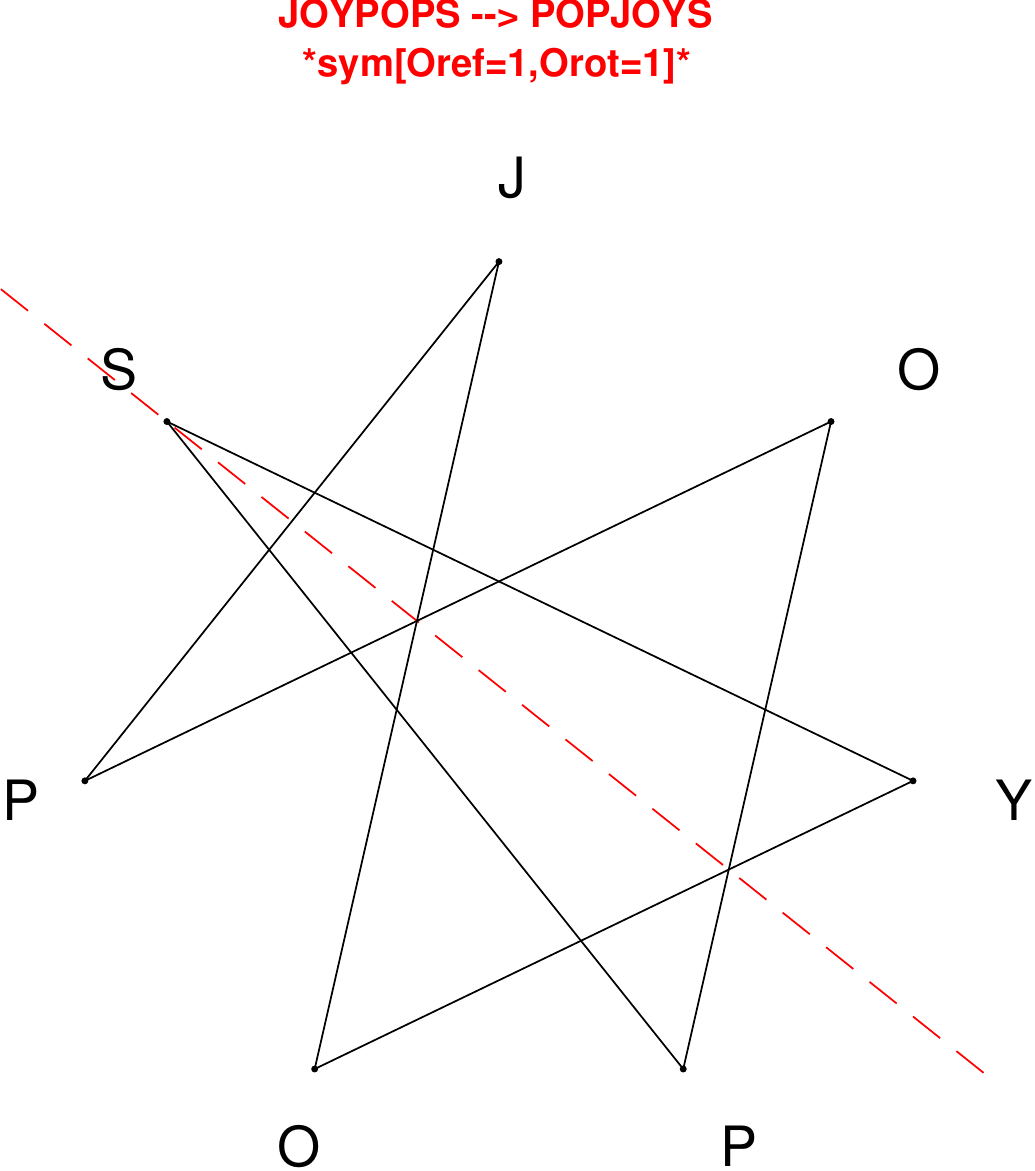}
\end{subfigure}
\hfill
\begin{subfigure}[T]{0.19\textwidth}
\centering
\includegraphics[width=\textwidth]{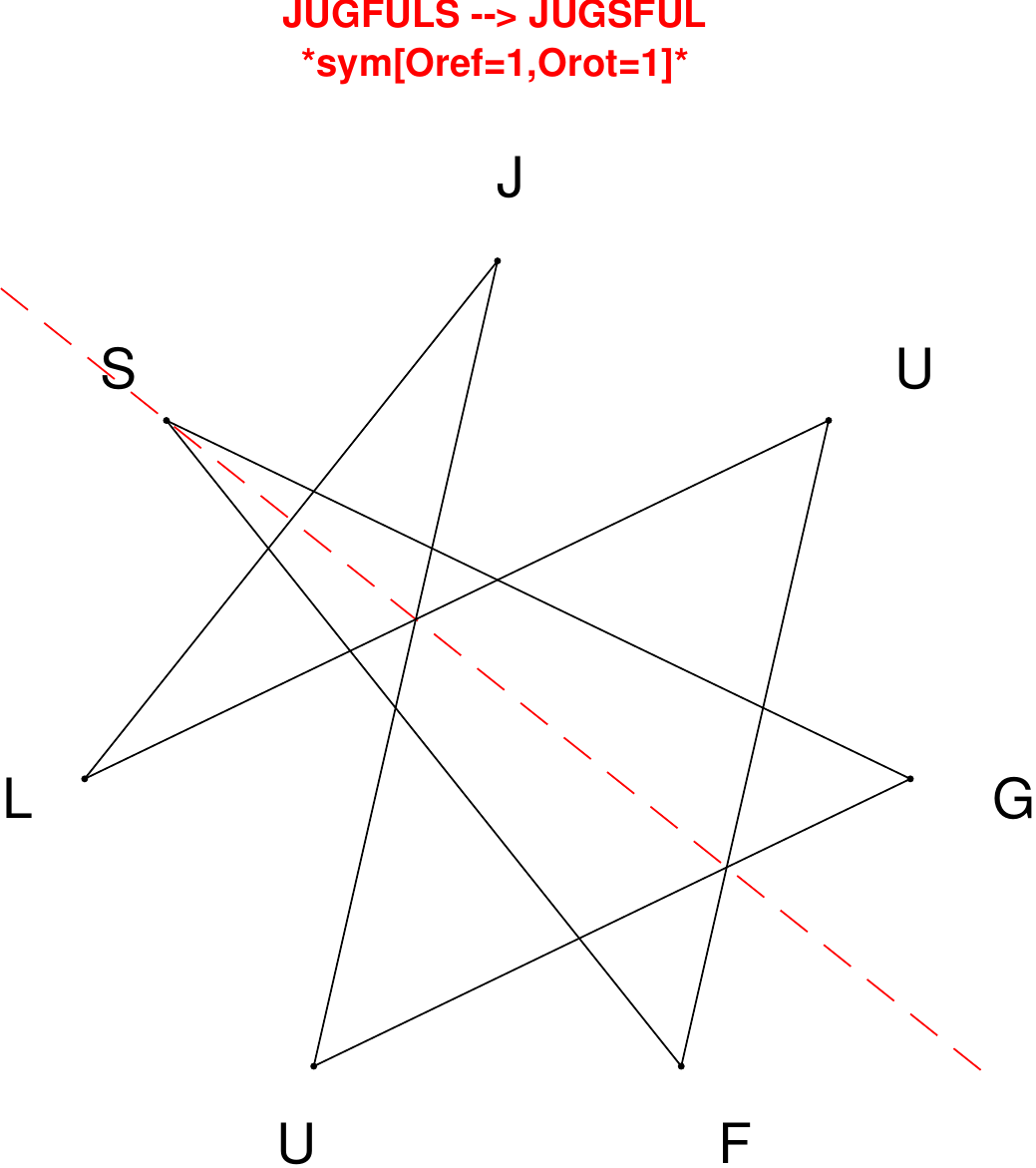}
\end{subfigure}
\end{figure}

\begin{figure}[H]
\centering
\begin{subfigure}[T]{0.19\textwidth}
\centering
\includegraphics[width=\textwidth]{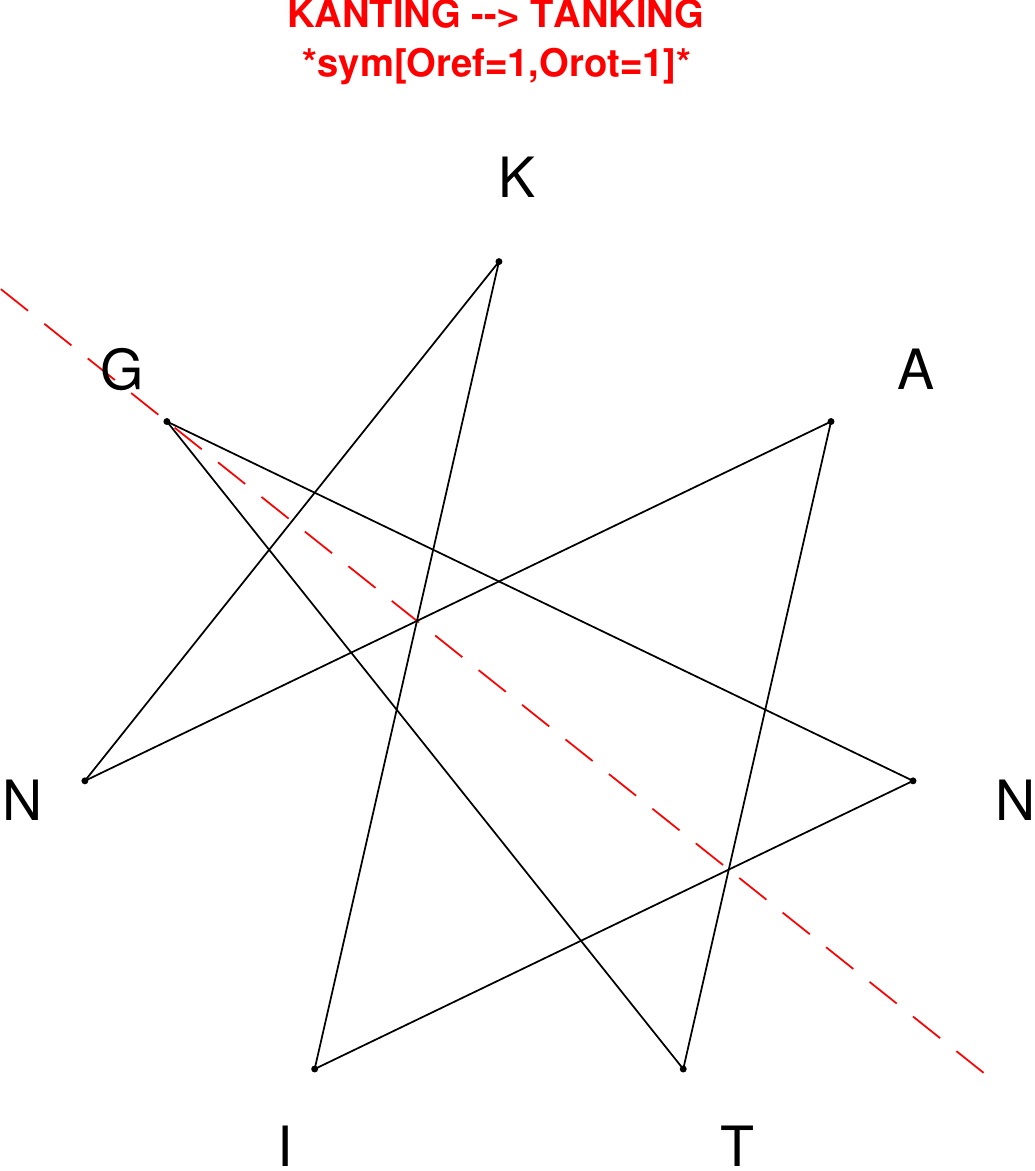}
\end{subfigure}
\hfill
\begin{subfigure}[T]{0.19\textwidth}
\centering
\includegraphics[width=\textwidth]{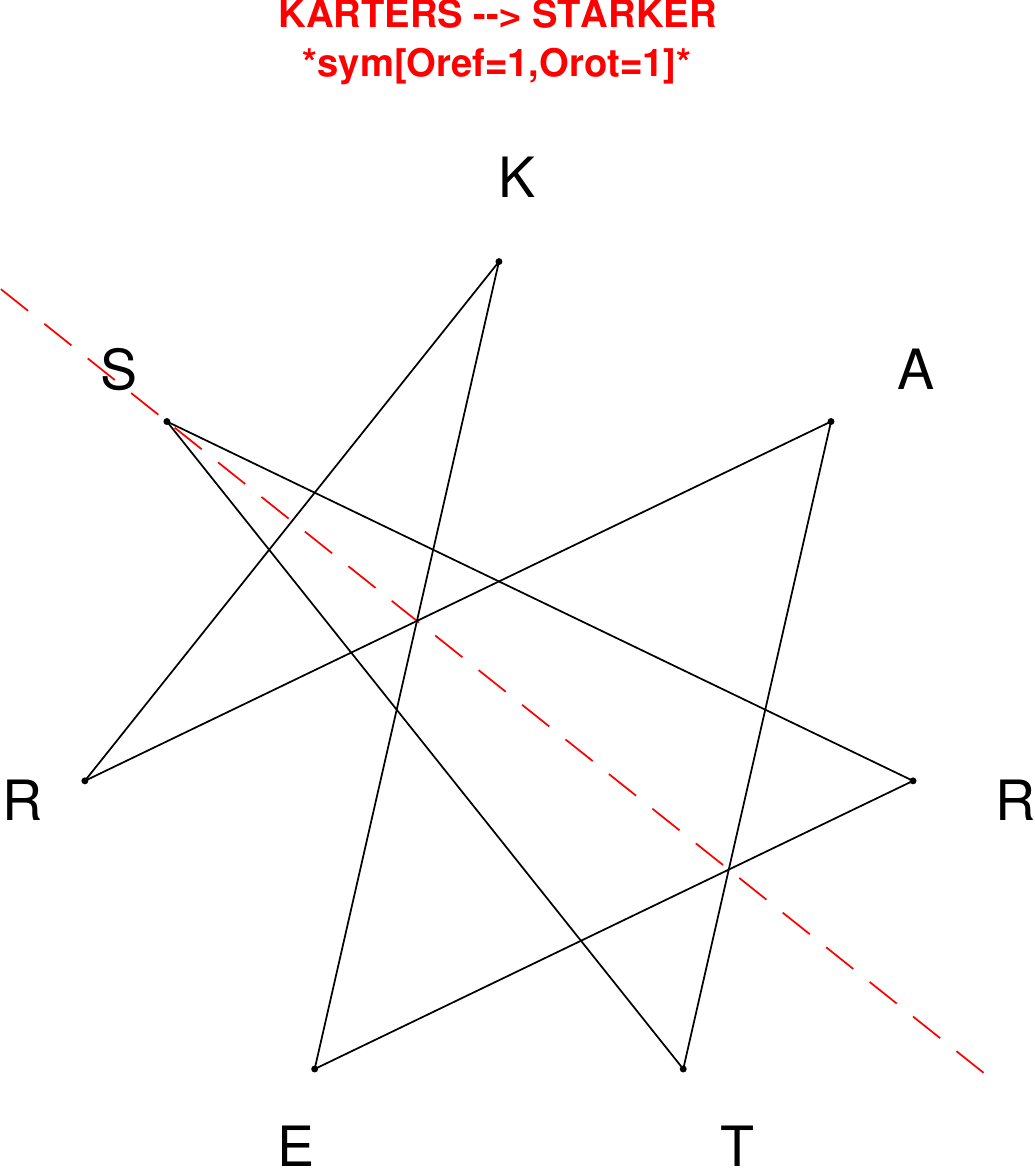}
\end{subfigure}
\hfill
\begin{subfigure}[T]{0.19\textwidth}
\centering
\includegraphics[width=\textwidth]{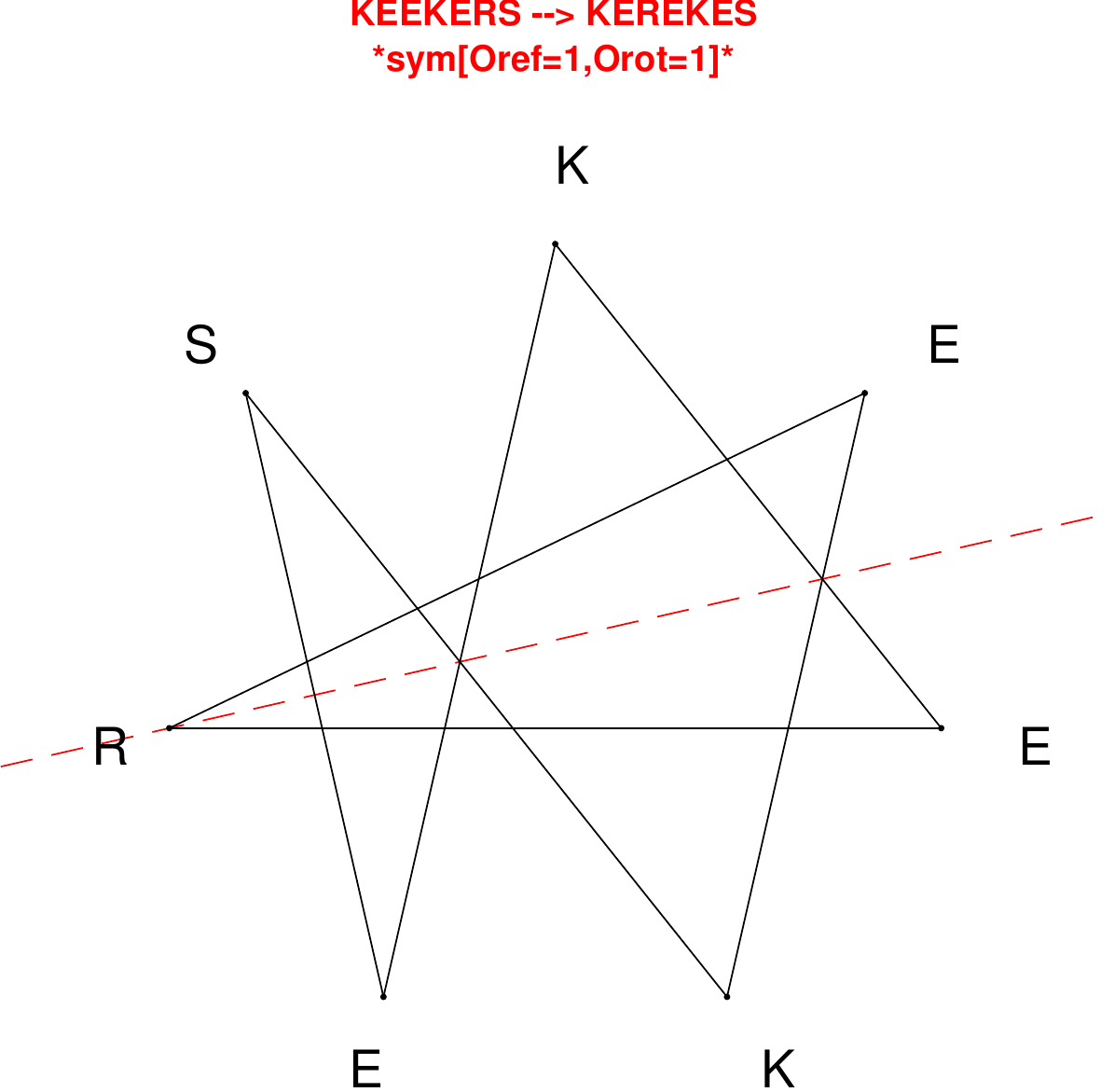}
\end{subfigure}
\hfill
\begin{subfigure}[T]{0.19\textwidth}
\centering
\includegraphics[width=\textwidth]{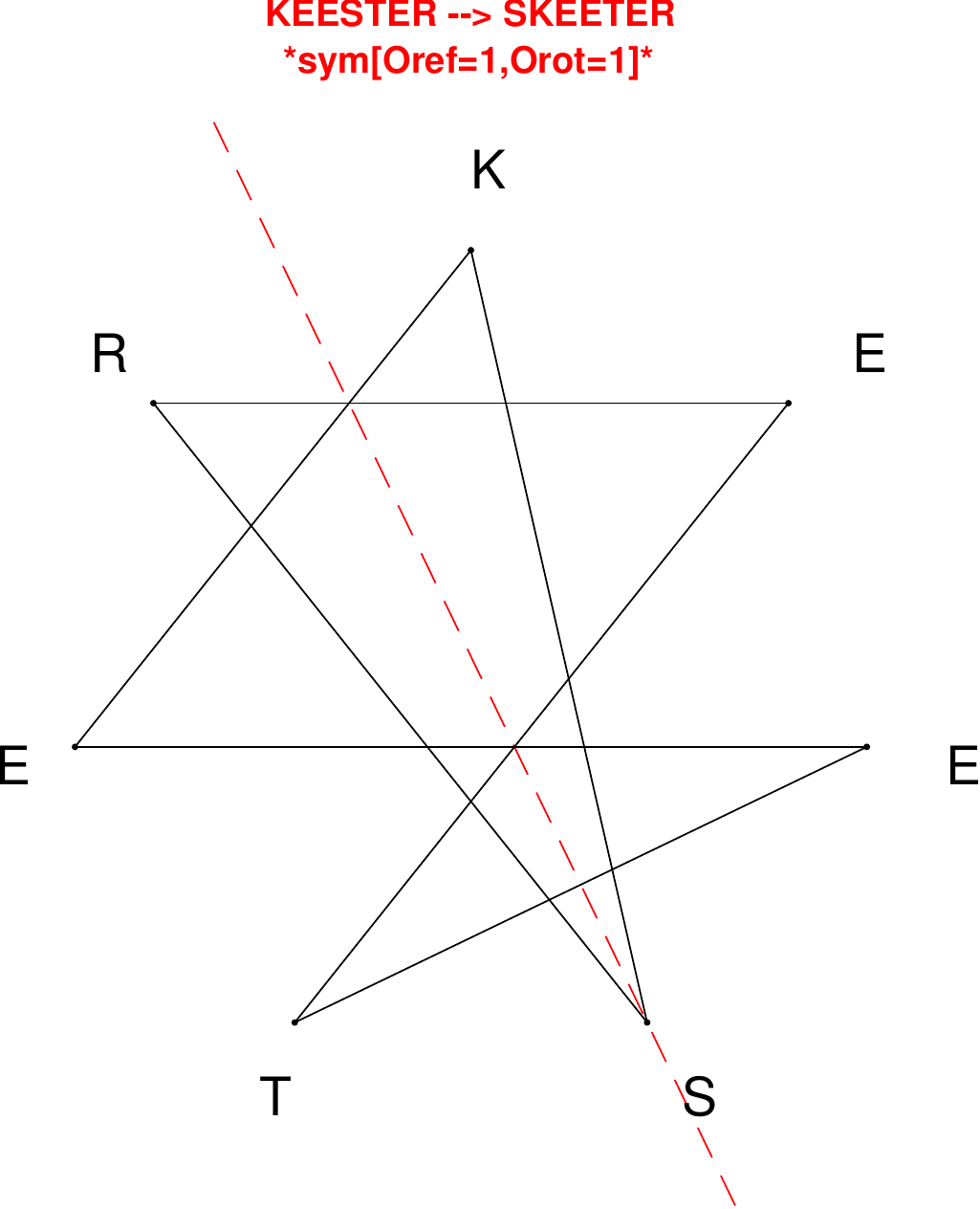}
\end{subfigure}
\hfill
\begin{subfigure}[T]{0.19\textwidth}
\centering
\includegraphics[width=\textwidth]{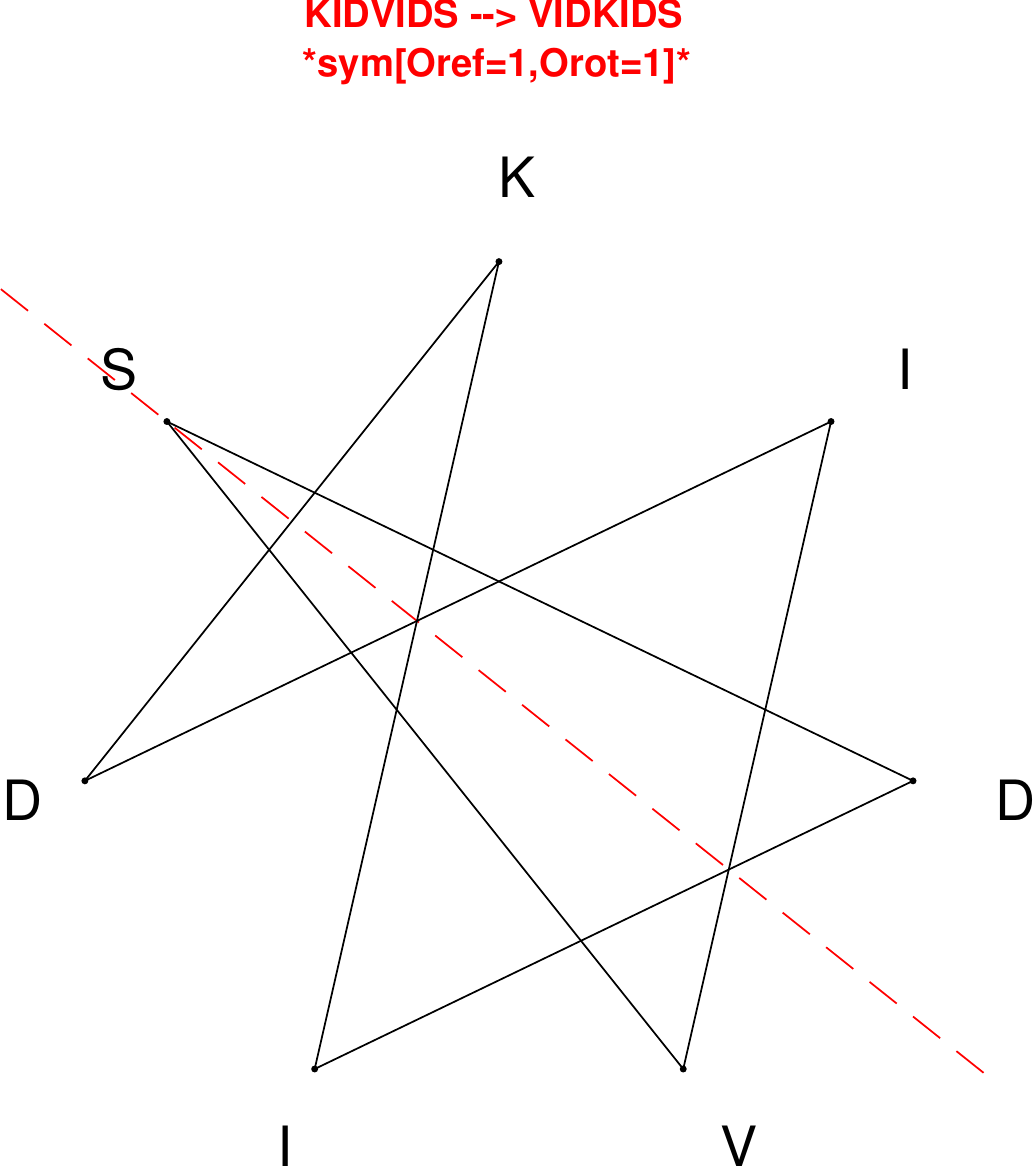}
\end{subfigure}
\end{figure}

\begin{figure}[H]
\centering
\begin{subfigure}[T]{0.19\textwidth}
\centering
\includegraphics[width=\textwidth]{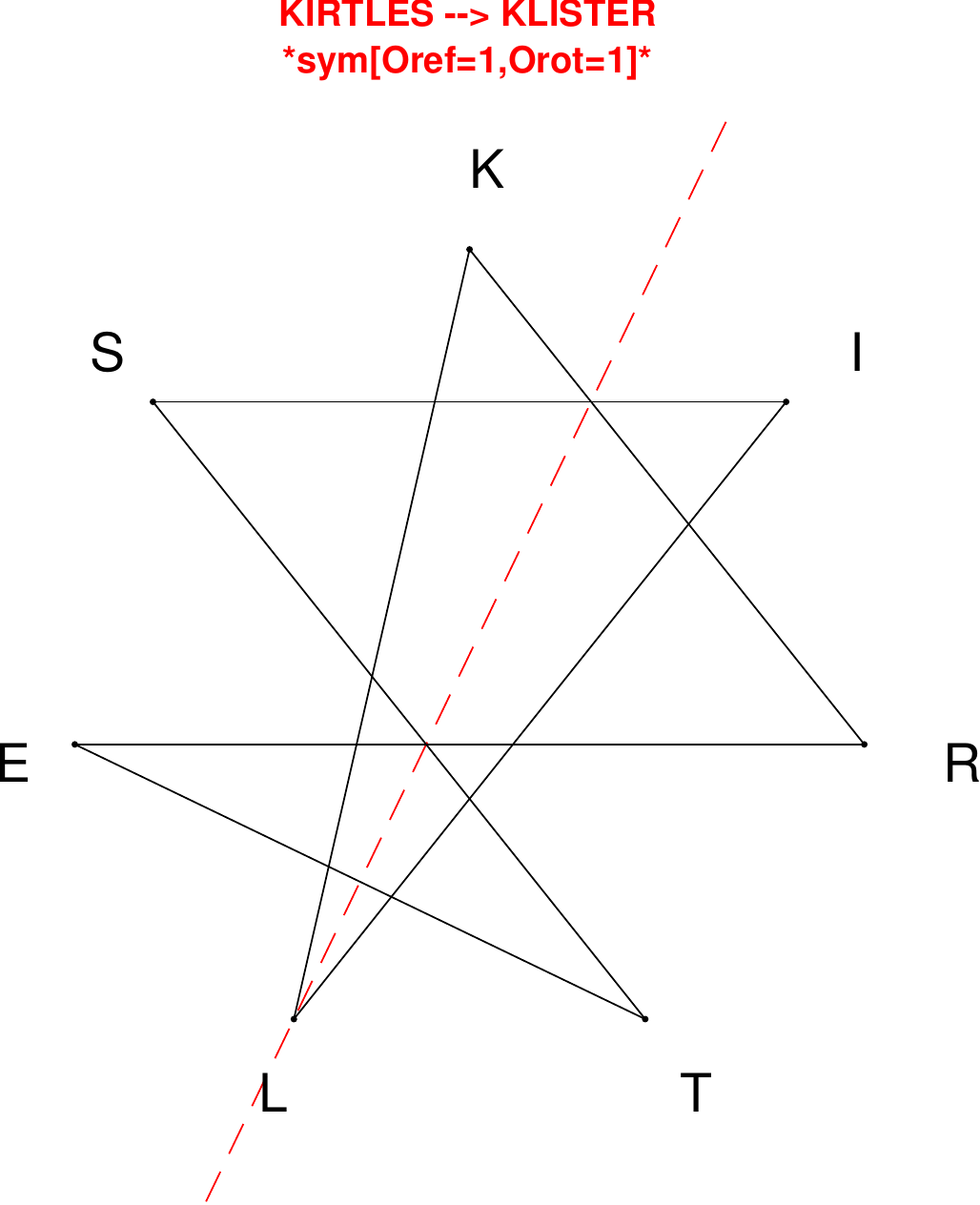}
\end{subfigure}
\hfill
\begin{subfigure}[T]{0.19\textwidth}
\centering
\includegraphics[width=\textwidth]{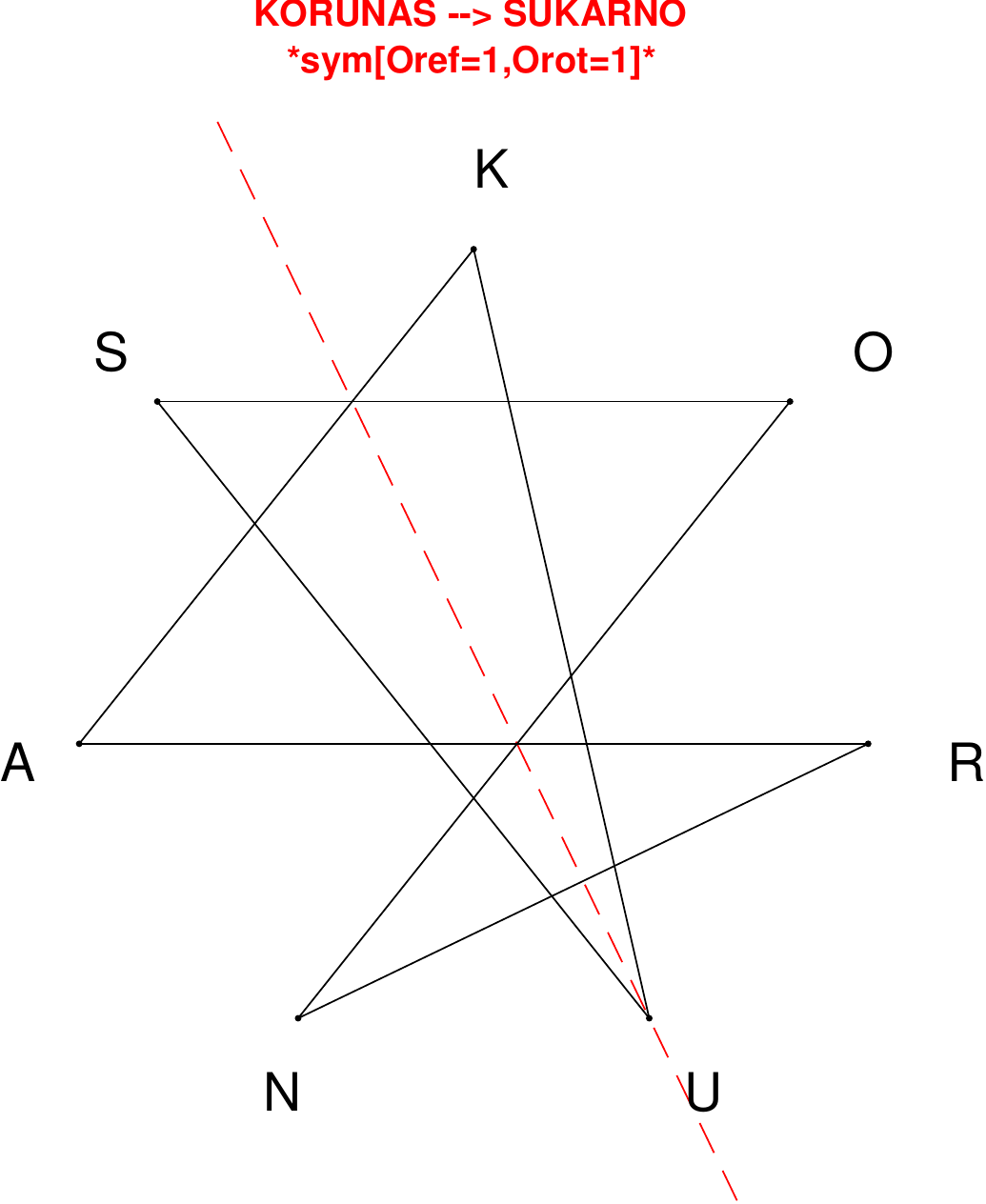}
\end{subfigure}
\hfill
\begin{subfigure}[T]{0.19\textwidth}
\centering
\includegraphics[width=\textwidth]{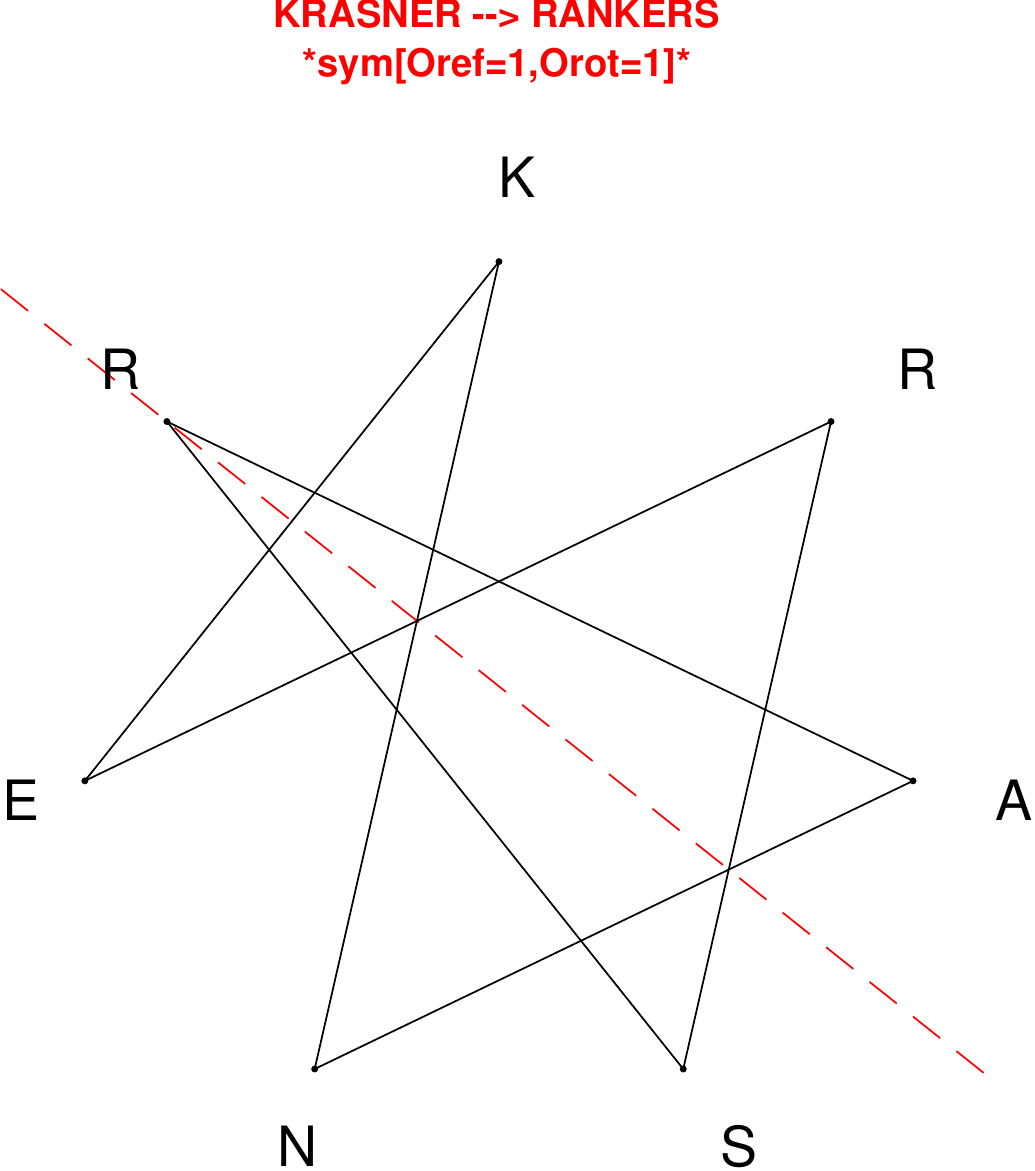}
\end{subfigure}
\hfill
\begin{subfigure}[T]{0.19\textwidth}
\centering
\includegraphics[width=\textwidth]{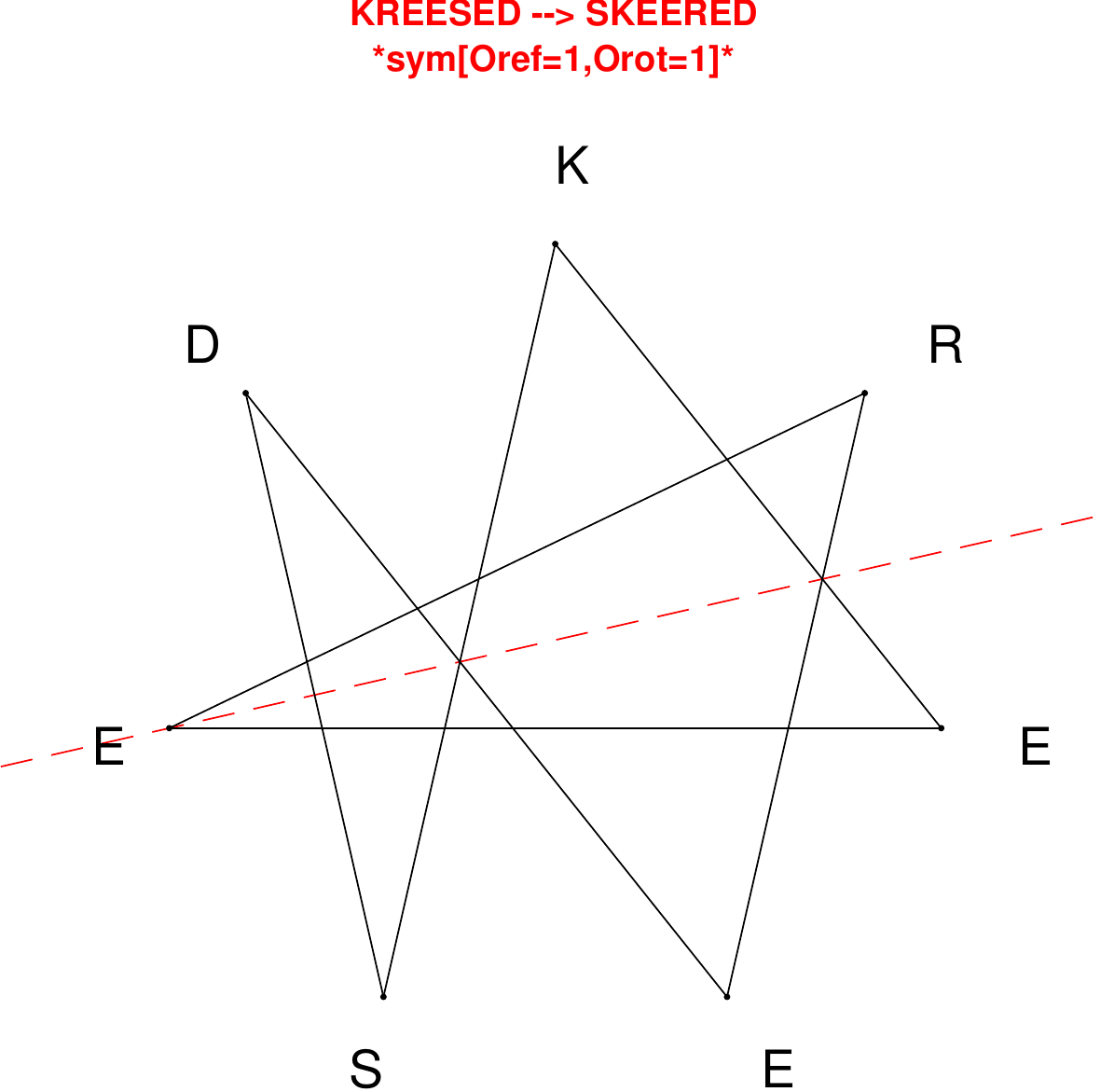}
\end{subfigure}
\hfill
\begin{subfigure}[T]{0.19\textwidth}
\centering
\includegraphics[width=\textwidth]{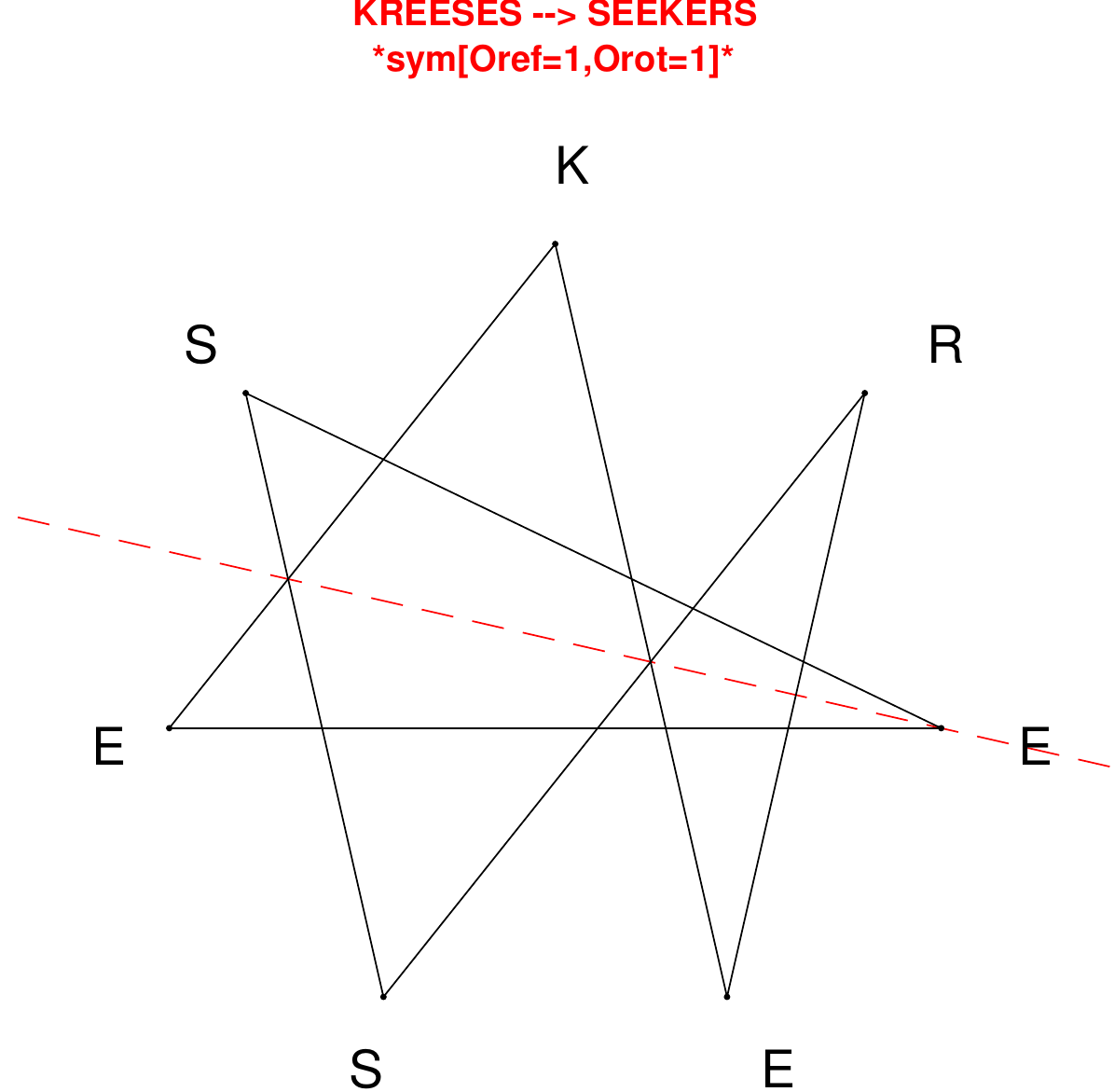}
\end{subfigure}
\end{figure}

\begin{figure}[H]
\centering
\begin{subfigure}[T]{0.19\textwidth}
\centering
\includegraphics[width=\textwidth]{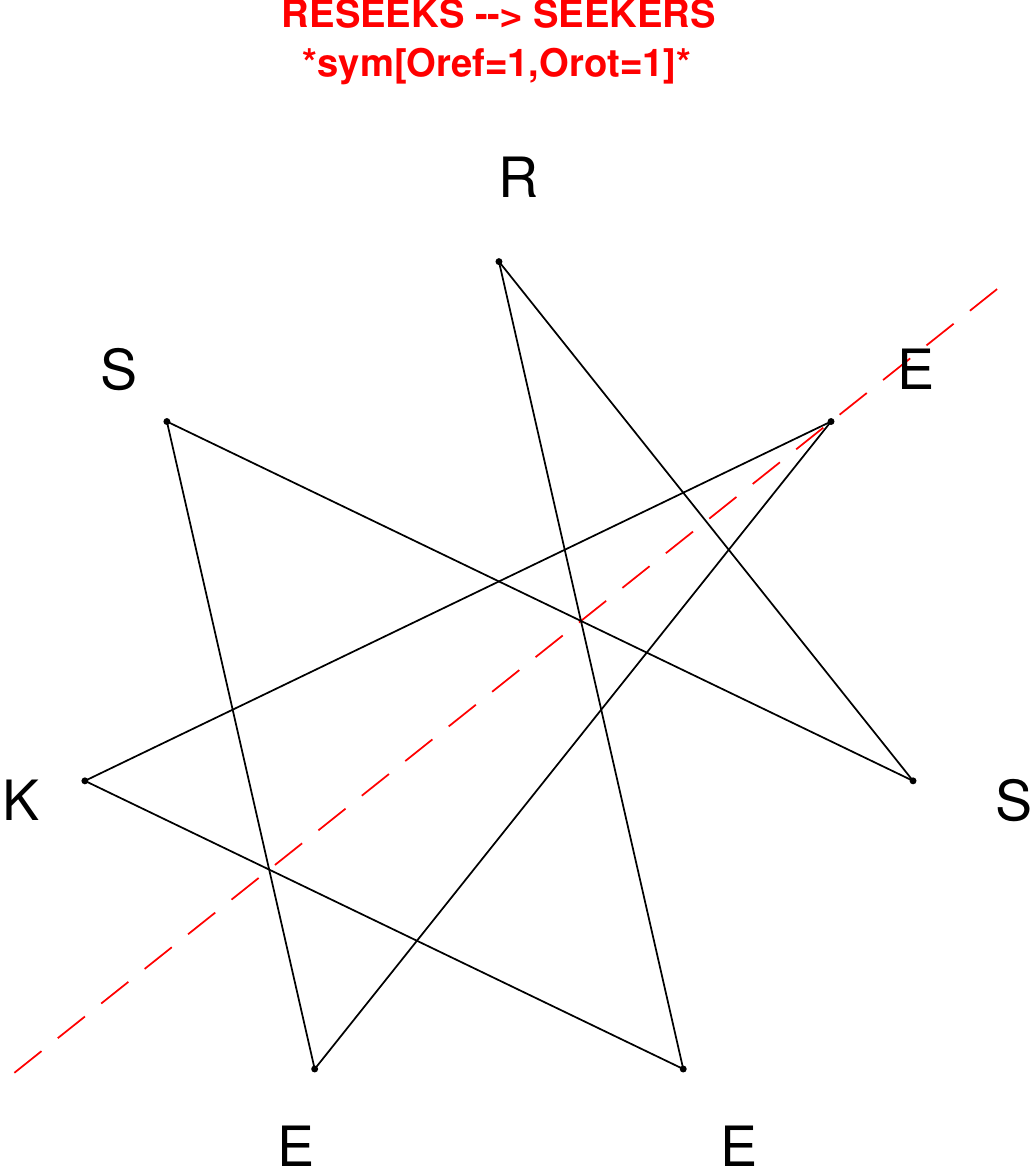}
\end{subfigure}
\hfill
\begin{subfigure}[T]{0.19\textwidth}
\centering
\includegraphics[width=\textwidth]{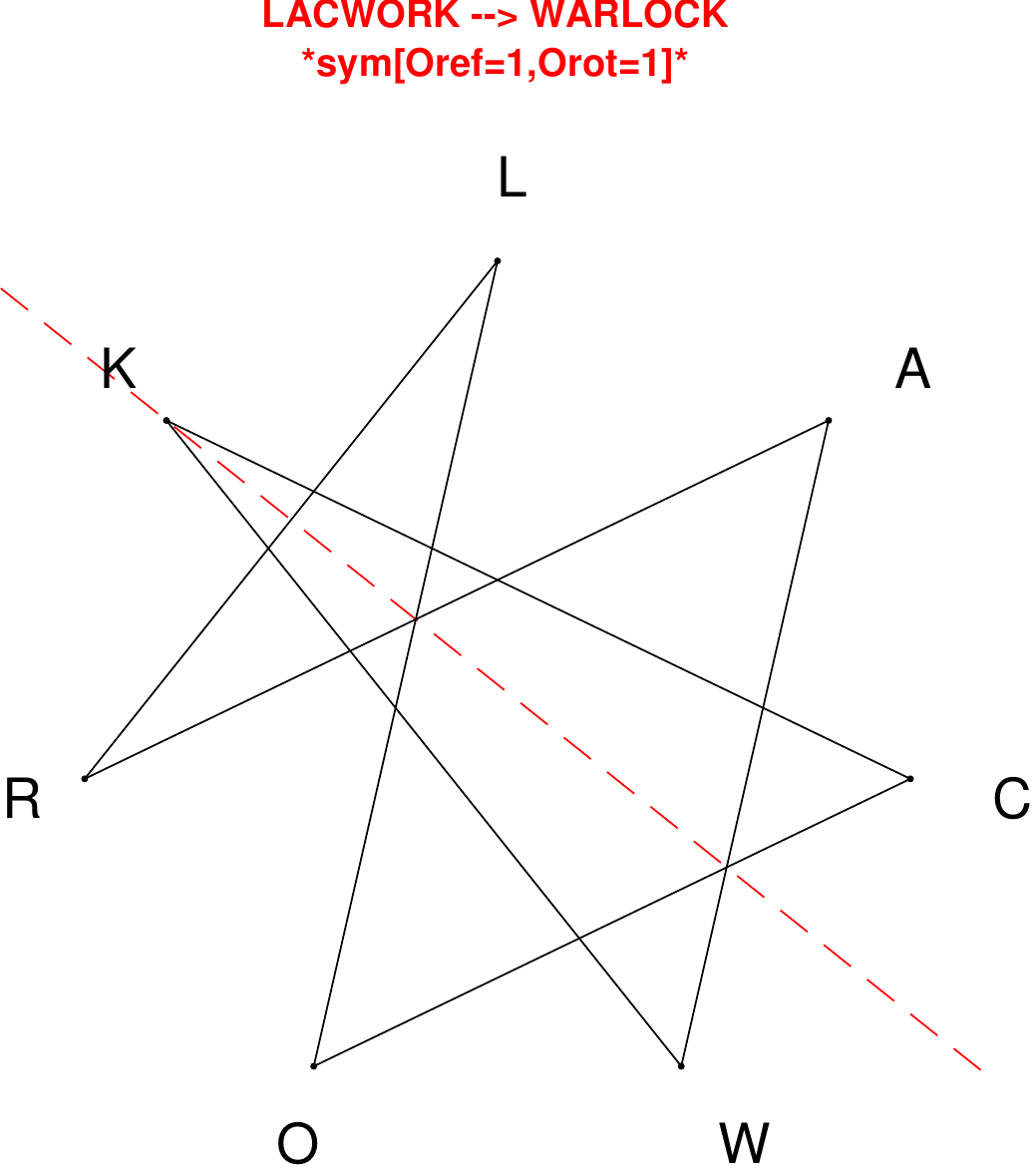}
\end{subfigure}
\hfill
\begin{subfigure}[T]{0.19\textwidth}
\centering
\includegraphics[width=\textwidth]{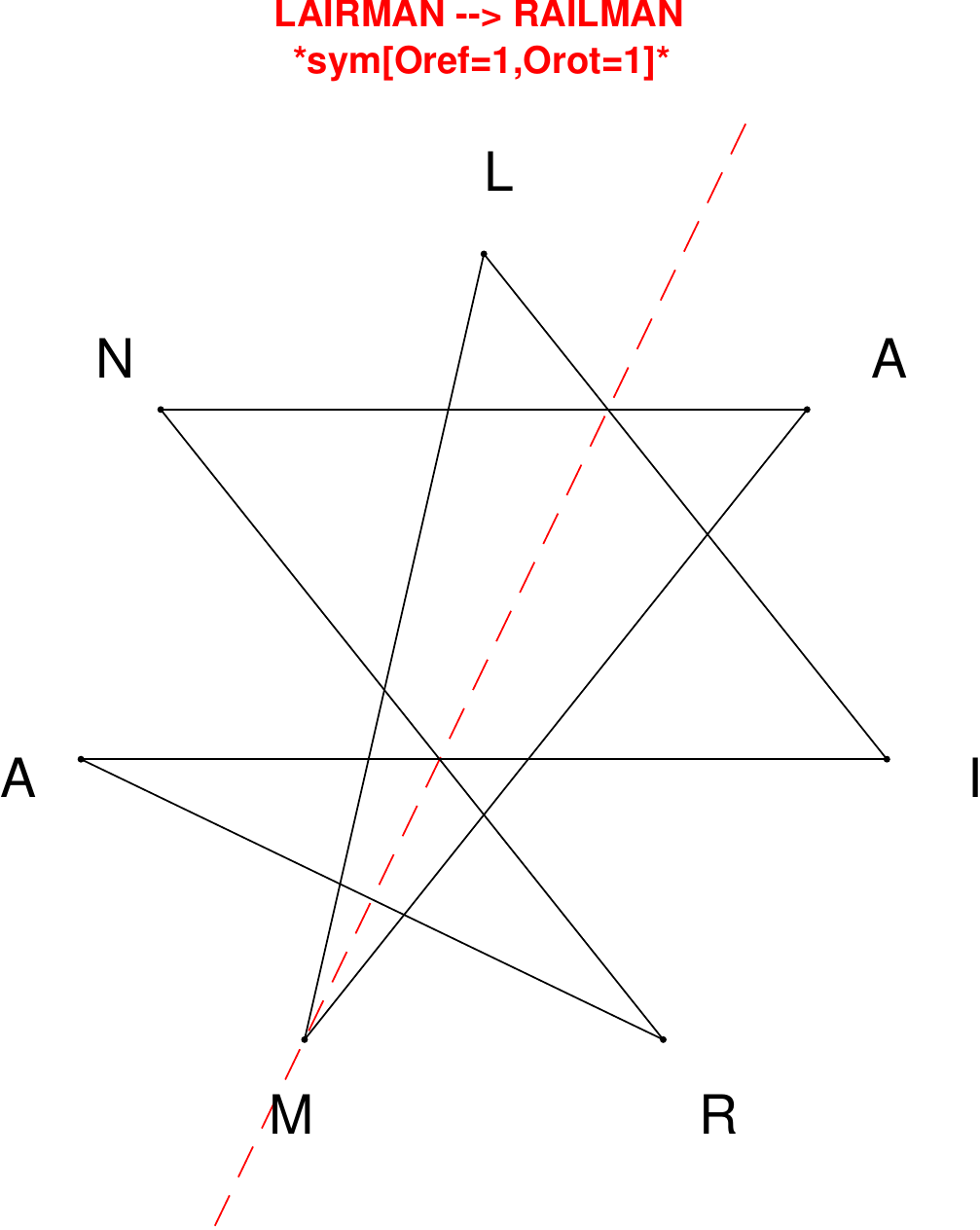}
\end{subfigure}
\hfill
\begin{subfigure}[T]{0.19\textwidth}
\centering
\includegraphics[width=\textwidth]{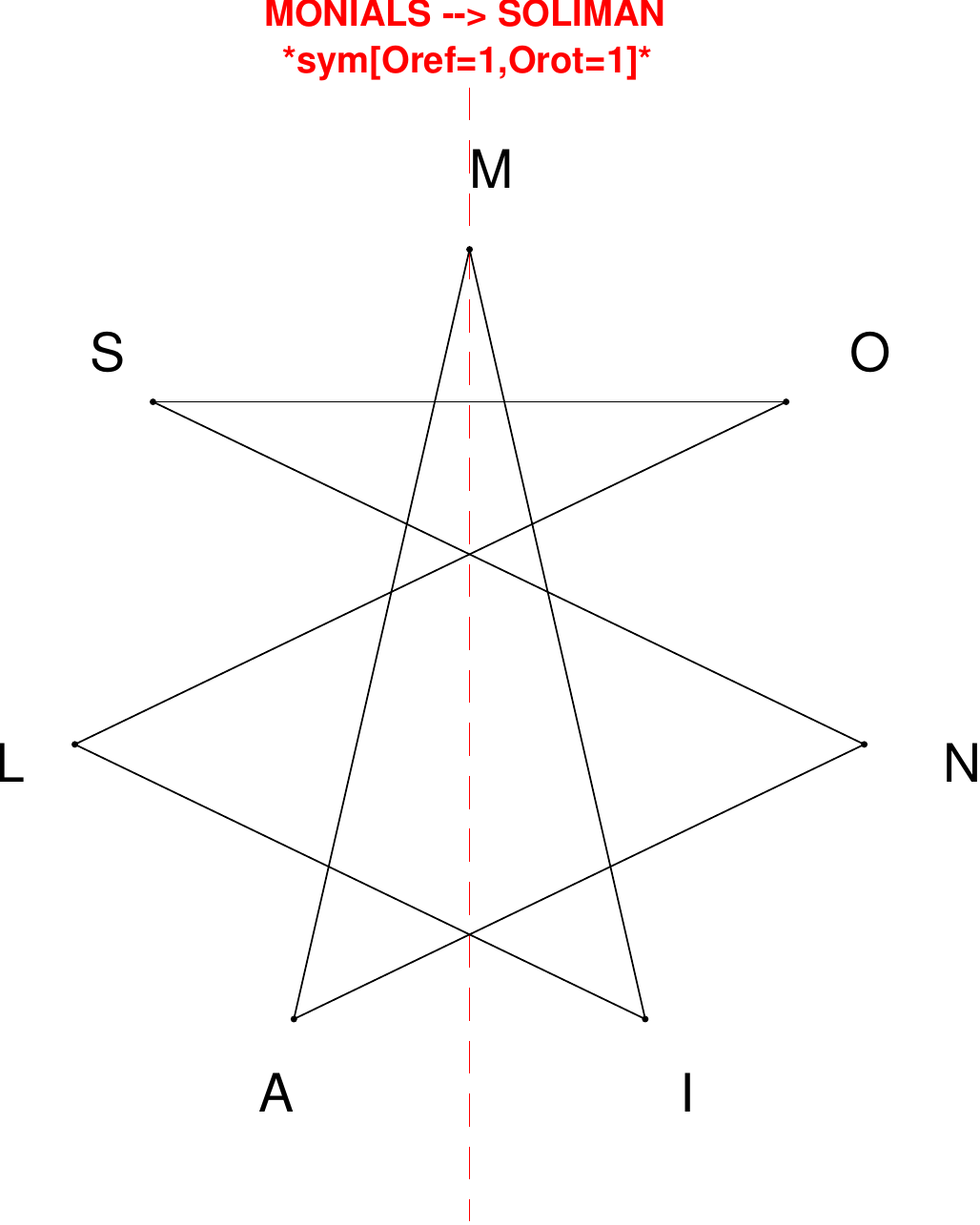}
\end{subfigure}
\hfill
\begin{subfigure}[T]{0.19\textwidth}
\centering
\includegraphics[width=\textwidth]{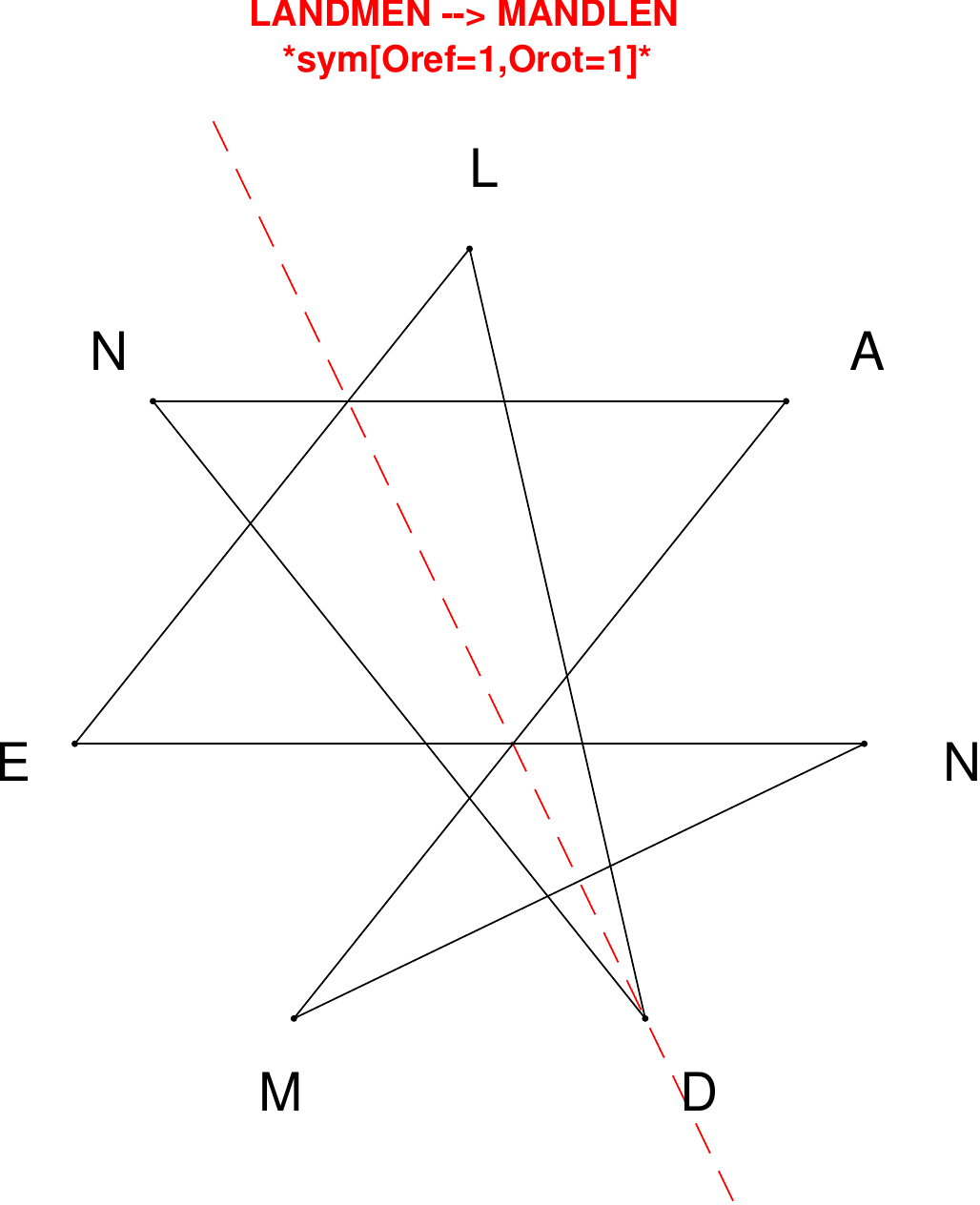}
\end{subfigure}
\end{figure}

\begin{figure}[H]
\centering
\begin{subfigure}[T]{0.19\textwidth}
\centering
\includegraphics[width=\textwidth]{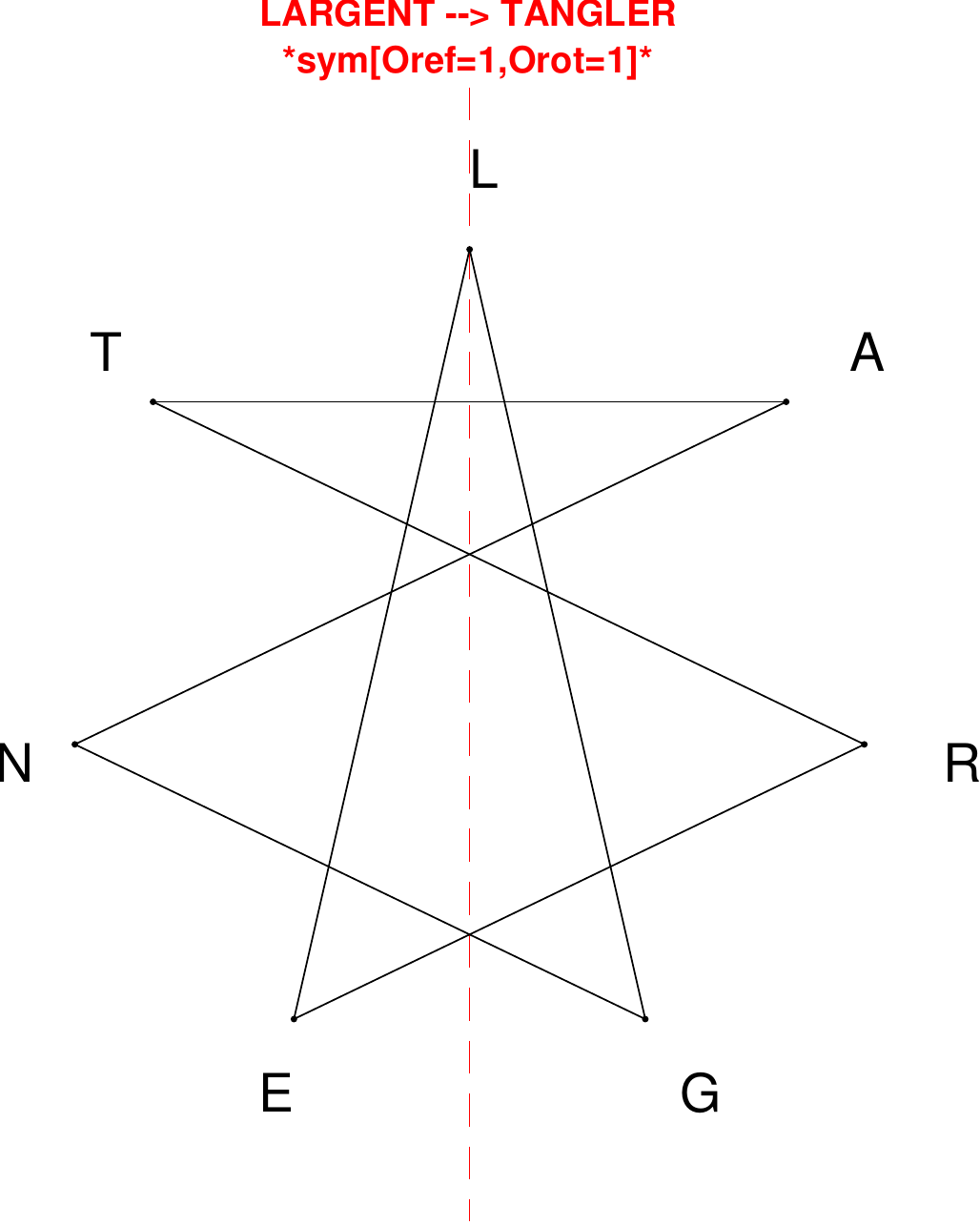}
\end{subfigure}
\hfill
\begin{subfigure}[T]{0.19\textwidth}
\centering
\includegraphics[width=\textwidth]{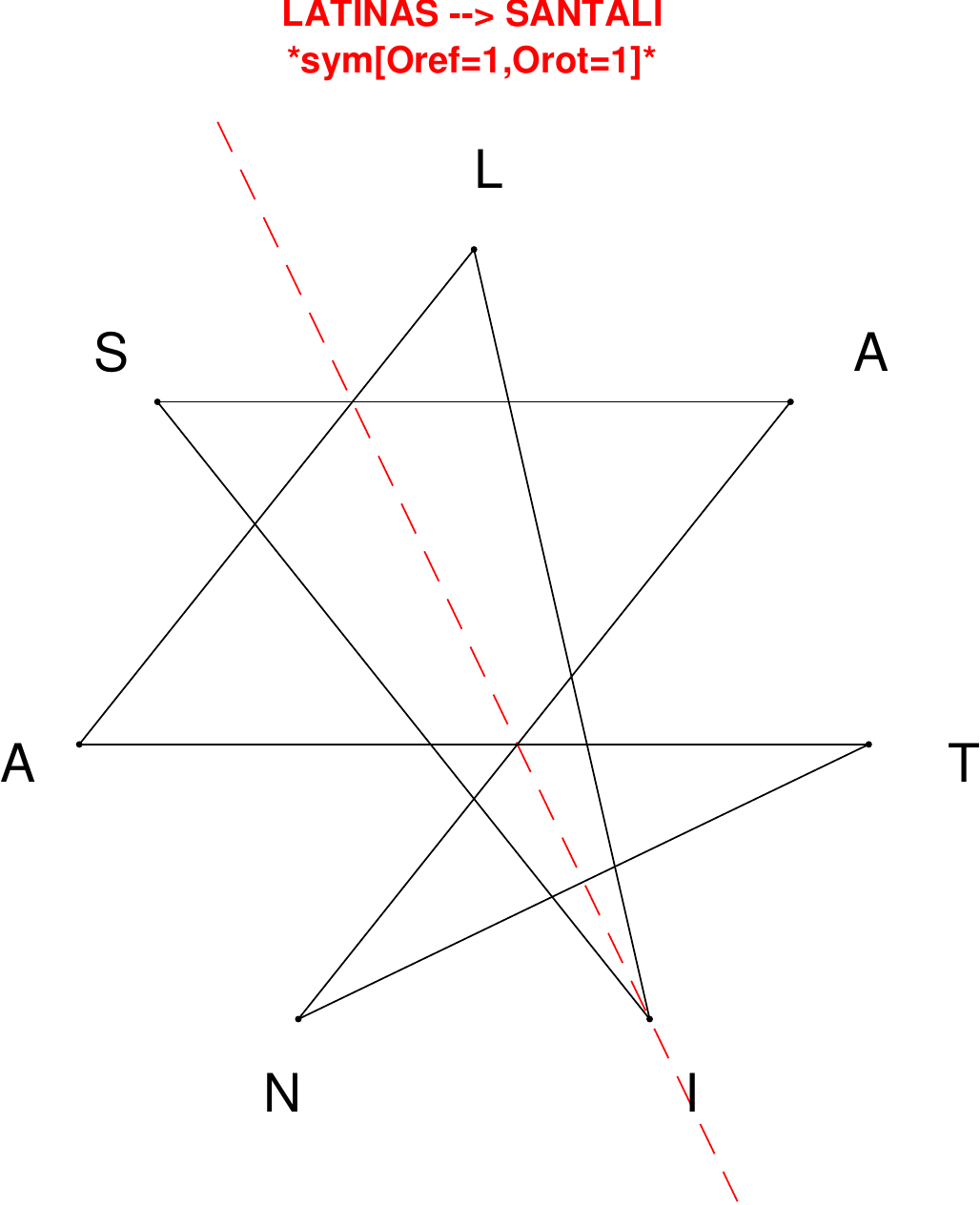}
\end{subfigure}
\hfill
\begin{subfigure}[T]{0.19\textwidth}
\centering
\includegraphics[width=\textwidth]{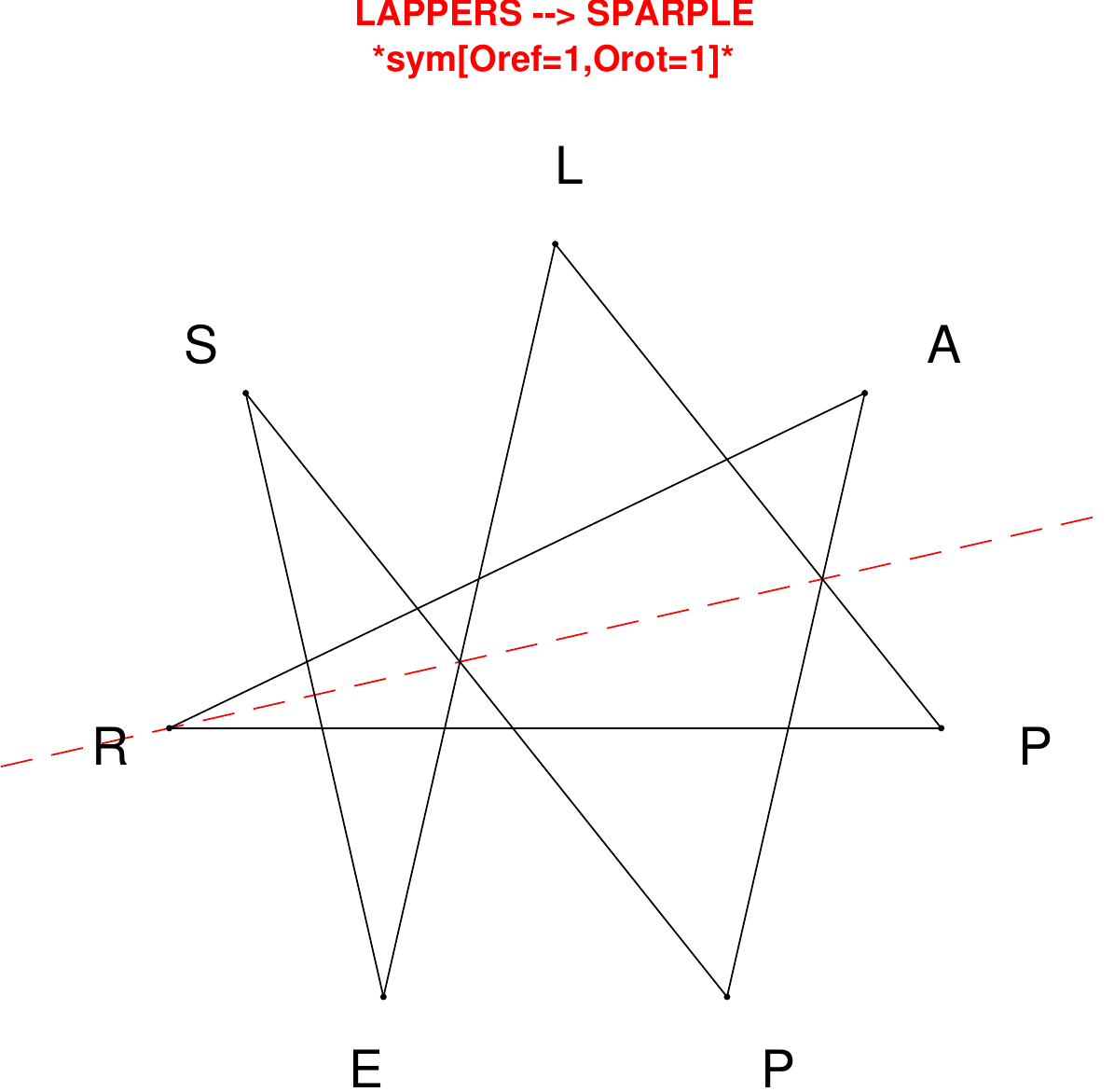}
\end{subfigure}
\hfill
\begin{subfigure}[T]{0.19\textwidth}
\centering
\includegraphics[width=\textwidth]{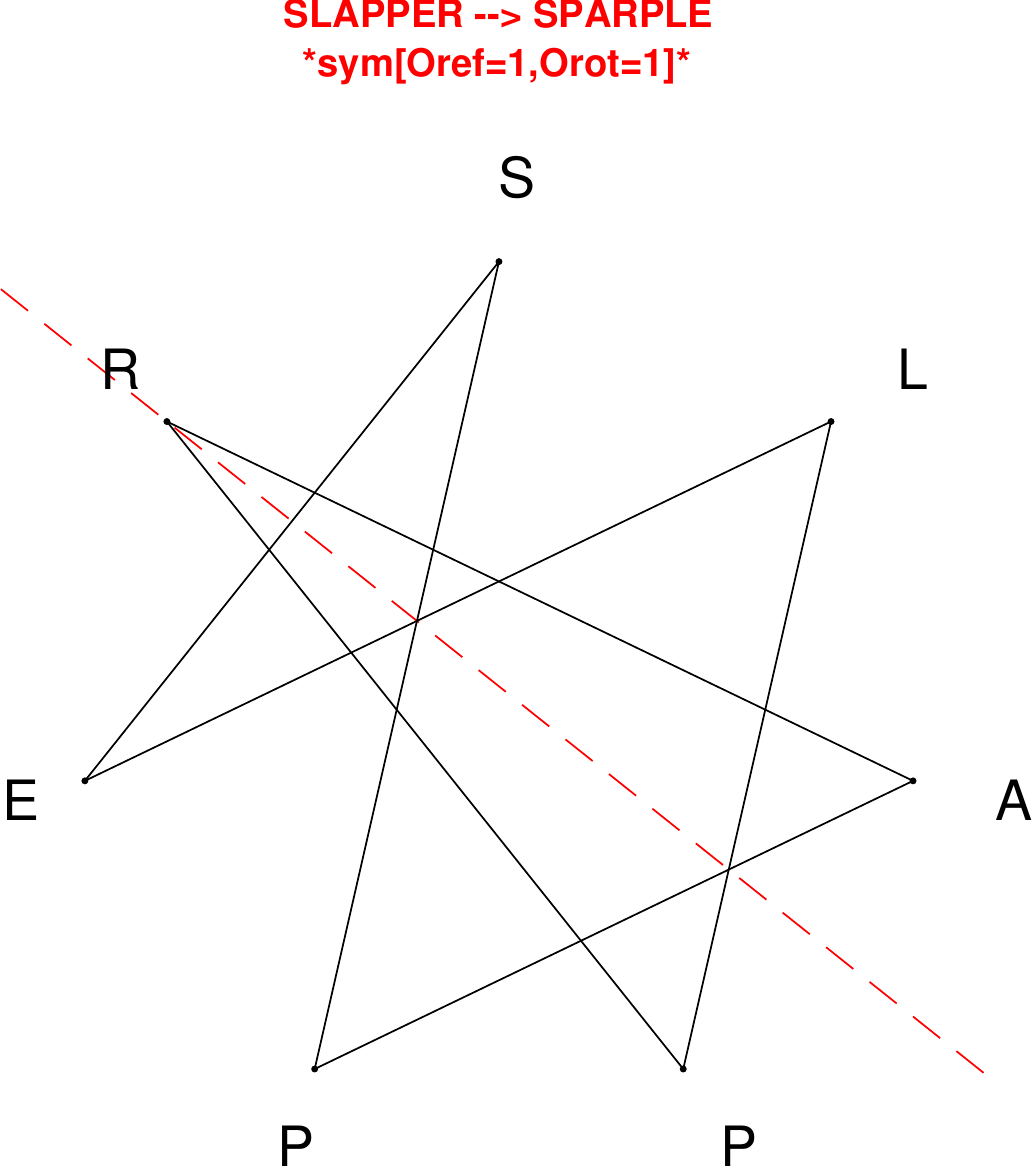}
\end{subfigure}
\hfill
\begin{subfigure}[T]{0.19\textwidth}
\centering
\includegraphics[width=\textwidth]{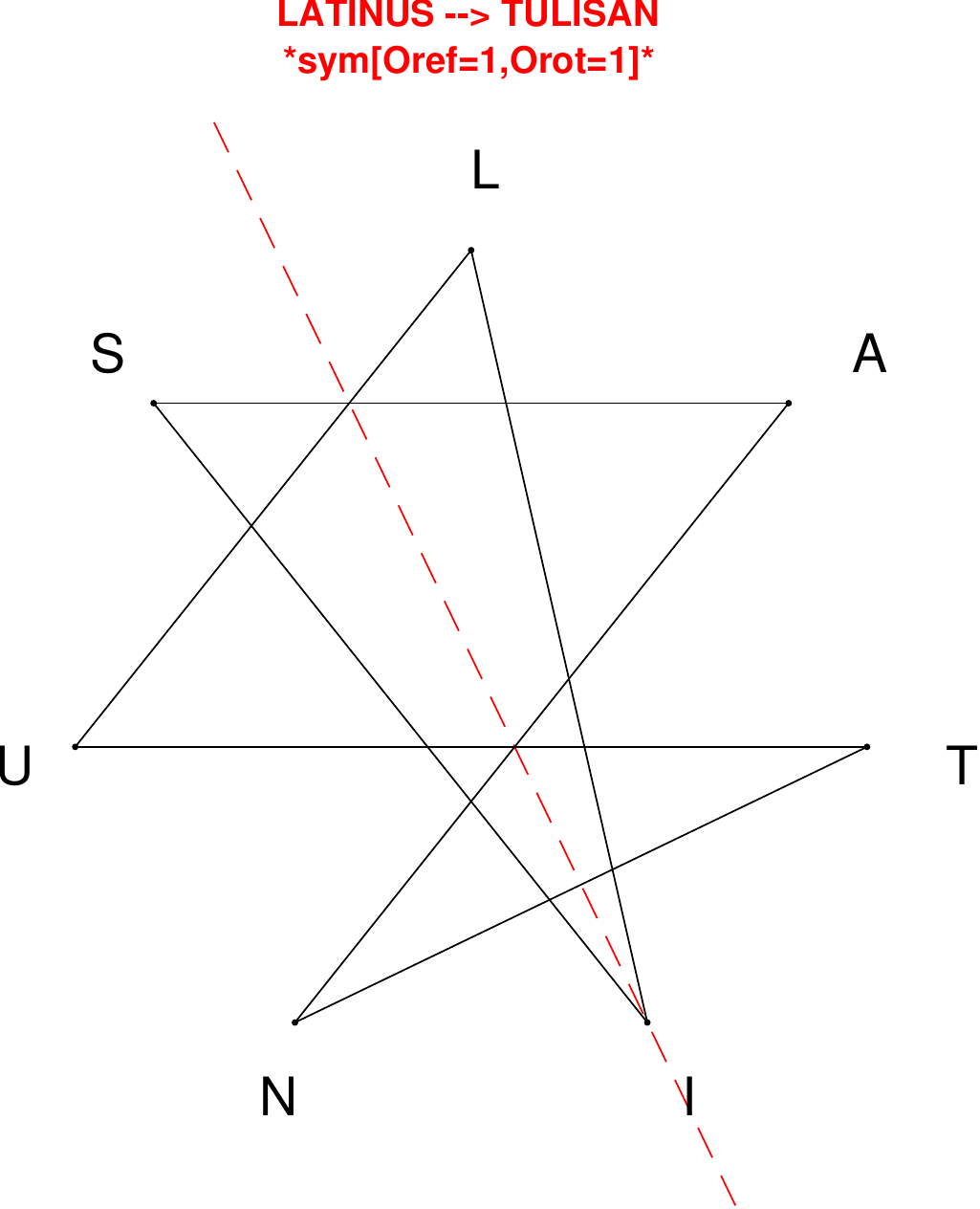}
\end{subfigure}
\end{figure}

\begin{figure}[H]
\centering
\begin{subfigure}[T]{0.19\textwidth}
\centering
\includegraphics[width=\textwidth]{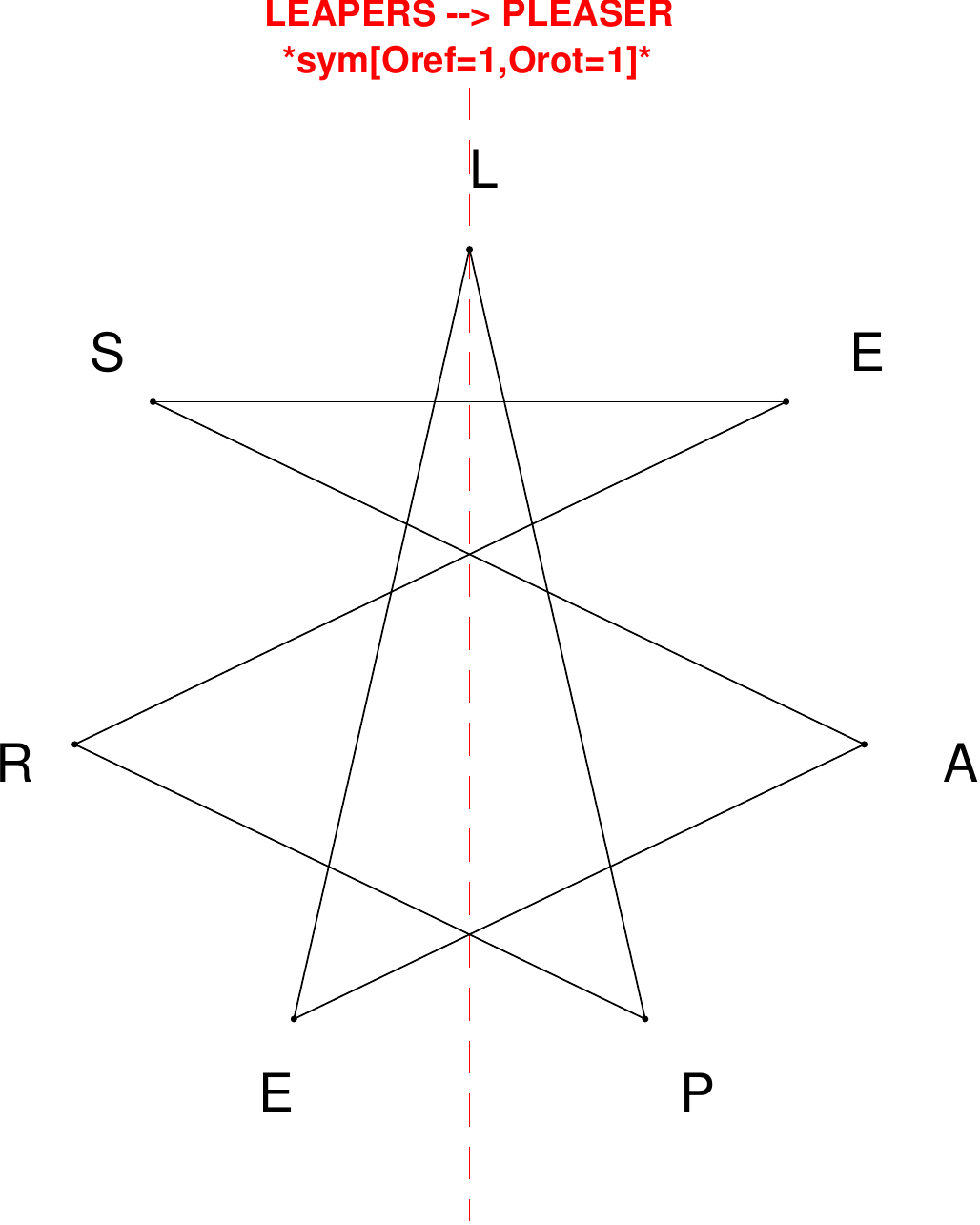}
\end{subfigure}
\hfill
\begin{subfigure}[T]{0.19\textwidth}
\centering
\includegraphics[width=\textwidth]{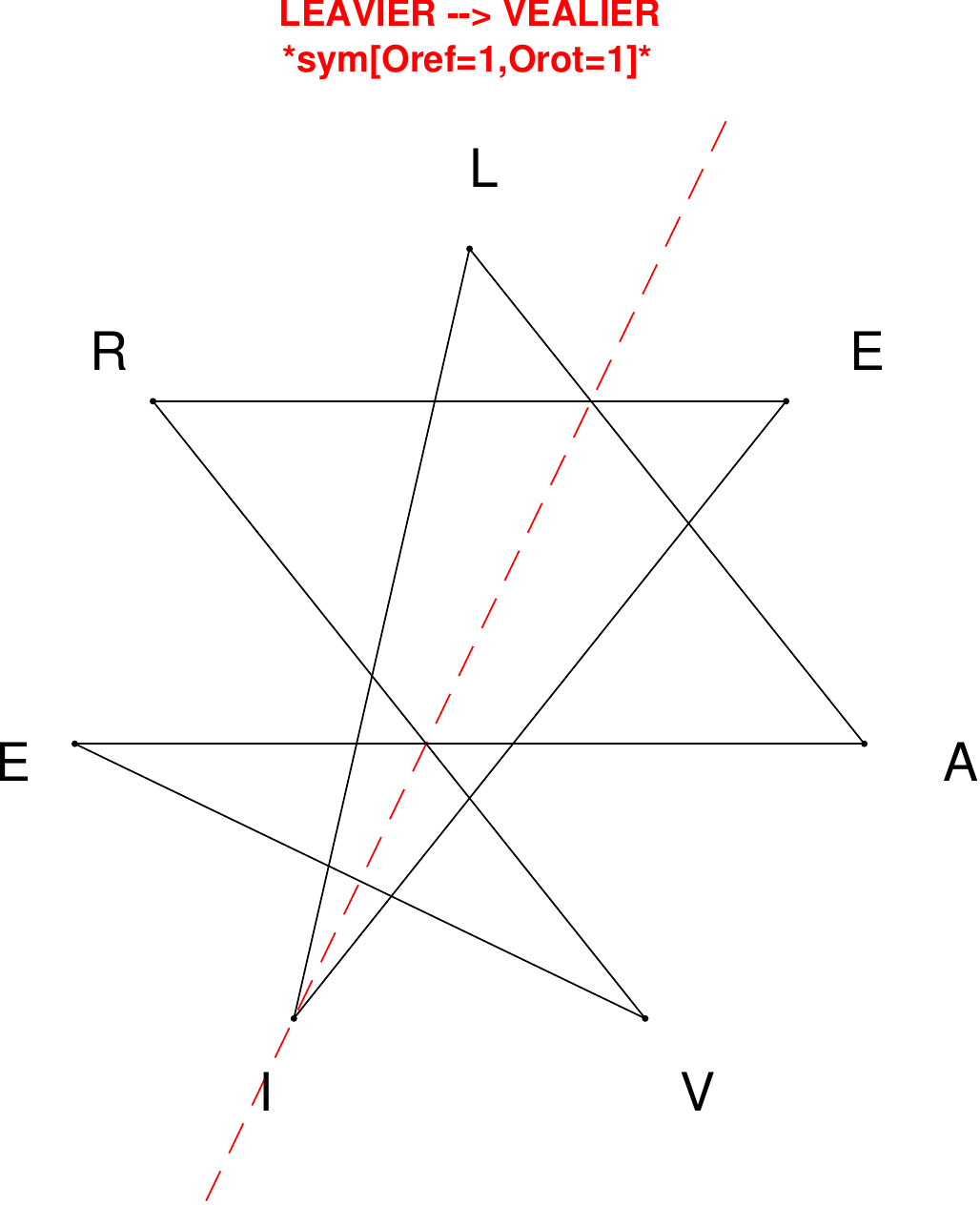}
\end{subfigure}
\hfill
\begin{subfigure}[T]{0.19\textwidth}
\centering
\includegraphics[width=\textwidth]{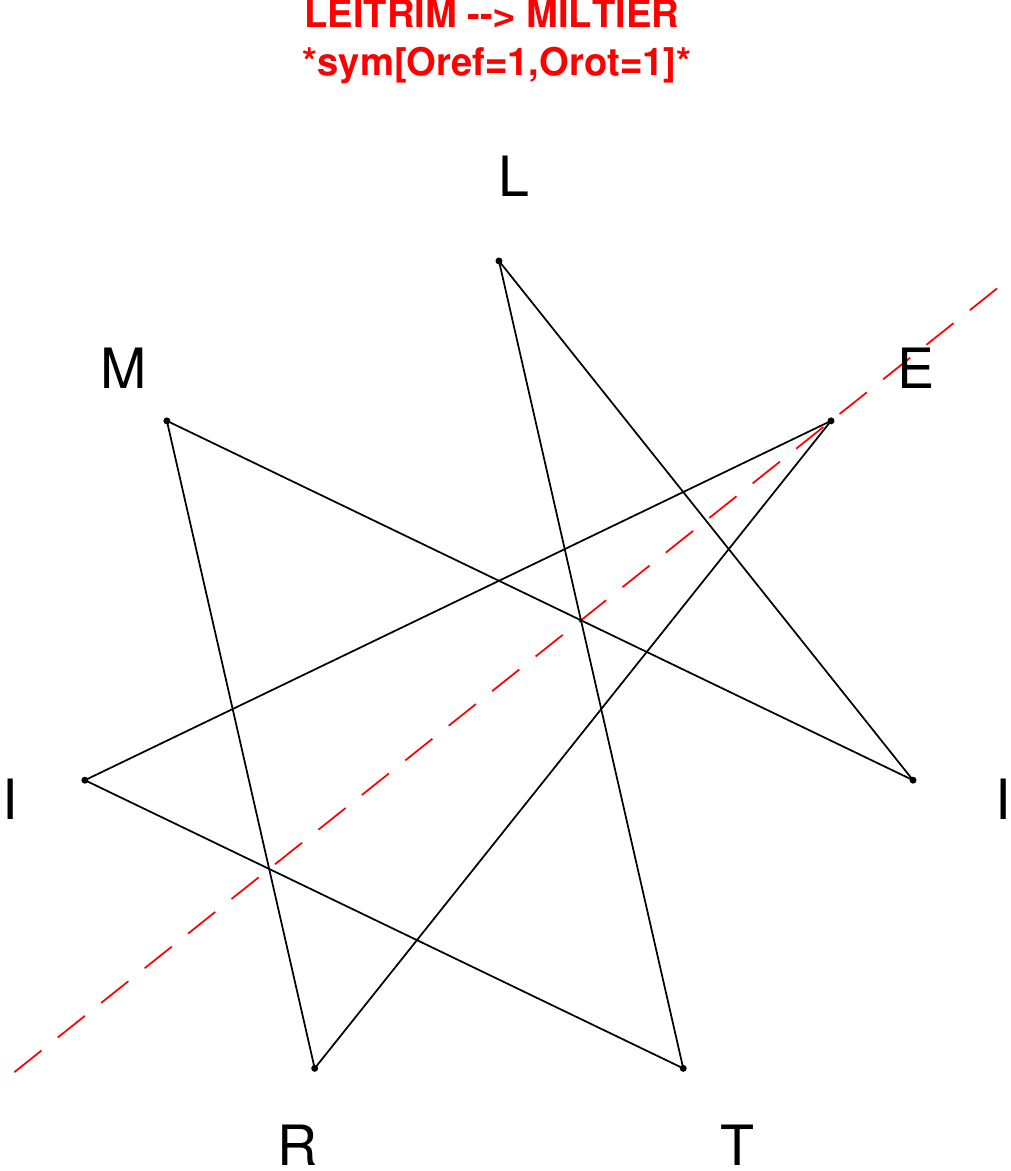}
\end{subfigure}
\hfill
\begin{subfigure}[T]{0.19\textwidth}
\centering
\includegraphics[width=\textwidth]{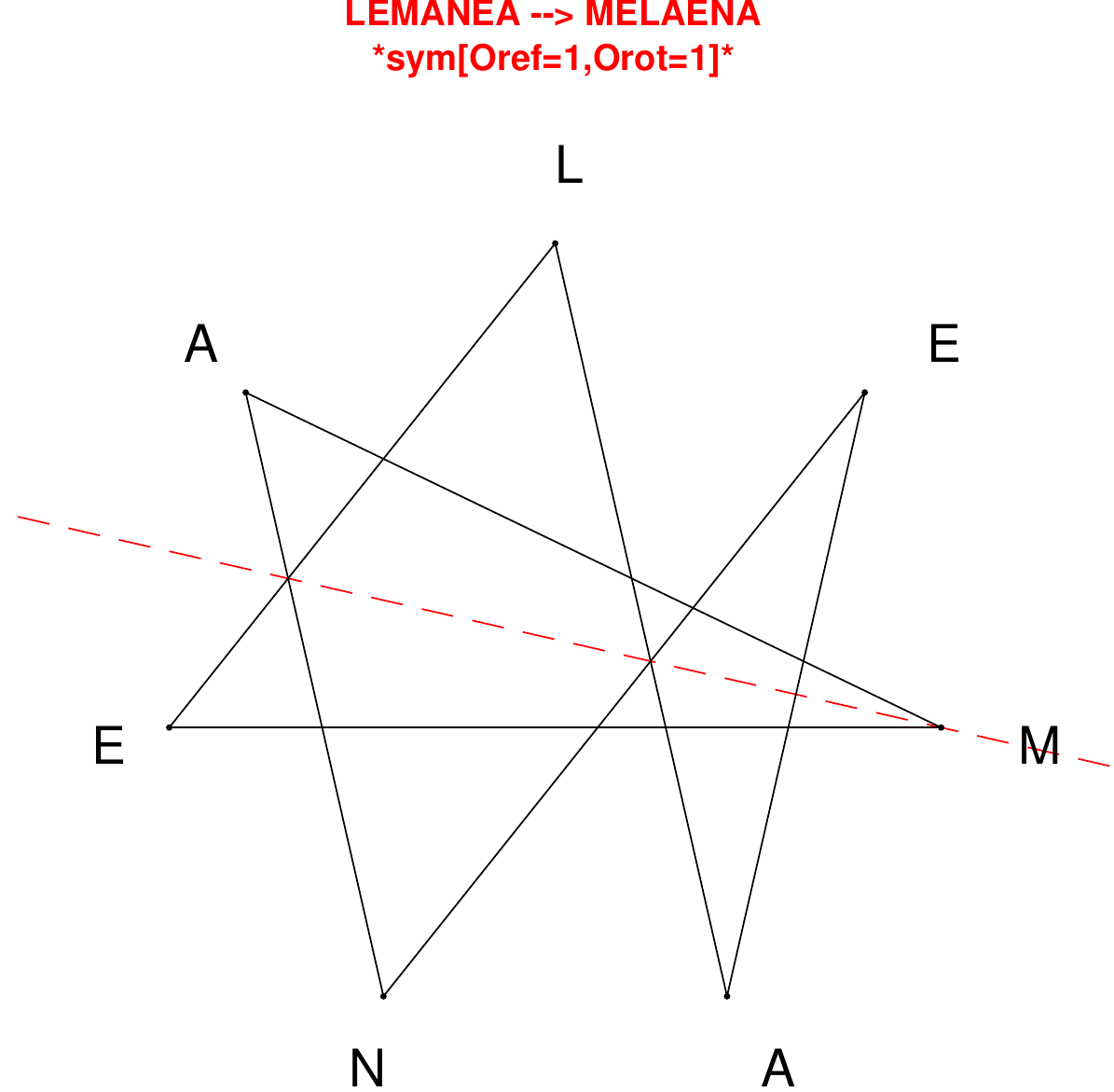}
\end{subfigure}
\hfill
\begin{subfigure}[T]{0.19\textwidth}
\centering
\includegraphics[width=\textwidth]{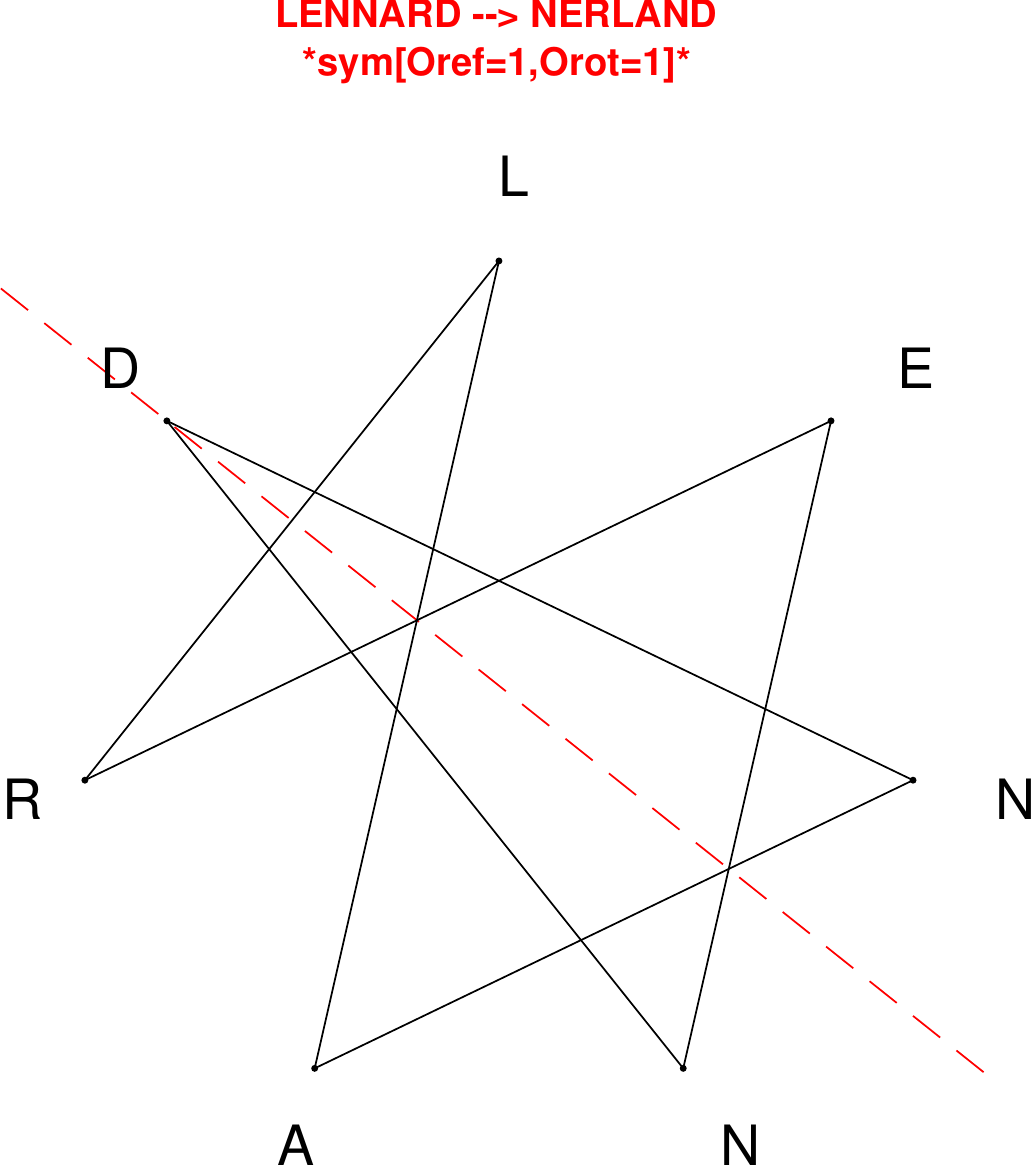}
\end{subfigure}
\end{figure}

\begin{figure}[H]
\centering
\begin{subfigure}[T]{0.19\textwidth}
\centering
\includegraphics[width=\textwidth]{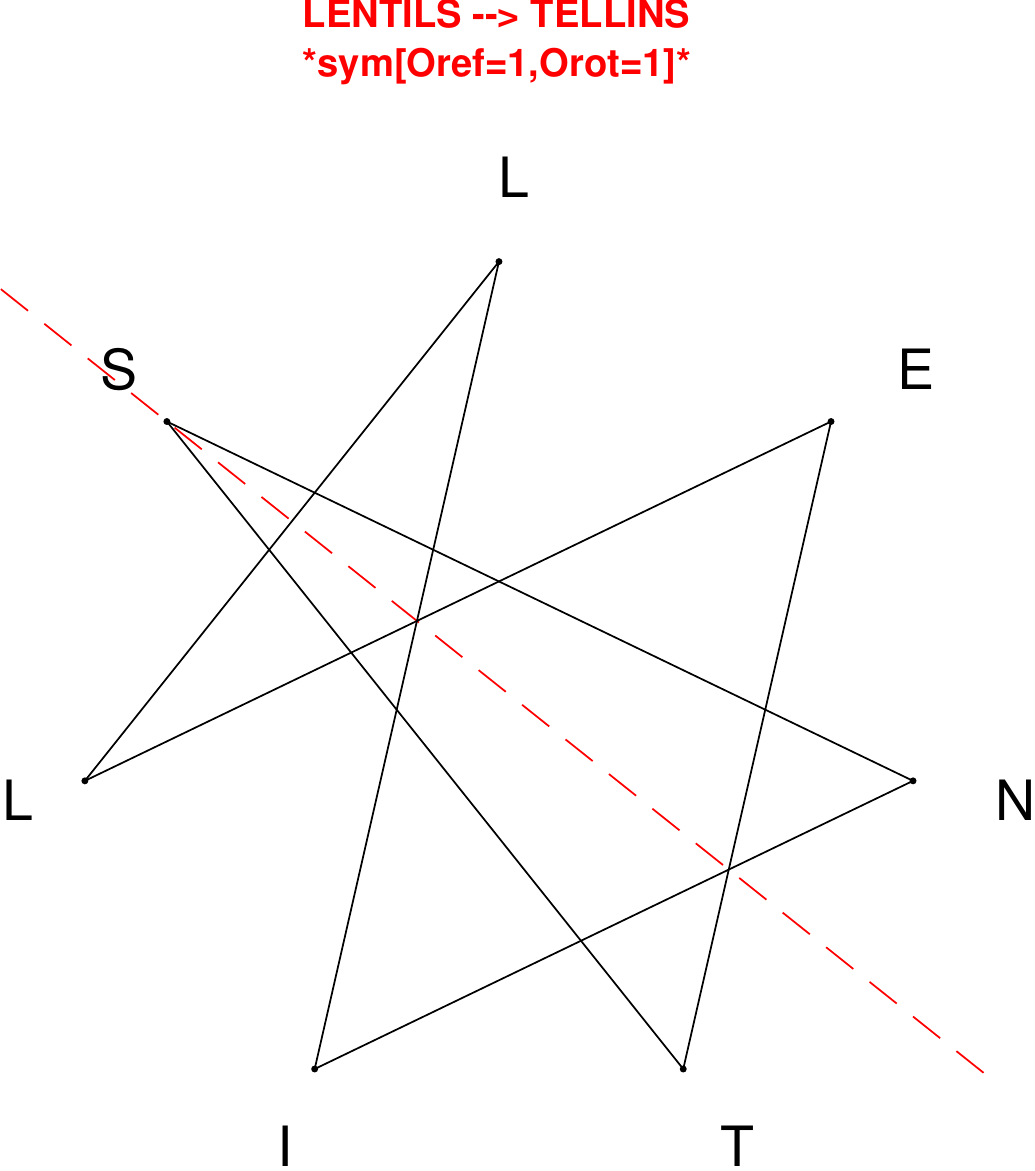}
\end{subfigure}
\hfill
\begin{subfigure}[T]{0.19\textwidth}
\centering
\includegraphics[width=\textwidth]{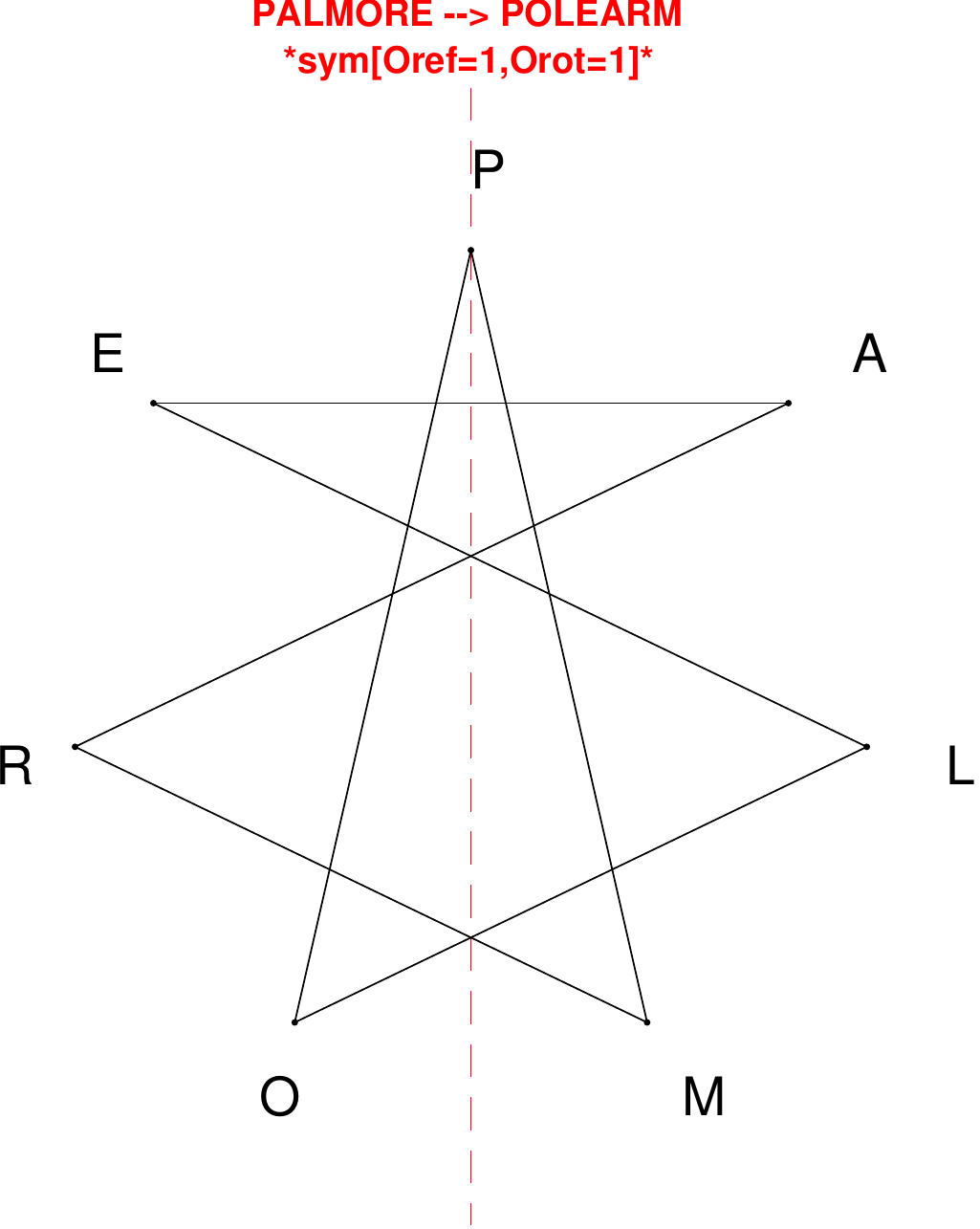}
\end{subfigure}
\hfill
\begin{subfigure}[T]{0.19\textwidth}
\centering
\includegraphics[width=\textwidth]{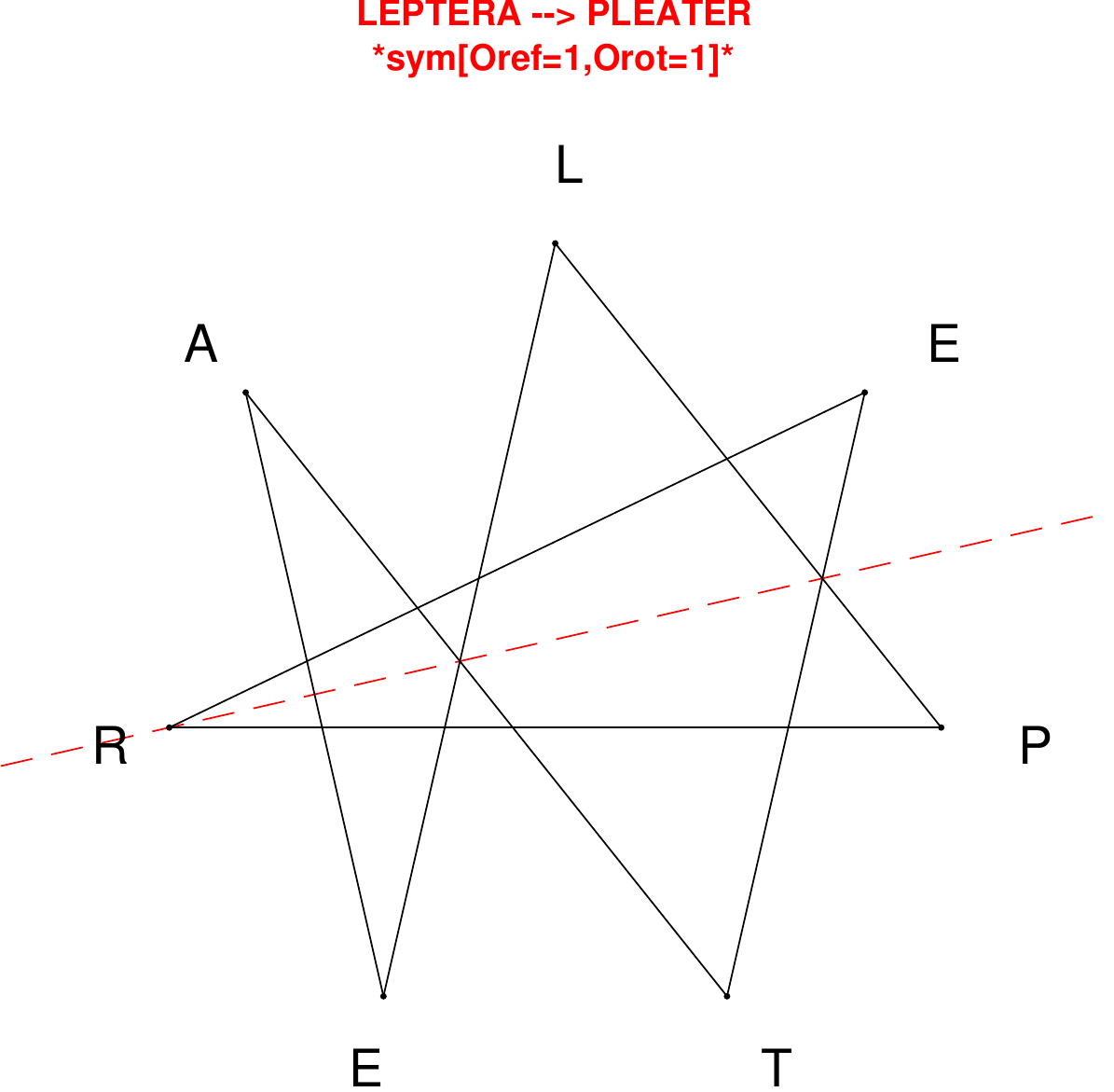}
\end{subfigure}
\hfill
\begin{subfigure}[T]{0.19\textwidth}
\centering
\includegraphics[width=\textwidth]{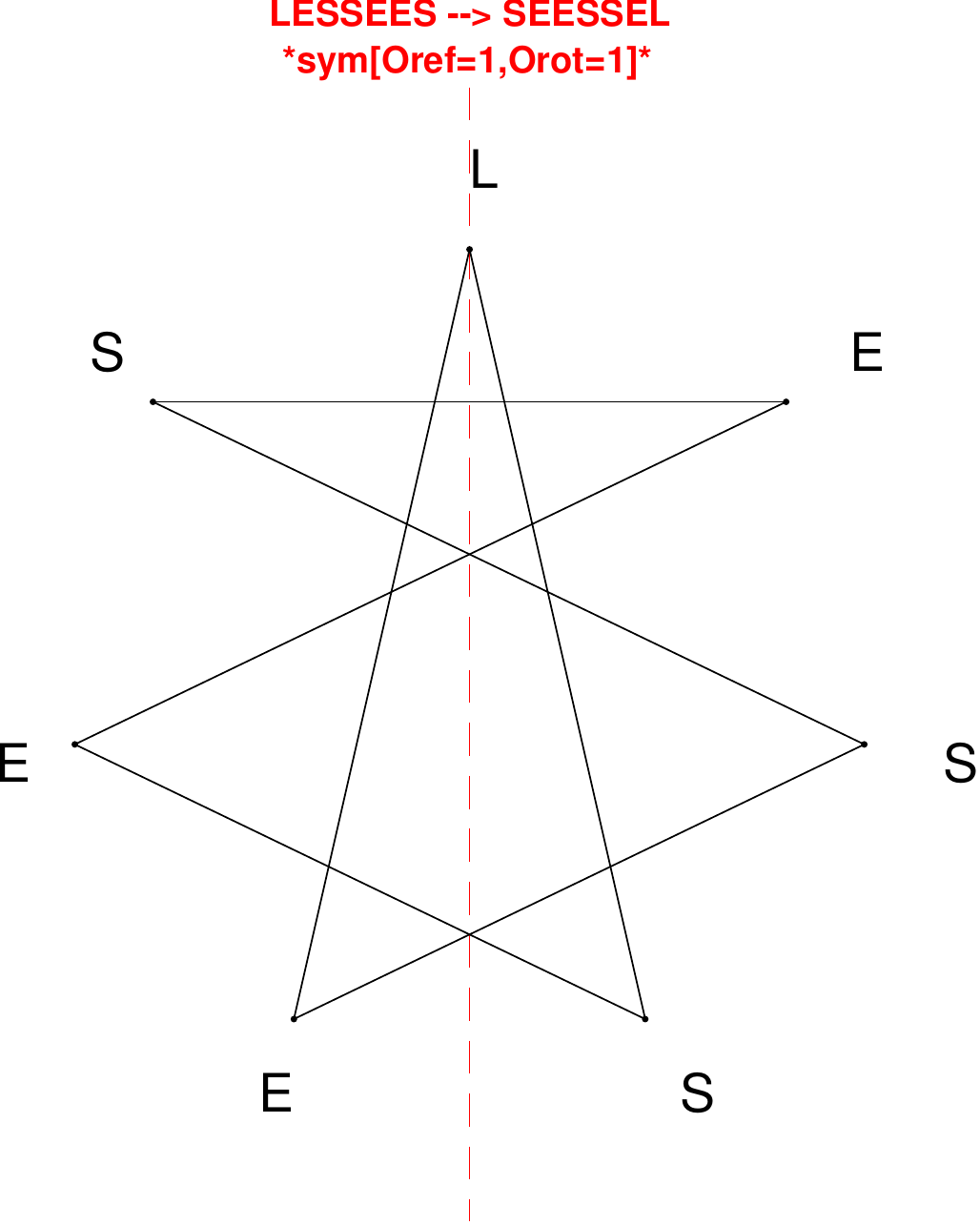}
\end{subfigure}
\hfill
\begin{subfigure}[T]{0.19\textwidth}
\centering
\includegraphics[width=\textwidth]{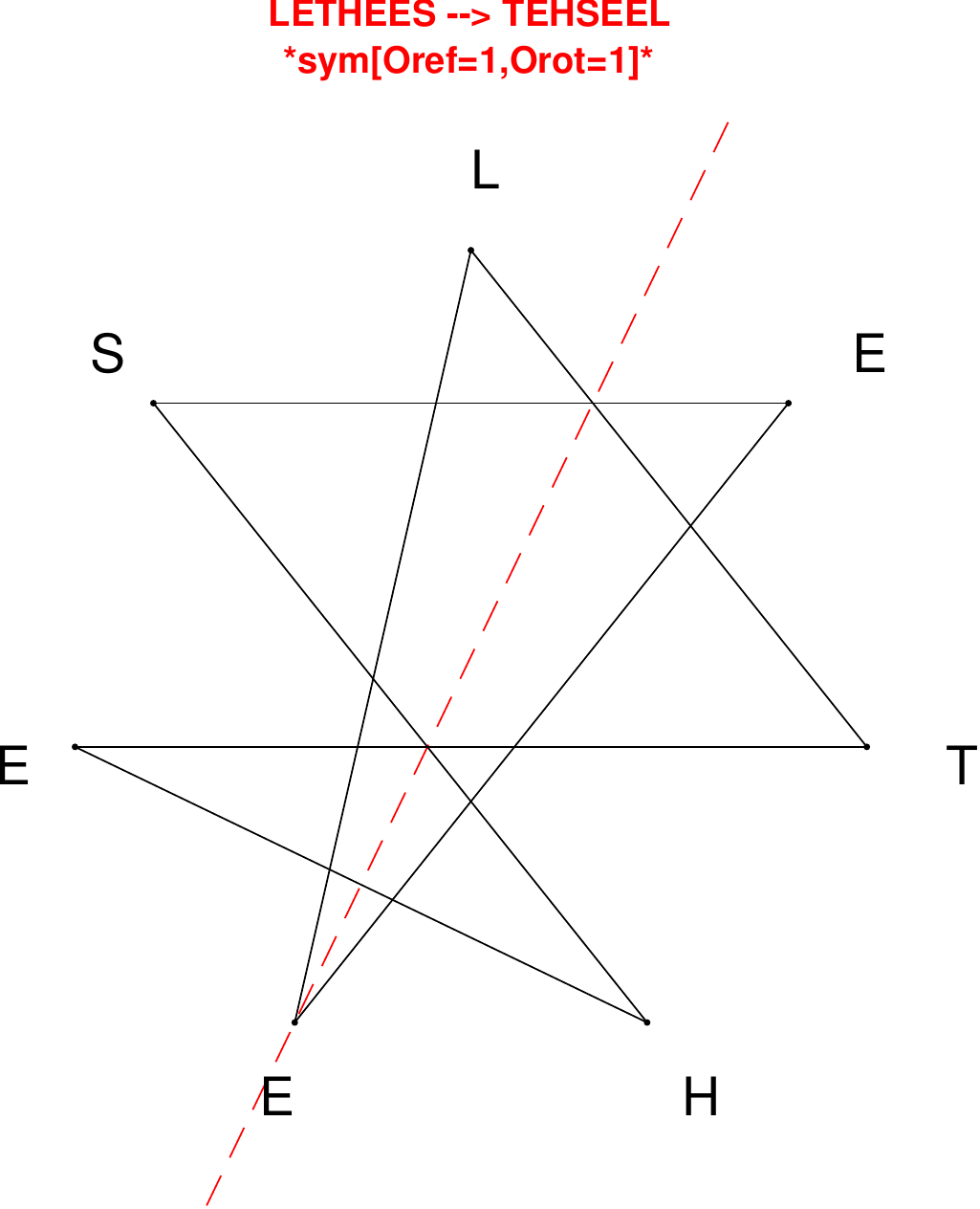}
\end{subfigure}
\end{figure}

\begin{figure}[H]
\centering
\begin{subfigure}[T]{0.19\textwidth}
\centering
\includegraphics[width=\textwidth]{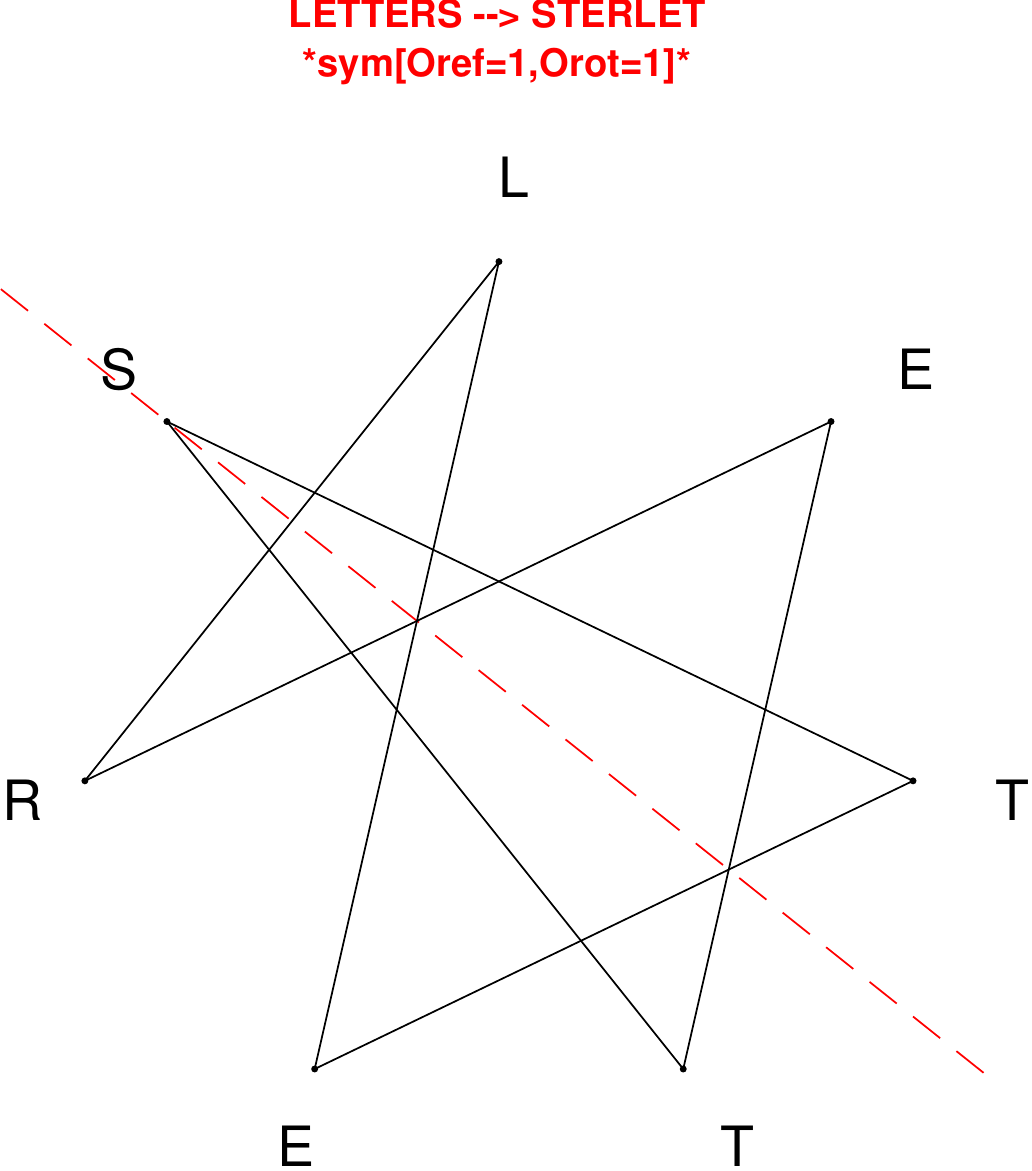}
\end{subfigure}
\hfill
\begin{subfigure}[T]{0.19\textwidth}
\centering
\includegraphics[width=\textwidth]{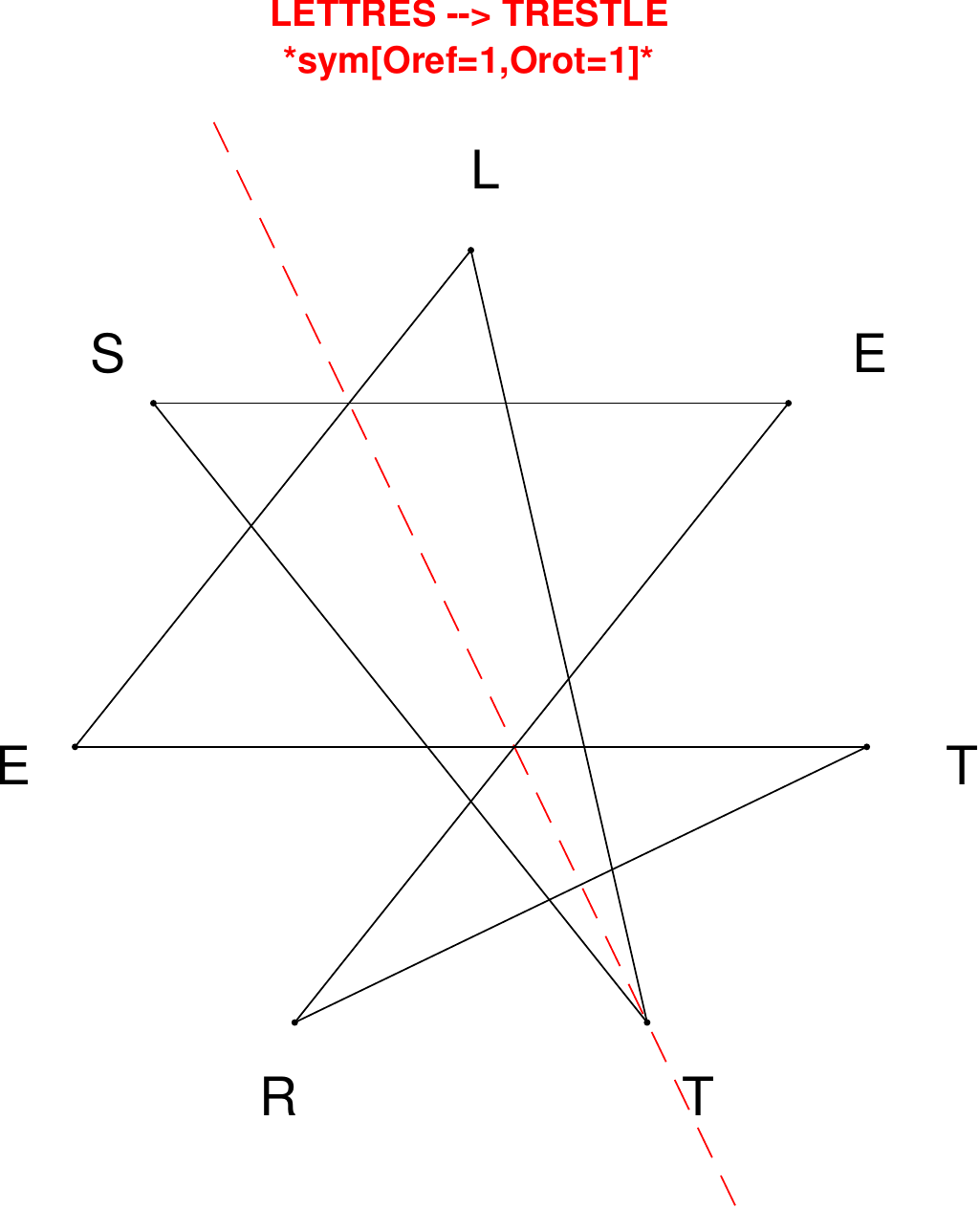}
\end{subfigure}
\hfill
\begin{subfigure}[T]{0.19\textwidth}
\centering
\includegraphics[width=\textwidth]{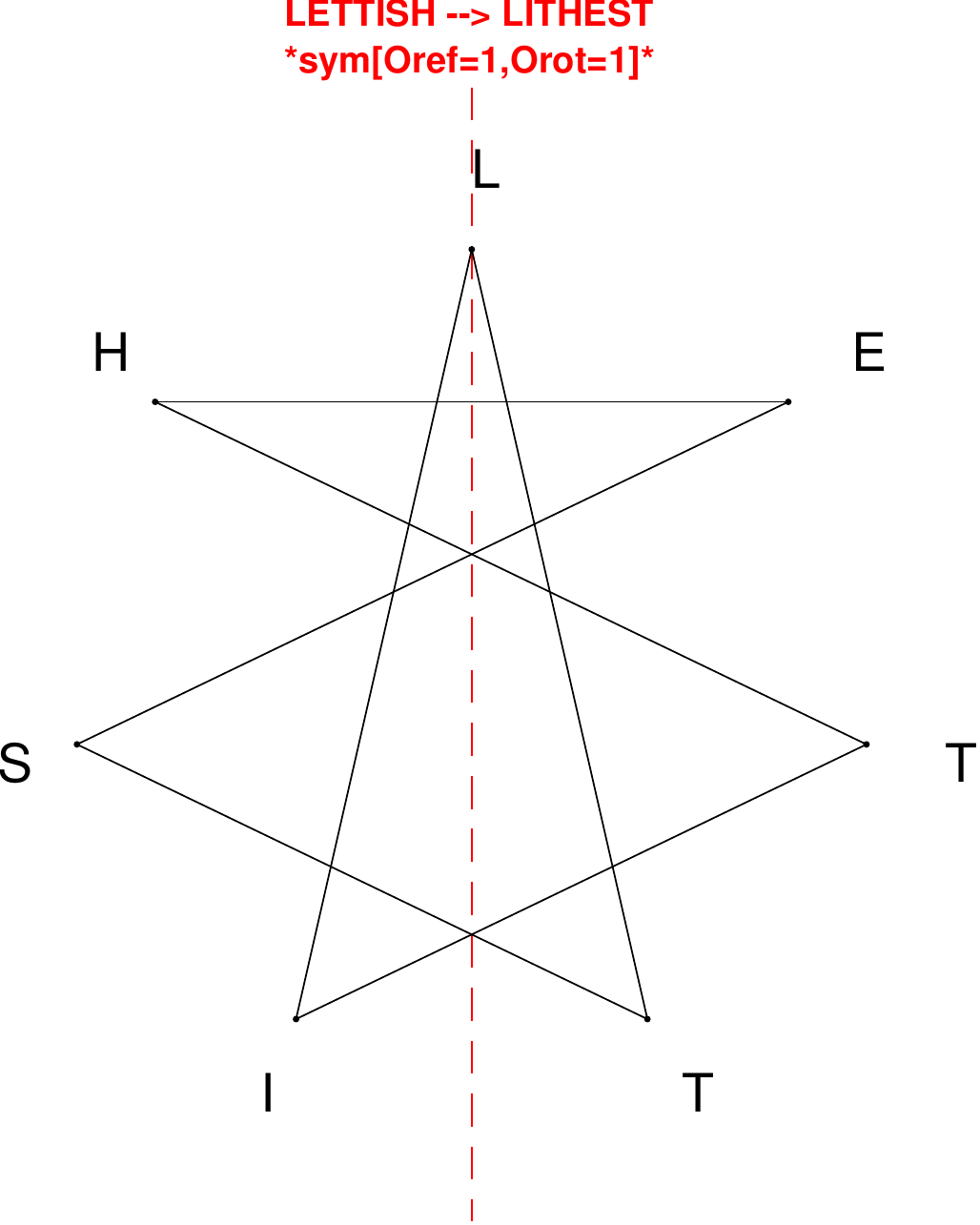}
\end{subfigure}
\hfill
\begin{subfigure}[T]{0.19\textwidth}
\centering
\includegraphics[width=\textwidth]{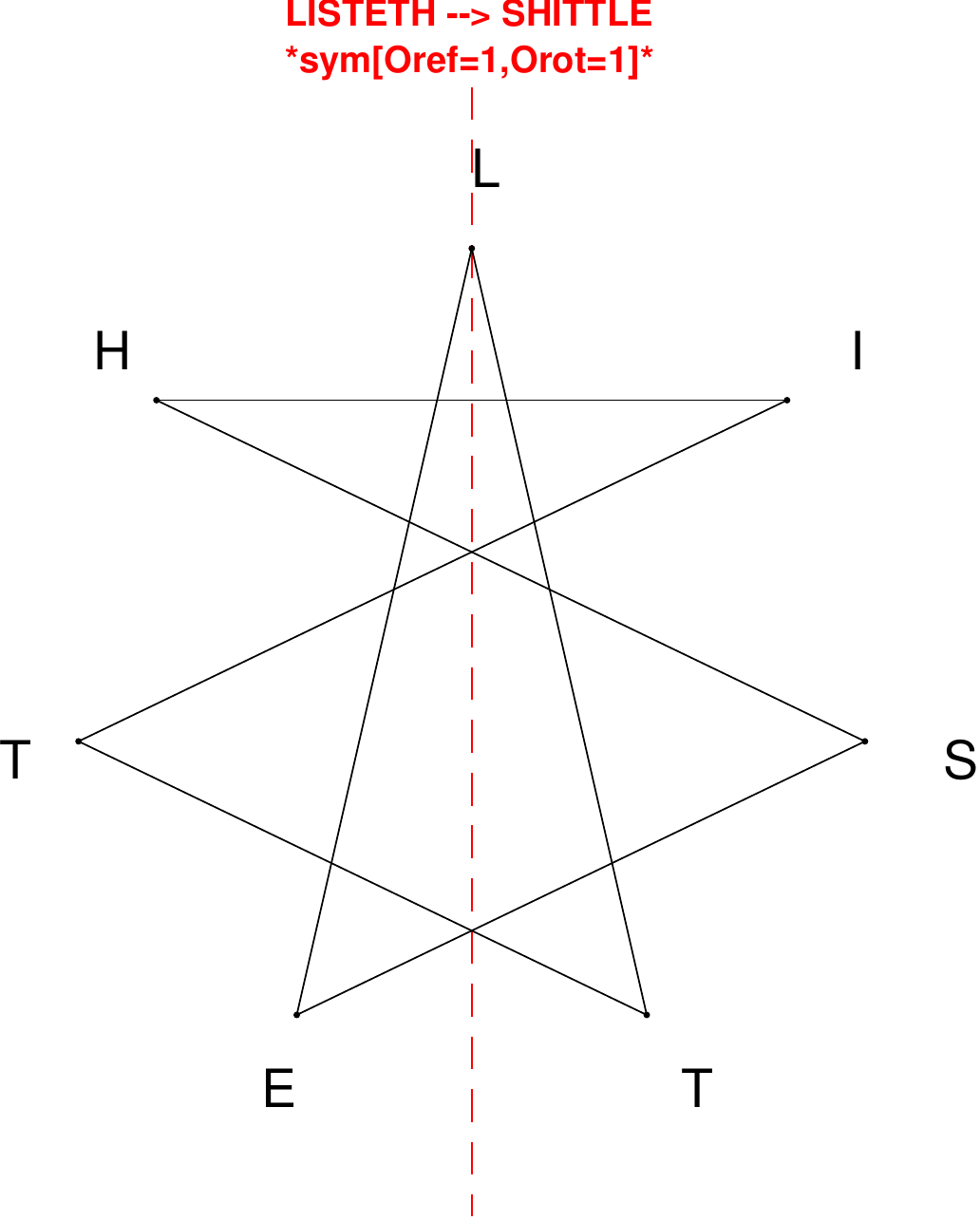}
\end{subfigure}
\hfill
\begin{subfigure}[T]{0.19\textwidth}
\centering
\includegraphics[width=\textwidth]{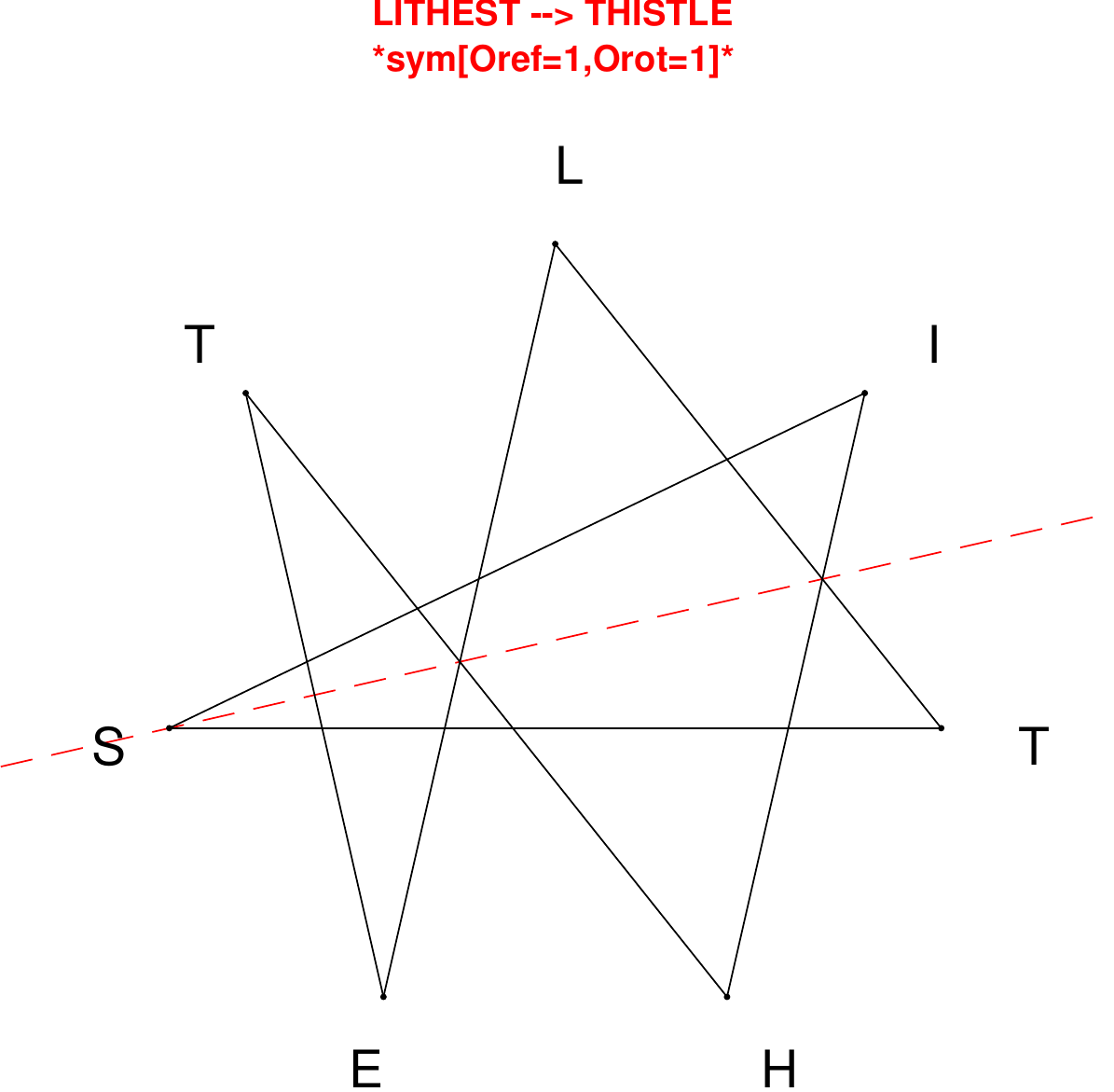}
\end{subfigure}
\end{figure}

\begin{figure}[H]
\centering
\begin{subfigure}[T]{0.19\textwidth}
\centering
\includegraphics[width=\textwidth]{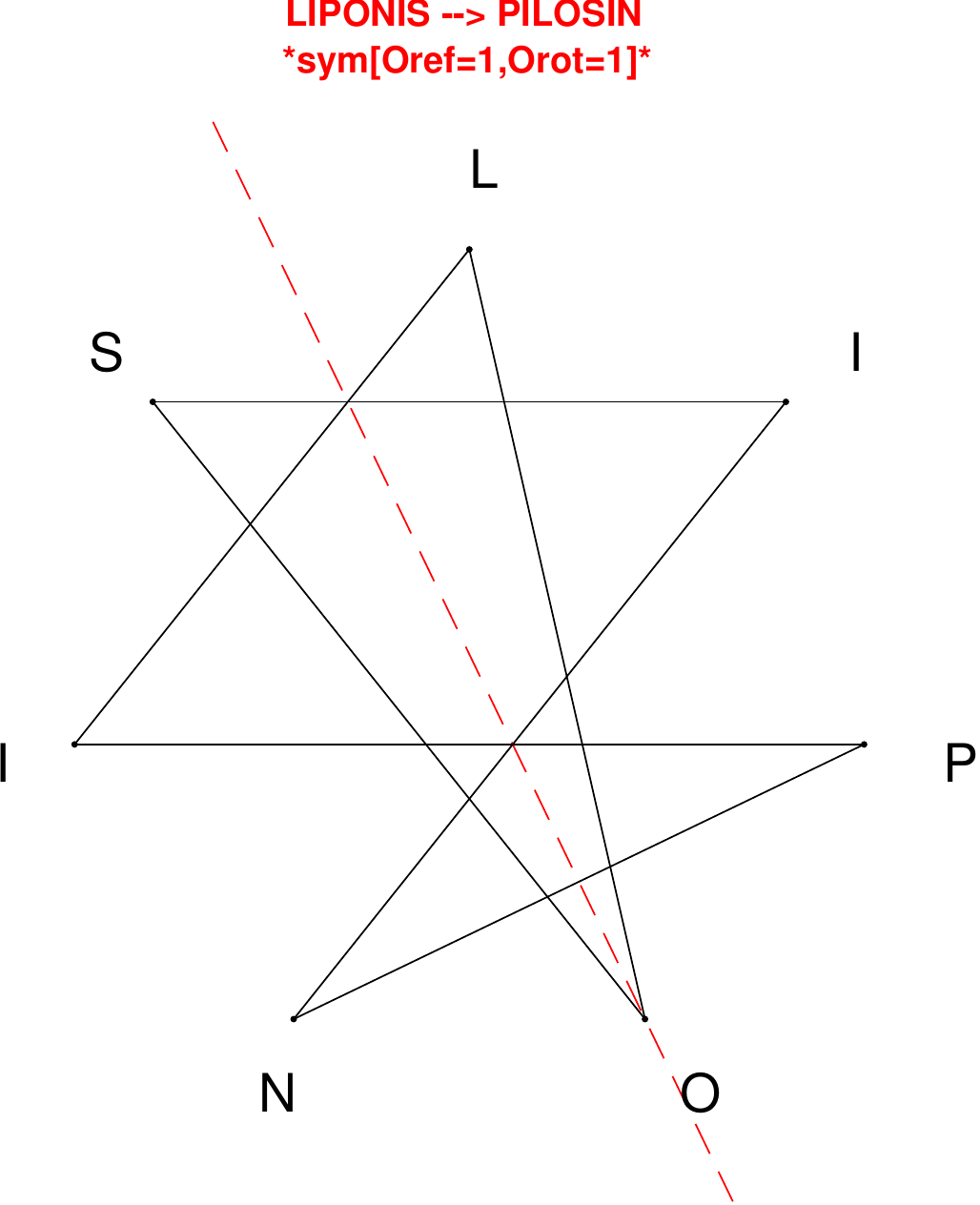}
\end{subfigure}
\hfill
\begin{subfigure}[T]{0.19\textwidth}
\centering
\includegraphics[width=\textwidth]{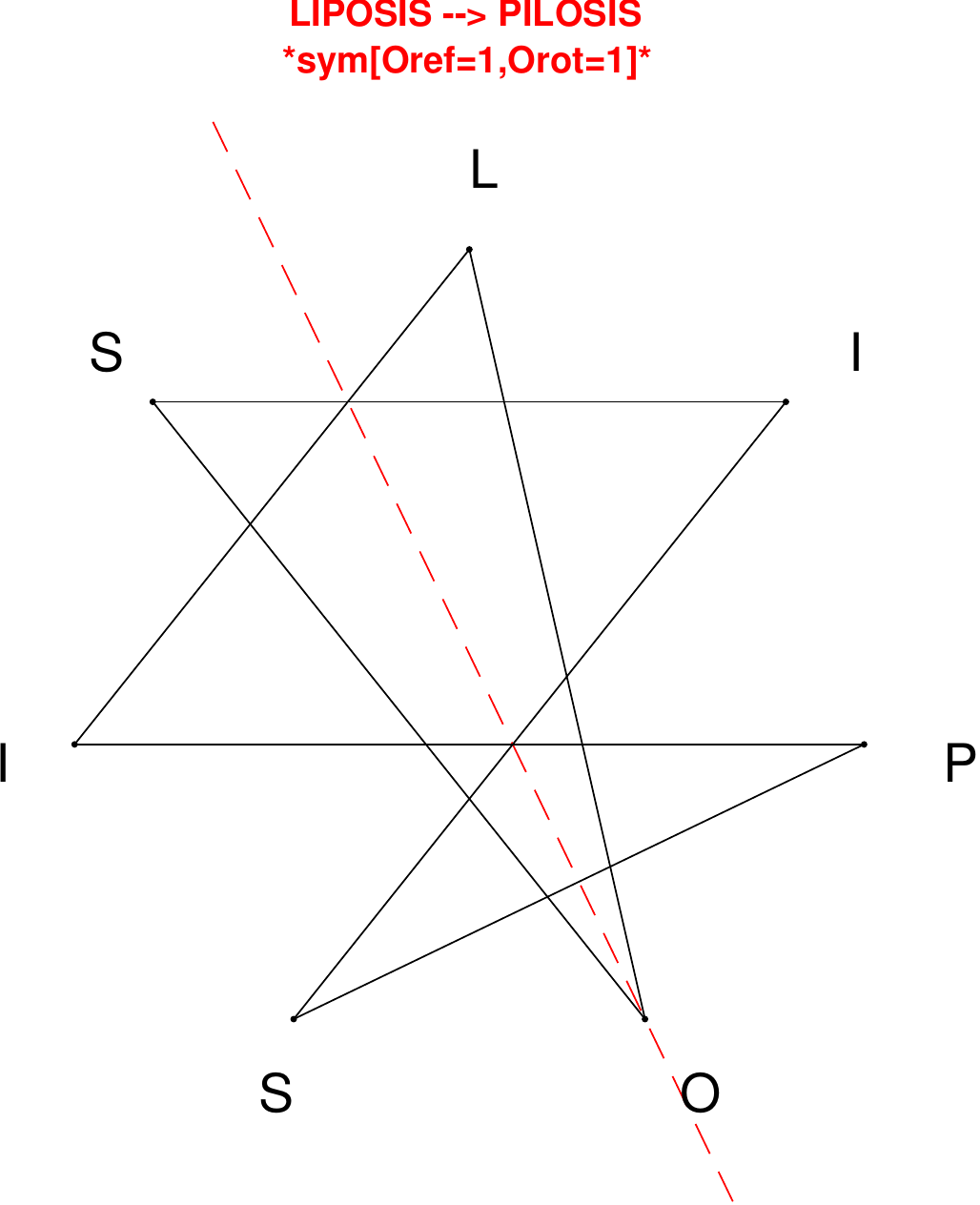}
\end{subfigure}
\hfill
\begin{subfigure}[T]{0.19\textwidth}
\centering
\includegraphics[width=\textwidth]{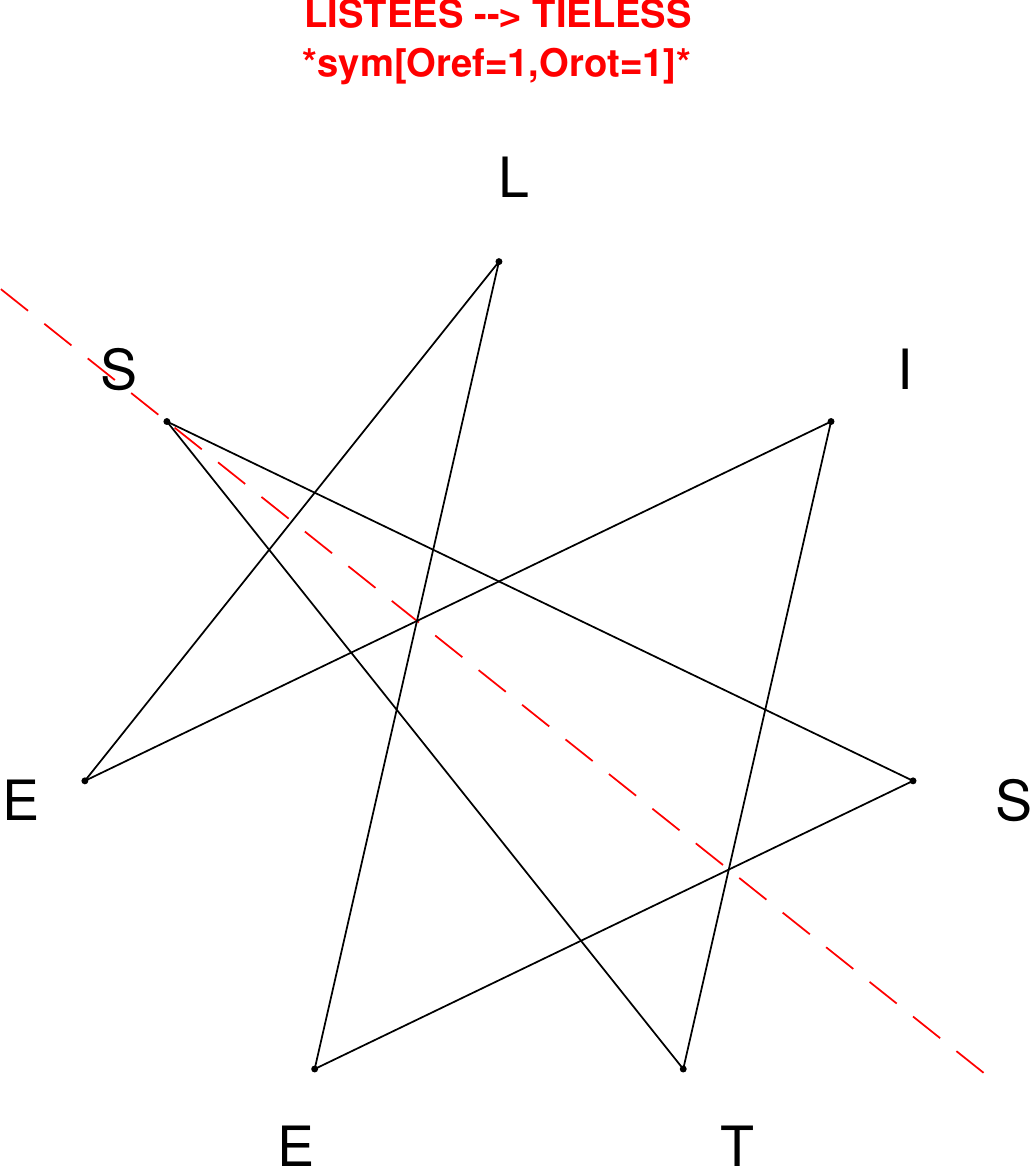}
\end{subfigure}
\hfill
\begin{subfigure}[T]{0.19\textwidth}
\centering
\includegraphics[width=\textwidth]{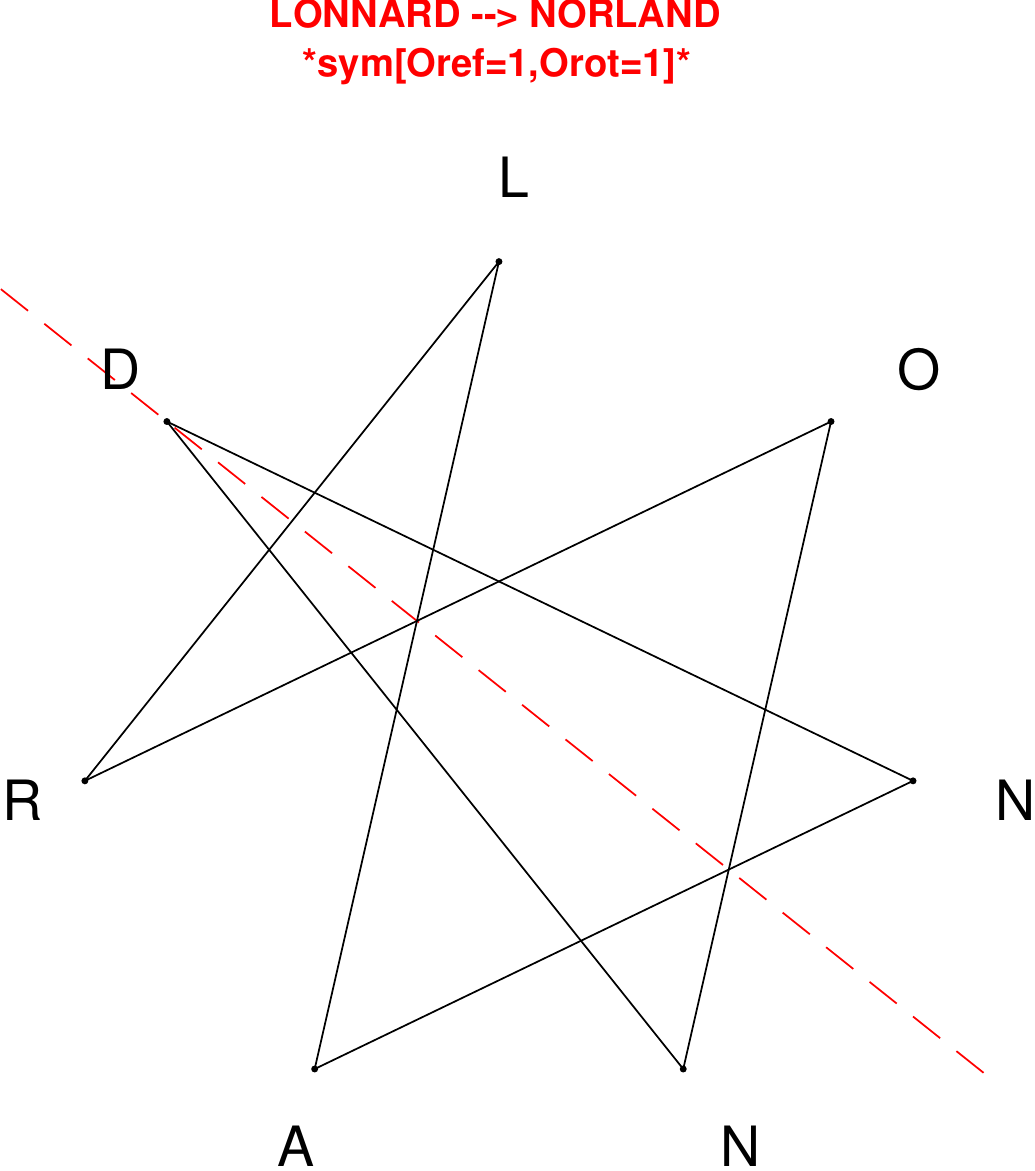}
\end{subfigure}
\hfill
\begin{subfigure}[T]{0.19\textwidth}
\centering
\includegraphics[width=\textwidth]{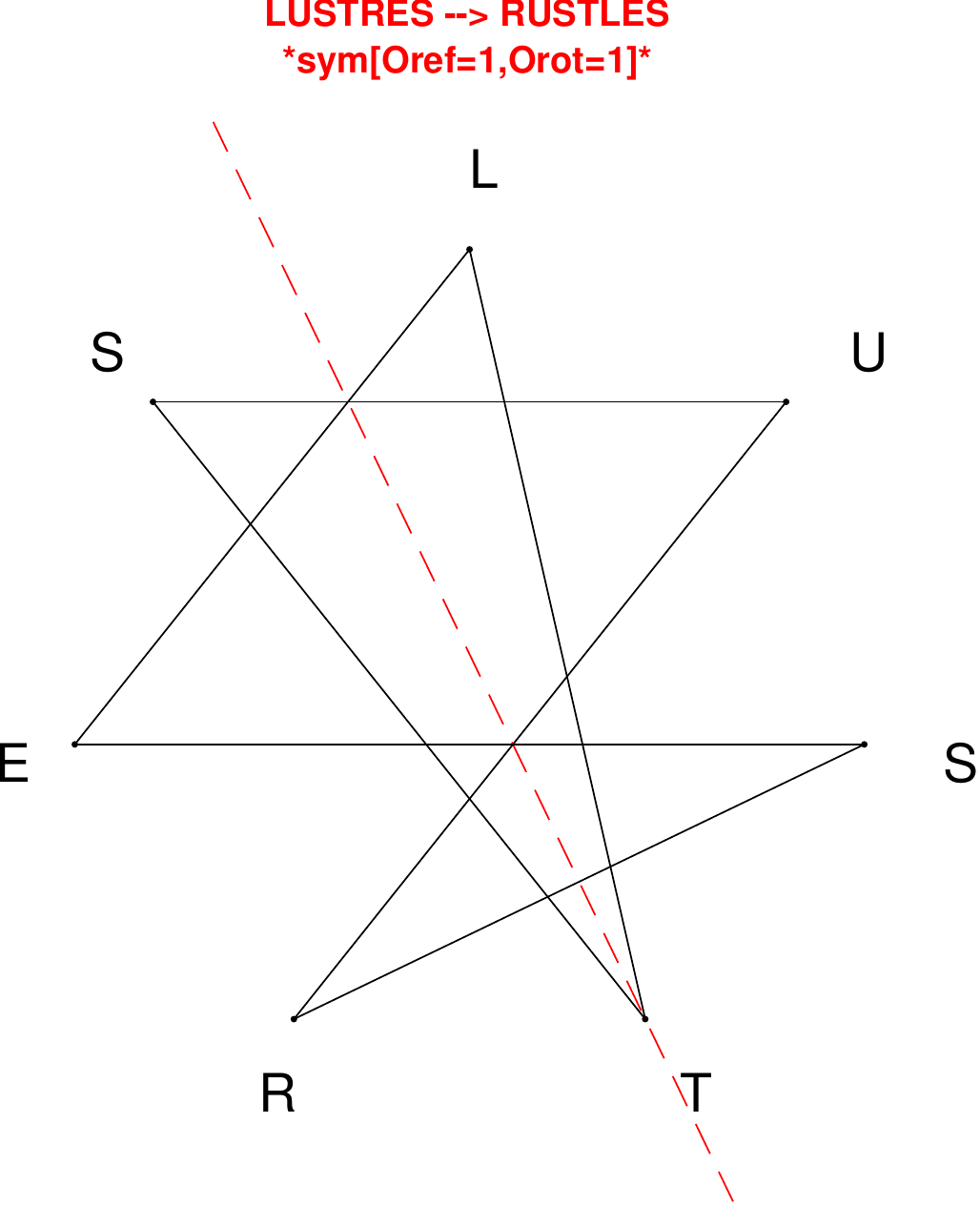}
\end{subfigure}
\end{figure}

\begin{figure}[H]
\centering
\begin{subfigure}[T]{0.19\textwidth}
\centering
\includegraphics[width=\textwidth]{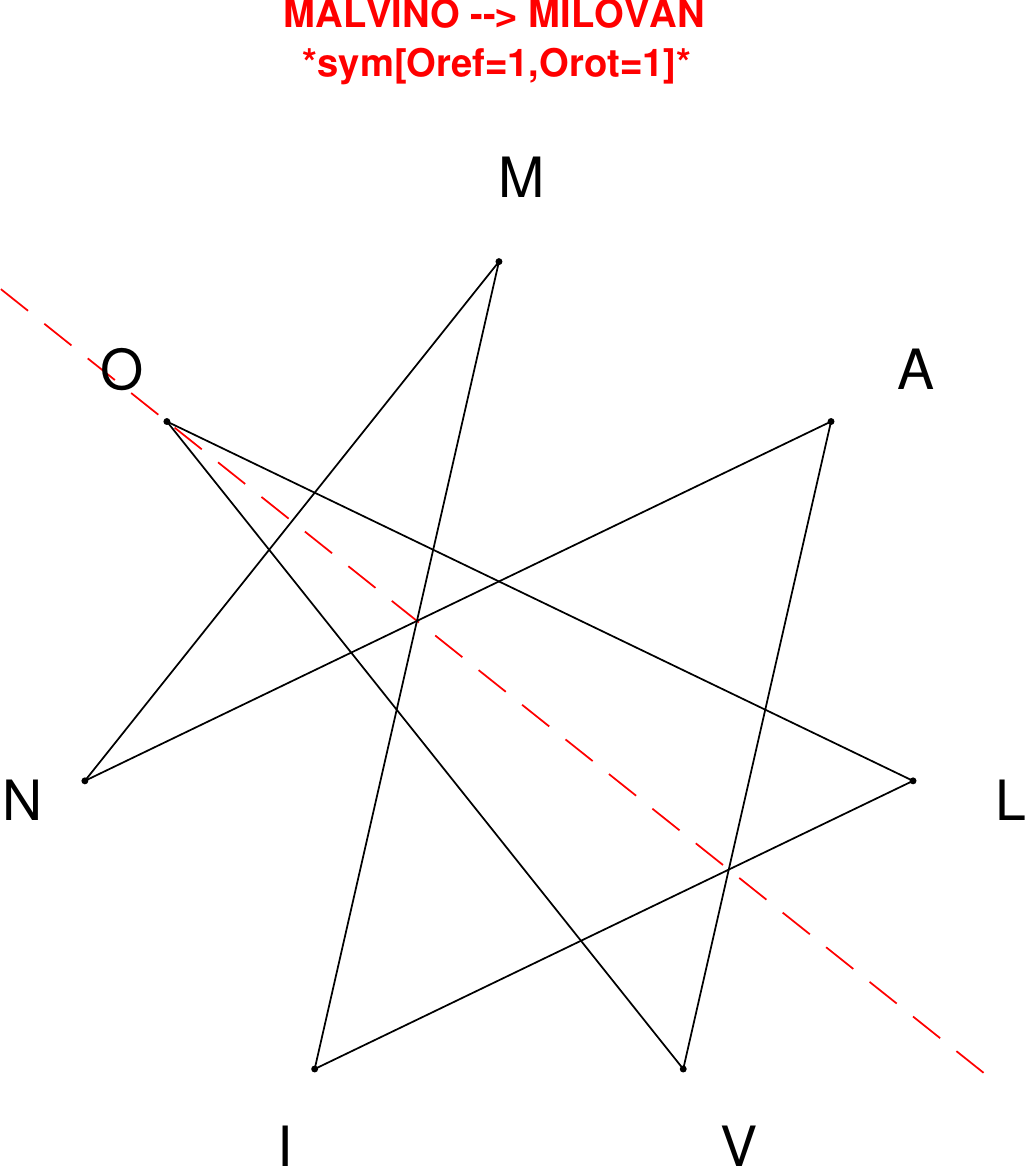}
\end{subfigure}
\hfill
\begin{subfigure}[T]{0.19\textwidth}
\centering
\includegraphics[width=\textwidth]{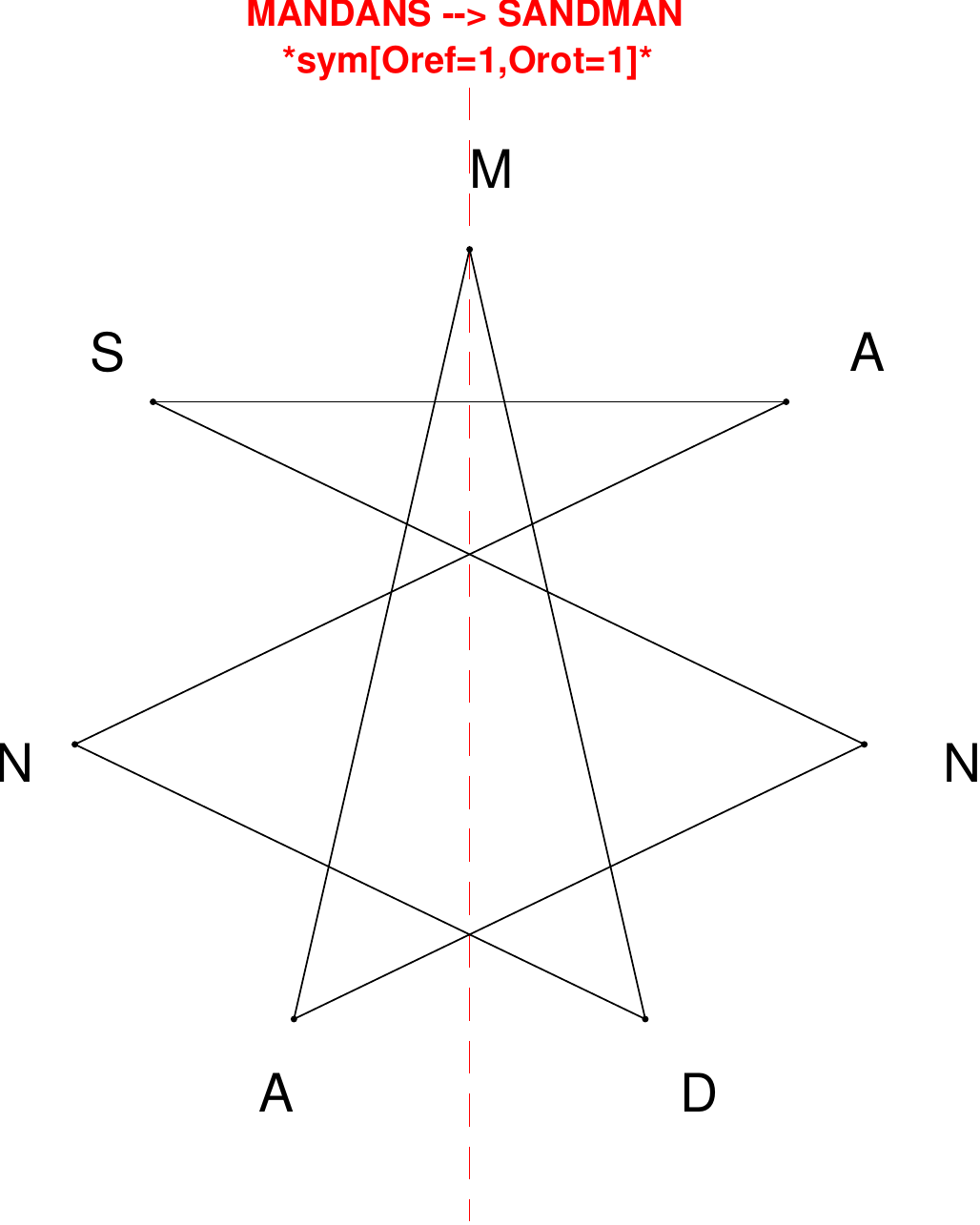}
\end{subfigure}
\hfill
\begin{subfigure}[T]{0.19\textwidth}
\centering
\includegraphics[width=\textwidth]{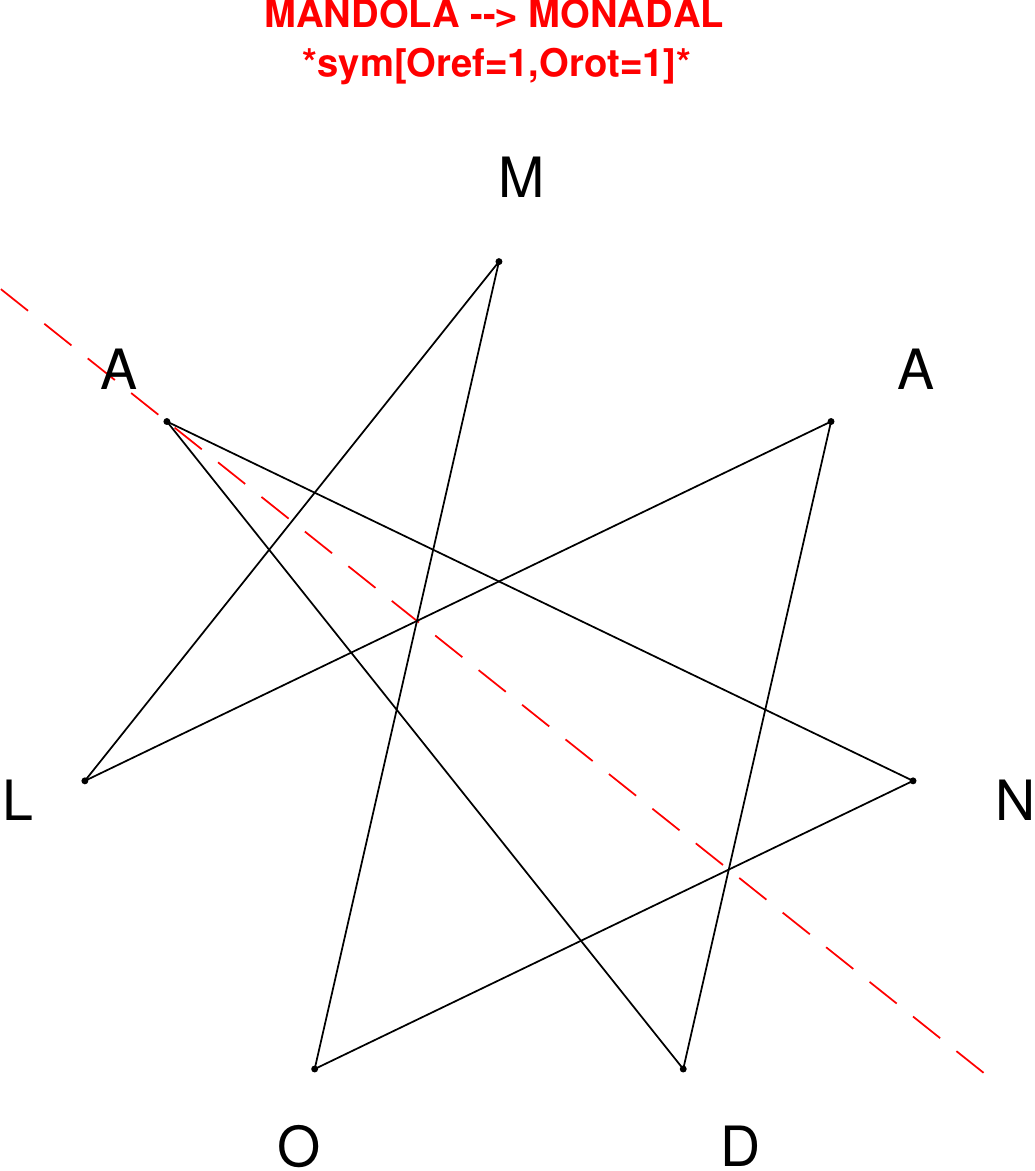}
\end{subfigure}
\hfill
\begin{subfigure}[T]{0.19\textwidth}
\centering
\includegraphics[width=\textwidth]{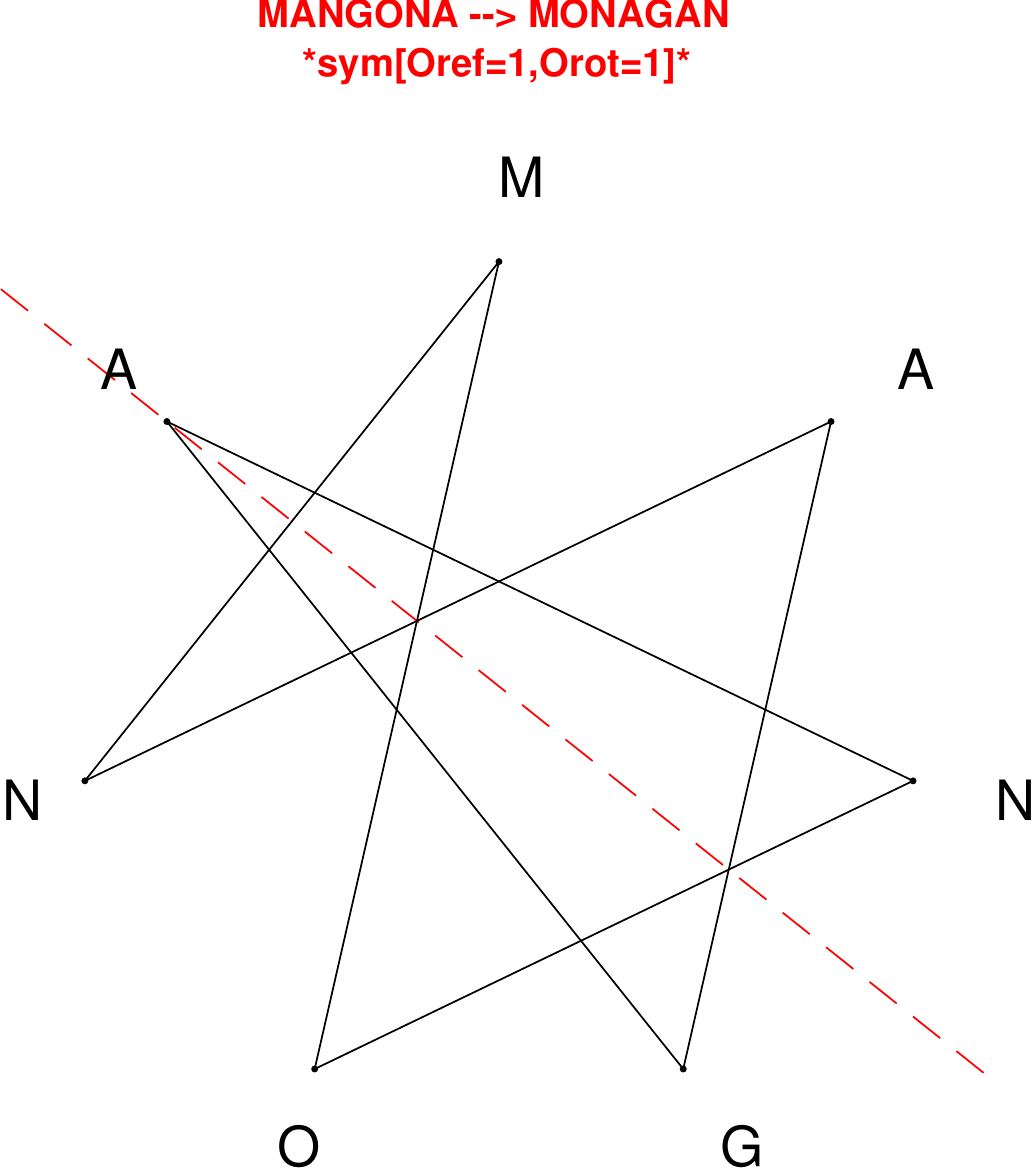}
\end{subfigure}
\hfill
\begin{subfigure}[T]{0.19\textwidth}
\centering
\includegraphics[width=\textwidth]{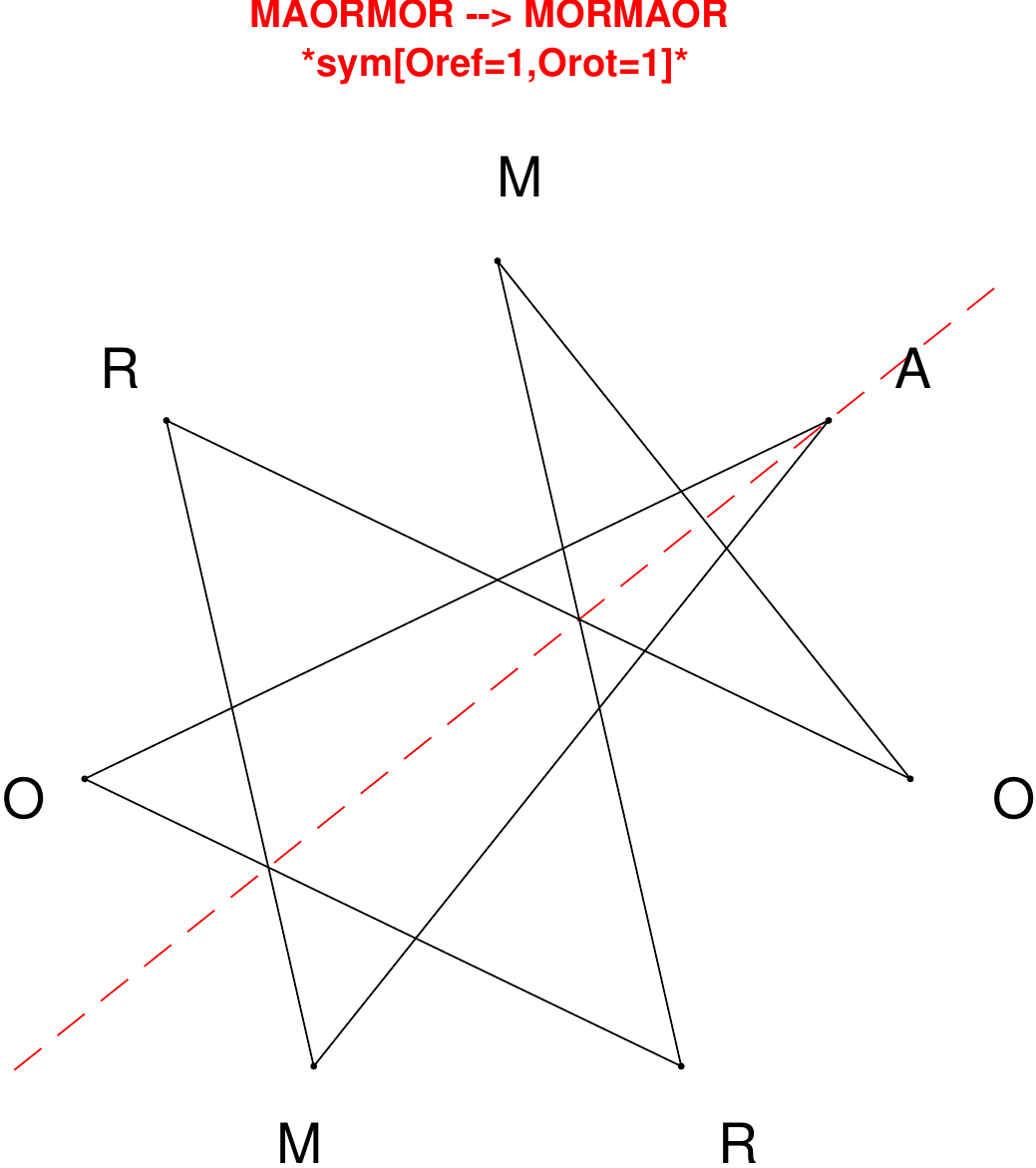}
\end{subfigure}
\end{figure}

\begin{figure}[H]
\centering
\begin{subfigure}[T]{0.19\textwidth}
\centering
\includegraphics[width=\textwidth]{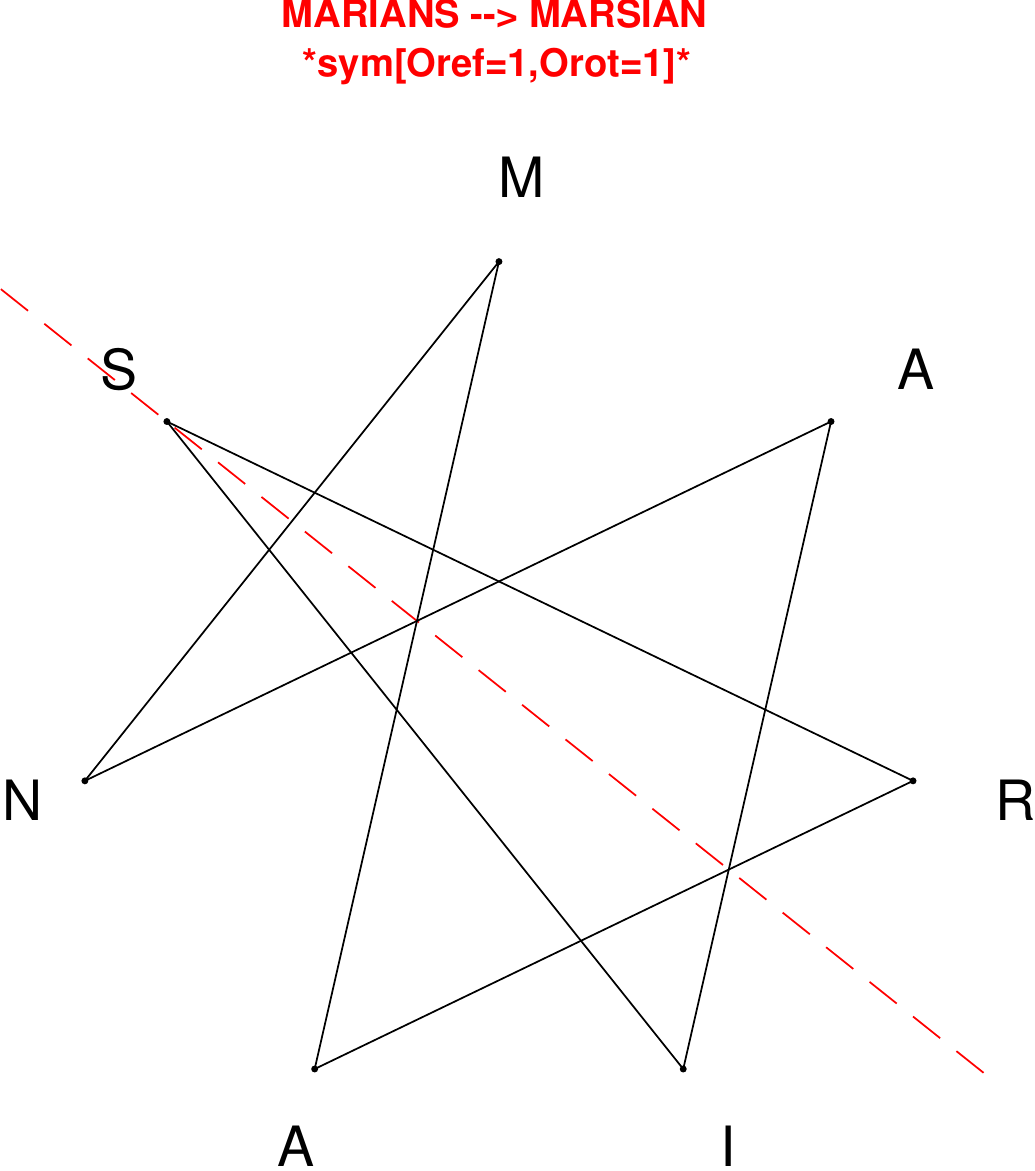}
\end{subfigure}
\hfill
\begin{subfigure}[T]{0.19\textwidth}
\centering
\includegraphics[width=\textwidth]{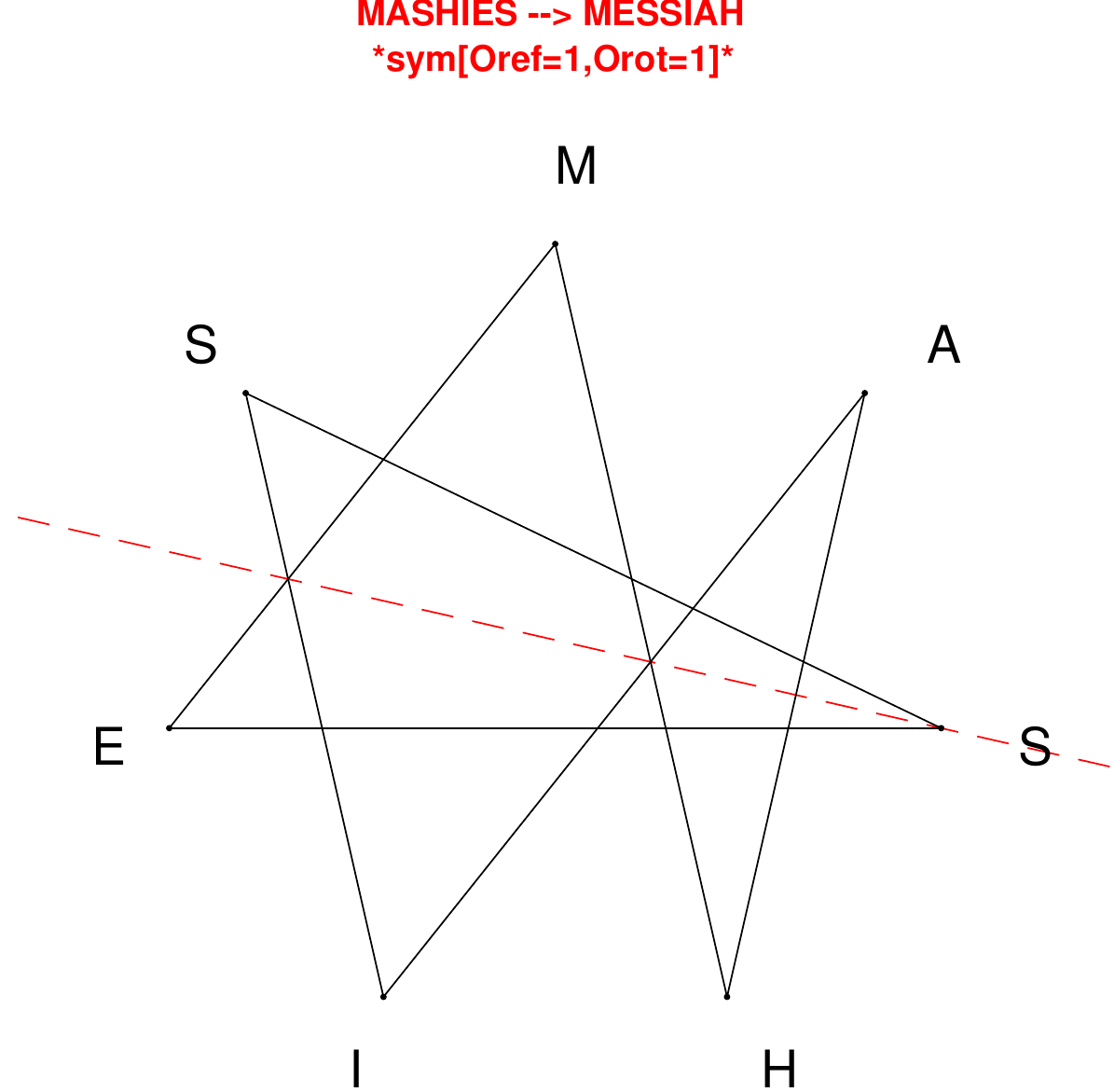}
\end{subfigure}
\hfill
\begin{subfigure}[T]{0.19\textwidth}
\centering
\includegraphics[width=\textwidth]{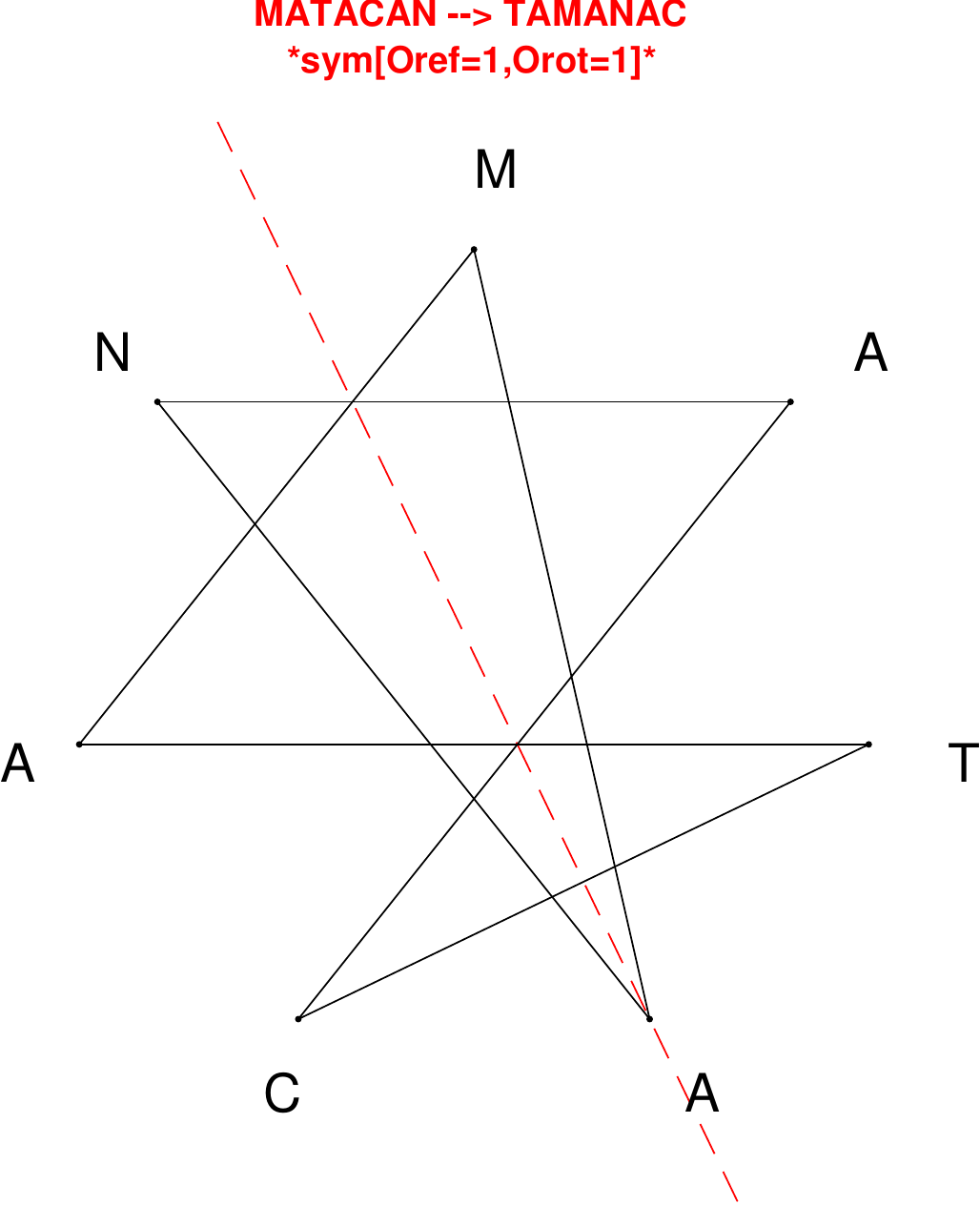}
\end{subfigure}
\hfill
\begin{subfigure}[T]{0.19\textwidth}
\centering
\includegraphics[width=\textwidth]{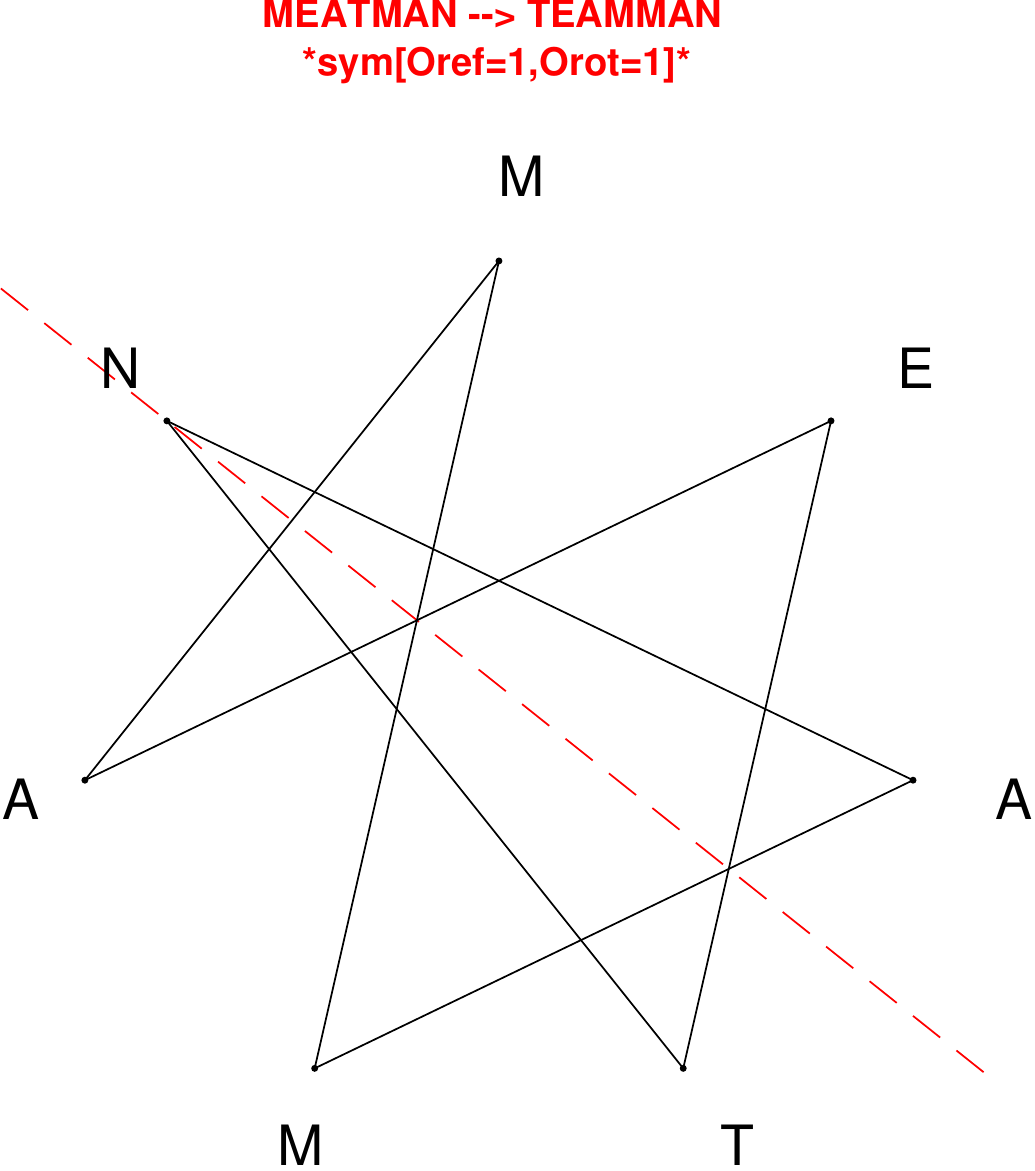}
\end{subfigure}
\hfill
\begin{subfigure}[T]{0.19\textwidth}
\centering
\includegraphics[width=\textwidth]{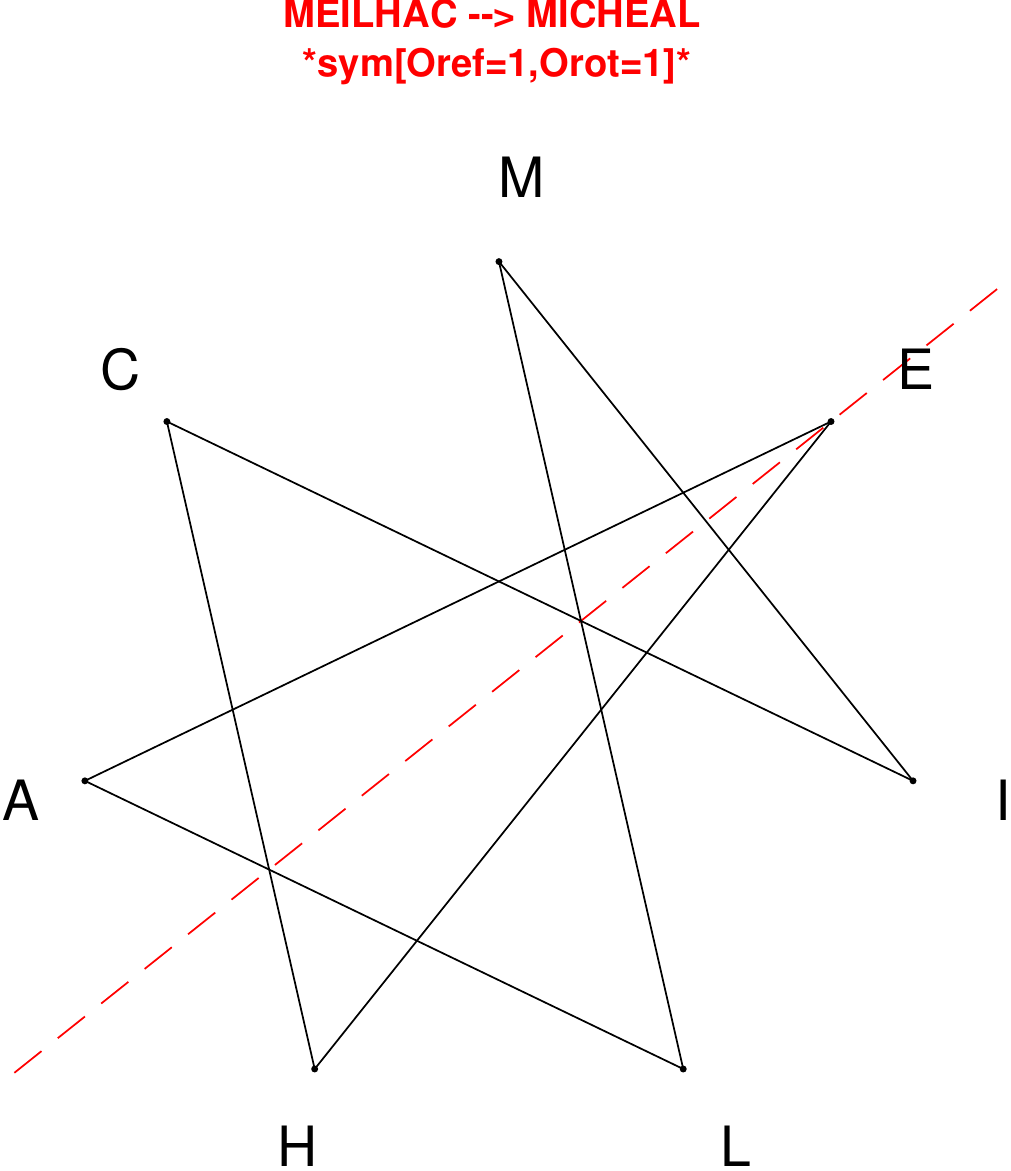}
\end{subfigure}
\end{figure}

\begin{figure}[H]
\centering
\begin{subfigure}[T]{0.19\textwidth}
\centering
\includegraphics[width=\textwidth]{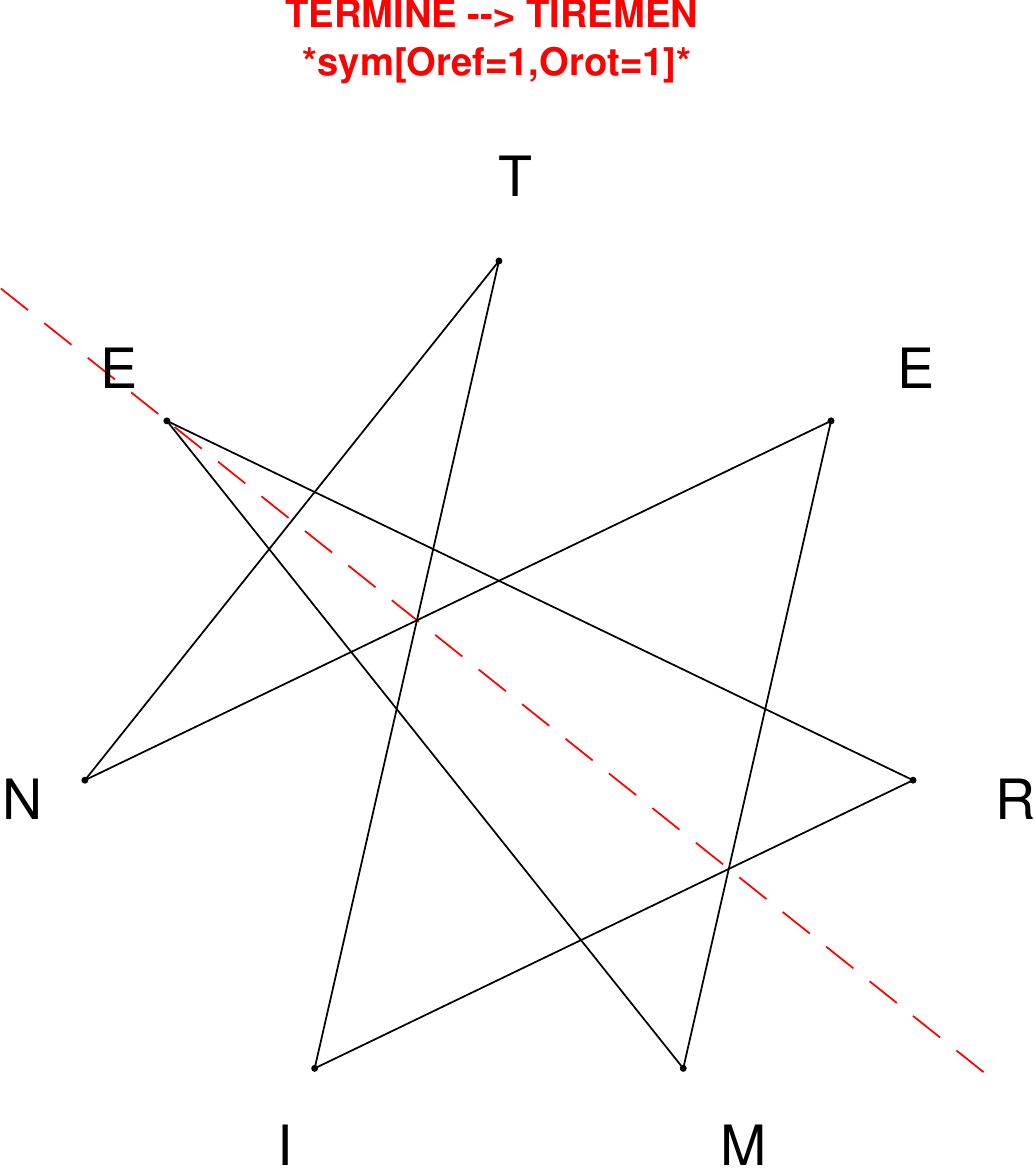}
\end{subfigure}
\hfill
\begin{subfigure}[T]{0.19\textwidth}
\centering
\includegraphics[width=\textwidth]{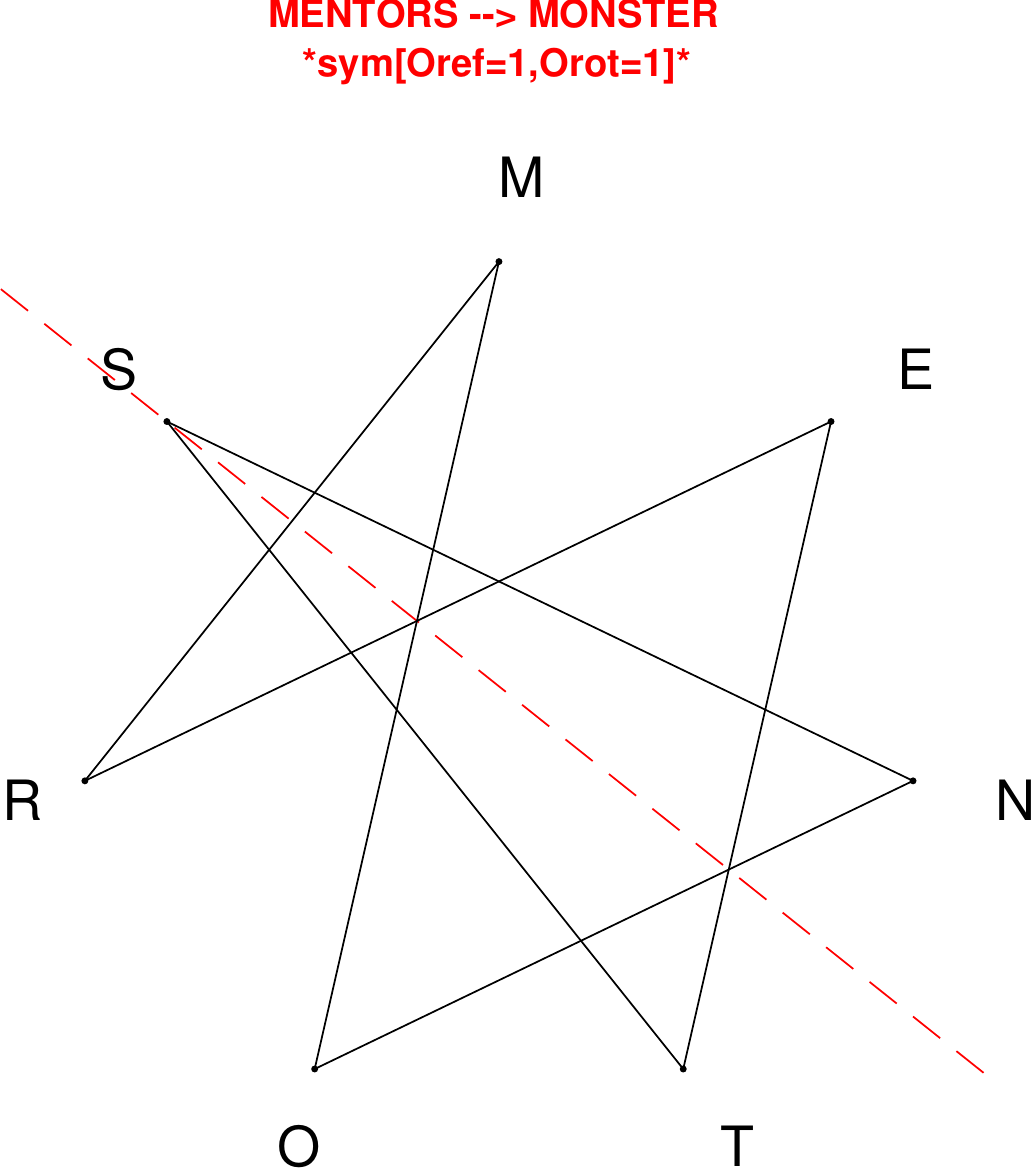}
\end{subfigure}
\hfill
\begin{subfigure}[T]{0.19\textwidth}
\centering
\includegraphics[width=\textwidth]{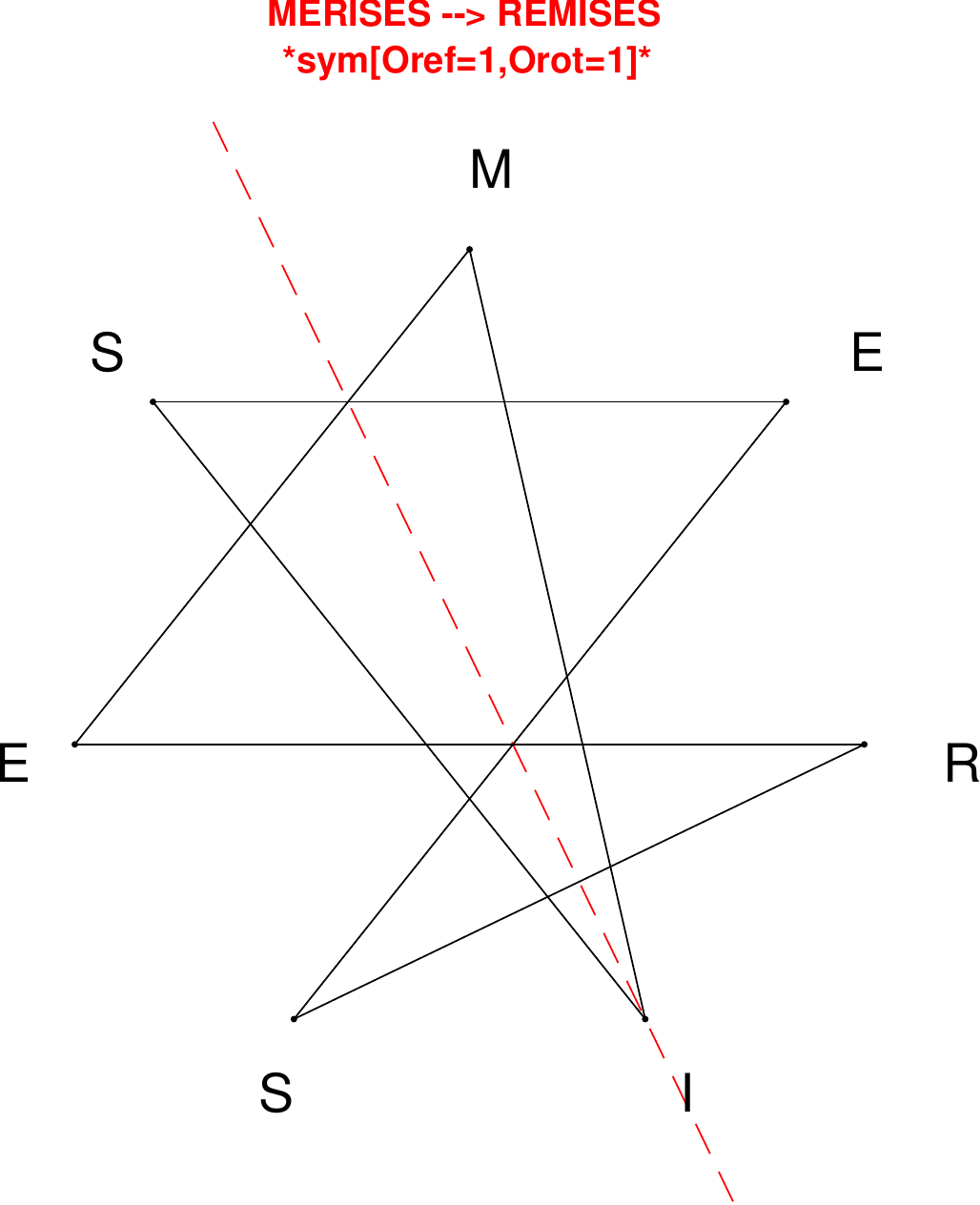}
\end{subfigure}
\hfill
\begin{subfigure}[T]{0.19\textwidth}
\centering
\includegraphics[width=\textwidth]{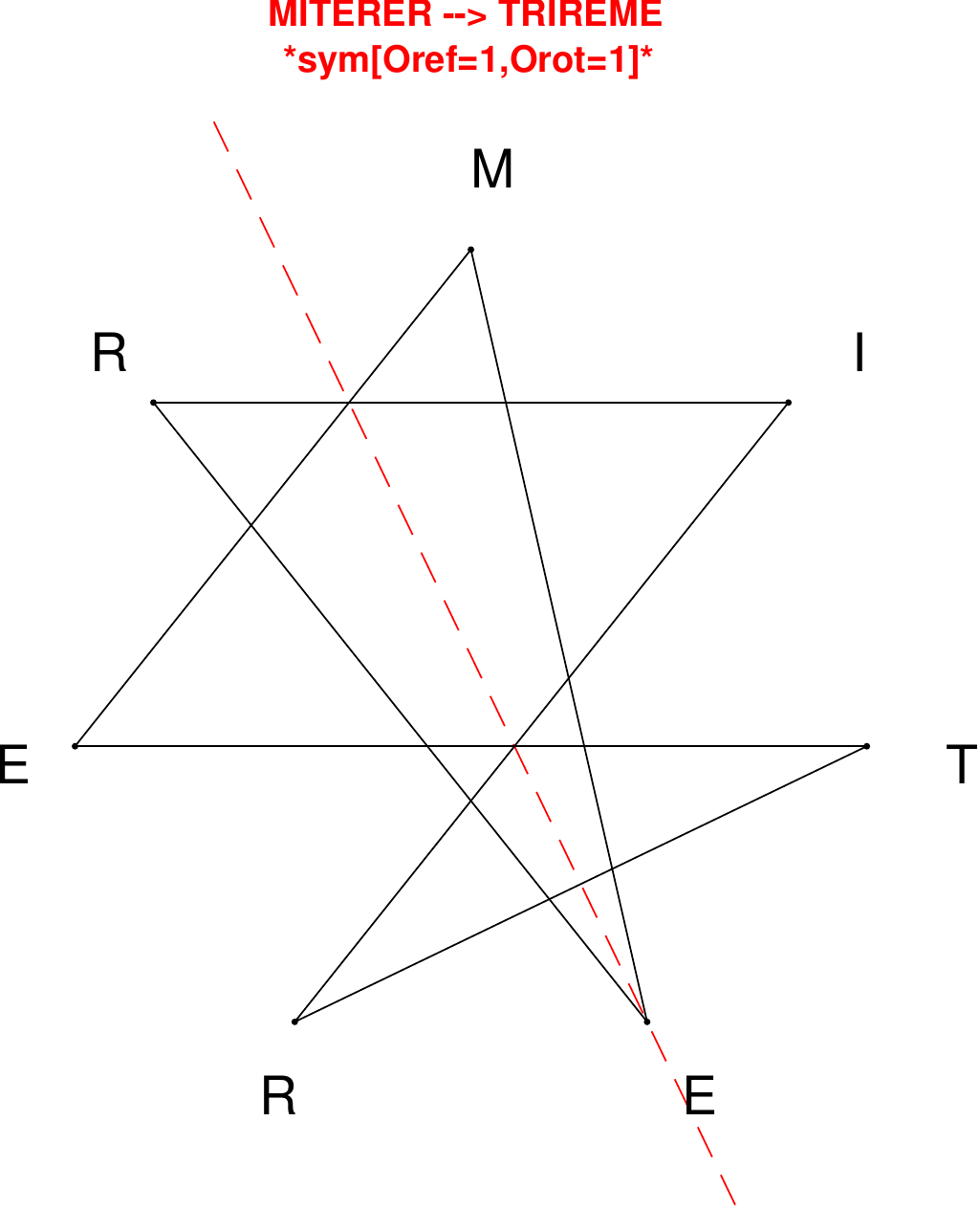}
\end{subfigure}
\hfill
\begin{subfigure}[T]{0.19\textwidth}
\centering
\includegraphics[width=\textwidth]{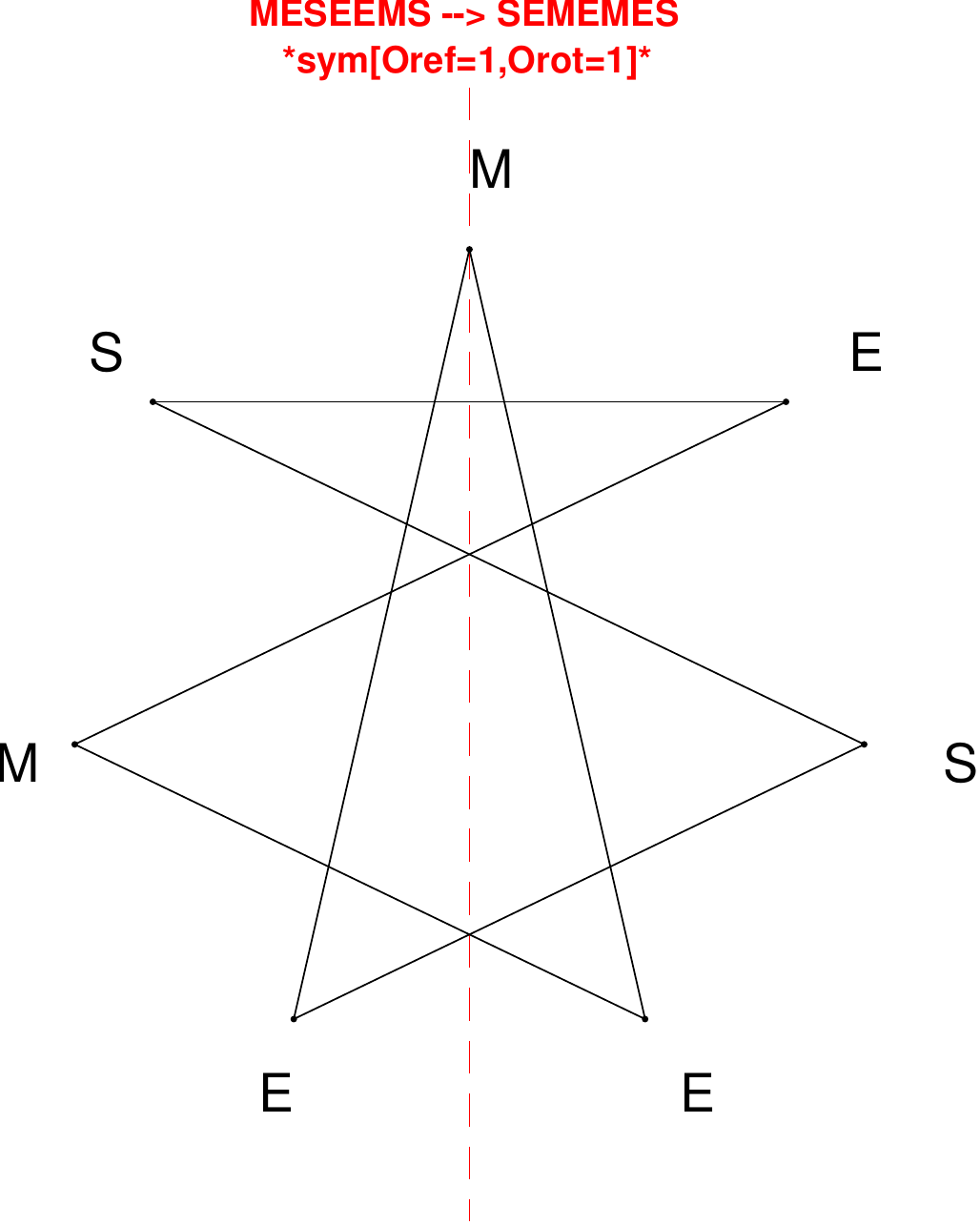}
\end{subfigure}
\end{figure}

\begin{figure}[H]
\centering
\begin{subfigure}[T]{0.19\textwidth}
\centering
\includegraphics[width=\textwidth]{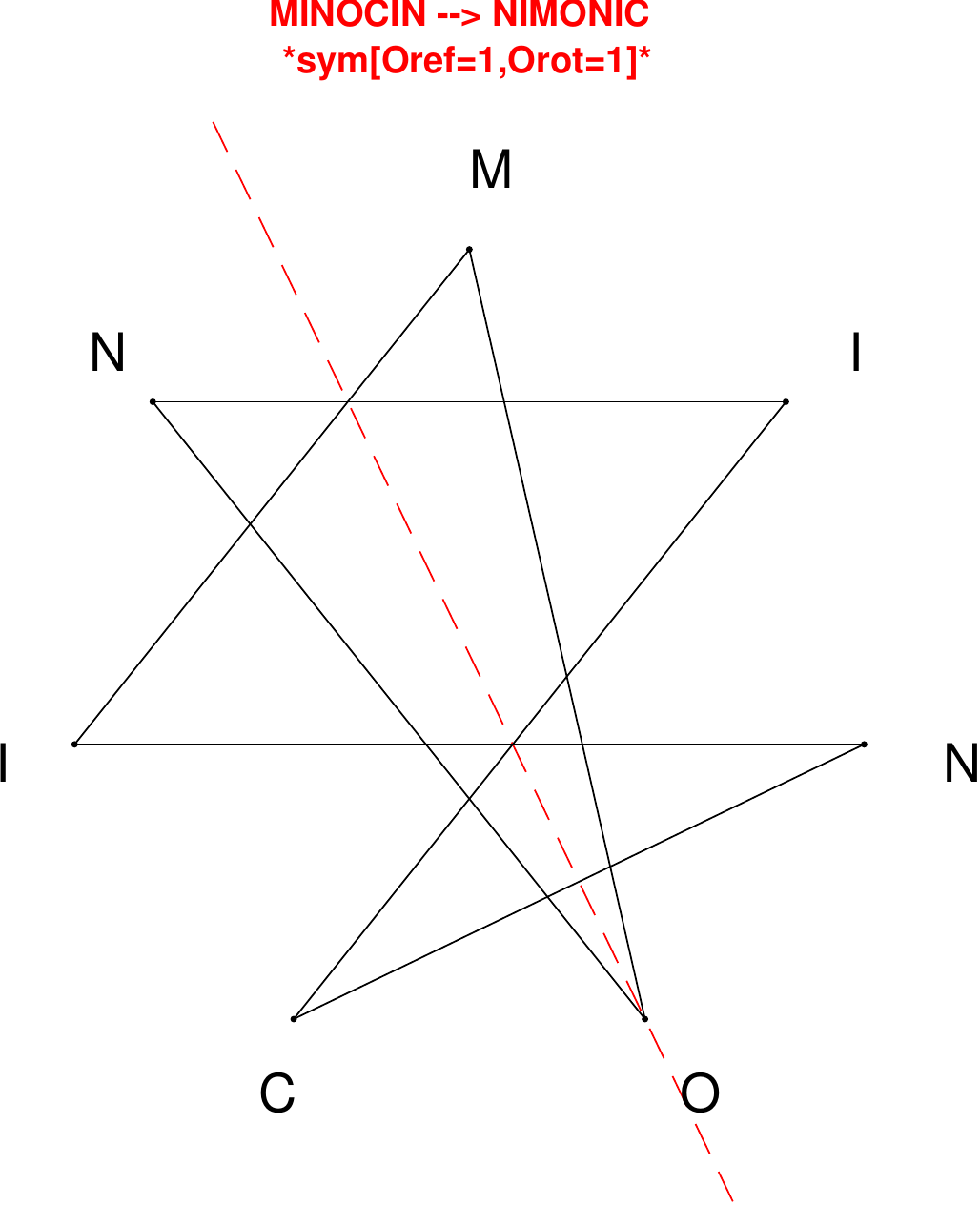}
\end{subfigure}
\hfill
\begin{subfigure}[T]{0.19\textwidth}
\centering
\includegraphics[width=\textwidth]{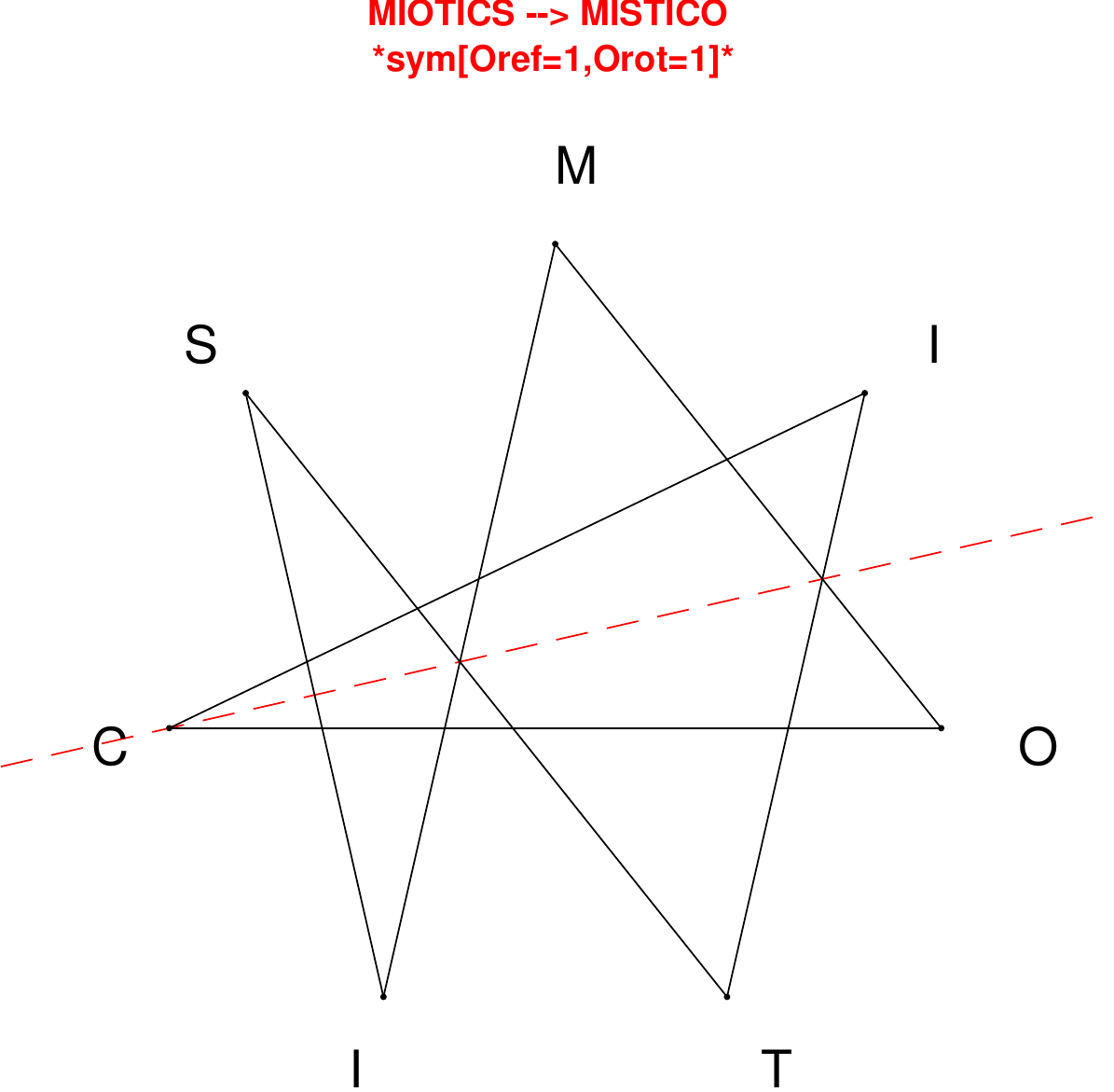}
\end{subfigure}
\hfill
\begin{subfigure}[T]{0.19\textwidth}
\centering
\includegraphics[width=\textwidth]{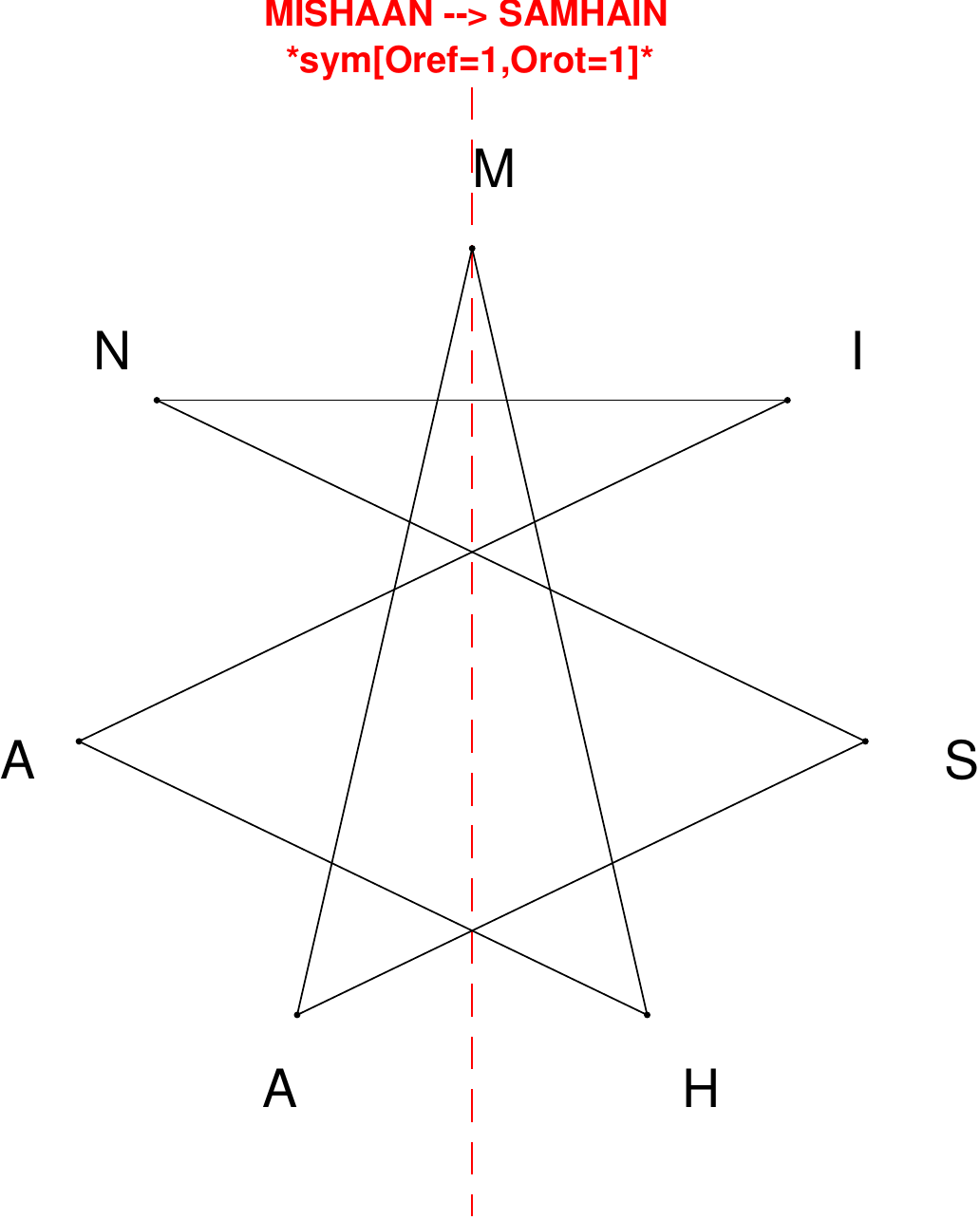}
\end{subfigure}
\hfill
\begin{subfigure}[T]{0.19\textwidth}
\centering
\includegraphics[width=\textwidth]{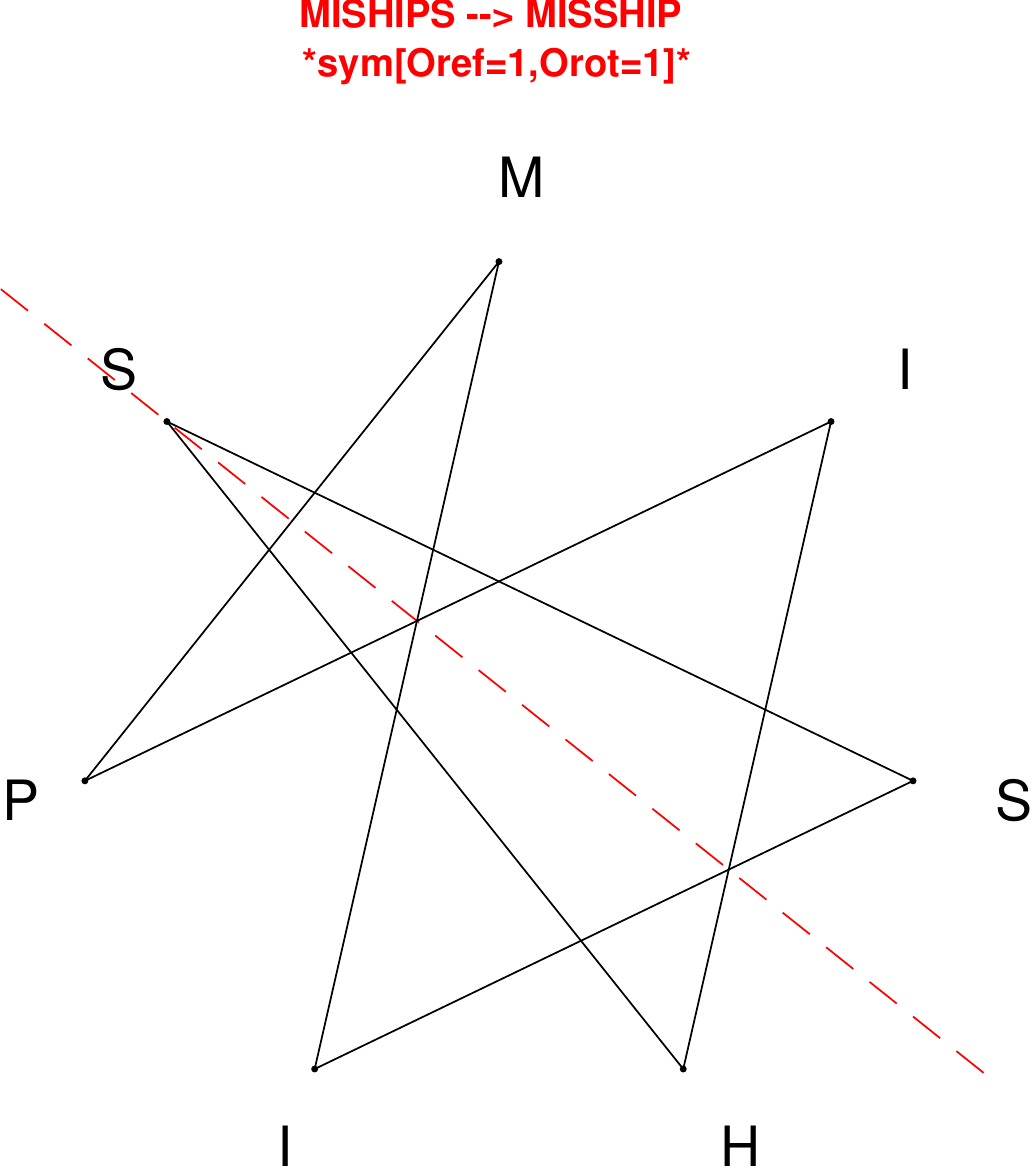}
\end{subfigure}
\hfill
\begin{subfigure}[T]{0.19\textwidth}
\centering
\includegraphics[width=\textwidth]{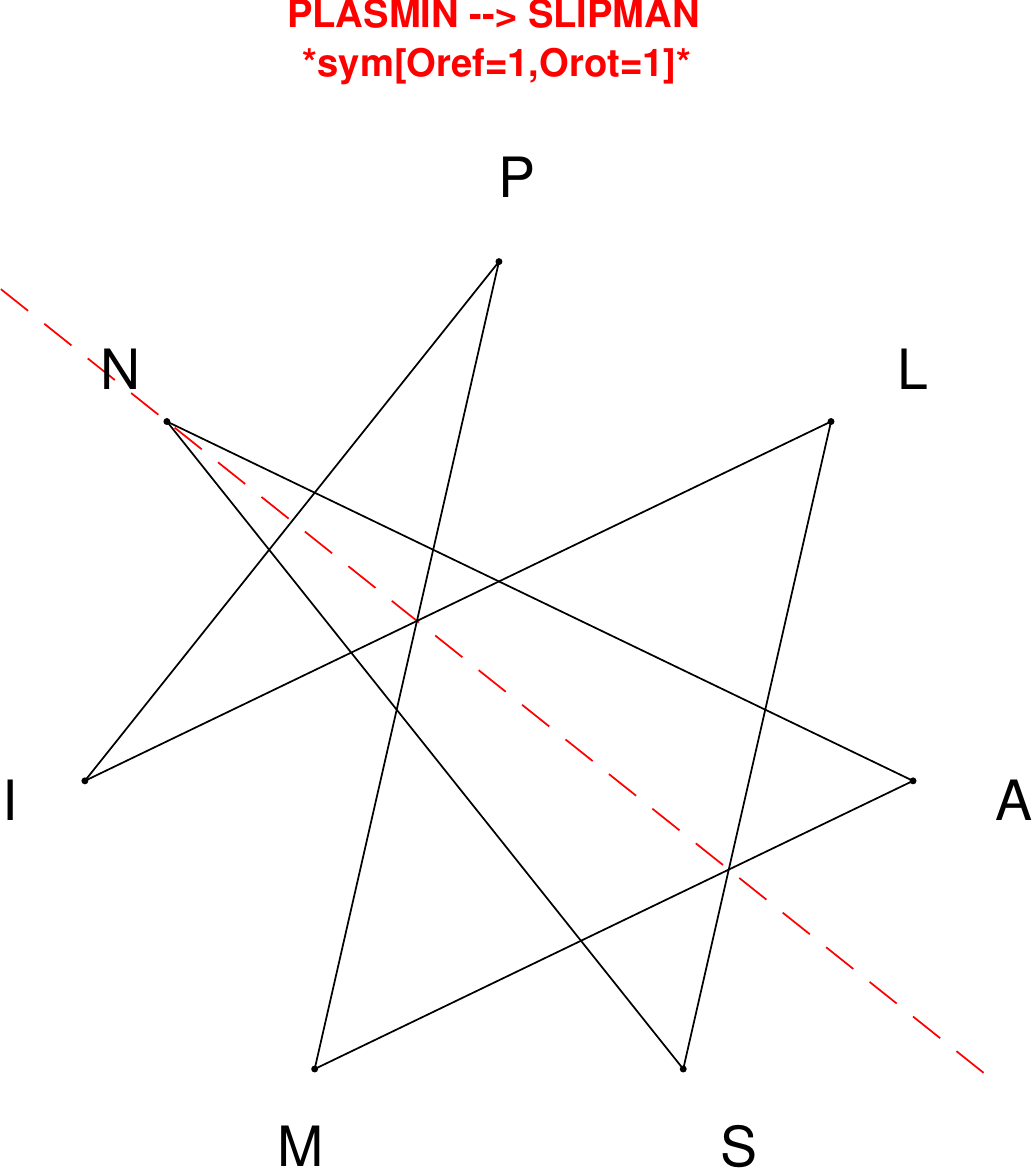}
\end{subfigure}
\end{figure}

\begin{figure}[H]
\centering
\begin{subfigure}[T]{0.19\textwidth}
\centering
\includegraphics[width=\textwidth]{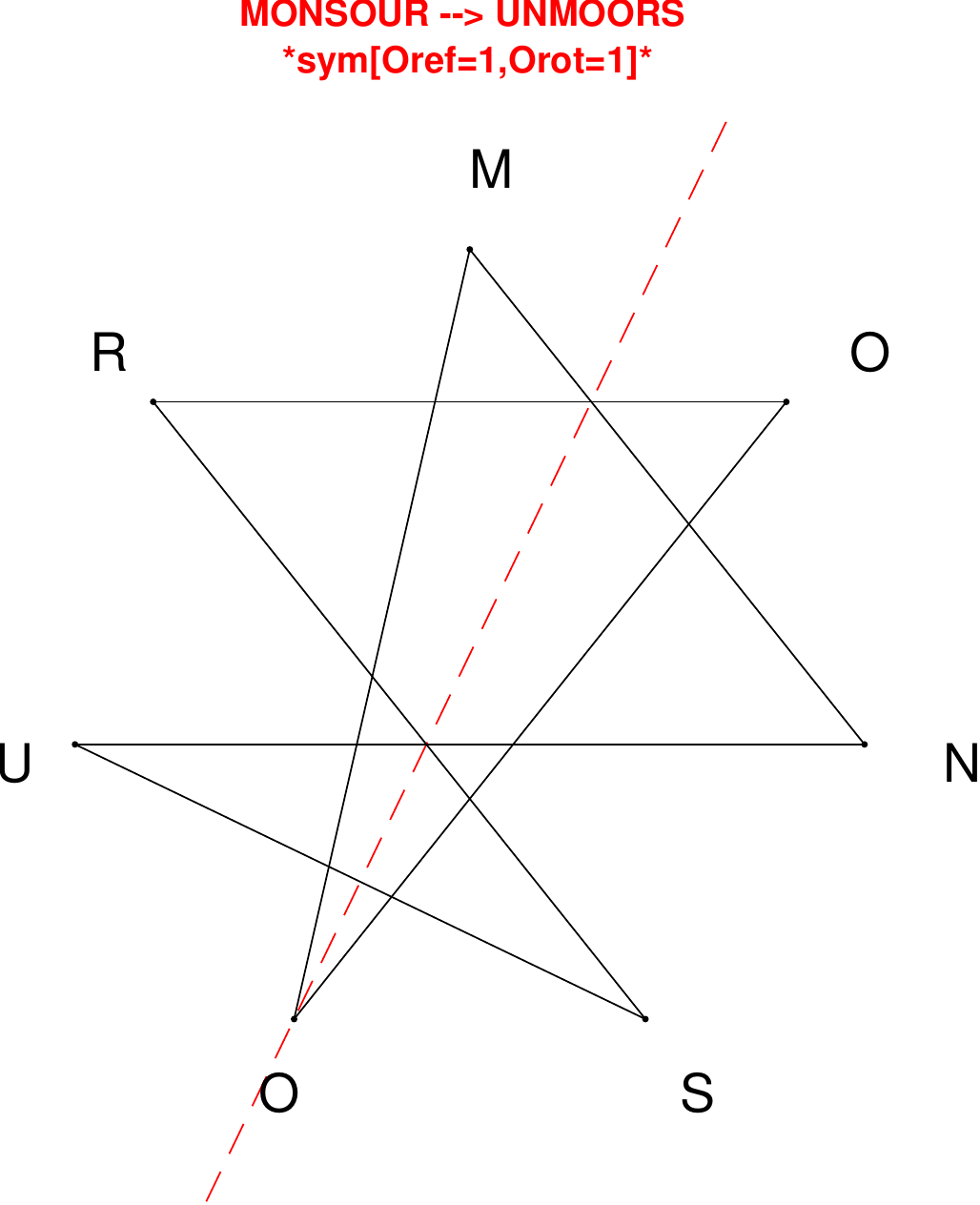}
\end{subfigure}
\hfill
\begin{subfigure}[T]{0.19\textwidth}
\centering
\includegraphics[width=\textwidth]{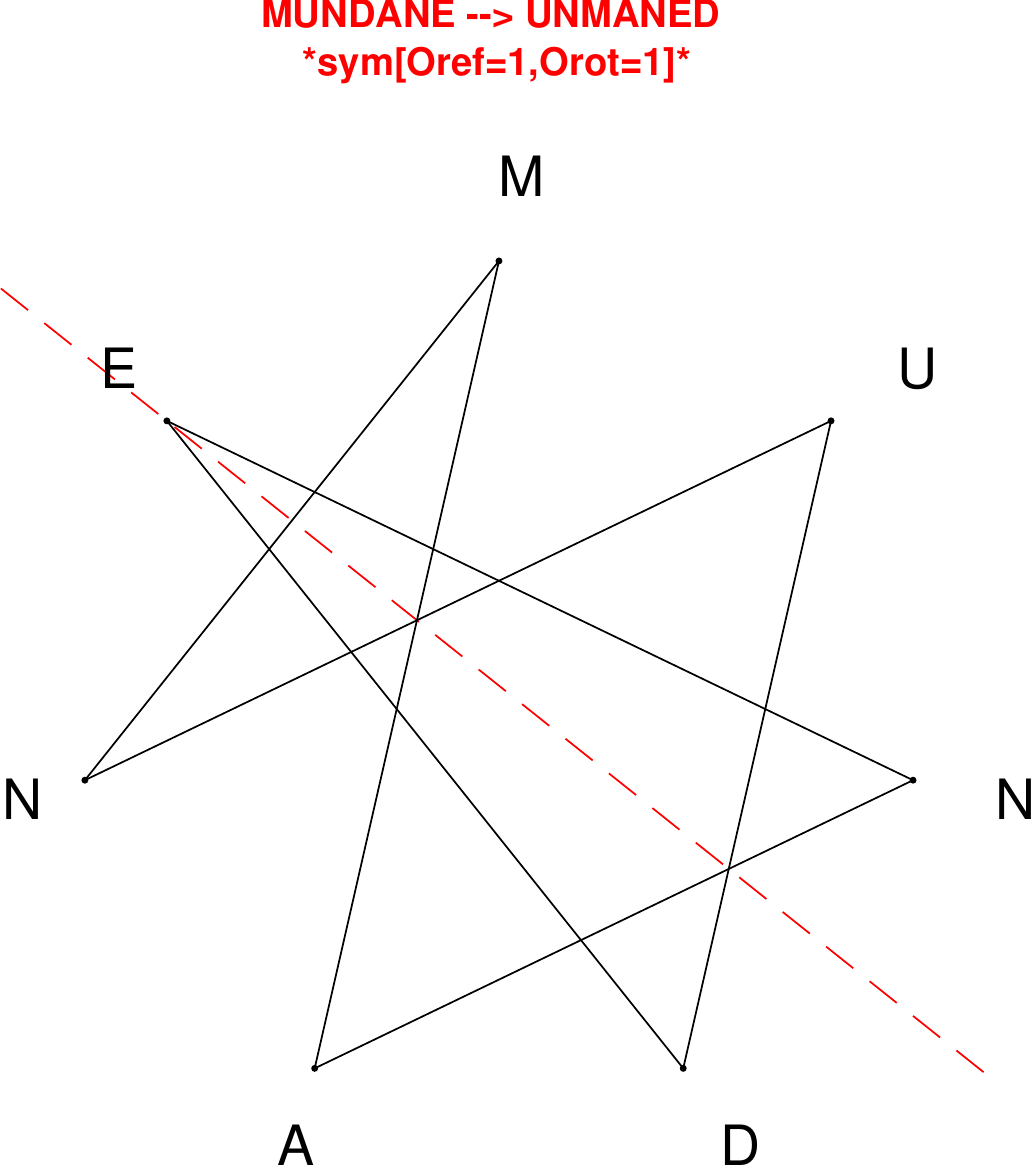}
\end{subfigure}
\hfill
\begin{subfigure}[T]{0.19\textwidth}
\centering
\includegraphics[width=\textwidth]{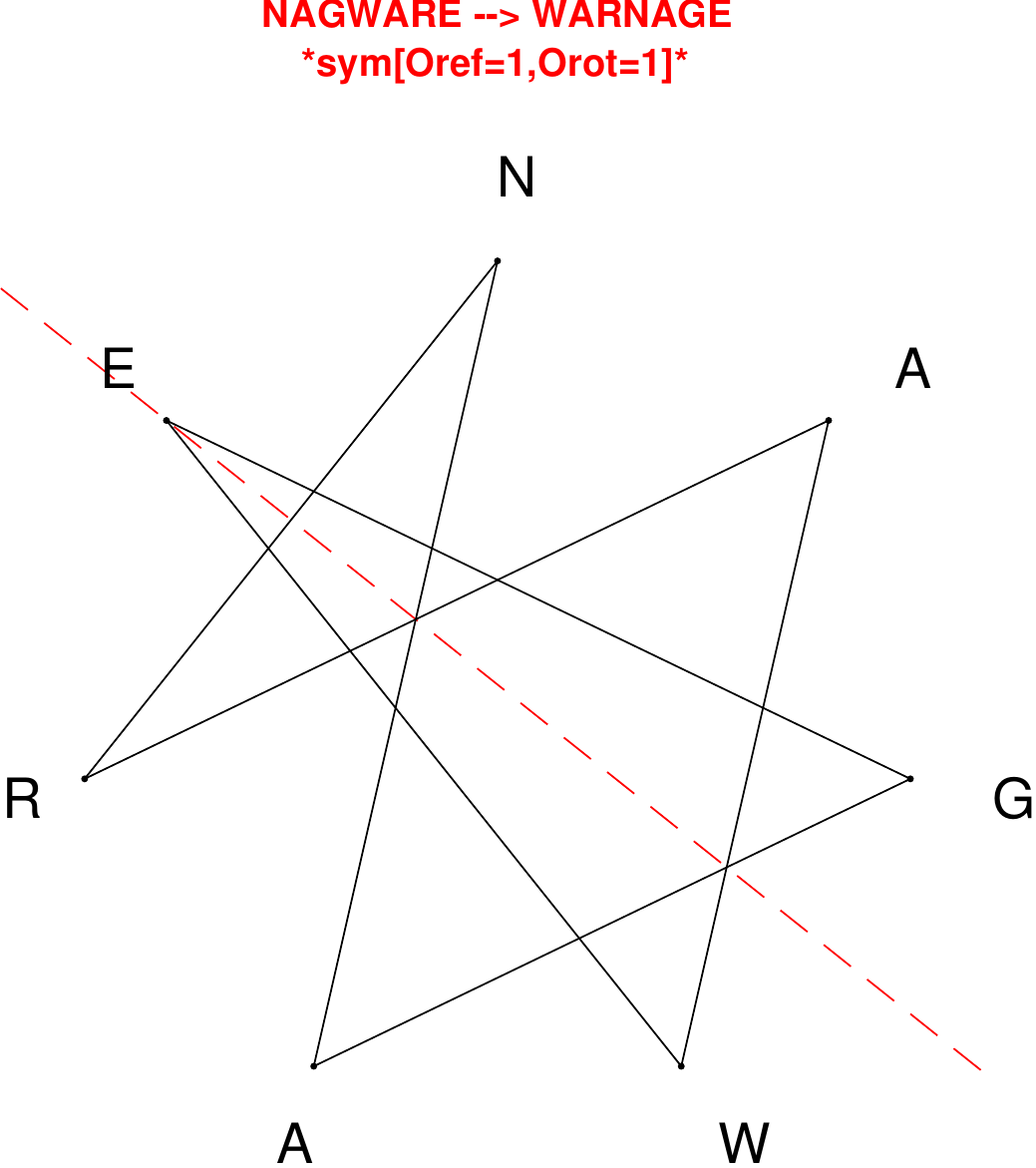}
\end{subfigure}
\hfill
\begin{subfigure}[T]{0.19\textwidth}
\centering
\includegraphics[width=\textwidth]{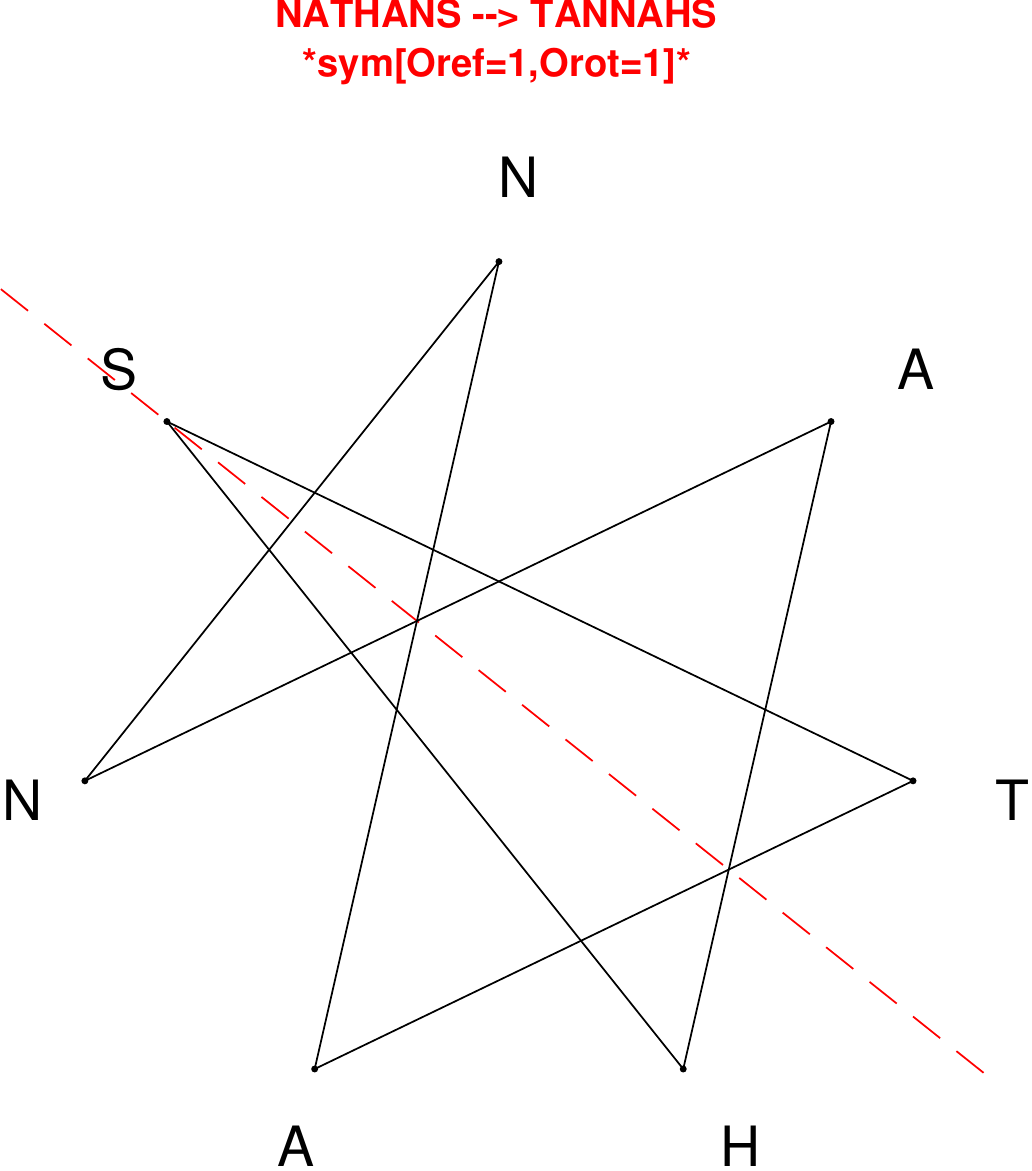}
\end{subfigure}
\hfill
\begin{subfigure}[T]{0.19\textwidth}
\centering
\includegraphics[width=\textwidth]{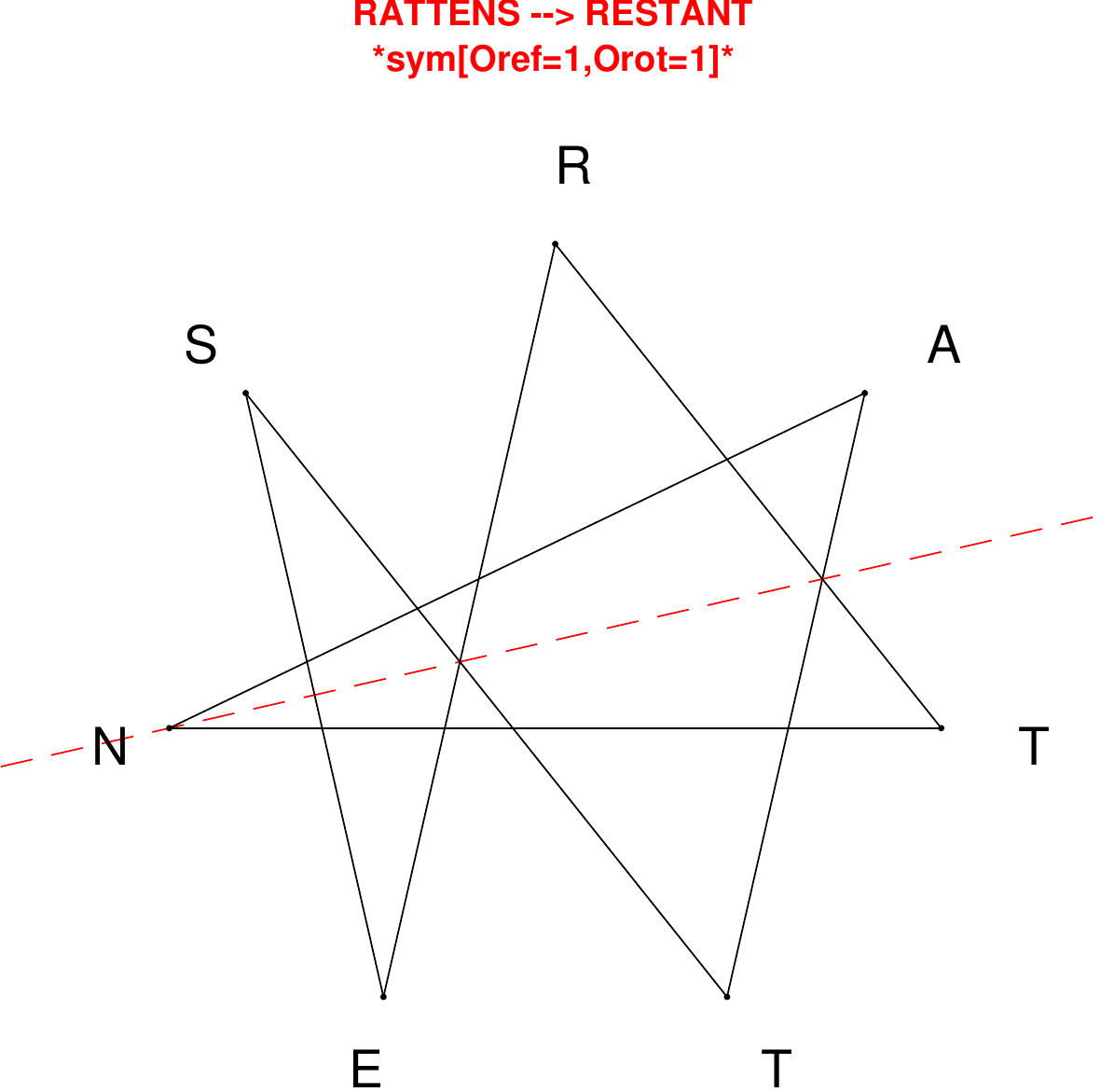}
\end{subfigure}
\end{figure}

\begin{figure}[H]
\centering
\begin{subfigure}[T]{0.19\textwidth}
\centering
\includegraphics[width=\textwidth]{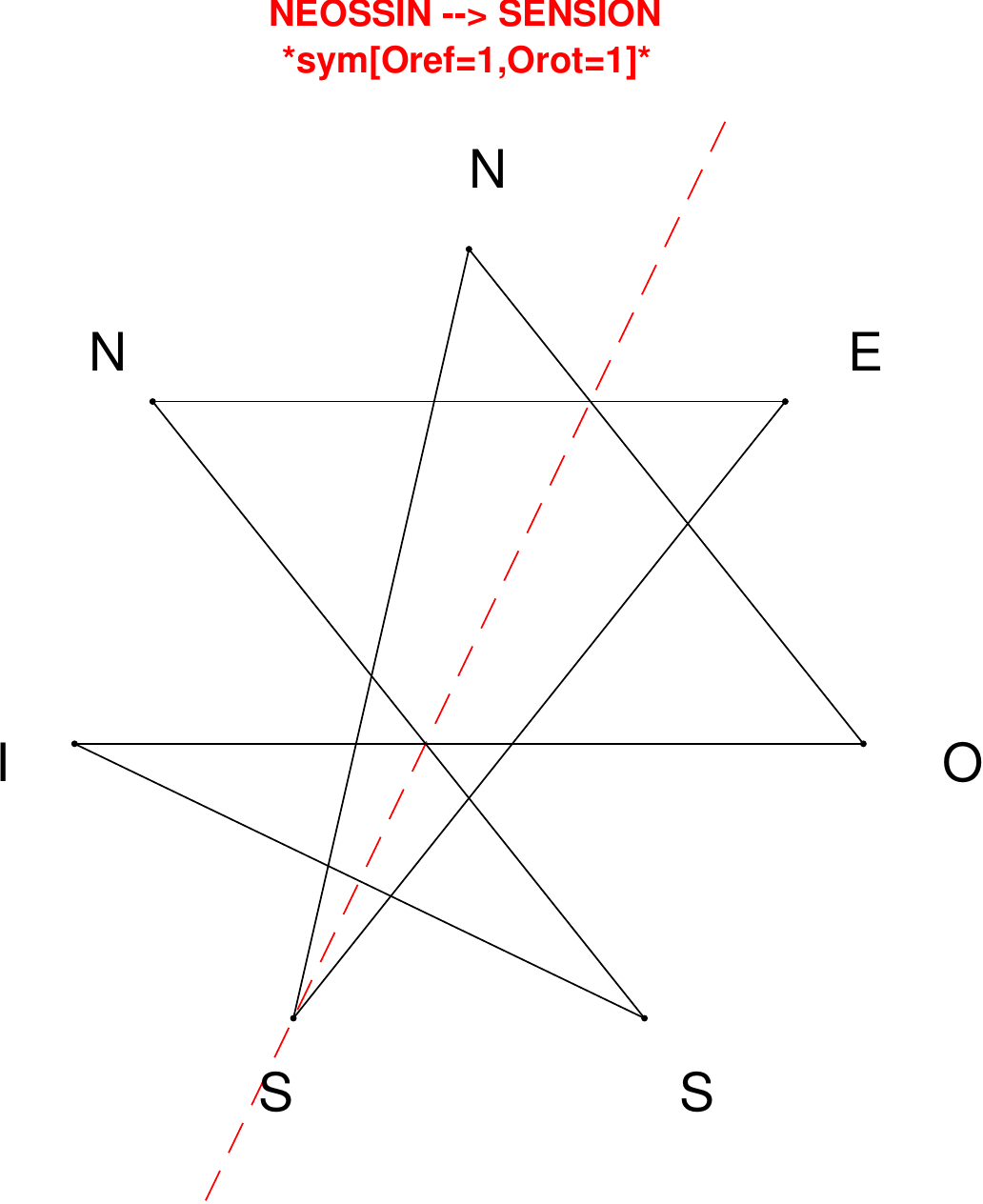}
\end{subfigure}
\hfill
\begin{subfigure}[T]{0.19\textwidth}
\centering
\includegraphics[width=\textwidth]{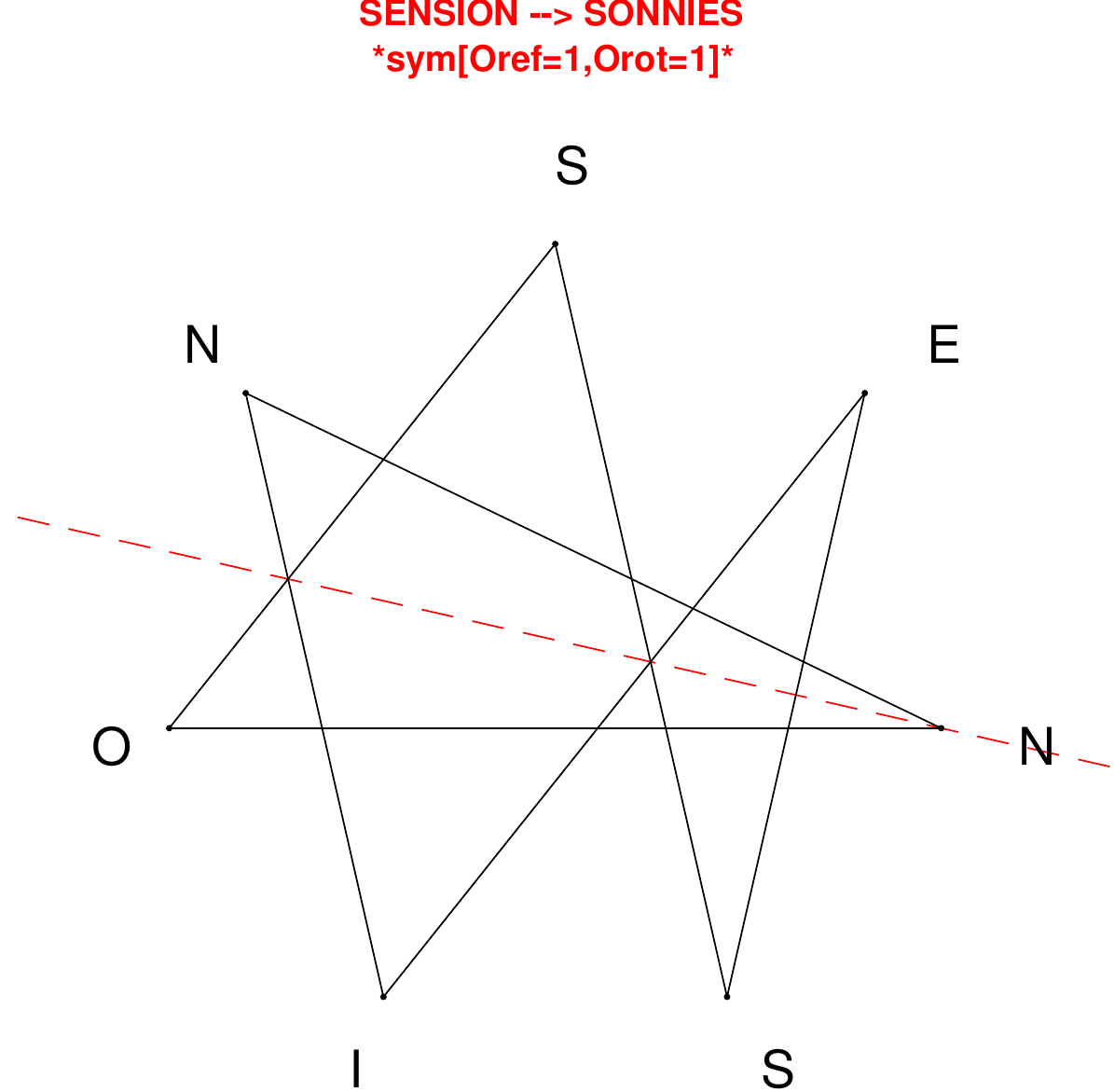}
\end{subfigure}
\hfill
\begin{subfigure}[T]{0.19\textwidth}
\centering
\includegraphics[width=\textwidth]{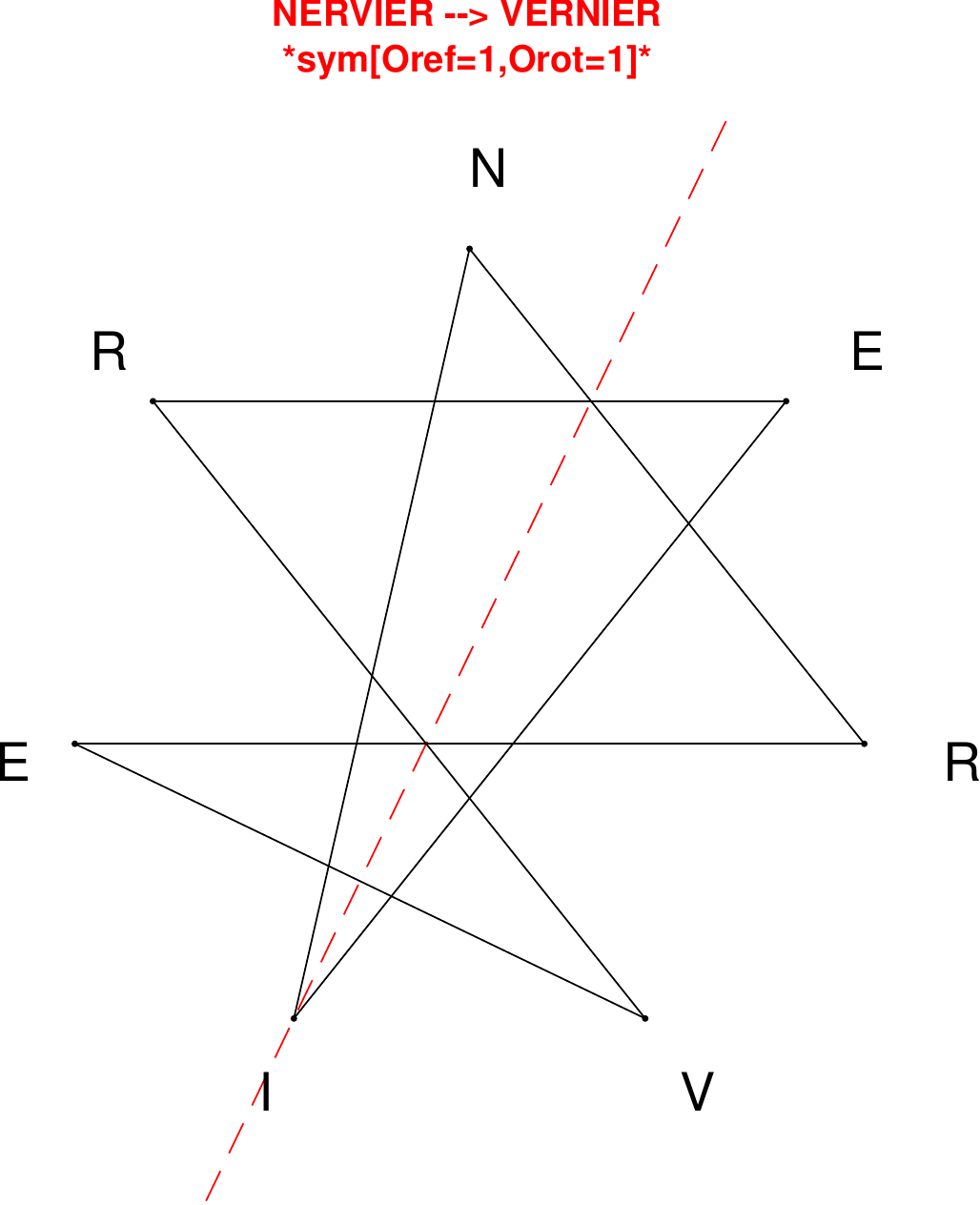}
\end{subfigure}
\hfill
\begin{subfigure}[T]{0.19\textwidth}
\centering
\includegraphics[width=\textwidth]{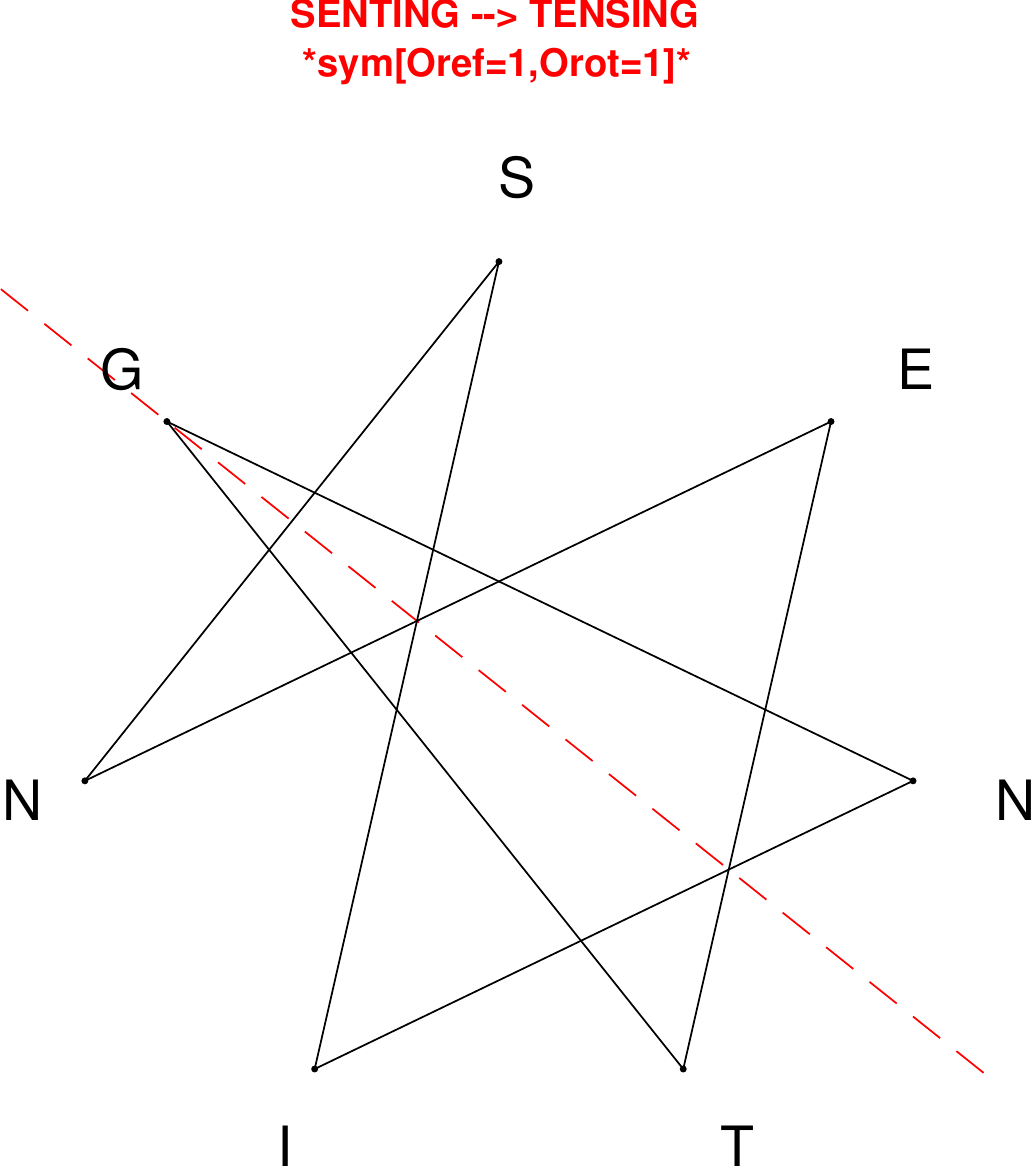}
\end{subfigure}
\hfill
\begin{subfigure}[T]{0.19\textwidth}
\centering
\includegraphics[width=\textwidth]{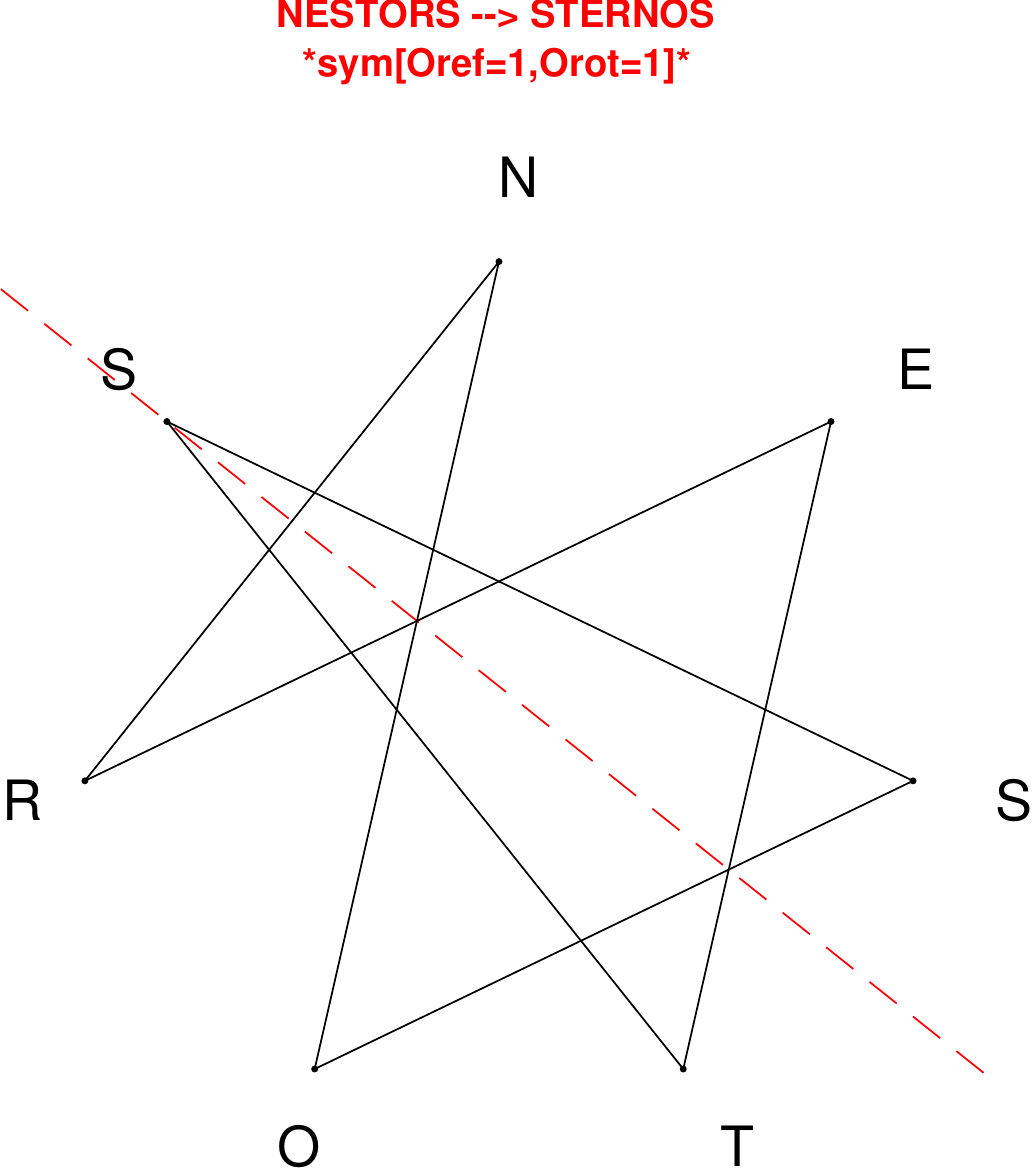}
\end{subfigure}
\end{figure}

\begin{figure}[H]
\centering
\begin{subfigure}[T]{0.19\textwidth}
\centering
\includegraphics[width=\textwidth]{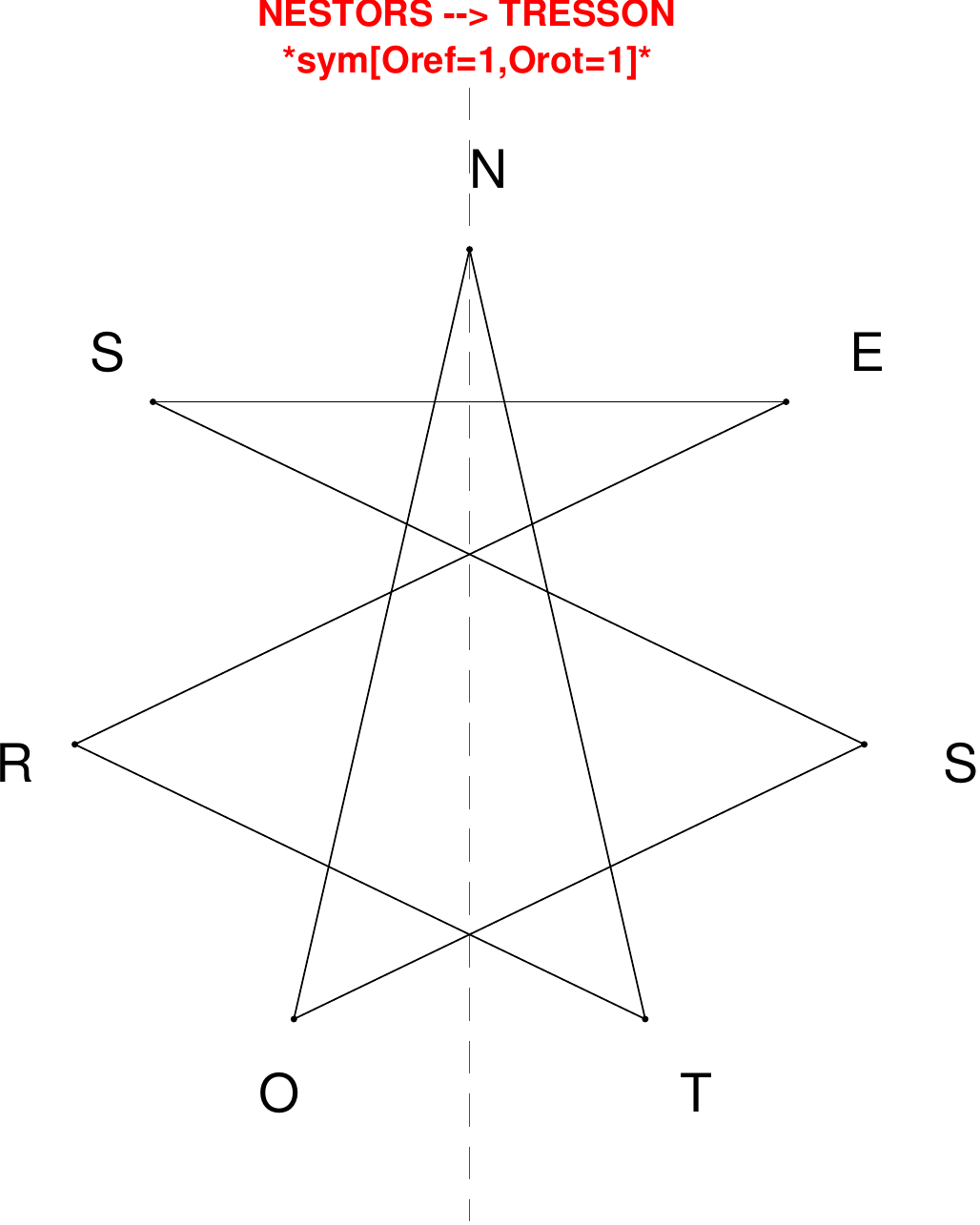}
\end{subfigure}
\hfill
\begin{subfigure}[T]{0.19\textwidth}
\centering
\includegraphics[width=\textwidth]{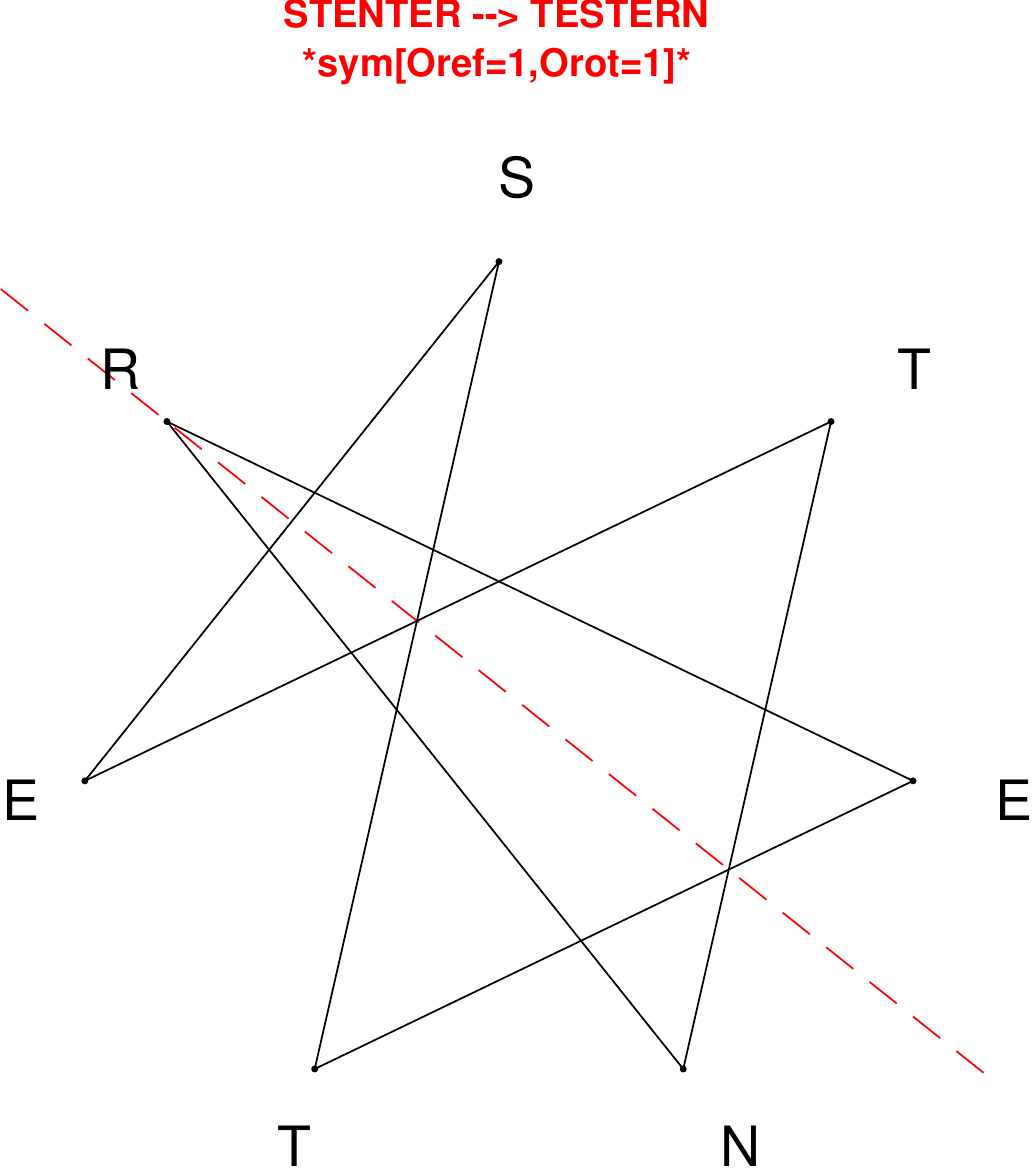}
\end{subfigure}
\hfill
\begin{subfigure}[T]{0.19\textwidth}
\centering
\includegraphics[width=\textwidth]{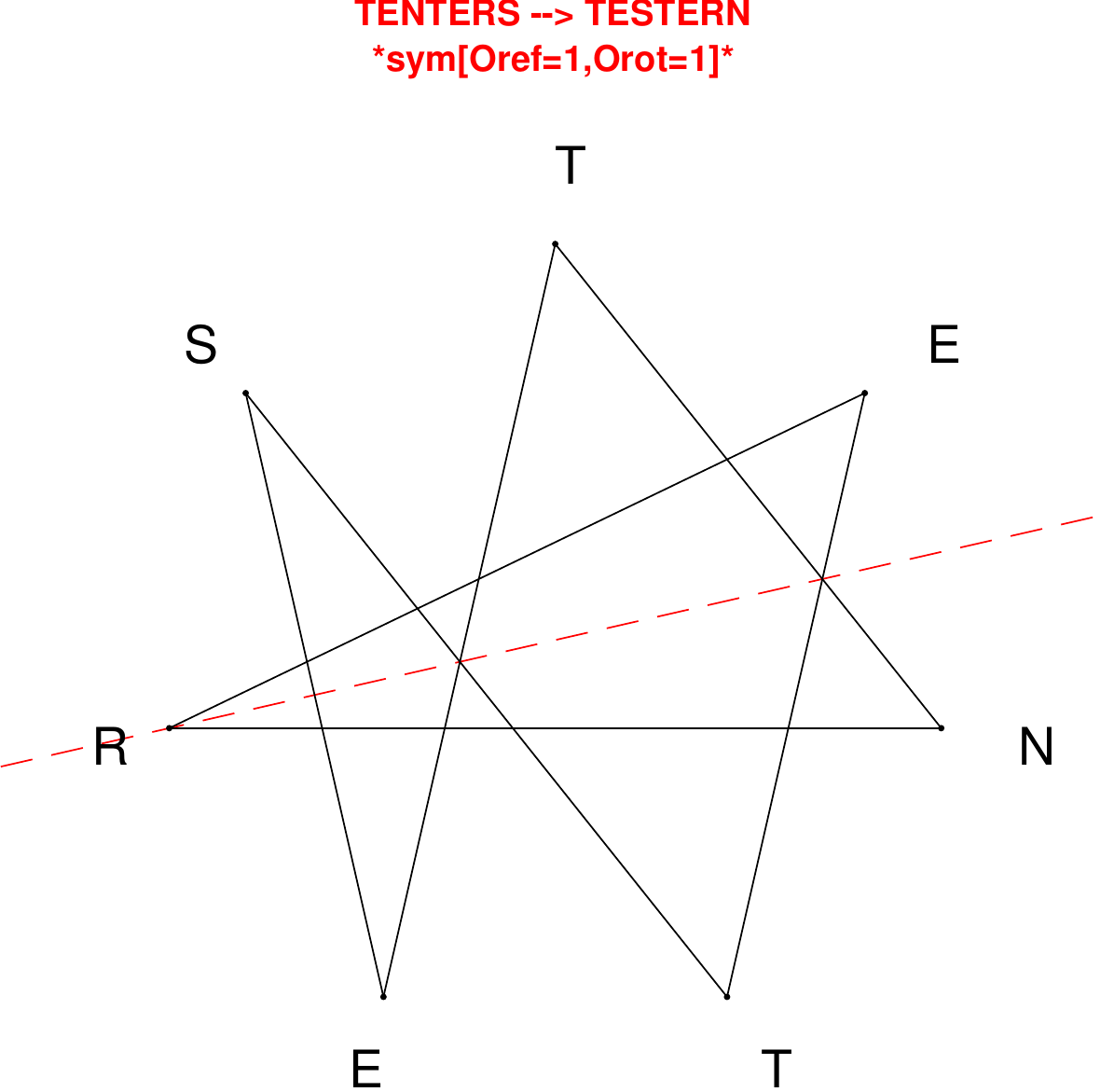}
\end{subfigure}
\hfill
\begin{subfigure}[T]{0.19\textwidth}
\centering
\includegraphics[width=\textwidth]{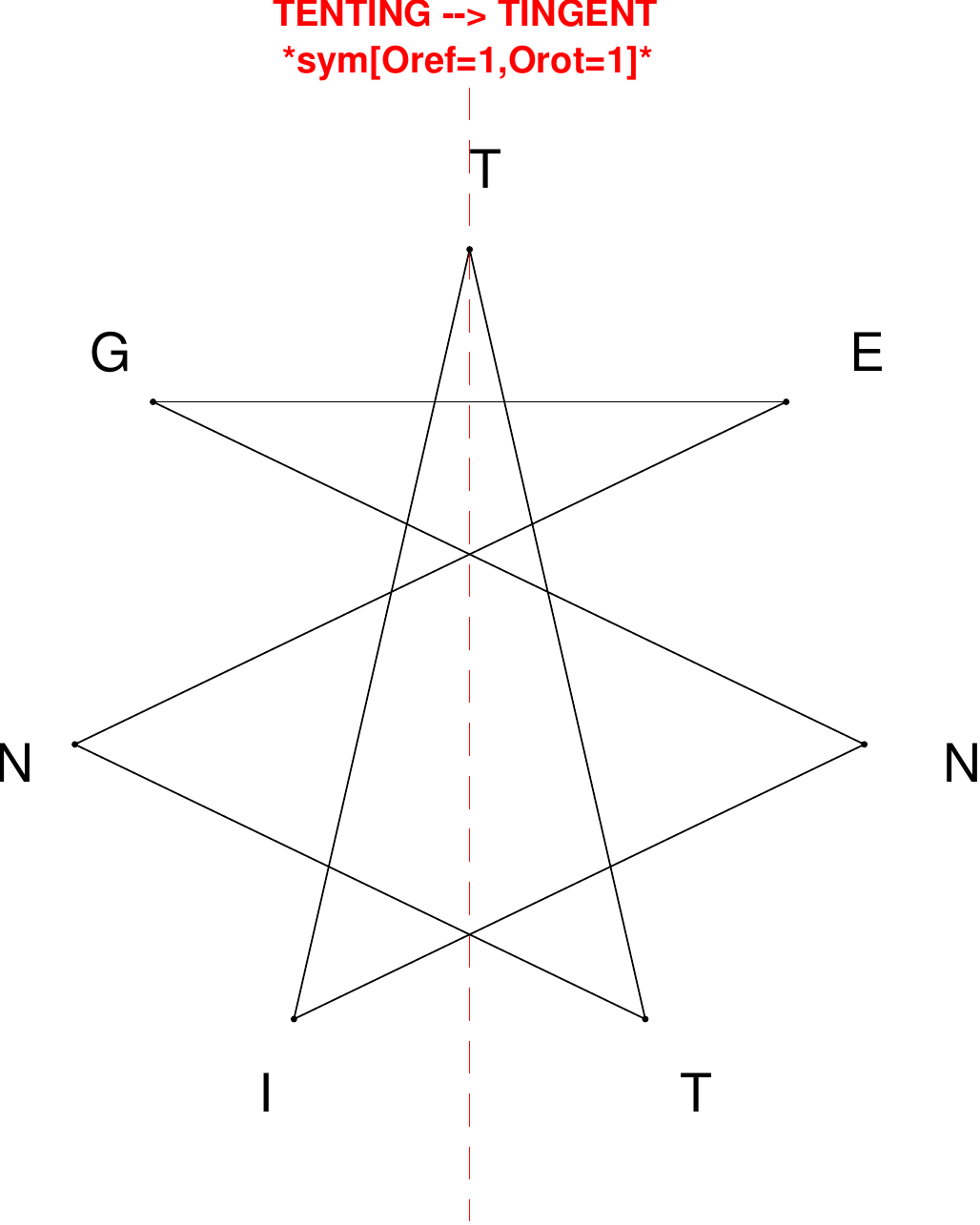}
\end{subfigure}
\hfill
\begin{subfigure}[T]{0.19\textwidth}
\centering
\includegraphics[width=\textwidth]{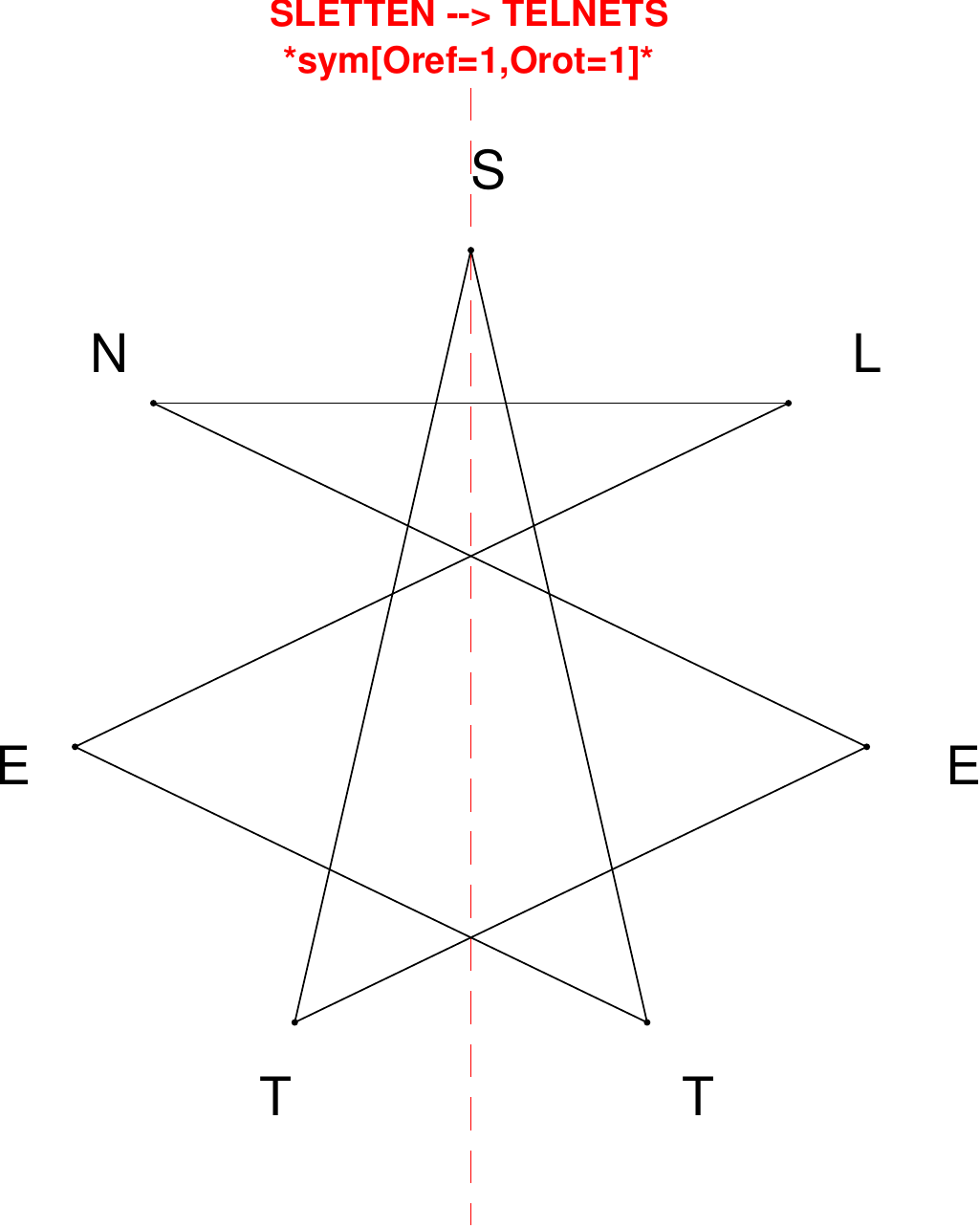}
\end{subfigure}
\end{figure}

\begin{figure}[H]
\centering
\begin{subfigure}[T]{0.19\textwidth}
\centering
\includegraphics[width=\textwidth]{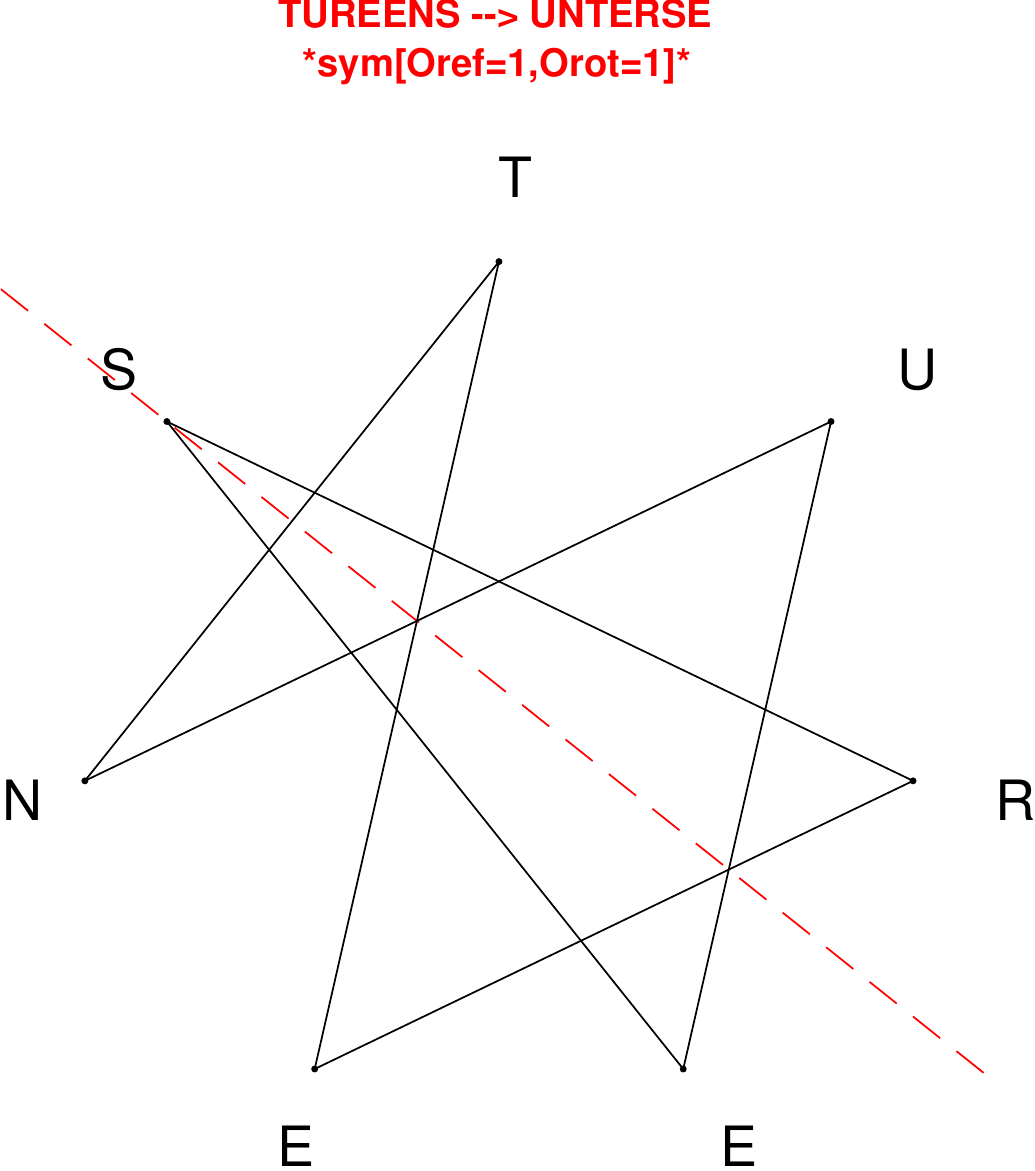}
\end{subfigure}
\hfill
\begin{subfigure}[T]{0.19\textwidth}
\centering
\includegraphics[width=\textwidth]{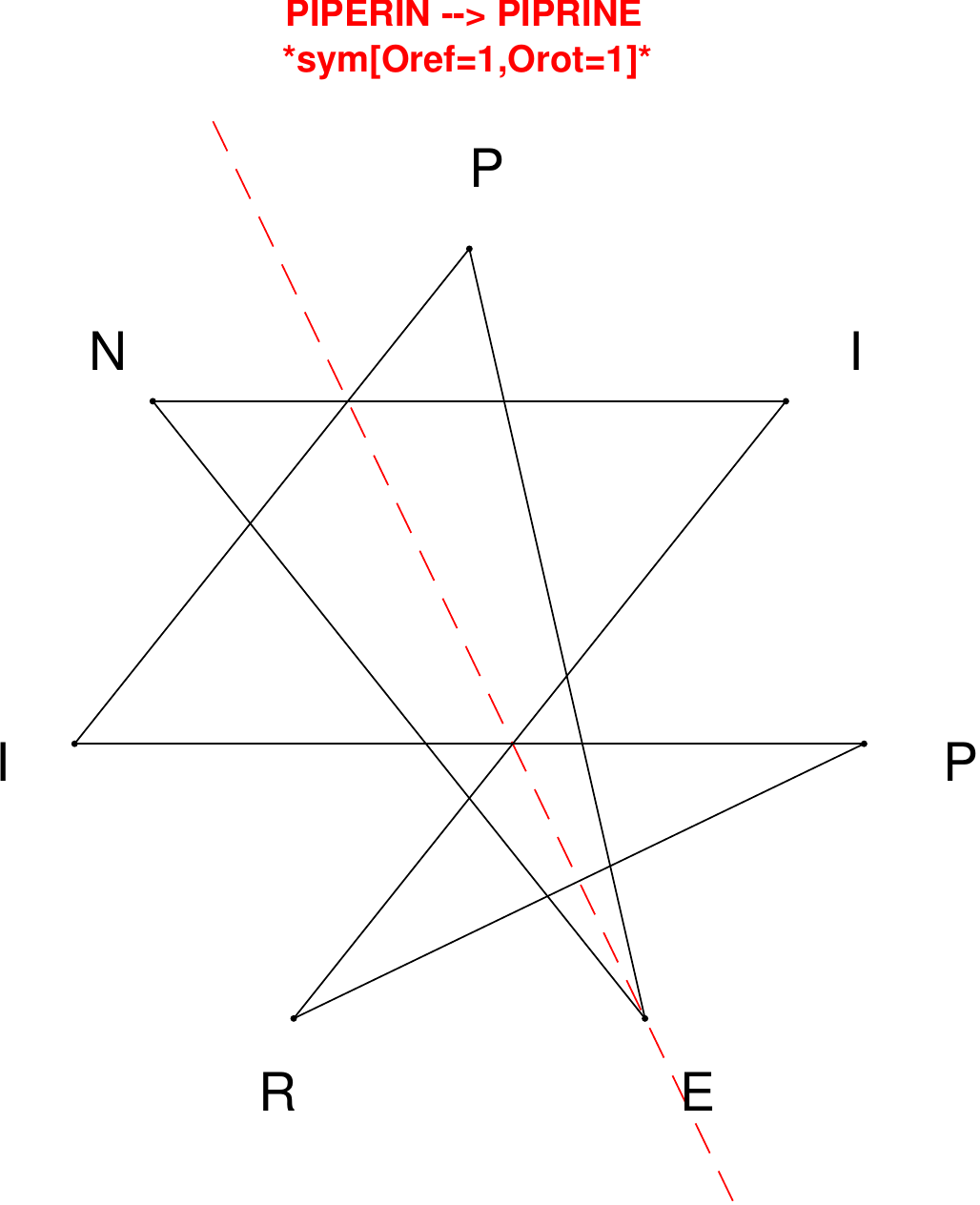}
\end{subfigure}
\hfill
\begin{subfigure}[T]{0.19\textwidth}
\centering
\includegraphics[width=\textwidth]{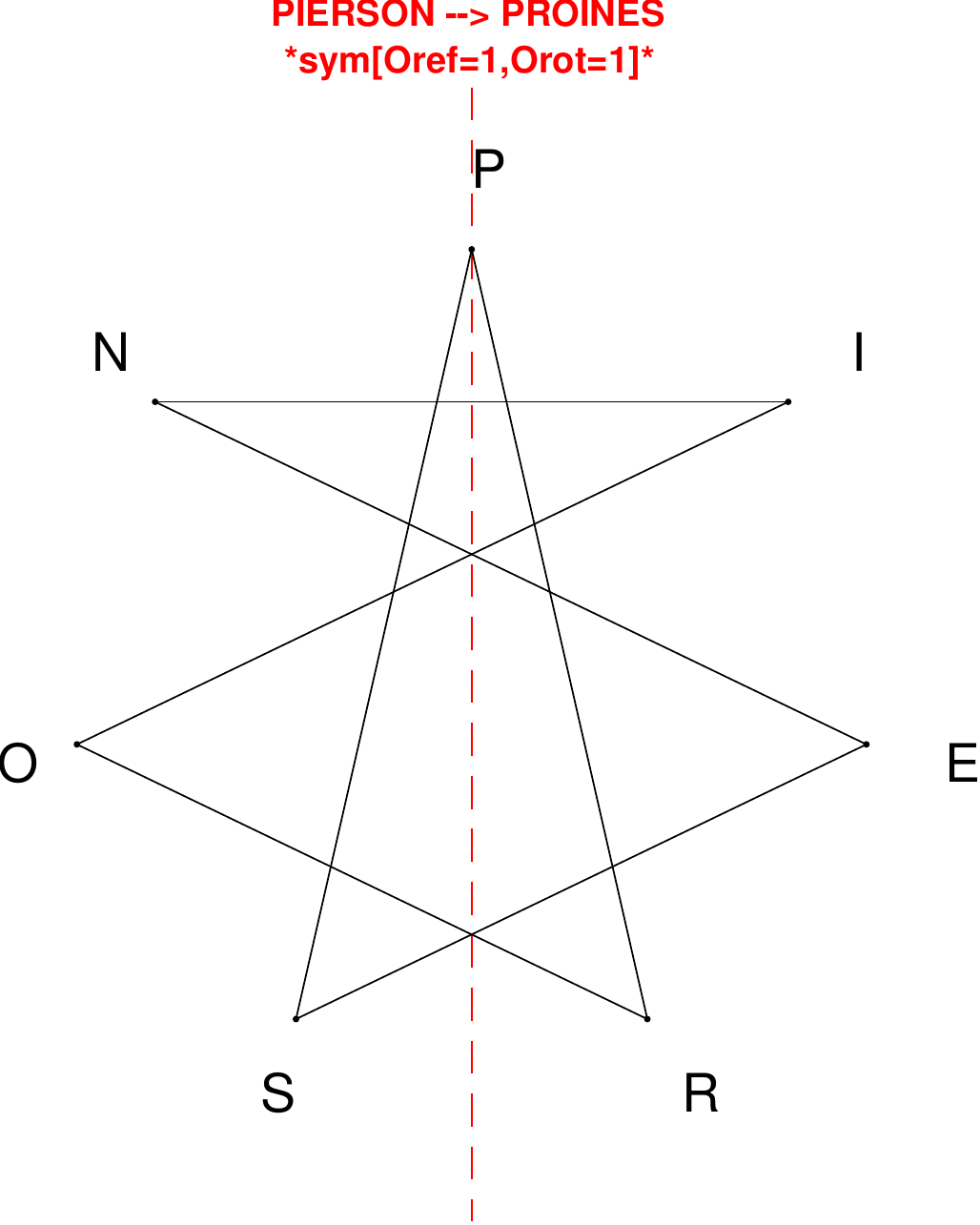}
\end{subfigure}
\hfill
\begin{subfigure}[T]{0.19\textwidth}
\centering
\includegraphics[width=\textwidth]{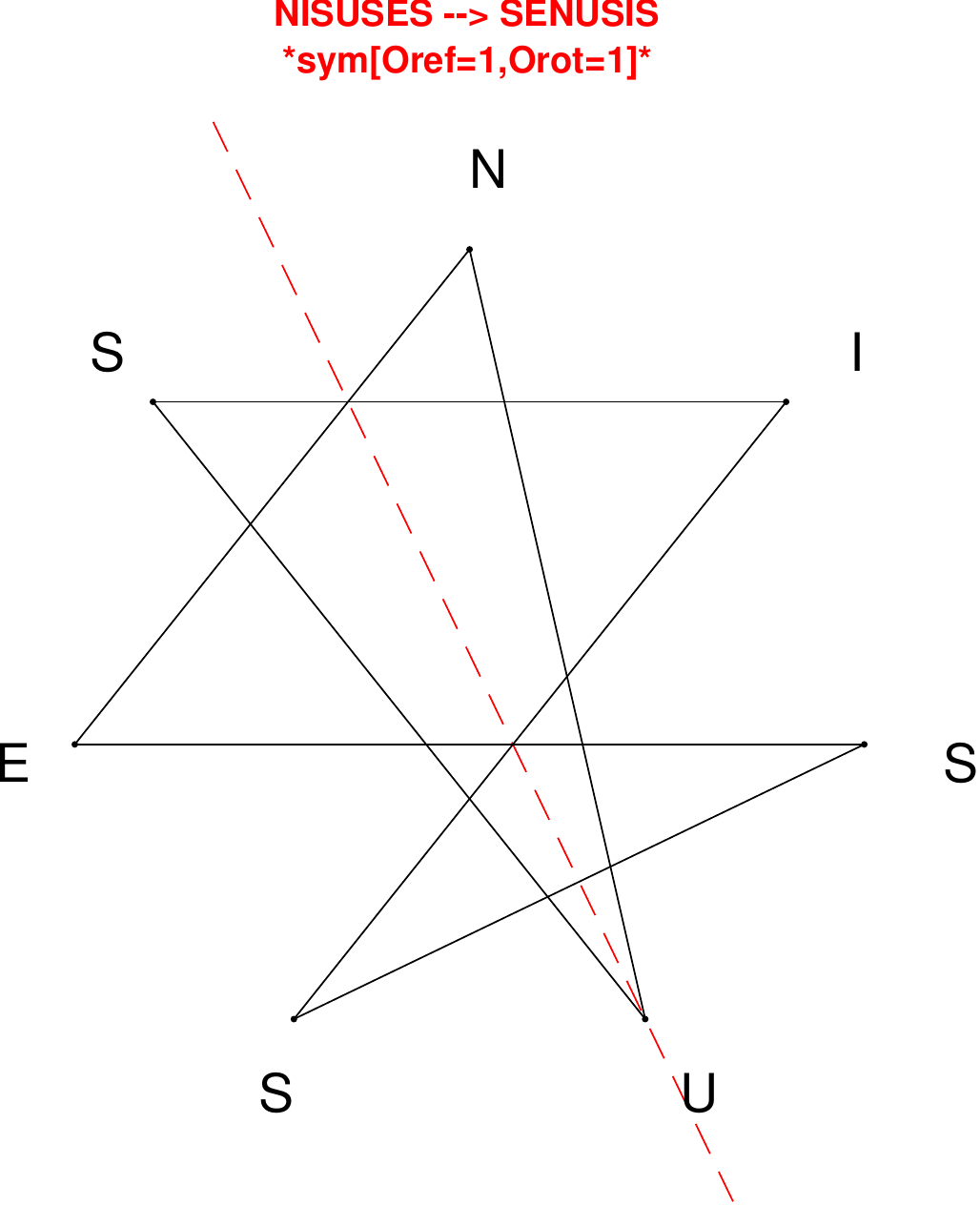}
\end{subfigure}
\hfill
\begin{subfigure}[T]{0.19\textwidth}
\centering
\includegraphics[width=\textwidth]{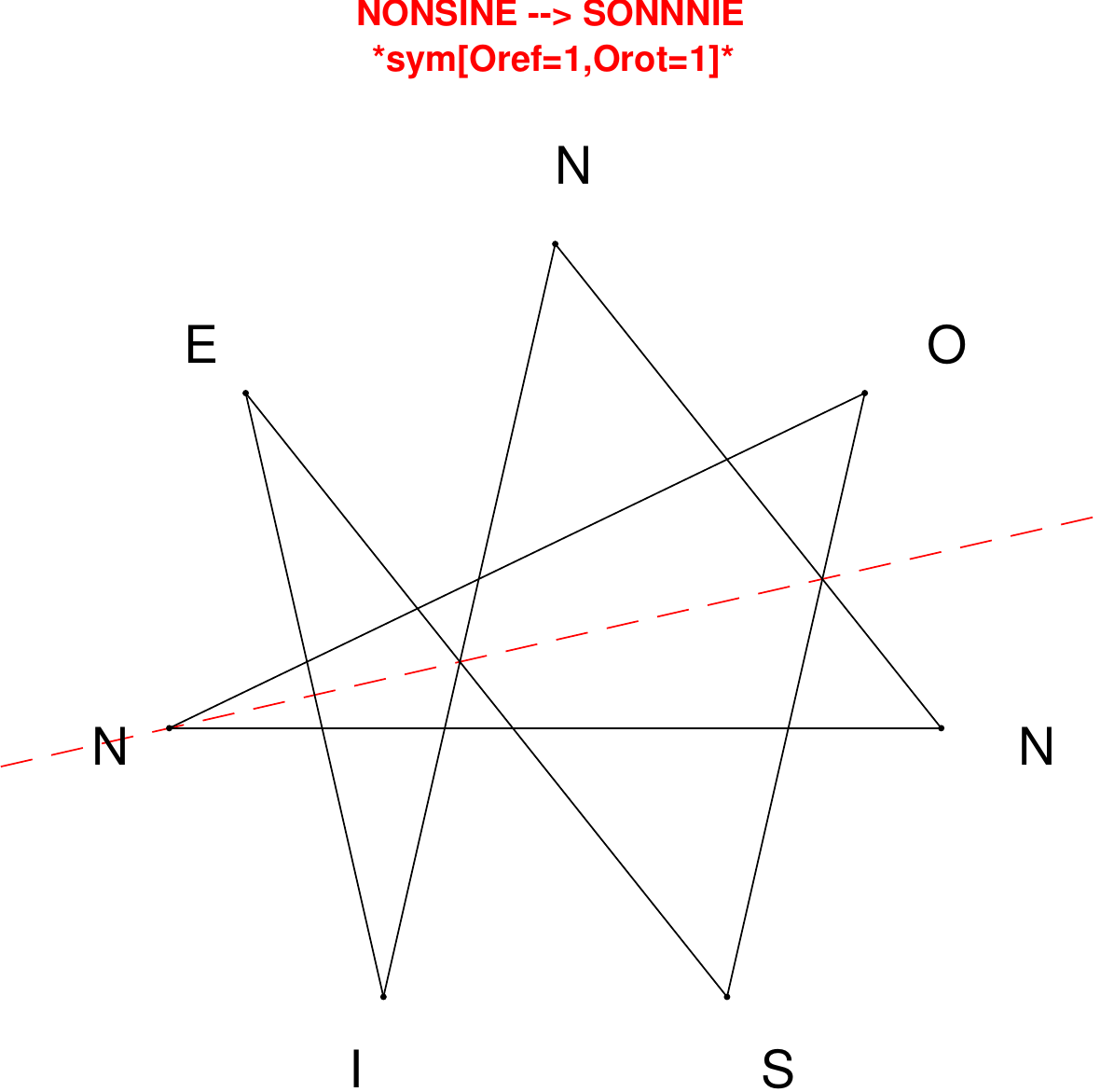}
\end{subfigure}
\end{figure}

\begin{figure}[H]
\centering
\begin{subfigure}[T]{0.19\textwidth}
\centering
\includegraphics[width=\textwidth]{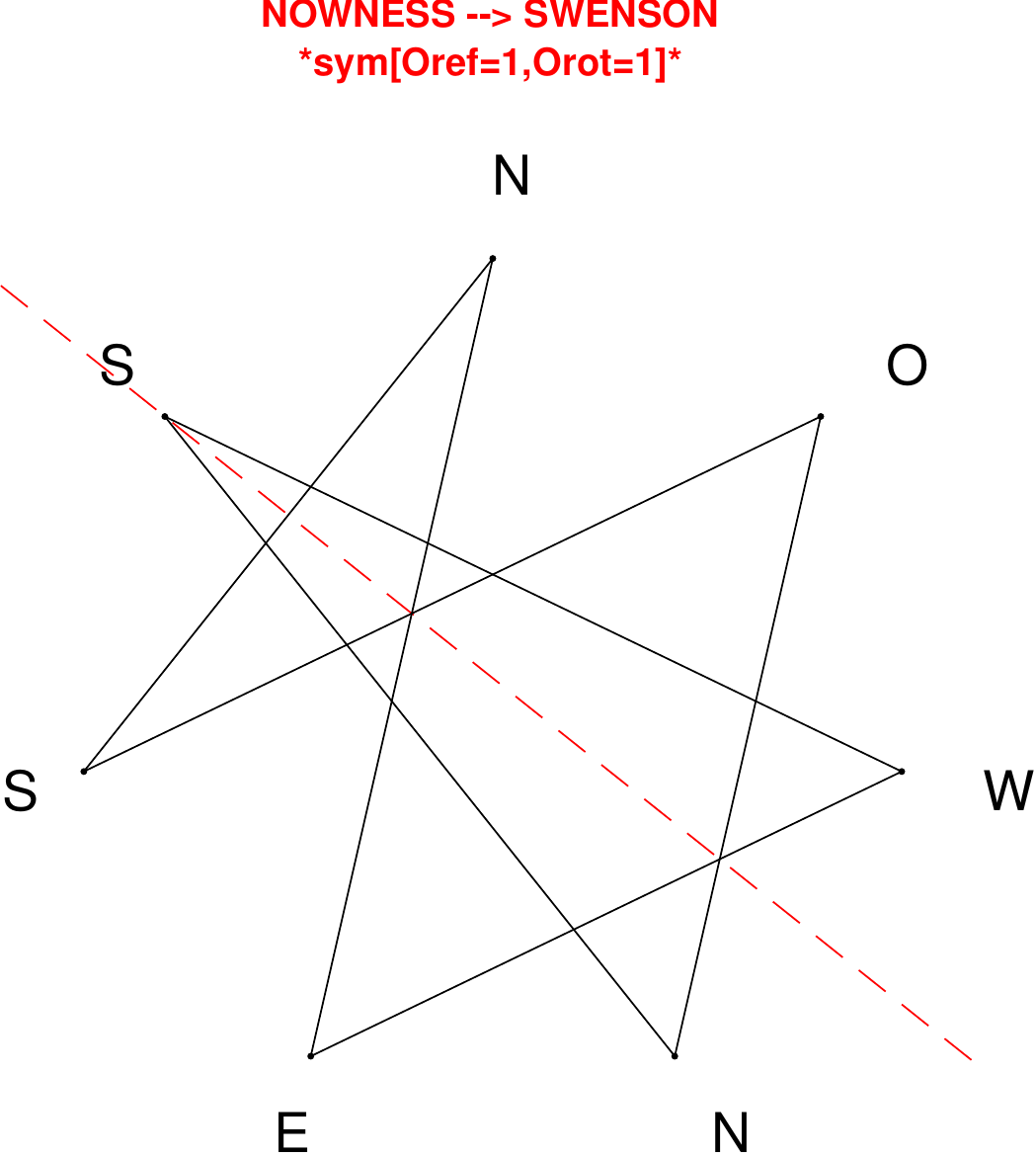}
\end{subfigure}
\hfill
\begin{subfigure}[T]{0.19\textwidth}
\centering
\includegraphics[width=\textwidth]{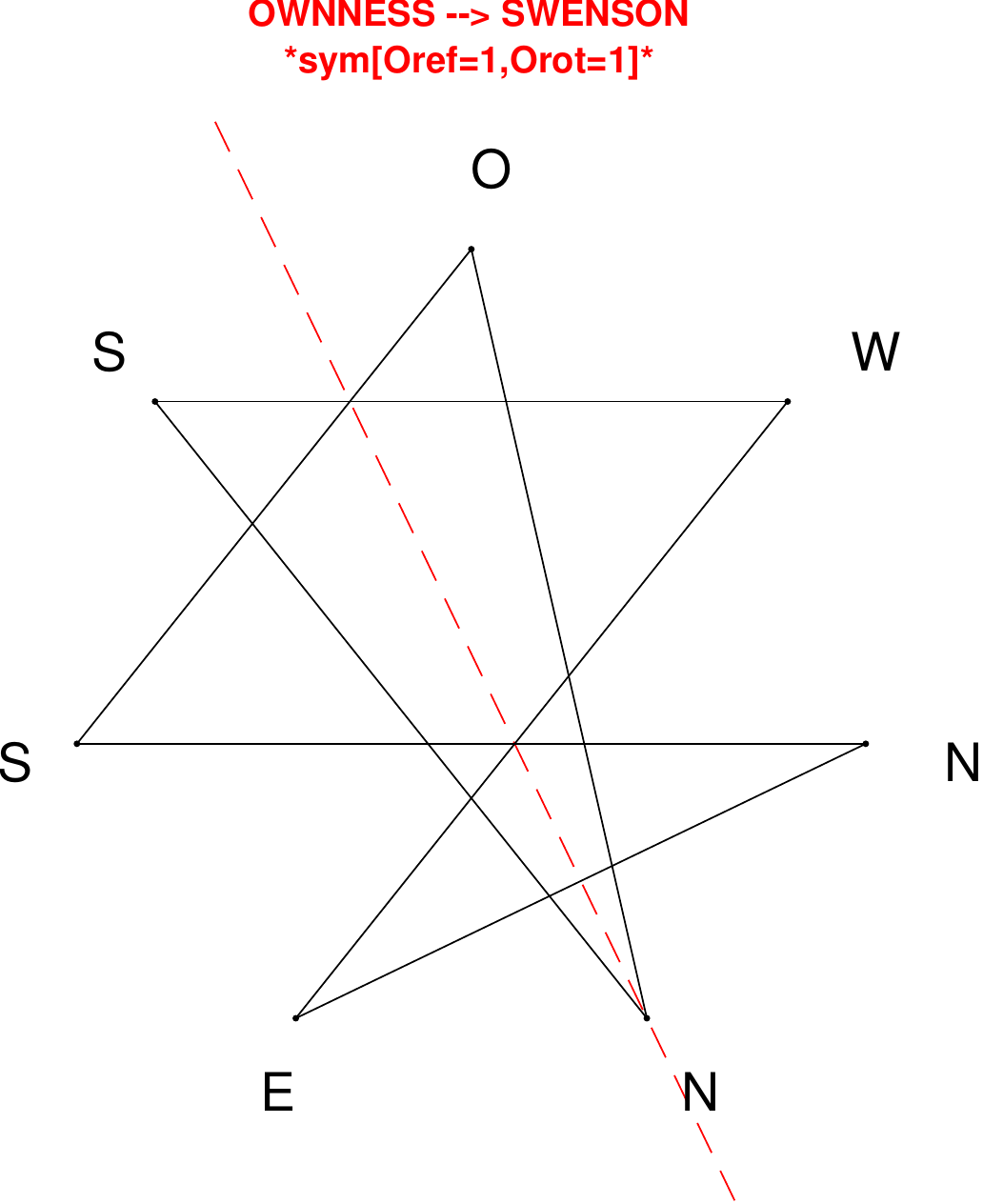}
\end{subfigure}
\hfill
\begin{subfigure}[T]{0.19\textwidth}
\centering
\includegraphics[width=\textwidth]{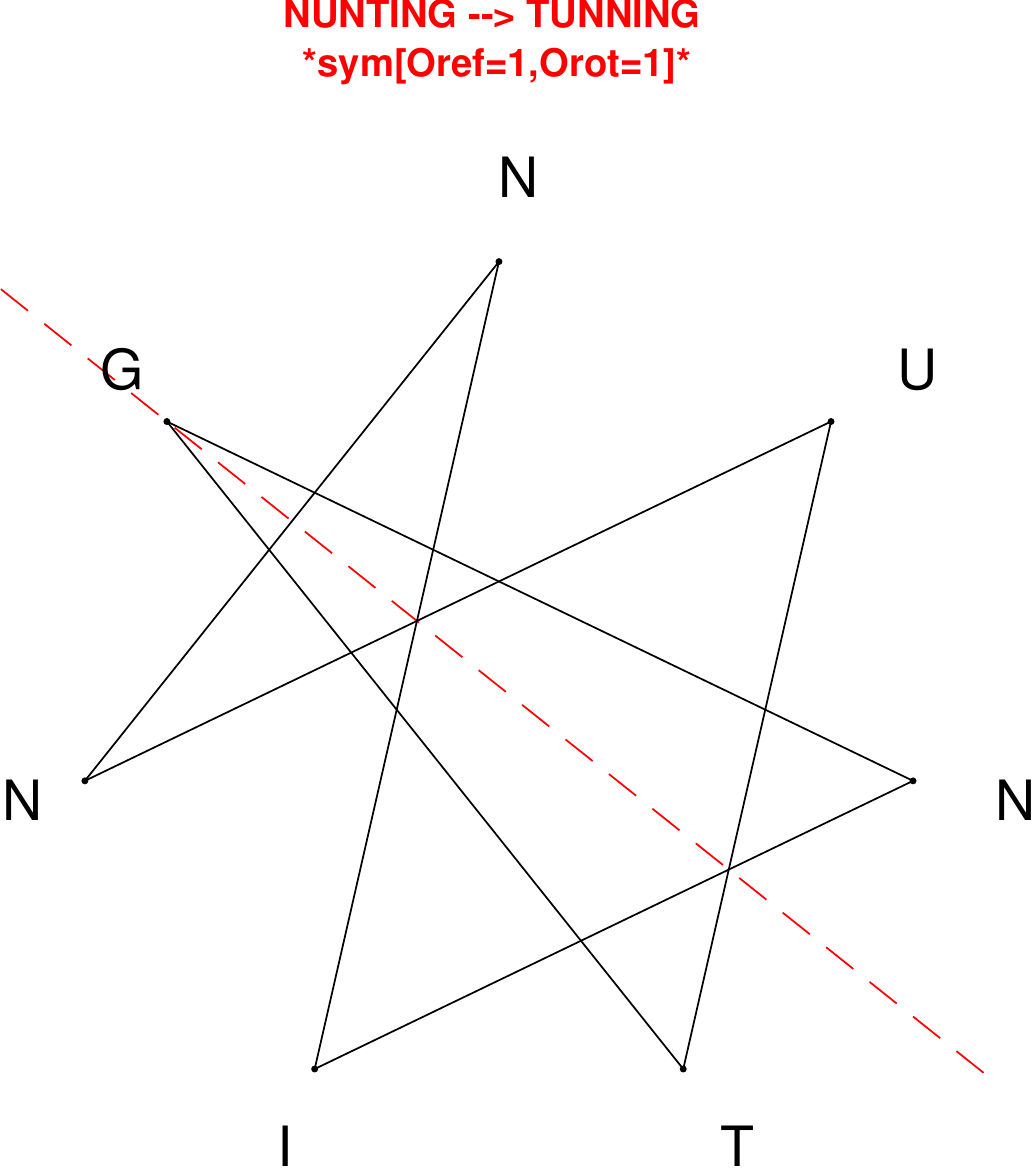}
\end{subfigure}
\hfill
\begin{subfigure}[T]{0.19\textwidth}
\centering
\includegraphics[width=\textwidth]{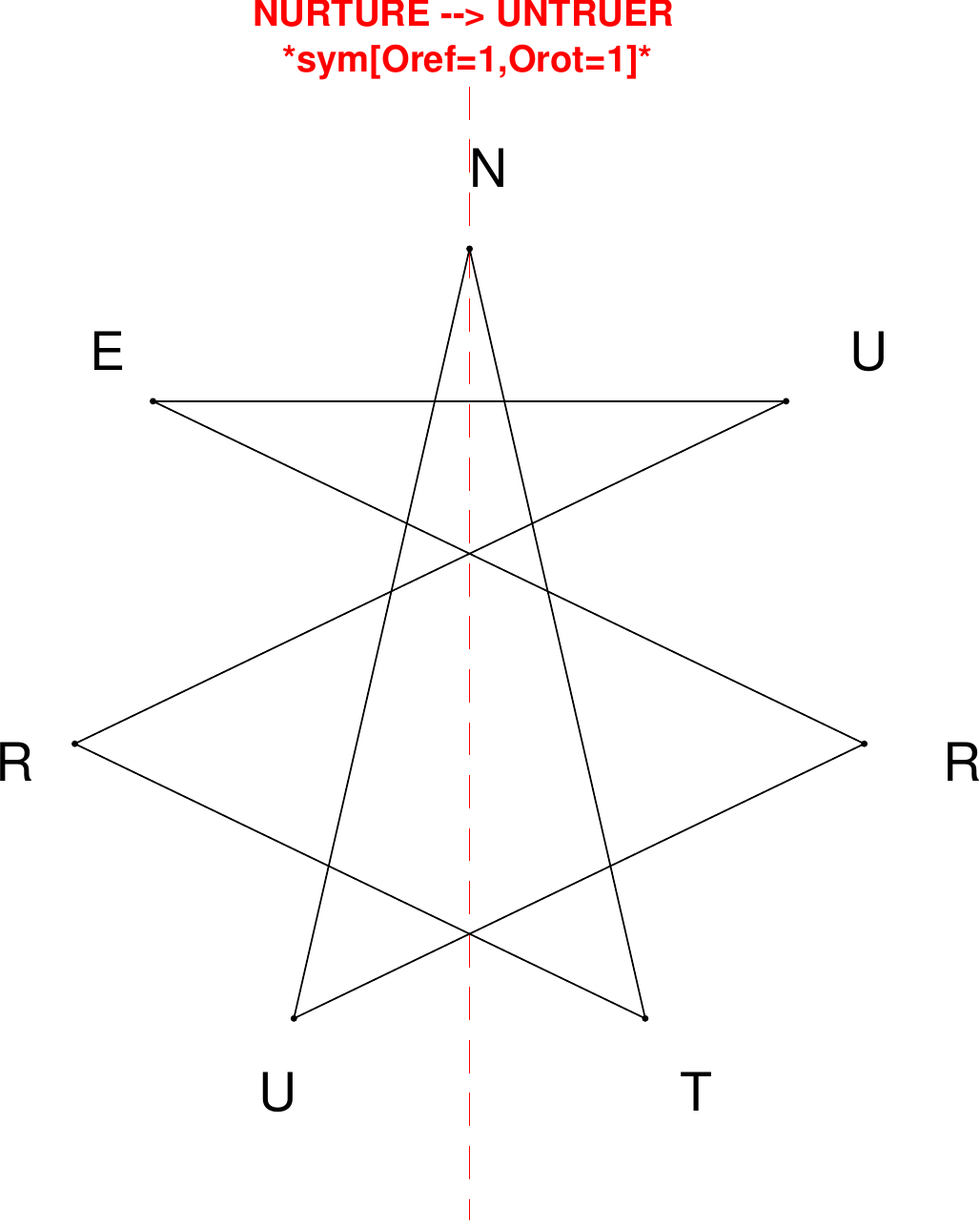}
\end{subfigure}
\hfill
\begin{subfigure}[T]{0.19\textwidth}
\centering
\includegraphics[width=\textwidth]{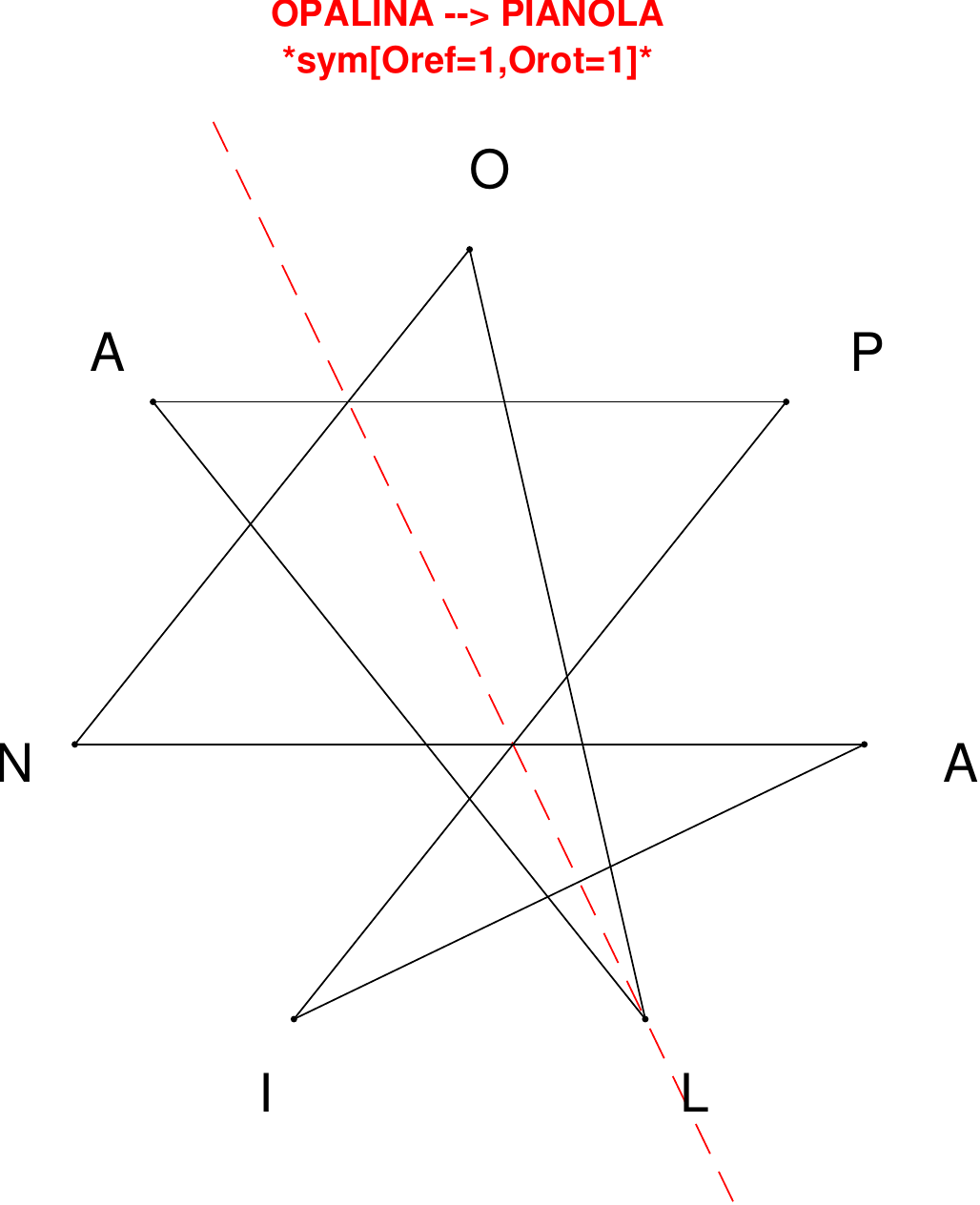}
\end{subfigure}
\end{figure}

\begin{figure}[H]
\centering
\begin{subfigure}[T]{0.19\textwidth}
\centering
\includegraphics[width=\textwidth]{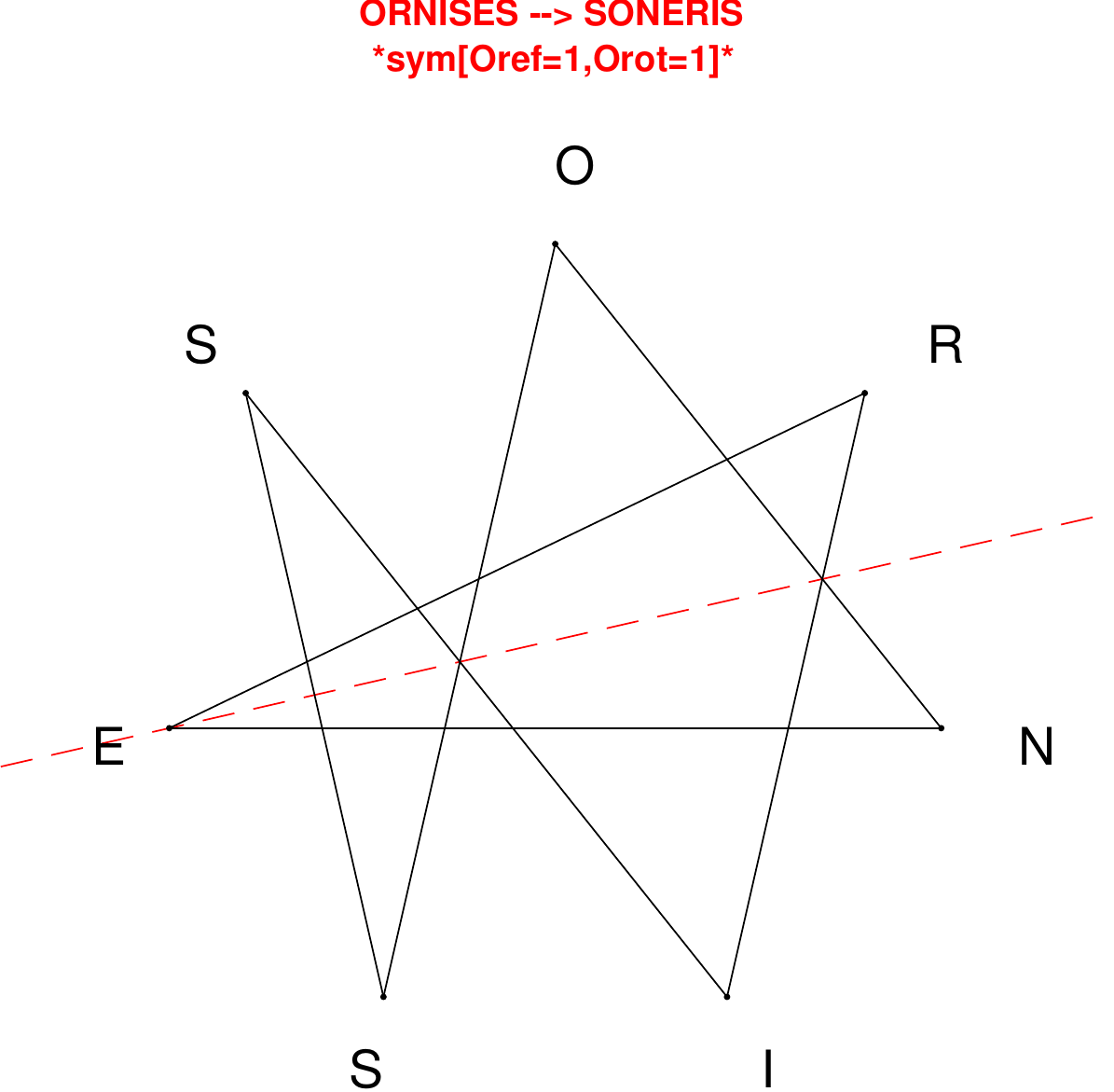}
\end{subfigure}
\hfill
\begin{subfigure}[T]{0.19\textwidth}
\centering
\includegraphics[width=\textwidth]{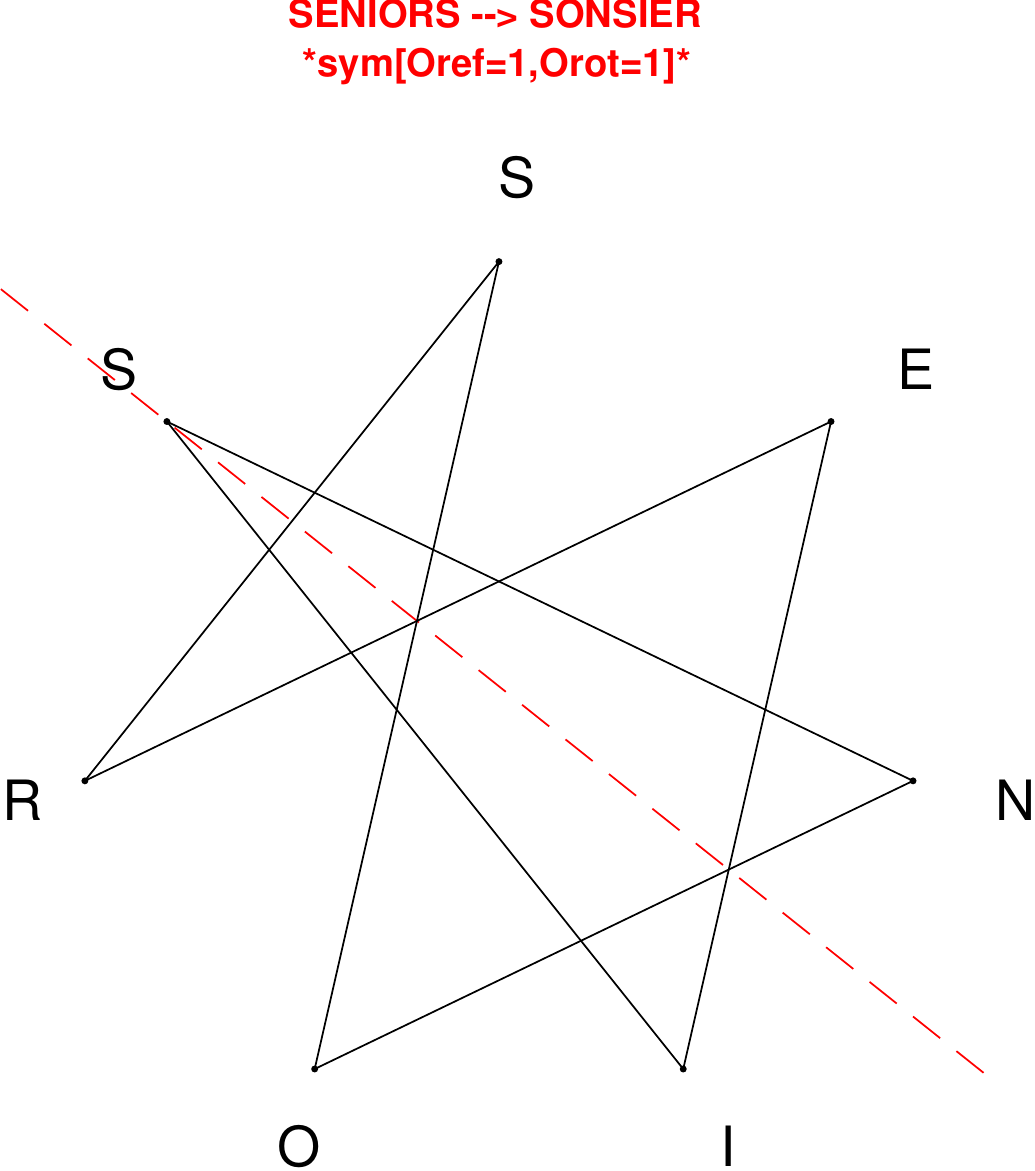}
\end{subfigure}
\hfill
\begin{subfigure}[T]{0.19\textwidth}
\centering
\includegraphics[width=\textwidth]{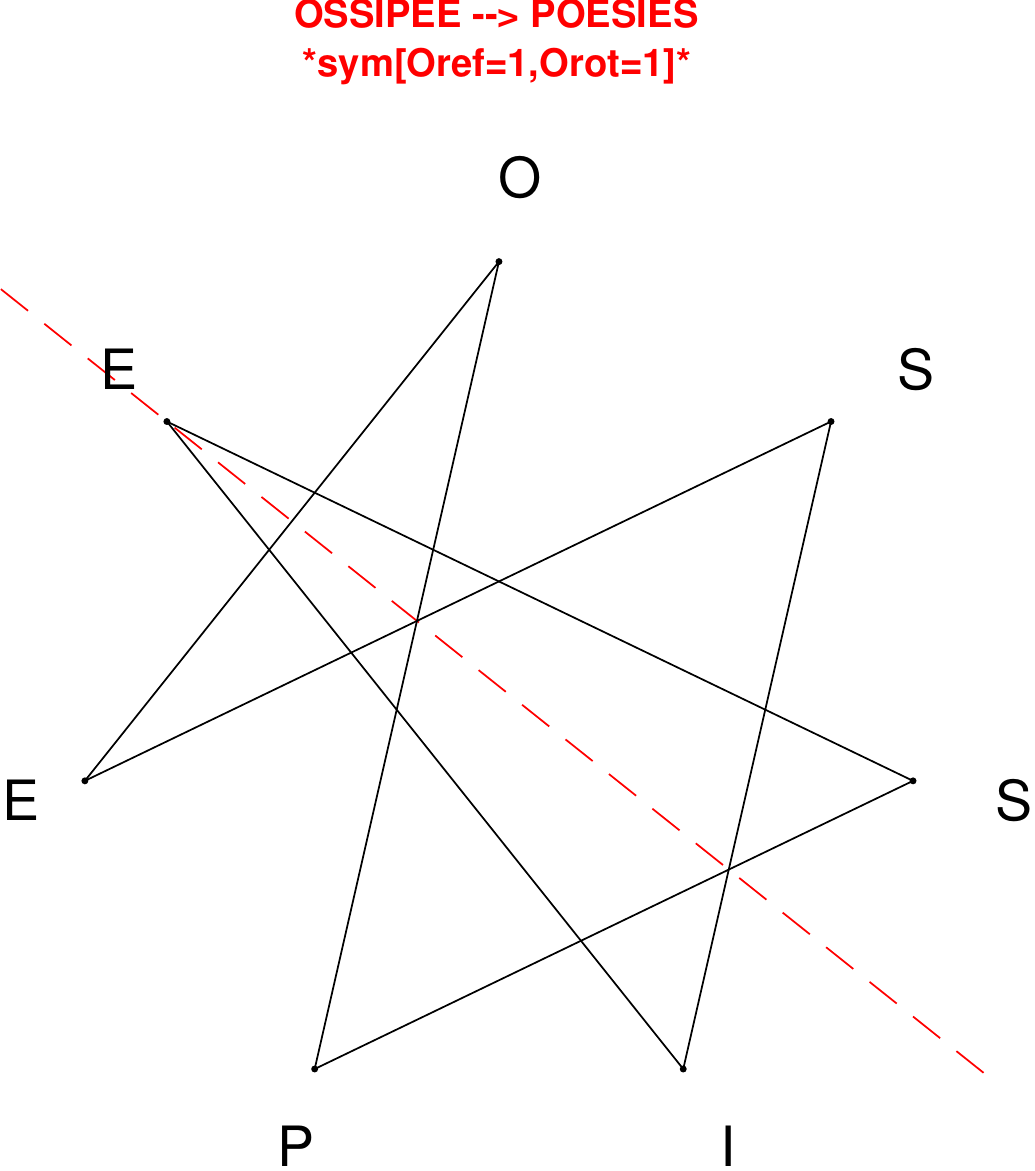}
\end{subfigure}
\hfill
\begin{subfigure}[T]{0.19\textwidth}
\centering
\includegraphics[width=\textwidth]{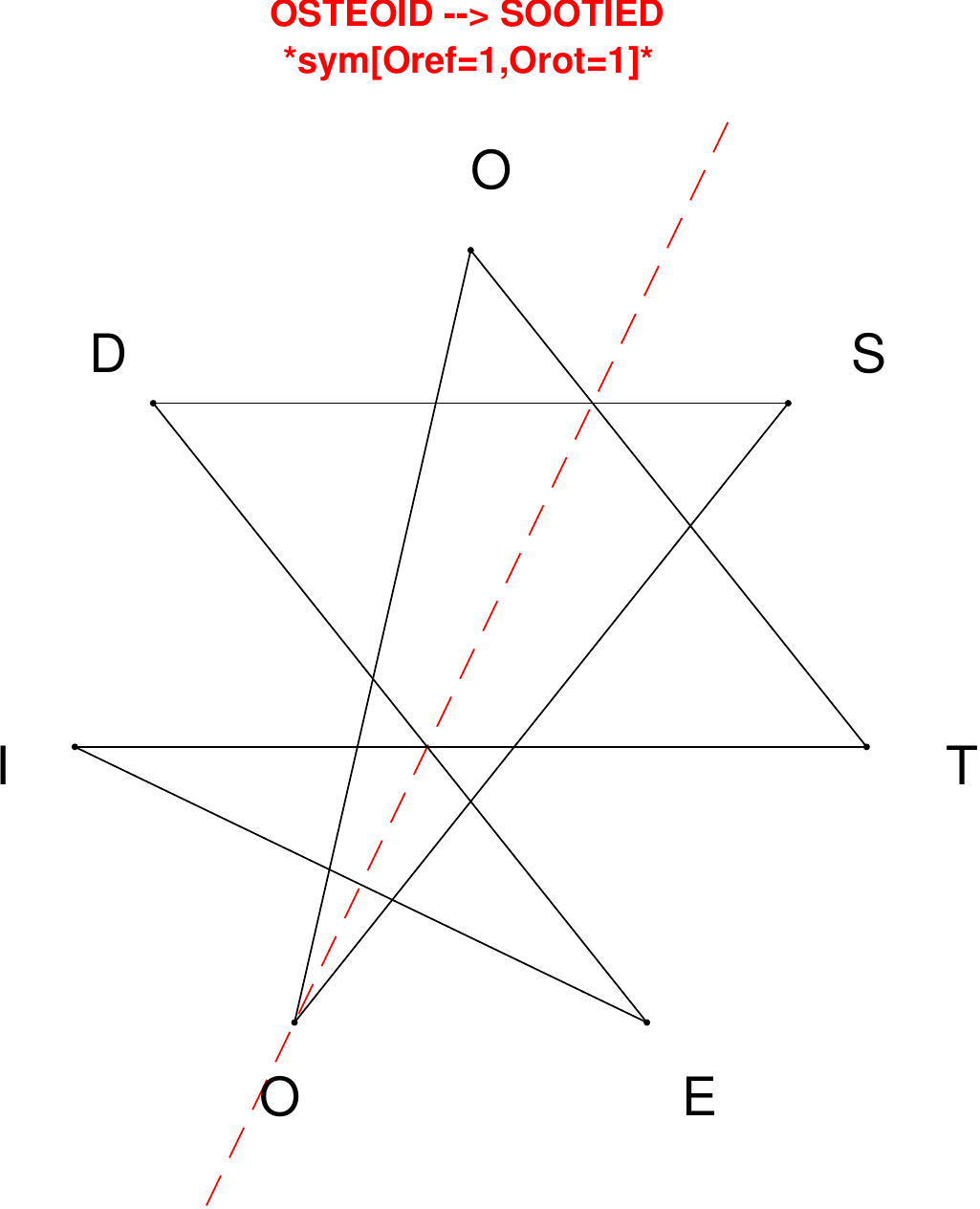}
\end{subfigure}
\hfill
\begin{subfigure}[T]{0.19\textwidth}
\centering
\includegraphics[width=\textwidth]{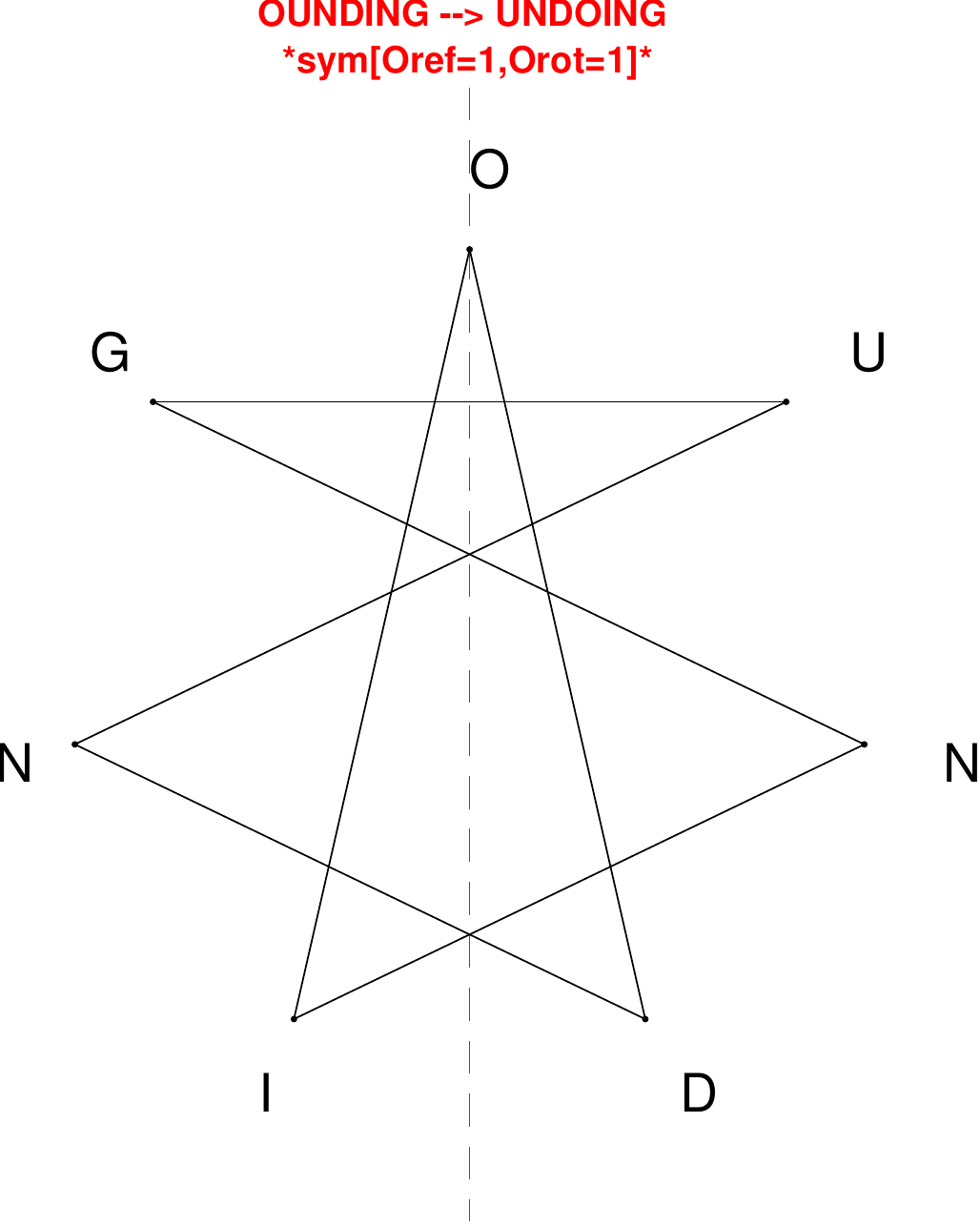}
\end{subfigure}
\end{figure}

\begin{figure}[H]
\centering
\begin{subfigure}[T]{0.19\textwidth}
\centering
\includegraphics[width=\textwidth]{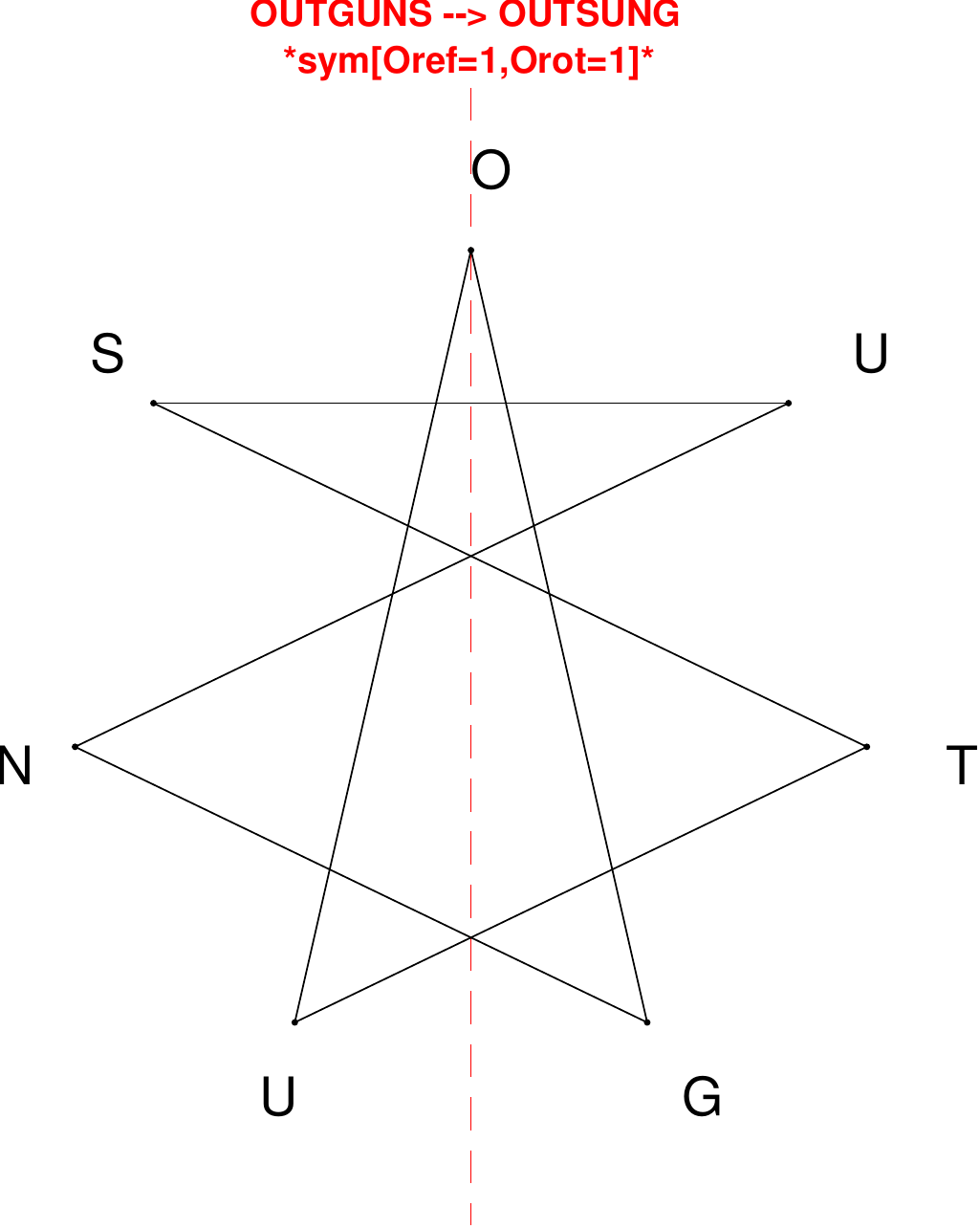}
\end{subfigure}
\hfill
\begin{subfigure}[T]{0.19\textwidth}
\centering
\includegraphics[width=\textwidth]{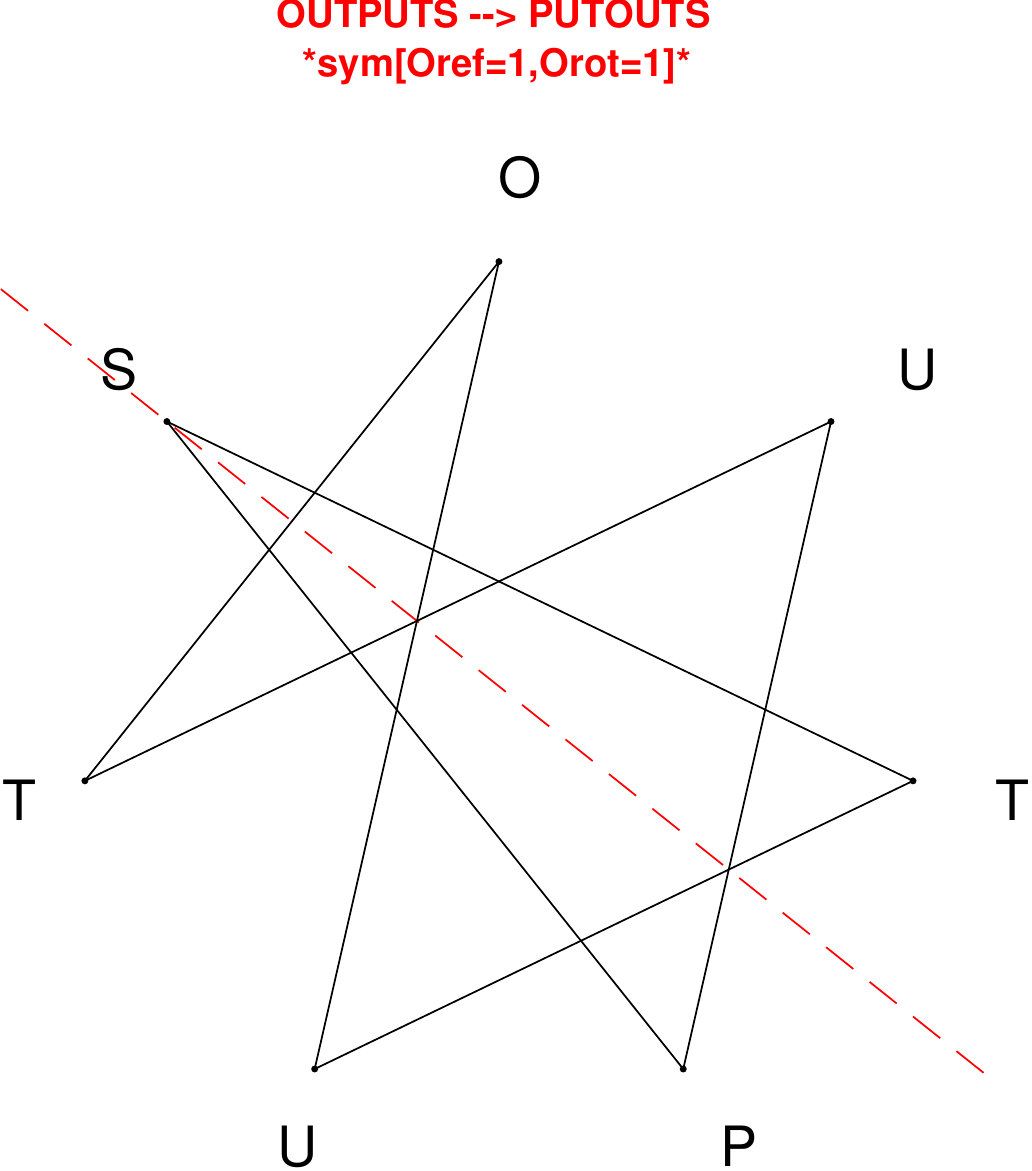}
\end{subfigure}
\hfill
\begin{subfigure}[T]{0.19\textwidth}
\centering
\includegraphics[width=\textwidth]{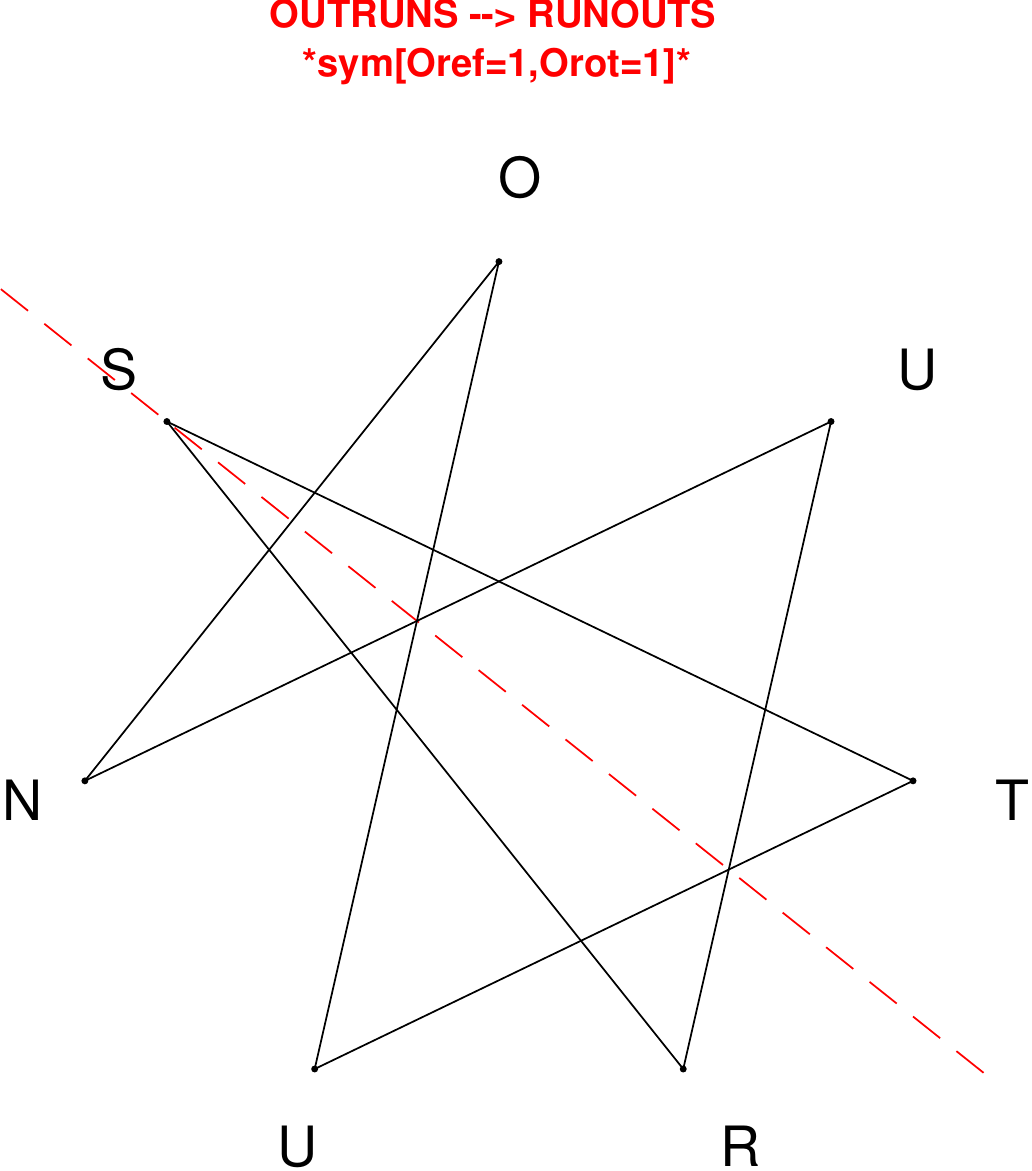}
\end{subfigure}
\hfill
\begin{subfigure}[T]{0.19\textwidth}
\centering
\includegraphics[width=\textwidth]{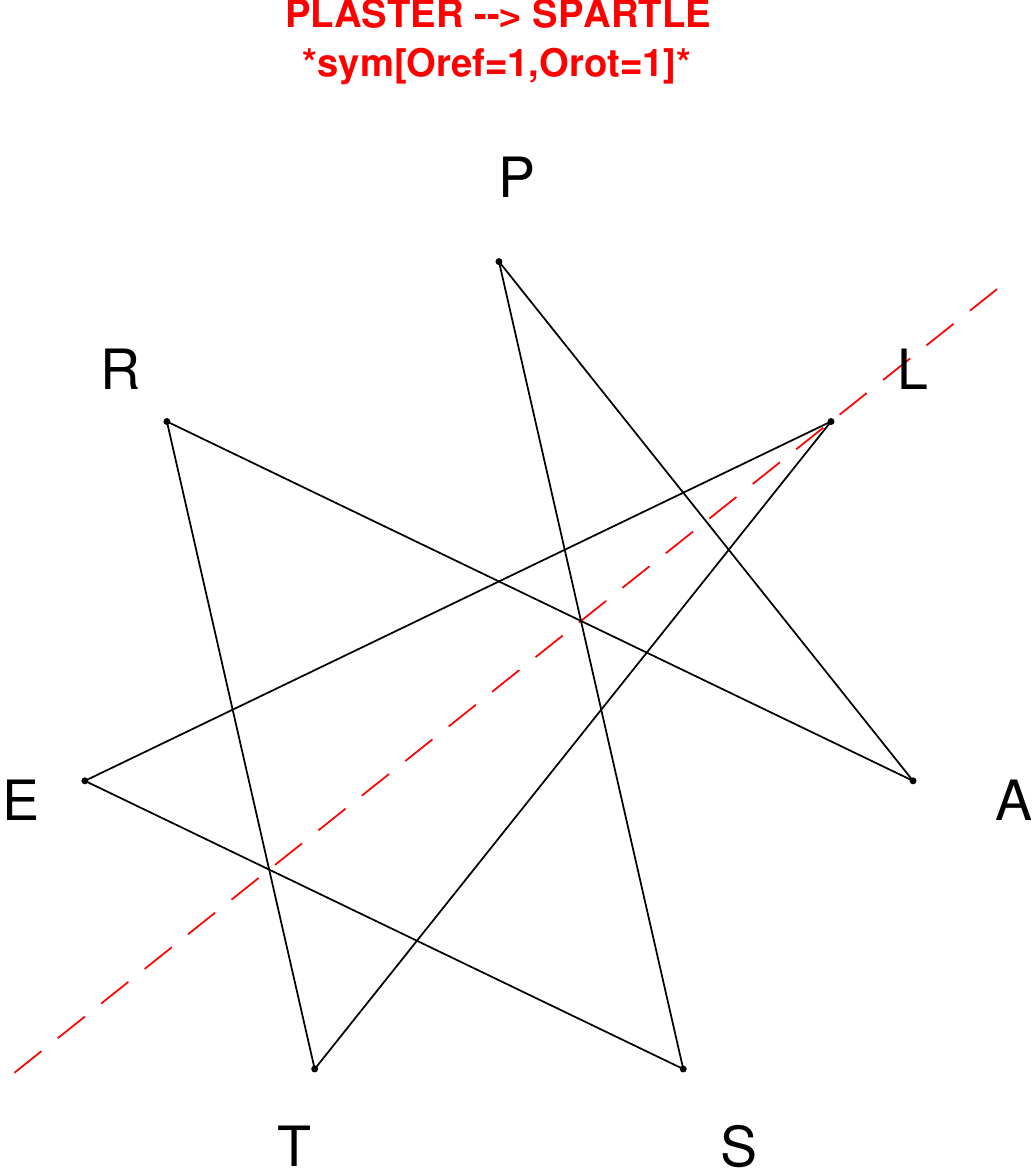}
\end{subfigure}
\hfill
\begin{subfigure}[T]{0.19\textwidth}
\centering
\includegraphics[width=\textwidth]{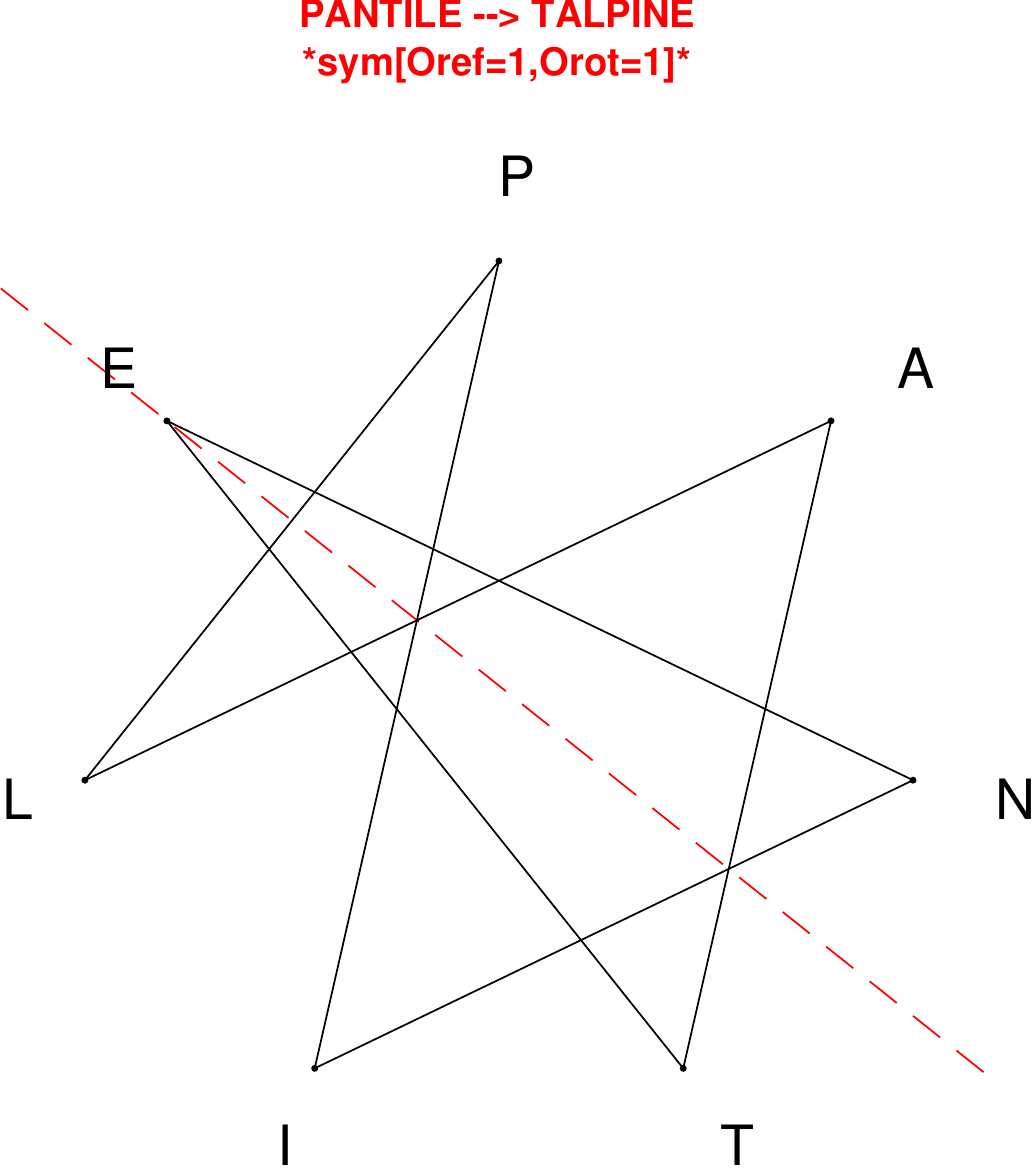}
\end{subfigure}
\end{figure}

\begin{figure}[H]
\centering
\begin{subfigure}[T]{0.19\textwidth}
\centering
\includegraphics[width=\textwidth]{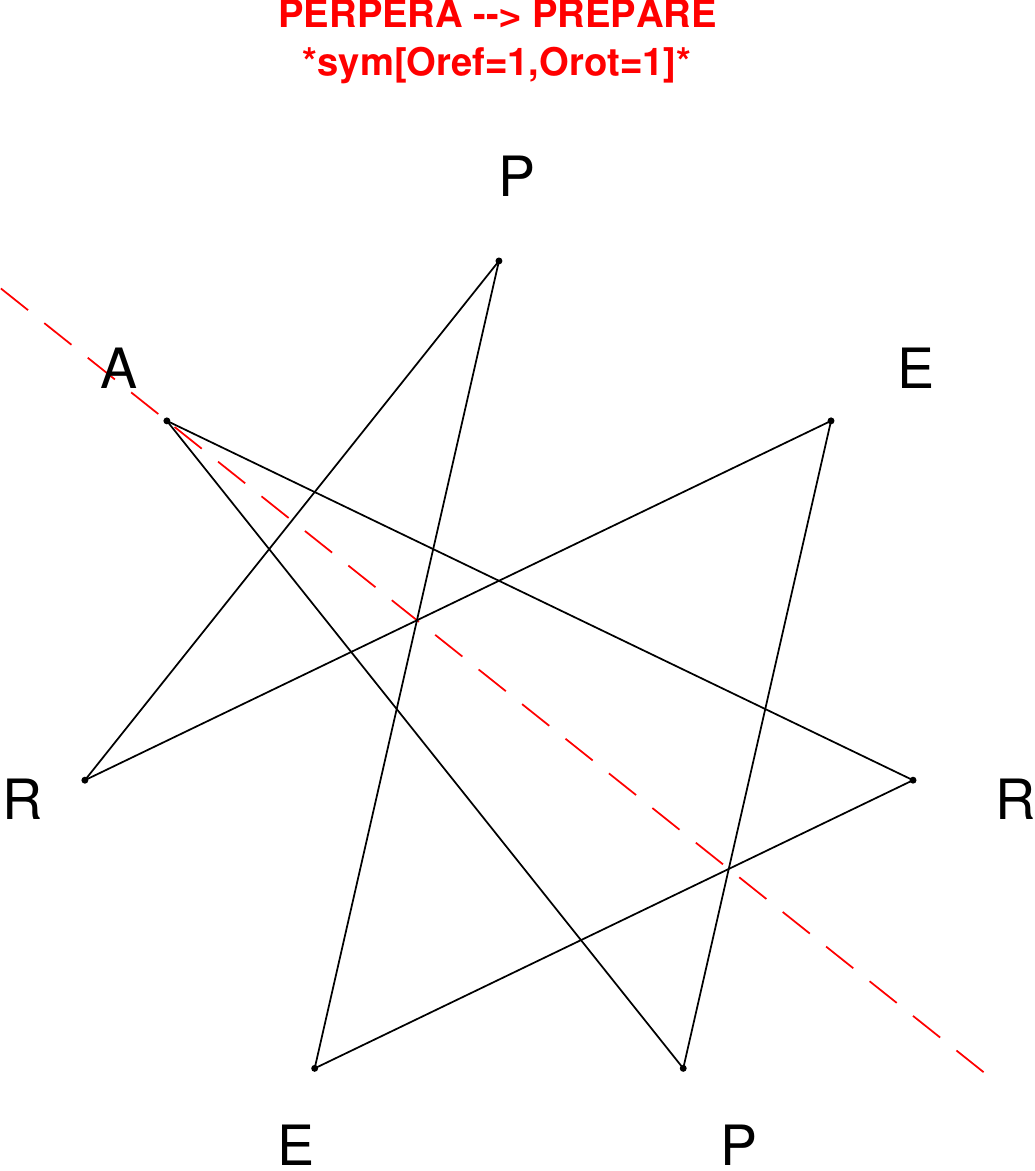}
\end{subfigure}
\hfill
\begin{subfigure}[T]{0.19\textwidth}
\centering
\includegraphics[width=\textwidth]{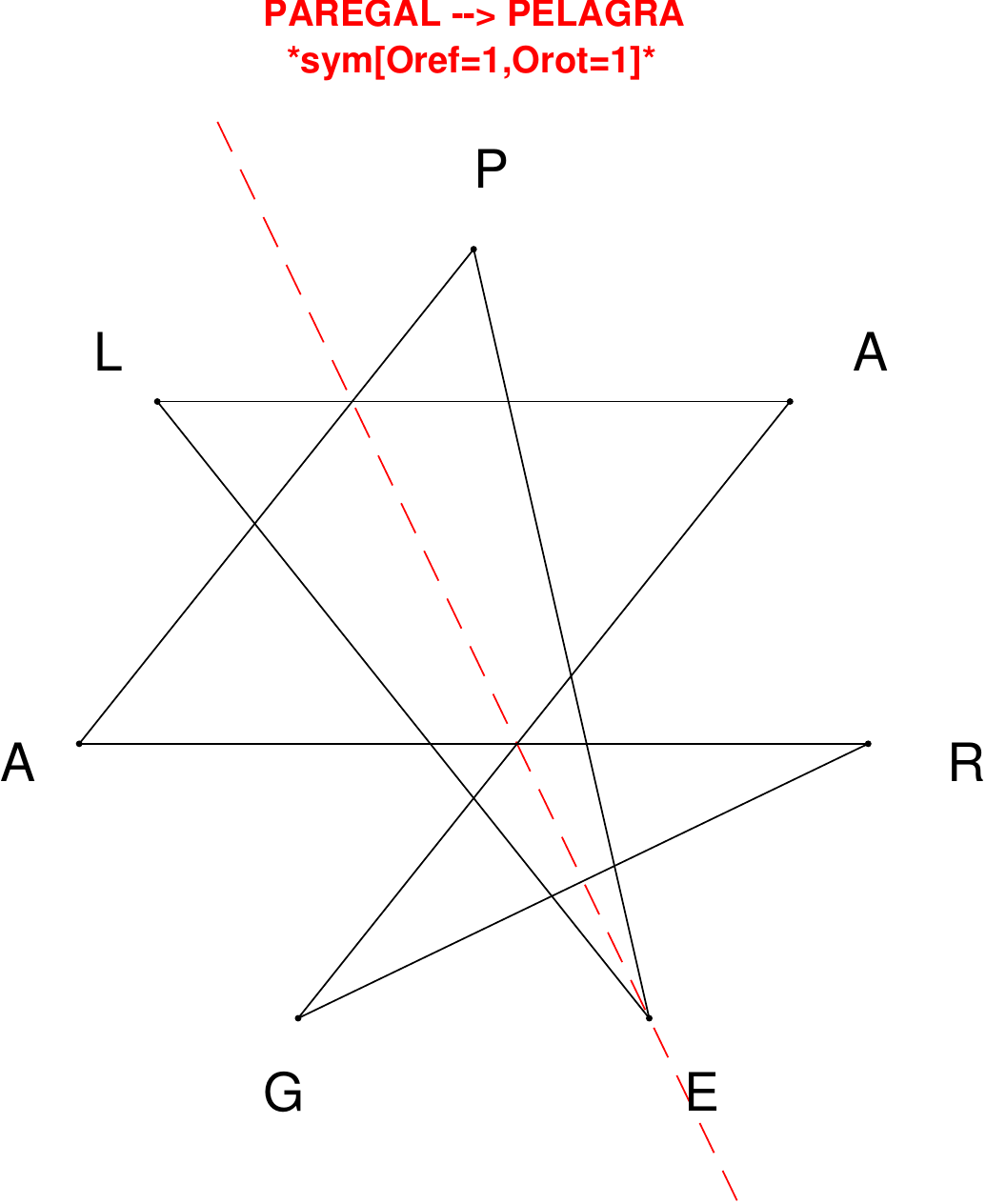}
\end{subfigure}
\hfill
\begin{subfigure}[T]{0.19\textwidth}
\centering
\includegraphics[width=\textwidth]{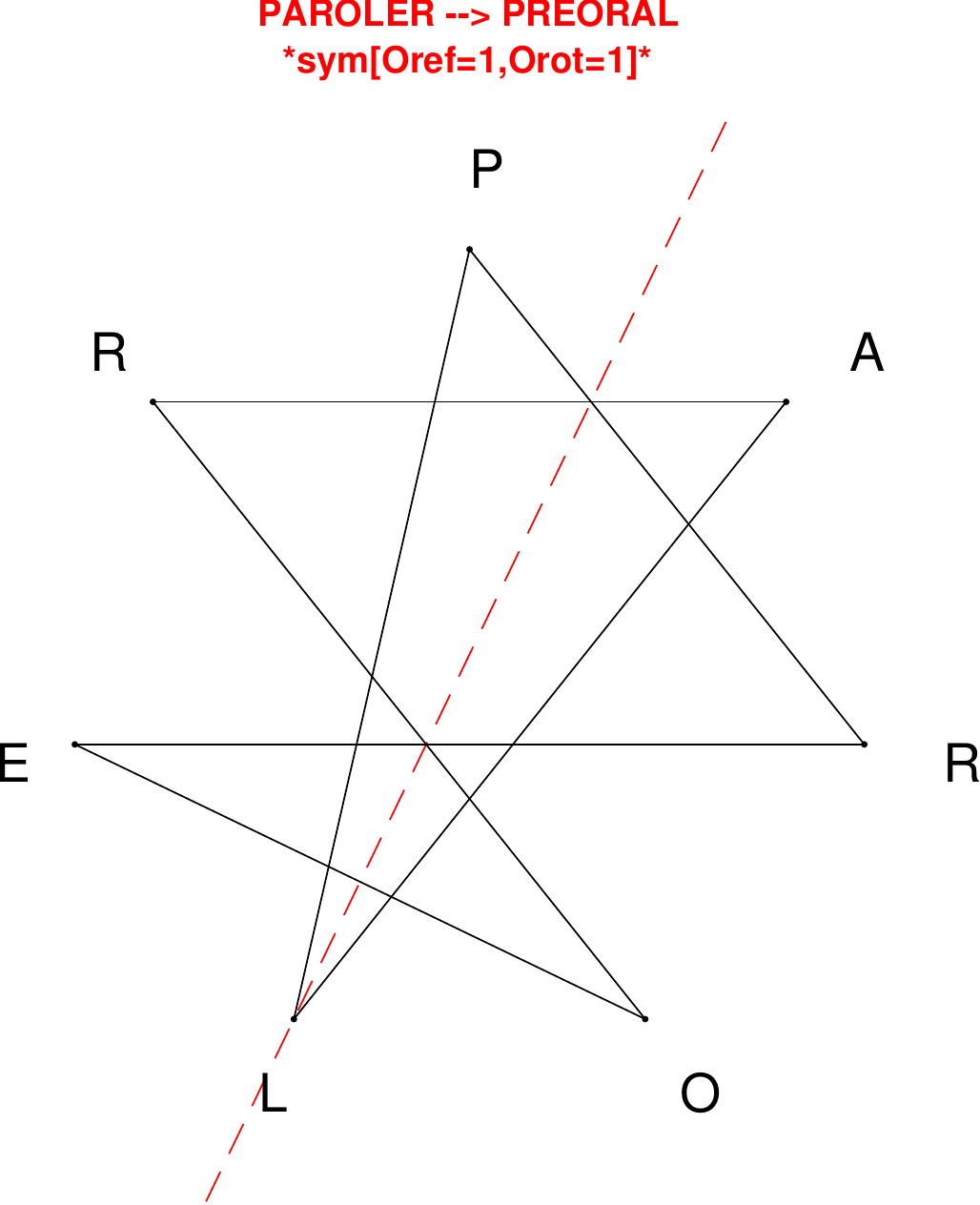}
\end{subfigure}
\hfill
\begin{subfigure}[T]{0.19\textwidth}
\centering
\includegraphics[width=\textwidth]{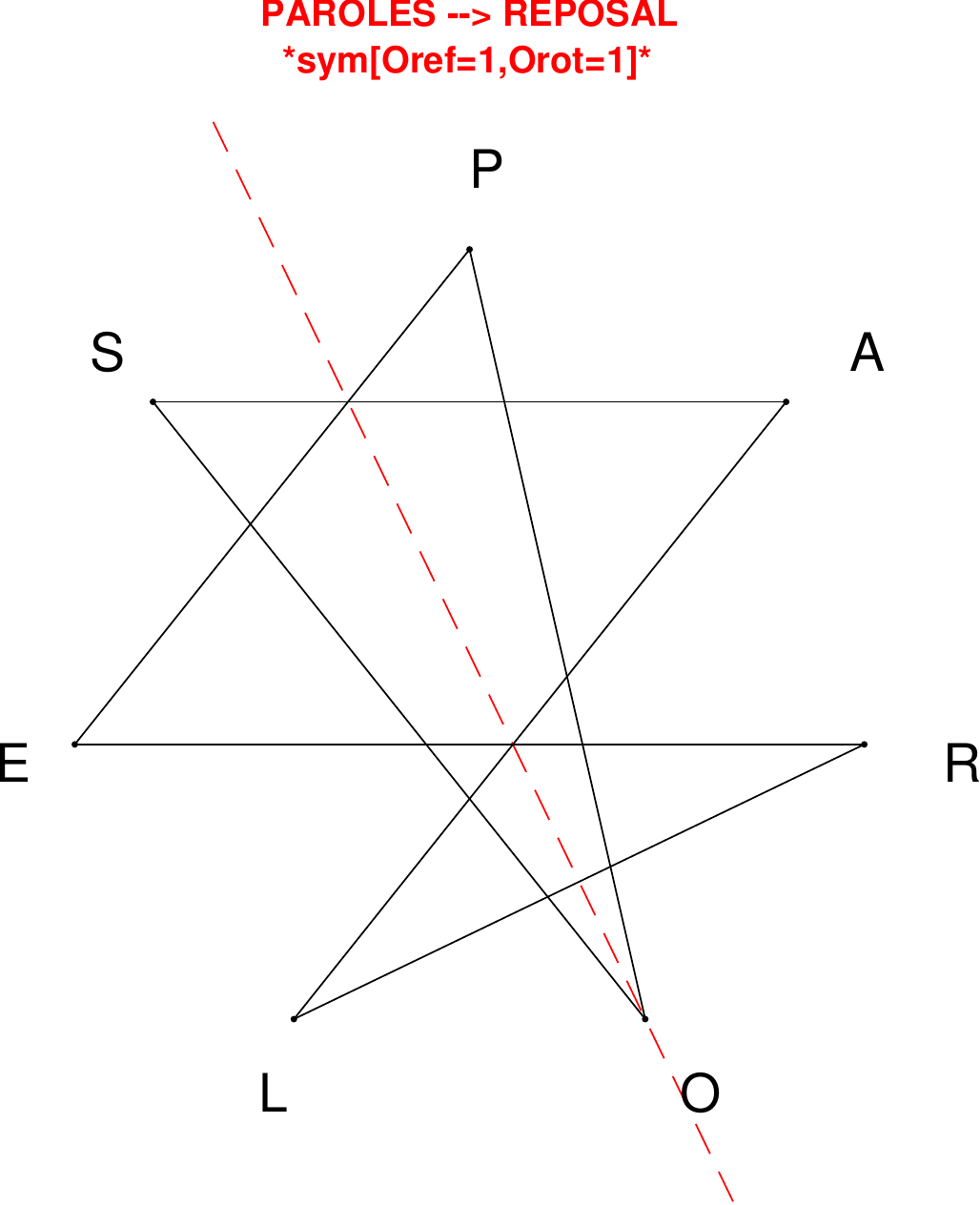}
\end{subfigure}
\hfill
\begin{subfigure}[T]{0.19\textwidth}
\centering
\includegraphics[width=\textwidth]{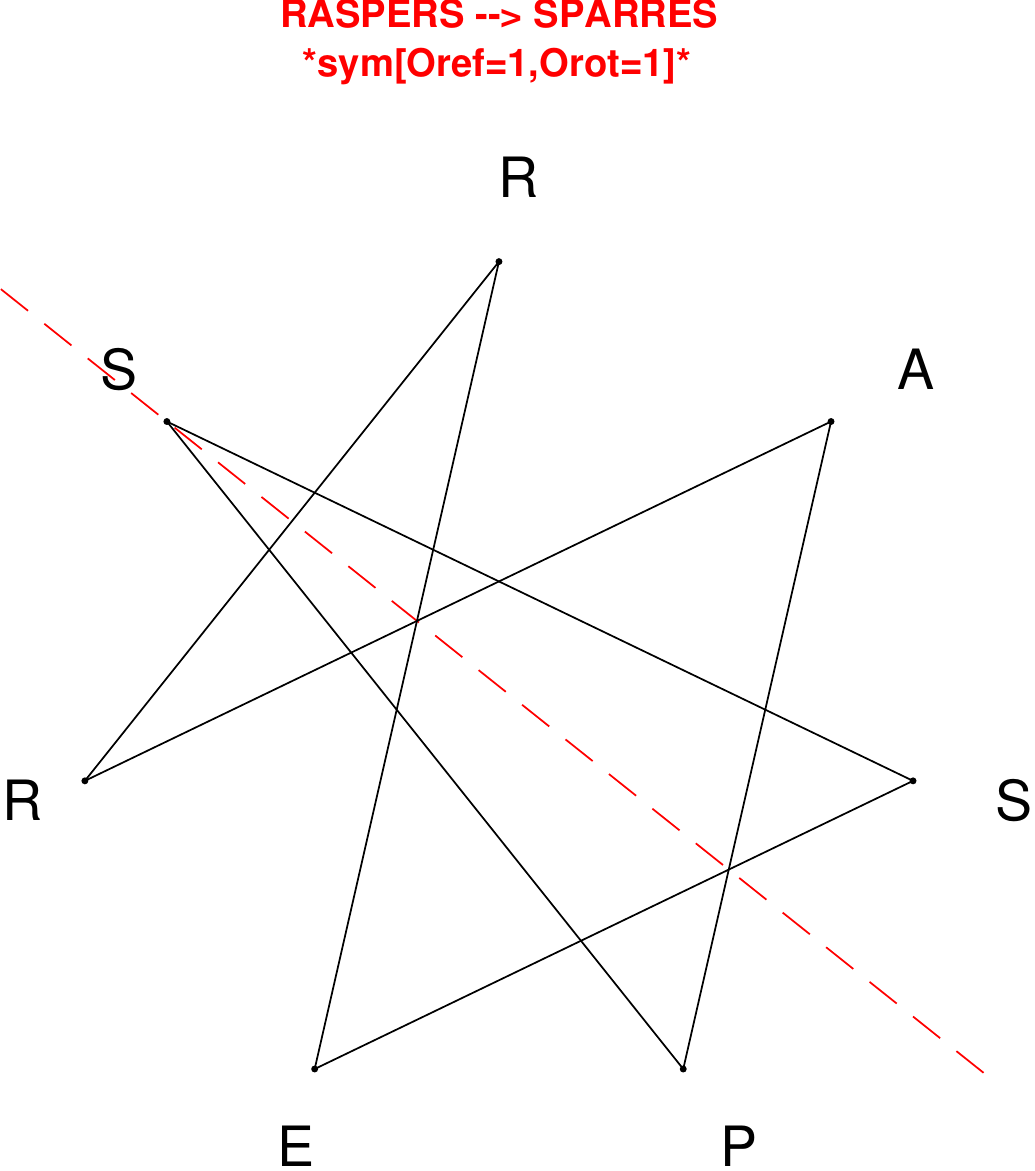}
\end{subfigure}
\end{figure}

\begin{figure}[H]
\centering
\begin{subfigure}[T]{0.19\textwidth}
\centering
\includegraphics[width=\textwidth]{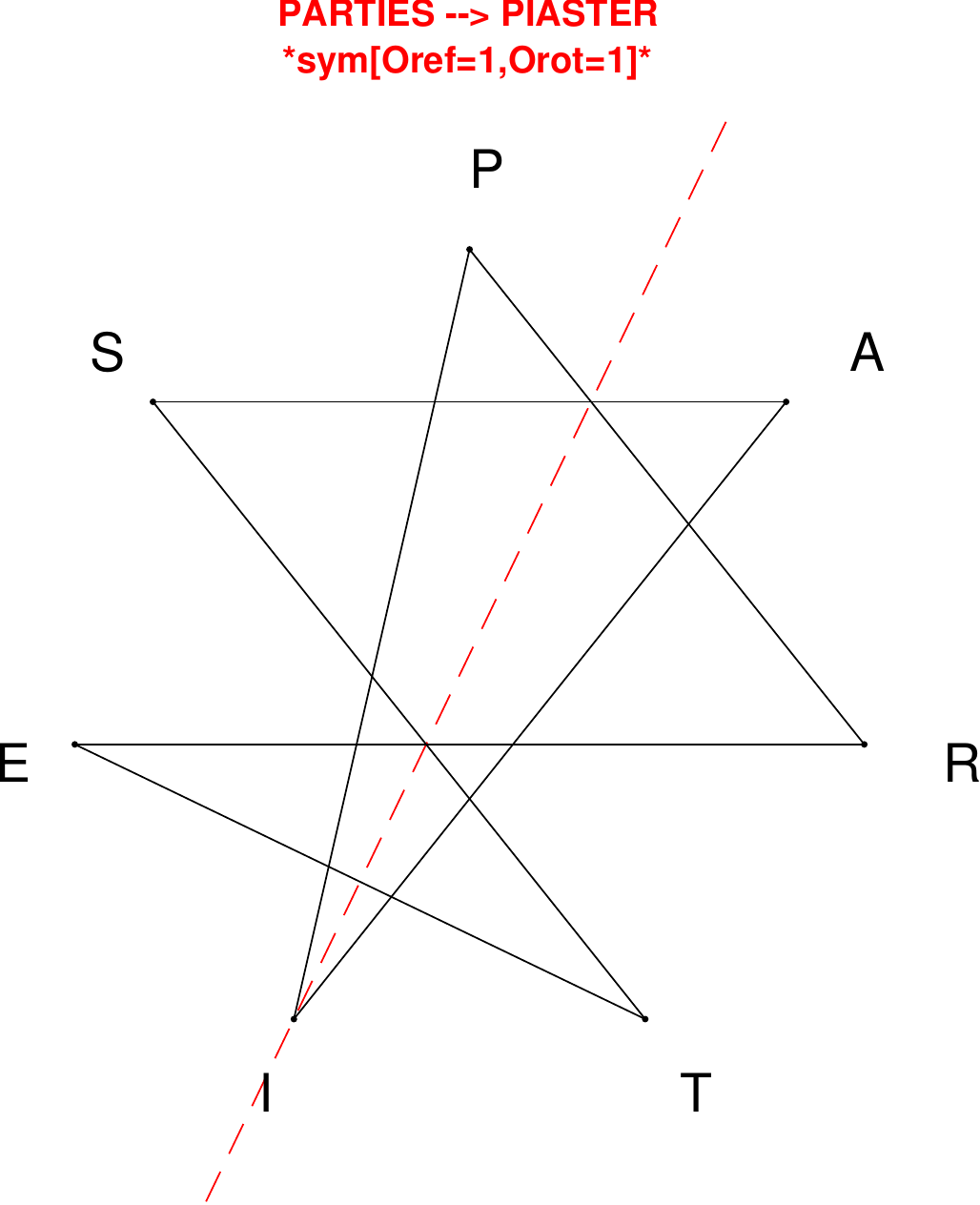}
\end{subfigure}
\hfill
\begin{subfigure}[T]{0.19\textwidth}
\centering
\includegraphics[width=\textwidth]{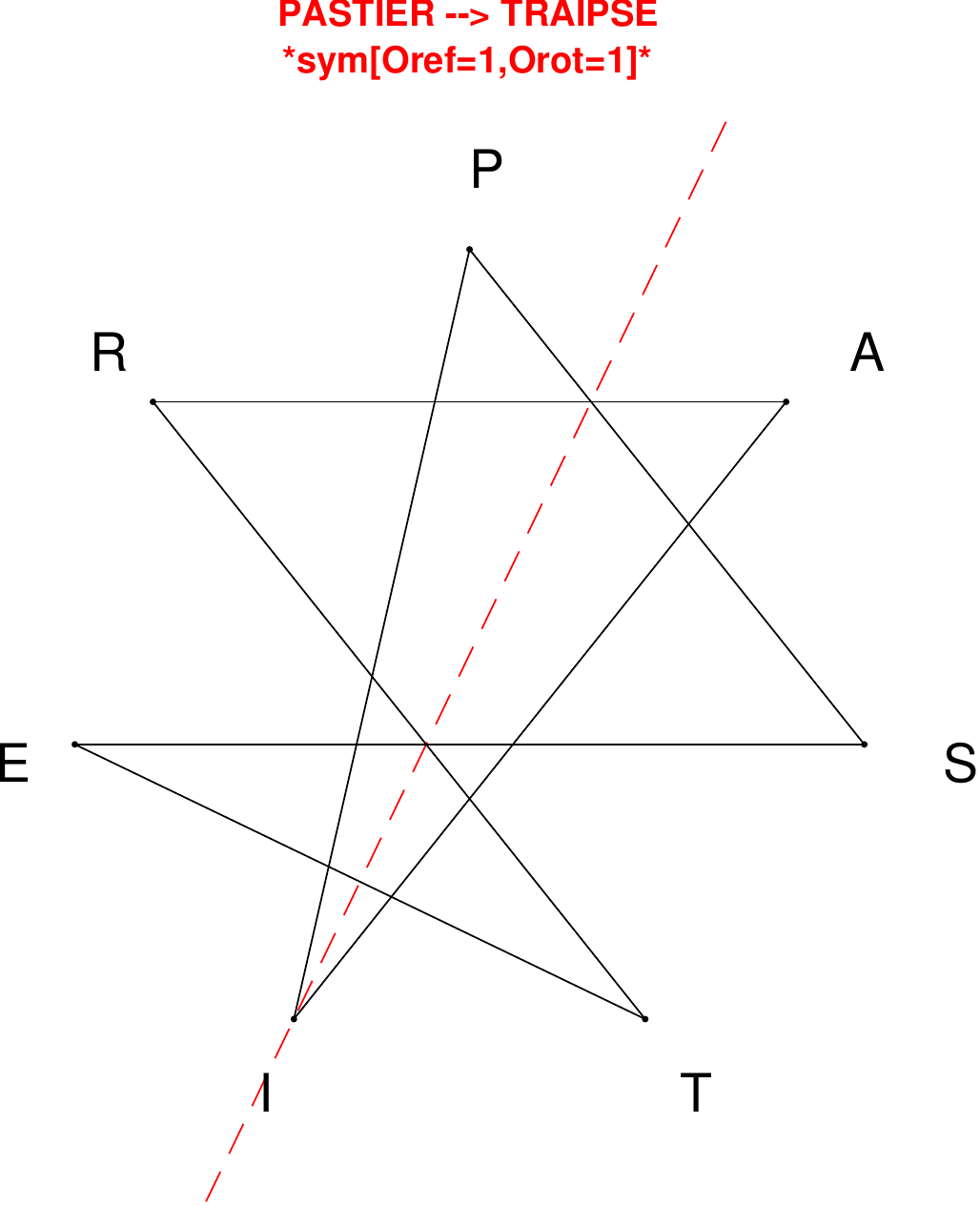}
\end{subfigure}
\hfill
\begin{subfigure}[T]{0.19\textwidth}
\centering
\includegraphics[width=\textwidth]{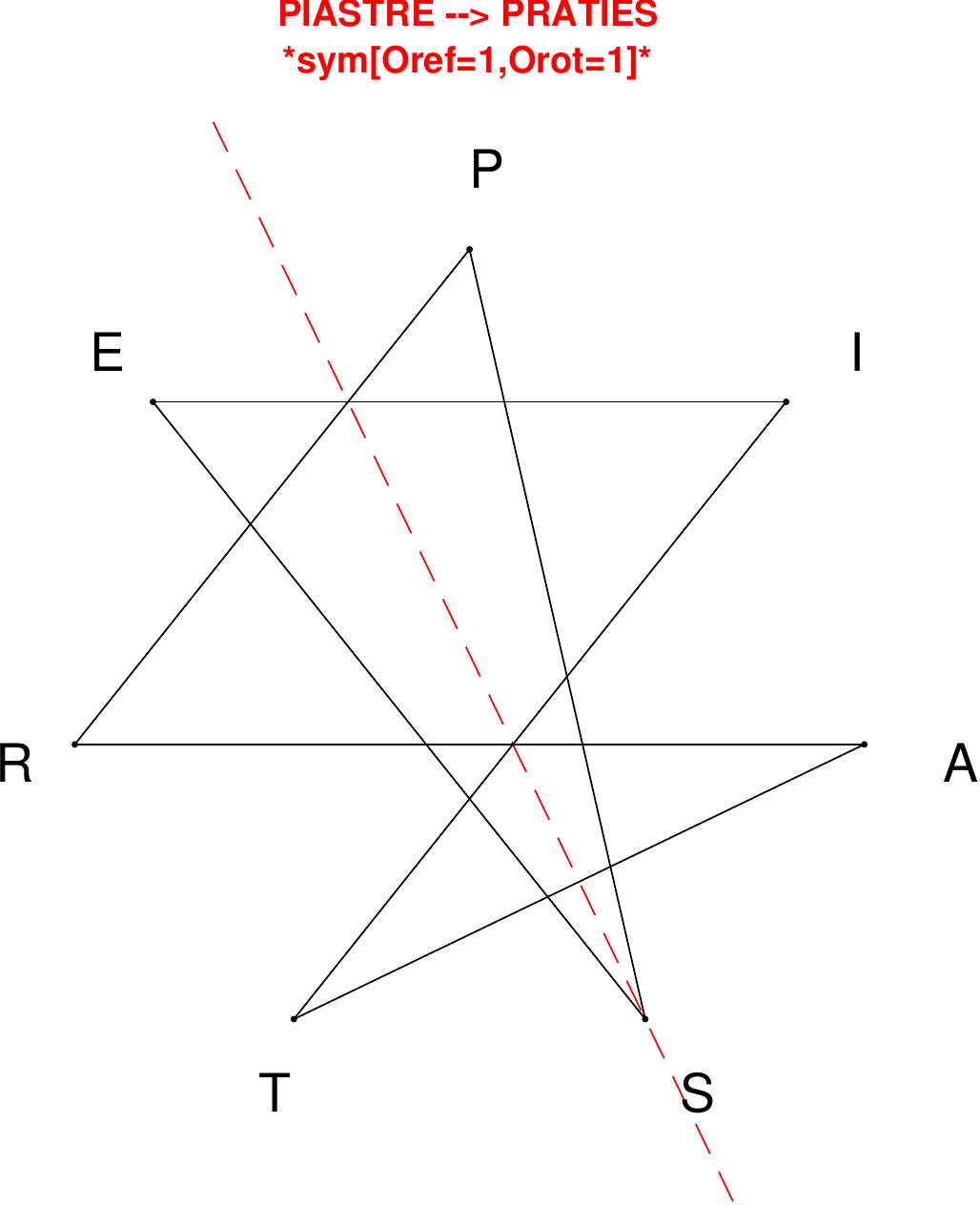}
\end{subfigure}
\hfill
\begin{subfigure}[T]{0.19\textwidth}
\centering
\includegraphics[width=\textwidth]{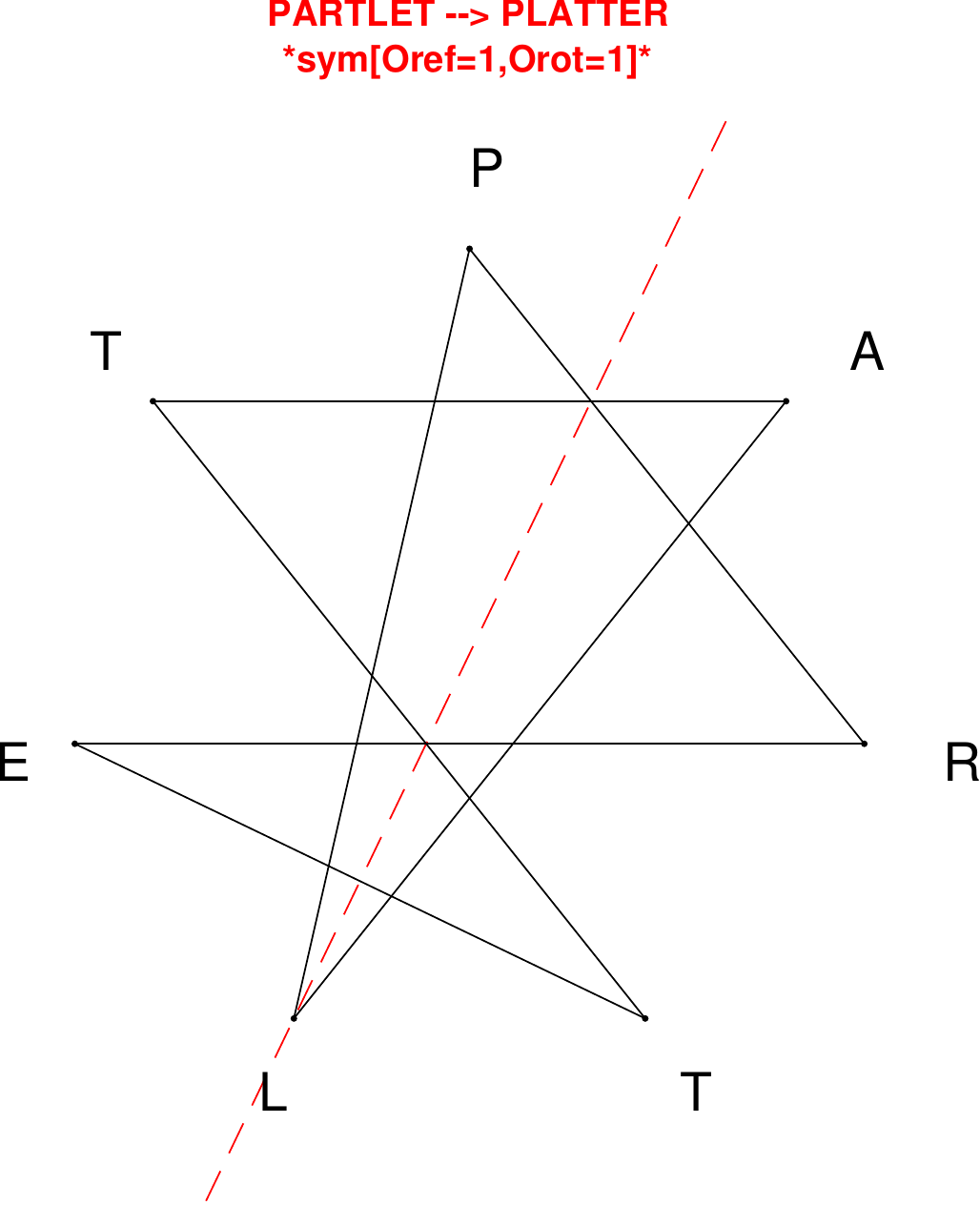}
\end{subfigure}
\hfill
\begin{subfigure}[T]{0.19\textwidth}
\centering
\includegraphics[width=\textwidth]{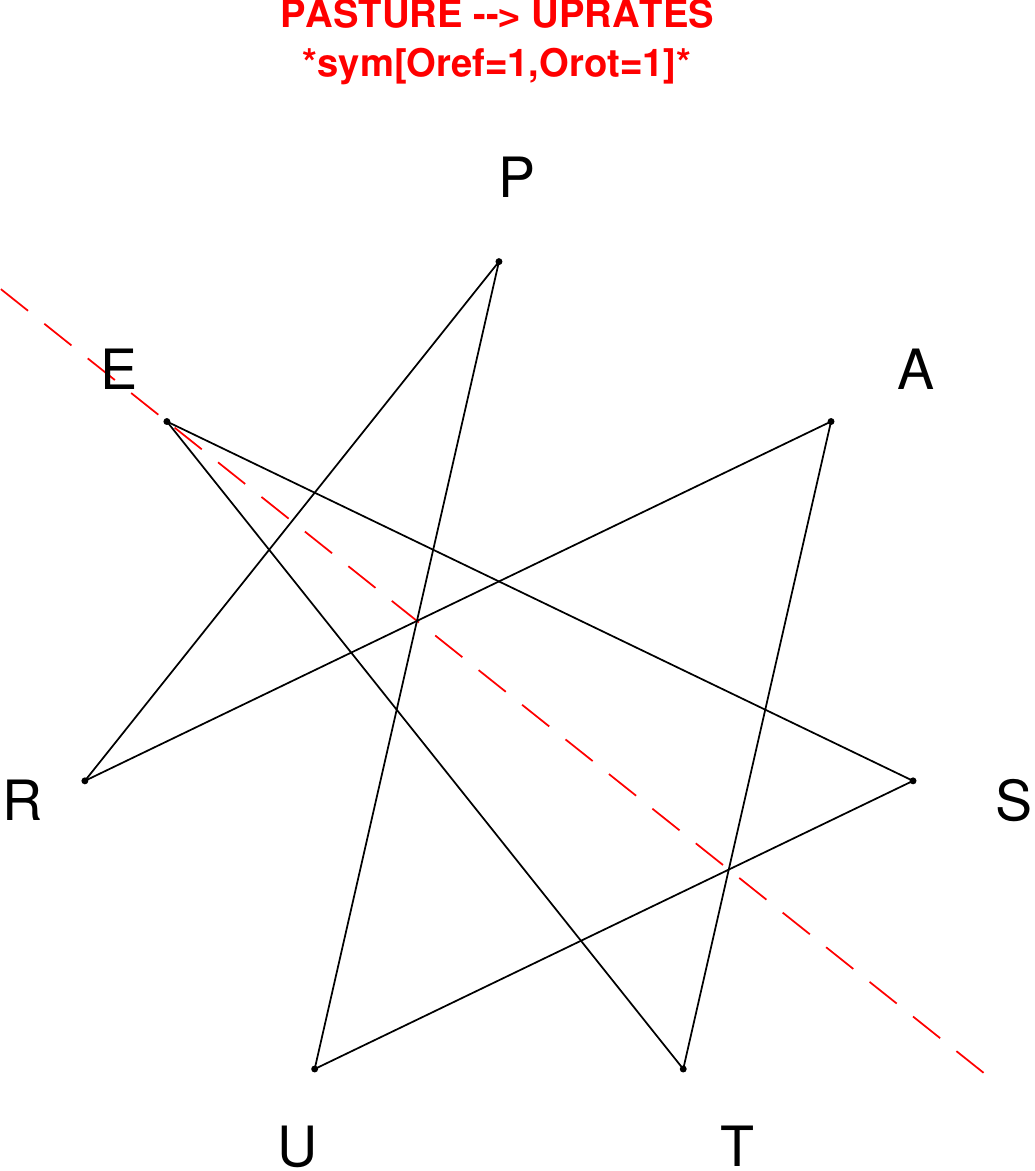}
\end{subfigure}
\end{figure}

\begin{figure}[H]
\centering
\begin{subfigure}[T]{0.19\textwidth}
\centering
\includegraphics[width=\textwidth]{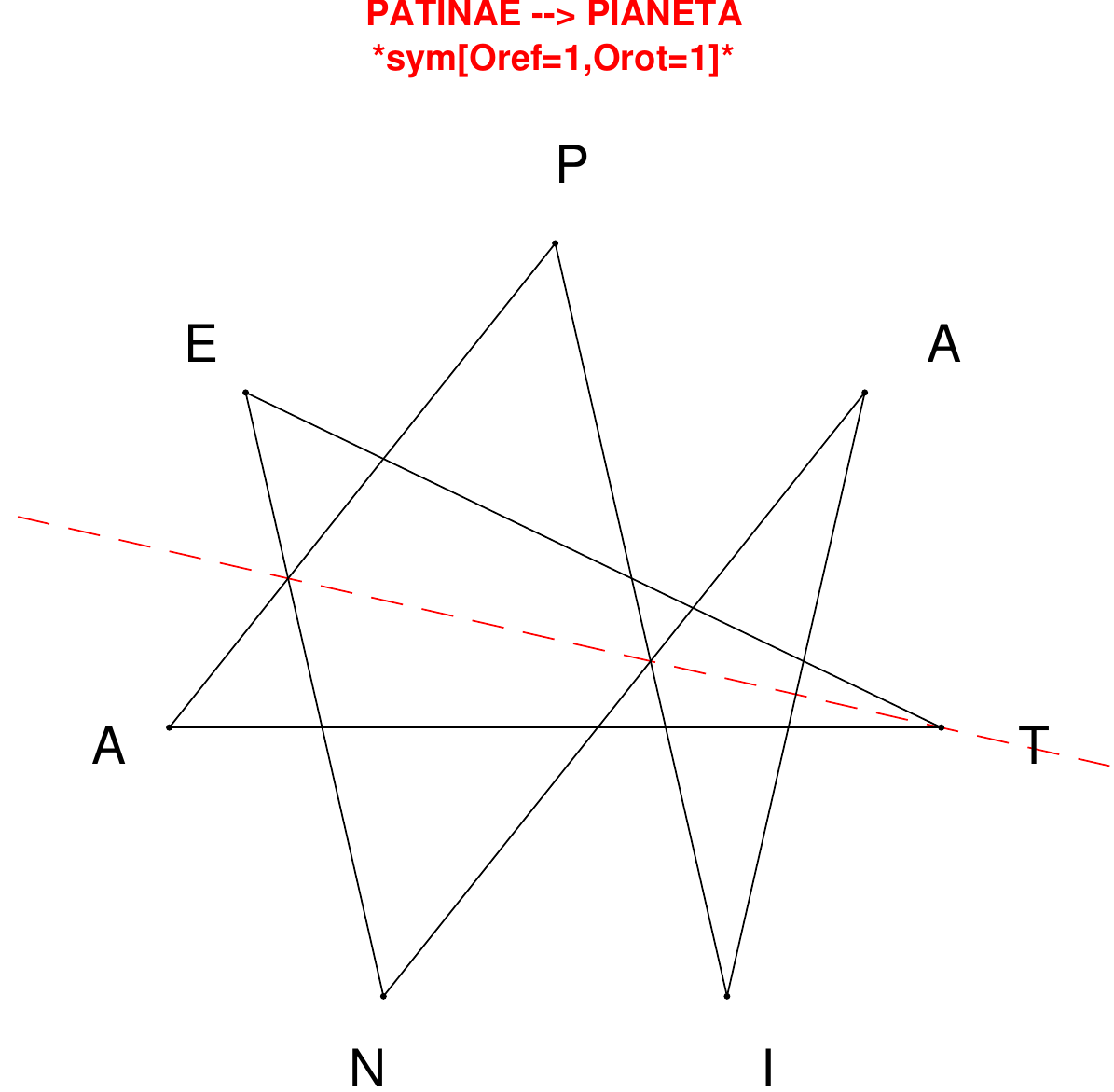}
\end{subfigure}
\hfill
\begin{subfigure}[T]{0.19\textwidth}
\centering
\includegraphics[width=\textwidth]{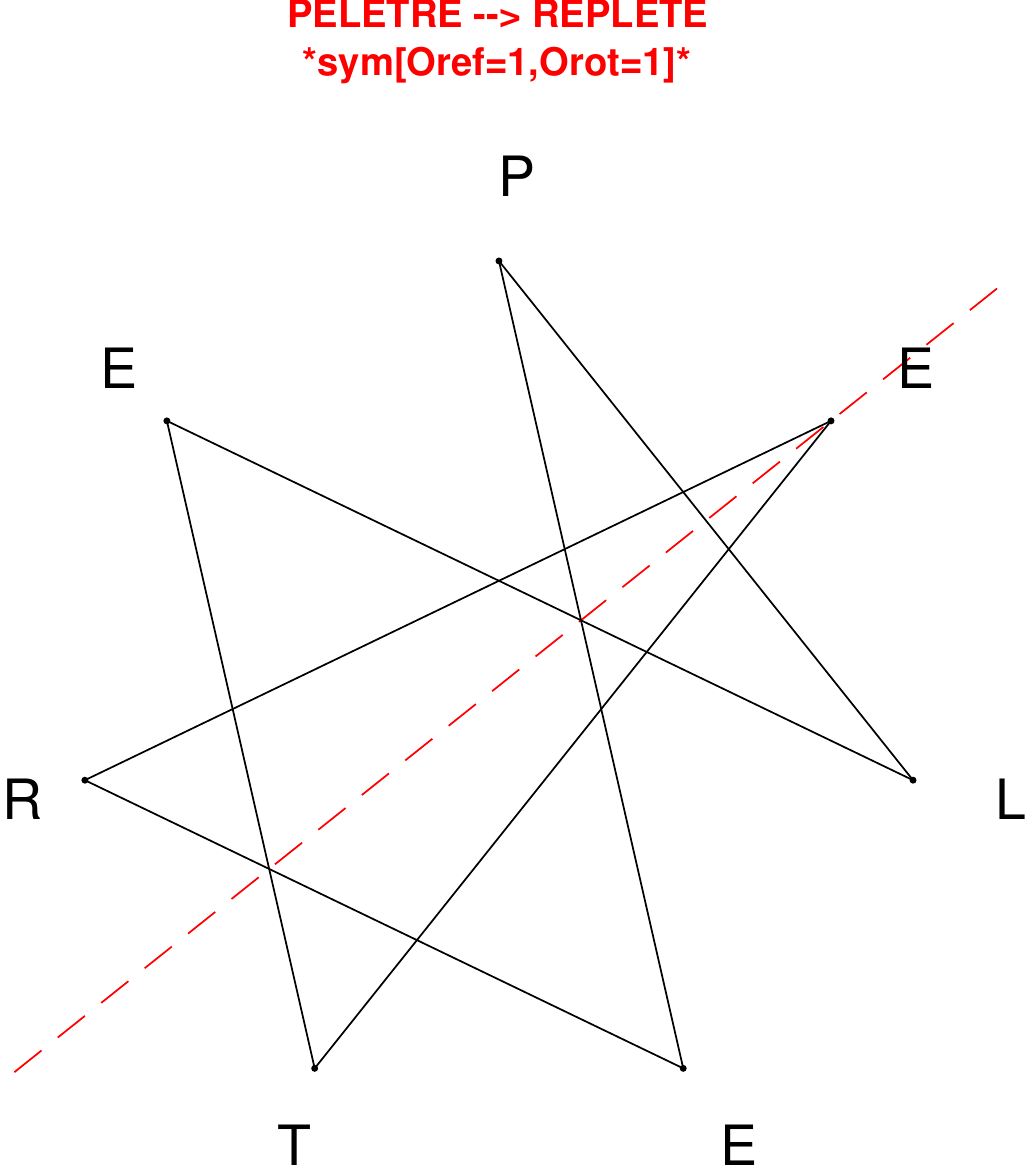}
\end{subfigure}
\hfill
\begin{subfigure}[T]{0.19\textwidth}
\centering
\includegraphics[width=\textwidth]{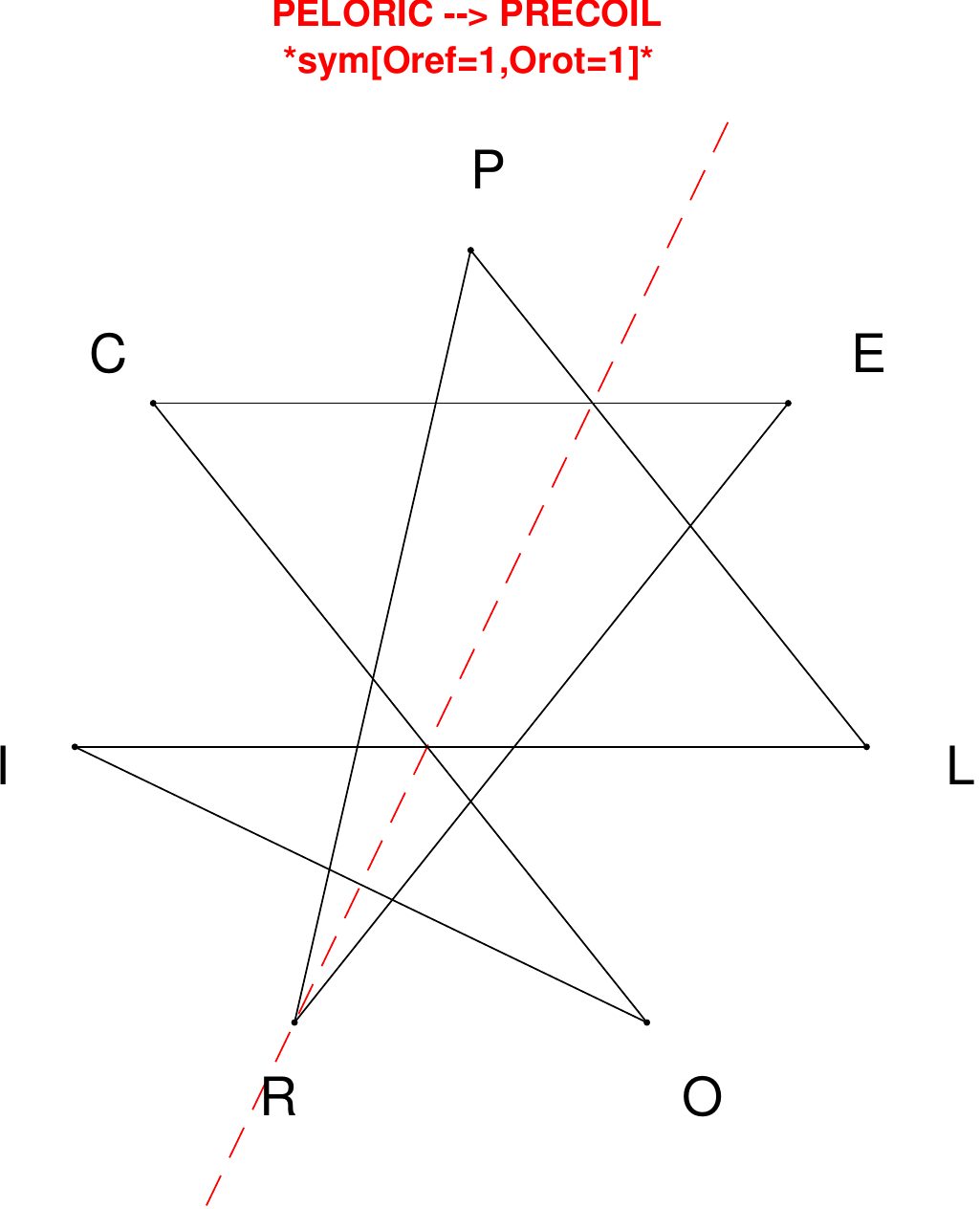}
\end{subfigure}
\hfill
\begin{subfigure}[T]{0.19\textwidth}
\centering
\includegraphics[width=\textwidth]{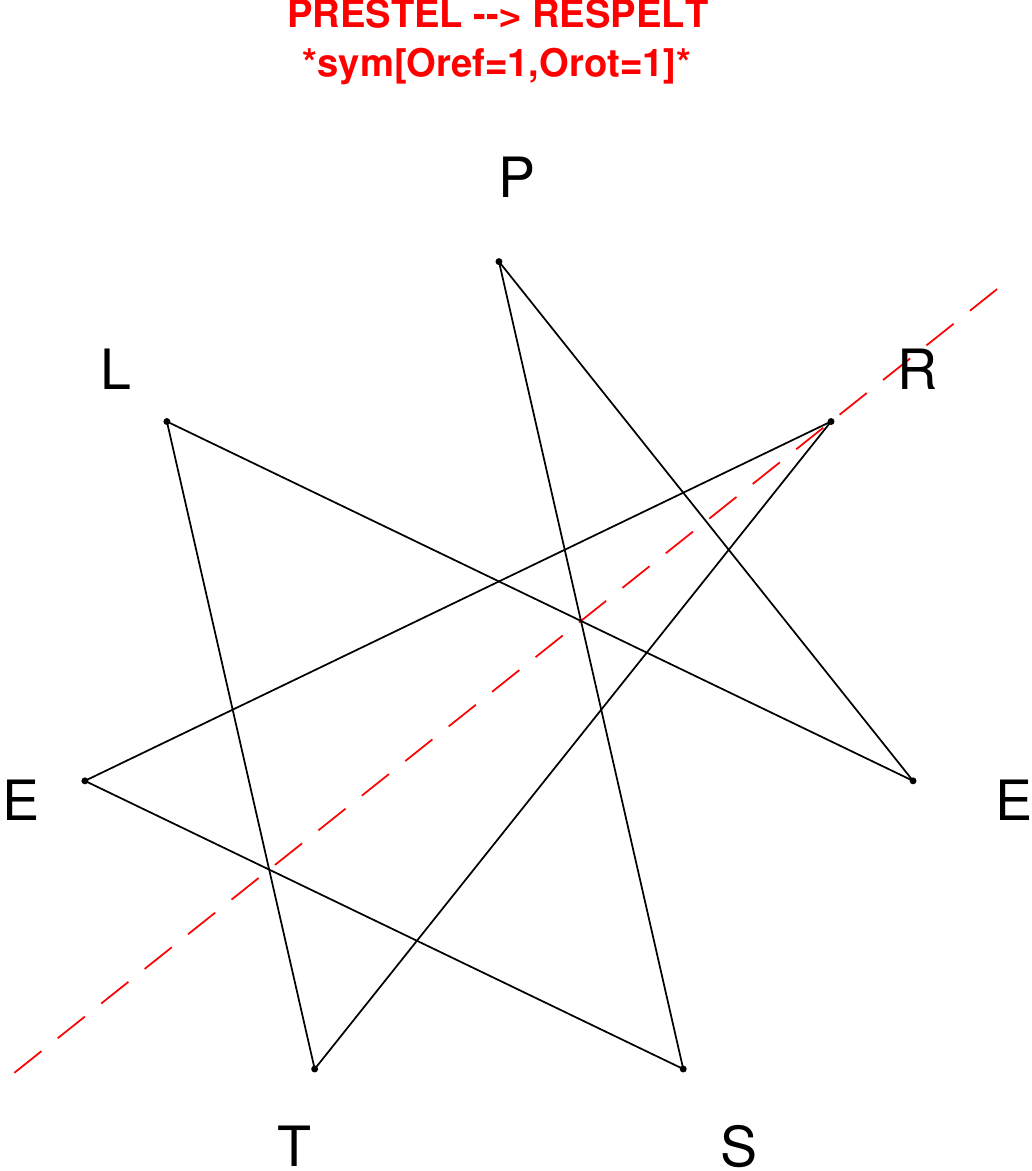}
\end{subfigure}
\hfill
\begin{subfigure}[T]{0.19\textwidth}
\centering
\includegraphics[width=\textwidth]{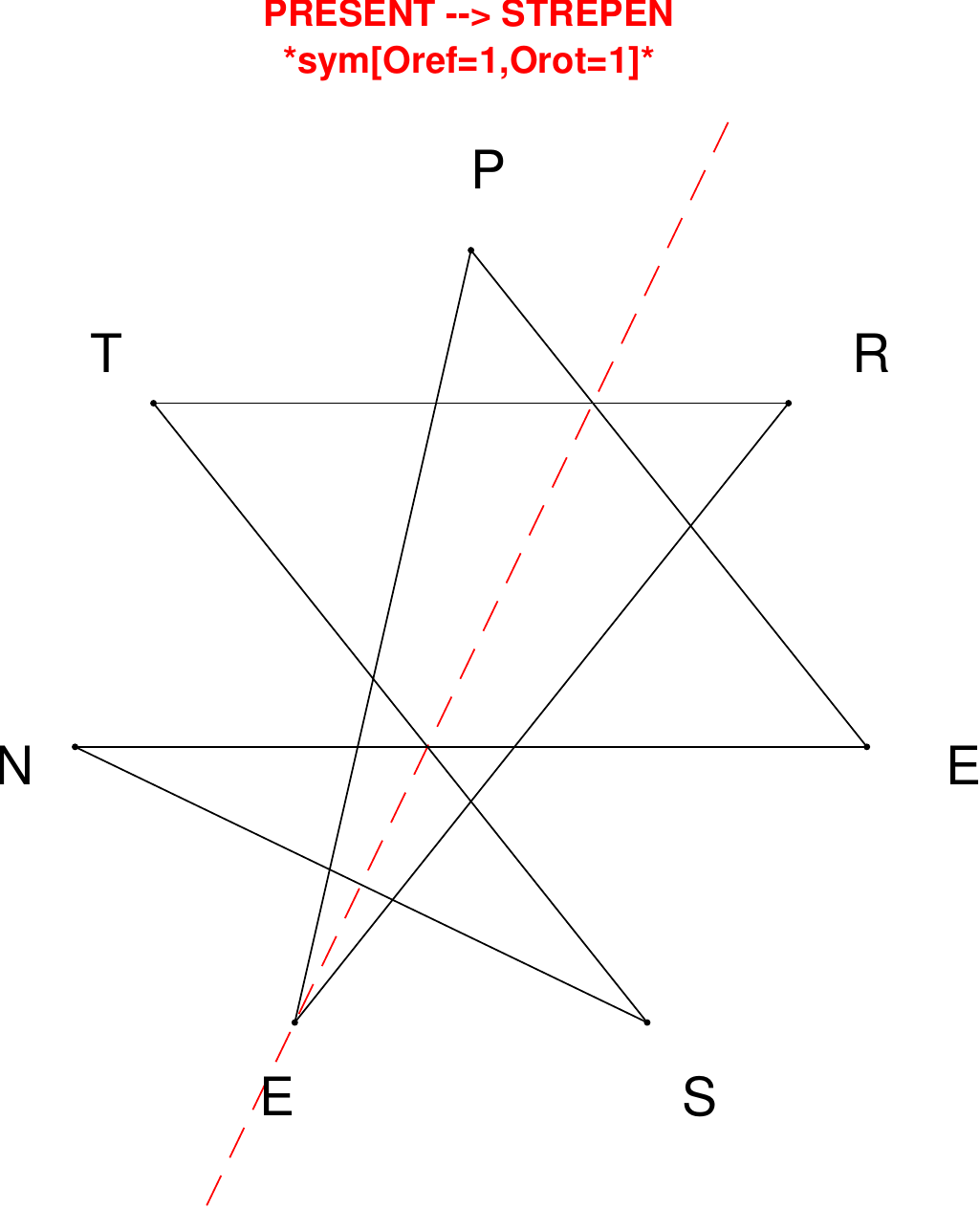}
\end{subfigure}
\end{figure}

\begin{figure}[H]
\centering
\begin{subfigure}[T]{0.19\textwidth}
\centering
\includegraphics[width=\textwidth]{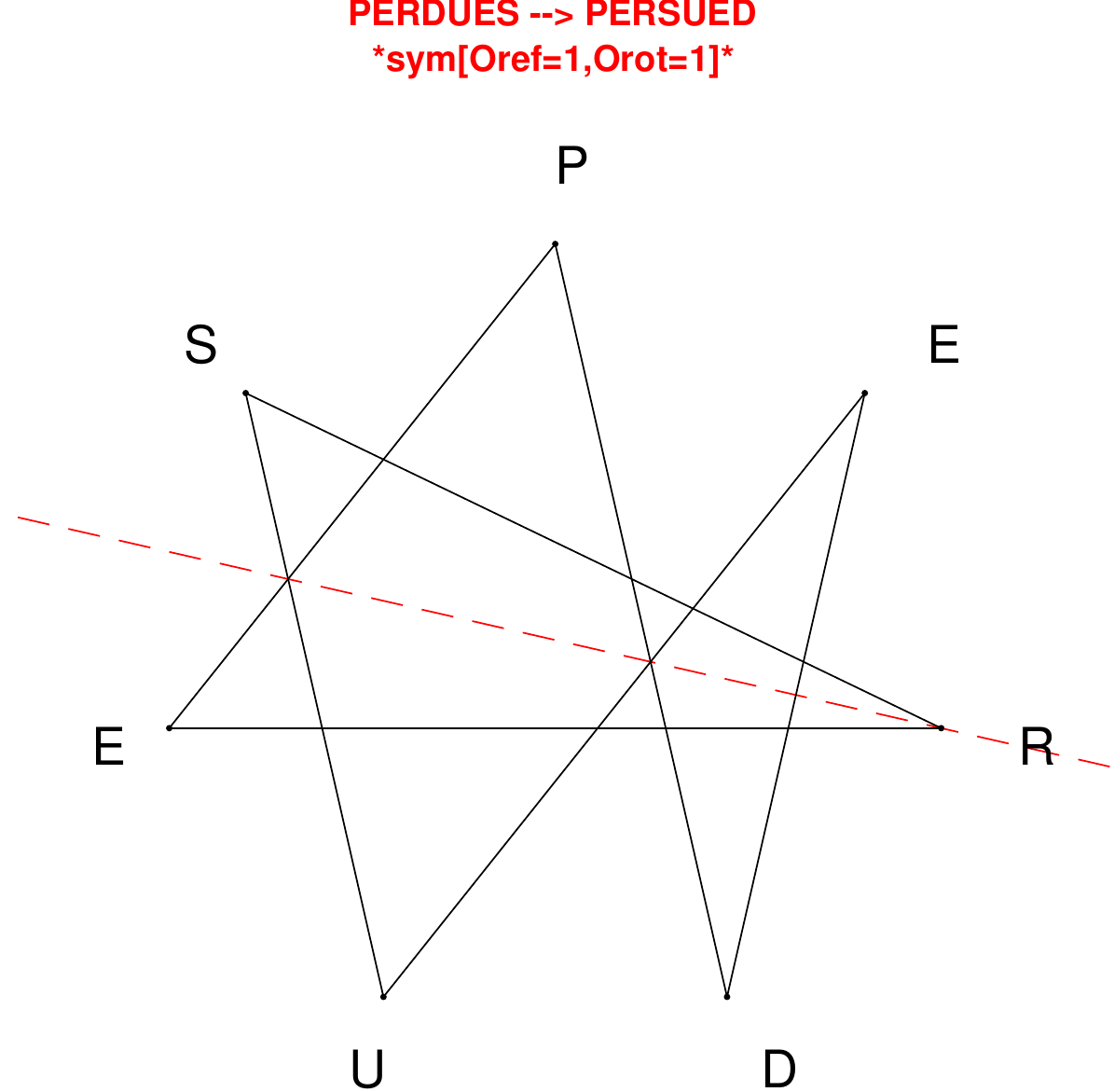}
\end{subfigure}
\hfill
\begin{subfigure}[T]{0.19\textwidth}
\centering
\includegraphics[width=\textwidth]{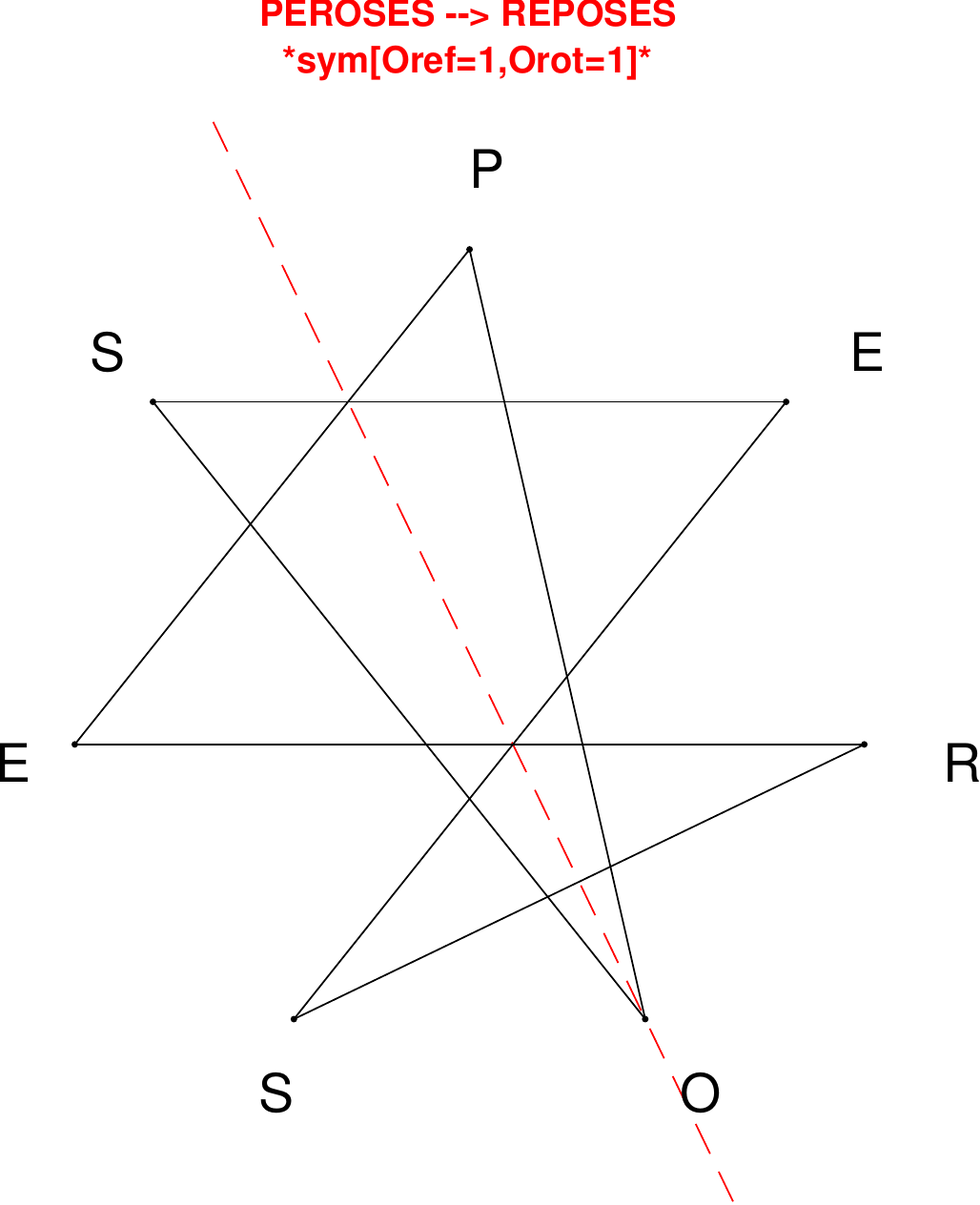}
\end{subfigure}
\hfill
\begin{subfigure}[T]{0.19\textwidth}
\centering
\includegraphics[width=\textwidth]{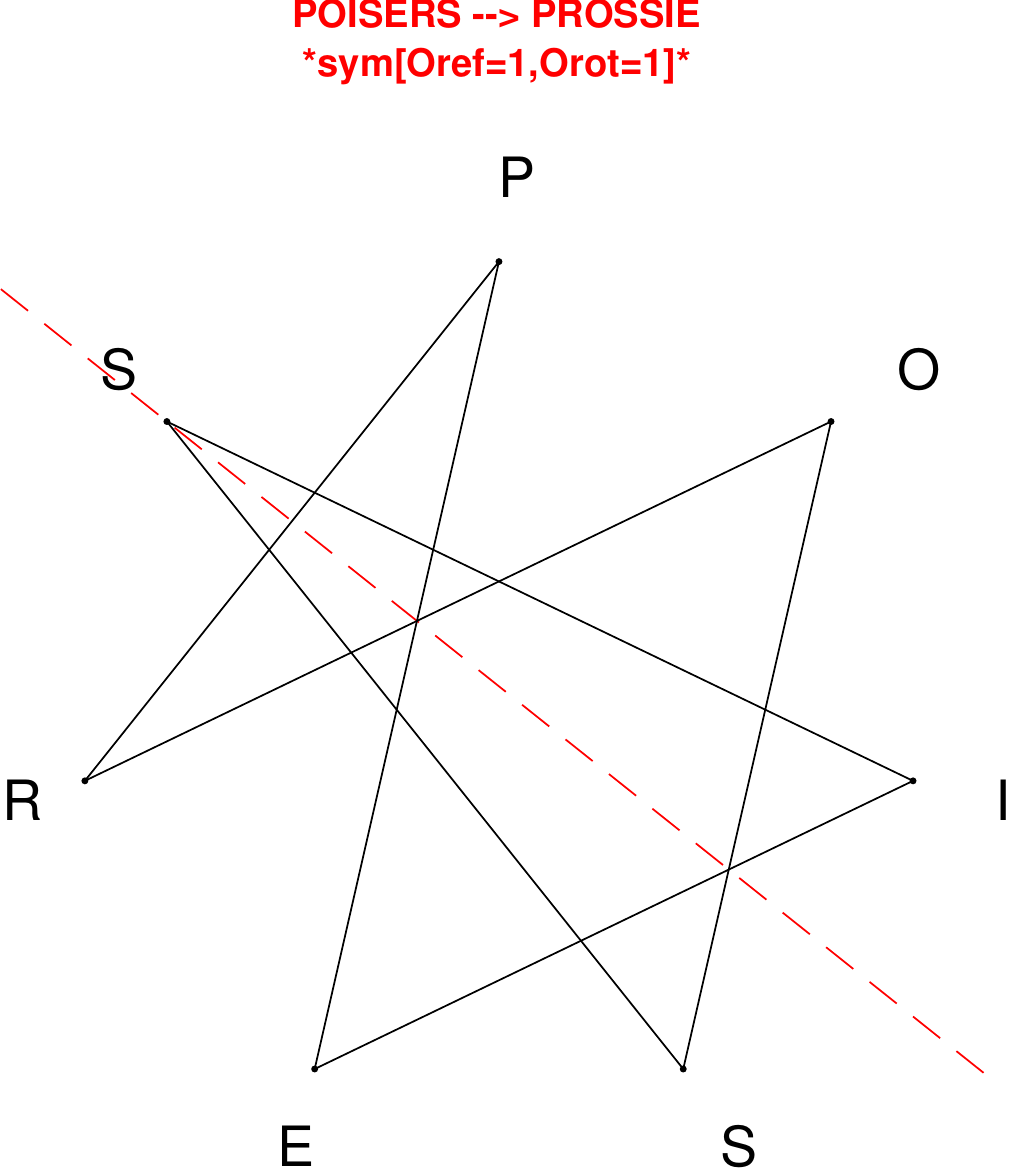}
\end{subfigure}
\hfill
\begin{subfigure}[T]{0.19\textwidth}
\centering
\includegraphics[width=\textwidth]{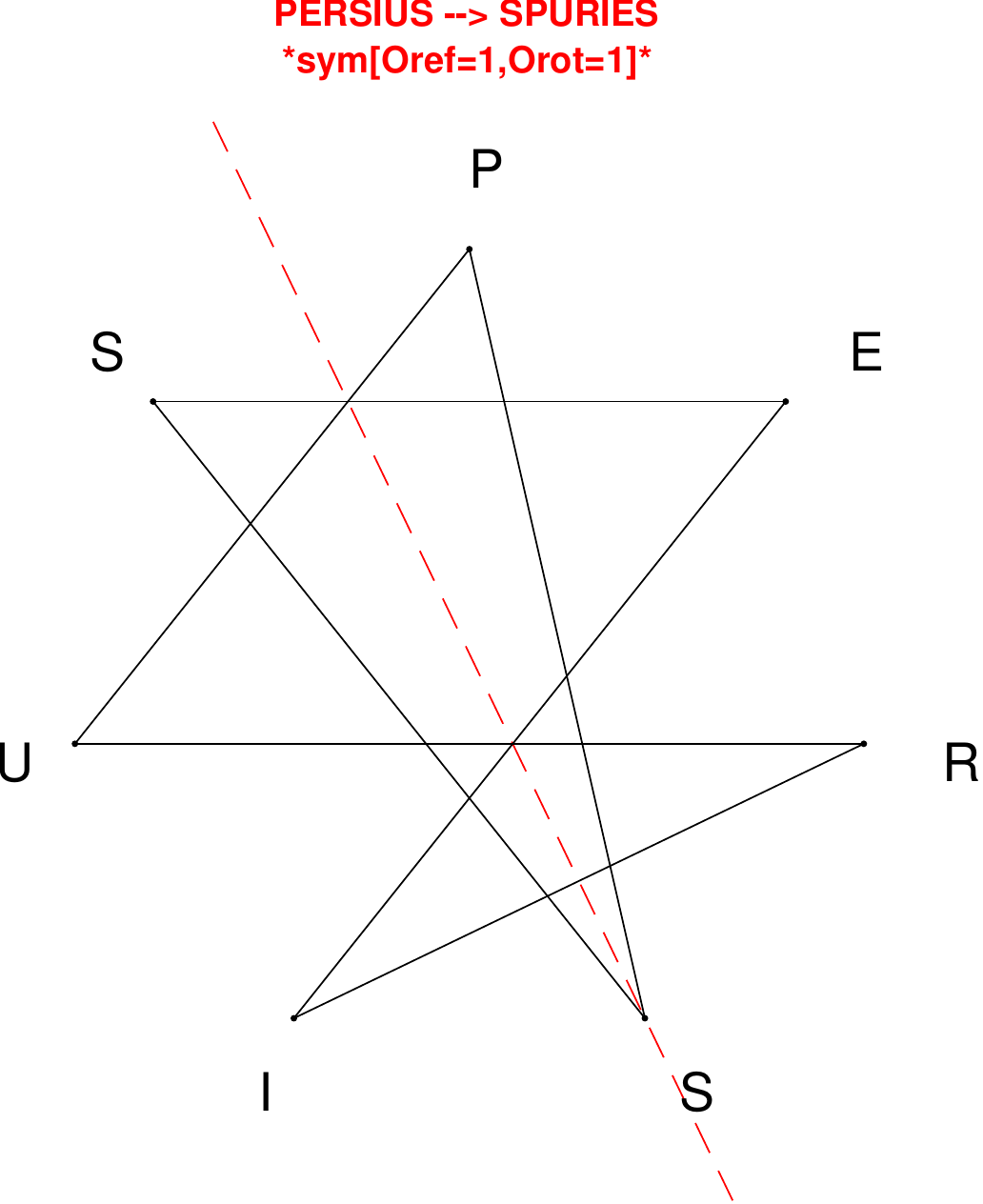}
\end{subfigure}
\hfill
\begin{subfigure}[T]{0.19\textwidth}
\centering
\includegraphics[width=\textwidth]{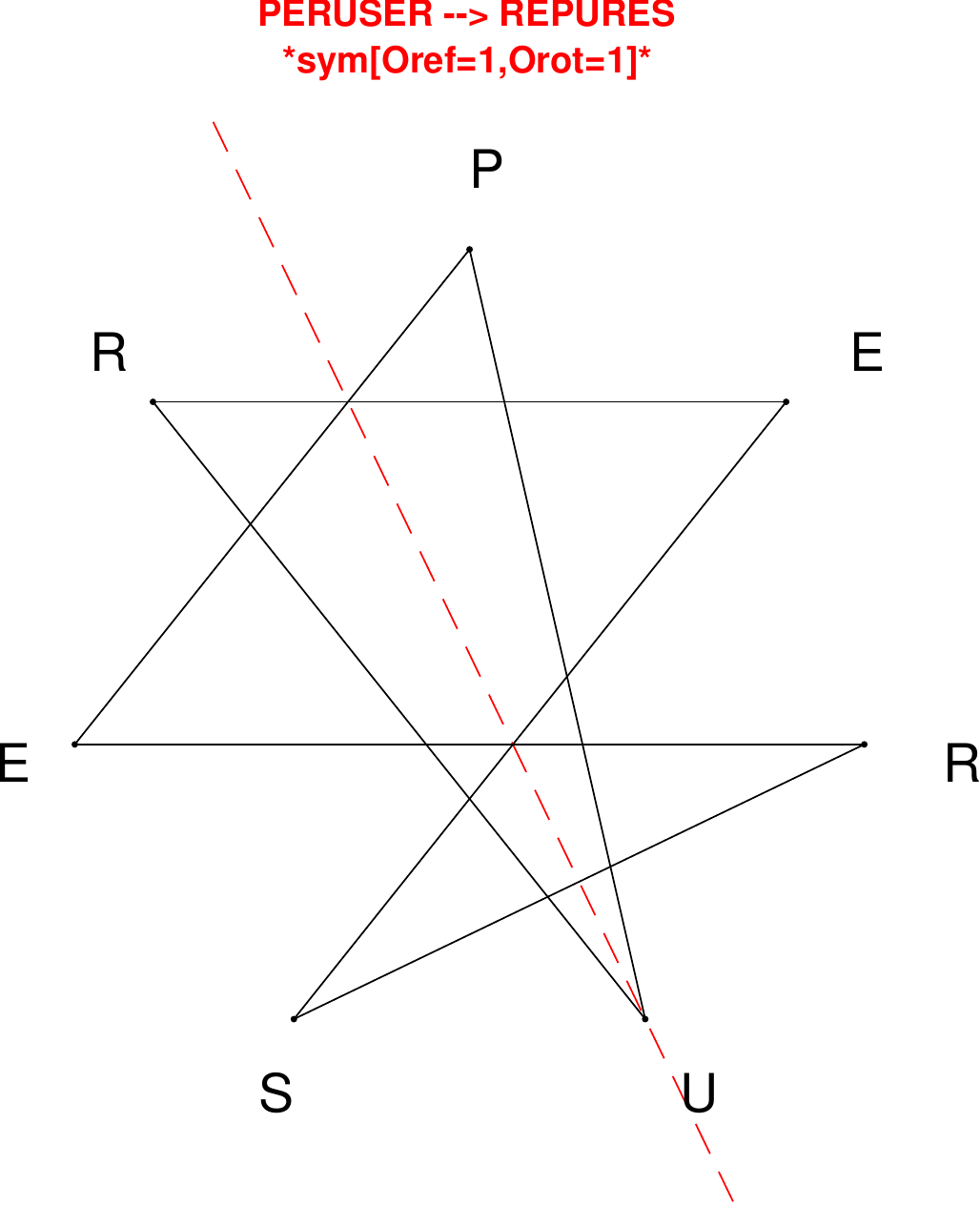}
\end{subfigure}
\end{figure}

\begin{figure}[H]
\centering
\begin{subfigure}[T]{0.19\textwidth}
\centering
\includegraphics[width=\textwidth]{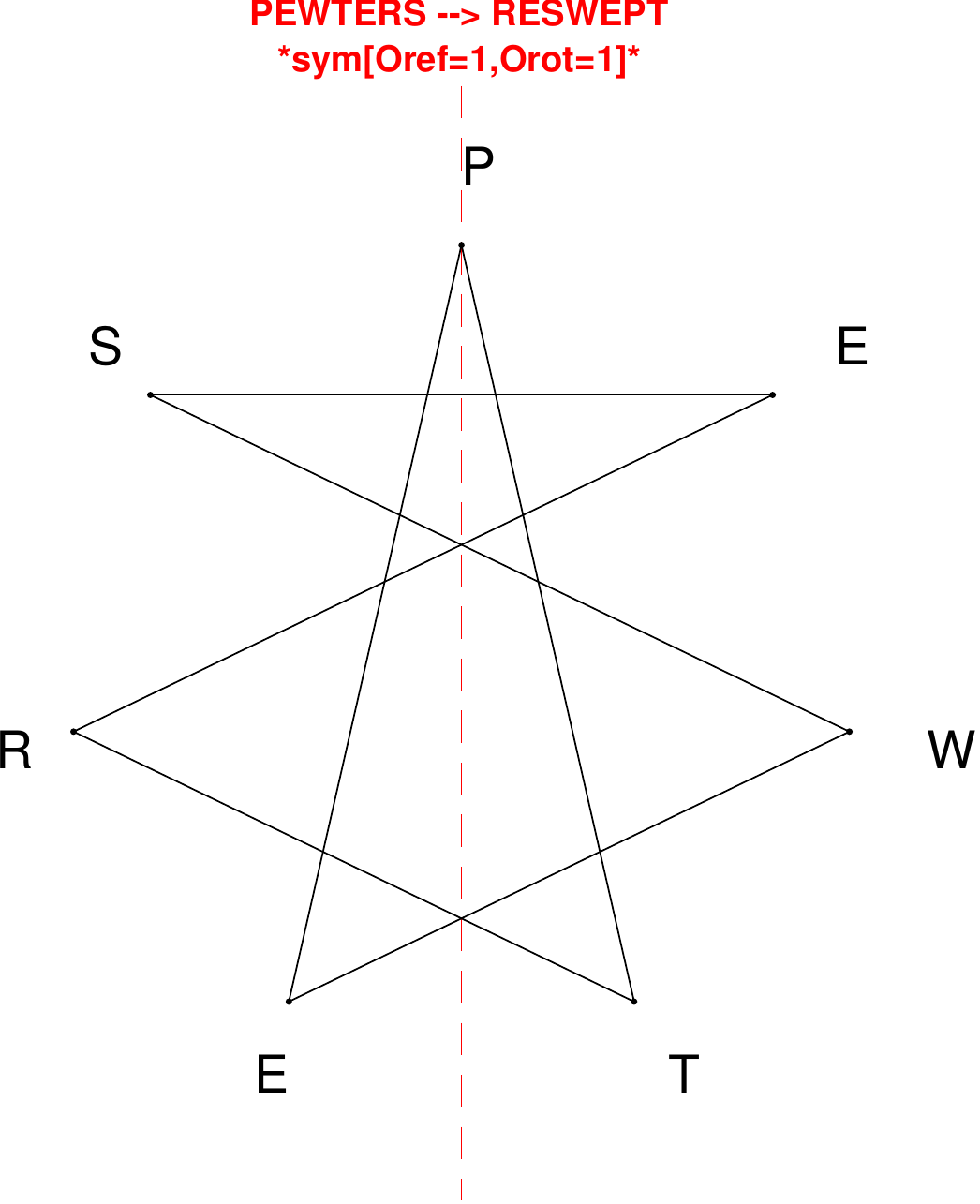}
\end{subfigure}
\hfill
\begin{subfigure}[T]{0.19\textwidth}
\centering
\includegraphics[width=\textwidth]{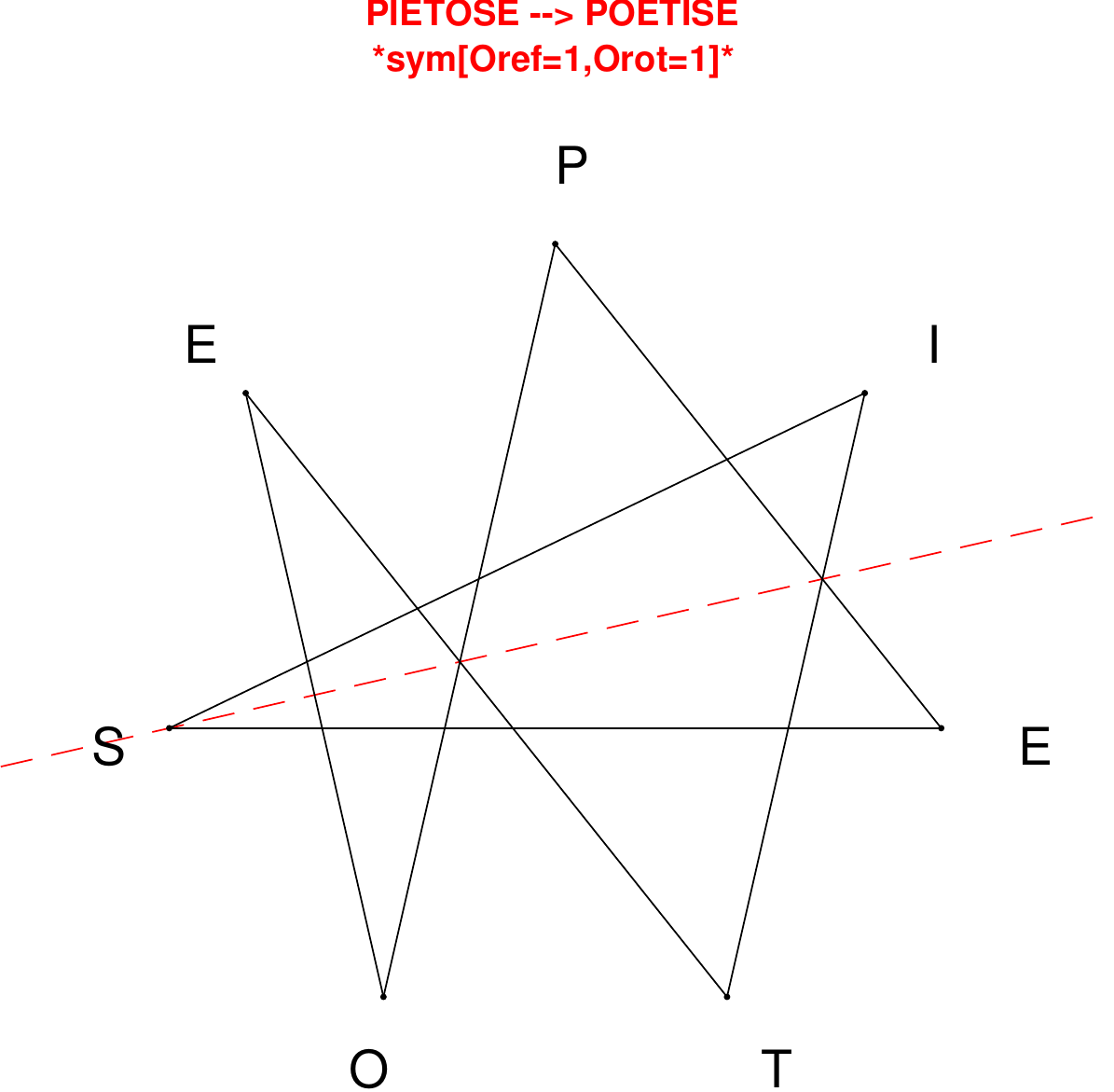}
\end{subfigure}
\hfill
\begin{subfigure}[T]{0.19\textwidth}
\centering
\includegraphics[width=\textwidth]{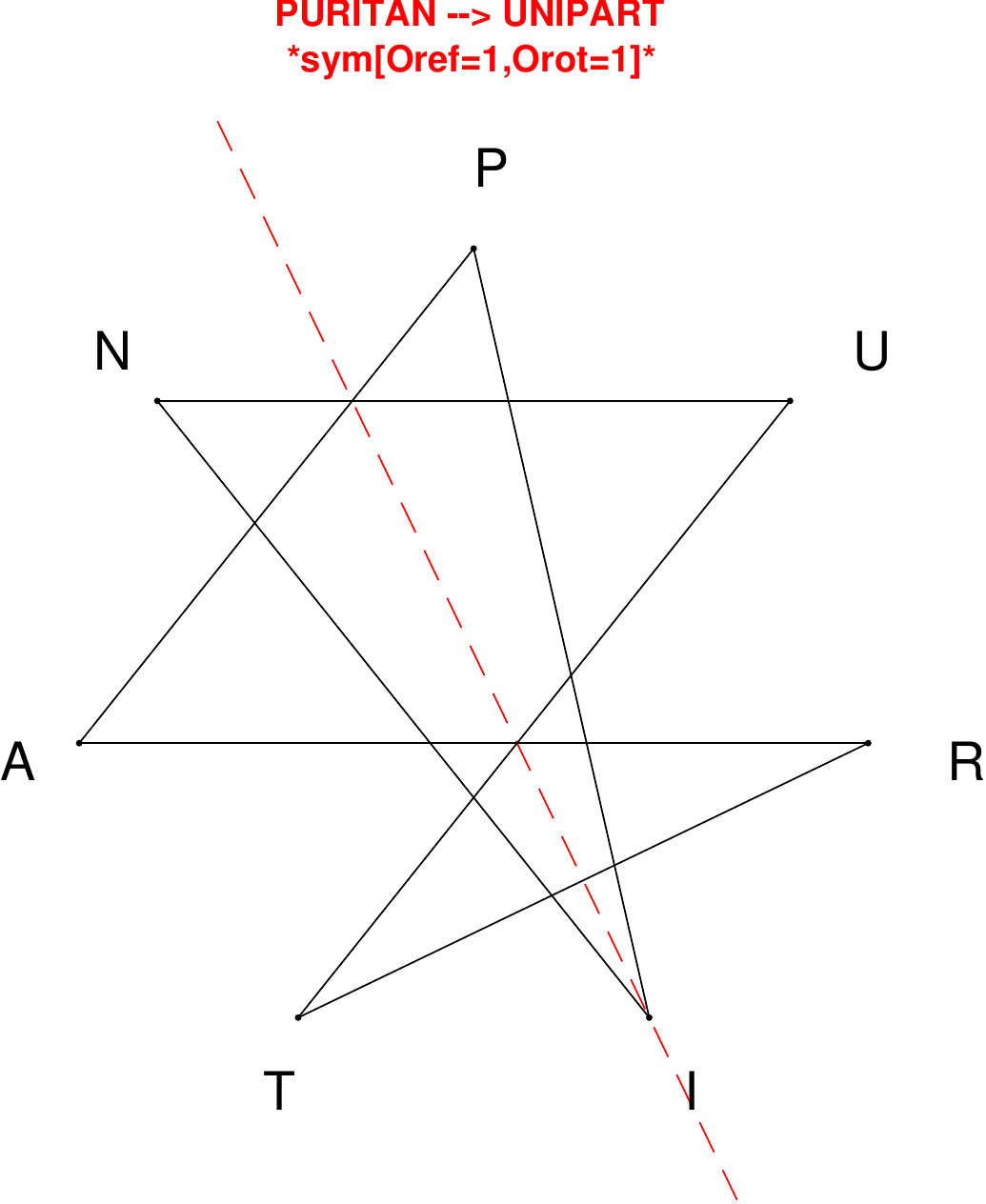}
\end{subfigure}
\hfill
\begin{subfigure}[T]{0.19\textwidth}
\centering
\includegraphics[width=\textwidth]{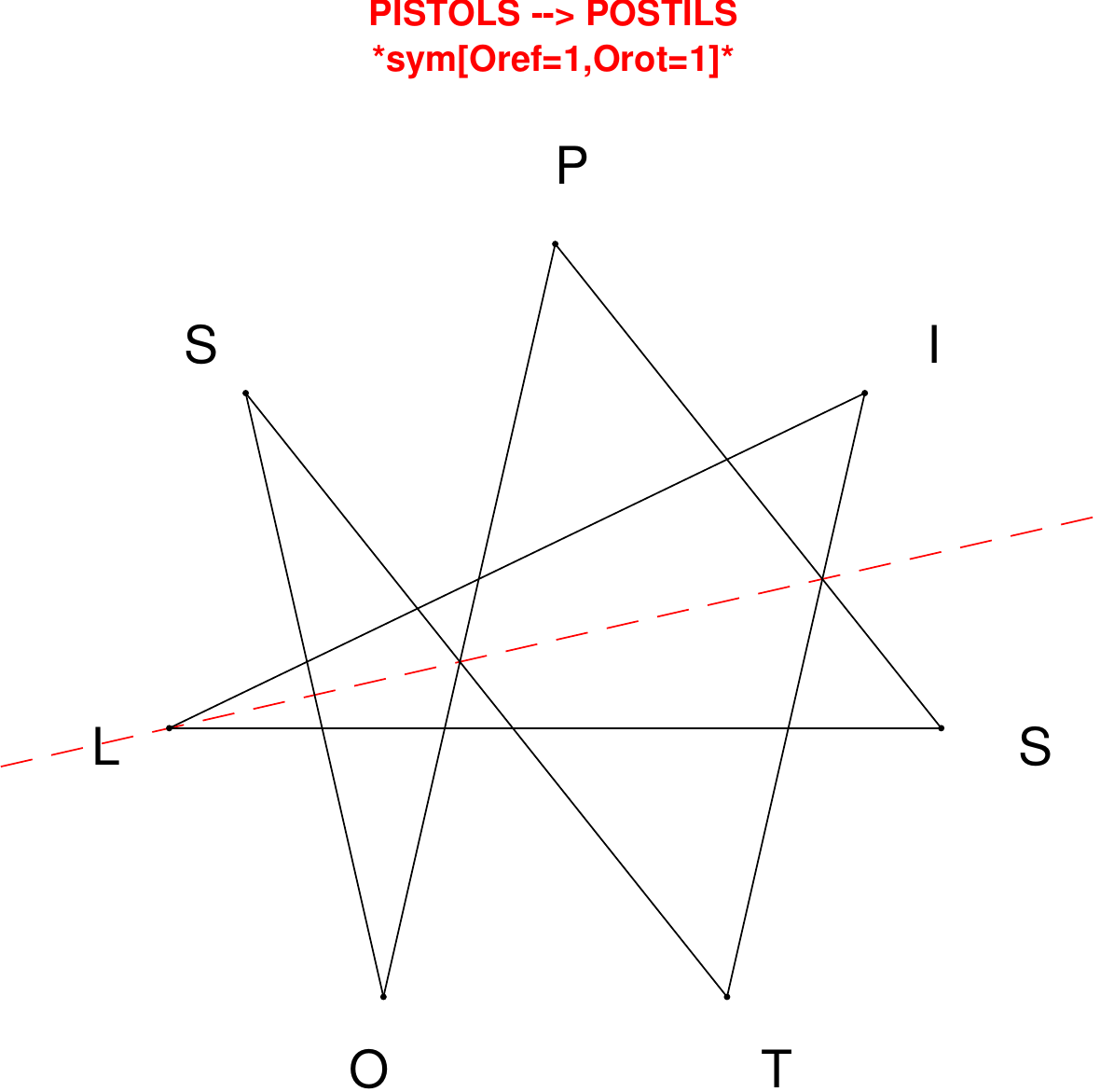}
\end{subfigure}
\hfill
\begin{subfigure}[T]{0.19\textwidth}
\centering
\includegraphics[width=\textwidth]{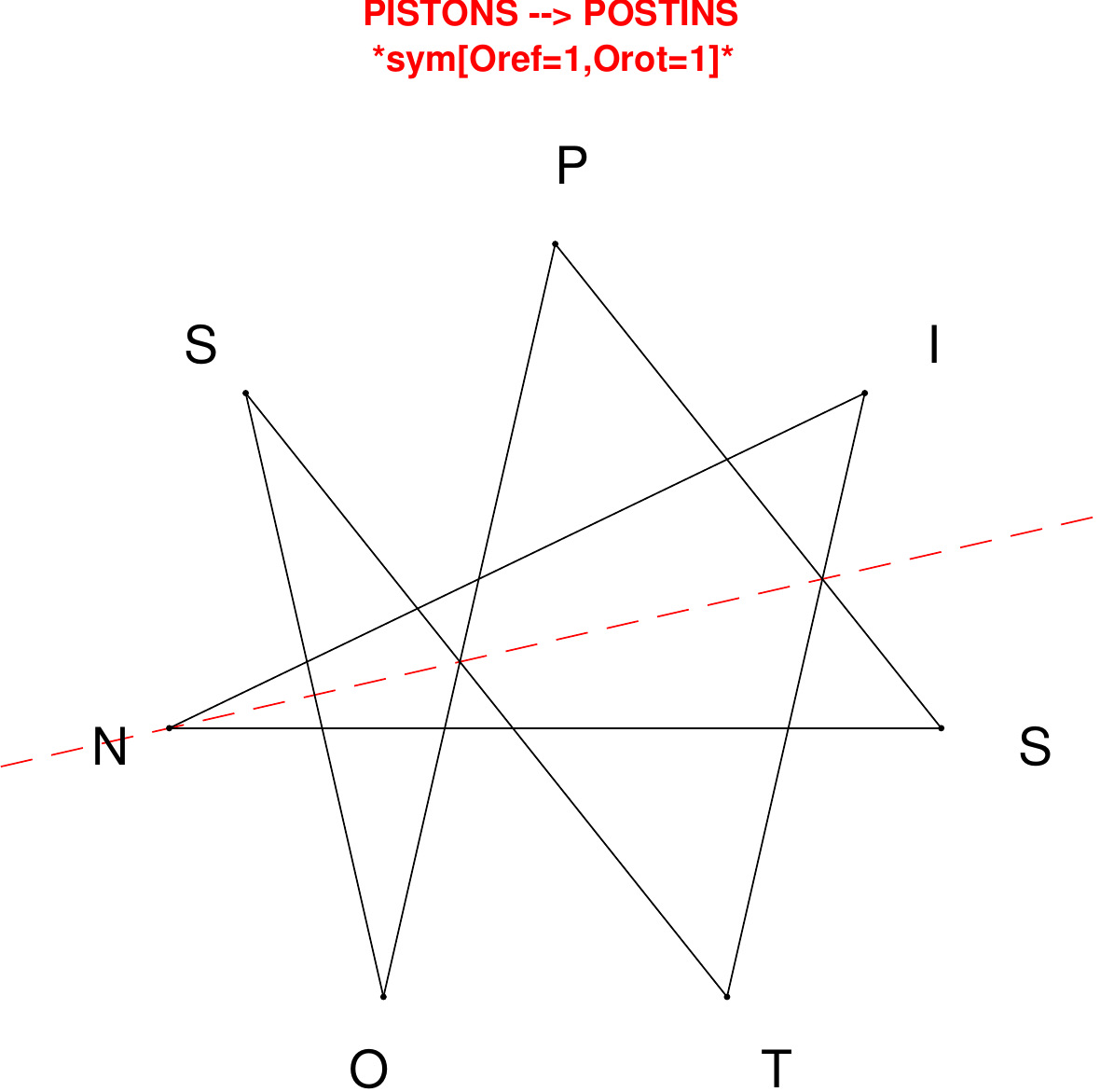}
\end{subfigure}
\end{figure}

\begin{figure}[H]
\centering
\begin{subfigure}[T]{0.19\textwidth}
\centering
\includegraphics[width=\textwidth]{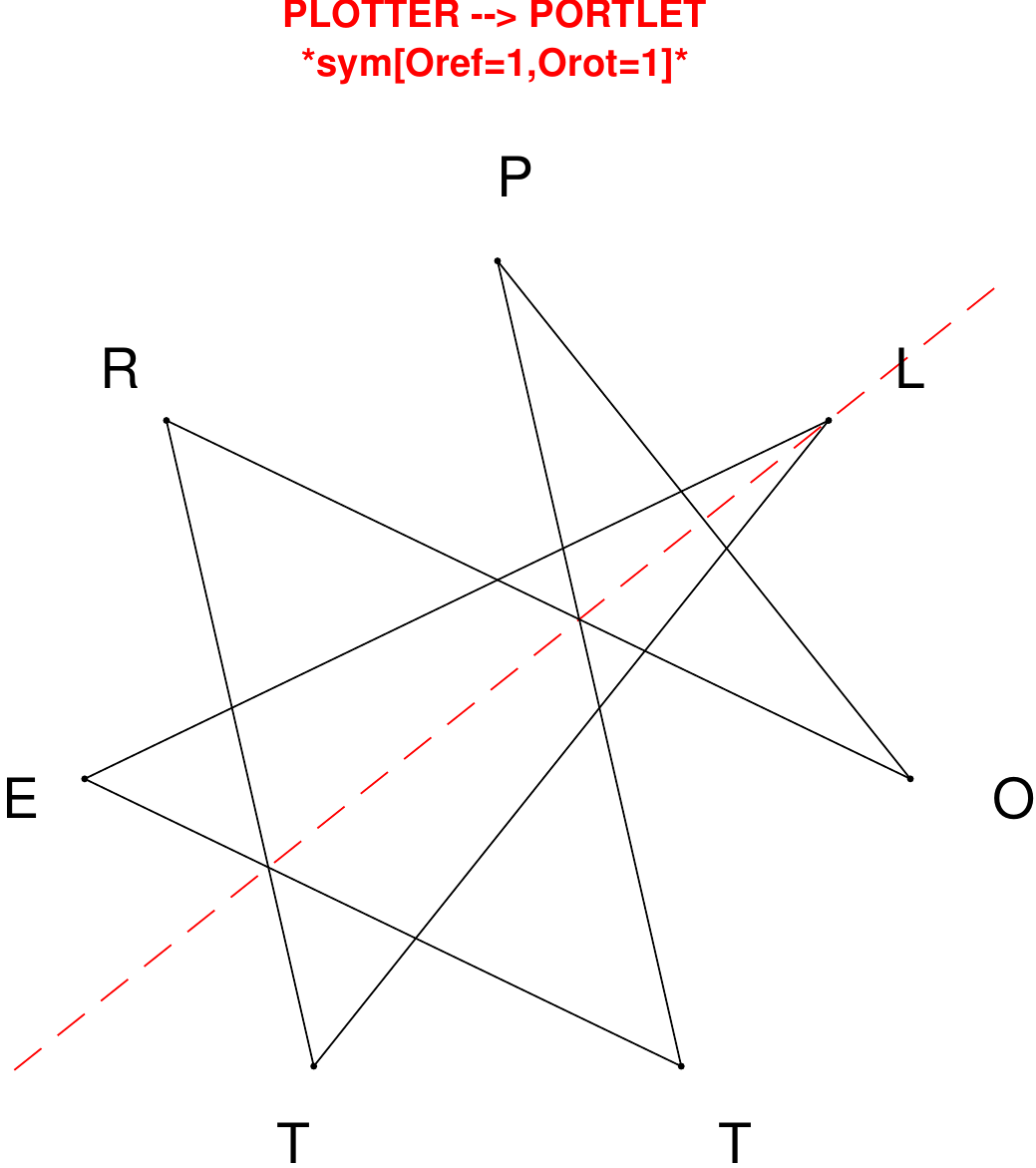}
\end{subfigure}
\hfill
\begin{subfigure}[T]{0.19\textwidth}
\centering
\includegraphics[width=\textwidth]{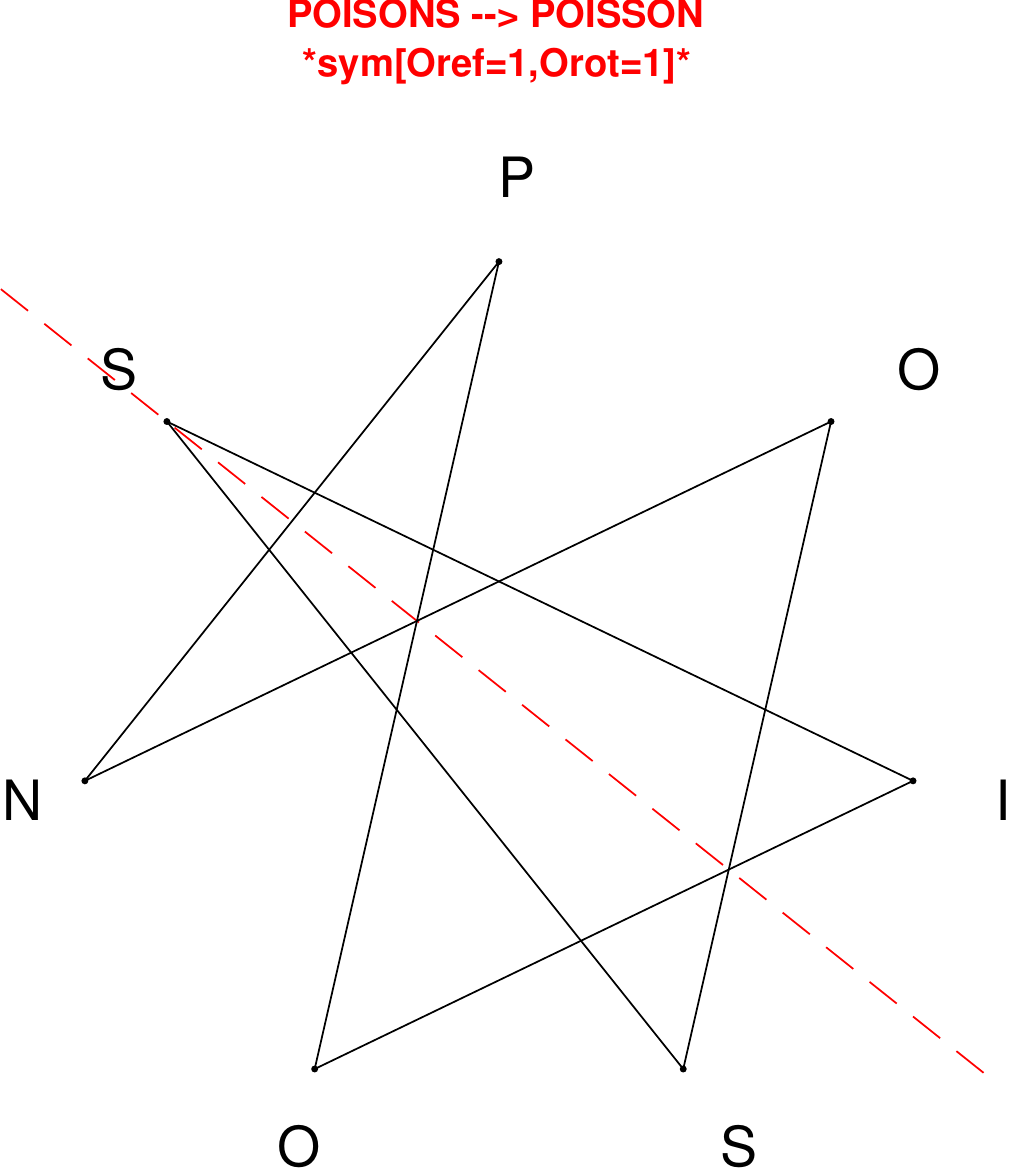}
\end{subfigure}
\hfill
\begin{subfigure}[T]{0.19\textwidth}
\centering
\includegraphics[width=\textwidth]{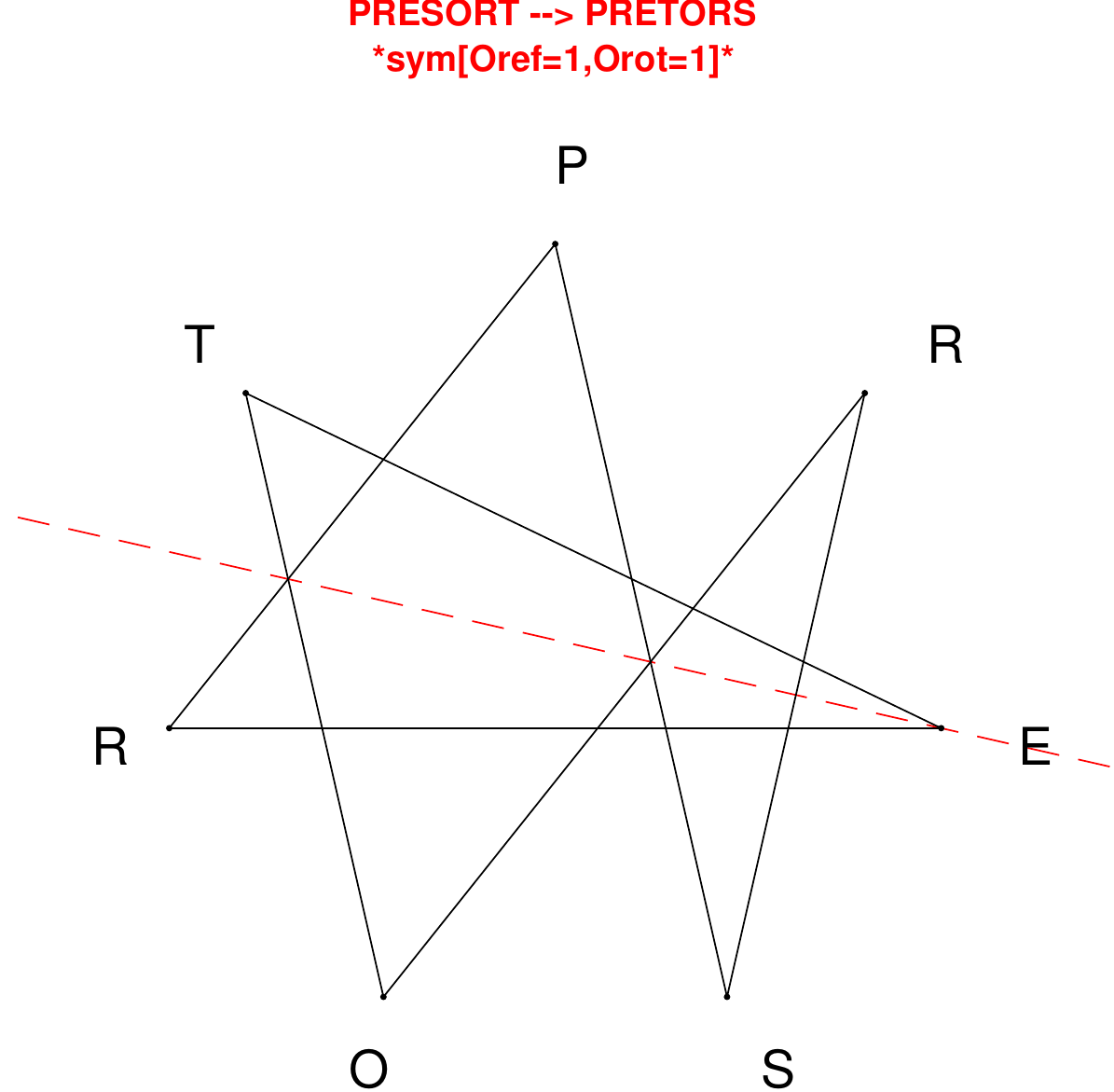}
\end{subfigure}
\hfill
\begin{subfigure}[T]{0.19\textwidth}
\centering
\includegraphics[width=\textwidth]{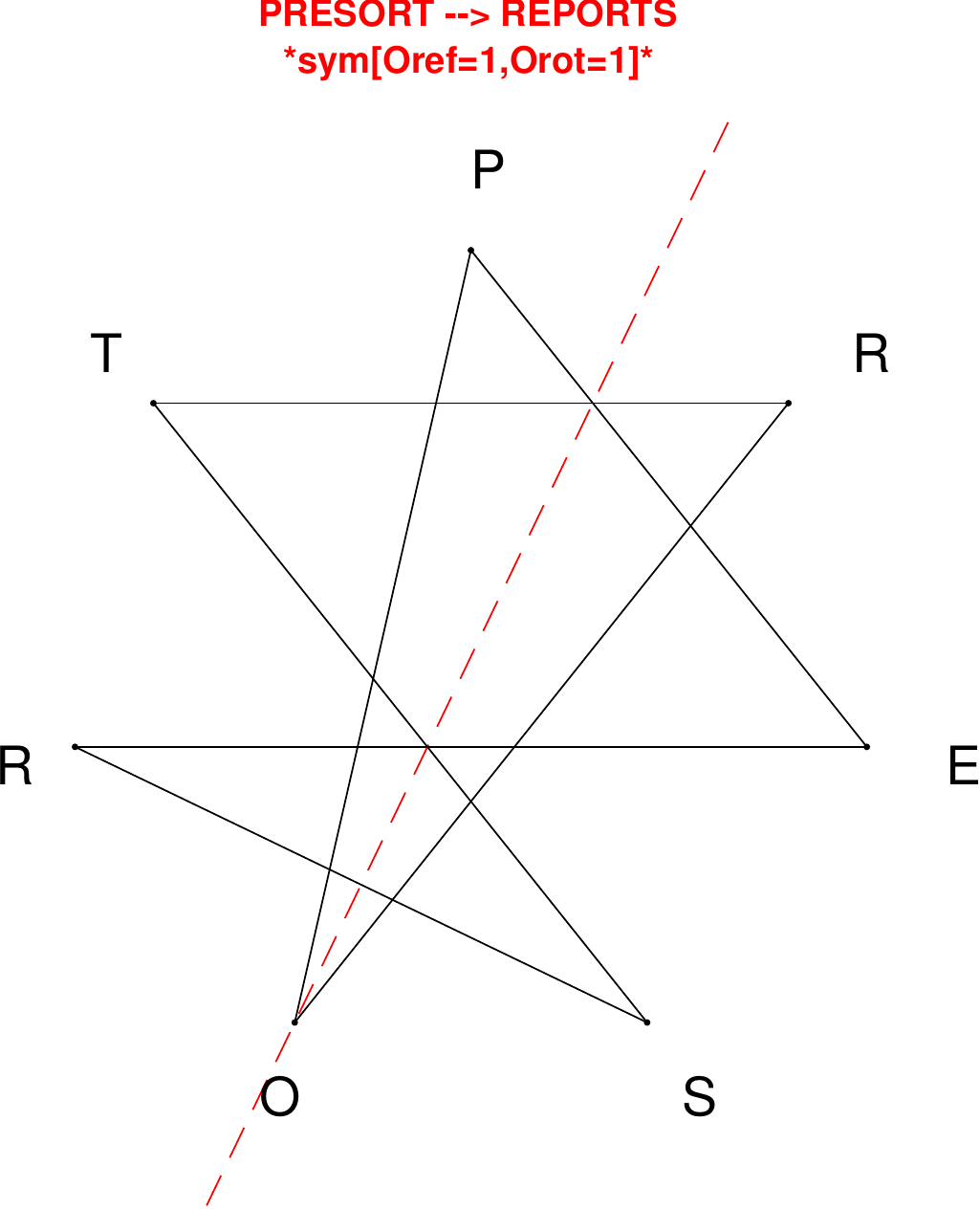}
\end{subfigure}
\hfill
\begin{subfigure}[T]{0.19\textwidth}
\centering
\includegraphics[width=\textwidth]{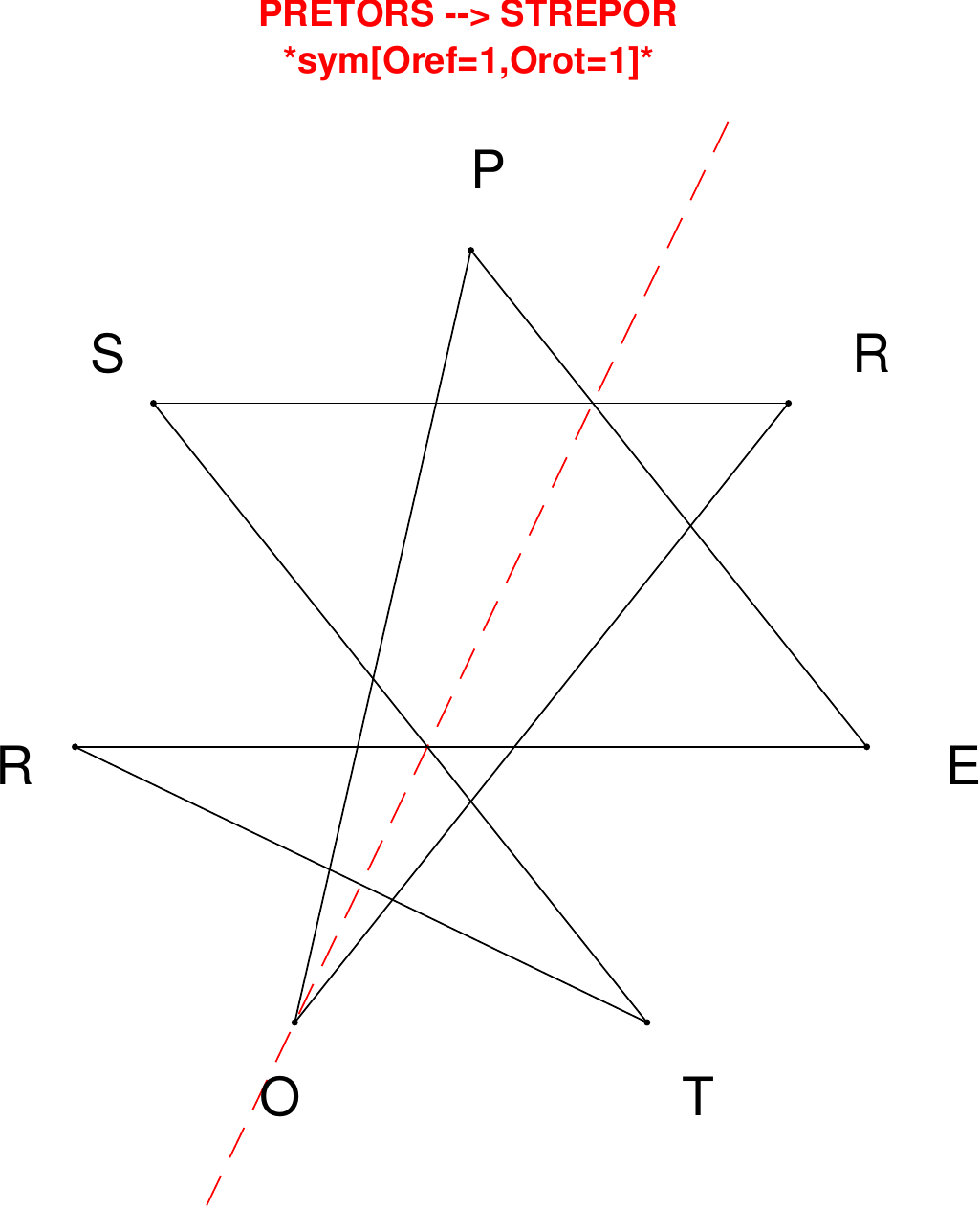}
\end{subfigure}
\end{figure}

\begin{figure}[H]
\centering
\begin{subfigure}[T]{0.19\textwidth}
\centering
\includegraphics[width=\textwidth]{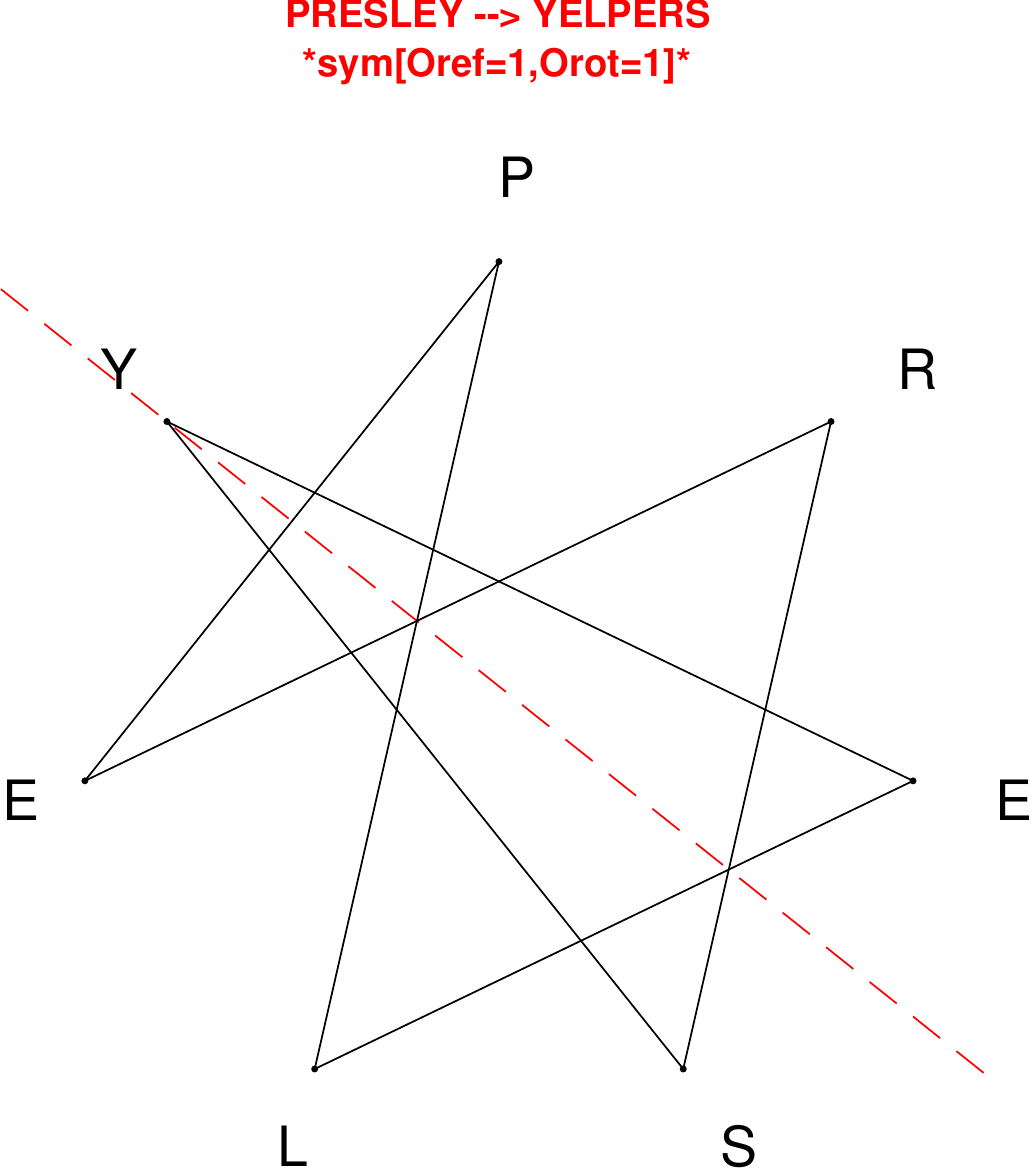}
\end{subfigure}
\hfill
\begin{subfigure}[T]{0.19\textwidth}
\centering
\includegraphics[width=\textwidth]{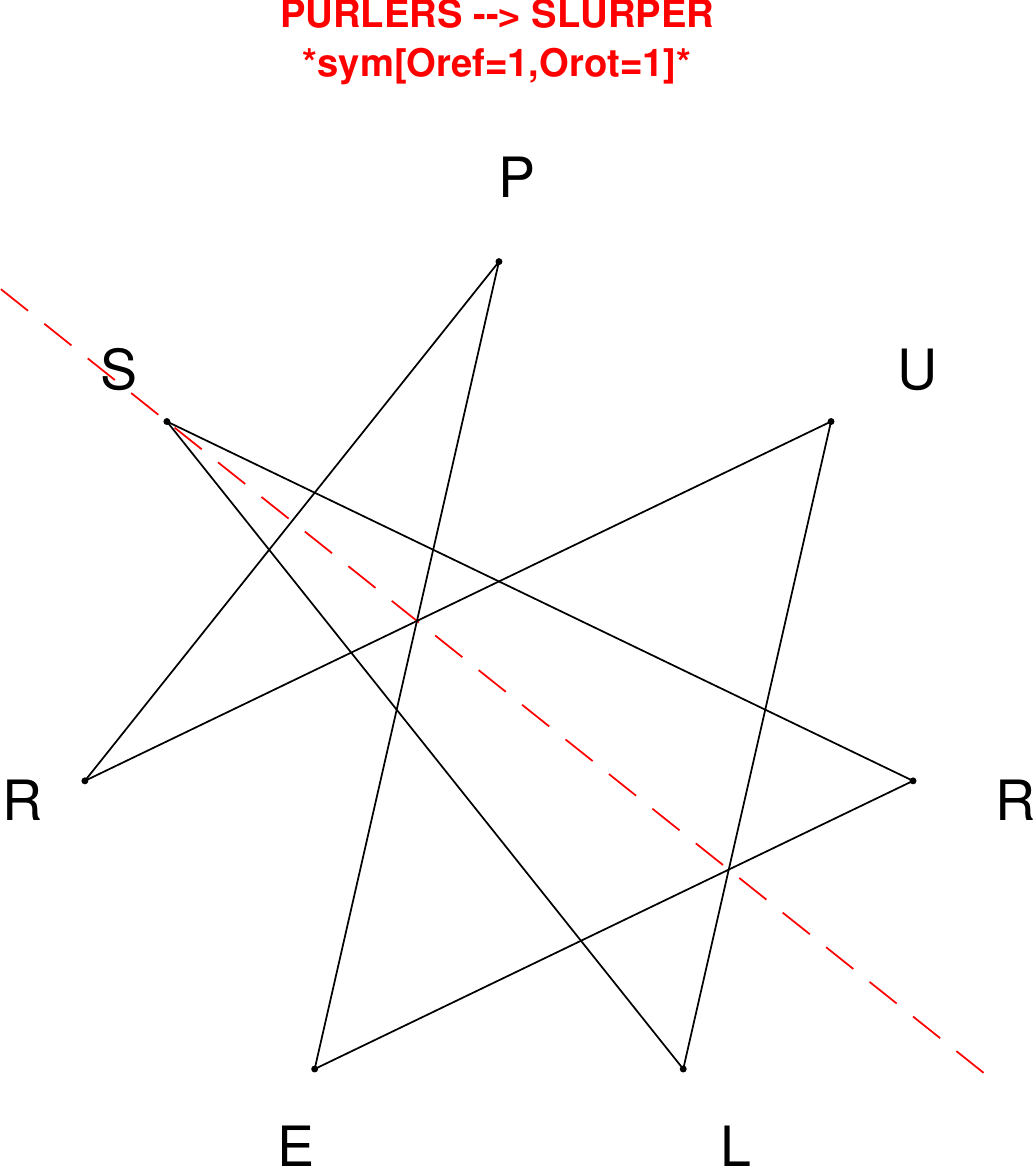}
\end{subfigure}
\hfill
\begin{subfigure}[T]{0.19\textwidth}
\centering
\includegraphics[width=\textwidth]{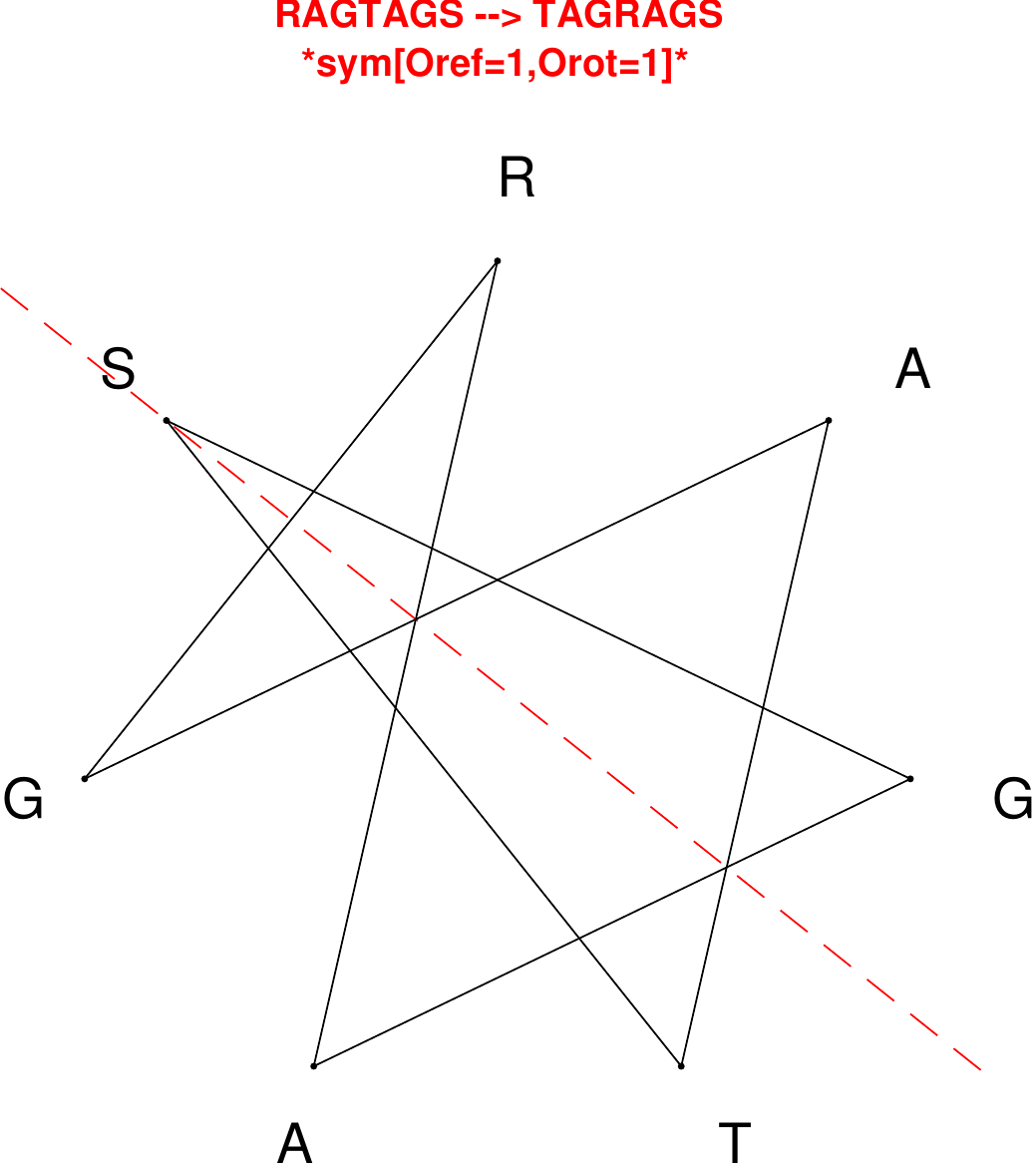}
\end{subfigure}
\hfill
\begin{subfigure}[T]{0.19\textwidth}
\centering
\includegraphics[width=\textwidth]{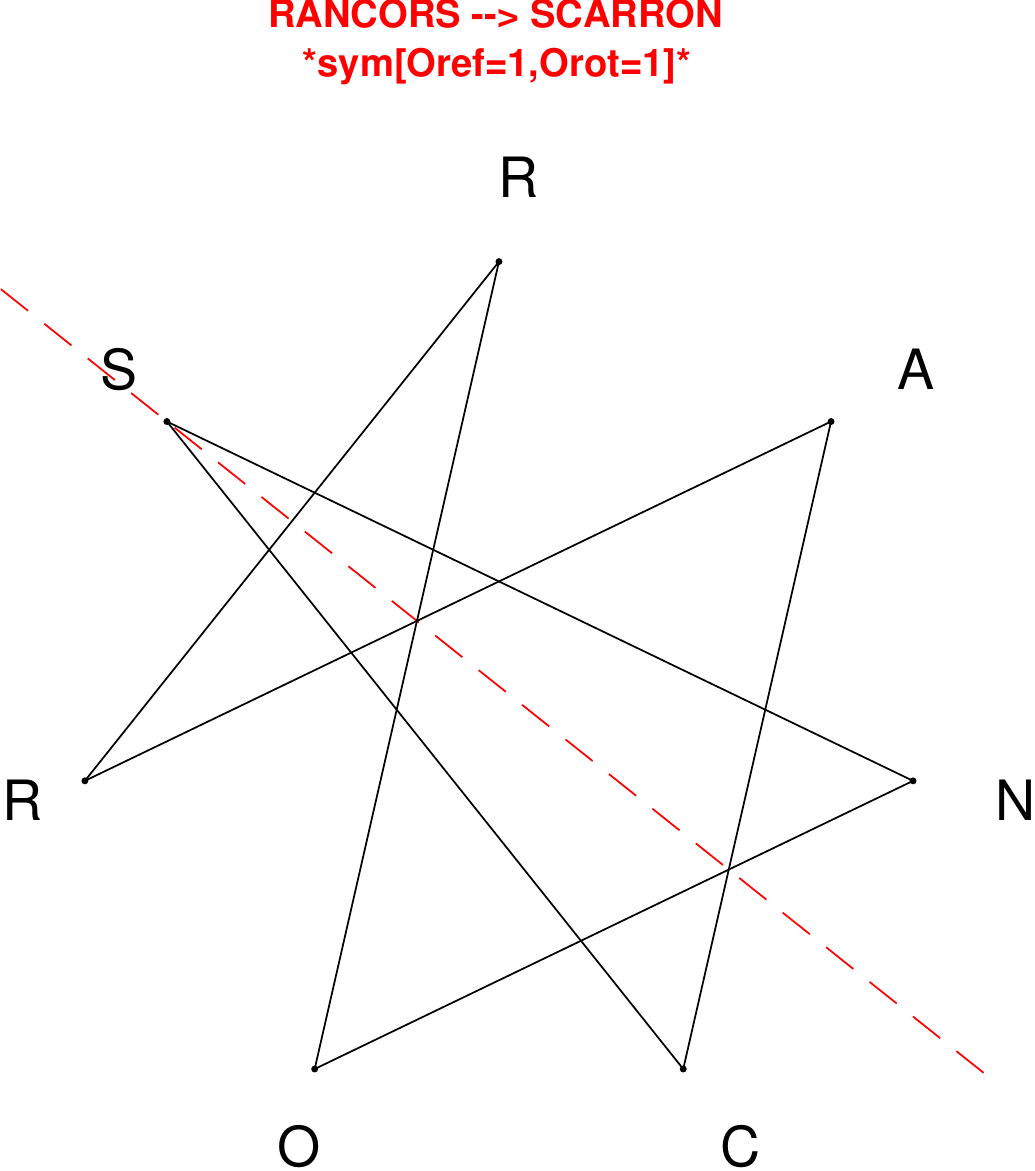}
\end{subfigure}
\hfill
\begin{subfigure}[T]{0.19\textwidth}
\centering
\includegraphics[width=\textwidth]{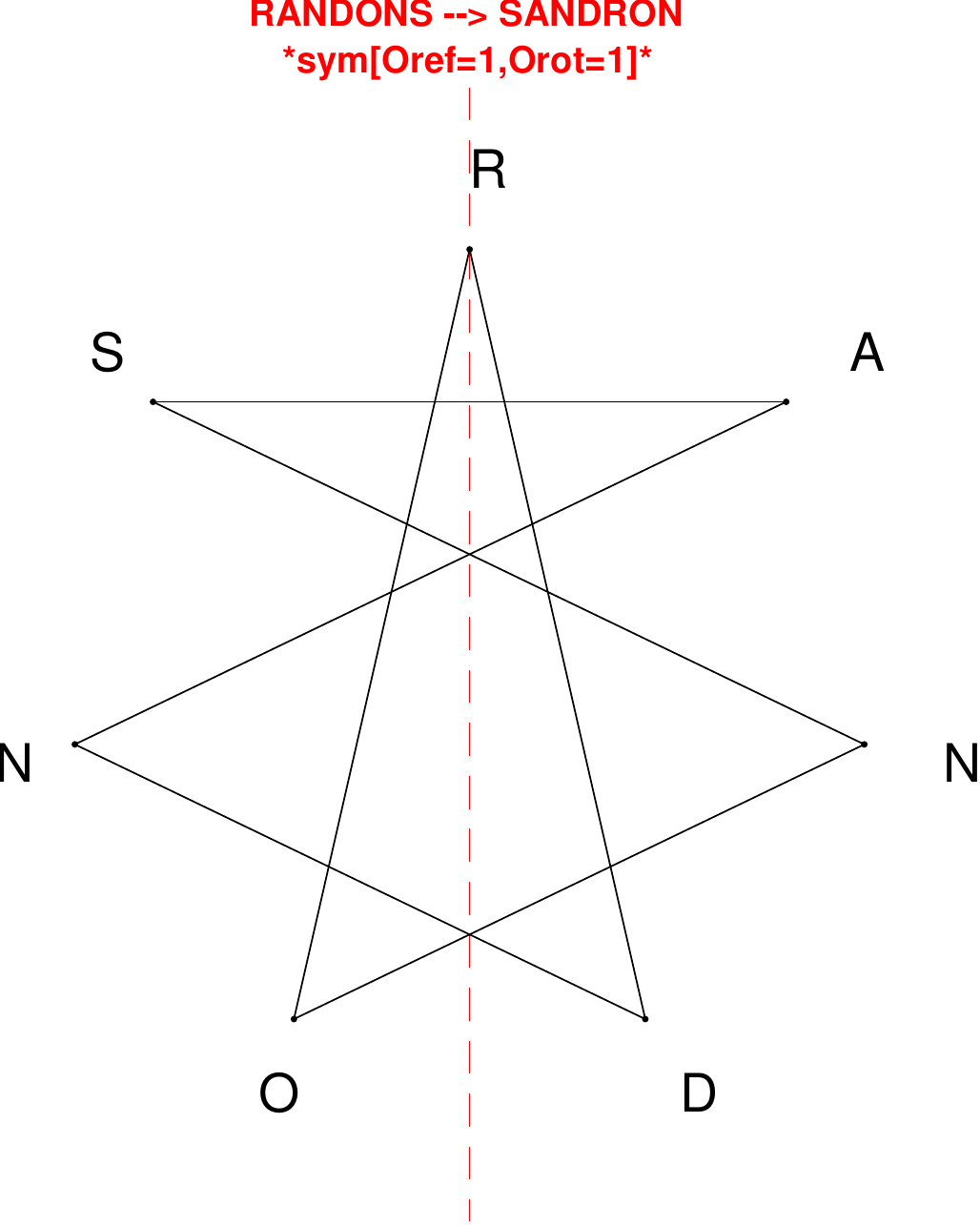}
\end{subfigure}
\end{figure}

\begin{figure}[H]
\centering
\begin{subfigure}[T]{0.19\textwidth}
\centering
\includegraphics[width=\textwidth]{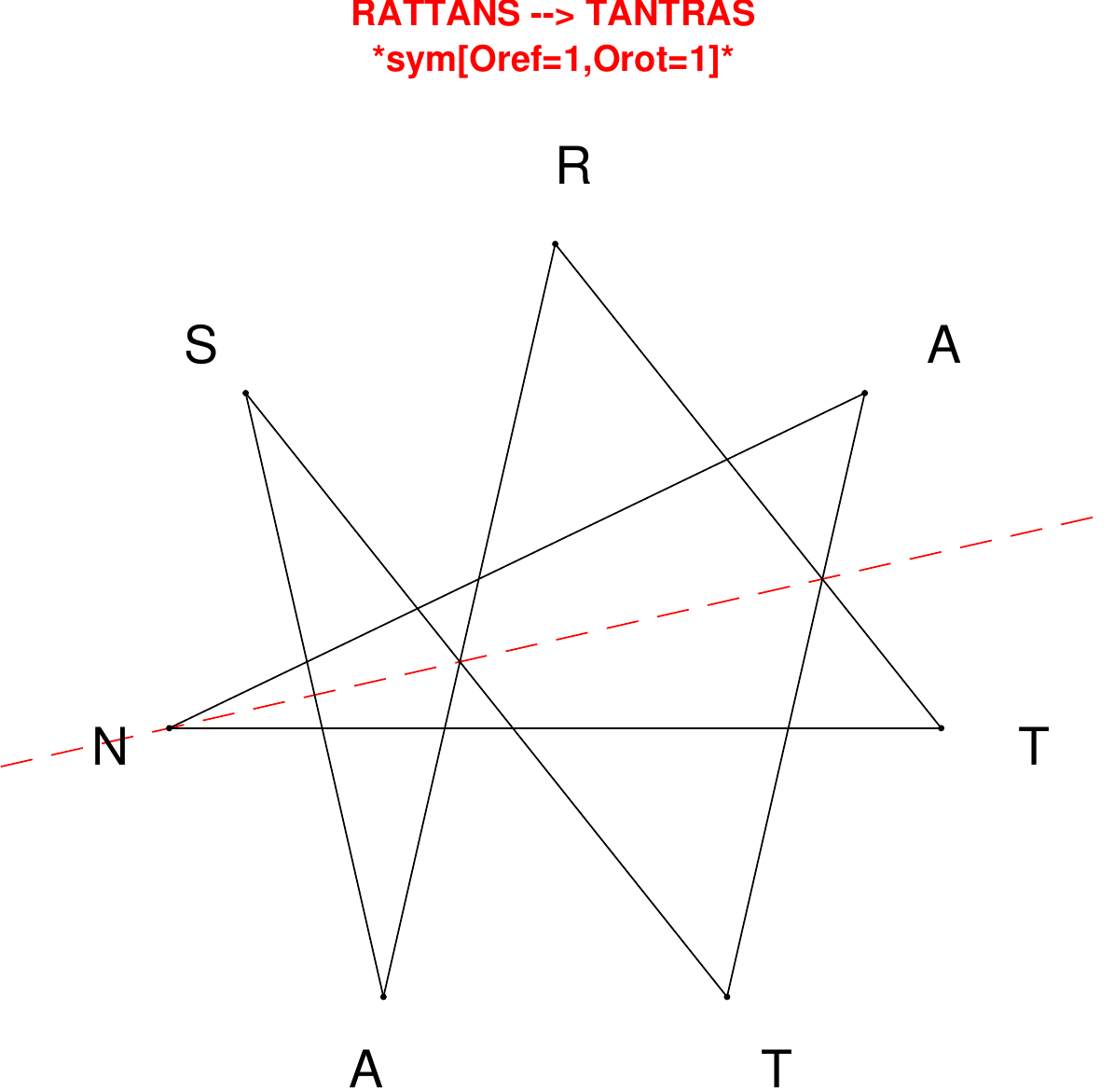}
\end{subfigure}
\hfill
\begin{subfigure}[T]{0.19\textwidth}
\centering
\includegraphics[width=\textwidth]{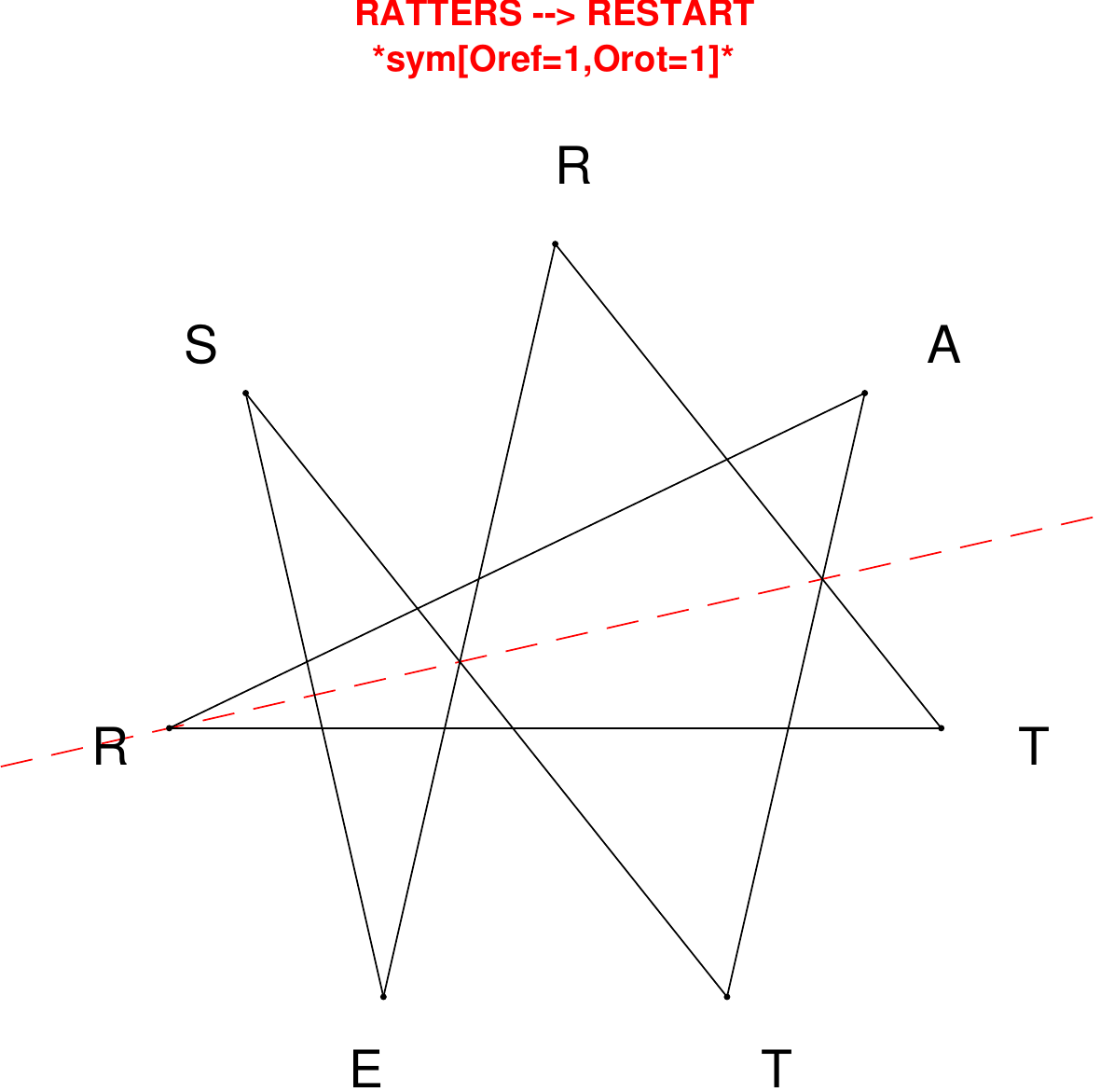}
\end{subfigure}
\hfill
\begin{subfigure}[T]{0.19\textwidth}
\centering
\includegraphics[width=\textwidth]{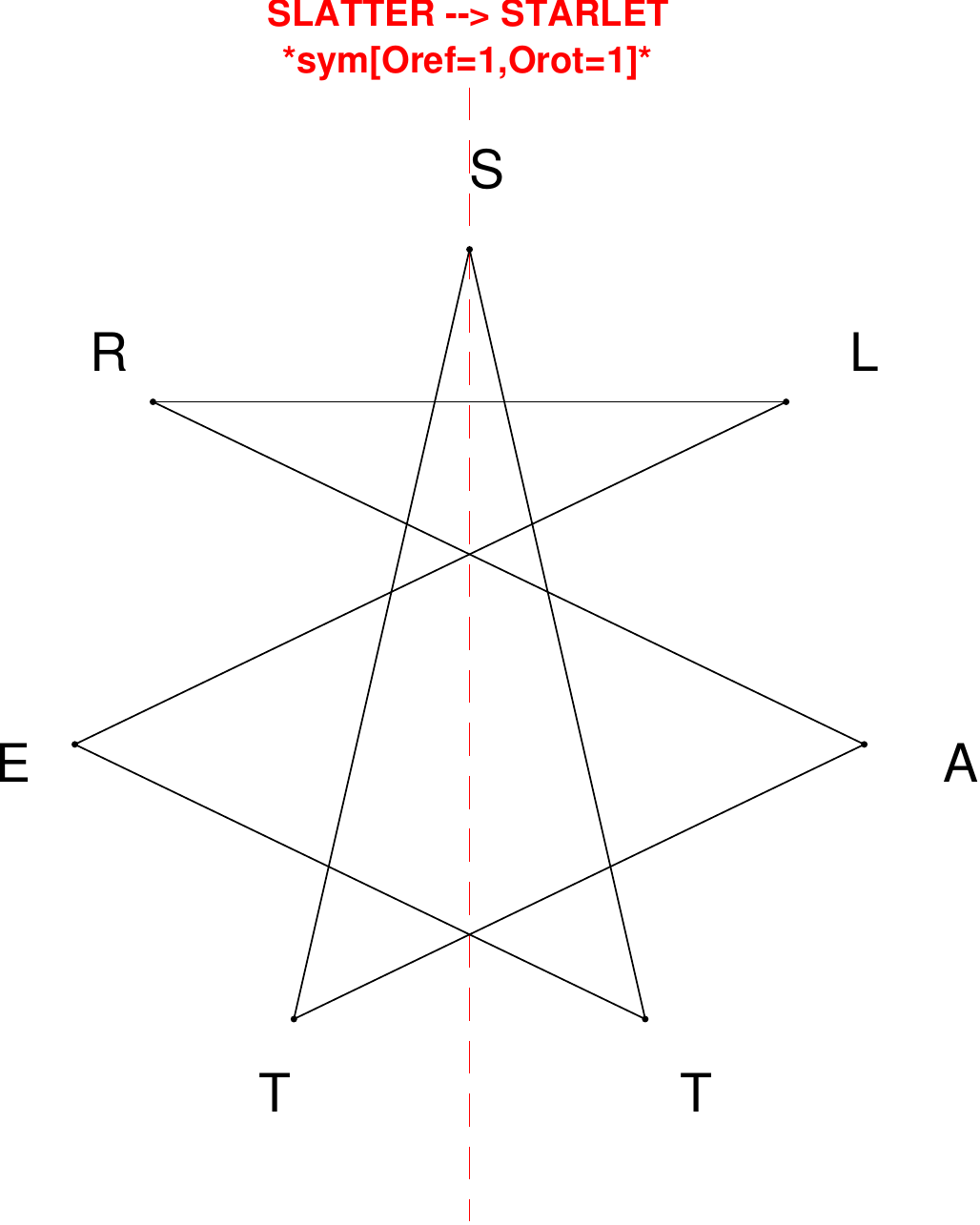}
\end{subfigure}
\hfill
\begin{subfigure}[T]{0.19\textwidth}
\centering
\includegraphics[width=\textwidth]{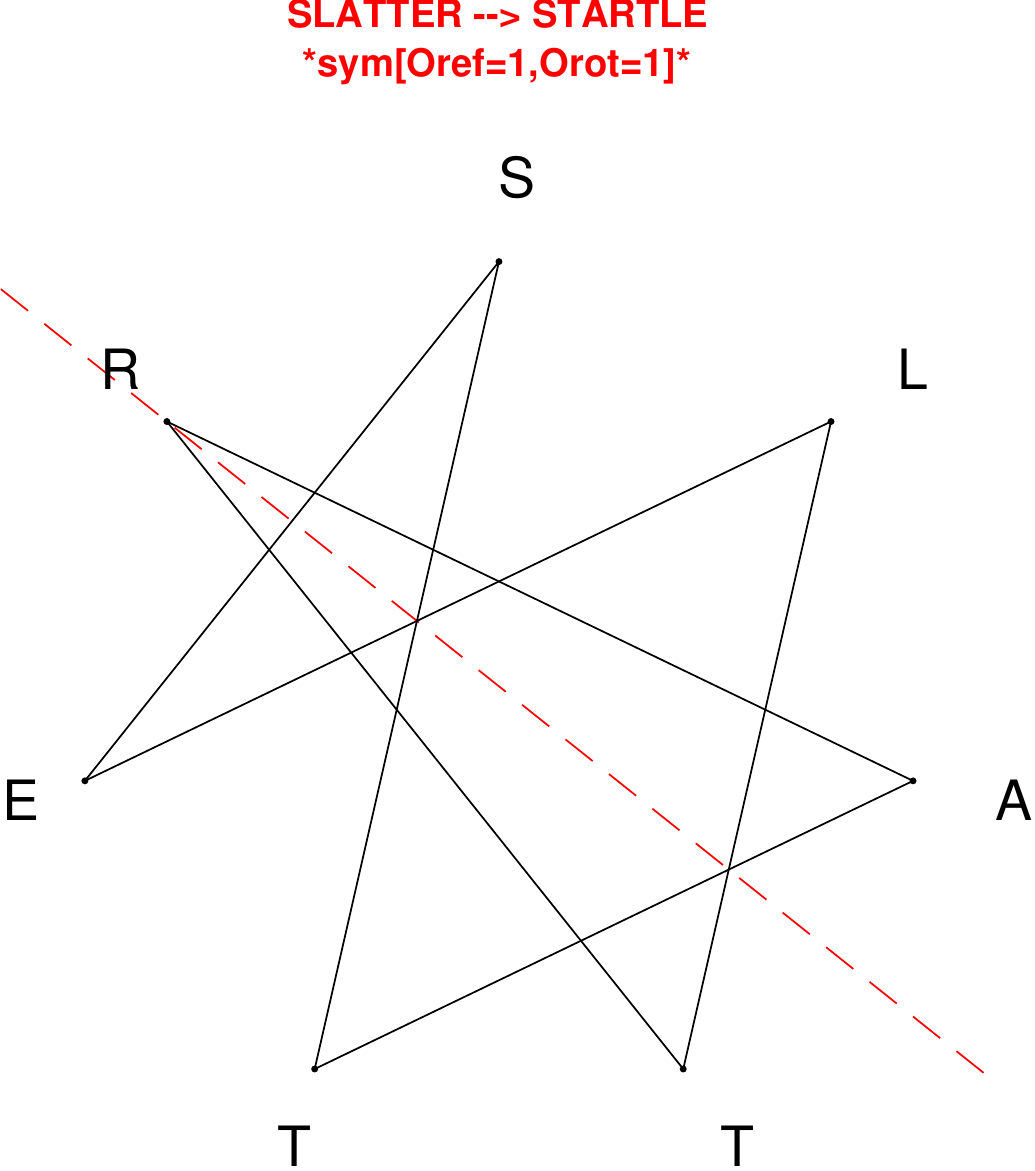}
\end{subfigure}
\hfill
\begin{subfigure}[T]{0.19\textwidth}
\centering
\includegraphics[width=\textwidth]{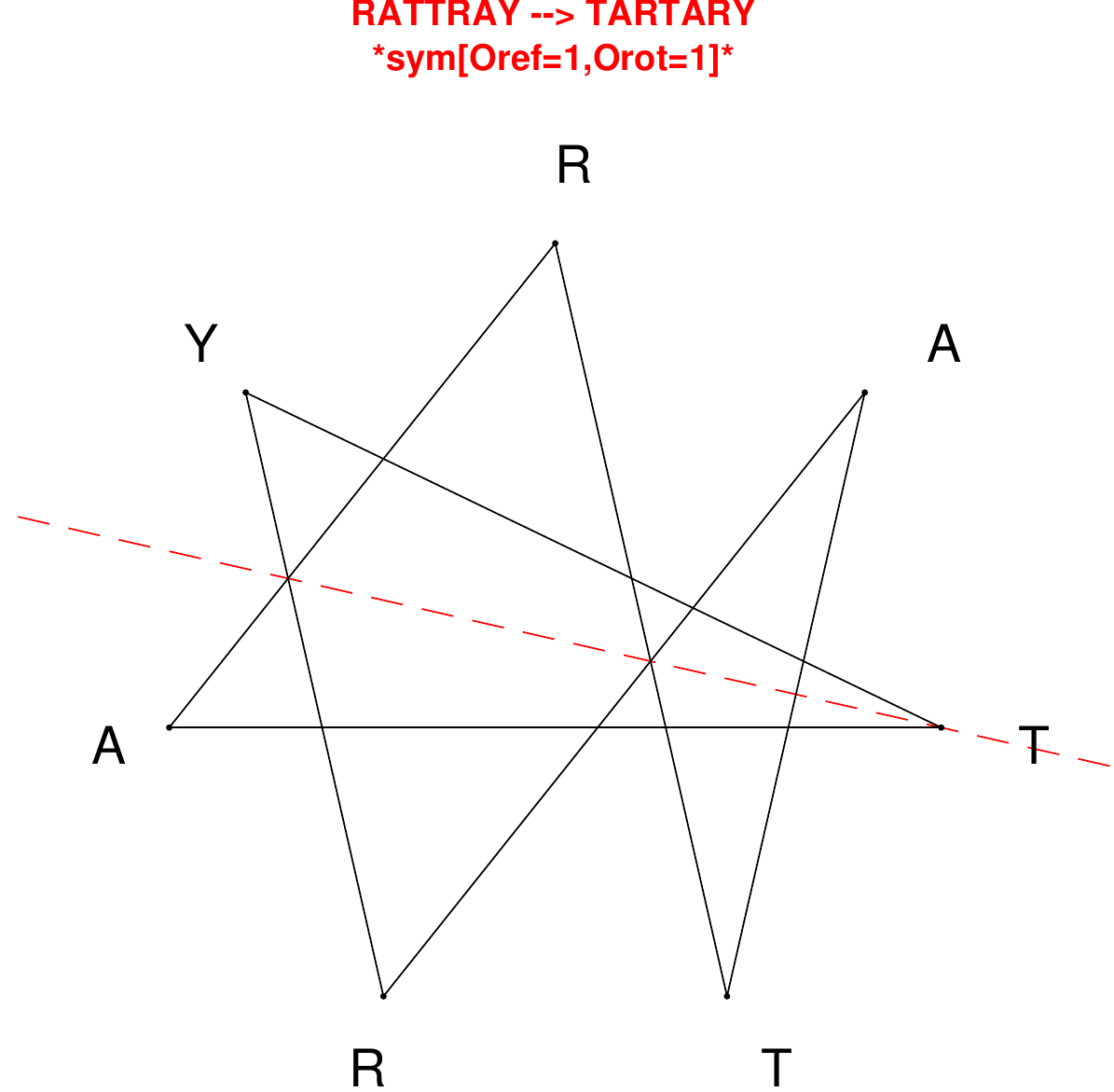}
\end{subfigure}
\end{figure}

\begin{figure}[H]
\centering
\begin{subfigure}[T]{0.19\textwidth}
\centering
\includegraphics[width=\textwidth]{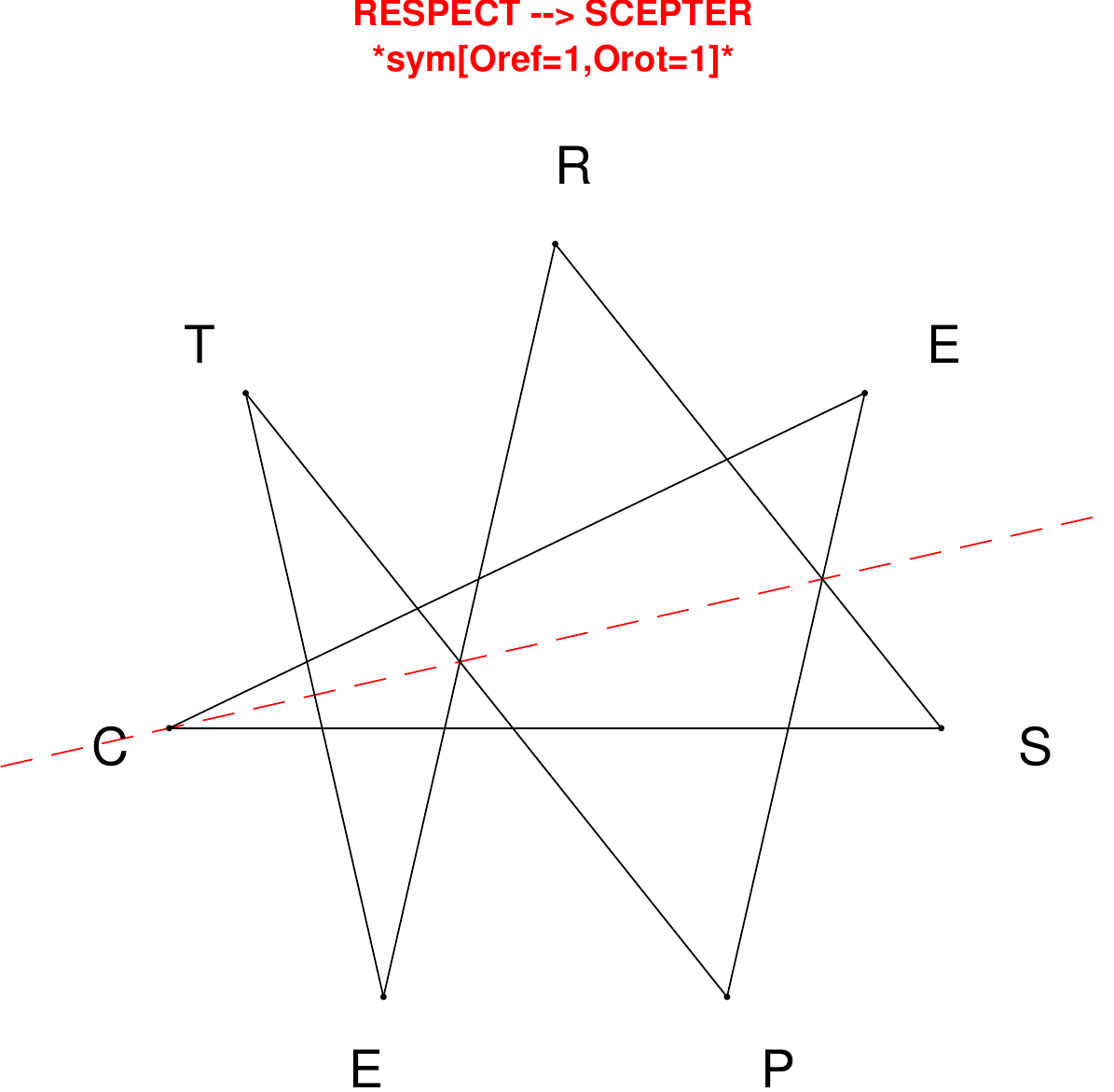}
\end{subfigure}
\hfill
\begin{subfigure}[T]{0.19\textwidth}
\centering
\includegraphics[width=\textwidth]{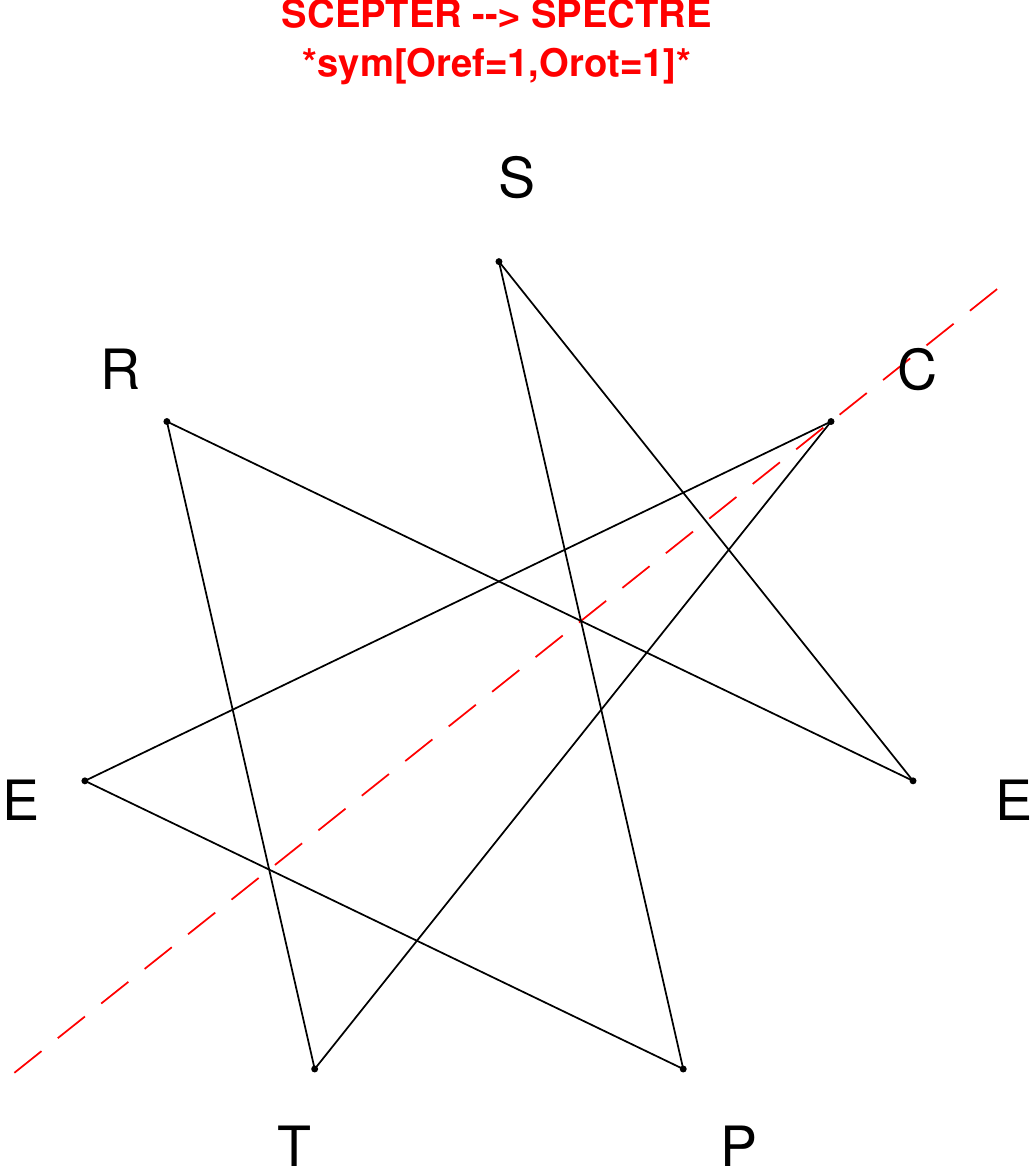}
\end{subfigure}
\hfill
\begin{subfigure}[T]{0.19\textwidth}
\centering
\includegraphics[width=\textwidth]{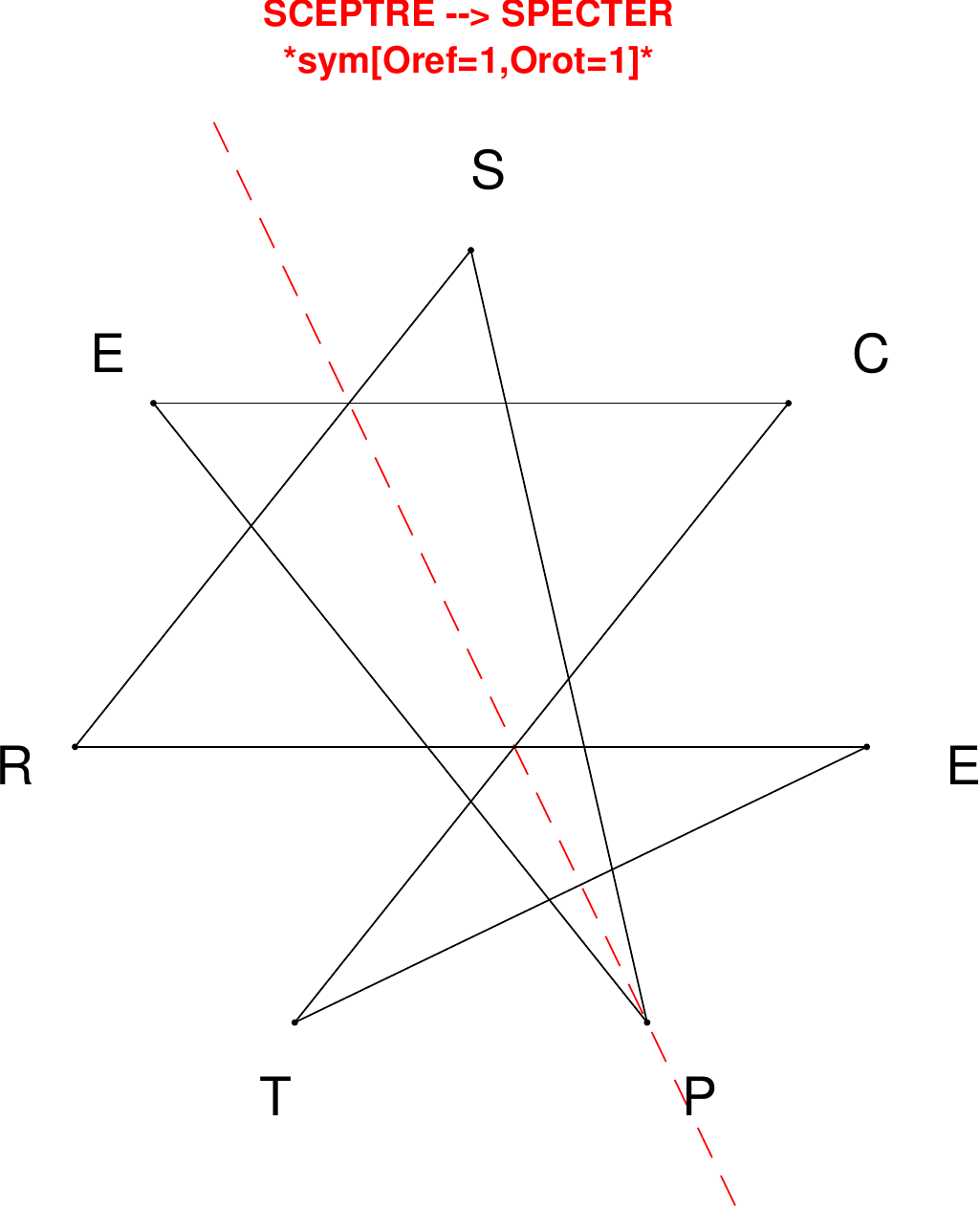}
\end{subfigure}
\hfill
\begin{subfigure}[T]{0.19\textwidth}
\centering
\includegraphics[width=\textwidth]{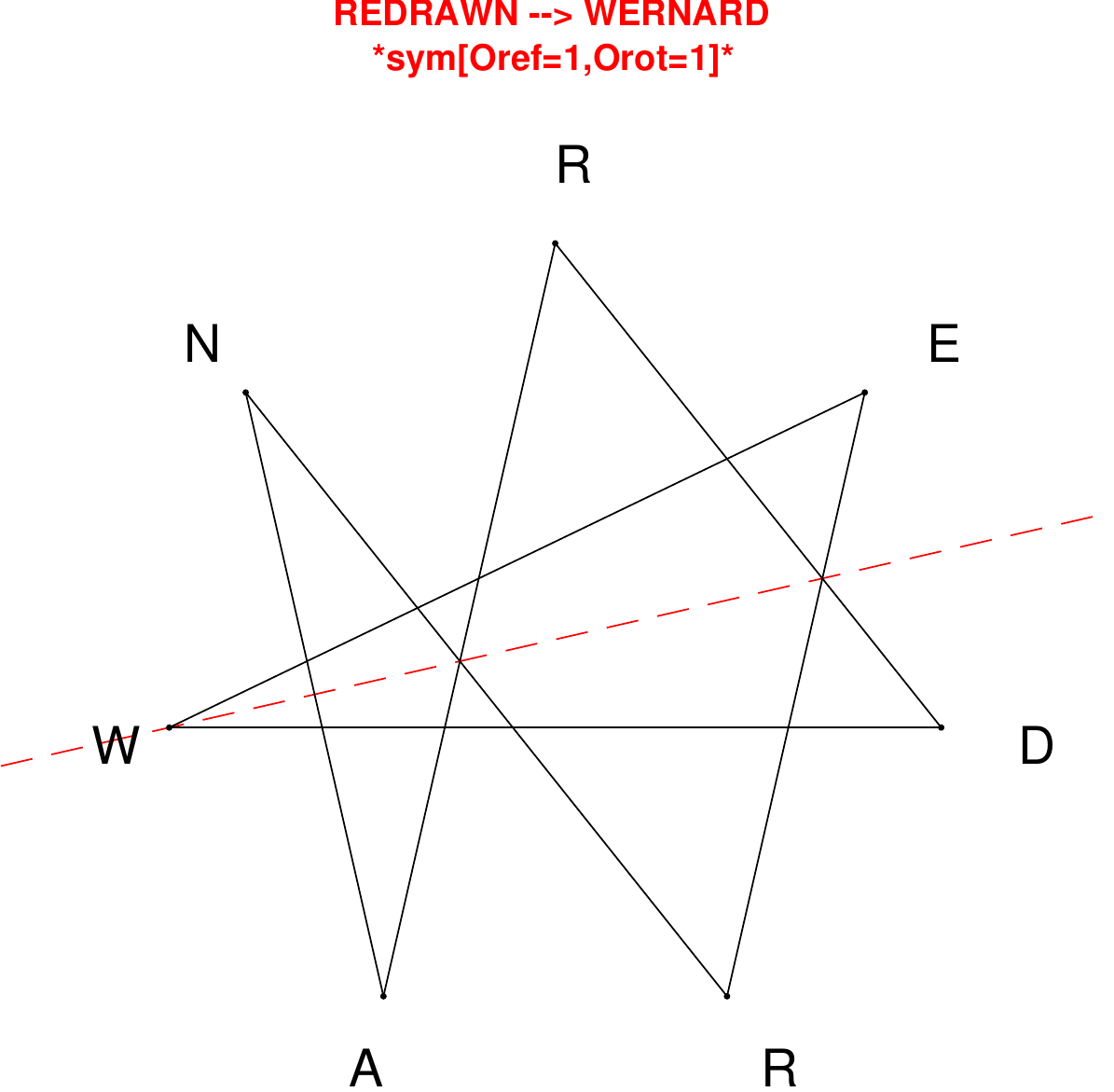}
\end{subfigure}
\hfill
\begin{subfigure}[T]{0.19\textwidth}
\centering
\includegraphics[width=\textwidth]{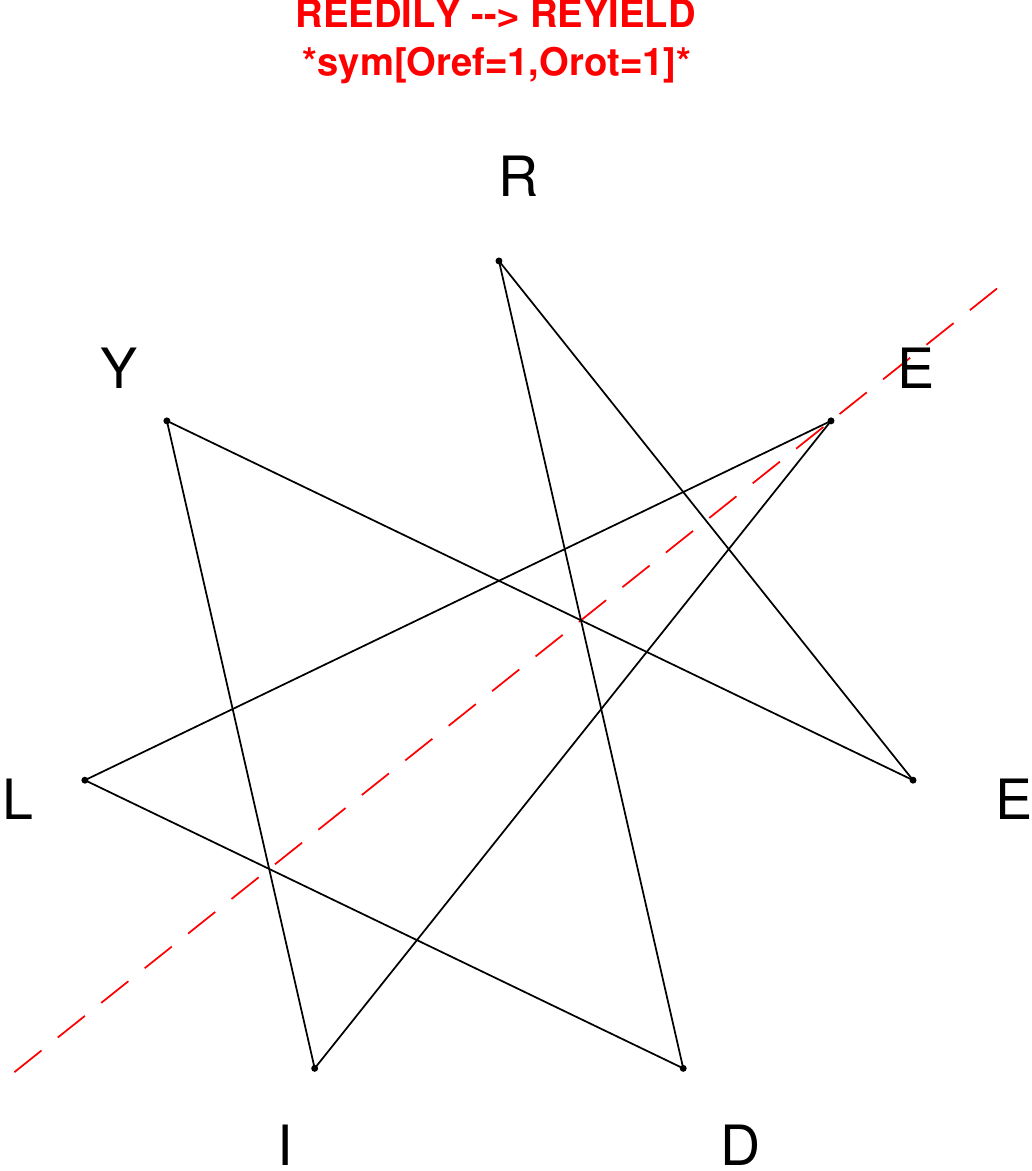}
\end{subfigure}
\end{figure}

\begin{figure}[H]
\centering
\begin{subfigure}[T]{0.19\textwidth}
\centering
\includegraphics[width=\textwidth]{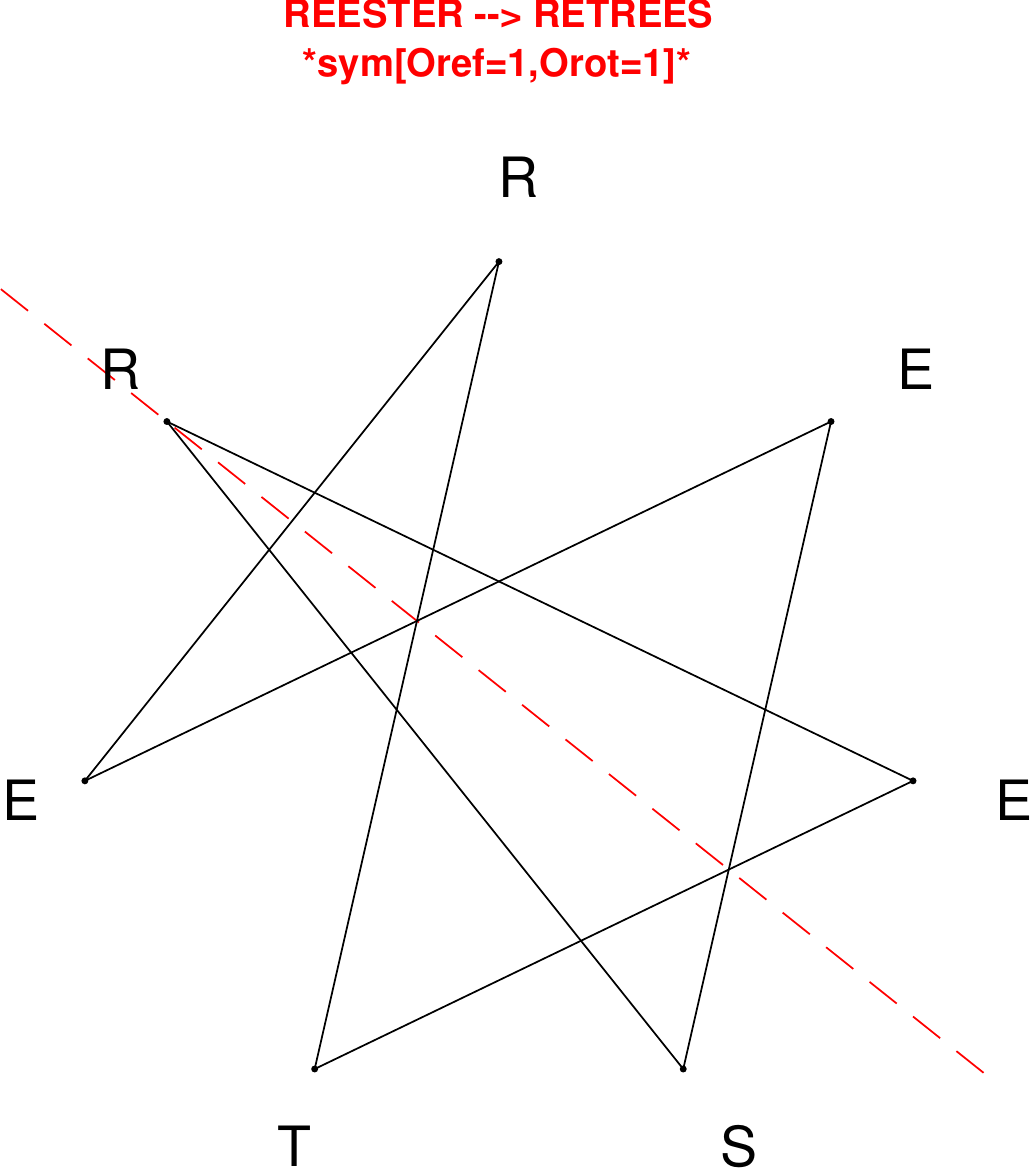}
\end{subfigure}
\hfill
\begin{subfigure}[T]{0.19\textwidth}
\centering
\includegraphics[width=\textwidth]{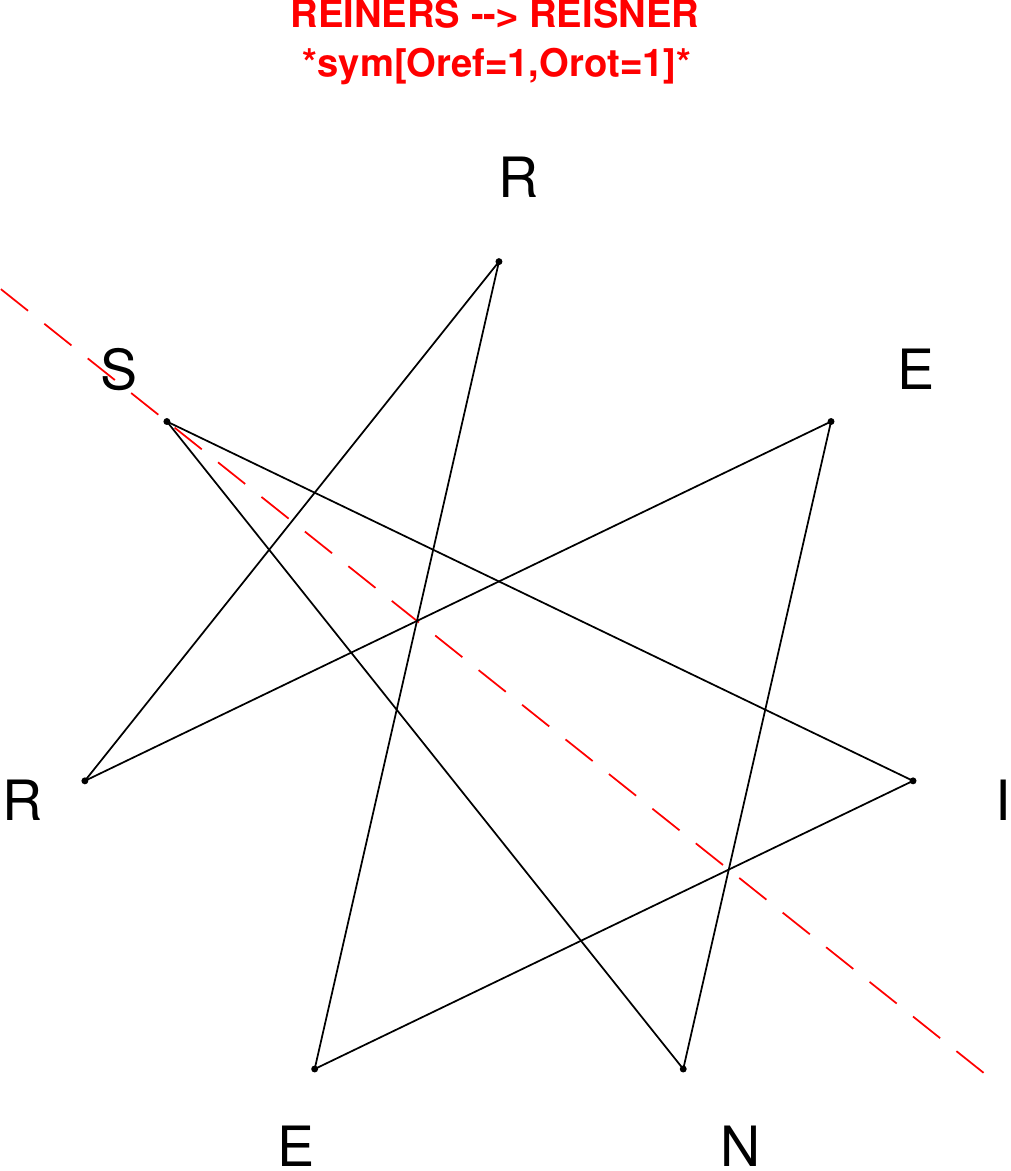}
\end{subfigure}
\hfill
\begin{subfigure}[T]{0.19\textwidth}
\centering
\includegraphics[width=\textwidth]{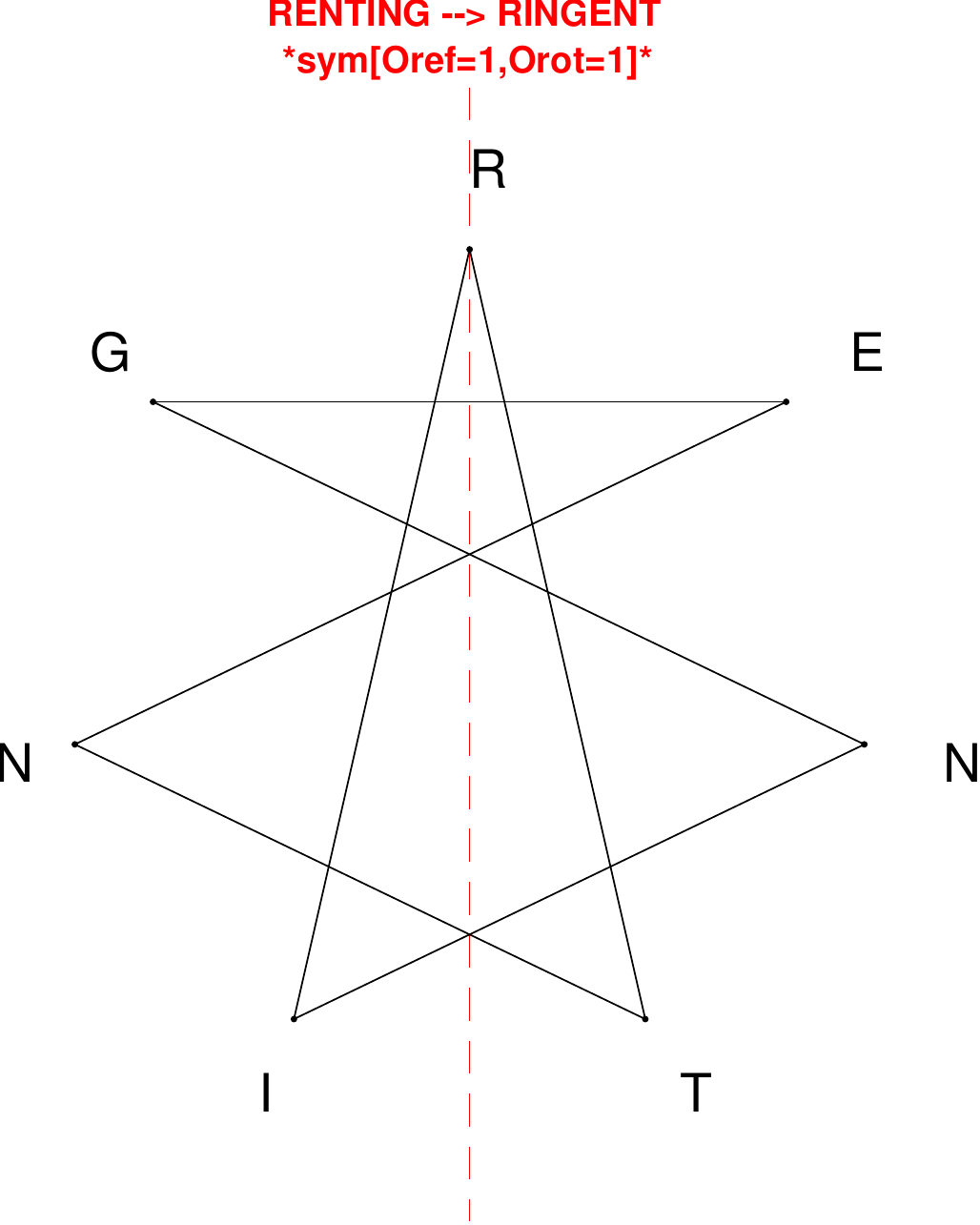}
\end{subfigure}
\hfill
\begin{subfigure}[T]{0.19\textwidth}
\centering
\includegraphics[width=\textwidth]{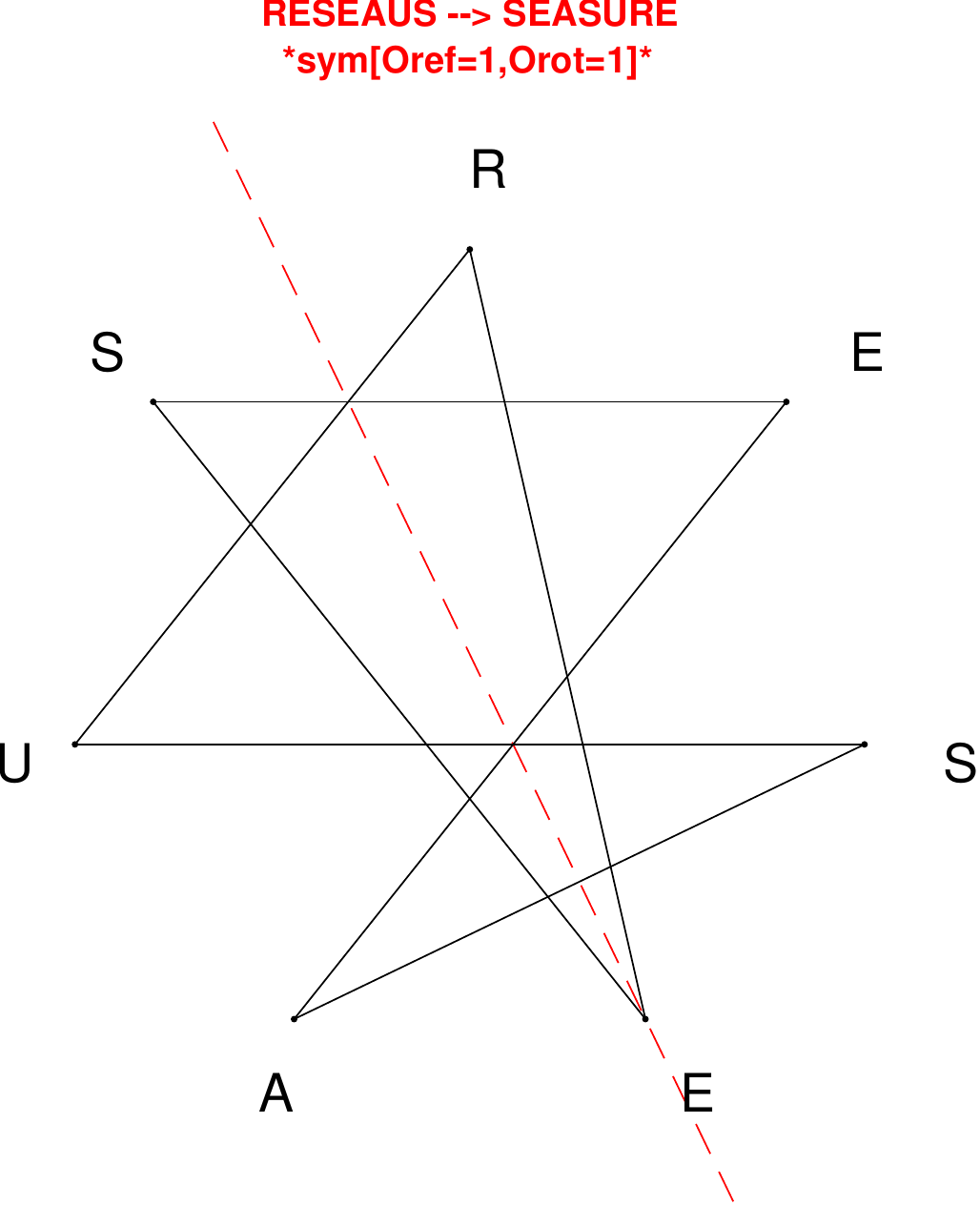}
\end{subfigure}
\hfill
\begin{subfigure}[T]{0.19\textwidth}
\centering
\includegraphics[width=\textwidth]{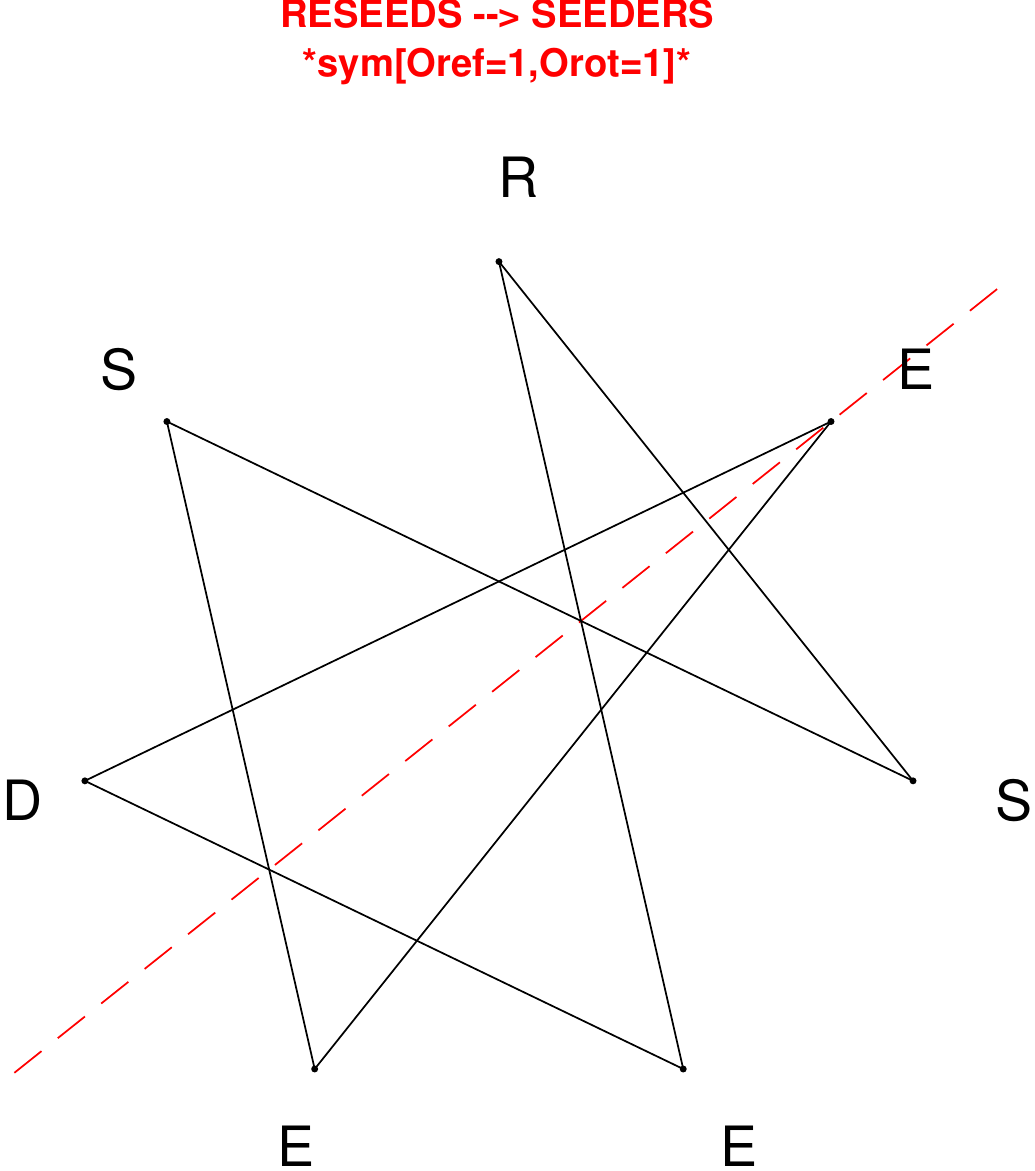}
\end{subfigure}
\end{figure}

\begin{figure}[H]
\centering
\begin{subfigure}[T]{0.19\textwidth}
\centering
\includegraphics[width=\textwidth]{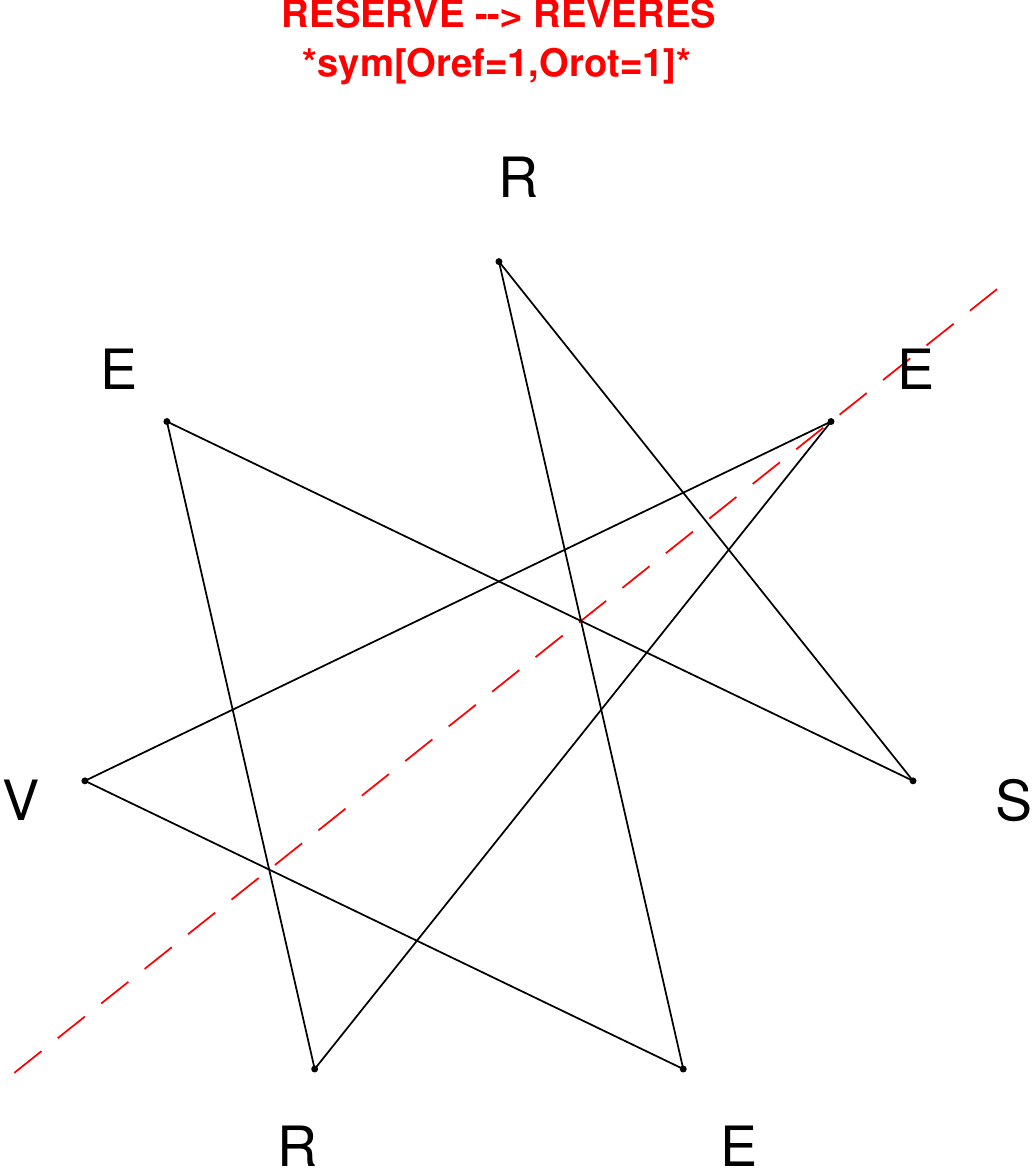}
\end{subfigure}
\hfill
\begin{subfigure}[T]{0.19\textwidth}
\centering
\includegraphics[width=\textwidth]{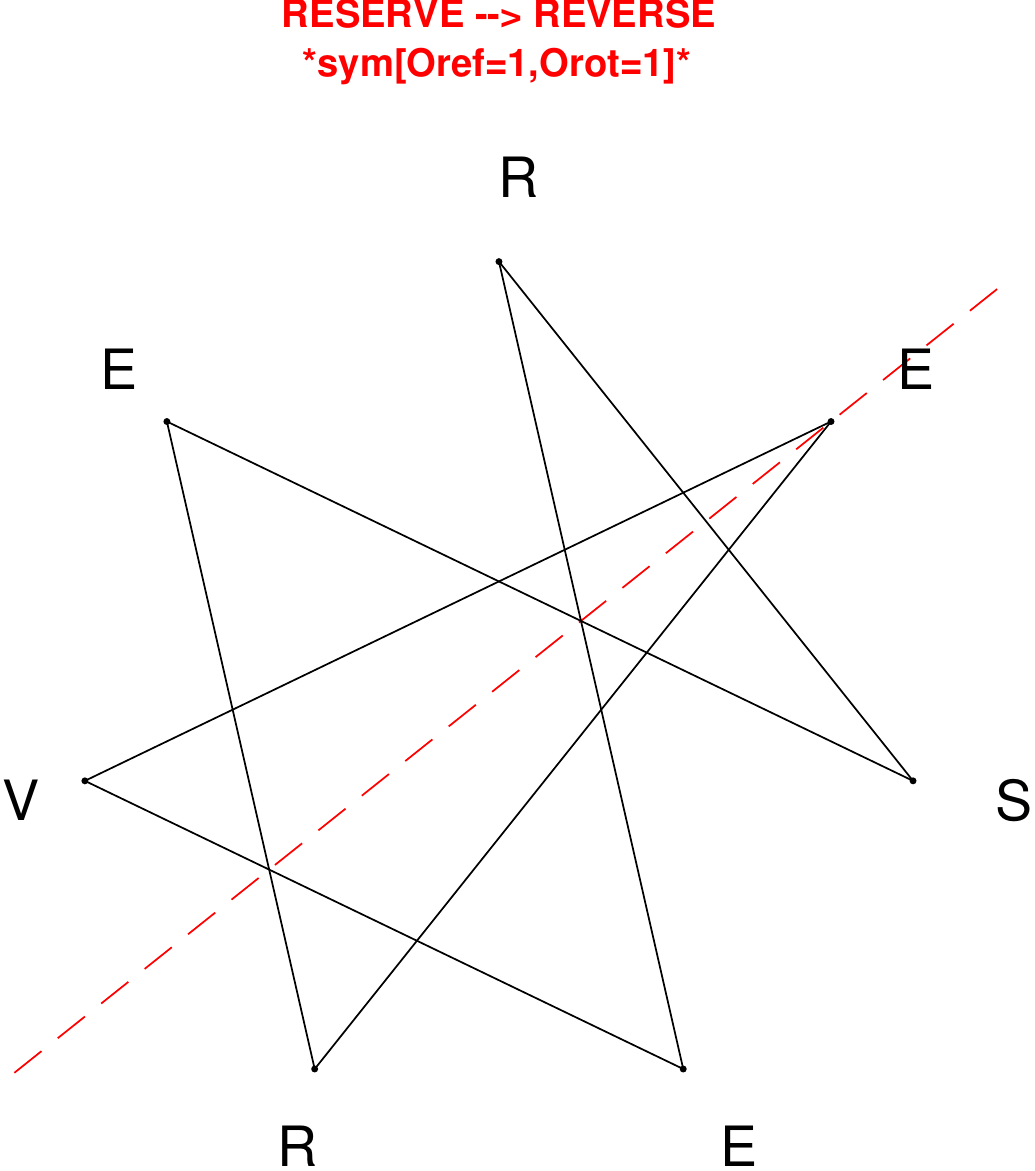}
\end{subfigure}
\hfill
\begin{subfigure}[T]{0.19\textwidth}
\centering
\includegraphics[width=\textwidth]{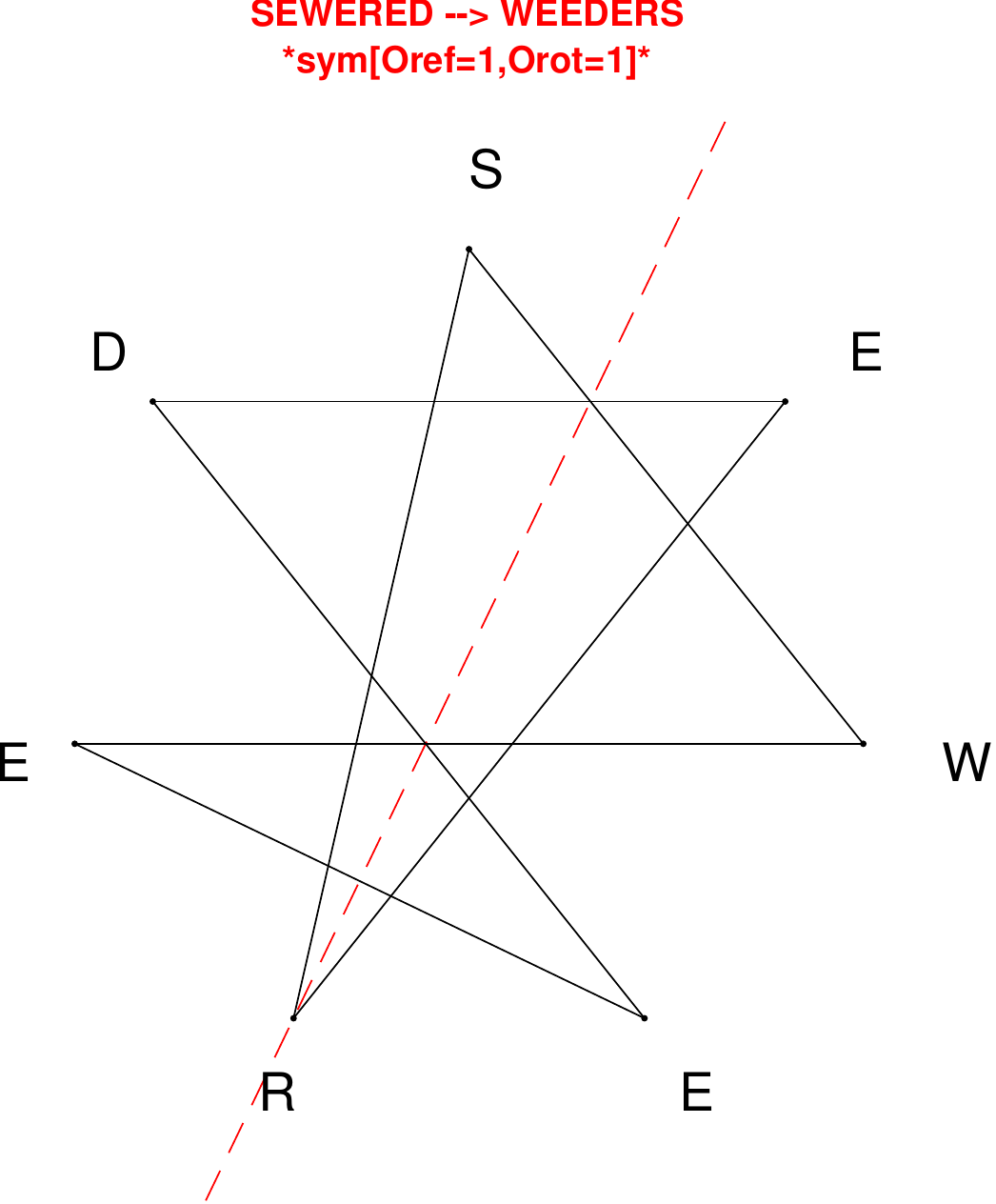}
\end{subfigure}
\hfill
\begin{subfigure}[T]{0.19\textwidth}
\centering
\includegraphics[width=\textwidth]{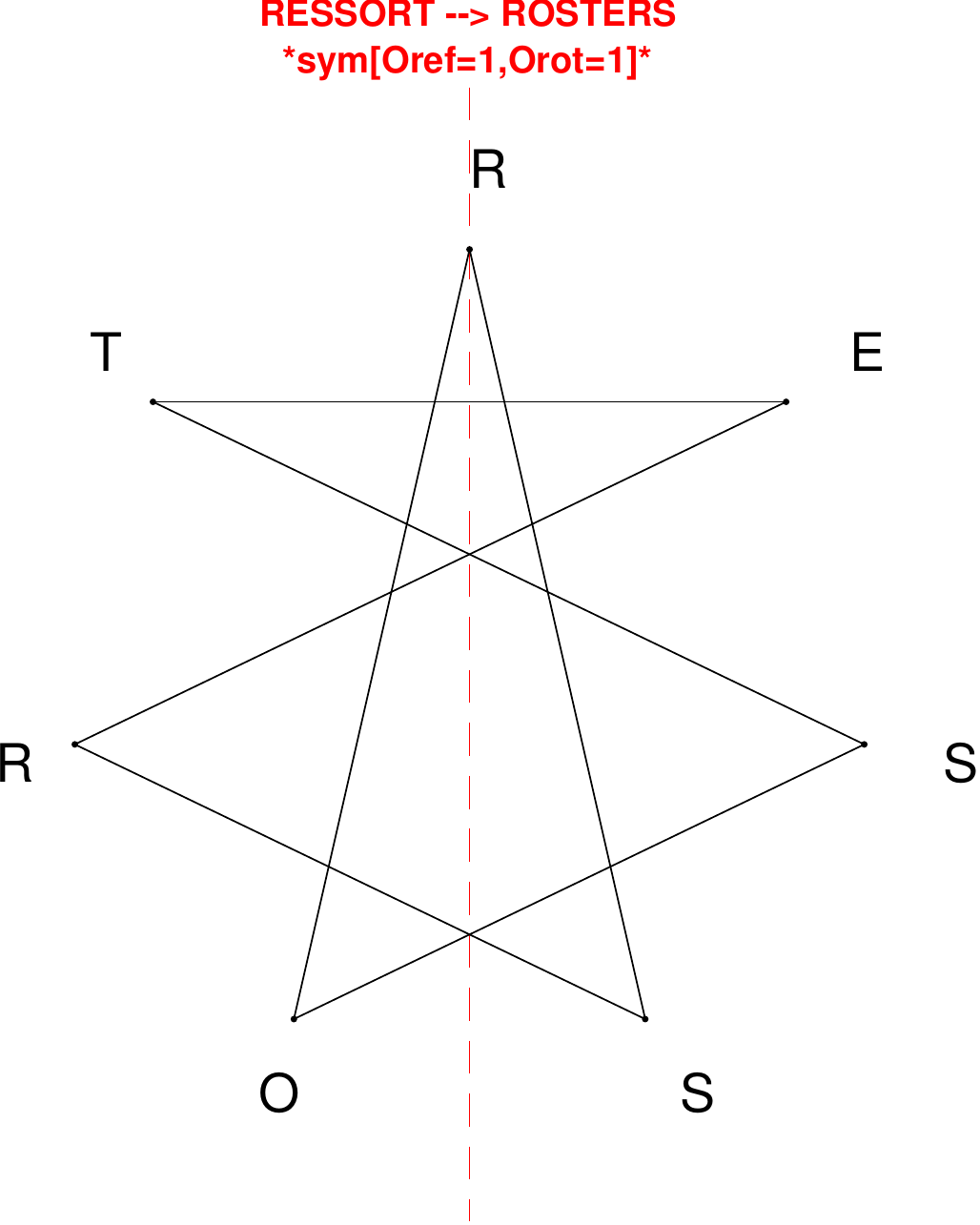}
\end{subfigure}
\hfill
\begin{subfigure}[T]{0.19\textwidth}
\centering
\includegraphics[width=\textwidth]{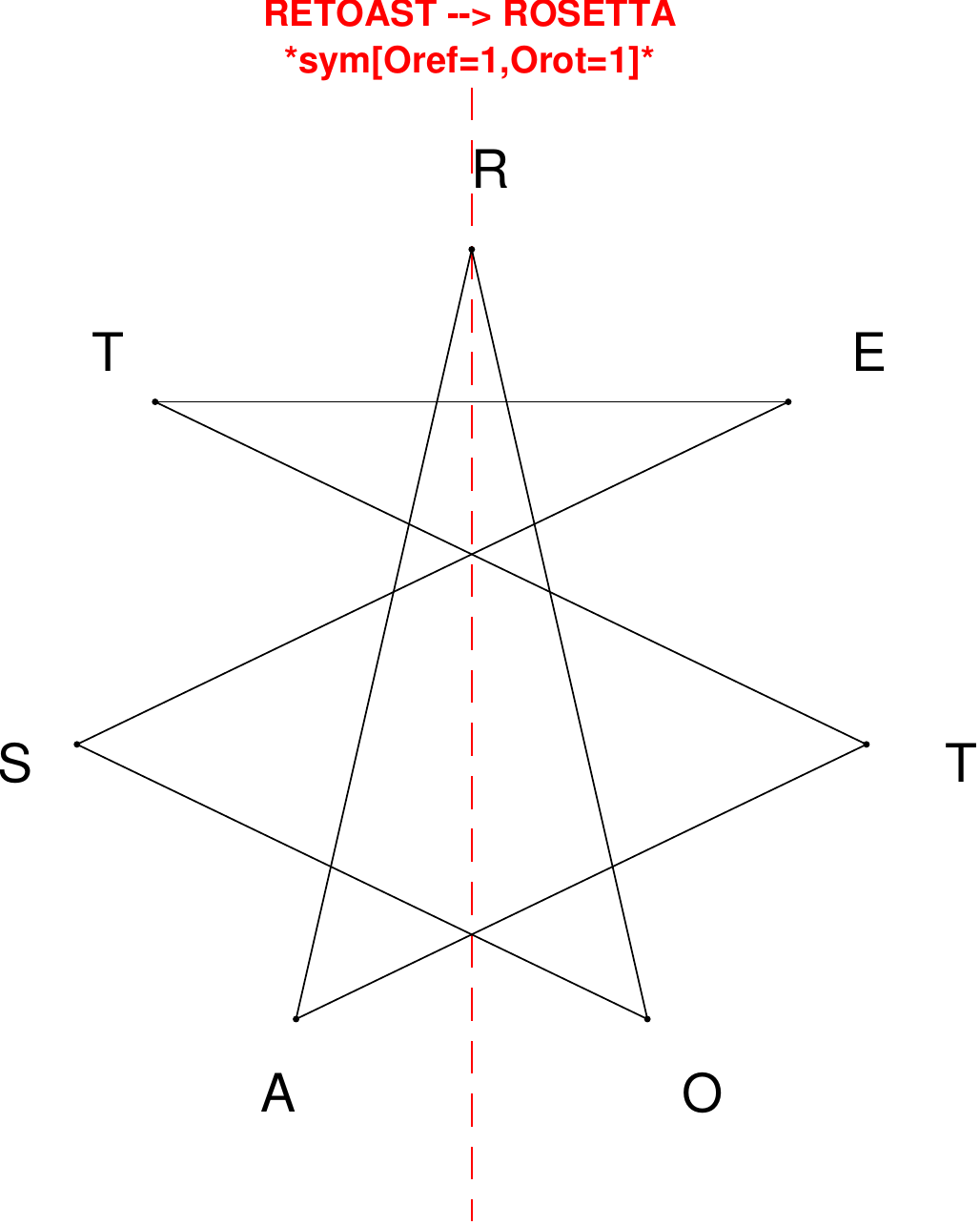}
\end{subfigure}
\end{figure}

\begin{figure}[H]
\centering
\begin{subfigure}[T]{0.19\textwidth}
\centering
\includegraphics[width=\textwidth]{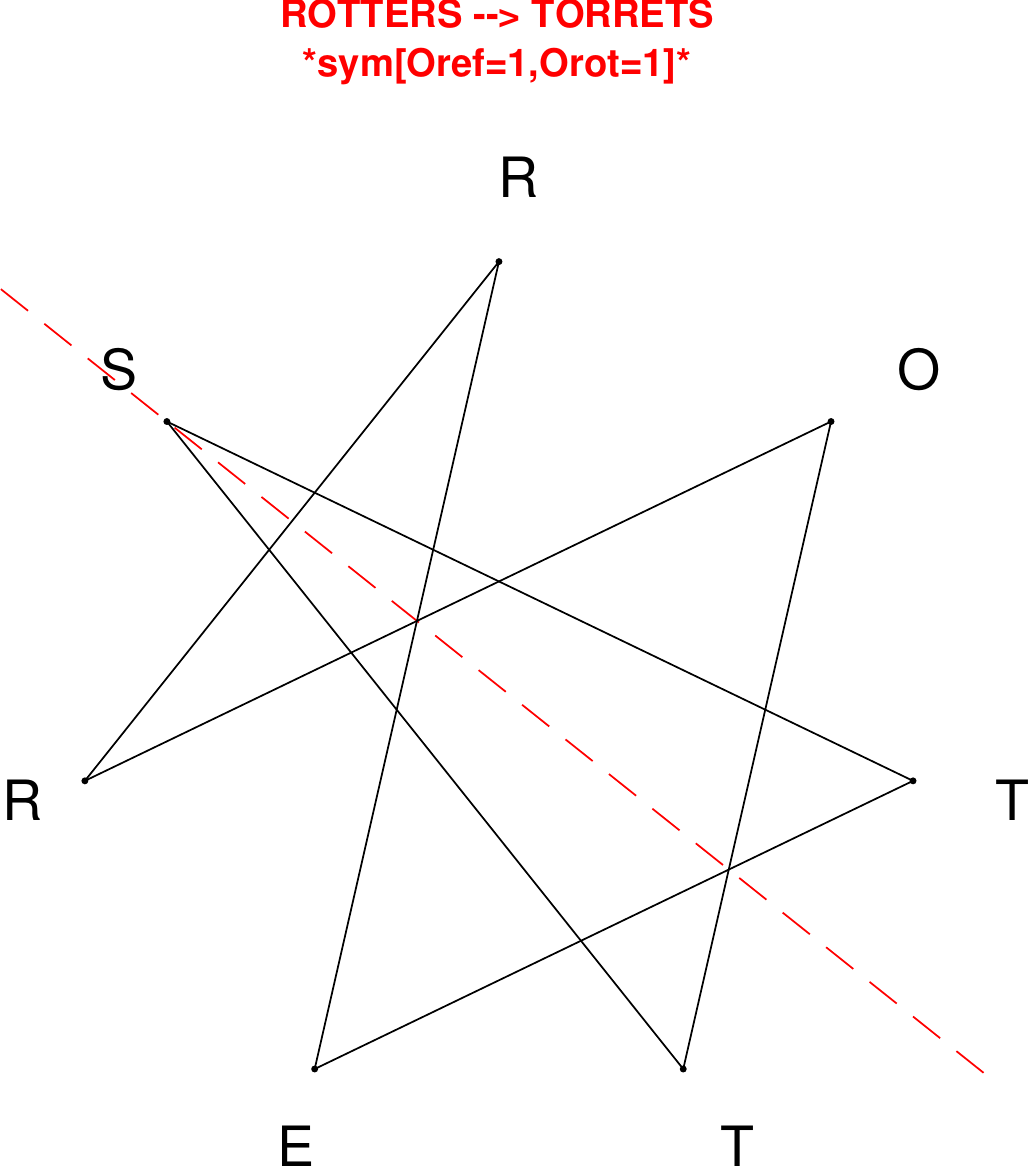}
\end{subfigure}
\hfill
\begin{subfigure}[T]{0.19\textwidth}
\centering
\includegraphics[width=\textwidth]{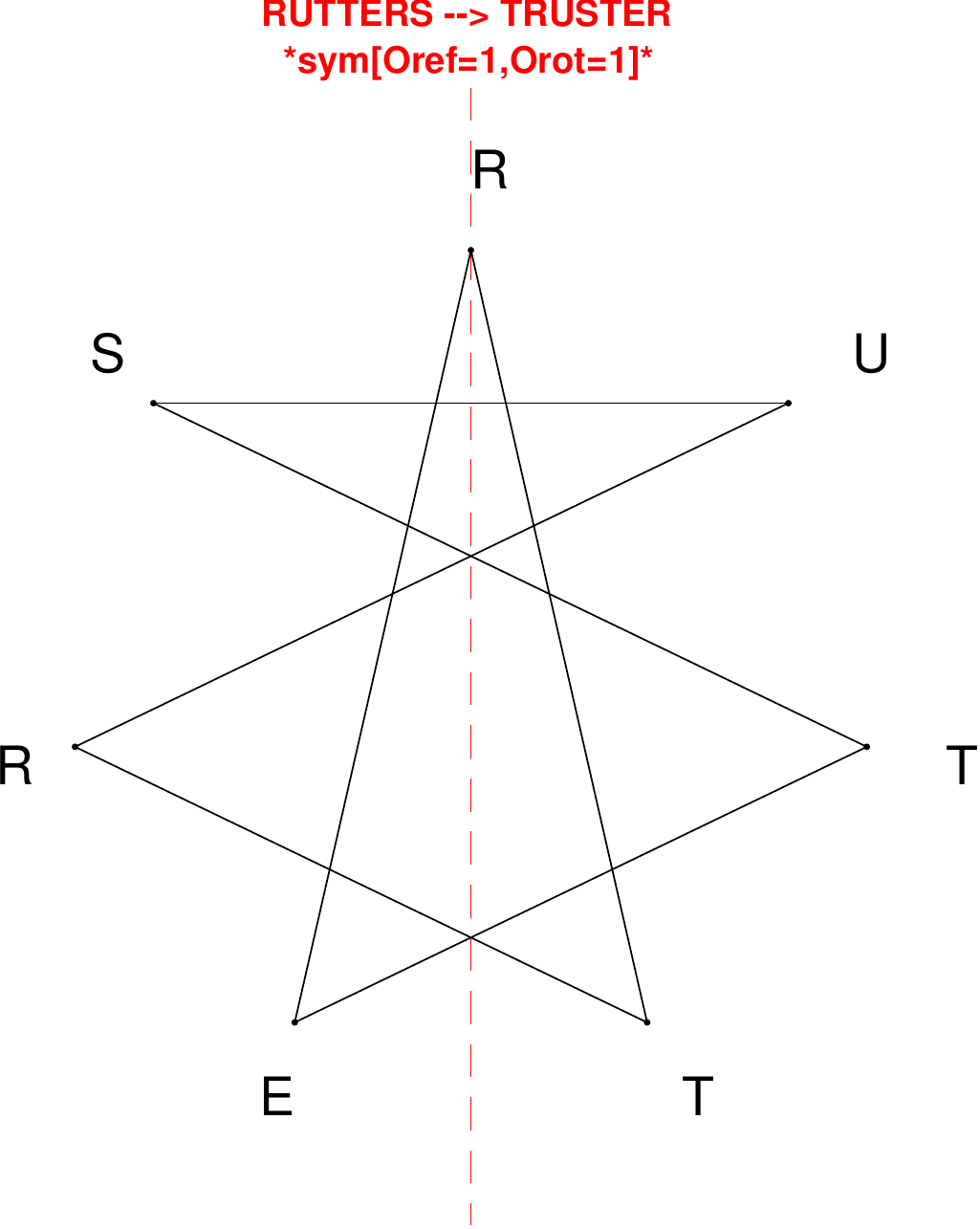}
\end{subfigure}
\hfill
\begin{subfigure}[T]{0.19\textwidth}
\centering
\includegraphics[width=\textwidth]{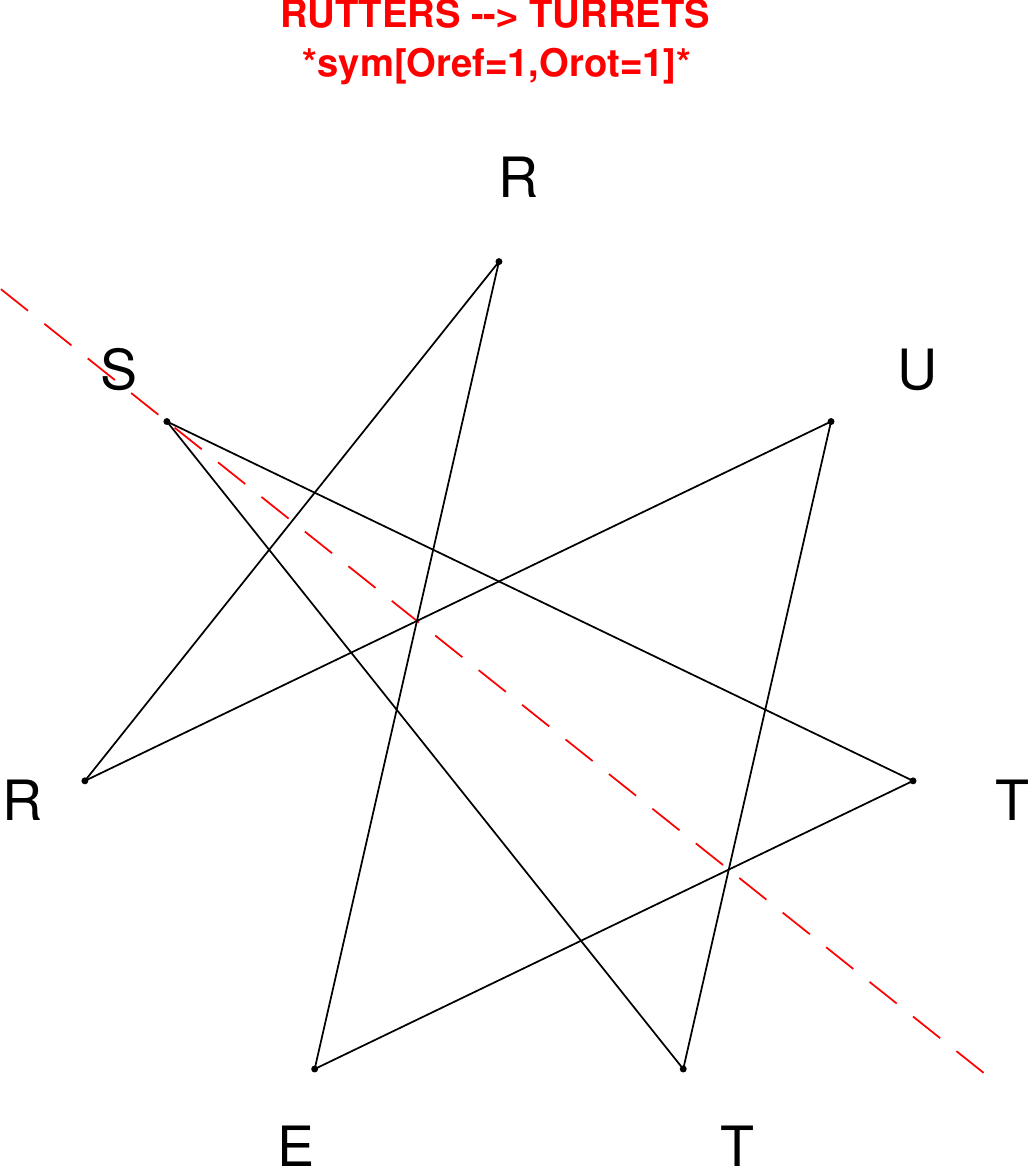}
\end{subfigure}
\hfill
\begin{subfigure}[T]{0.19\textwidth}
\centering
\includegraphics[width=\textwidth]{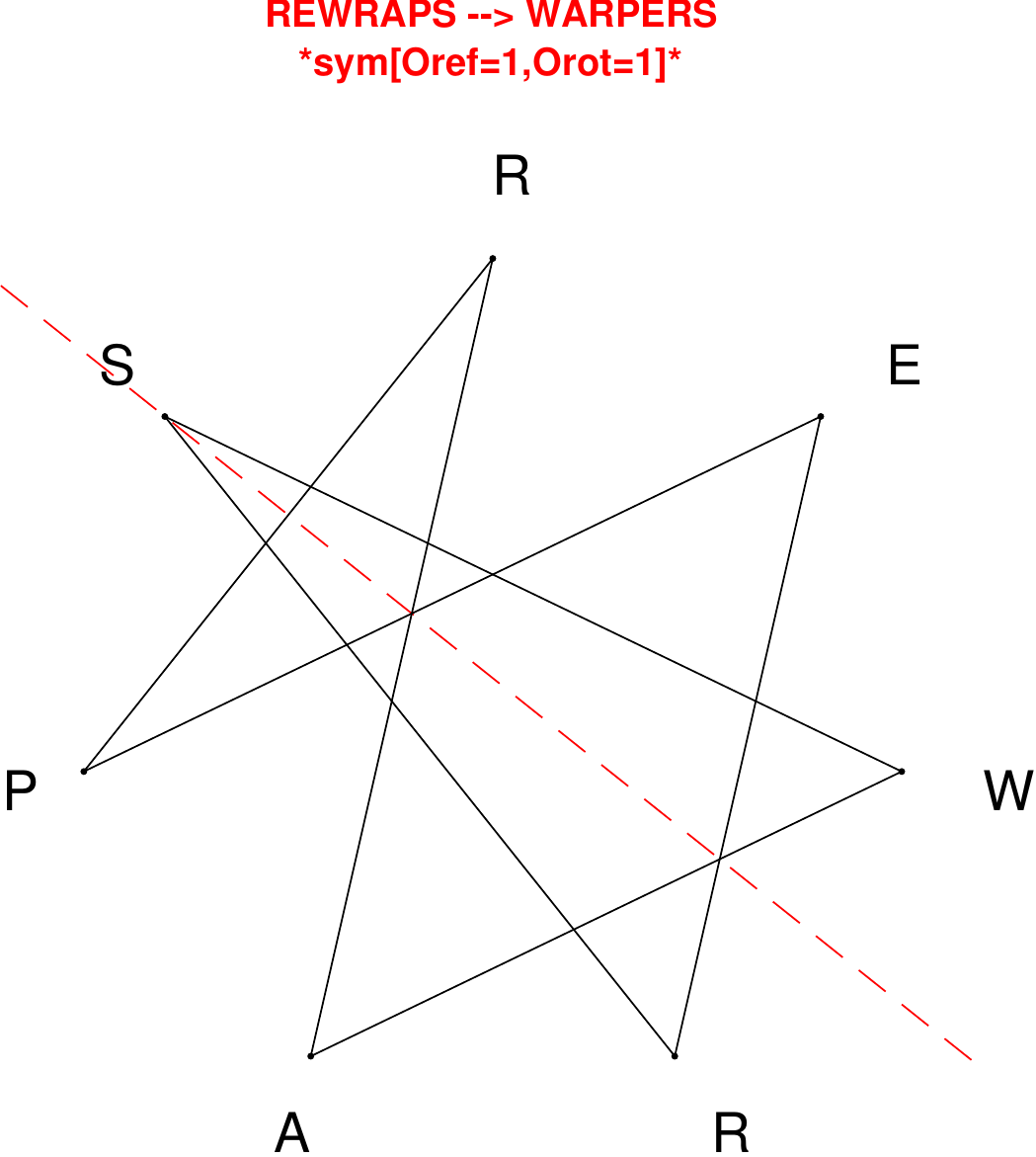}
\end{subfigure}
\hfill
\begin{subfigure}[T]{0.19\textwidth}
\centering
\includegraphics[width=\textwidth]{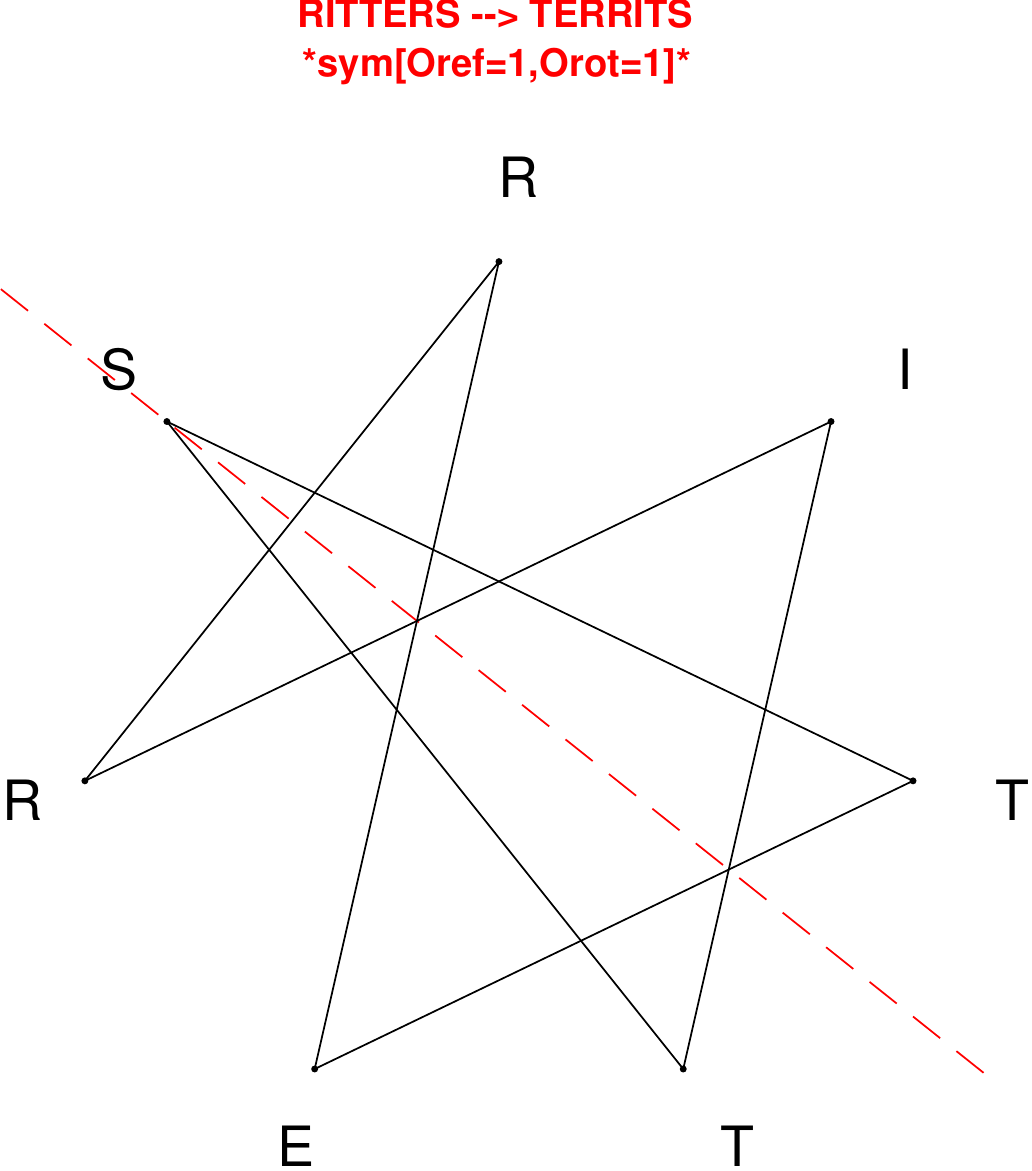}
\end{subfigure}
\end{figure}

\begin{figure}[H]
\centering
\begin{subfigure}[T]{0.19\textwidth}
\centering
\includegraphics[width=\textwidth]{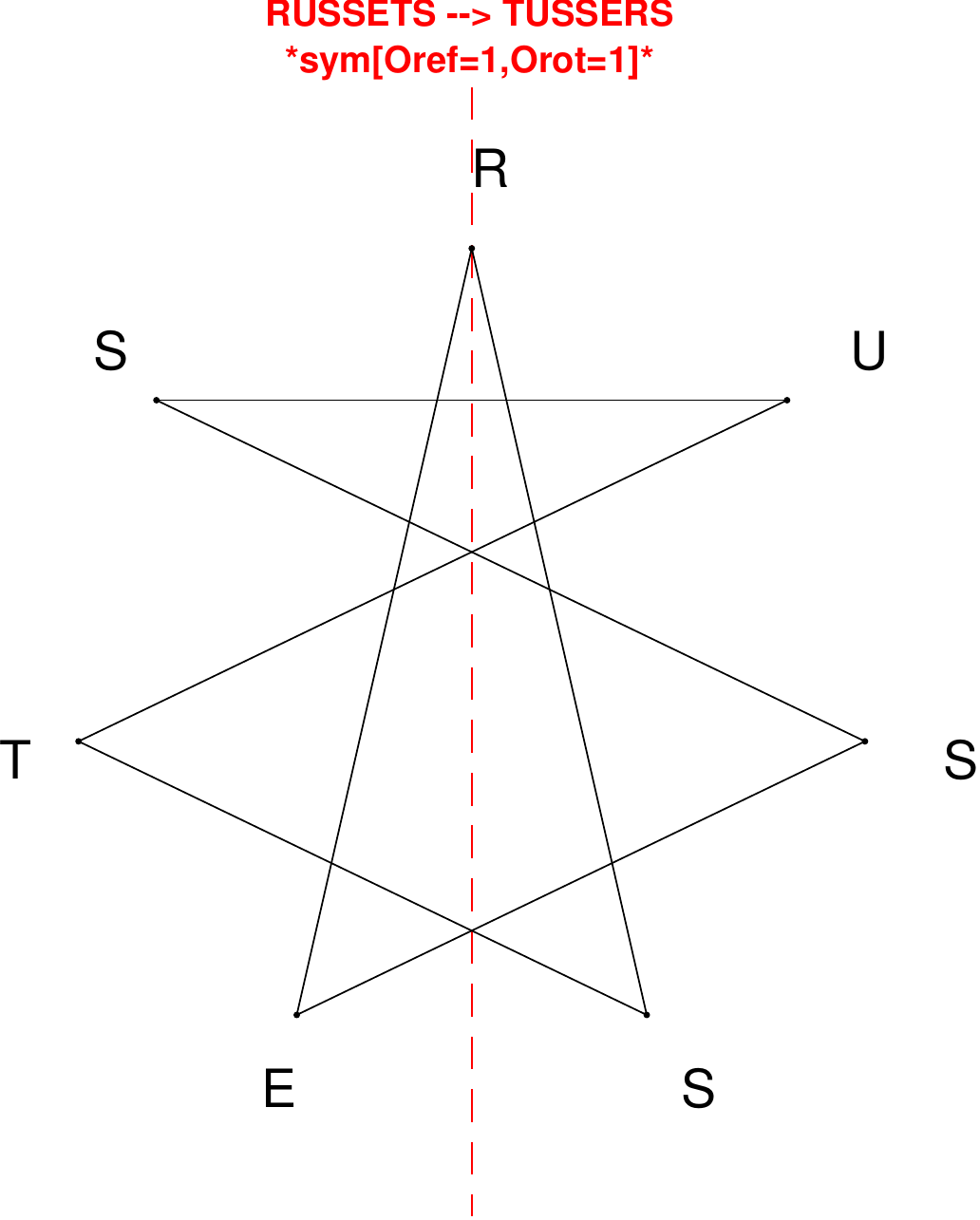}
\end{subfigure}
\hfill
\begin{subfigure}[T]{0.19\textwidth}
\centering
\includegraphics[width=\textwidth]{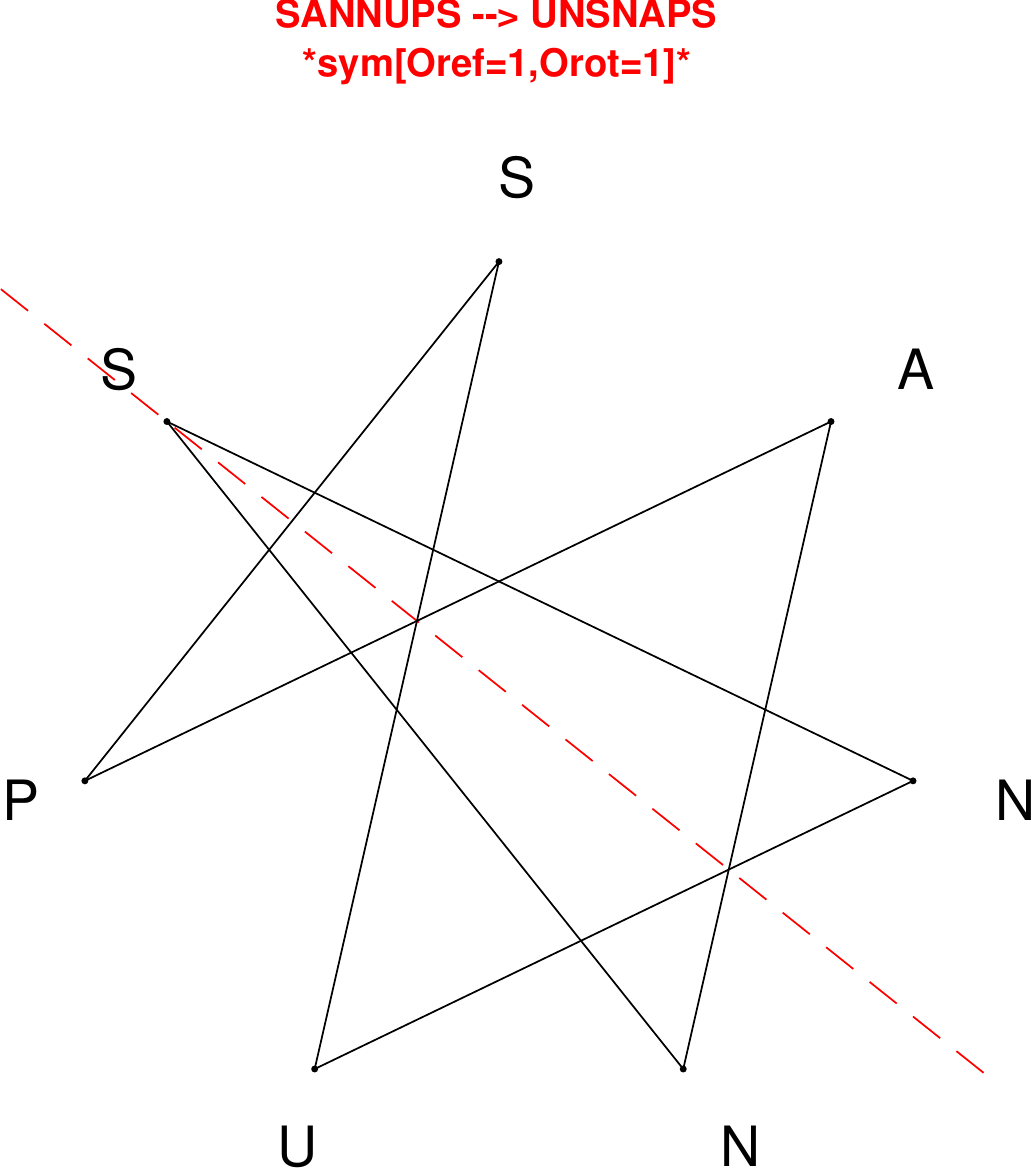}
\end{subfigure}
\hfill
\begin{subfigure}[T]{0.19\textwidth}
\centering
\includegraphics[width=\textwidth]{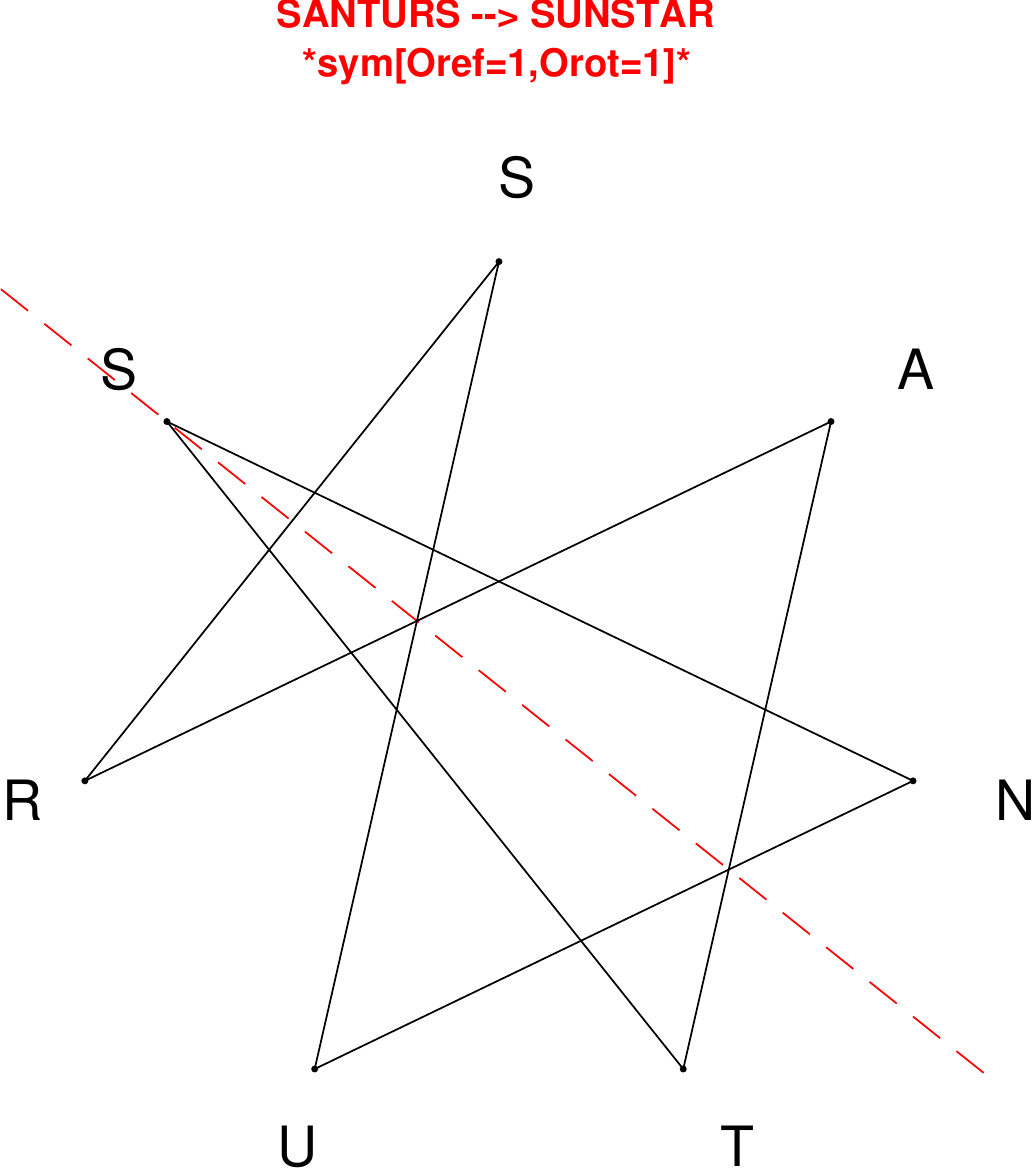}
\end{subfigure}
\hfill
\begin{subfigure}[T]{0.19\textwidth}
\centering
\includegraphics[width=\textwidth]{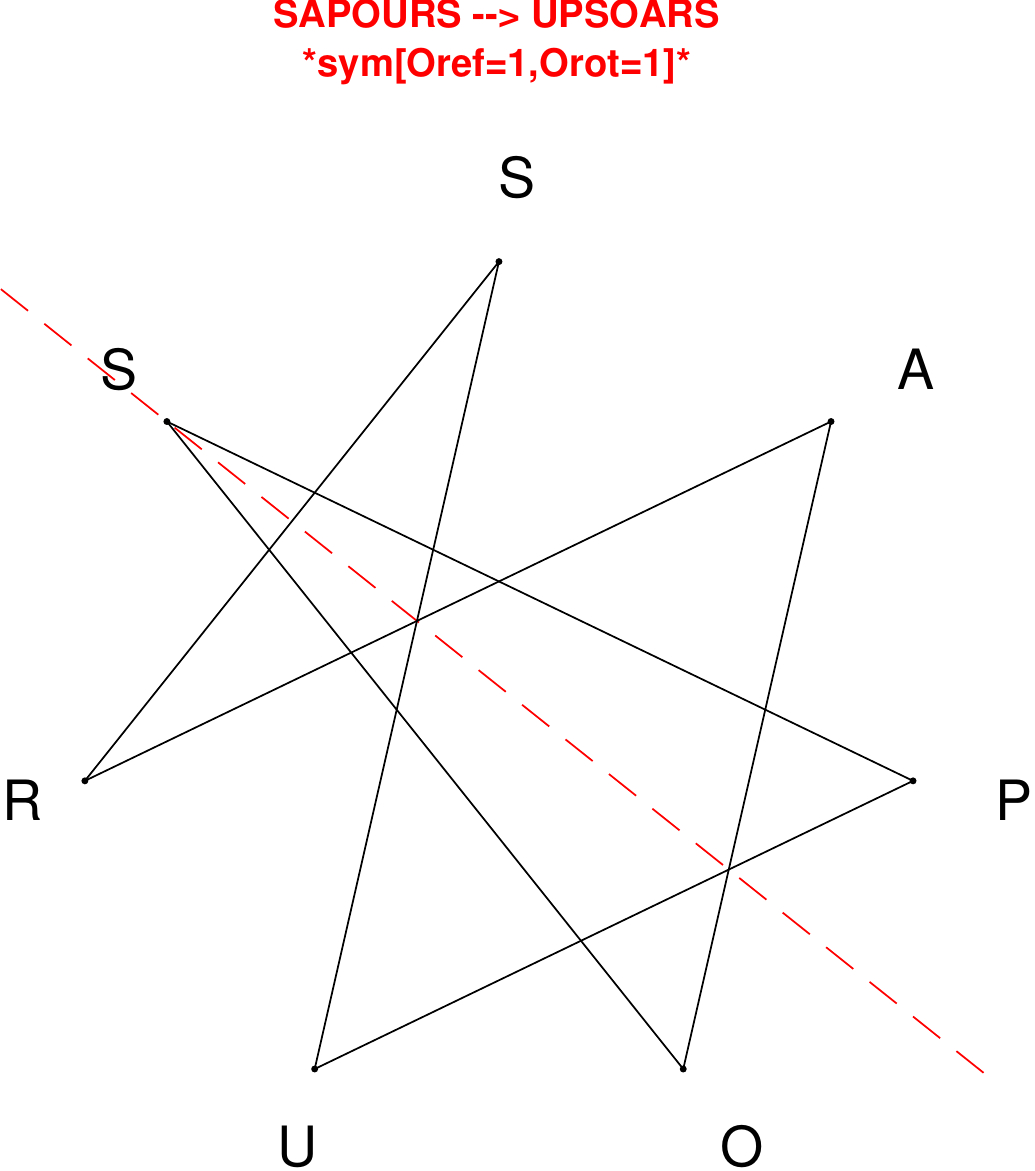}
\end{subfigure}
\hfill
\begin{subfigure}[T]{0.19\textwidth}
\centering
\includegraphics[width=\textwidth]{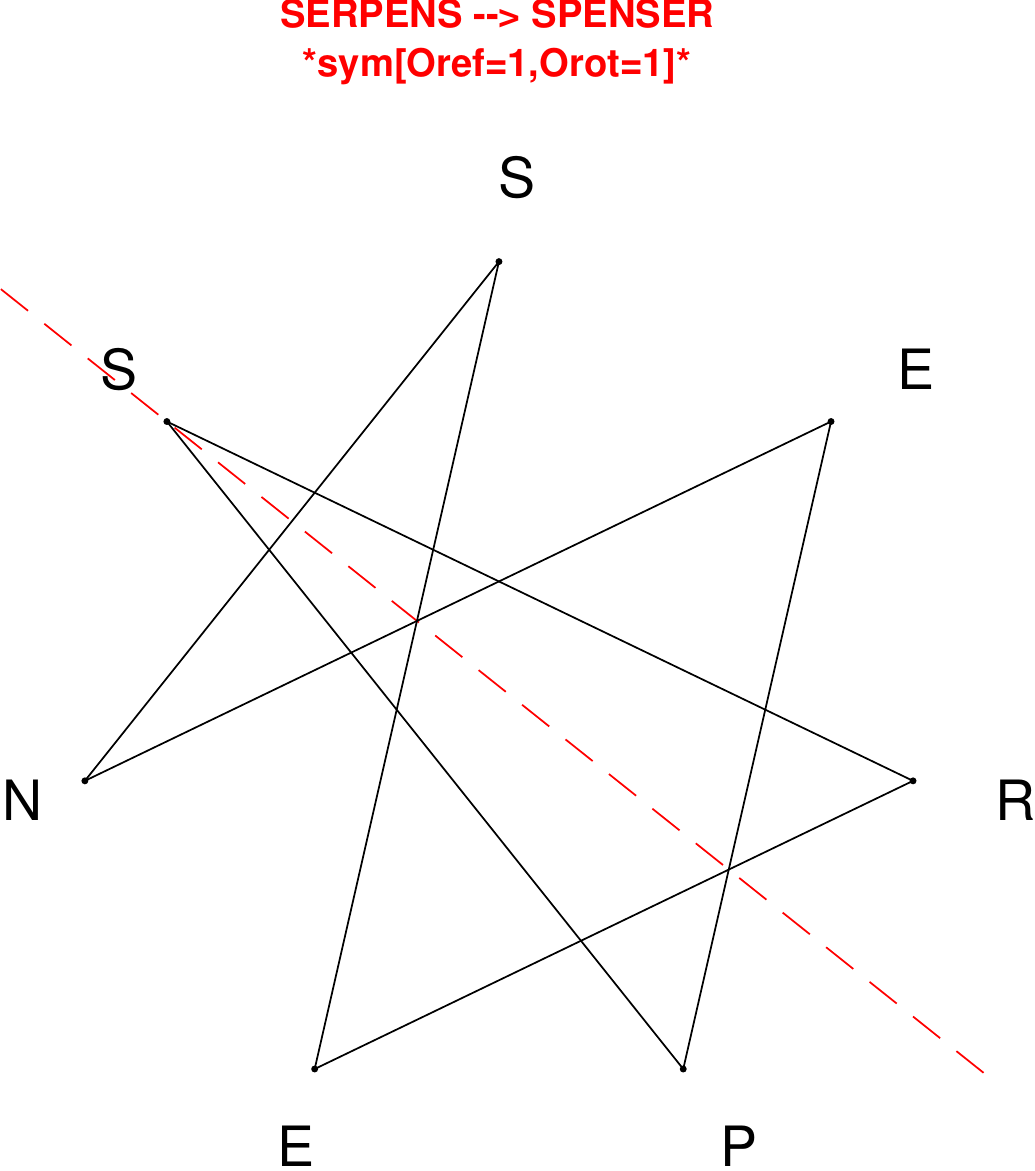}
\end{subfigure}
\end{figure}

\begin{figure}[H]
\centering
\begin{subfigure}[T]{0.19\textwidth}
\centering
\includegraphics[width=\textwidth]{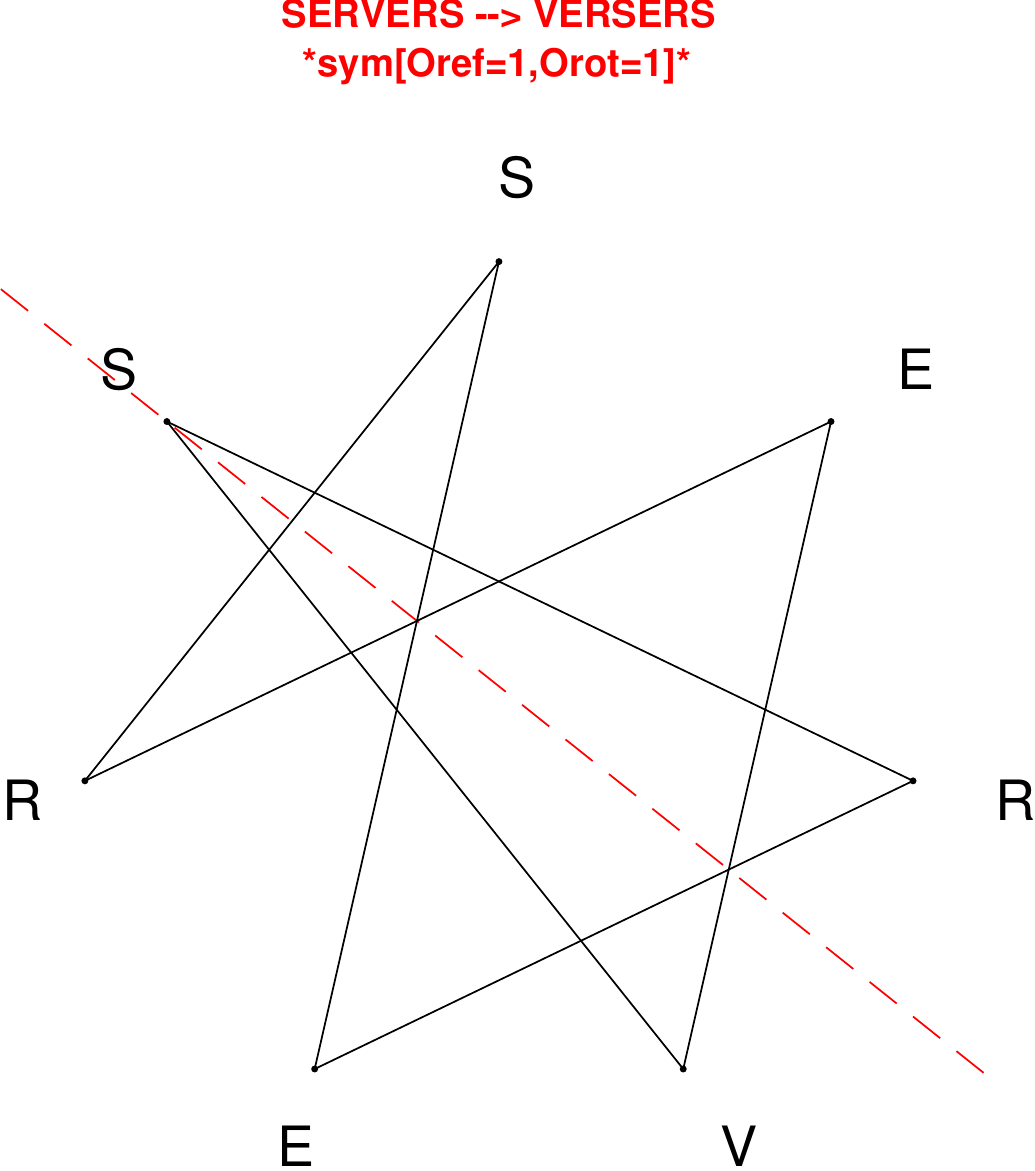}
\end{subfigure}
\hfill
\begin{subfigure}[T]{0.19\textwidth}
\centering
\includegraphics[width=\textwidth]{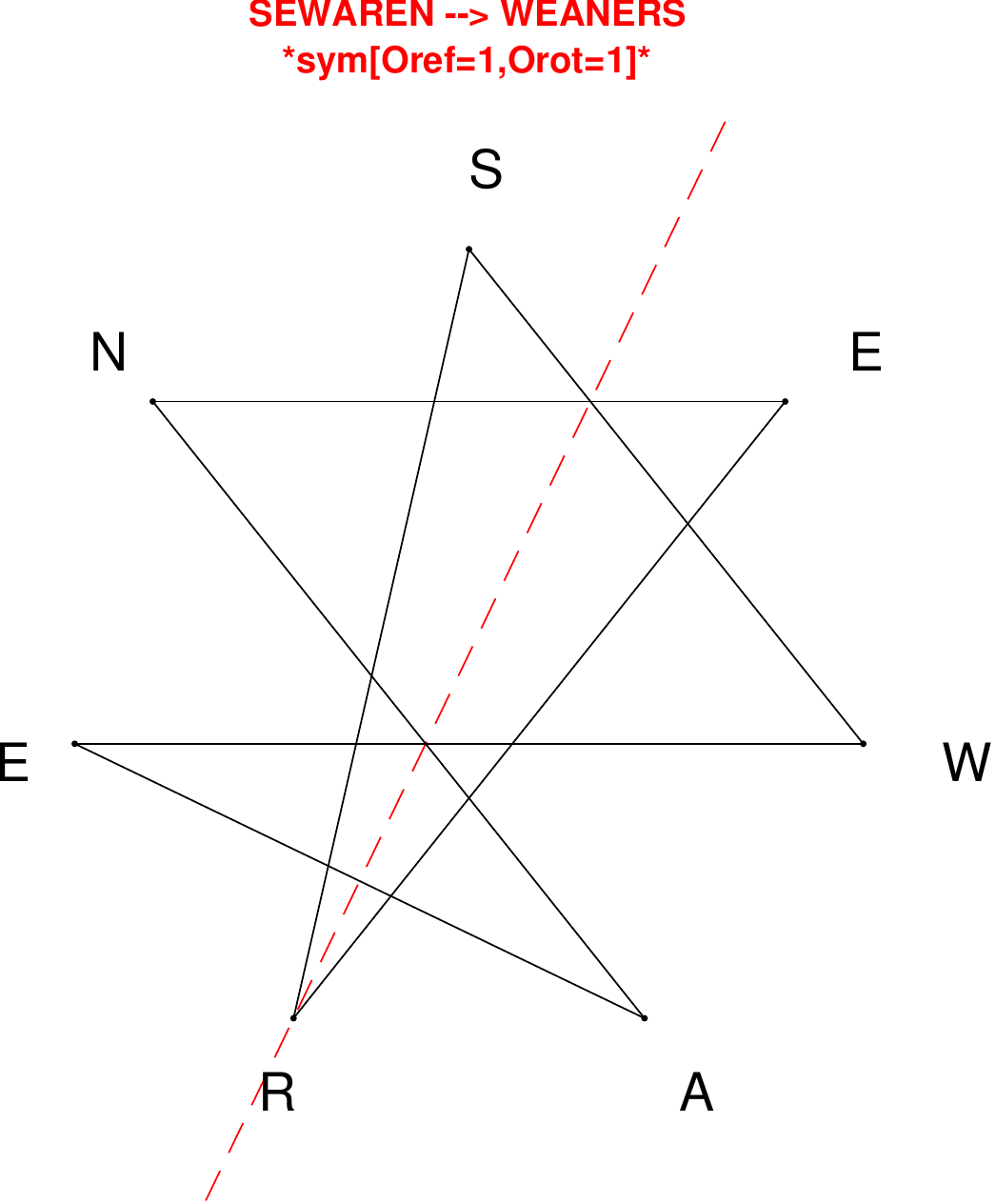}
\end{subfigure}
\hfill
\begin{subfigure}[T]{0.19\textwidth}
\centering
\includegraphics[width=\textwidth]{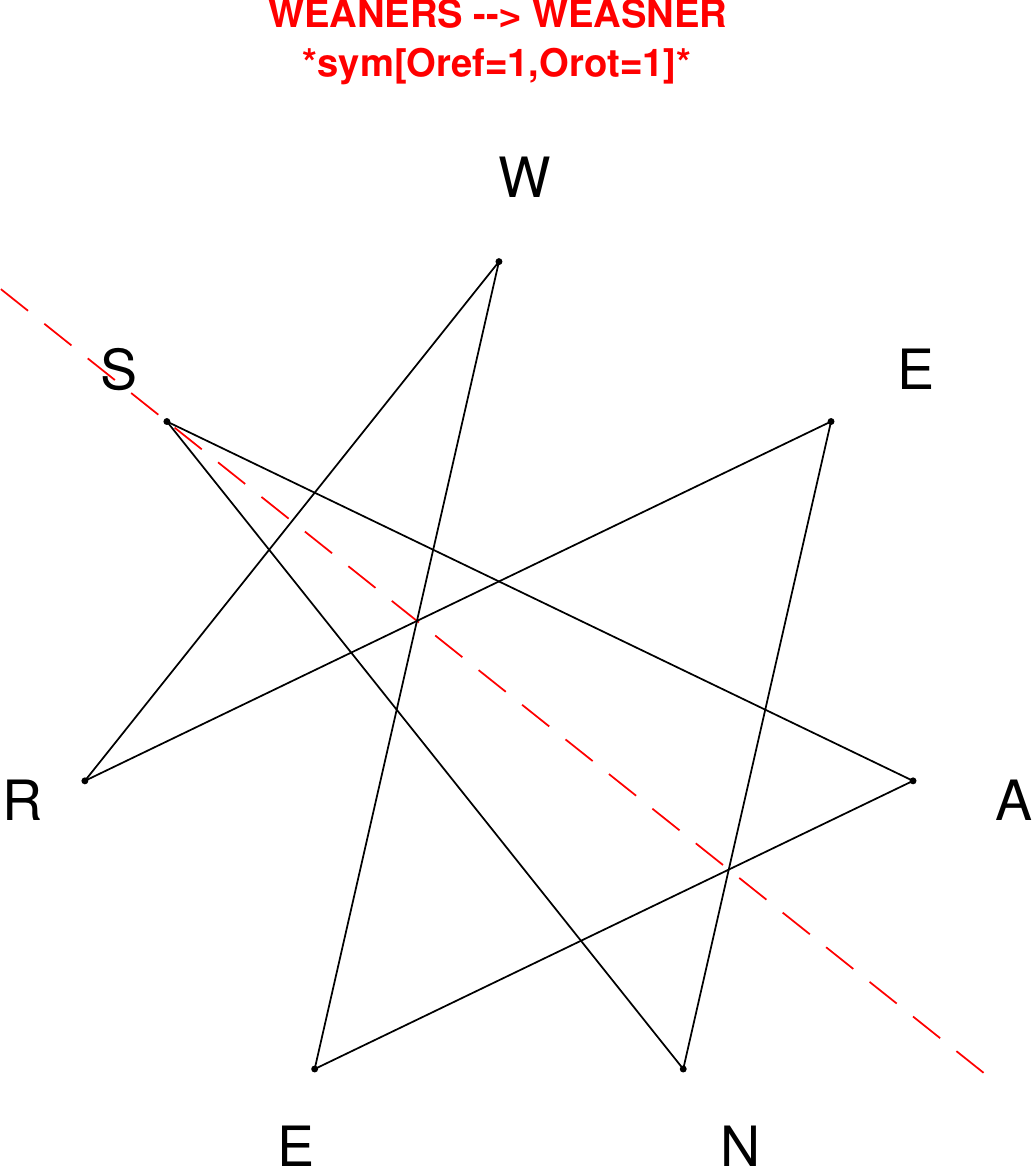}
\end{subfigure}
\hfill
\begin{subfigure}[T]{0.19\textwidth}
\centering
\includegraphics[width=\textwidth]{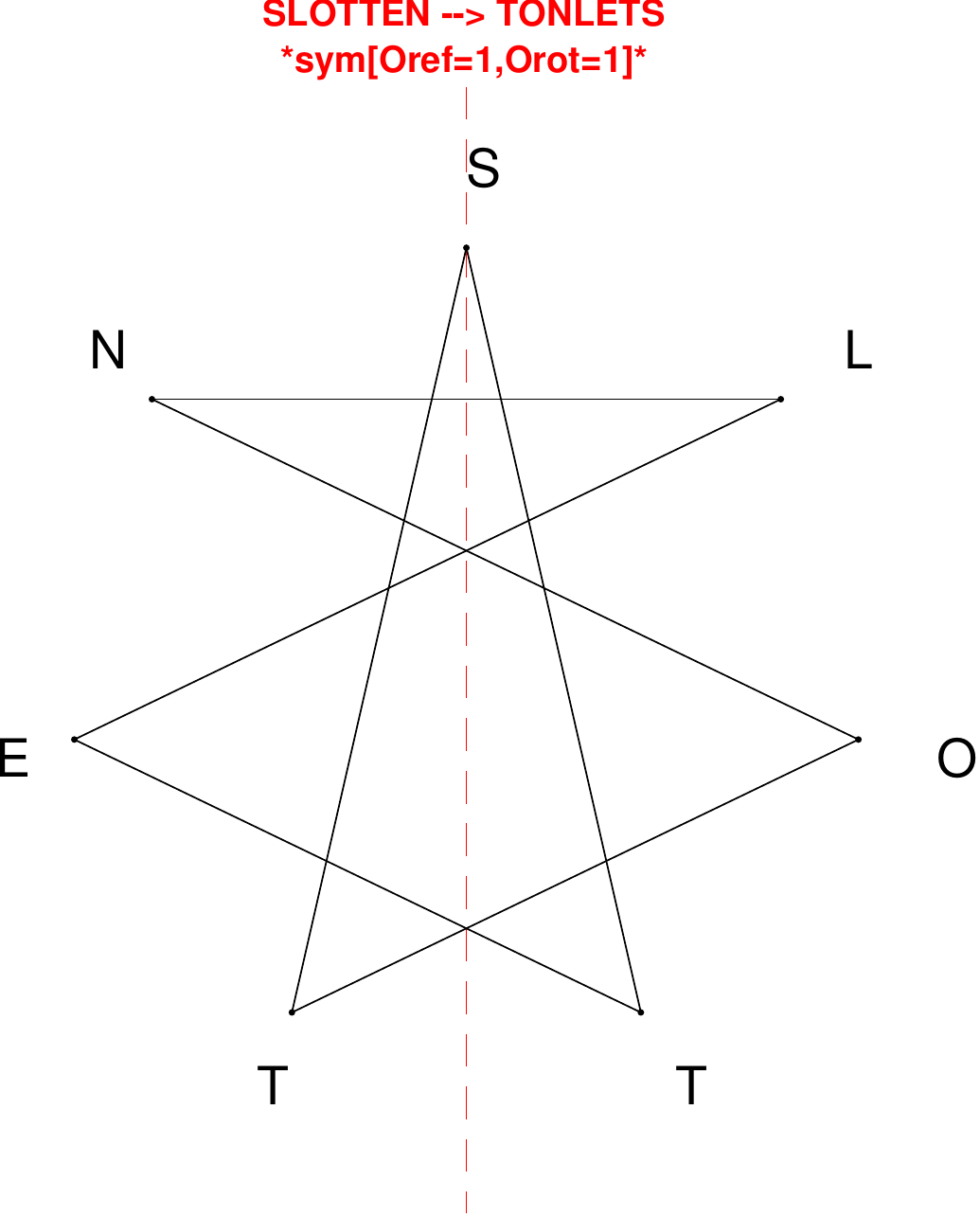}
\end{subfigure}
\hfill
\begin{subfigure}[T]{0.19\textwidth}
\centering
\includegraphics[width=\textwidth]{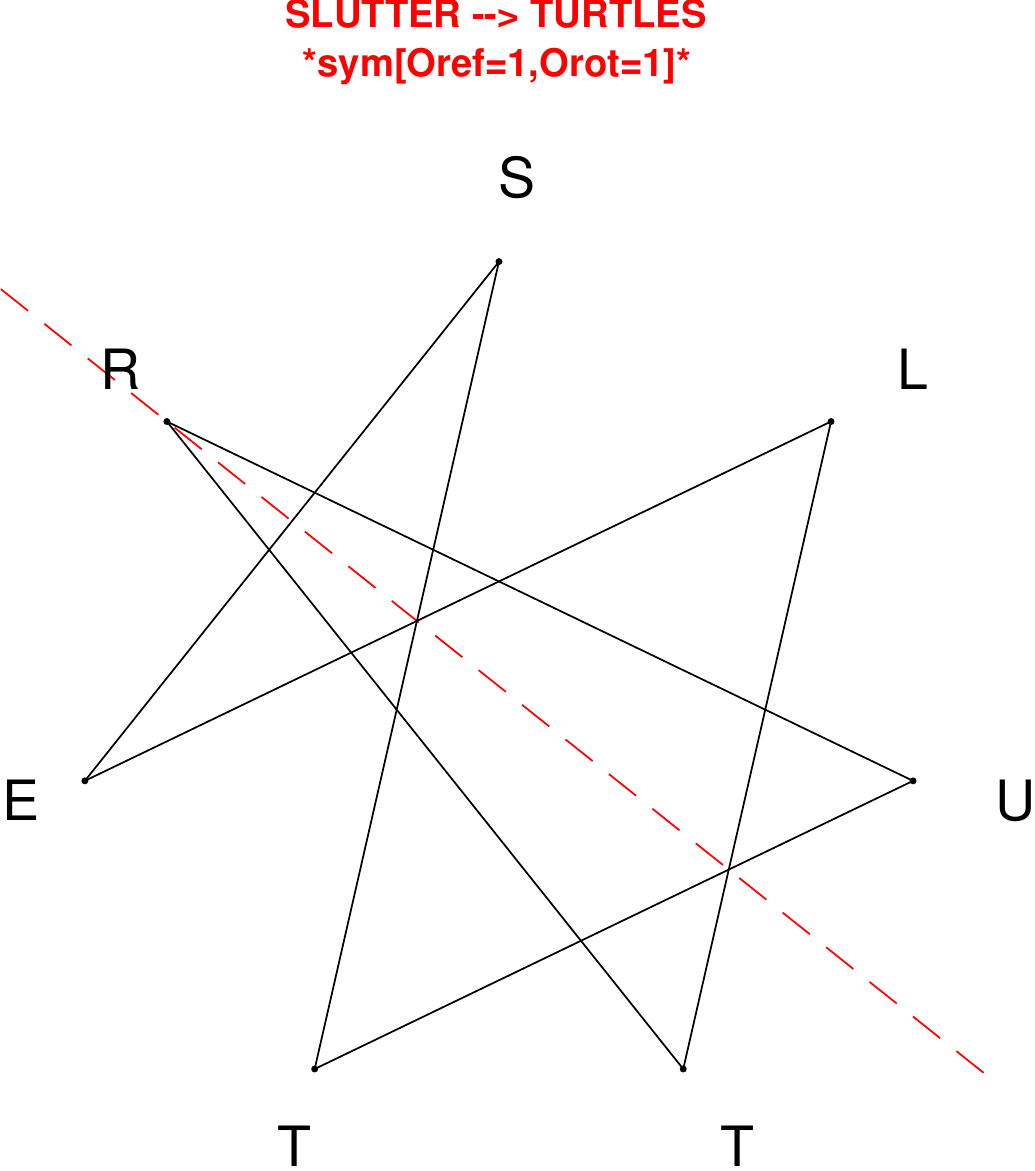}
\end{subfigure}
\end{figure}

\begin{figure}[H]
\centering
\begin{subfigure}[T]{0.19\textwidth}
\centering
\includegraphics[width=\textwidth]{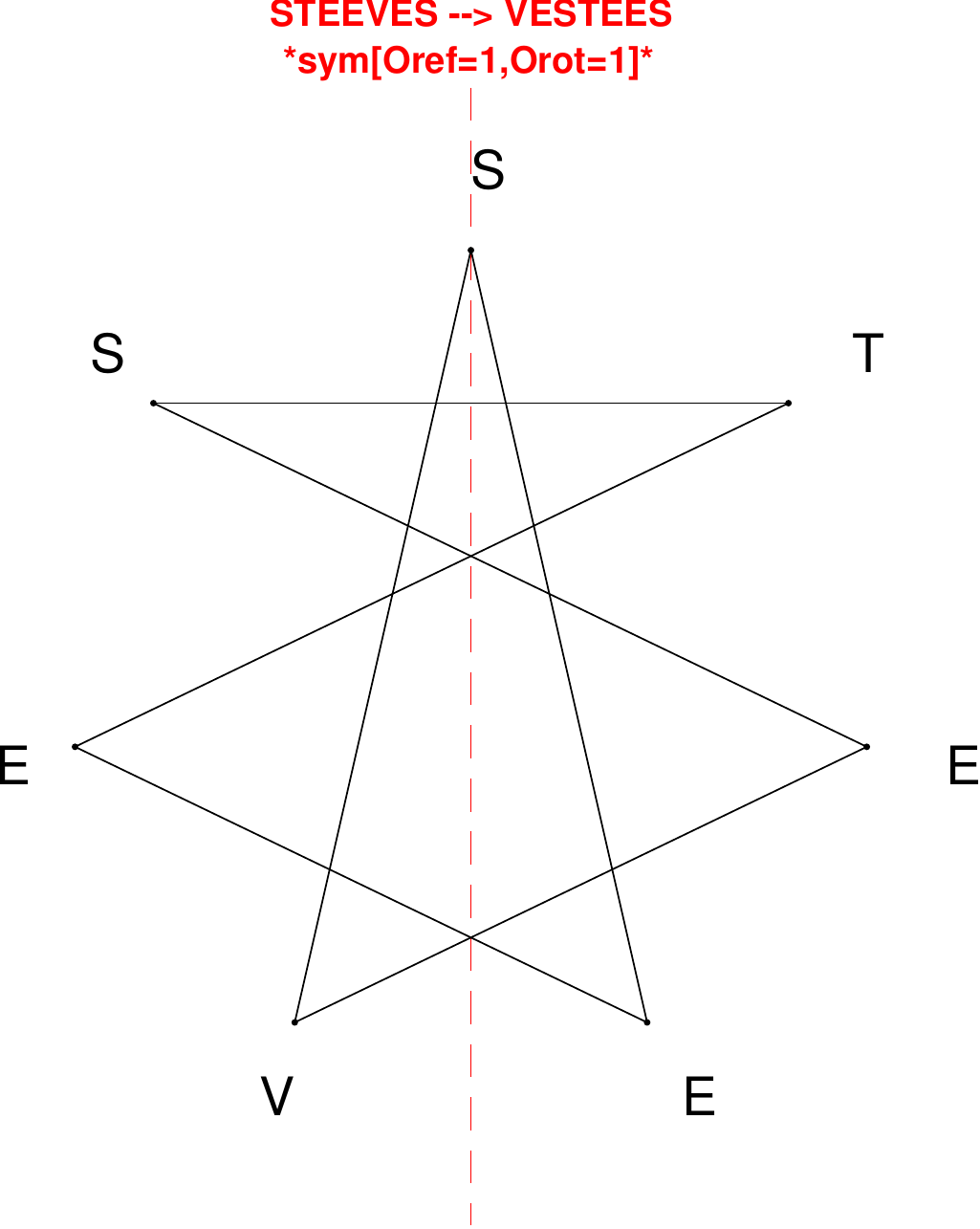}
\end{subfigure}
\hfill
\begin{subfigure}[T]{0.19\textwidth}
\centering
\includegraphics[width=\textwidth]{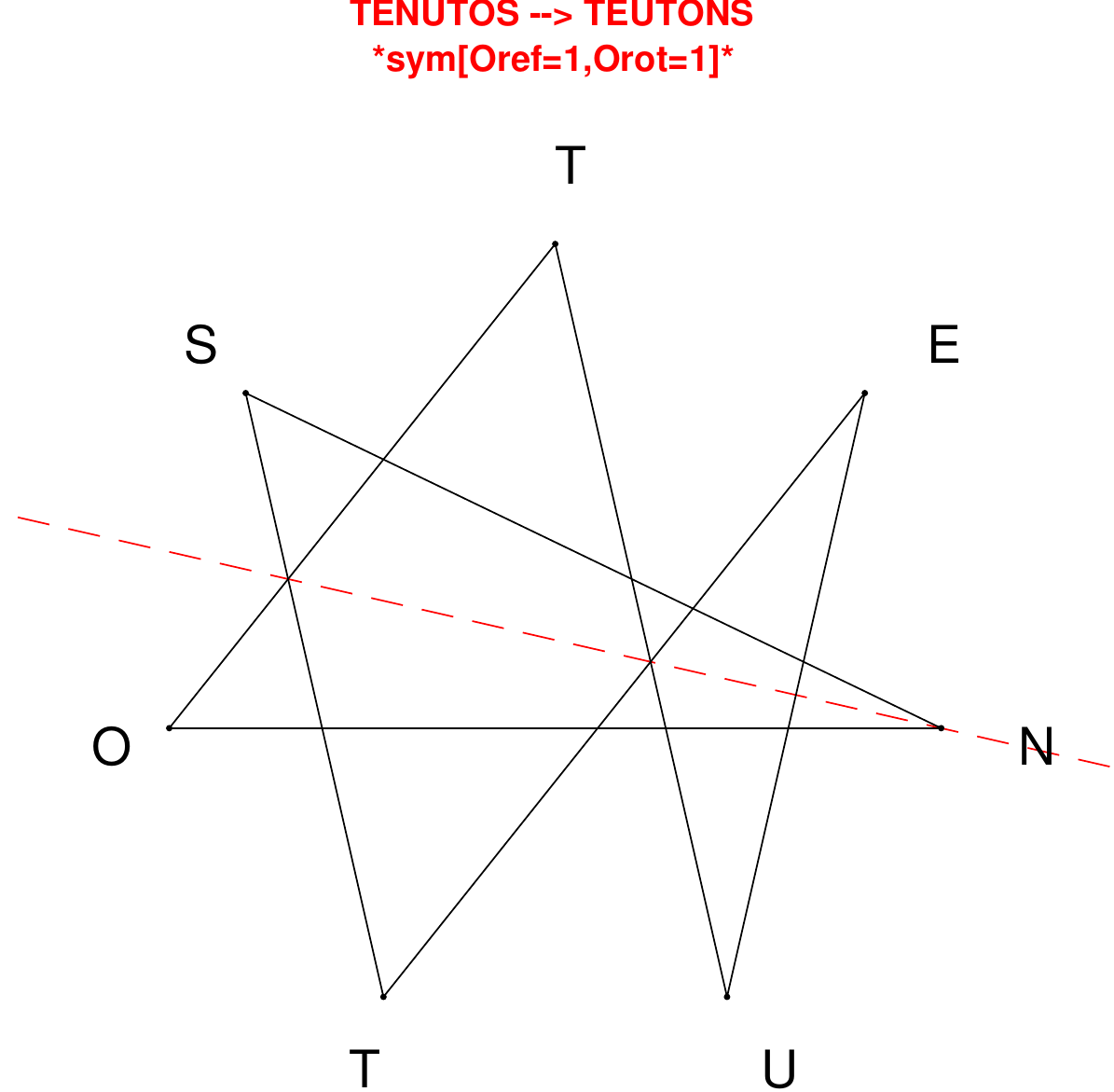}
\end{subfigure}
\hfill
\begin{subfigure}[T]{0.19\textwidth}
\centering
\includegraphics[width=\textwidth]{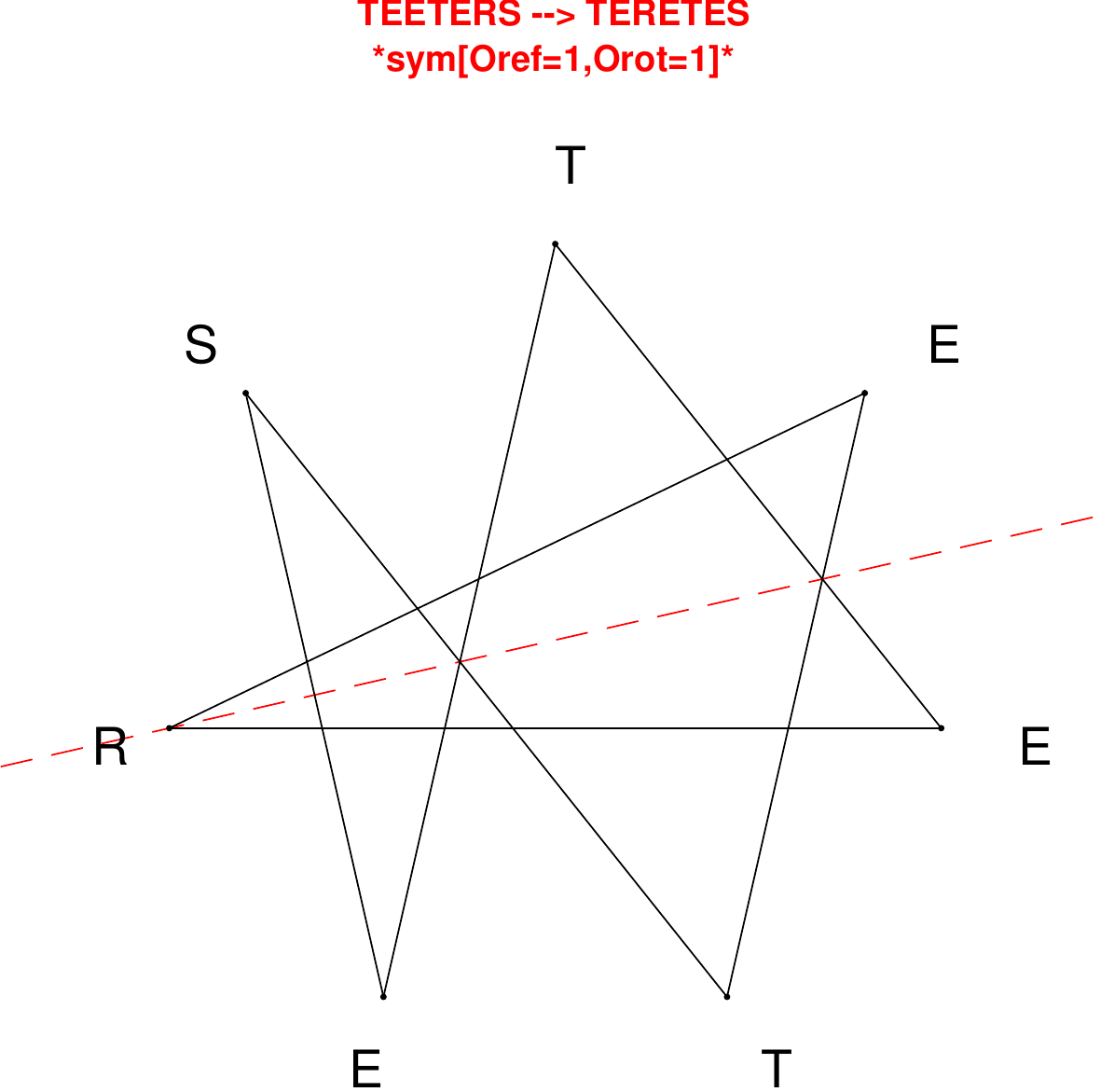}
\end{subfigure}
\hfill
\begin{subfigure}[T]{0.19\textwidth}
\centering
\includegraphics[width=\textwidth]{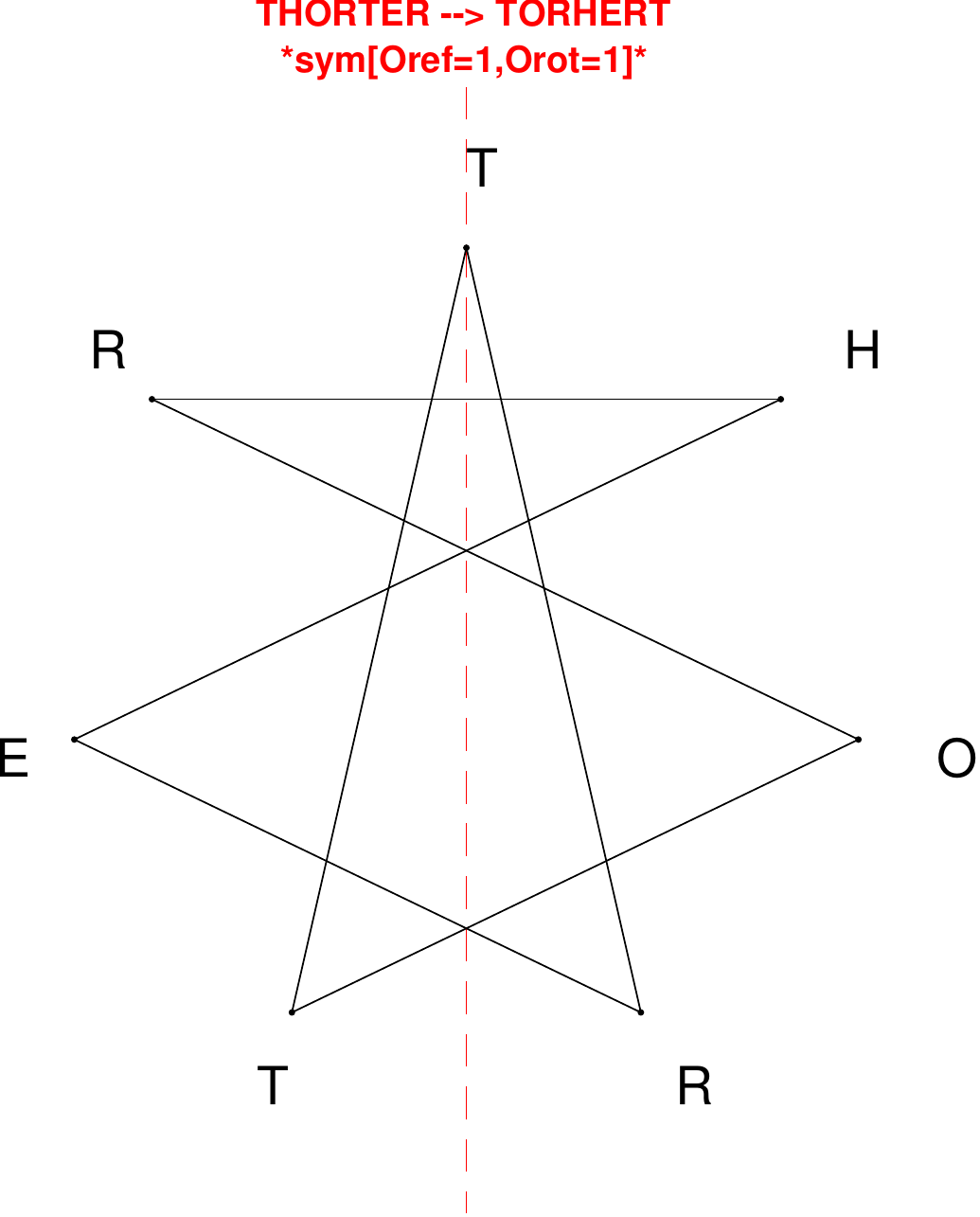}
\end{subfigure}
\hfill
\begin{subfigure}[T]{0.19\textwidth}
\centering
\includegraphics[width=\textwidth]{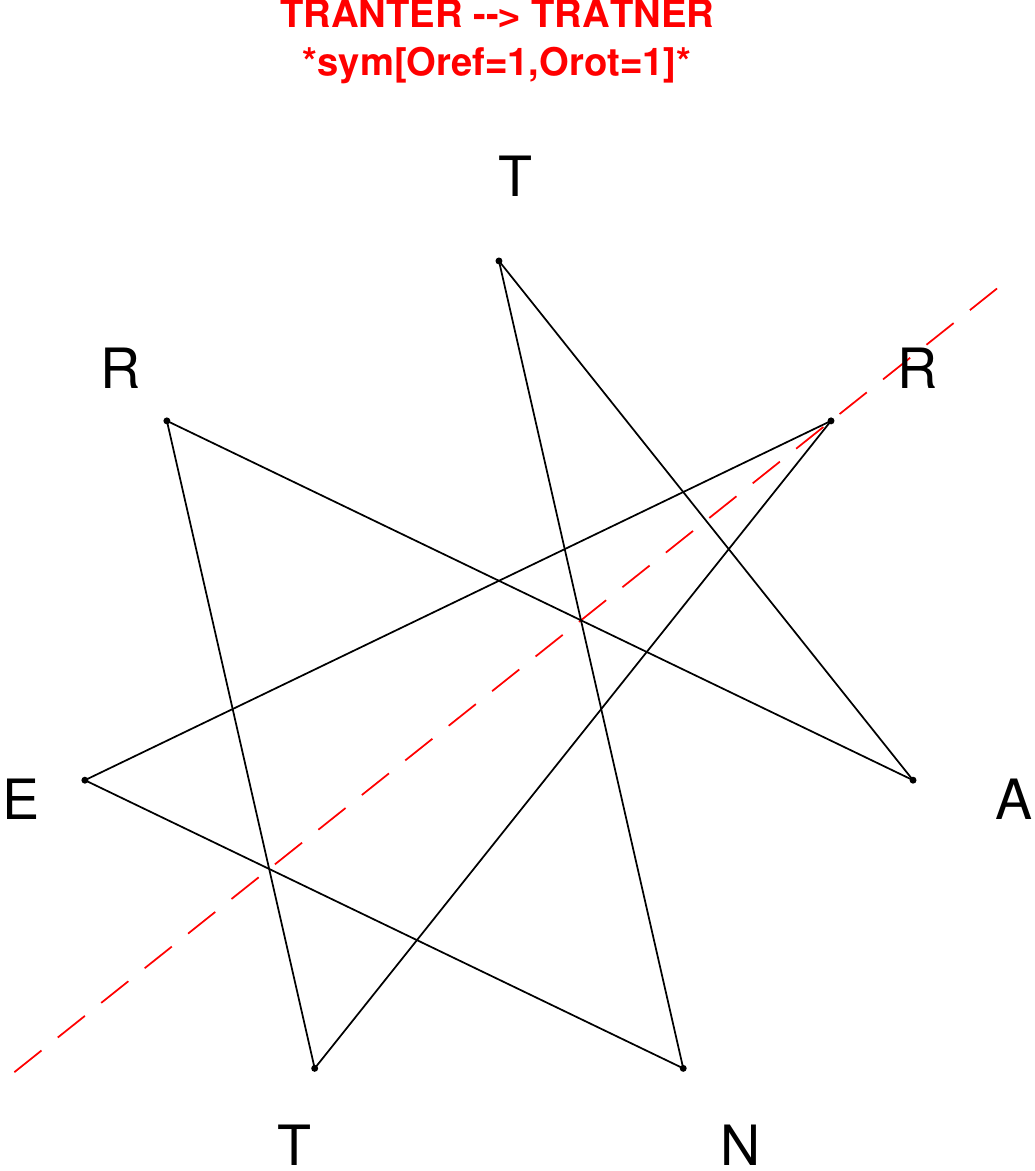}
\end{subfigure}
\end{figure}

\begin{figure}[H]
\centering
\begin{subfigure}[T]{0.19\textwidth}
\centering
\includegraphics[width=\textwidth]{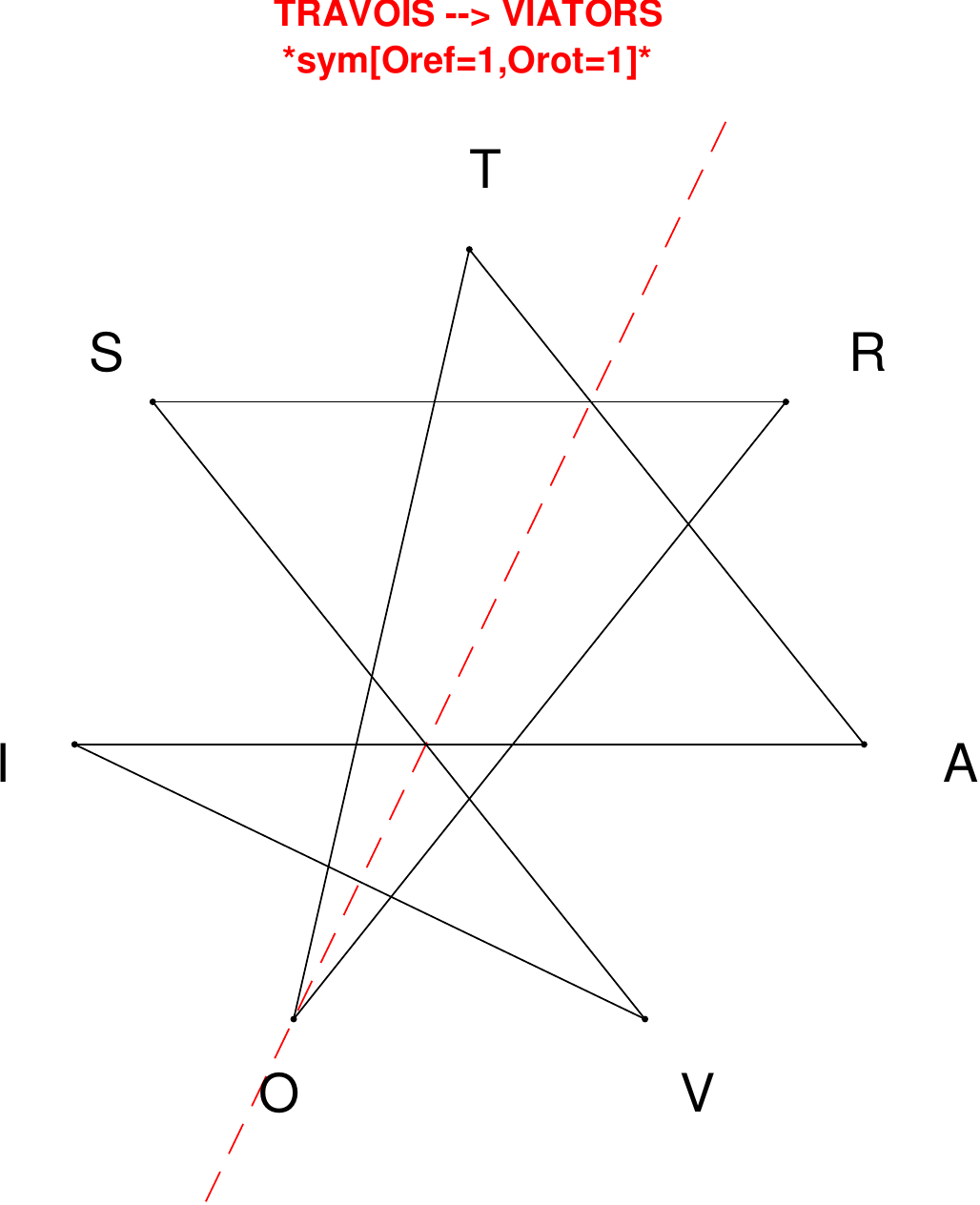}
\end{subfigure}
\hfill
\begin{subfigure}[T]{0.19\textwidth}
\centering
\includegraphics[width=\textwidth]{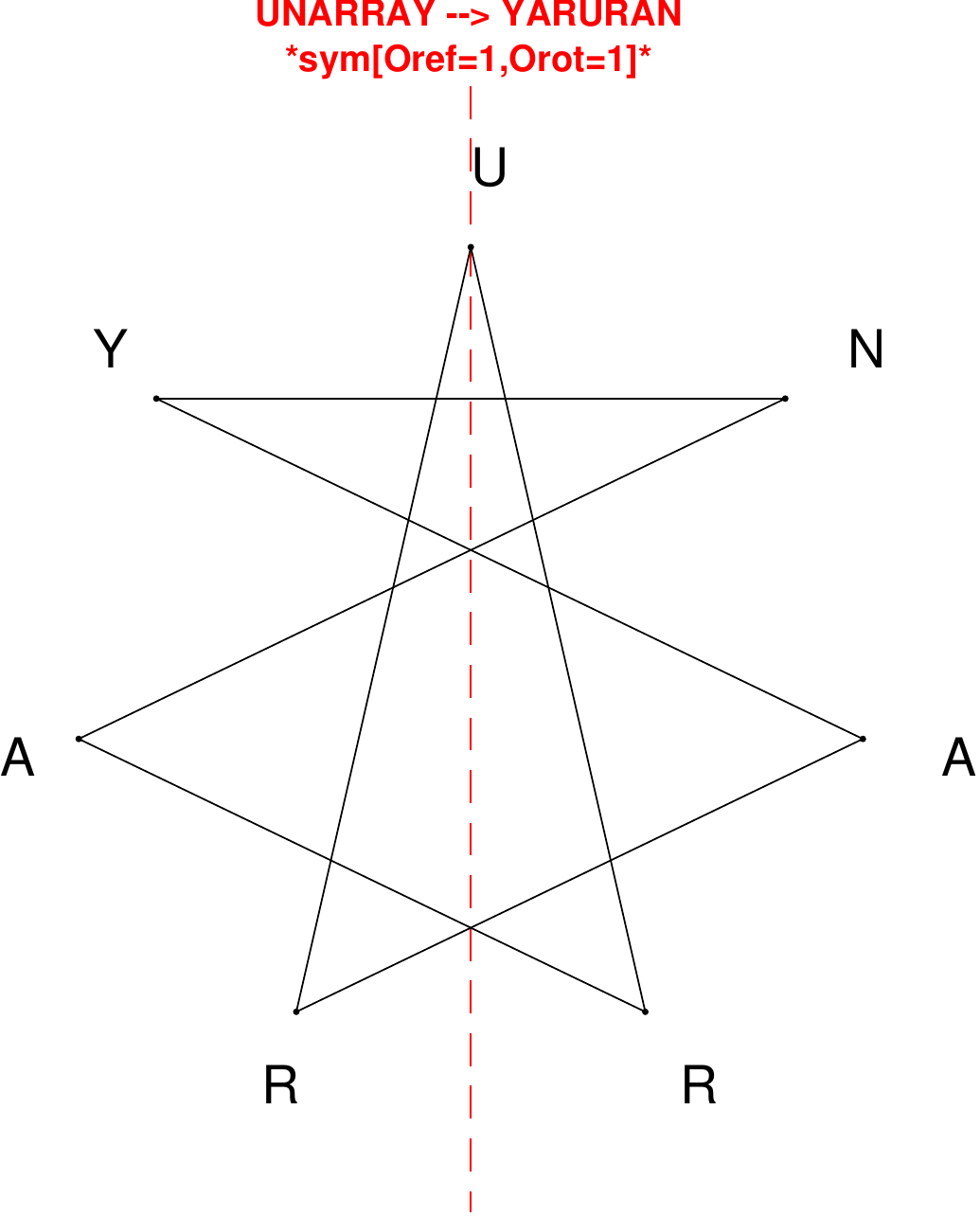}
\end{subfigure}
\hfill
\begin{subfigure}[T]{0.19\textwidth}
\centering
\includegraphics[width=\textwidth]{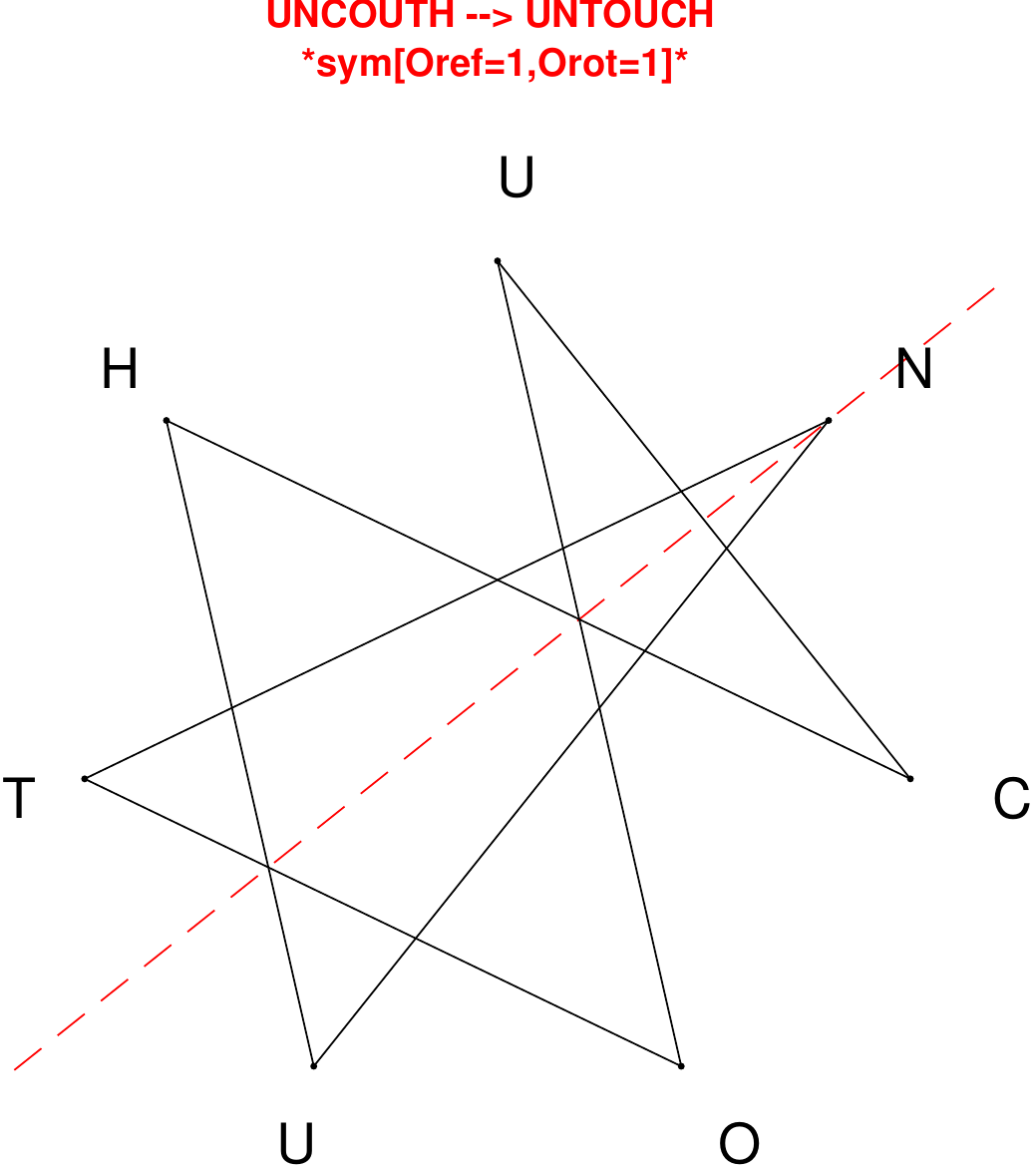}
\end{subfigure}
\hfill
\begin{subfigure}[T]{0.19\textwidth}
\centering
\includegraphics[width=\textwidth]{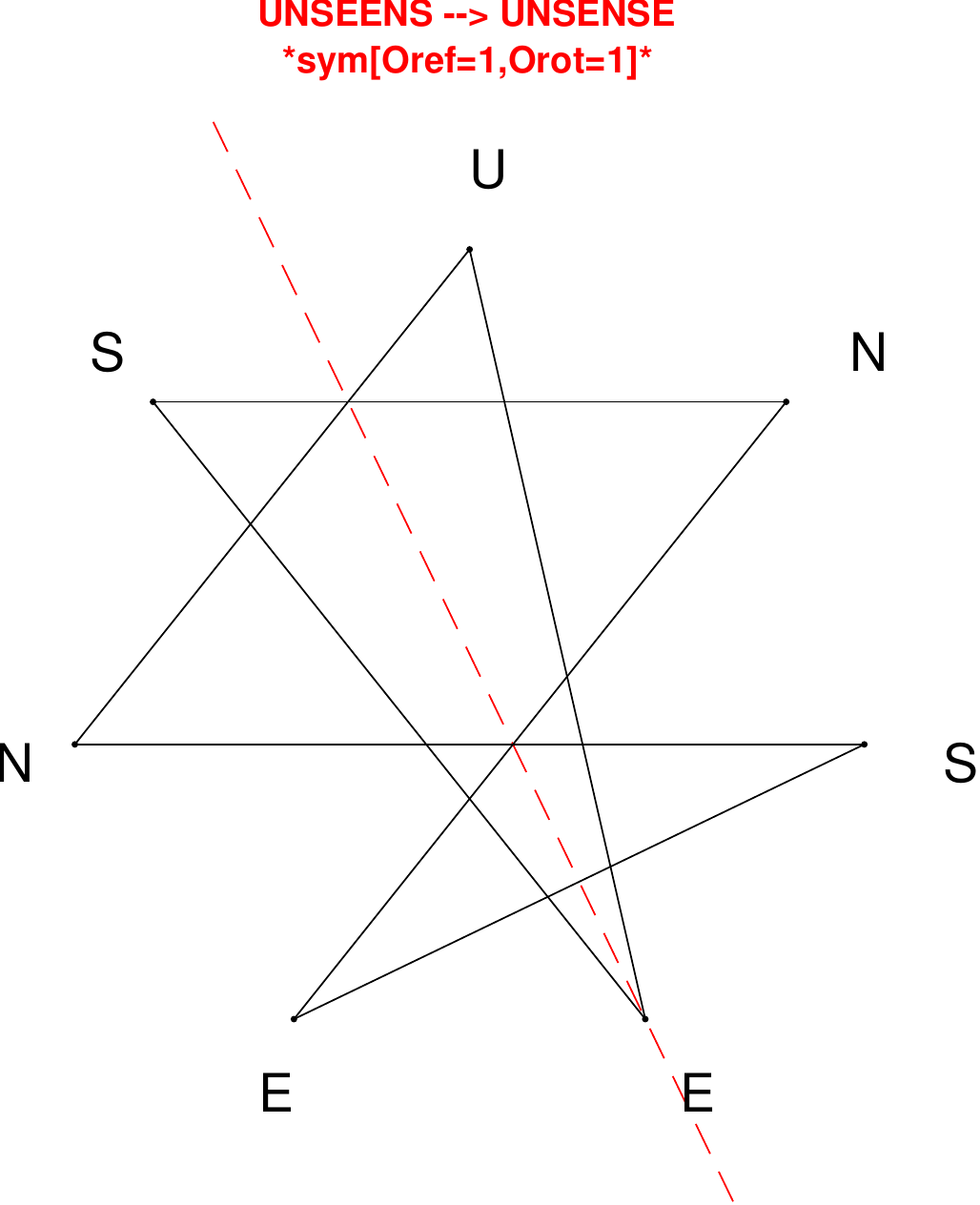}
\end{subfigure}
\hfill
\begin{subfigure}[T]{0.19\textwidth}
\centering
\includegraphics[width=\textwidth]{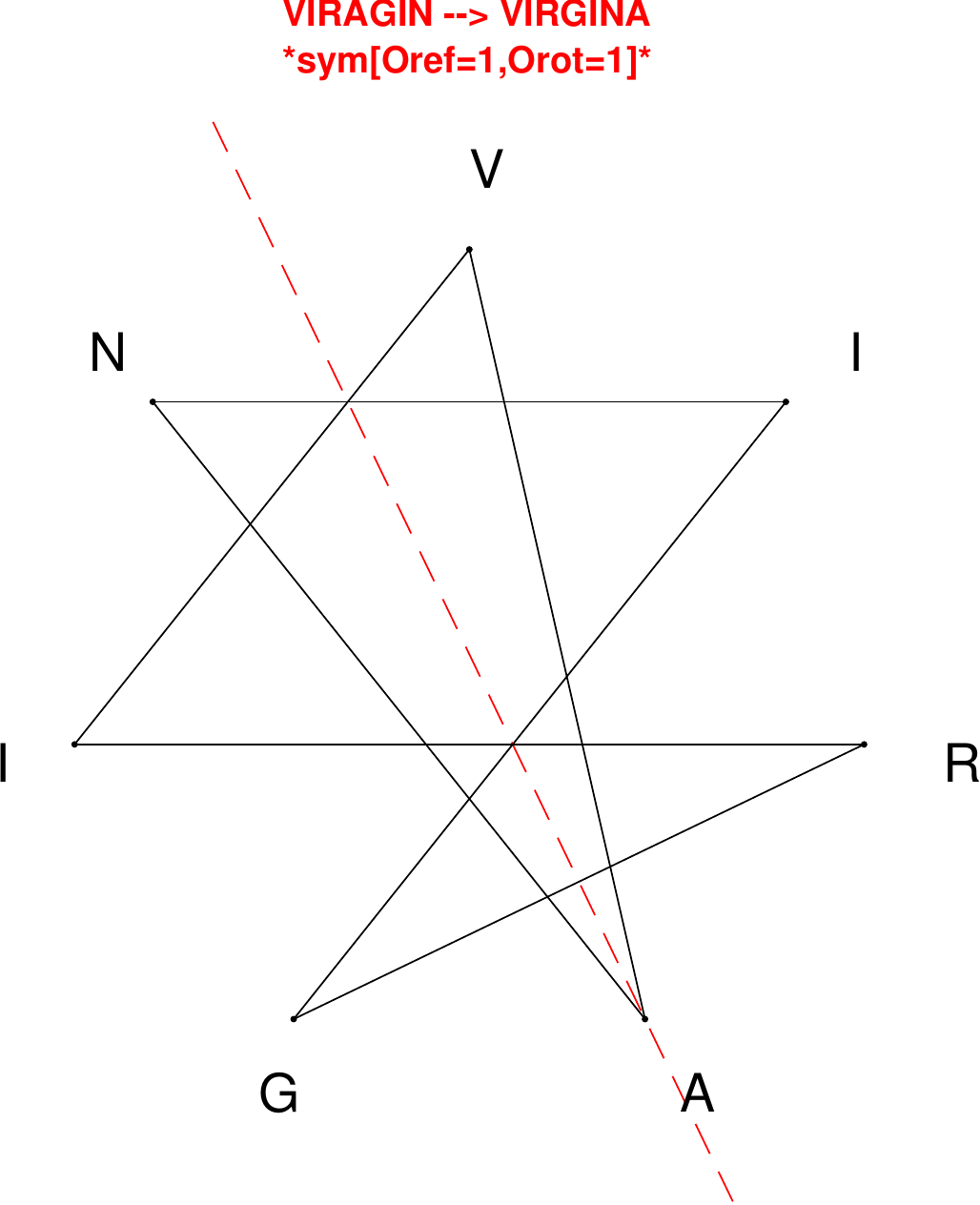}
\end{subfigure}
\end{figure}

\begin{figure}[H]
\centering
\begin{subfigure}[T]{0.19\textwidth}
\centering
\includegraphics[width=\textwidth]{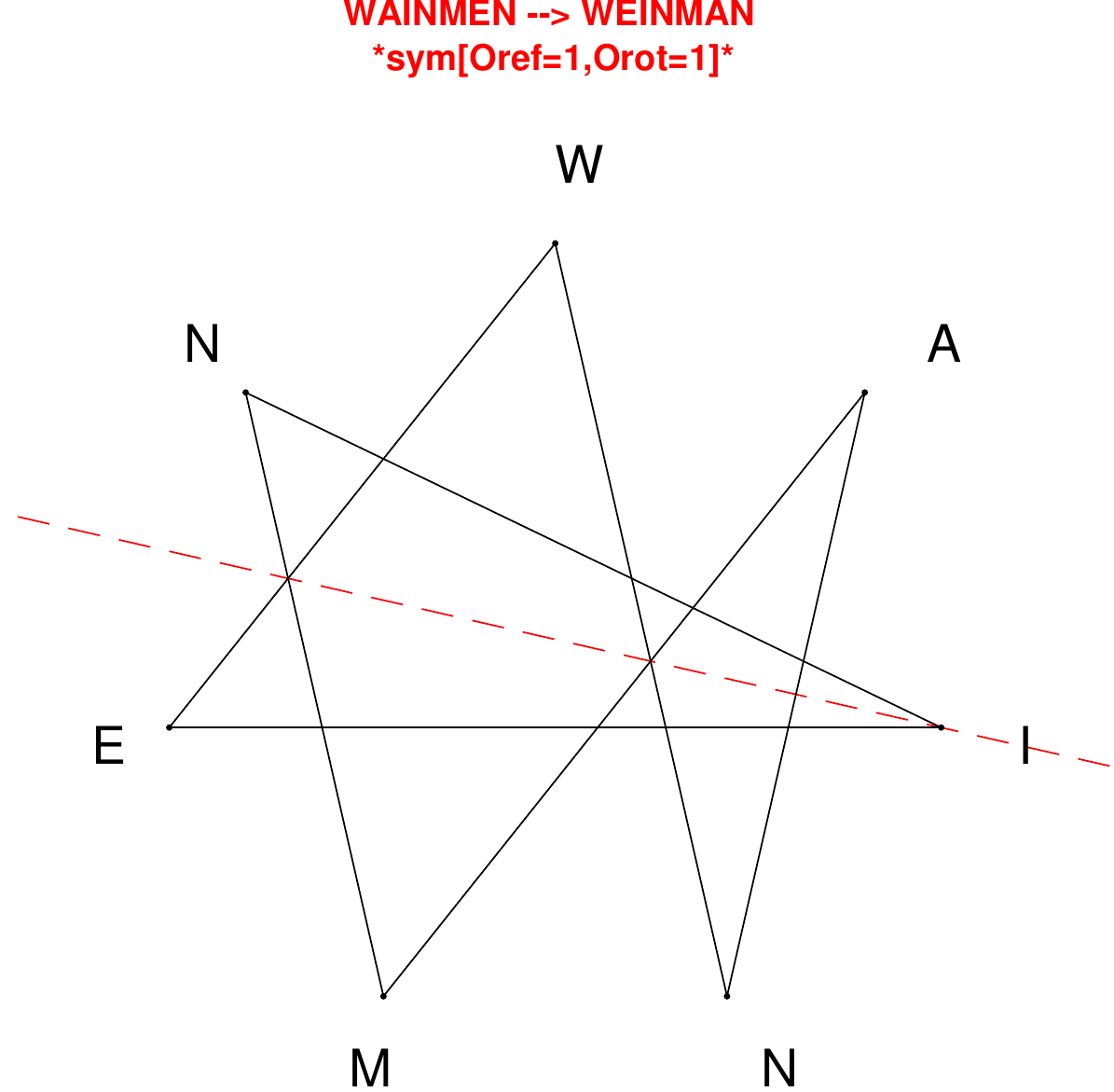}
\end{subfigure}
\hfill
\begin{subfigure}[T]{0.19\textwidth}
\centering
\includegraphics[width=\textwidth]{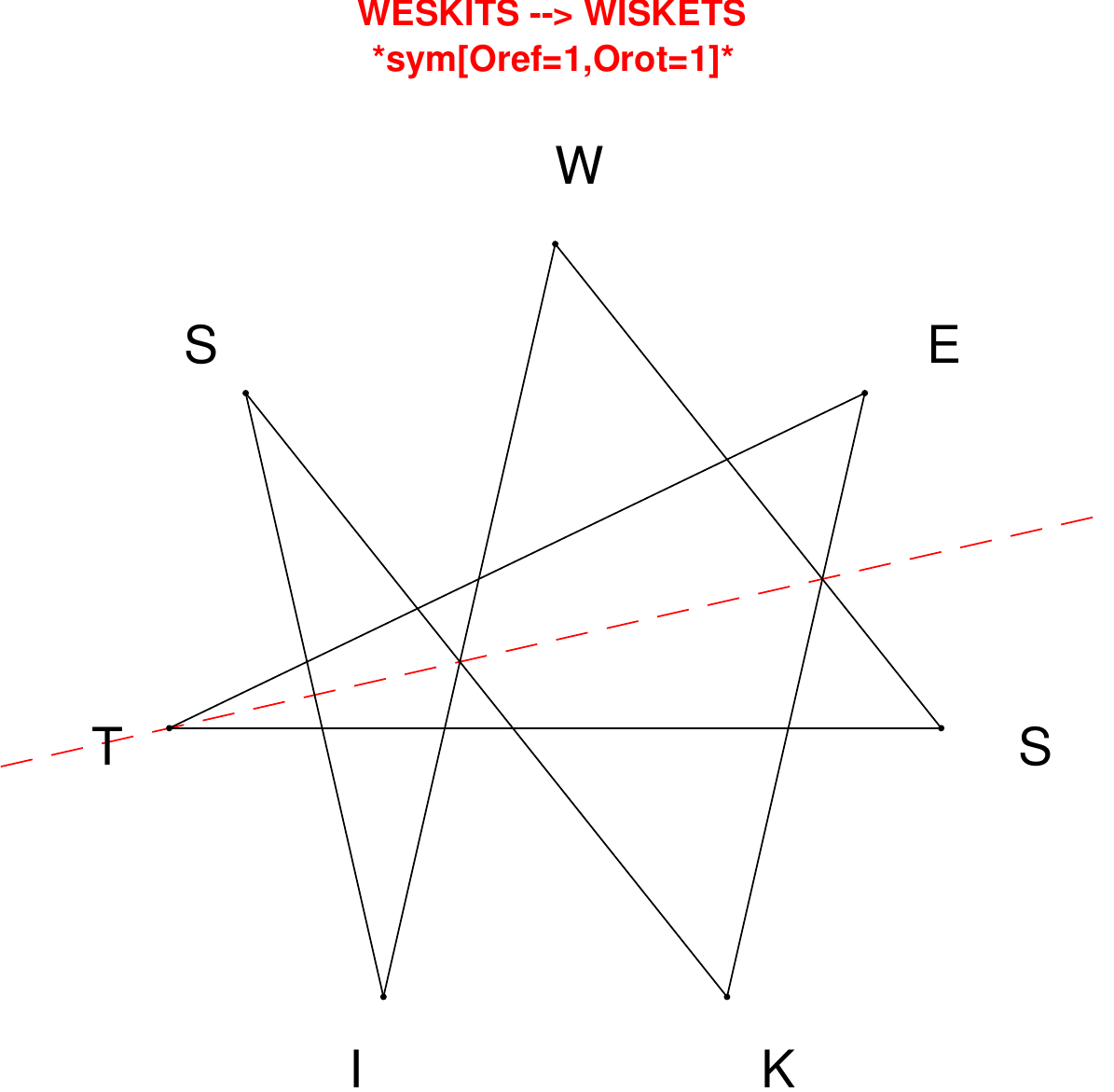}
\end{subfigure}
\hfill
\begin{subfigure}[T]{0.19\textwidth}
\centering
\includegraphics[width=\textwidth]{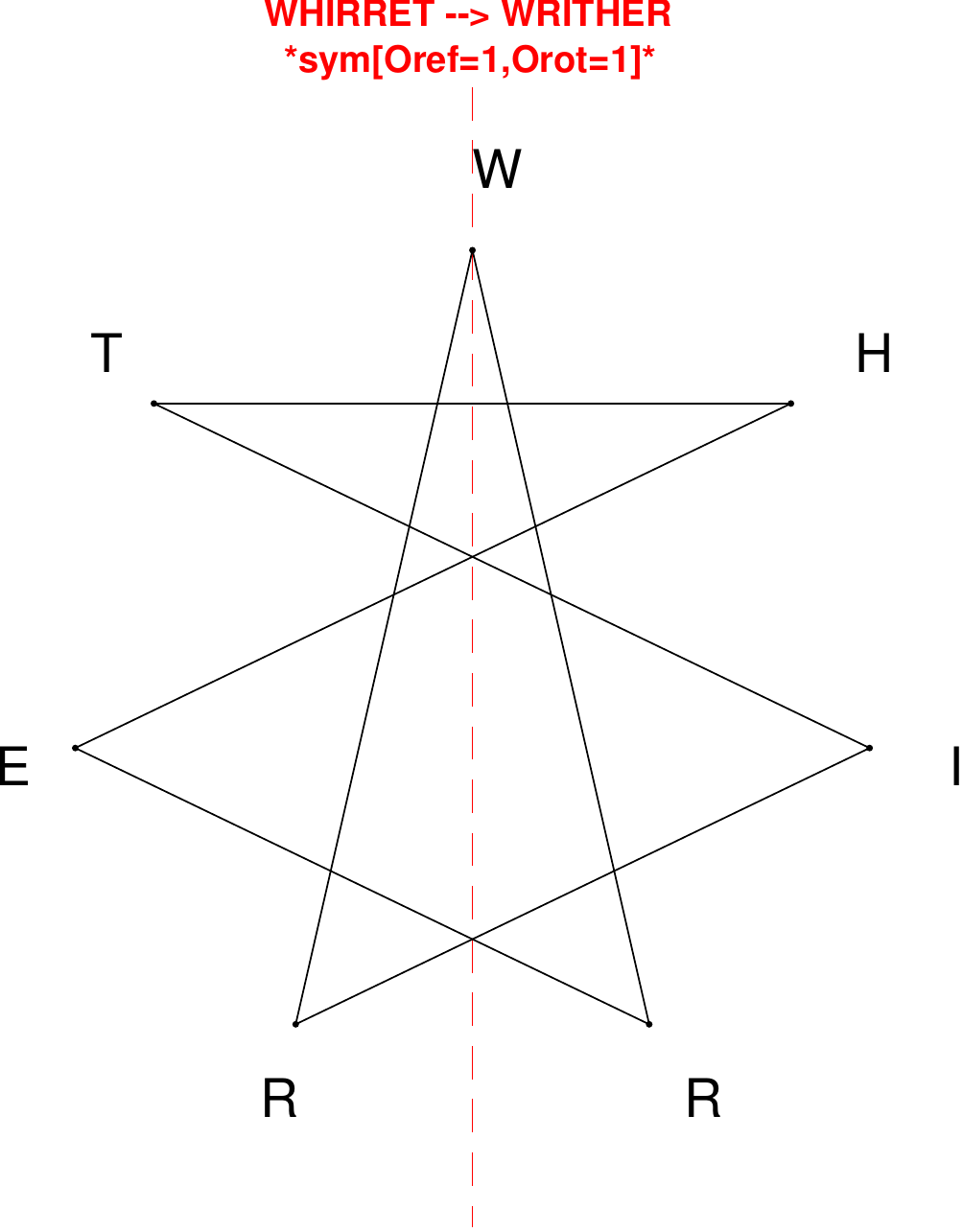}
\end{subfigure}
\hfill
\begin{subfigure}[T]{0.19\textwidth}
\centering
\includegraphics[width=\textwidth]{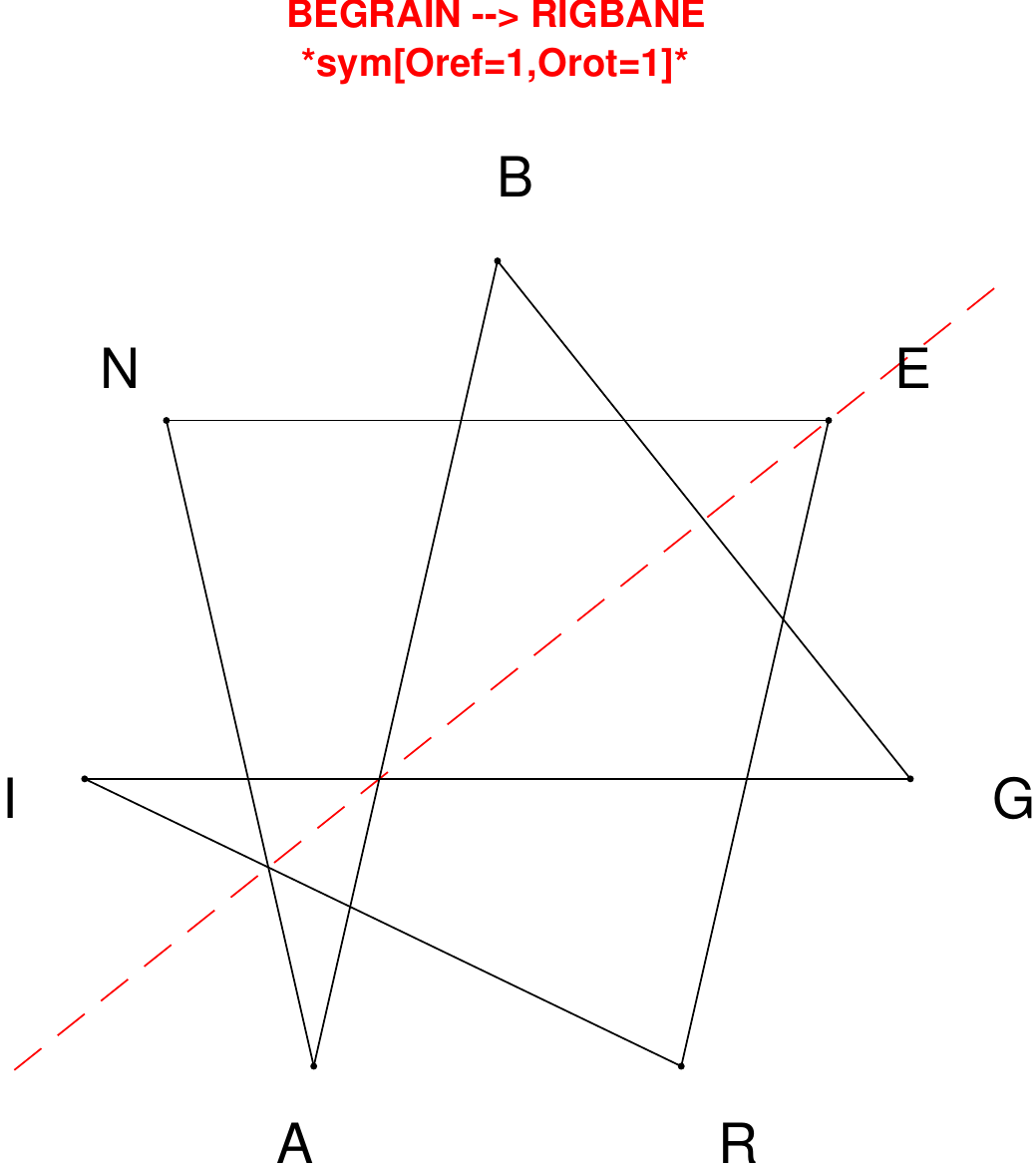}
\end{subfigure}
\hfill
\begin{subfigure}[T]{0.19\textwidth}
\centering
\includegraphics[width=\textwidth]{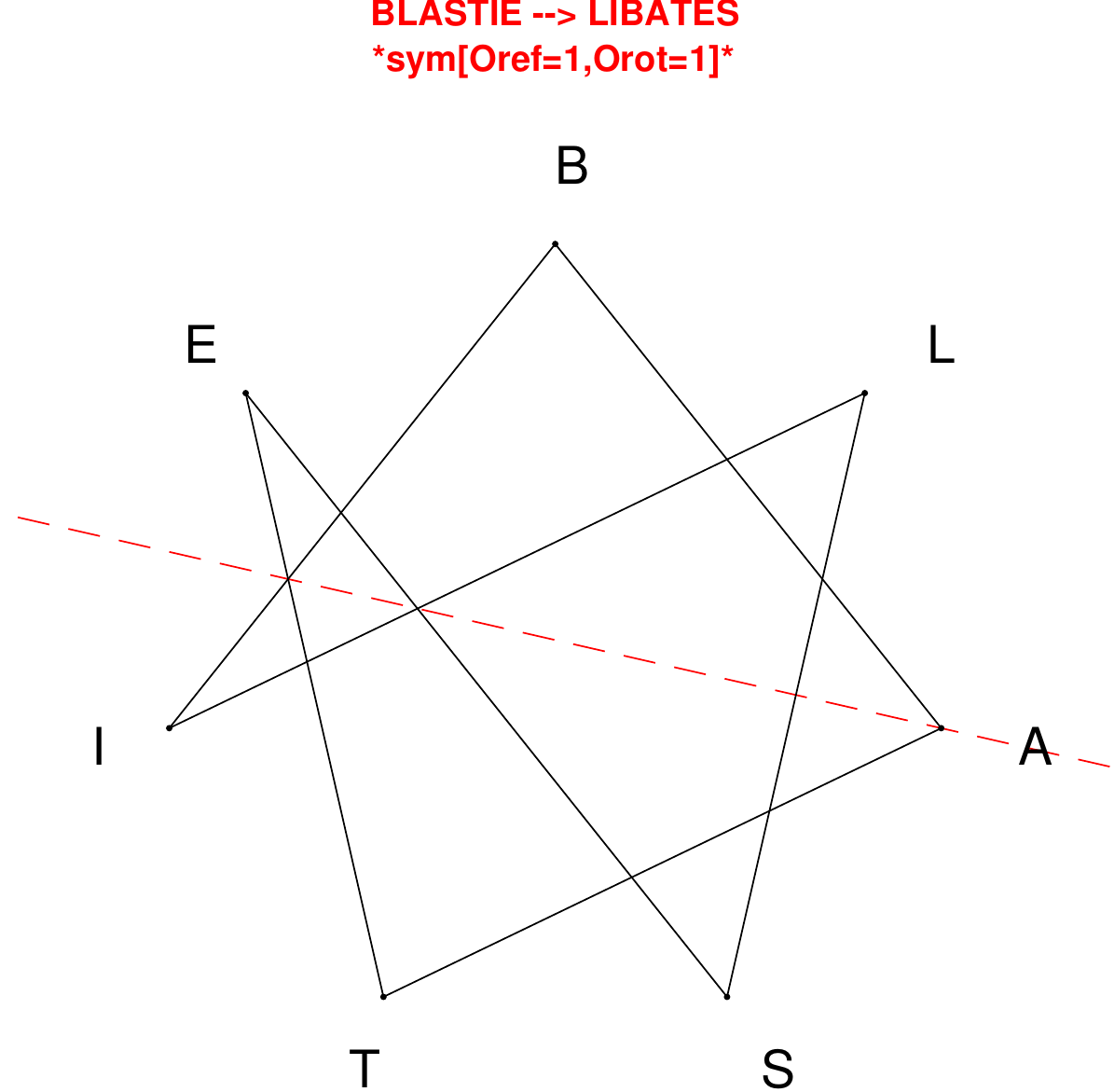}
\end{subfigure}
\end{figure}

\begin{figure}[H]
\centering
\begin{subfigure}[T]{0.19\textwidth}
\centering
\includegraphics[width=\textwidth]{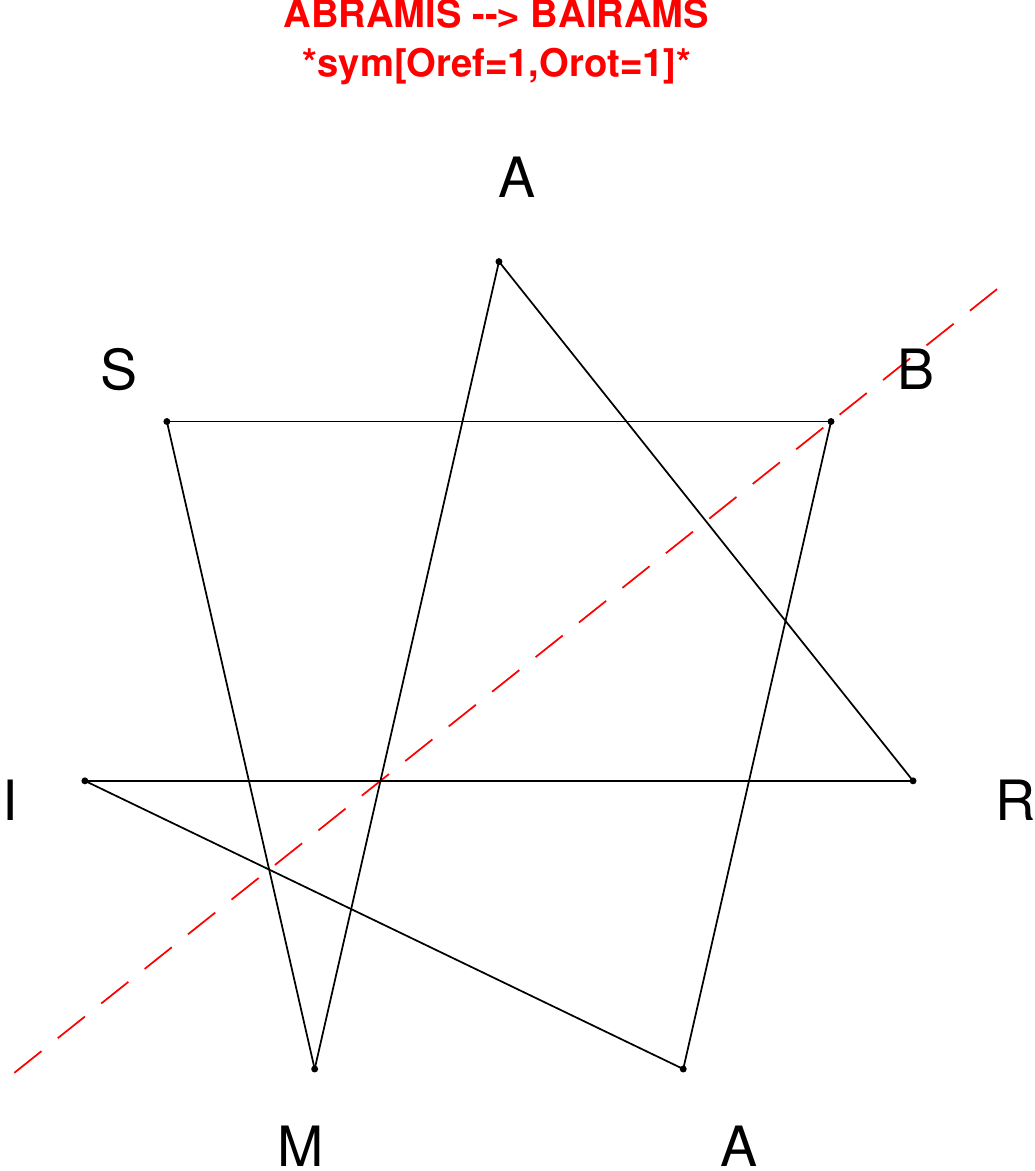}
\end{subfigure}
\hfill
\begin{subfigure}[T]{0.19\textwidth}
\centering
\includegraphics[width=\textwidth]{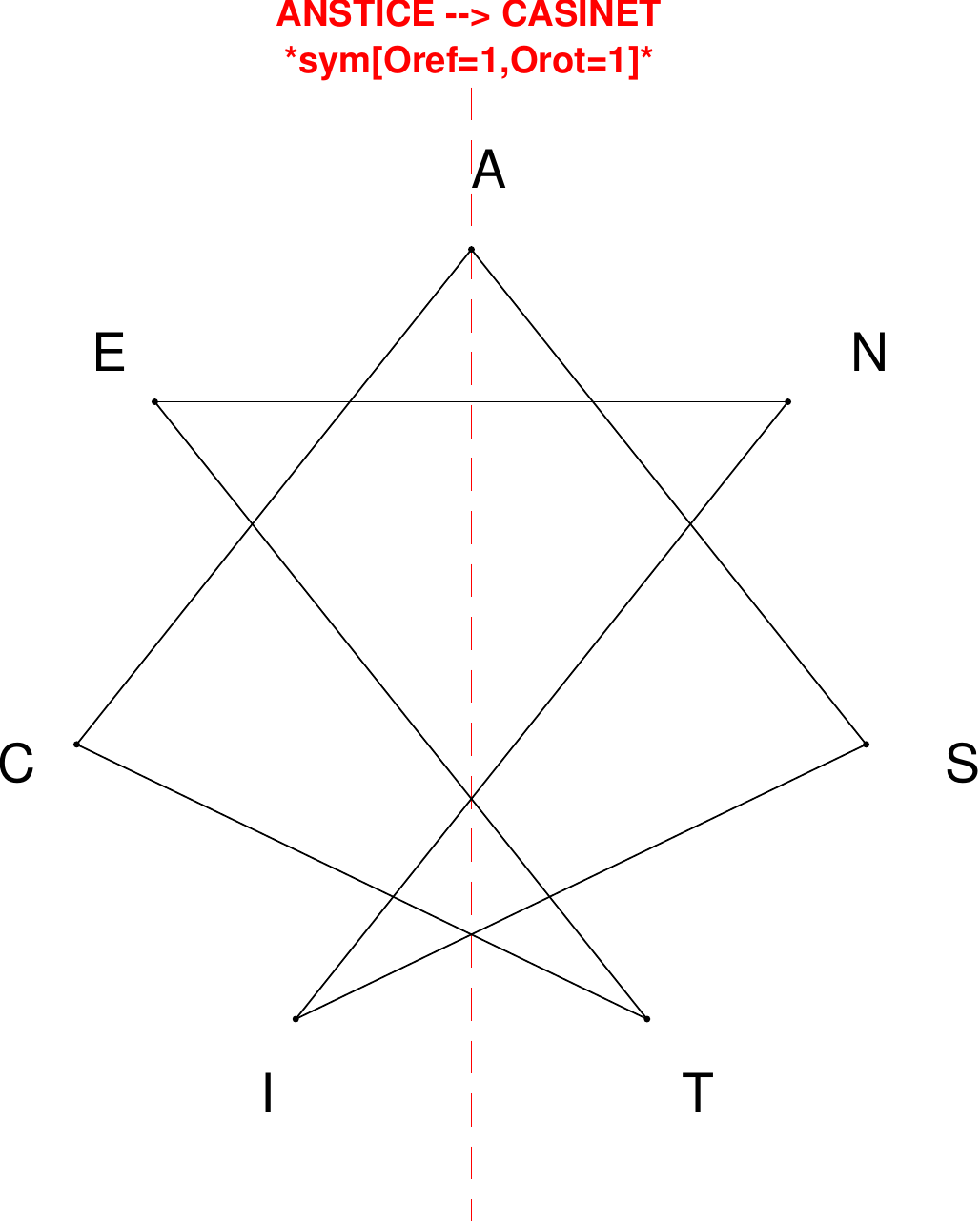}
\end{subfigure}
\hfill
\begin{subfigure}[T]{0.19\textwidth}
\centering
\includegraphics[width=\textwidth]{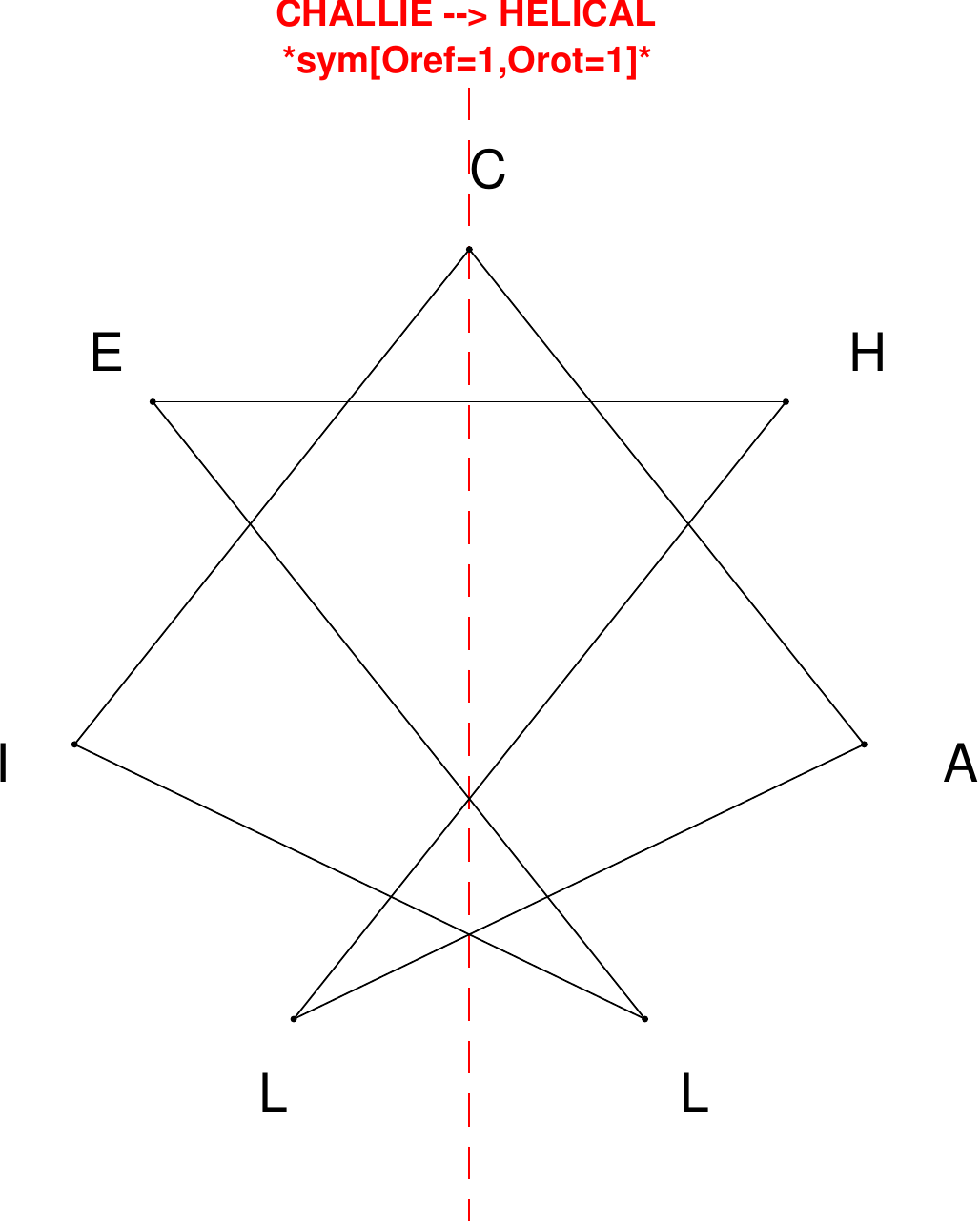}
\end{subfigure}
\hfill
\begin{subfigure}[T]{0.19\textwidth}
\centering
\includegraphics[width=\textwidth]{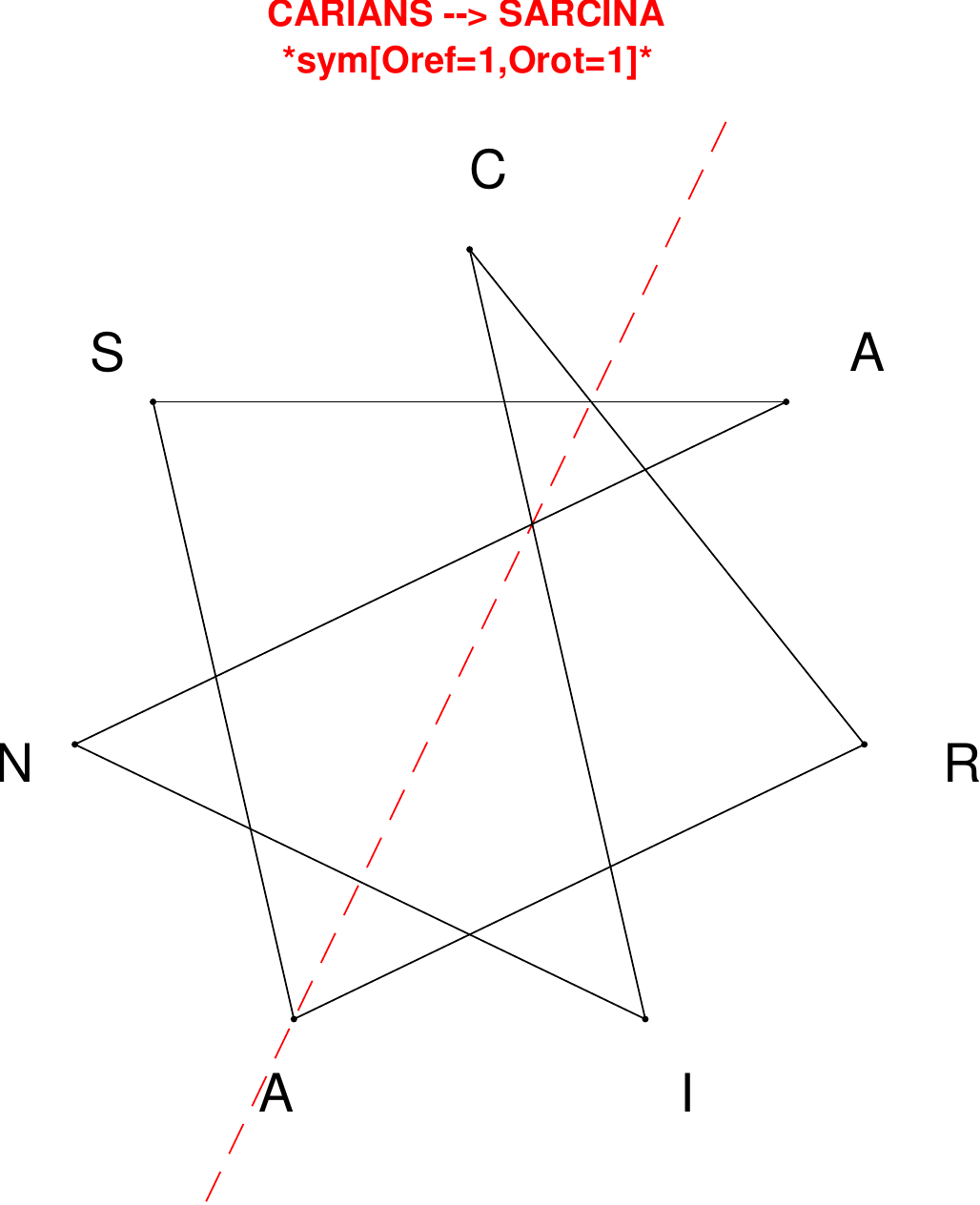}
\end{subfigure}
\hfill
\begin{subfigure}[T]{0.19\textwidth}
\centering
\includegraphics[width=\textwidth]{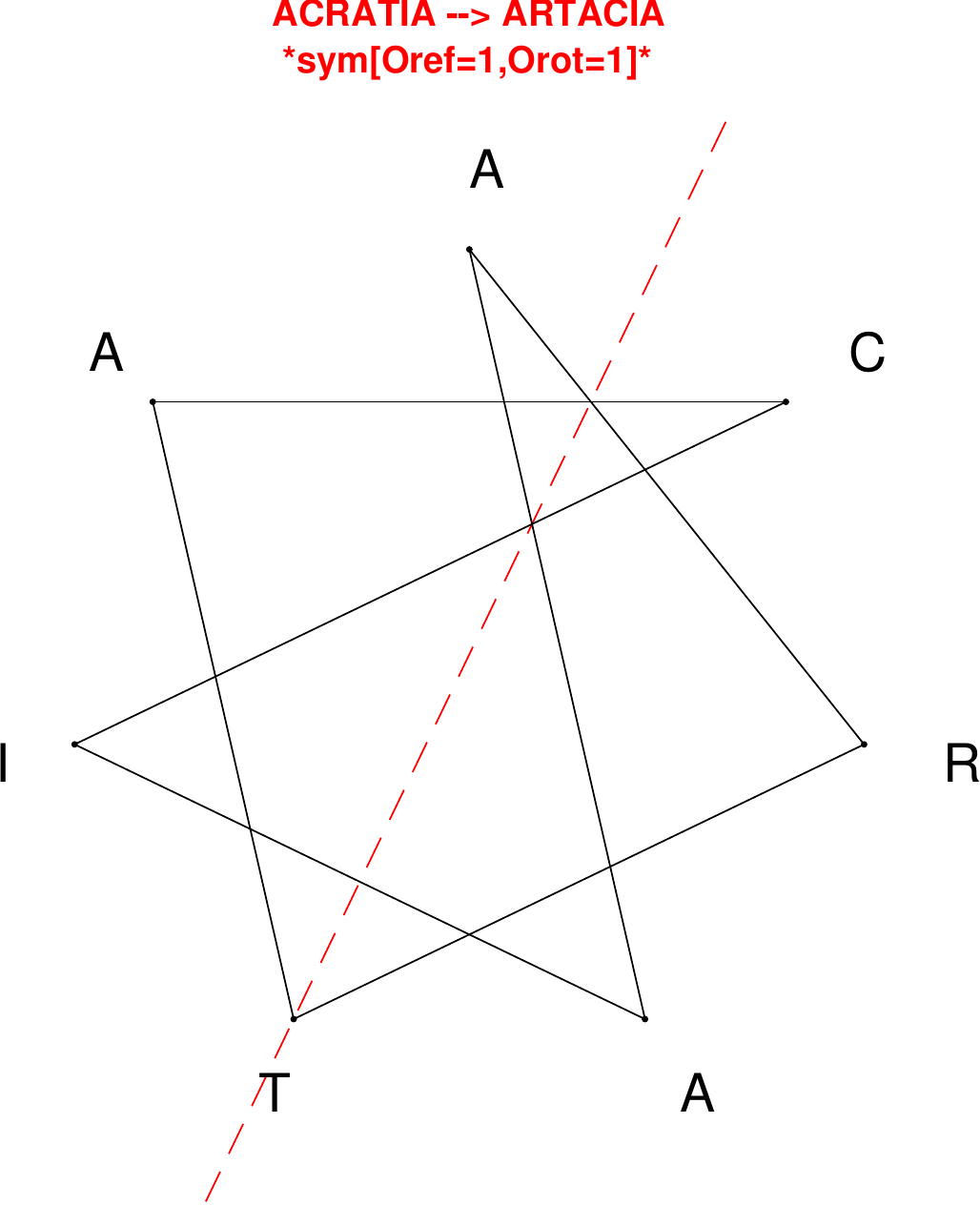}
\end{subfigure}
\end{figure}

\begin{figure}[H]
\centering
\begin{subfigure}[T]{0.19\textwidth}
\centering
\includegraphics[width=\textwidth]{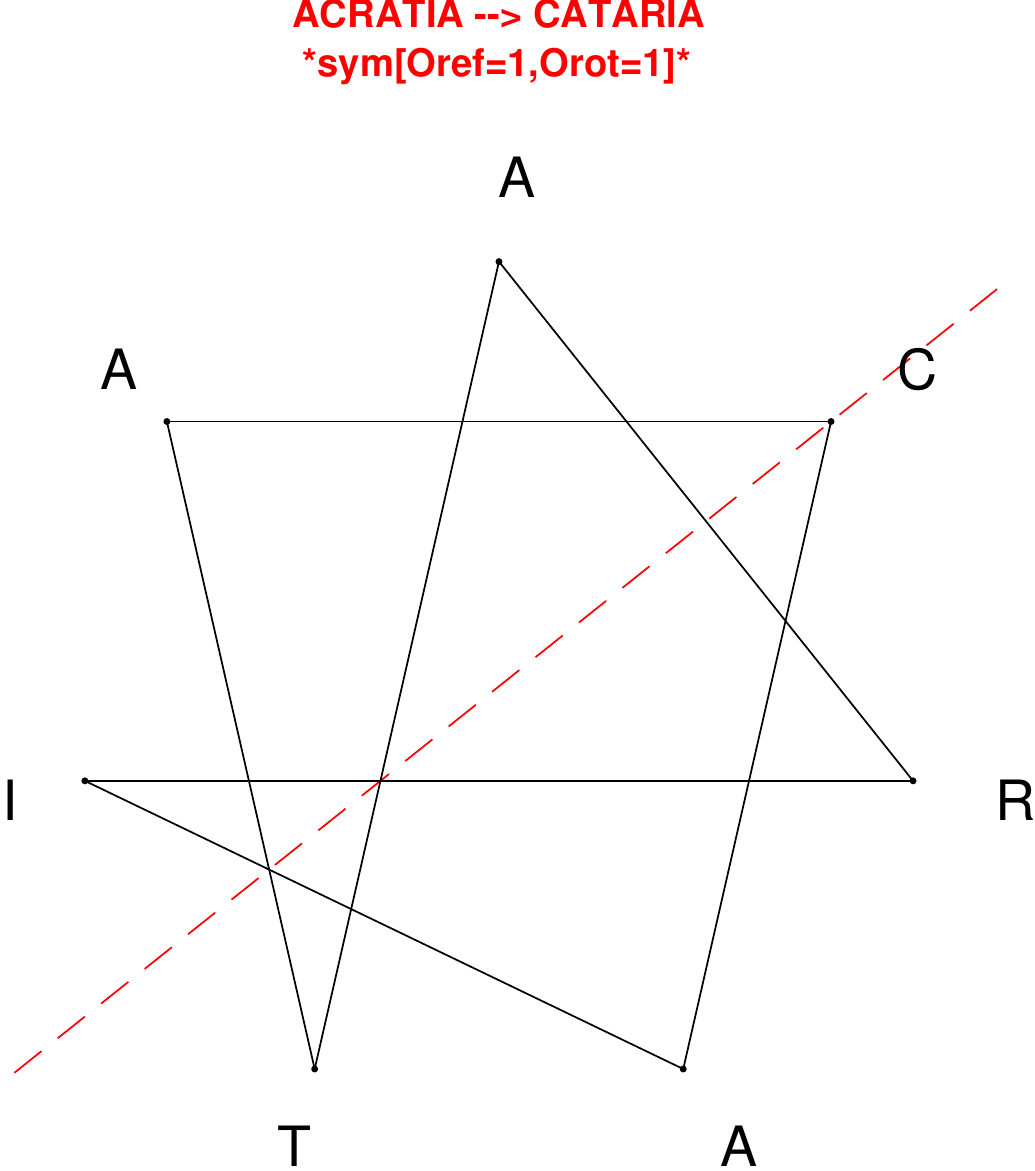}
\end{subfigure}
\hfill
\begin{subfigure}[T]{0.19\textwidth}
\centering
\includegraphics[width=\textwidth]{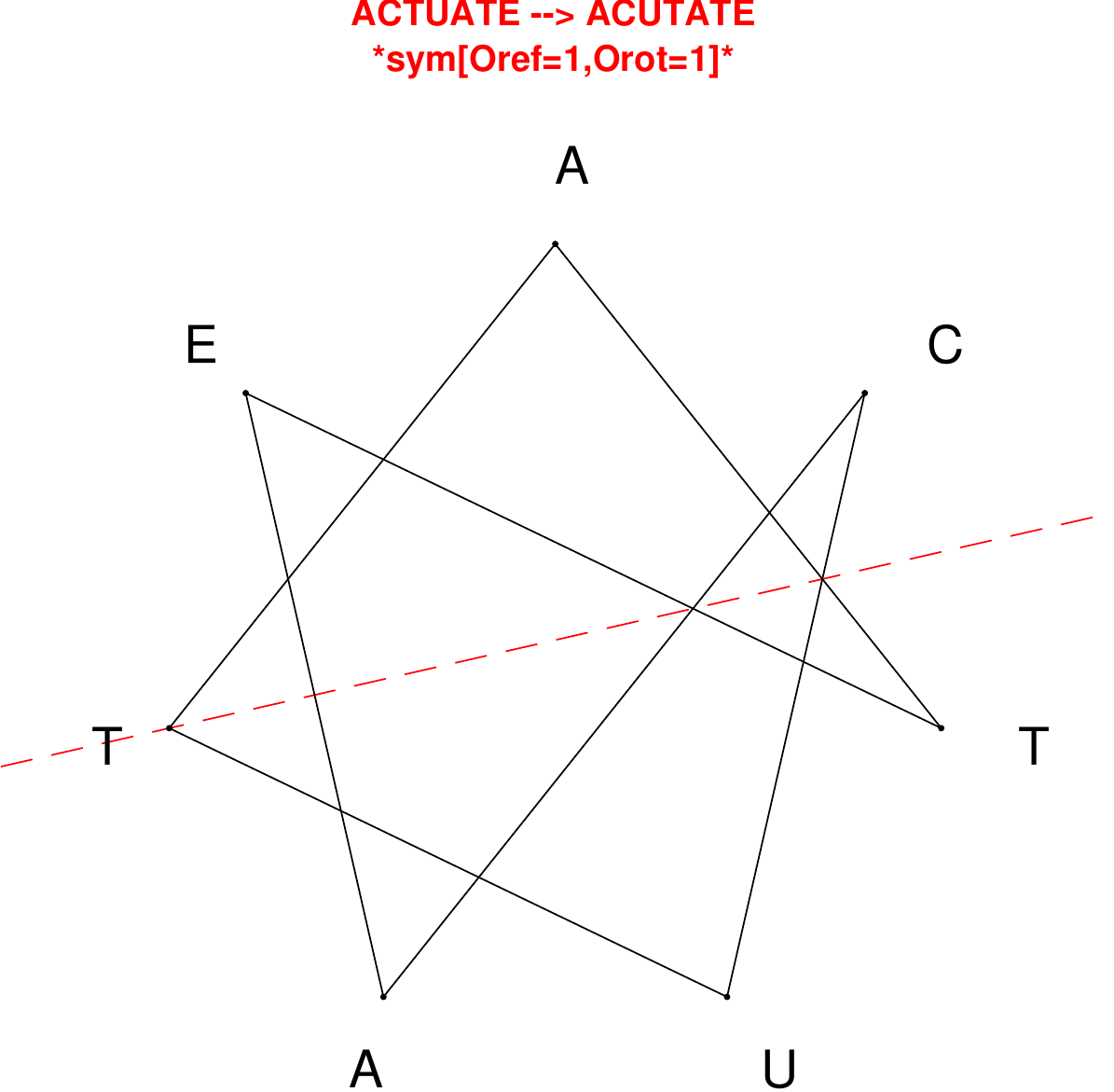}
\end{subfigure}
\hfill
\begin{subfigure}[T]{0.19\textwidth}
\centering
\includegraphics[width=\textwidth]{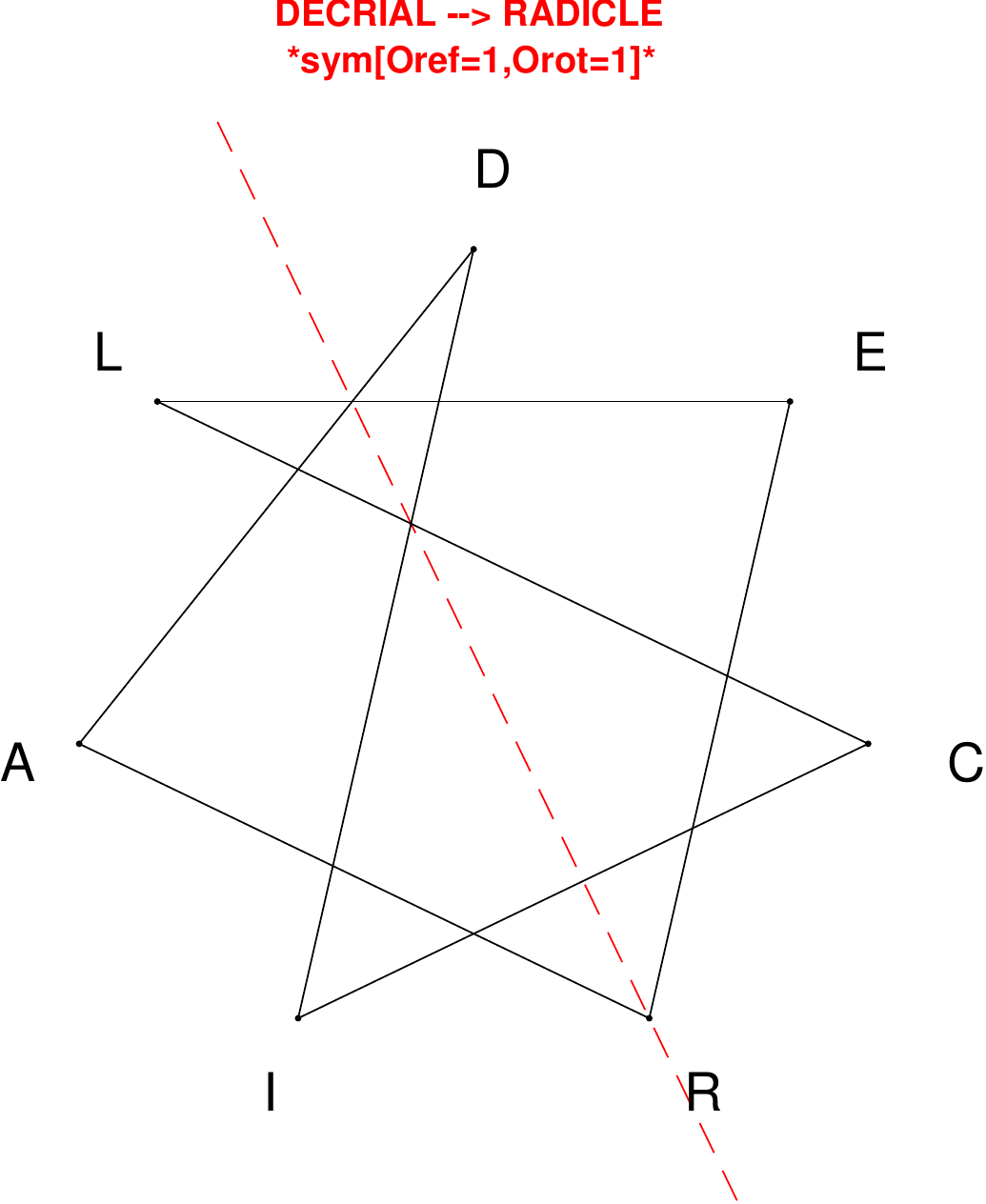}
\end{subfigure}
\hfill
\begin{subfigure}[T]{0.19\textwidth}
\centering
\includegraphics[width=\textwidth]{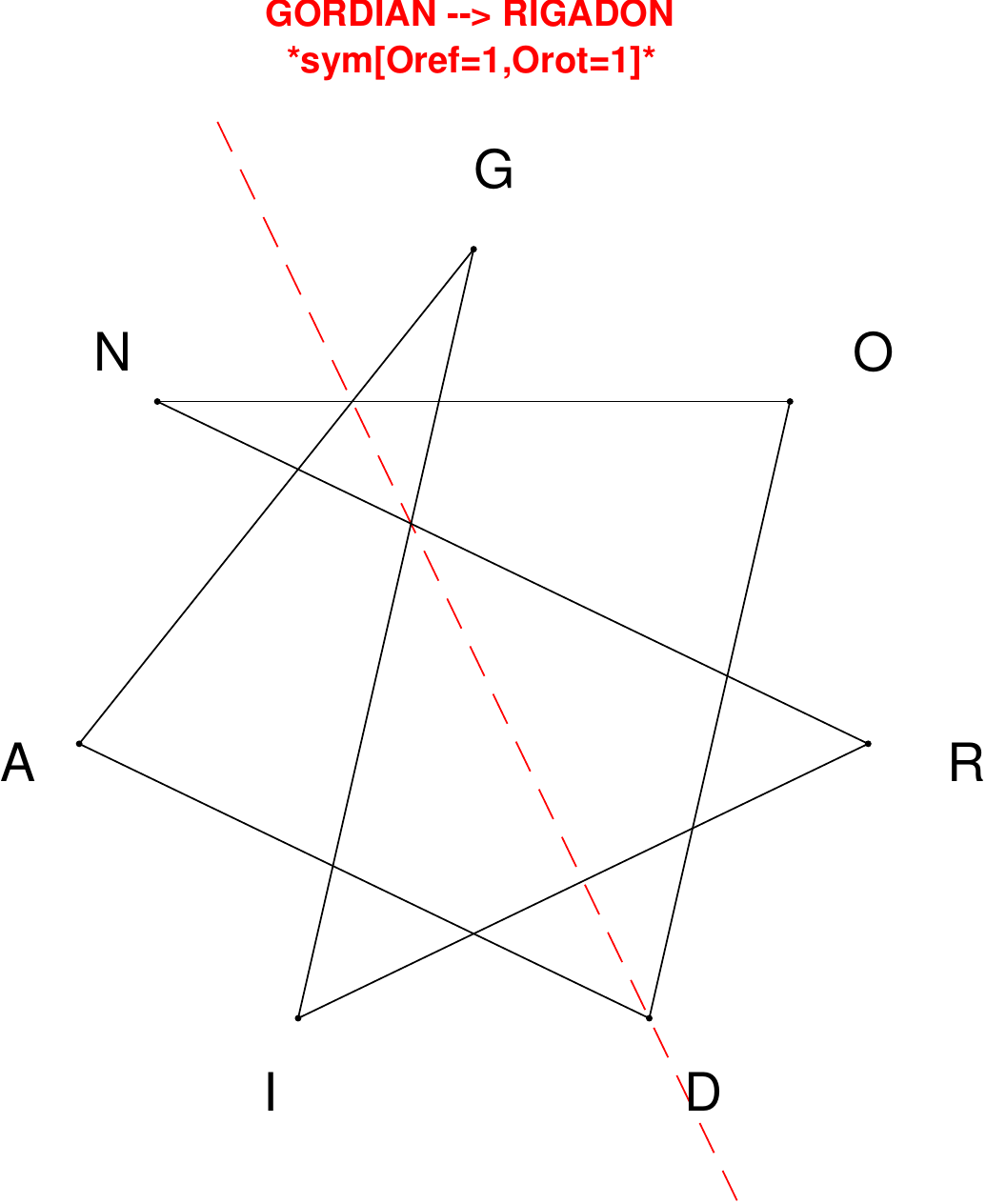}
\end{subfigure}
\hfill
\begin{subfigure}[T]{0.19\textwidth}
\centering
\includegraphics[width=\textwidth]{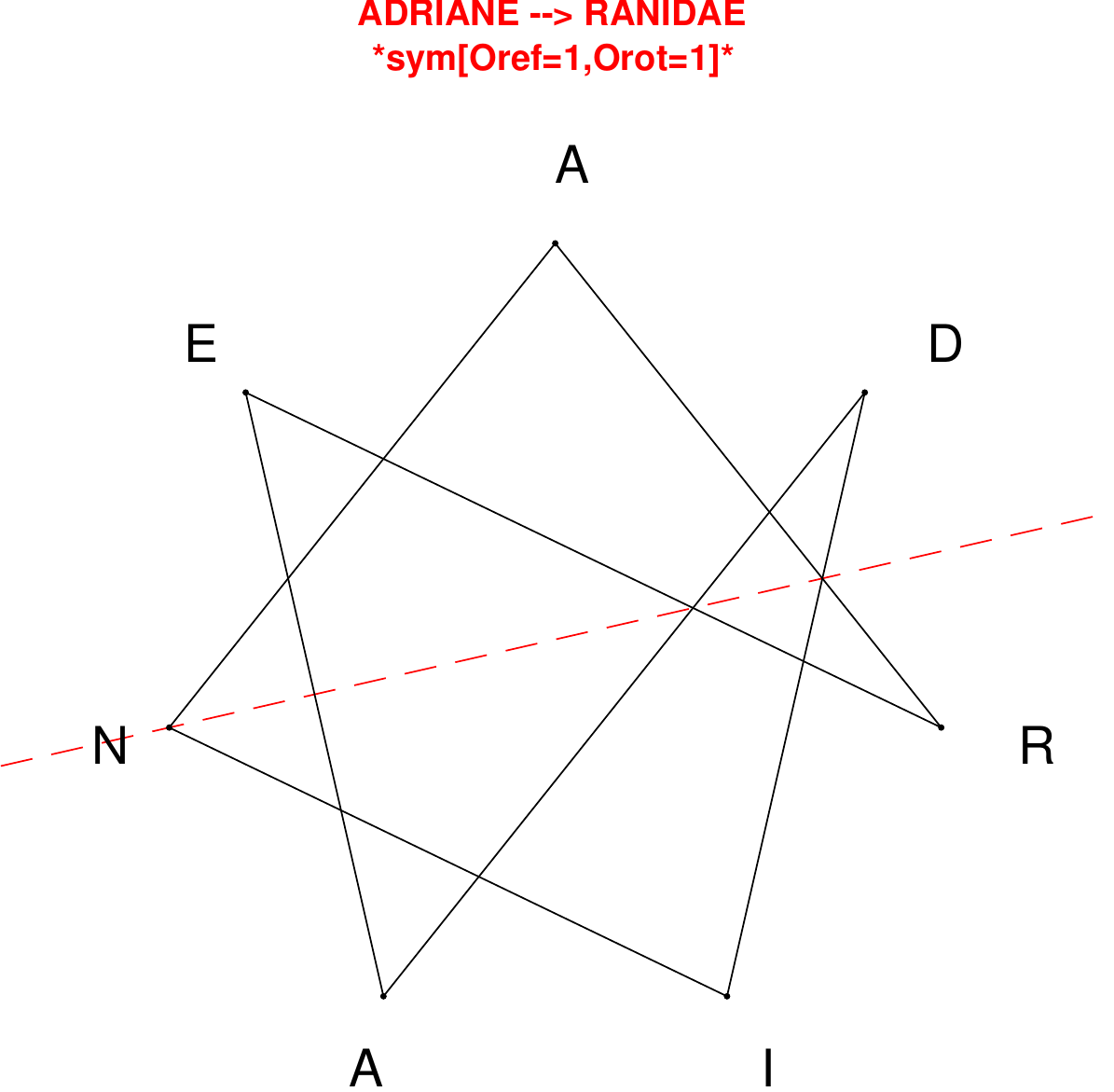}
\end{subfigure}
\end{figure}

\begin{figure}[H]
\centering
\begin{subfigure}[T]{0.19\textwidth}
\centering
\includegraphics[width=\textwidth]{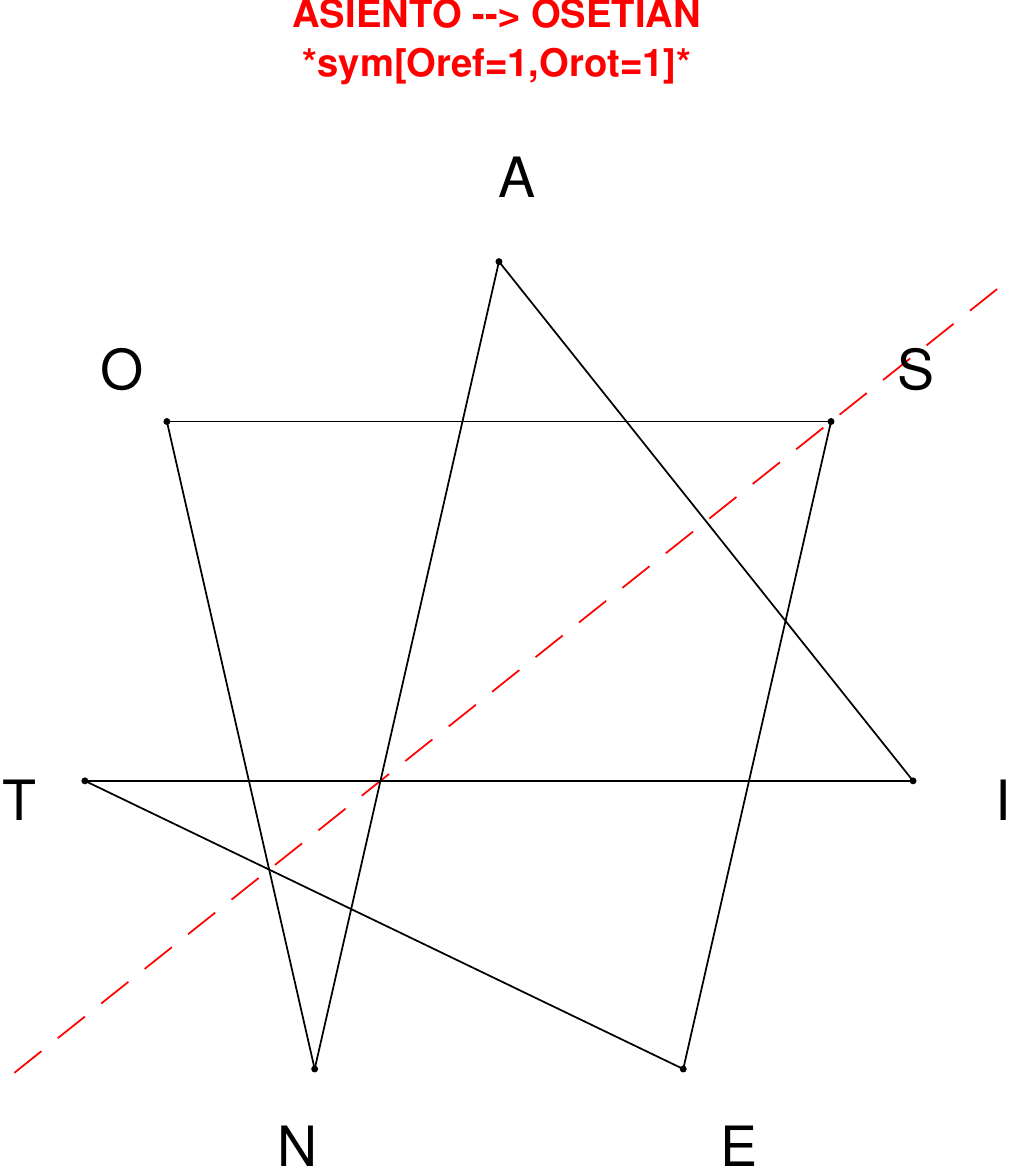}
\end{subfigure}
\hfill
\begin{subfigure}[T]{0.19\textwidth}
\centering
\includegraphics[width=\textwidth]{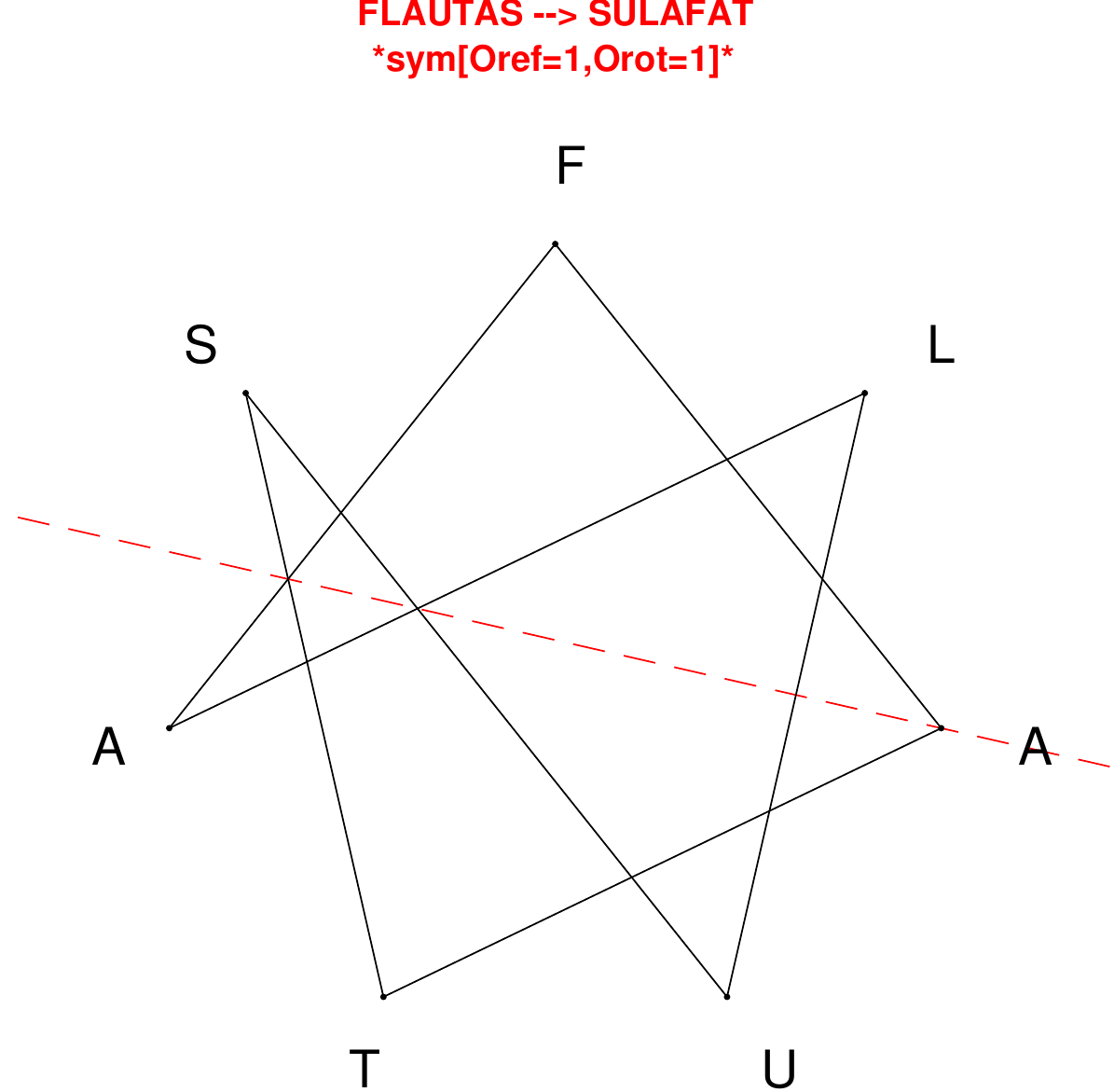}
\end{subfigure}
\hfill
\begin{subfigure}[T]{0.19\textwidth}
\centering
\includegraphics[width=\textwidth]{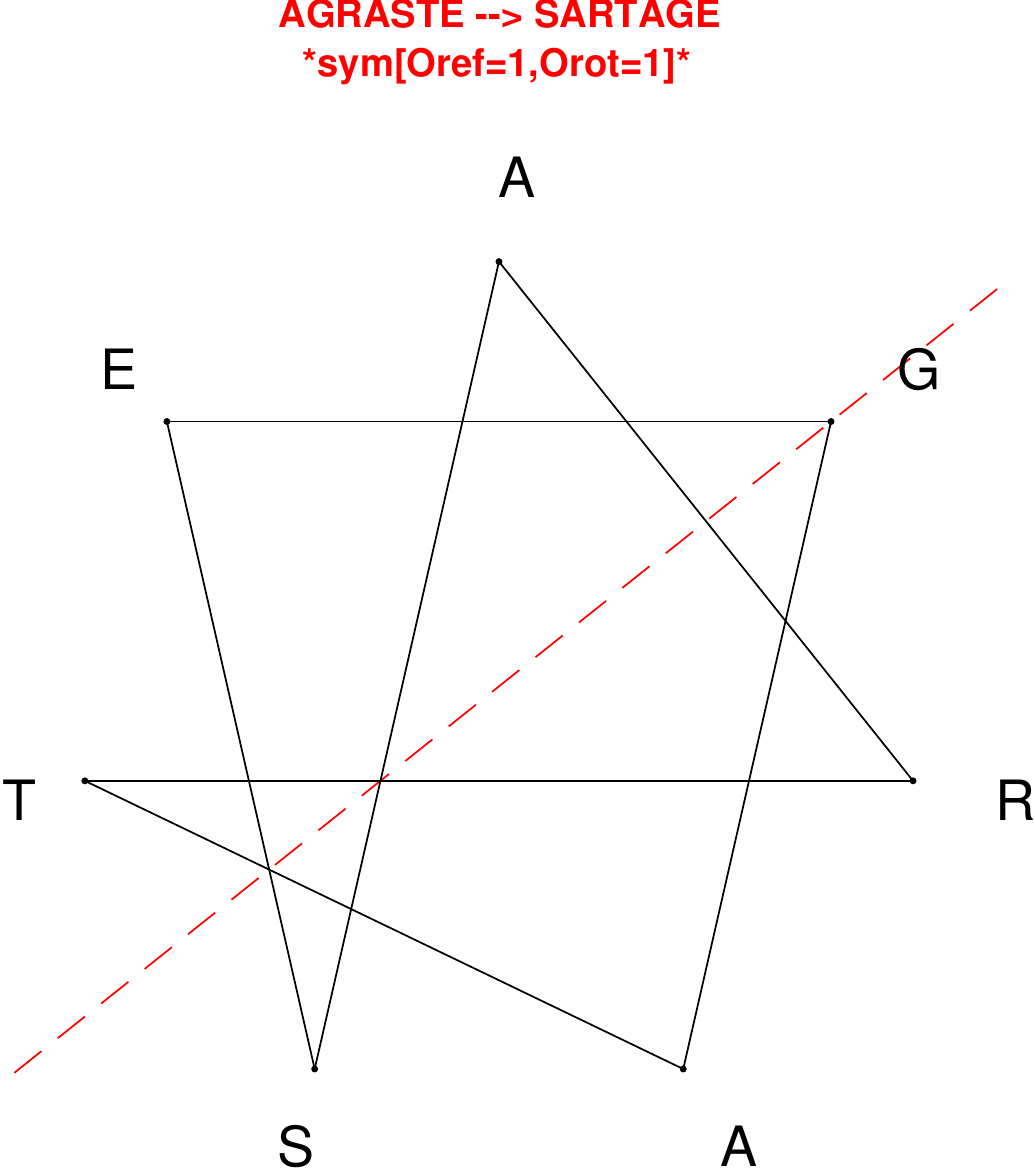}
\end{subfigure}
\hfill
\begin{subfigure}[T]{0.19\textwidth}
\centering
\includegraphics[width=\textwidth]{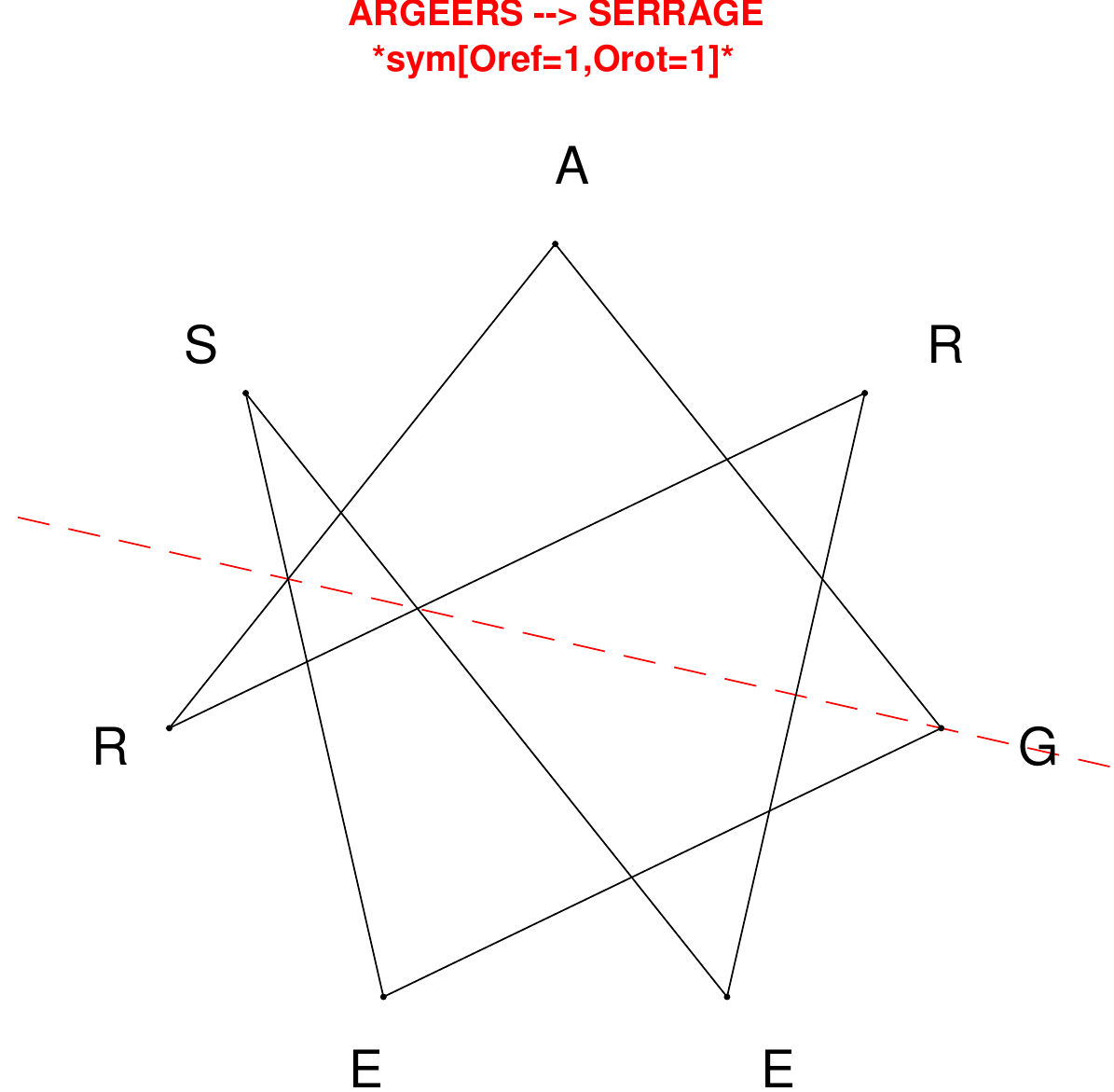}
\end{subfigure}
\hfill
\begin{subfigure}[T]{0.19\textwidth}
\centering
\includegraphics[width=\textwidth]{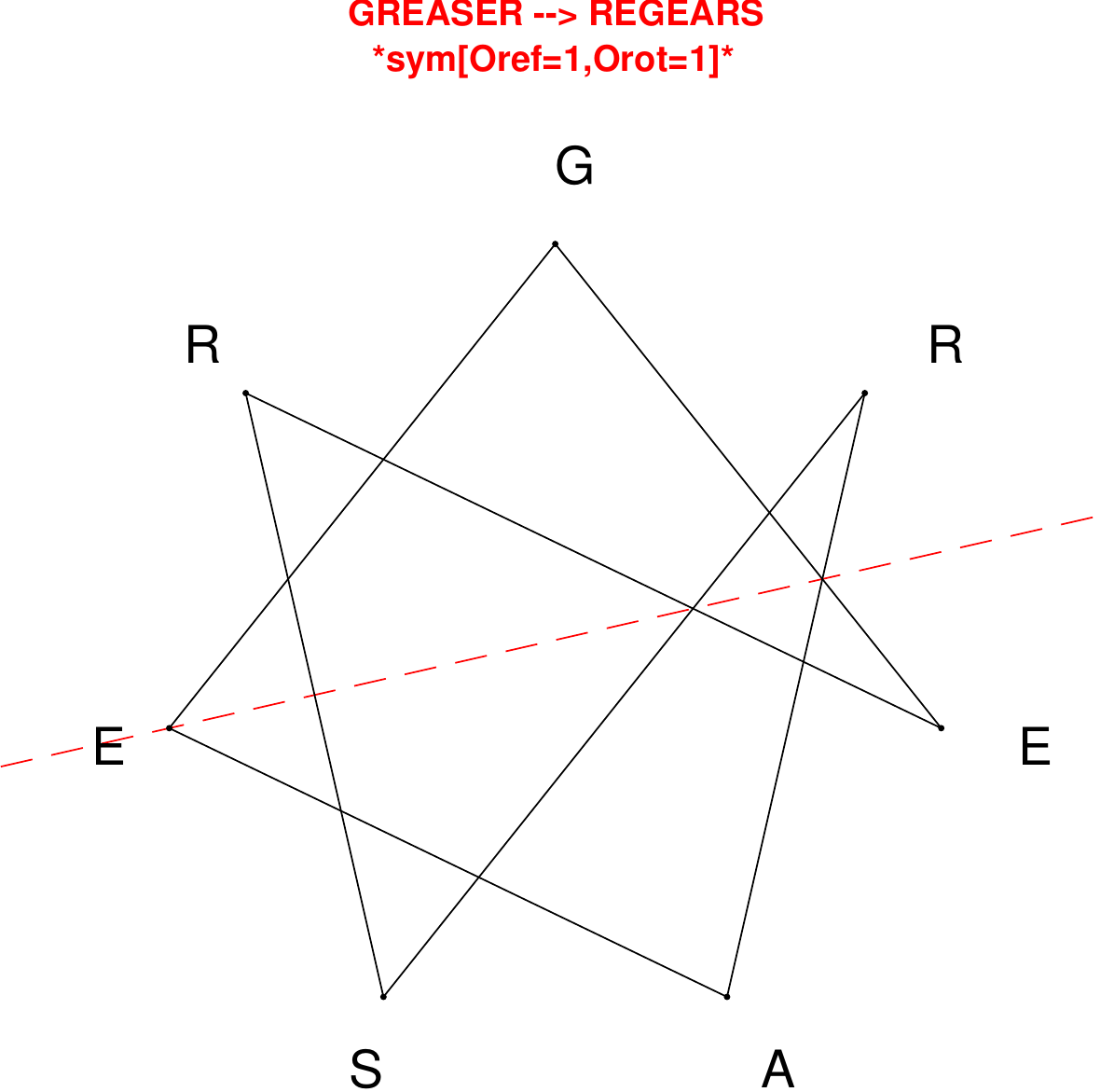}
\end{subfigure}
\end{figure}

\begin{figure}[H]
\centering
\begin{subfigure}[T]{0.19\textwidth}
\centering
\includegraphics[width=\textwidth]{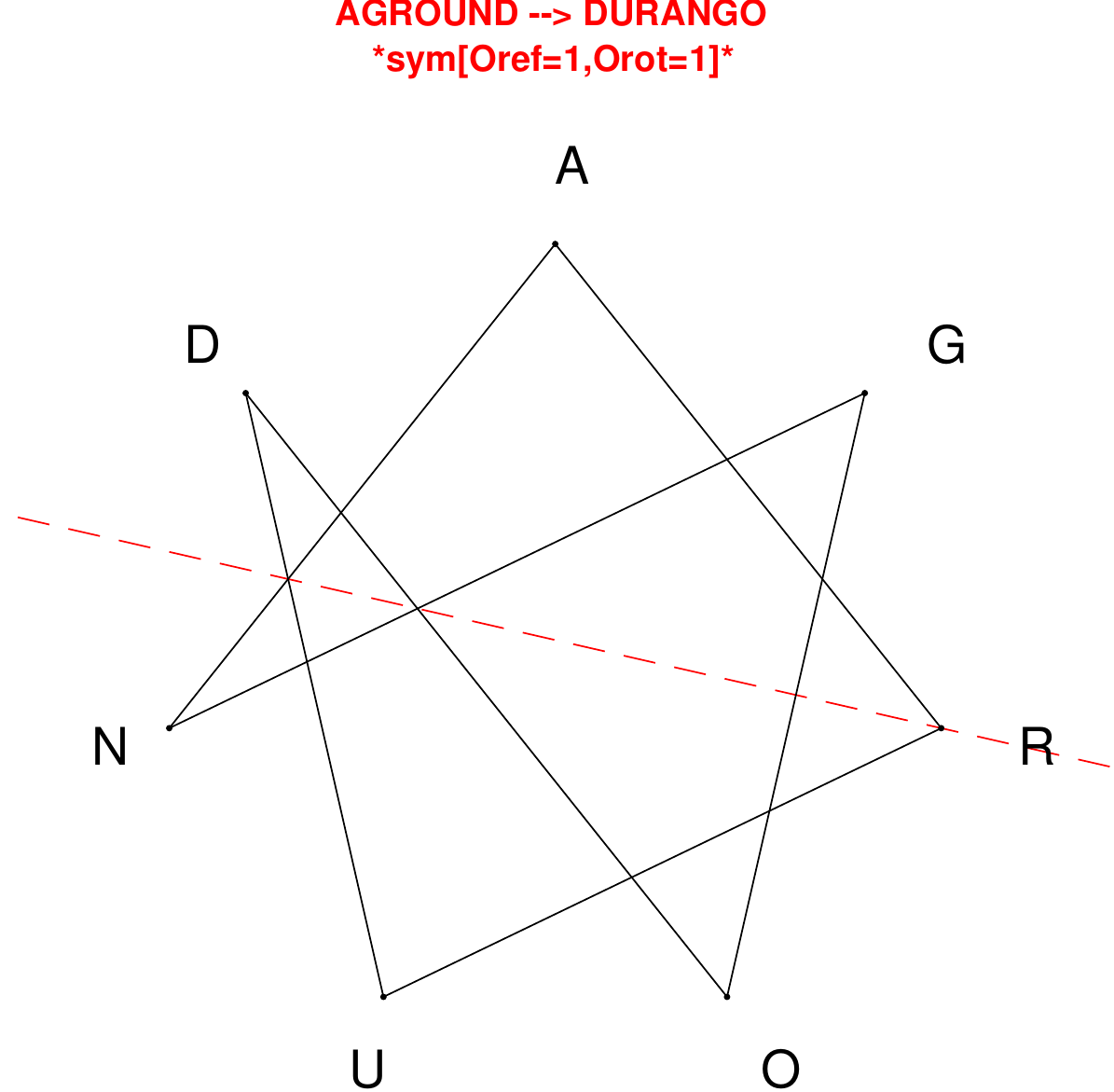}
\end{subfigure}
\hfill
\begin{subfigure}[T]{0.19\textwidth}
\centering
\includegraphics[width=\textwidth]{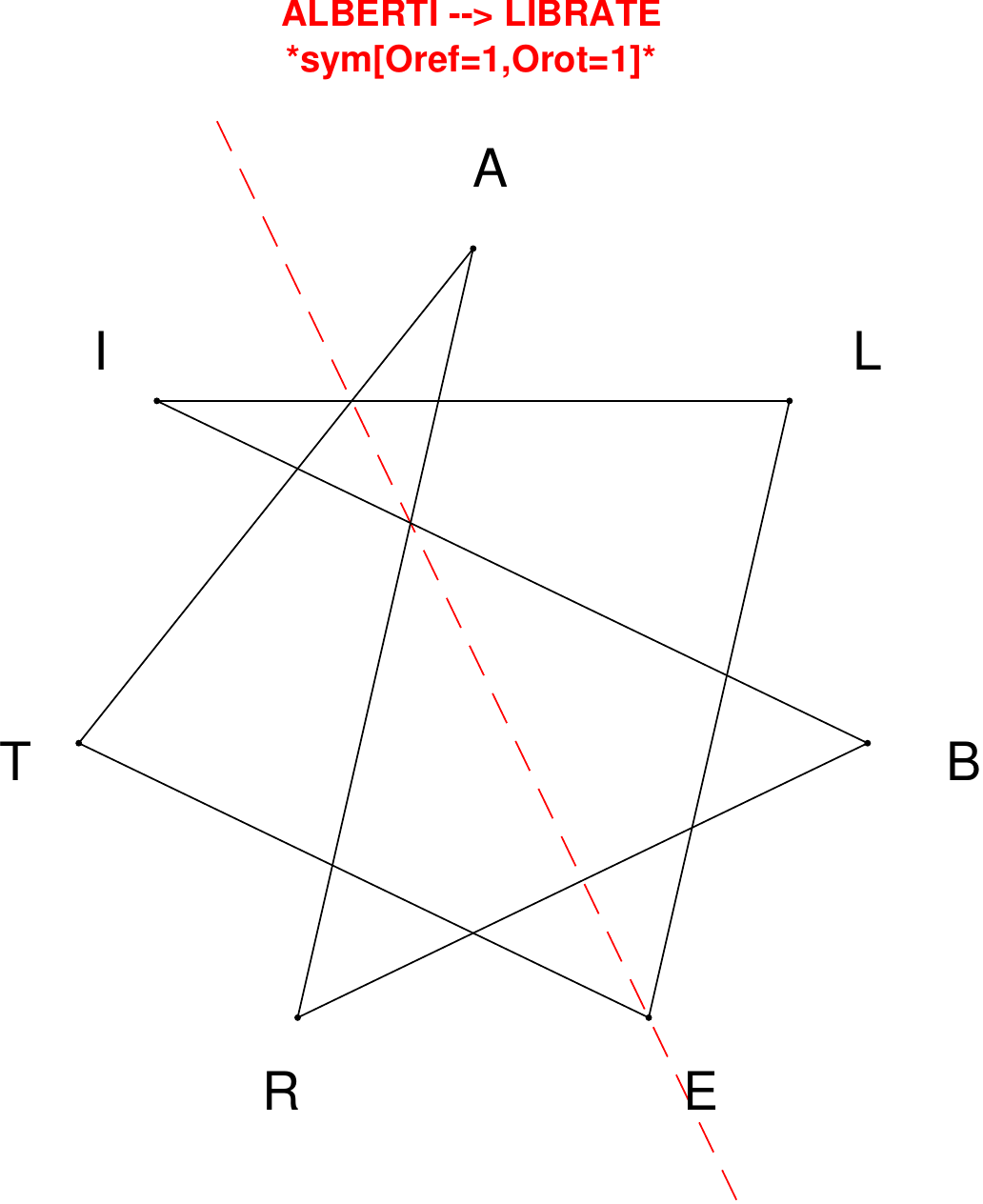}
\end{subfigure}
\hfill
\begin{subfigure}[T]{0.19\textwidth}
\centering
\includegraphics[width=\textwidth]{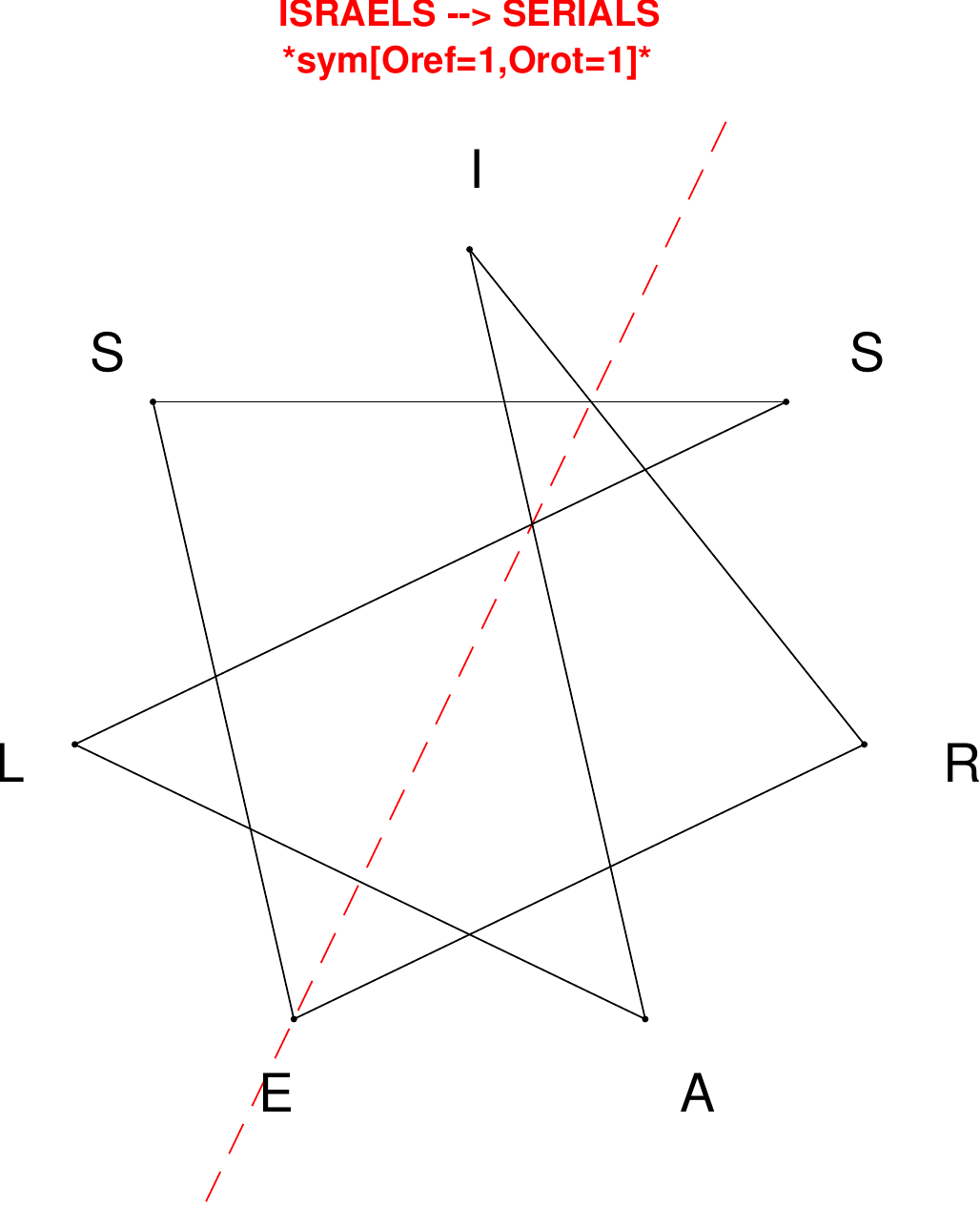}
\end{subfigure}
\hfill
\begin{subfigure}[T]{0.19\textwidth}
\centering
\includegraphics[width=\textwidth]{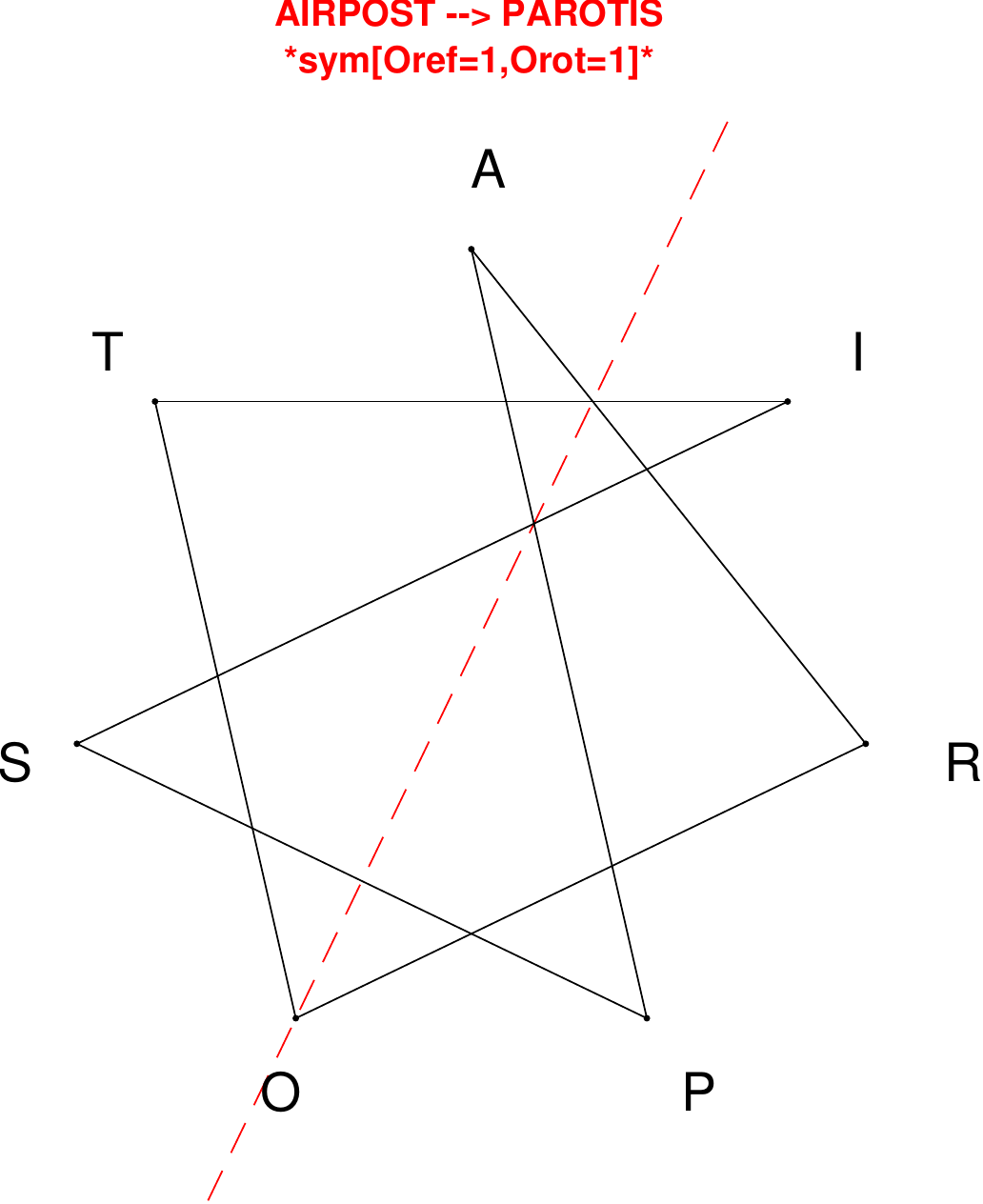}
\end{subfigure}
\hfill
\begin{subfigure}[T]{0.19\textwidth}
\centering
\includegraphics[width=\textwidth]{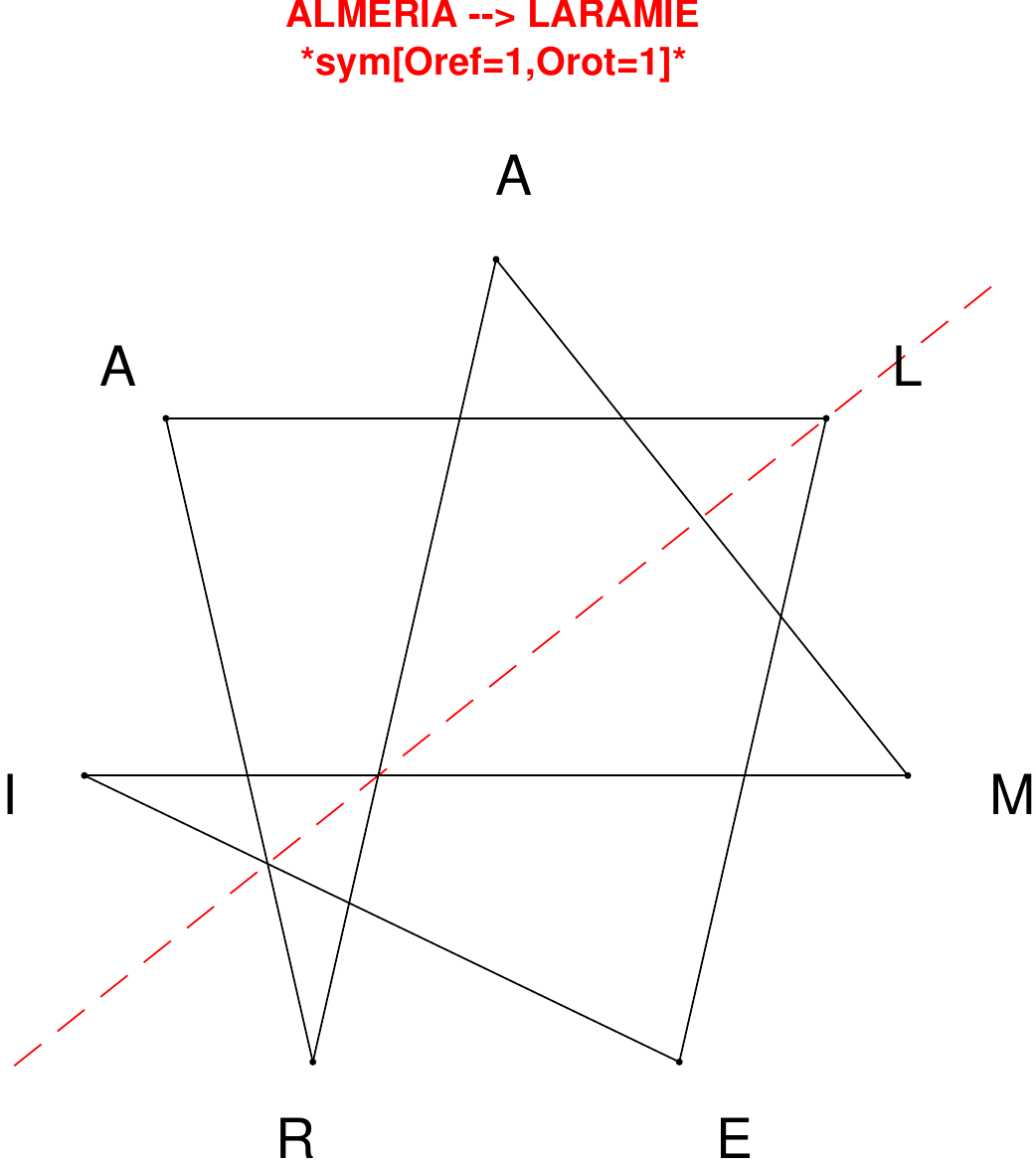}
\end{subfigure}
\end{figure}

\begin{figure}[H]
\centering
\begin{subfigure}[T]{0.19\textwidth}
\centering
\includegraphics[width=\textwidth]{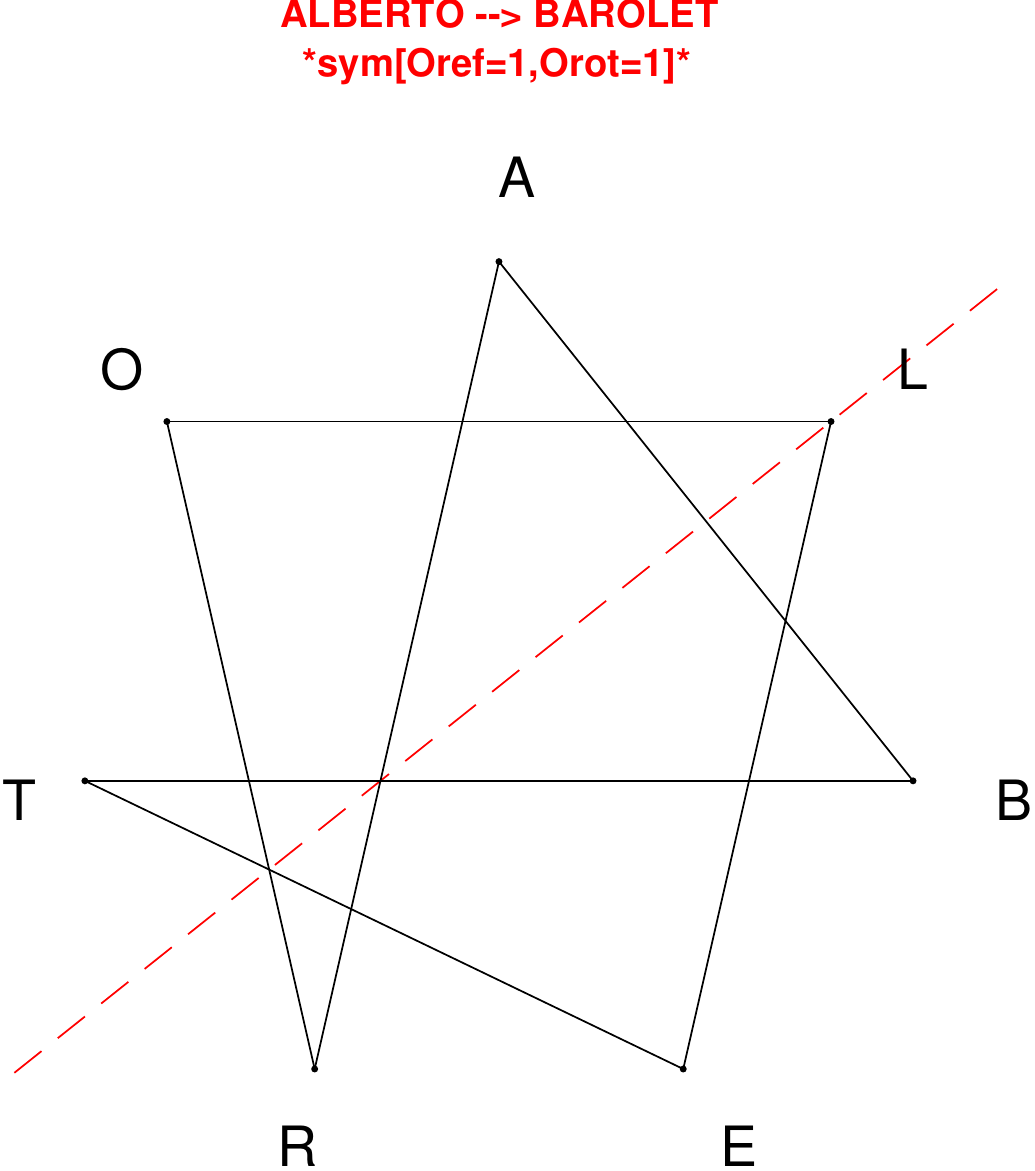}
\end{subfigure}
\hfill
\begin{subfigure}[T]{0.19\textwidth}
\centering
\includegraphics[width=\textwidth]{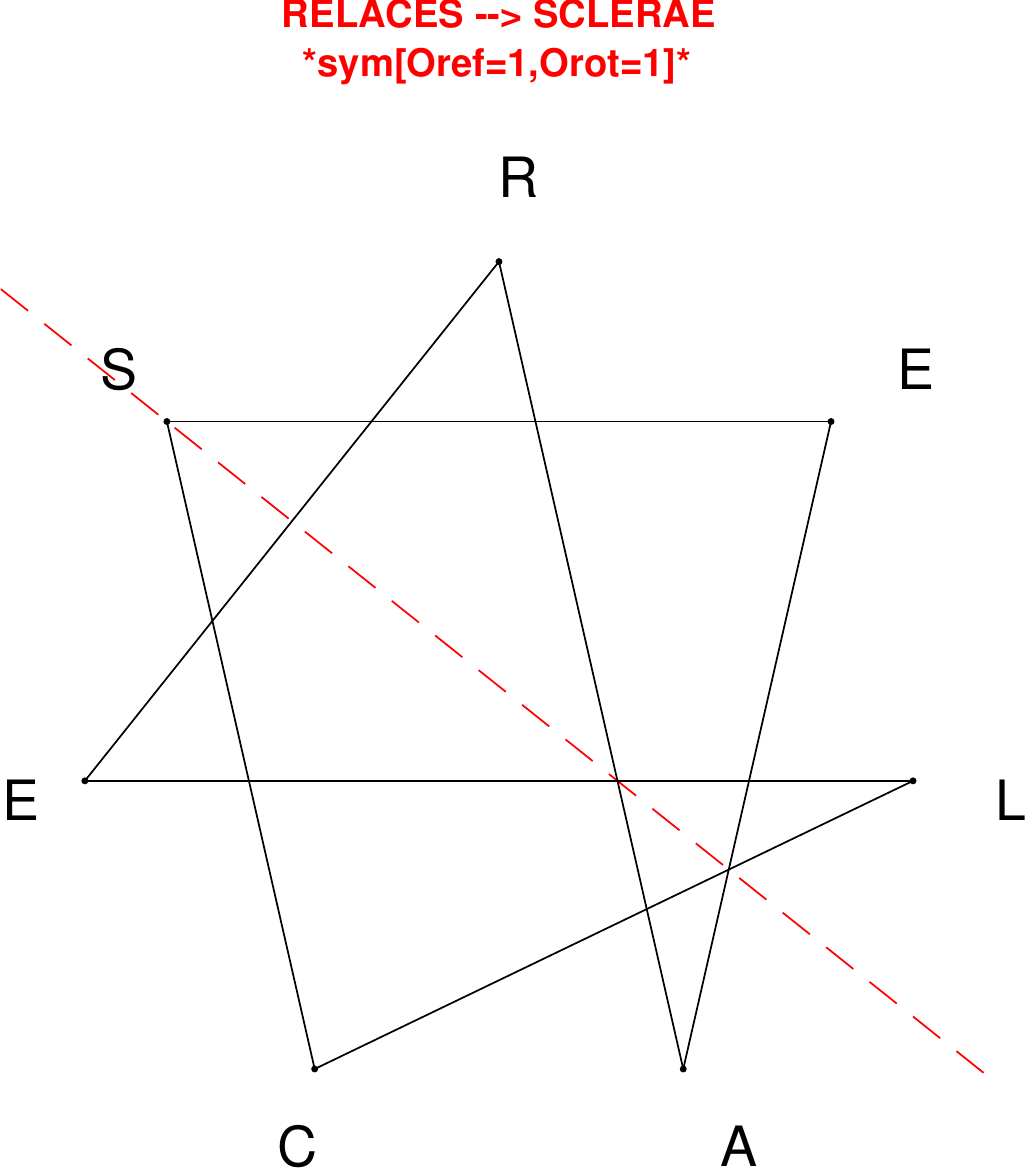}
\end{subfigure}
\hfill
\begin{subfigure}[T]{0.19\textwidth}
\centering
\includegraphics[width=\textwidth]{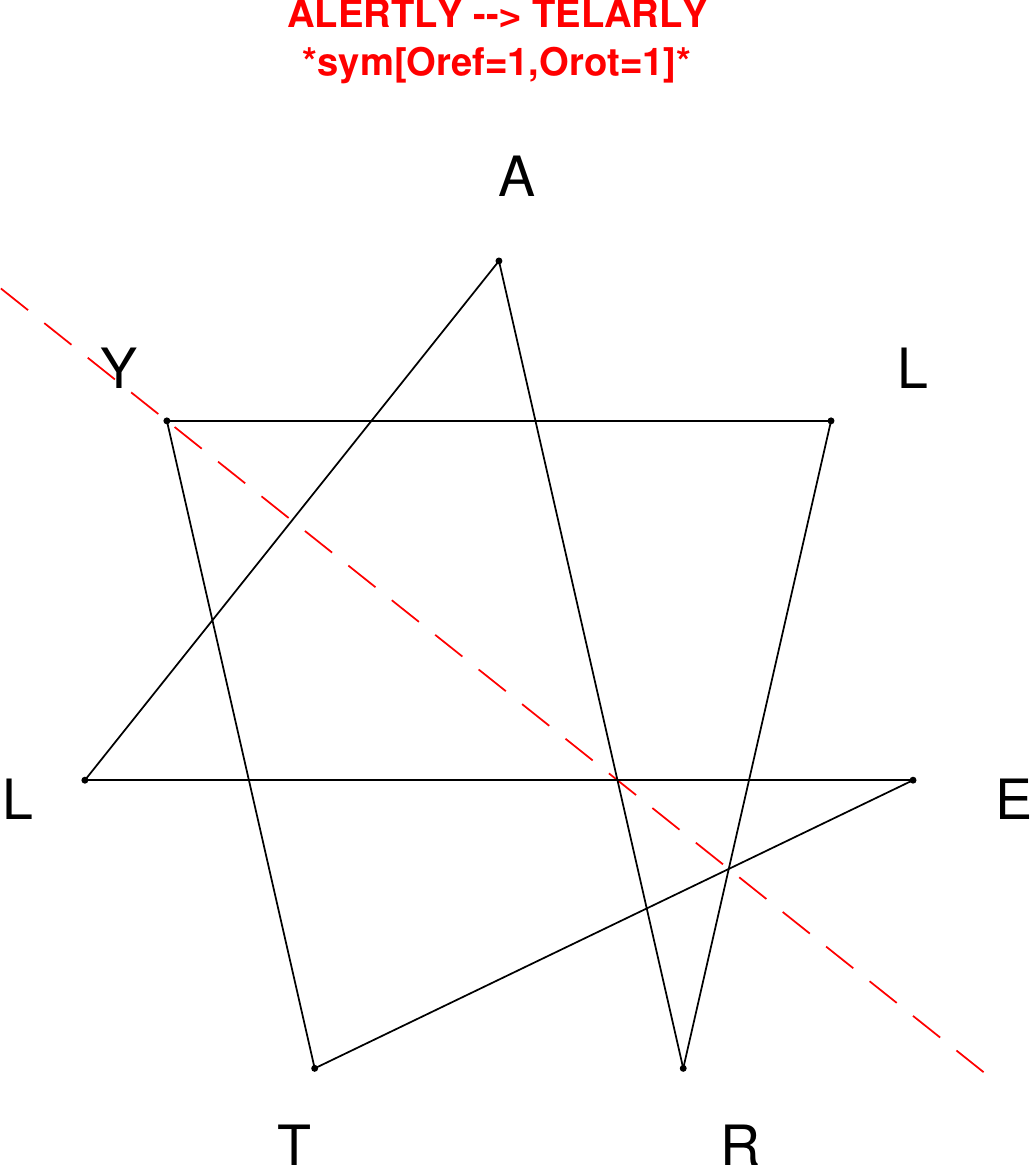}
\end{subfigure}
\hfill
\begin{subfigure}[T]{0.19\textwidth}
\centering
\includegraphics[width=\textwidth]{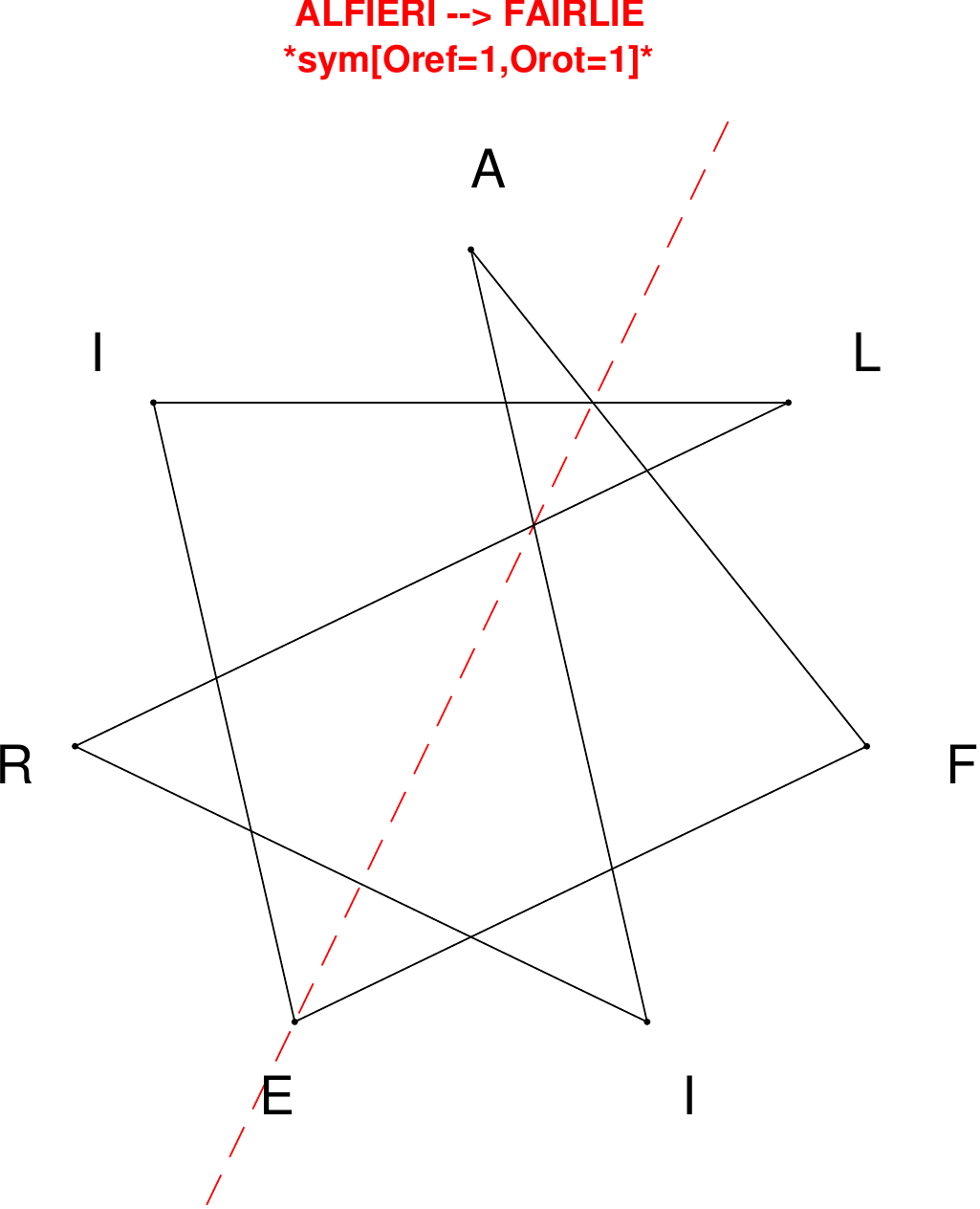}
\end{subfigure}
\hfill
\begin{subfigure}[T]{0.19\textwidth}
\centering
\includegraphics[width=\textwidth]{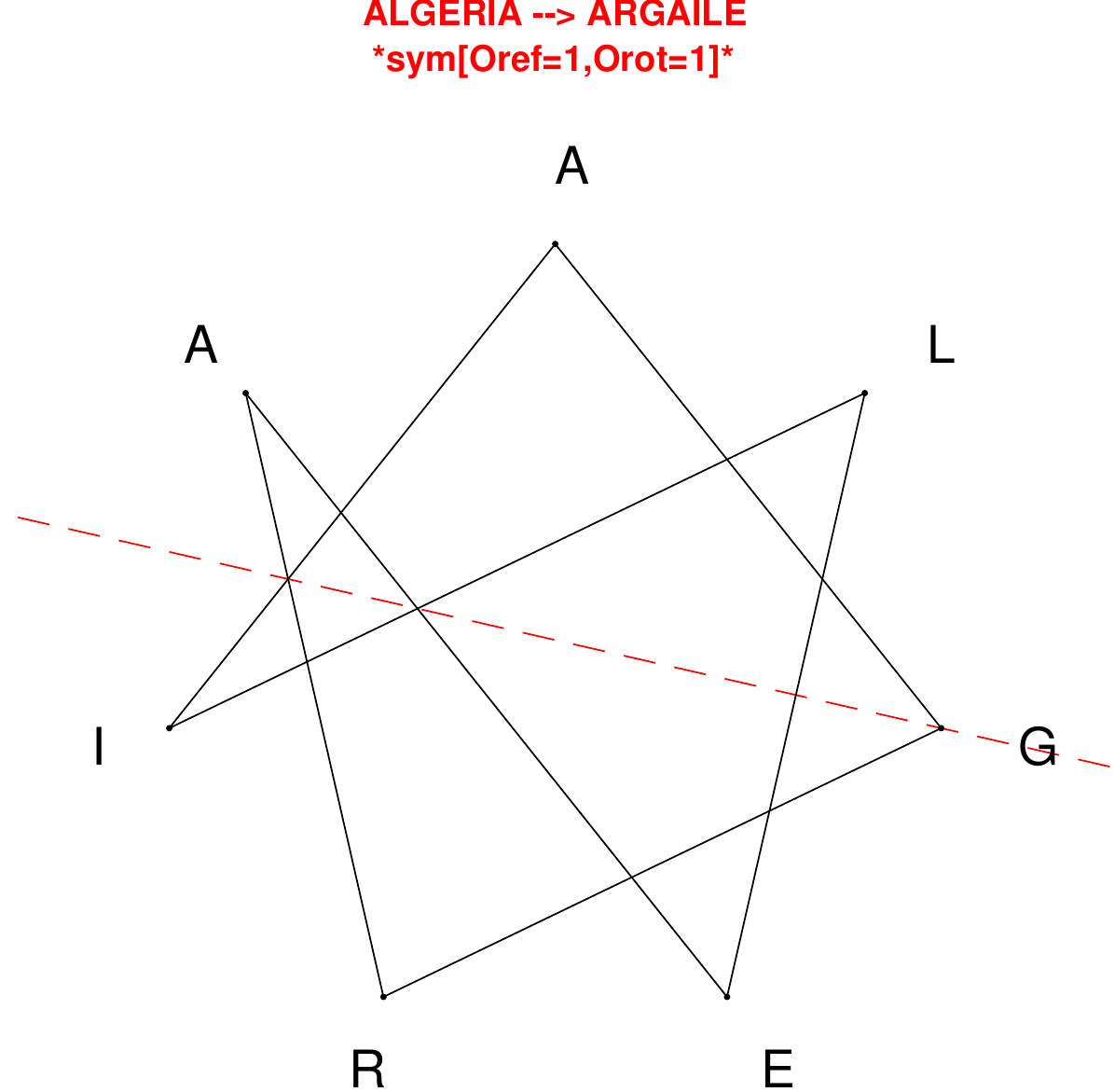}
\end{subfigure}
\end{figure}

\begin{figure}[H]
\centering
\begin{subfigure}[T]{0.19\textwidth}
\centering
\includegraphics[width=\textwidth]{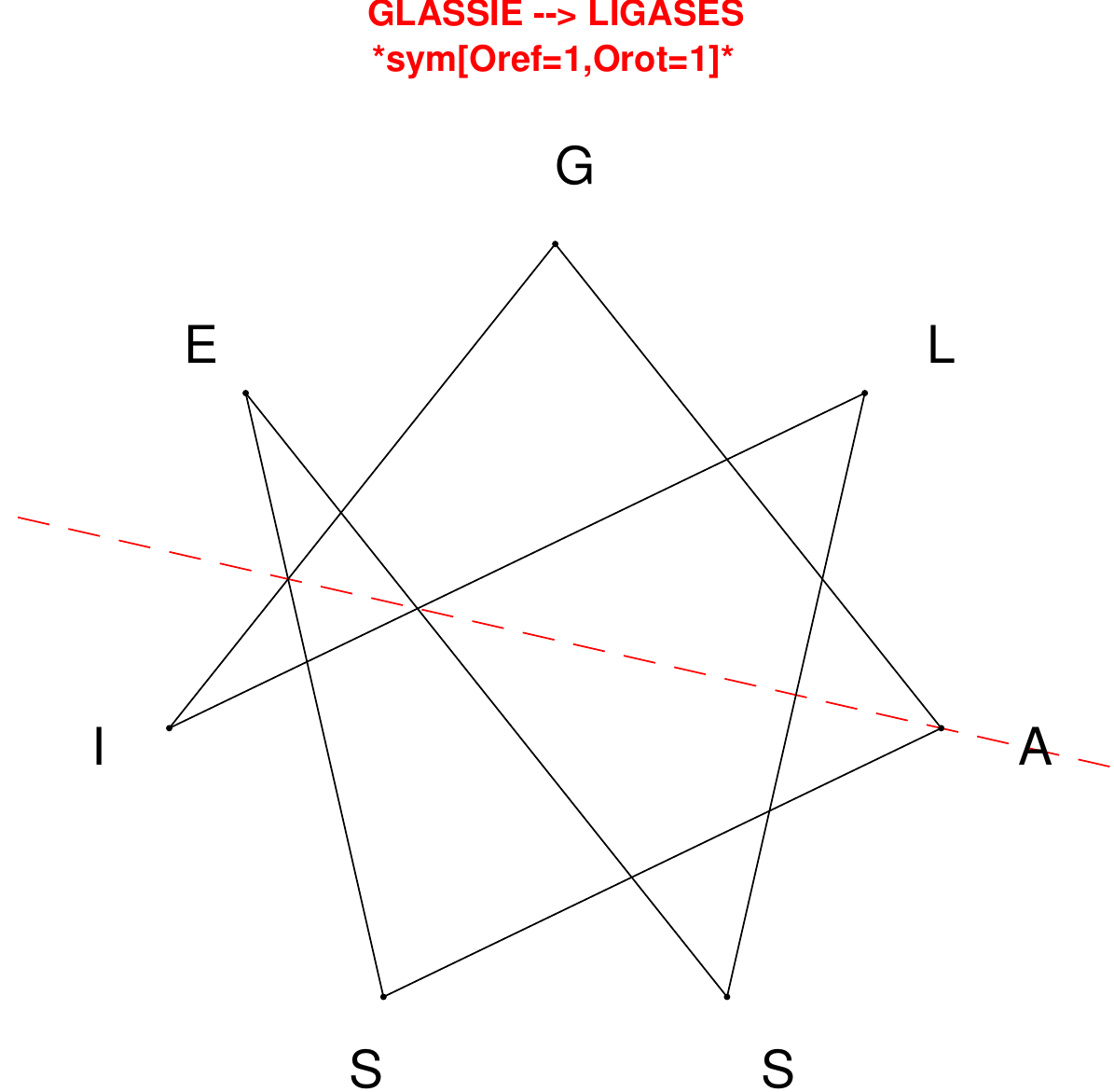}
\end{subfigure}
\hfill
\begin{subfigure}[T]{0.19\textwidth}
\centering
\includegraphics[width=\textwidth]{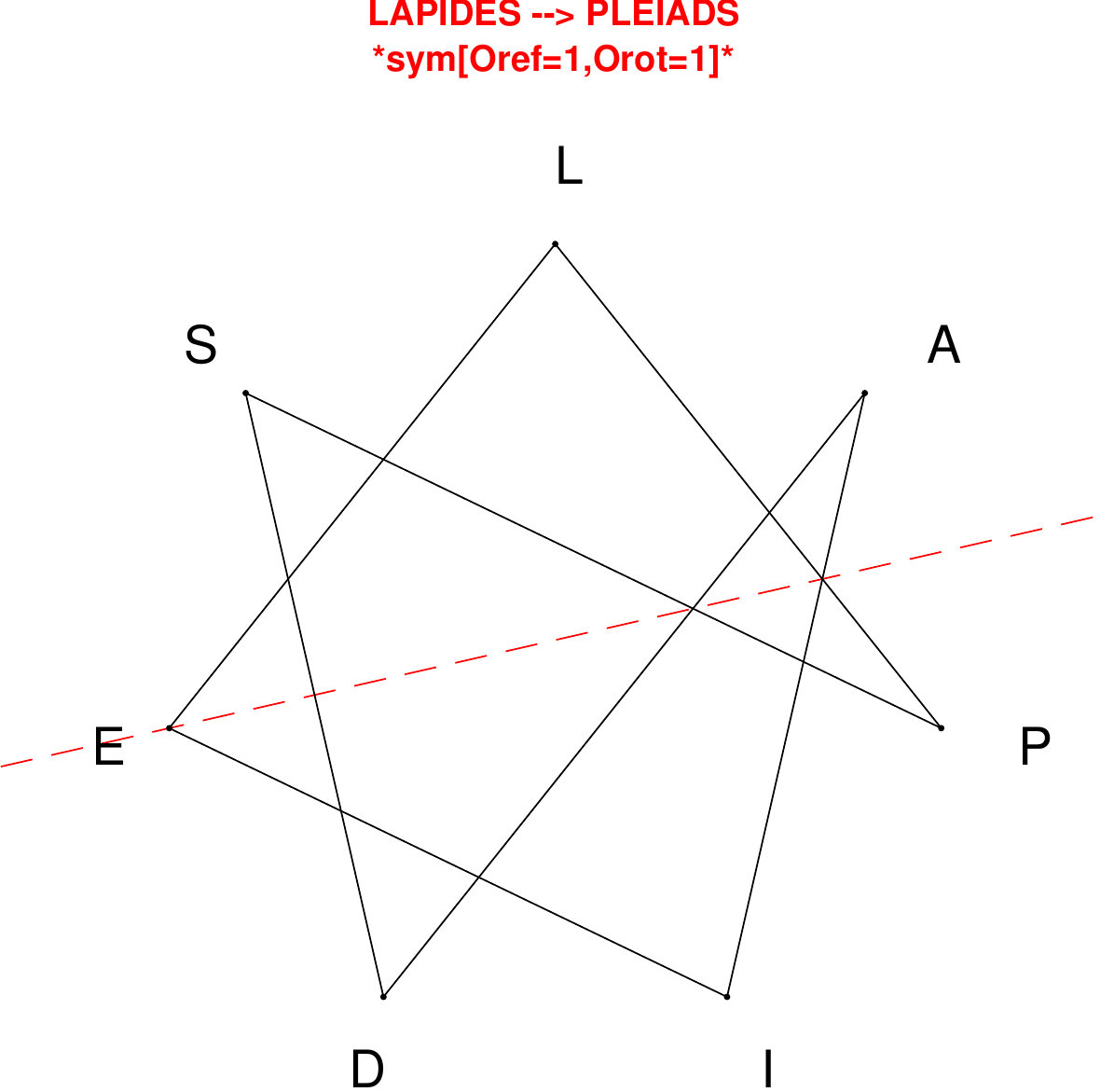}
\end{subfigure}
\hfill
\begin{subfigure}[T]{0.19\textwidth}
\centering
\includegraphics[width=\textwidth]{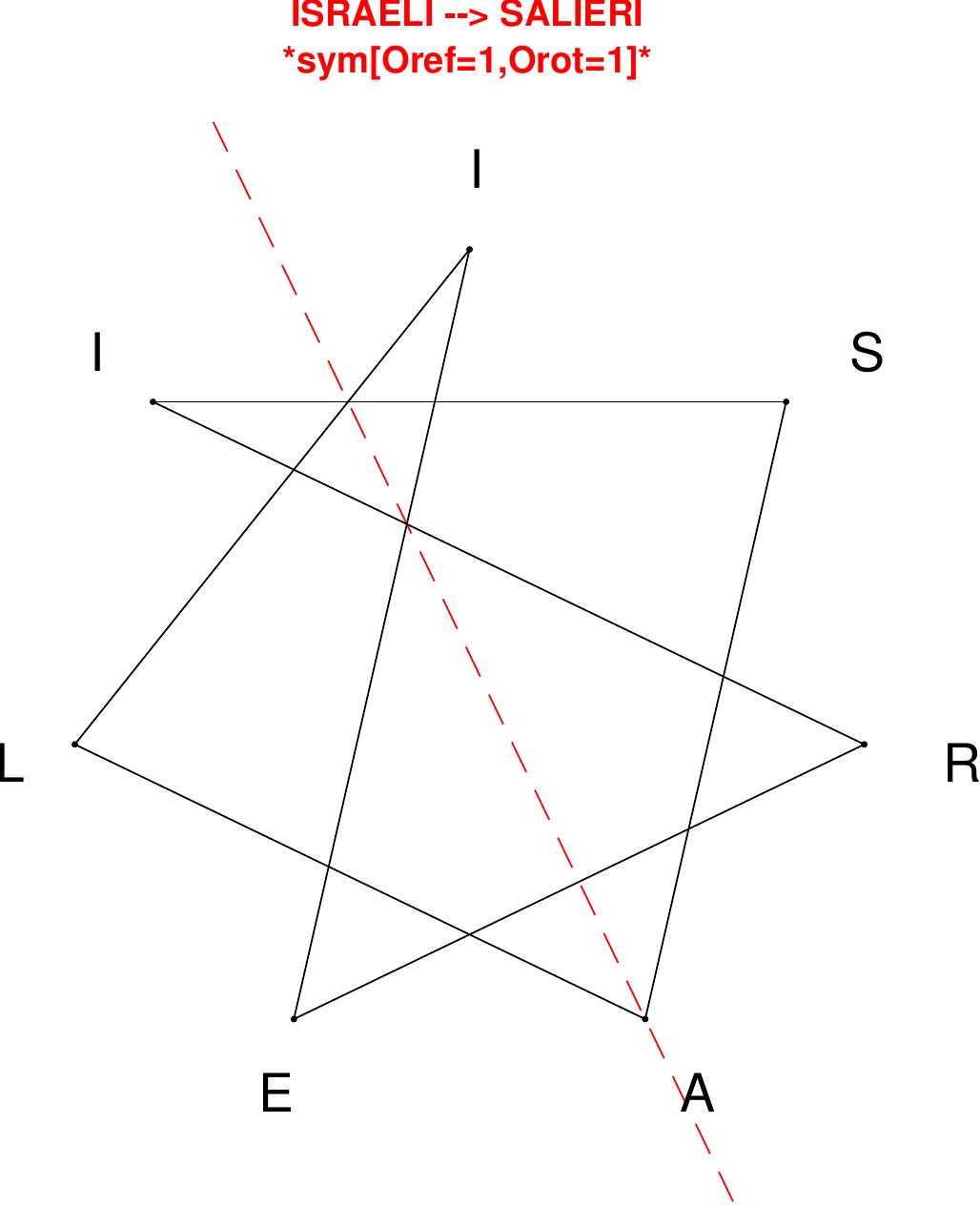}
\end{subfigure}
\hfill
\begin{subfigure}[T]{0.19\textwidth}
\centering
\includegraphics[width=\textwidth]{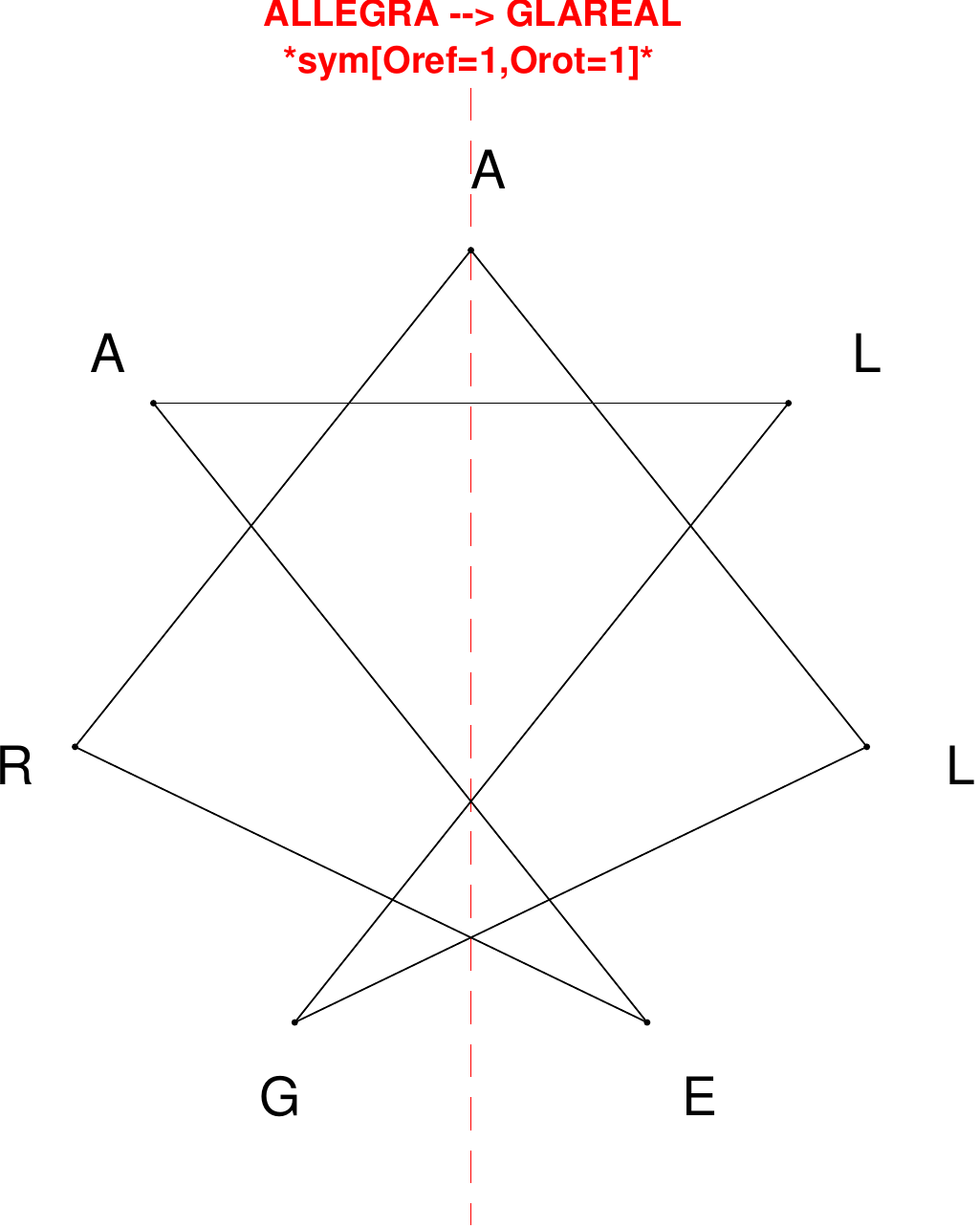}
\end{subfigure}
\hfill
\begin{subfigure}[T]{0.19\textwidth}
\centering
\includegraphics[width=\textwidth]{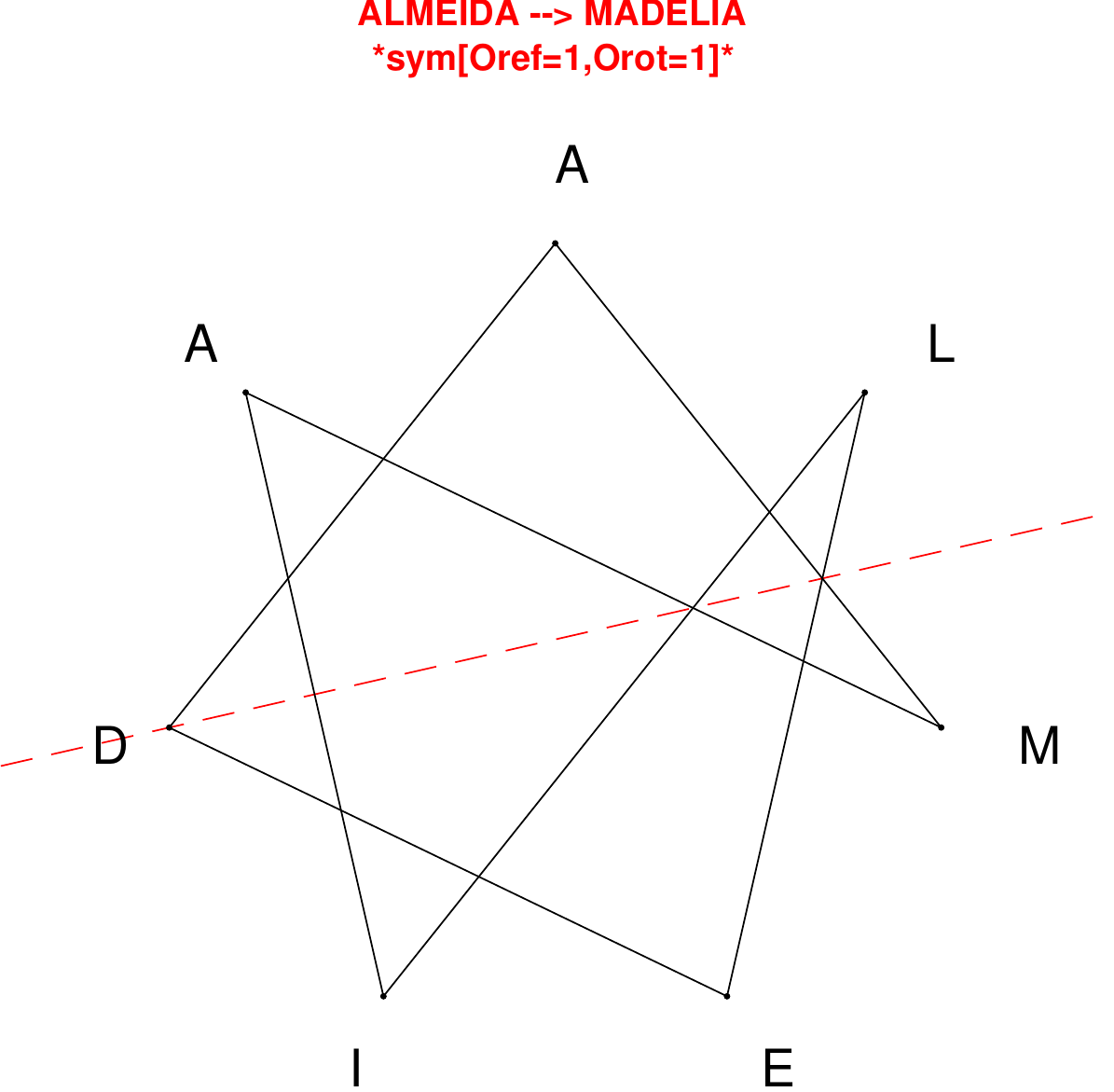}
\end{subfigure}
\end{figure}

\begin{figure}[H]
\centering
\begin{subfigure}[T]{0.19\textwidth}
\centering
\includegraphics[width=\textwidth]{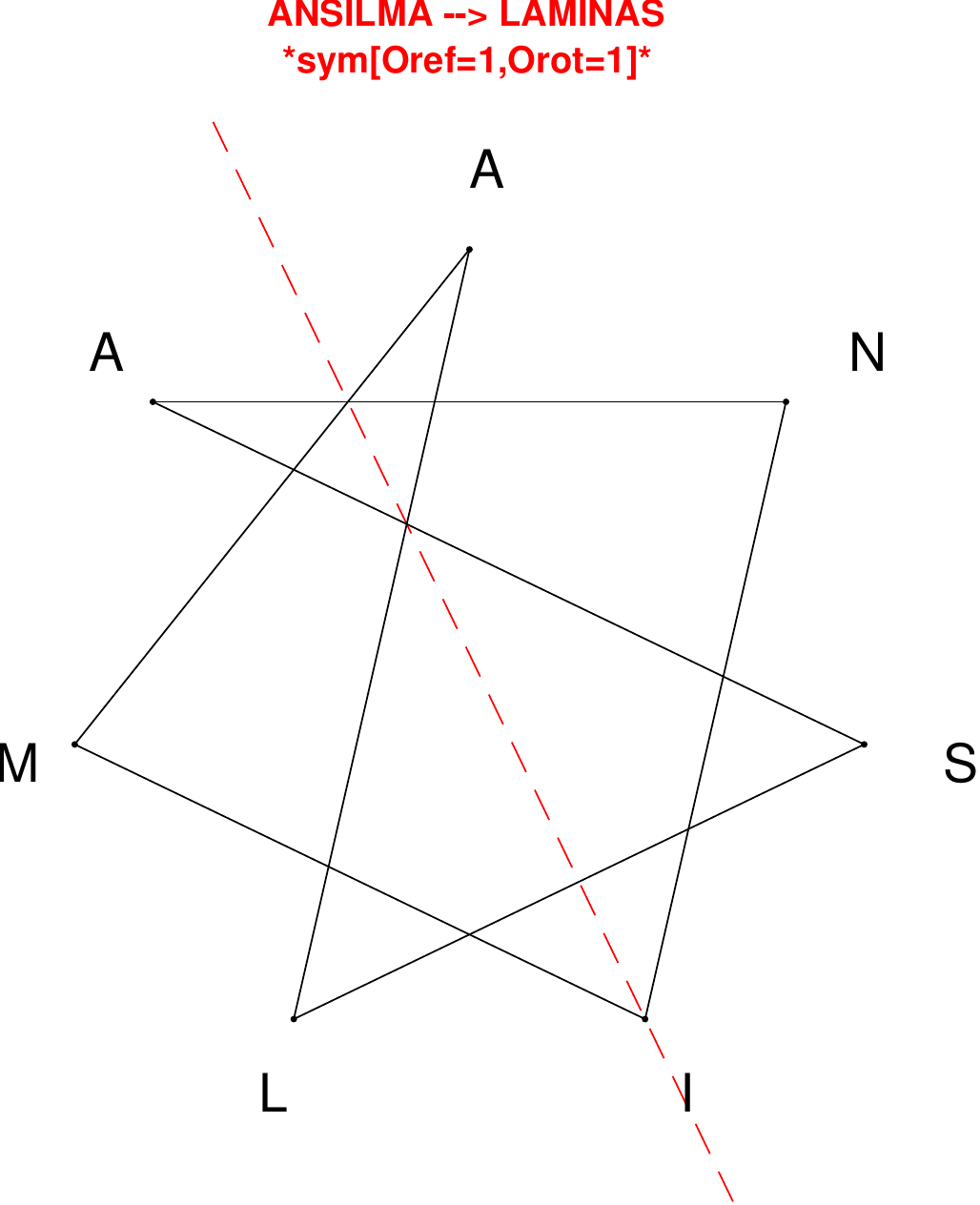}
\end{subfigure}
\hfill
\begin{subfigure}[T]{0.19\textwidth}
\centering
\includegraphics[width=\textwidth]{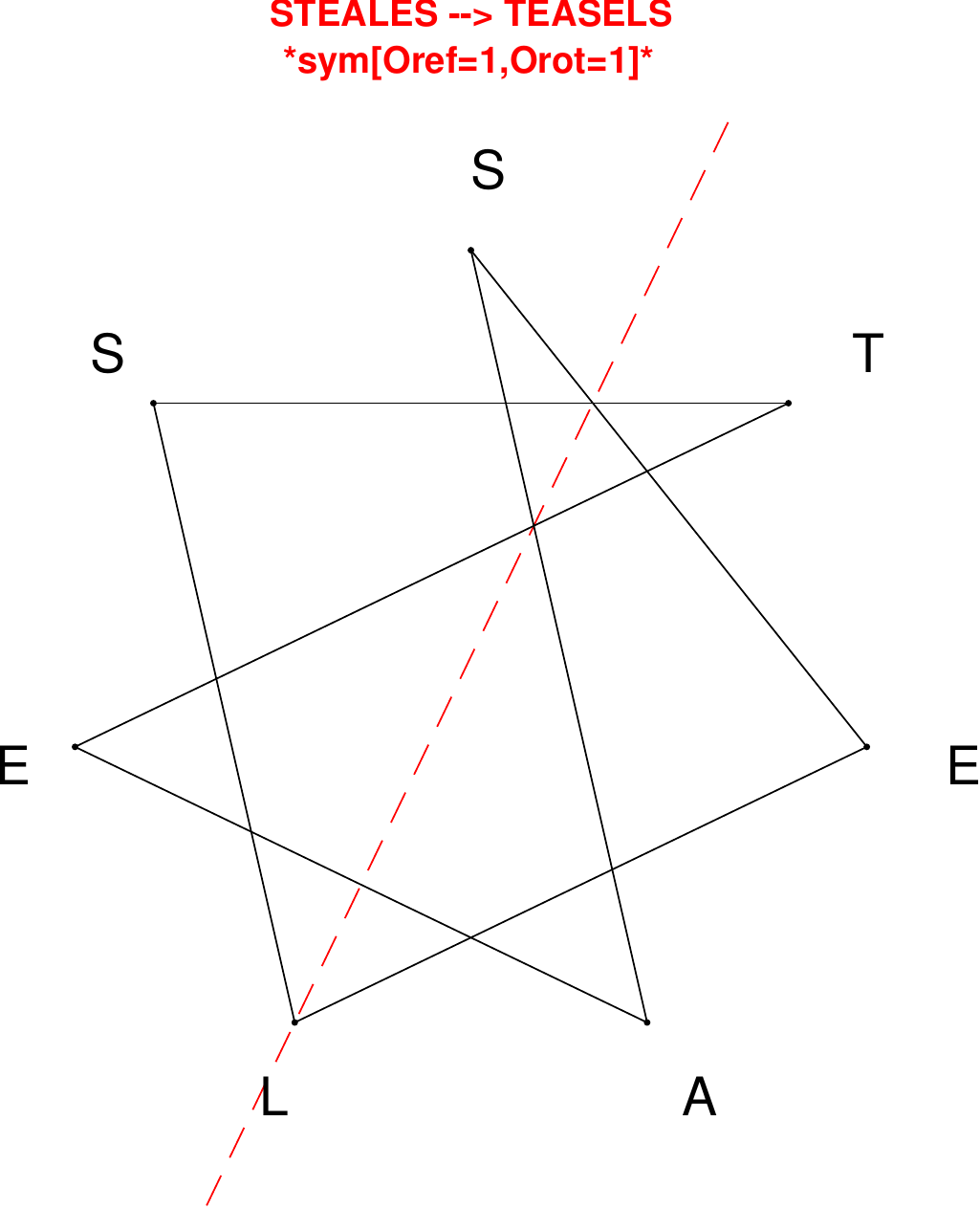}
\end{subfigure}
\hfill
\begin{subfigure}[T]{0.19\textwidth}
\centering
\includegraphics[width=\textwidth]{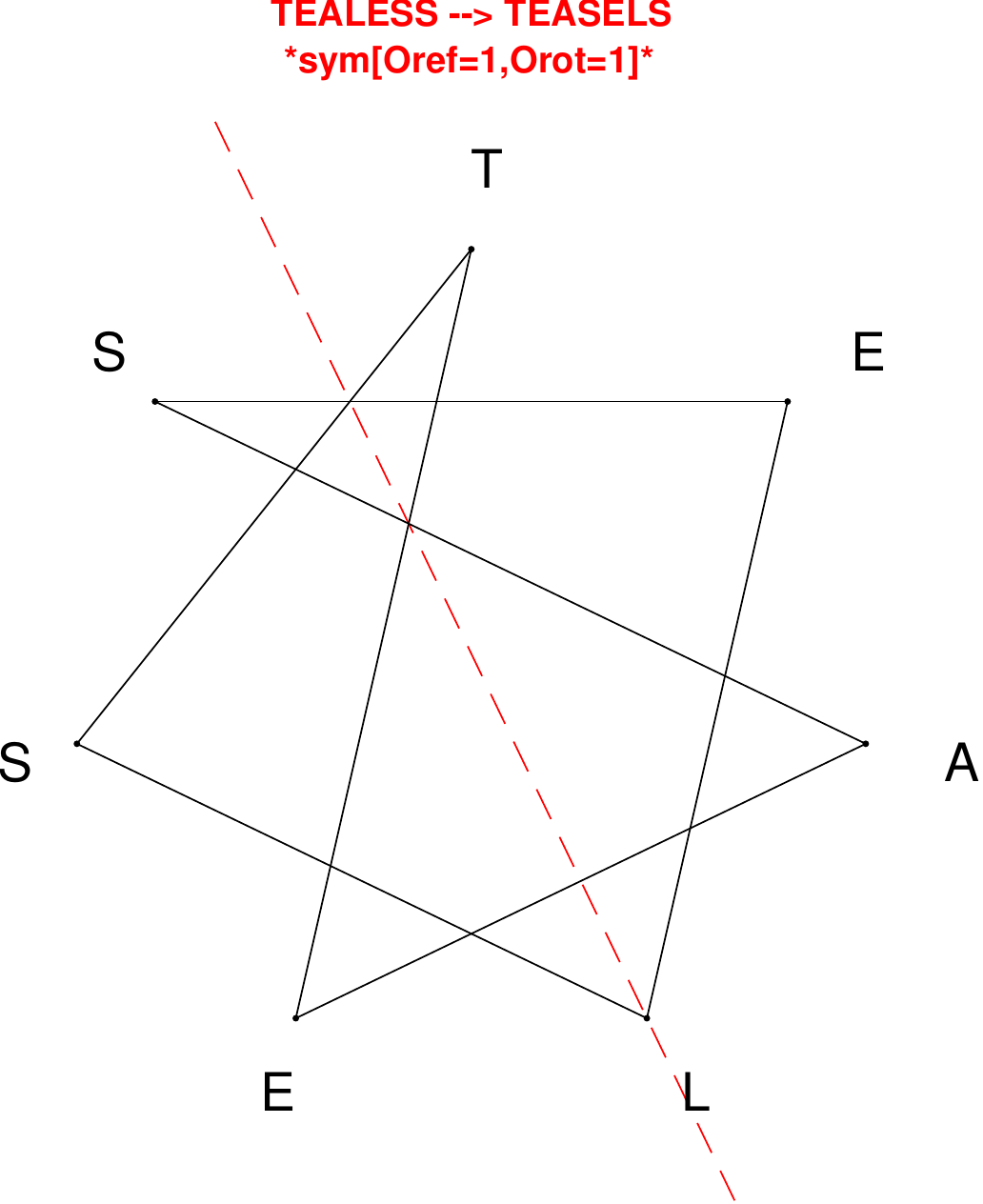}
\end{subfigure}
\hfill
\begin{subfigure}[T]{0.19\textwidth}
\centering
\includegraphics[width=\textwidth]{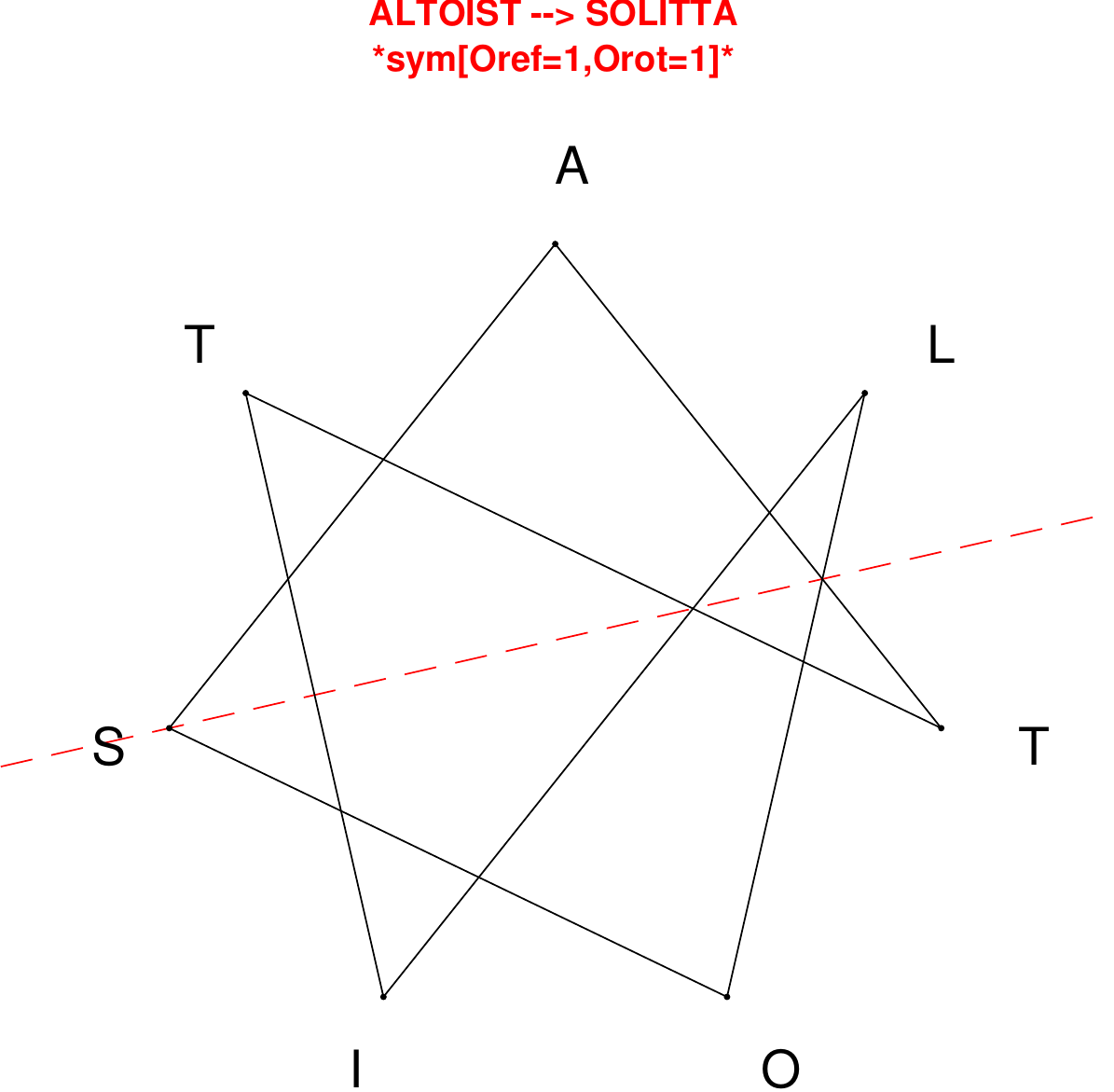}
\end{subfigure}
\hfill
\begin{subfigure}[T]{0.19\textwidth}
\centering
\includegraphics[width=\textwidth]{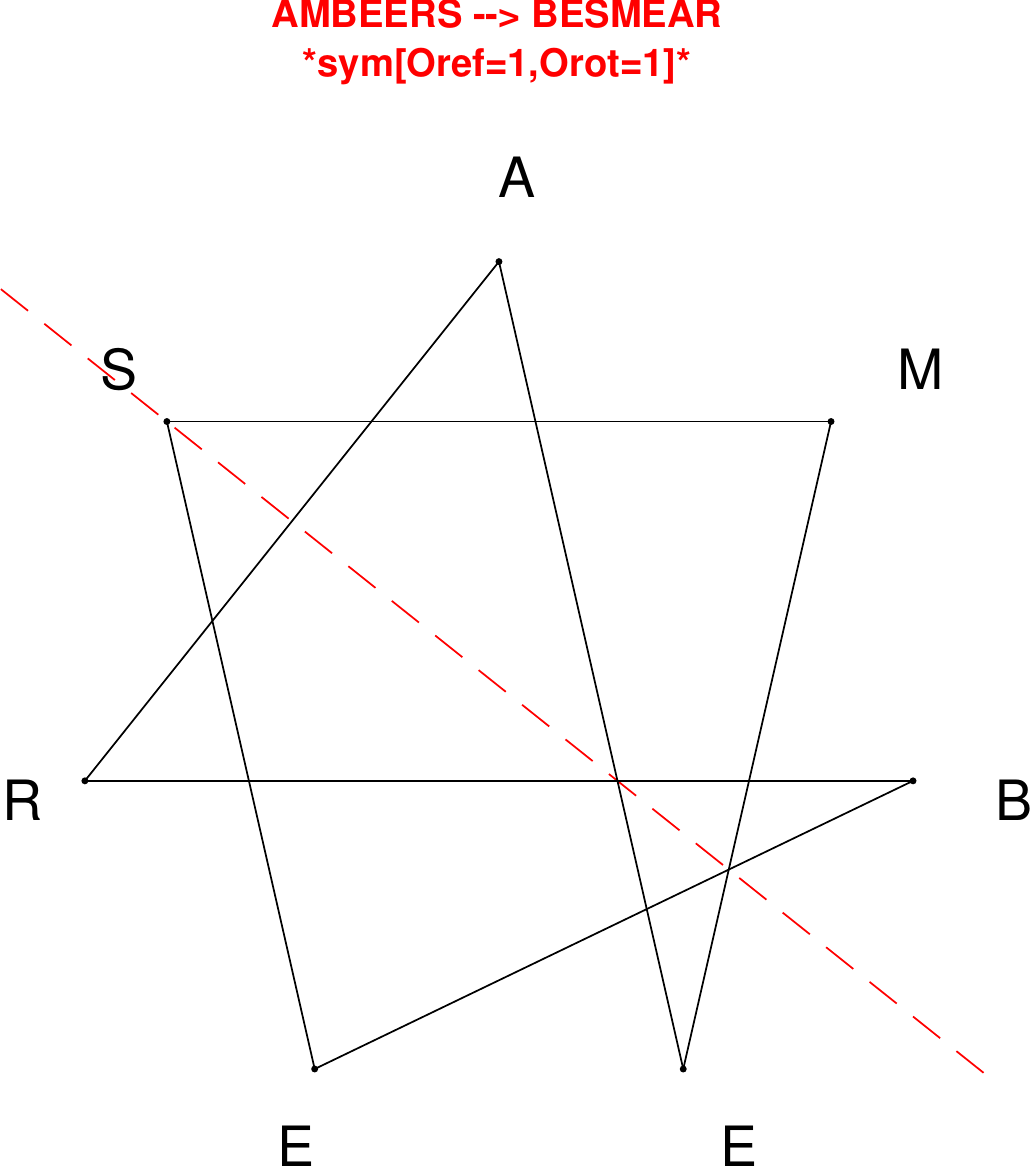}
\end{subfigure}
\end{figure}

\begin{figure}[H]
\centering
\begin{subfigure}[T]{0.19\textwidth}
\centering
\includegraphics[width=\textwidth]{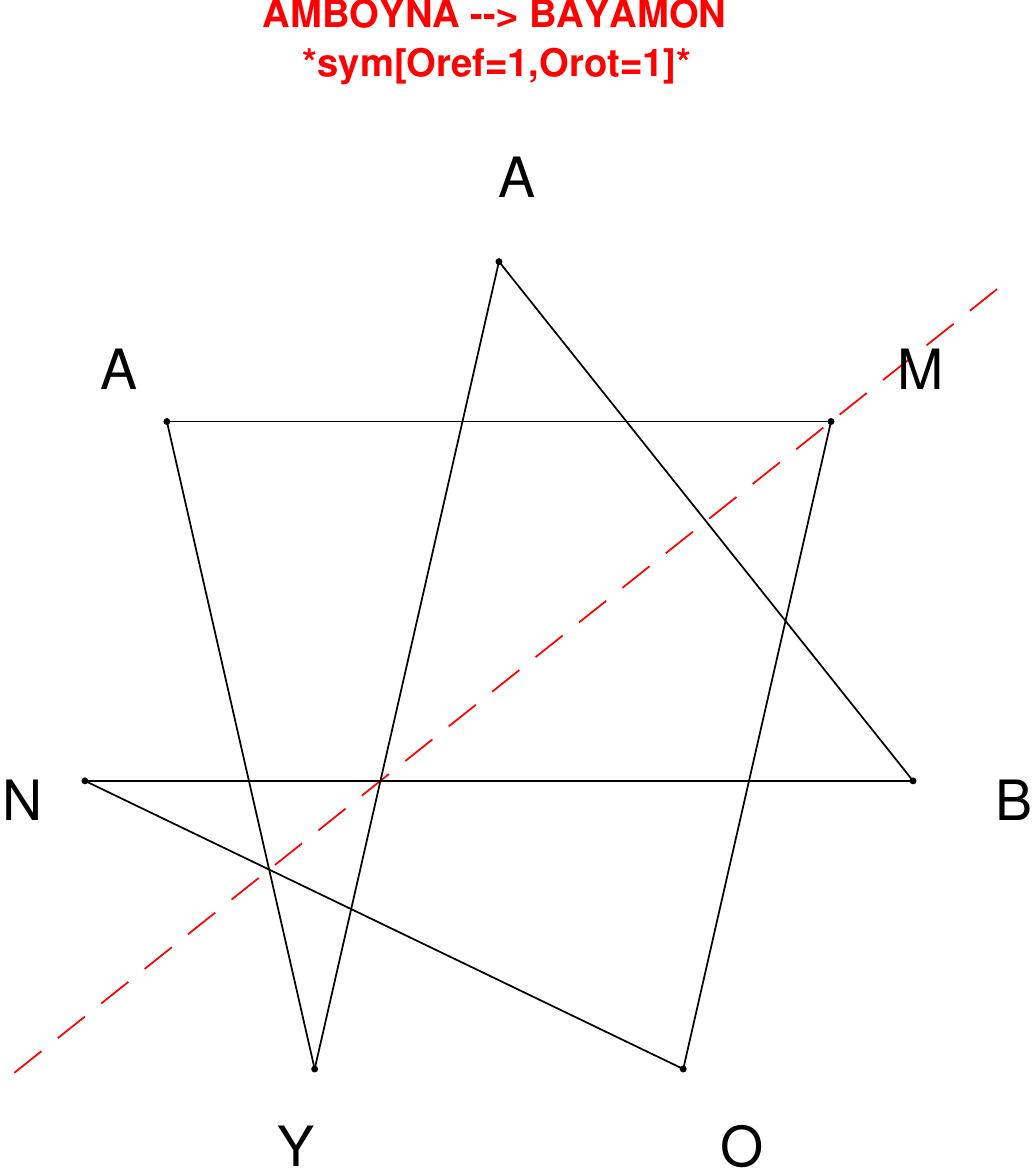}
\end{subfigure}
\hfill
\begin{subfigure}[T]{0.19\textwidth}
\centering
\includegraphics[width=\textwidth]{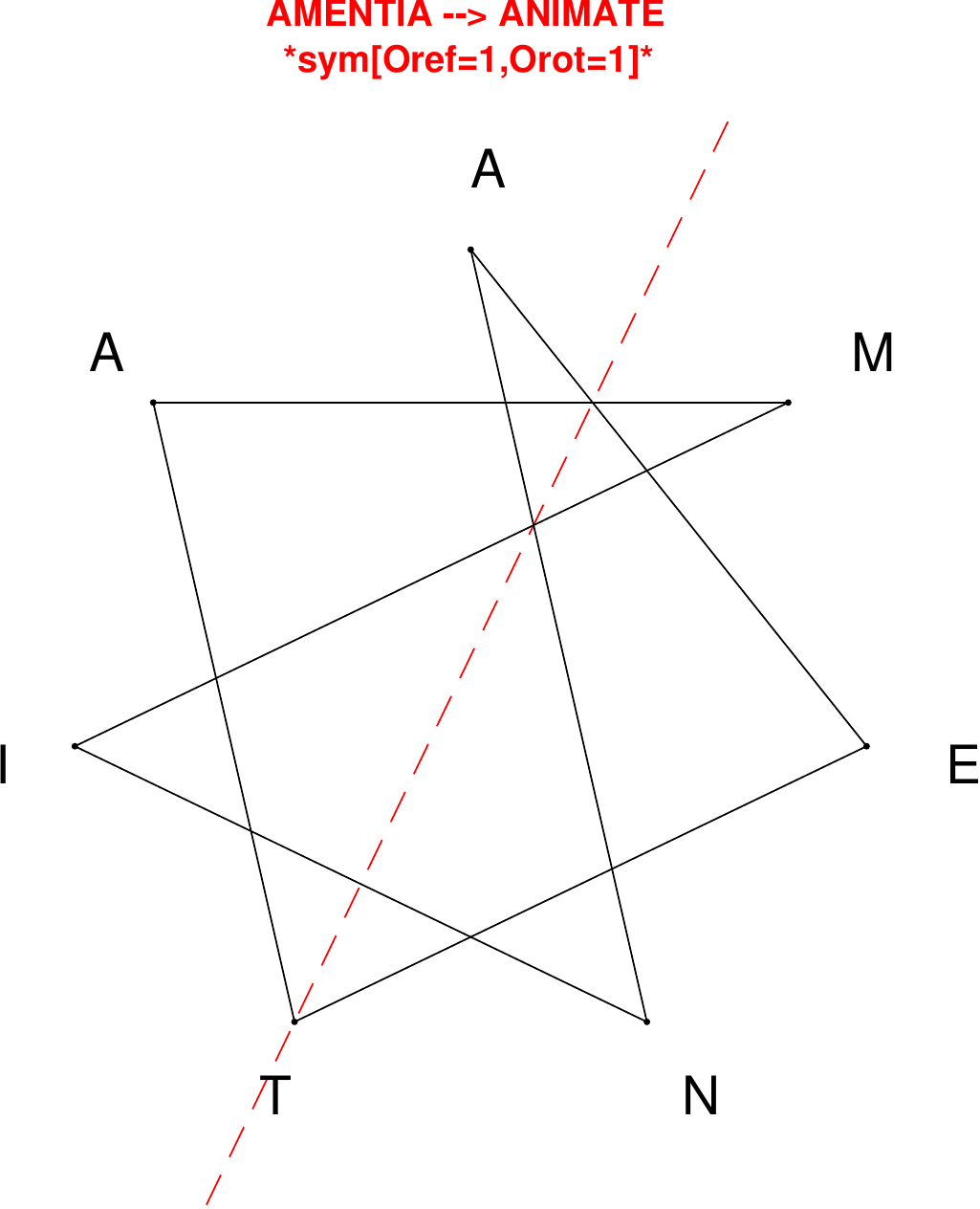}
\end{subfigure}
\hfill
\begin{subfigure}[T]{0.19\textwidth}
\centering
\includegraphics[width=\textwidth]{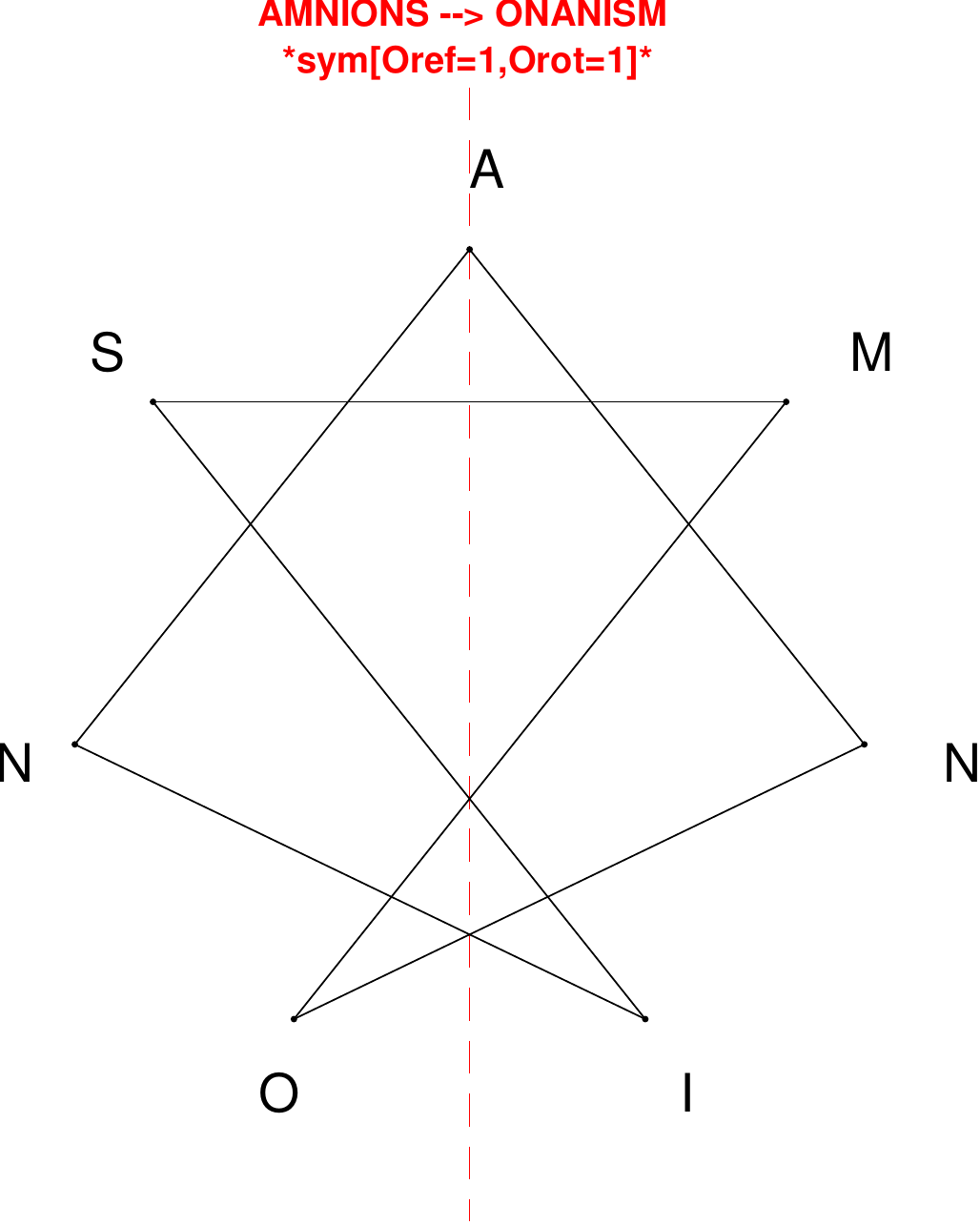}
\end{subfigure}
\hfill
\begin{subfigure}[T]{0.19\textwidth}
\centering
\includegraphics[width=\textwidth]{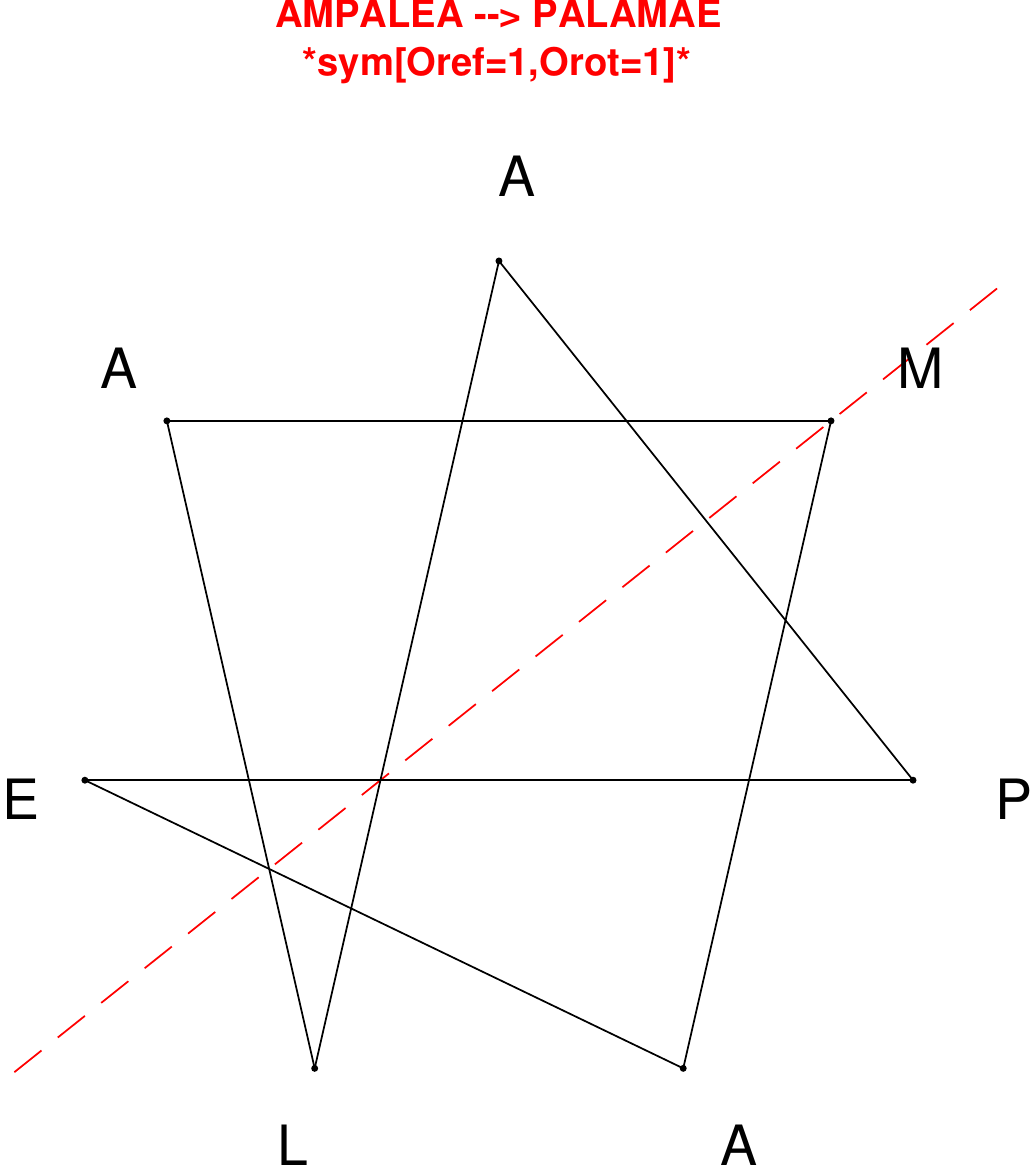}
\end{subfigure}
\hfill
\begin{subfigure}[T]{0.19\textwidth}
\centering
\includegraphics[width=\textwidth]{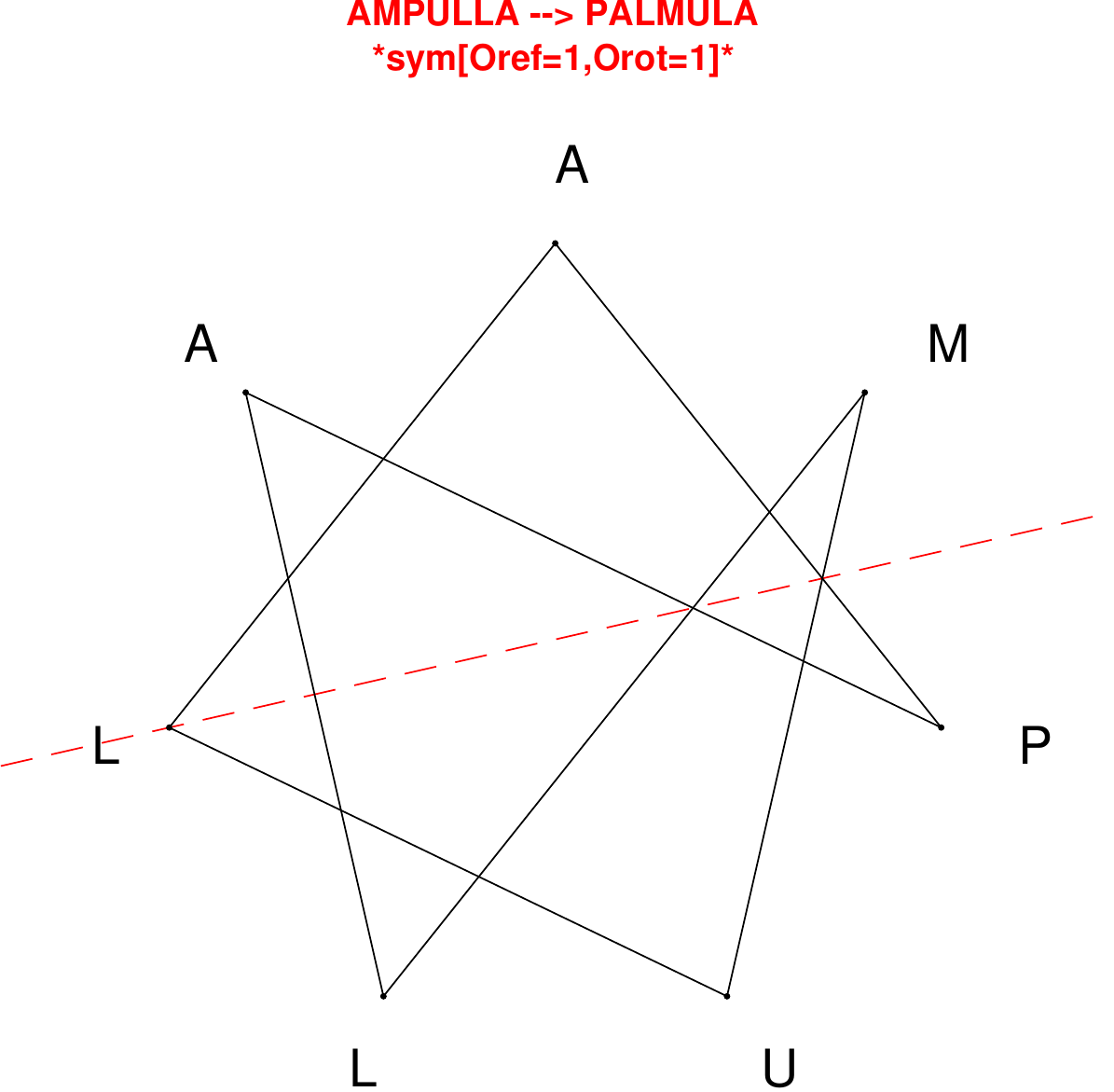}
\end{subfigure}
\end{figure}

\begin{figure}[H]
\centering
\begin{subfigure}[T]{0.19\textwidth}
\centering
\includegraphics[width=\textwidth]{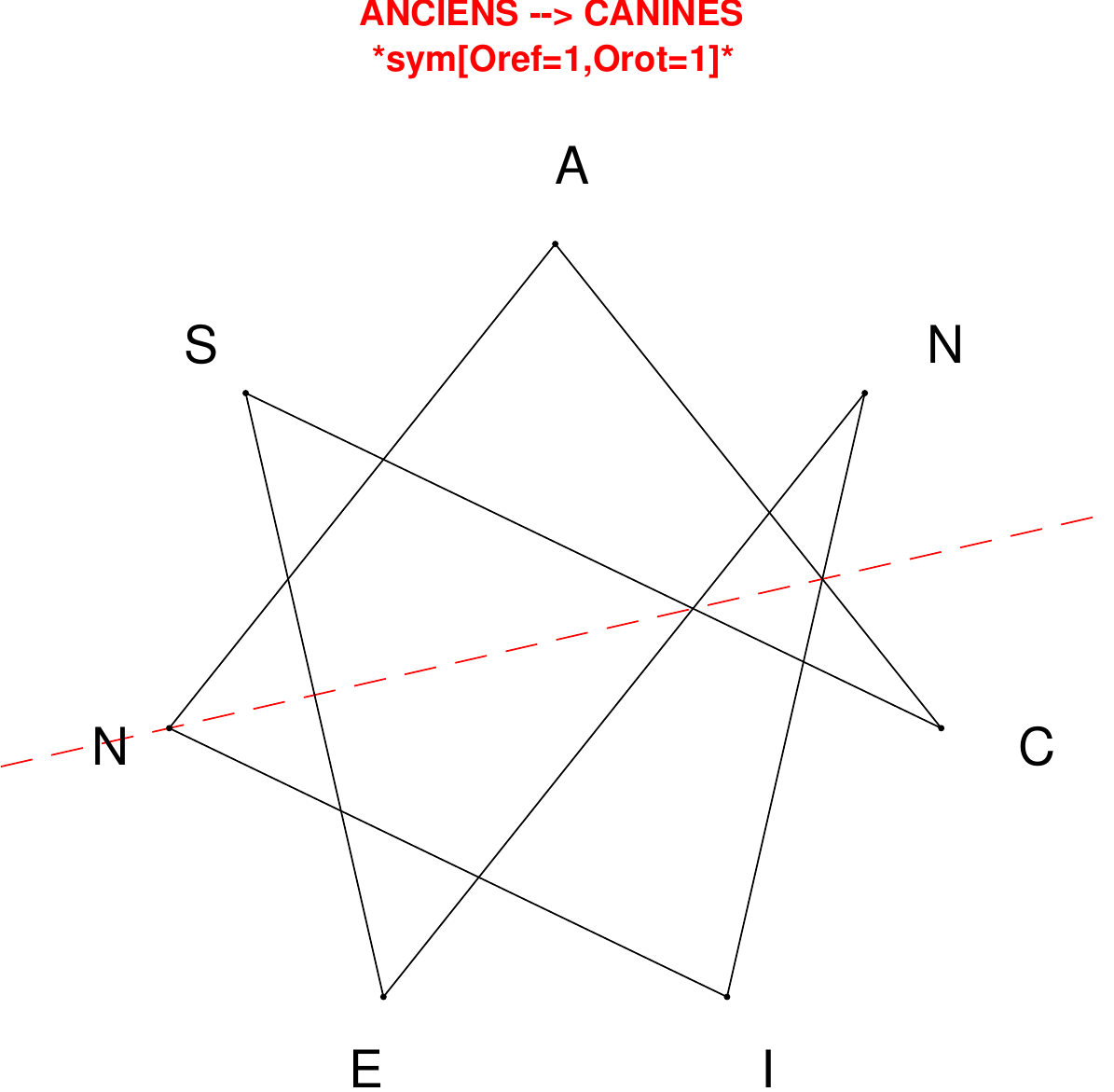}
\end{subfigure}
\hfill
\begin{subfigure}[T]{0.19\textwidth}
\centering
\includegraphics[width=\textwidth]{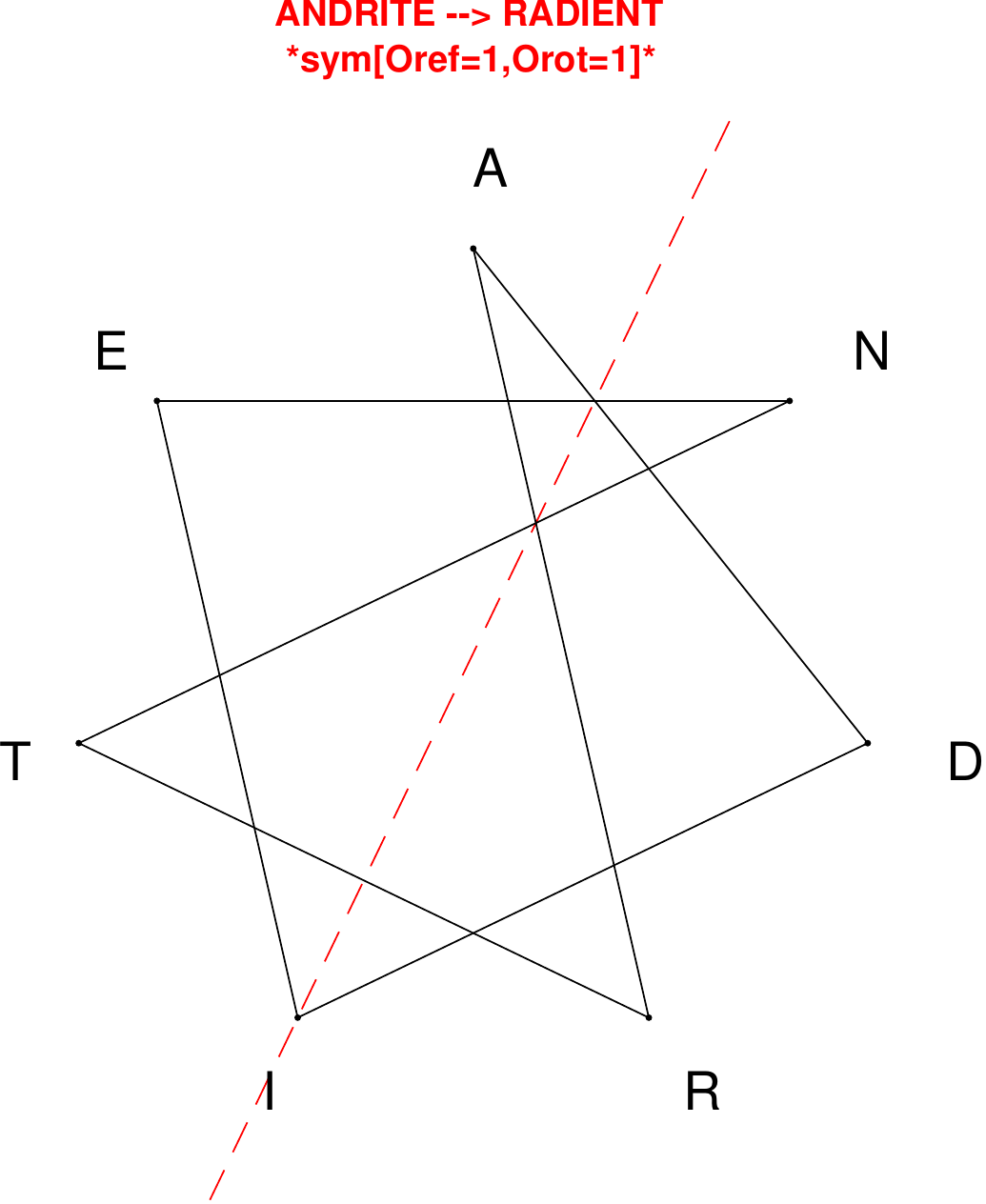}
\end{subfigure}
\hfill
\begin{subfigure}[T]{0.19\textwidth}
\centering
\includegraphics[width=\textwidth]{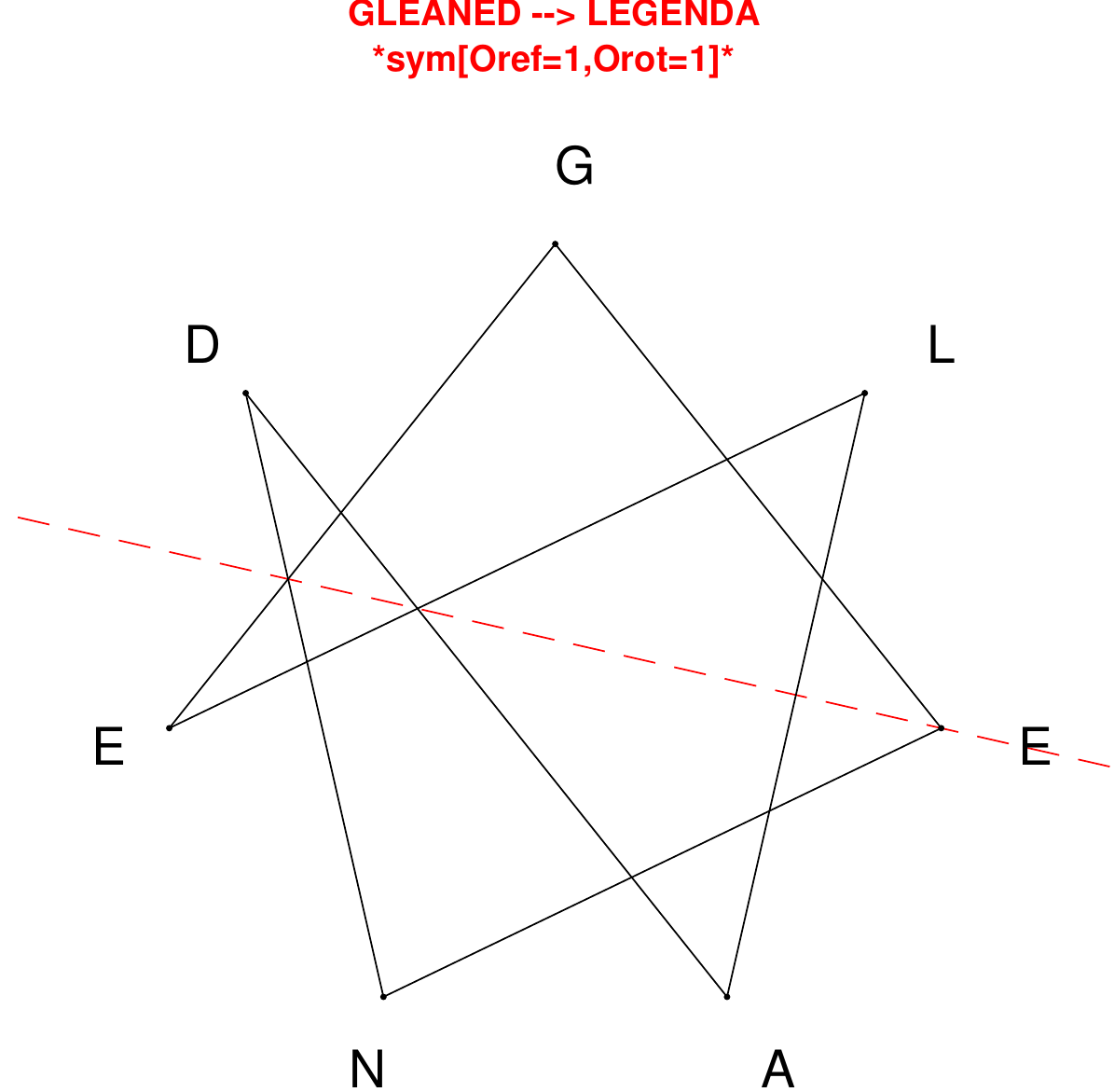}
\end{subfigure}
\hfill
\begin{subfigure}[T]{0.19\textwidth}
\centering
\includegraphics[width=\textwidth]{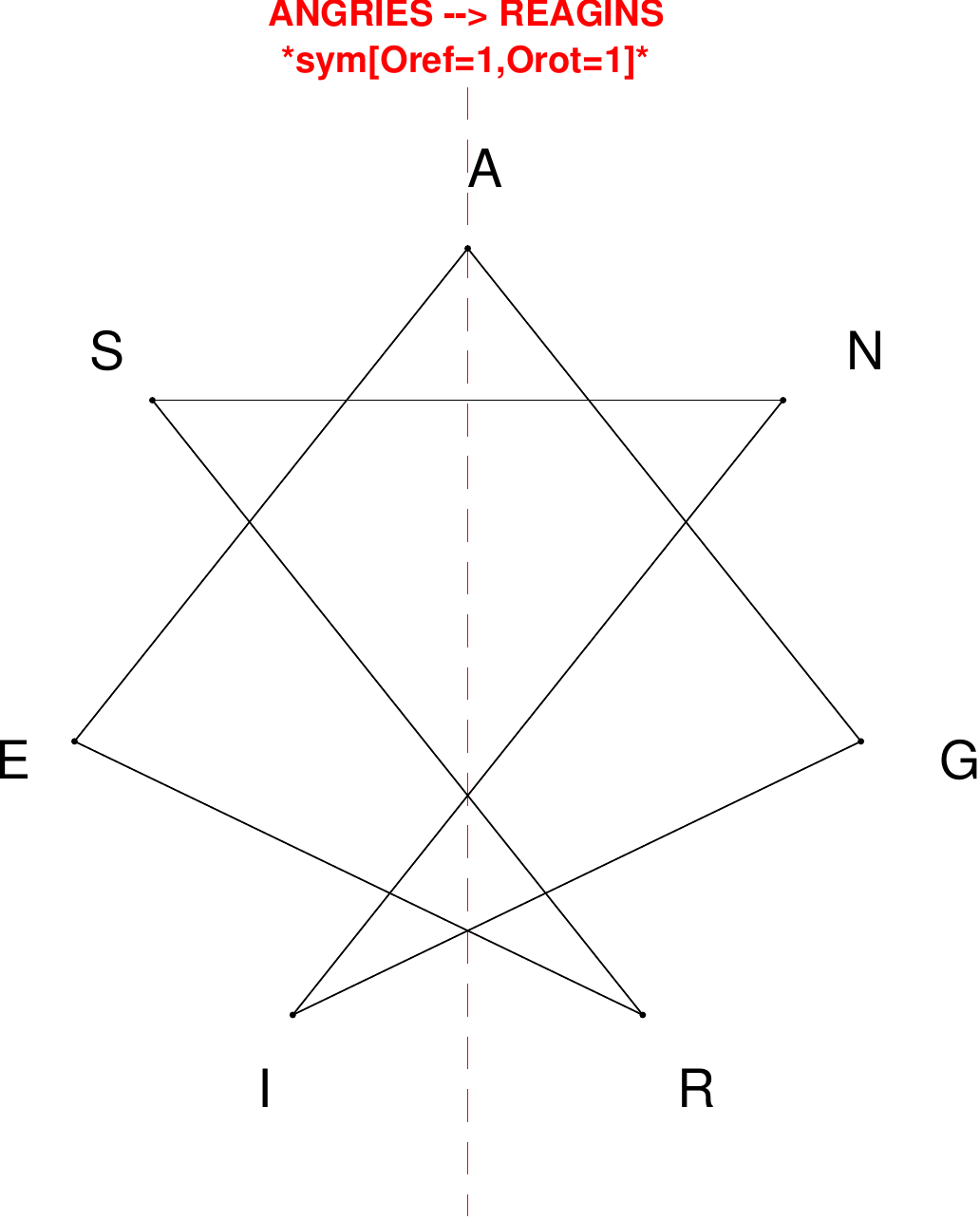}
\end{subfigure}
\hfill
\begin{subfigure}[T]{0.19\textwidth}
\centering
\includegraphics[width=\textwidth]{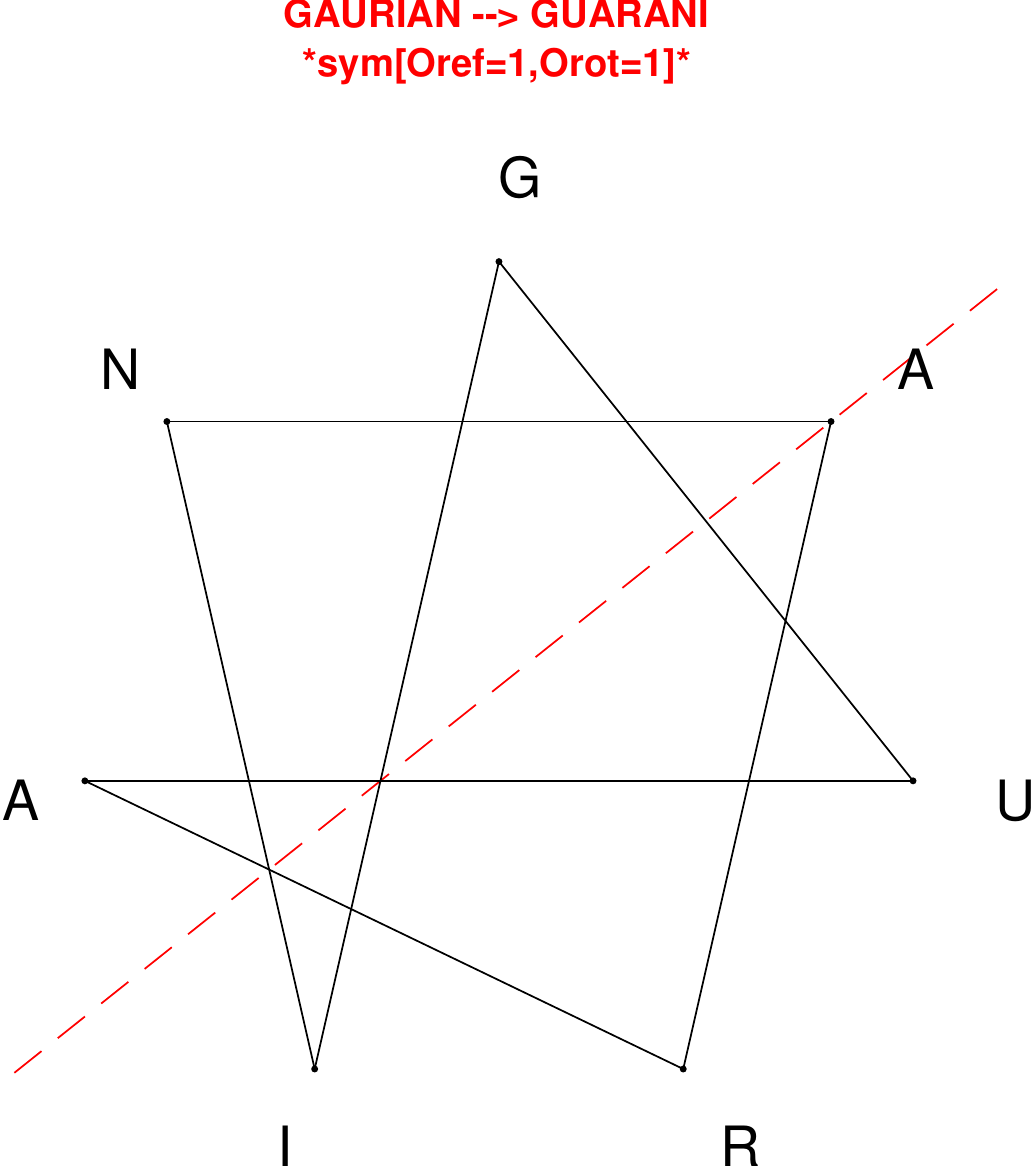}
\end{subfigure}
\end{figure}

\begin{figure}[H]
\centering
\begin{subfigure}[T]{0.19\textwidth}
\centering
\includegraphics[width=\textwidth]{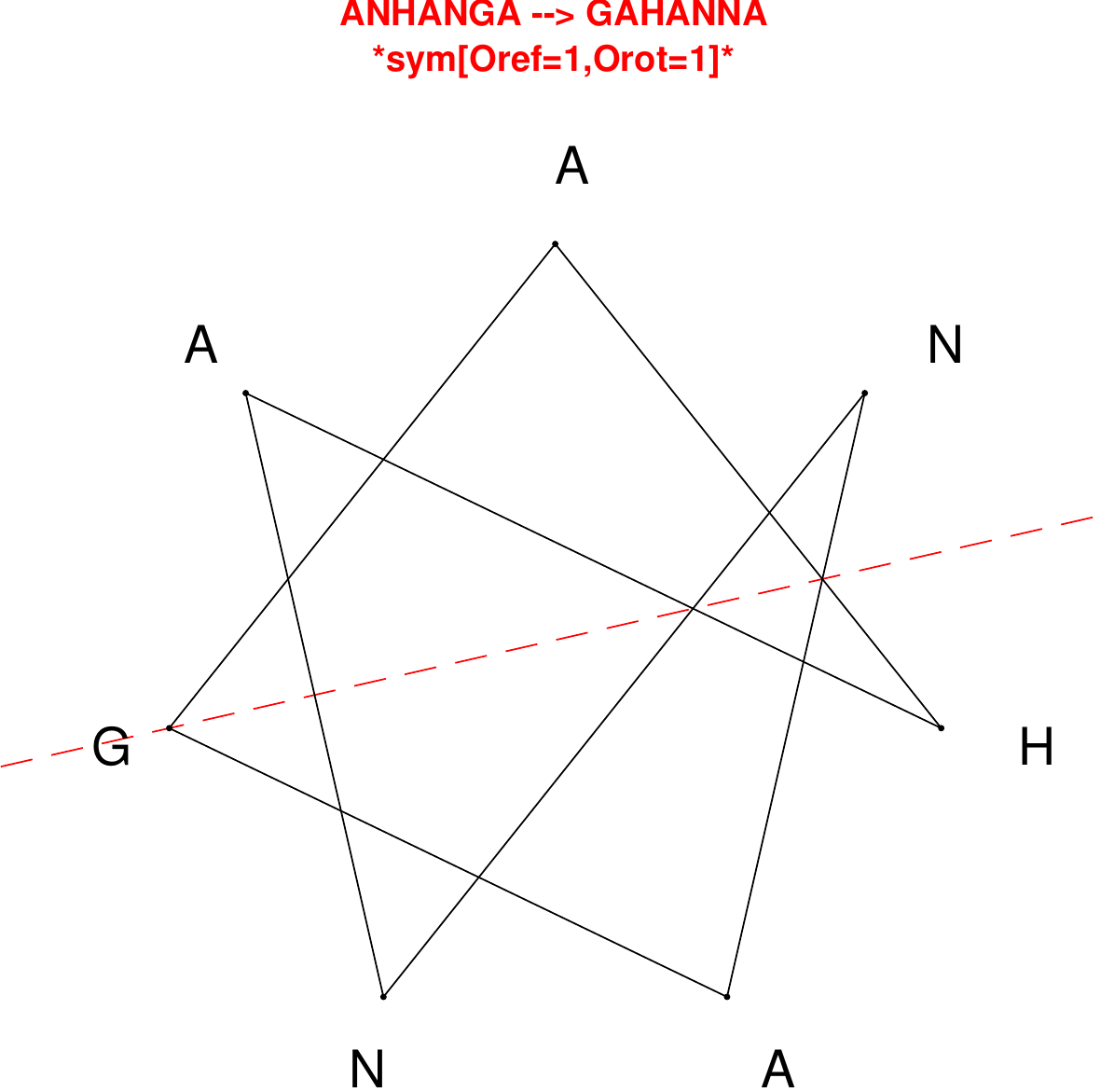}
\end{subfigure}
\hfill
\begin{subfigure}[T]{0.19\textwidth}
\centering
\includegraphics[width=\textwidth]{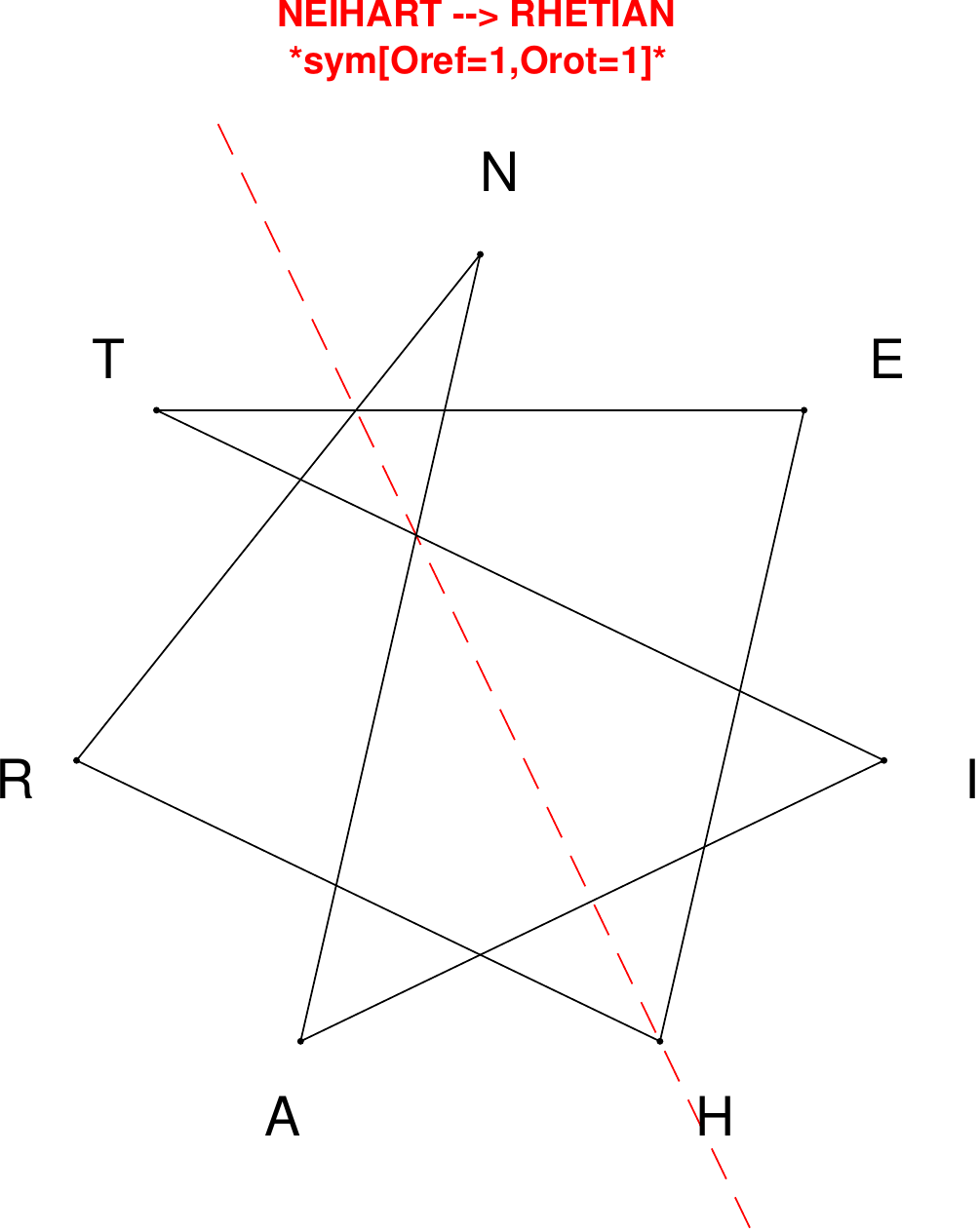}
\end{subfigure}
\hfill
\begin{subfigure}[T]{0.19\textwidth}
\centering
\includegraphics[width=\textwidth]{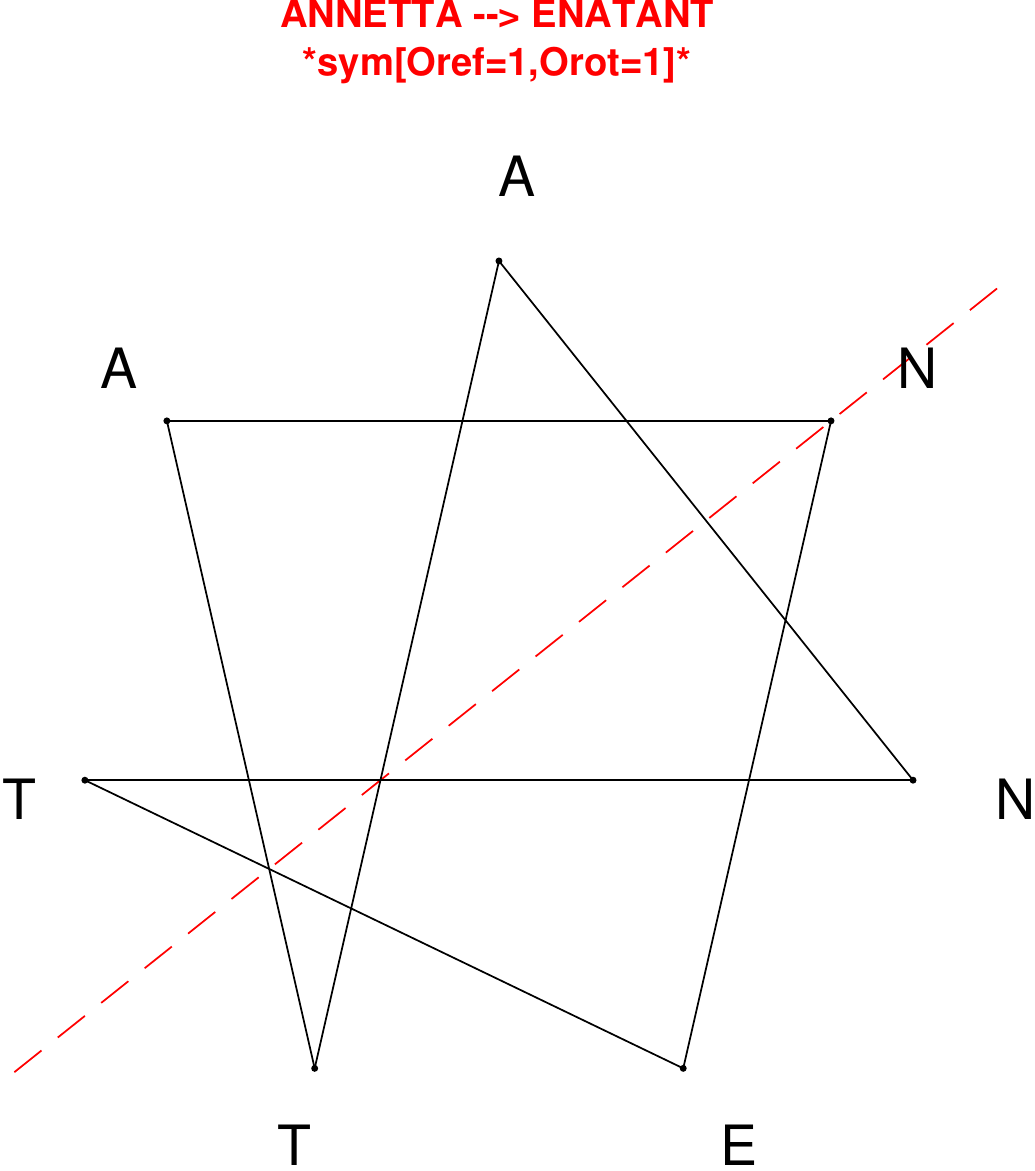}
\end{subfigure}
\hfill
\begin{subfigure}[T]{0.19\textwidth}
\centering
\includegraphics[width=\textwidth]{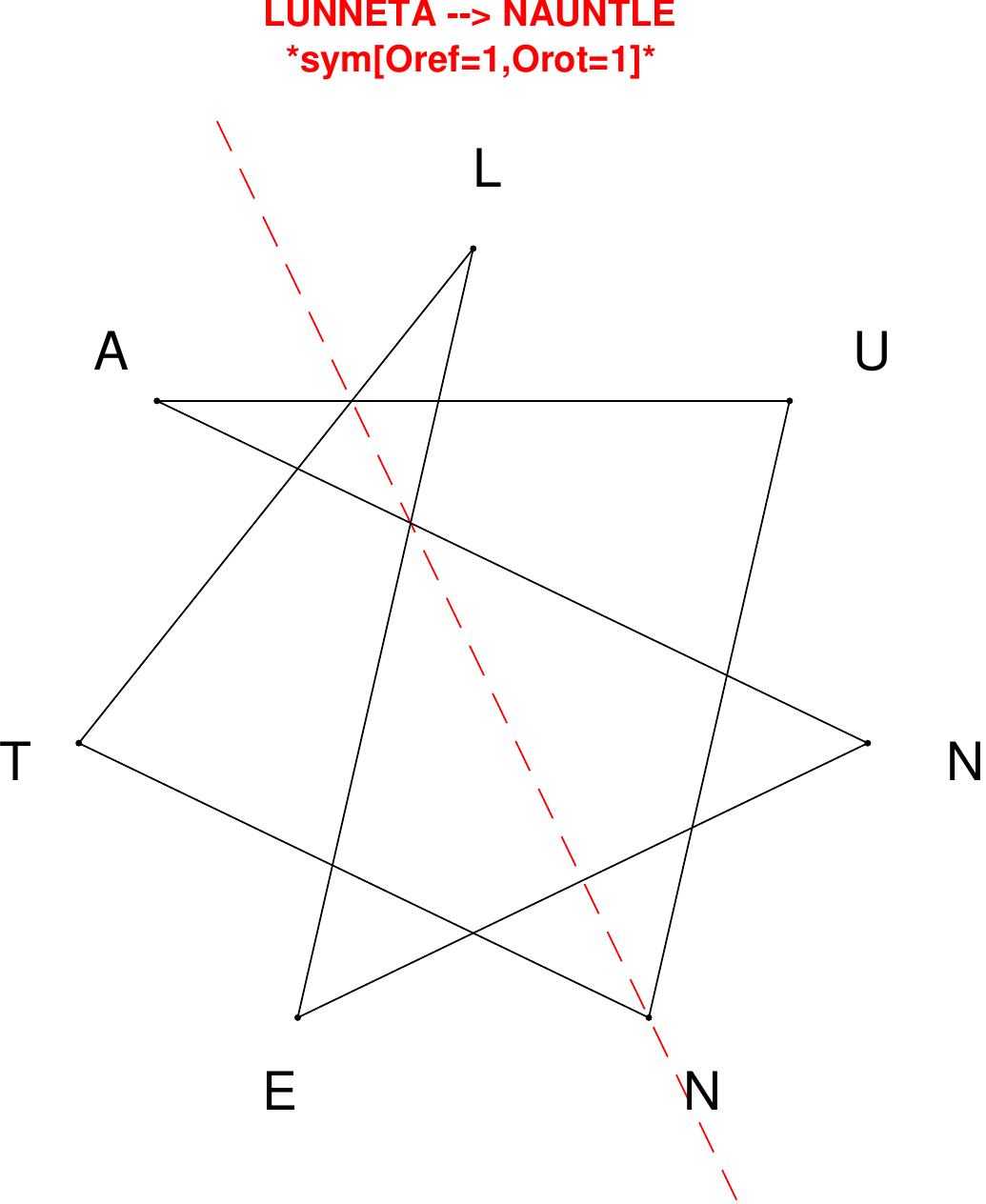}
\end{subfigure}
\hfill
\begin{subfigure}[T]{0.19\textwidth}
\centering
\includegraphics[width=\textwidth]{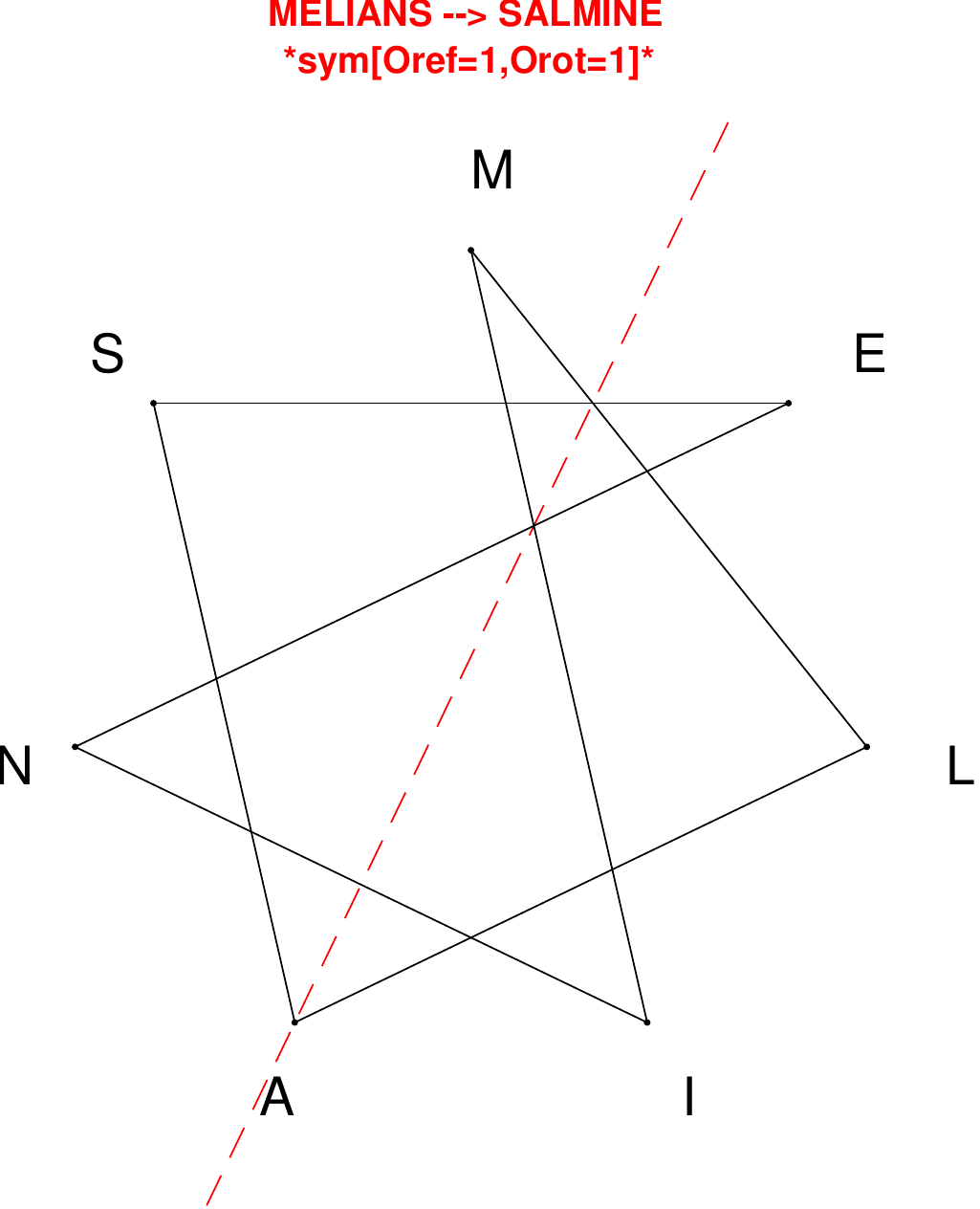}
\end{subfigure}
\end{figure}

\begin{figure}[H]
\centering
\begin{subfigure}[T]{0.19\textwidth}
\centering
\includegraphics[width=\textwidth]{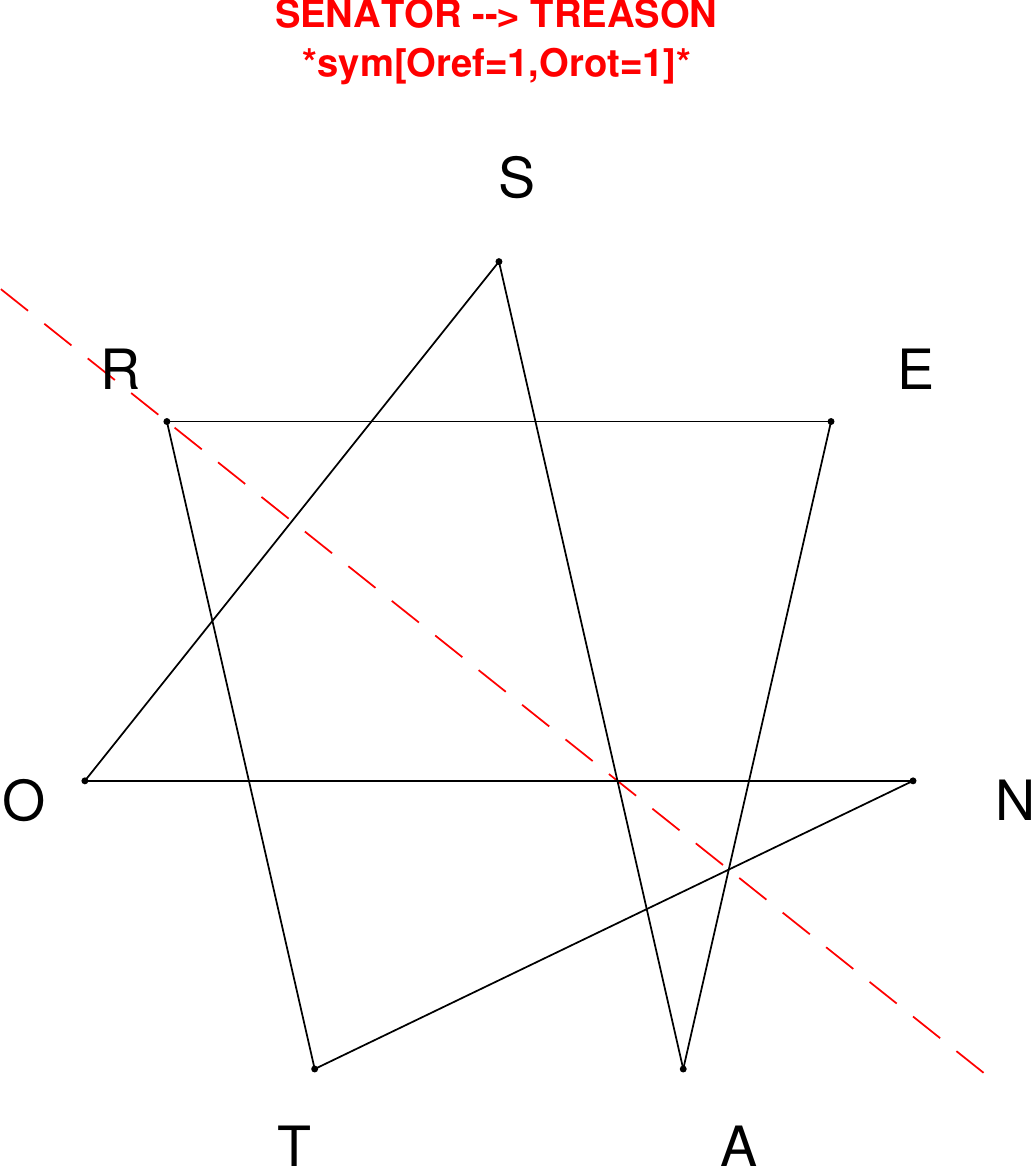}
\end{subfigure}
\hfill
\begin{subfigure}[T]{0.19\textwidth}
\centering
\includegraphics[width=\textwidth]{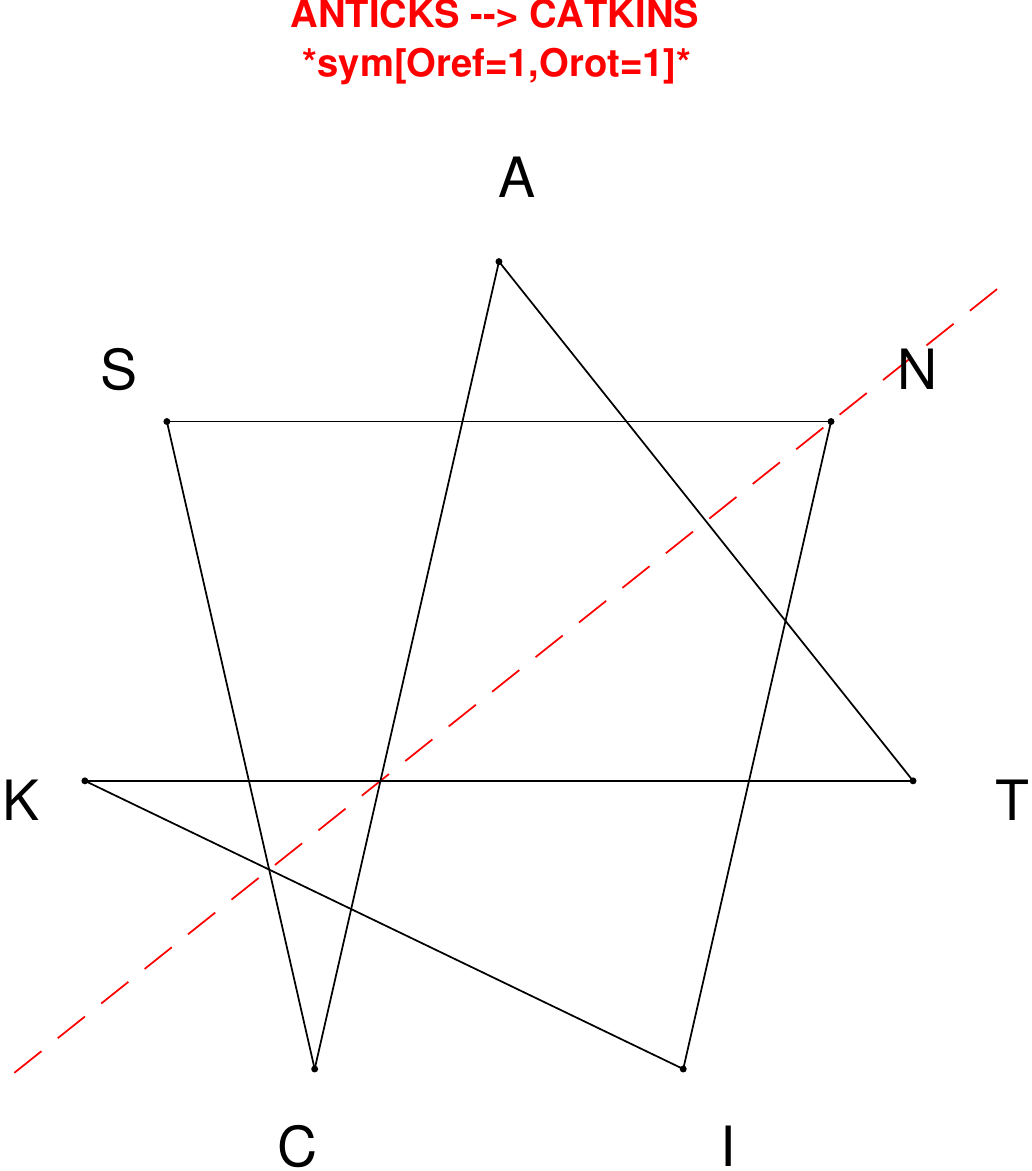}
\end{subfigure}
\hfill
\begin{subfigure}[T]{0.19\textwidth}
\centering
\includegraphics[width=\textwidth]{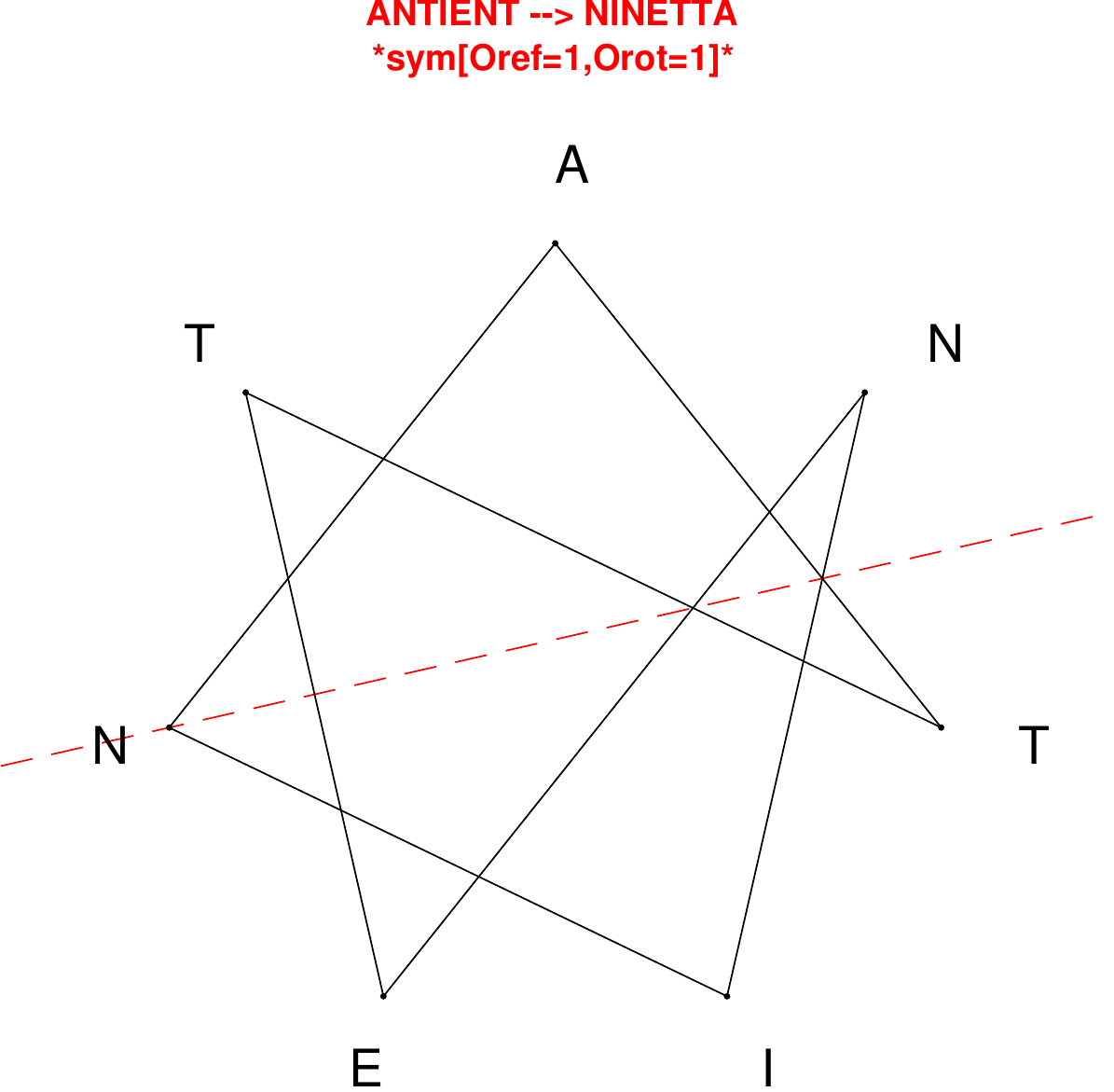}
\end{subfigure}
\hfill
\begin{subfigure}[T]{0.19\textwidth}
\centering
\includegraphics[width=\textwidth]{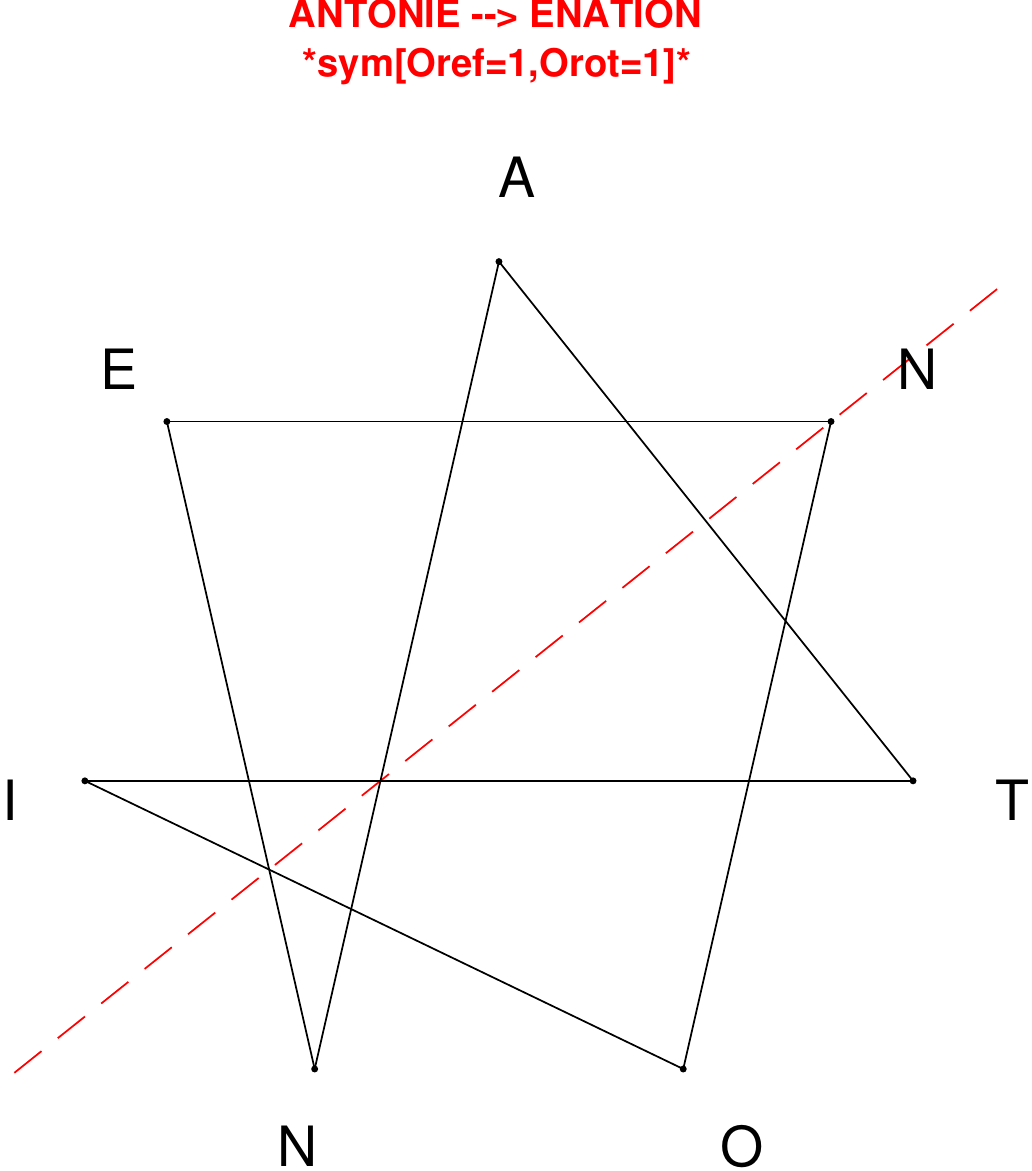}
\end{subfigure}
\hfill
\begin{subfigure}[T]{0.19\textwidth}
\centering
\includegraphics[width=\textwidth]{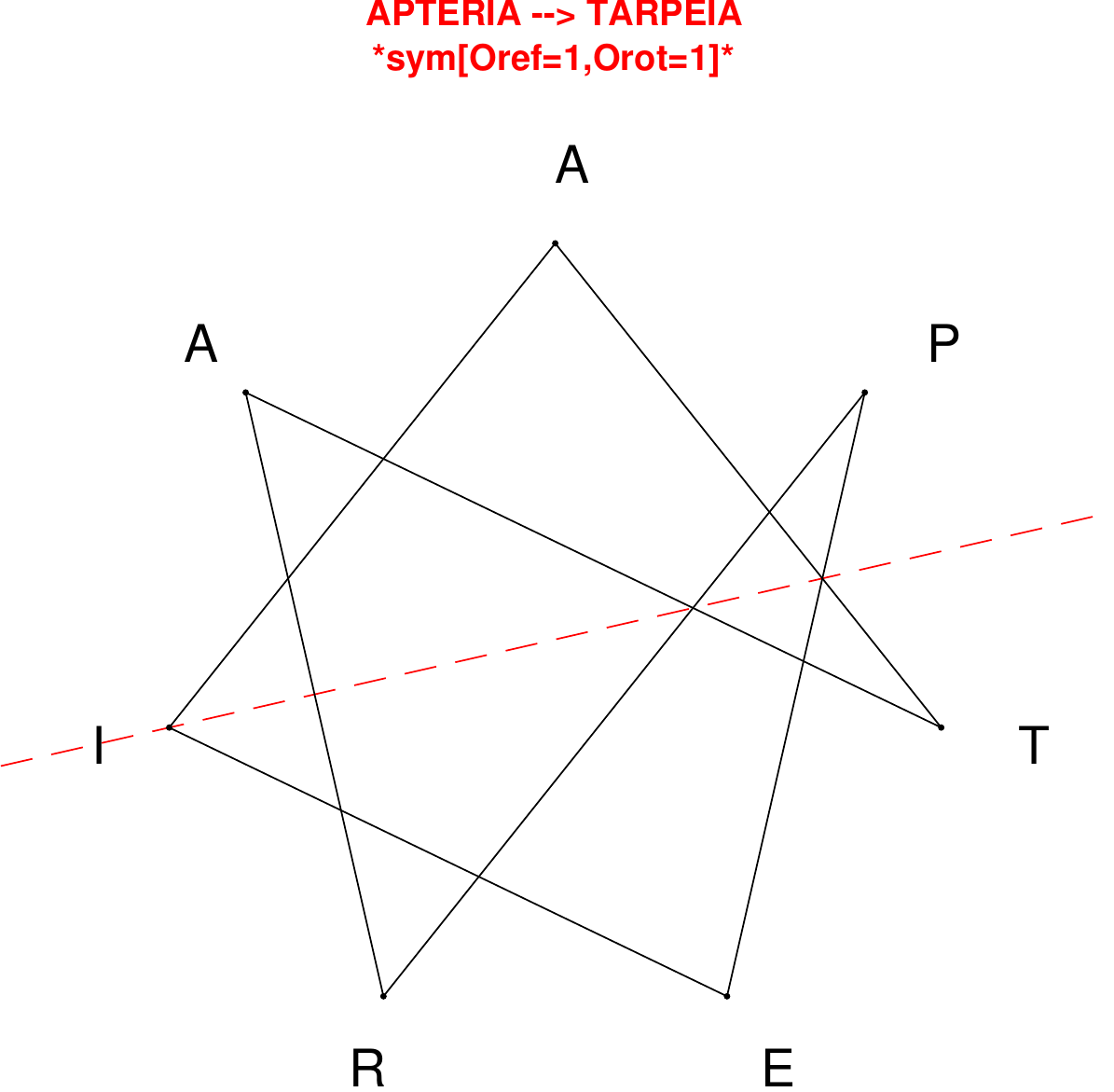}
\end{subfigure}
\end{figure}

\begin{figure}[H]
\centering
\begin{subfigure}[T]{0.19\textwidth}
\centering
\includegraphics[width=\textwidth]{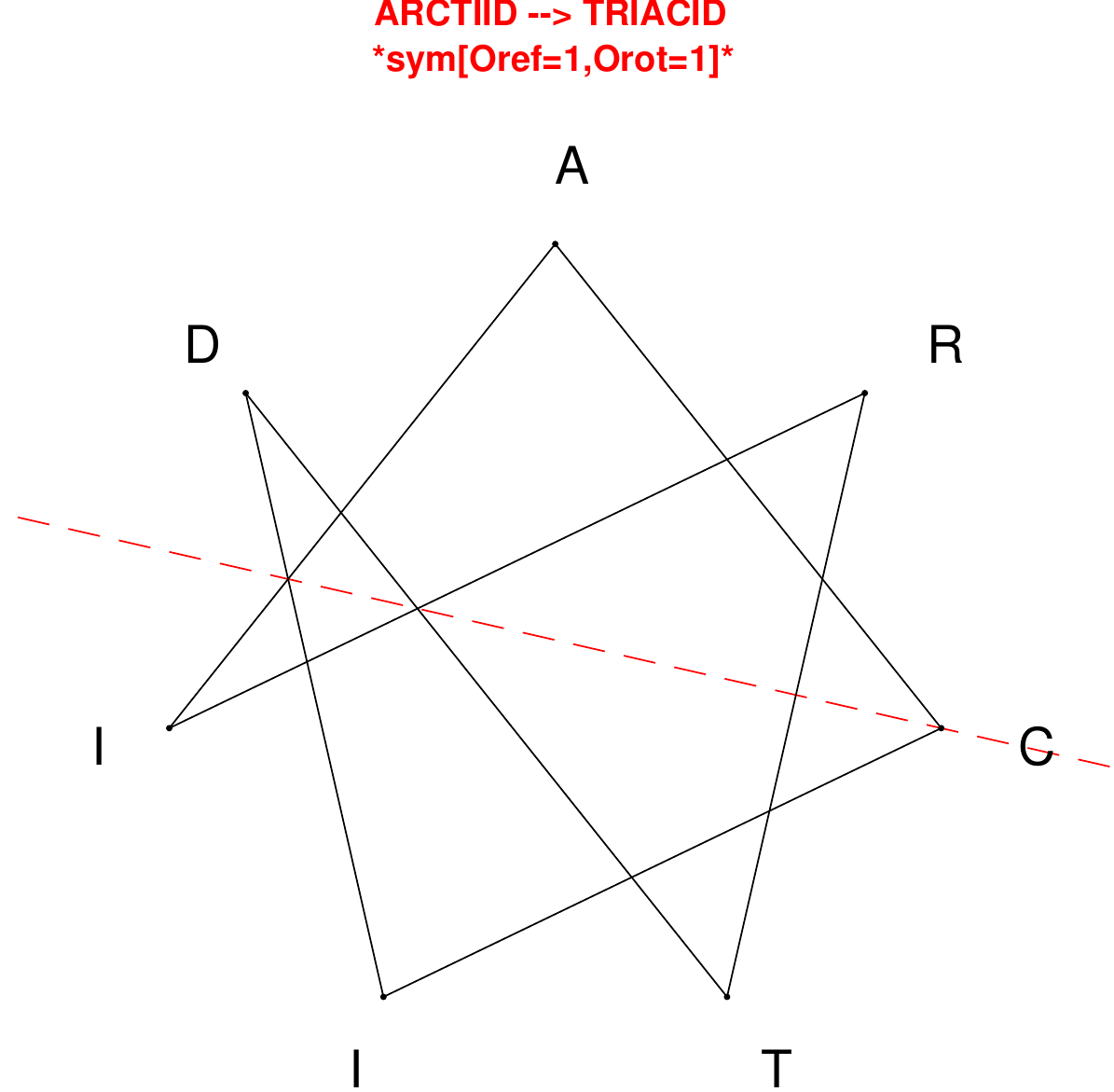}
\end{subfigure}
\hfill
\begin{subfigure}[T]{0.19\textwidth}
\centering
\includegraphics[width=\textwidth]{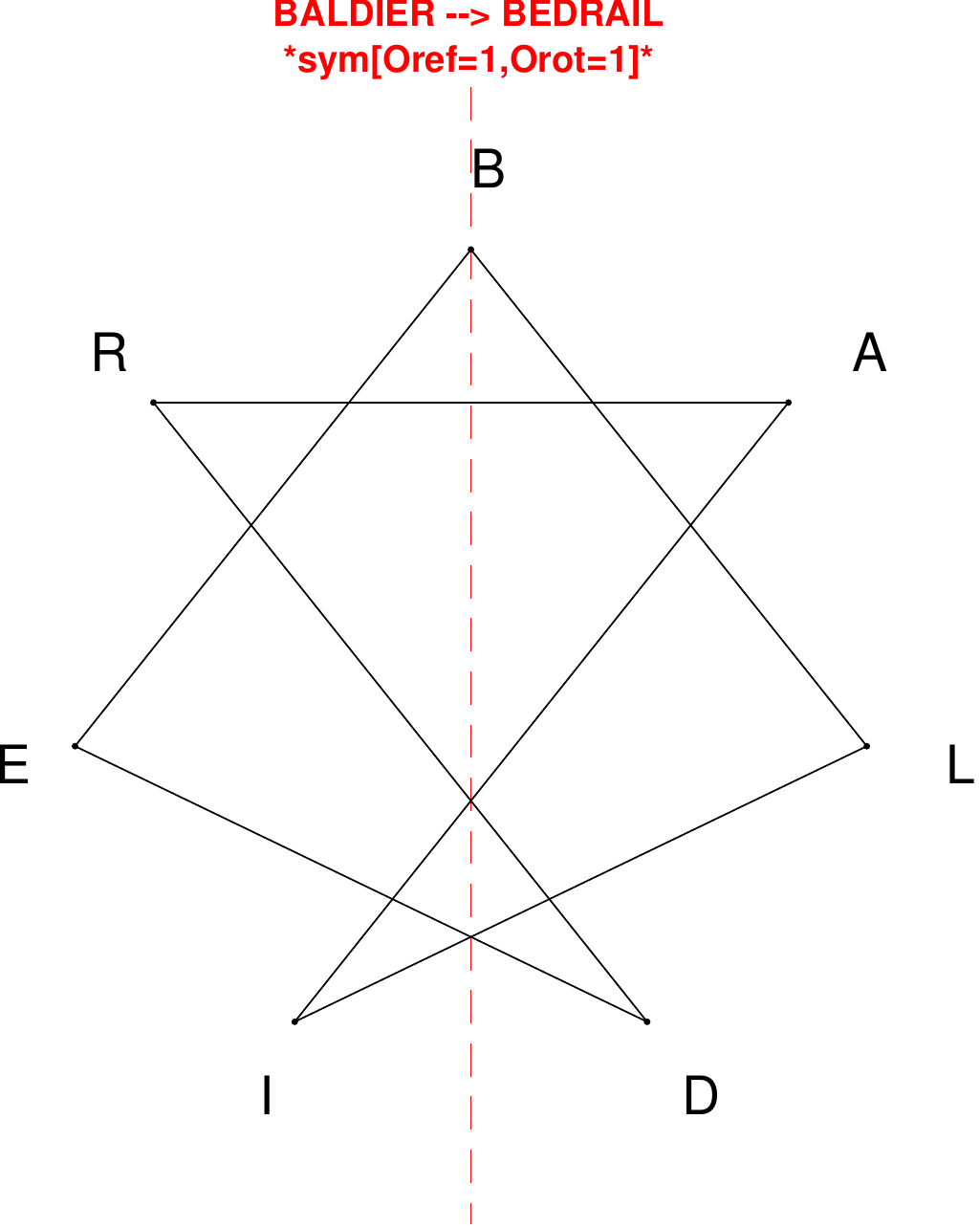}
\end{subfigure}
\hfill
\begin{subfigure}[T]{0.19\textwidth}
\centering
\includegraphics[width=\textwidth]{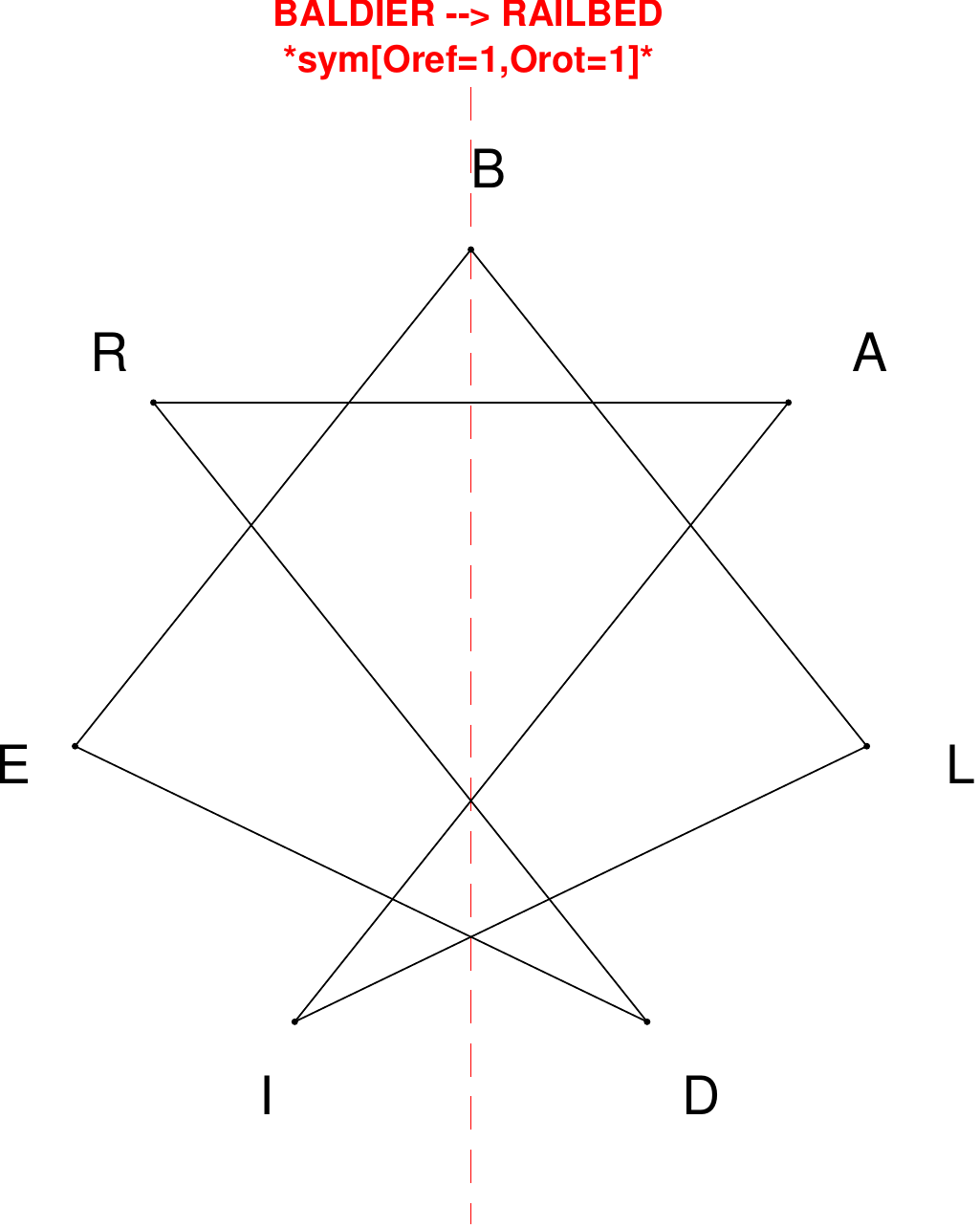}
\end{subfigure}
\hfill
\begin{subfigure}[T]{0.19\textwidth}
\centering
\includegraphics[width=\textwidth]{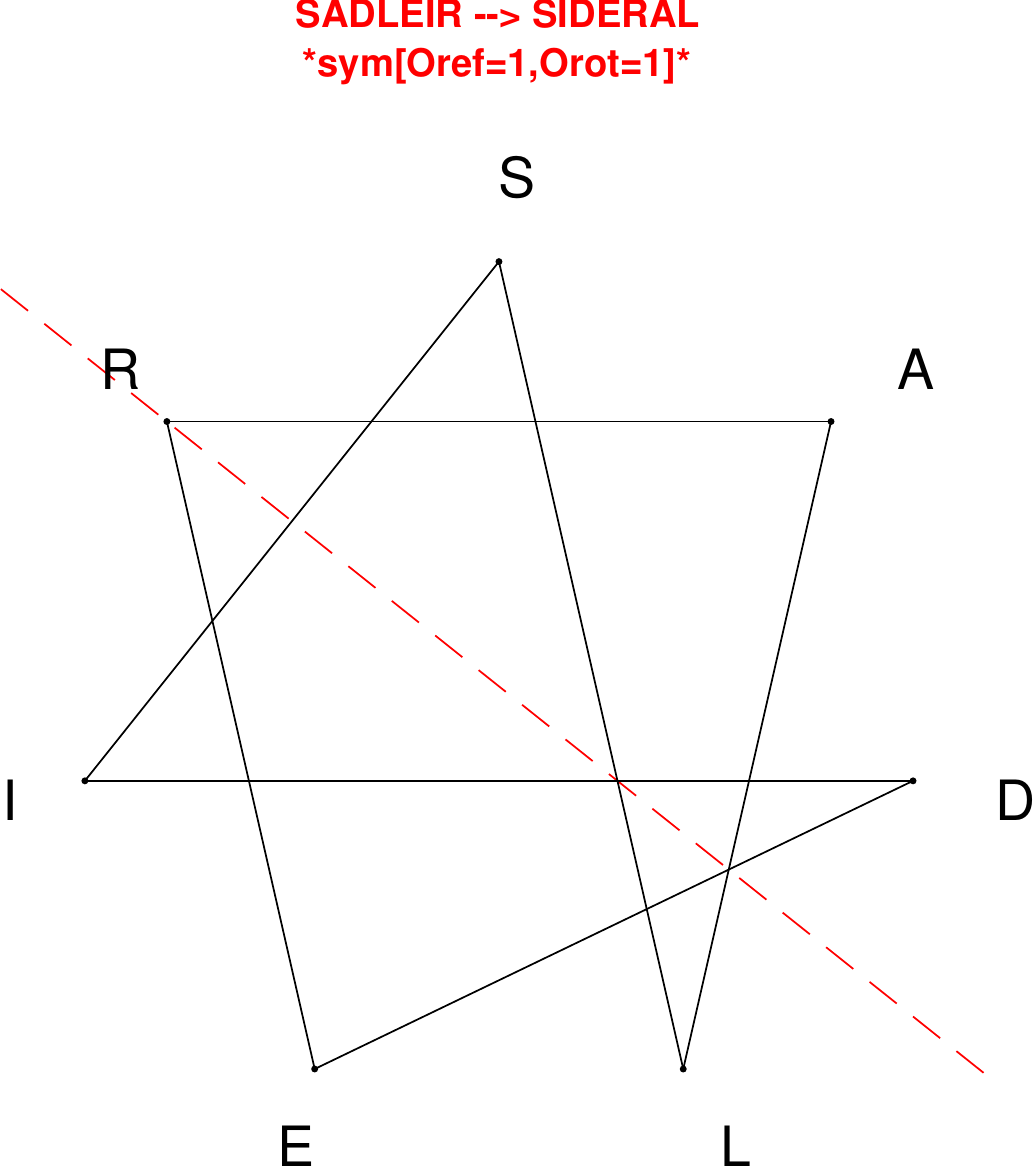}
\end{subfigure}
\hfill
\begin{subfigure}[T]{0.19\textwidth}
\centering
\includegraphics[width=\textwidth]{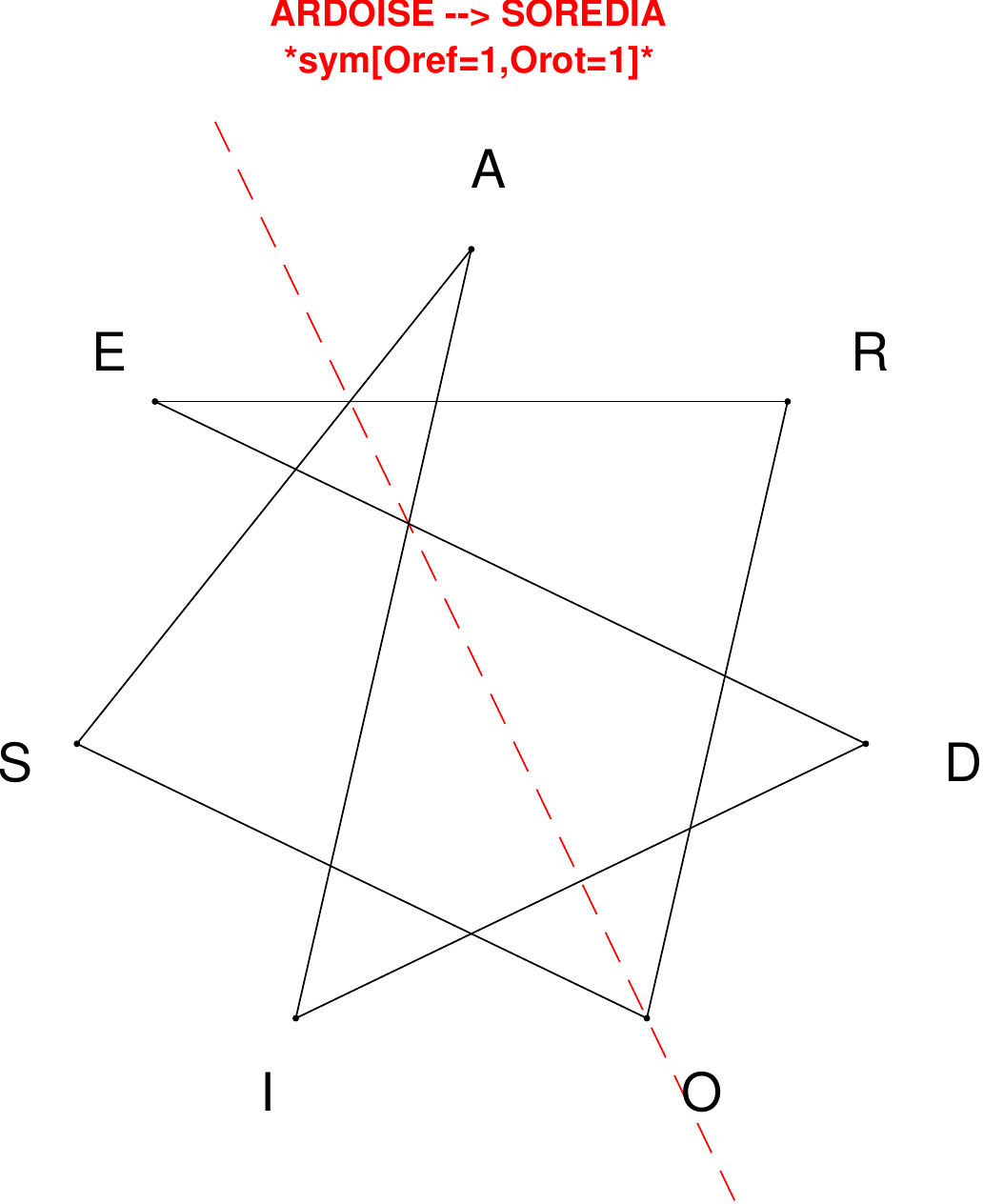}
\end{subfigure}
\end{figure}

\begin{figure}[H]
\centering
\begin{subfigure}[T]{0.19\textwidth}
\centering
\includegraphics[width=\textwidth]{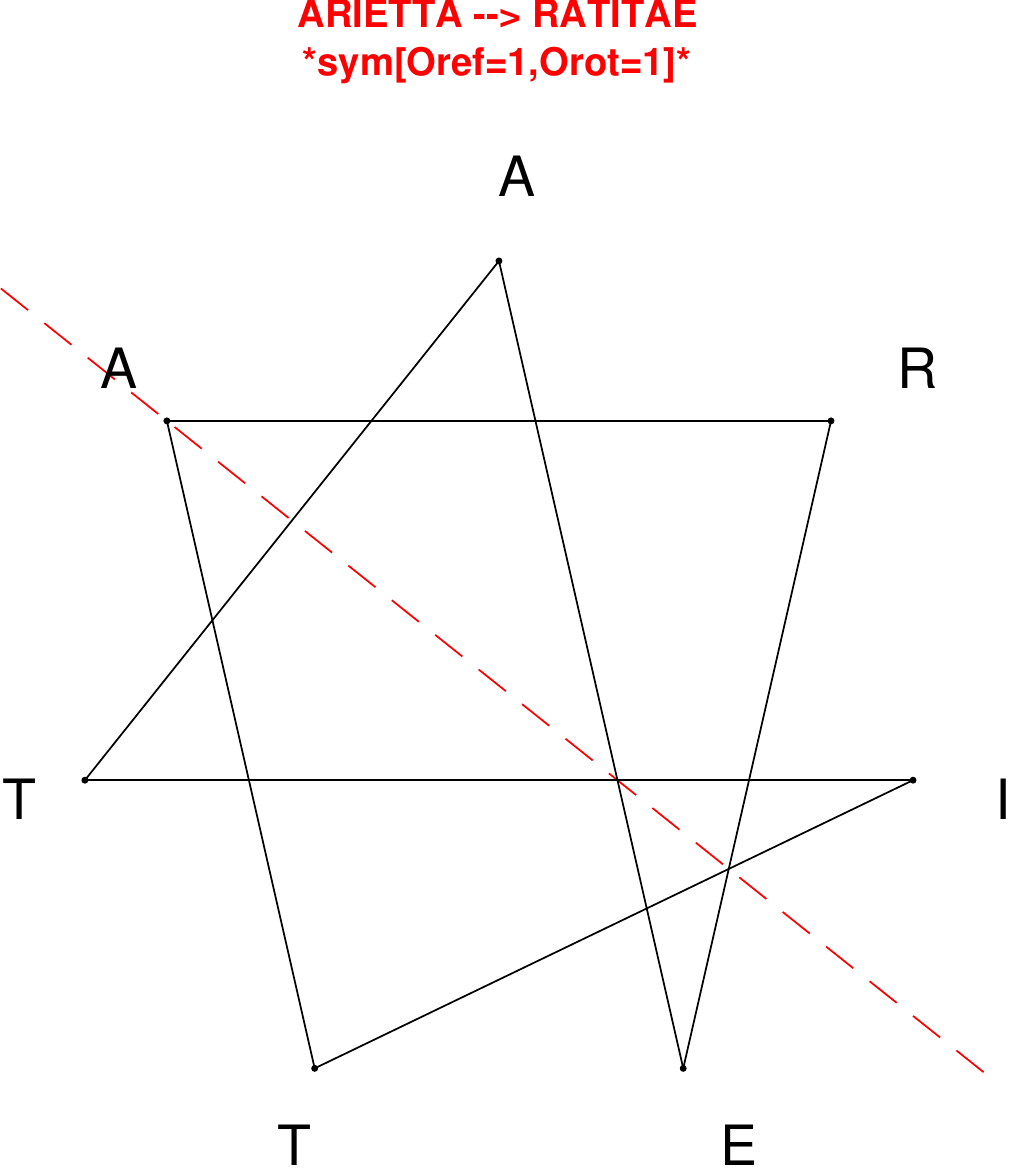}
\end{subfigure}
\hfill
\begin{subfigure}[T]{0.19\textwidth}
\centering
\includegraphics[width=\textwidth]{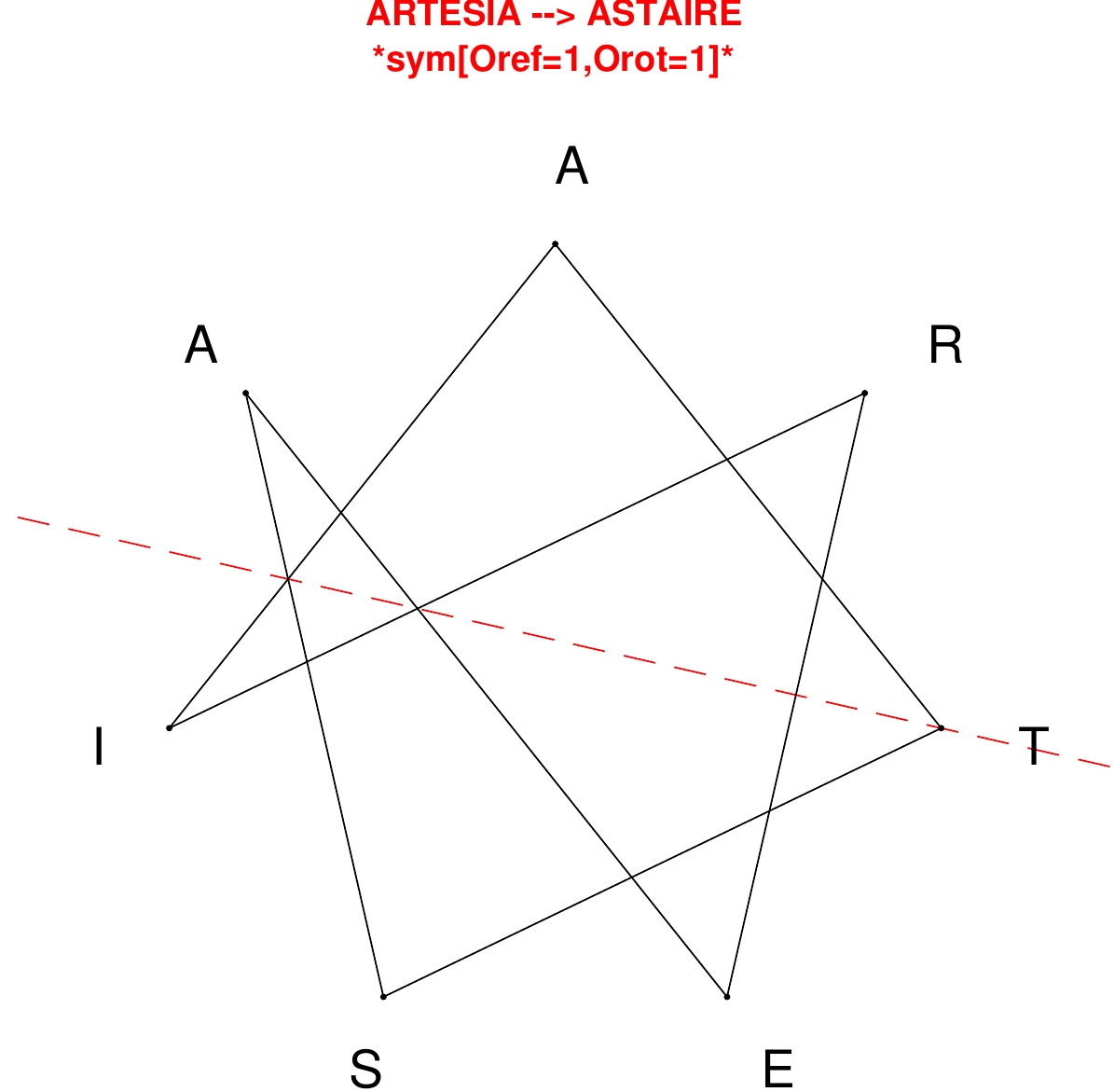}
\end{subfigure}
\hfill
\begin{subfigure}[T]{0.19\textwidth}
\centering
\includegraphics[width=\textwidth]{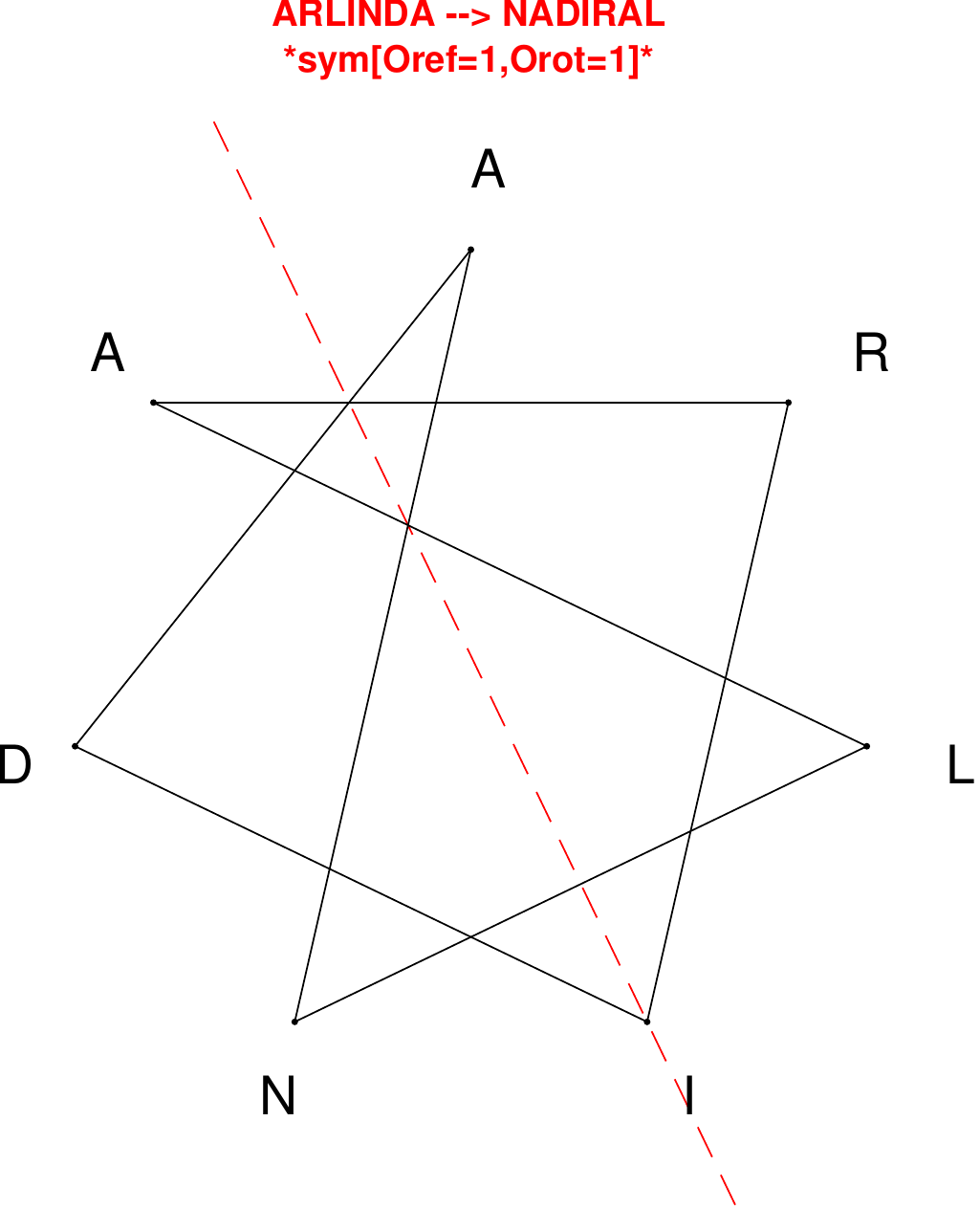}
\end{subfigure}
\hfill
\begin{subfigure}[T]{0.19\textwidth}
\centering
\includegraphics[width=\textwidth]{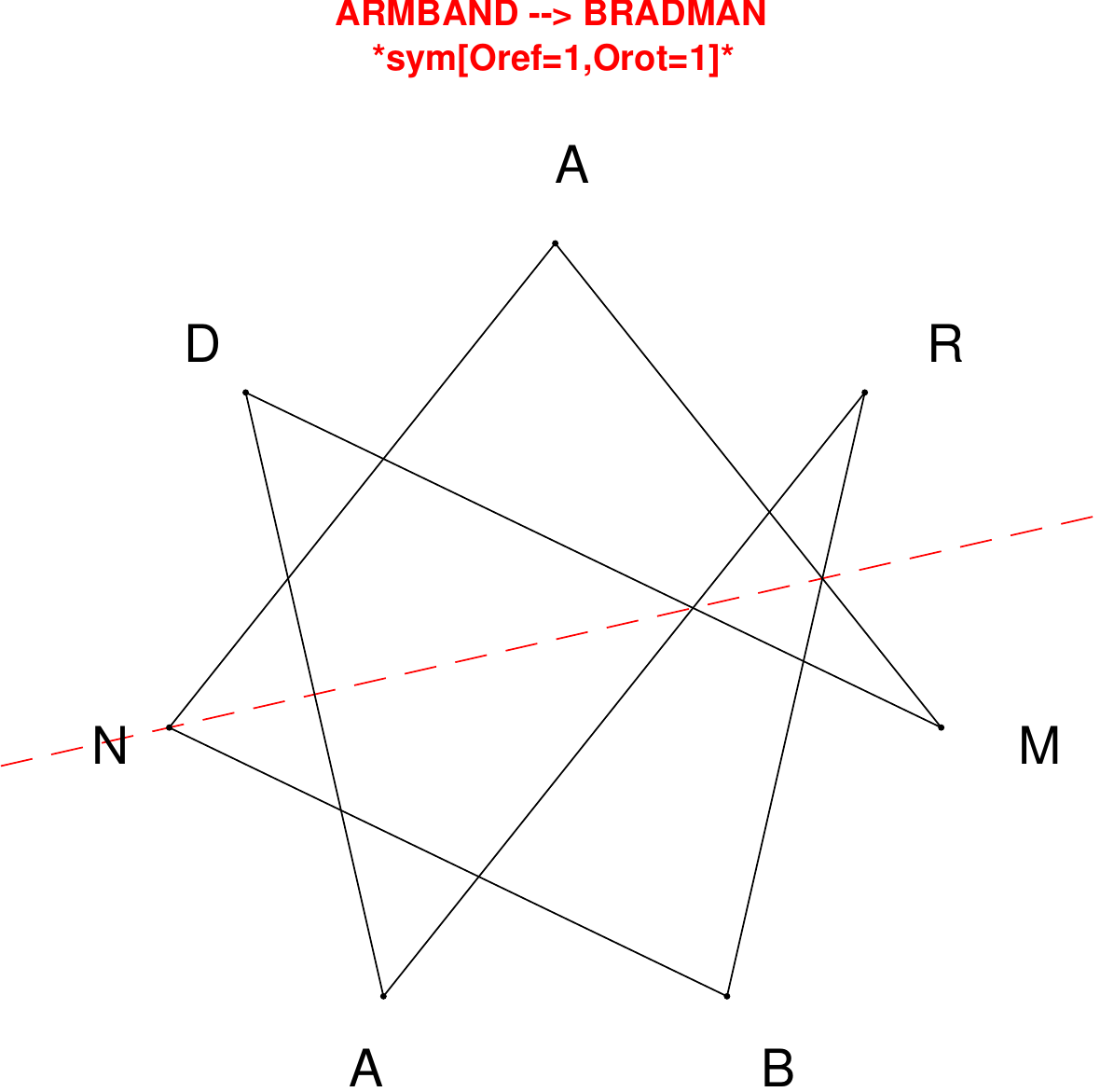}
\end{subfigure}
\hfill
\begin{subfigure}[T]{0.19\textwidth}
\centering
\includegraphics[width=\textwidth]{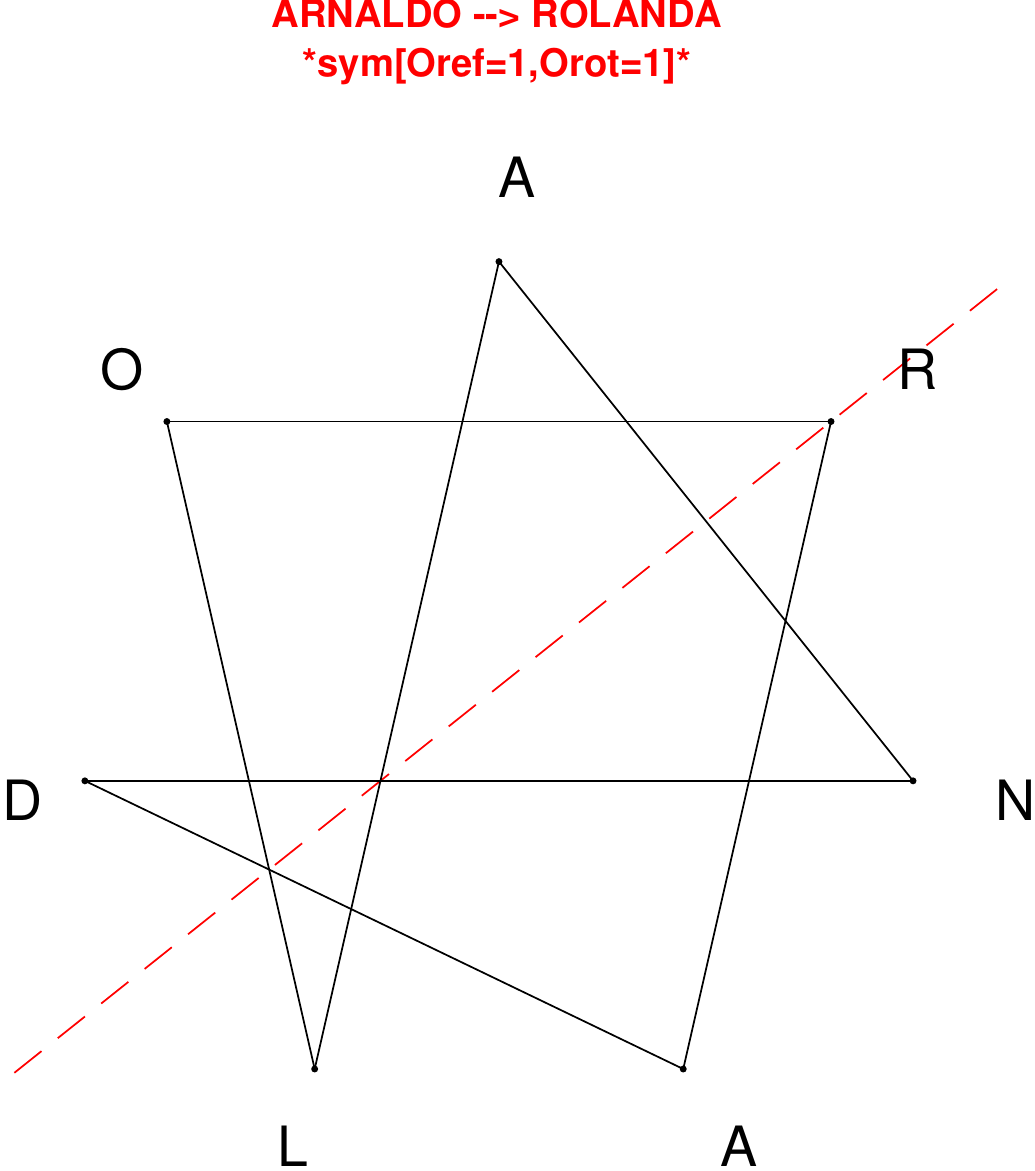}
\end{subfigure}
\end{figure}

\begin{figure}[H]
\centering
\begin{subfigure}[T]{0.19\textwidth}
\centering
\includegraphics[width=\textwidth]{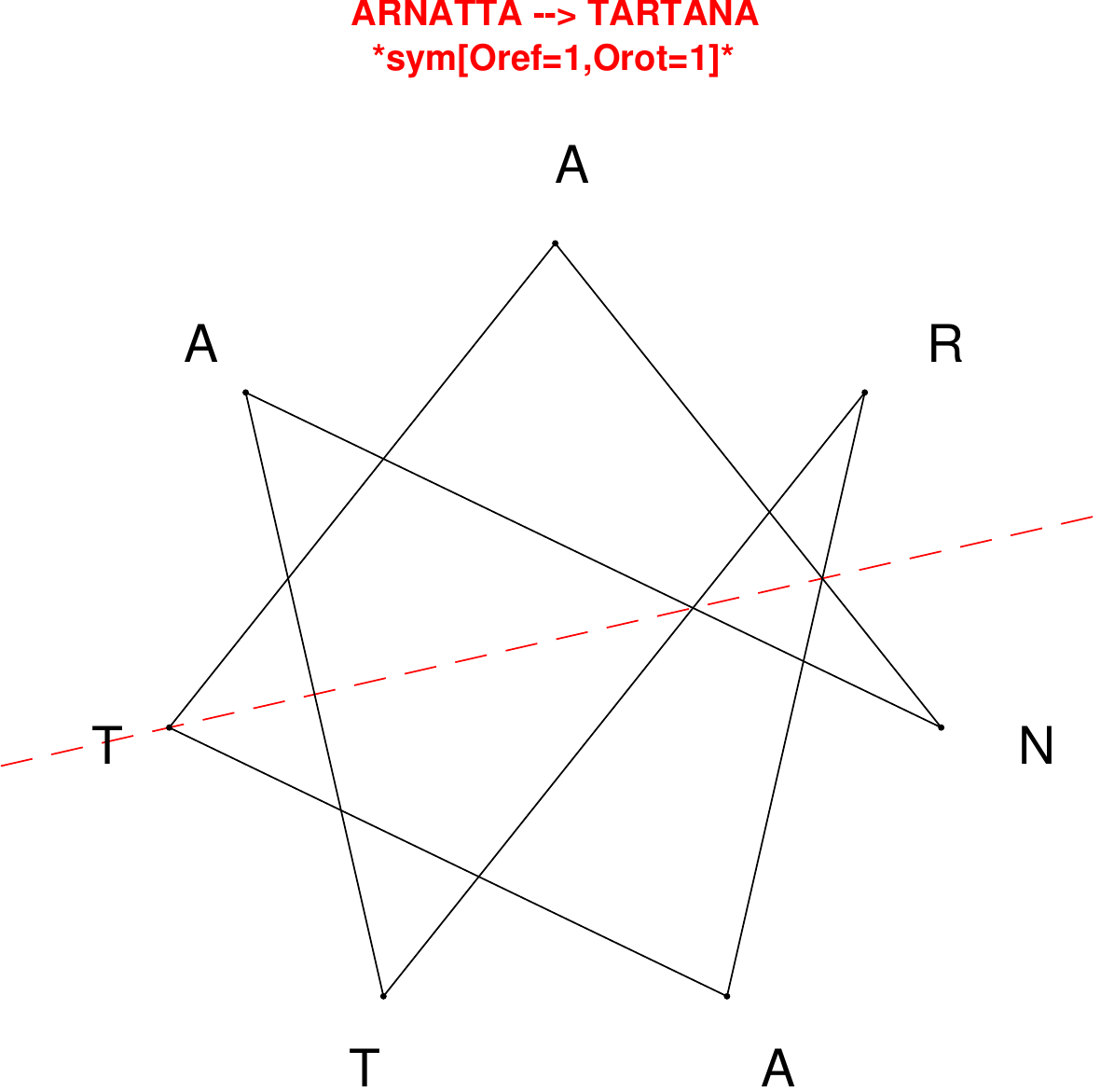}
\end{subfigure}
\hfill
\begin{subfigure}[T]{0.19\textwidth}
\centering
\includegraphics[width=\textwidth]{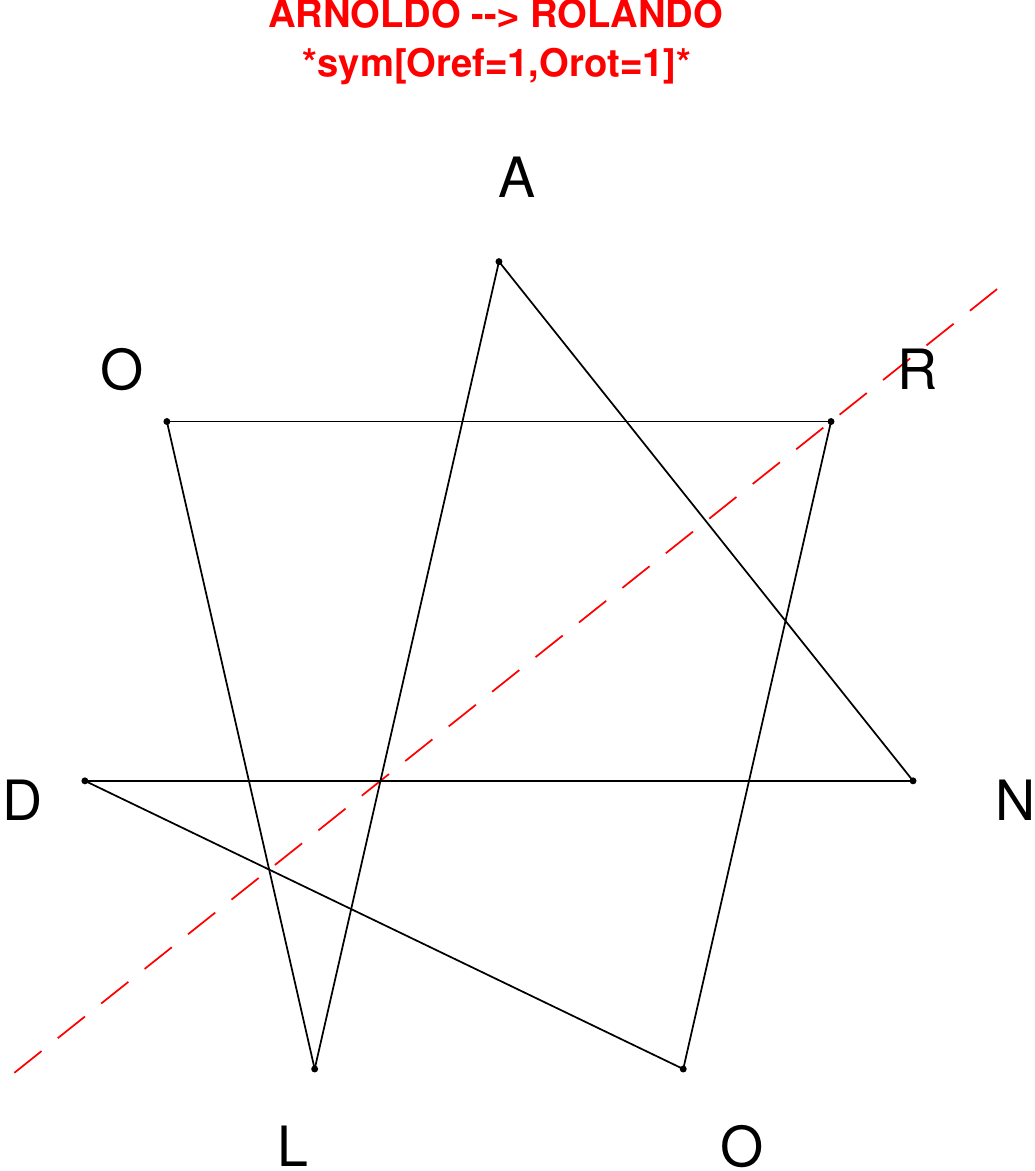}
\end{subfigure}
\hfill
\begin{subfigure}[T]{0.19\textwidth}
\centering
\includegraphics[width=\textwidth]{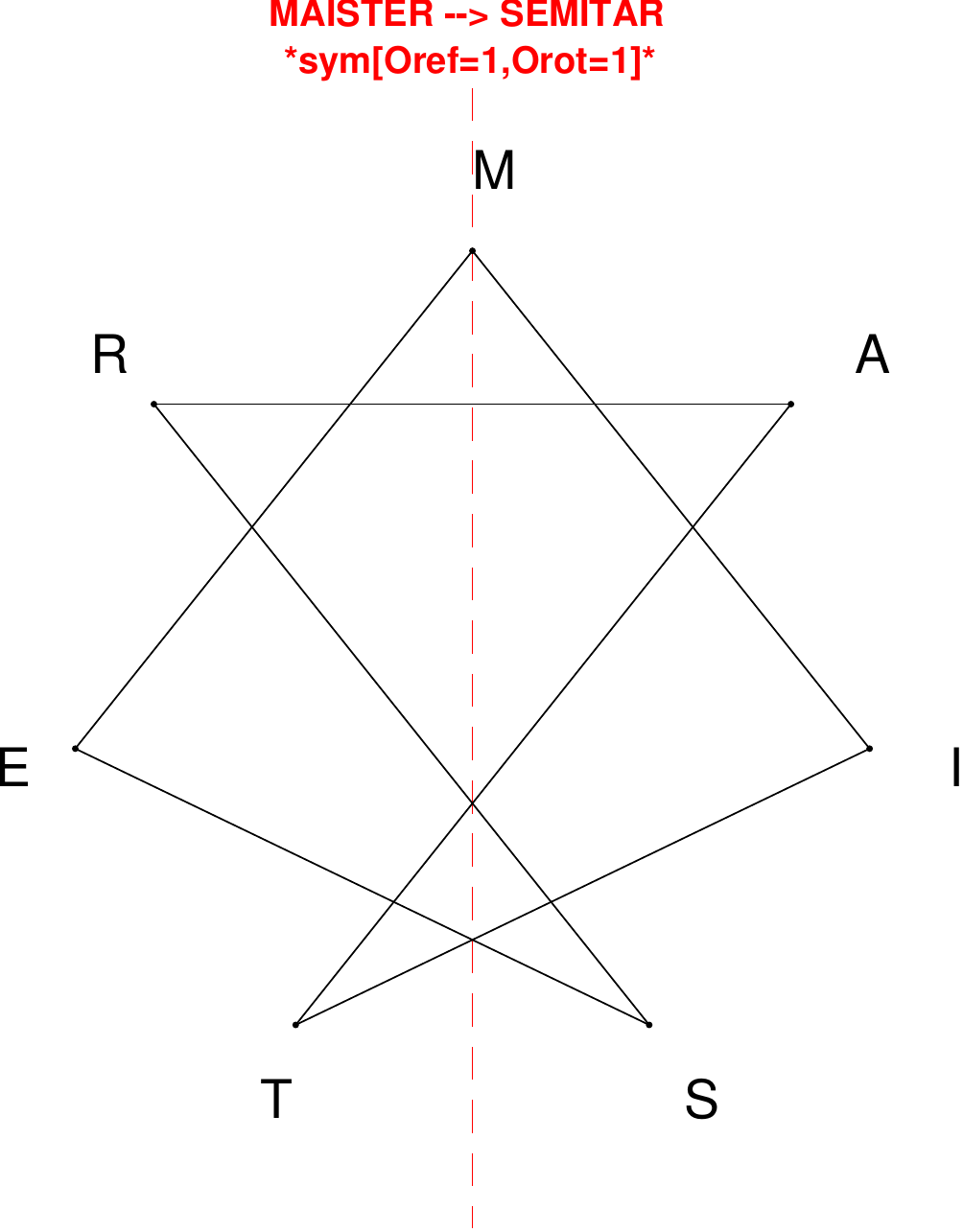}
\end{subfigure}
\hfill
\begin{subfigure}[T]{0.19\textwidth}
\centering
\includegraphics[width=\textwidth]{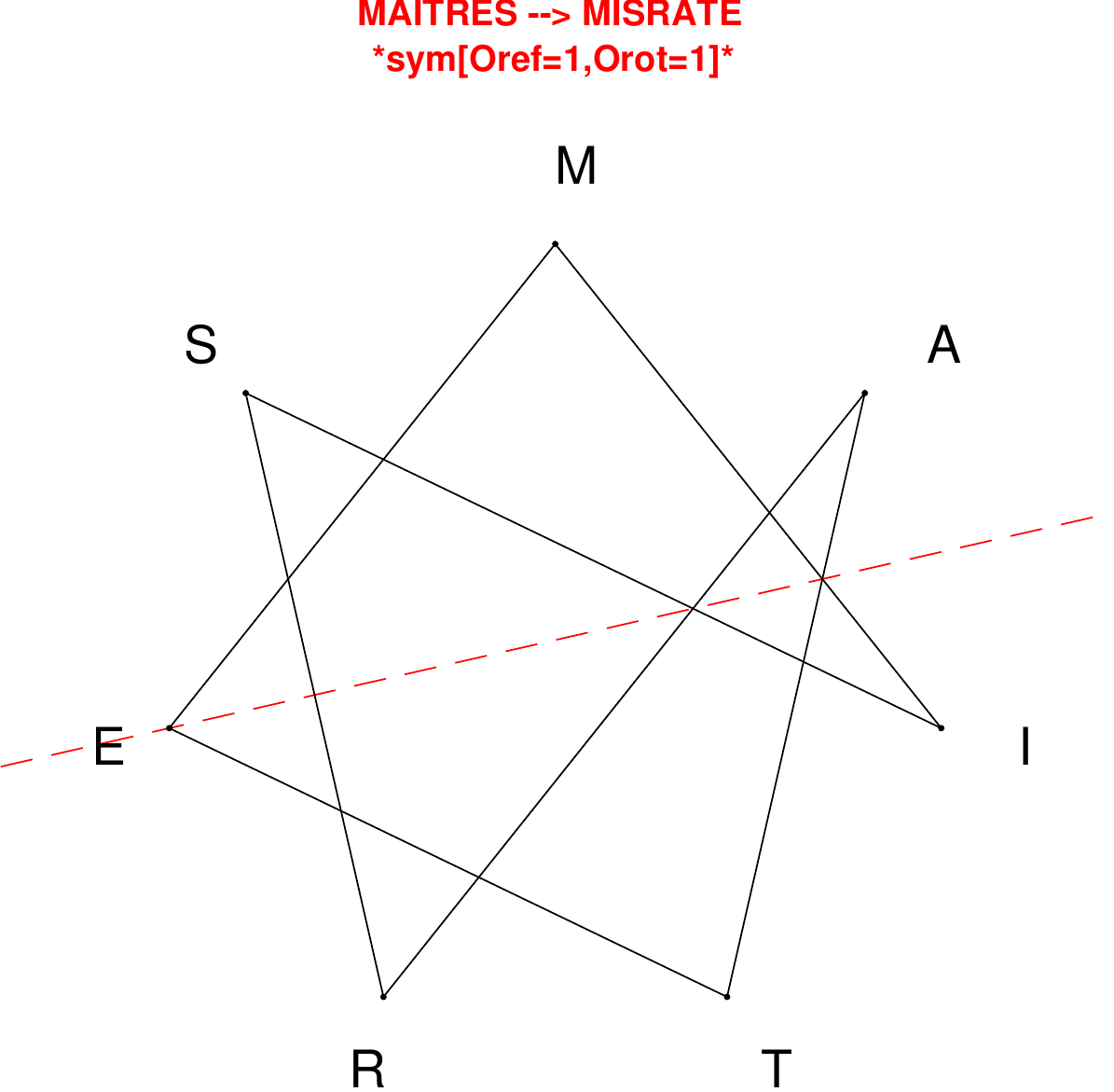}
\end{subfigure}
\hfill
\begin{subfigure}[T]{0.19\textwidth}
\centering
\includegraphics[width=\textwidth]{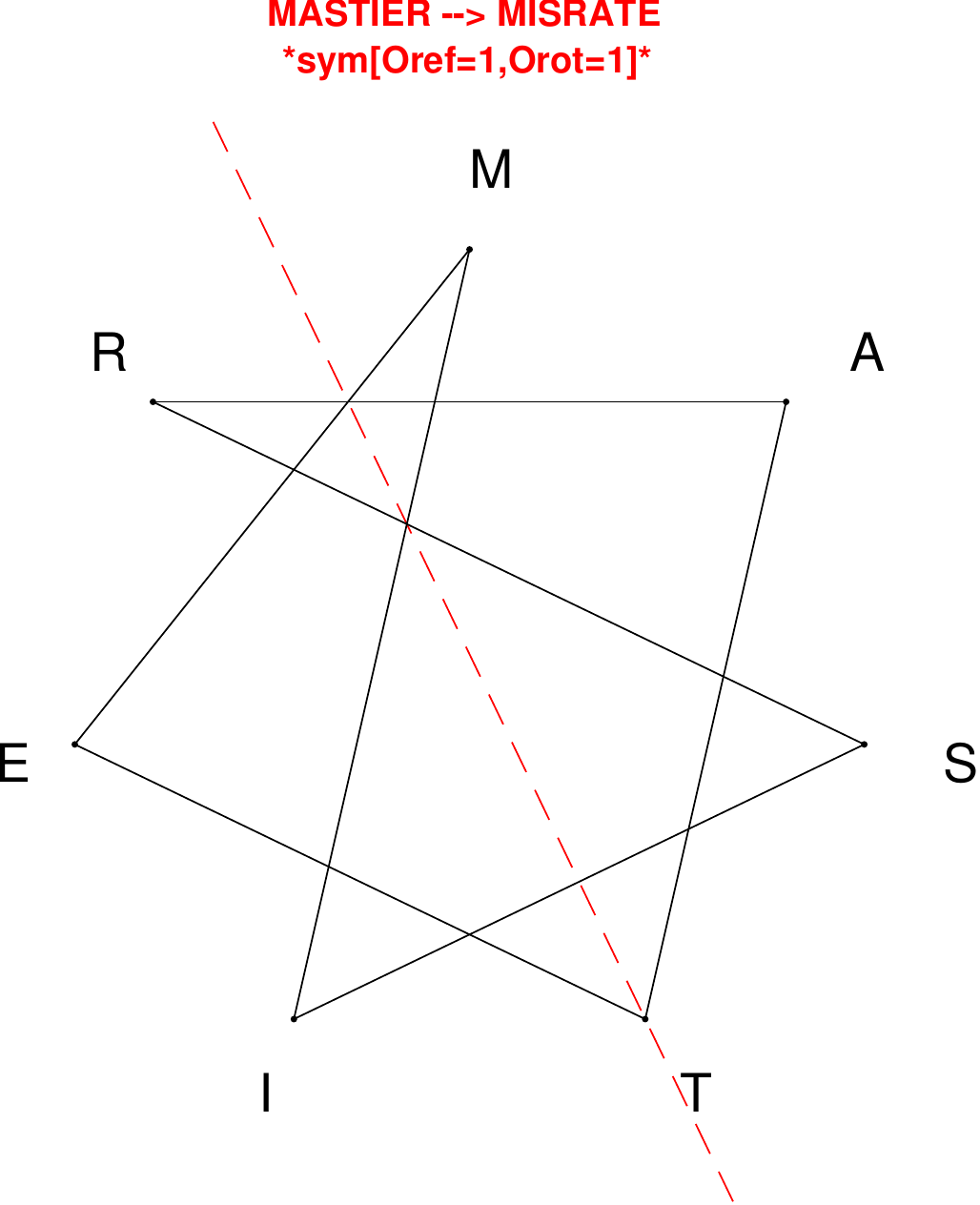}
\end{subfigure}
\end{figure}

\begin{figure}[H]
\centering
\begin{subfigure}[T]{0.19\textwidth}
\centering
\includegraphics[width=\textwidth]{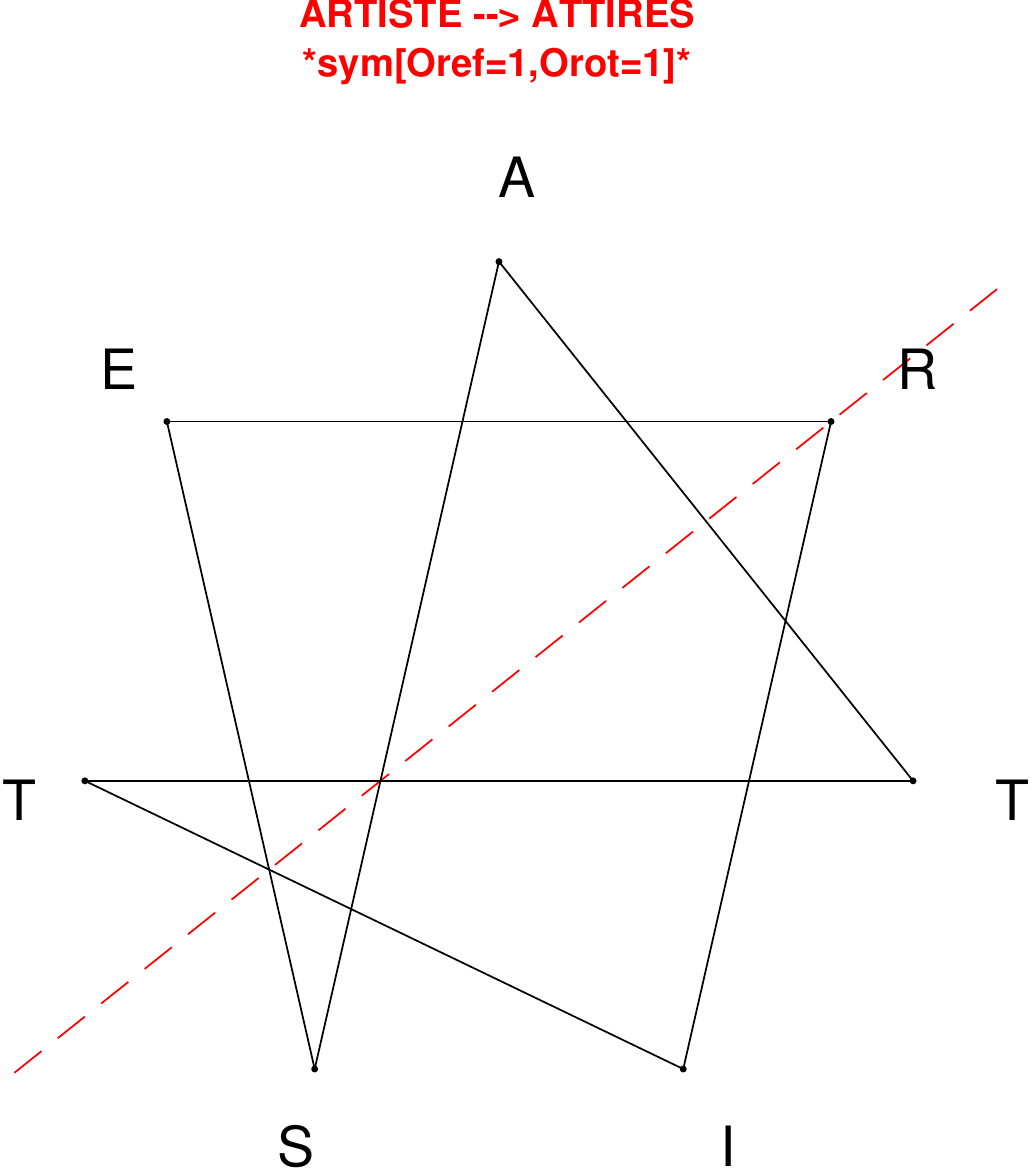}
\end{subfigure}
\hfill
\begin{subfigure}[T]{0.19\textwidth}
\centering
\includegraphics[width=\textwidth]{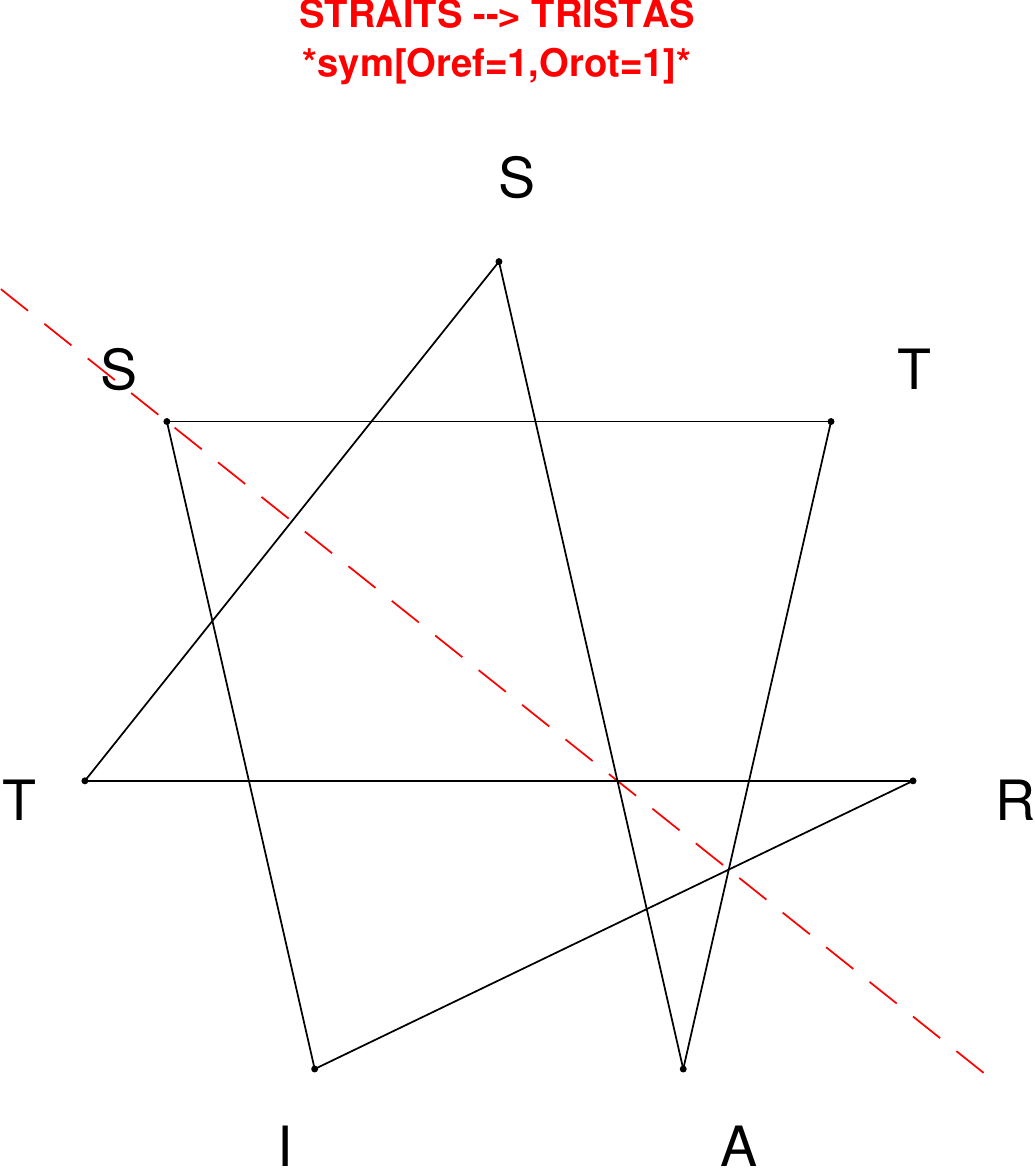}
\end{subfigure}
\hfill
\begin{subfigure}[T]{0.19\textwidth}
\centering
\includegraphics[width=\textwidth]{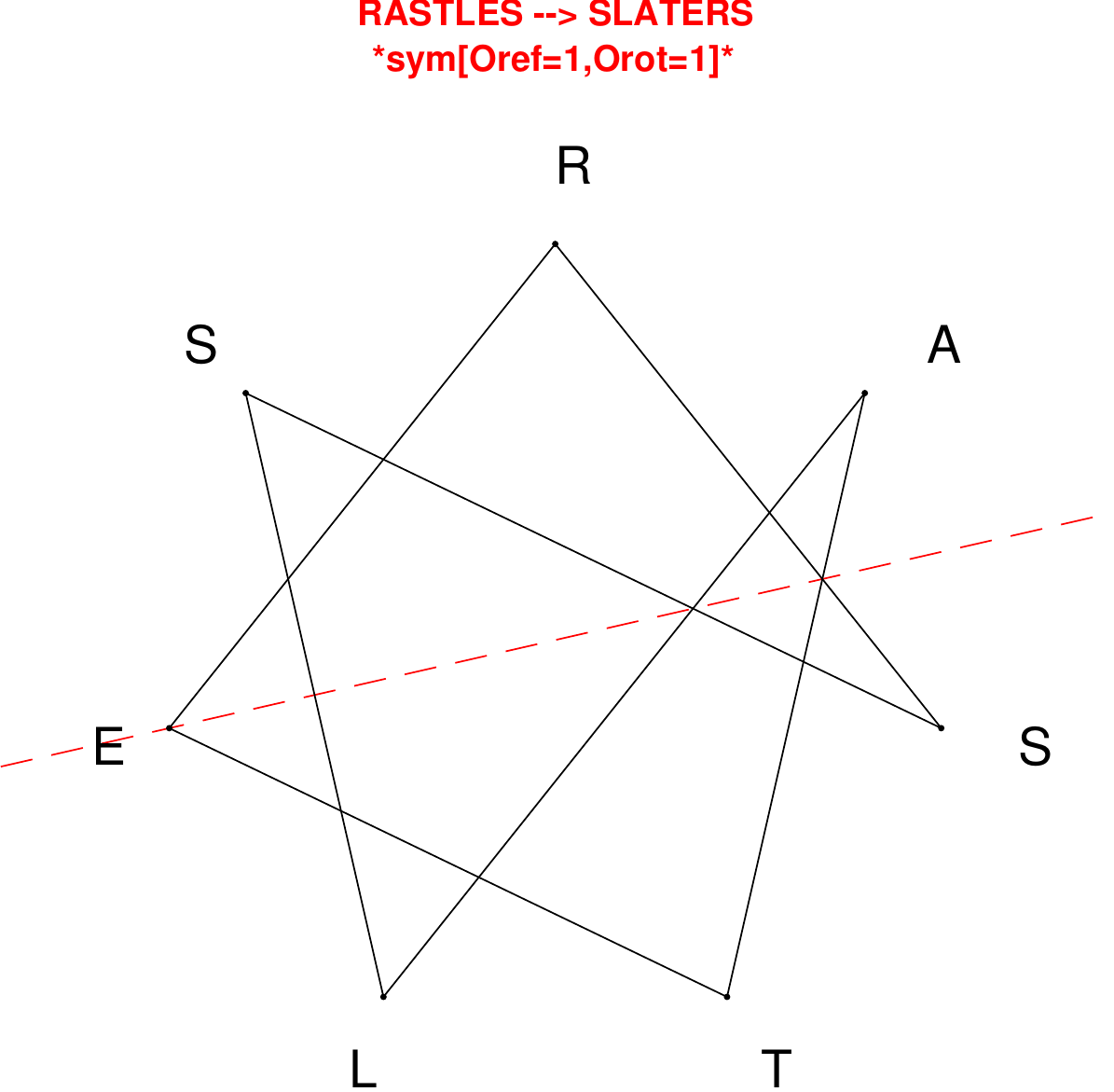}
\end{subfigure}
\hfill
\begin{subfigure}[T]{0.19\textwidth}
\centering
\includegraphics[width=\textwidth]{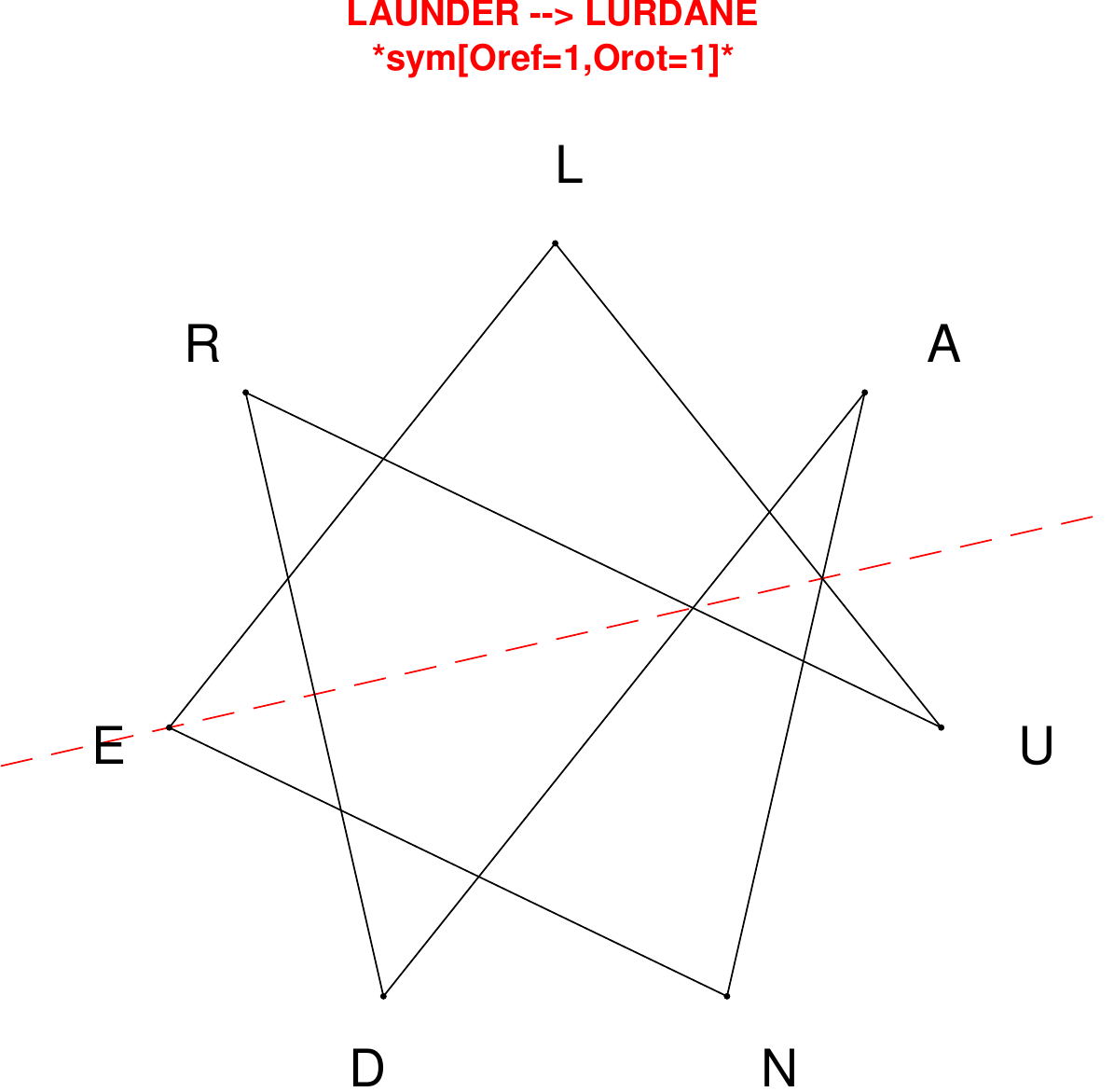}
\end{subfigure}
\hfill
\begin{subfigure}[T]{0.19\textwidth}
\centering
\includegraphics[width=\textwidth]{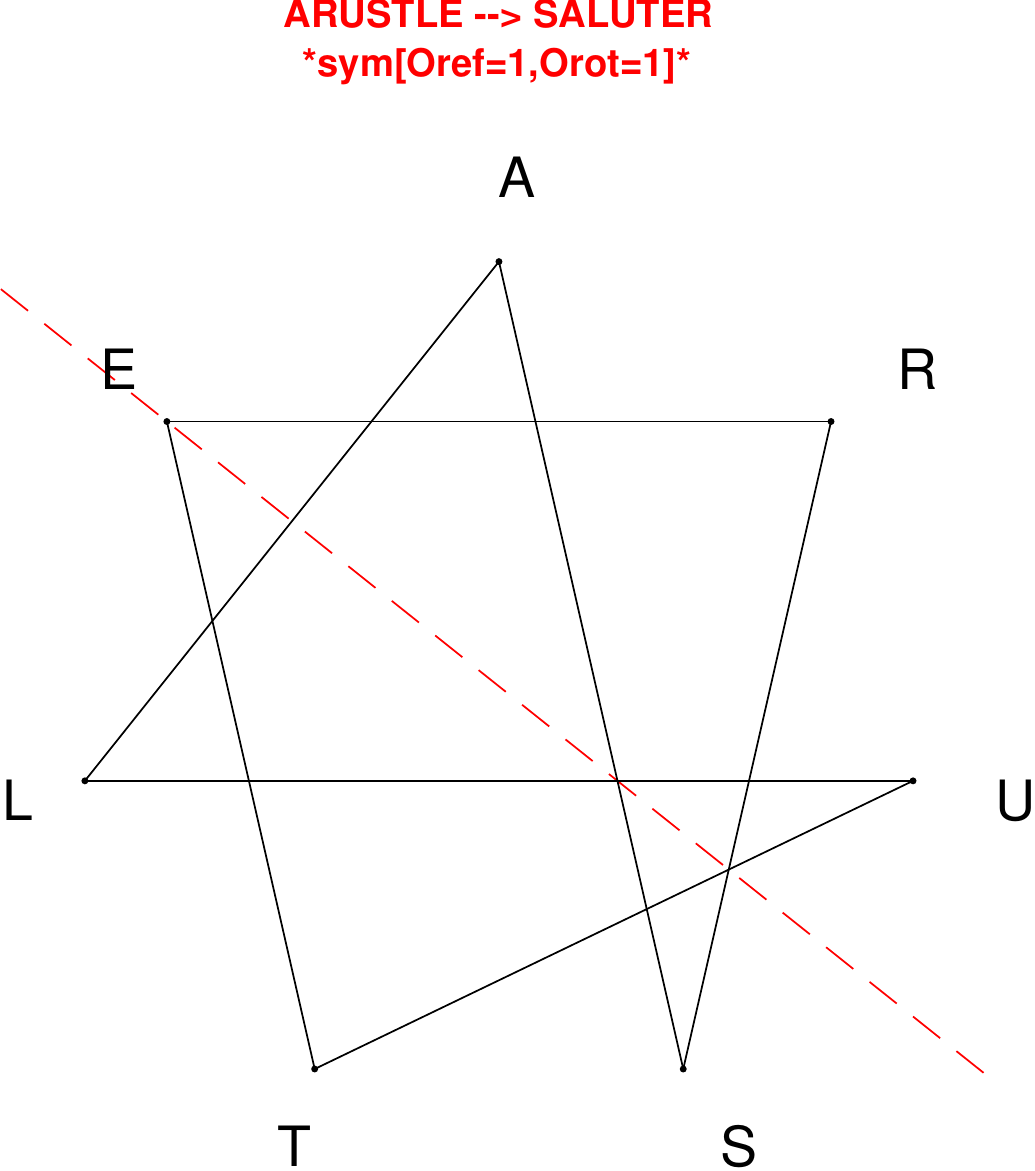}
\end{subfigure}
\end{figure}

\begin{figure}[H]
\centering
\begin{subfigure}[T]{0.19\textwidth}
\centering
\includegraphics[width=\textwidth]{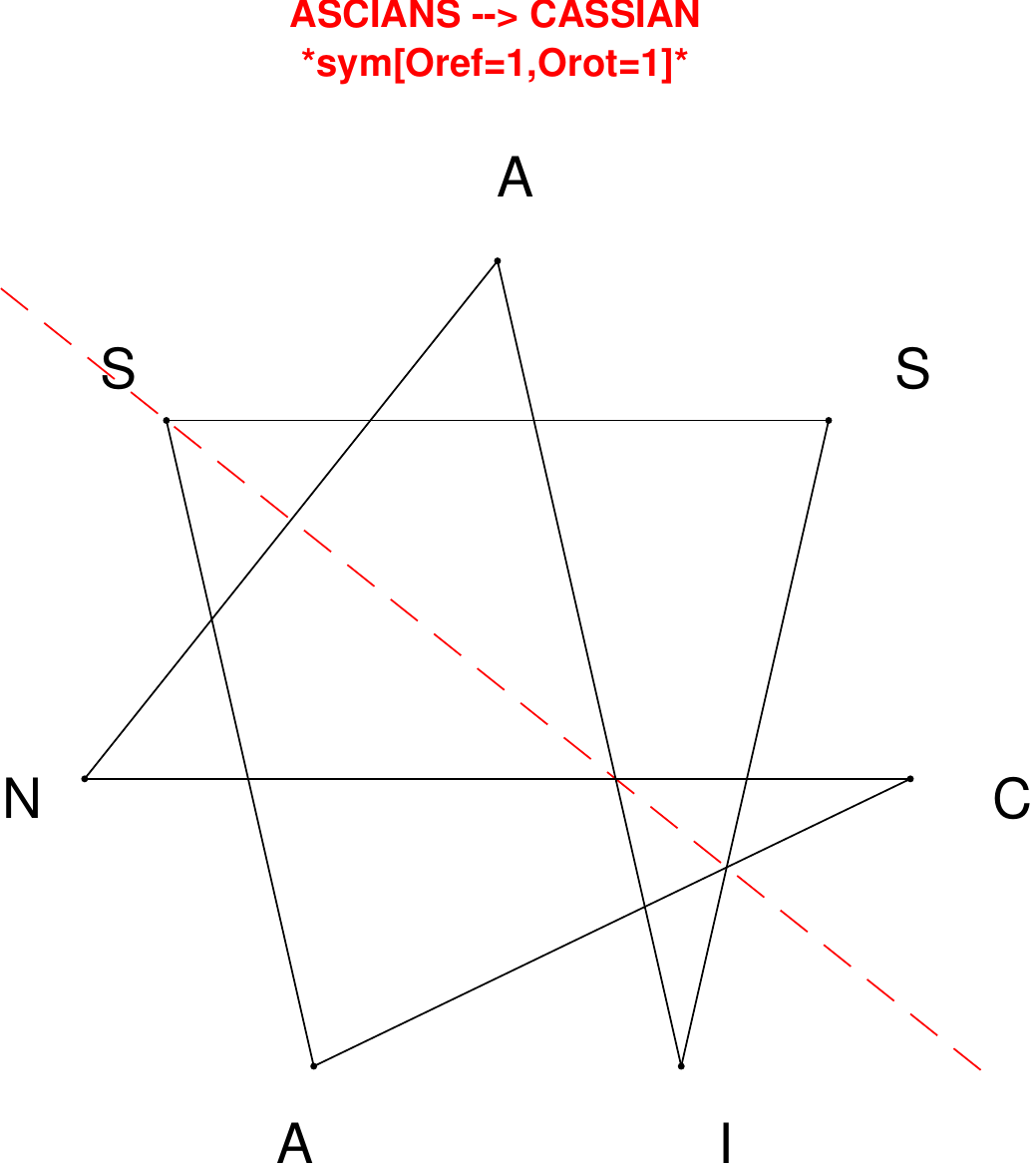}
\end{subfigure}
\hfill
\begin{subfigure}[T]{0.19\textwidth}
\centering
\includegraphics[width=\textwidth]{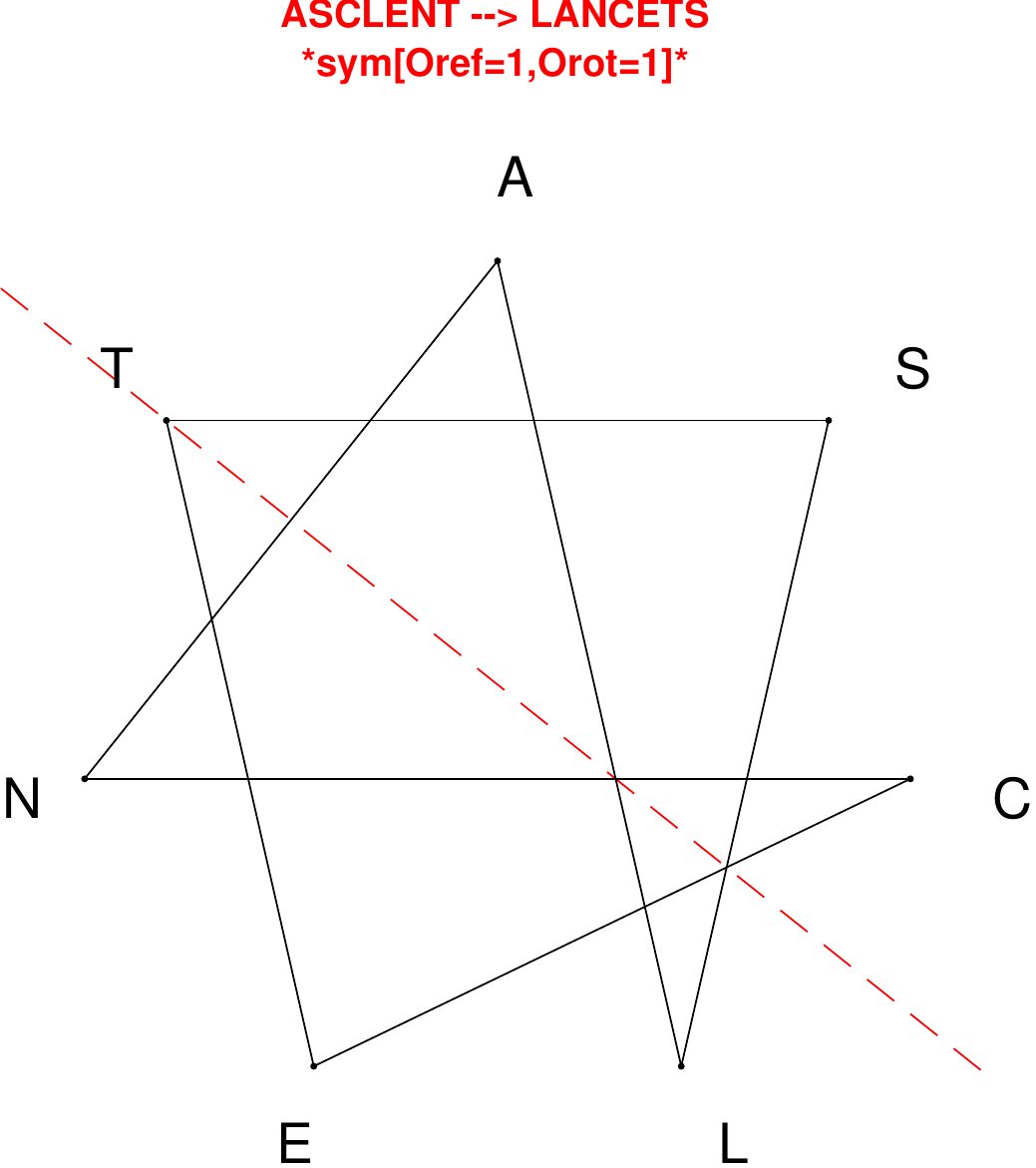}
\end{subfigure}
\hfill
\begin{subfigure}[T]{0.19\textwidth}
\centering
\includegraphics[width=\textwidth]{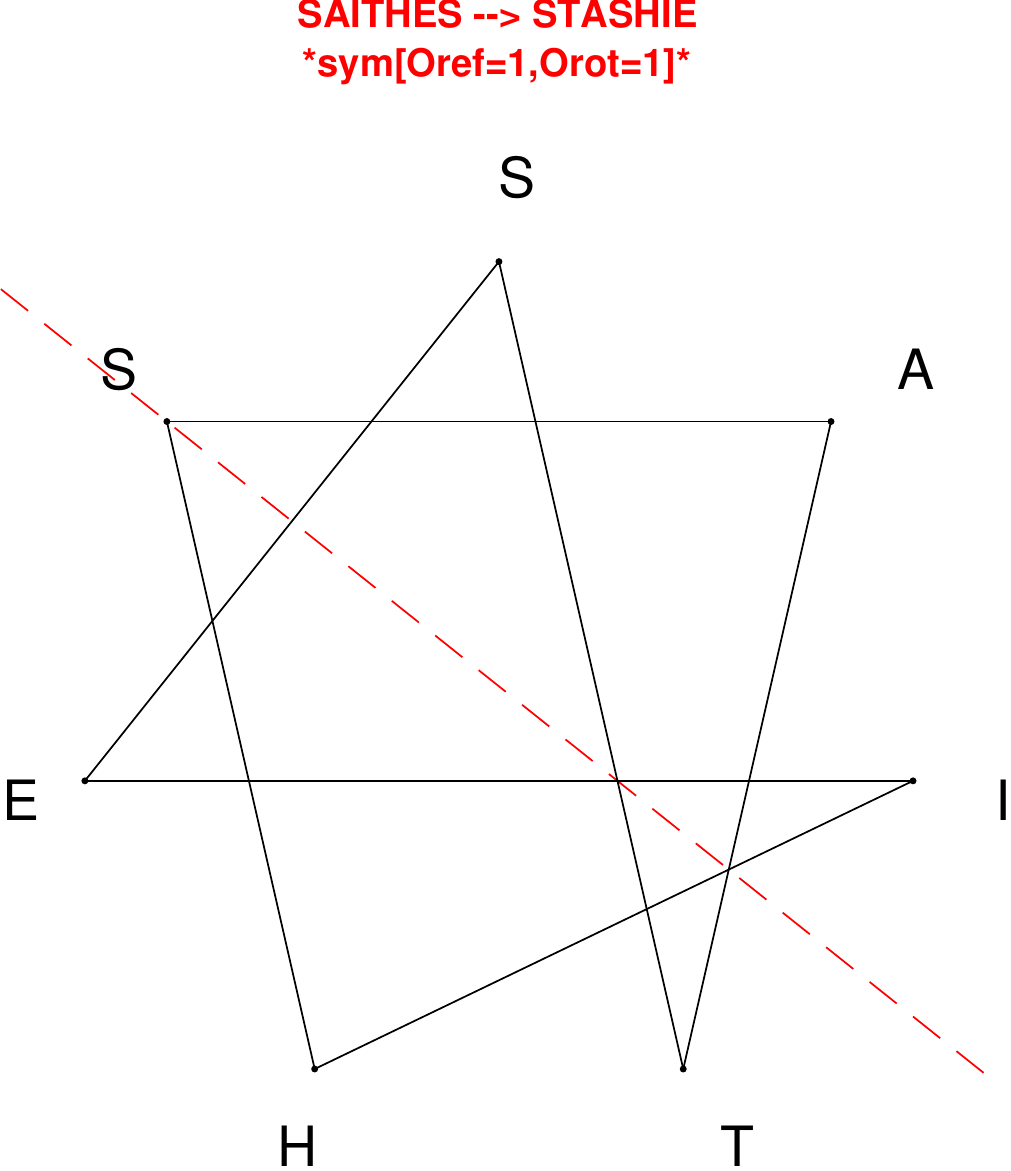}
\end{subfigure}
\hfill
\begin{subfigure}[T]{0.19\textwidth}
\centering
\includegraphics[width=\textwidth]{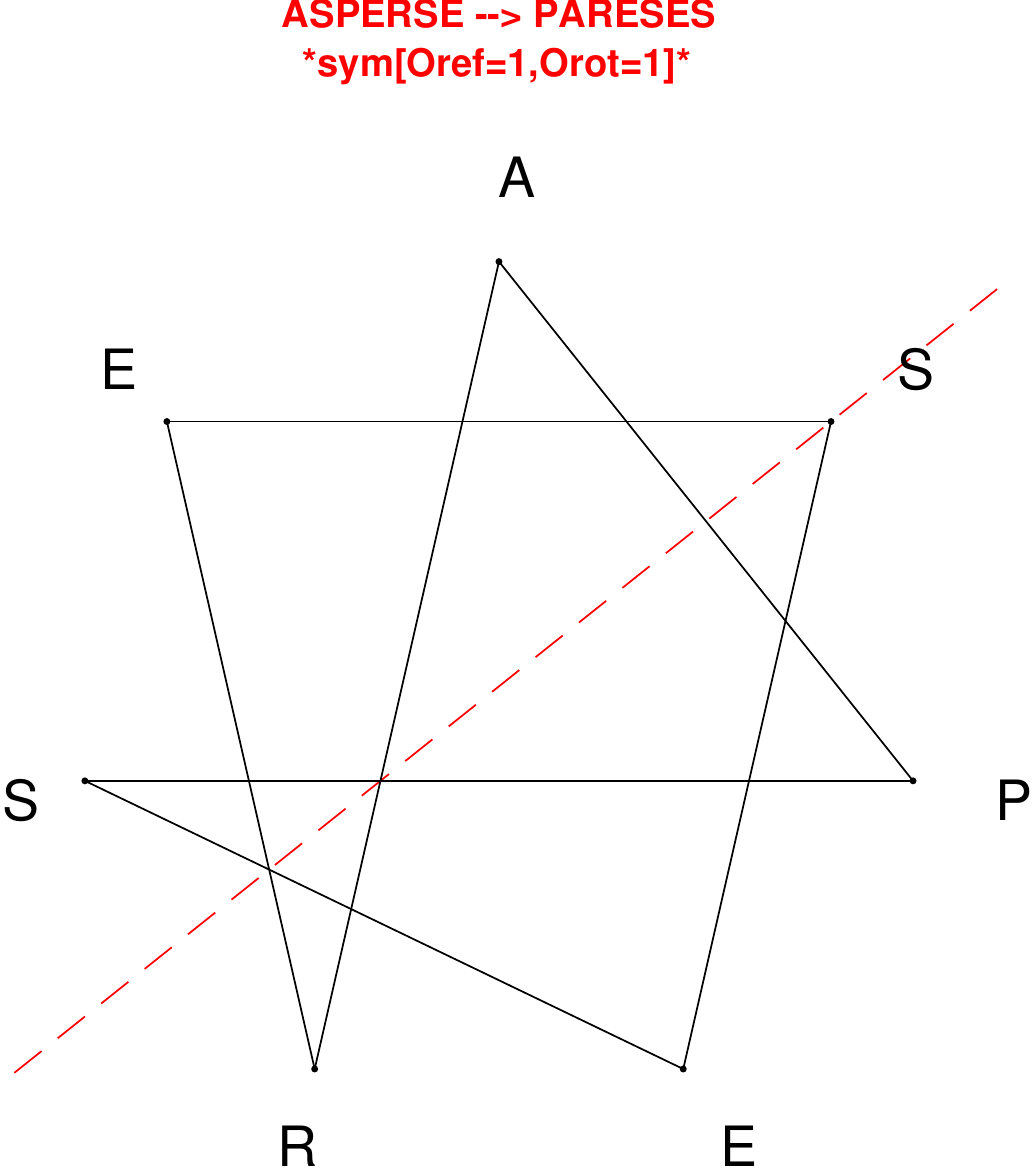}
\end{subfigure}
\hfill
\begin{subfigure}[T]{0.19\textwidth}
\centering
\includegraphics[width=\textwidth]{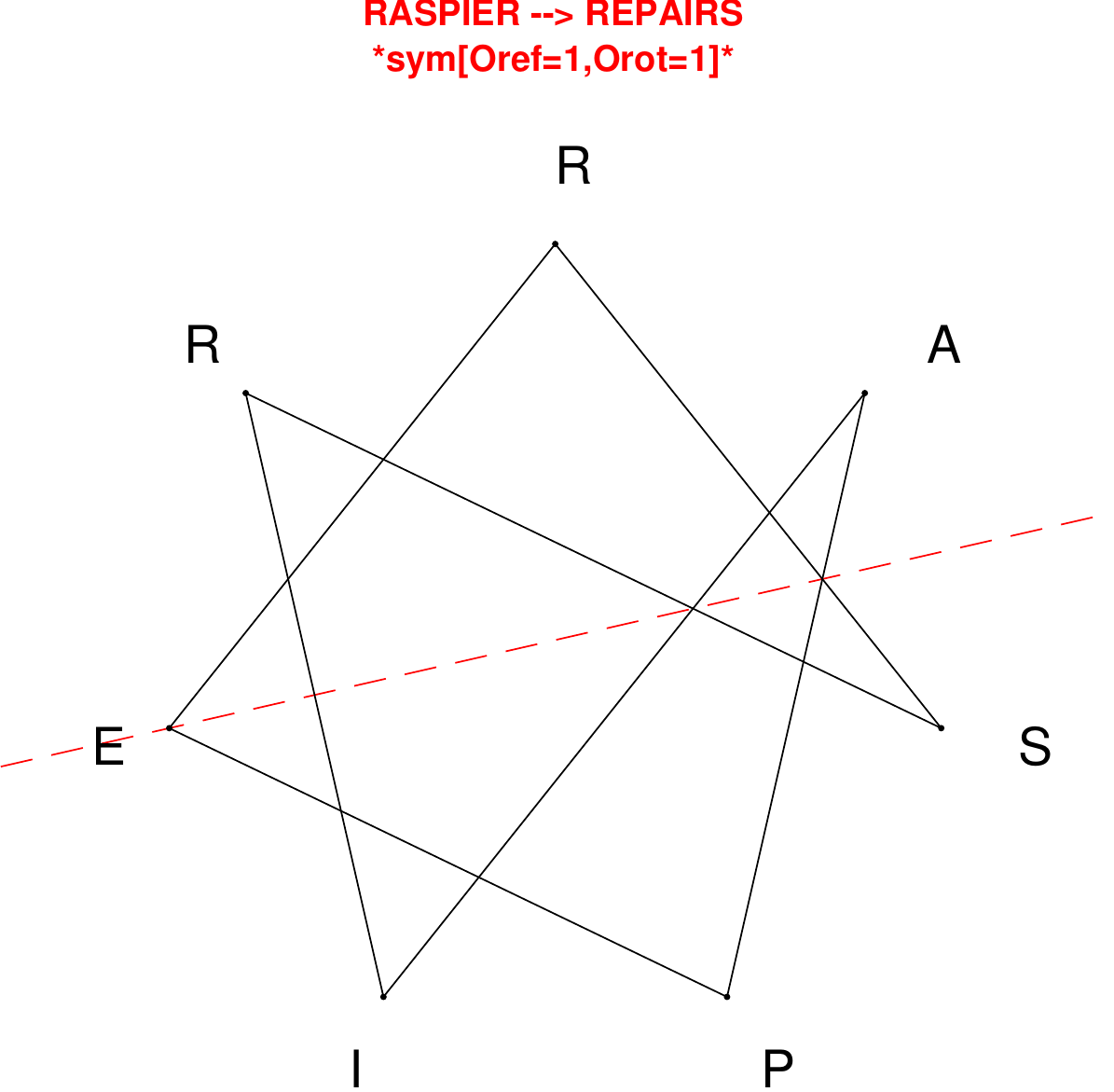}
\end{subfigure}
\end{figure}

\begin{figure}[H]
\centering
\begin{subfigure}[T]{0.19\textwidth}
\centering
\includegraphics[width=\textwidth]{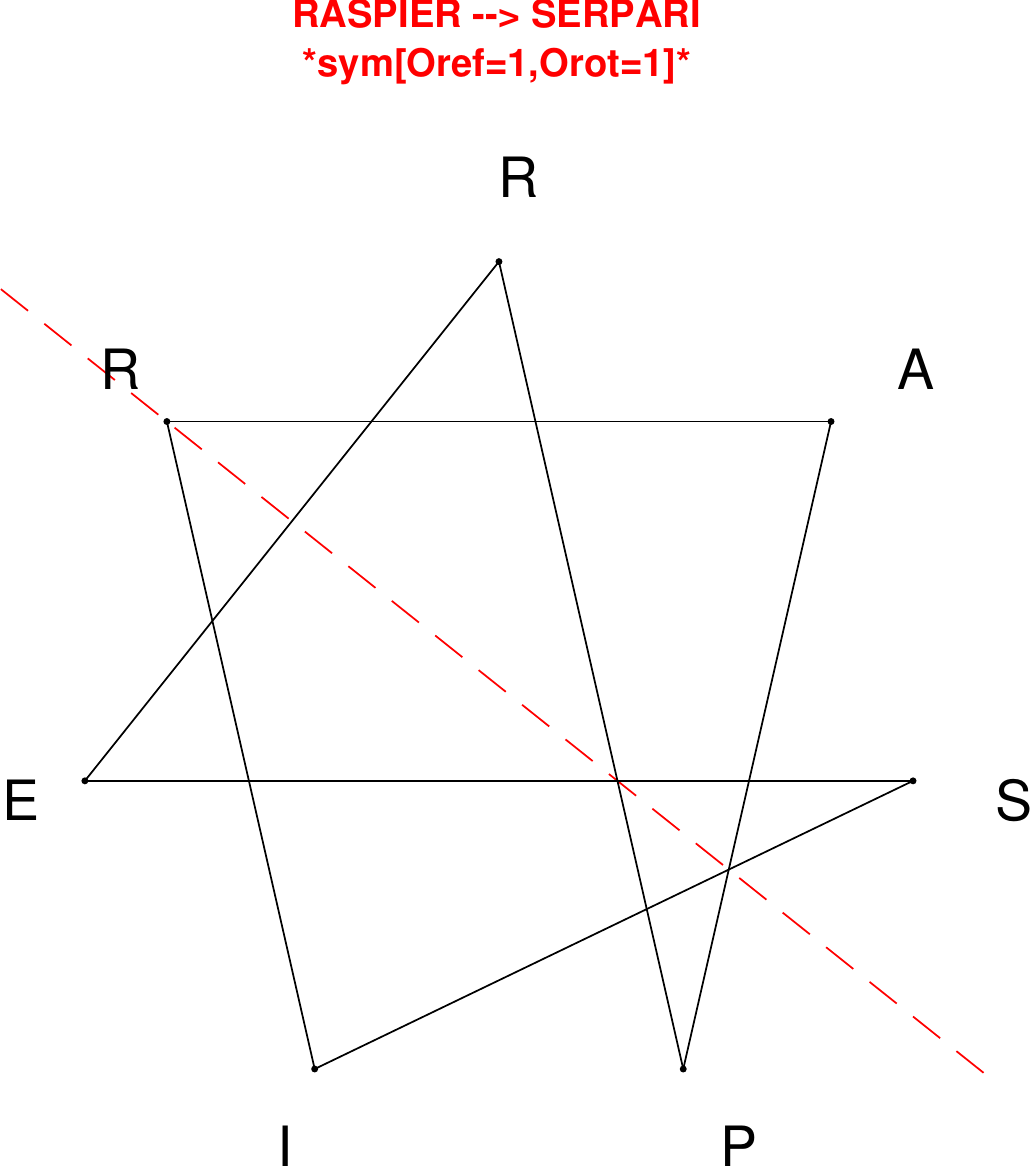}
\end{subfigure}
\hfill
\begin{subfigure}[T]{0.19\textwidth}
\centering
\includegraphics[width=\textwidth]{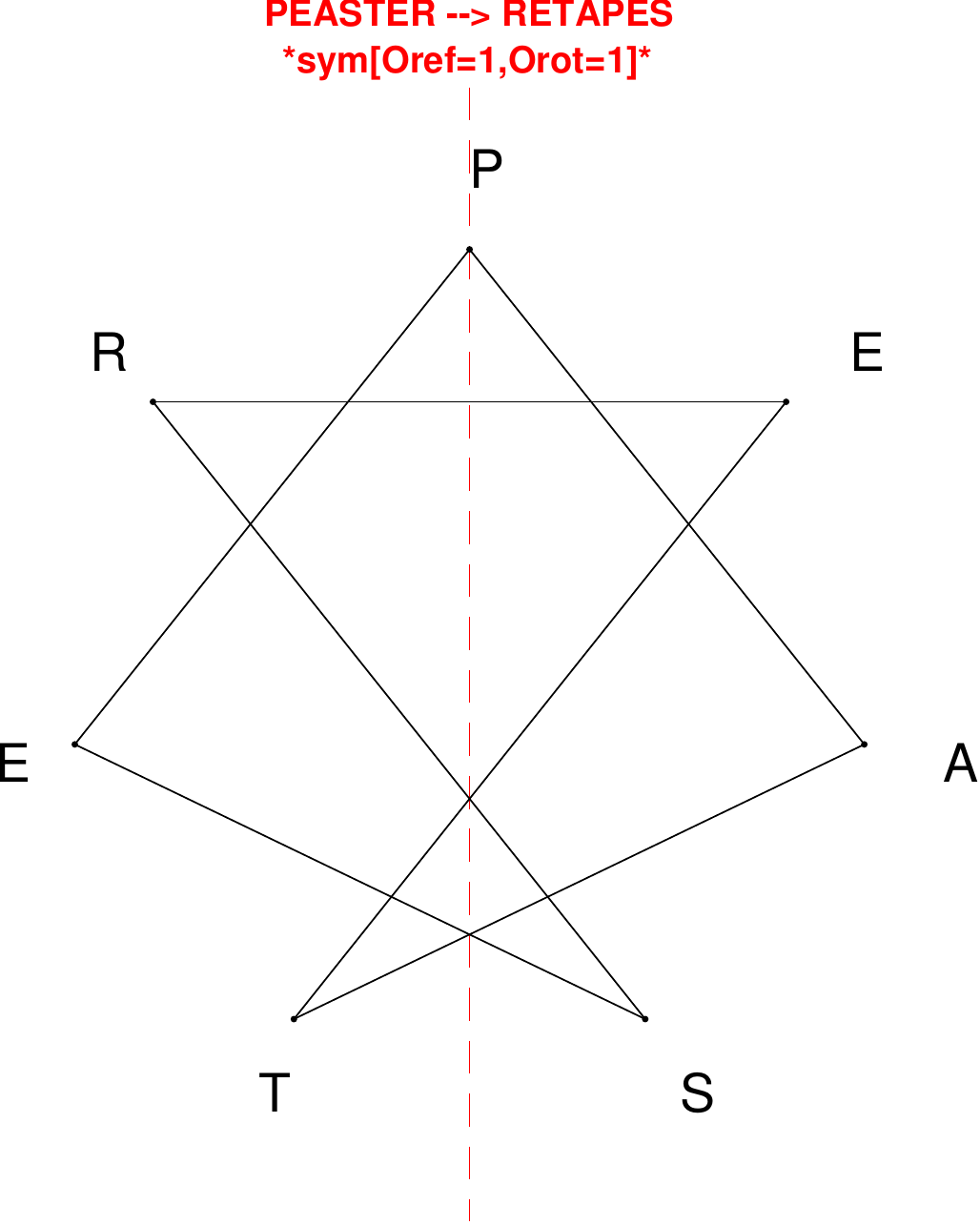}
\end{subfigure}
\hfill
\begin{subfigure}[T]{0.19\textwidth}
\centering
\includegraphics[width=\textwidth]{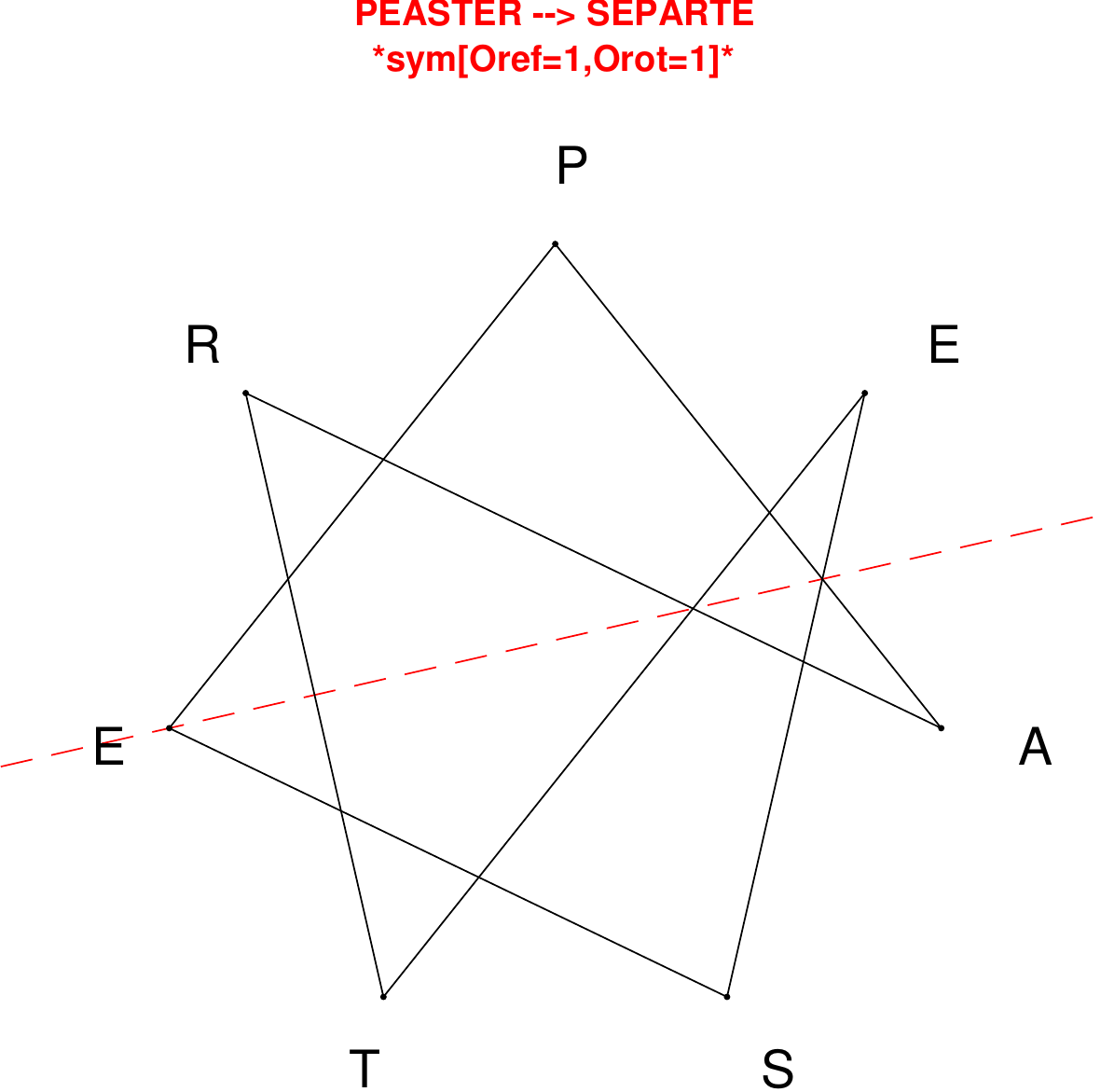}
\end{subfigure}
\hfill
\begin{subfigure}[T]{0.19\textwidth}
\centering
\includegraphics[width=\textwidth]{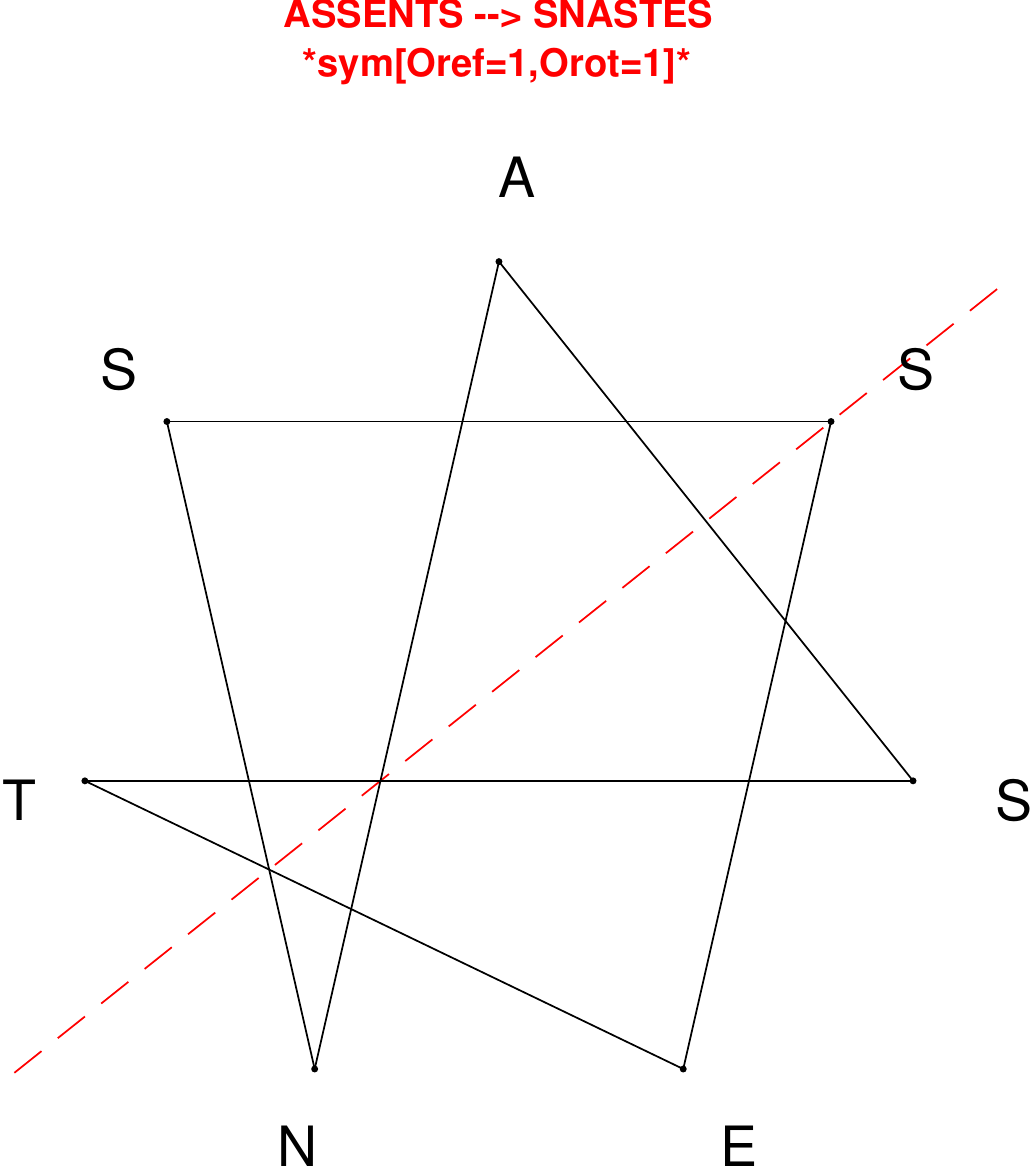}
\end{subfigure}
\hfill
\begin{subfigure}[T]{0.19\textwidth}
\centering
\includegraphics[width=\textwidth]{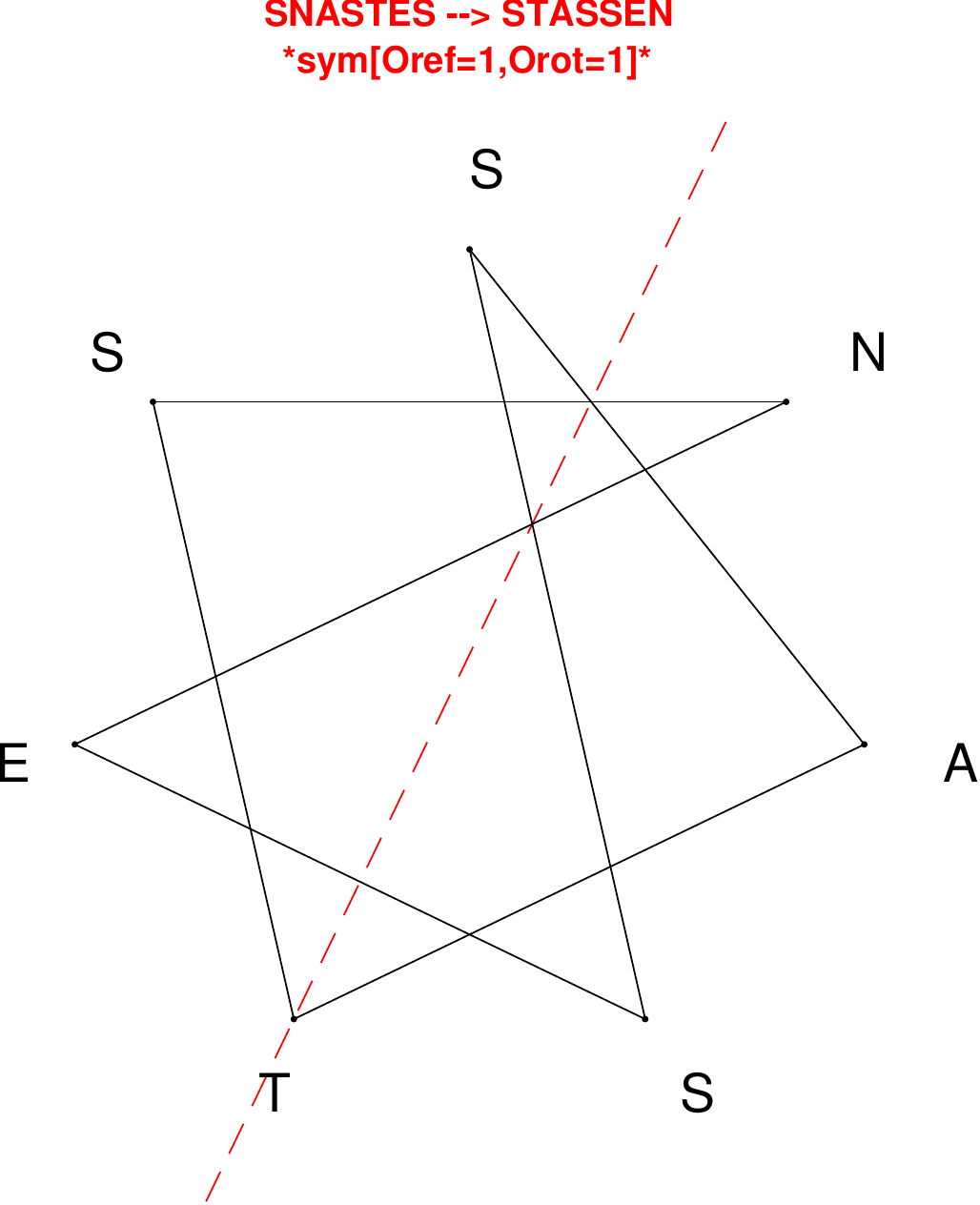}
\end{subfigure}
\end{figure}

\begin{figure}[H]
\centering
\begin{subfigure}[T]{0.19\textwidth}
\centering
\includegraphics[width=\textwidth]{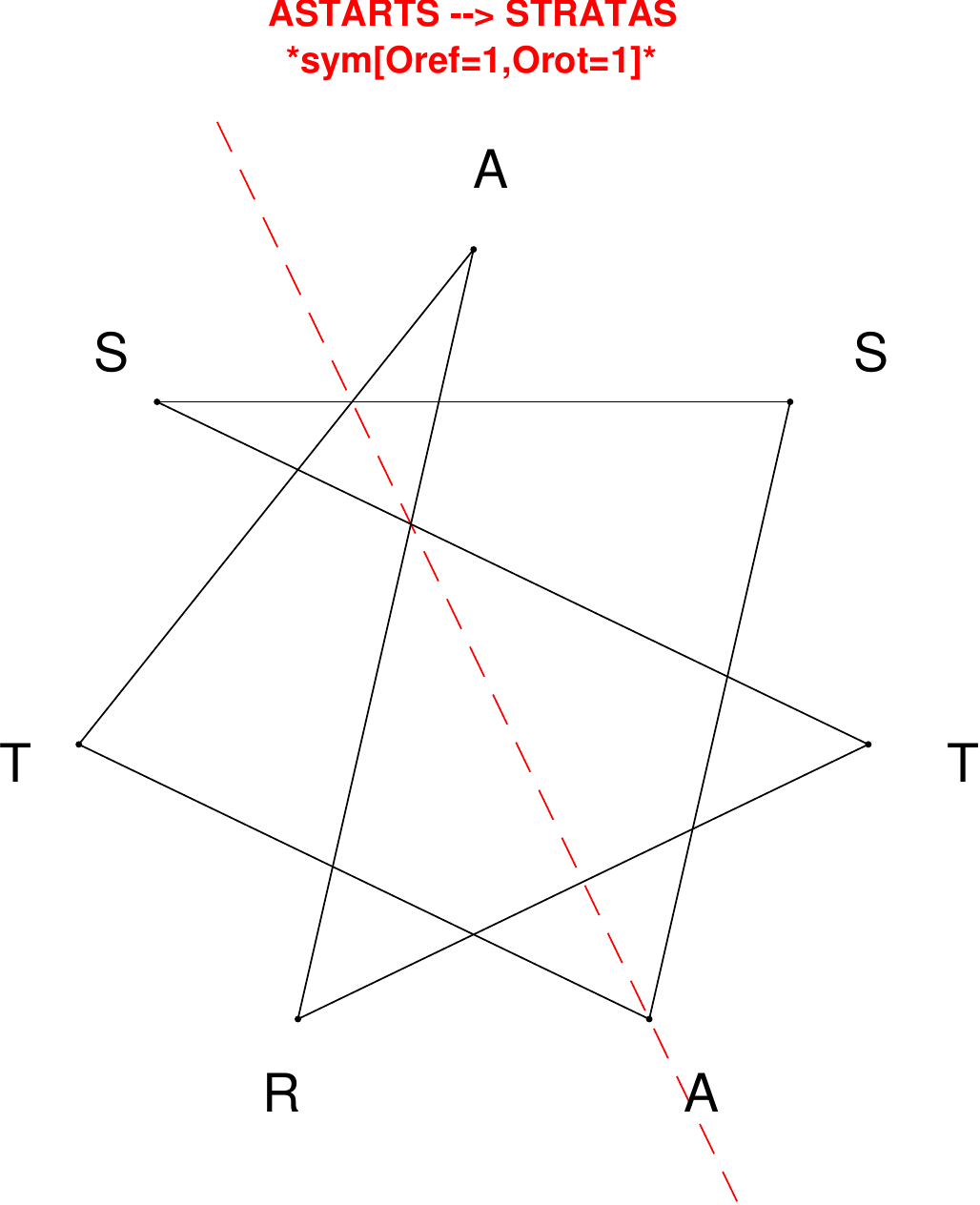}
\end{subfigure}
\hfill
\begin{subfigure}[T]{0.19\textwidth}
\centering
\includegraphics[width=\textwidth]{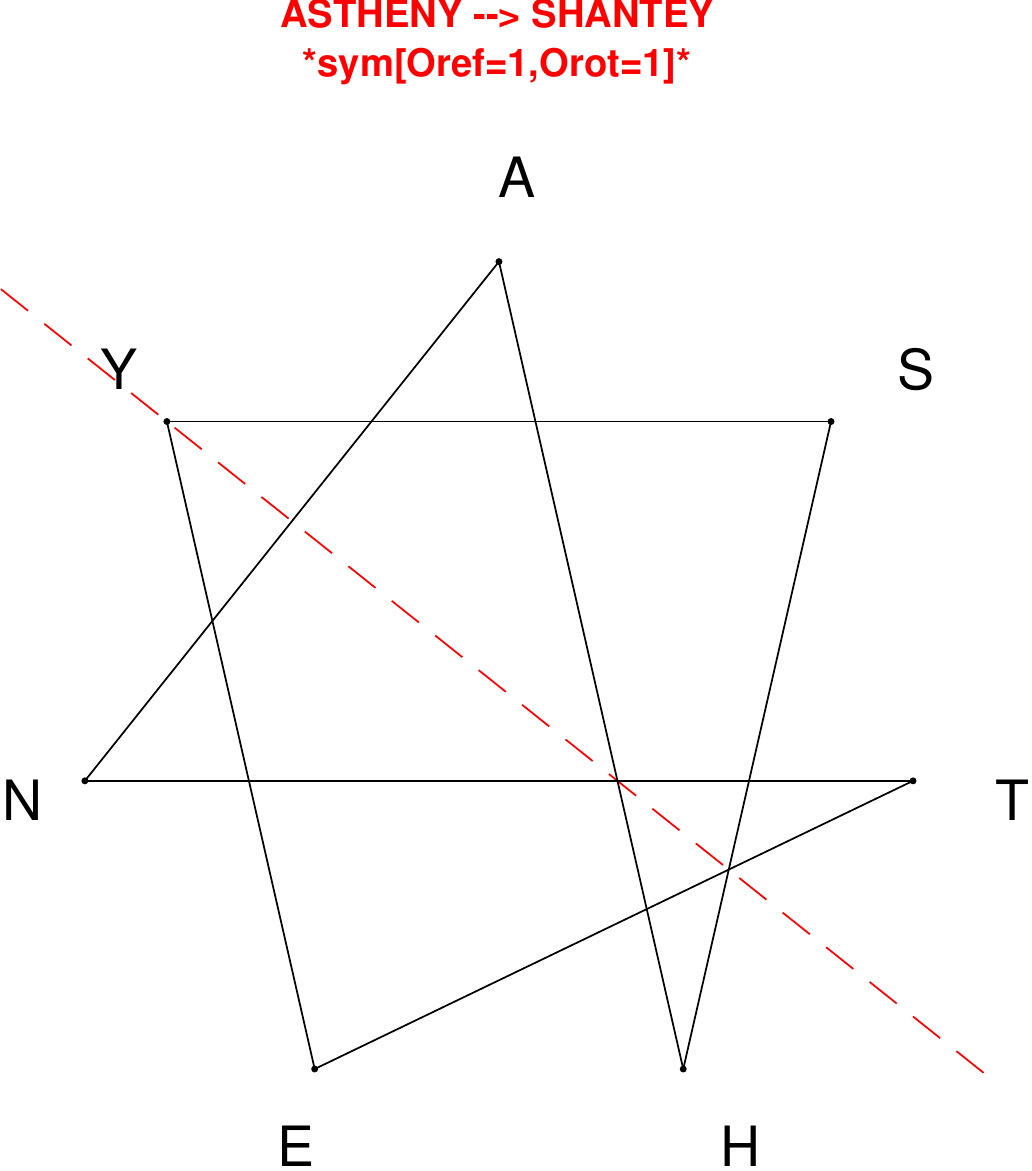}
\end{subfigure}
\hfill
\begin{subfigure}[T]{0.19\textwidth}
\centering
\includegraphics[width=\textwidth]{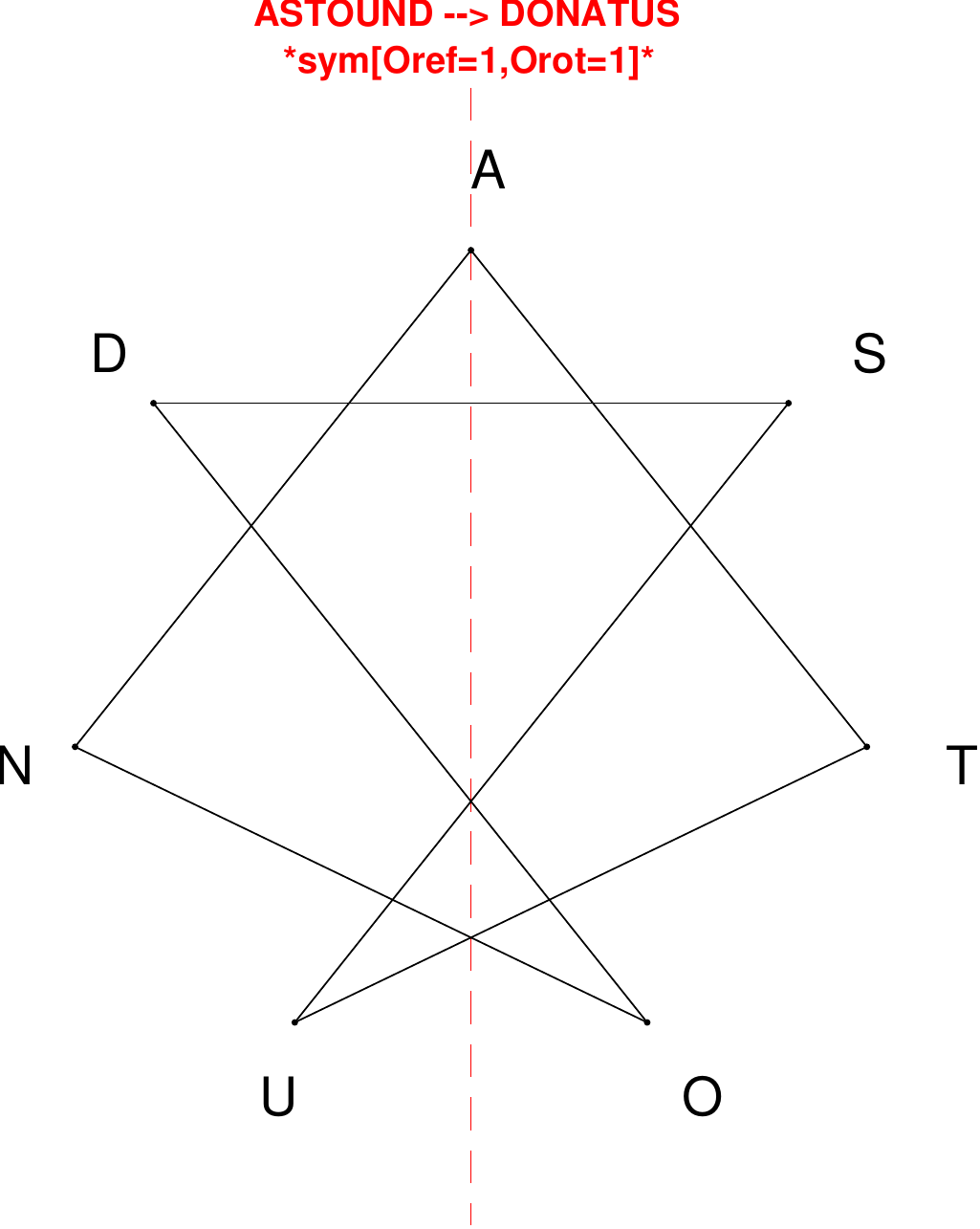}
\end{subfigure}
\hfill
\begin{subfigure}[T]{0.19\textwidth}
\centering
\includegraphics[width=\textwidth]{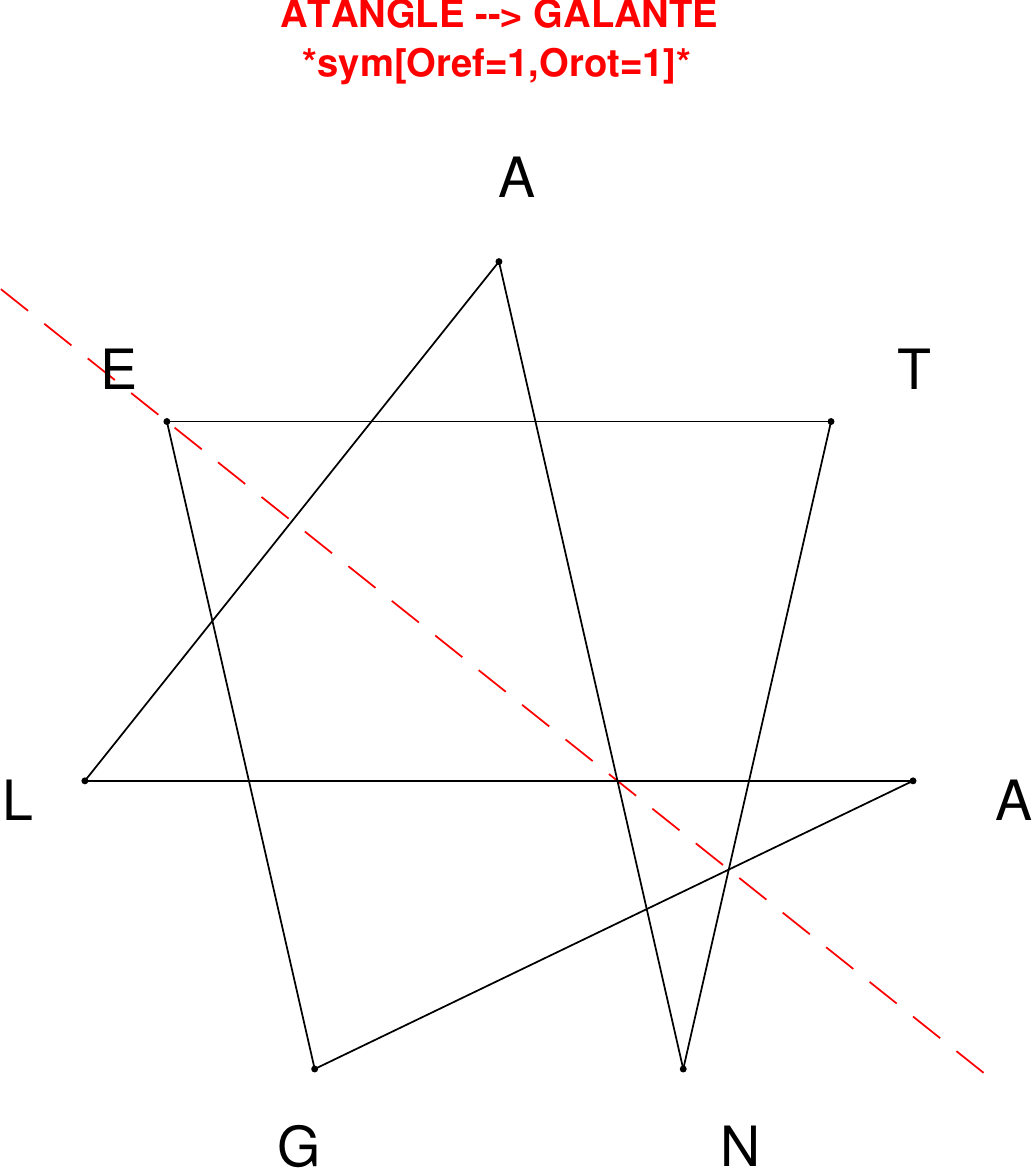}
\end{subfigure}
\hfill
\begin{subfigure}[T]{0.19\textwidth}
\centering
\includegraphics[width=\textwidth]{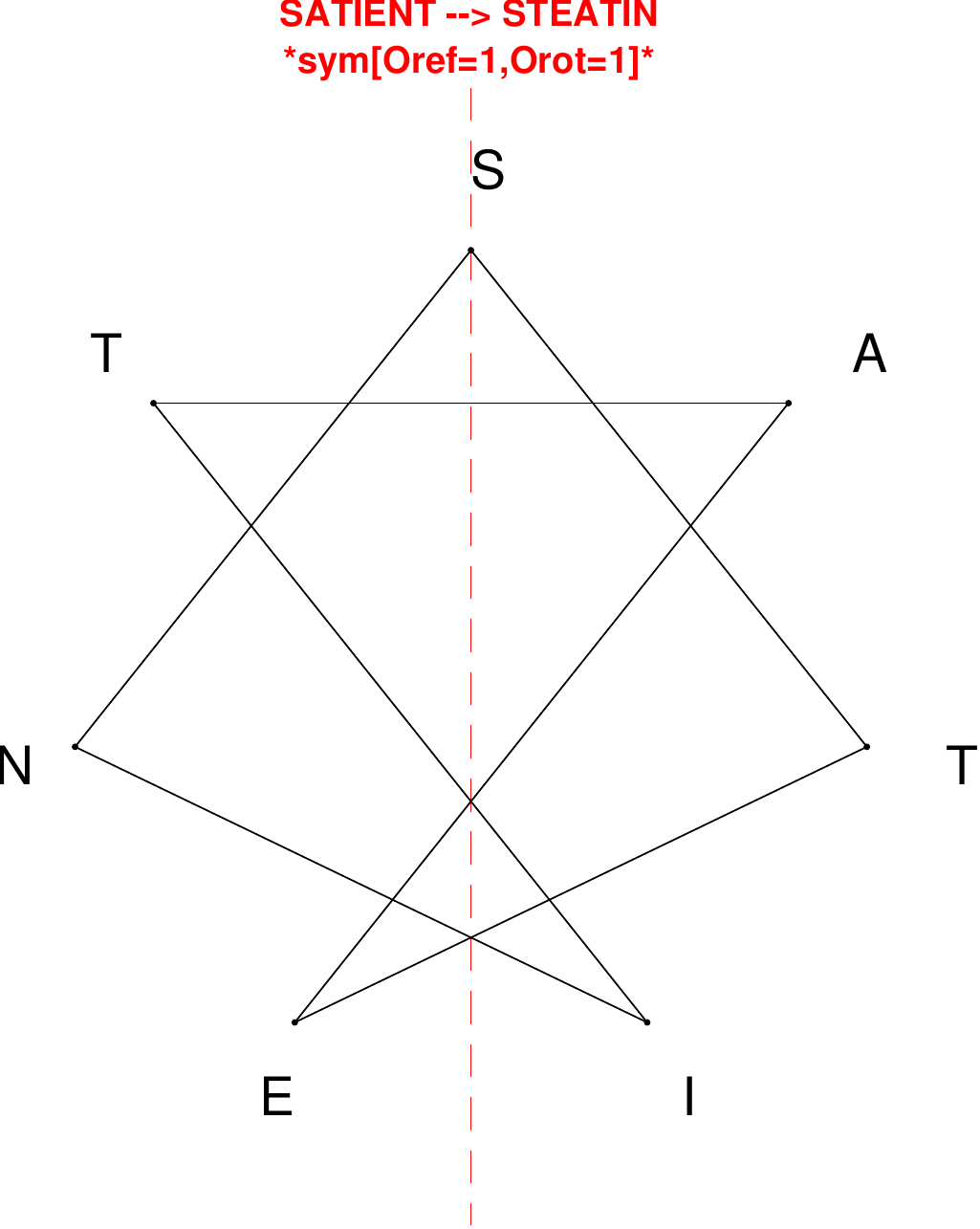}
\end{subfigure}
\end{figure}

\begin{figure}[H]
\centering
\begin{subfigure}[T]{0.19\textwidth}
\centering
\includegraphics[width=\textwidth]{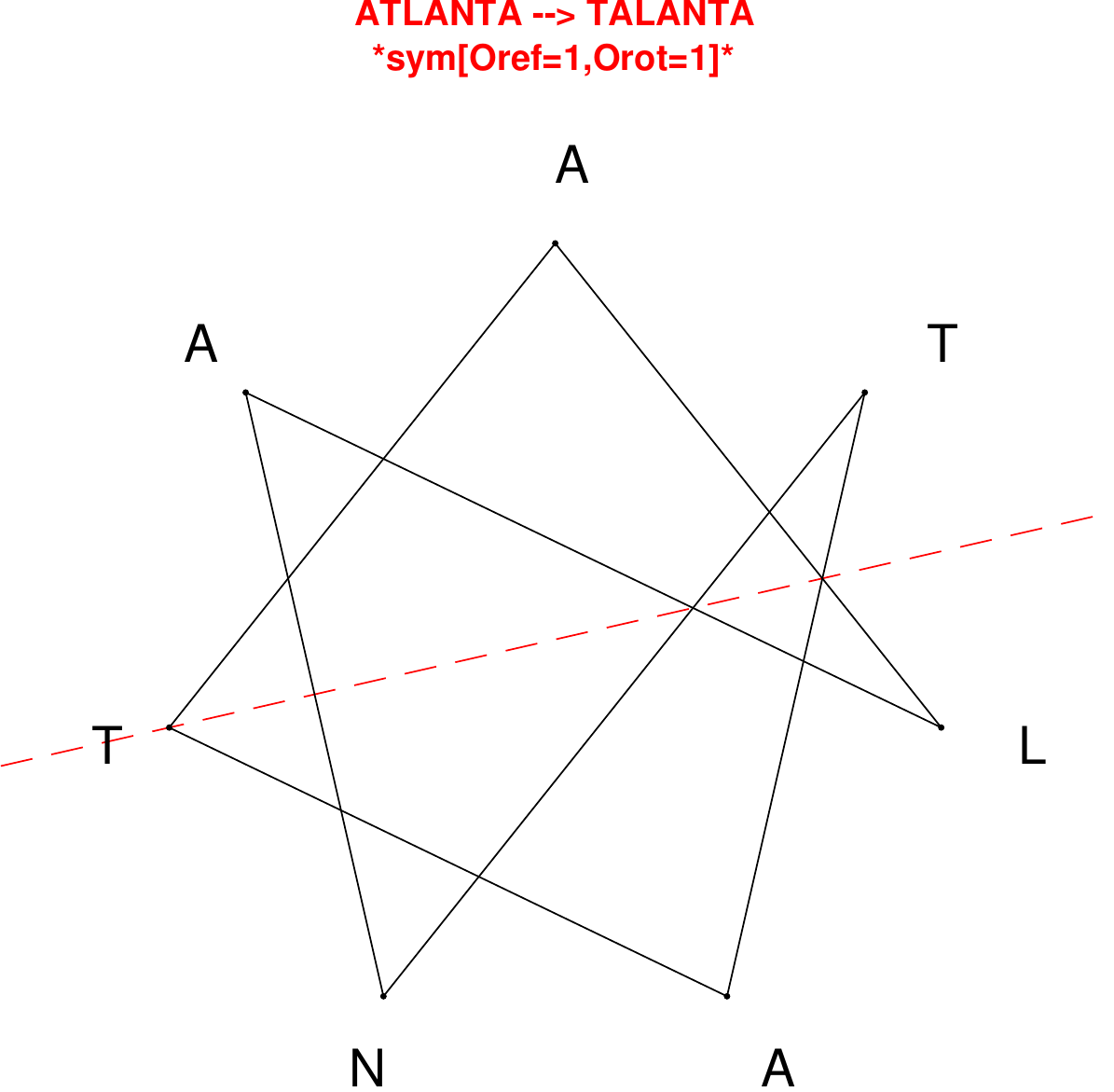}
\end{subfigure}
\hfill
\begin{subfigure}[T]{0.19\textwidth}
\centering
\includegraphics[width=\textwidth]{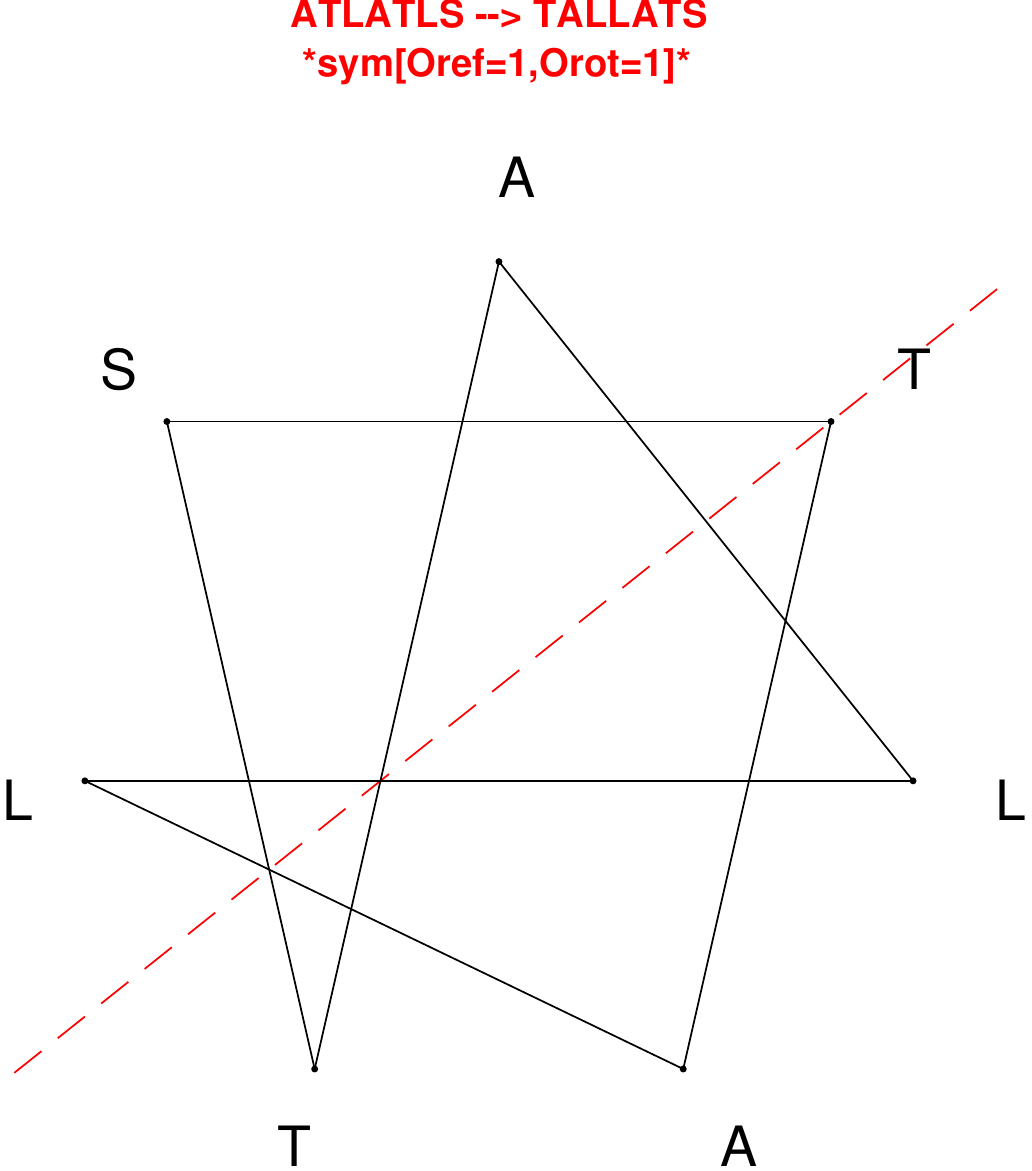}
\end{subfigure}
\hfill
\begin{subfigure}[T]{0.19\textwidth}
\centering
\includegraphics[width=\textwidth]{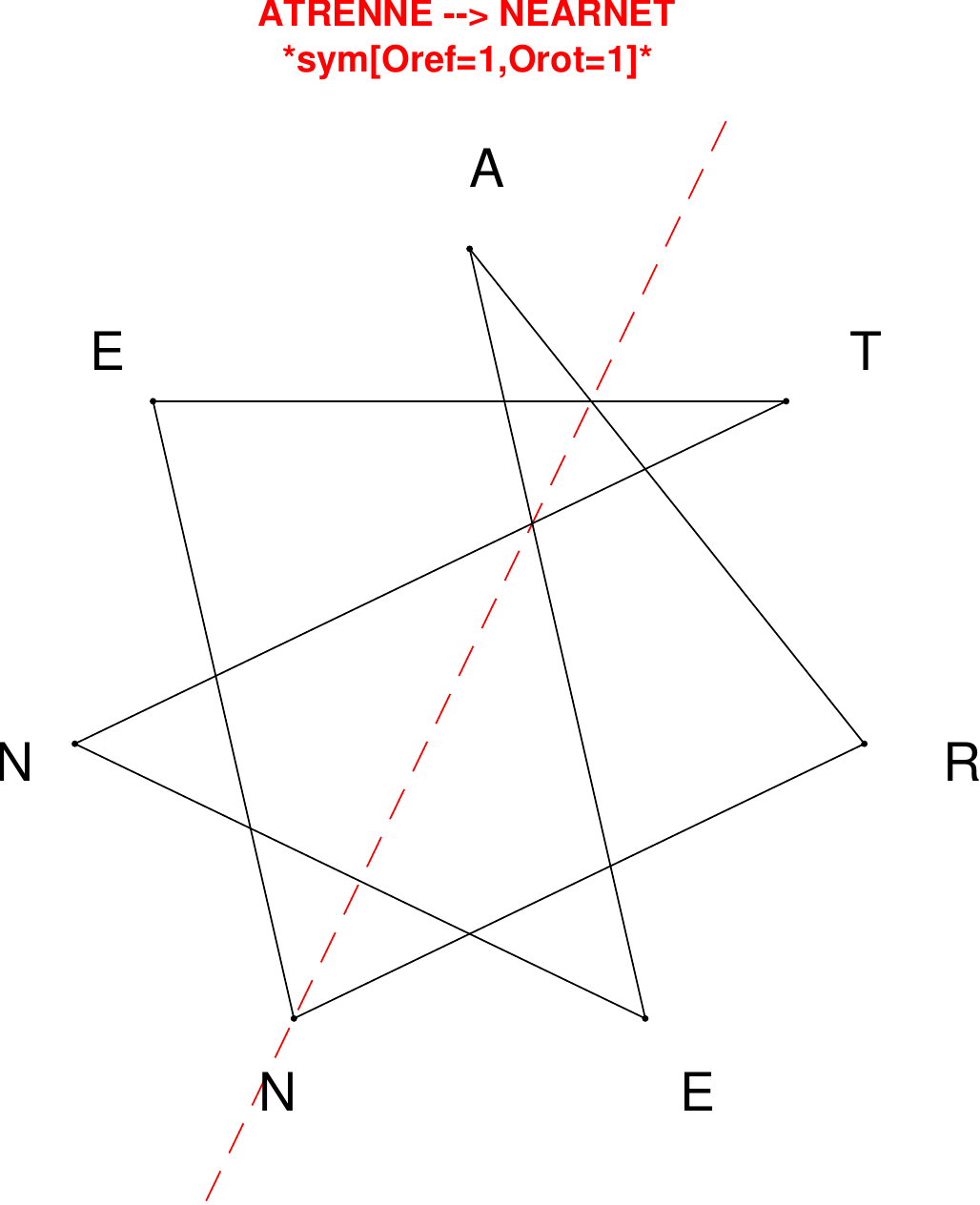}
\end{subfigure}
\hfill
\begin{subfigure}[T]{0.19\textwidth}
\centering
\includegraphics[width=\textwidth]{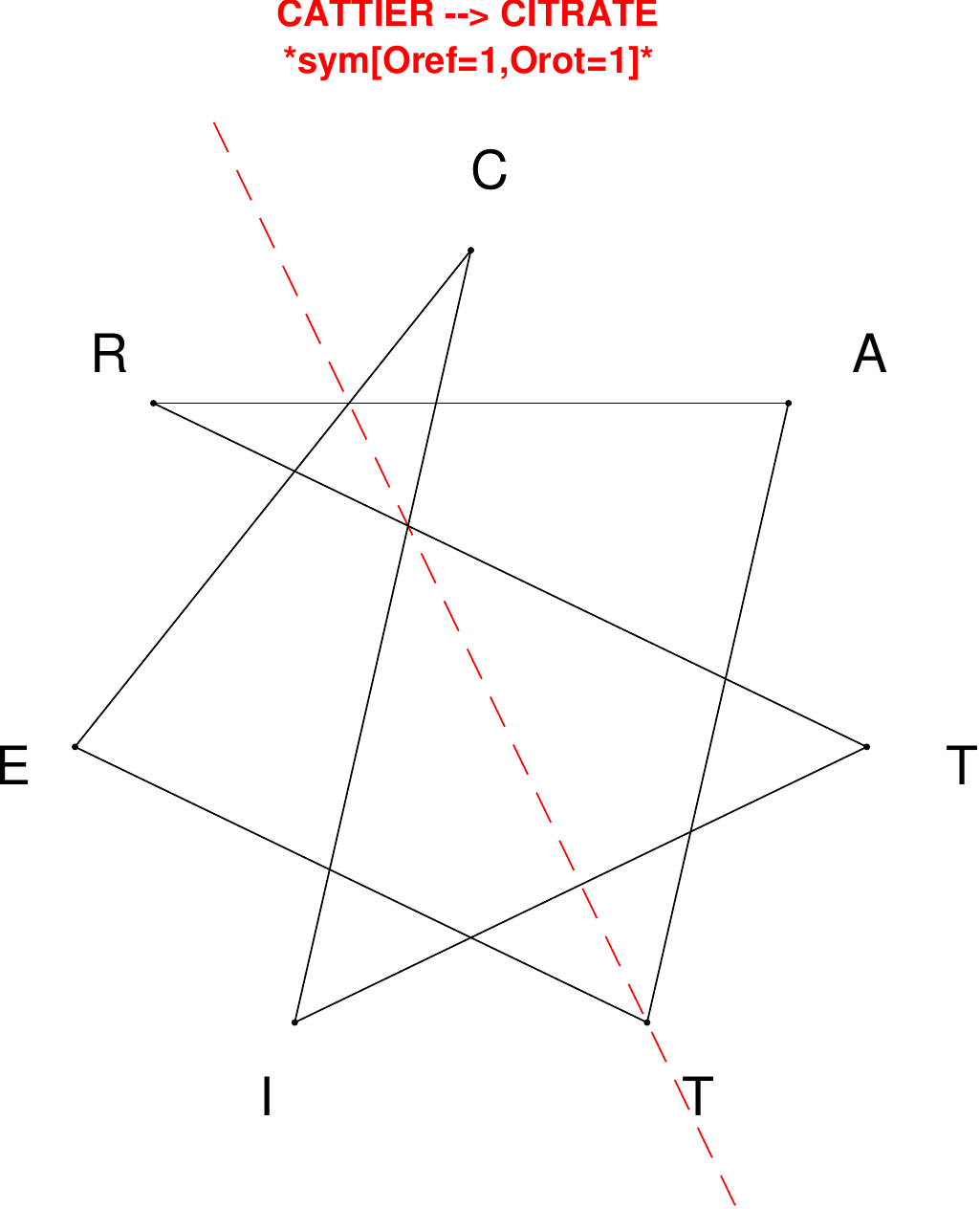}
\end{subfigure}
\hfill
\begin{subfigure}[T]{0.19\textwidth}
\centering
\includegraphics[width=\textwidth]{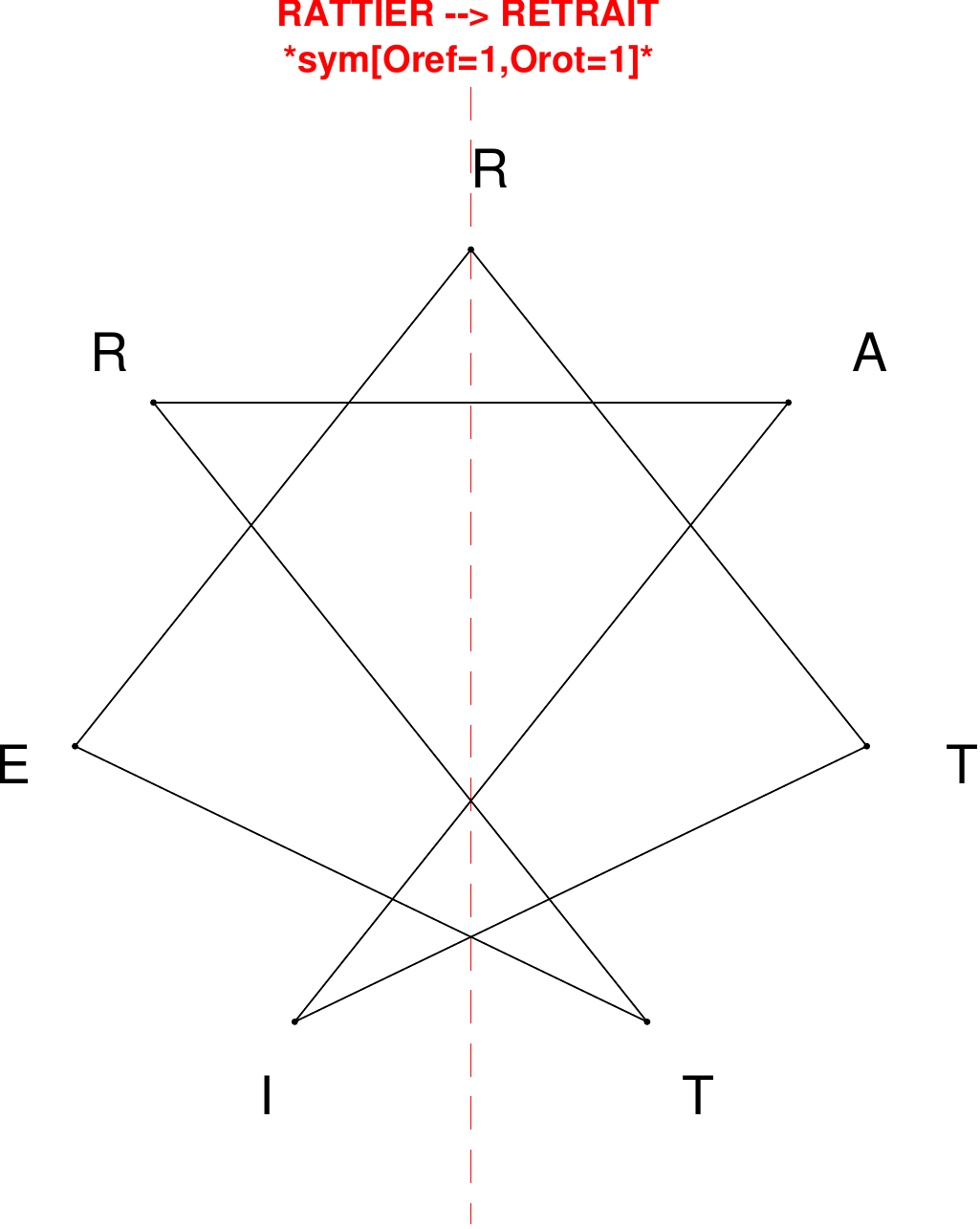}
\end{subfigure}
\end{figure}

\begin{figure}[H]
\centering
\begin{subfigure}[T]{0.19\textwidth}
\centering
\includegraphics[width=\textwidth]{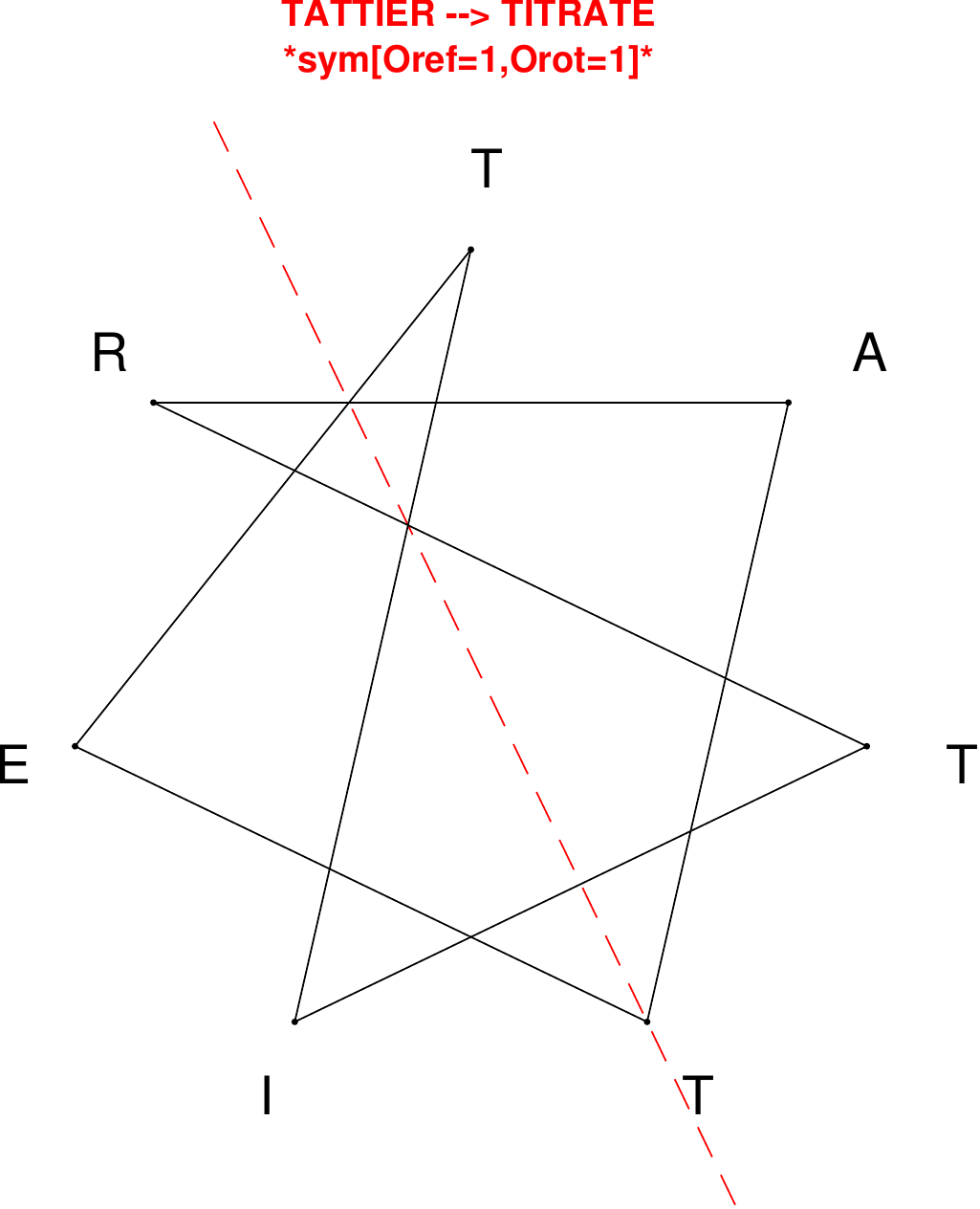}
\end{subfigure}
\hfill
\begin{subfigure}[T]{0.19\textwidth}
\centering
\includegraphics[width=\textwidth]{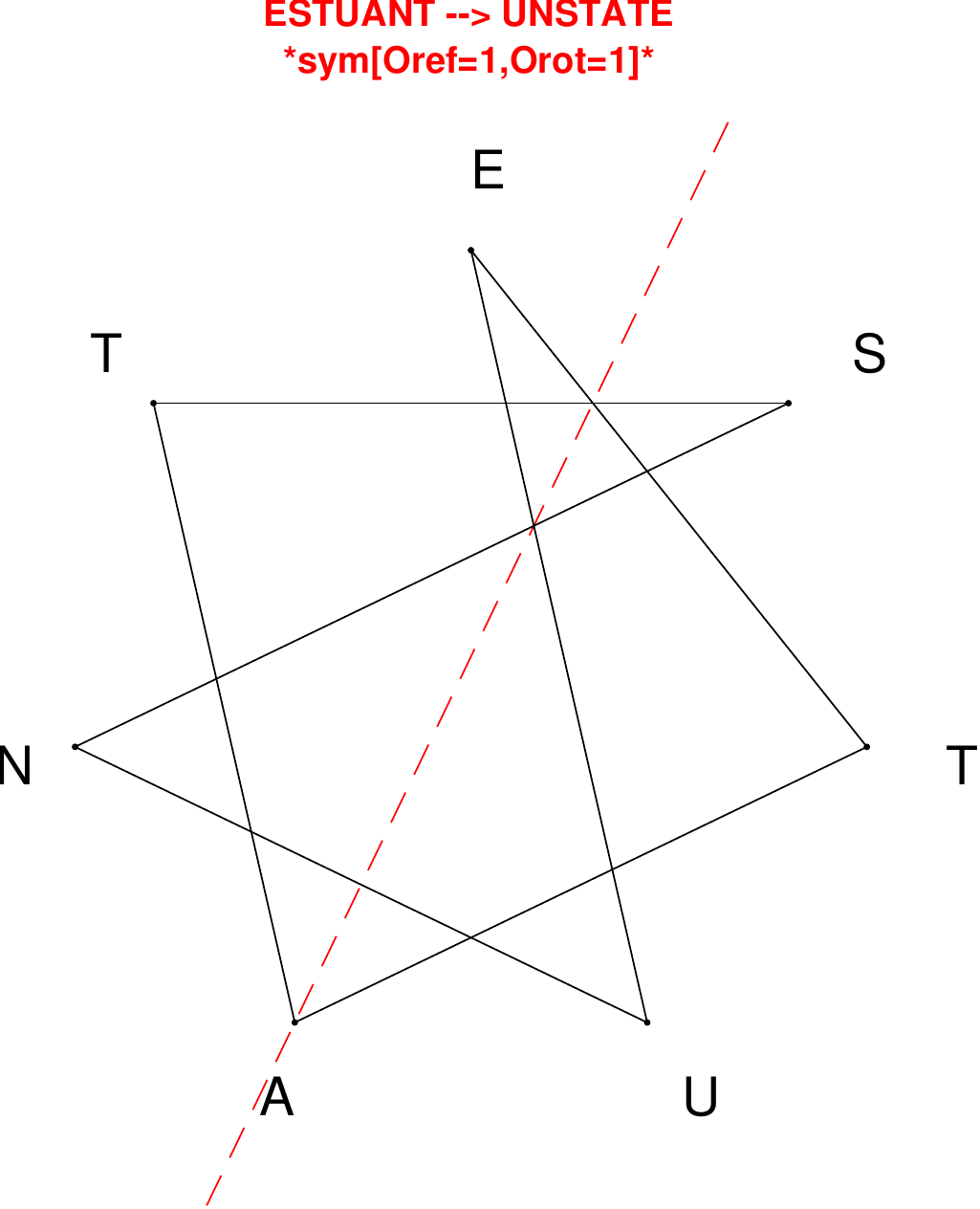}
\end{subfigure}
\hfill
\begin{subfigure}[T]{0.19\textwidth}
\centering
\includegraphics[width=\textwidth]{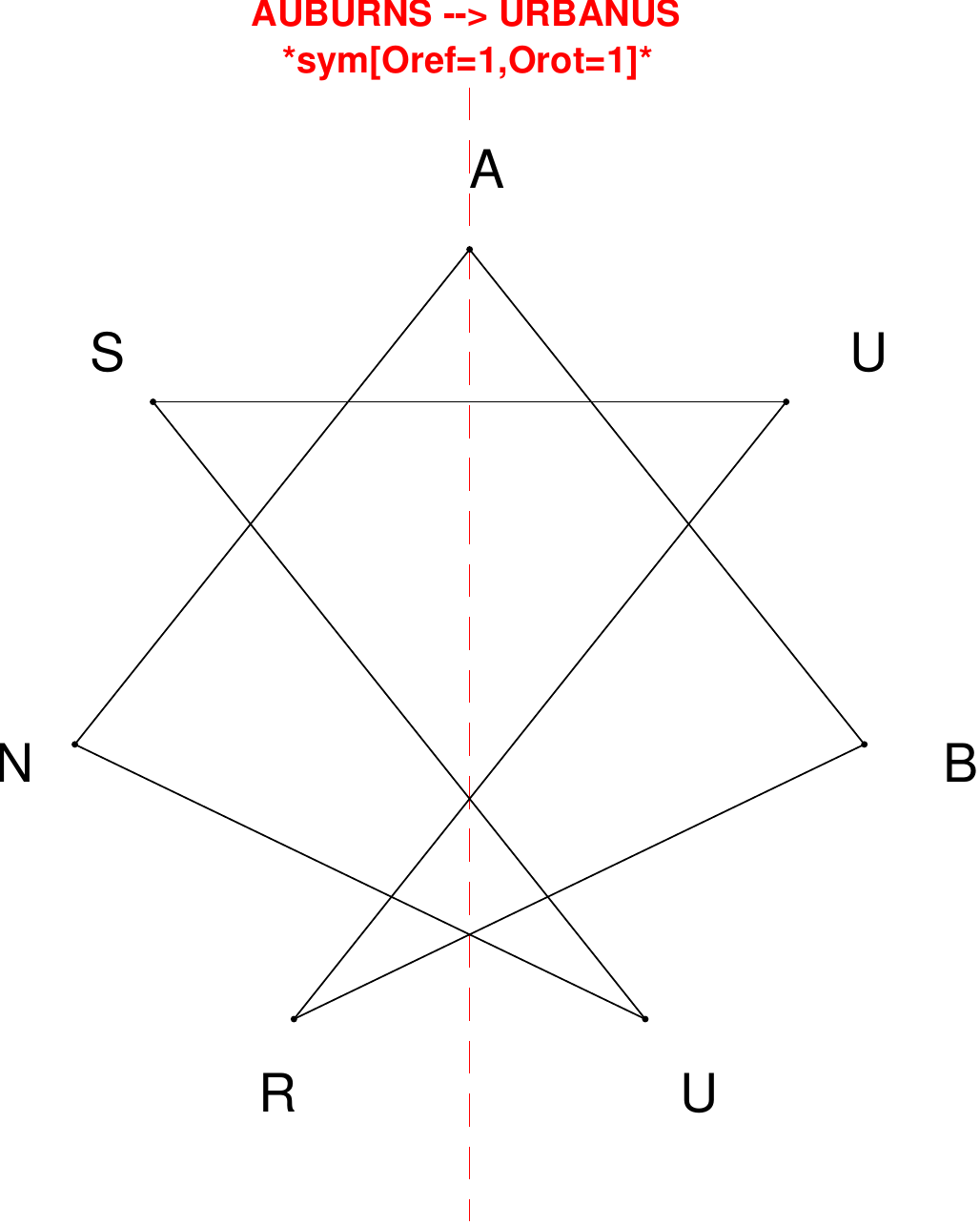}
\end{subfigure}
\hfill
\begin{subfigure}[T]{0.19\textwidth}
\centering
\includegraphics[width=\textwidth]{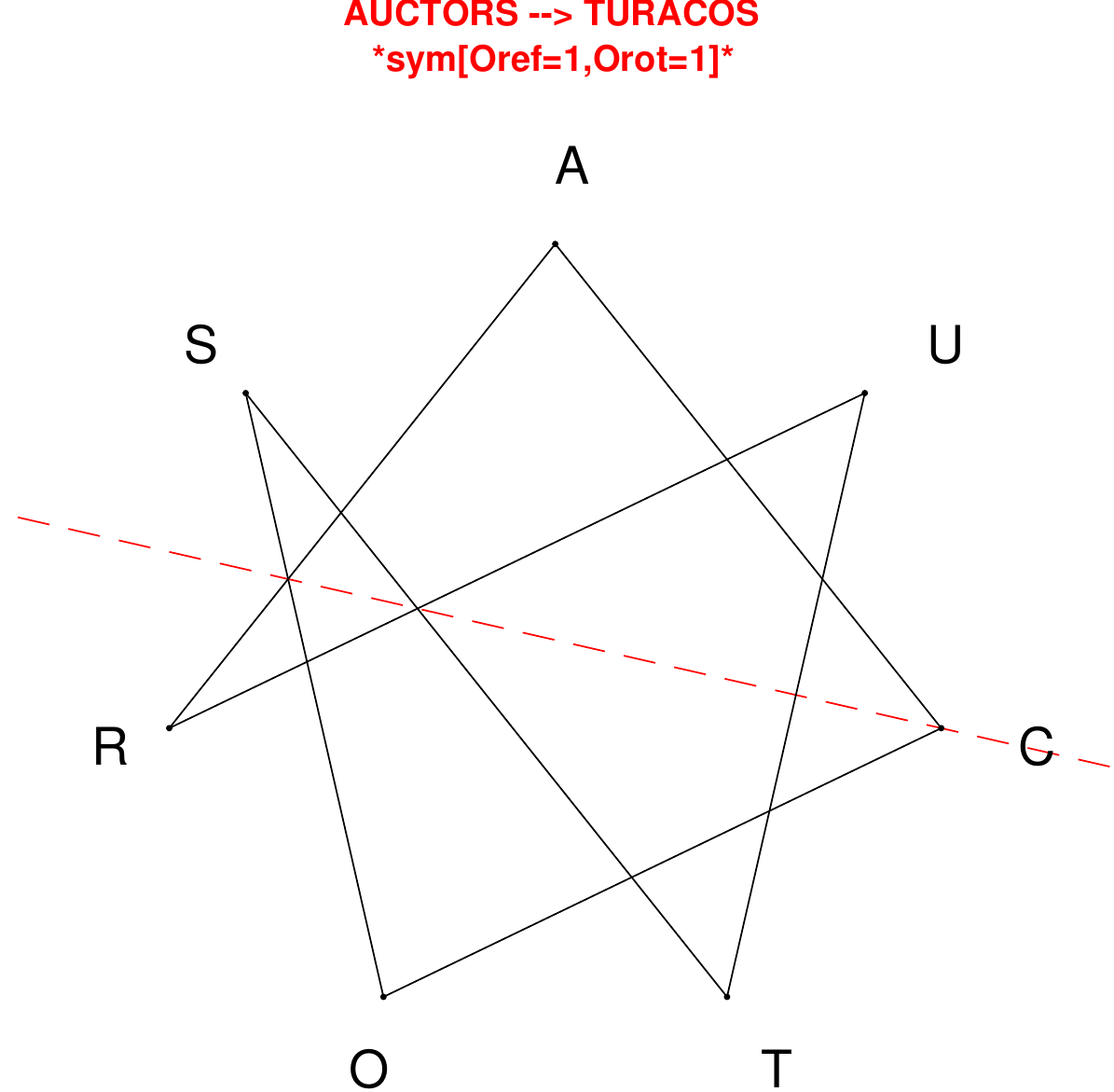}
\end{subfigure}
\hfill
\begin{subfigure}[T]{0.19\textwidth}
\centering
\includegraphics[width=\textwidth]{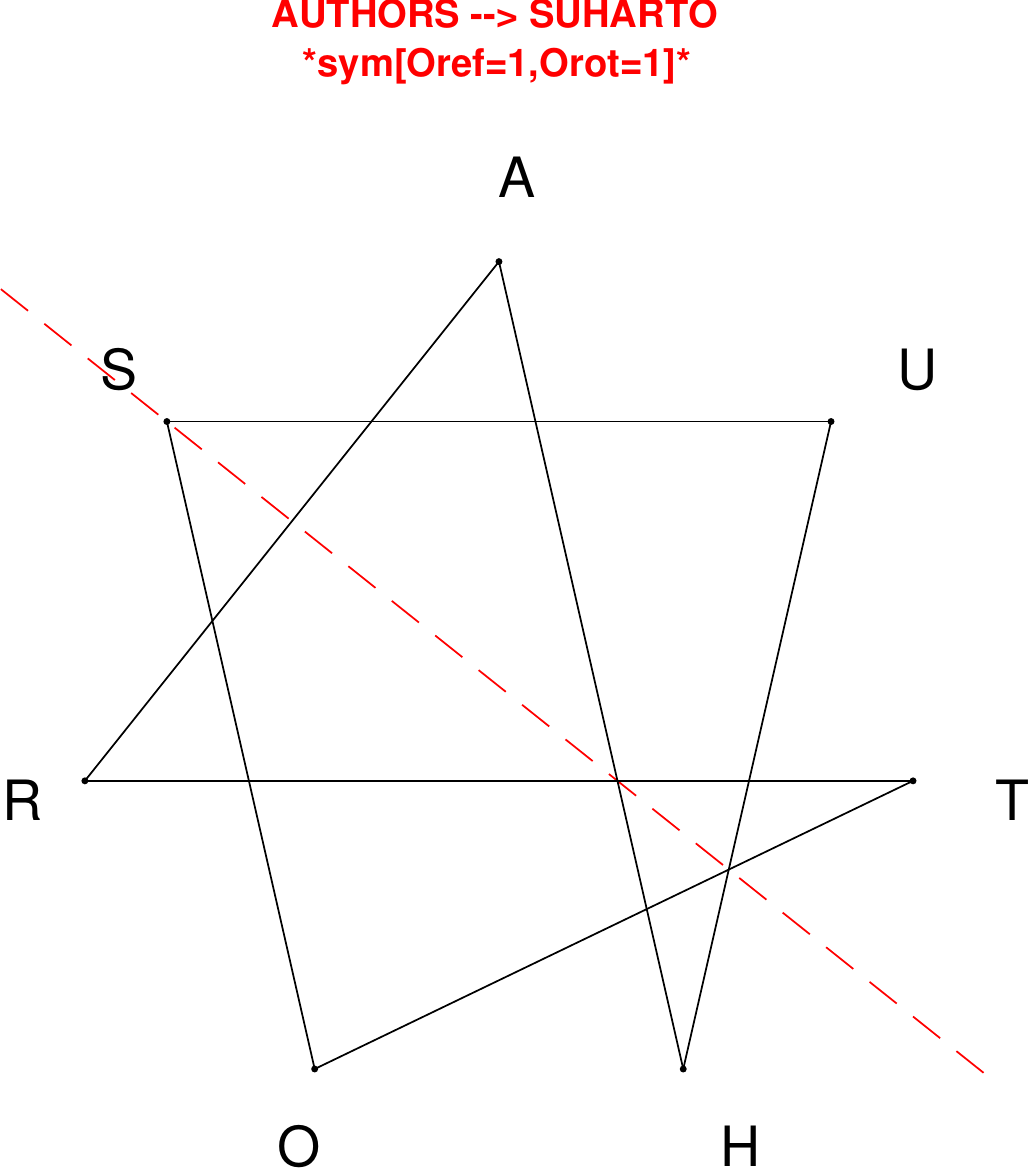}
\end{subfigure}
\end{figure}

\begin{figure}[H]
\centering
\begin{subfigure}[T]{0.19\textwidth}
\centering
\includegraphics[width=\textwidth]{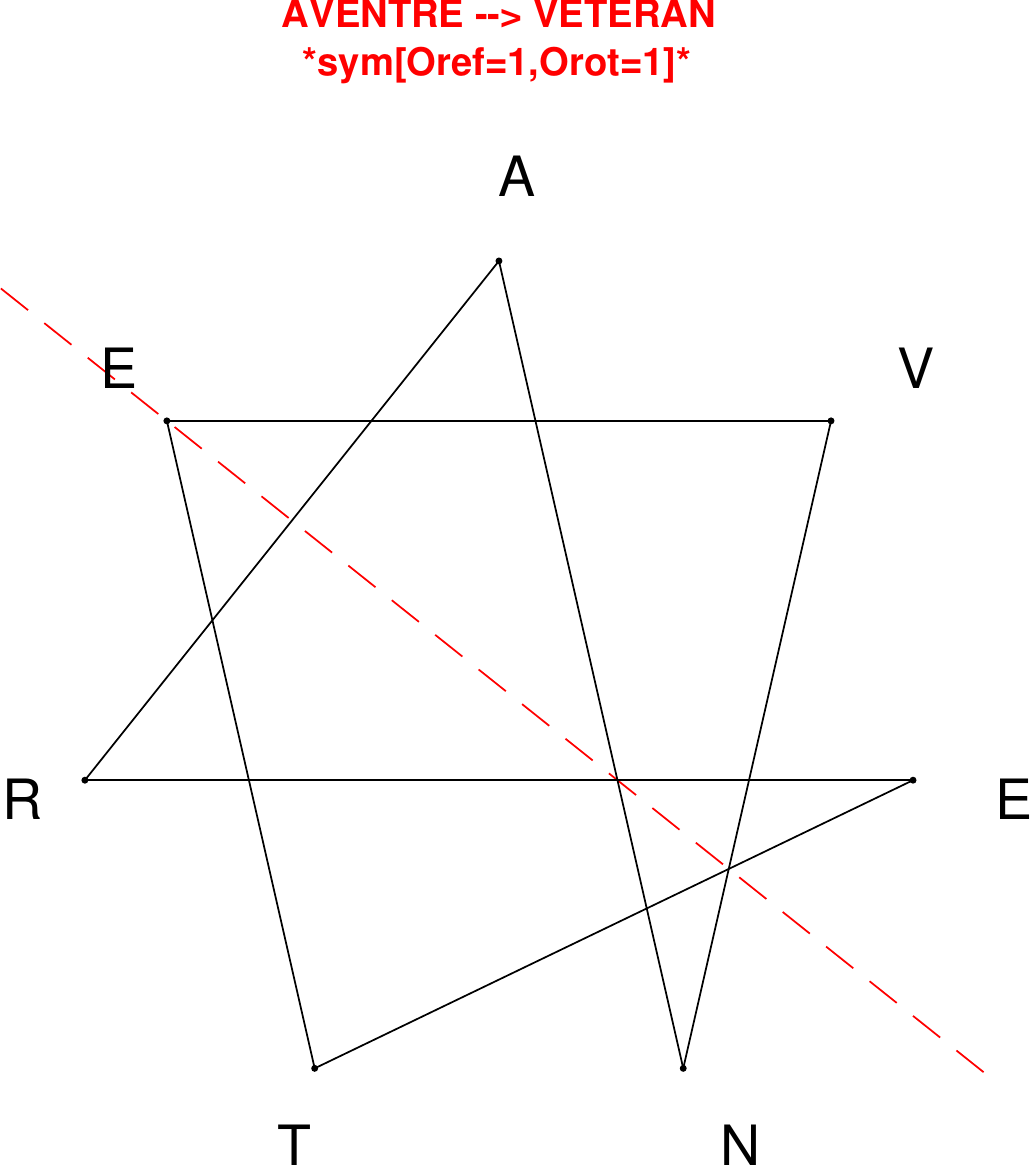}
\end{subfigure}
\hfill
\begin{subfigure}[T]{0.19\textwidth}
\centering
\includegraphics[width=\textwidth]{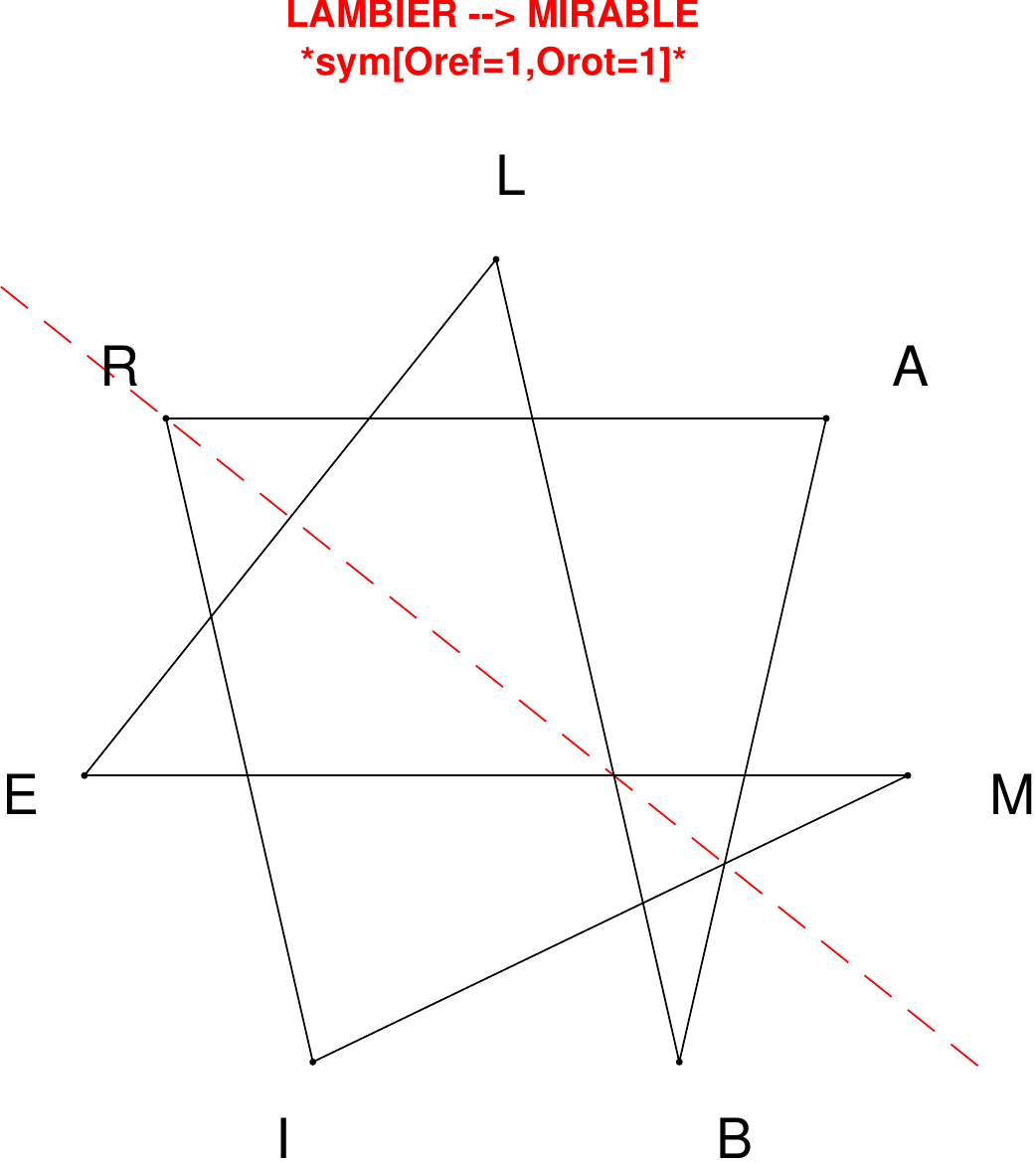}
\end{subfigure}
\hfill
\begin{subfigure}[T]{0.19\textwidth}
\centering
\includegraphics[width=\textwidth]{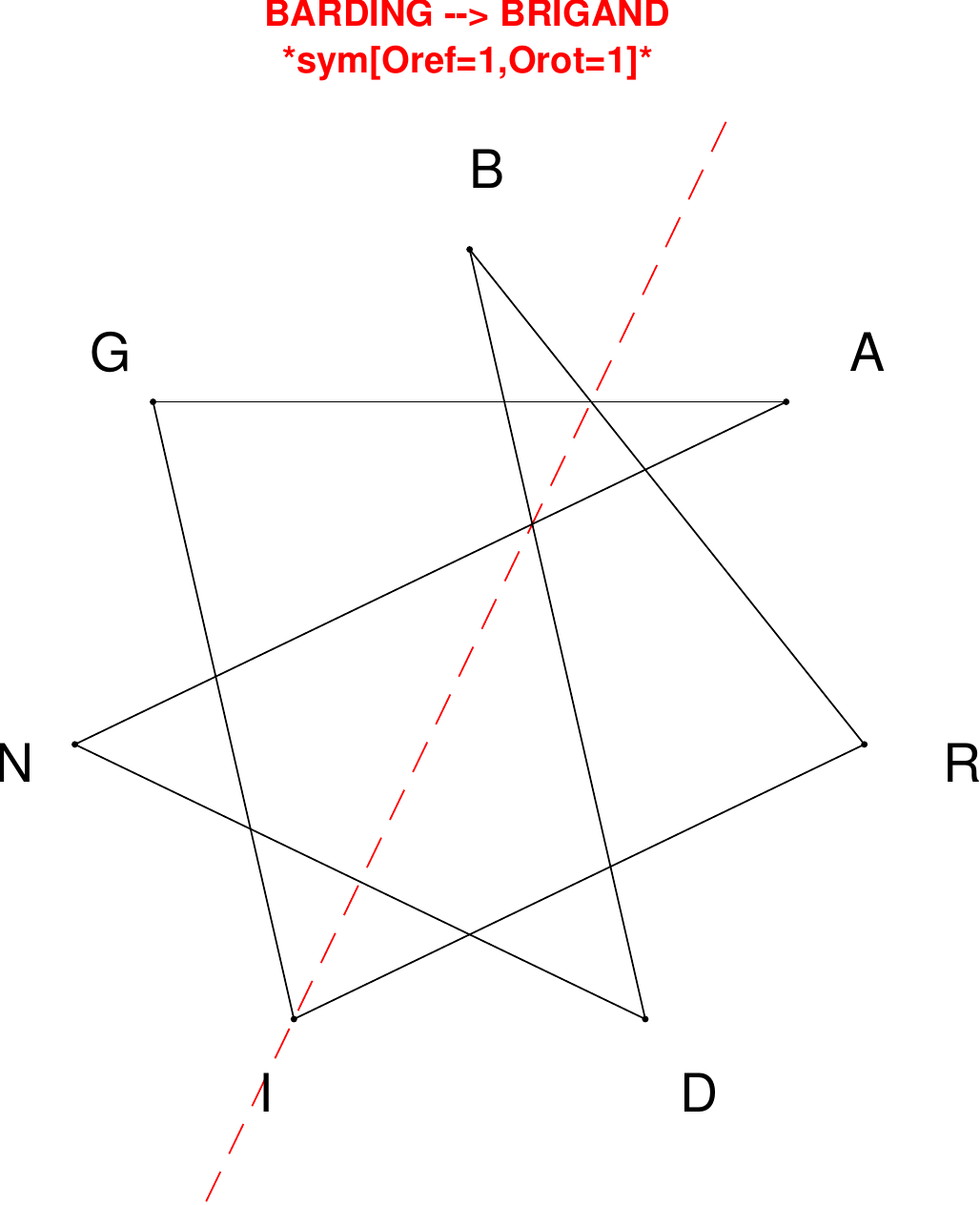}
\end{subfigure}
\hfill
\begin{subfigure}[T]{0.19\textwidth}
\centering
\includegraphics[width=\textwidth]{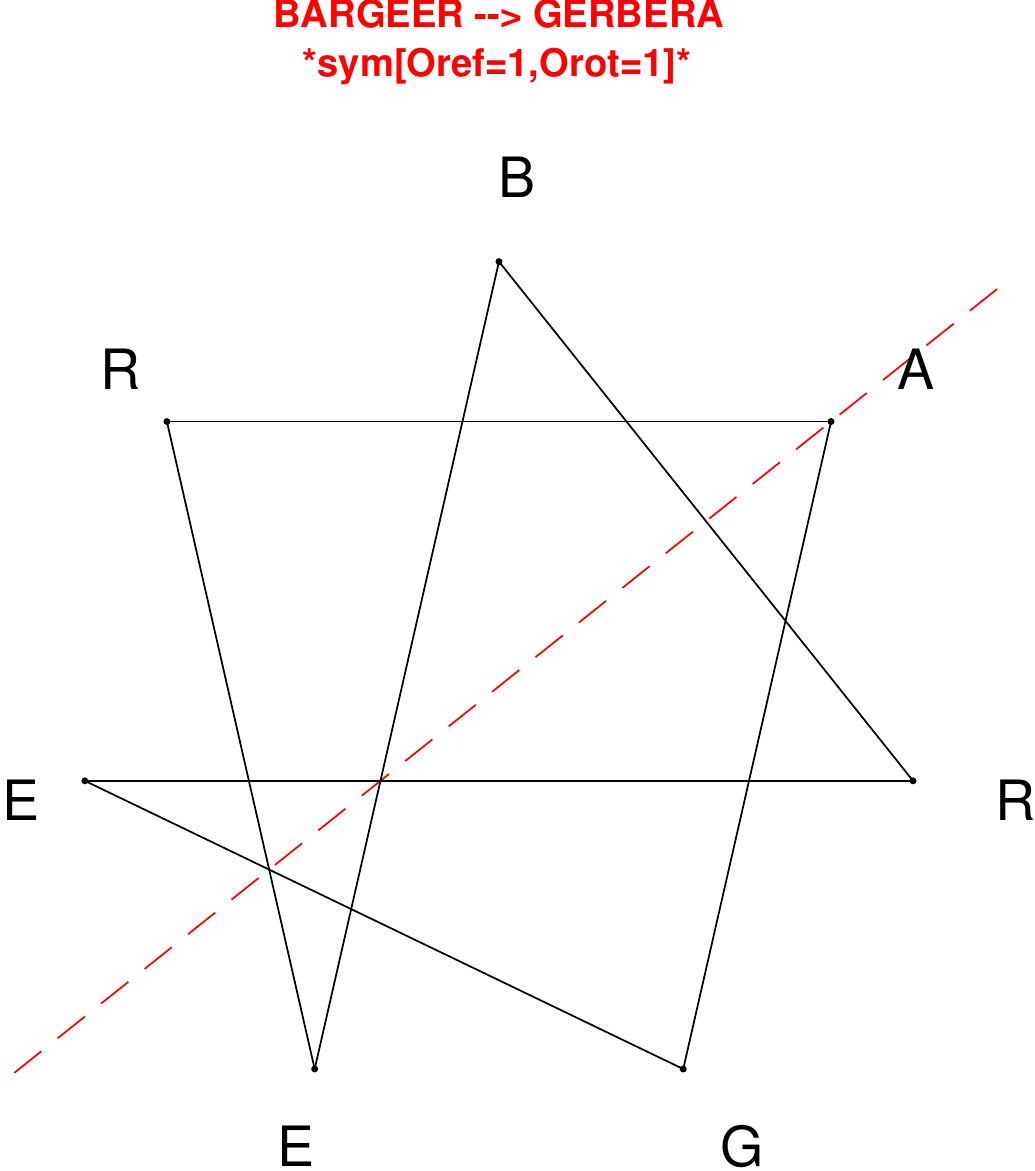}
\end{subfigure}
\hfill
\begin{subfigure}[T]{0.19\textwidth}
\centering
\includegraphics[width=\textwidth]{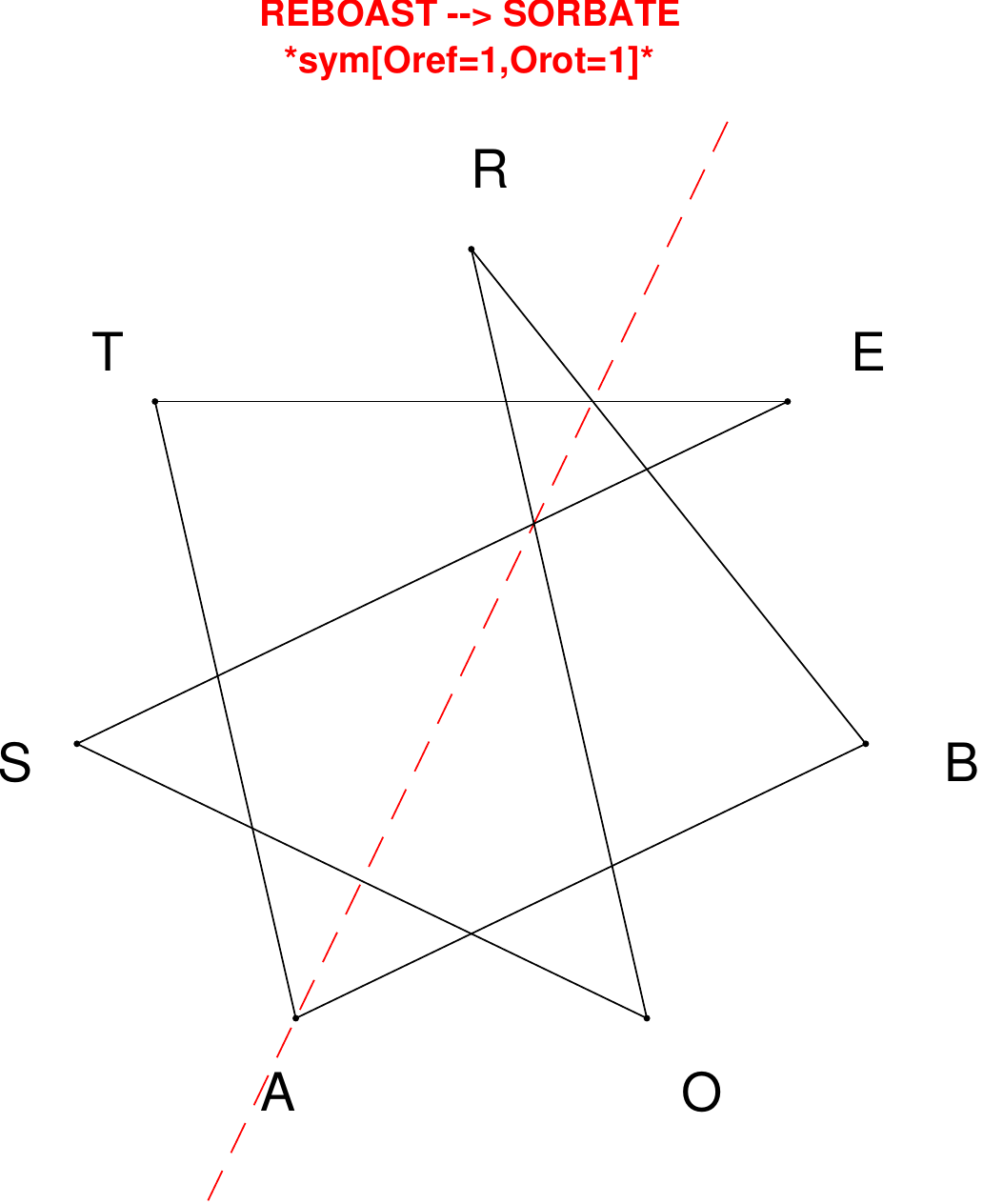}
\end{subfigure}
\end{figure}

\begin{figure}[H]
\centering
\begin{subfigure}[T]{0.19\textwidth}
\centering
\includegraphics[width=\textwidth]{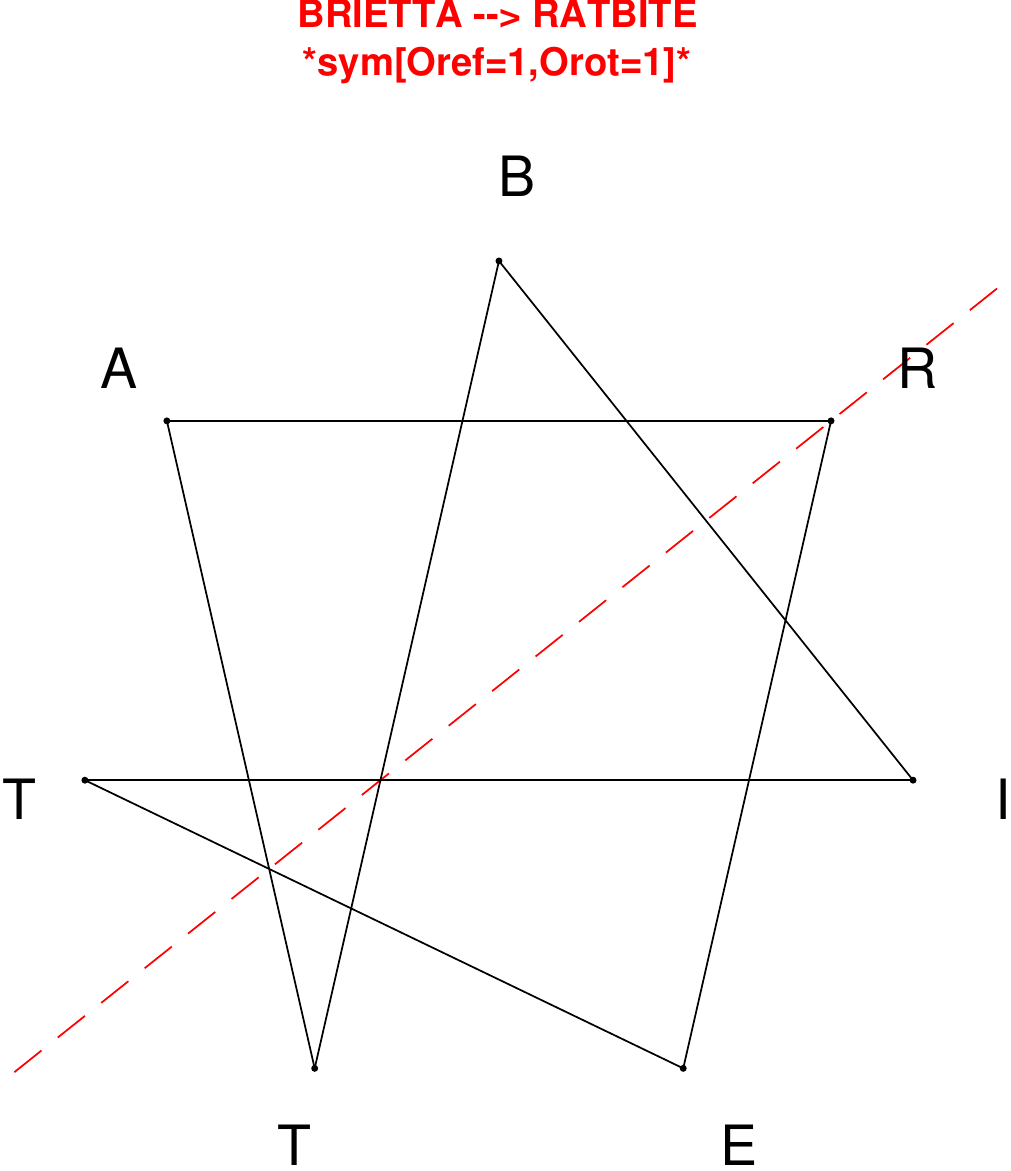}
\end{subfigure}
\hfill
\begin{subfigure}[T]{0.19\textwidth}
\centering
\includegraphics[width=\textwidth]{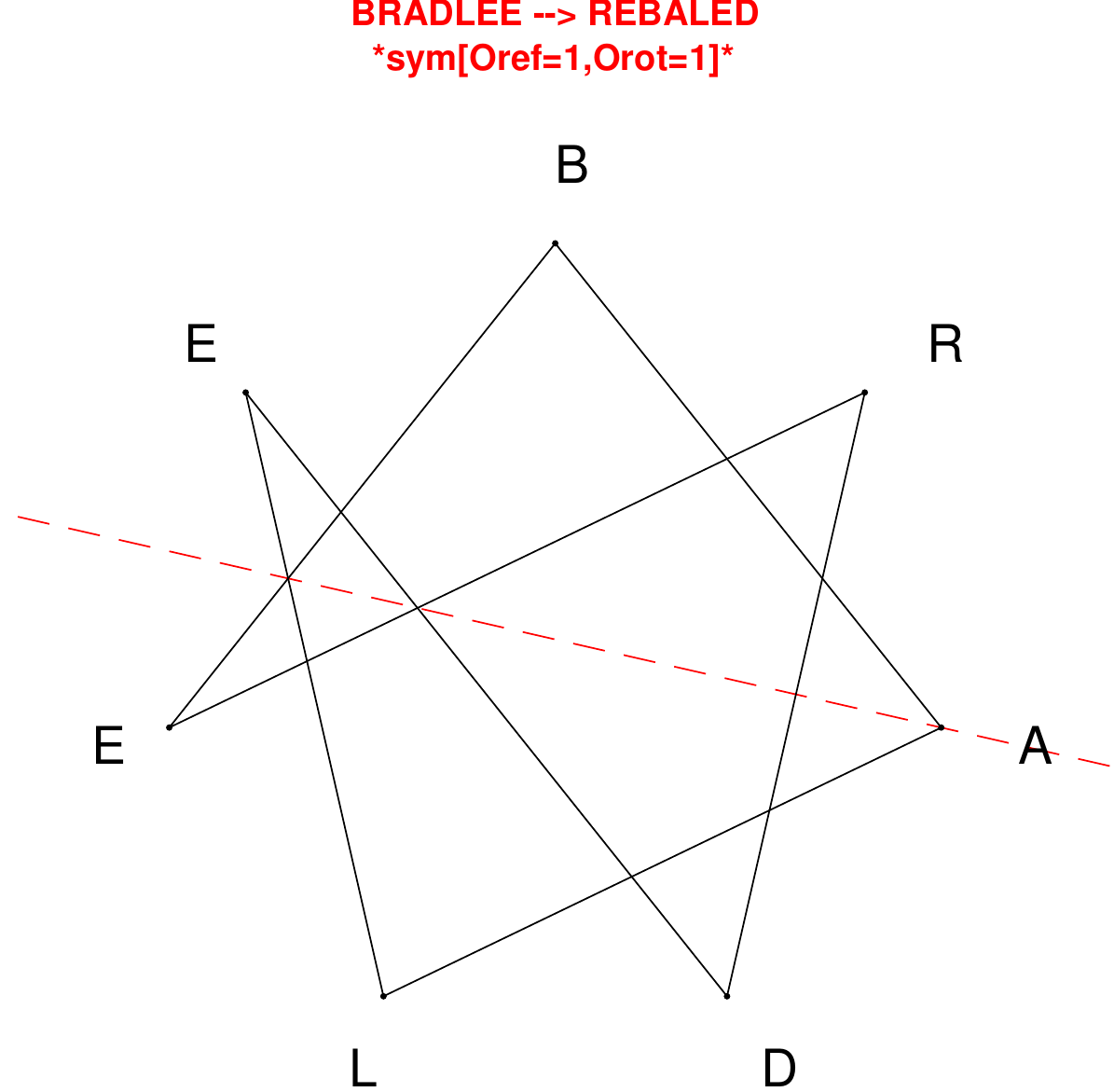}
\end{subfigure}
\hfill
\begin{subfigure}[T]{0.19\textwidth}
\centering
\includegraphics[width=\textwidth]{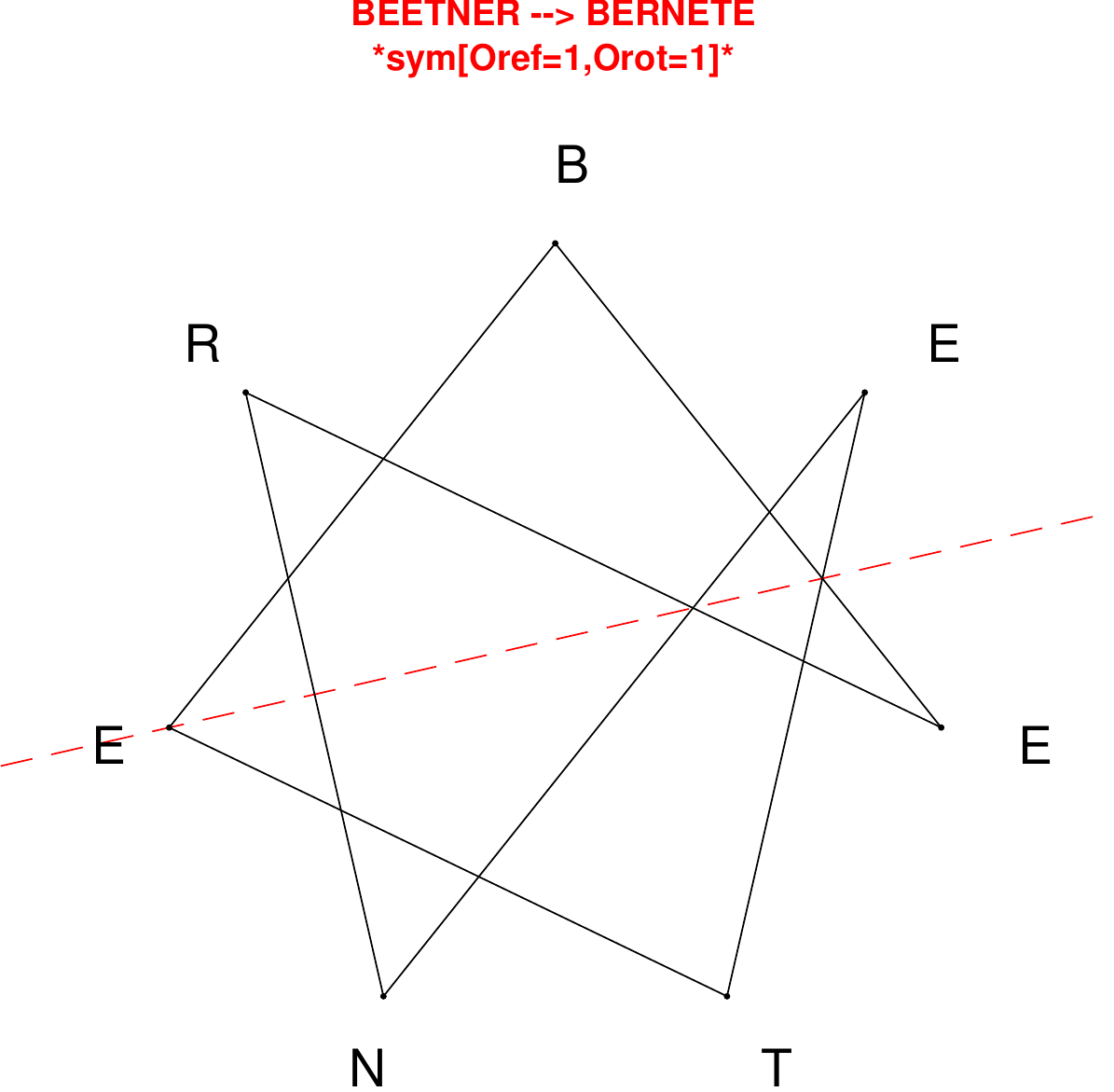}
\end{subfigure}
\hfill
\begin{subfigure}[T]{0.19\textwidth}
\centering
\includegraphics[width=\textwidth]{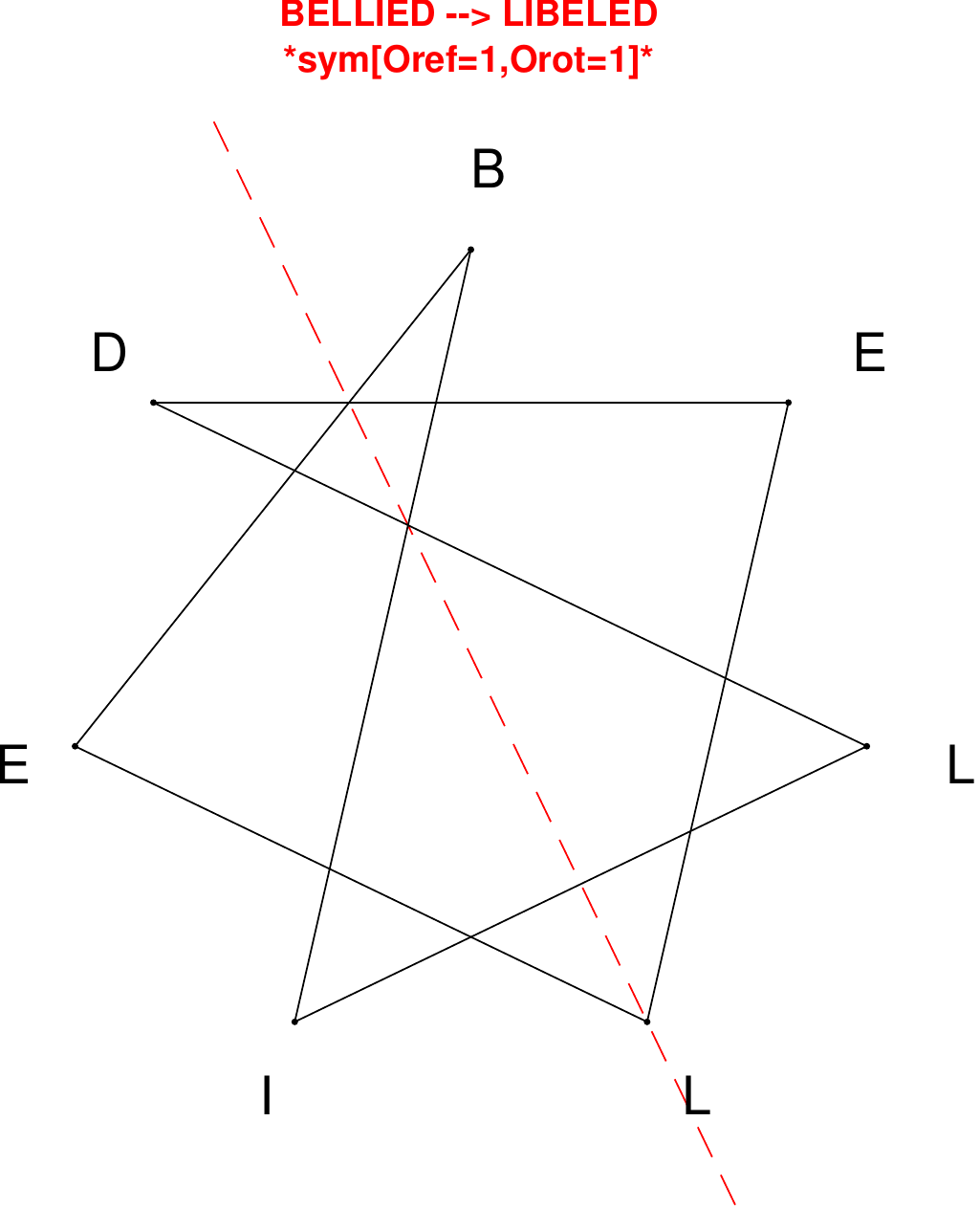}
\end{subfigure}
\hfill
\begin{subfigure}[T]{0.19\textwidth}
\centering
\includegraphics[width=\textwidth]{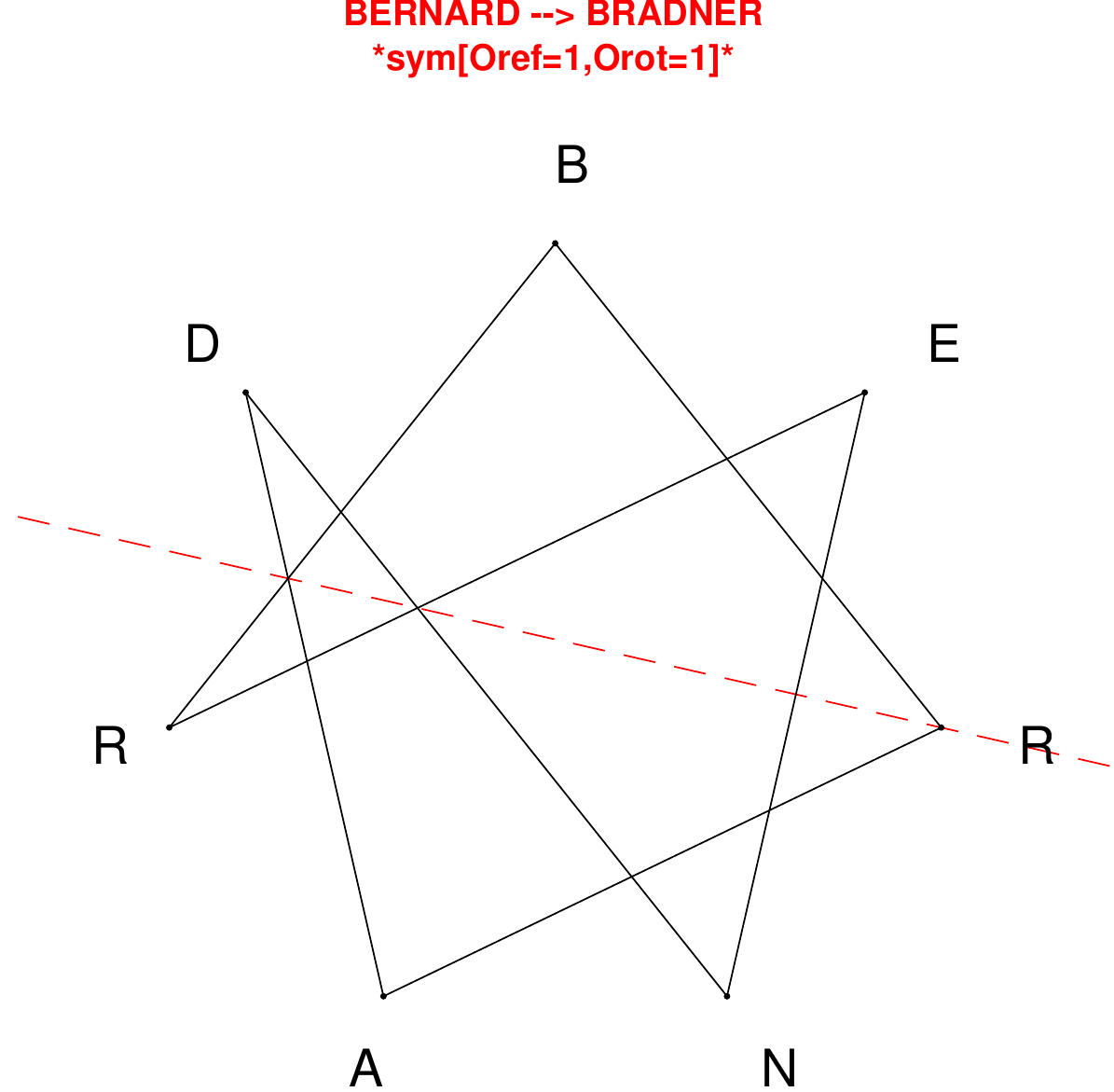}
\end{subfigure}
\end{figure}

\begin{figure}[H]
\centering
\begin{subfigure}[T]{0.19\textwidth}
\centering
\includegraphics[width=\textwidth]{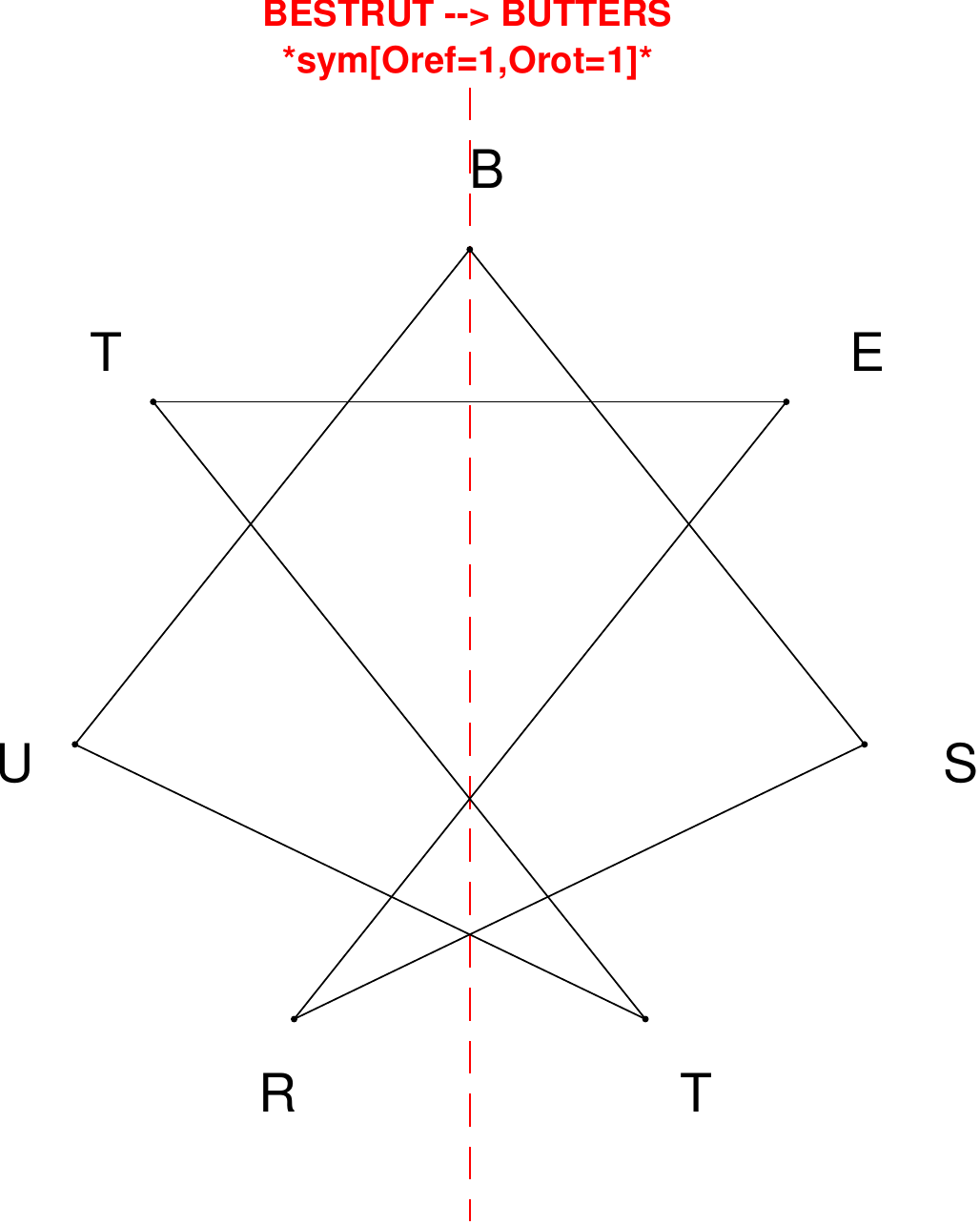}
\end{subfigure}
\hfill
\begin{subfigure}[T]{0.19\textwidth}
\centering
\includegraphics[width=\textwidth]{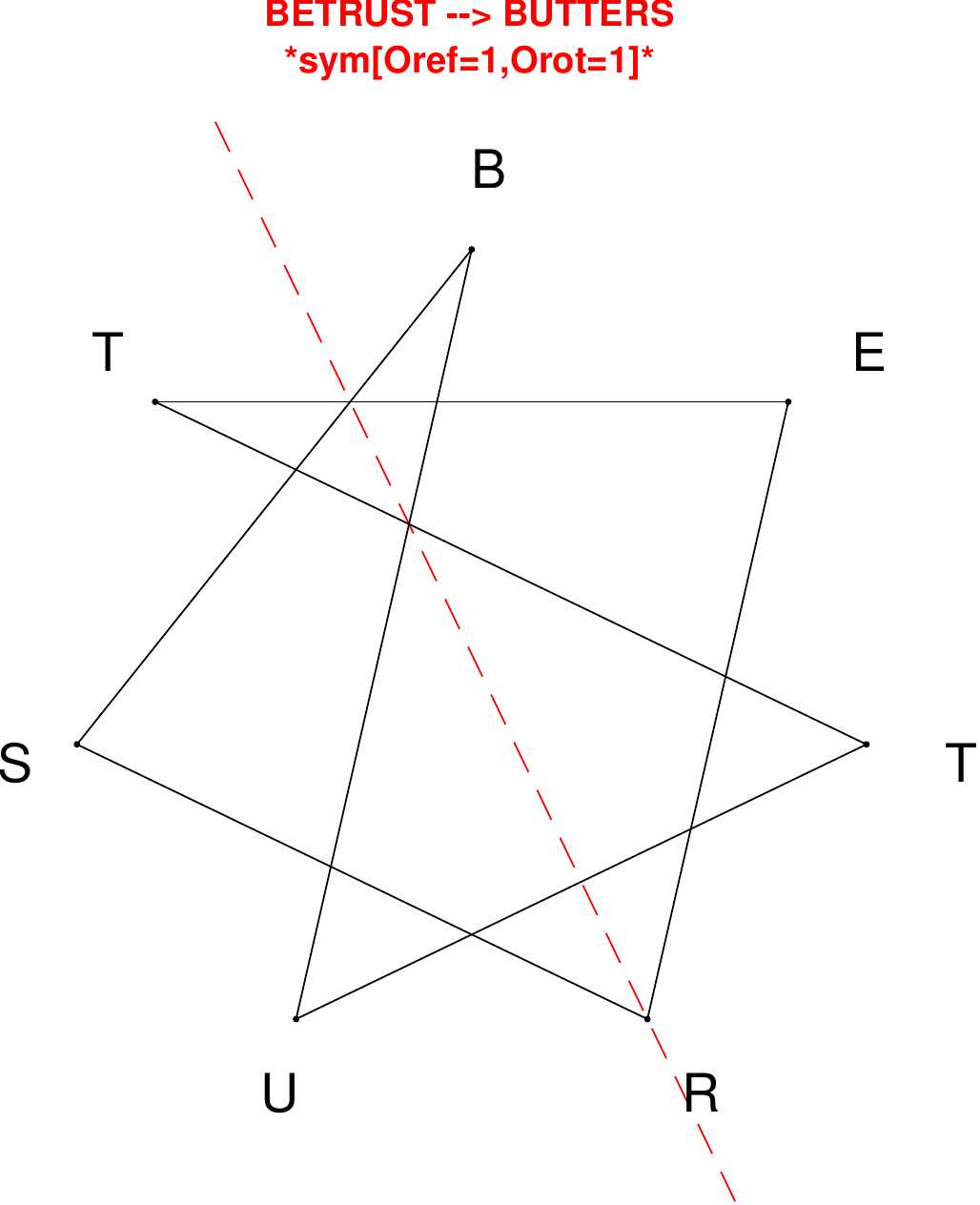}
\end{subfigure}
\hfill
\begin{subfigure}[T]{0.19\textwidth}
\centering
\includegraphics[width=\textwidth]{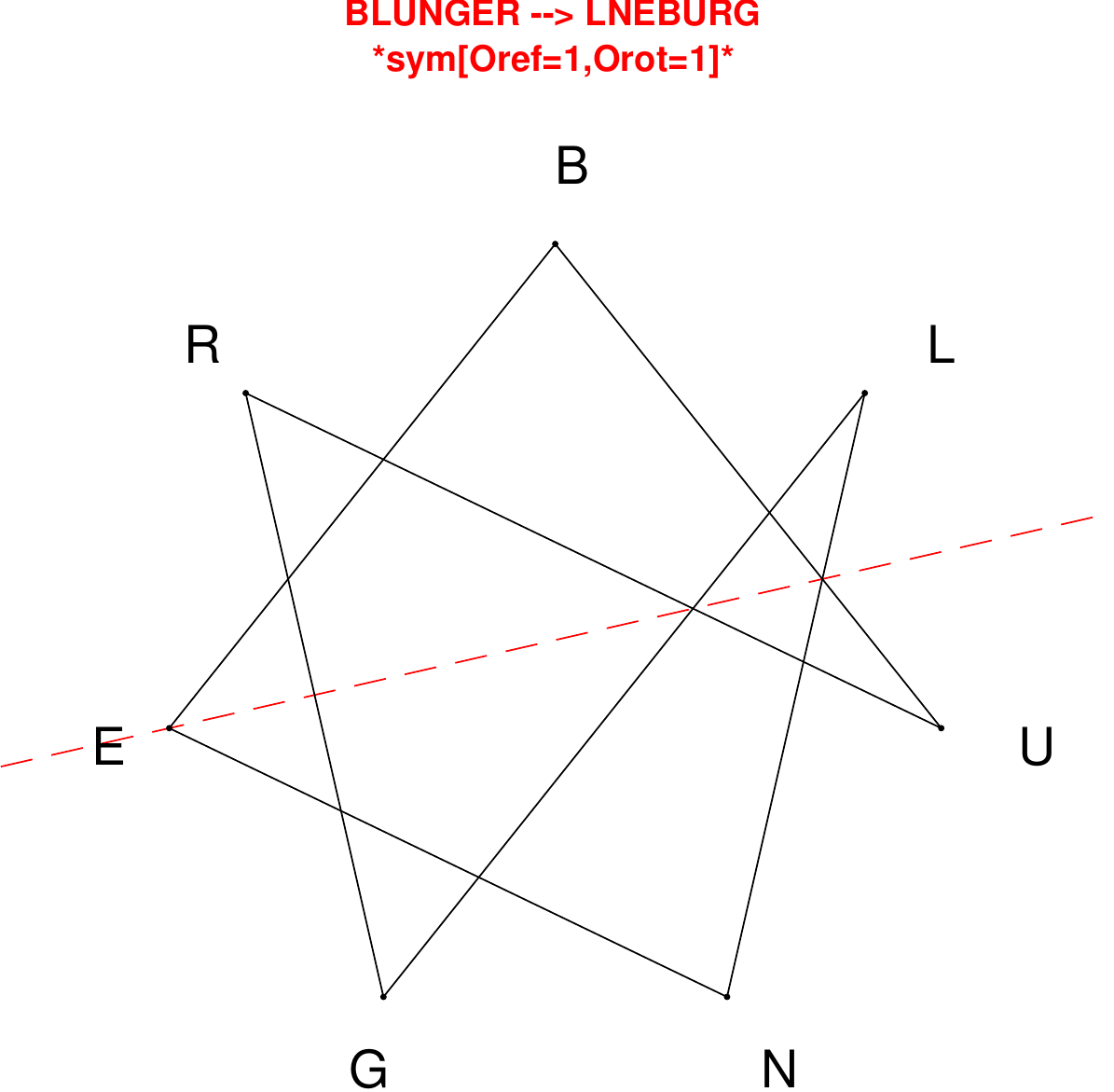}
\end{subfigure}
\hfill
\begin{subfigure}[T]{0.19\textwidth}
\centering
\includegraphics[width=\textwidth]{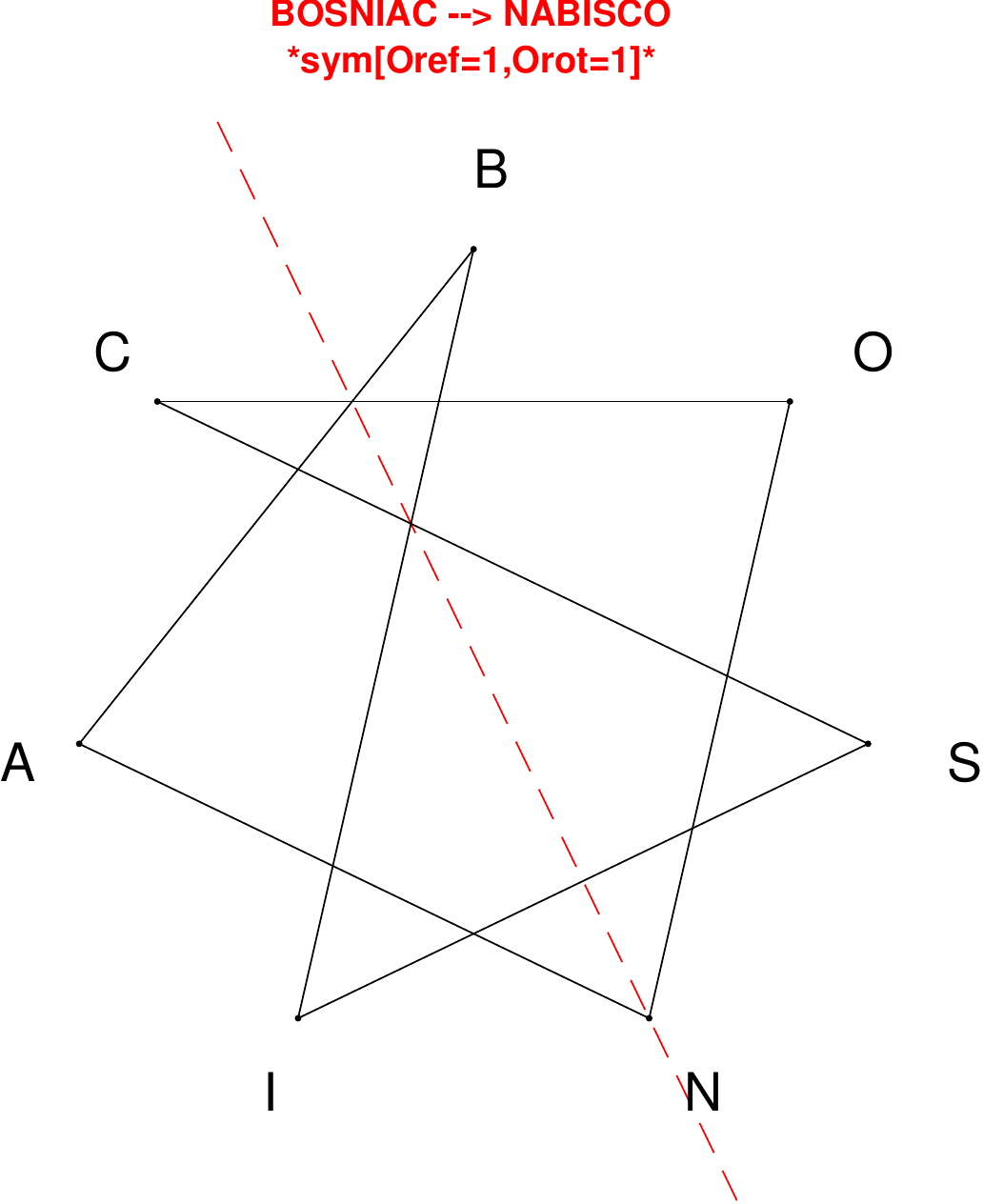}
\end{subfigure}
\hfill
\begin{subfigure}[T]{0.19\textwidth}
\centering
\includegraphics[width=\textwidth]{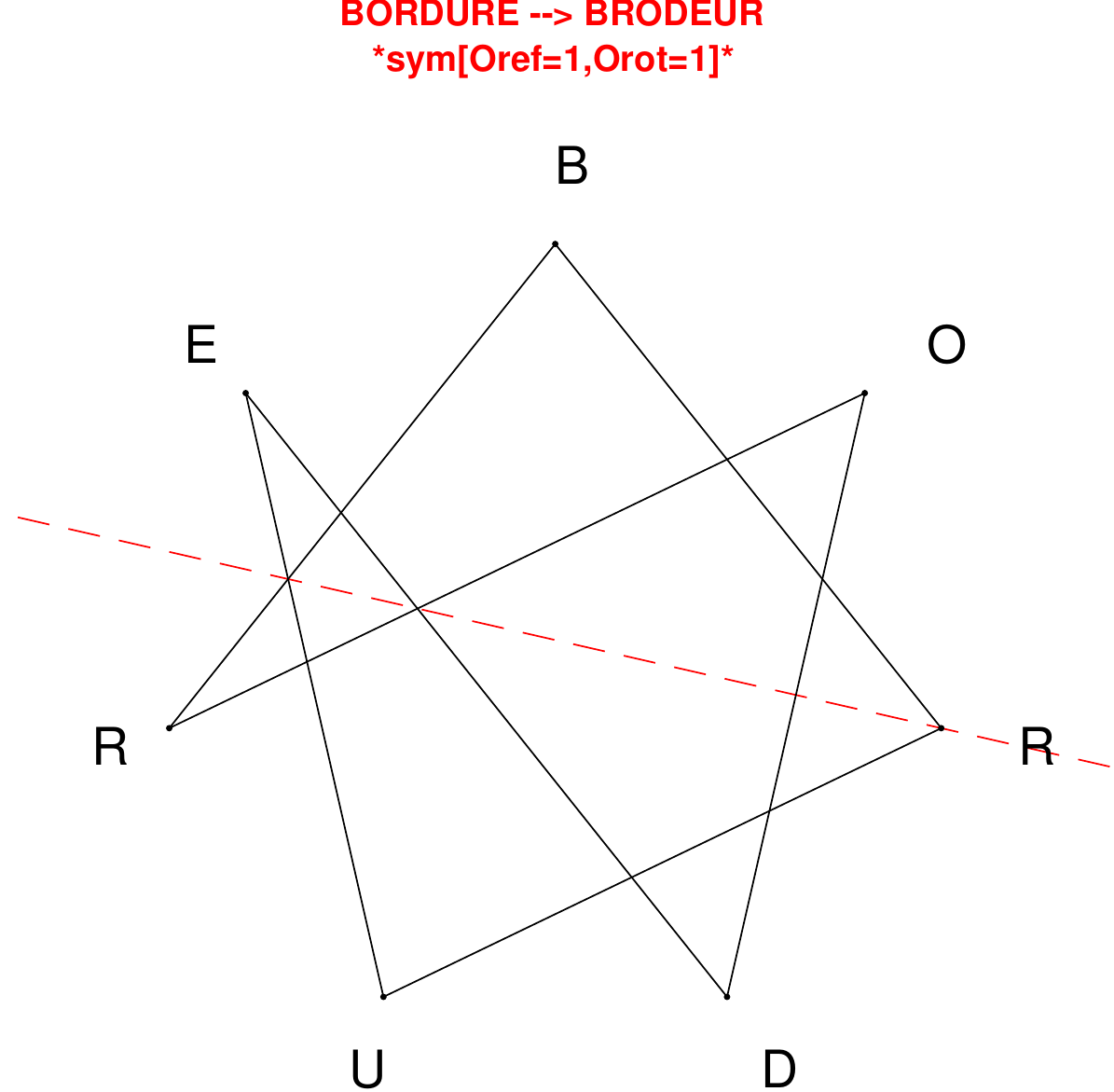}
\end{subfigure}
\end{figure}

\begin{figure}[H]
\centering
\begin{subfigure}[T]{0.19\textwidth}
\centering
\includegraphics[width=\textwidth]{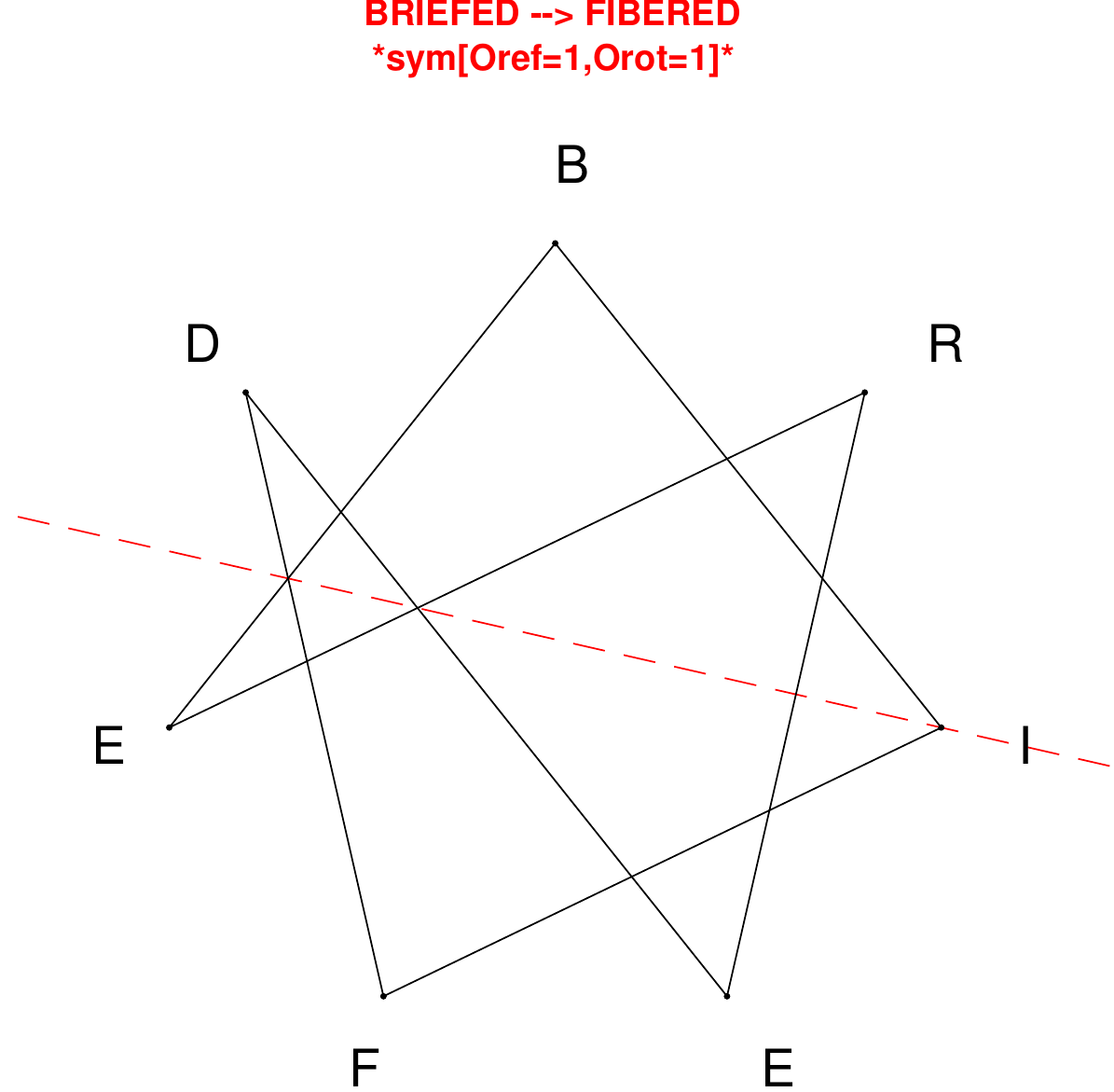}
\end{subfigure}
\hfill
\begin{subfigure}[T]{0.19\textwidth}
\centering
\includegraphics[width=\textwidth]{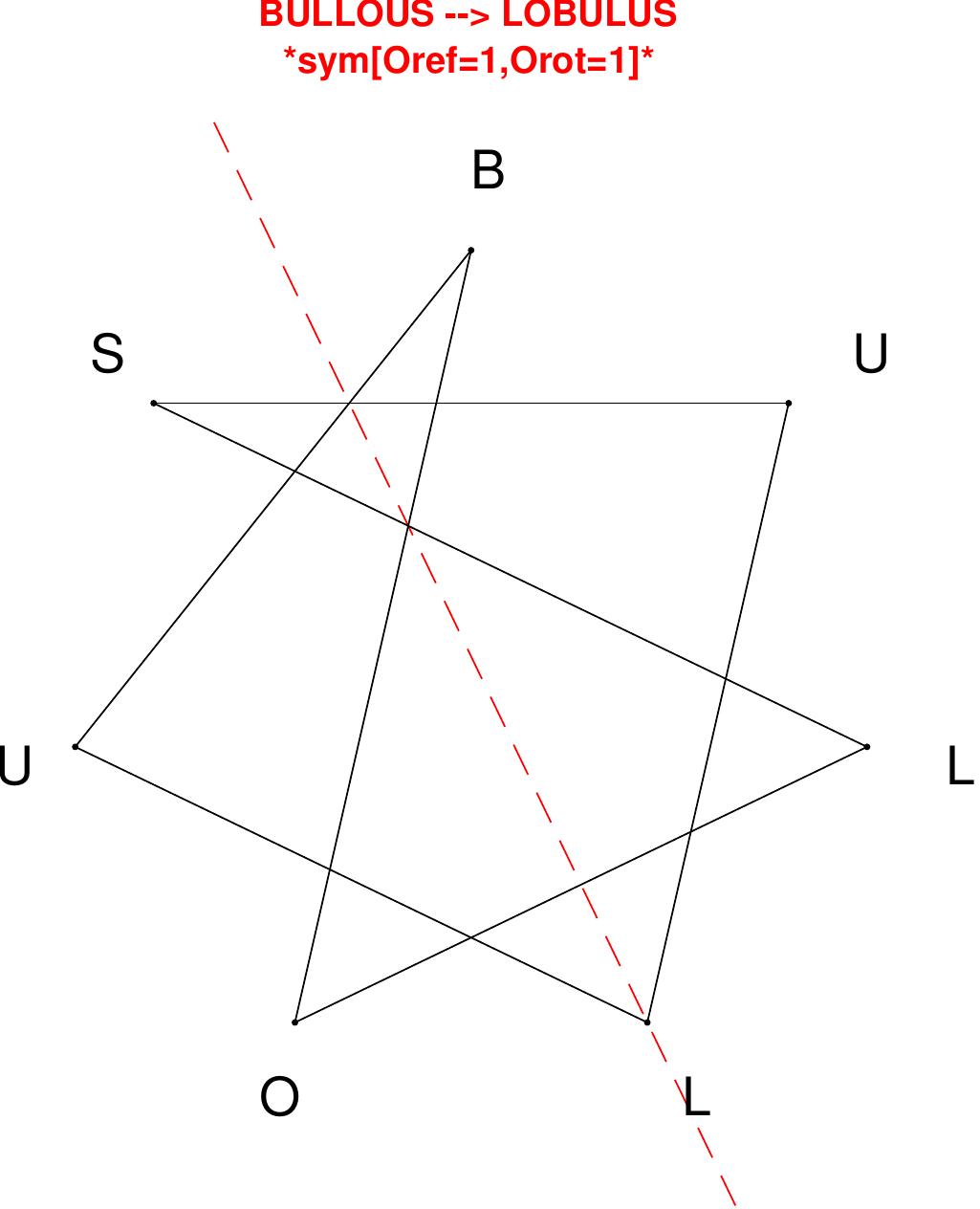}
\end{subfigure}
\hfill
\begin{subfigure}[T]{0.19\textwidth}
\centering
\includegraphics[width=\textwidth]{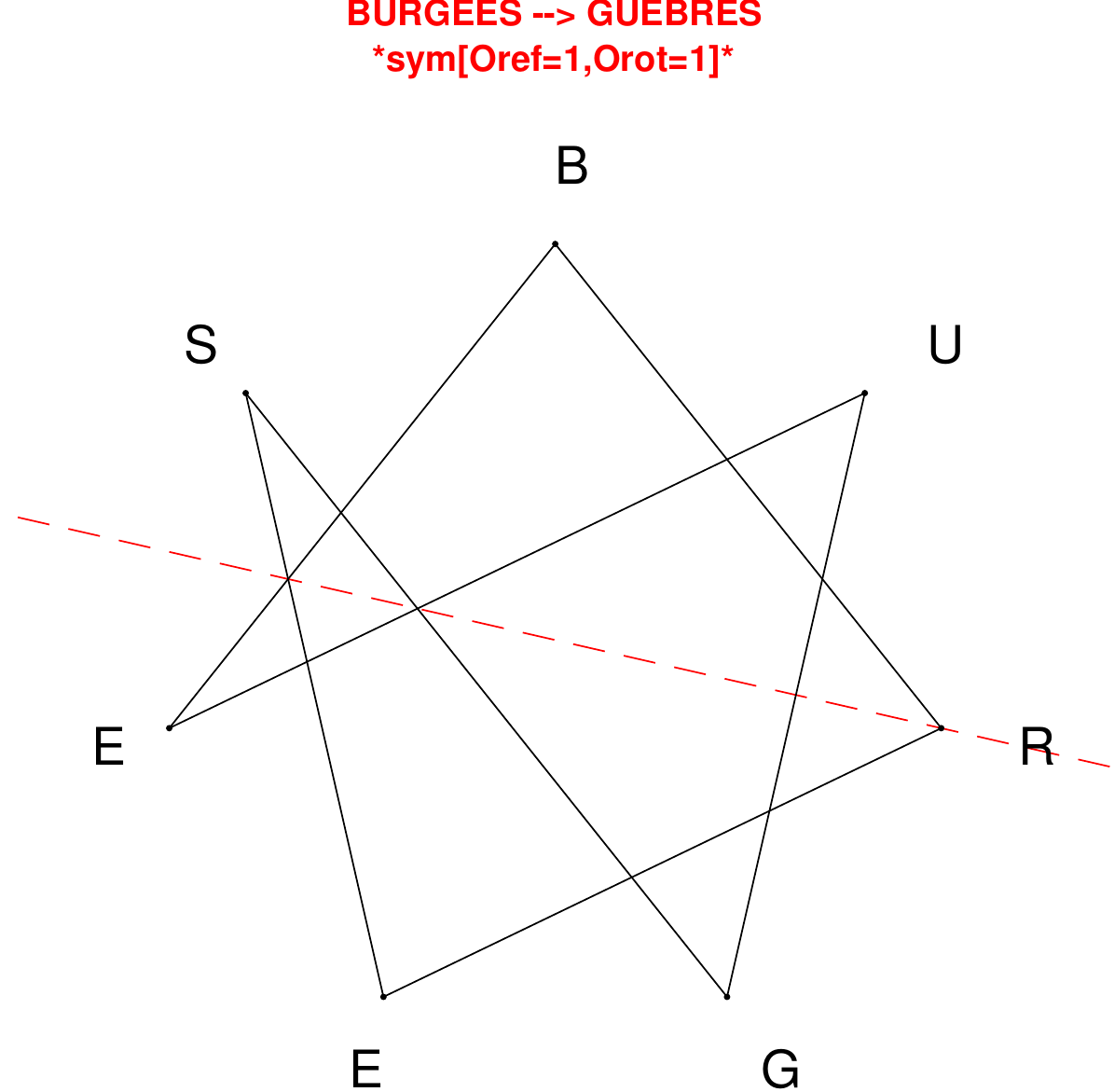}
\end{subfigure}
\hfill
\begin{subfigure}[T]{0.19\textwidth}
\centering
\includegraphics[width=\textwidth]{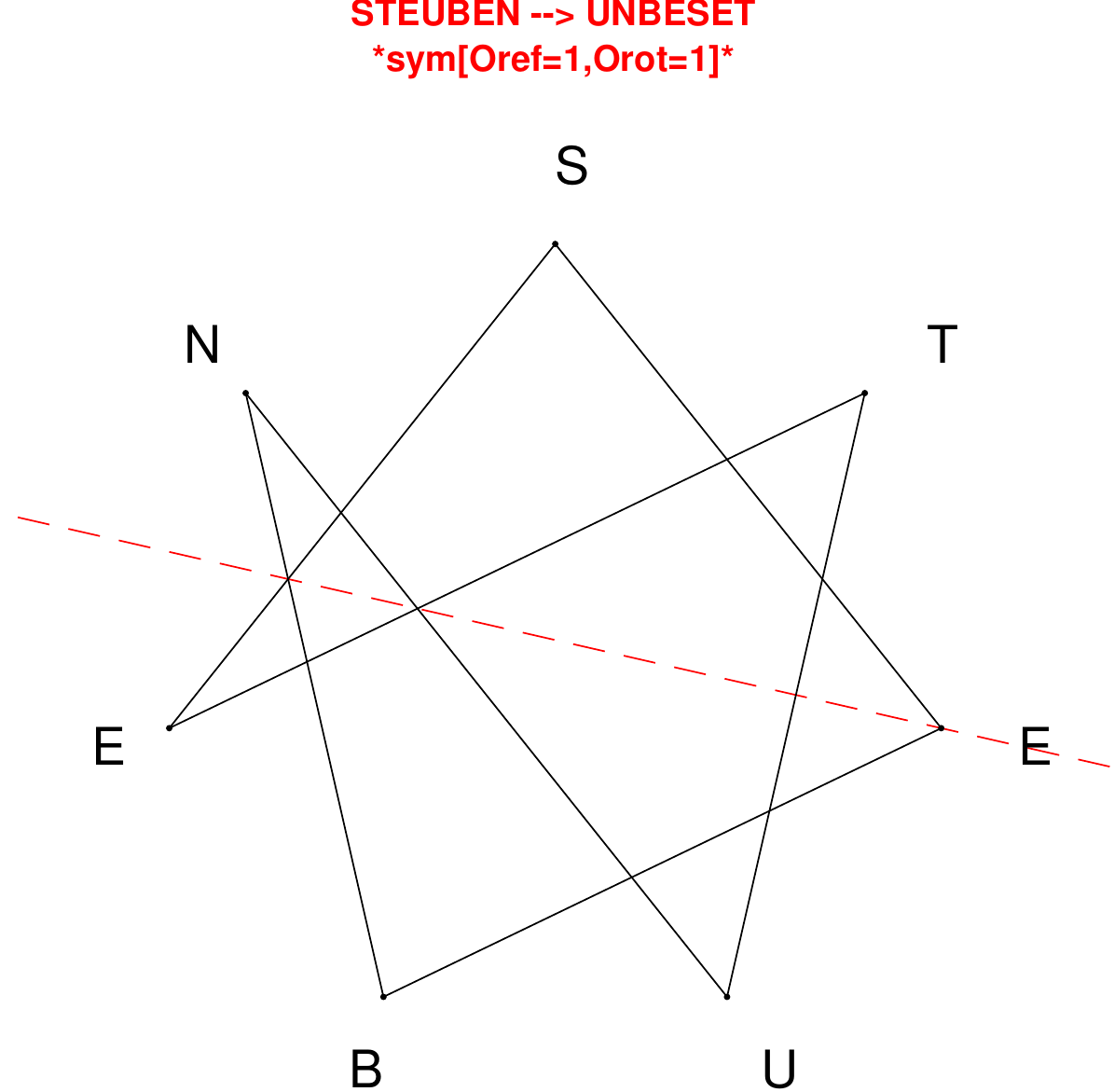}
\end{subfigure}
\hfill
\begin{subfigure}[T]{0.19\textwidth}
\centering
\includegraphics[width=\textwidth]{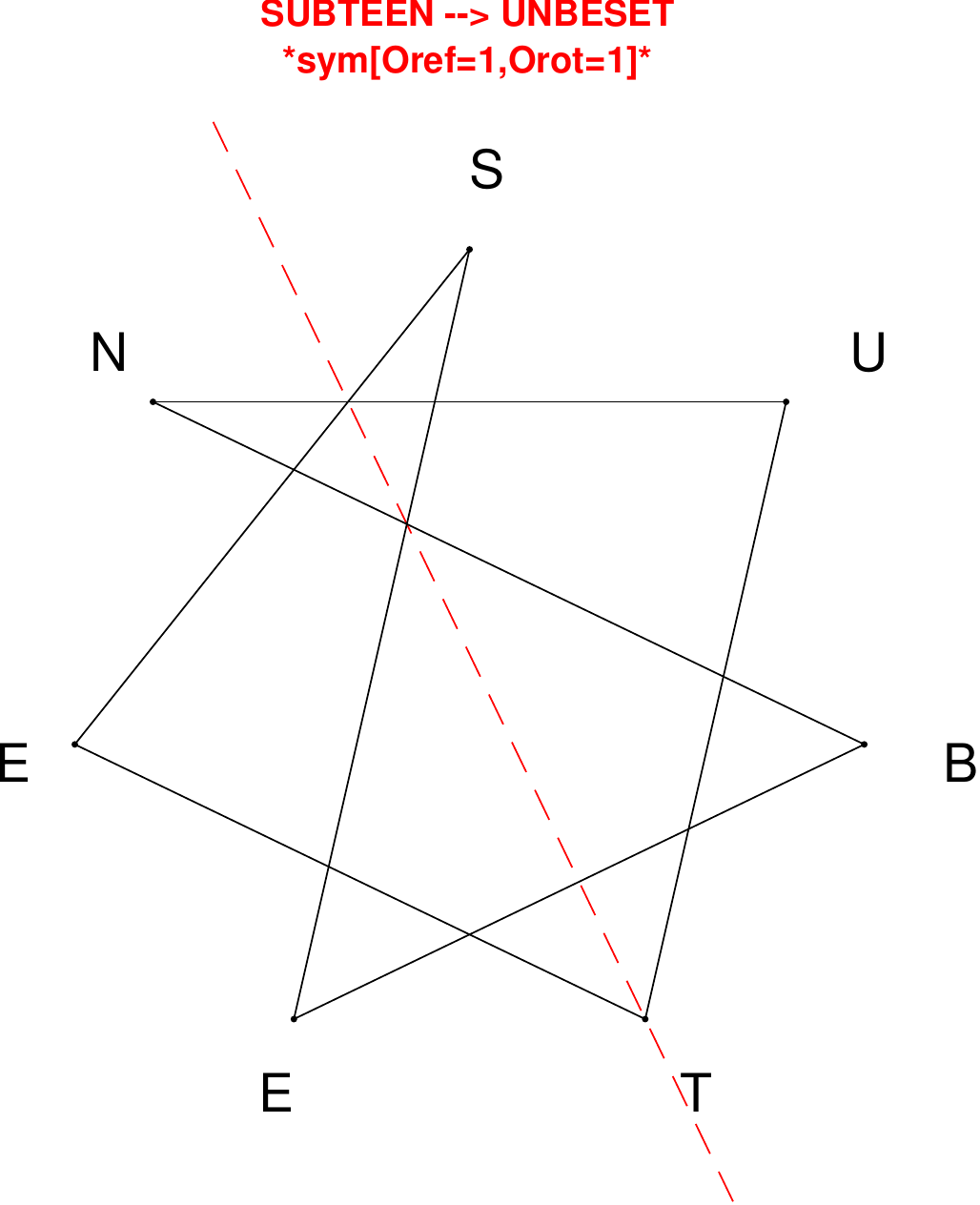}
\end{subfigure}
\end{figure}

\begin{figure}[H]
\centering
\begin{subfigure}[T]{0.19\textwidth}
\centering
\includegraphics[width=\textwidth]{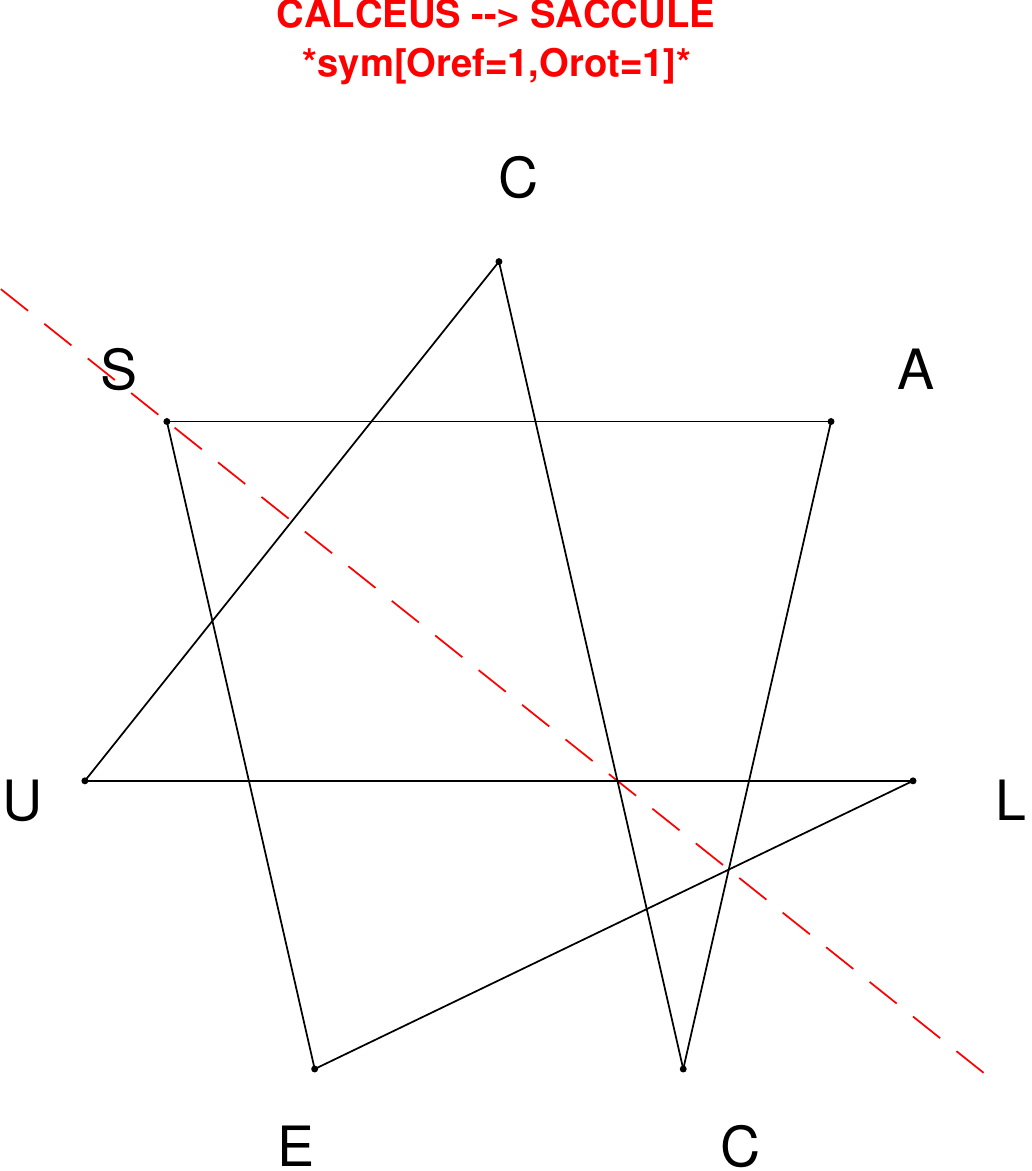}
\end{subfigure}
\hfill
\begin{subfigure}[T]{0.19\textwidth}
\centering
\includegraphics[width=\textwidth]{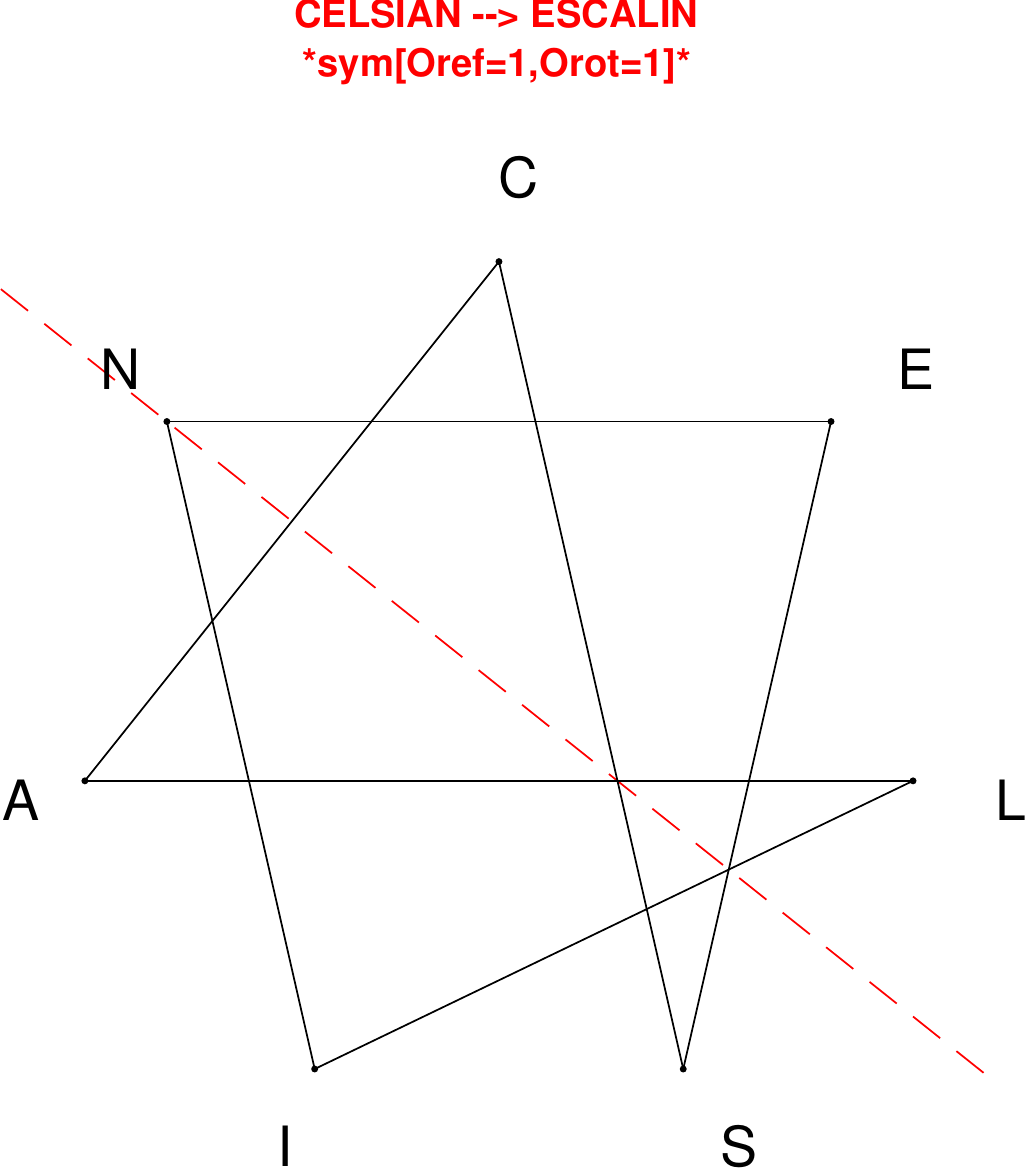}
\end{subfigure}
\hfill
\begin{subfigure}[T]{0.19\textwidth}
\centering
\includegraphics[width=\textwidth]{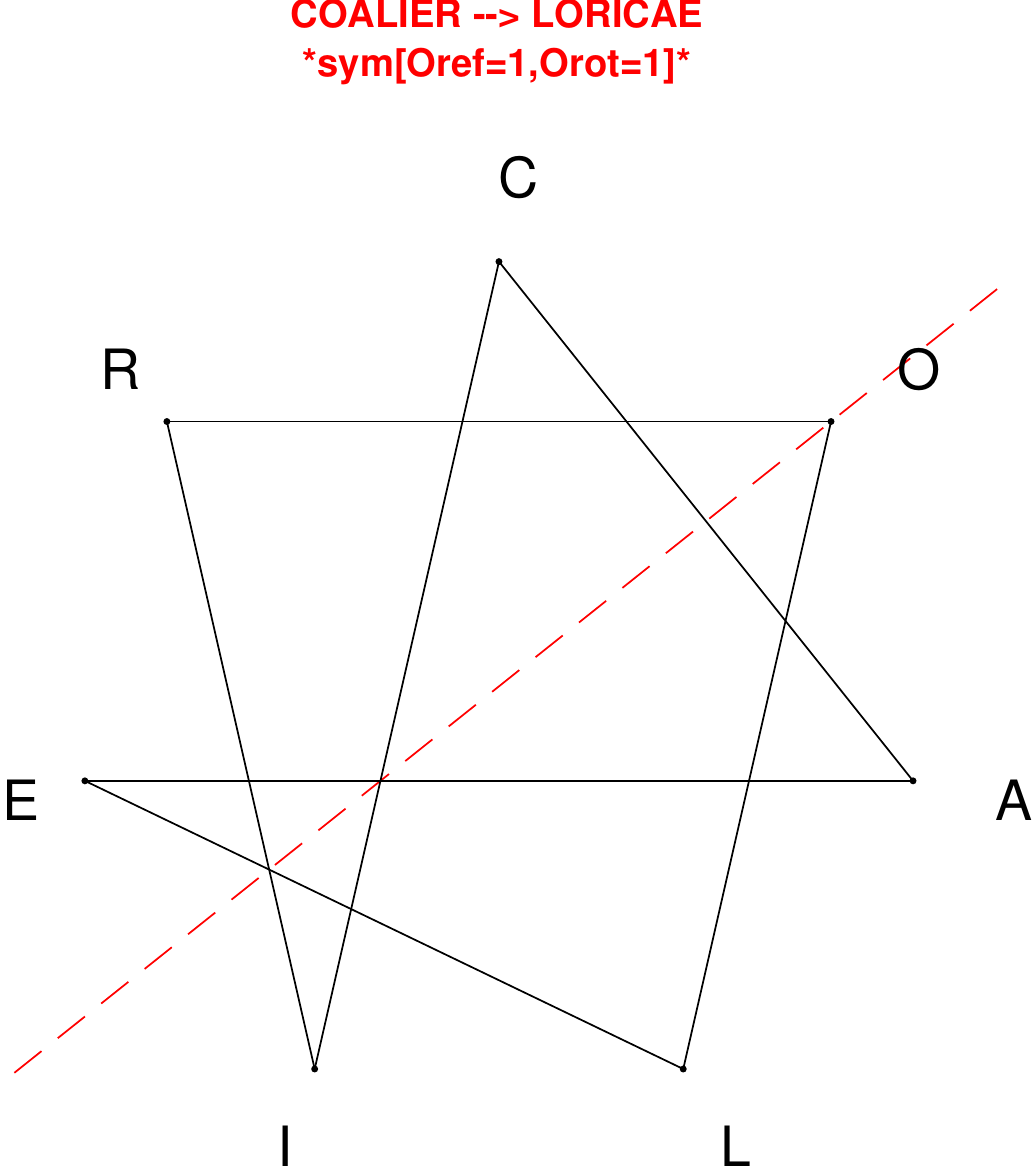}
\end{subfigure}
\hfill
\begin{subfigure}[T]{0.19\textwidth}
\centering
\includegraphics[width=\textwidth]{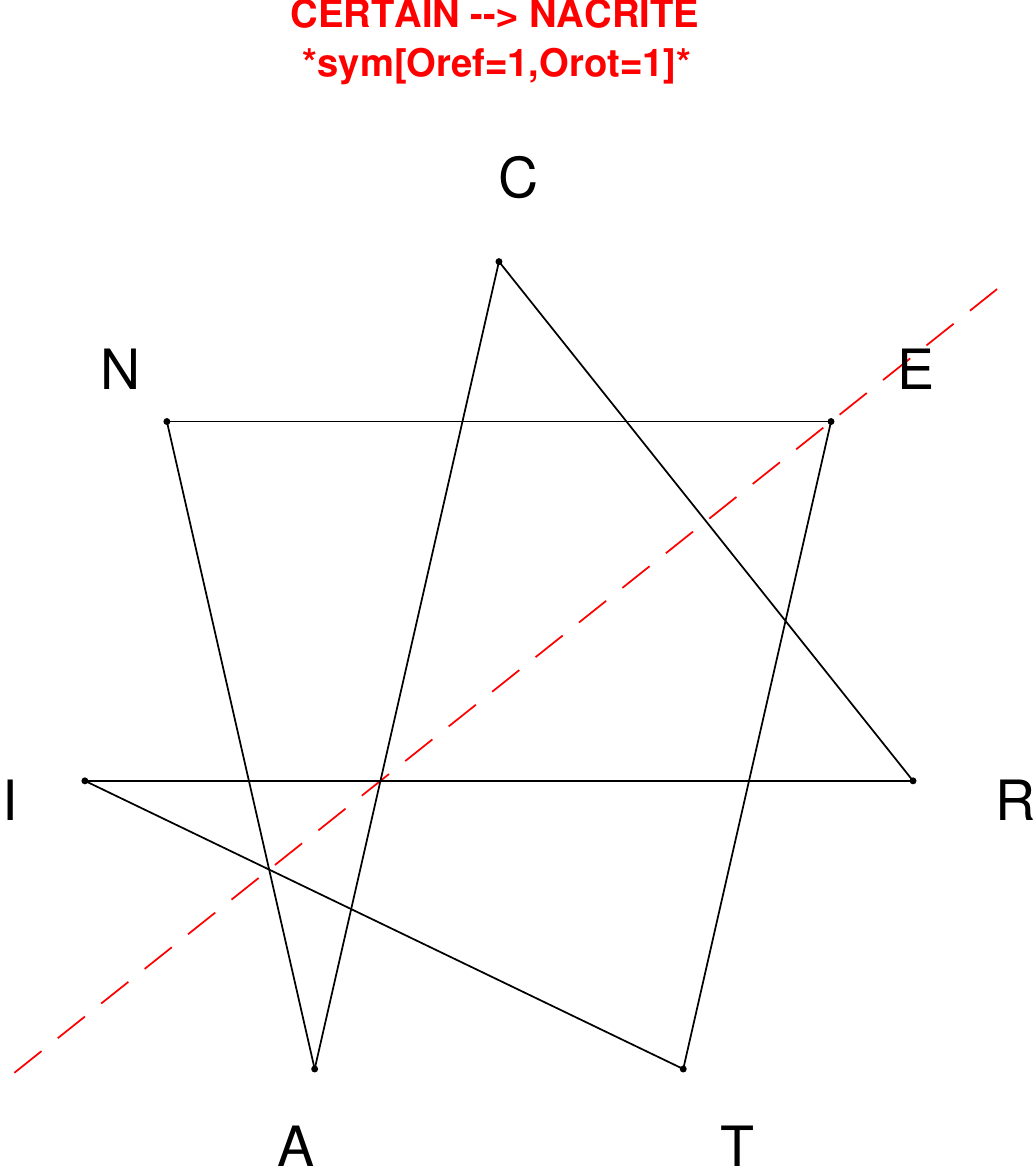}
\end{subfigure}
\hfill
\begin{subfigure}[T]{0.19\textwidth}
\centering
\includegraphics[width=\textwidth]{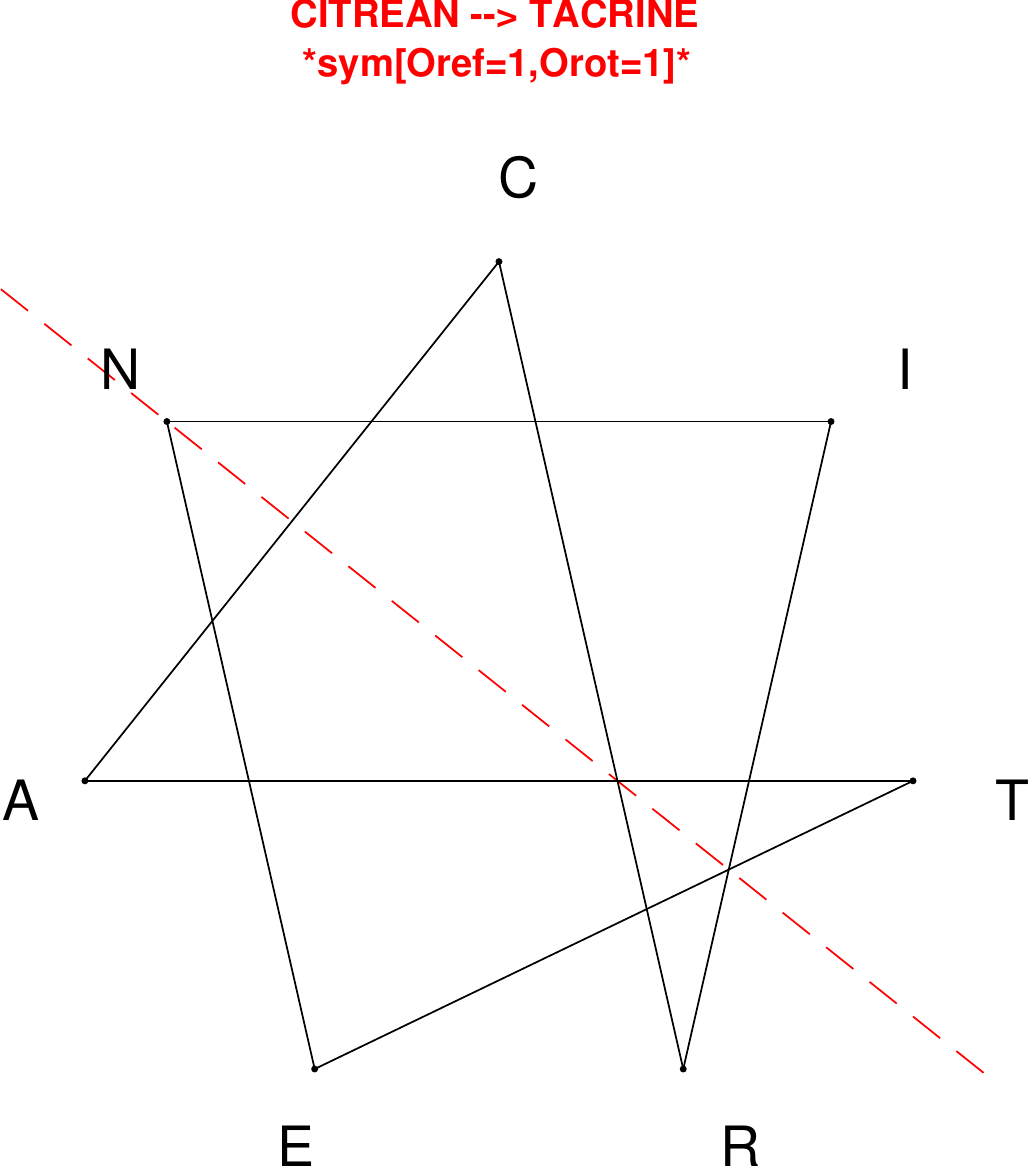}
\end{subfigure}
\end{figure}

\begin{figure}[H]
\centering
\begin{subfigure}[T]{0.19\textwidth}
\centering
\includegraphics[width=\textwidth]{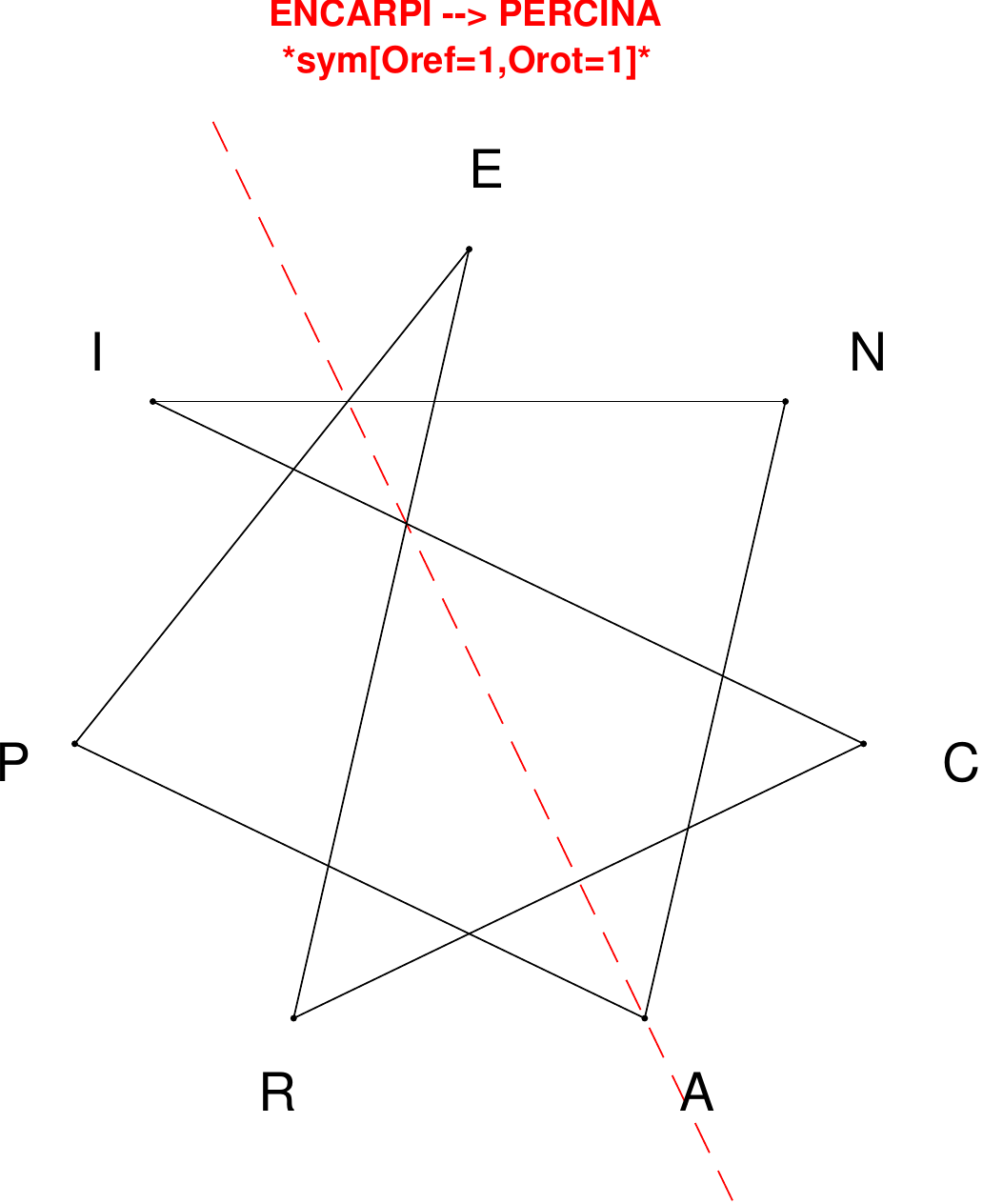}
\end{subfigure}
\hfill
\begin{subfigure}[T]{0.19\textwidth}
\centering
\includegraphics[width=\textwidth]{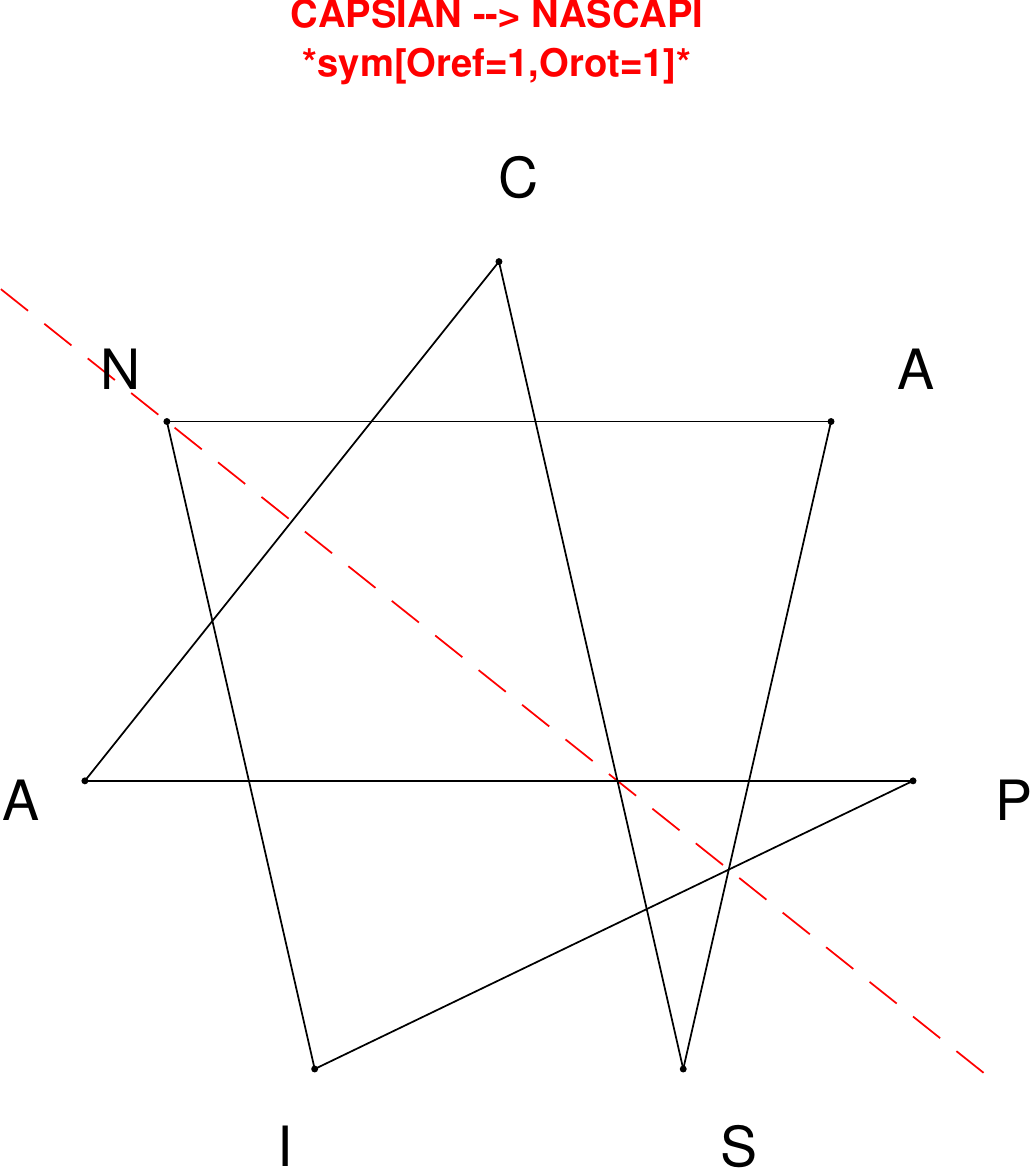}
\end{subfigure}
\hfill
\begin{subfigure}[T]{0.19\textwidth}
\centering
\includegraphics[width=\textwidth]{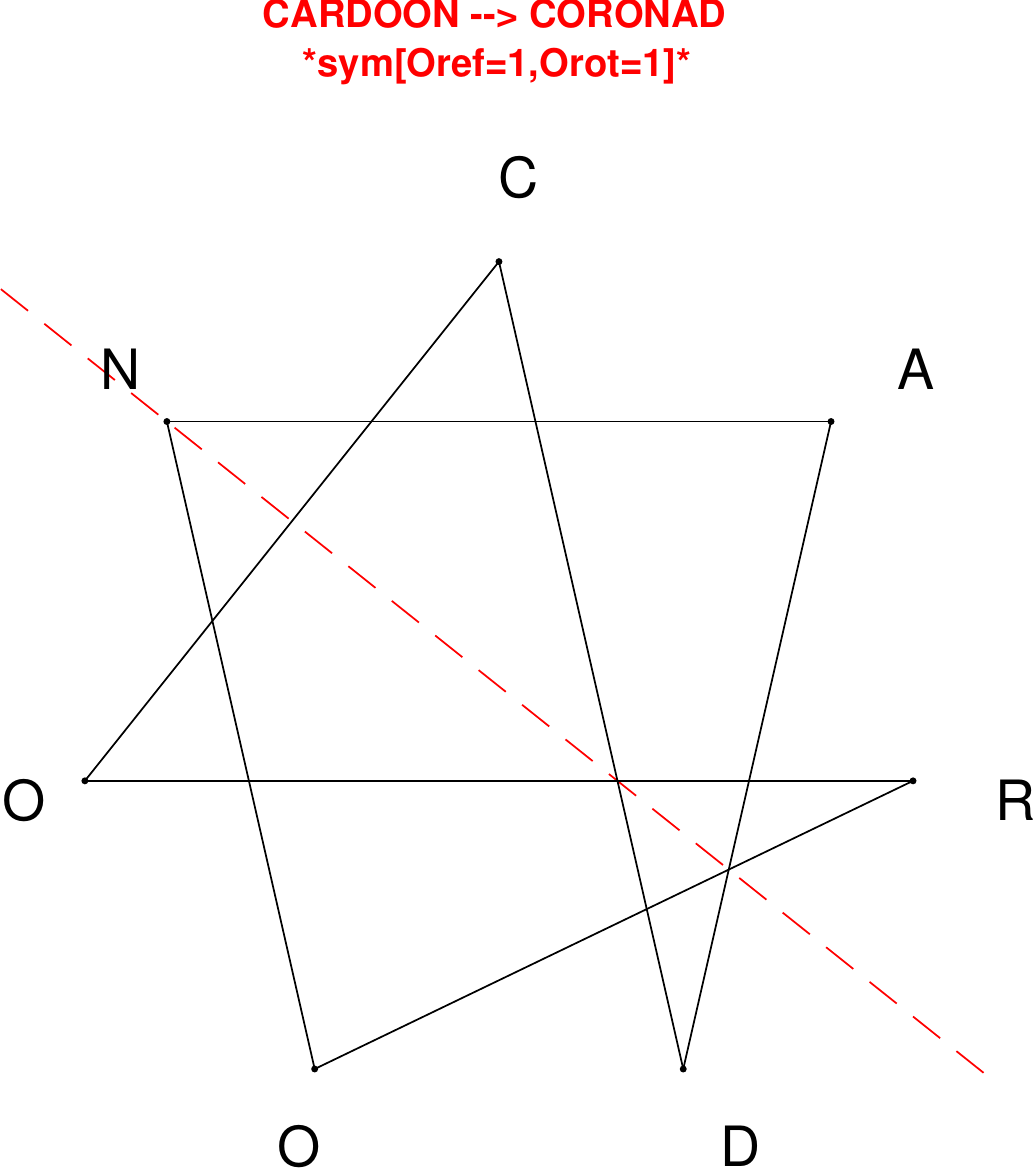}
\end{subfigure}
\hfill
\begin{subfigure}[T]{0.19\textwidth}
\centering
\includegraphics[width=\textwidth]{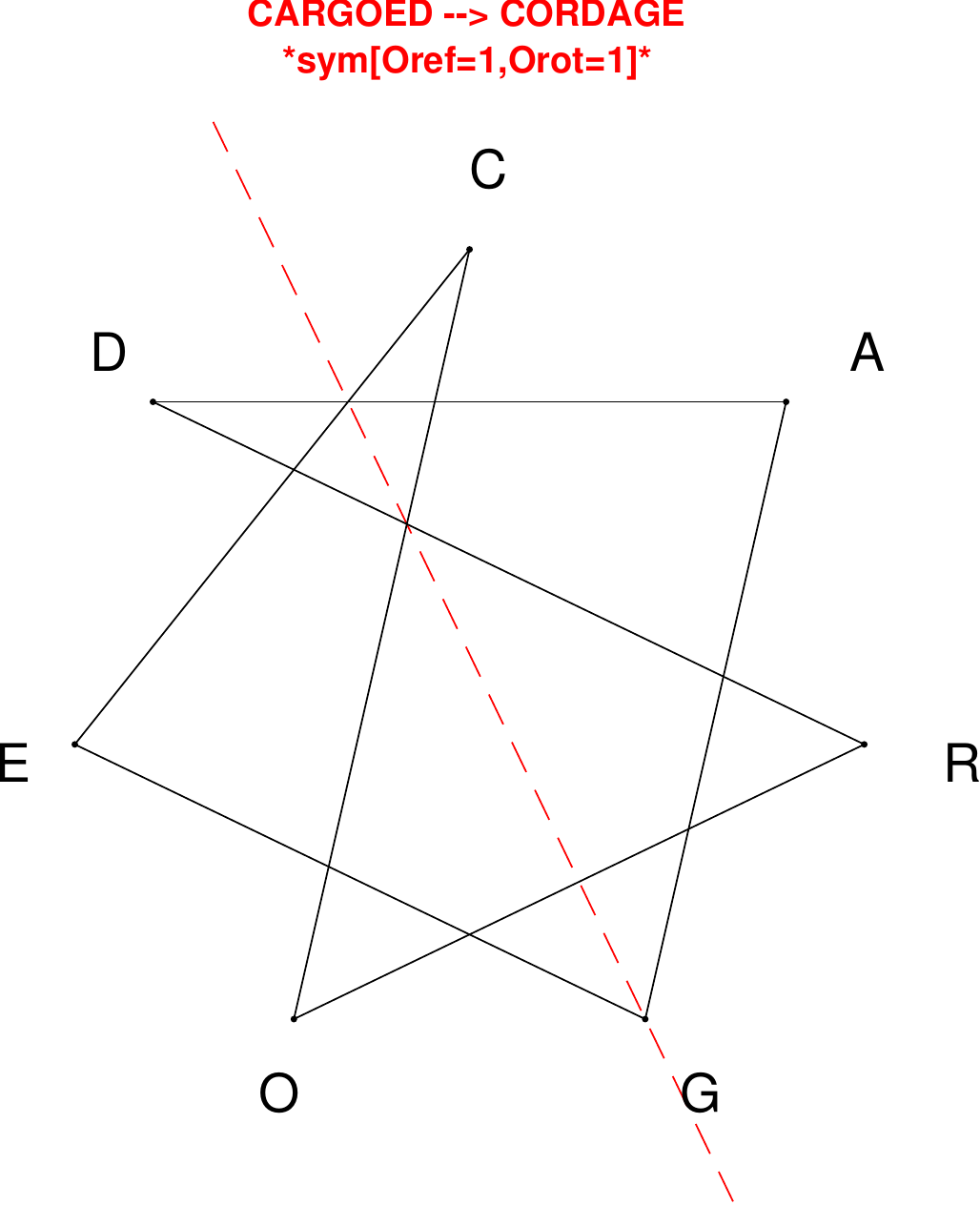}
\end{subfigure}
\hfill
\begin{subfigure}[T]{0.19\textwidth}
\centering
\includegraphics[width=\textwidth]{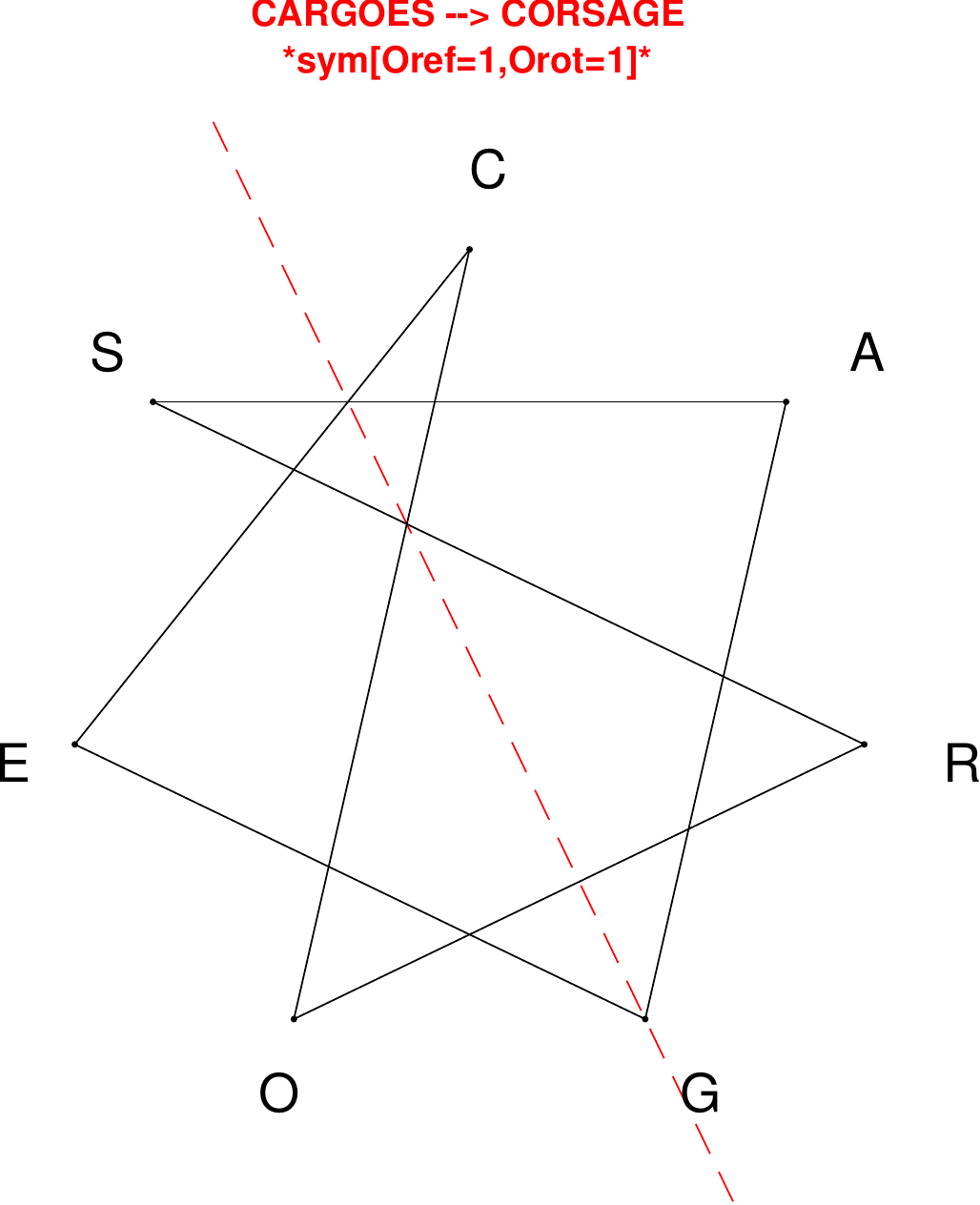}
\end{subfigure}
\end{figure}

\begin{figure}[H]
\centering
\begin{subfigure}[T]{0.19\textwidth}
\centering
\includegraphics[width=\textwidth]{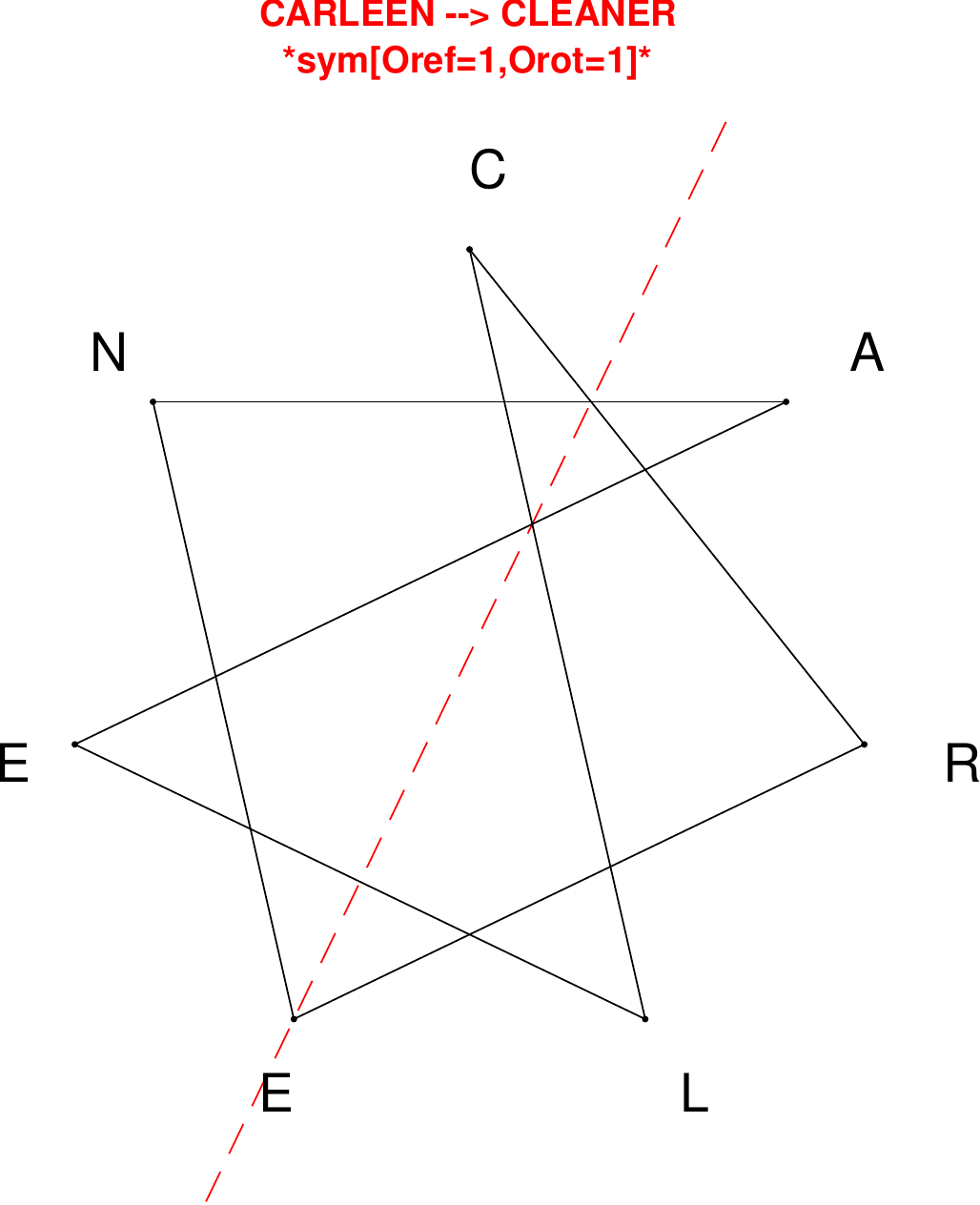}
\end{subfigure}
\hfill
\begin{subfigure}[T]{0.19\textwidth}
\centering
\includegraphics[width=\textwidth]{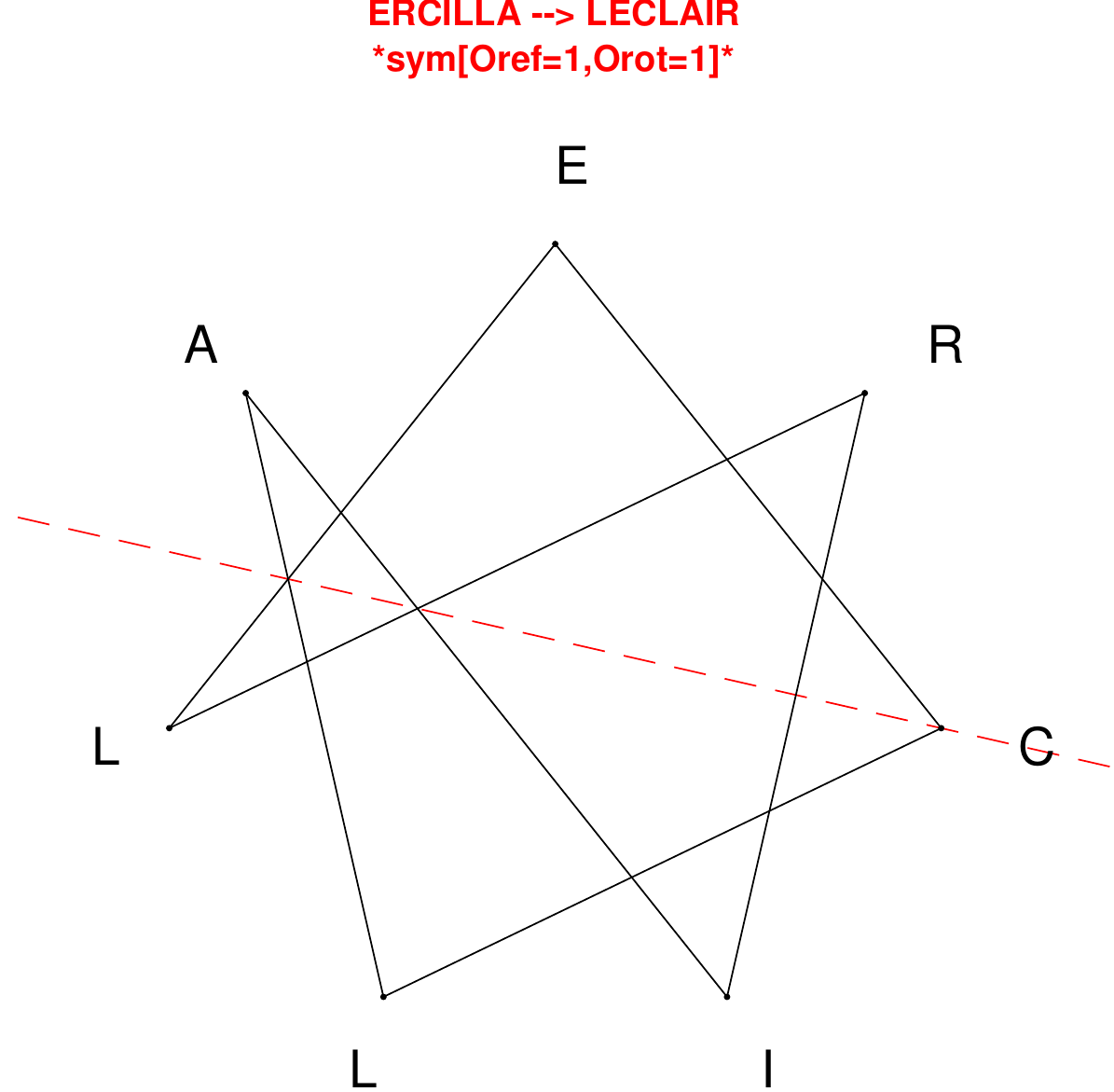}
\end{subfigure}
\hfill
\begin{subfigure}[T]{0.19\textwidth}
\centering
\includegraphics[width=\textwidth]{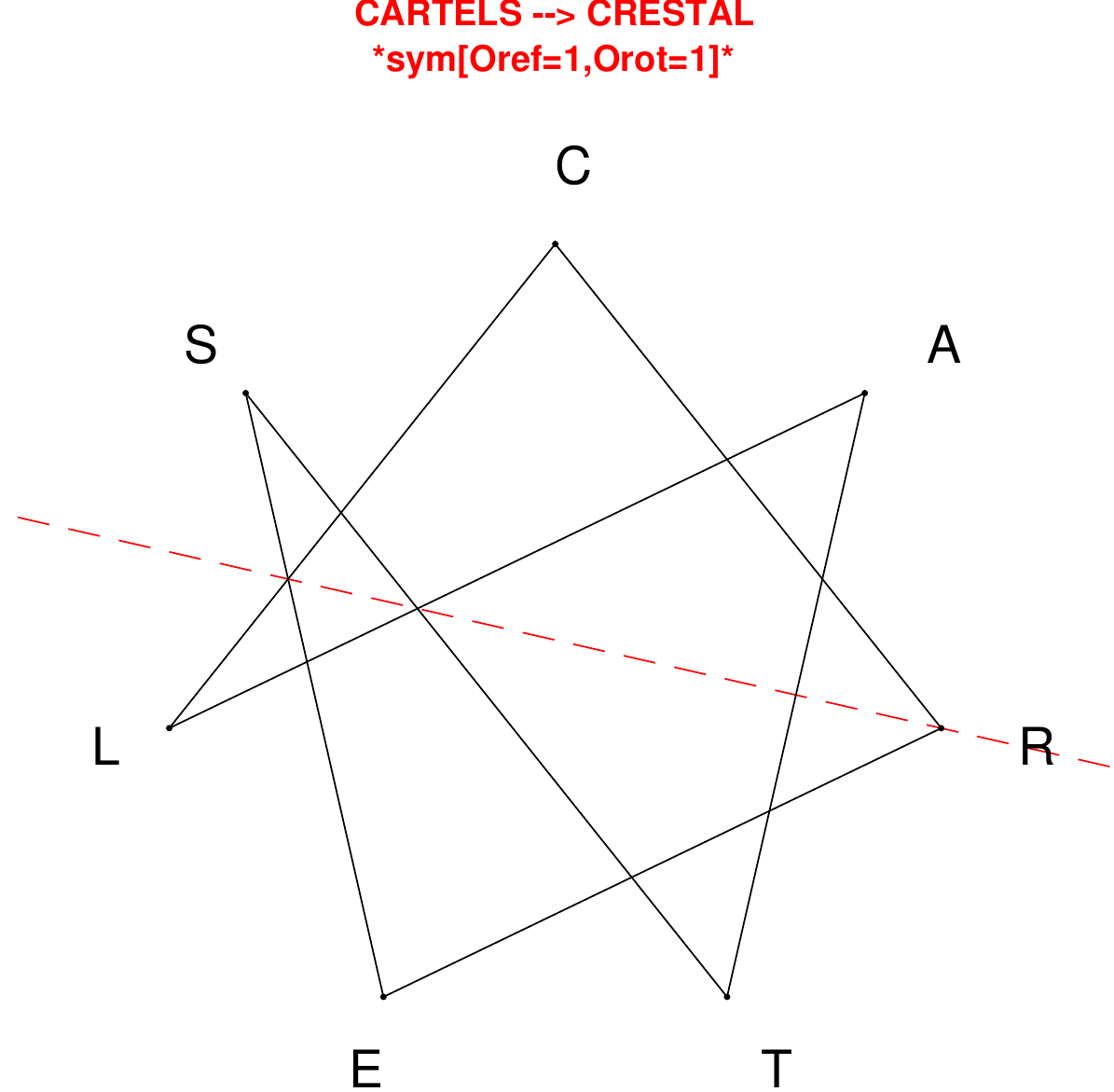}
\end{subfigure}
\hfill
\begin{subfigure}[T]{0.19\textwidth}
\centering
\includegraphics[width=\textwidth]{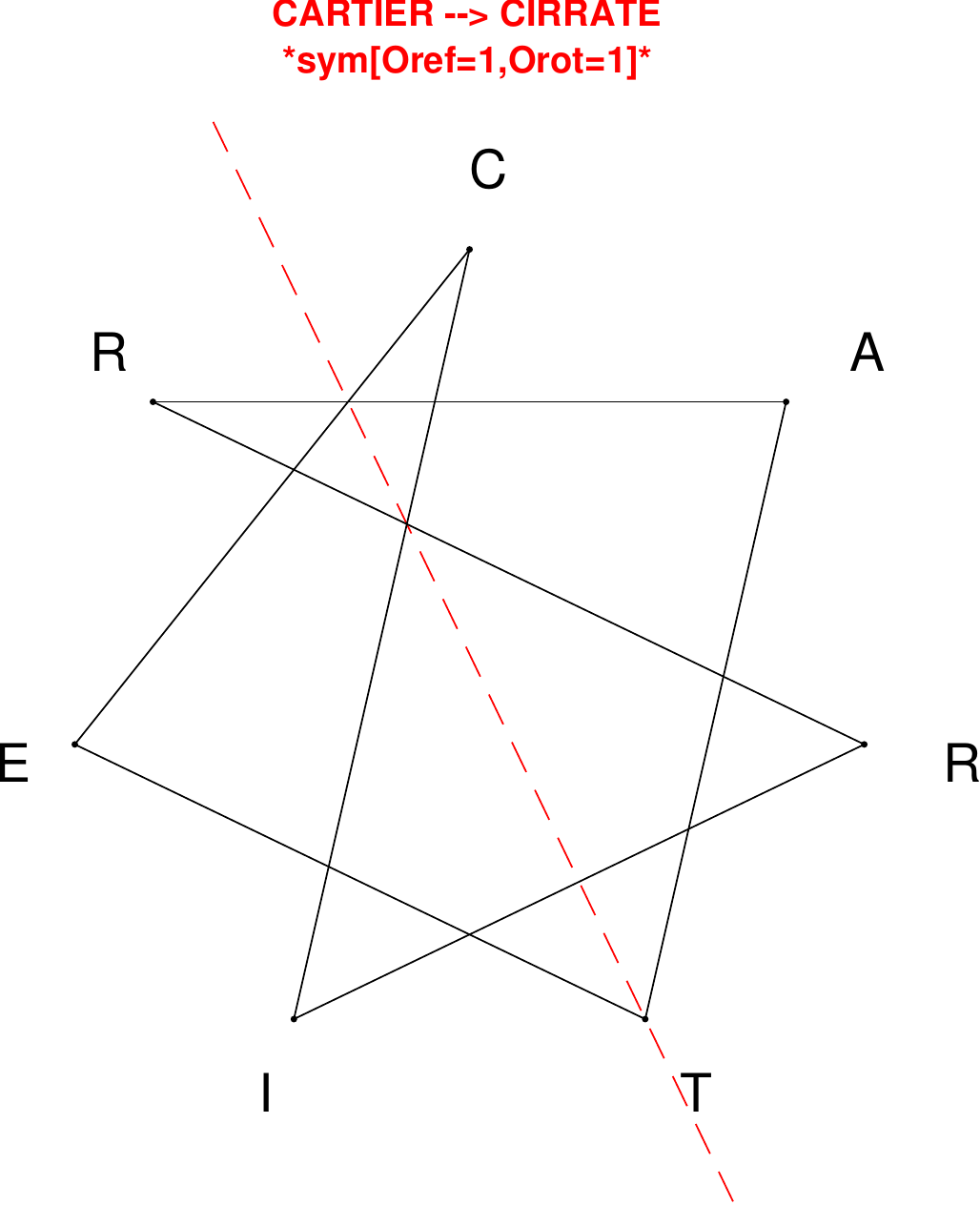}
\end{subfigure}
\hfill
\begin{subfigure}[T]{0.19\textwidth}
\centering
\includegraphics[width=\textwidth]{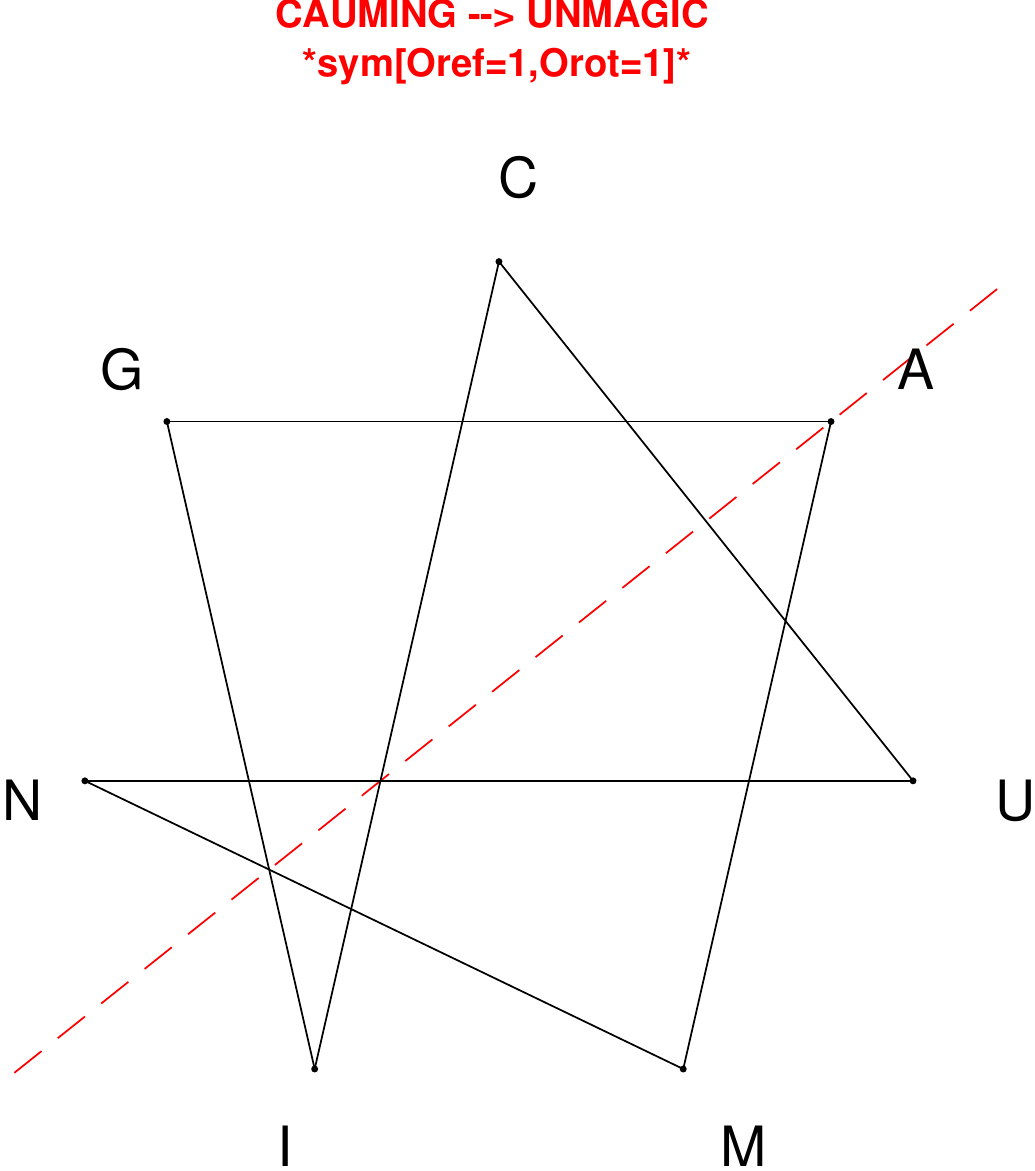}
\end{subfigure}
\end{figure}

\begin{figure}[H]
\centering
\begin{subfigure}[T]{0.19\textwidth}
\centering
\includegraphics[width=\textwidth]{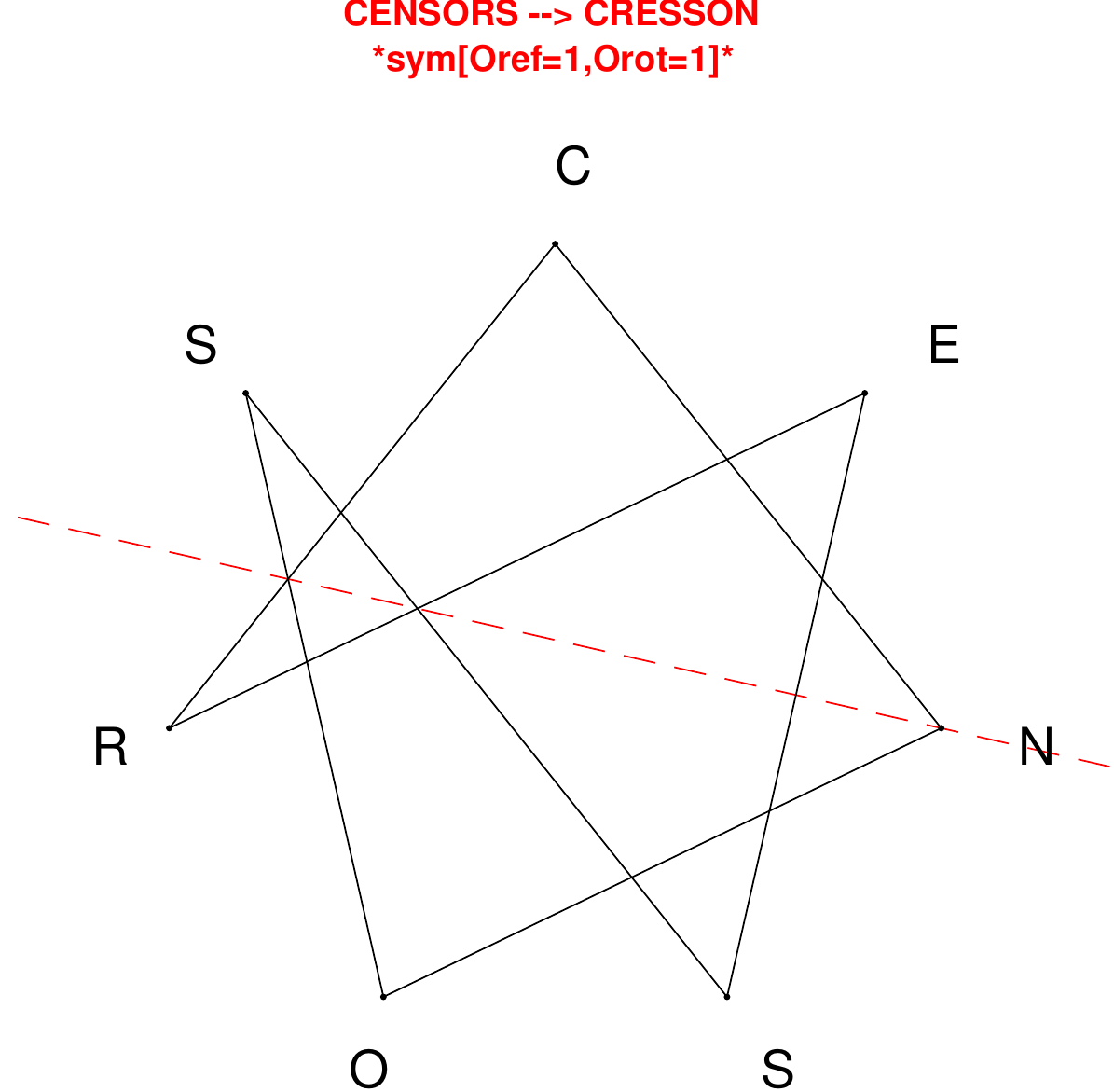}
\end{subfigure}
\hfill
\begin{subfigure}[T]{0.19\textwidth}
\centering
\includegraphics[width=\textwidth]{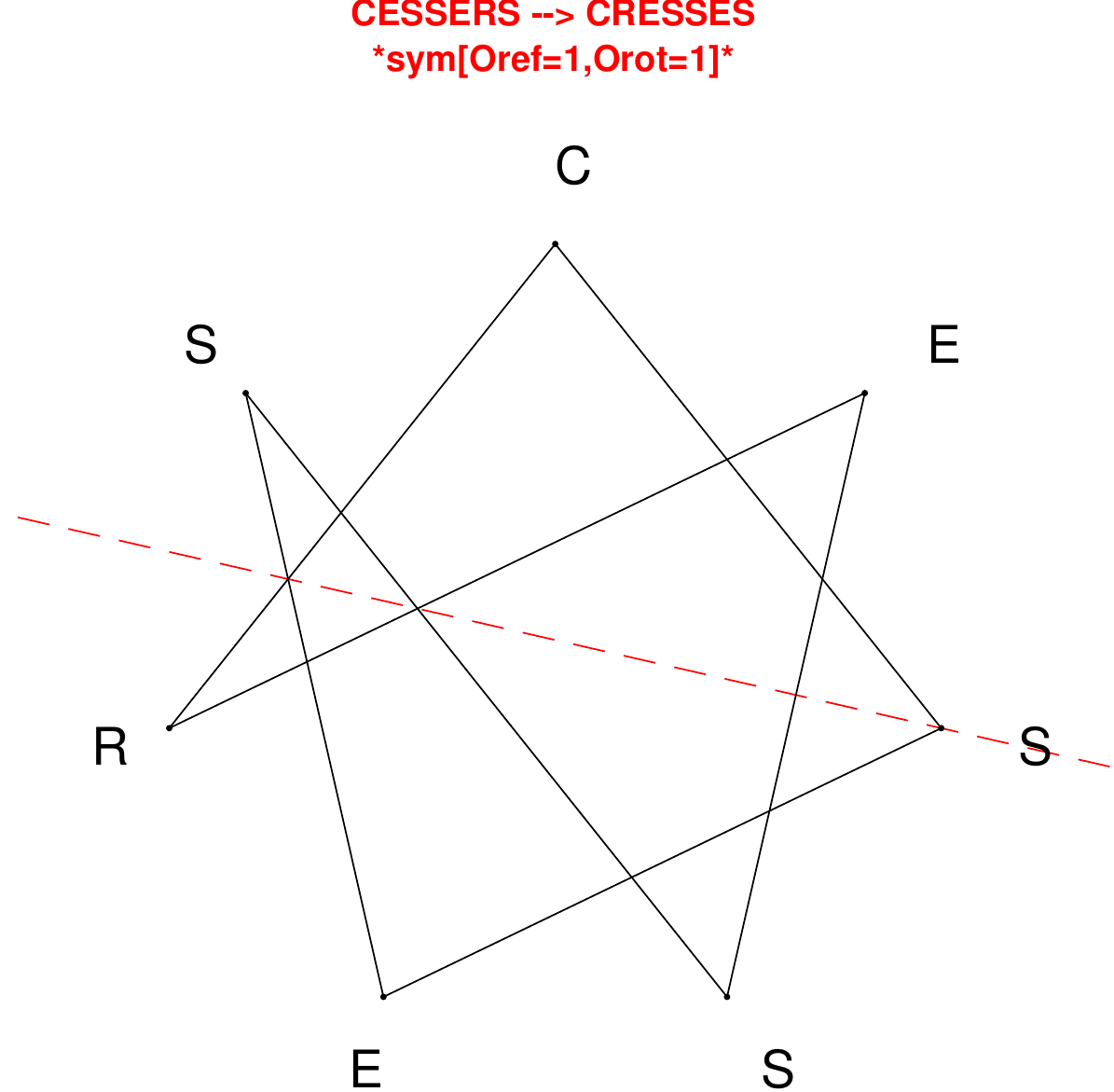}
\end{subfigure}
\hfill
\begin{subfigure}[T]{0.19\textwidth}
\centering
\includegraphics[width=\textwidth]{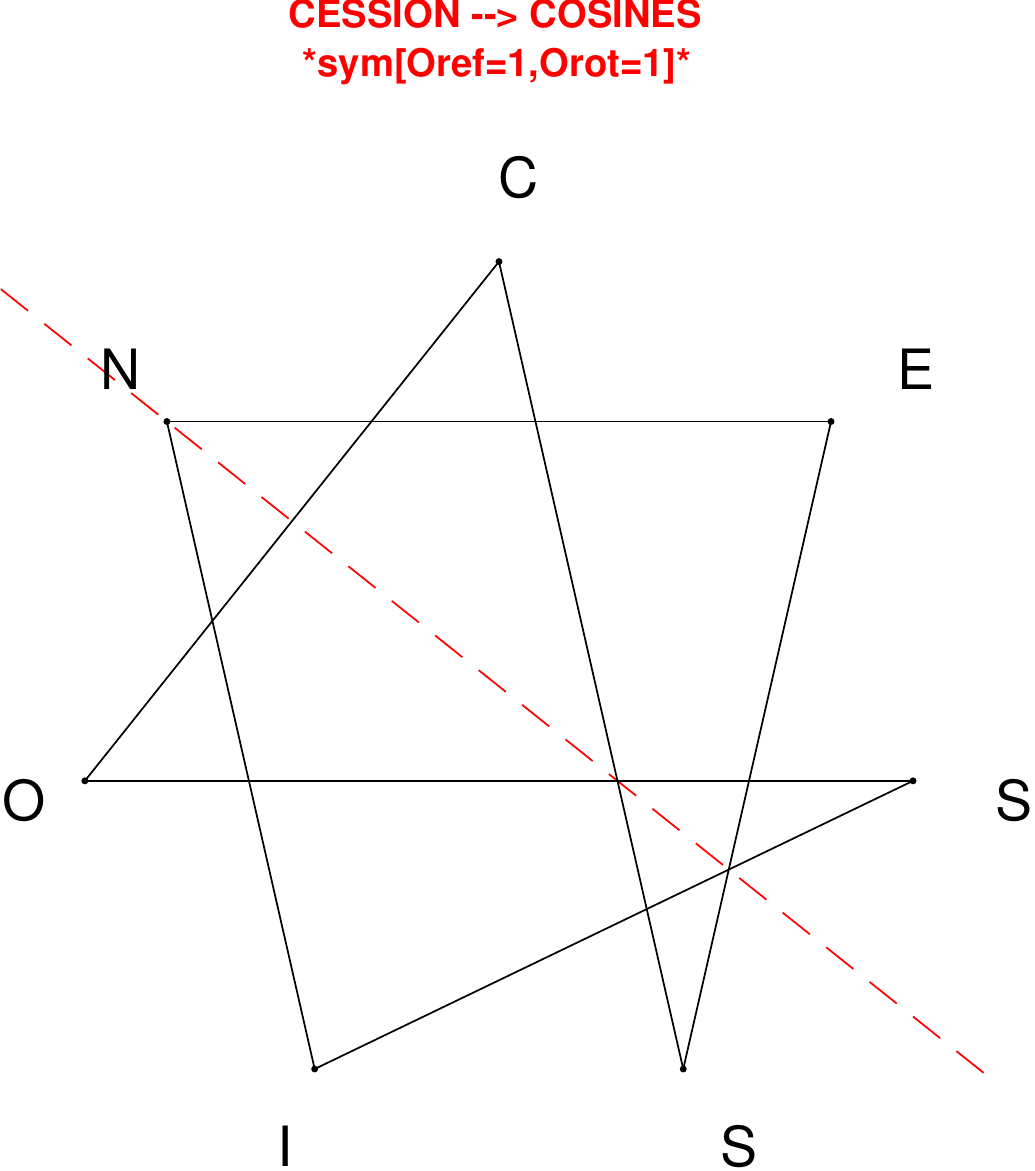}
\end{subfigure}
\hfill
\begin{subfigure}[T]{0.19\textwidth}
\centering
\includegraphics[width=\textwidth]{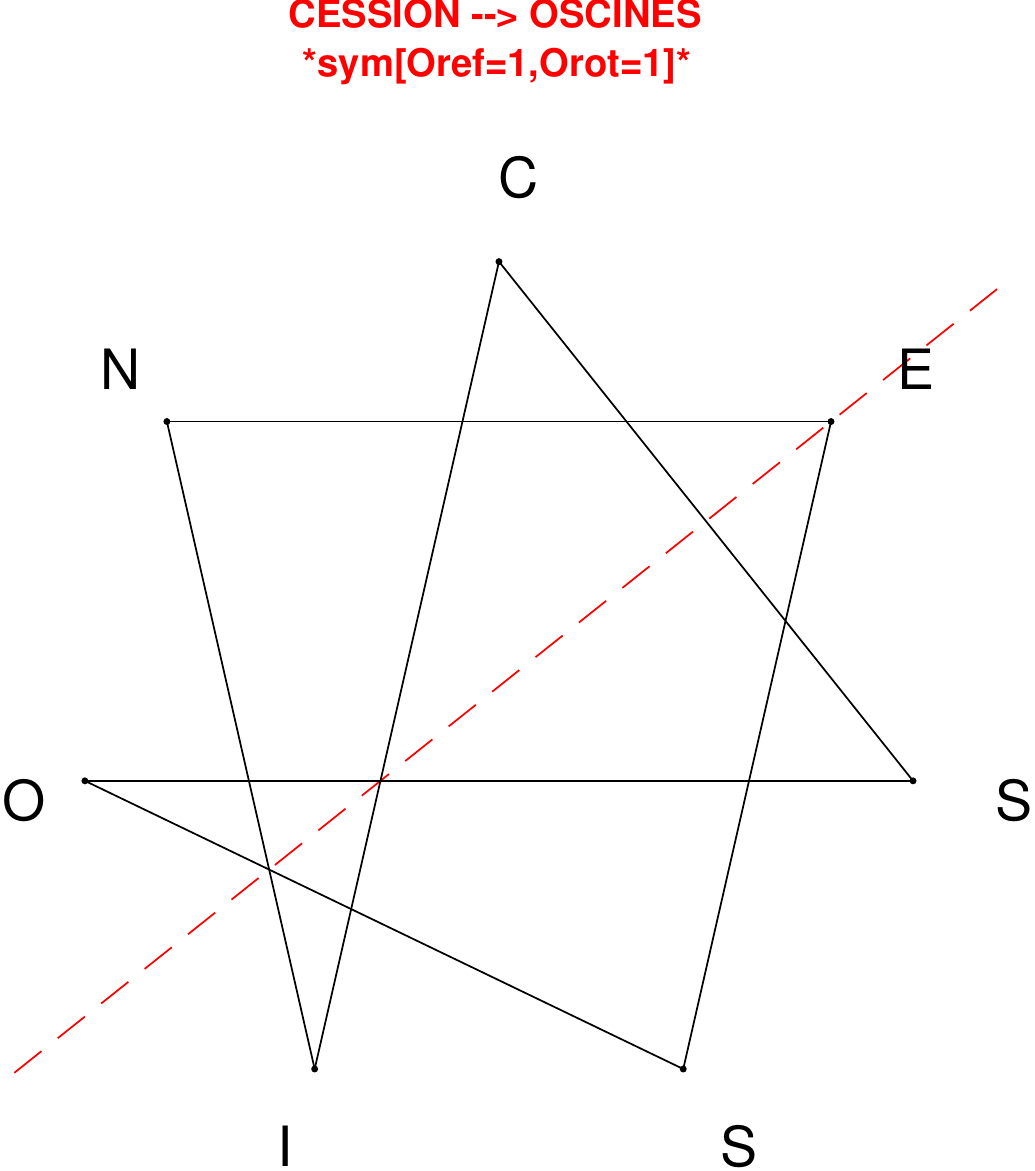}
\end{subfigure}
\hfill
\begin{subfigure}[T]{0.19\textwidth}
\centering
\includegraphics[width=\textwidth]{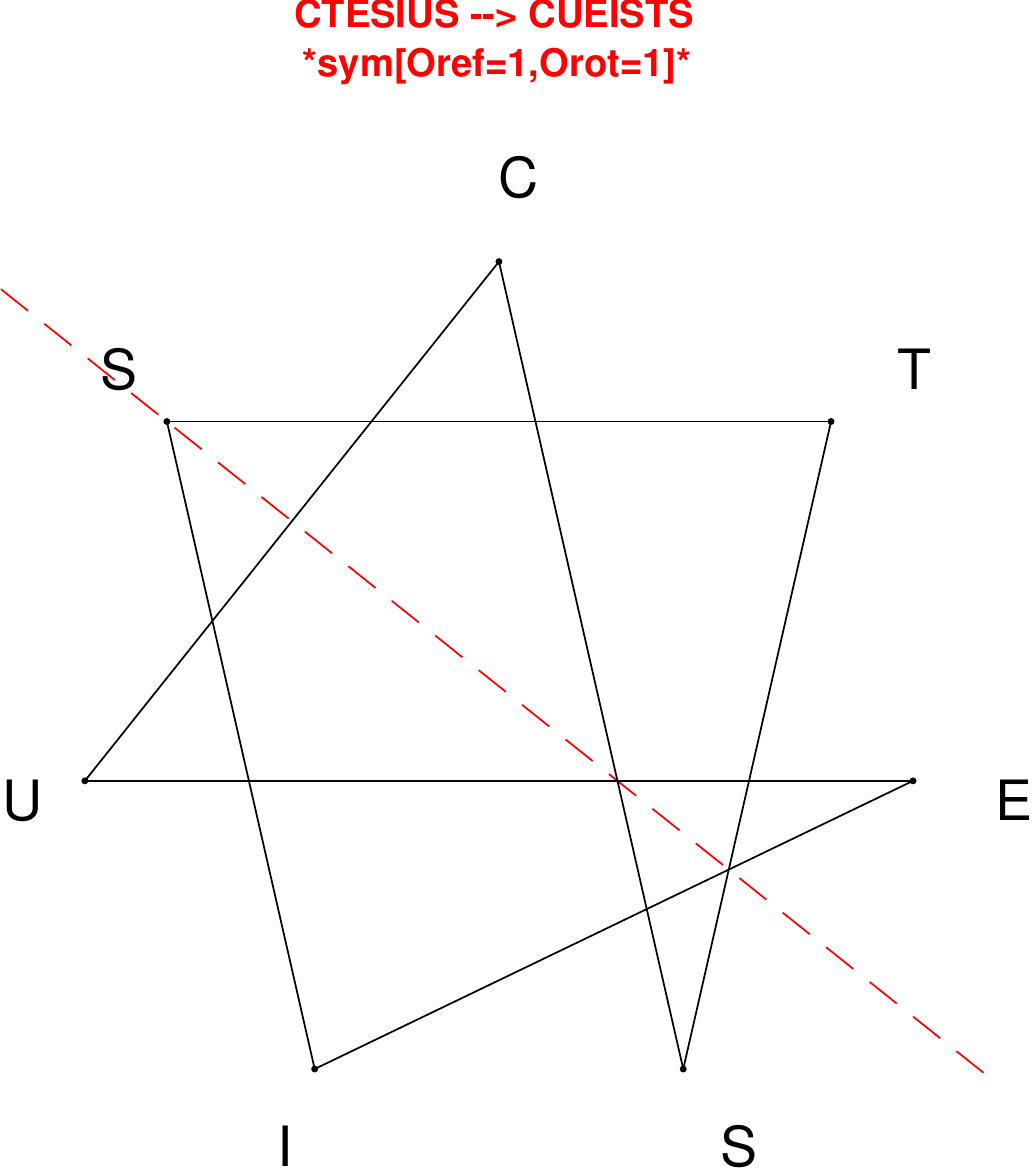}
\end{subfigure}
\end{figure}

\begin{figure}[H]
\centering
\begin{subfigure}[T]{0.19\textwidth}
\centering
\includegraphics[width=\textwidth]{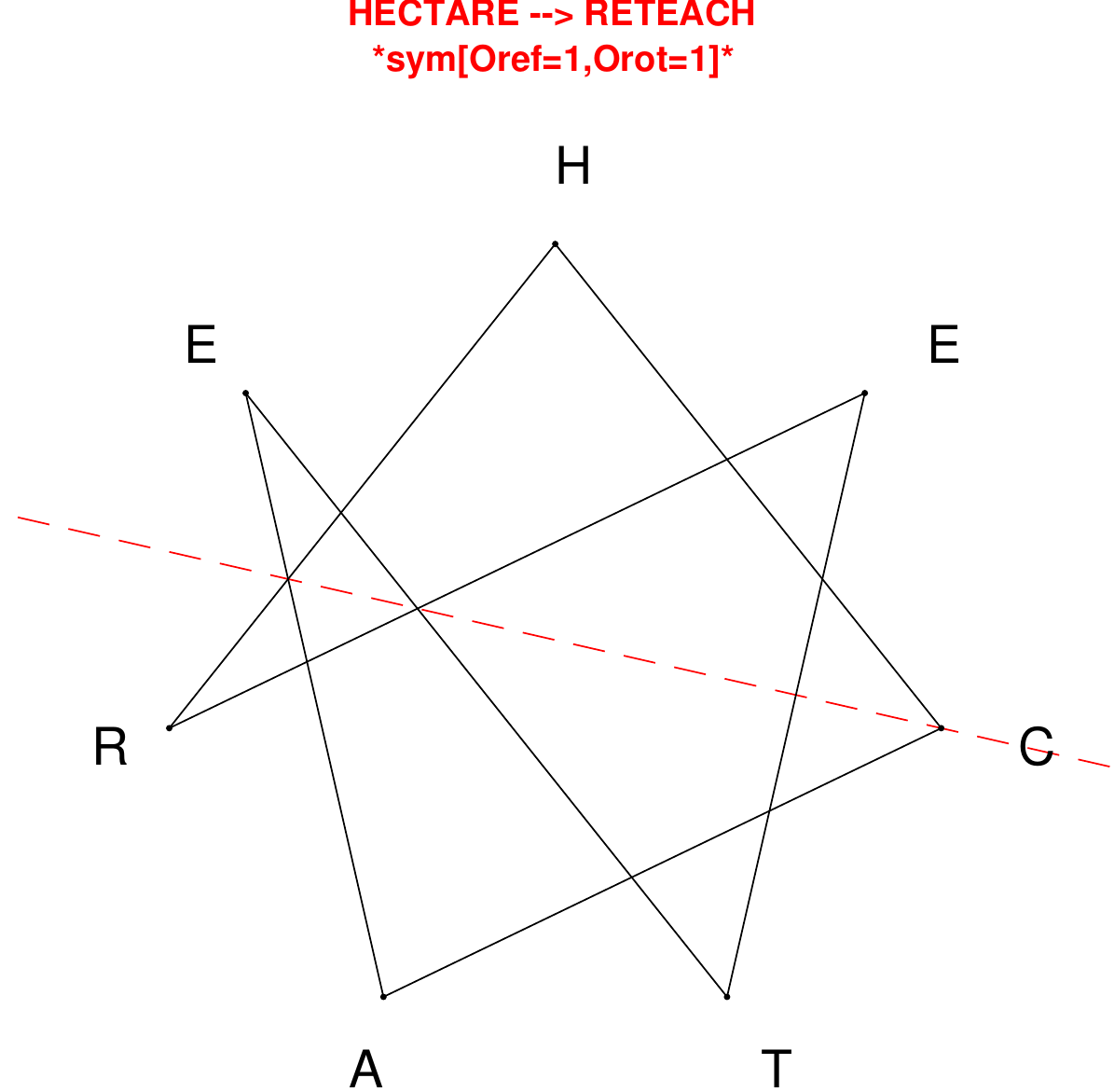}
\end{subfigure}
\hfill
\begin{subfigure}[T]{0.19\textwidth}
\centering
\includegraphics[width=\textwidth]{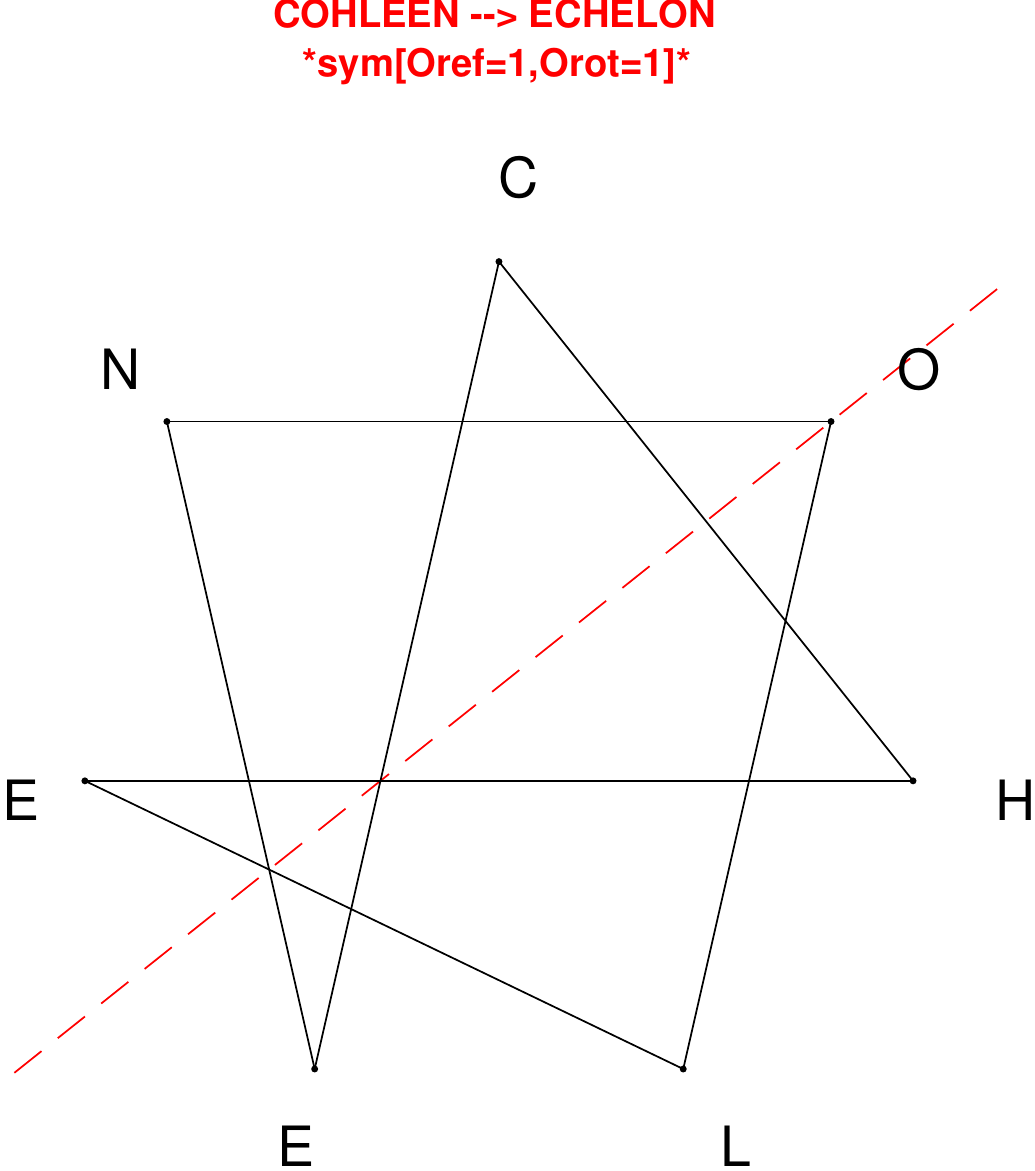}
\end{subfigure}
\hfill
\begin{subfigure}[T]{0.19\textwidth}
\centering
\includegraphics[width=\textwidth]{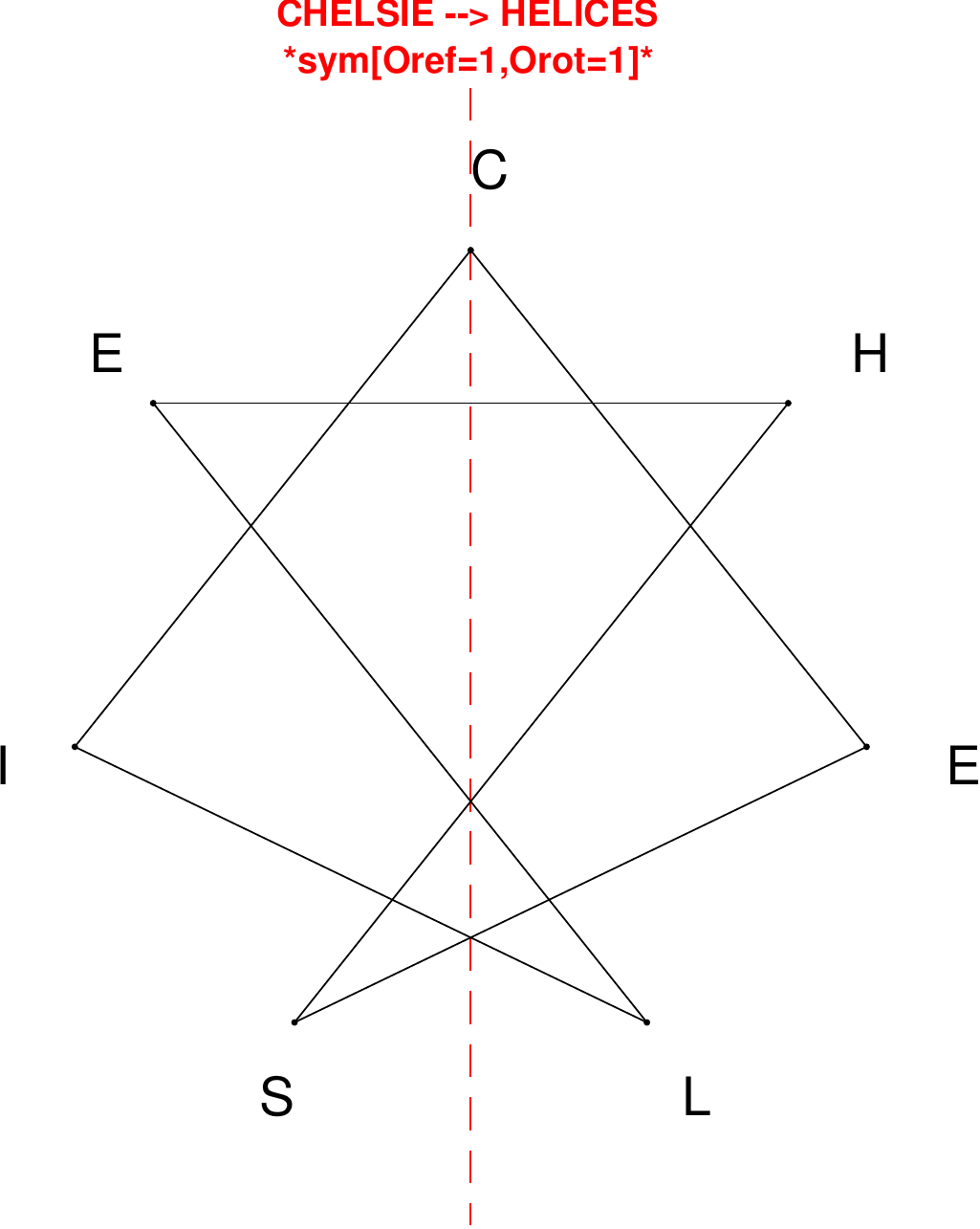}
\end{subfigure}
\hfill
\begin{subfigure}[T]{0.19\textwidth}
\centering
\includegraphics[width=\textwidth]{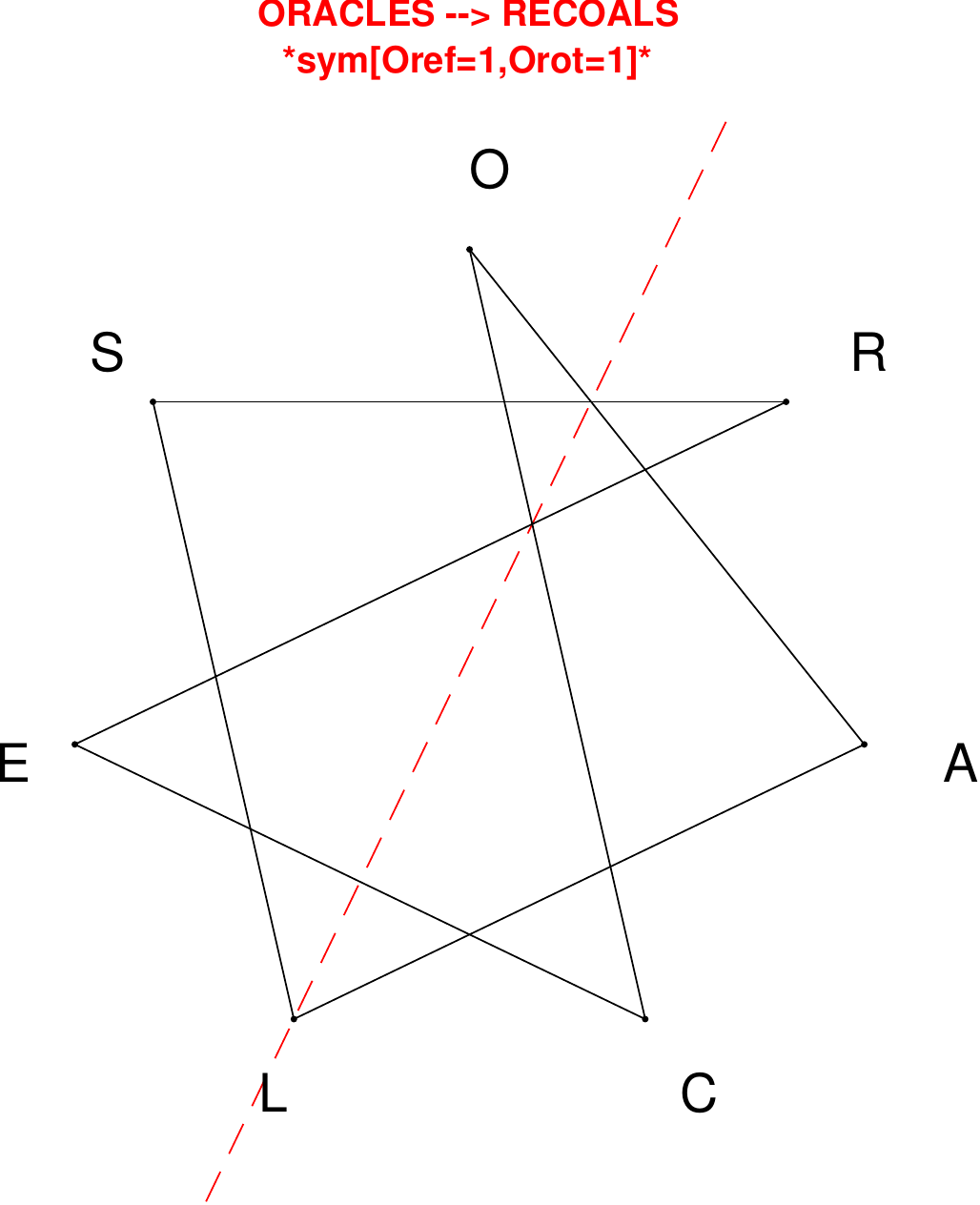}
\end{subfigure}
\hfill
\begin{subfigure}[T]{0.19\textwidth}
\centering
\includegraphics[width=\textwidth]{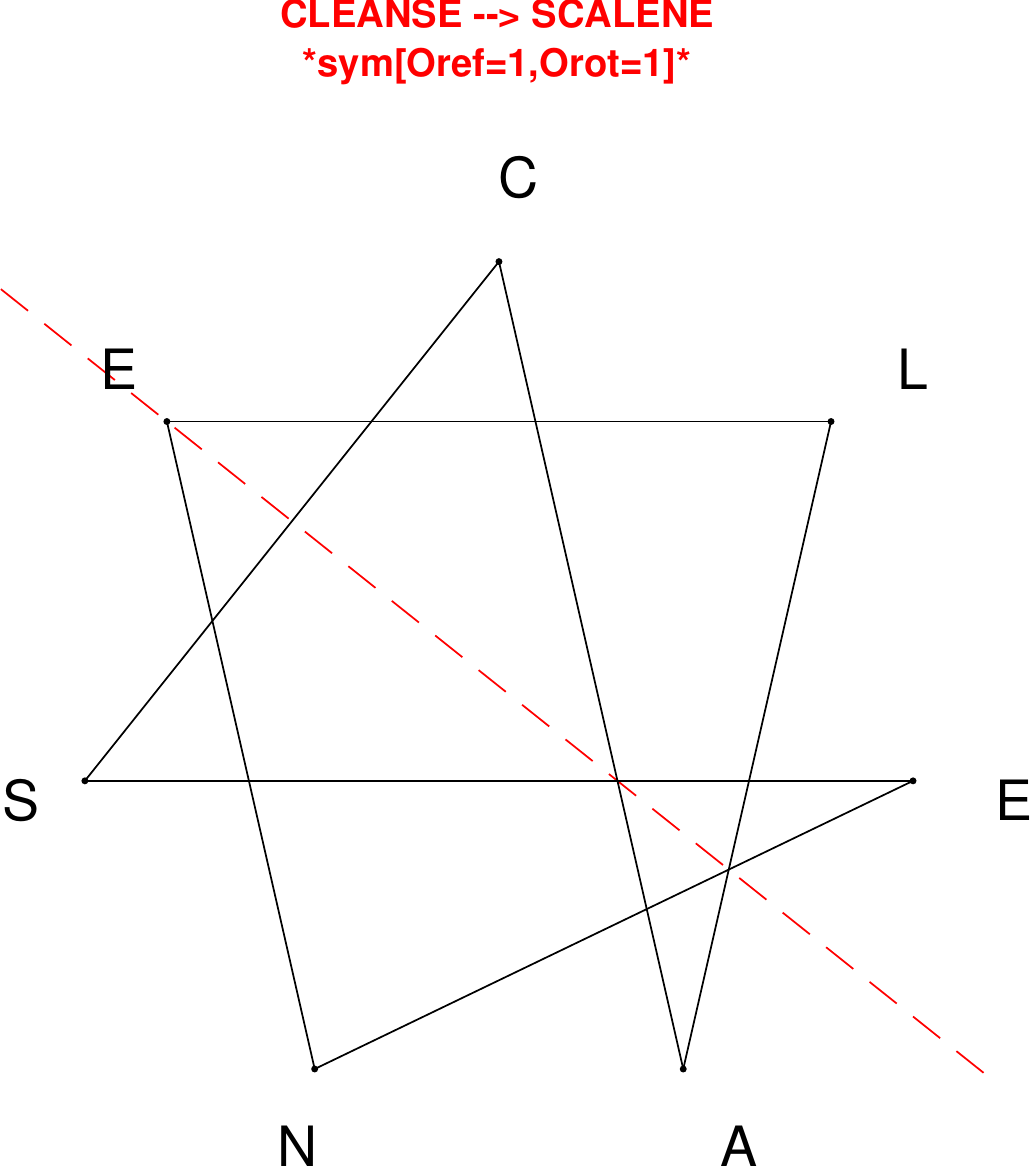}
\end{subfigure}
\end{figure}

\begin{figure}[H]
\centering
\begin{subfigure}[T]{0.19\textwidth}
\centering
\includegraphics[width=\textwidth]{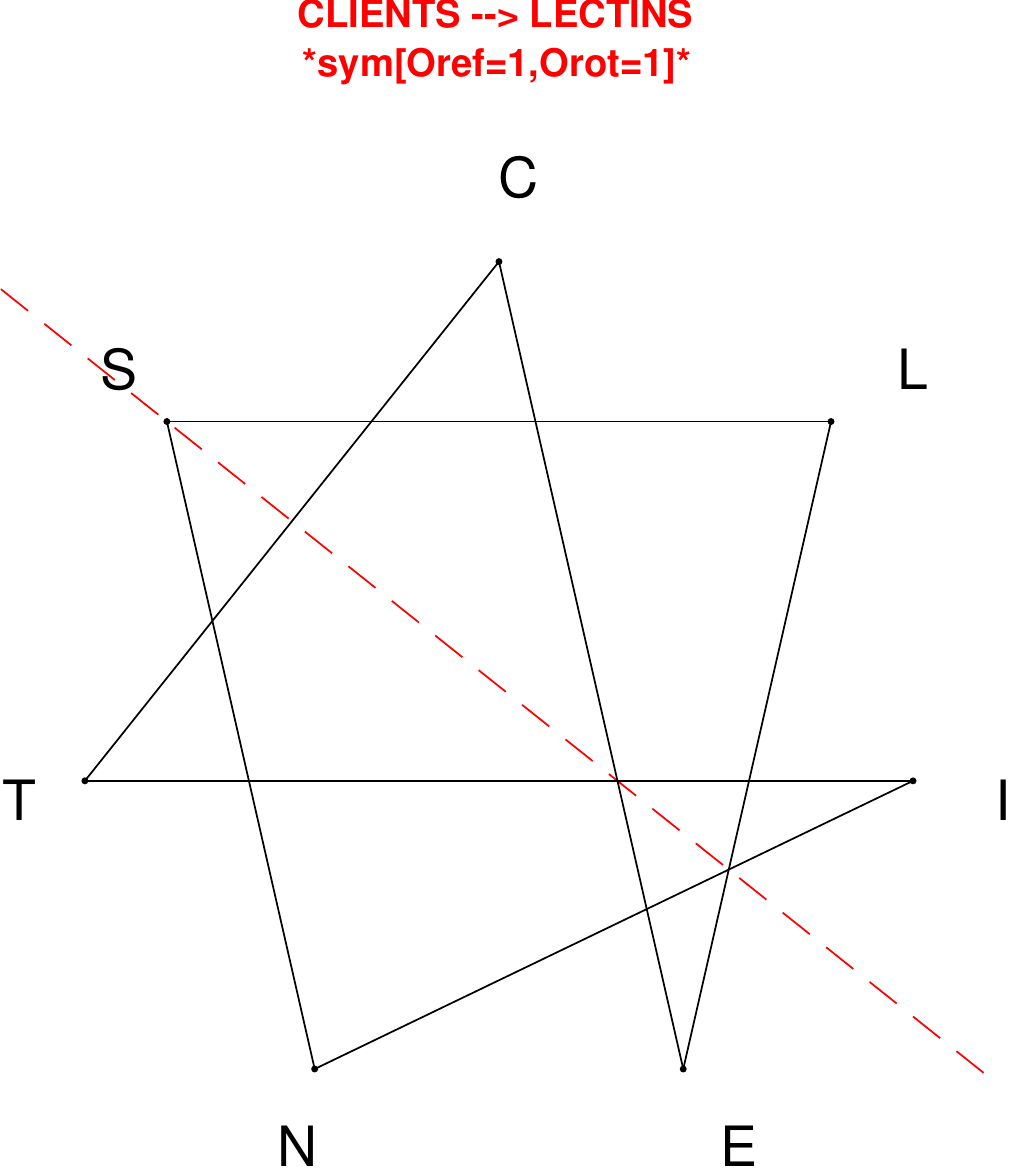}
\end{subfigure}
\hfill
\begin{subfigure}[T]{0.19\textwidth}
\centering
\includegraphics[width=\textwidth]{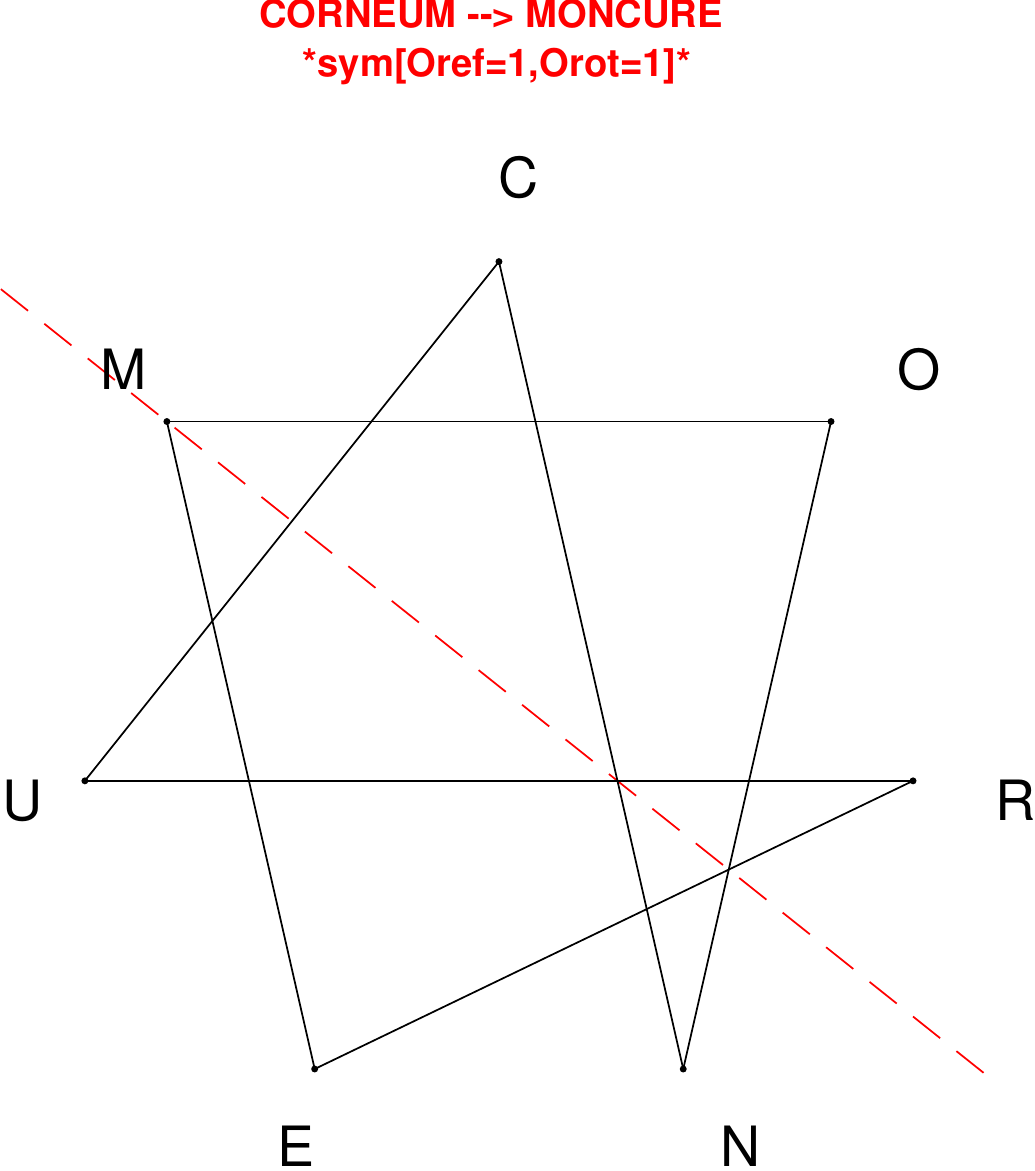}
\end{subfigure}
\hfill
\begin{subfigure}[T]{0.19\textwidth}
\centering
\includegraphics[width=\textwidth]{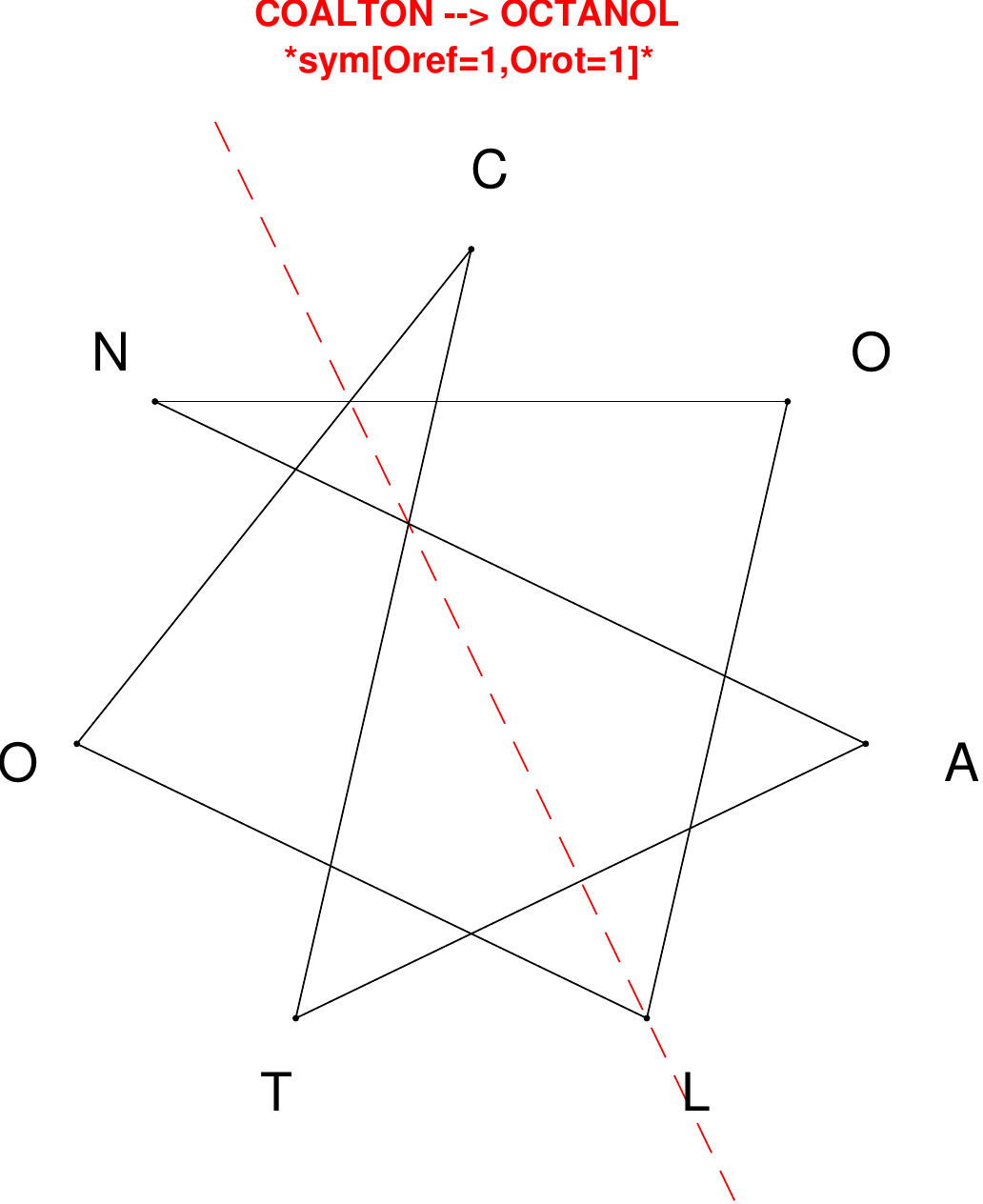}
\end{subfigure}
\hfill
\begin{subfigure}[T]{0.19\textwidth}
\centering
\includegraphics[width=\textwidth]{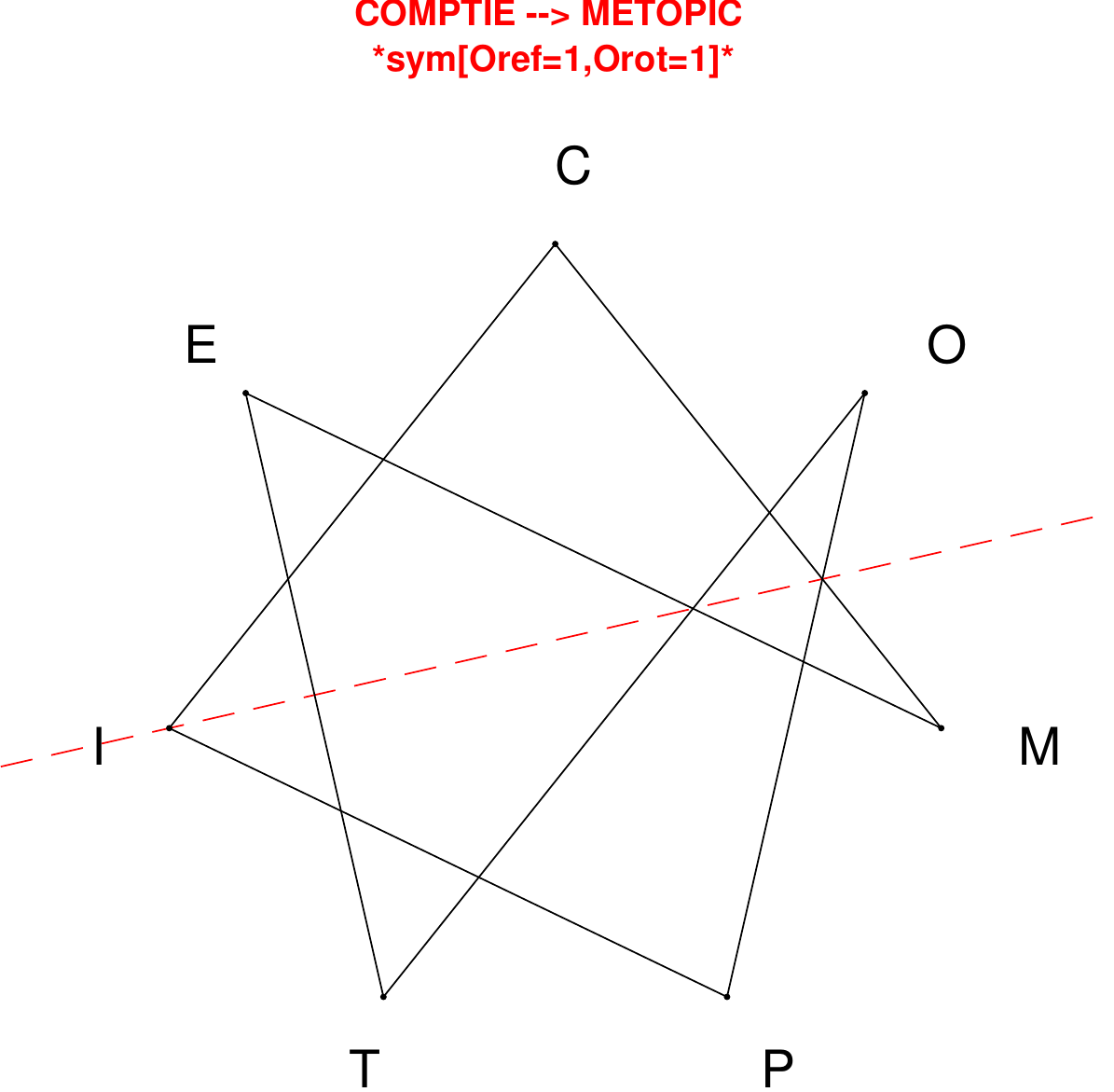}
\end{subfigure}
\hfill
\begin{subfigure}[T]{0.19\textwidth}
\centering
\includegraphics[width=\textwidth]{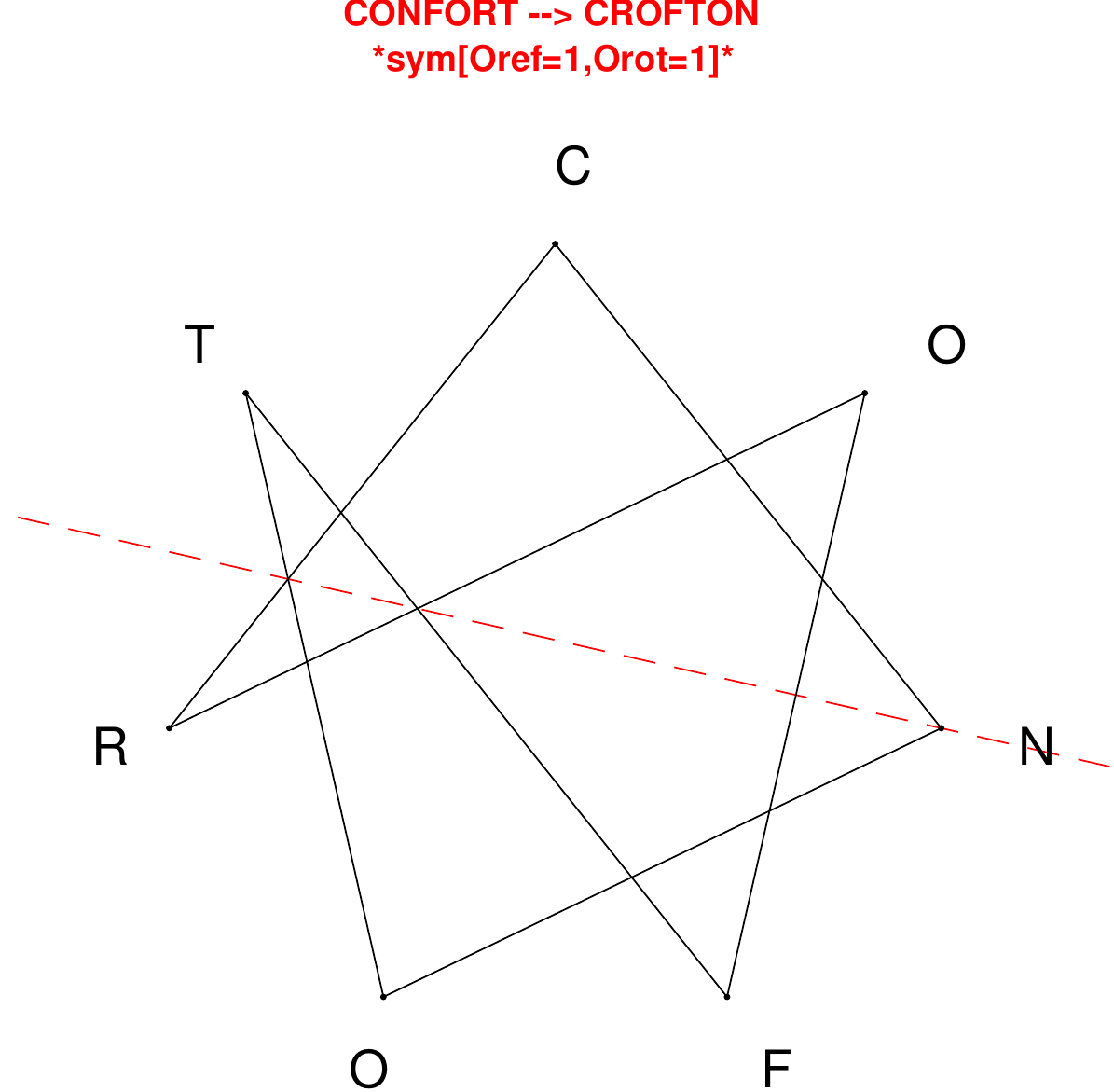}
\end{subfigure}
\end{figure}

\begin{figure}[H]
\centering
\begin{subfigure}[T]{0.19\textwidth}
\centering
\includegraphics[width=\textwidth]{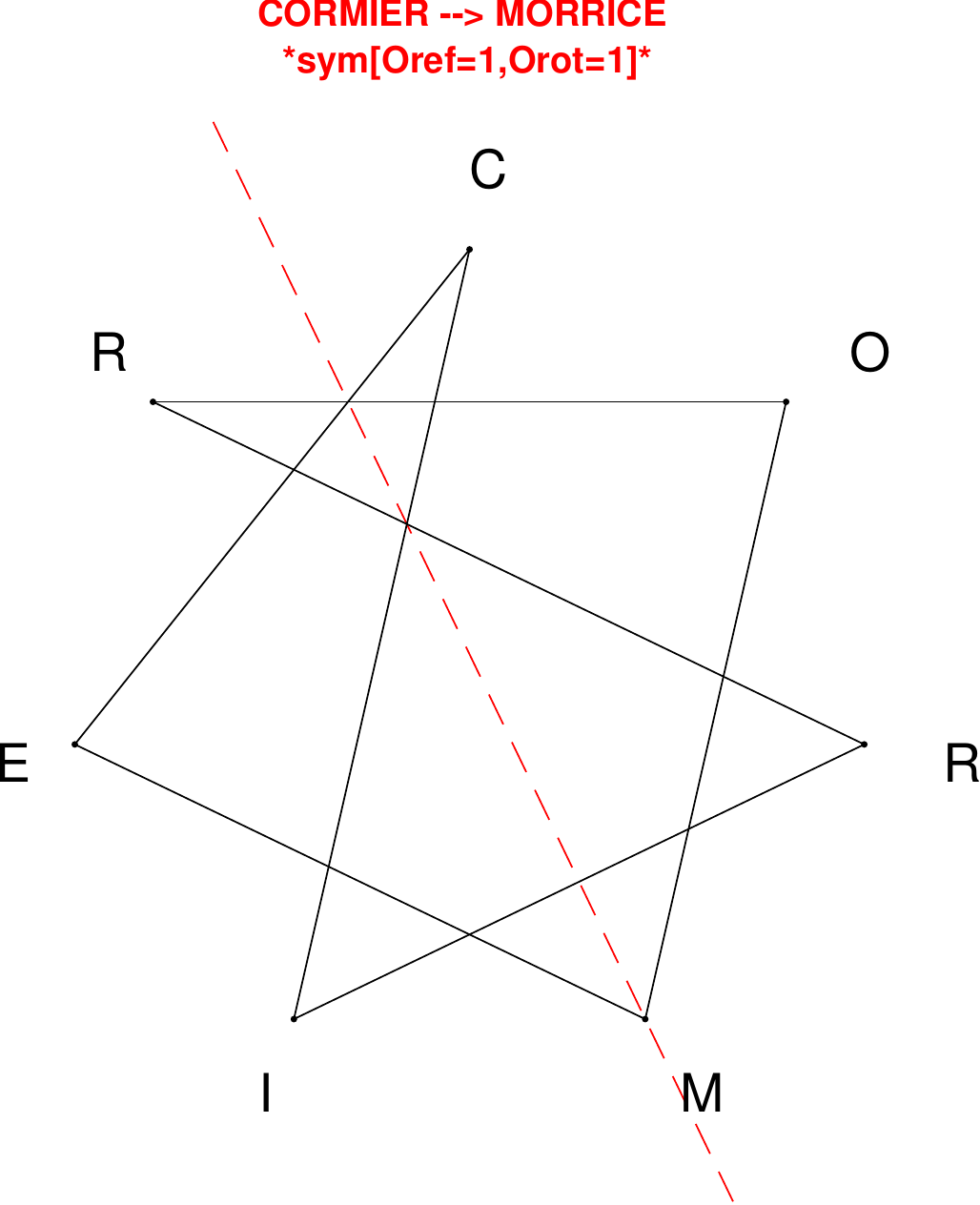}
\end{subfigure}
\hfill
\begin{subfigure}[T]{0.19\textwidth}
\centering
\includegraphics[width=\textwidth]{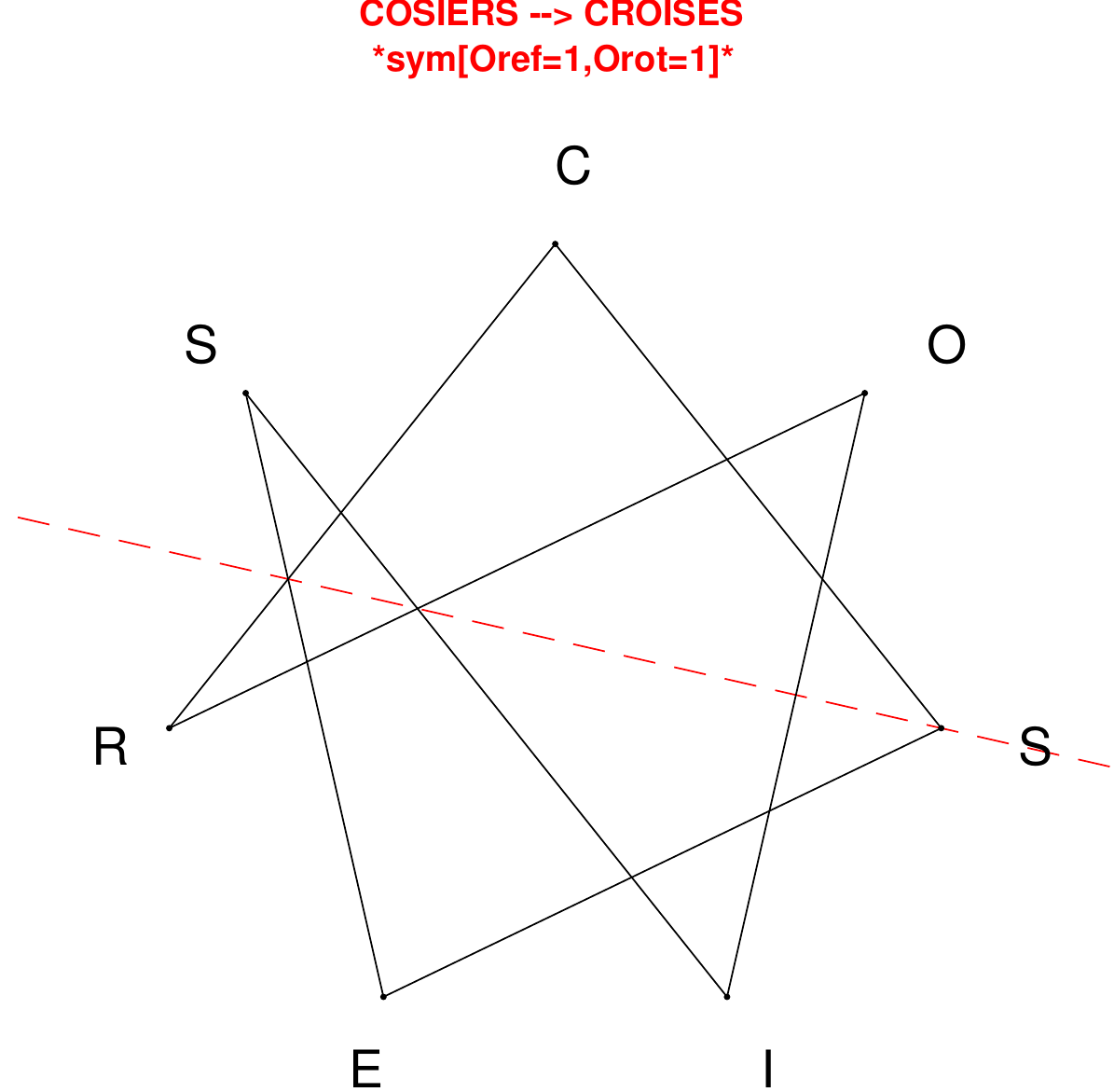}
\end{subfigure}
\hfill
\begin{subfigure}[T]{0.19\textwidth}
\centering
\includegraphics[width=\textwidth]{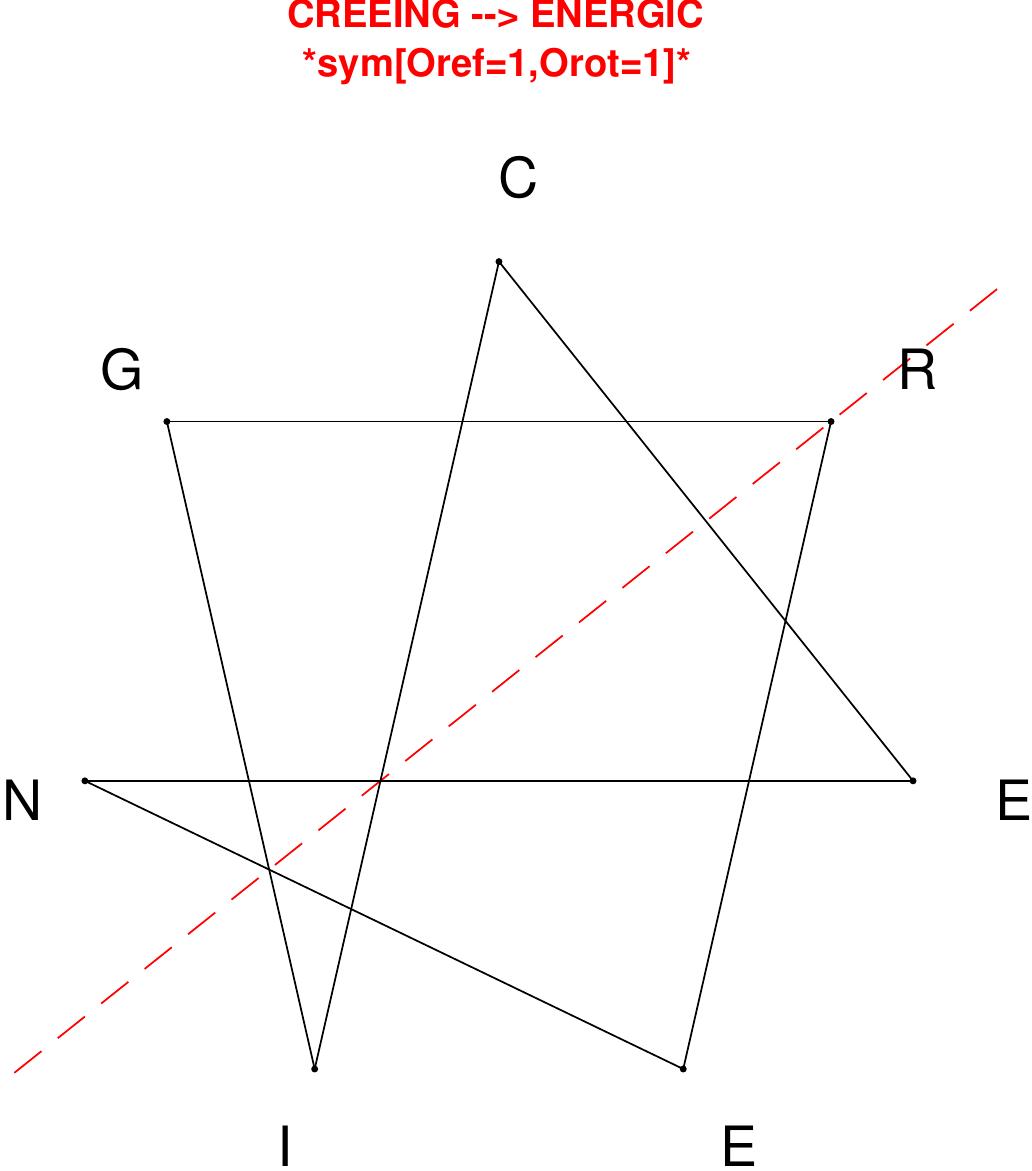}
\end{subfigure}
\hfill
\begin{subfigure}[T]{0.19\textwidth}
\centering
\includegraphics[width=\textwidth]{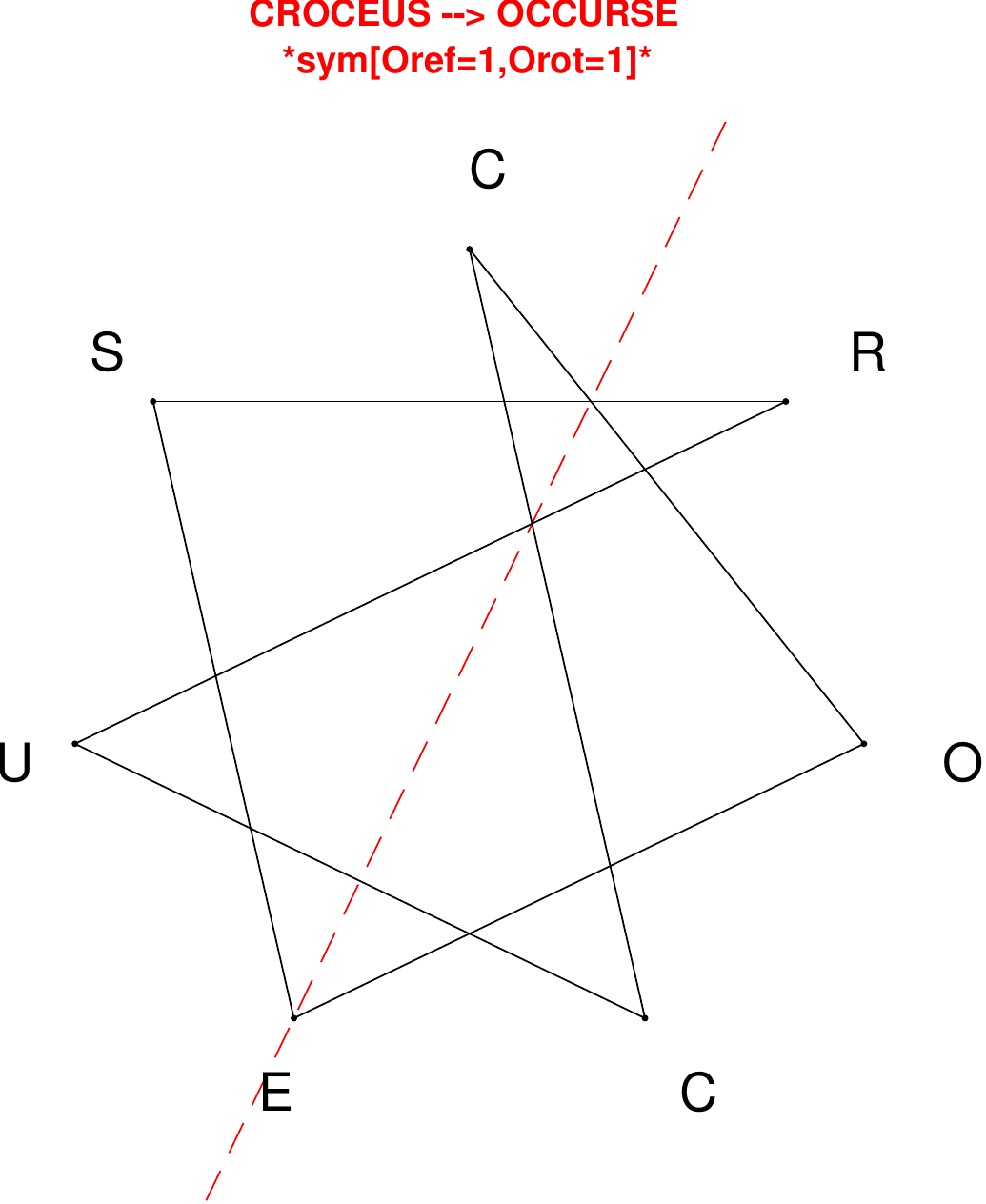}
\end{subfigure}
\hfill
\begin{subfigure}[T]{0.19\textwidth}
\centering
\includegraphics[width=\textwidth]{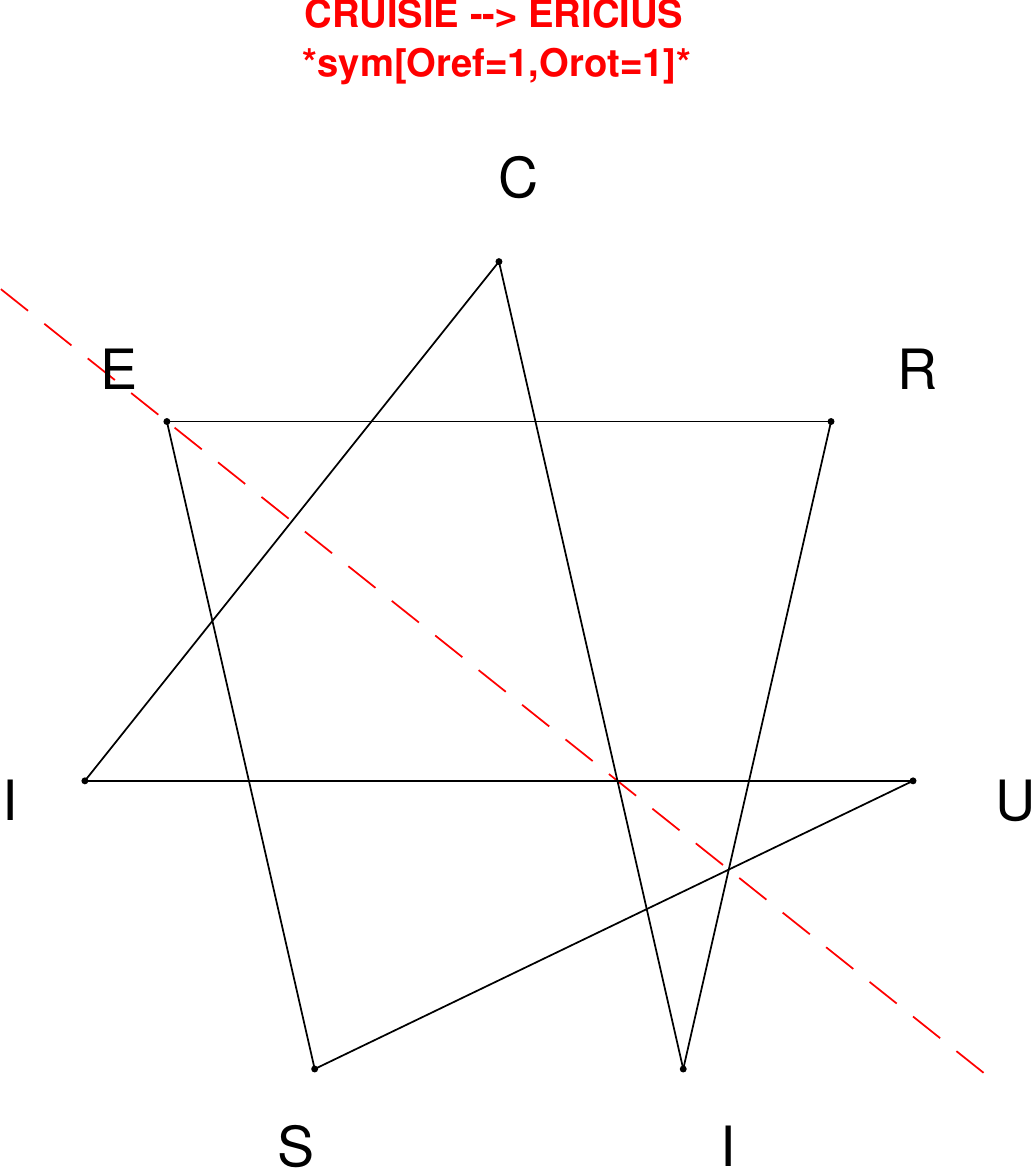}
\end{subfigure}
\end{figure}

\begin{figure}[H]
\centering
\begin{subfigure}[T]{0.19\textwidth}
\centering
\includegraphics[width=\textwidth]{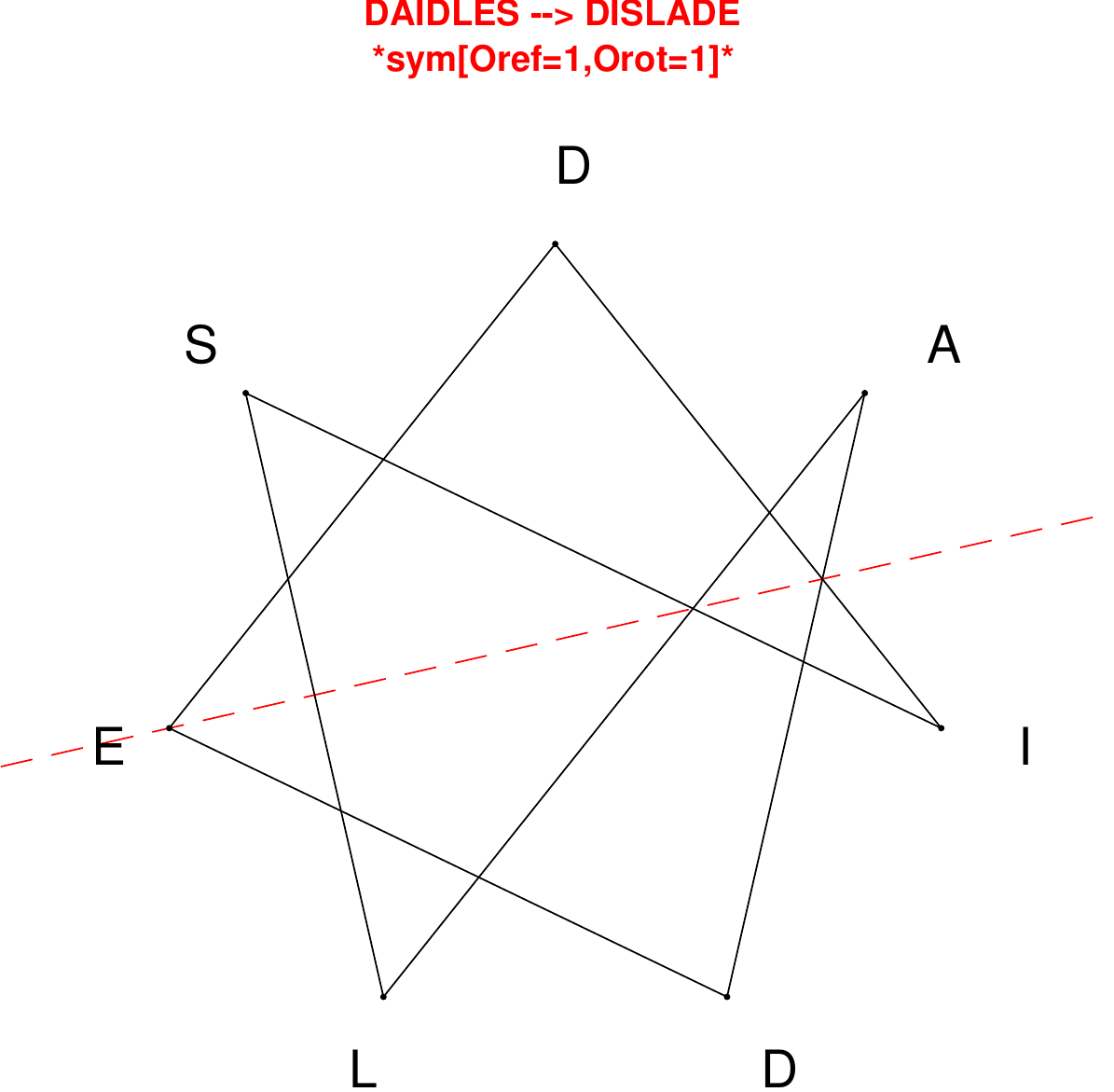}
\end{subfigure}
\hfill
\begin{subfigure}[T]{0.19\textwidth}
\centering
\includegraphics[width=\textwidth]{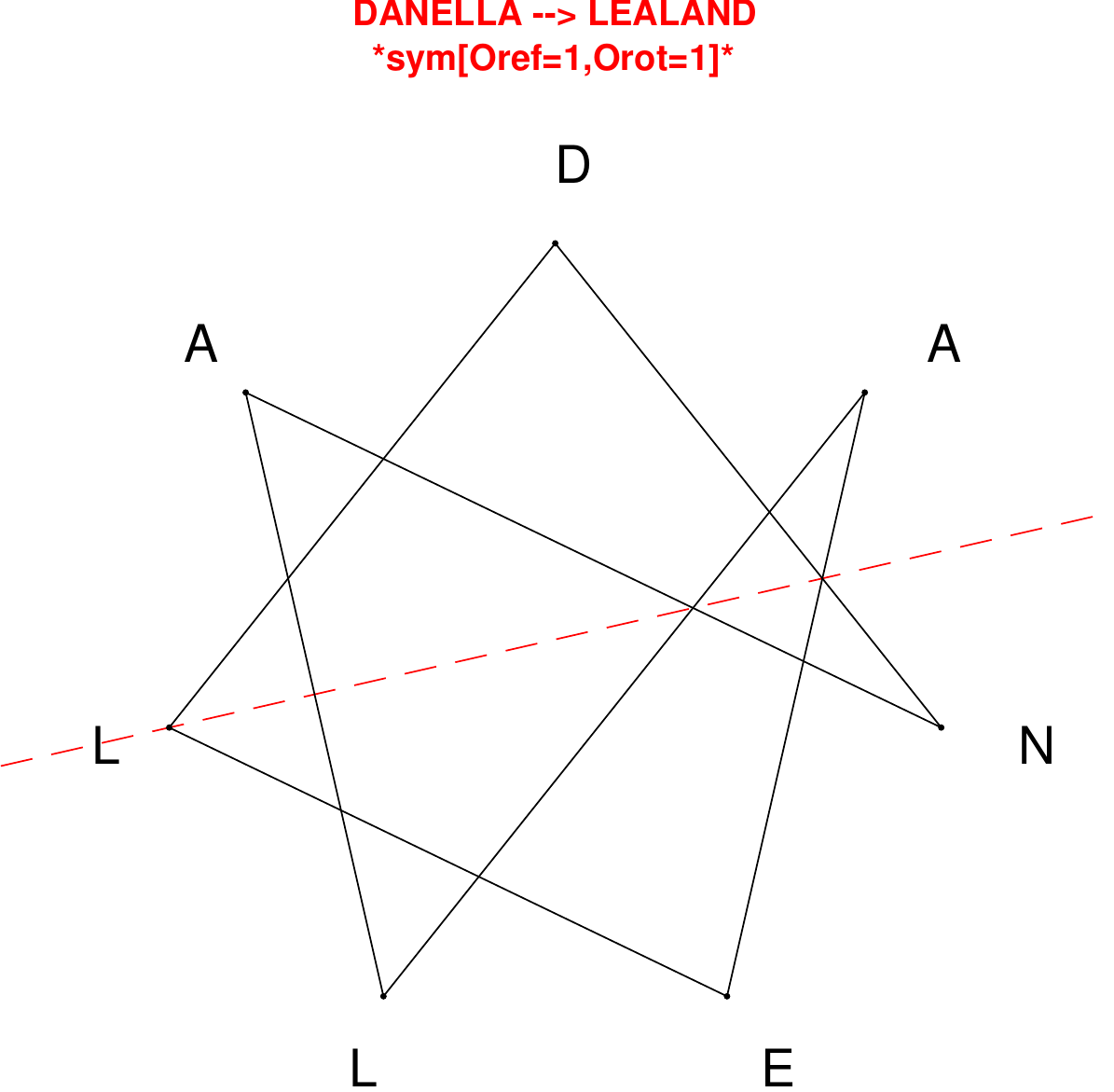}
\end{subfigure}
\hfill
\begin{subfigure}[T]{0.19\textwidth}
\centering
\includegraphics[width=\textwidth]{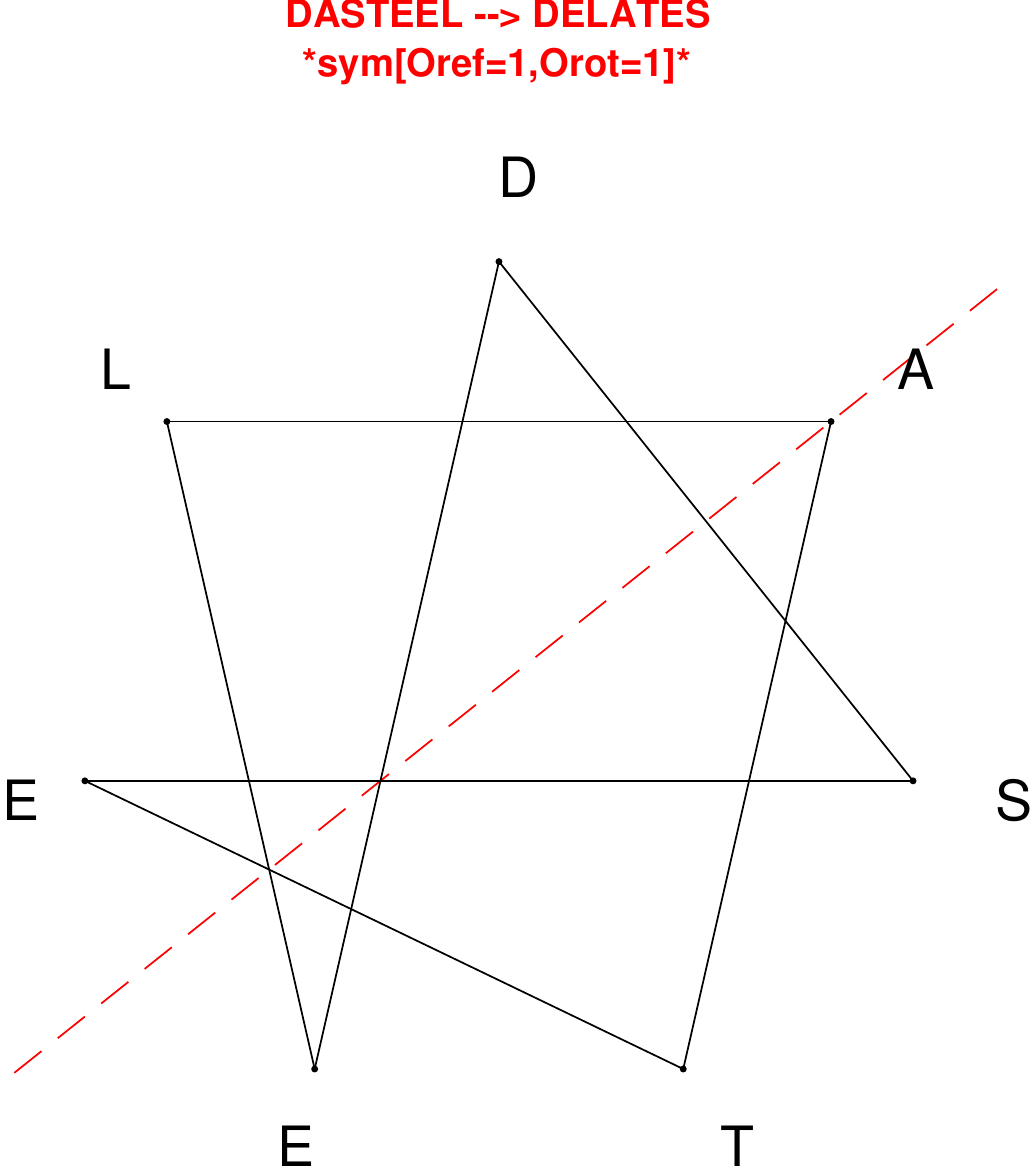}
\end{subfigure}
\hfill
\begin{subfigure}[T]{0.19\textwidth}
\centering
\includegraphics[width=\textwidth]{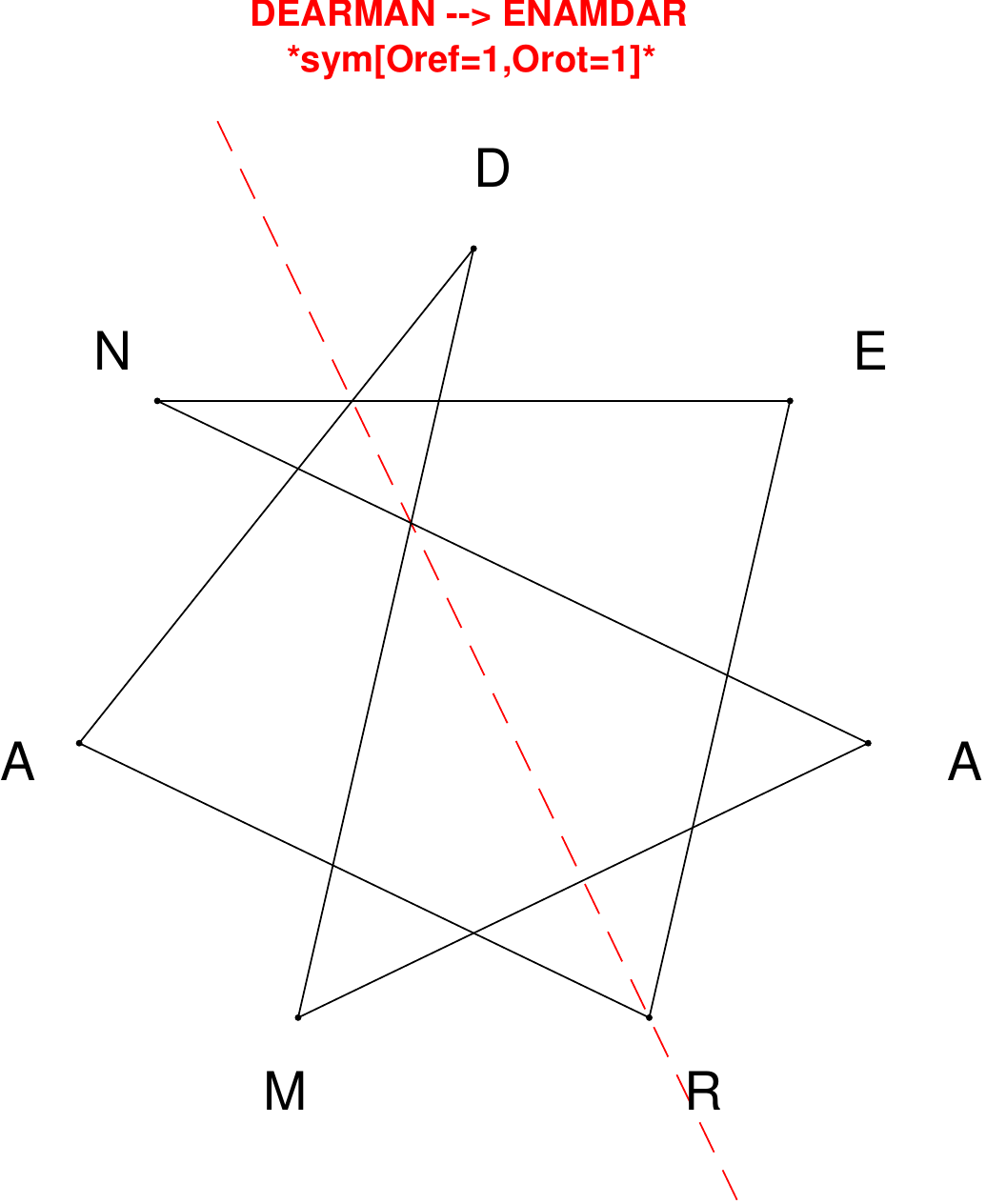}
\end{subfigure}
\hfill
\begin{subfigure}[T]{0.19\textwidth}
\centering
\includegraphics[width=\textwidth]{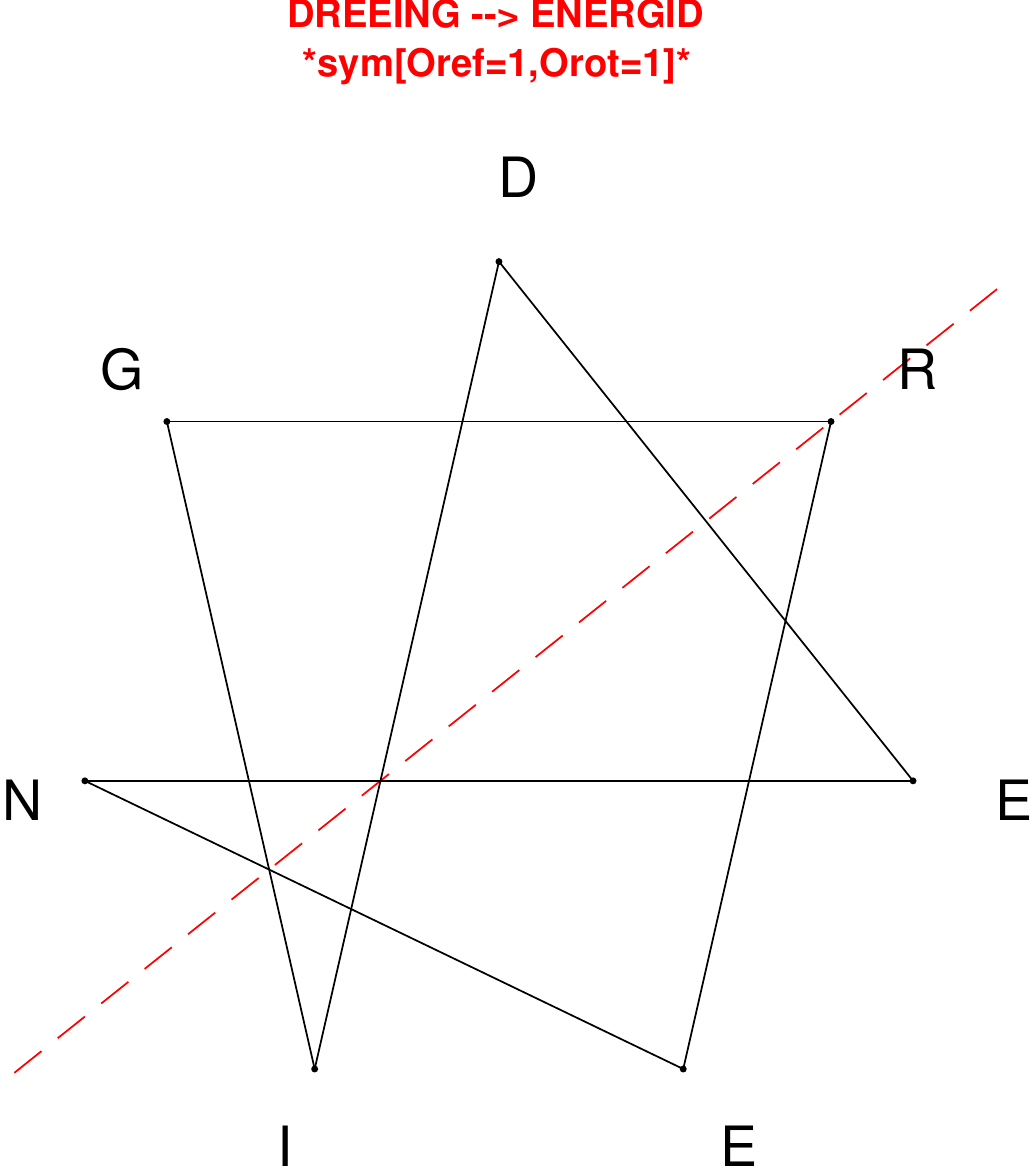}
\end{subfigure}
\end{figure}

\begin{figure}[H]
\centering
\begin{subfigure}[T]{0.19\textwidth}
\centering
\includegraphics[width=\textwidth]{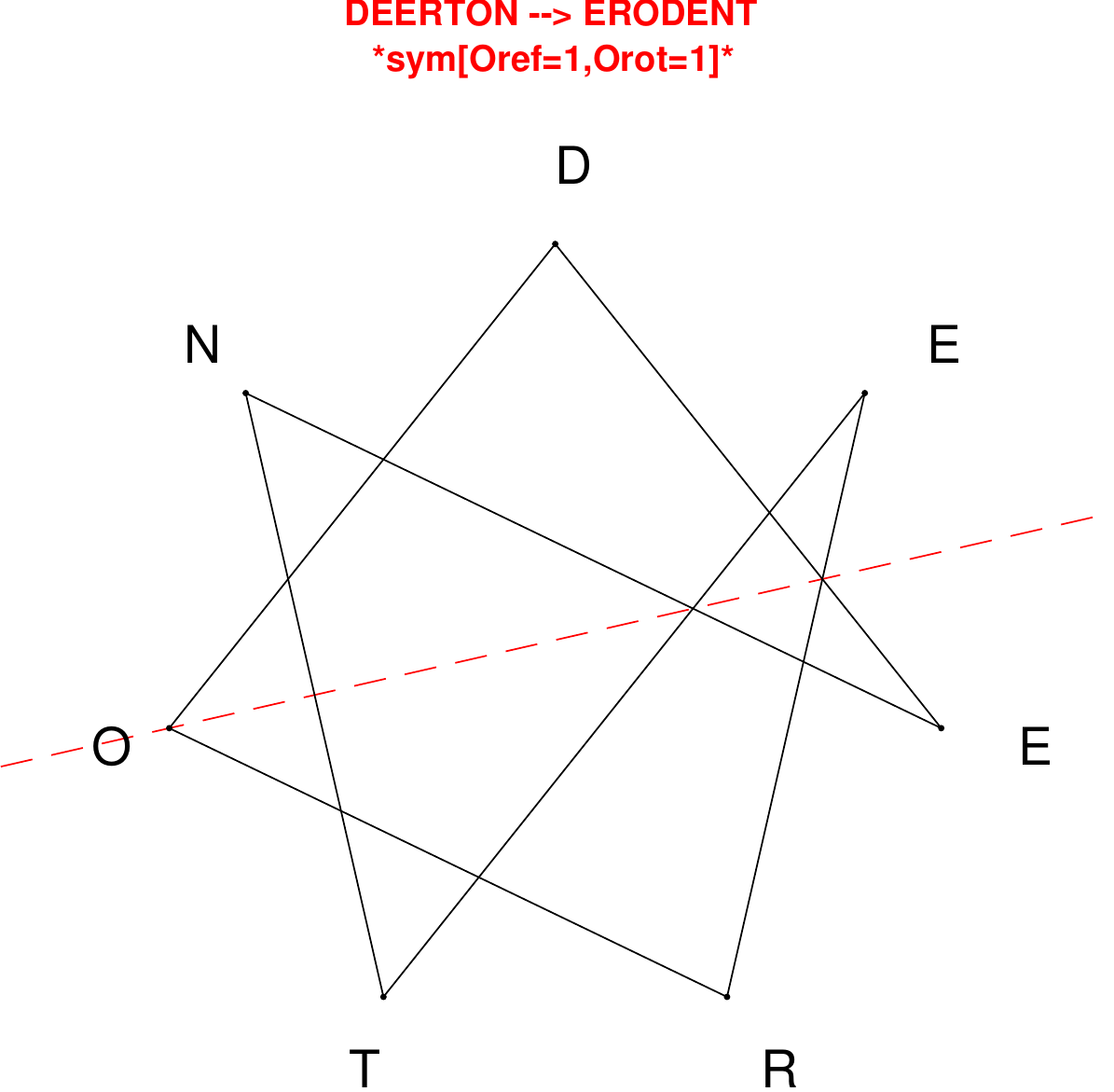}
\end{subfigure}
\hfill
\begin{subfigure}[T]{0.19\textwidth}
\centering
\includegraphics[width=\textwidth]{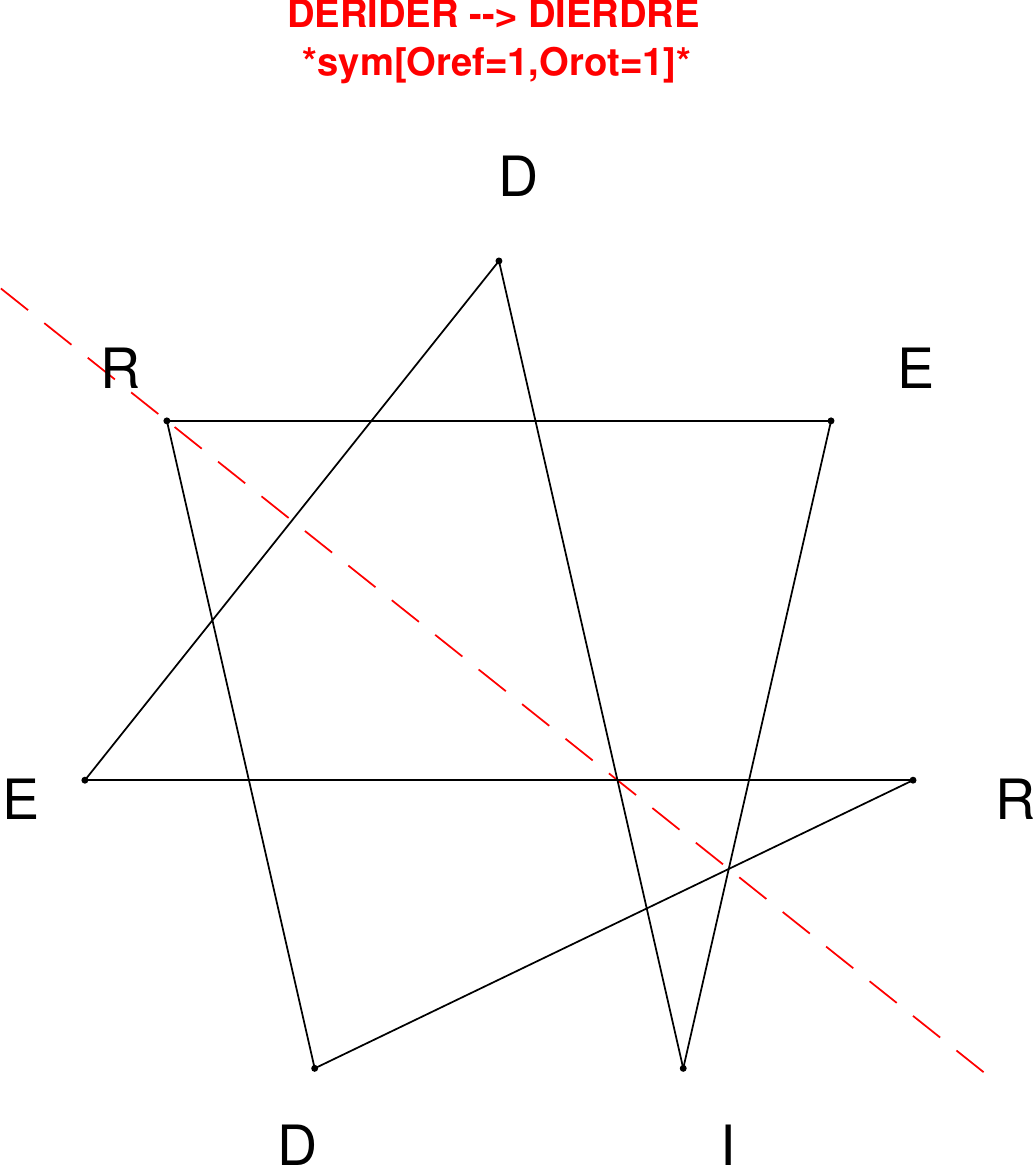}
\end{subfigure}
\hfill
\begin{subfigure}[T]{0.19\textwidth}
\centering
\includegraphics[width=\textwidth]{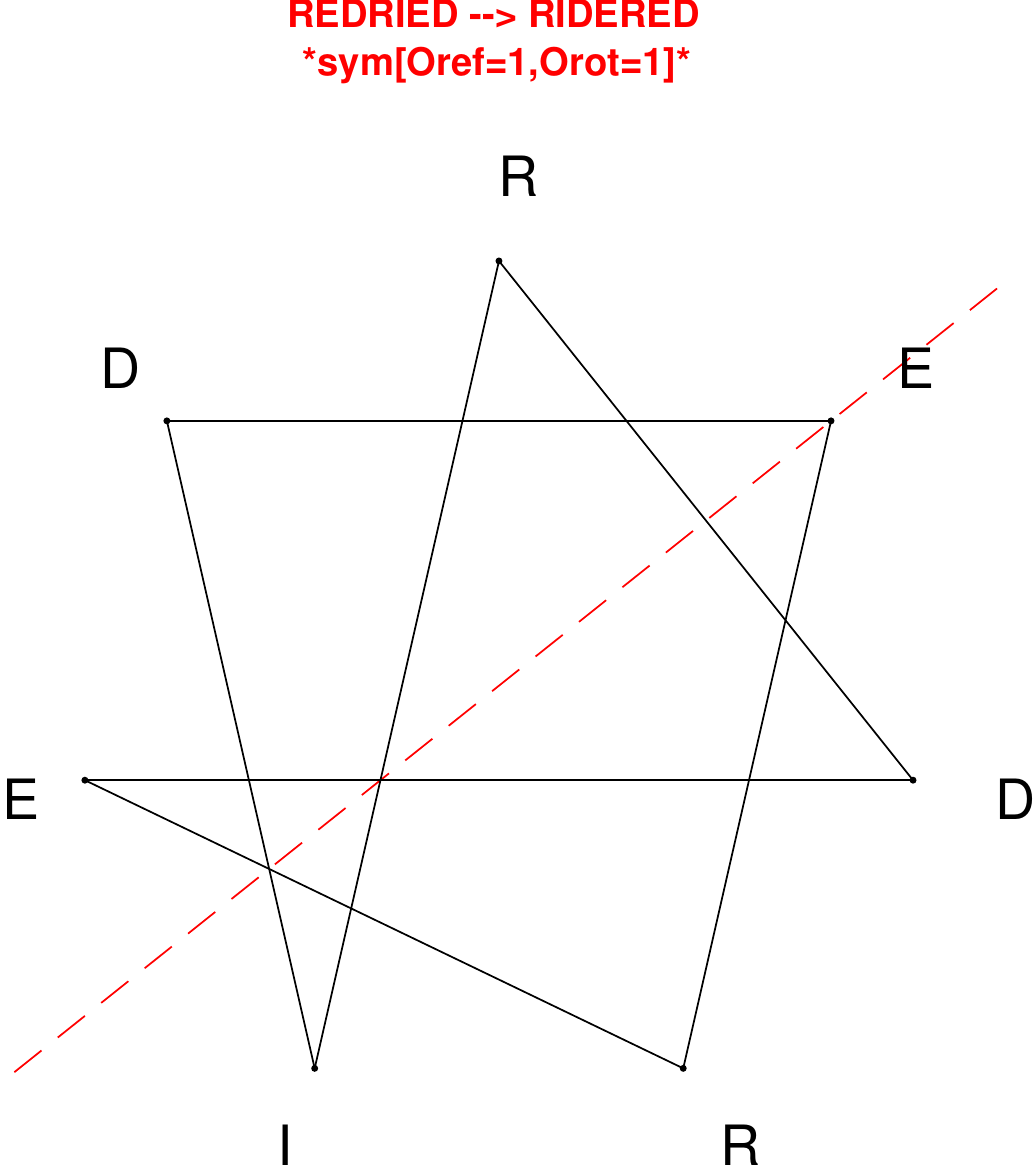}
\end{subfigure}
\hfill
\begin{subfigure}[T]{0.19\textwidth}
\centering
\includegraphics[width=\textwidth]{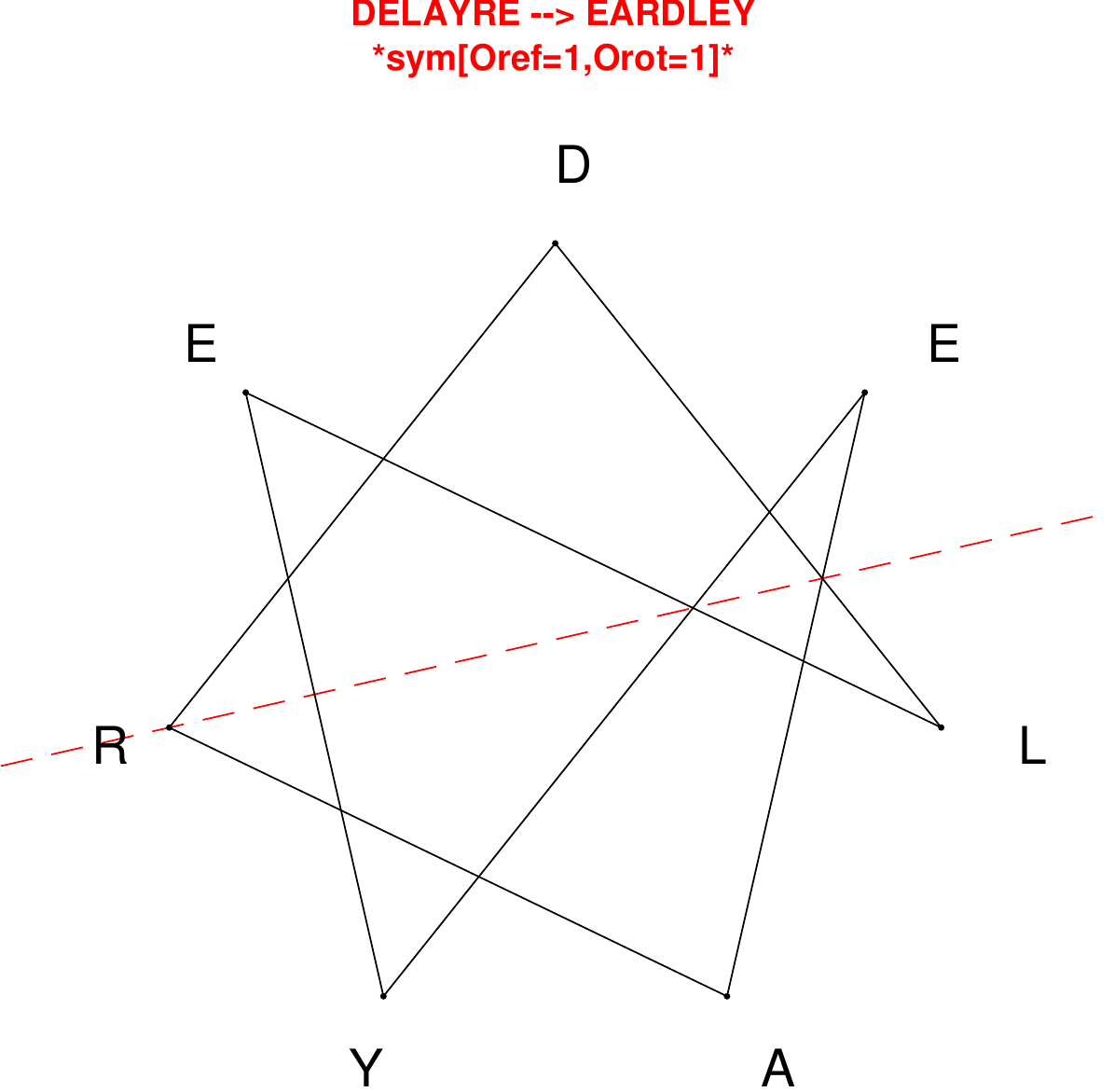}
\end{subfigure}
\hfill
\begin{subfigure}[T]{0.19\textwidth}
\centering
\includegraphics[width=\textwidth]{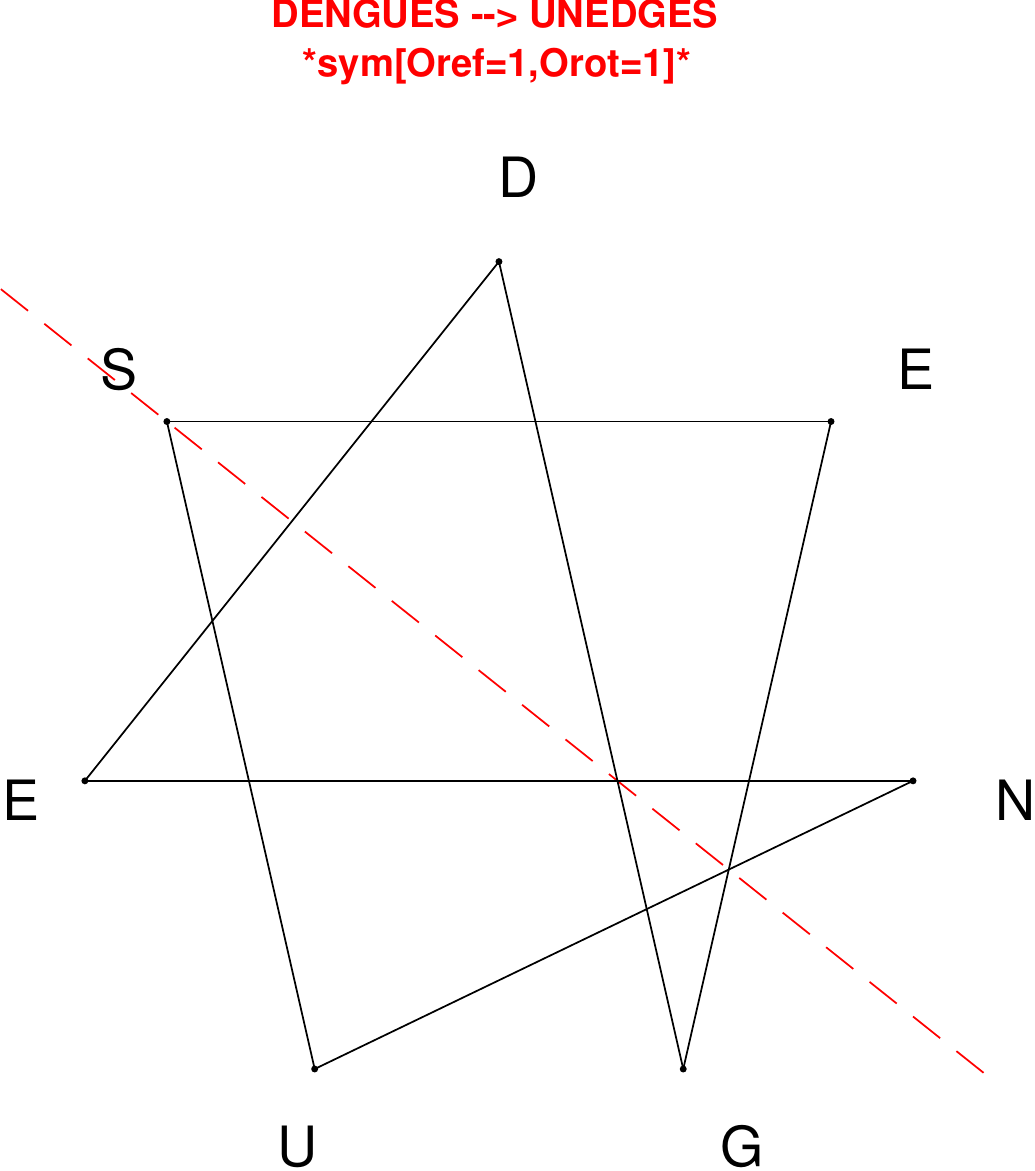}
\end{subfigure}
\end{figure}

\begin{figure}[H]
\centering
\begin{subfigure}[T]{0.19\textwidth}
\centering
\includegraphics[width=\textwidth]{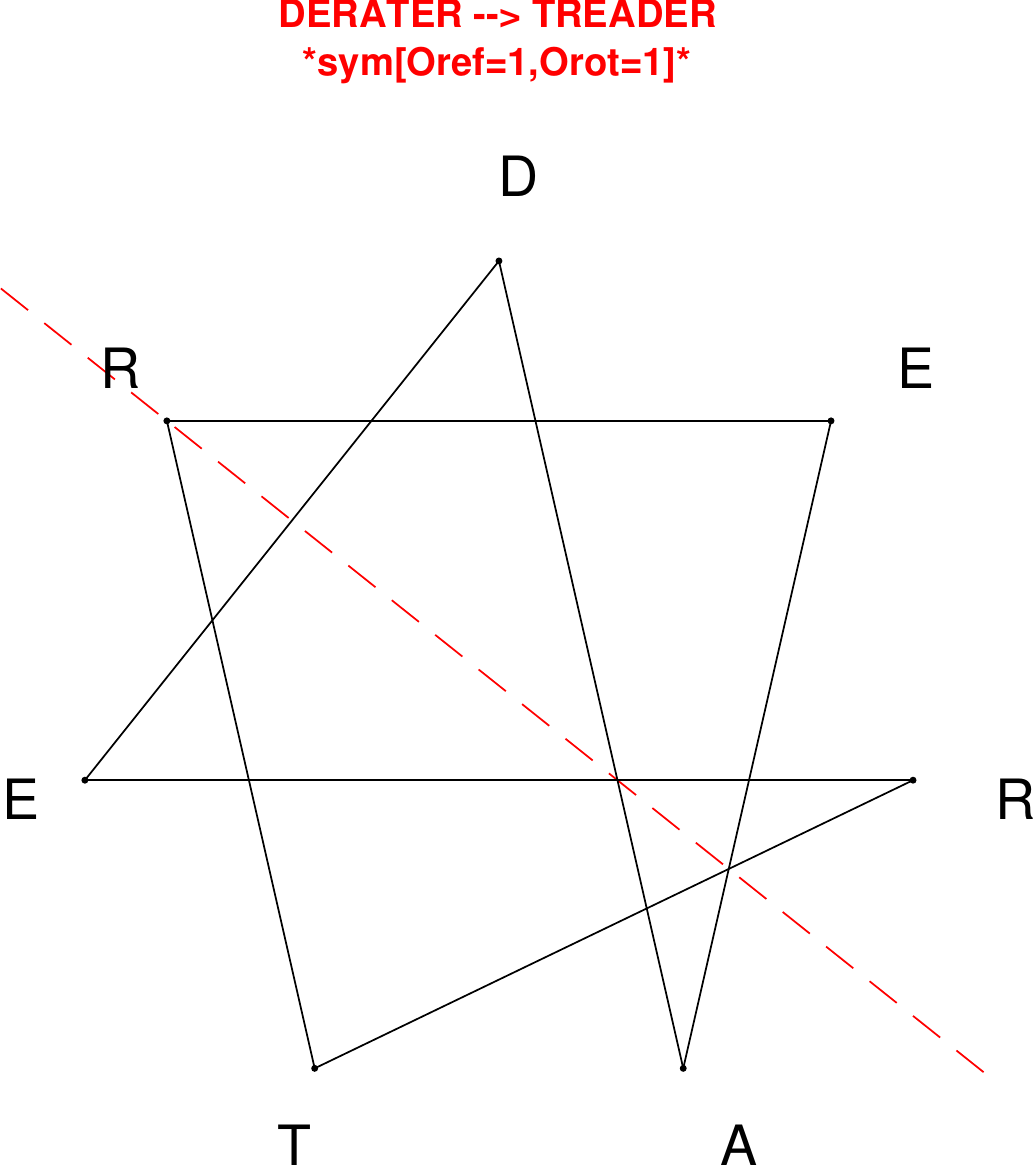}
\end{subfigure}
\hfill
\begin{subfigure}[T]{0.19\textwidth}
\centering
\includegraphics[width=\textwidth]{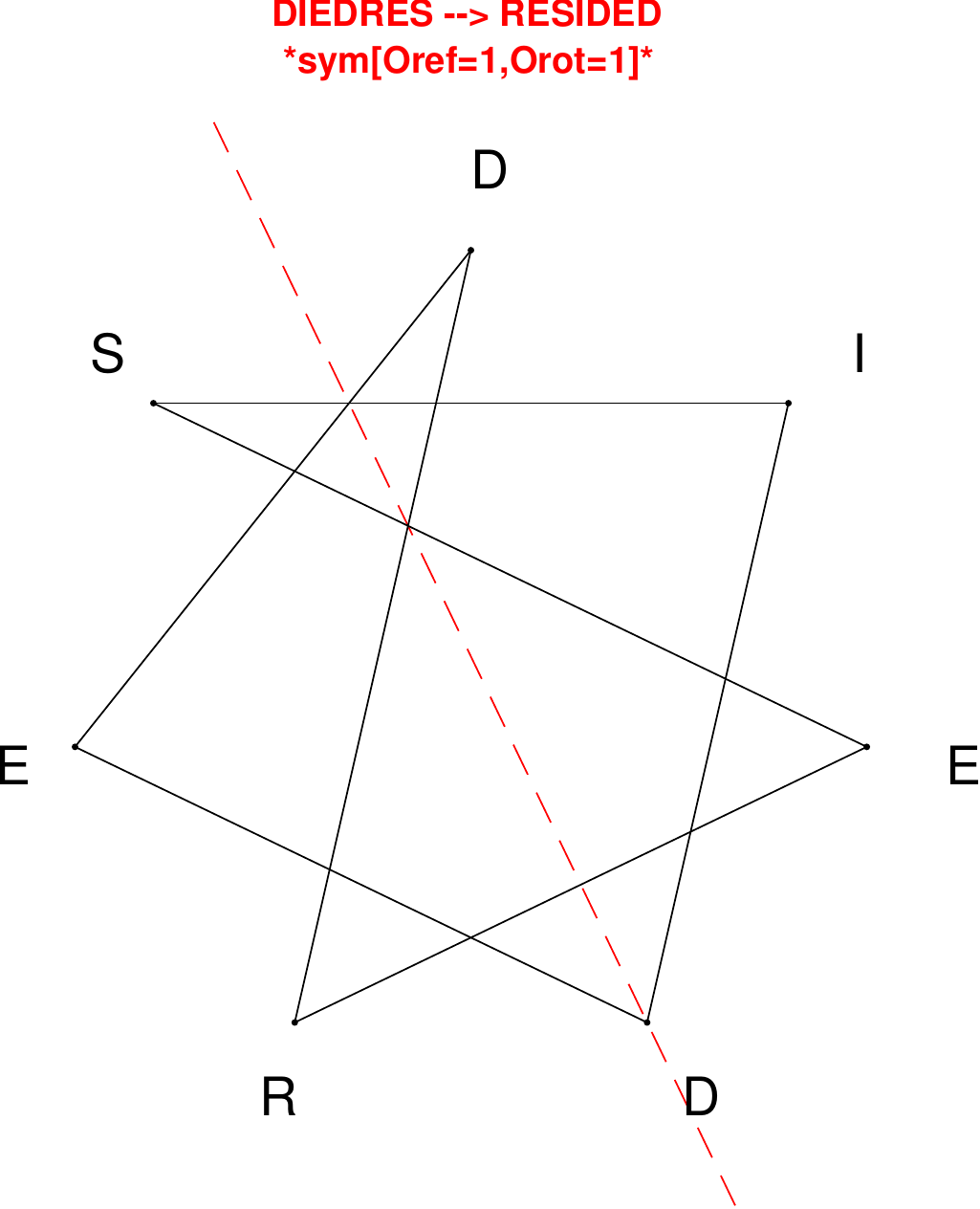}
\end{subfigure}
\hfill
\begin{subfigure}[T]{0.19\textwidth}
\centering
\includegraphics[width=\textwidth]{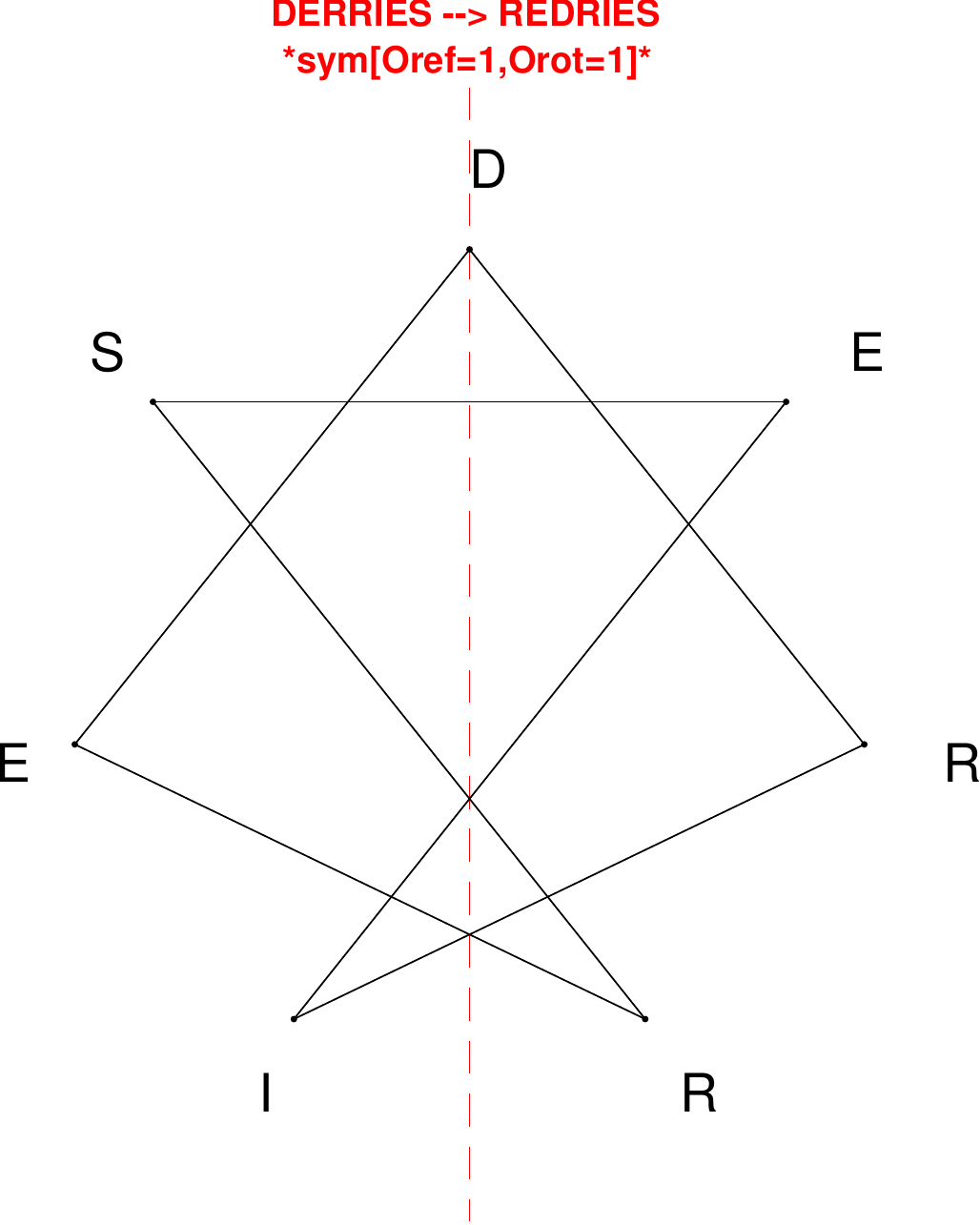}
\end{subfigure}
\hfill
\begin{subfigure}[T]{0.19\textwidth}
\centering
\includegraphics[width=\textwidth]{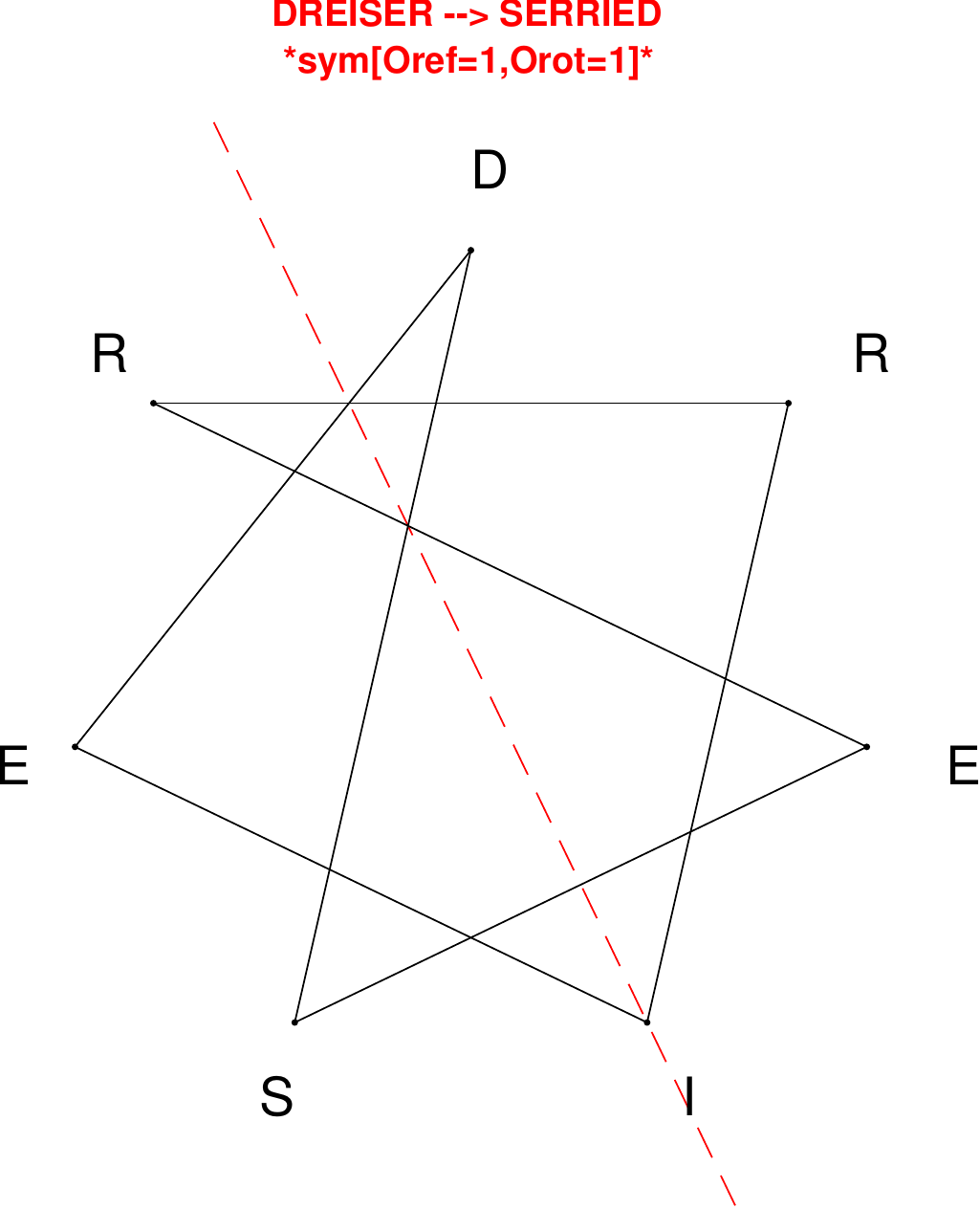}
\end{subfigure}
\hfill
\begin{subfigure}[T]{0.19\textwidth}
\centering
\includegraphics[width=\textwidth]{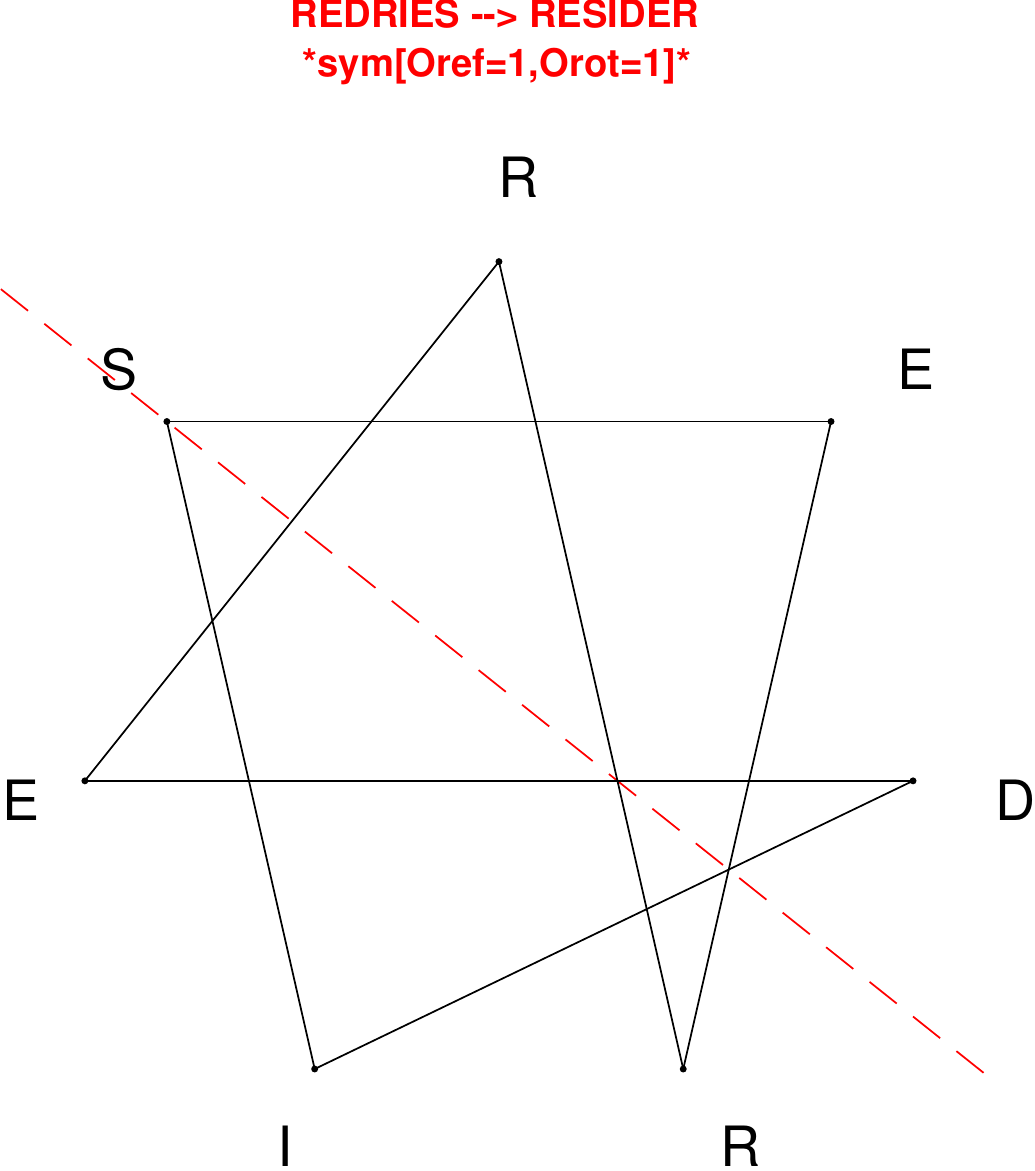}
\end{subfigure}
\end{figure}

\begin{figure}[H]
\centering
\begin{subfigure}[T]{0.19\textwidth}
\centering
\includegraphics[width=\textwidth]{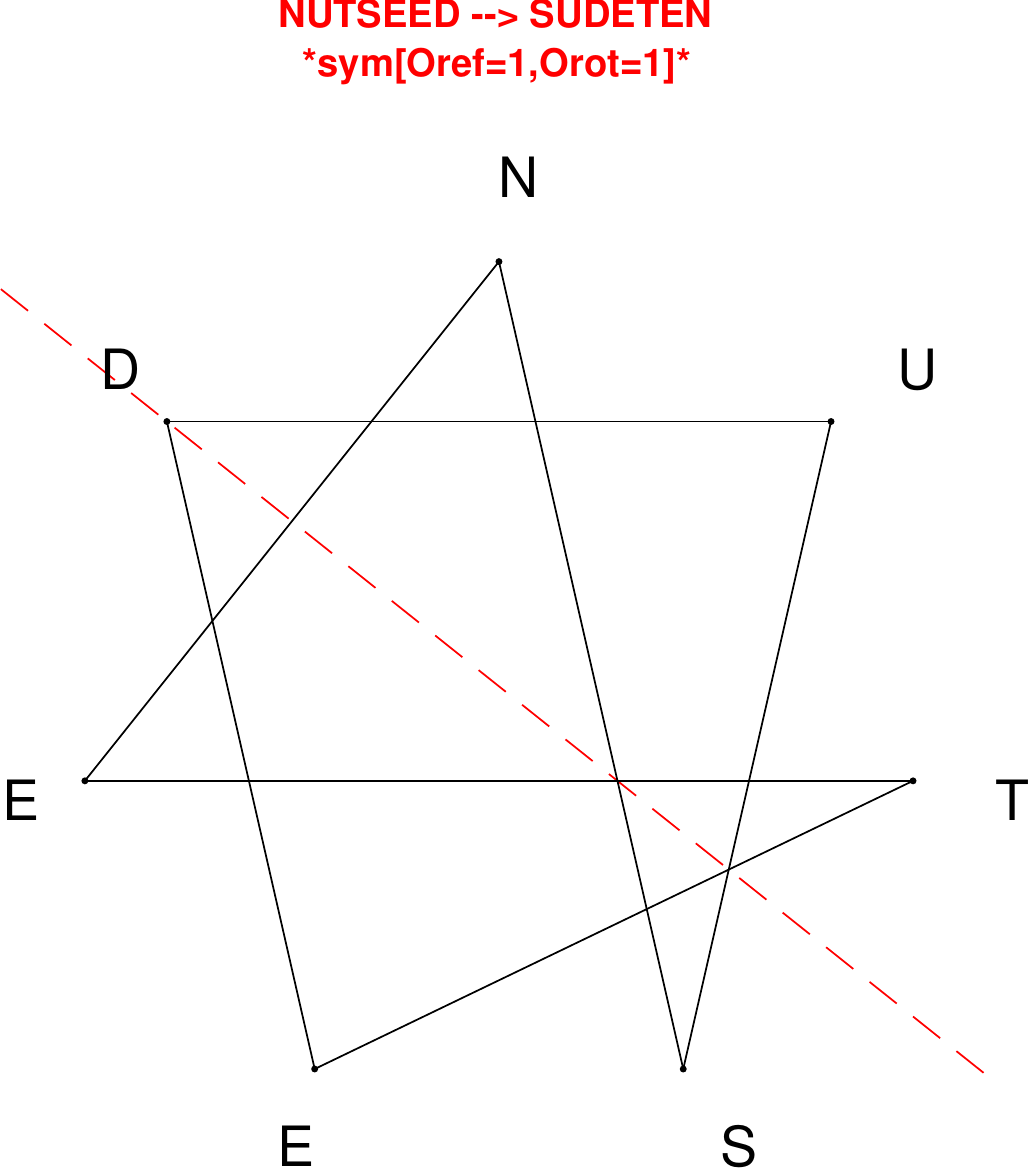}
\end{subfigure}
\hfill
\begin{subfigure}[T]{0.19\textwidth}
\centering
\includegraphics[width=\textwidth]{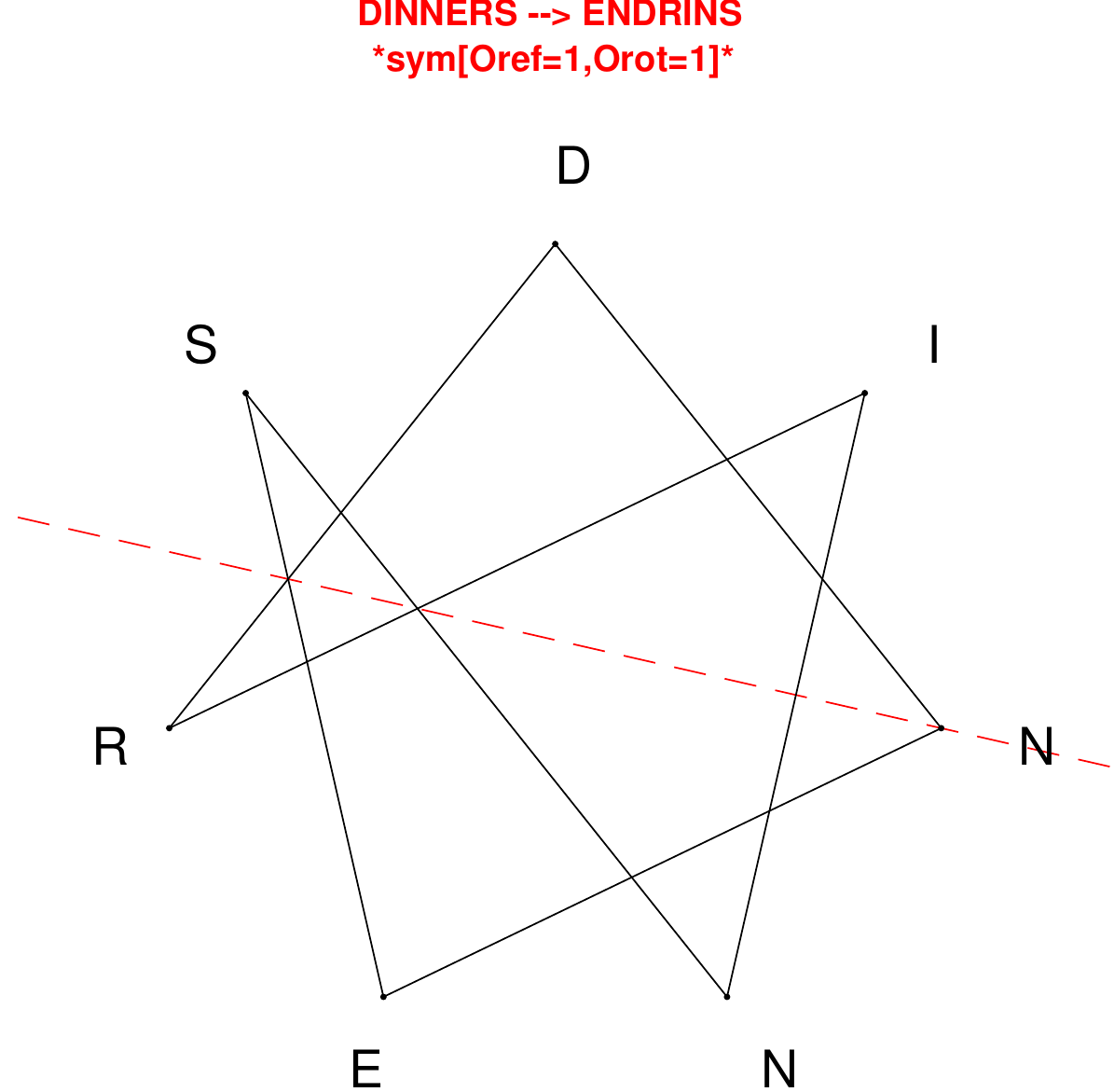}
\end{subfigure}
\hfill
\begin{subfigure}[T]{0.19\textwidth}
\centering
\includegraphics[width=\textwidth]{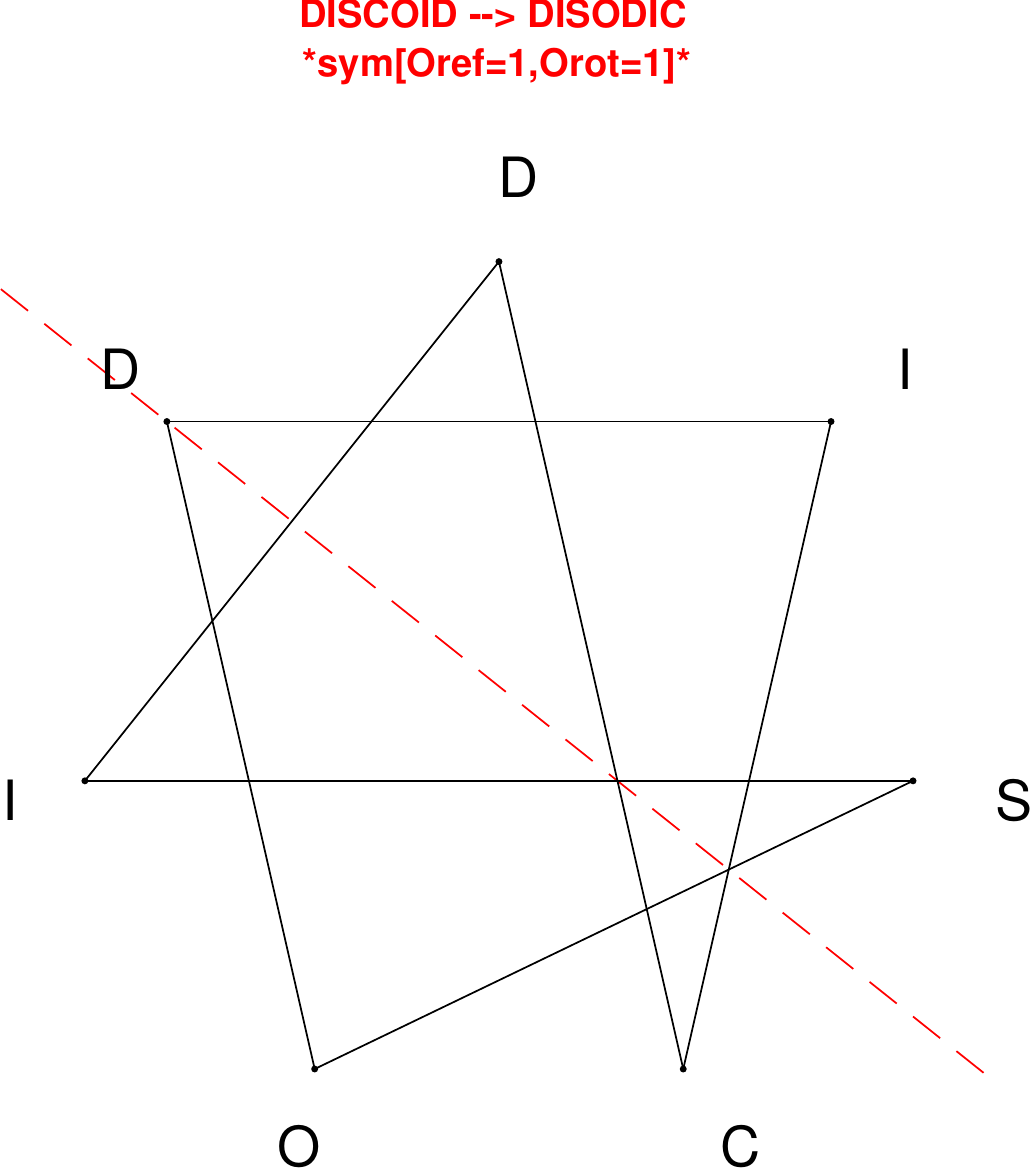}
\end{subfigure}
\hfill
\begin{subfigure}[T]{0.19\textwidth}
\centering
\includegraphics[width=\textwidth]{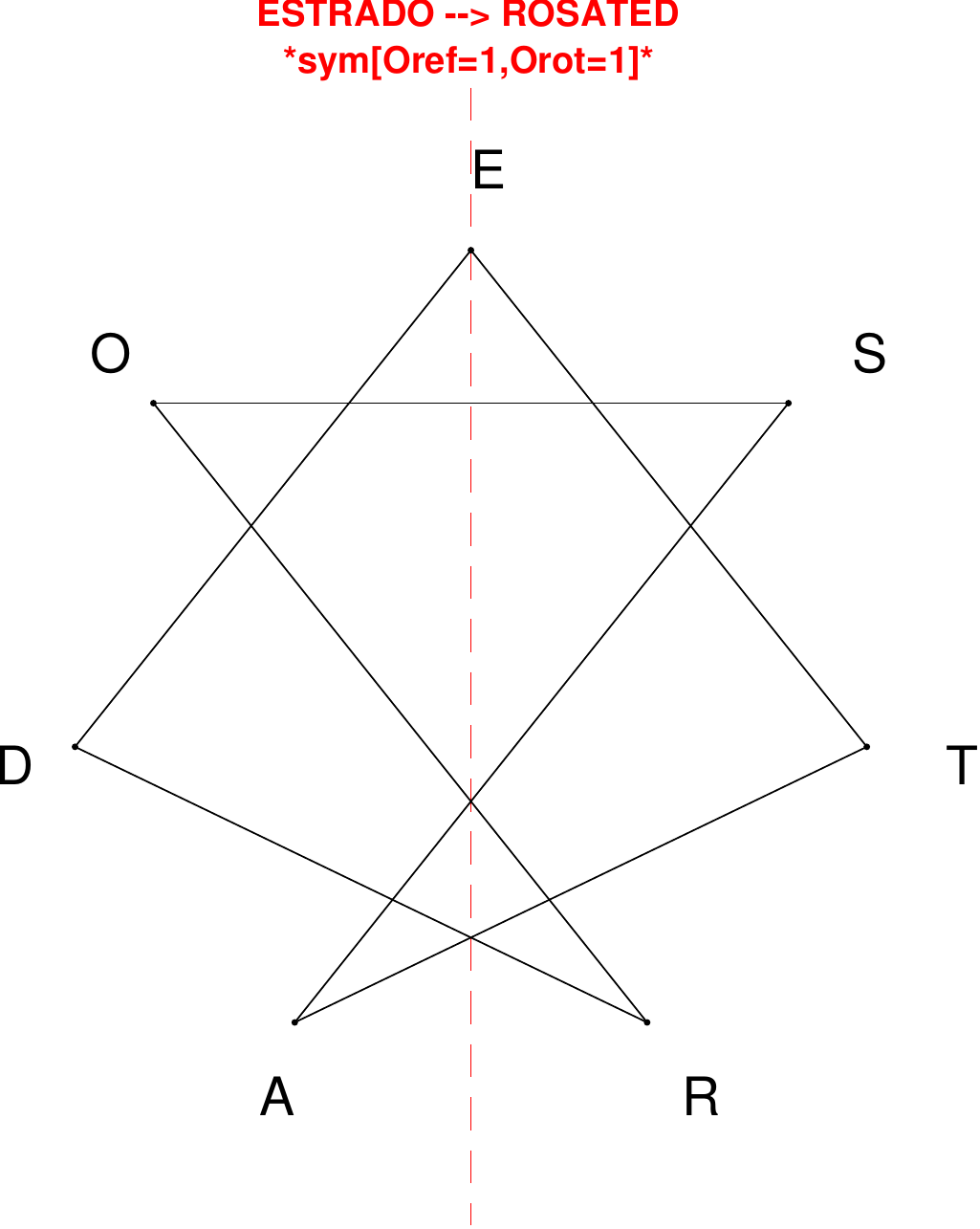}
\end{subfigure}
\hfill
\begin{subfigure}[T]{0.19\textwidth}
\centering
\includegraphics[width=\textwidth]{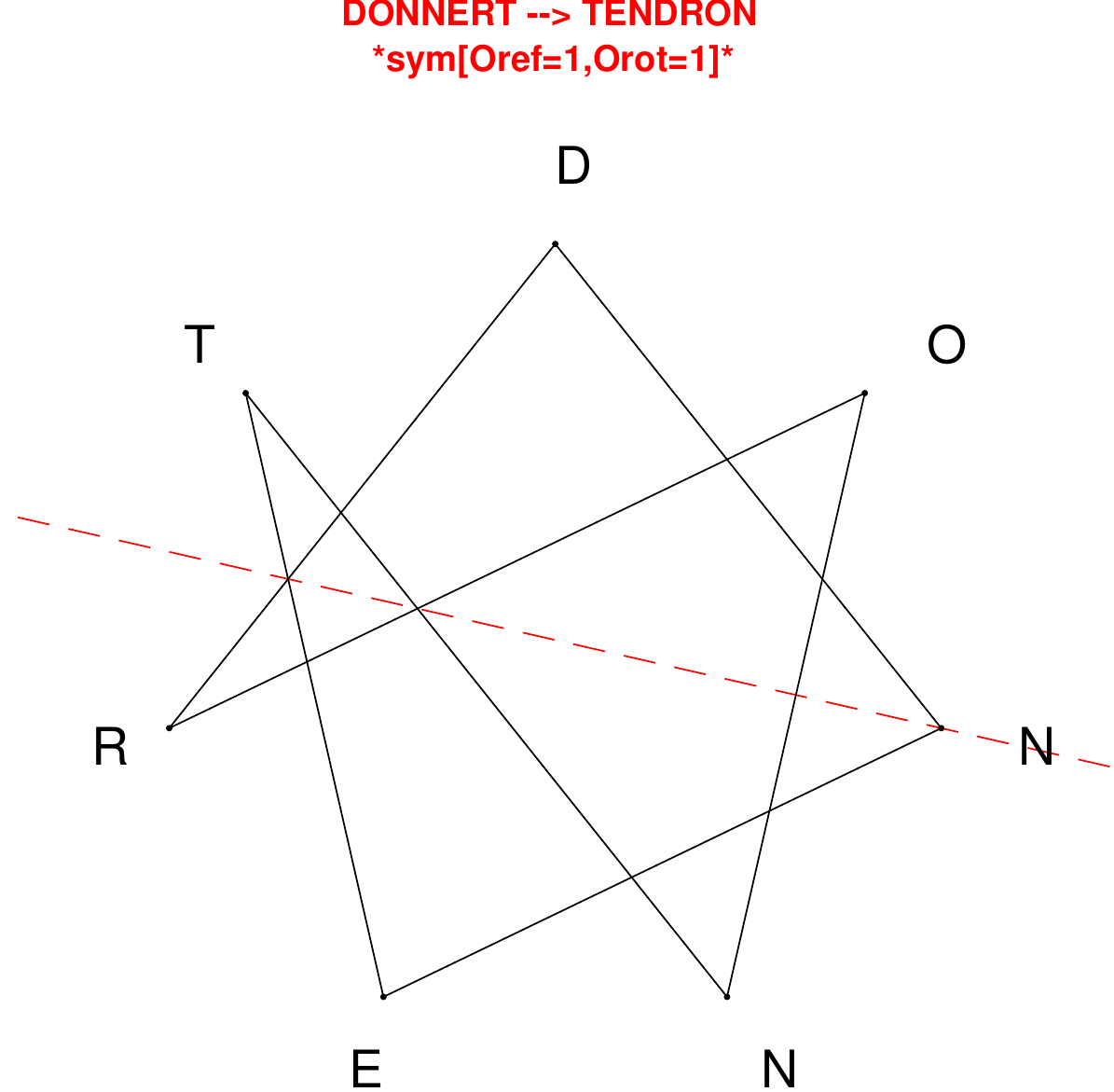}
\end{subfigure}
\end{figure}

\begin{figure}[H]
\centering
\begin{subfigure}[T]{0.19\textwidth}
\centering
\includegraphics[width=\textwidth]{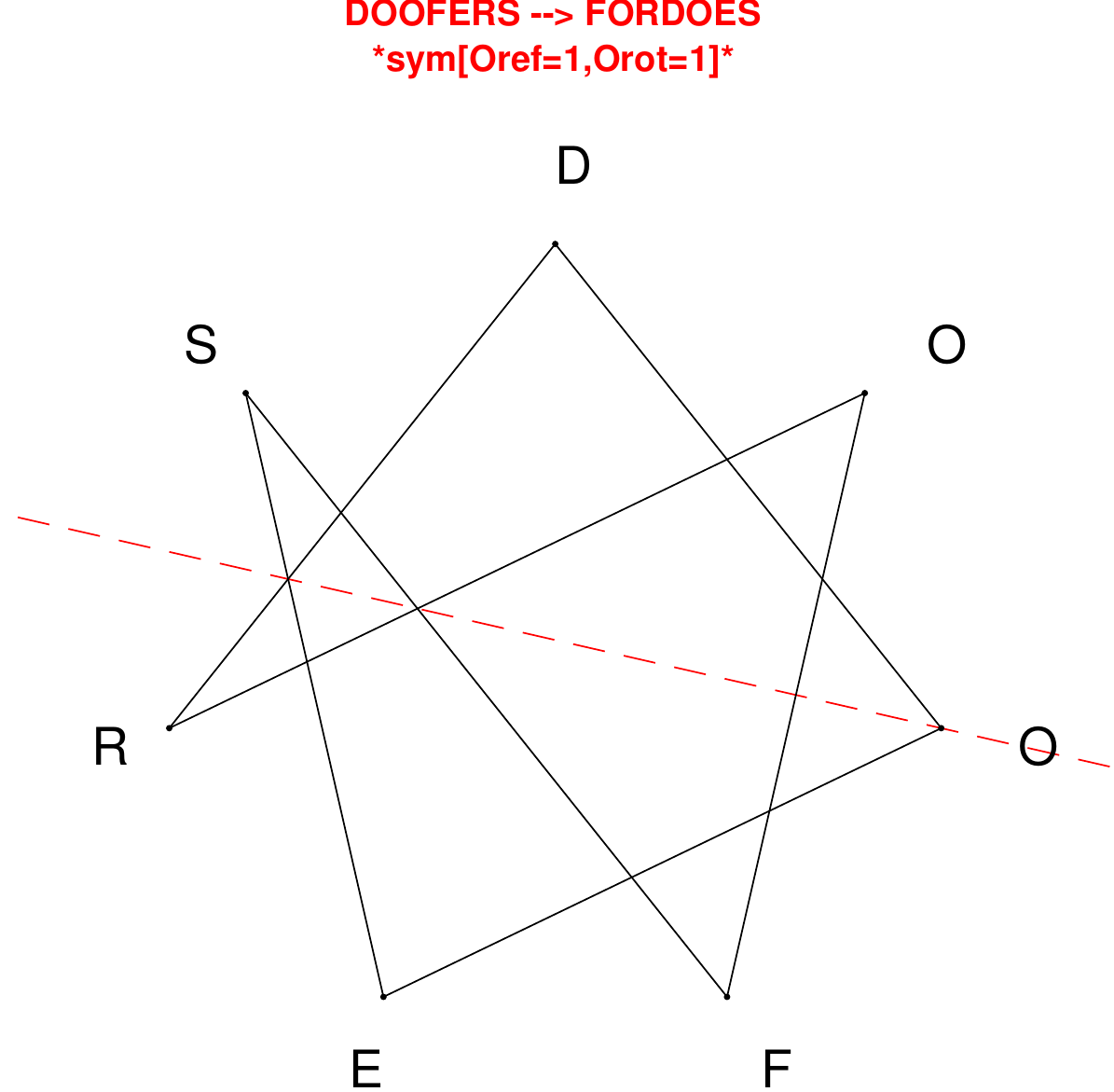}
\end{subfigure}
\hfill
\begin{subfigure}[T]{0.19\textwidth}
\centering
\includegraphics[width=\textwidth]{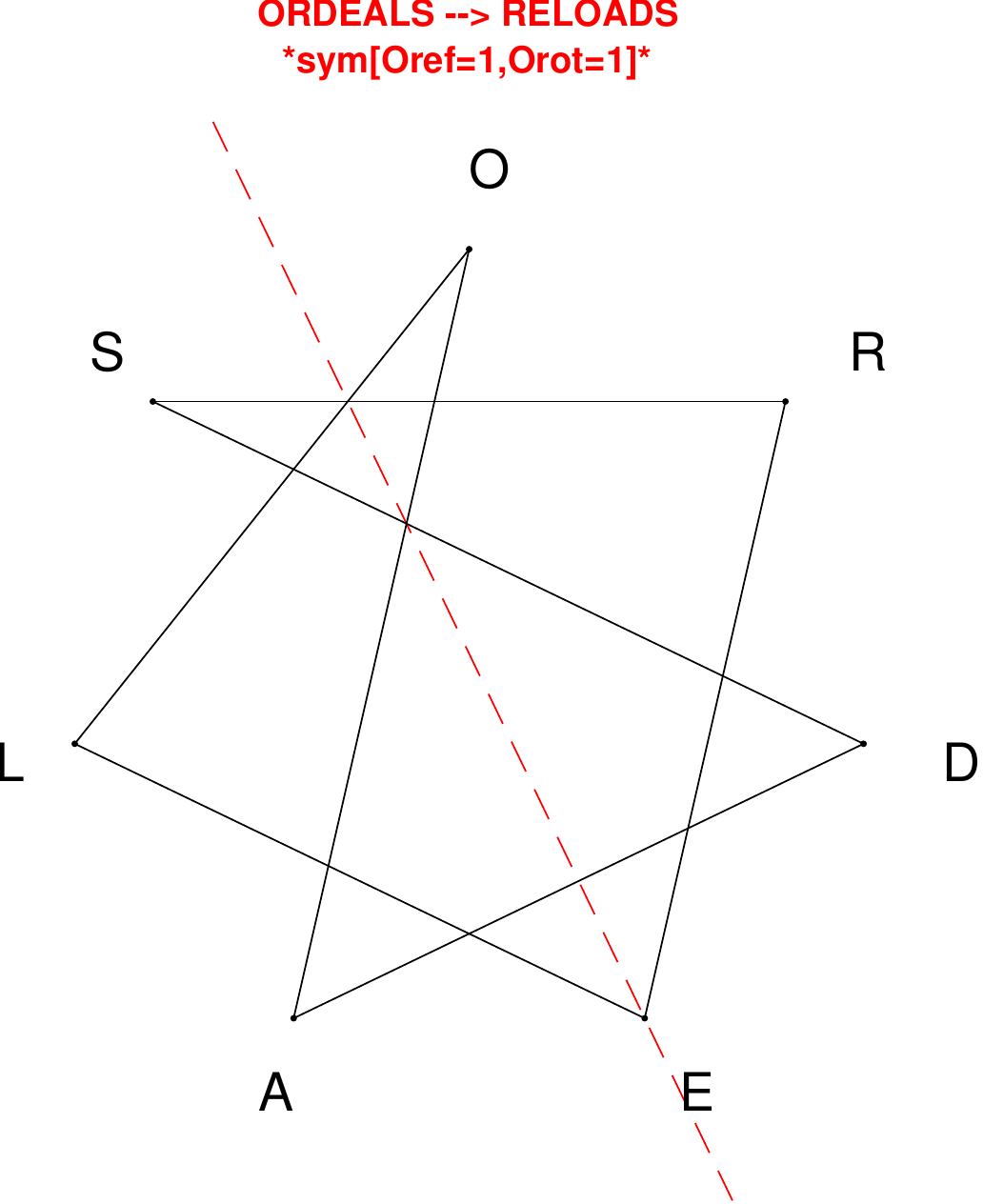}
\end{subfigure}
\hfill
\begin{subfigure}[T]{0.19\textwidth}
\centering
\includegraphics[width=\textwidth]{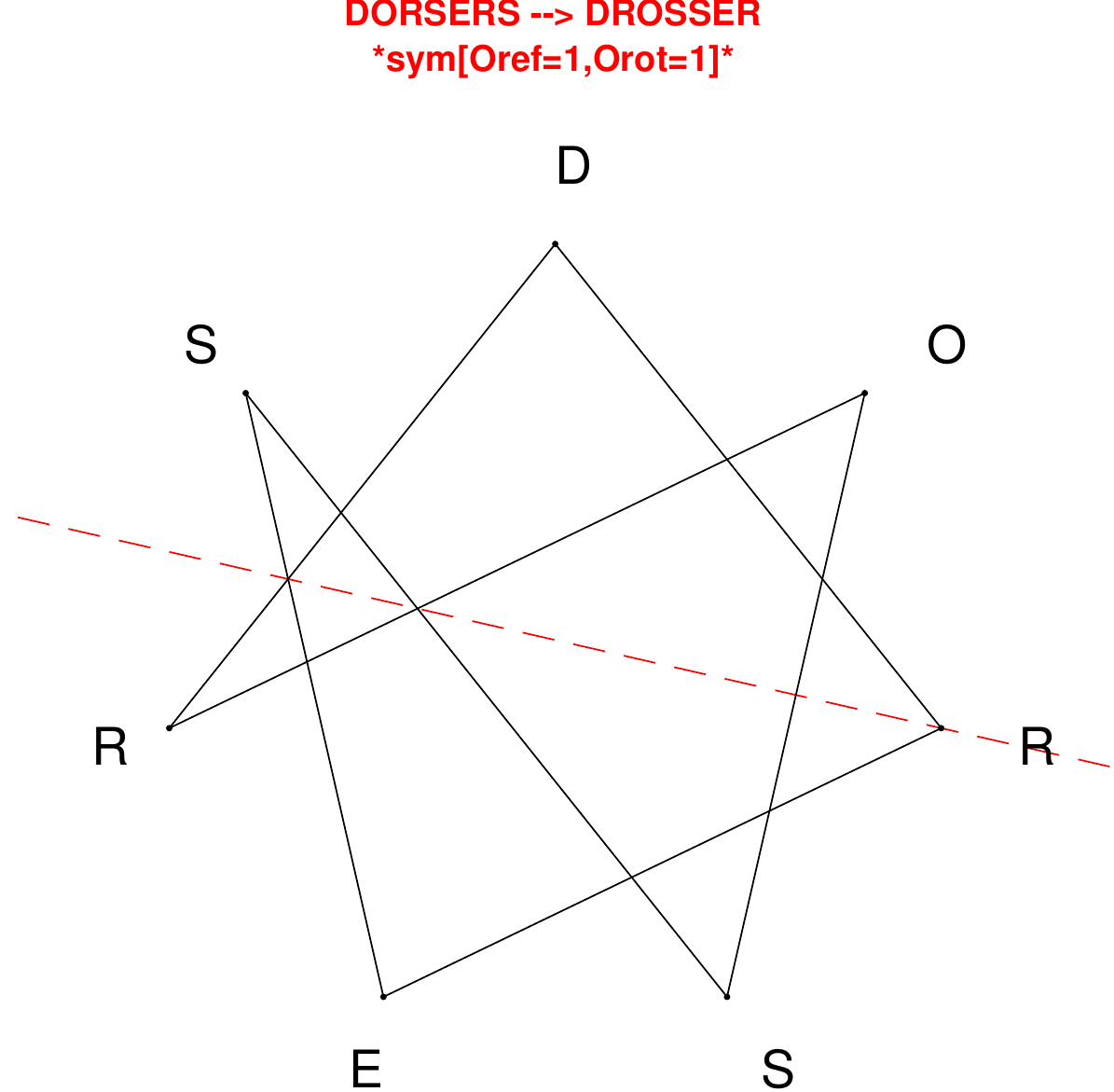}
\end{subfigure}
\hfill
\begin{subfigure}[T]{0.19\textwidth}
\centering
\includegraphics[width=\textwidth]{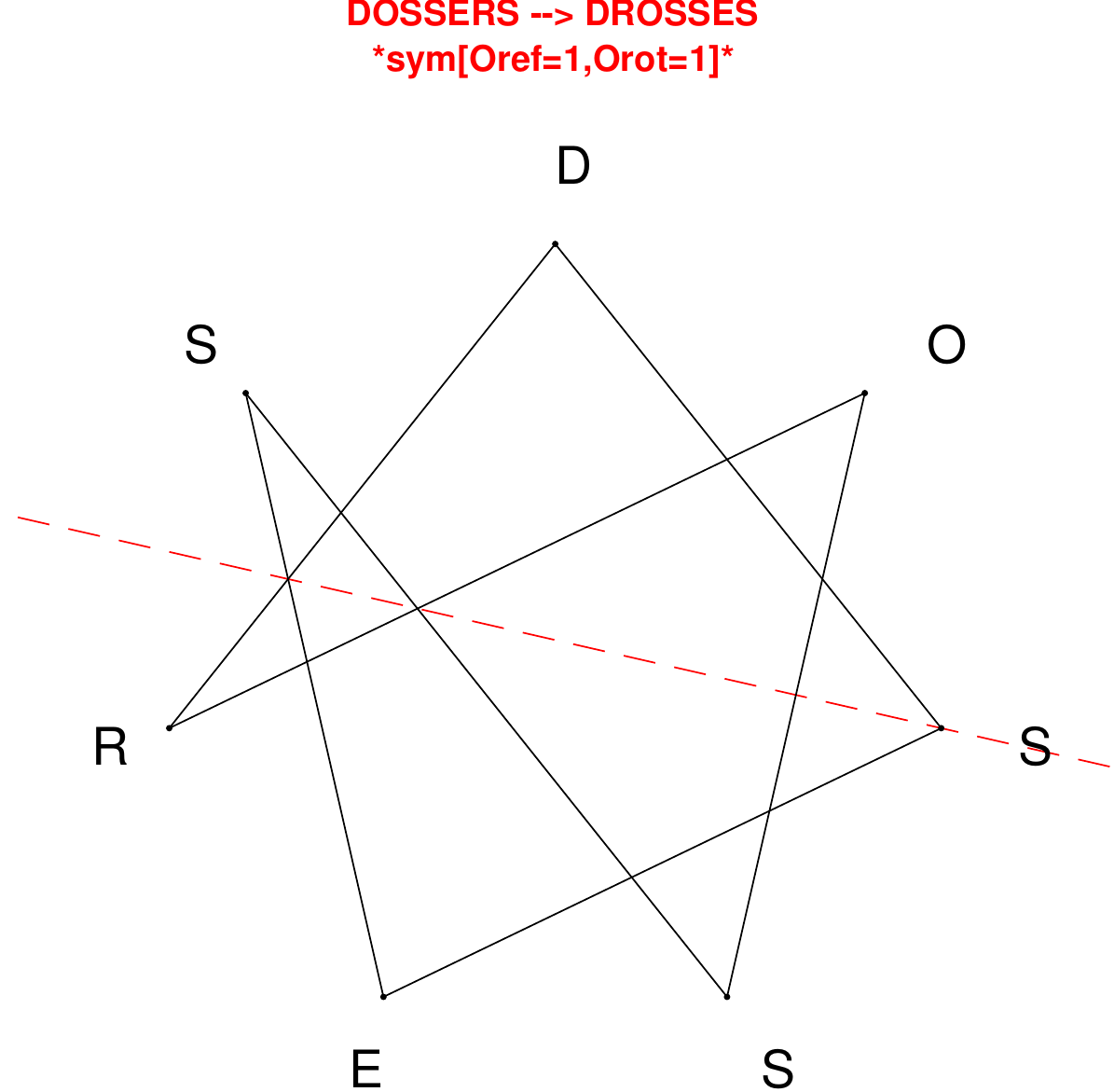}
\end{subfigure}
\hfill
\begin{subfigure}[T]{0.19\textwidth}
\centering
\includegraphics[width=\textwidth]{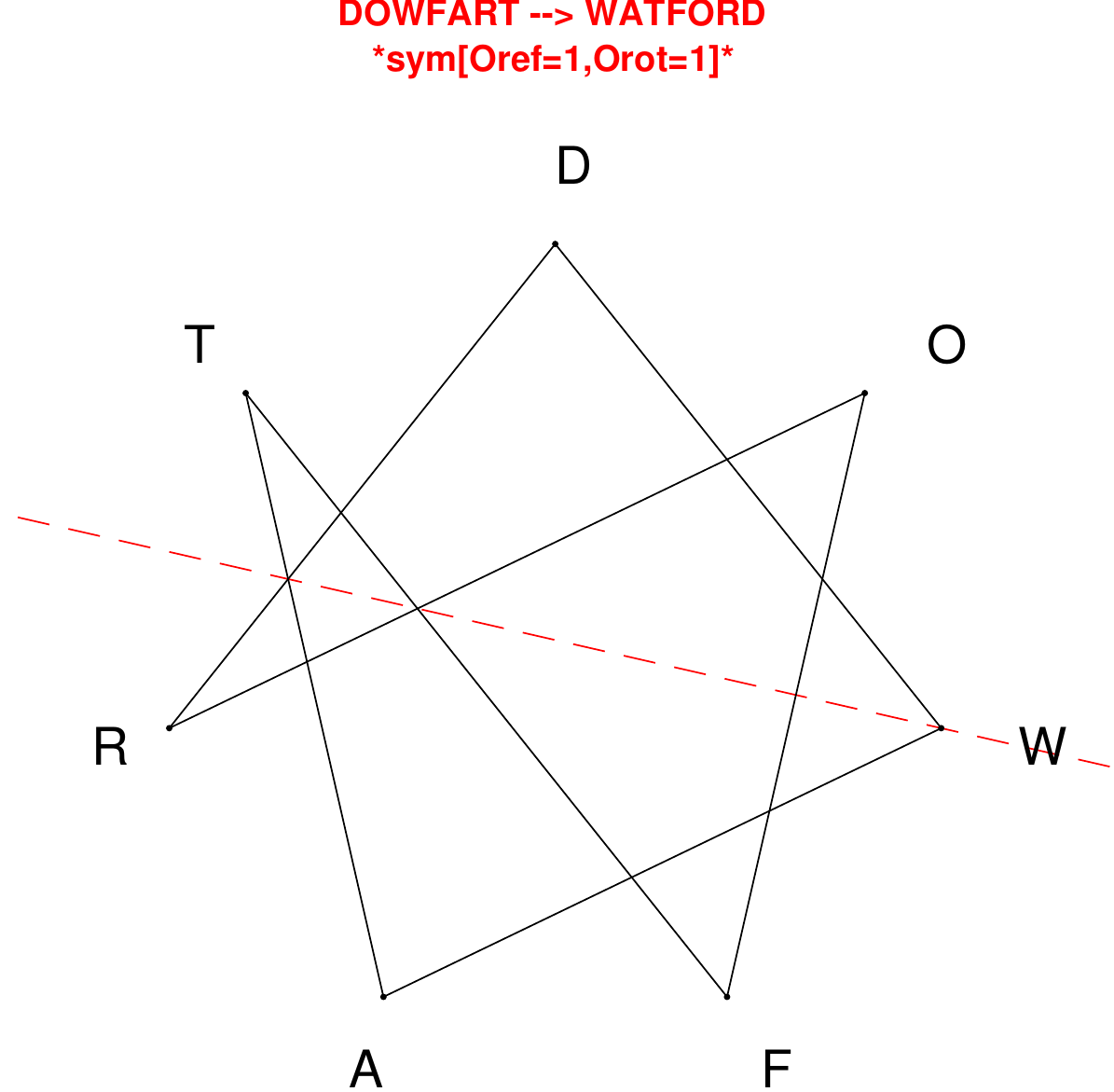}
\end{subfigure}
\end{figure}

\begin{figure}[H]
\centering
\begin{subfigure}[T]{0.19\textwidth}
\centering
\includegraphics[width=\textwidth]{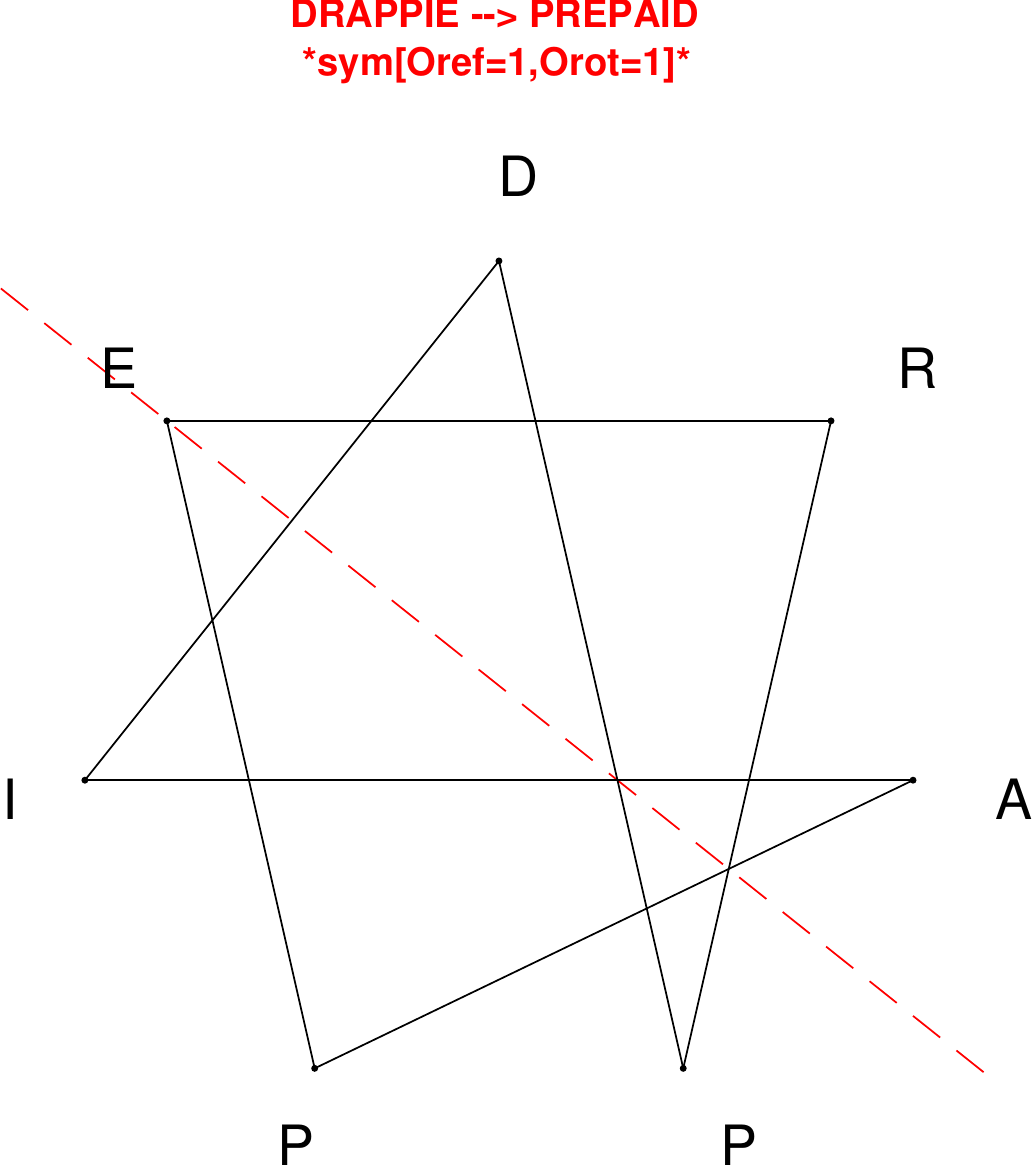}
\end{subfigure}
\hfill
\begin{subfigure}[T]{0.19\textwidth}
\centering
\includegraphics[width=\textwidth]{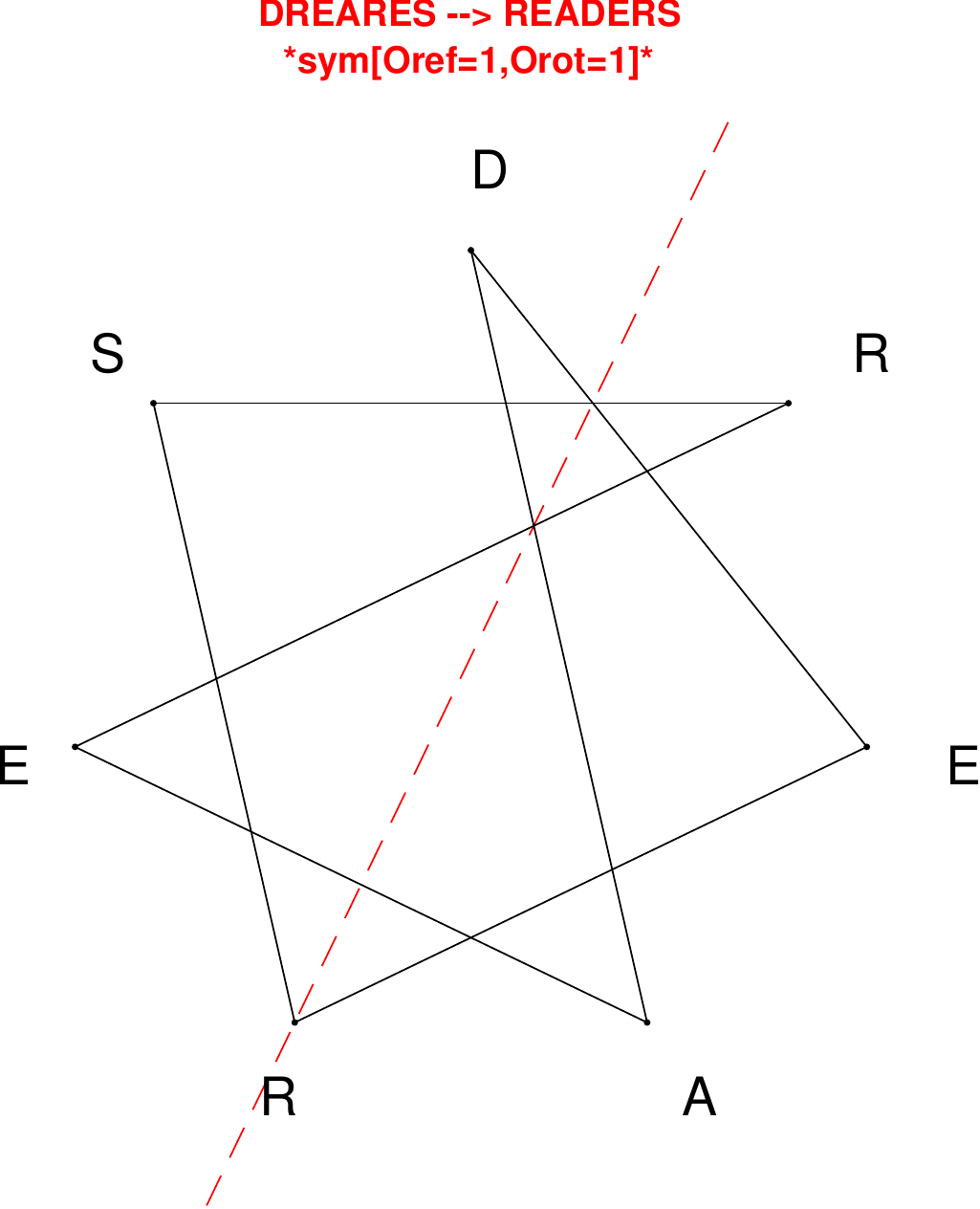}
\end{subfigure}
\hfill
\begin{subfigure}[T]{0.19\textwidth}
\centering
\includegraphics[width=\textwidth]{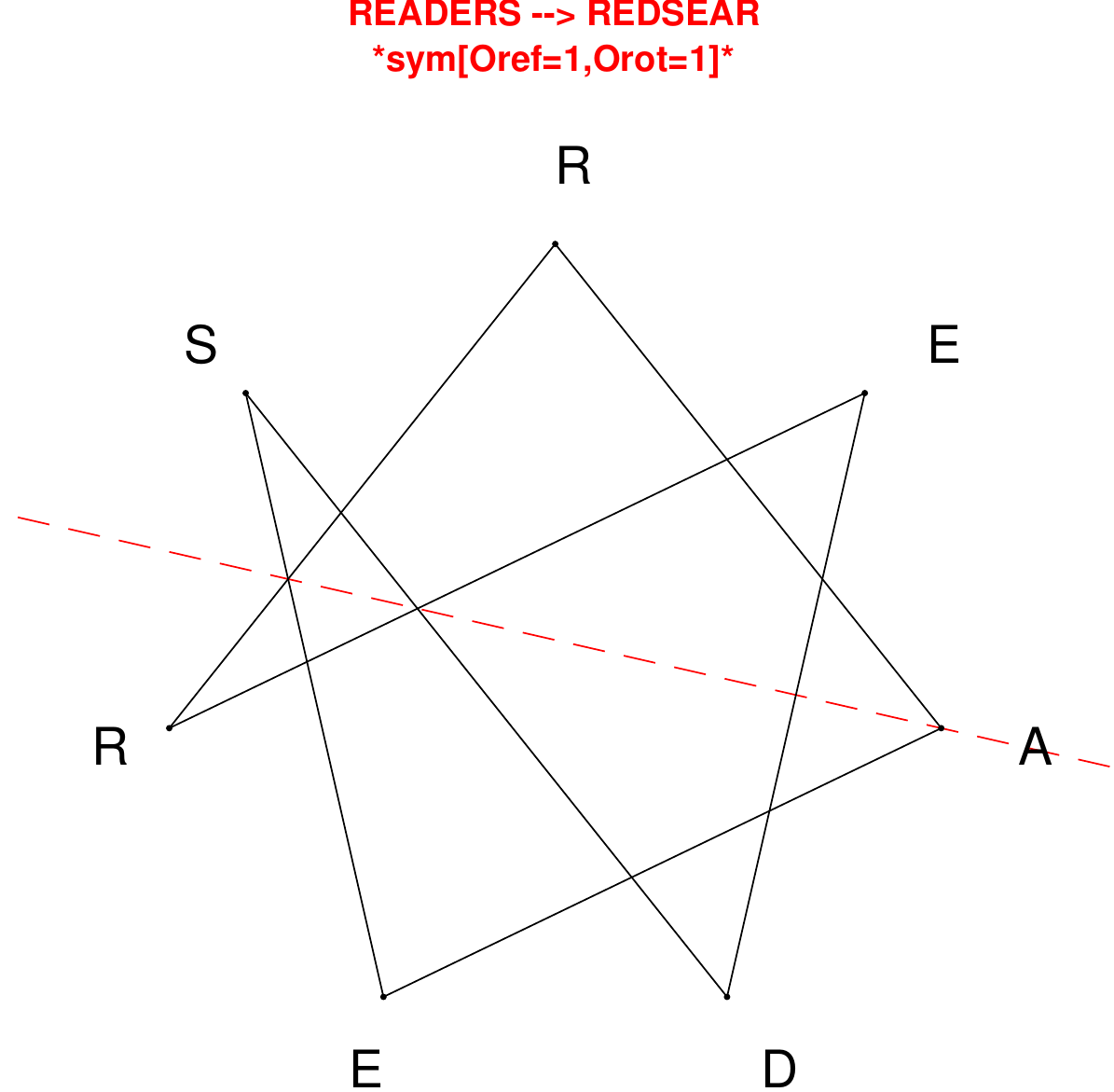}
\end{subfigure}
\hfill
\begin{subfigure}[T]{0.19\textwidth}
\centering
\includegraphics[width=\textwidth]{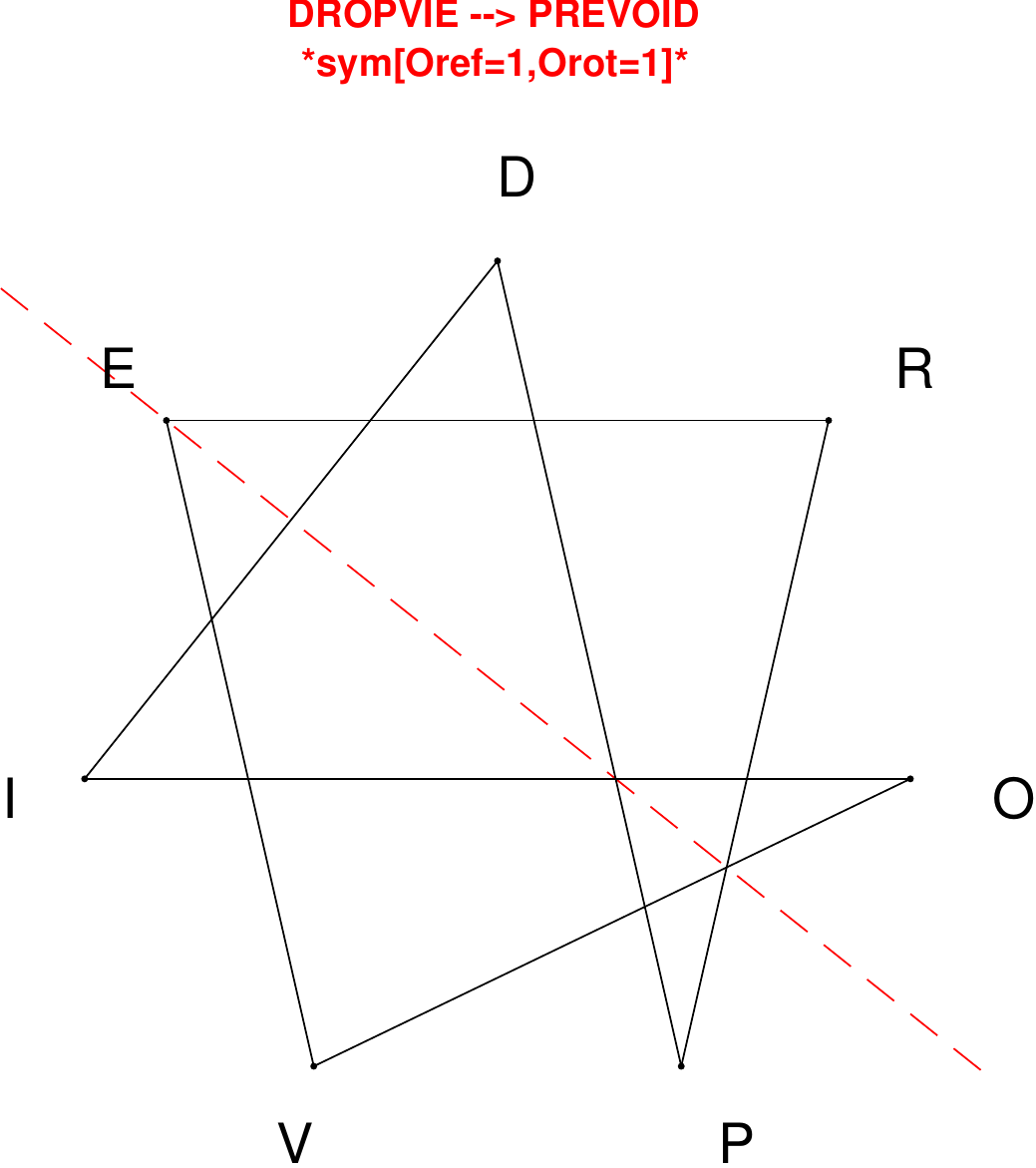}
\end{subfigure}
\hfill
\begin{subfigure}[T]{0.19\textwidth}
\centering
\includegraphics[width=\textwidth]{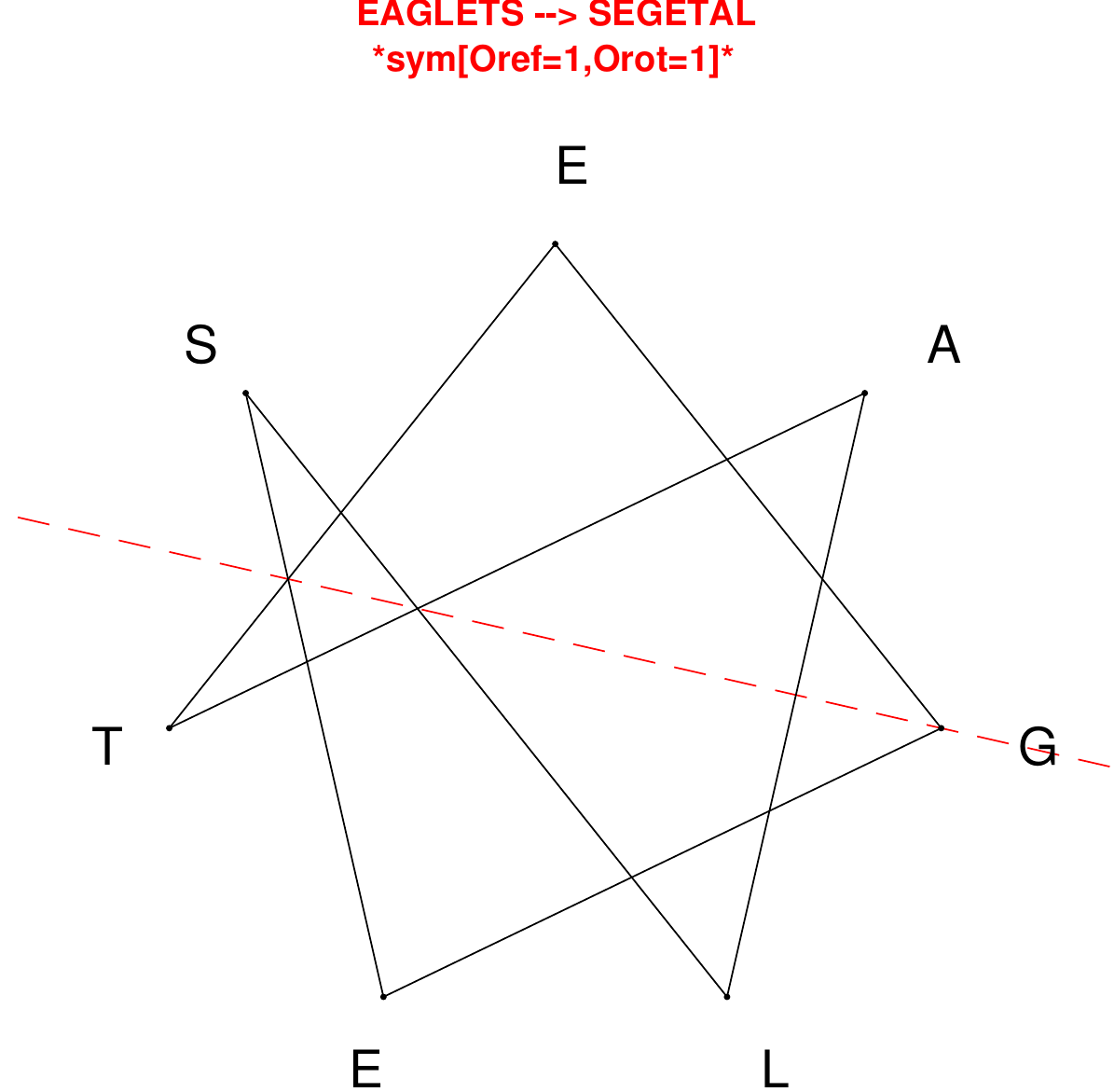}
\end{subfigure}
\end{figure}

\begin{figure}[H]
\centering
\begin{subfigure}[T]{0.19\textwidth}
\centering
\includegraphics[width=\textwidth]{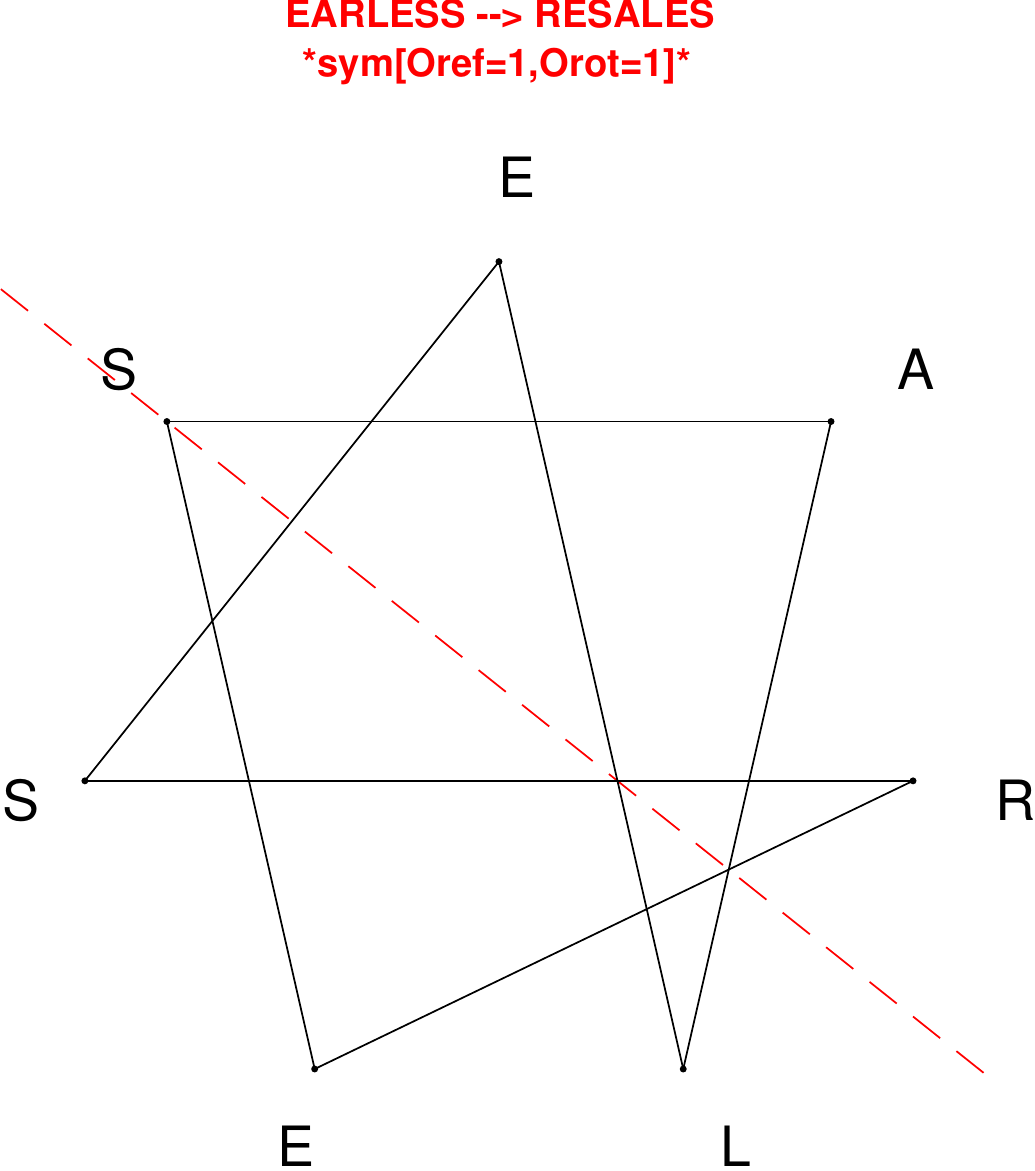}
\end{subfigure}
\hfill
\begin{subfigure}[T]{0.19\textwidth}
\centering
\includegraphics[width=\textwidth]{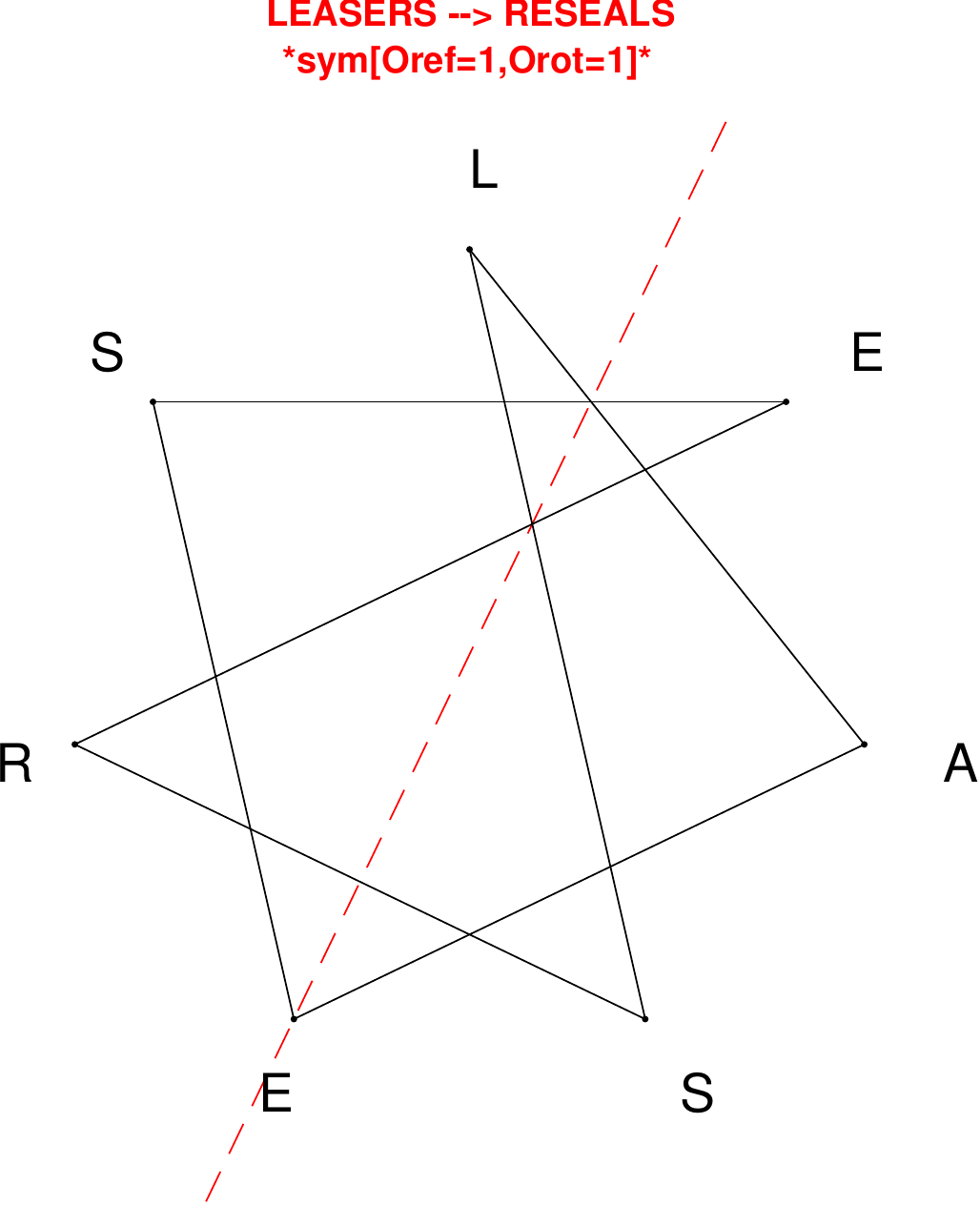}
\end{subfigure}
\hfill
\begin{subfigure}[T]{0.19\textwidth}
\centering
\includegraphics[width=\textwidth]{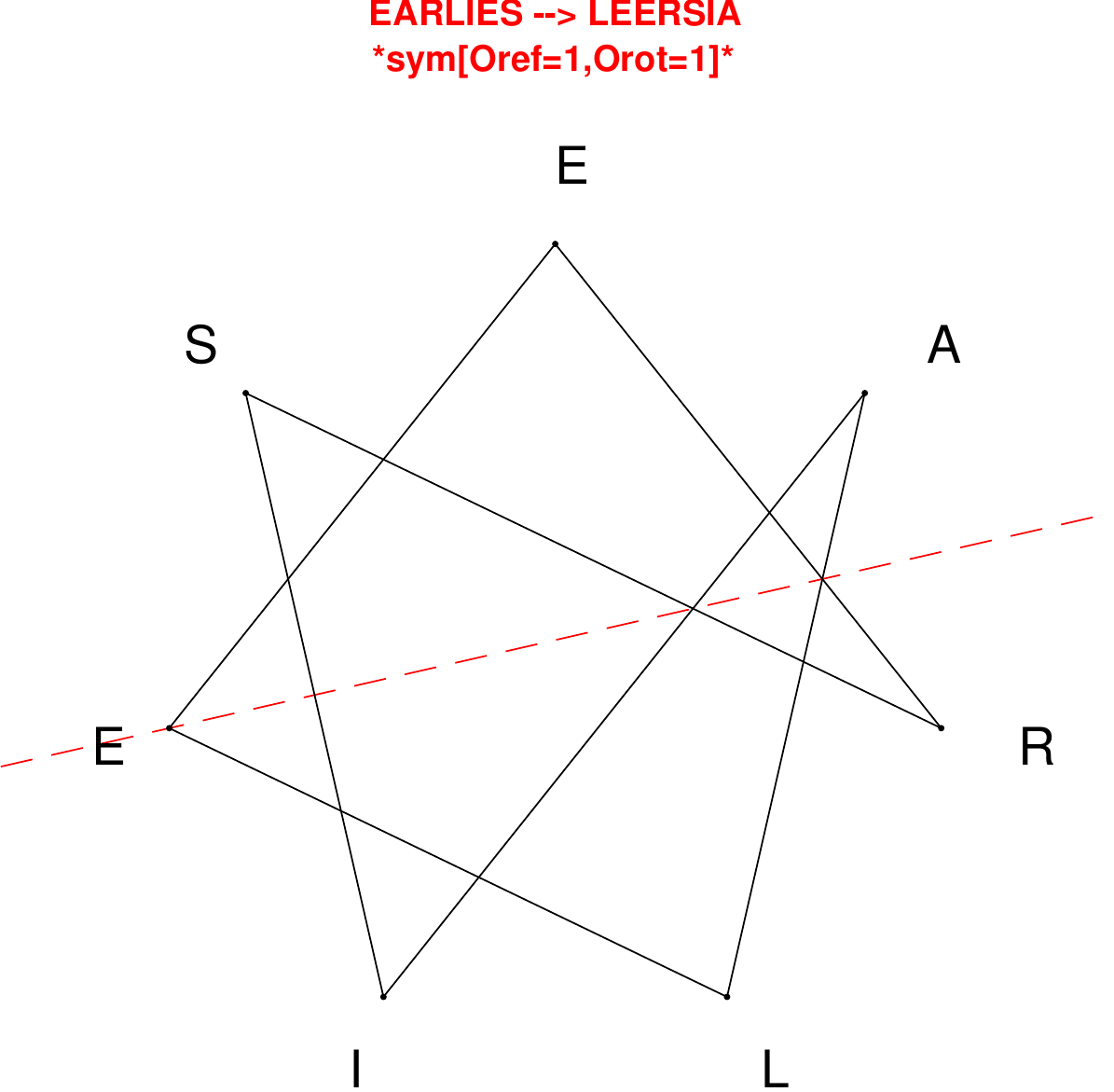}
\end{subfigure}
\hfill
\begin{subfigure}[T]{0.19\textwidth}
\centering
\includegraphics[width=\textwidth]{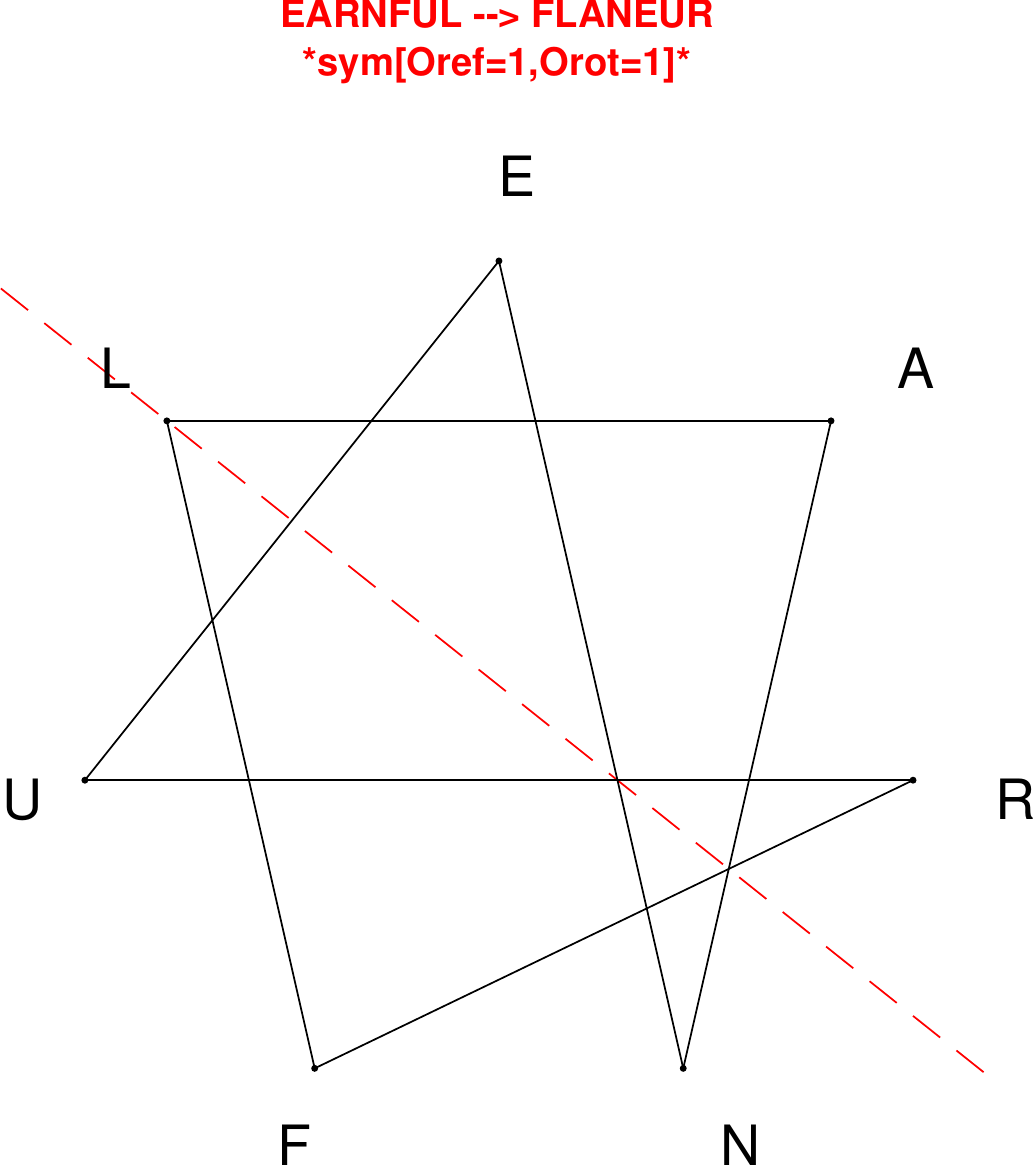}
\end{subfigure}
\hfill
\begin{subfigure}[T]{0.19\textwidth}
\centering
\includegraphics[width=\textwidth]{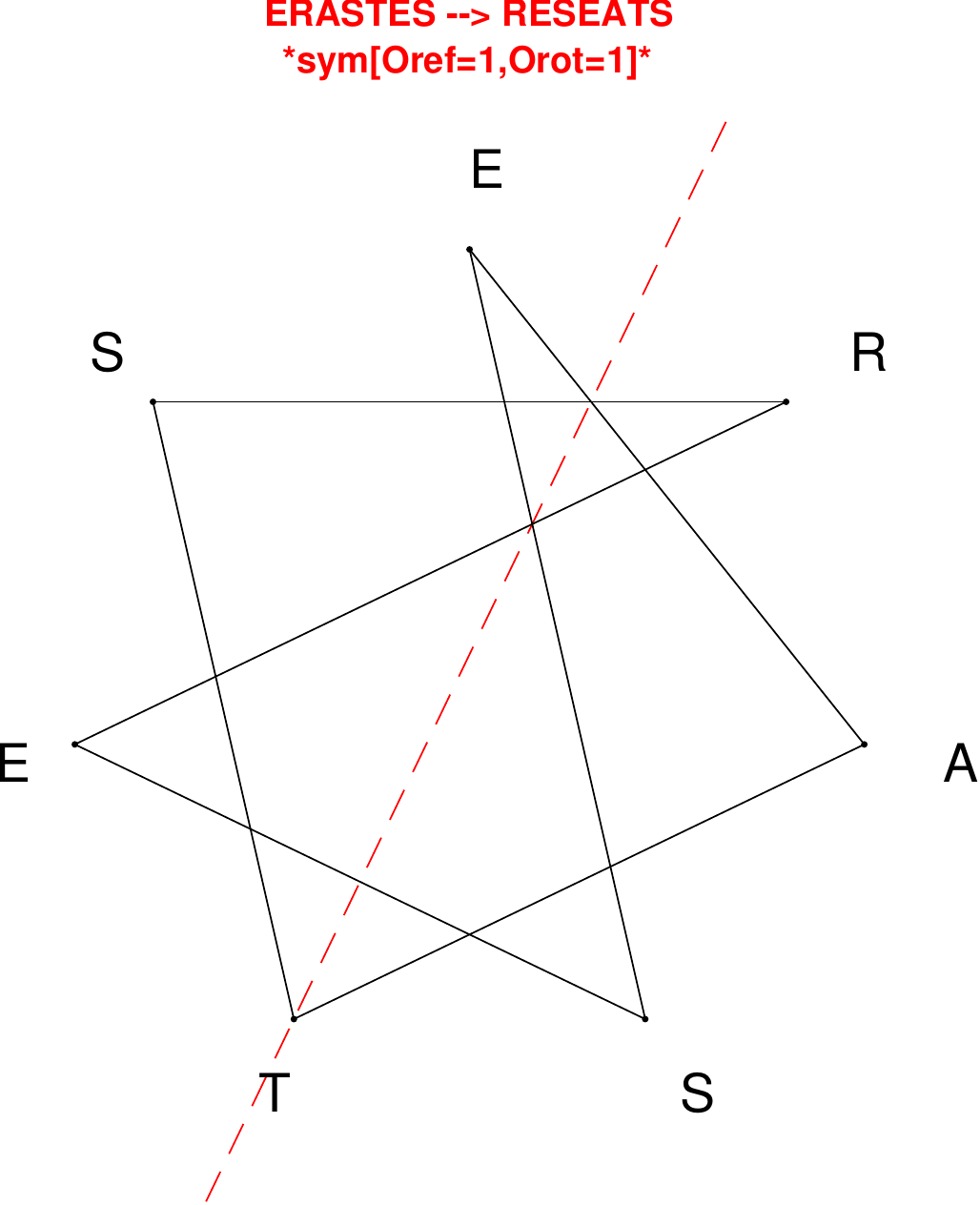}
\end{subfigure}
\end{figure}

\begin{figure}[H]
\centering
\begin{subfigure}[T]{0.19\textwidth}
\centering
\includegraphics[width=\textwidth]{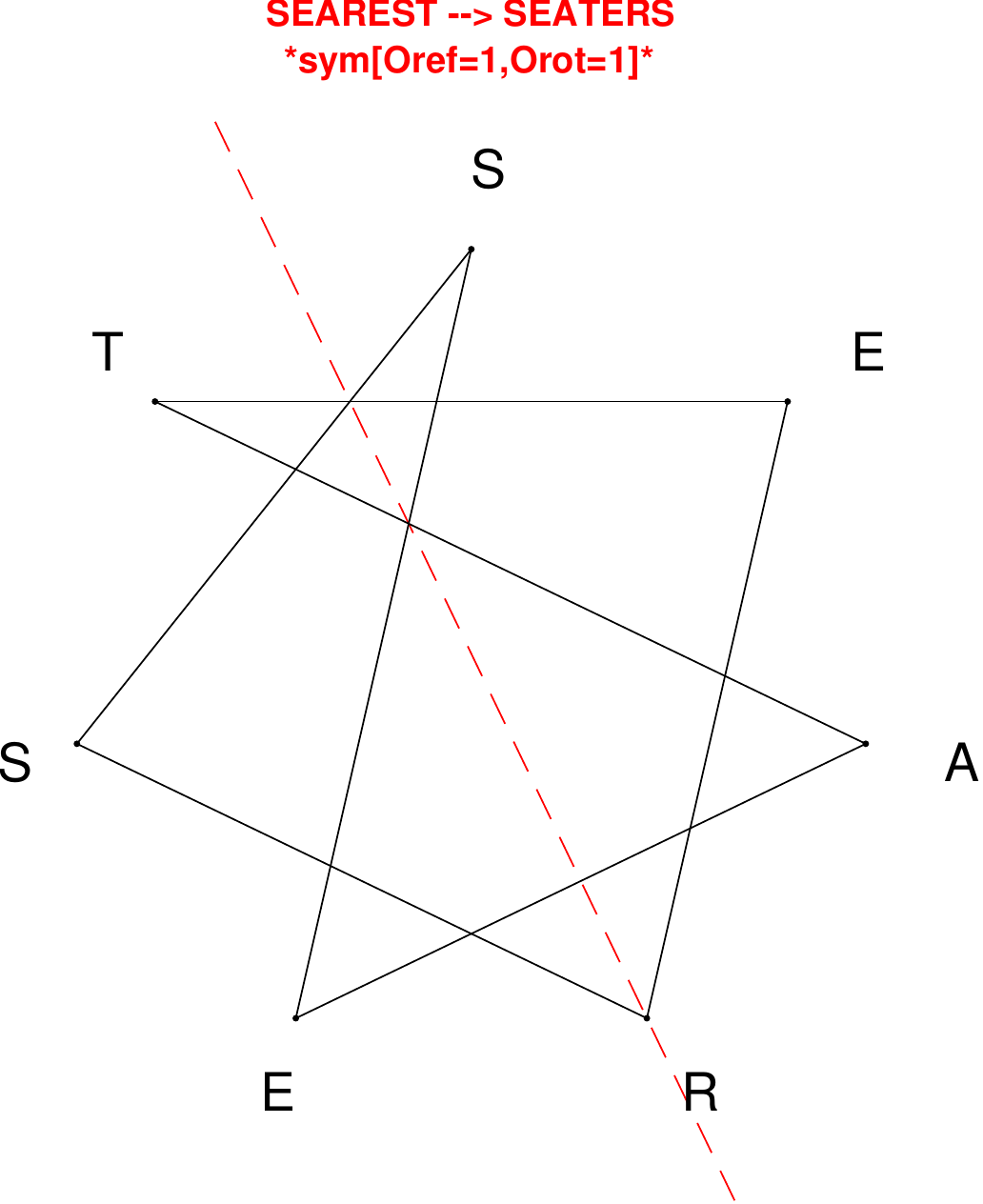}
\end{subfigure}
\hfill
\begin{subfigure}[T]{0.19\textwidth}
\centering
\includegraphics[width=\textwidth]{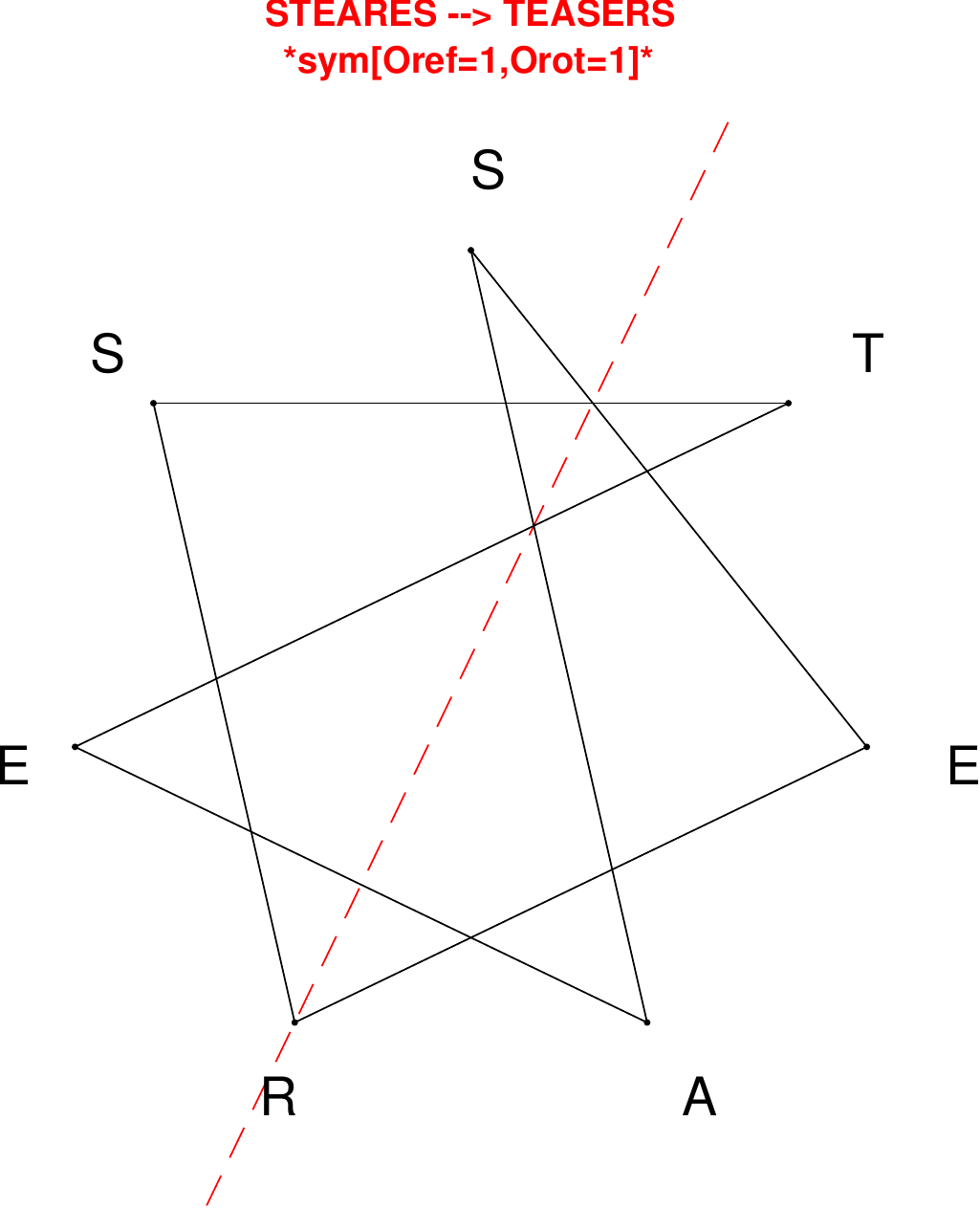}
\end{subfigure}
\hfill
\begin{subfigure}[T]{0.19\textwidth}
\centering
\includegraphics[width=\textwidth]{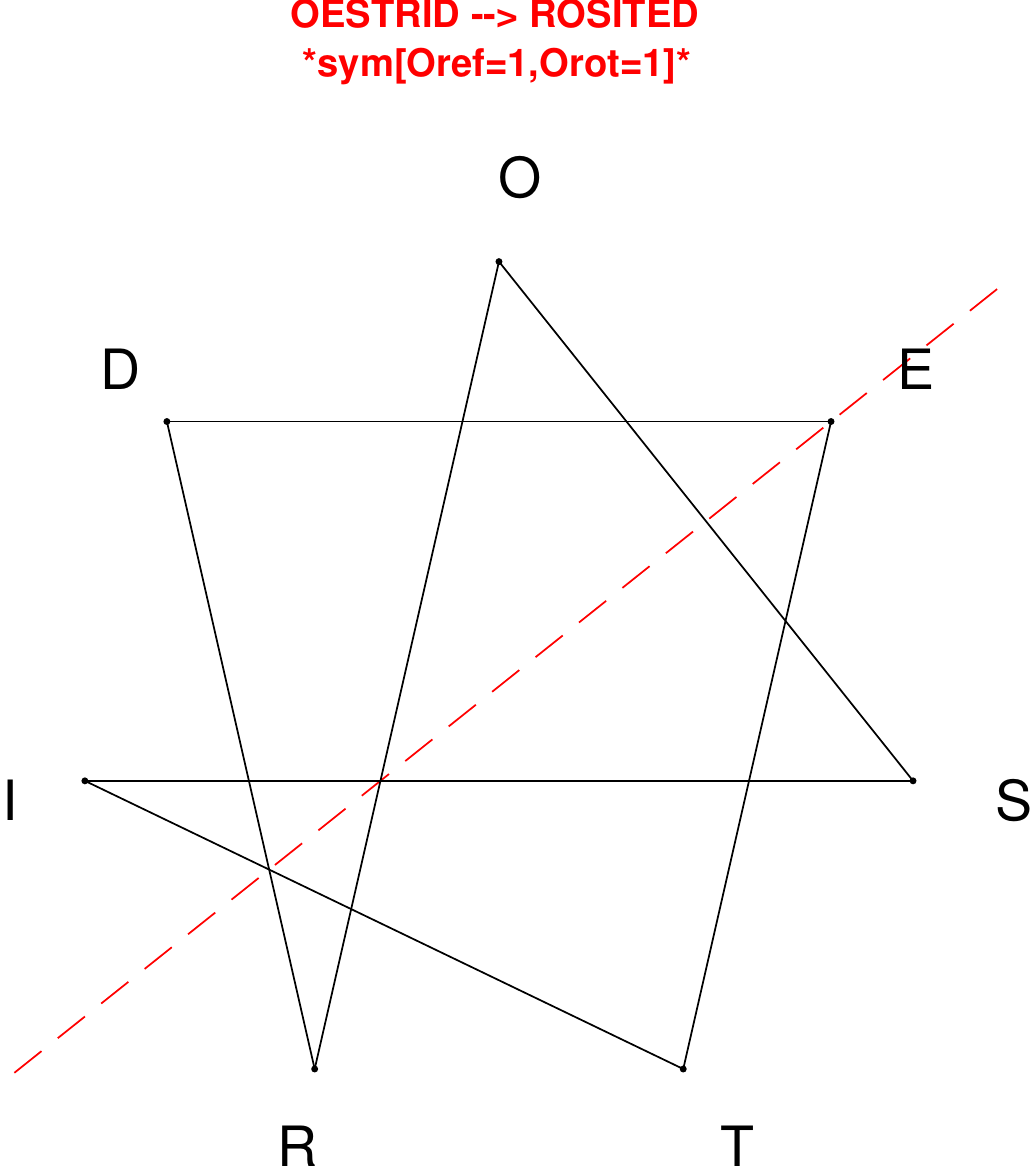}
\end{subfigure}
\hfill
\begin{subfigure}[T]{0.19\textwidth}
\centering
\includegraphics[width=\textwidth]{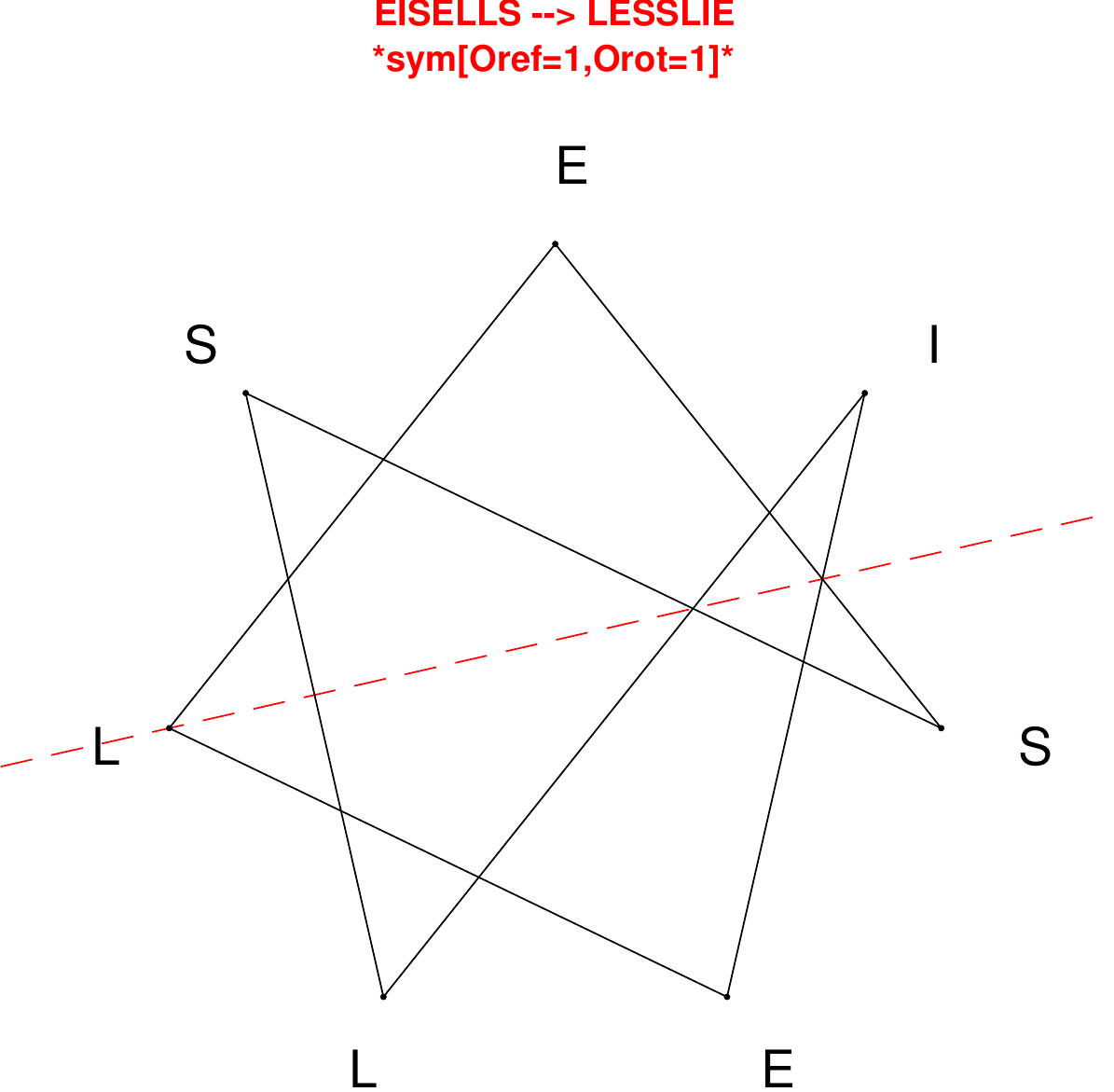}
\end{subfigure}
\hfill
\begin{subfigure}[T]{0.19\textwidth}
\centering
\includegraphics[width=\textwidth]{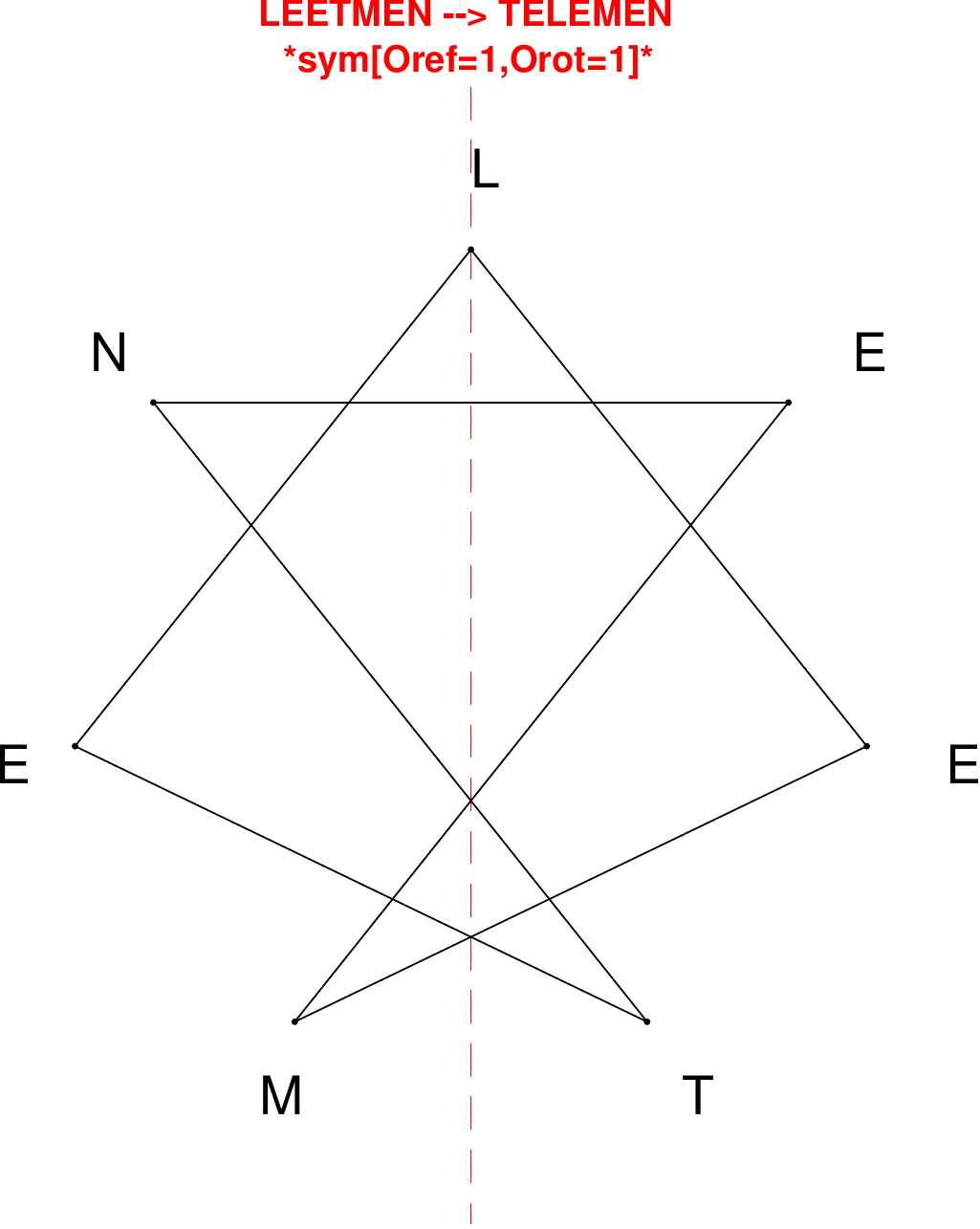}
\end{subfigure}
\end{figure}

\begin{figure}[H]
\centering
\begin{subfigure}[T]{0.19\textwidth}
\centering
\includegraphics[width=\textwidth]{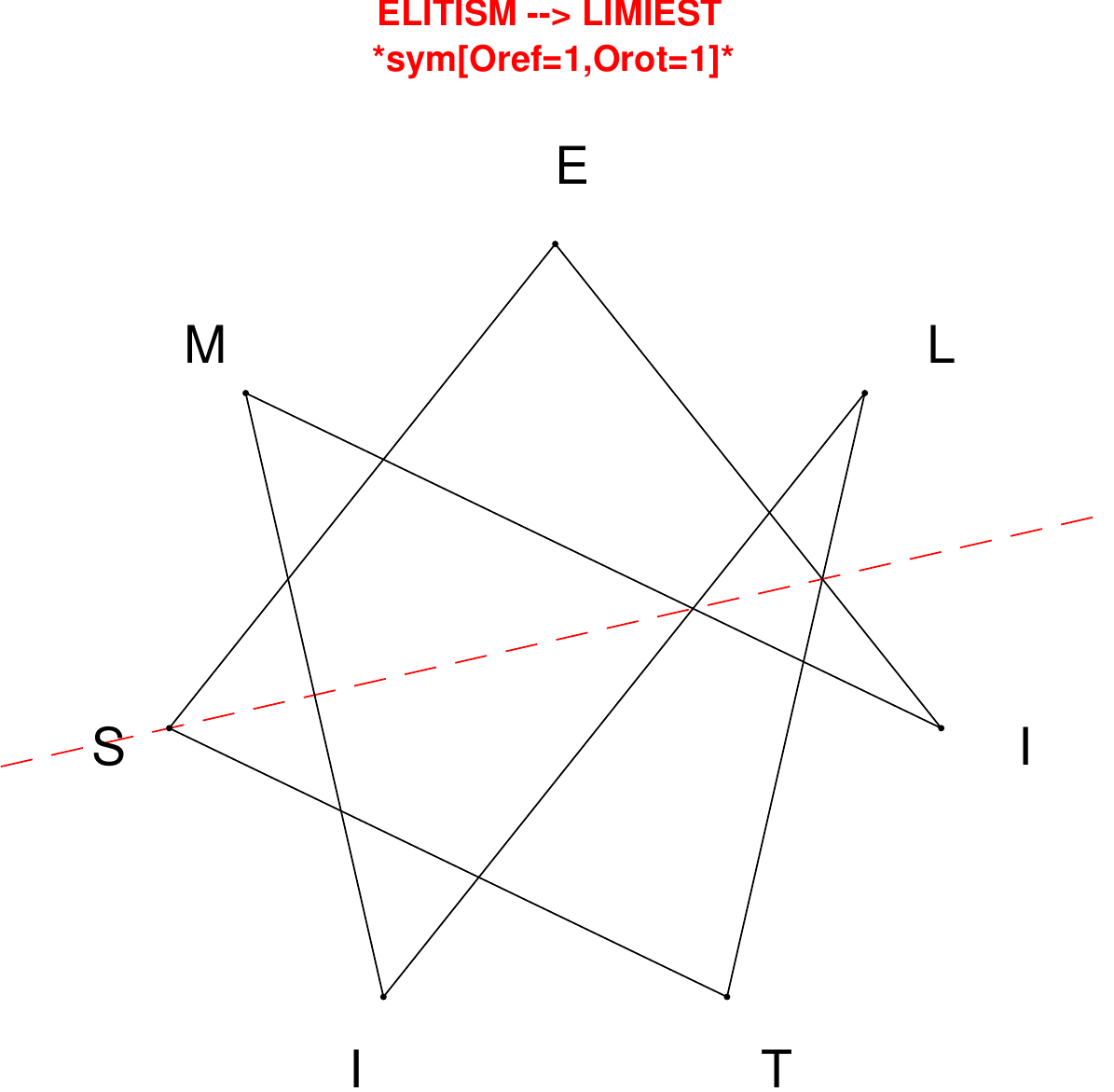}
\end{subfigure}
\hfill
\begin{subfigure}[T]{0.19\textwidth}
\centering
\includegraphics[width=\textwidth]{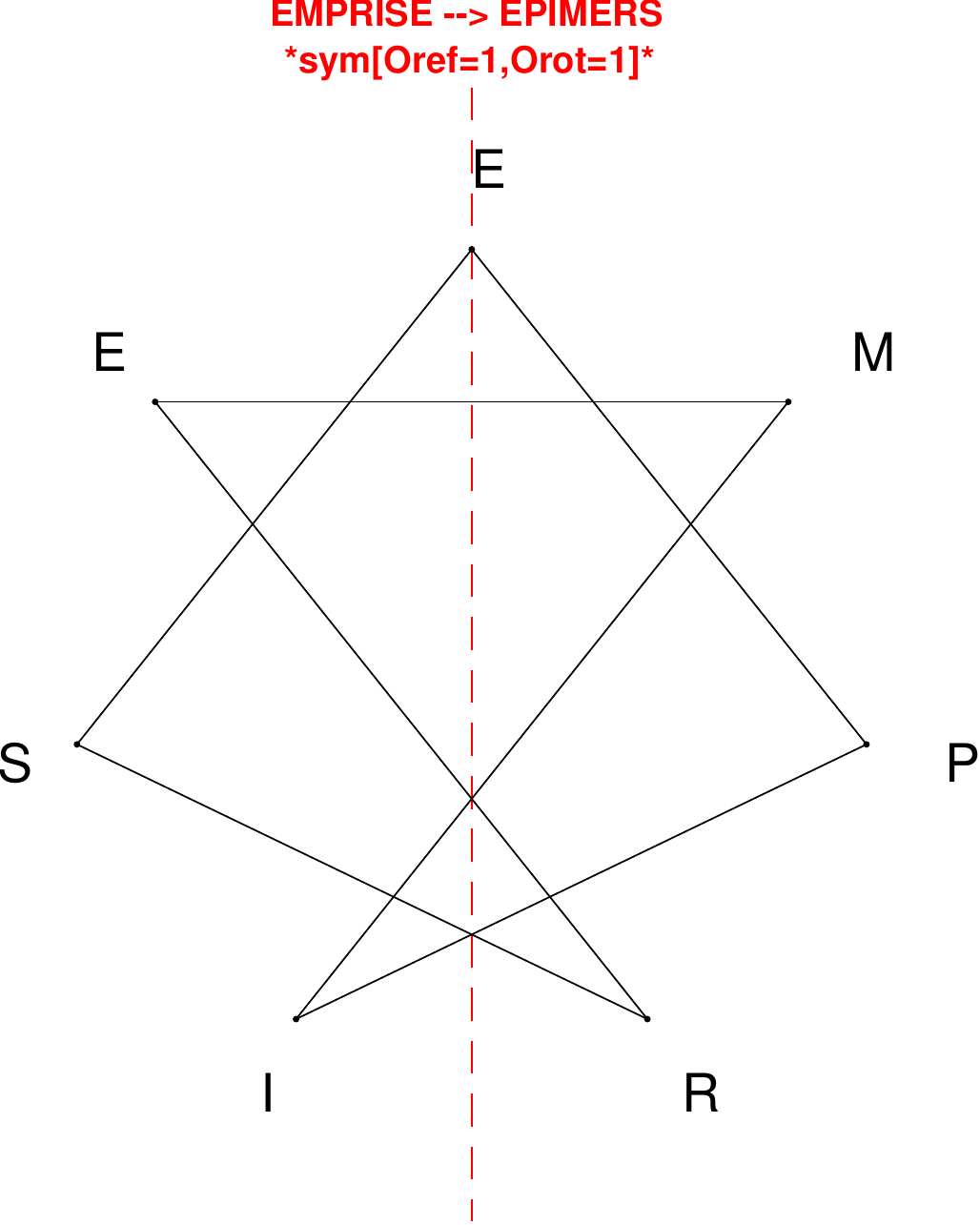}
\end{subfigure}
\hfill
\begin{subfigure}[T]{0.19\textwidth}
\centering
\includegraphics[width=\textwidth]{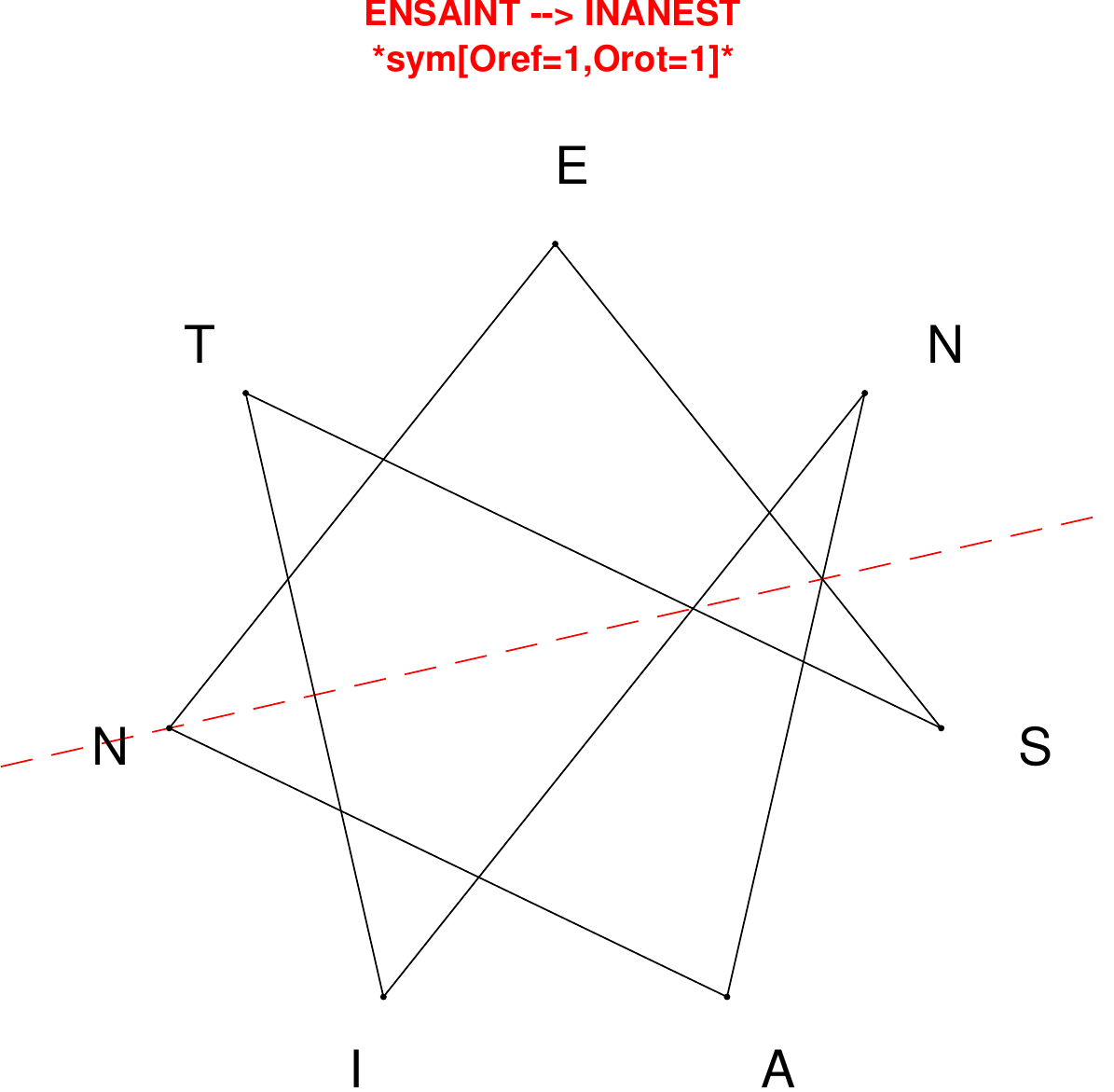}
\end{subfigure}
\hfill
\begin{subfigure}[T]{0.19\textwidth}
\centering
\includegraphics[width=\textwidth]{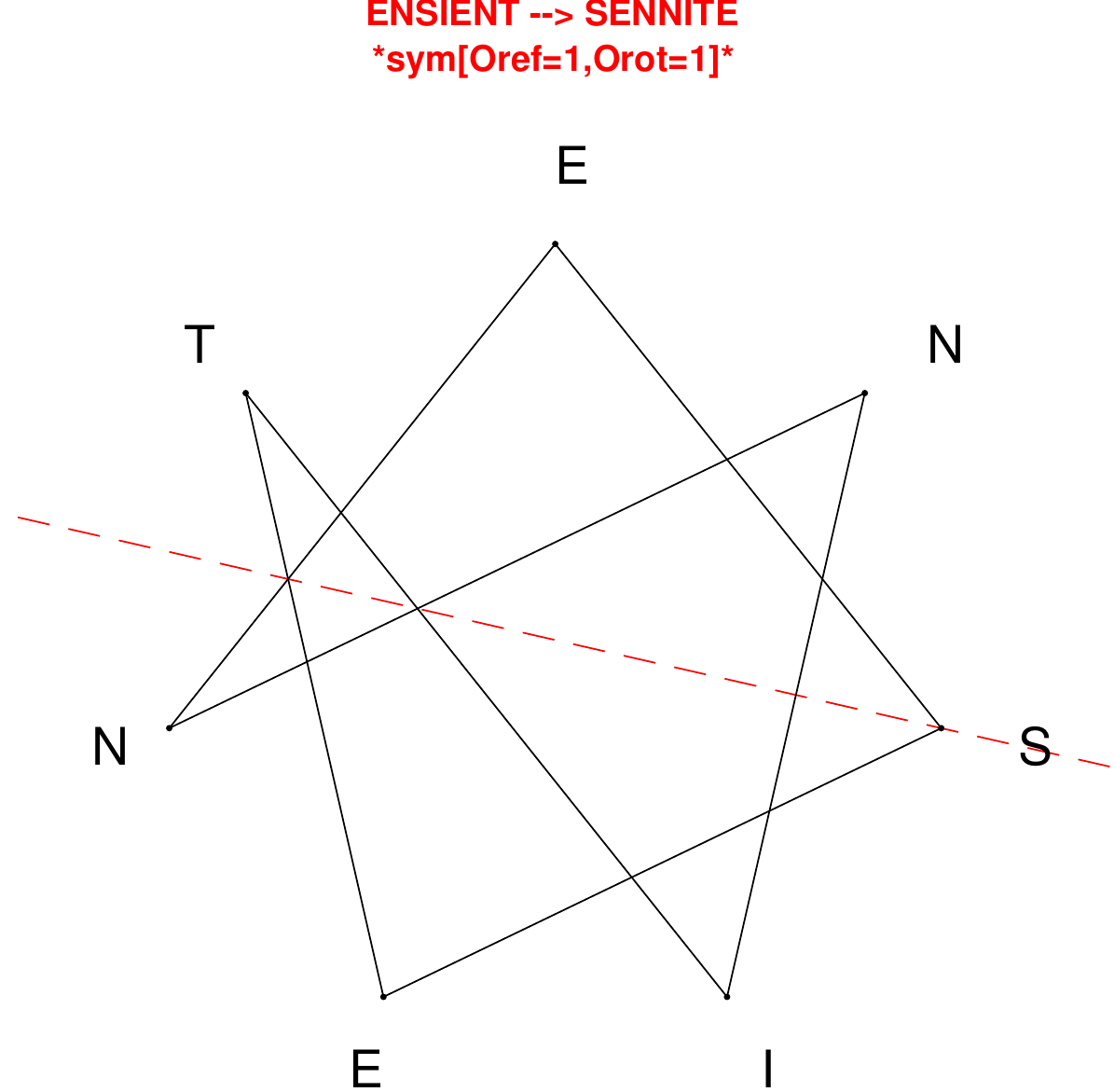}
\end{subfigure}
\hfill
\begin{subfigure}[T]{0.19\textwidth}
\centering
\includegraphics[width=\textwidth]{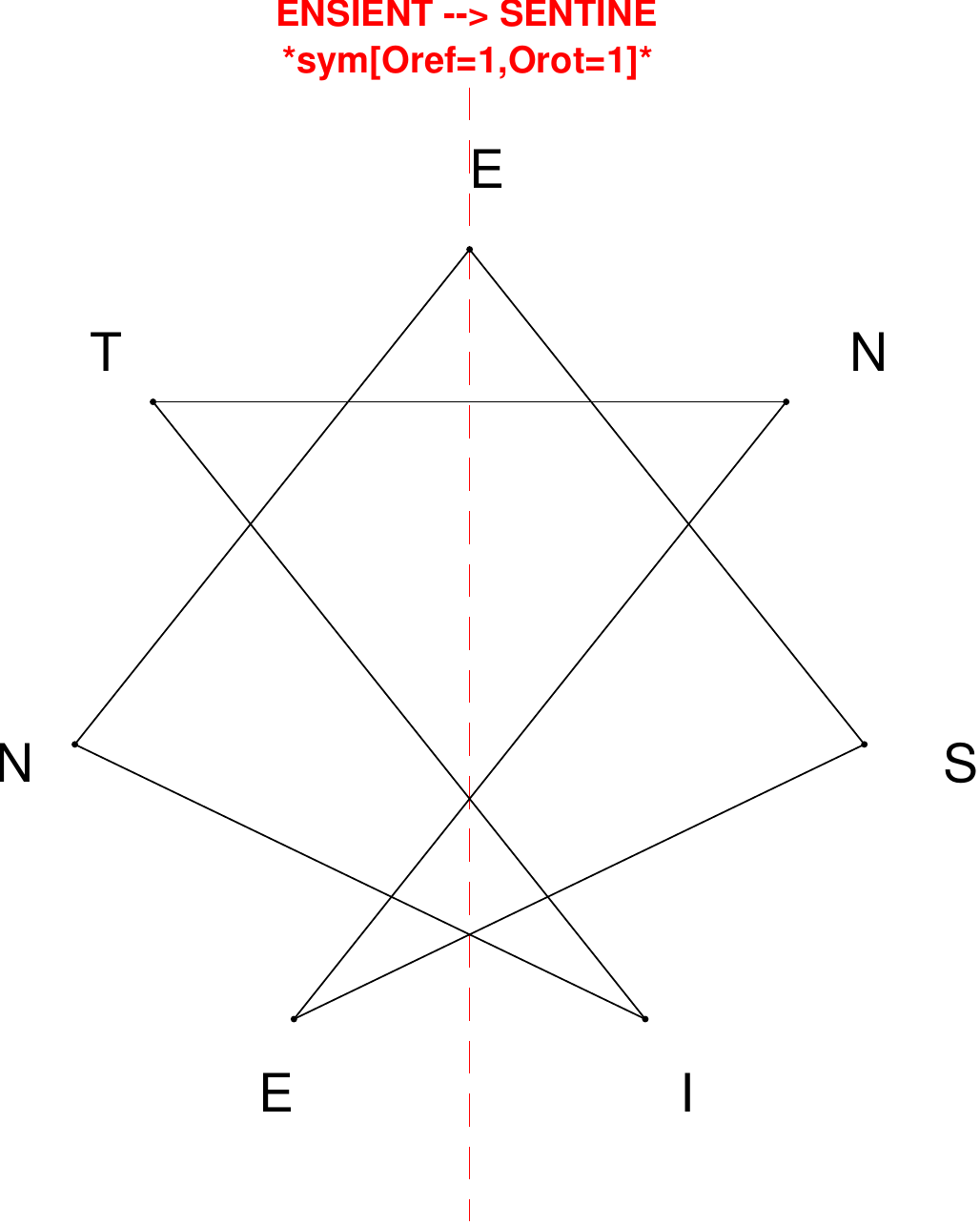}
\end{subfigure}
\end{figure}

\begin{figure}[H]
\centering
\begin{subfigure}[T]{0.19\textwidth}
\centering
\includegraphics[width=\textwidth]{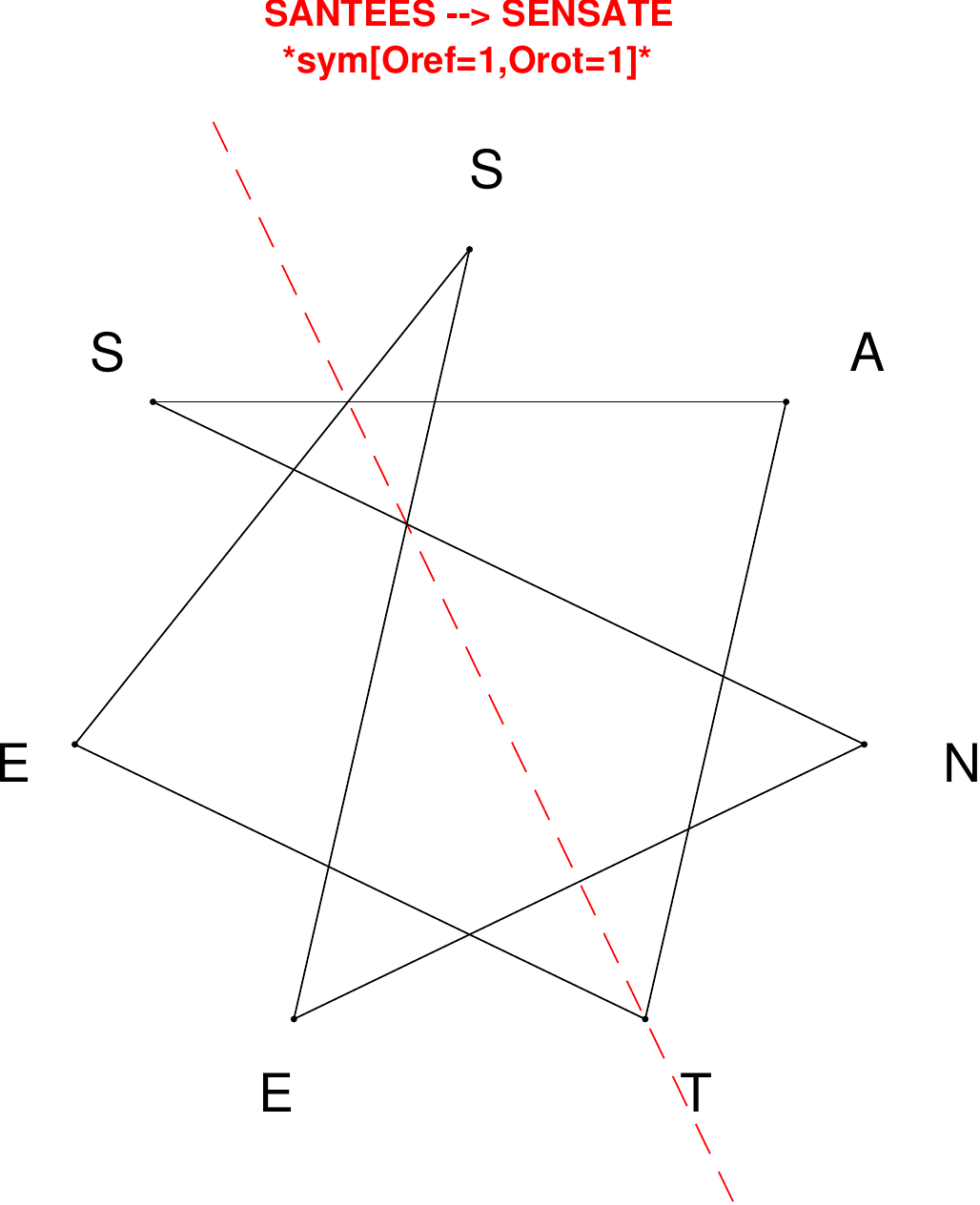}
\end{subfigure}
\hfill
\begin{subfigure}[T]{0.19\textwidth}
\centering
\includegraphics[width=\textwidth]{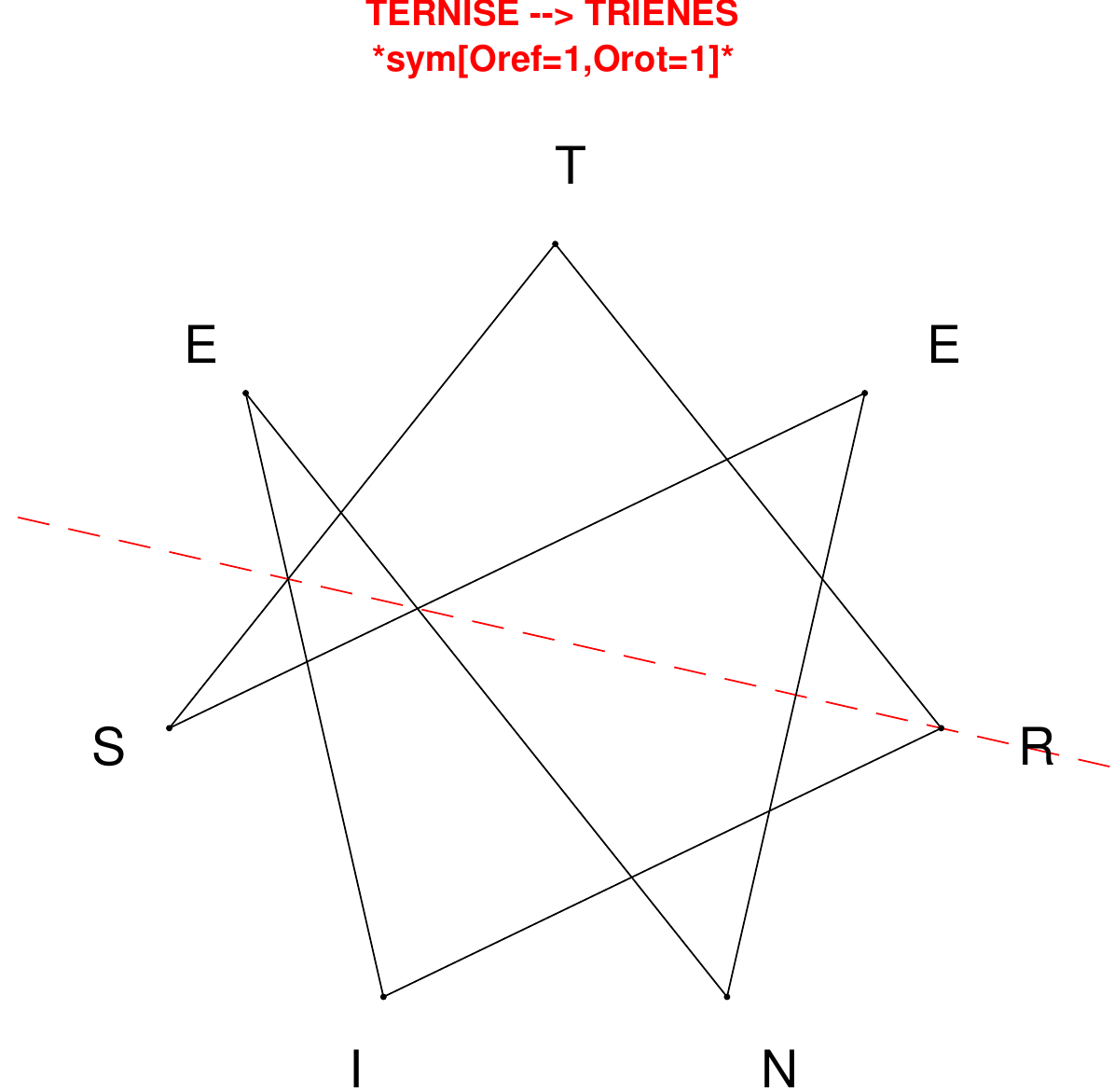}
\end{subfigure}
\hfill
\begin{subfigure}[T]{0.19\textwidth}
\centering
\includegraphics[width=\textwidth]{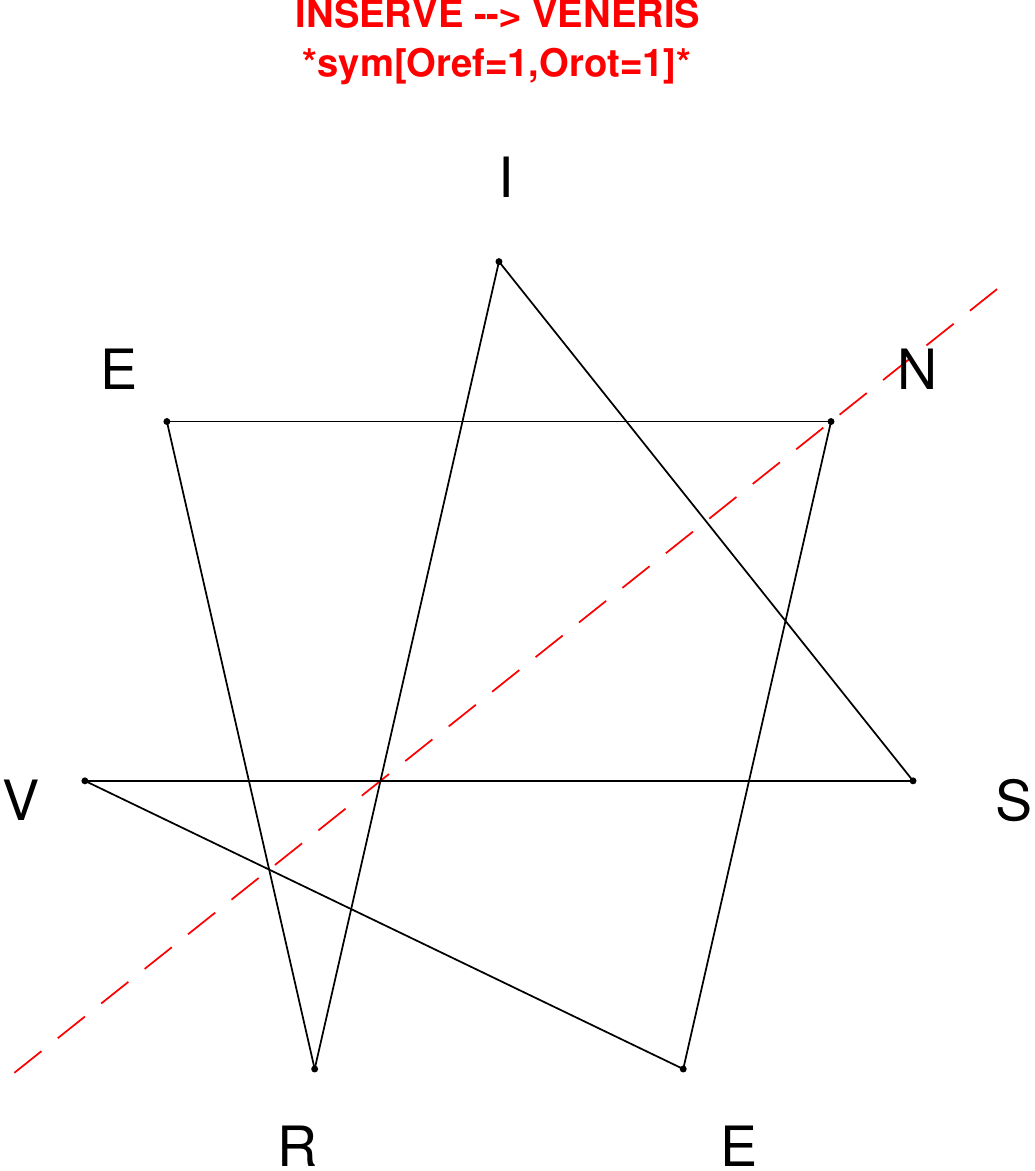}
\end{subfigure}
\hfill
\begin{subfigure}[T]{0.19\textwidth}
\centering
\includegraphics[width=\textwidth]{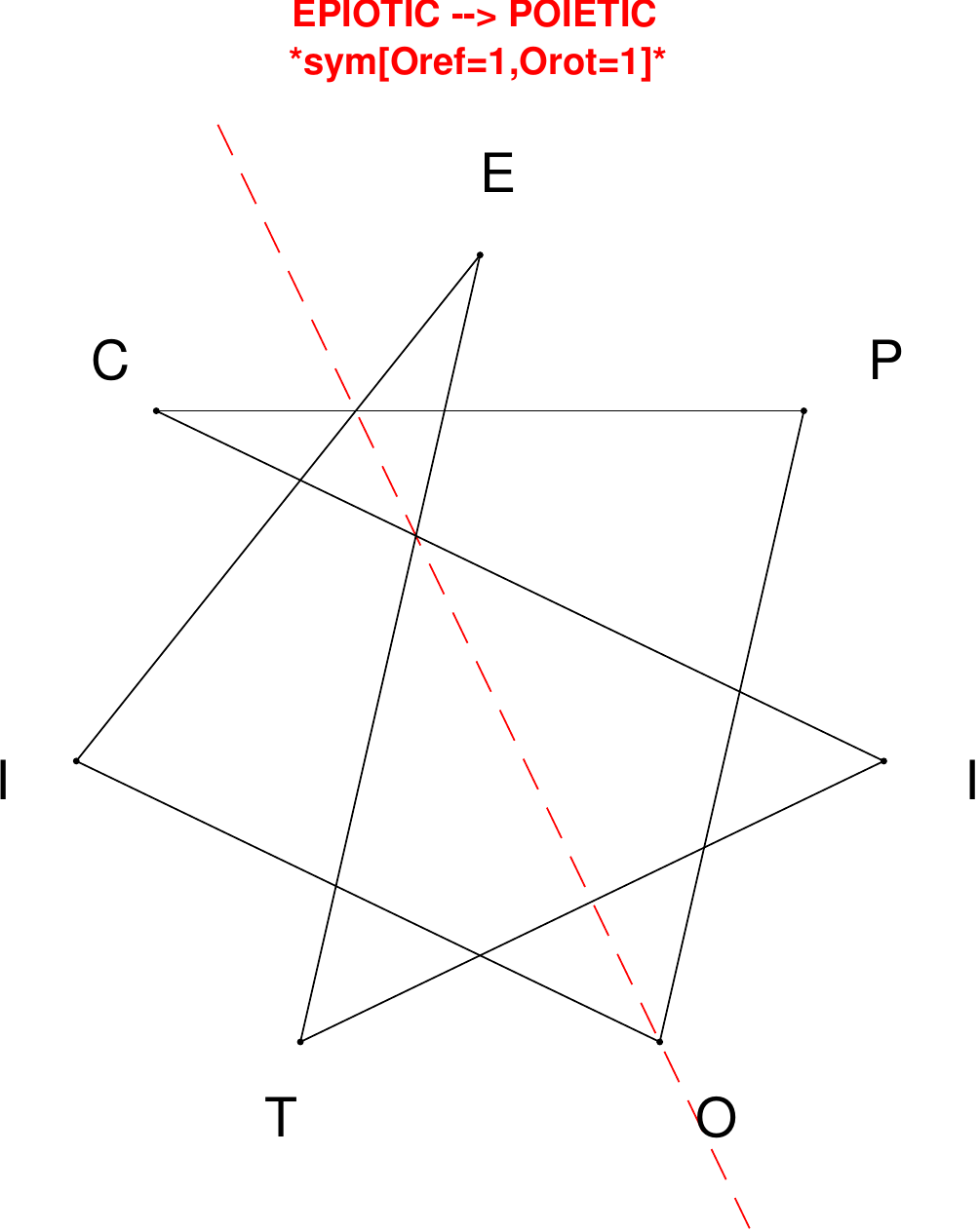}
\end{subfigure}
\hfill
\begin{subfigure}[T]{0.19\textwidth}
\centering
\includegraphics[width=\textwidth]{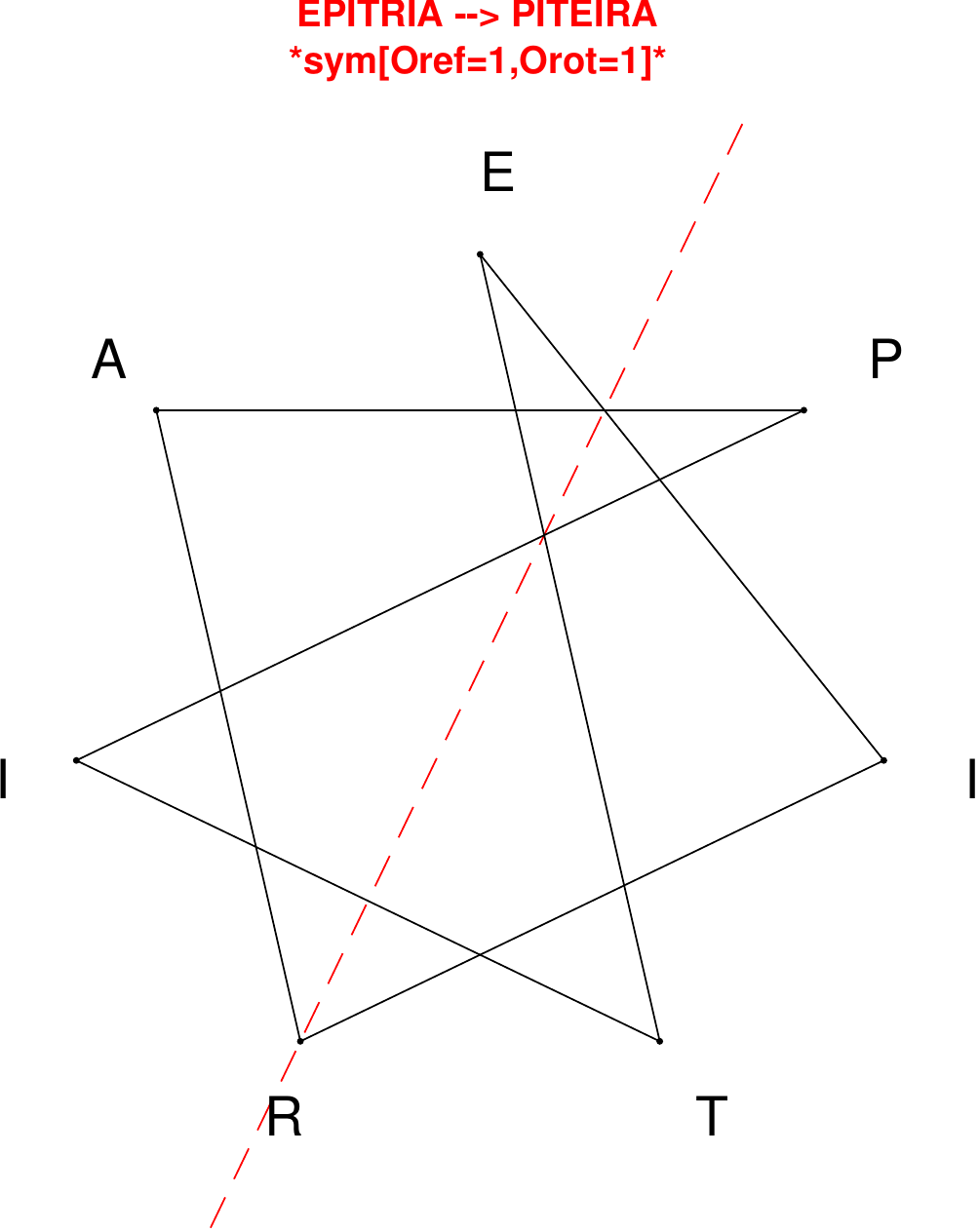}
\end{subfigure}
\end{figure}

\begin{figure}[H]
\centering
\begin{subfigure}[T]{0.19\textwidth}
\centering
\includegraphics[width=\textwidth]{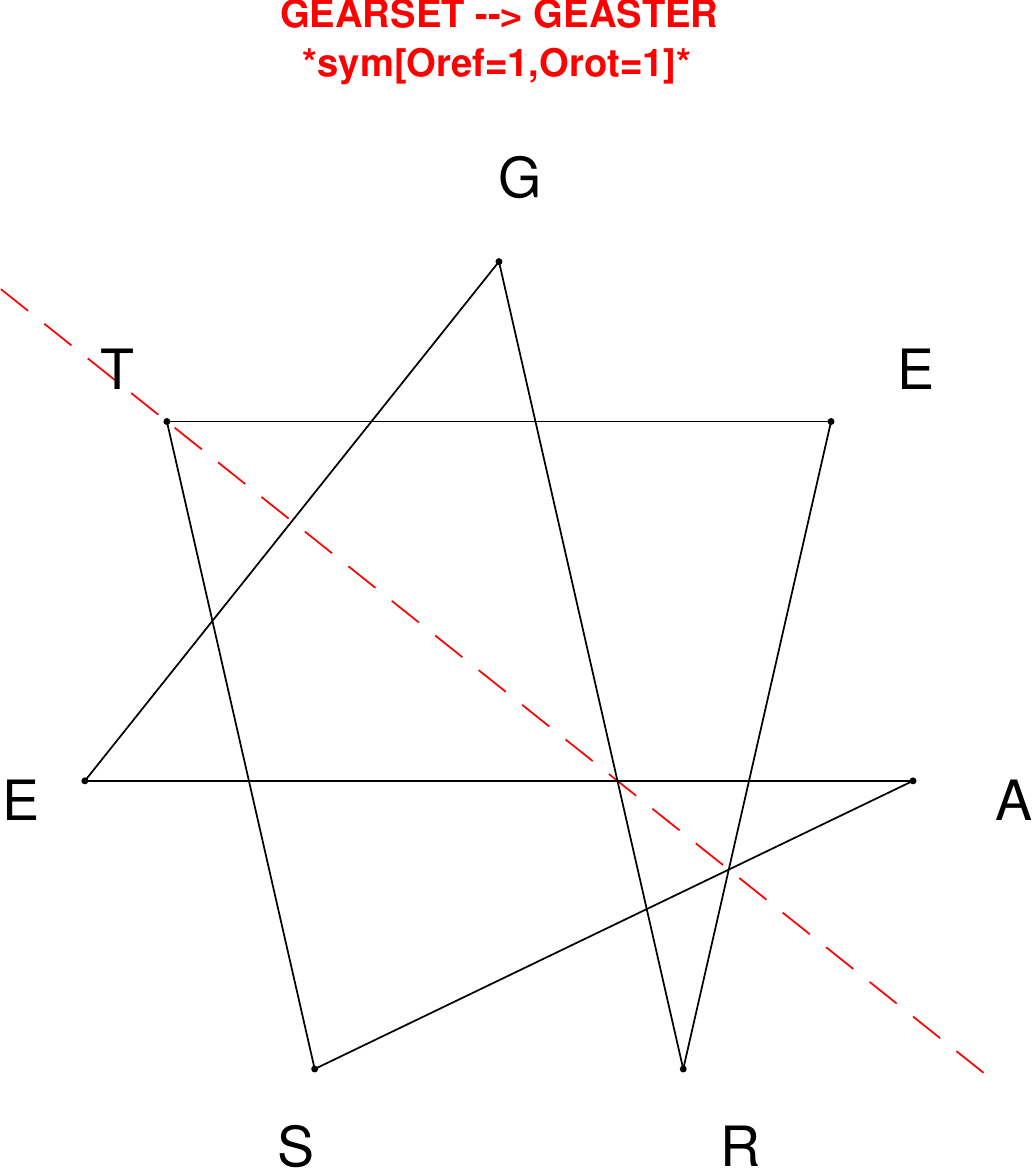}
\end{subfigure}
\hfill
\begin{subfigure}[T]{0.19\textwidth}
\centering
\includegraphics[width=\textwidth]{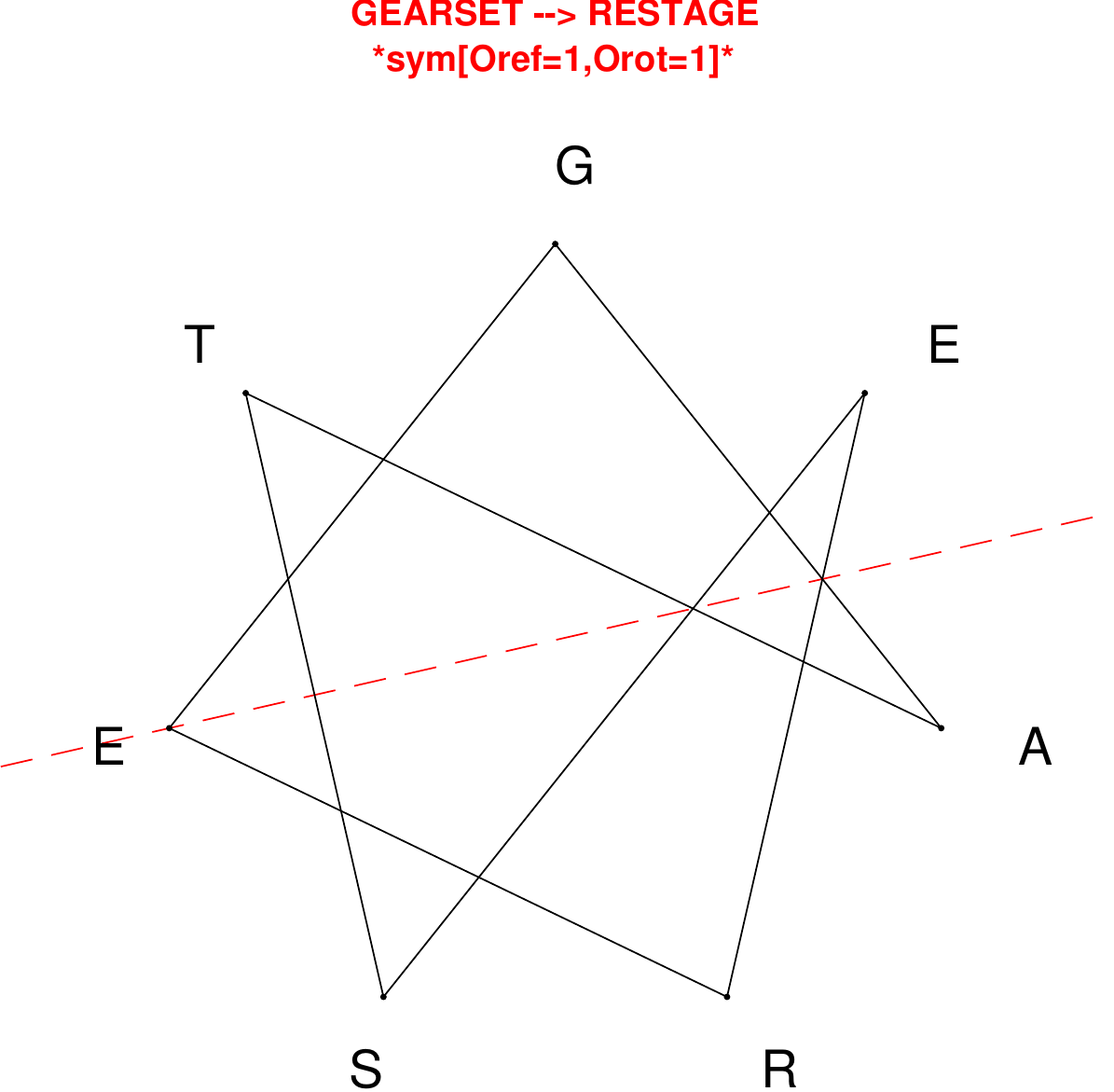}
\end{subfigure}
\hfill
\begin{subfigure}[T]{0.19\textwidth}
\centering
\includegraphics[width=\textwidth]{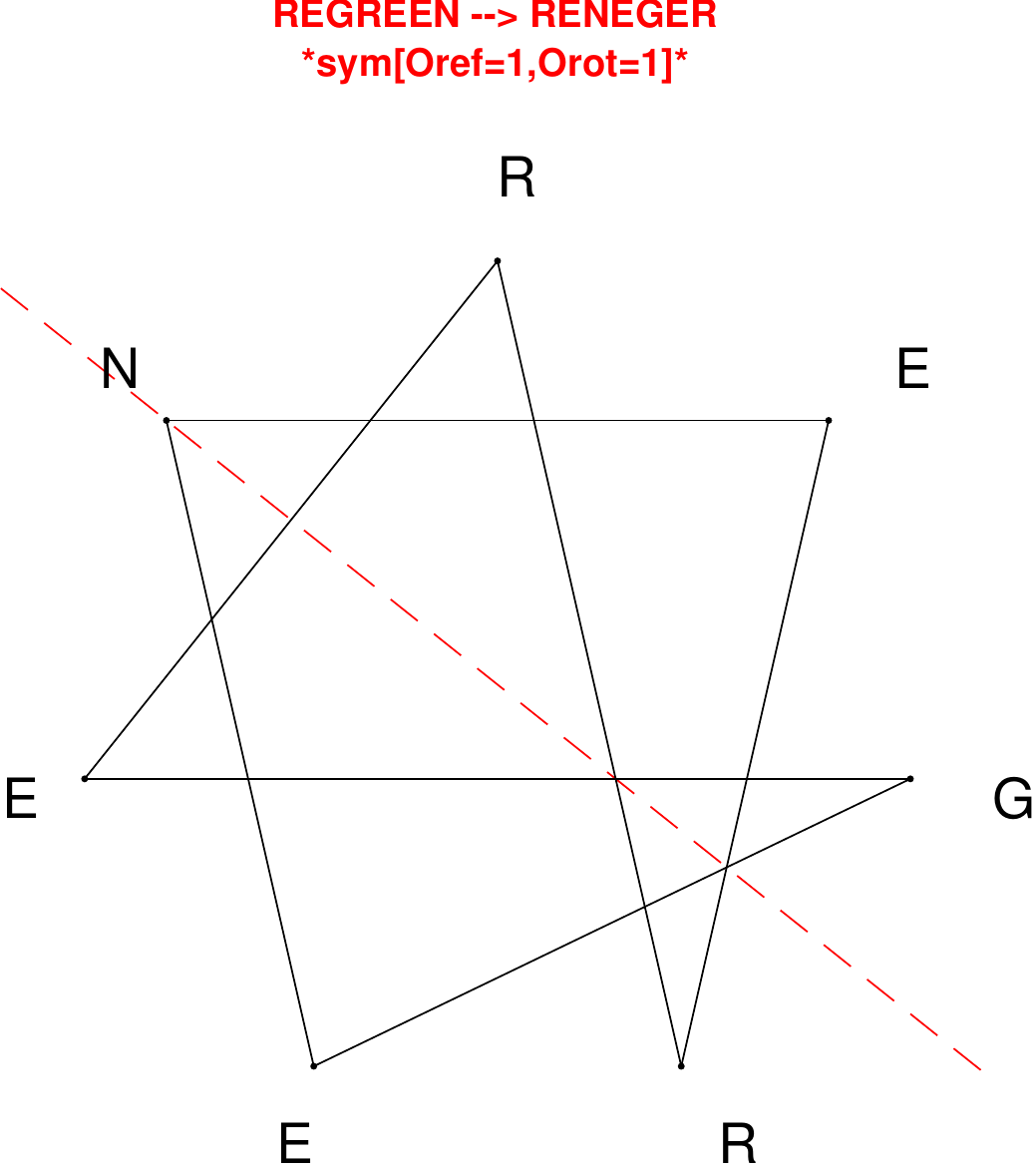}
\end{subfigure}
\hfill
\begin{subfigure}[T]{0.19\textwidth}
\centering
\includegraphics[width=\textwidth]{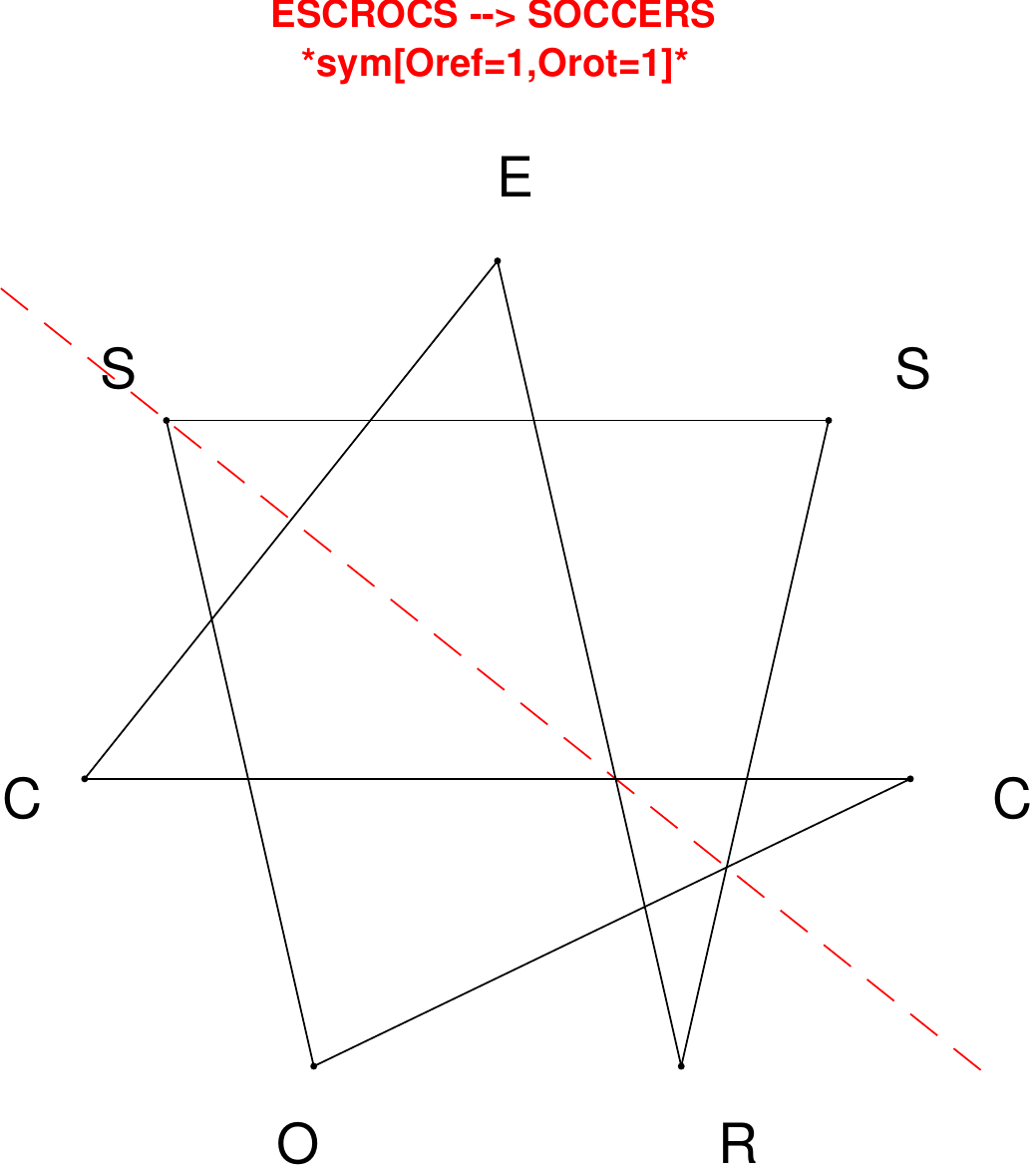}
\end{subfigure}
\hfill
\begin{subfigure}[T]{0.19\textwidth}
\centering
\includegraphics[width=\textwidth]{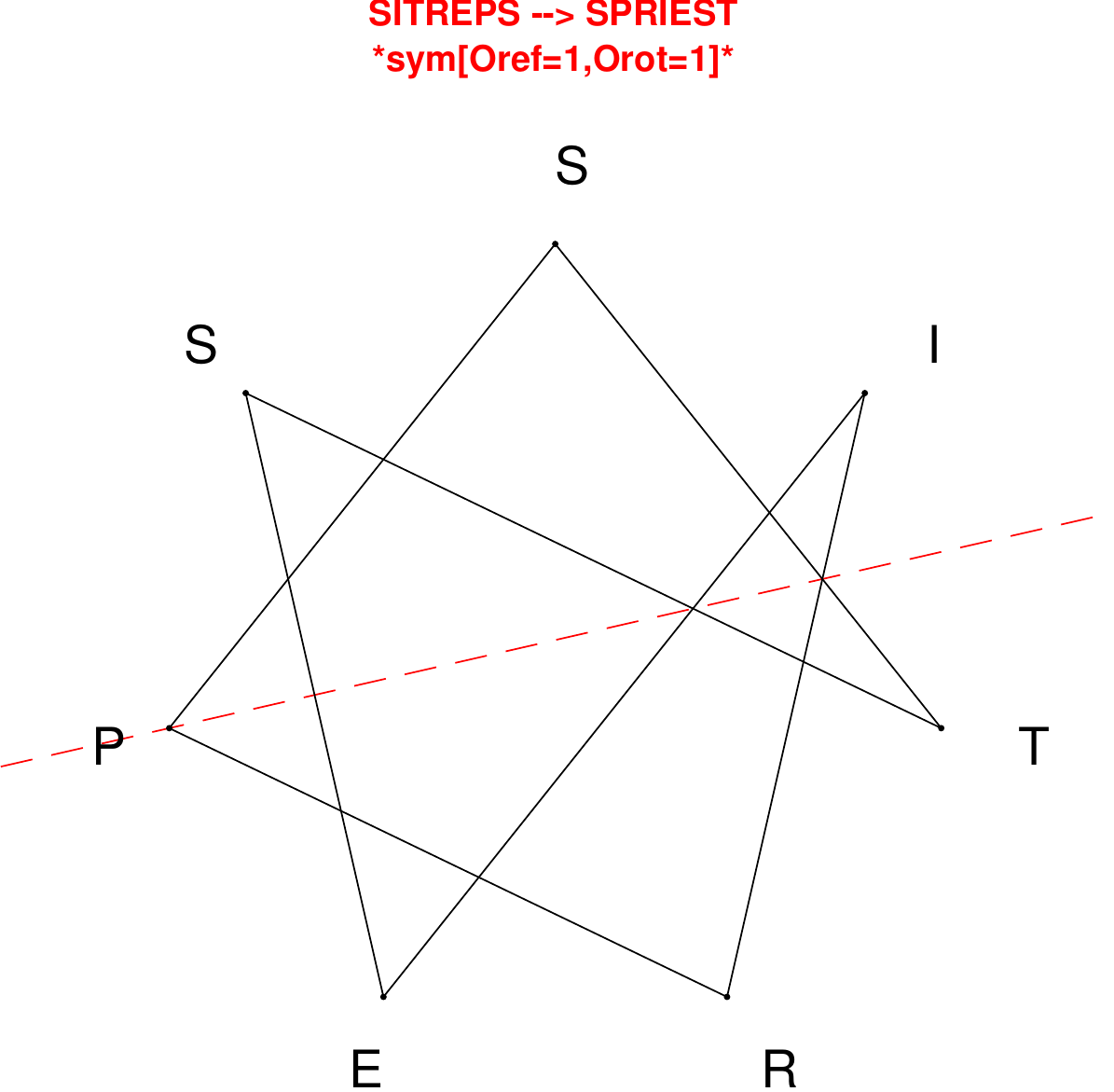}
\end{subfigure}
\end{figure}

\begin{figure}[H]
\centering
\begin{subfigure}[T]{0.19\textwidth}
\centering
\includegraphics[width=\textwidth]{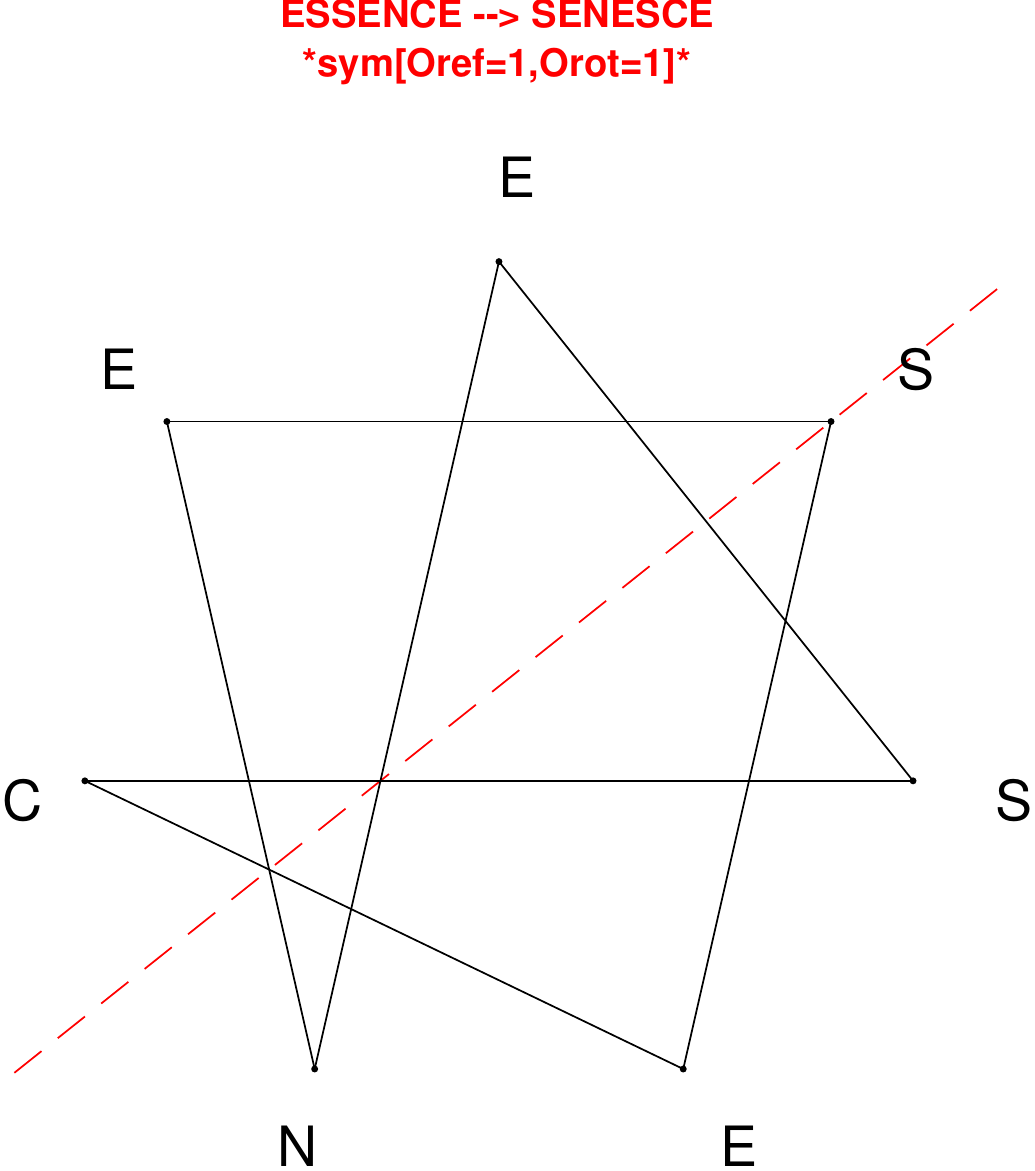}
\end{subfigure}
\hfill
\begin{subfigure}[T]{0.19\textwidth}
\centering
\includegraphics[width=\textwidth]{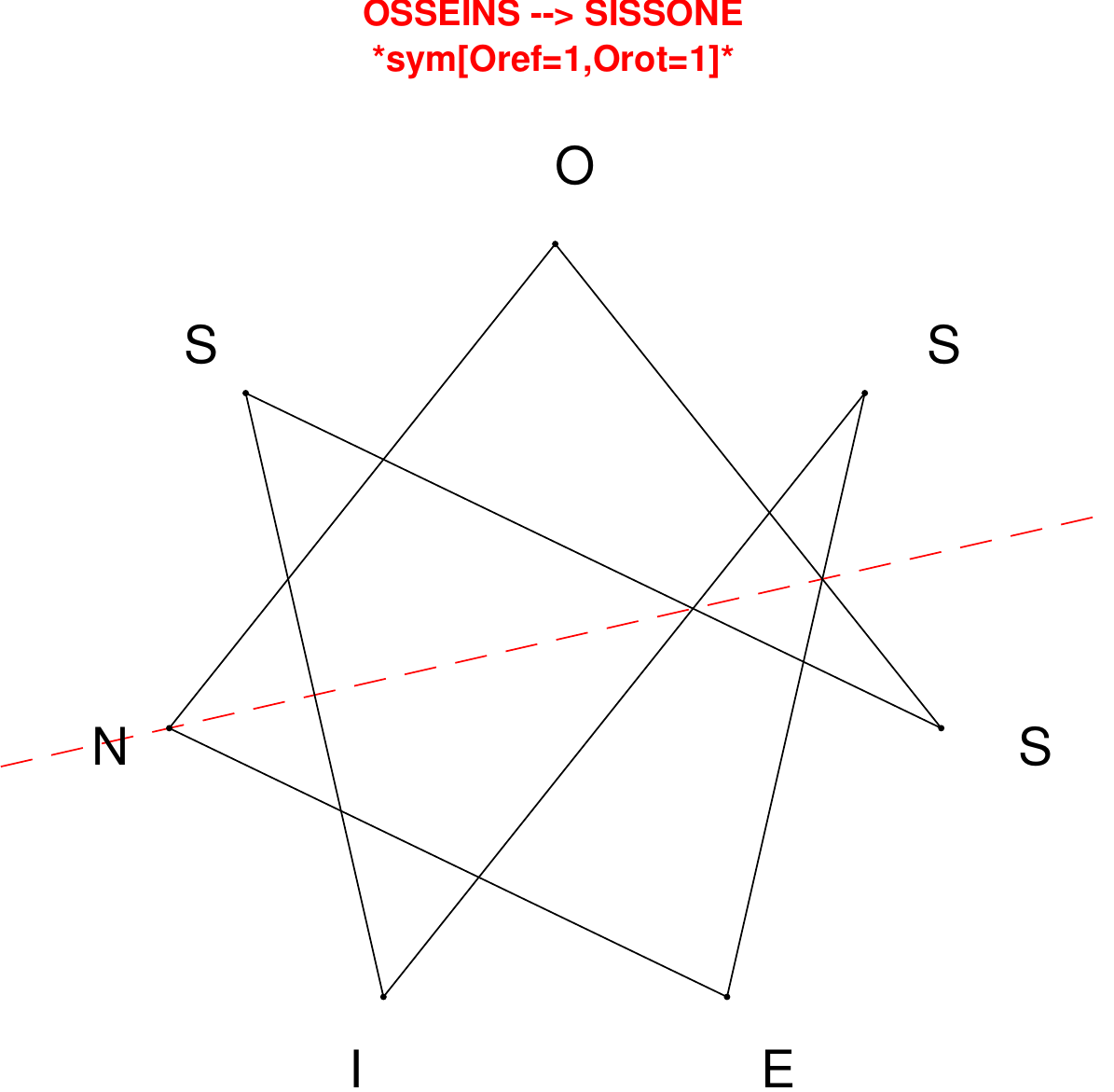}
\end{subfigure}
\hfill
\begin{subfigure}[T]{0.19\textwidth}
\centering
\includegraphics[width=\textwidth]{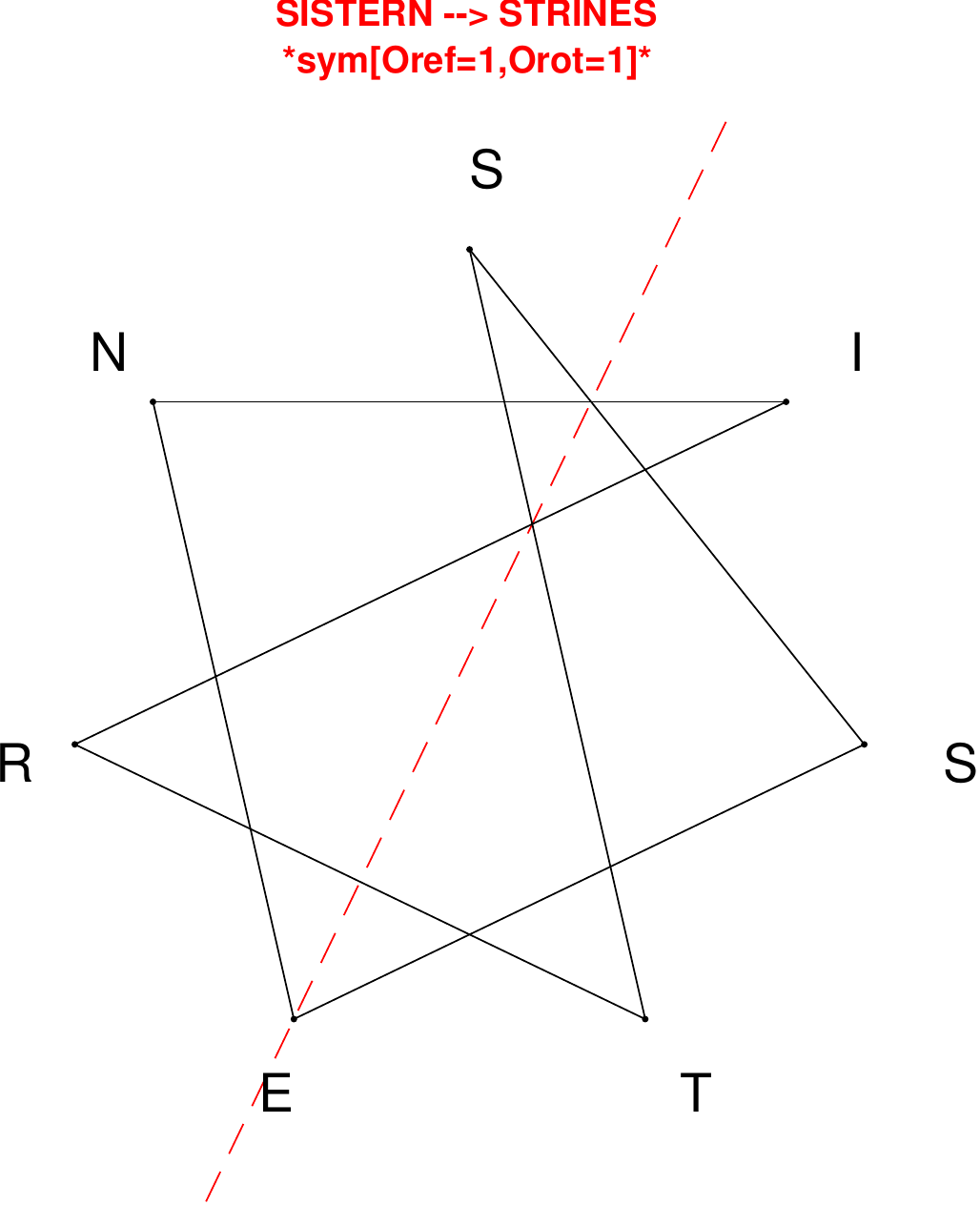}
\end{subfigure}
\hfill
\begin{subfigure}[T]{0.19\textwidth}
\centering
\includegraphics[width=\textwidth]{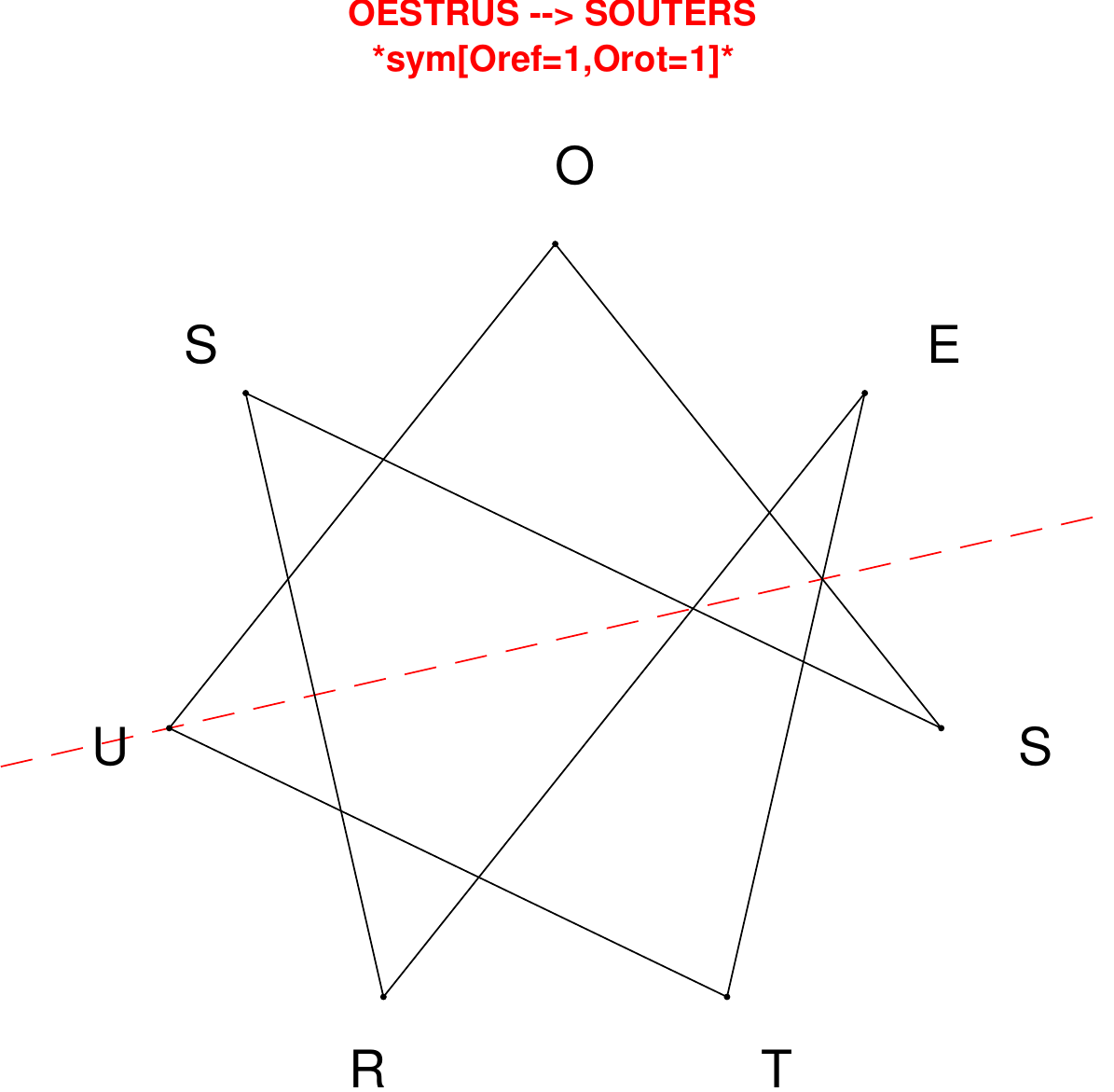}
\end{subfigure}
\hfill
\begin{subfigure}[T]{0.19\textwidth}
\centering
\includegraphics[width=\textwidth]{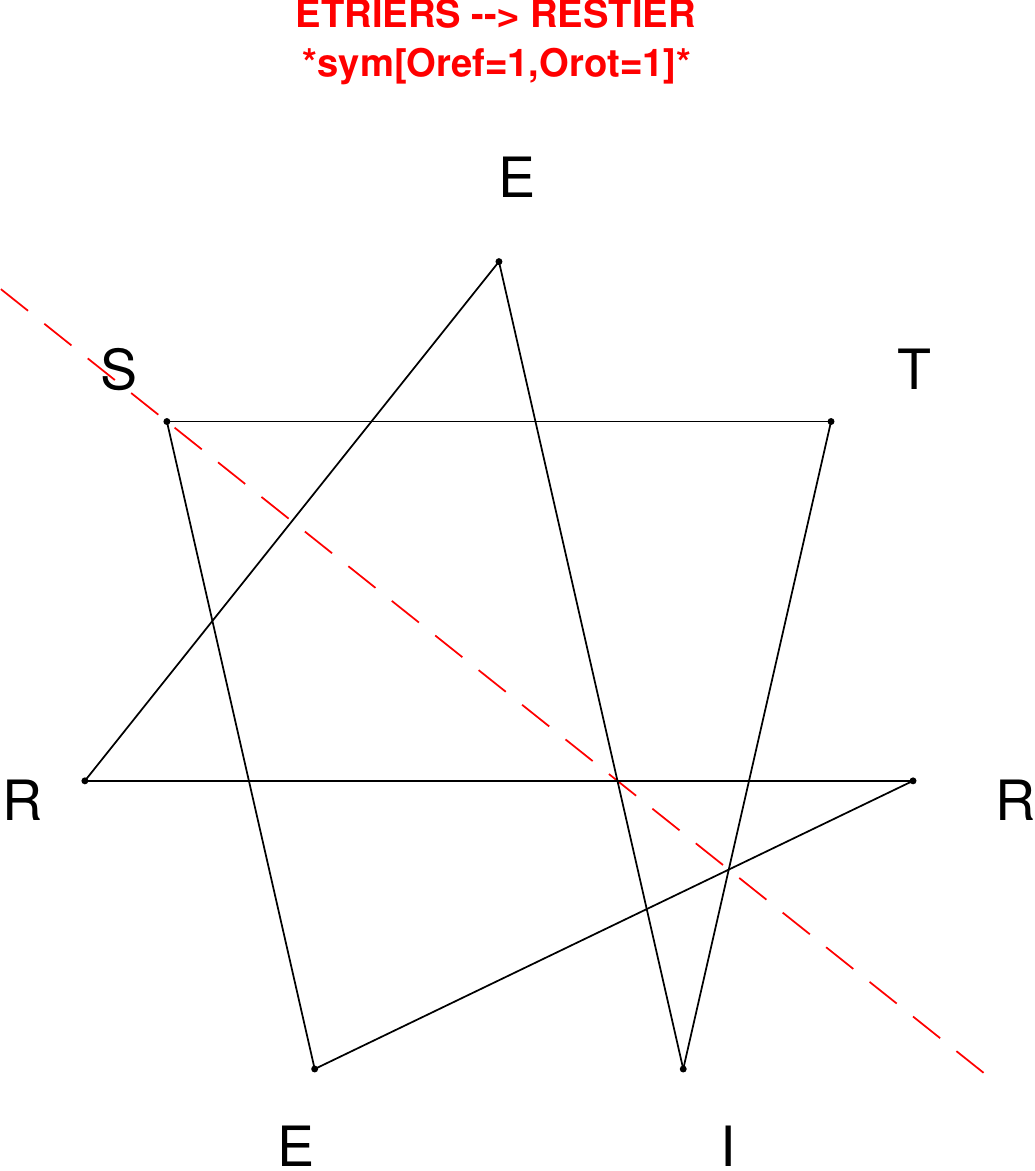}
\end{subfigure}
\end{figure}

\begin{figure}[H]
\centering
\begin{subfigure}[T]{0.19\textwidth}
\centering
\includegraphics[width=\textwidth]{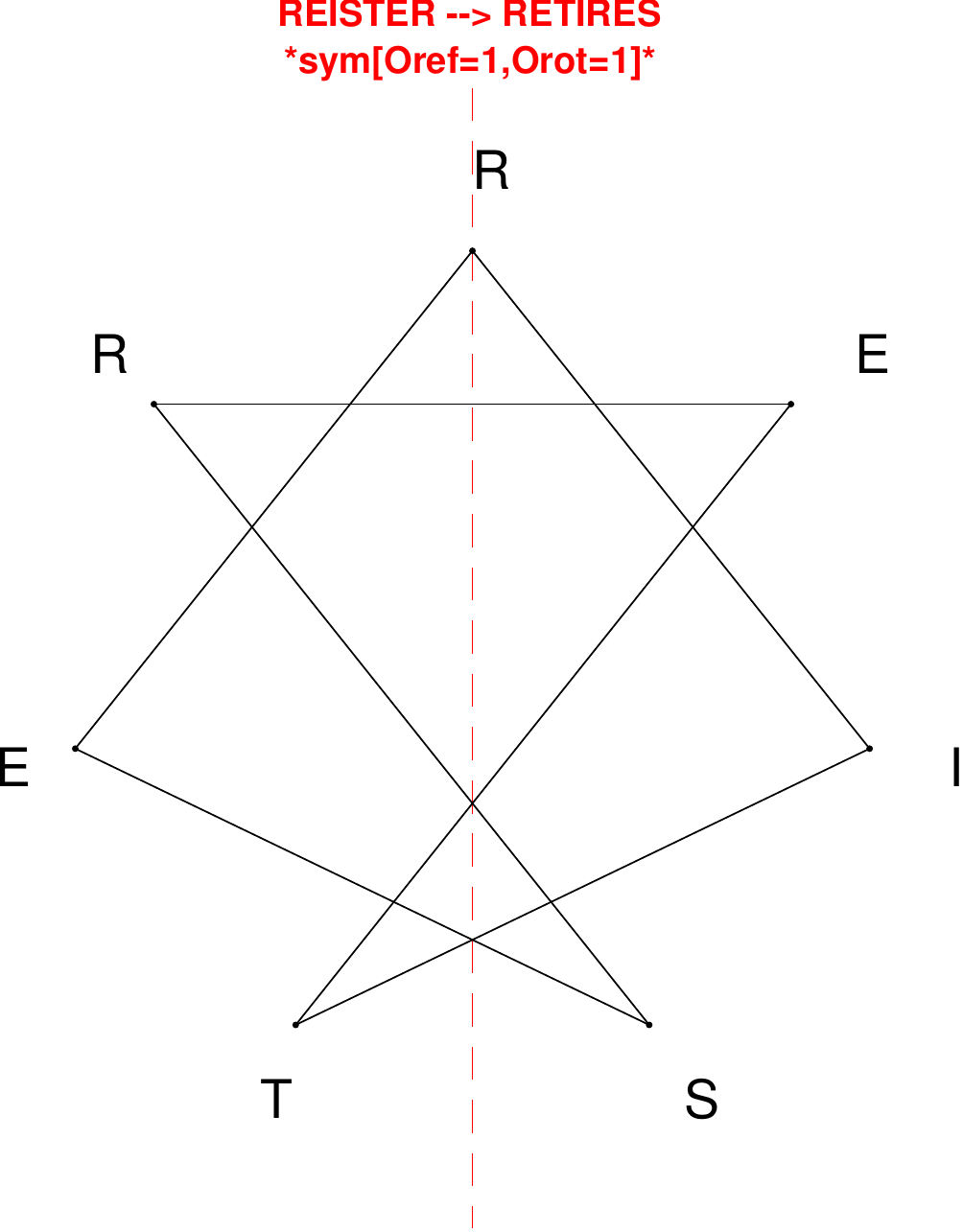}
\end{subfigure}
\hfill
\begin{subfigure}[T]{0.19\textwidth}
\centering
\includegraphics[width=\textwidth]{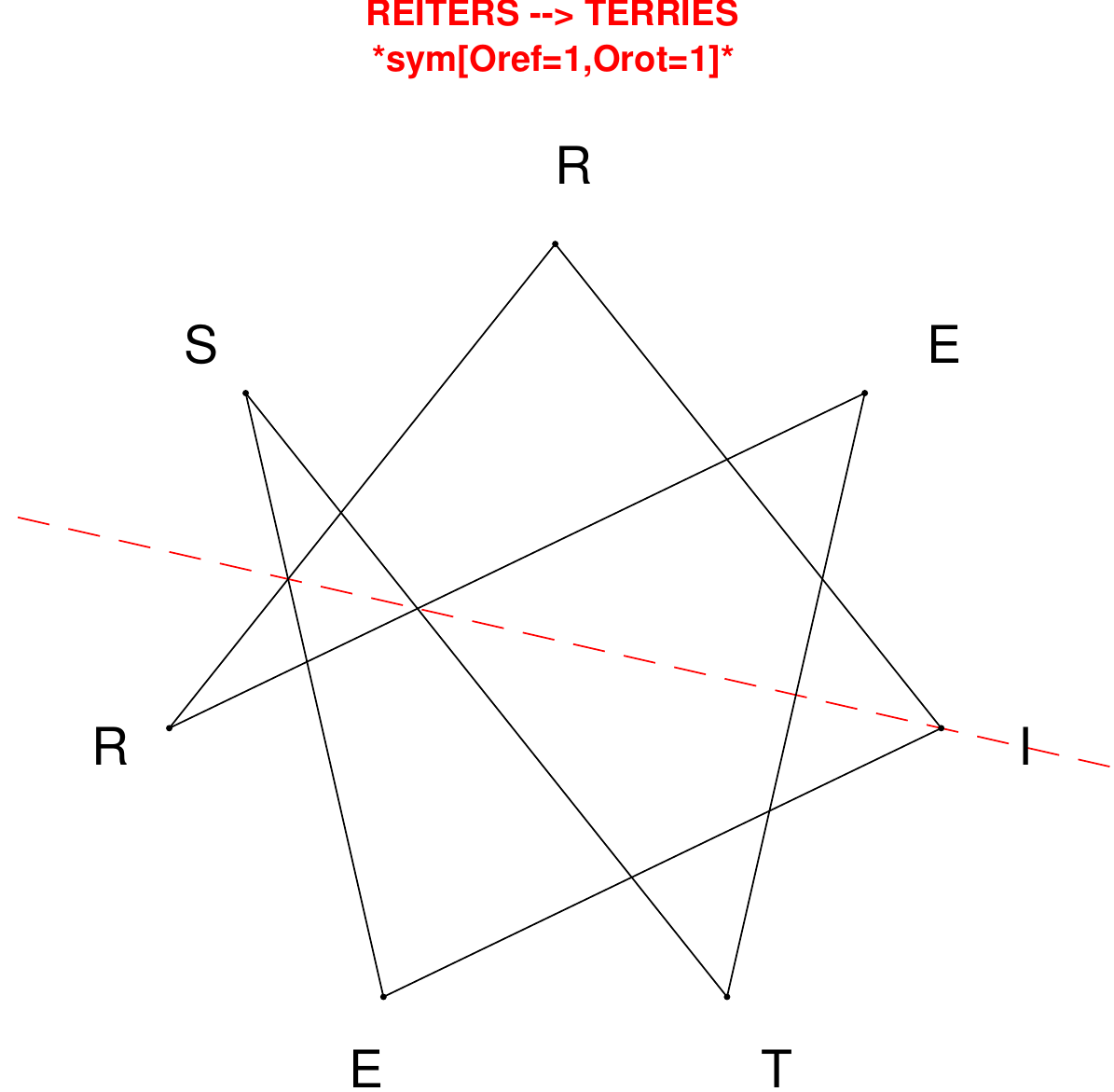}
\end{subfigure}
\hfill
\begin{subfigure}[T]{0.19\textwidth}
\centering
\includegraphics[width=\textwidth]{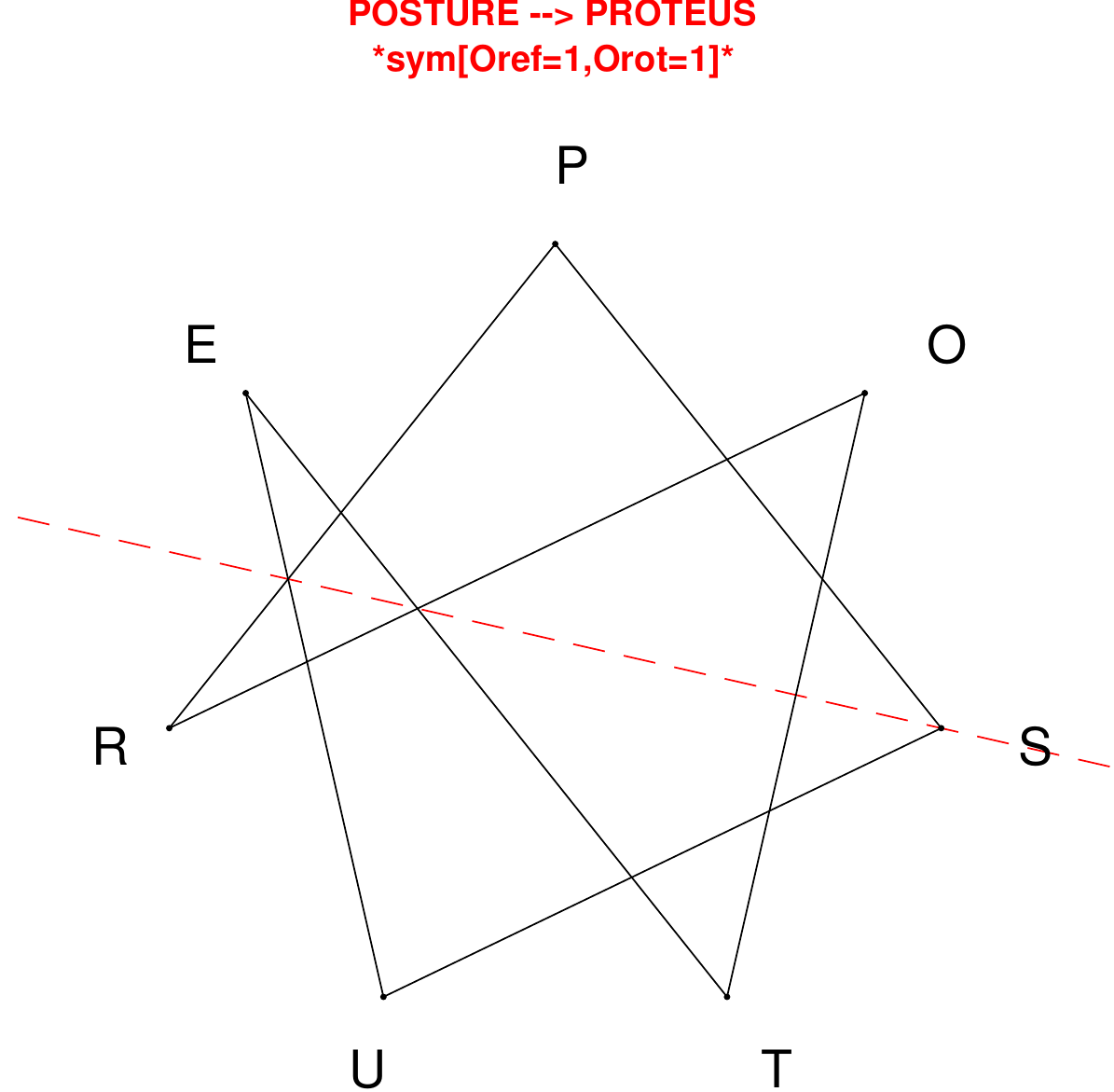}
\end{subfigure}
\hfill
\begin{subfigure}[T]{0.19\textwidth}
\centering
\includegraphics[width=\textwidth]{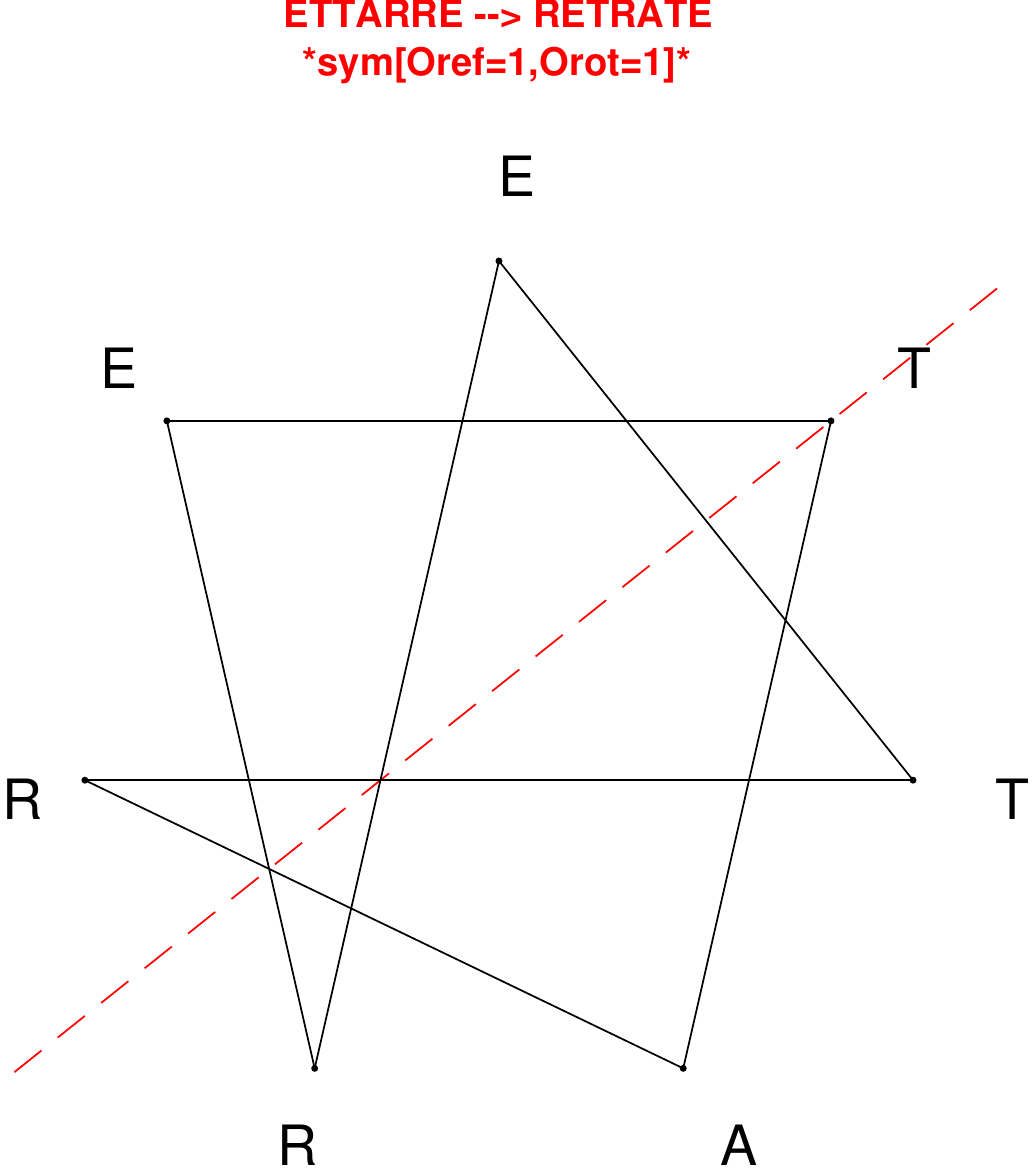}
\end{subfigure}
\hfill
\begin{subfigure}[T]{0.19\textwidth}
\centering
\includegraphics[width=\textwidth]{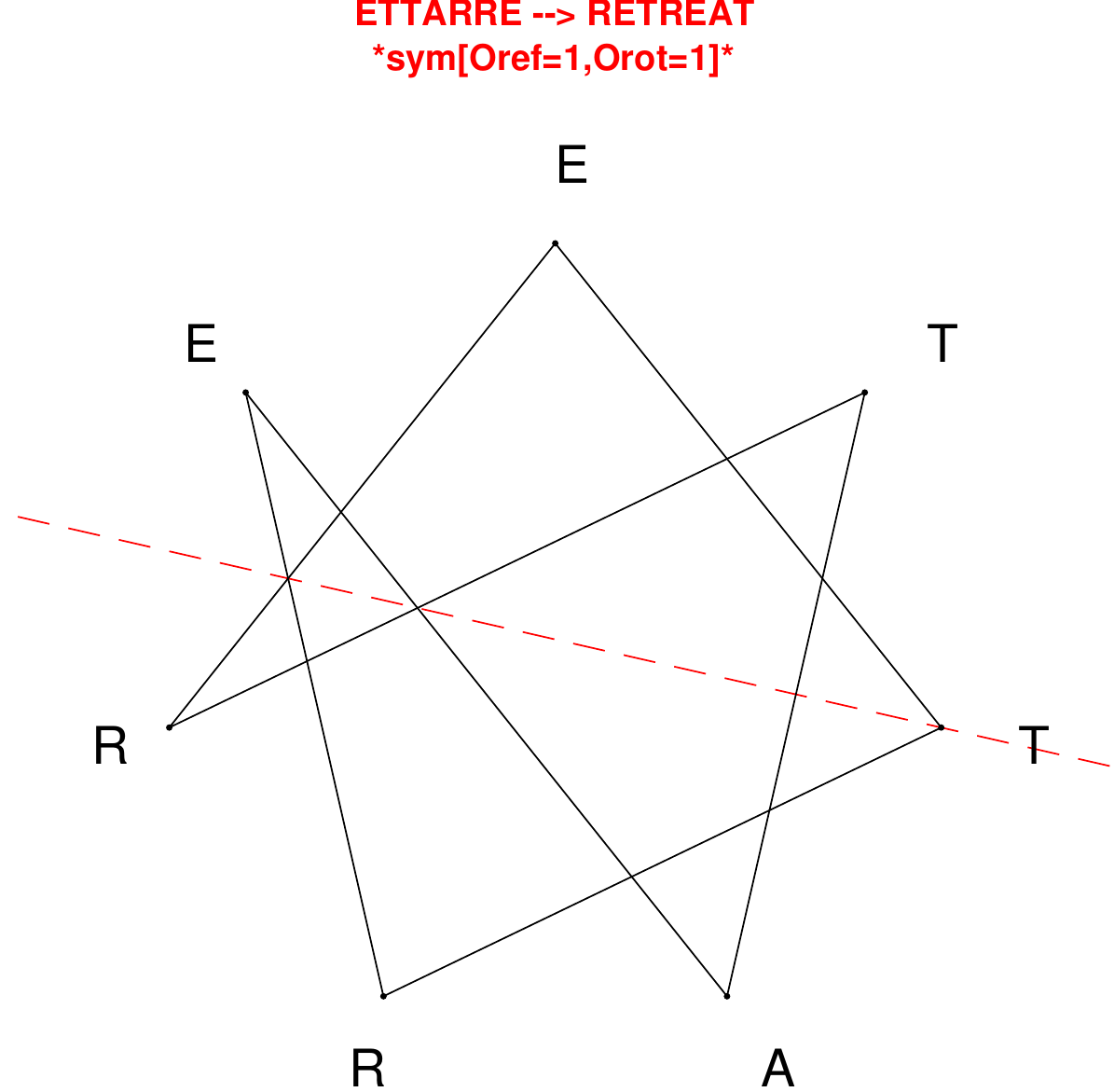}
\end{subfigure}
\end{figure}

\begin{figure}[H]
\centering
\begin{subfigure}[T]{0.19\textwidth}
\centering
\includegraphics[width=\textwidth]{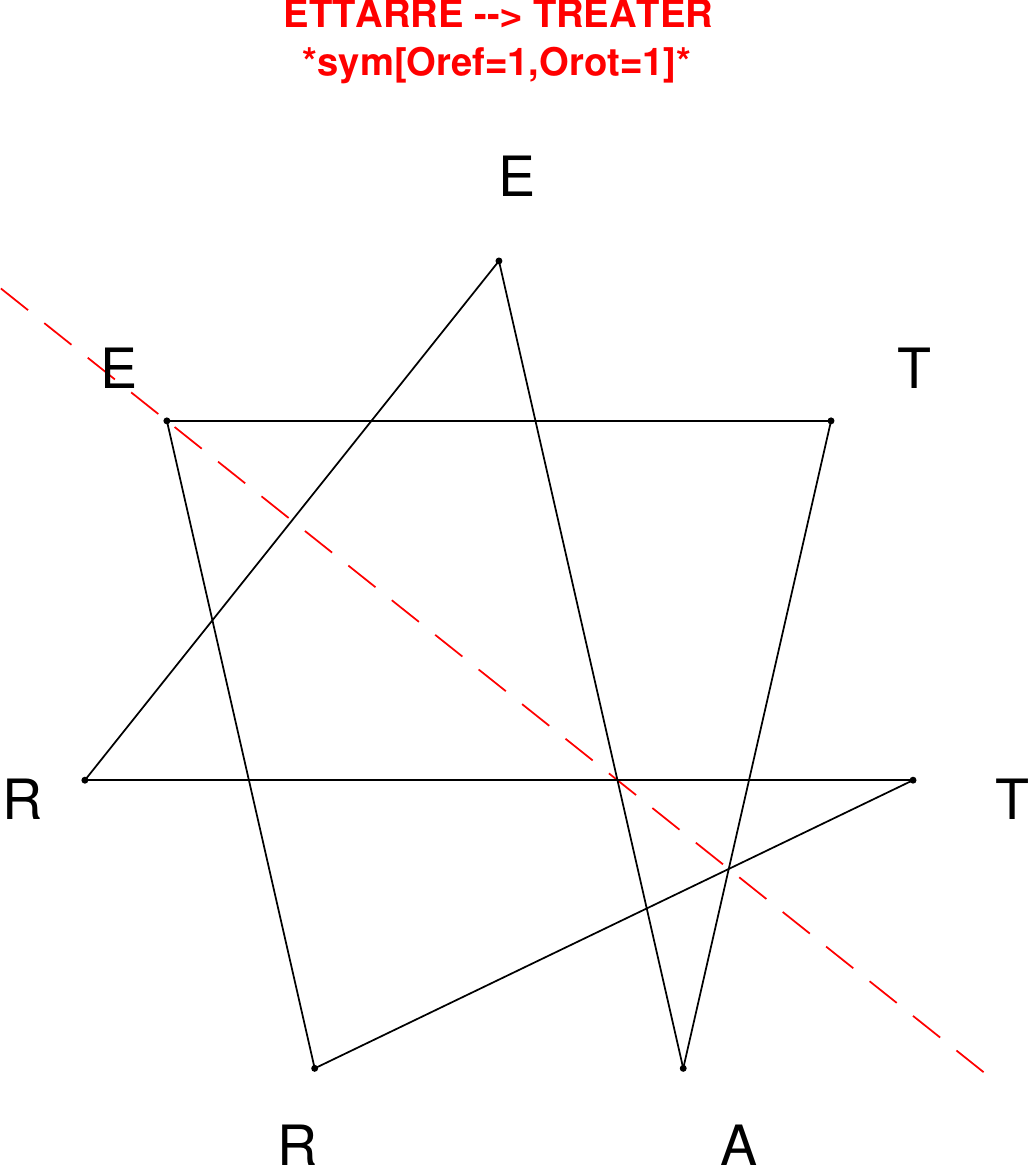}
\end{subfigure}
\hfill
\begin{subfigure}[T]{0.19\textwidth}
\centering
\includegraphics[width=\textwidth]{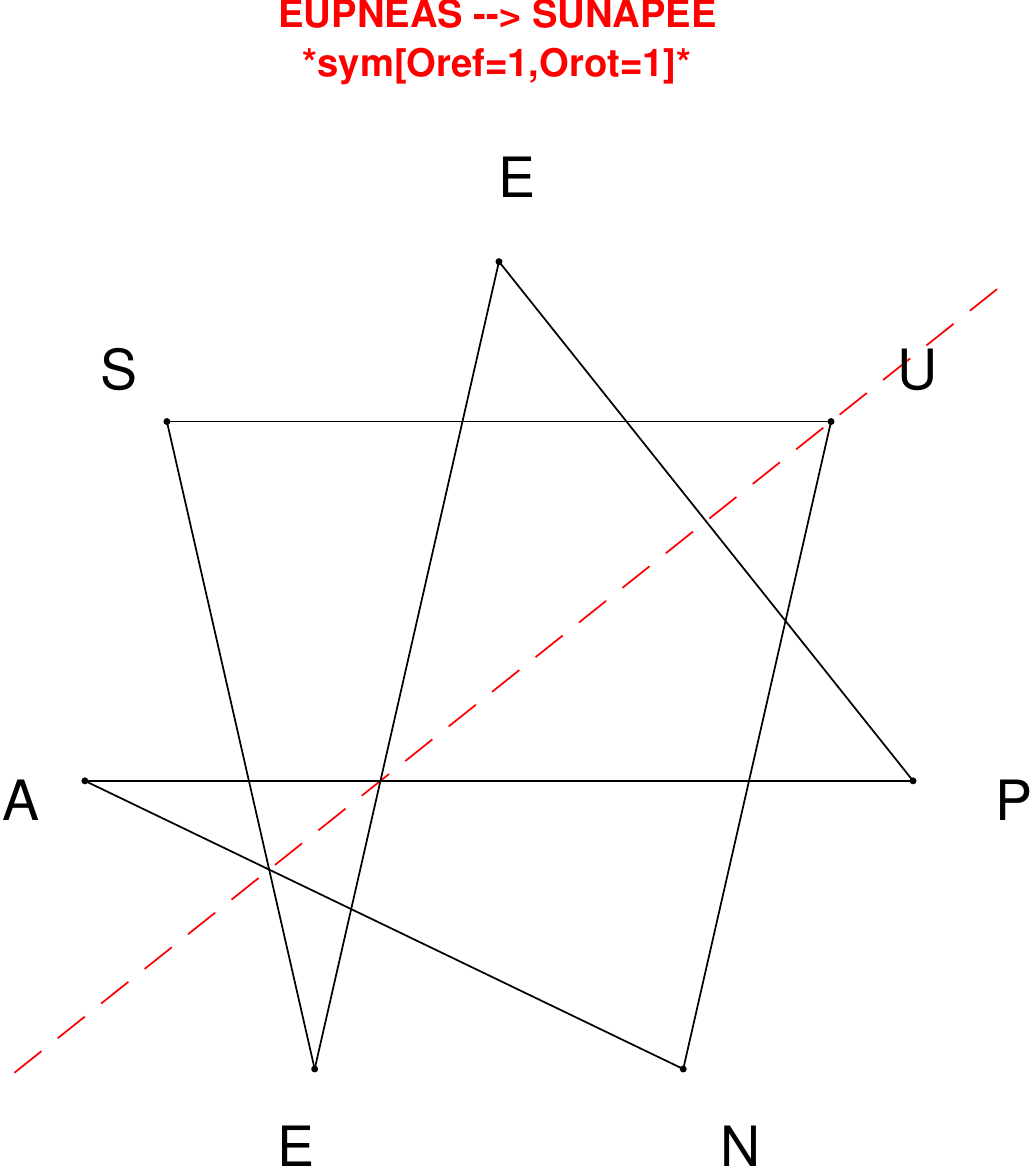}
\end{subfigure}
\hfill
\begin{subfigure}[T]{0.19\textwidth}
\centering
\includegraphics[width=\textwidth]{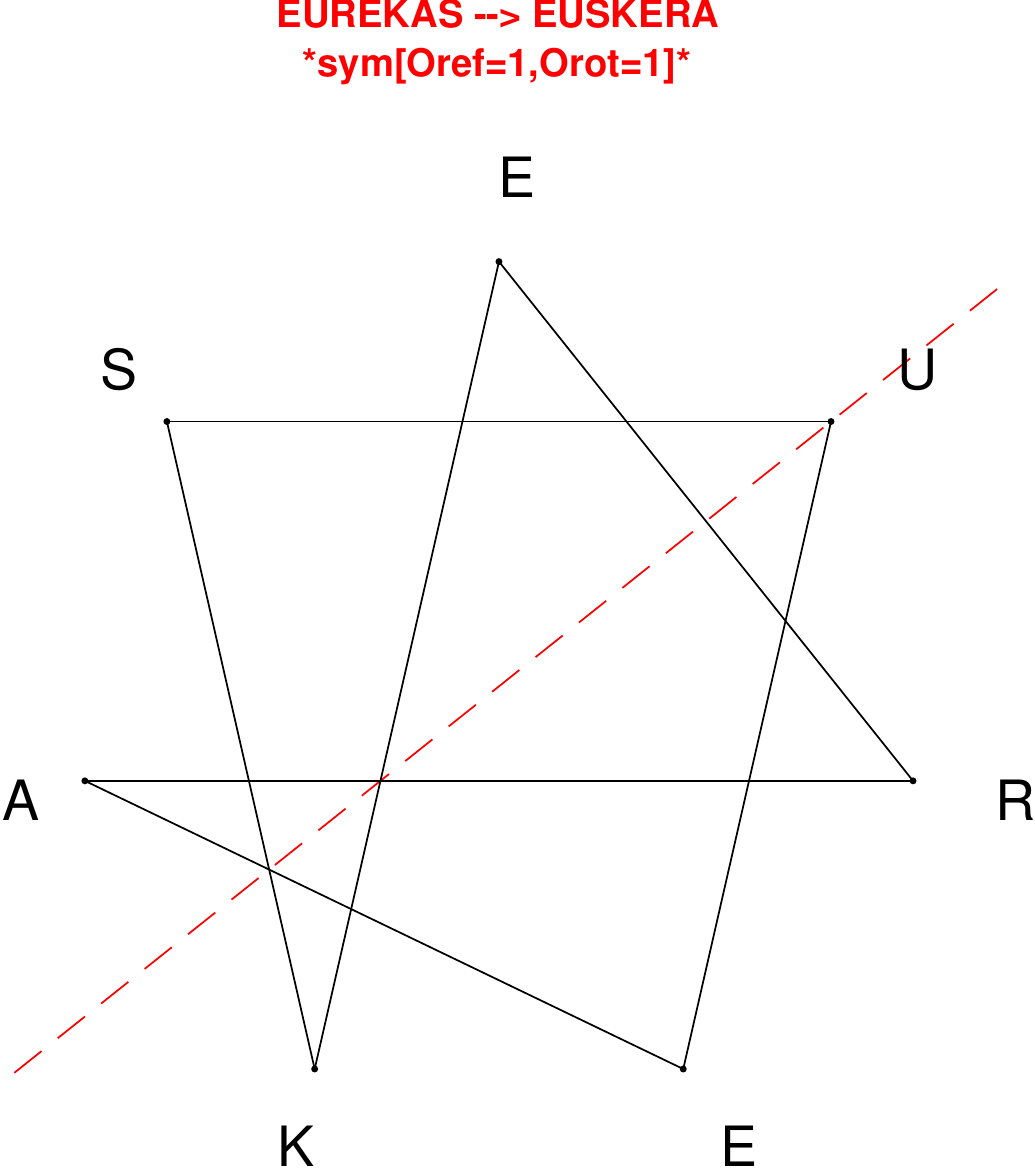}
\end{subfigure}
\hfill
\begin{subfigure}[T]{0.19\textwidth}
\centering
\includegraphics[width=\textwidth]{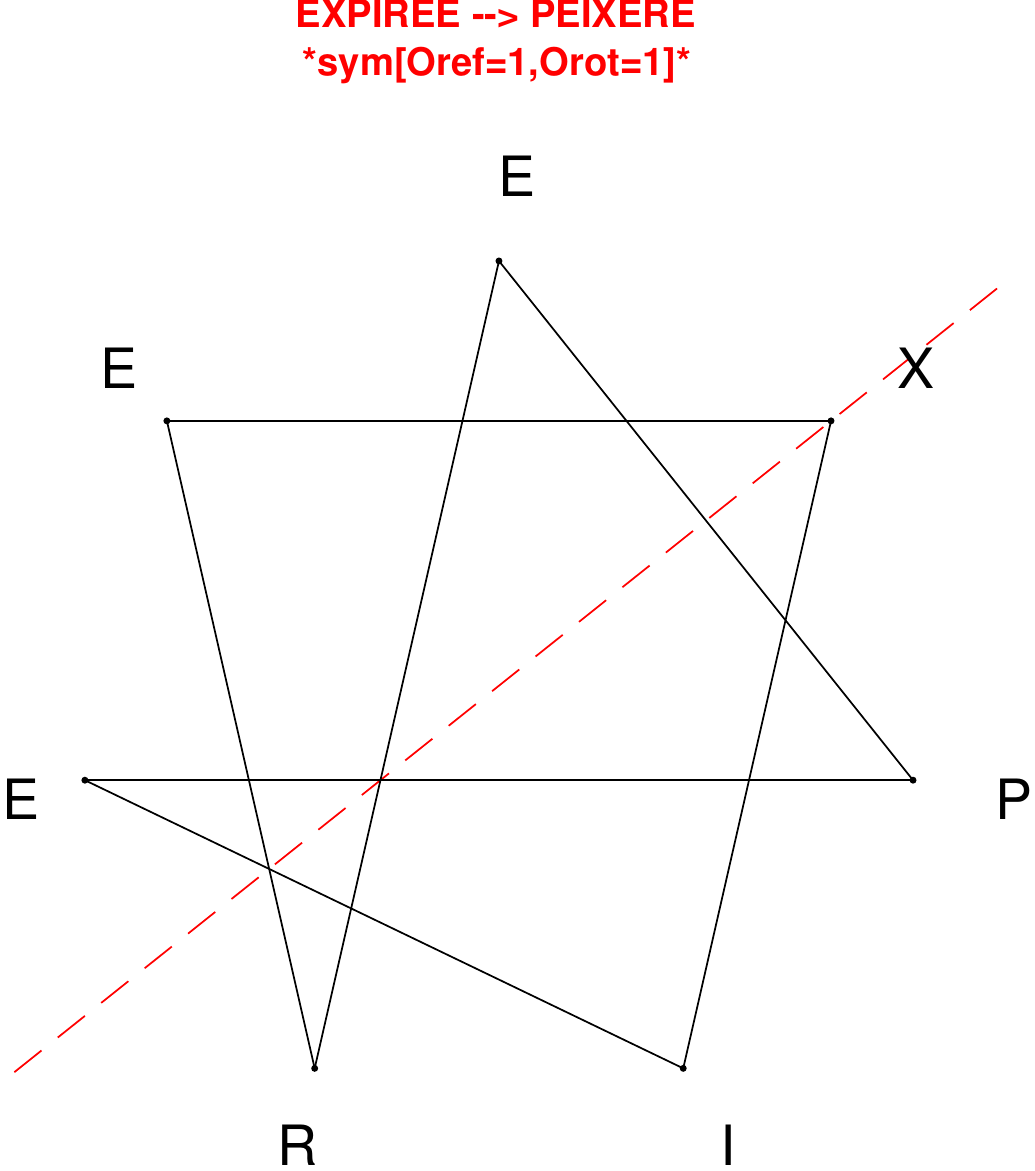}
\end{subfigure}
\hfill
\begin{subfigure}[T]{0.19\textwidth}
\centering
\includegraphics[width=\textwidth]{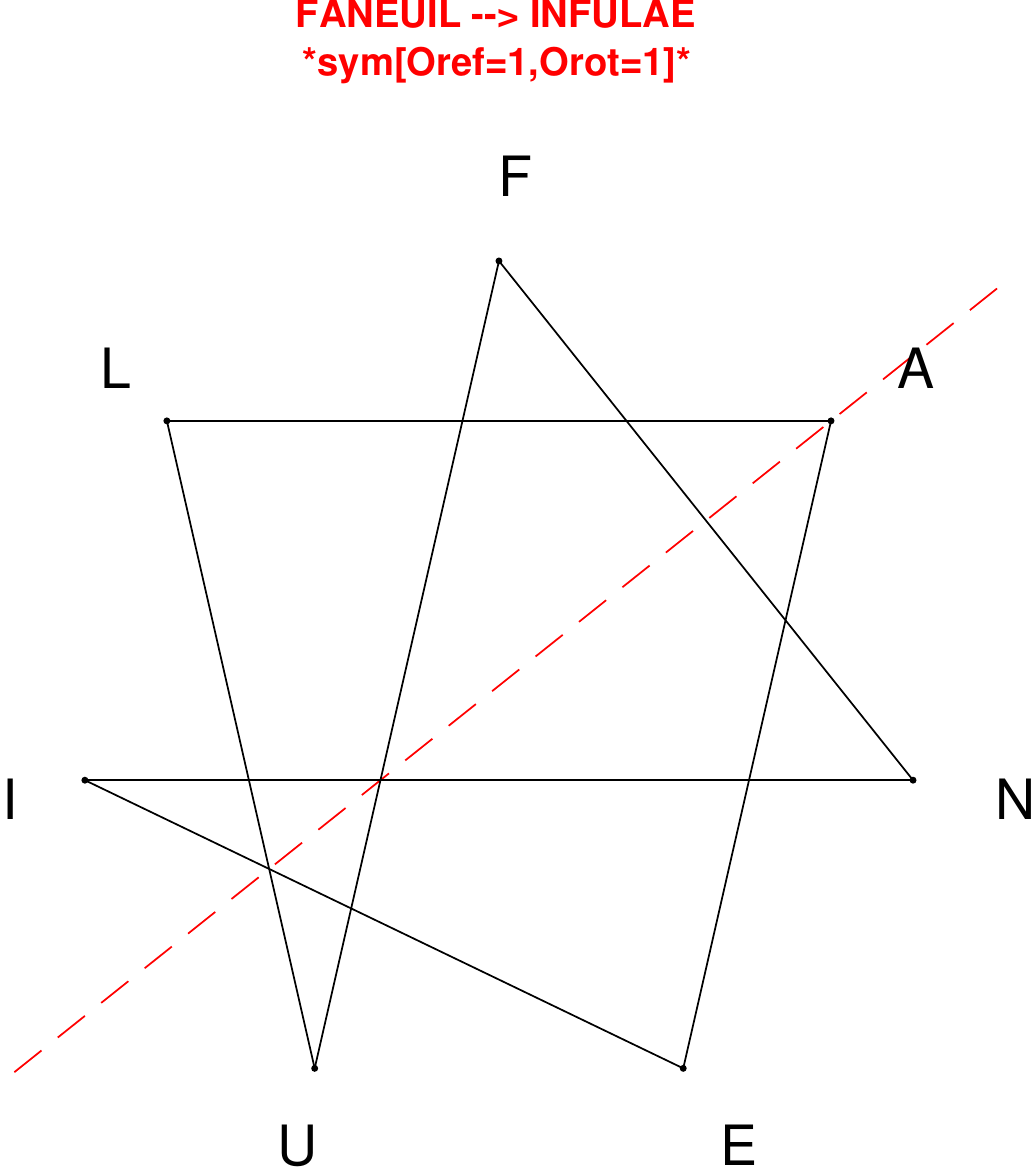}
\end{subfigure}
\end{figure}

\begin{figure}[H]
\centering
\begin{subfigure}[T]{0.19\textwidth}
\centering
\includegraphics[width=\textwidth]{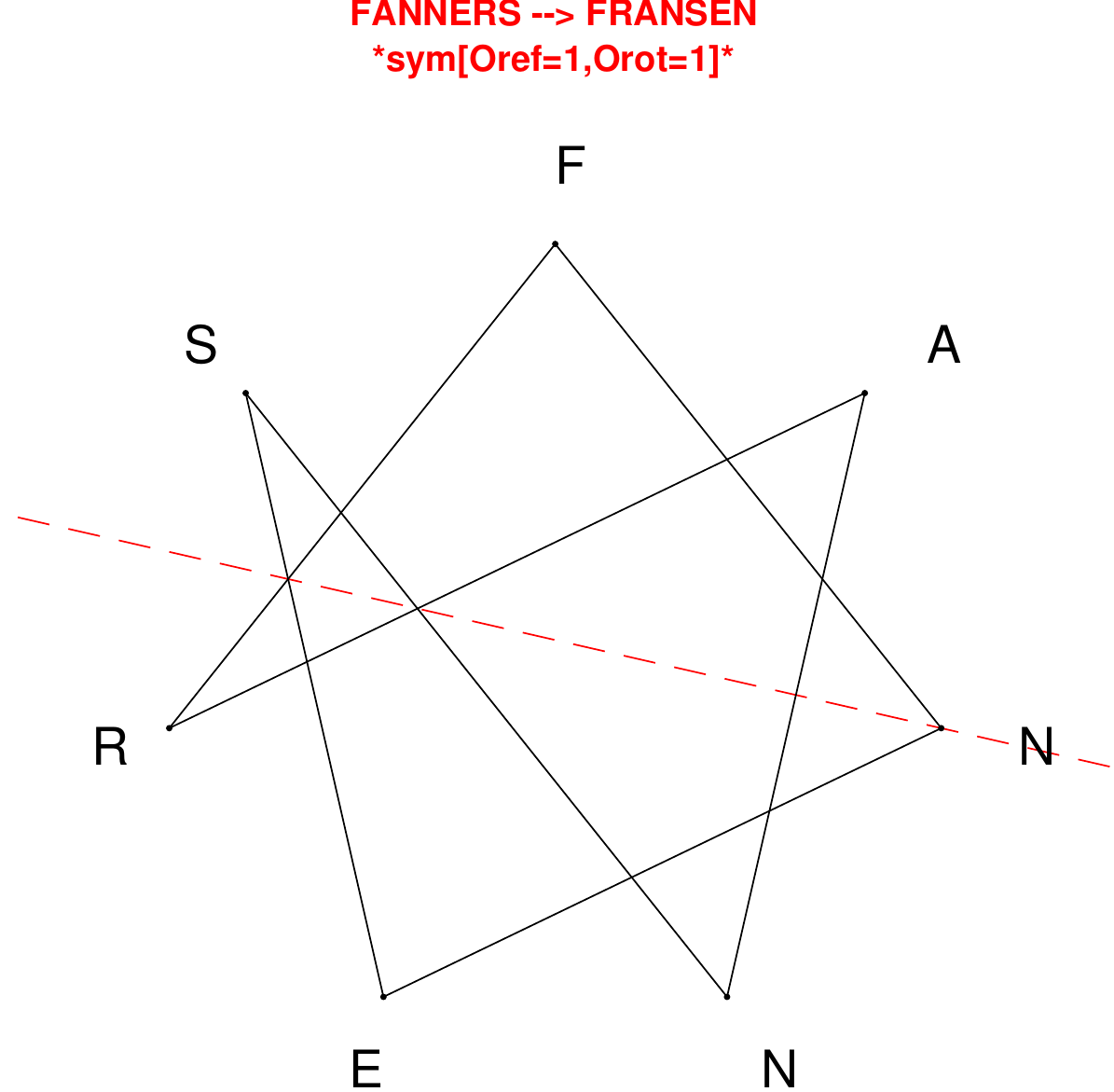}
\end{subfigure}
\hfill
\begin{subfigure}[T]{0.19\textwidth}
\centering
\includegraphics[width=\textwidth]{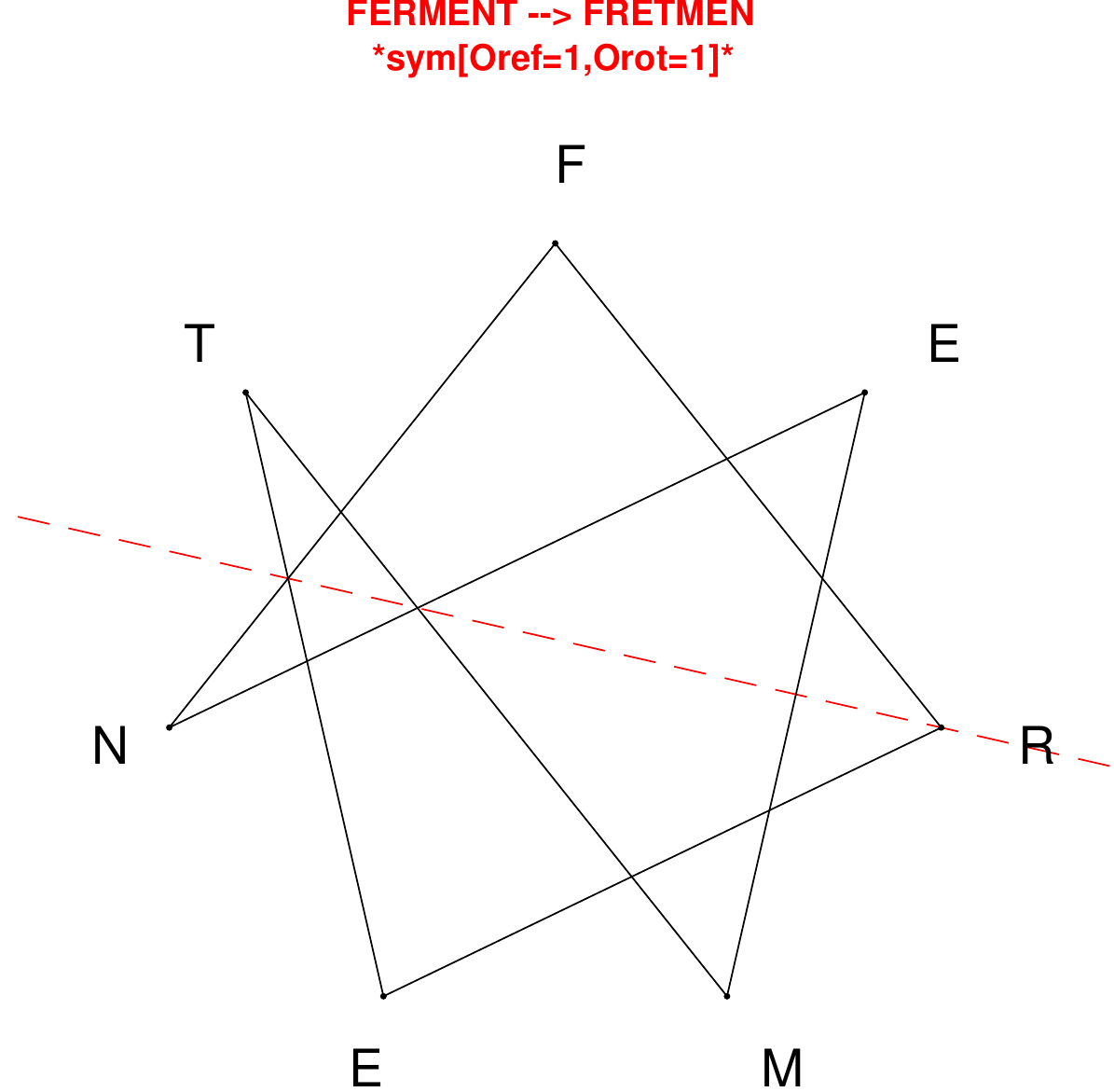}
\end{subfigure}
\hfill
\begin{subfigure}[T]{0.19\textwidth}
\centering
\includegraphics[width=\textwidth]{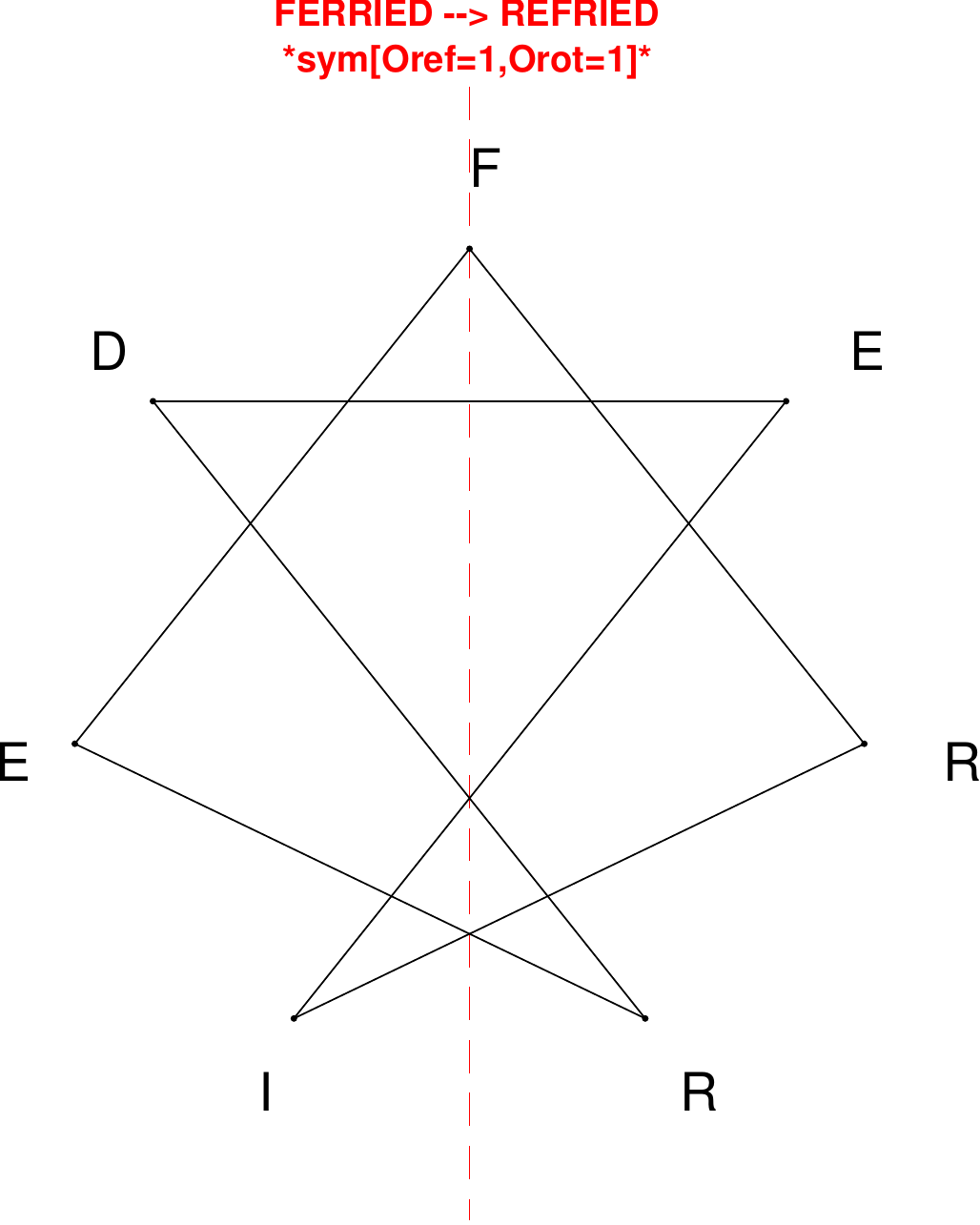}
\end{subfigure}
\hfill
\begin{subfigure}[T]{0.19\textwidth}
\centering
\includegraphics[width=\textwidth]{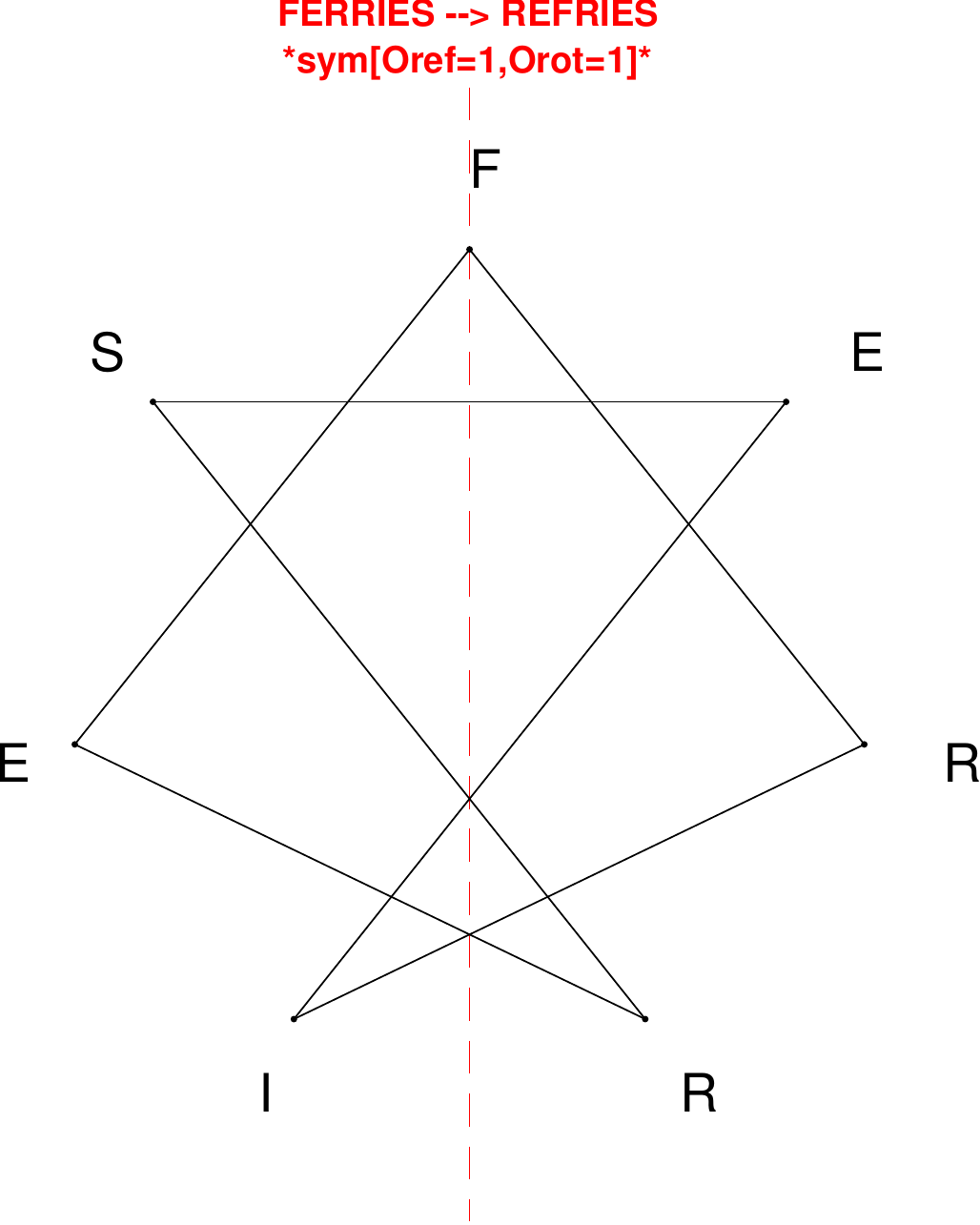}
\end{subfigure}
\hfill
\begin{subfigure}[T]{0.19\textwidth}
\centering
\includegraphics[width=\textwidth]{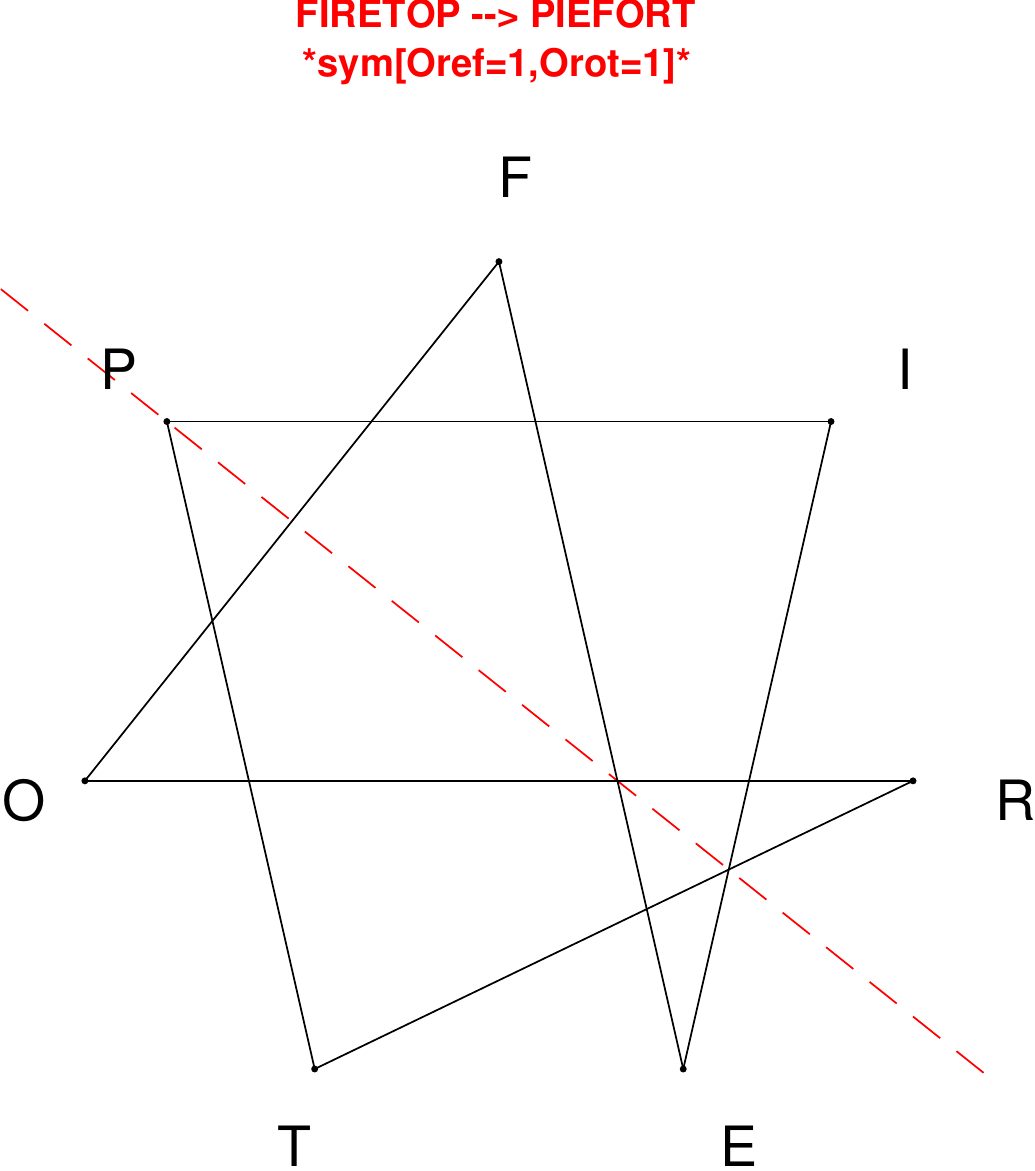}
\end{subfigure}
\end{figure}

\begin{figure}[H]
\centering
\begin{subfigure}[T]{0.19\textwidth}
\centering
\includegraphics[width=\textwidth]{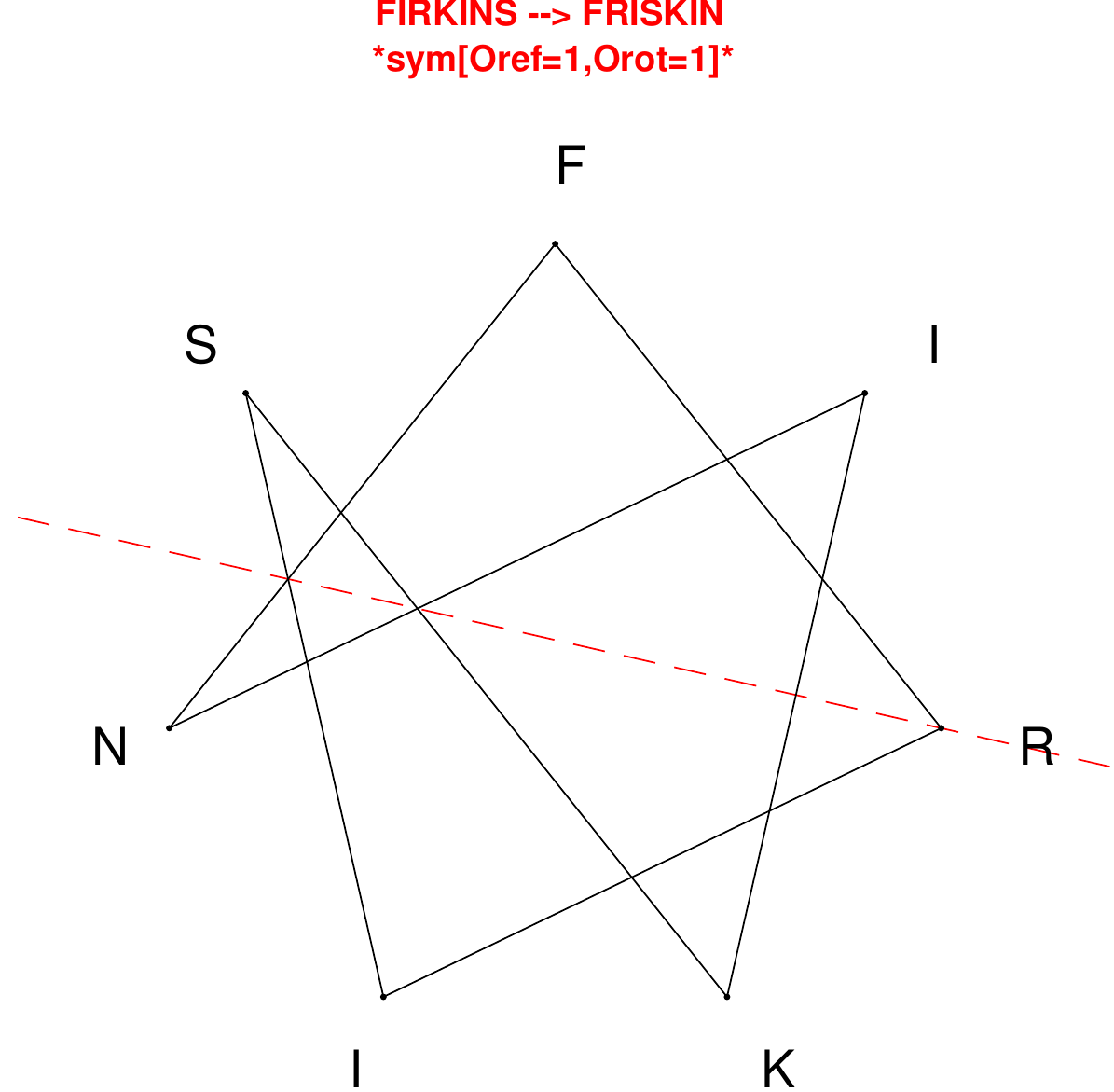}
\end{subfigure}
\hfill
\begin{subfigure}[T]{0.19\textwidth}
\centering
\includegraphics[width=\textwidth]{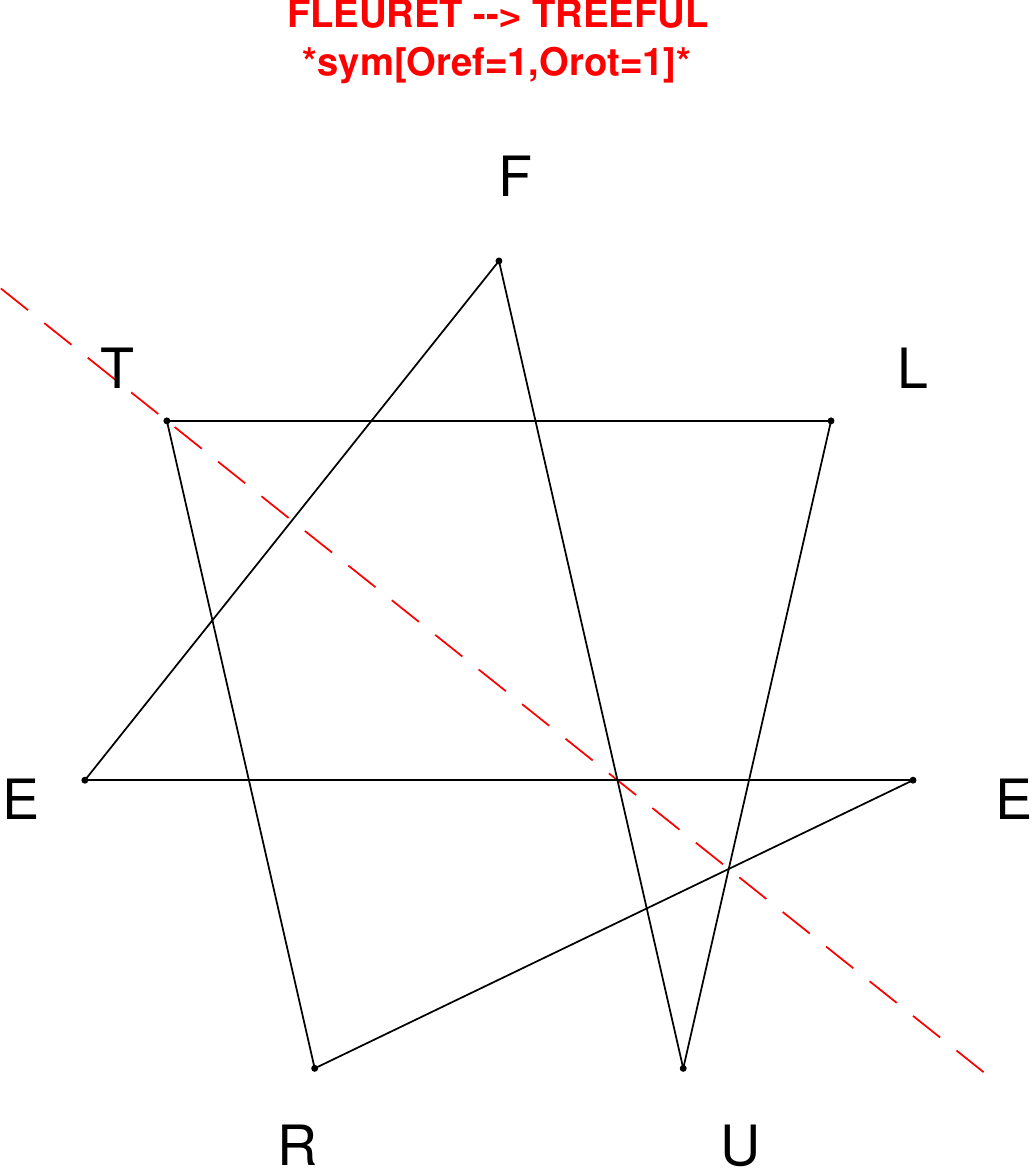}
\end{subfigure}
\hfill
\begin{subfigure}[T]{0.19\textwidth}
\centering
\includegraphics[width=\textwidth]{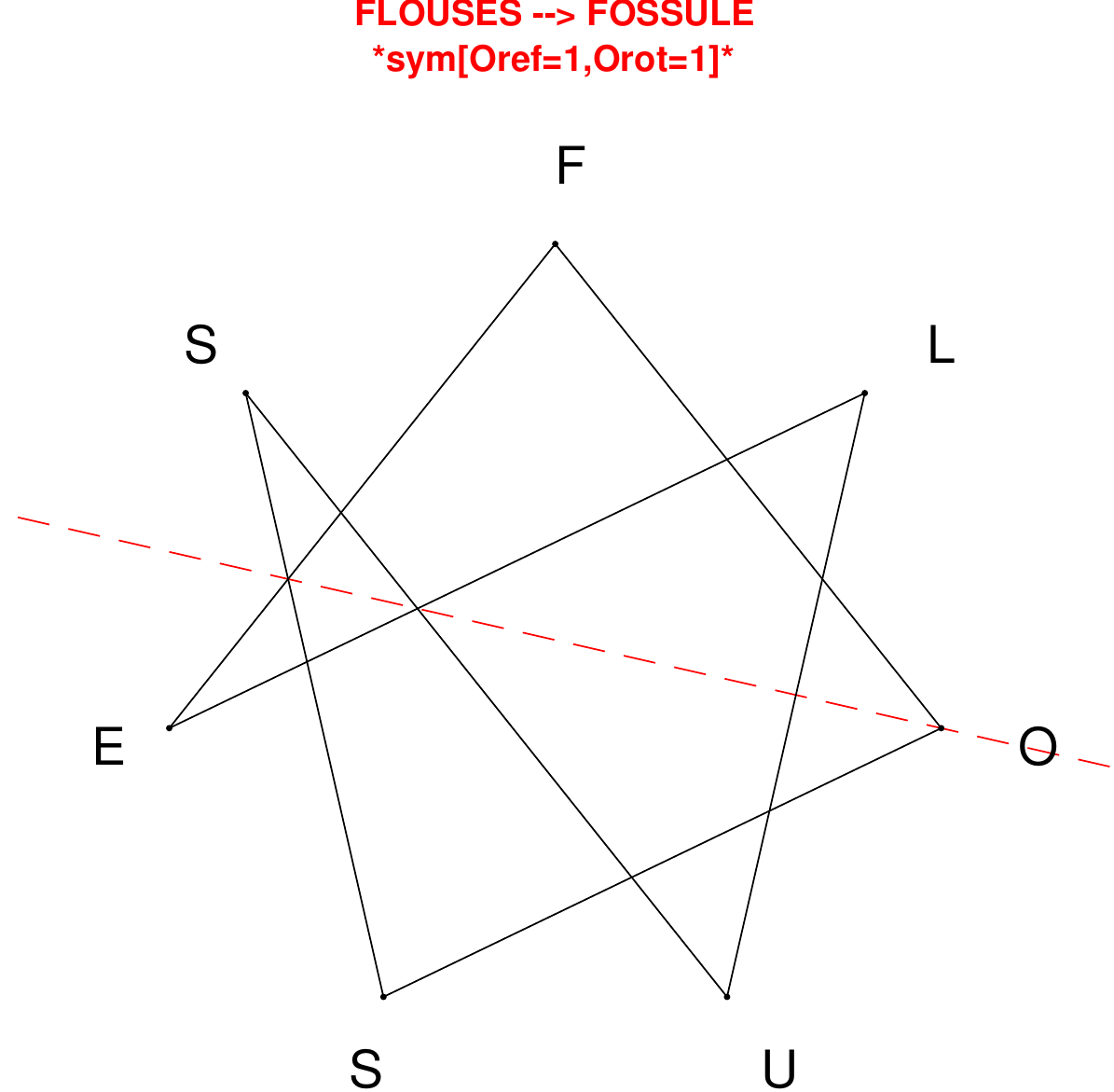}
\end{subfigure}
\hfill
\begin{subfigure}[T]{0.19\textwidth}
\centering
\includegraphics[width=\textwidth]{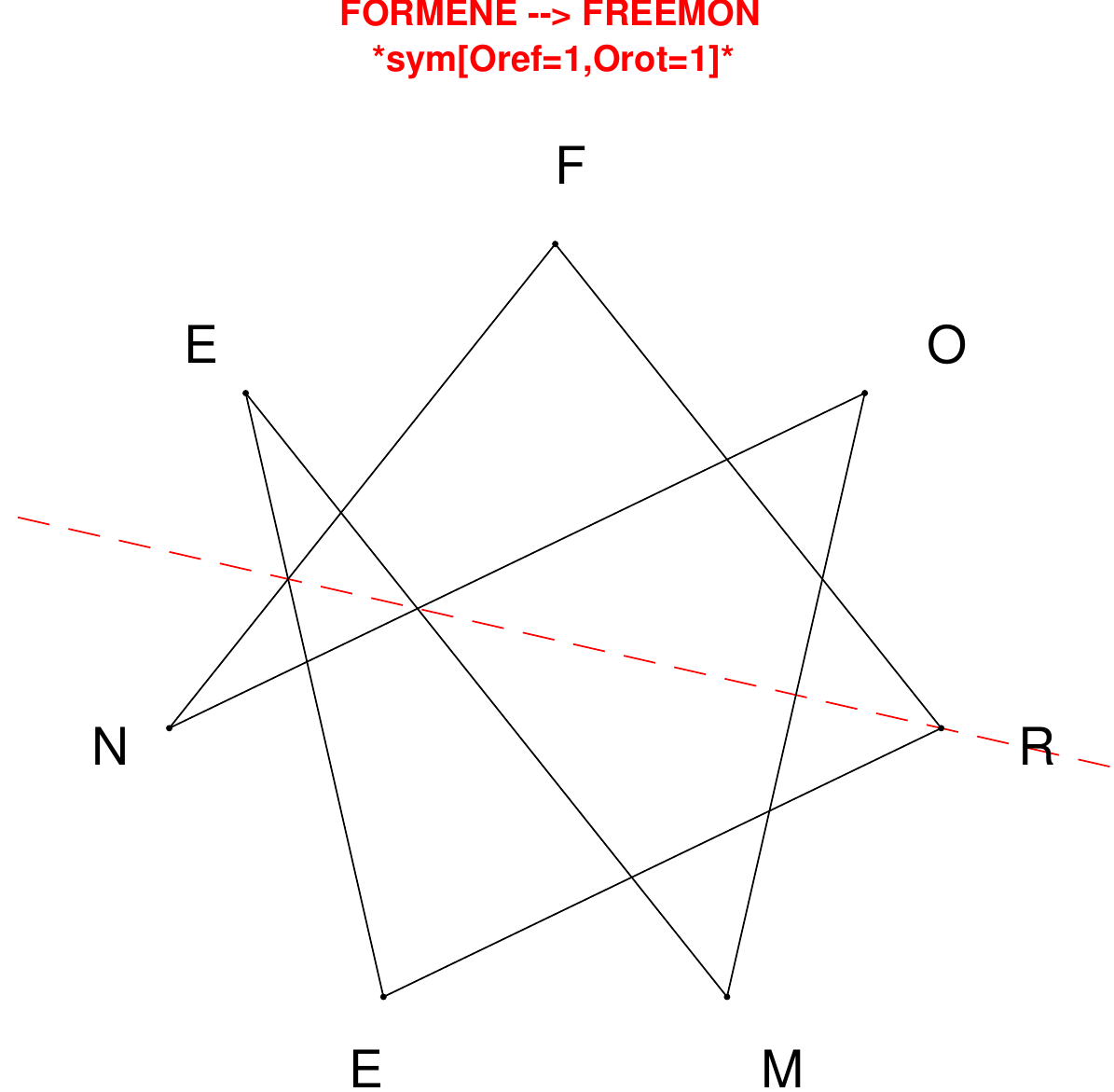}
\end{subfigure}
\hfill
\begin{subfigure}[T]{0.19\textwidth}
\centering
\includegraphics[width=\textwidth]{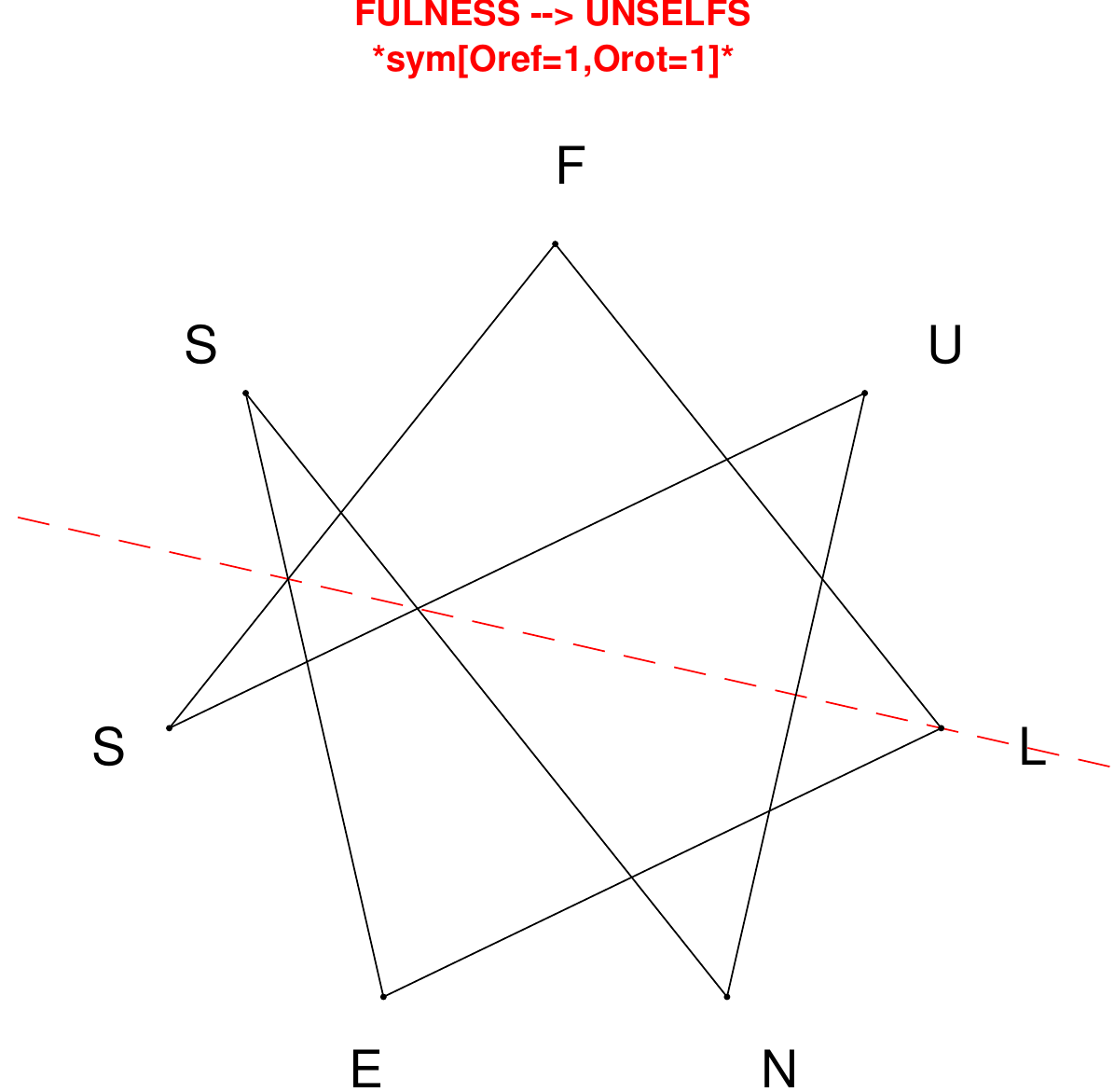}
\end{subfigure}
\end{figure}

\begin{figure}[H]
\centering
\begin{subfigure}[T]{0.19\textwidth}
\centering
\includegraphics[width=\textwidth]{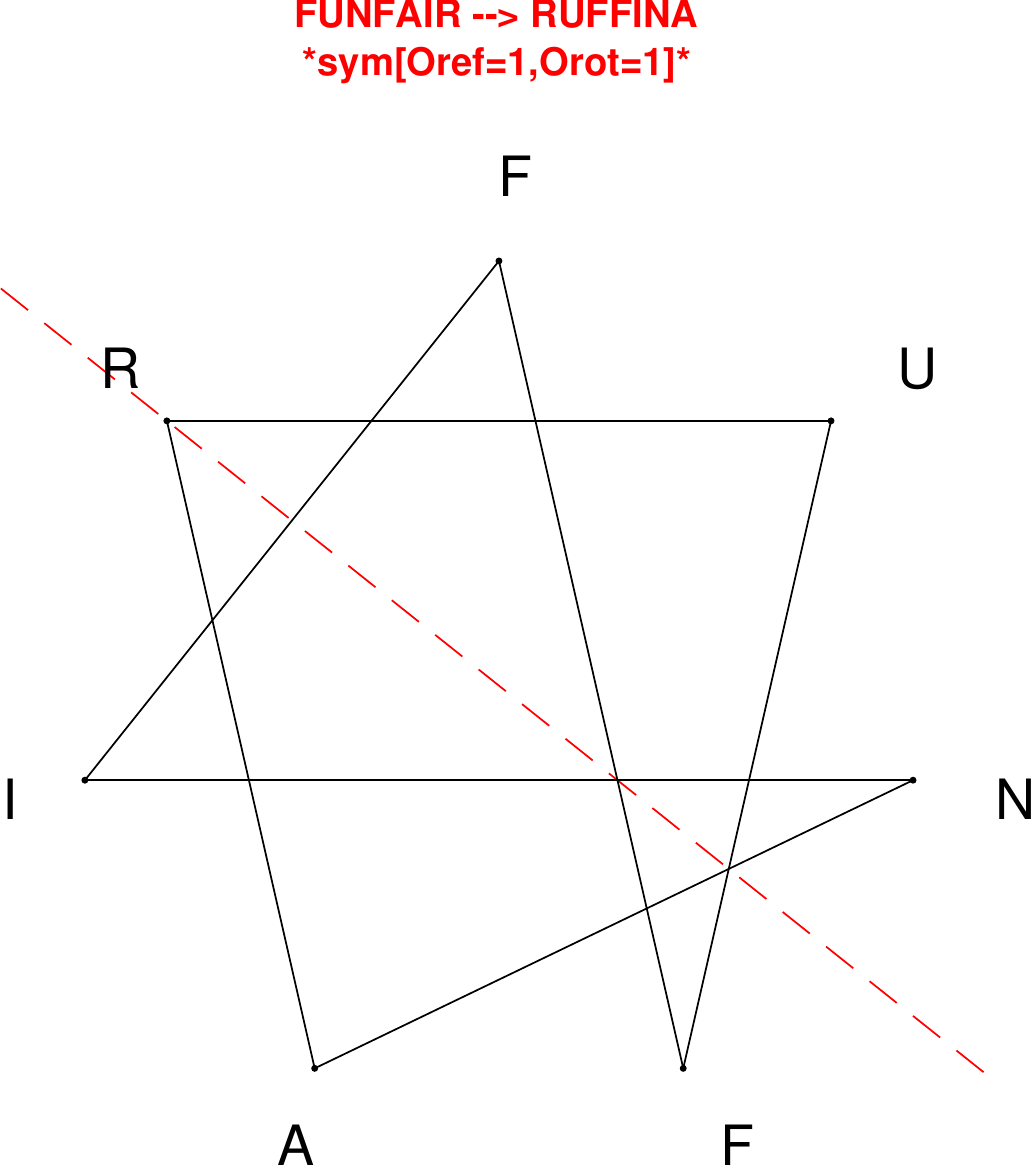}
\end{subfigure}
\hfill
\begin{subfigure}[T]{0.19\textwidth}
\centering
\includegraphics[width=\textwidth]{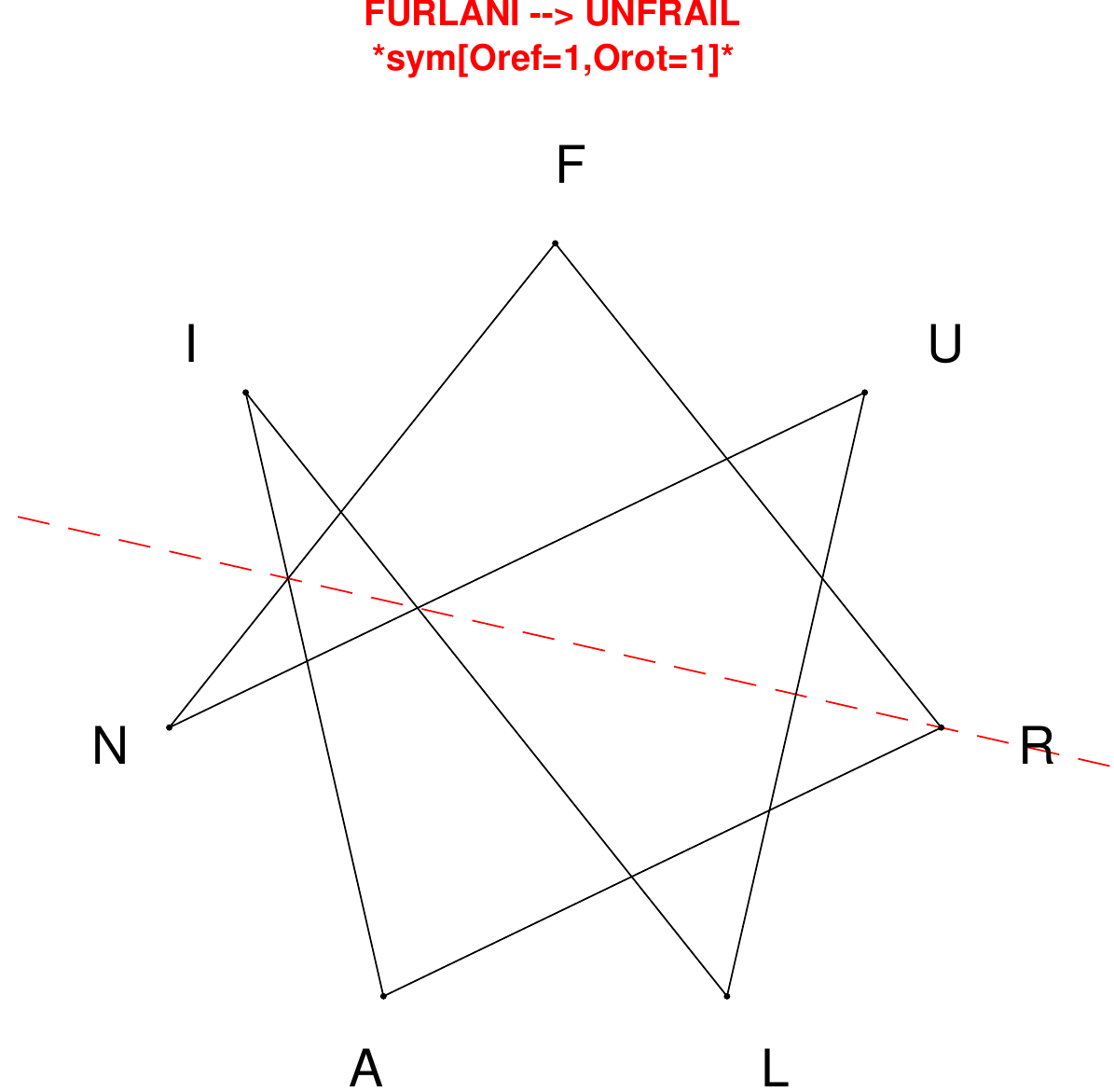}
\end{subfigure}
\hfill
\begin{subfigure}[T]{0.19\textwidth}
\centering
\includegraphics[width=\textwidth]{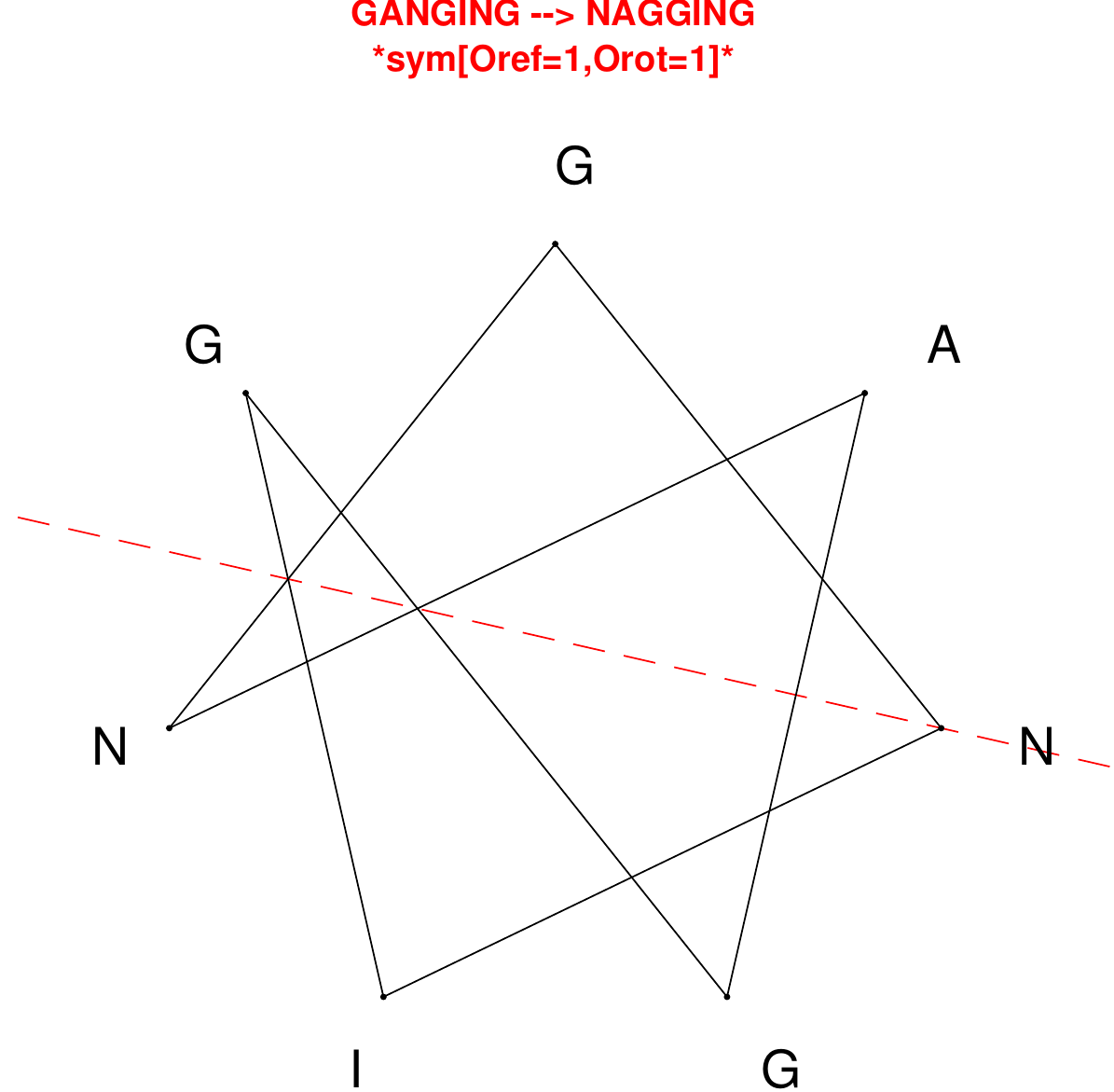}
\end{subfigure}
\hfill
\begin{subfigure}[T]{0.19\textwidth}
\centering
\includegraphics[width=\textwidth]{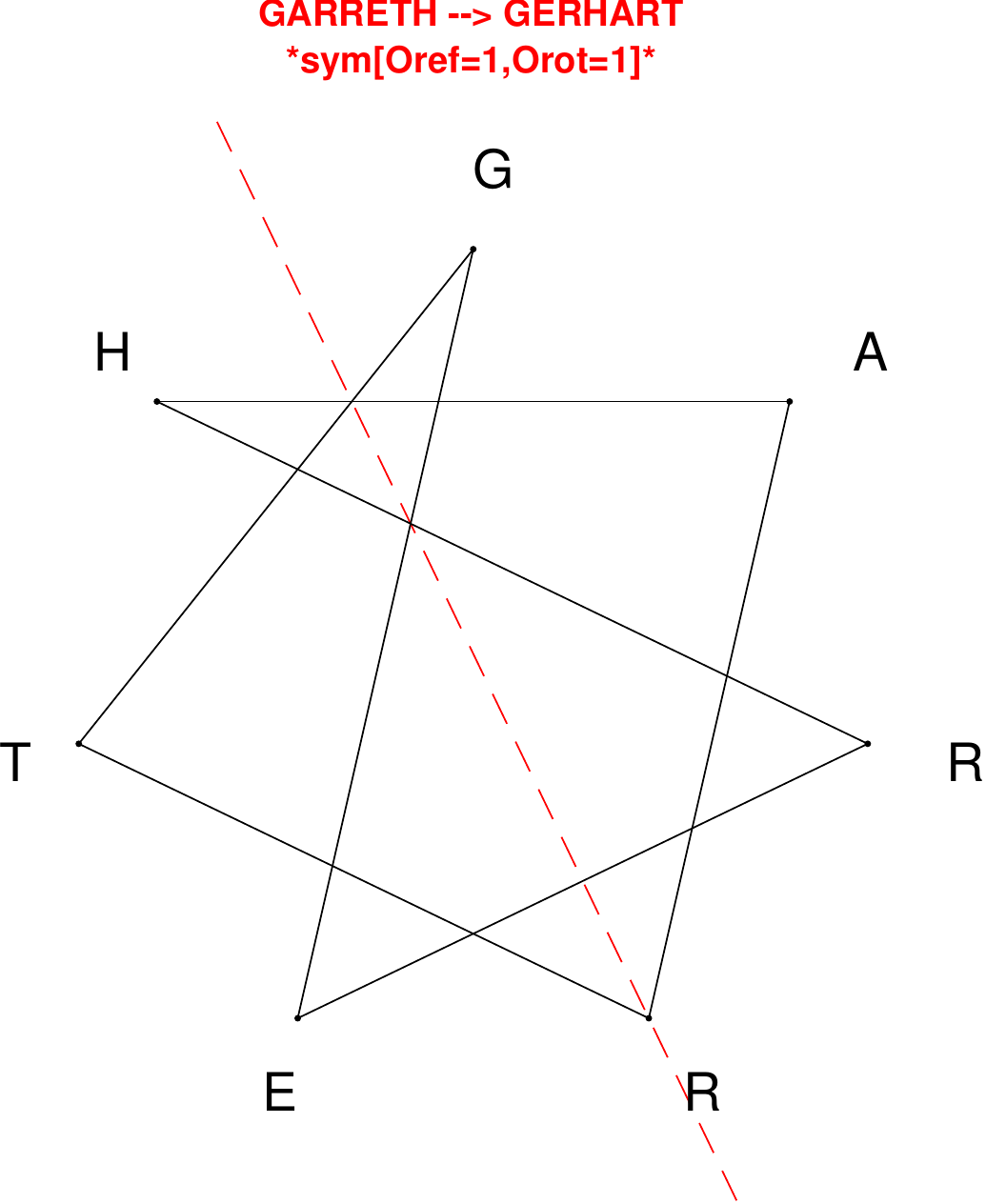}
\end{subfigure}
\hfill
\begin{subfigure}[T]{0.19\textwidth}
\centering
\includegraphics[width=\textwidth]{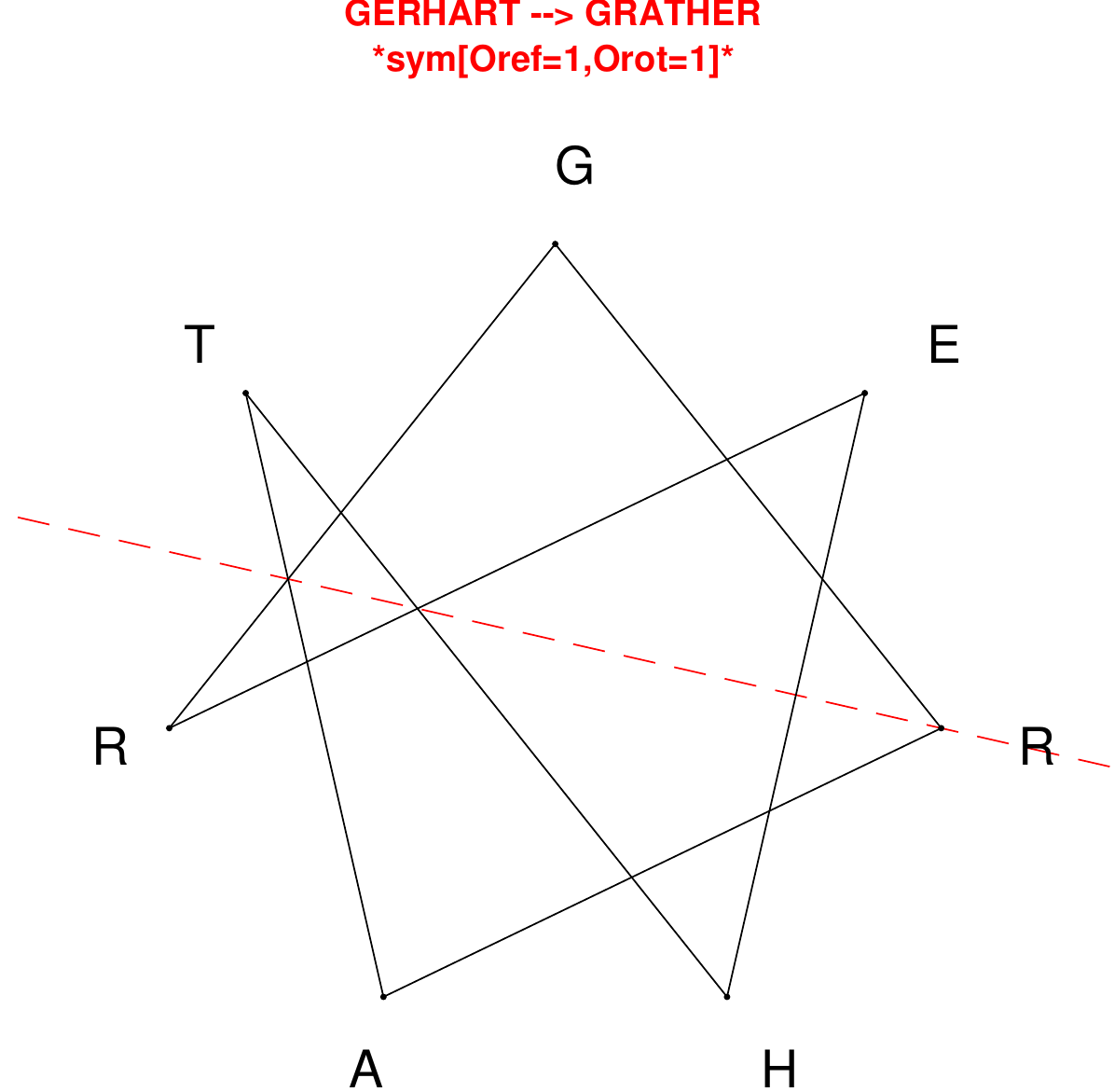}
\end{subfigure}
\end{figure}

\begin{figure}[H]
\centering
\begin{subfigure}[T]{0.19\textwidth}
\centering
\includegraphics[width=\textwidth]{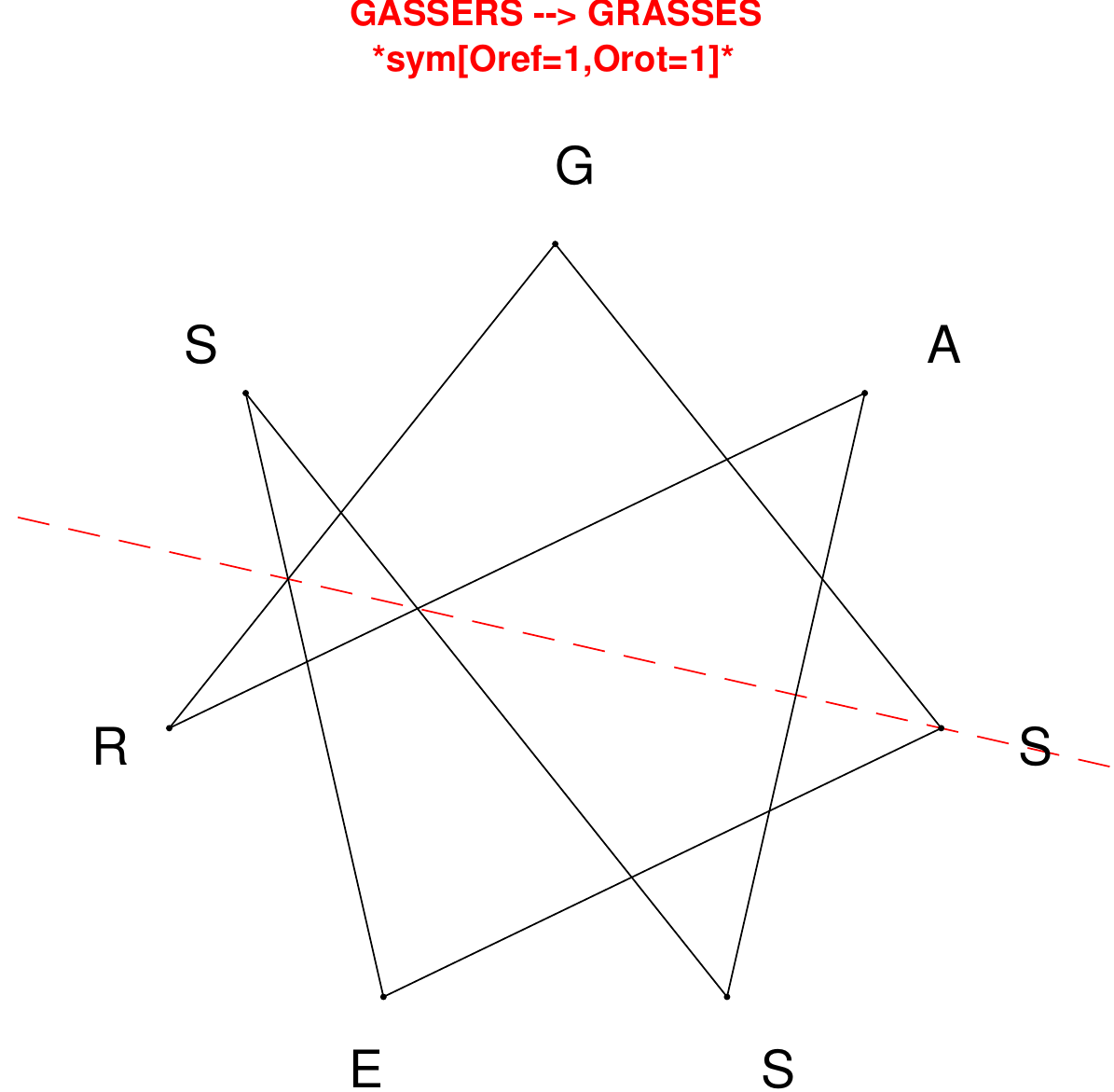}
\end{subfigure}
\hfill
\begin{subfigure}[T]{0.19\textwidth}
\centering
\includegraphics[width=\textwidth]{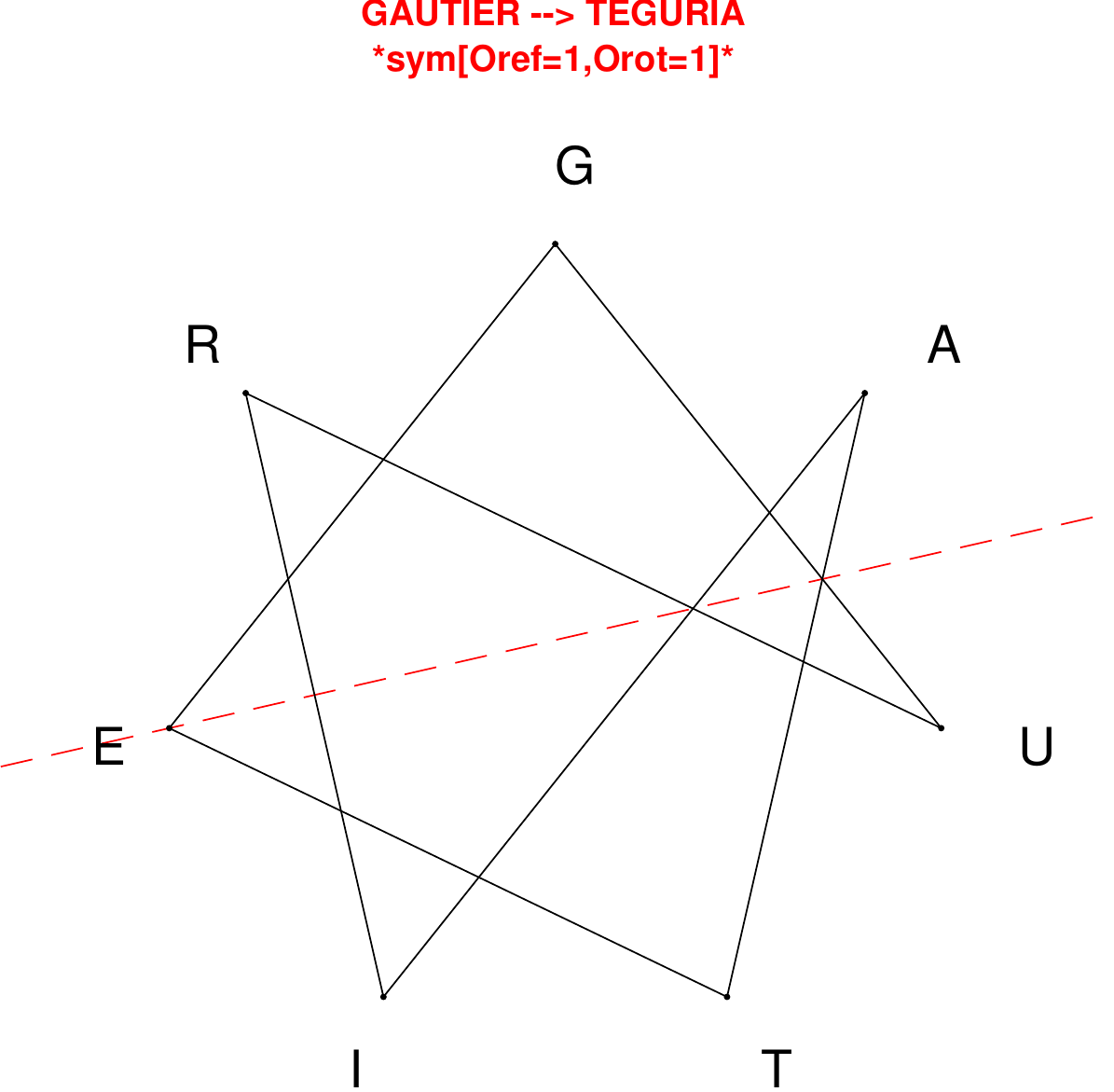}
\end{subfigure}
\hfill
\begin{subfigure}[T]{0.19\textwidth}
\centering
\includegraphics[width=\textwidth]{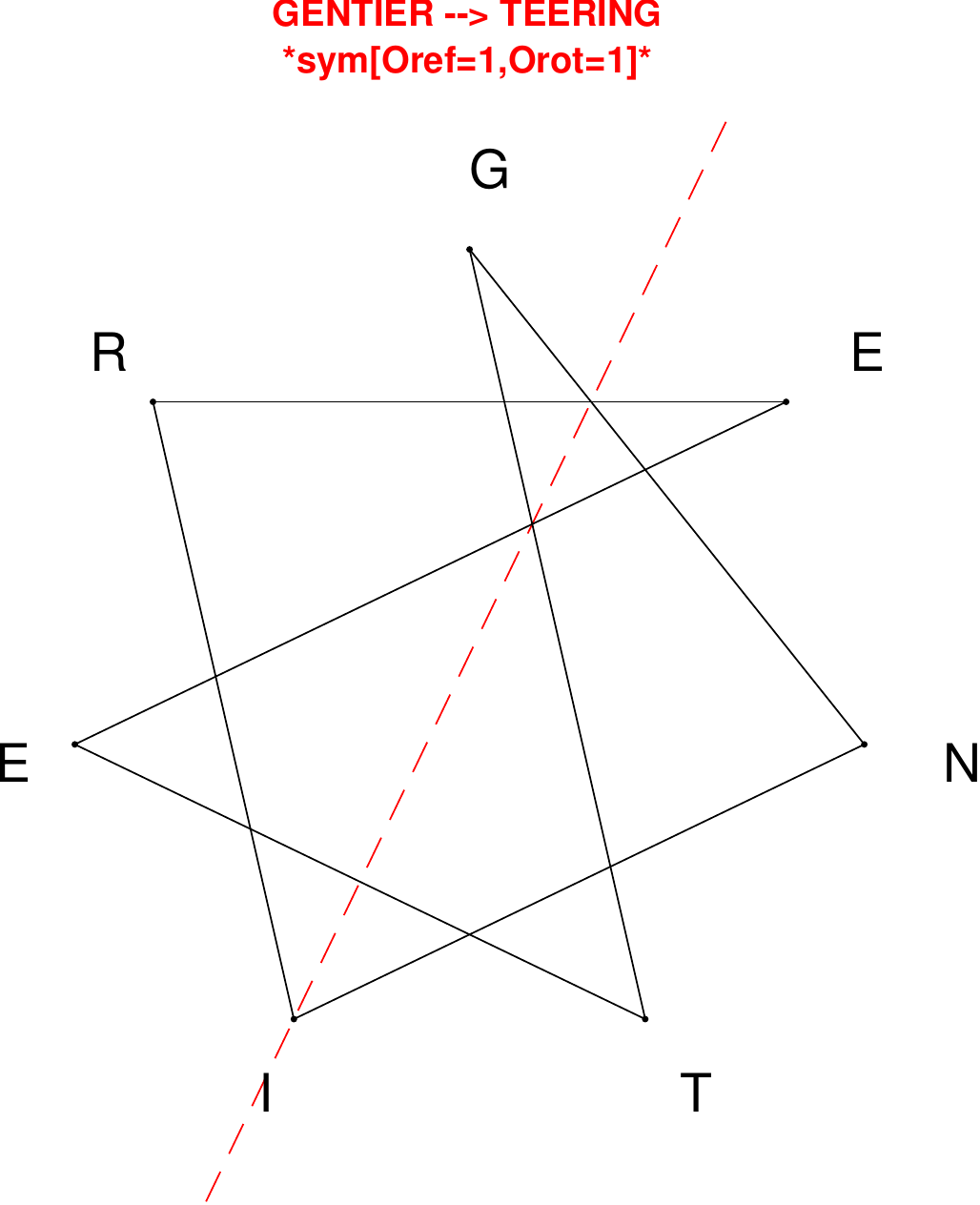}
\end{subfigure}
\hfill
\begin{subfigure}[T]{0.19\textwidth}
\centering
\includegraphics[width=\textwidth]{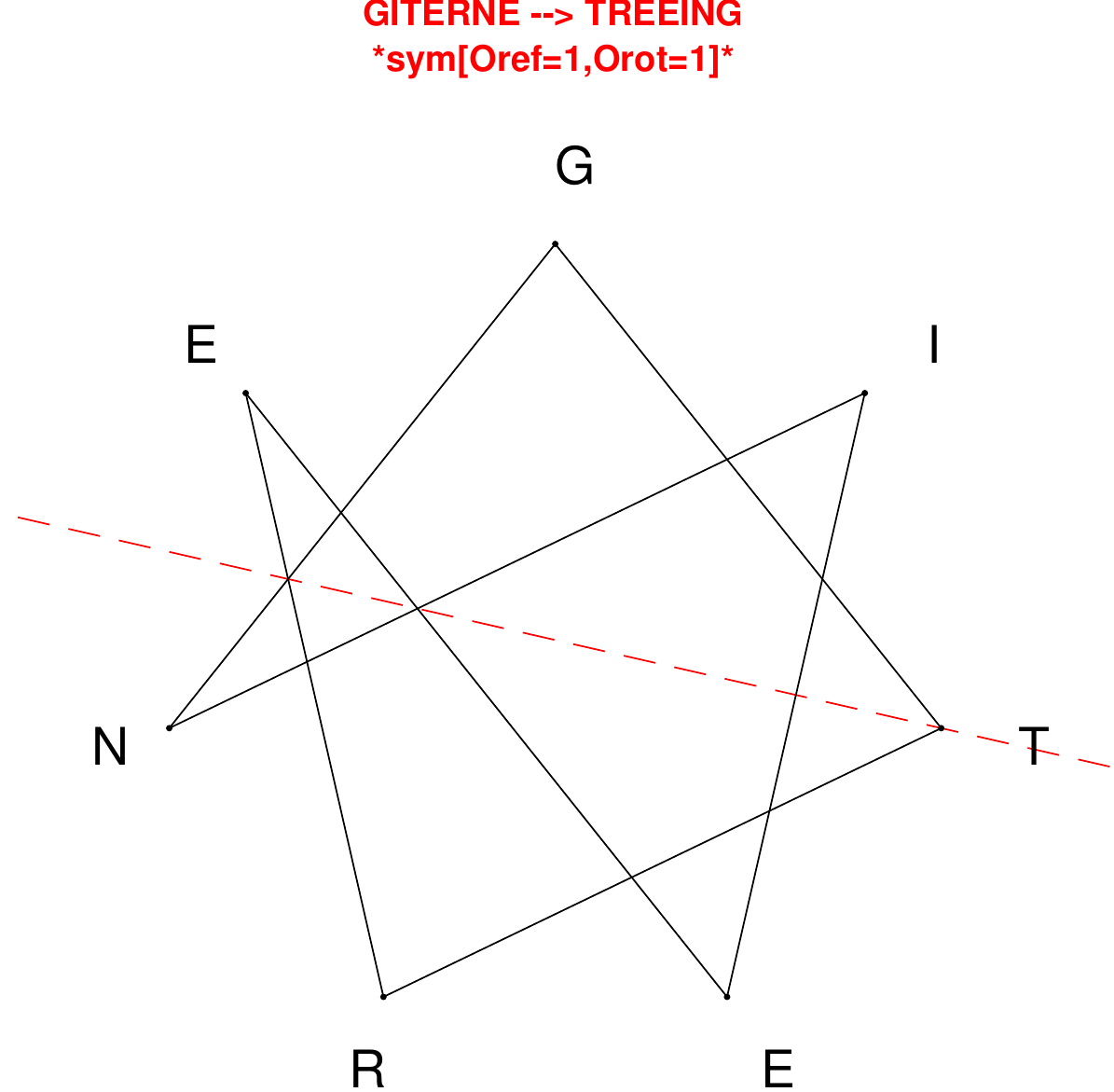}
\end{subfigure}
\hfill
\begin{subfigure}[T]{0.19\textwidth}
\centering
\includegraphics[width=\textwidth]{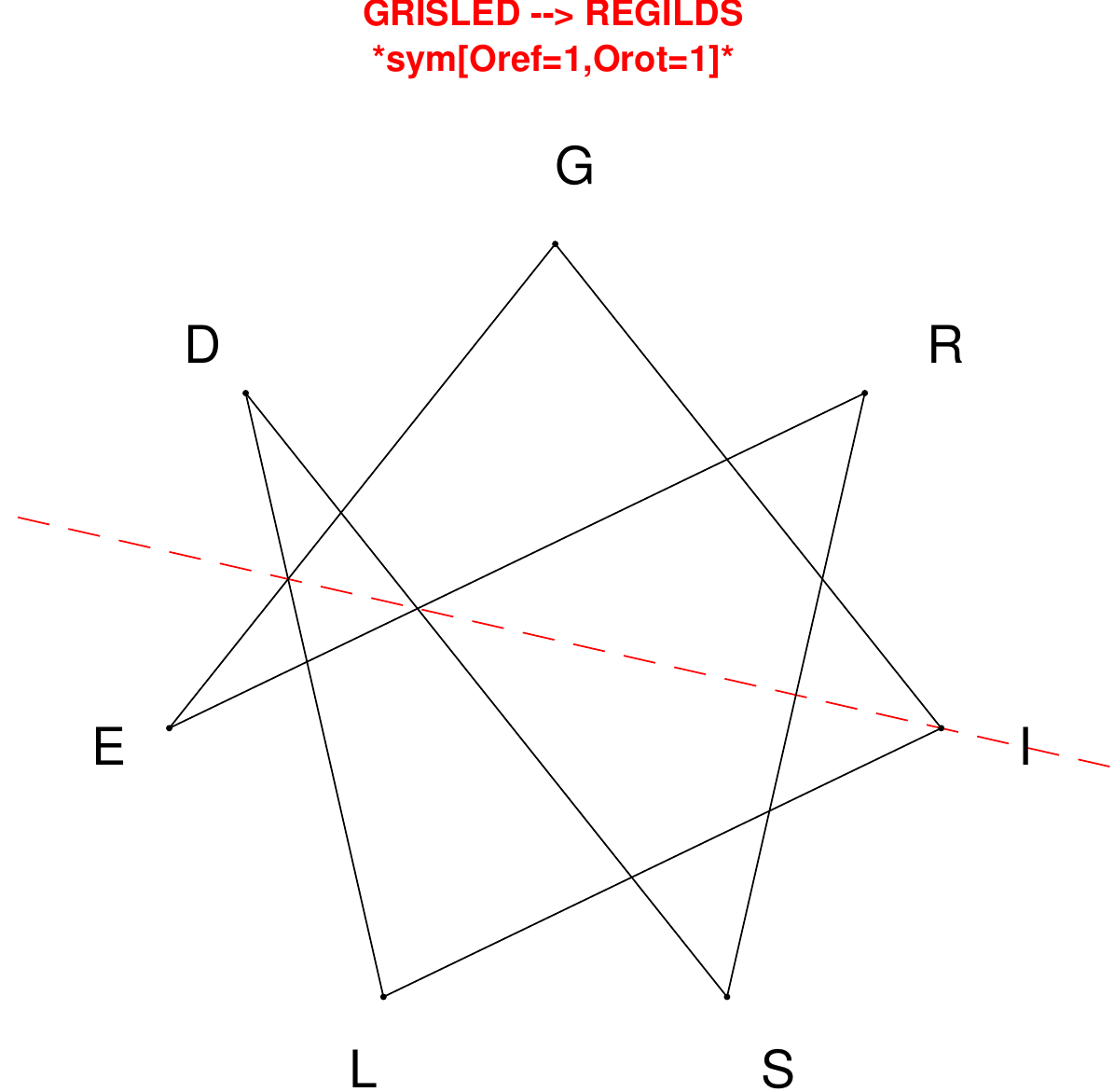}
\end{subfigure}
\end{figure}

\begin{figure}[H]
\centering
\begin{subfigure}[T]{0.19\textwidth}
\centering
\includegraphics[width=\textwidth]{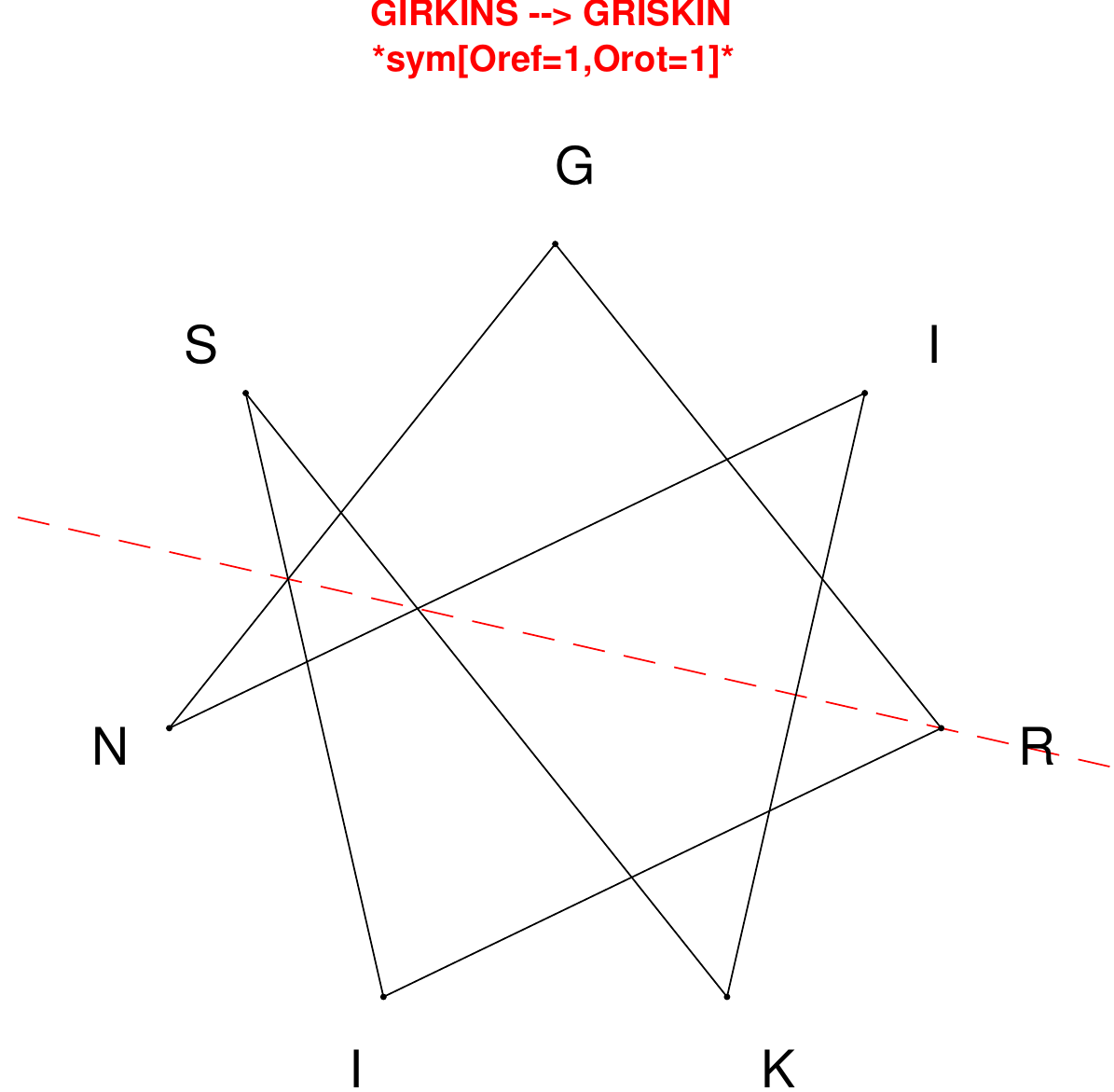}
\end{subfigure}
\hfill
\begin{subfigure}[T]{0.19\textwidth}
\centering
\includegraphics[width=\textwidth]{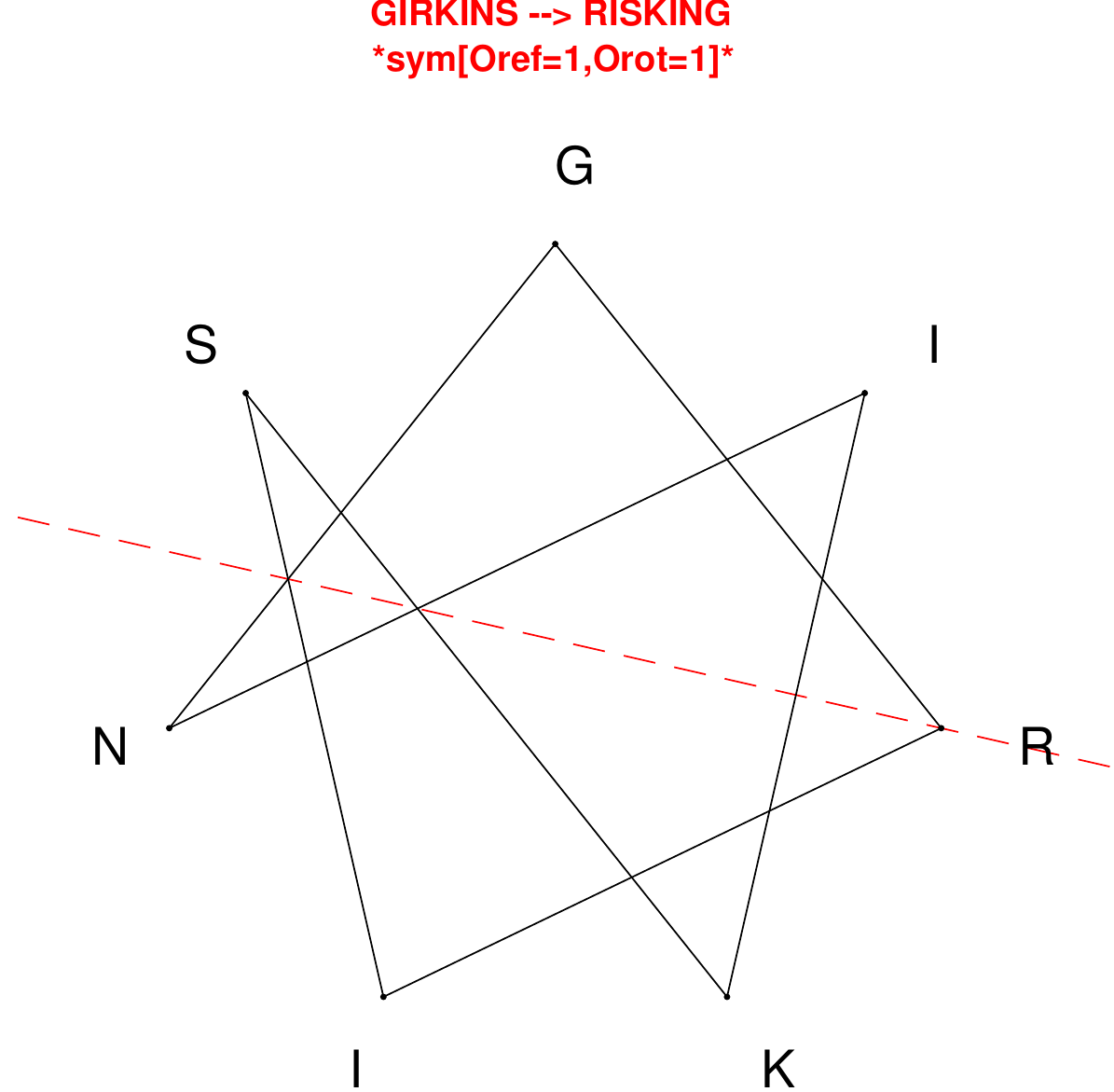}
\end{subfigure}
\hfill
\begin{subfigure}[T]{0.19\textwidth}
\centering
\includegraphics[width=\textwidth]{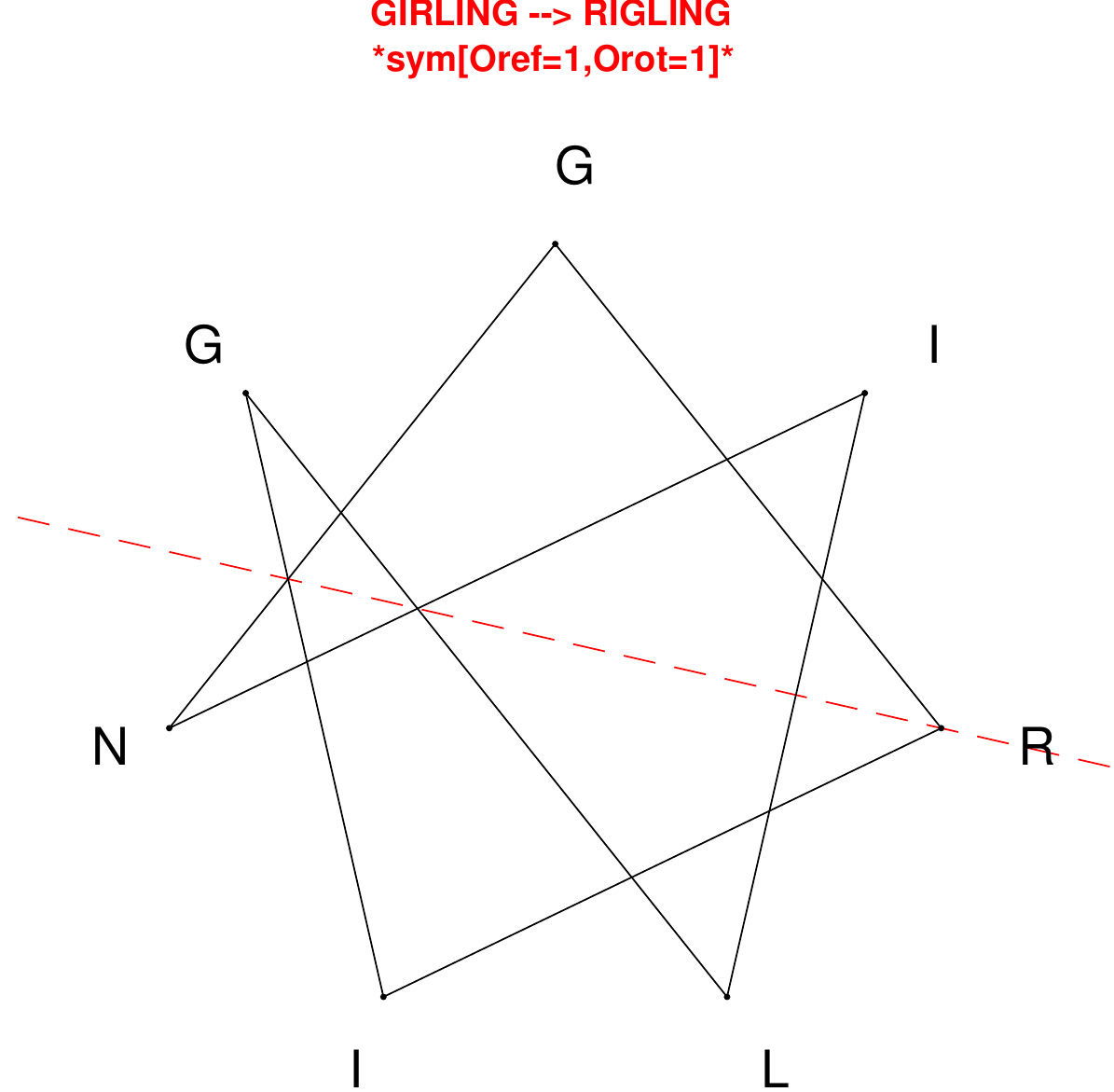}
\end{subfigure}
\hfill
\begin{subfigure}[T]{0.19\textwidth}
\centering
\includegraphics[width=\textwidth]{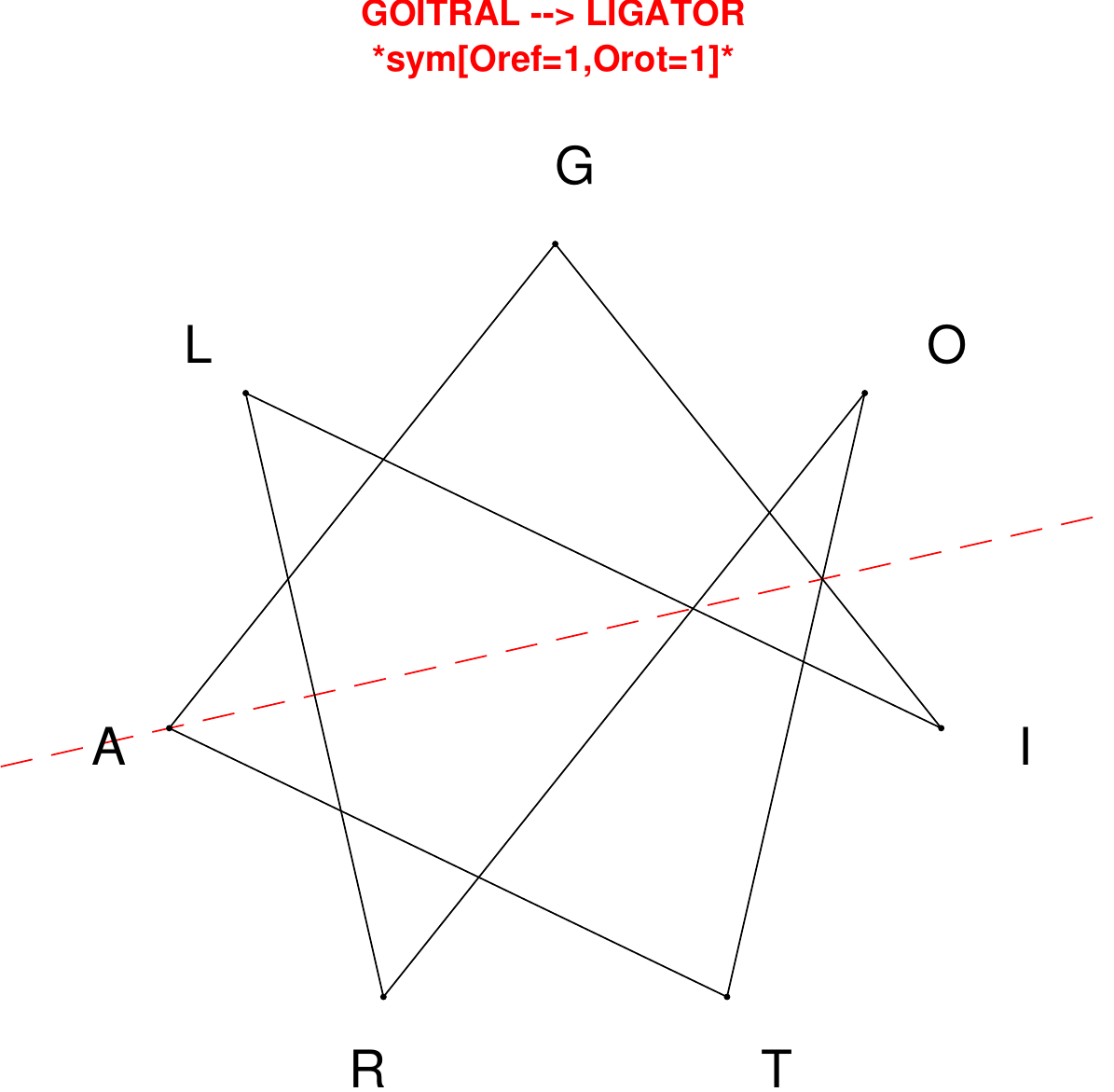}
\end{subfigure}
\hfill
\begin{subfigure}[T]{0.19\textwidth}
\centering
\includegraphics[width=\textwidth]{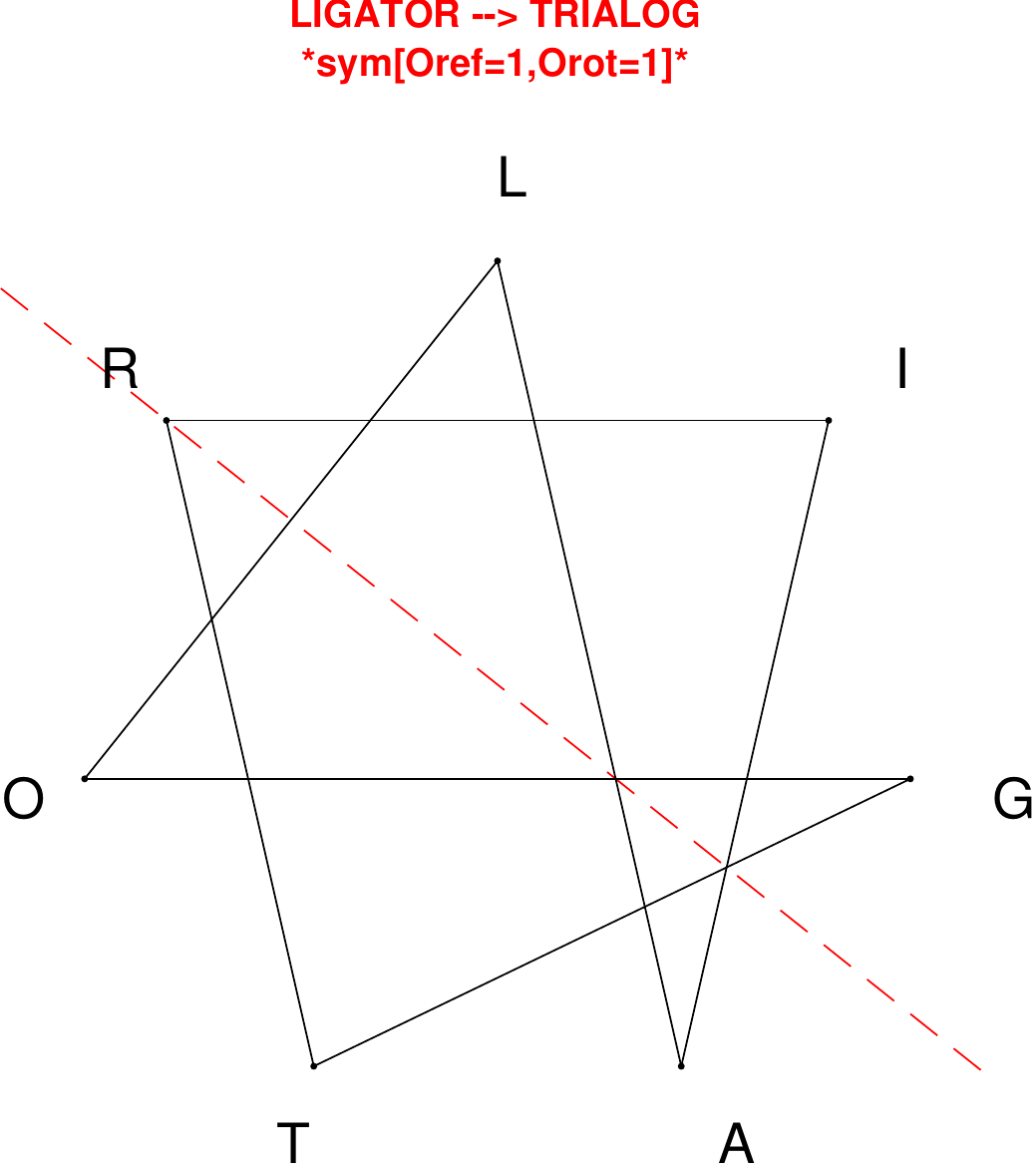}
\end{subfigure}
\end{figure}

\begin{figure}[H]
\centering
\begin{subfigure}[T]{0.19\textwidth}
\centering
\includegraphics[width=\textwidth]{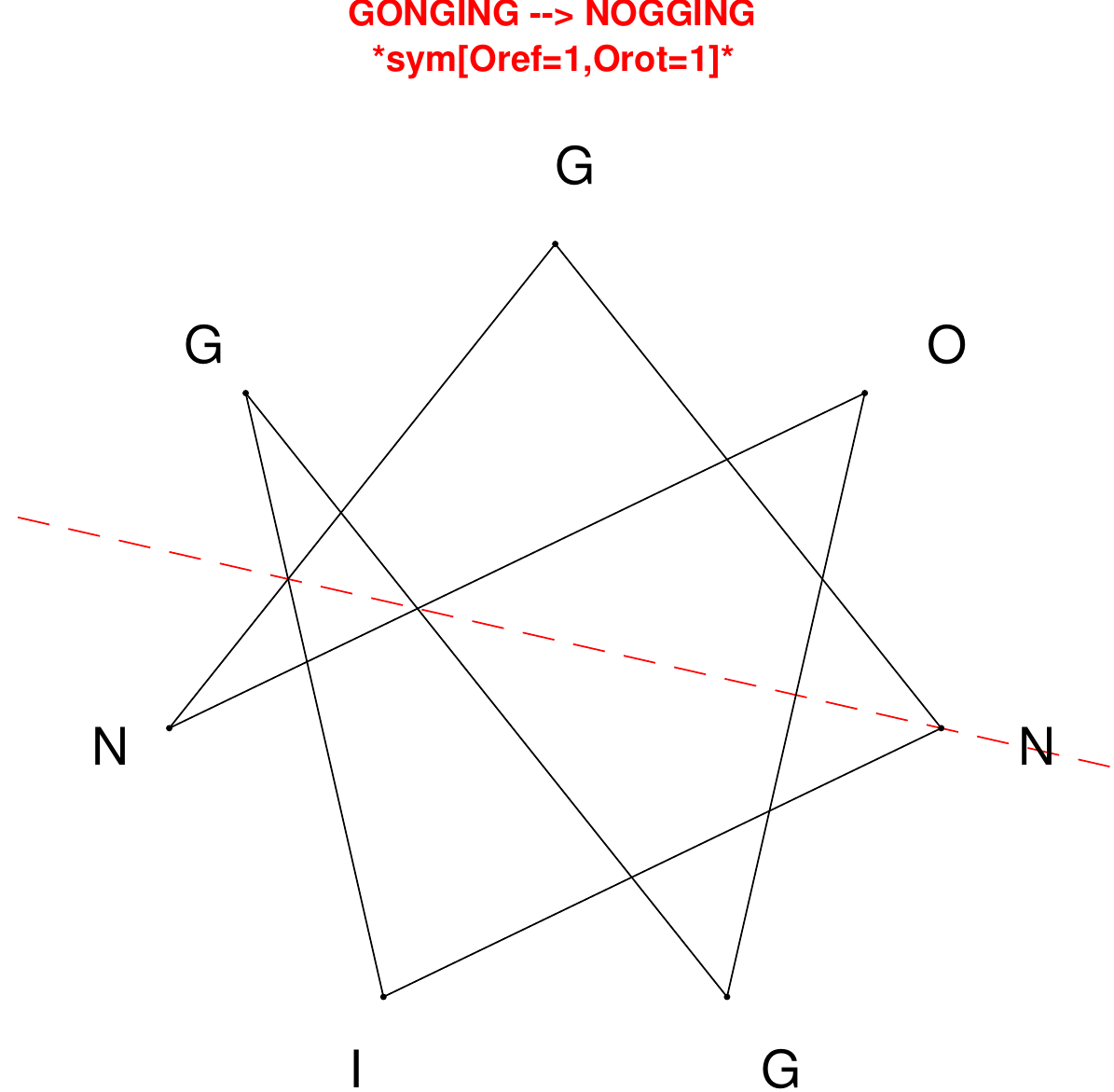}
\end{subfigure}
\hfill
\begin{subfigure}[T]{0.19\textwidth}
\centering
\includegraphics[width=\textwidth]{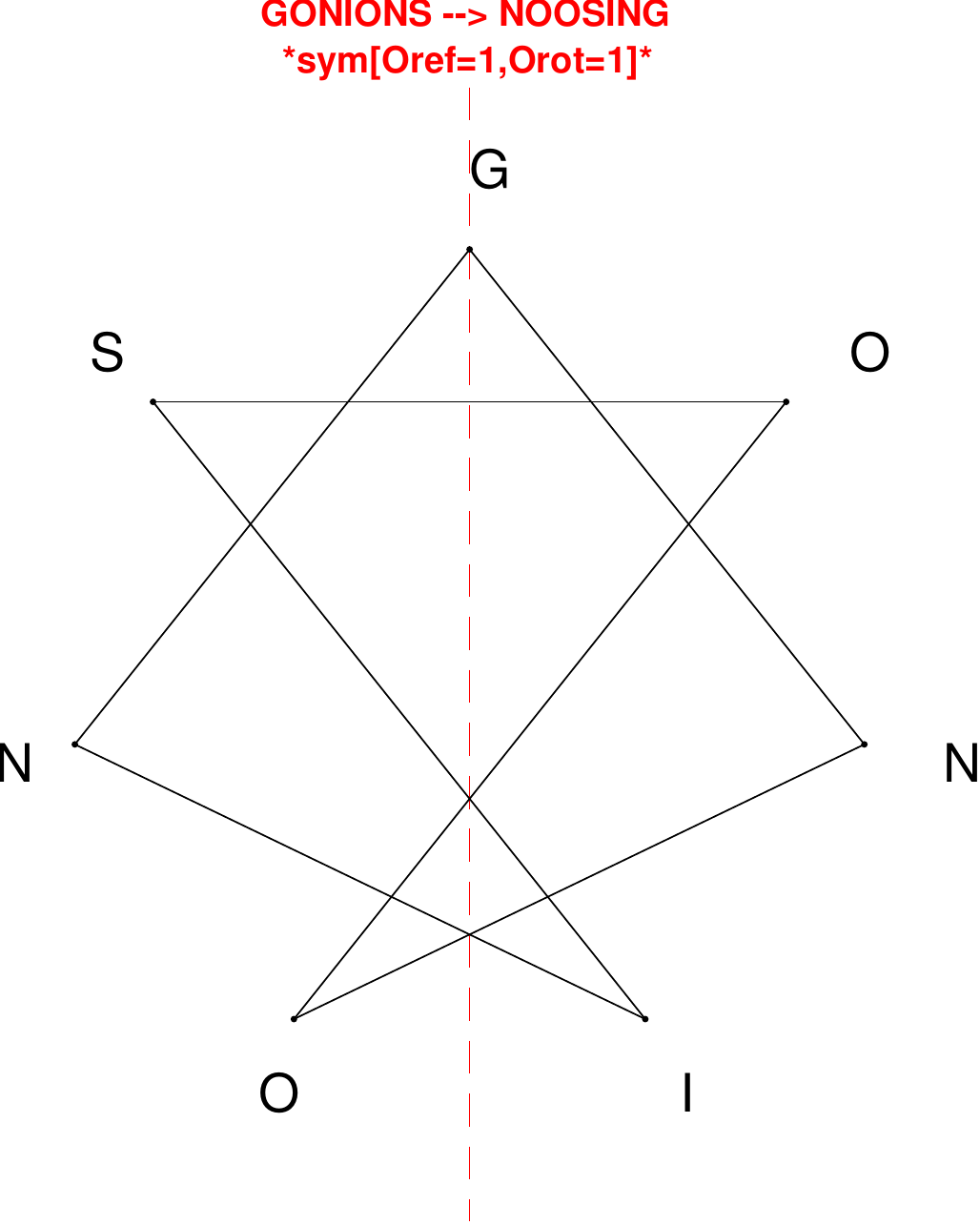}
\end{subfigure}
\hfill
\begin{subfigure}[T]{0.19\textwidth}
\centering
\includegraphics[width=\textwidth]{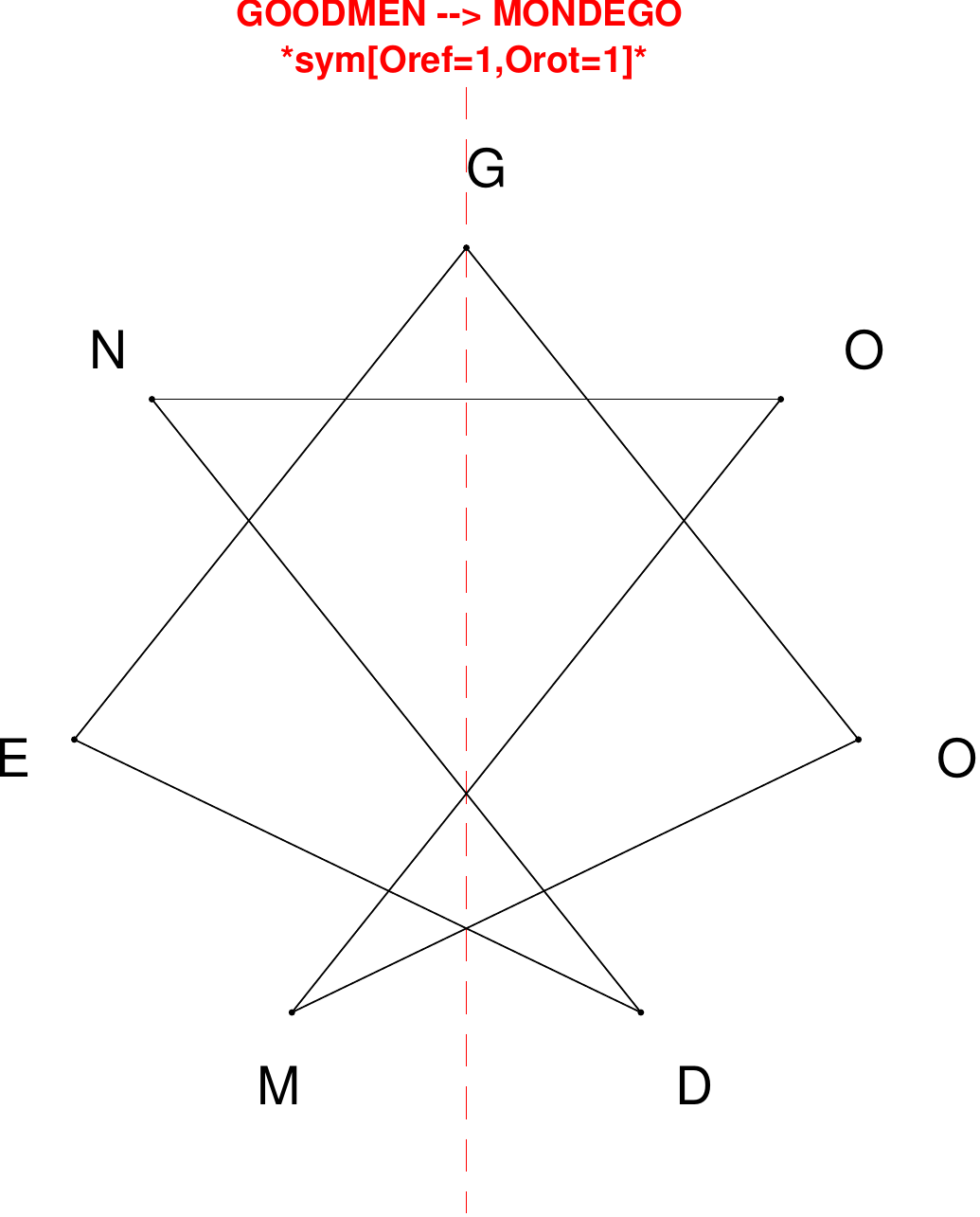}
\end{subfigure}
\hfill
\begin{subfigure}[T]{0.19\textwidth}
\centering
\includegraphics[width=\textwidth]{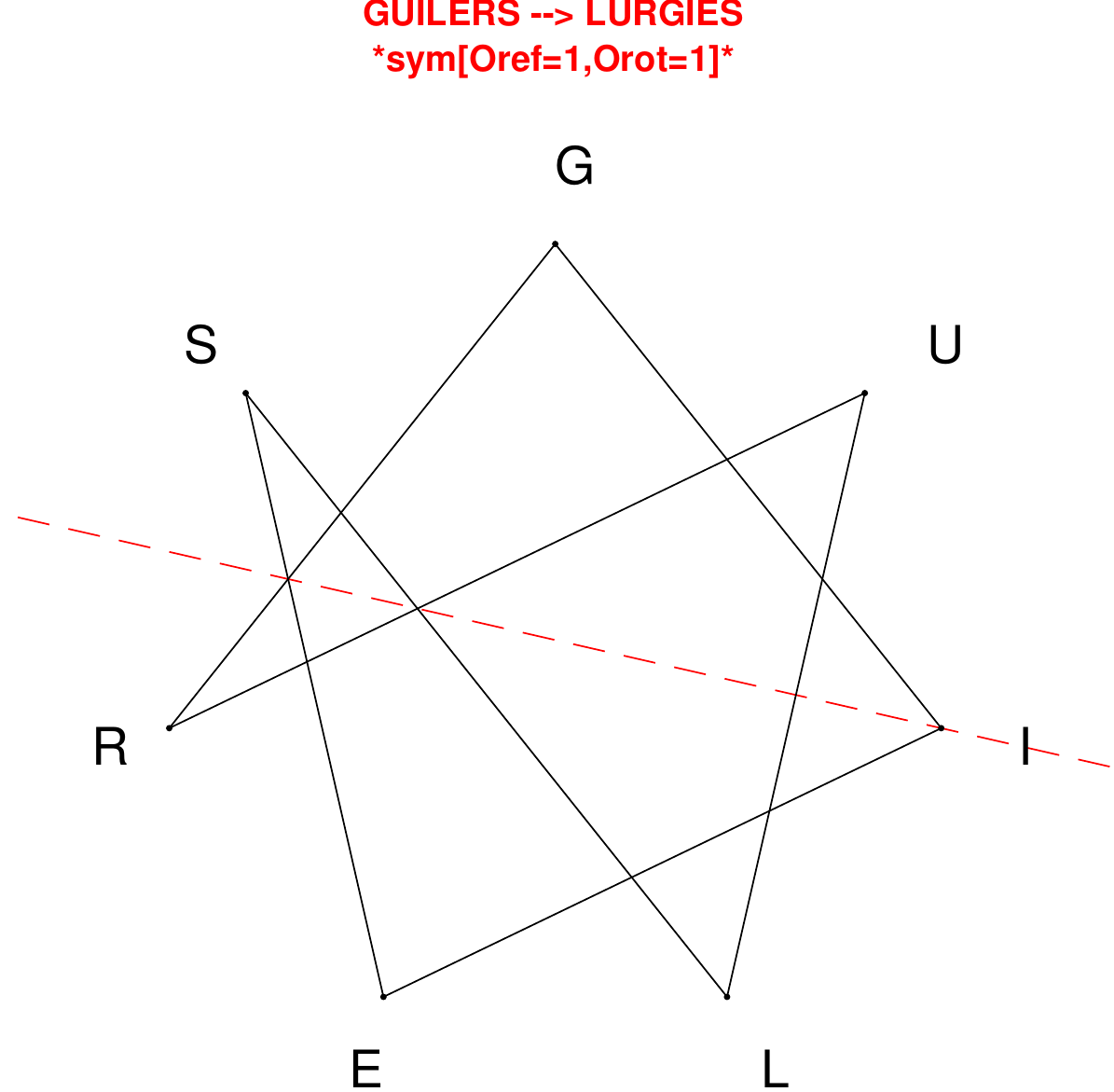}
\end{subfigure}
\hfill
\begin{subfigure}[T]{0.19\textwidth}
\centering
\includegraphics[width=\textwidth]{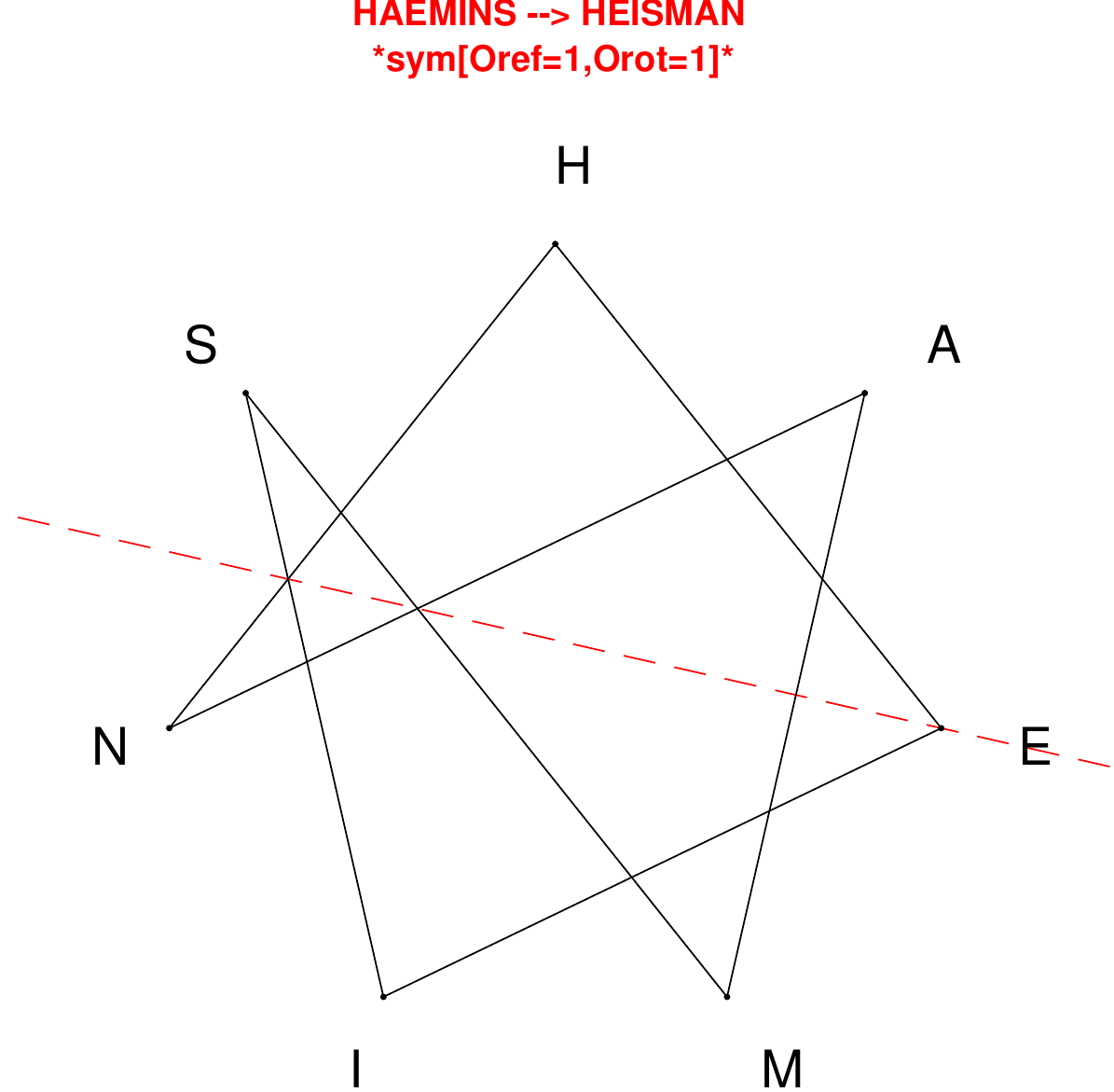}
\end{subfigure}
\end{figure}

\begin{figure}[H]
\centering
\begin{subfigure}[T]{0.19\textwidth}
\centering
\includegraphics[width=\textwidth]{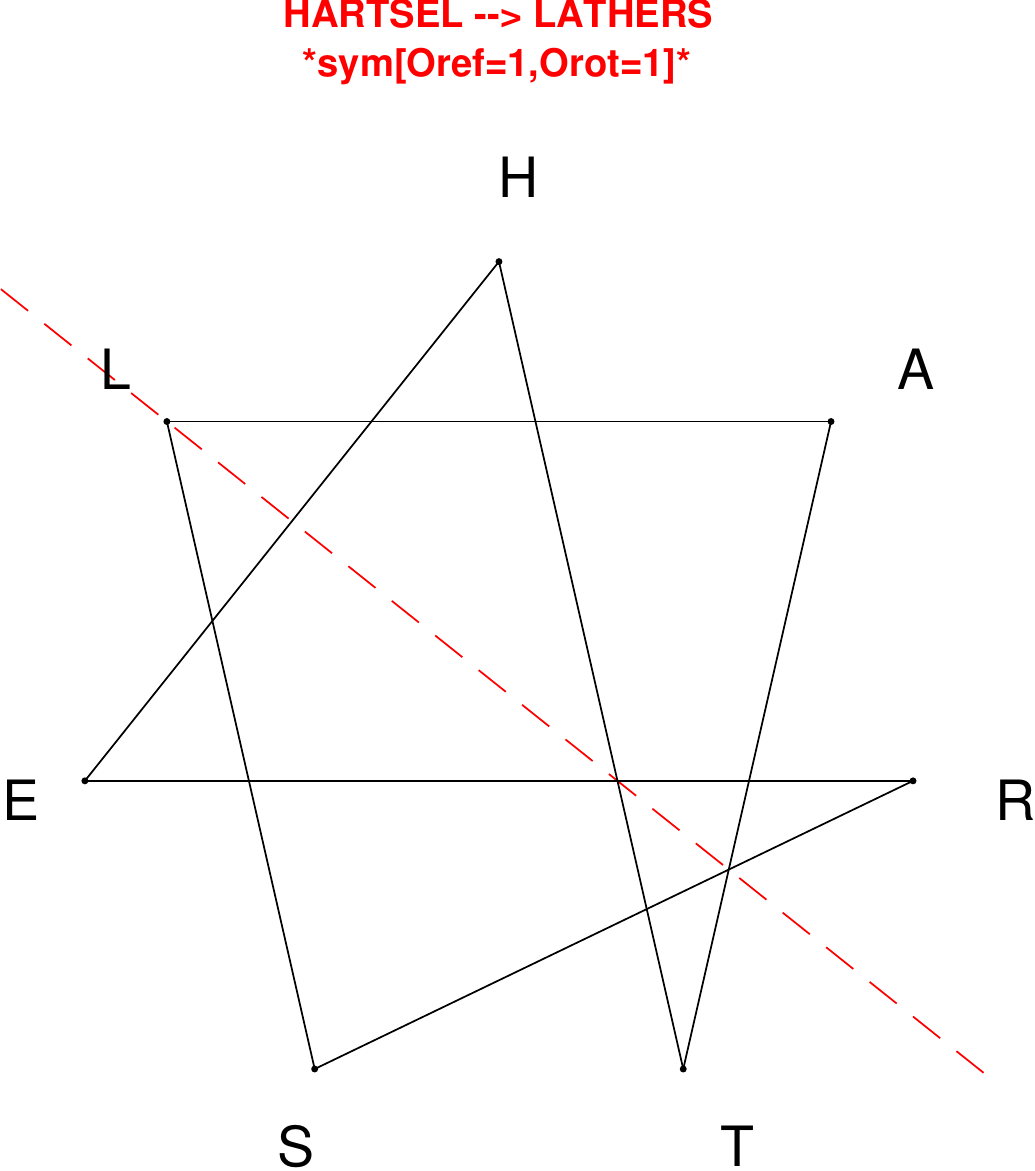}
\end{subfigure}
\hfill
\begin{subfigure}[T]{0.19\textwidth}
\centering
\includegraphics[width=\textwidth]{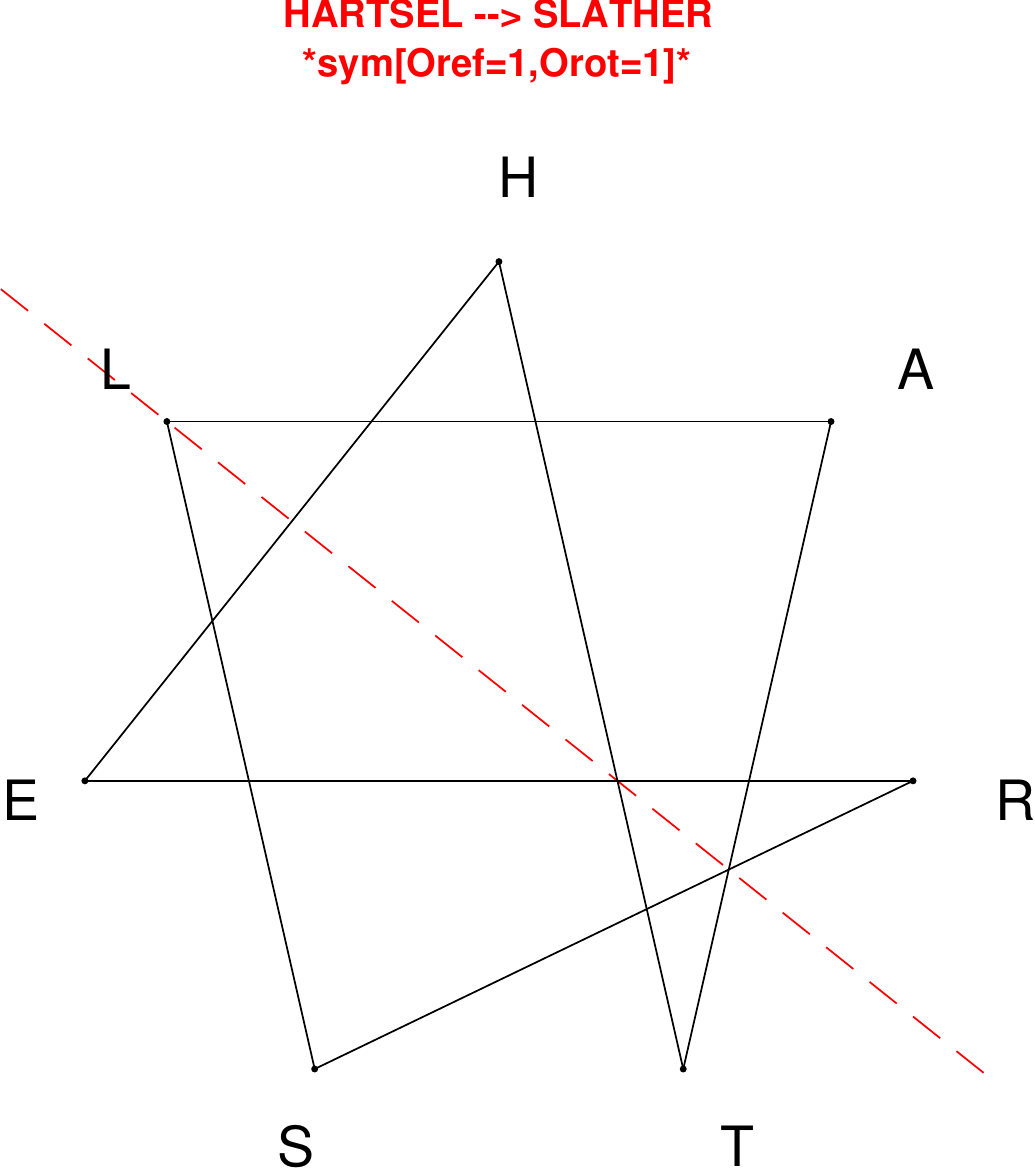}
\end{subfigure}
\hfill
\begin{subfigure}[T]{0.19\textwidth}
\centering
\includegraphics[width=\textwidth]{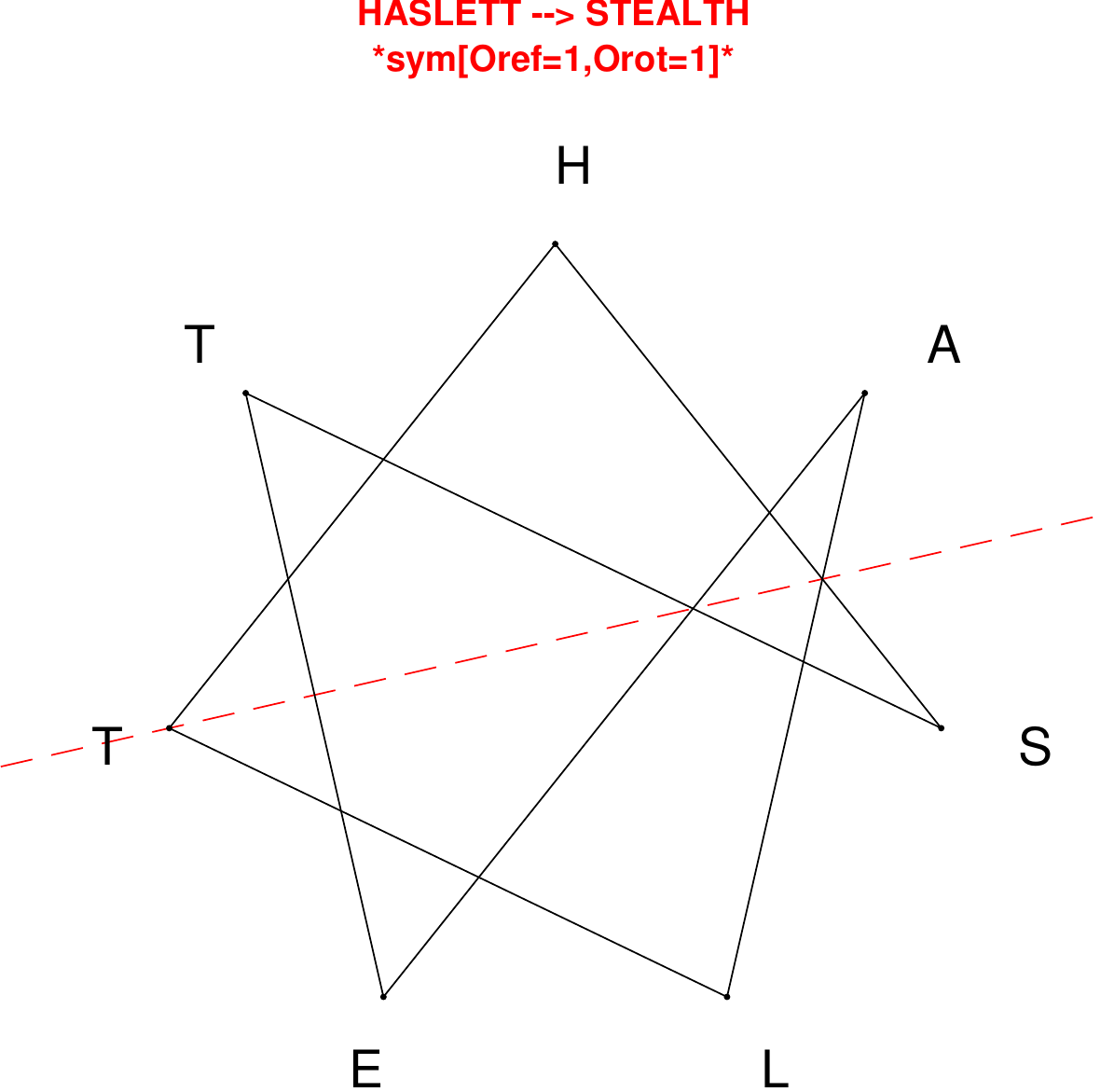}
\end{subfigure}
\hfill
\begin{subfigure}[T]{0.19\textwidth}
\centering
\includegraphics[width=\textwidth]{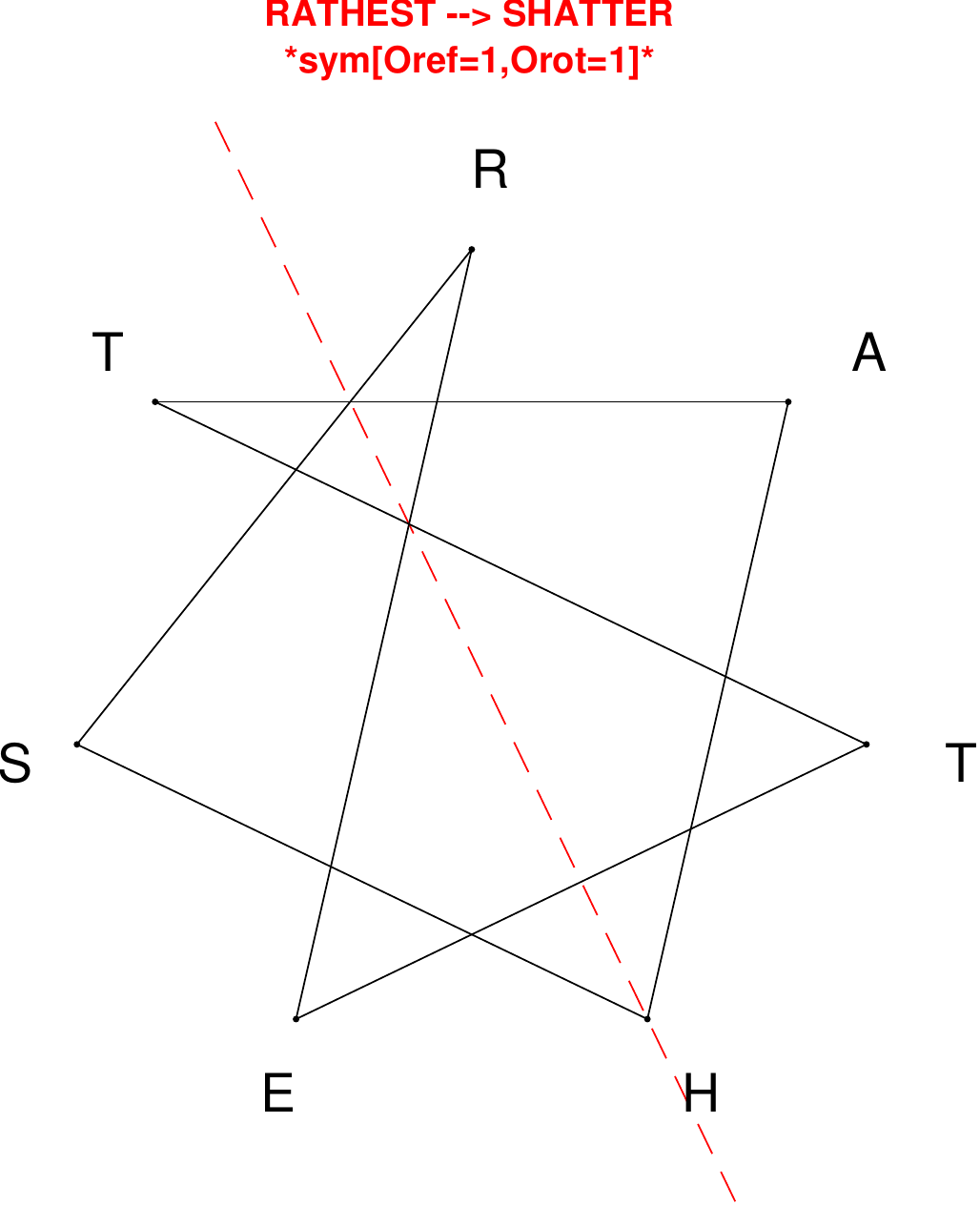}
\end{subfigure}
\hfill
\begin{subfigure}[T]{0.19\textwidth}
\centering
\includegraphics[width=\textwidth]{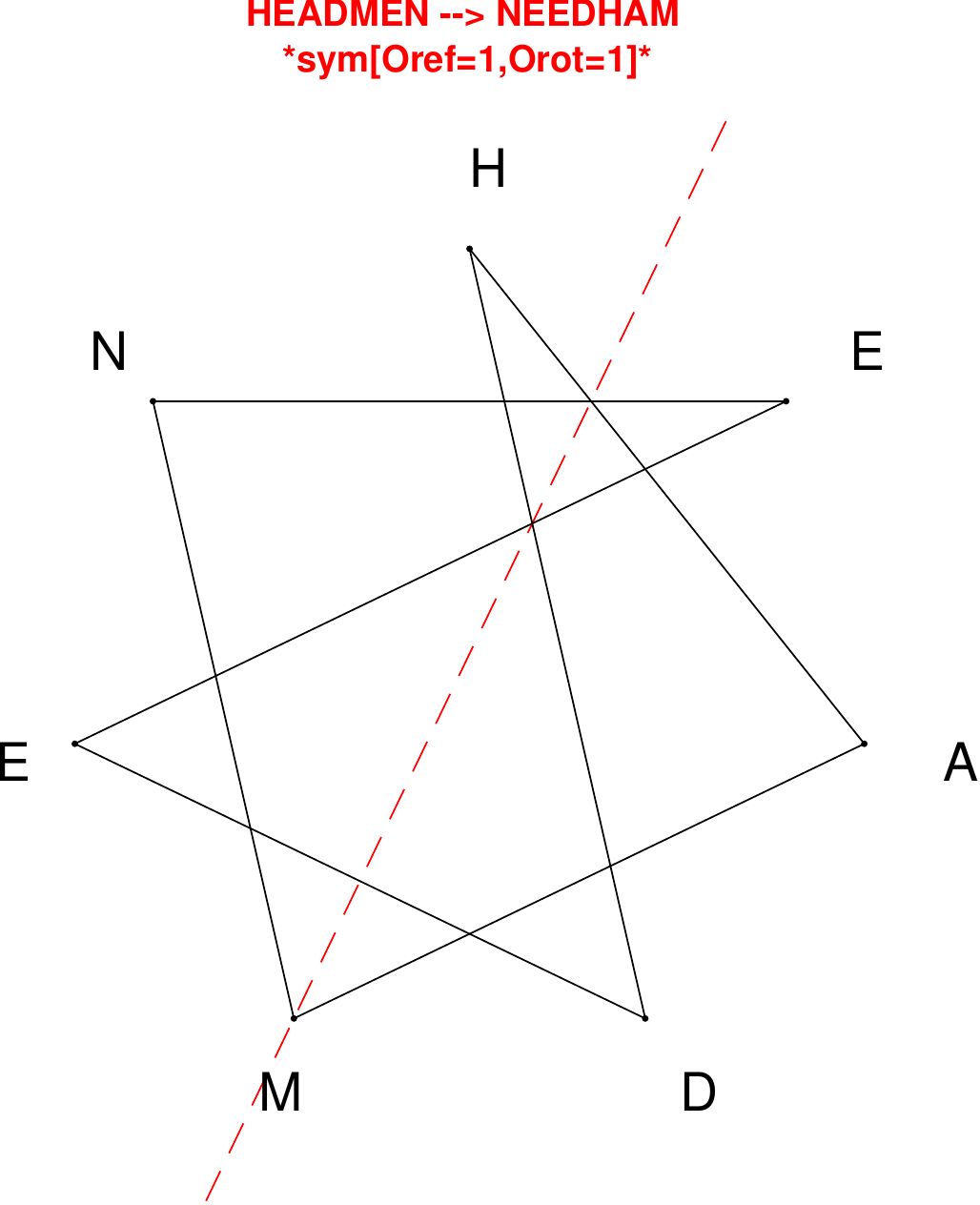}
\end{subfigure}
\end{figure}

\begin{figure}[H]
\centering
\begin{subfigure}[T]{0.19\textwidth}
\centering
\includegraphics[width=\textwidth]{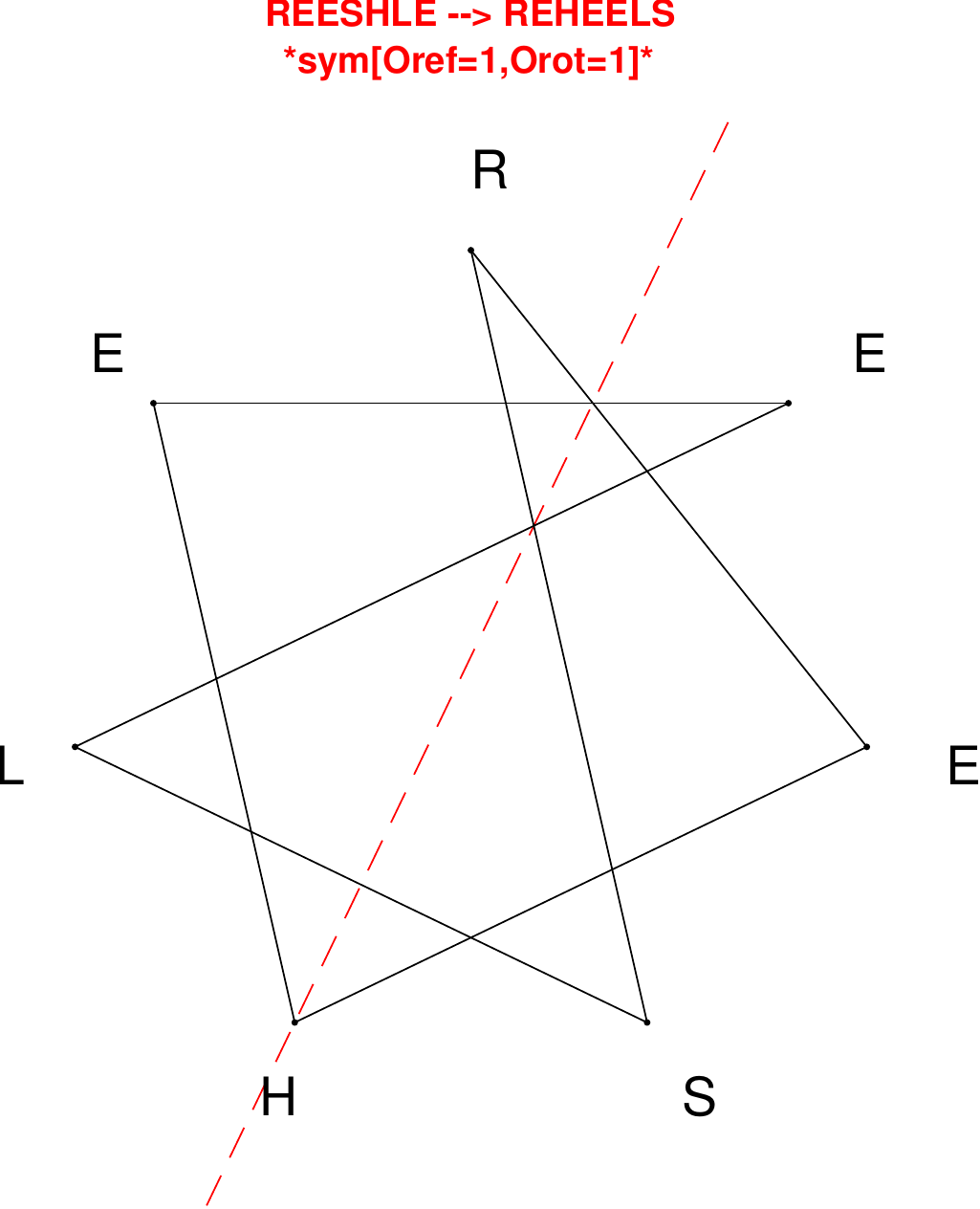}
\end{subfigure}
\hfill
\begin{subfigure}[T]{0.19\textwidth}
\centering
\includegraphics[width=\textwidth]{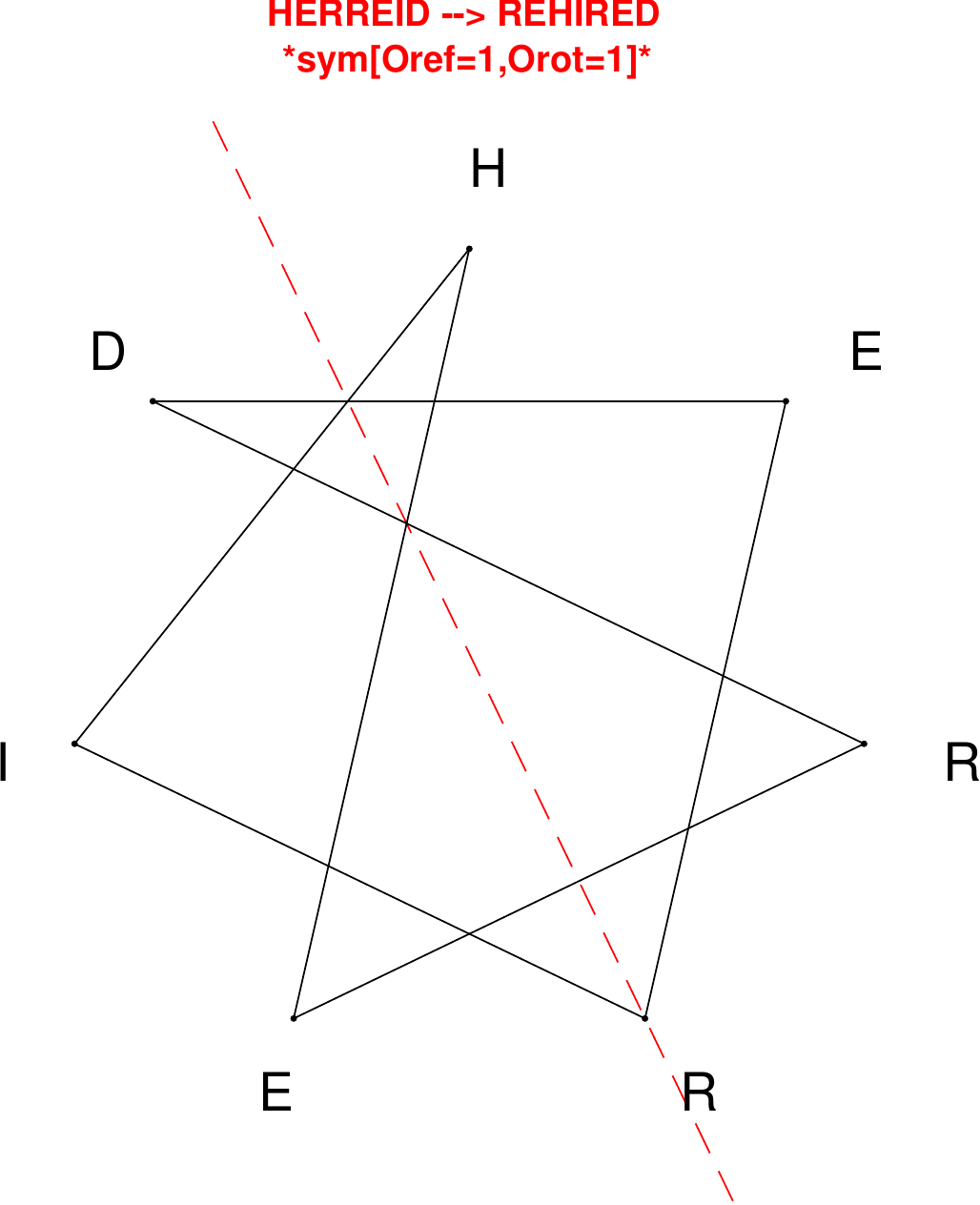}
\end{subfigure}
\hfill
\begin{subfigure}[T]{0.19\textwidth}
\centering
\includegraphics[width=\textwidth]{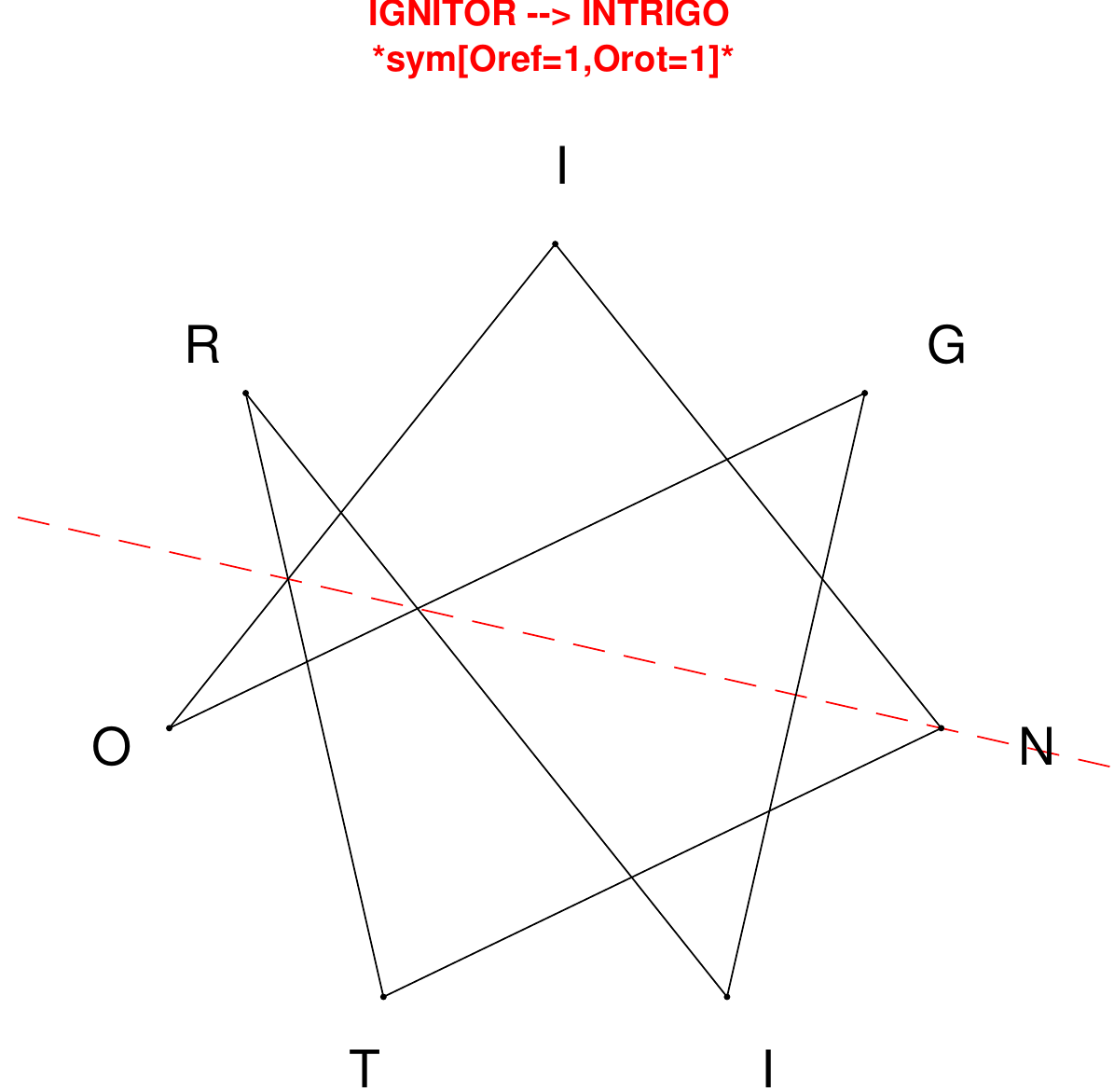}
\end{subfigure}
\hfill
\begin{subfigure}[T]{0.19\textwidth}
\centering
\includegraphics[width=\textwidth]{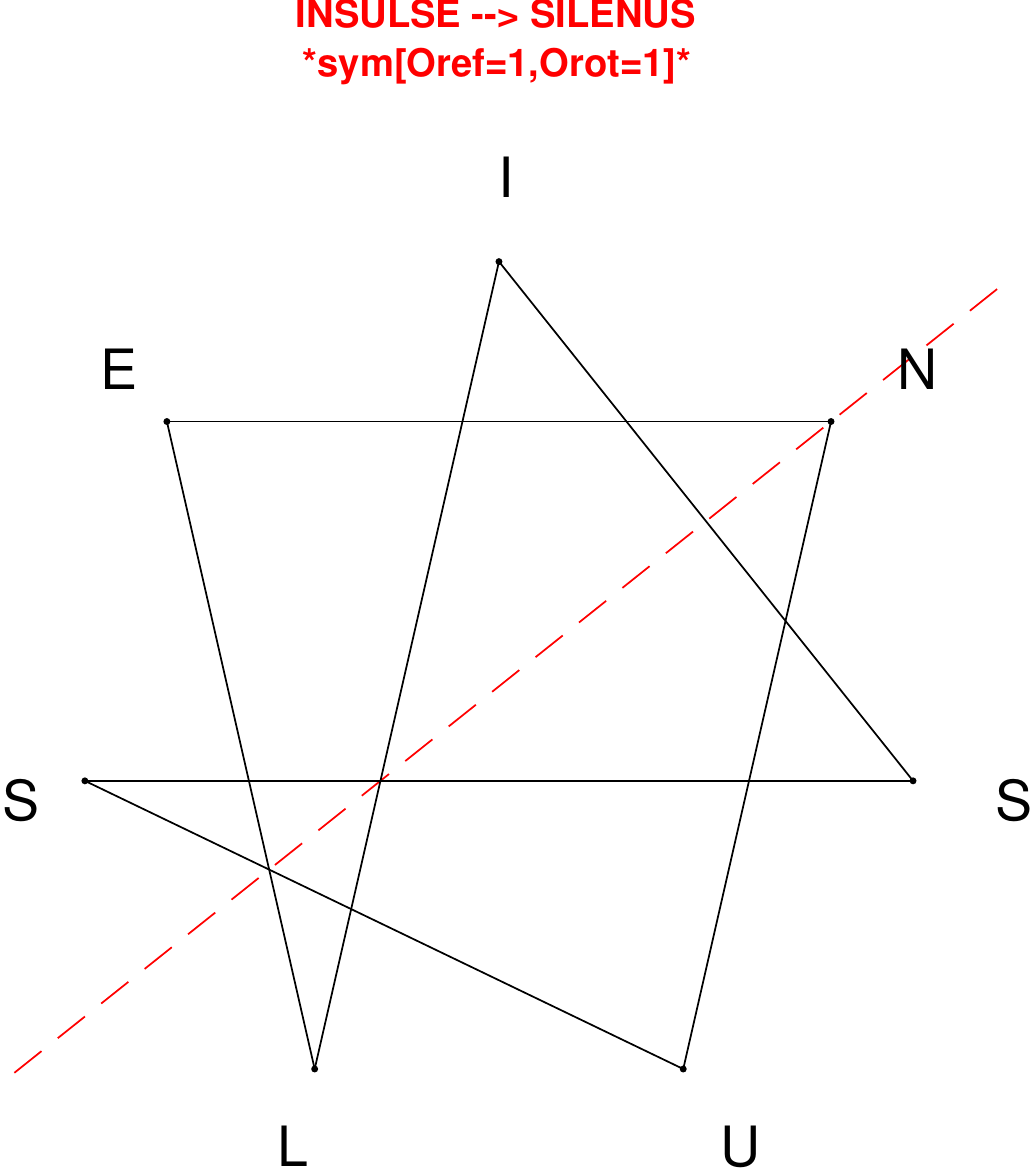}
\end{subfigure}
\hfill
\begin{subfigure}[T]{0.19\textwidth}
\centering
\includegraphics[width=\textwidth]{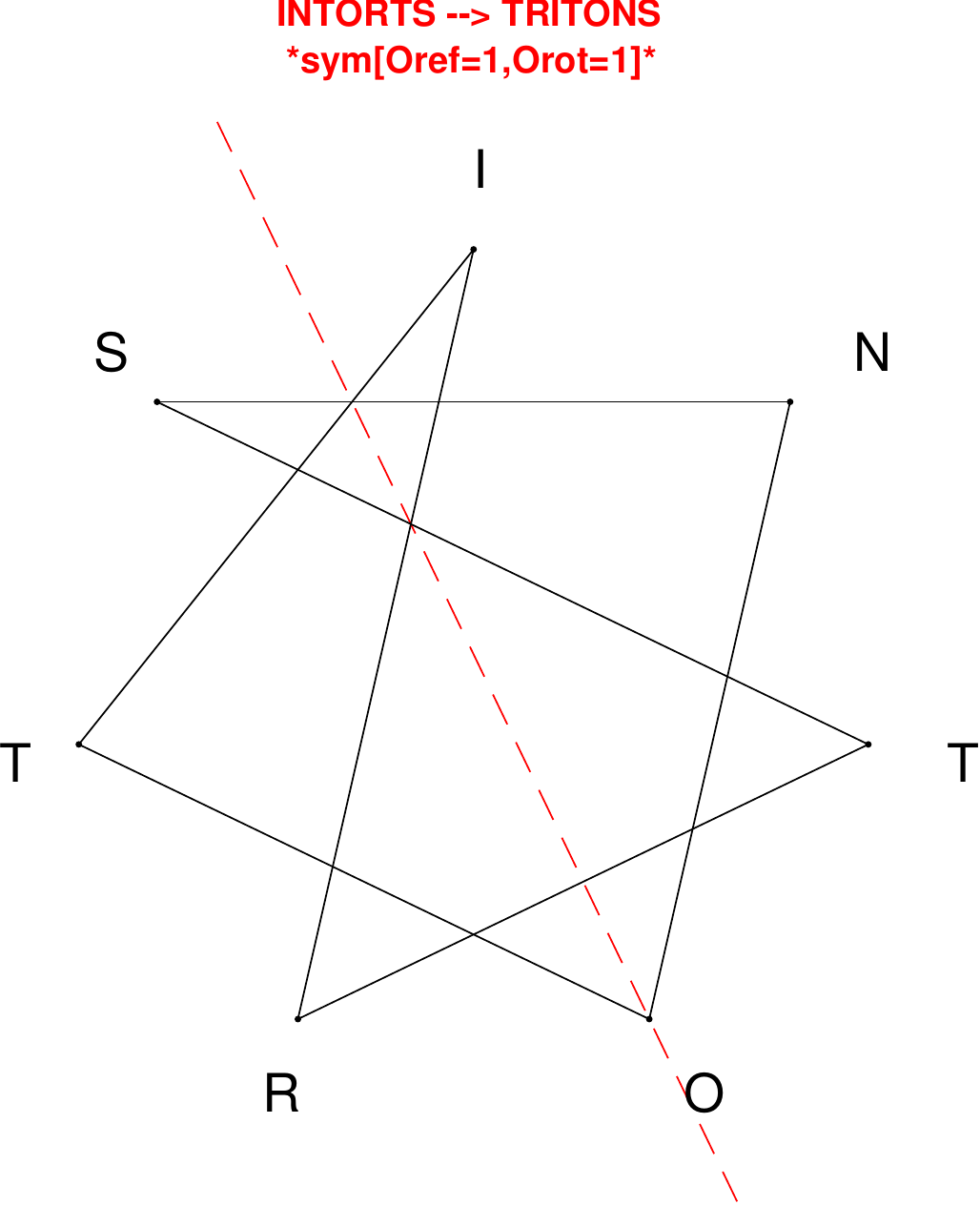}
\end{subfigure}
\end{figure}

\begin{figure}[H]
\centering
\begin{subfigure}[T]{0.19\textwidth}
\centering
\includegraphics[width=\textwidth]{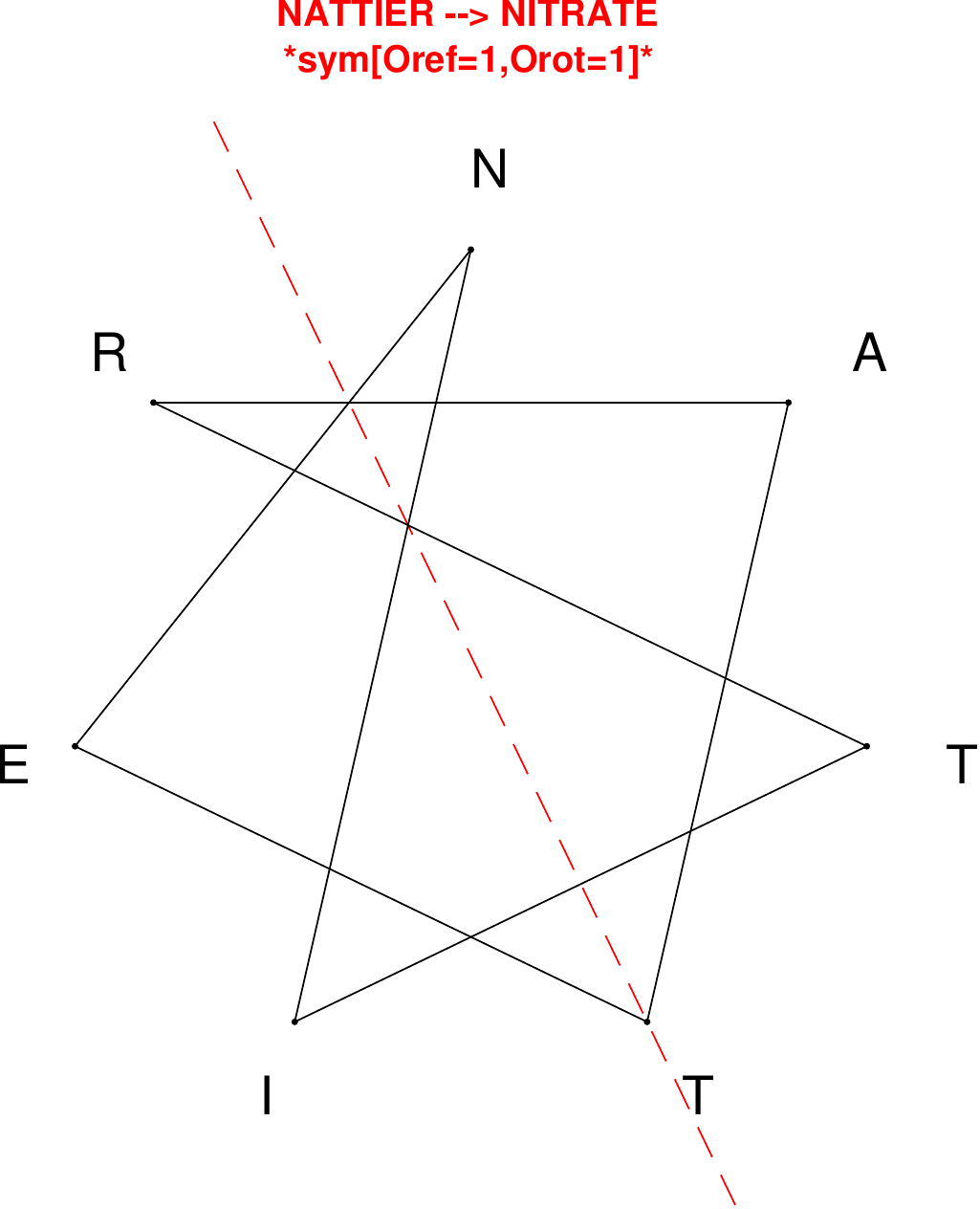}
\end{subfigure}
\hfill
\begin{subfigure}[T]{0.19\textwidth}
\centering
\includegraphics[width=\textwidth]{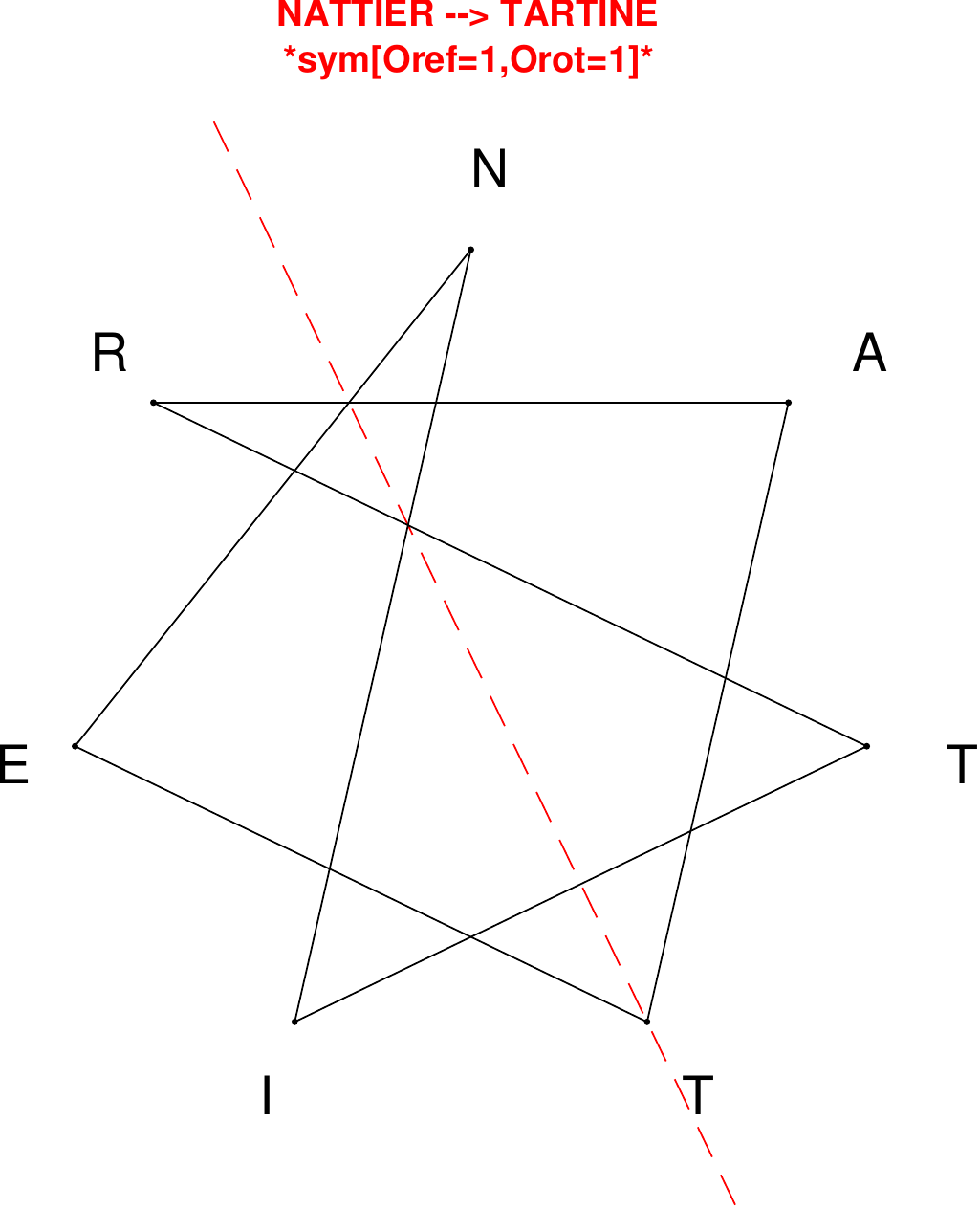}
\end{subfigure}
\hfill
\begin{subfigure}[T]{0.19\textwidth}
\centering
\includegraphics[width=\textwidth]{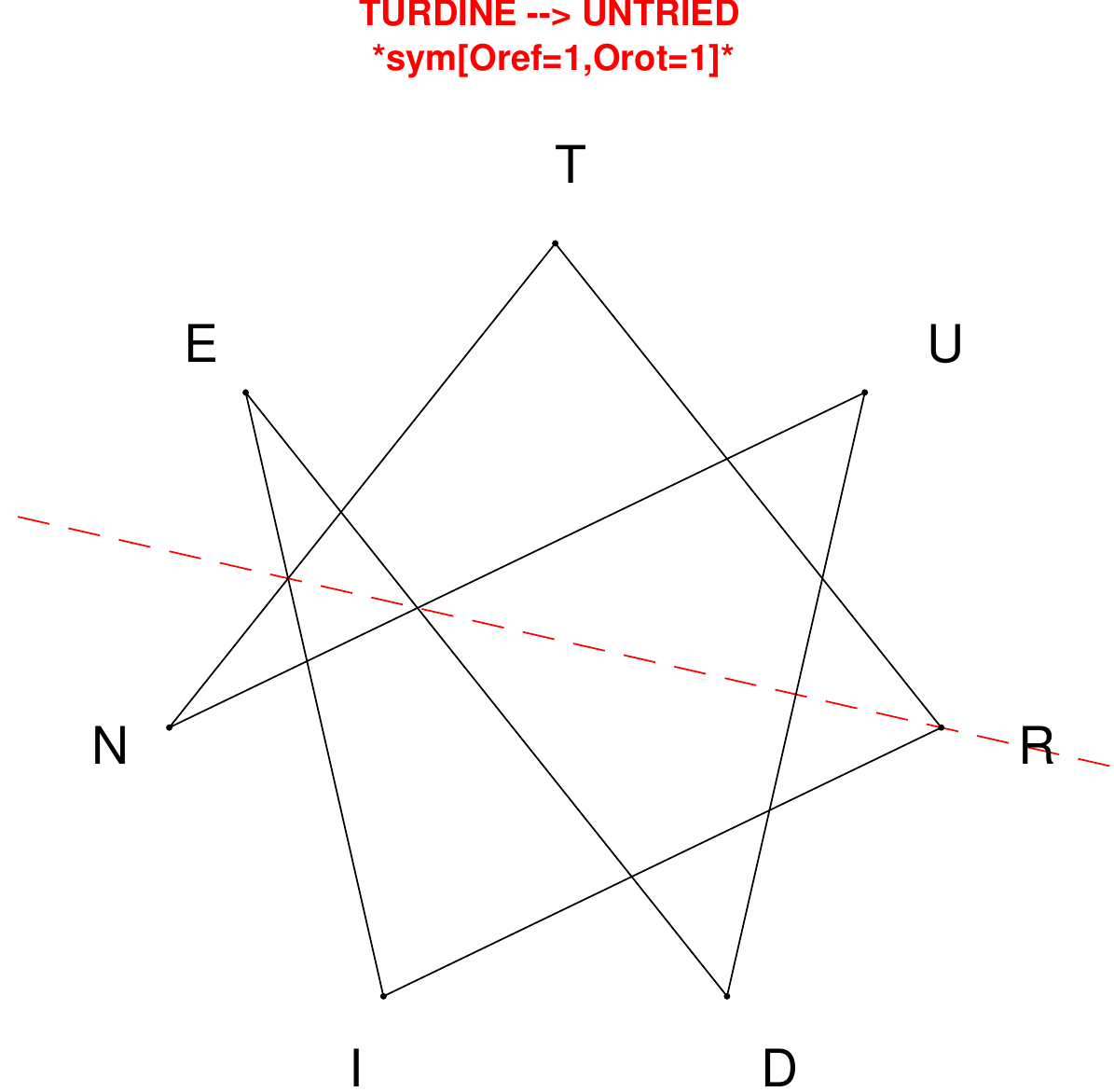}
\end{subfigure}
\hfill
\begin{subfigure}[T]{0.19\textwidth}
\centering
\includegraphics[width=\textwidth]{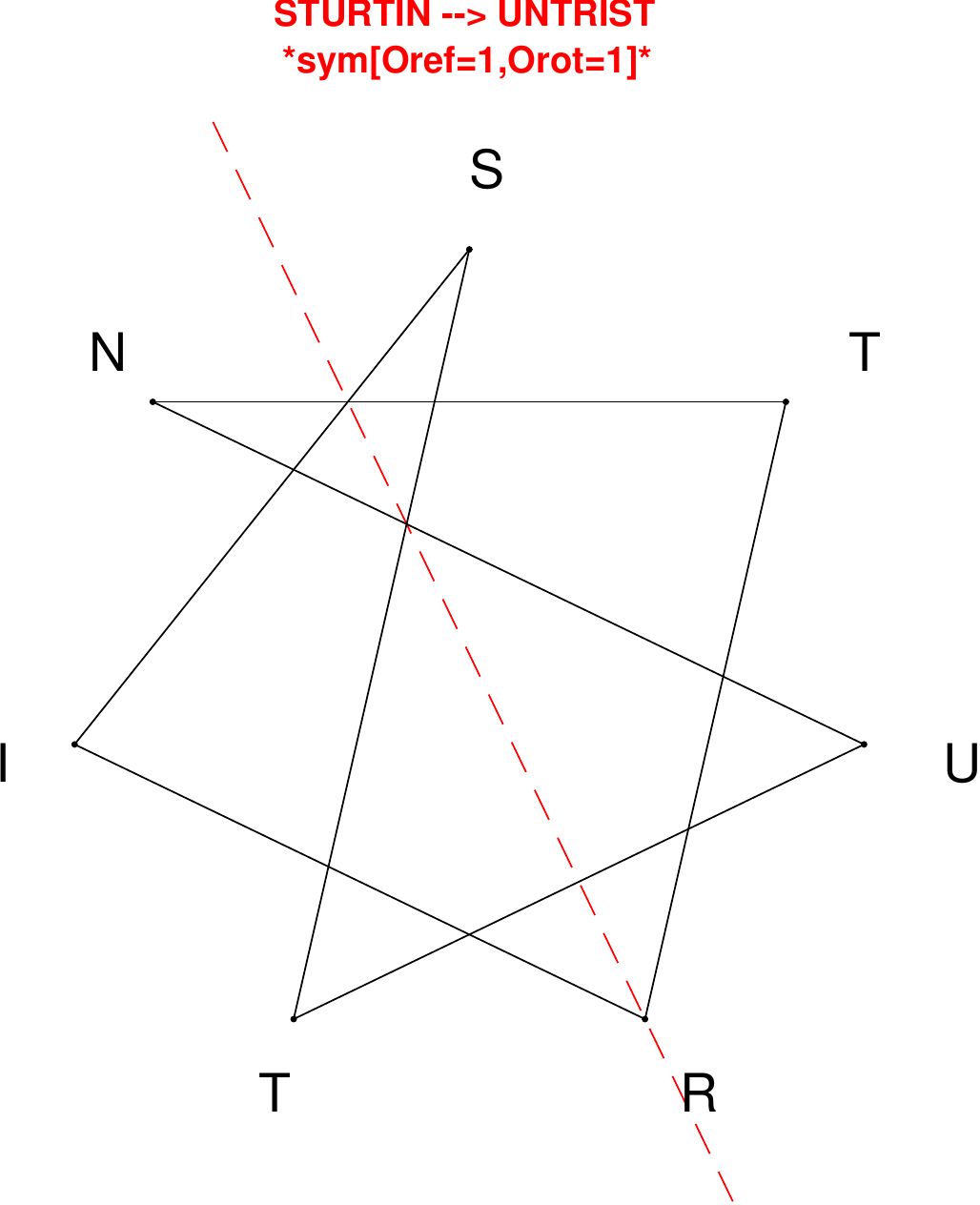}
\end{subfigure}
\hfill
\begin{subfigure}[T]{0.19\textwidth}
\centering
\includegraphics[width=\textwidth]{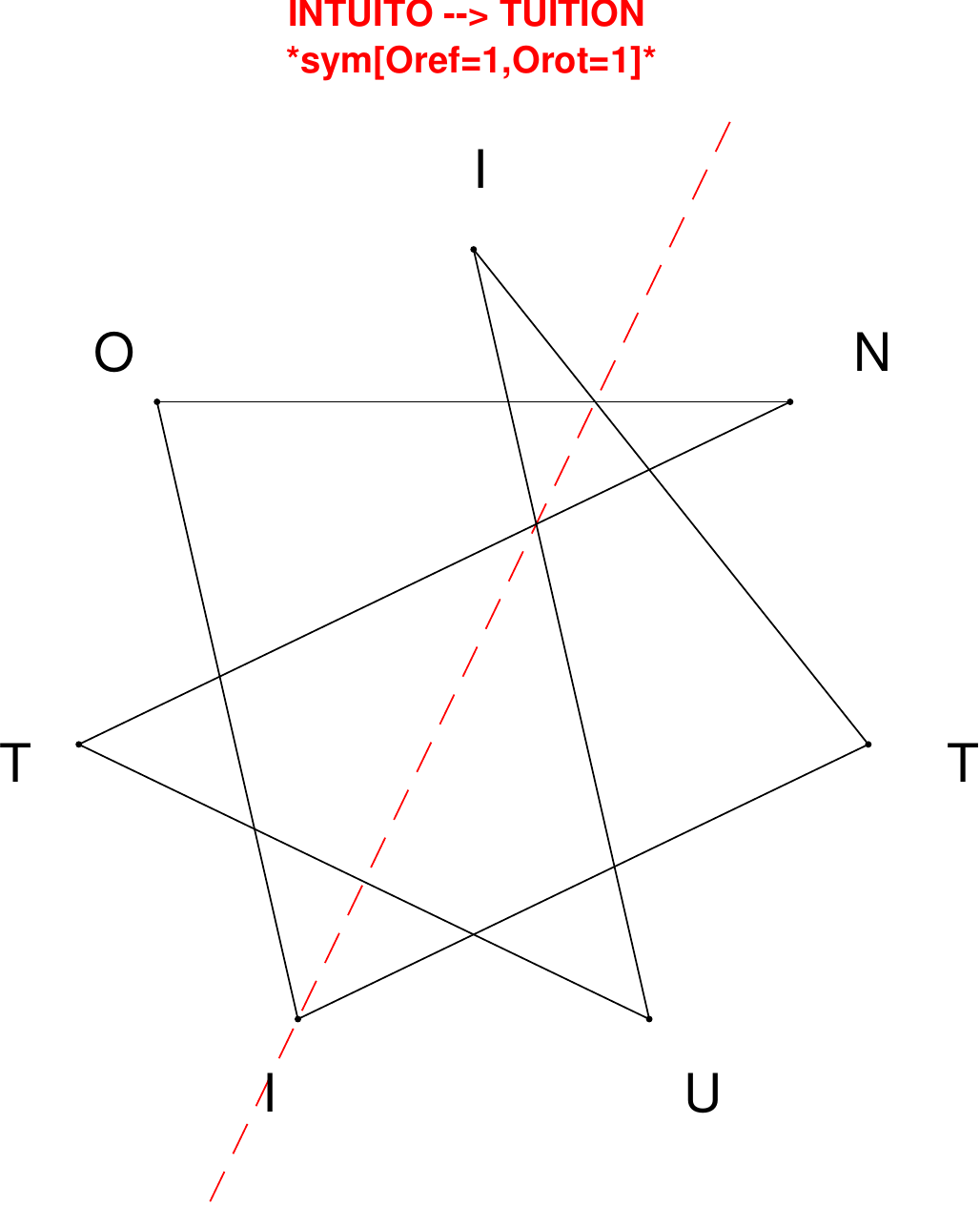}
\end{subfigure}
\end{figure}

\begin{figure}[H]
\centering
\begin{subfigure}[T]{0.19\textwidth}
\centering
\includegraphics[width=\textwidth]{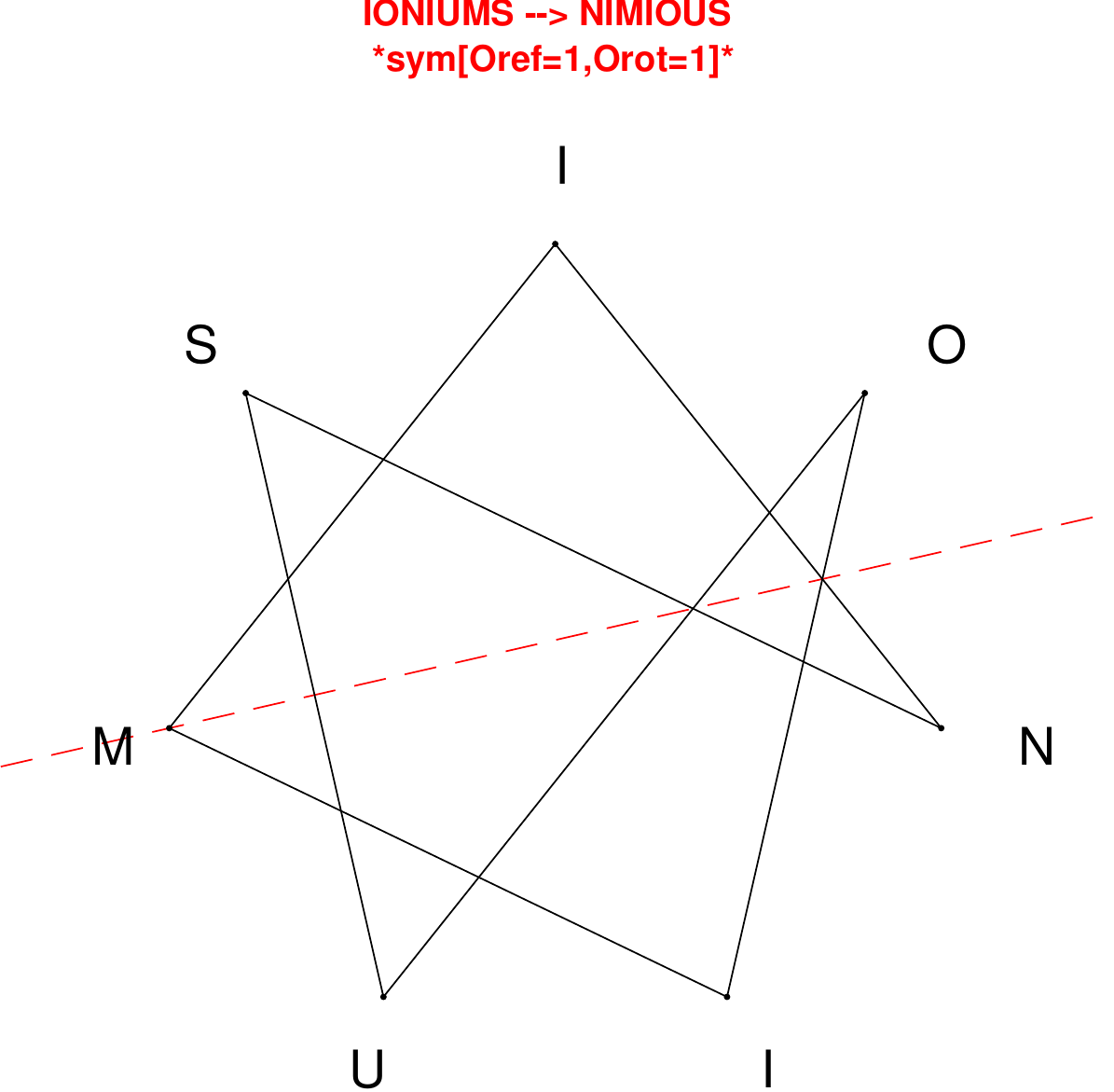}
\end{subfigure}
\hfill
\begin{subfigure}[T]{0.19\textwidth}
\centering
\includegraphics[width=\textwidth]{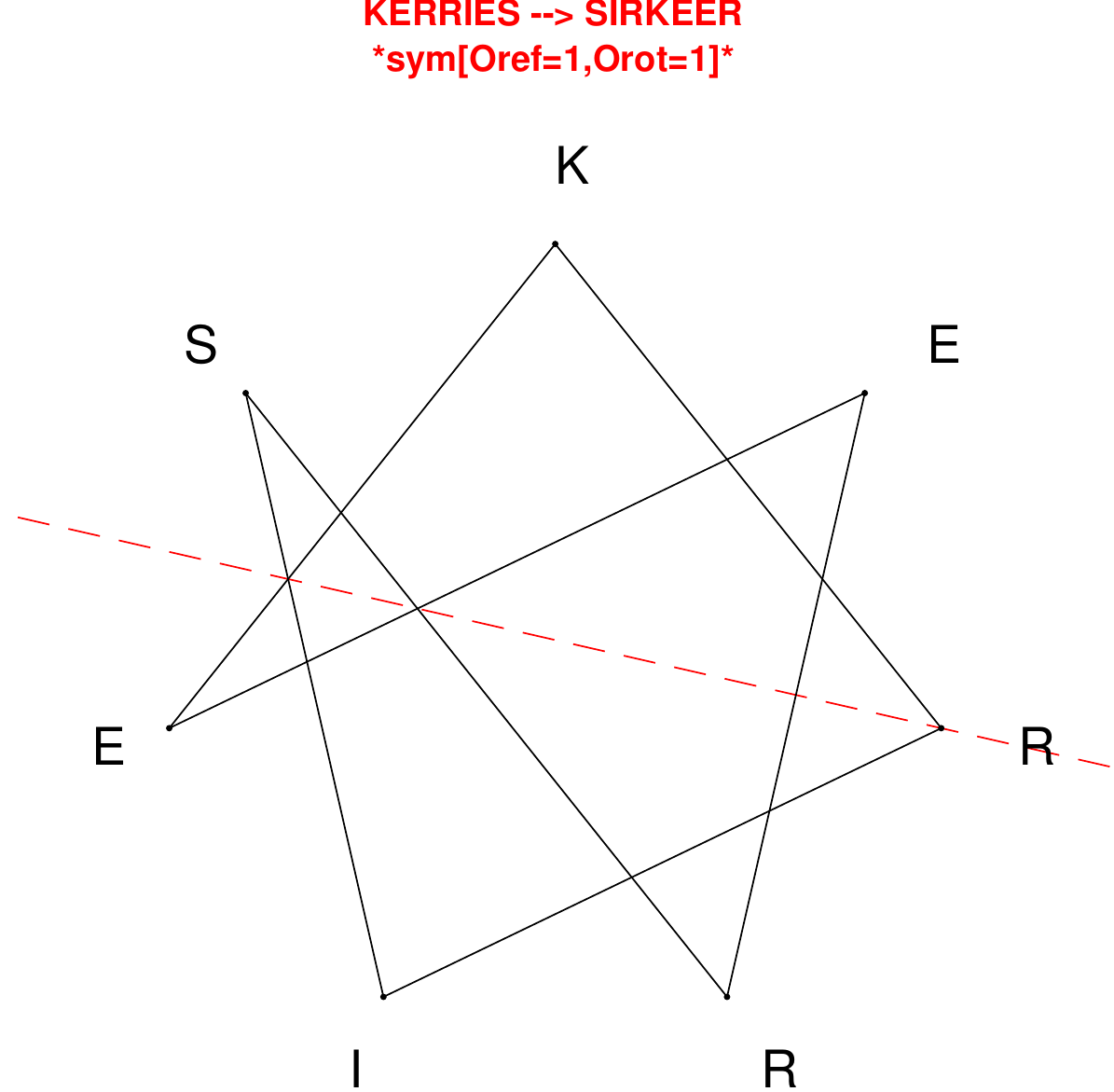}
\end{subfigure}
\hfill
\begin{subfigure}[T]{0.19\textwidth}
\centering
\includegraphics[width=\textwidth]{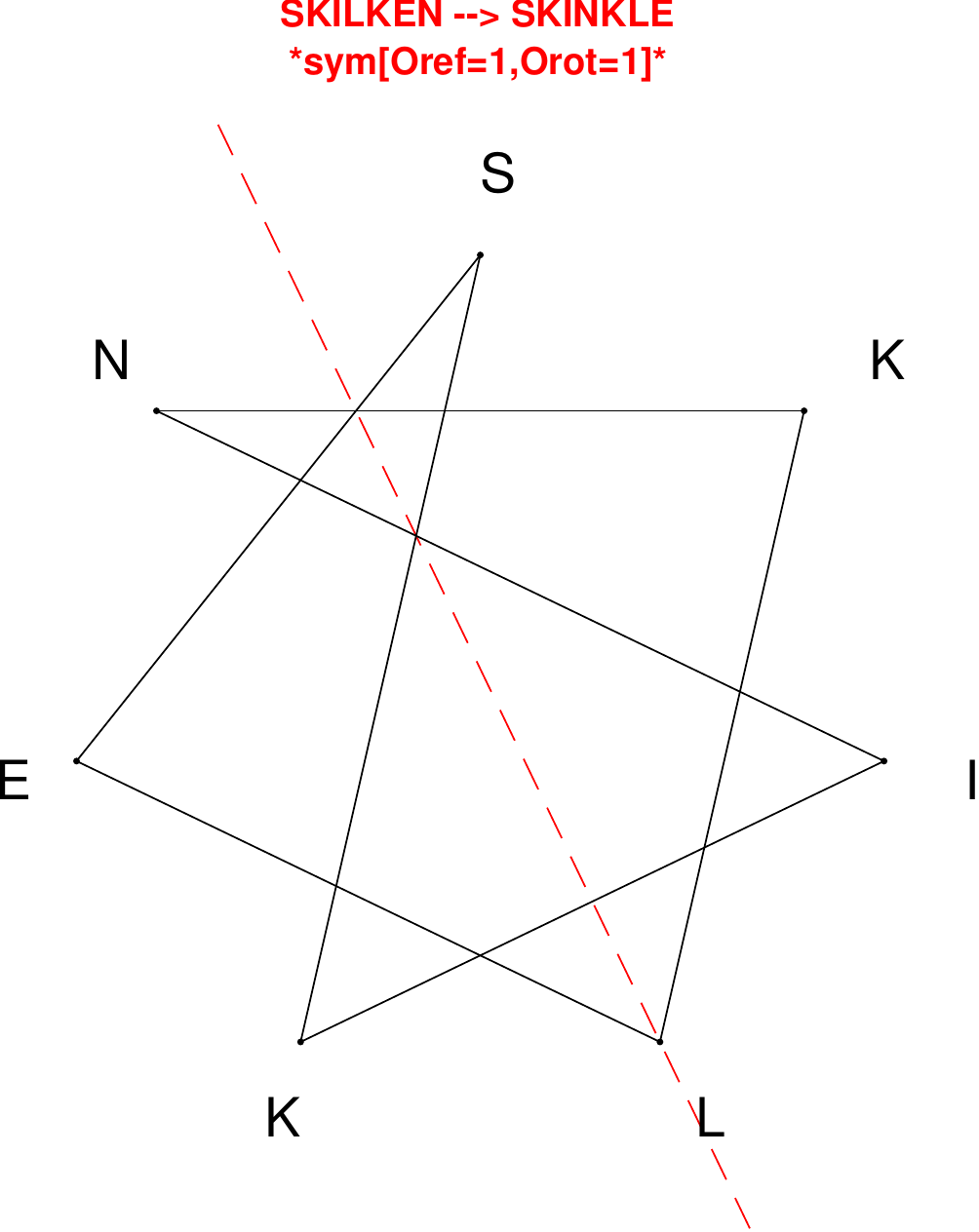}
\end{subfigure}
\hfill
\begin{subfigure}[T]{0.19\textwidth}
\centering
\includegraphics[width=\textwidth]{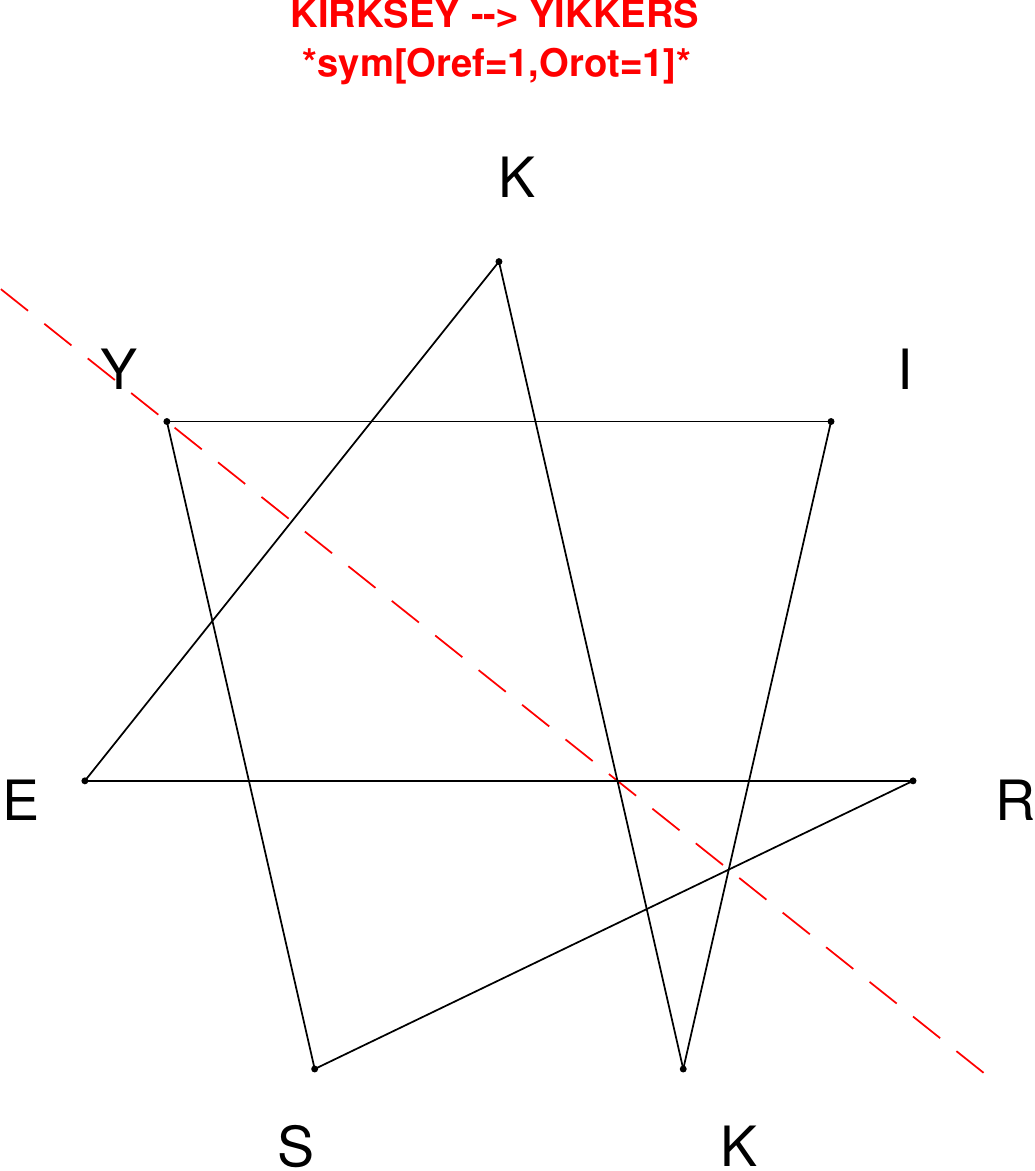}
\end{subfigure}
\hfill
\begin{subfigure}[T]{0.19\textwidth}
\centering
\includegraphics[width=\textwidth]{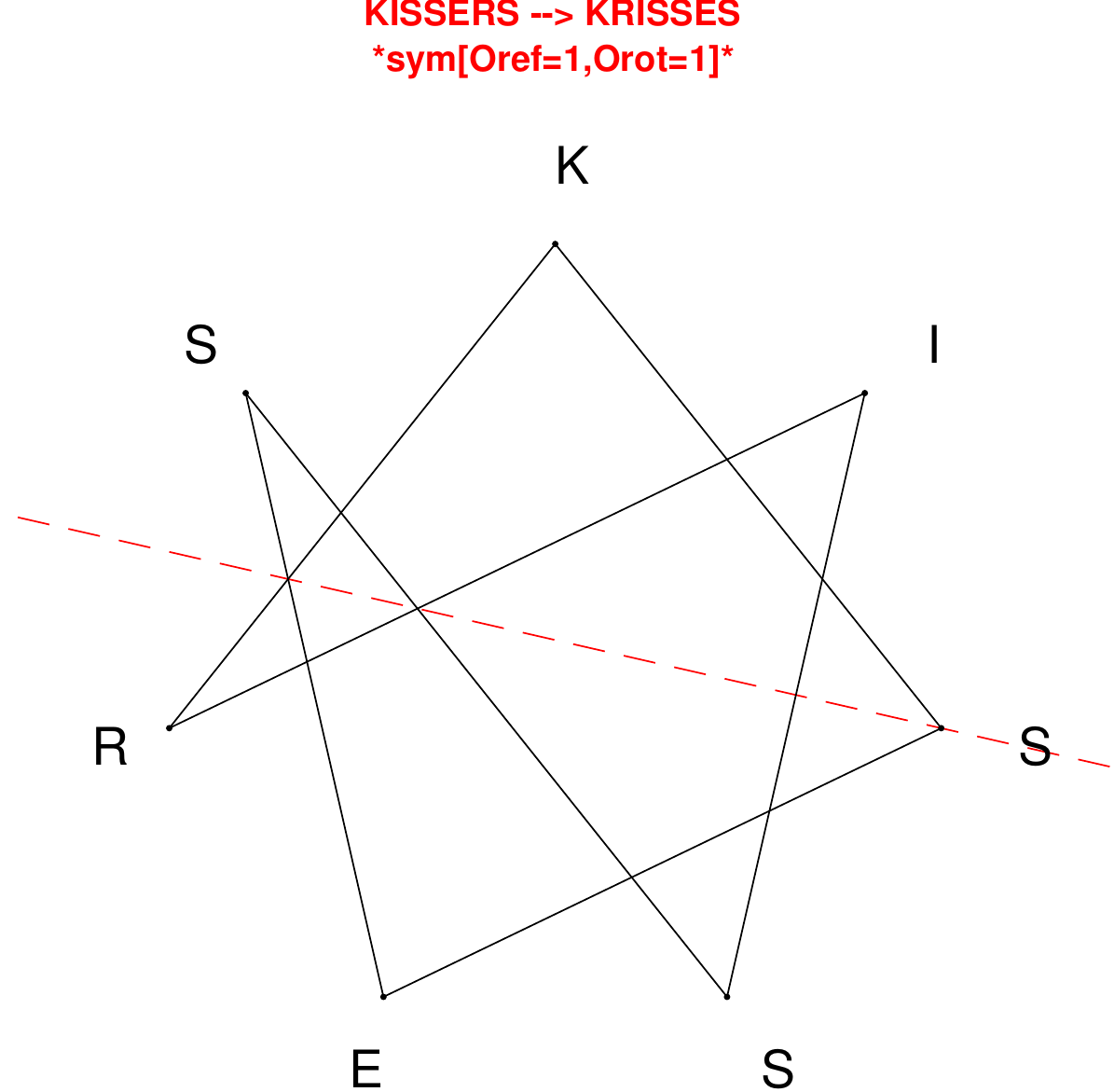}
\end{subfigure}
\end{figure}

\begin{figure}[H]
\centering
\begin{subfigure}[T]{0.19\textwidth}
\centering
\includegraphics[width=\textwidth]{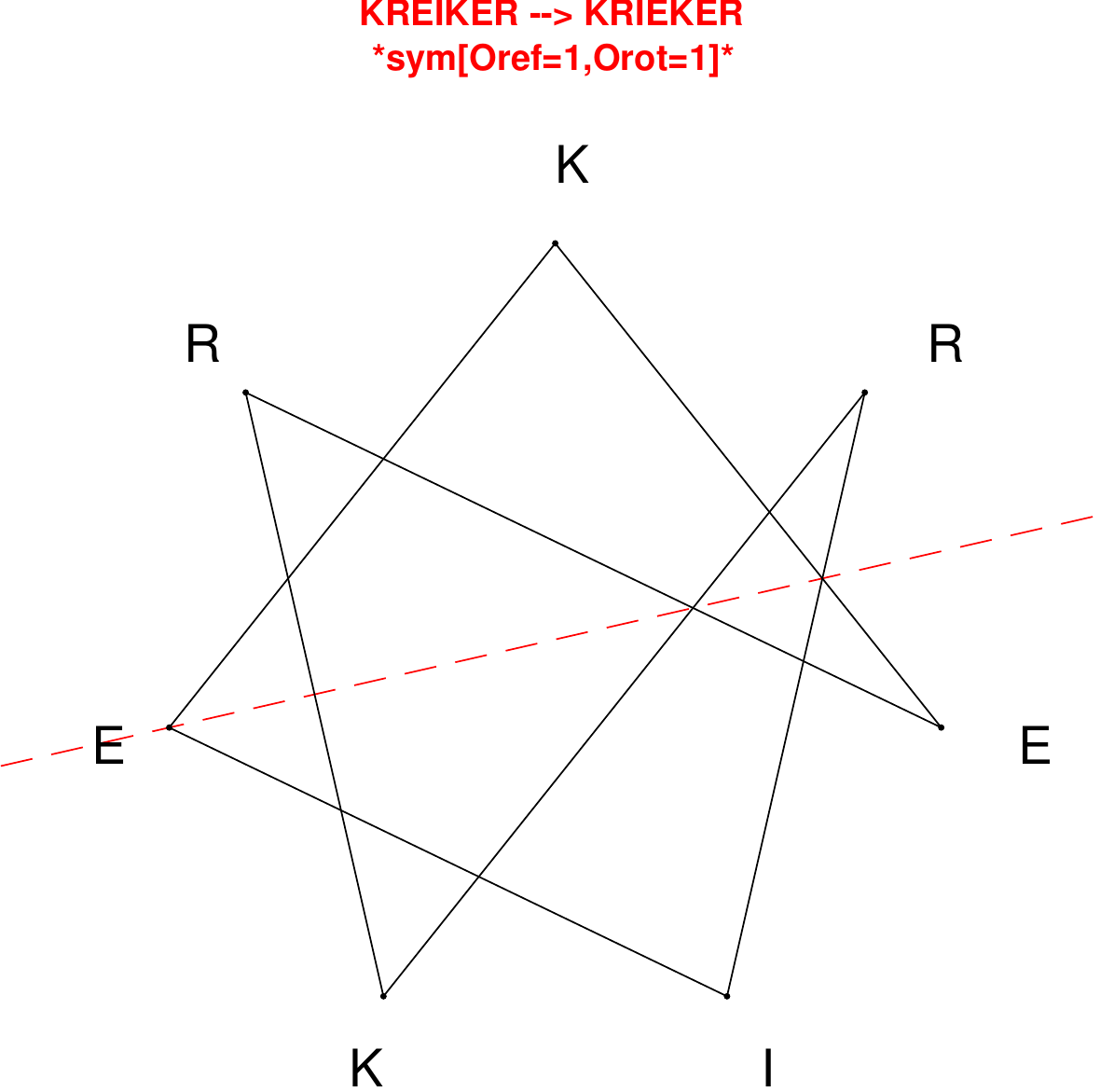}
\end{subfigure}
\hfill
\begin{subfigure}[T]{0.19\textwidth}
\centering
\includegraphics[width=\textwidth]{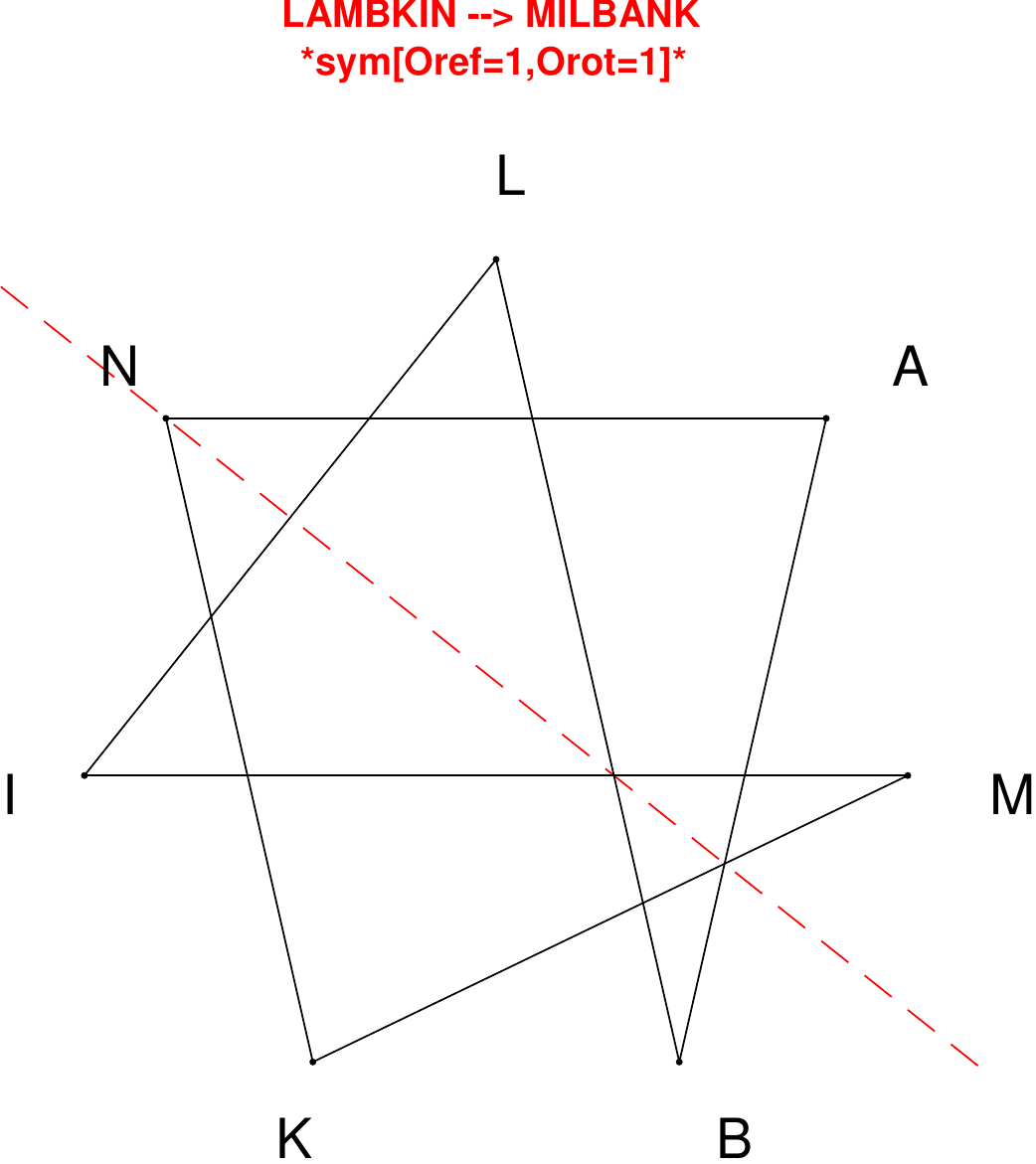}
\end{subfigure}
\hfill
\begin{subfigure}[T]{0.19\textwidth}
\centering
\includegraphics[width=\textwidth]{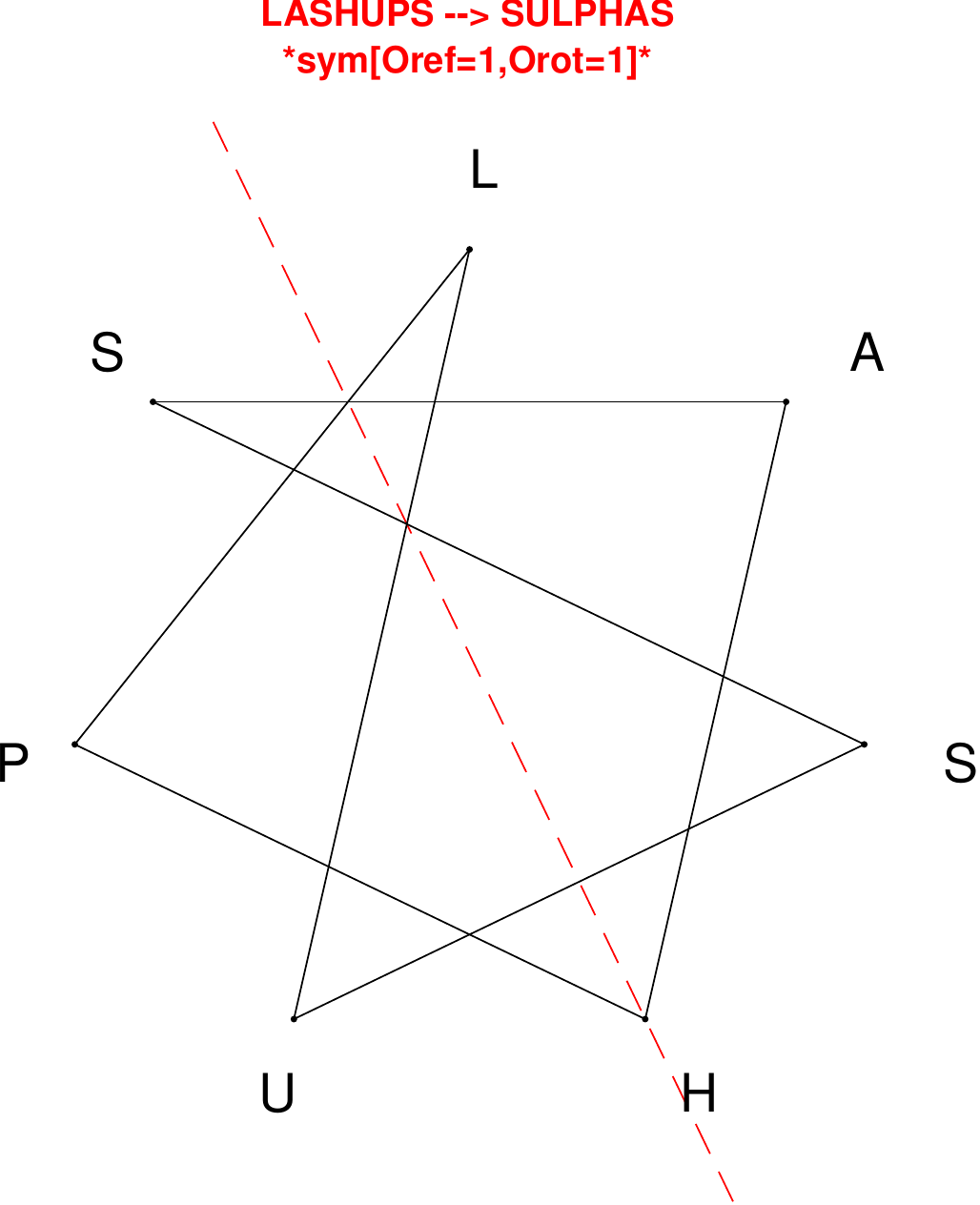}
\end{subfigure}
\hfill
\begin{subfigure}[T]{0.19\textwidth}
\centering
\includegraphics[width=\textwidth]{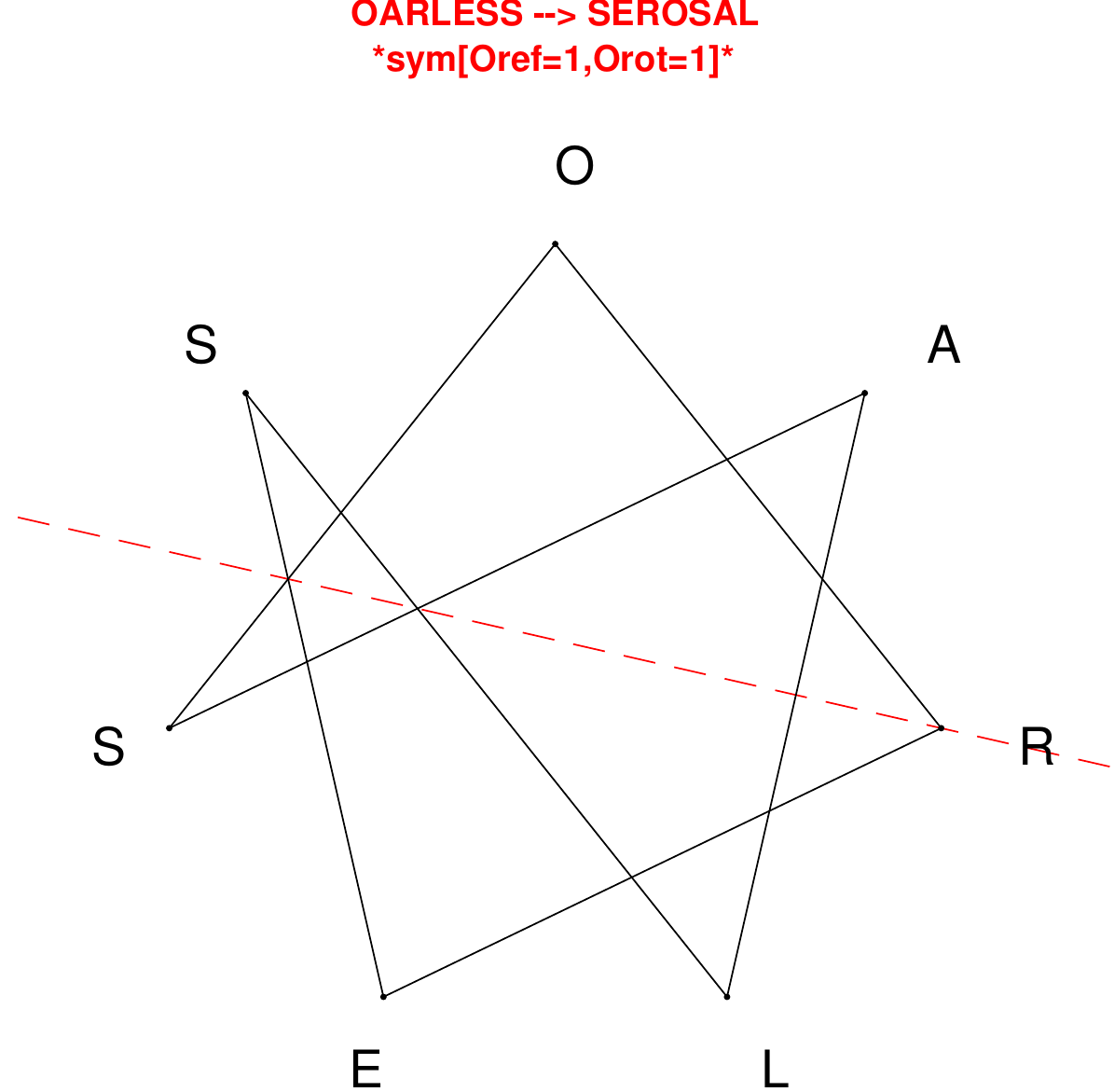}
\end{subfigure}
\hfill
\begin{subfigure}[T]{0.19\textwidth}
\centering
\includegraphics[width=\textwidth]{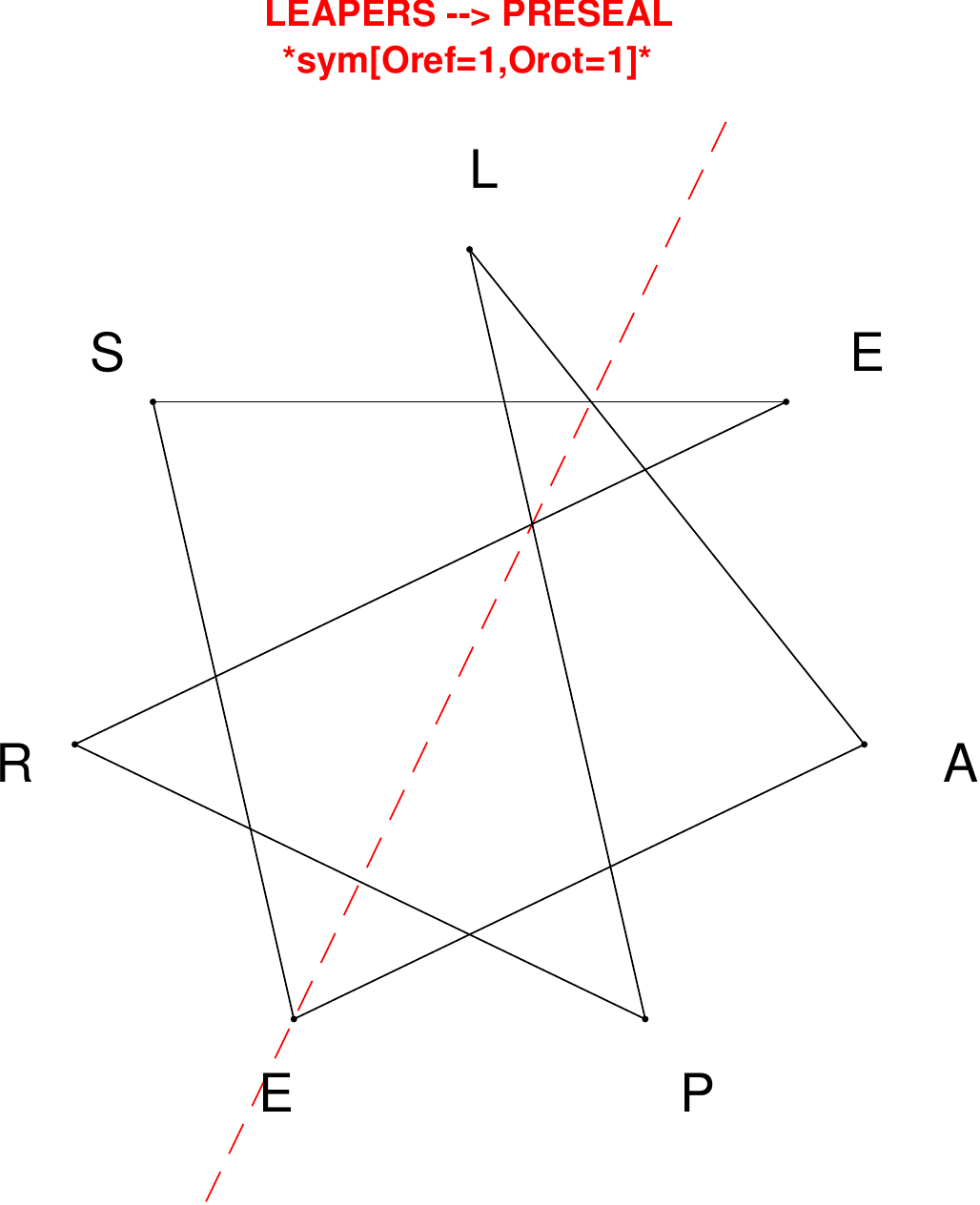}
\end{subfigure}
\end{figure}

\begin{figure}[H]
\centering
\begin{subfigure}[T]{0.19\textwidth}
\centering
\includegraphics[width=\textwidth]{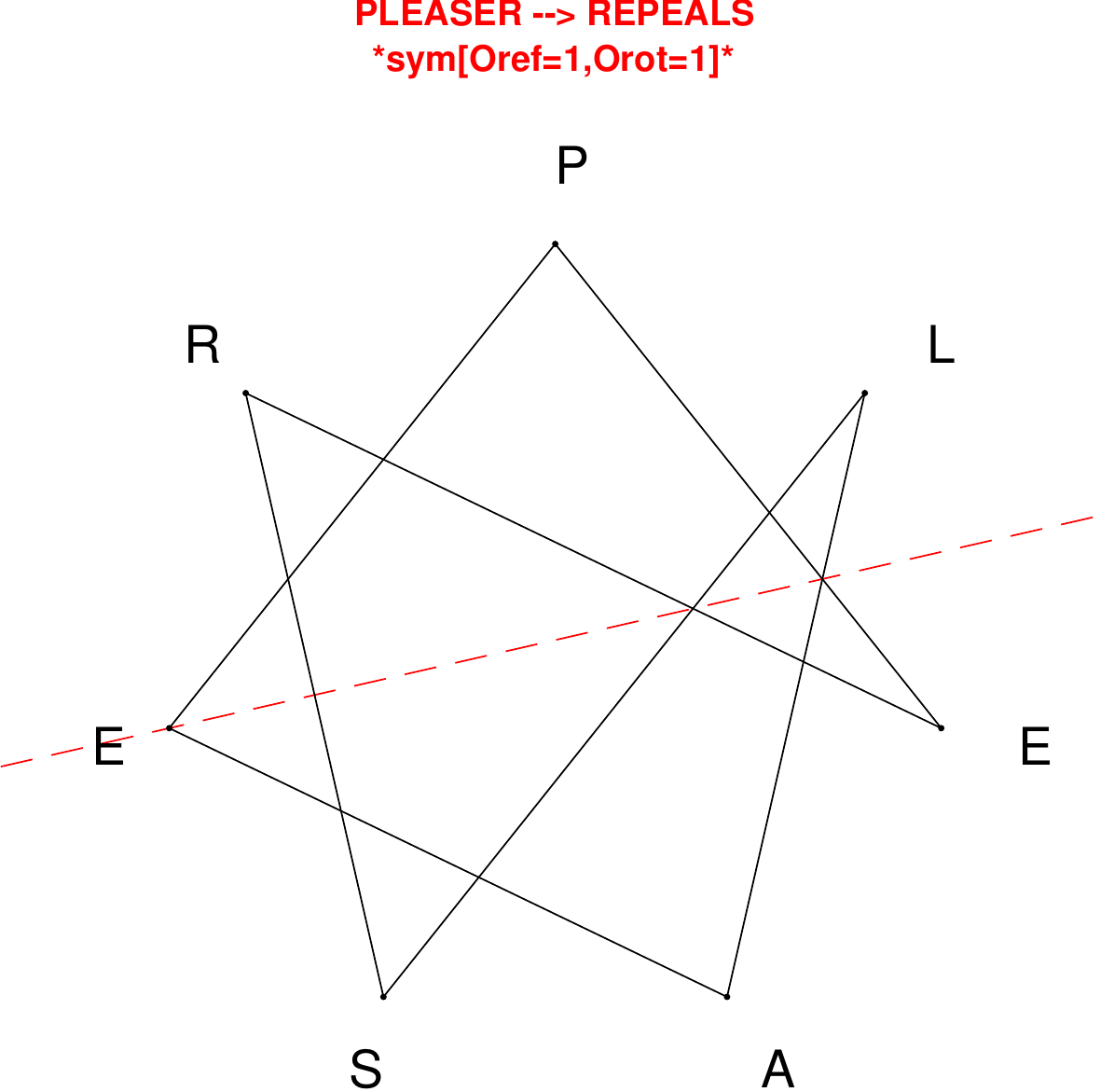}
\end{subfigure}
\hfill
\begin{subfigure}[T]{0.19\textwidth}
\centering
\includegraphics[width=\textwidth]{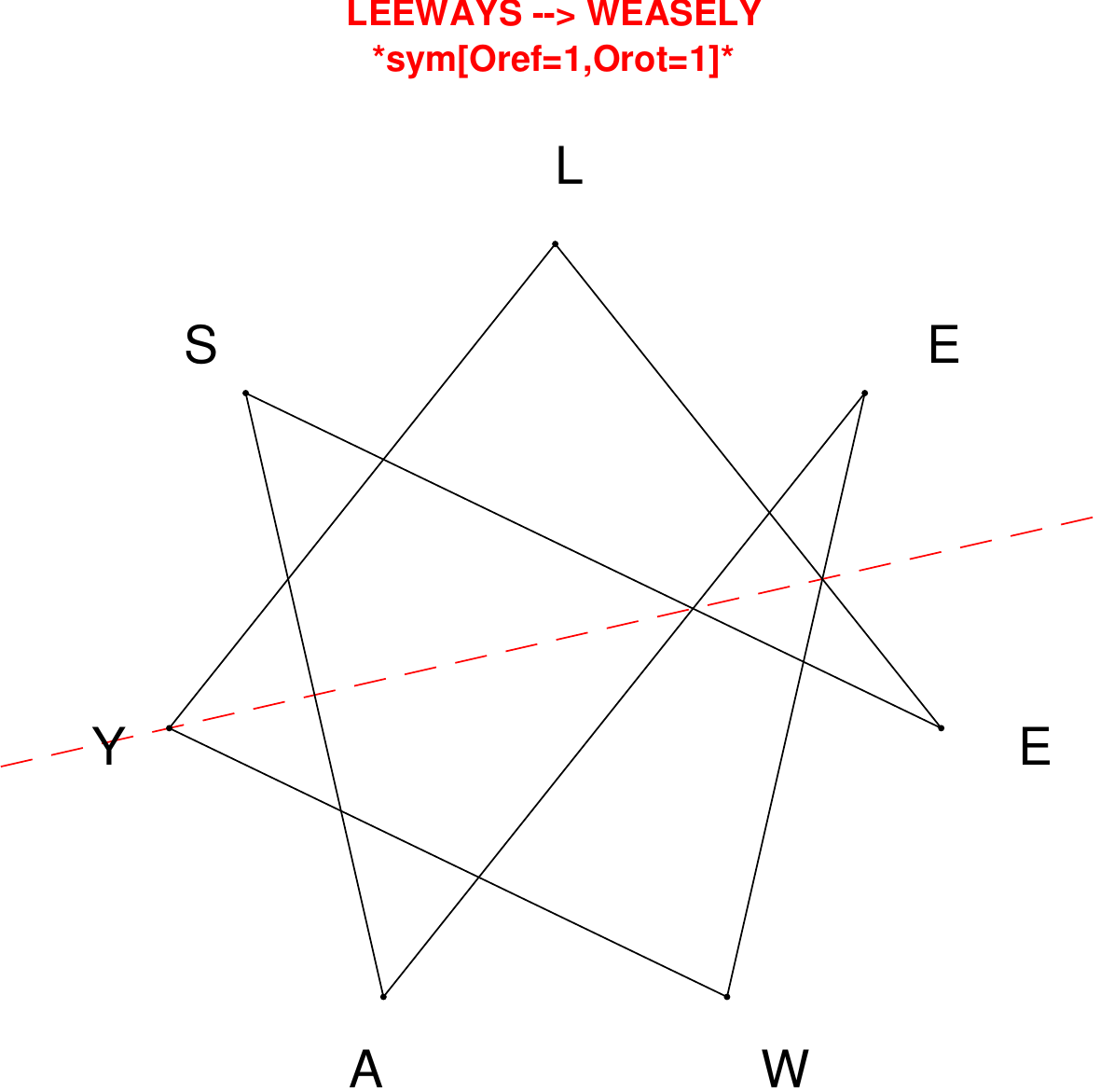}
\end{subfigure}
\hfill
\begin{subfigure}[T]{0.19\textwidth}
\centering
\includegraphics[width=\textwidth]{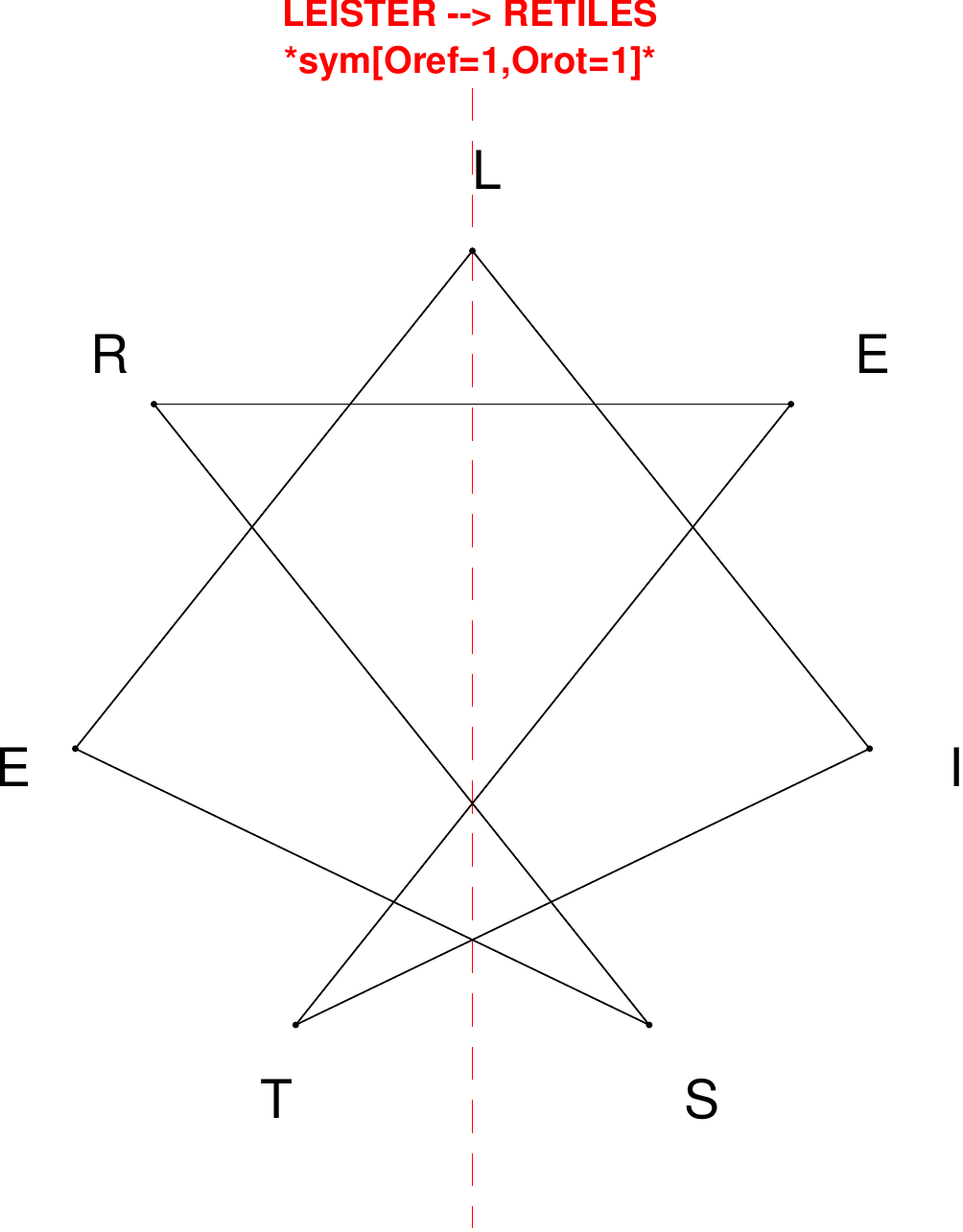}
\end{subfigure}
\hfill
\begin{subfigure}[T]{0.19\textwidth}
\centering
\includegraphics[width=\textwidth]{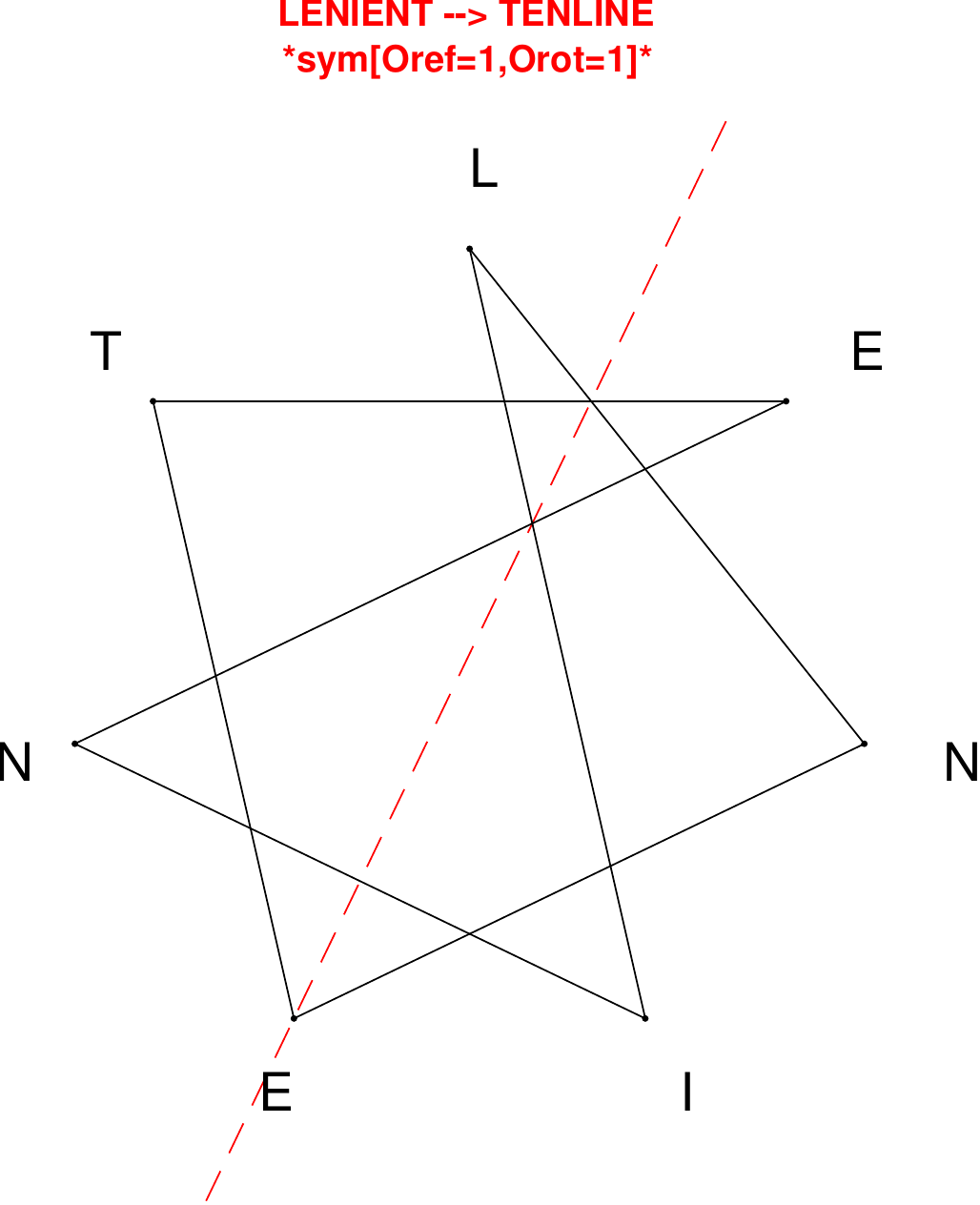}
\end{subfigure}
\hfill
\begin{subfigure}[T]{0.19\textwidth}
\centering
\includegraphics[width=\textwidth]{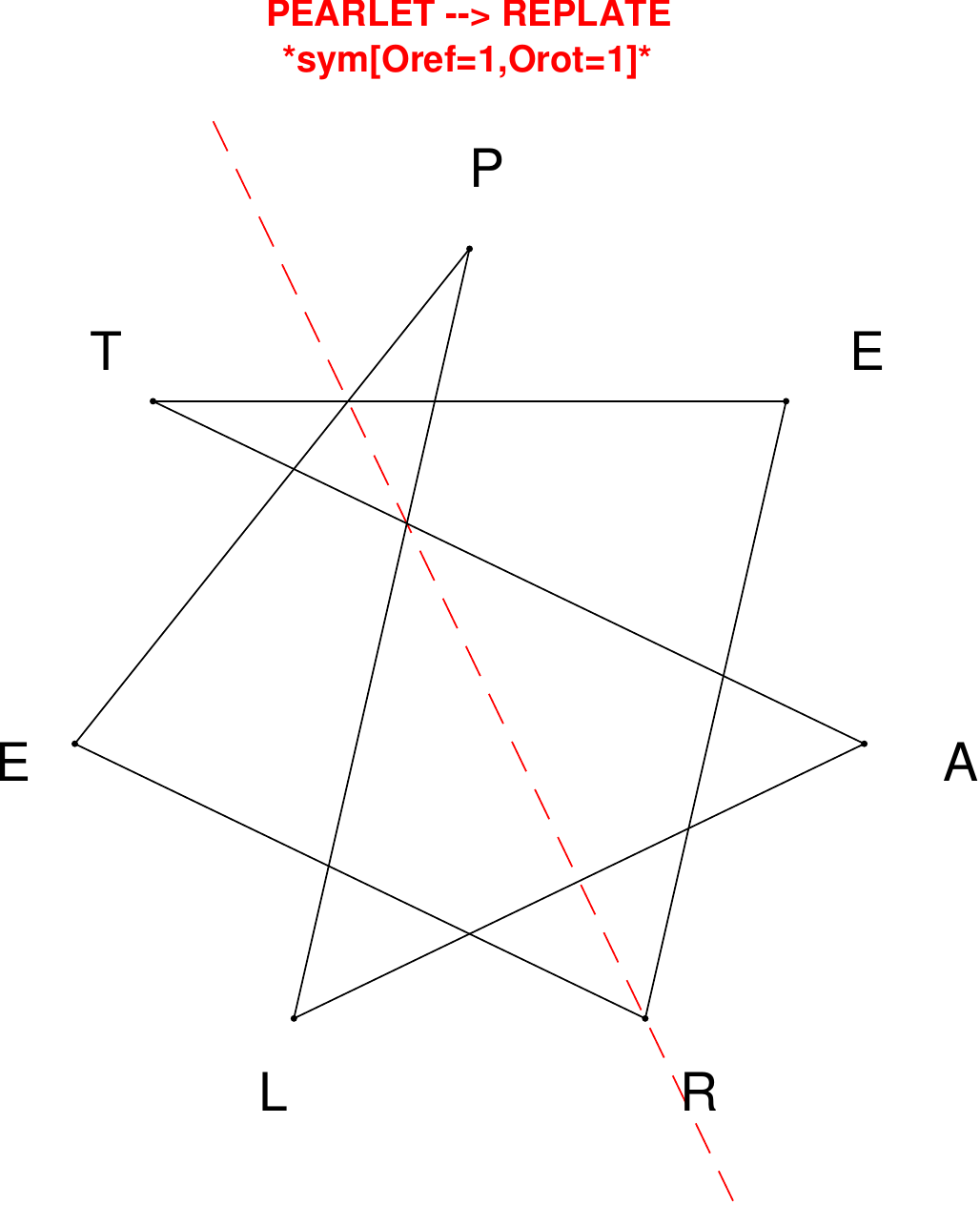}
\end{subfigure}
\end{figure}

\begin{figure}[H]
\centering
\begin{subfigure}[T]{0.19\textwidth}
\centering
\includegraphics[width=\textwidth]{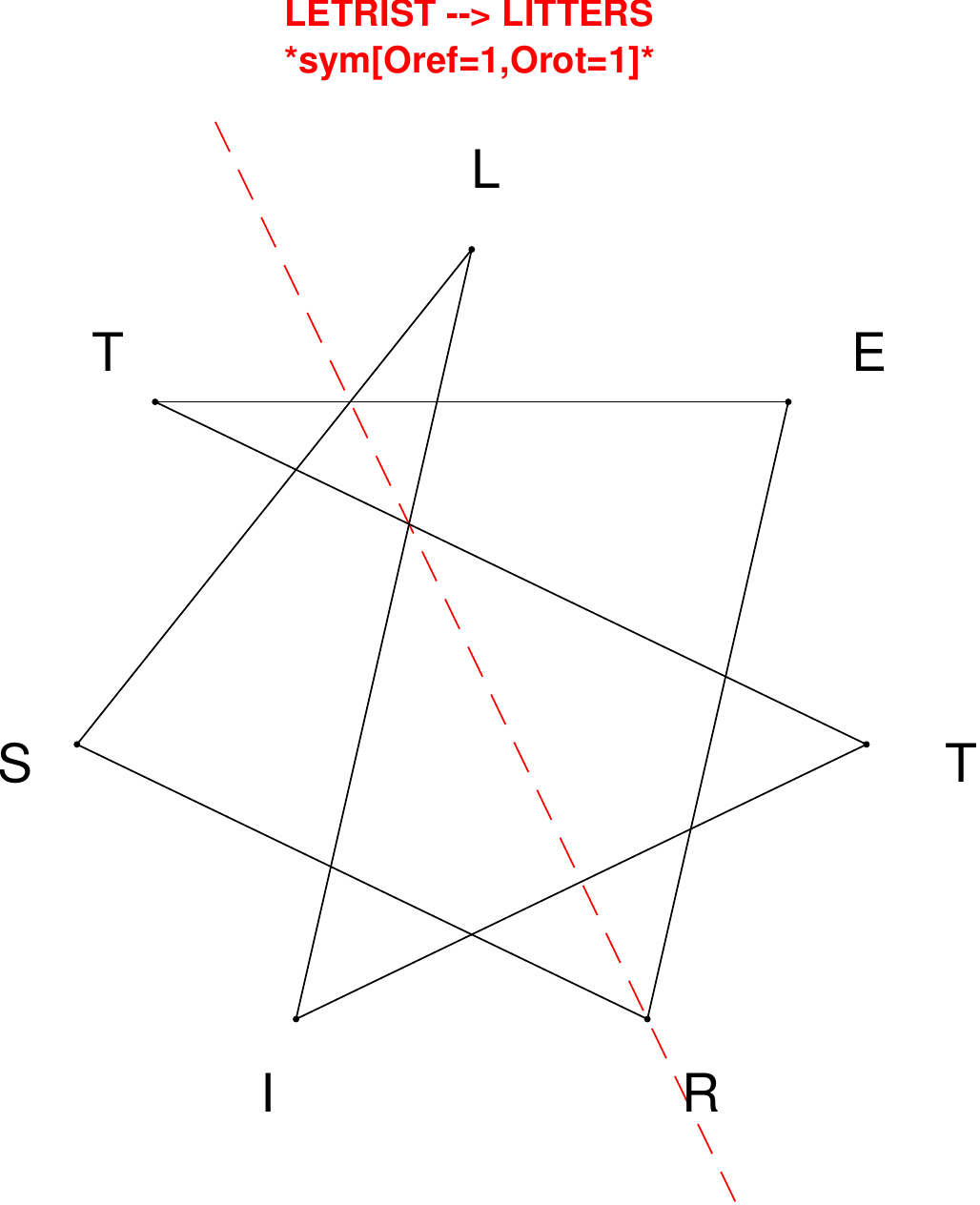}
\end{subfigure}
\hfill
\begin{subfigure}[T]{0.19\textwidth}
\centering
\includegraphics[width=\textwidth]{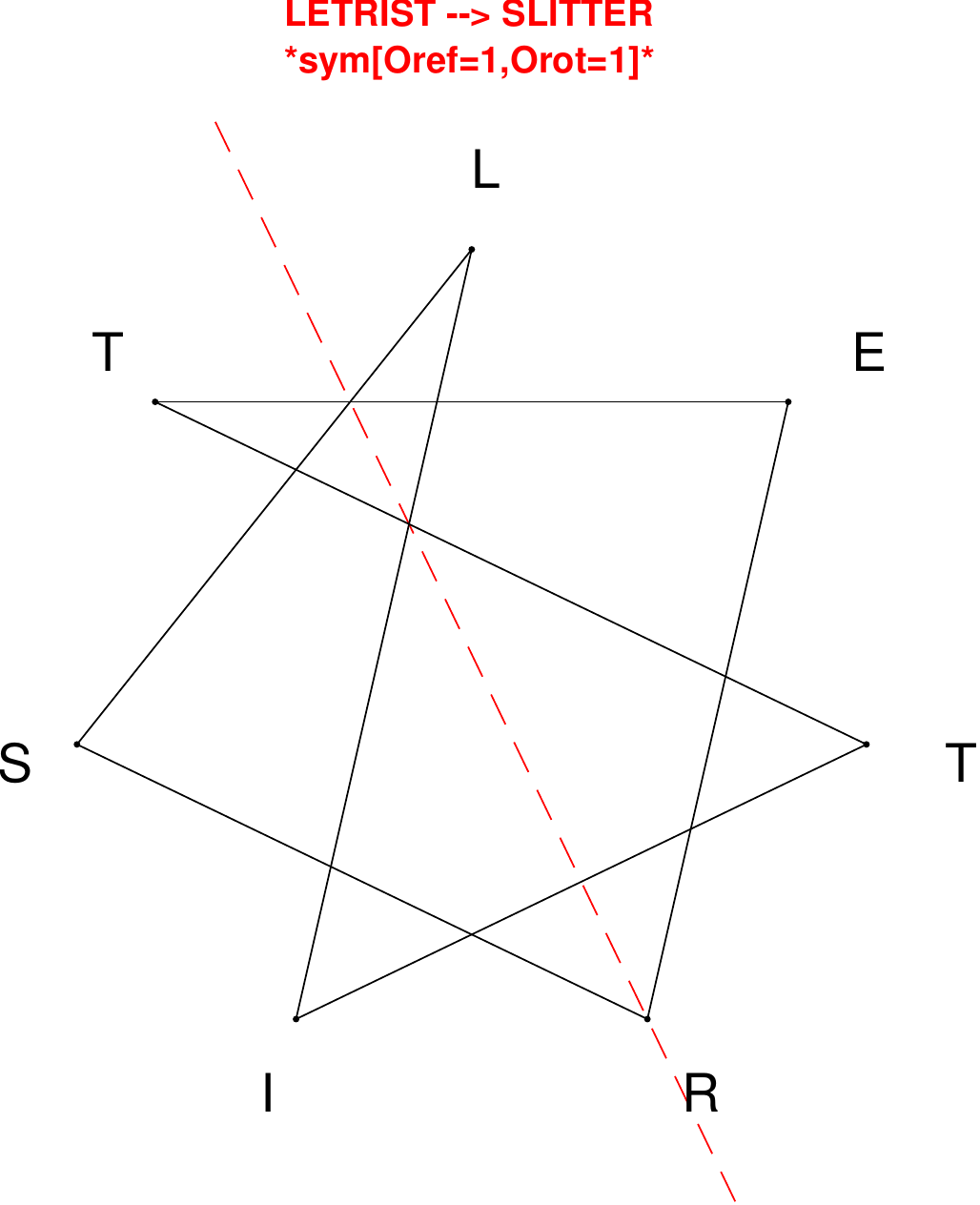}
\end{subfigure}
\hfill
\begin{subfigure}[T]{0.19\textwidth}
\centering
\includegraphics[width=\textwidth]{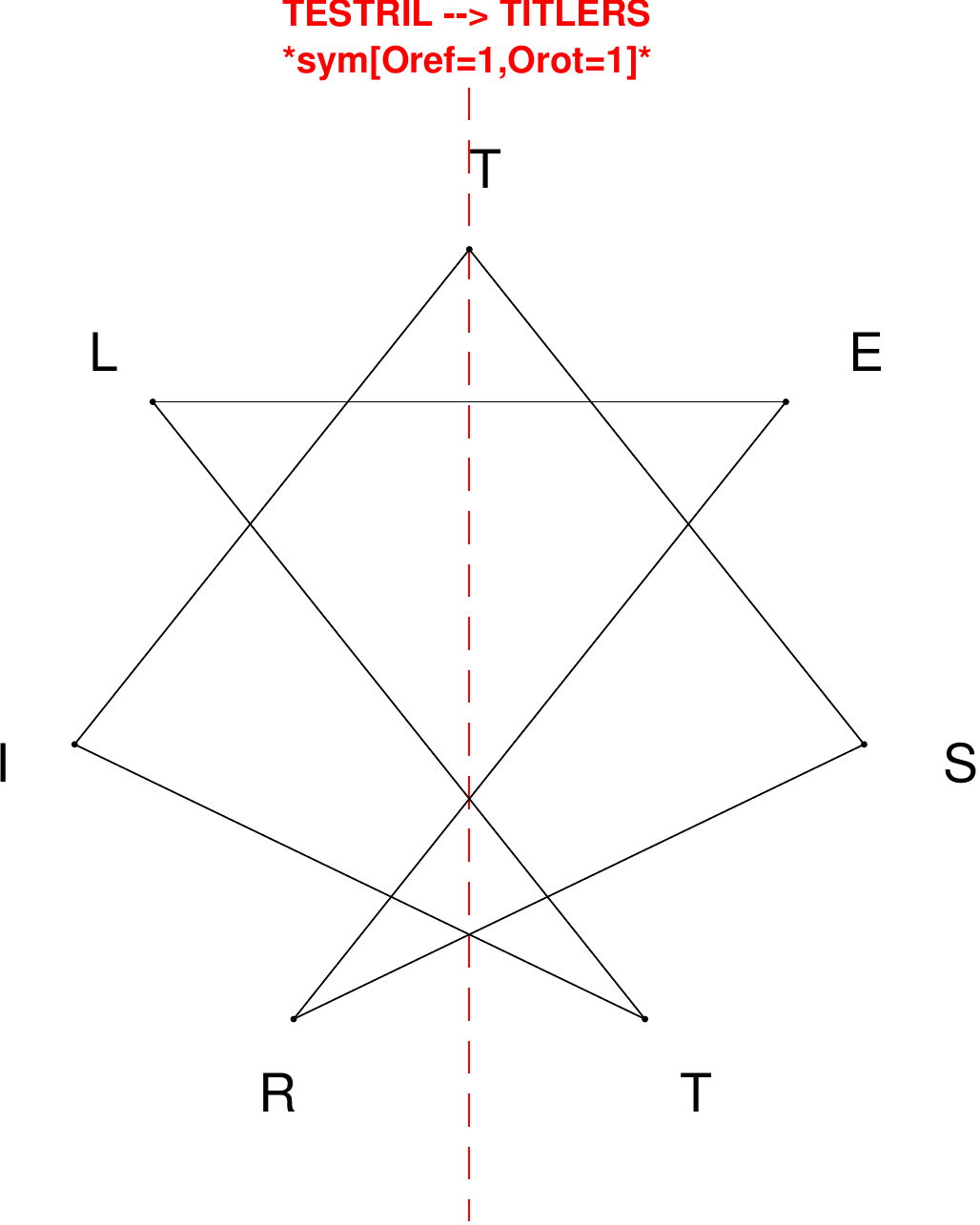}
\end{subfigure}
\hfill
\begin{subfigure}[T]{0.19\textwidth}
\centering
\includegraphics[width=\textwidth]{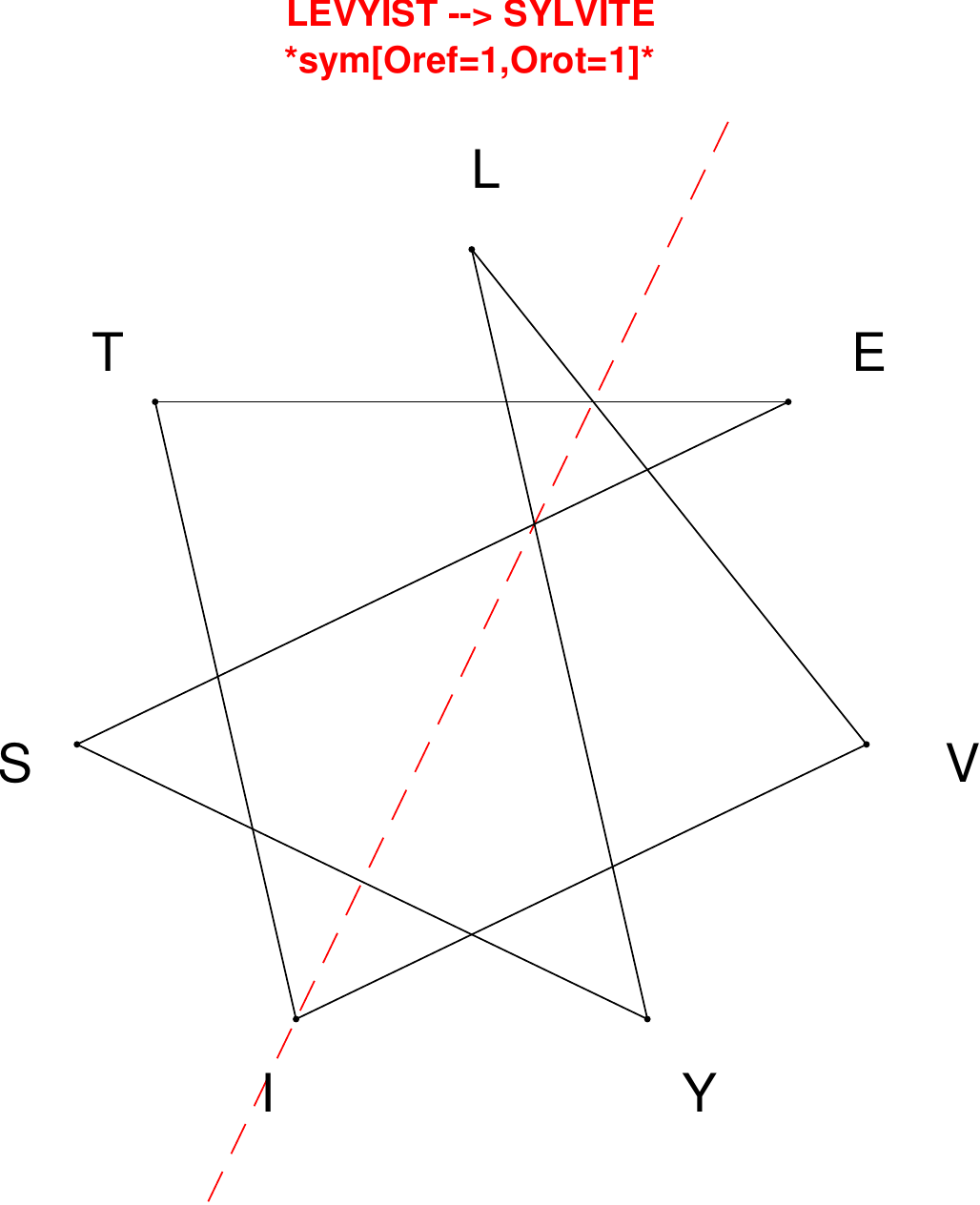}
\end{subfigure}
\hfill
\begin{subfigure}[T]{0.19\textwidth}
\centering
\includegraphics[width=\textwidth]{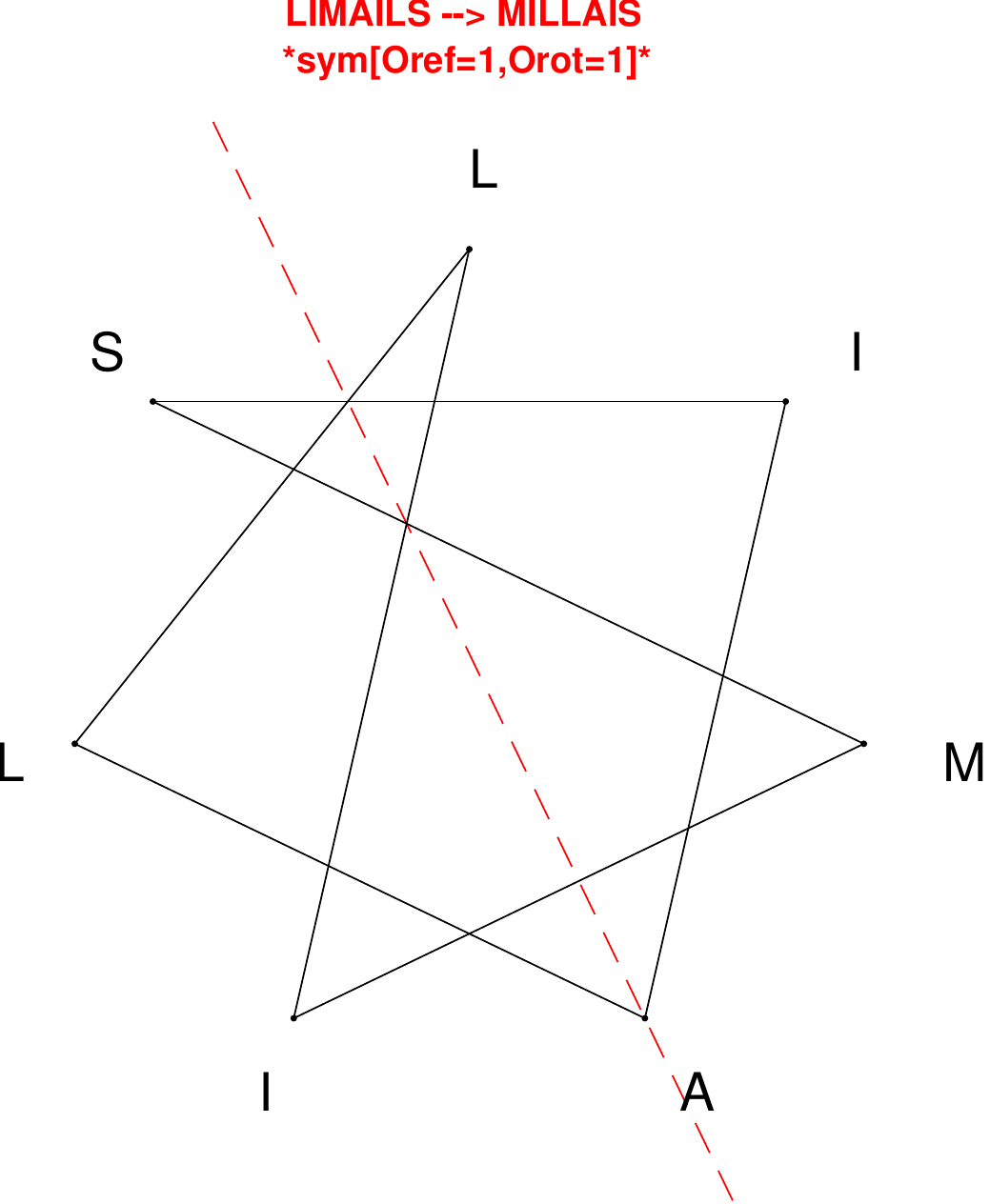}
\end{subfigure}
\end{figure}

\begin{figure}[H]
\centering
\begin{subfigure}[T]{0.19\textwidth}
\centering
\includegraphics[width=\textwidth]{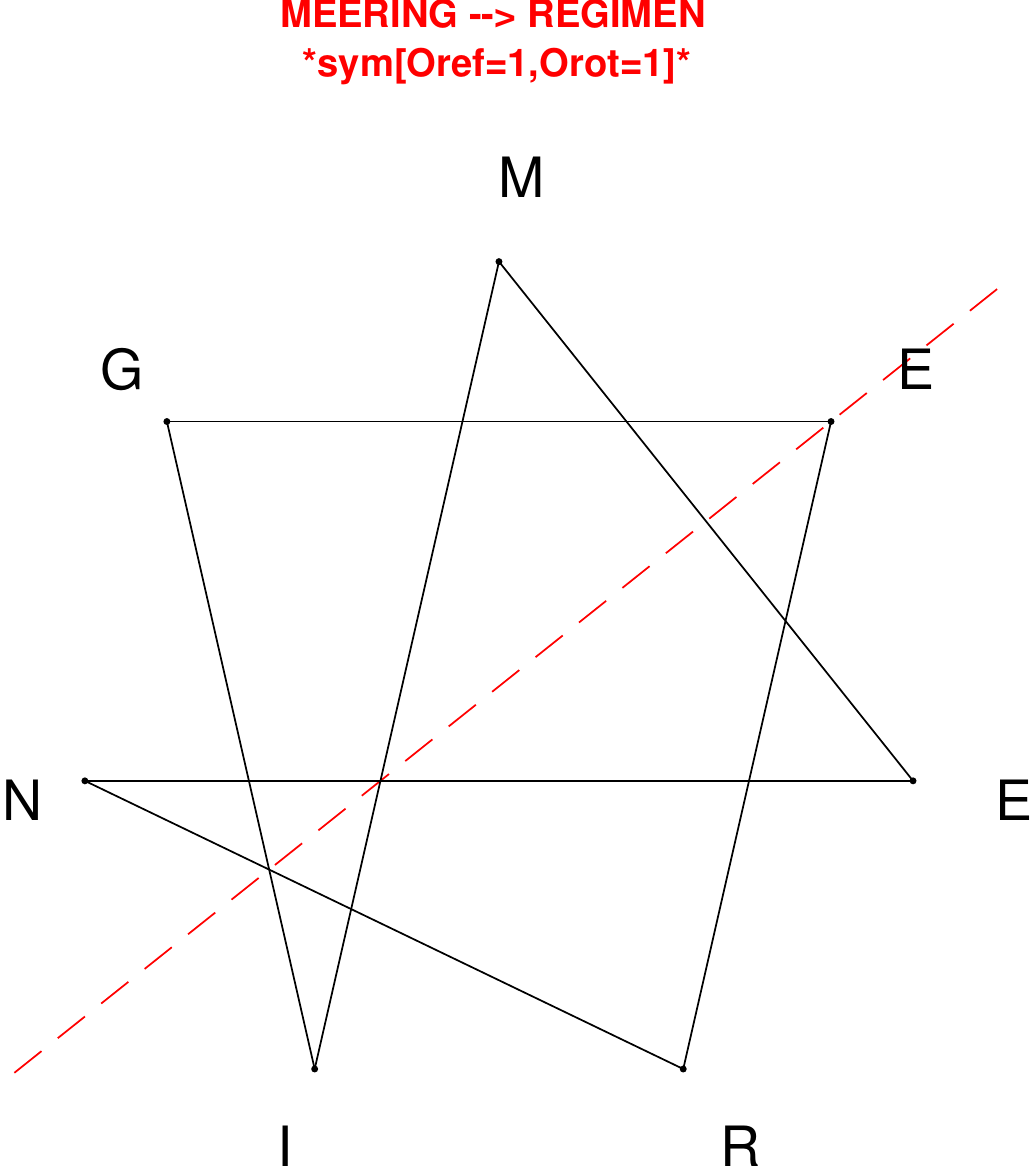}
\end{subfigure}
\hfill
\begin{subfigure}[T]{0.19\textwidth}
\centering
\includegraphics[width=\textwidth]{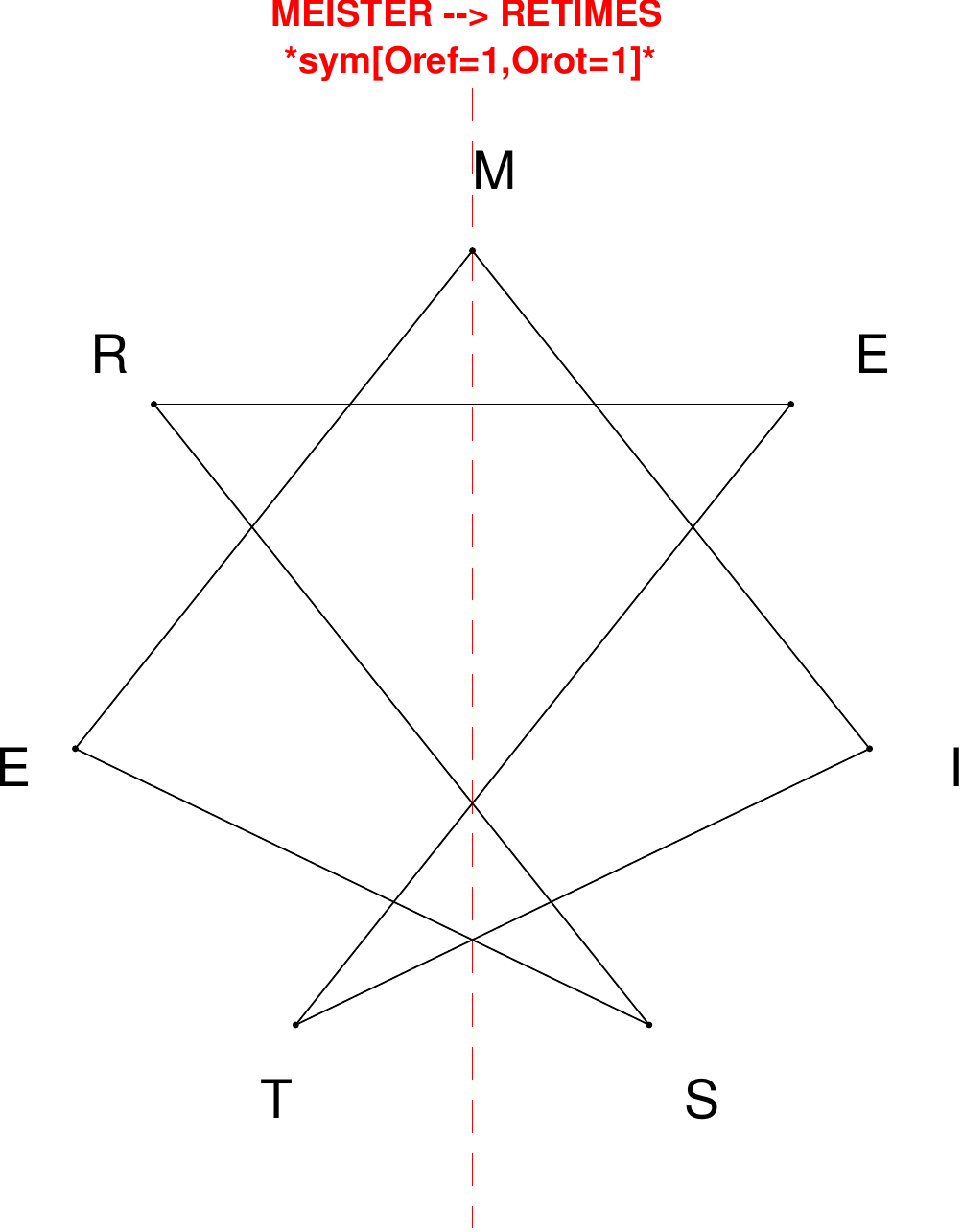}
\end{subfigure}
\hfill
\begin{subfigure}[T]{0.19\textwidth}
\centering
\includegraphics[width=\textwidth]{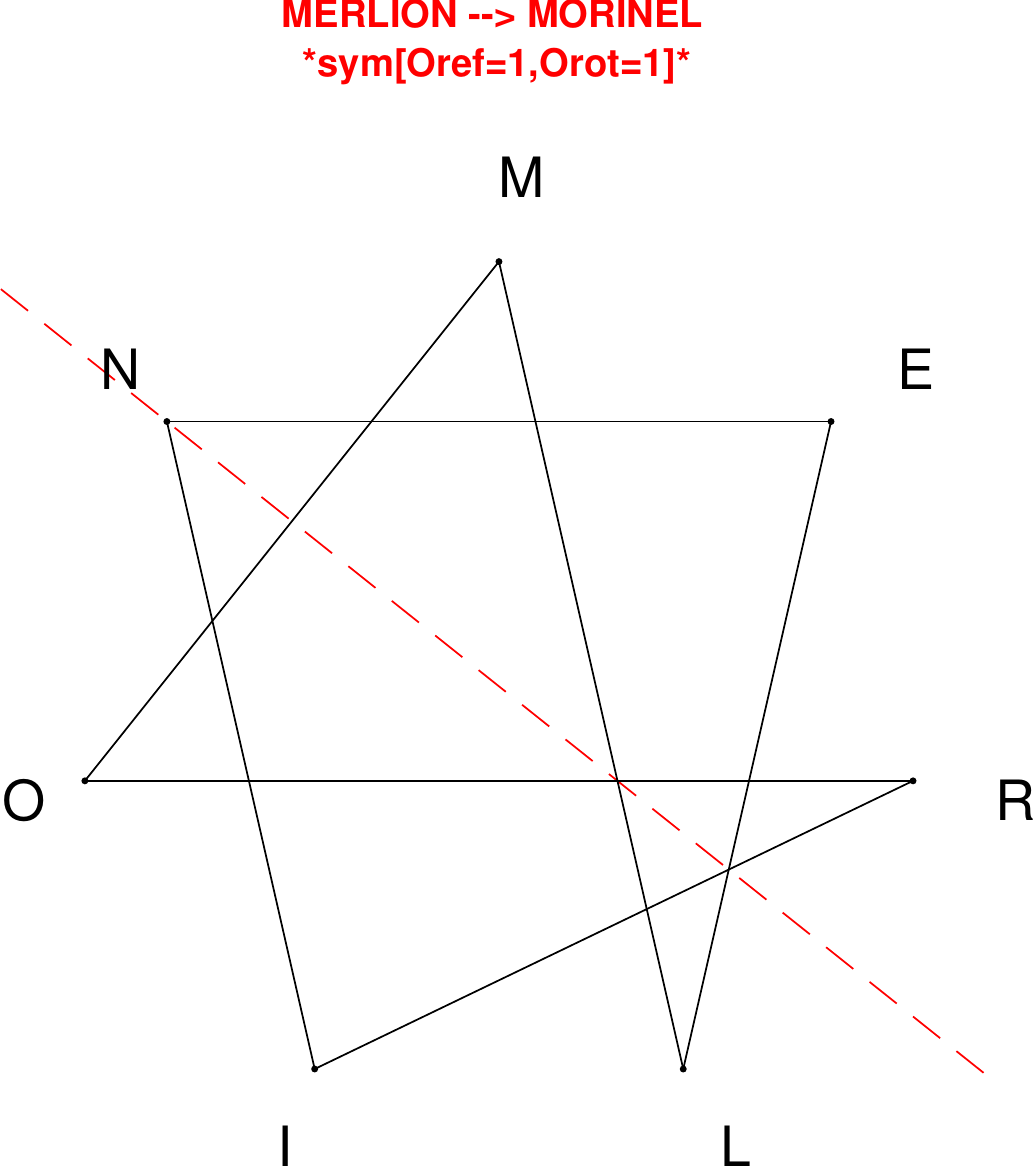}
\end{subfigure}
\hfill
\begin{subfigure}[T]{0.19\textwidth}
\centering
\includegraphics[width=\textwidth]{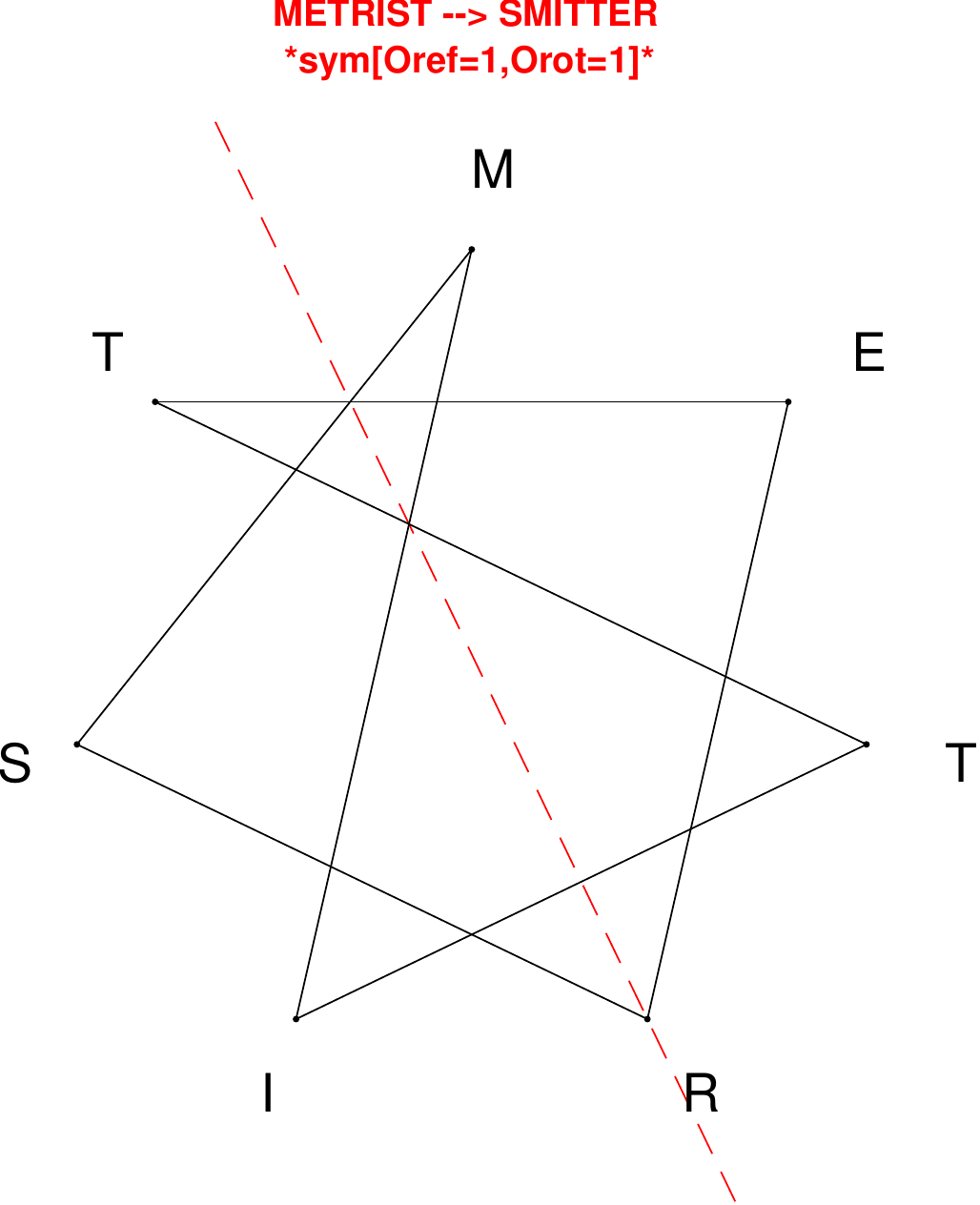}
\end{subfigure}
\hfill
\begin{subfigure}[T]{0.19\textwidth}
\centering
\includegraphics[width=\textwidth]{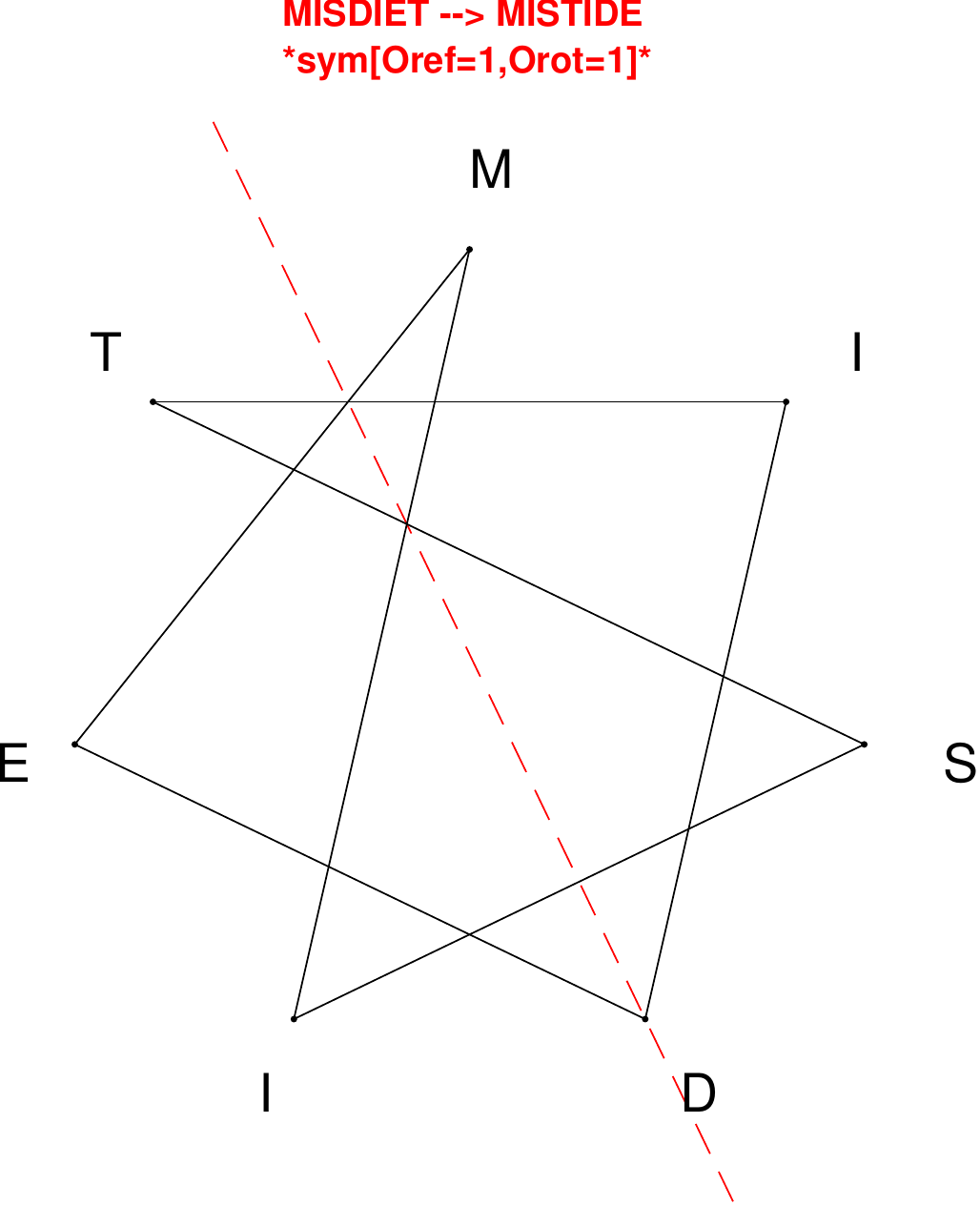}
\end{subfigure}
\end{figure}

\begin{figure}[H]
\centering
\begin{subfigure}[T]{0.19\textwidth}
\centering
\includegraphics[width=\textwidth]{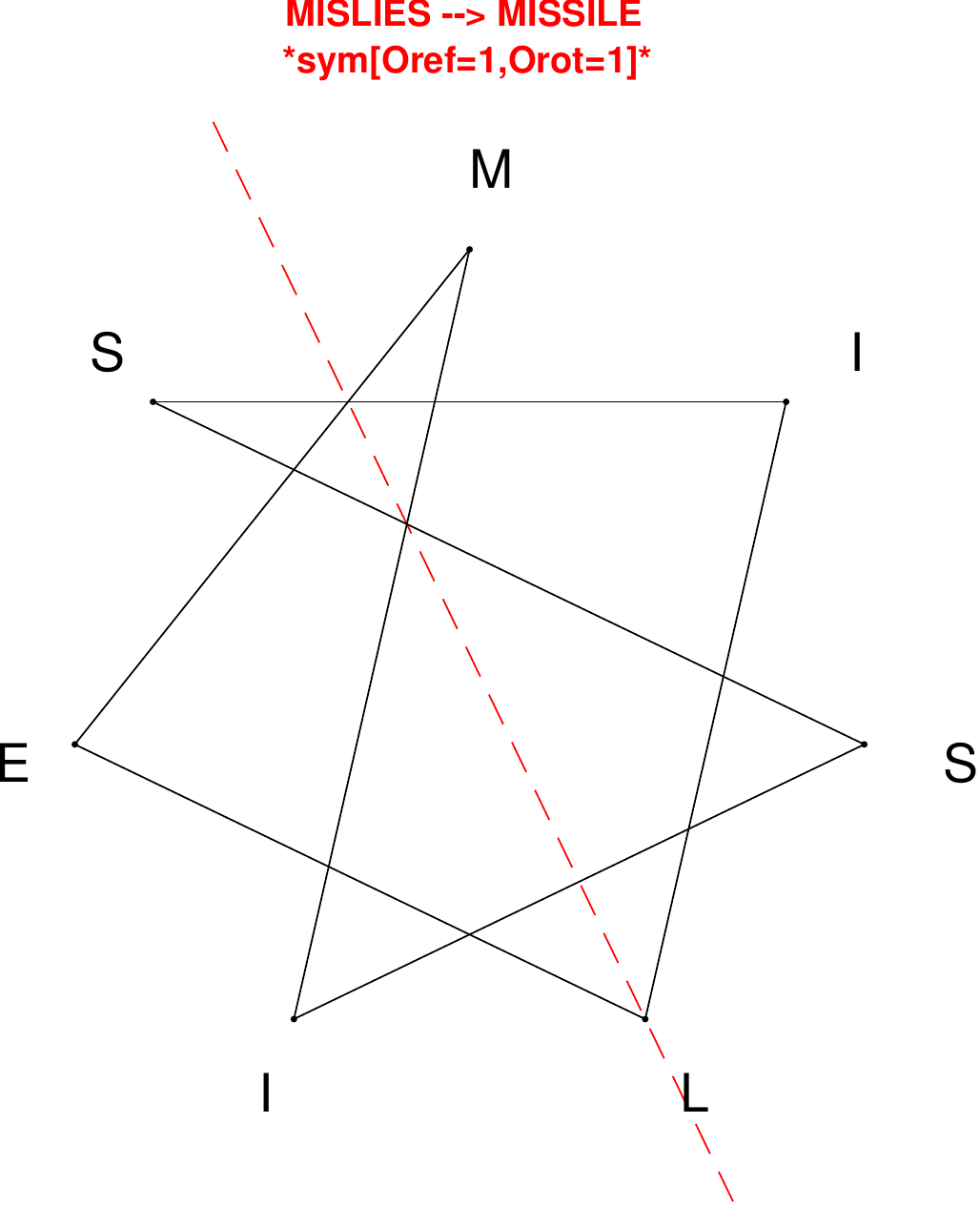}
\end{subfigure}
\hfill
\begin{subfigure}[T]{0.19\textwidth}
\centering
\includegraphics[width=\textwidth]{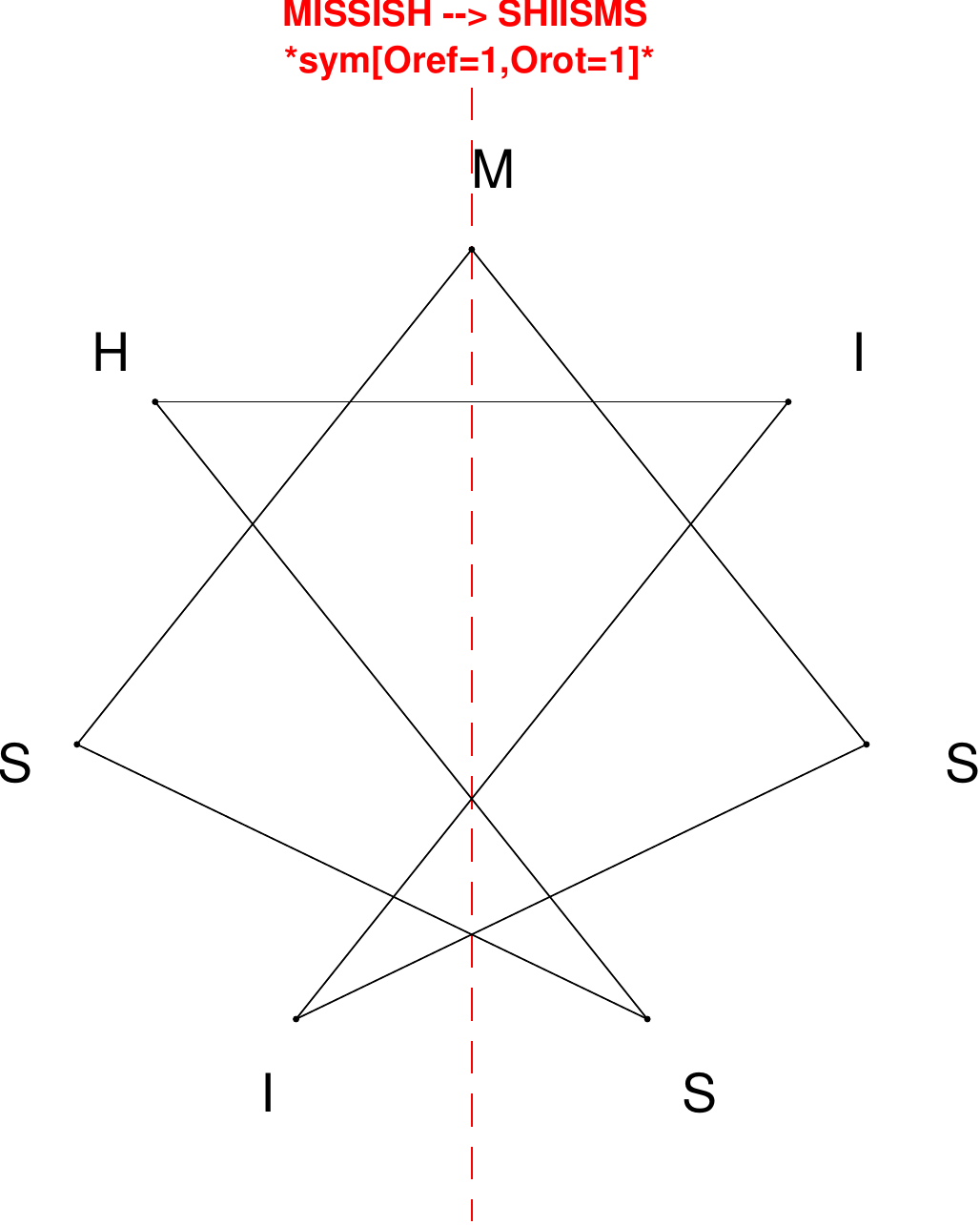}
\end{subfigure}
\hfill
\begin{subfigure}[T]{0.19\textwidth}
\centering
\includegraphics[width=\textwidth]{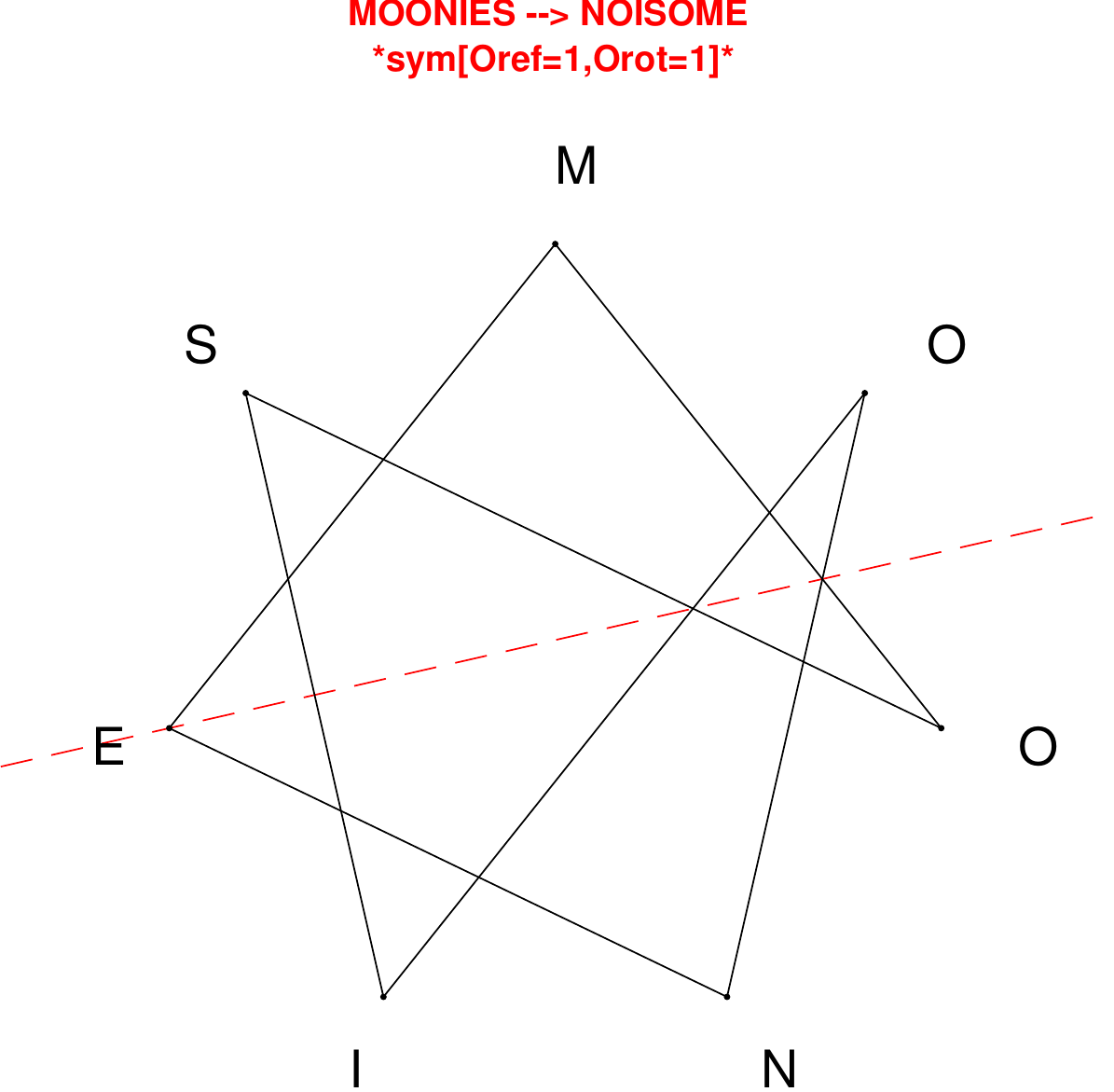}
\end{subfigure}
\hfill
\begin{subfigure}[T]{0.19\textwidth}
\centering
\includegraphics[width=\textwidth]{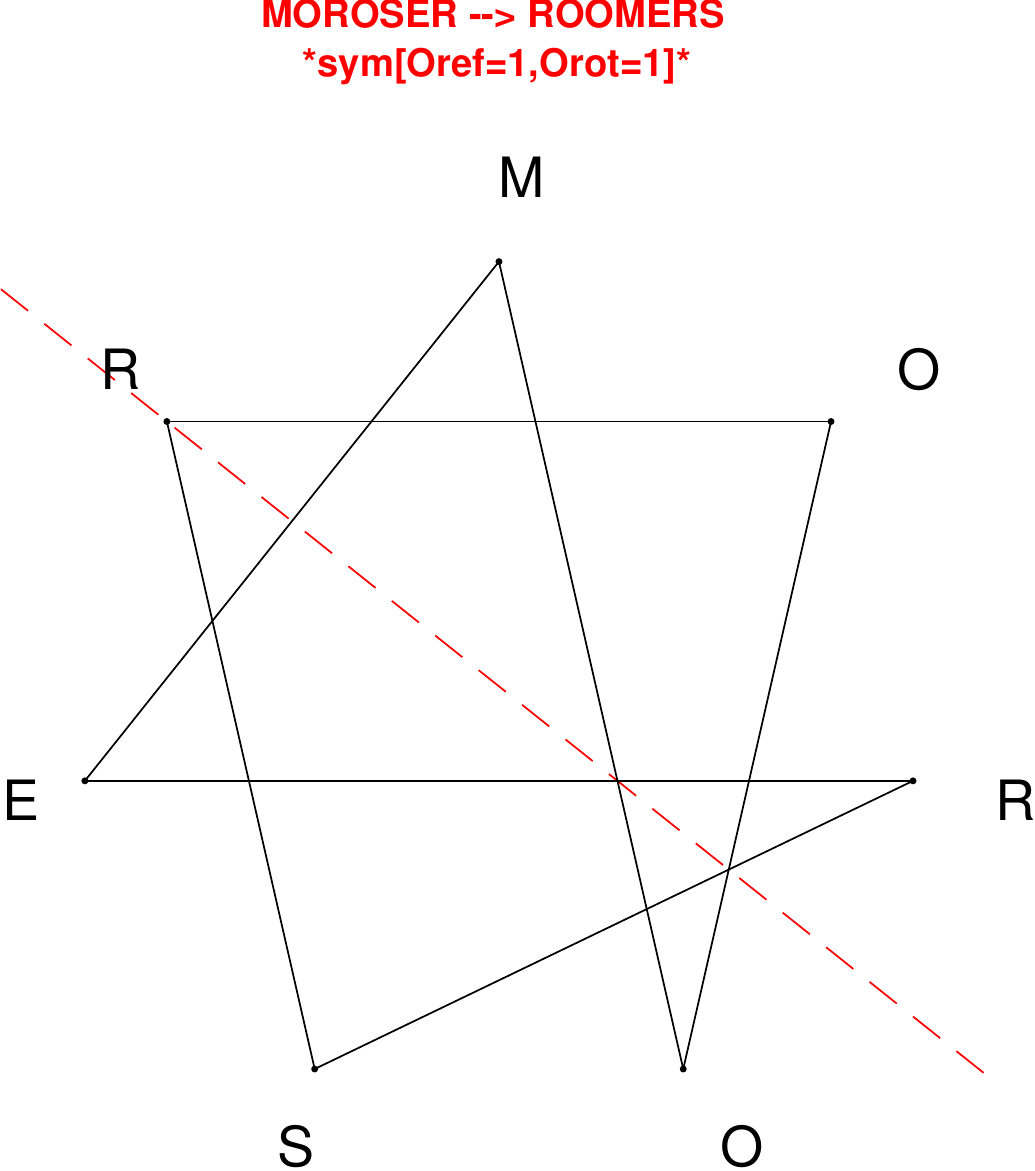}
\end{subfigure}
\hfill
\begin{subfigure}[T]{0.19\textwidth}
\centering
\includegraphics[width=\textwidth]{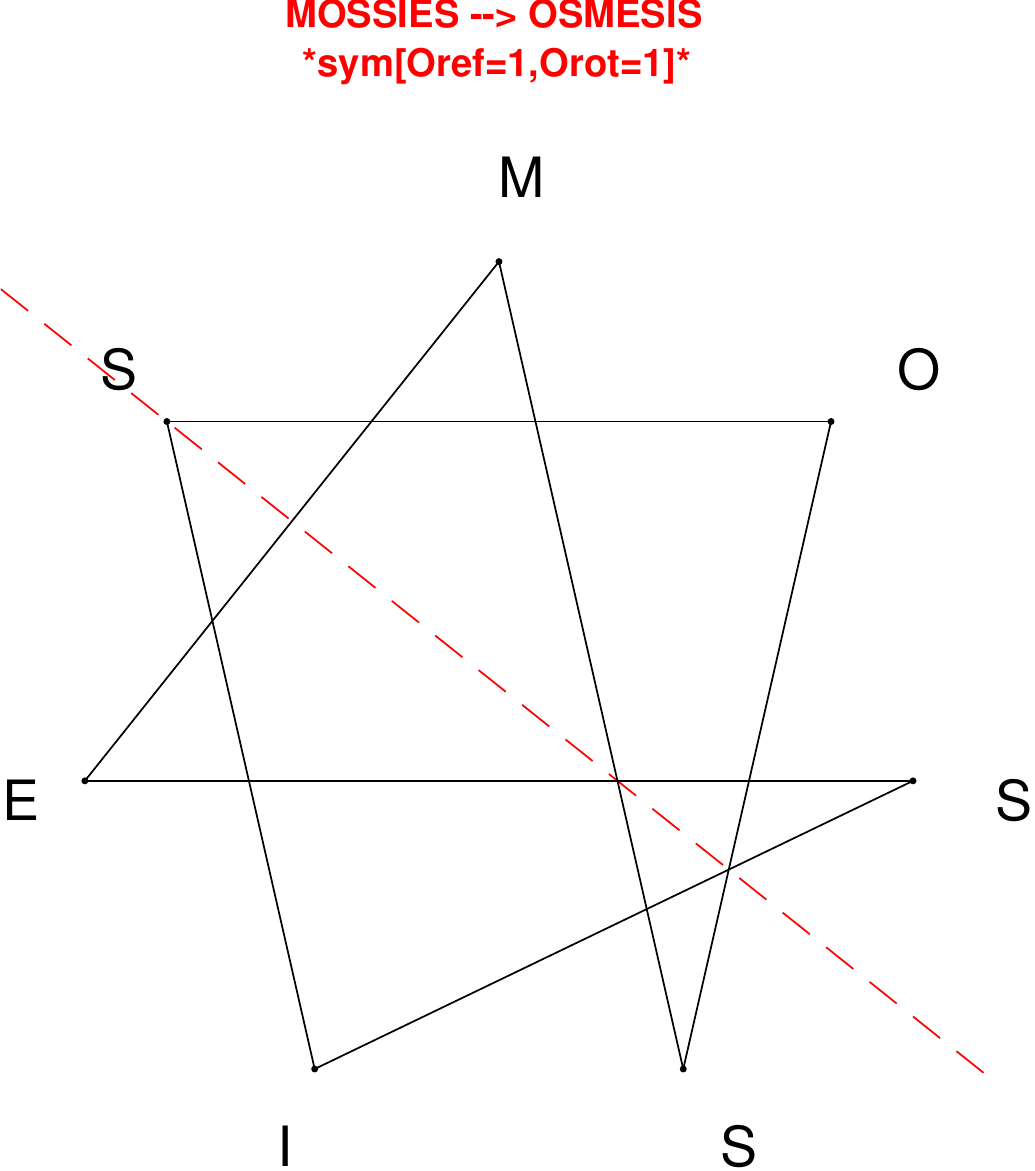}
\end{subfigure}
\end{figure}

\begin{figure}[H]
\centering
\begin{subfigure}[T]{0.19\textwidth}
\centering
\includegraphics[width=\textwidth]{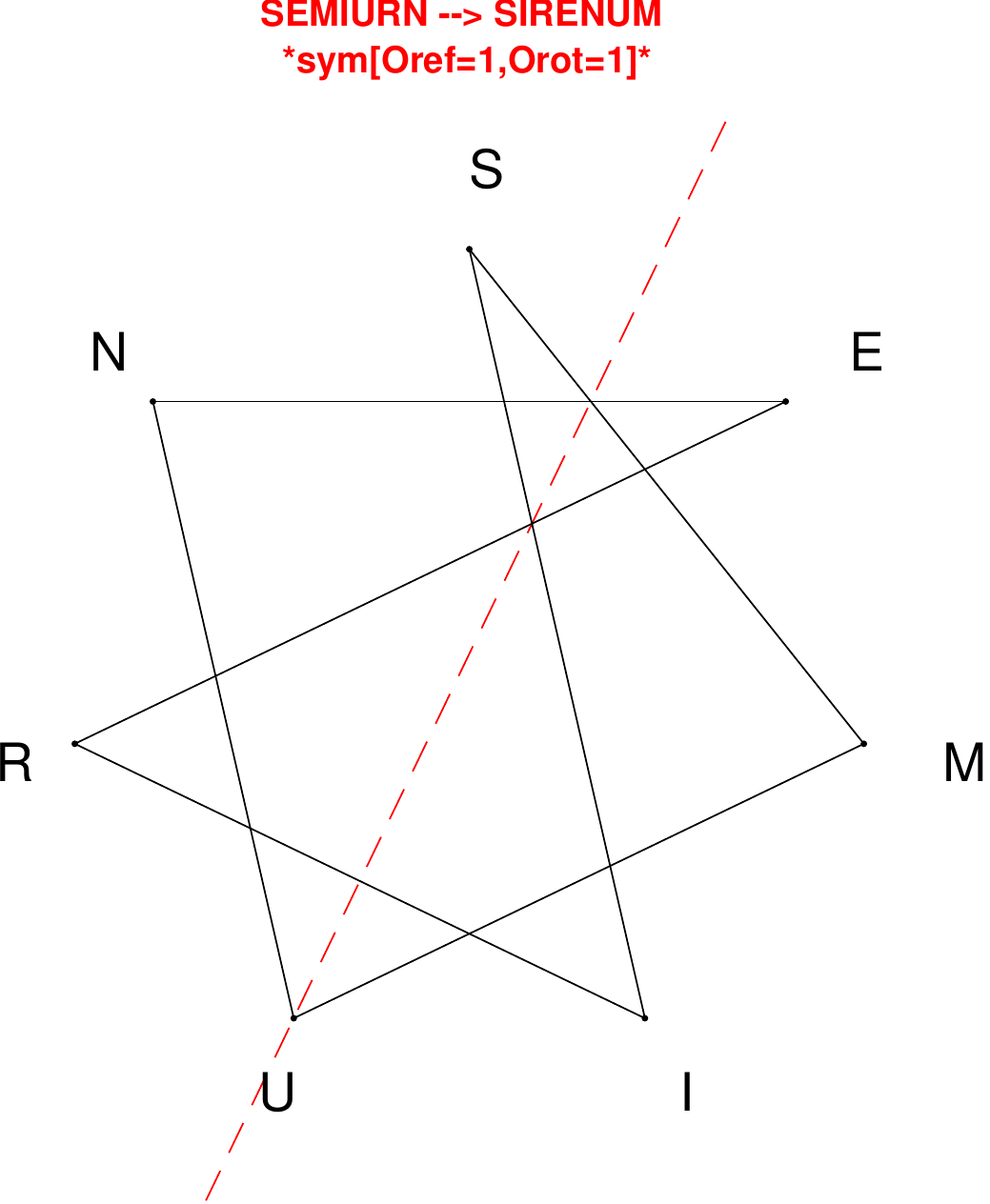}
\end{subfigure}
\hfill
\begin{subfigure}[T]{0.19\textwidth}
\centering
\includegraphics[width=\textwidth]{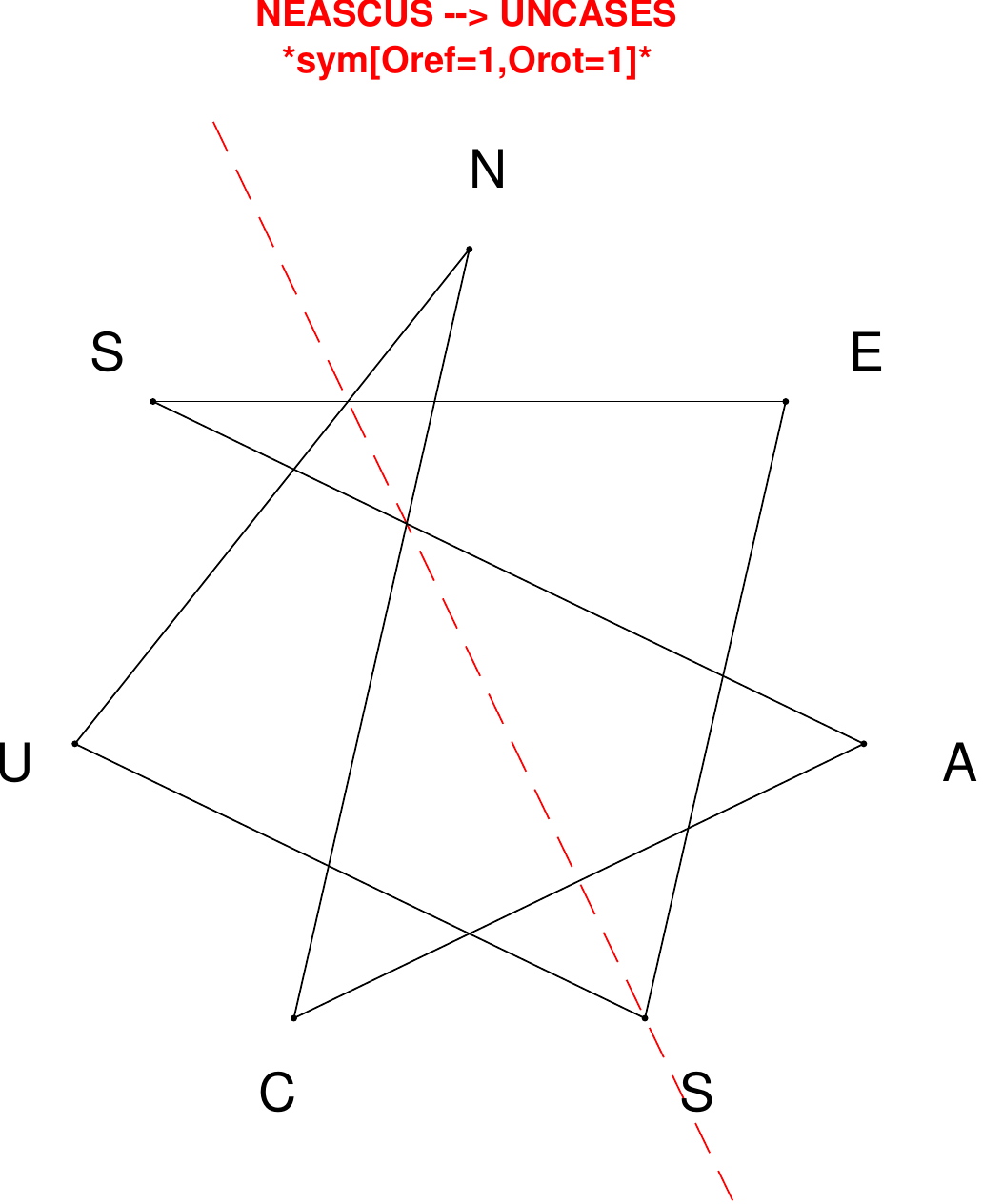}
\end{subfigure}
\hfill
\begin{subfigure}[T]{0.19\textwidth}
\centering
\includegraphics[width=\textwidth]{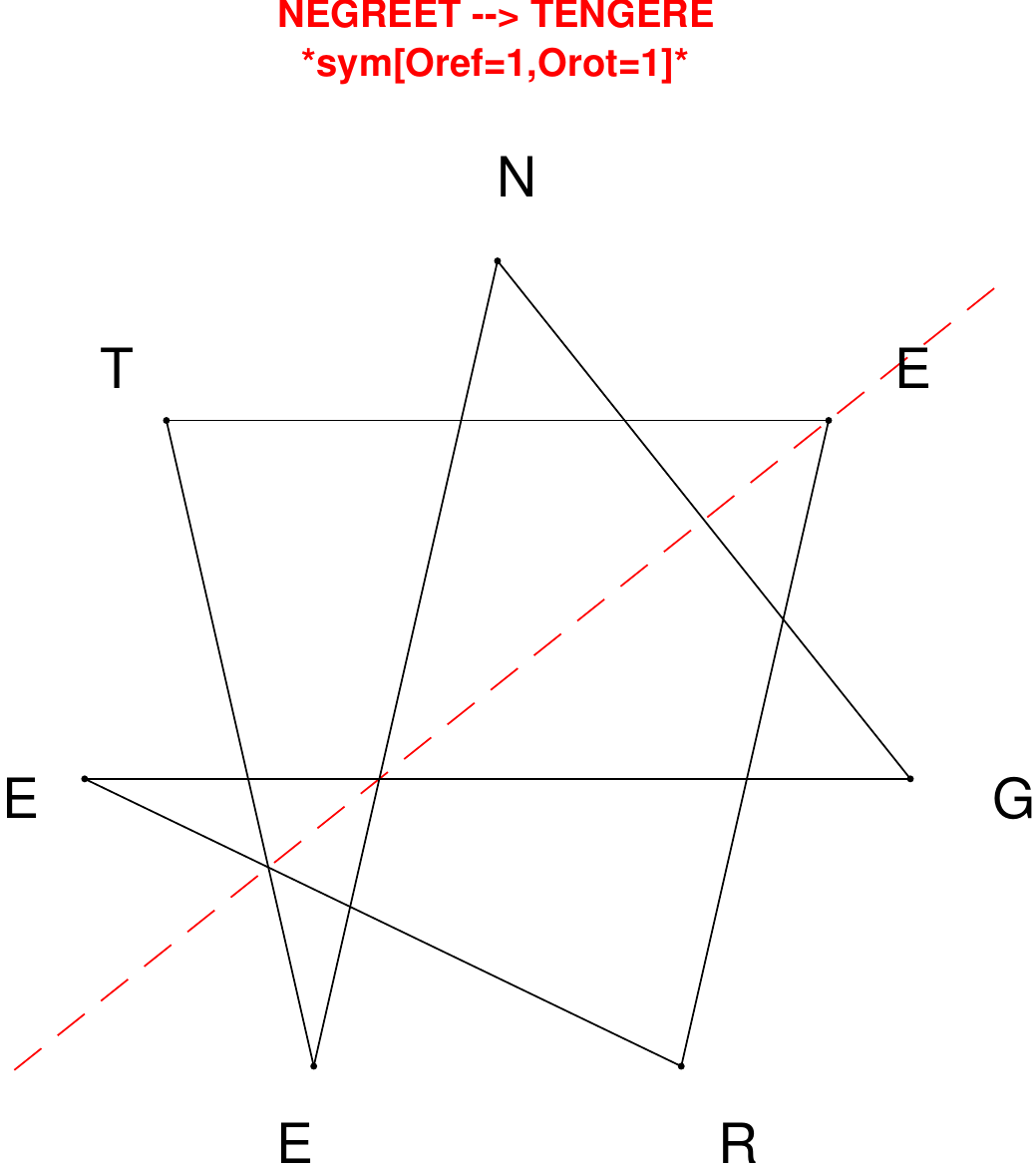}
\end{subfigure}
\hfill
\begin{subfigure}[T]{0.19\textwidth}
\centering
\includegraphics[width=\textwidth]{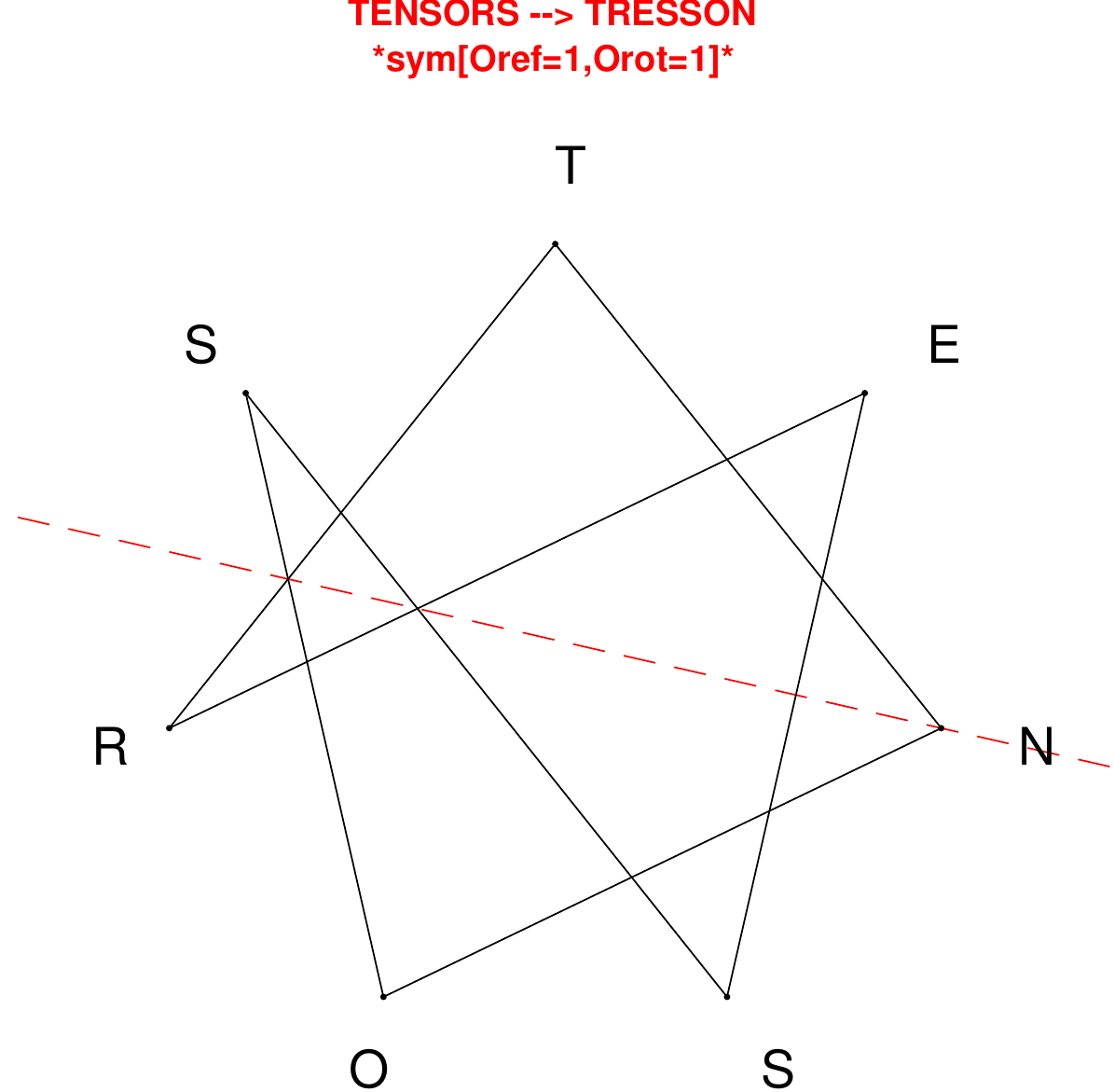}
\end{subfigure}
\hfill
\begin{subfigure}[T]{0.19\textwidth}
\centering
\includegraphics[width=\textwidth]{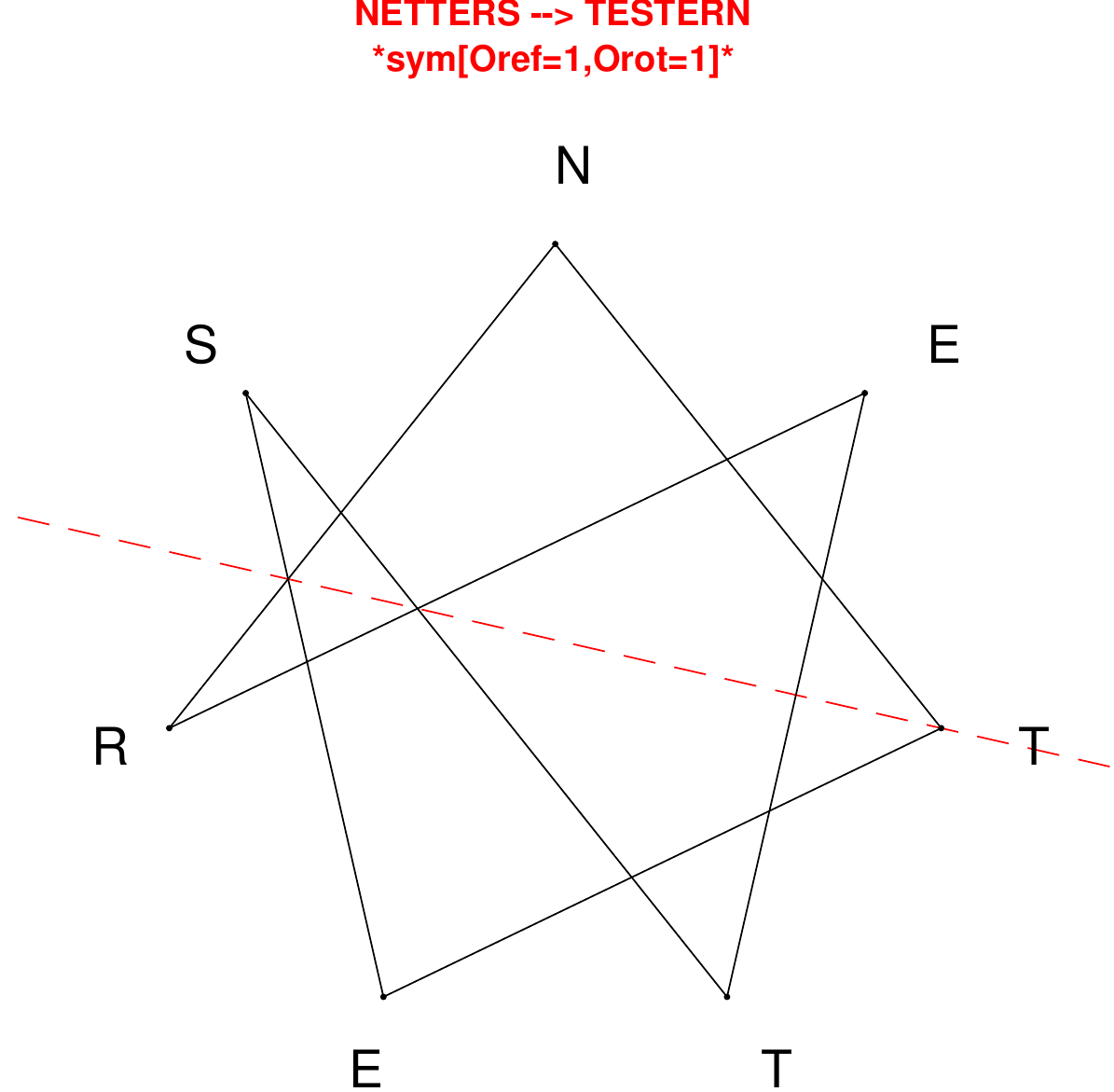}
\end{subfigure}
\end{figure}

\begin{figure}[H]
\centering
\begin{subfigure}[T]{0.19\textwidth}
\centering
\includegraphics[width=\textwidth]{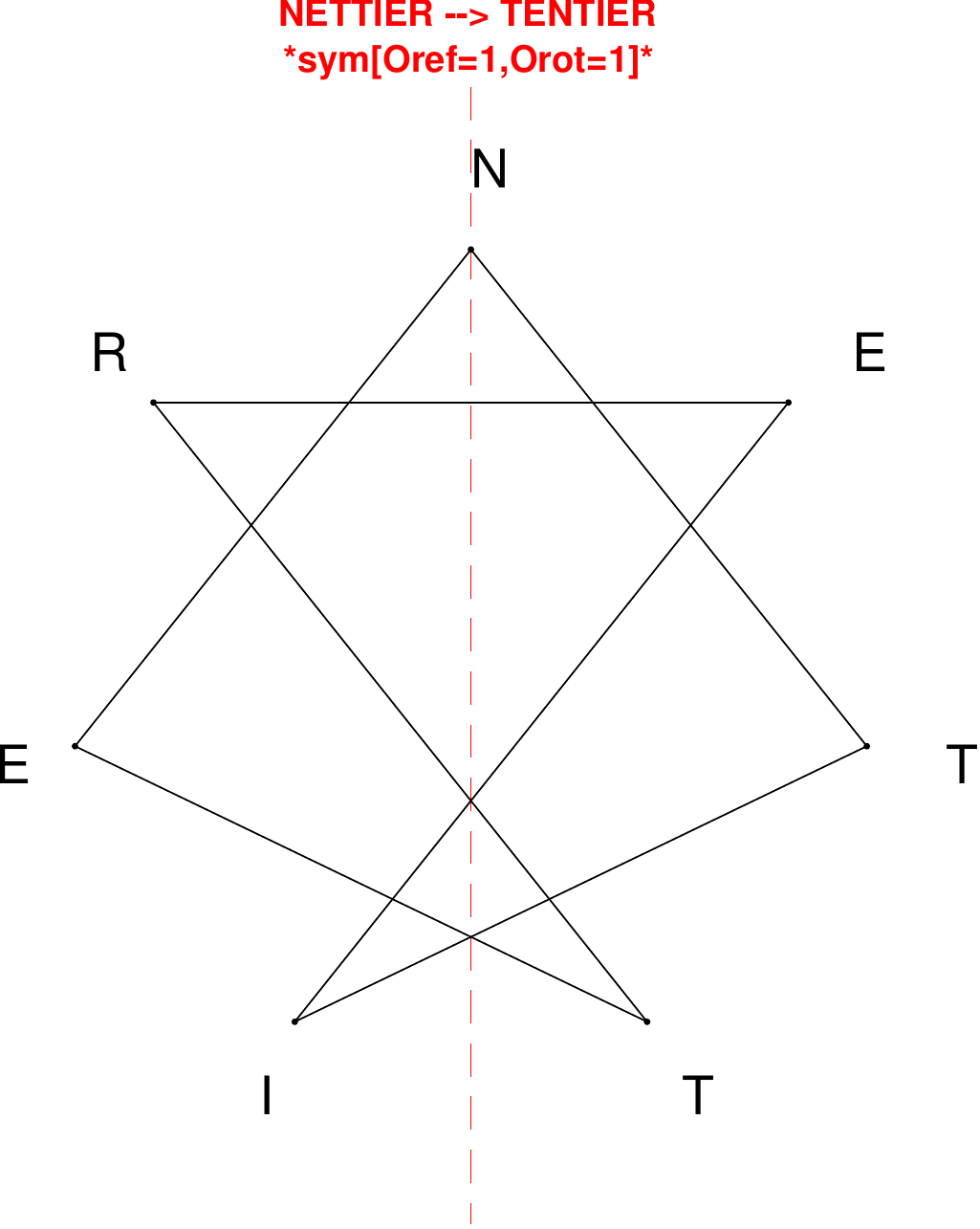}
\end{subfigure}
\hfill
\begin{subfigure}[T]{0.19\textwidth}
\centering
\includegraphics[width=\textwidth]{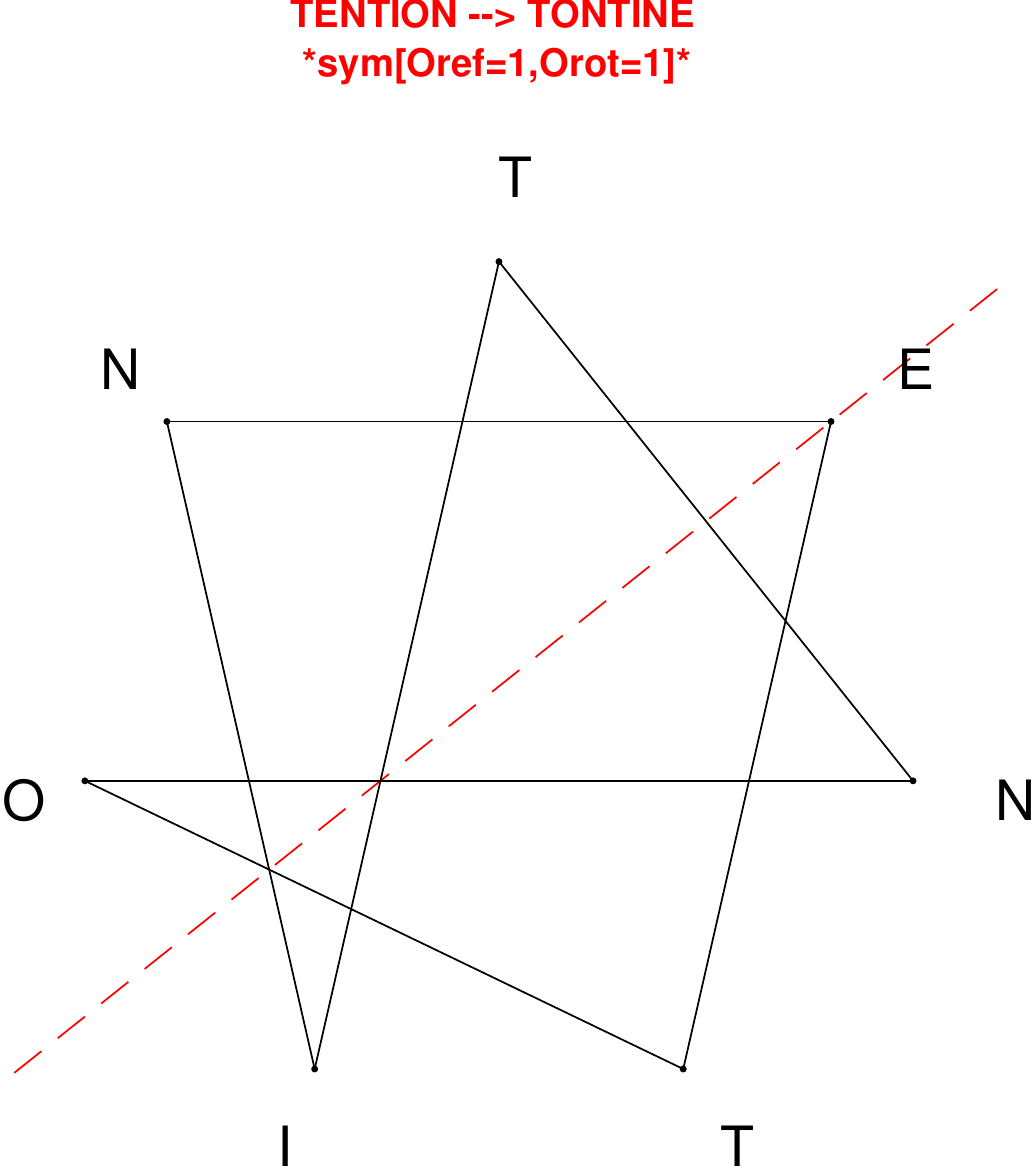}
\end{subfigure}
\hfill
\begin{subfigure}[T]{0.19\textwidth}
\centering
\includegraphics[width=\textwidth]{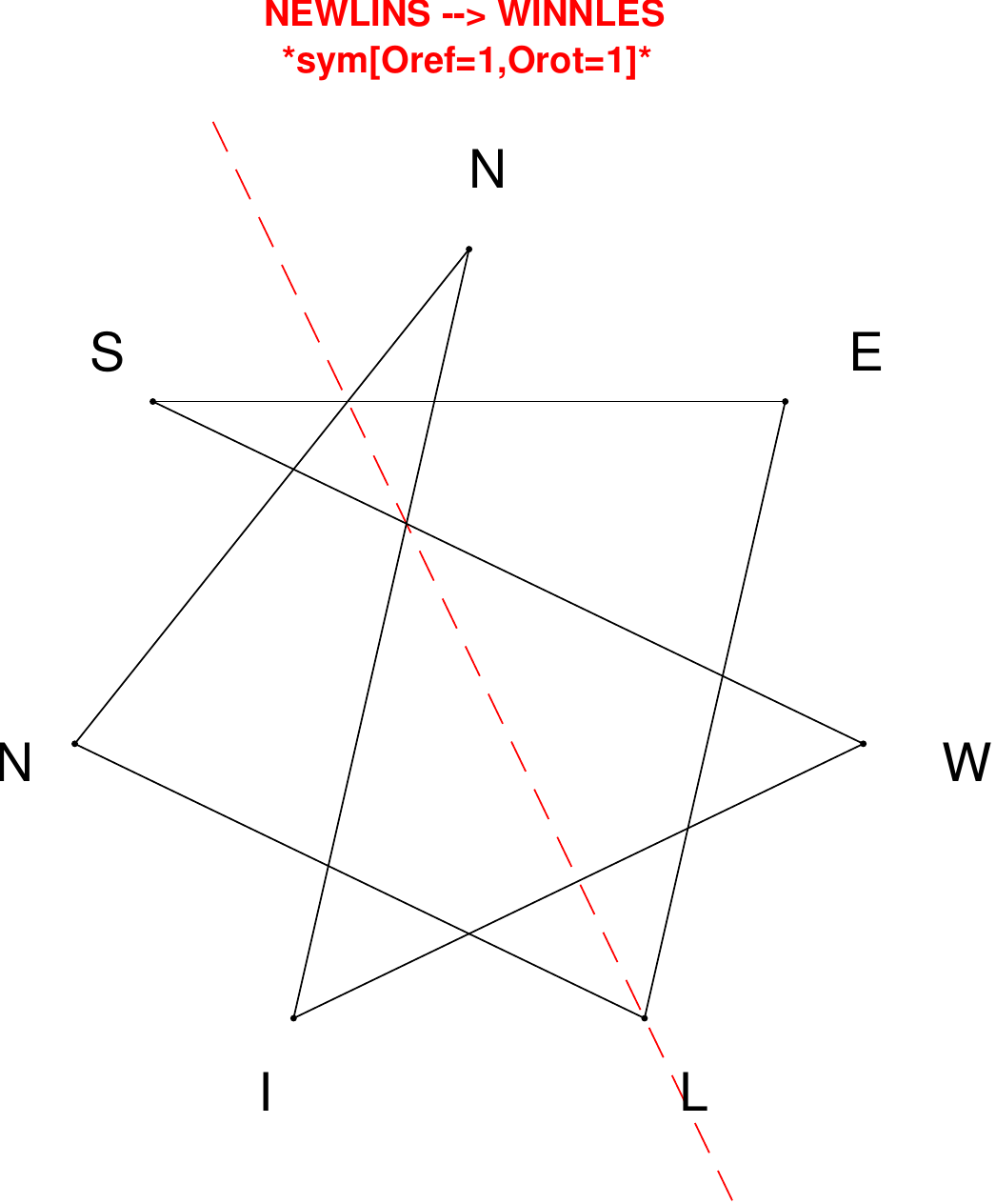}
\end{subfigure}
\hfill
\begin{subfigure}[T]{0.19\textwidth}
\centering
\includegraphics[width=\textwidth]{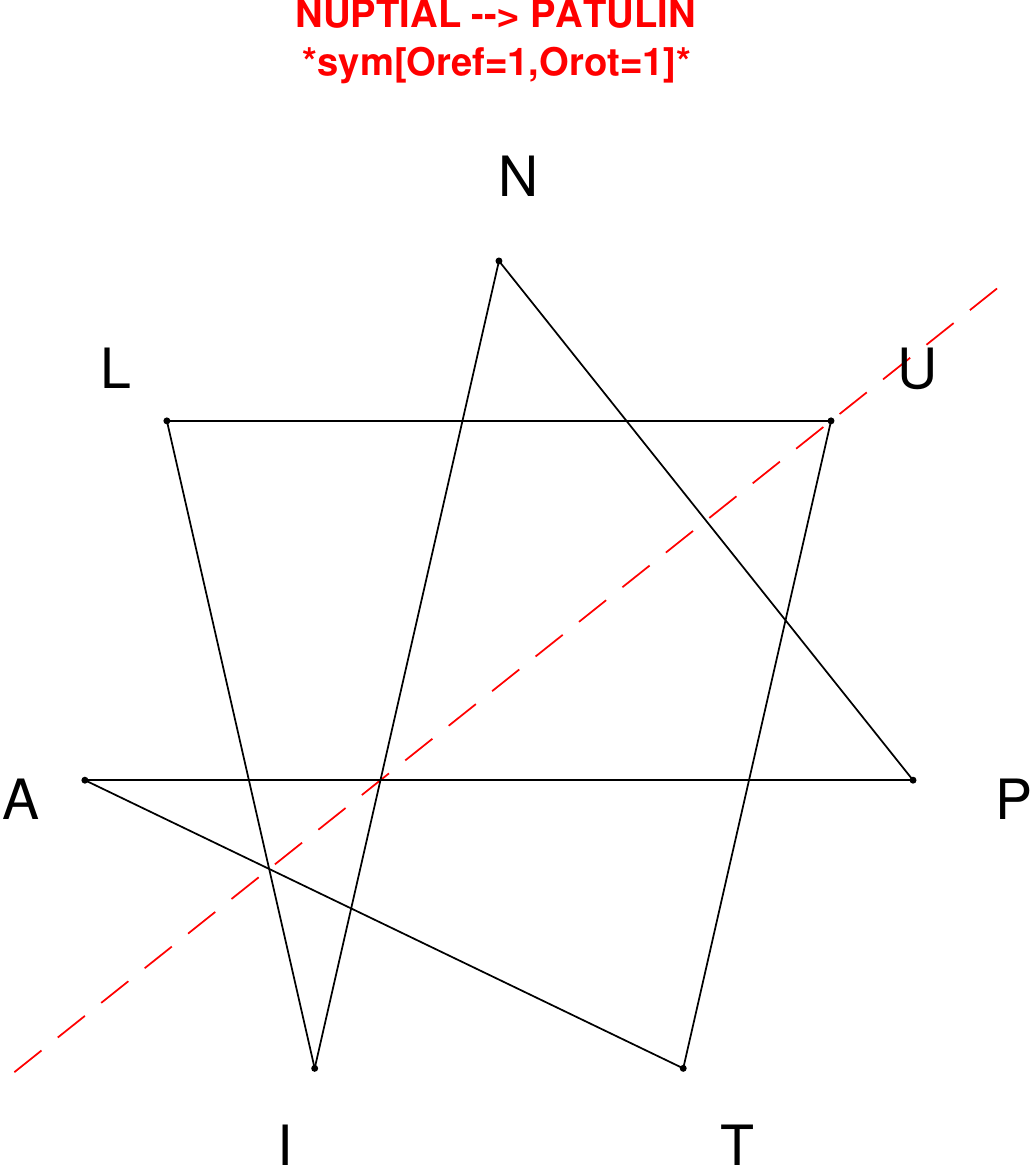}
\end{subfigure}
\hfill
\begin{subfigure}[T]{0.19\textwidth}
\centering
\includegraphics[width=\textwidth]{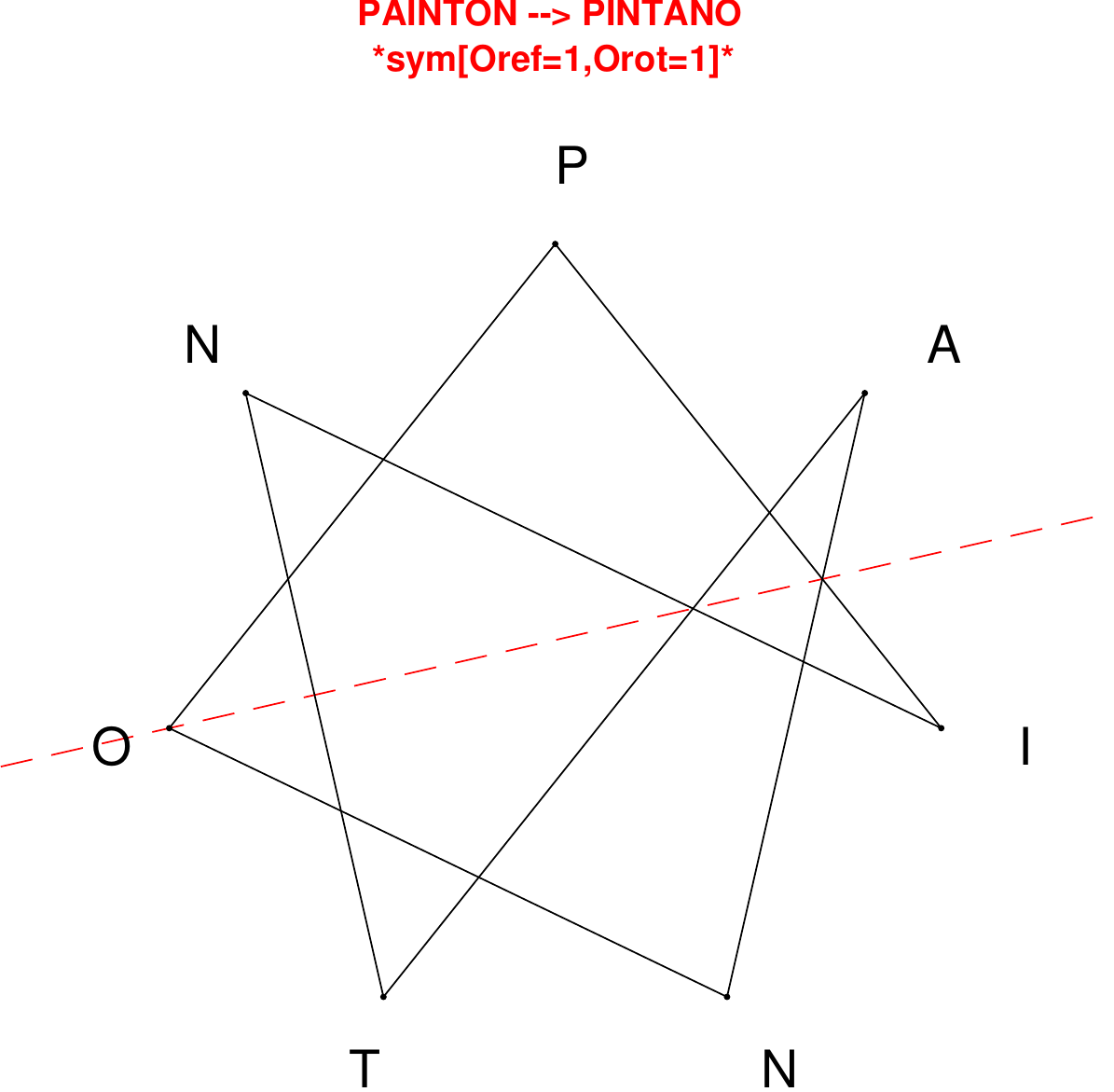}
\end{subfigure}
\end{figure}

\begin{figure}[H]
\centering
\begin{subfigure}[T]{0.19\textwidth}
\centering
\includegraphics[width=\textwidth]{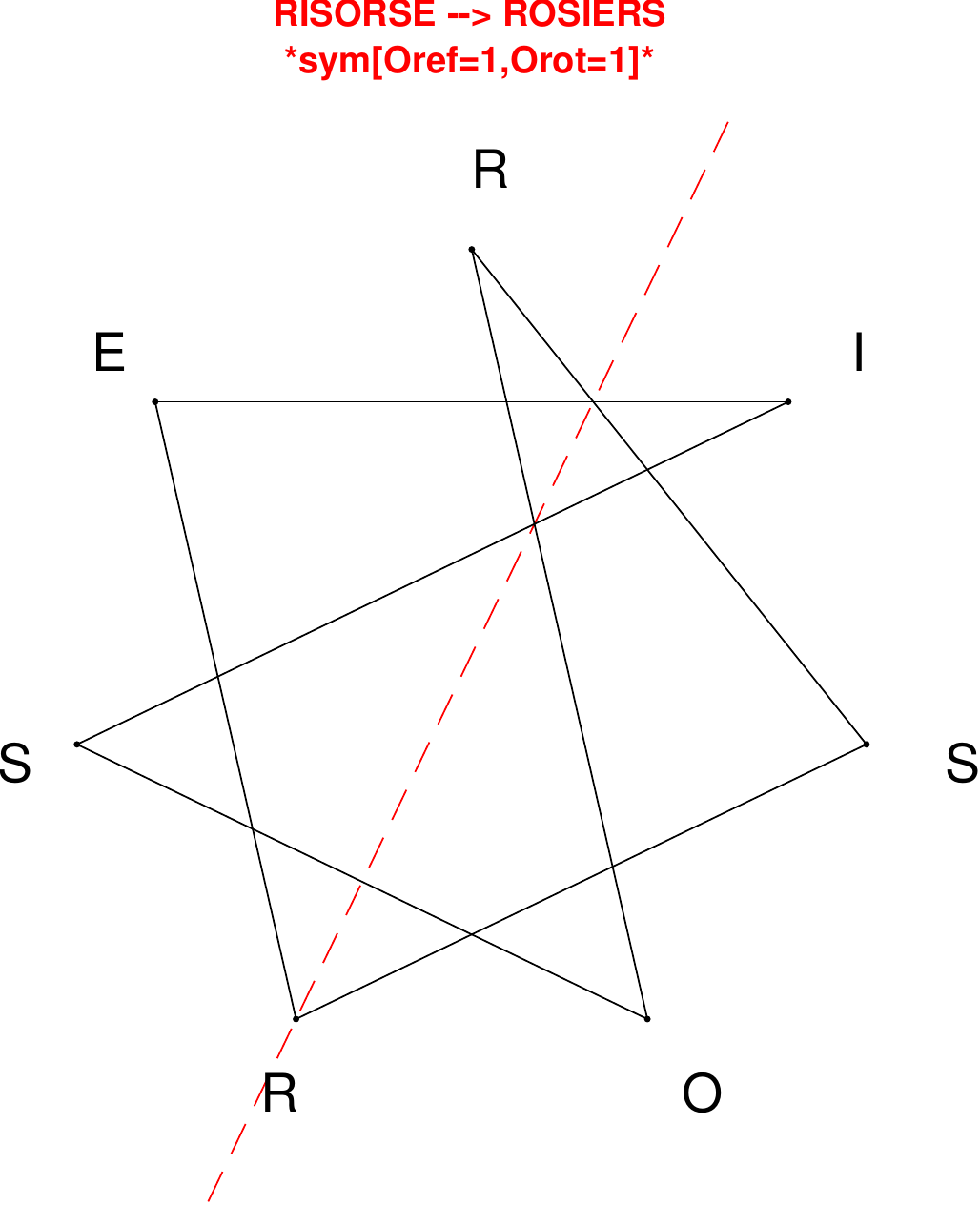}
\end{subfigure}
\hfill
\begin{subfigure}[T]{0.19\textwidth}
\centering
\includegraphics[width=\textwidth]{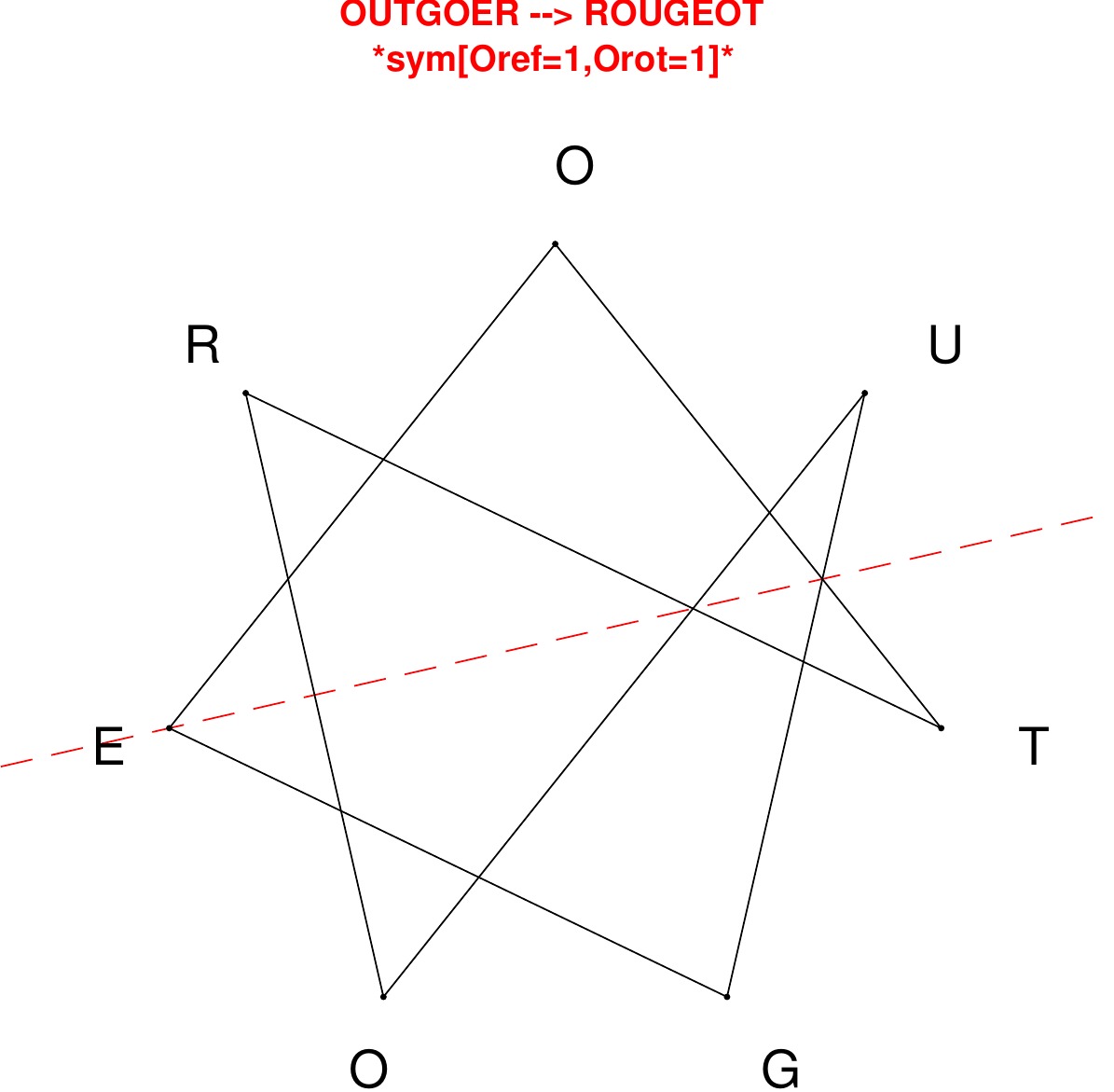}
\end{subfigure}
\hfill
\begin{subfigure}[T]{0.19\textwidth}
\centering
\includegraphics[width=\textwidth]{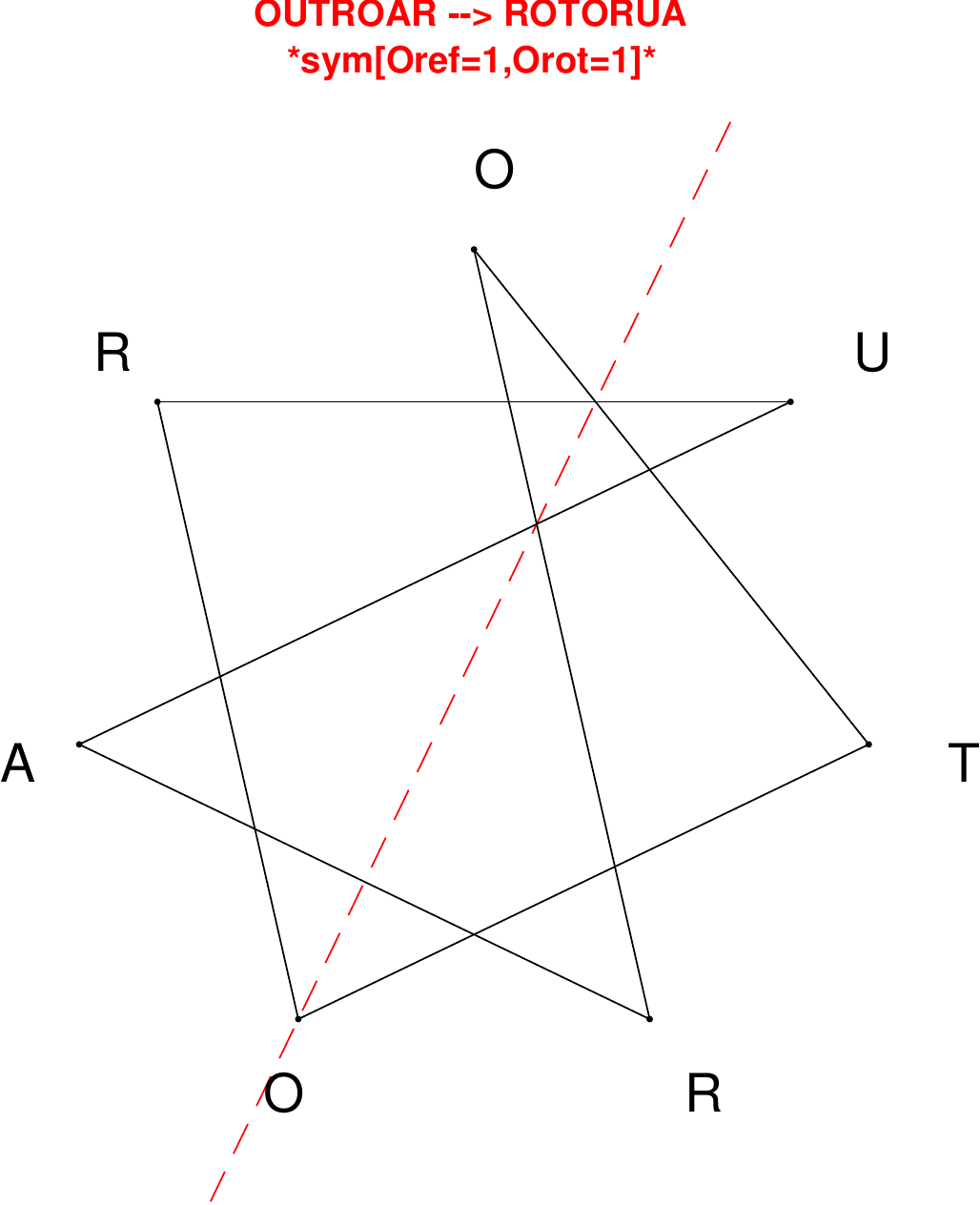}
\end{subfigure}
\hfill
\begin{subfigure}[T]{0.19\textwidth}
\centering
\includegraphics[width=\textwidth]{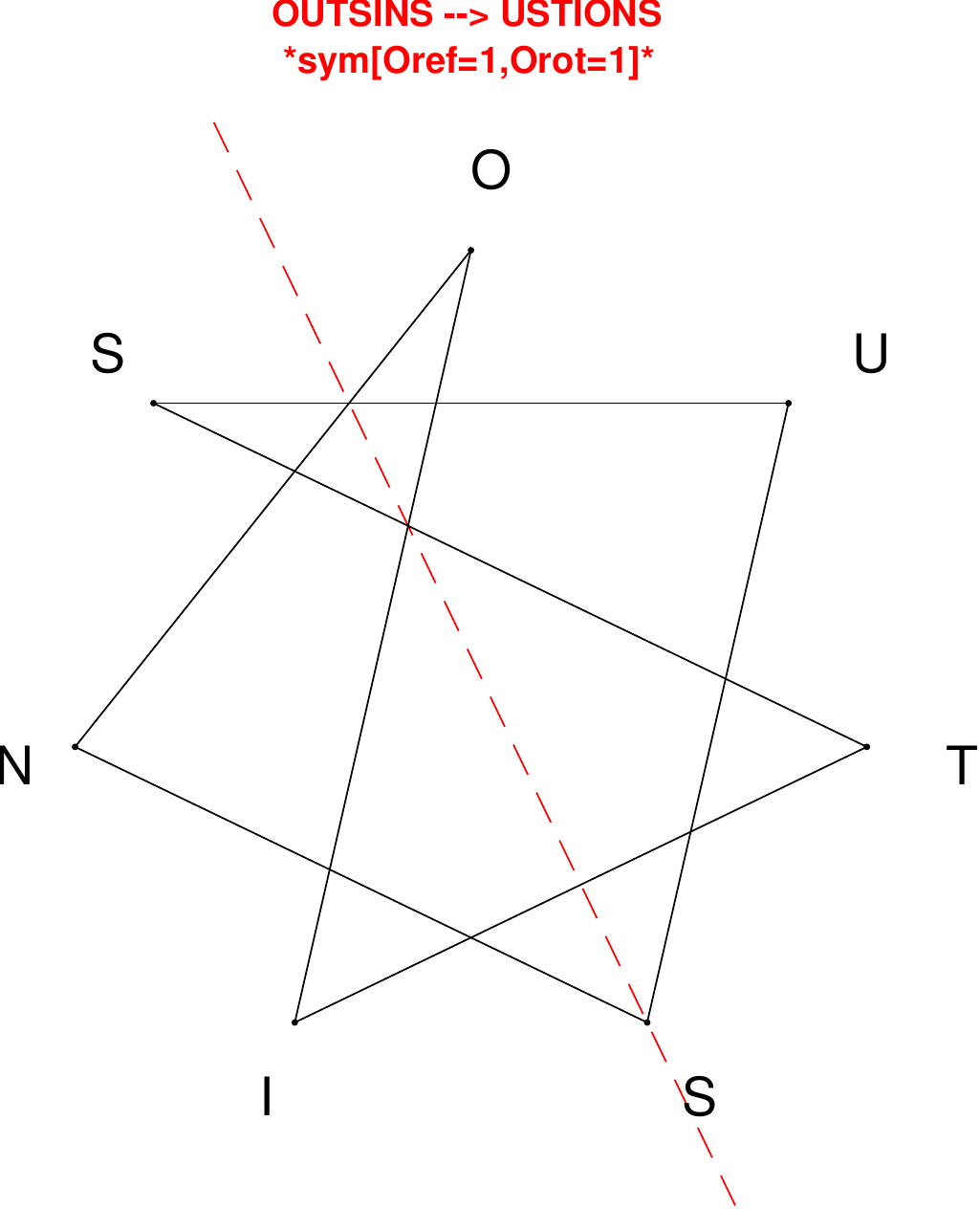}
\end{subfigure}
\hfill
\begin{subfigure}[T]{0.19\textwidth}
\centering
\includegraphics[width=\textwidth]{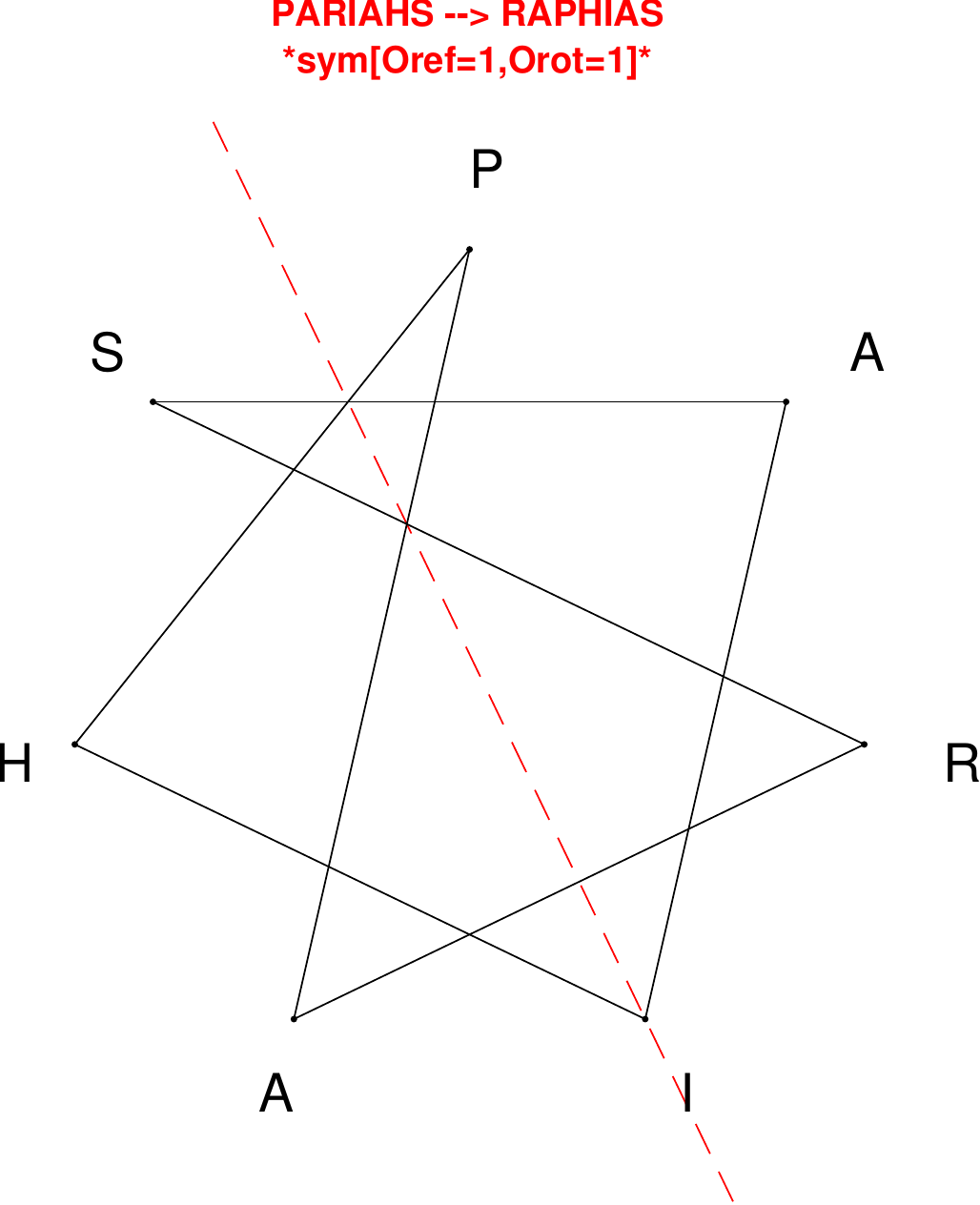}
\end{subfigure}
\end{figure}

\begin{figure}[H]
\centering
\begin{subfigure}[T]{0.19\textwidth}
\centering
\includegraphics[width=\textwidth]{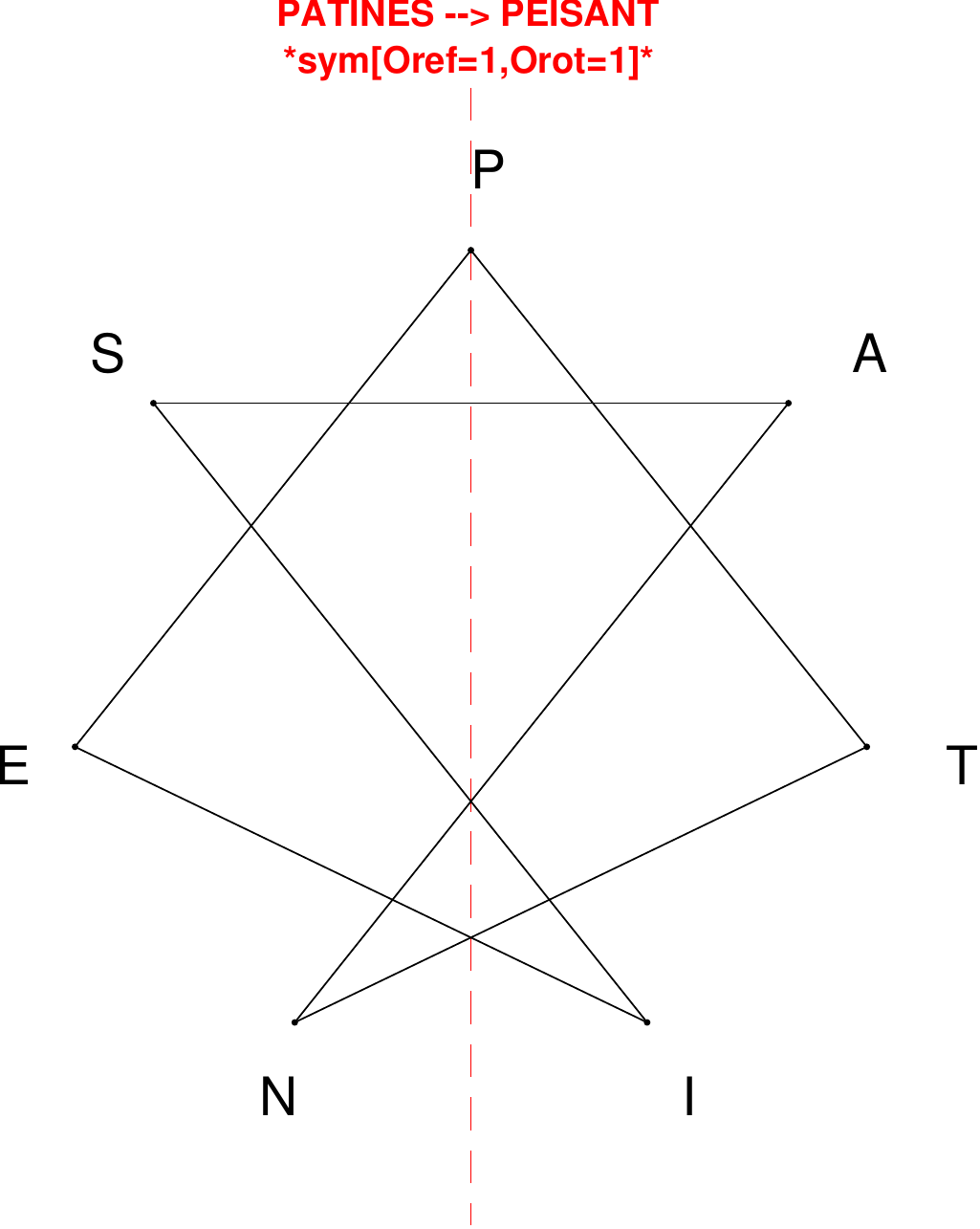}
\end{subfigure}
\hfill
\begin{subfigure}[T]{0.19\textwidth}
\centering
\includegraphics[width=\textwidth]{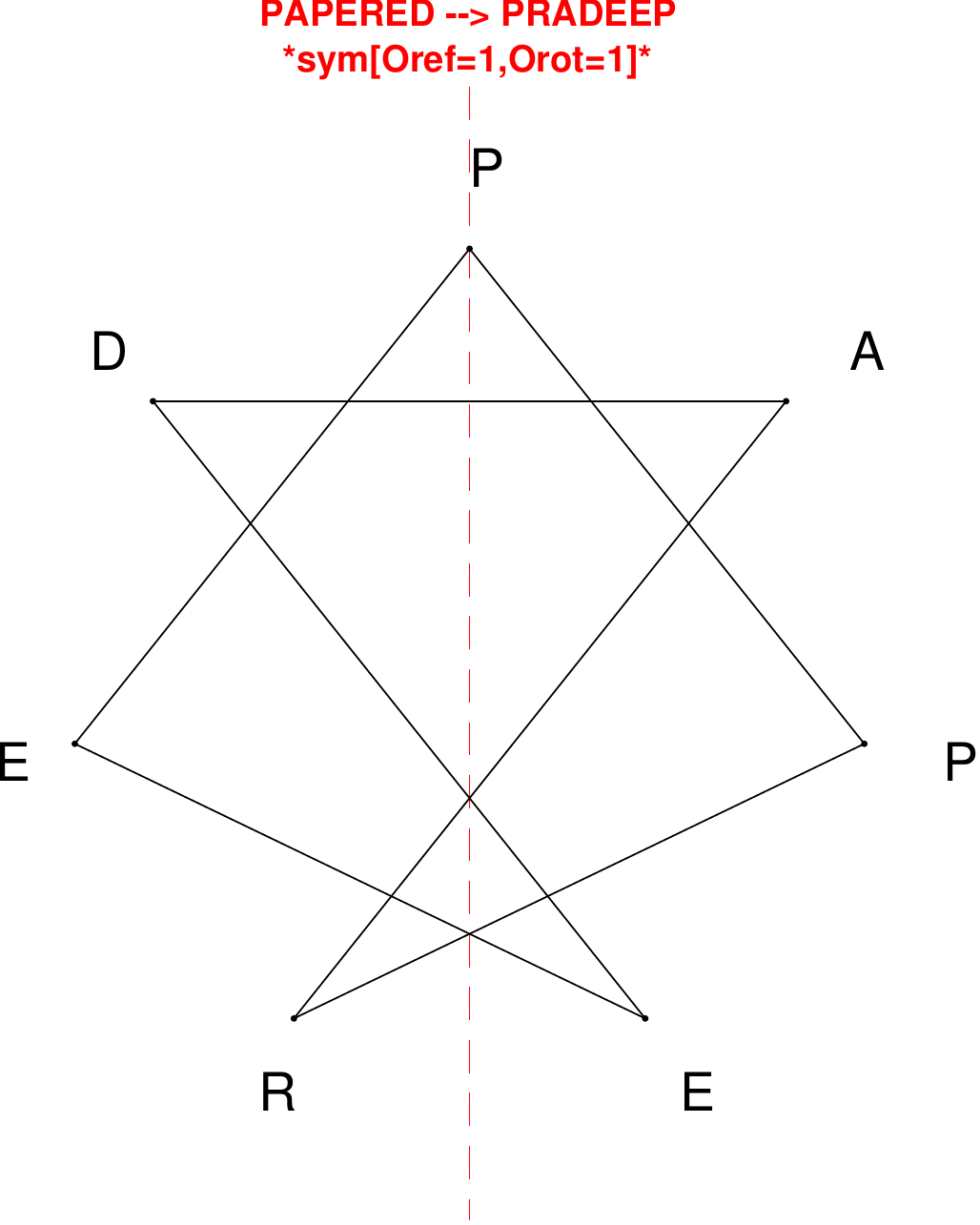}
\end{subfigure}
\hfill
\begin{subfigure}[T]{0.19\textwidth}
\centering
\includegraphics[width=\textwidth]{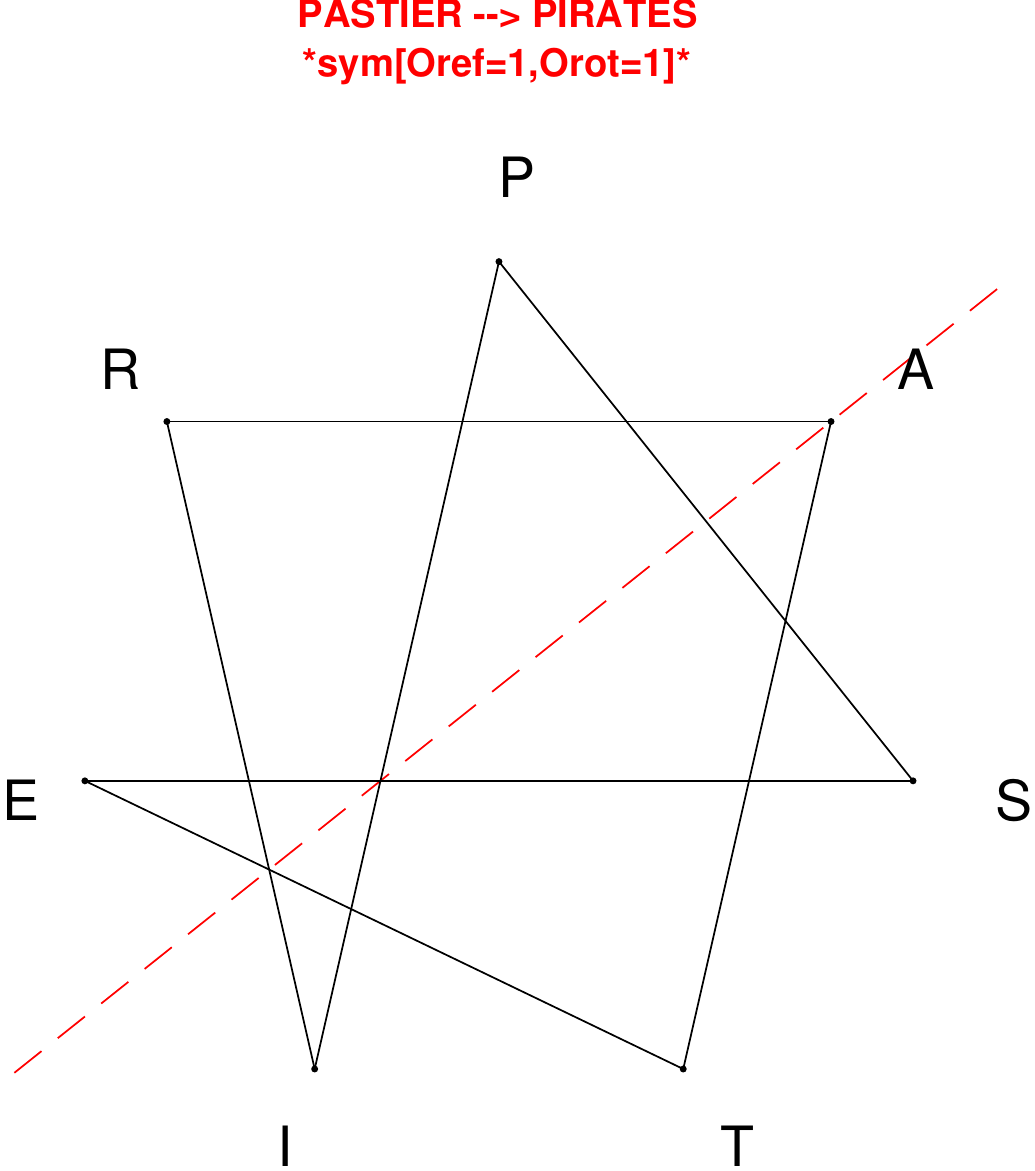}
\end{subfigure}
\hfill
\begin{subfigure}[T]{0.19\textwidth}
\centering
\includegraphics[width=\textwidth]{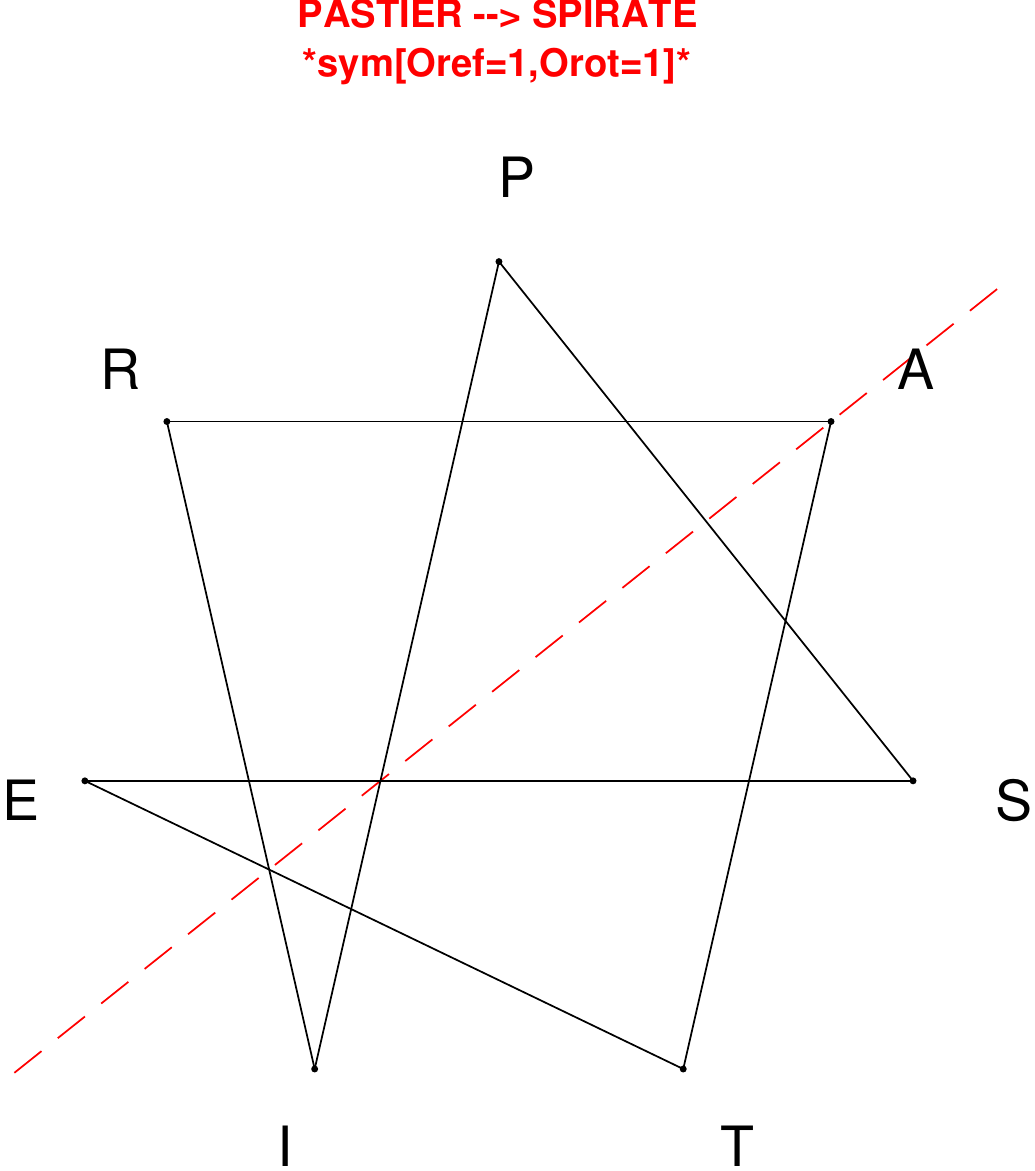}
\end{subfigure}
\hfill
\begin{subfigure}[T]{0.19\textwidth}
\centering
\includegraphics[width=\textwidth]{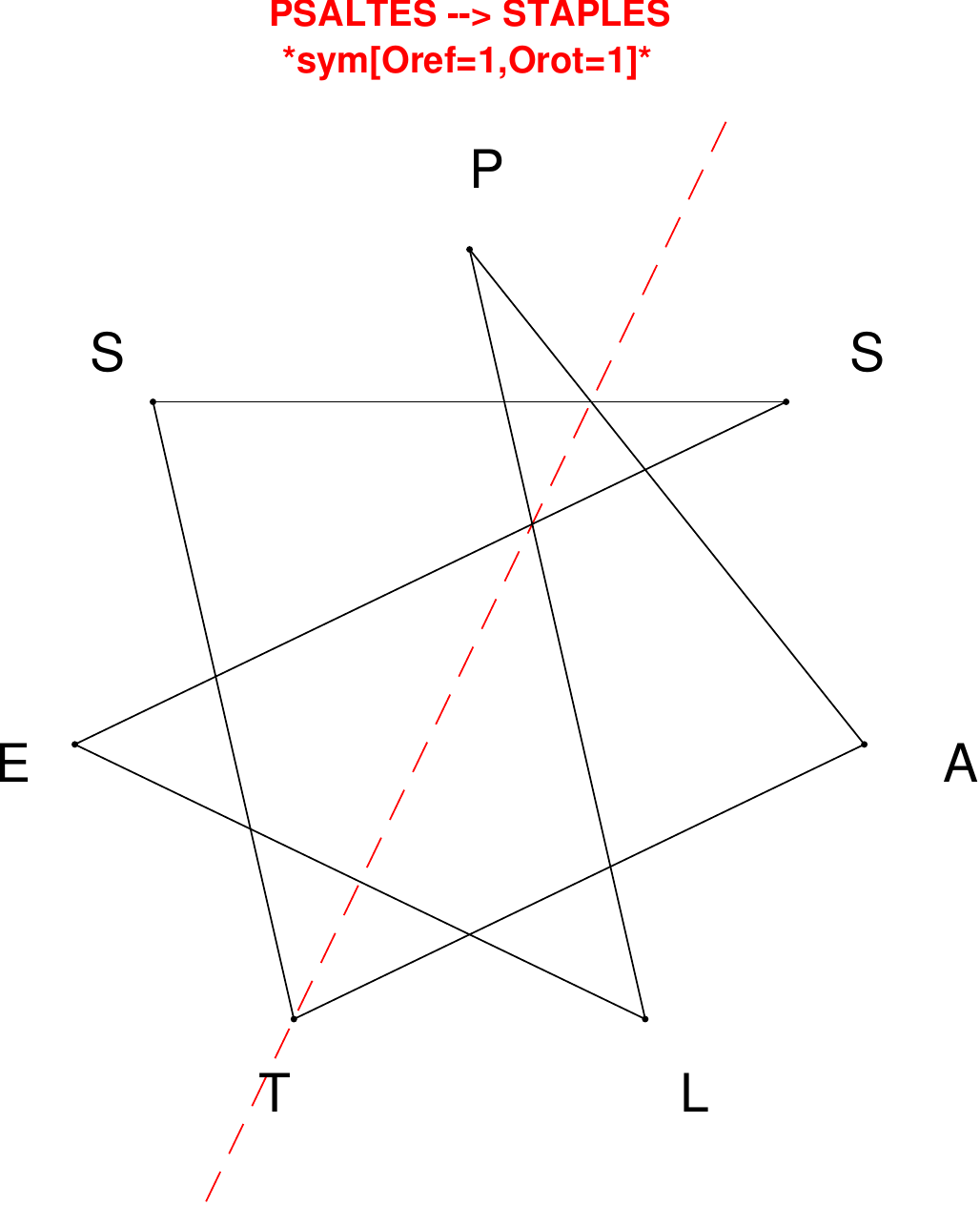}
\end{subfigure}
\end{figure}

\begin{figure}[H]
\centering
\begin{subfigure}[T]{0.19\textwidth}
\centering
\includegraphics[width=\textwidth]{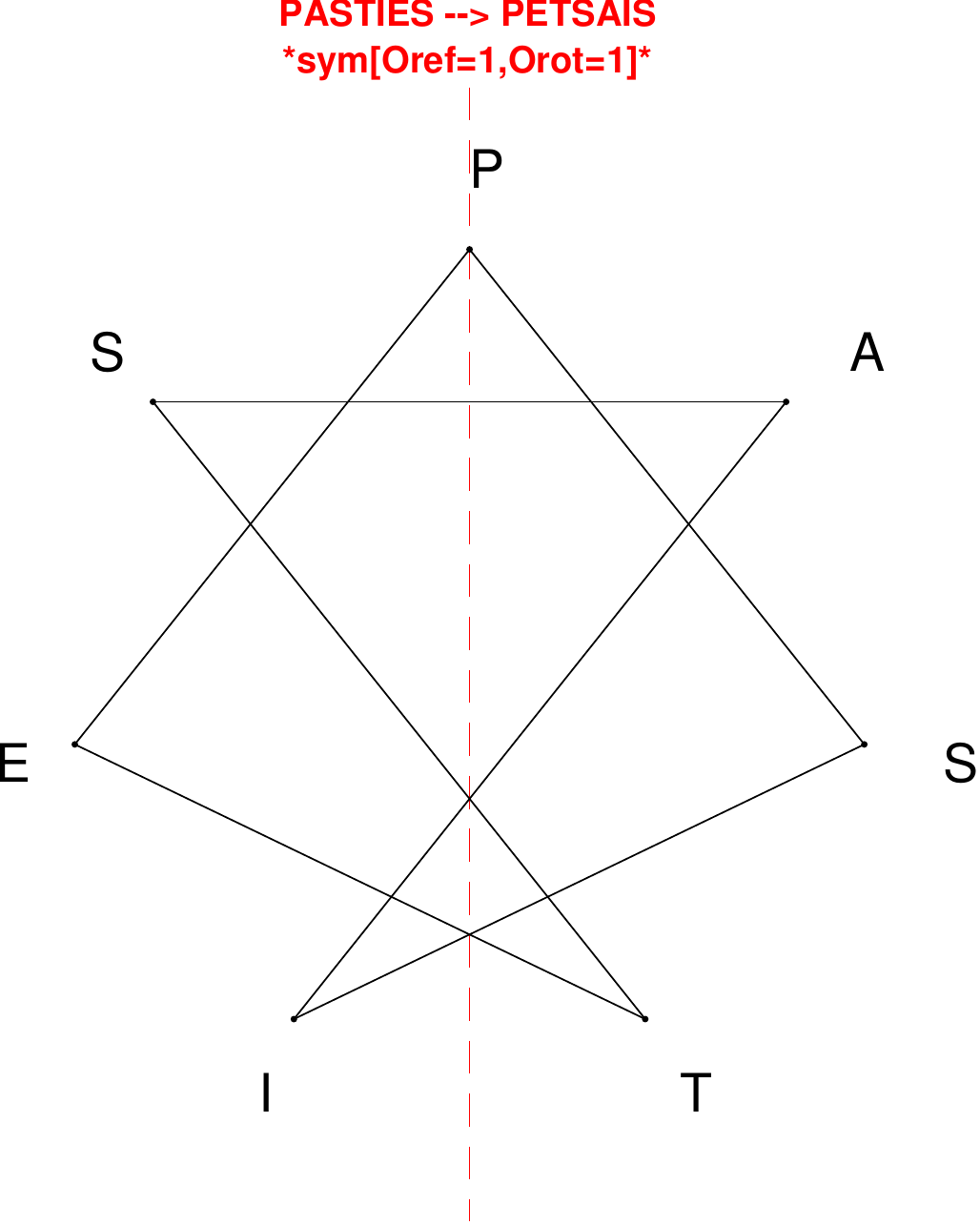}
\end{subfigure}
\hfill
\begin{subfigure}[T]{0.19\textwidth}
\centering
\includegraphics[width=\textwidth]{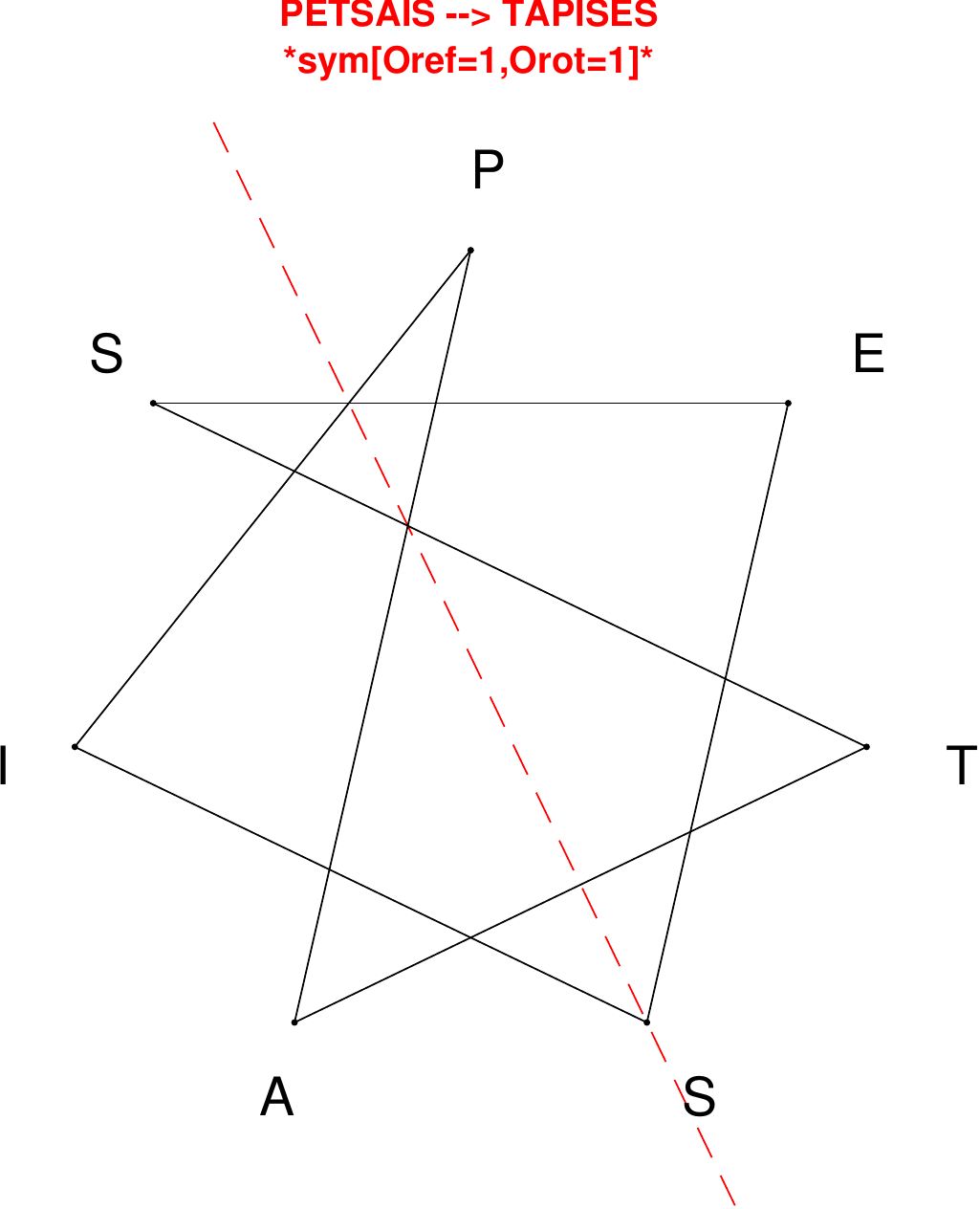}
\end{subfigure}
\hfill
\begin{subfigure}[T]{0.19\textwidth}
\centering
\includegraphics[width=\textwidth]{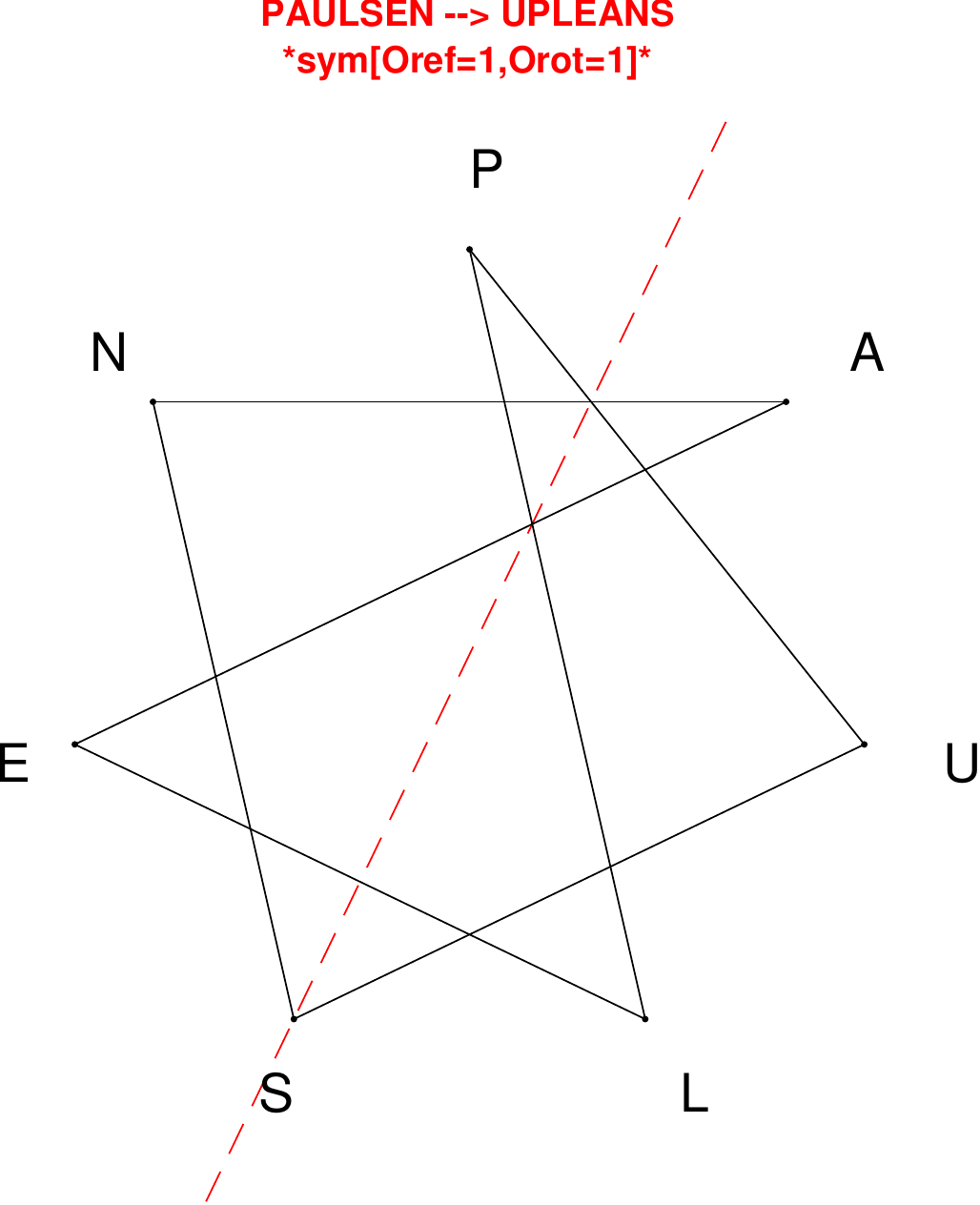}
\end{subfigure}
\hfill
\begin{subfigure}[T]{0.19\textwidth}
\centering
\includegraphics[width=\textwidth]{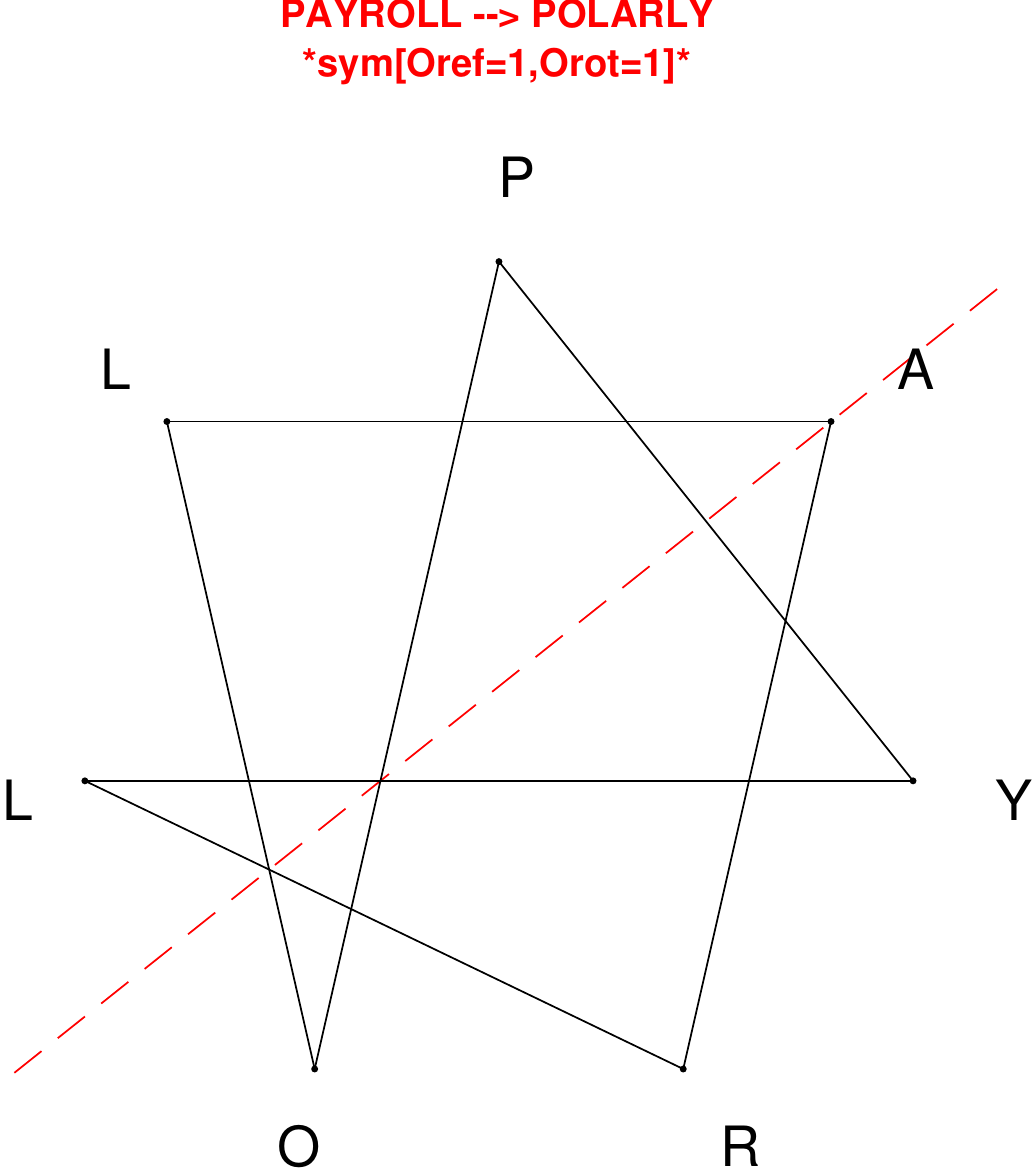}
\end{subfigure}
\hfill
\begin{subfigure}[T]{0.19\textwidth}
\centering
\includegraphics[width=\textwidth]{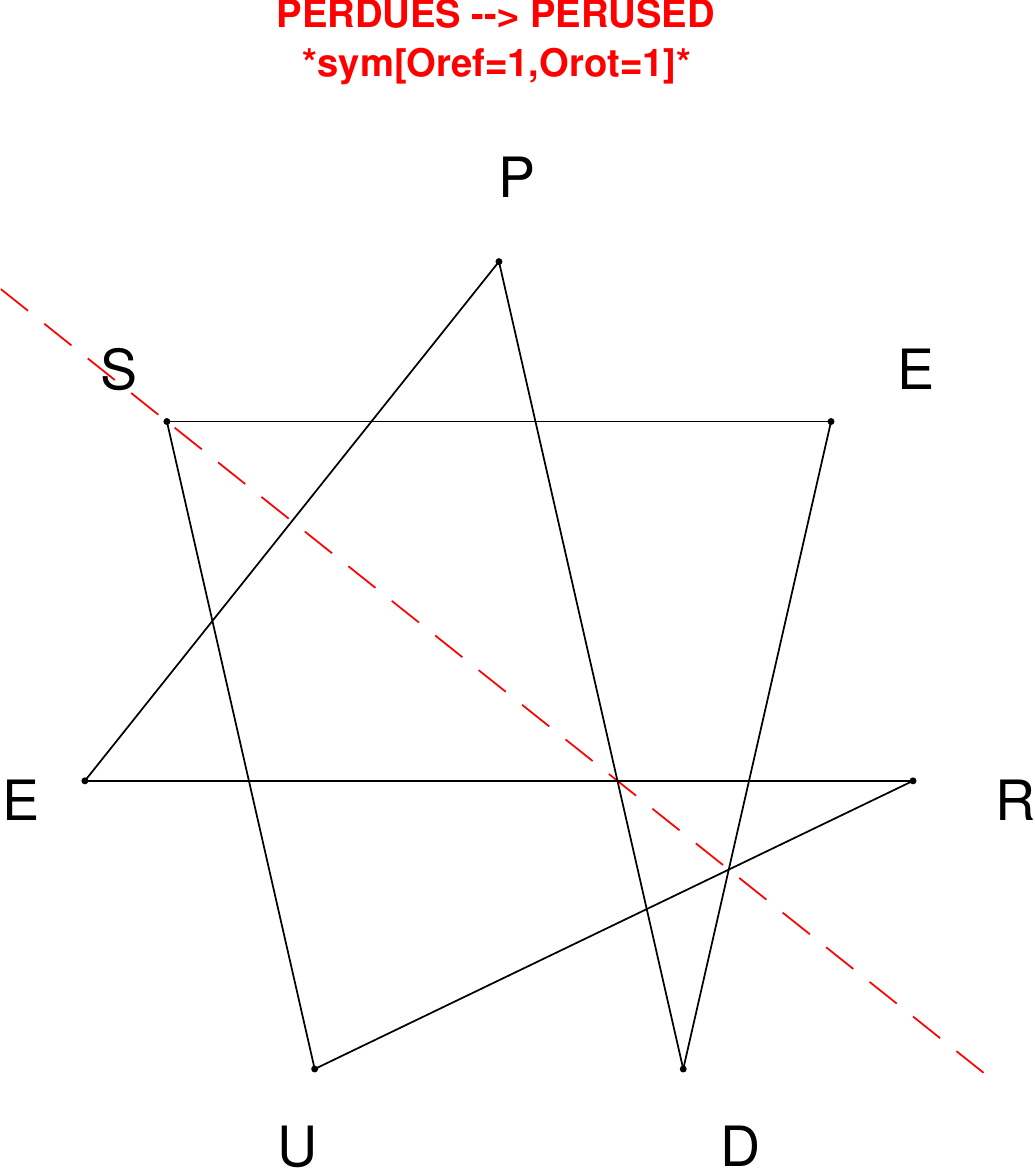}
\end{subfigure}
\end{figure}

\begin{figure}[H]
\centering
\begin{subfigure}[T]{0.19\textwidth}
\centering
\includegraphics[width=\textwidth]{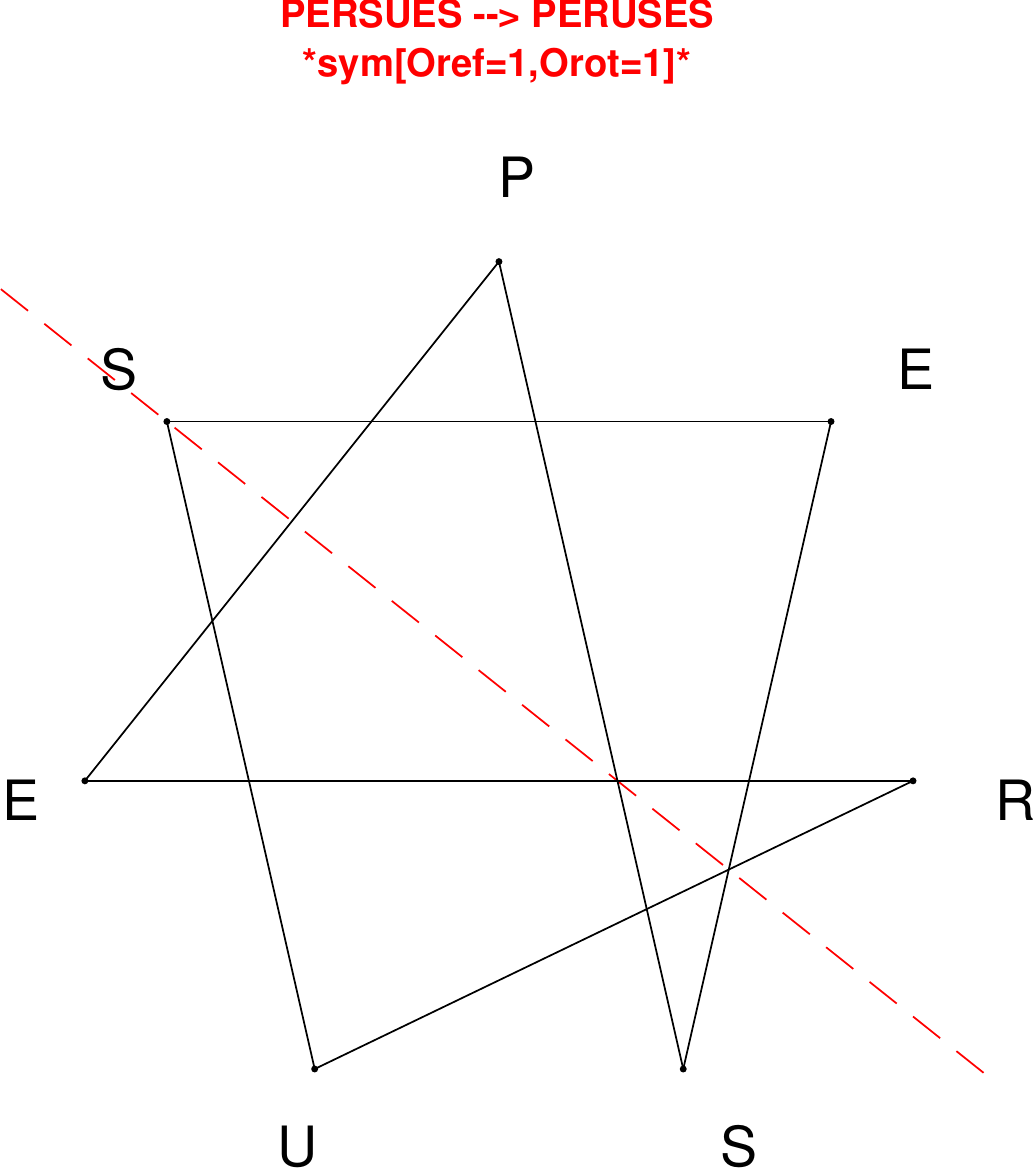}
\end{subfigure}
\hfill
\begin{subfigure}[T]{0.19\textwidth}
\centering
\includegraphics[width=\textwidth]{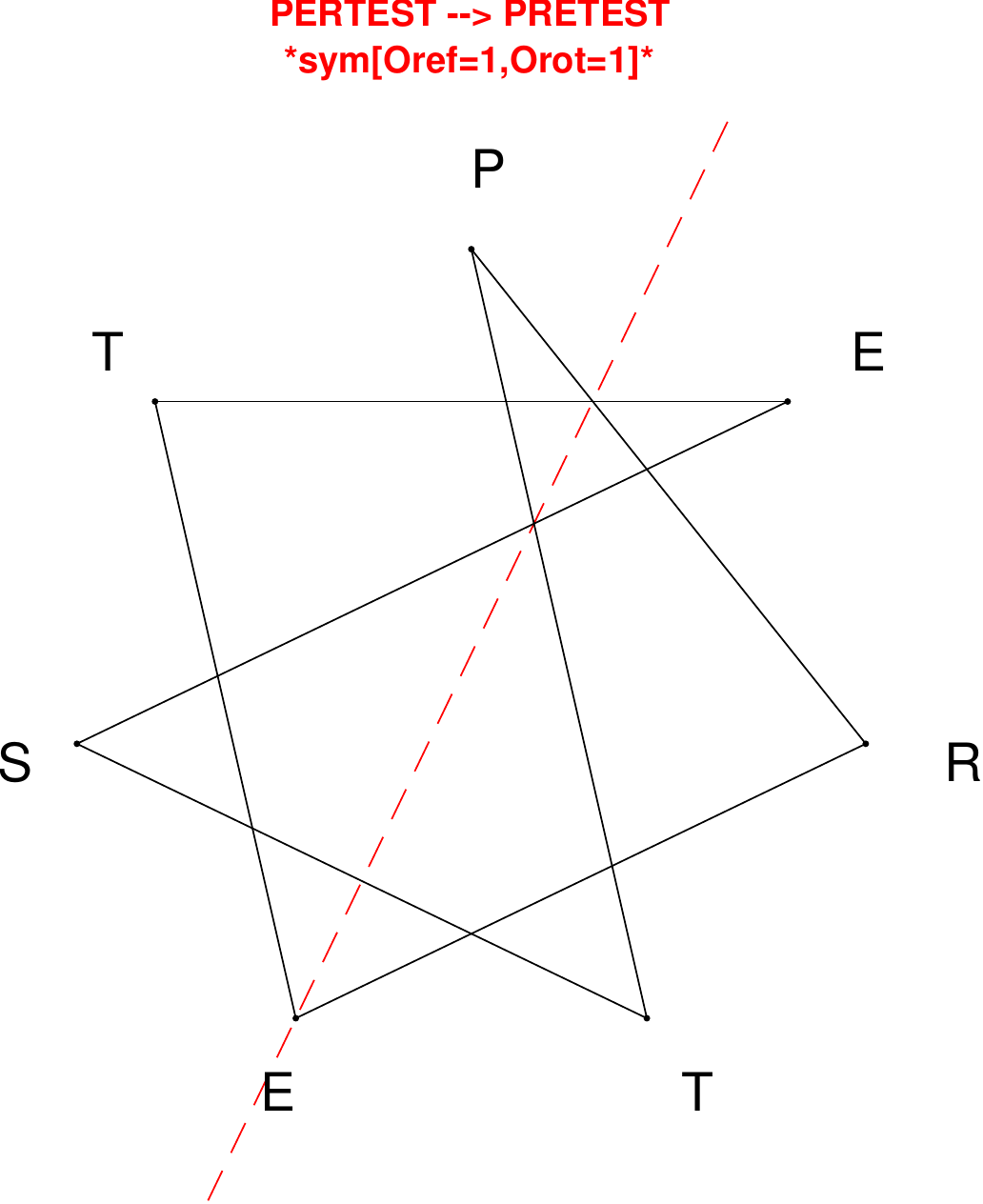}
\end{subfigure}
\hfill
\begin{subfigure}[T]{0.19\textwidth}
\centering
\includegraphics[width=\textwidth]{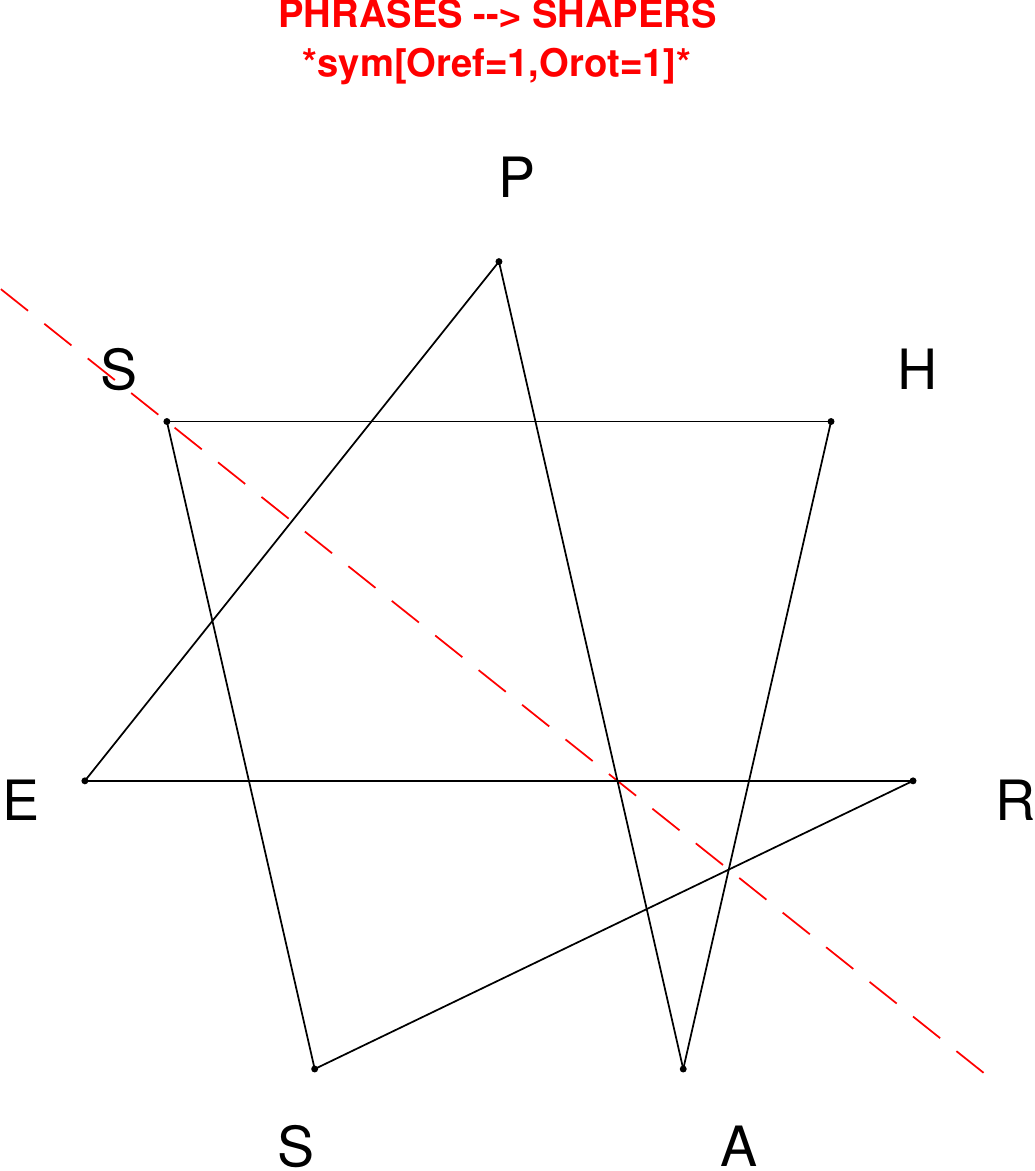}
\end{subfigure}
\hfill
\begin{subfigure}[T]{0.19\textwidth}
\centering
\includegraphics[width=\textwidth]{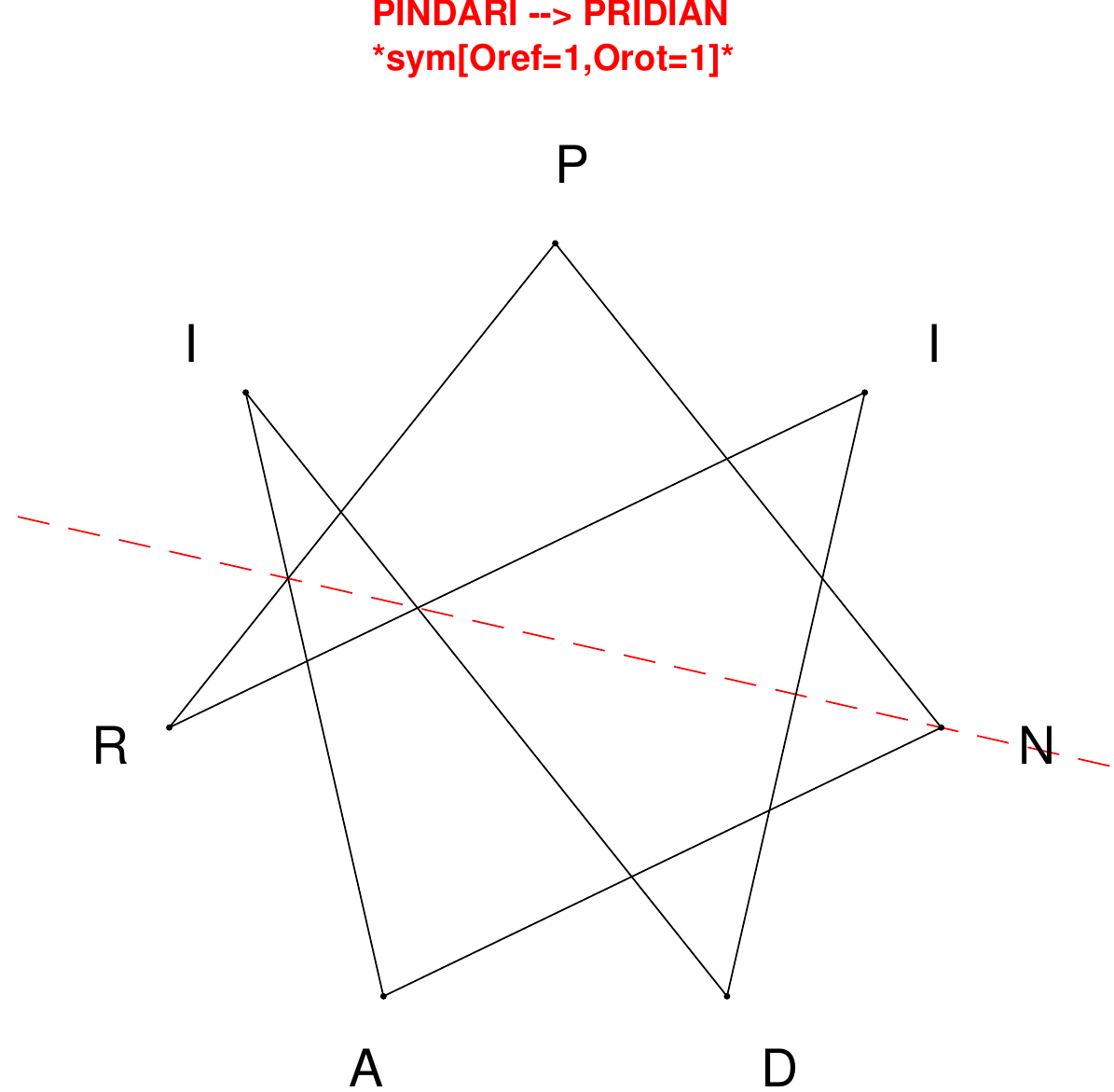}
\end{subfigure}
\hfill
\begin{subfigure}[T]{0.19\textwidth}
\centering
\includegraphics[width=\textwidth]{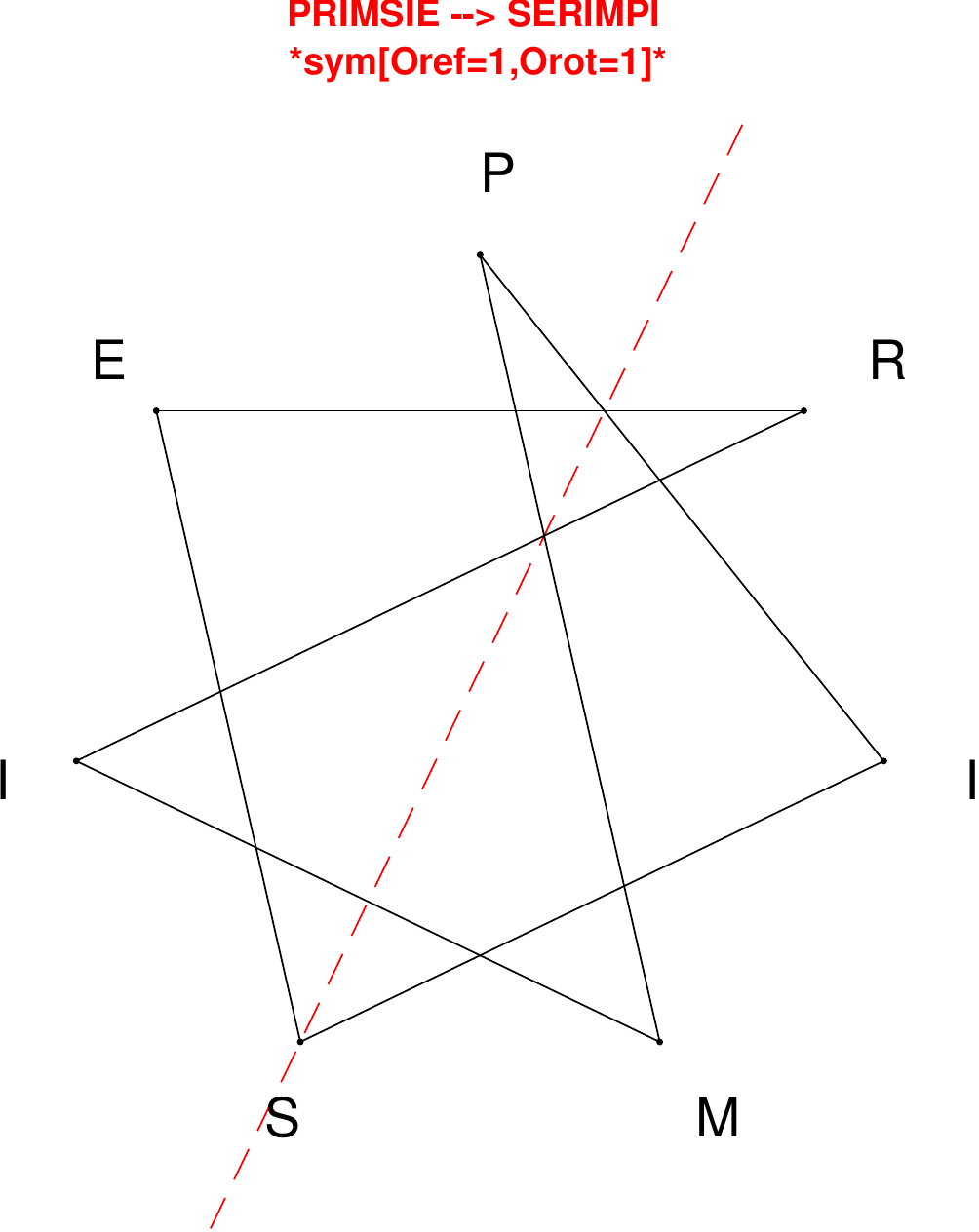}
\end{subfigure}
\end{figure}

\begin{figure}[H]
\centering
\begin{subfigure}[T]{0.19\textwidth}
\centering
\includegraphics[width=\textwidth]{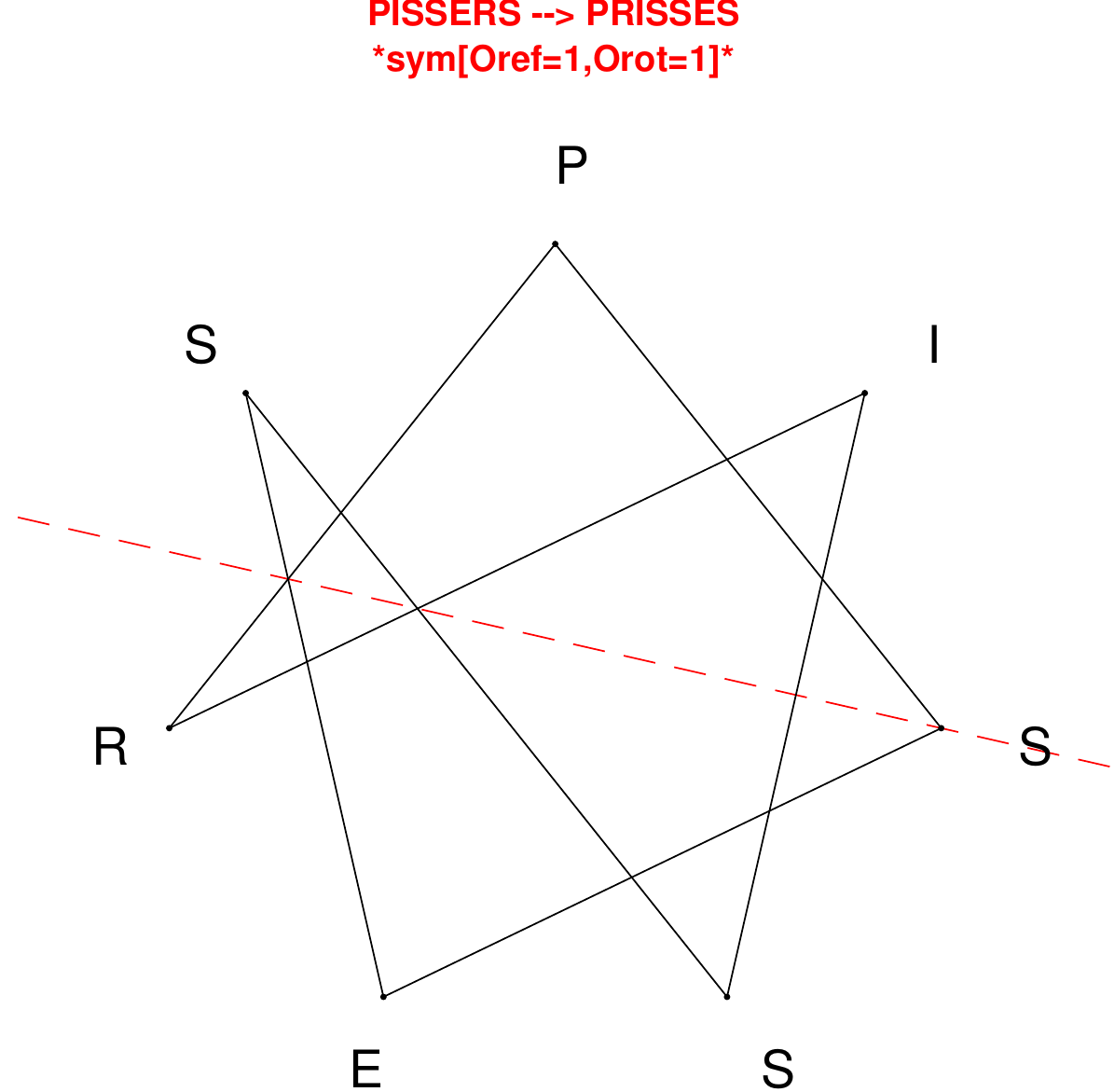}
\end{subfigure}
\hfill
\begin{subfigure}[T]{0.19\textwidth}
\centering
\includegraphics[width=\textwidth]{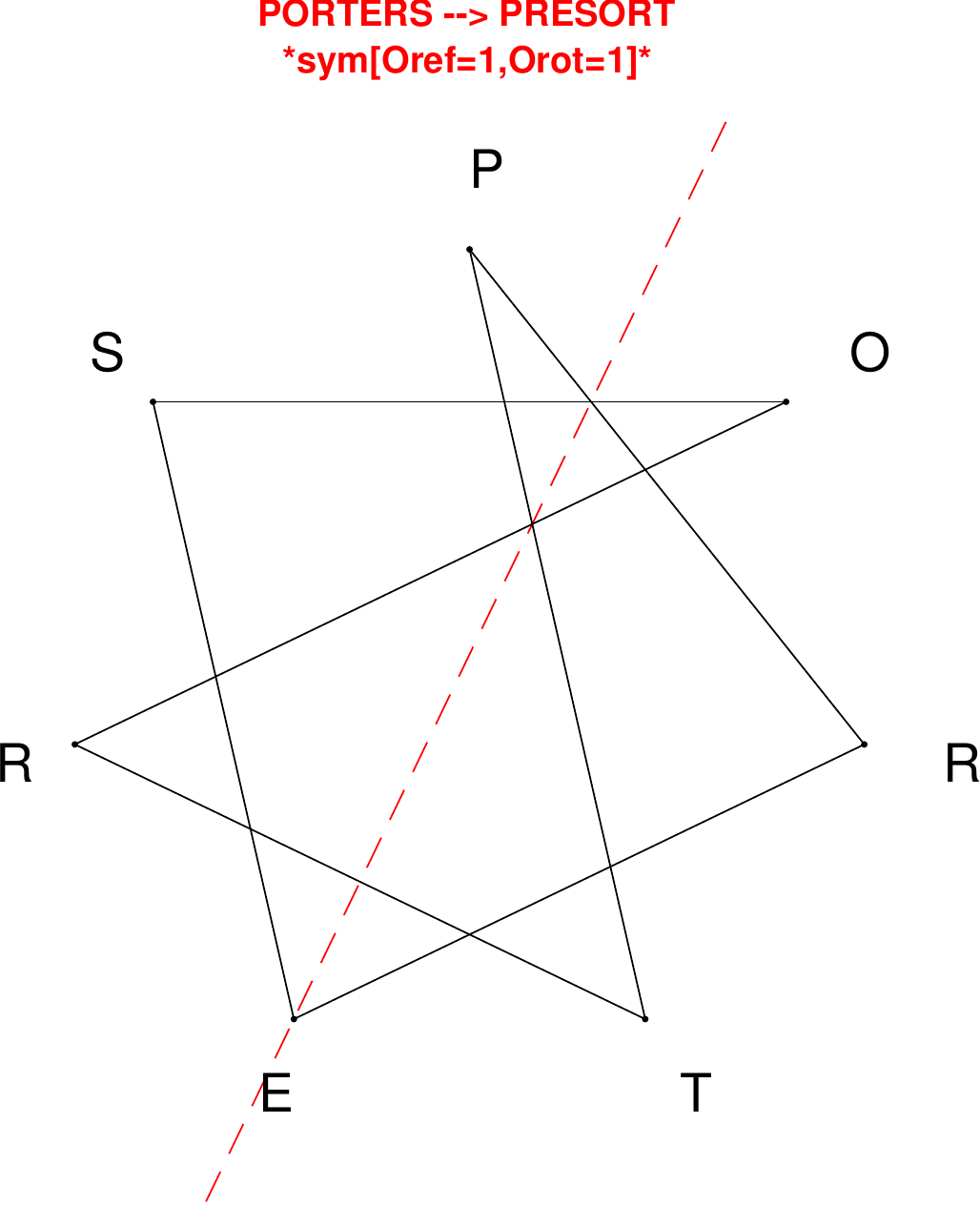}
\end{subfigure}
\hfill
\begin{subfigure}[T]{0.19\textwidth}
\centering
\includegraphics[width=\textwidth]{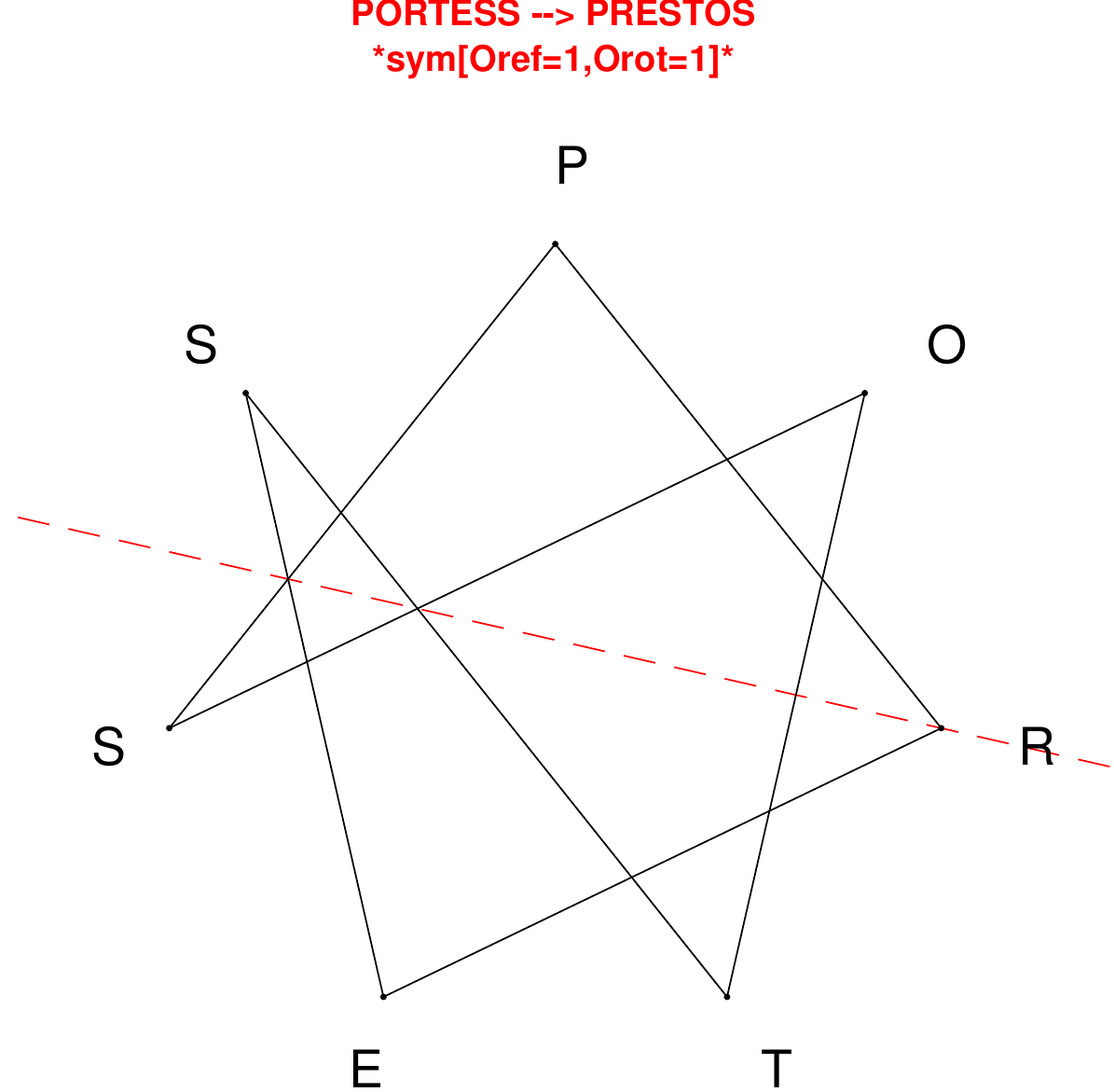}
\end{subfigure}
\hfill
\begin{subfigure}[T]{0.19\textwidth}
\centering
\includegraphics[width=\textwidth]{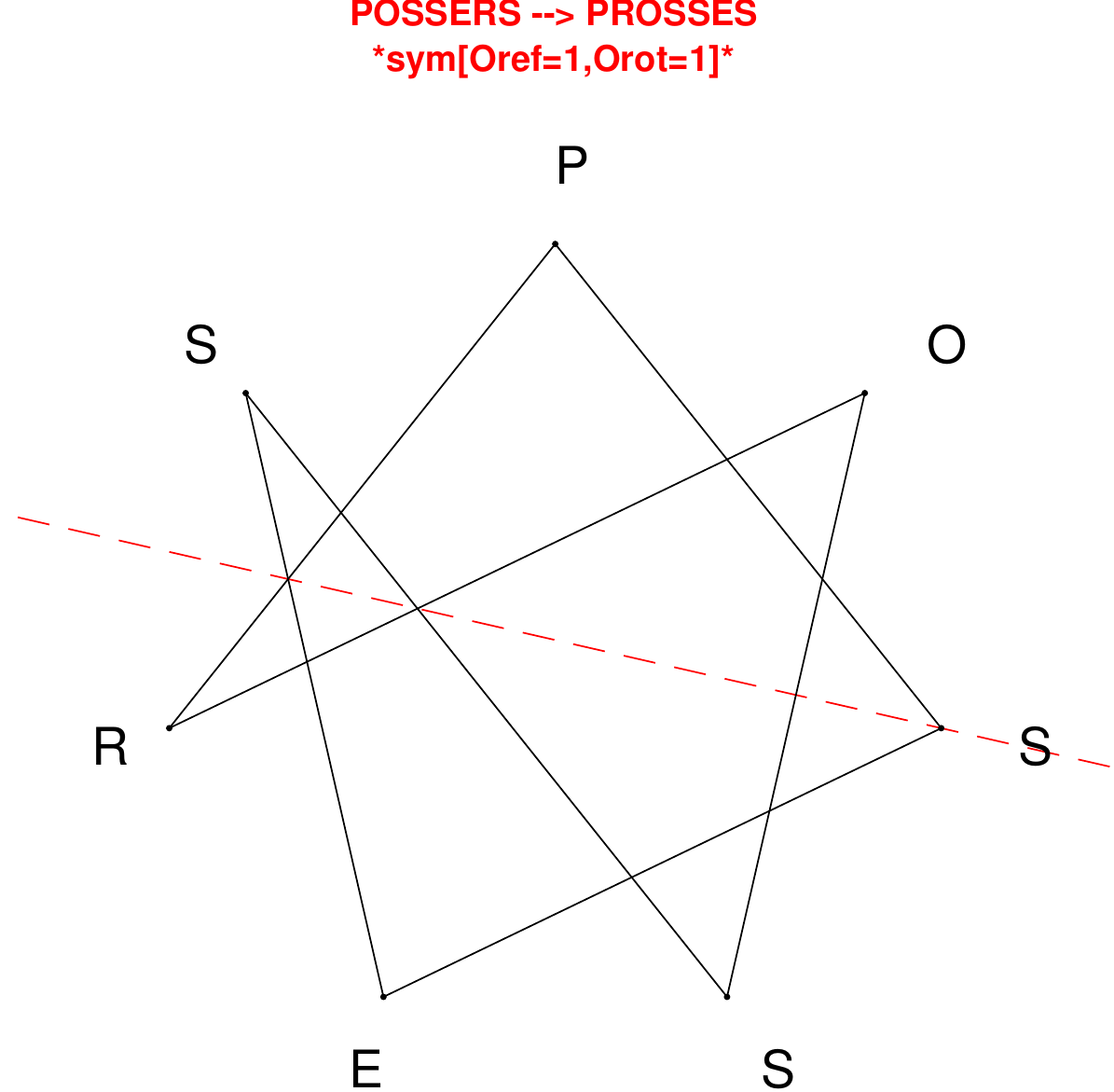}
\end{subfigure}
\hfill
\begin{subfigure}[T]{0.19\textwidth}
\centering
\includegraphics[width=\textwidth]{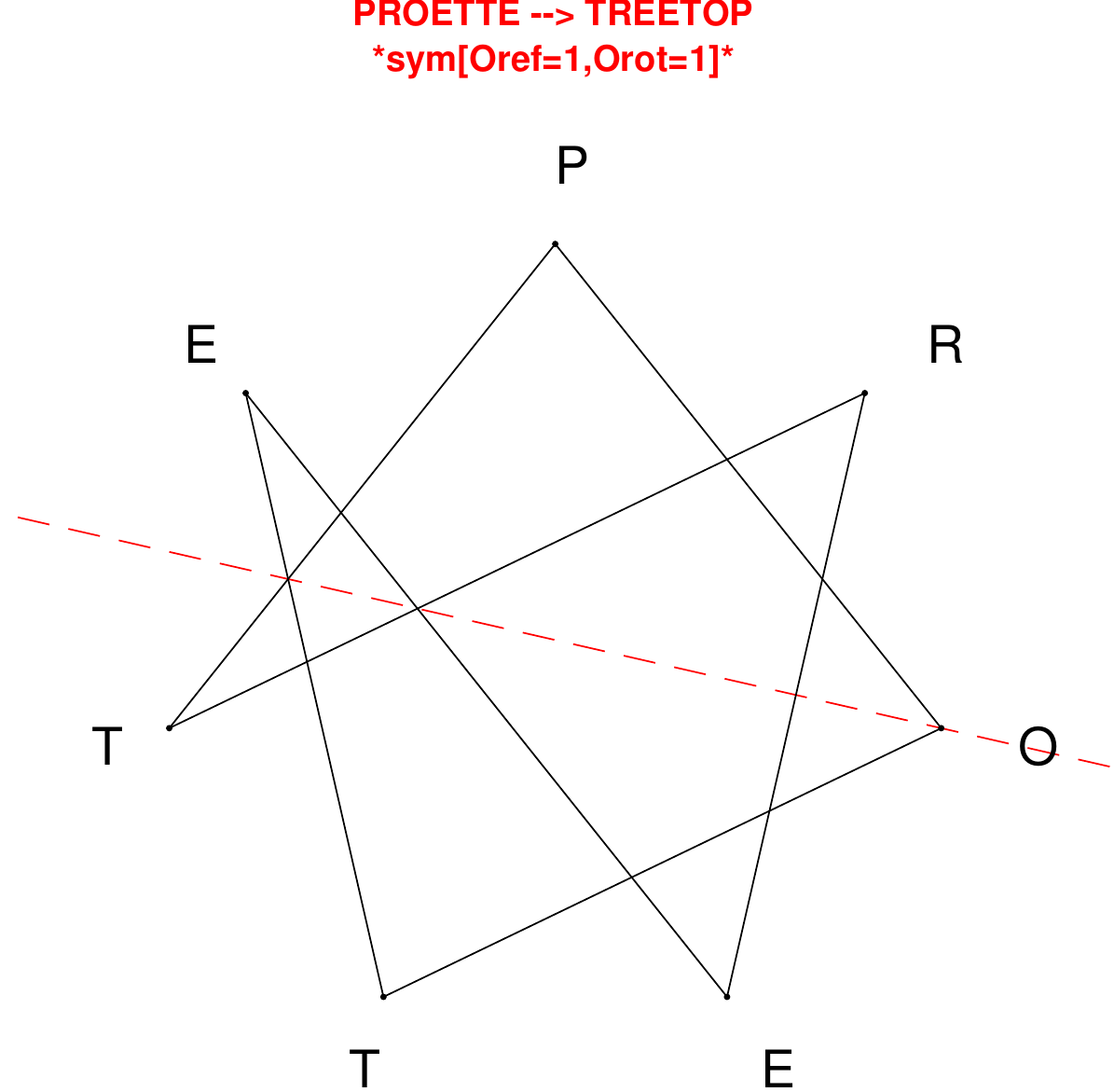}
\end{subfigure}
\end{figure}

\begin{figure}[H]
\centering
\begin{subfigure}[T]{0.19\textwidth}
\centering
\includegraphics[width=\textwidth]{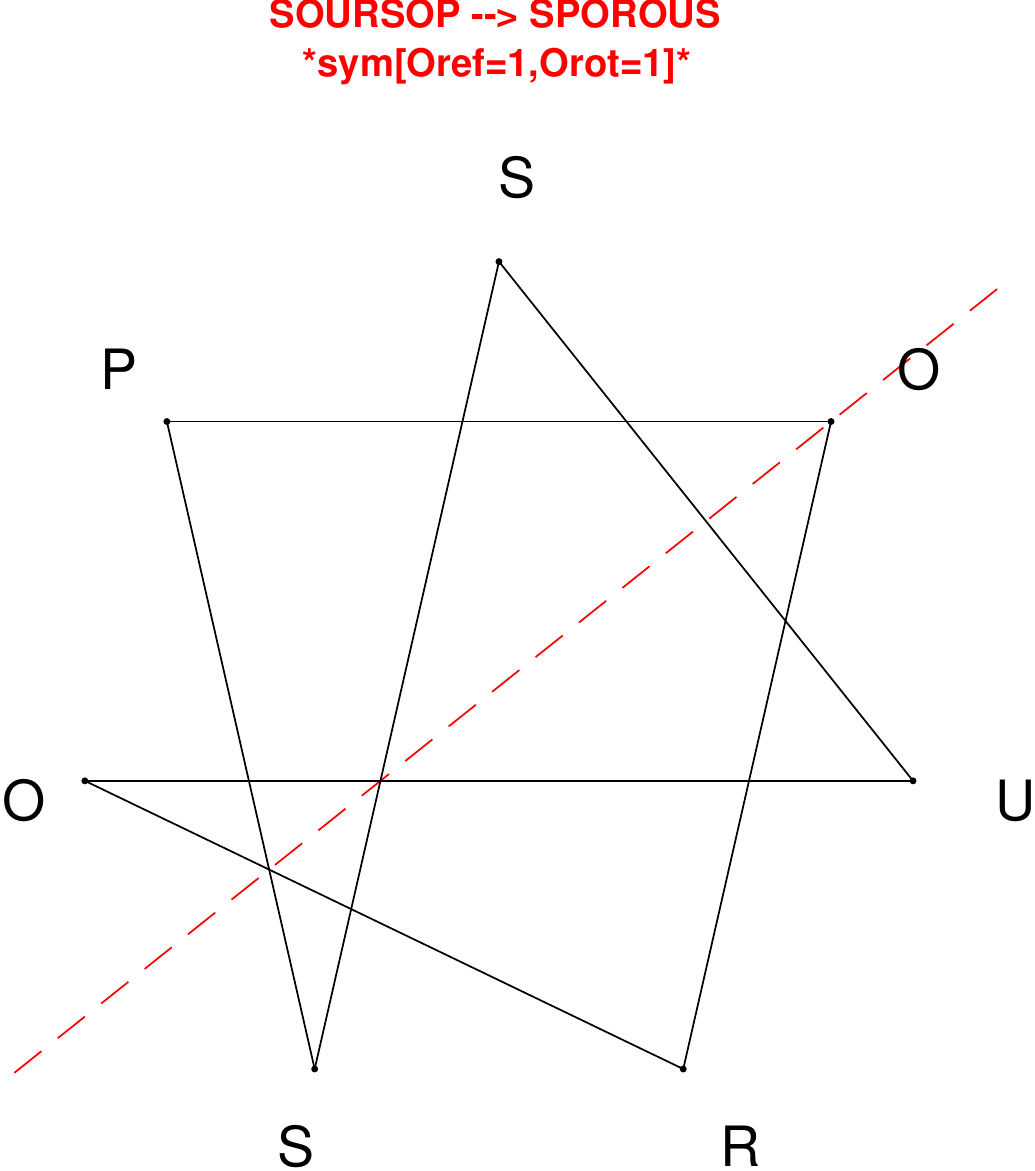}
\end{subfigure}
\hfill
\begin{subfigure}[T]{0.19\textwidth}
\centering
\includegraphics[width=\textwidth]{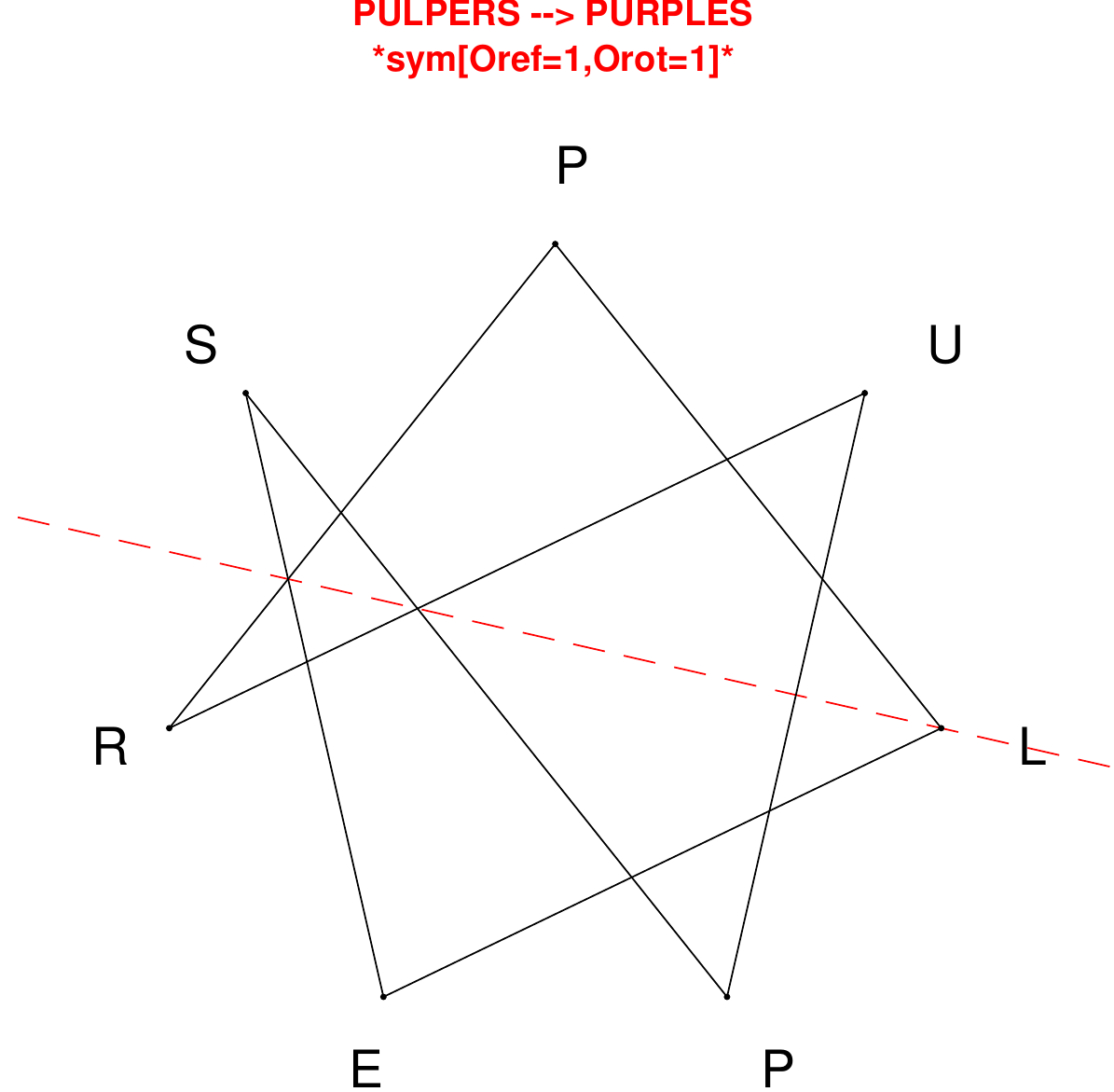}
\end{subfigure}
\hfill
\begin{subfigure}[T]{0.19\textwidth}
\centering
\includegraphics[width=\textwidth]{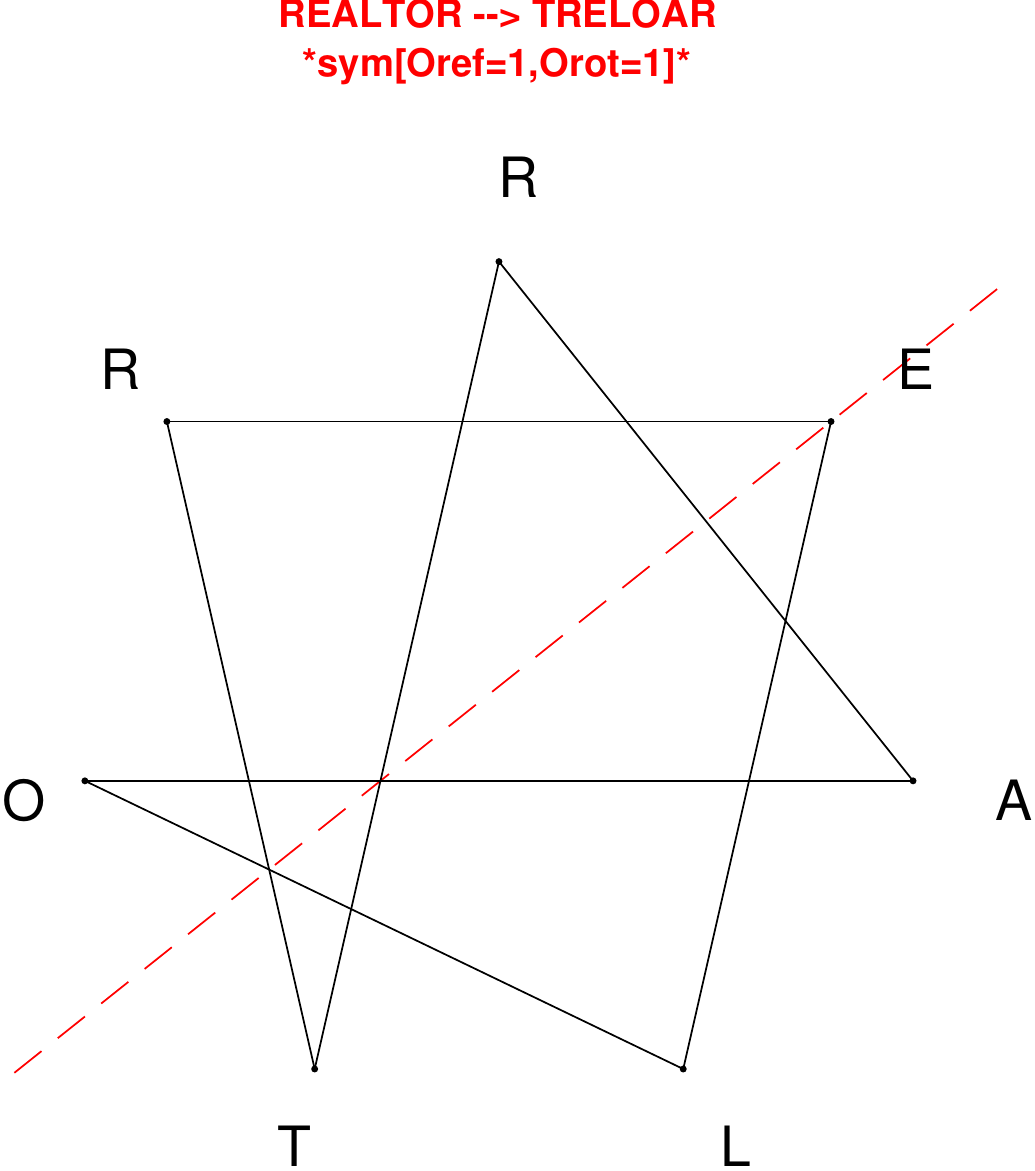}
\end{subfigure}
\hfill
\begin{subfigure}[T]{0.19\textwidth}
\centering
\includegraphics[width=\textwidth]{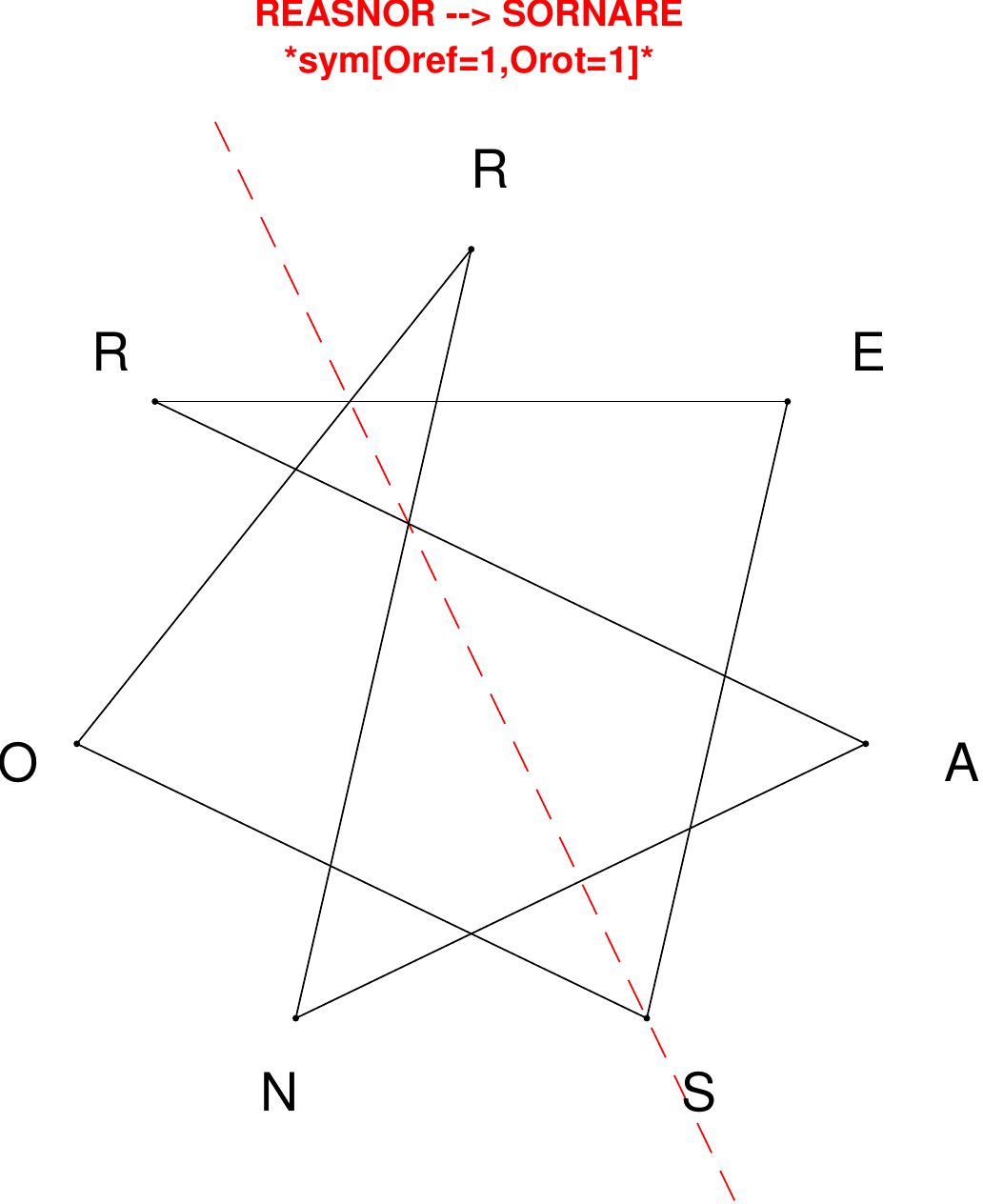}
\end{subfigure}
\hfill
\begin{subfigure}[T]{0.19\textwidth}
\centering
\includegraphics[width=\textwidth]{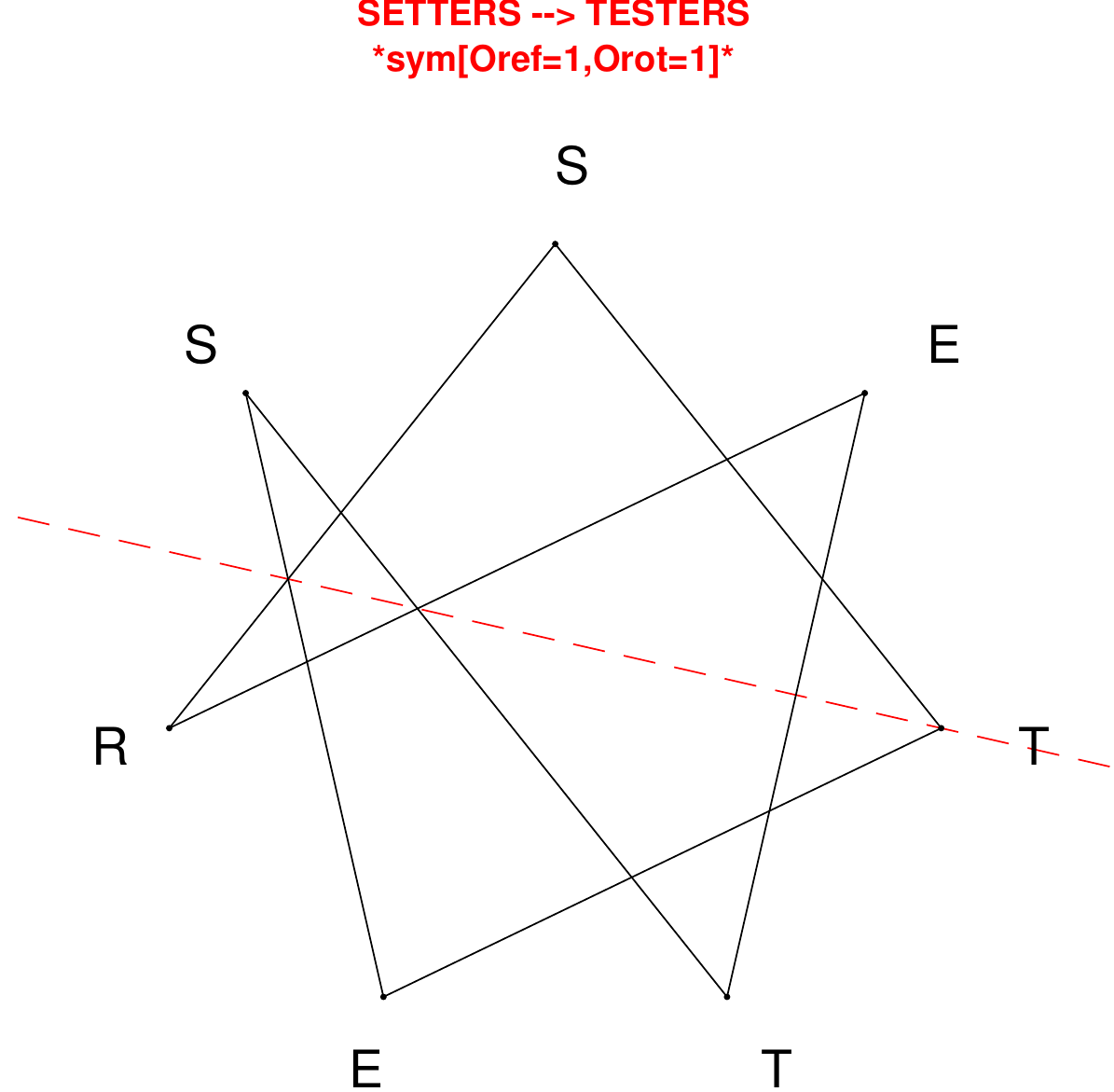}
\end{subfigure}
\end{figure}

\begin{figure}[H]
\centering
\begin{subfigure}[T]{0.19\textwidth}
\centering
\includegraphics[width=\textwidth]{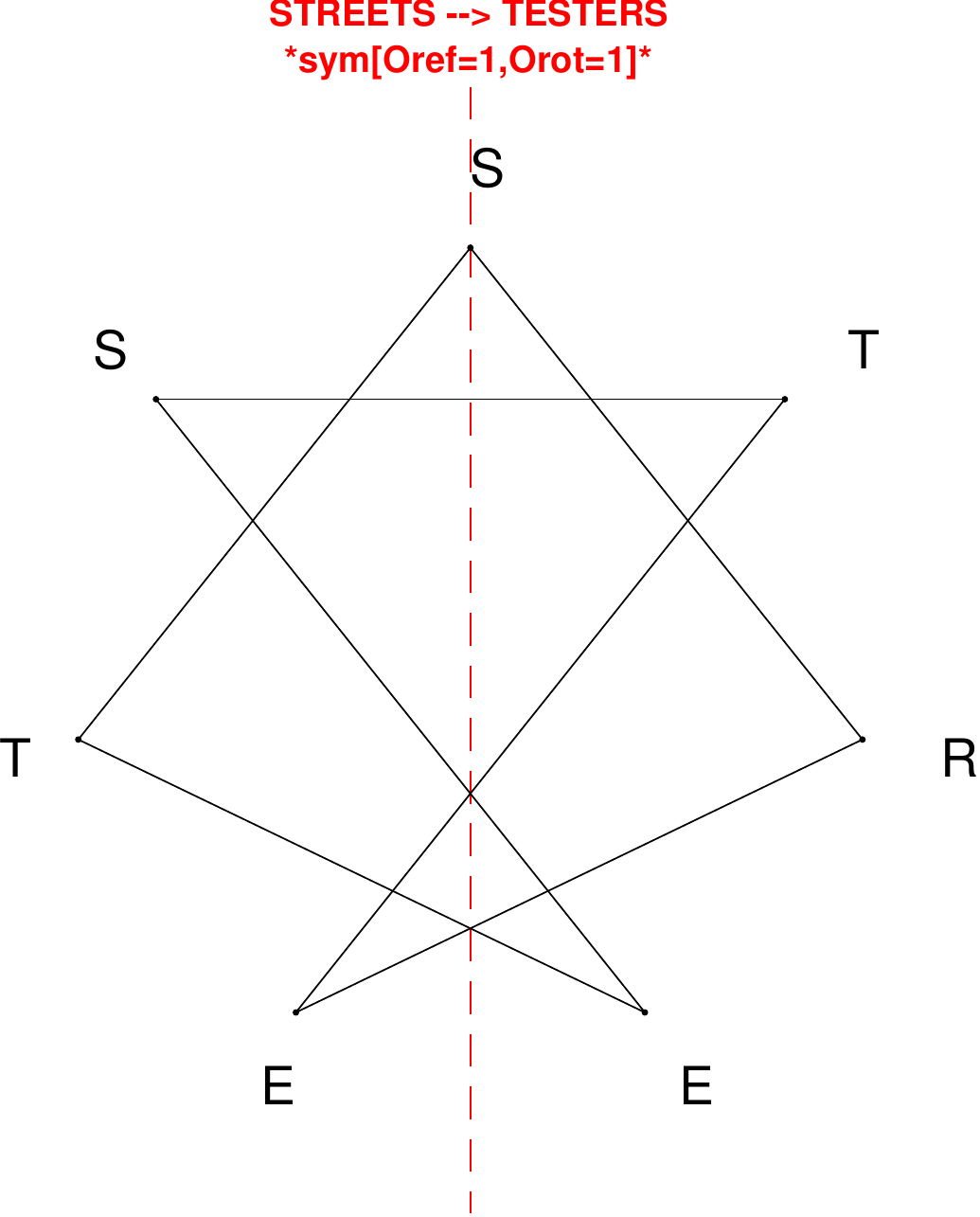}
\end{subfigure}
\hfill
\begin{subfigure}[T]{0.19\textwidth}
\centering
\includegraphics[width=\textwidth]{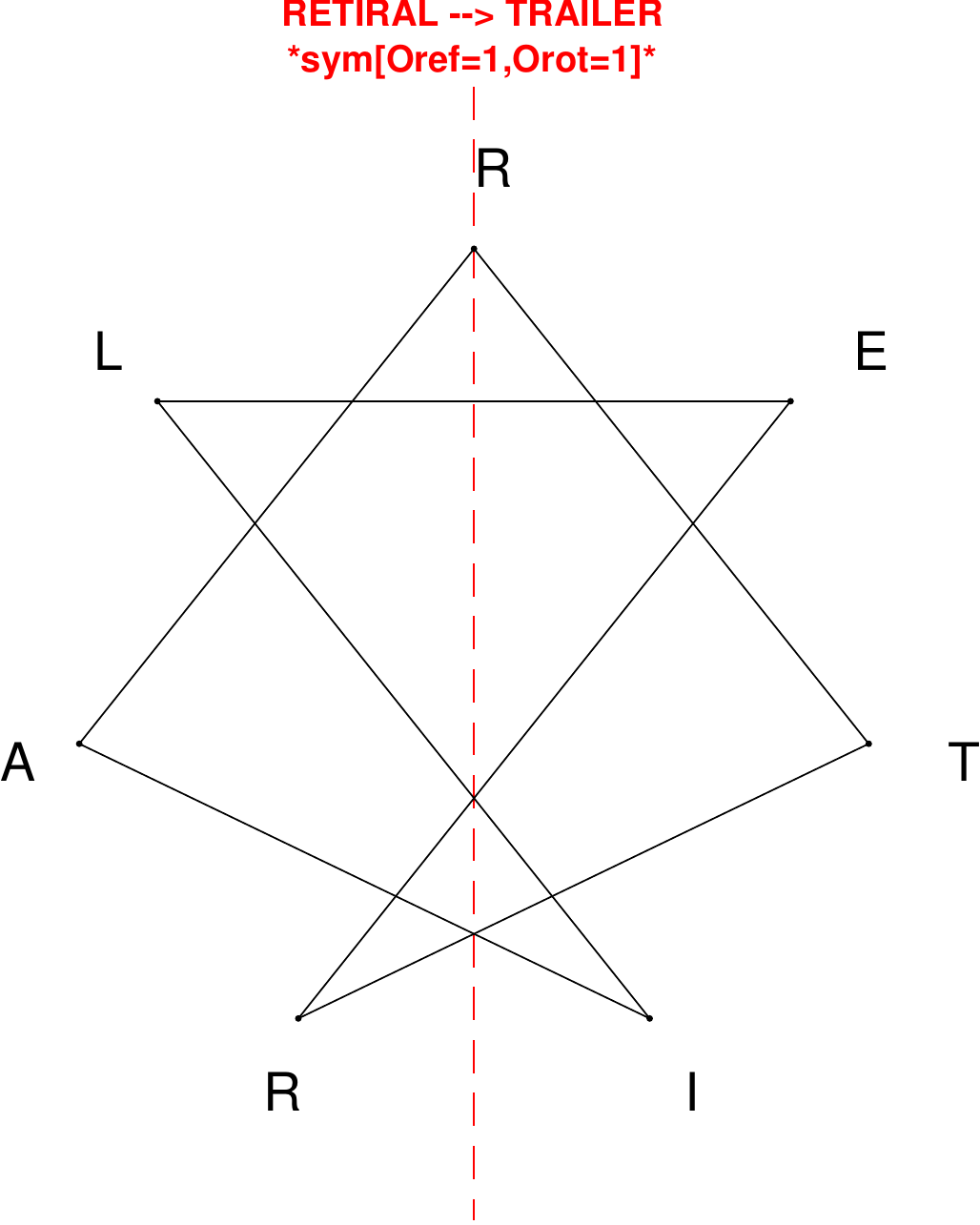}
\end{subfigure}
\hfill
\begin{subfigure}[T]{0.19\textwidth}
\centering
\includegraphics[width=\textwidth]{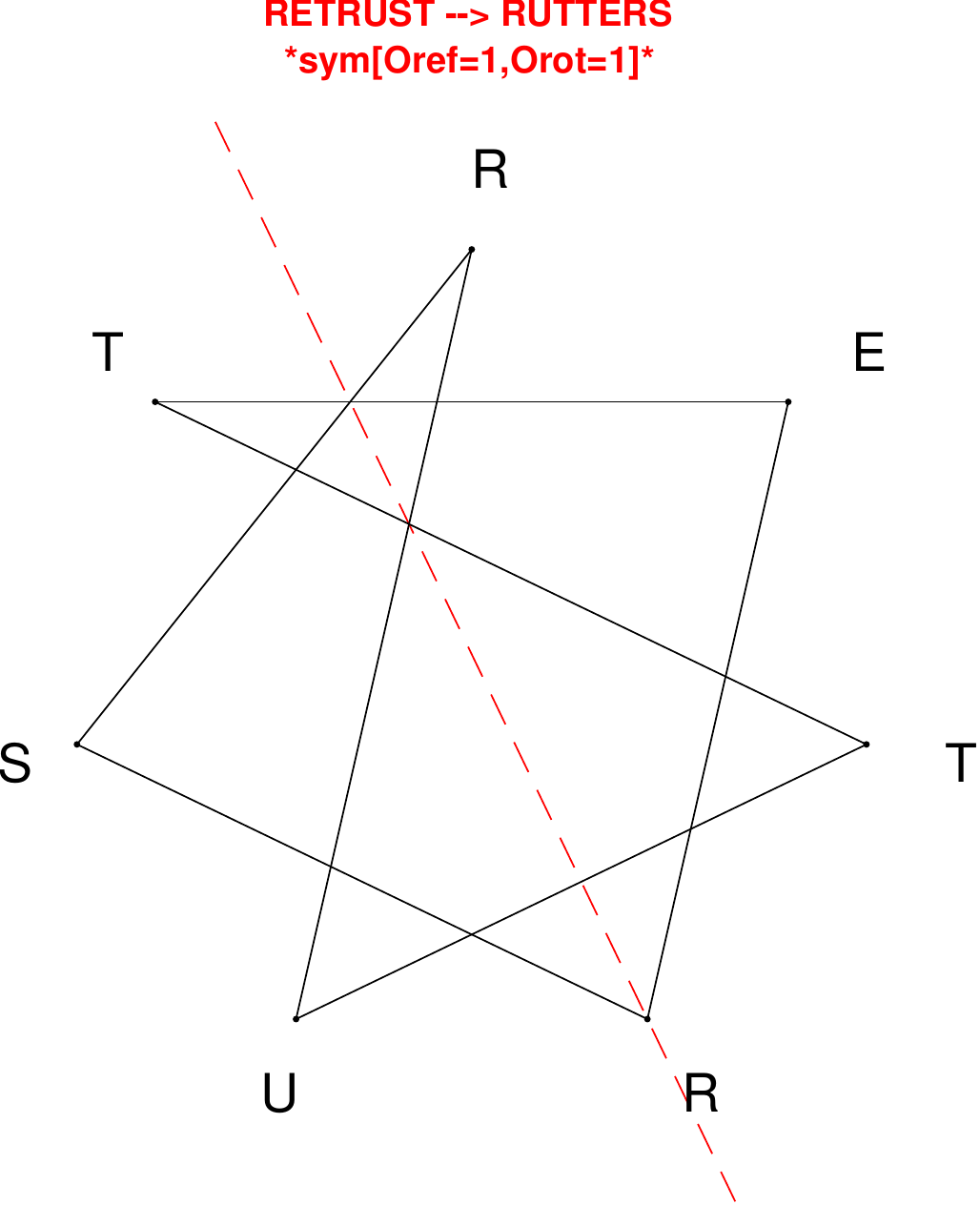}
\end{subfigure}
\hfill
\begin{subfigure}[T]{0.19\textwidth}
\centering
\includegraphics[width=\textwidth]{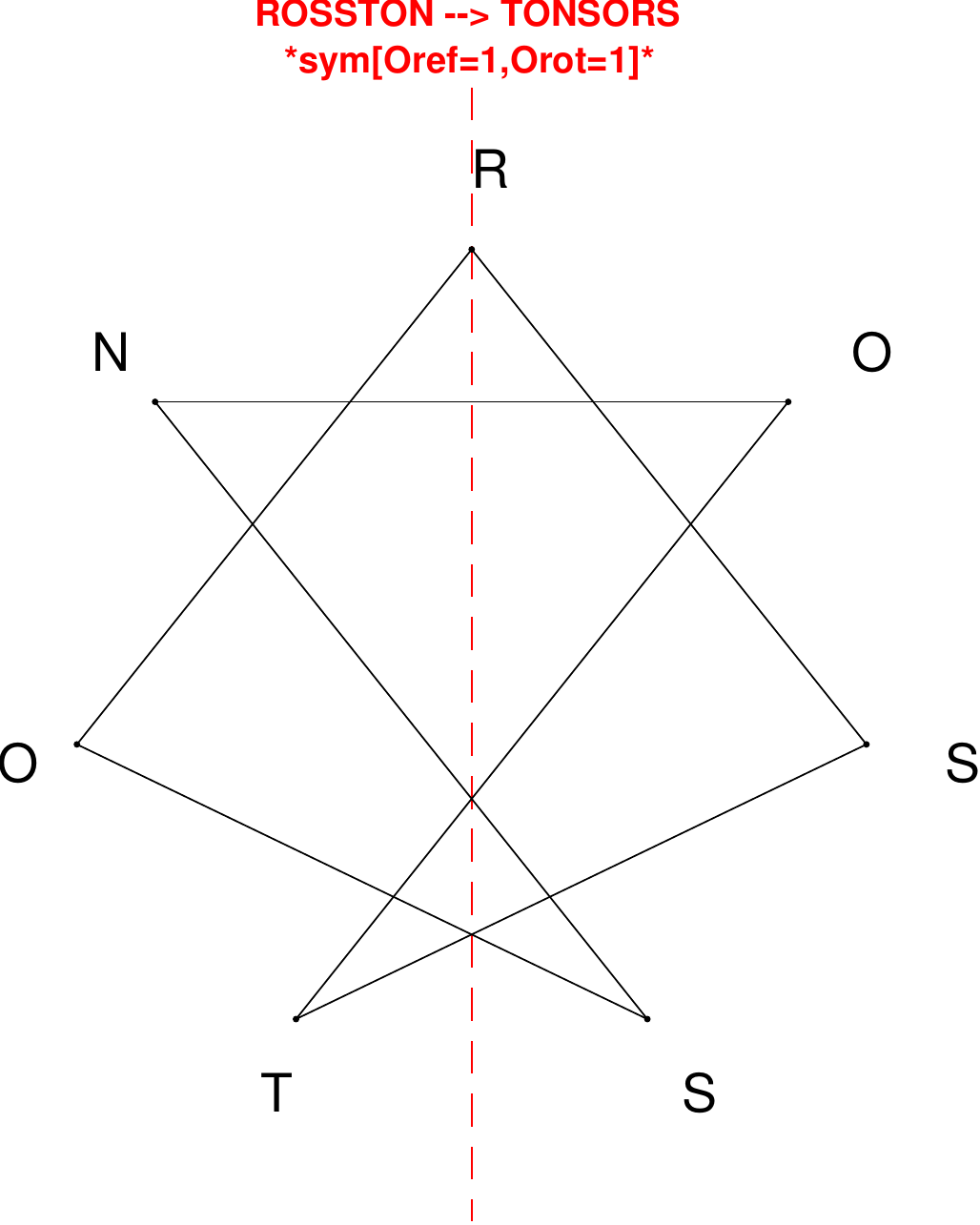}
\end{subfigure}
\hfill
\begin{subfigure}[T]{0.19\textwidth}
\centering
\includegraphics[width=\textwidth]{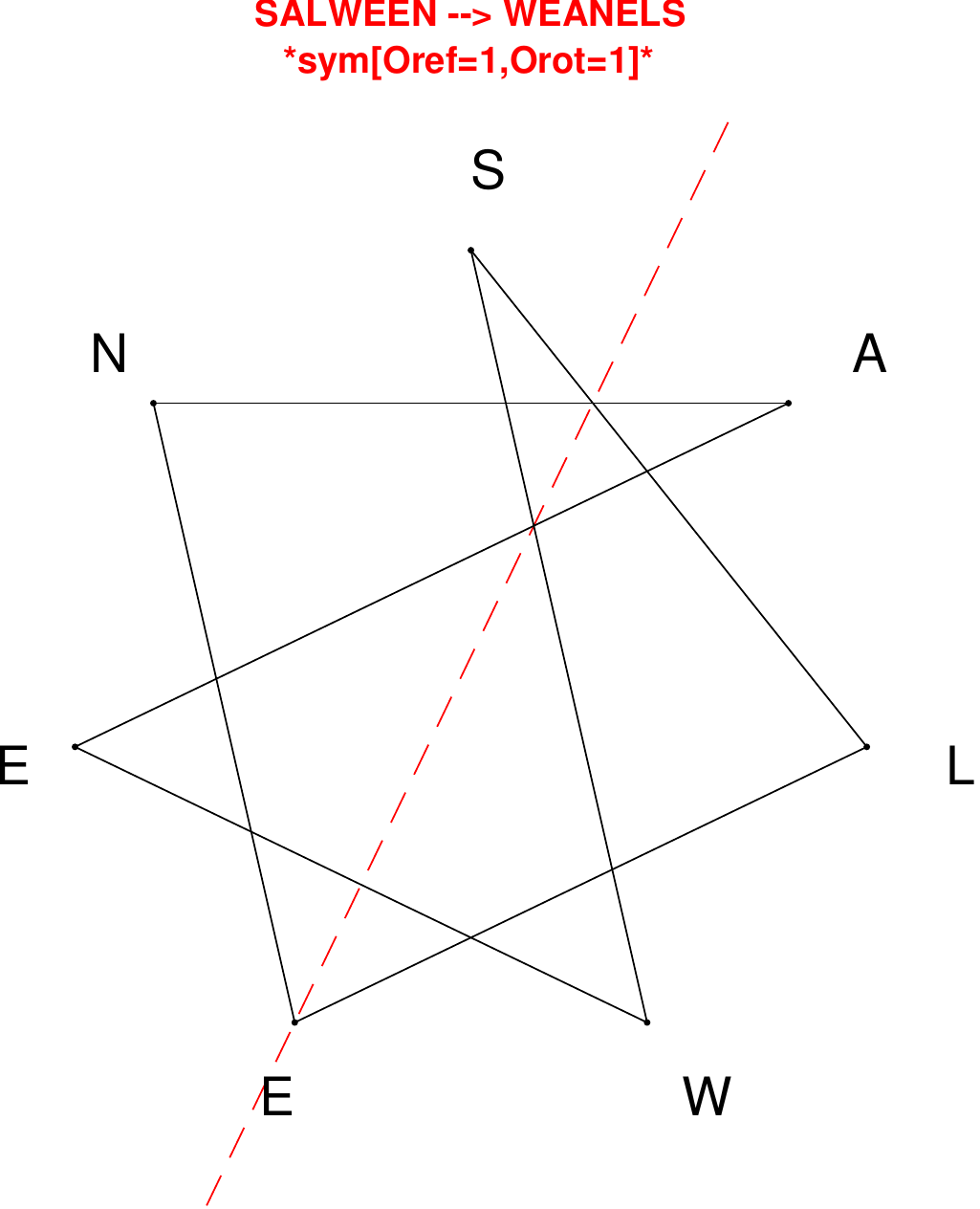}
\end{subfigure}
\end{figure}

\begin{figure}[H]
\centering
\begin{subfigure}[T]{0.19\textwidth}
\centering
\includegraphics[width=\textwidth]{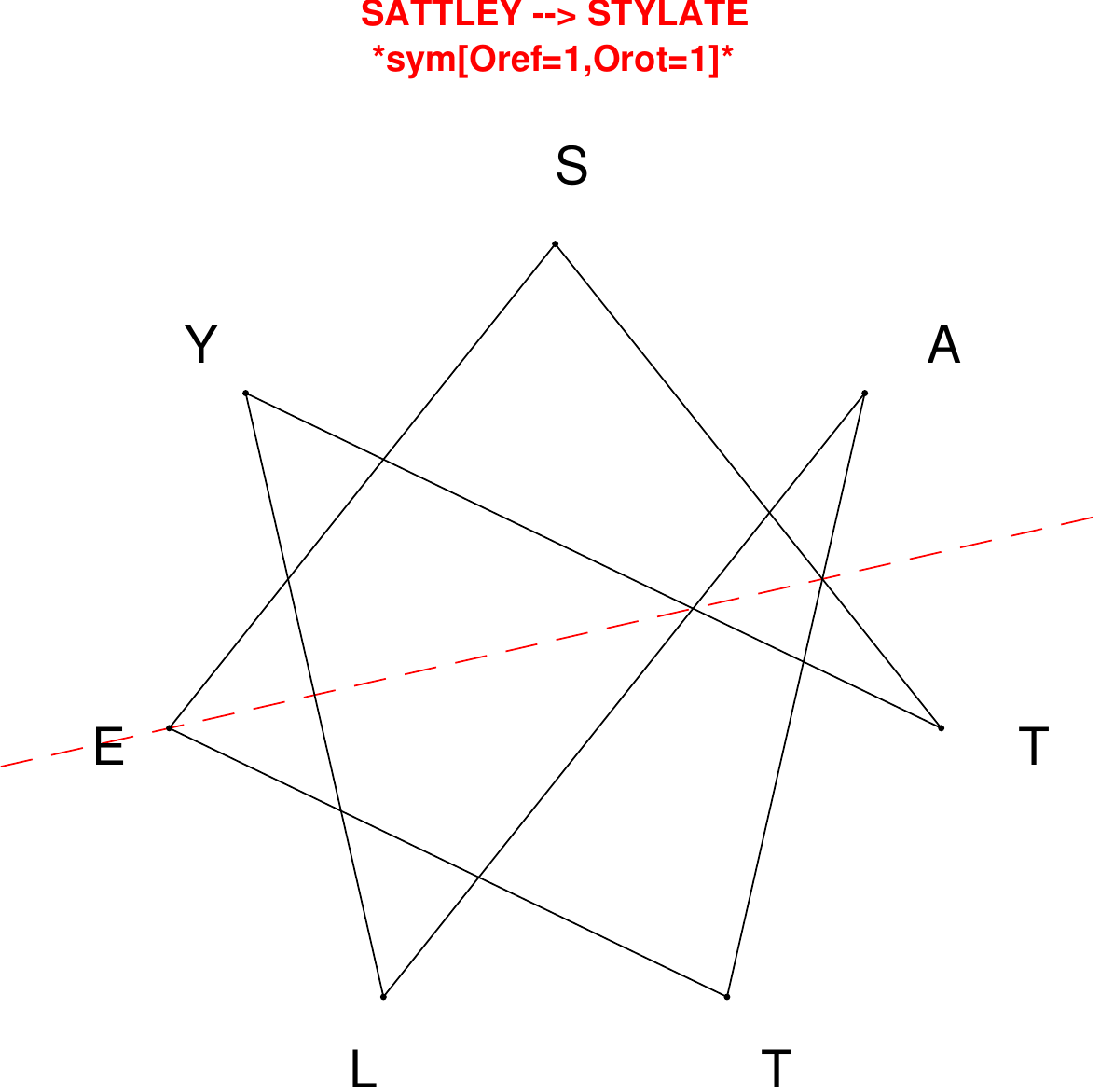}
\end{subfigure}
\hfill
\begin{subfigure}[T]{0.19\textwidth}
\centering
\includegraphics[width=\textwidth]{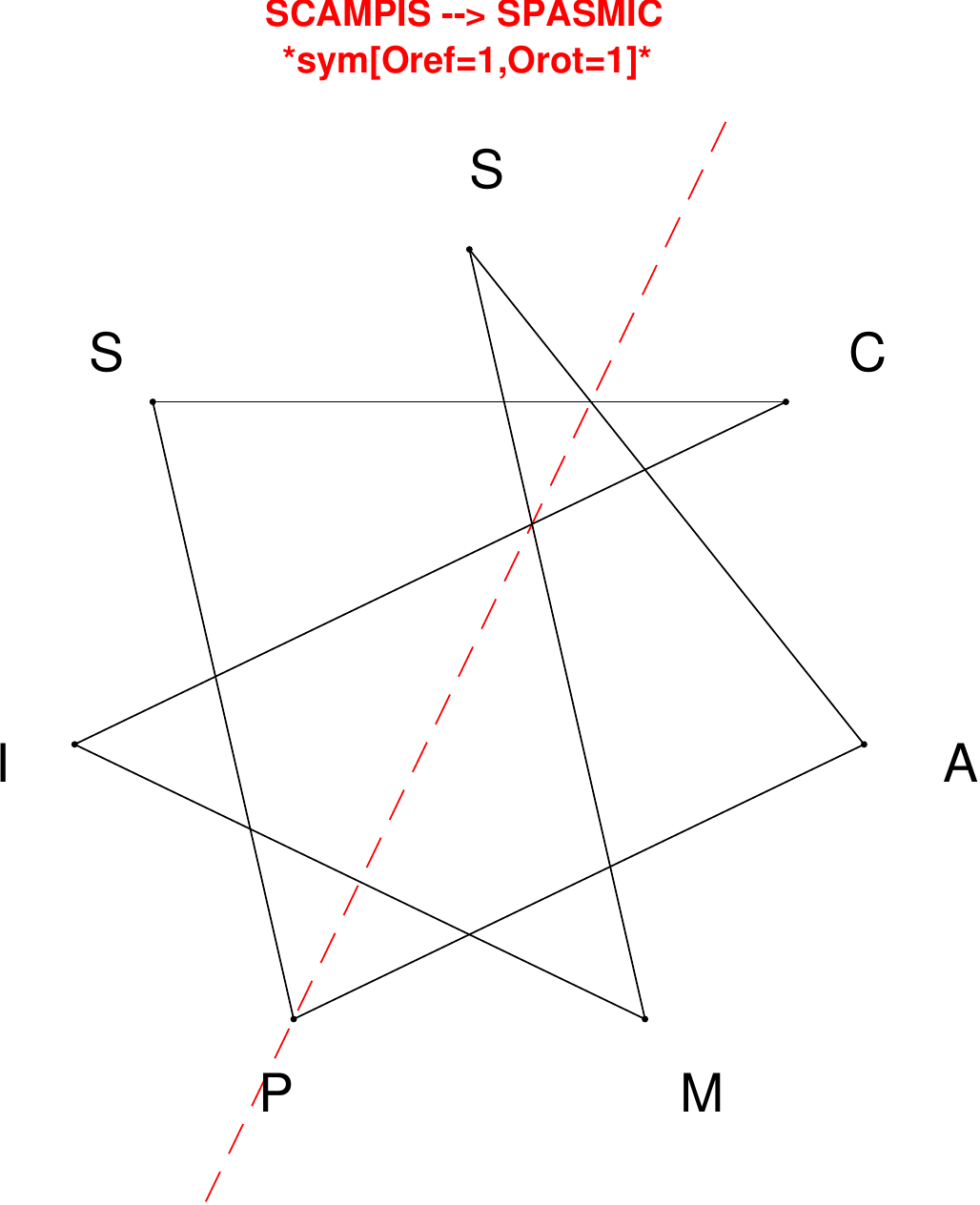}
\end{subfigure}
\hfill
\begin{subfigure}[T]{0.19\textwidth}
\centering
\includegraphics[width=\textwidth]{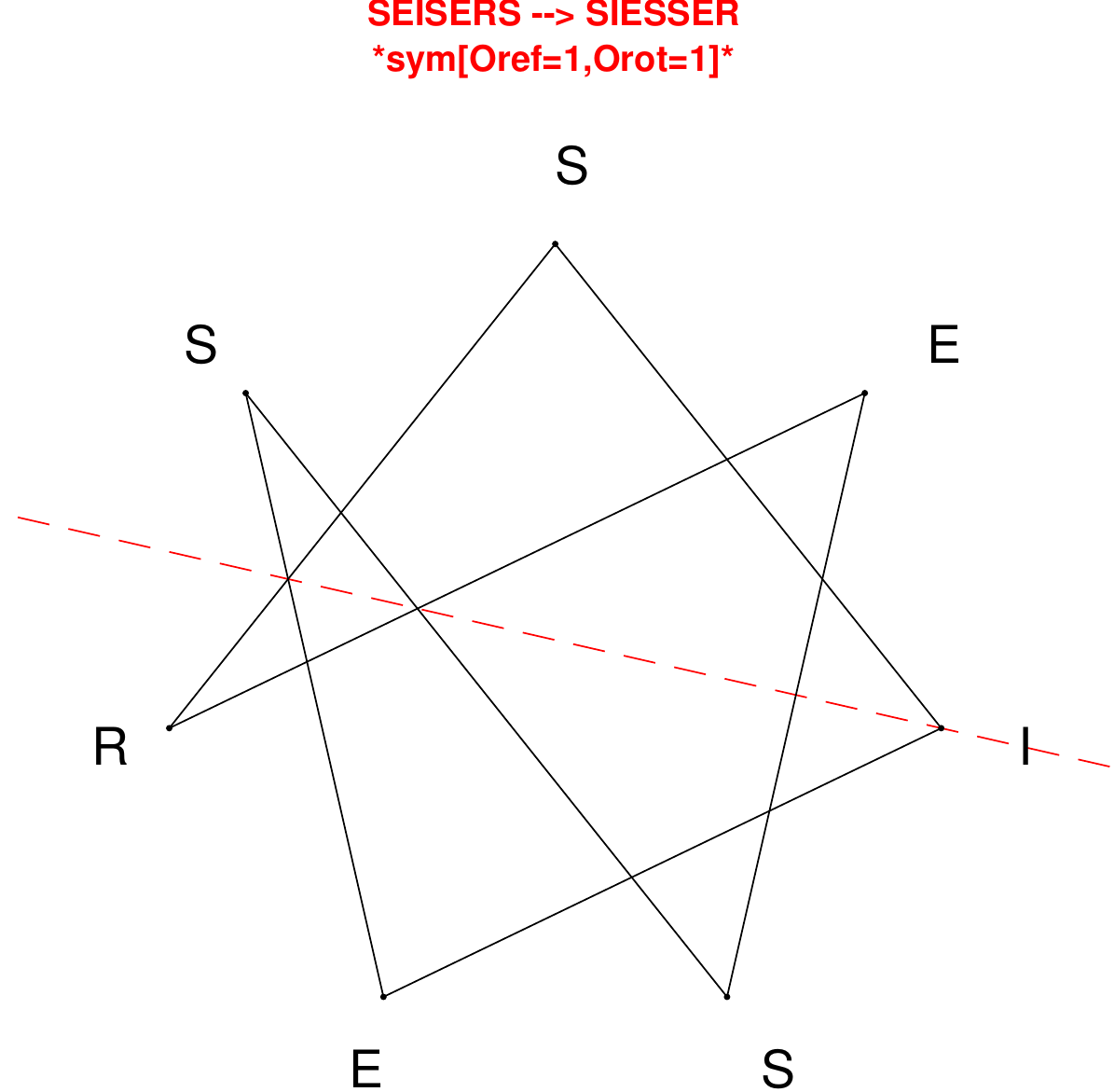}
\end{subfigure}
\hfill
\begin{subfigure}[T]{0.19\textwidth}
\centering
\includegraphics[width=\textwidth]{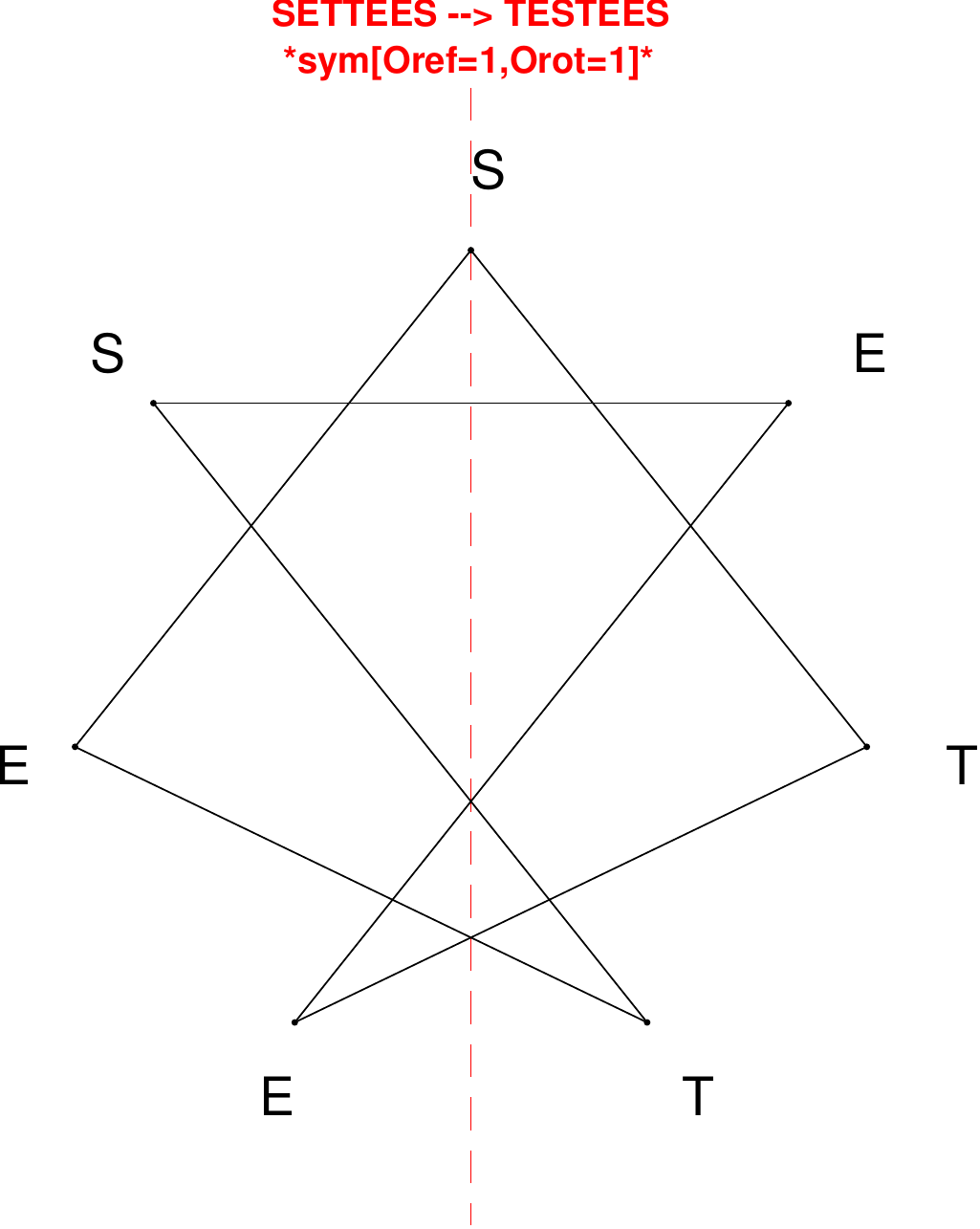}
\end{subfigure}
\hfill
\begin{subfigure}[T]{0.19\textwidth}
\centering
\includegraphics[width=\textwidth]{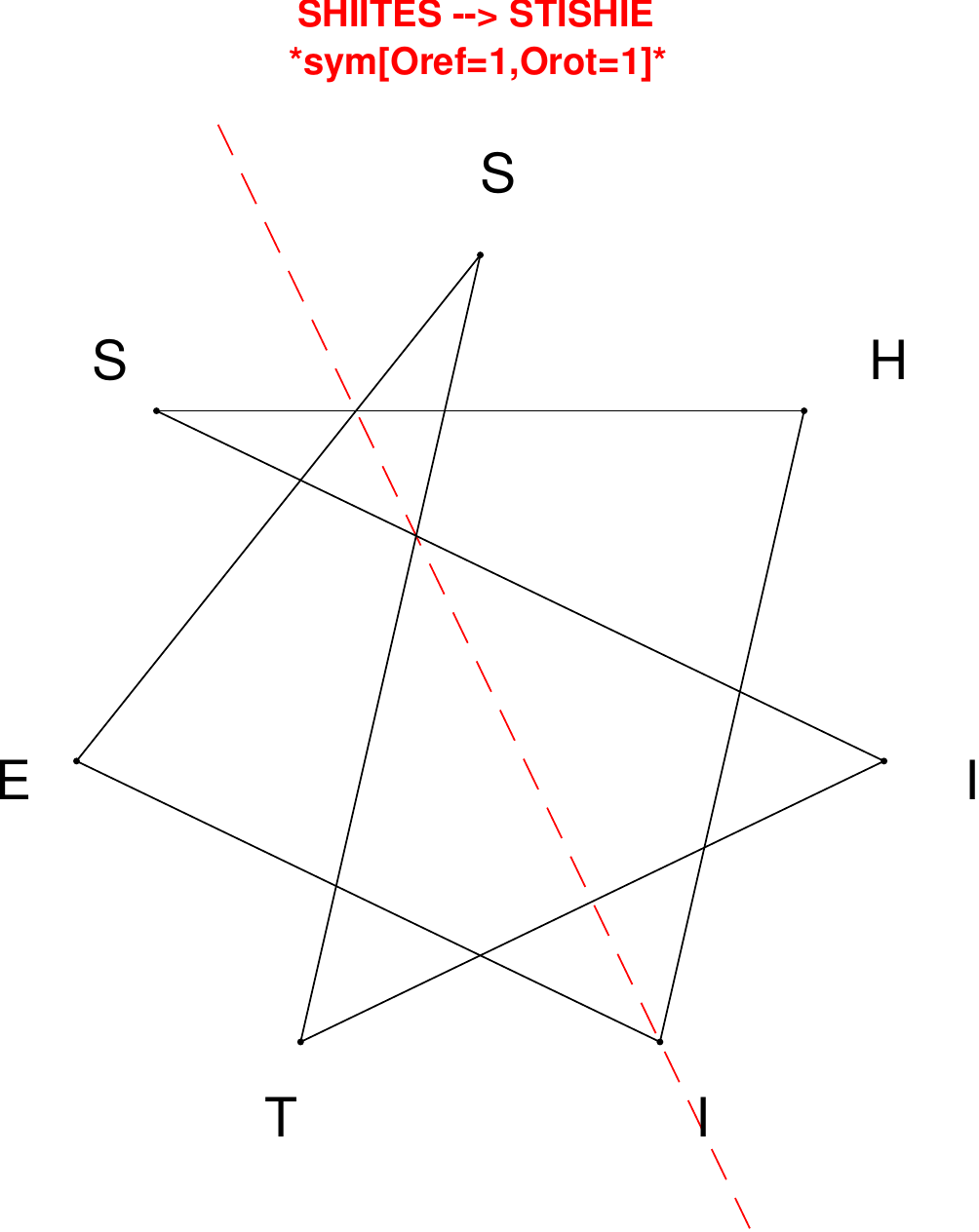}
\end{subfigure}
\end{figure}

\begin{figure}[H]
\centering
\begin{subfigure}[T]{0.19\textwidth}
\centering
\includegraphics[width=\textwidth]{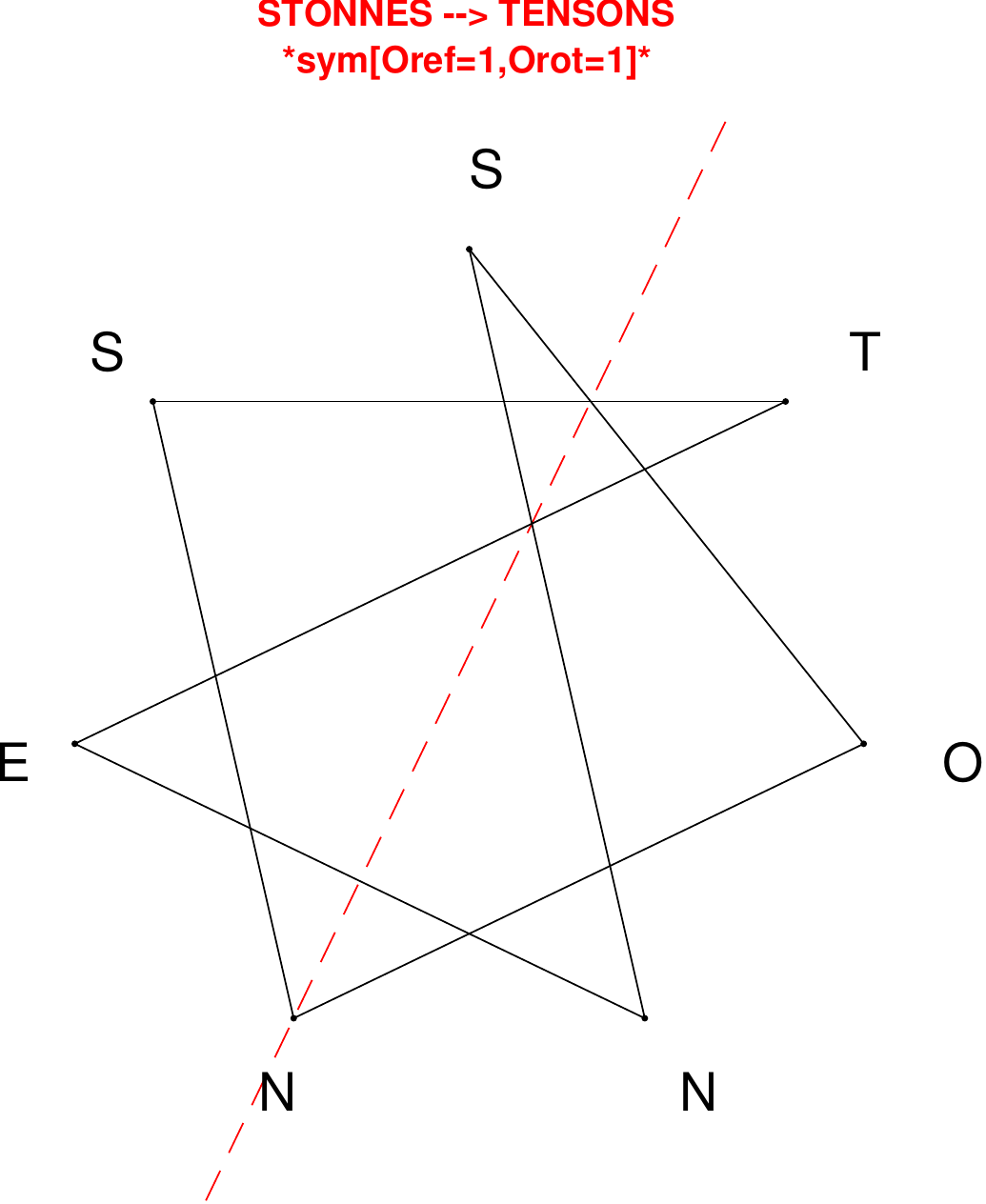}
\end{subfigure}
\hfill
\begin{subfigure}[T]{0.19\textwidth}
\centering
\includegraphics[width=\textwidth]{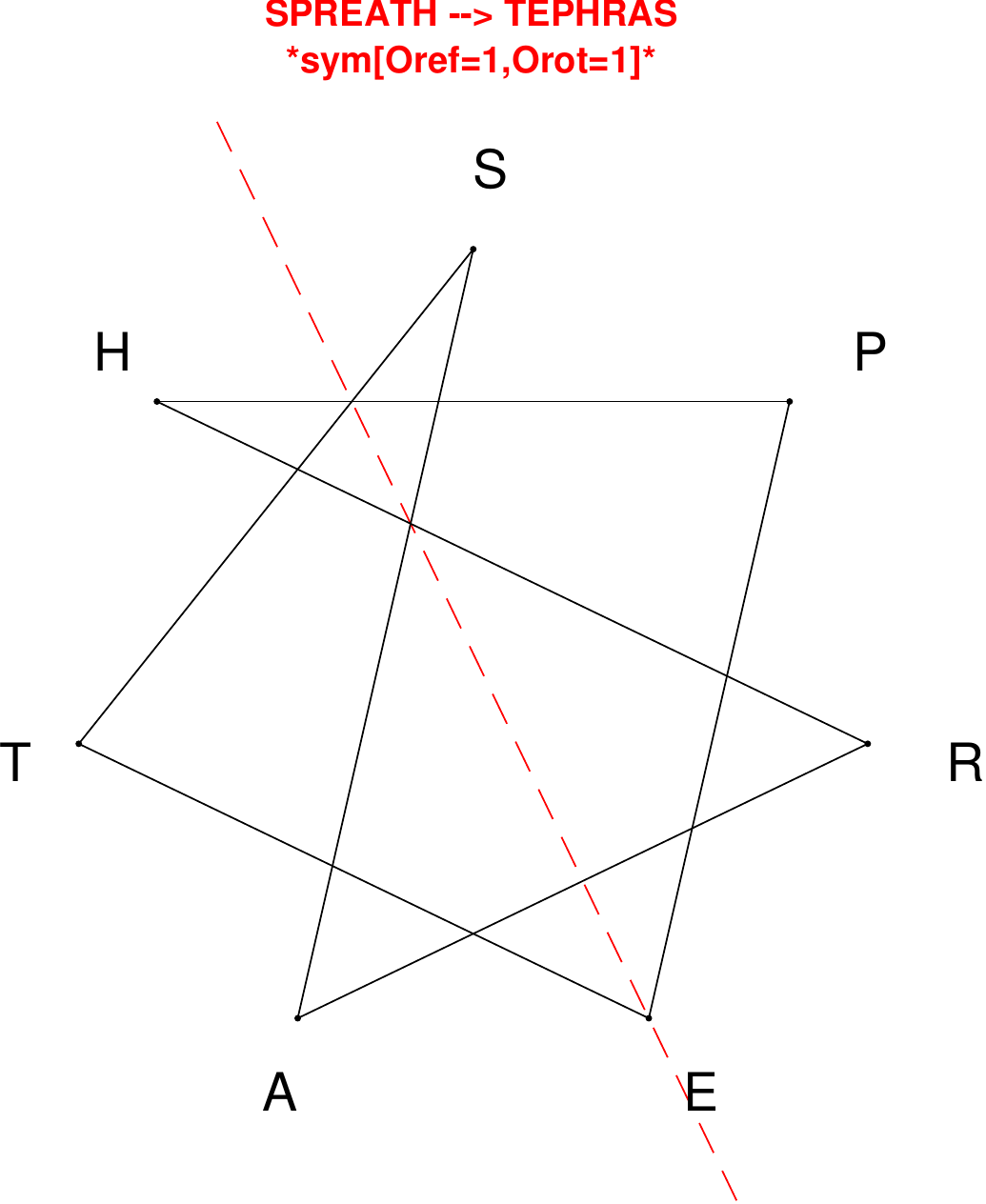}
\end{subfigure}
\hfill
\begin{subfigure}[T]{0.19\textwidth}
\centering
\includegraphics[width=\textwidth]{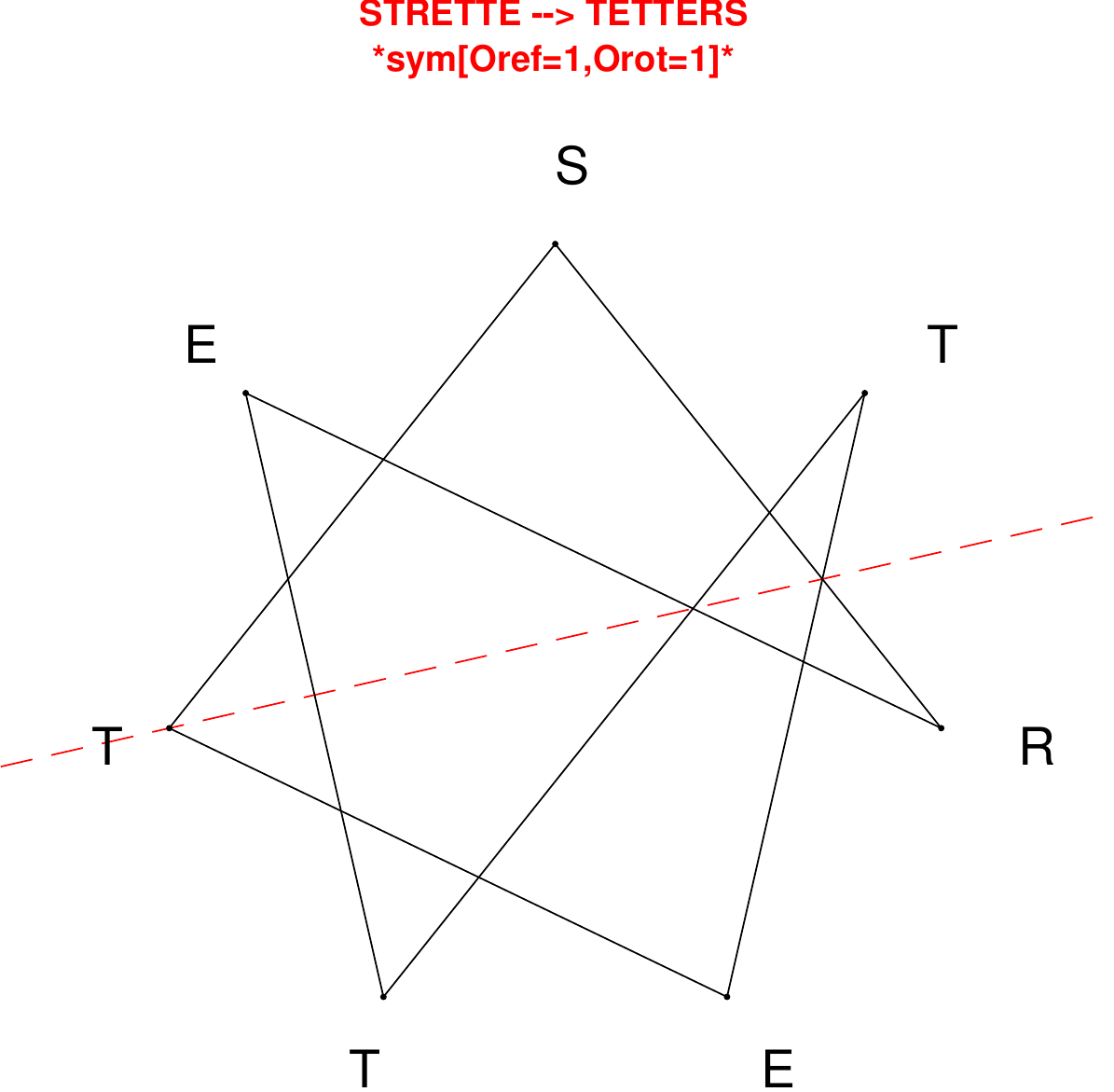}
\end{subfigure}
\hfill
\begin{subfigure}[T]{0.19\textwidth}
\centering
\includegraphics[width=\textwidth]{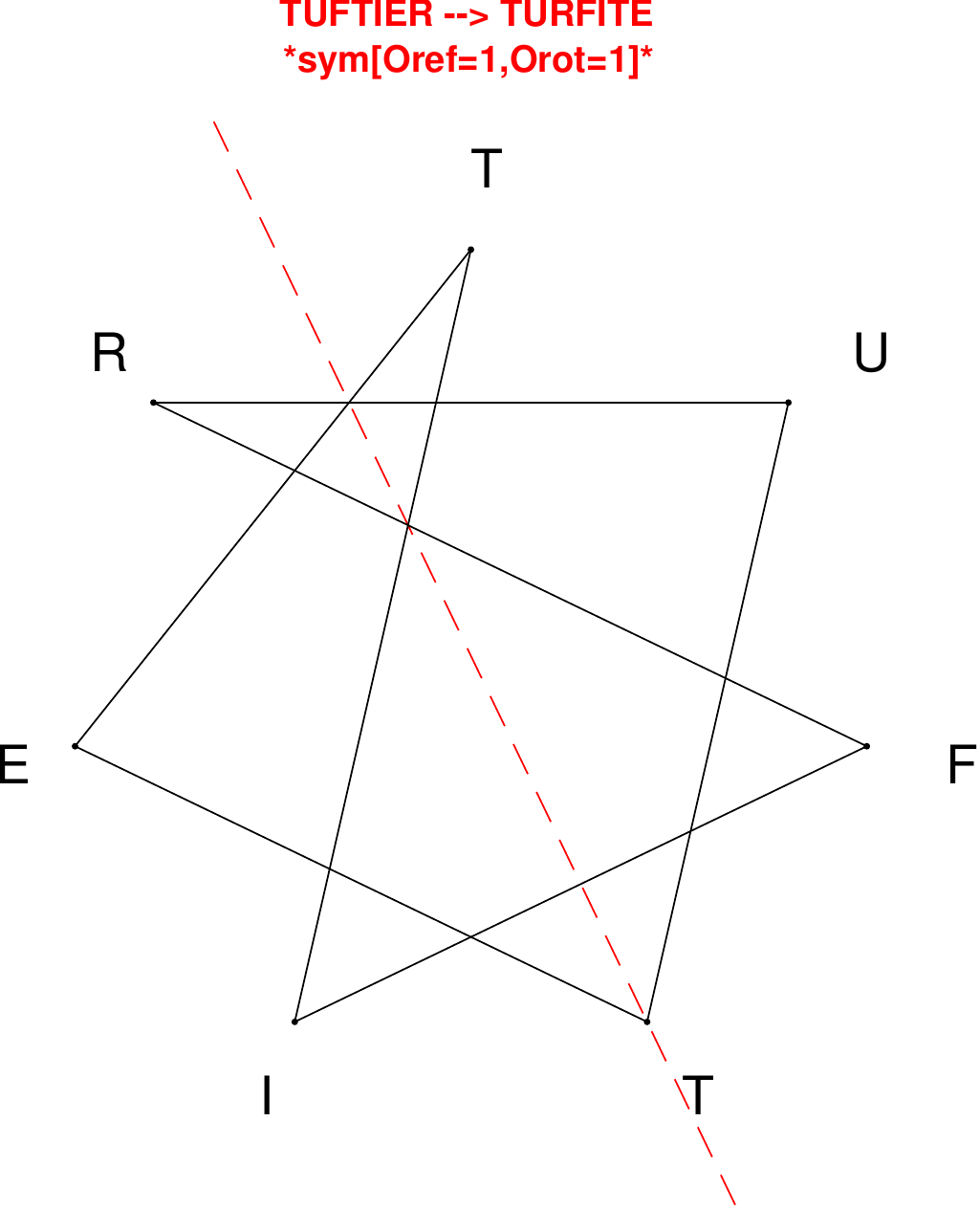}
\end{subfigure}
\hfill
\begin{subfigure}[T]{0.19\textwidth}
\centering
\includegraphics[width=\textwidth]{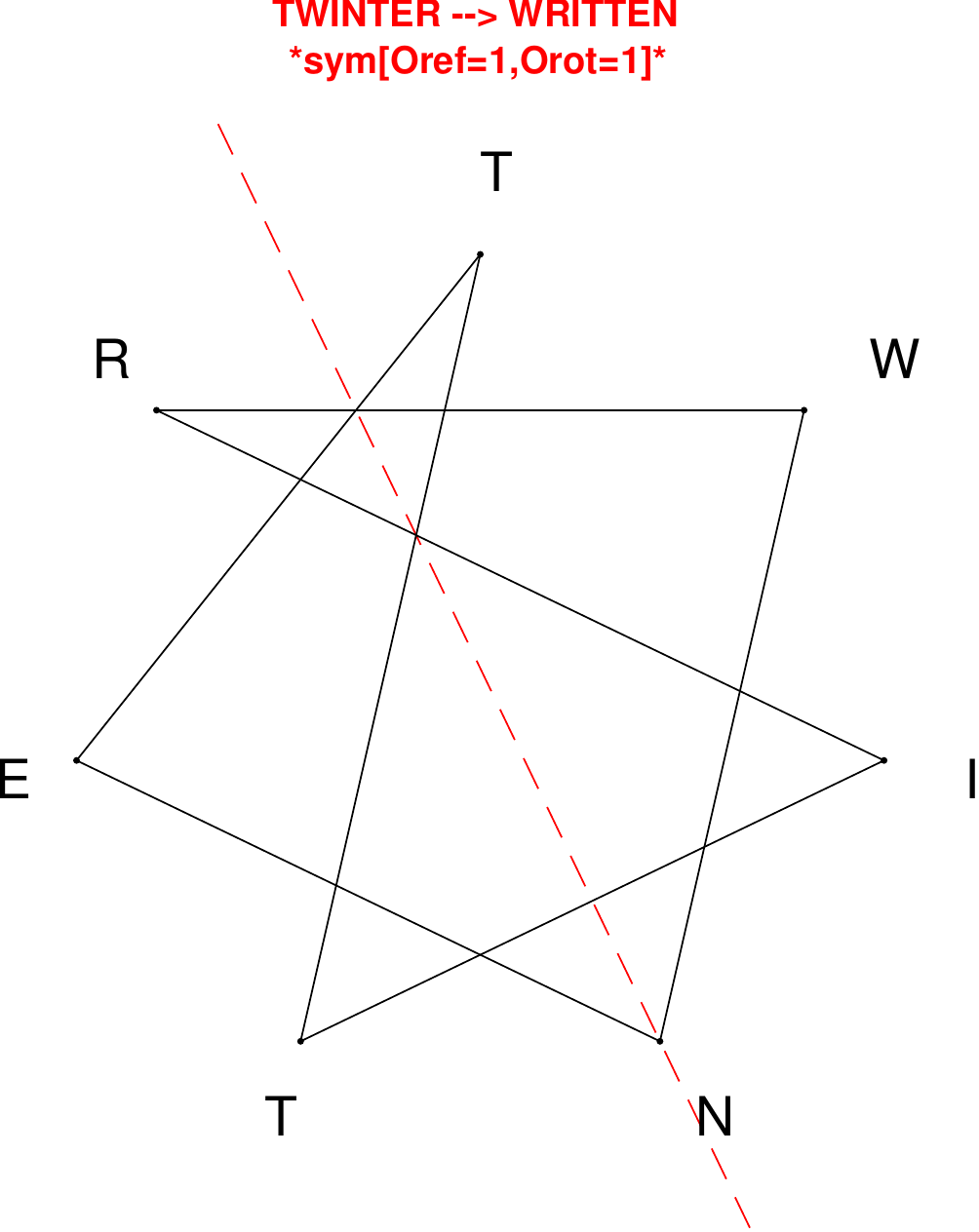}
\end{subfigure}
\end{figure}

\begin{figure}[H]
\centering
\begin{subfigure}[T]{0.19\textwidth}
\centering
\includegraphics[width=\textwidth]{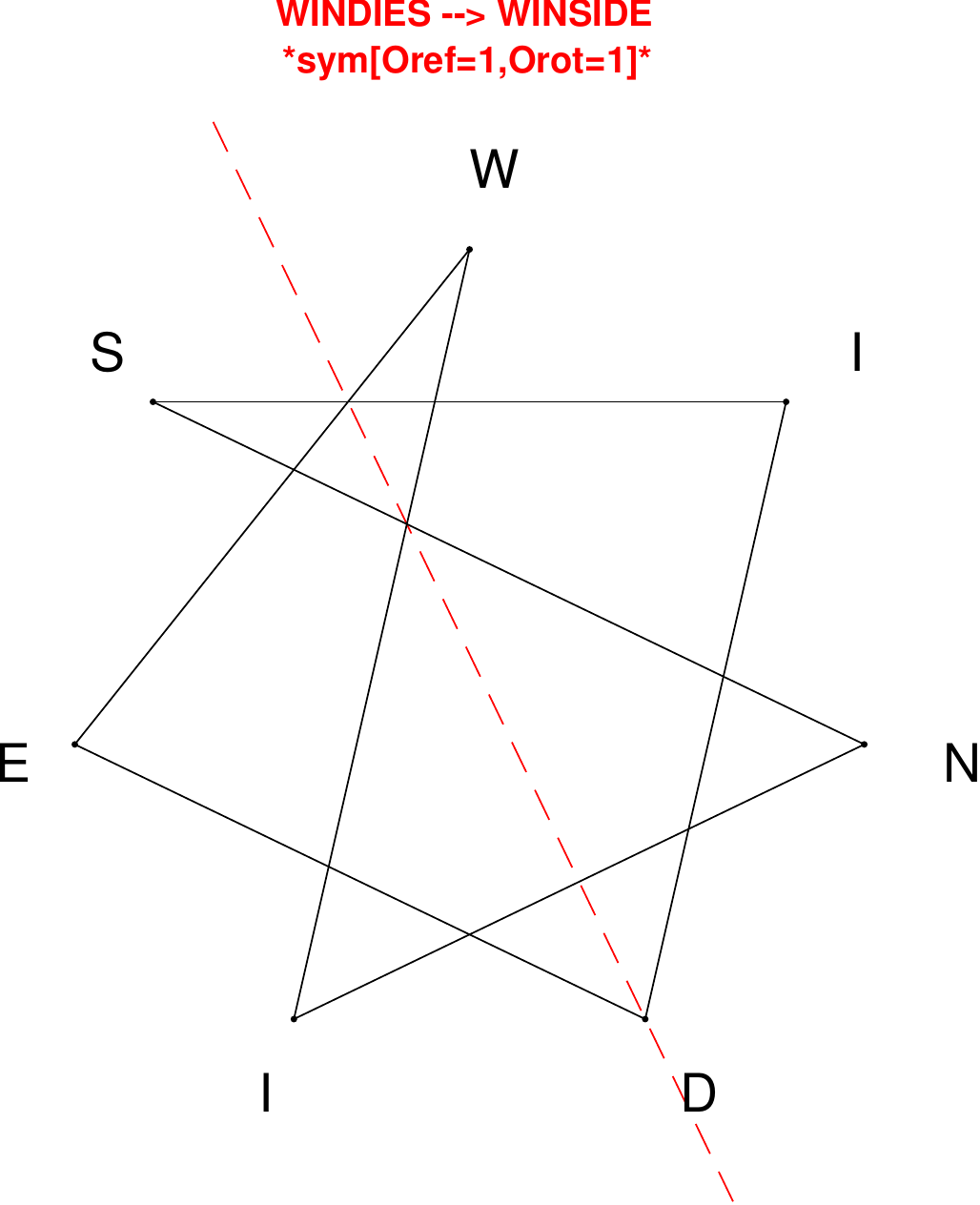}
\end{subfigure}
\hfill
\end{figure}

%%%%%%%%%%%%%%%%%%
\clearpage
\subsection{Star Anagrams $N = 6$}
All of the stars for $N=6$ are symmetric. As we expected from the analysis above, none of them are perfect.

\begin{figure}[H]
\centering
\begin{subfigure}[T]{0.19\textwidth}
\centering
\includegraphics[width=\textwidth]{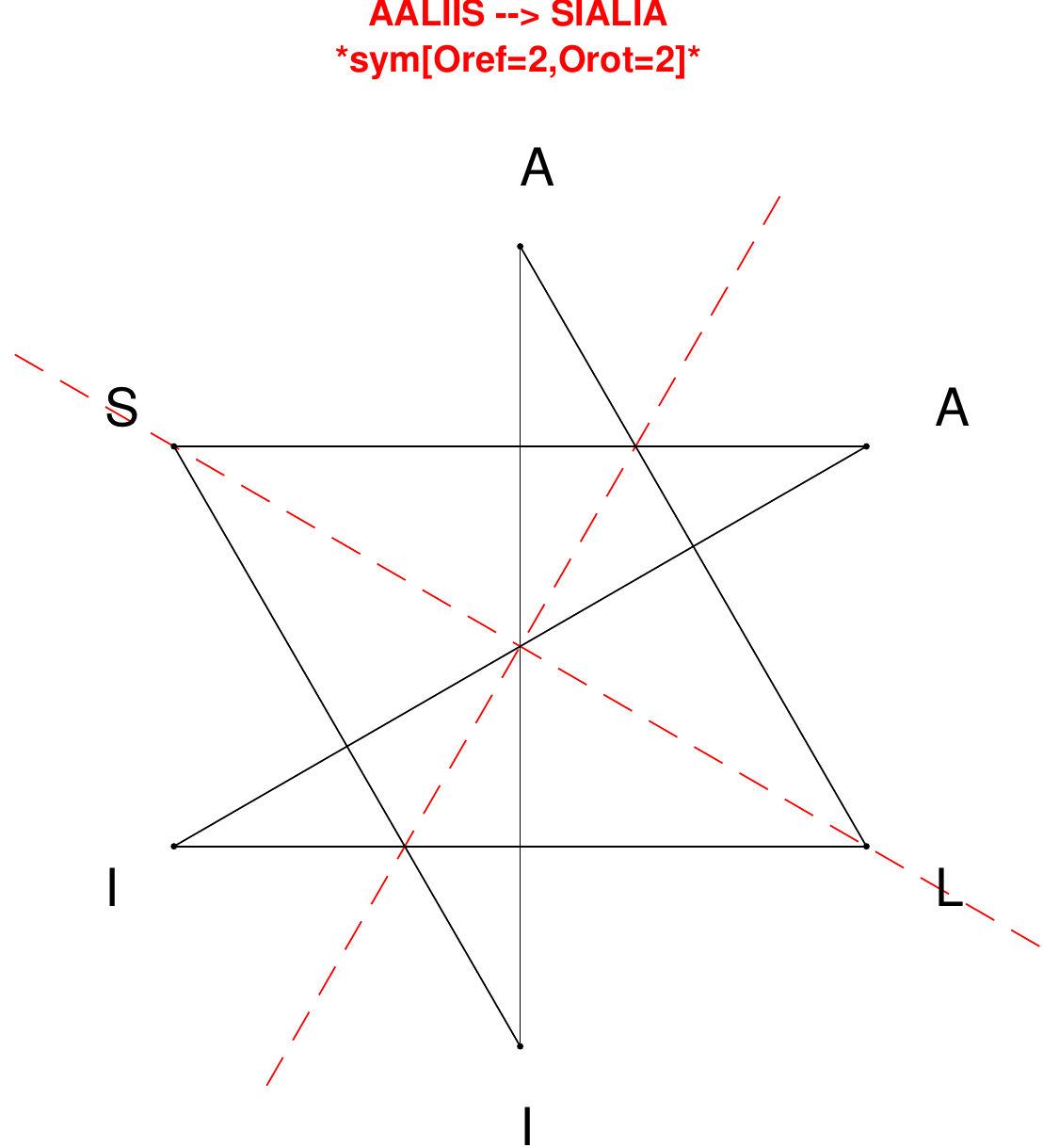}
\end{subfigure}
\hfill
\begin{subfigure}[T]{0.19\textwidth}
\centering
\includegraphics[width=\textwidth]{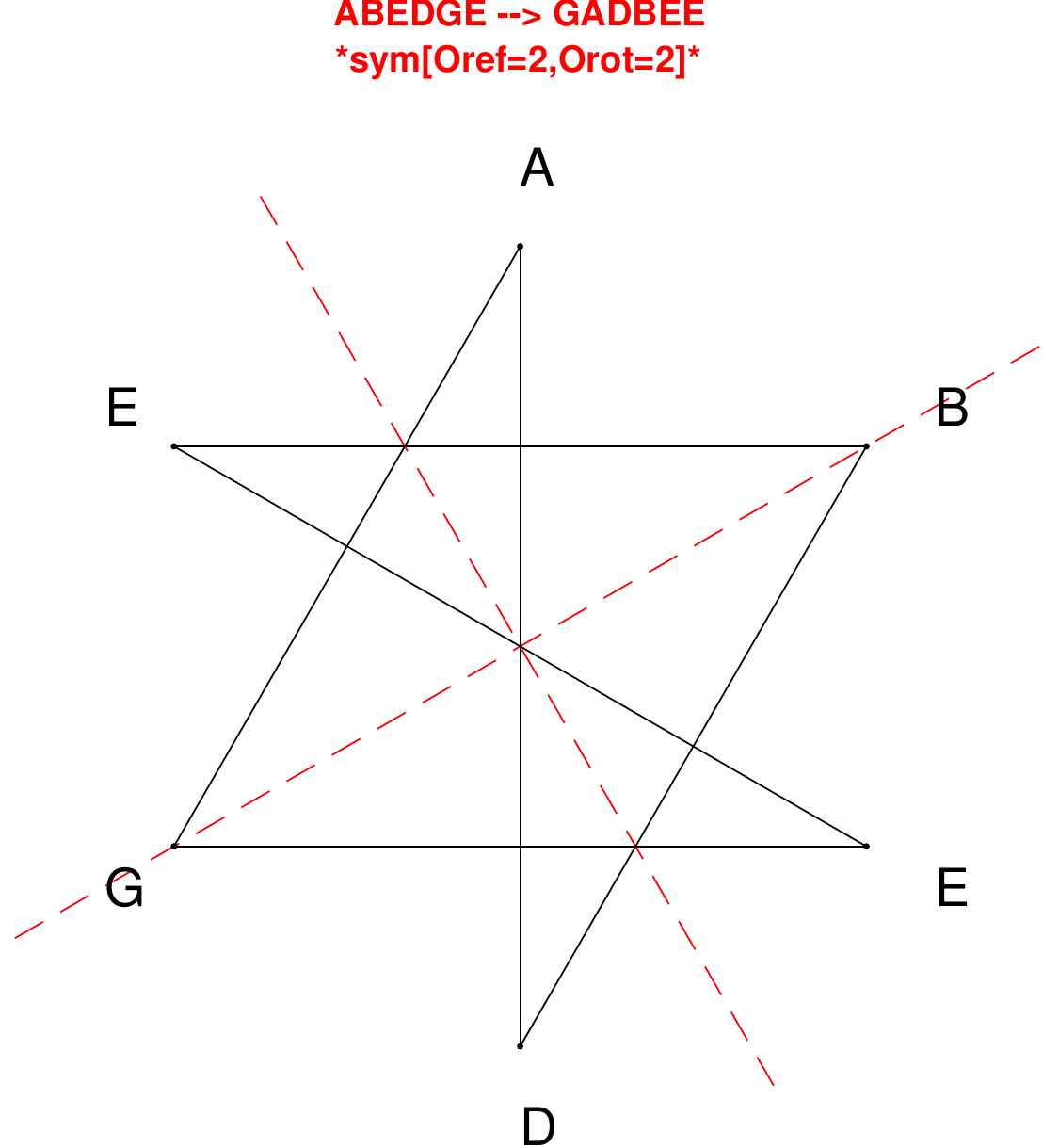}
\end{subfigure}
\hfill
\begin{subfigure}[T]{0.19\textwidth}
\centering
\includegraphics[width=\textwidth]{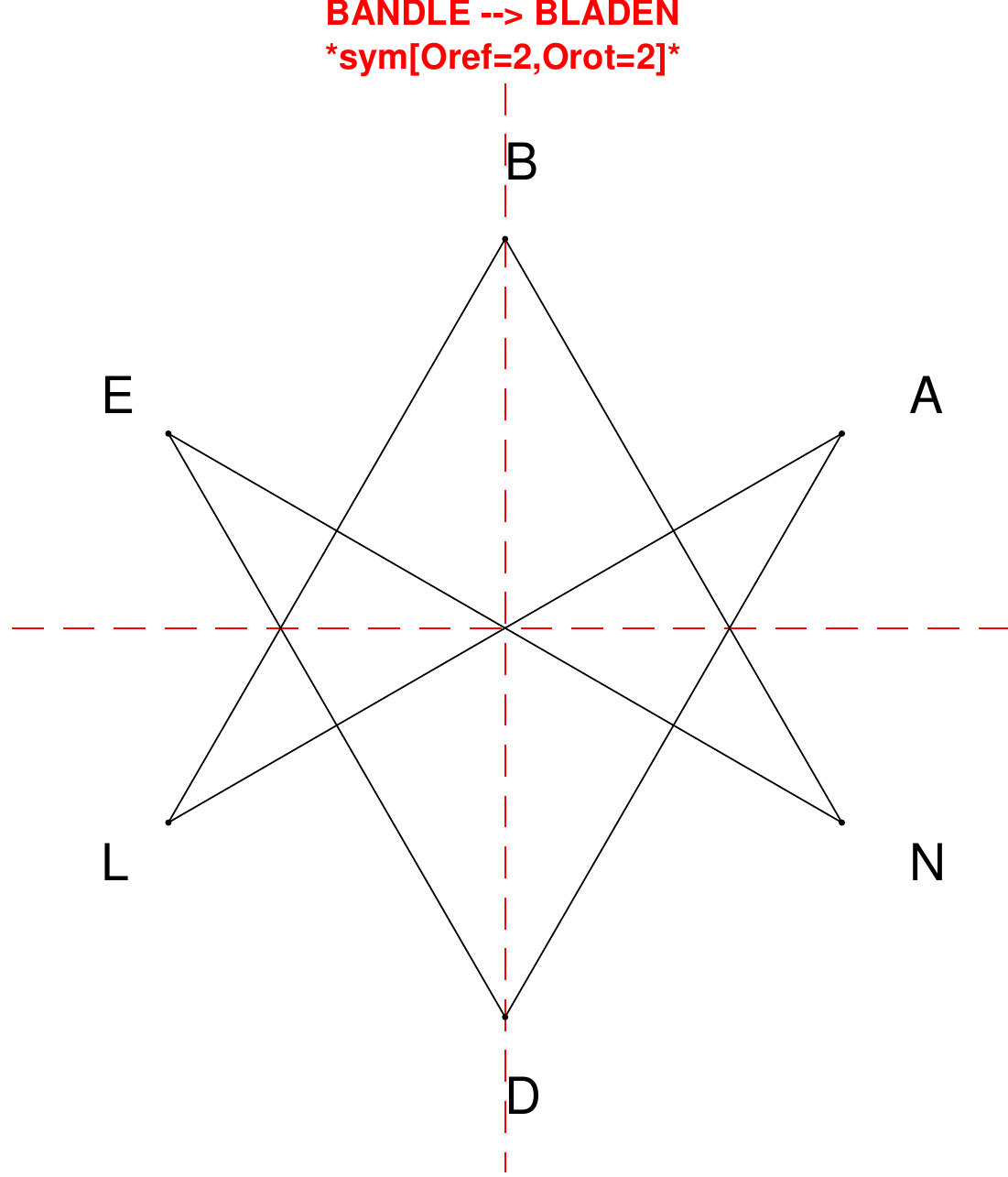}
\end{subfigure}
\hfill
\begin{subfigure}[T]{0.19\textwidth}
\centering
\includegraphics[width=\textwidth]{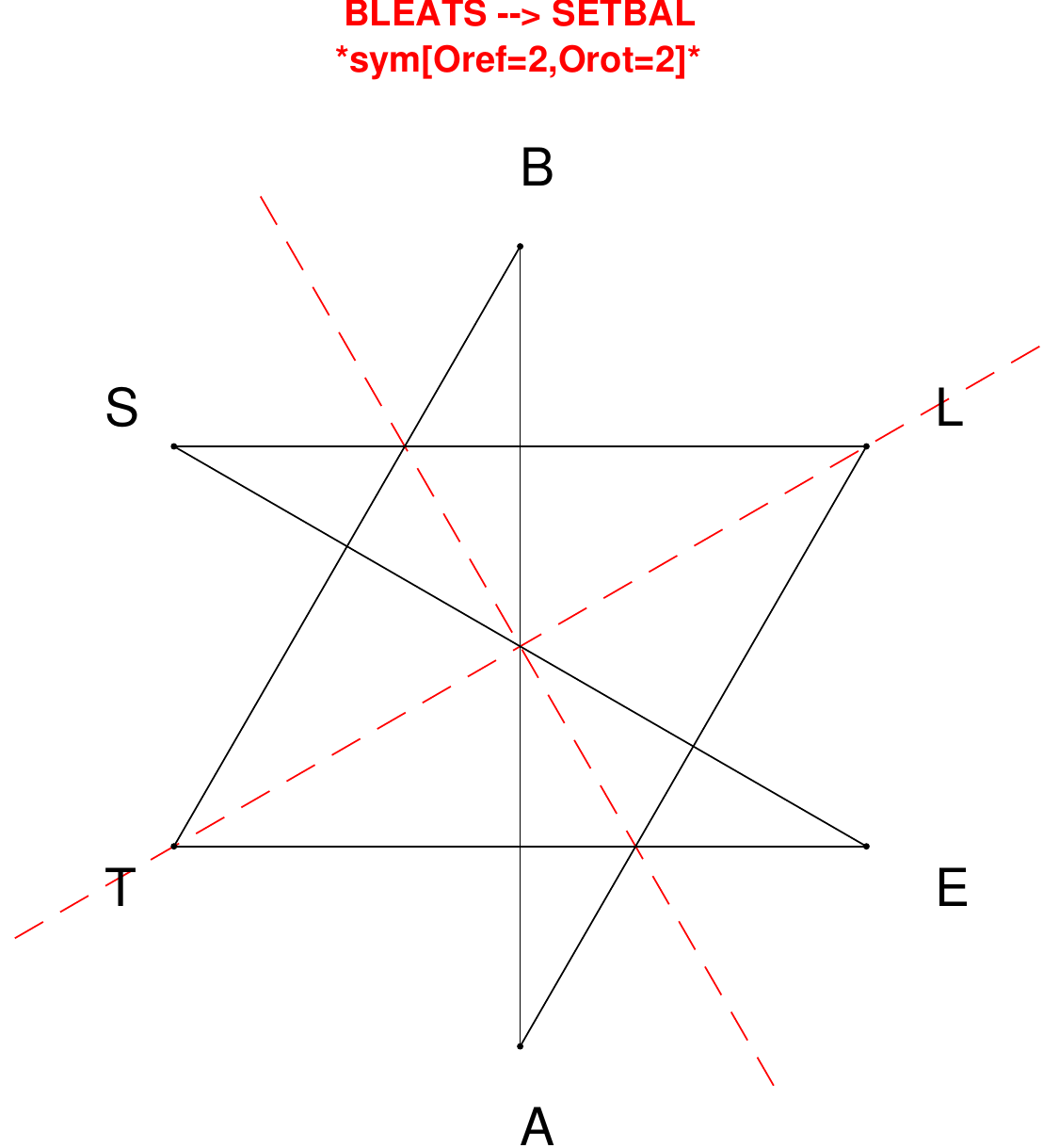}
\end{subfigure}
\hfill
\begin{subfigure}[T]{0.19\textwidth}
\centering
\includegraphics[width=\textwidth]{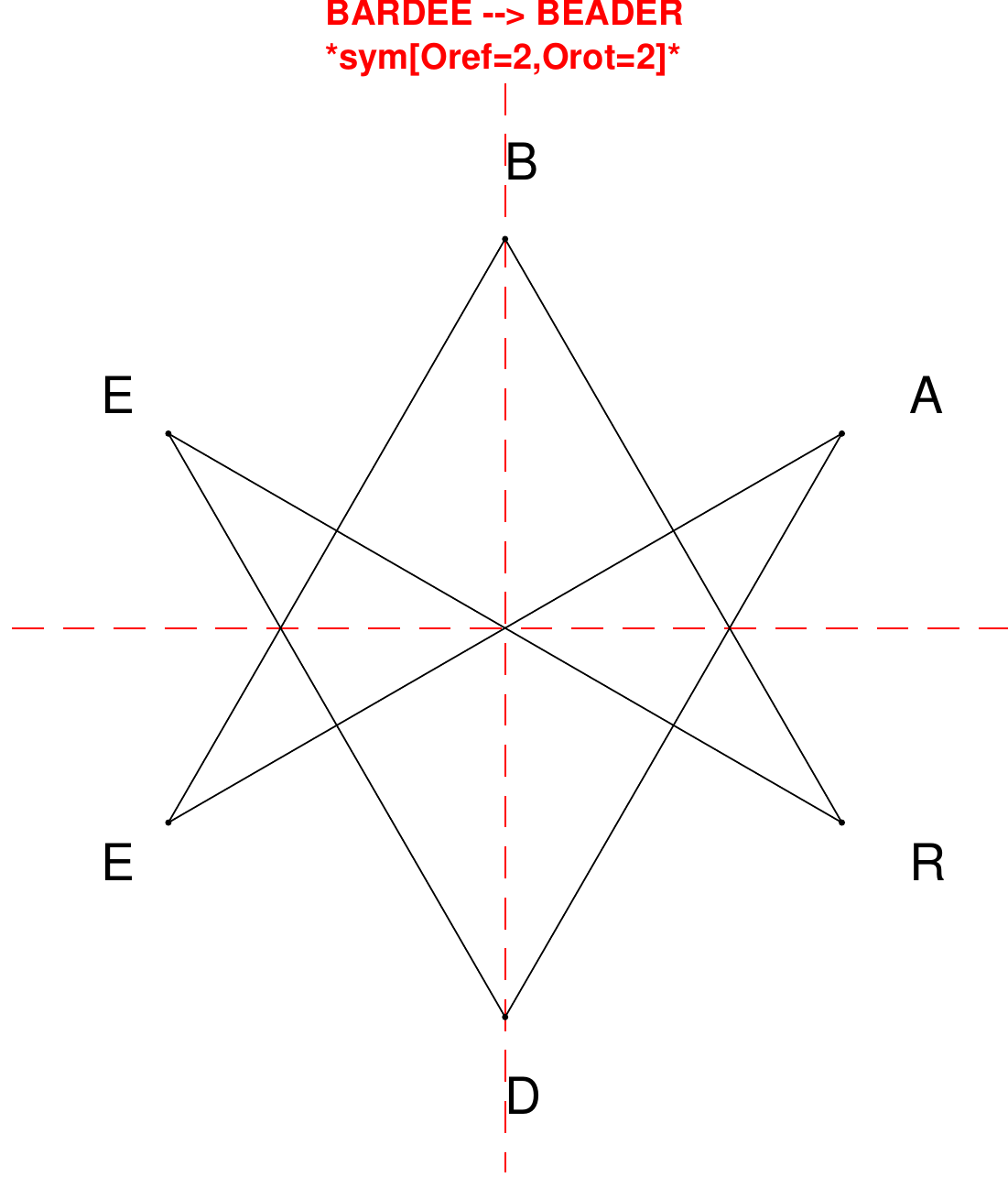}
\end{subfigure}
\end{figure}

\begin{figure}[H]
\centering
\begin{subfigure}[T]{0.19\textwidth}
\centering
\includegraphics[width=\textwidth]{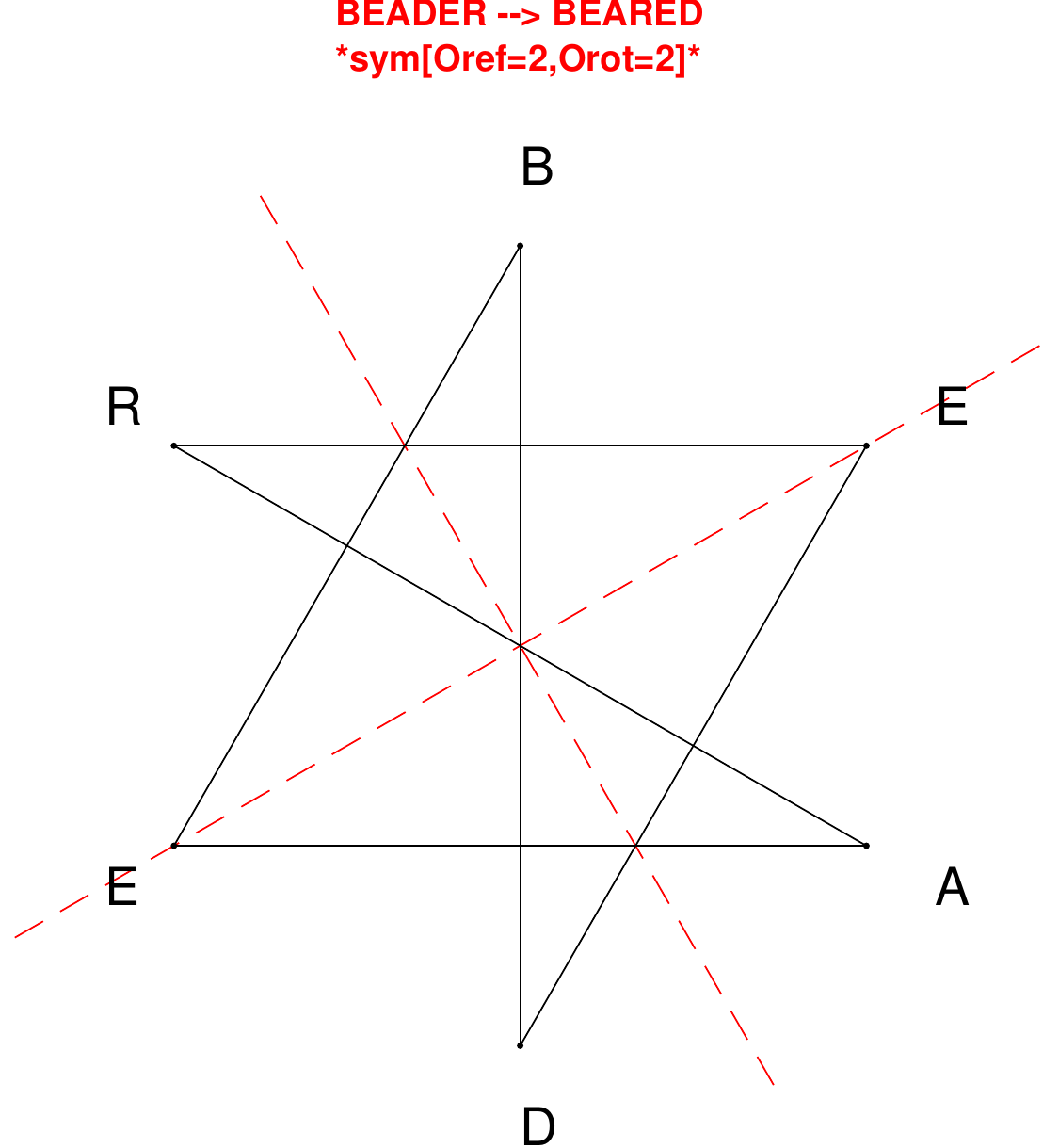}
\end{subfigure}
\hfill
\begin{subfigure}[T]{0.19\textwidth}
\centering
\includegraphics[width=\textwidth]{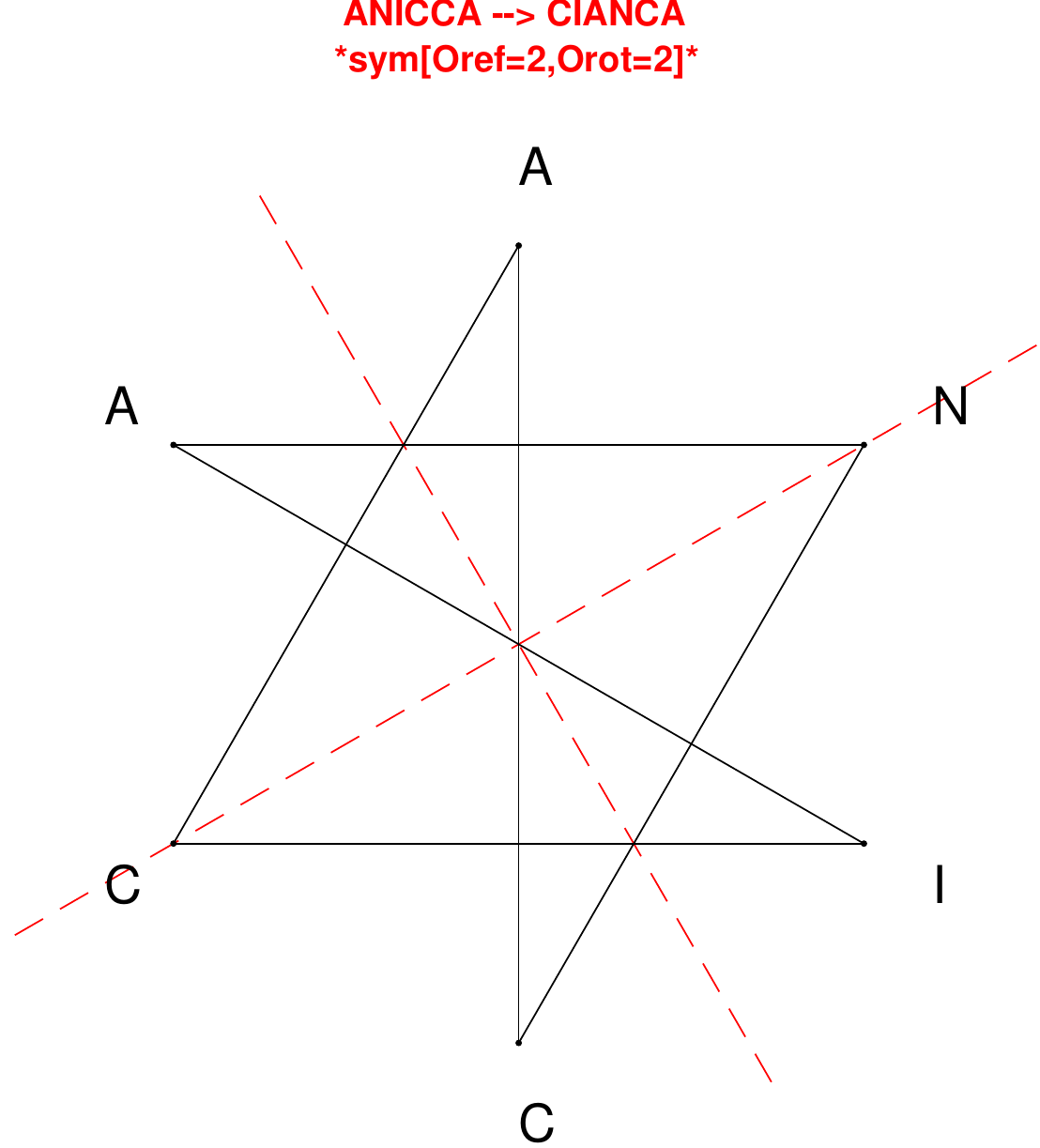}
\end{subfigure}
\hfill
\begin{subfigure}[T]{0.19\textwidth}
\centering
\includegraphics[width=\textwidth]{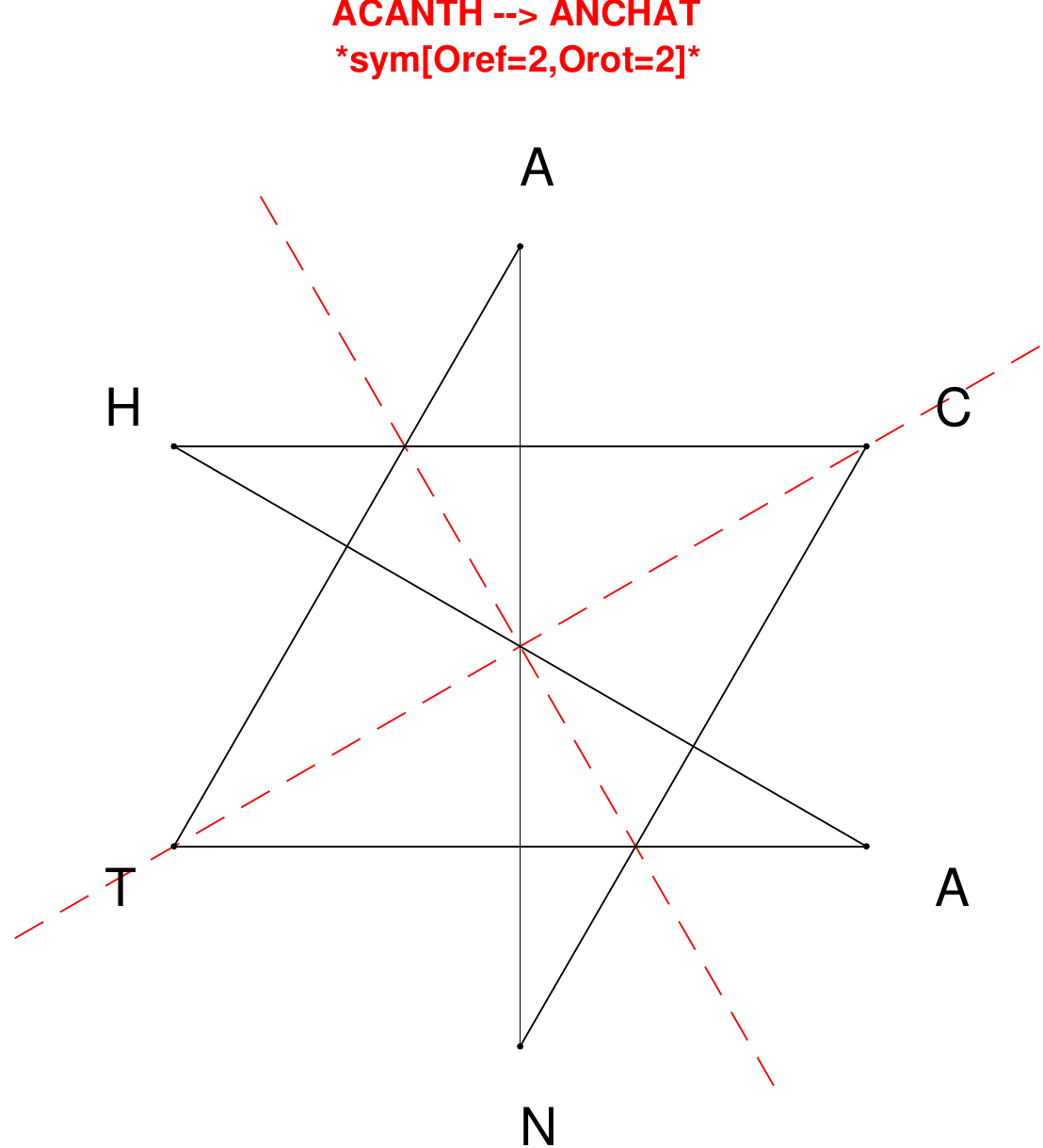}
\end{subfigure}
\hfill
\begin{subfigure}[T]{0.19\textwidth}
\centering
\includegraphics[width=\textwidth]{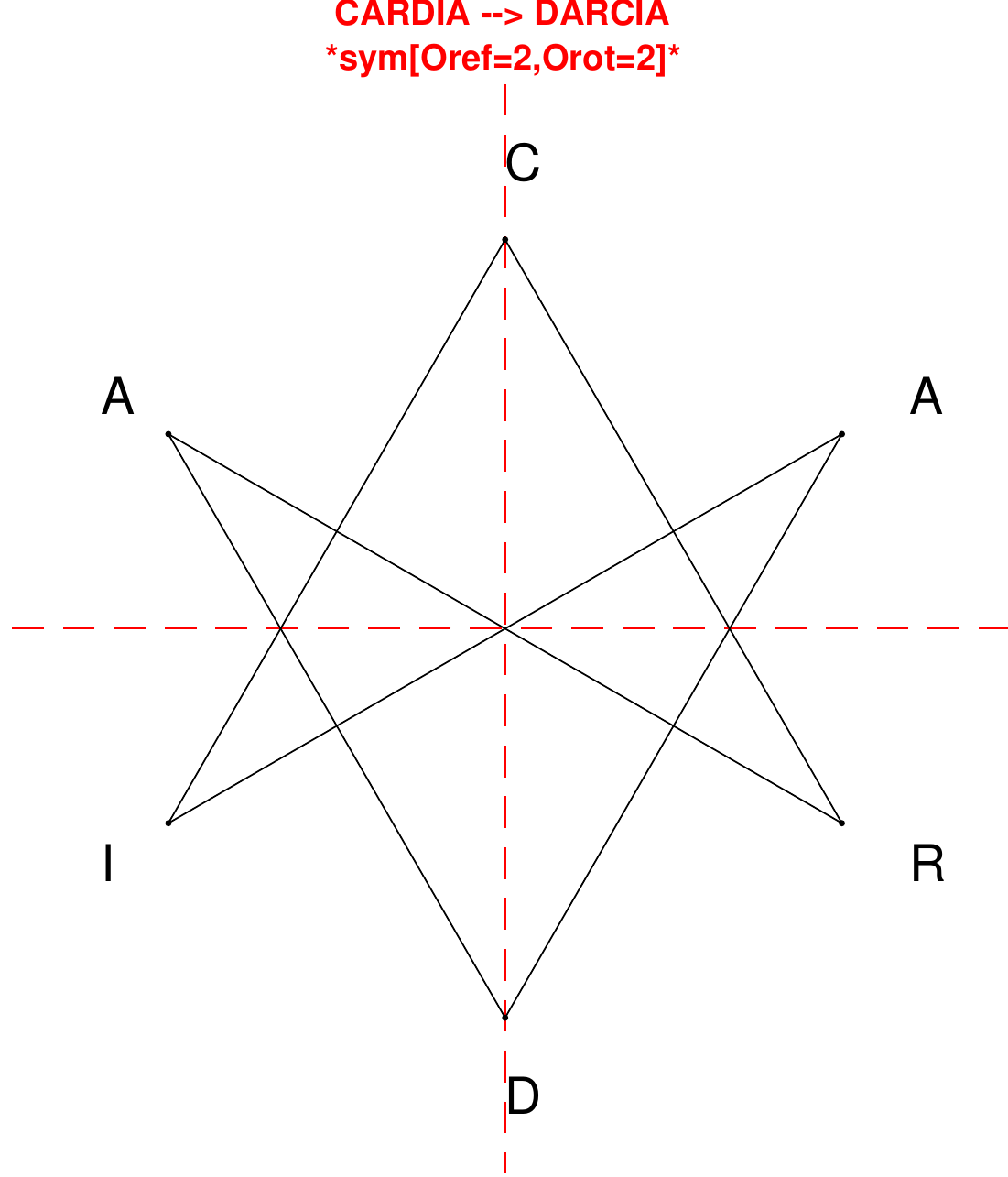}
\end{subfigure}
\hfill
\begin{subfigure}[T]{0.19\textwidth}
\centering
\includegraphics[width=\textwidth]{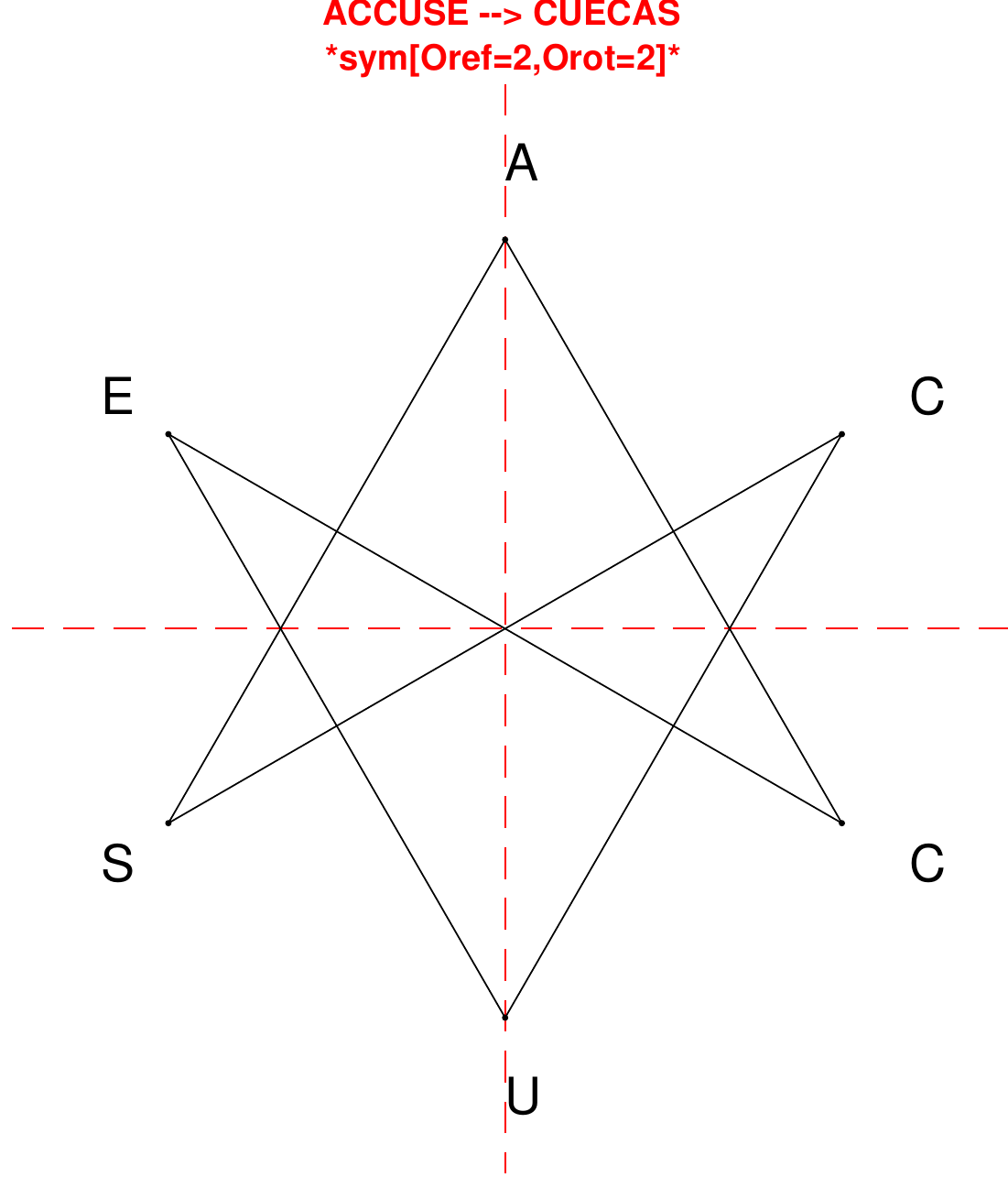}
\end{subfigure}
\end{figure}

\begin{figure}[H]
\centering
\begin{subfigure}[T]{0.19\textwidth}
\centering
\includegraphics[width=\textwidth]{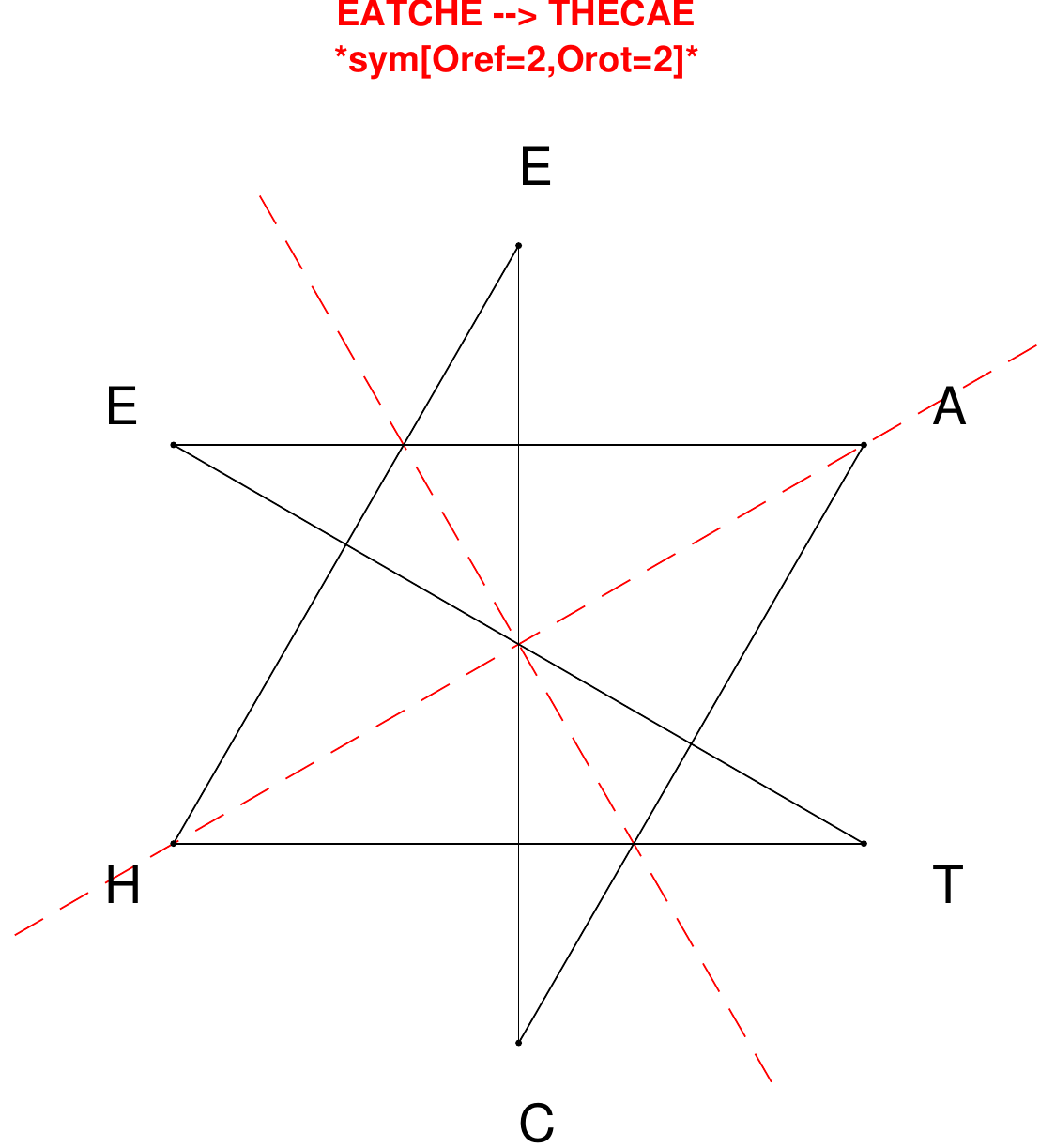}
\end{subfigure}
\hfill
\begin{subfigure}[T]{0.19\textwidth}
\centering
\includegraphics[width=\textwidth]{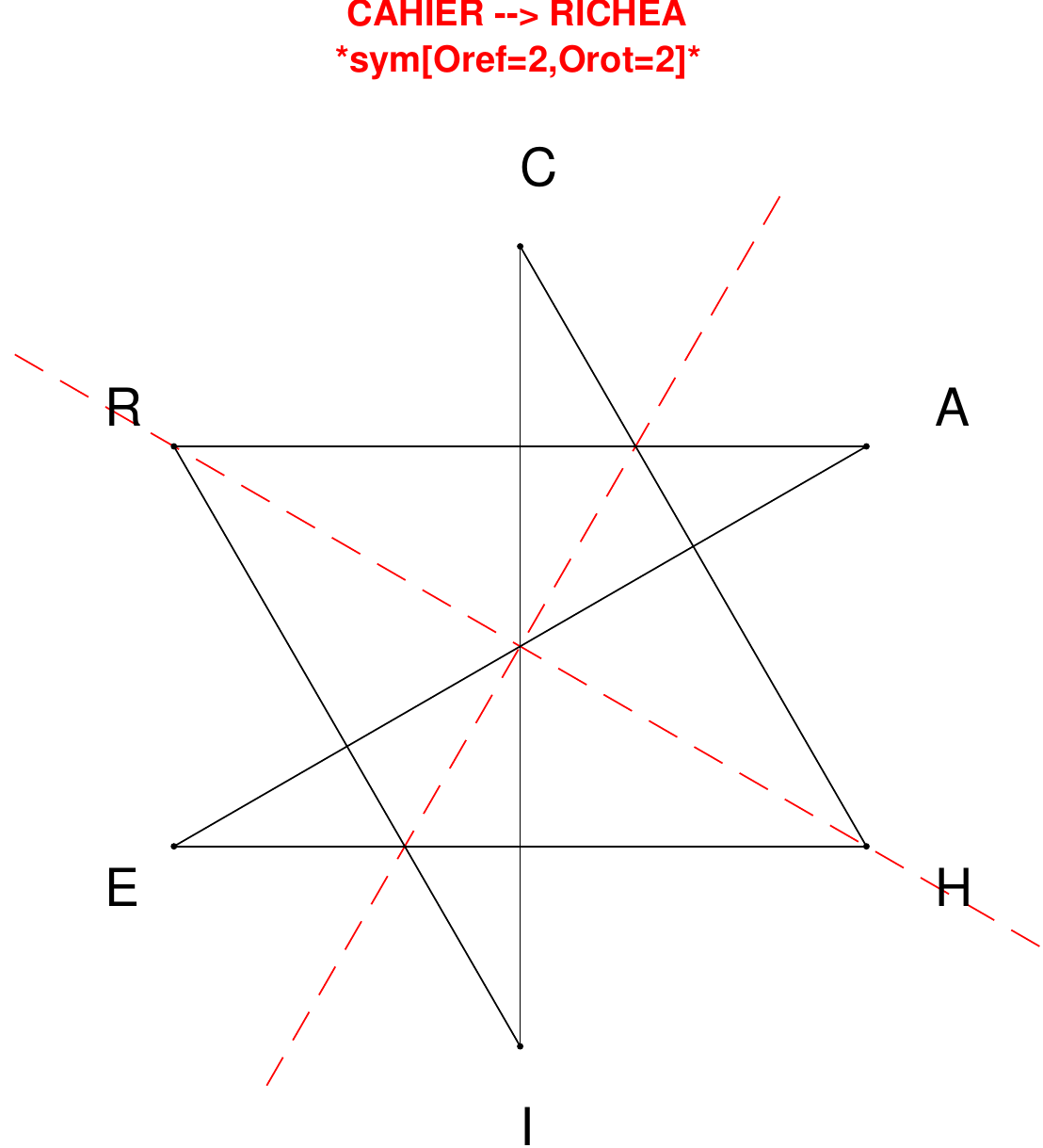}
\end{subfigure}
\hfill
\begin{subfigure}[T]{0.19\textwidth}
\centering
\includegraphics[width=\textwidth]{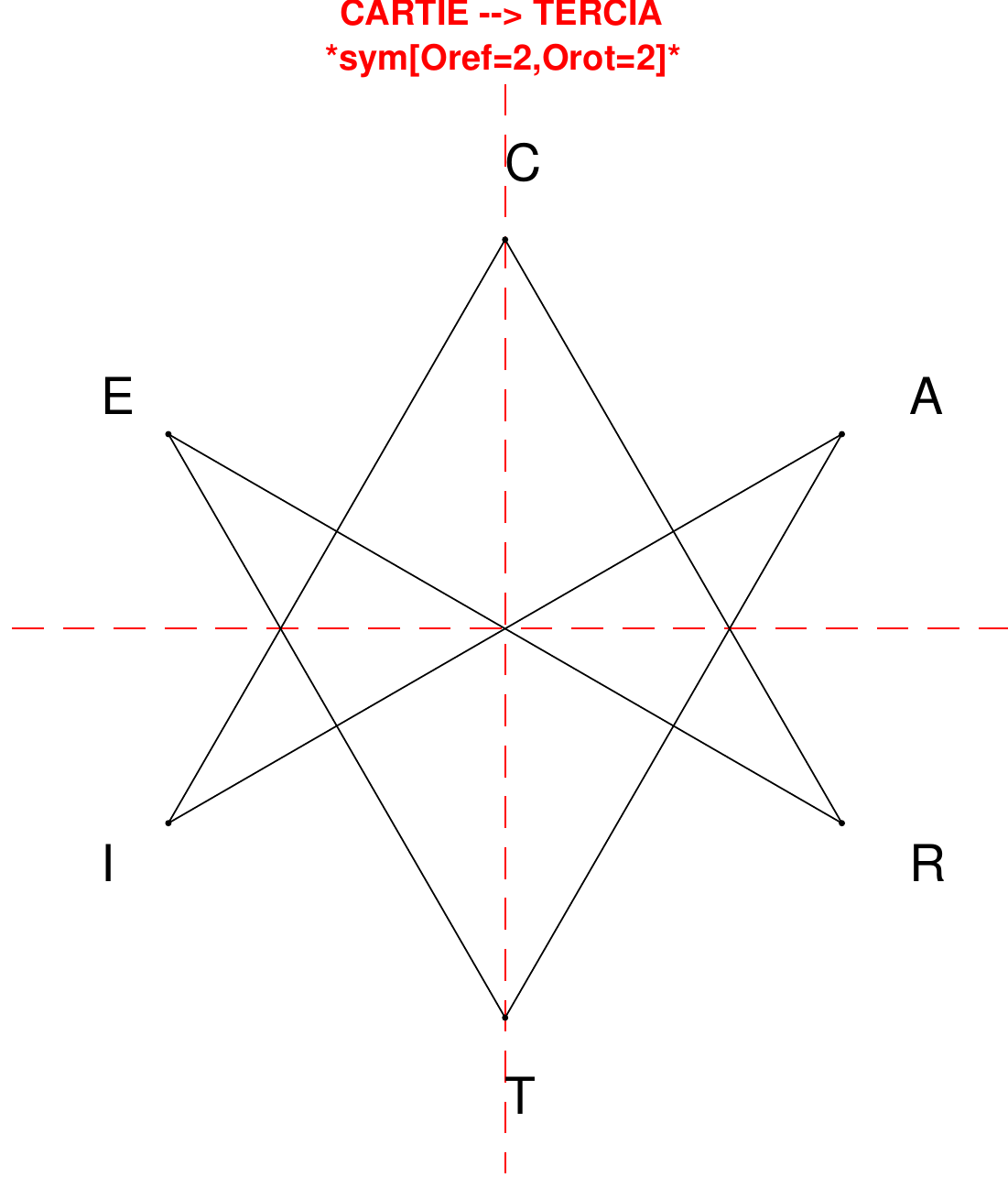}
\end{subfigure}
\hfill
\begin{subfigure}[T]{0.19\textwidth}
\centering
\includegraphics[width=\textwidth]{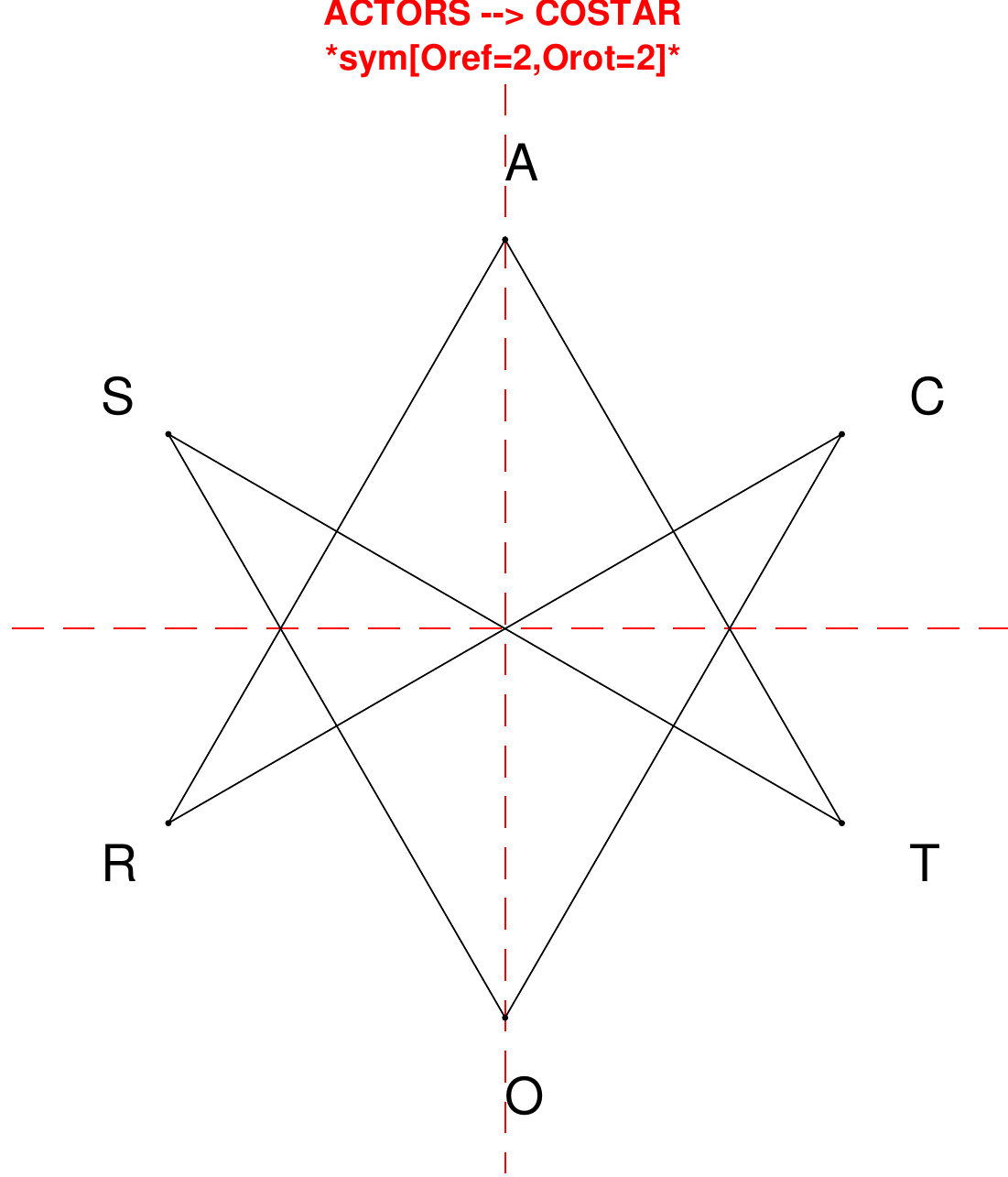}
\end{subfigure}
\hfill
\begin{subfigure}[T]{0.19\textwidth}
\centering
\includegraphics[width=\textwidth]{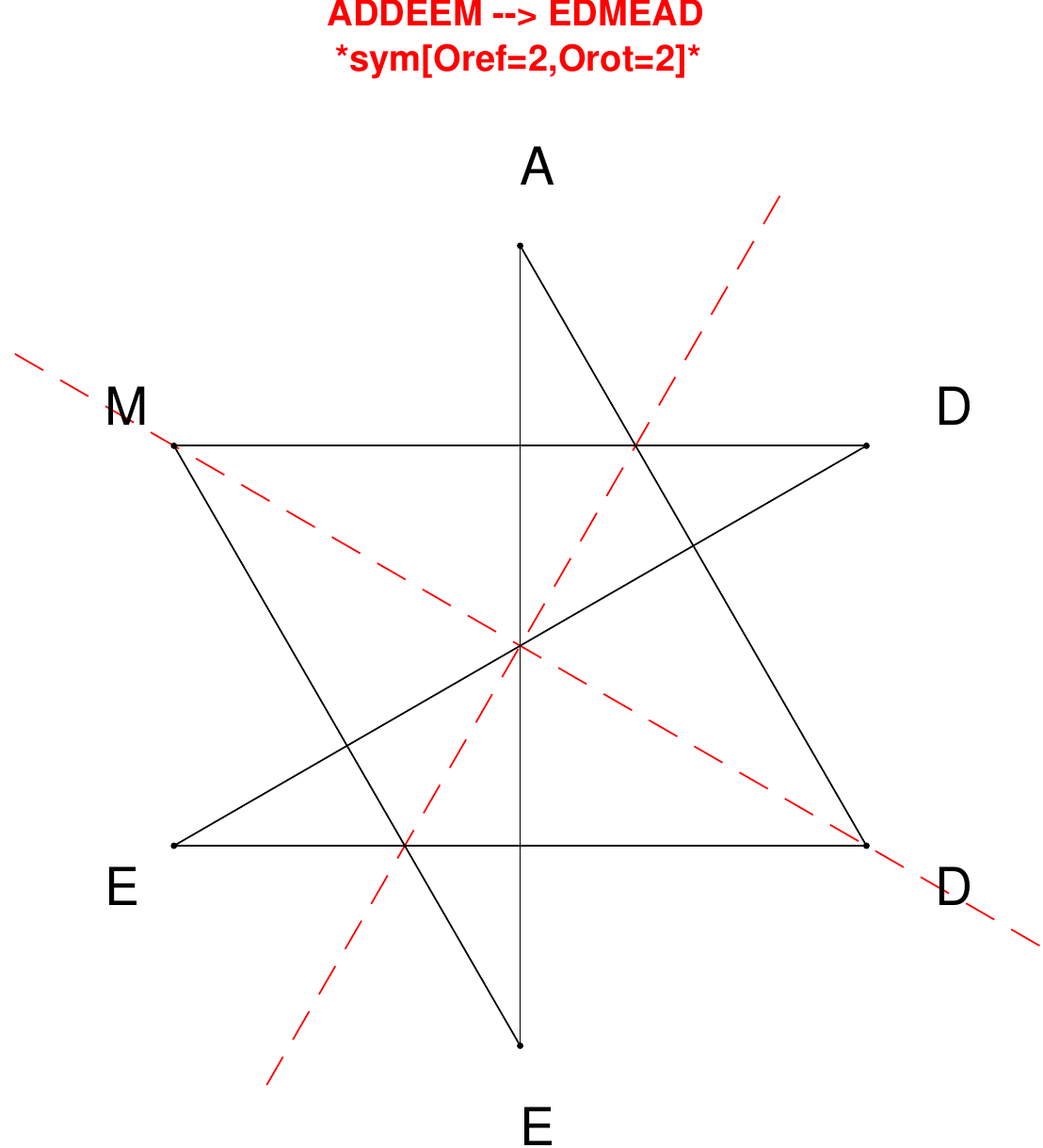}
\end{subfigure}
\end{figure}

\begin{figure}[H]
\centering
\begin{subfigure}[T]{0.19\textwidth}
\centering
\includegraphics[width=\textwidth]{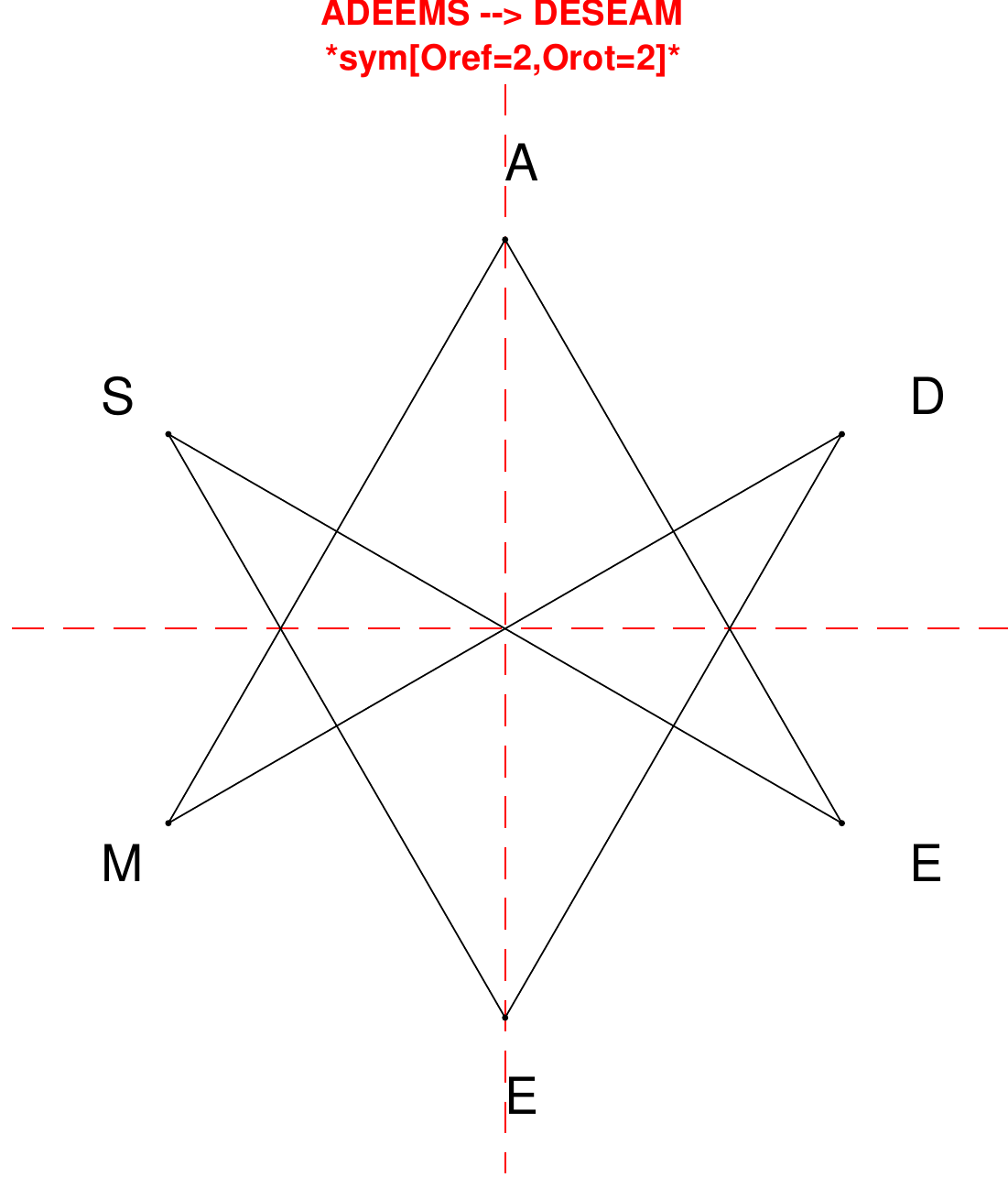}
\end{subfigure}
\hfill
\begin{subfigure}[T]{0.19\textwidth}
\centering
\includegraphics[width=\textwidth]{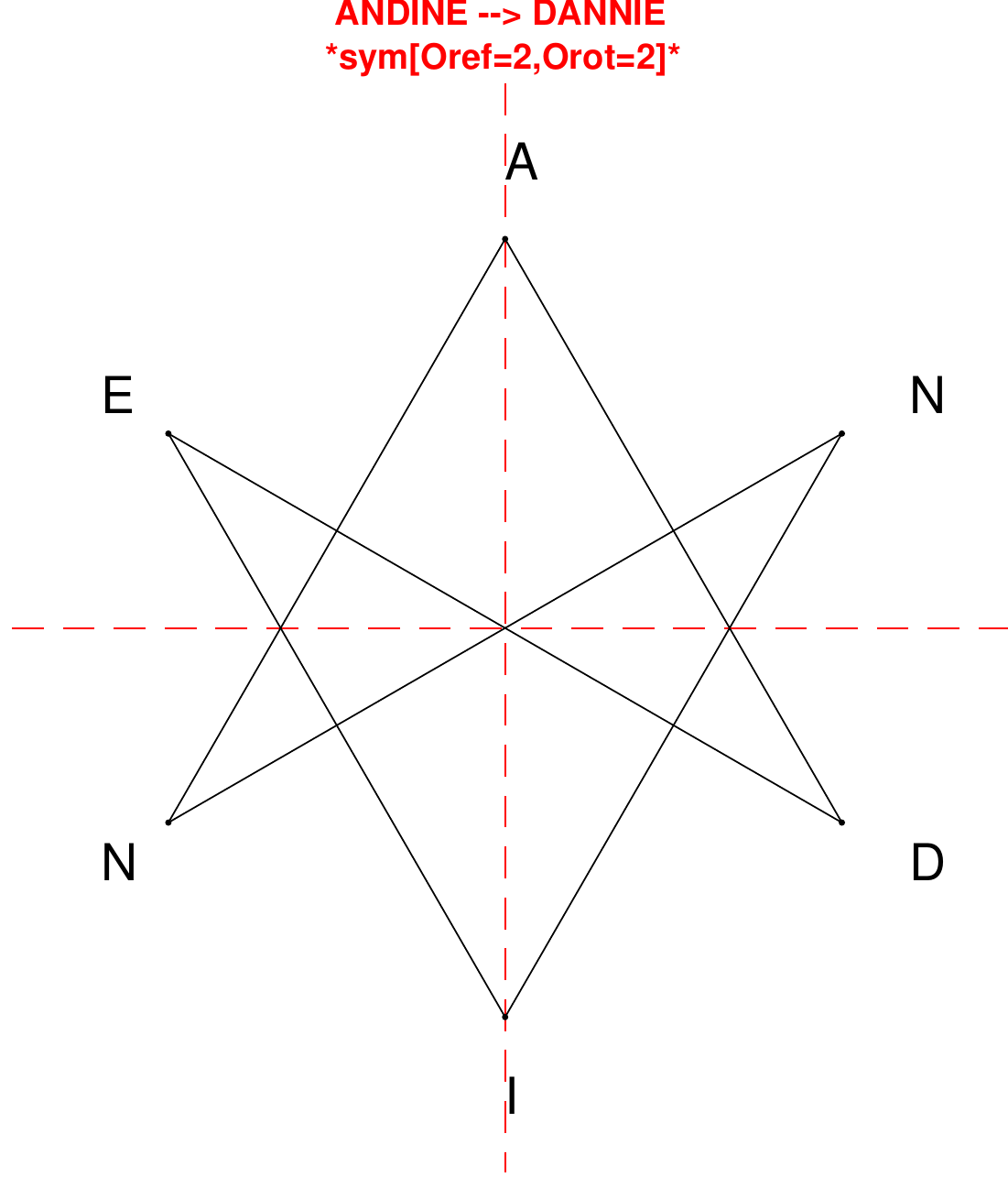}
\end{subfigure}
\hfill
\begin{subfigure}[T]{0.19\textwidth}
\centering
\includegraphics[width=\textwidth]{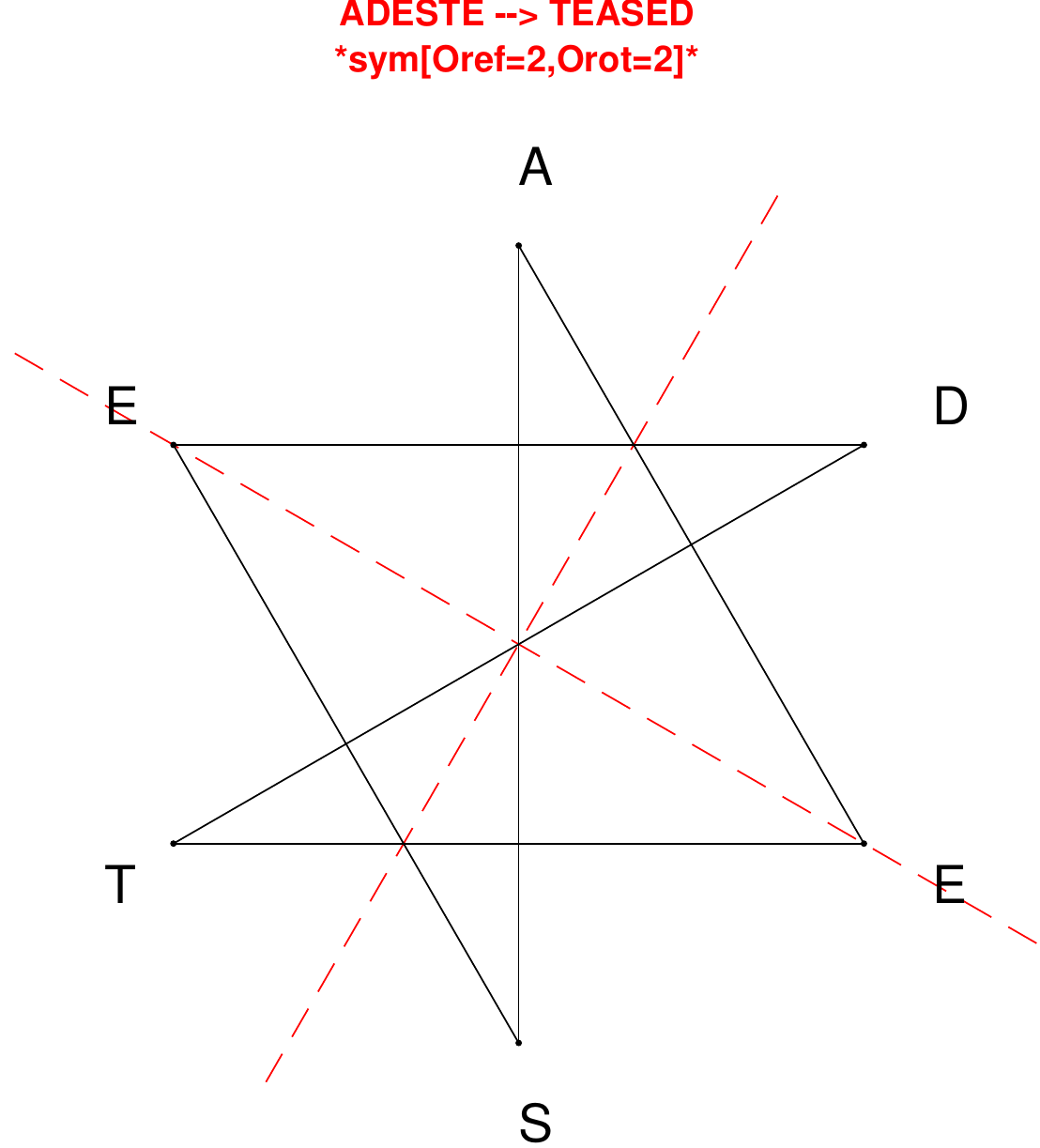}
\end{subfigure}
\hfill
\begin{subfigure}[T]{0.19\textwidth}
\centering
\includegraphics[width=\textwidth]{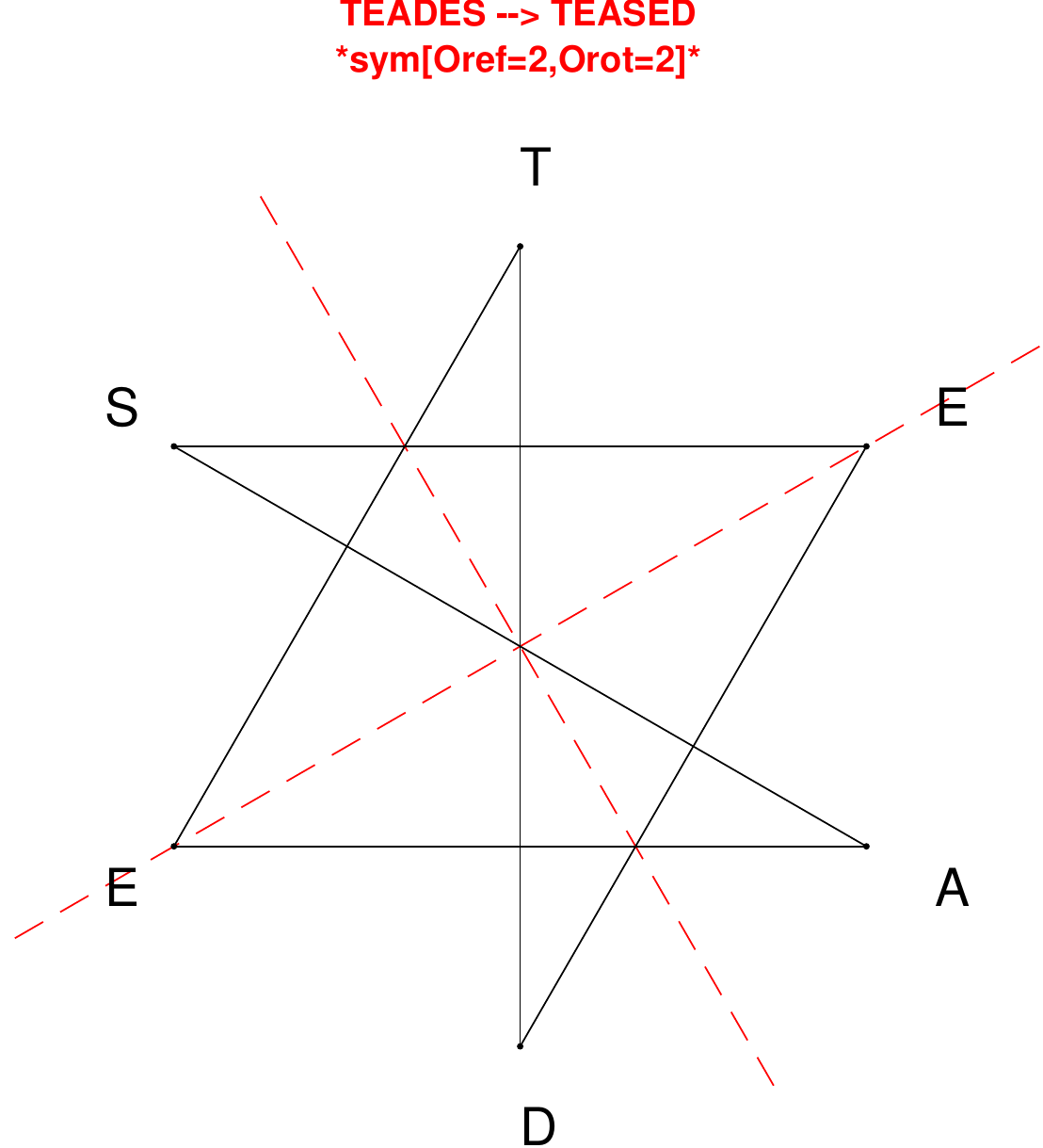}
\end{subfigure}
\hfill
\begin{subfigure}[T]{0.19\textwidth}
\centering
\includegraphics[width=\textwidth]{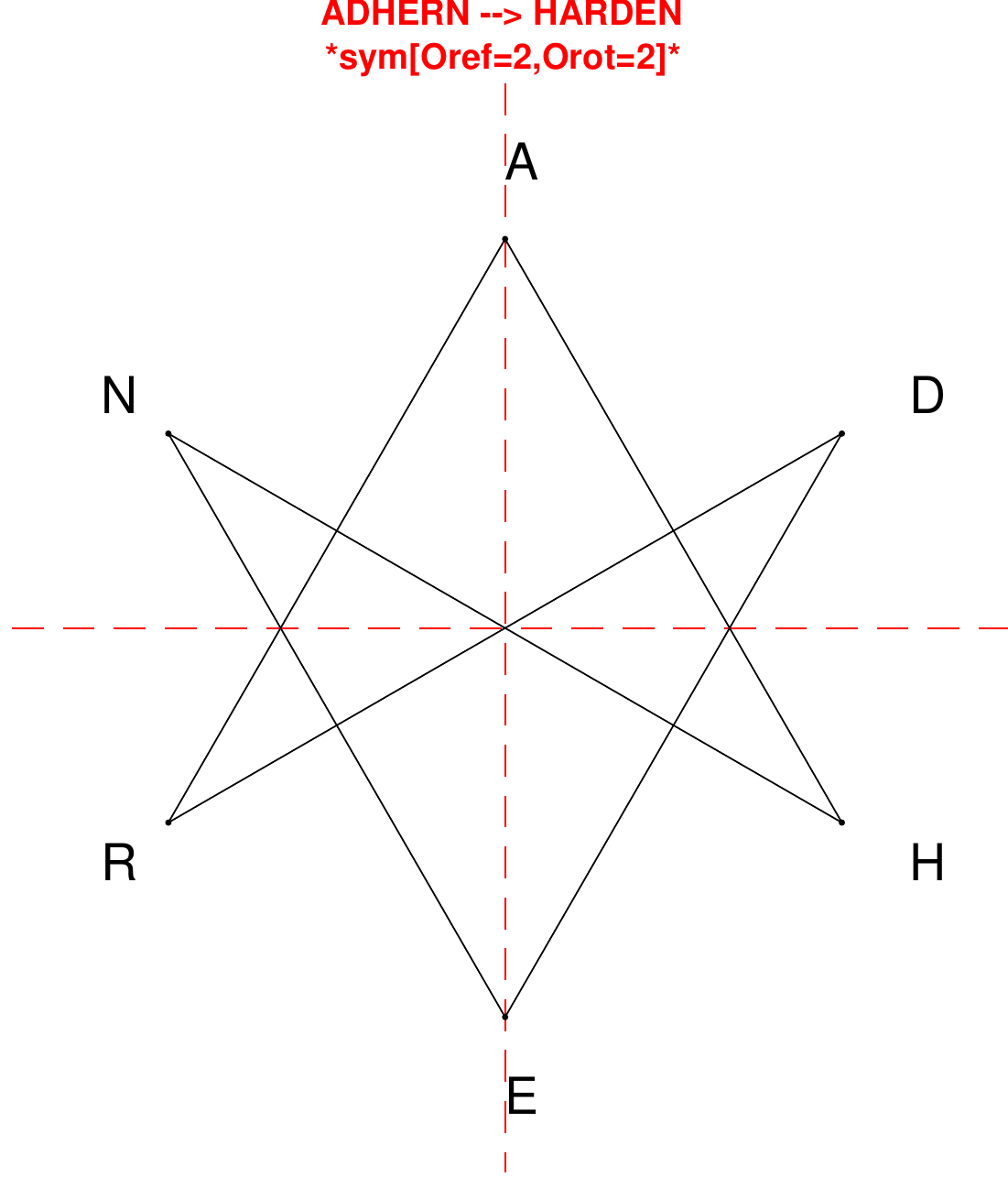}
\end{subfigure}
\end{figure}

\begin{figure}[H]
\centering
\begin{subfigure}[T]{0.19\textwidth}
\centering
\includegraphics[width=\textwidth]{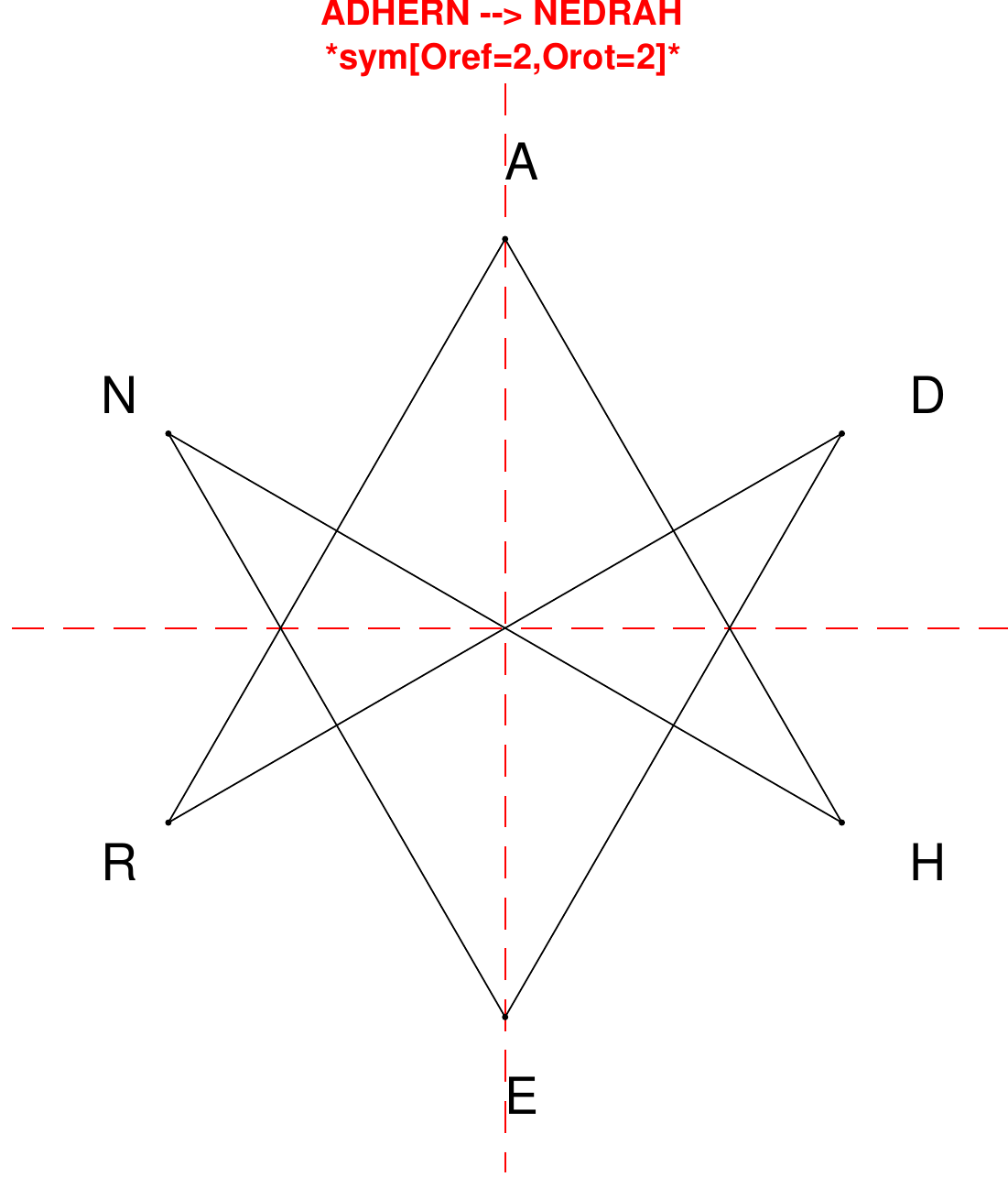}
\end{subfigure}
\hfill
\begin{subfigure}[T]{0.19\textwidth}
\centering
\includegraphics[width=\textwidth]{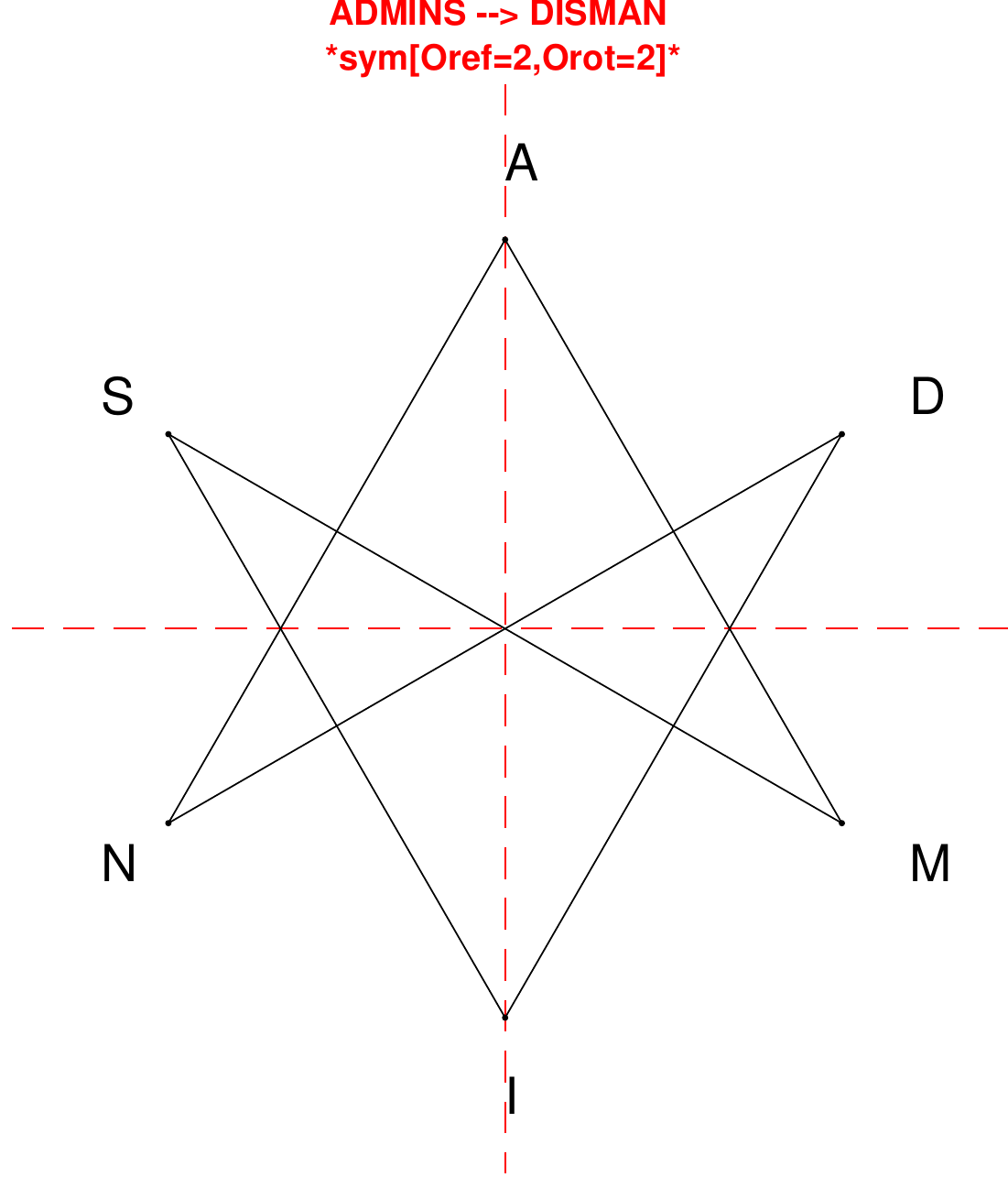}
\end{subfigure}
\hfill
\begin{subfigure}[T]{0.19\textwidth}
\centering
\includegraphics[width=\textwidth]{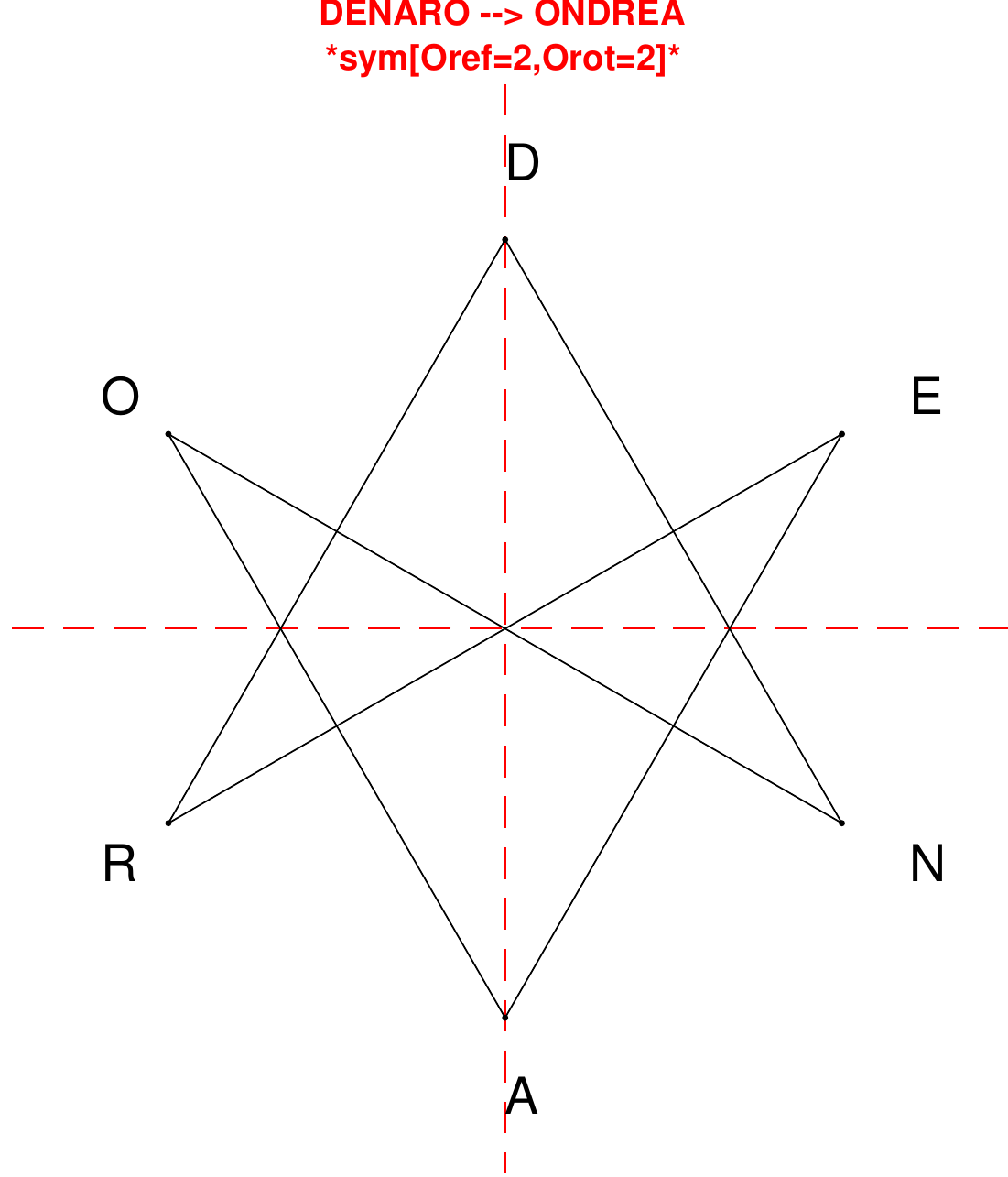}
\end{subfigure}
\hfill
\begin{subfigure}[T]{0.19\textwidth}
\centering
\includegraphics[width=\textwidth]{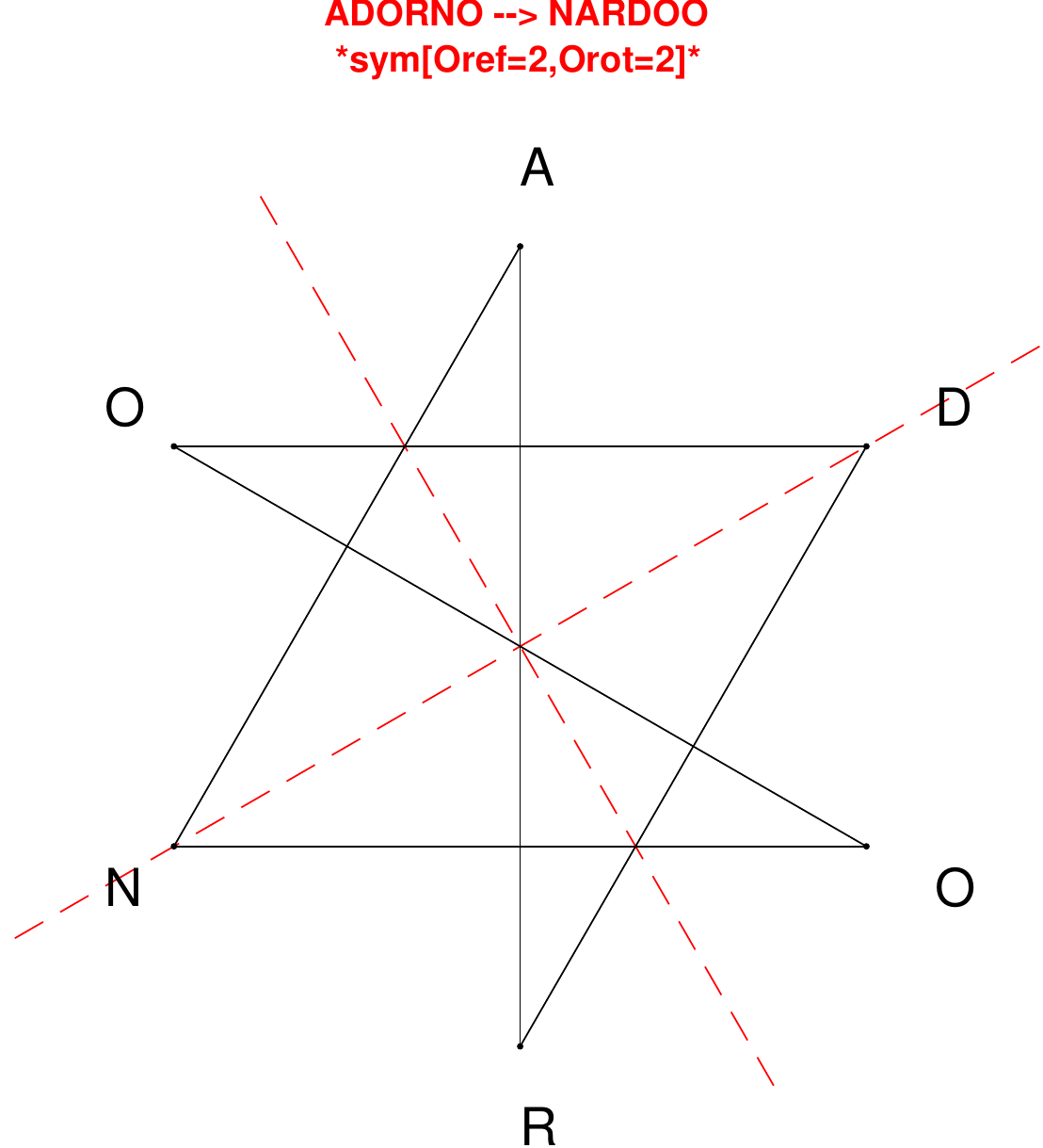}
\end{subfigure}
\hfill
\begin{subfigure}[T]{0.19\textwidth}
\centering
\includegraphics[width=\textwidth]{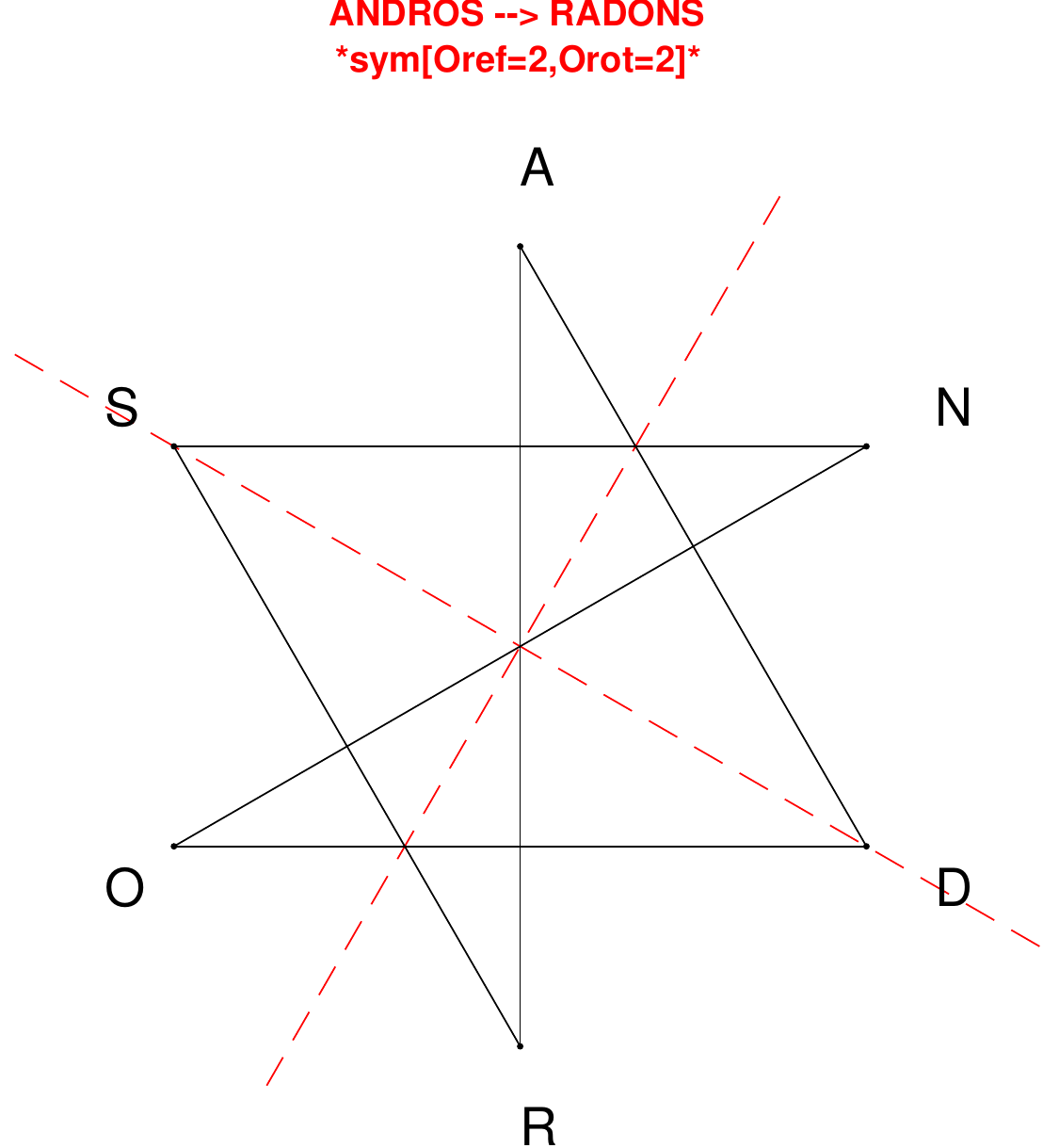}
\end{subfigure}
\end{figure}

\begin{figure}[H]
\centering
\begin{subfigure}[T]{0.19\textwidth}
\centering
\includegraphics[width=\textwidth]{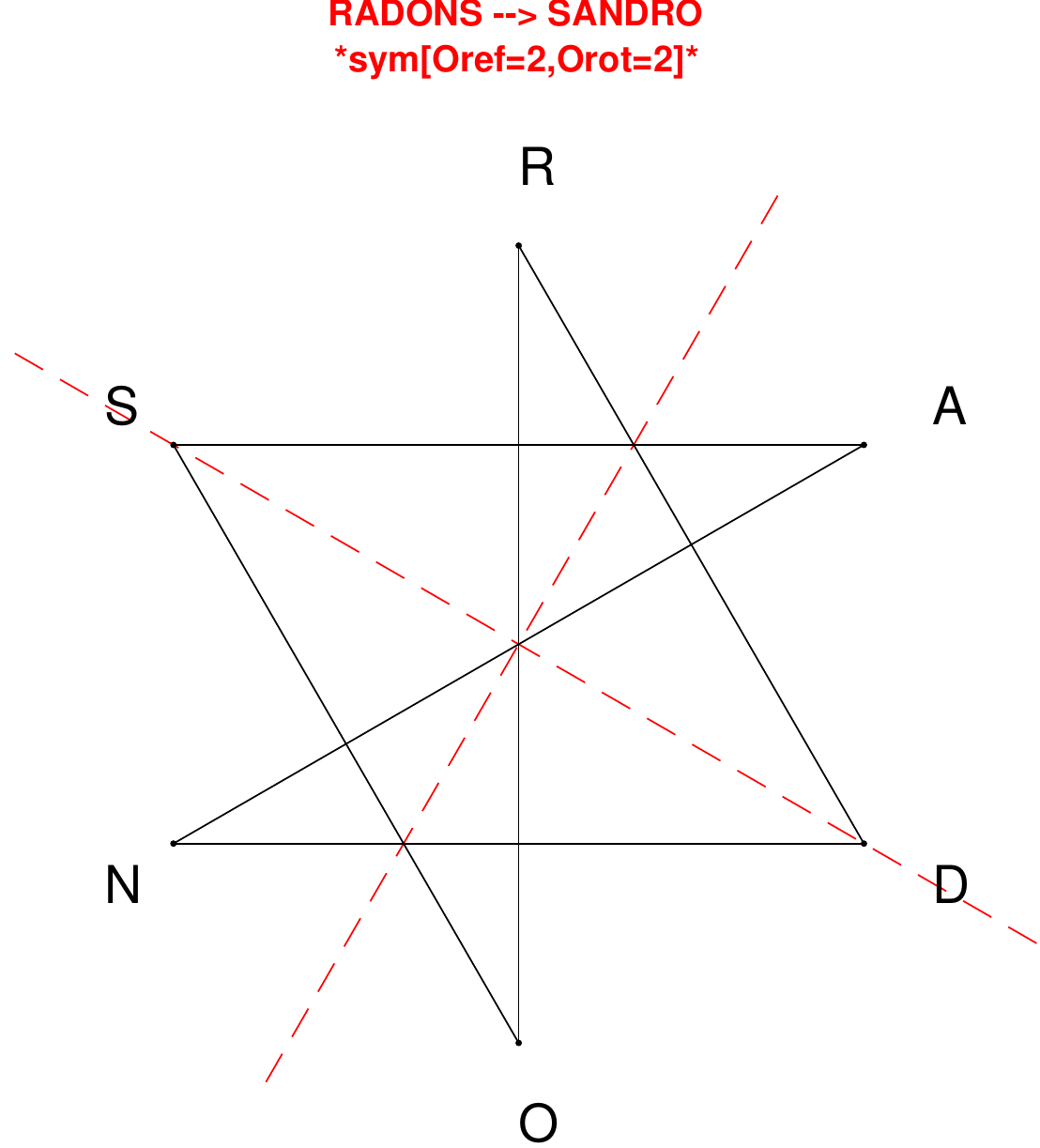}
\end{subfigure}
\hfill
\begin{subfigure}[T]{0.19\textwidth}
\centering
\includegraphics[width=\textwidth]{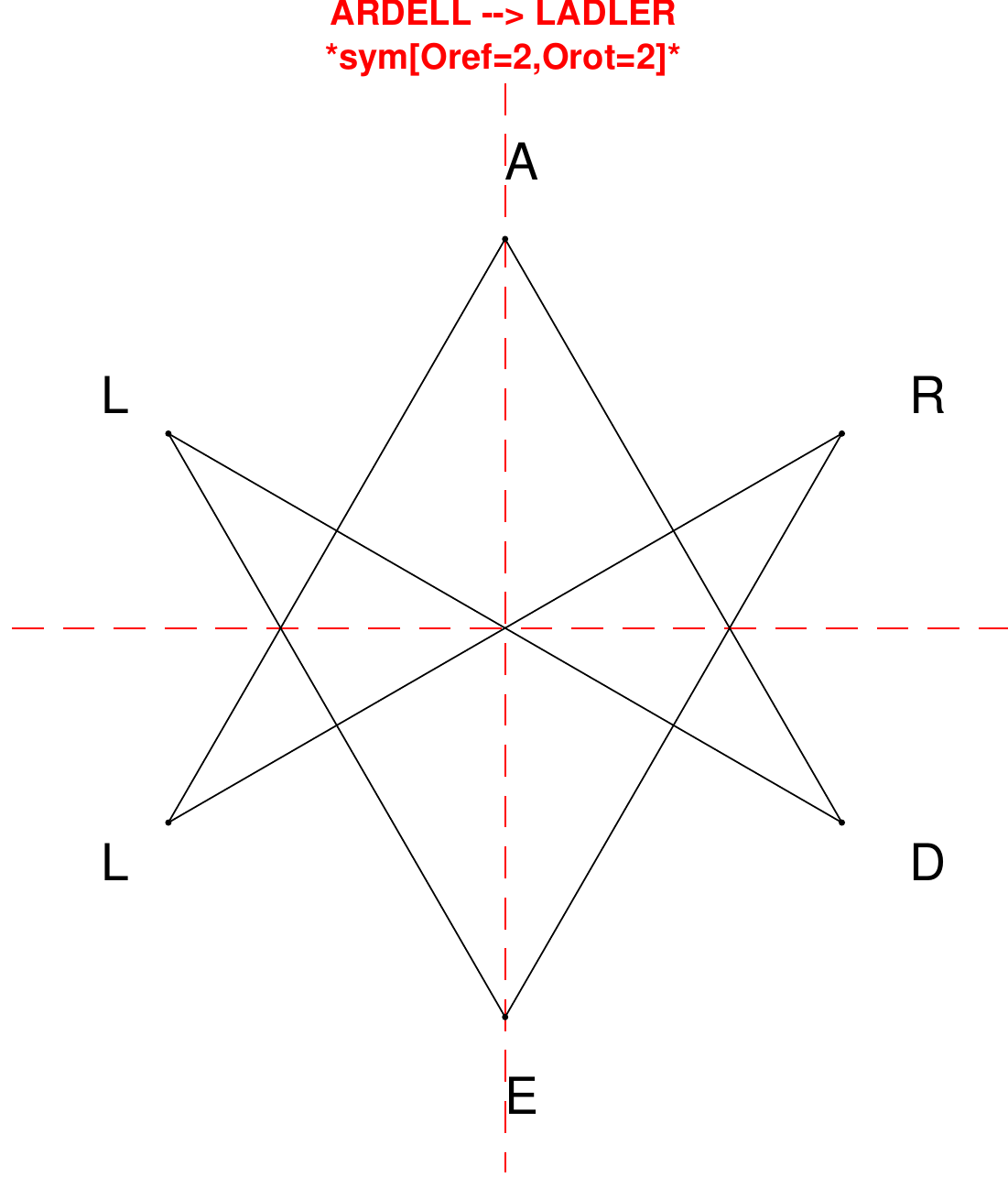}
\end{subfigure}
\hfill
\begin{subfigure}[T]{0.19\textwidth}
\centering
\includegraphics[width=\textwidth]{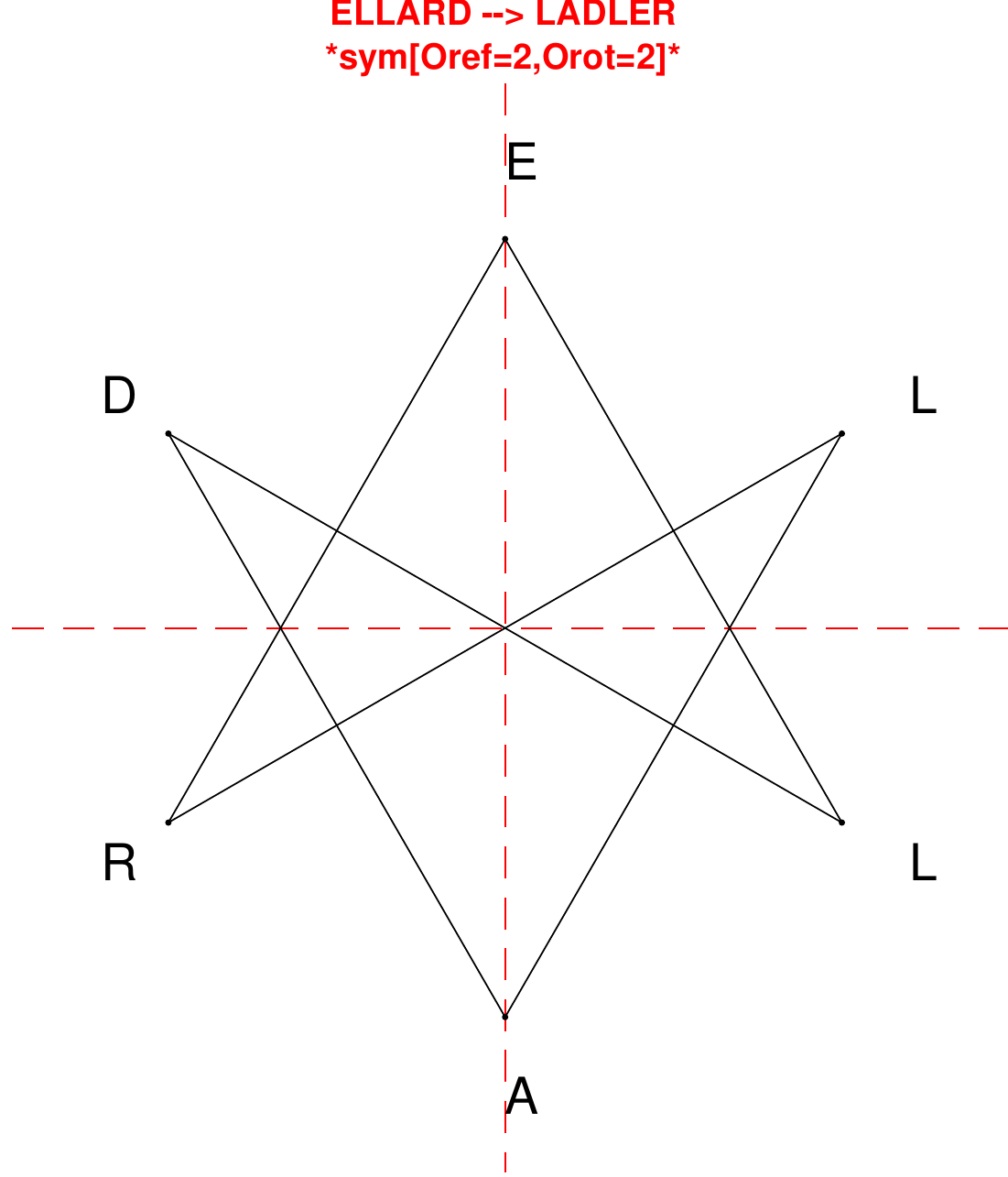}
\end{subfigure}
\hfill
\begin{subfigure}[T]{0.19\textwidth}
\centering
\includegraphics[width=\textwidth]{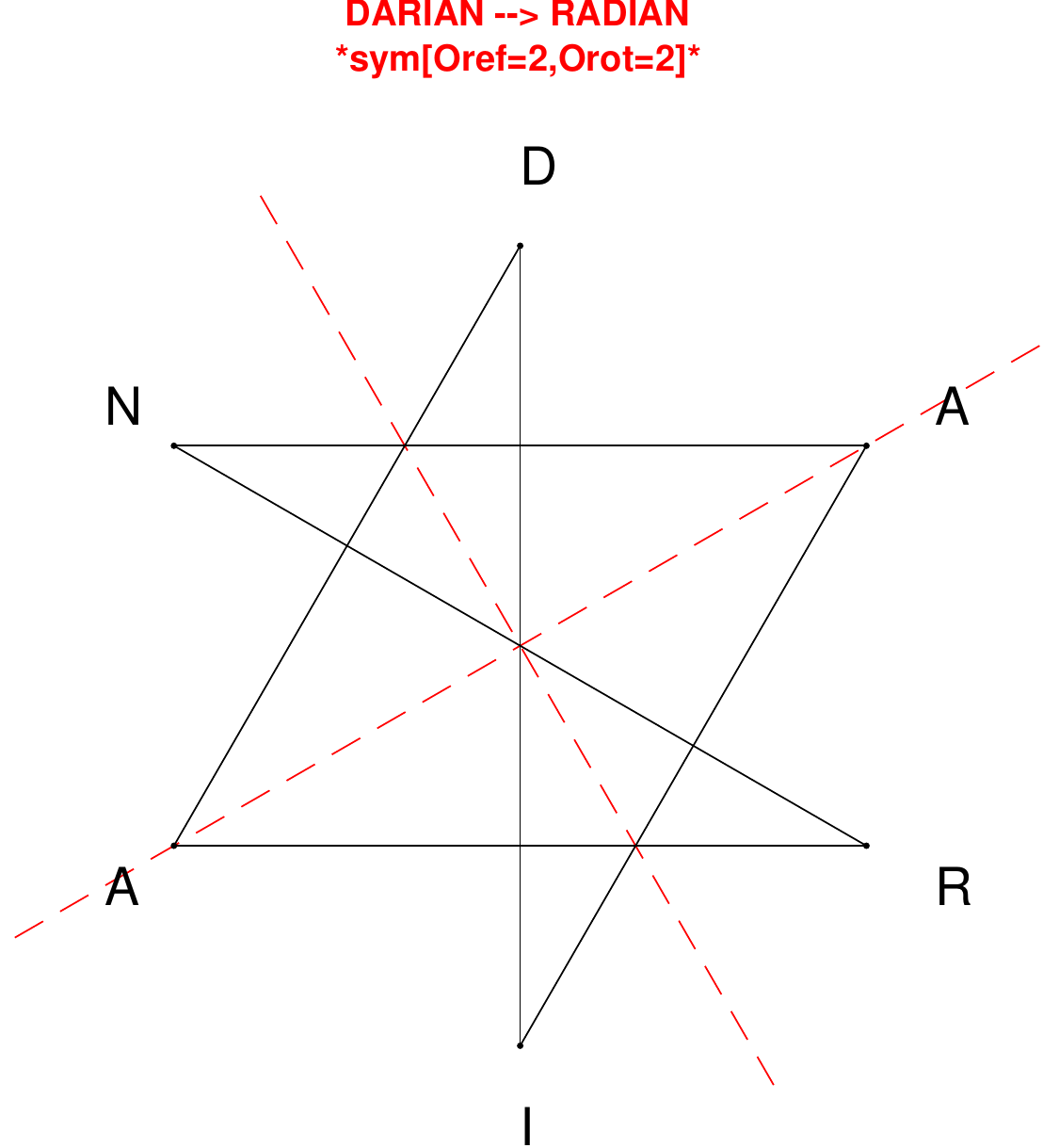}
\end{subfigure}
\hfill
\begin{subfigure}[T]{0.19\textwidth}
\centering
\includegraphics[width=\textwidth]{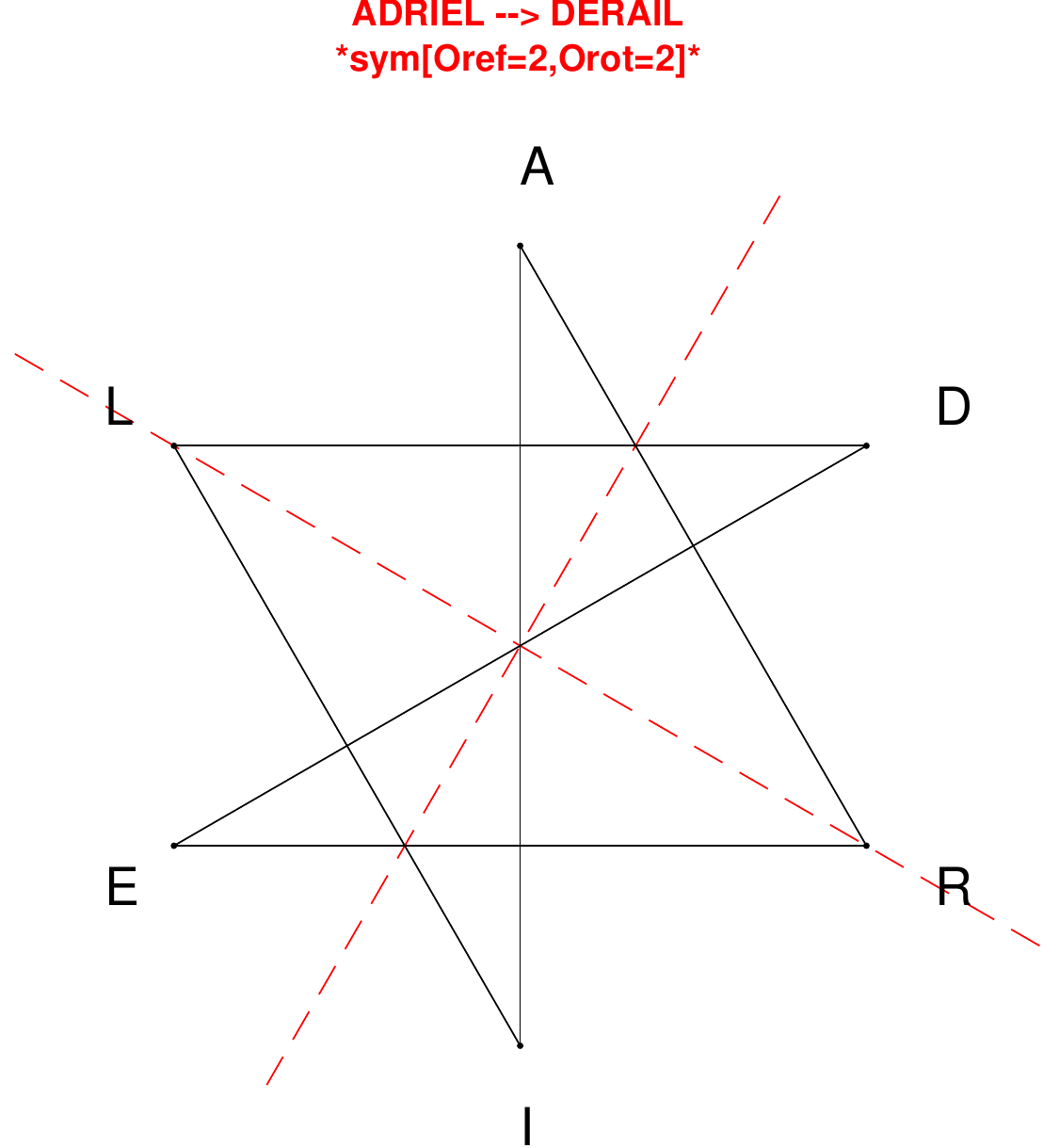}
\end{subfigure}
\end{figure}

\begin{figure}[H]
\centering
\begin{subfigure}[T]{0.19\textwidth}
\centering
\includegraphics[width=\textwidth]{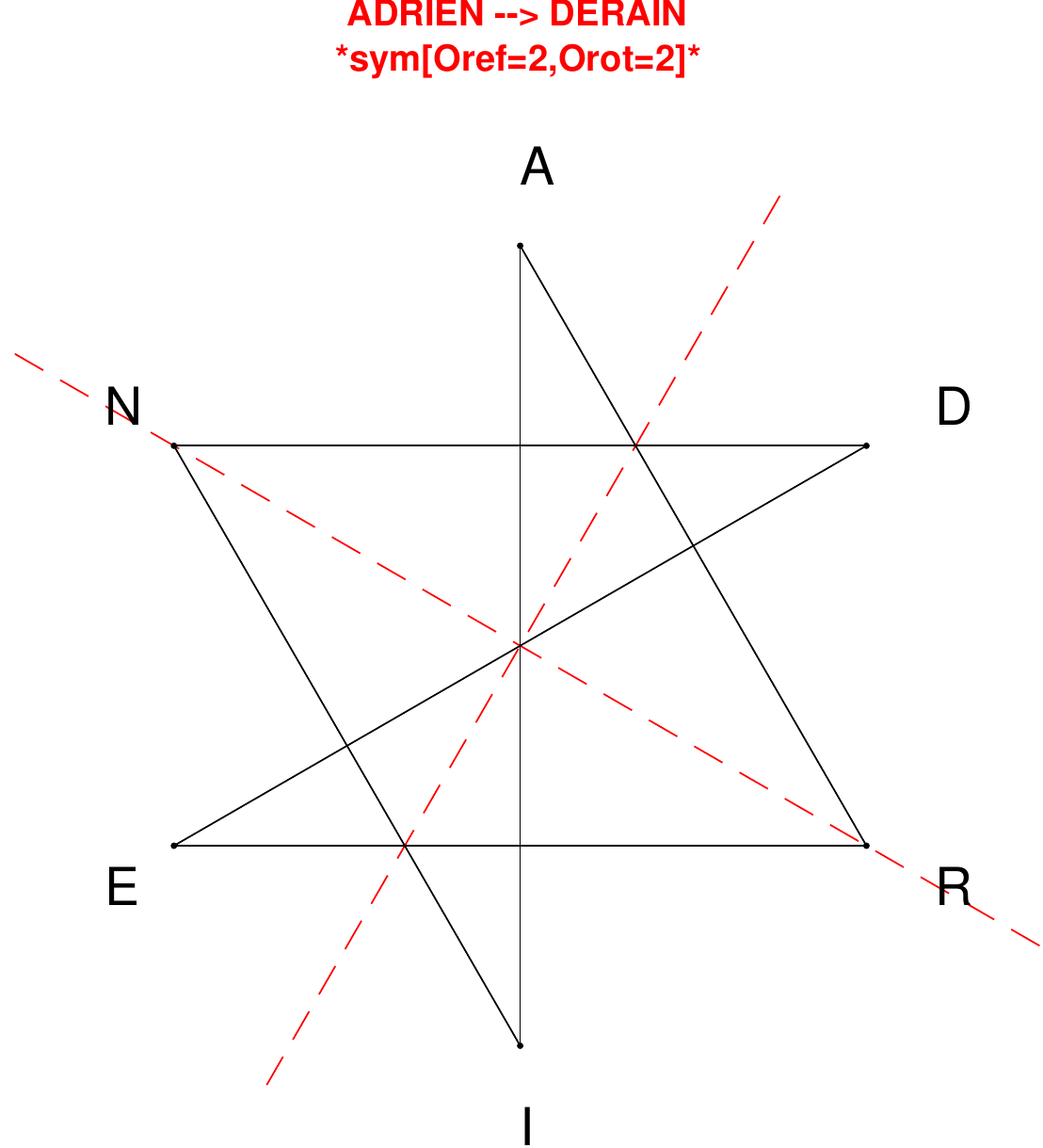}
\end{subfigure}
\hfill
\begin{subfigure}[T]{0.19\textwidth}
\centering
\includegraphics[width=\textwidth]{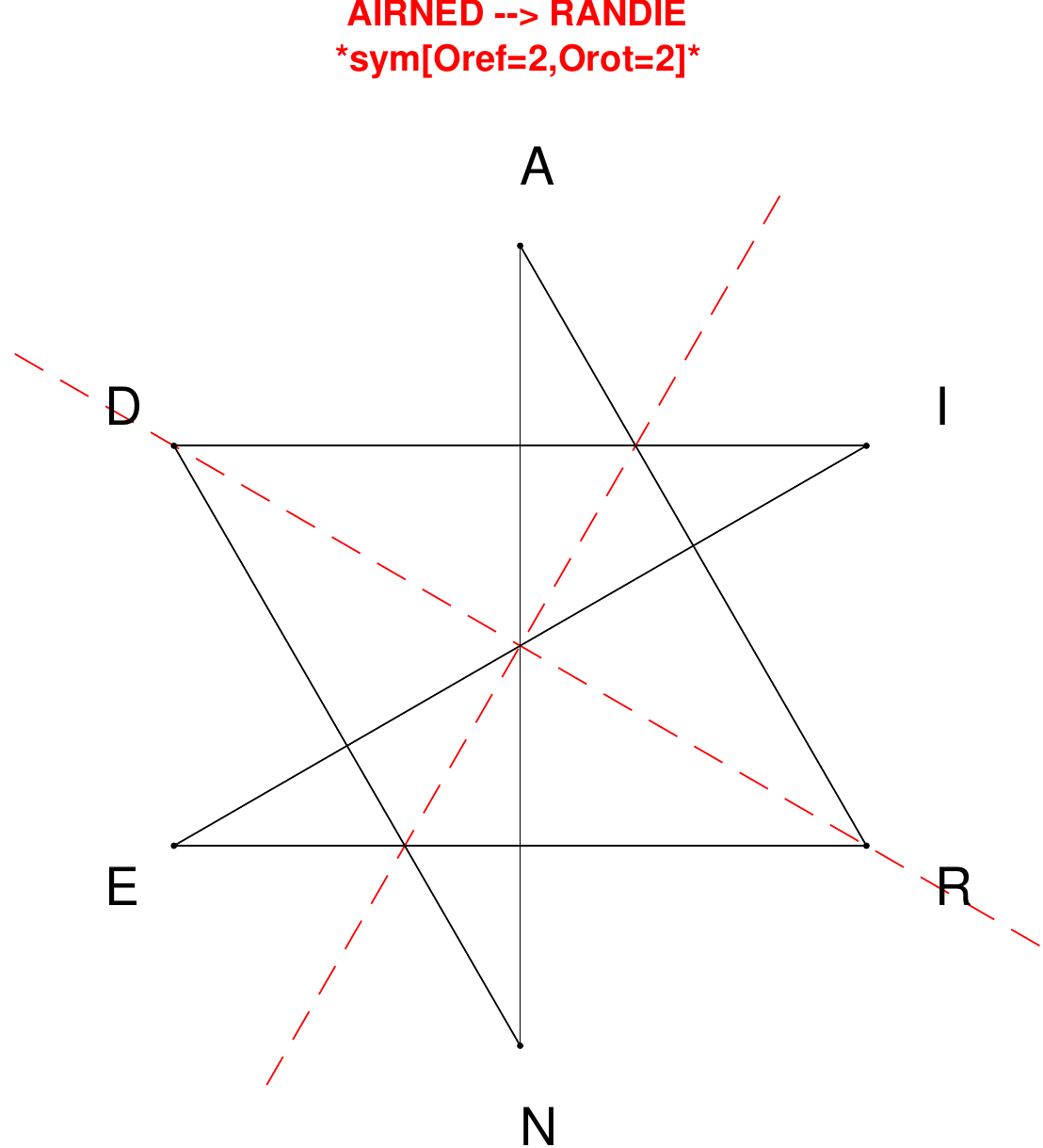}
\end{subfigure}
\hfill
\begin{subfigure}[T]{0.19\textwidth}
\centering
\includegraphics[width=\textwidth]{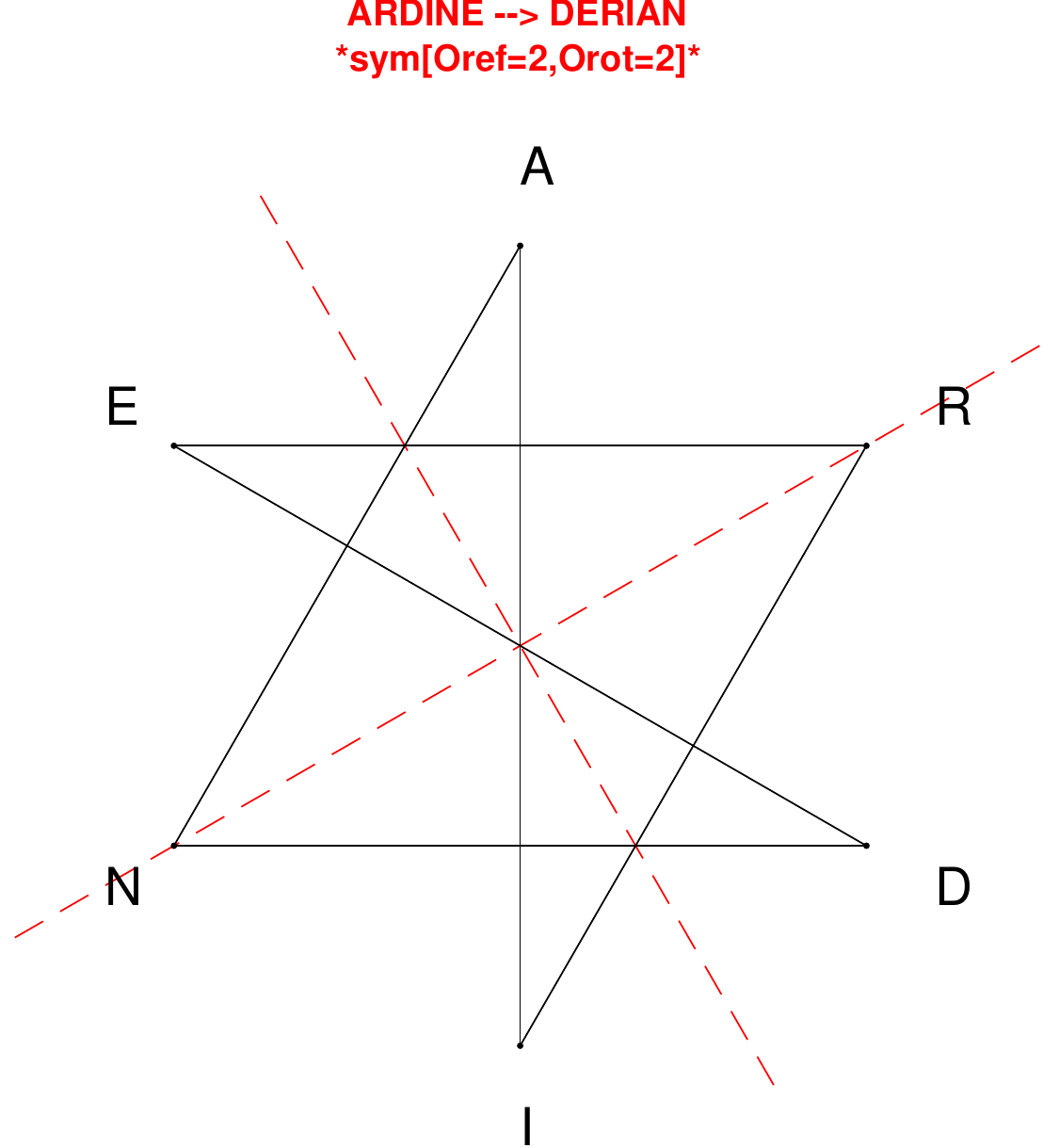}
\end{subfigure}
\hfill
\begin{subfigure}[T]{0.19\textwidth}
\centering
\includegraphics[width=\textwidth]{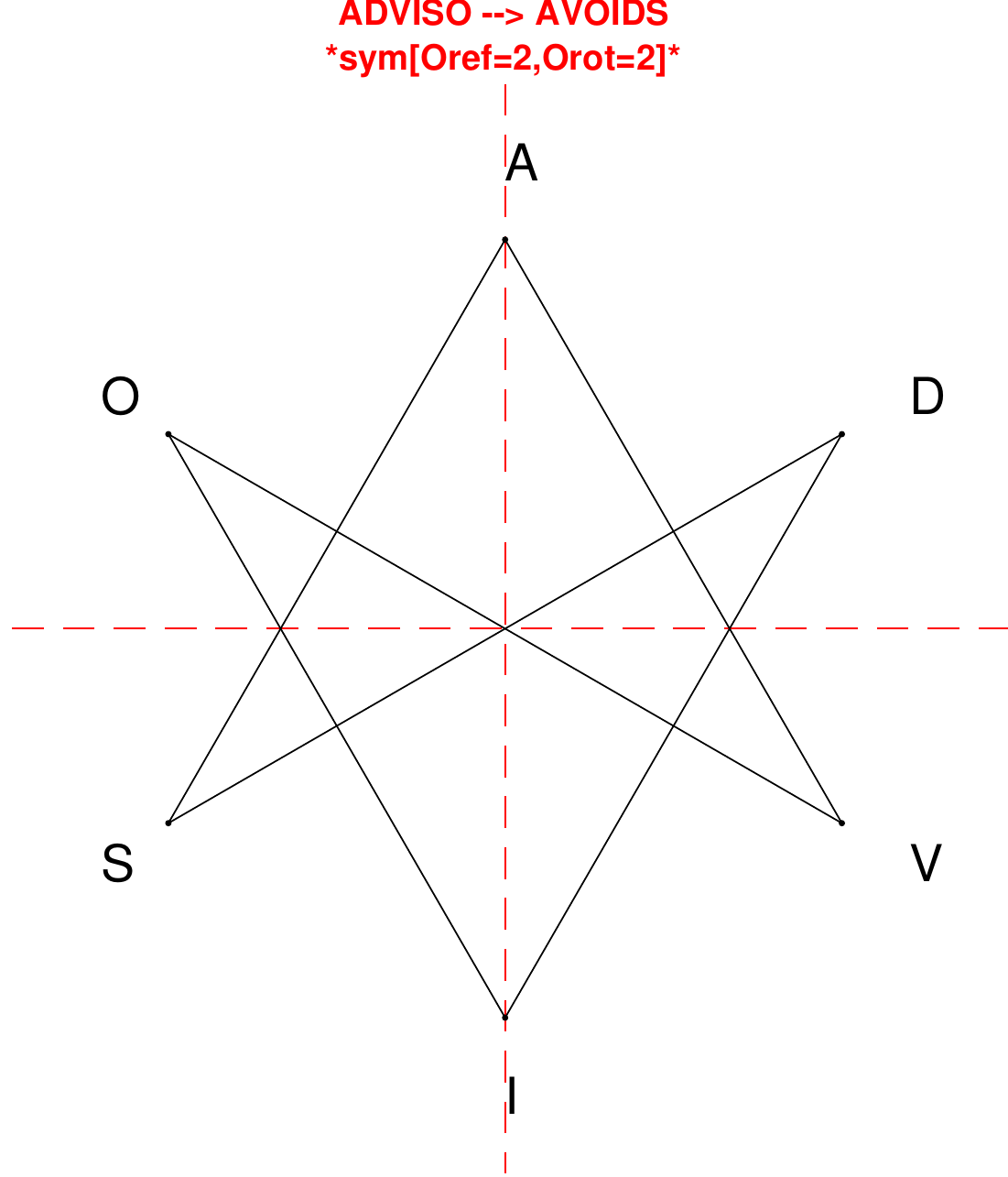}
\end{subfigure}
\hfill
\begin{subfigure}[T]{0.19\textwidth}
\centering
\includegraphics[width=\textwidth]{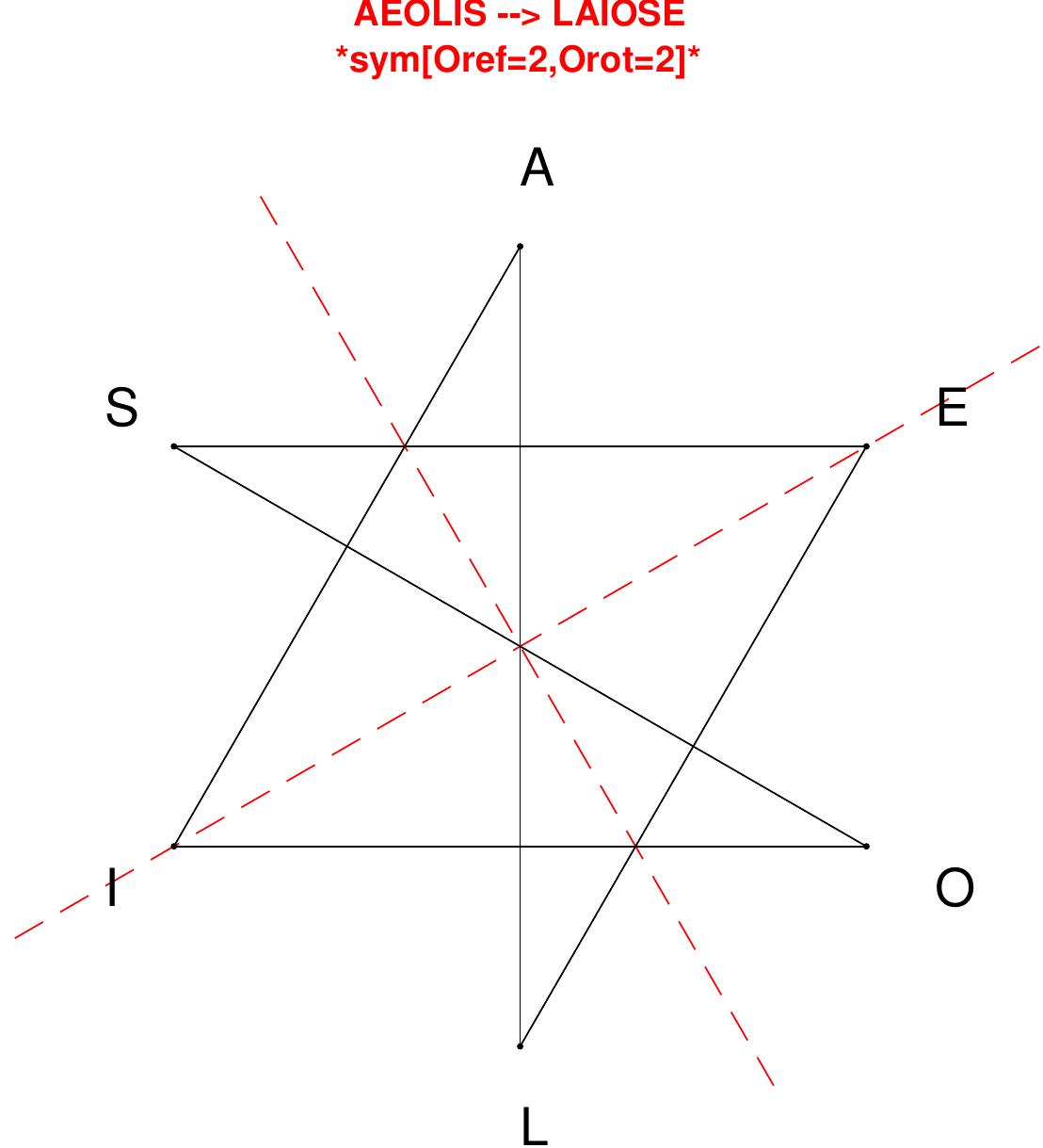}
\end{subfigure}
\end{figure}

\begin{figure}[H]
\centering
\begin{subfigure}[T]{0.19\textwidth}
\centering
\includegraphics[width=\textwidth]{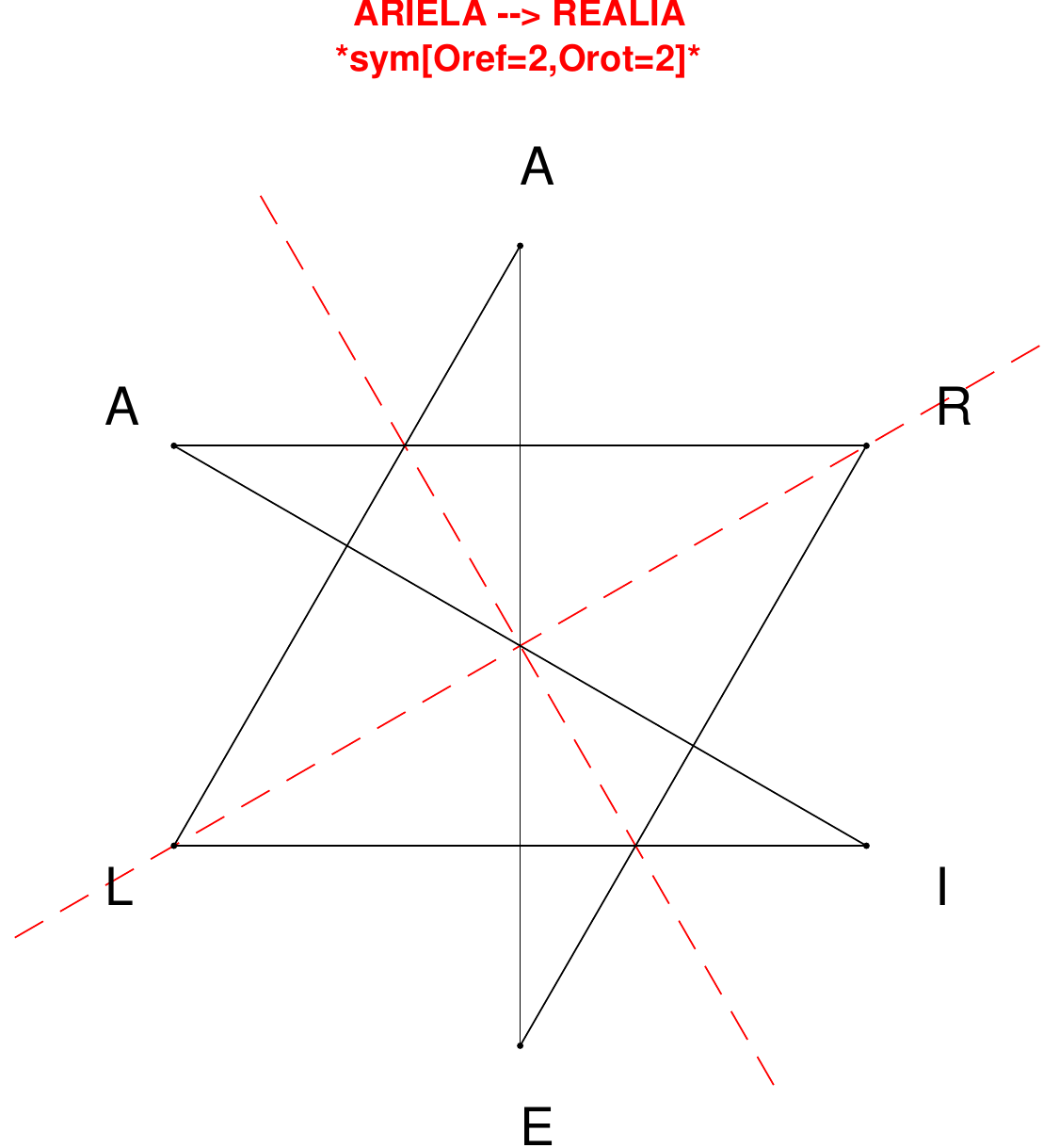}
\end{subfigure}
\hfill
\begin{subfigure}[T]{0.19\textwidth}
\centering
\includegraphics[width=\textwidth]{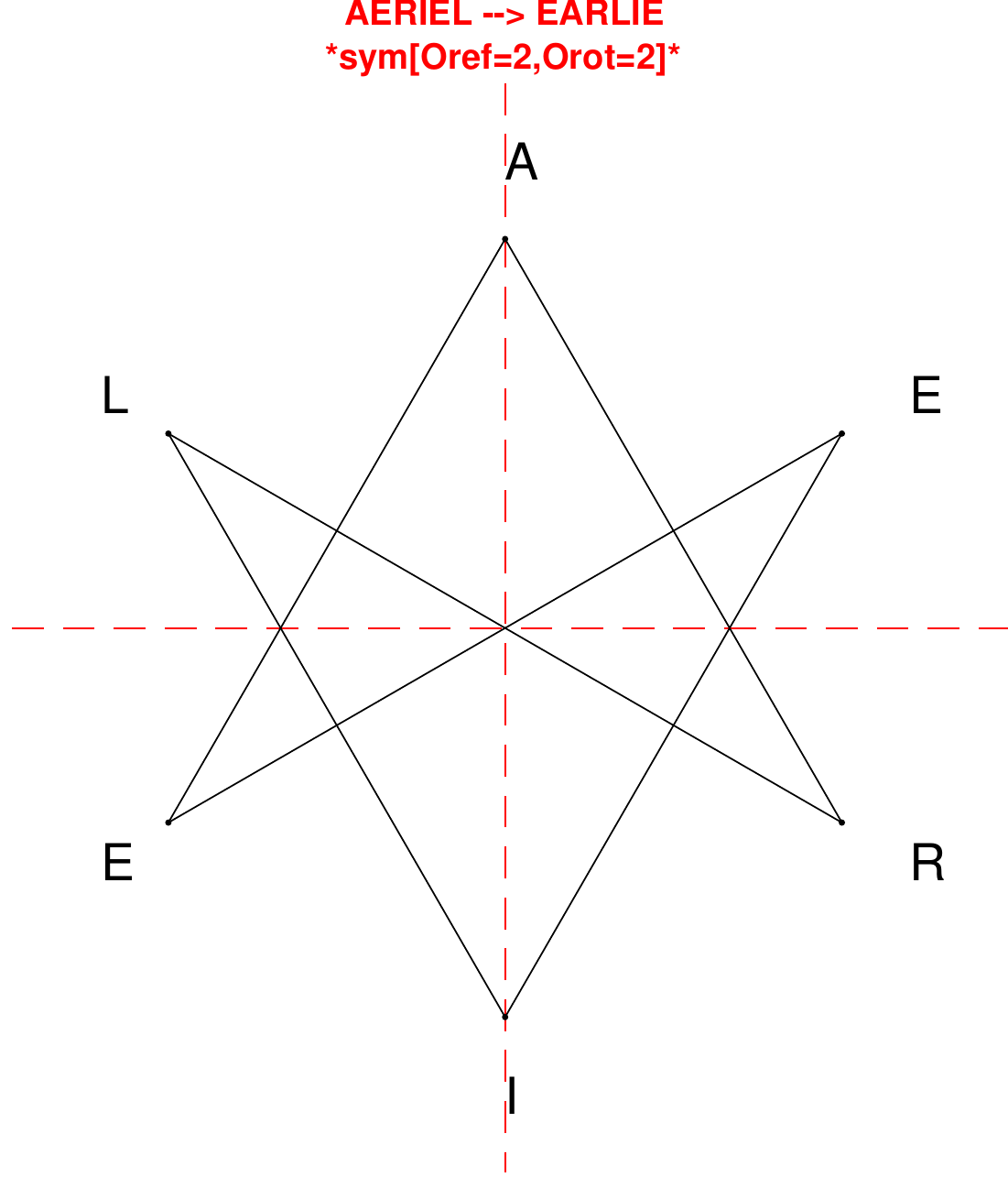}
\end{subfigure}
\hfill
\begin{subfigure}[T]{0.19\textwidth}
\centering
\includegraphics[width=\textwidth]{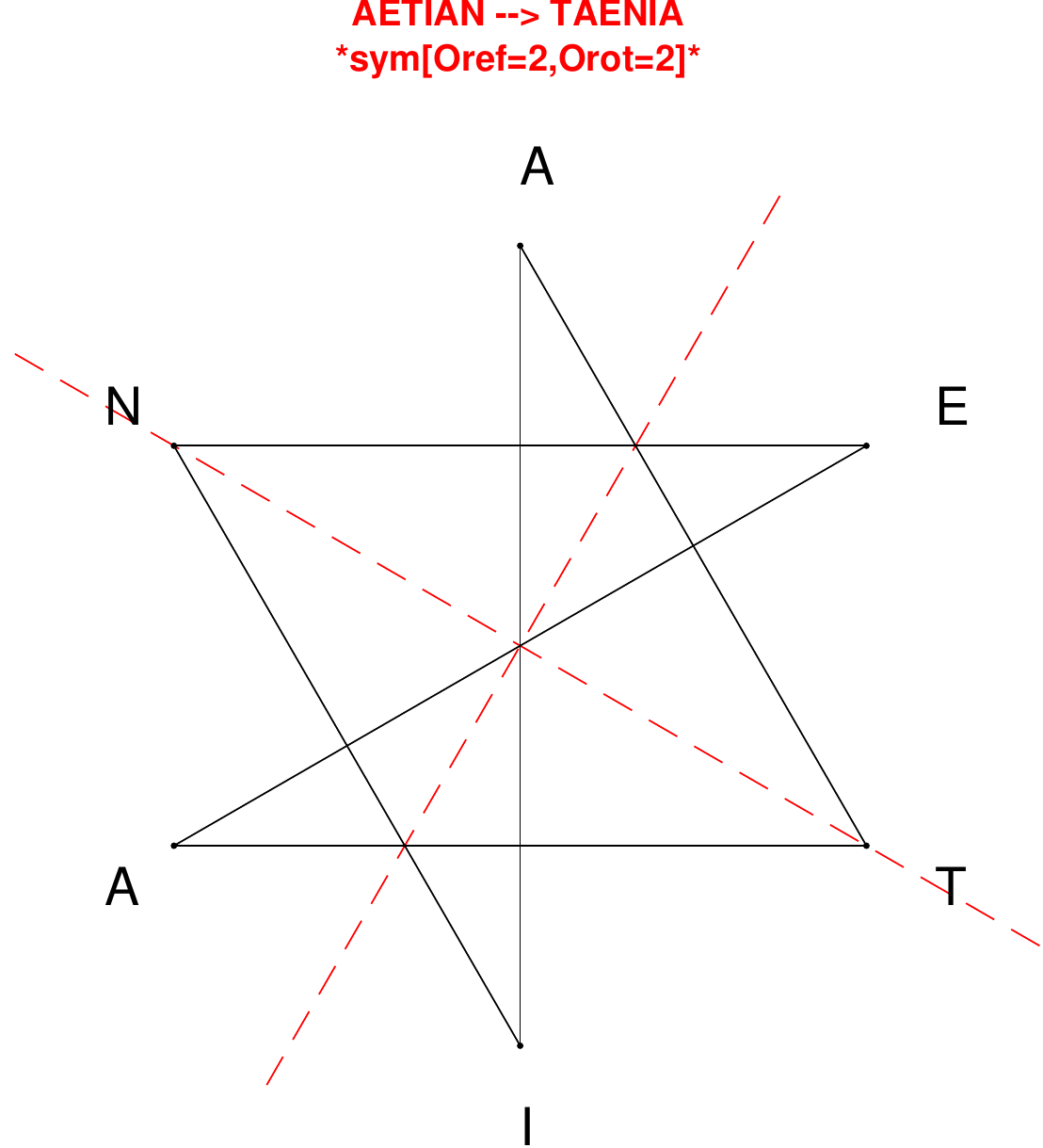}
\end{subfigure}
\hfill
\begin{subfigure}[T]{0.19\textwidth}
\centering
\includegraphics[width=\textwidth]{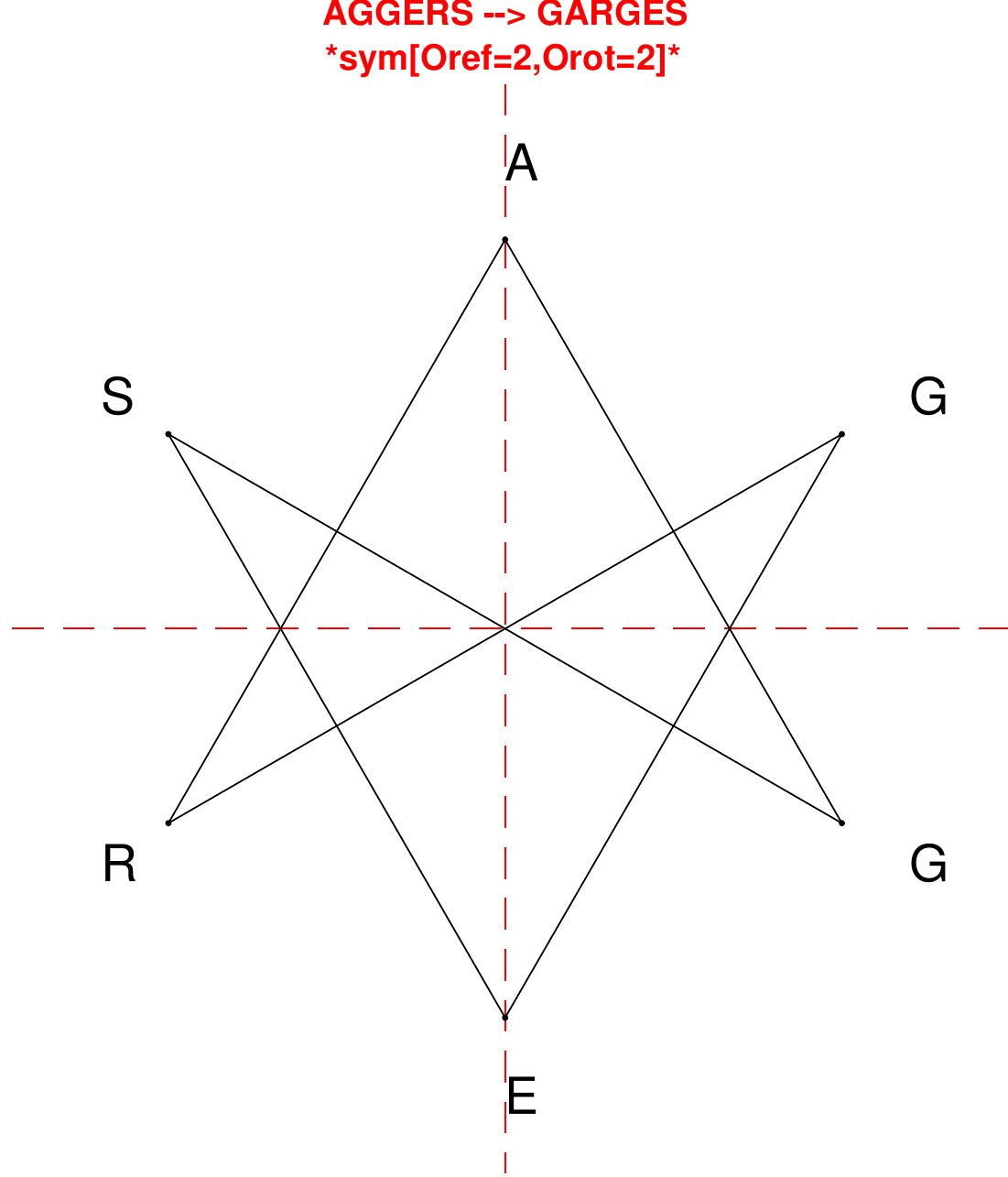}
\end{subfigure}
\hfill
\begin{subfigure}[T]{0.19\textwidth}
\centering
\includegraphics[width=\textwidth]{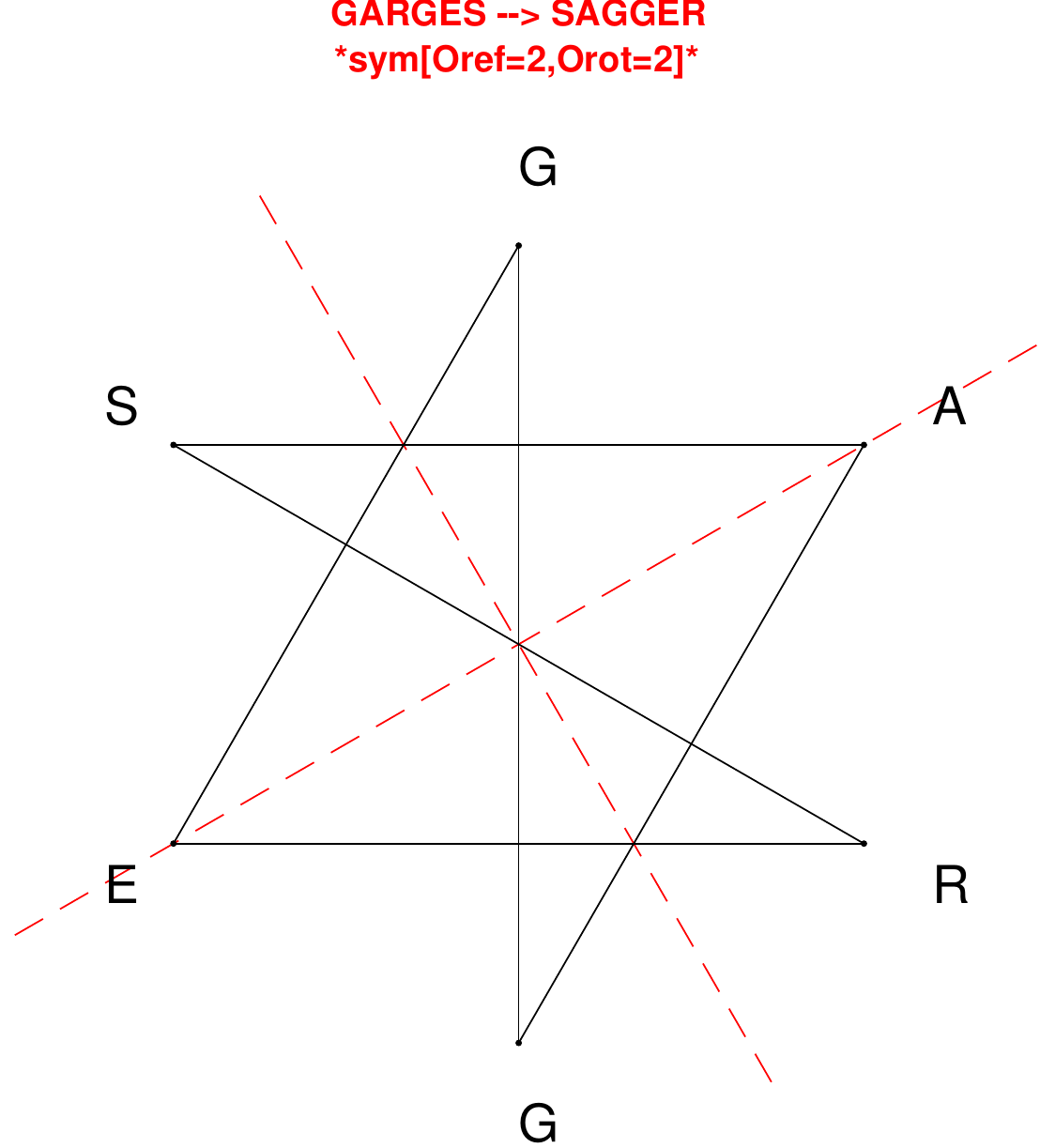}
\end{subfigure}
\end{figure}

\begin{figure}[H]
\centering
\begin{subfigure}[T]{0.19\textwidth}
\centering
\includegraphics[width=\textwidth]{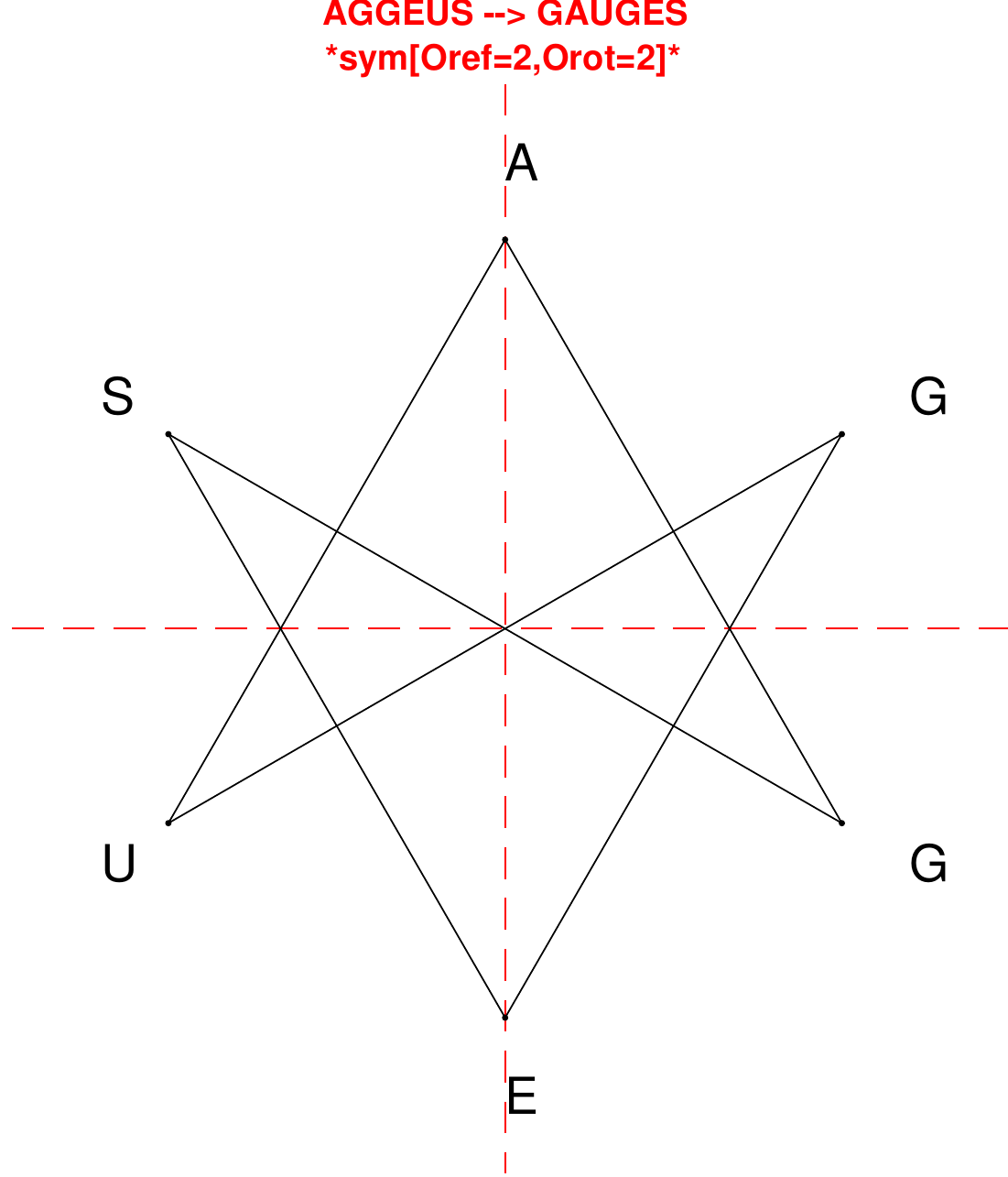}
\end{subfigure}
\hfill
\begin{subfigure}[T]{0.19\textwidth}
\centering
\includegraphics[width=\textwidth]{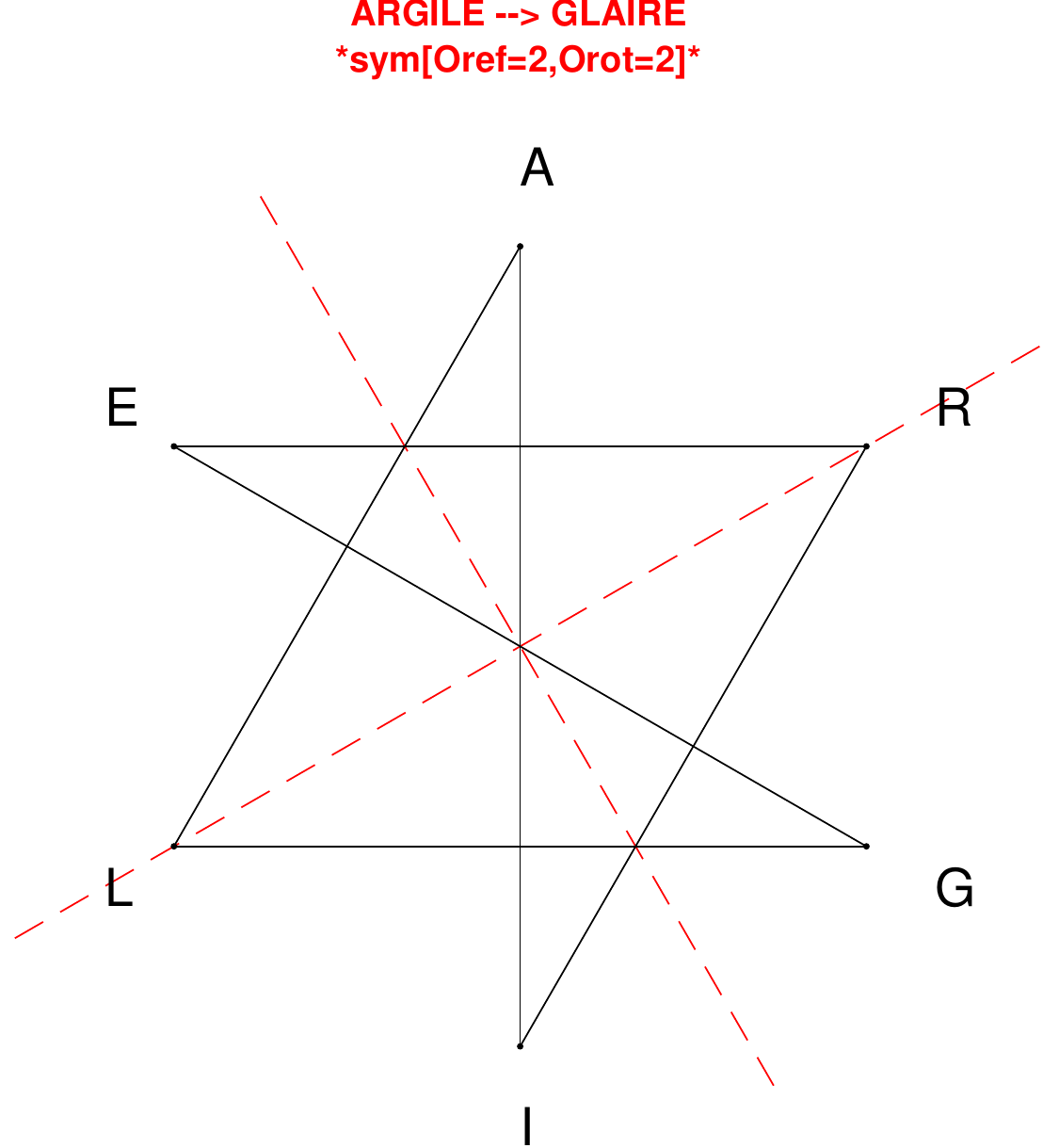}
\end{subfigure}
\hfill
\begin{subfigure}[T]{0.19\textwidth}
\centering
\includegraphics[width=\textwidth]{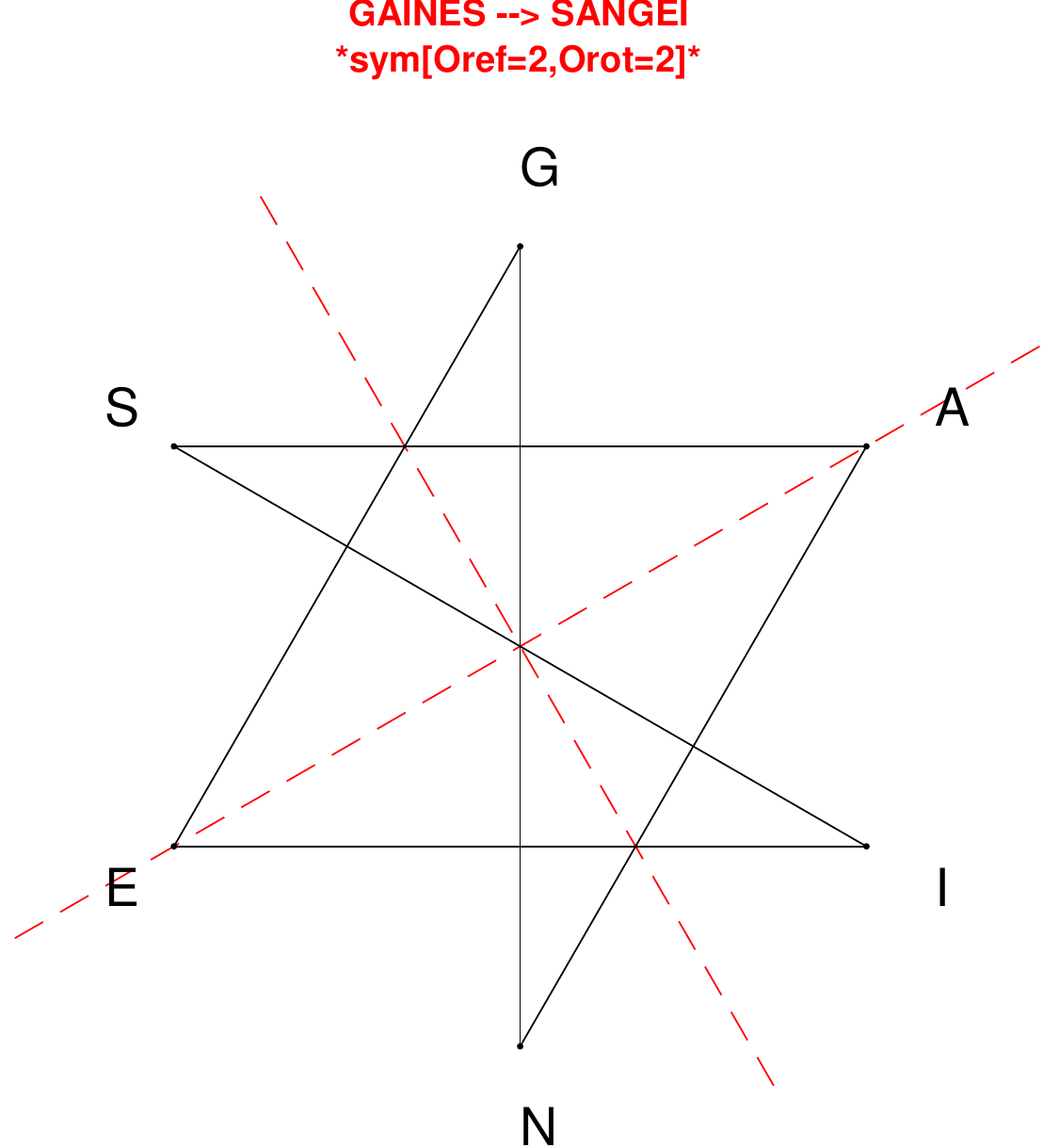}
\end{subfigure}
\hfill
\begin{subfigure}[T]{0.19\textwidth}
\centering
\includegraphics[width=\textwidth]{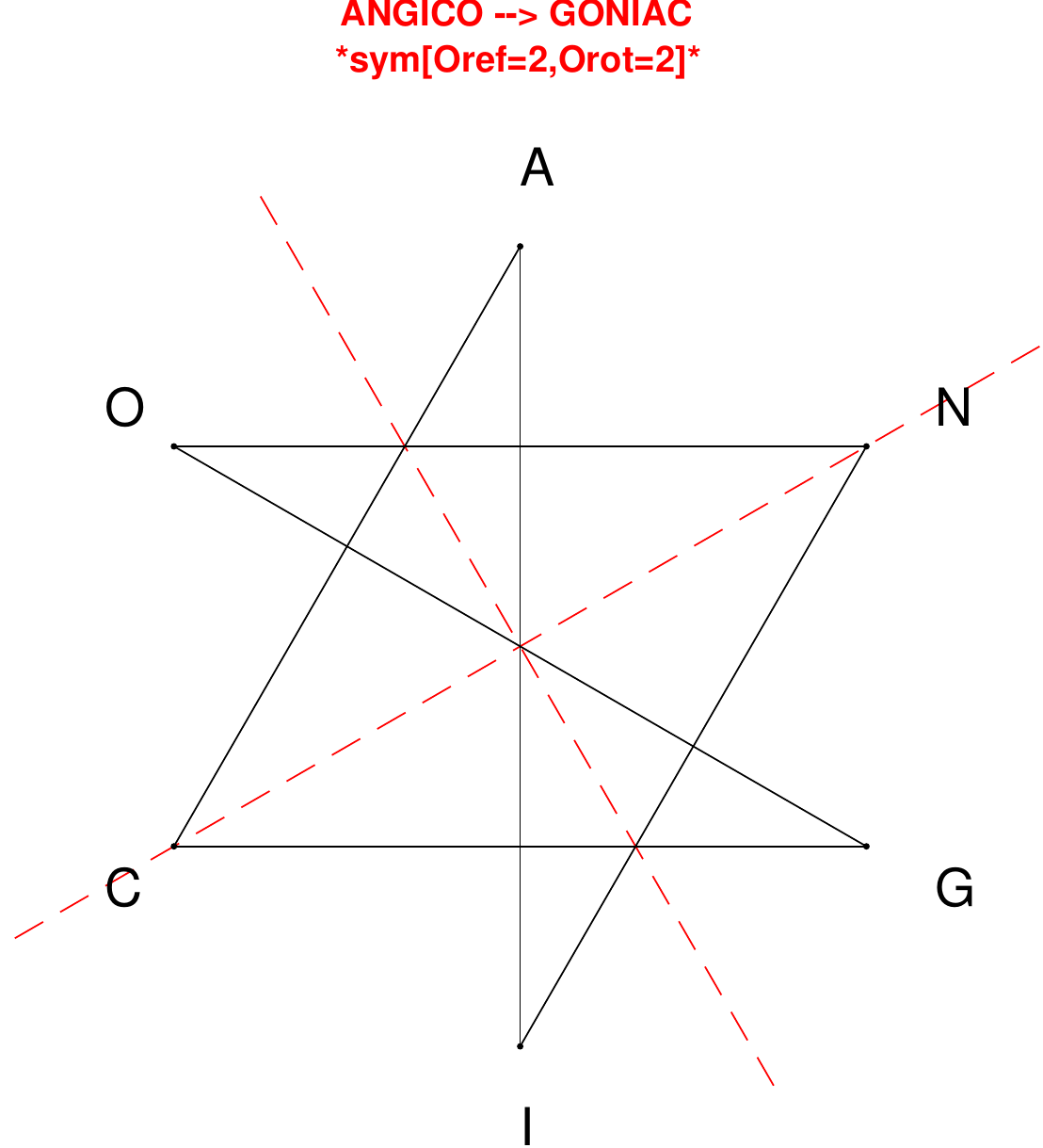}
\end{subfigure}
\hfill
\begin{subfigure}[T]{0.19\textwidth}
\centering
\includegraphics[width=\textwidth]{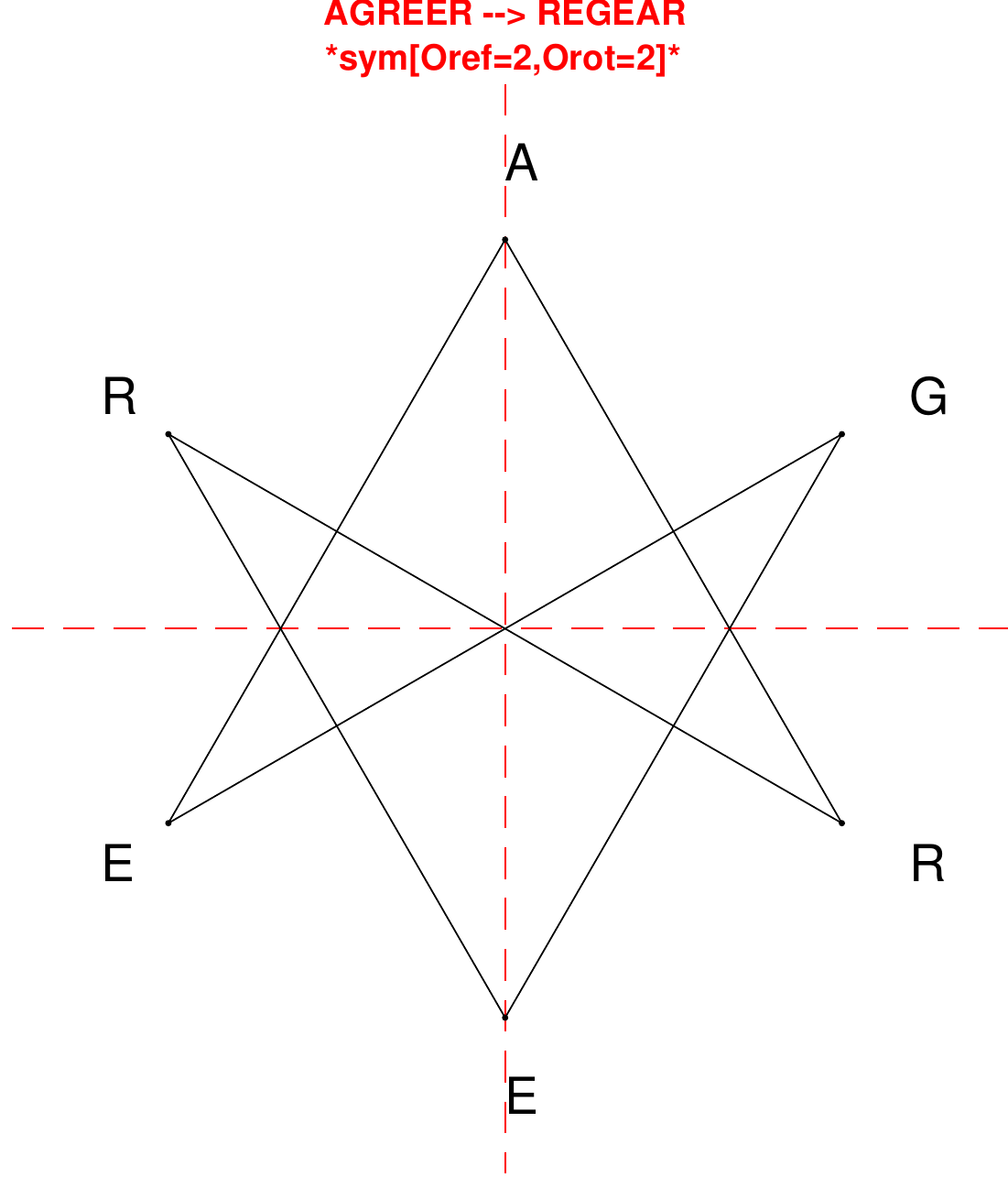}
\end{subfigure}
\end{figure}

\begin{figure}[H]
\centering
\begin{subfigure}[T]{0.19\textwidth}
\centering
\includegraphics[width=\textwidth]{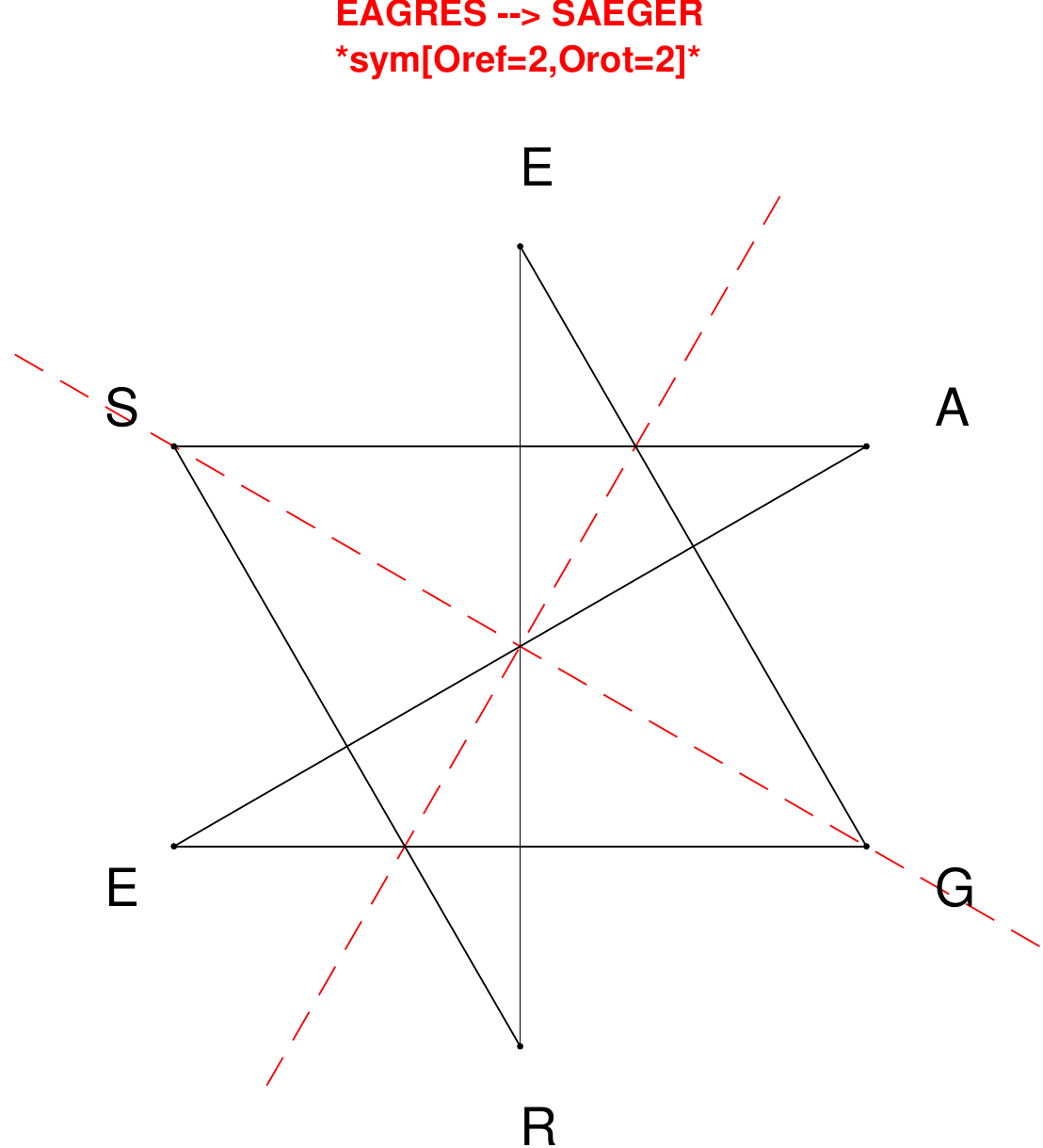}
\end{subfigure}
\hfill
\begin{subfigure}[T]{0.19\textwidth}
\centering
\includegraphics[width=\textwidth]{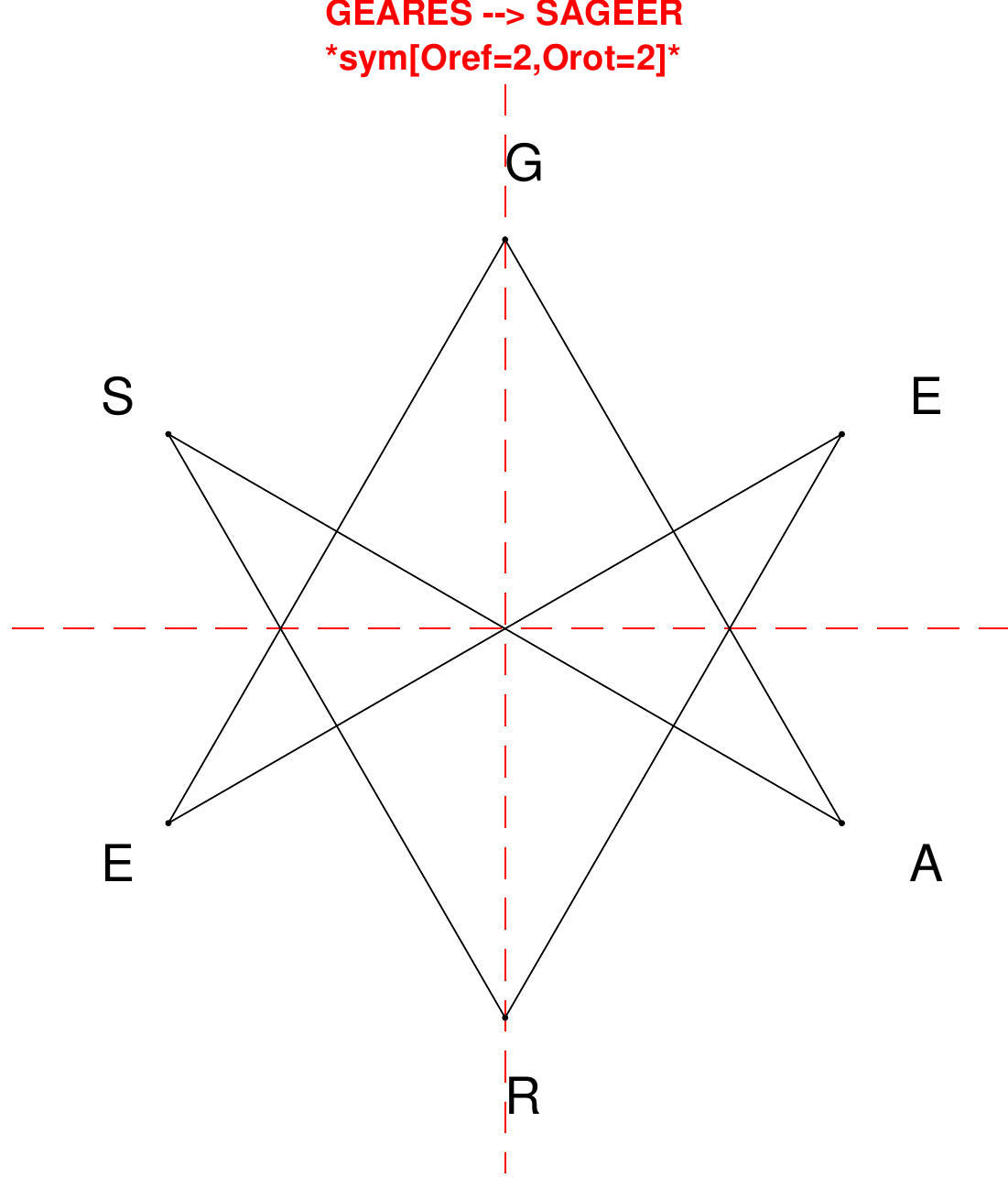}
\end{subfigure}
\hfill
\begin{subfigure}[T]{0.19\textwidth}
\centering
\includegraphics[width=\textwidth]{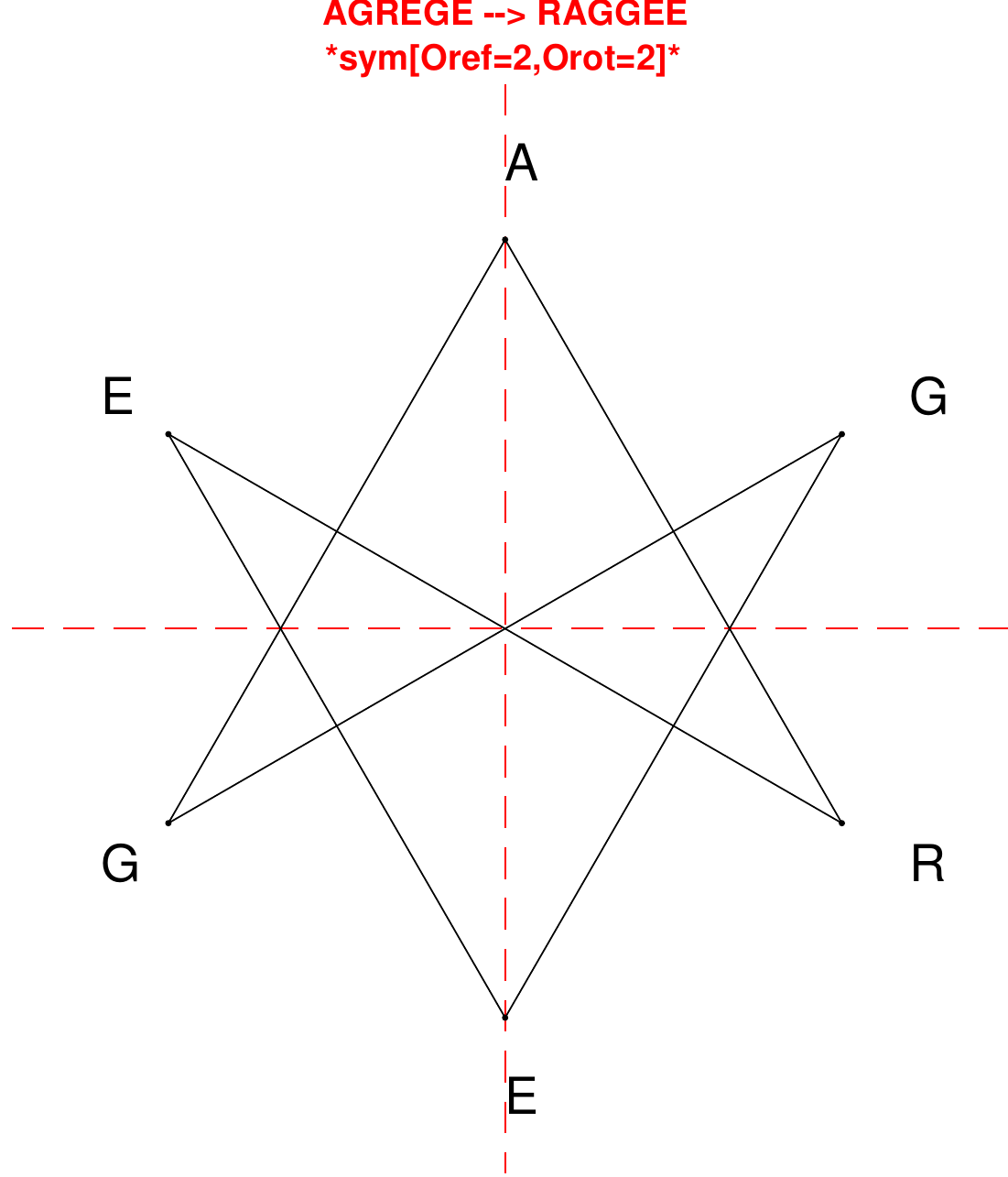}
\end{subfigure}
\hfill
\begin{subfigure}[T]{0.19\textwidth}
\centering
\includegraphics[width=\textwidth]{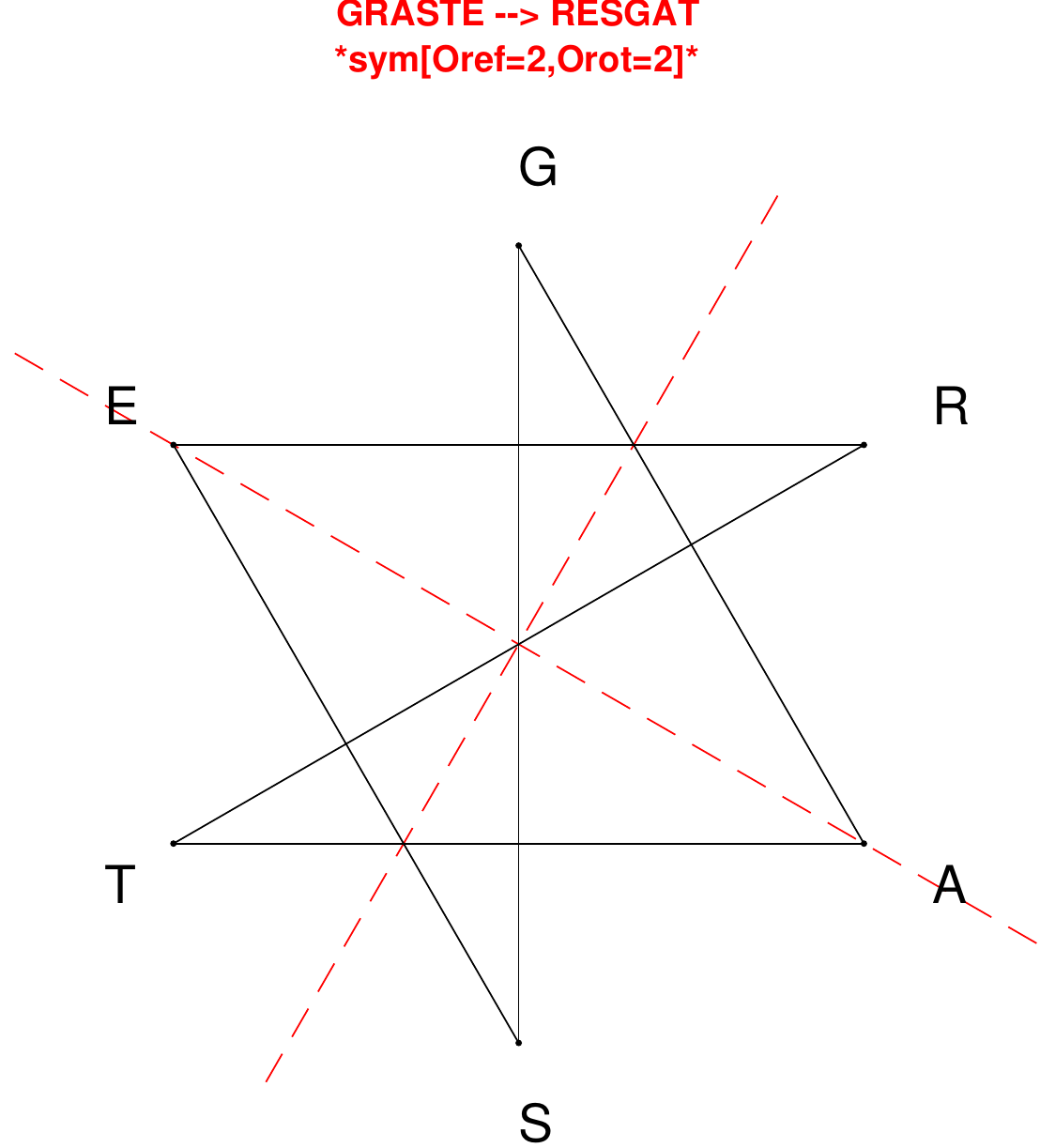}
\end{subfigure}
\hfill
\begin{subfigure}[T]{0.19\textwidth}
\centering
\includegraphics[width=\textwidth]{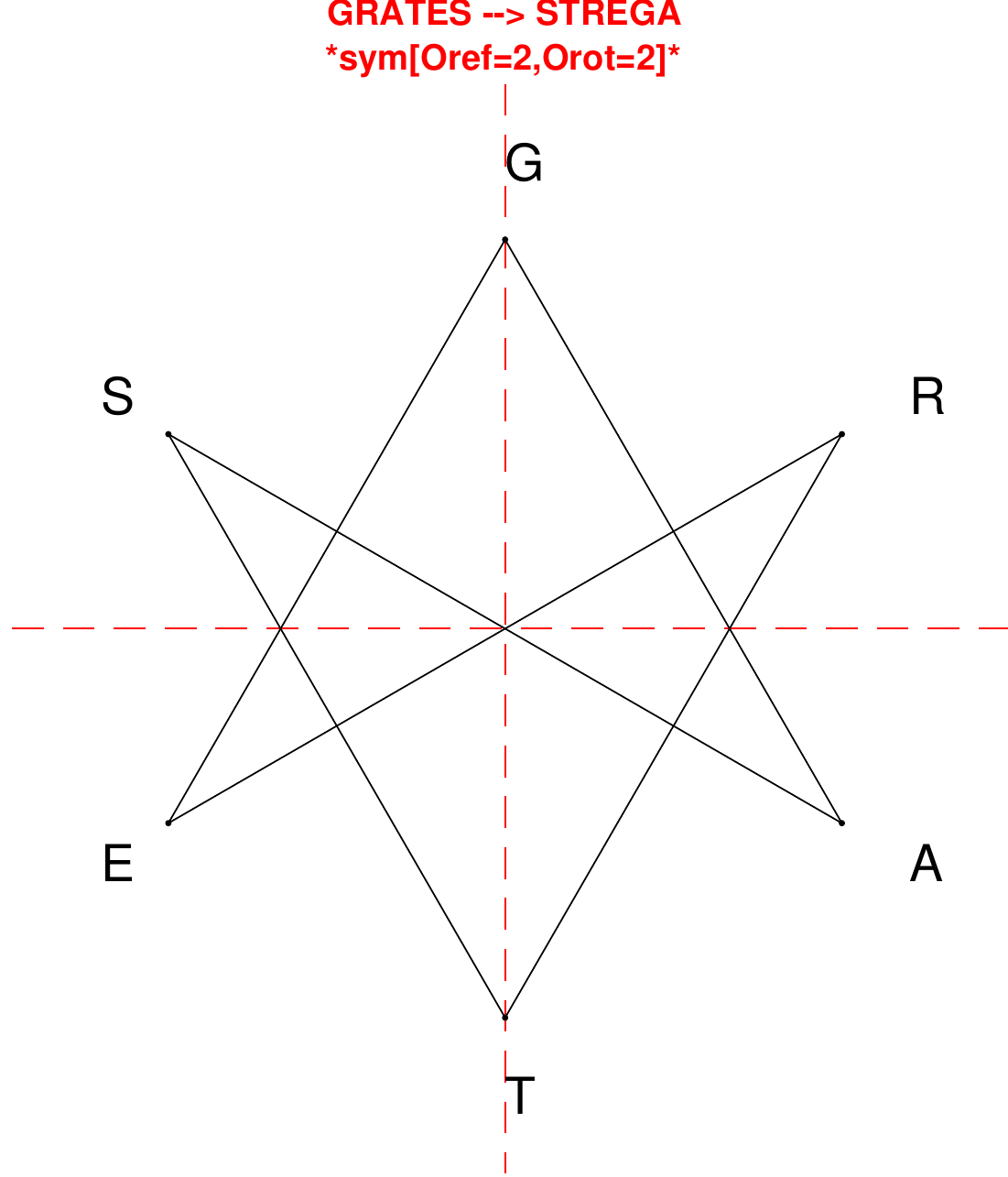}
\end{subfigure}
\end{figure}

\begin{figure}[H]
\centering
\begin{subfigure}[T]{0.19\textwidth}
\centering
\includegraphics[width=\textwidth]{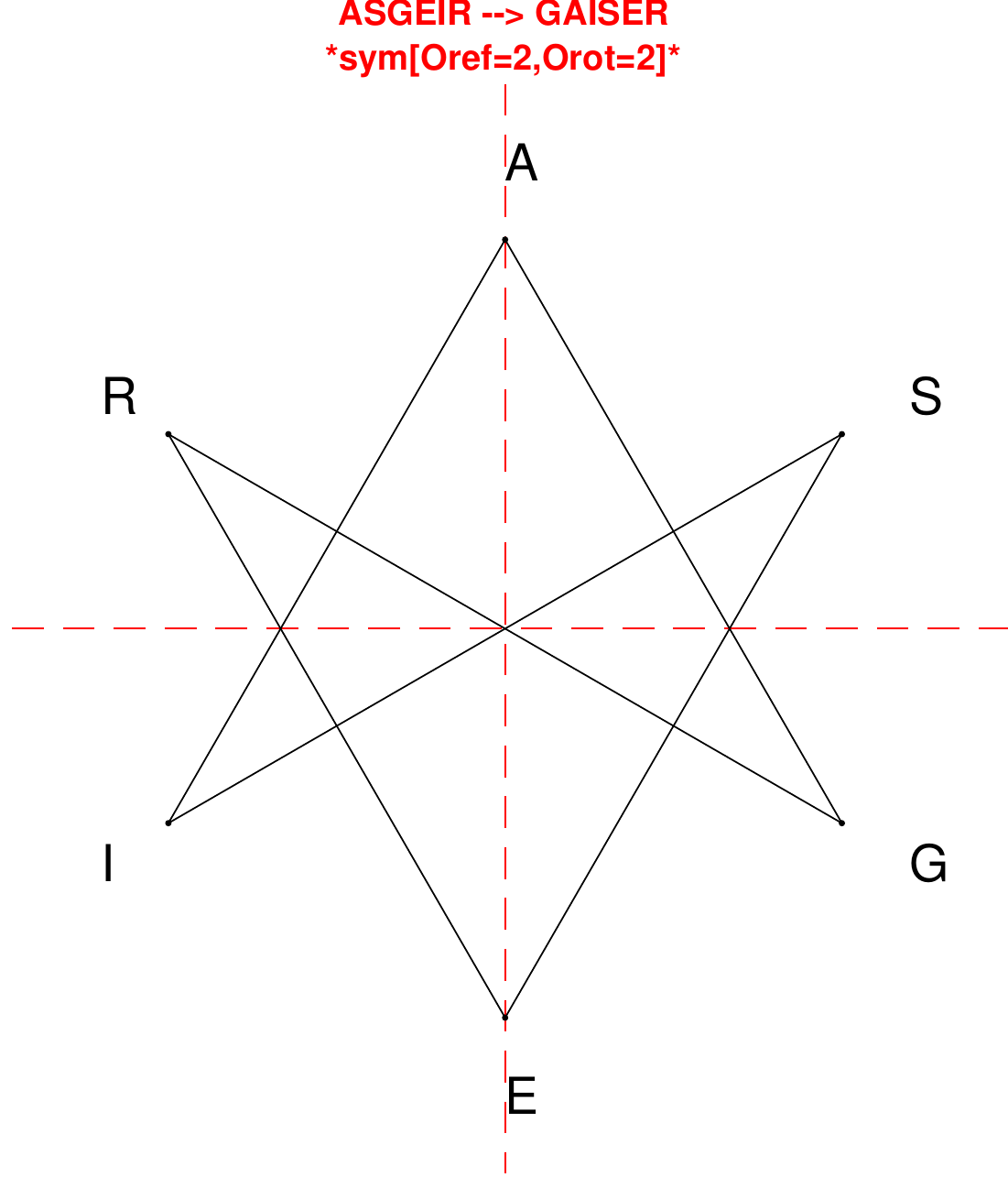}
\end{subfigure}
\hfill
\begin{subfigure}[T]{0.19\textwidth}
\centering
\includegraphics[width=\textwidth]{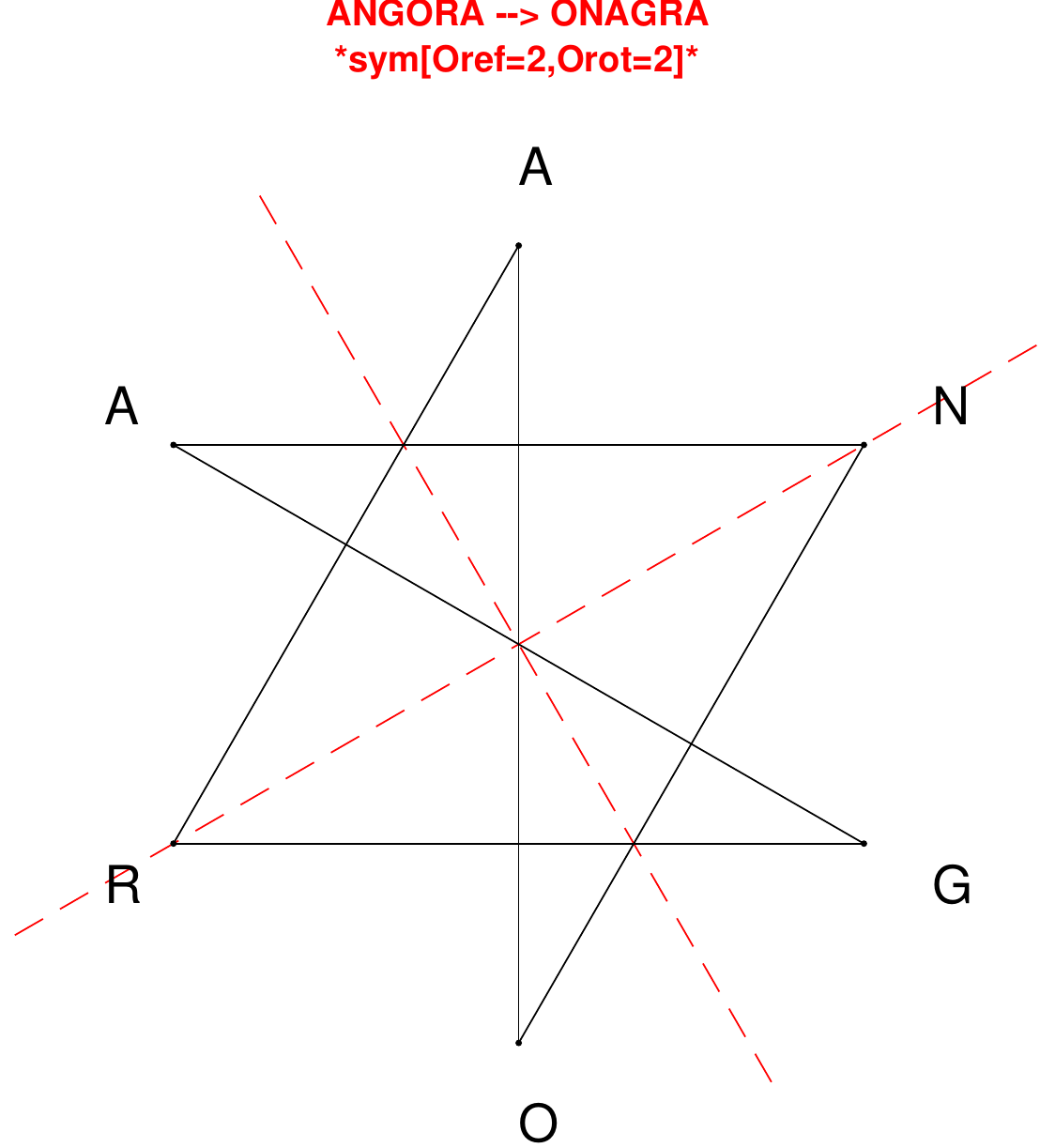}
\end{subfigure}
\hfill
\begin{subfigure}[T]{0.19\textwidth}
\centering
\includegraphics[width=\textwidth]{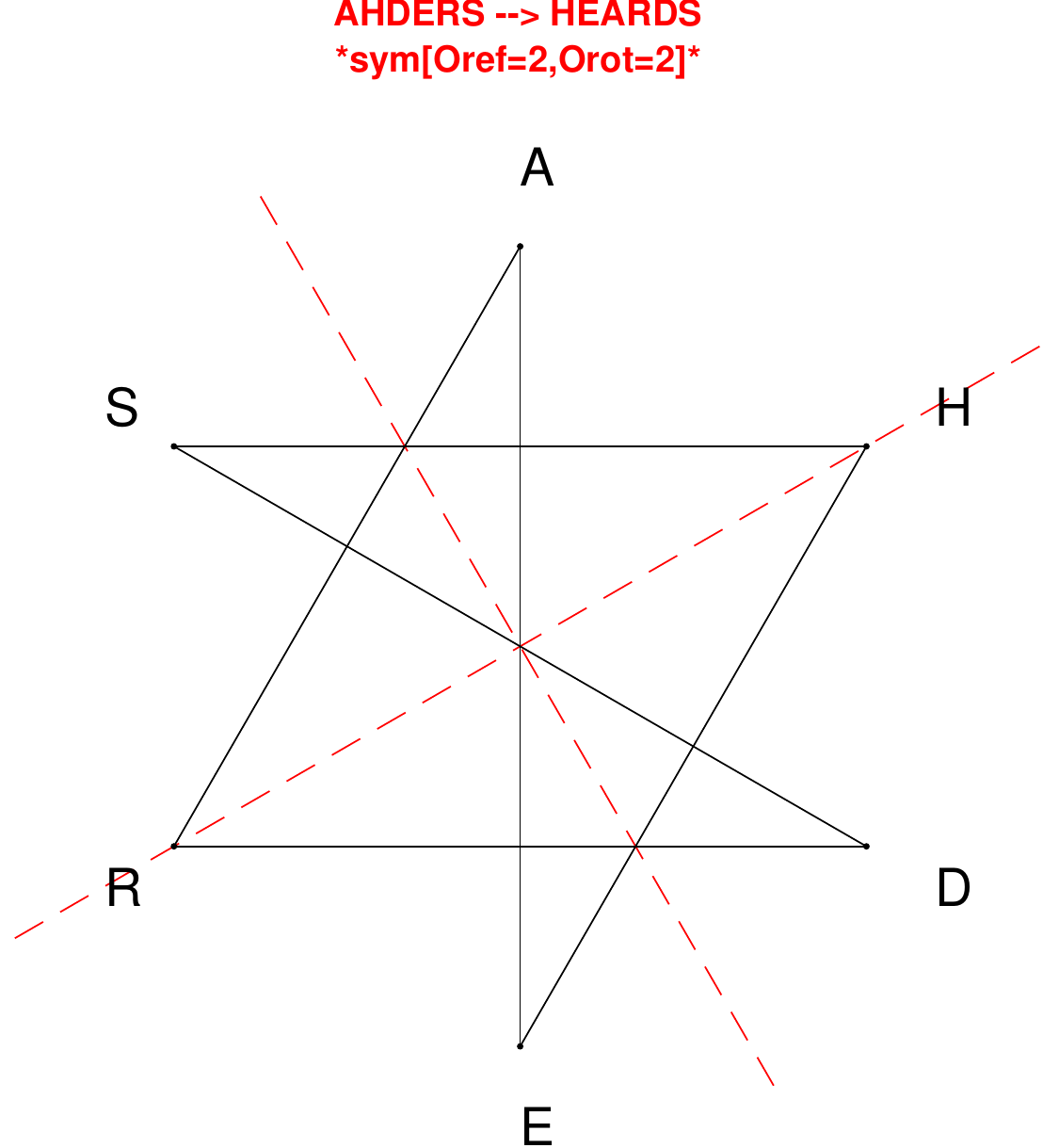}
\end{subfigure}
\hfill
\begin{subfigure}[T]{0.19\textwidth}
\centering
\includegraphics[width=\textwidth]{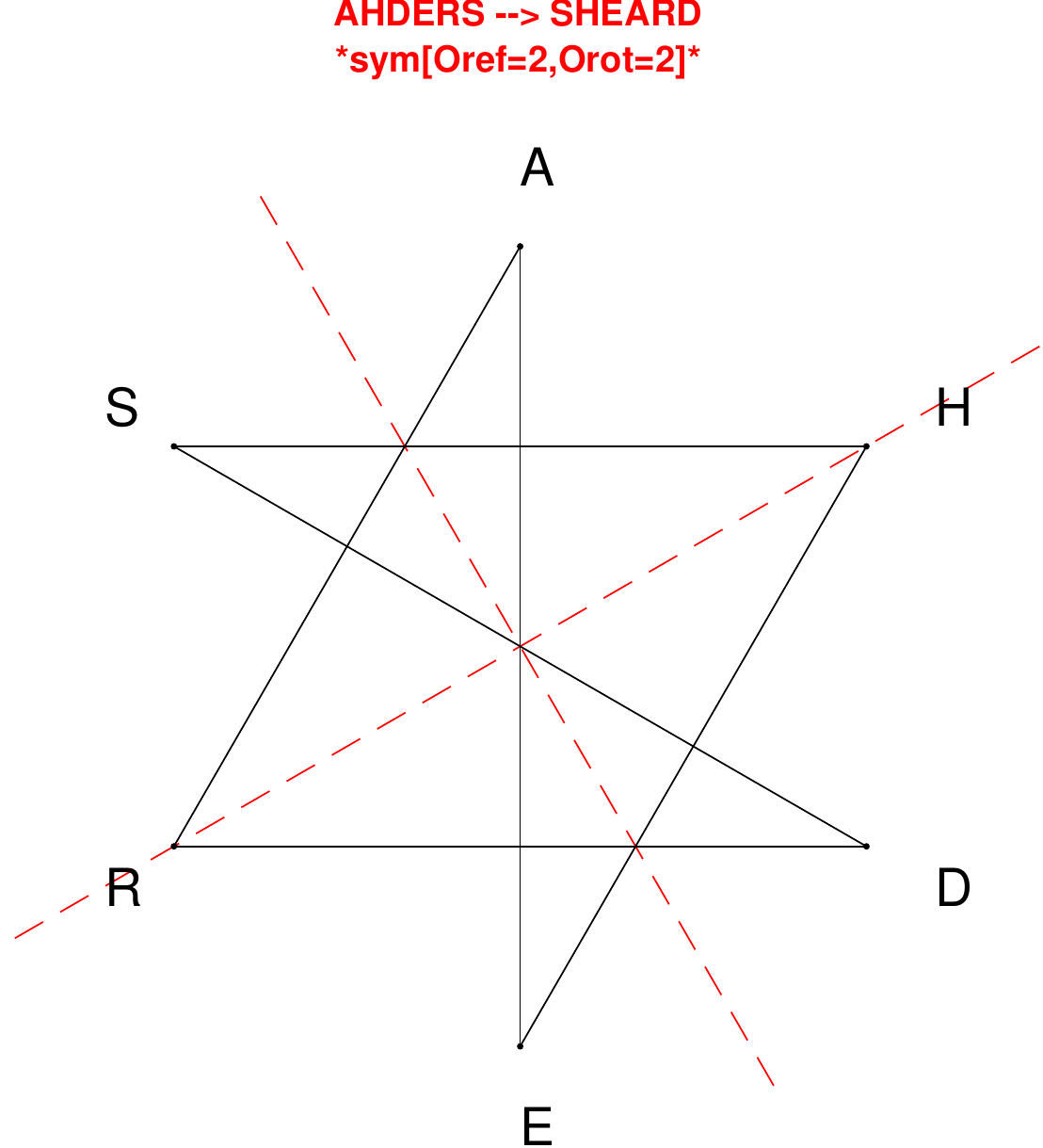}
\end{subfigure}
\hfill
\begin{subfigure}[T]{0.19\textwidth}
\centering
\includegraphics[width=\textwidth]{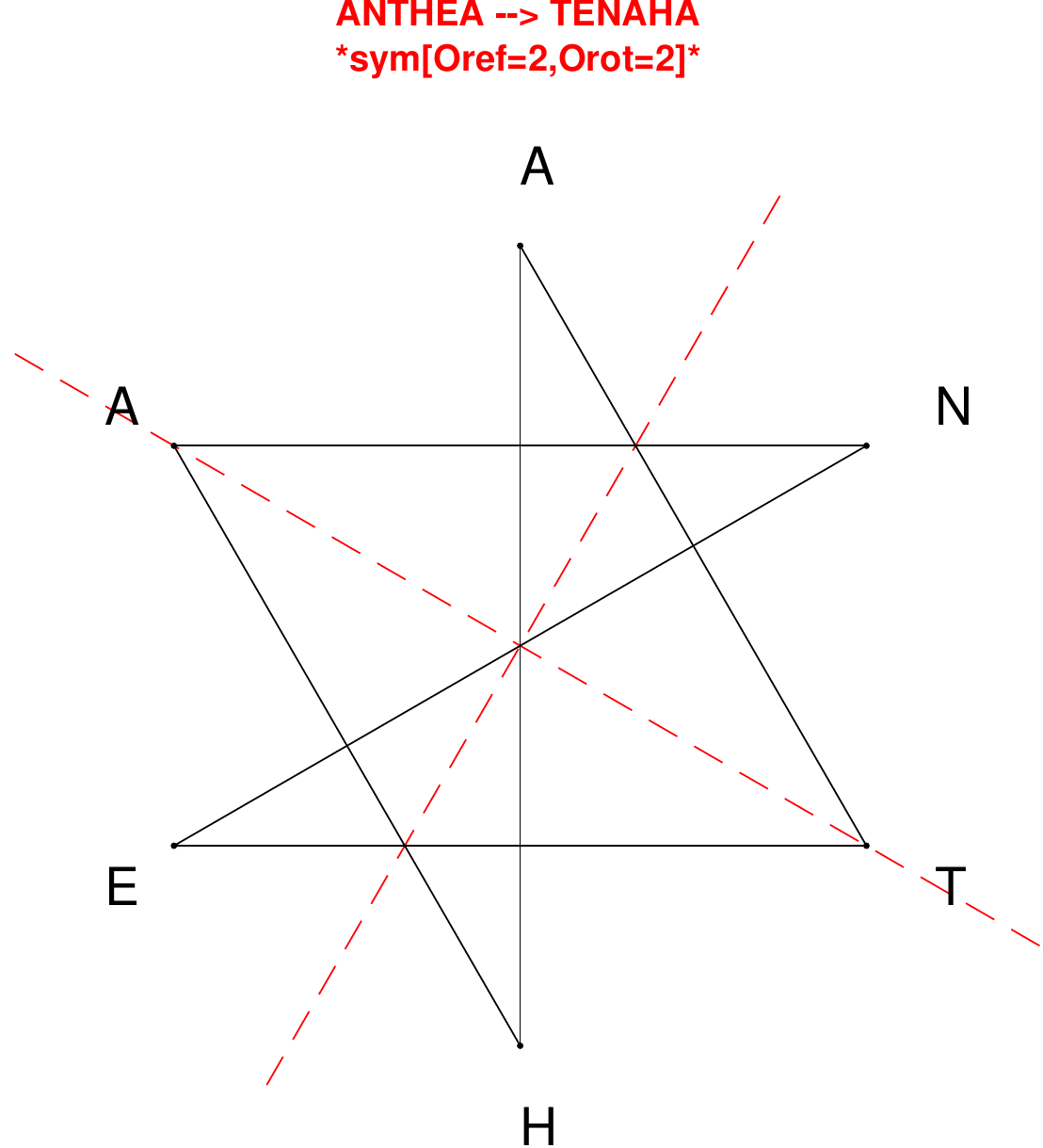}
\end{subfigure}
\end{figure}

\begin{figure}[H]
\centering
\begin{subfigure}[T]{0.19\textwidth}
\centering
\includegraphics[width=\textwidth]{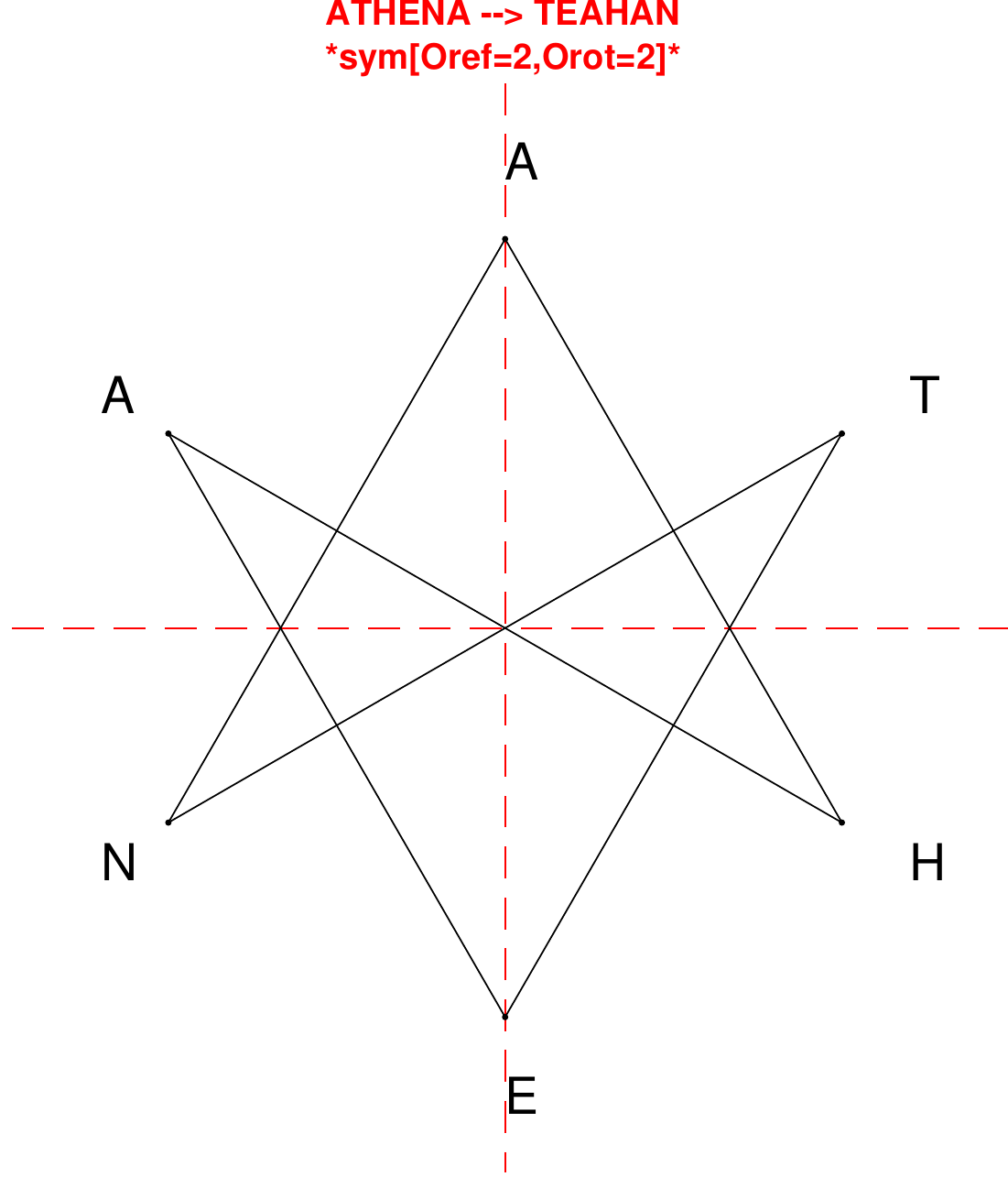}
\end{subfigure}
\hfill
\begin{subfigure}[T]{0.19\textwidth}
\centering
\includegraphics[width=\textwidth]{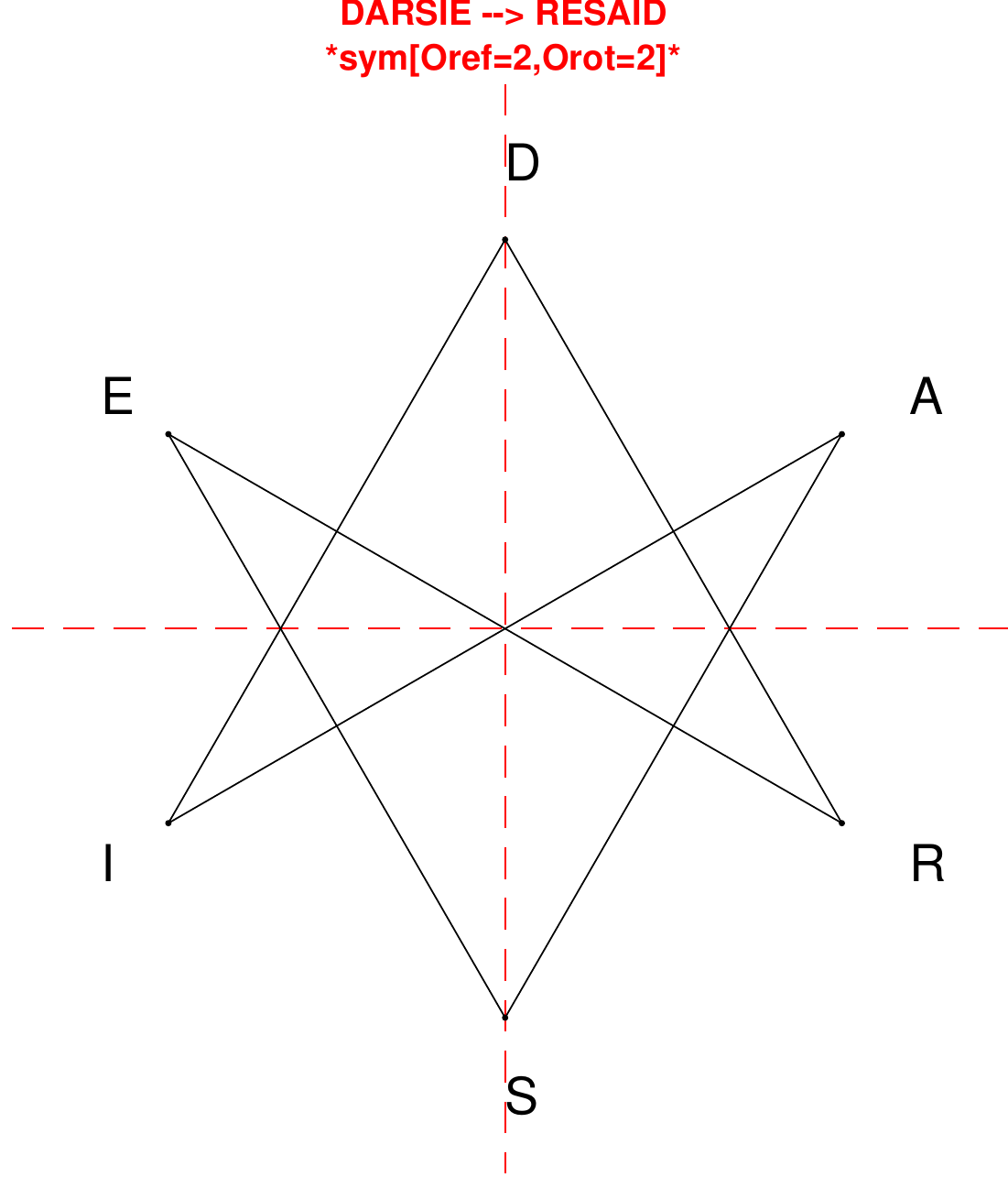}
\end{subfigure}
\hfill
\begin{subfigure}[T]{0.19\textwidth}
\centering
\includegraphics[width=\textwidth]{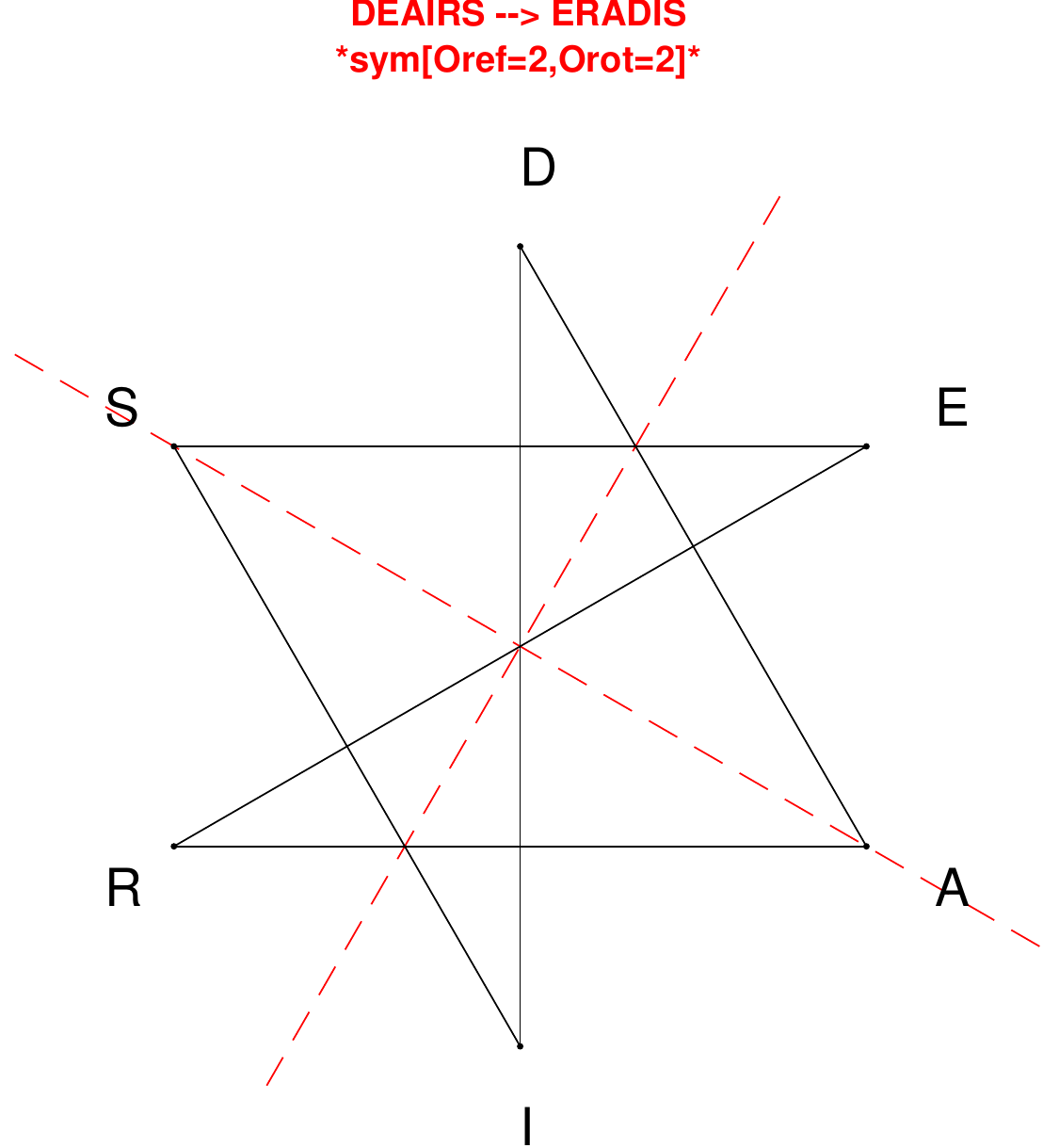}
\end{subfigure}
\hfill
\begin{subfigure}[T]{0.19\textwidth}
\centering
\includegraphics[width=\textwidth]{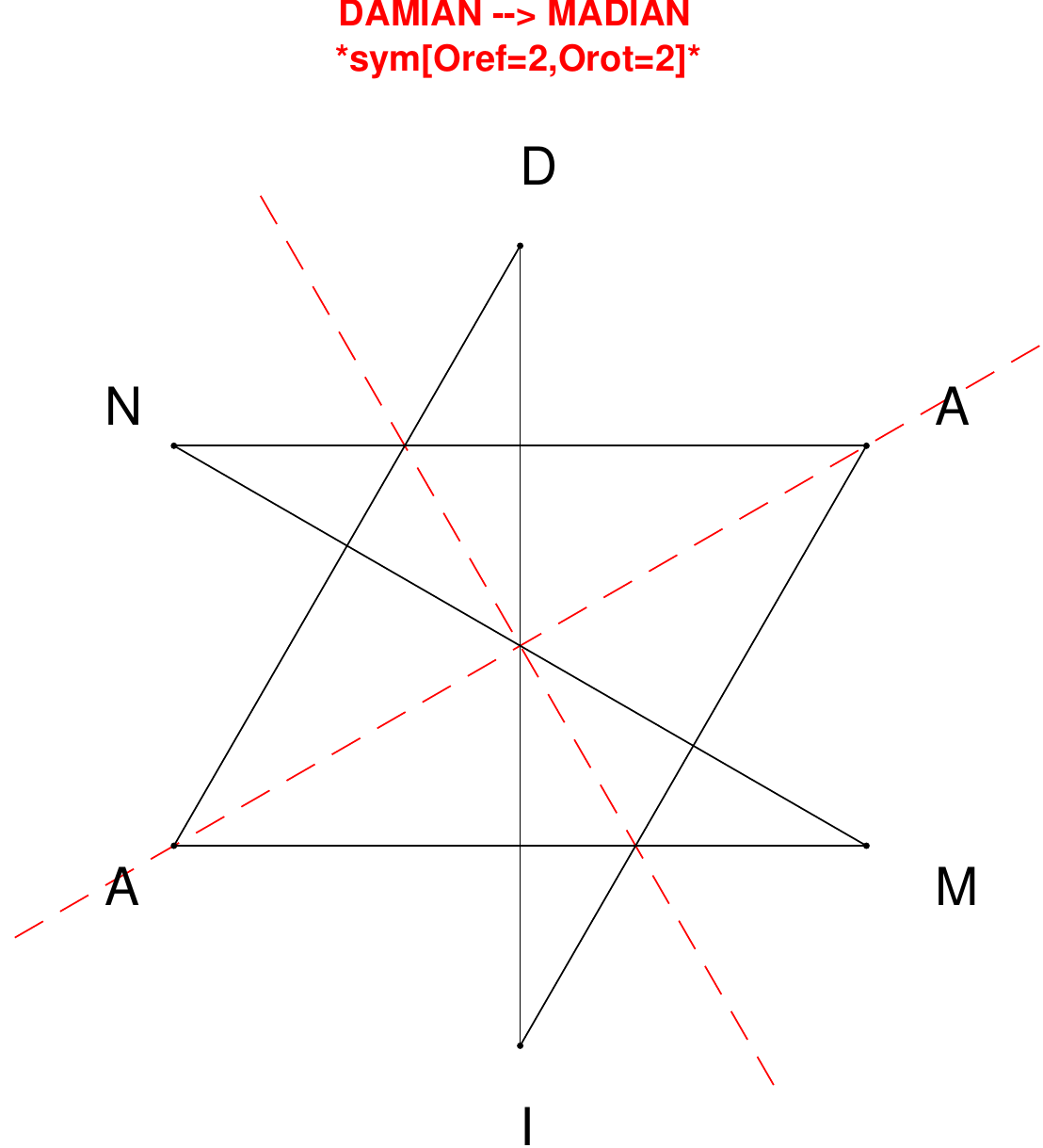}
\end{subfigure}
\hfill
\begin{subfigure}[T]{0.19\textwidth}
\centering
\includegraphics[width=\textwidth]{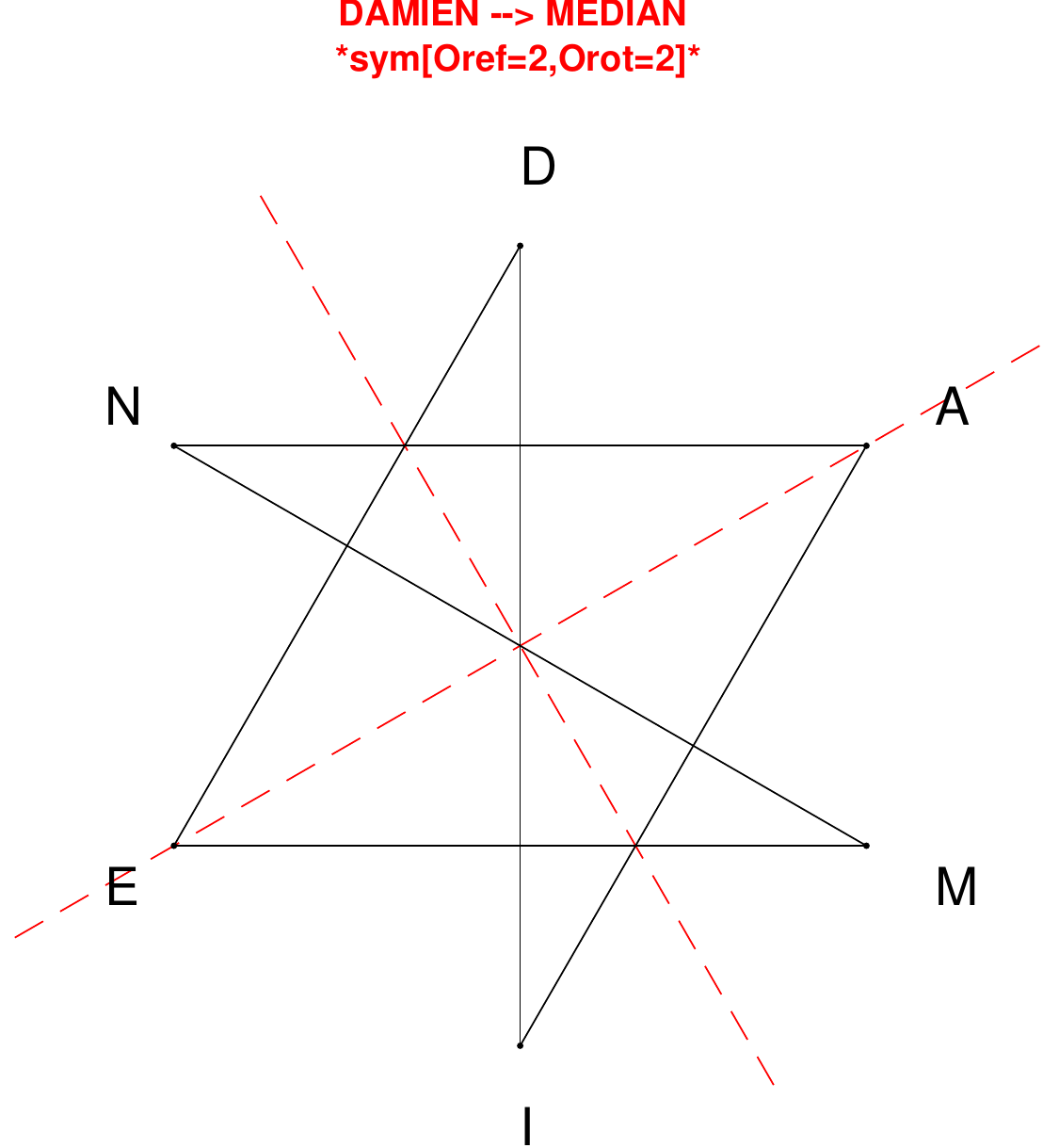}
\end{subfigure}
\end{figure}

\begin{figure}[H]
\centering
\begin{subfigure}[T]{0.19\textwidth}
\centering
\includegraphics[width=\textwidth]{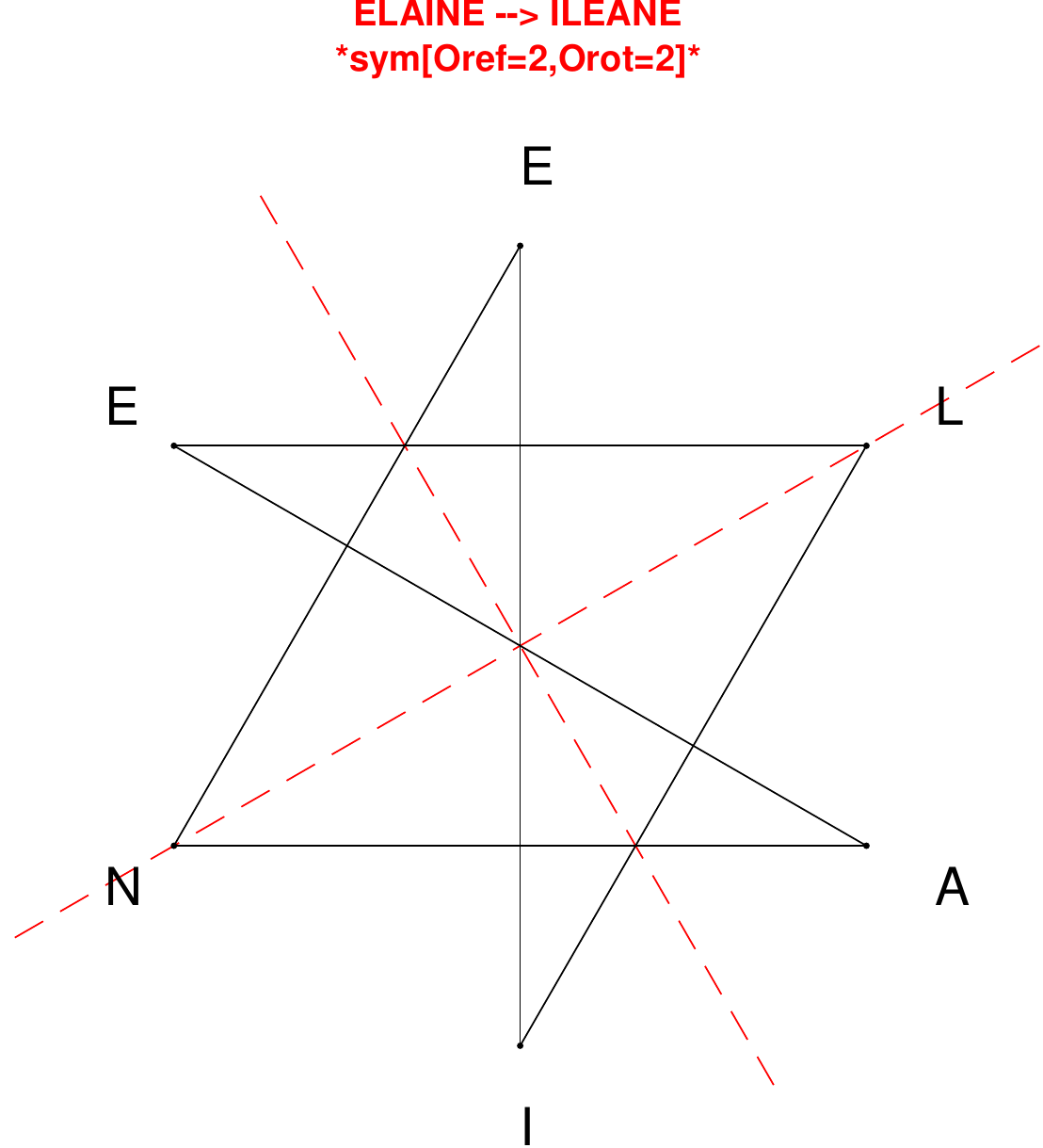}
\end{subfigure}
\hfill
\begin{subfigure}[T]{0.19\textwidth}
\centering
\includegraphics[width=\textwidth]{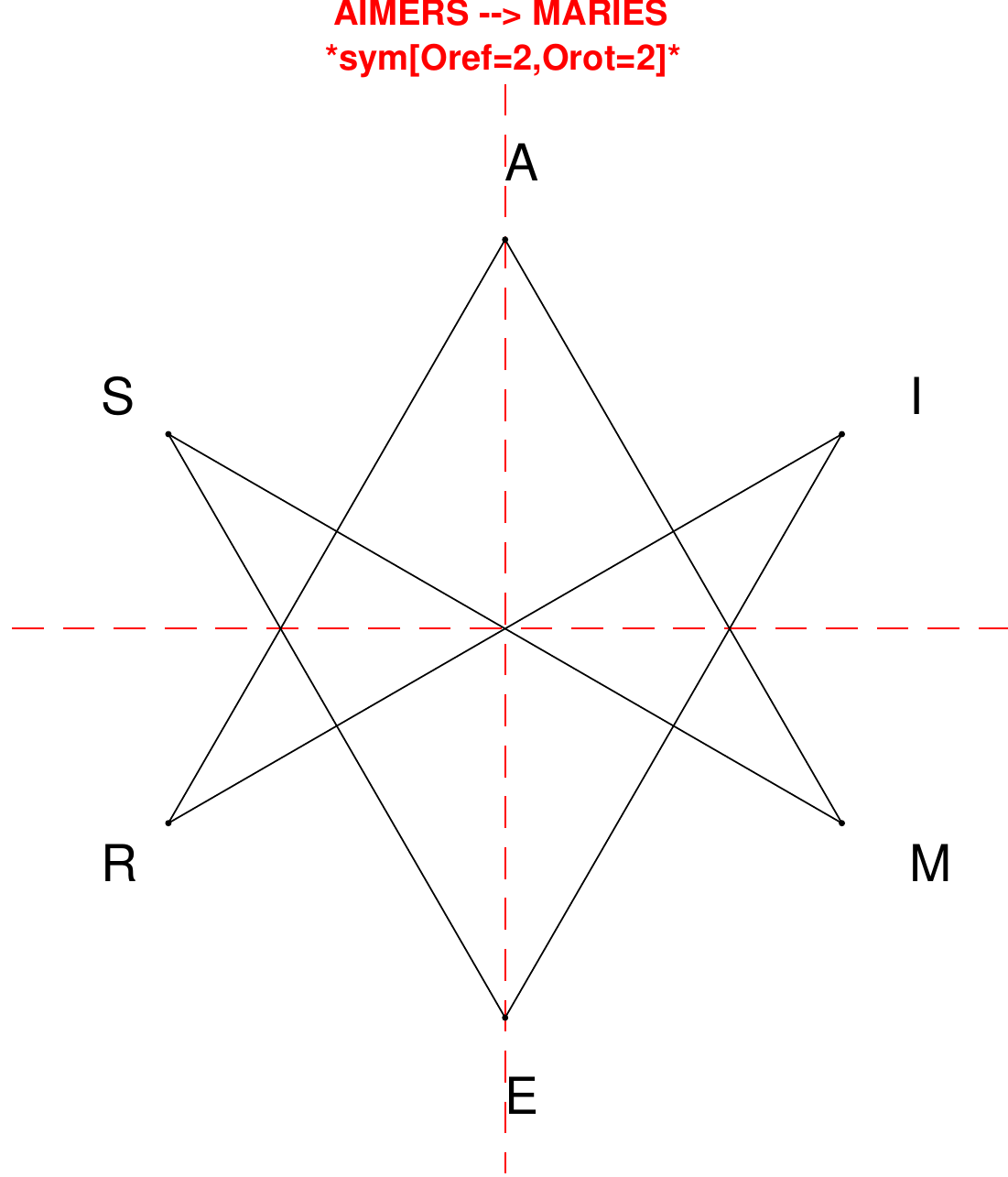}
\end{subfigure}
\hfill
\begin{subfigure}[T]{0.19\textwidth}
\centering
\includegraphics[width=\textwidth]{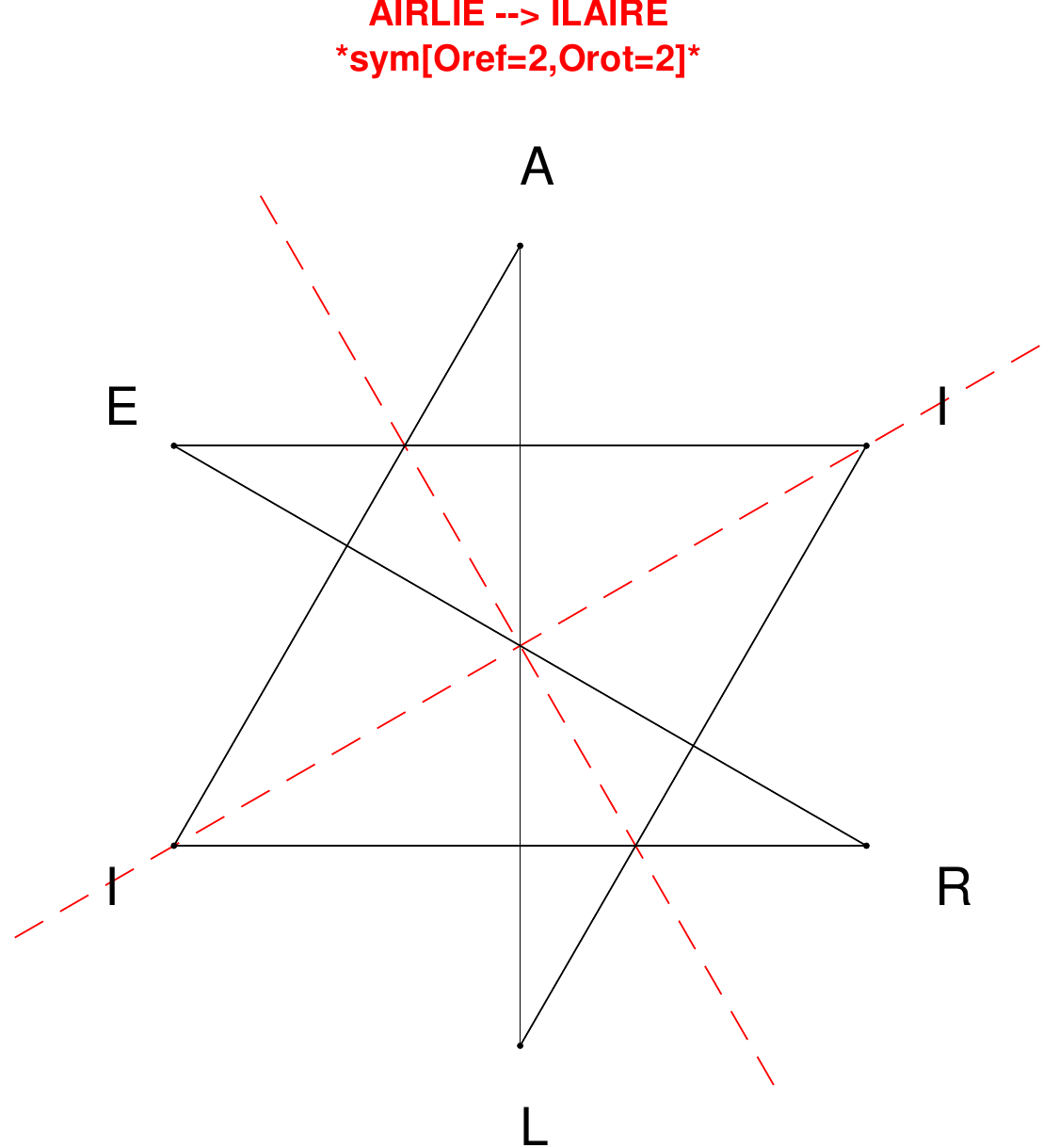}
\end{subfigure}
\hfill
\begin{subfigure}[T]{0.19\textwidth}
\centering
\includegraphics[width=\textwidth]{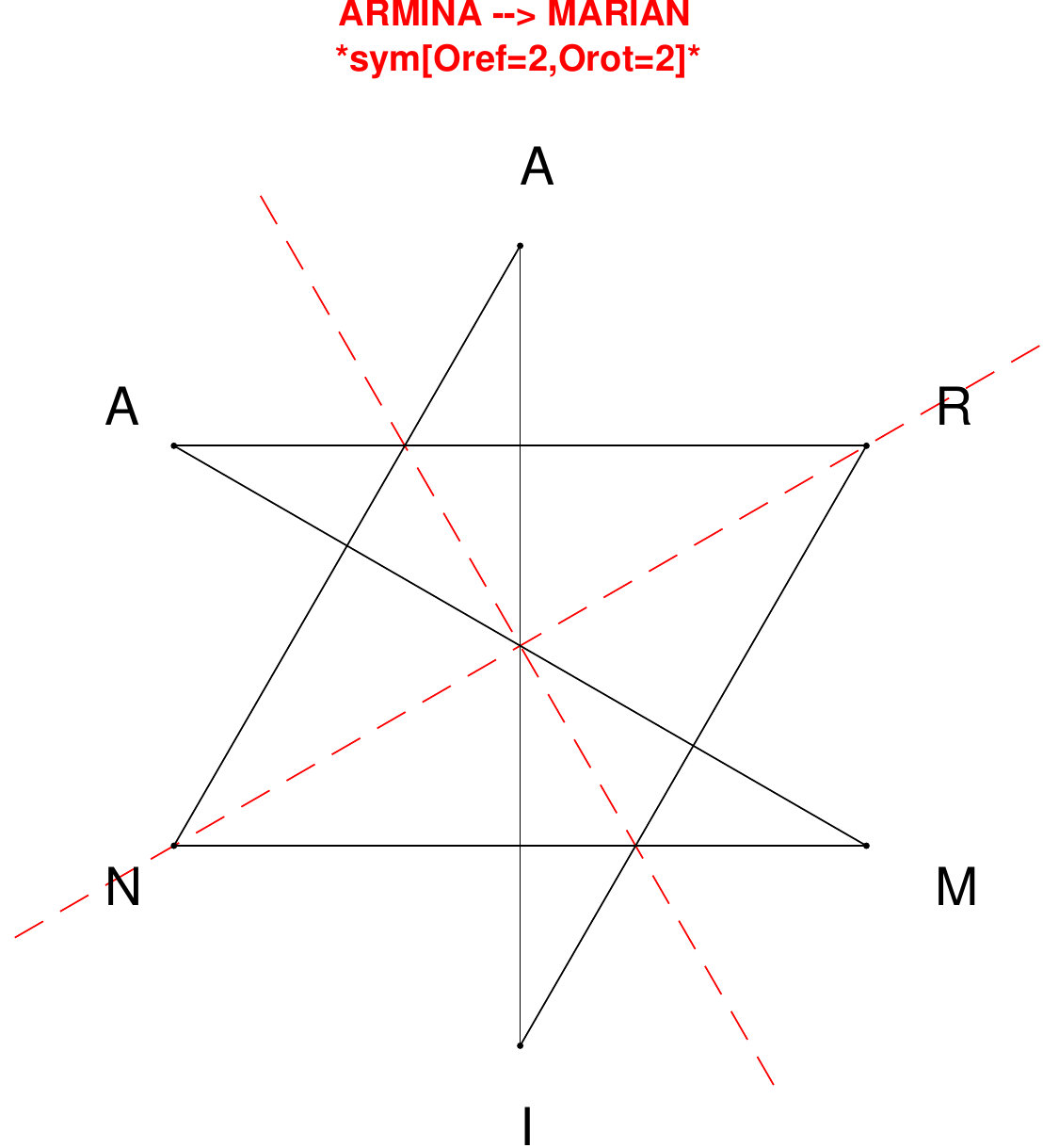}
\end{subfigure}
\hfill
\begin{subfigure}[T]{0.19\textwidth}
\centering
\includegraphics[width=\textwidth]{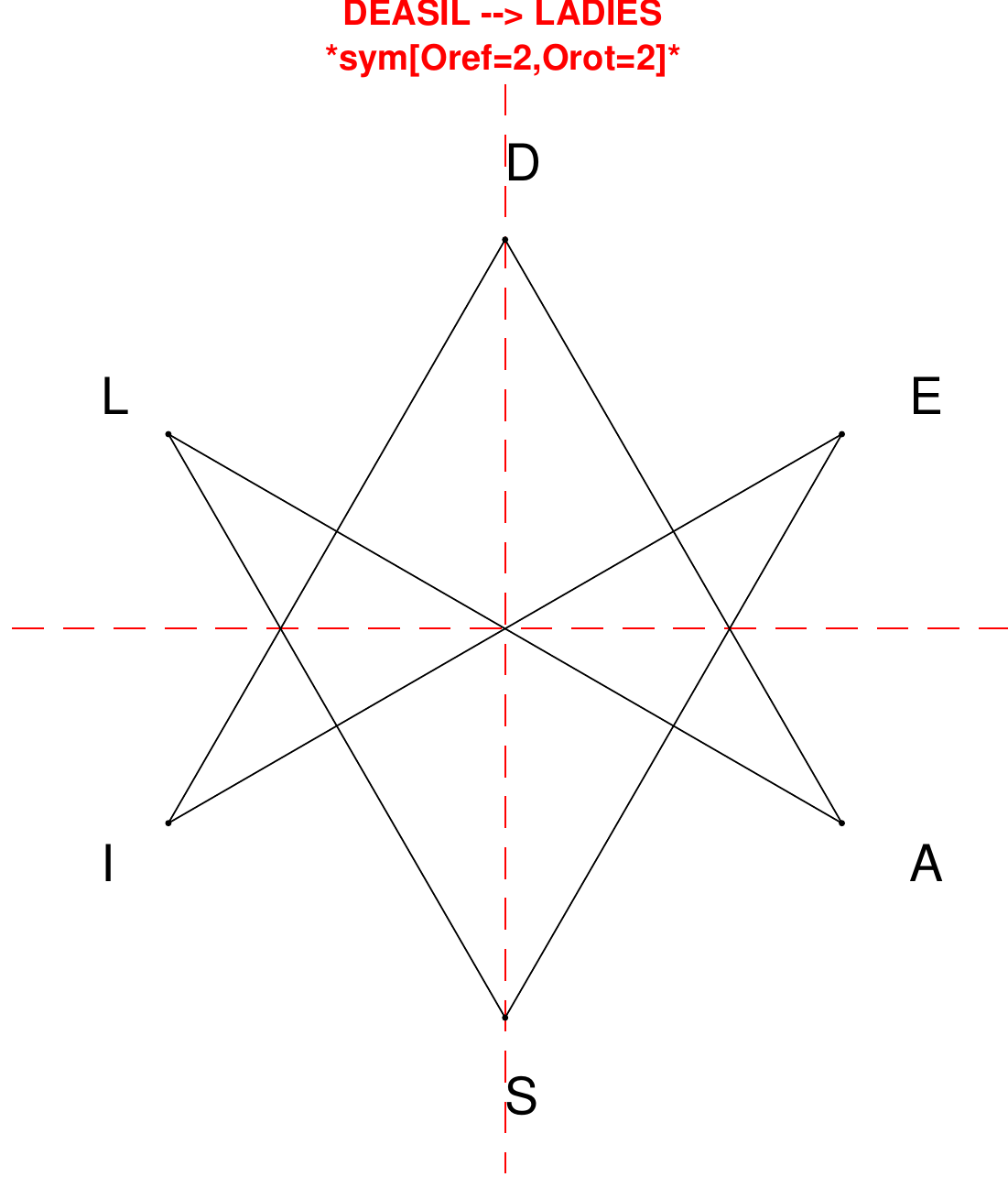}
\end{subfigure}
\end{figure}

\begin{figure}[H]
\centering
\begin{subfigure}[T]{0.19\textwidth}
\centering
\includegraphics[width=\textwidth]{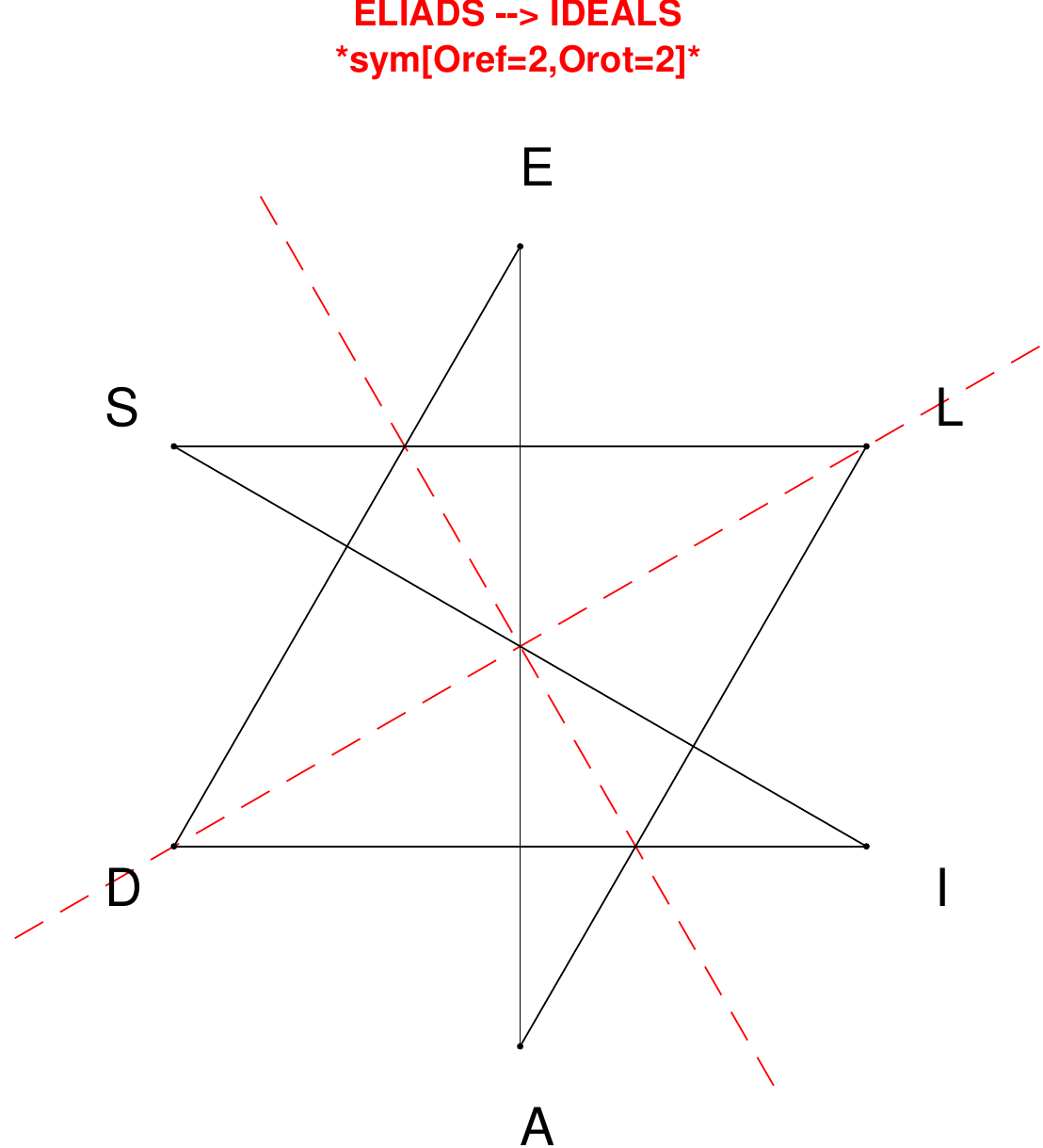}
\end{subfigure}
\hfill
\begin{subfigure}[T]{0.19\textwidth}
\centering
\includegraphics[width=\textwidth]{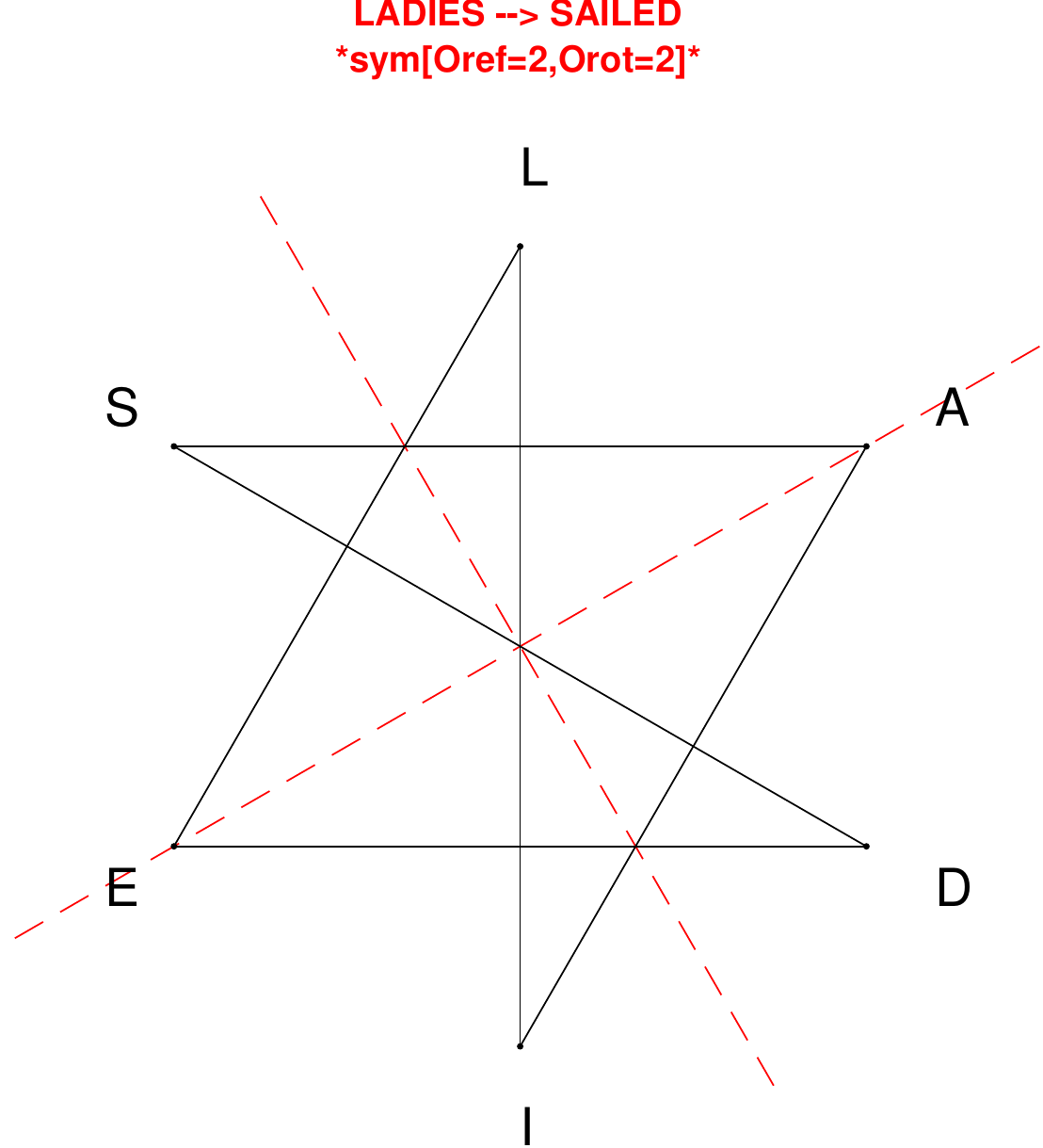}
\end{subfigure}
\hfill
\begin{subfigure}[T]{0.19\textwidth}
\centering
\includegraphics[width=\textwidth]{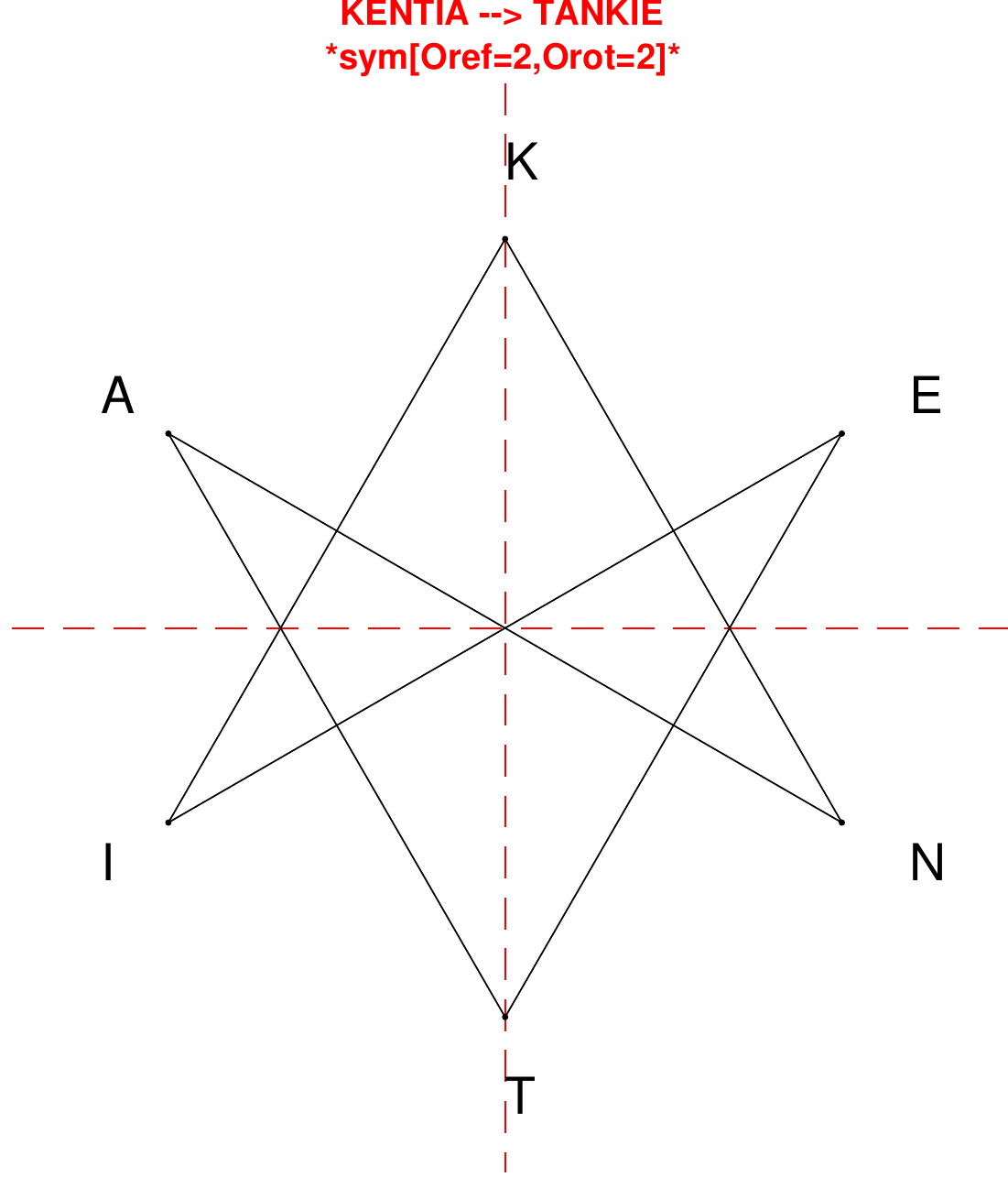}
\end{subfigure}
\hfill
\begin{subfigure}[T]{0.19\textwidth}
\centering
\includegraphics[width=\textwidth]{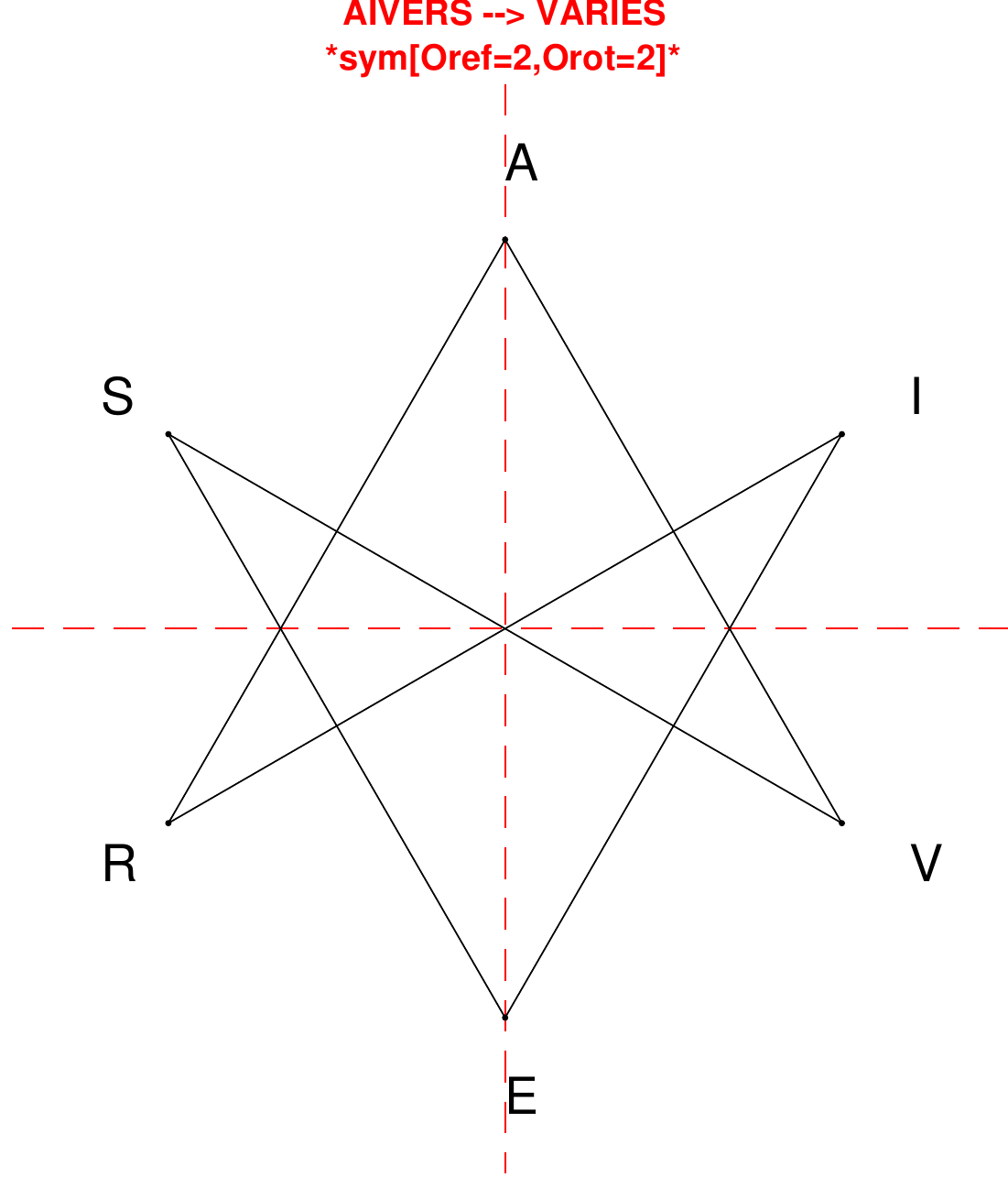}
\end{subfigure}
\hfill
\begin{subfigure}[T]{0.19\textwidth}
\centering
\includegraphics[width=\textwidth]{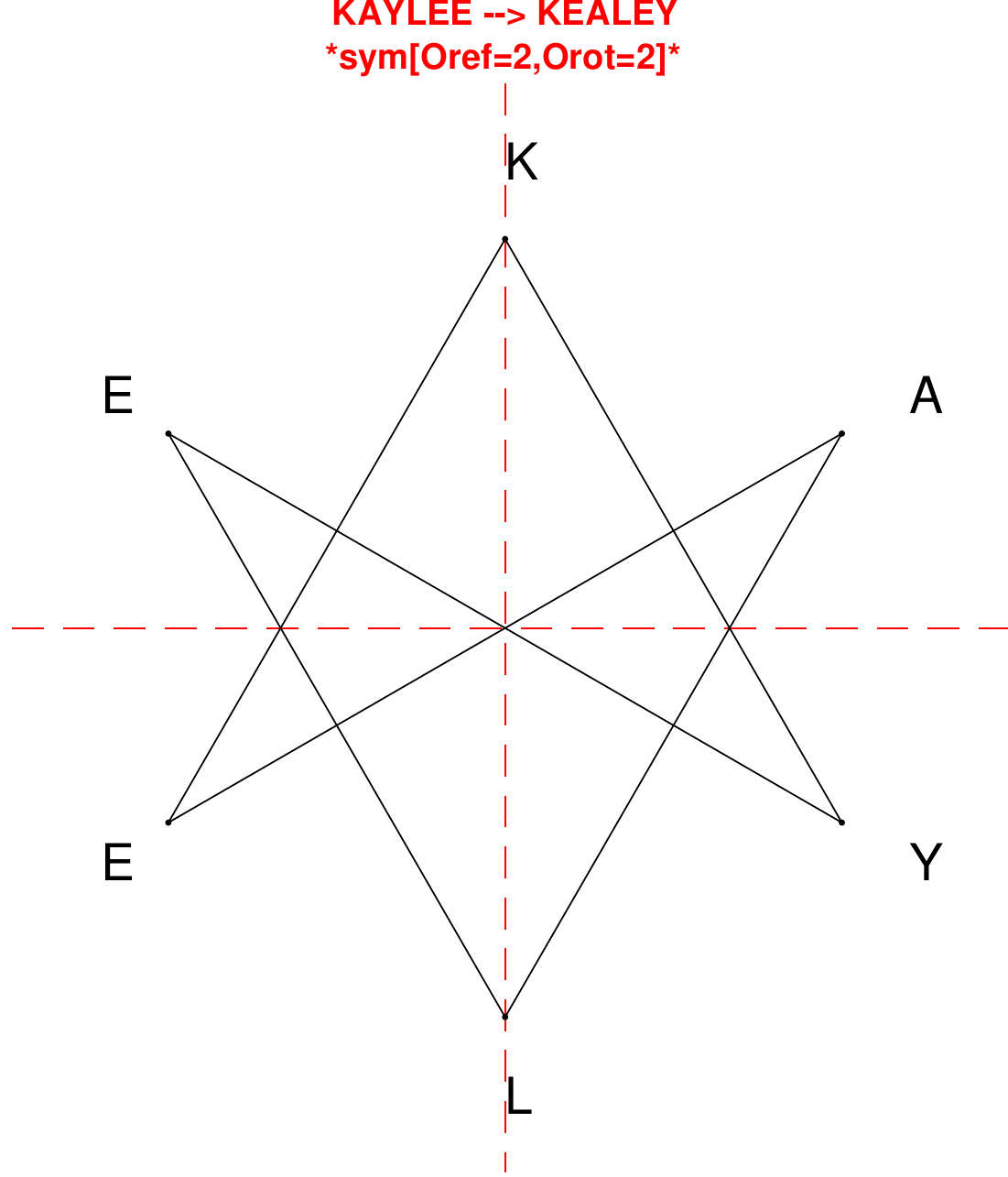}
\end{subfigure}
\end{figure}

\begin{figure}[H]
\centering
\begin{subfigure}[T]{0.19\textwidth}
\centering
\includegraphics[width=\textwidth]{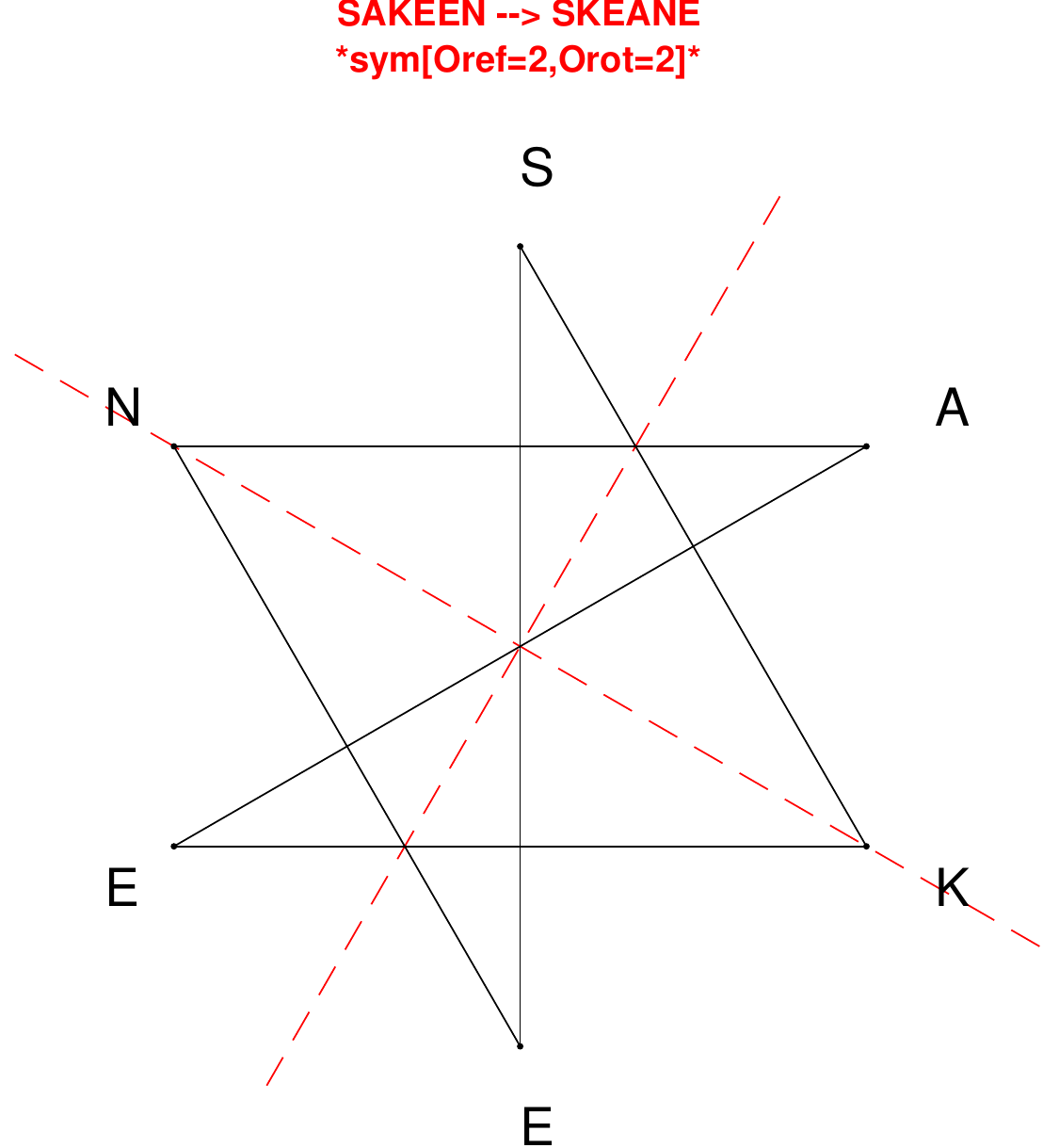}
\end{subfigure}
\hfill
\begin{subfigure}[T]{0.19\textwidth}
\centering
\includegraphics[width=\textwidth]{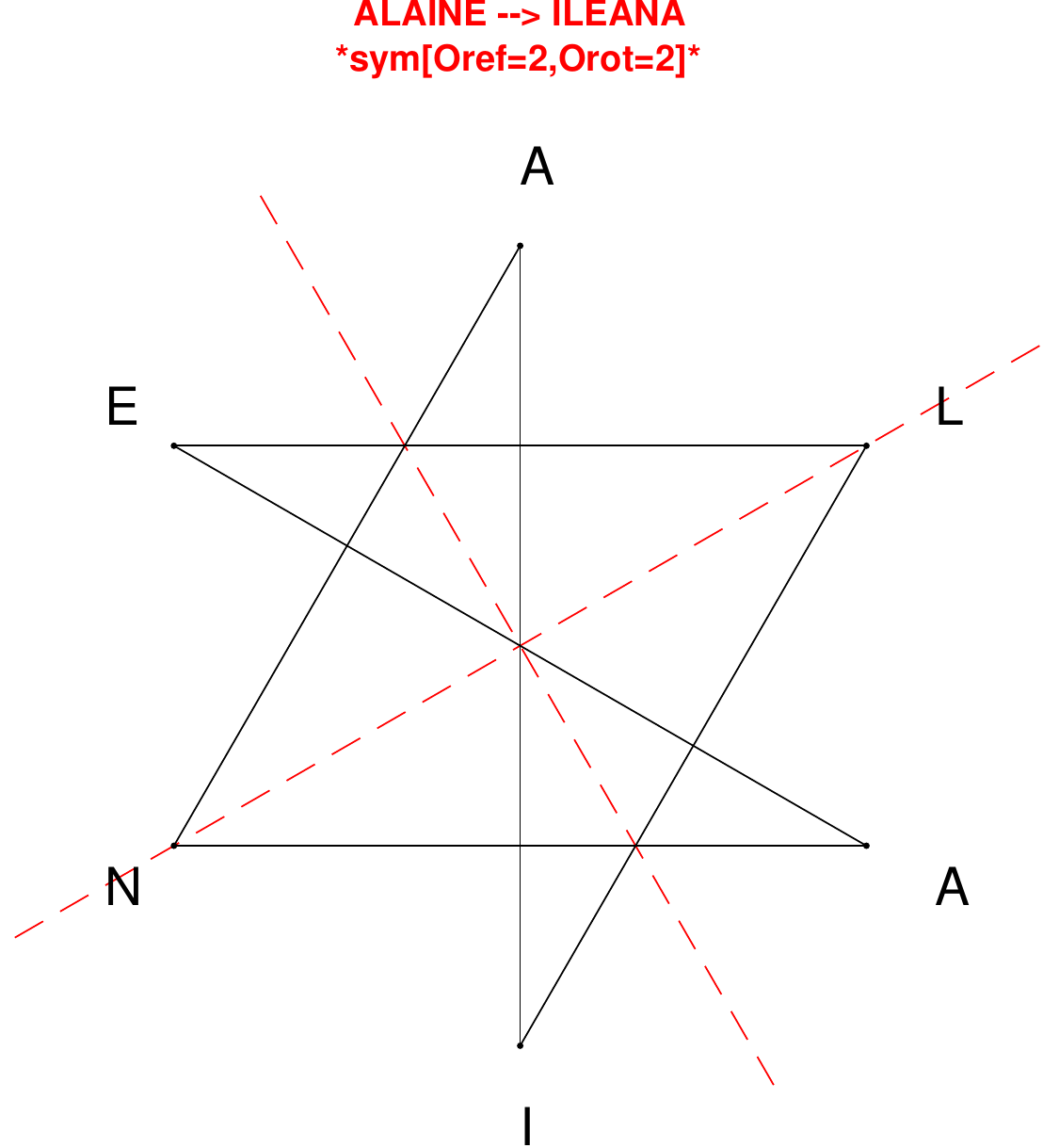}
\end{subfigure}
\hfill
\begin{subfigure}[T]{0.19\textwidth}
\centering
\includegraphics[width=\textwidth]{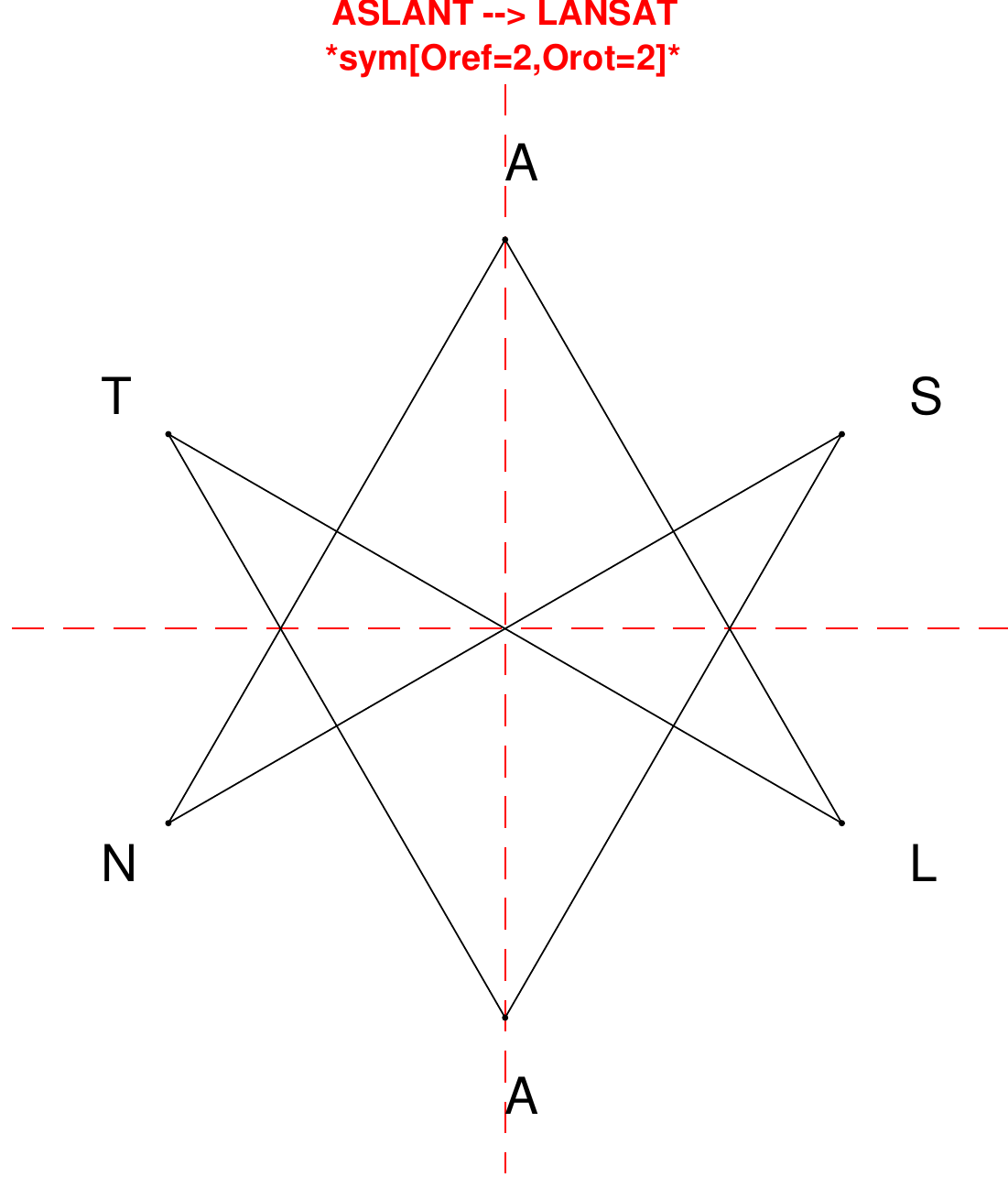}
\end{subfigure}
\hfill
\begin{subfigure}[T]{0.19\textwidth}
\centering
\includegraphics[width=\textwidth]{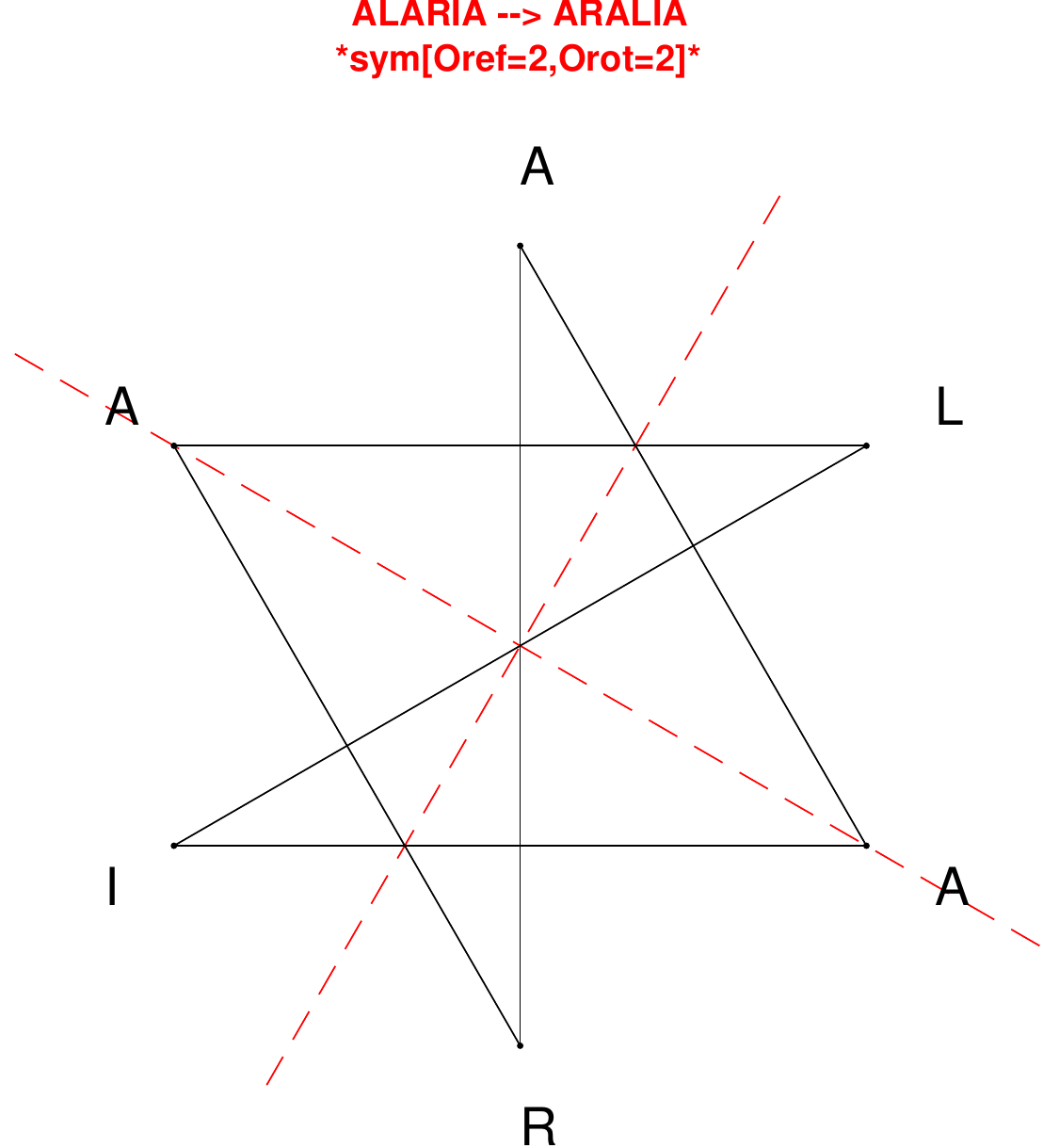}
\end{subfigure}
\hfill
\begin{subfigure}[T]{0.19\textwidth}
\centering
\includegraphics[width=\textwidth]{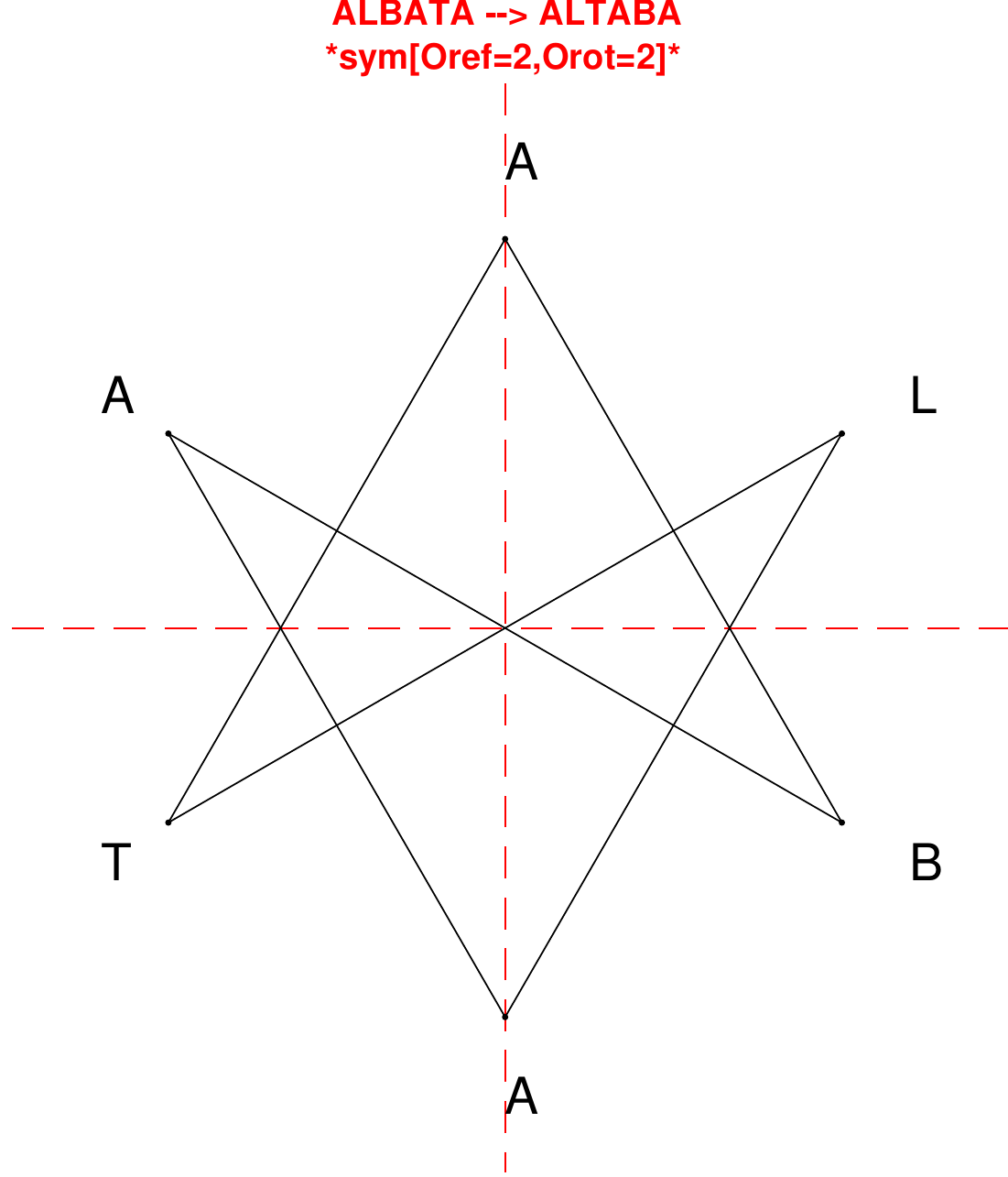}
\end{subfigure}
\end{figure}

\begin{figure}[H]
\centering
\begin{subfigure}[T]{0.19\textwidth}
\centering
\includegraphics[width=\textwidth]{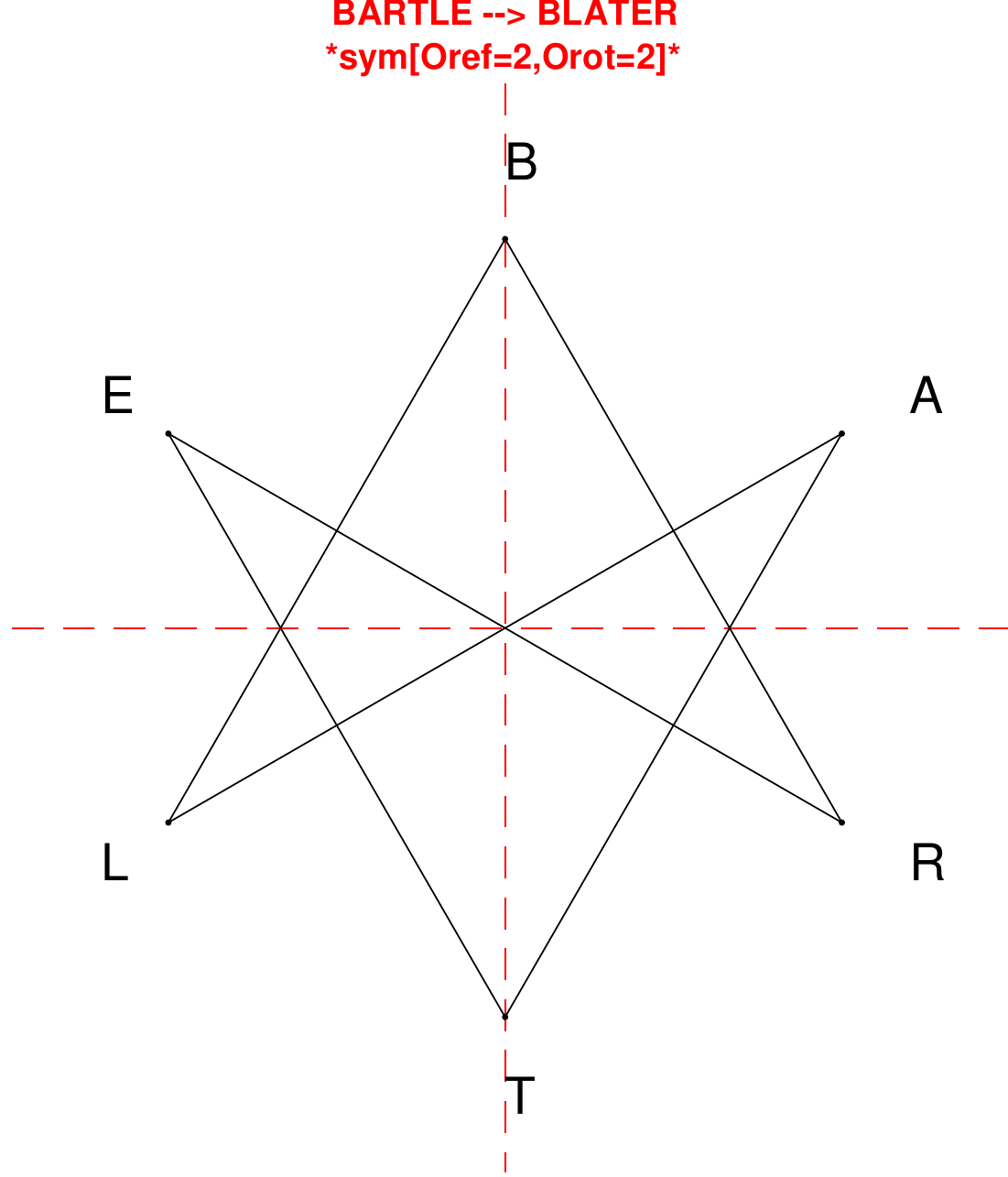}
\end{subfigure}
\hfill
\begin{subfigure}[T]{0.19\textwidth}
\centering
\includegraphics[width=\textwidth]{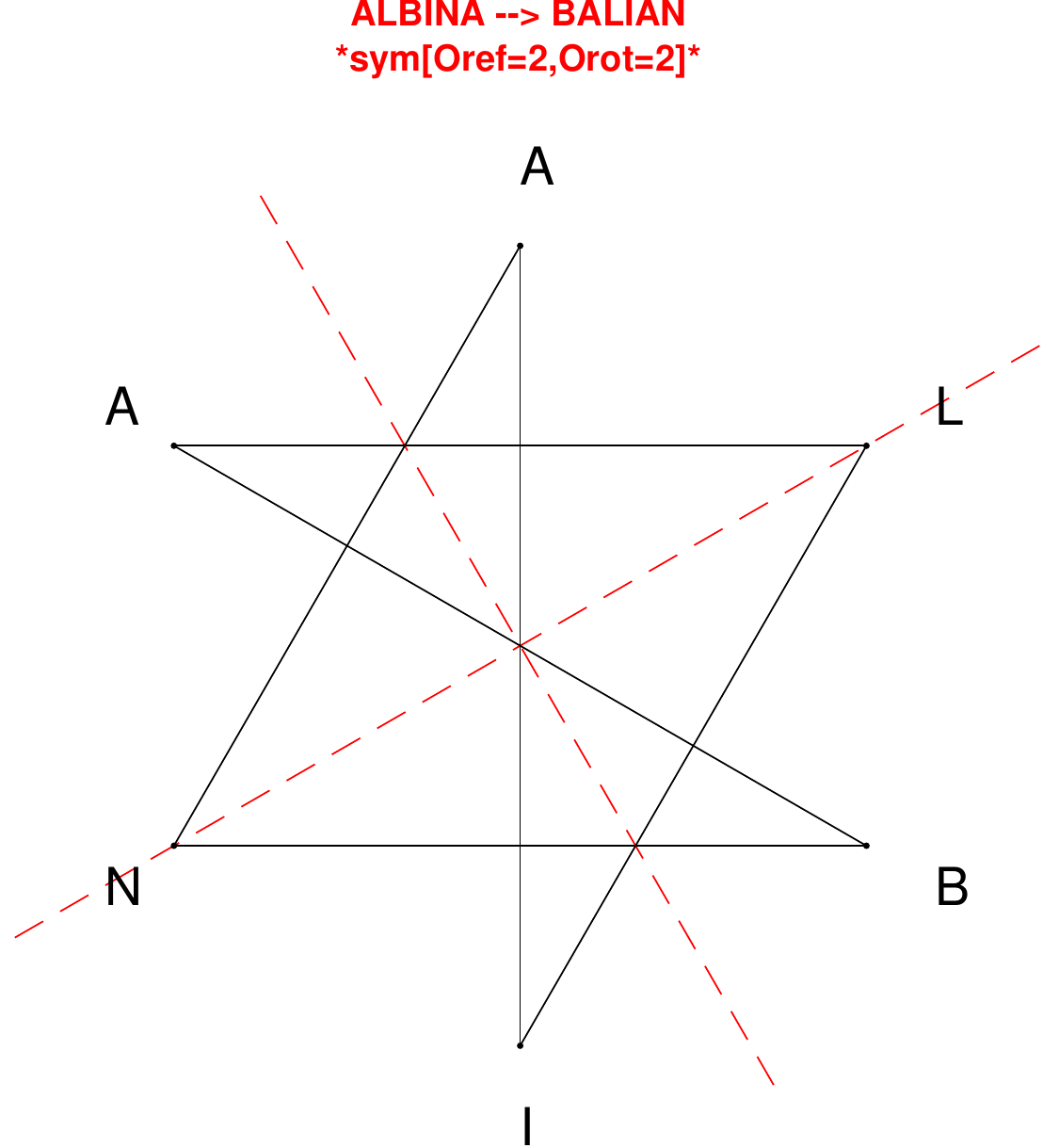}
\end{subfigure}
\hfill
\begin{subfigure}[T]{0.19\textwidth}
\centering
\includegraphics[width=\textwidth]{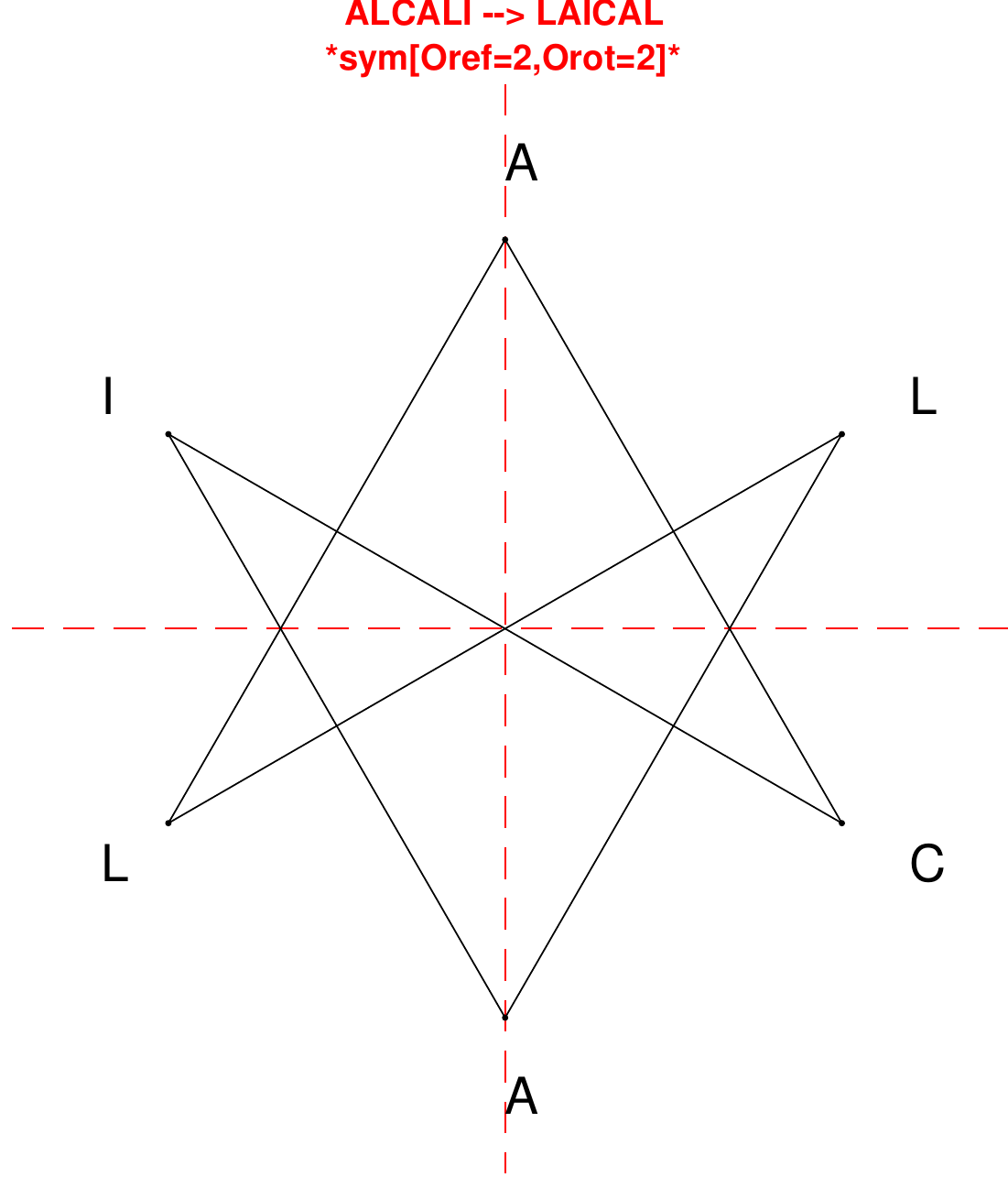}
\end{subfigure}
\hfill
\begin{subfigure}[T]{0.19\textwidth}
\centering
\includegraphics[width=\textwidth]{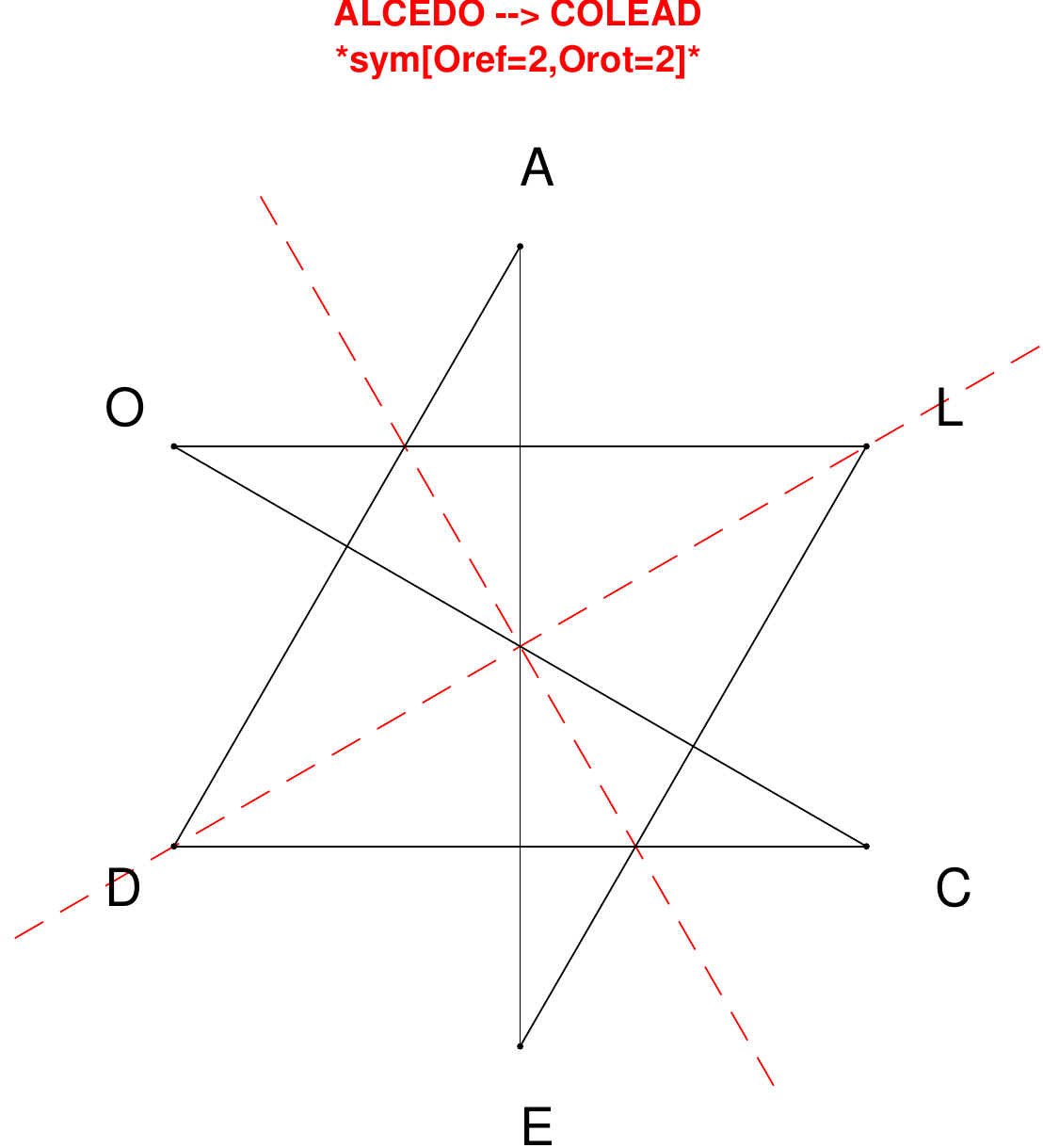}
\end{subfigure}
\hfill
\begin{subfigure}[T]{0.19\textwidth}
\centering
\includegraphics[width=\textwidth]{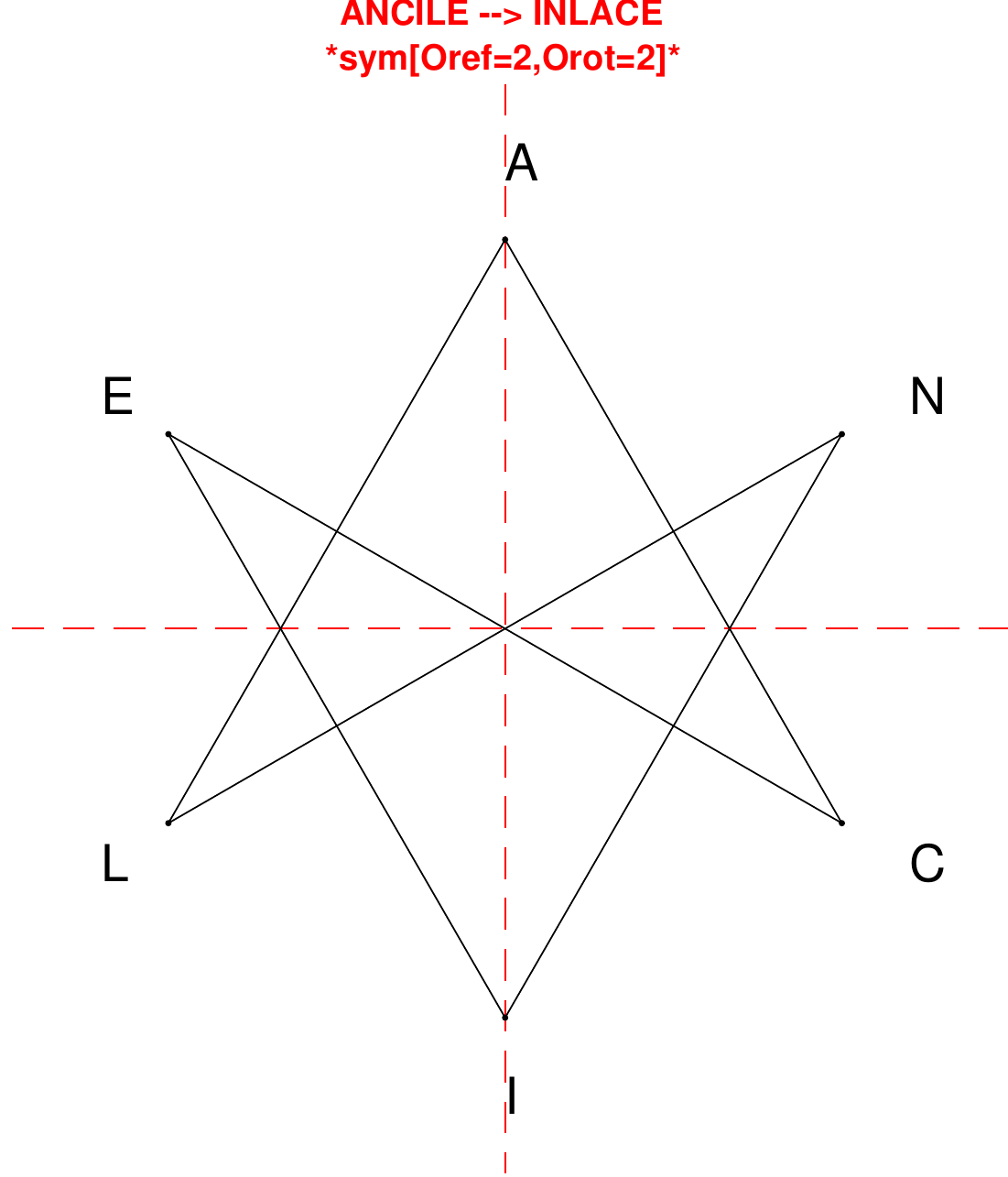}
\end{subfigure}
\end{figure}

\begin{figure}[H]
\centering
\begin{subfigure}[T]{0.19\textwidth}
\centering
\includegraphics[width=\textwidth]{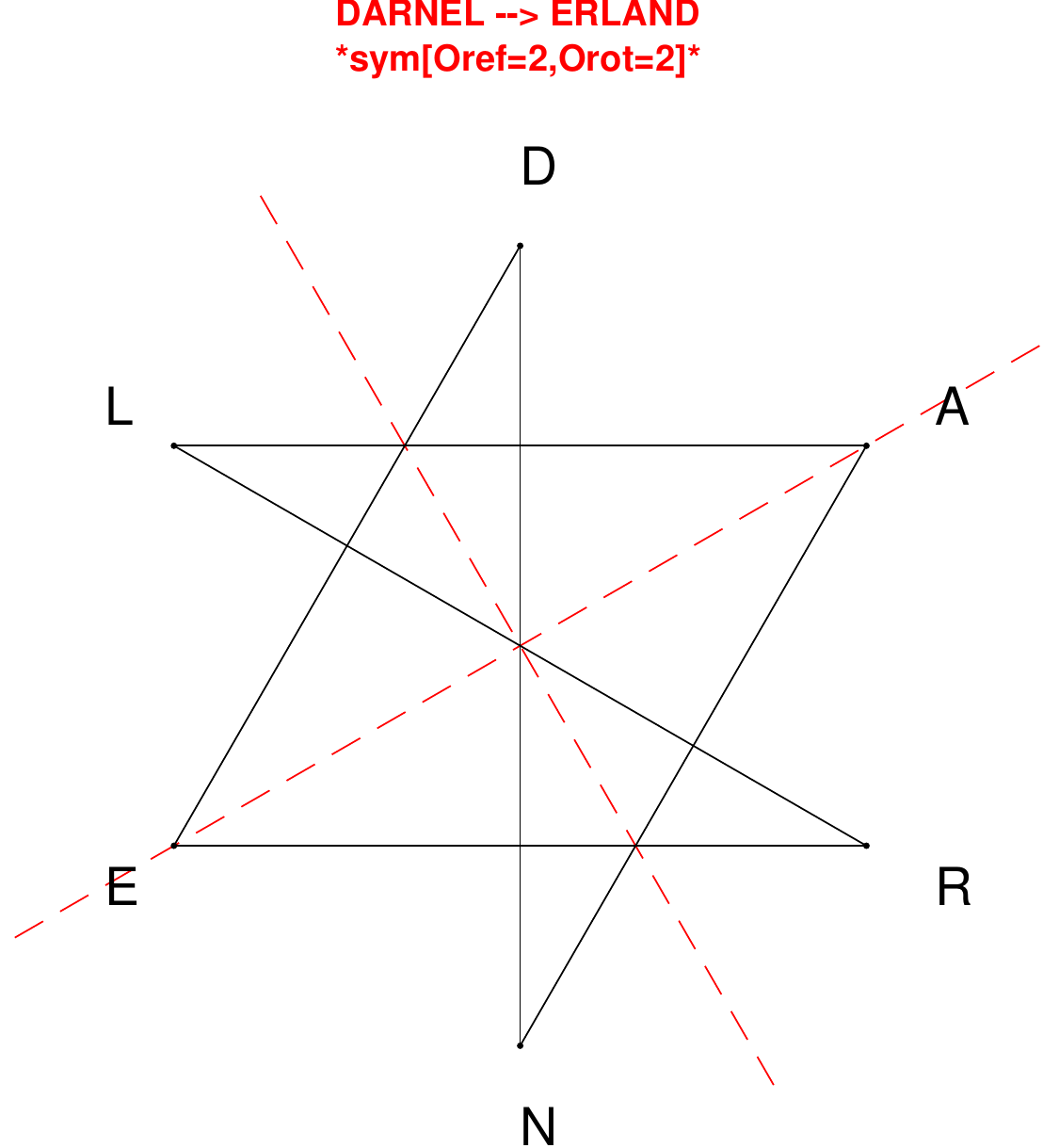}
\end{subfigure}
\hfill
\begin{subfigure}[T]{0.19\textwidth}
\centering
\includegraphics[width=\textwidth]{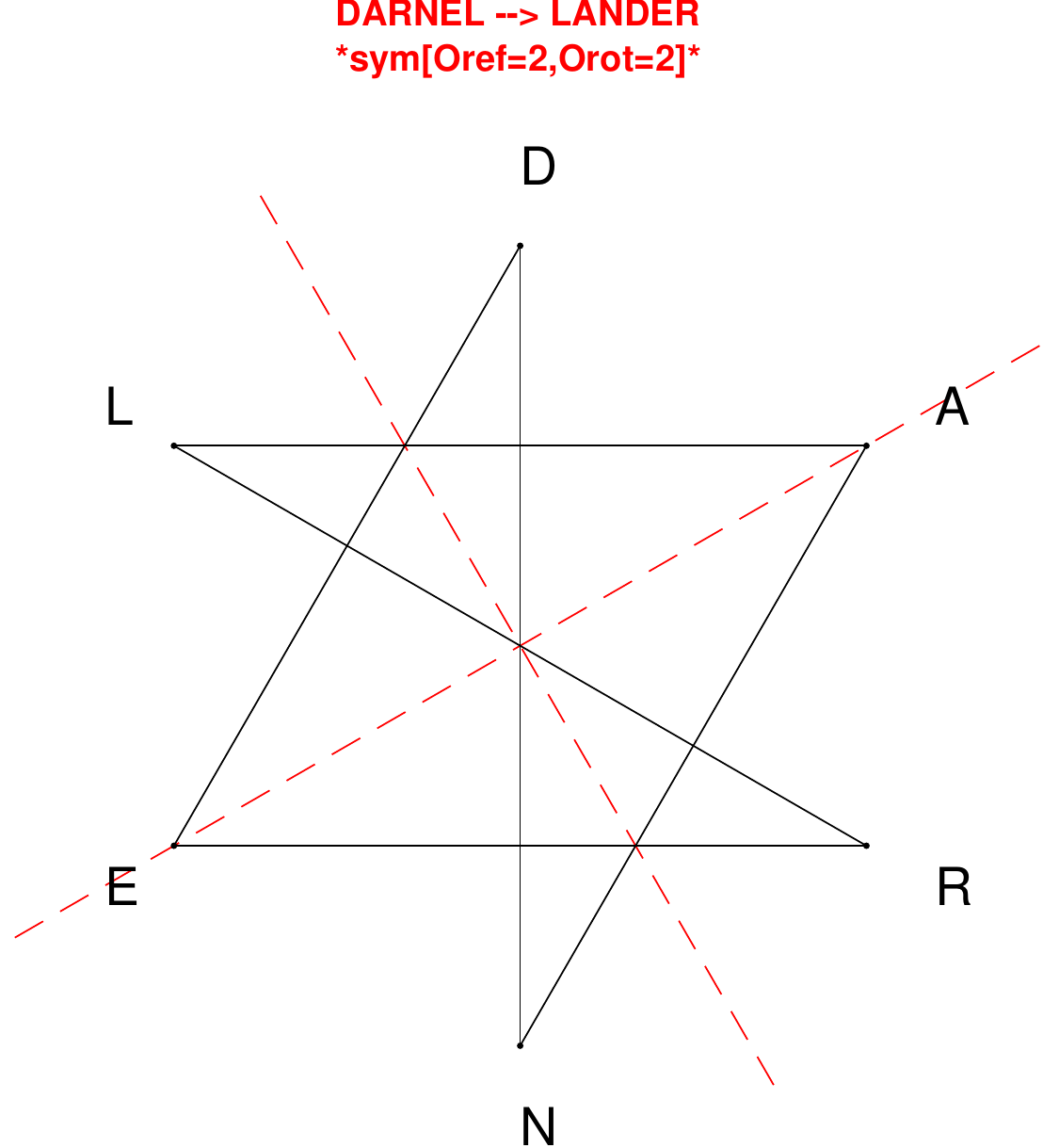}
\end{subfigure}
\hfill
\begin{subfigure}[T]{0.19\textwidth}
\centering
\includegraphics[width=\textwidth]{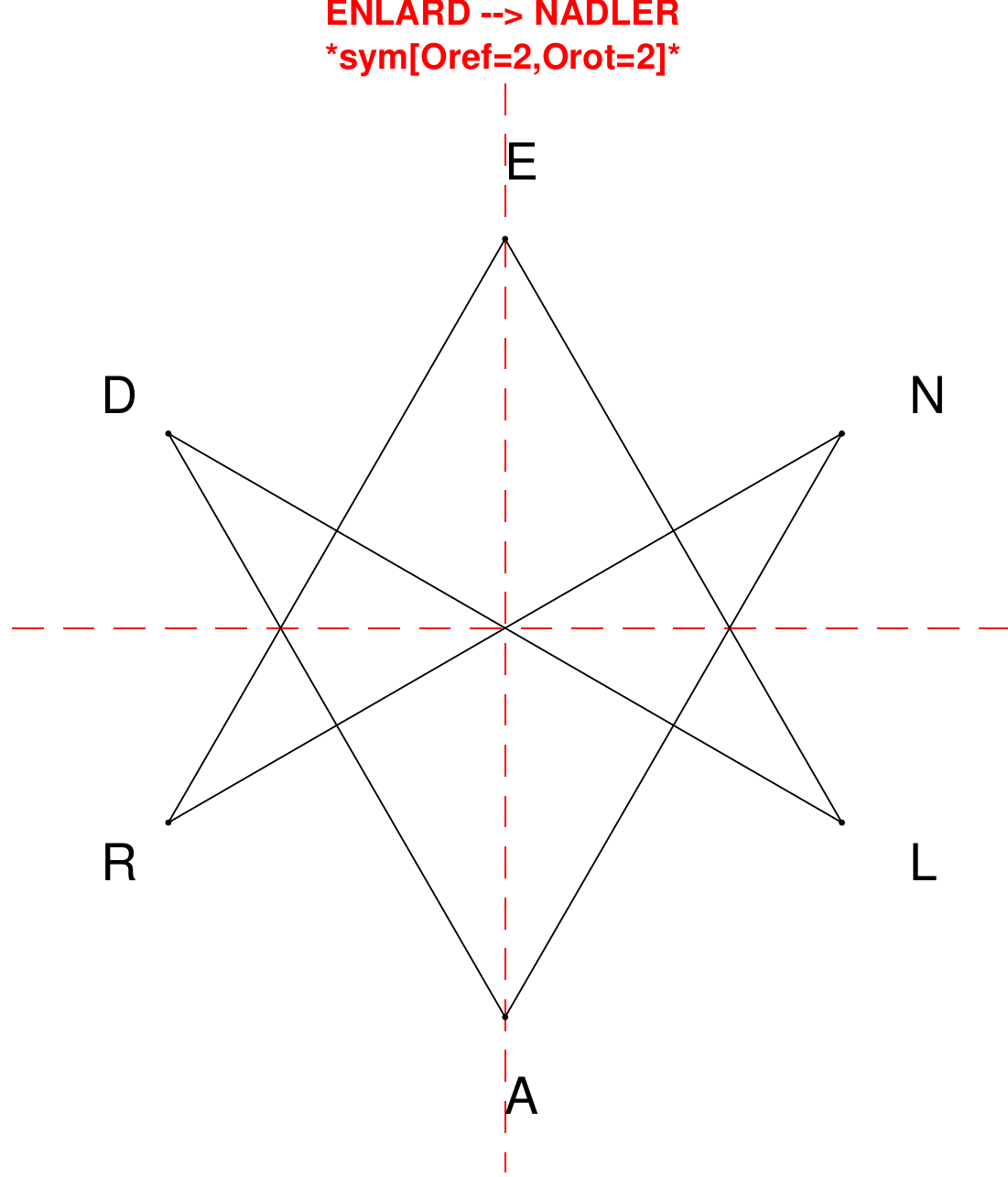}
\end{subfigure}
\hfill
\begin{subfigure}[T]{0.19\textwidth}
\centering
\includegraphics[width=\textwidth]{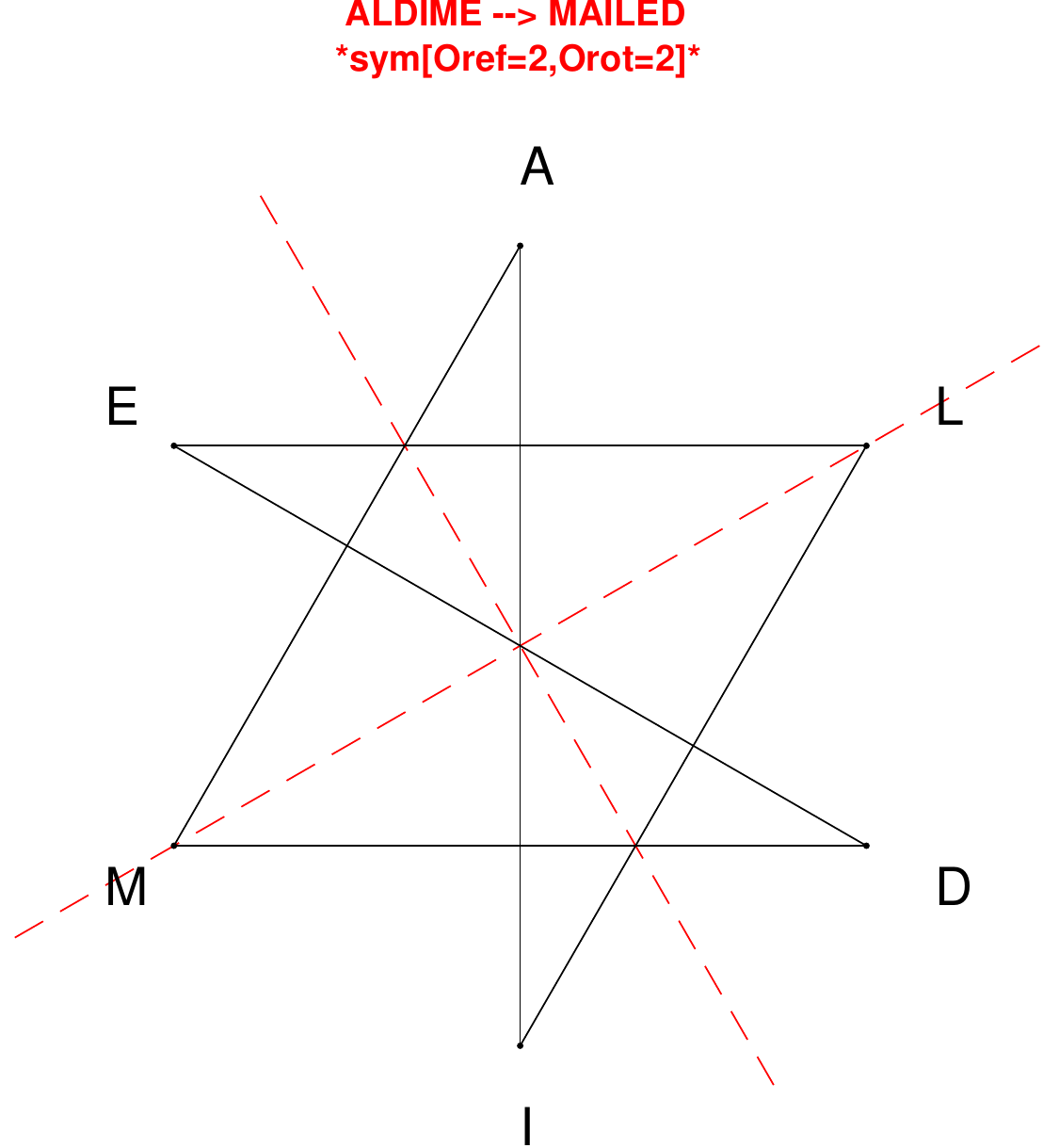}
\end{subfigure}
\hfill
\begin{subfigure}[T]{0.19\textwidth}
\centering
\includegraphics[width=\textwidth]{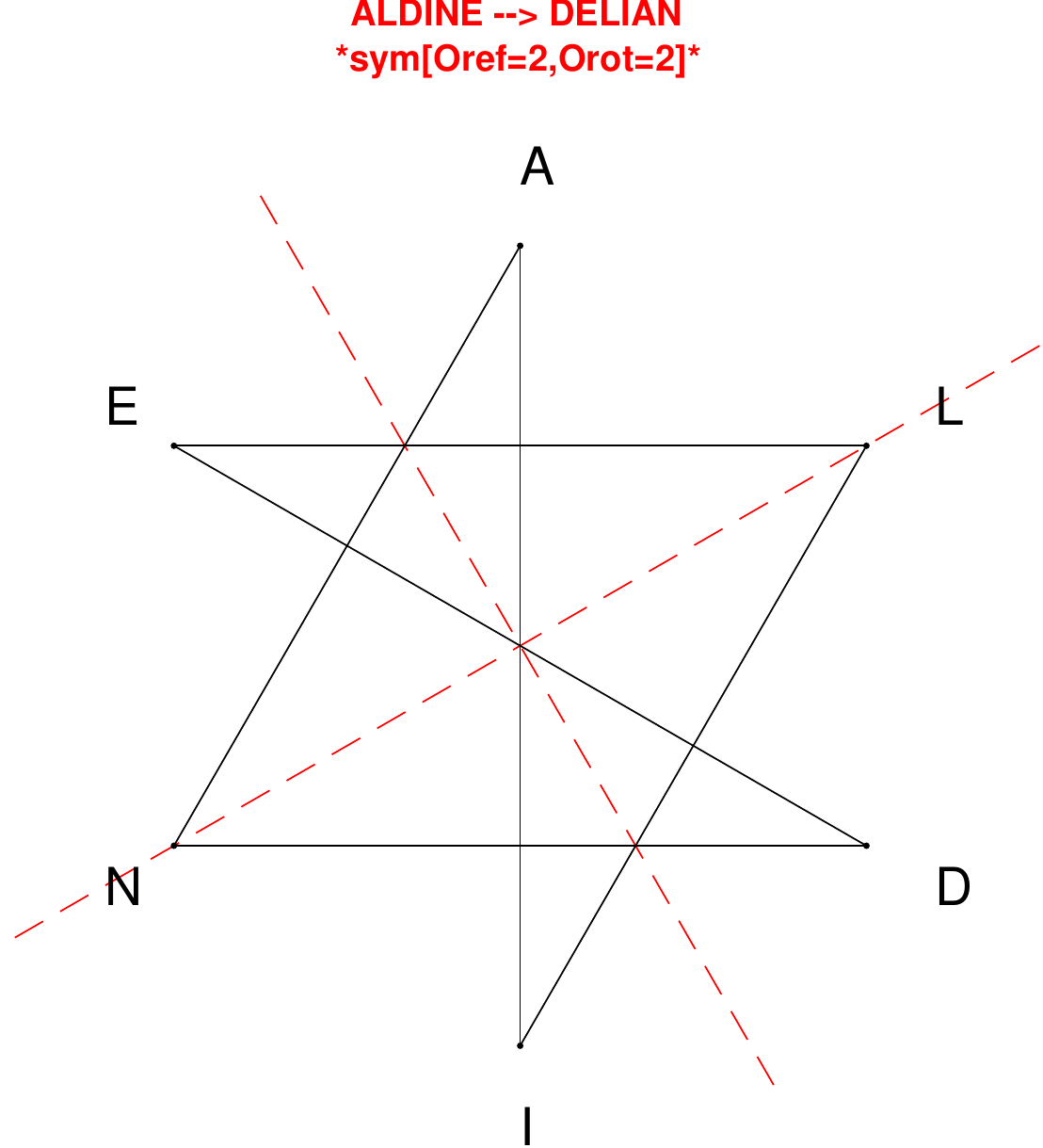}
\end{subfigure}
\end{figure}

\begin{figure}[H]
\centering
\begin{subfigure}[T]{0.19\textwidth}
\centering
\includegraphics[width=\textwidth]{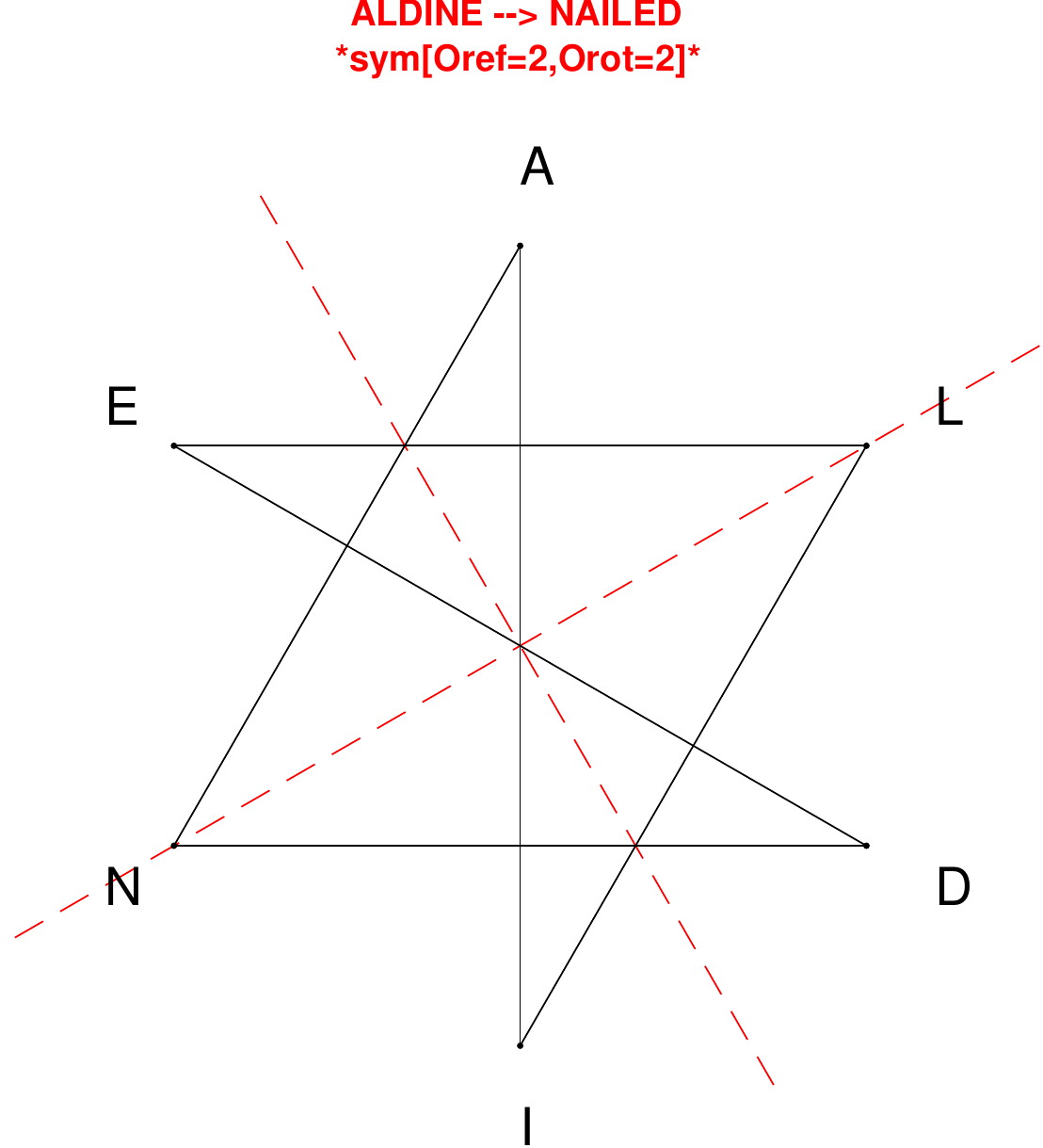}
\end{subfigure}
\hfill
\begin{subfigure}[T]{0.19\textwidth}
\centering
\includegraphics[width=\textwidth]{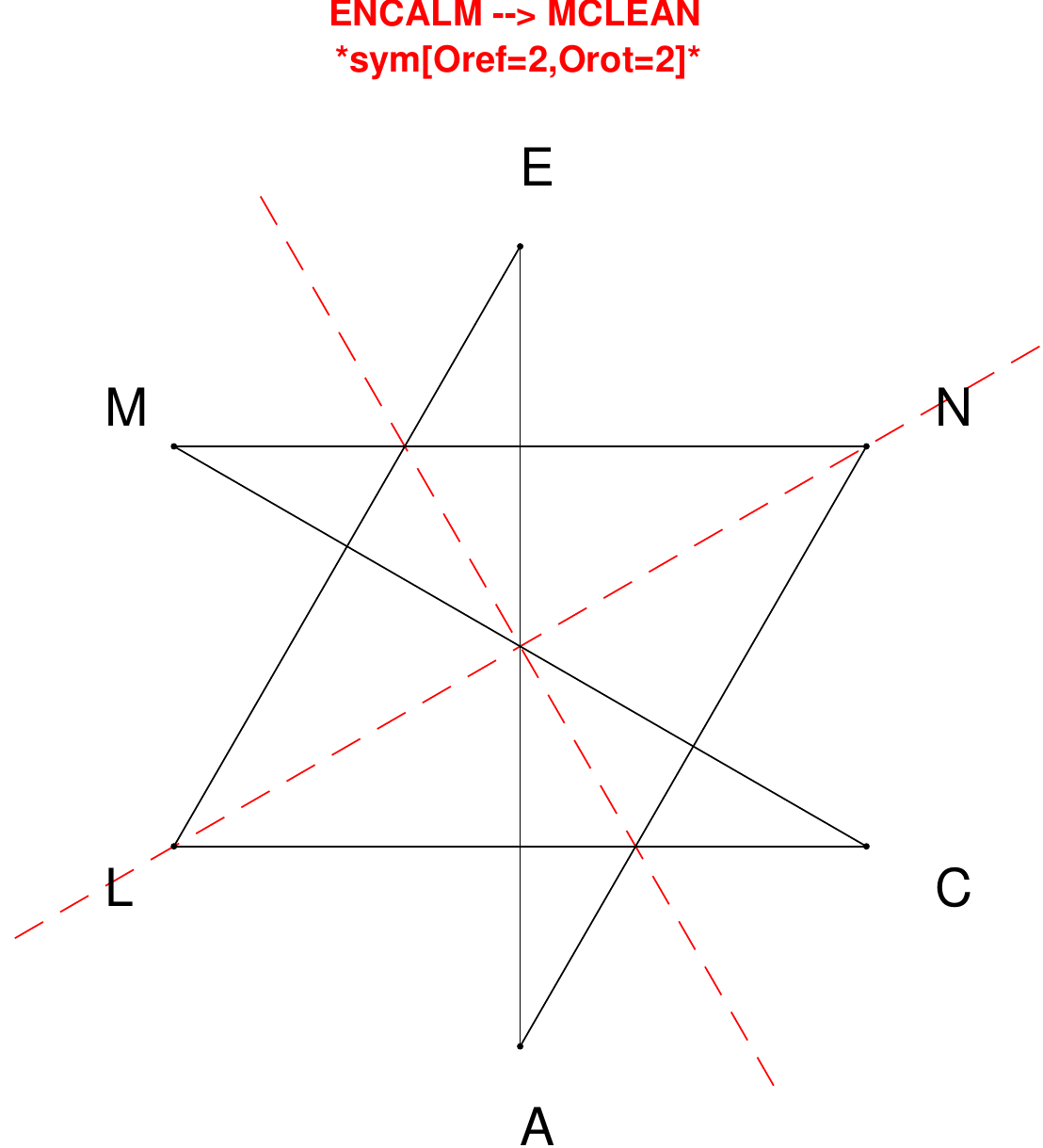}
\end{subfigure}
\hfill
\begin{subfigure}[T]{0.19\textwidth}
\centering
\includegraphics[width=\textwidth]{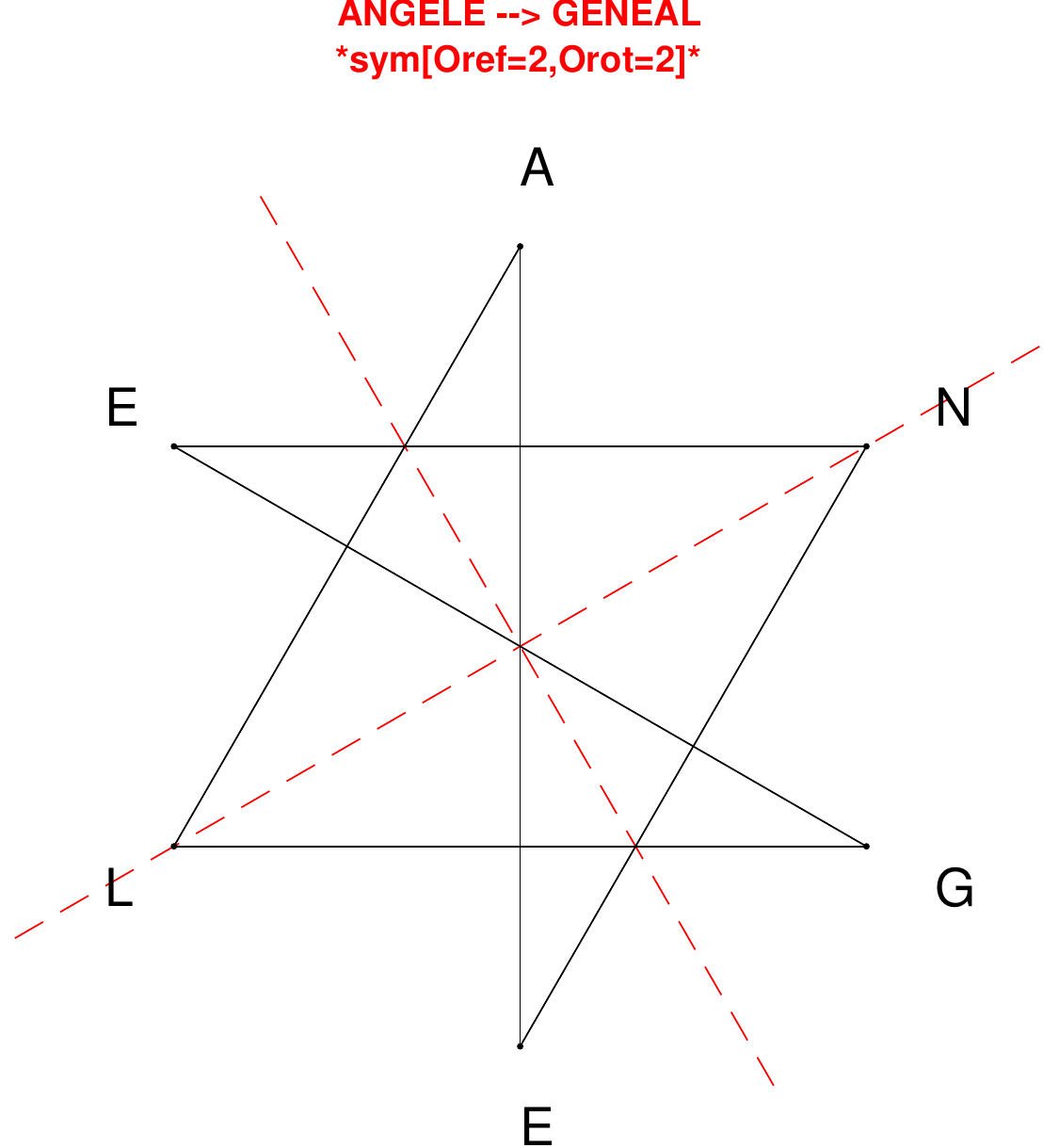}
\end{subfigure}
\hfill
\begin{subfigure}[T]{0.19\textwidth}
\centering
\includegraphics[width=\textwidth]{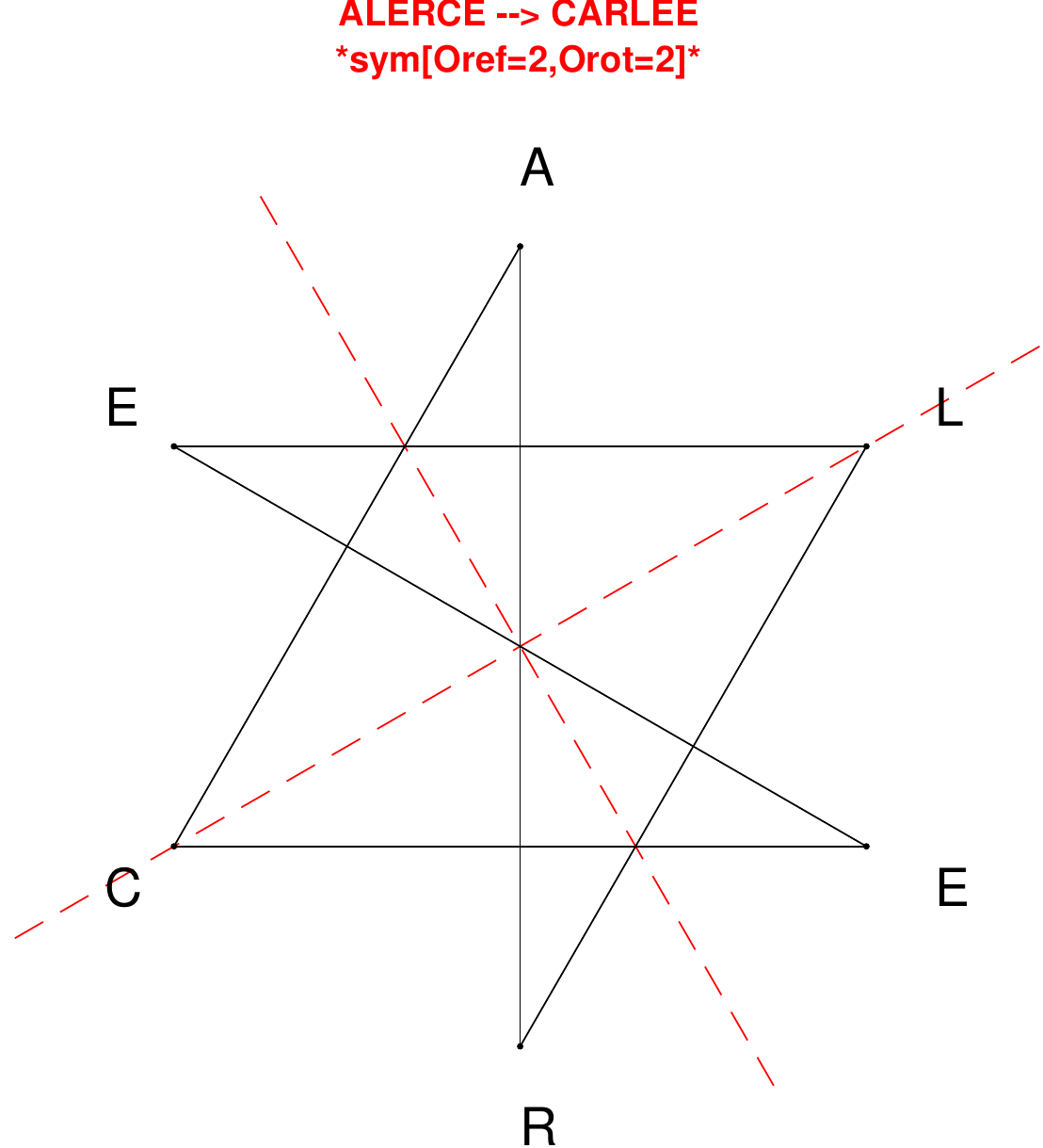}
\end{subfigure}
\hfill
\begin{subfigure}[T]{0.19\textwidth}
\centering
\includegraphics[width=\textwidth]{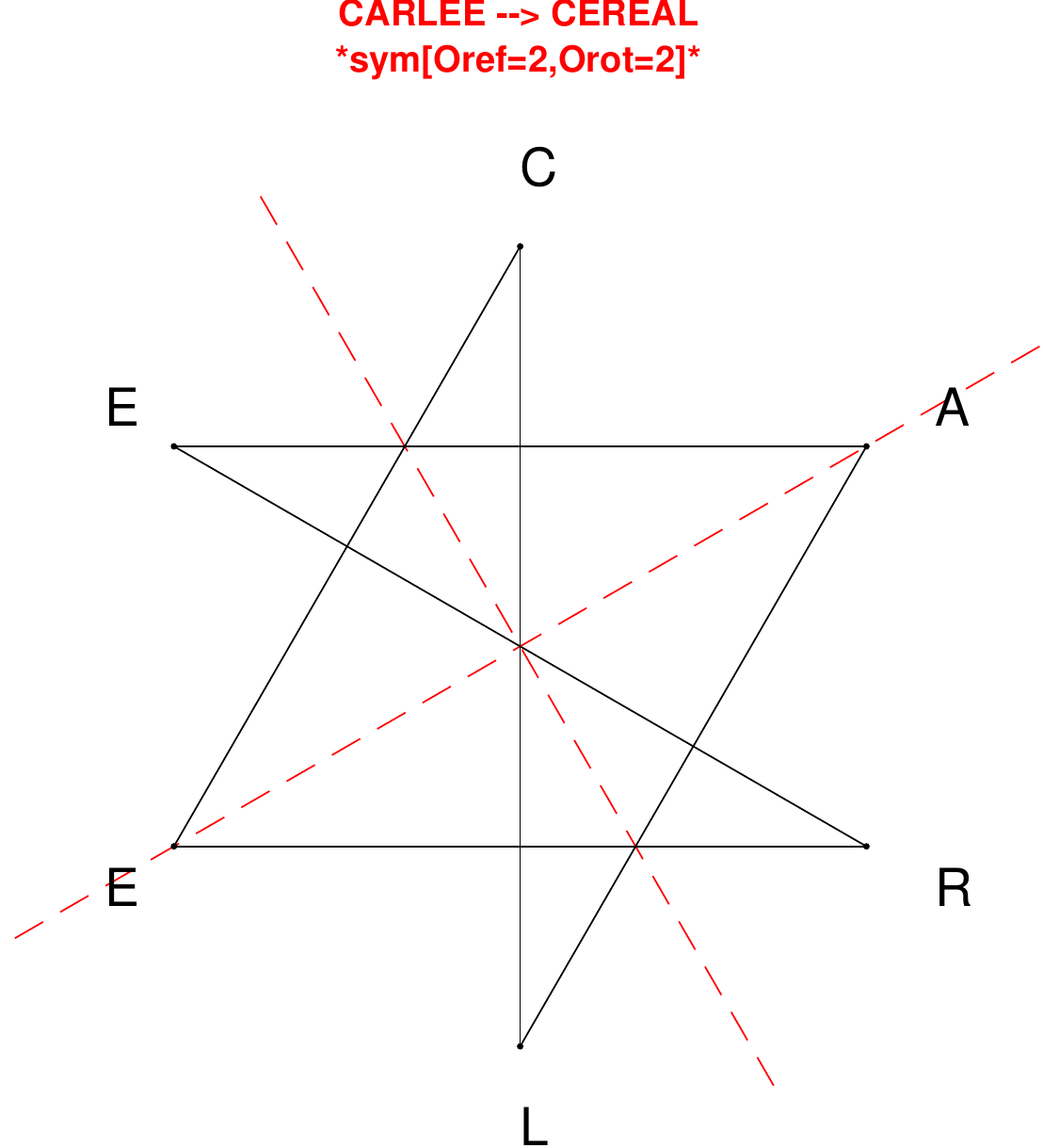}
\end{subfigure}
\end{figure}

\begin{figure}[H]
\centering
\begin{subfigure}[T]{0.19\textwidth}
\centering
\includegraphics[width=\textwidth]{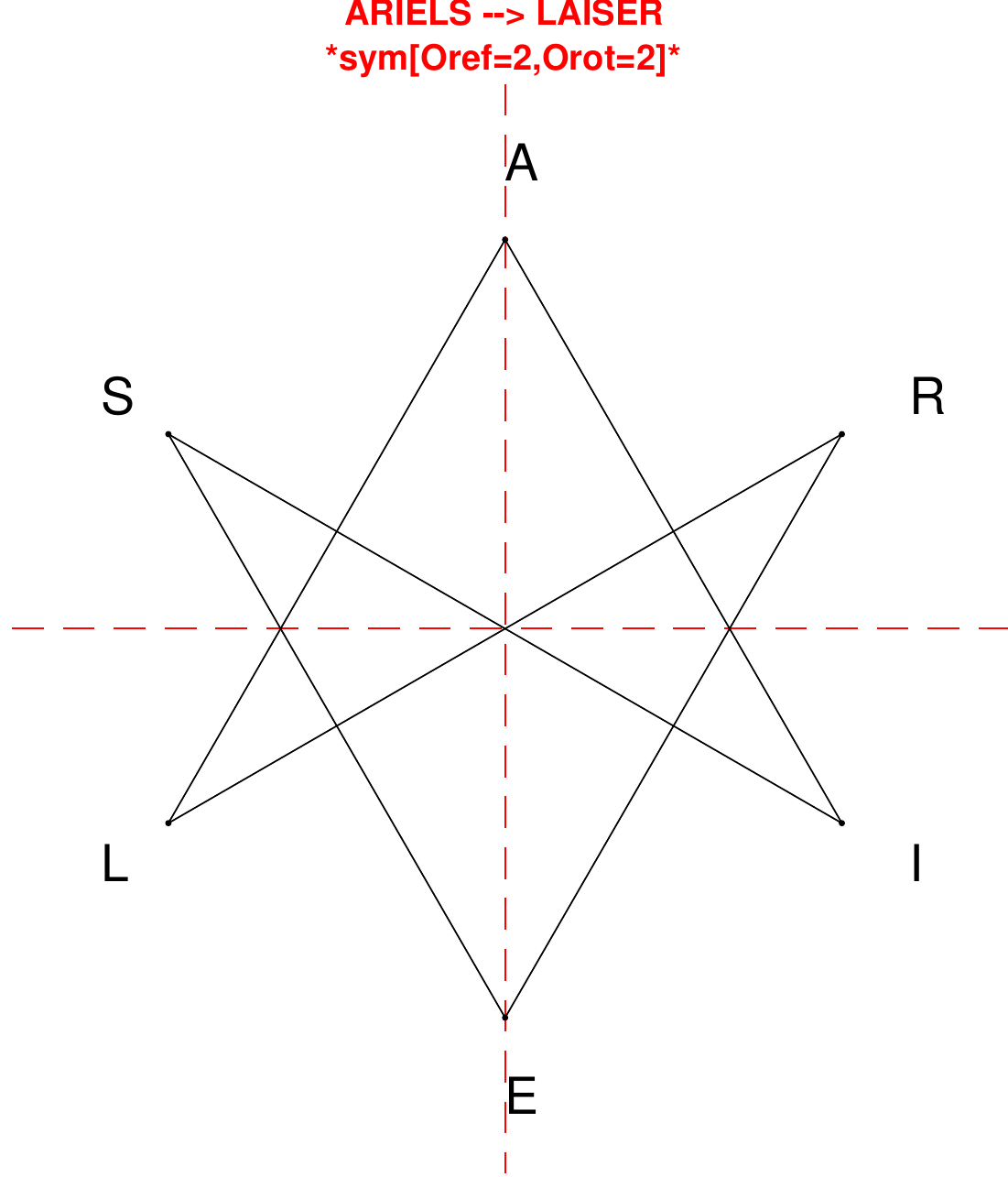}
\end{subfigure}
\hfill
\begin{subfigure}[T]{0.19\textwidth}
\centering
\includegraphics[width=\textwidth]{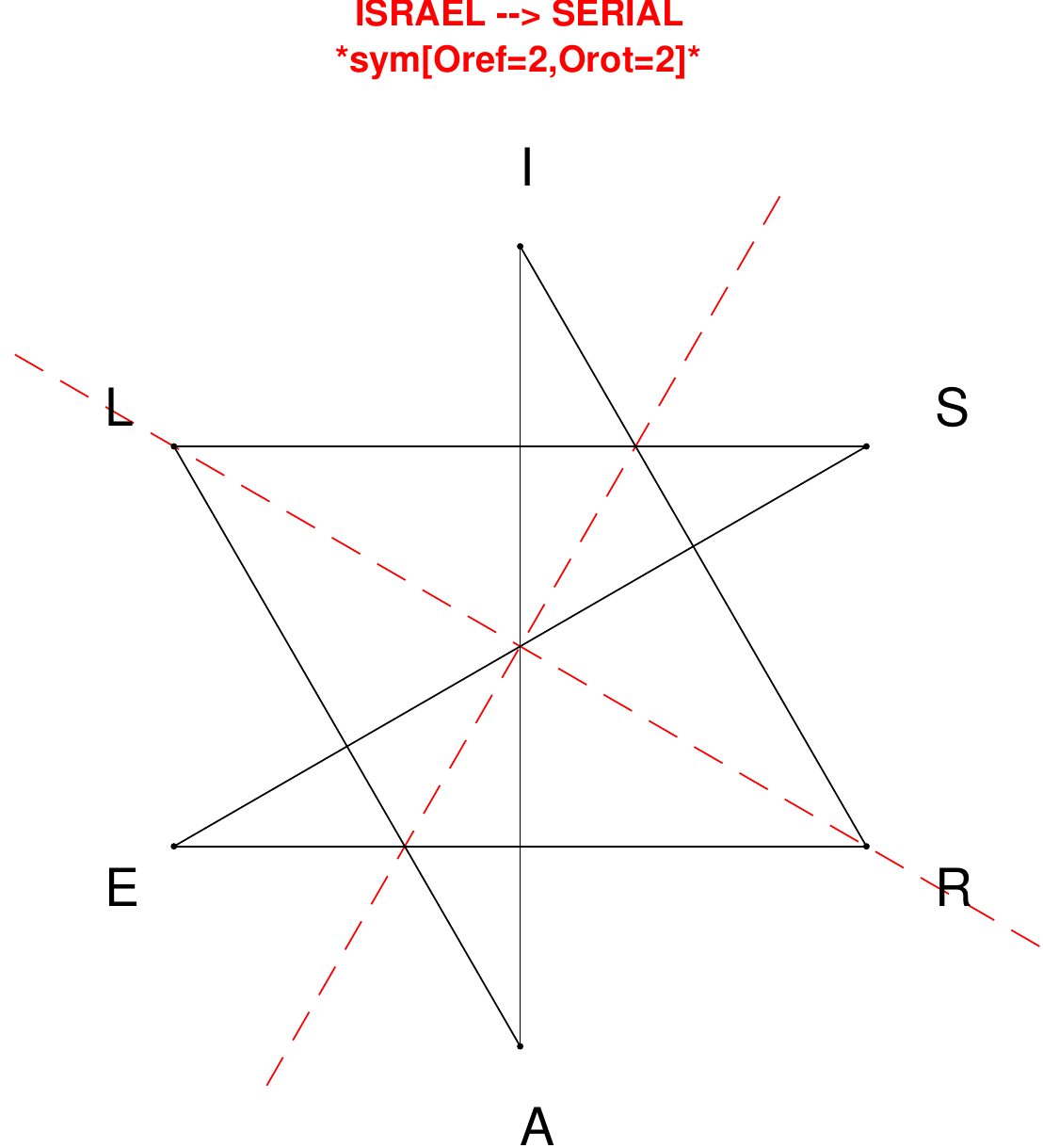}
\end{subfigure}
\hfill
\begin{subfigure}[T]{0.19\textwidth}
\centering
\includegraphics[width=\textwidth]{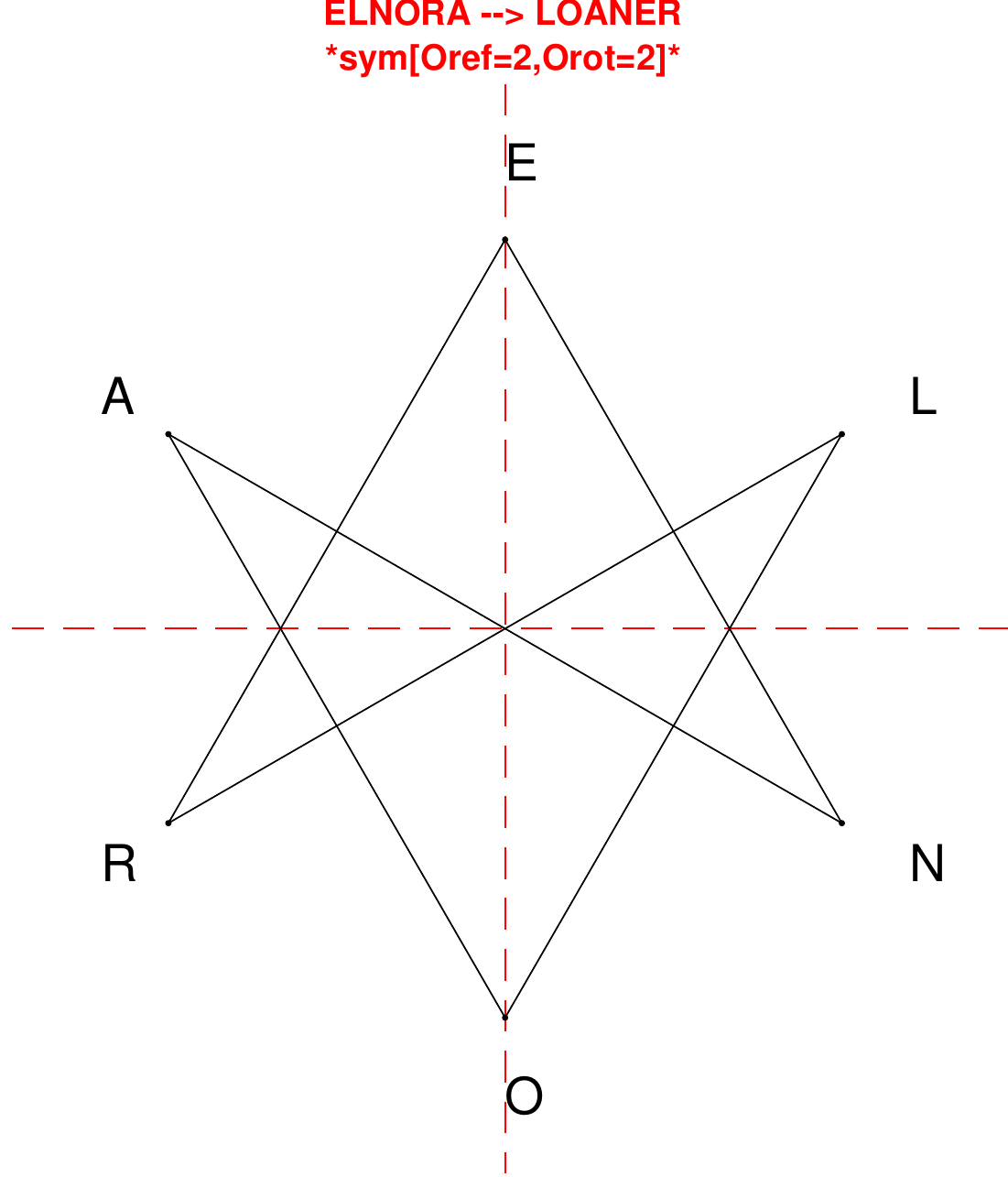}
\end{subfigure}
\hfill
\begin{subfigure}[T]{0.19\textwidth}
\centering
\includegraphics[width=\textwidth]{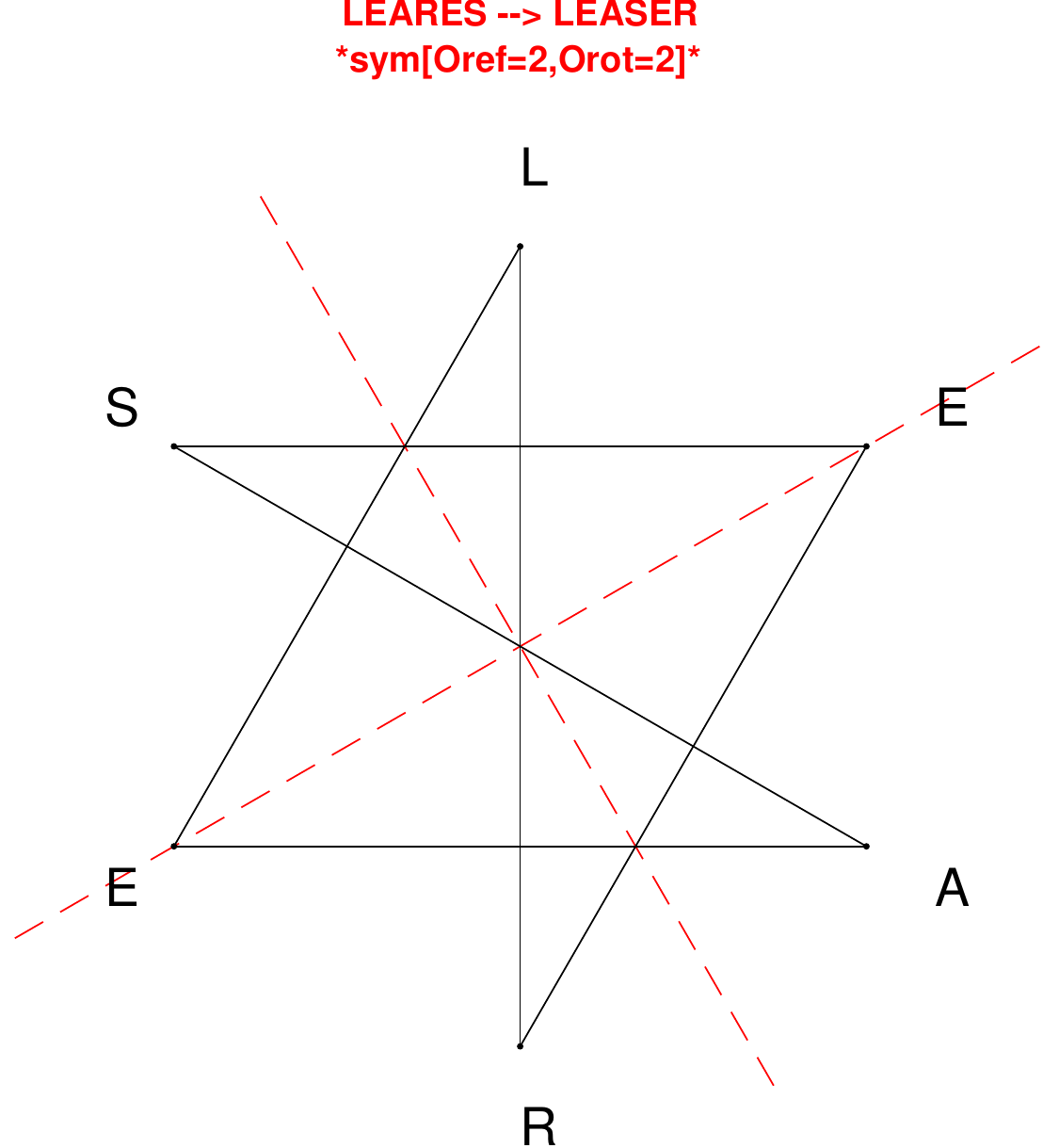}
\end{subfigure}
\hfill
\begin{subfigure}[T]{0.19\textwidth}
\centering
\includegraphics[width=\textwidth]{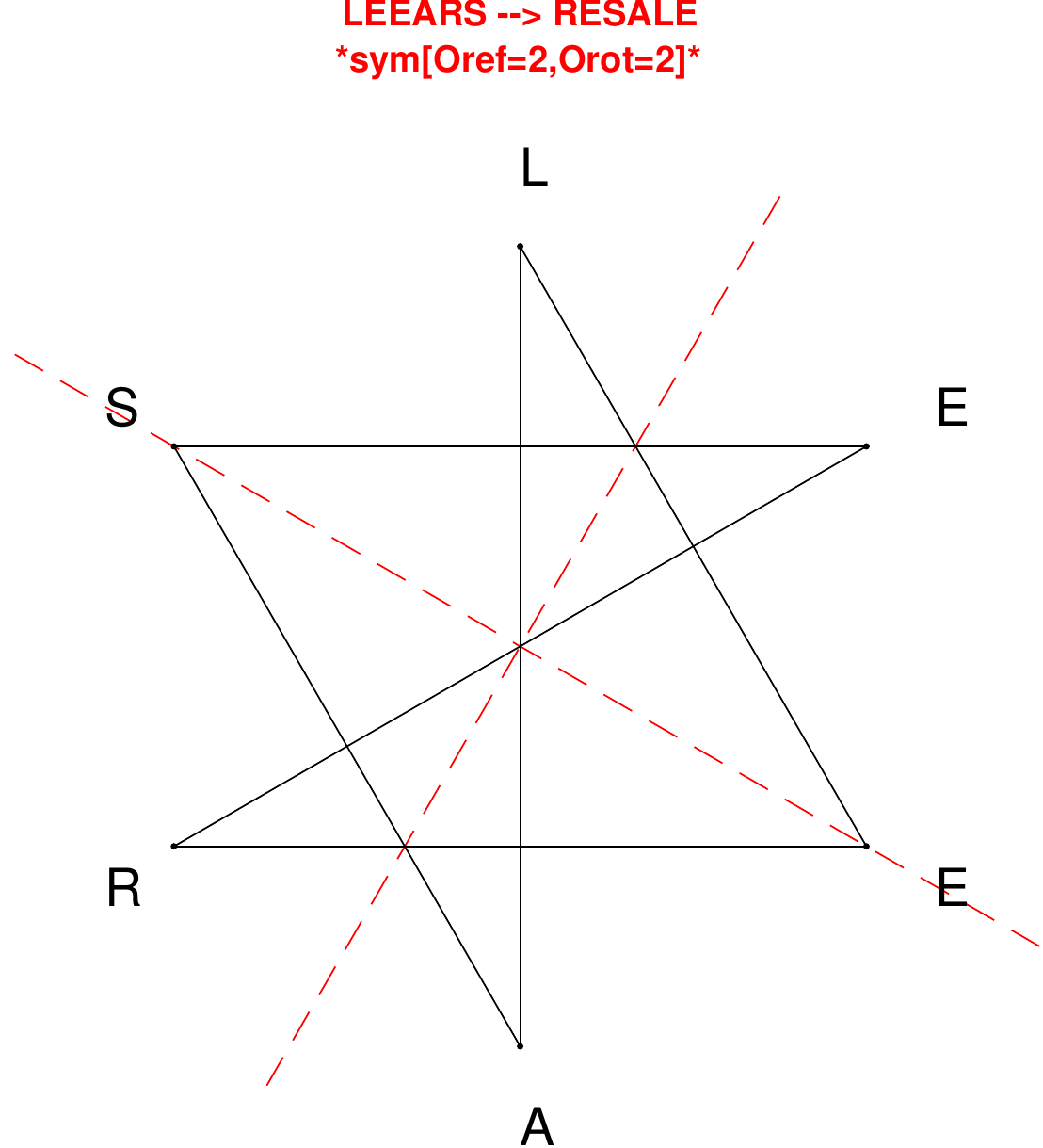}
\end{subfigure}
\end{figure}

\begin{figure}[H]
\centering
\begin{subfigure}[T]{0.19\textwidth}
\centering
\includegraphics[width=\textwidth]{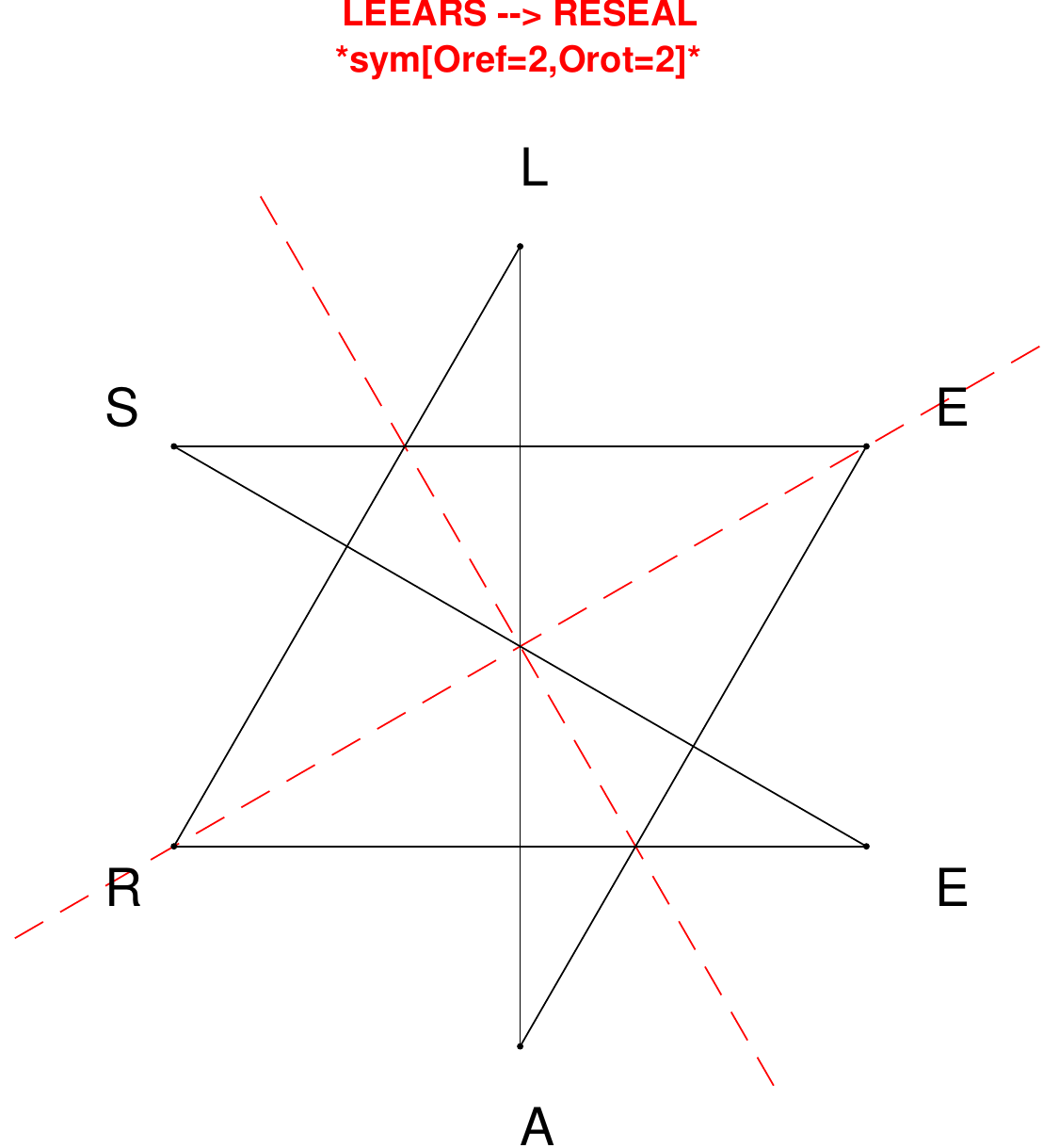}
\end{subfigure}
\hfill
\begin{subfigure}[T]{0.19\textwidth}
\centering
\includegraphics[width=\textwidth]{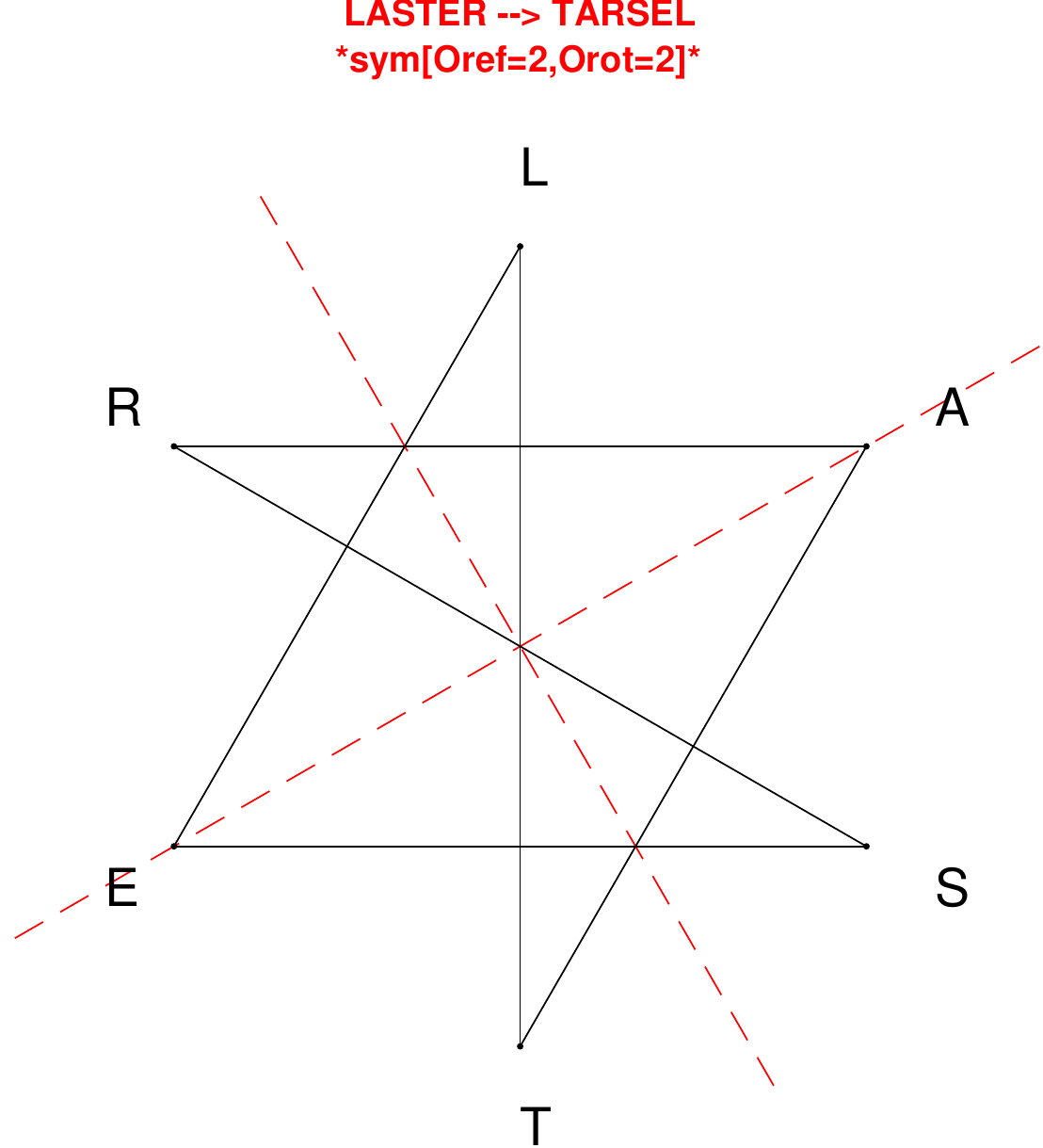}
\end{subfigure}
\hfill
\begin{subfigure}[T]{0.19\textwidth}
\centering
\includegraphics[width=\textwidth]{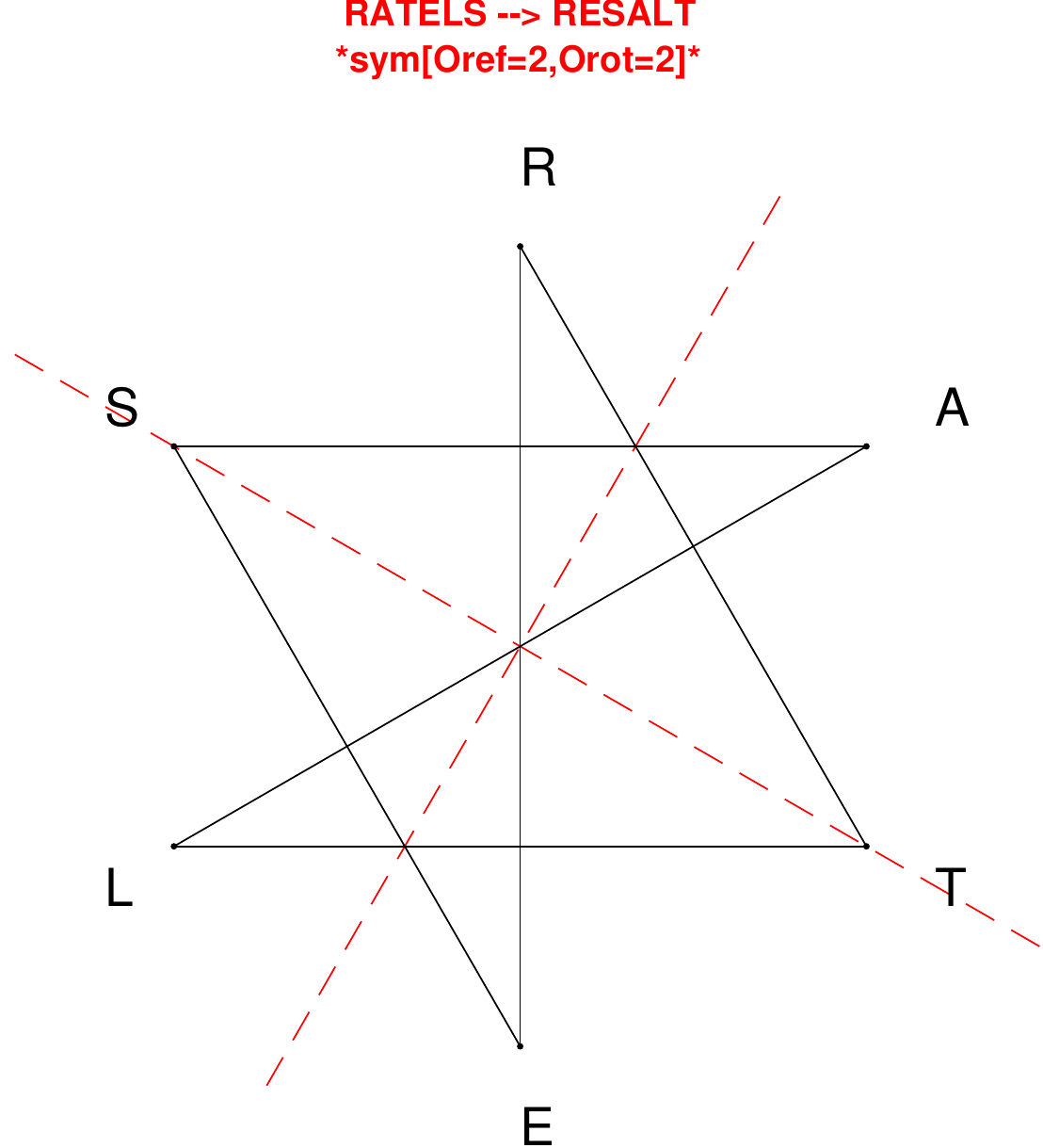}
\end{subfigure}
\hfill
\begin{subfigure}[T]{0.19\textwidth}
\centering
\includegraphics[width=\textwidth]{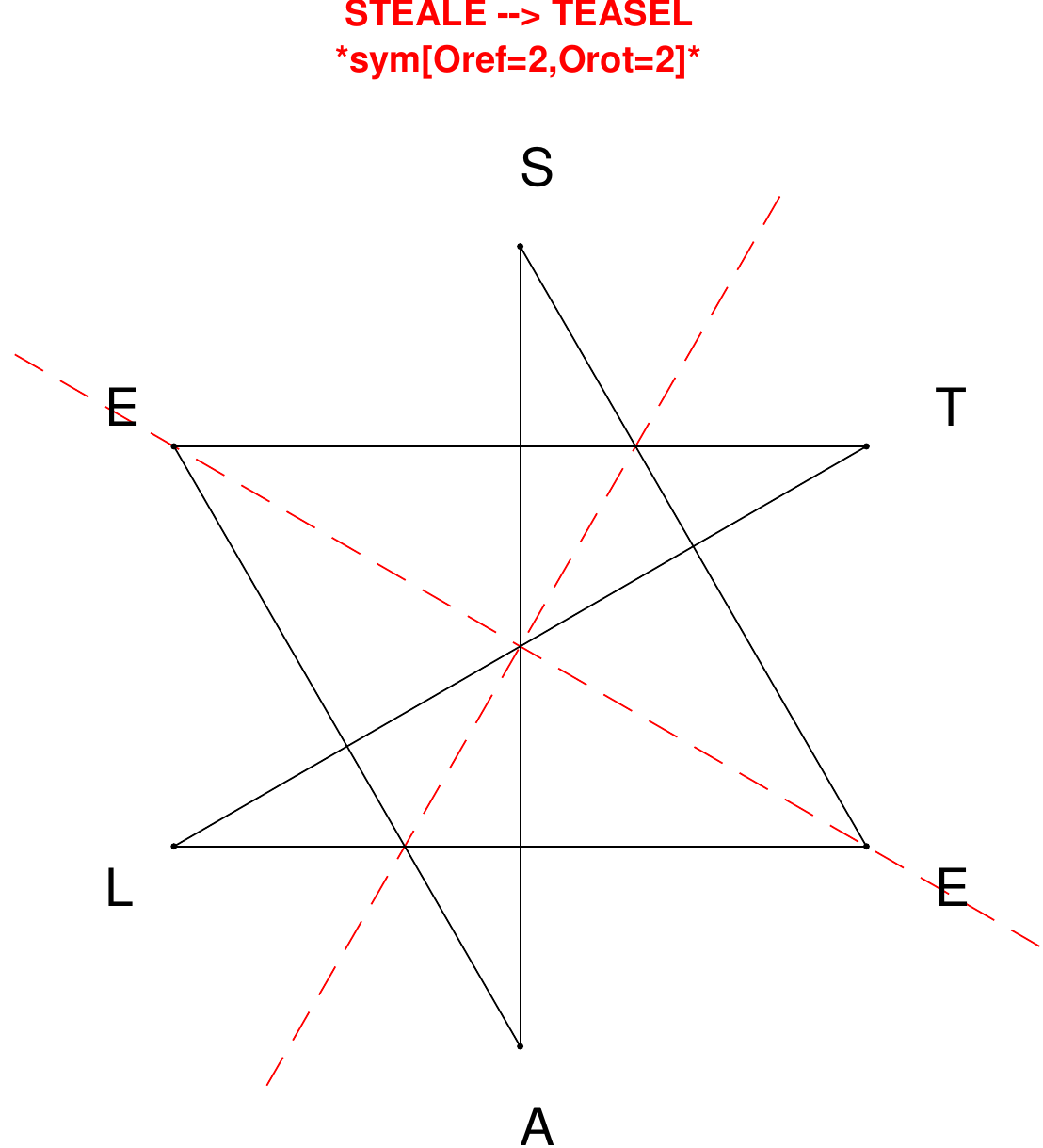}
\end{subfigure}
\hfill
\begin{subfigure}[T]{0.19\textwidth}
\centering
\includegraphics[width=\textwidth]{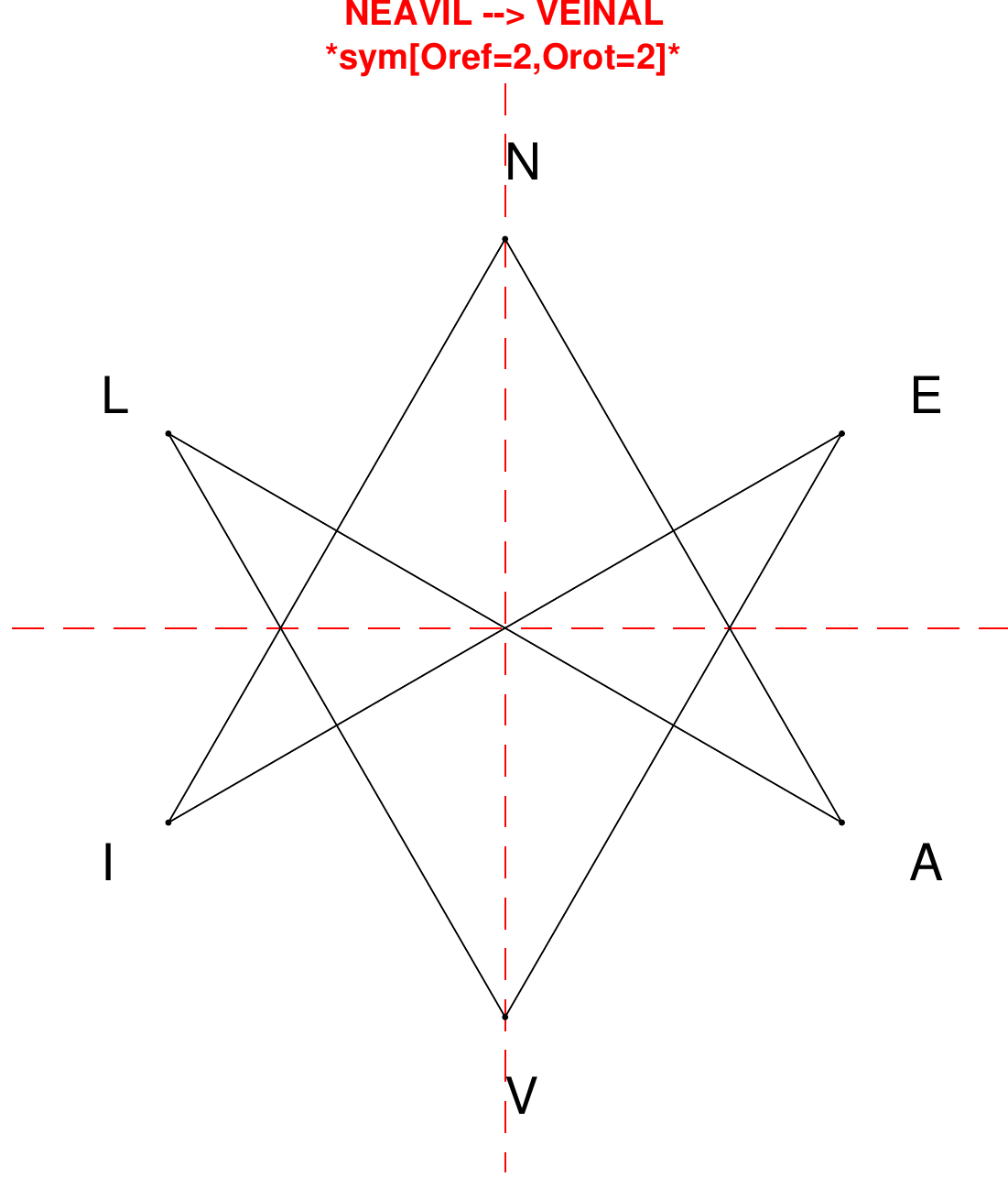}
\end{subfigure}
\end{figure}

\begin{figure}[H]
\centering
\begin{subfigure}[T]{0.19\textwidth}
\centering
\includegraphics[width=\textwidth]{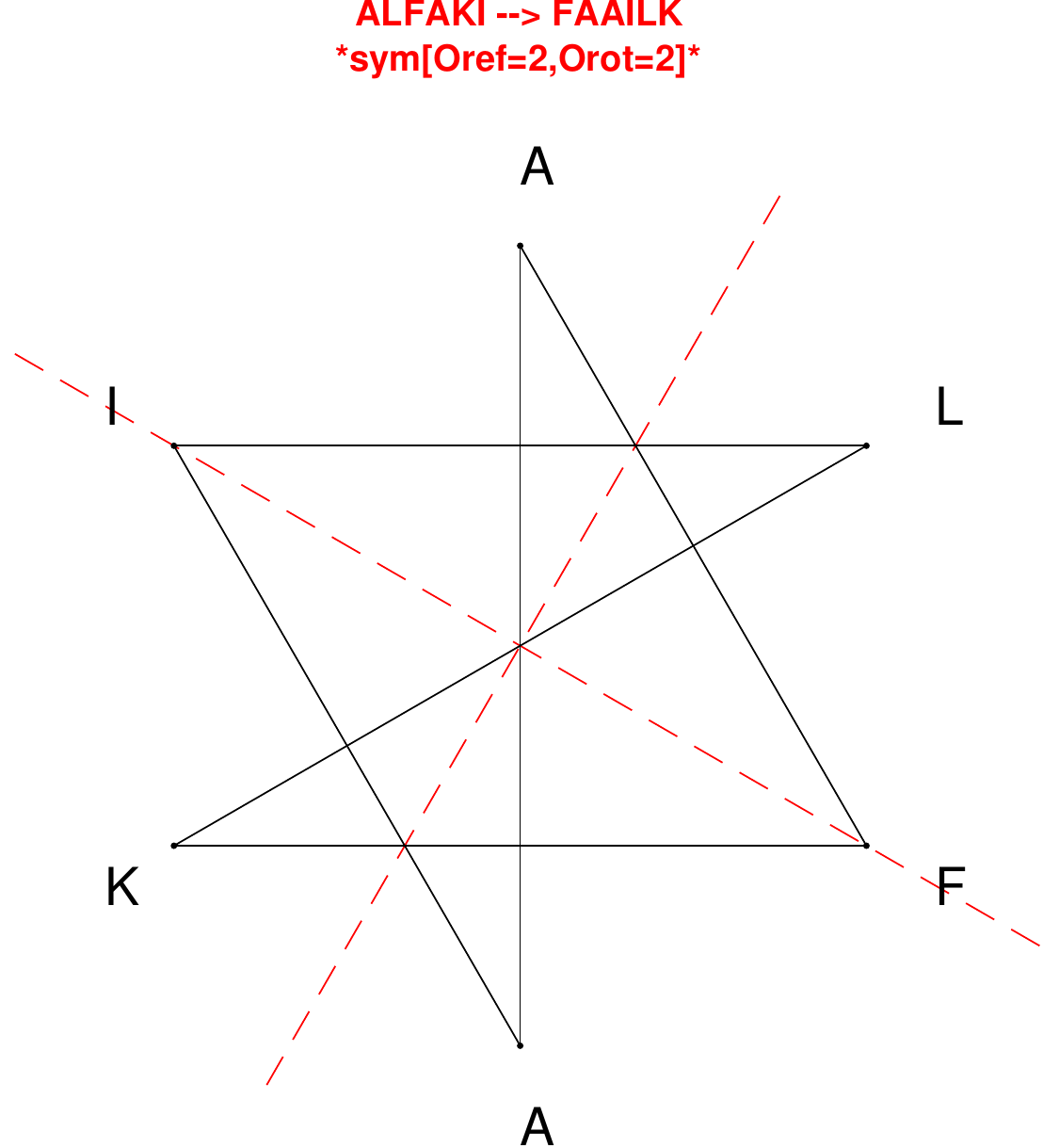}
\end{subfigure}
\hfill
\begin{subfigure}[T]{0.19\textwidth}
\centering
\includegraphics[width=\textwidth]{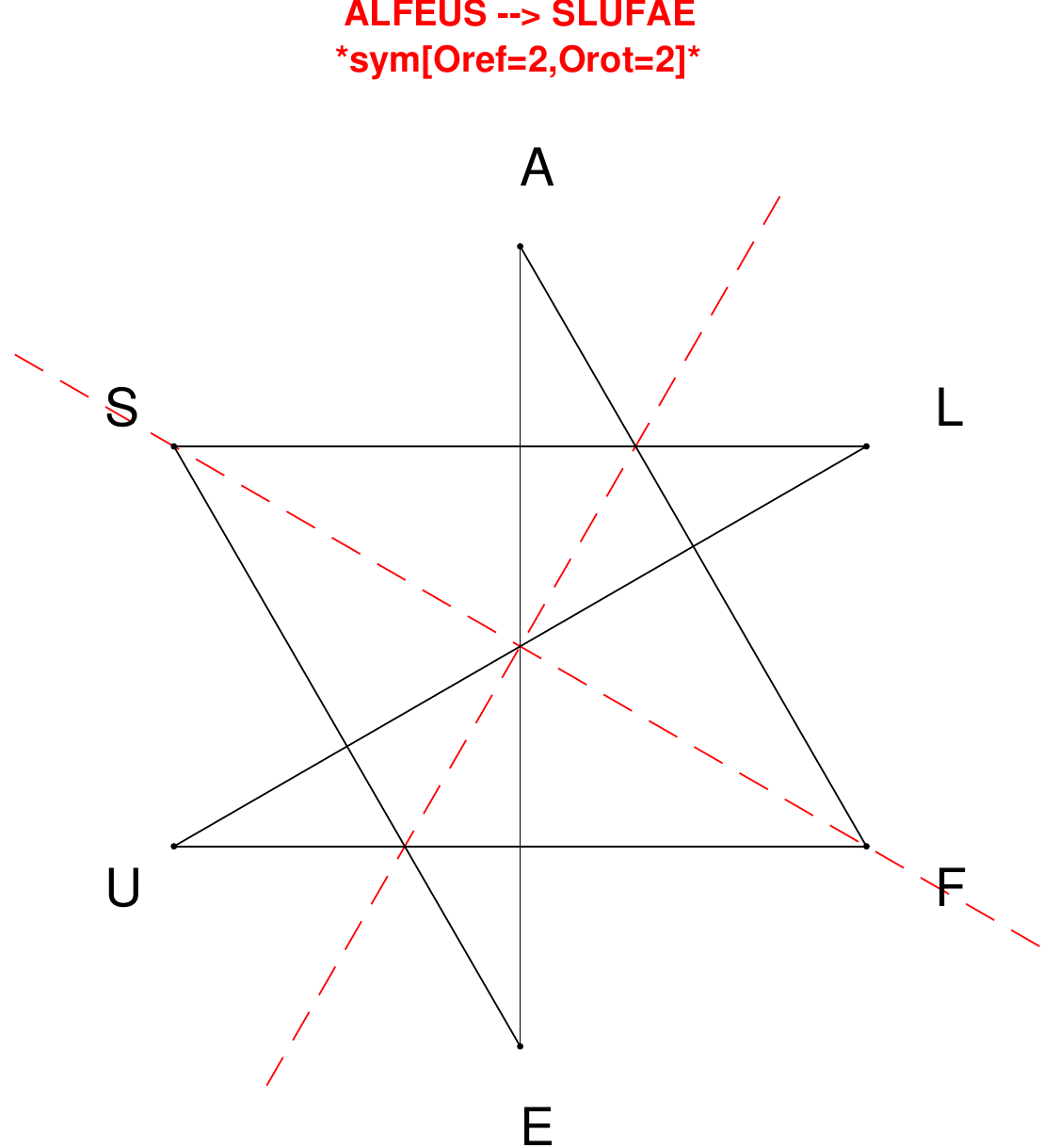}
\end{subfigure}
\hfill
\begin{subfigure}[T]{0.19\textwidth}
\centering
\includegraphics[width=\textwidth]{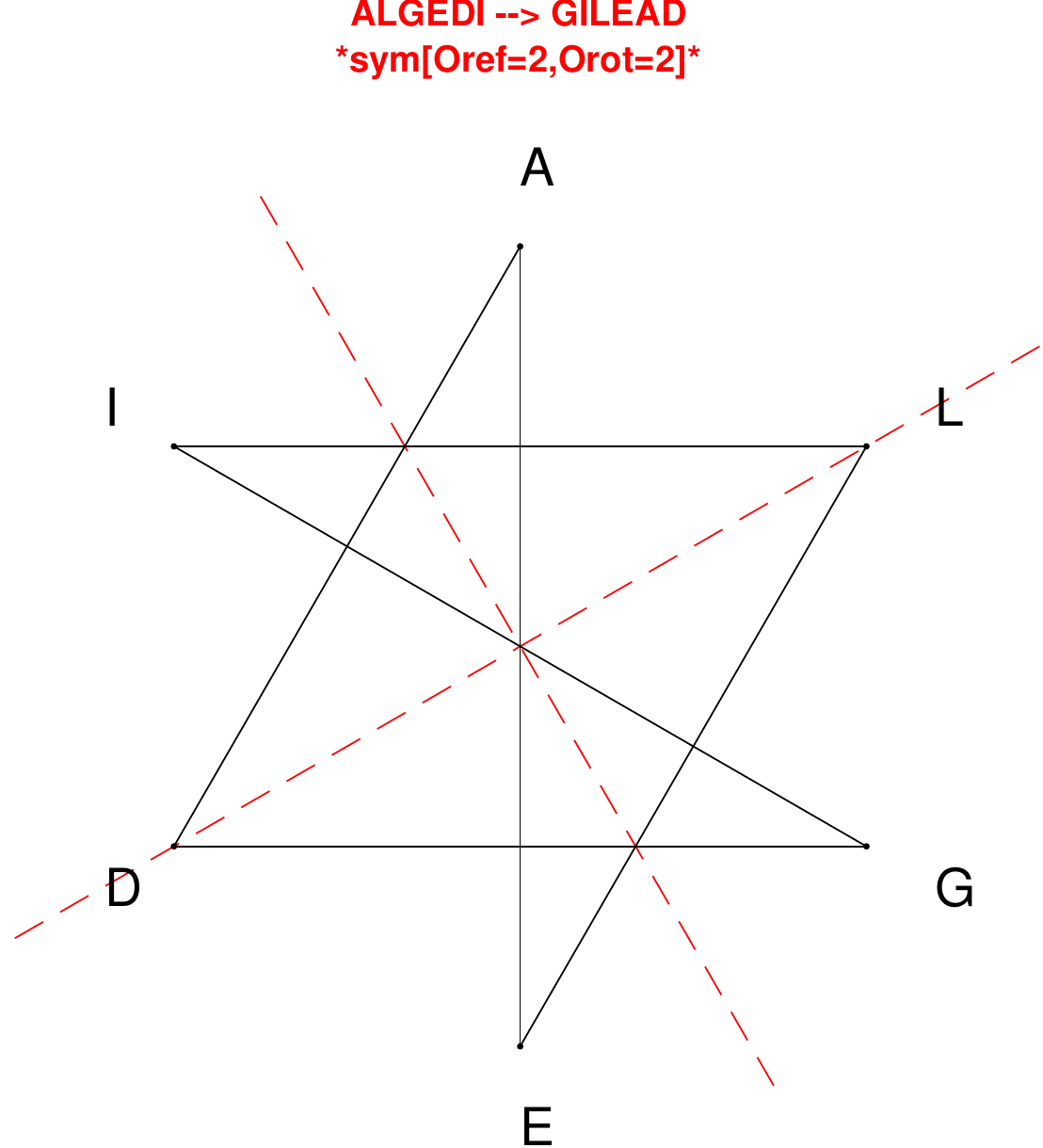}
\end{subfigure}
\hfill
\begin{subfigure}[T]{0.19\textwidth}
\centering
\includegraphics[width=\textwidth]{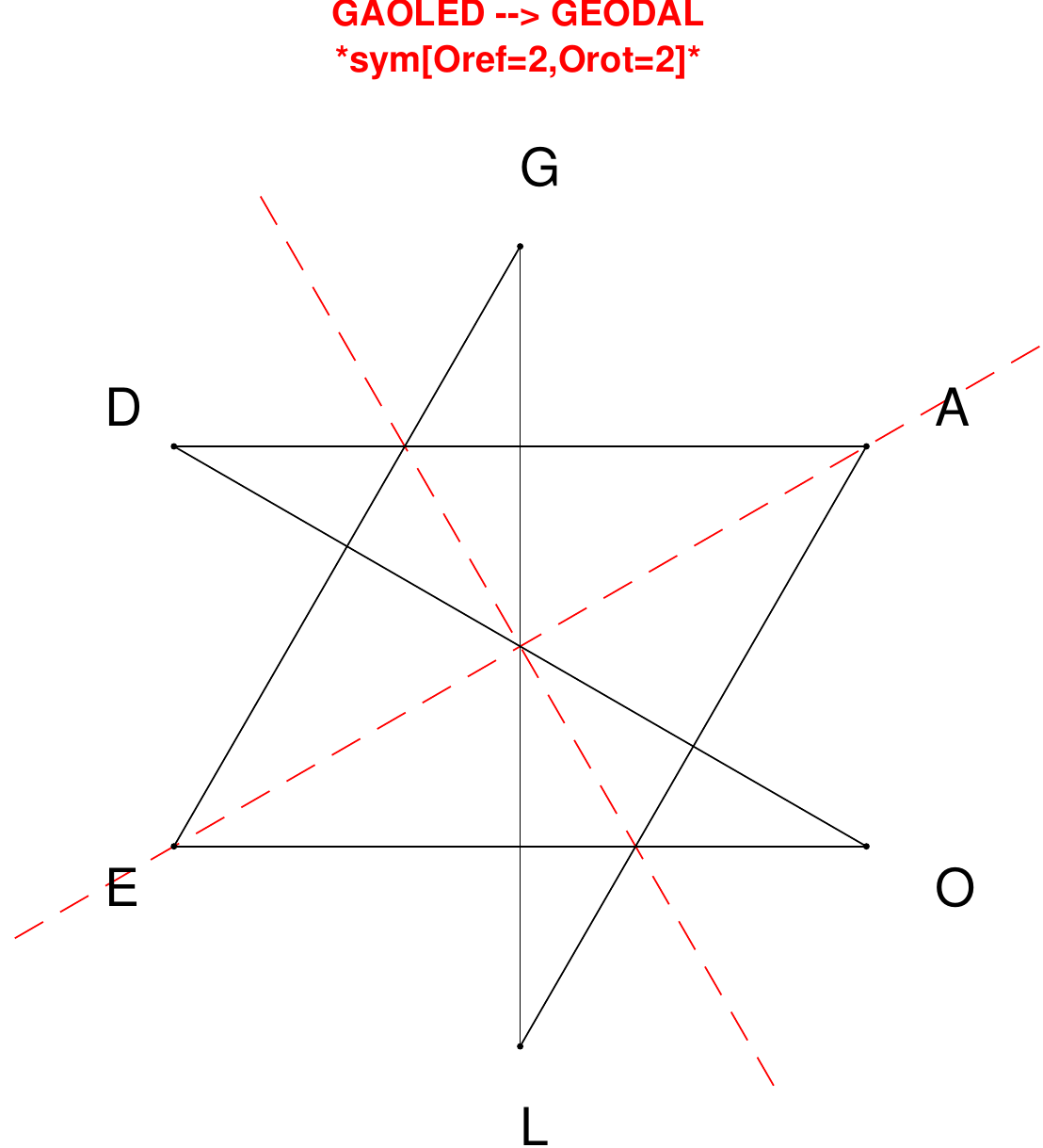}
\end{subfigure}
\hfill
\begin{subfigure}[T]{0.19\textwidth}
\centering
\includegraphics[width=\textwidth]{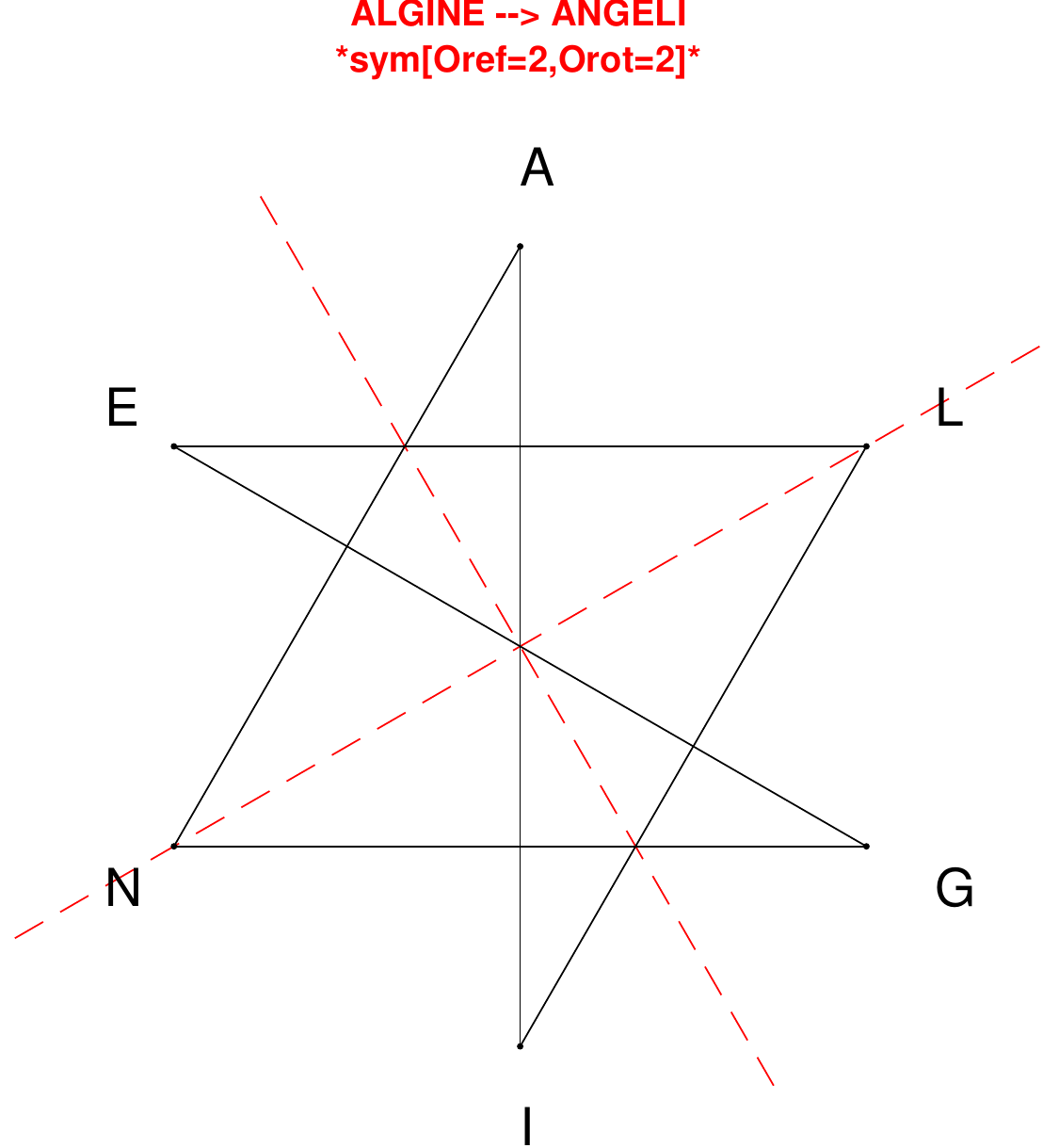}
\end{subfigure}
\end{figure}

\begin{figure}[H]
\centering
\begin{subfigure}[T]{0.19\textwidth}
\centering
\includegraphics[width=\textwidth]{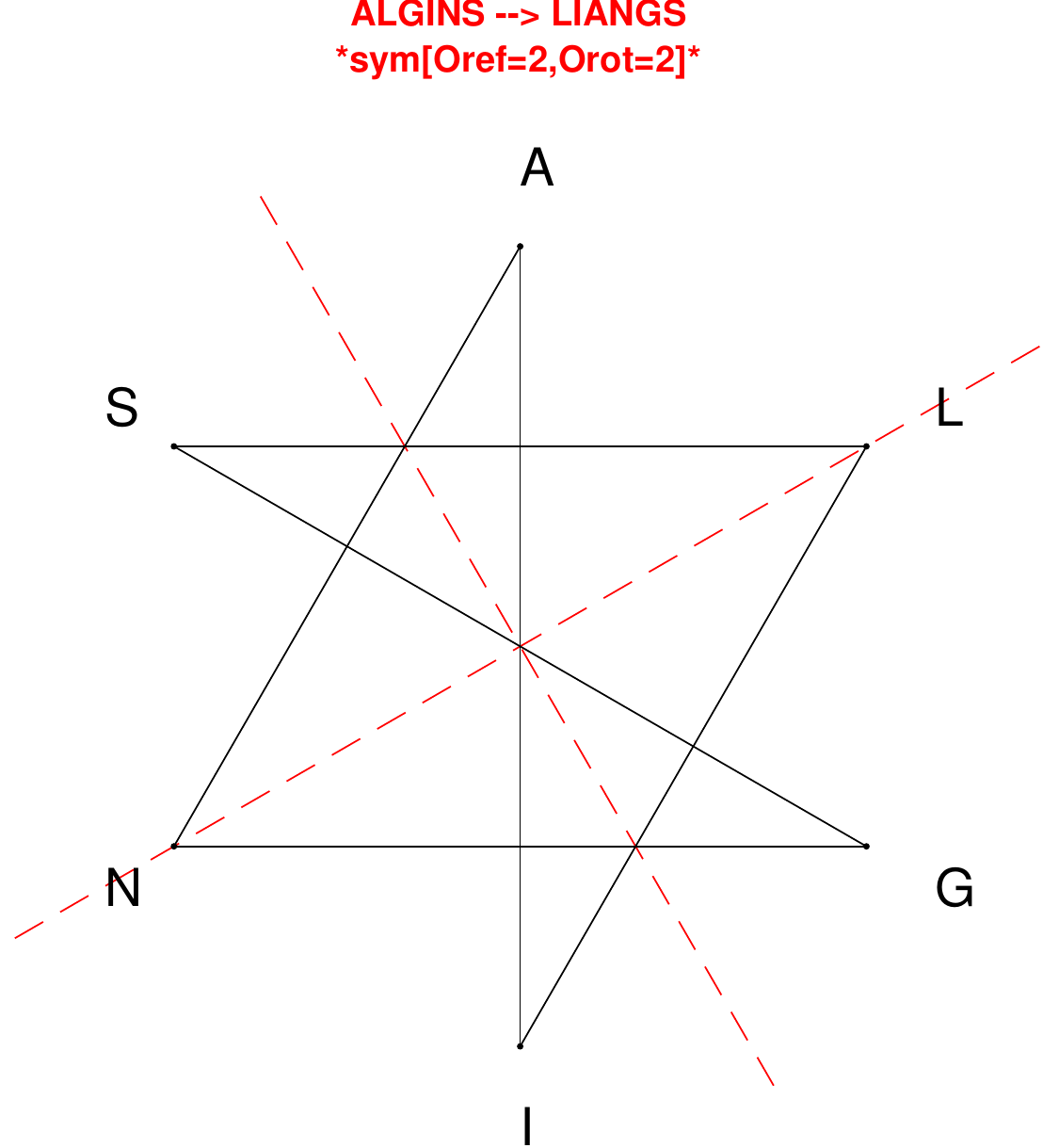}
\end{subfigure}
\hfill
\begin{subfigure}[T]{0.19\textwidth}
\centering
\includegraphics[width=\textwidth]{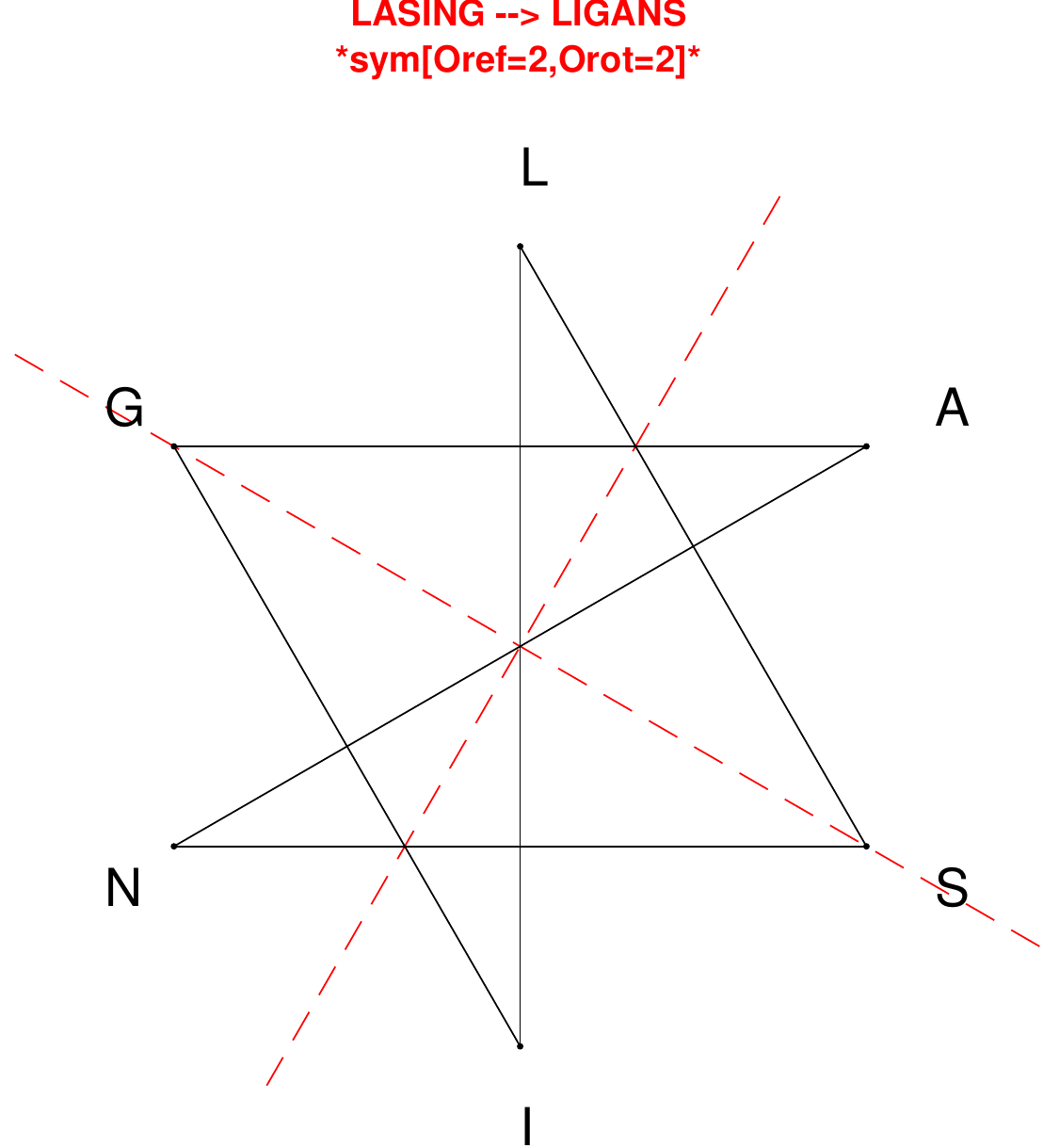}
\end{subfigure}
\hfill
\begin{subfigure}[T]{0.19\textwidth}
\centering
\includegraphics[width=\textwidth]{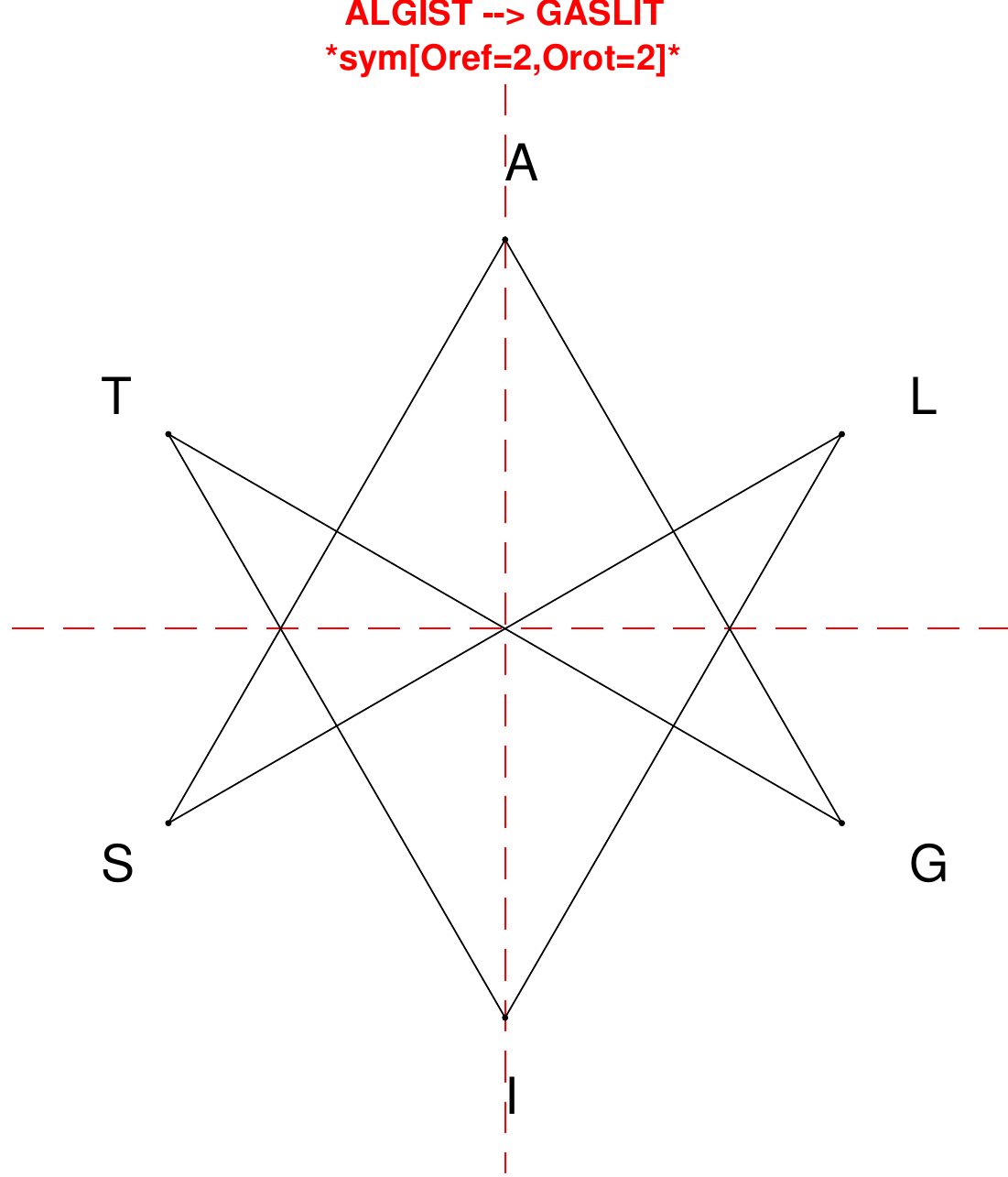}
\end{subfigure}
\hfill
\begin{subfigure}[T]{0.19\textwidth}
\centering
\includegraphics[width=\textwidth]{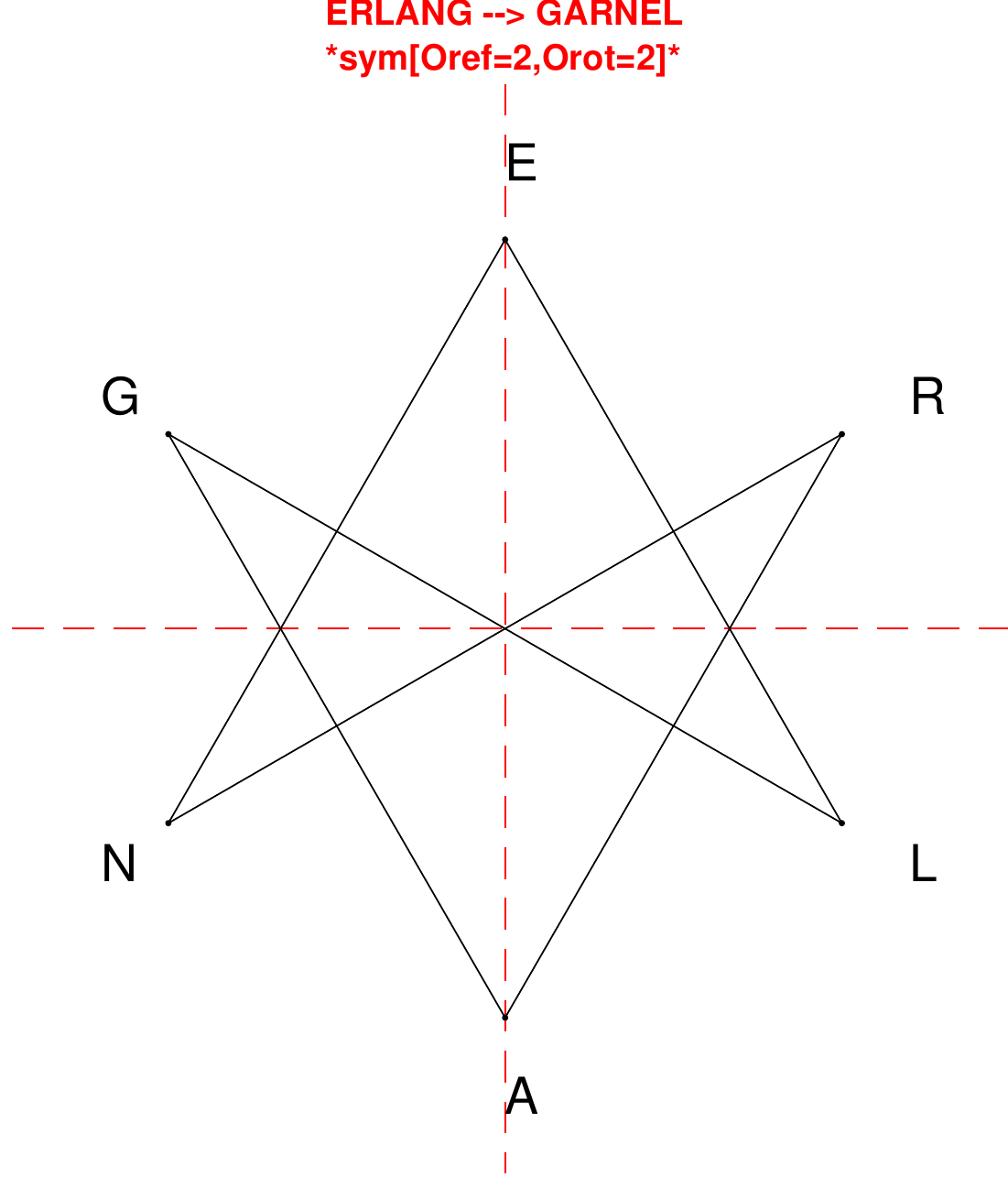}
\end{subfigure}
\hfill
\begin{subfigure}[T]{0.19\textwidth}
\centering
\includegraphics[width=\textwidth]{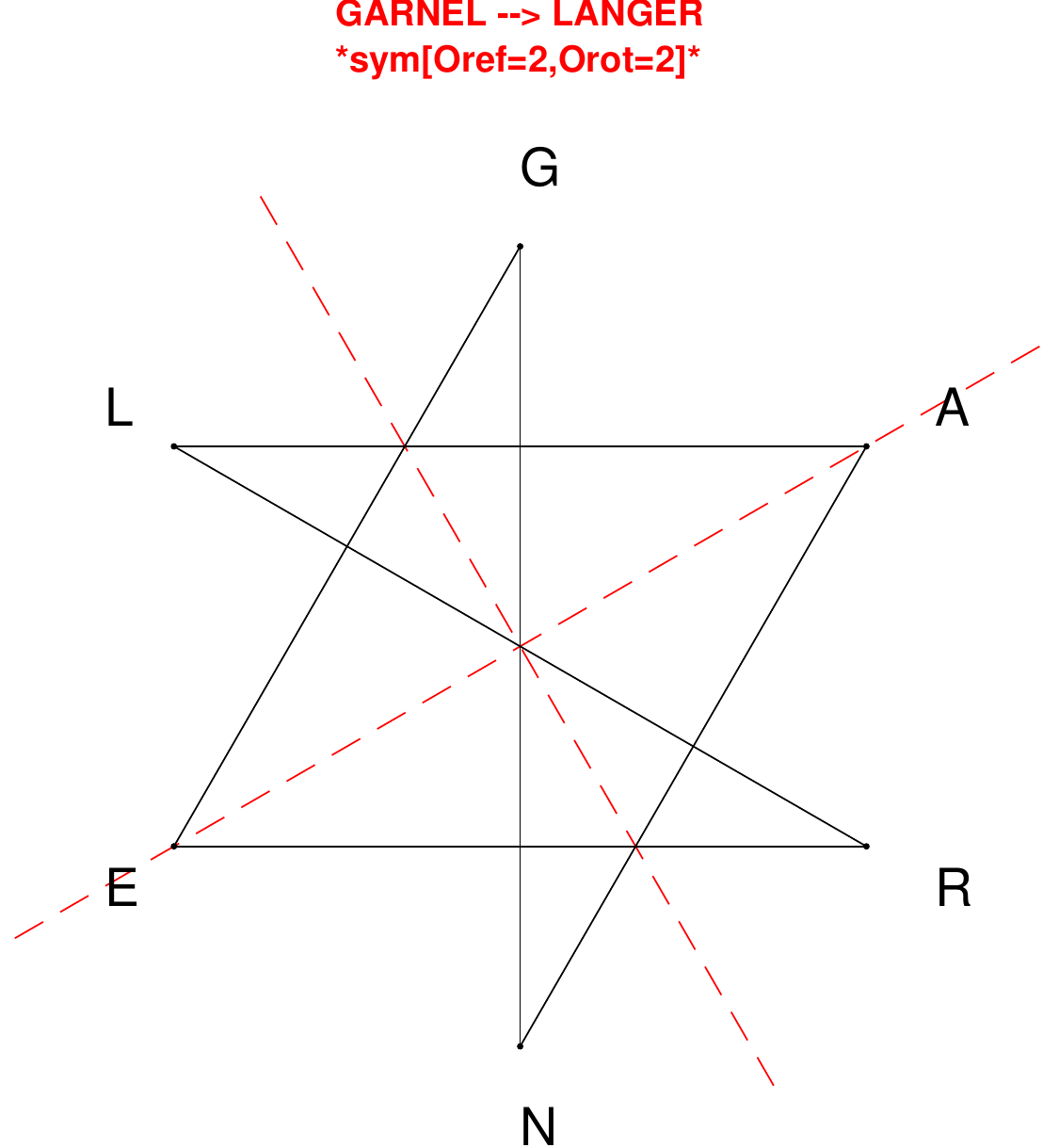}
\end{subfigure}
\end{figure}

\begin{figure}[H]
\centering
\begin{subfigure}[T]{0.19\textwidth}
\centering
\includegraphics[width=\textwidth]{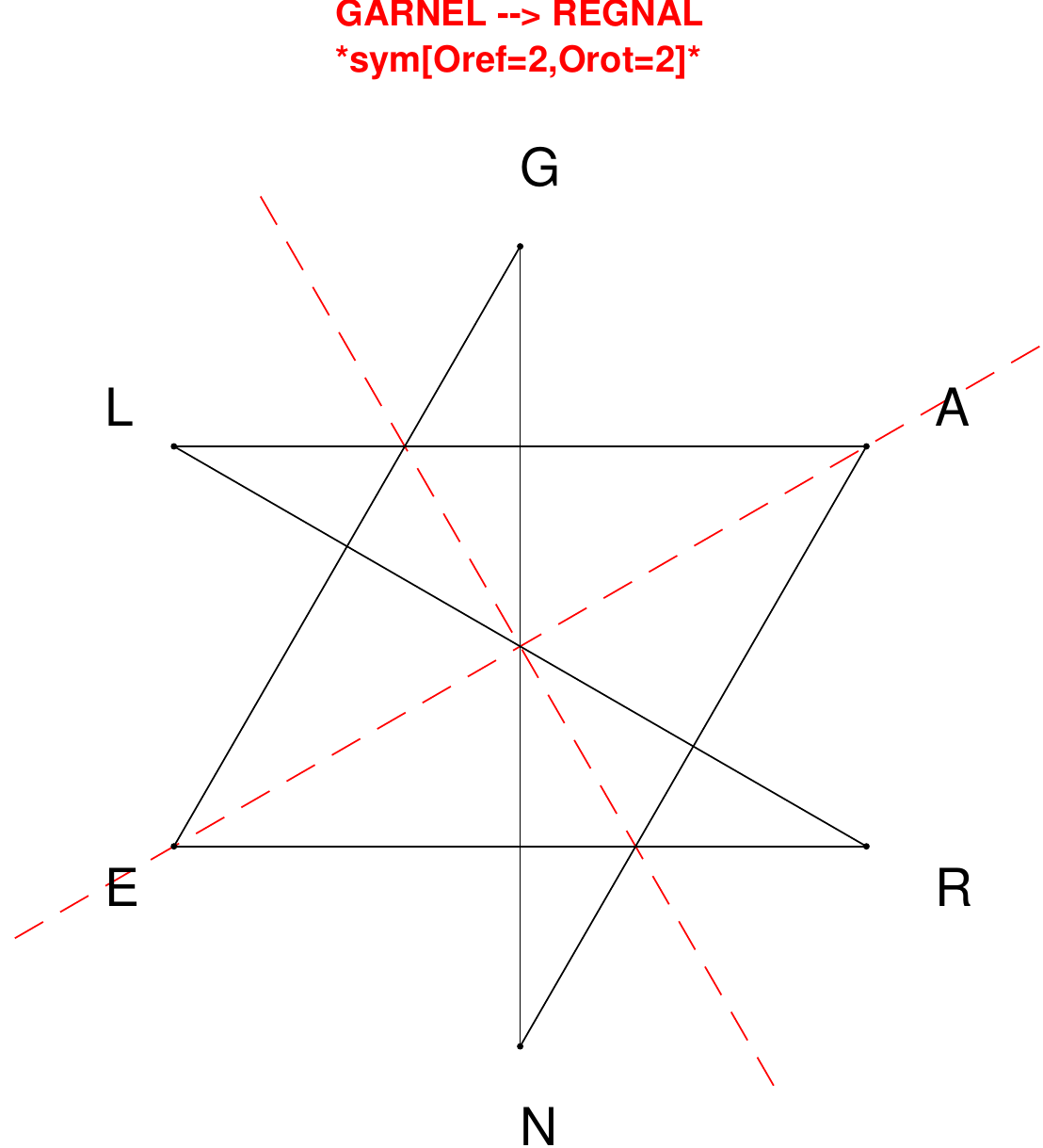}
\end{subfigure}
\hfill
\begin{subfigure}[T]{0.19\textwidth}
\centering
\includegraphics[width=\textwidth]{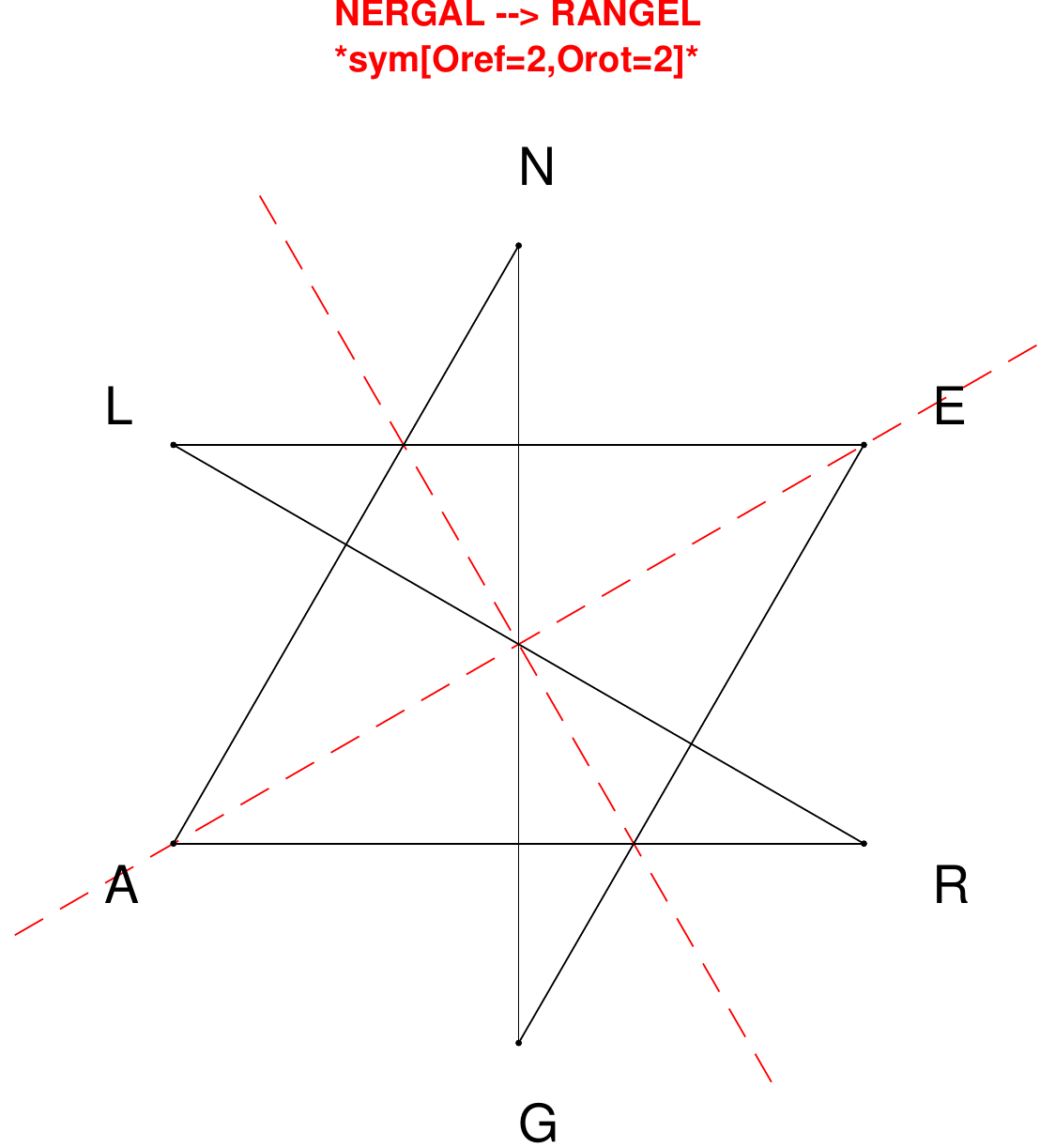}
\end{subfigure}
\hfill
\begin{subfigure}[T]{0.19\textwidth}
\centering
\includegraphics[width=\textwidth]{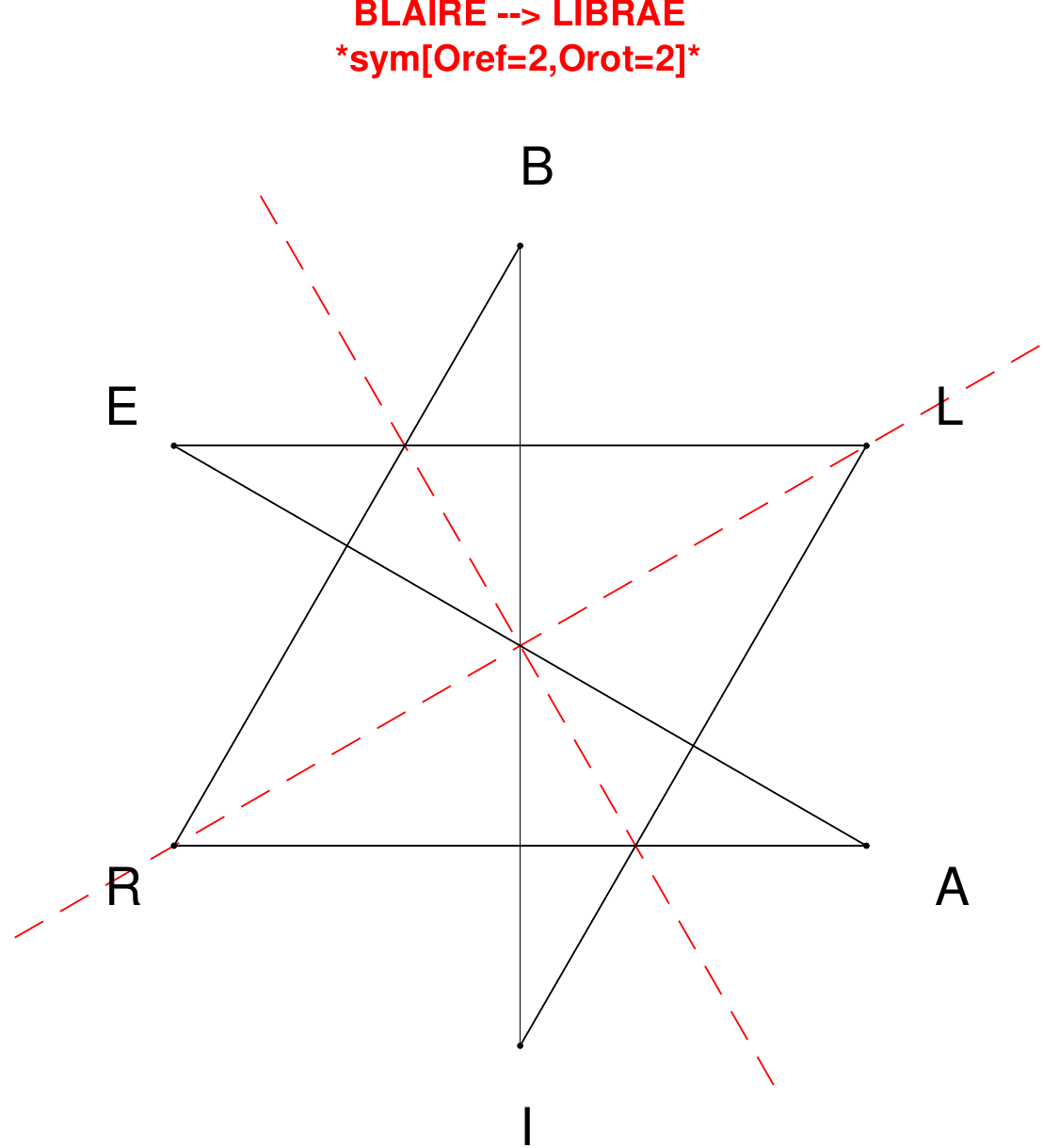}
\end{subfigure}
\hfill
\begin{subfigure}[T]{0.19\textwidth}
\centering
\includegraphics[width=\textwidth]{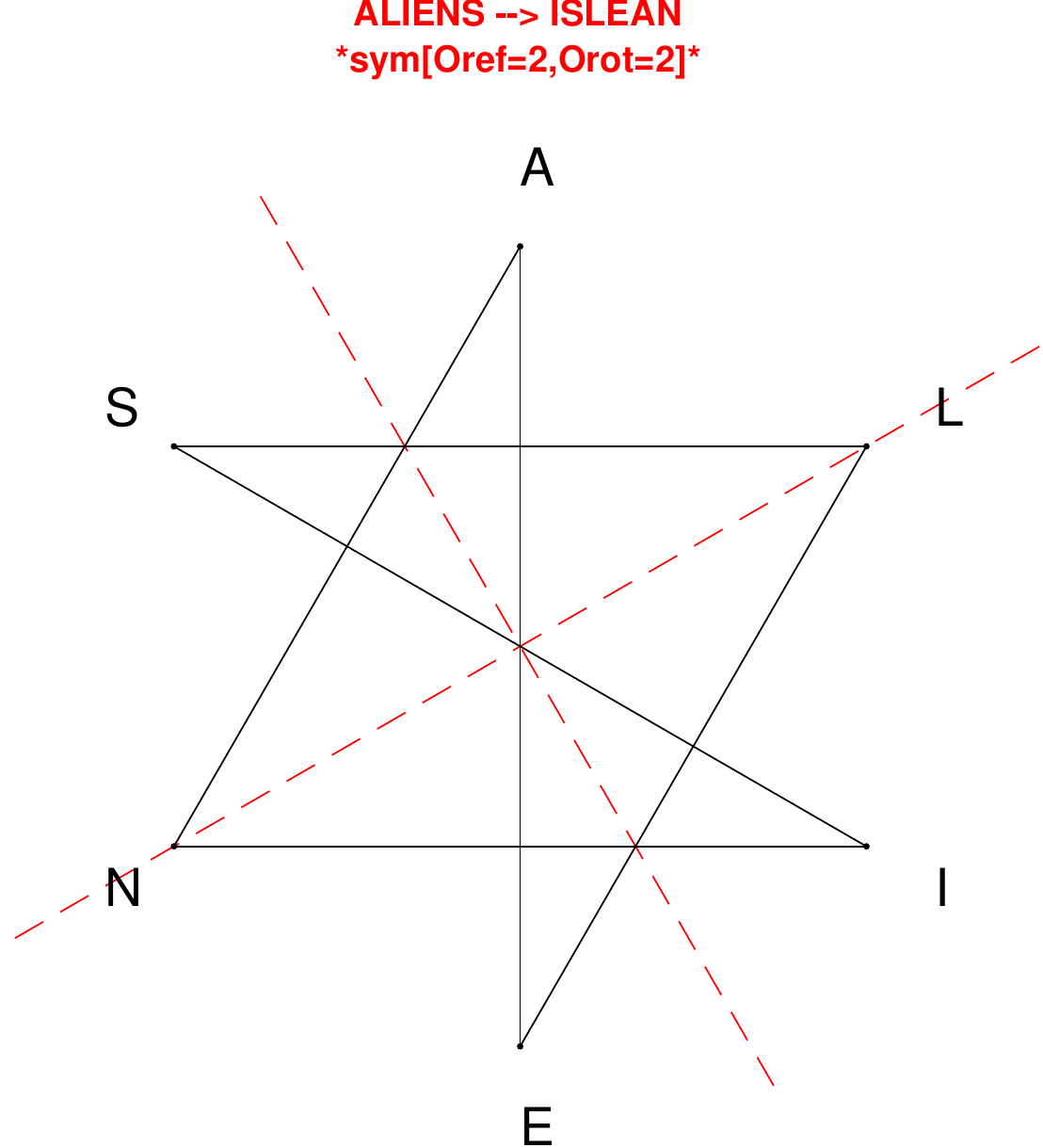}
\end{subfigure}
\hfill
\begin{subfigure}[T]{0.19\textwidth}
\centering
\includegraphics[width=\textwidth]{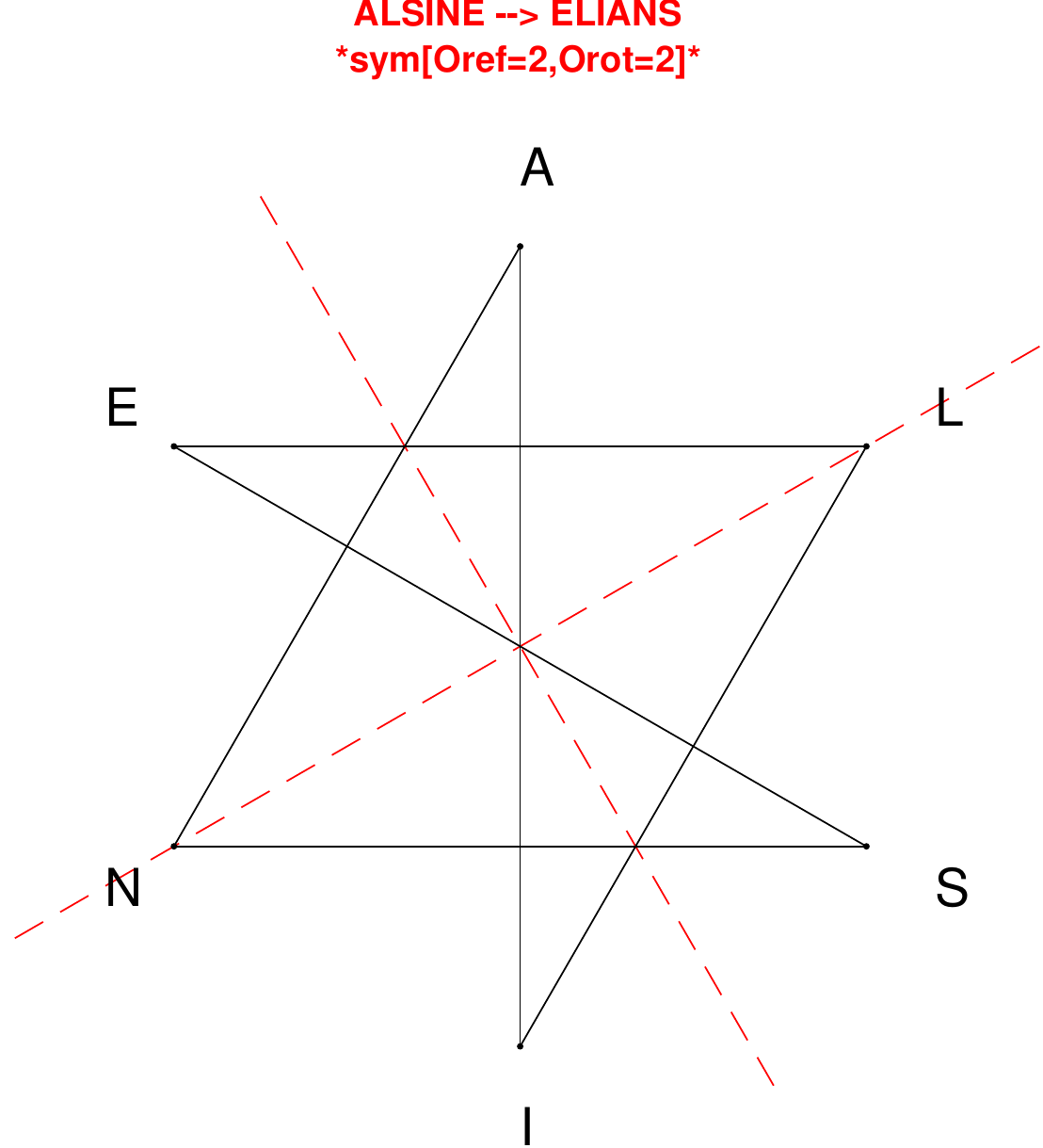}
\end{subfigure}
\end{figure}

\begin{figure}[H]
\centering
\begin{subfigure}[T]{0.19\textwidth}
\centering
\includegraphics[width=\textwidth]{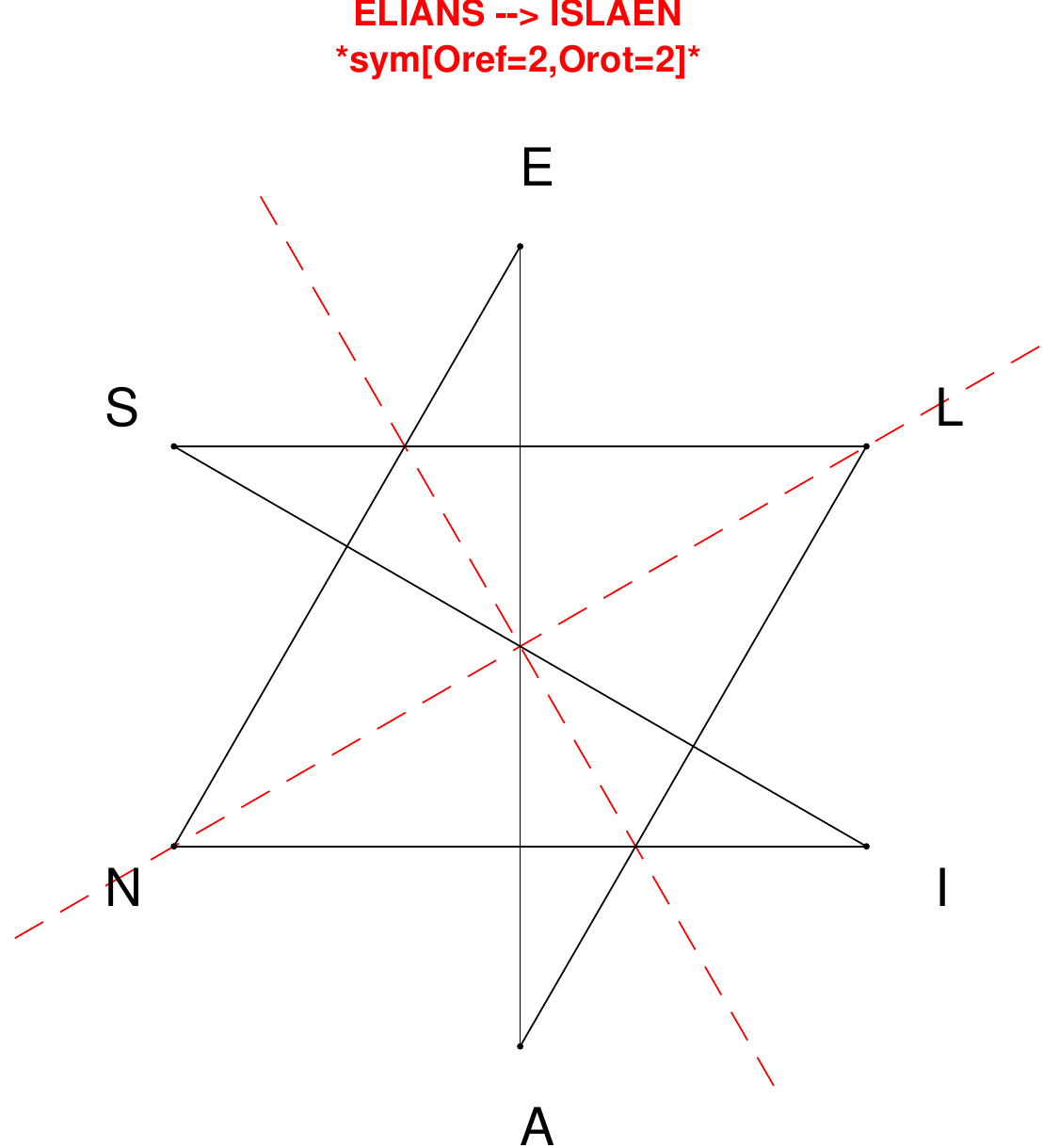}
\end{subfigure}
\hfill
\begin{subfigure}[T]{0.19\textwidth}
\centering
\includegraphics[width=\textwidth]{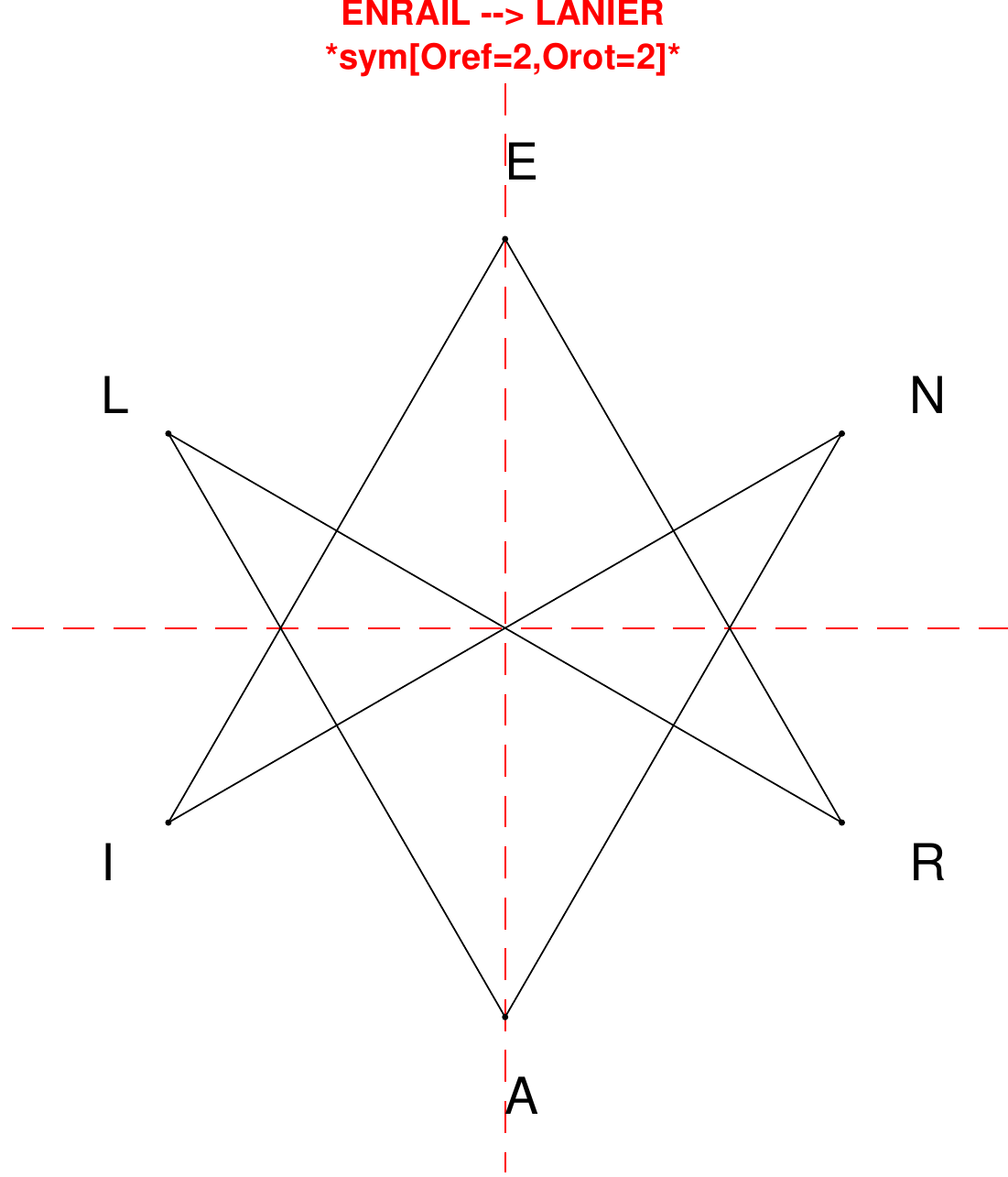}
\end{subfigure}
\hfill
\begin{subfigure}[T]{0.19\textwidth}
\centering
\includegraphics[width=\textwidth]{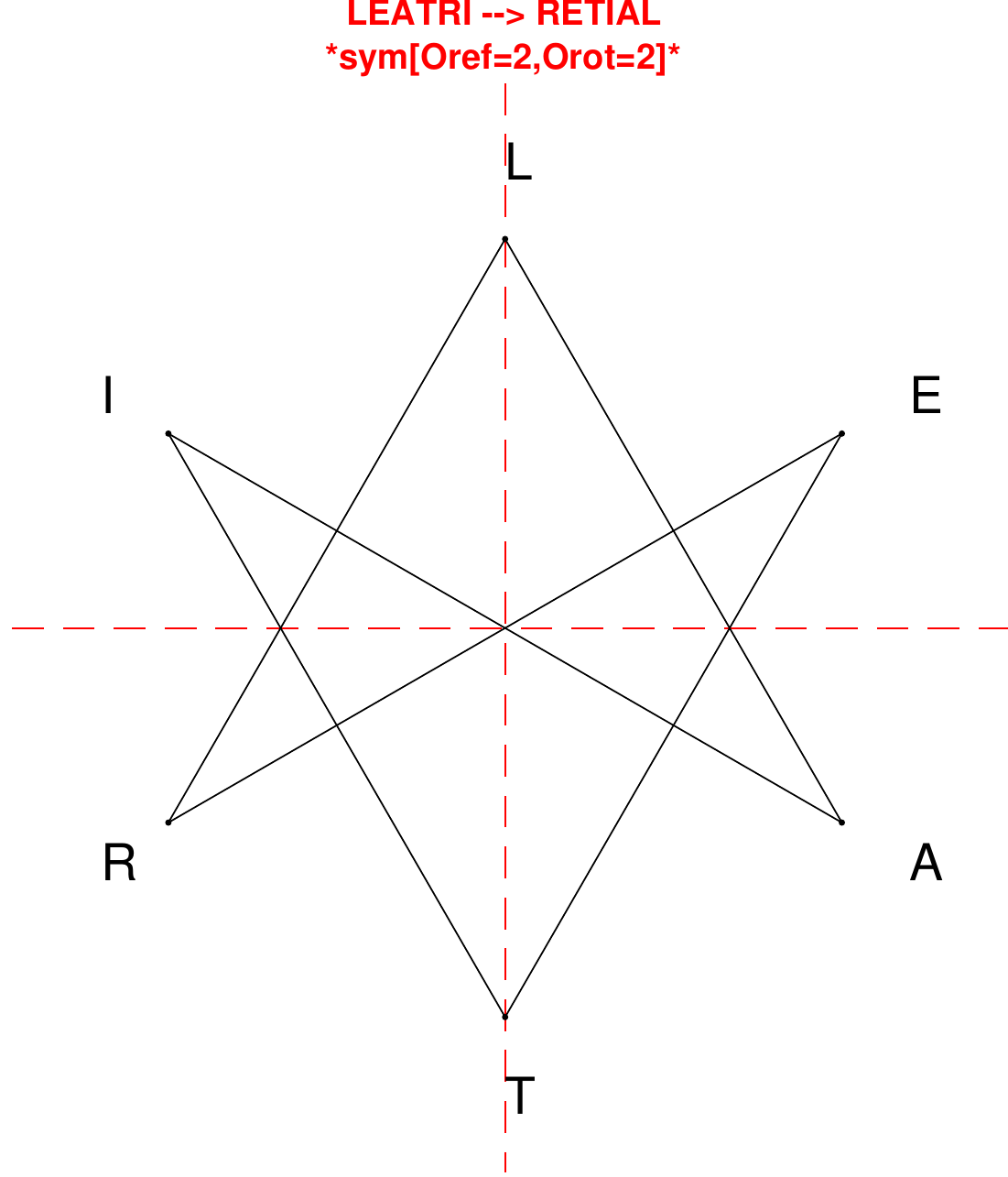}
\end{subfigure}
\hfill
\begin{subfigure}[T]{0.19\textwidth}
\centering
\includegraphics[width=\textwidth]{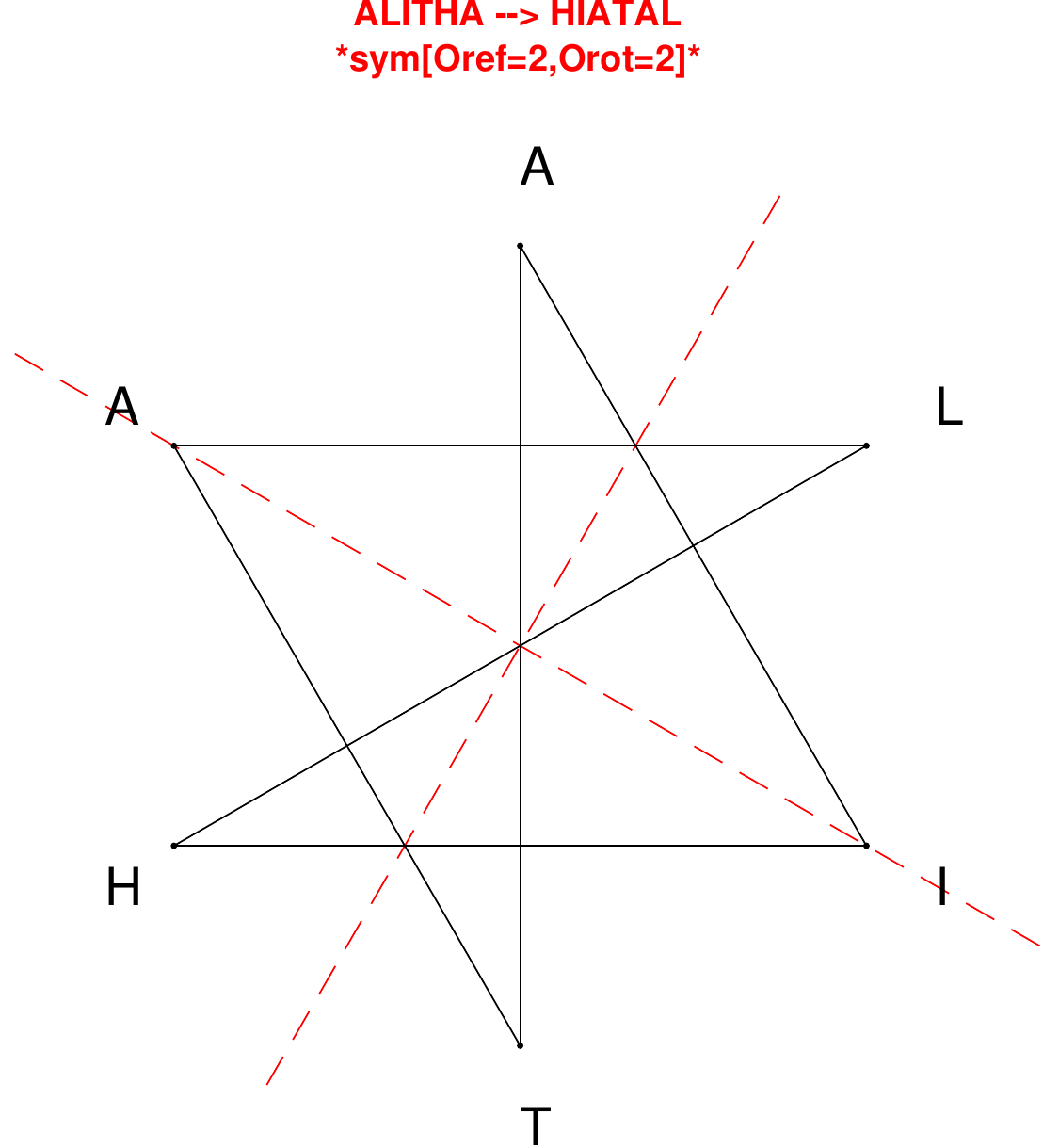}
\end{subfigure}
\hfill
\begin{subfigure}[T]{0.19\textwidth}
\centering
\includegraphics[width=\textwidth]{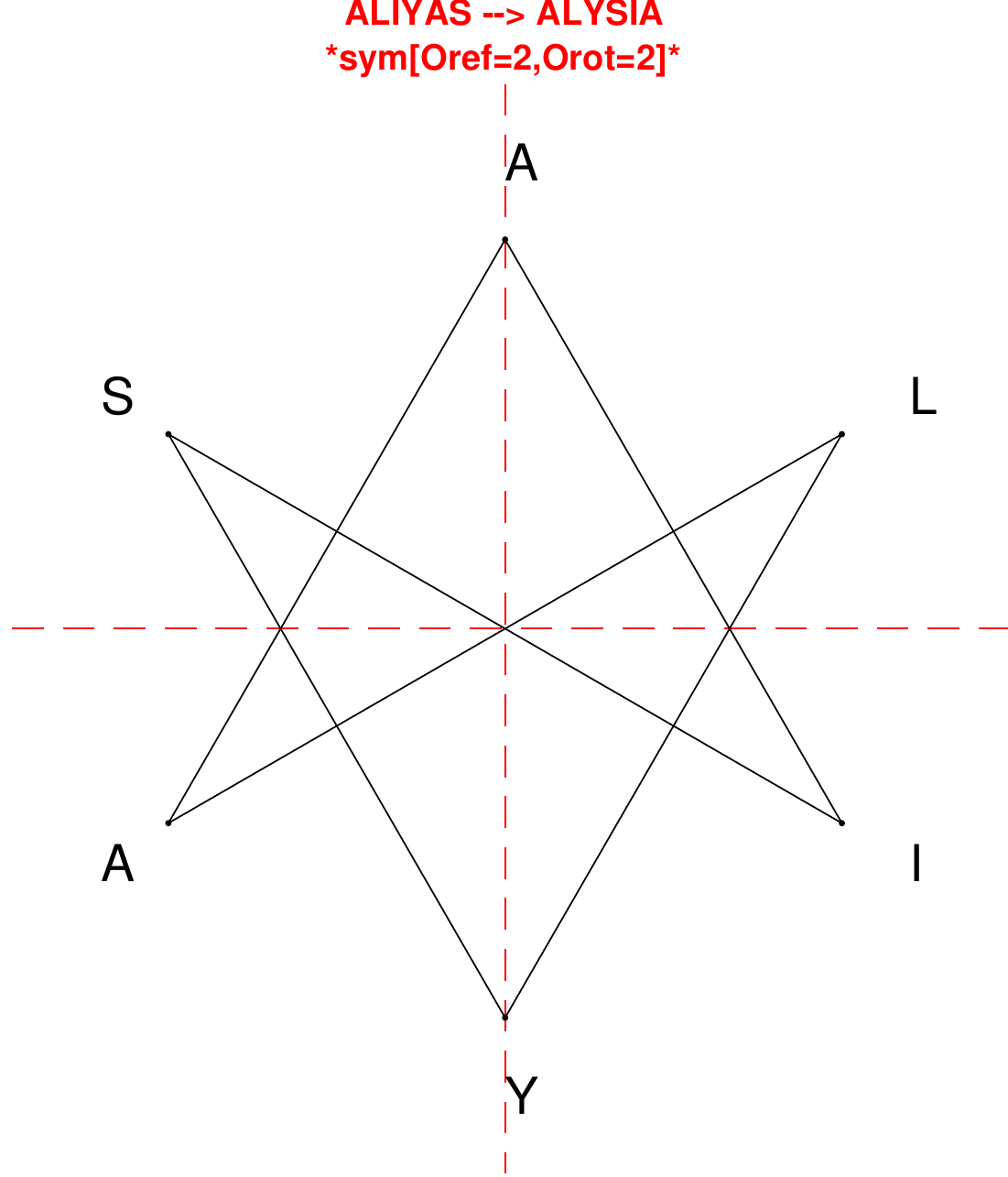}
\end{subfigure}
\end{figure}

\begin{figure}[H]
\centering
\begin{subfigure}[T]{0.19\textwidth}
\centering
\includegraphics[width=\textwidth]{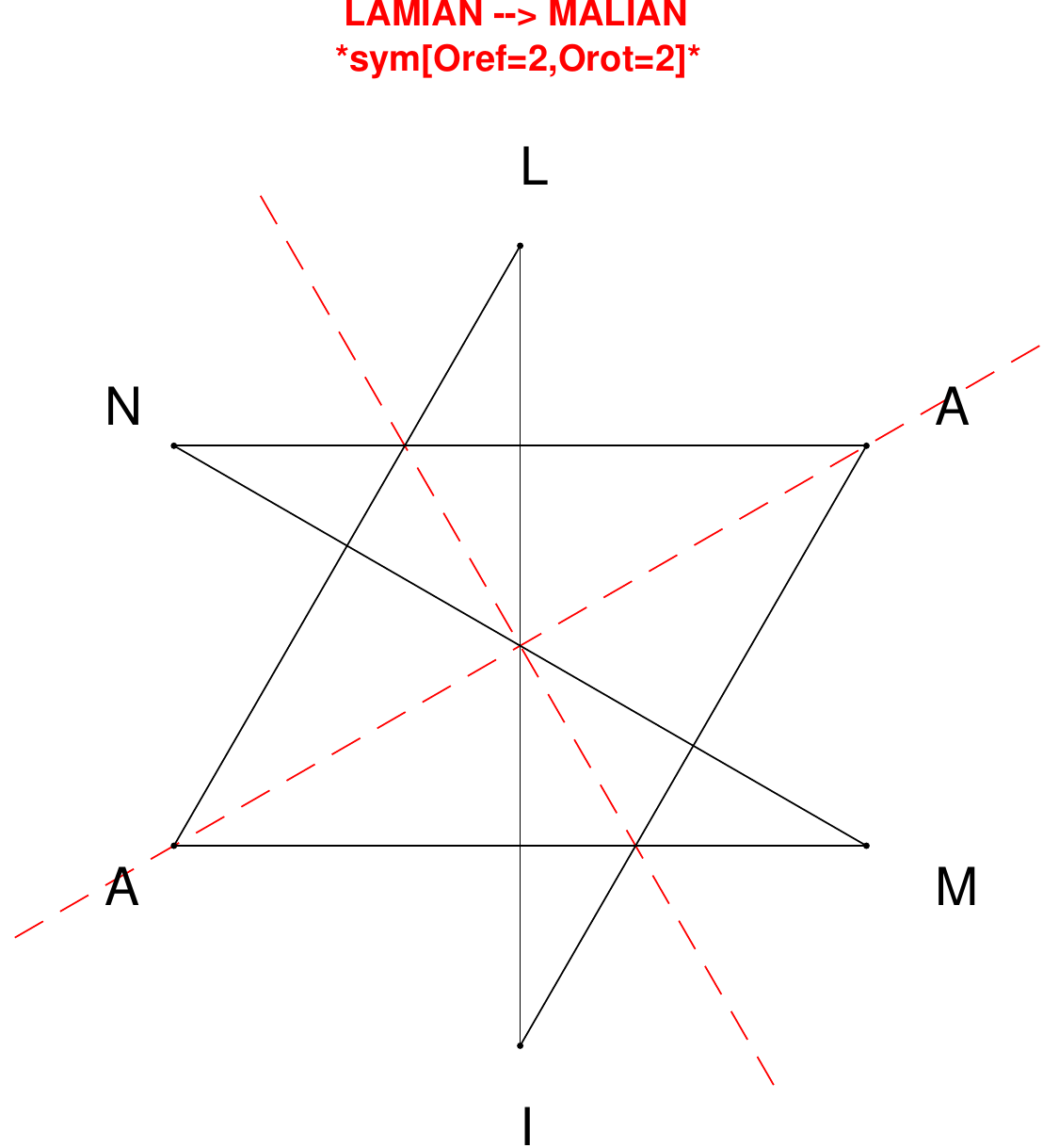}
\end{subfigure}
\hfill
\begin{subfigure}[T]{0.19\textwidth}
\centering
\includegraphics[width=\textwidth]{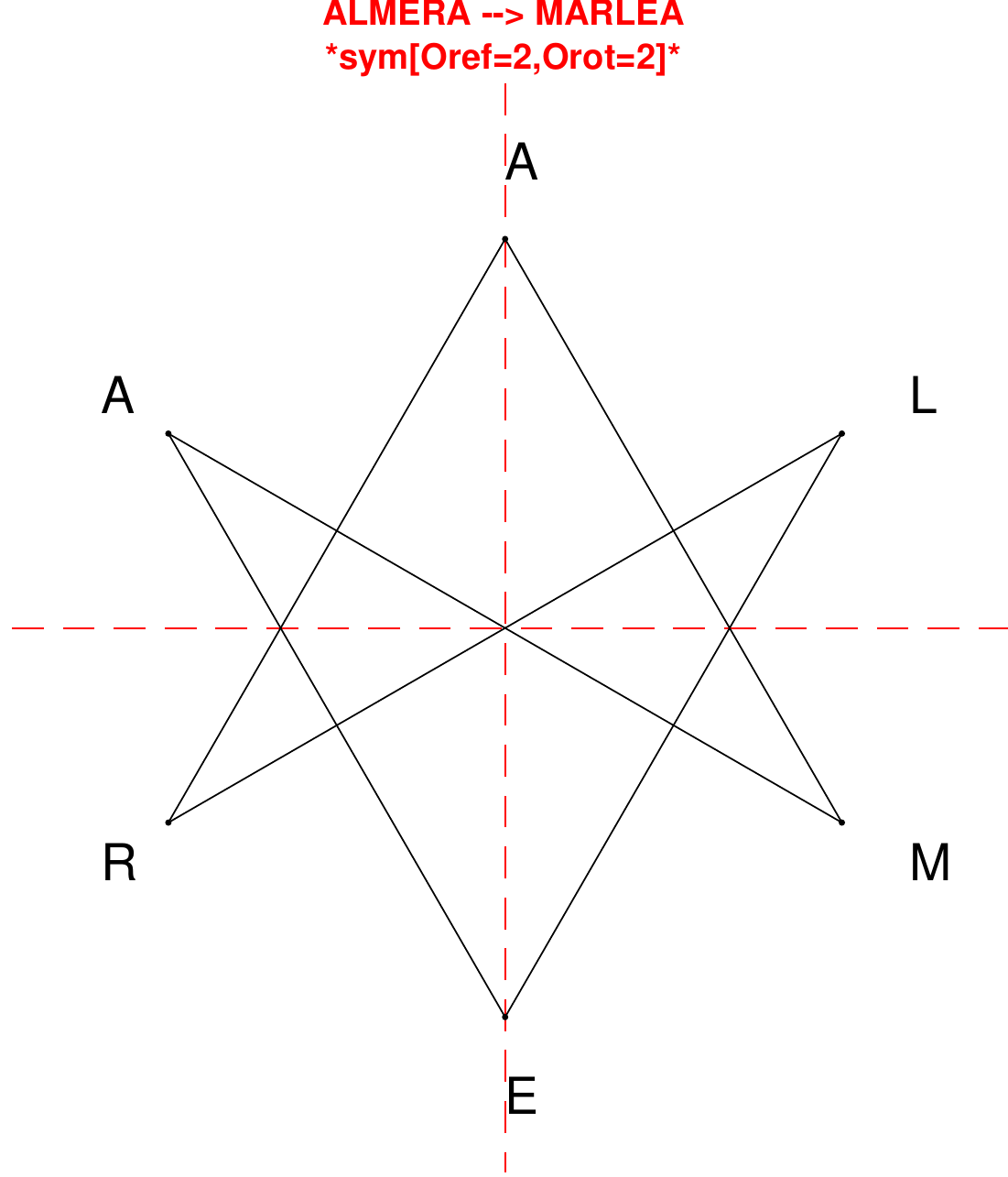}
\end{subfigure}
\hfill
\begin{subfigure}[T]{0.19\textwidth}
\centering
\includegraphics[width=\textwidth]{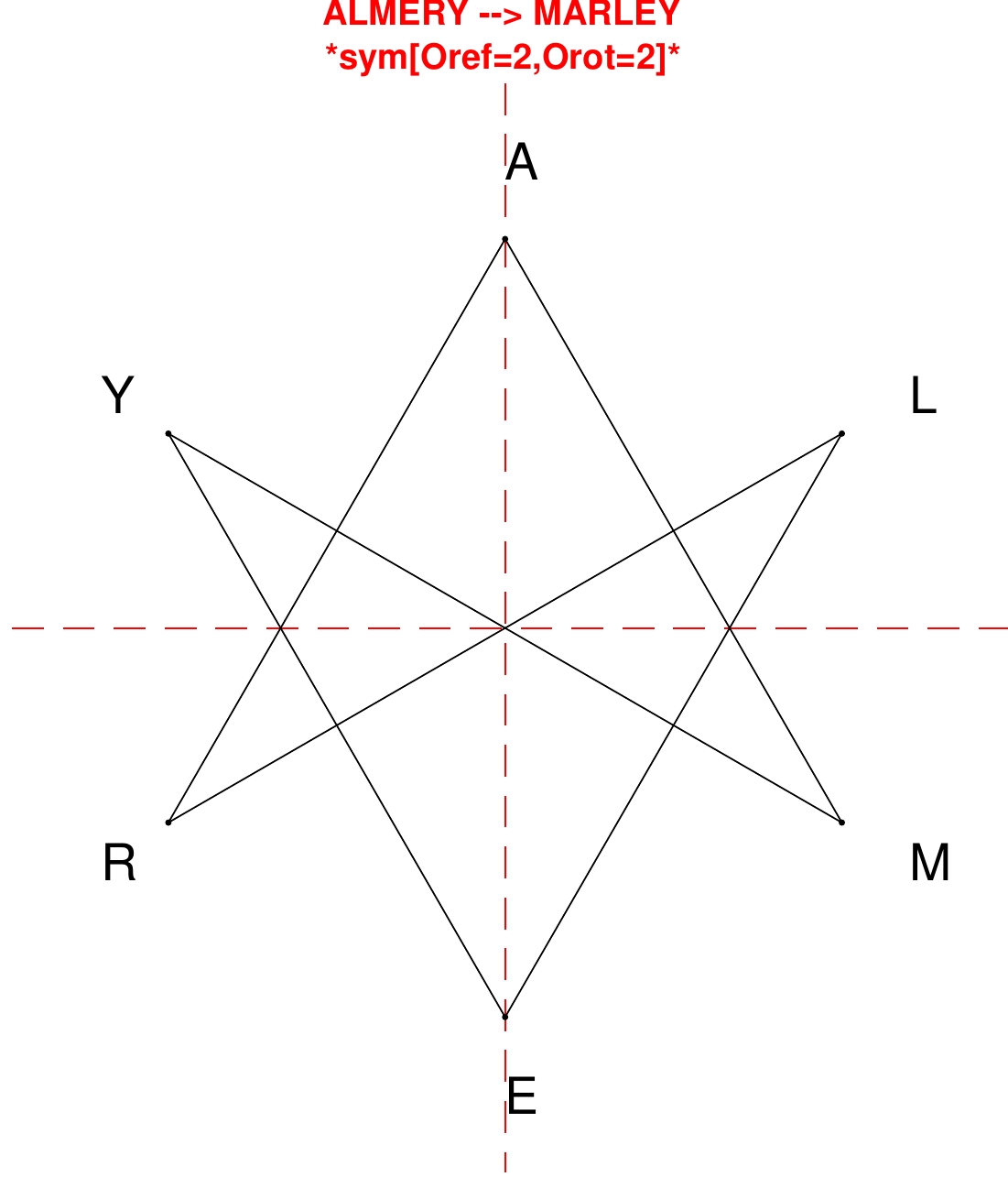}
\end{subfigure}
\hfill
\begin{subfigure}[T]{0.19\textwidth}
\centering
\includegraphics[width=\textwidth]{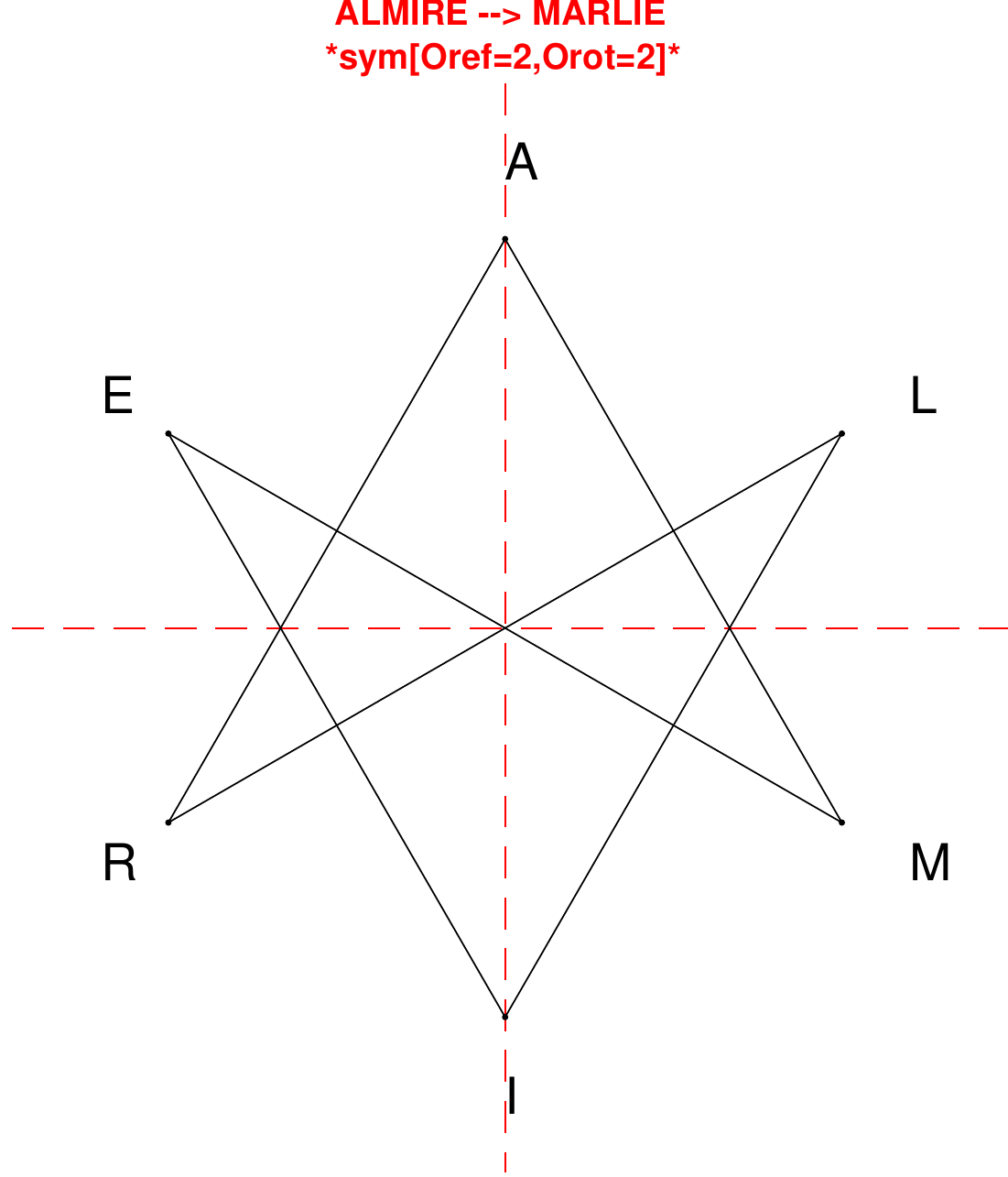}
\end{subfigure}
\hfill
\begin{subfigure}[T]{0.19\textwidth}
\centering
\includegraphics[width=\textwidth]{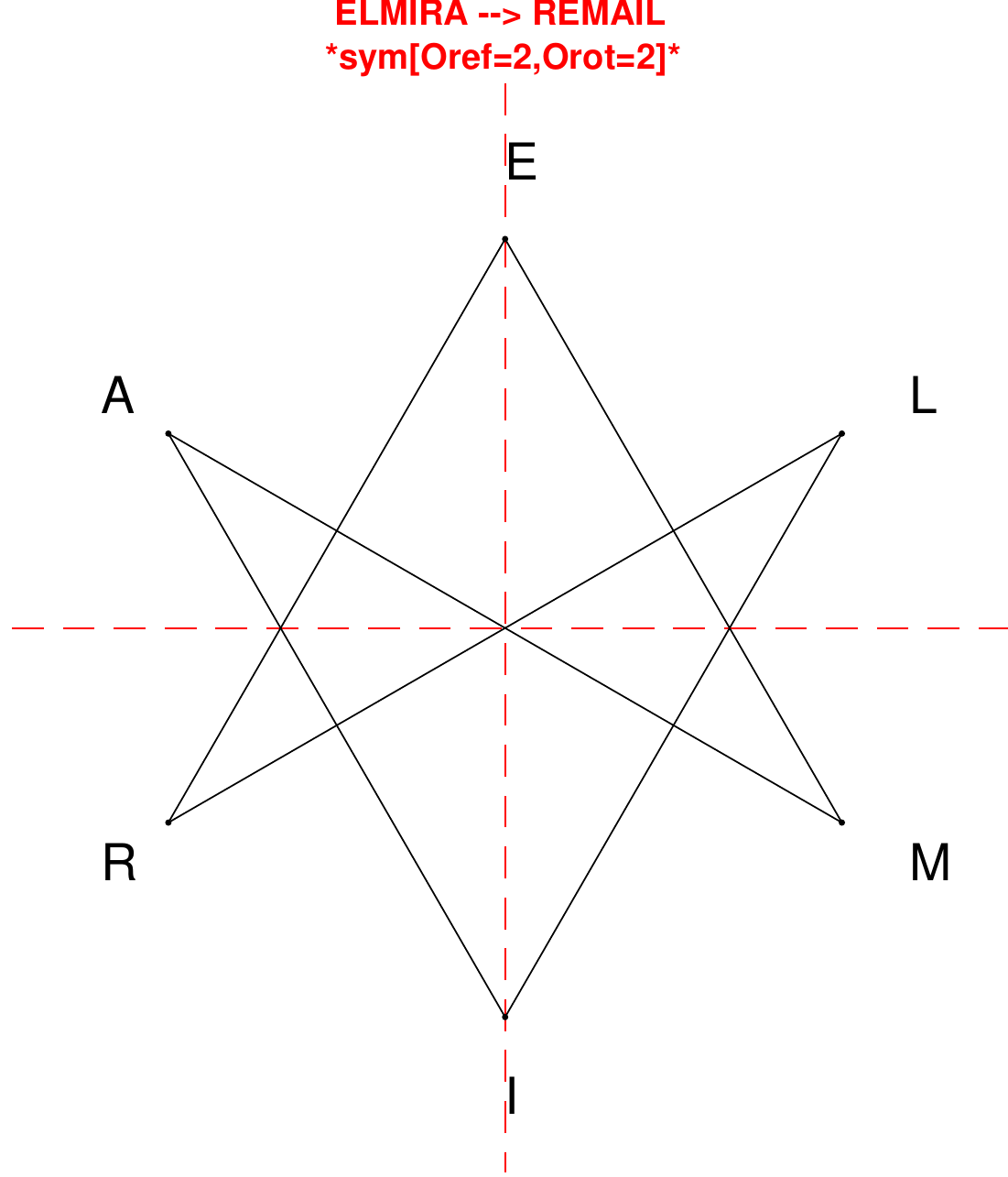}
\end{subfigure}
\end{figure}

\begin{figure}[H]
\centering
\begin{subfigure}[T]{0.19\textwidth}
\centering
\includegraphics[width=\textwidth]{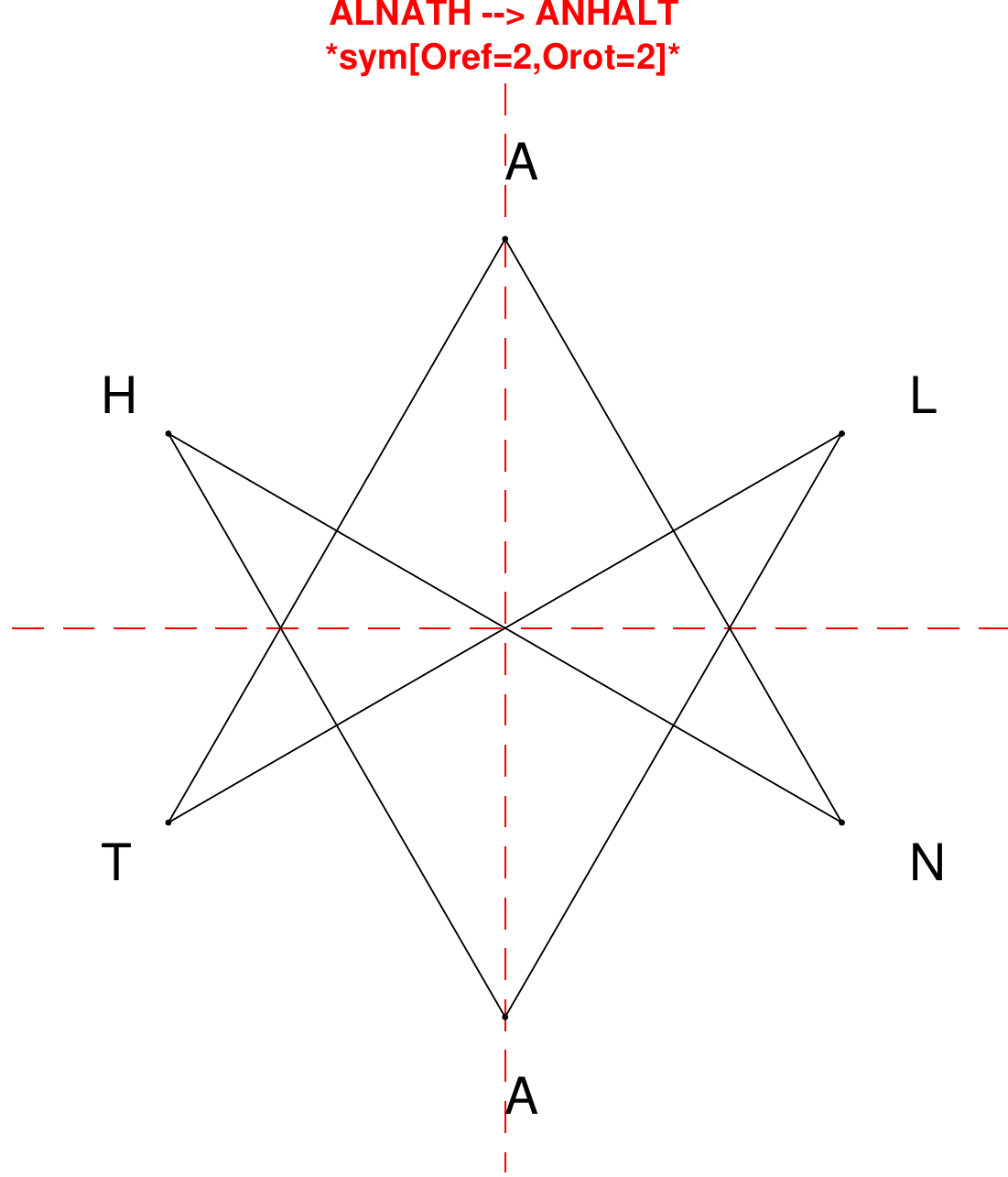}
\end{subfigure}
\hfill
\begin{subfigure}[T]{0.19\textwidth}
\centering
\includegraphics[width=\textwidth]{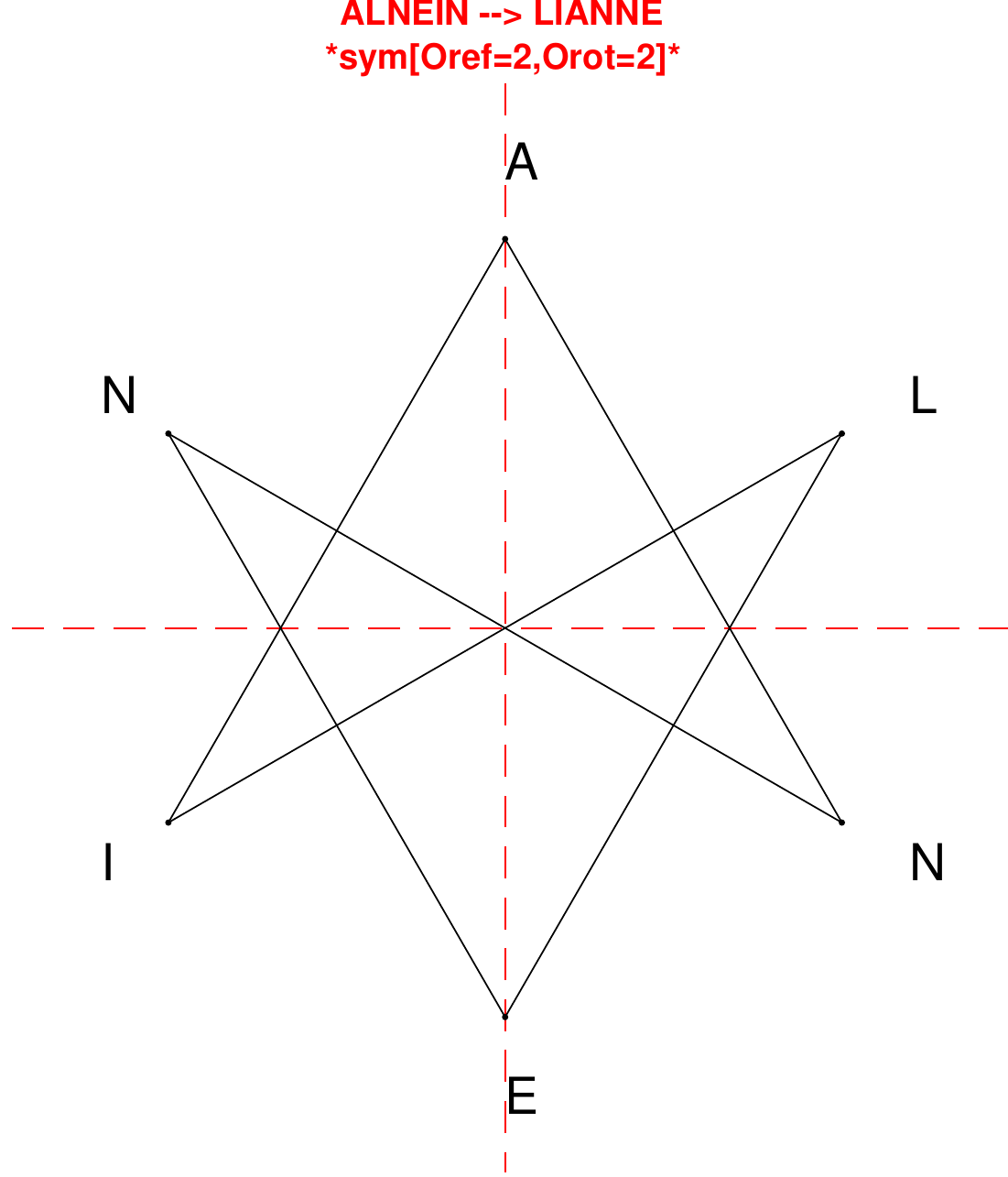}
\end{subfigure}
\hfill
\begin{subfigure}[T]{0.19\textwidth}
\centering
\includegraphics[width=\textwidth]{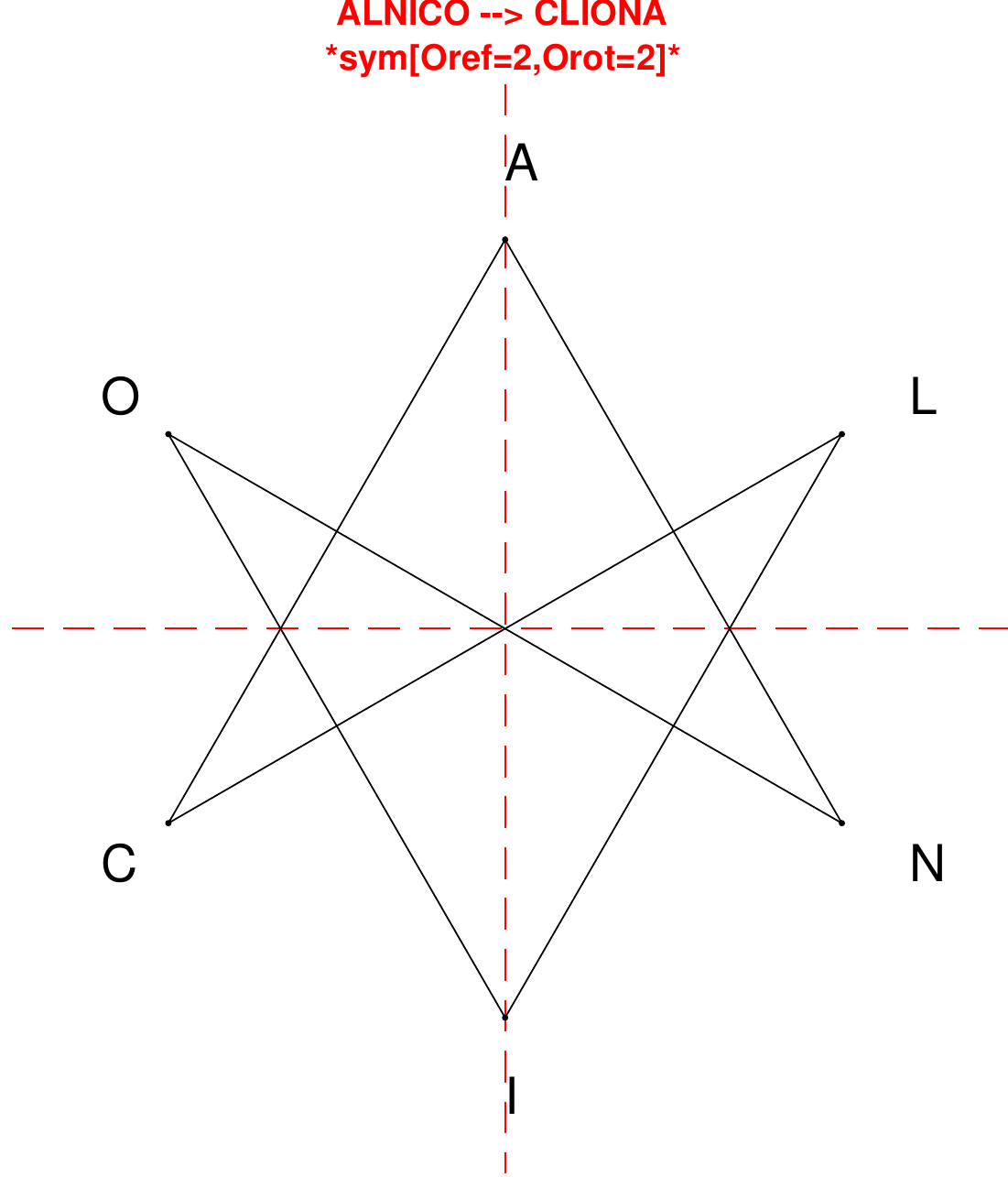}
\end{subfigure}
\hfill
\begin{subfigure}[T]{0.19\textwidth}
\centering
\includegraphics[width=\textwidth]{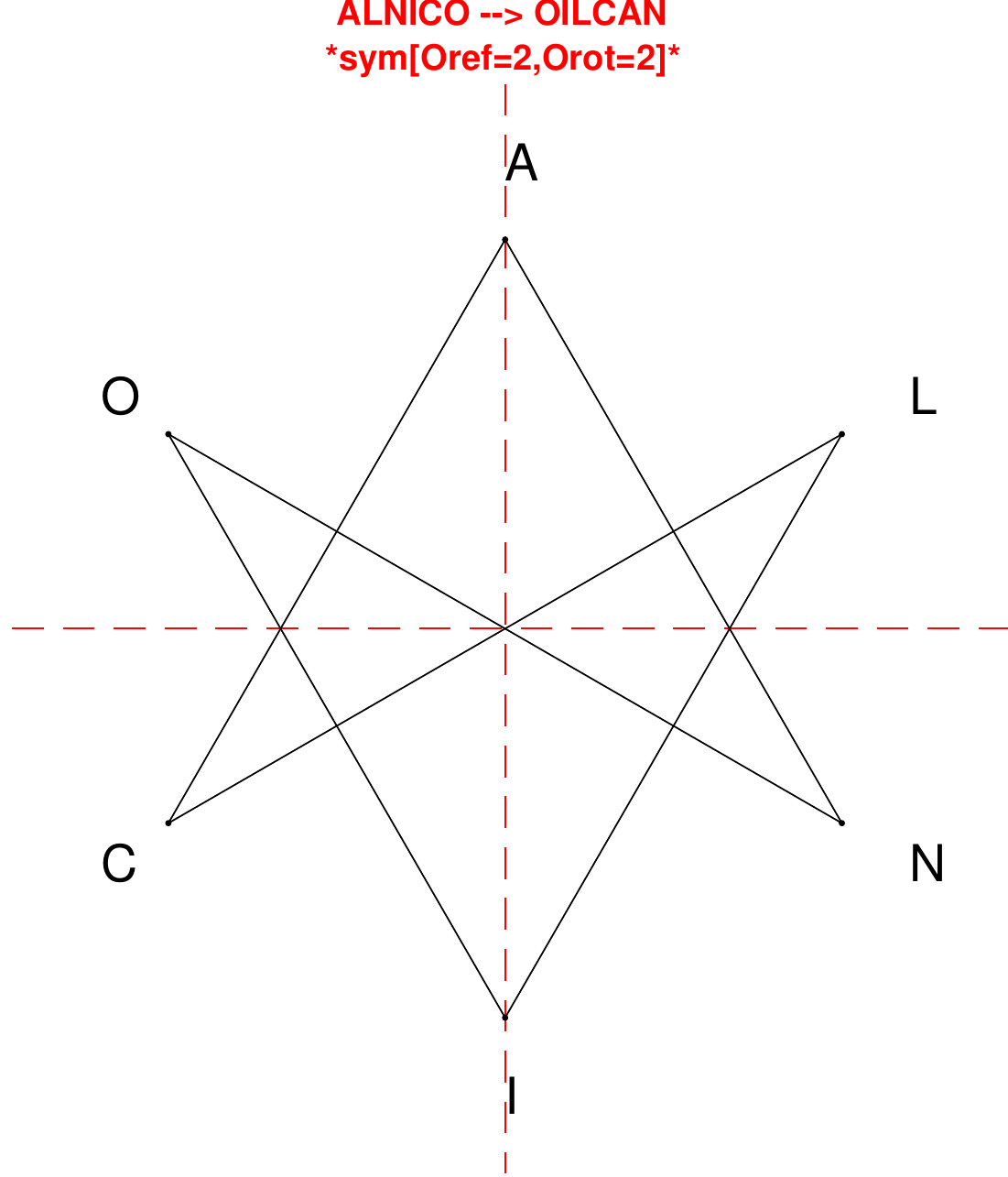}
\end{subfigure}
\hfill
\begin{subfigure}[T]{0.19\textwidth}
\centering
\includegraphics[width=\textwidth]{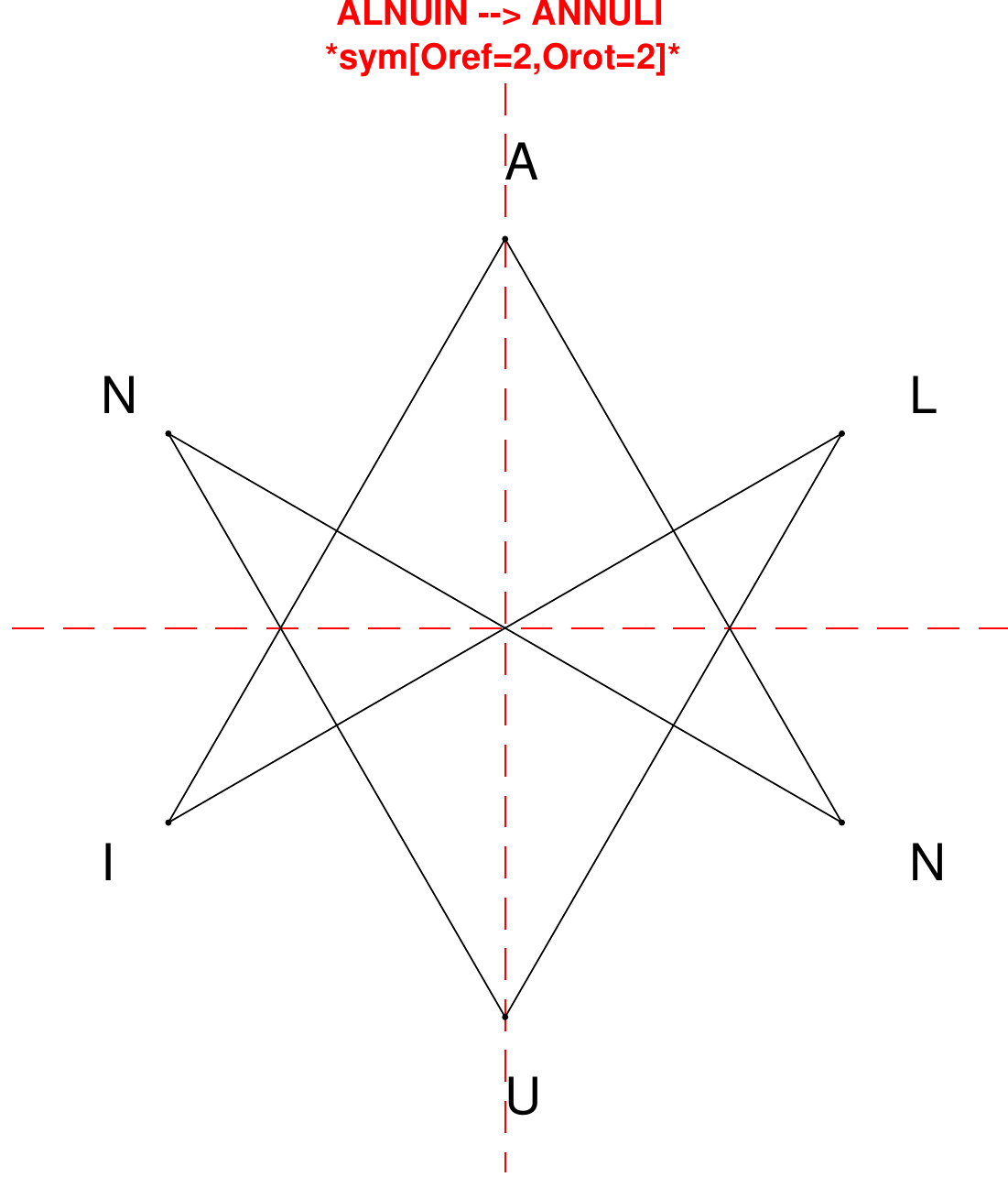}
\end{subfigure}
\end{figure}

\begin{figure}[H]
\centering
\begin{subfigure}[T]{0.19\textwidth}
\centering
\includegraphics[width=\textwidth]{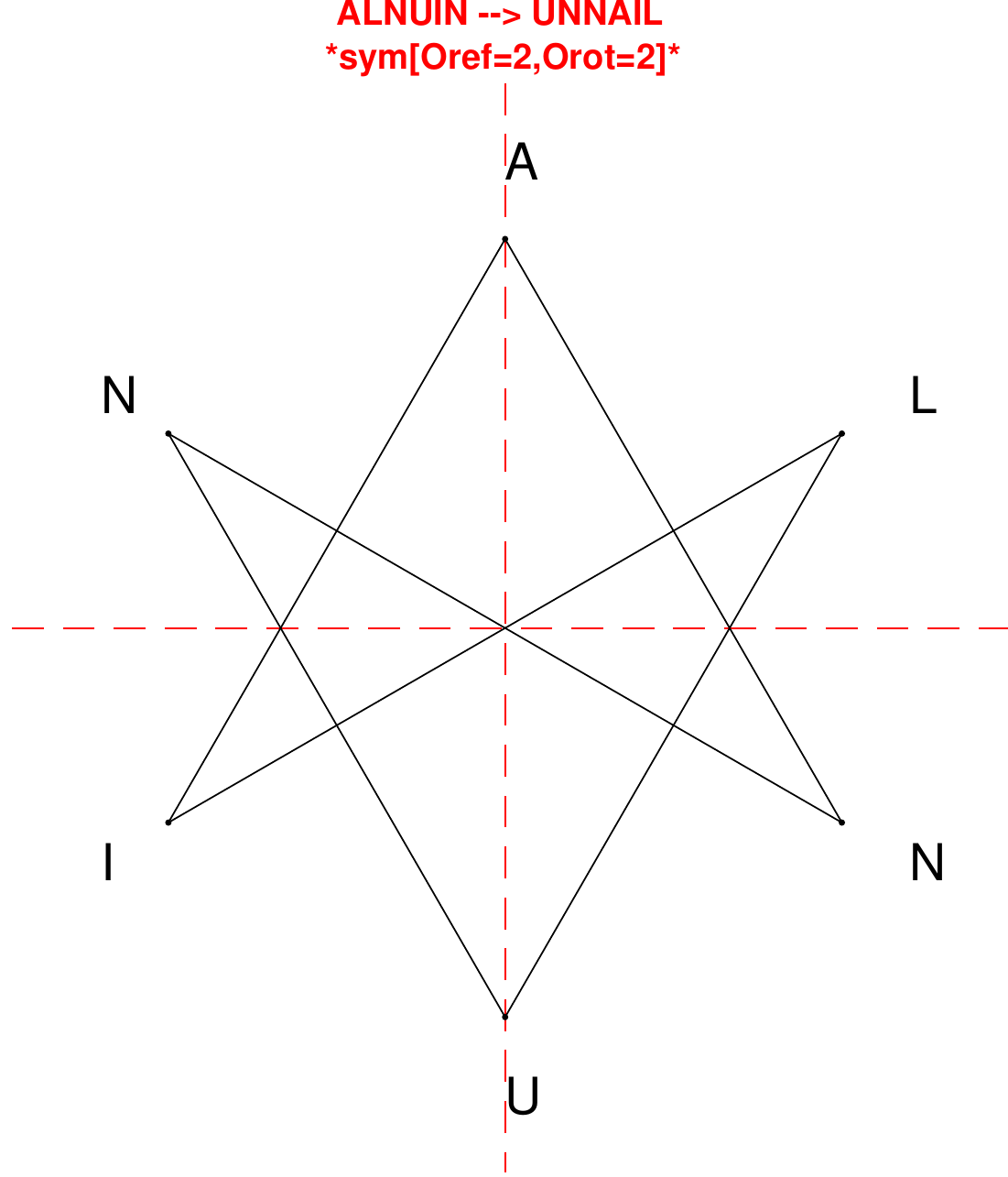}
\end{subfigure}
\hfill
\begin{subfigure}[T]{0.19\textwidth}
\centering
\includegraphics[width=\textwidth]{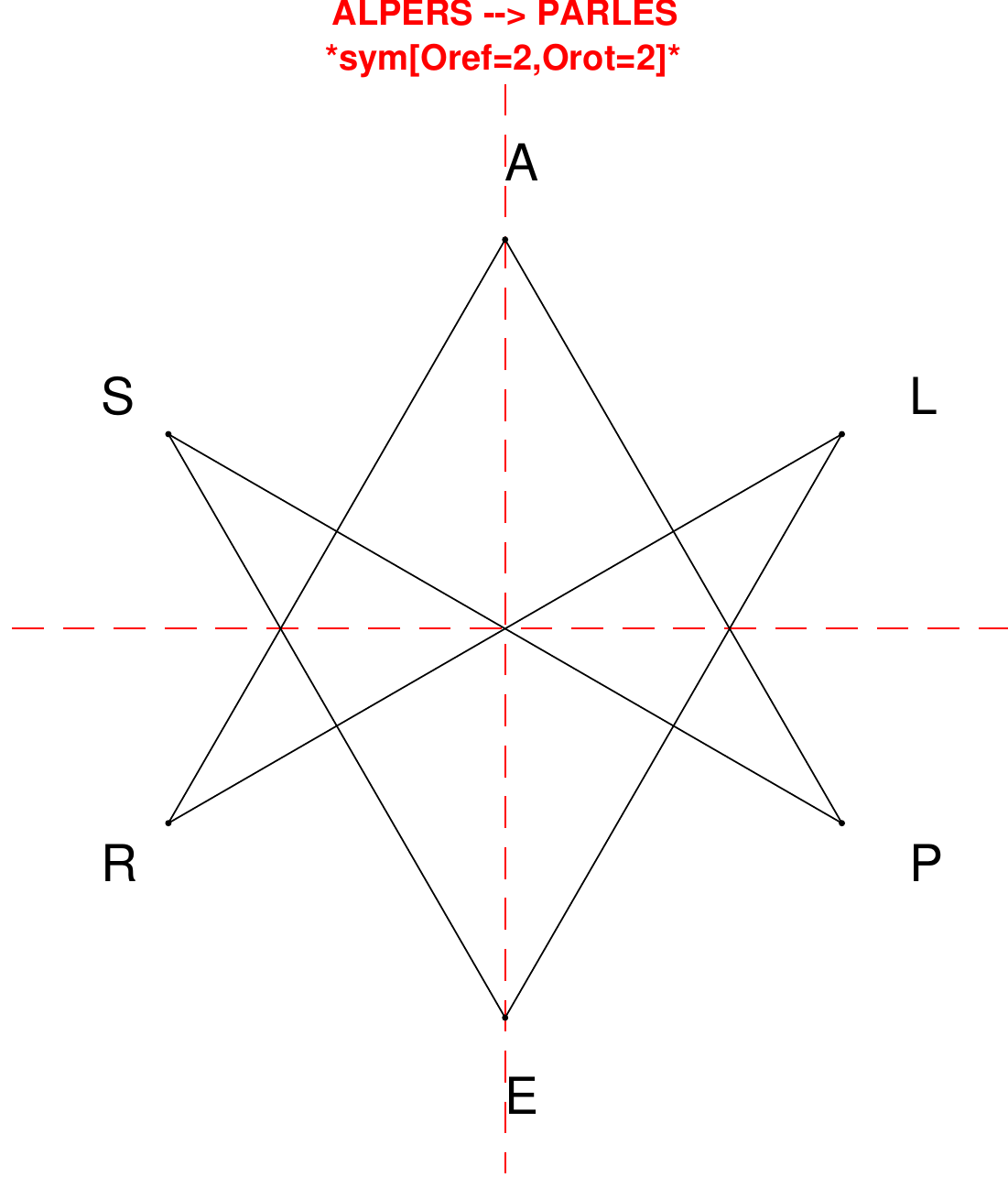}
\end{subfigure}
\hfill
\begin{subfigure}[T]{0.19\textwidth}
\centering
\includegraphics[width=\textwidth]{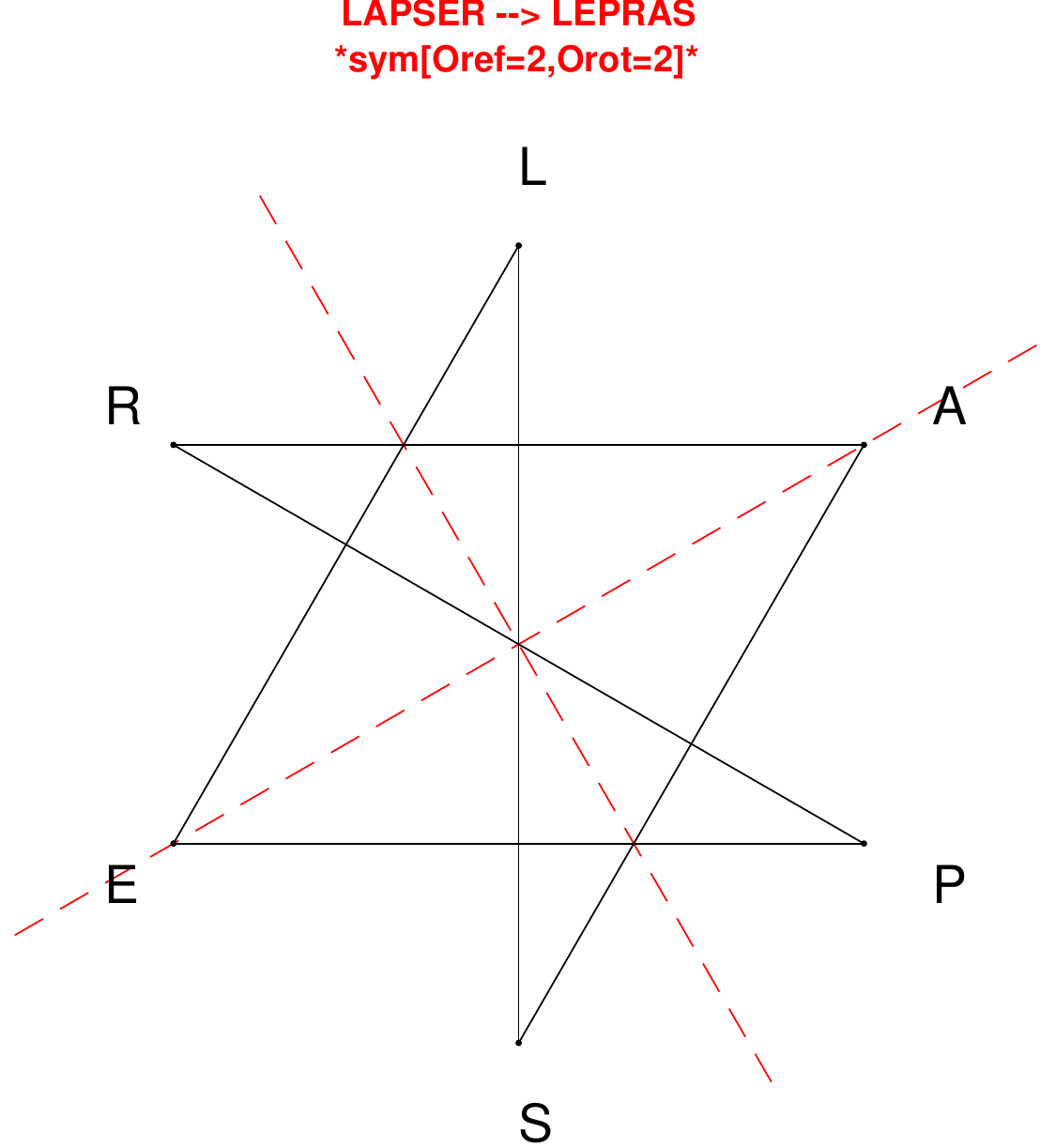}
\end{subfigure}
\hfill
\begin{subfigure}[T]{0.19\textwidth}
\centering
\includegraphics[width=\textwidth]{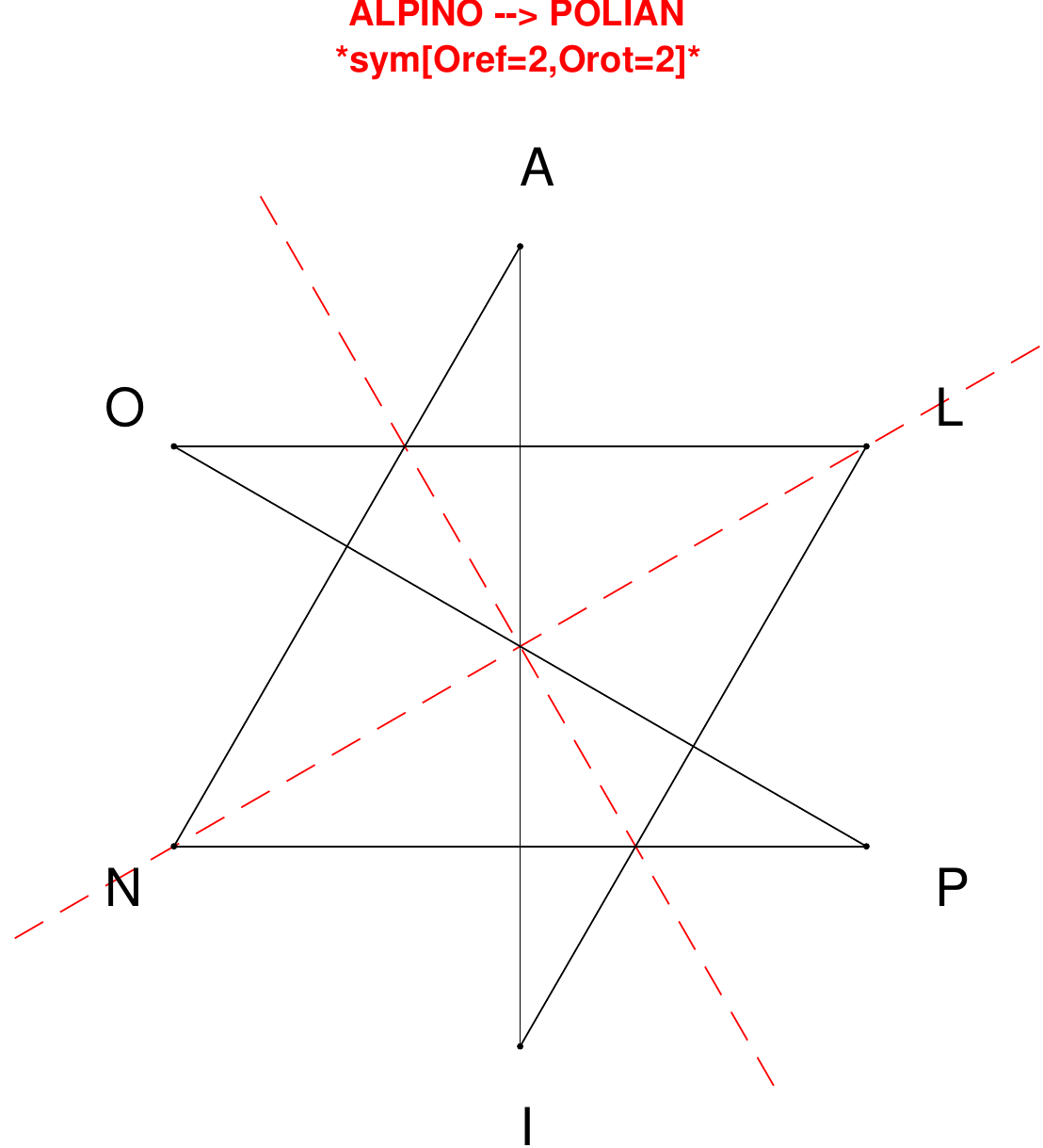}
\end{subfigure}
\hfill
\begin{subfigure}[T]{0.19\textwidth}
\centering
\includegraphics[width=\textwidth]{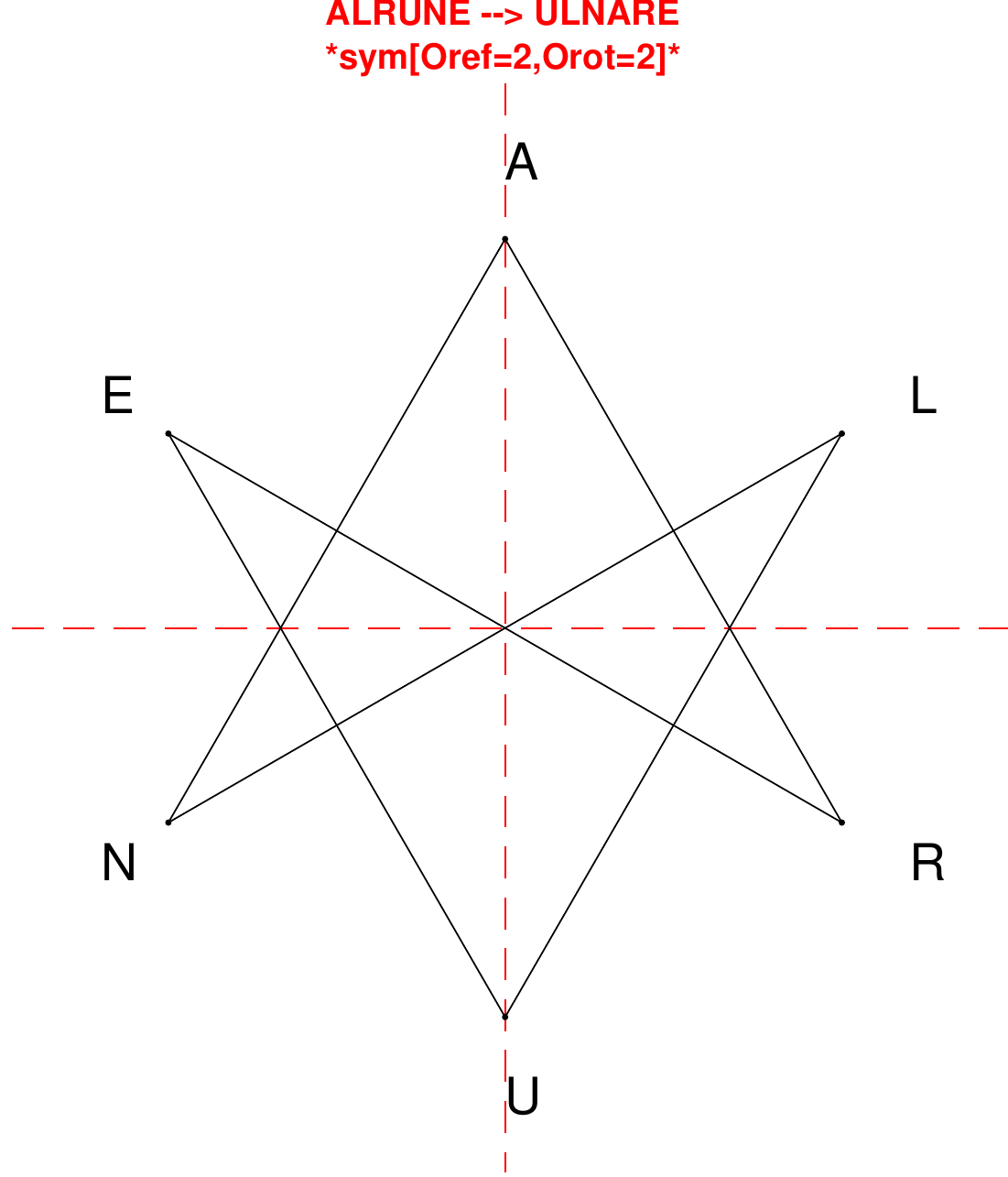}
\end{subfigure}
\end{figure}

\begin{figure}[H]
\centering
\begin{subfigure}[T]{0.19\textwidth}
\centering
\includegraphics[width=\textwidth]{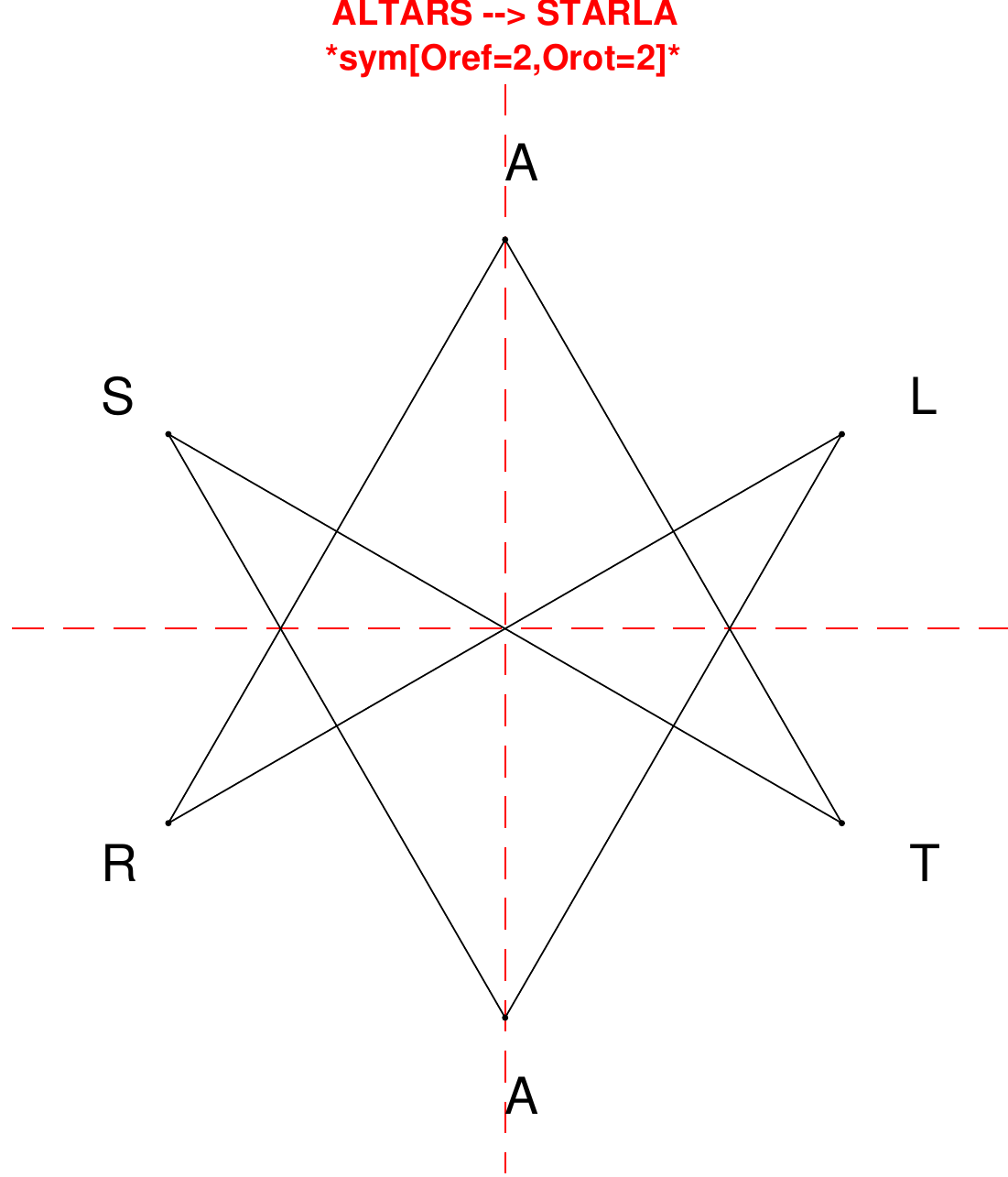}
\end{subfigure}
\hfill
\begin{subfigure}[T]{0.19\textwidth}
\centering
\includegraphics[width=\textwidth]{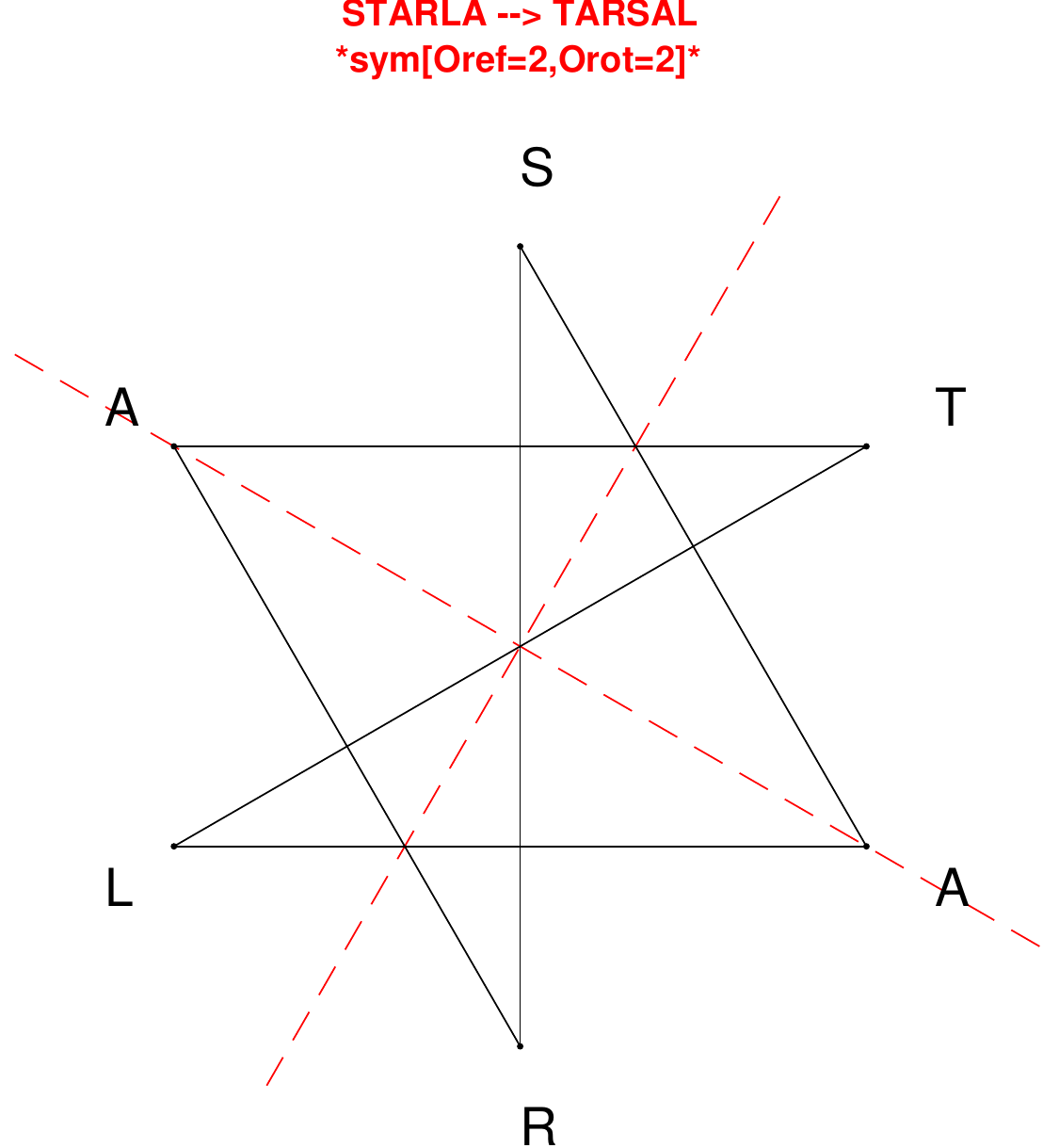}
\end{subfigure}
\hfill
\begin{subfigure}[T]{0.19\textwidth}
\centering
\includegraphics[width=\textwidth]{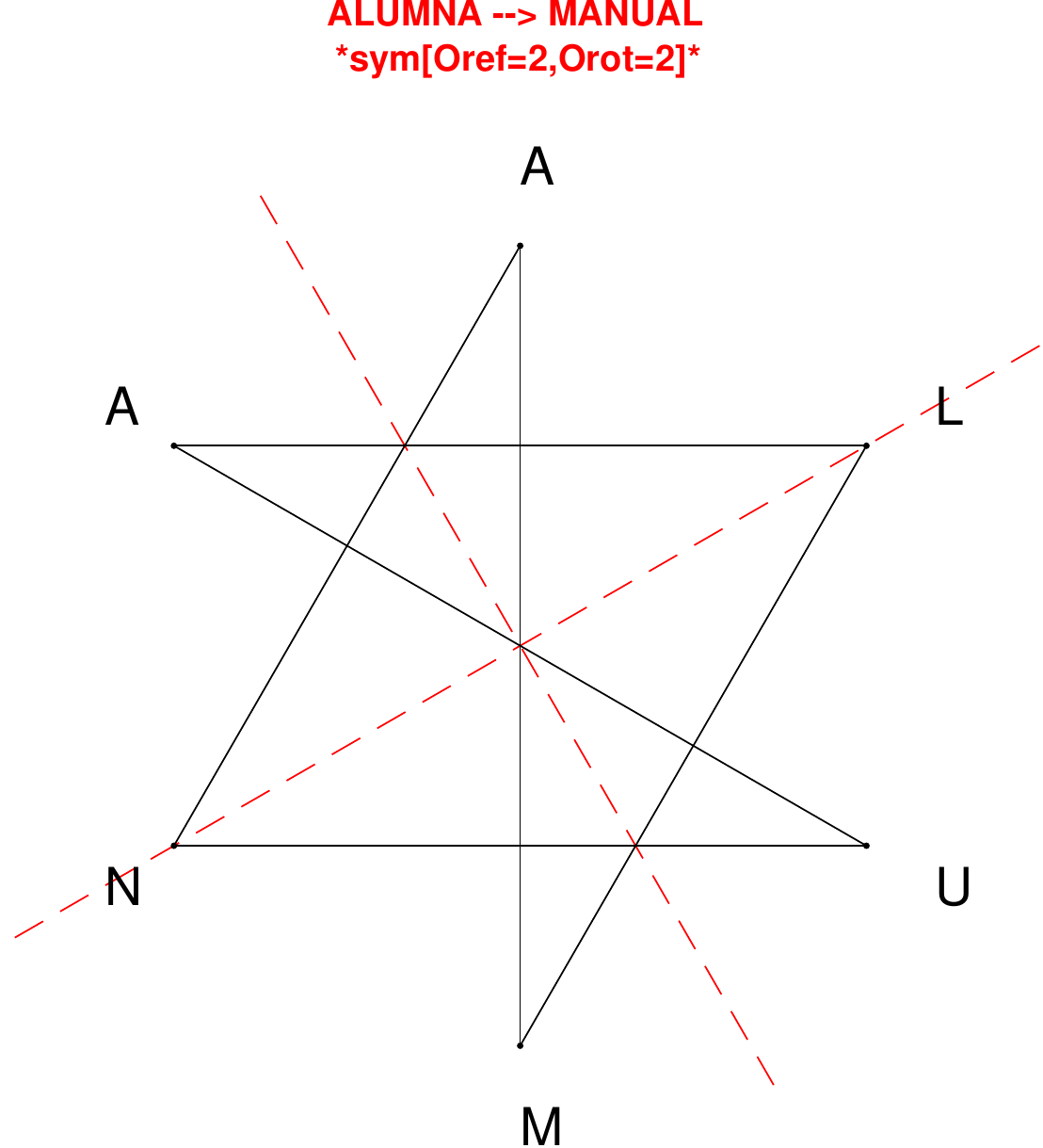}
\end{subfigure}
\hfill
\begin{subfigure}[T]{0.19\textwidth}
\centering
\includegraphics[width=\textwidth]{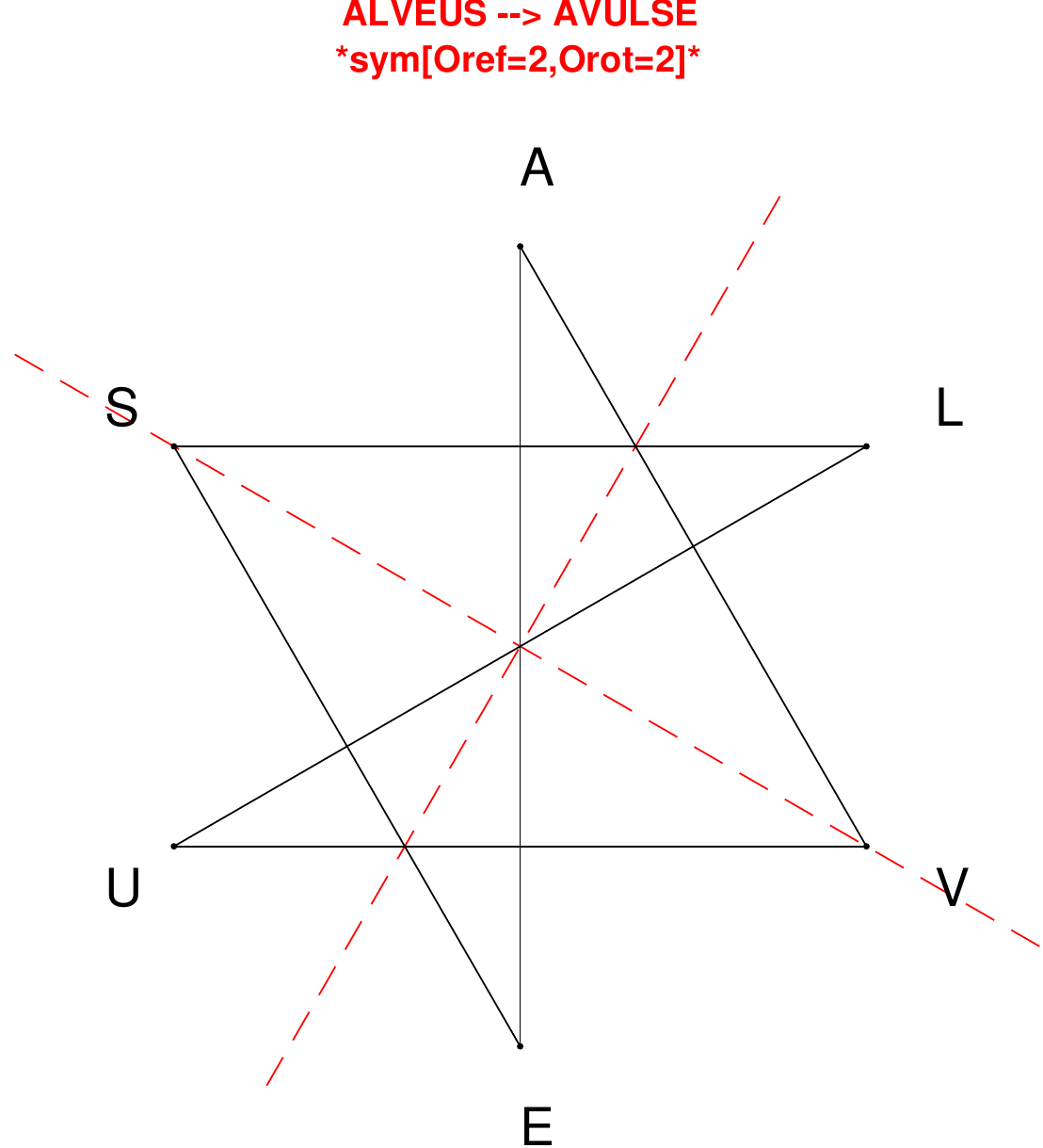}
\end{subfigure}
\hfill
\begin{subfigure}[T]{0.19\textwidth}
\centering
\includegraphics[width=\textwidth]{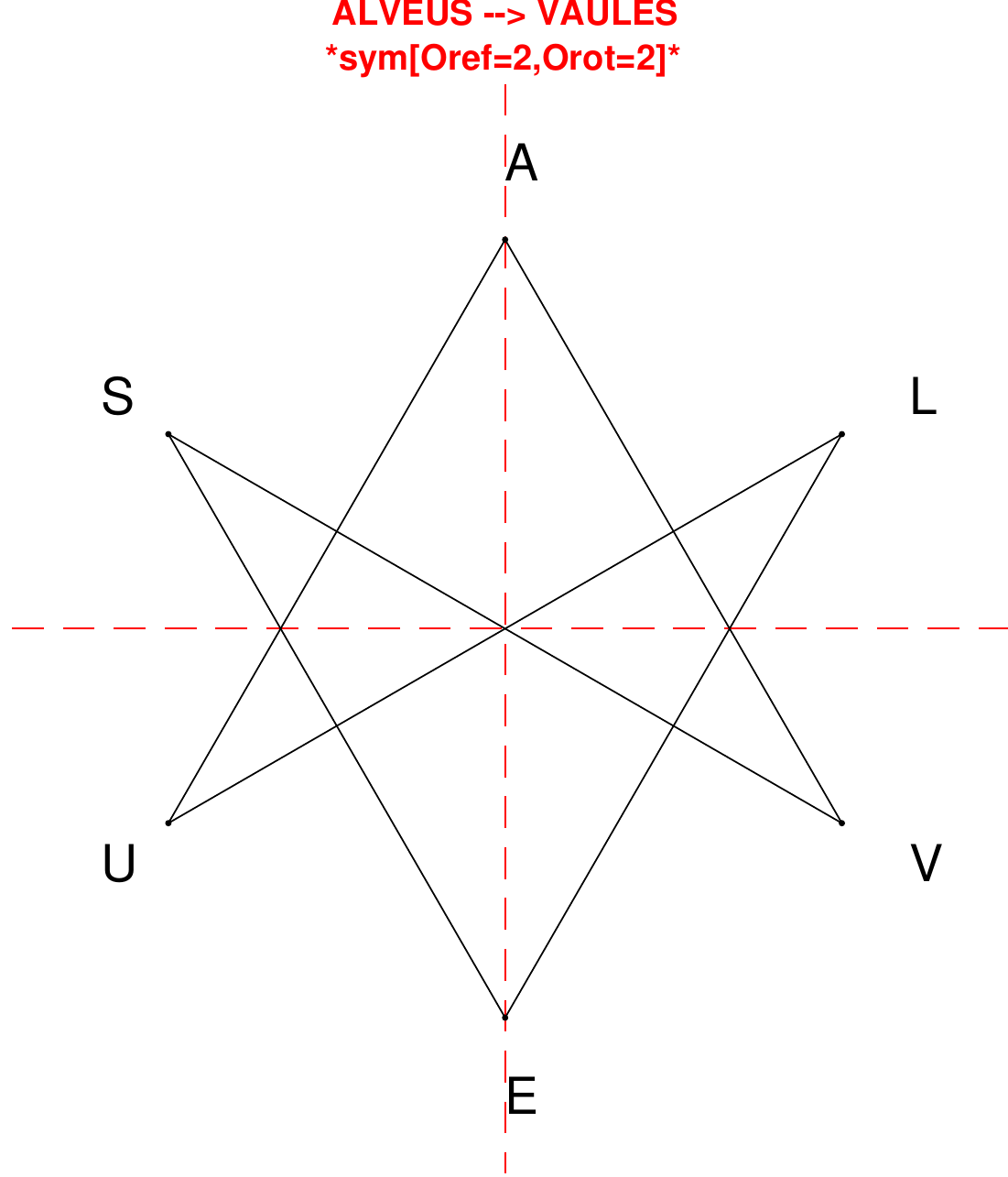}
\end{subfigure}
\end{figure}

\begin{figure}[H]
\centering
\begin{subfigure}[T]{0.19\textwidth}
\centering
\includegraphics[width=\textwidth]{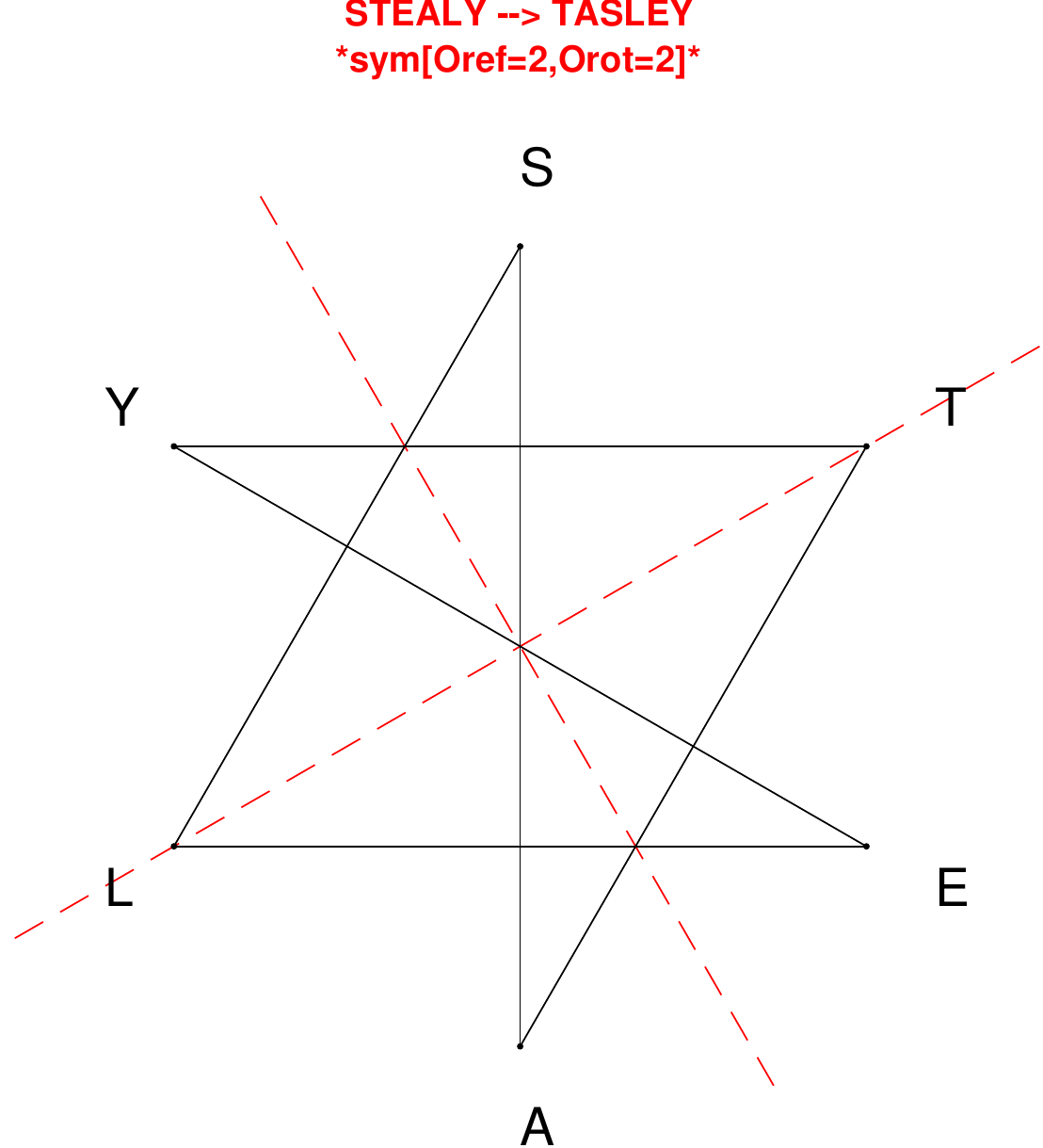}
\end{subfigure}
\hfill
\begin{subfigure}[T]{0.19\textwidth}
\centering
\includegraphics[width=\textwidth]{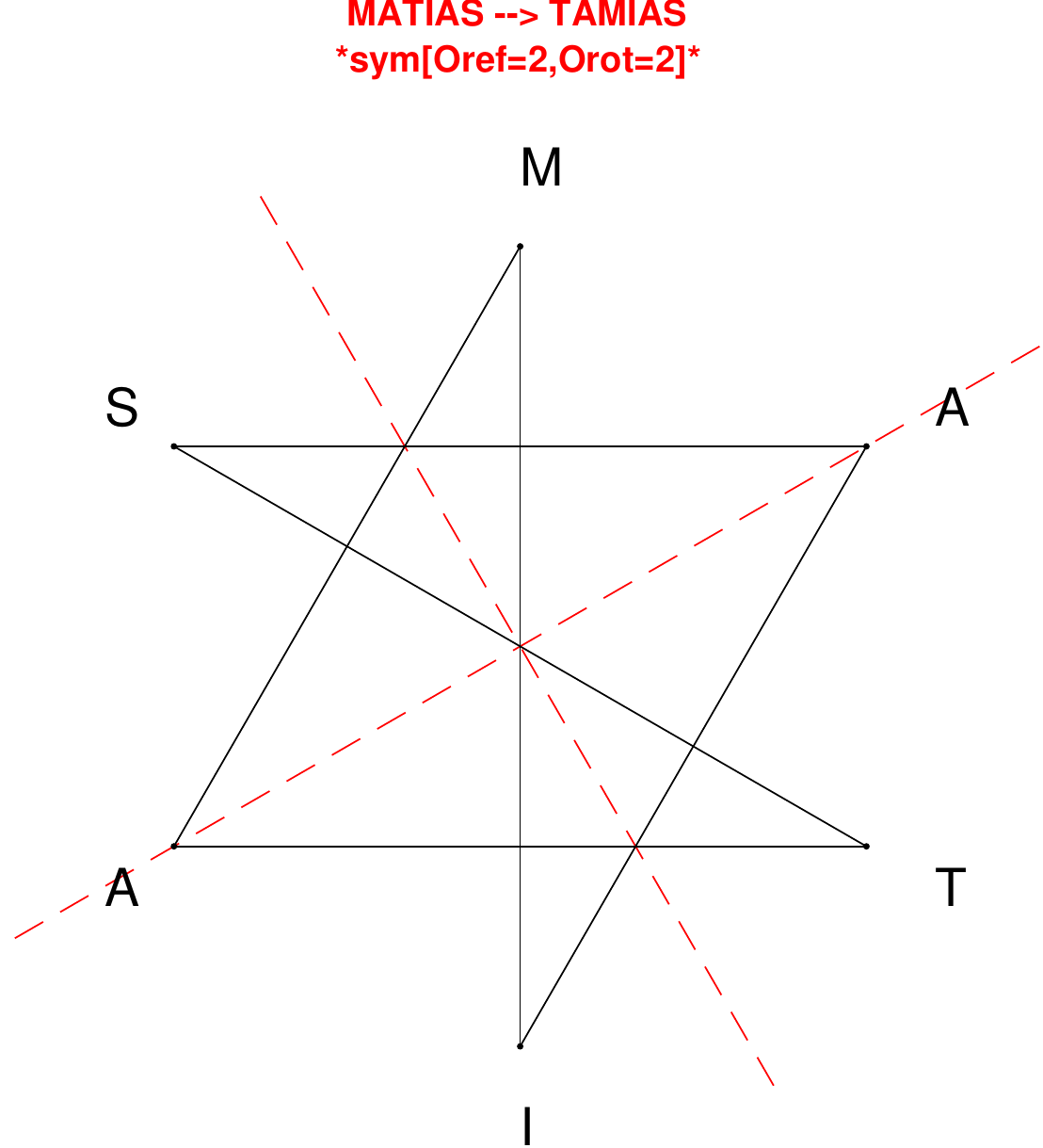}
\end{subfigure}
\hfill
\begin{subfigure}[T]{0.19\textwidth}
\centering
\includegraphics[width=\textwidth]{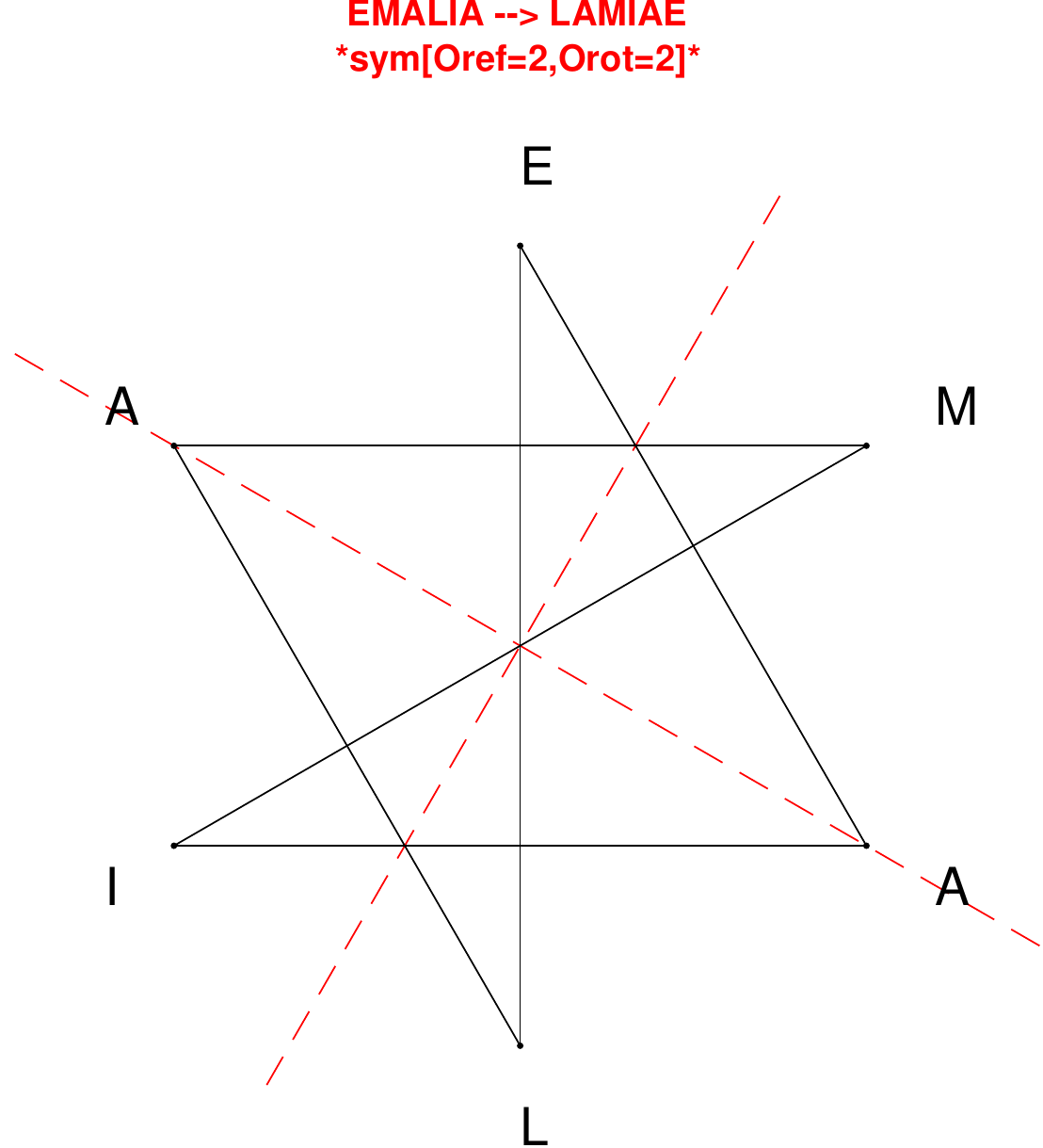}
\end{subfigure}
\hfill
\begin{subfigure}[T]{0.19\textwidth}
\centering
\includegraphics[width=\textwidth]{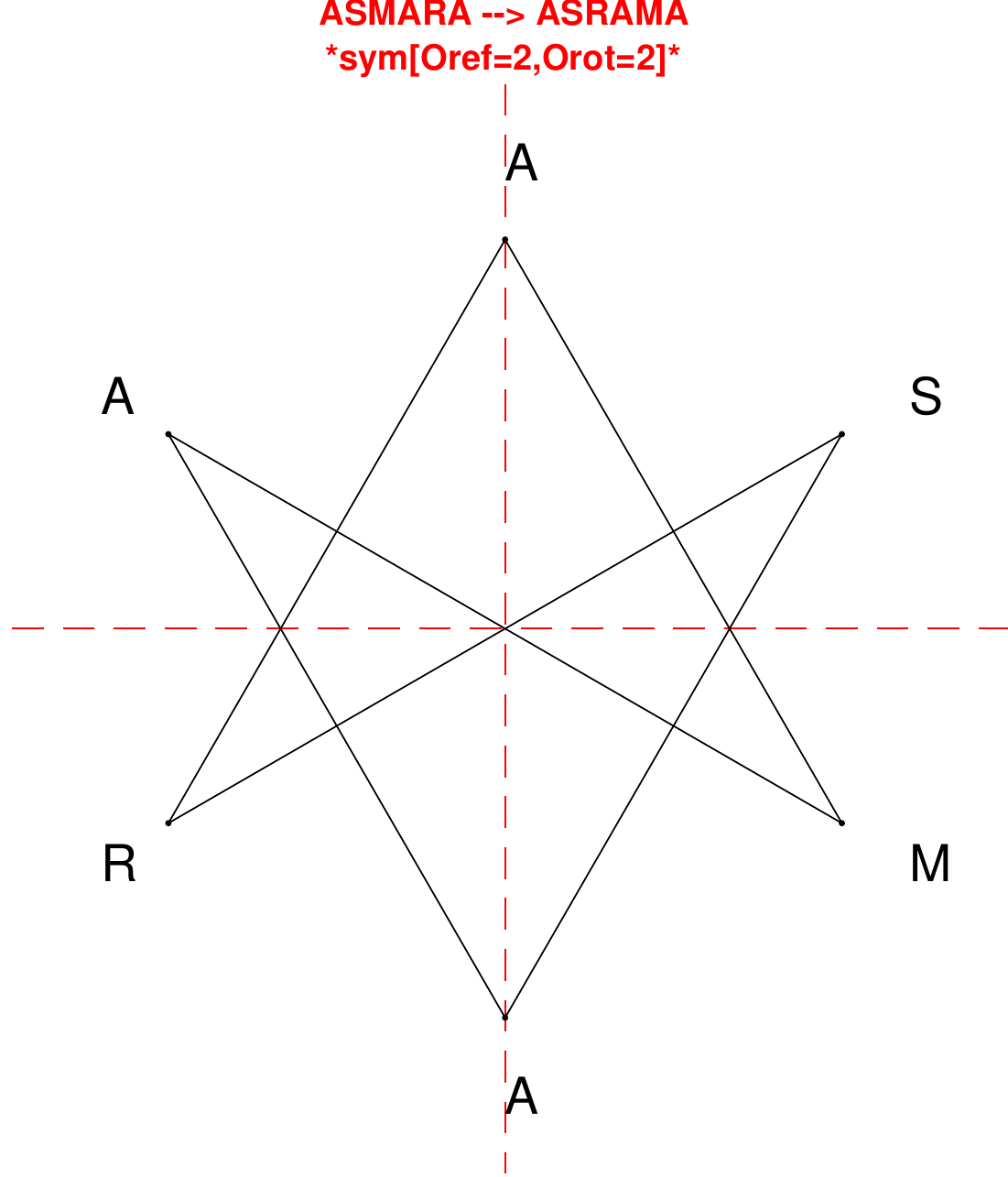}
\end{subfigure}
\hfill
\begin{subfigure}[T]{0.19\textwidth}
\centering
\includegraphics[width=\textwidth]{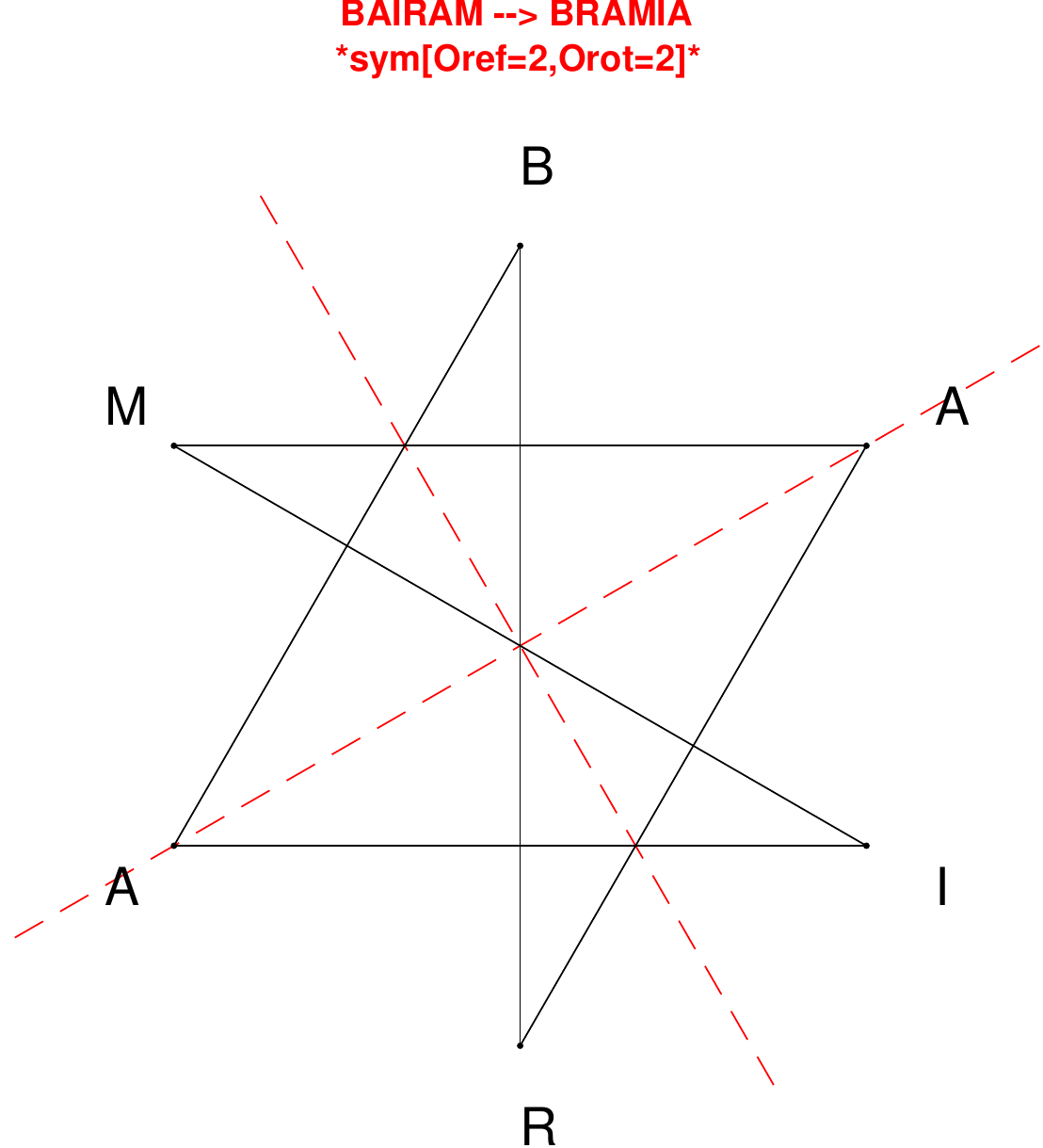}
\end{subfigure}
\end{figure}

\begin{figure}[H]
\centering
\begin{subfigure}[T]{0.19\textwidth}
\centering
\includegraphics[width=\textwidth]{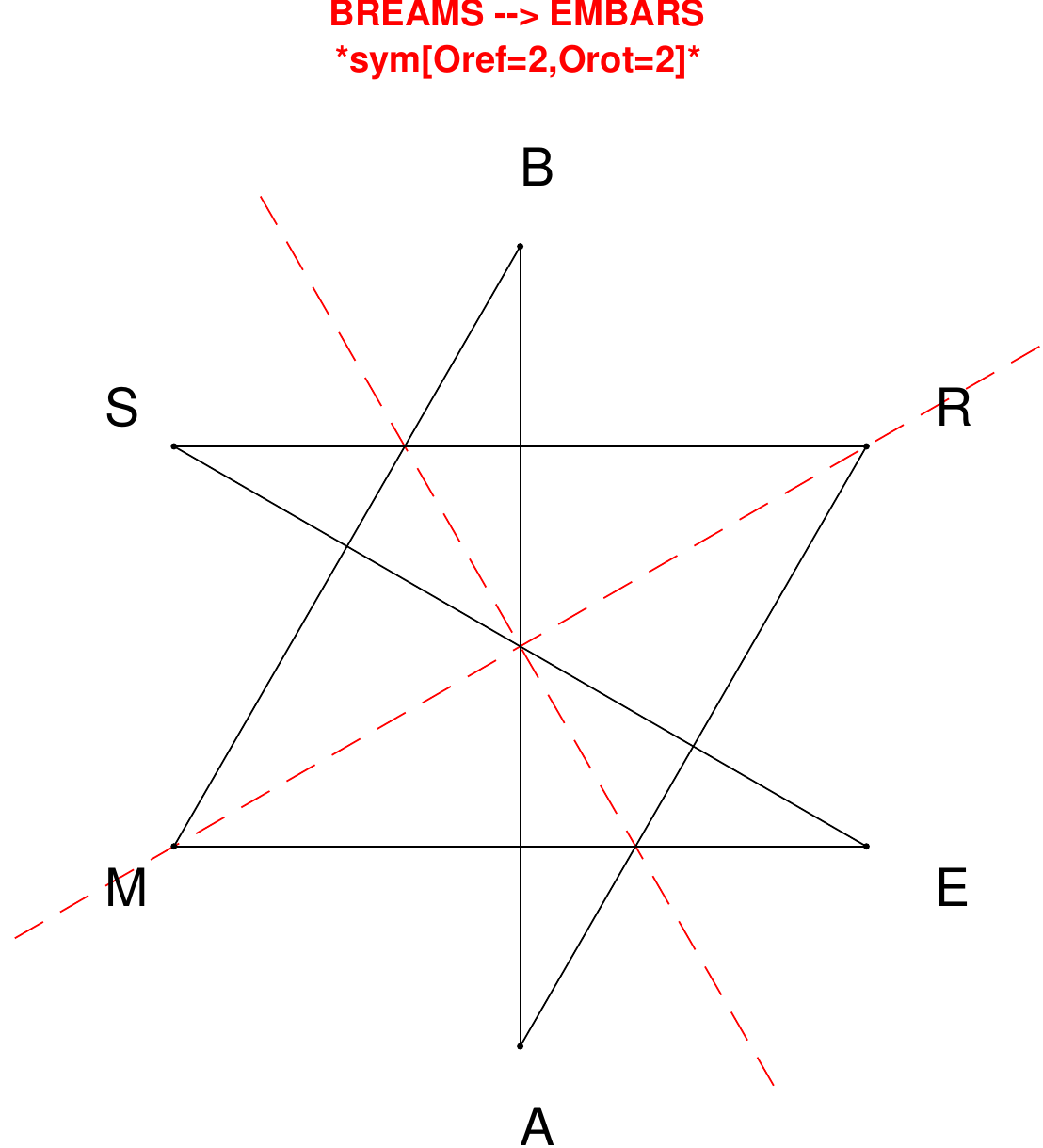}
\end{subfigure}
\hfill
\begin{subfigure}[T]{0.19\textwidth}
\centering
\includegraphics[width=\textwidth]{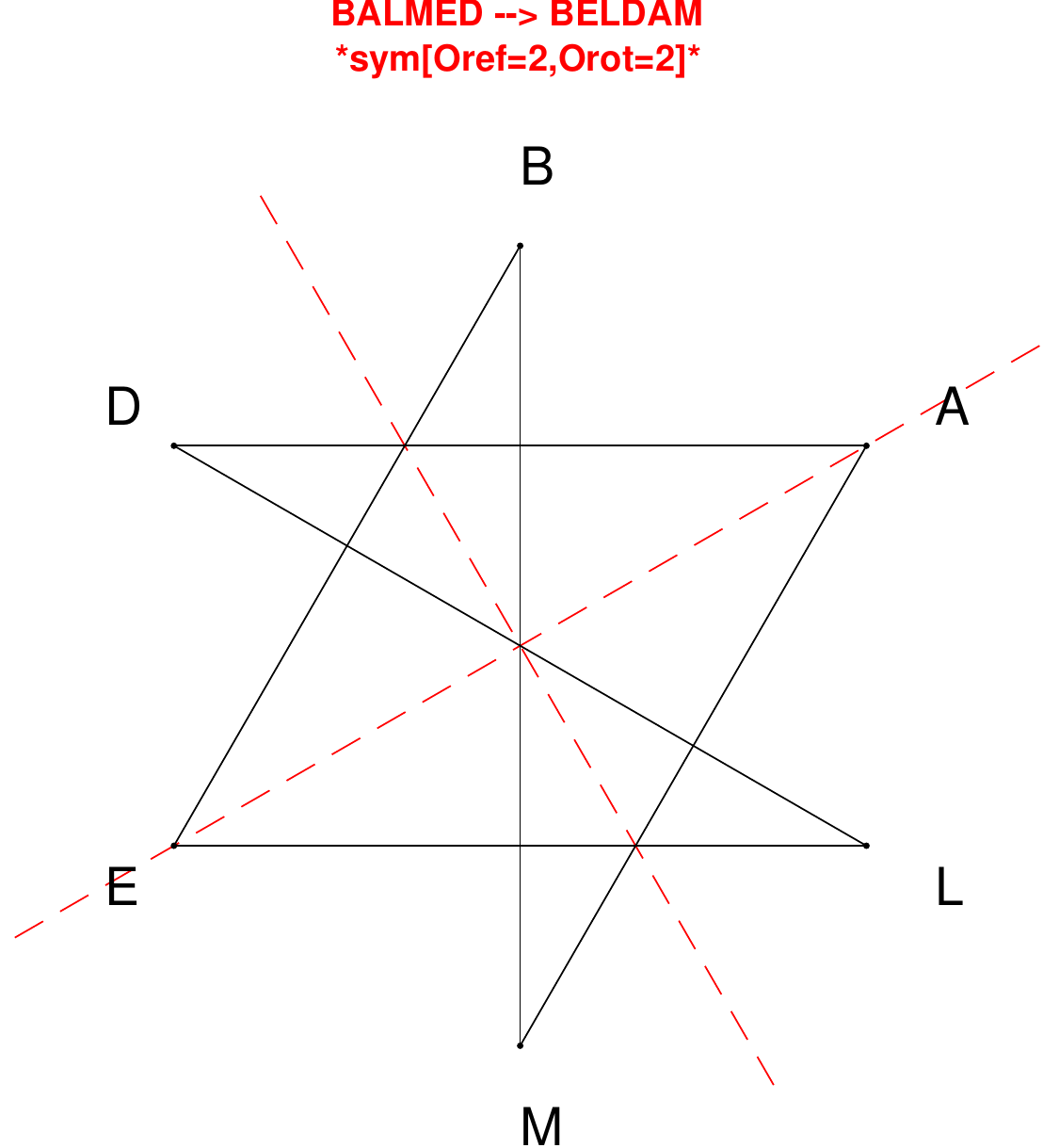}
\end{subfigure}
\hfill
\begin{subfigure}[T]{0.19\textwidth}
\centering
\includegraphics[width=\textwidth]{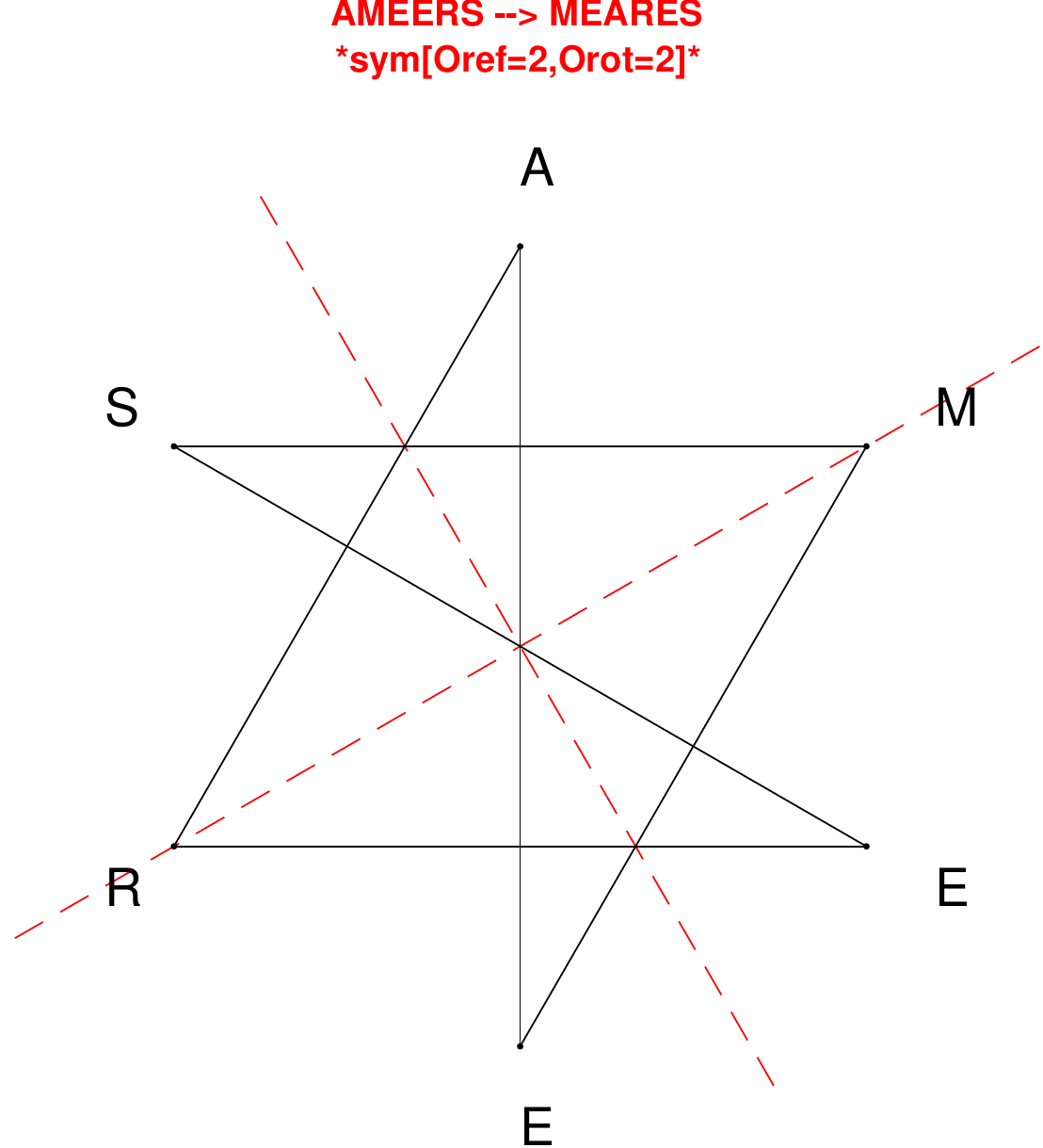}
\end{subfigure}
\hfill
\begin{subfigure}[T]{0.19\textwidth}
\centering
\includegraphics[width=\textwidth]{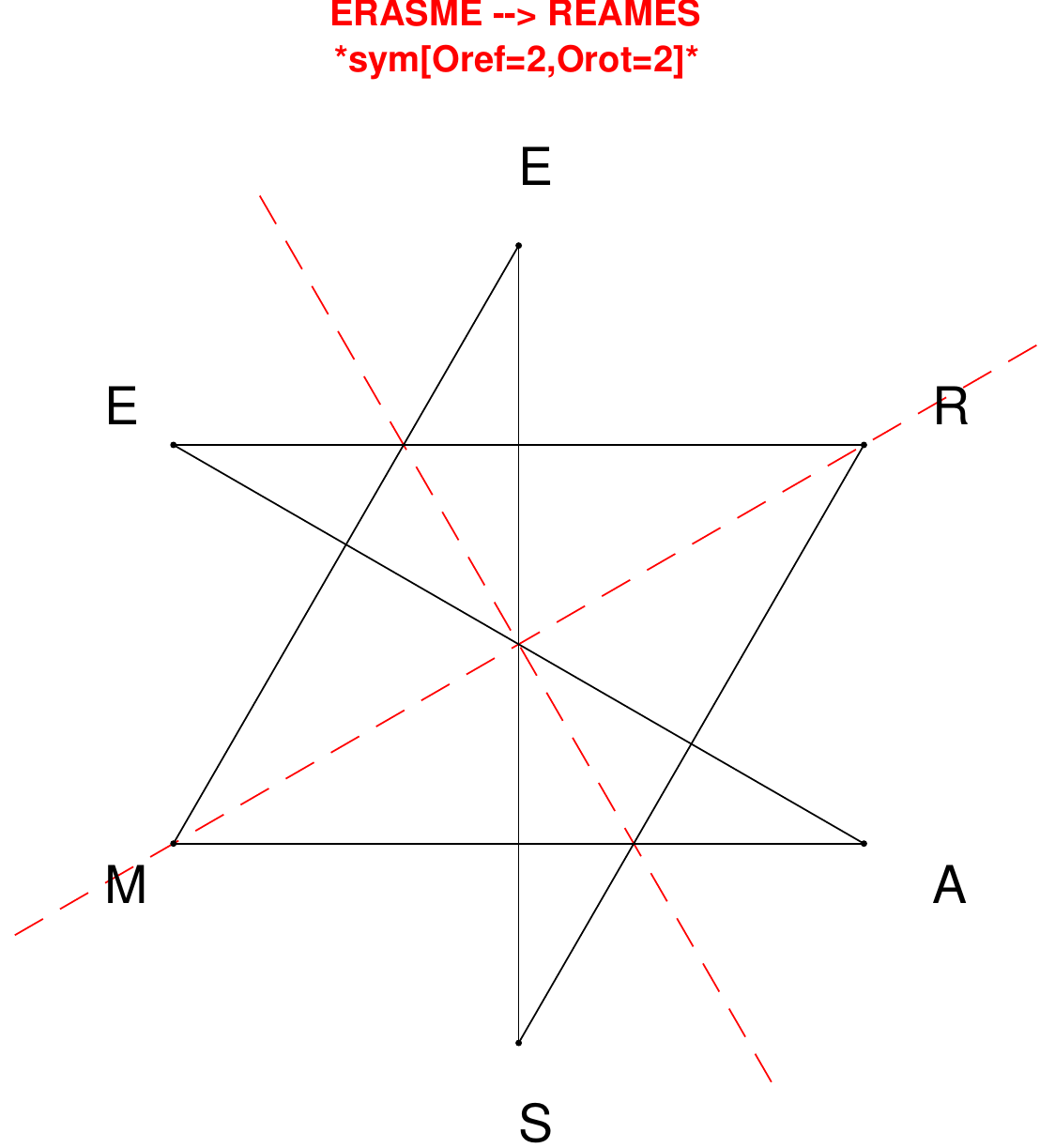}
\end{subfigure}
\hfill
\begin{subfigure}[T]{0.19\textwidth}
\centering
\includegraphics[width=\textwidth]{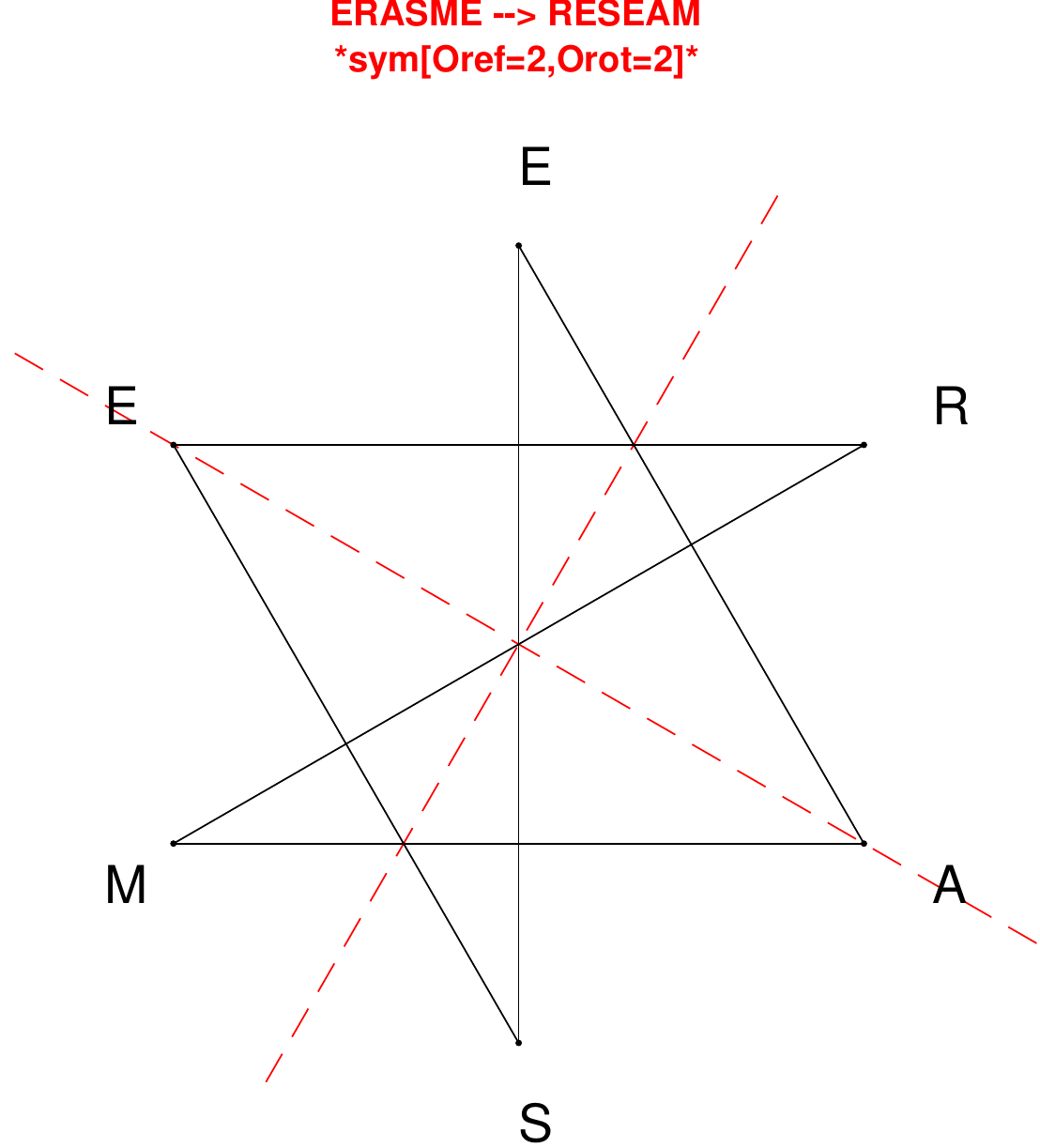}
\end{subfigure}
\end{figure}

\begin{figure}[H]
\centering
\begin{subfigure}[T]{0.19\textwidth}
\centering
\includegraphics[width=\textwidth]{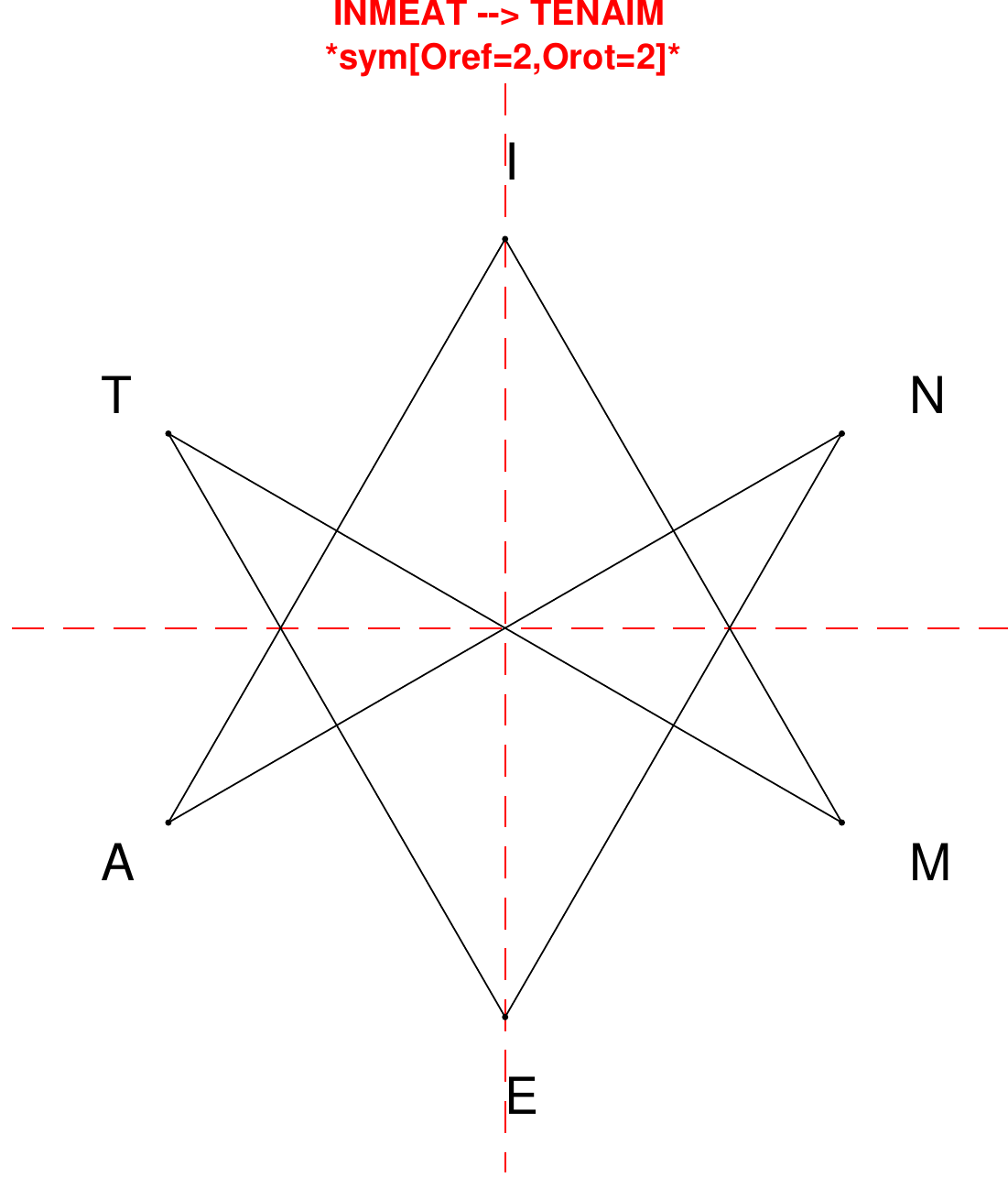}
\end{subfigure}
\hfill
\begin{subfigure}[T]{0.19\textwidth}
\centering
\includegraphics[width=\textwidth]{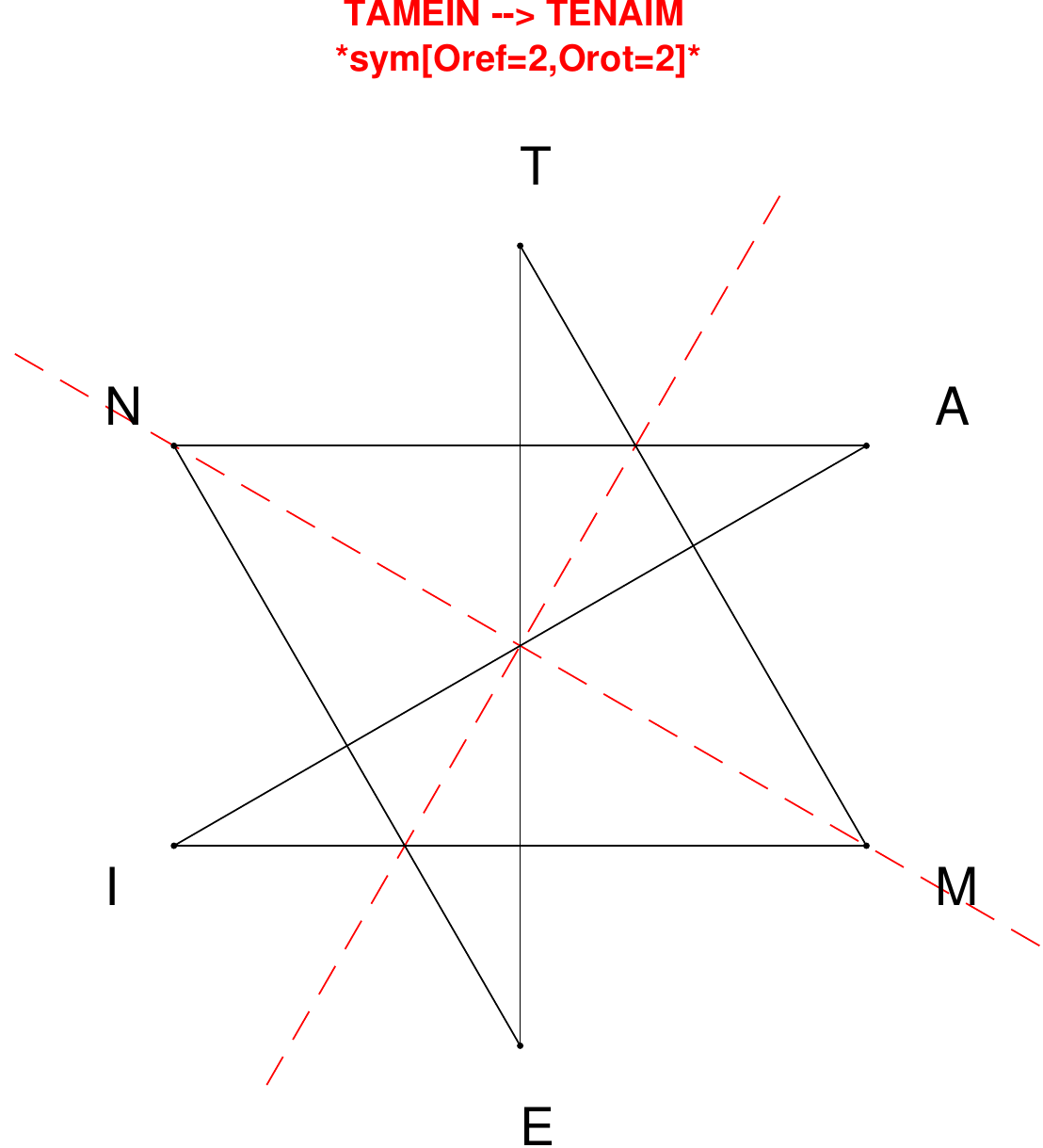}
\end{subfigure}
\hfill
\begin{subfigure}[T]{0.19\textwidth}
\centering
\includegraphics[width=\textwidth]{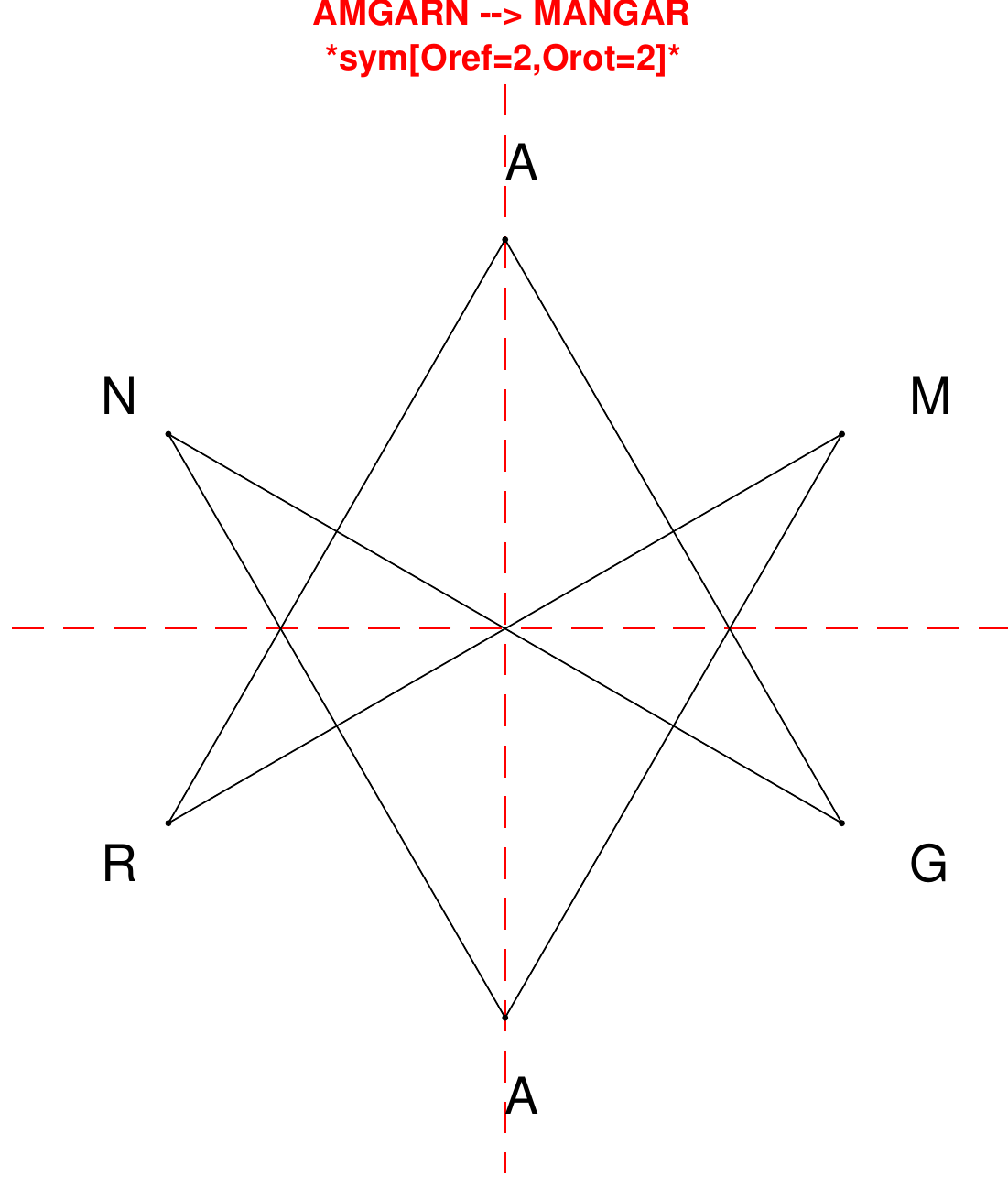}
\end{subfigure}
\hfill
\begin{subfigure}[T]{0.19\textwidth}
\centering
\includegraphics[width=\textwidth]{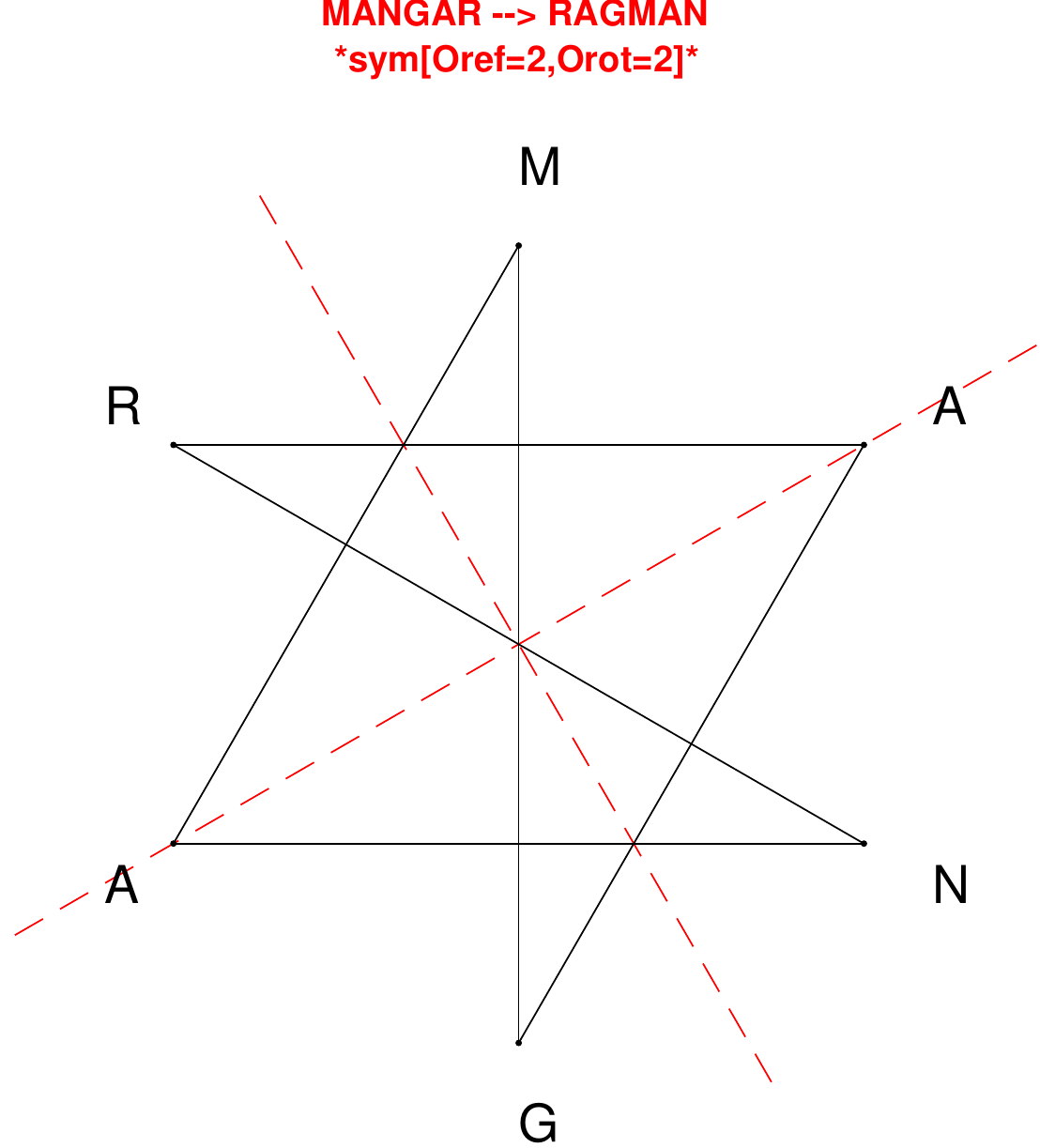}
\end{subfigure}
\hfill
\begin{subfigure}[T]{0.19\textwidth}
\centering
\includegraphics[width=\textwidth]{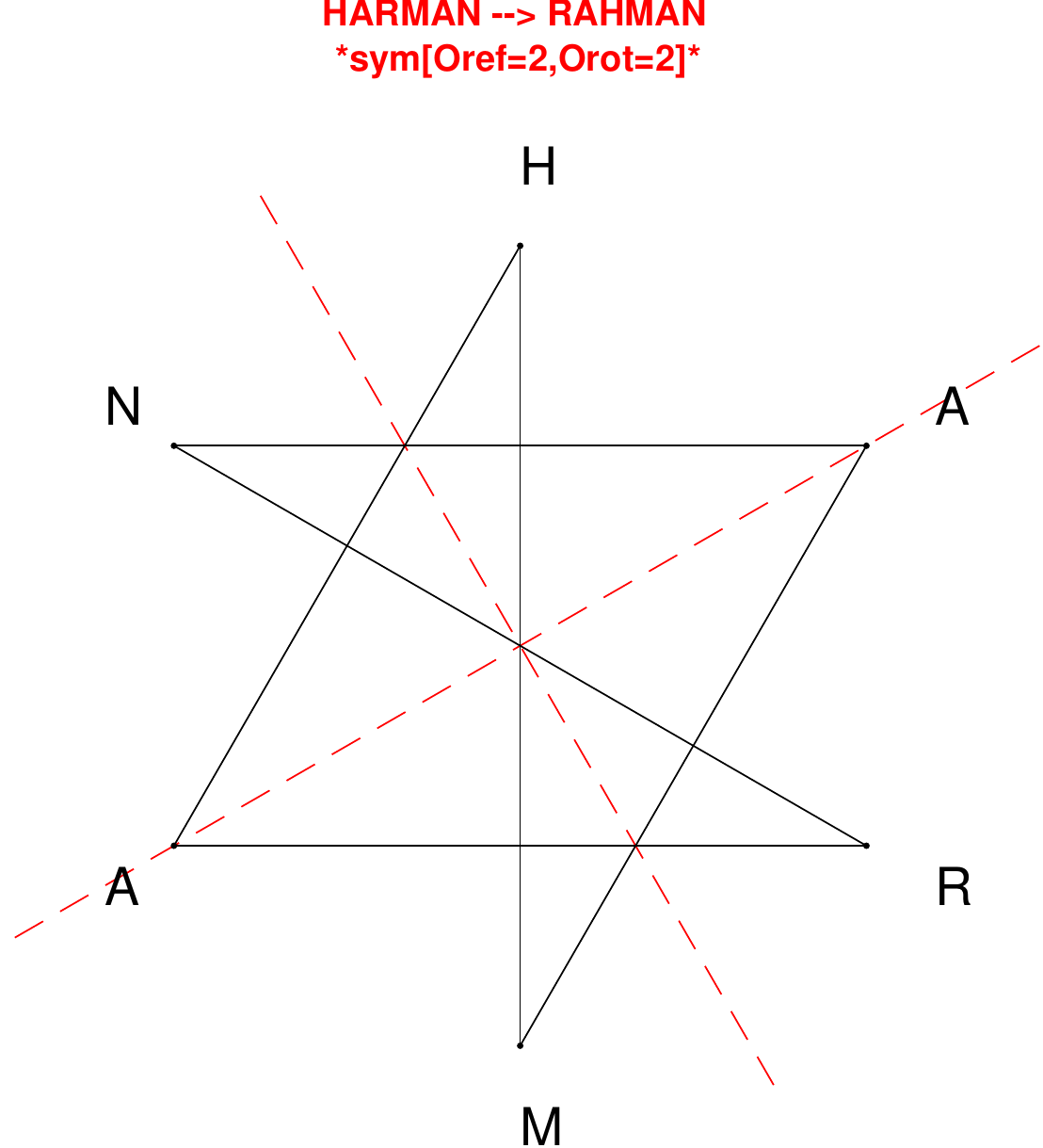}
\end{subfigure}
\end{figure}

\begin{figure}[H]
\centering
\begin{subfigure}[T]{0.19\textwidth}
\centering
\includegraphics[width=\textwidth]{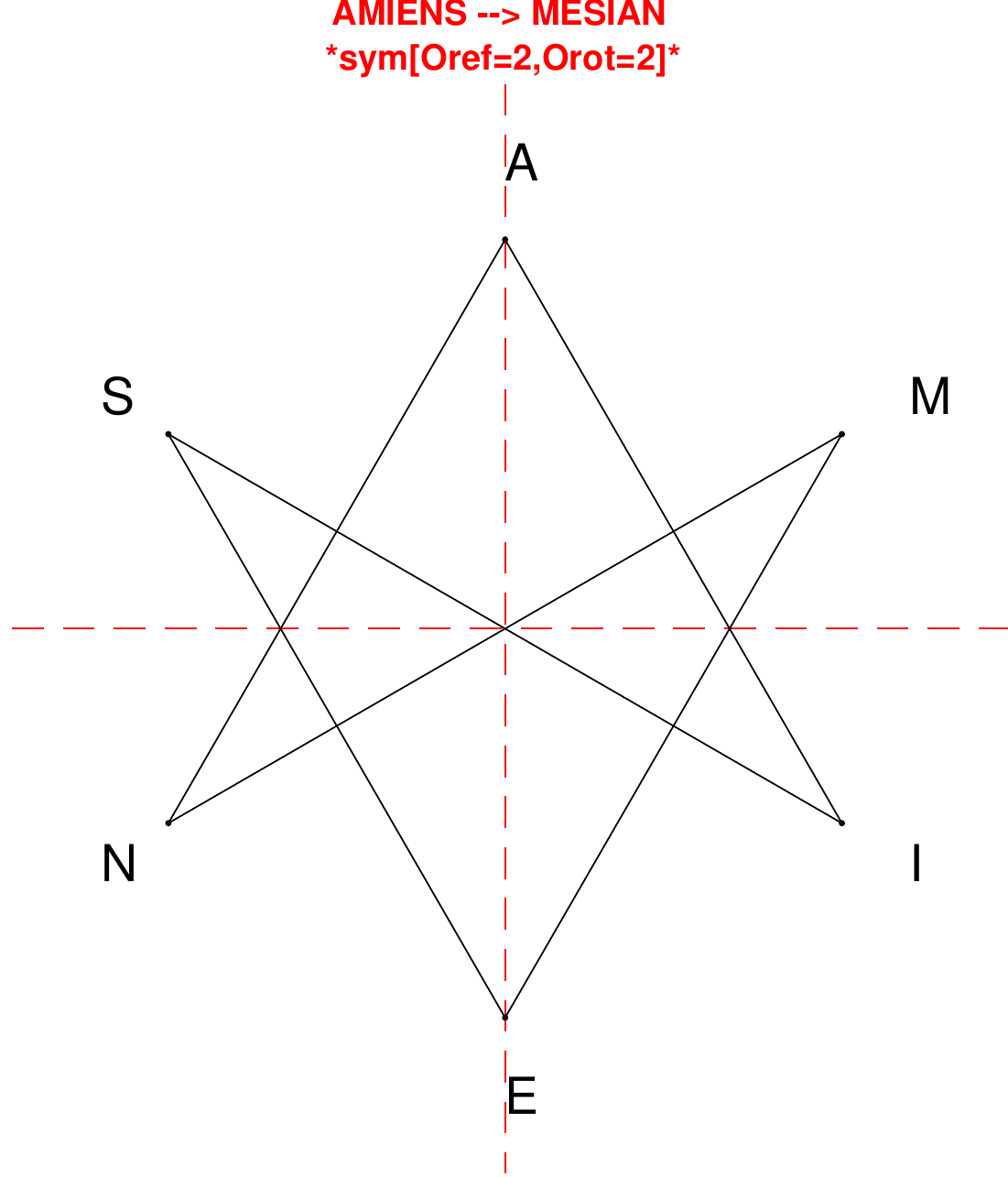}
\end{subfigure}
\hfill
\begin{subfigure}[T]{0.19\textwidth}
\centering
\includegraphics[width=\textwidth]{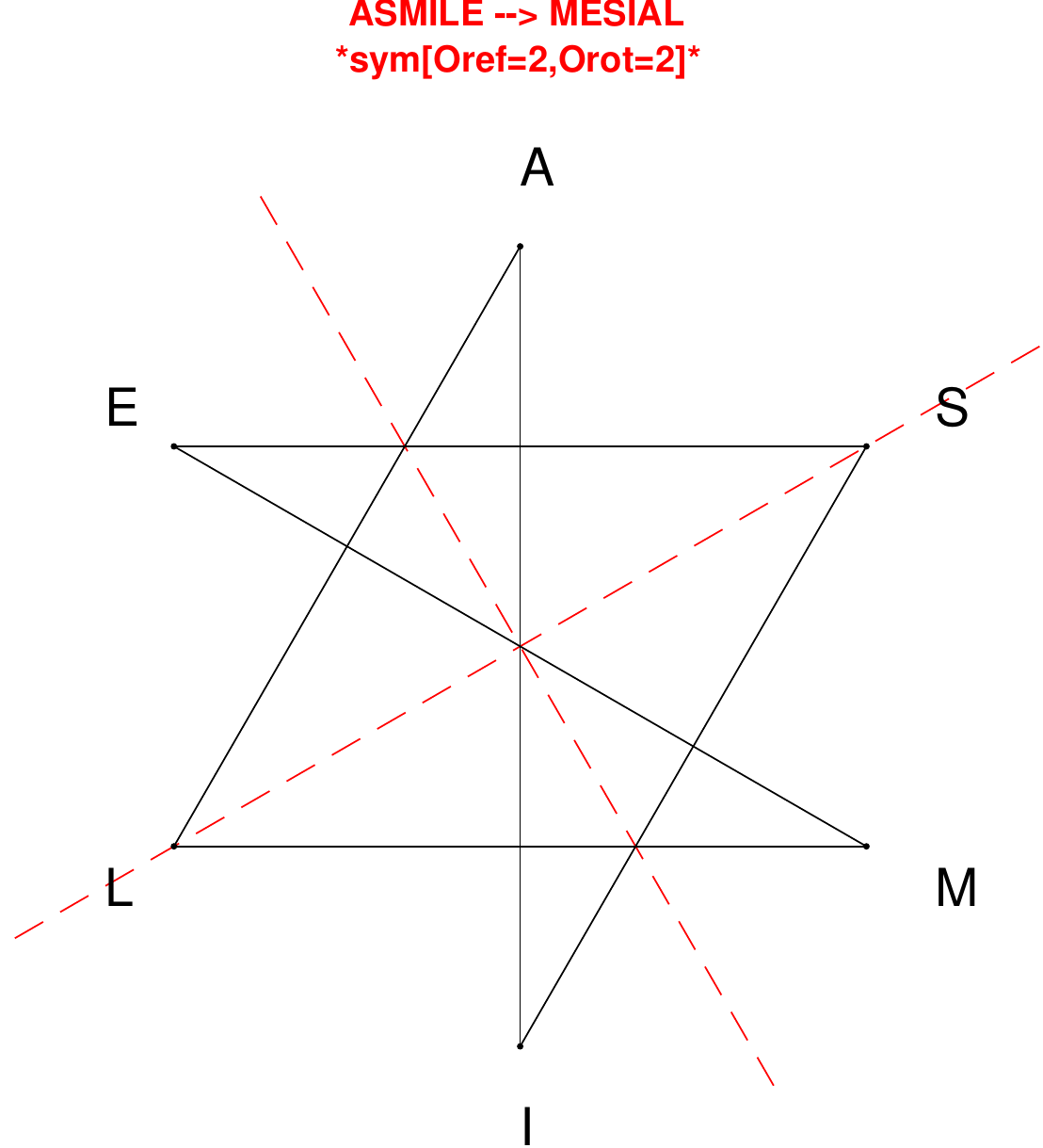}
\end{subfigure}
\hfill
\begin{subfigure}[T]{0.19\textwidth}
\centering
\includegraphics[width=\textwidth]{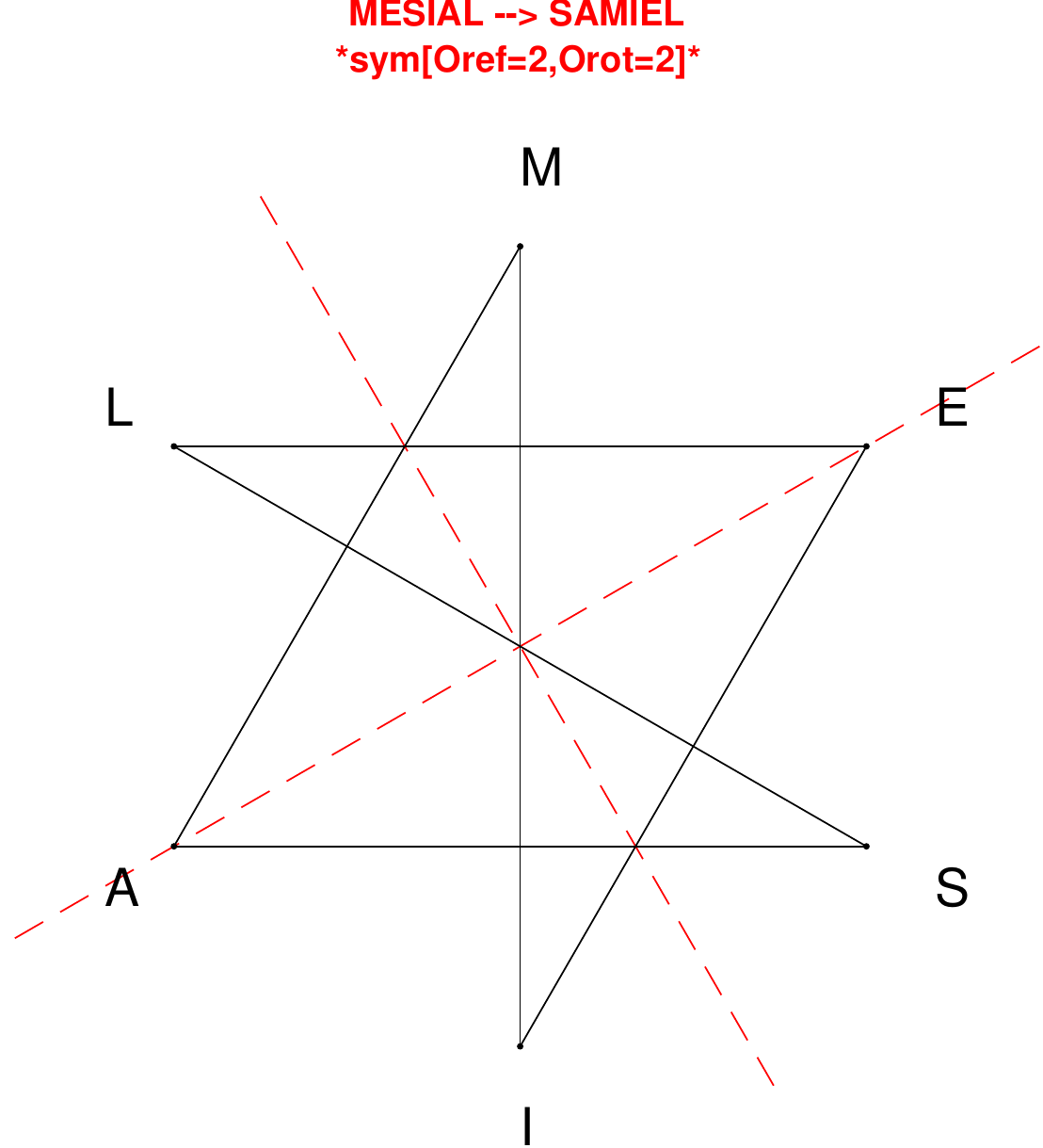}
\end{subfigure}
\hfill
\begin{subfigure}[T]{0.19\textwidth}
\centering
\includegraphics[width=\textwidth]{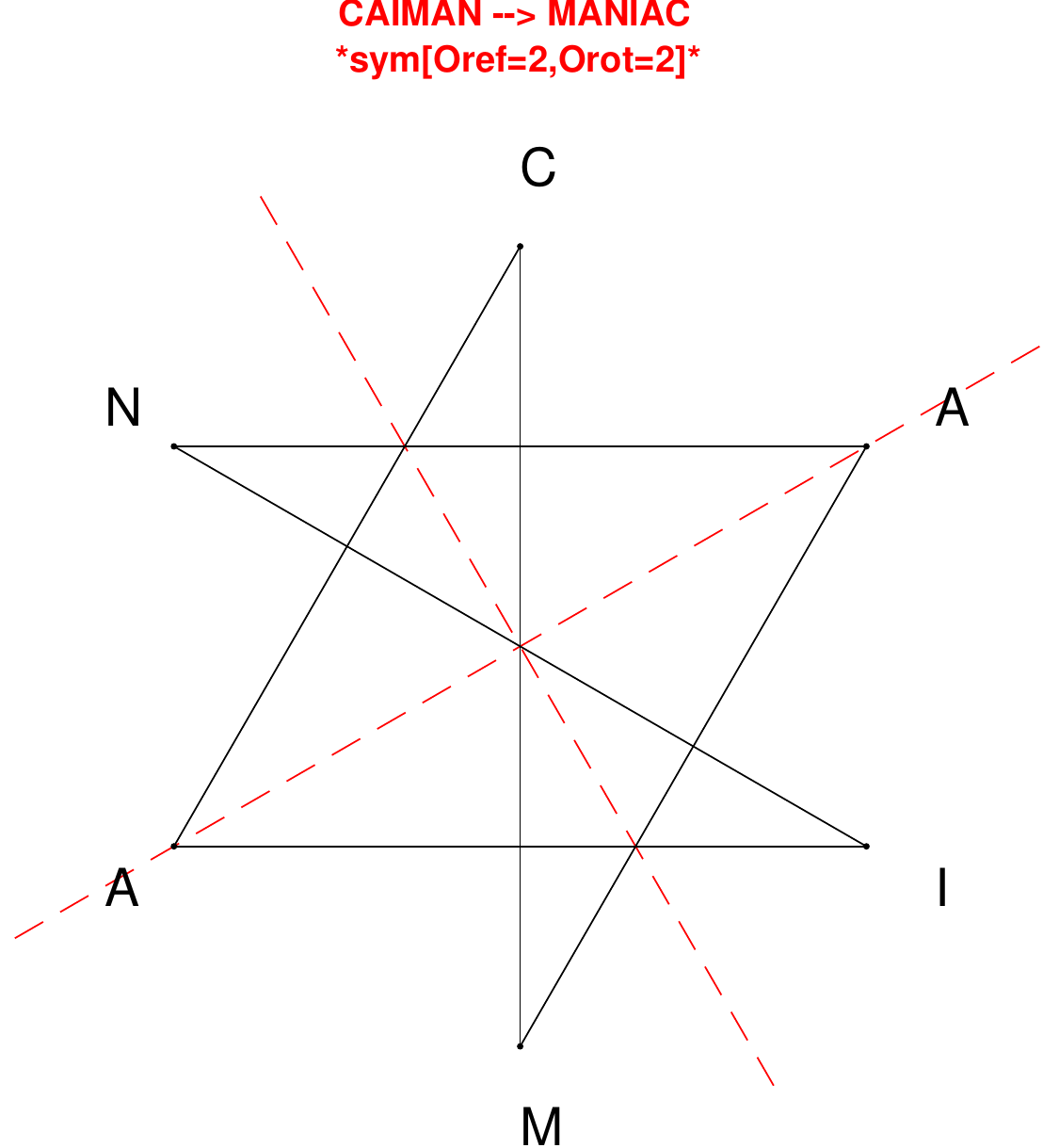}
\end{subfigure}
\hfill
\begin{subfigure}[T]{0.19\textwidth}
\centering
\includegraphics[width=\textwidth]{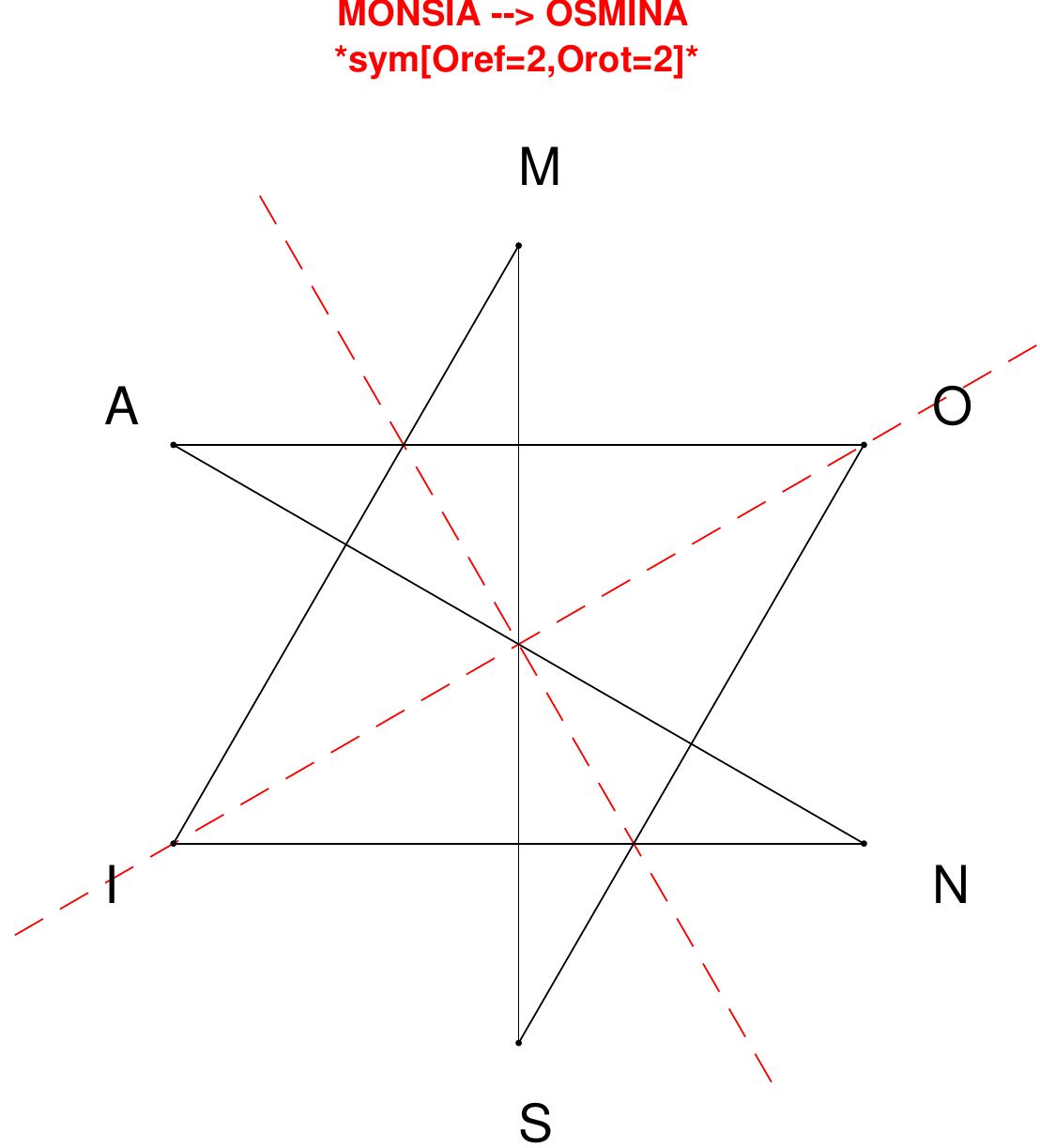}
\end{subfigure}
\end{figure}

\begin{figure}[H]
\centering
\begin{subfigure}[T]{0.19\textwidth}
\centering
\includegraphics[width=\textwidth]{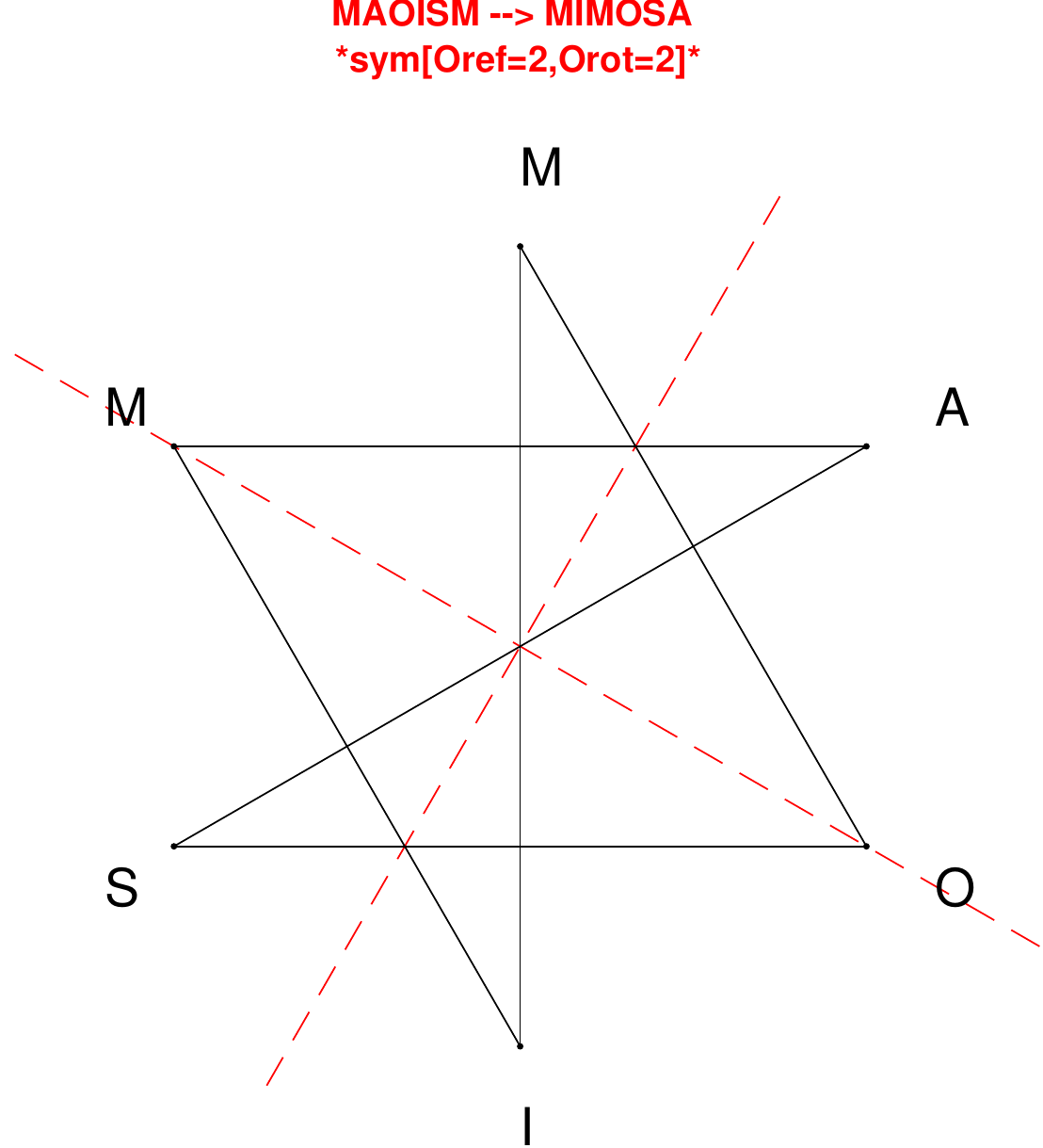}
\end{subfigure}
\hfill
\begin{subfigure}[T]{0.19\textwidth}
\centering
\includegraphics[width=\textwidth]{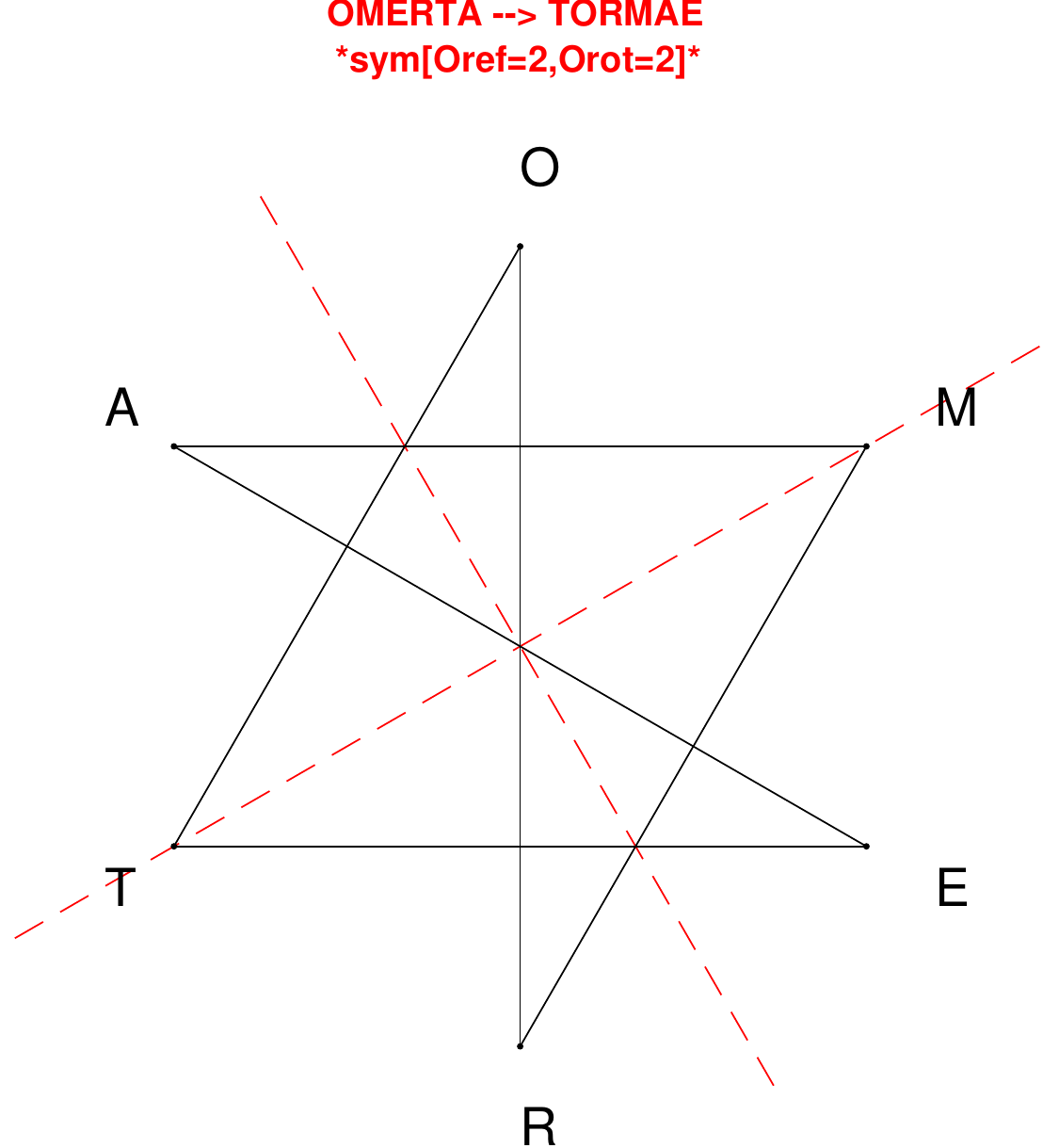}
\end{subfigure}
\hfill
\begin{subfigure}[T]{0.19\textwidth}
\centering
\includegraphics[width=\textwidth]{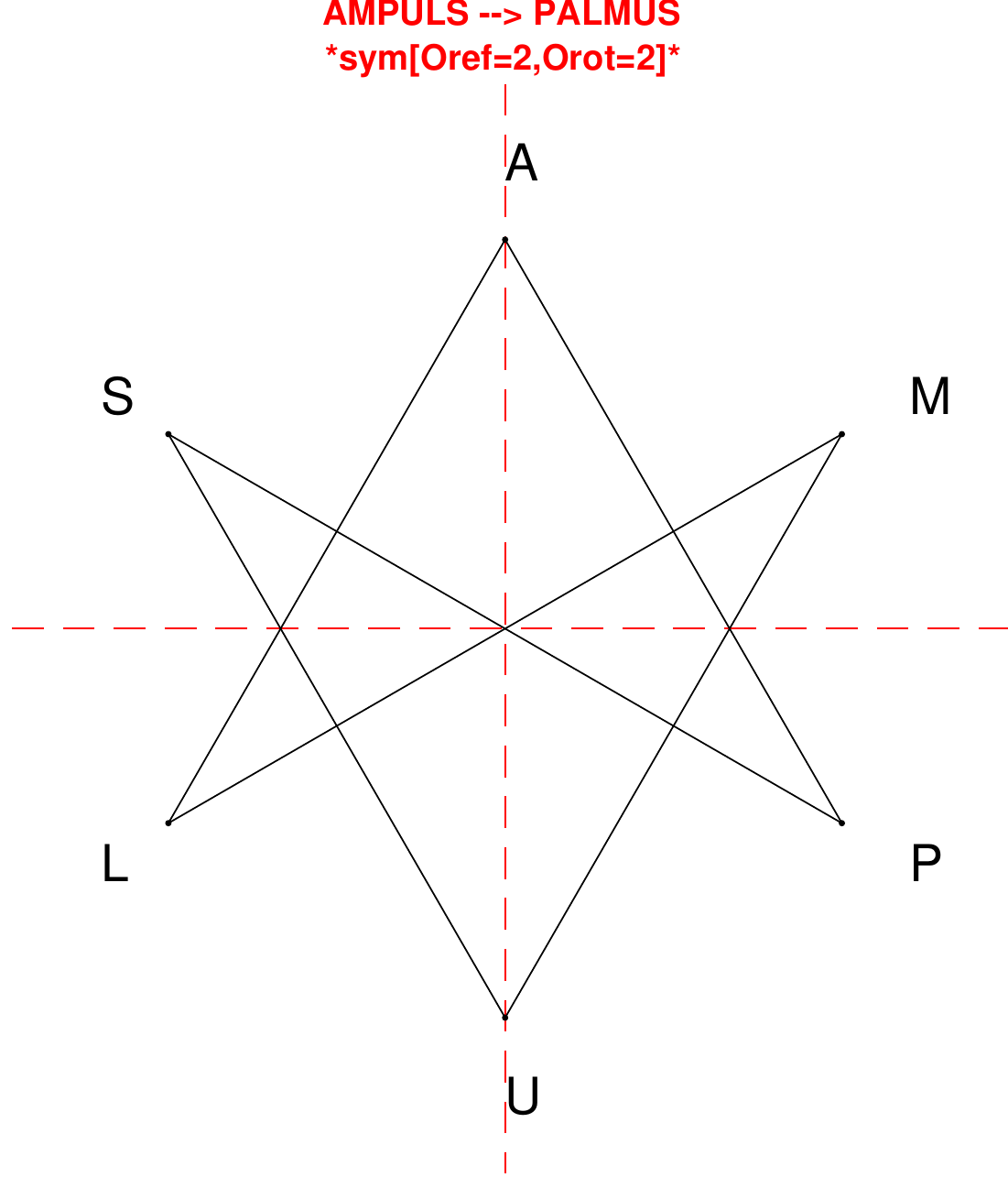}
\end{subfigure}
\hfill
\begin{subfigure}[T]{0.19\textwidth}
\centering
\includegraphics[width=\textwidth]{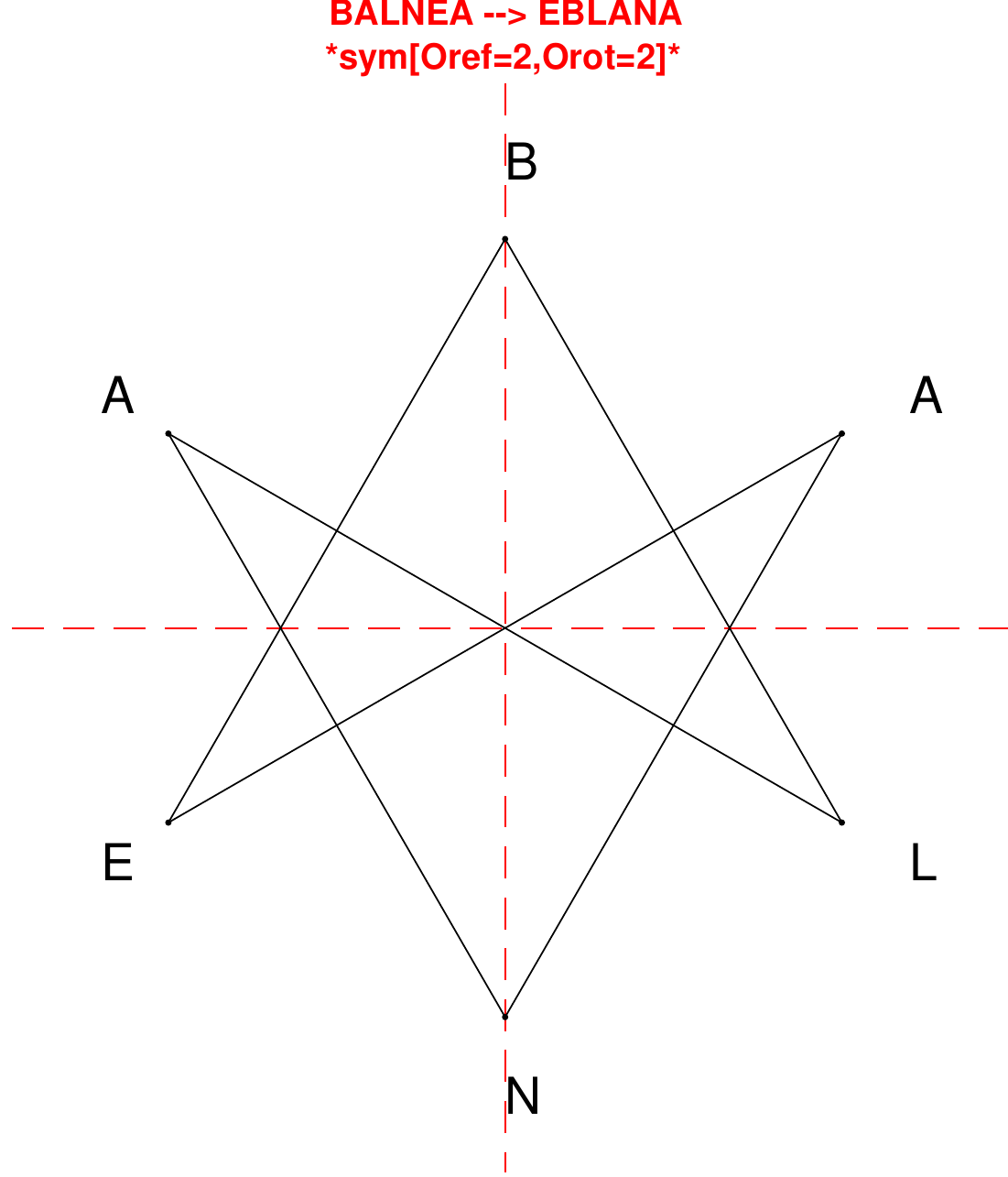}
\end{subfigure}
\hfill
\begin{subfigure}[T]{0.19\textwidth}
\centering
\includegraphics[width=\textwidth]{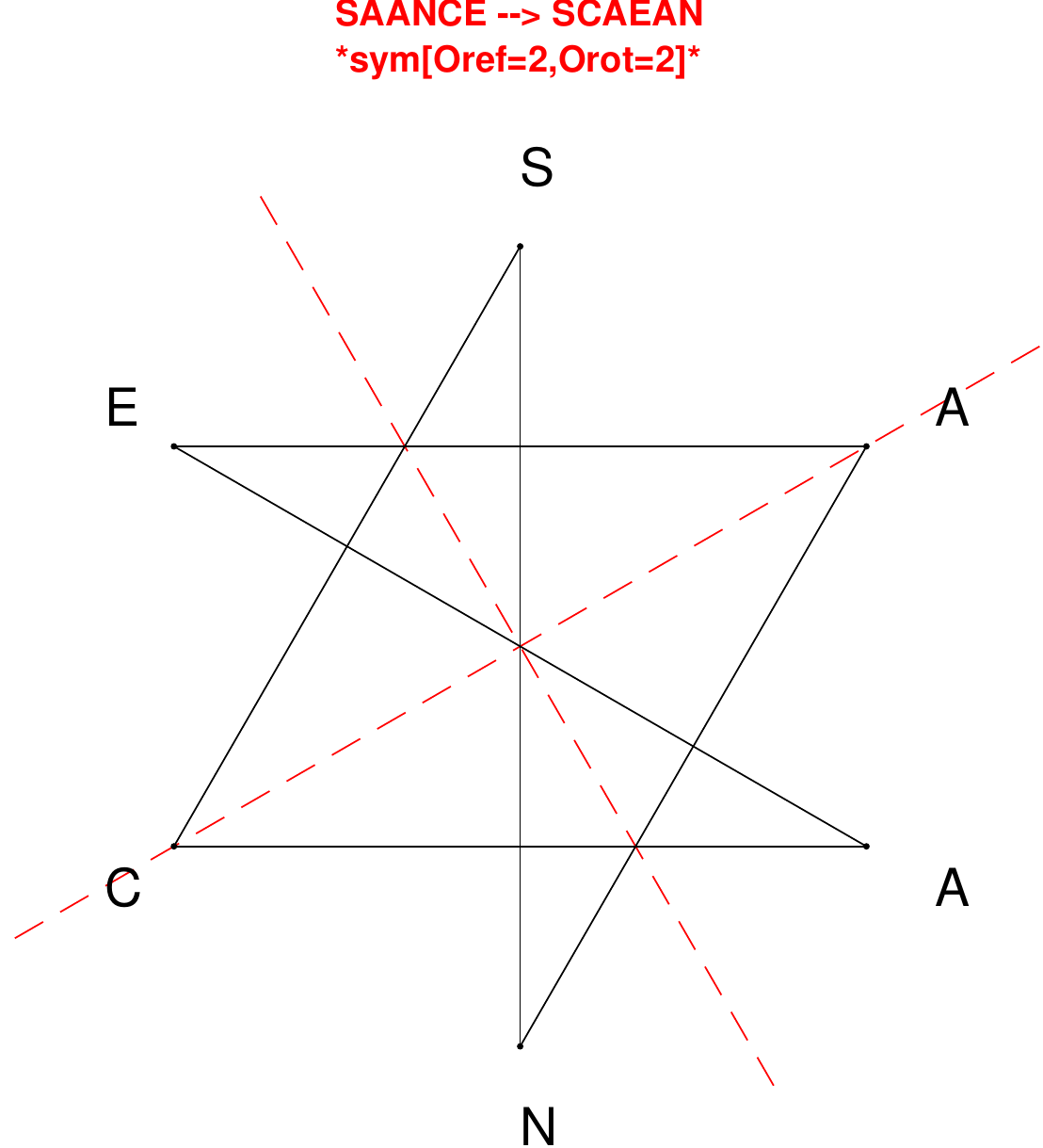}
\end{subfigure}
\end{figure}

\begin{figure}[H]
\centering
\begin{subfigure}[T]{0.19\textwidth}
\centering
\includegraphics[width=\textwidth]{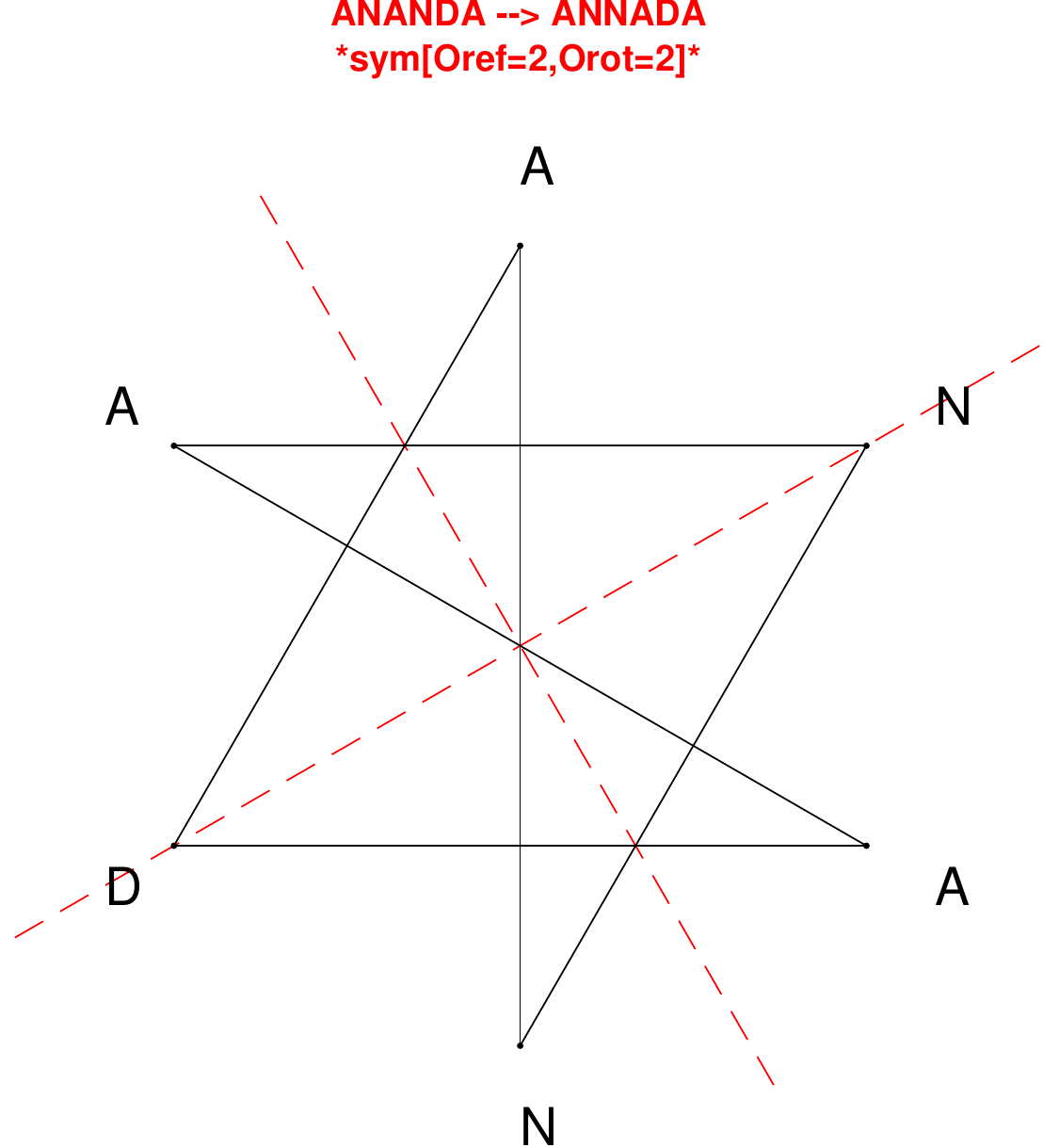}
\end{subfigure}
\hfill
\begin{subfigure}[T]{0.19\textwidth}
\centering
\includegraphics[width=\textwidth]{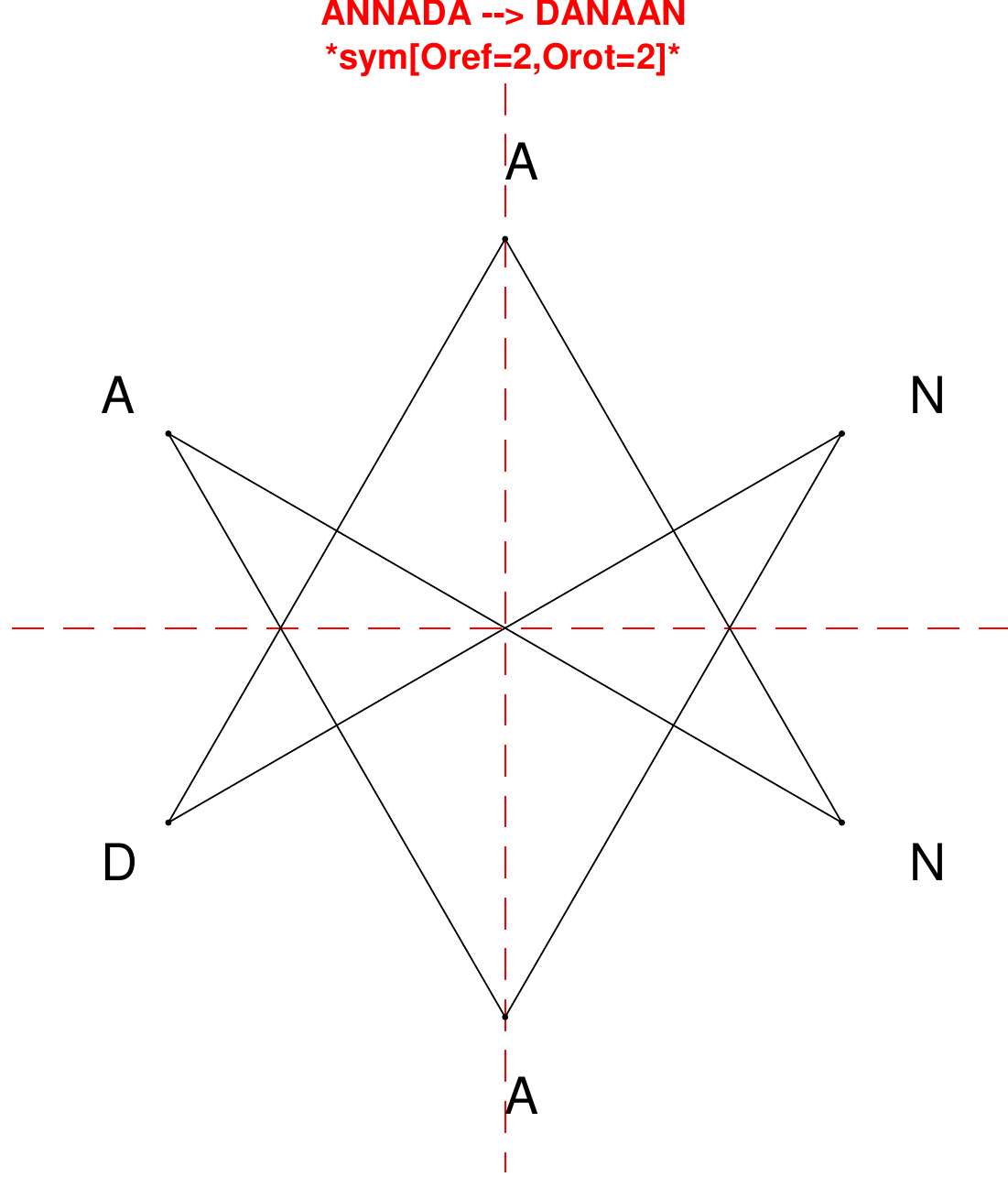}
\end{subfigure}
\hfill
\begin{subfigure}[T]{0.19\textwidth}
\centering
\includegraphics[width=\textwidth]{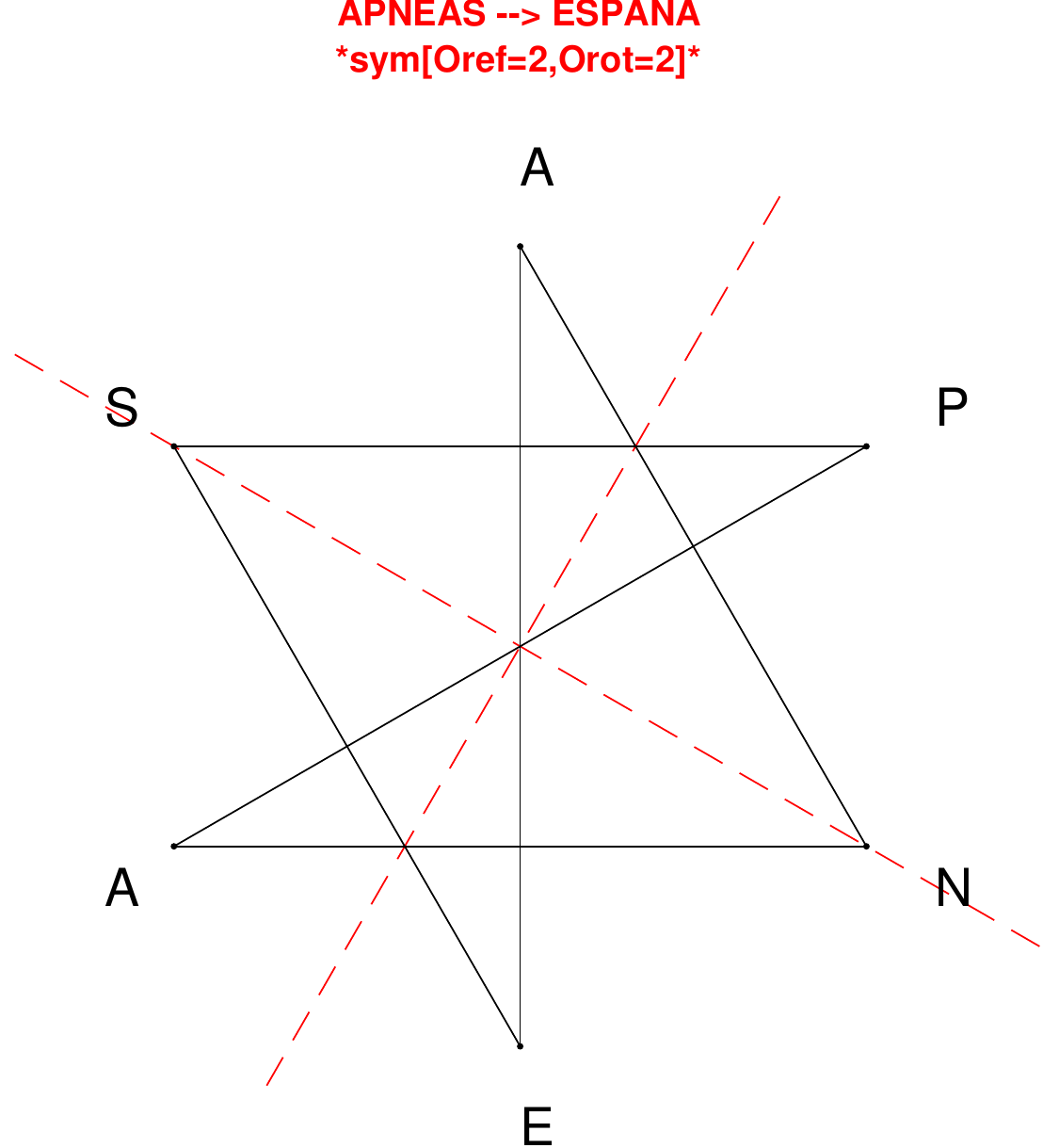}
\end{subfigure}
\hfill
\begin{subfigure}[T]{0.19\textwidth}
\centering
\includegraphics[width=\textwidth]{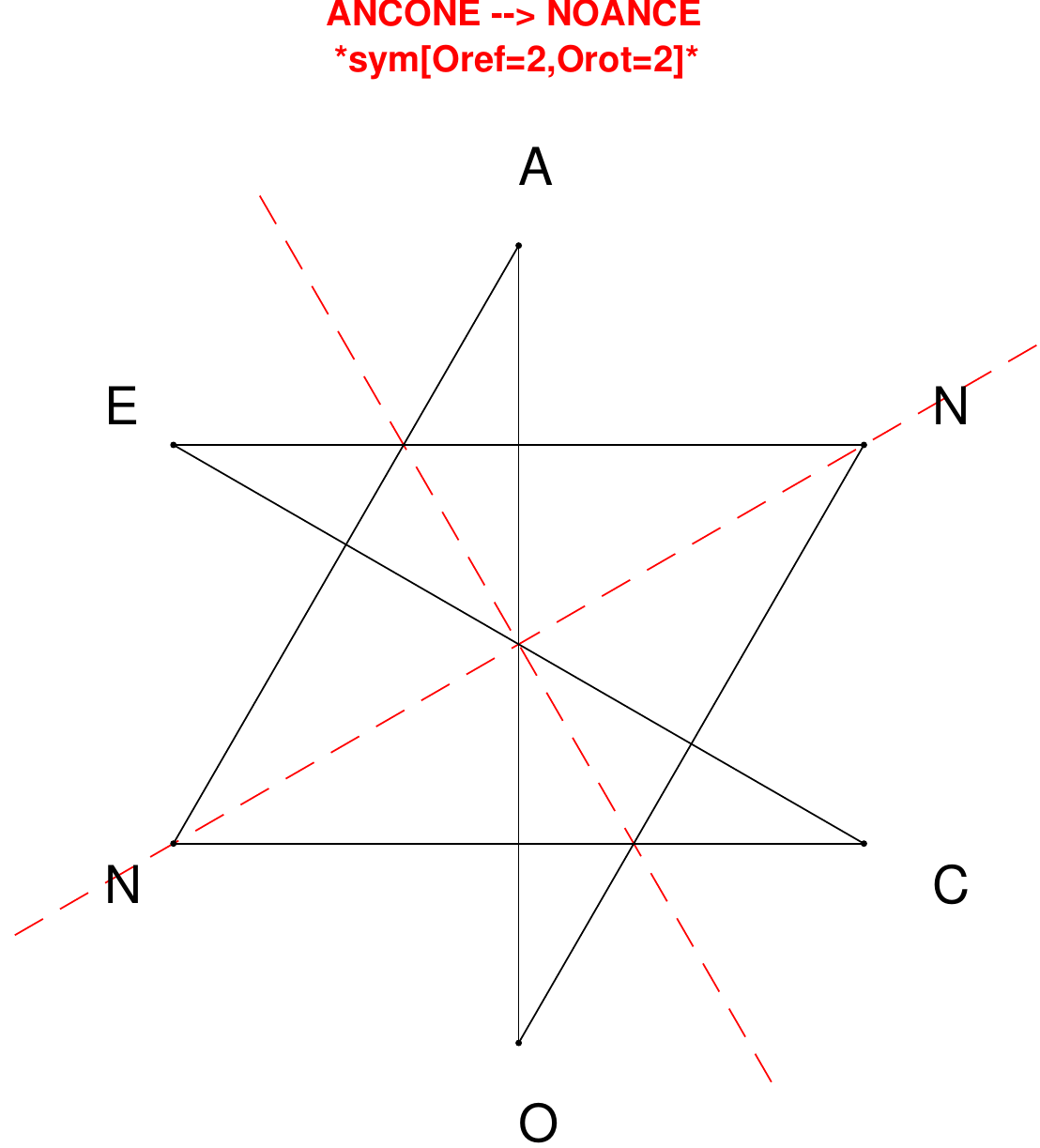}
\end{subfigure}
\hfill
\begin{subfigure}[T]{0.19\textwidth}
\centering
\includegraphics[width=\textwidth]{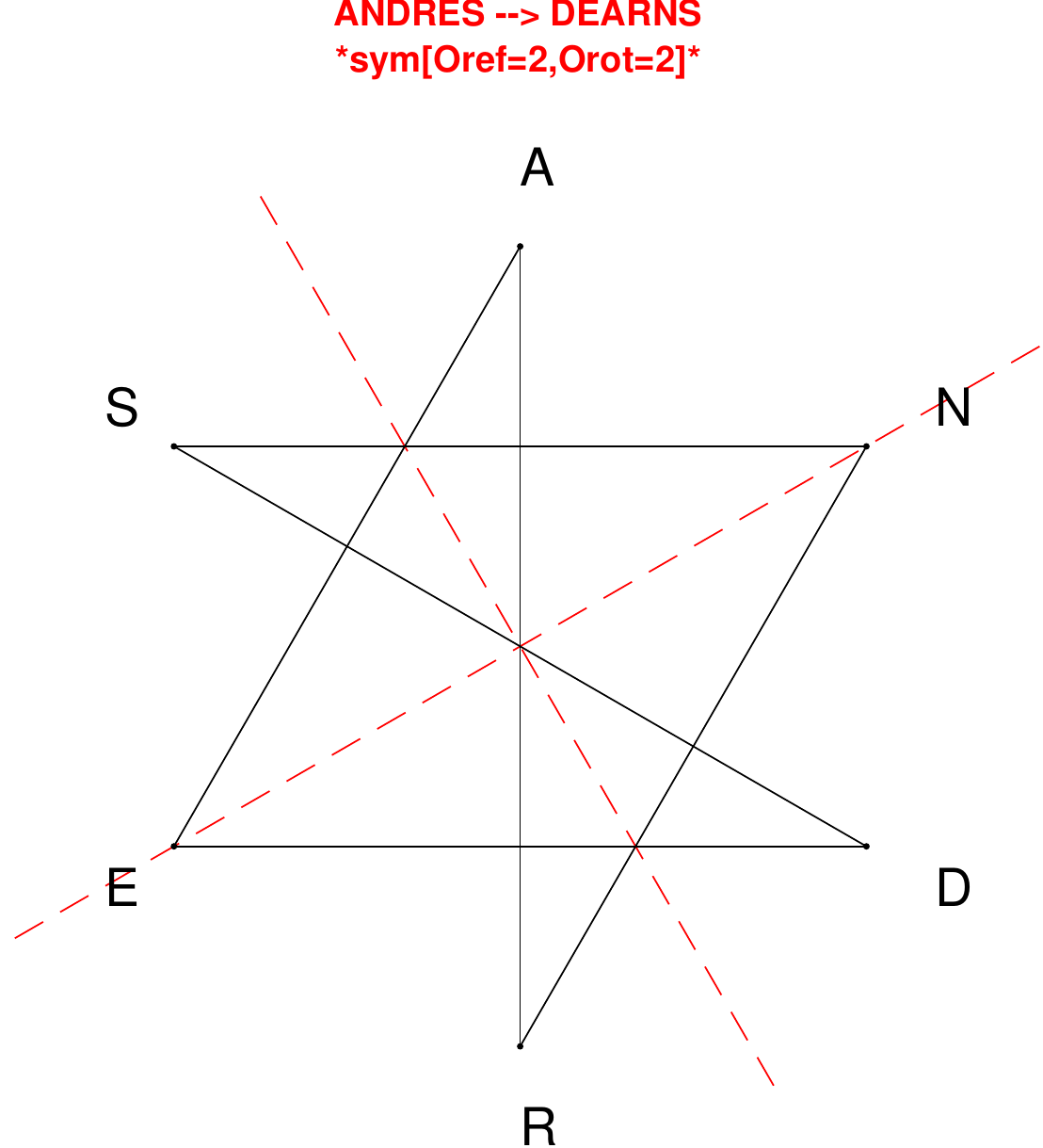}
\end{subfigure}
\end{figure}

\begin{figure}[H]
\centering
\begin{subfigure}[T]{0.19\textwidth}
\centering
\includegraphics[width=\textwidth]{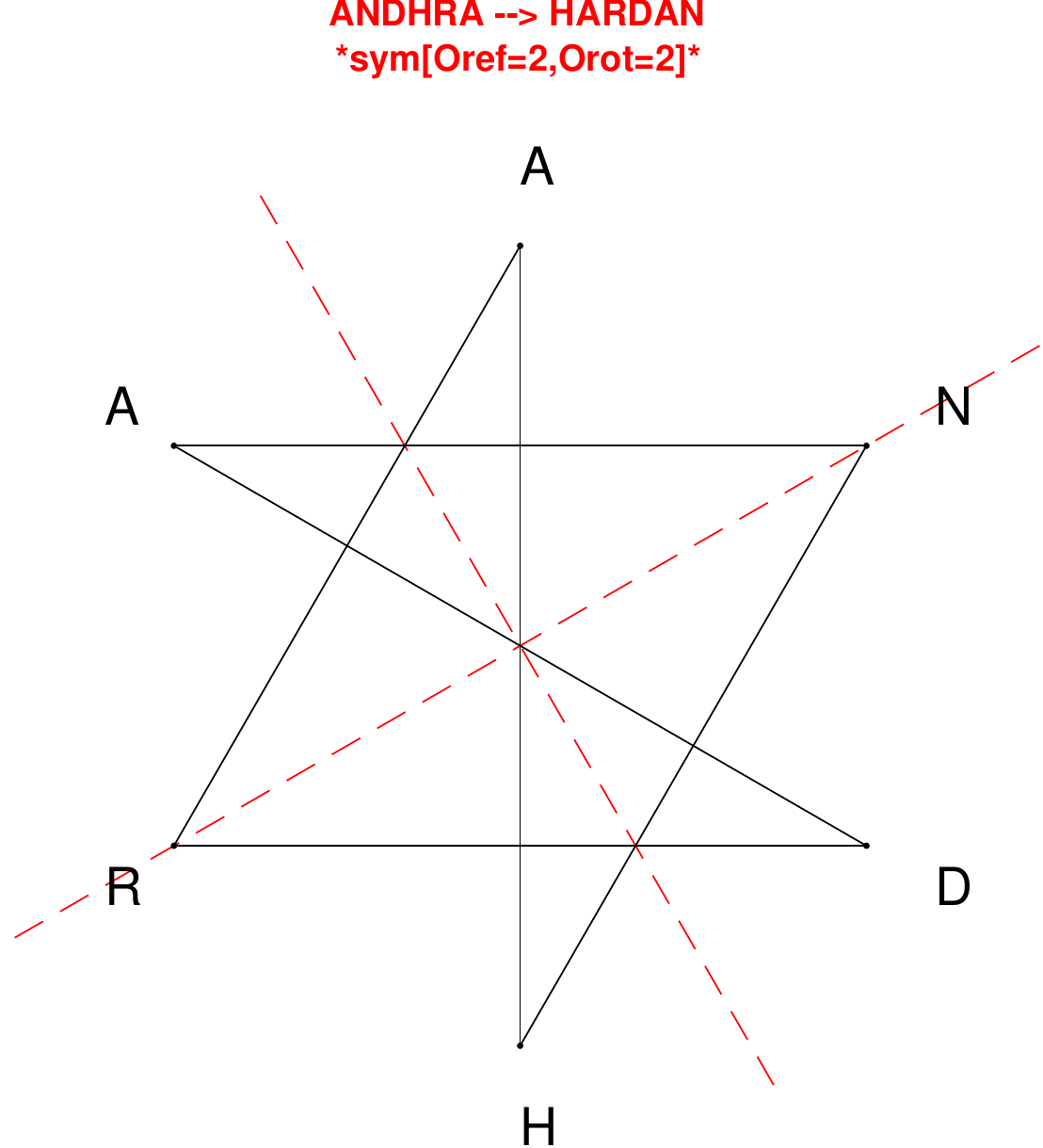}
\end{subfigure}
\hfill
\begin{subfigure}[T]{0.19\textwidth}
\centering
\includegraphics[width=\textwidth]{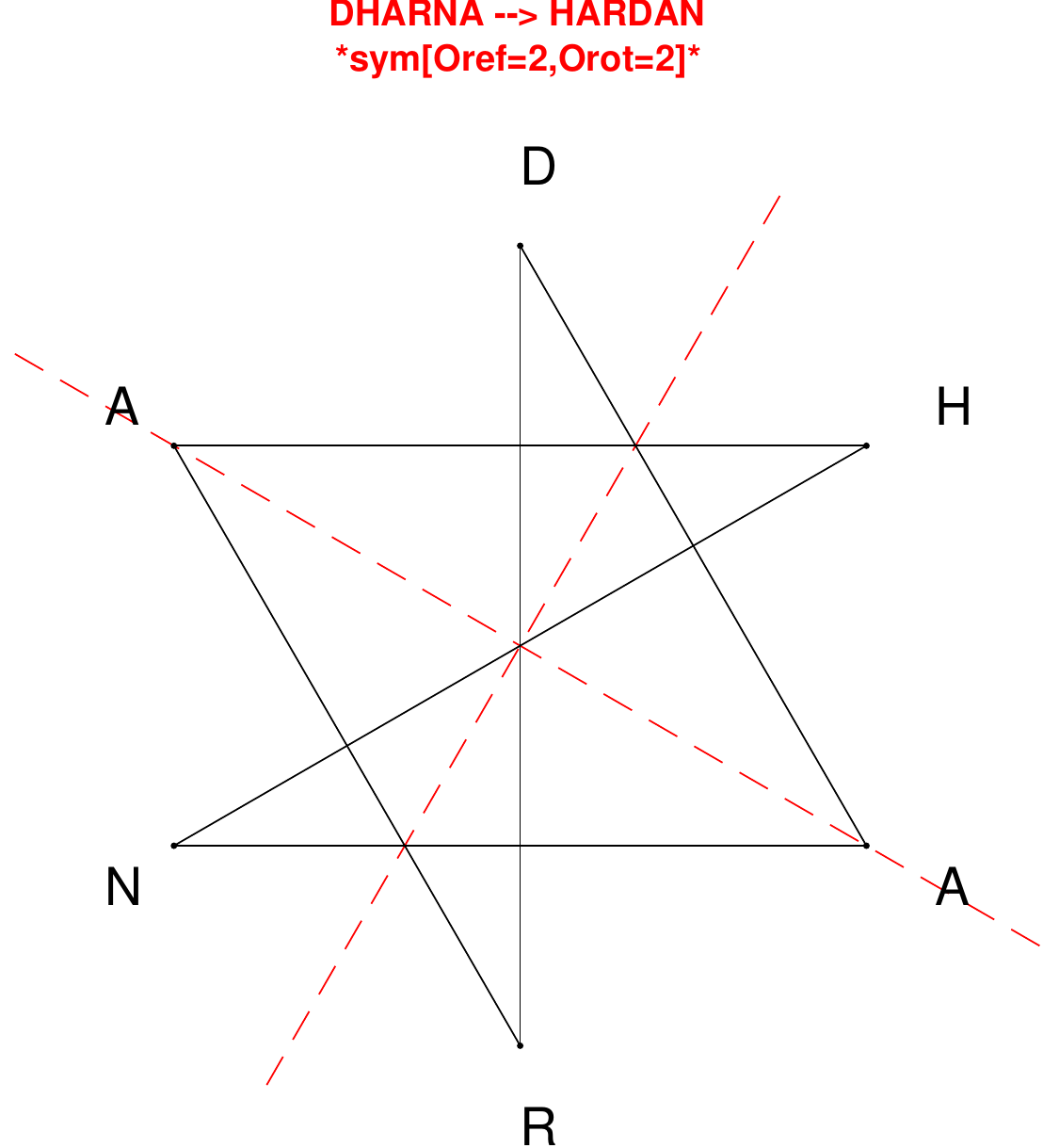}
\end{subfigure}
\hfill
\begin{subfigure}[T]{0.19\textwidth}
\centering
\includegraphics[width=\textwidth]{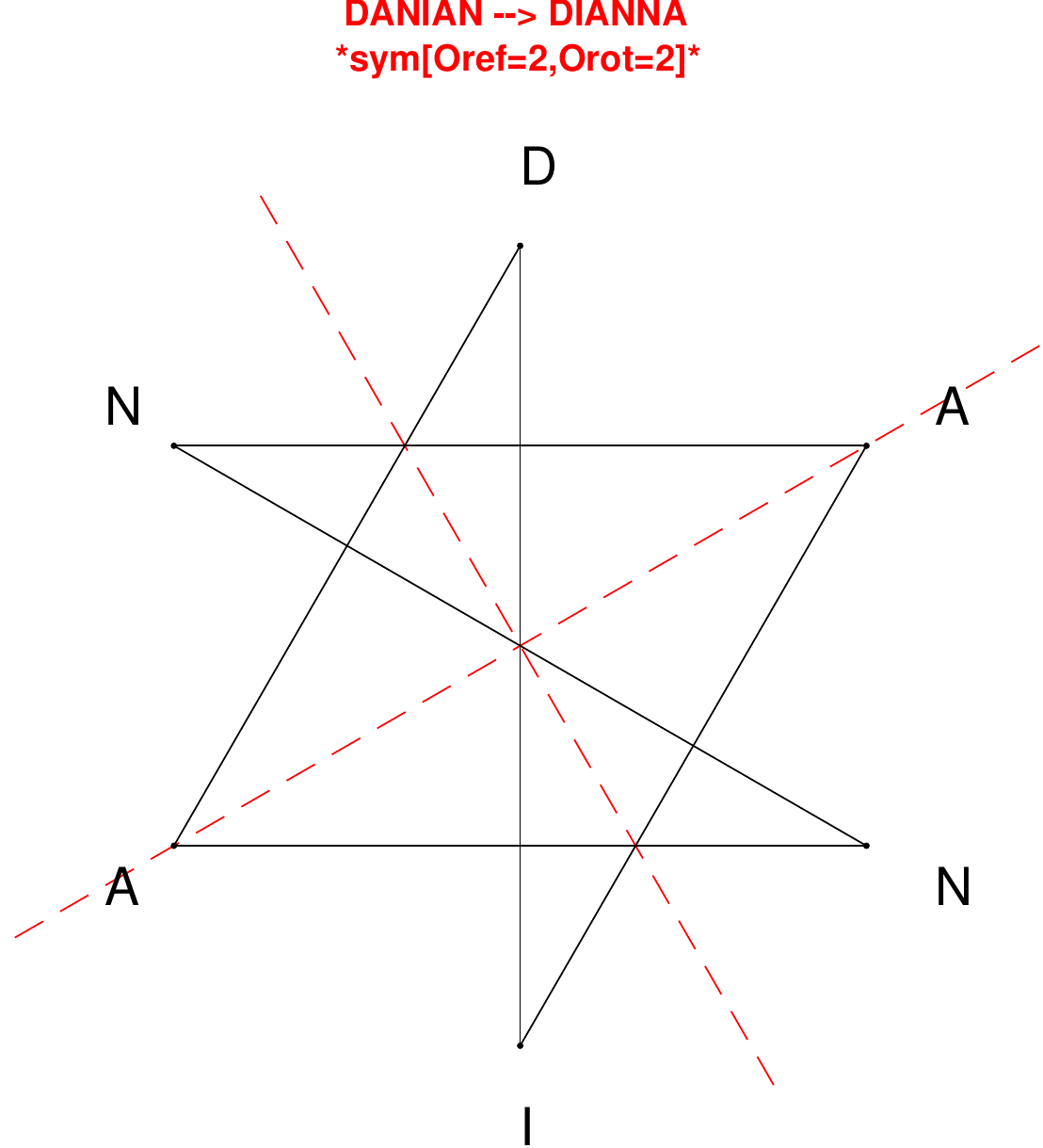}
\end{subfigure}
\hfill
\begin{subfigure}[T]{0.19\textwidth}
\centering
\includegraphics[width=\textwidth]{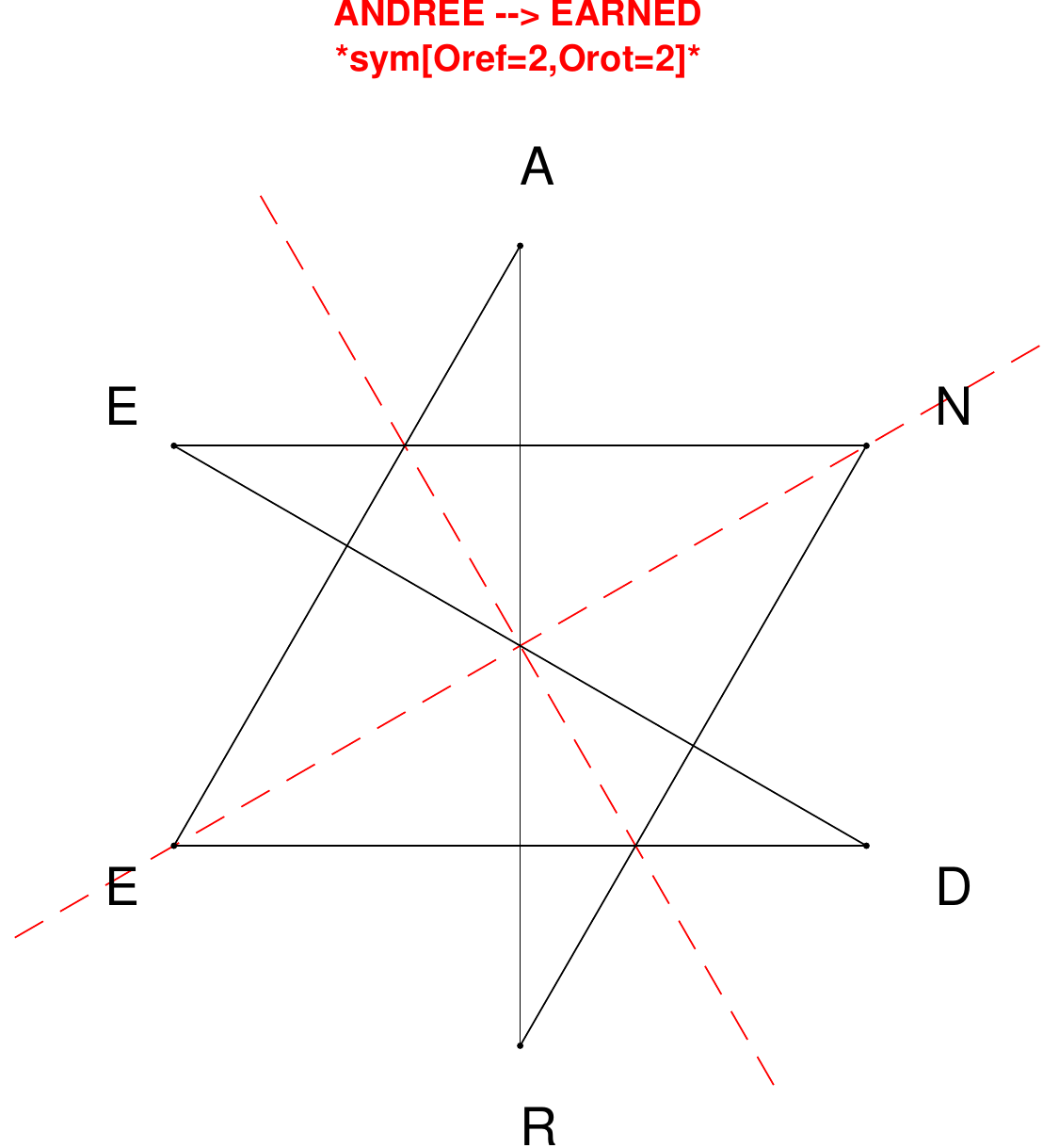}
\end{subfigure}
\hfill
\begin{subfigure}[T]{0.19\textwidth}
\centering
\includegraphics[width=\textwidth]{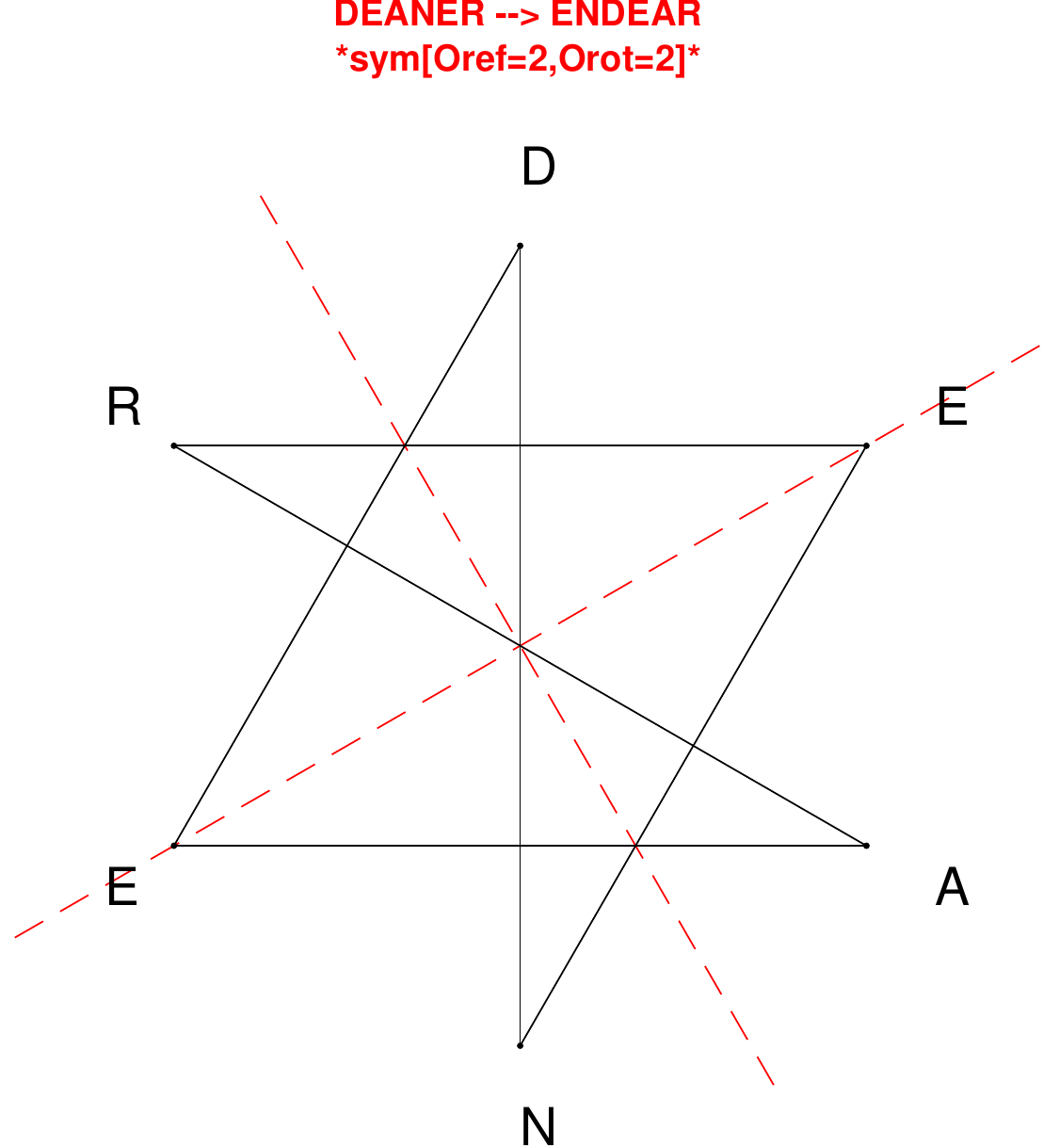}
\end{subfigure}
\end{figure}

\begin{figure}[H]
\centering
\begin{subfigure}[T]{0.19\textwidth}
\centering
\includegraphics[width=\textwidth]{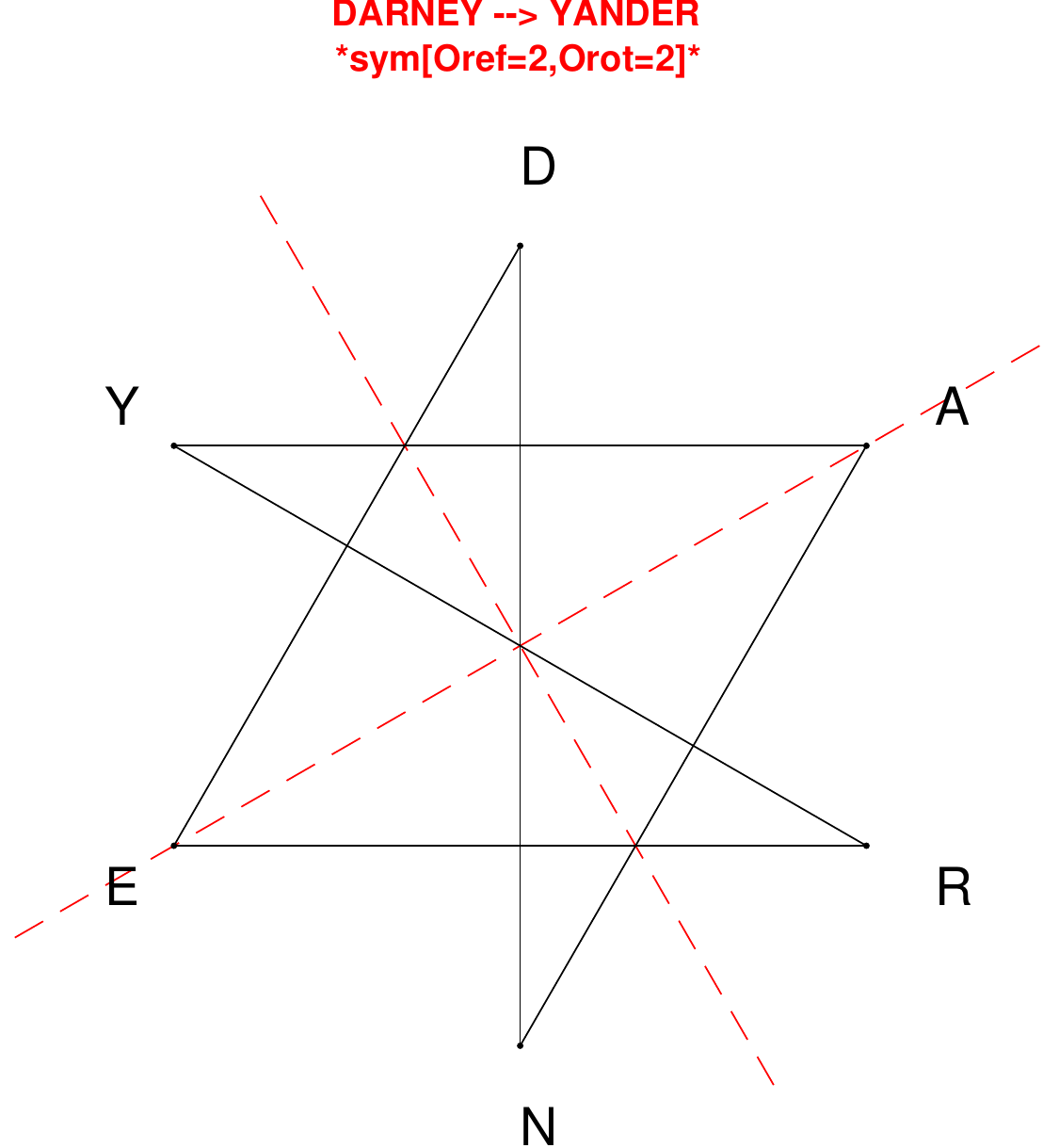}
\end{subfigure}
\hfill
\begin{subfigure}[T]{0.19\textwidth}
\centering
\includegraphics[width=\textwidth]{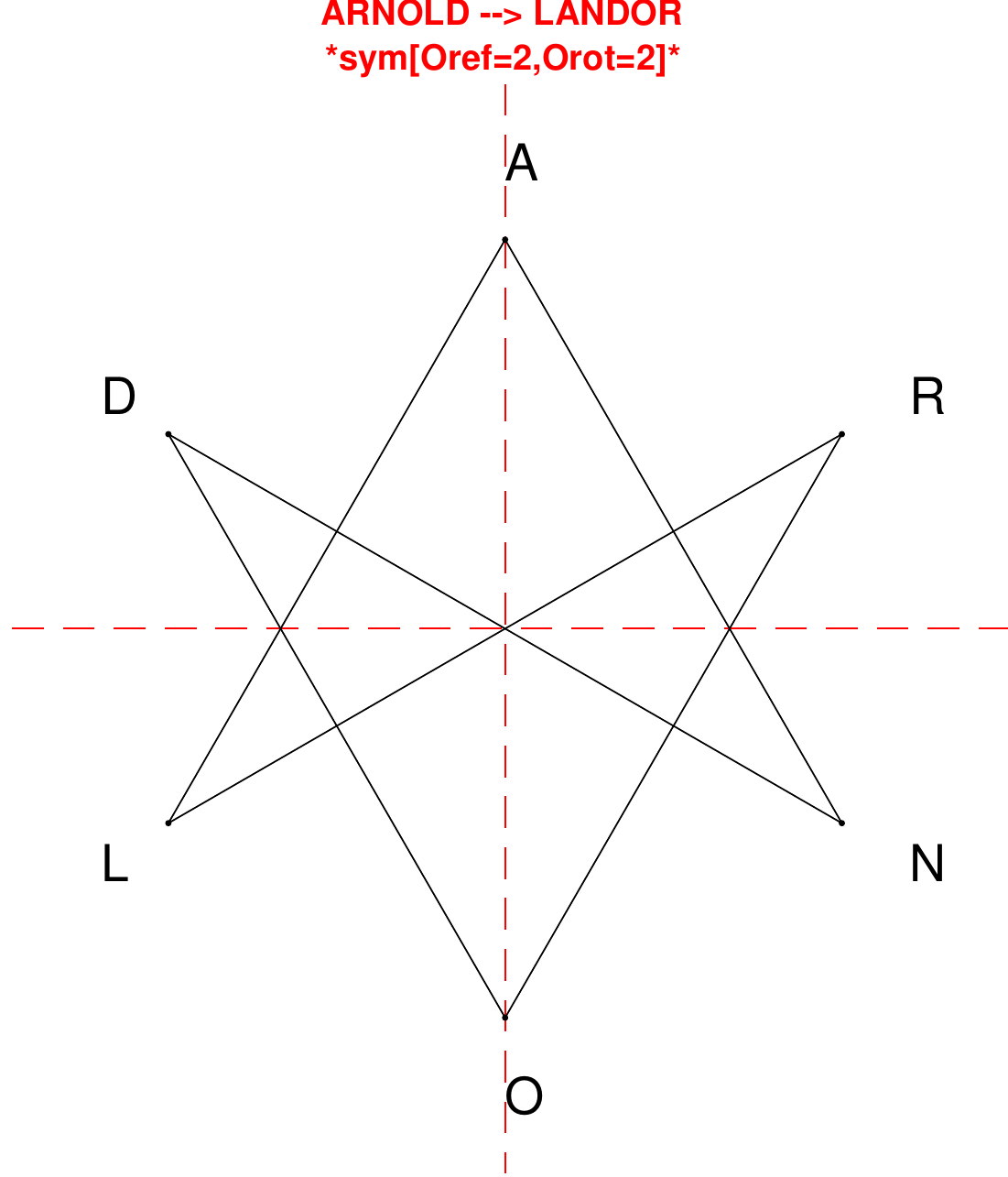}
\end{subfigure}
\hfill
\begin{subfigure}[T]{0.19\textwidth}
\centering
\includegraphics[width=\textwidth]{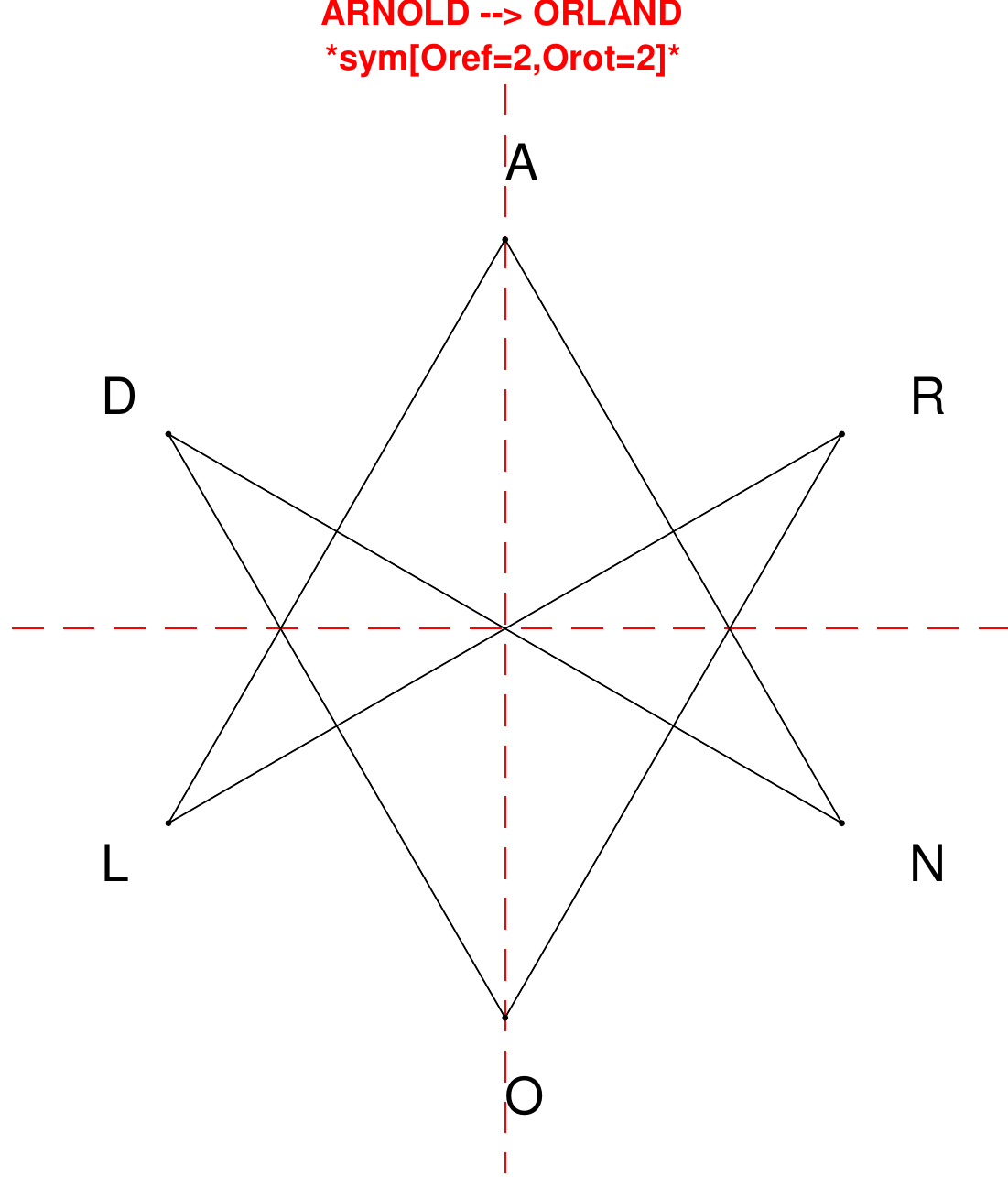}
\end{subfigure}
\hfill
\begin{subfigure}[T]{0.19\textwidth}
\centering
\includegraphics[width=\textwidth]{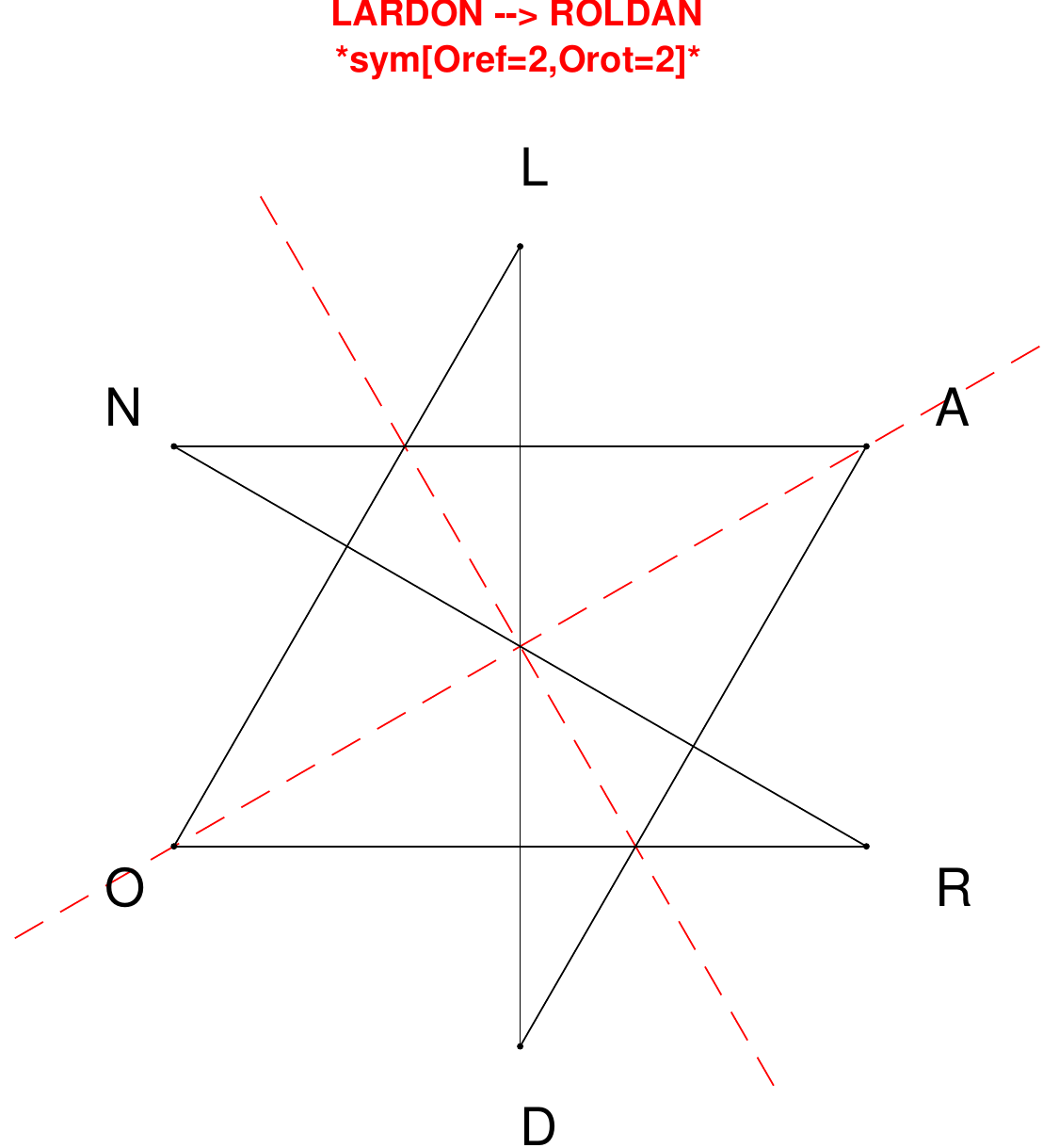}
\end{subfigure}
\hfill
\begin{subfigure}[T]{0.19\textwidth}
\centering
\includegraphics[width=\textwidth]{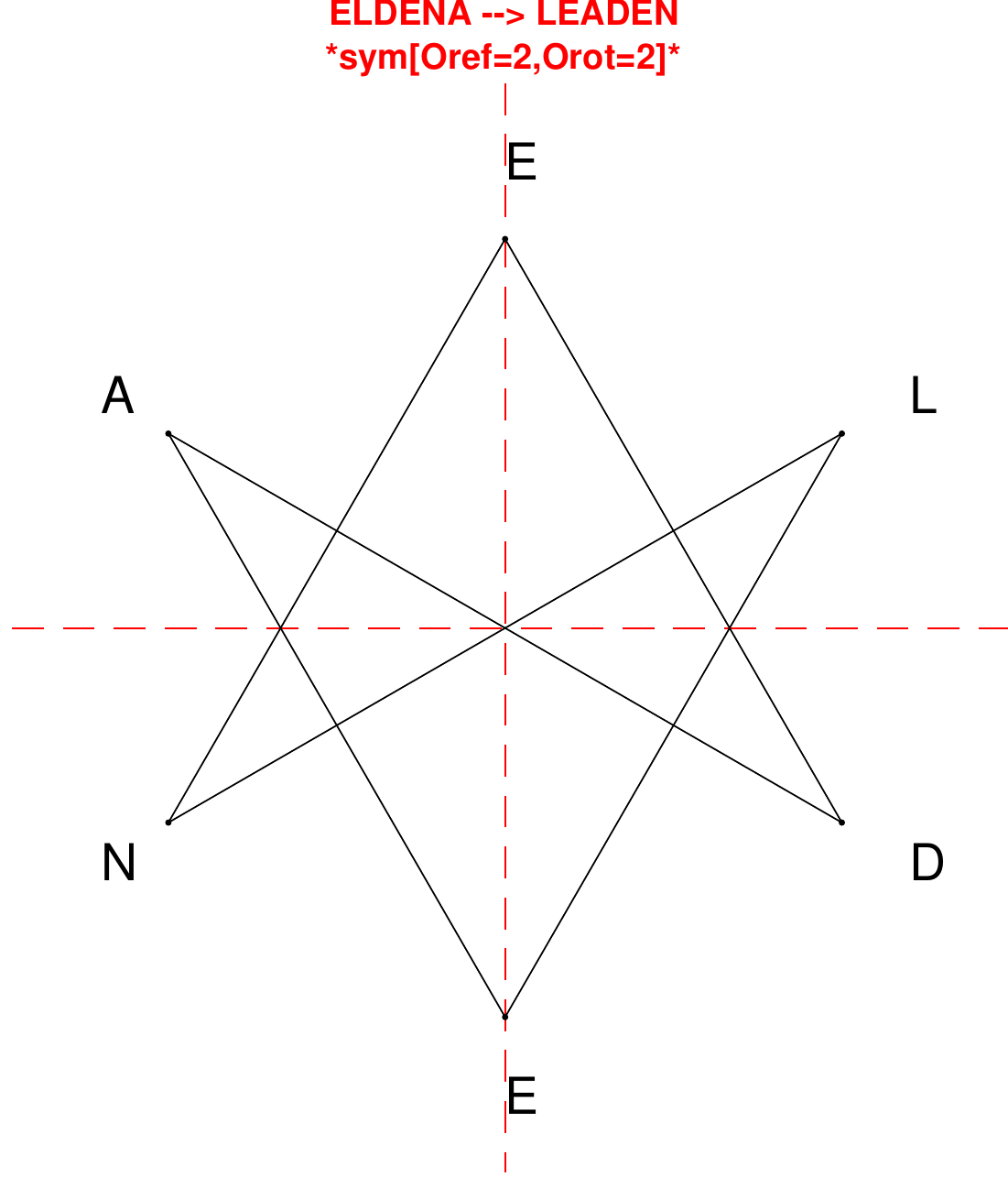}
\end{subfigure}
\end{figure}

\begin{figure}[H]
\centering
\begin{subfigure}[T]{0.19\textwidth}
\centering
\includegraphics[width=\textwidth]{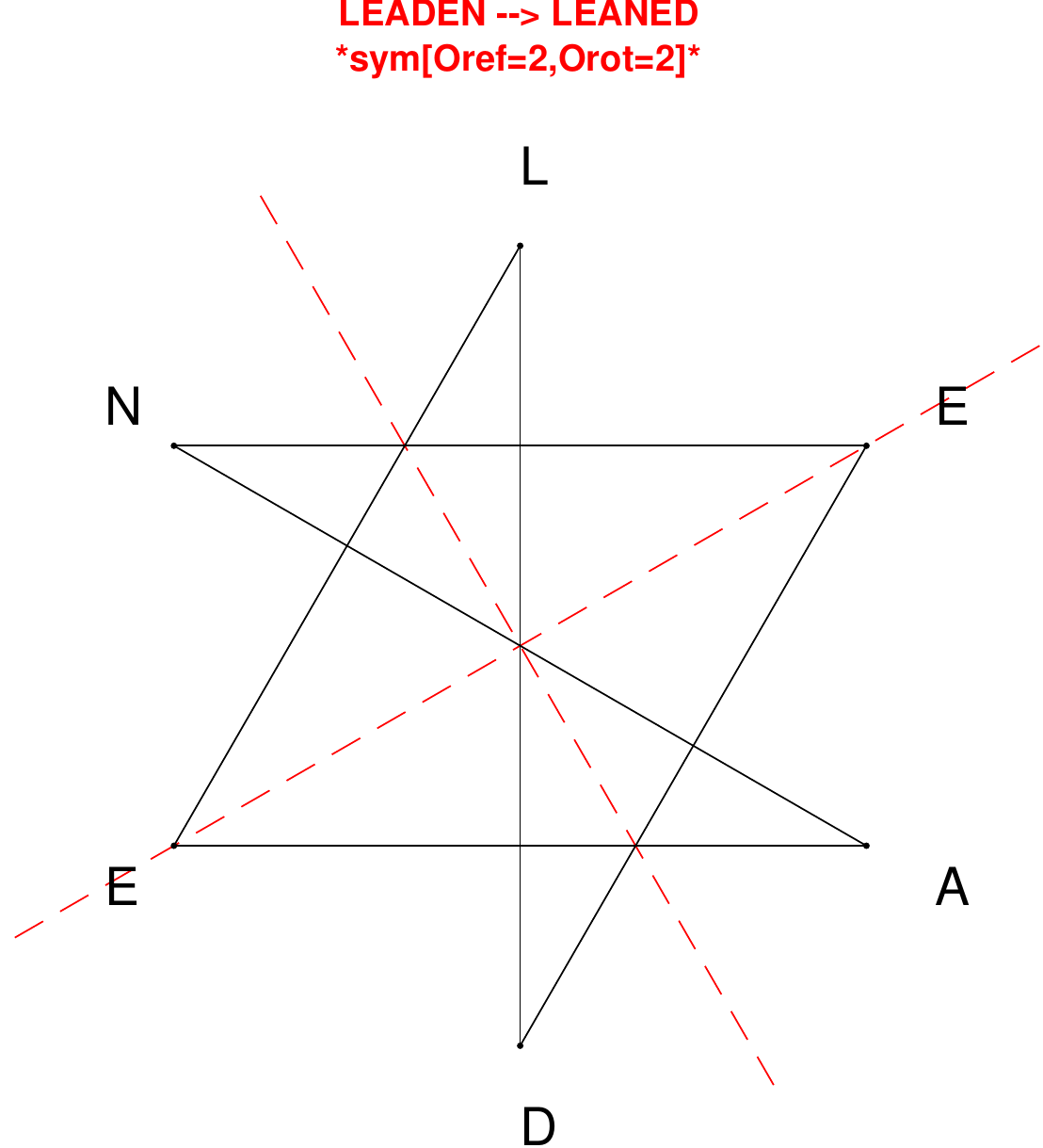}
\end{subfigure}
\hfill
\begin{subfigure}[T]{0.19\textwidth}
\centering
\includegraphics[width=\textwidth]{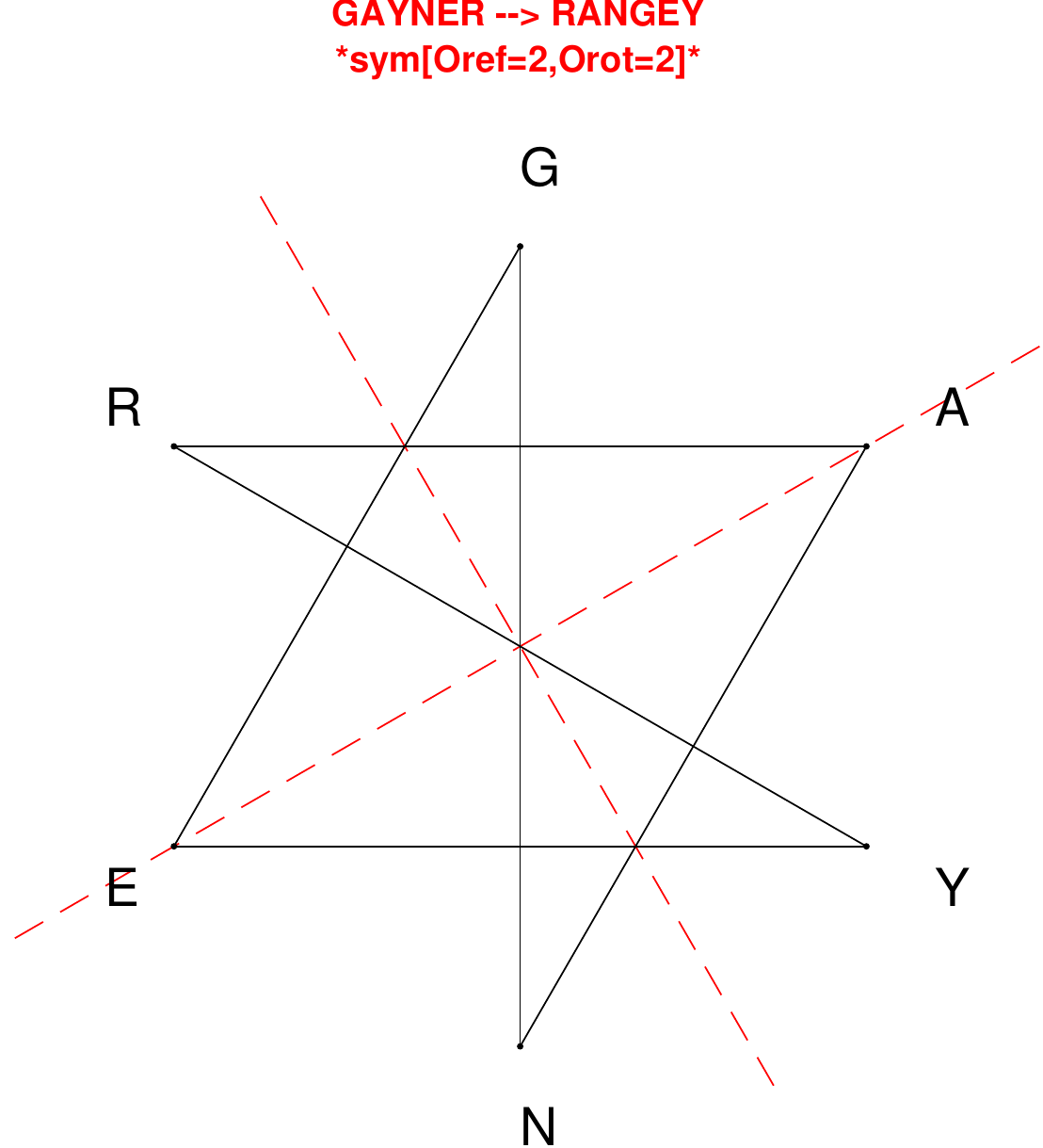}
\end{subfigure}
\hfill
\begin{subfigure}[T]{0.19\textwidth}
\centering
\includegraphics[width=\textwidth]{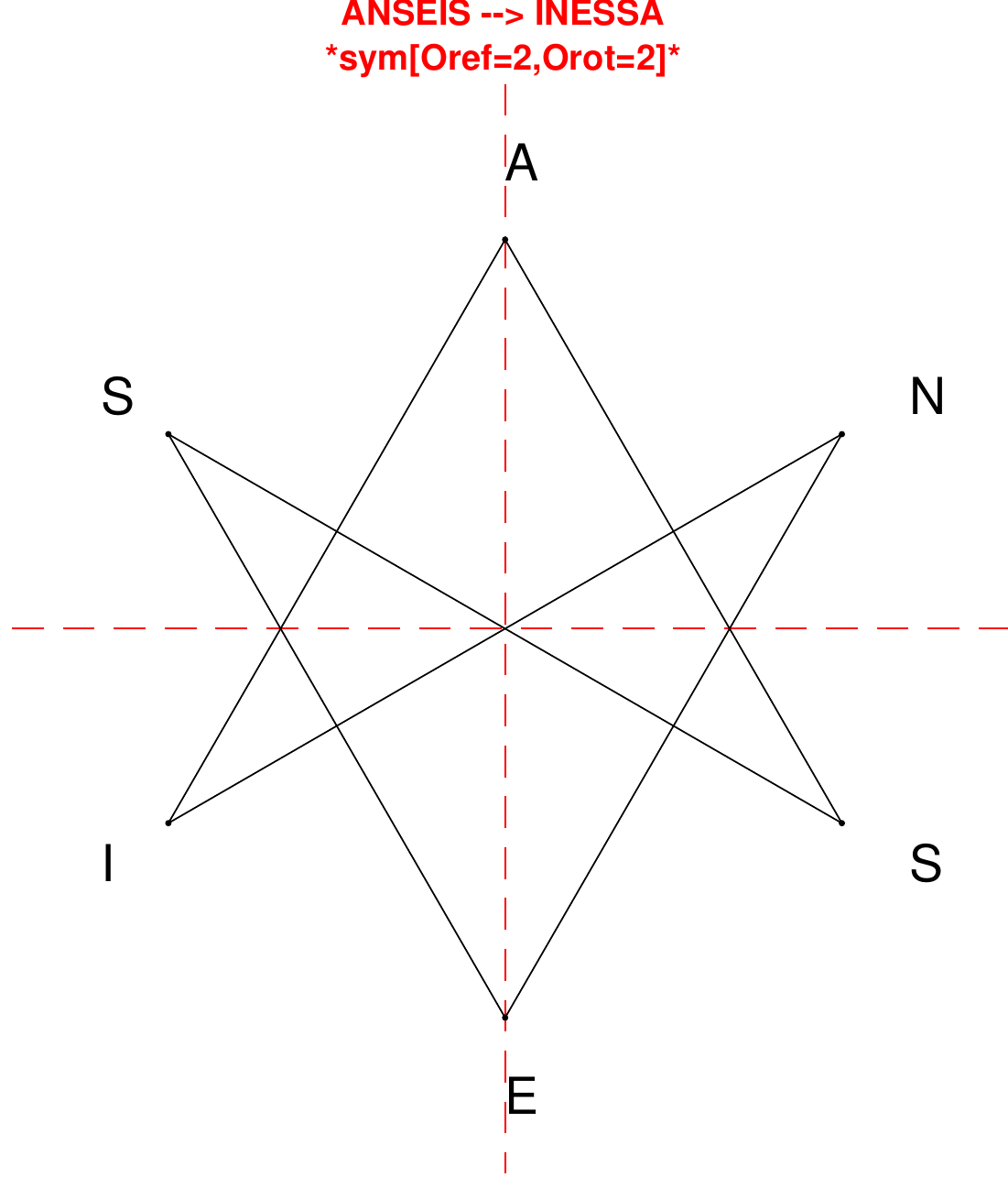}
\end{subfigure}
\hfill
\begin{subfigure}[T]{0.19\textwidth}
\centering
\includegraphics[width=\textwidth]{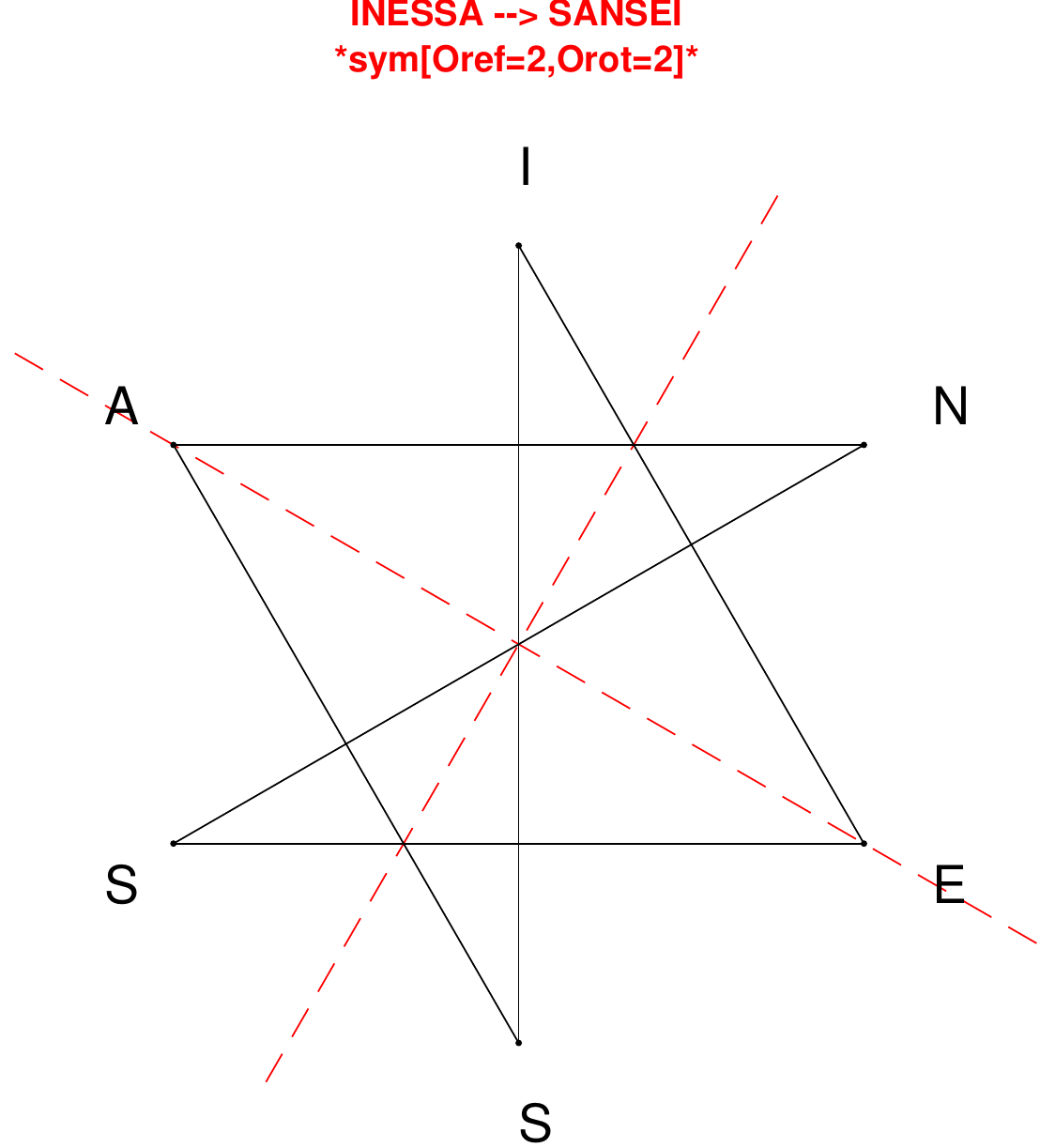}
\end{subfigure}
\hfill
\begin{subfigure}[T]{0.19\textwidth}
\centering
\includegraphics[width=\textwidth]{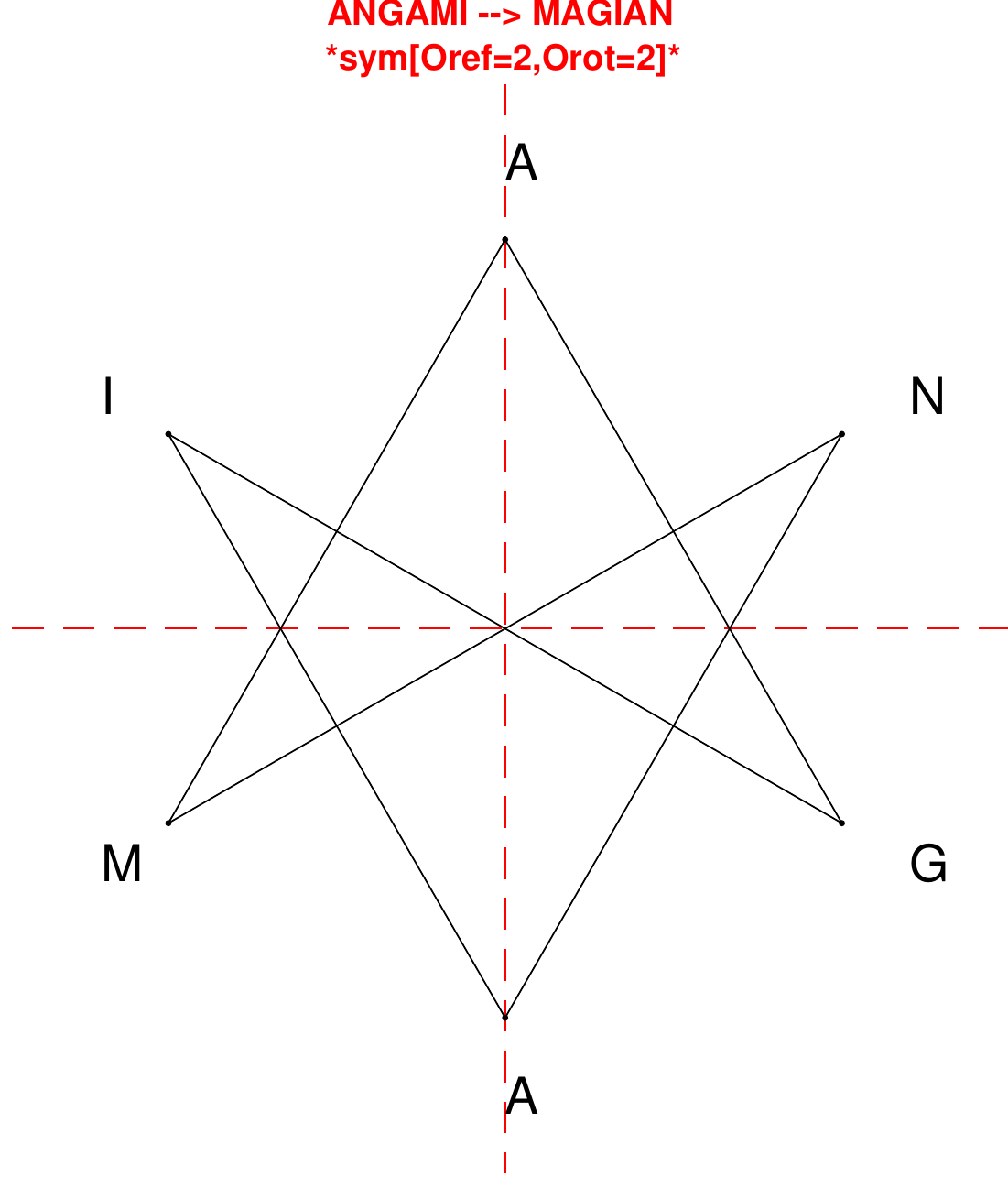}
\end{subfigure}
\end{figure}

\begin{figure}[H]
\centering
\begin{subfigure}[T]{0.19\textwidth}
\centering
\includegraphics[width=\textwidth]{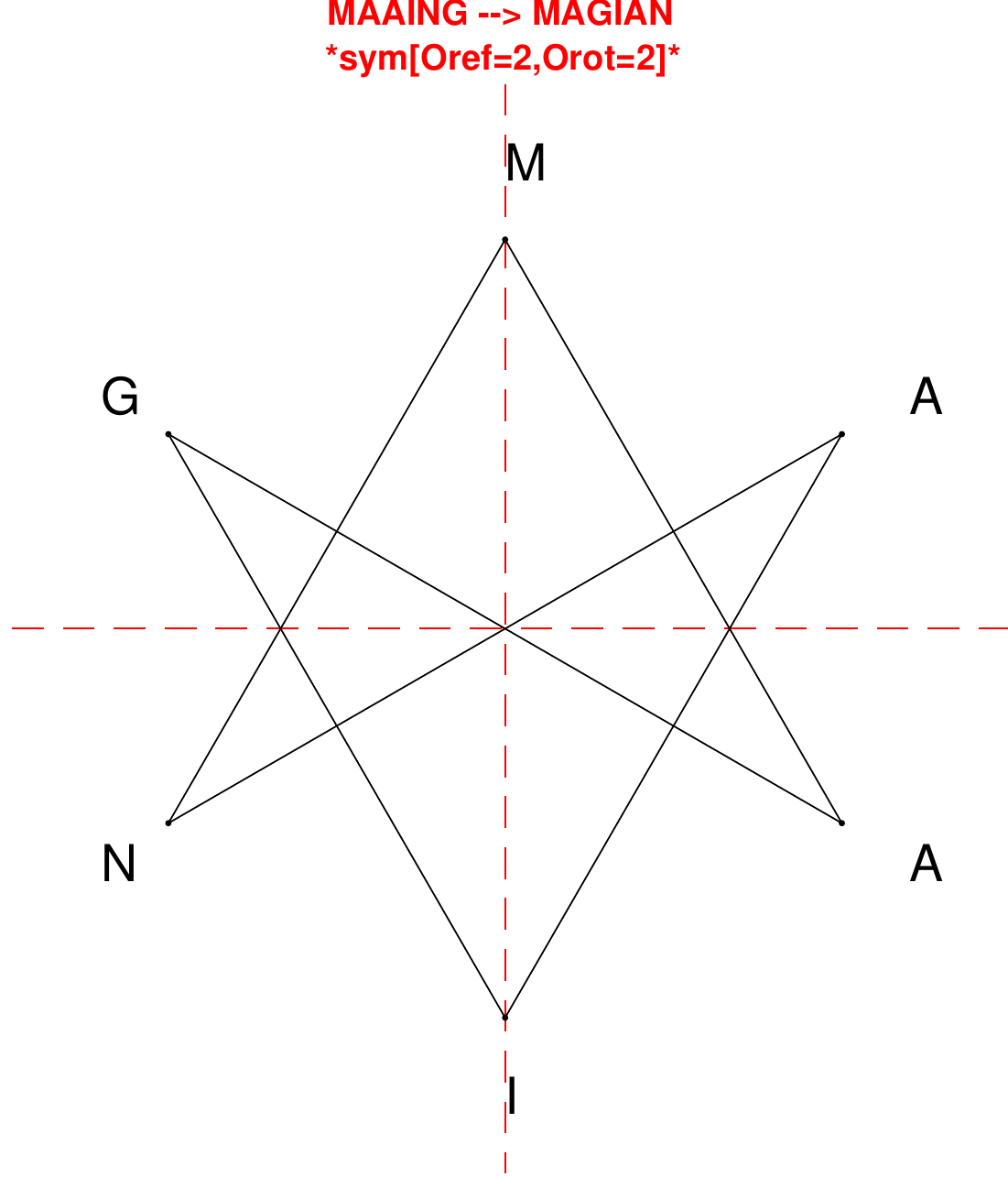}
\end{subfigure}
\hfill
\begin{subfigure}[T]{0.19\textwidth}
\centering
\includegraphics[width=\textwidth]{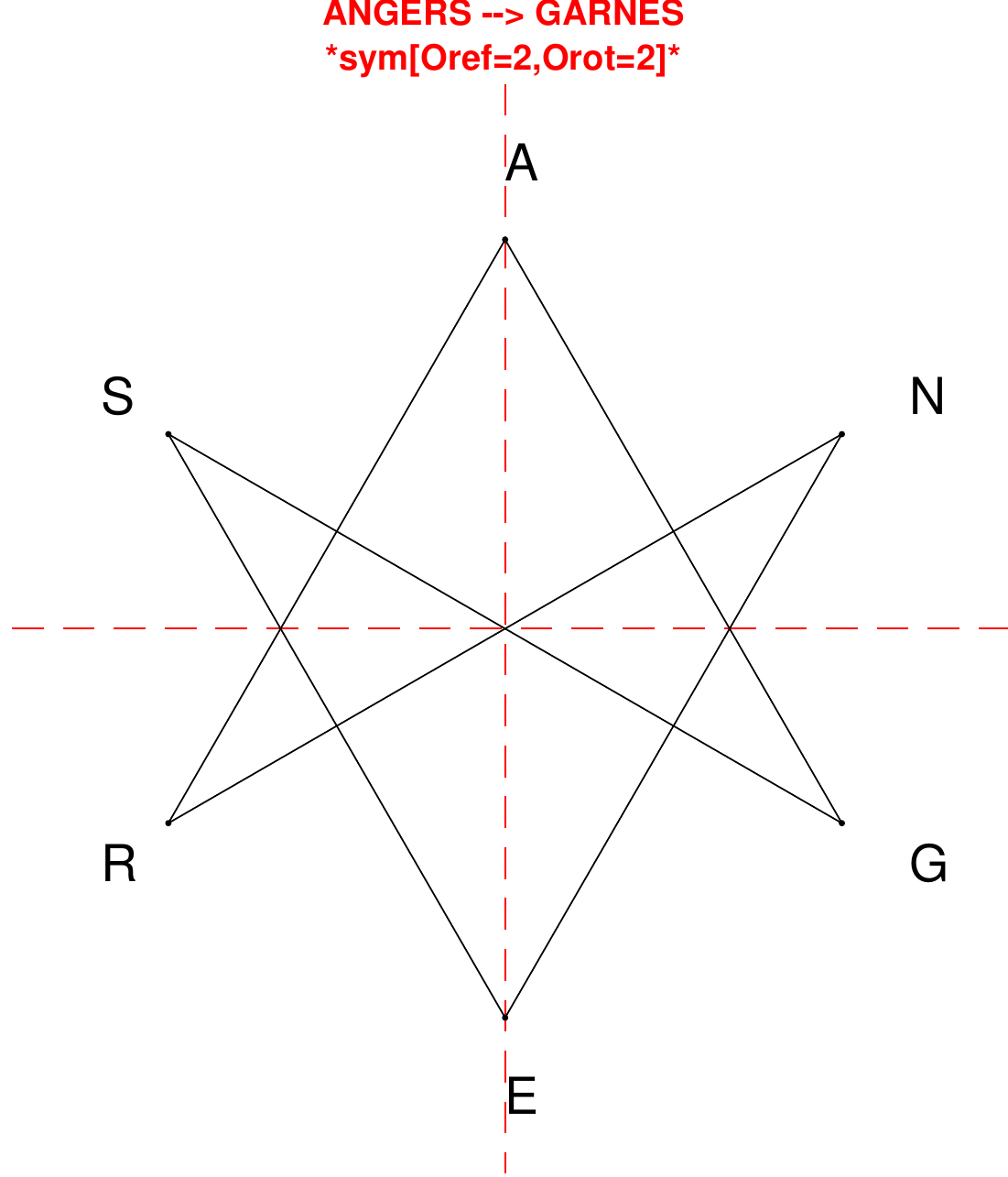}
\end{subfigure}
\hfill
\begin{subfigure}[T]{0.19\textwidth}
\centering
\includegraphics[width=\textwidth]{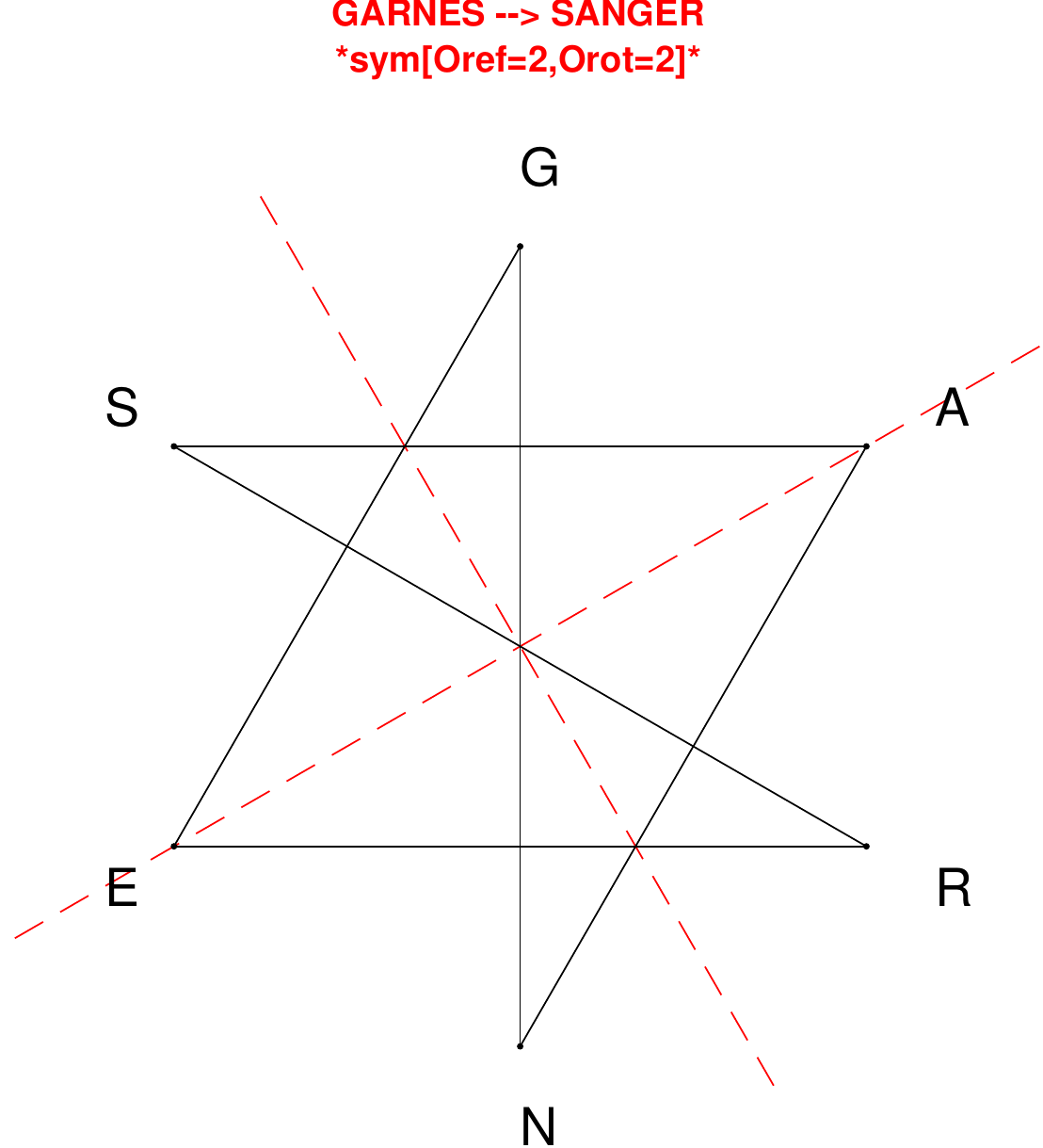}
\end{subfigure}
\hfill
\begin{subfigure}[T]{0.19\textwidth}
\centering
\includegraphics[width=\textwidth]{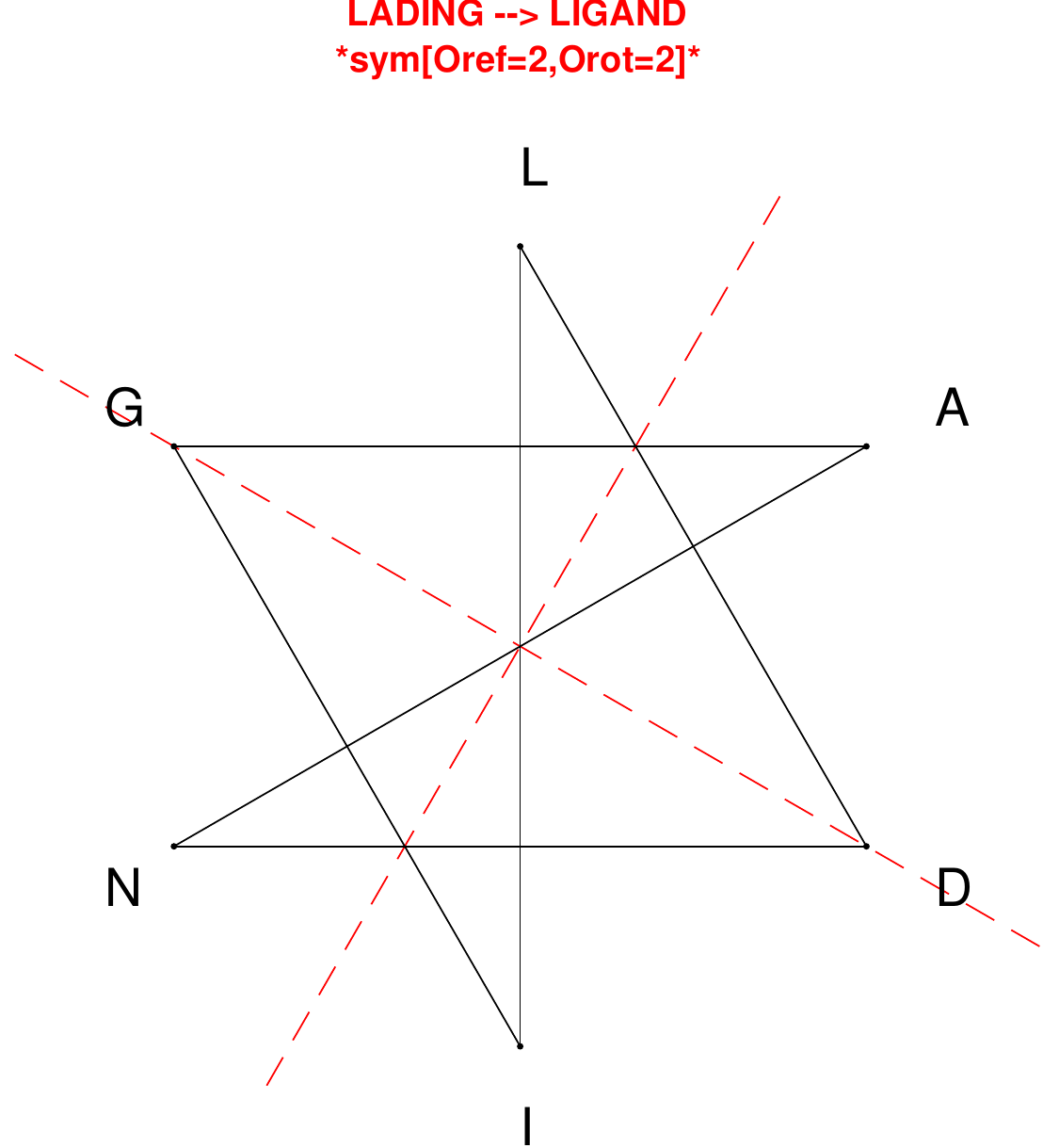}
\end{subfigure}
\hfill
\begin{subfigure}[T]{0.19\textwidth}
\centering
\includegraphics[width=\textwidth]{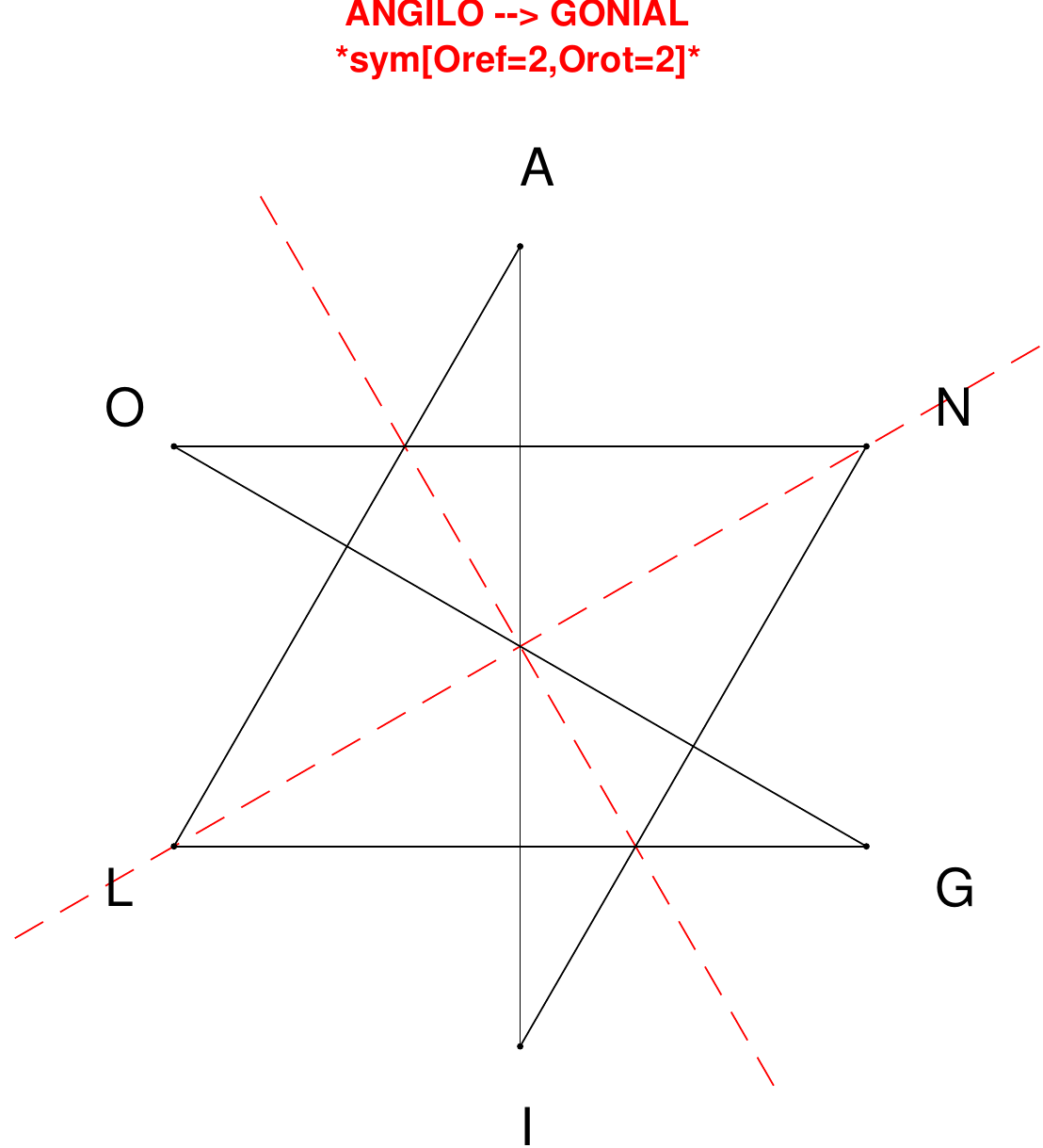}
\end{subfigure}
\end{figure}

\begin{figure}[H]
\centering
\begin{subfigure}[T]{0.19\textwidth}
\centering
\includegraphics[width=\textwidth]{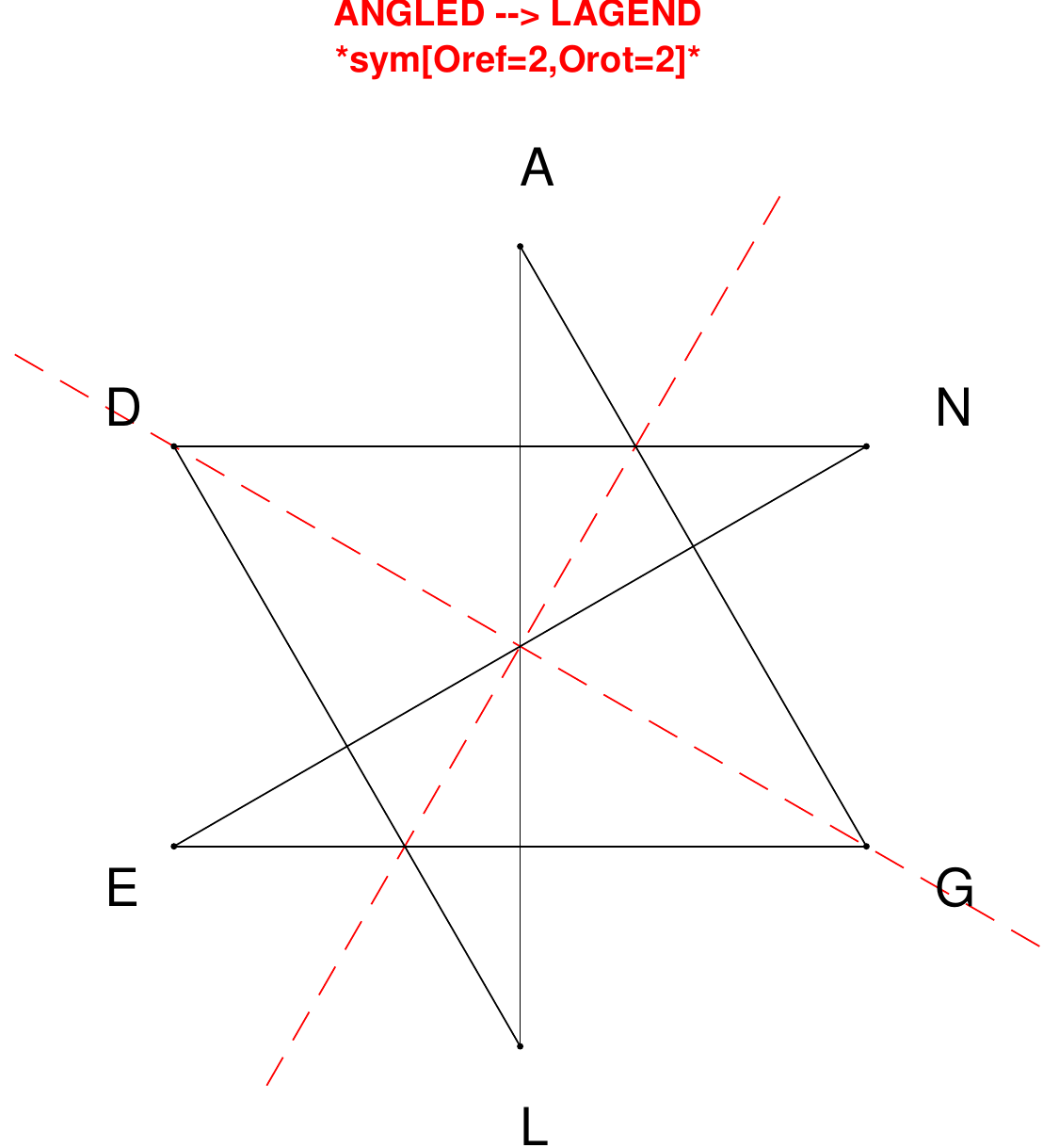}
\end{subfigure}
\hfill
\begin{subfigure}[T]{0.19\textwidth}
\centering
\includegraphics[width=\textwidth]{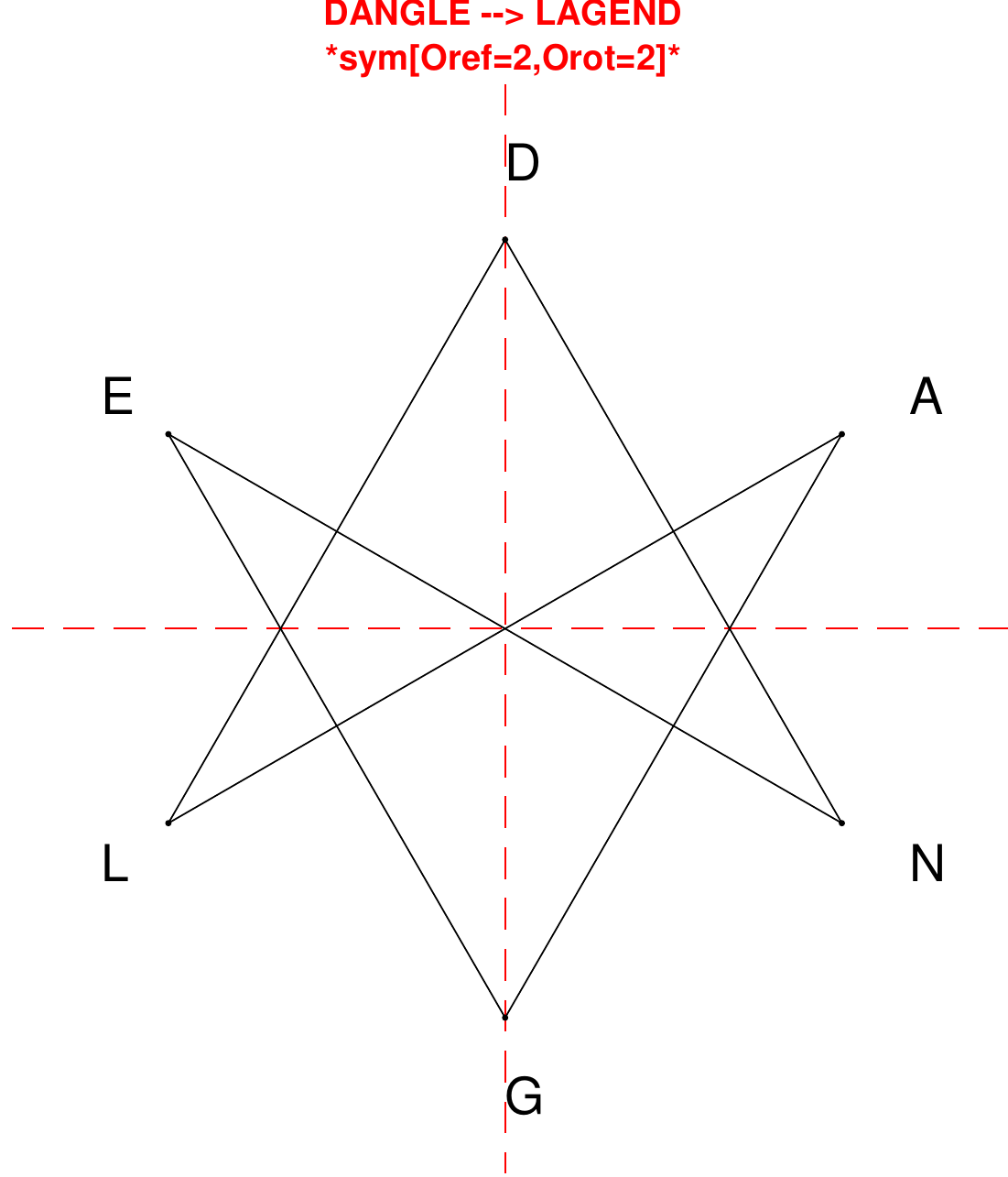}
\end{subfigure}
\hfill
\begin{subfigure}[T]{0.19\textwidth}
\centering
\includegraphics[width=\textwidth]{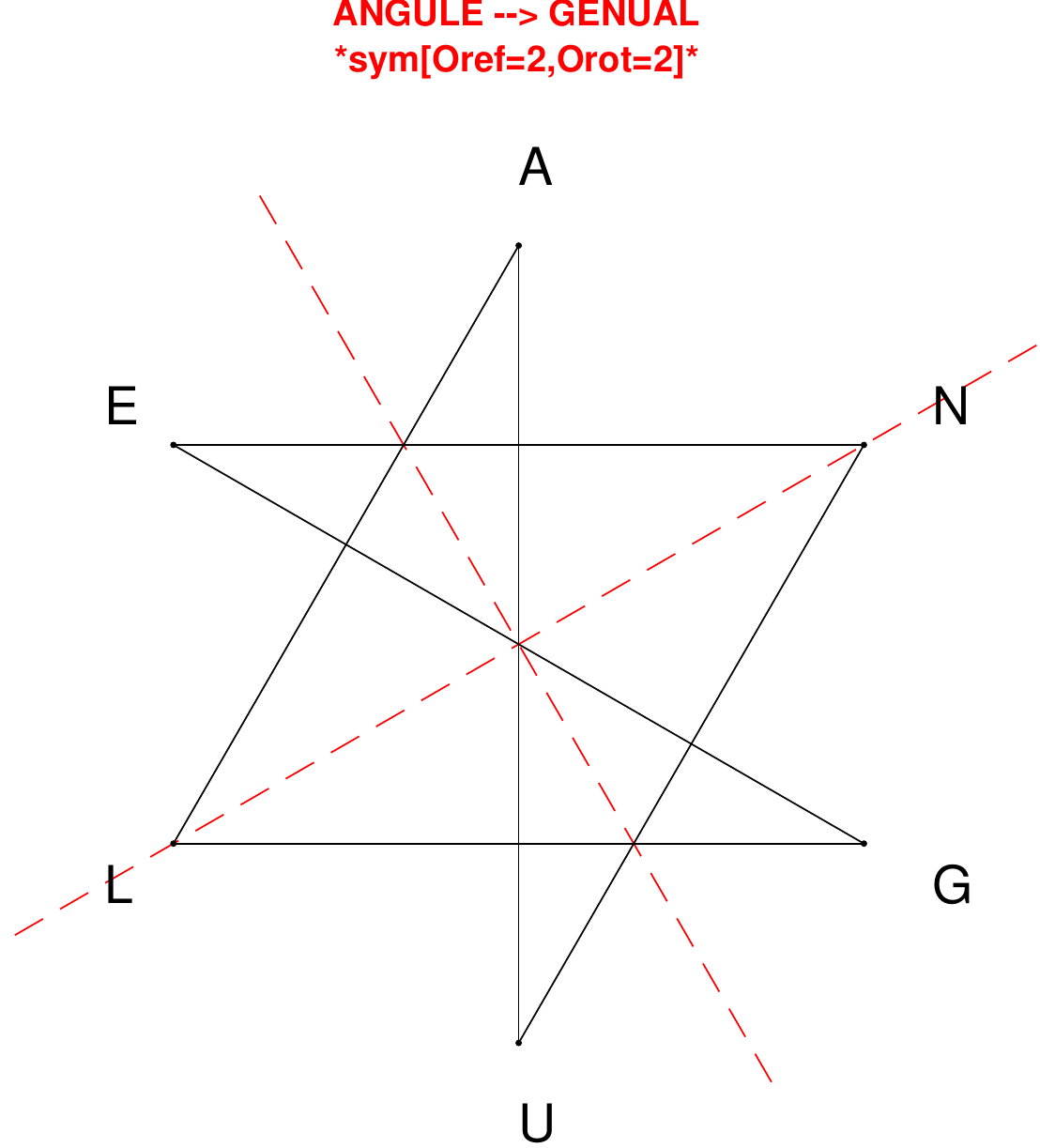}
\end{subfigure}
\hfill
\begin{subfigure}[T]{0.19\textwidth}
\centering
\includegraphics[width=\textwidth]{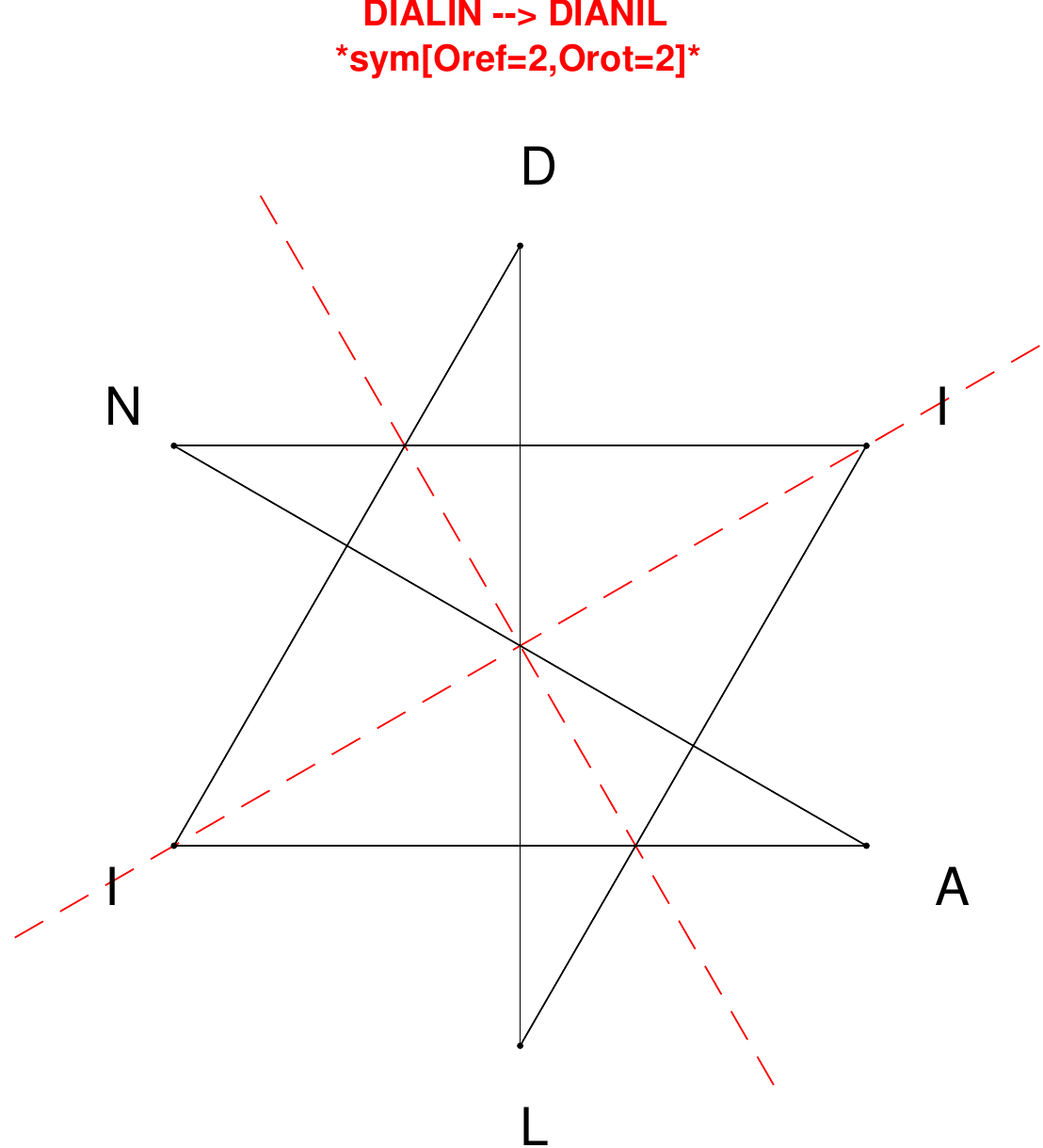}
\end{subfigure}
\hfill
\begin{subfigure}[T]{0.19\textwidth}
\centering
\includegraphics[width=\textwidth]{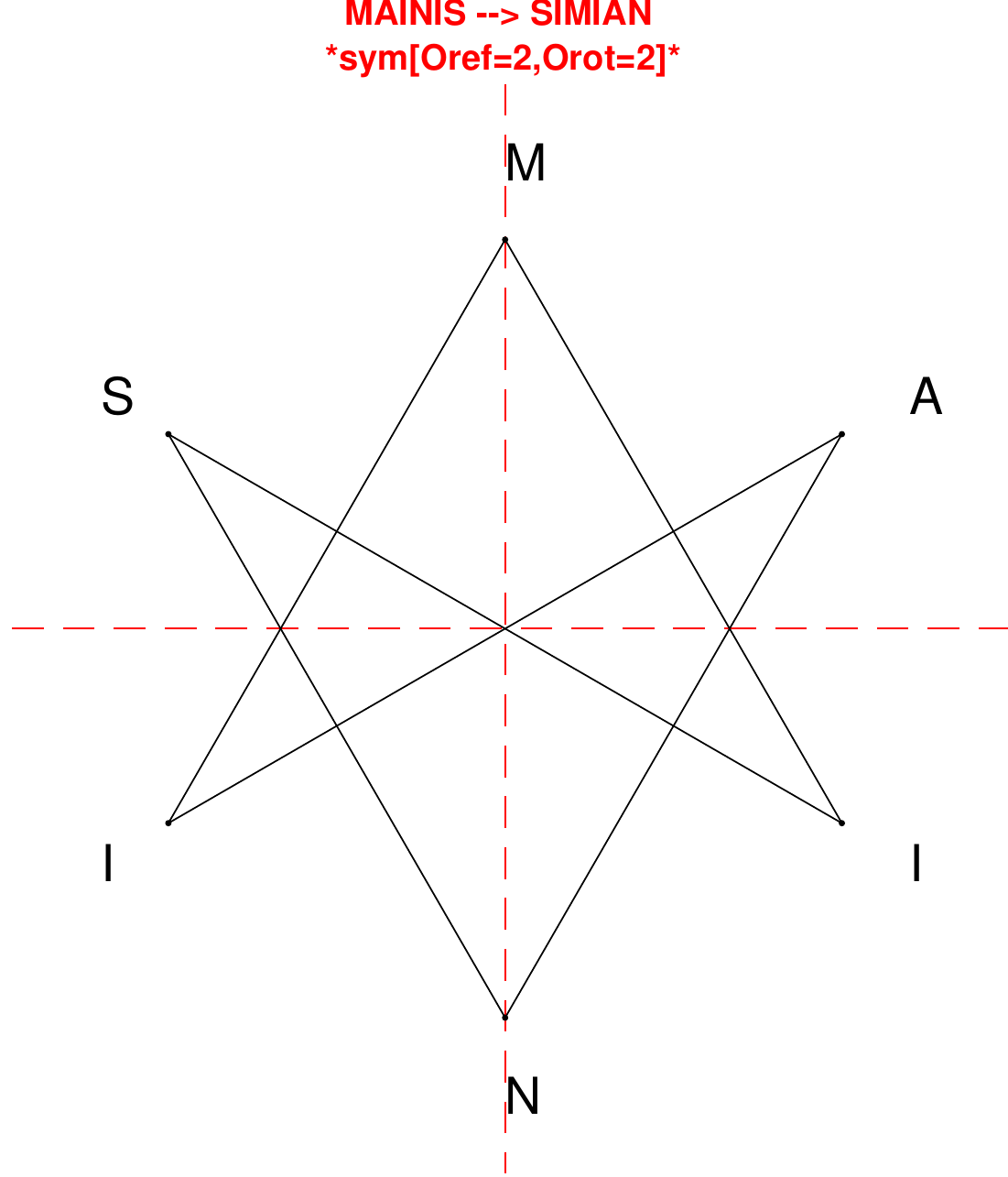}
\end{subfigure}
\end{figure}

\begin{figure}[H]
\centering
\begin{subfigure}[T]{0.19\textwidth}
\centering
\includegraphics[width=\textwidth]{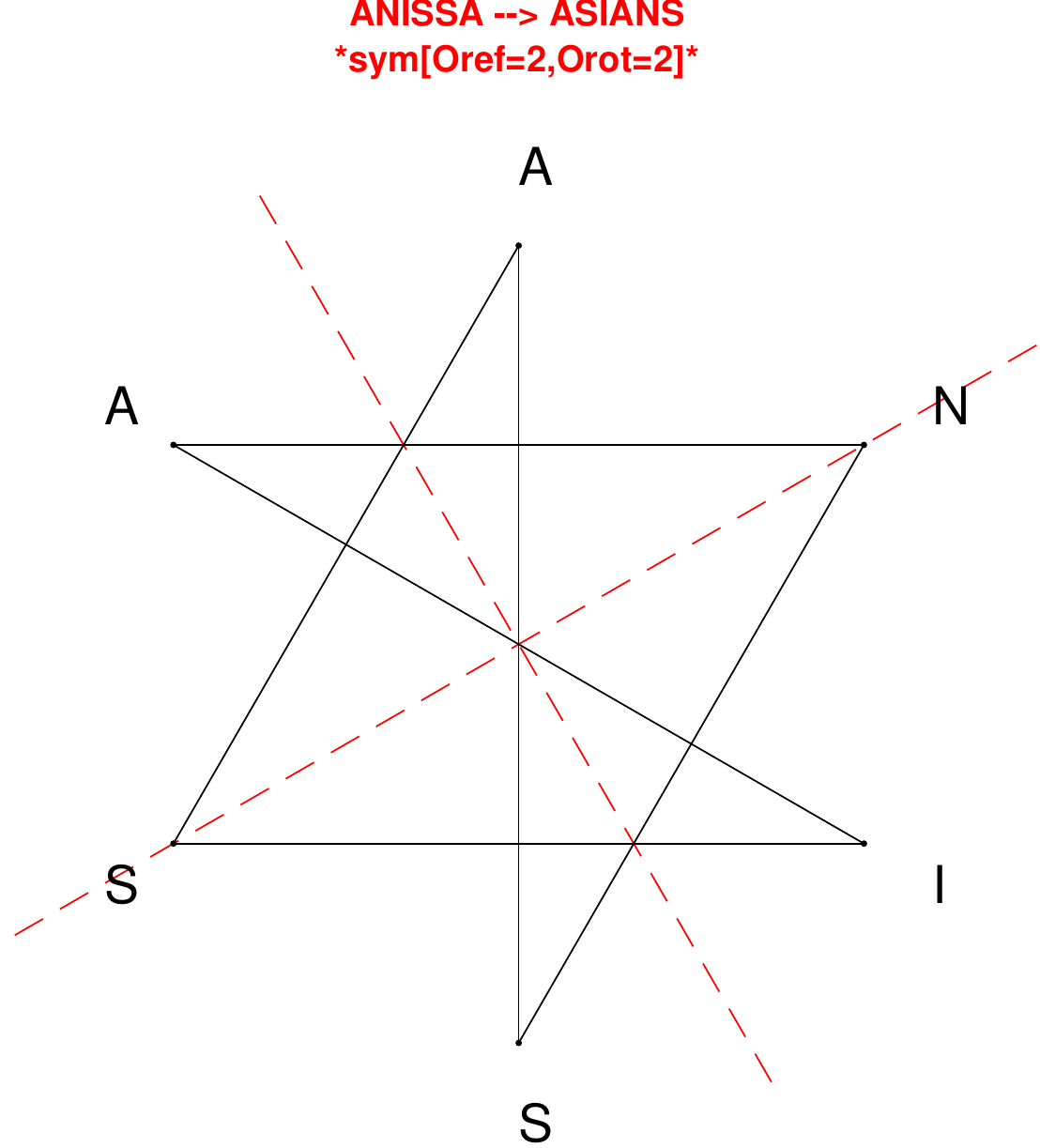}
\end{subfigure}
\hfill
\begin{subfigure}[T]{0.19\textwidth}
\centering
\includegraphics[width=\textwidth]{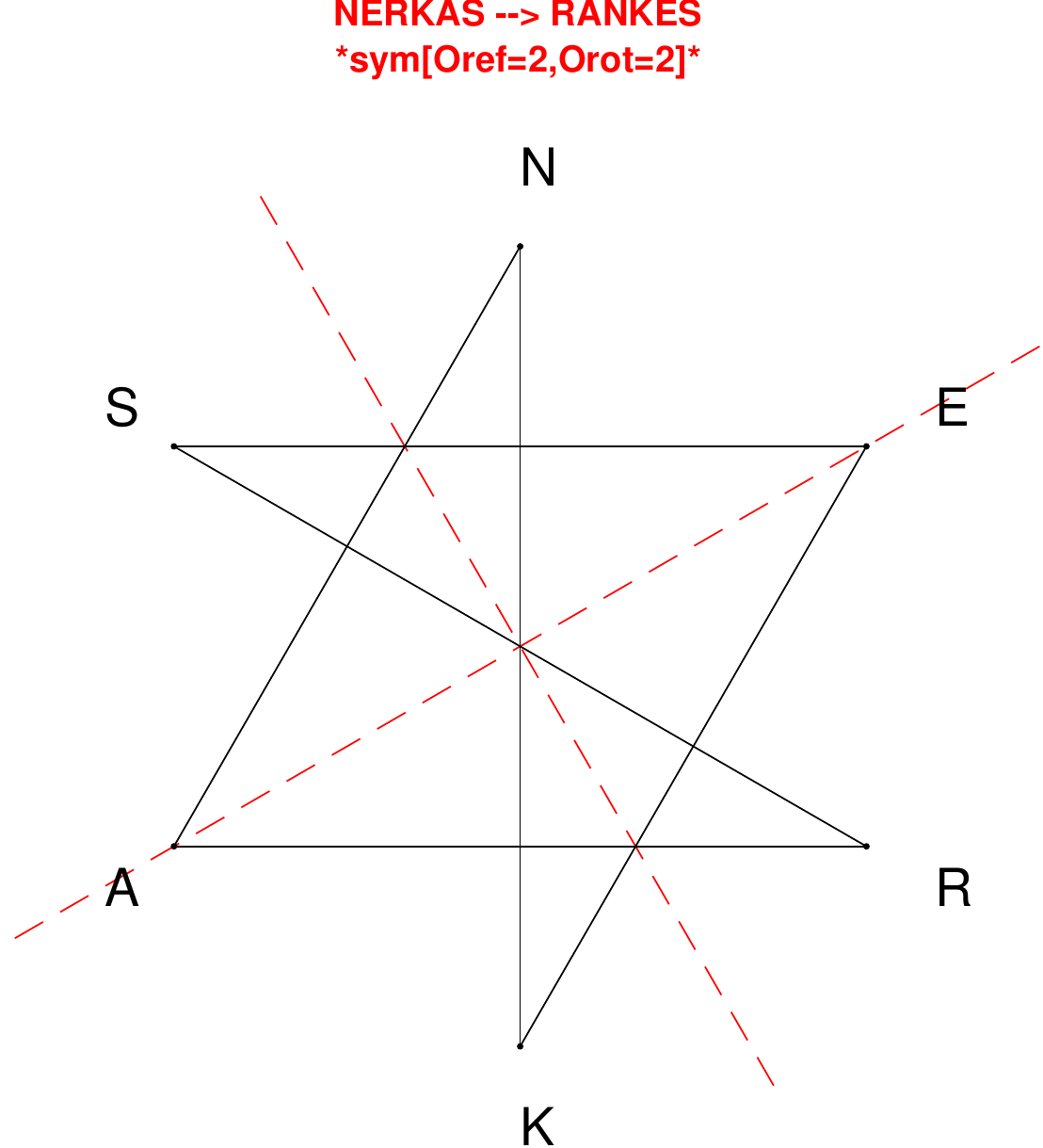}
\end{subfigure}
\hfill
\begin{subfigure}[T]{0.19\textwidth}
\centering
\includegraphics[width=\textwidth]{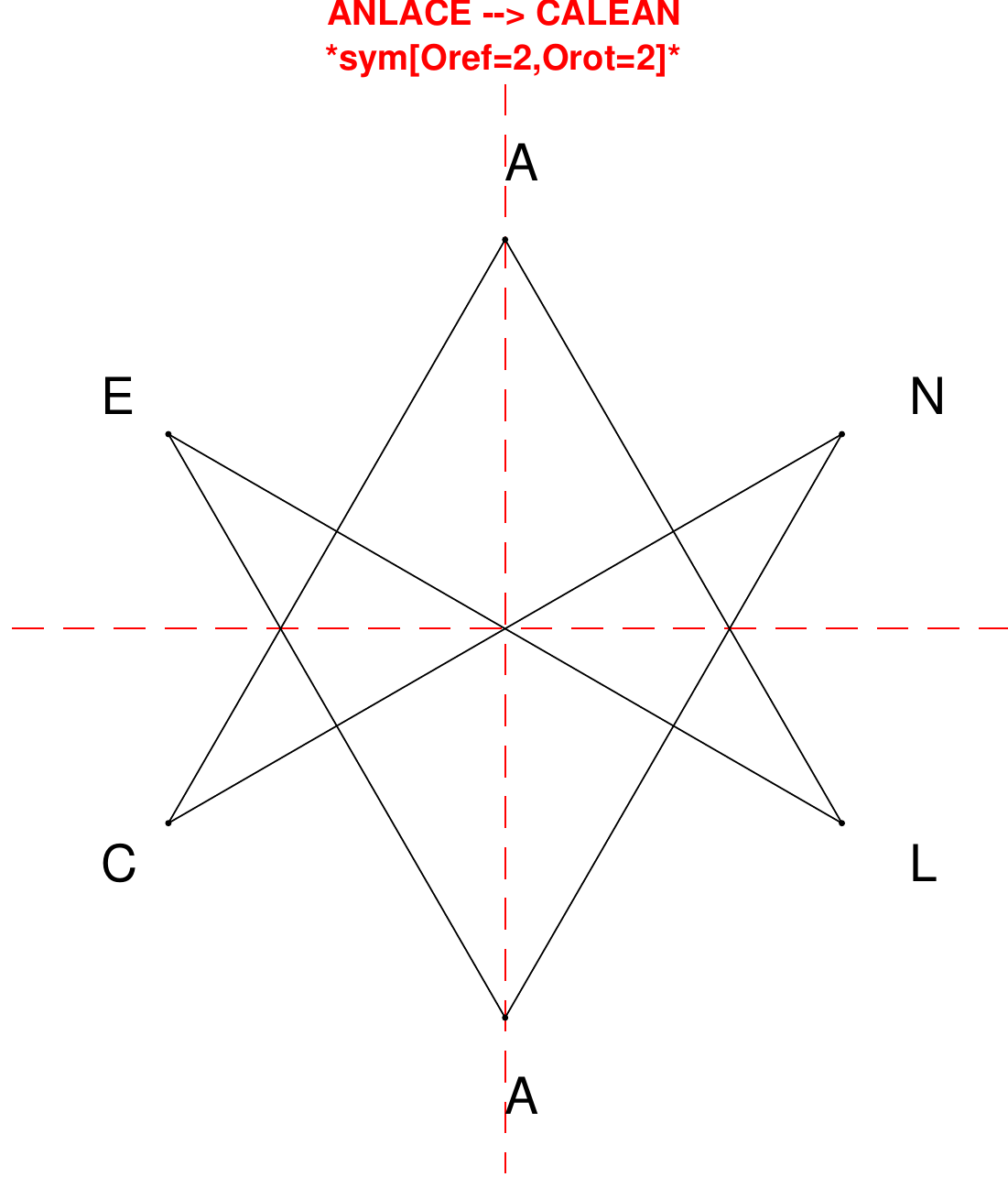}
\end{subfigure}
\hfill
\begin{subfigure}[T]{0.19\textwidth}
\centering
\includegraphics[width=\textwidth]{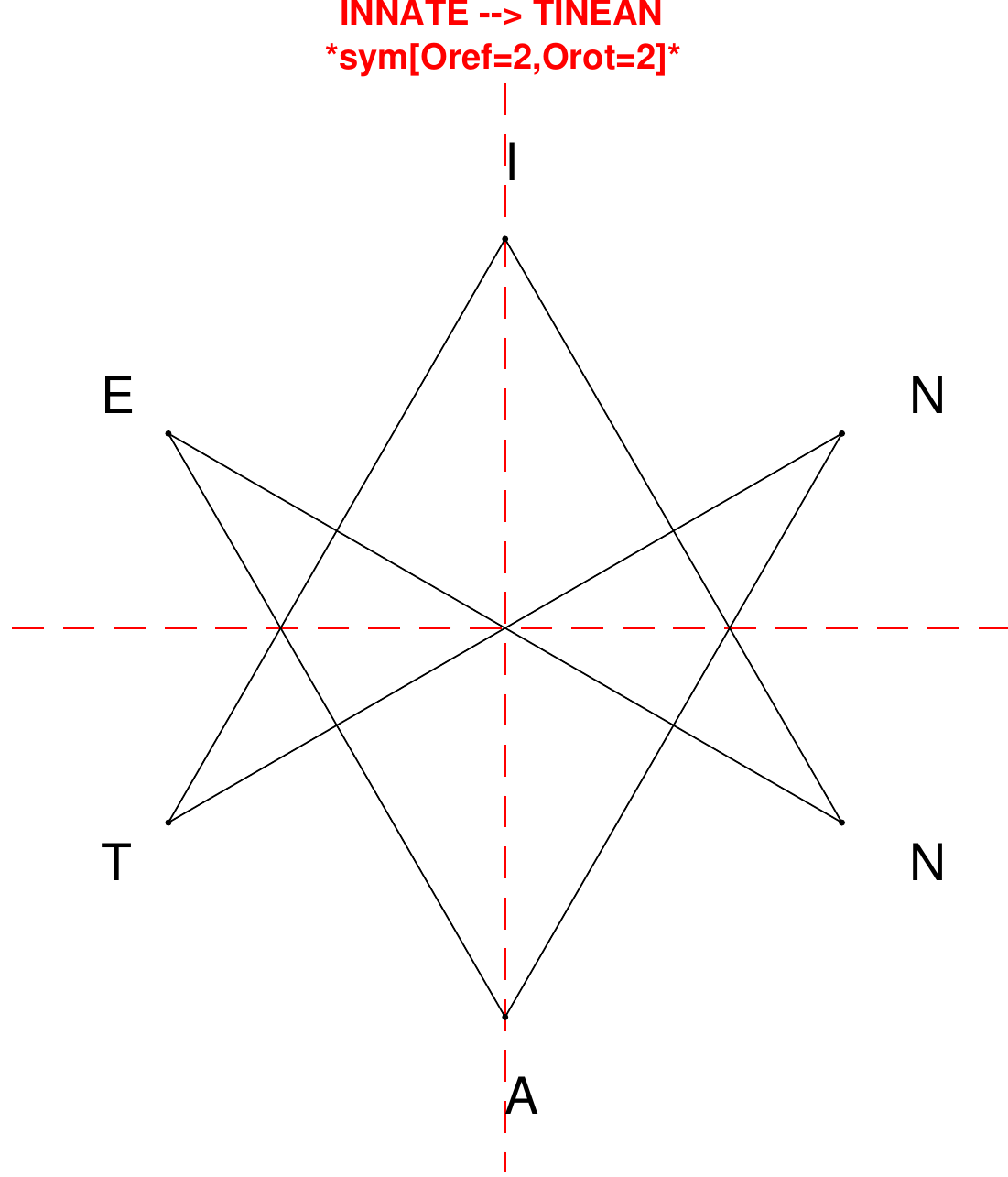}
\end{subfigure}
\hfill
\begin{subfigure}[T]{0.19\textwidth}
\centering
\includegraphics[width=\textwidth]{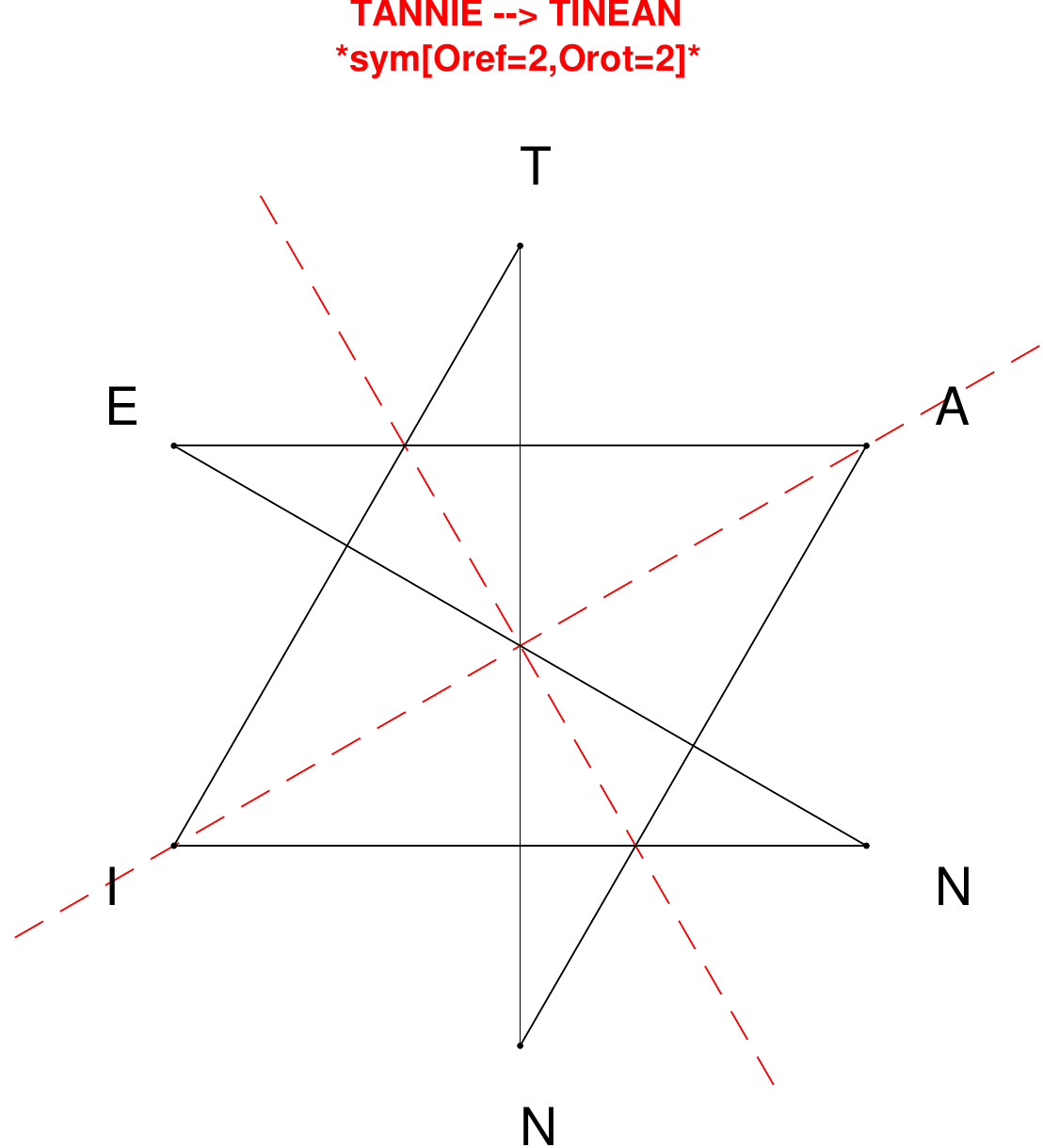}
\end{subfigure}
\end{figure}

\begin{figure}[H]
\centering
\begin{subfigure}[T]{0.19\textwidth}
\centering
\includegraphics[width=\textwidth]{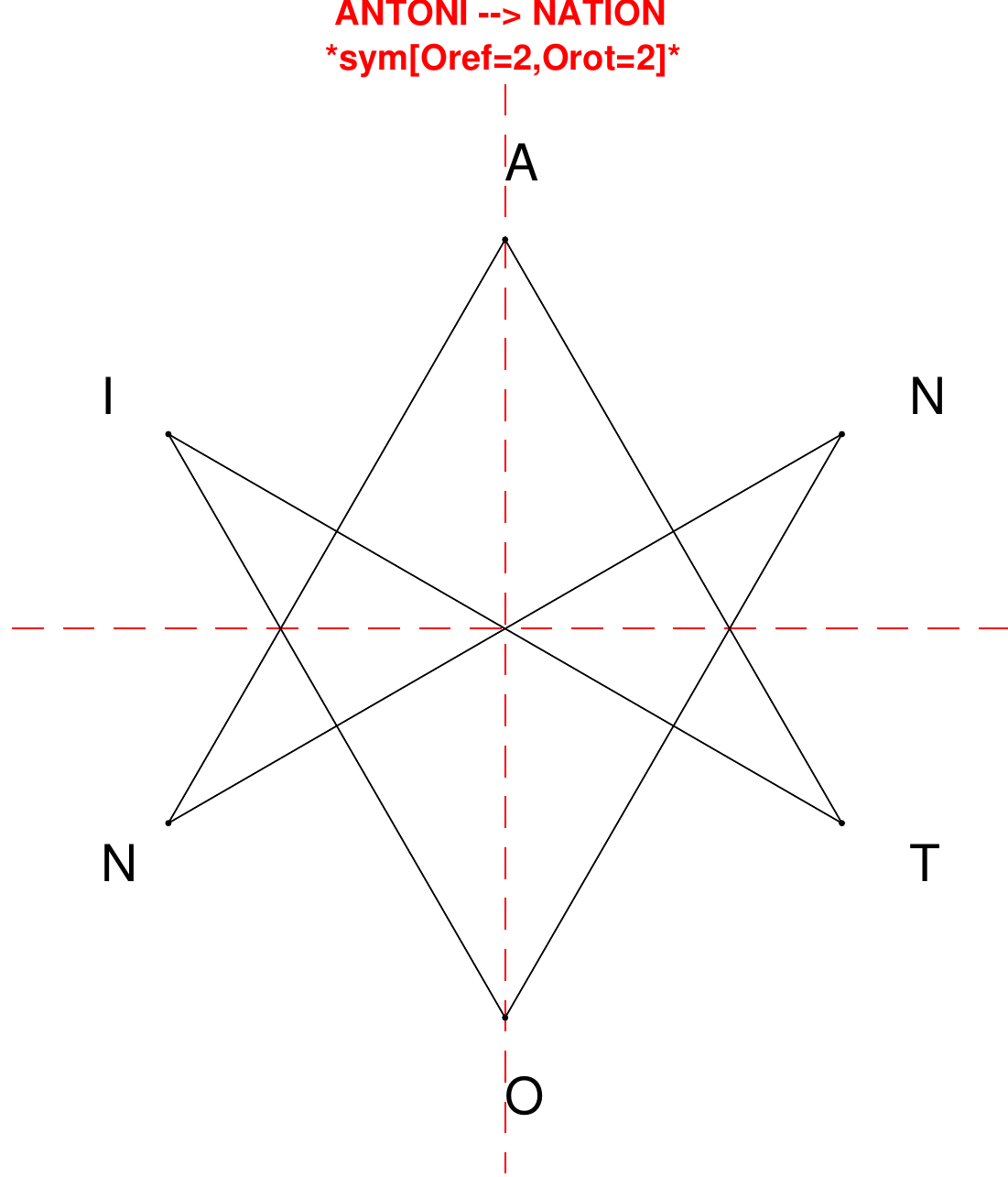}
\end{subfigure}
\hfill
\begin{subfigure}[T]{0.19\textwidth}
\centering
\includegraphics[width=\textwidth]{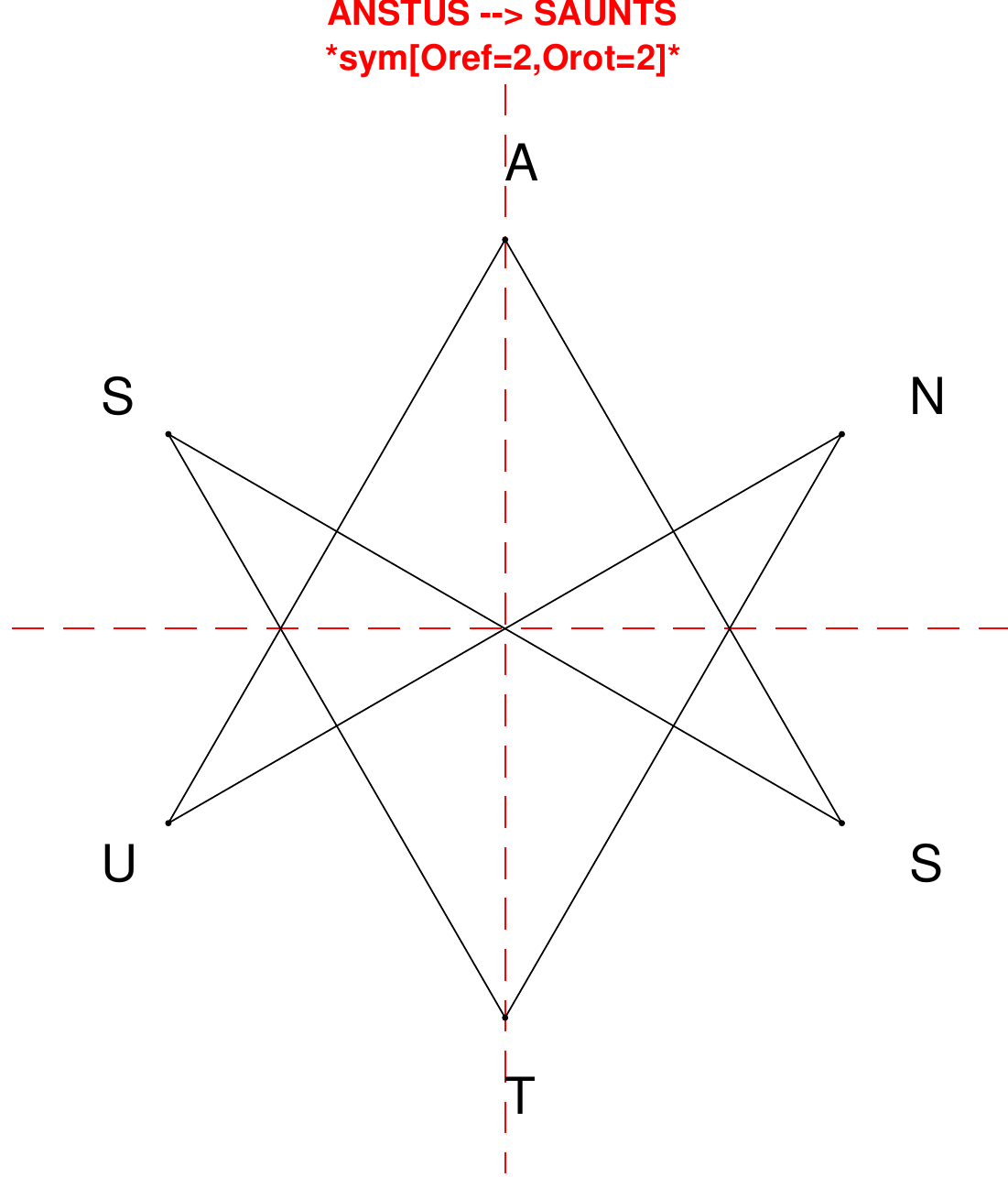}
\end{subfigure}
\hfill
\begin{subfigure}[T]{0.19\textwidth}
\centering
\includegraphics[width=\textwidth]{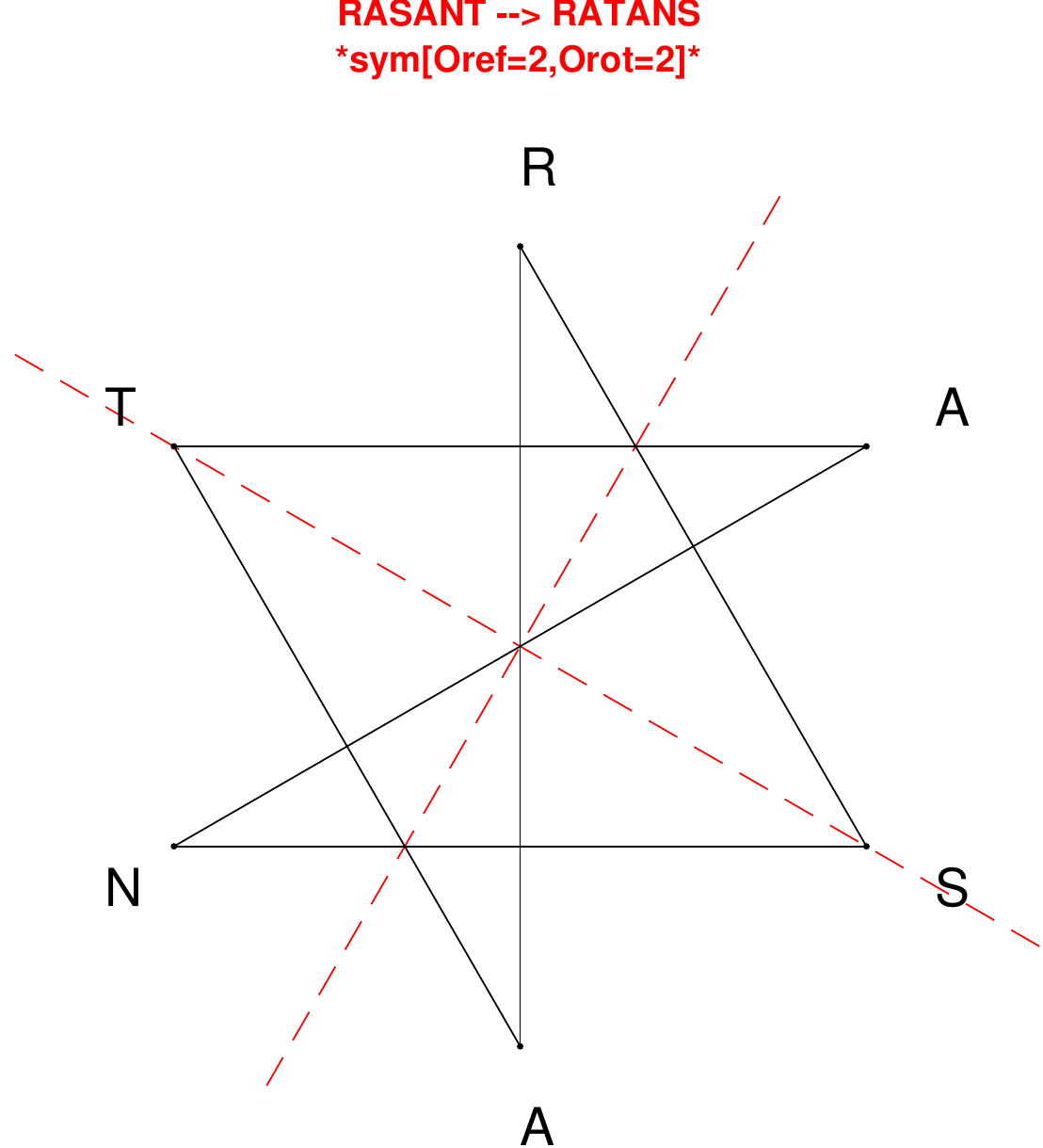}
\end{subfigure}
\hfill
\begin{subfigure}[T]{0.19\textwidth}
\centering
\includegraphics[width=\textwidth]{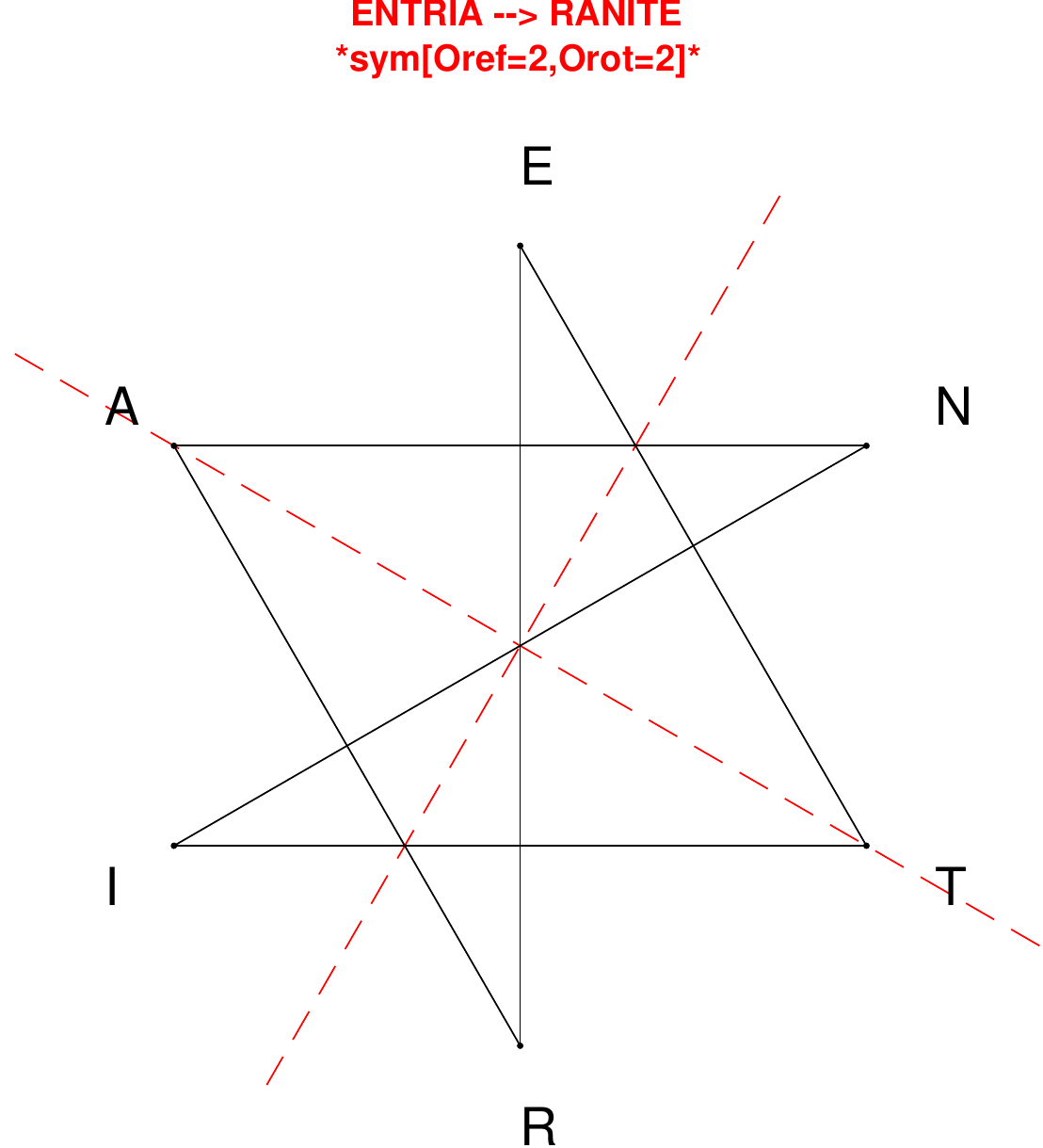}
\end{subfigure}
\hfill
\begin{subfigure}[T]{0.19\textwidth}
\centering
\includegraphics[width=\textwidth]{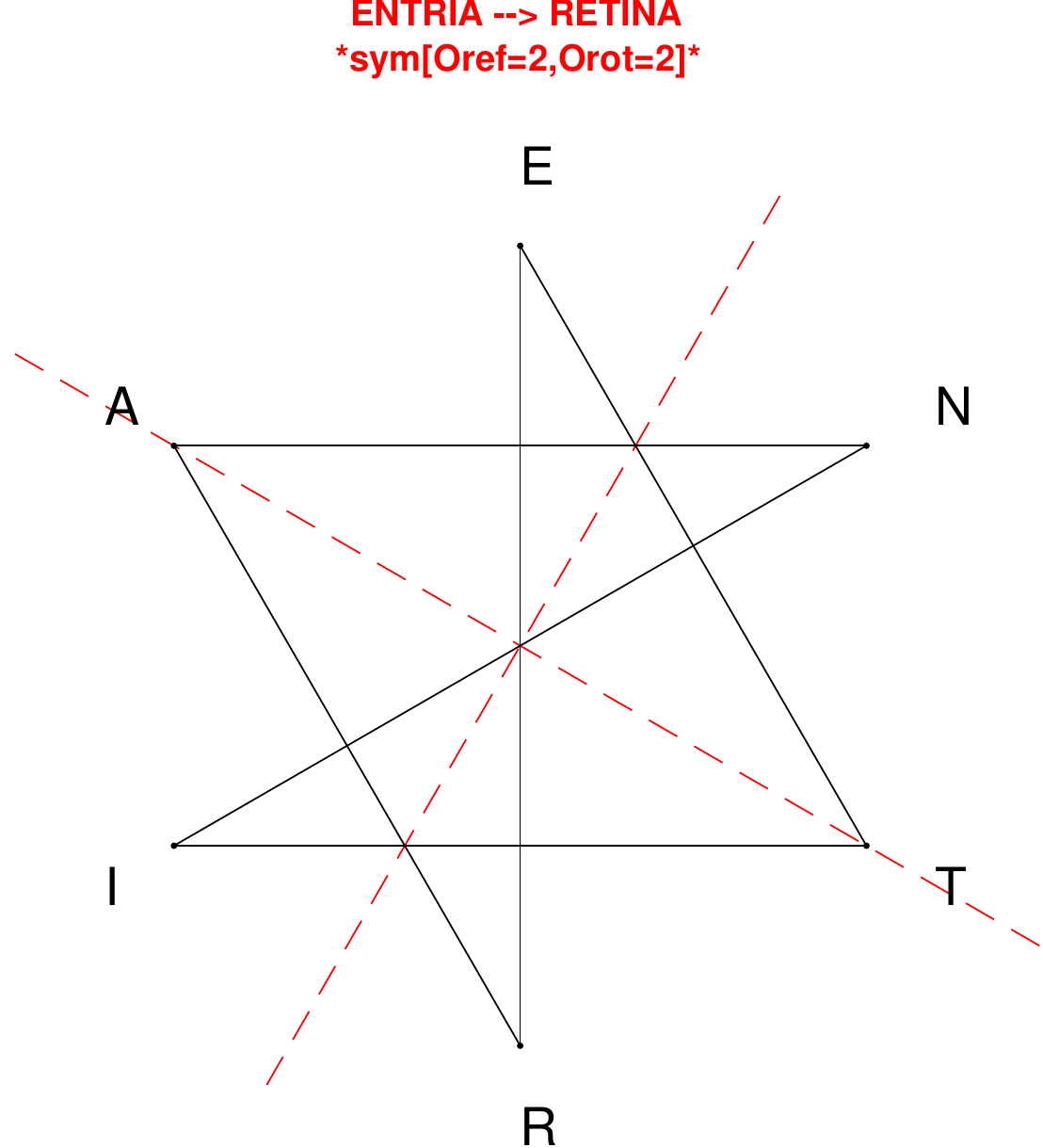}
\end{subfigure}
\end{figure}

\begin{figure}[H]
\centering
\begin{subfigure}[T]{0.19\textwidth}
\centering
\includegraphics[width=\textwidth]{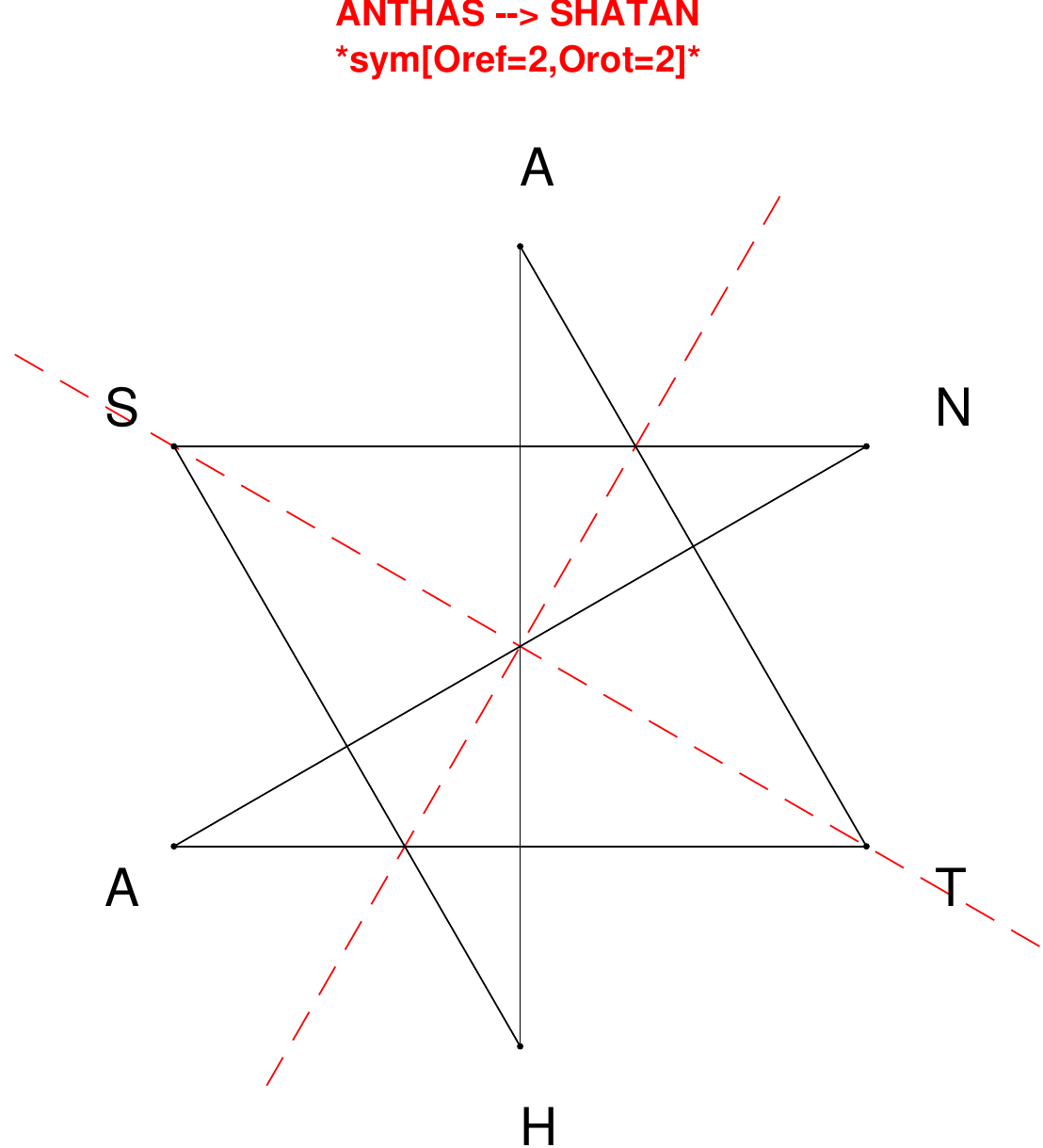}
\end{subfigure}
\hfill
\begin{subfigure}[T]{0.19\textwidth}
\centering
\includegraphics[width=\textwidth]{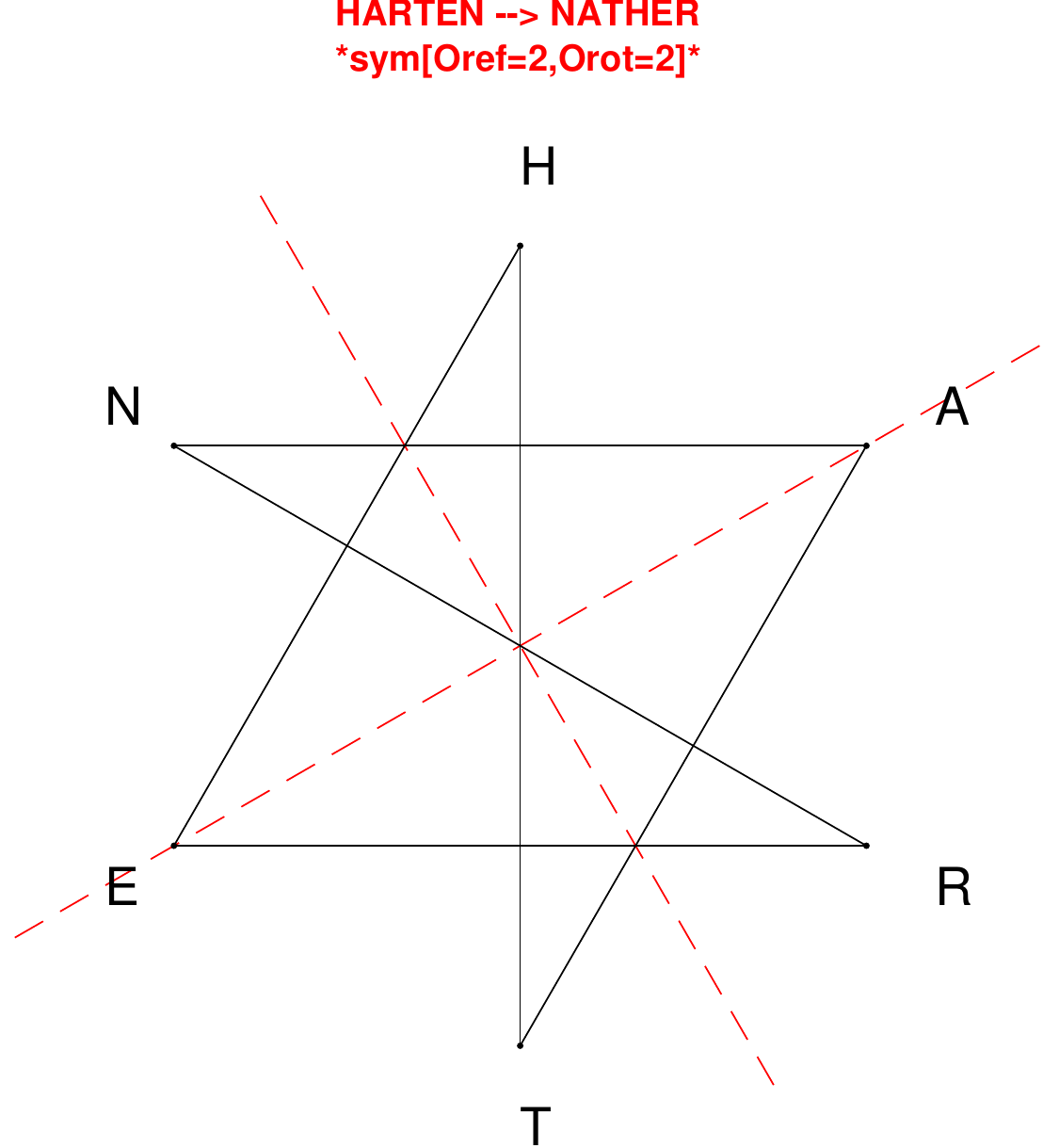}
\end{subfigure}
\hfill
\begin{subfigure}[T]{0.19\textwidth}
\centering
\includegraphics[width=\textwidth]{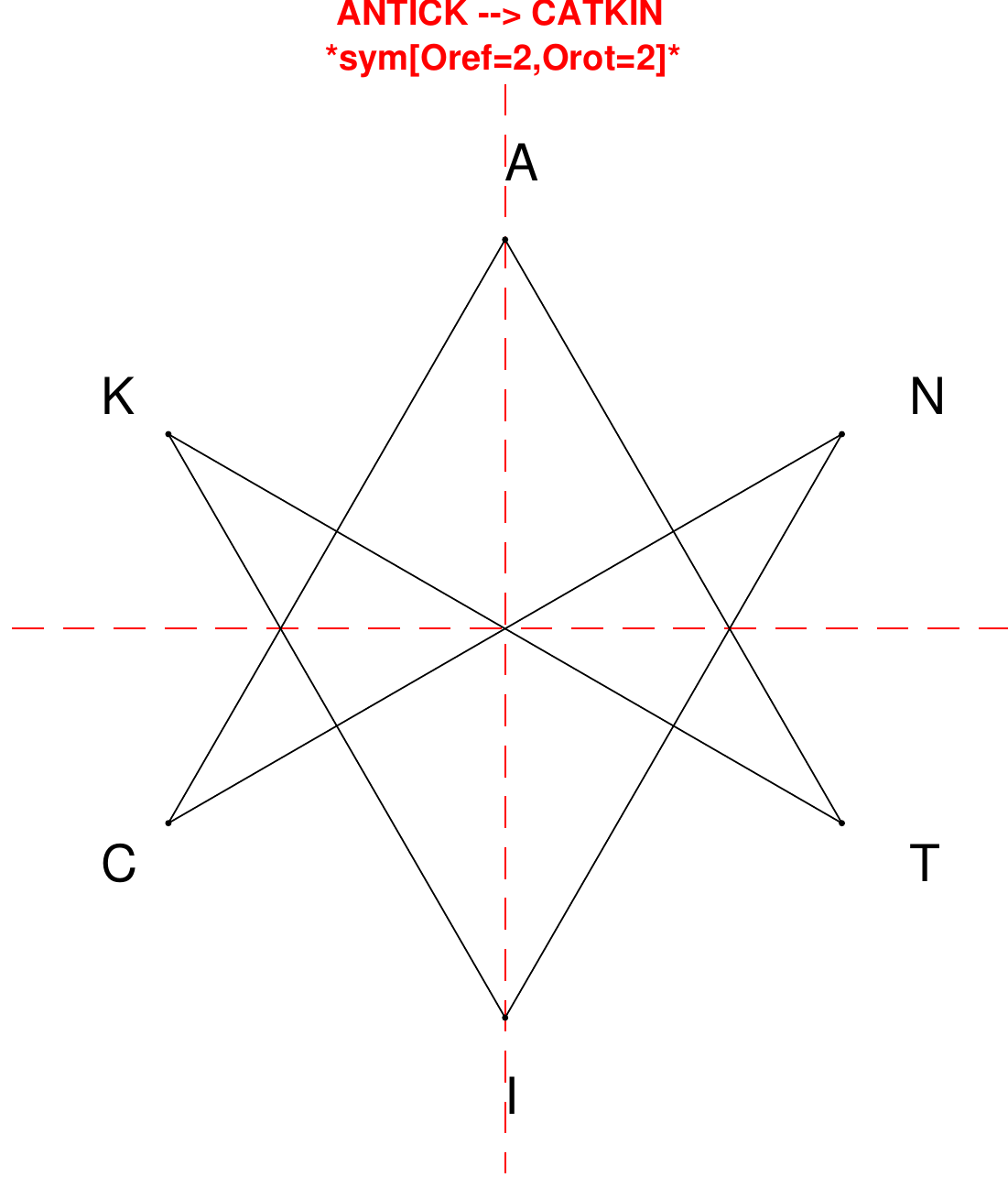}
\end{subfigure}
\hfill
\begin{subfigure}[T]{0.19\textwidth}
\centering
\includegraphics[width=\textwidth]{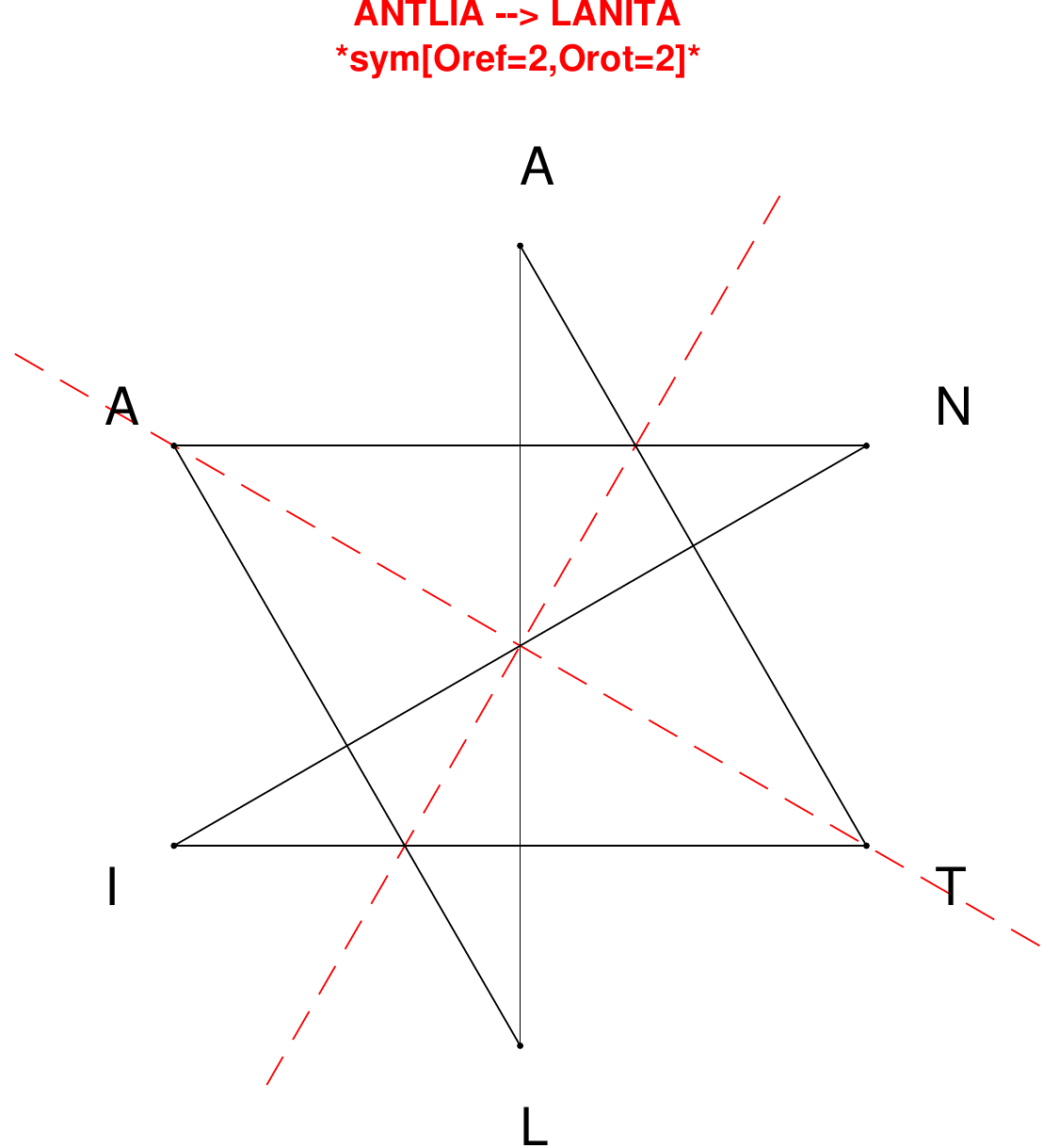}
\end{subfigure}
\hfill
\begin{subfigure}[T]{0.19\textwidth}
\centering
\includegraphics[width=\textwidth]{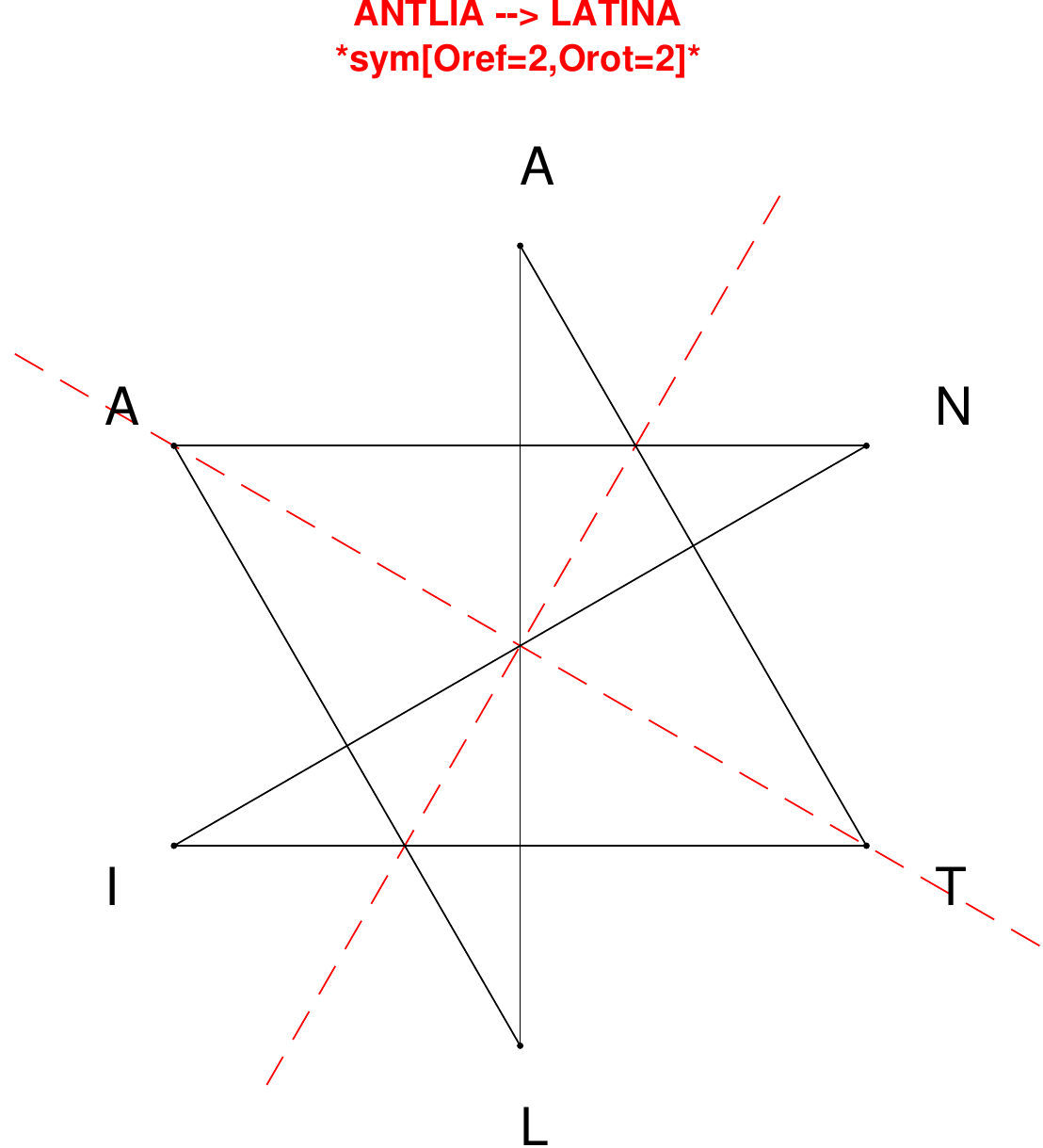}
\end{subfigure}
\end{figure}

\begin{figure}[H]
\centering
\begin{subfigure}[T]{0.19\textwidth}
\centering
\includegraphics[width=\textwidth]{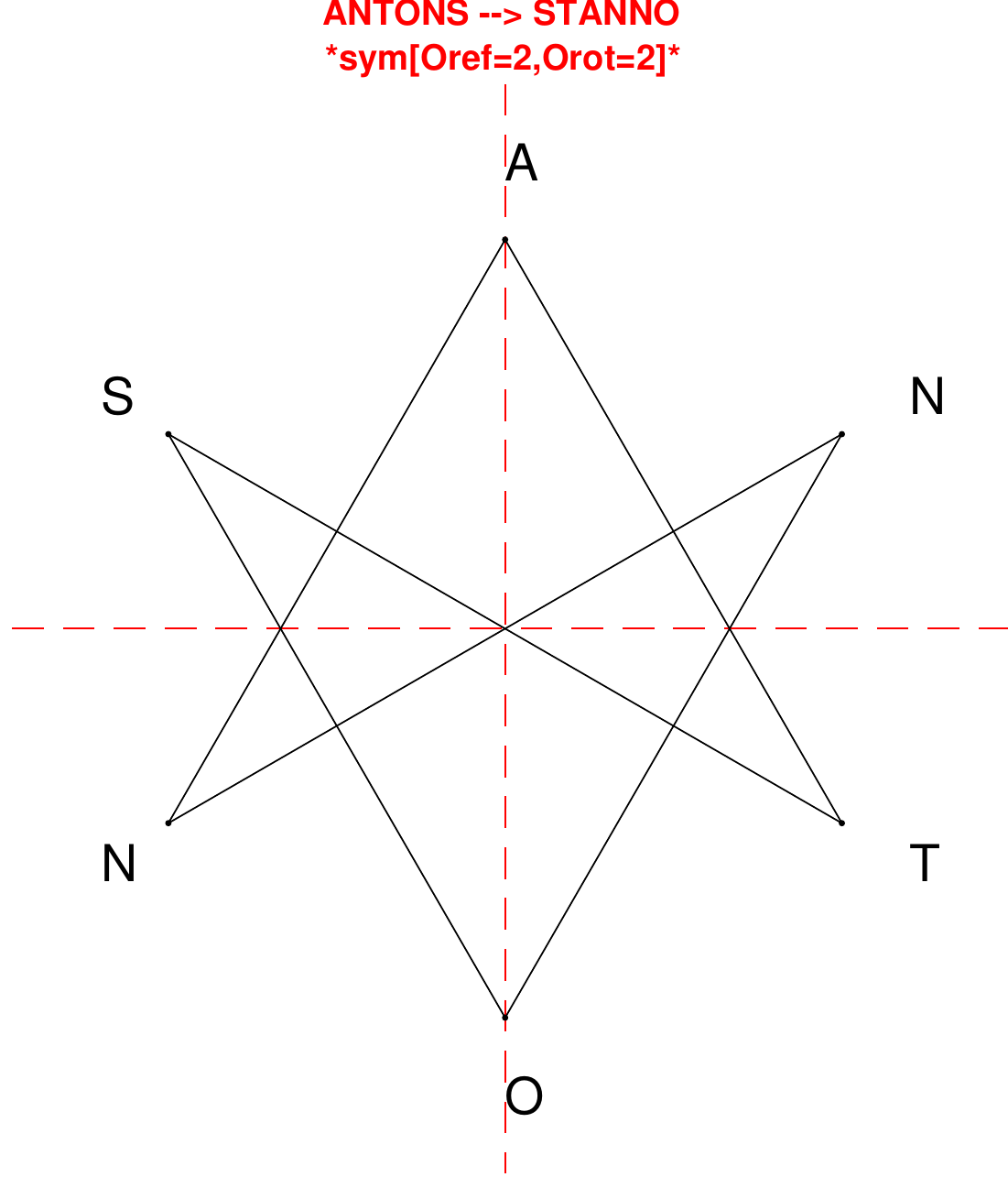}
\end{subfigure}
\hfill
\begin{subfigure}[T]{0.19\textwidth}
\centering
\includegraphics[width=\textwidth]{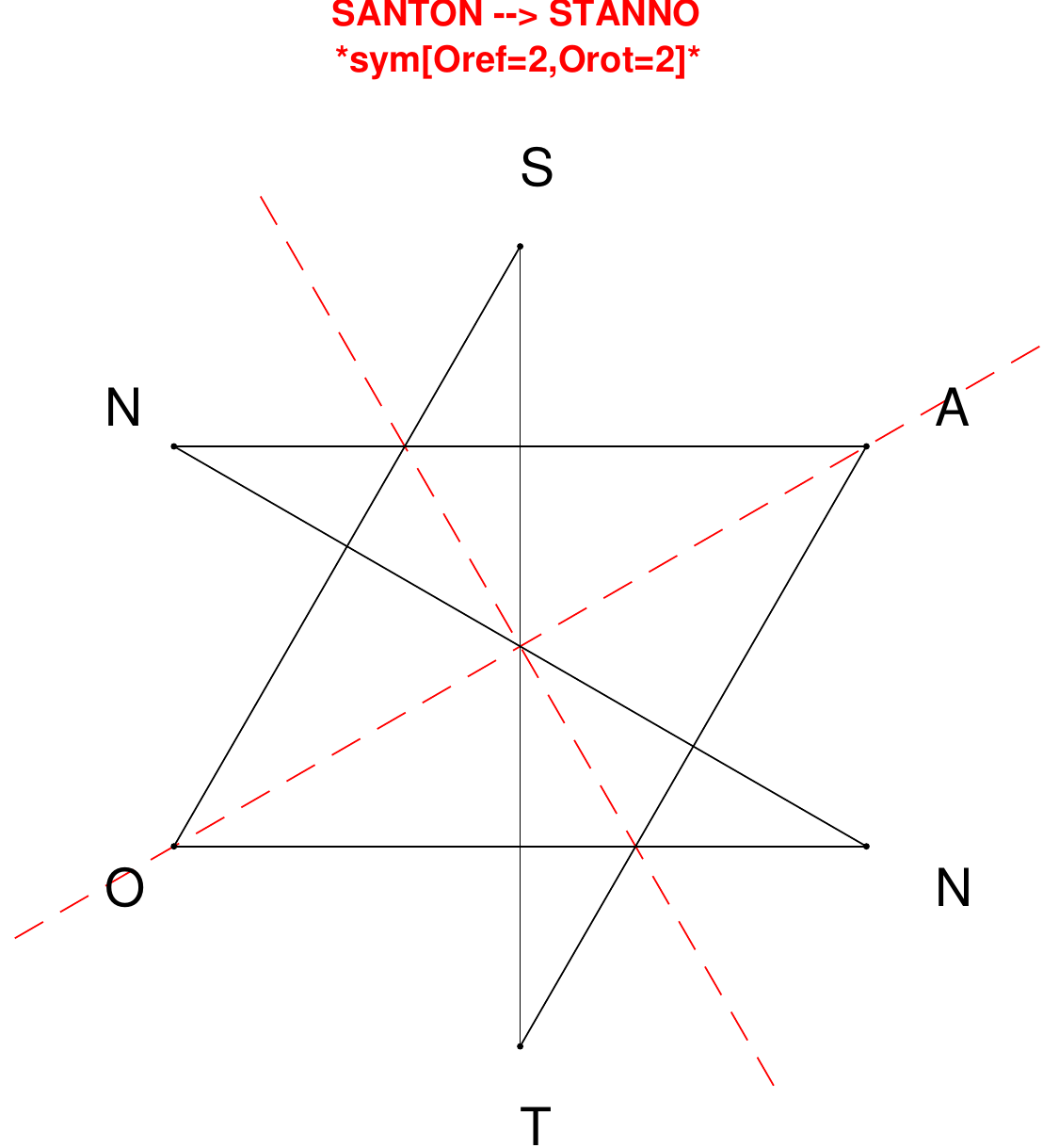}
\end{subfigure}
\hfill
\begin{subfigure}[T]{0.19\textwidth}
\centering
\includegraphics[width=\textwidth]{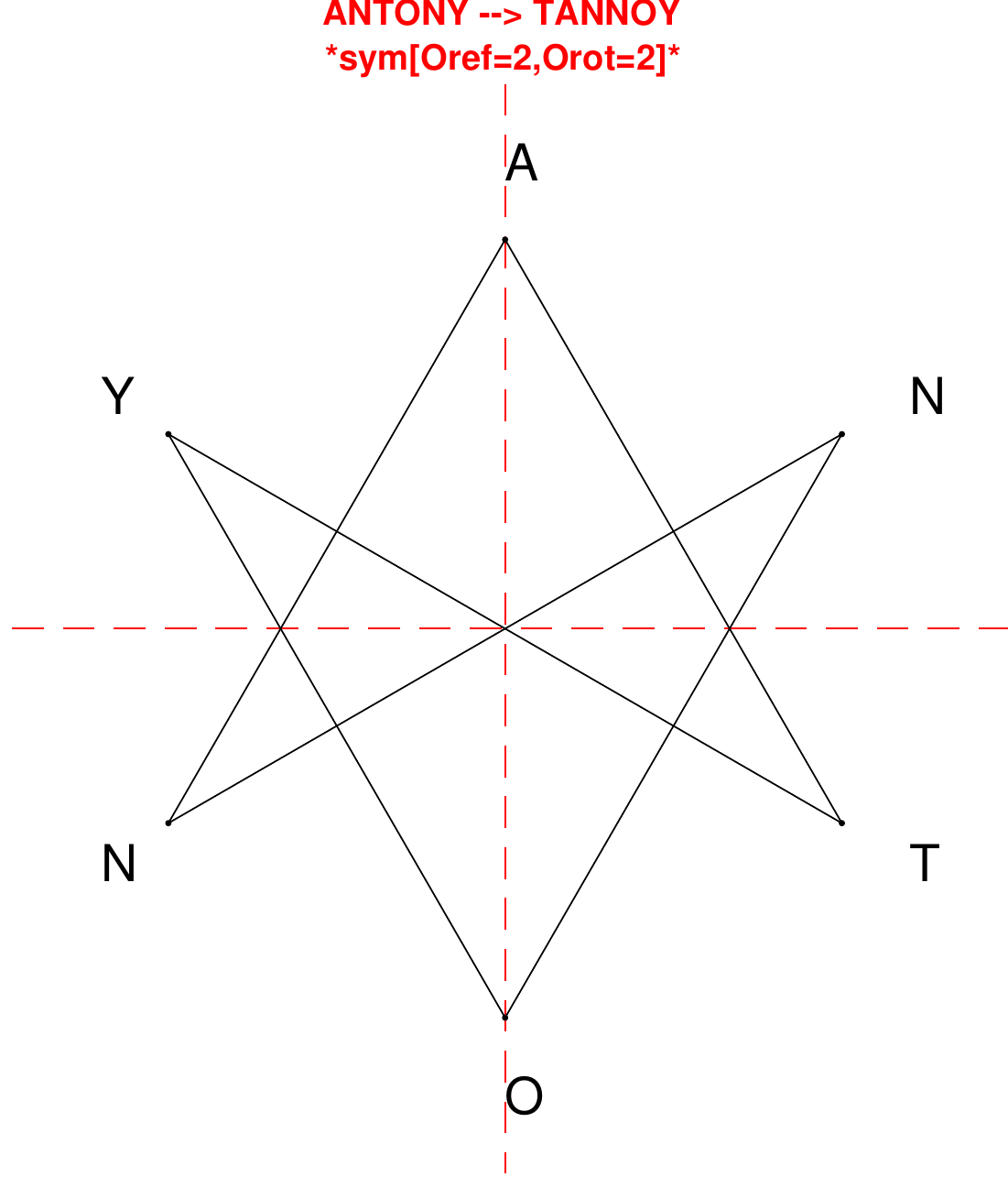}
\end{subfigure}
\hfill
\begin{subfigure}[T]{0.19\textwidth}
\centering
\includegraphics[width=\textwidth]{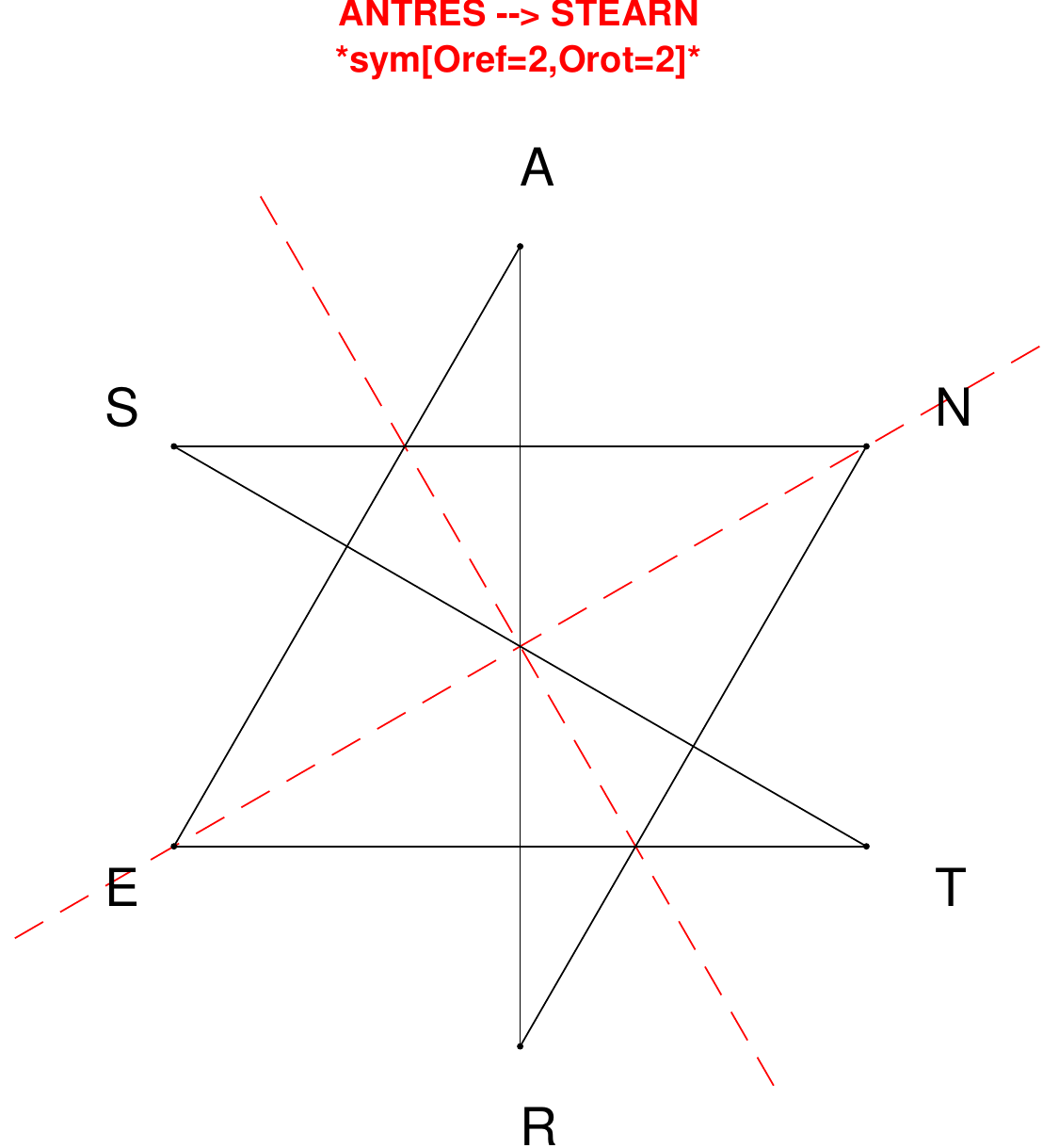}
\end{subfigure}
\hfill
\begin{subfigure}[T]{0.19\textwidth}
\centering
\includegraphics[width=\textwidth]{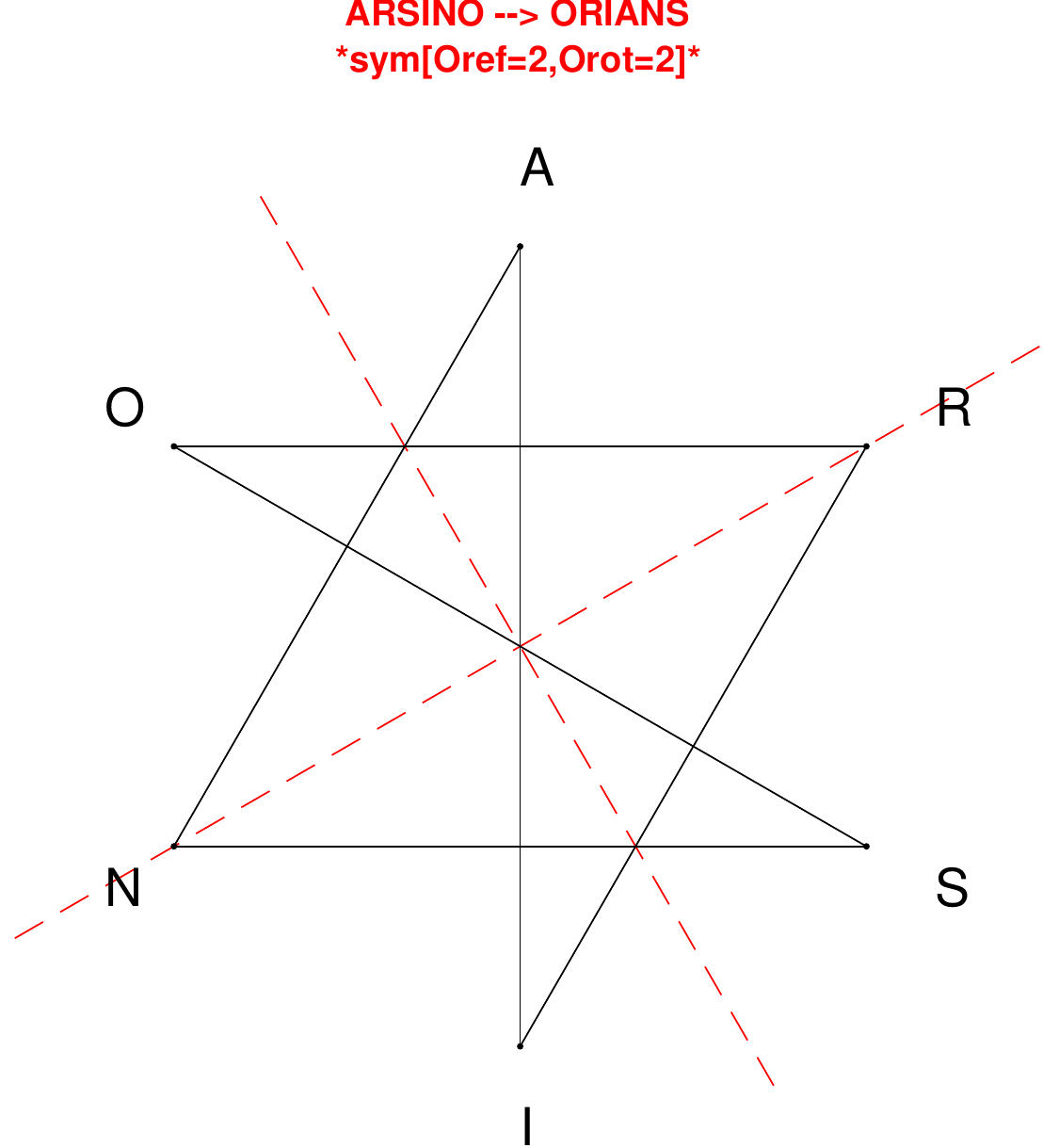}
\end{subfigure}
\end{figure}

\begin{figure}[H]
\centering
\begin{subfigure}[T]{0.19\textwidth}
\centering
\includegraphics[width=\textwidth]{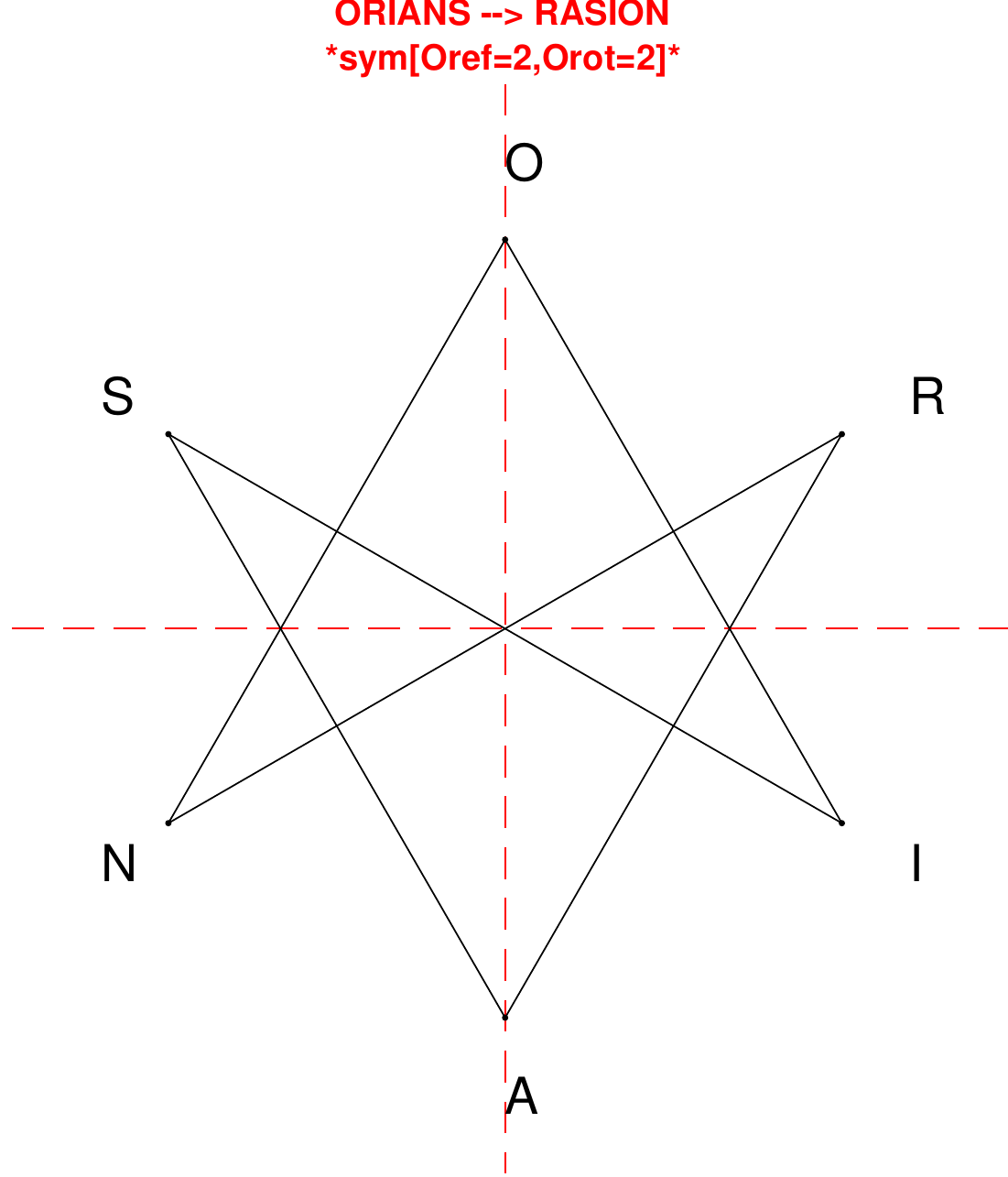}
\end{subfigure}
\hfill
\begin{subfigure}[T]{0.19\textwidth}
\centering
\includegraphics[width=\textwidth]{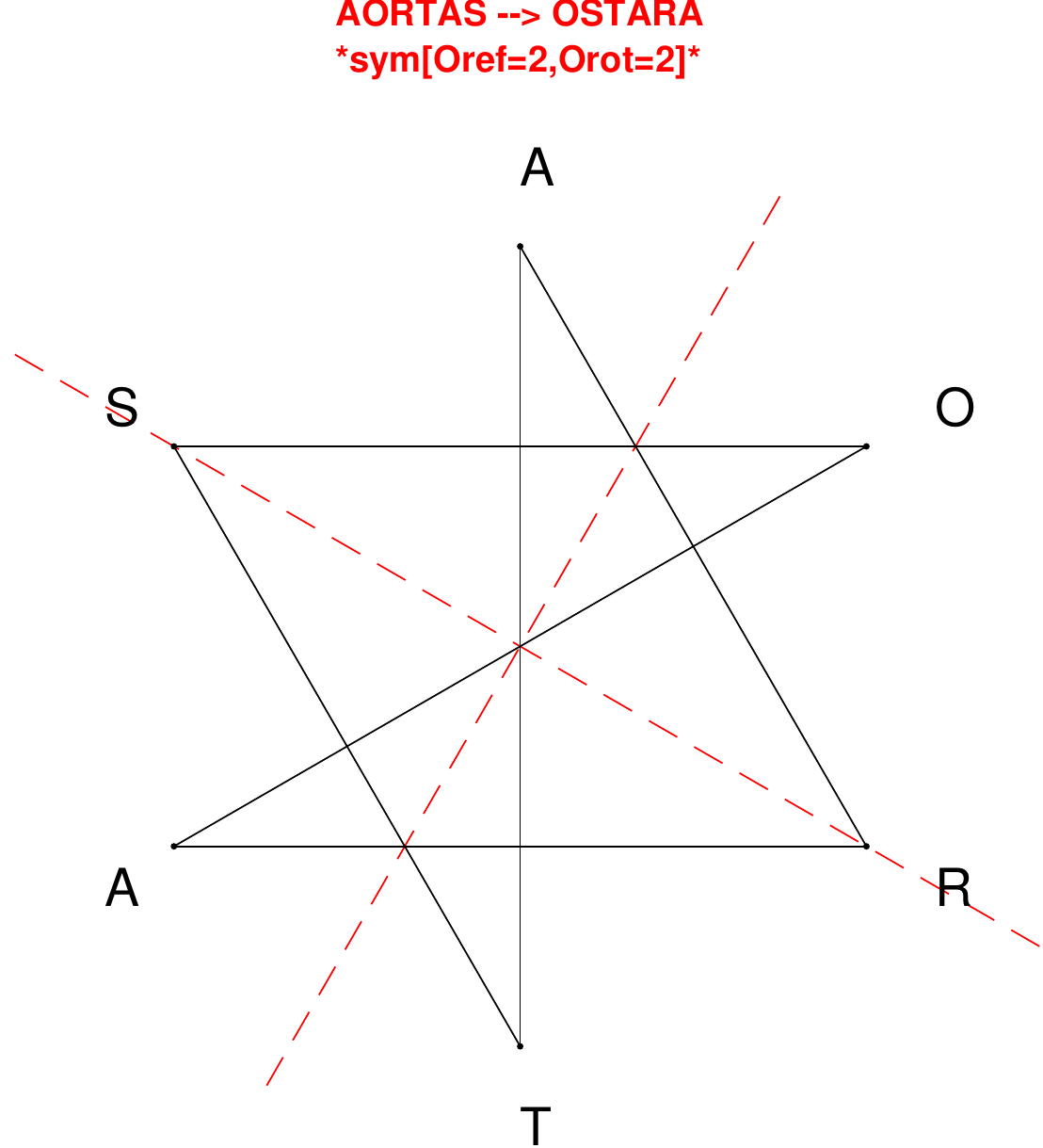}
\end{subfigure}
\hfill
\begin{subfigure}[T]{0.19\textwidth}
\centering
\includegraphics[width=\textwidth]{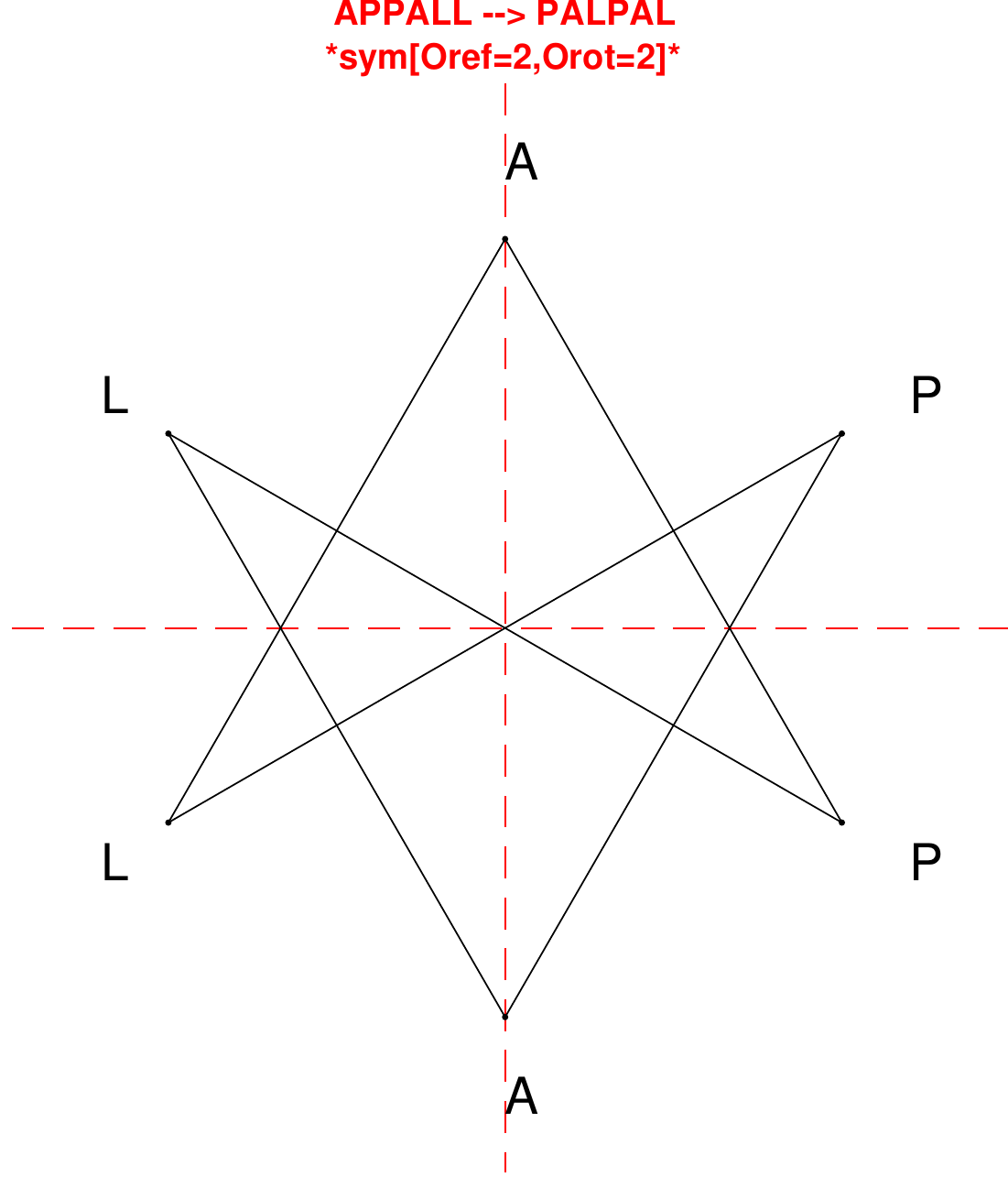}
\end{subfigure}
\hfill
\begin{subfigure}[T]{0.19\textwidth}
\centering
\includegraphics[width=\textwidth]{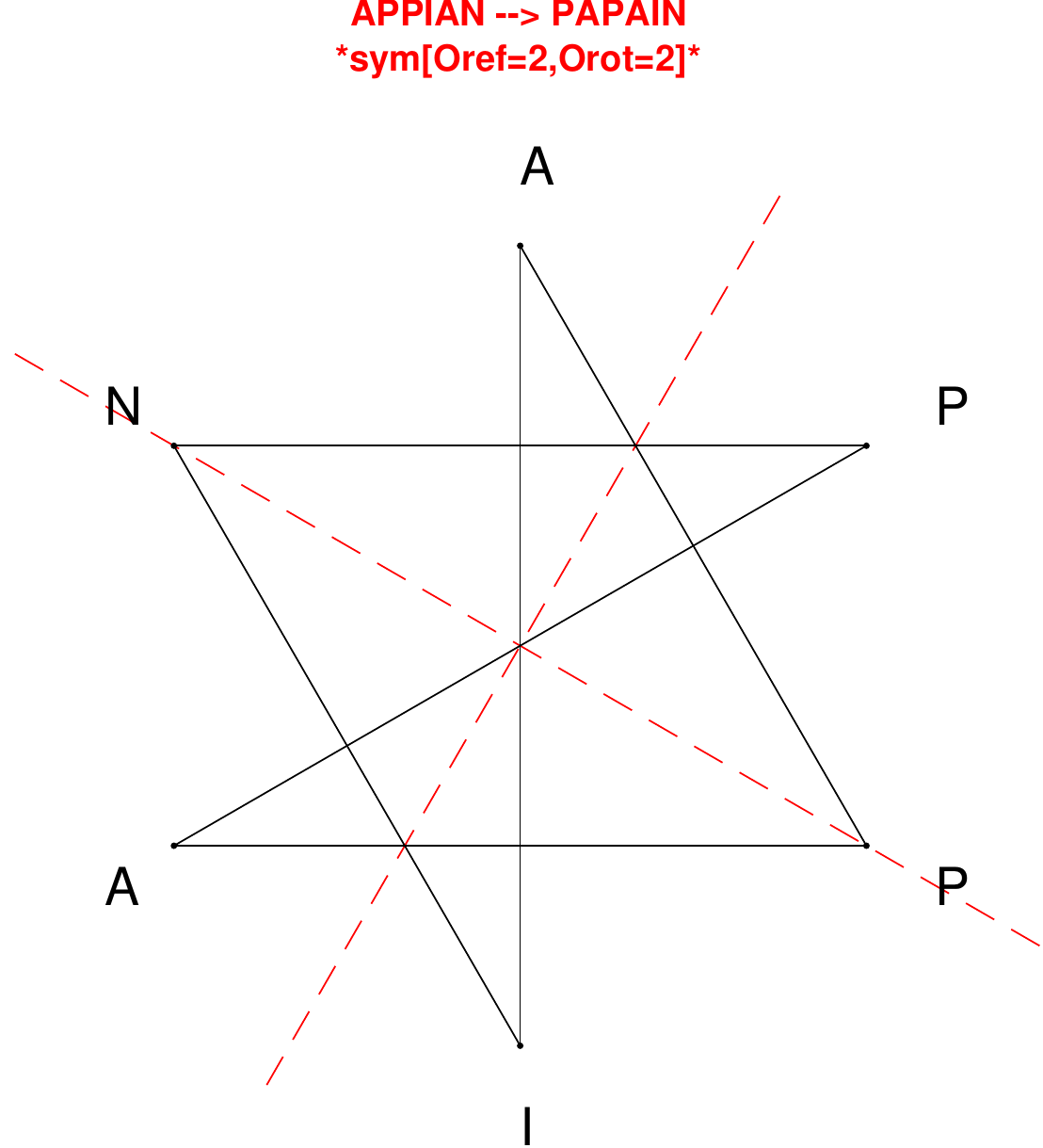}
\end{subfigure}
\hfill
\begin{subfigure}[T]{0.19\textwidth}
\centering
\includegraphics[width=\textwidth]{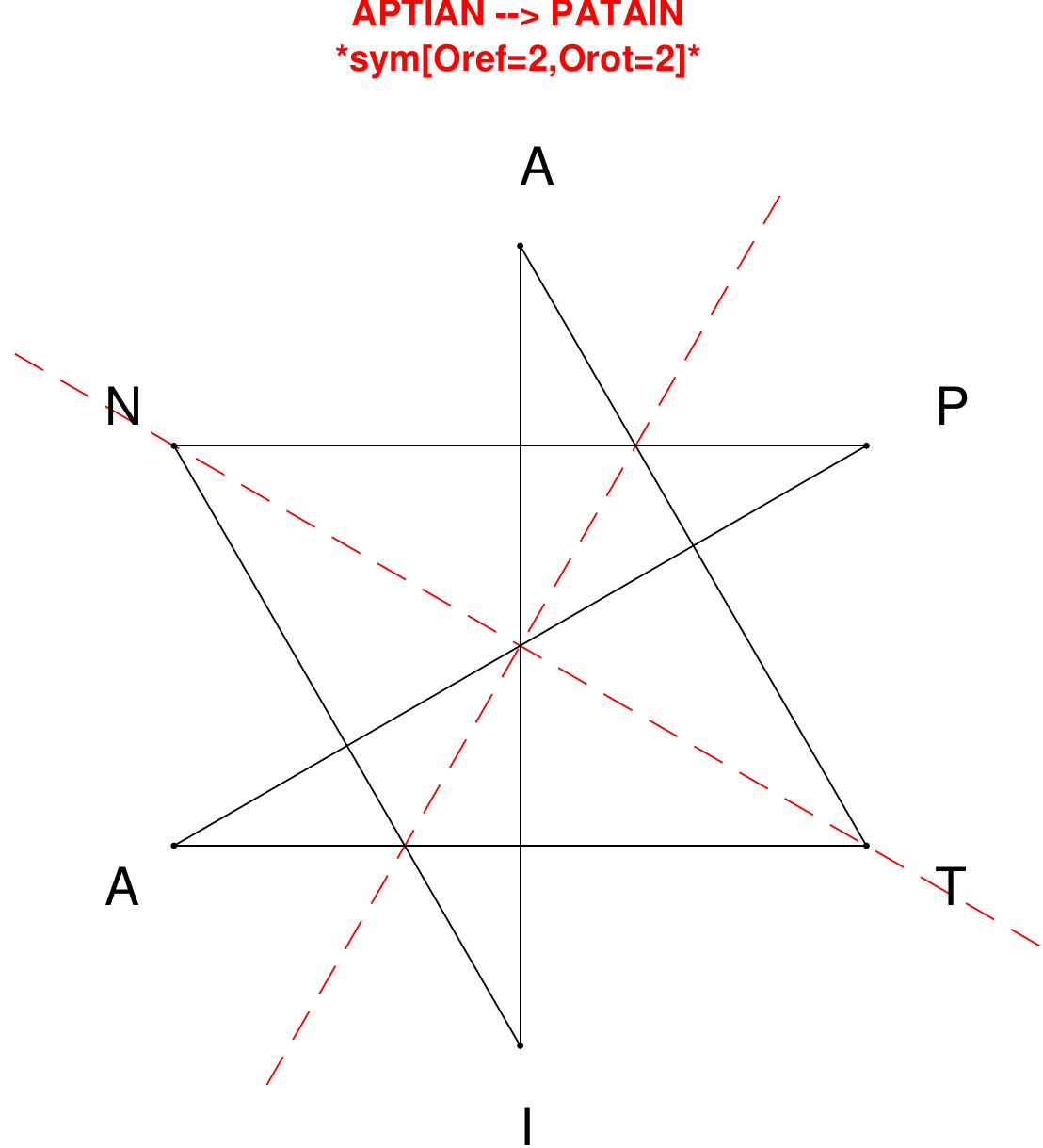}
\end{subfigure}
\end{figure}

\begin{figure}[H]
\centering
\begin{subfigure}[T]{0.19\textwidth}
\centering
\includegraphics[width=\textwidth]{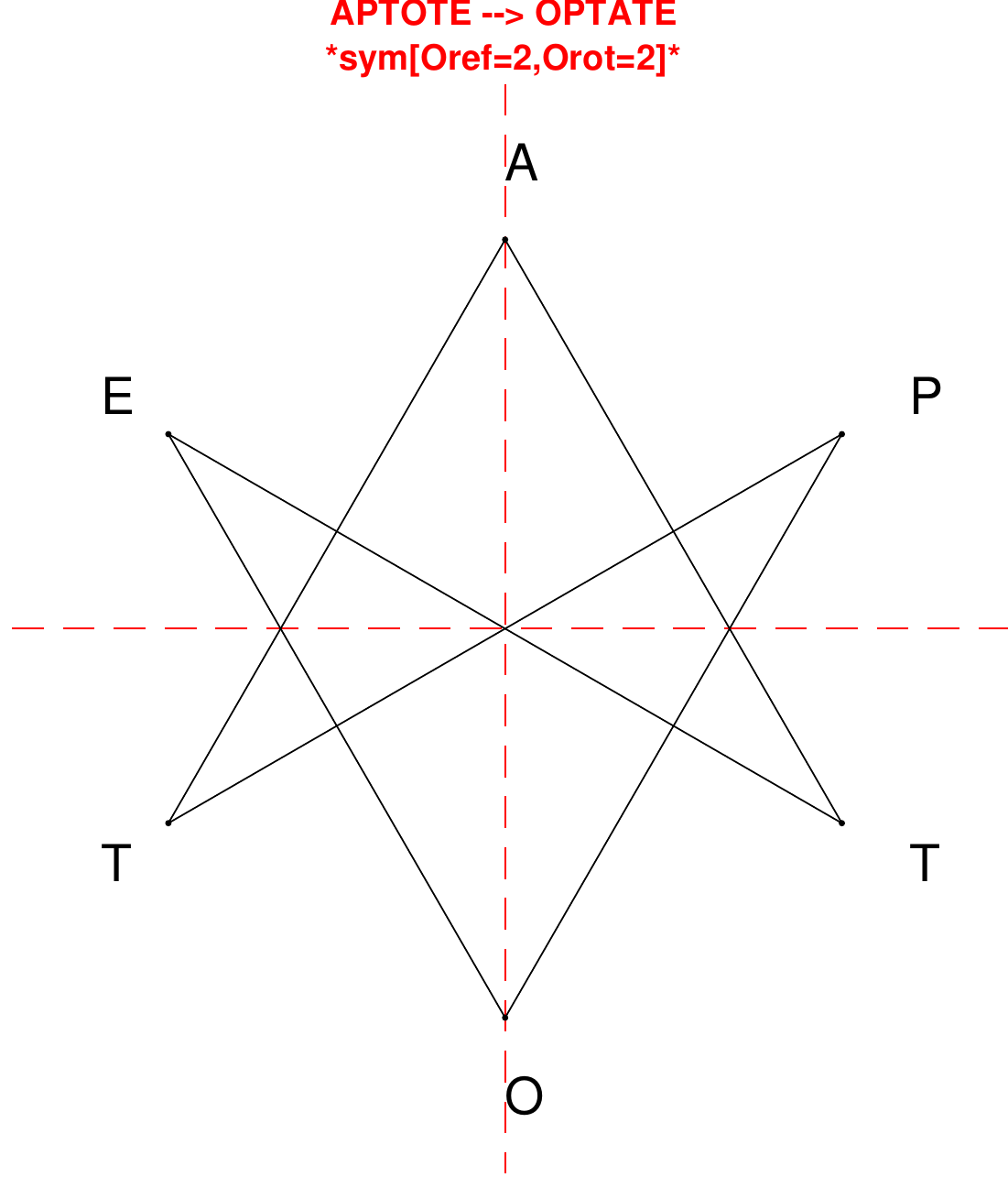}
\end{subfigure}
\hfill
\begin{subfigure}[T]{0.19\textwidth}
\centering
\includegraphics[width=\textwidth]{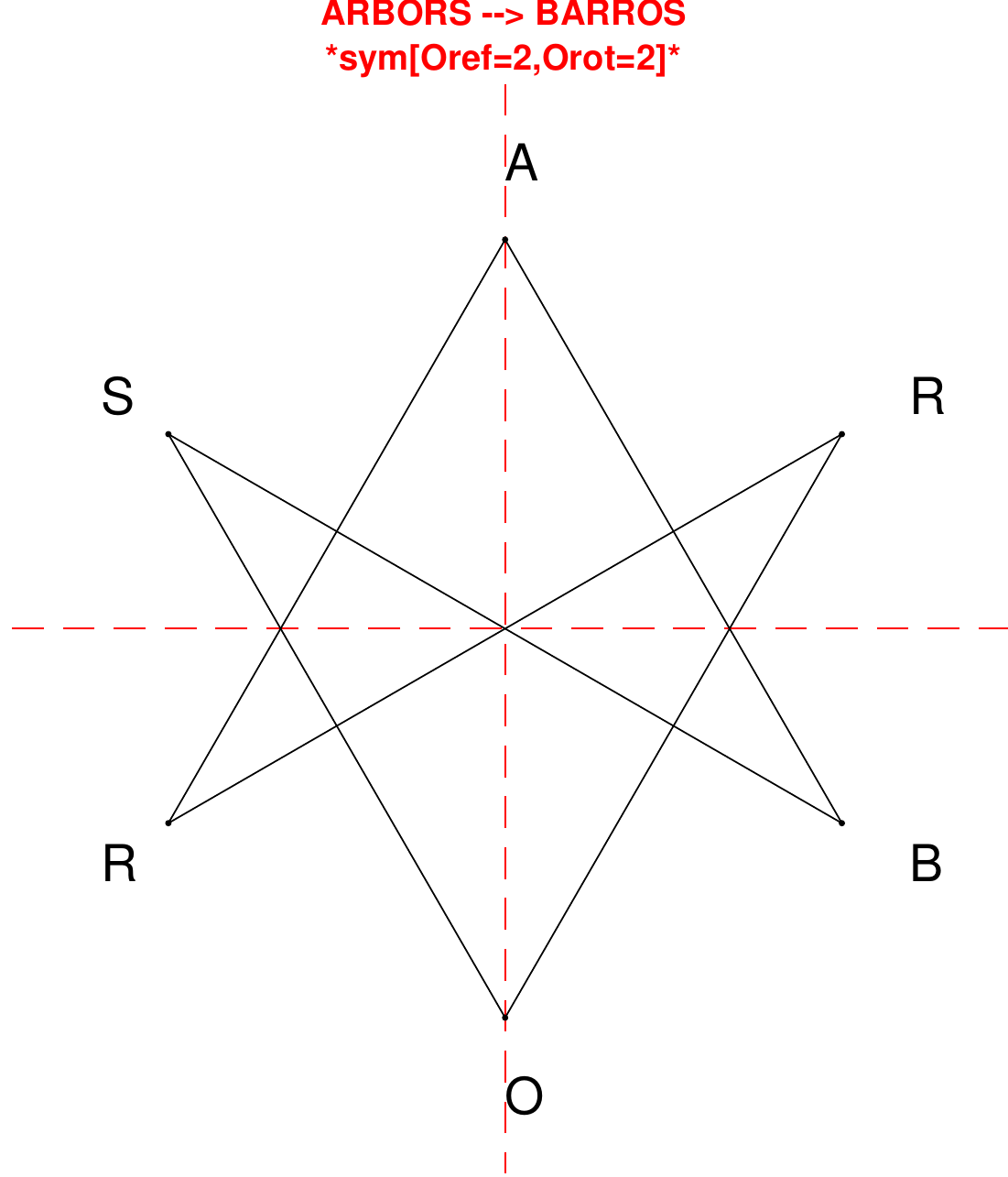}
\end{subfigure}
\hfill
\begin{subfigure}[T]{0.19\textwidth}
\centering
\includegraphics[width=\textwidth]{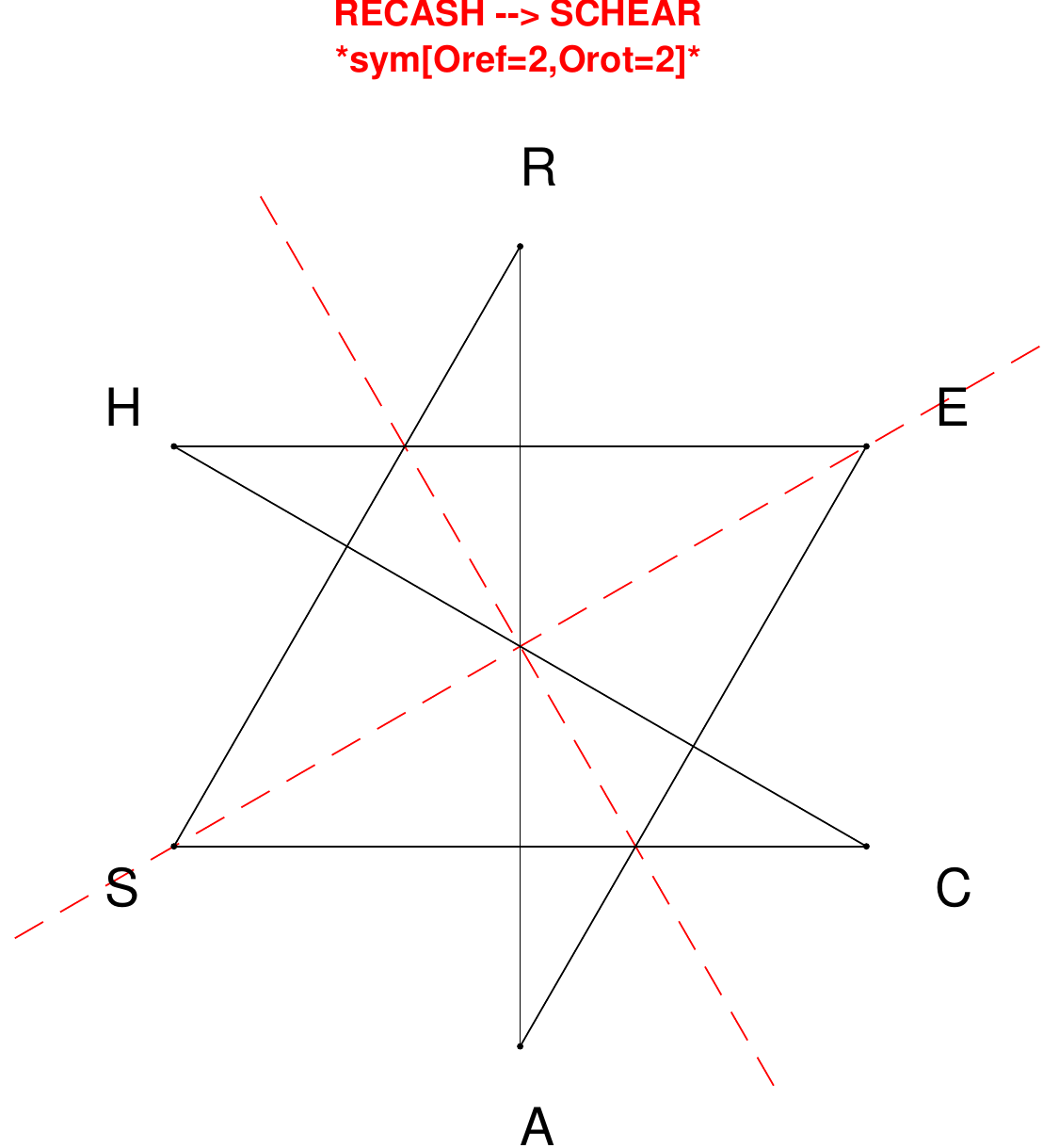}
\end{subfigure}
\hfill
\begin{subfigure}[T]{0.19\textwidth}
\centering
\includegraphics[width=\textwidth]{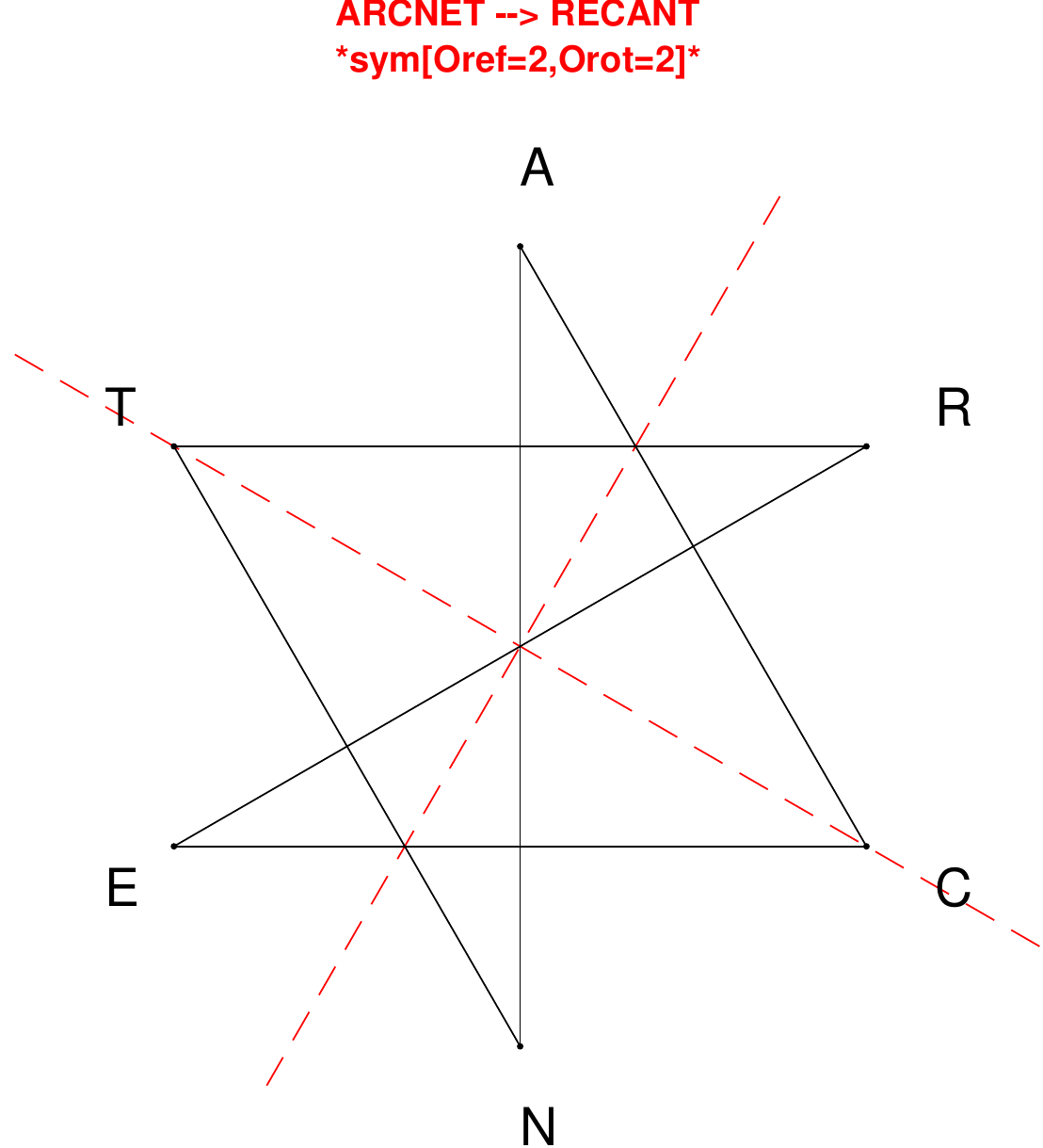}
\end{subfigure}
\hfill
\begin{subfigure}[T]{0.19\textwidth}
\centering
\includegraphics[width=\textwidth]{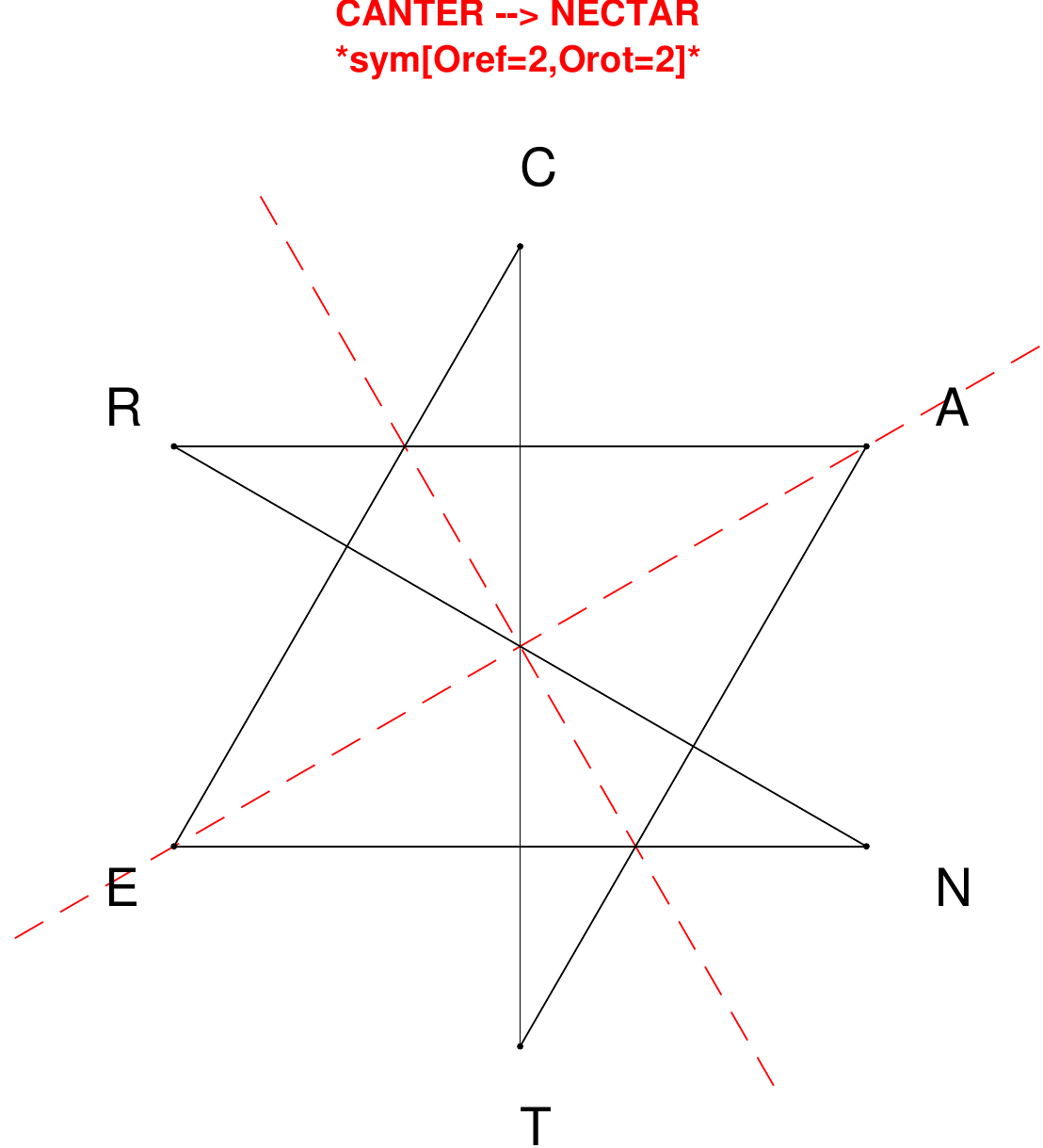}
\end{subfigure}
\end{figure}

\begin{figure}[H]
\centering
\begin{subfigure}[T]{0.19\textwidth}
\centering
\includegraphics[width=\textwidth]{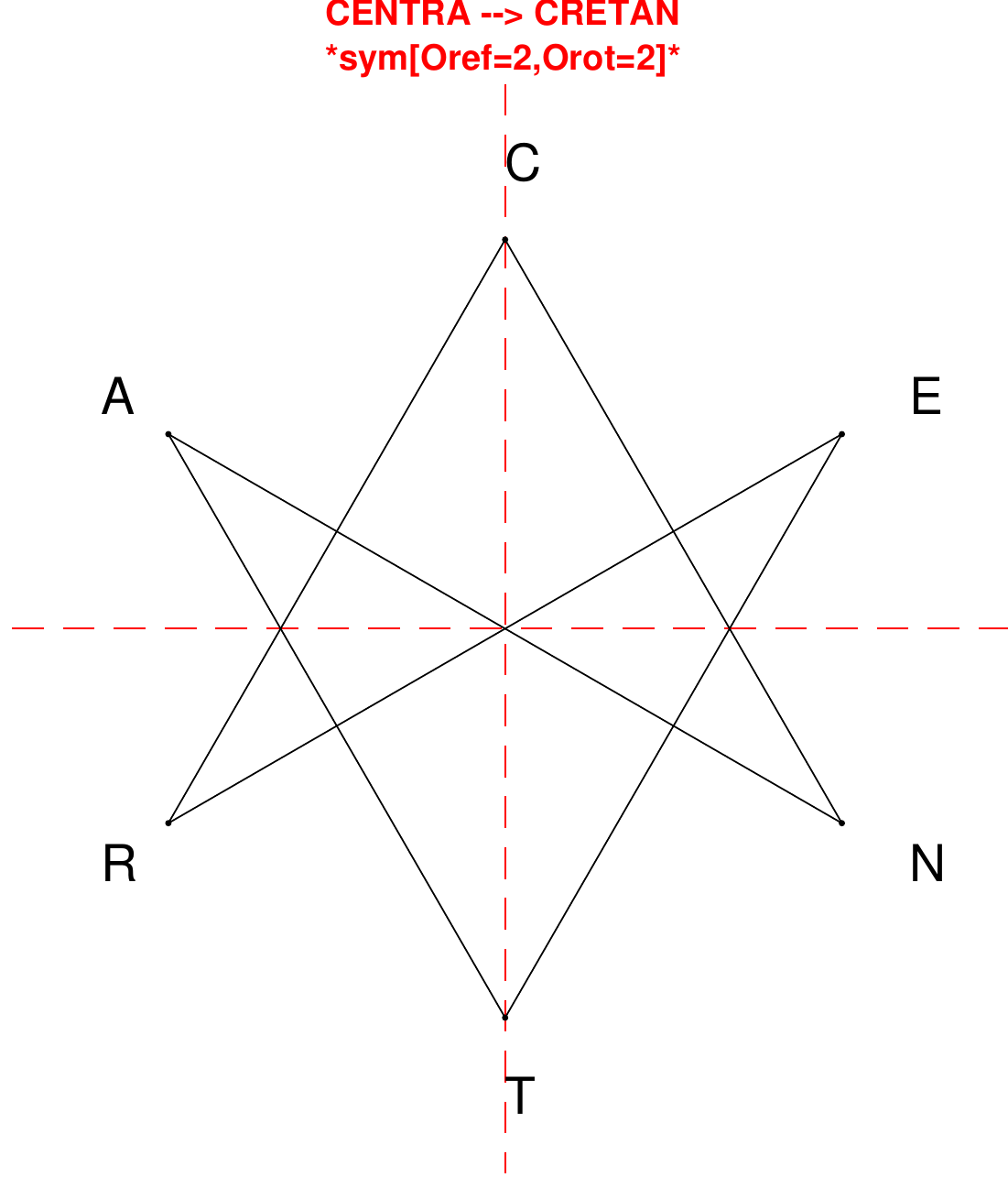}
\end{subfigure}
\hfill
\begin{subfigure}[T]{0.19\textwidth}
\centering
\includegraphics[width=\textwidth]{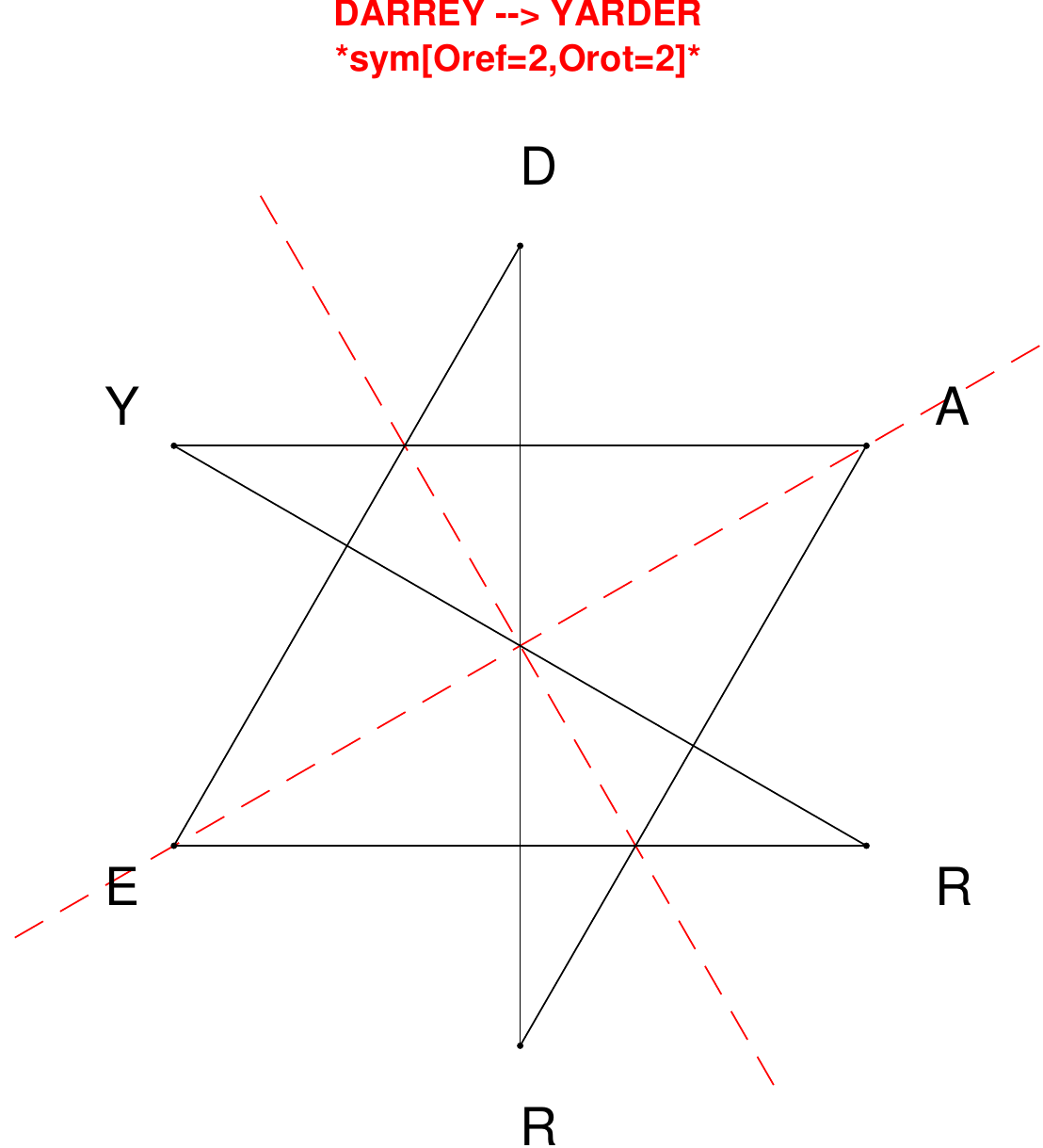}
\end{subfigure}
\hfill
\begin{subfigure}[T]{0.19\textwidth}
\centering
\includegraphics[width=\textwidth]{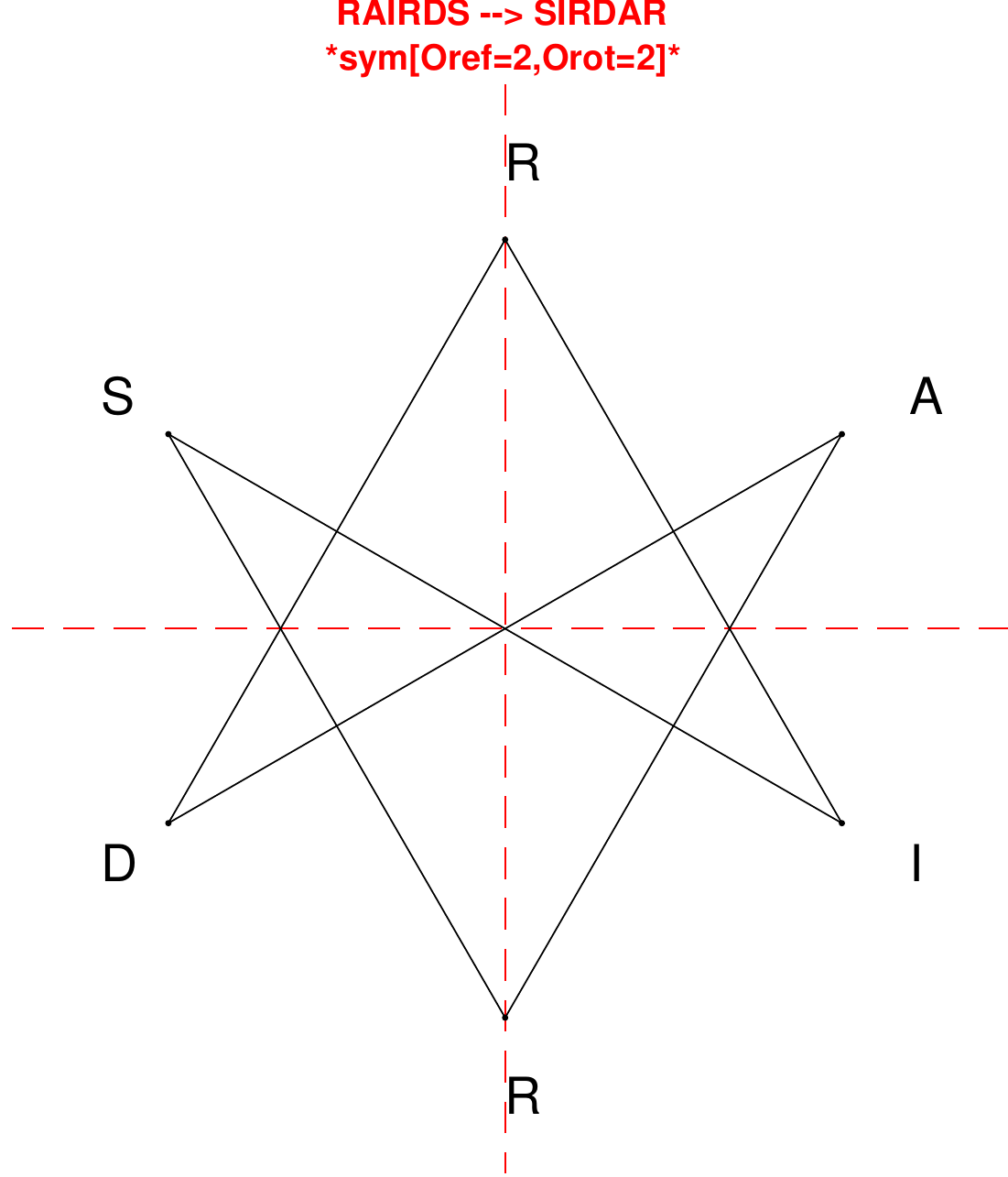}
\end{subfigure}
\hfill
\begin{subfigure}[T]{0.19\textwidth}
\centering
\includegraphics[width=\textwidth]{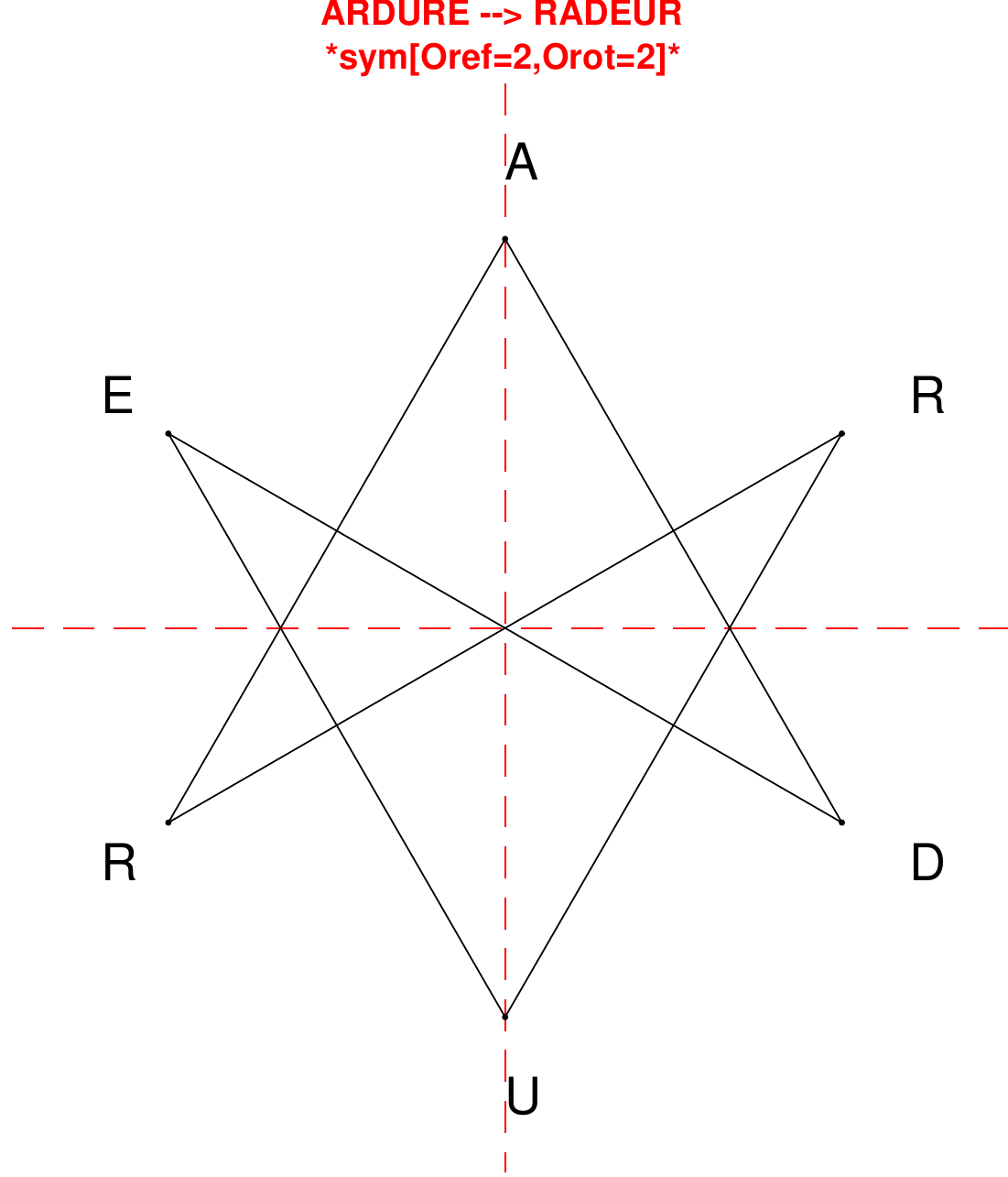}
\end{subfigure}
\hfill
\begin{subfigure}[T]{0.19\textwidth}
\centering
\includegraphics[width=\textwidth]{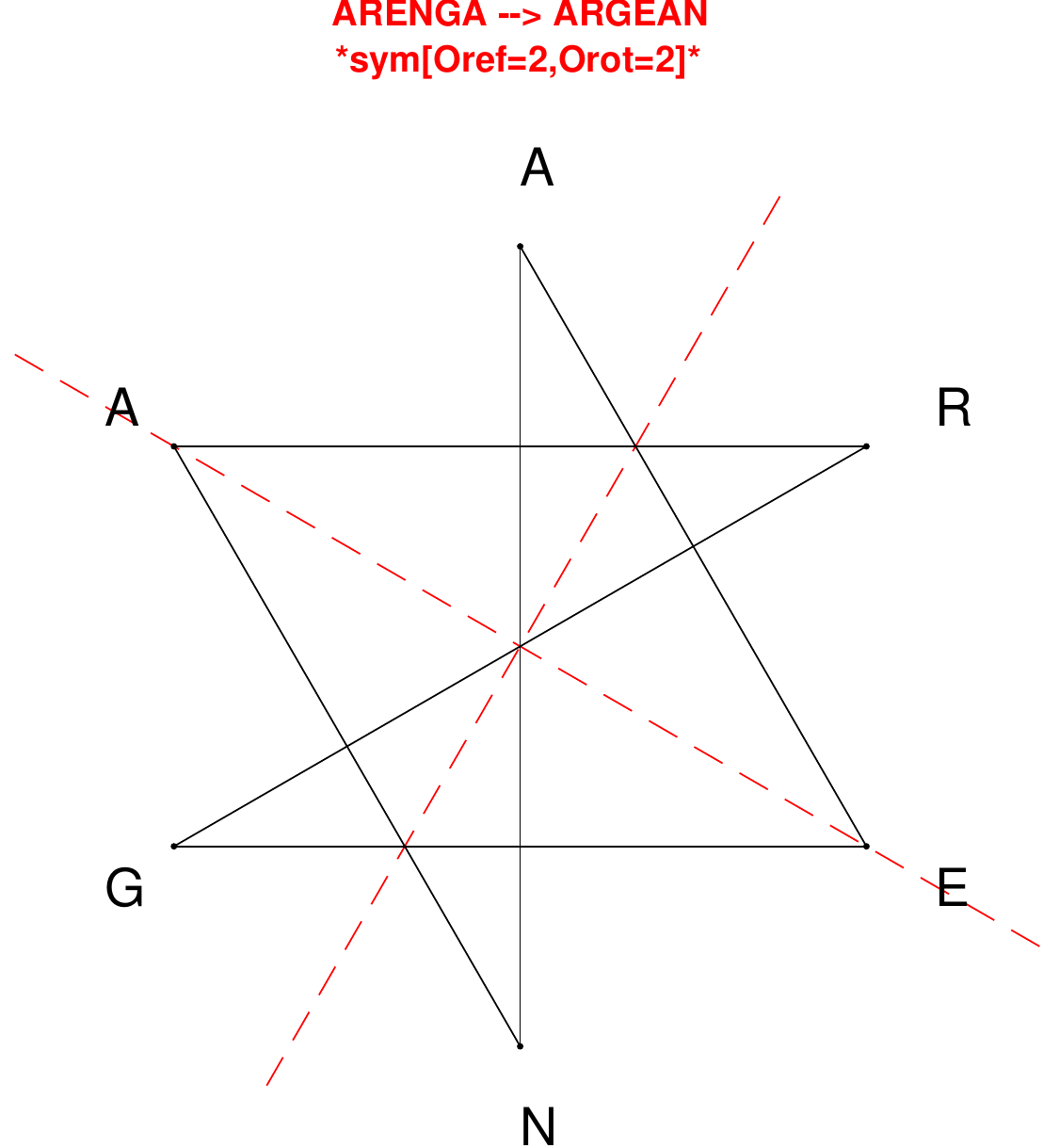}
\end{subfigure}
\end{figure}

\begin{figure}[H]
\centering
\begin{subfigure}[T]{0.19\textwidth}
\centering
\includegraphics[width=\textwidth]{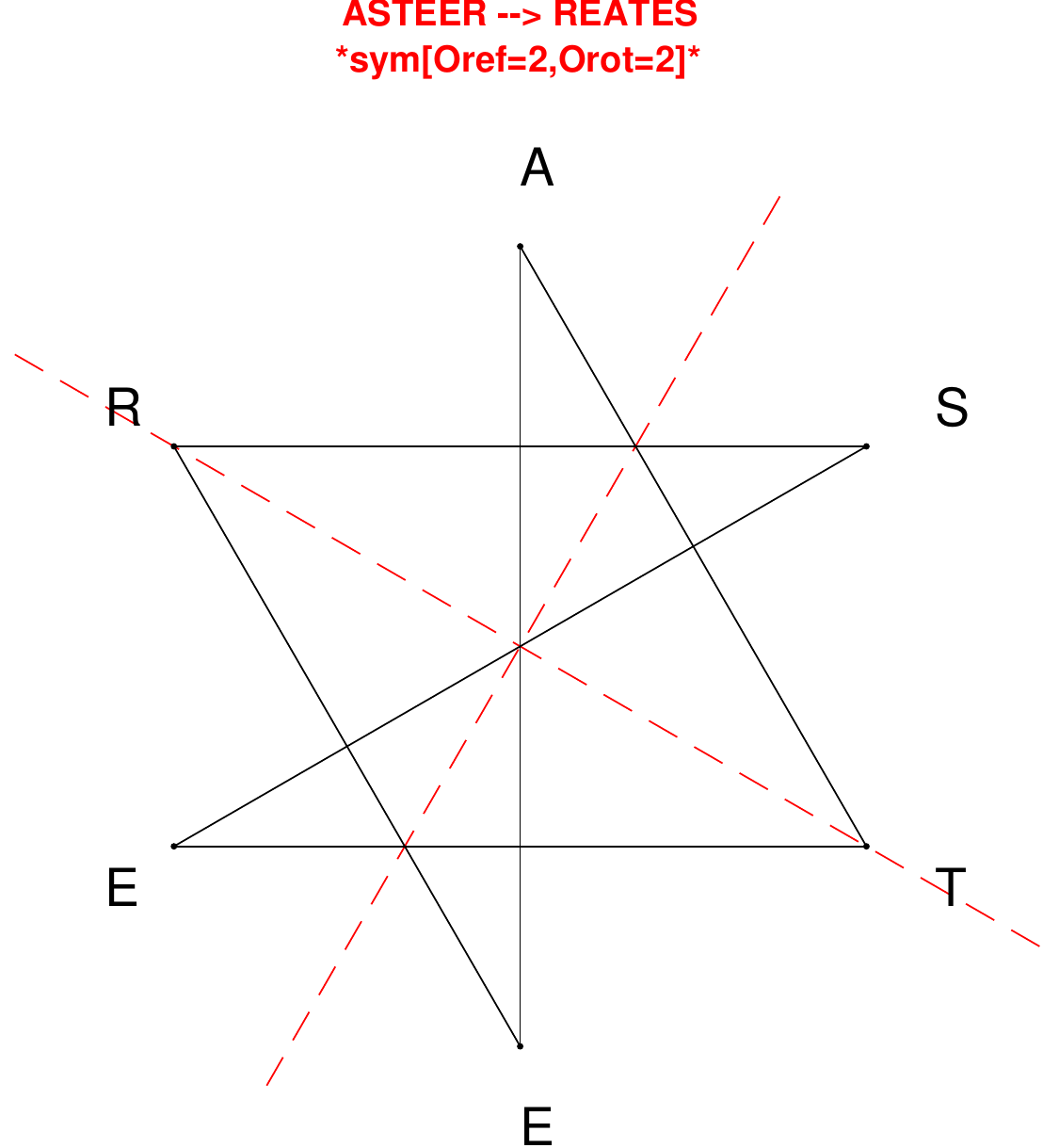}
\end{subfigure}
\hfill
\begin{subfigure}[T]{0.19\textwidth}
\centering
\includegraphics[width=\textwidth]{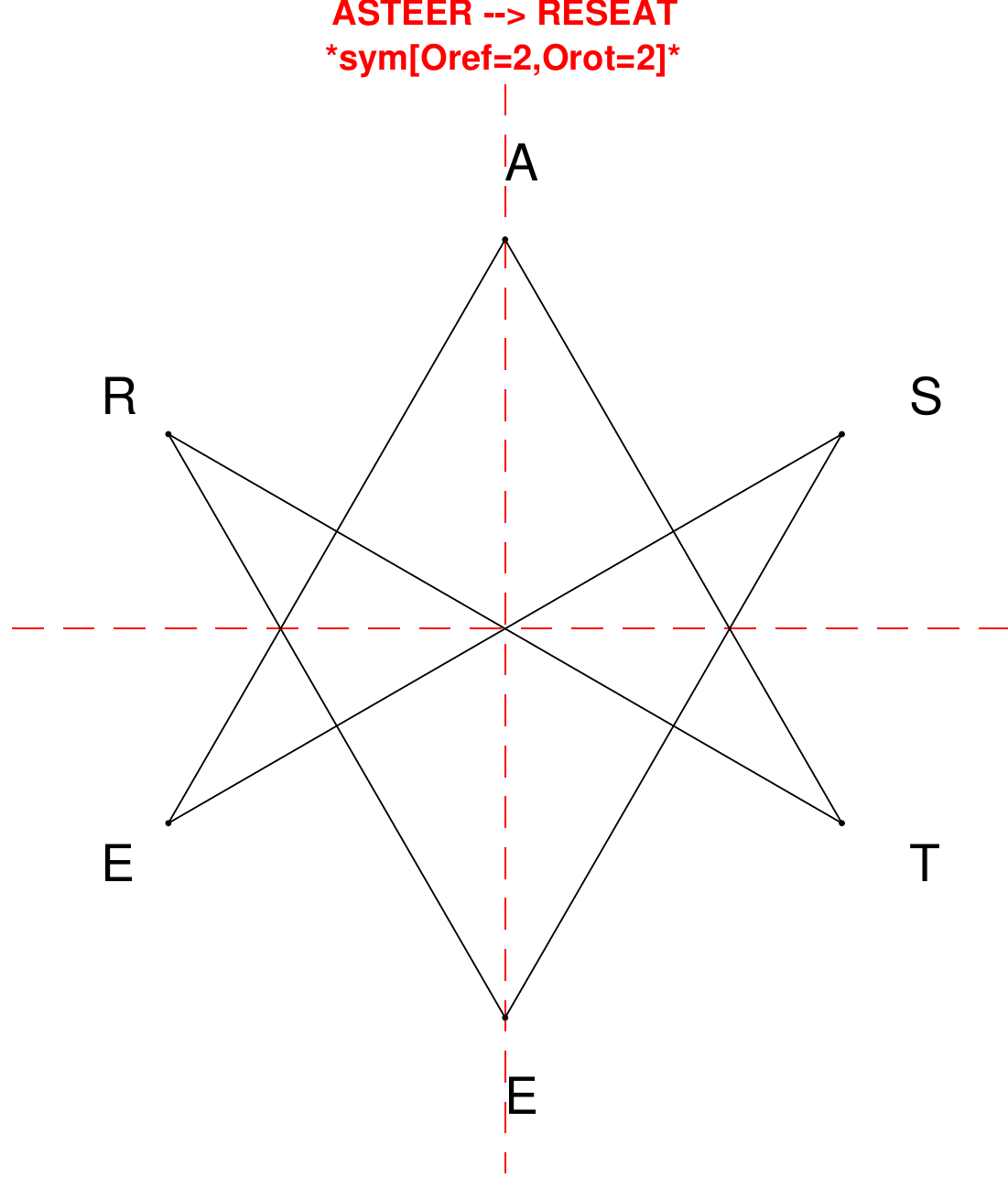}
\end{subfigure}
\hfill
\begin{subfigure}[T]{0.19\textwidth}
\centering
\includegraphics[width=\textwidth]{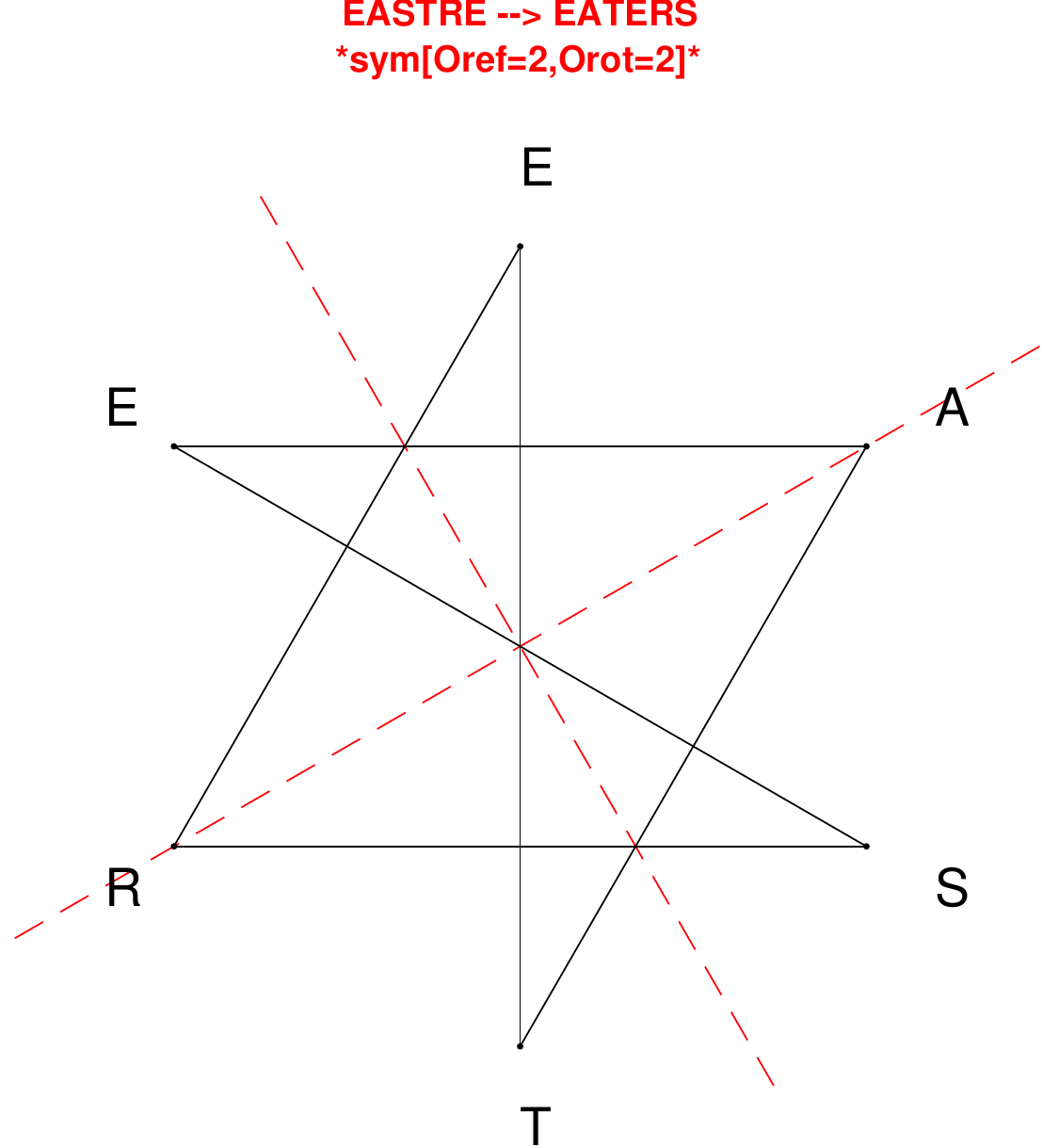}
\end{subfigure}
\hfill
\begin{subfigure}[T]{0.19\textwidth}
\centering
\includegraphics[width=\textwidth]{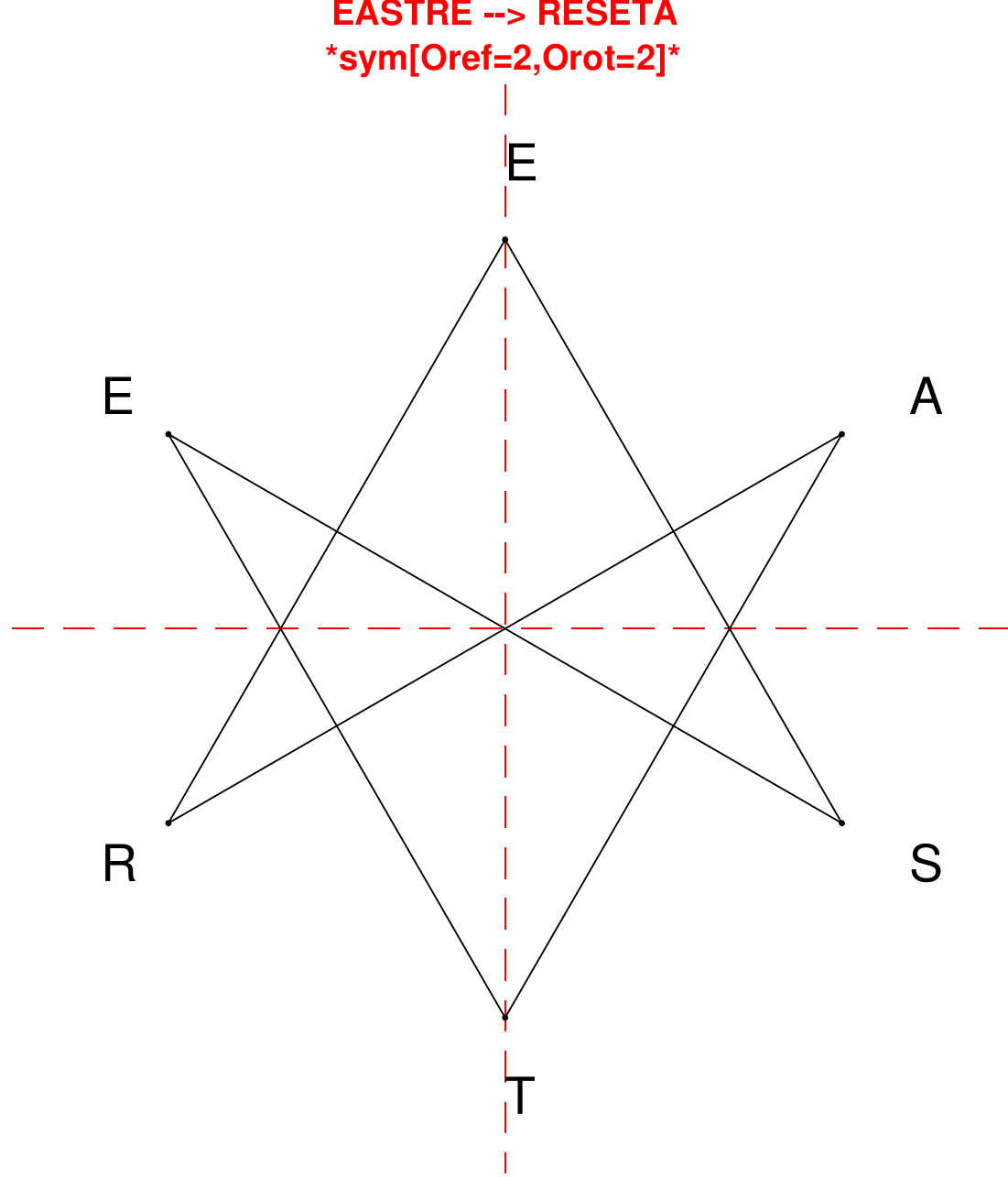}
\end{subfigure}
\hfill
\begin{subfigure}[T]{0.19\textwidth}
\centering
\includegraphics[width=\textwidth]{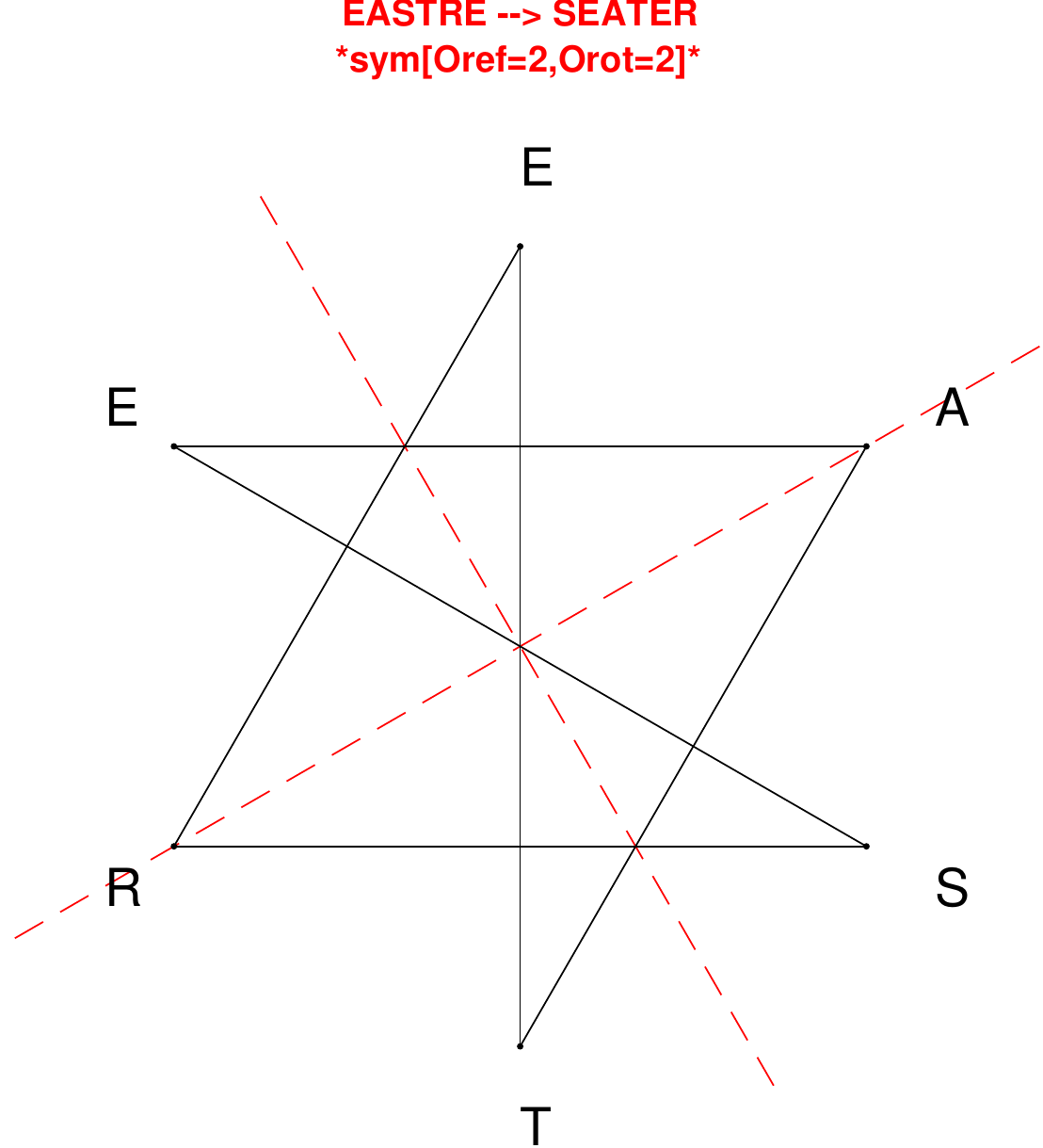}
\end{subfigure}
\end{figure}

\begin{figure}[H]
\centering
\begin{subfigure}[T]{0.19\textwidth}
\centering
\includegraphics[width=\textwidth]{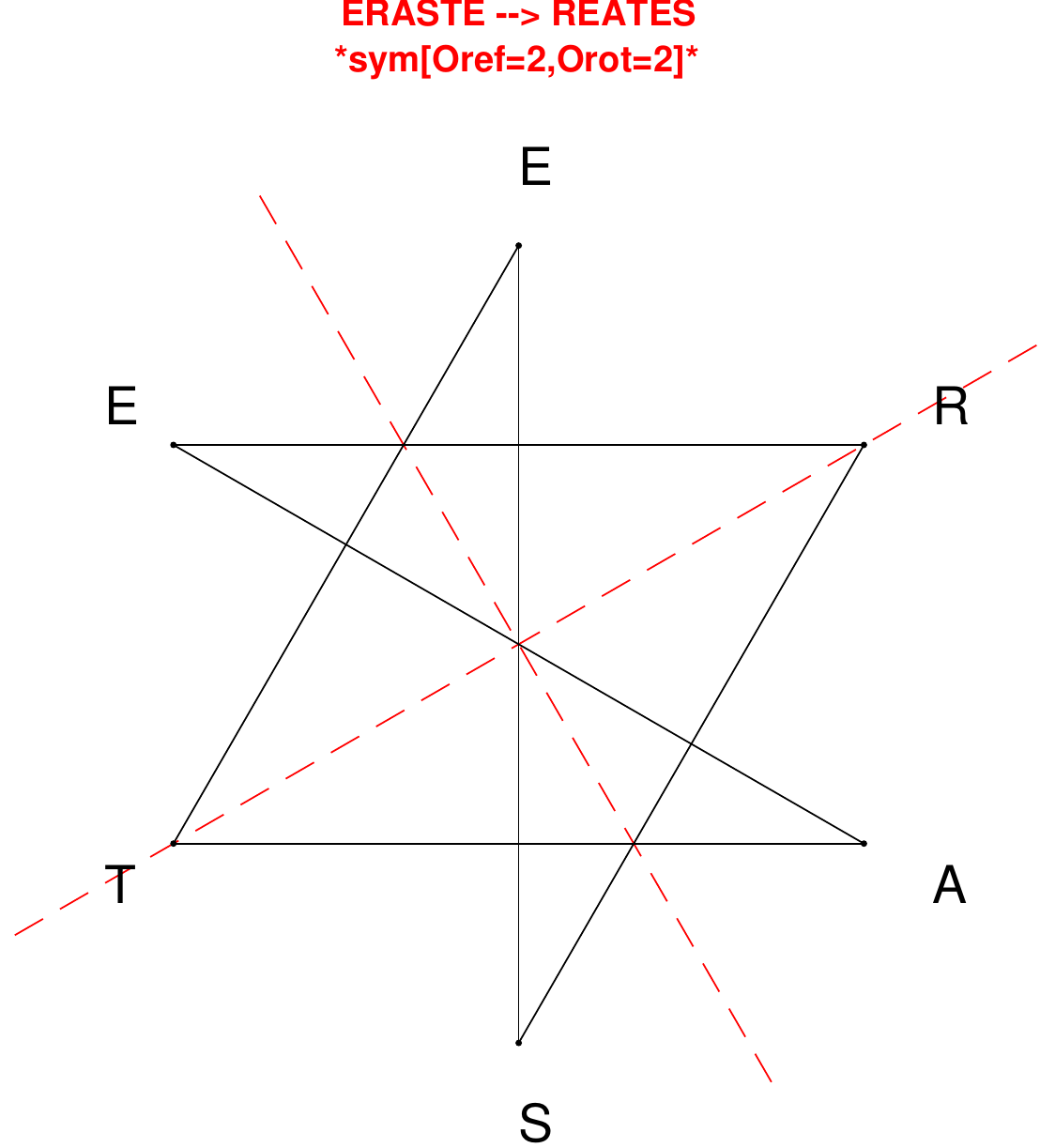}
\end{subfigure}
\hfill
\begin{subfigure}[T]{0.19\textwidth}
\centering
\includegraphics[width=\textwidth]{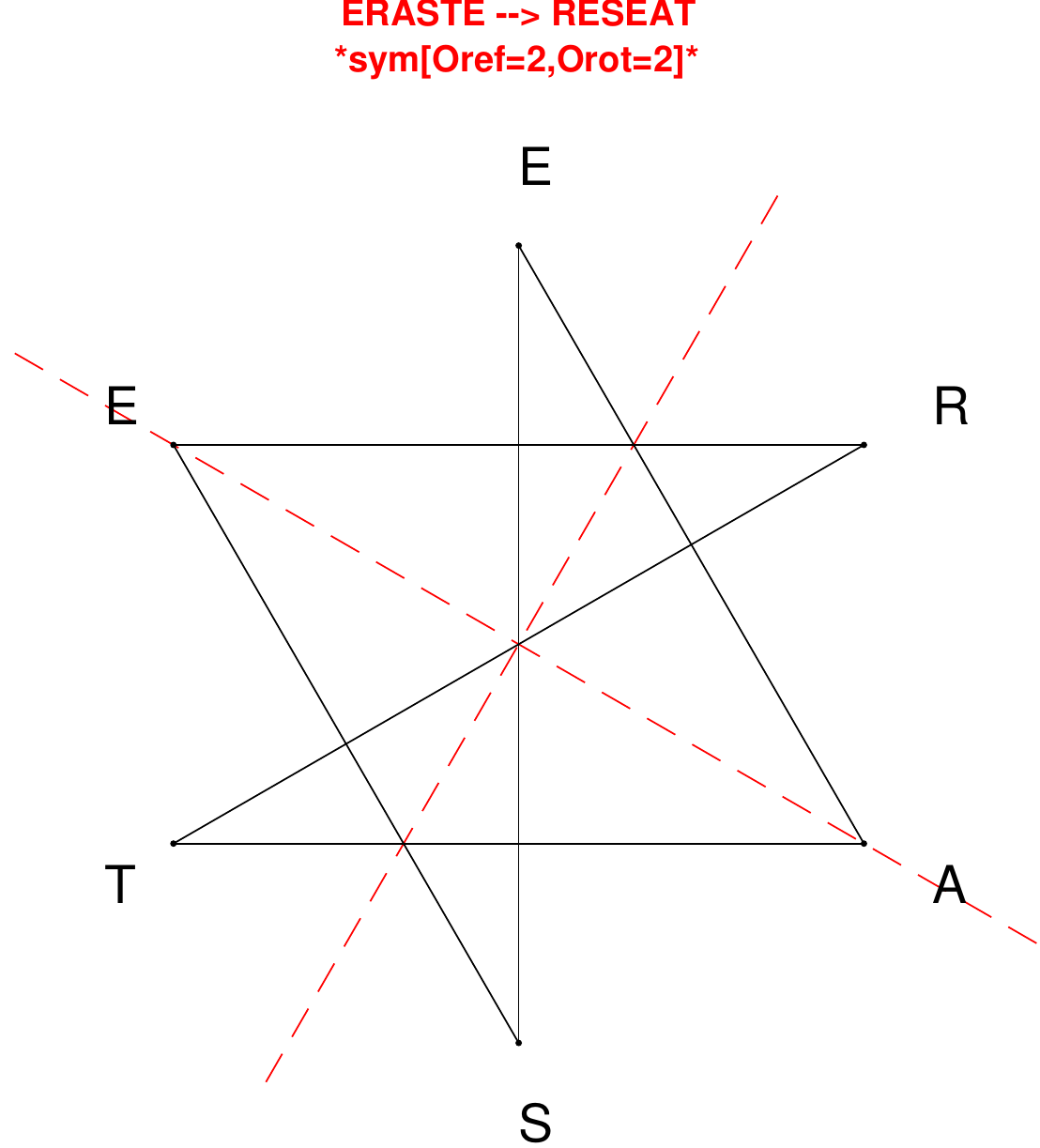}
\end{subfigure}
\hfill
\begin{subfigure}[T]{0.19\textwidth}
\centering
\includegraphics[width=\textwidth]{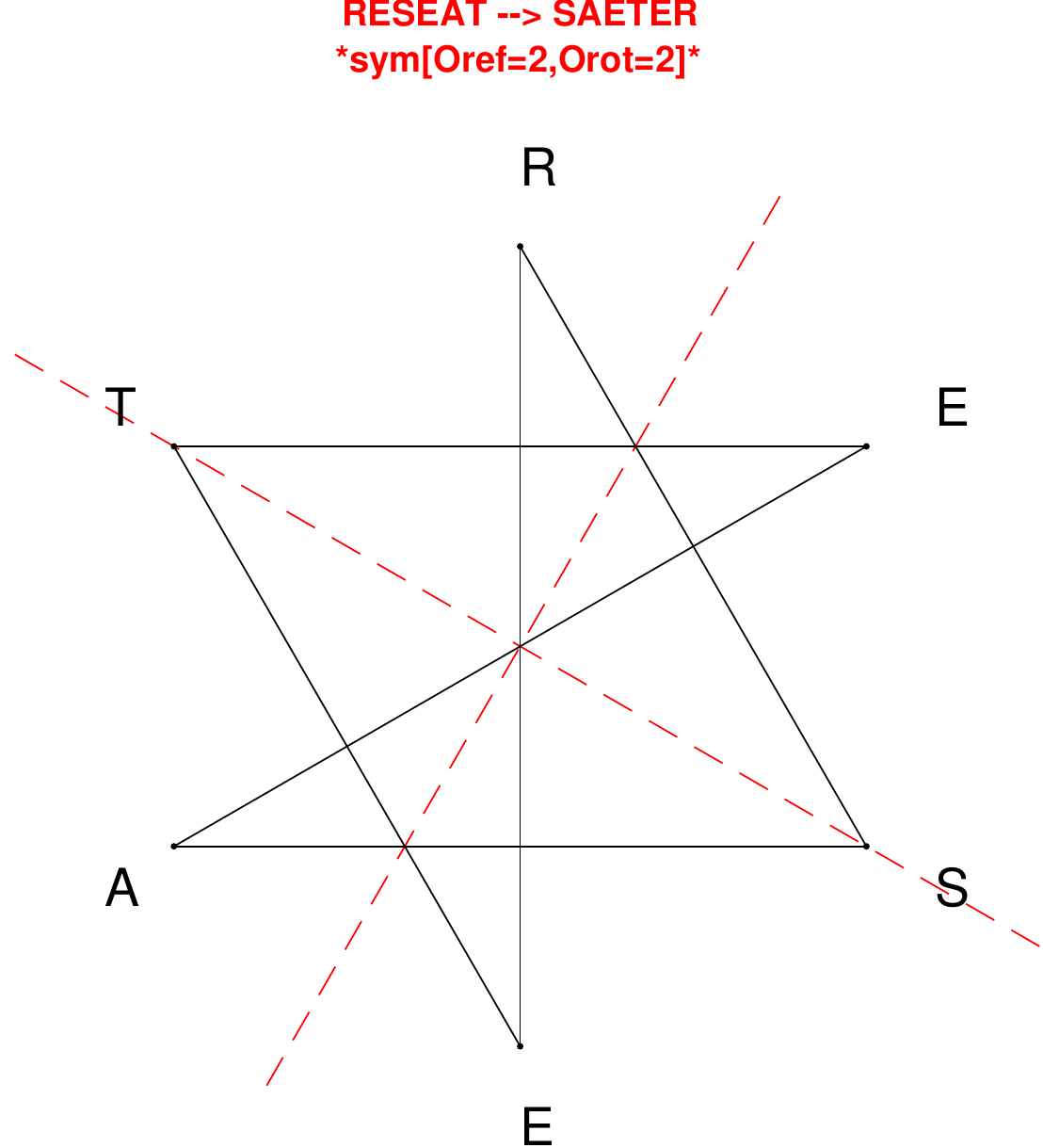}
\end{subfigure}
\hfill
\begin{subfigure}[T]{0.19\textwidth}
\centering
\includegraphics[width=\textwidth]{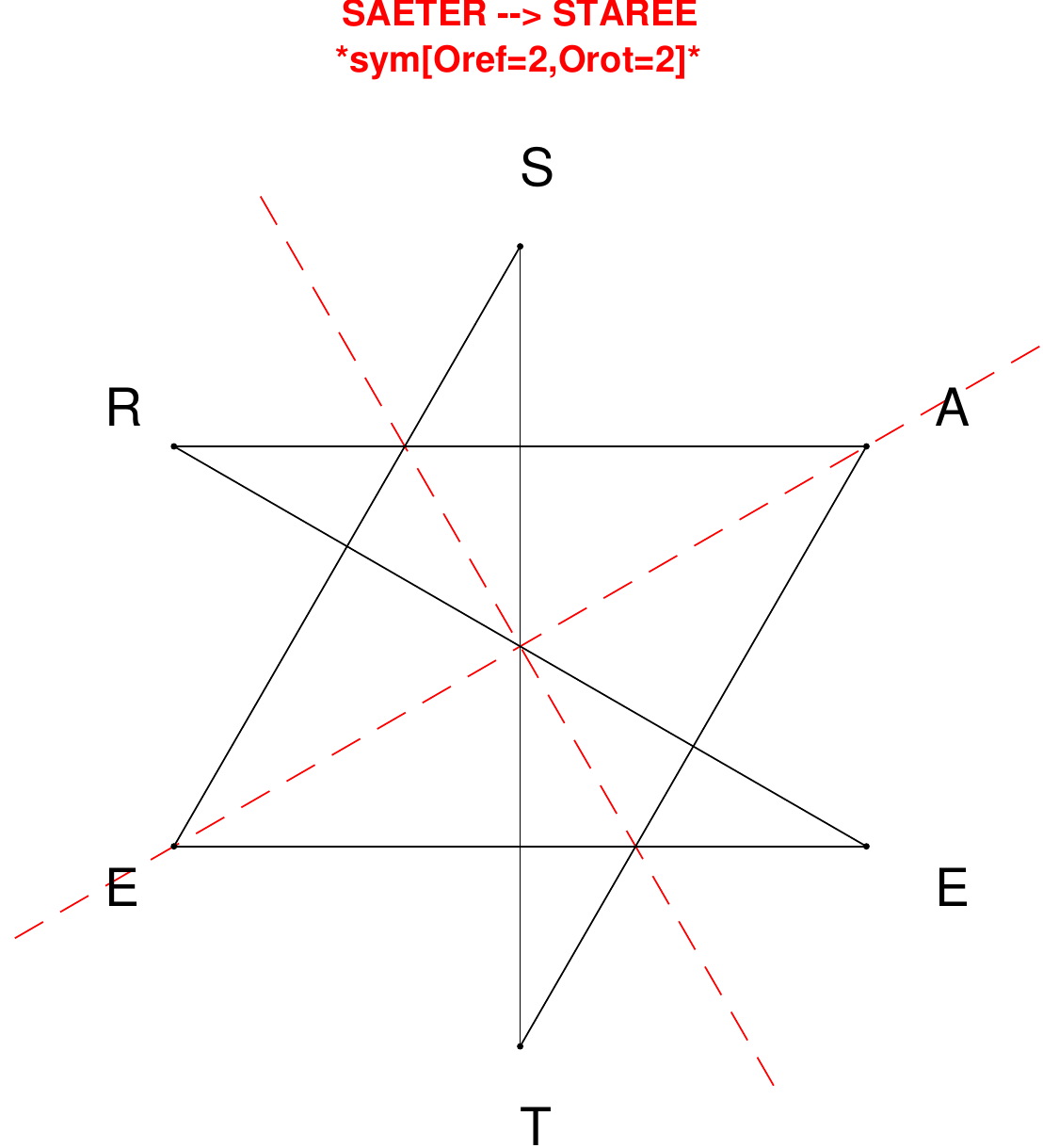}
\end{subfigure}
\hfill
\begin{subfigure}[T]{0.19\textwidth}
\centering
\includegraphics[width=\textwidth]{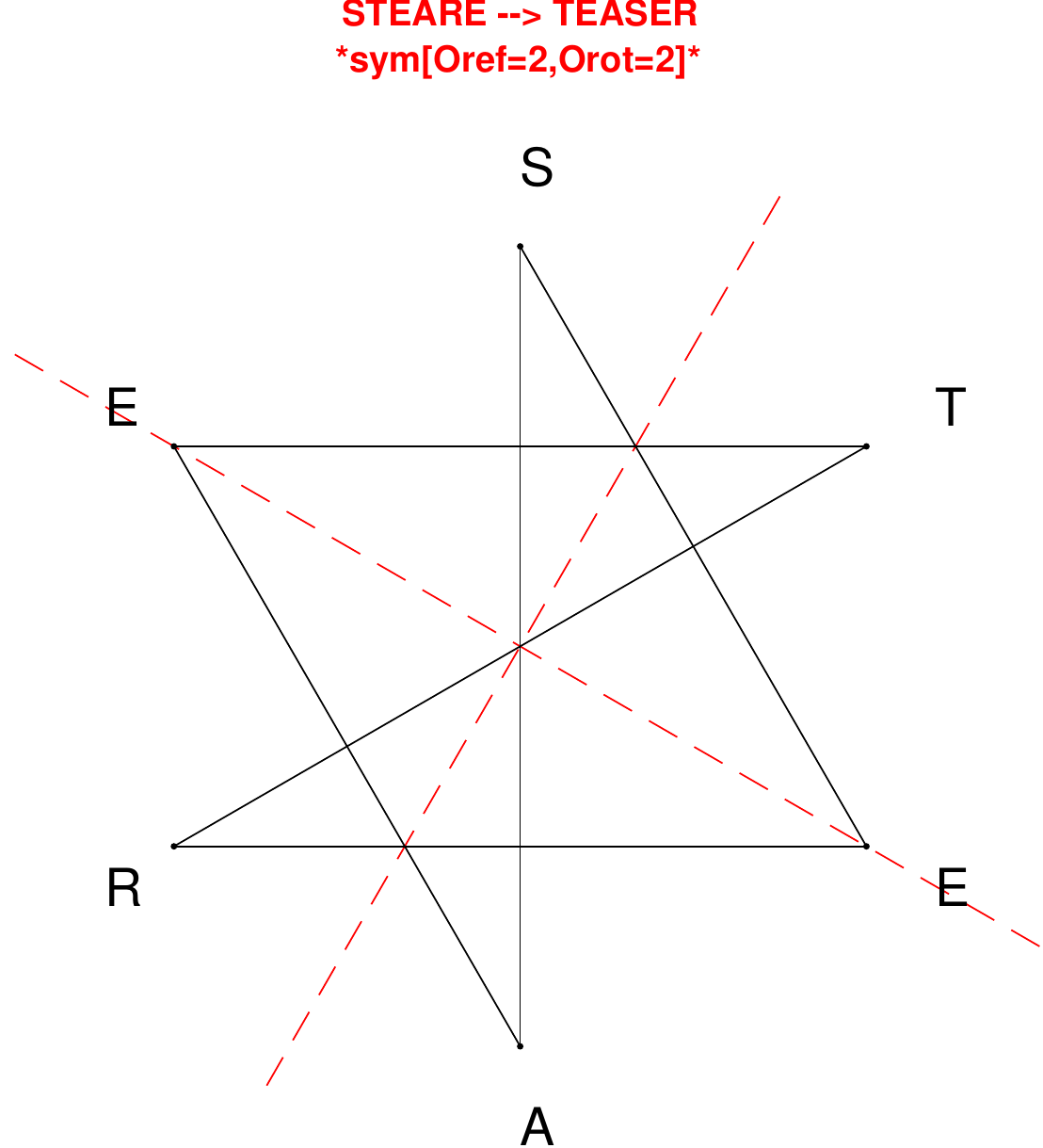}
\end{subfigure}
\end{figure}

\begin{figure}[H]
\centering
\begin{subfigure}[T]{0.19\textwidth}
\centering
\includegraphics[width=\textwidth]{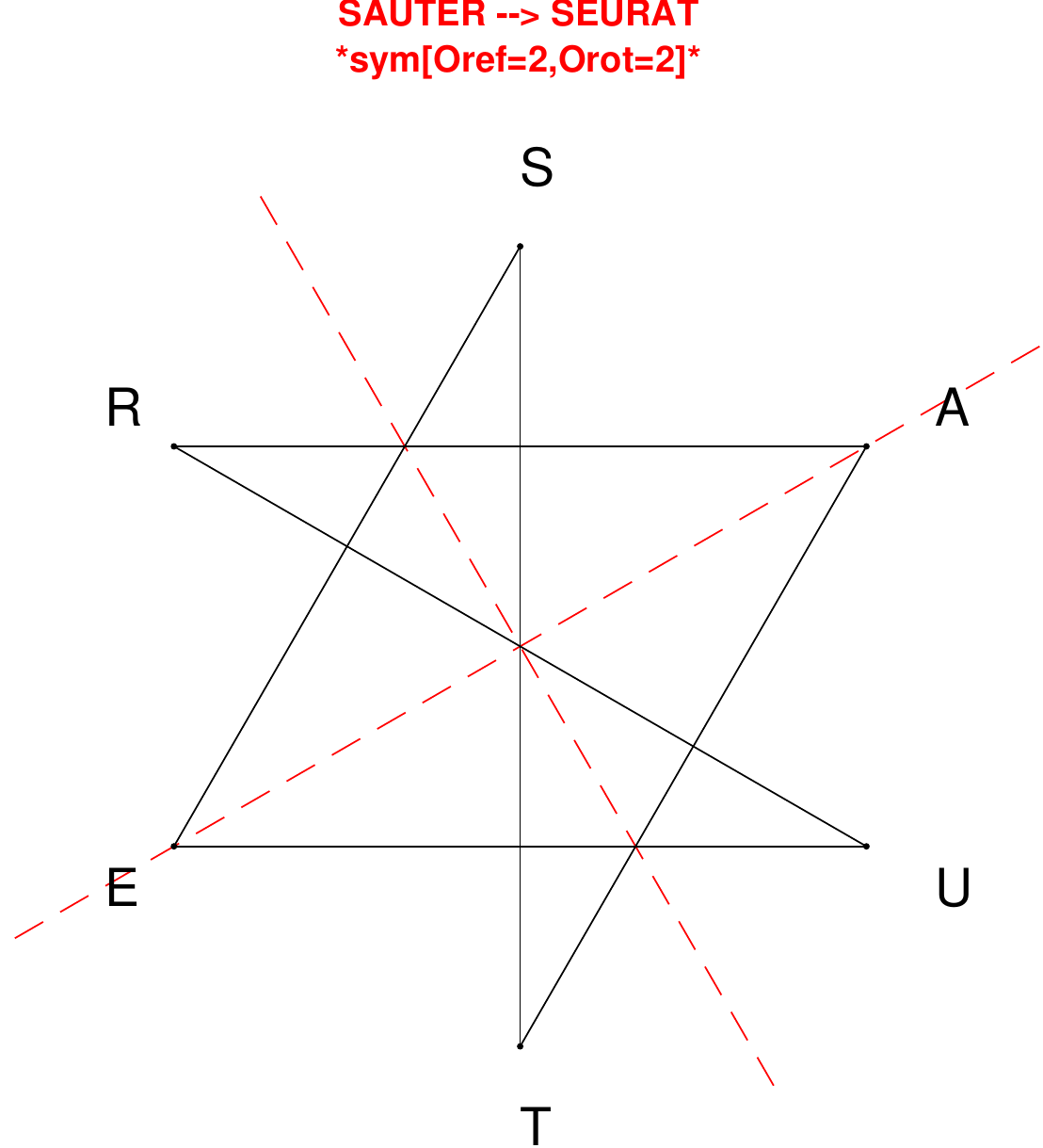}
\end{subfigure}
\hfill
\begin{subfigure}[T]{0.19\textwidth}
\centering
\includegraphics[width=\textwidth]{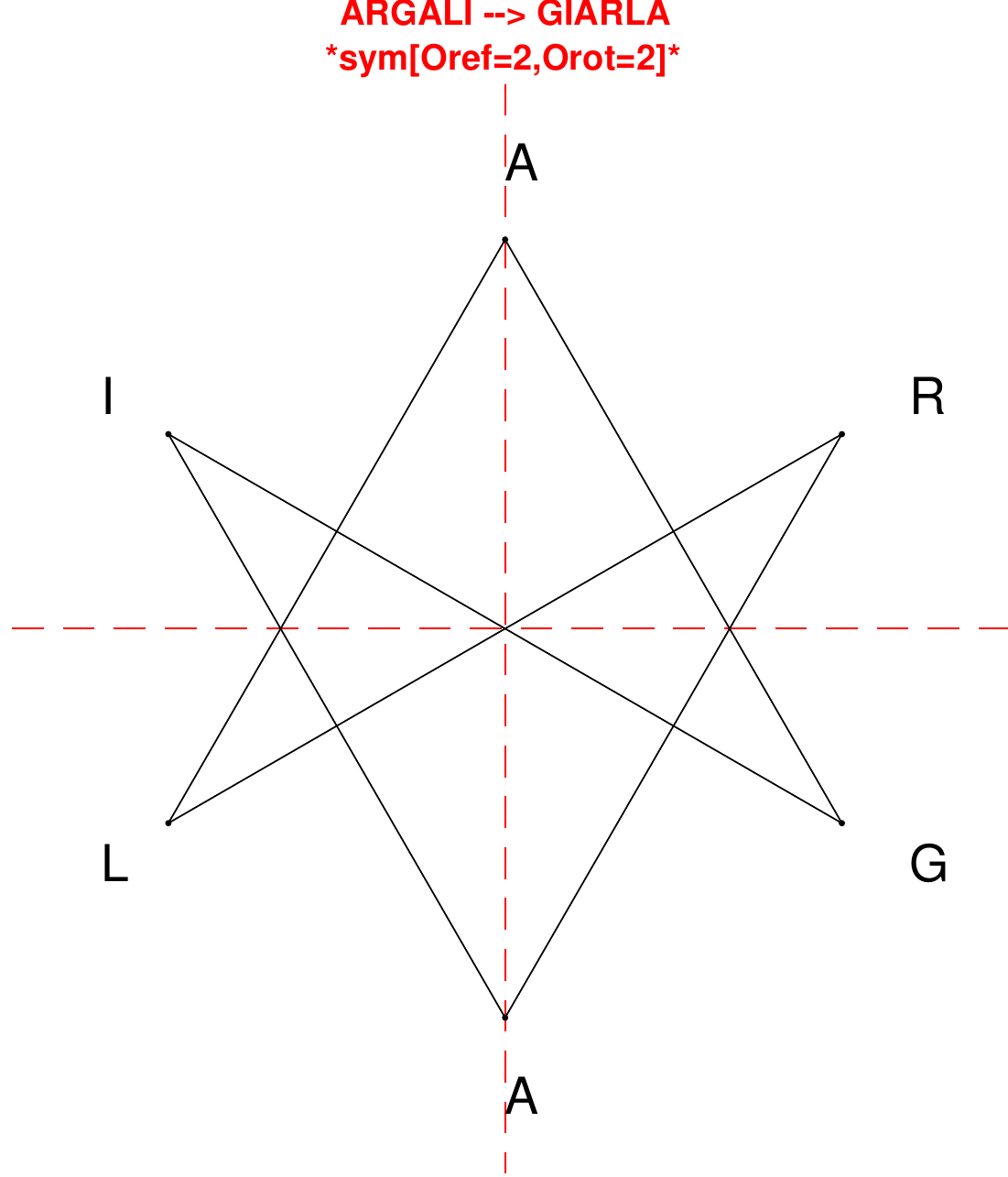}
\end{subfigure}
\hfill
\begin{subfigure}[T]{0.19\textwidth}
\centering
\includegraphics[width=\textwidth]{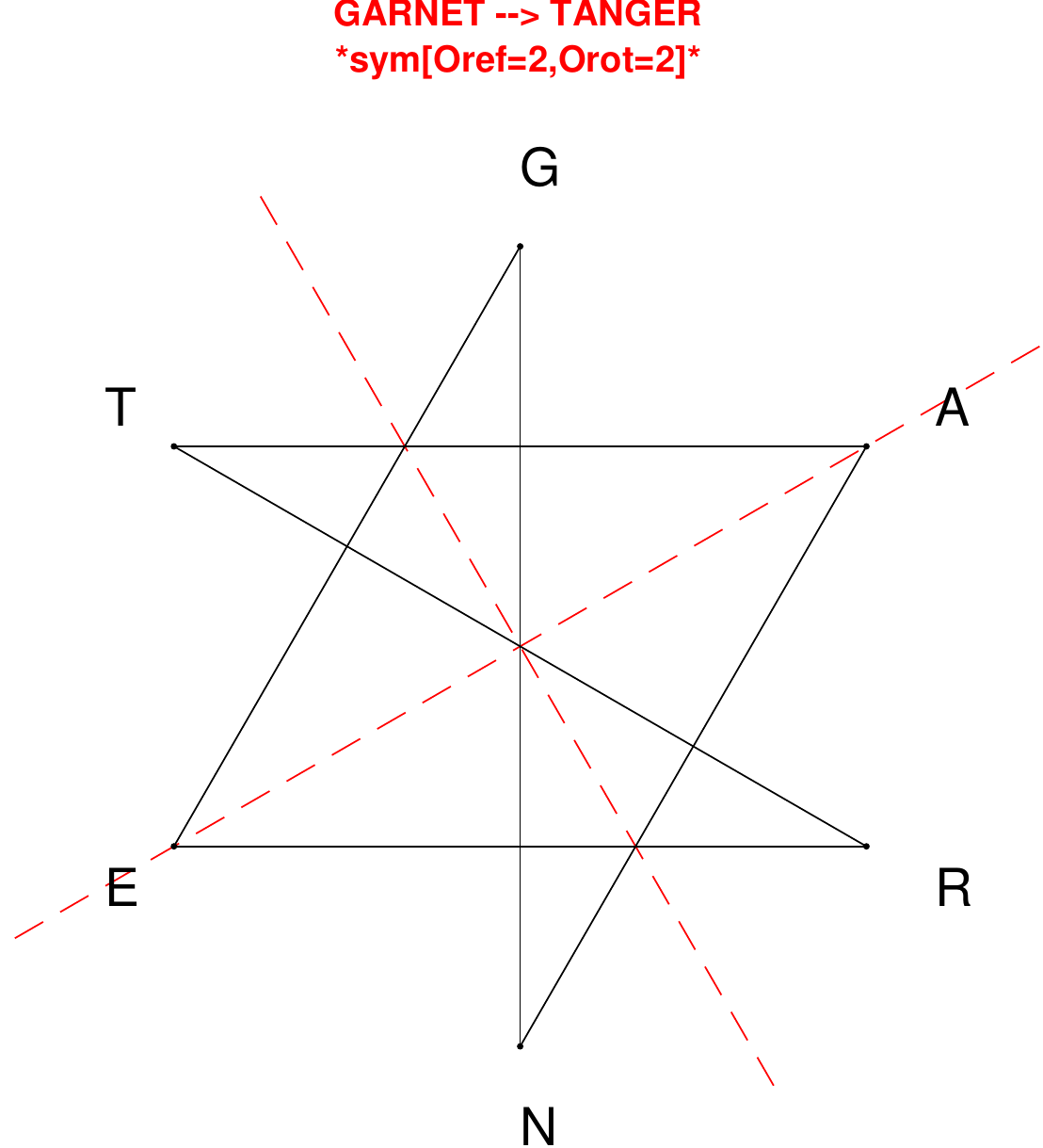}
\end{subfigure}
\hfill
\begin{subfigure}[T]{0.19\textwidth}
\centering
\includegraphics[width=\textwidth]{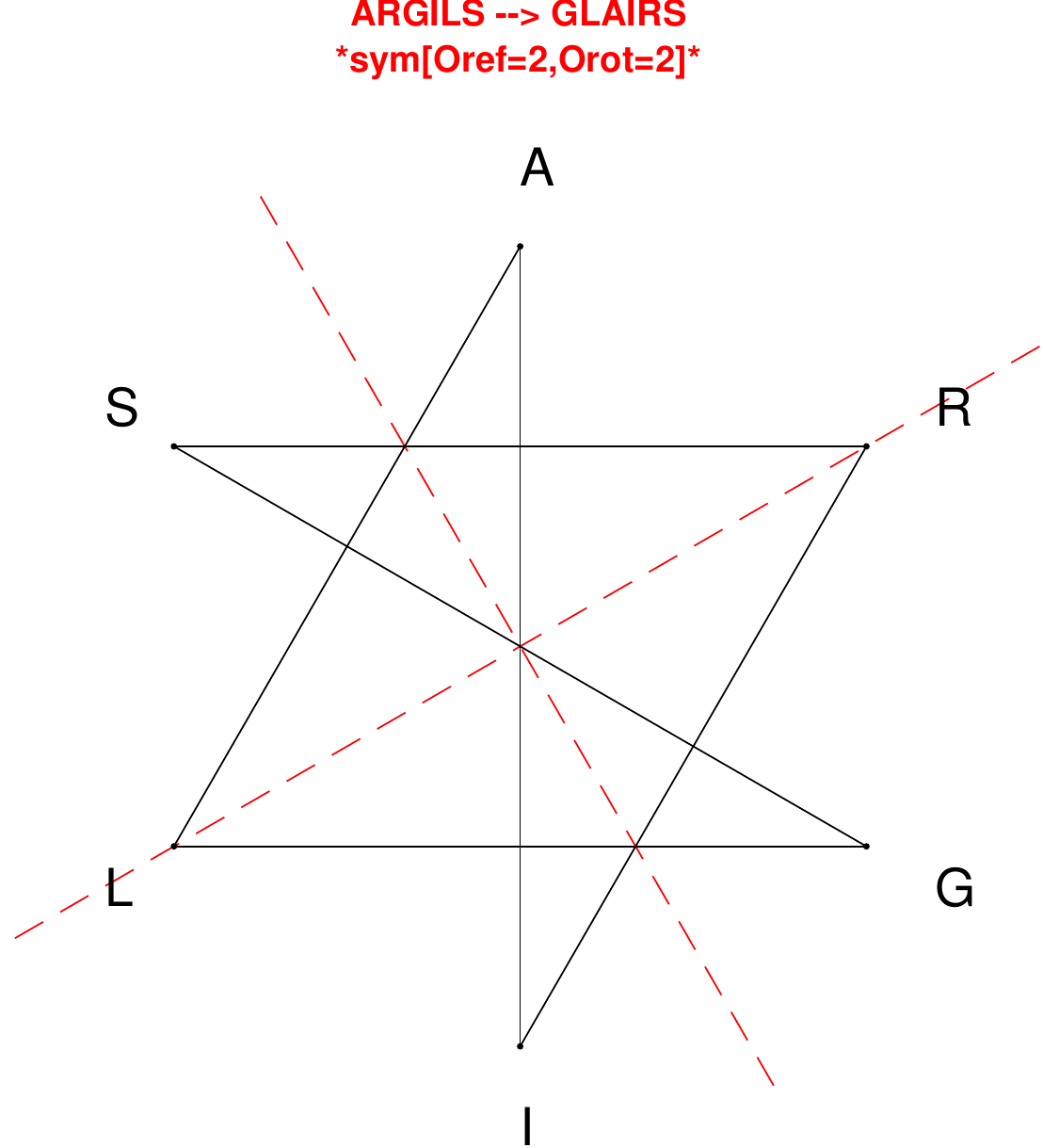}
\end{subfigure}
\hfill
\begin{subfigure}[T]{0.19\textwidth}
\centering
\includegraphics[width=\textwidth]{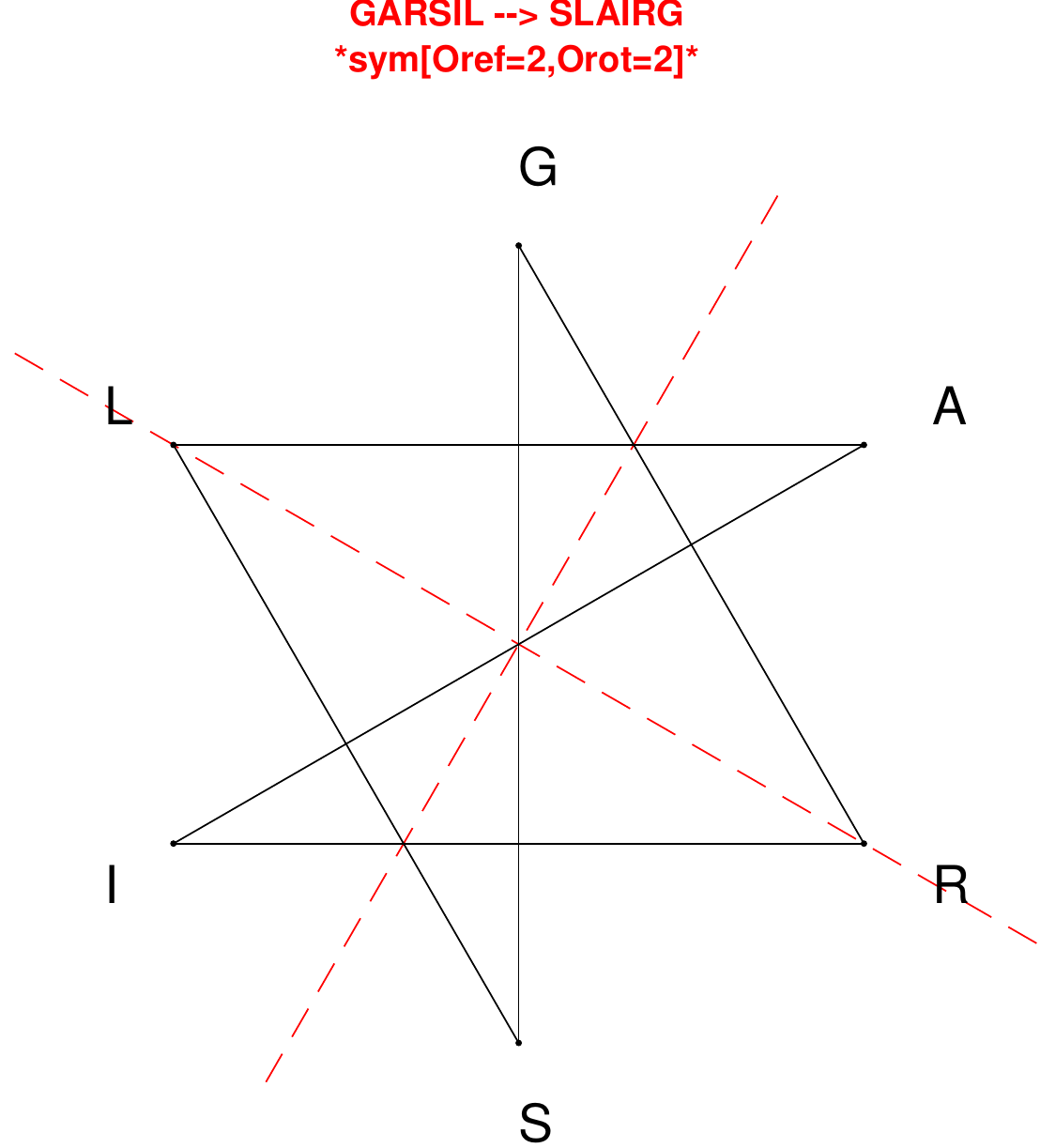}
\end{subfigure}
\end{figure}

\begin{figure}[H]
\centering
\begin{subfigure}[T]{0.19\textwidth}
\centering
\includegraphics[width=\textwidth]{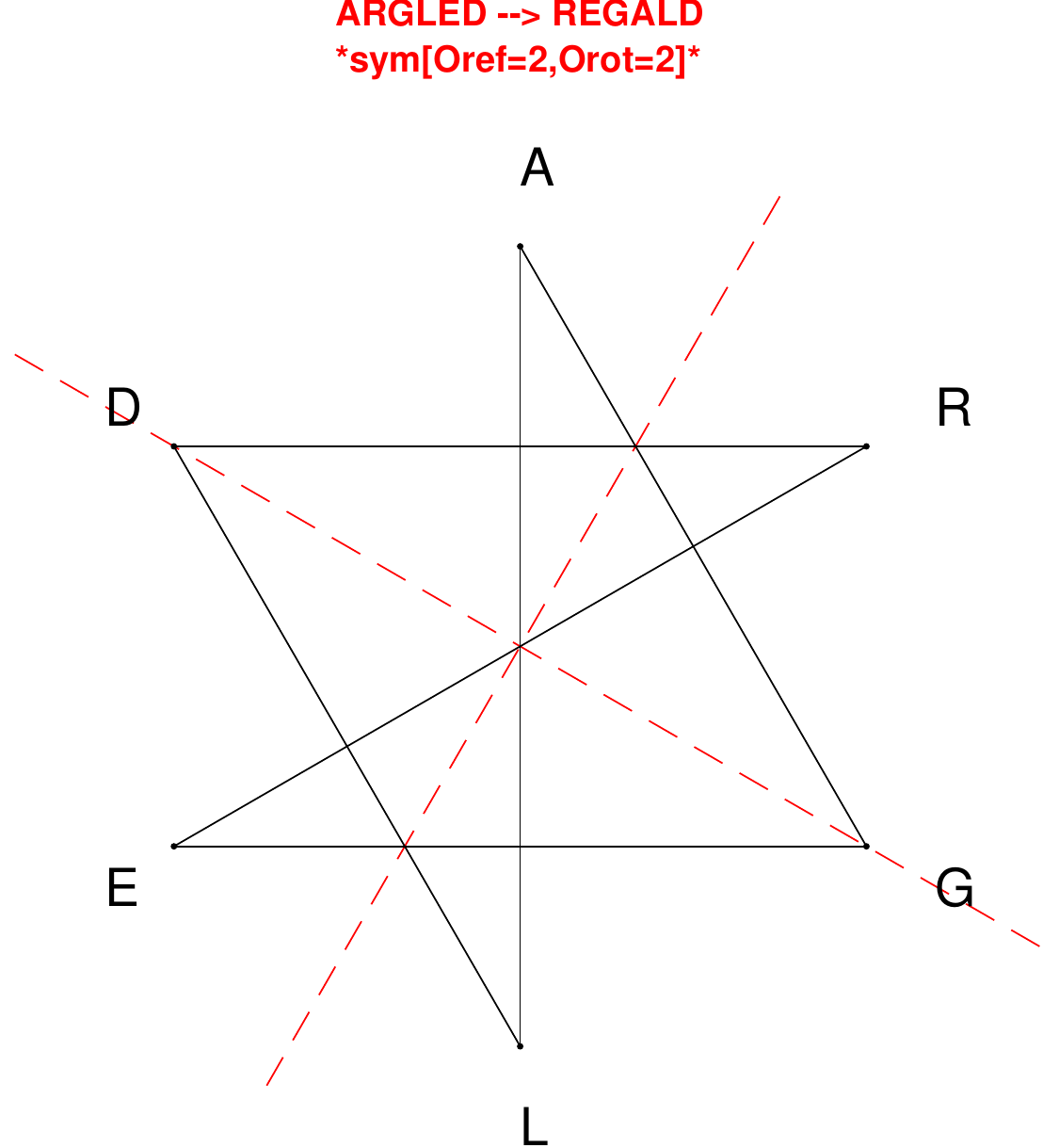}
\end{subfigure}
\hfill
\begin{subfigure}[T]{0.19\textwidth}
\centering
\includegraphics[width=\textwidth]{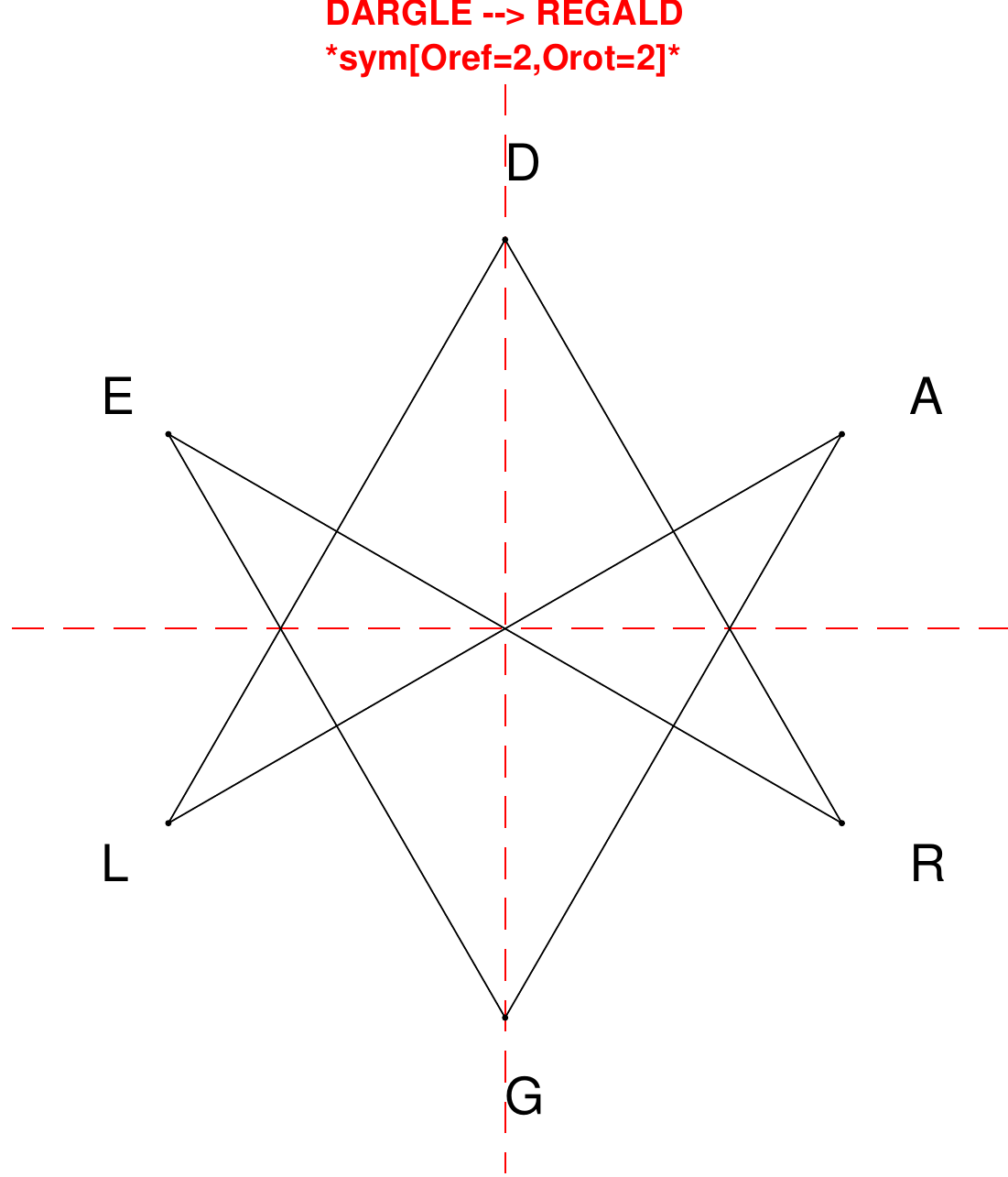}
\end{subfigure}
\hfill
\begin{subfigure}[T]{0.19\textwidth}
\centering
\includegraphics[width=\textwidth]{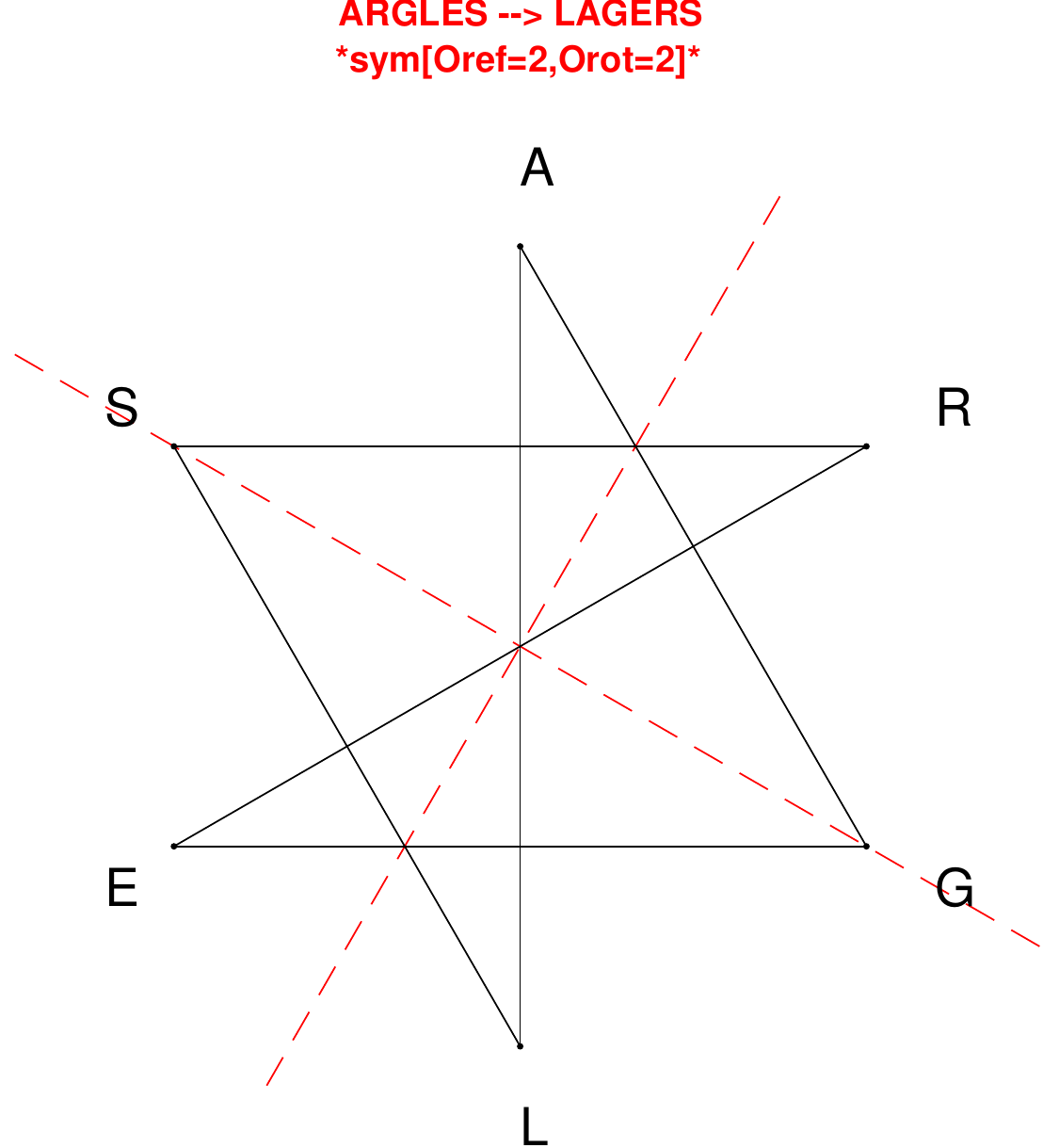}
\end{subfigure}
\hfill
\begin{subfigure}[T]{0.19\textwidth}
\centering
\includegraphics[width=\textwidth]{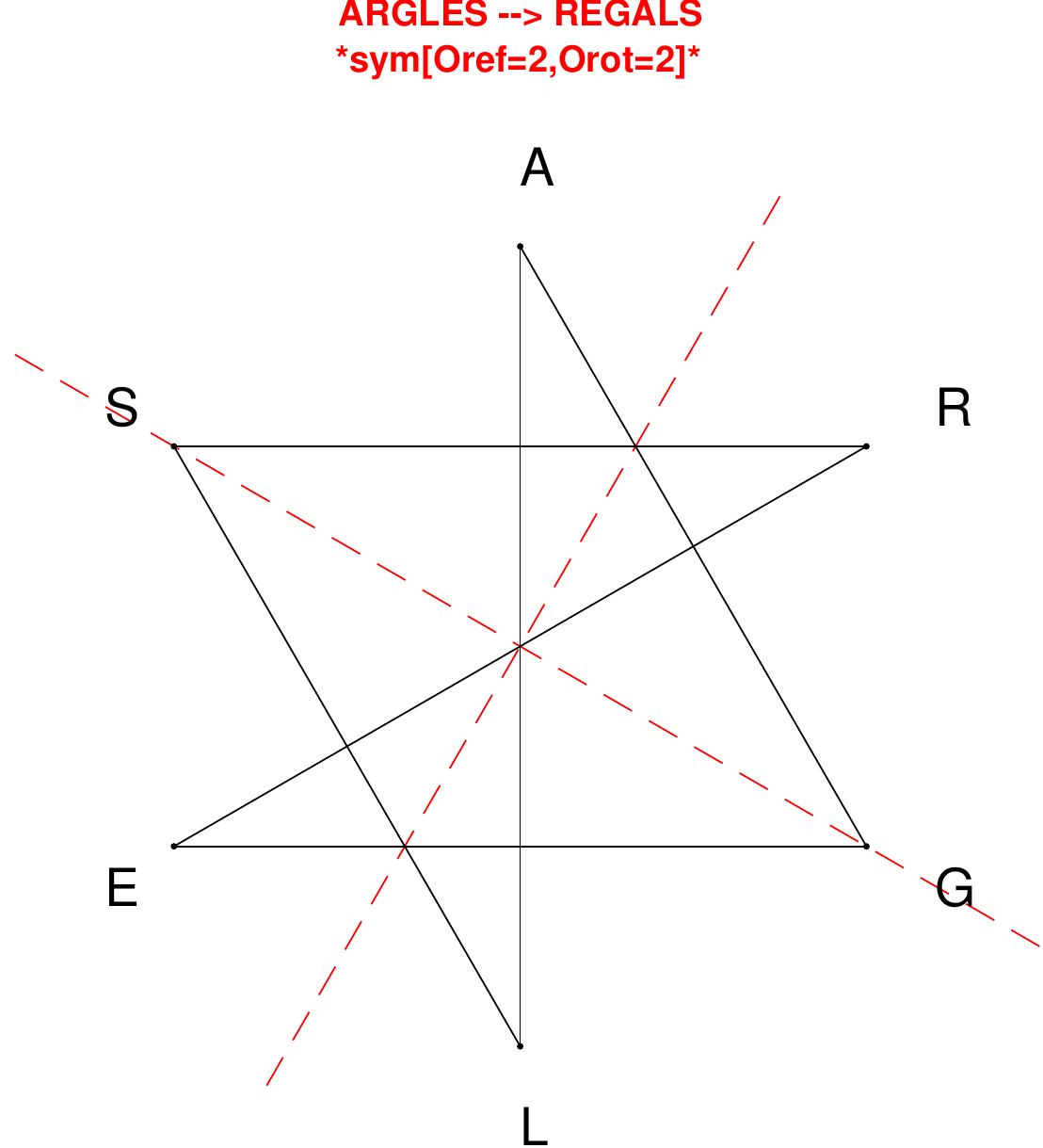}
\end{subfigure}
\hfill
\begin{subfigure}[T]{0.19\textwidth}
\centering
\includegraphics[width=\textwidth]{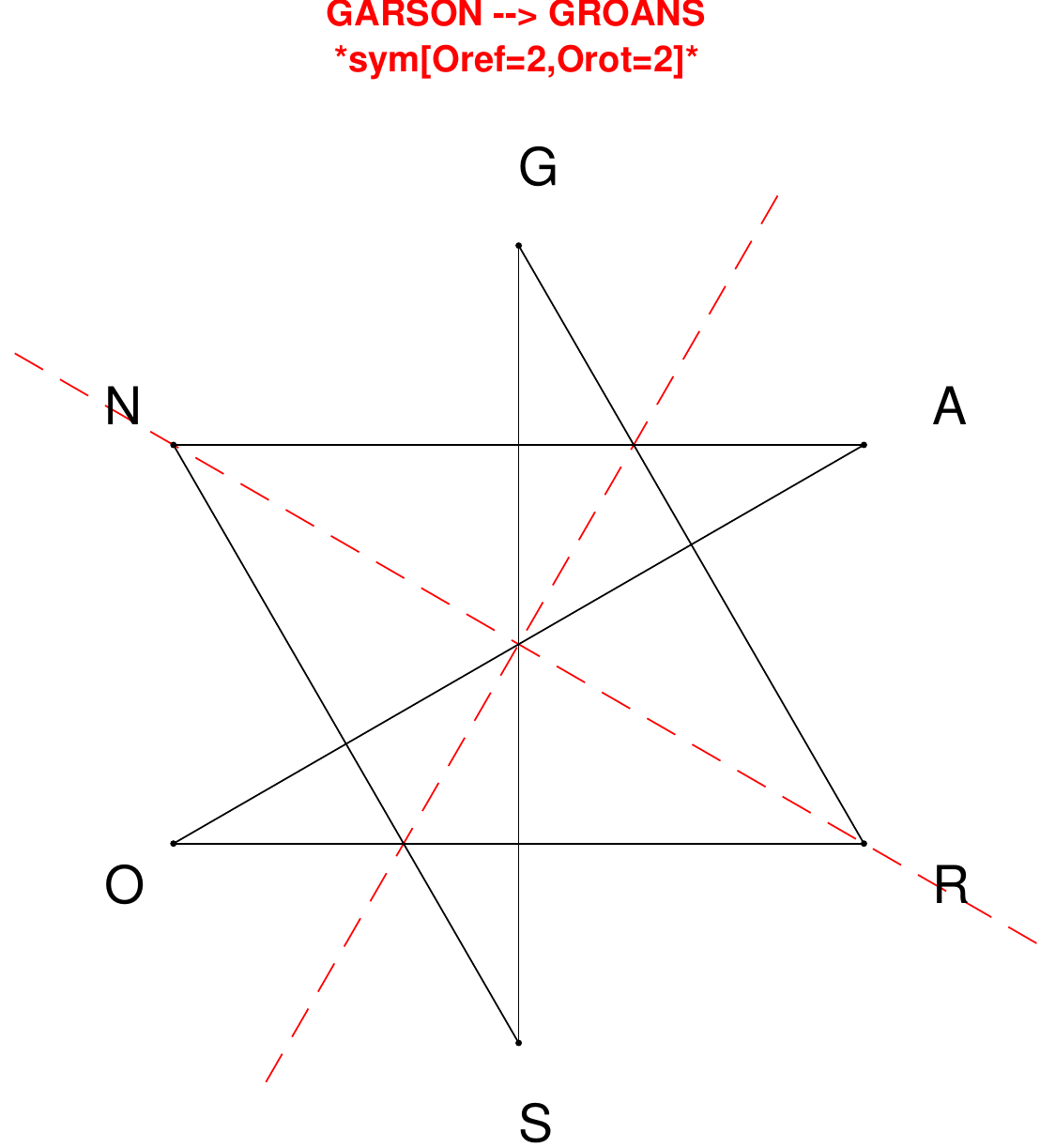}
\end{subfigure}
\end{figure}

\begin{figure}[H]
\centering
\begin{subfigure}[T]{0.19\textwidth}
\centering
\includegraphics[width=\textwidth]{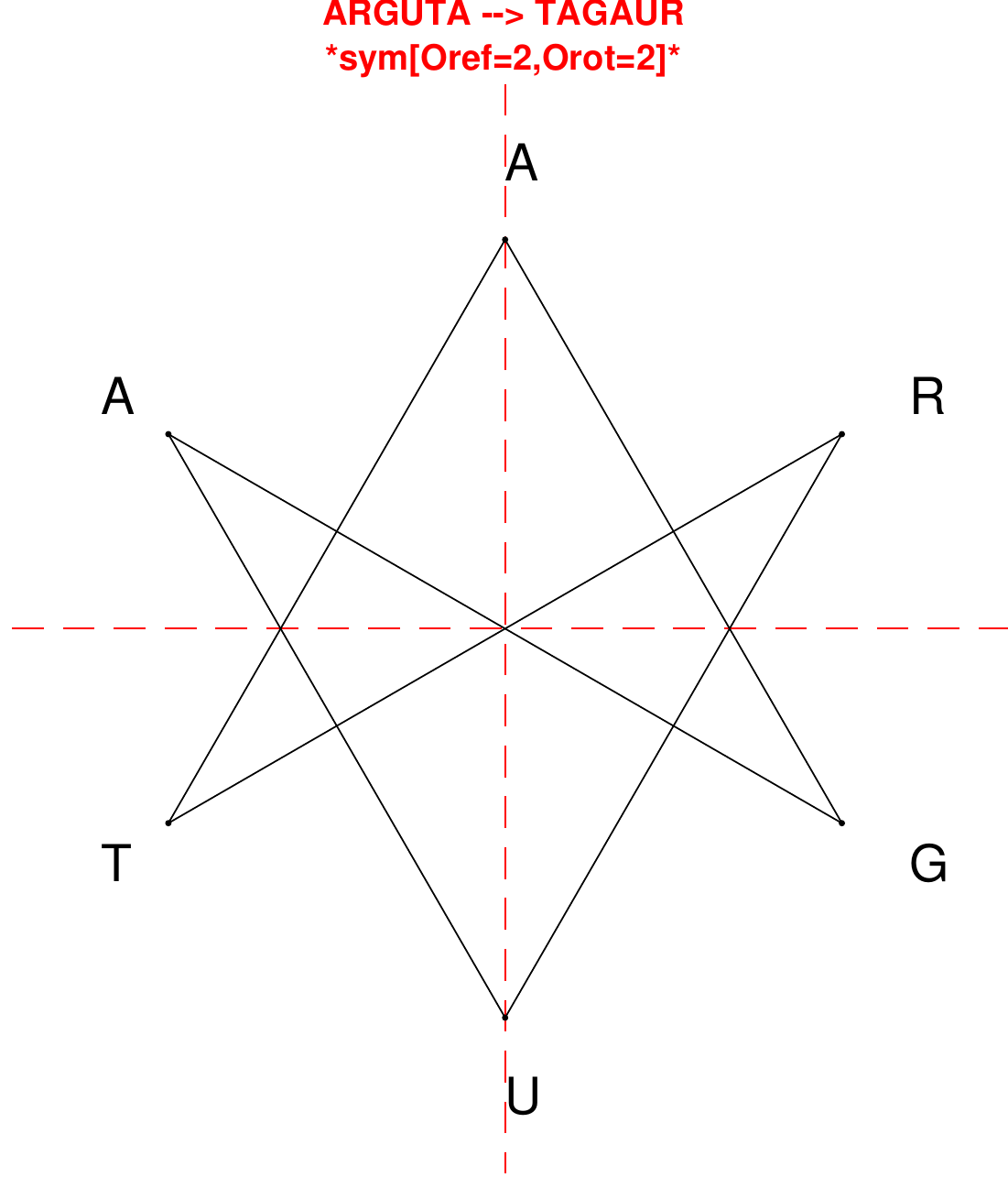}
\end{subfigure}
\hfill
\begin{subfigure}[T]{0.19\textwidth}
\centering
\includegraphics[width=\textwidth]{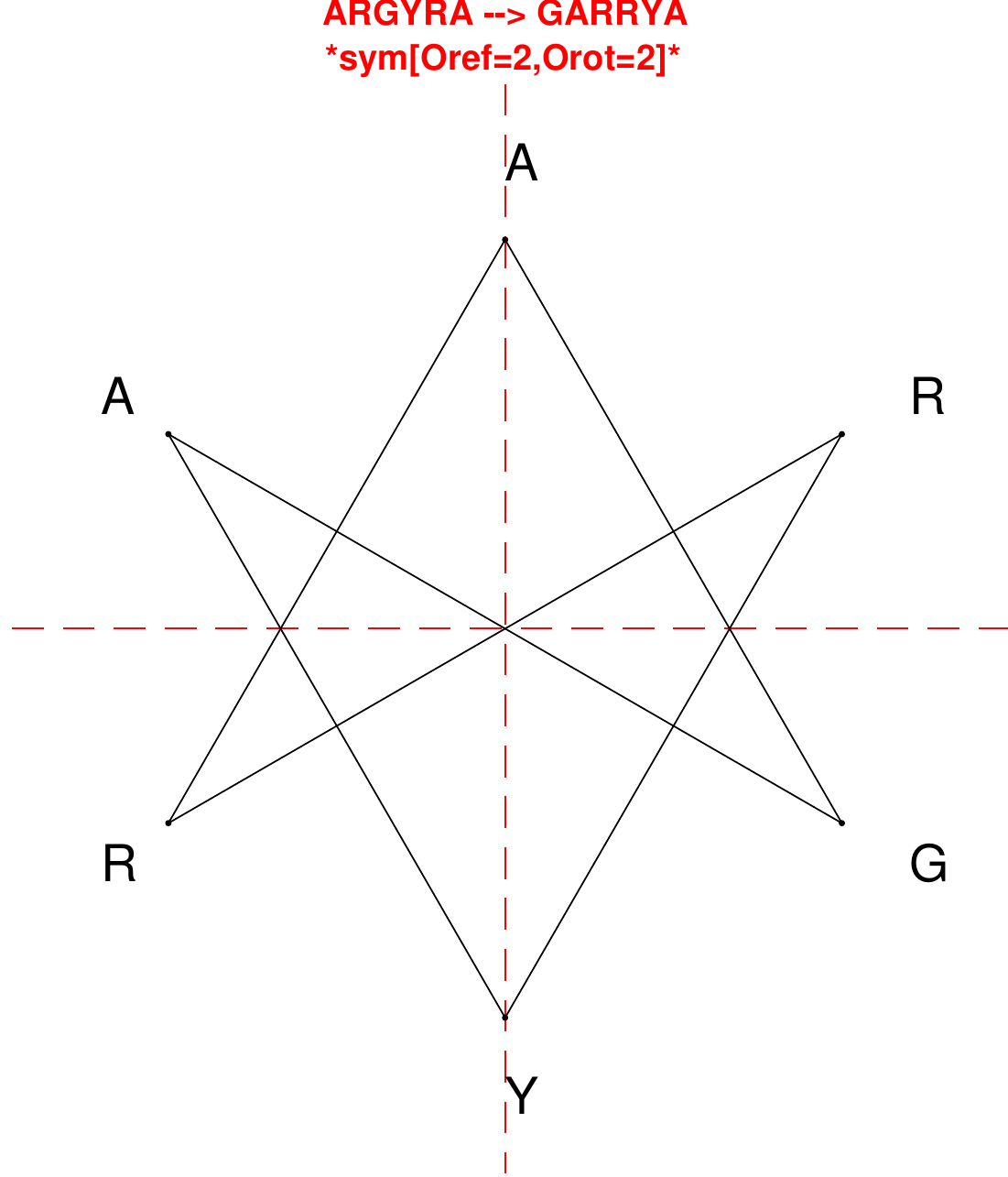}
\end{subfigure}
\hfill
\begin{subfigure}[T]{0.19\textwidth}
\centering
\includegraphics[width=\textwidth]{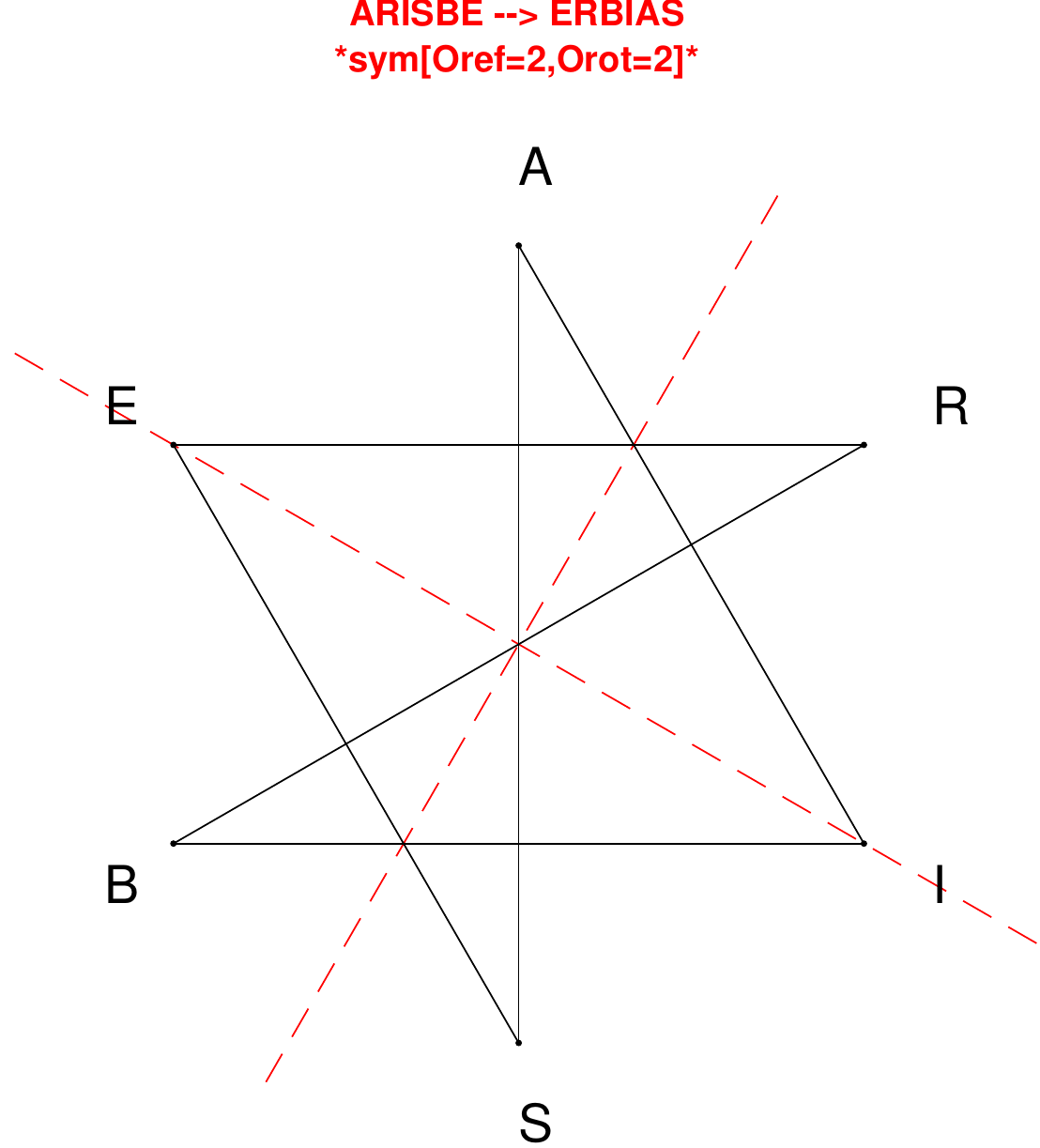}
\end{subfigure}
\hfill
\begin{subfigure}[T]{0.19\textwidth}
\centering
\includegraphics[width=\textwidth]{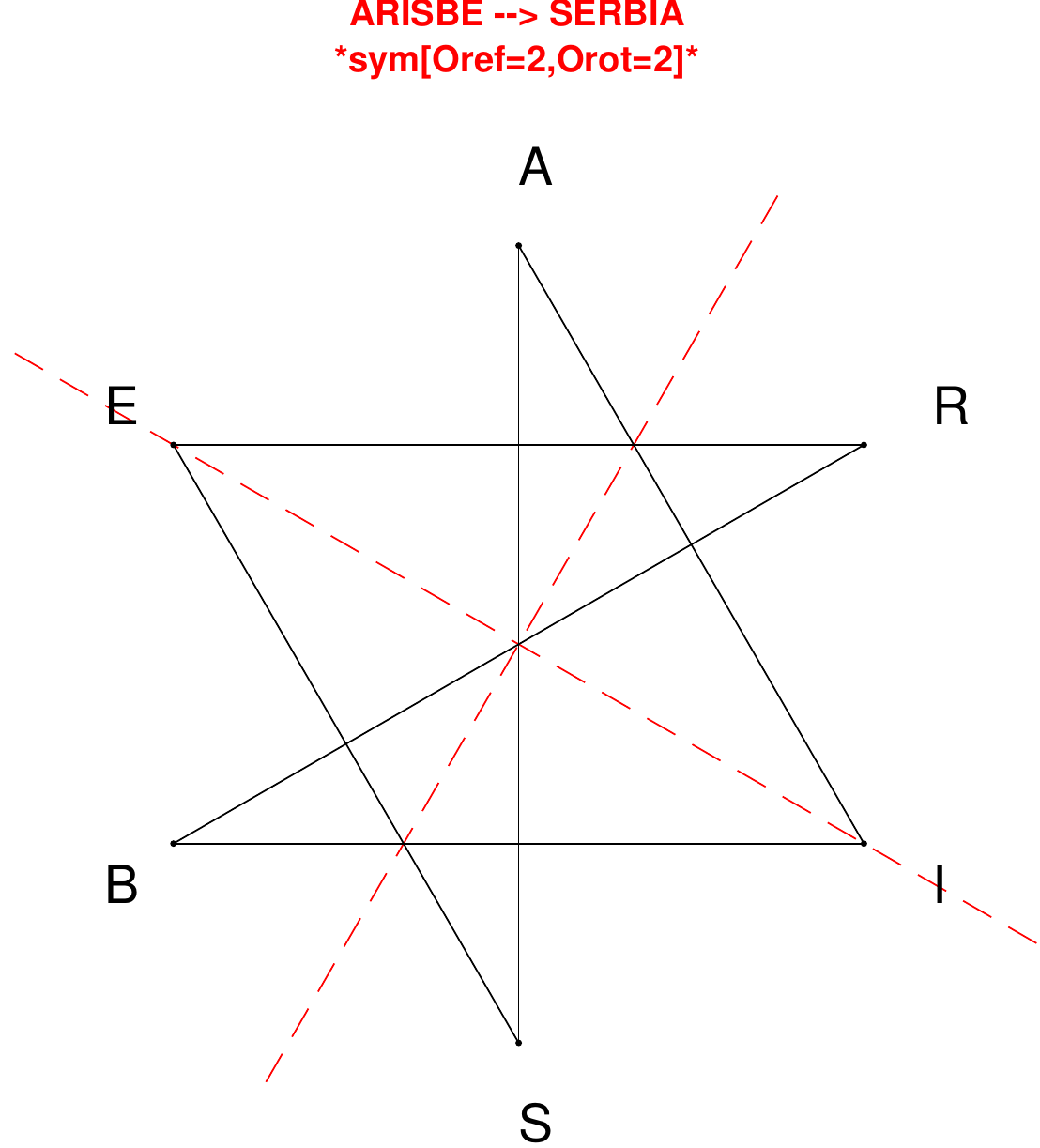}
\end{subfigure}
\hfill
\begin{subfigure}[T]{0.19\textwidth}
\centering
\includegraphics[width=\textwidth]{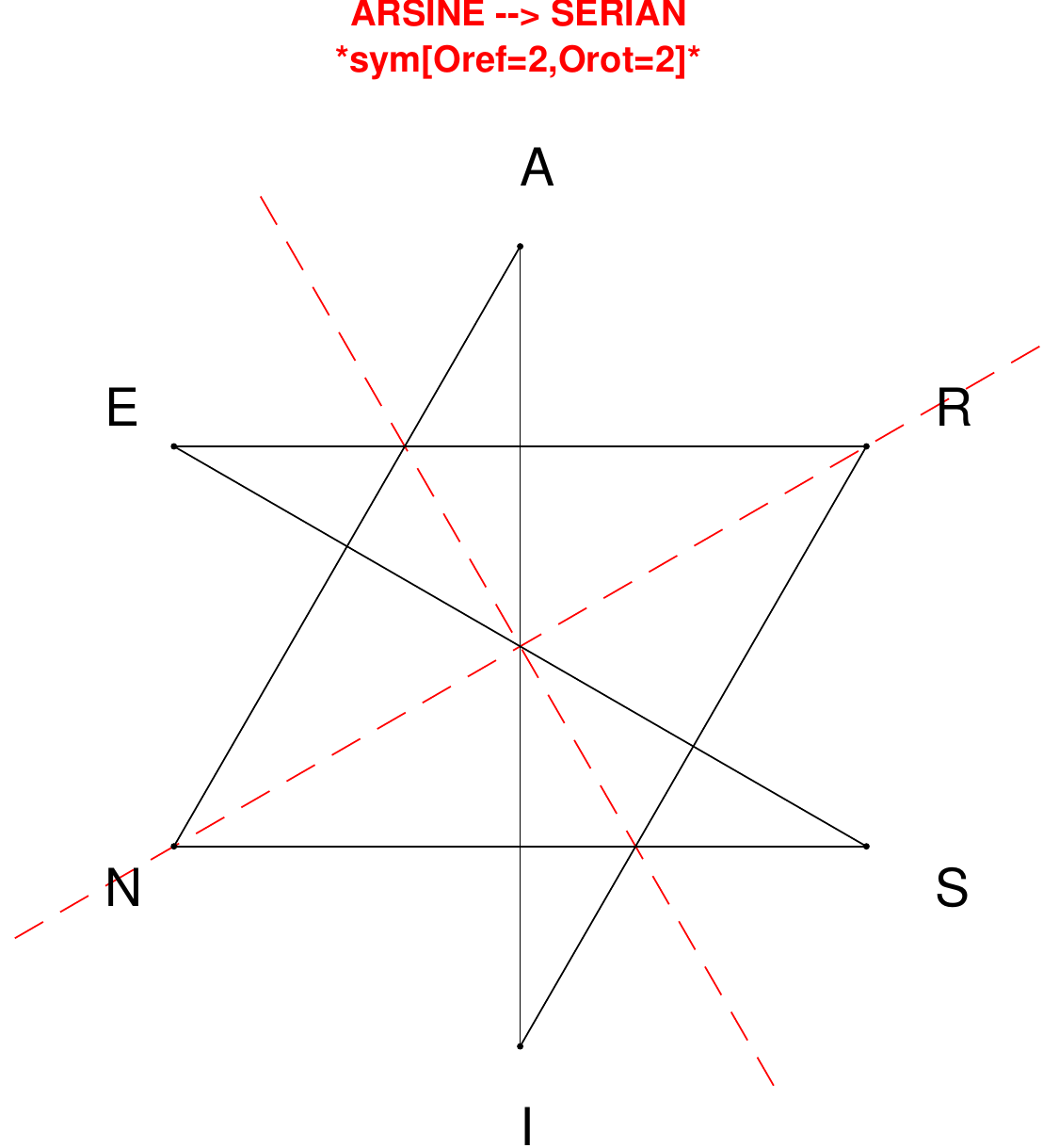}
\end{subfigure}
\end{figure}

\begin{figure}[H]
\centering
\begin{subfigure}[T]{0.19\textwidth}
\centering
\includegraphics[width=\textwidth]{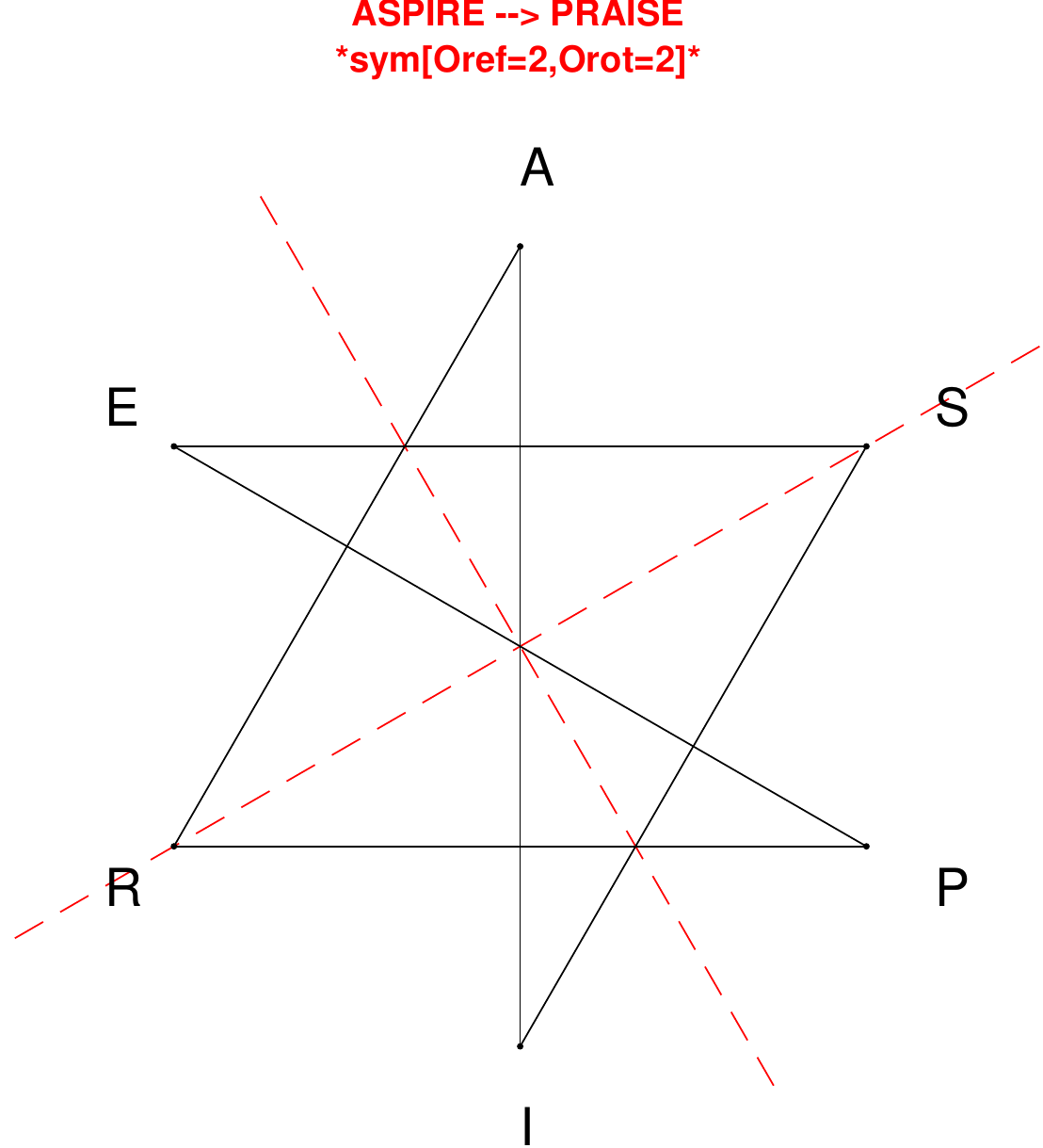}
\end{subfigure}
\hfill
\begin{subfigure}[T]{0.19\textwidth}
\centering
\includegraphics[width=\textwidth]{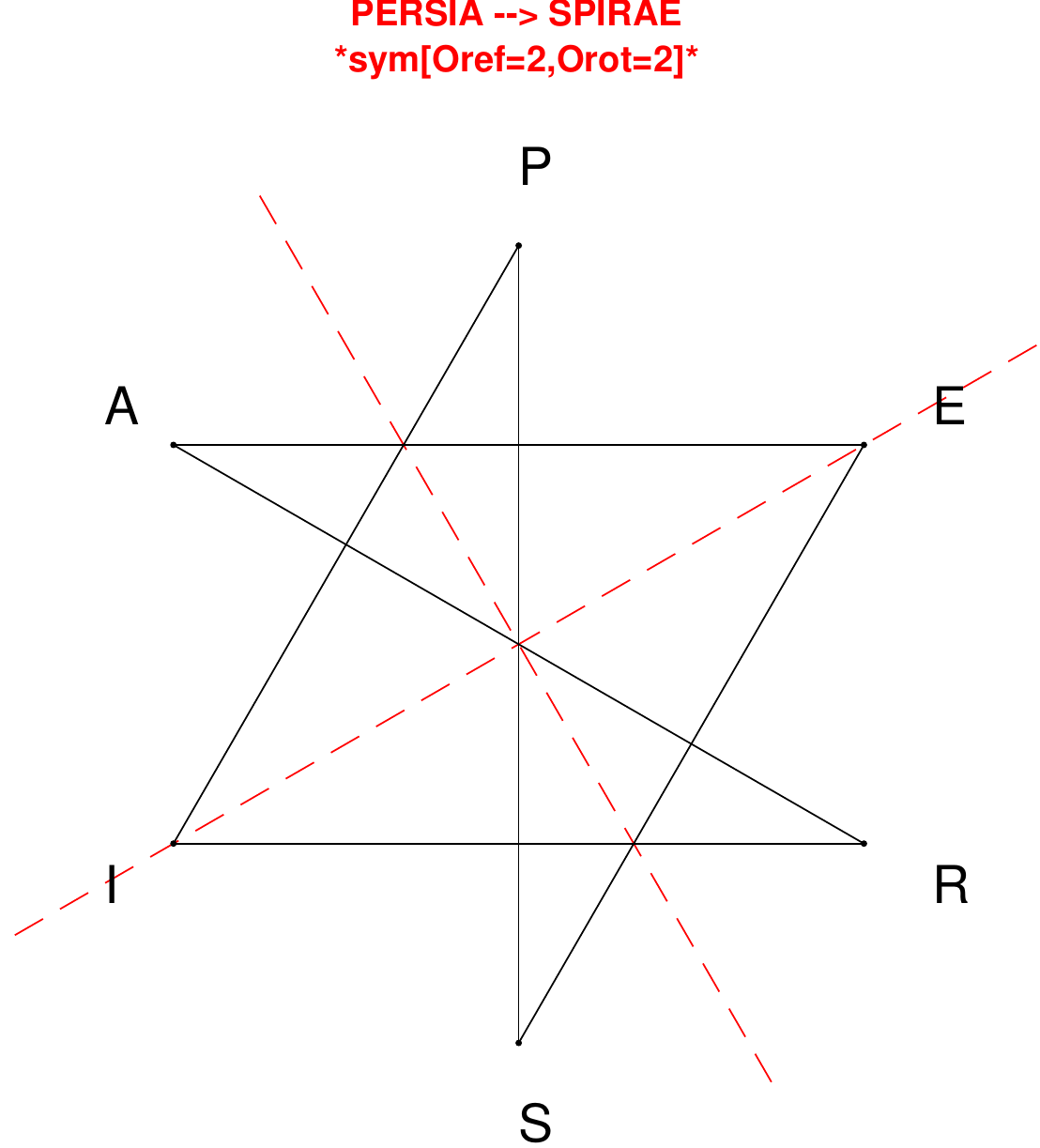}
\end{subfigure}
\hfill
\begin{subfigure}[T]{0.19\textwidth}
\centering
\includegraphics[width=\textwidth]{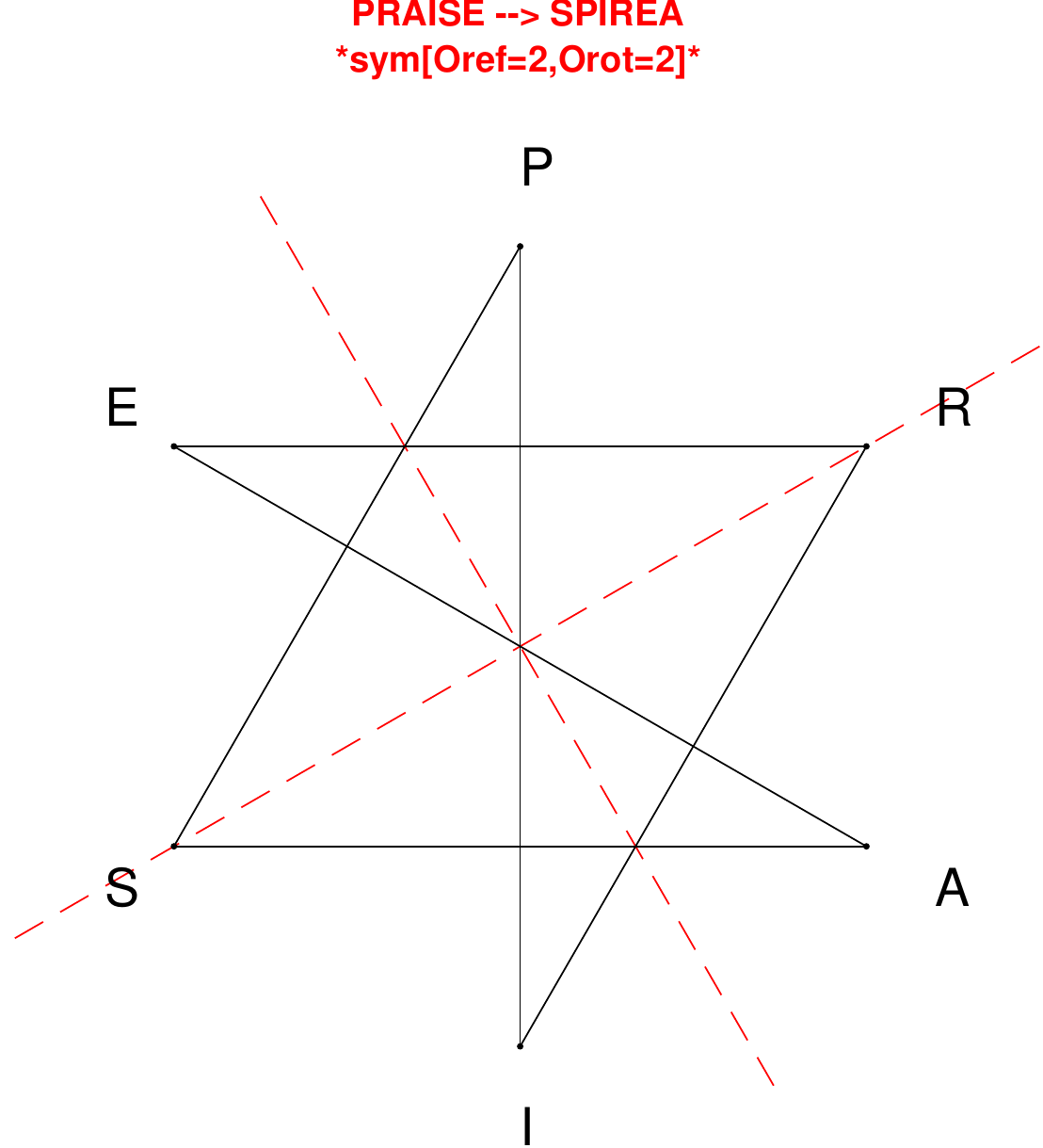}
\end{subfigure}
\hfill
\begin{subfigure}[T]{0.19\textwidth}
\centering
\includegraphics[width=\textwidth]{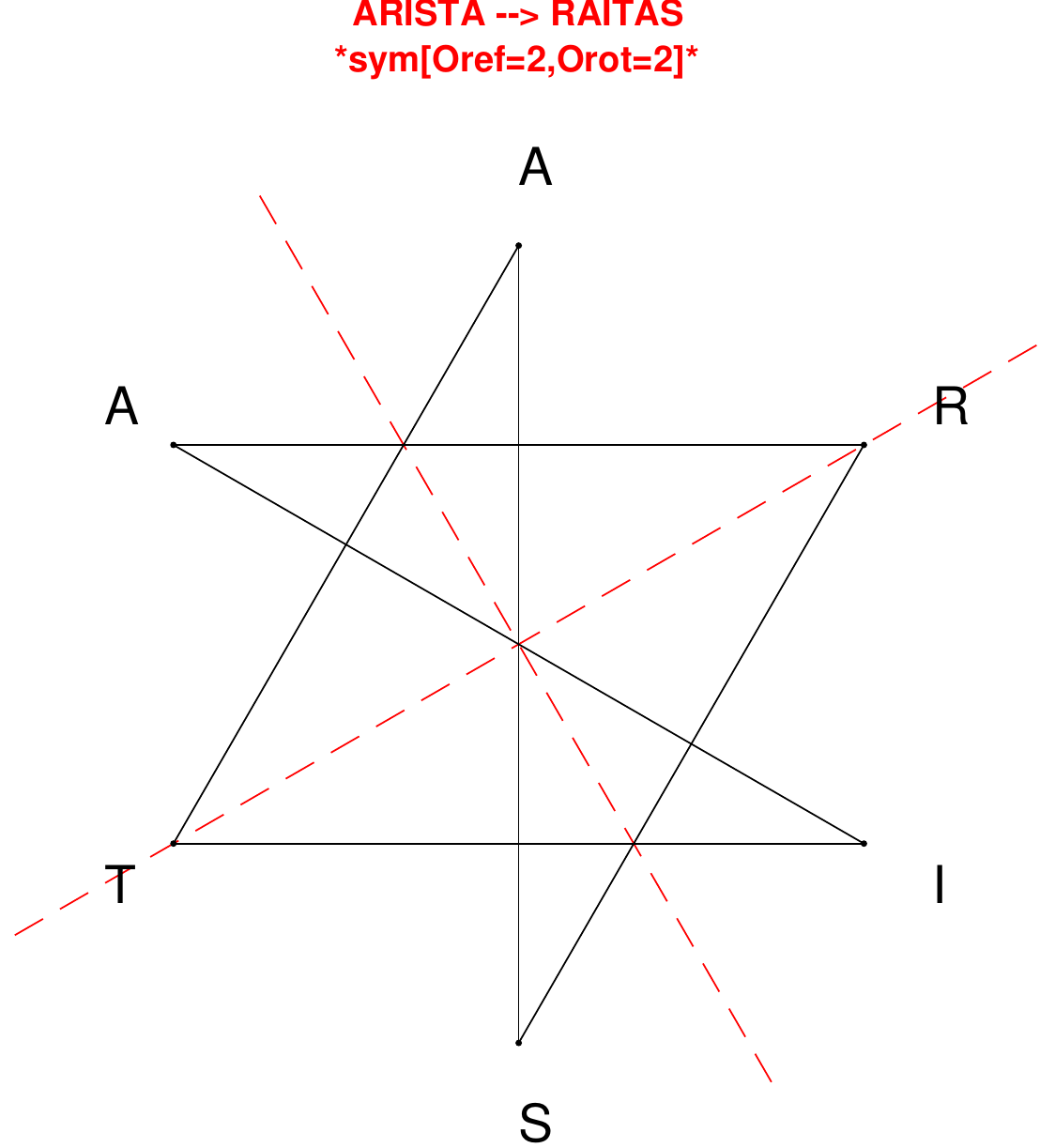}
\end{subfigure}
\hfill
\begin{subfigure}[T]{0.19\textwidth}
\centering
\includegraphics[width=\textwidth]{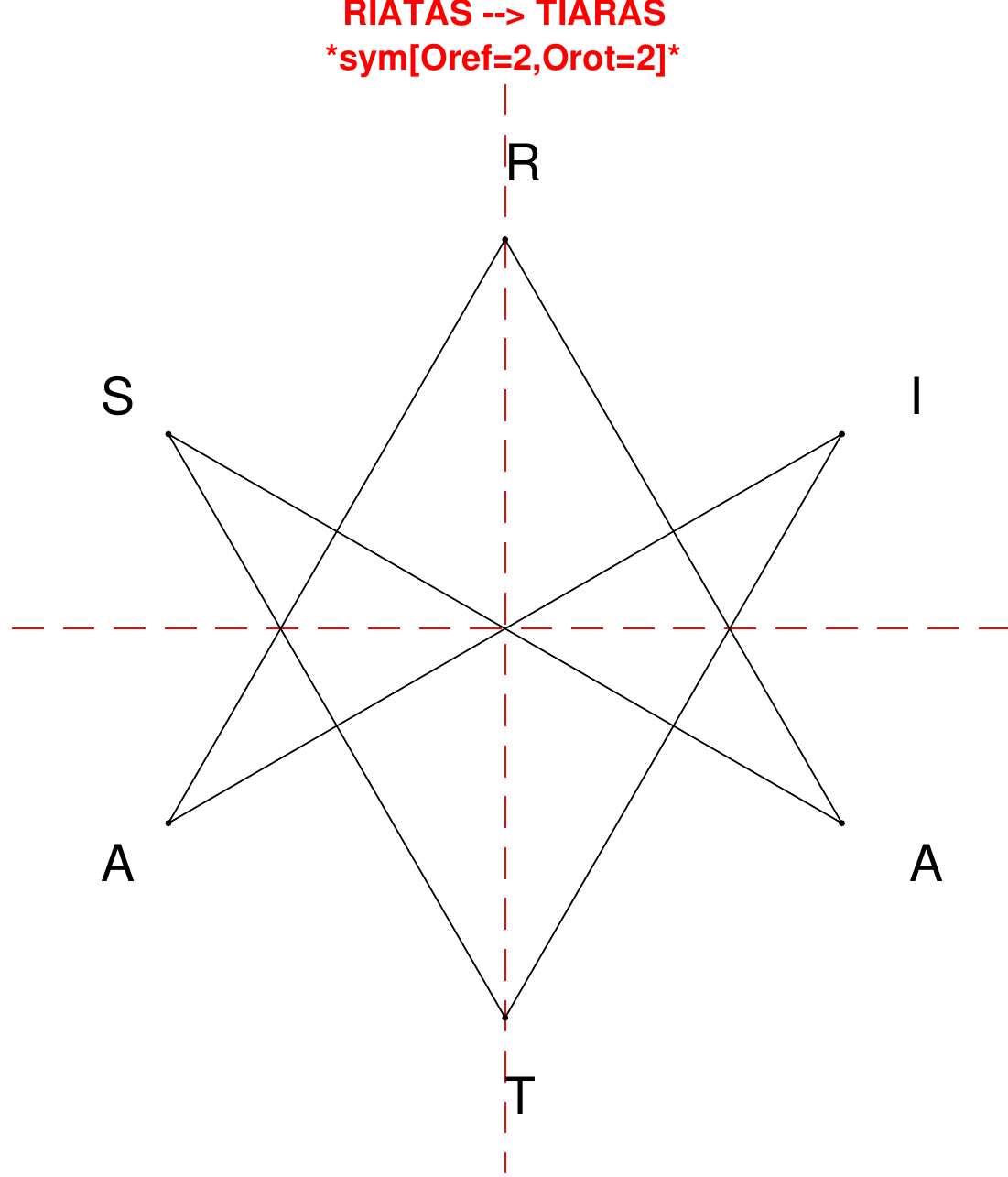}
\end{subfigure}
\end{figure}

\begin{figure}[H]
\centering
\begin{subfigure}[T]{0.19\textwidth}
\centering
\includegraphics[width=\textwidth]{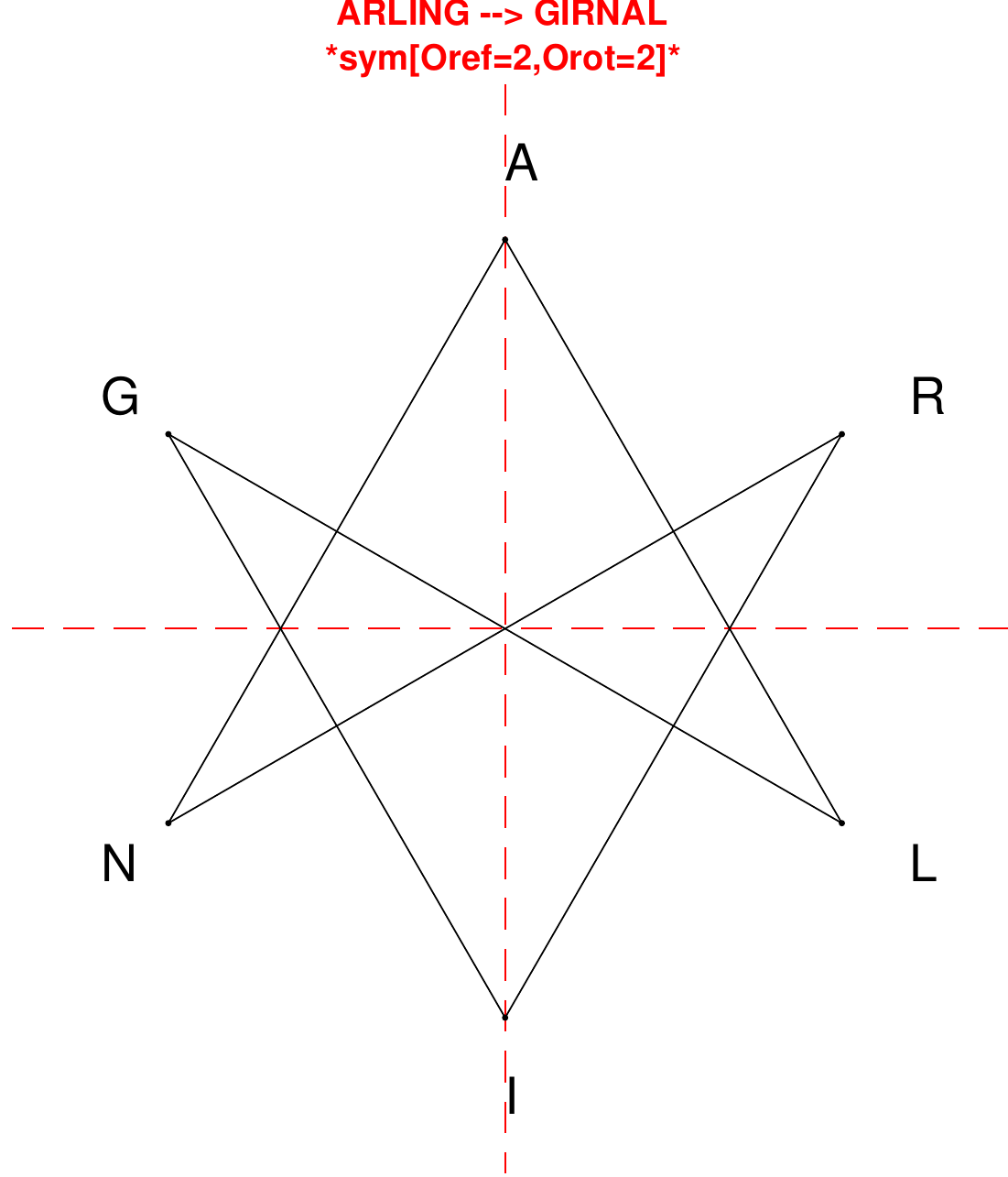}
\end{subfigure}
\hfill
\begin{subfigure}[T]{0.19\textwidth}
\centering
\includegraphics[width=\textwidth]{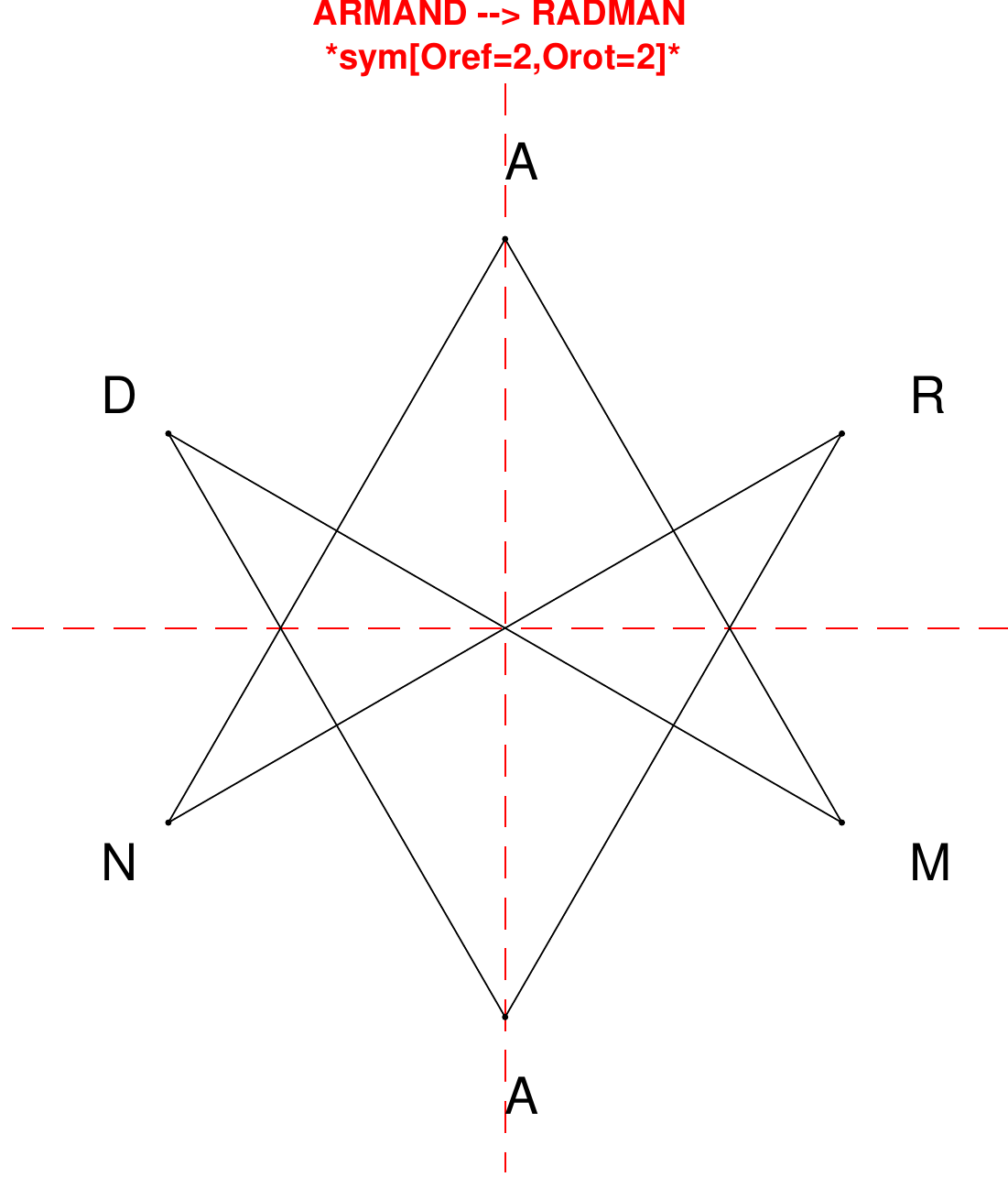}
\end{subfigure}
\hfill
\begin{subfigure}[T]{0.19\textwidth}
\centering
\includegraphics[width=\textwidth]{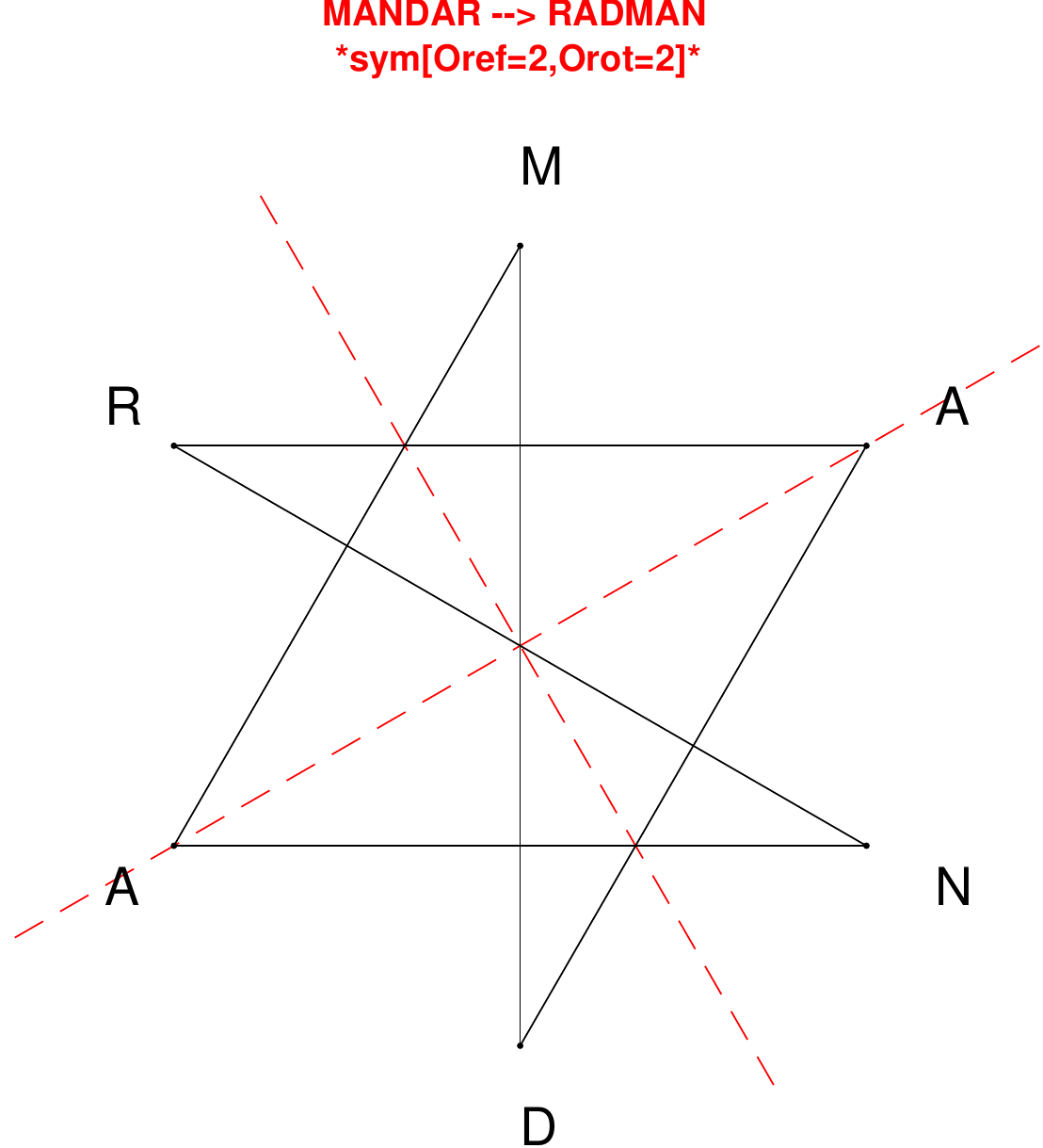}
\end{subfigure}
\hfill
\begin{subfigure}[T]{0.19\textwidth}
\centering
\includegraphics[width=\textwidth]{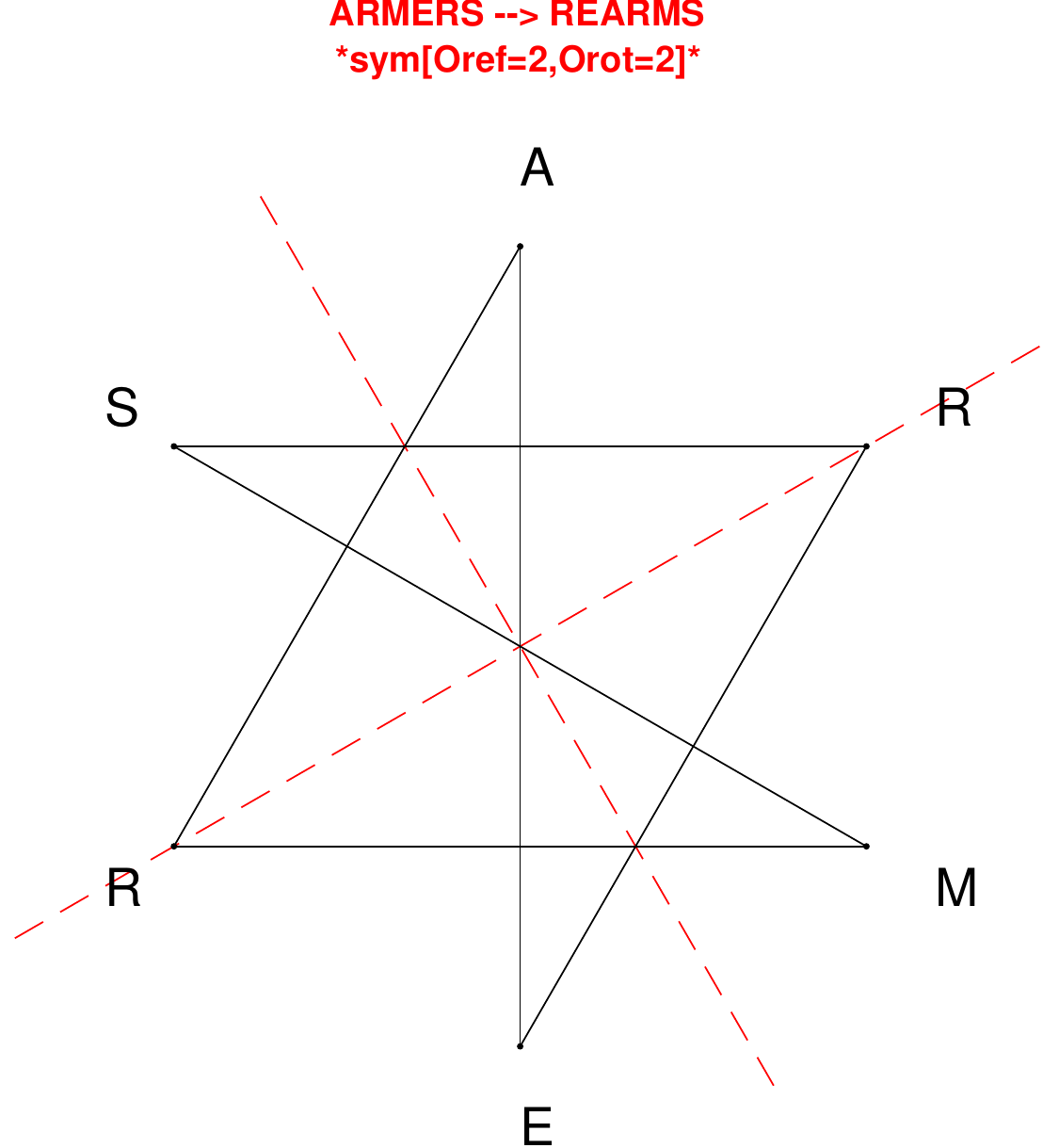}
\end{subfigure}
\hfill
\begin{subfigure}[T]{0.19\textwidth}
\centering
\includegraphics[width=\textwidth]{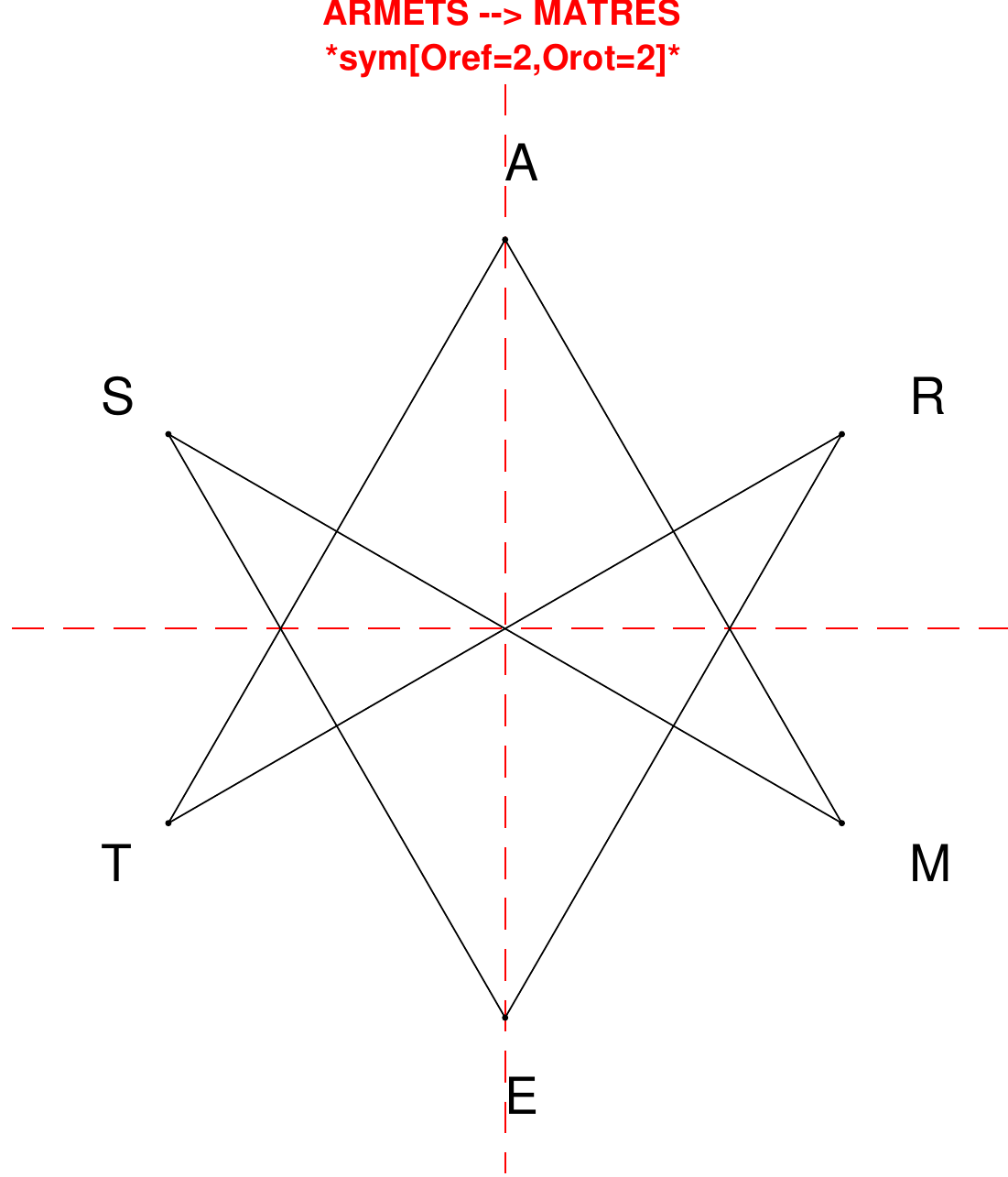}
\end{subfigure}
\end{figure}

\begin{figure}[H]
\centering
\begin{subfigure}[T]{0.19\textwidth}
\centering
\includegraphics[width=\textwidth]{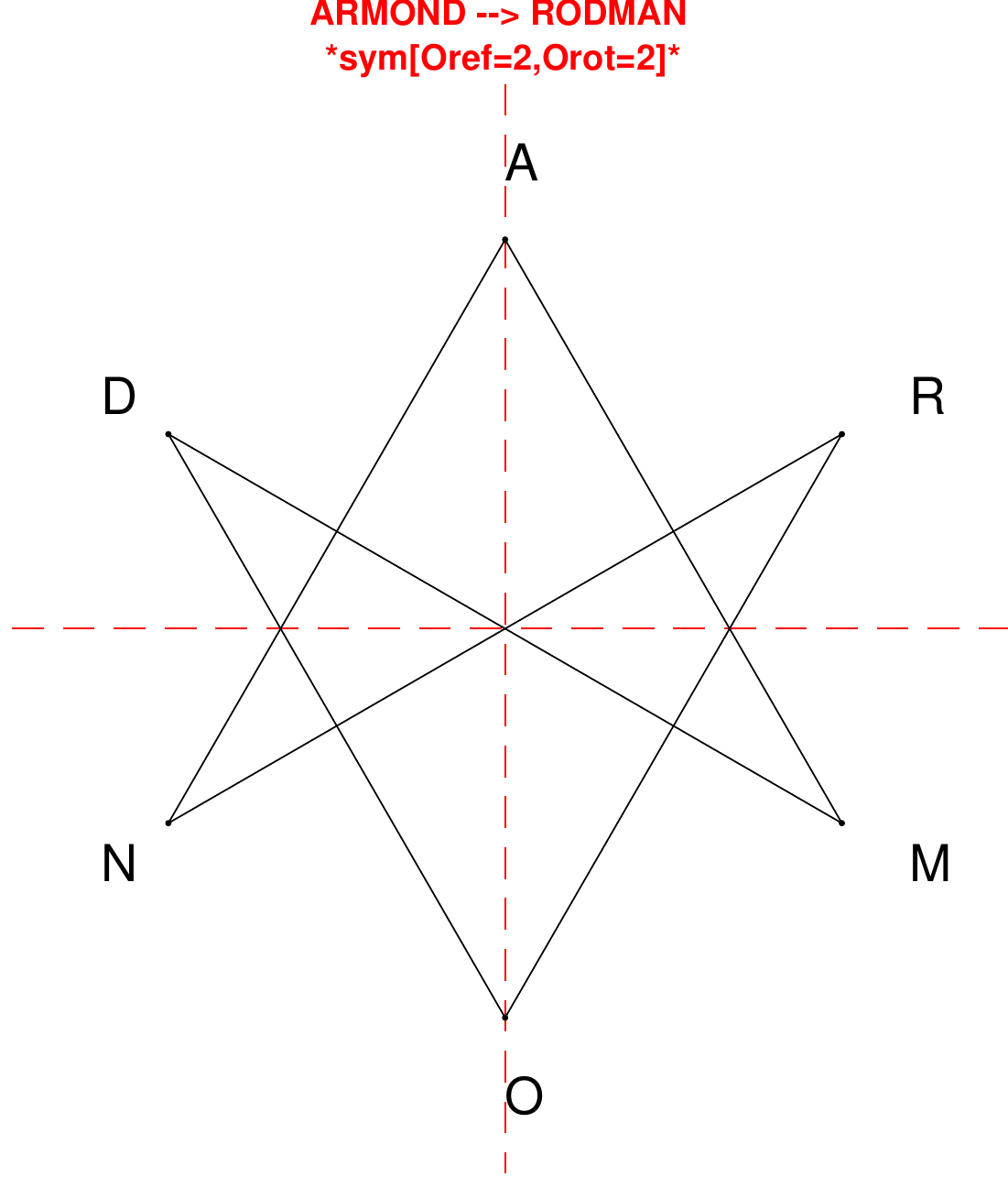}
\end{subfigure}
\hfill
\begin{subfigure}[T]{0.19\textwidth}
\centering
\includegraphics[width=\textwidth]{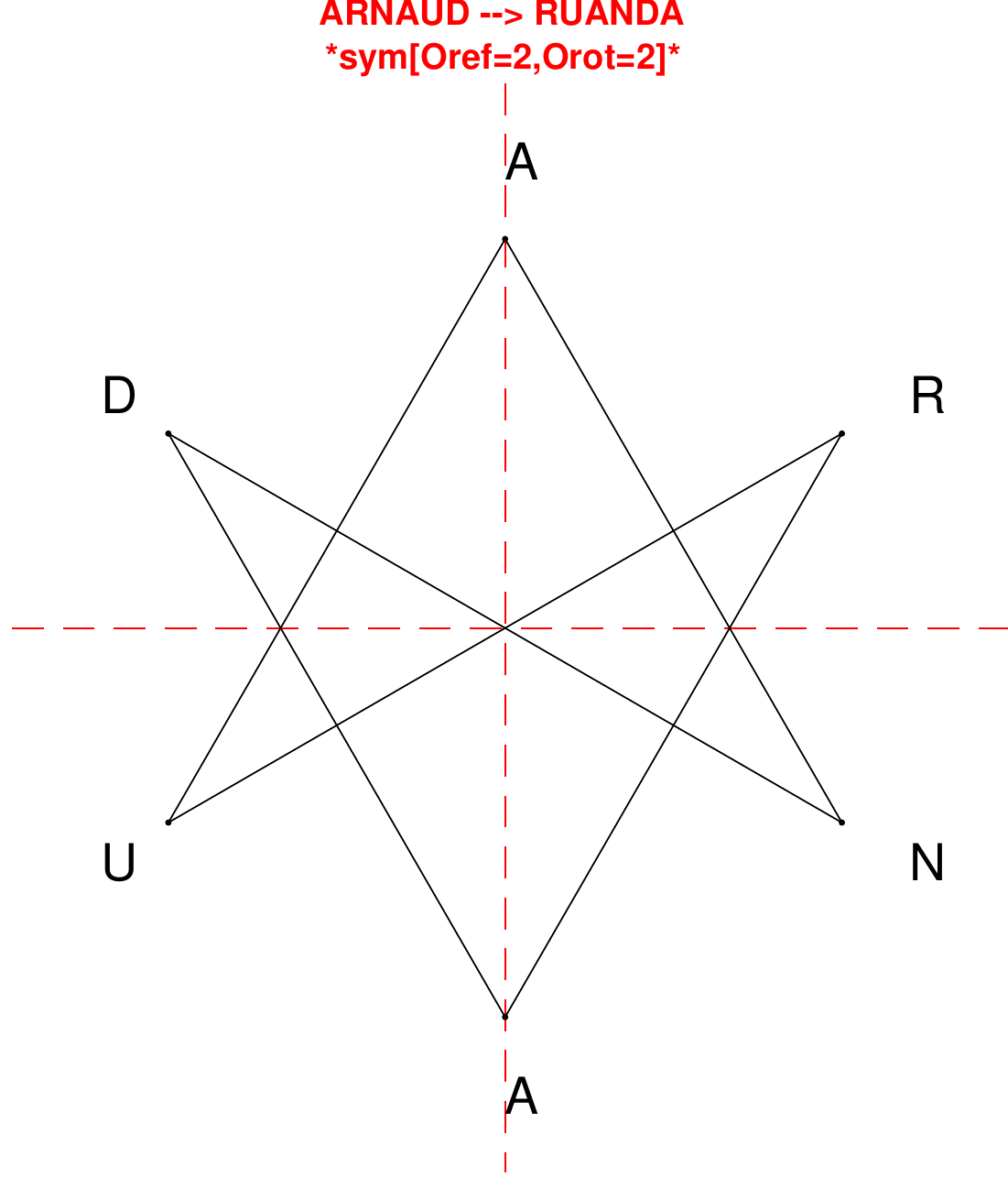}
\end{subfigure}
\hfill
\begin{subfigure}[T]{0.19\textwidth}
\centering
\includegraphics[width=\textwidth]{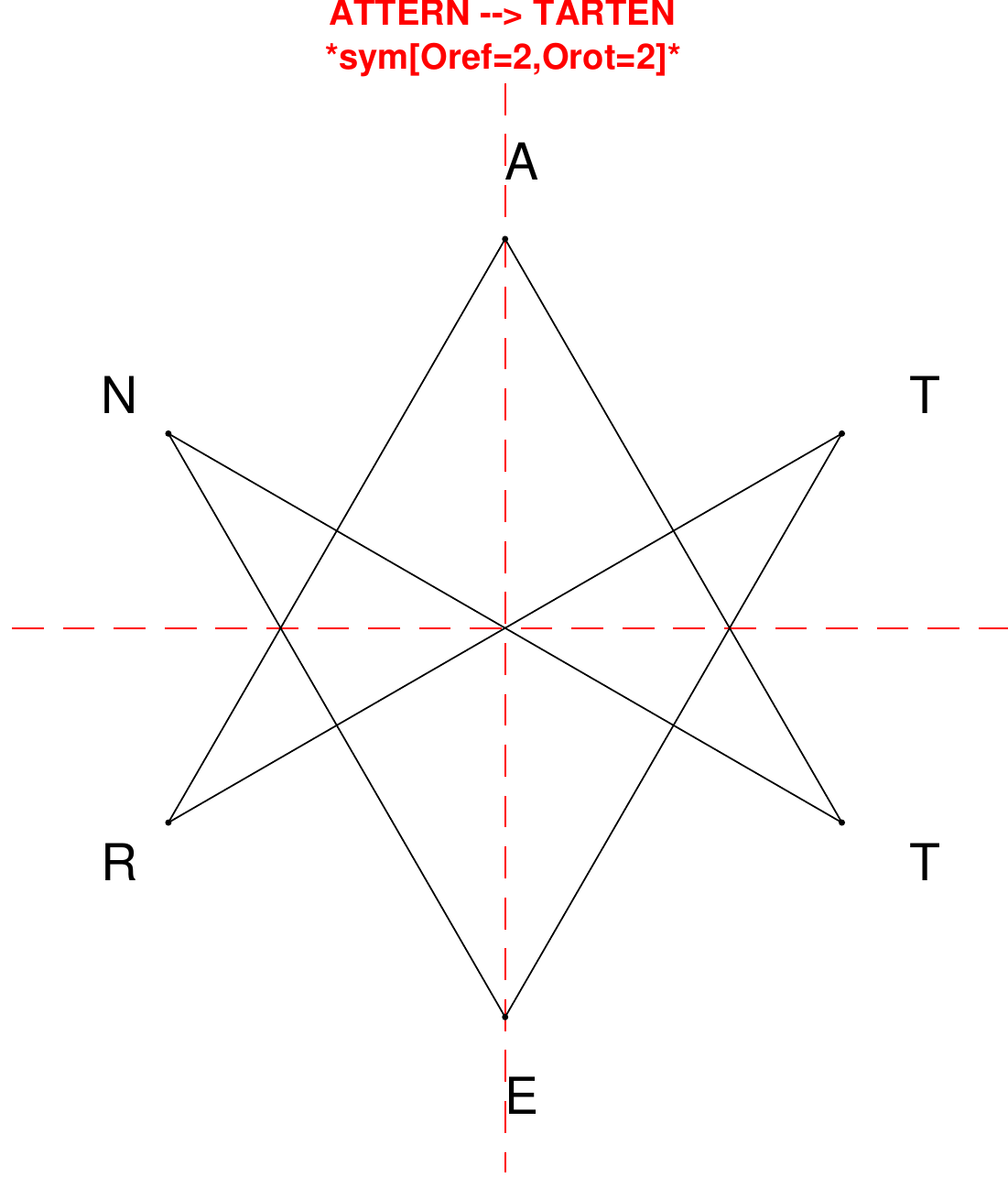}
\end{subfigure}
\hfill
\begin{subfigure}[T]{0.19\textwidth}
\centering
\includegraphics[width=\textwidth]{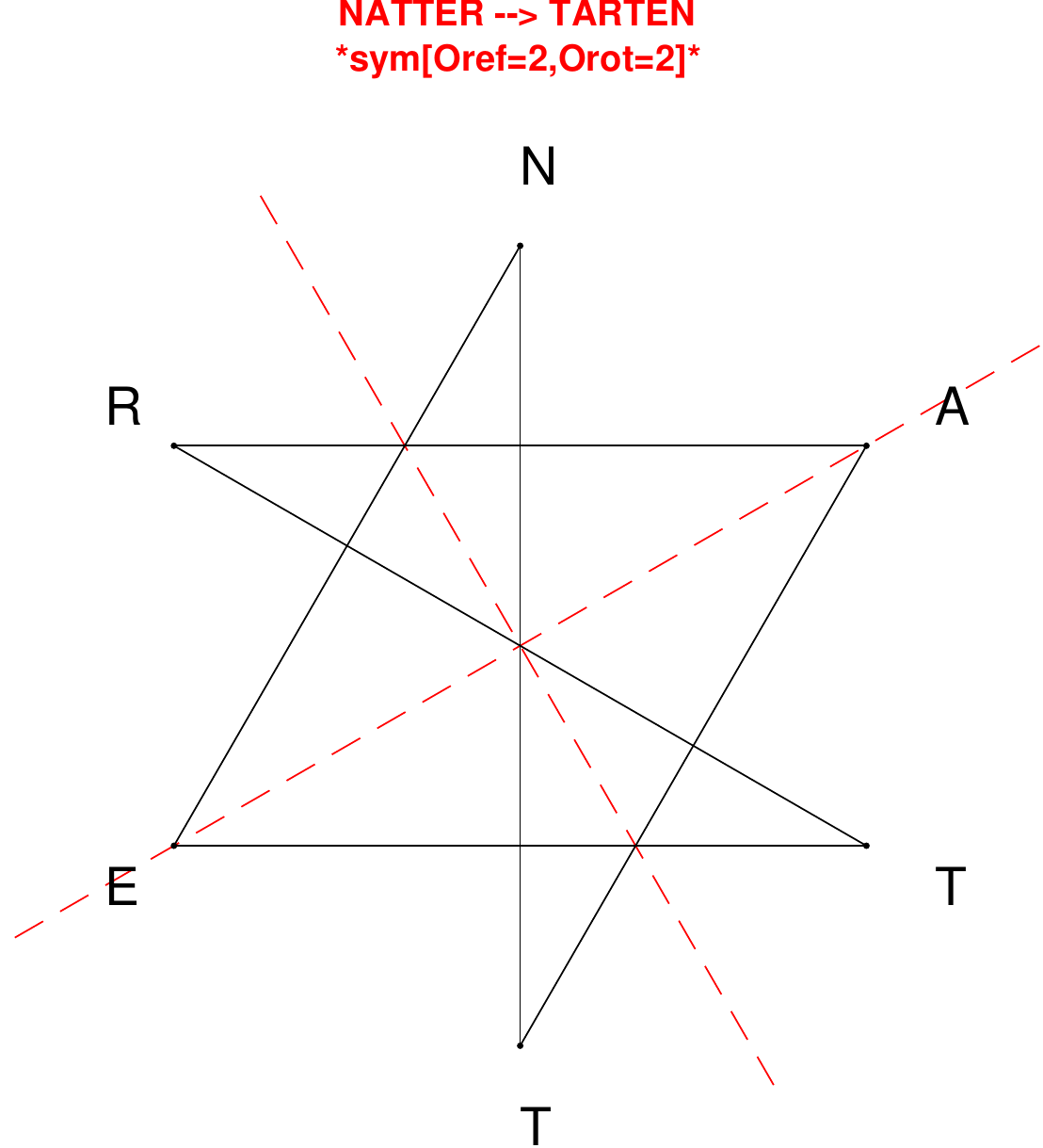}
\end{subfigure}
\hfill
\begin{subfigure}[T]{0.19\textwidth}
\centering
\includegraphics[width=\textwidth]{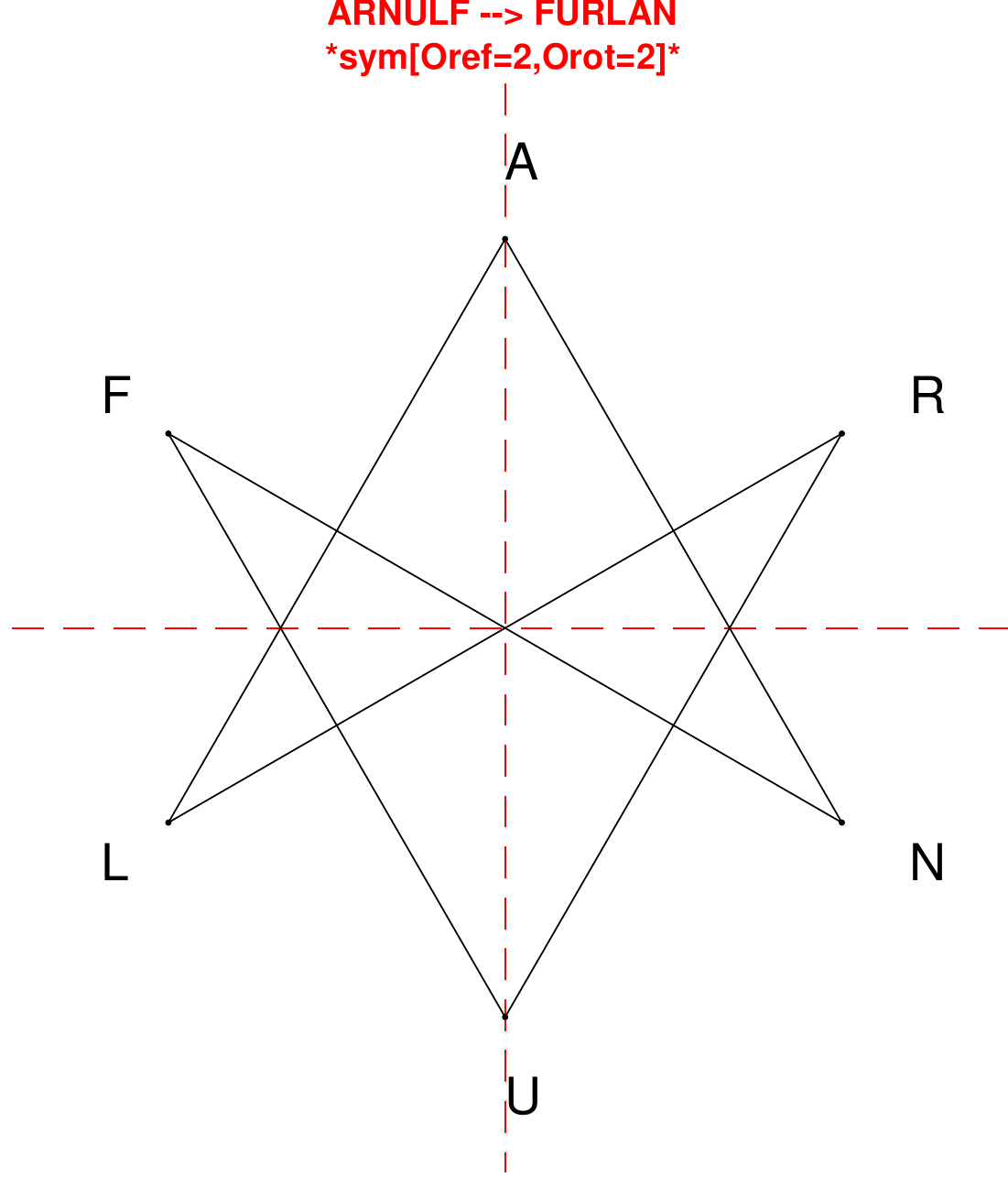}
\end{subfigure}
\end{figure}

\begin{figure}[H]
\centering
\begin{subfigure}[T]{0.19\textwidth}
\centering
\includegraphics[width=\textwidth]{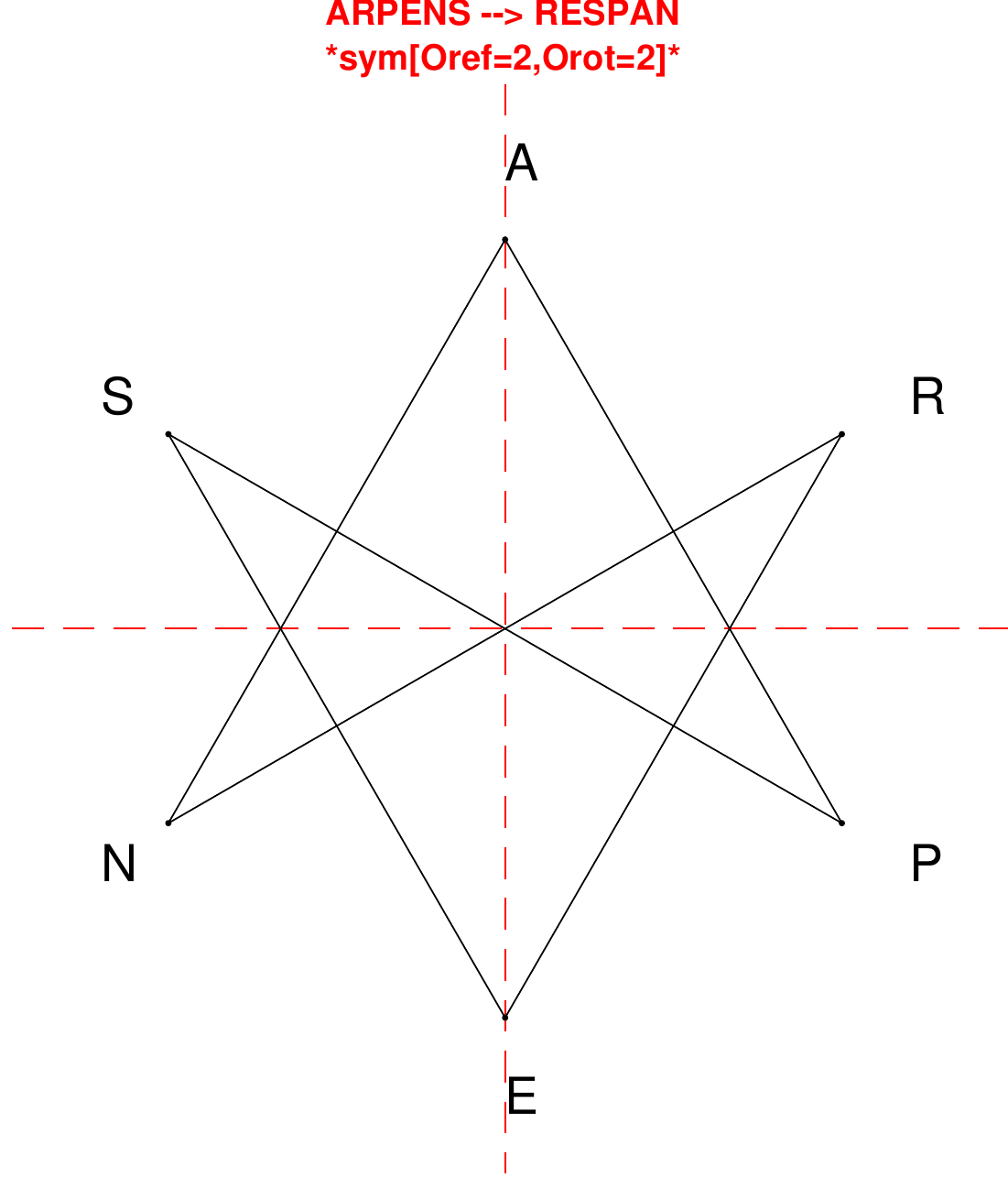}
\end{subfigure}
\hfill
\begin{subfigure}[T]{0.19\textwidth}
\centering
\includegraphics[width=\textwidth]{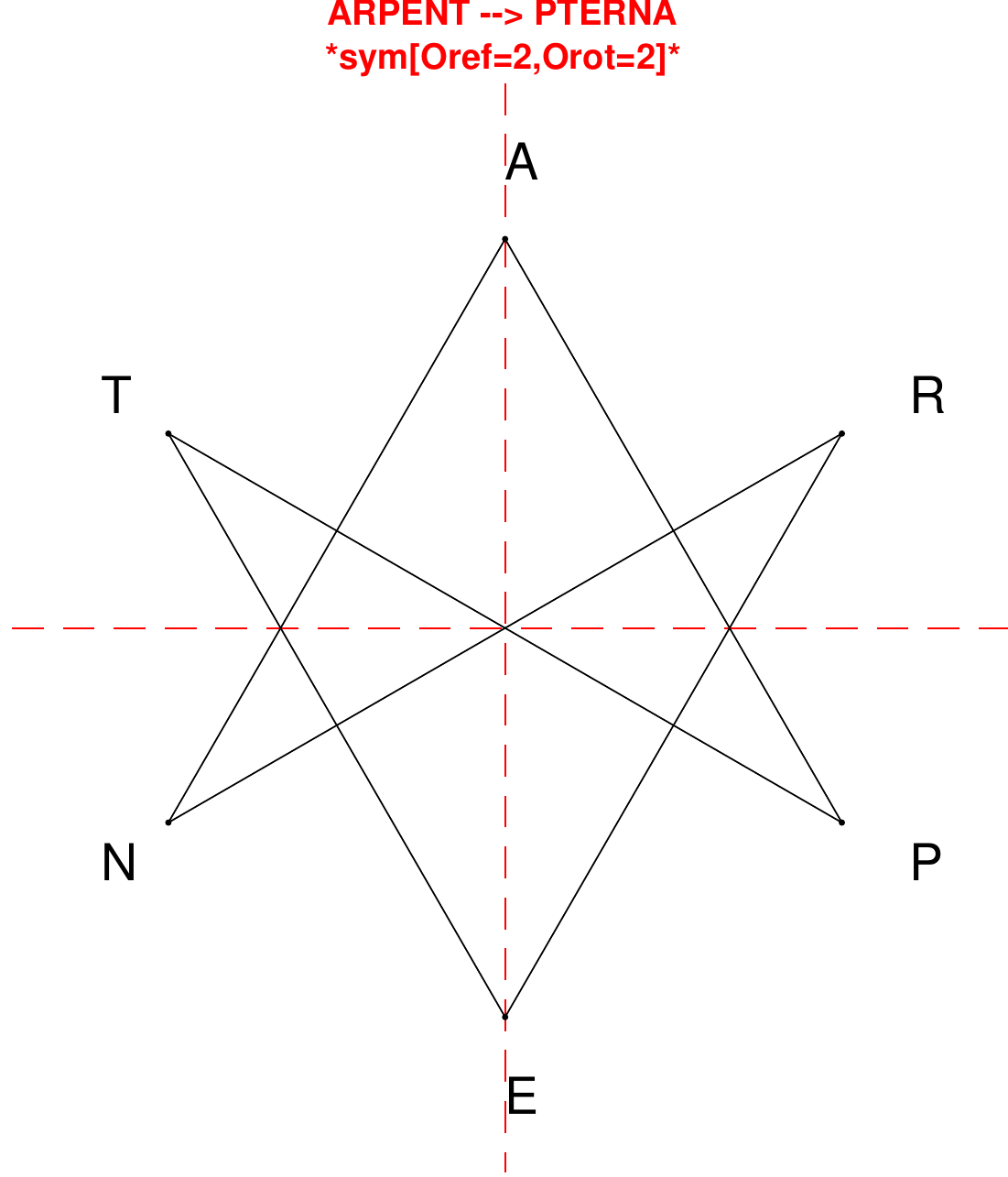}
\end{subfigure}
\hfill
\begin{subfigure}[T]{0.19\textwidth}
\centering
\includegraphics[width=\textwidth]{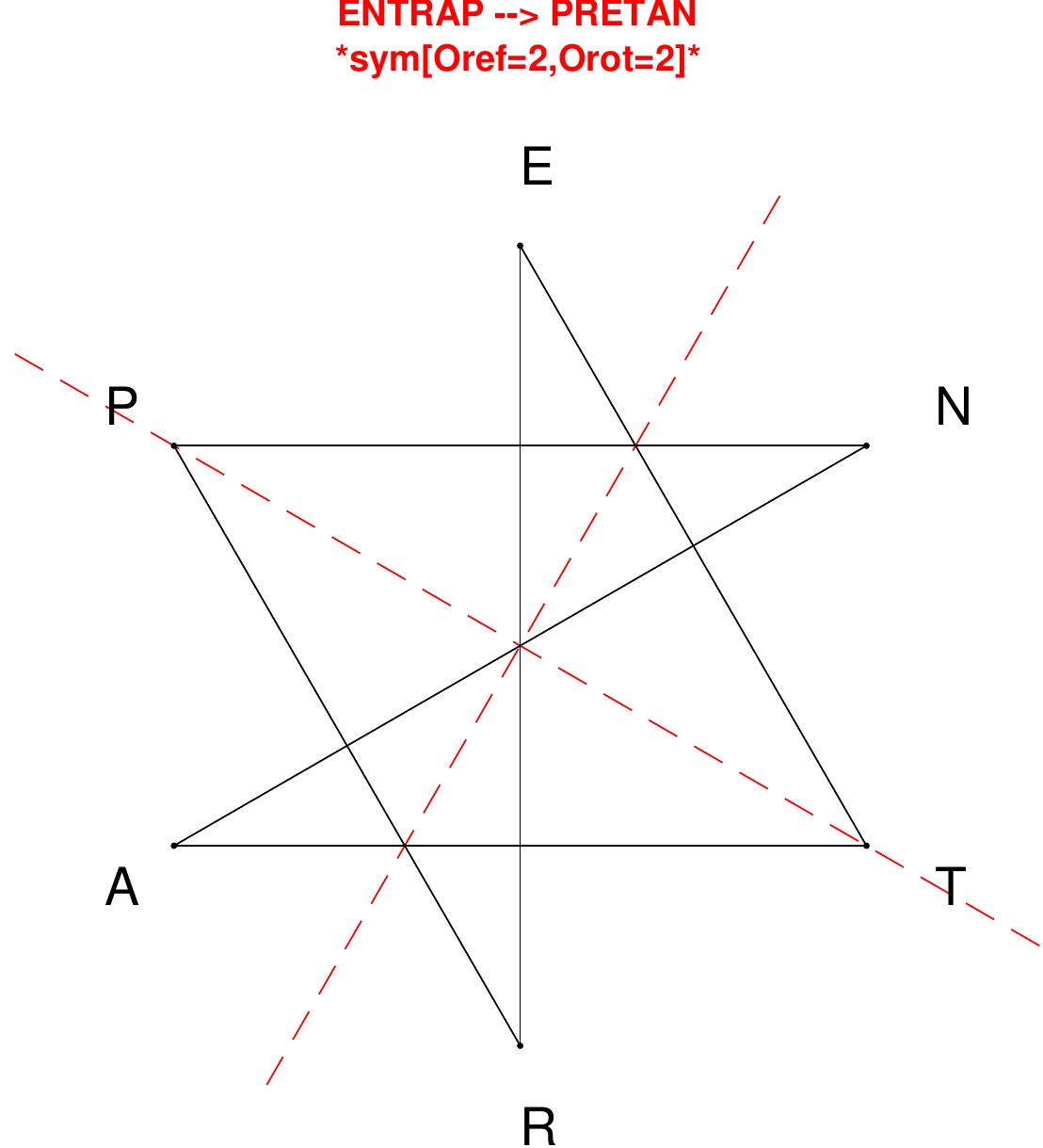}
\end{subfigure}
\hfill
\begin{subfigure}[T]{0.19\textwidth}
\centering
\includegraphics[width=\textwidth]{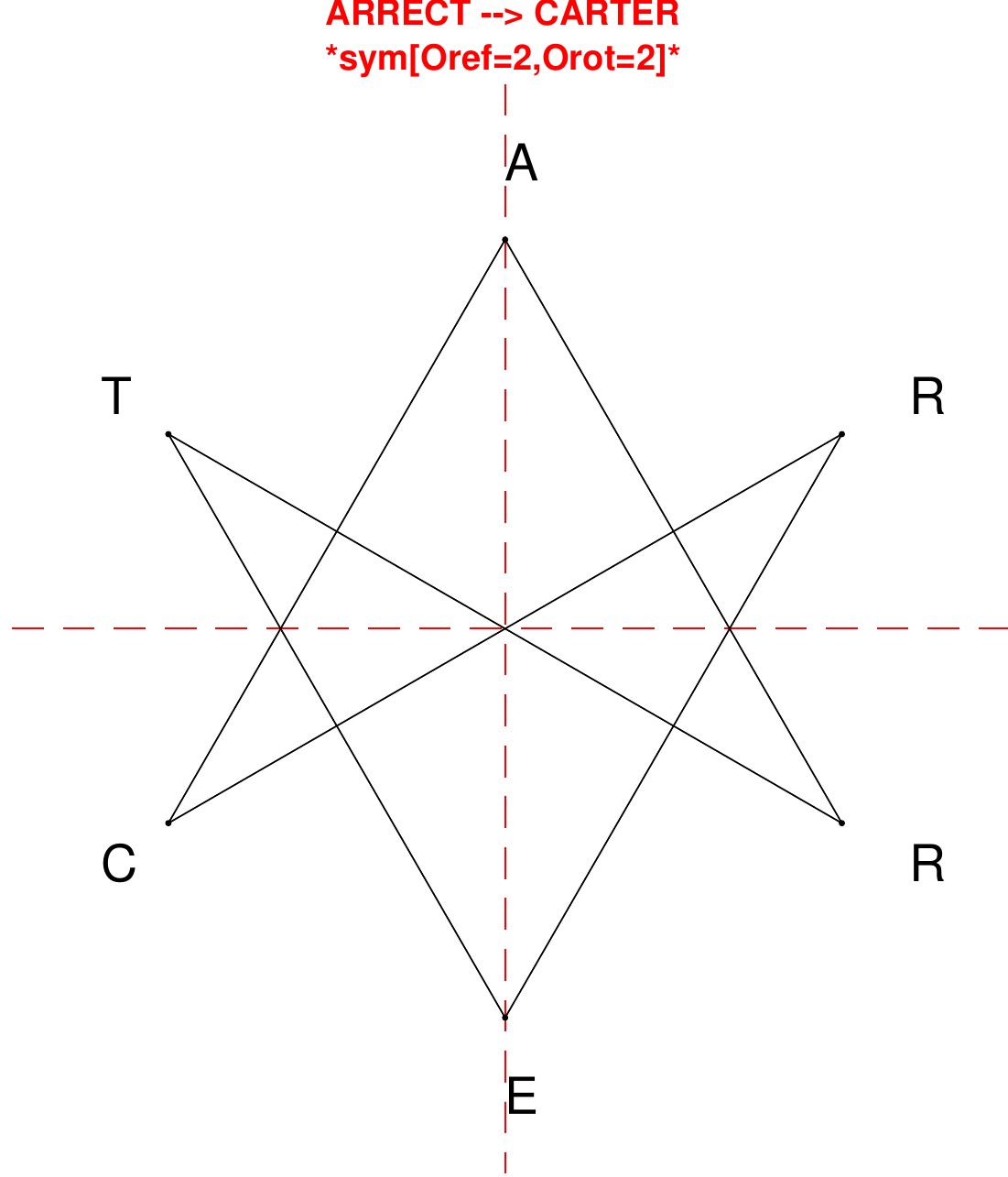}
\end{subfigure}
\hfill
\begin{subfigure}[T]{0.19\textwidth}
\centering
\includegraphics[width=\textwidth]{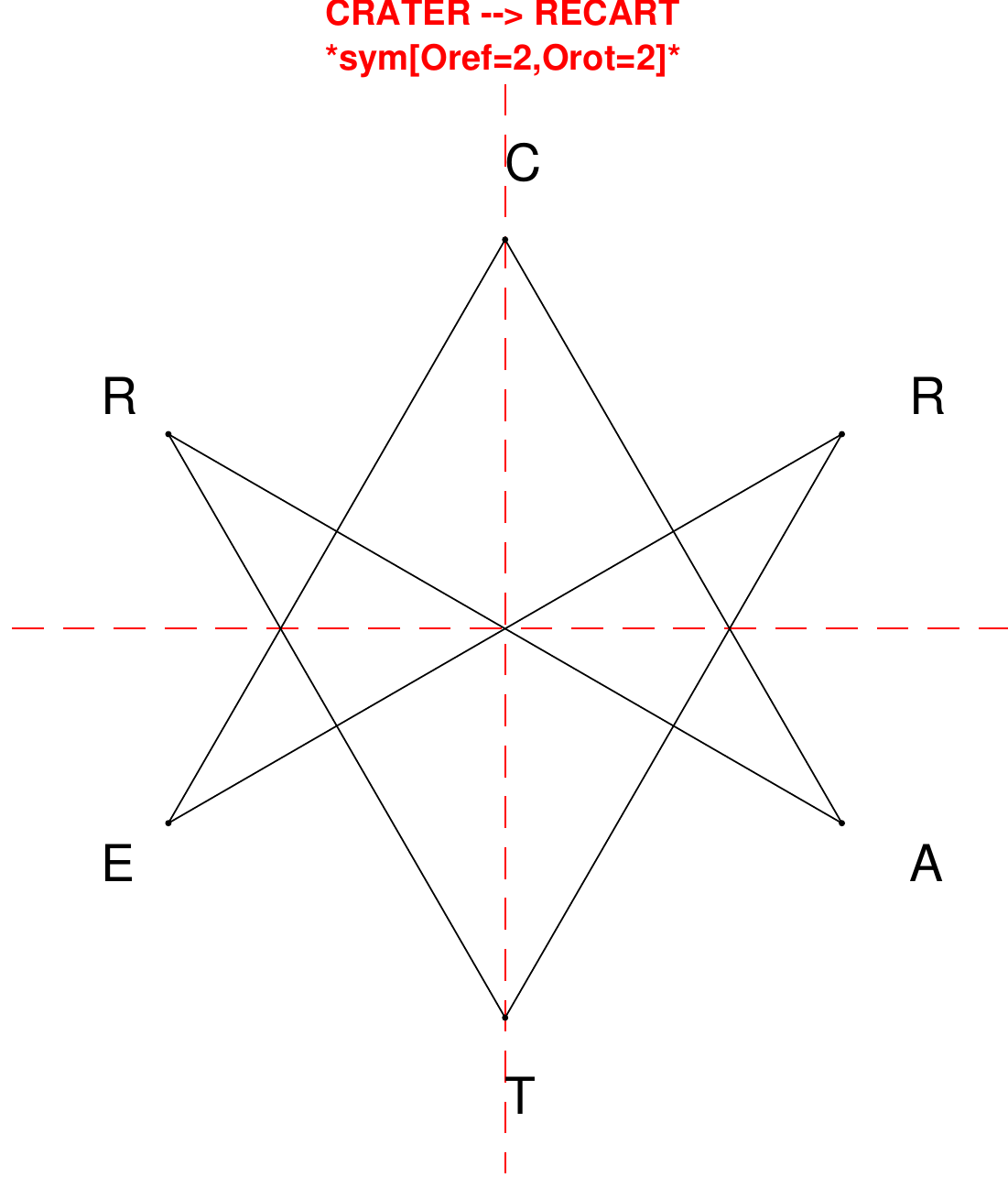}
\end{subfigure}
\end{figure}

\begin{figure}[H]
\centering
\begin{subfigure}[T]{0.19\textwidth}
\centering
\includegraphics[width=\textwidth]{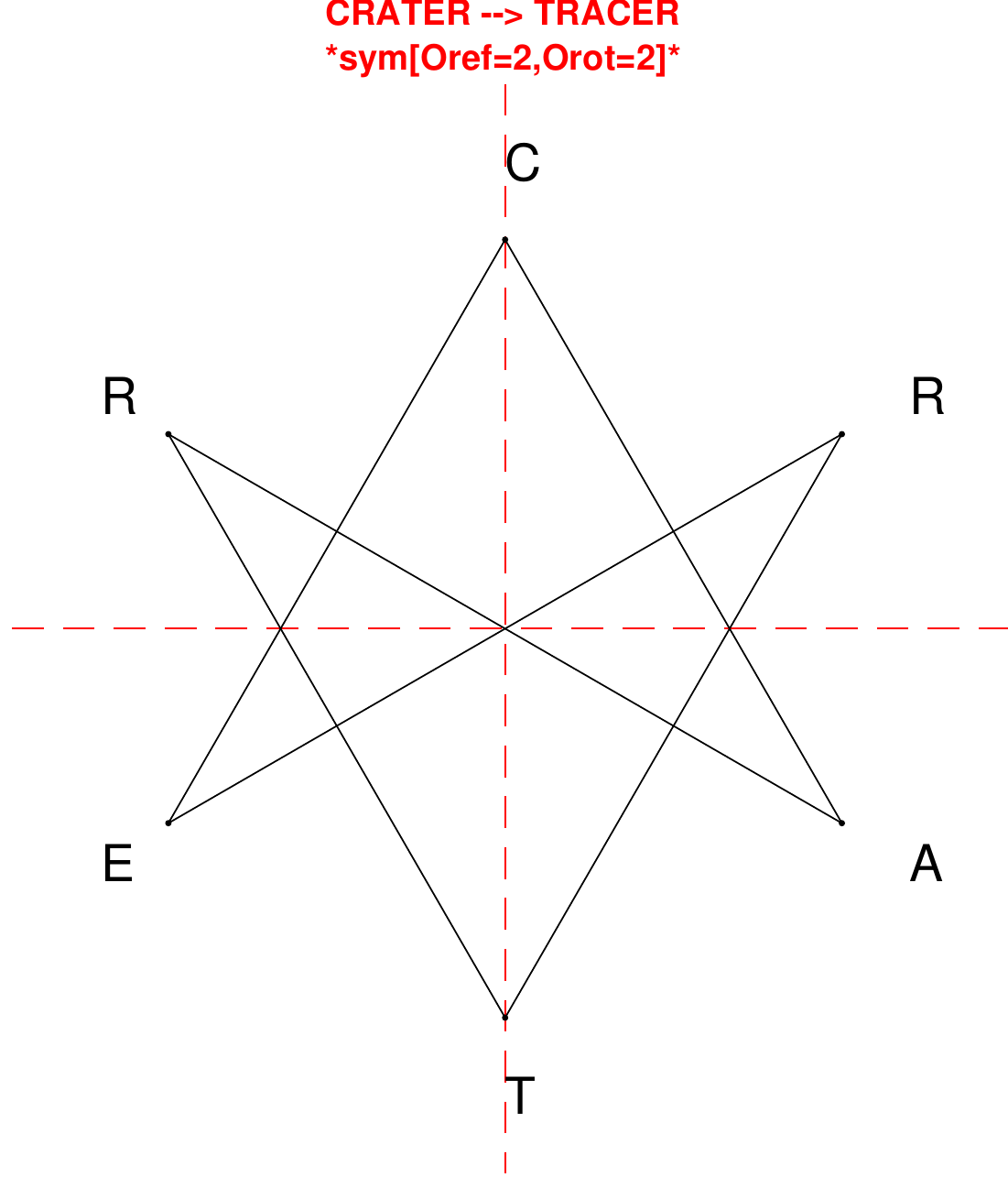}
\end{subfigure}
\hfill
\begin{subfigure}[T]{0.19\textwidth}
\centering
\includegraphics[width=\textwidth]{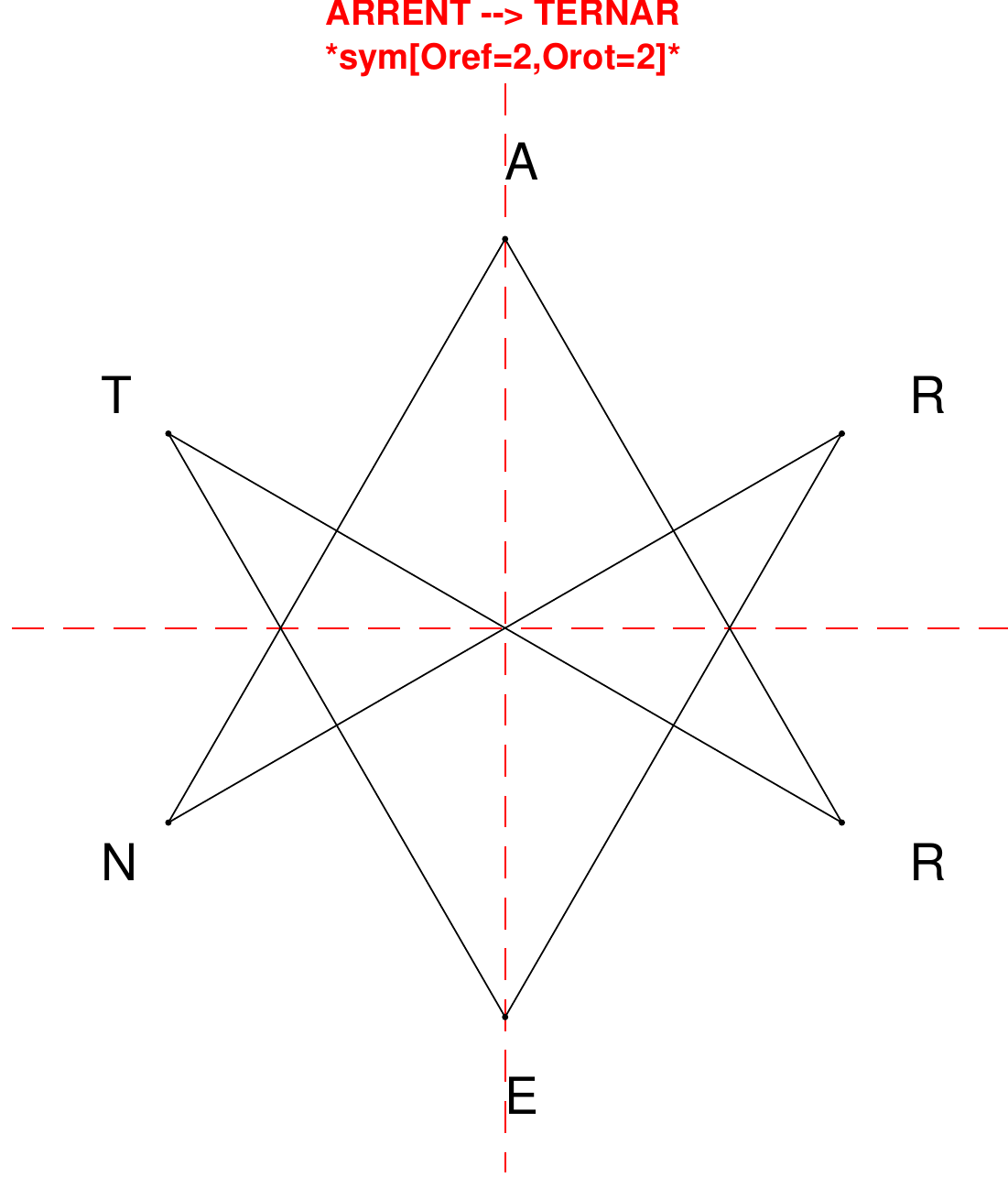}
\end{subfigure}
\hfill
\begin{subfigure}[T]{0.19\textwidth}
\centering
\includegraphics[width=\textwidth]{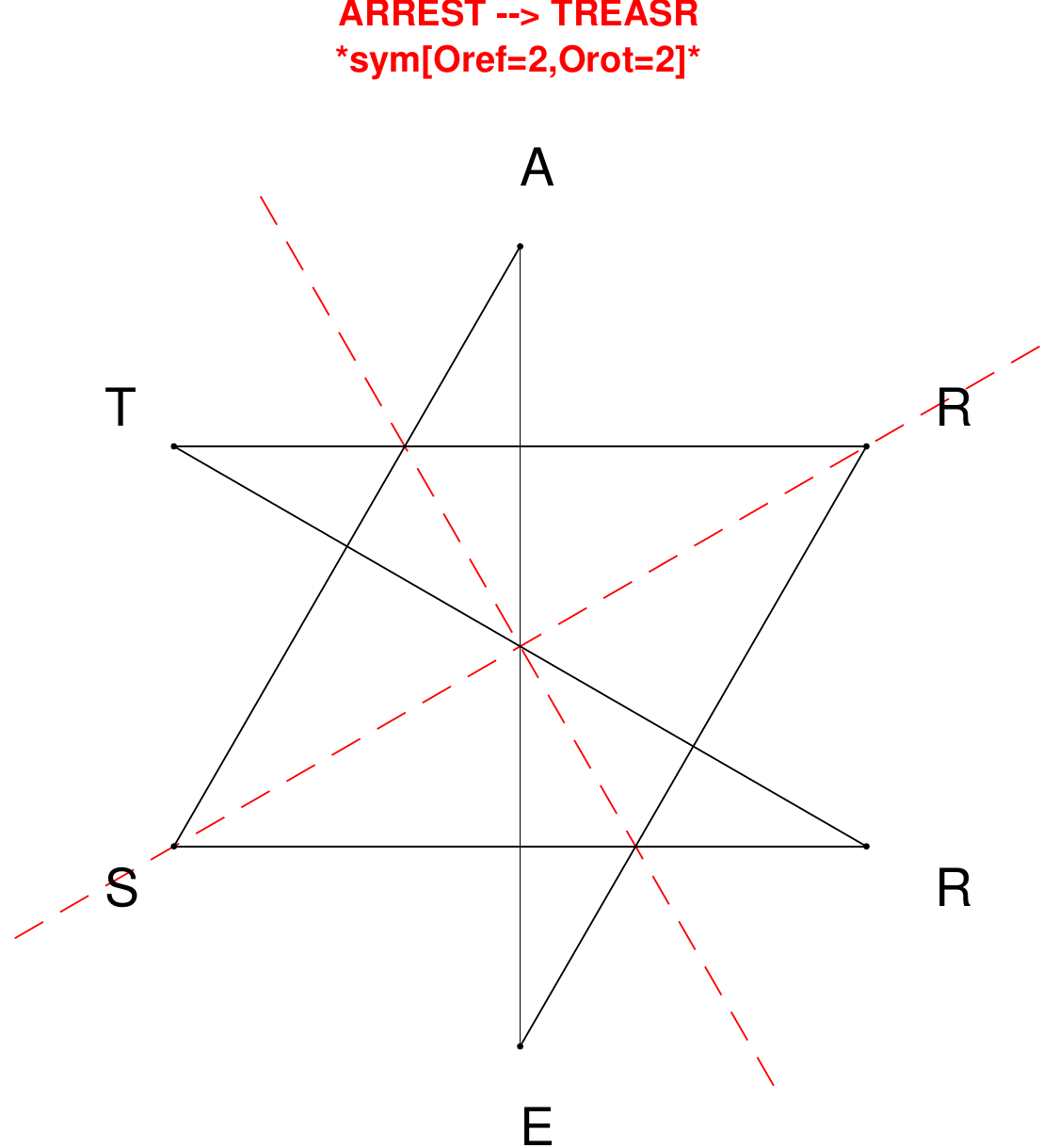}
\end{subfigure}
\hfill
\begin{subfigure}[T]{0.19\textwidth}
\centering
\includegraphics[width=\textwidth]{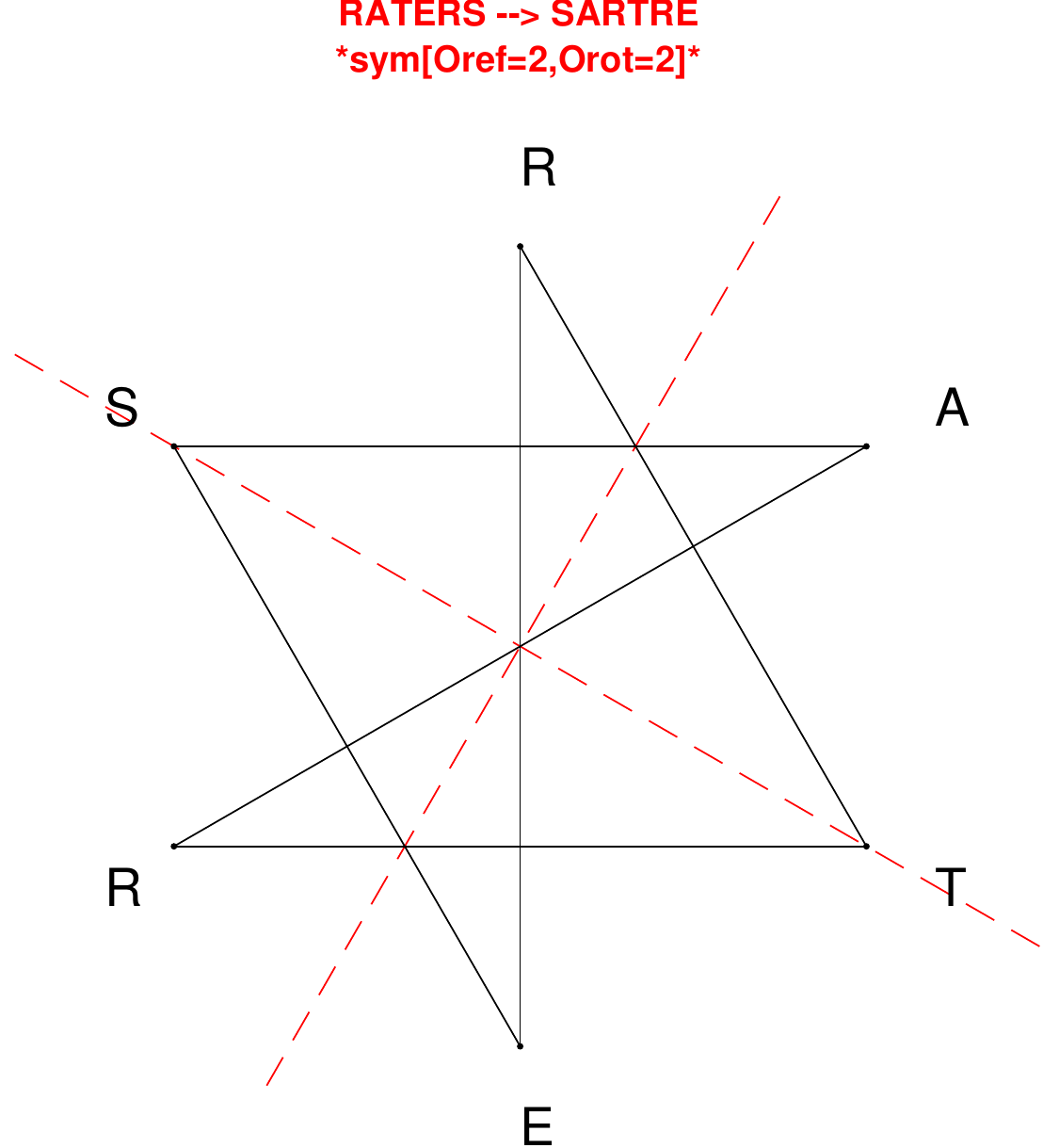}
\end{subfigure}
\hfill
\begin{subfigure}[T]{0.19\textwidth}
\centering
\includegraphics[width=\textwidth]{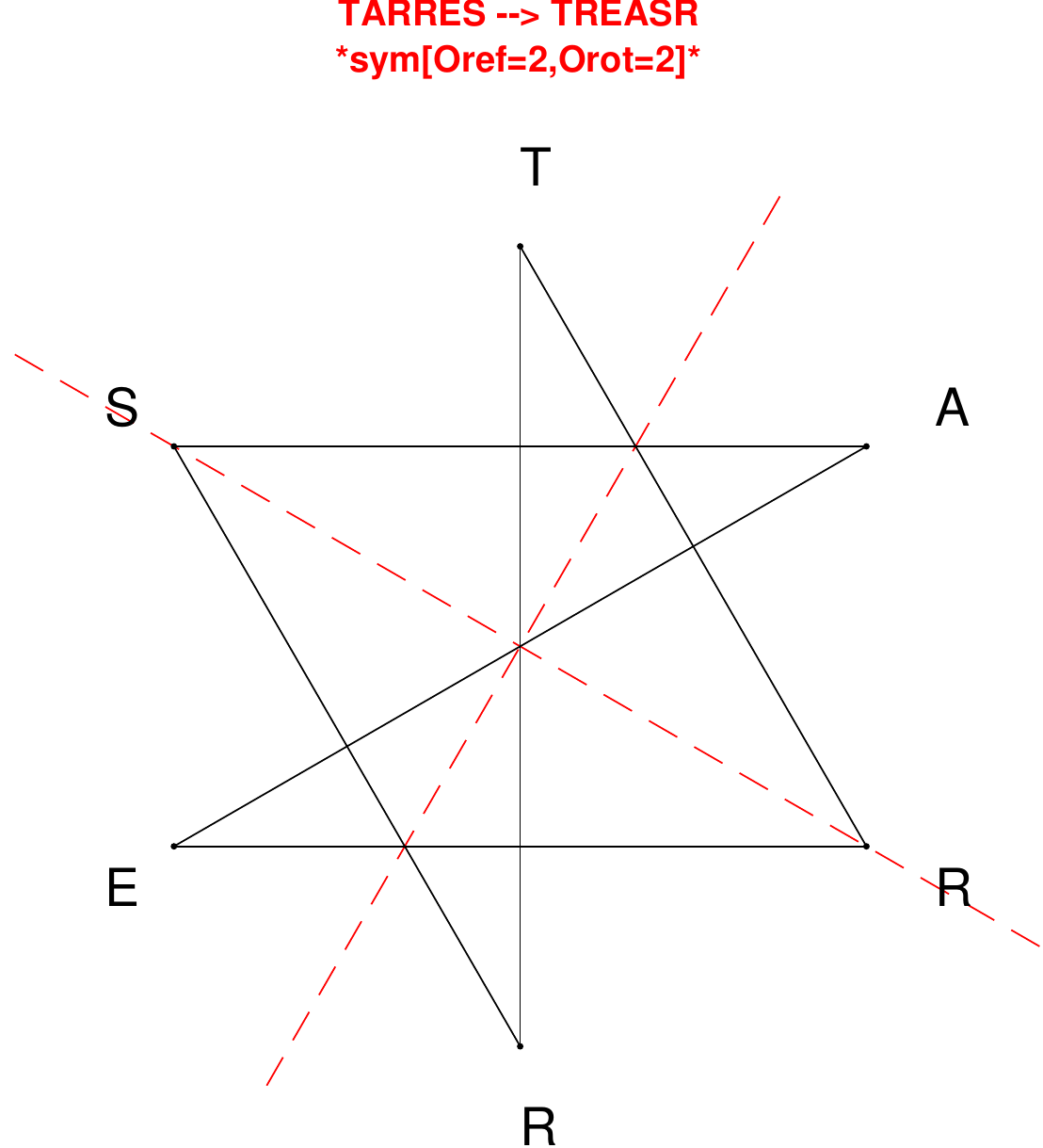}
\end{subfigure}
\end{figure}

\begin{figure}[H]
\centering
\begin{subfigure}[T]{0.19\textwidth}
\centering
\includegraphics[width=\textwidth]{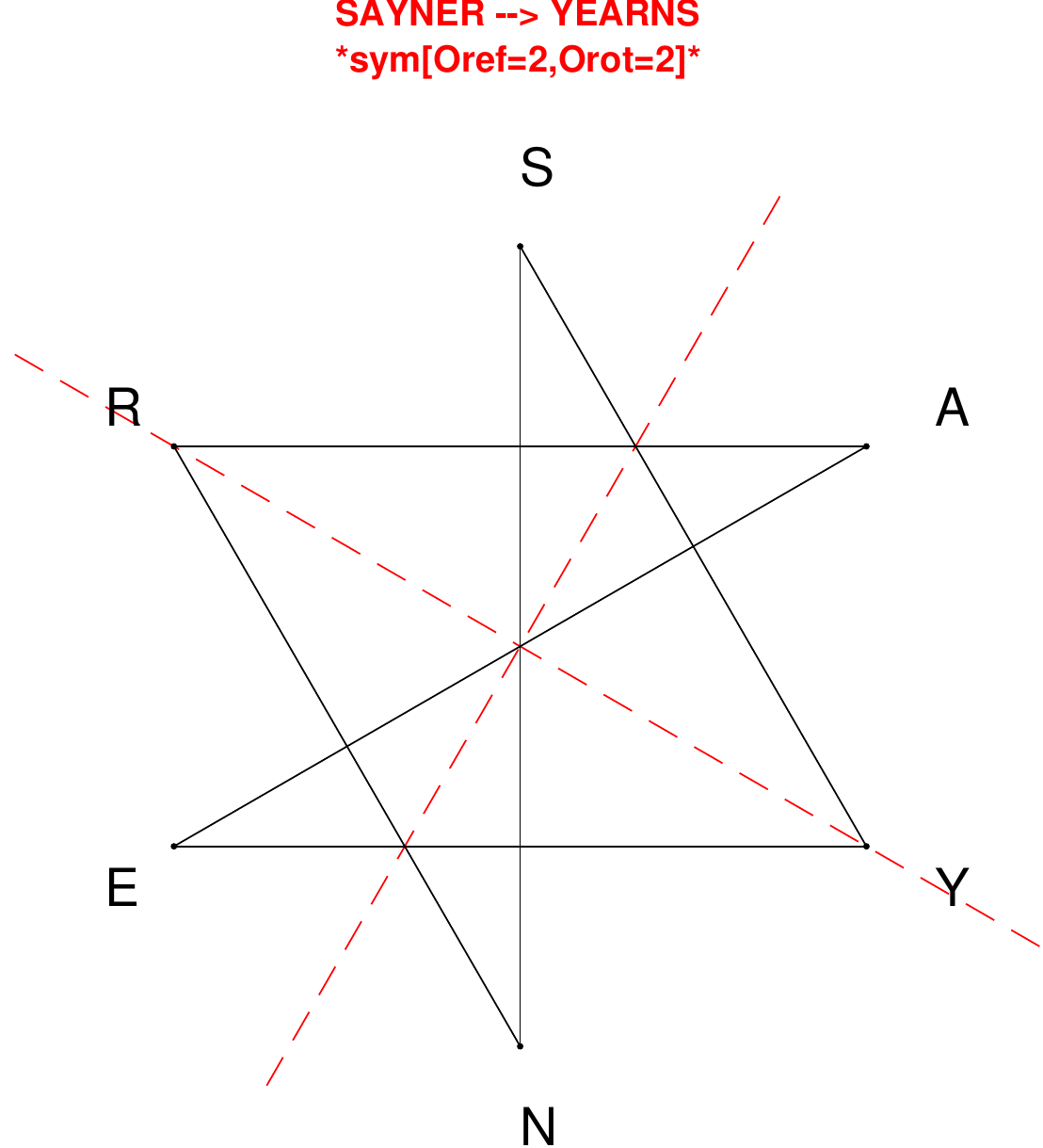}
\end{subfigure}
\hfill
\begin{subfigure}[T]{0.19\textwidth}
\centering
\includegraphics[width=\textwidth]{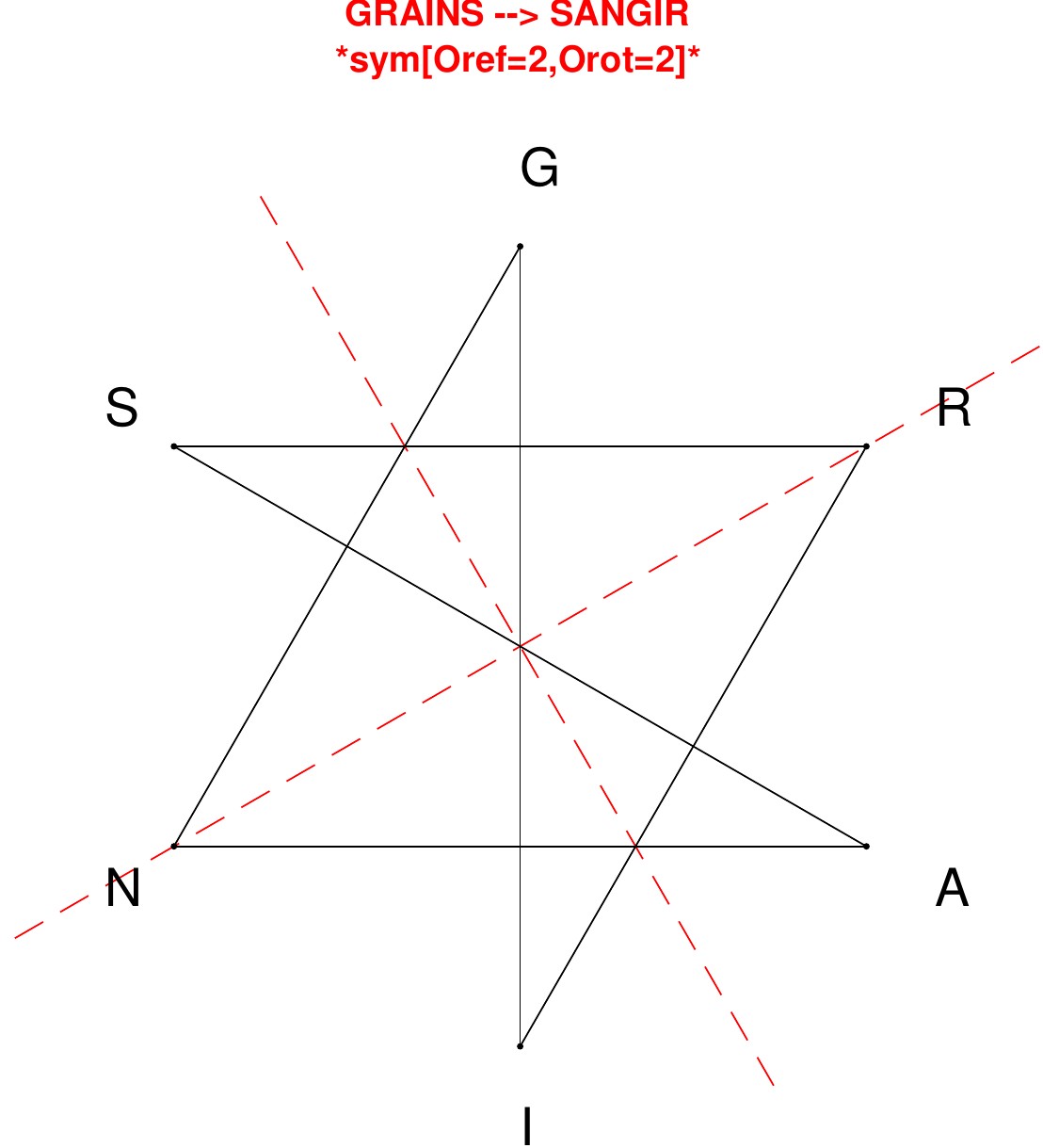}
\end{subfigure}
\hfill
\begin{subfigure}[T]{0.19\textwidth}
\centering
\includegraphics[width=\textwidth]{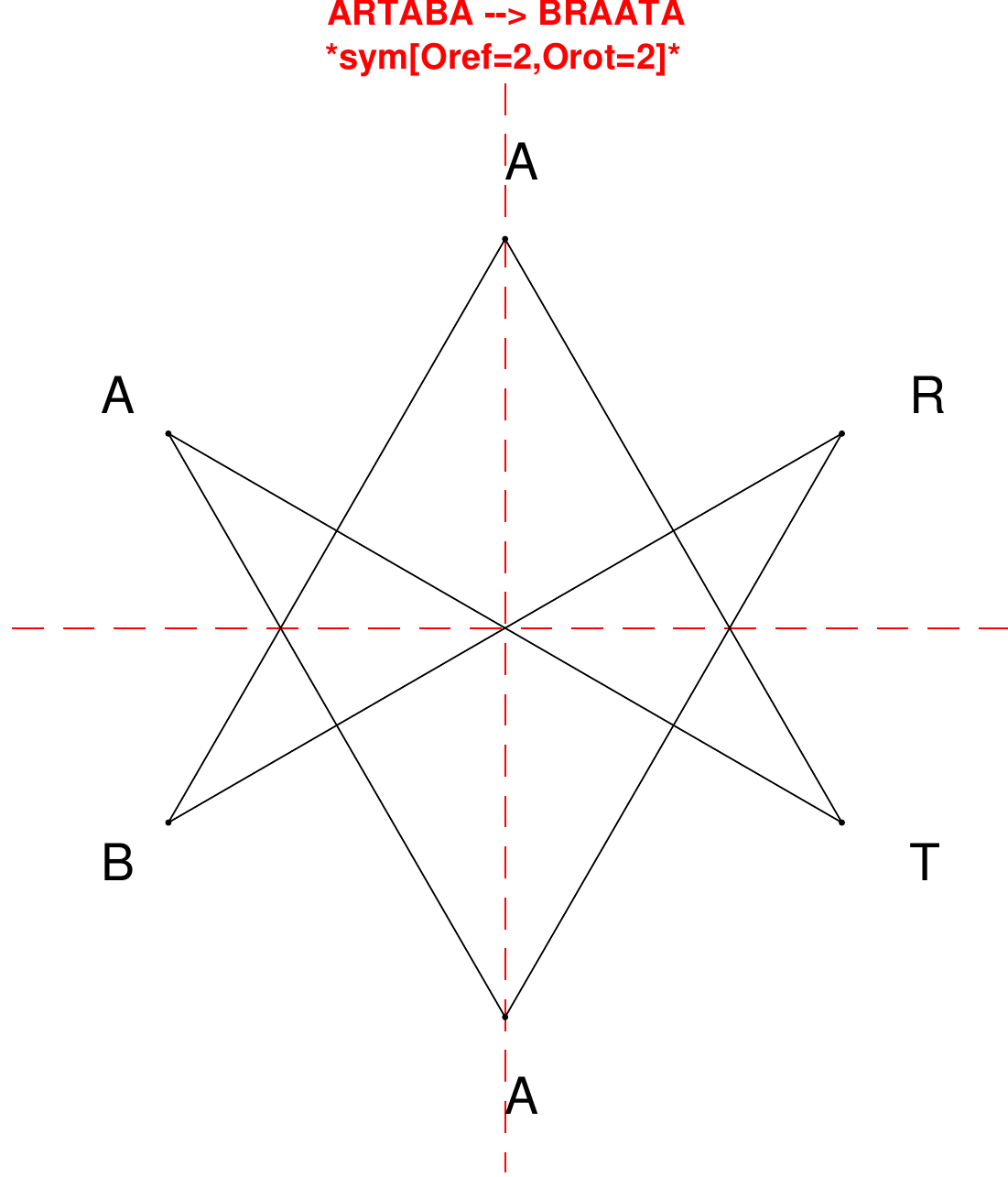}
\end{subfigure}
\hfill
\begin{subfigure}[T]{0.19\textwidth}
\centering
\includegraphics[width=\textwidth]{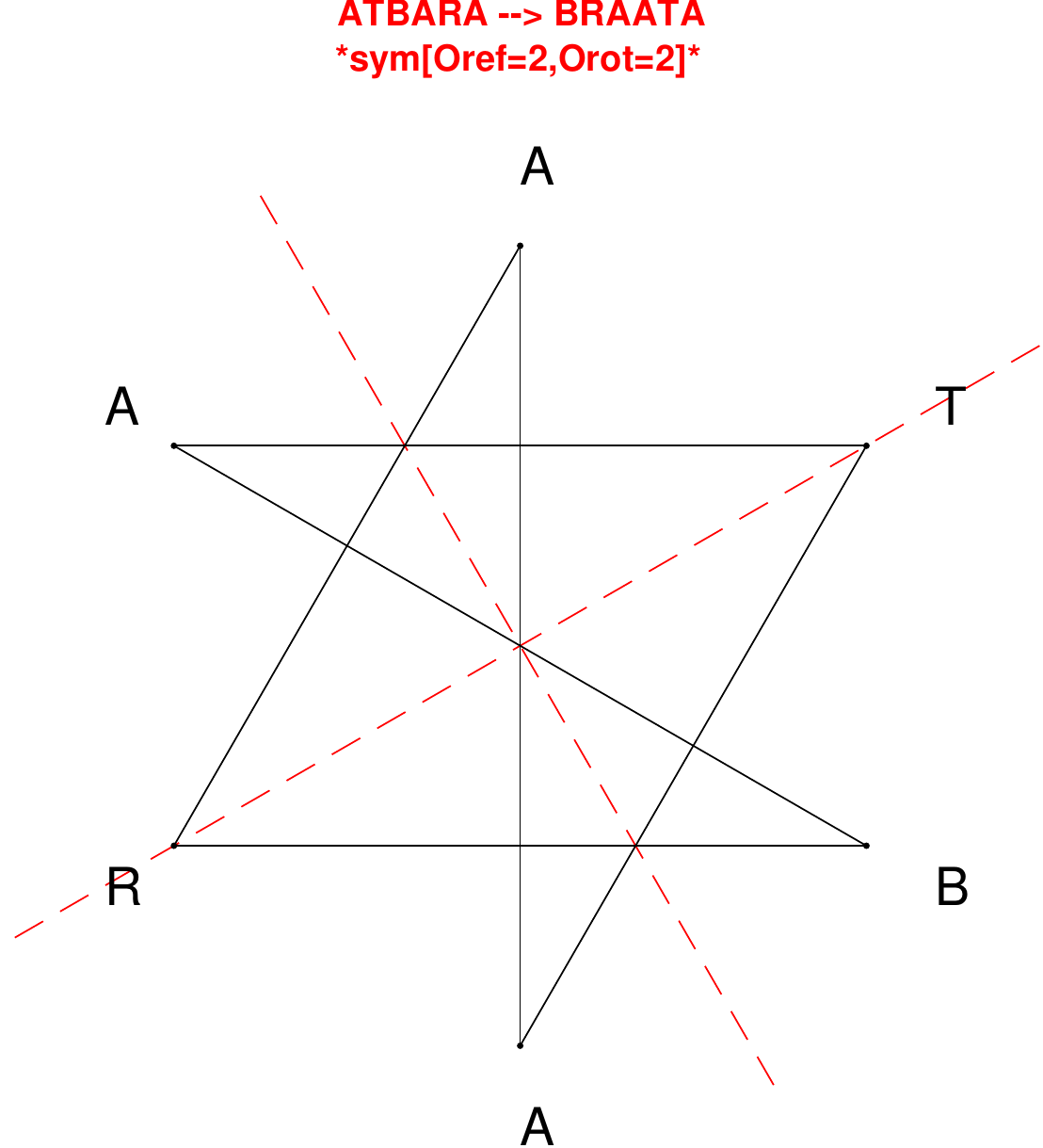}
\end{subfigure}
\hfill
\begin{subfigure}[T]{0.19\textwidth}
\centering
\includegraphics[width=\textwidth]{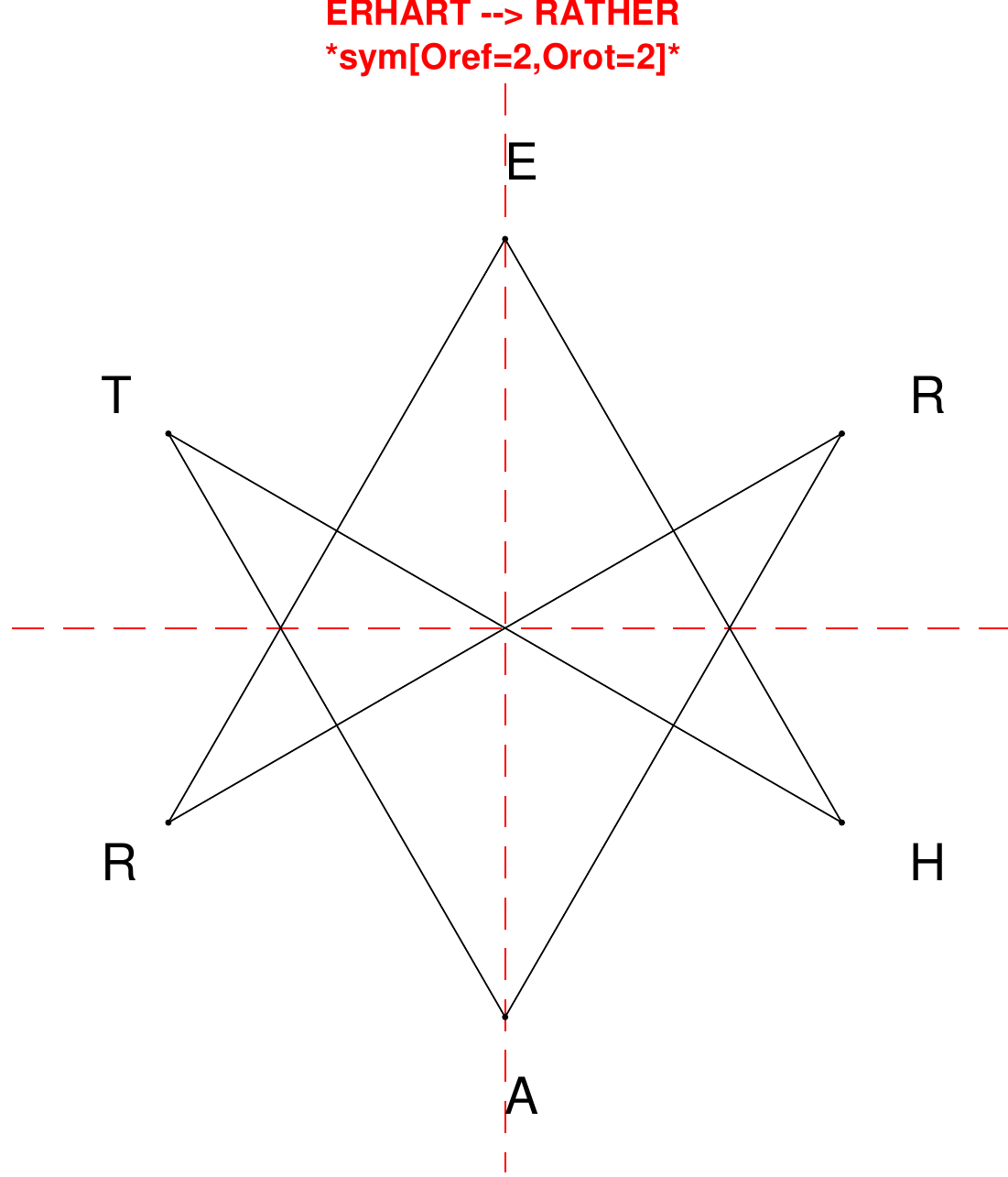}
\end{subfigure}
\end{figure}

\begin{figure}[H]
\centering
\begin{subfigure}[T]{0.19\textwidth}
\centering
\includegraphics[width=\textwidth]{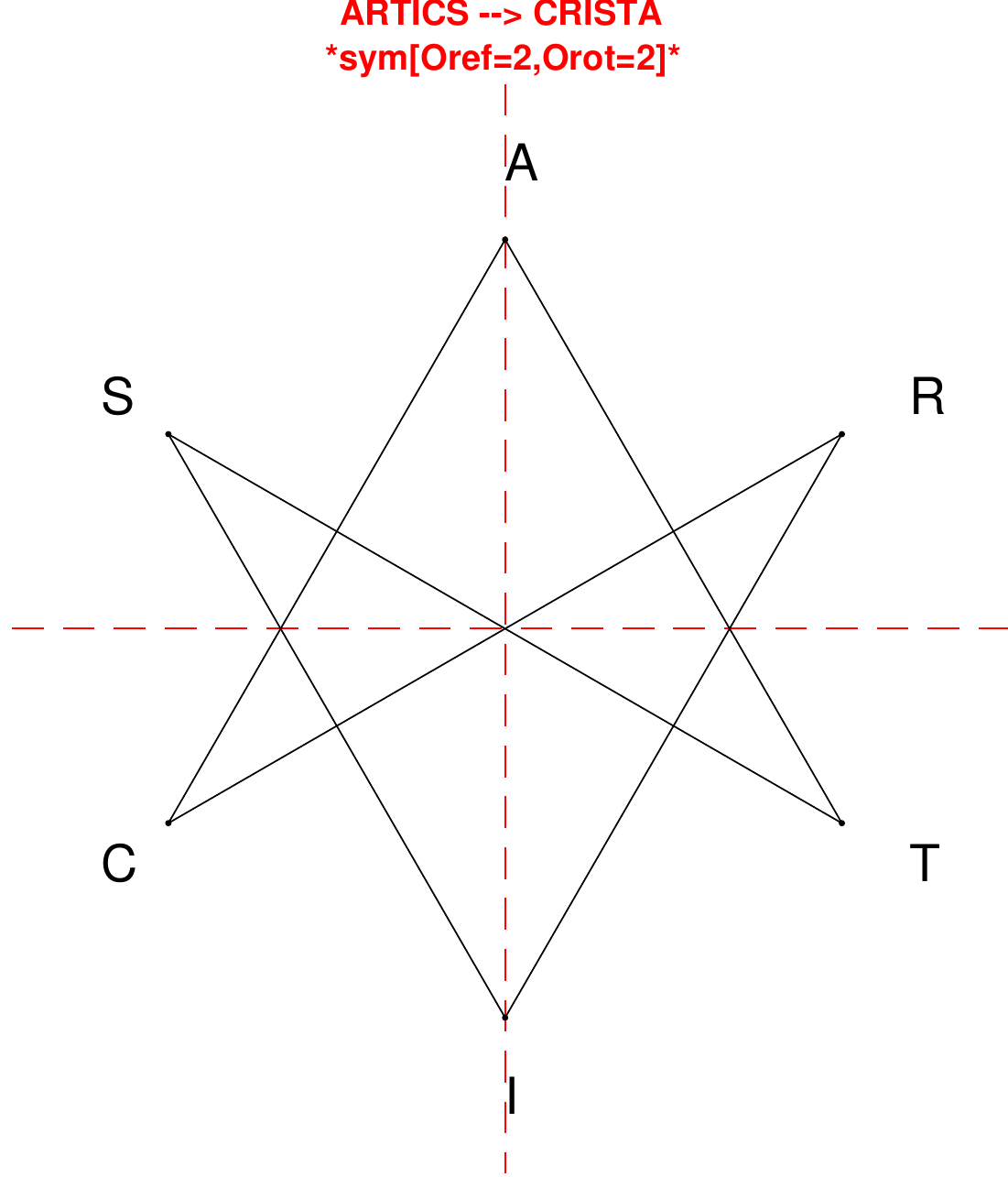}
\end{subfigure}
\hfill
\begin{subfigure}[T]{0.19\textwidth}
\centering
\includegraphics[width=\textwidth]{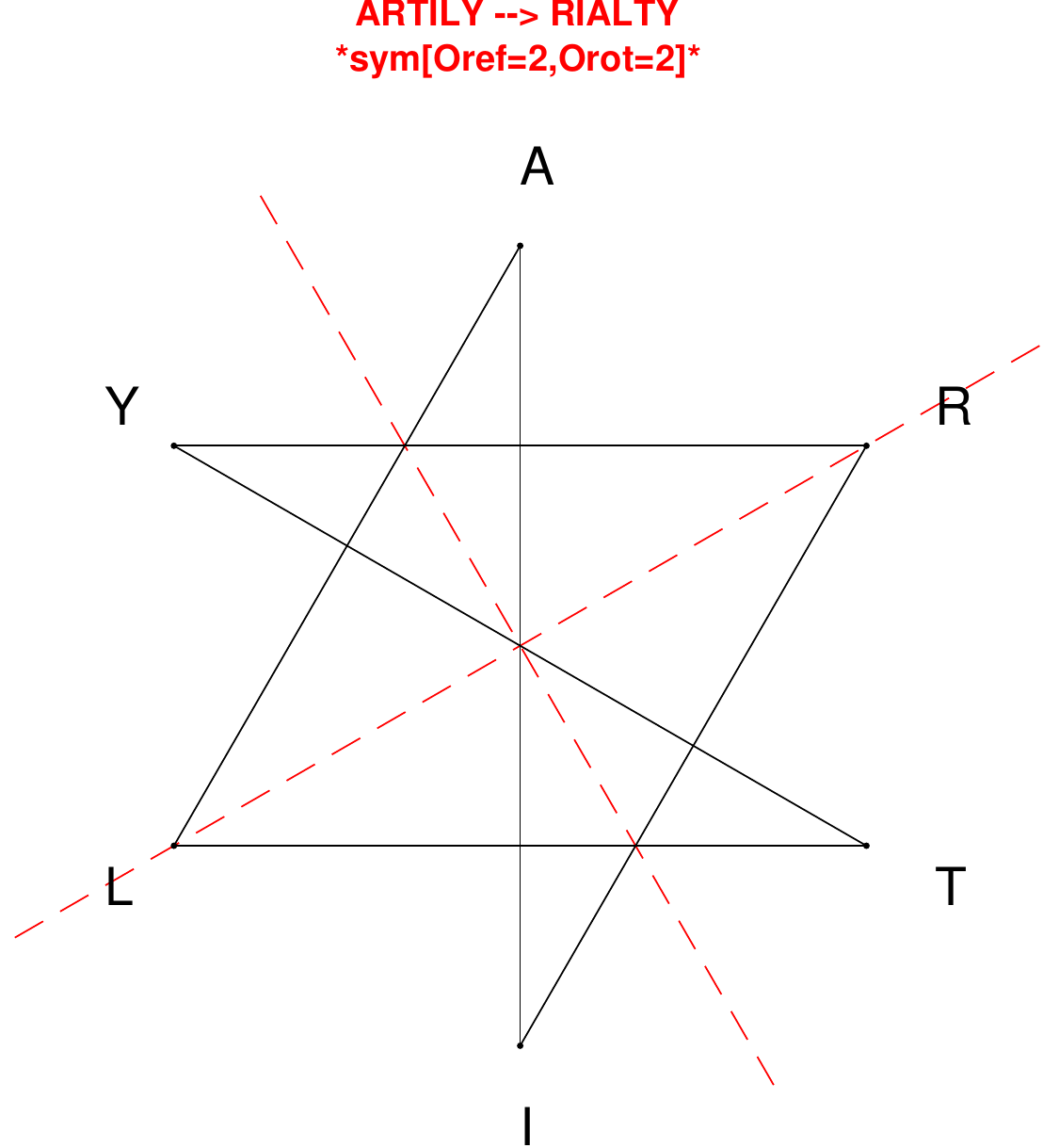}
\end{subfigure}
\hfill
\begin{subfigure}[T]{0.19\textwidth}
\centering
\includegraphics[width=\textwidth]{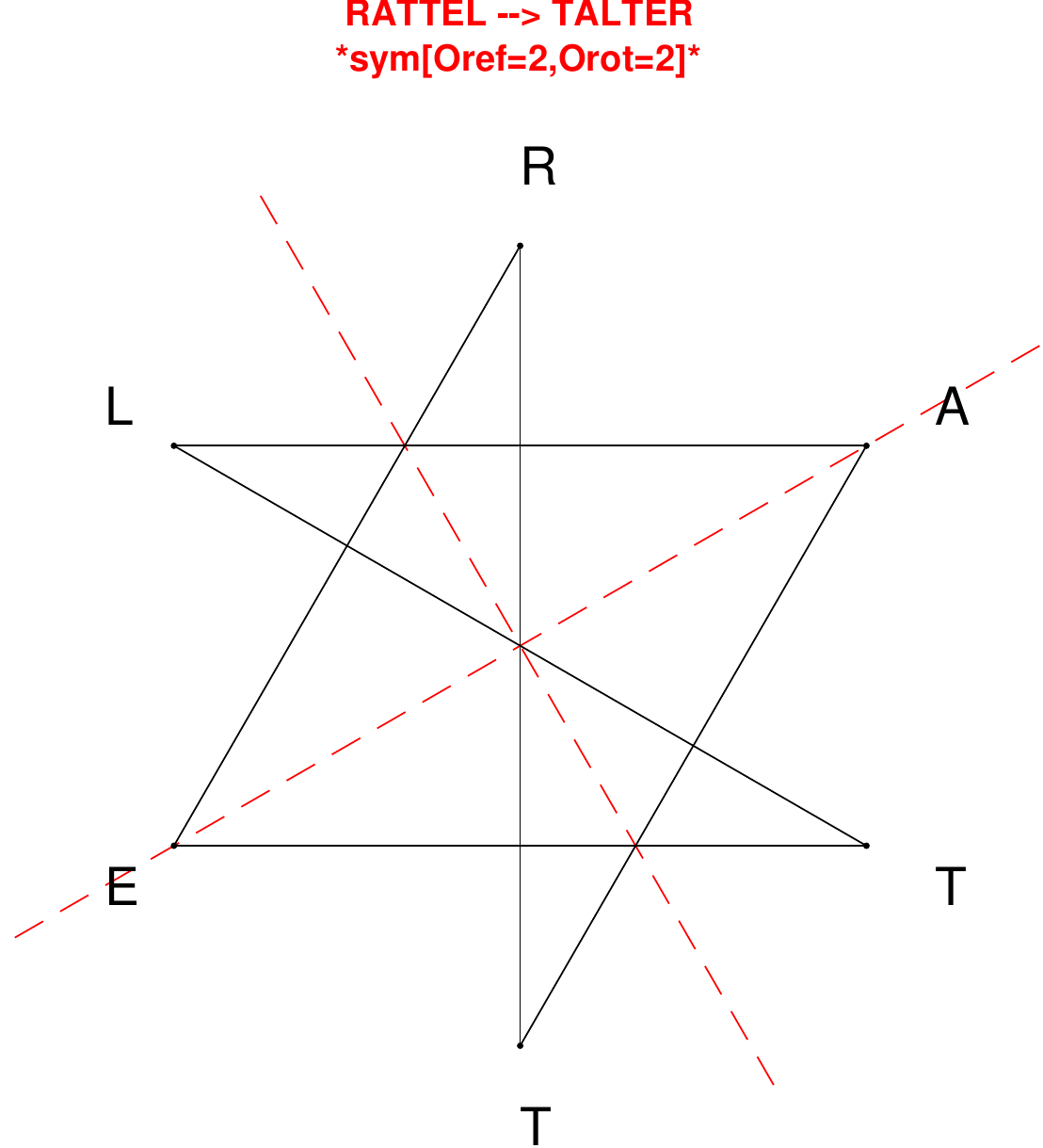}
\end{subfigure}
\hfill
\begin{subfigure}[T]{0.19\textwidth}
\centering
\includegraphics[width=\textwidth]{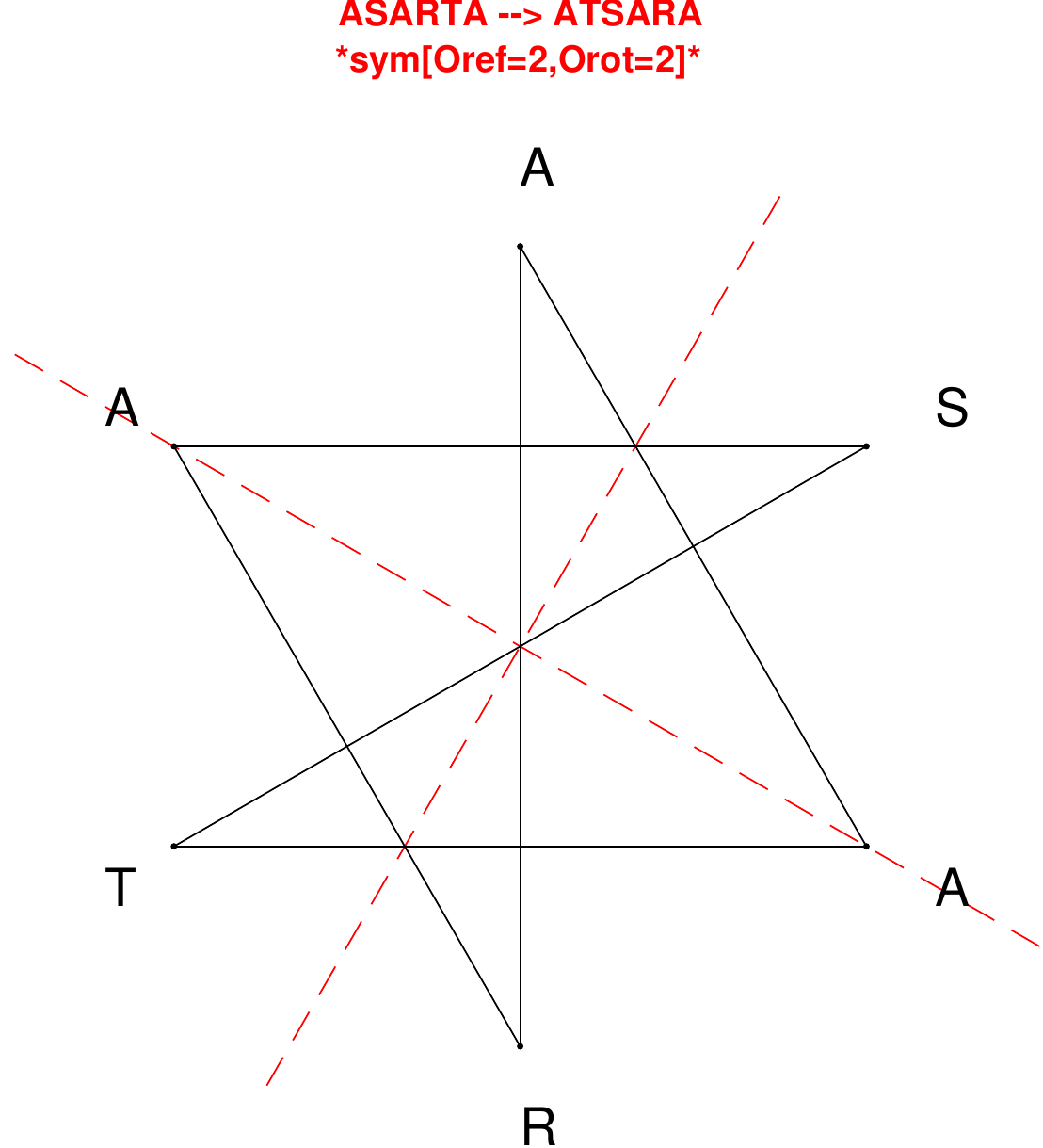}
\end{subfigure}
\hfill
\begin{subfigure}[T]{0.19\textwidth}
\centering
\includegraphics[width=\textwidth]{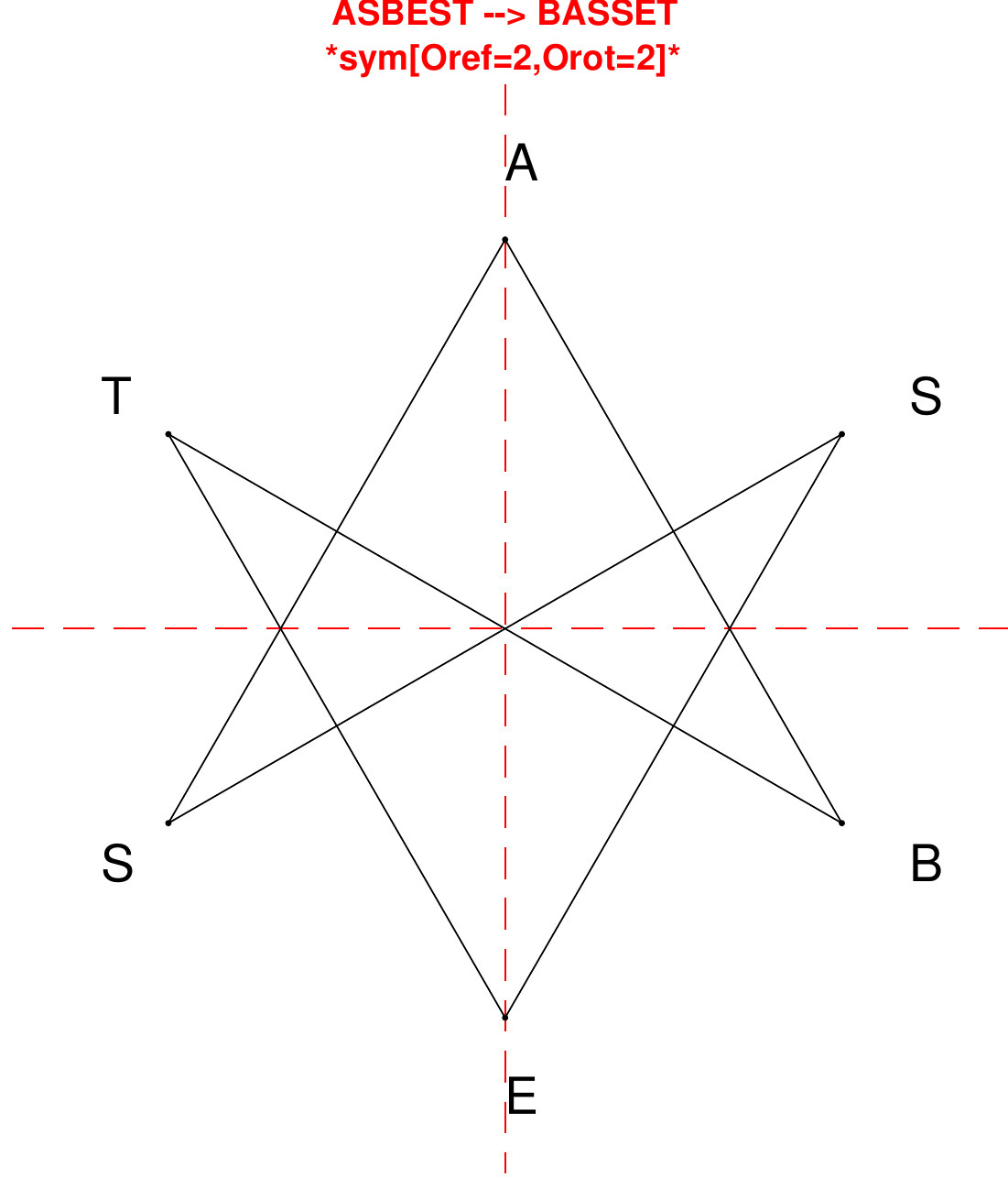}
\end{subfigure}
\end{figure}

\begin{figure}[H]
\centering
\begin{subfigure}[T]{0.19\textwidth}
\centering
\includegraphics[width=\textwidth]{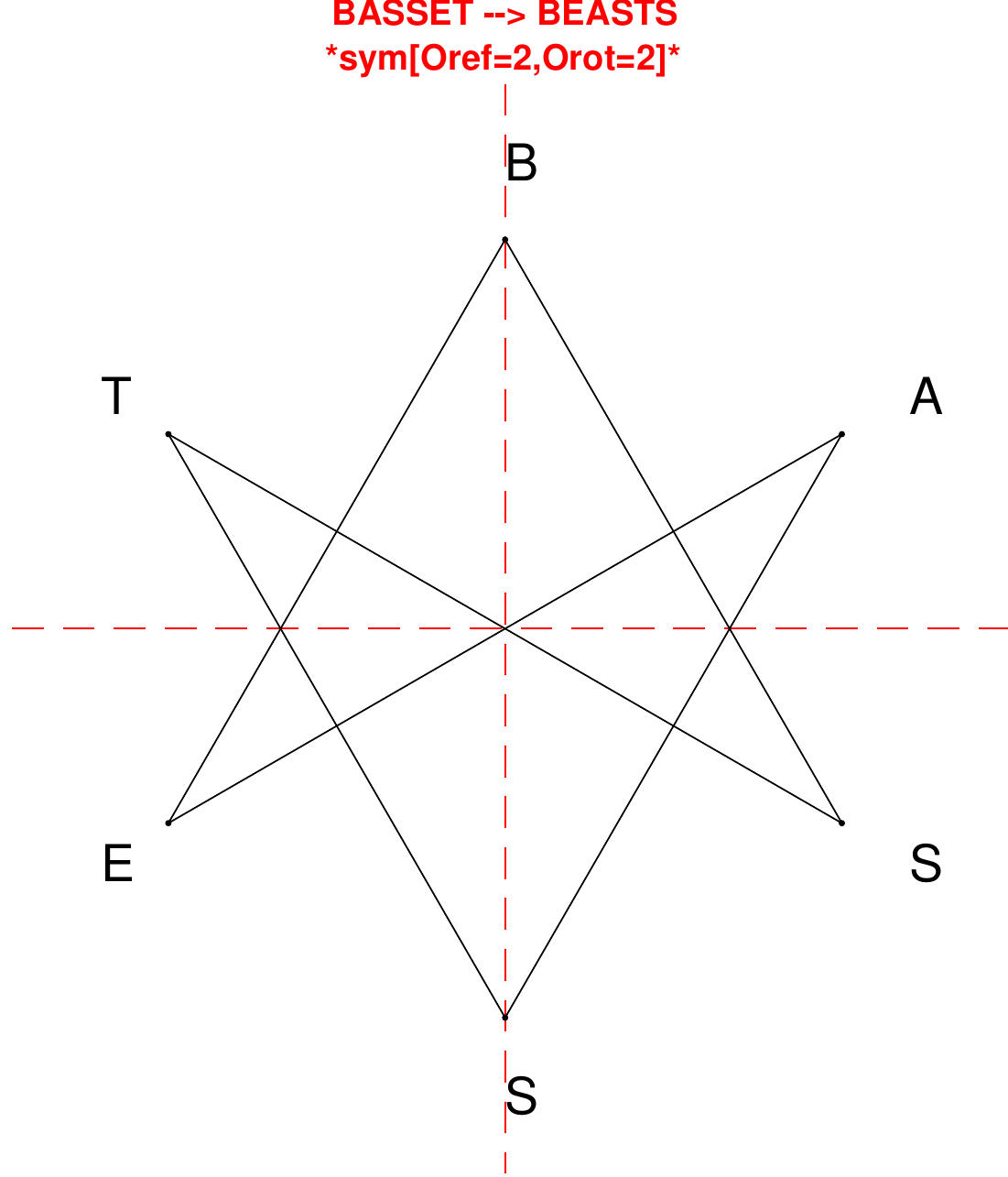}
\end{subfigure}
\hfill
\begin{subfigure}[T]{0.19\textwidth}
\centering
\includegraphics[width=\textwidth]{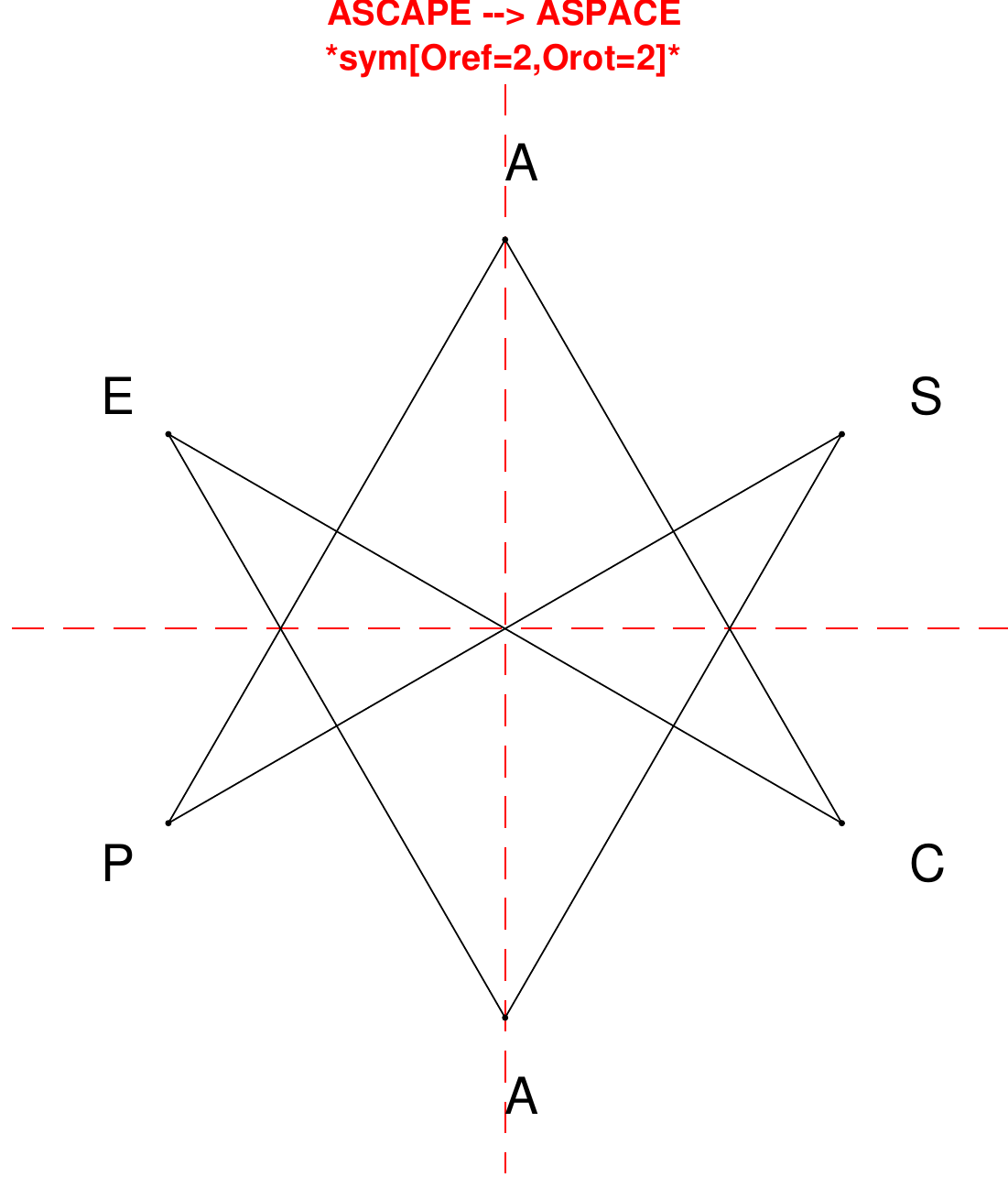}
\end{subfigure}
\hfill
\begin{subfigure}[T]{0.19\textwidth}
\centering
\includegraphics[width=\textwidth]{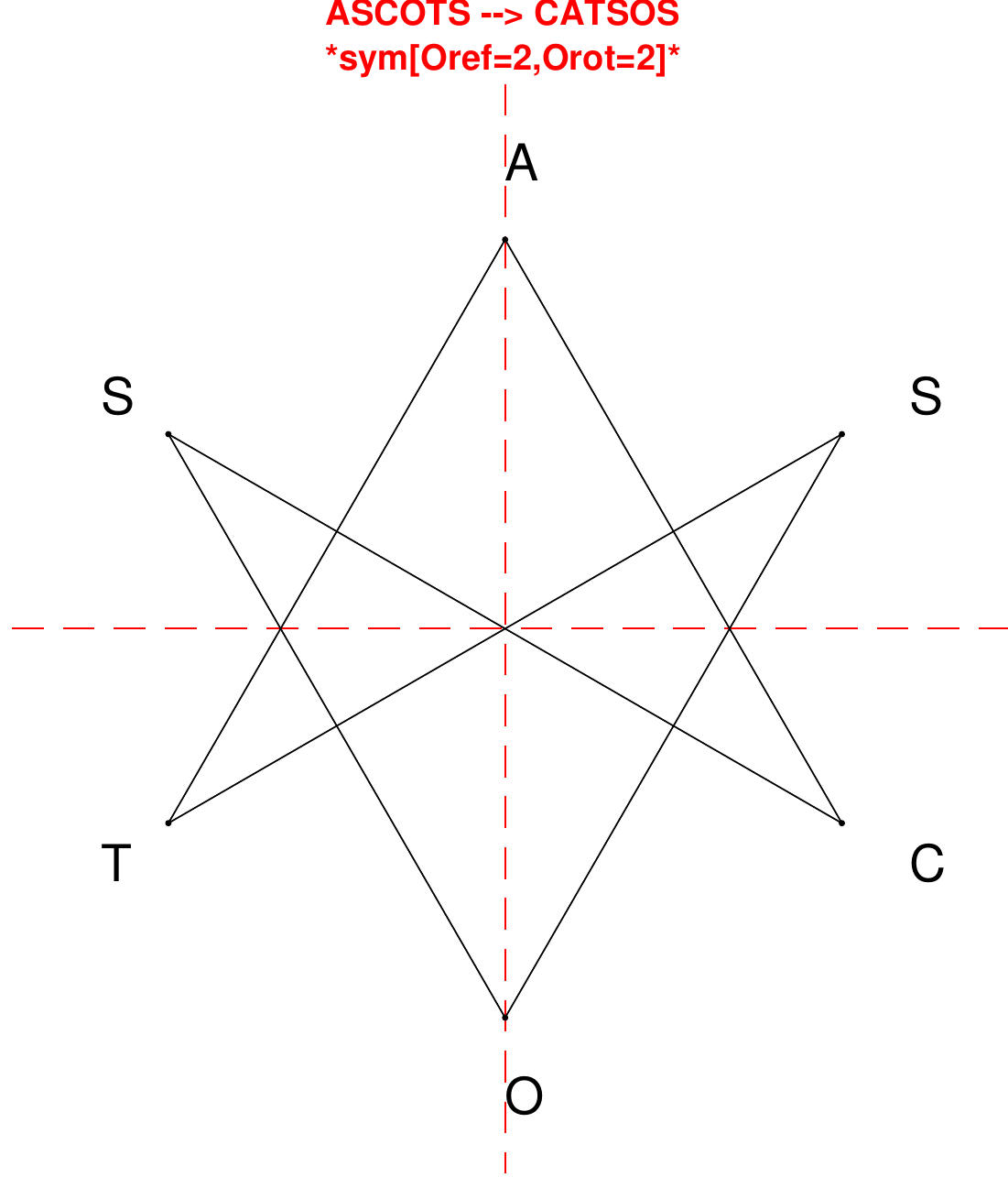}
\end{subfigure}
\hfill
\begin{subfigure}[T]{0.19\textwidth}
\centering
\includegraphics[width=\textwidth]{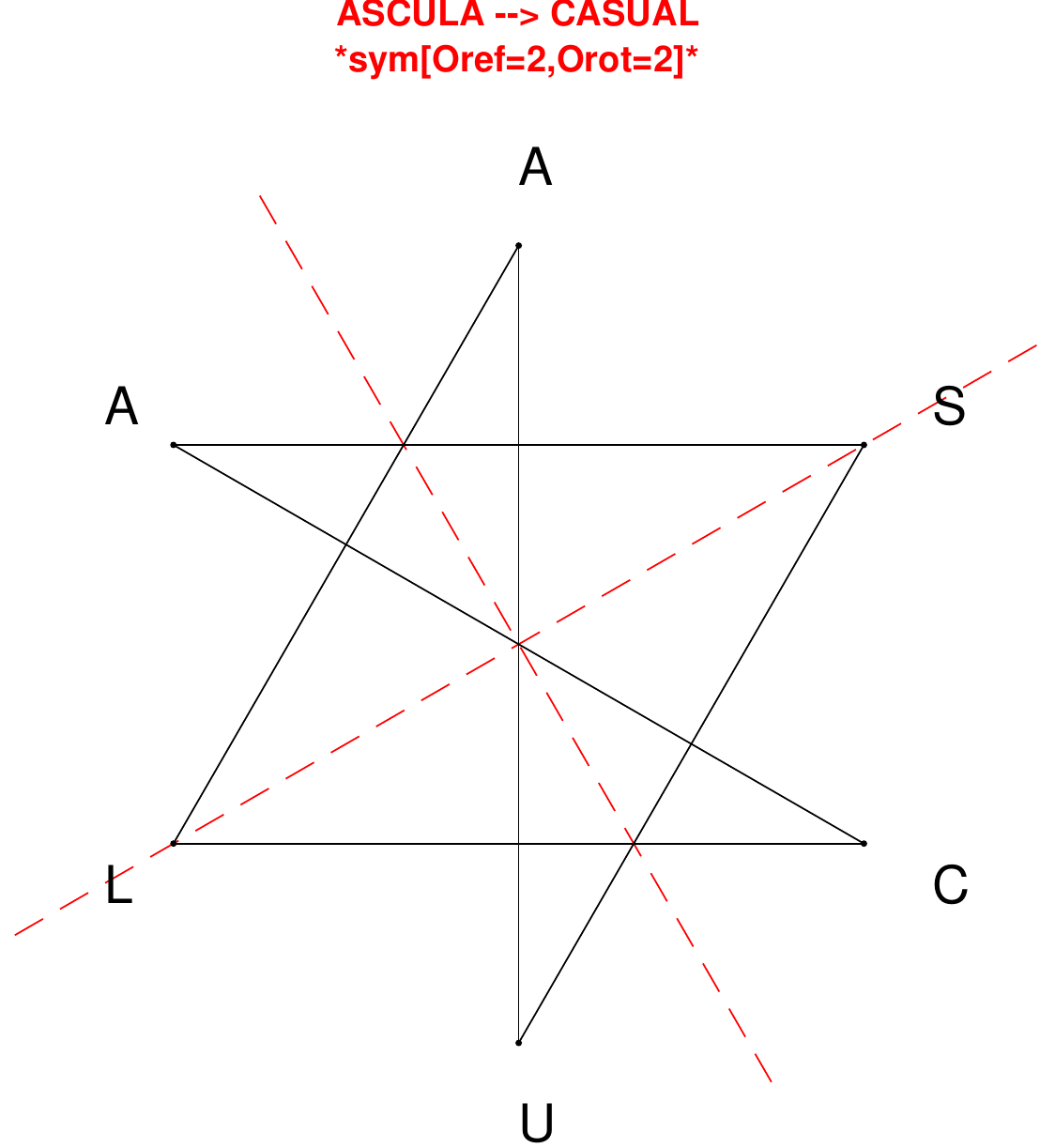}
\end{subfigure}
\hfill
\begin{subfigure}[T]{0.19\textwidth}
\centering
\includegraphics[width=\textwidth]{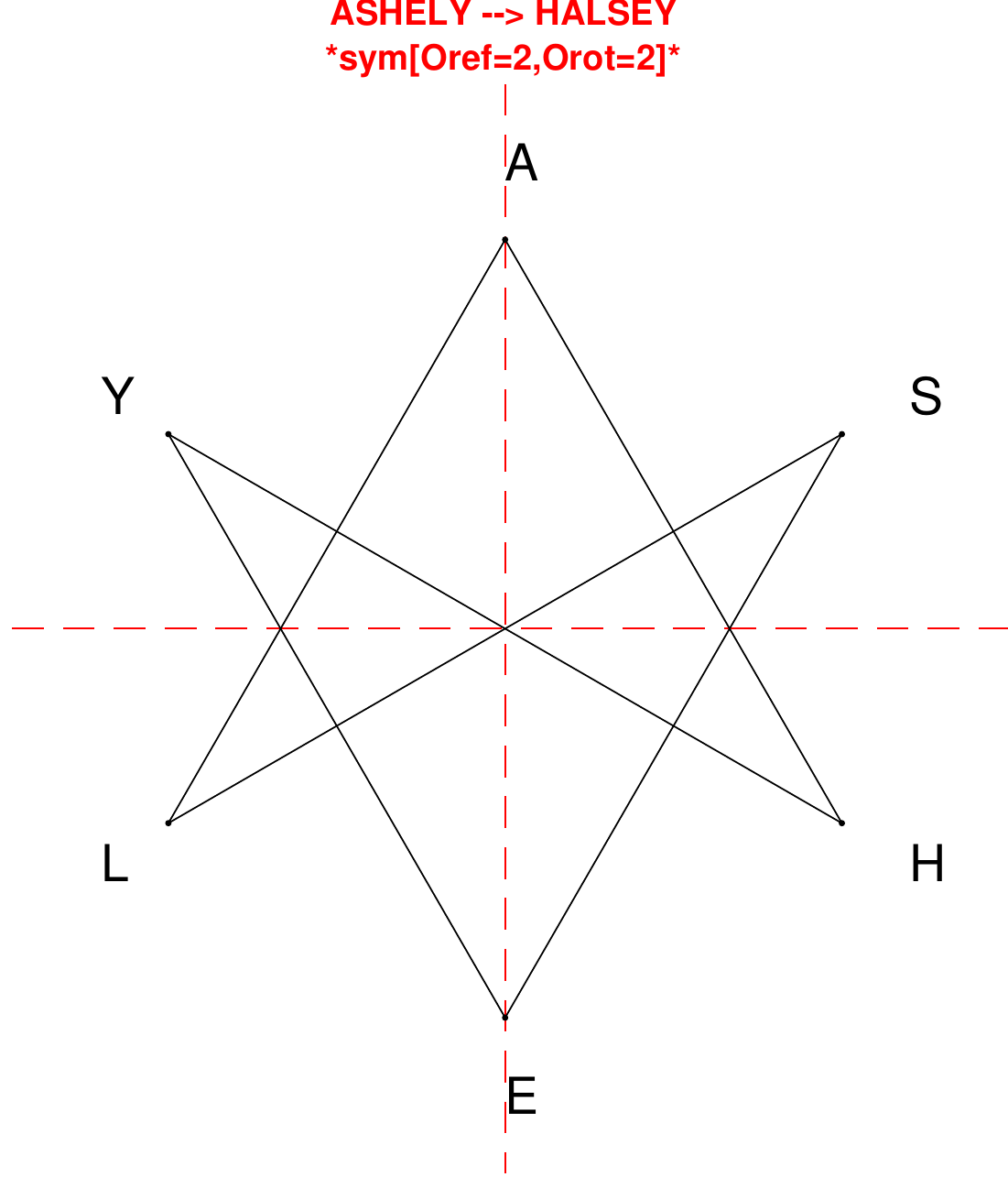}
\end{subfigure}
\end{figure}

\begin{figure}[H]
\centering
\begin{subfigure}[T]{0.19\textwidth}
\centering
\includegraphics[width=\textwidth]{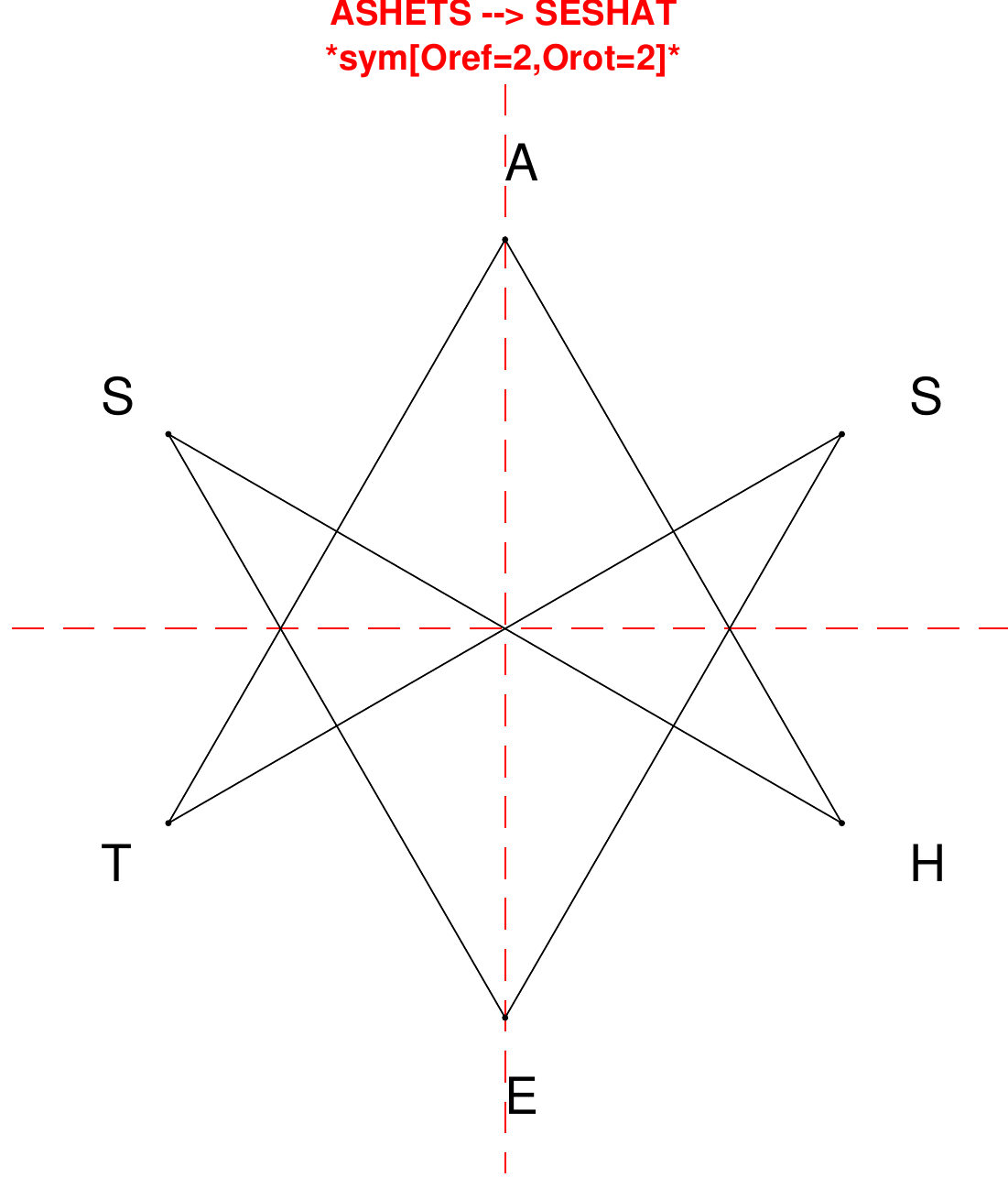}
\end{subfigure}
\hfill
\begin{subfigure}[T]{0.19\textwidth}
\centering
\includegraphics[width=\textwidth]{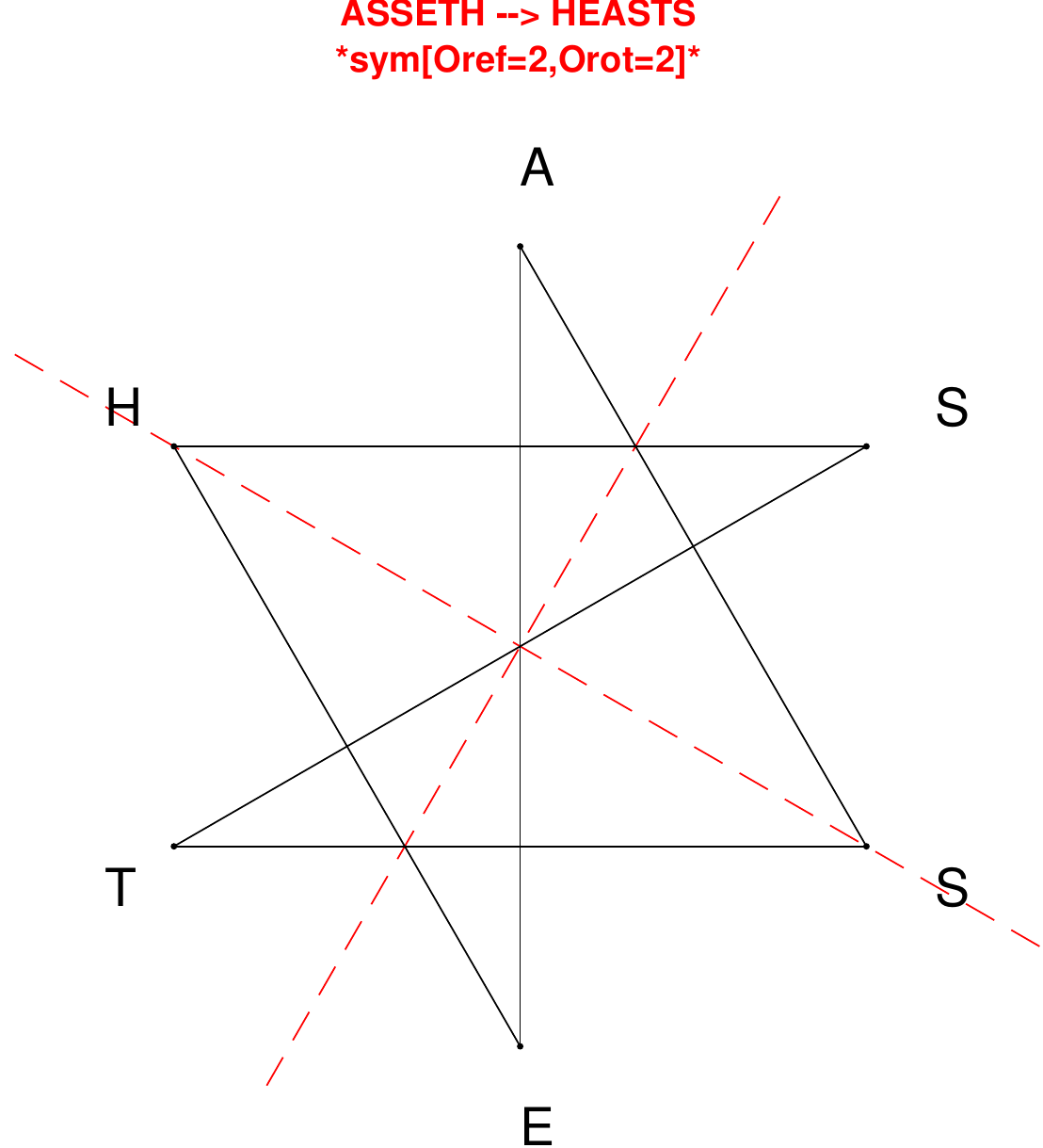}
\end{subfigure}
\hfill
\begin{subfigure}[T]{0.19\textwidth}
\centering
\includegraphics[width=\textwidth]{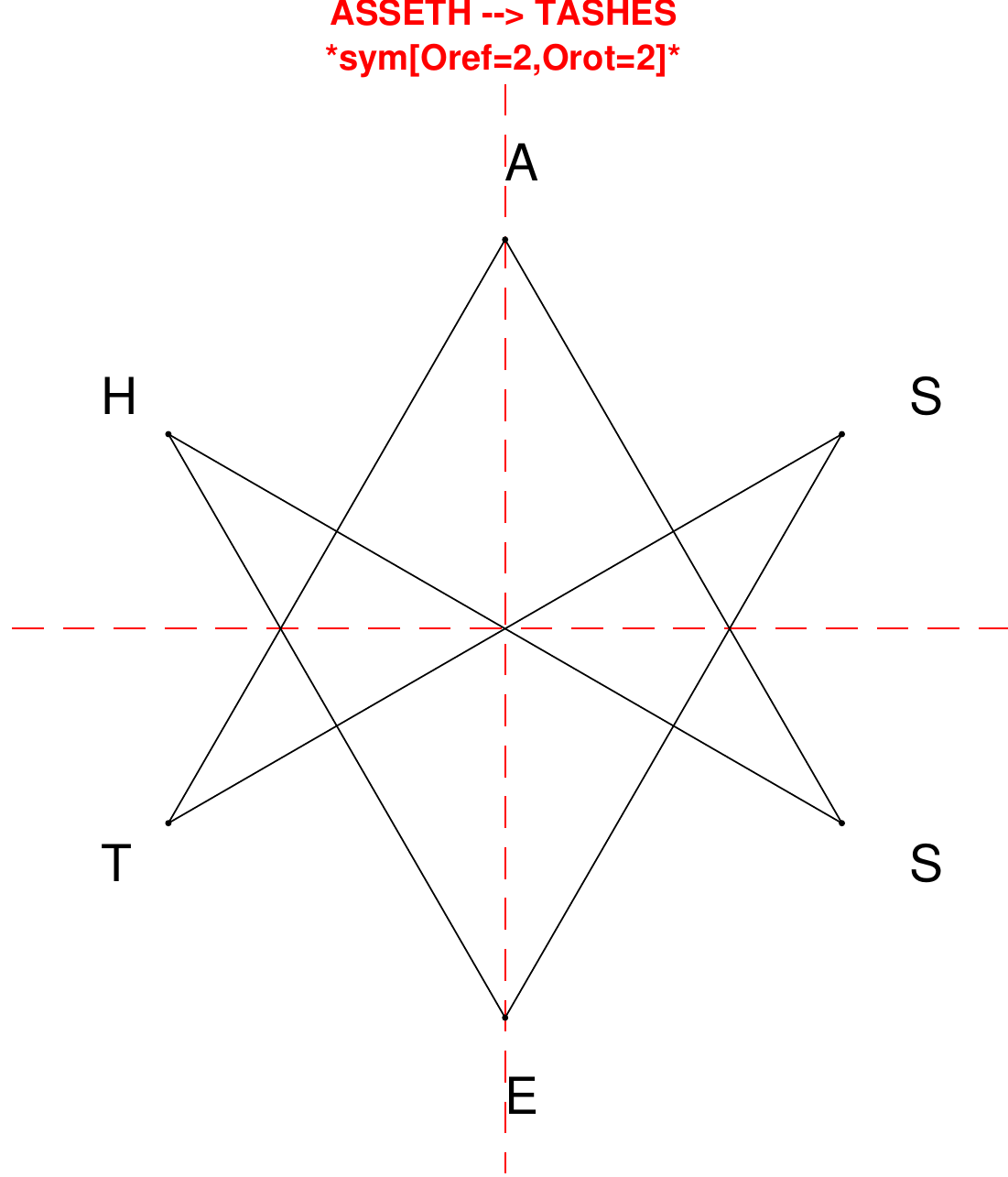}
\end{subfigure}
\hfill
\begin{subfigure}[T]{0.19\textwidth}
\centering
\includegraphics[width=\textwidth]{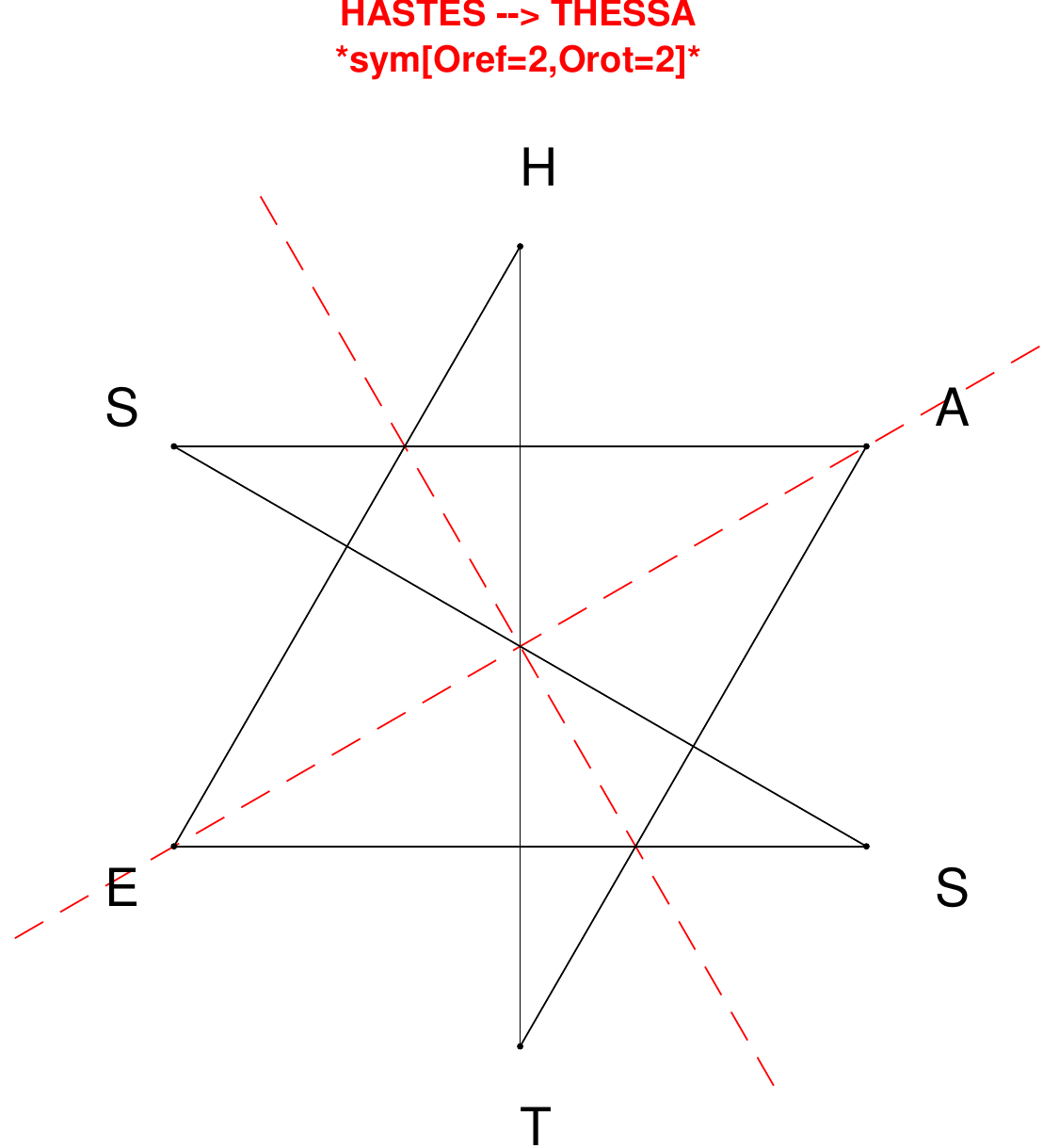}
\end{subfigure}
\hfill
\begin{subfigure}[T]{0.19\textwidth}
\centering
\includegraphics[width=\textwidth]{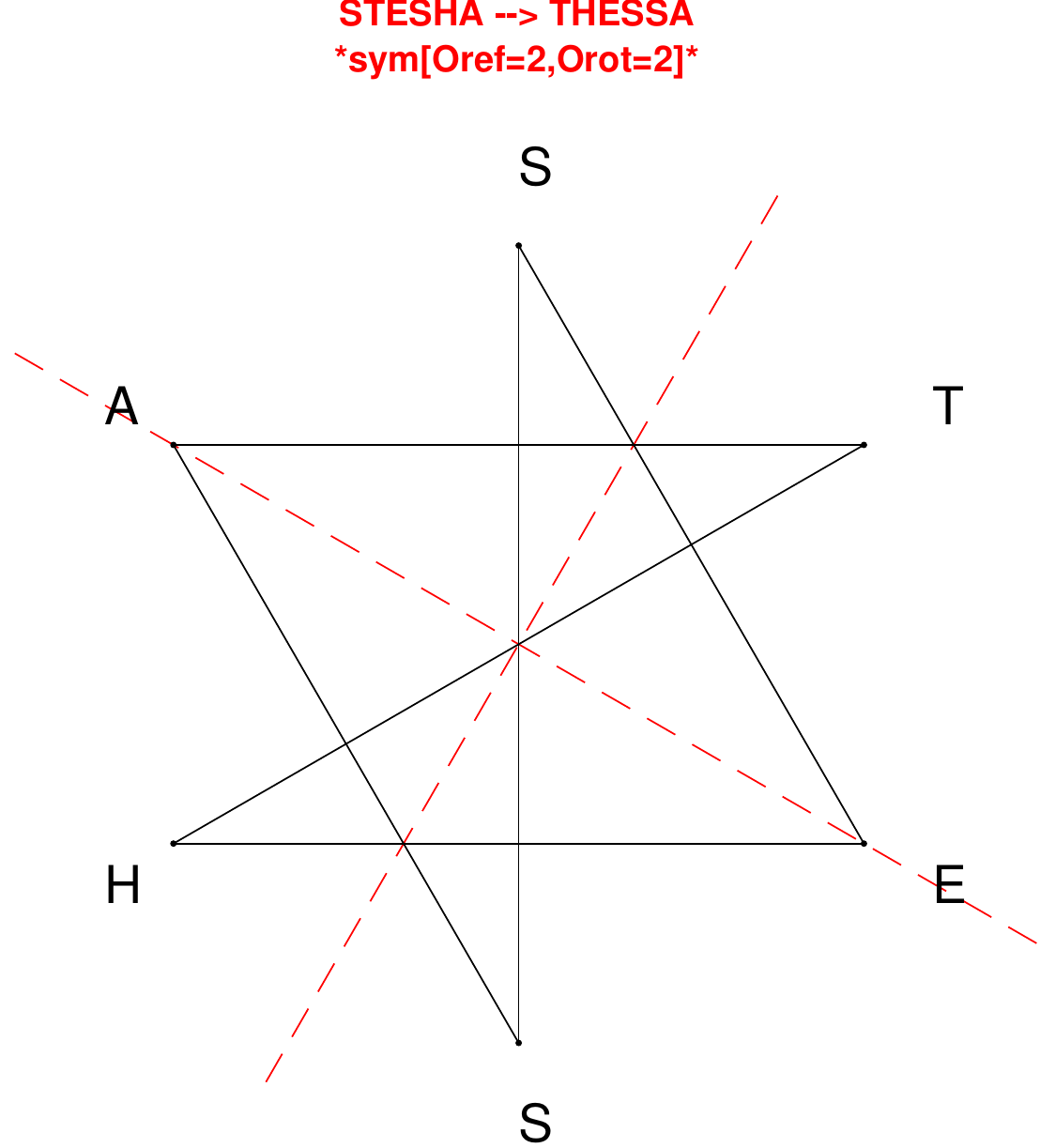}
\end{subfigure}
\end{figure}

\begin{figure}[H]
\centering
\begin{subfigure}[T]{0.19\textwidth}
\centering
\includegraphics[width=\textwidth]{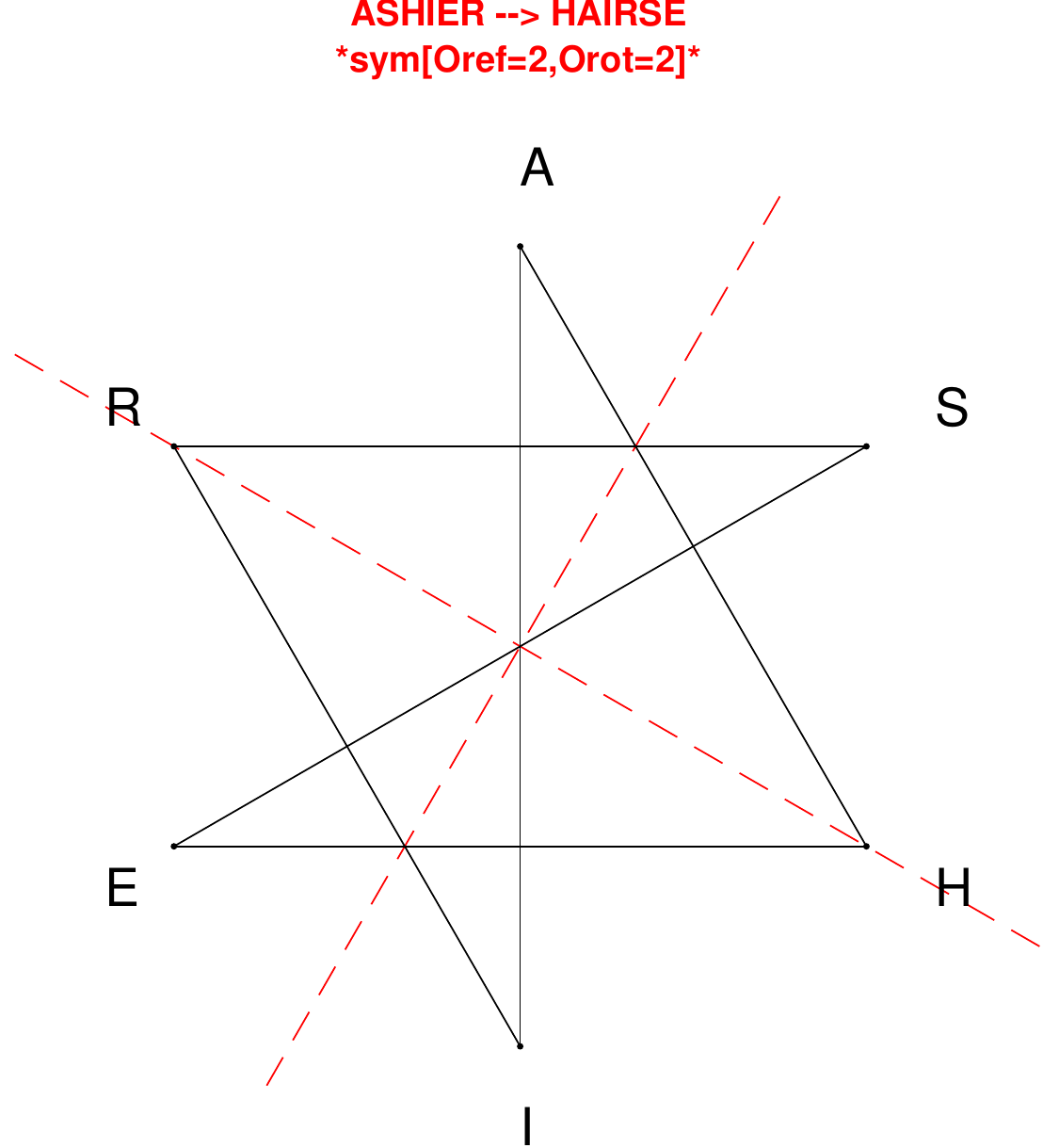}
\end{subfigure}
\hfill
\begin{subfigure}[T]{0.19\textwidth}
\centering
\includegraphics[width=\textwidth]{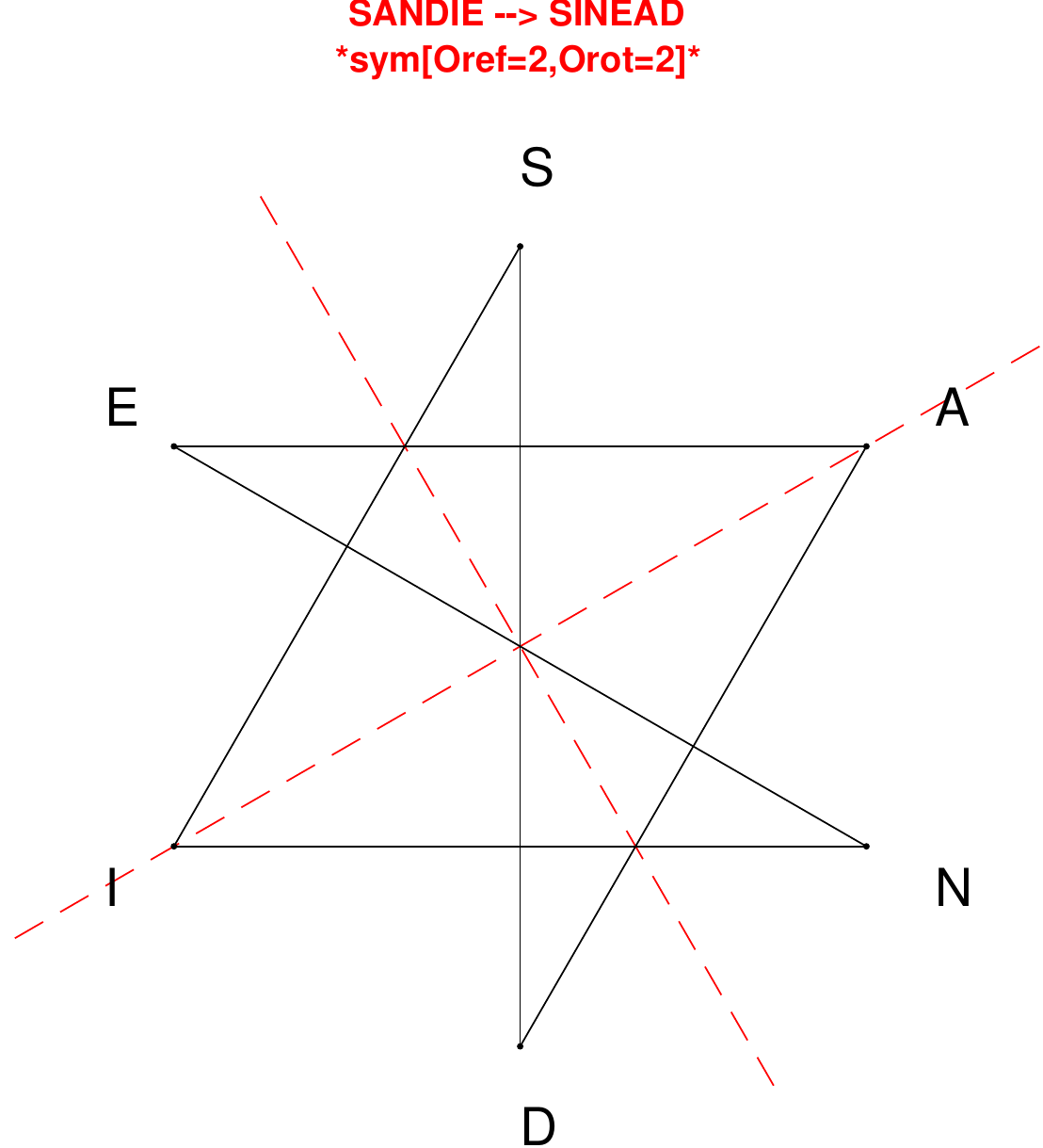}
\end{subfigure}
\hfill
\begin{subfigure}[T]{0.19\textwidth}
\centering
\includegraphics[width=\textwidth]{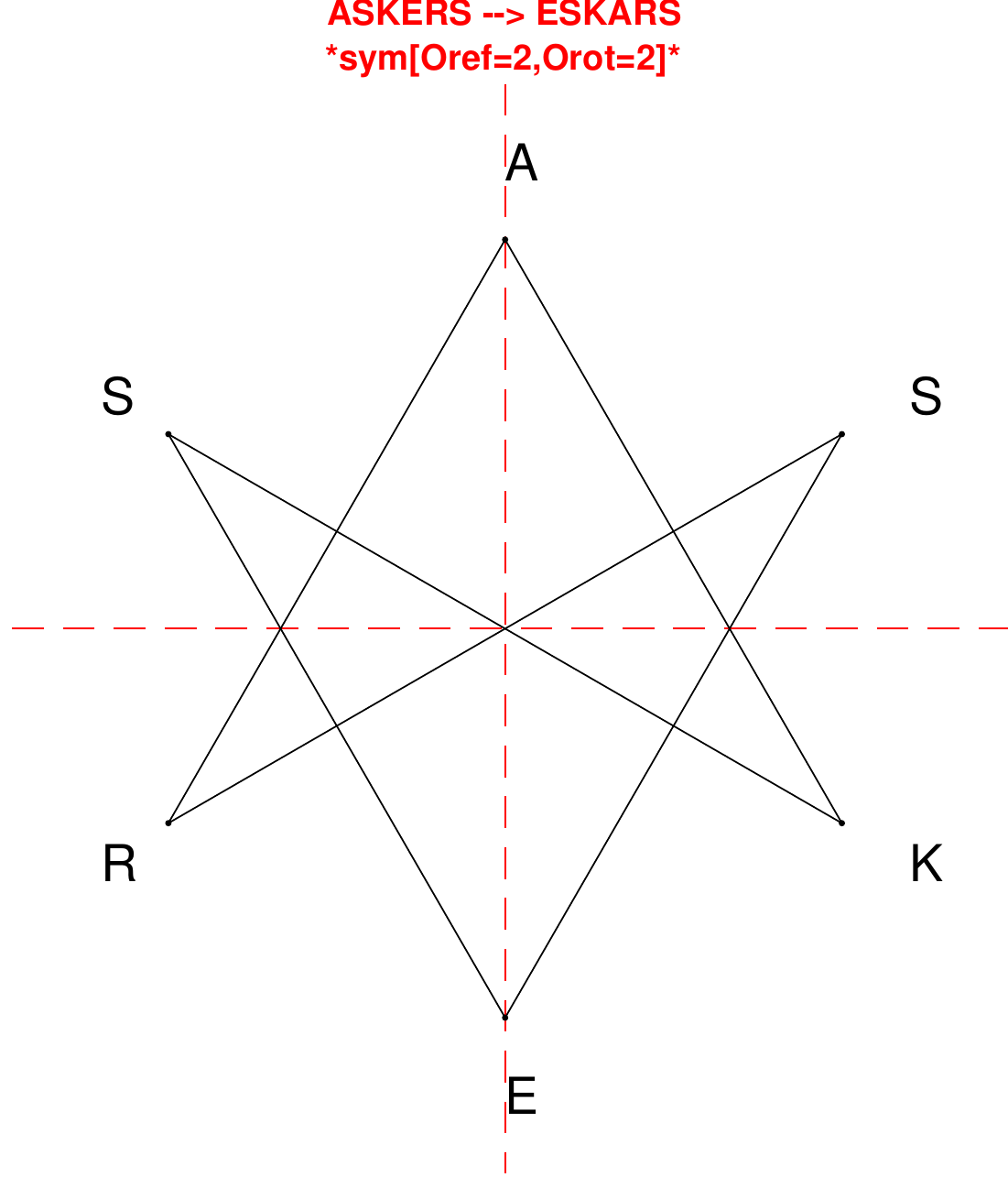}
\end{subfigure}
\hfill
\begin{subfigure}[T]{0.19\textwidth}
\centering
\includegraphics[width=\textwidth]{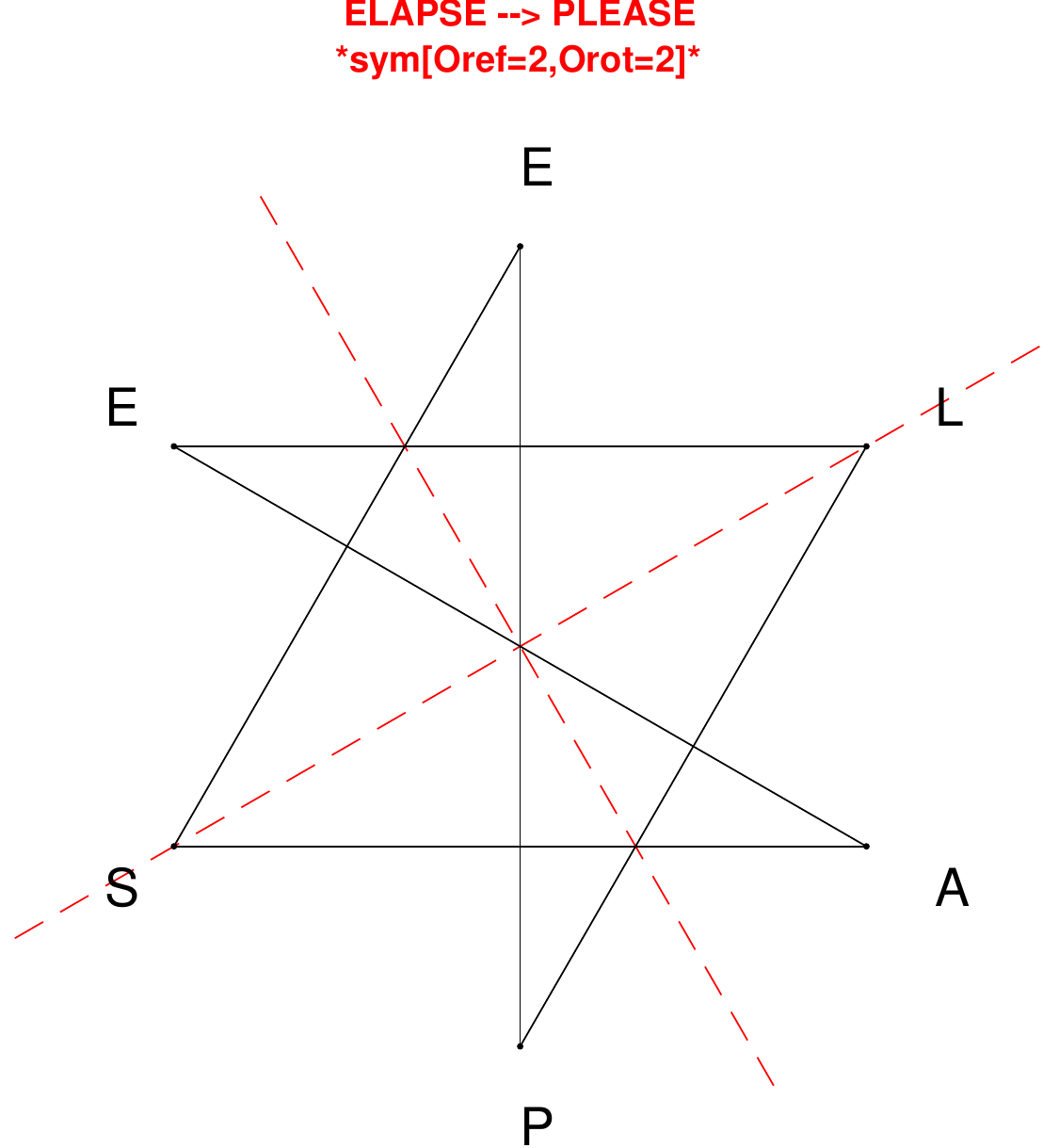}
\end{subfigure}
\hfill
\begin{subfigure}[T]{0.19\textwidth}
\centering
\includegraphics[width=\textwidth]{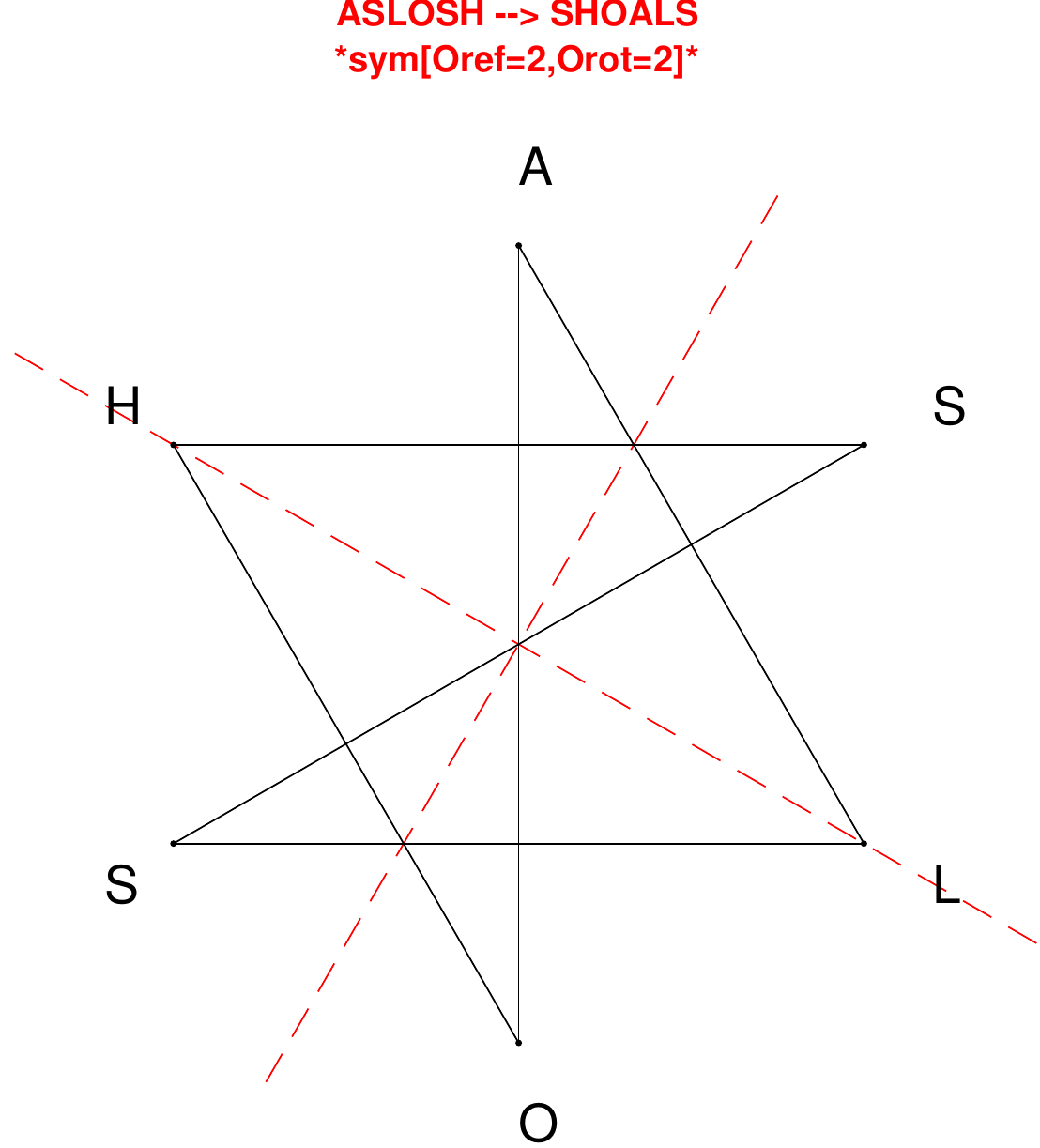}
\end{subfigure}
\end{figure}

\begin{figure}[H]
\centering
\begin{subfigure}[T]{0.19\textwidth}
\centering
\includegraphics[width=\textwidth]{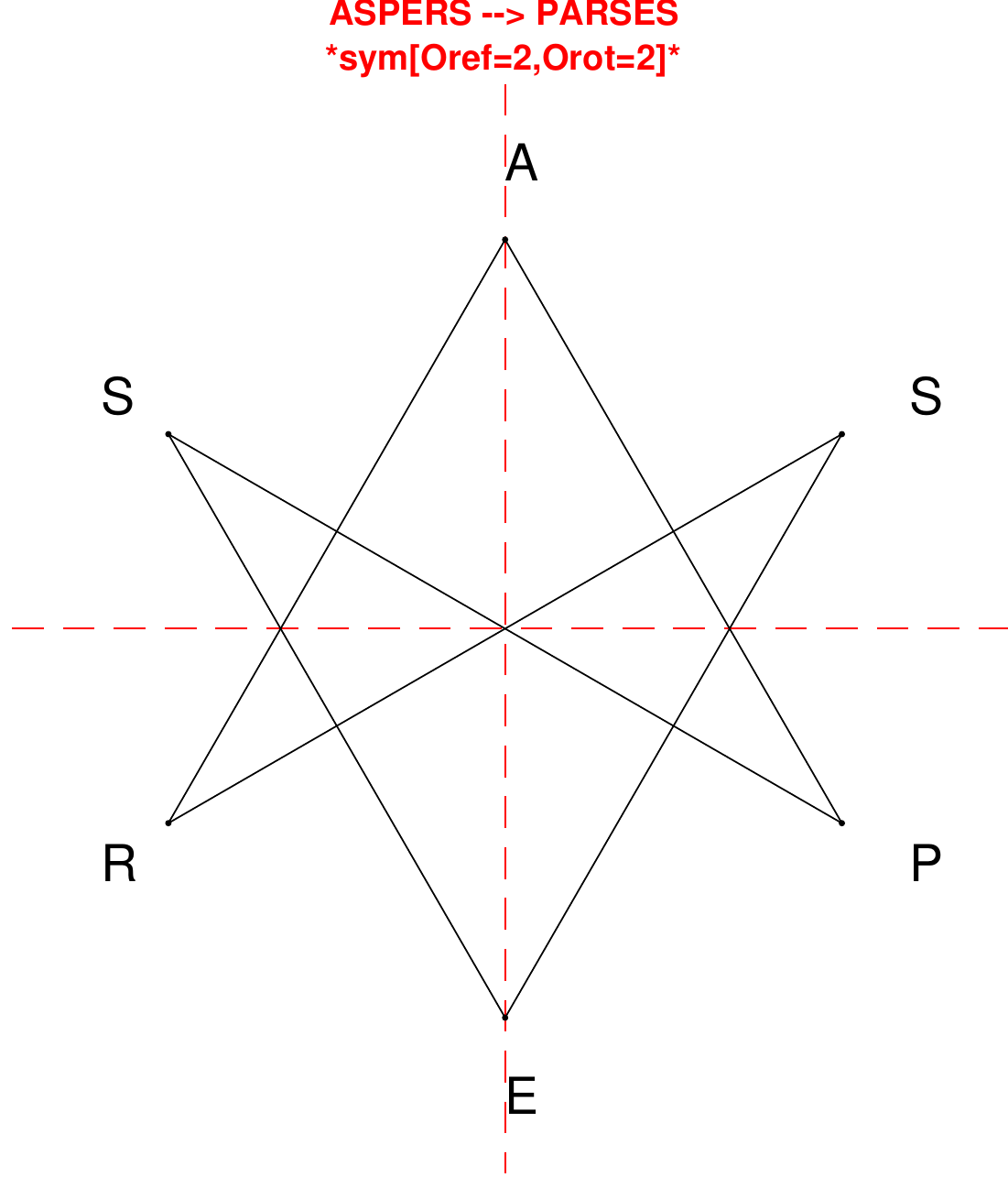}
\end{subfigure}
\hfill
\begin{subfigure}[T]{0.19\textwidth}
\centering
\includegraphics[width=\textwidth]{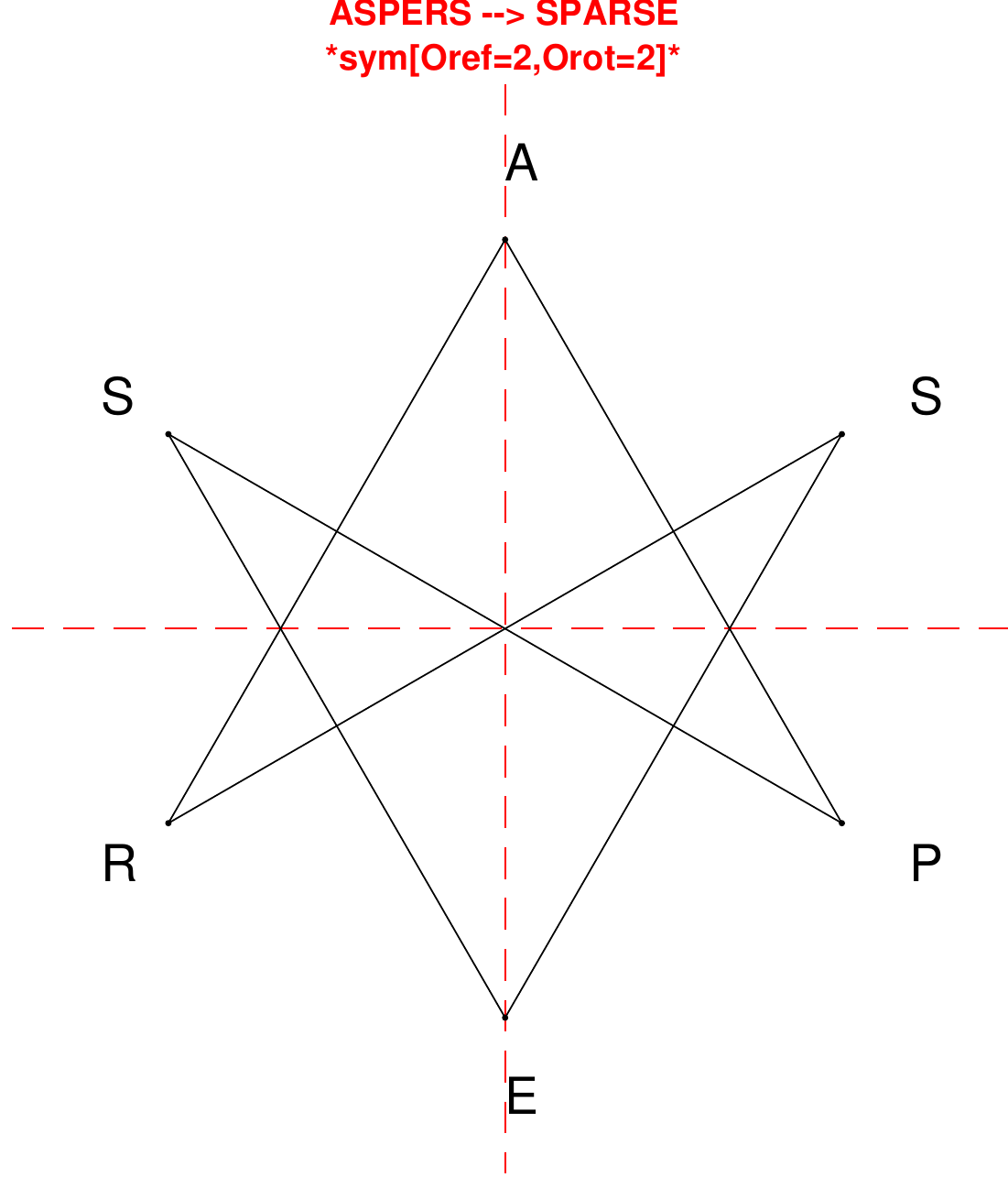}
\end{subfigure}
\hfill
\begin{subfigure}[T]{0.19\textwidth}
\centering
\includegraphics[width=\textwidth]{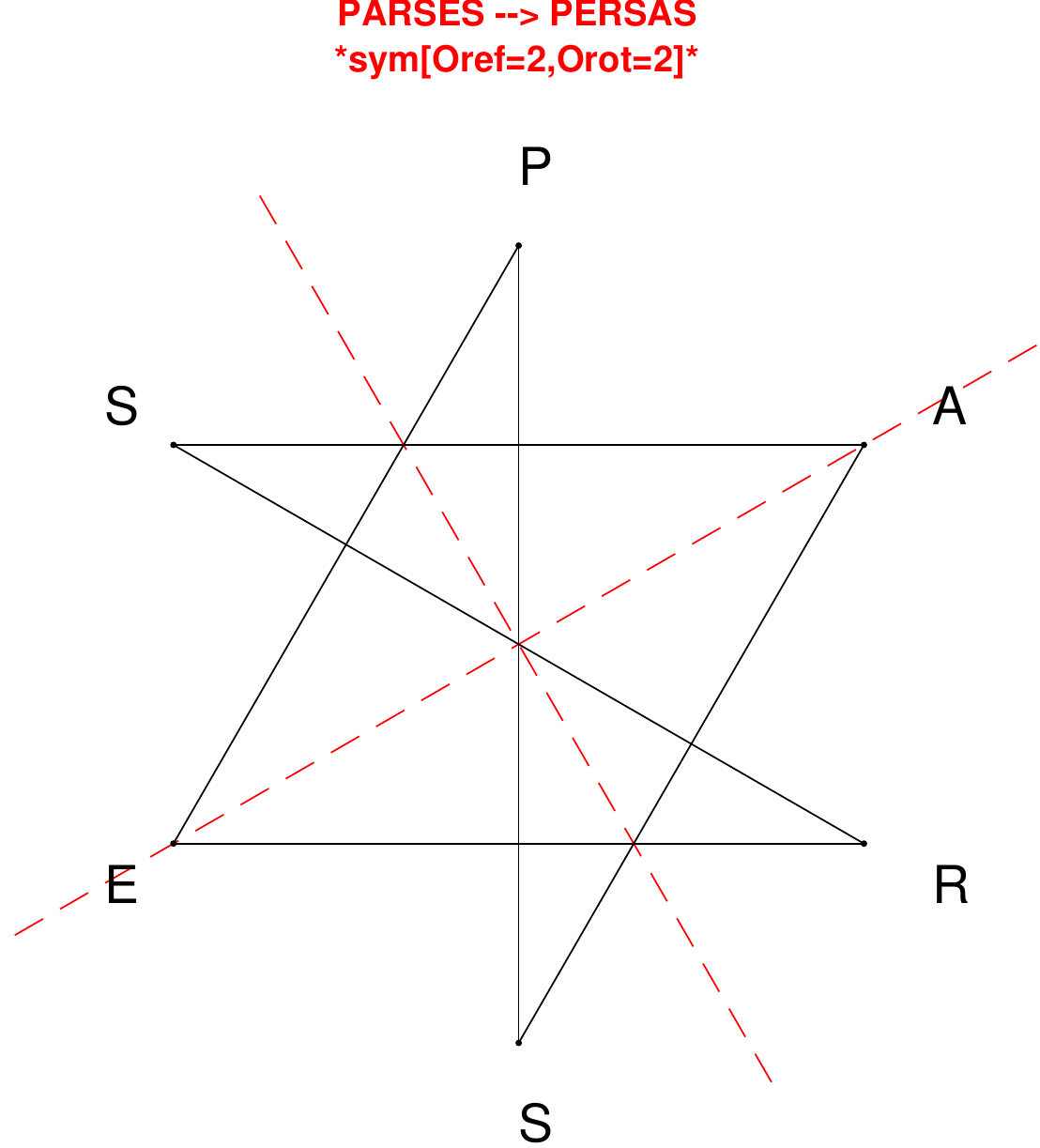}
\end{subfigure}
\hfill
\begin{subfigure}[T]{0.19\textwidth}
\centering
\includegraphics[width=\textwidth]{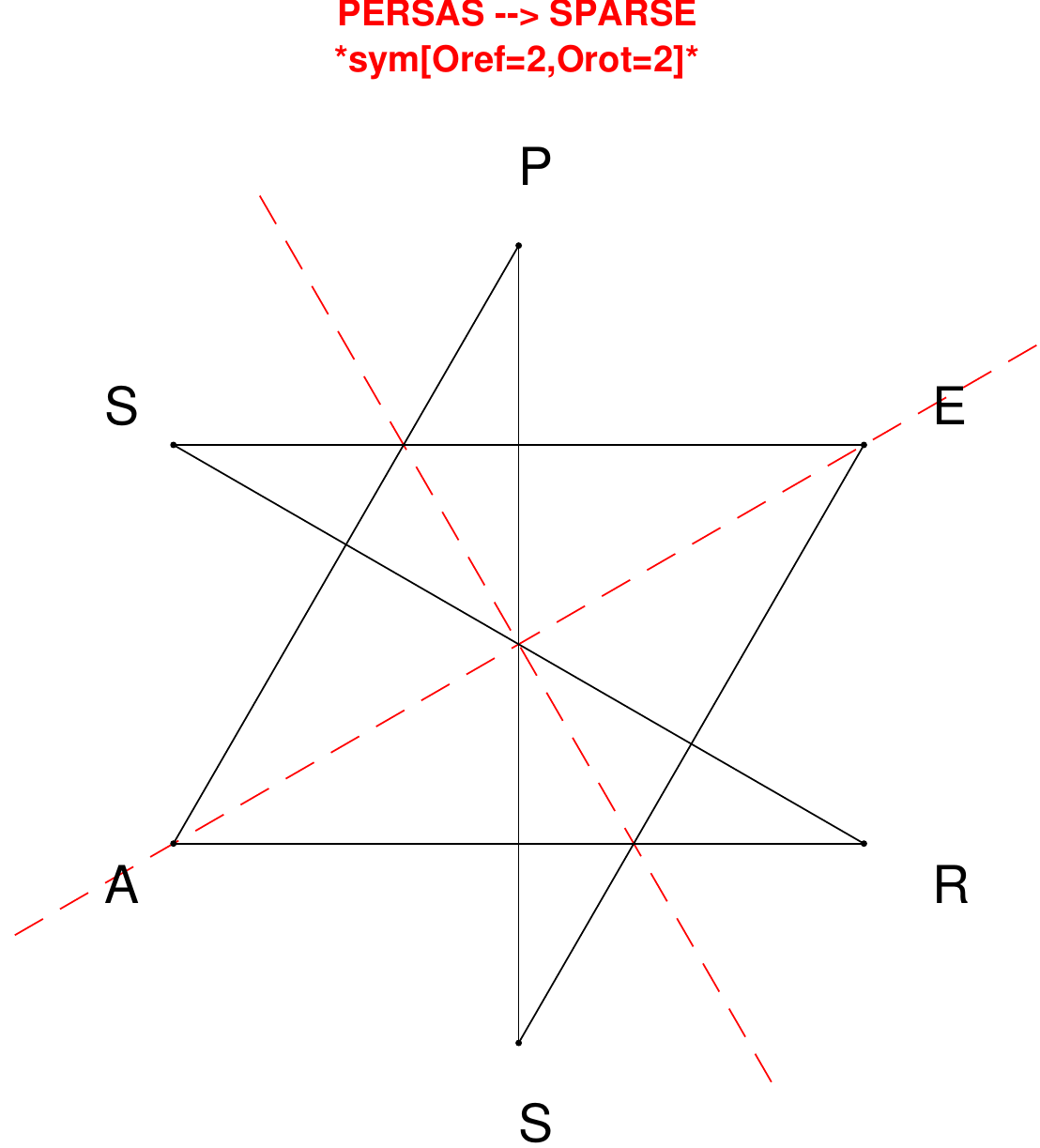}
\end{subfigure}
\hfill
\begin{subfigure}[T]{0.19\textwidth}
\centering
\includegraphics[width=\textwidth]{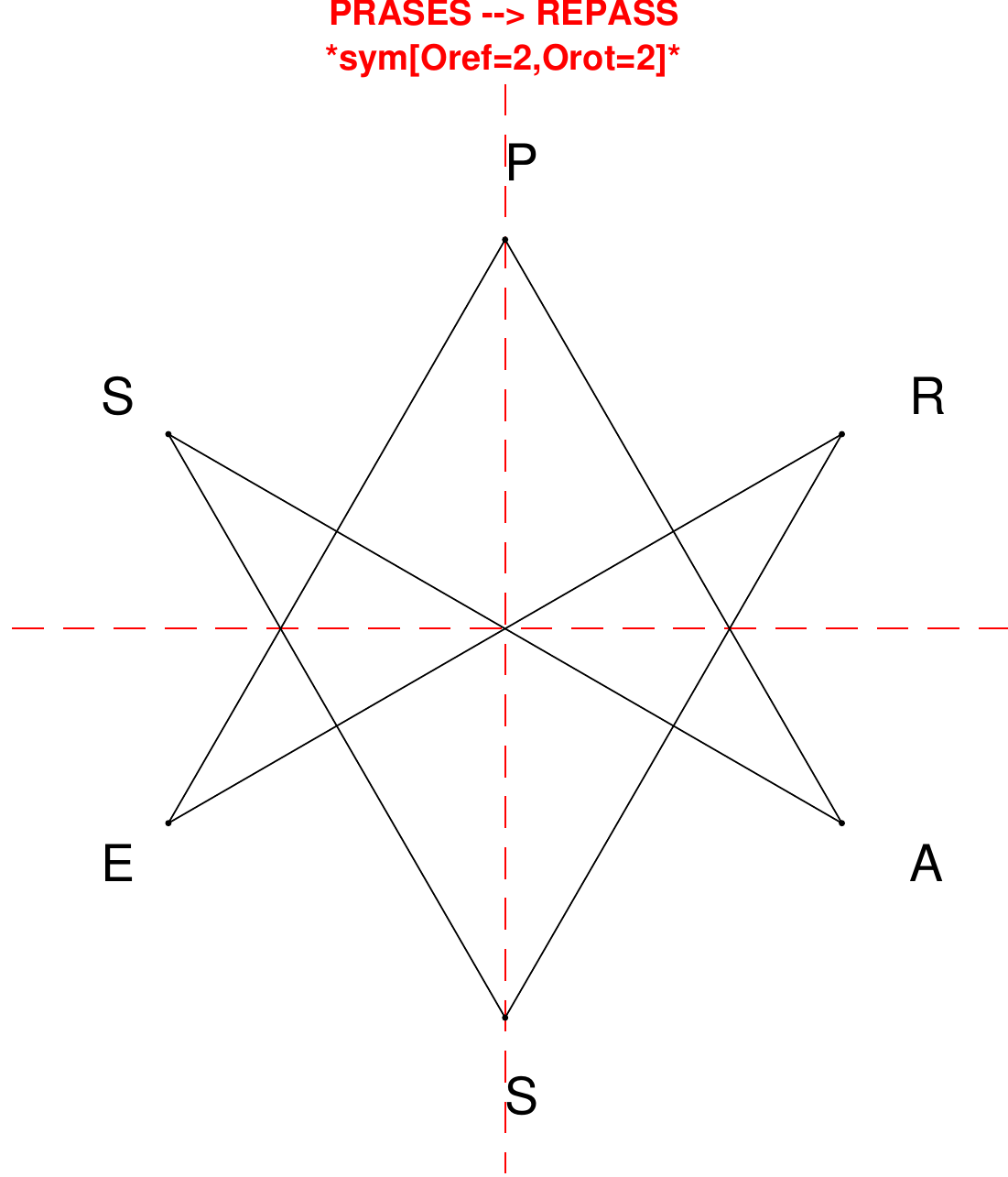}
\end{subfigure}
\end{figure}

\begin{figure}[H]
\centering
\begin{subfigure}[T]{0.19\textwidth}
\centering
\includegraphics[width=\textwidth]{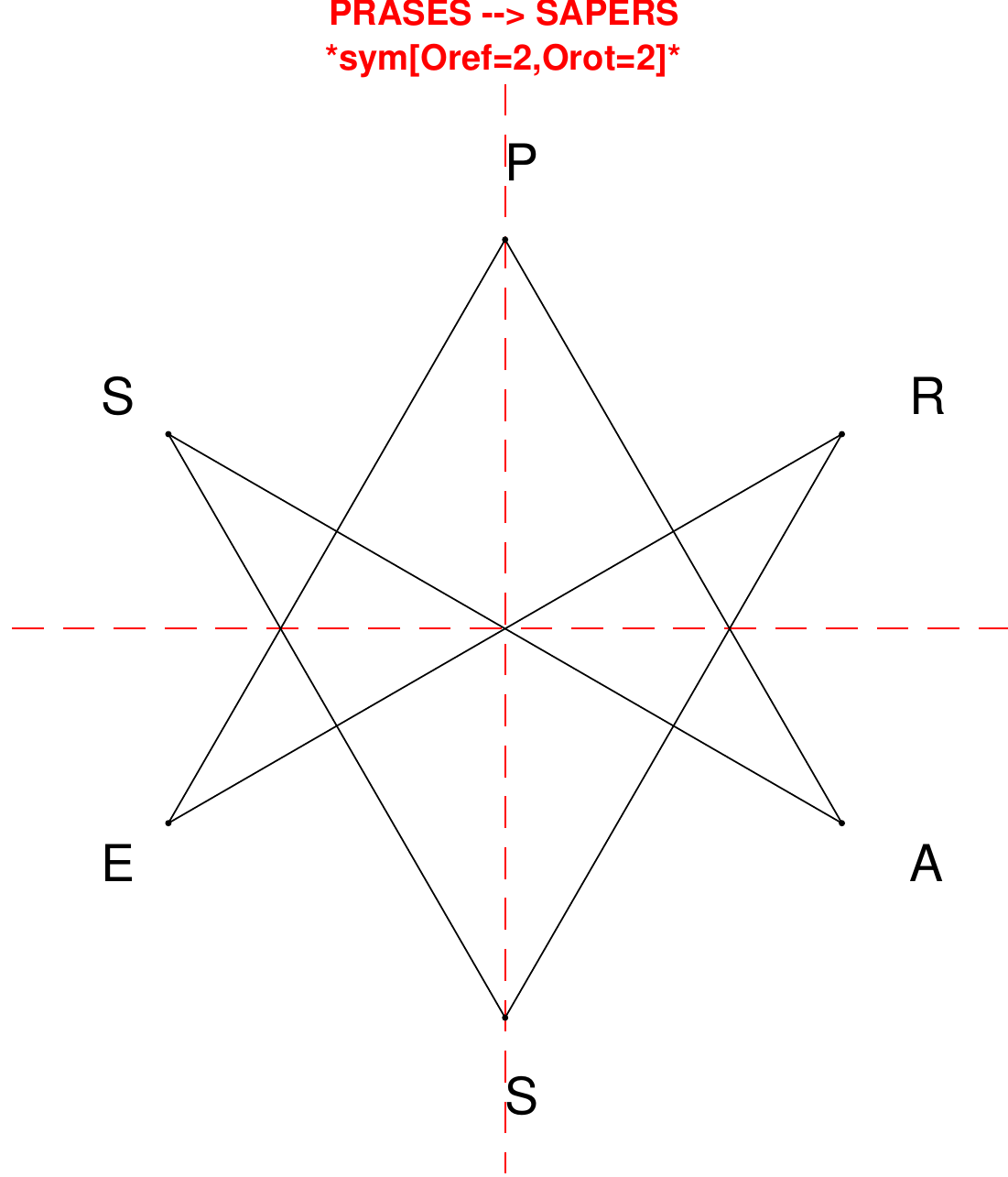}
\end{subfigure}
\hfill
\begin{subfigure}[T]{0.19\textwidth}
\centering
\includegraphics[width=\textwidth]{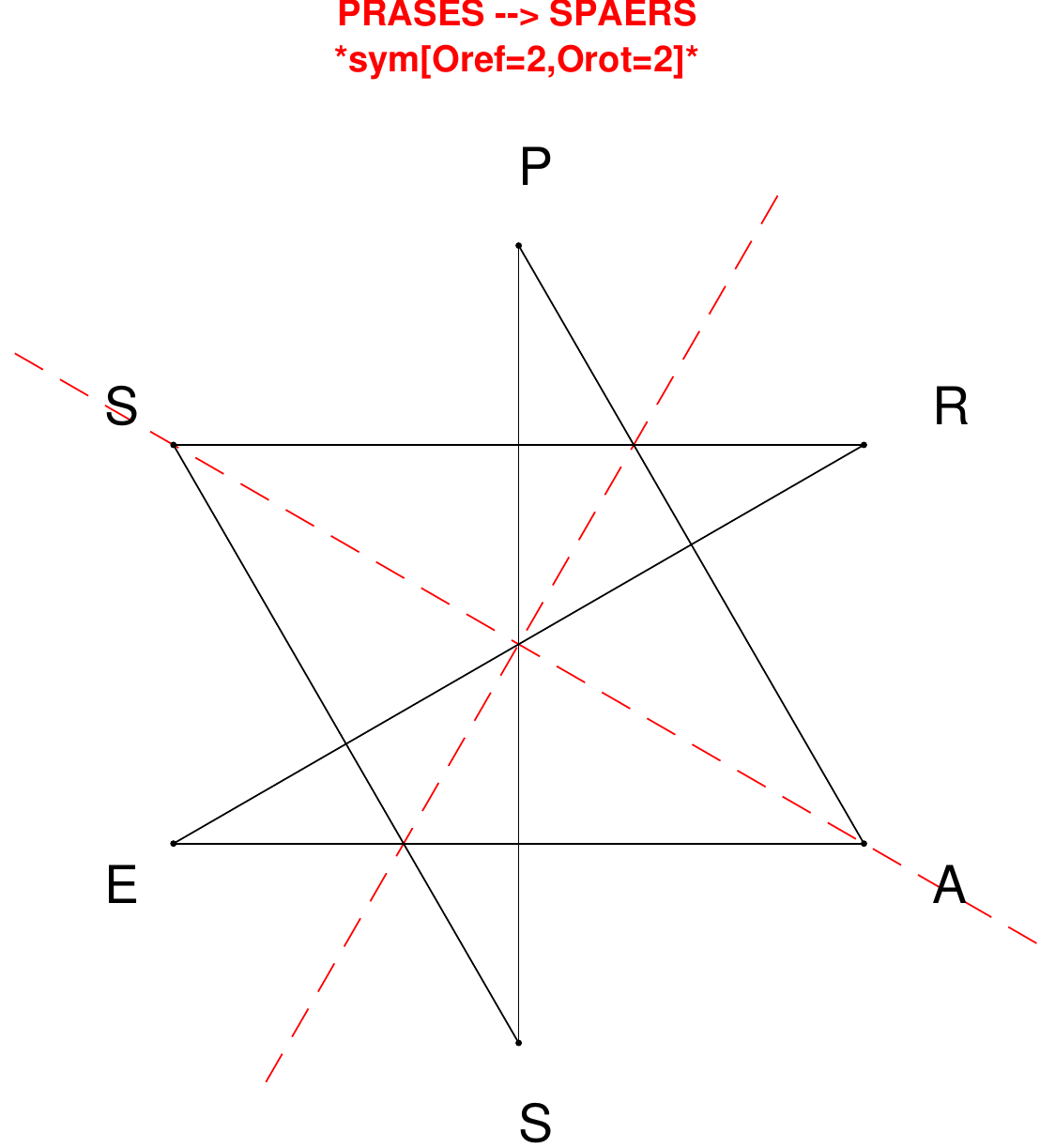}
\end{subfigure}
\hfill
\begin{subfigure}[T]{0.19\textwidth}
\centering
\includegraphics[width=\textwidth]{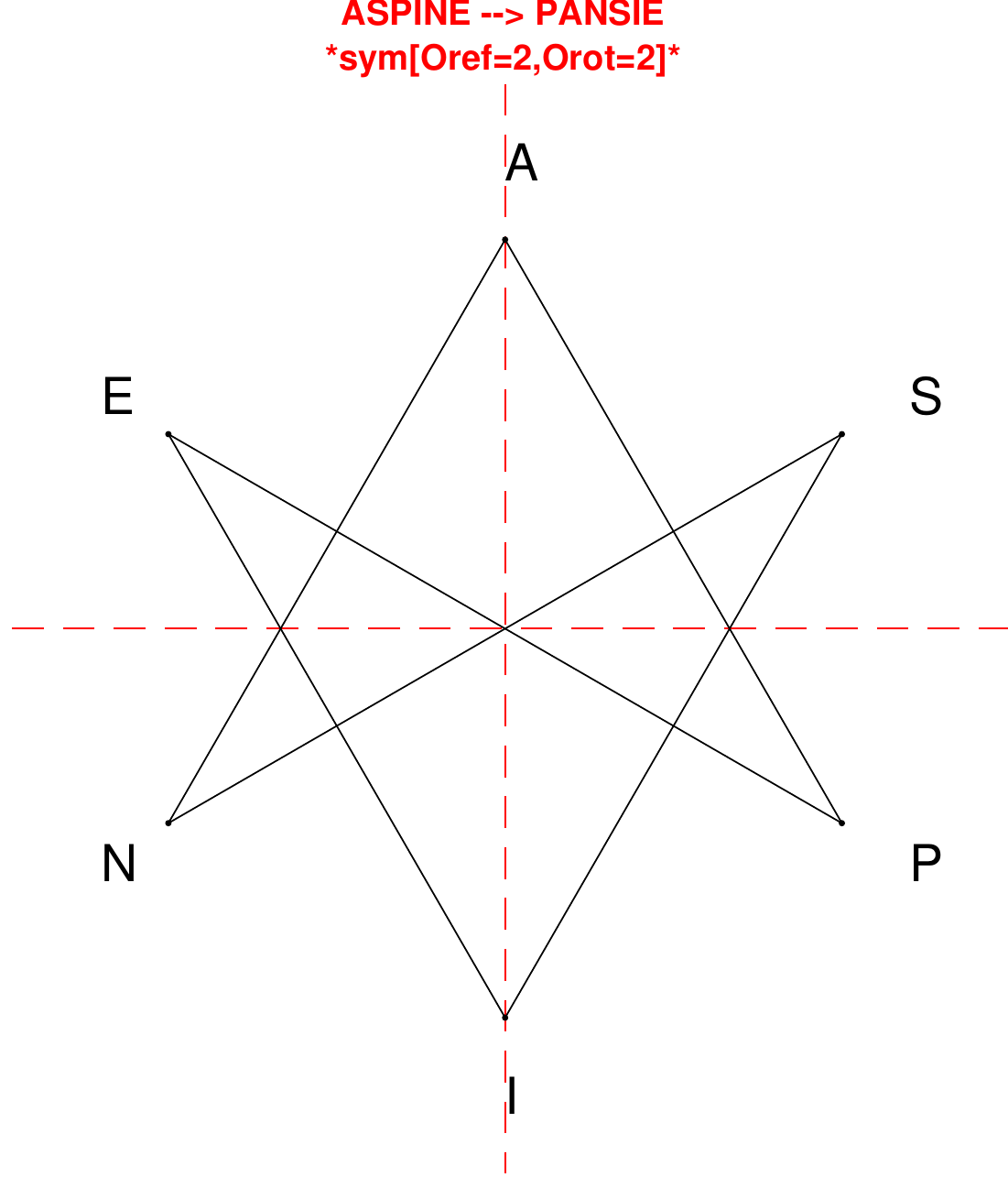}
\end{subfigure}
\hfill
\begin{subfigure}[T]{0.19\textwidth}
\centering
\includegraphics[width=\textwidth]{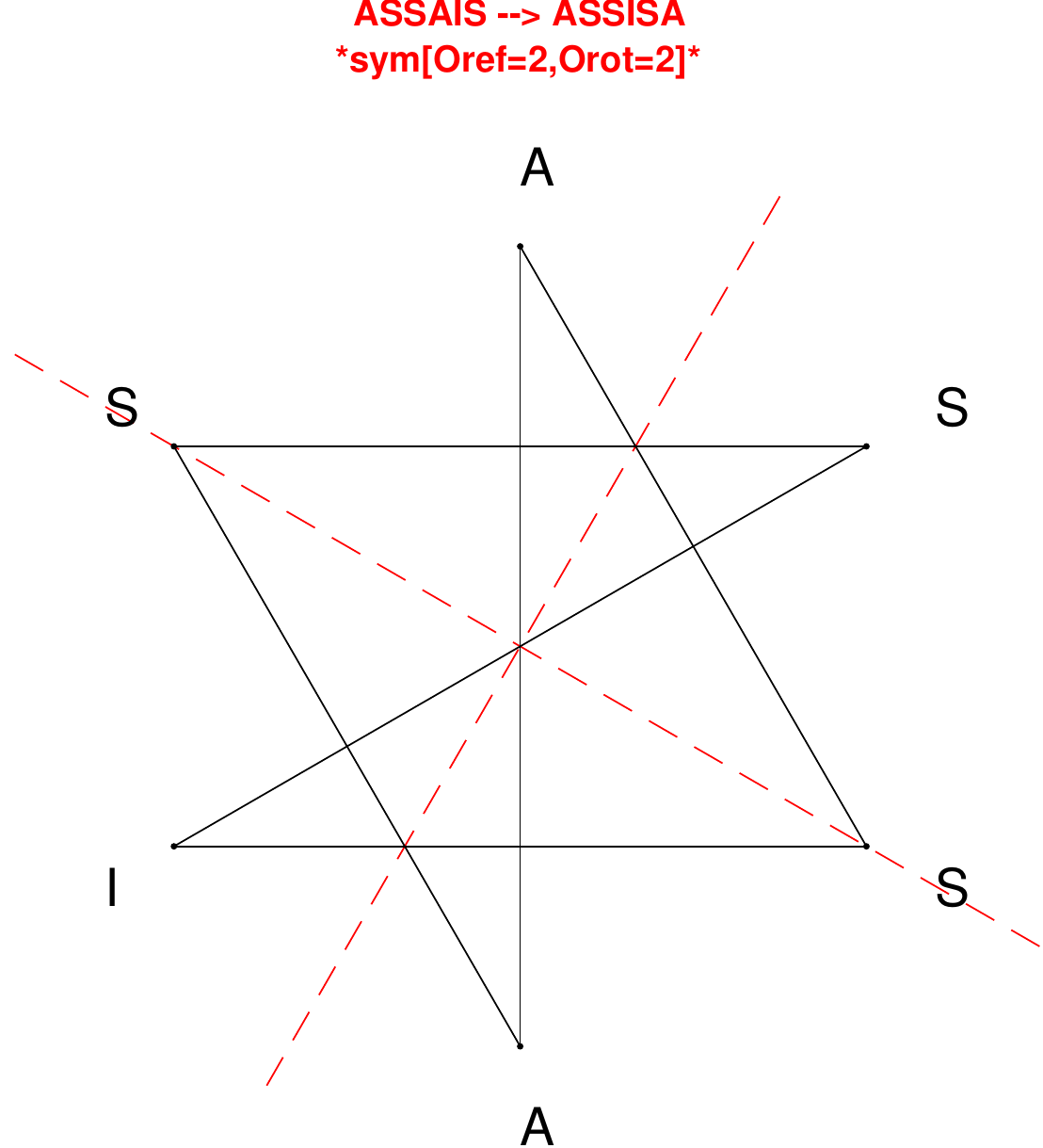}
\end{subfigure}
\hfill
\begin{subfigure}[T]{0.19\textwidth}
\centering
\includegraphics[width=\textwidth]{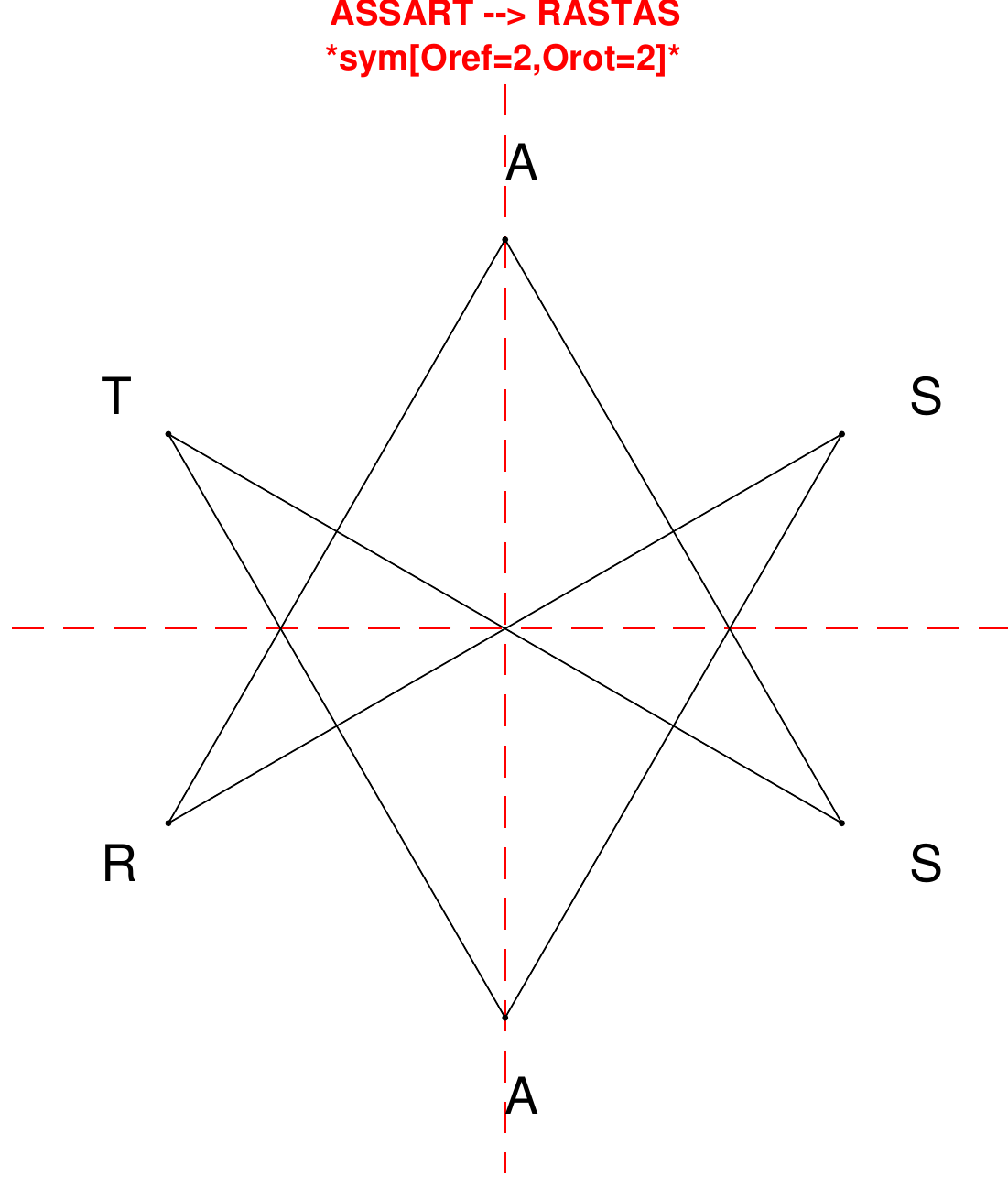}
\end{subfigure}
\end{figure}

\begin{figure}[H]
\centering
\begin{subfigure}[T]{0.19\textwidth}
\centering
\includegraphics[width=\textwidth]{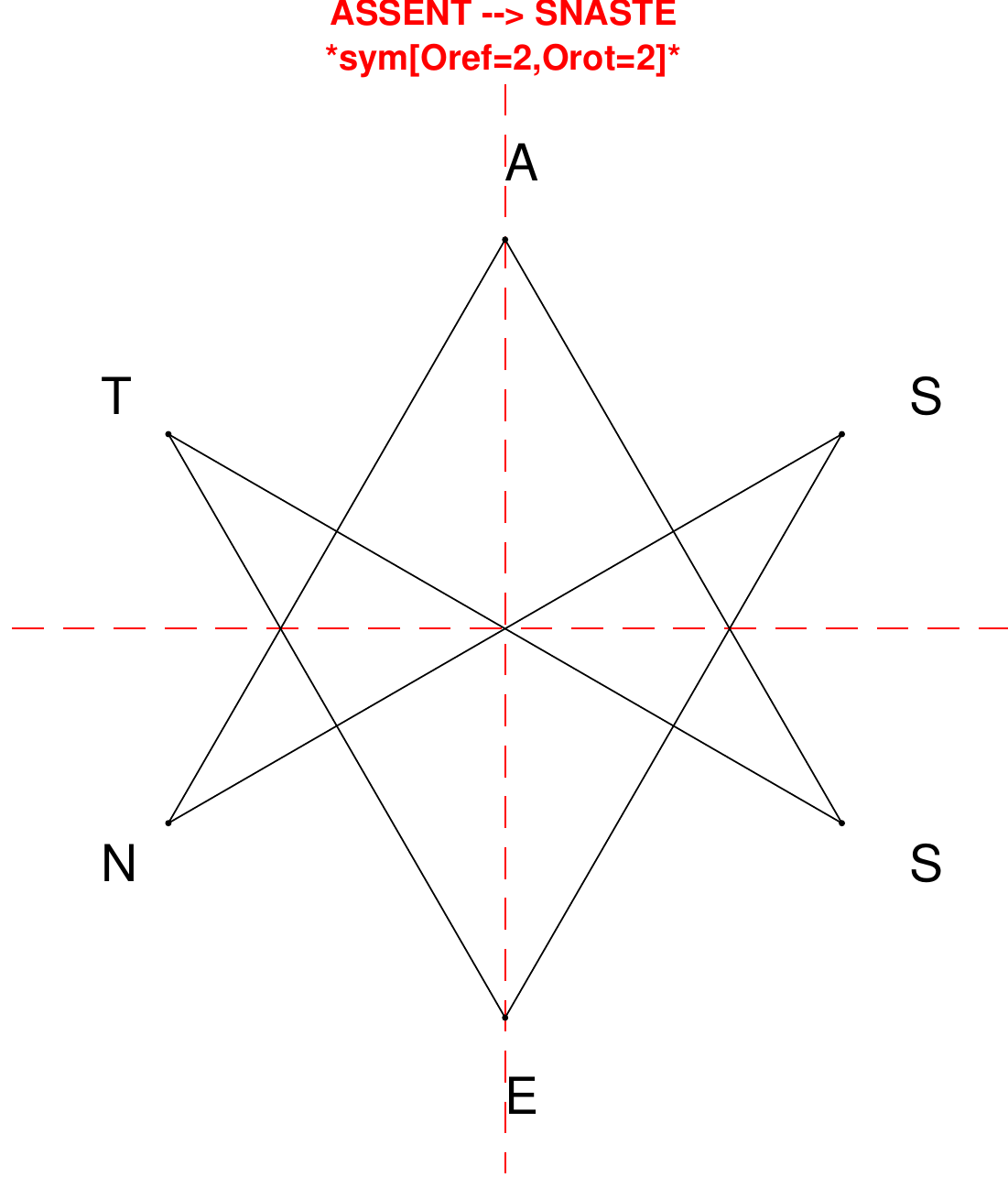}
\end{subfigure}
\hfill
\begin{subfigure}[T]{0.19\textwidth}
\centering
\includegraphics[width=\textwidth]{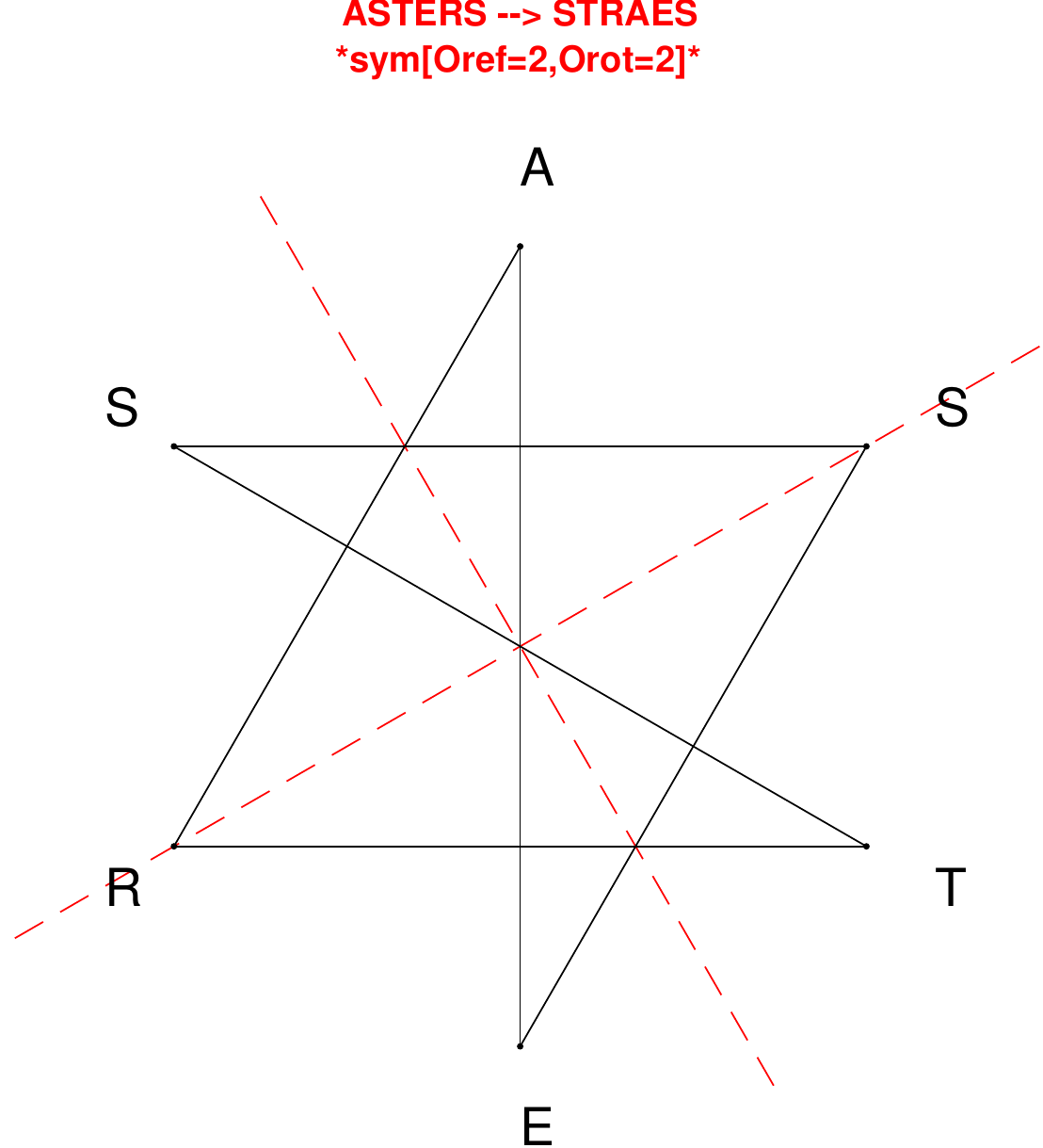}
\end{subfigure}
\hfill
\begin{subfigure}[T]{0.19\textwidth}
\centering
\includegraphics[width=\textwidth]{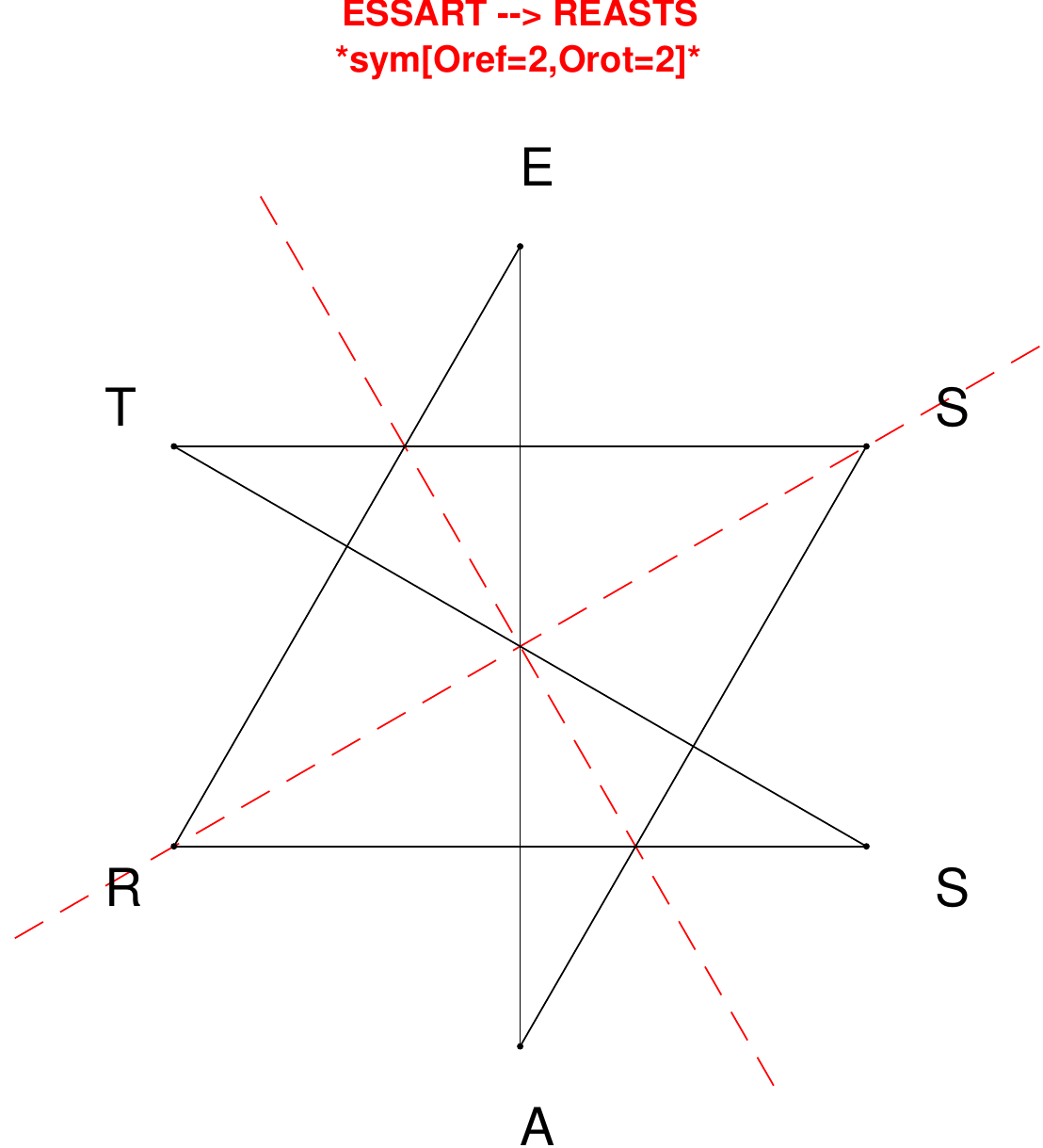}
\end{subfigure}
\hfill
\begin{subfigure}[T]{0.19\textwidth}
\centering
\includegraphics[width=\textwidth]{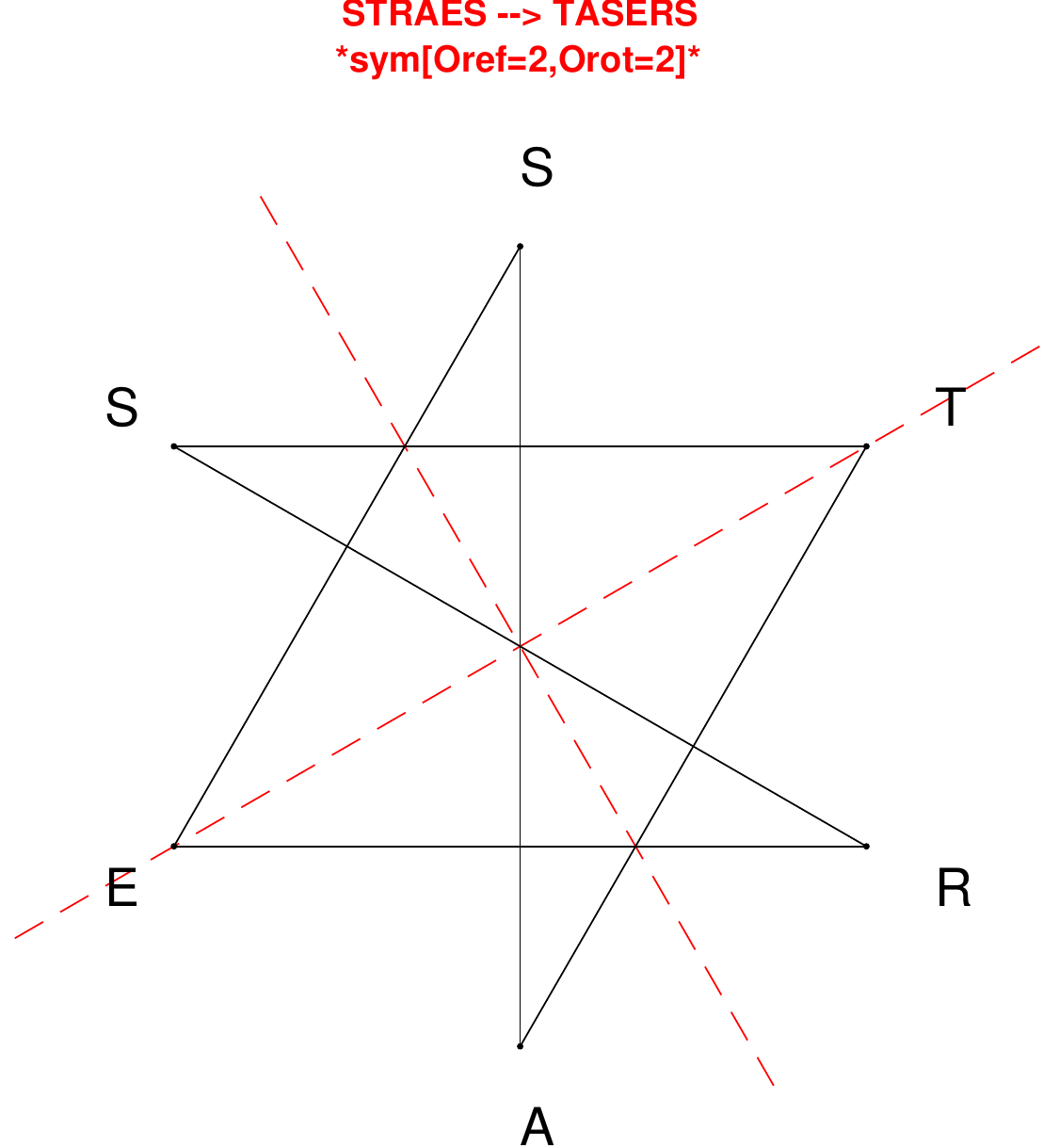}
\end{subfigure}
\hfill
\begin{subfigure}[T]{0.19\textwidth}
\centering
\includegraphics[width=\textwidth]{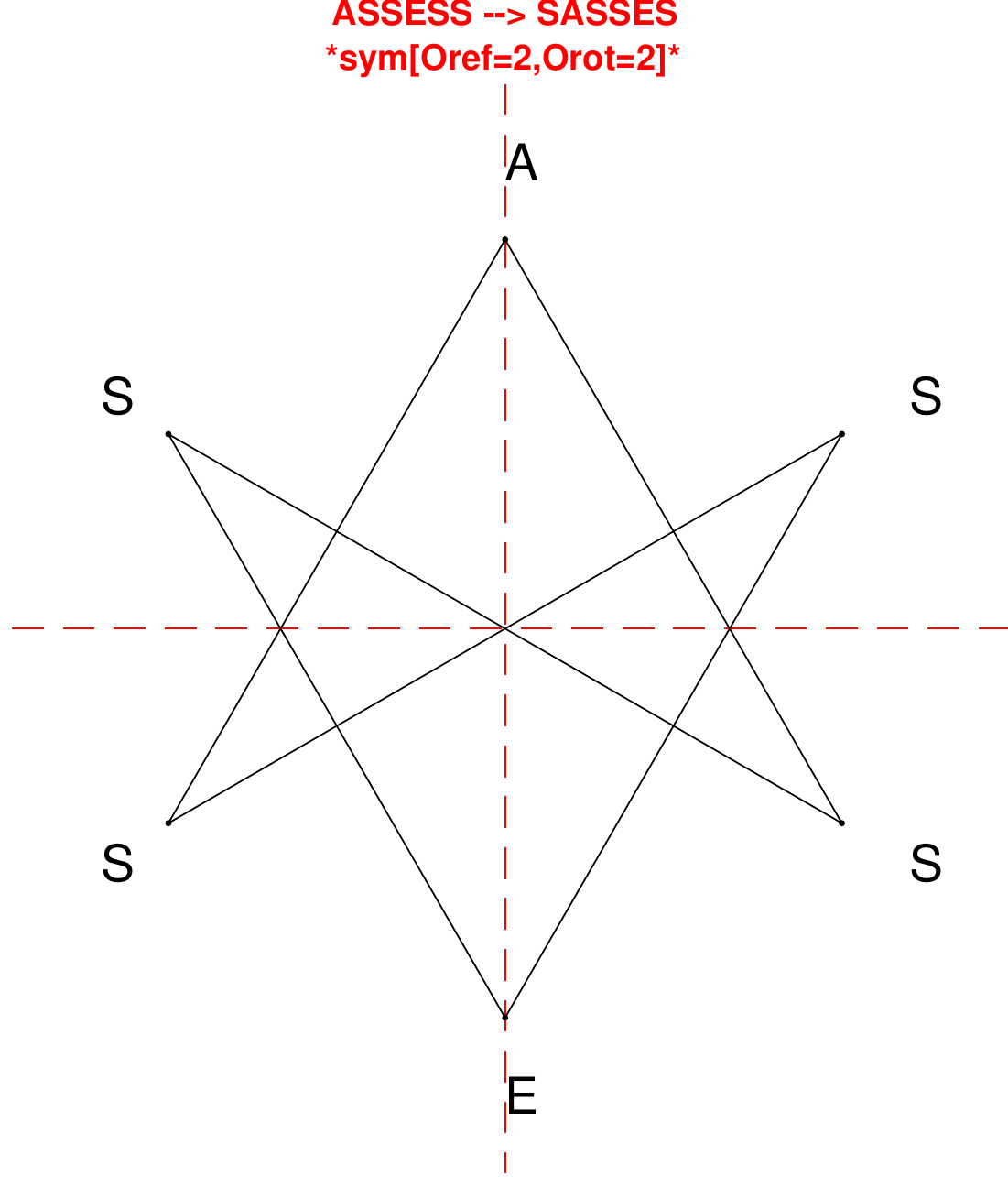}
\end{subfigure}
\end{figure}

\begin{figure}[H]
\centering
\begin{subfigure}[T]{0.19\textwidth}
\centering
\includegraphics[width=\textwidth]{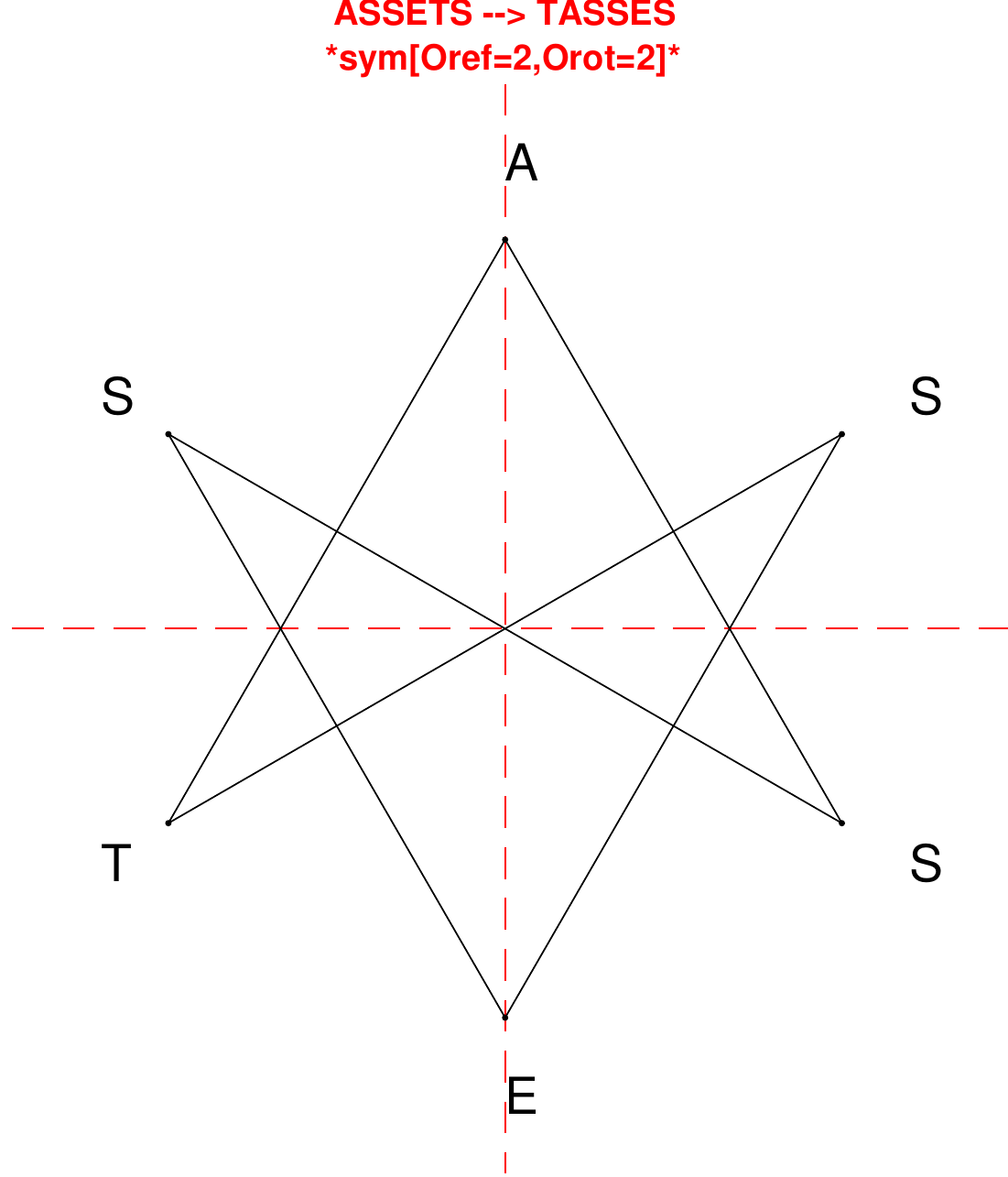}
\end{subfigure}
\hfill
\begin{subfigure}[T]{0.19\textwidth}
\centering
\includegraphics[width=\textwidth]{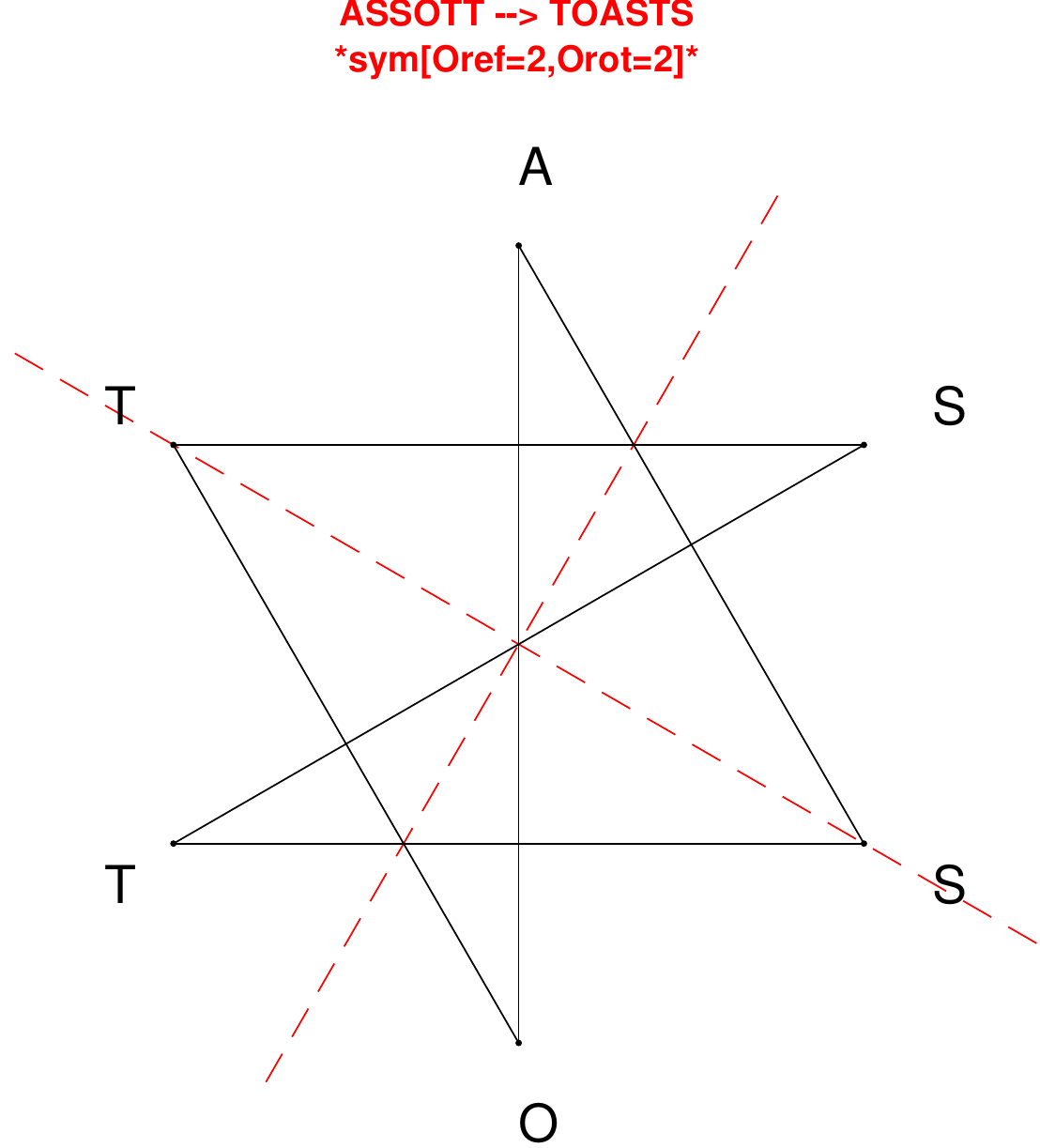}
\end{subfigure}
\hfill
\begin{subfigure}[T]{0.19\textwidth}
\centering
\includegraphics[width=\textwidth]{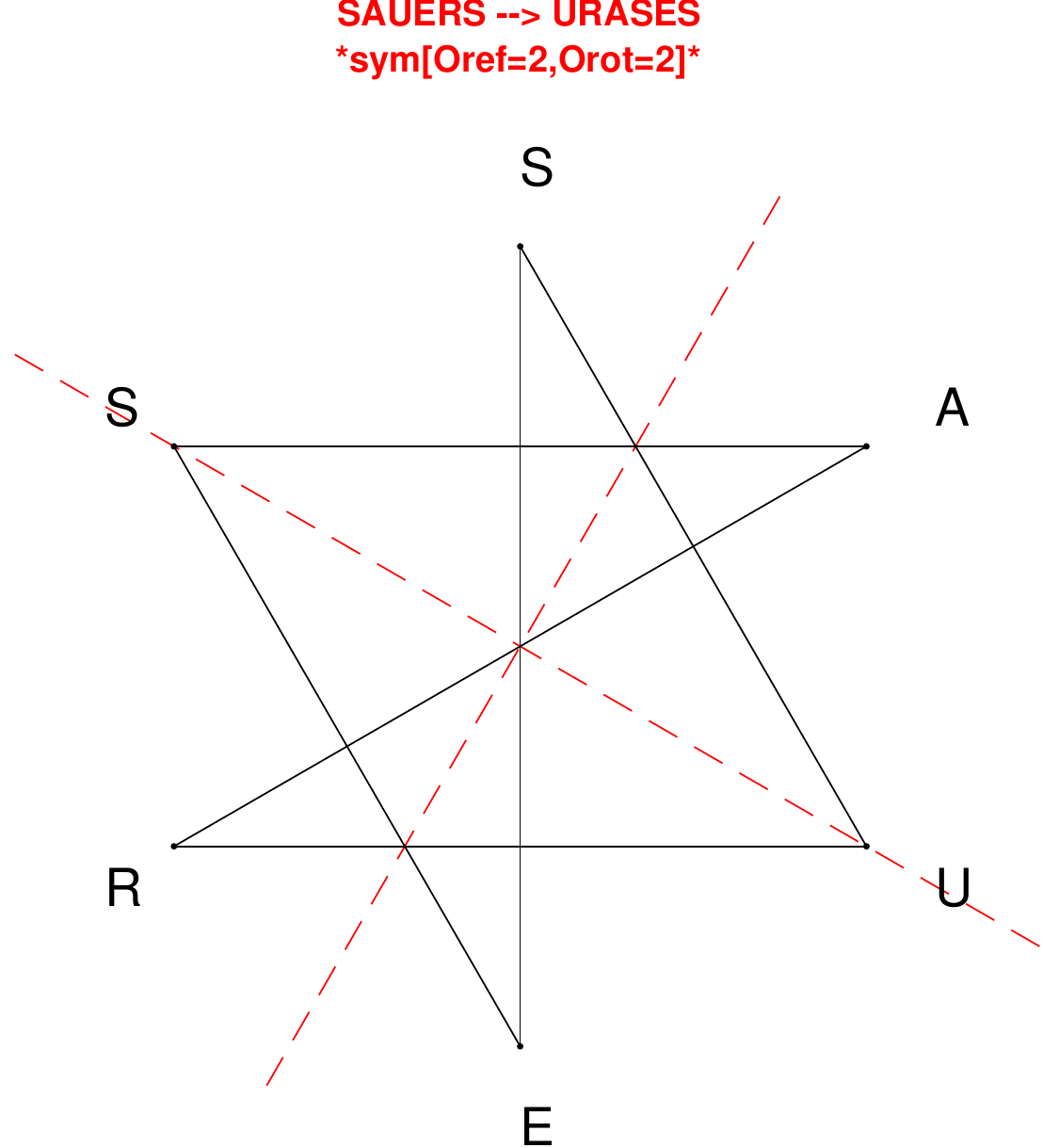}
\end{subfigure}
\hfill
\begin{subfigure}[T]{0.19\textwidth}
\centering
\includegraphics[width=\textwidth]{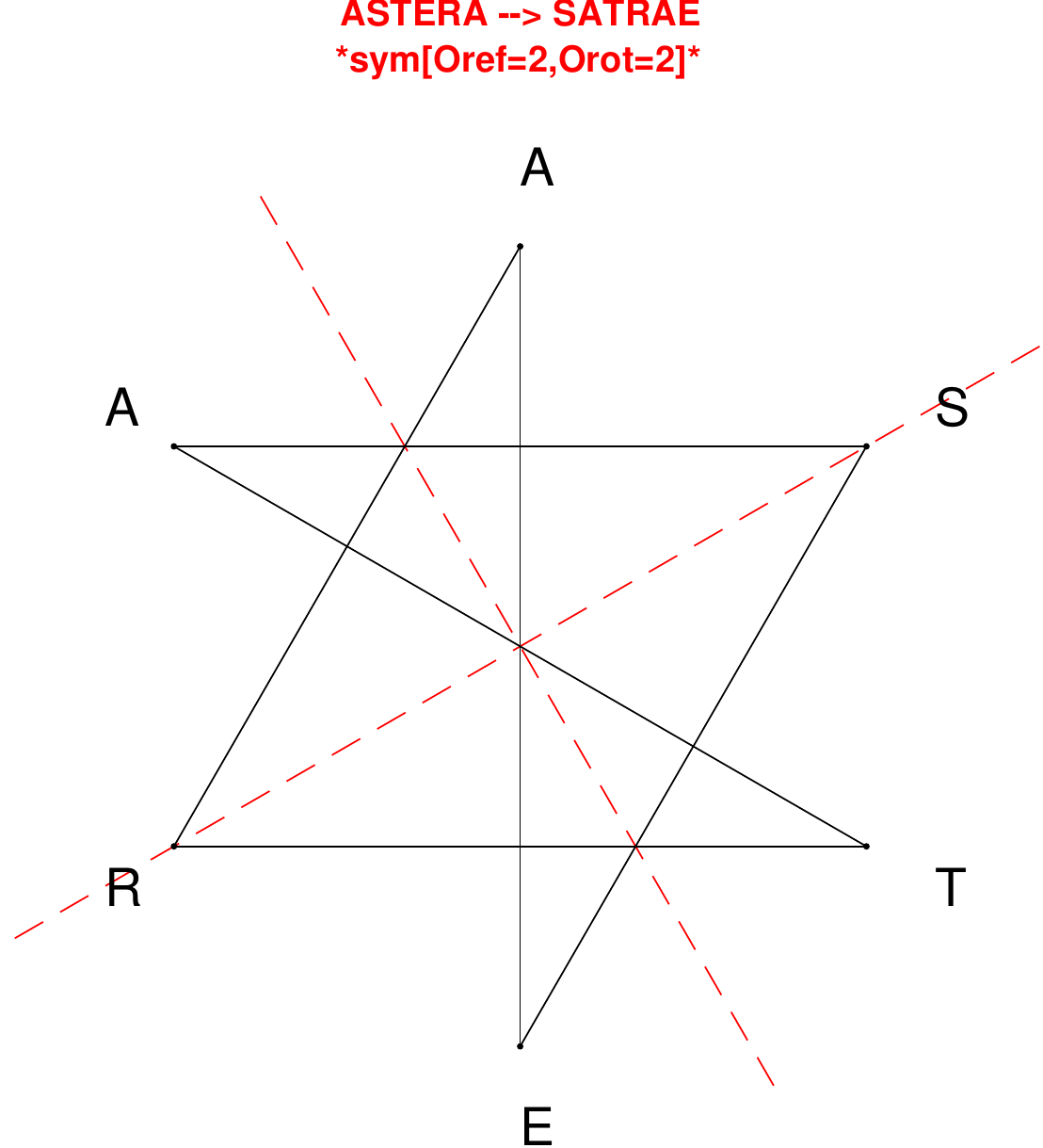}
\end{subfigure}
\hfill
\begin{subfigure}[T]{0.19\textwidth}
\centering
\includegraphics[width=\textwidth]{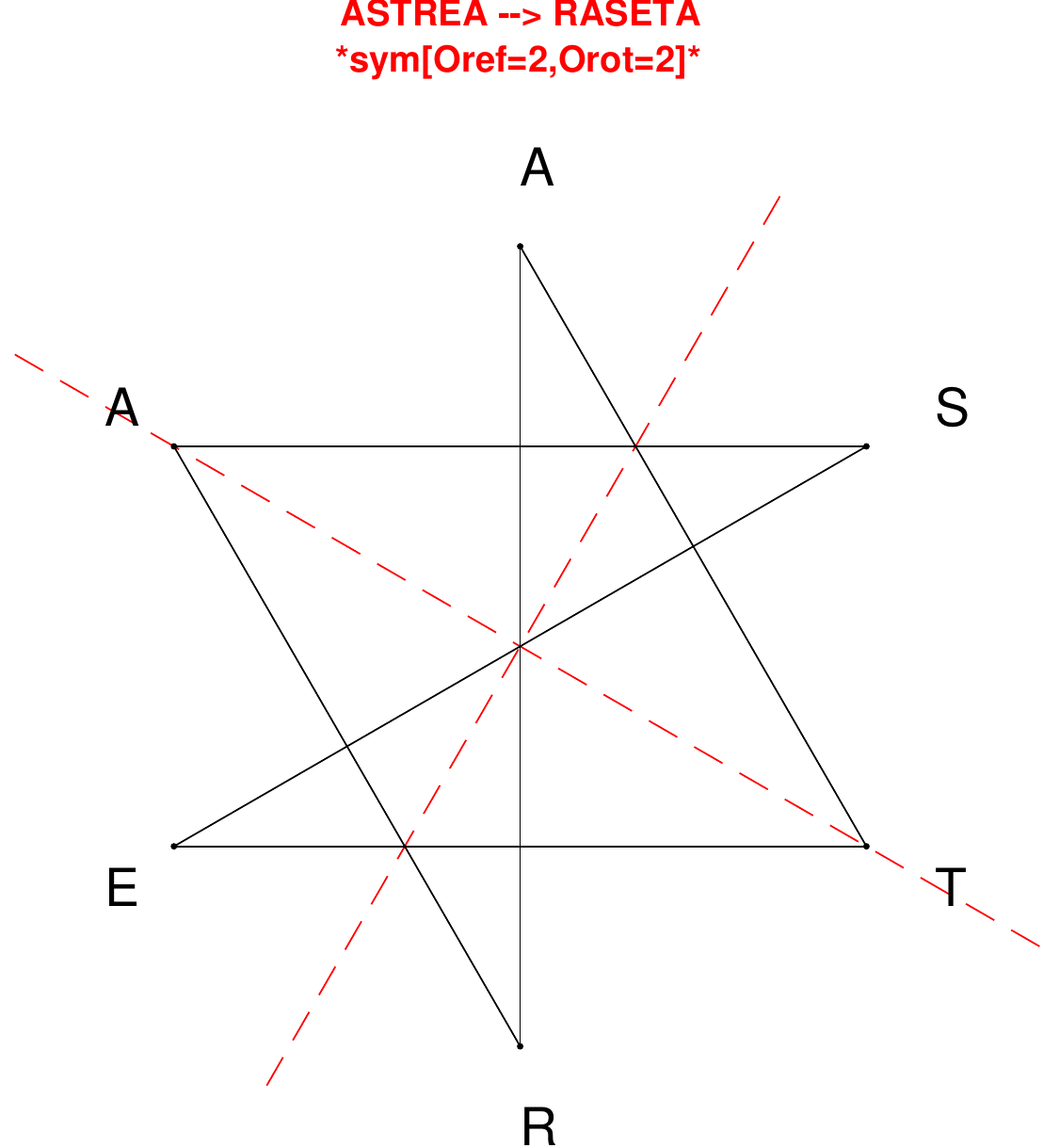}
\end{subfigure}
\end{figure}

\begin{figure}[H]
\centering
\begin{subfigure}[T]{0.19\textwidth}
\centering
\includegraphics[width=\textwidth]{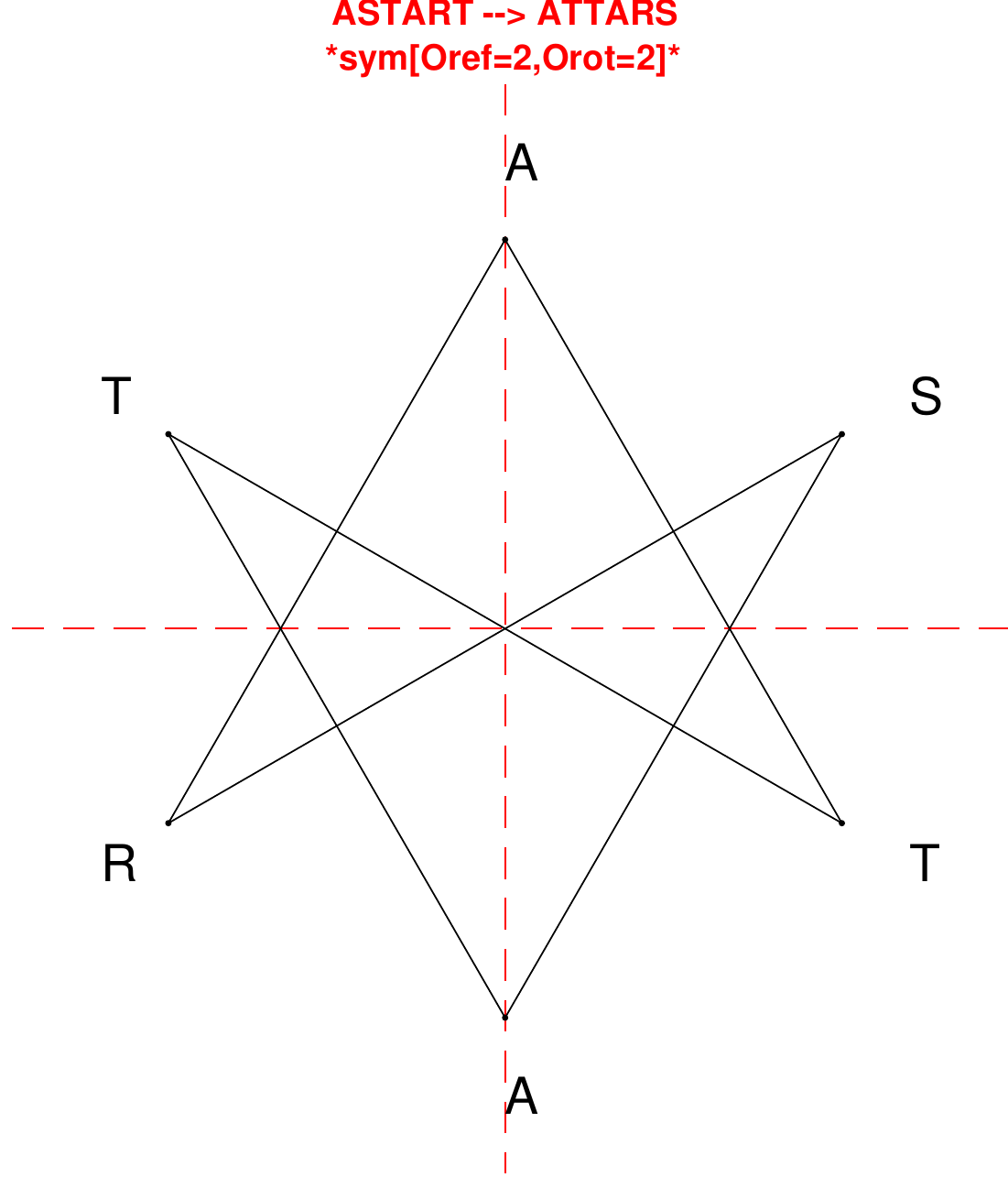}
\end{subfigure}
\hfill
\begin{subfigure}[T]{0.19\textwidth}
\centering
\includegraphics[width=\textwidth]{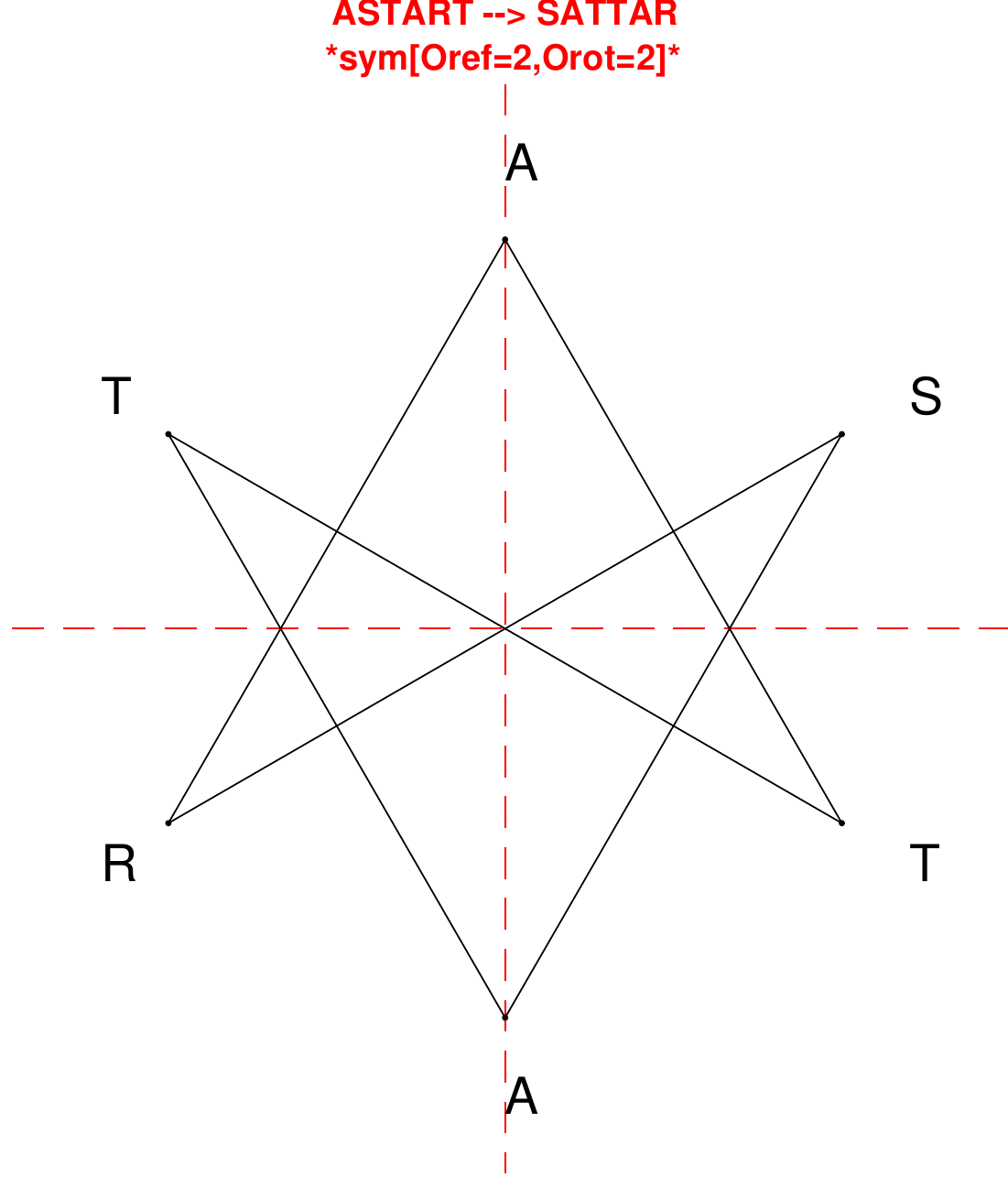}
\end{subfigure}
\hfill
\begin{subfigure}[T]{0.19\textwidth}
\centering
\includegraphics[width=\textwidth]{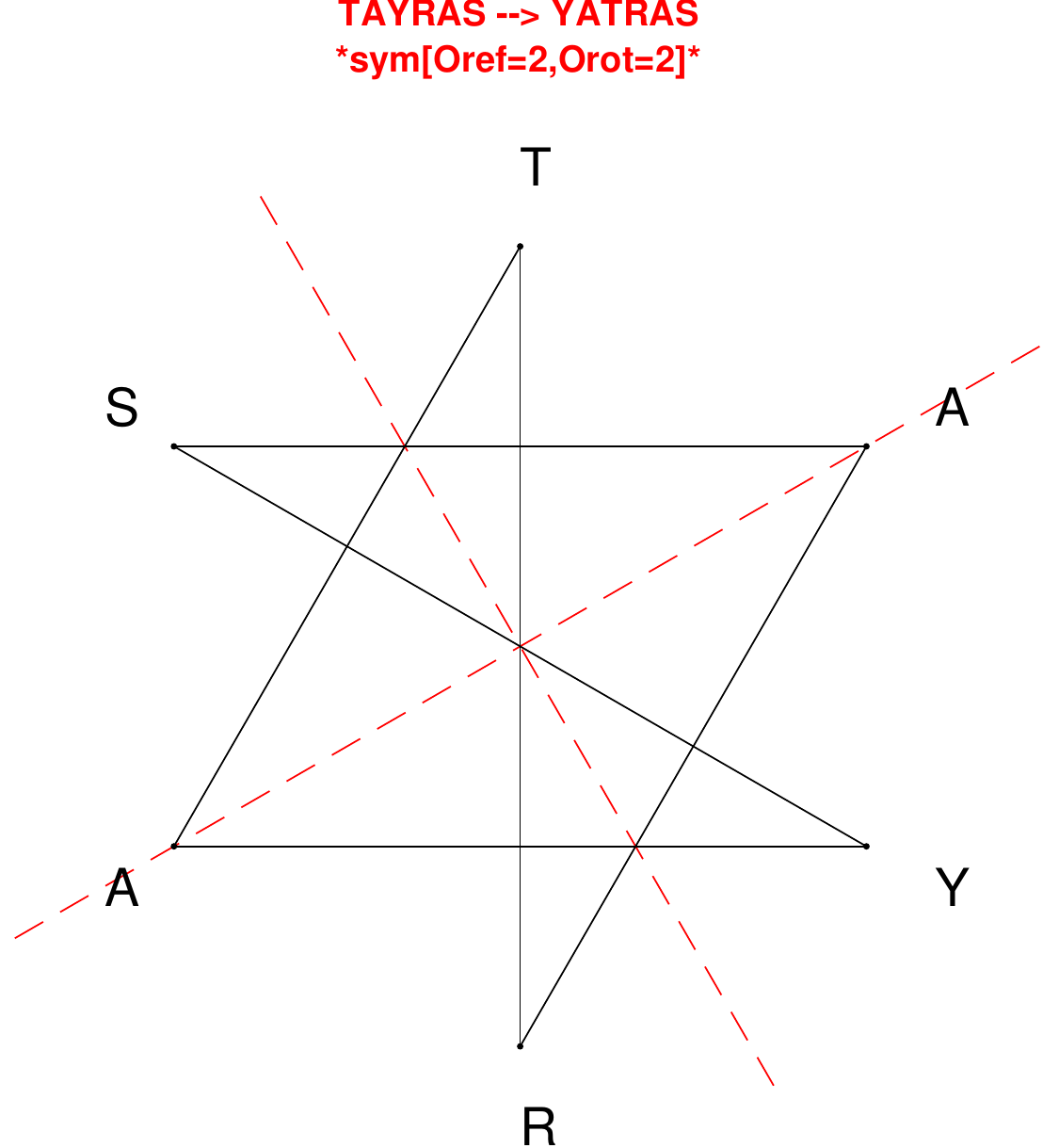}
\end{subfigure}
\hfill
\begin{subfigure}[T]{0.19\textwidth}
\centering
\includegraphics[width=\textwidth]{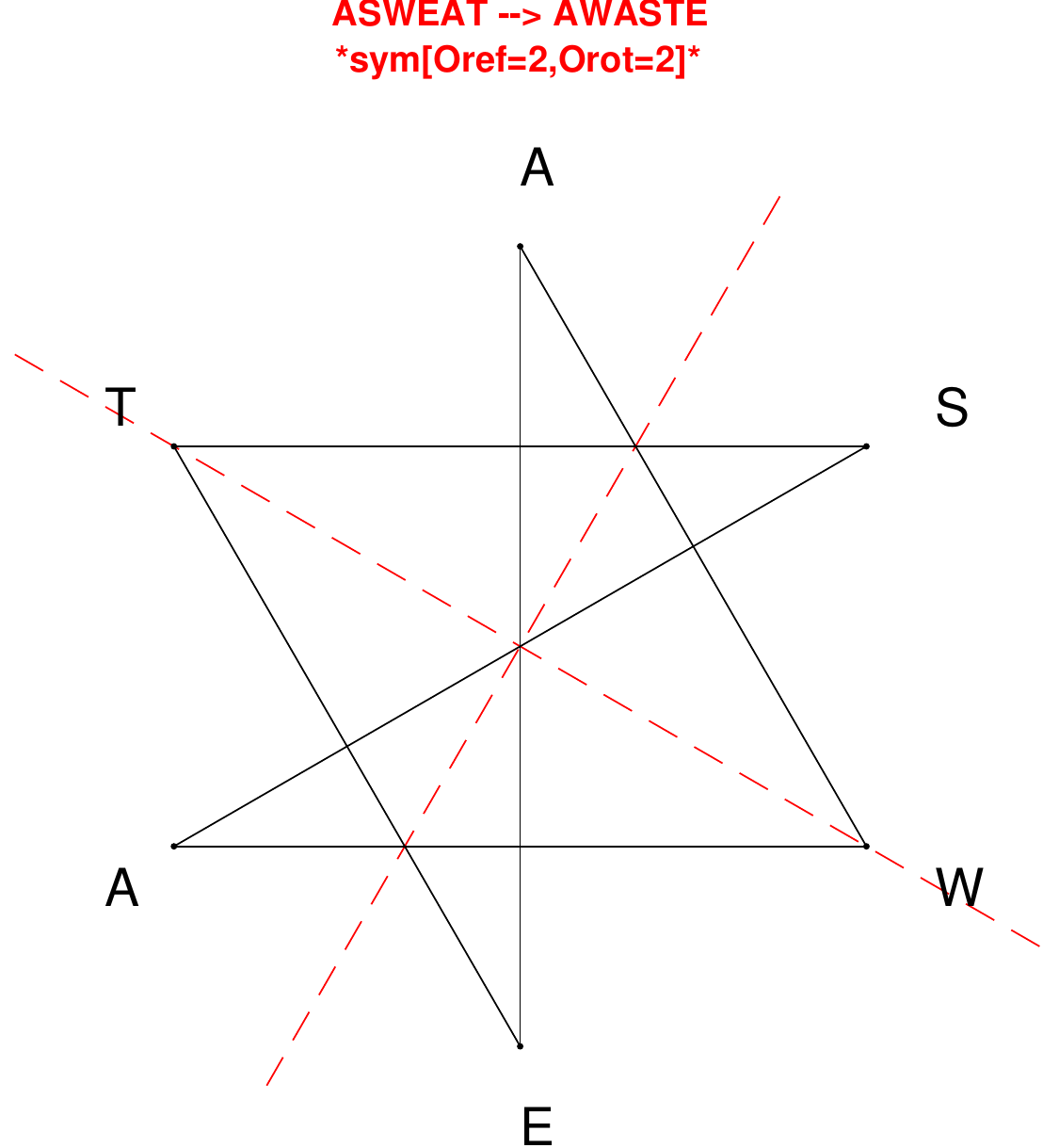}
\end{subfigure}
\hfill
\begin{subfigure}[T]{0.19\textwidth}
\centering
\includegraphics[width=\textwidth]{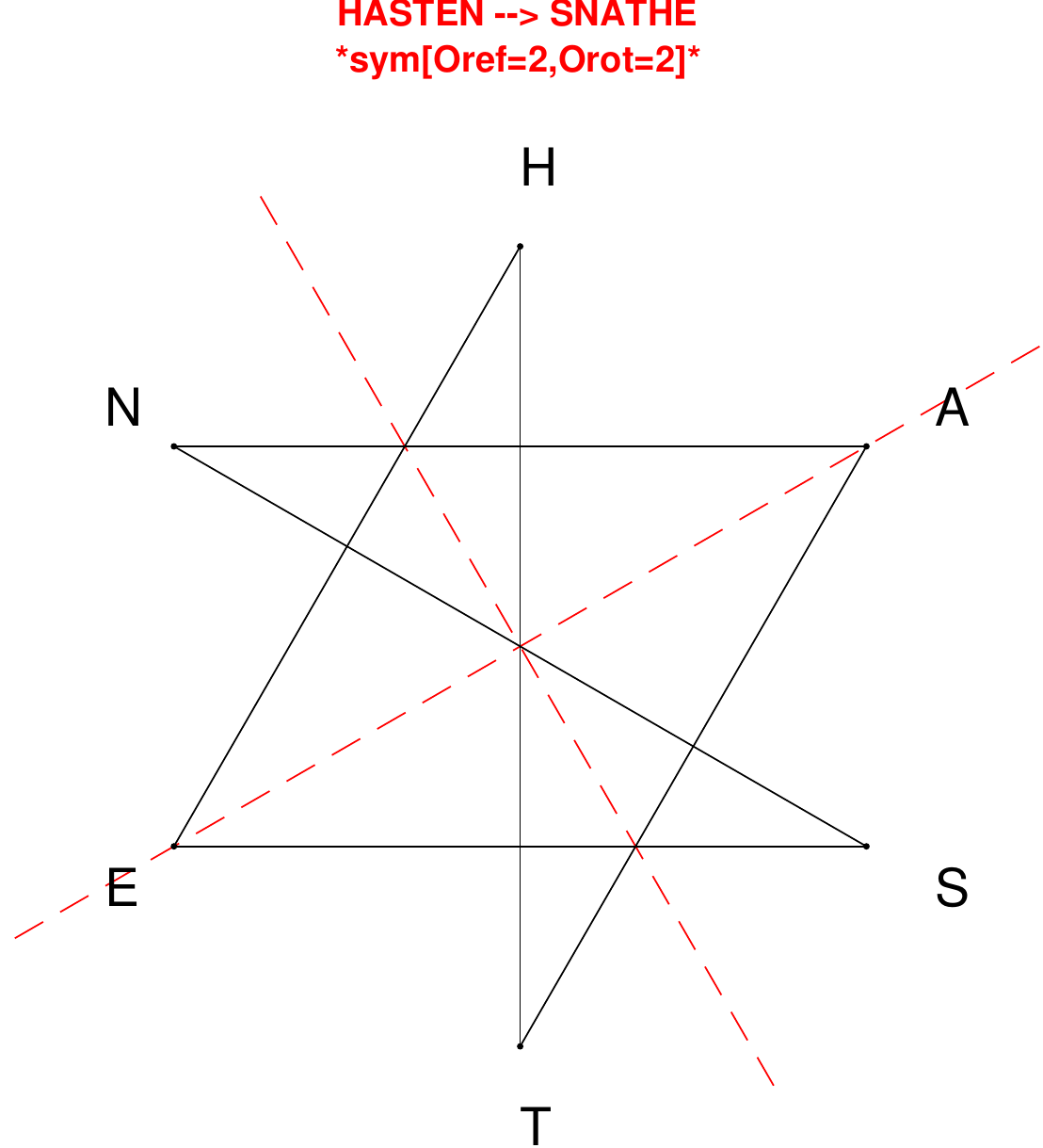}
\end{subfigure}
\end{figure}

\begin{figure}[H]
\centering
\begin{subfigure}[T]{0.19\textwidth}
\centering
\includegraphics[width=\textwidth]{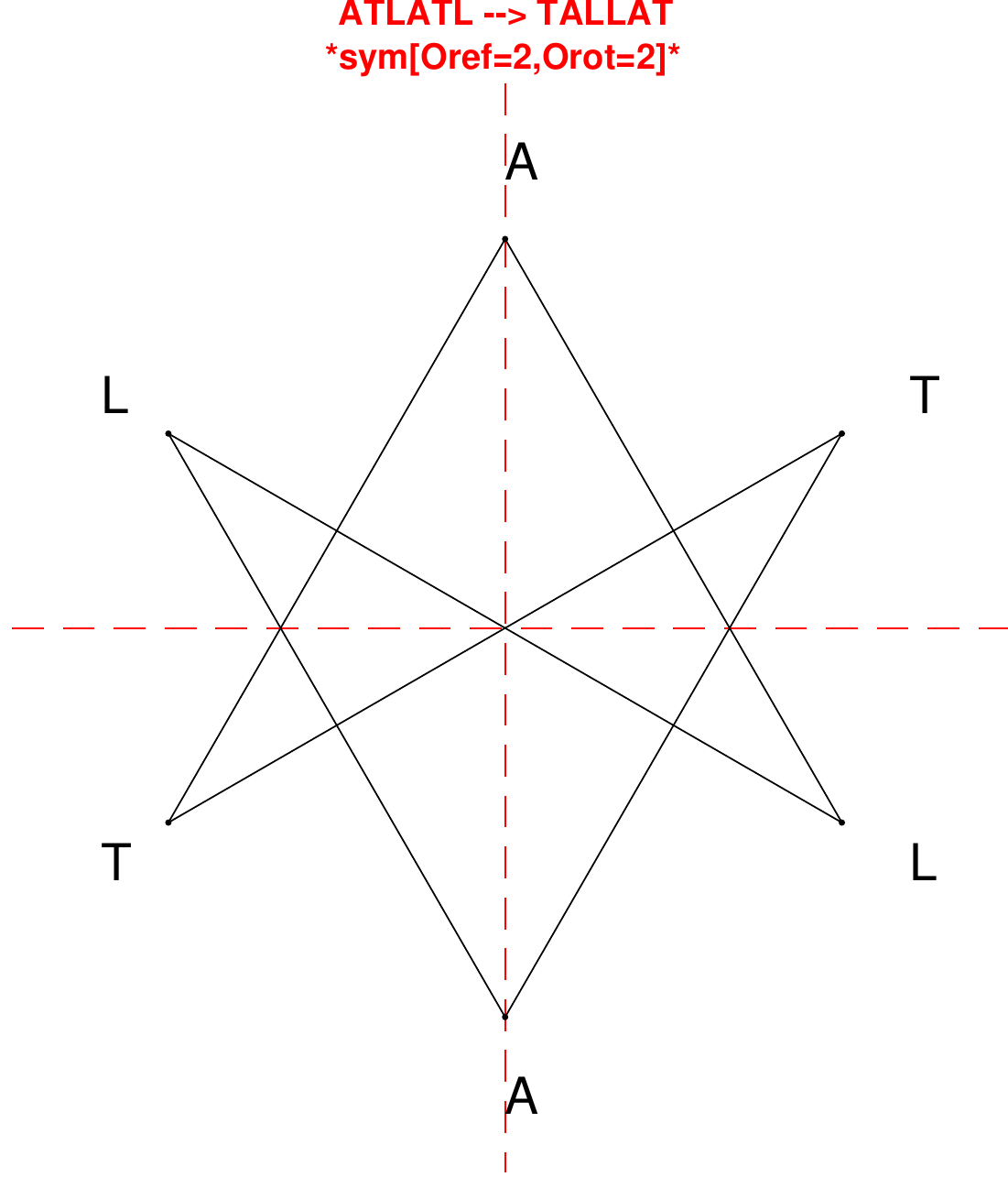}
\end{subfigure}
\hfill
\begin{subfigure}[T]{0.19\textwidth}
\centering
\includegraphics[width=\textwidth]{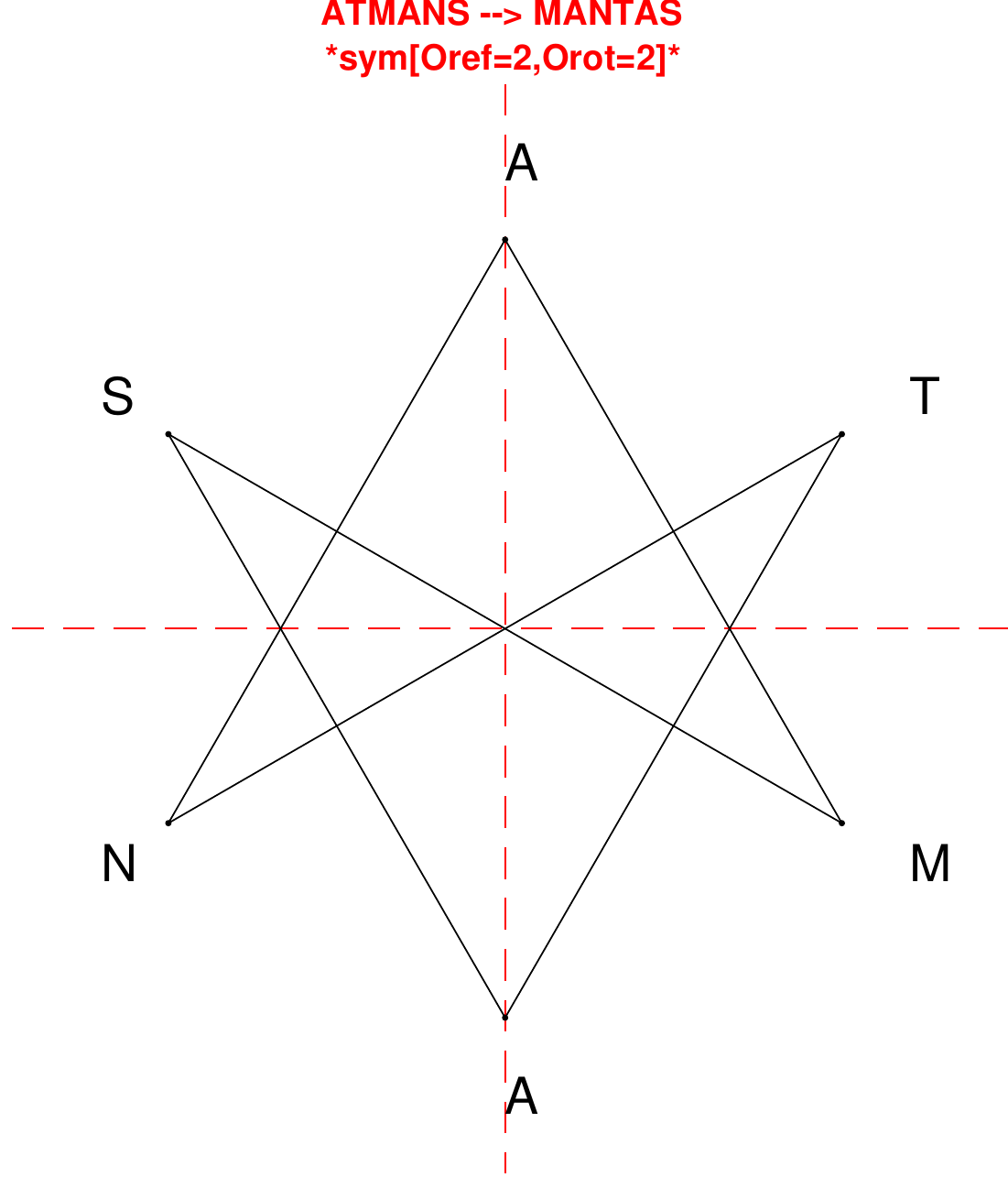}
\end{subfigure}
\hfill
\begin{subfigure}[T]{0.19\textwidth}
\centering
\includegraphics[width=\textwidth]{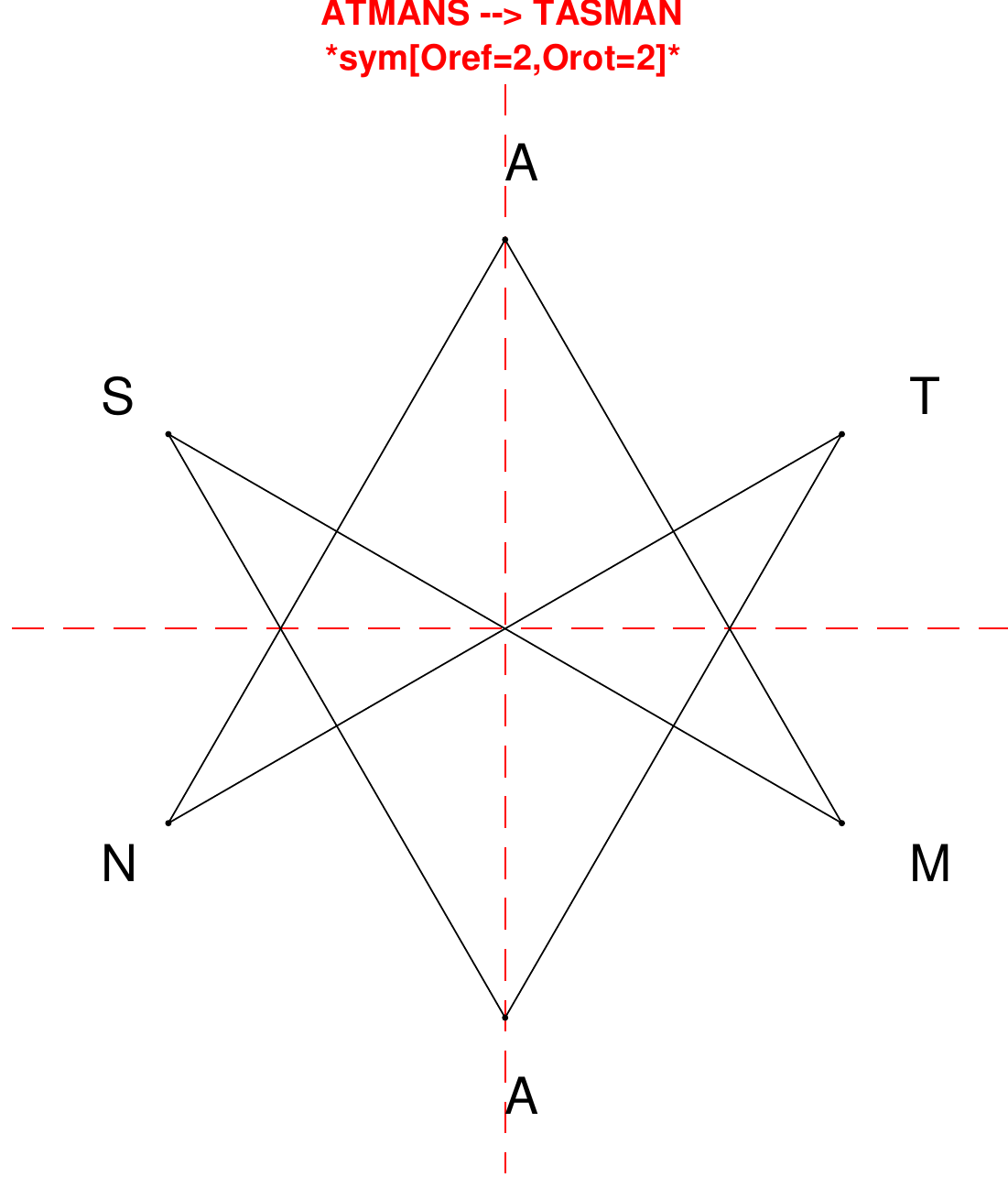}
\end{subfigure}
\hfill
\begin{subfigure}[T]{0.19\textwidth}
\centering
\includegraphics[width=\textwidth]{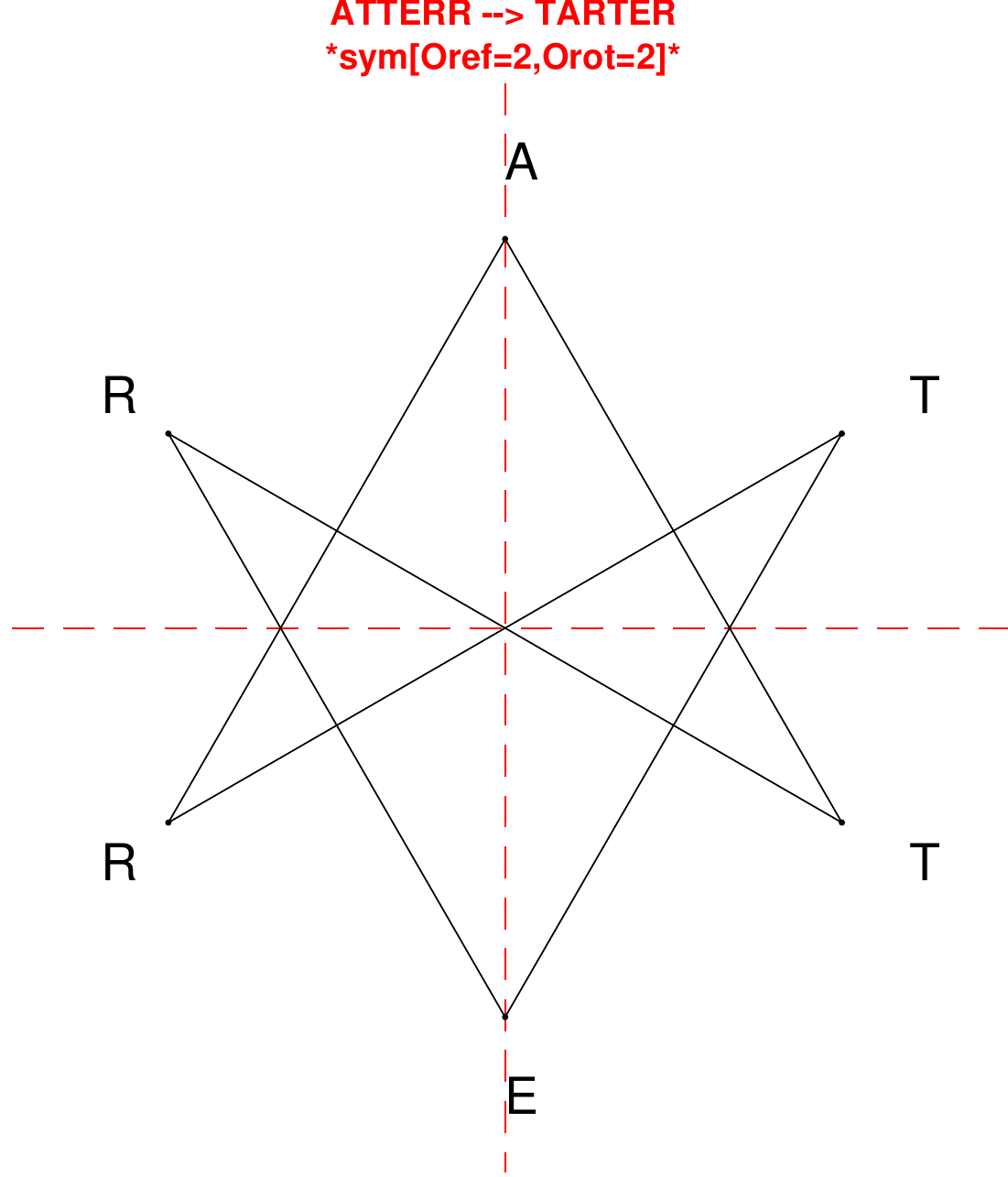}
\end{subfigure}
\hfill
\begin{subfigure}[T]{0.19\textwidth}
\centering
\includegraphics[width=\textwidth]{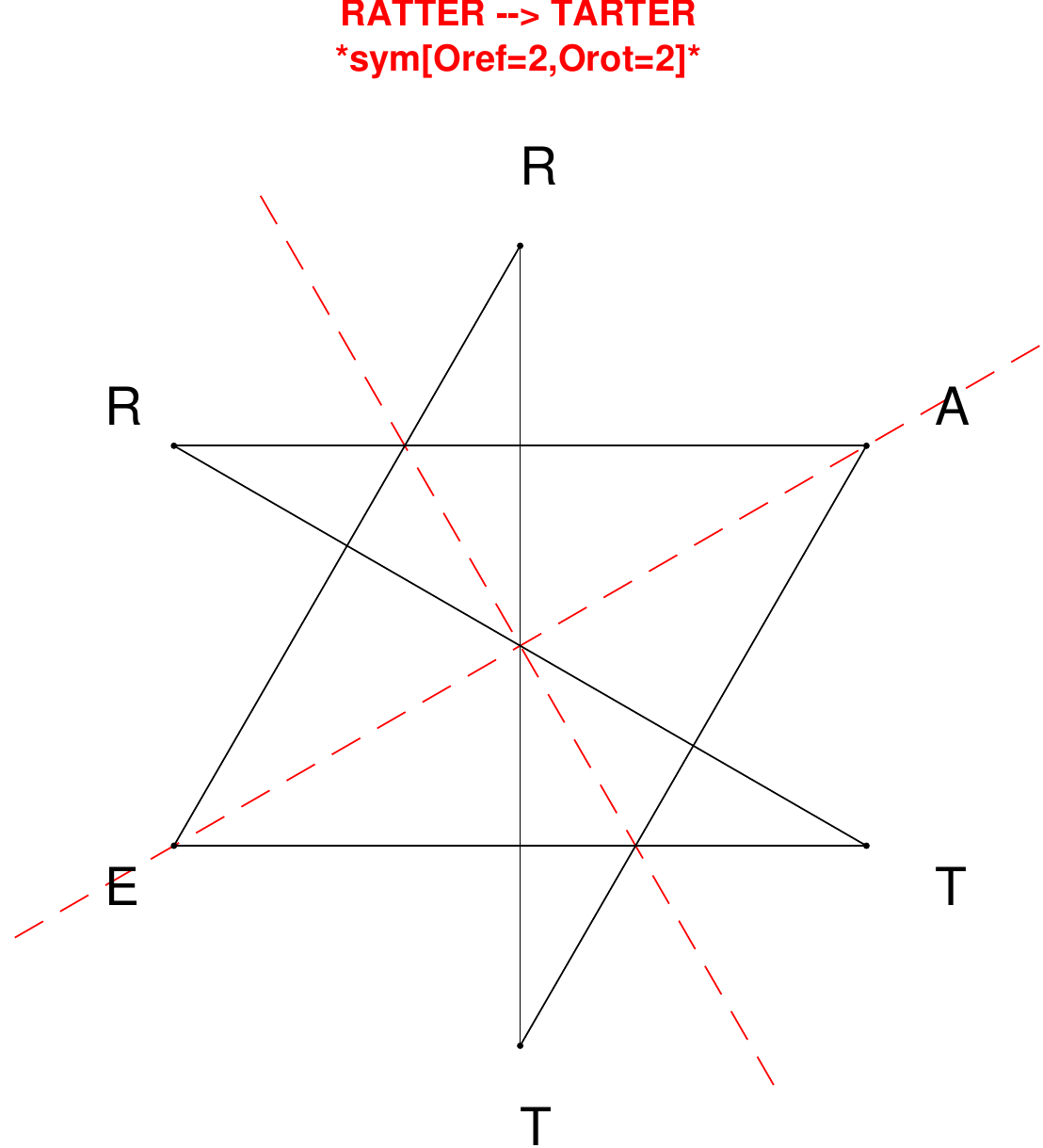}
\end{subfigure}
\end{figure}

\begin{figure}[H]
\centering
\begin{subfigure}[T]{0.19\textwidth}
\centering
\includegraphics[width=\textwidth]{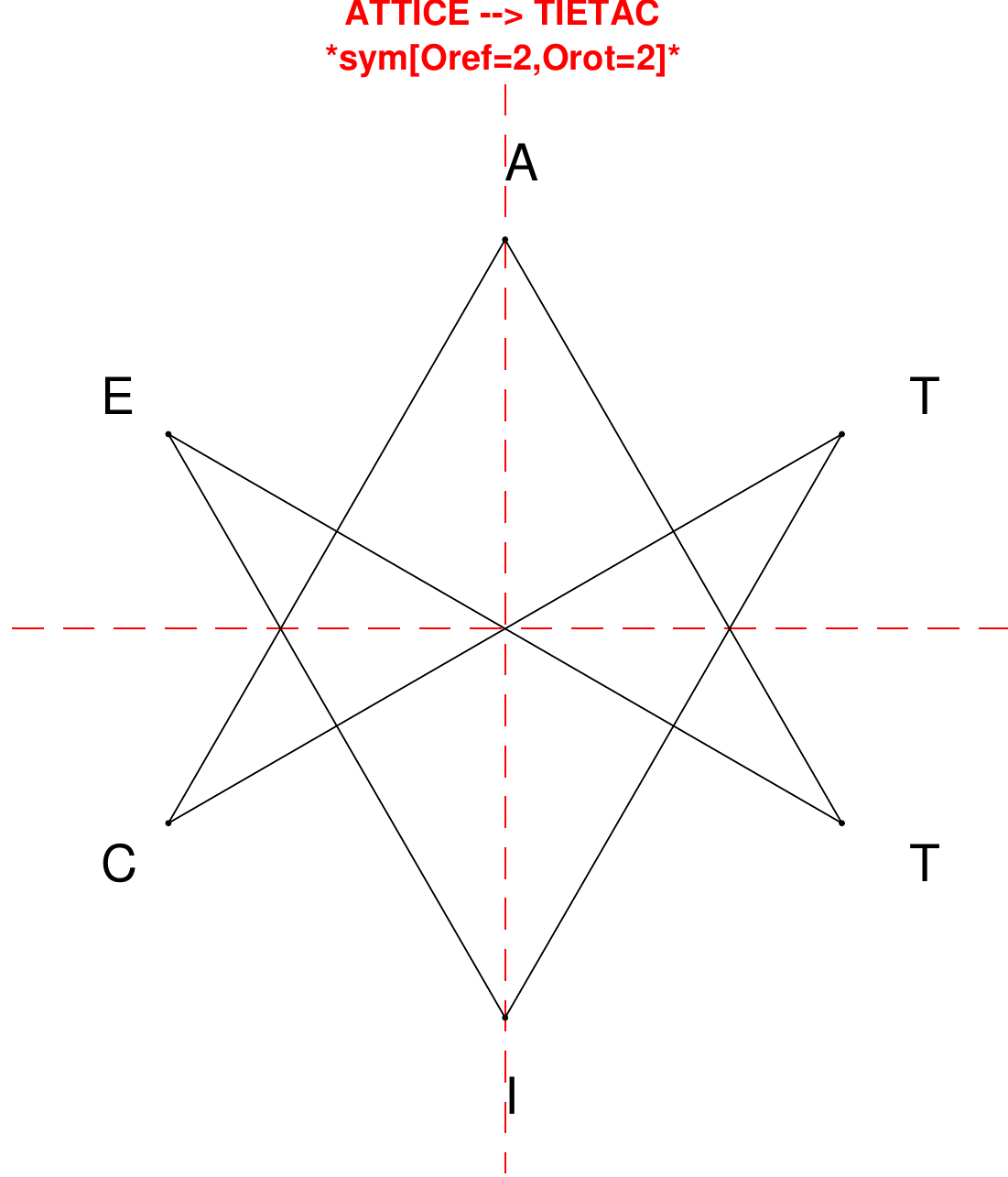}
\end{subfigure}
\hfill
\begin{subfigure}[T]{0.19\textwidth}
\centering
\includegraphics[width=\textwidth]{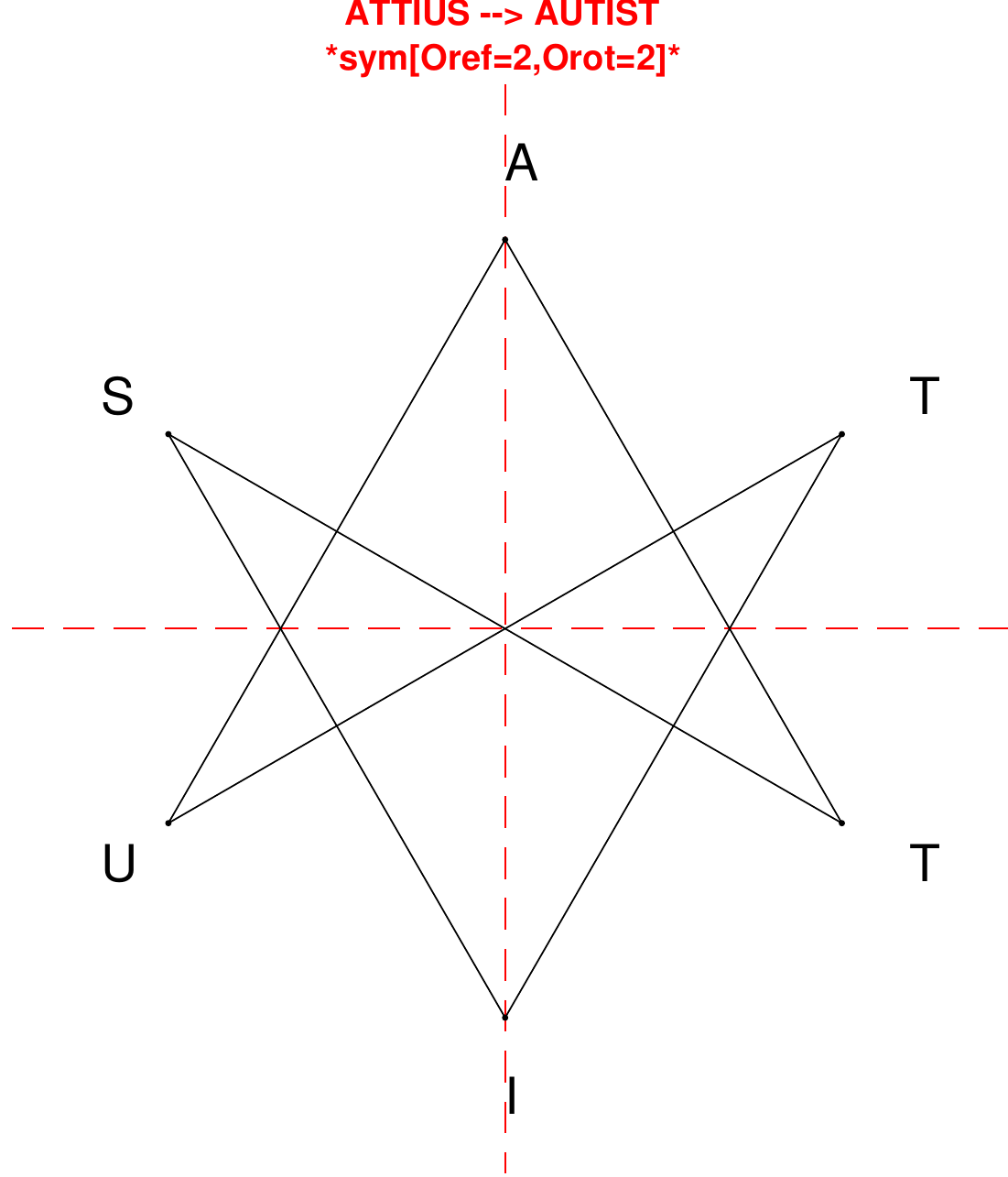}
\end{subfigure}
\hfill
\begin{subfigure}[T]{0.19\textwidth}
\centering
\includegraphics[width=\textwidth]{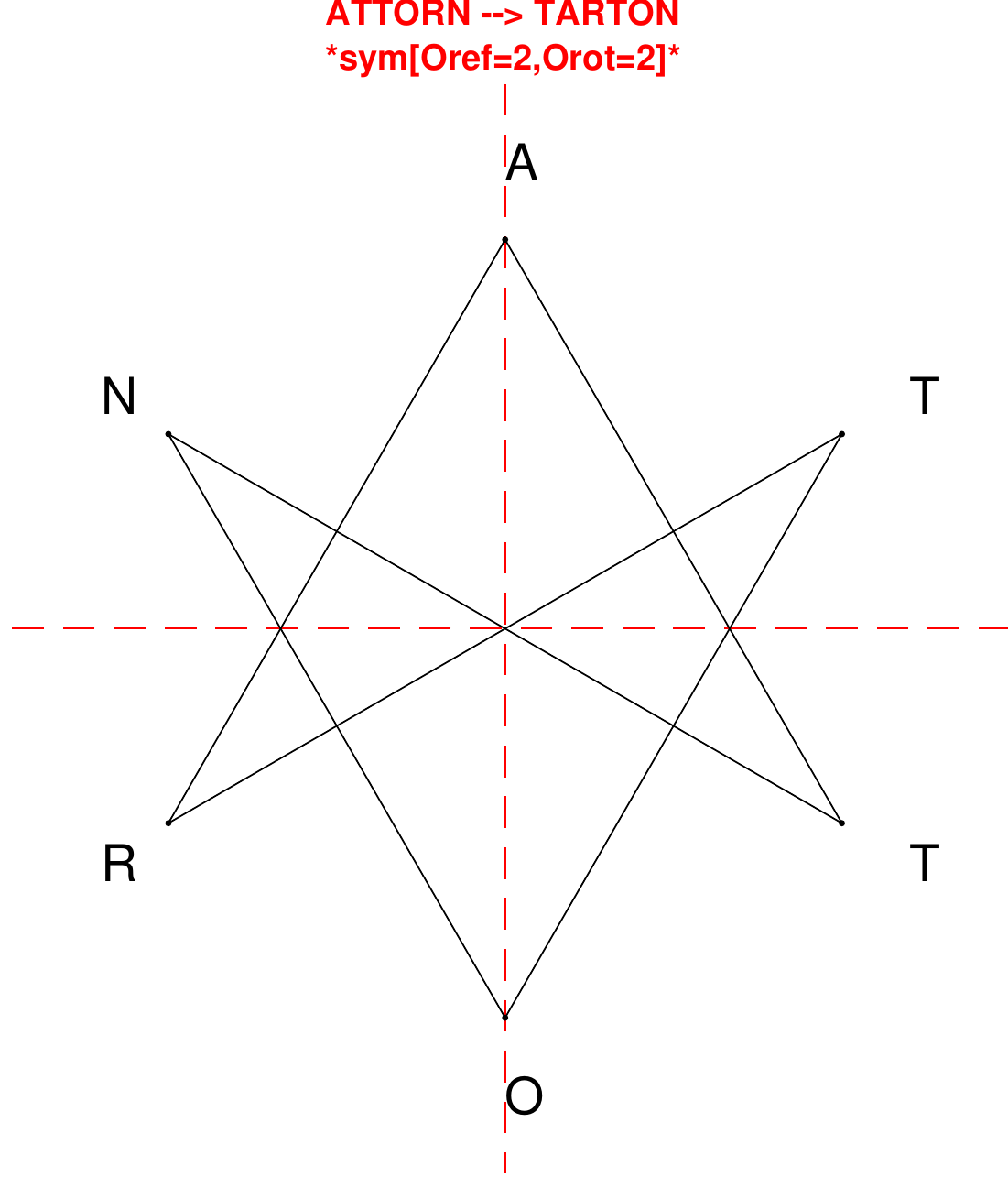}
\end{subfigure}
\hfill
\begin{subfigure}[T]{0.19\textwidth}
\centering
\includegraphics[width=\textwidth]{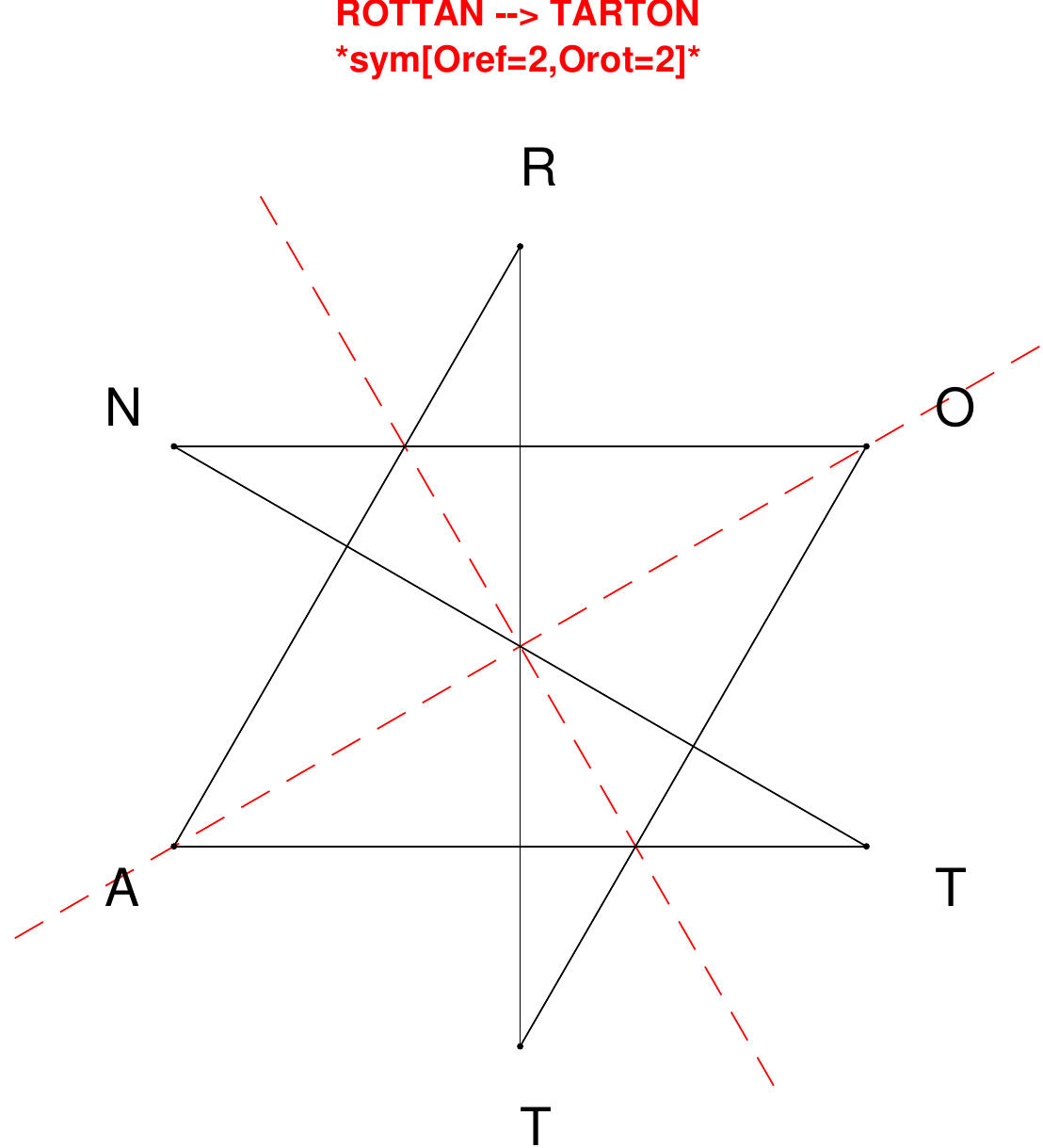}
\end{subfigure}
\hfill
\begin{subfigure}[T]{0.19\textwidth}
\centering
\includegraphics[width=\textwidth]{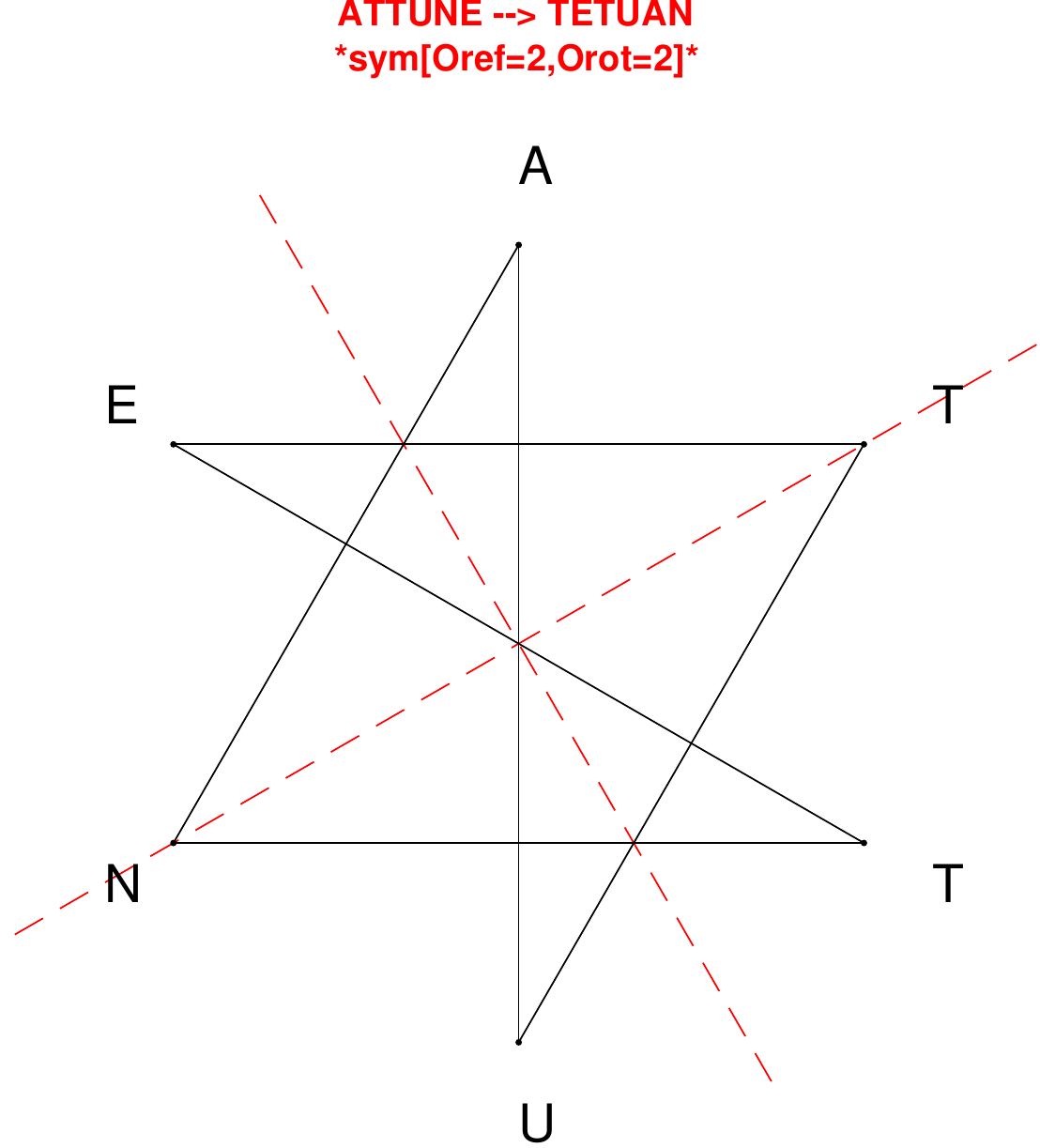}
\end{subfigure}
\end{figure}

\begin{figure}[H]
\centering
\begin{subfigure}[T]{0.19\textwidth}
\centering
\includegraphics[width=\textwidth]{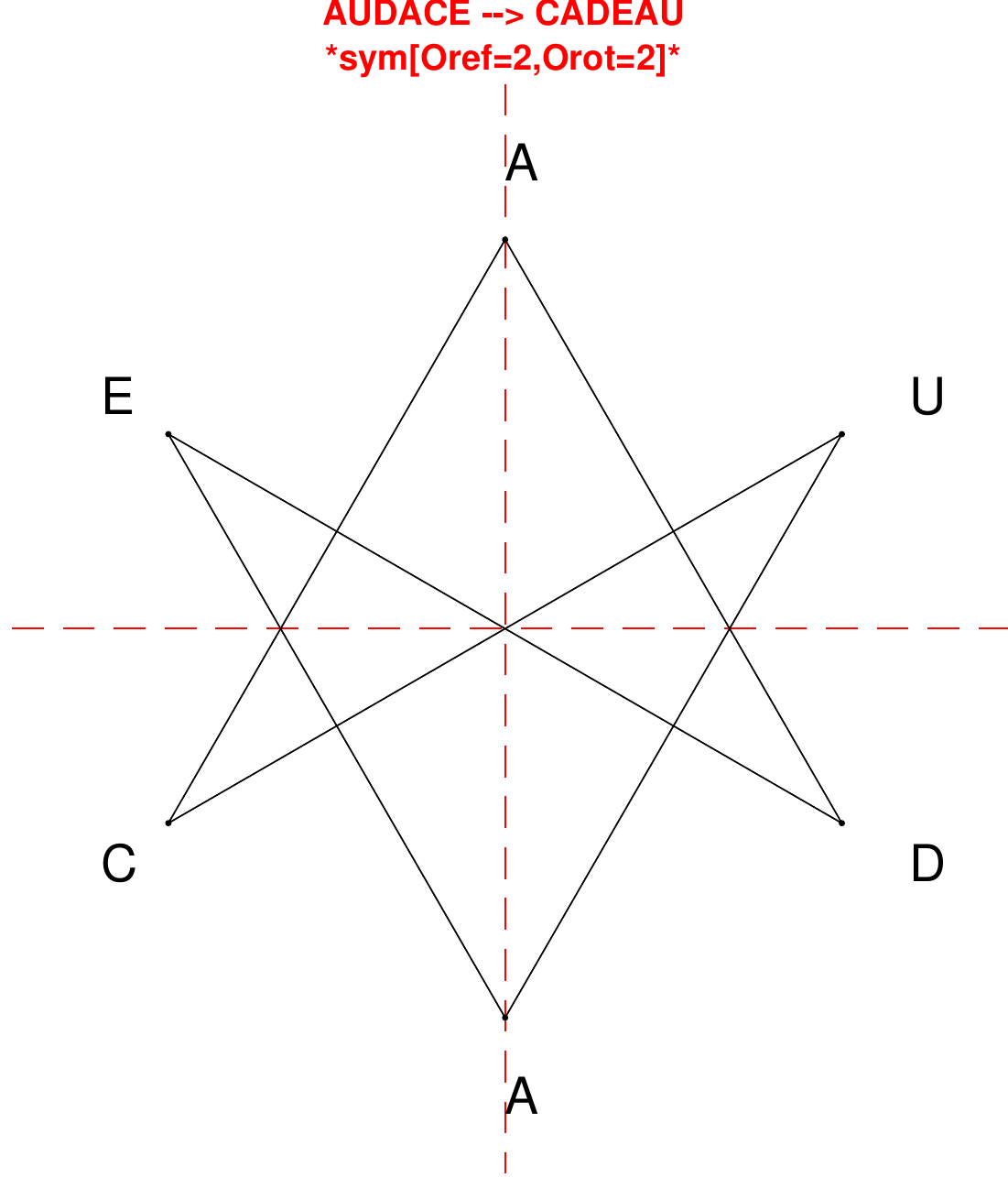}
\end{subfigure}
\hfill
\begin{subfigure}[T]{0.19\textwidth}
\centering
\includegraphics[width=\textwidth]{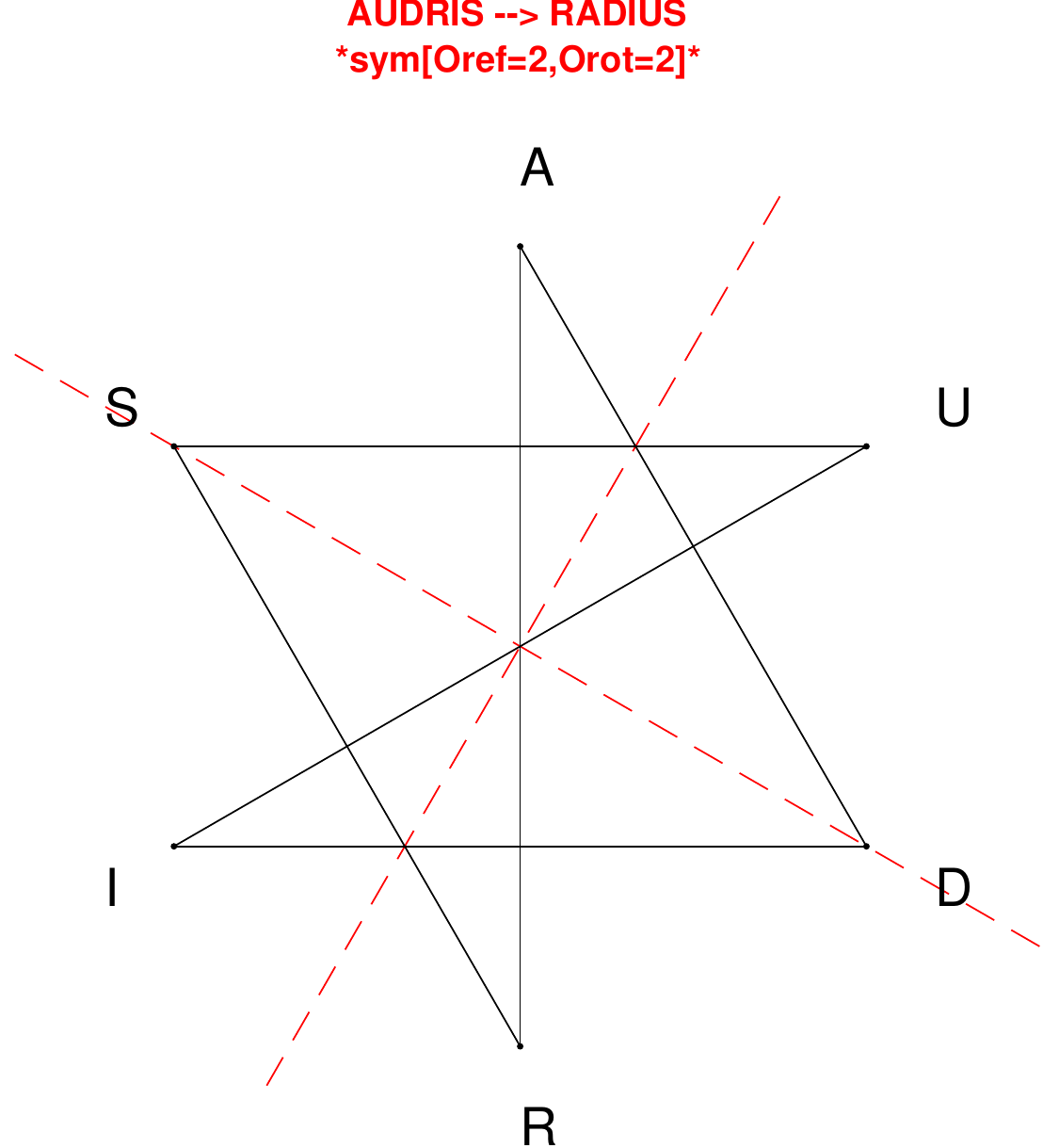}
\end{subfigure}
\hfill
\begin{subfigure}[T]{0.19\textwidth}
\centering
\includegraphics[width=\textwidth]{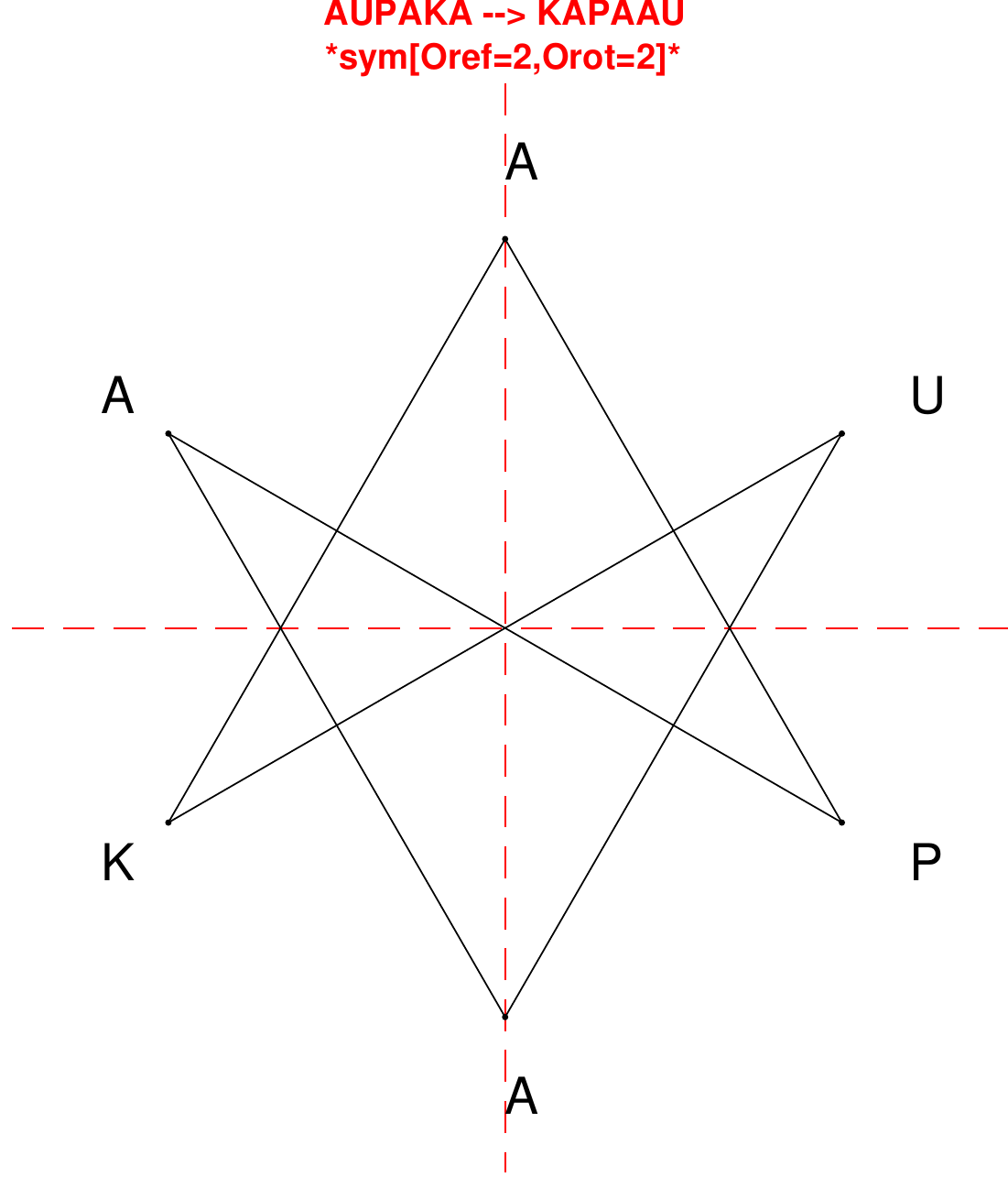}
\end{subfigure}
\hfill
\begin{subfigure}[T]{0.19\textwidth}
\centering
\includegraphics[width=\textwidth]{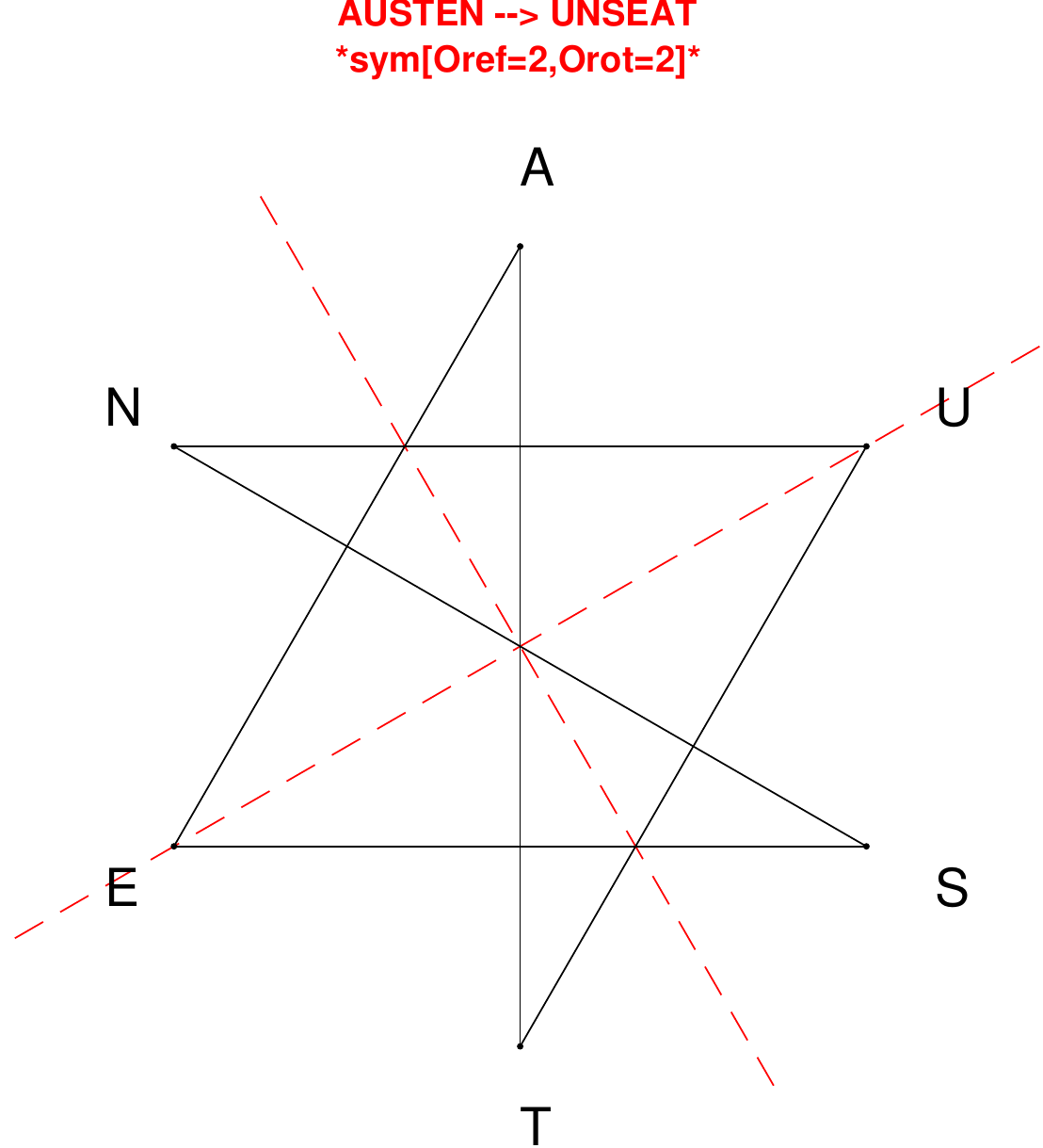}
\end{subfigure}
\hfill
\begin{subfigure}[T]{0.19\textwidth}
\centering
\includegraphics[width=\textwidth]{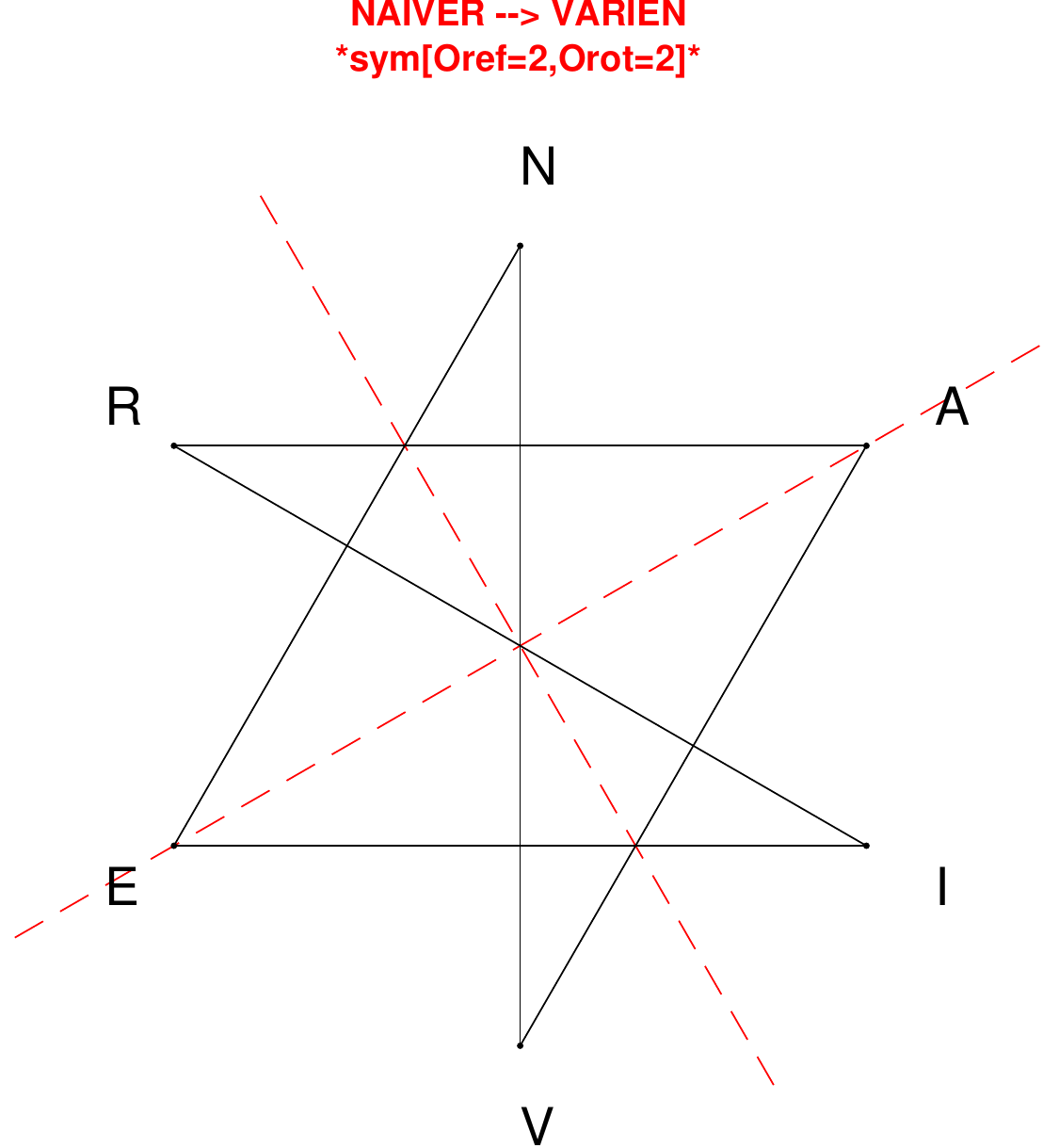}
\end{subfigure}
\end{figure}

\begin{figure}[H]
\centering
\begin{subfigure}[T]{0.19\textwidth}
\centering
\includegraphics[width=\textwidth]{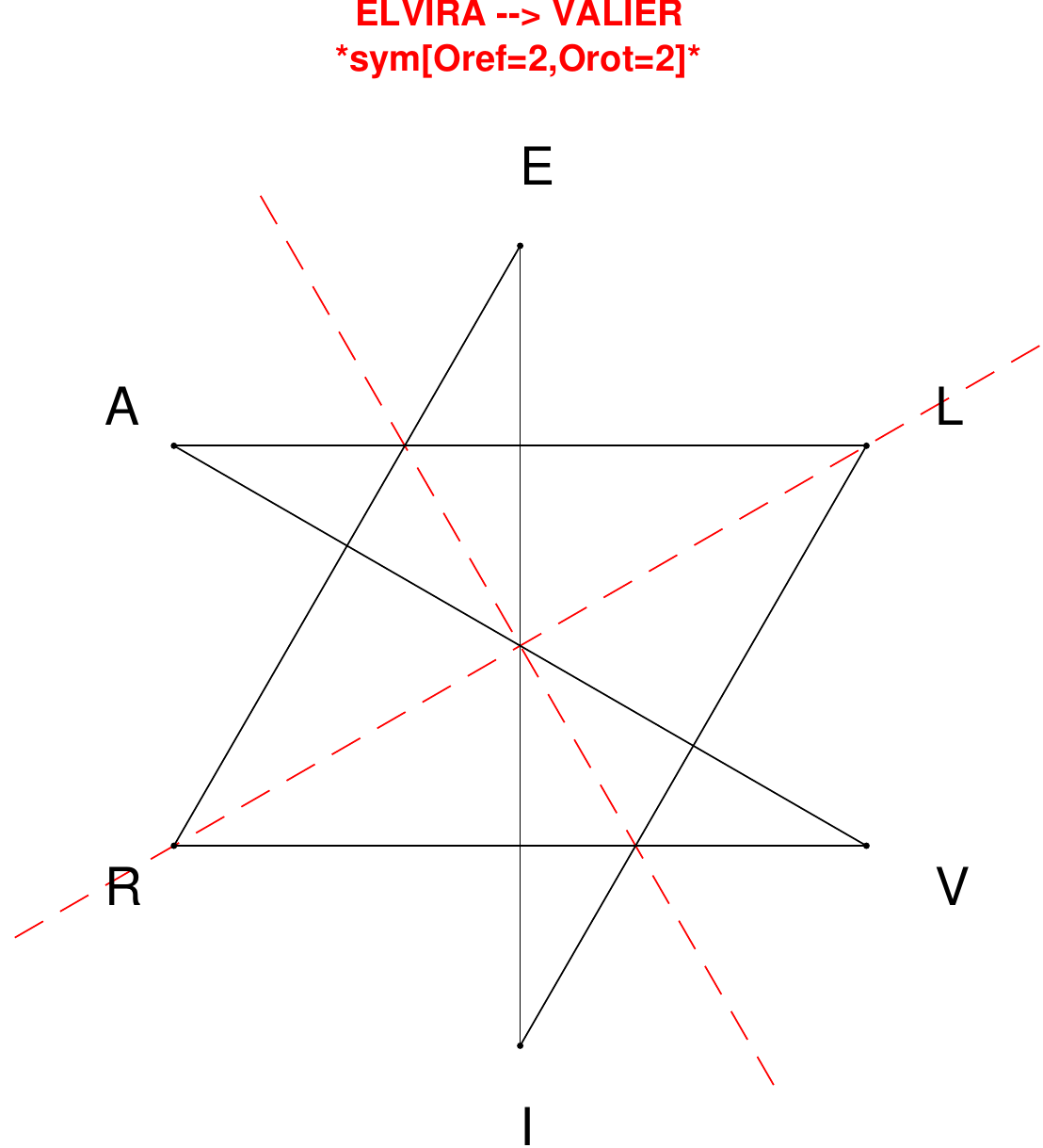}
\end{subfigure}
\hfill
\begin{subfigure}[T]{0.19\textwidth}
\centering
\includegraphics[width=\textwidth]{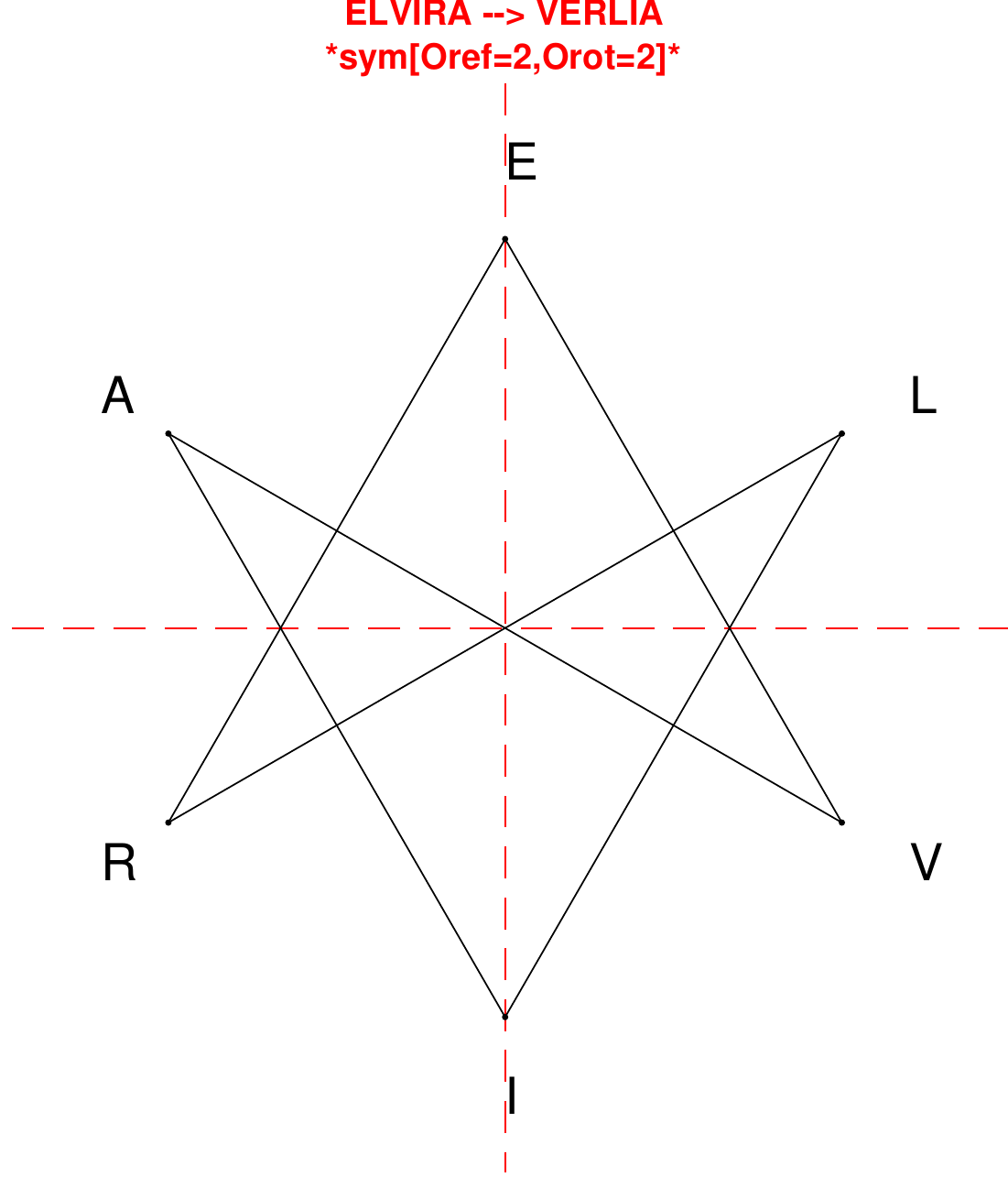}
\end{subfigure}
\hfill
\begin{subfigure}[T]{0.19\textwidth}
\centering
\includegraphics[width=\textwidth]{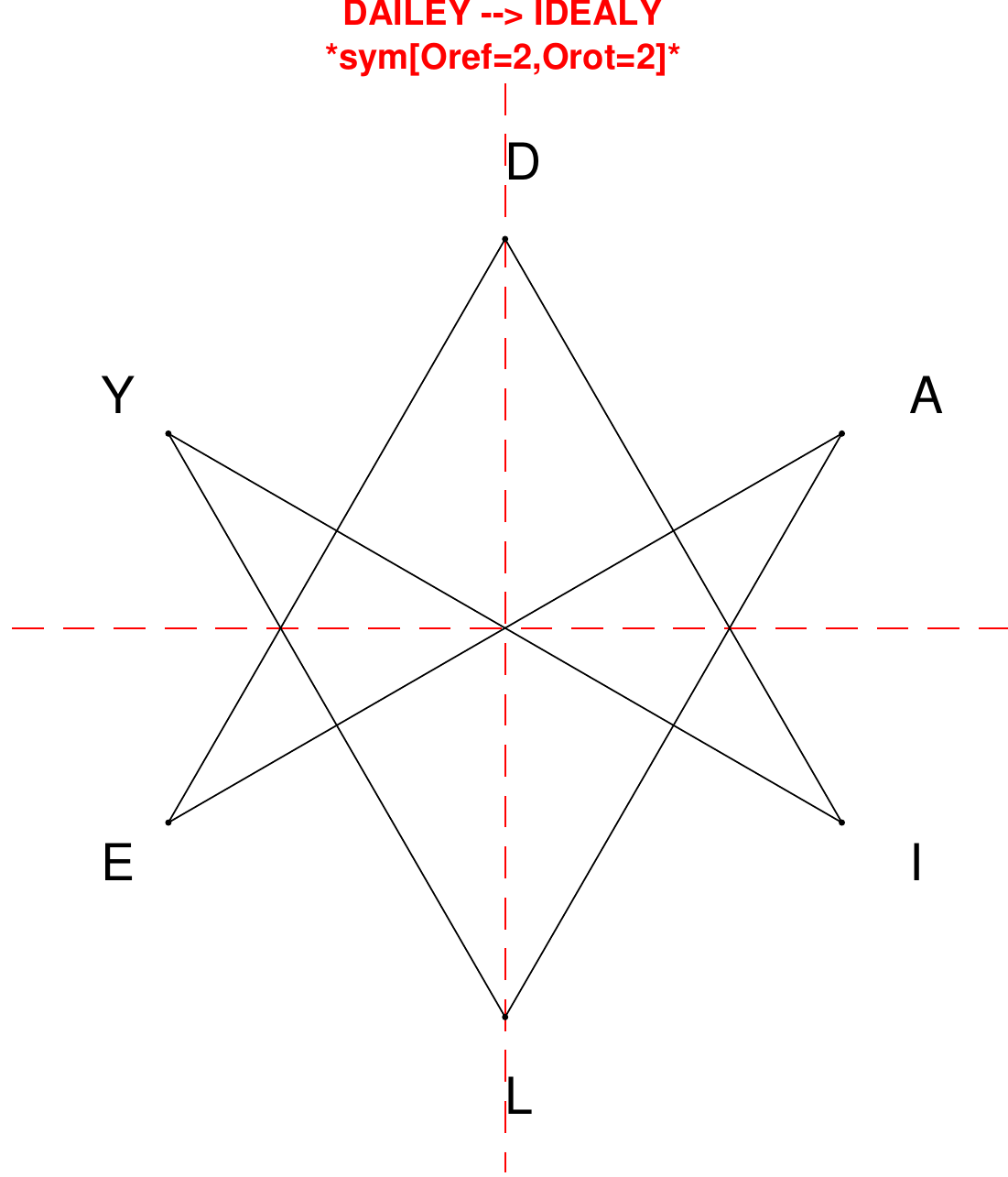}
\end{subfigure}
\hfill
\begin{subfigure}[T]{0.19\textwidth}
\centering
\includegraphics[width=\textwidth]{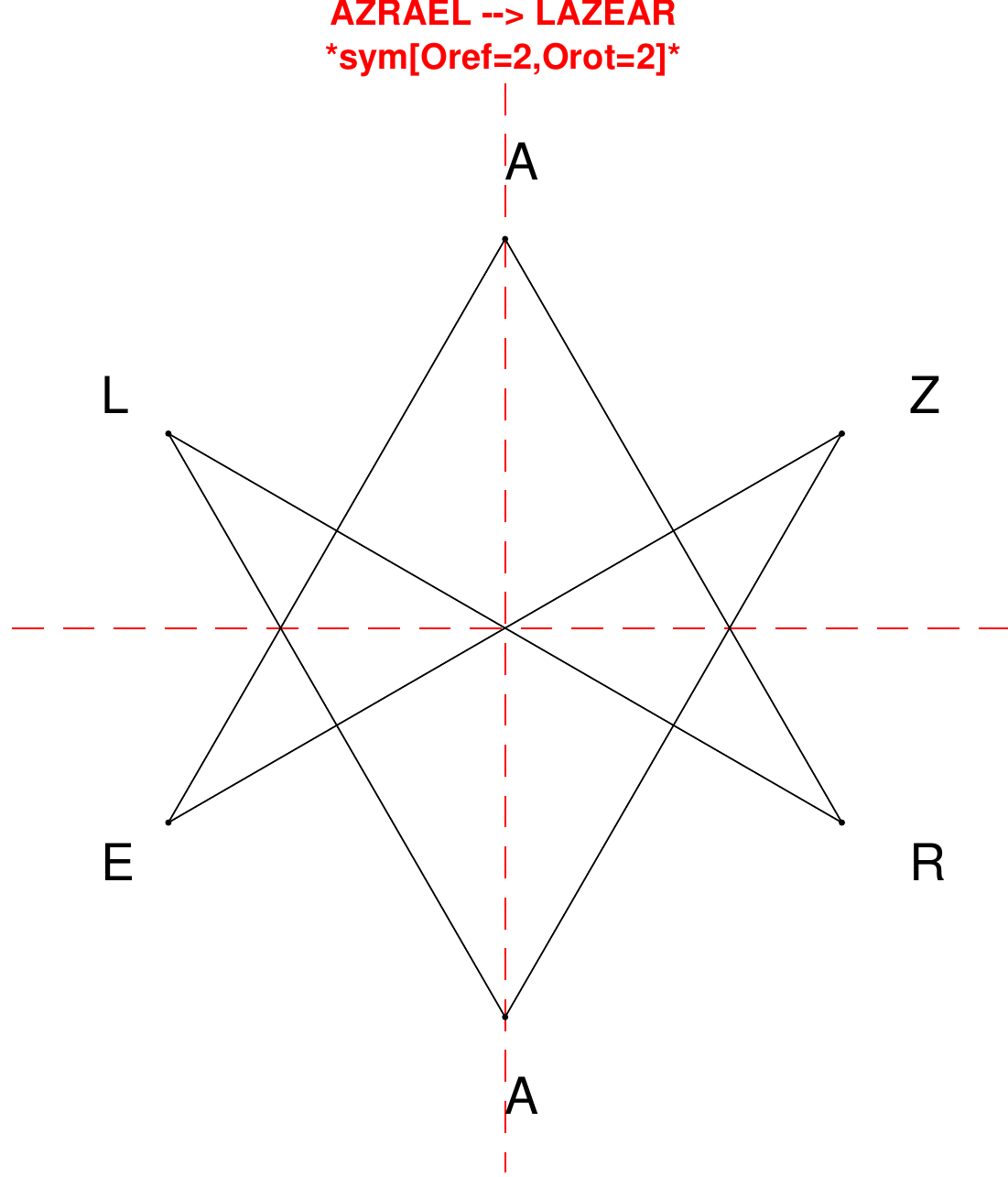}
\end{subfigure}
\hfill
\begin{subfigure}[T]{0.19\textwidth}
\centering
\includegraphics[width=\textwidth]{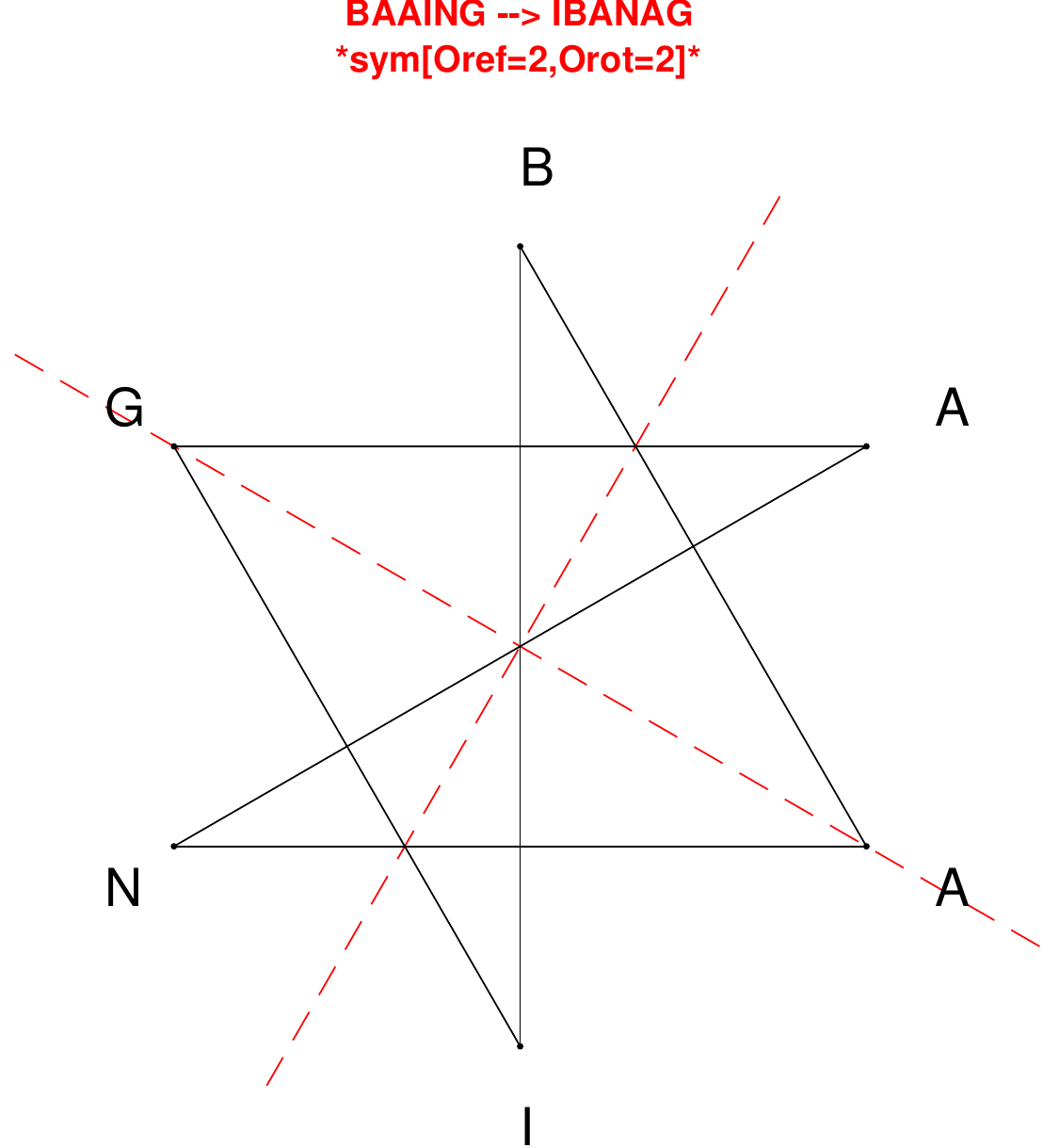}
\end{subfigure}
\end{figure}

\begin{figure}[H]
\centering
\begin{subfigure}[T]{0.19\textwidth}
\centering
\includegraphics[width=\textwidth]{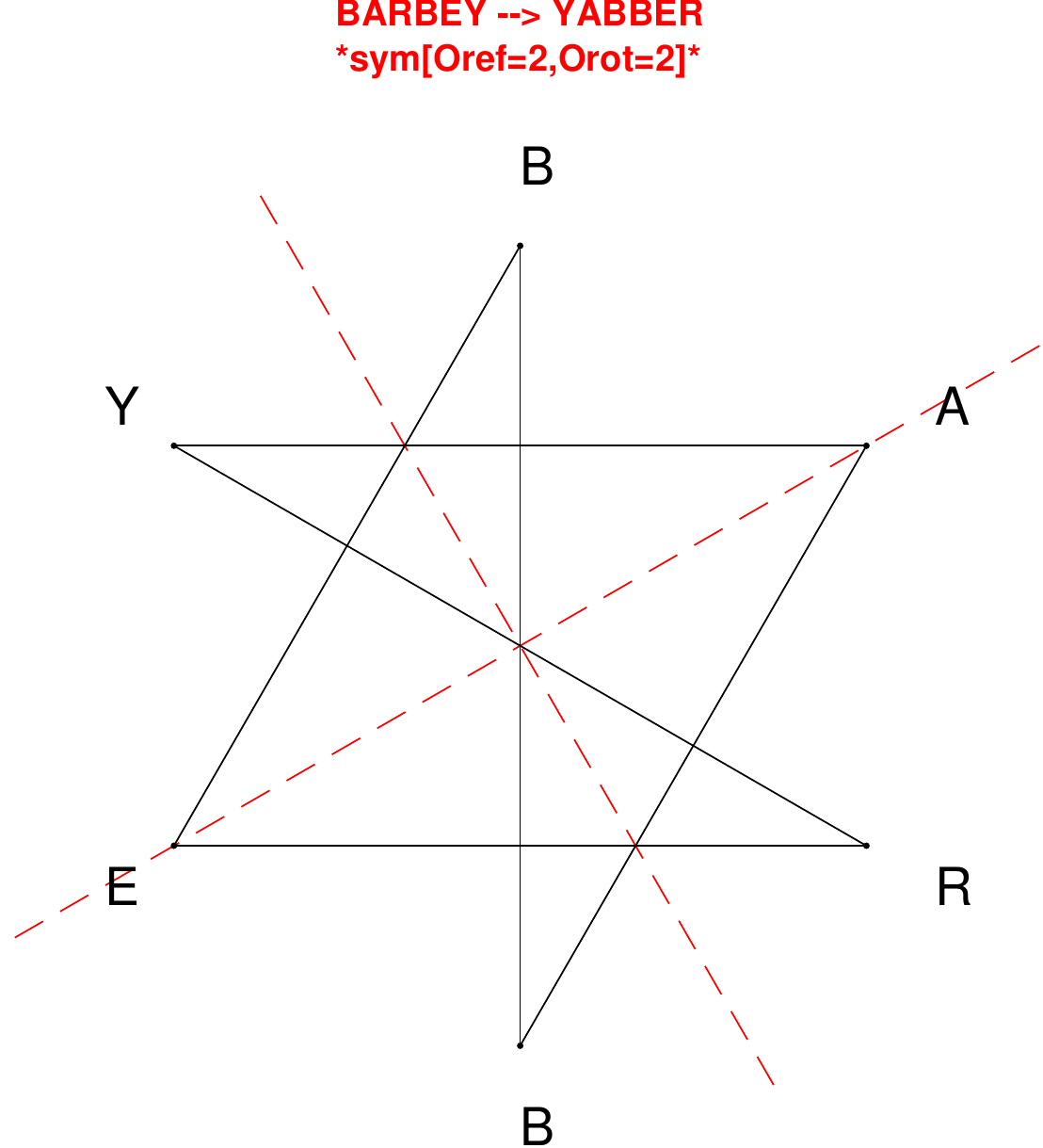}
\end{subfigure}
\hfill
\begin{subfigure}[T]{0.19\textwidth}
\centering
\includegraphics[width=\textwidth]{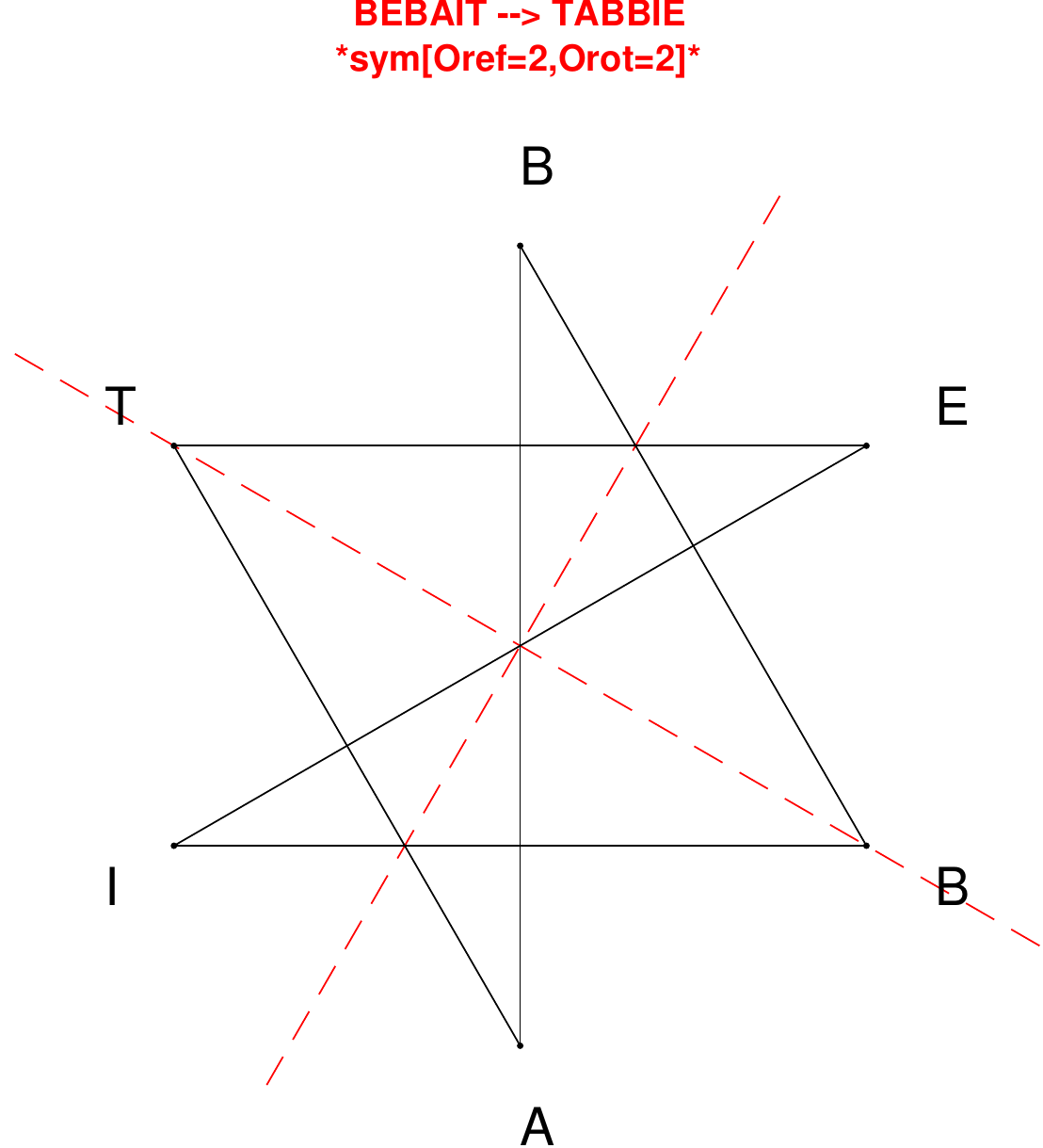}
\end{subfigure}
\hfill
\begin{subfigure}[T]{0.19\textwidth}
\centering
\includegraphics[width=\textwidth]{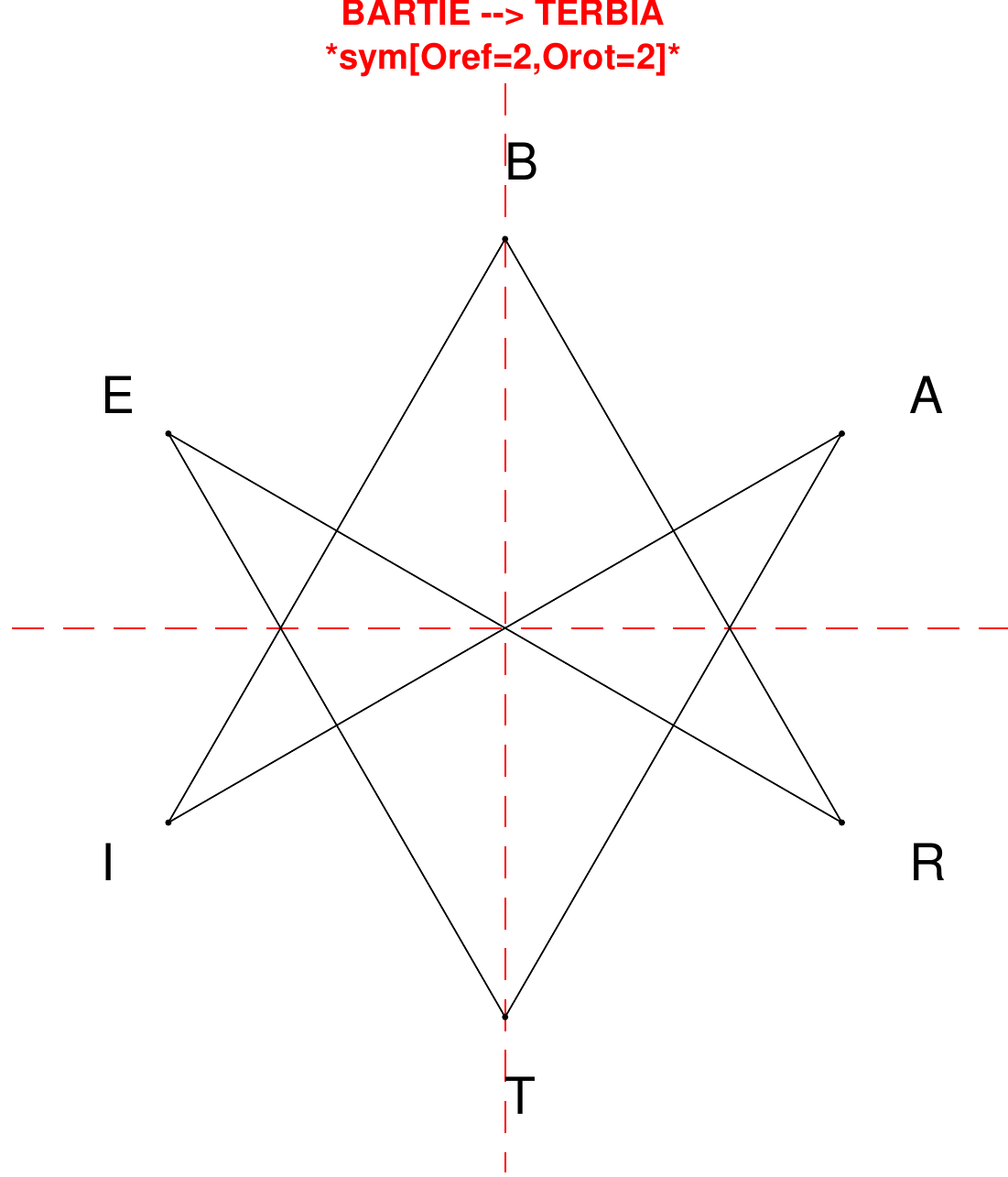}
\end{subfigure}
\hfill
\begin{subfigure}[T]{0.19\textwidth}
\centering
\includegraphics[width=\textwidth]{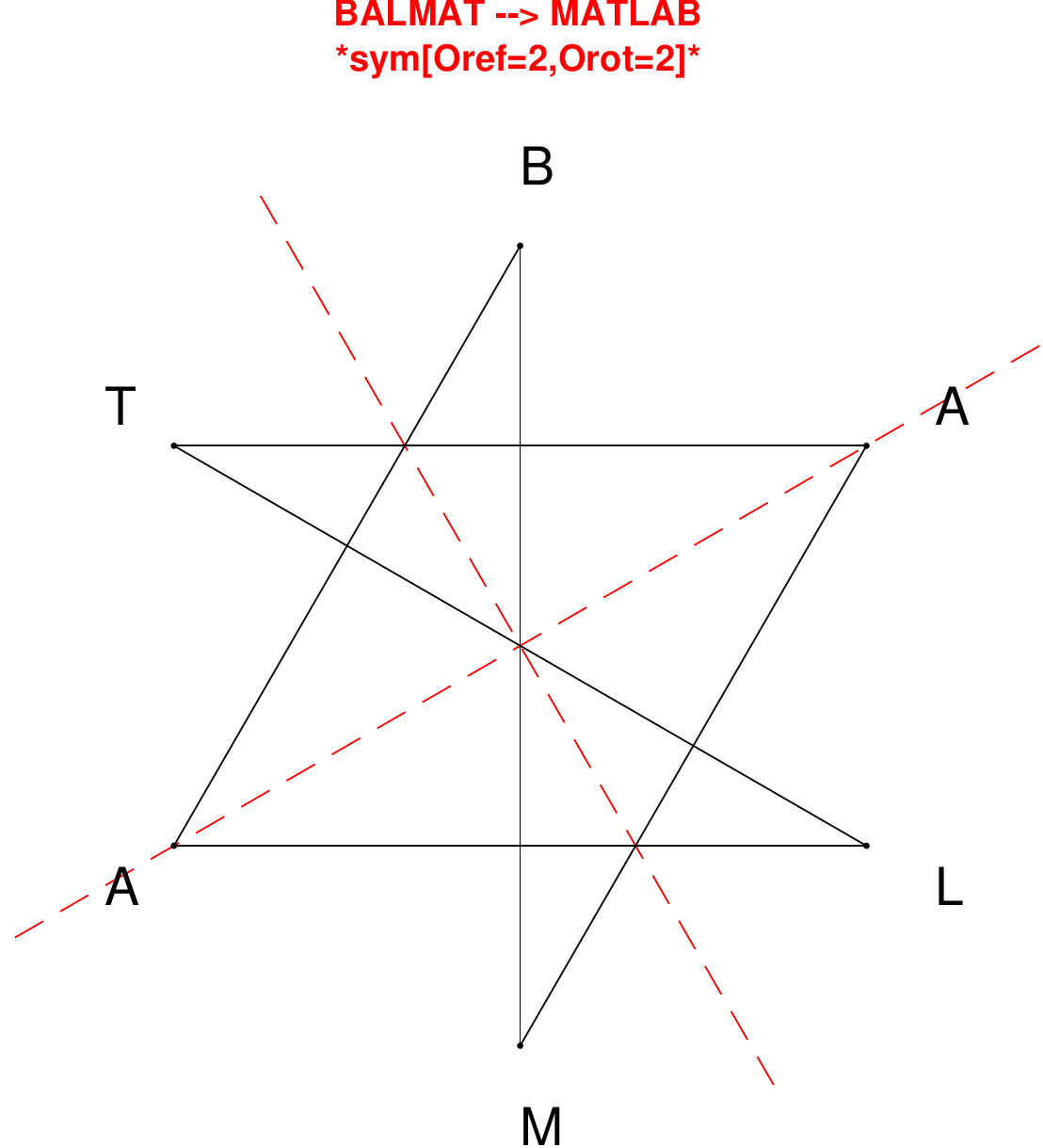}
\end{subfigure}
\hfill
\begin{subfigure}[T]{0.19\textwidth}
\centering
\includegraphics[width=\textwidth]{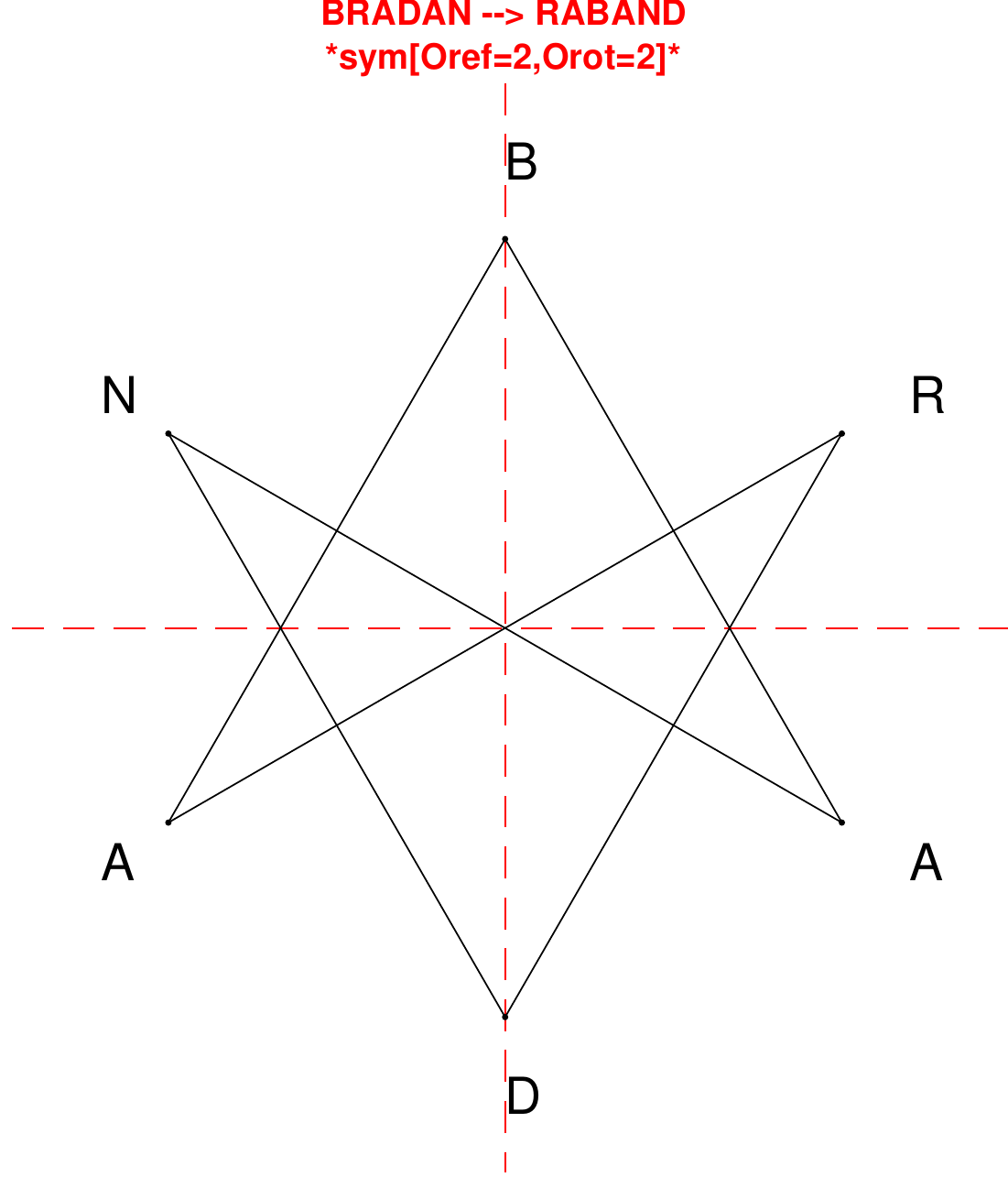}
\end{subfigure}
\end{figure}

\begin{figure}[H]
\centering
\begin{subfigure}[T]{0.19\textwidth}
\centering
\includegraphics[width=\textwidth]{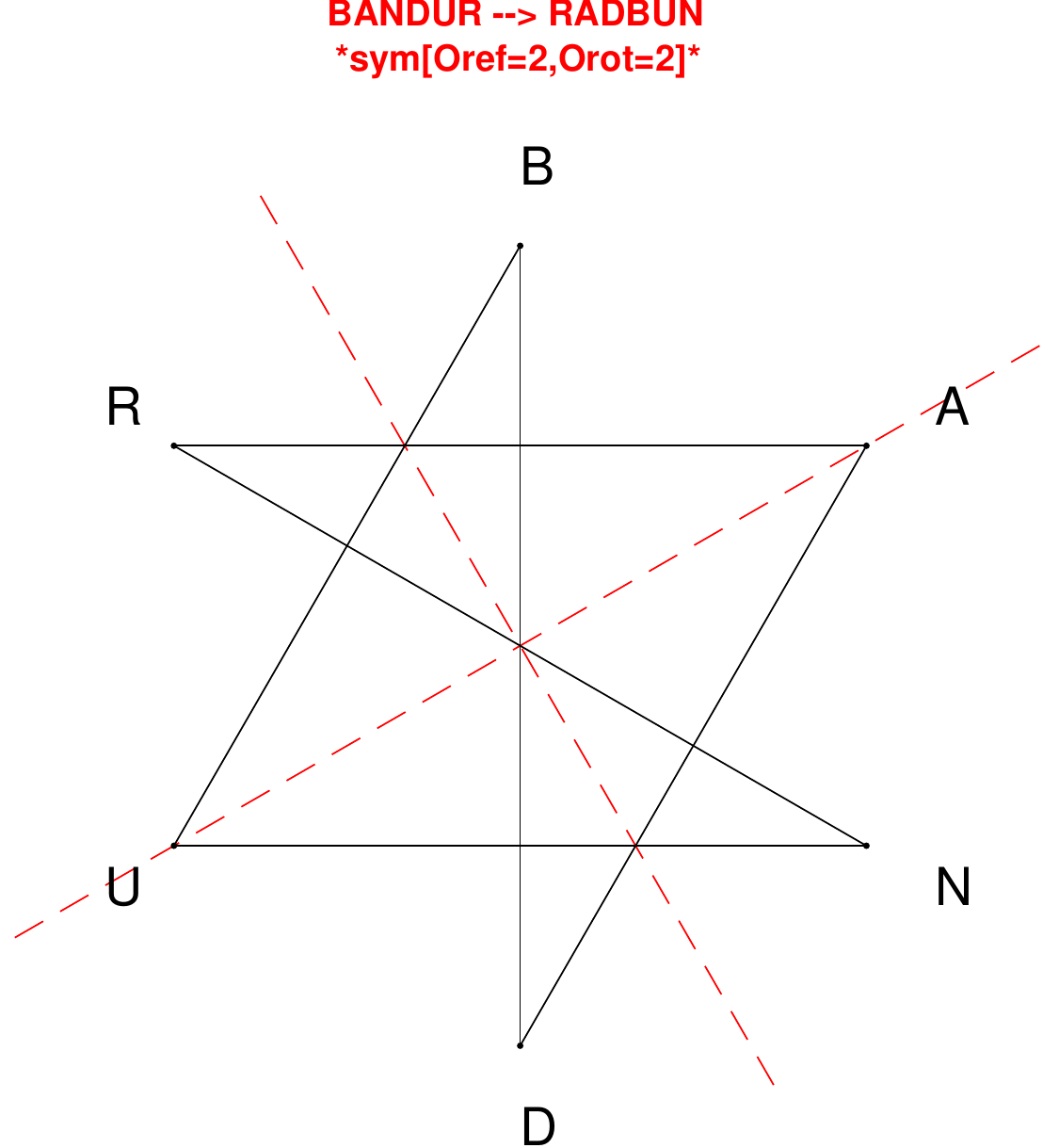}
\end{subfigure}
\hfill
\begin{subfigure}[T]{0.19\textwidth}
\centering
\includegraphics[width=\textwidth]{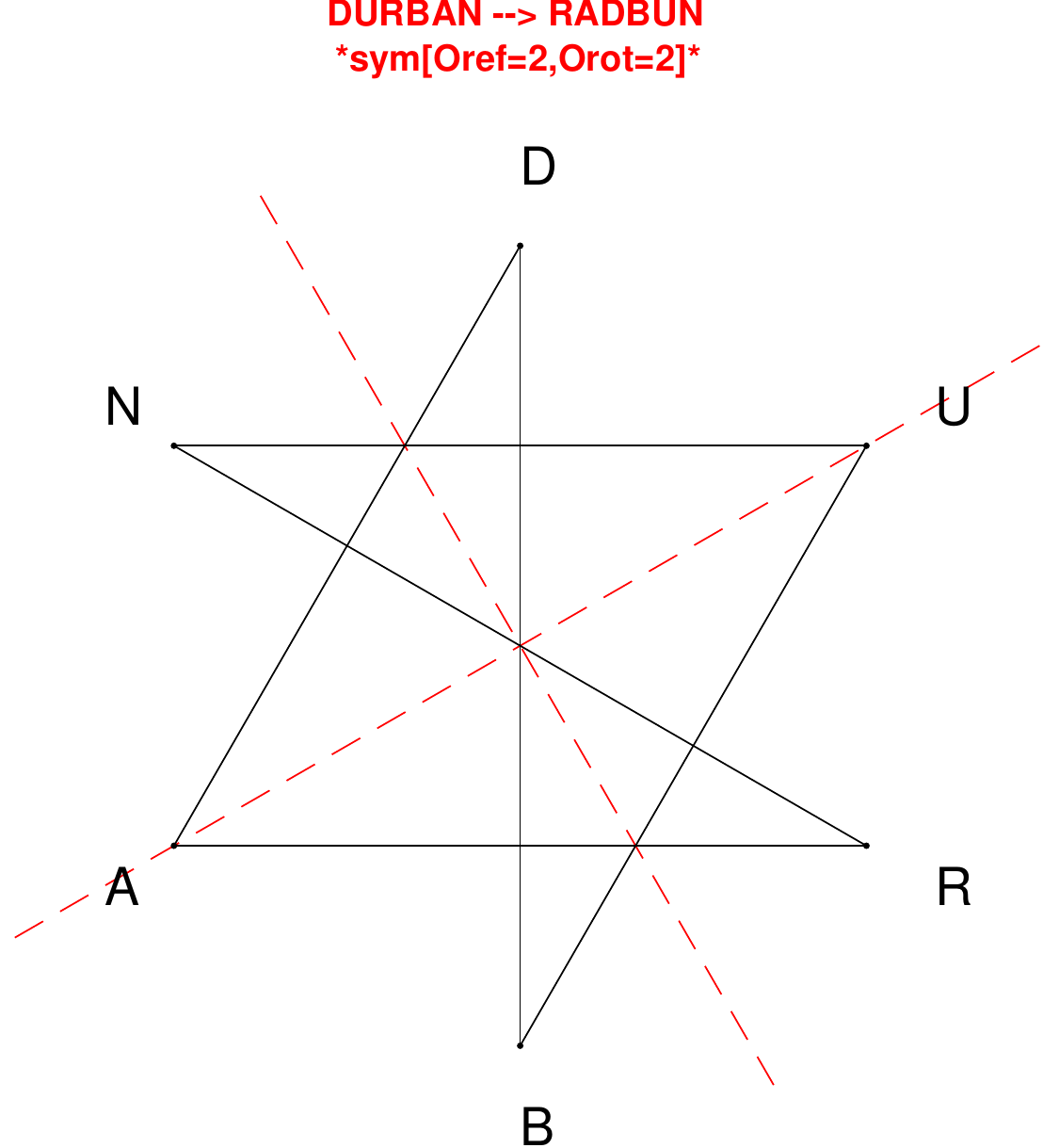}
\end{subfigure}
\hfill
\begin{subfigure}[T]{0.19\textwidth}
\centering
\includegraphics[width=\textwidth]{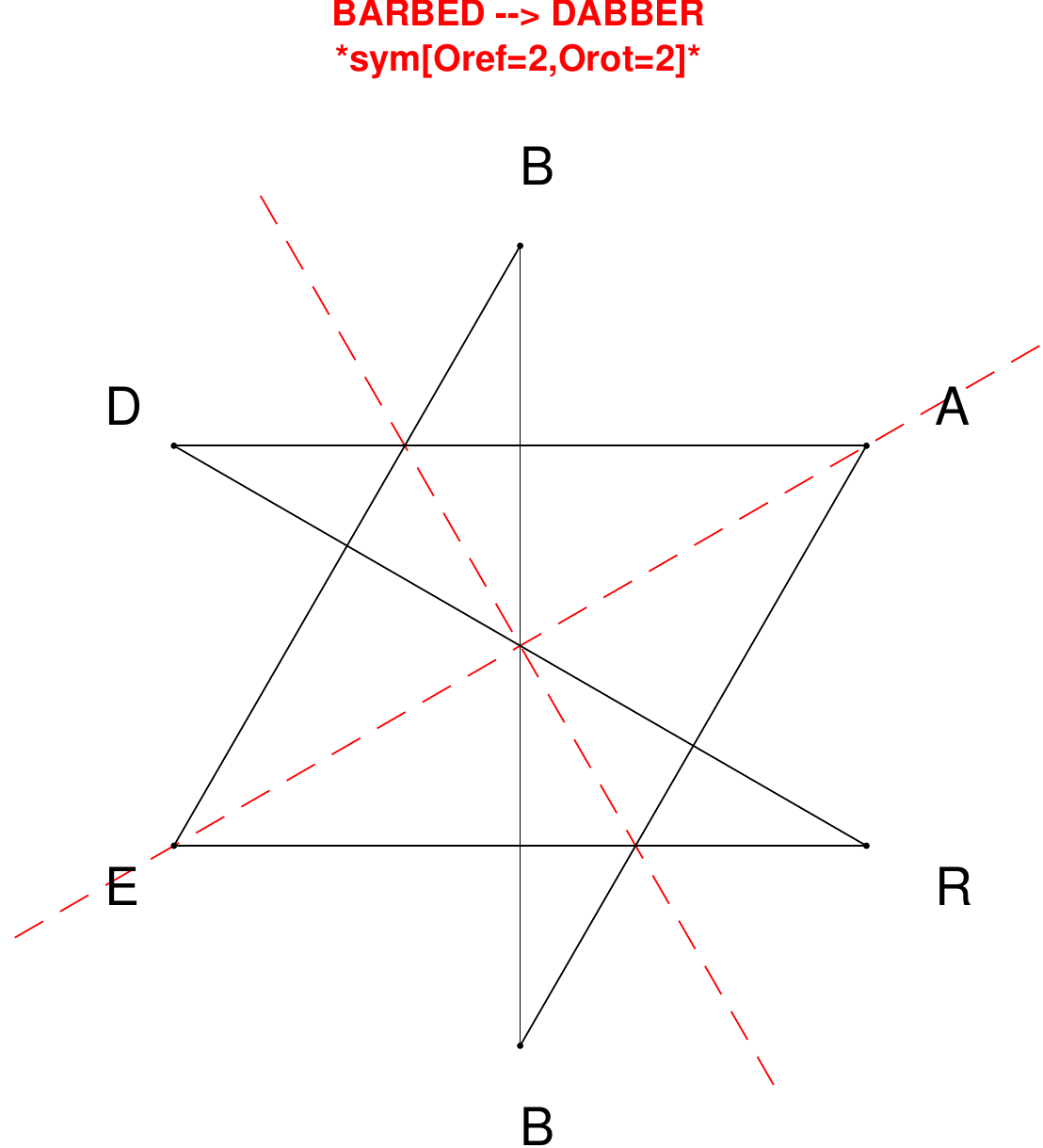}
\end{subfigure}
\hfill
\begin{subfigure}[T]{0.19\textwidth}
\centering
\includegraphics[width=\textwidth]{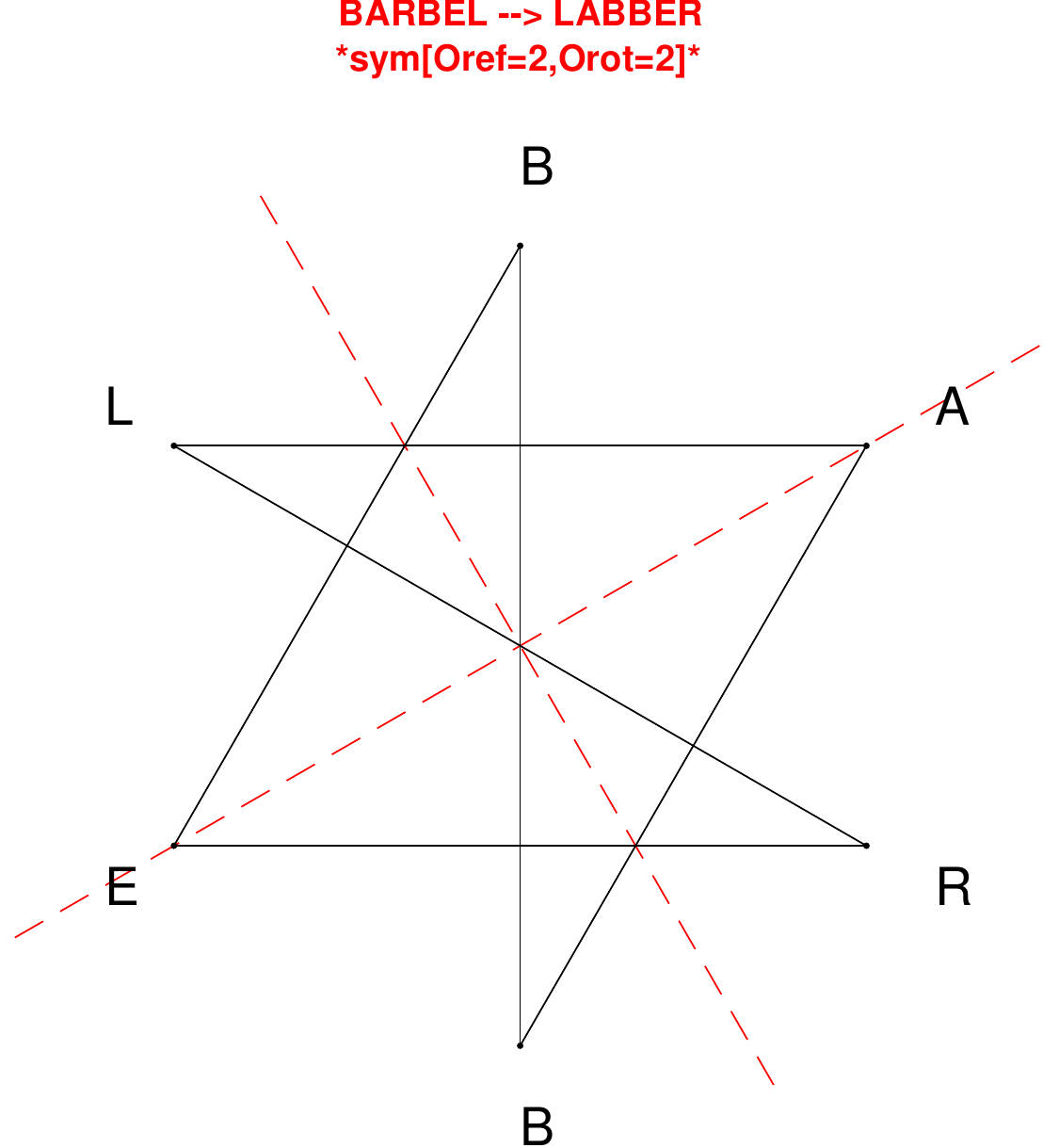}
\end{subfigure}
\hfill
\begin{subfigure}[T]{0.19\textwidth}
\centering
\includegraphics[width=\textwidth]{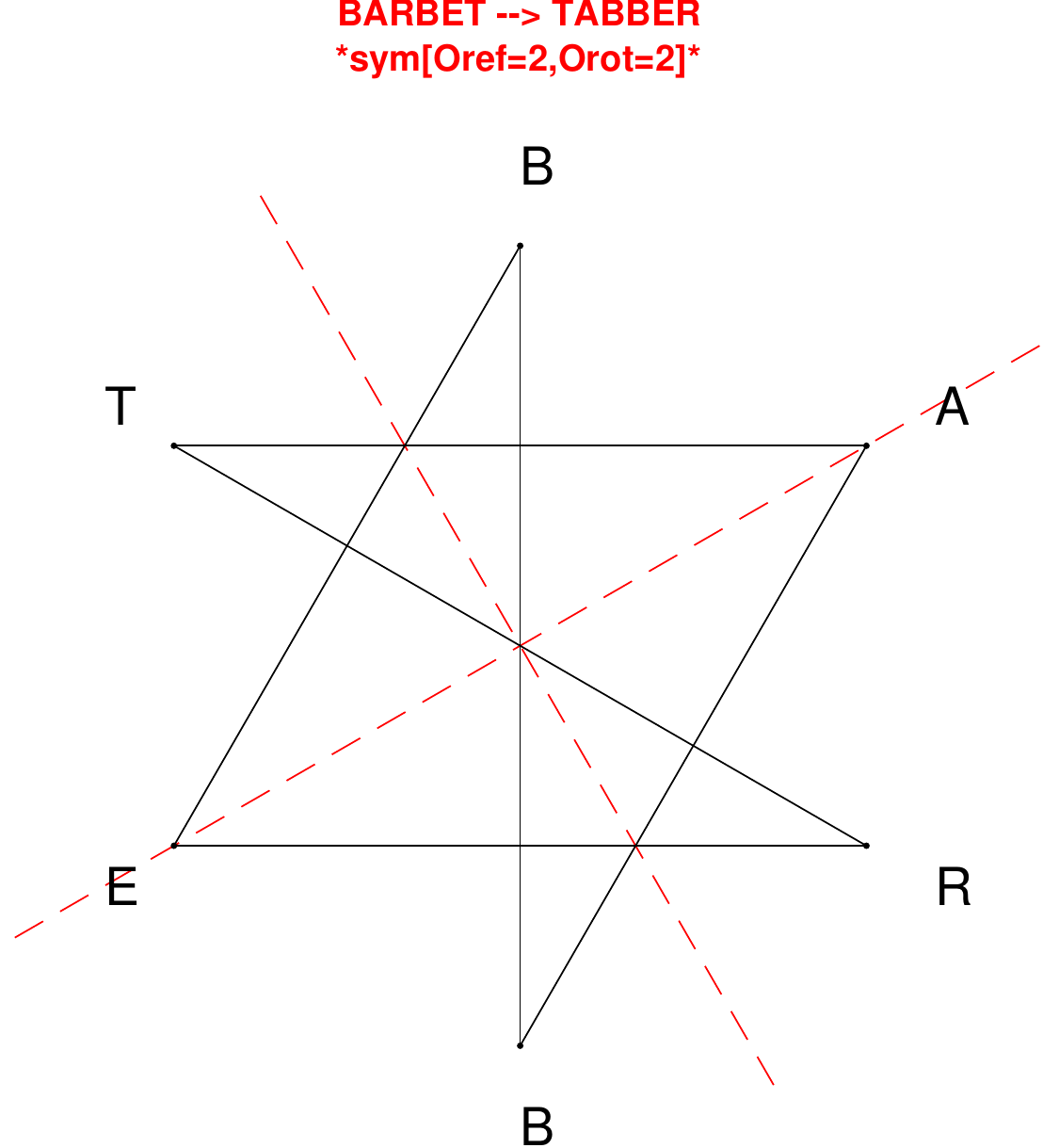}
\end{subfigure}
\end{figure}

\begin{figure}[H]
\centering
\begin{subfigure}[T]{0.19\textwidth}
\centering
\includegraphics[width=\textwidth]{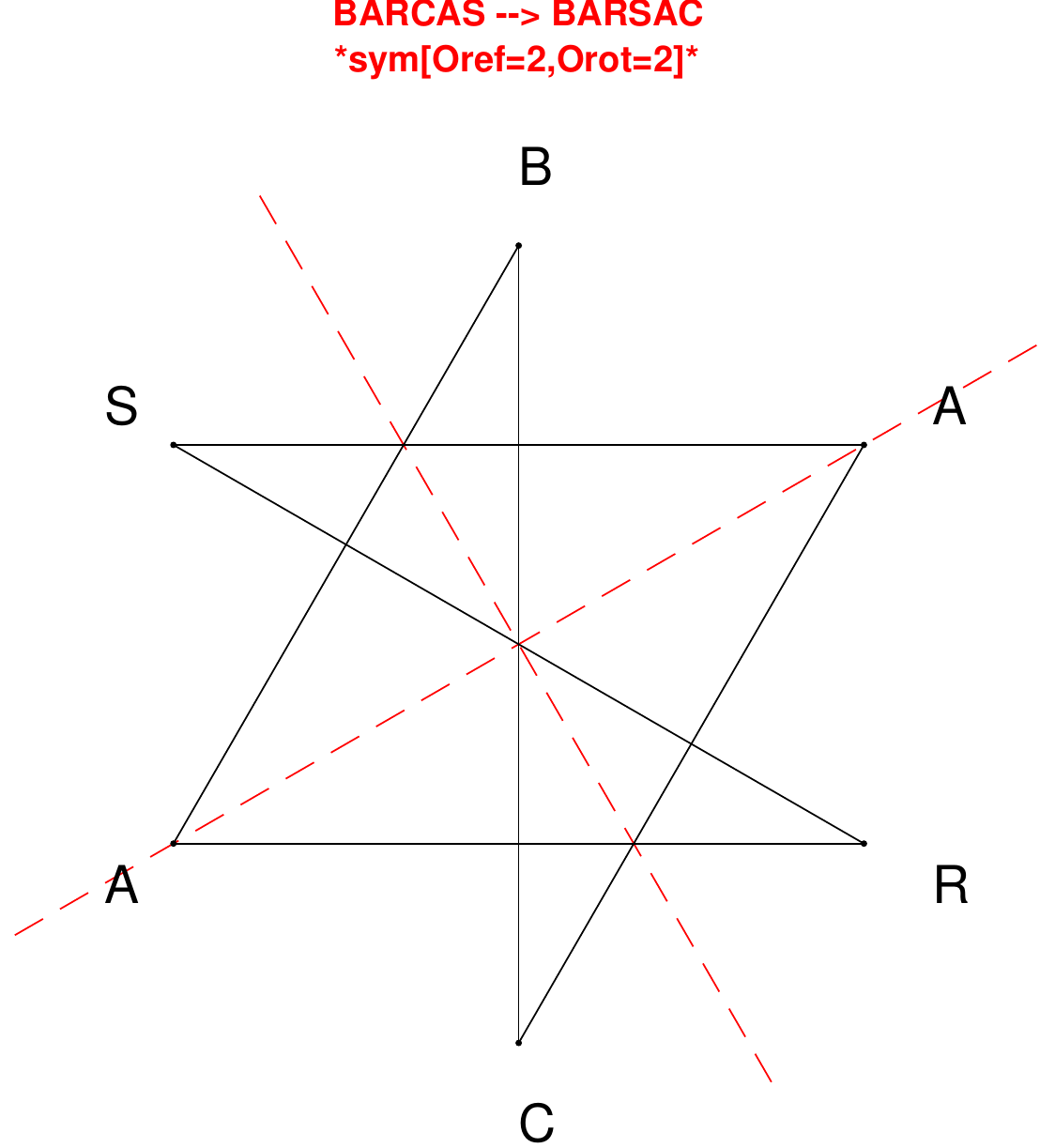}
\end{subfigure}
\hfill
\begin{subfigure}[T]{0.19\textwidth}
\centering
\includegraphics[width=\textwidth]{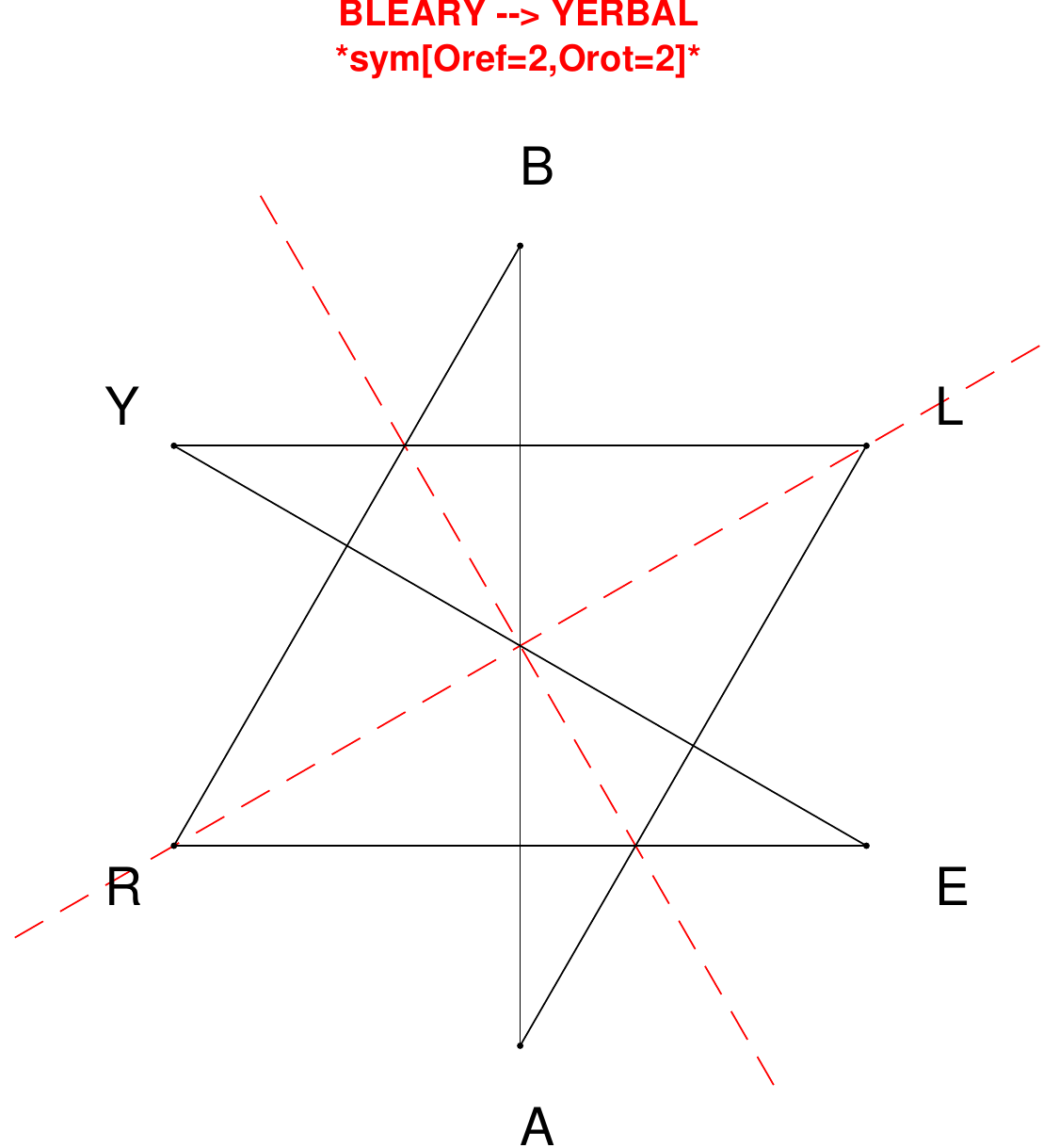}
\end{subfigure}
\hfill
\begin{subfigure}[T]{0.19\textwidth}
\centering
\includegraphics[width=\textwidth]{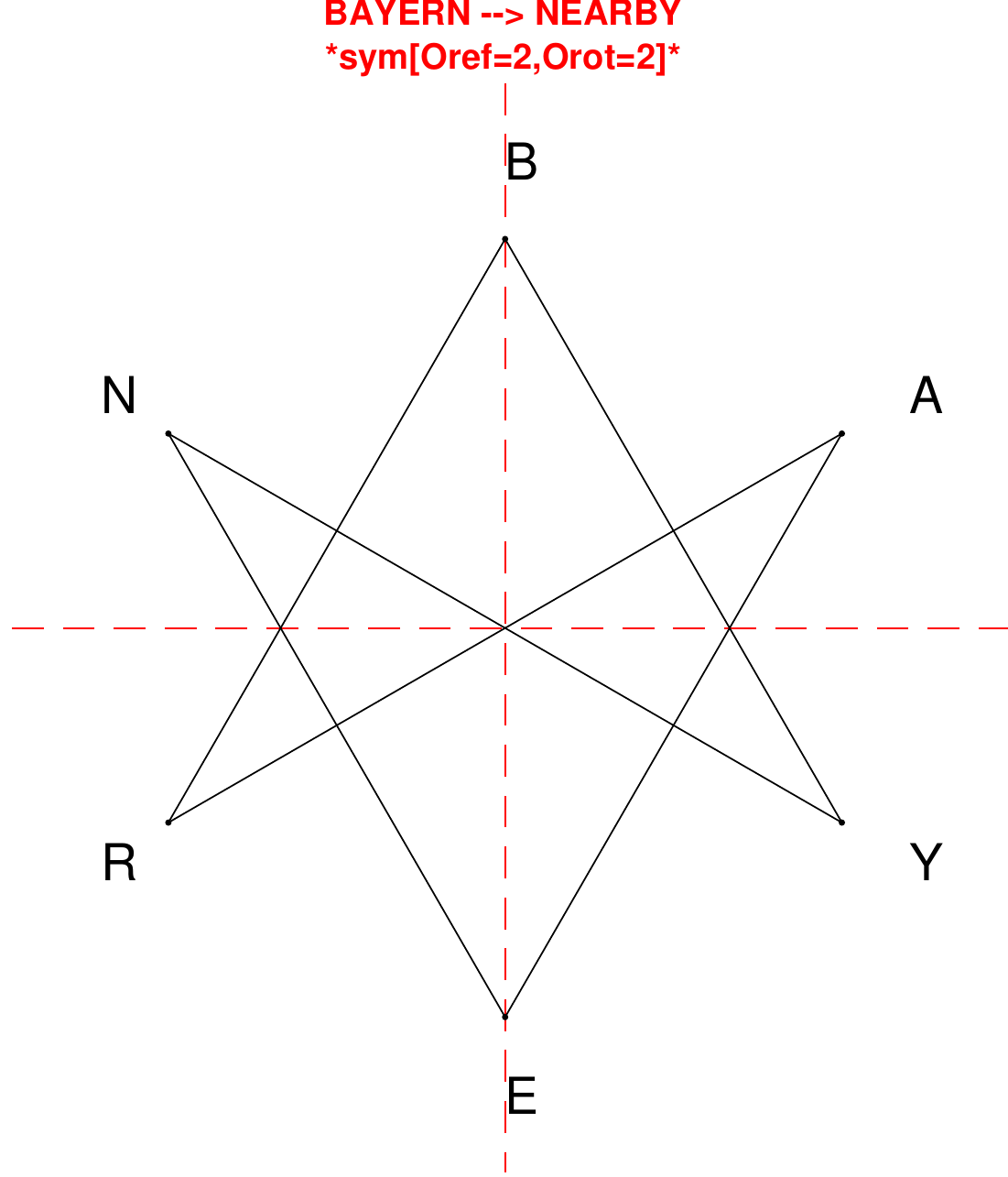}
\end{subfigure}
\hfill
\begin{subfigure}[T]{0.19\textwidth}
\centering
\includegraphics[width=\textwidth]{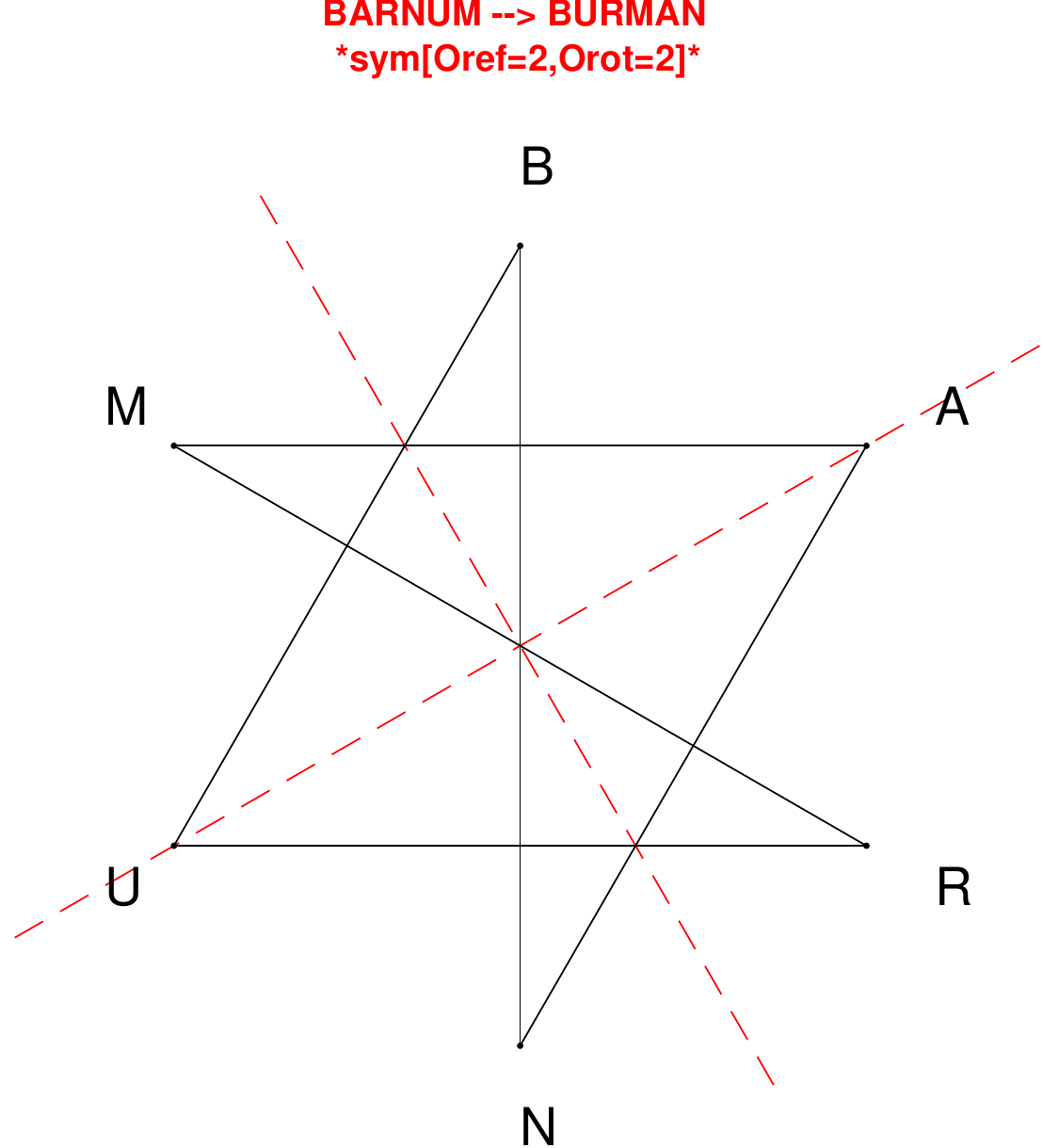}
\end{subfigure}
\hfill
\begin{subfigure}[T]{0.19\textwidth}
\centering
\includegraphics[width=\textwidth]{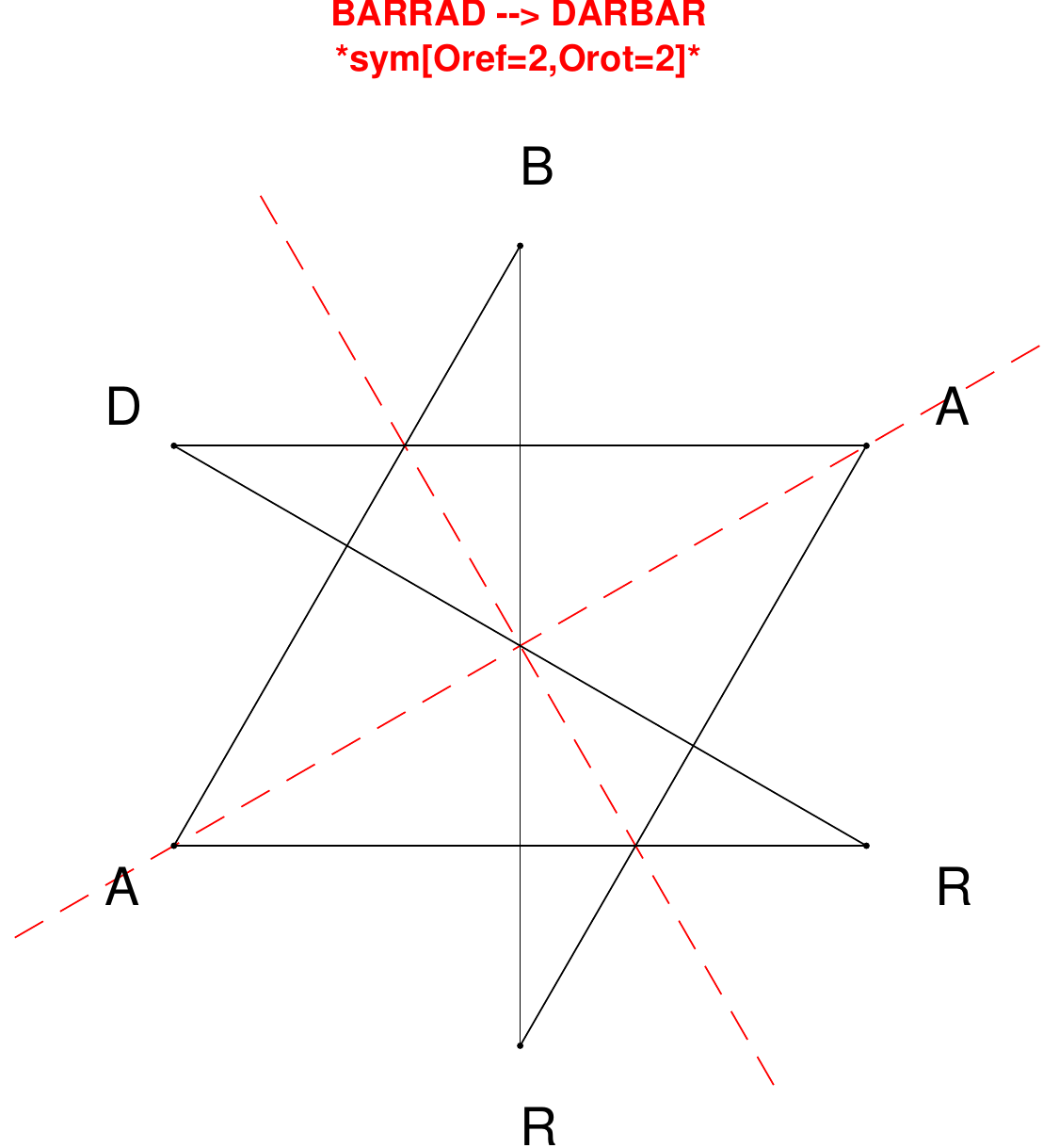}
\end{subfigure}
\end{figure}

\begin{figure}[H]
\centering
\begin{subfigure}[T]{0.19\textwidth}
\centering
\includegraphics[width=\textwidth]{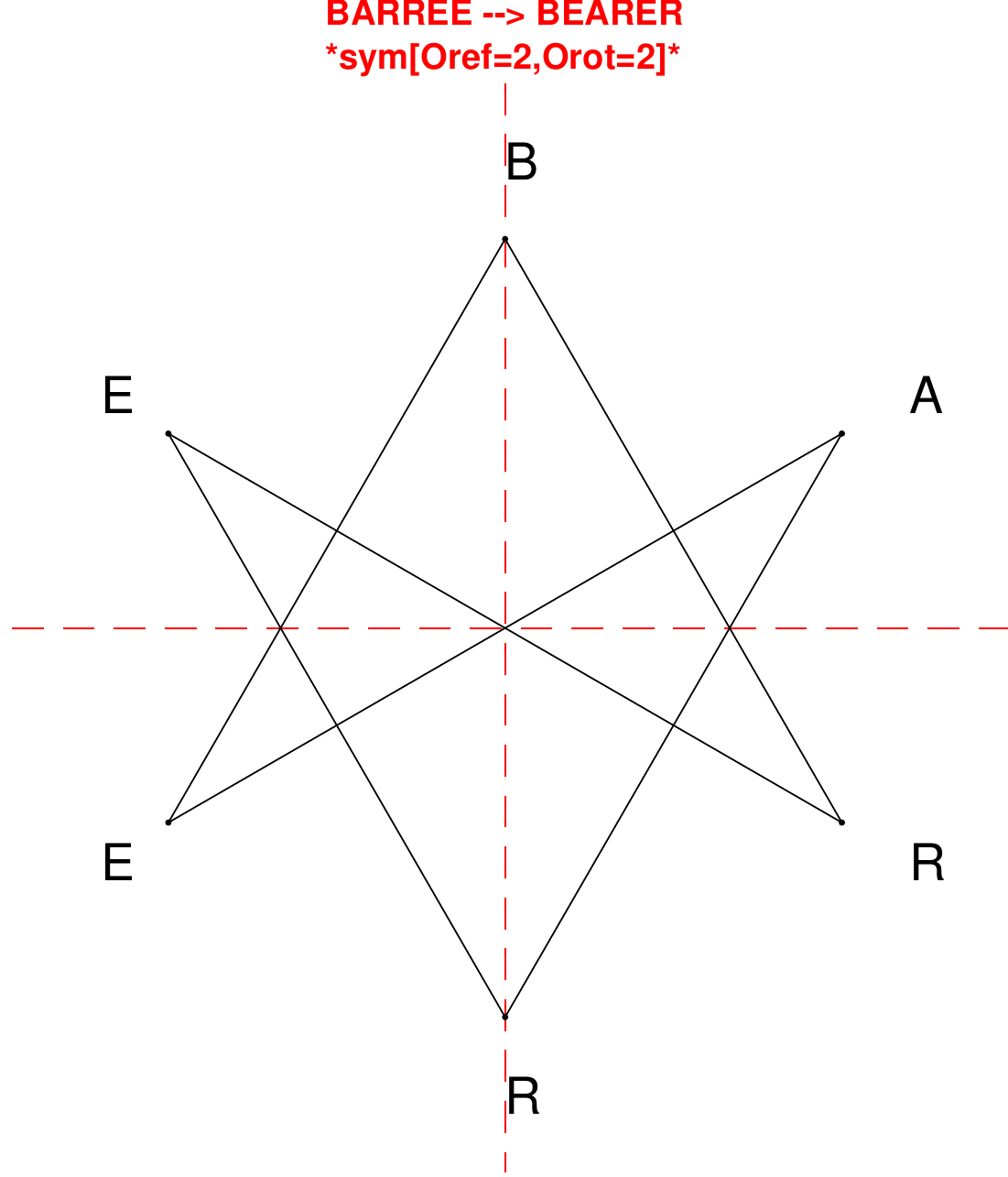}
\end{subfigure}
\hfill
\begin{subfigure}[T]{0.19\textwidth}
\centering
\includegraphics[width=\textwidth]{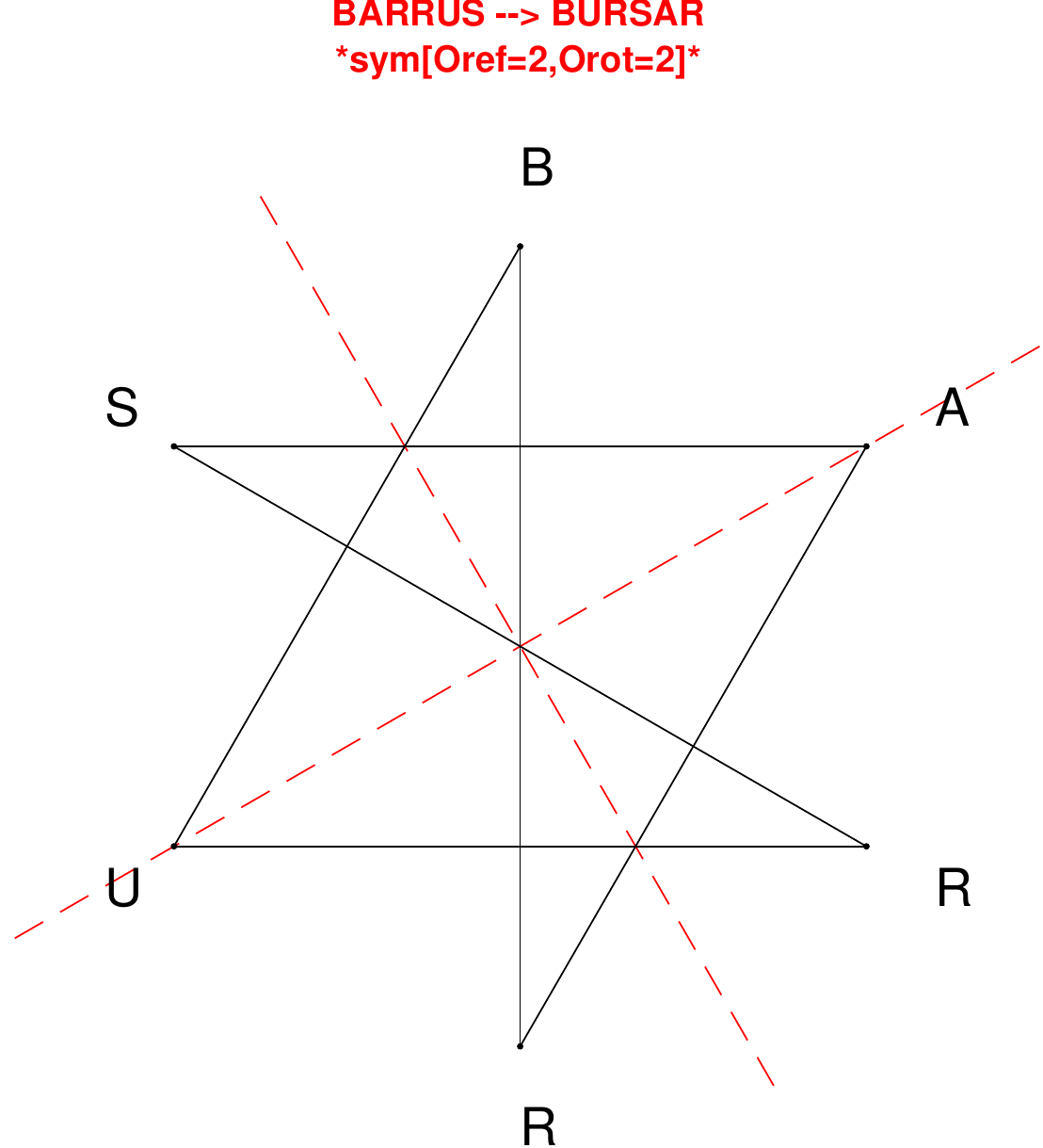}
\end{subfigure}
\hfill
\begin{subfigure}[T]{0.19\textwidth}
\centering
\includegraphics[width=\textwidth]{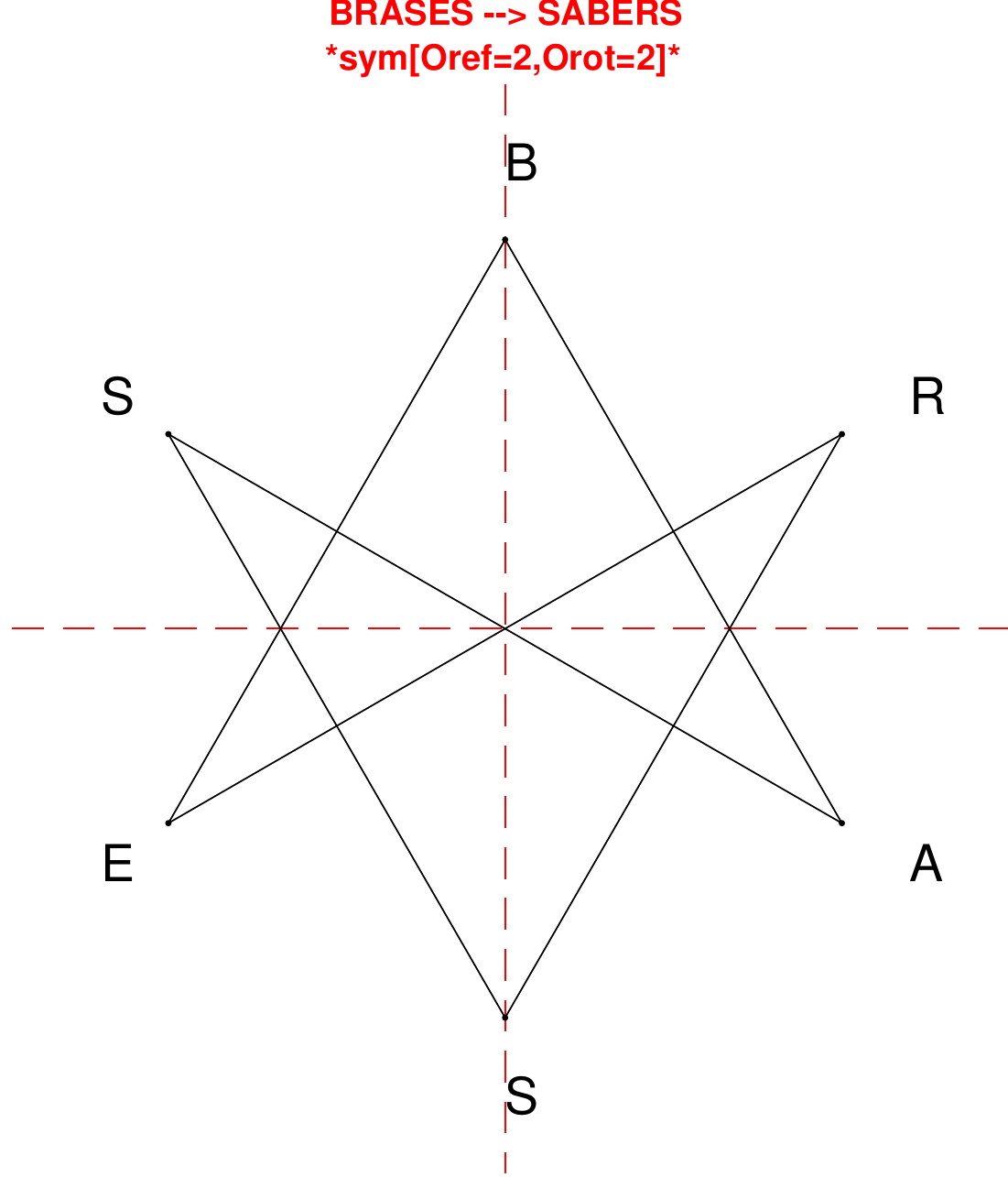}
\end{subfigure}
\hfill
\begin{subfigure}[T]{0.19\textwidth}
\centering
\includegraphics[width=\textwidth]{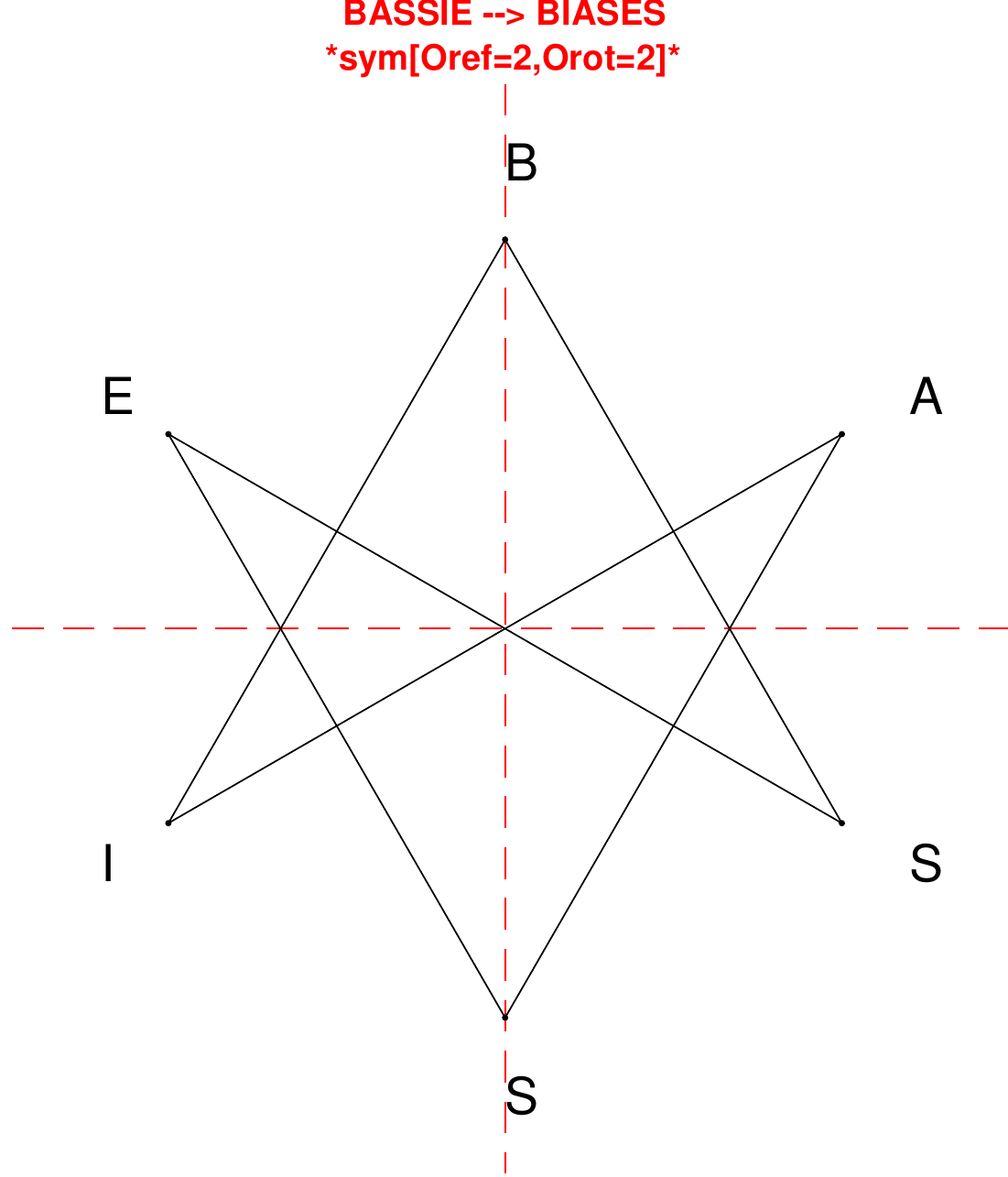}
\end{subfigure}
\hfill
\begin{subfigure}[T]{0.19\textwidth}
\centering
\includegraphics[width=\textwidth]{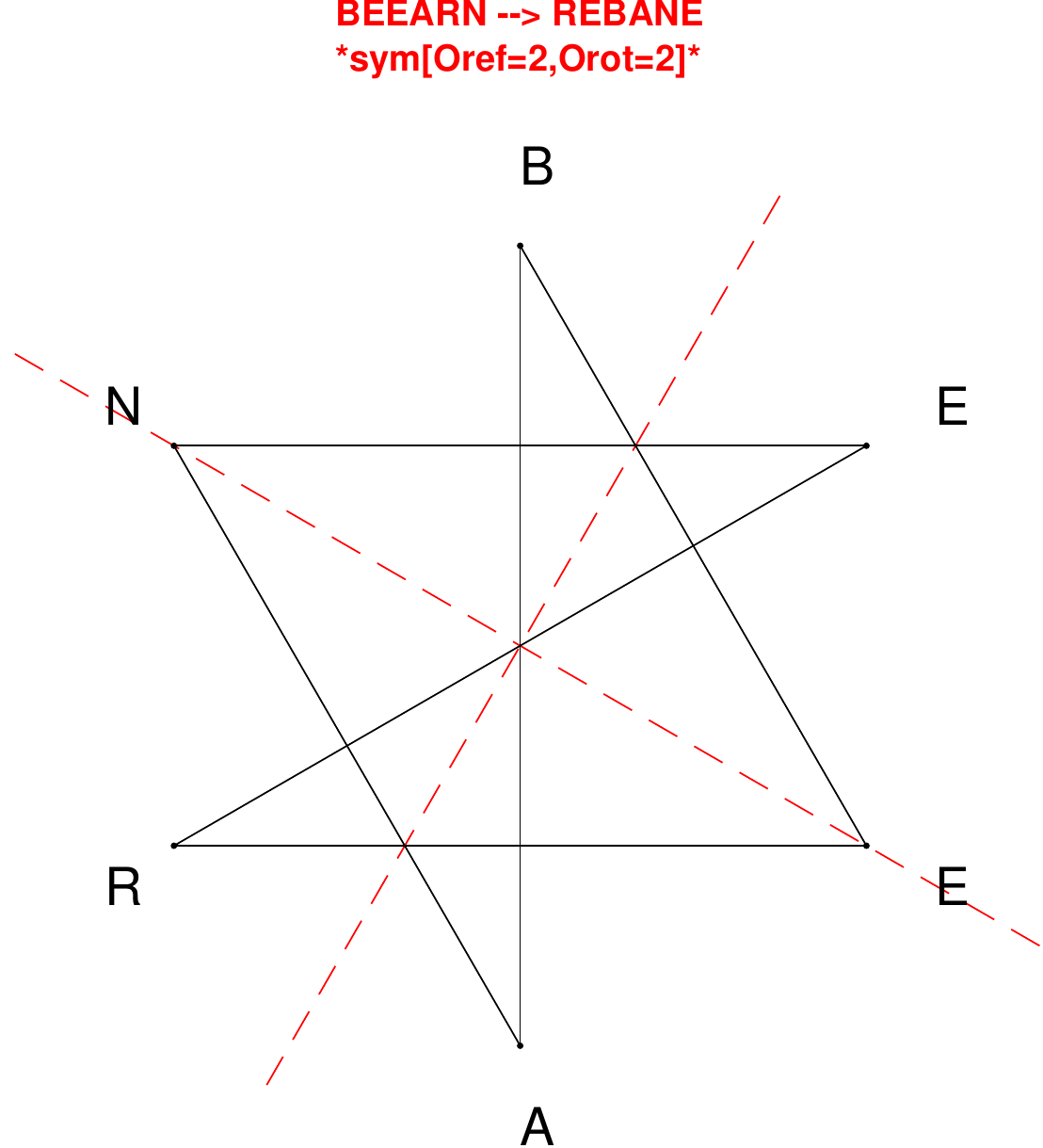}
\end{subfigure}
\end{figure}

\begin{figure}[H]
\centering
\begin{subfigure}[T]{0.19\textwidth}
\centering
\includegraphics[width=\textwidth]{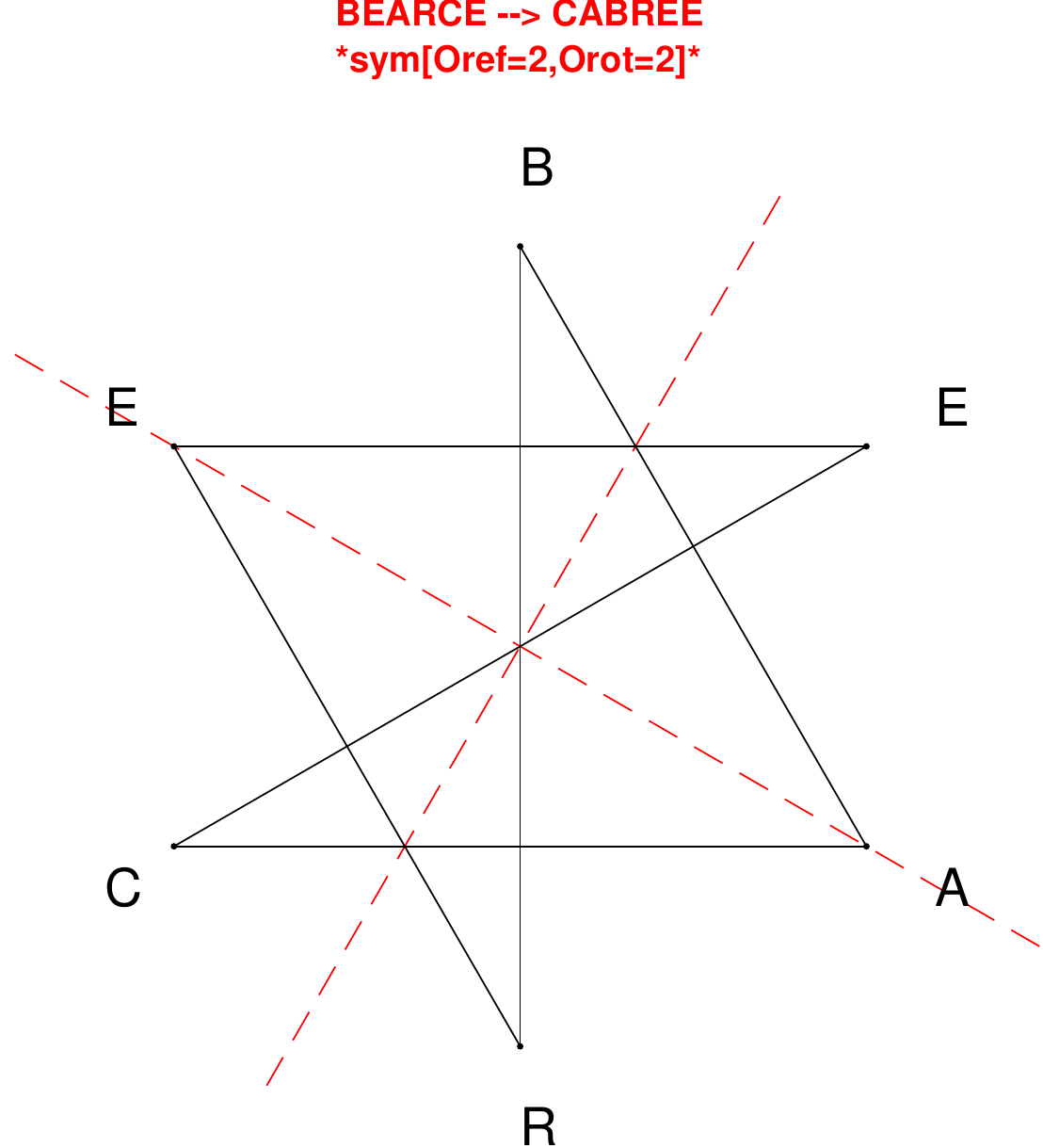}
\end{subfigure}
\hfill
\begin{subfigure}[T]{0.19\textwidth}
\centering
\includegraphics[width=\textwidth]{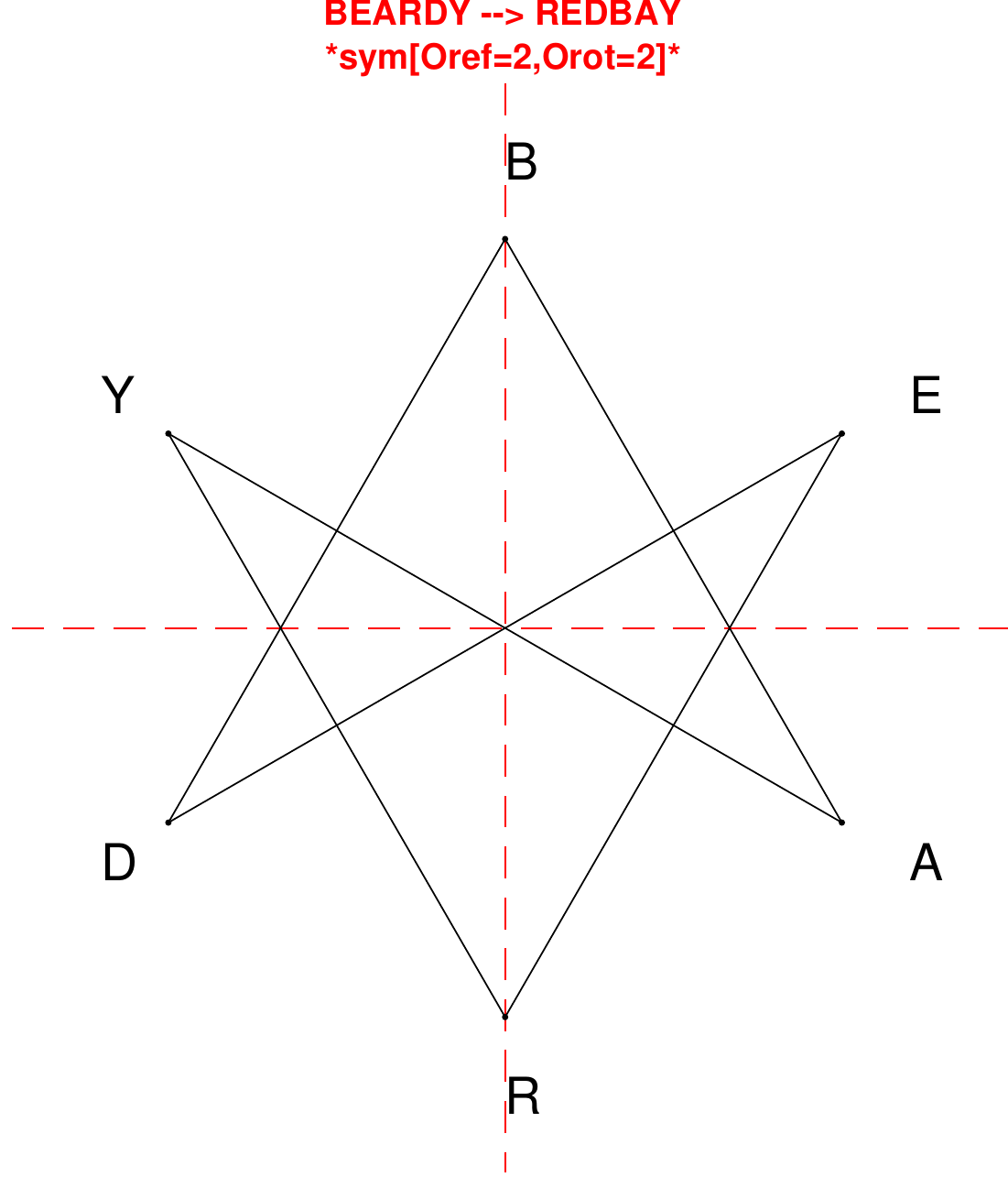}
\end{subfigure}
\hfill
\begin{subfigure}[T]{0.19\textwidth}
\centering
\includegraphics[width=\textwidth]{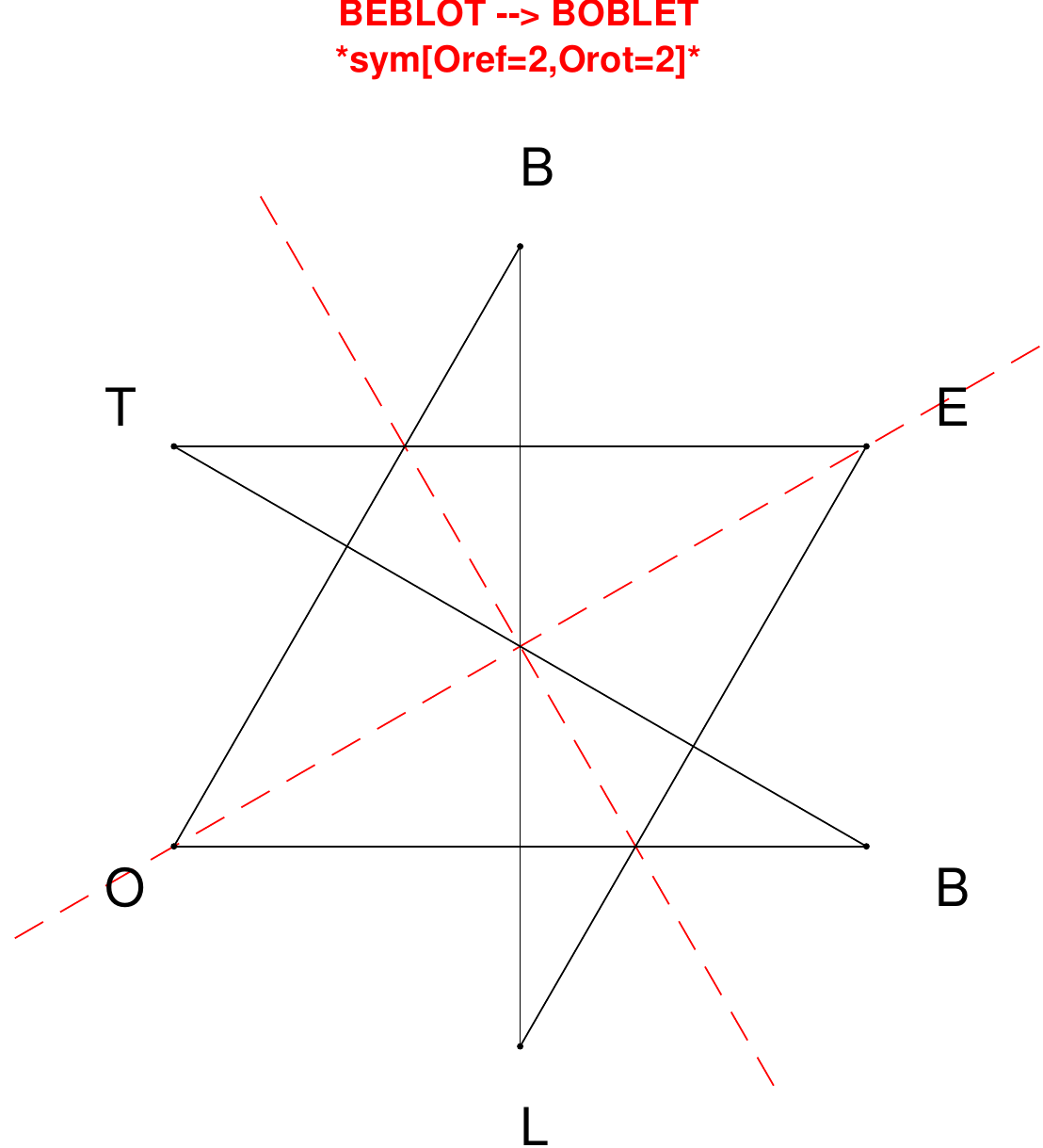}
\end{subfigure}
\hfill
\begin{subfigure}[T]{0.19\textwidth}
\centering
\includegraphics[width=\textwidth]{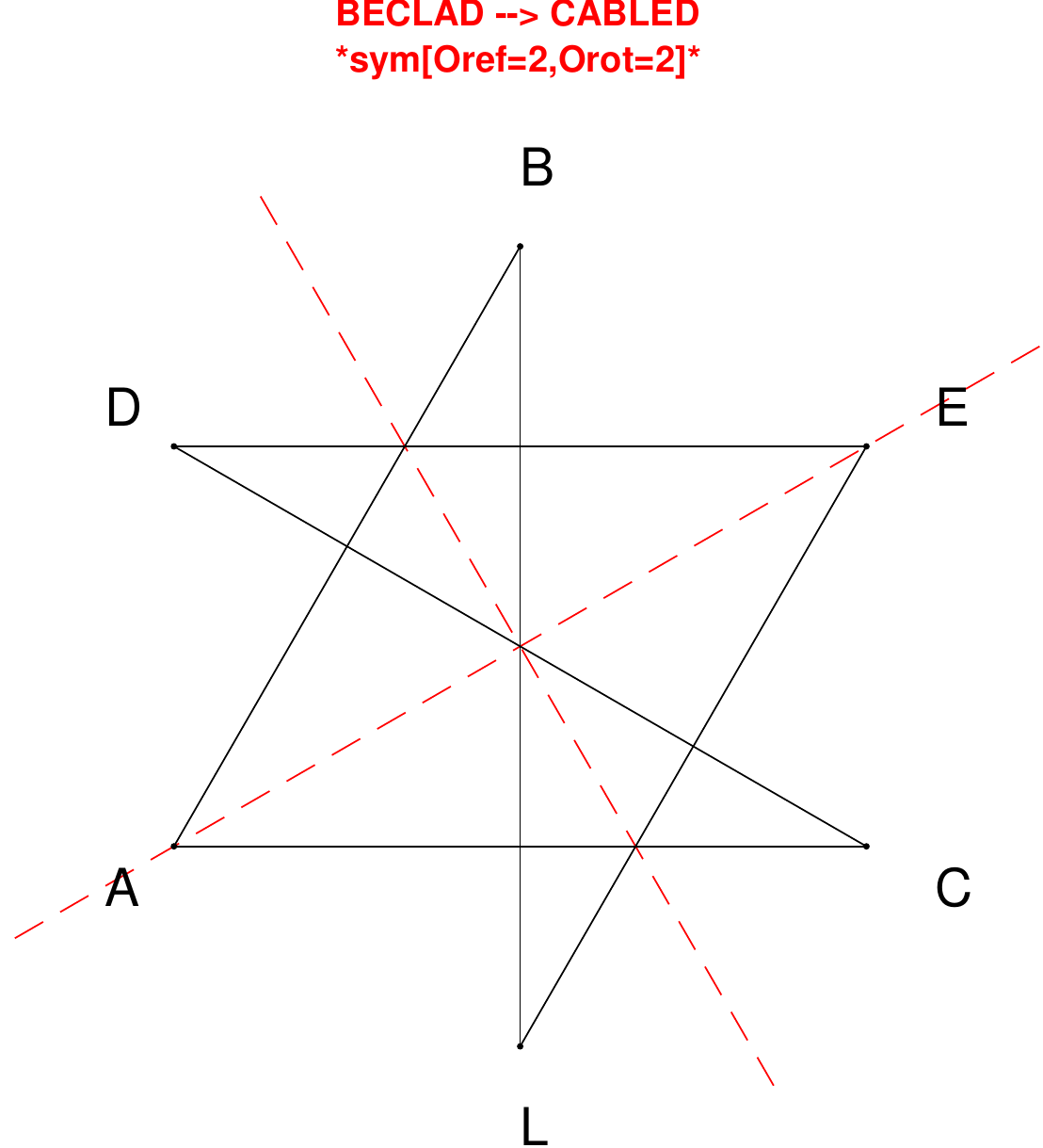}
\end{subfigure}
\hfill
\begin{subfigure}[T]{0.19\textwidth}
\centering
\includegraphics[width=\textwidth]{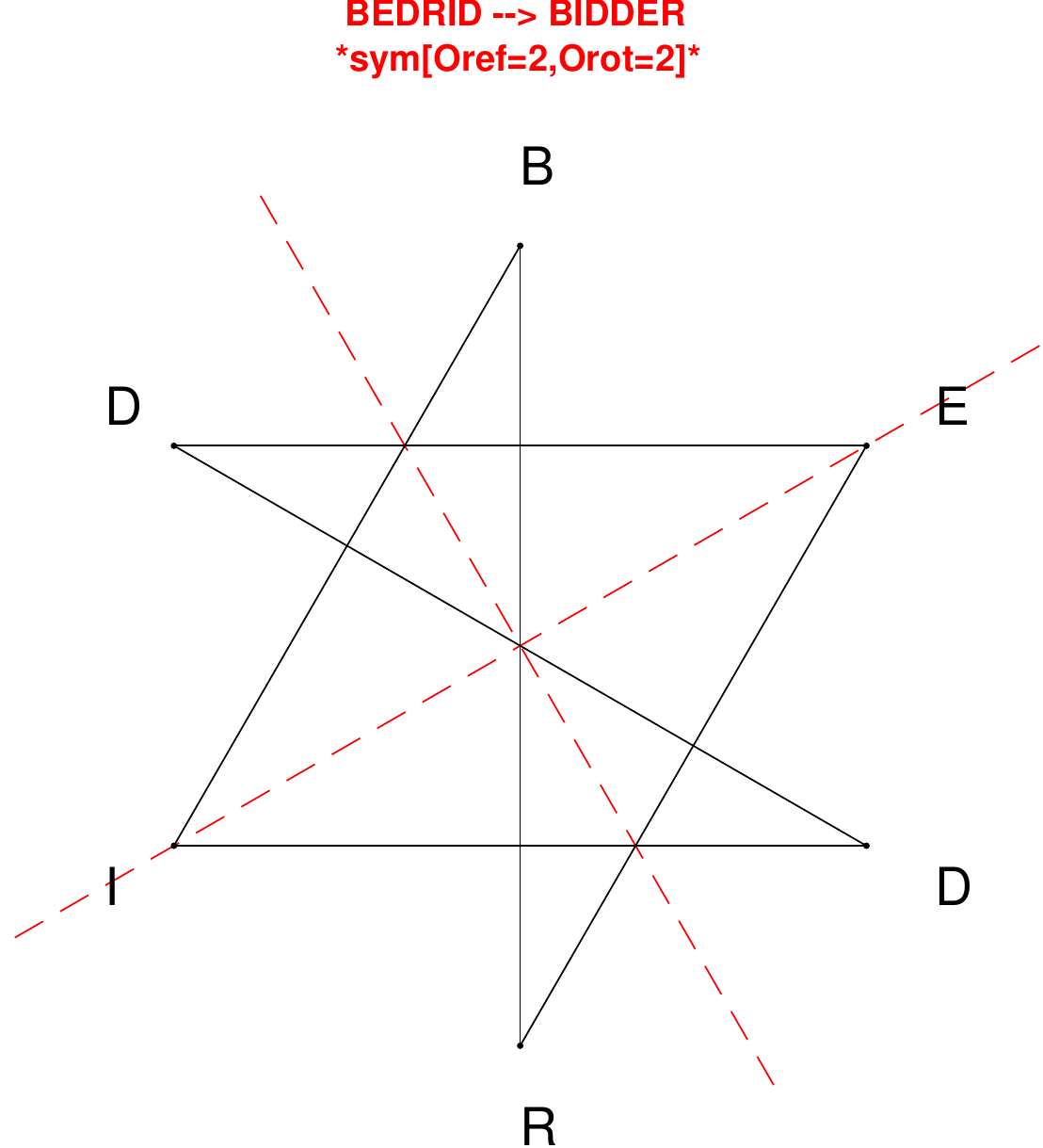}
\end{subfigure}
\end{figure}

\begin{figure}[H]
\centering
\begin{subfigure}[T]{0.19\textwidth}
\centering
\includegraphics[width=\textwidth]{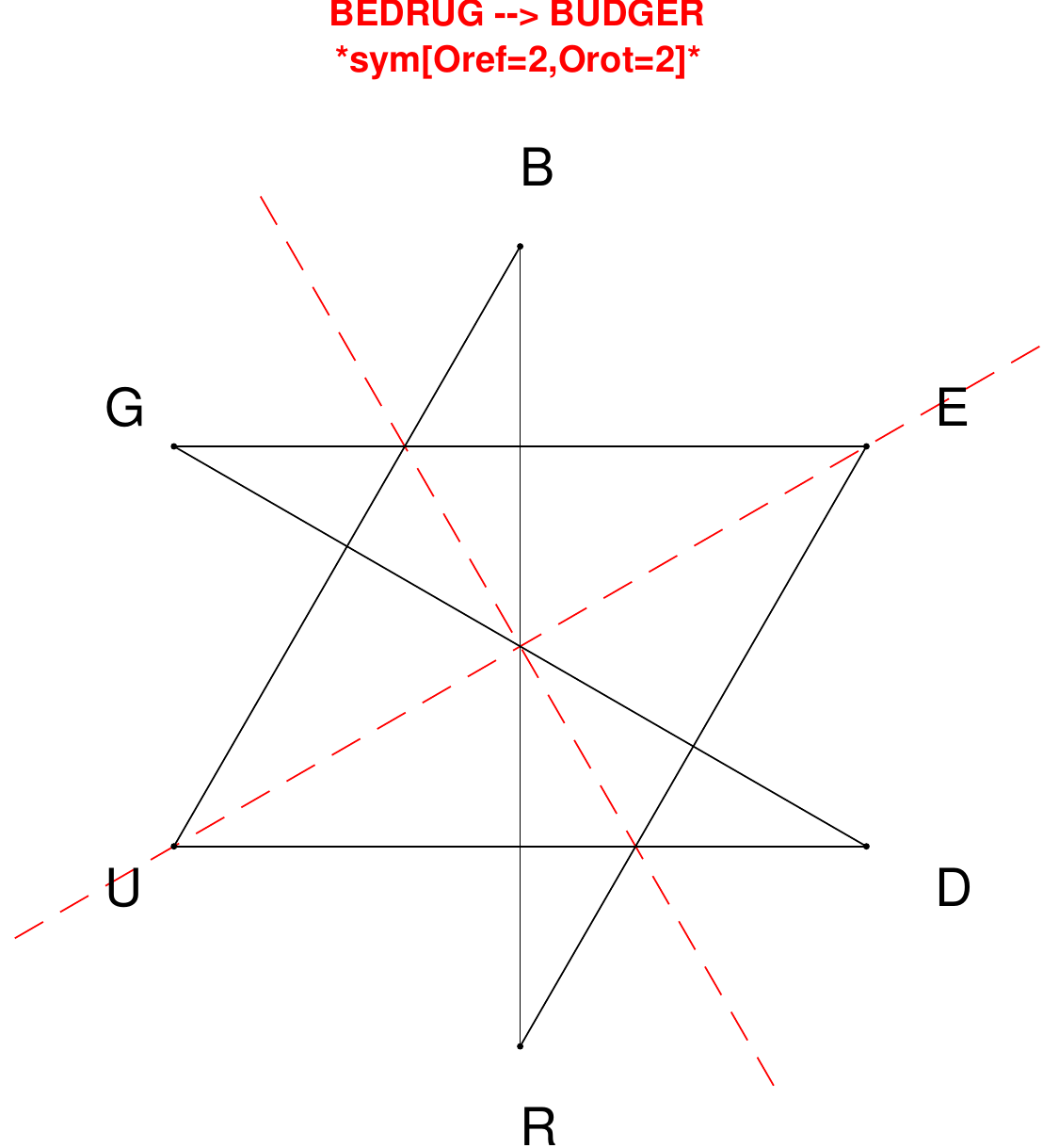}
\end{subfigure}
\hfill
\begin{subfigure}[T]{0.19\textwidth}
\centering
\includegraphics[width=\textwidth]{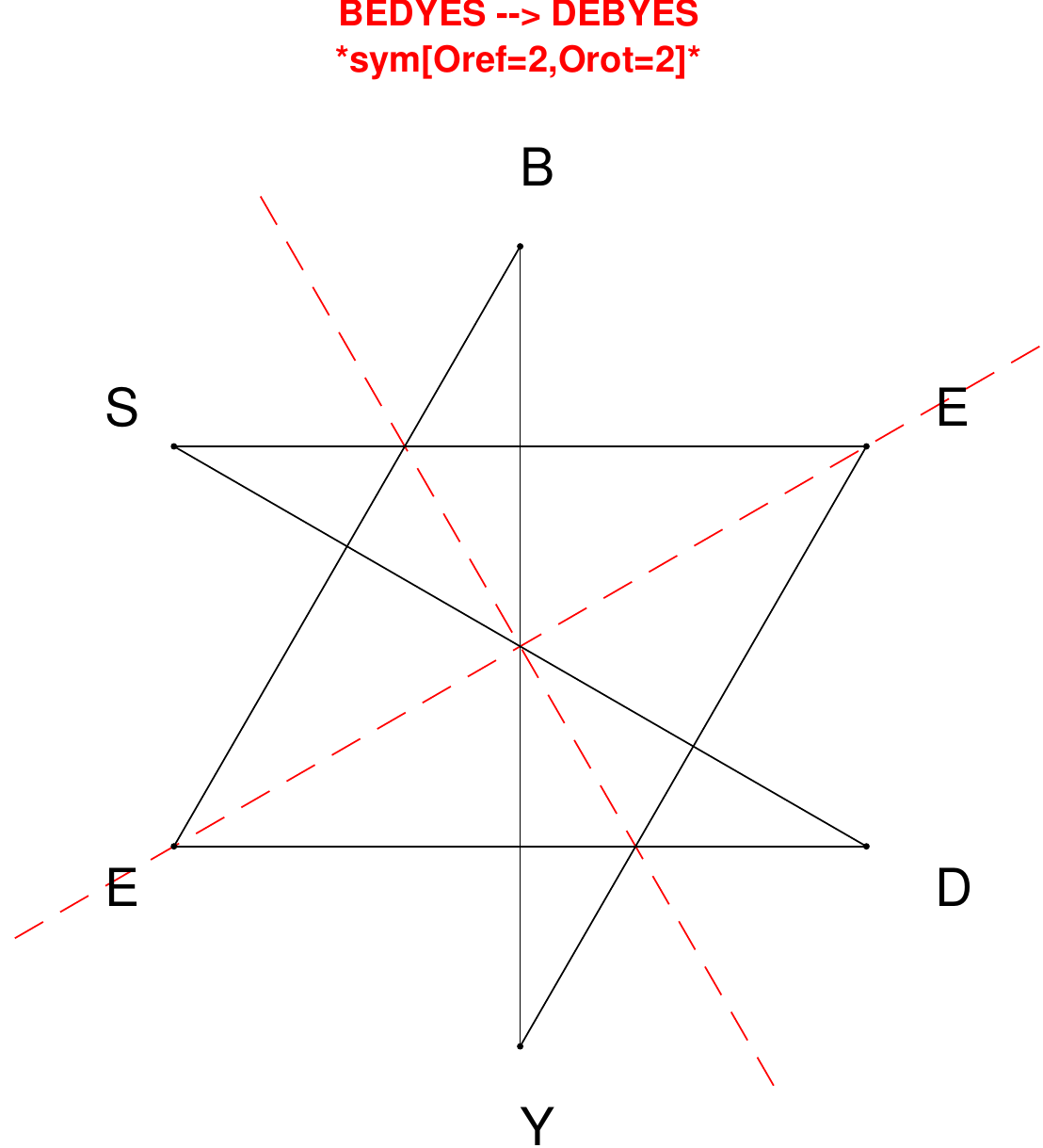}
\end{subfigure}
\hfill
\begin{subfigure}[T]{0.19\textwidth}
\centering
\includegraphics[width=\textwidth]{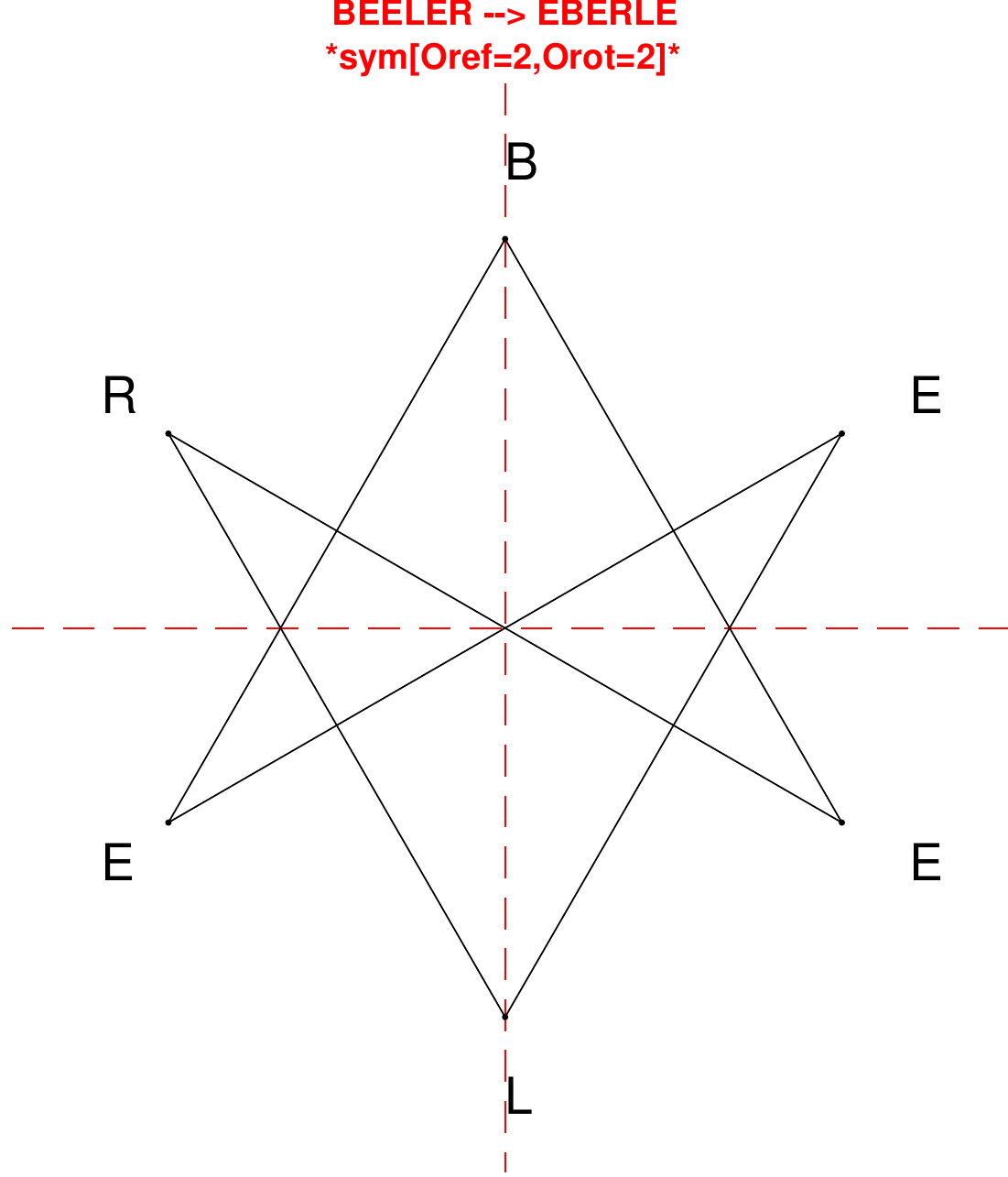}
\end{subfigure}
\hfill
\begin{subfigure}[T]{0.19\textwidth}
\centering
\includegraphics[width=\textwidth]{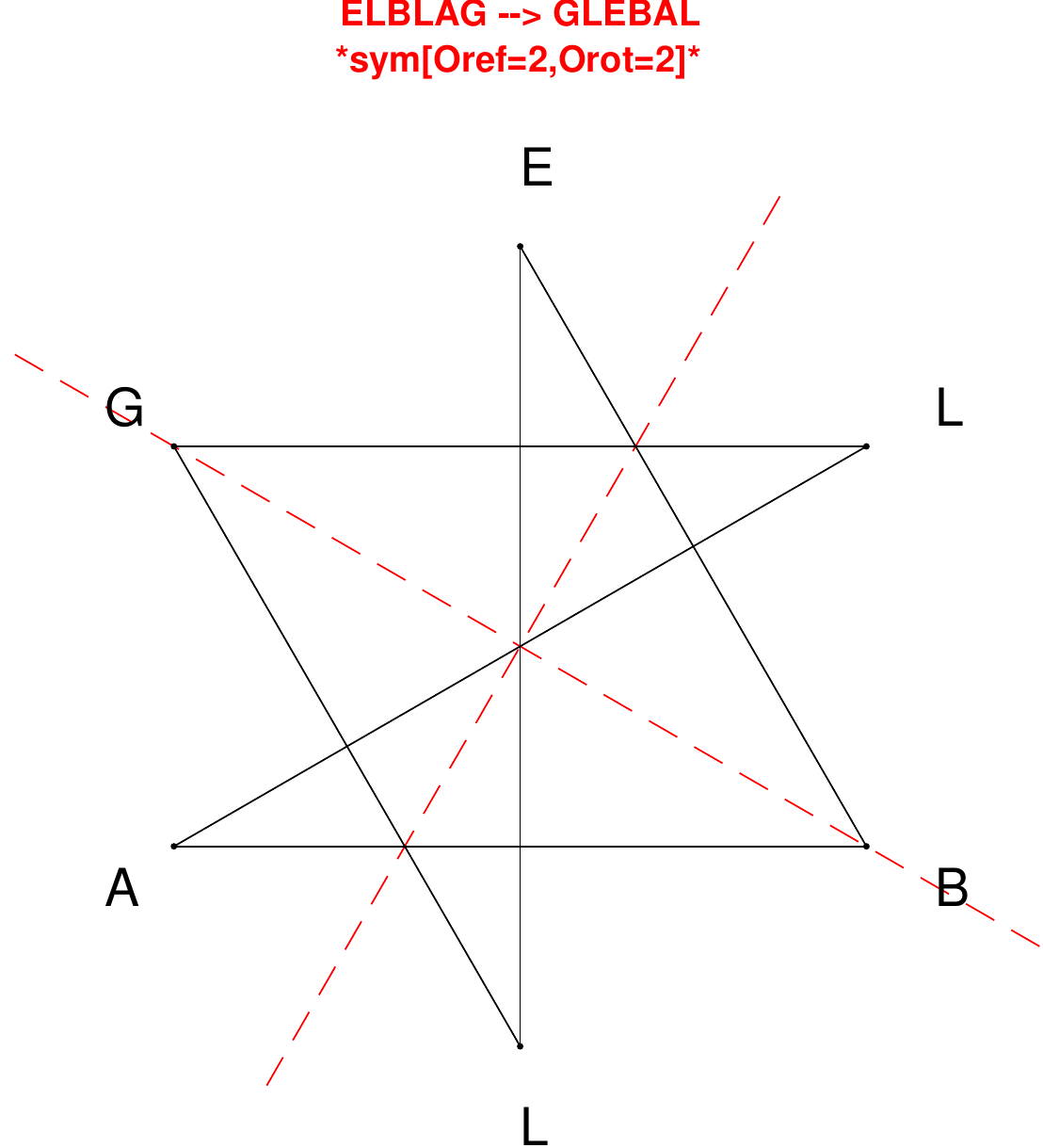}
\end{subfigure}
\hfill
\begin{subfigure}[T]{0.19\textwidth}
\centering
\includegraphics[width=\textwidth]{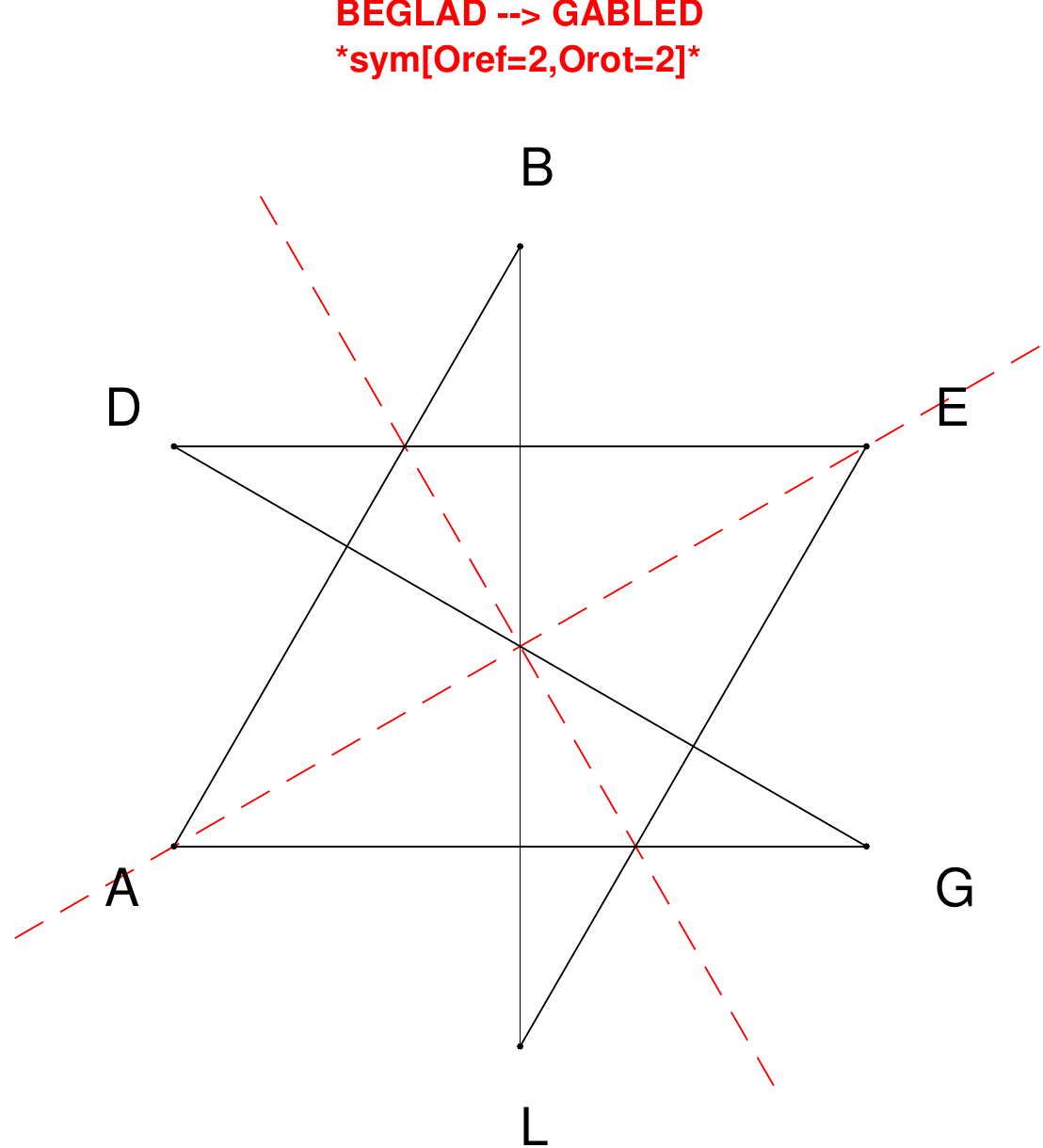}
\end{subfigure}
\end{figure}

\begin{figure}[H]
\centering
\begin{subfigure}[T]{0.19\textwidth}
\centering
\includegraphics[width=\textwidth]{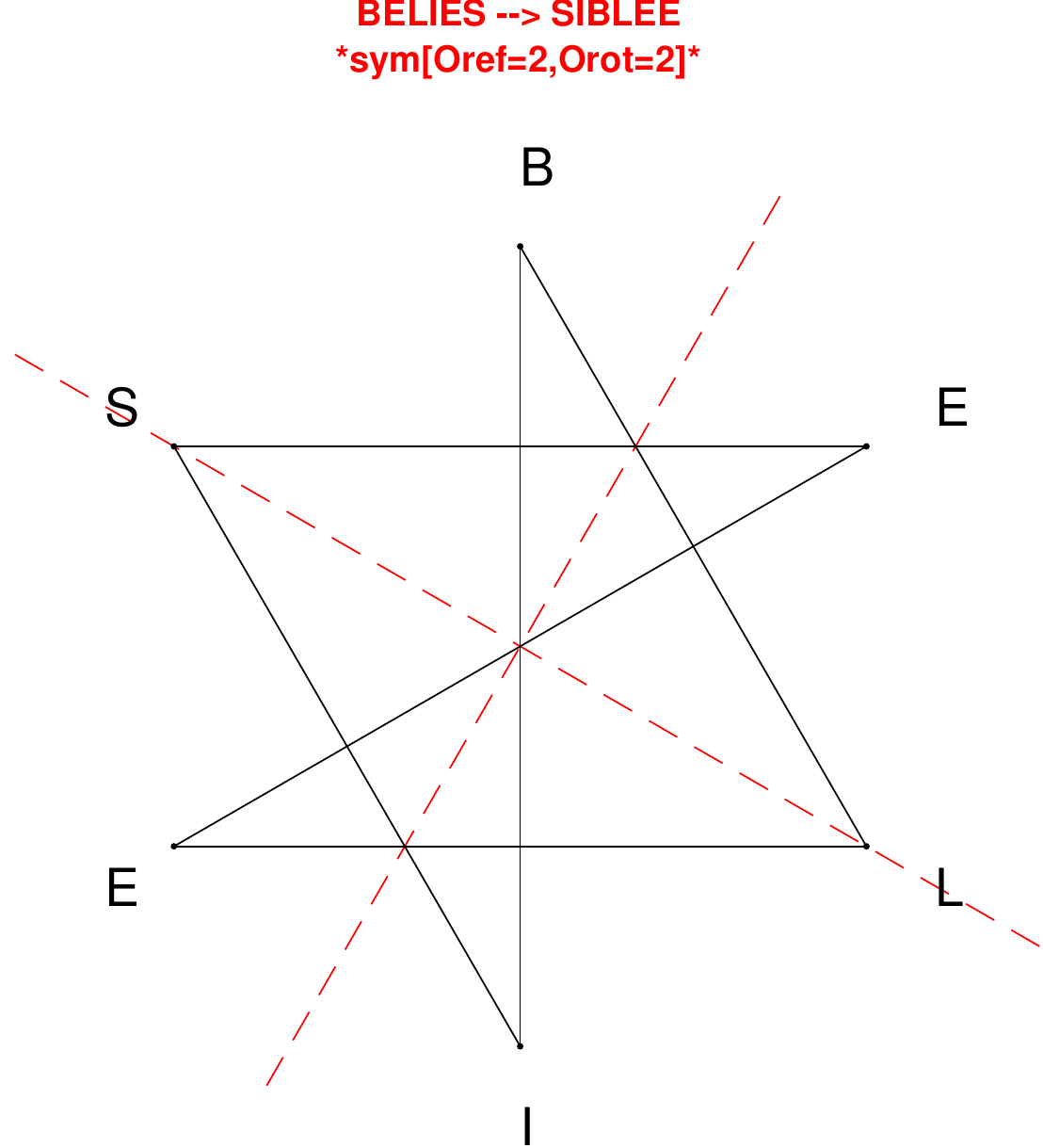}
\end{subfigure}
\hfill
\begin{subfigure}[T]{0.19\textwidth}
\centering
\includegraphics[width=\textwidth]{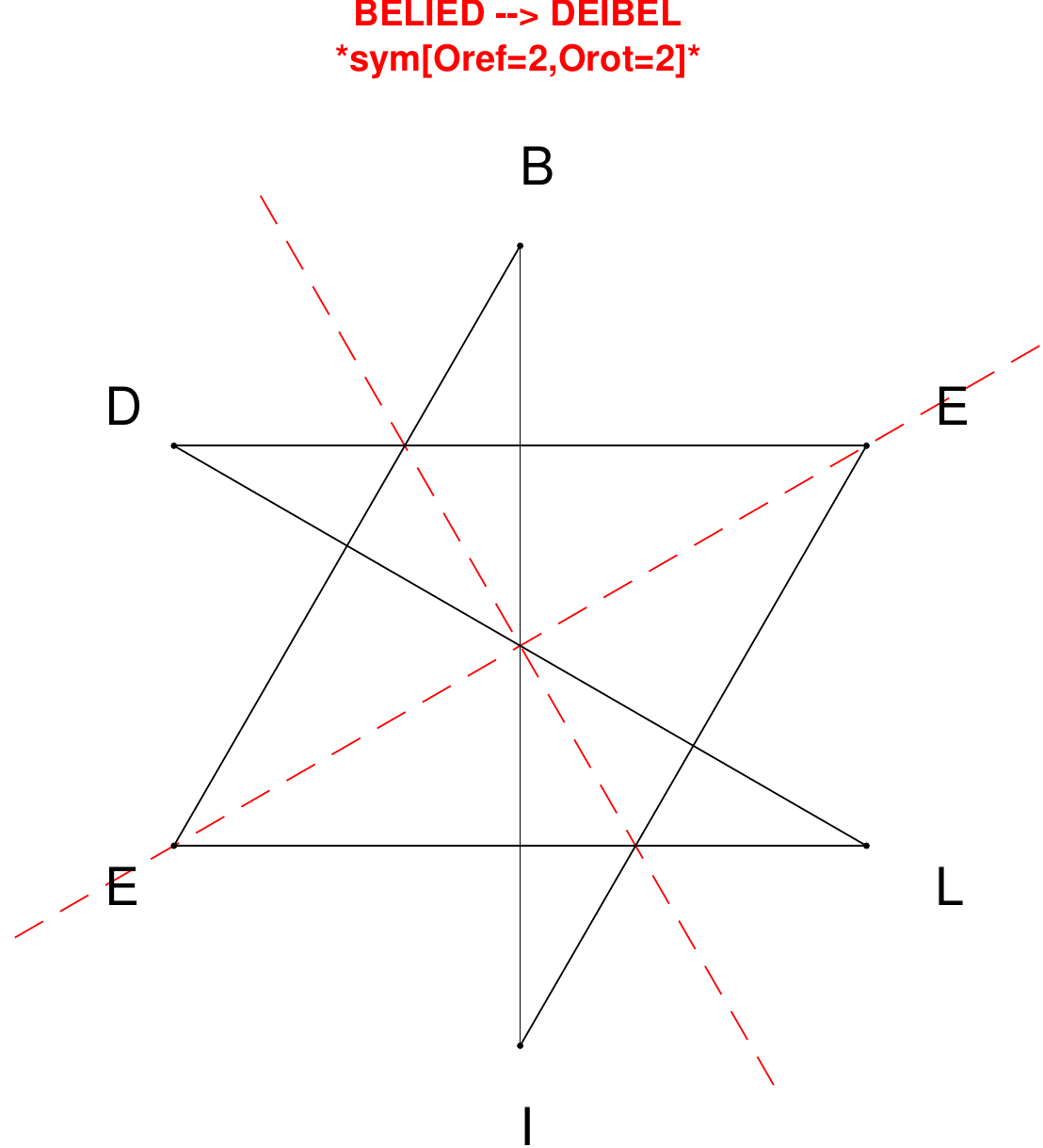}
\end{subfigure}
\hfill
\begin{subfigure}[T]{0.19\textwidth}
\centering
\includegraphics[width=\textwidth]{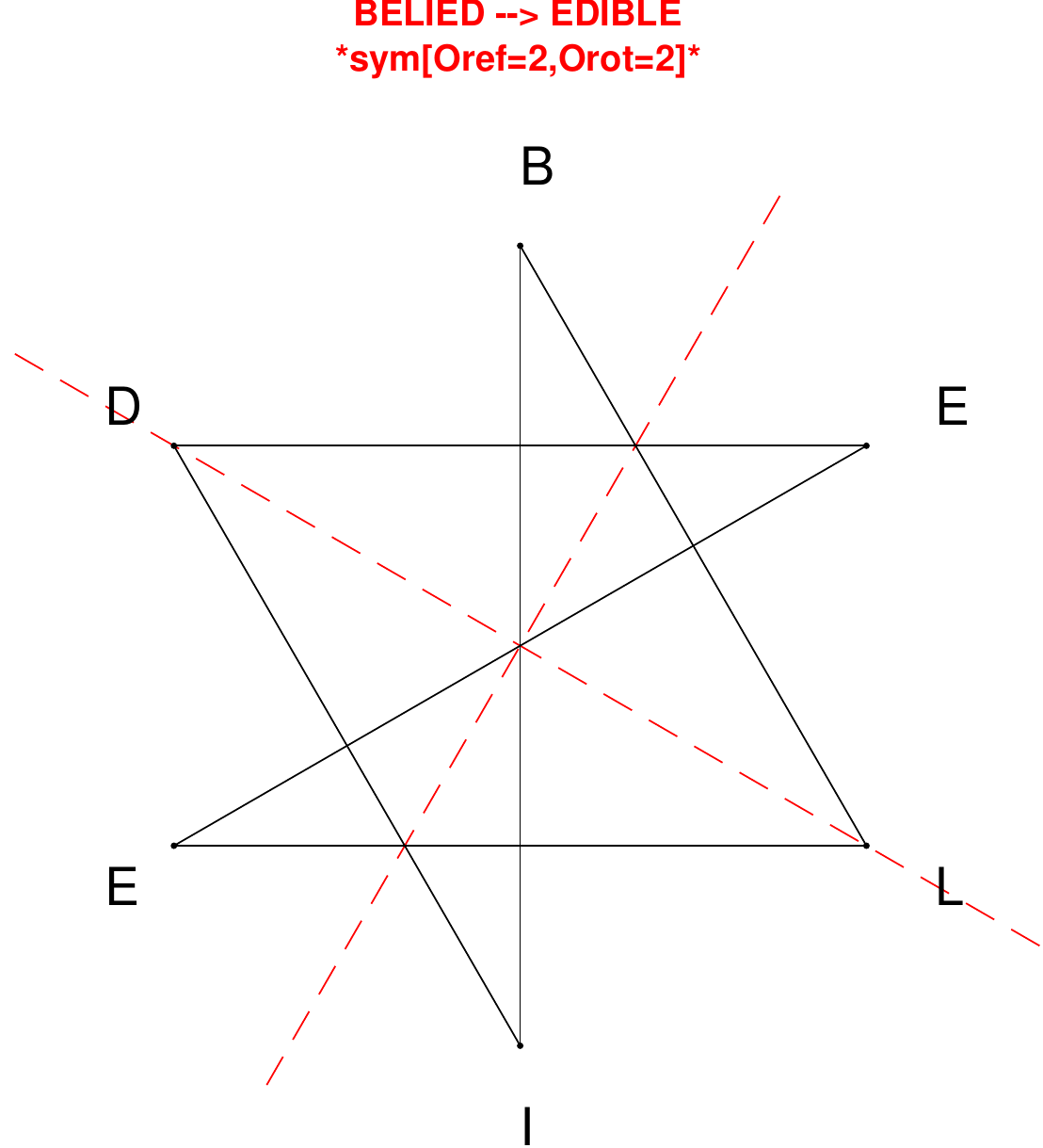}
\end{subfigure}
\hfill
\begin{subfigure}[T]{0.19\textwidth}
\centering
\includegraphics[width=\textwidth]{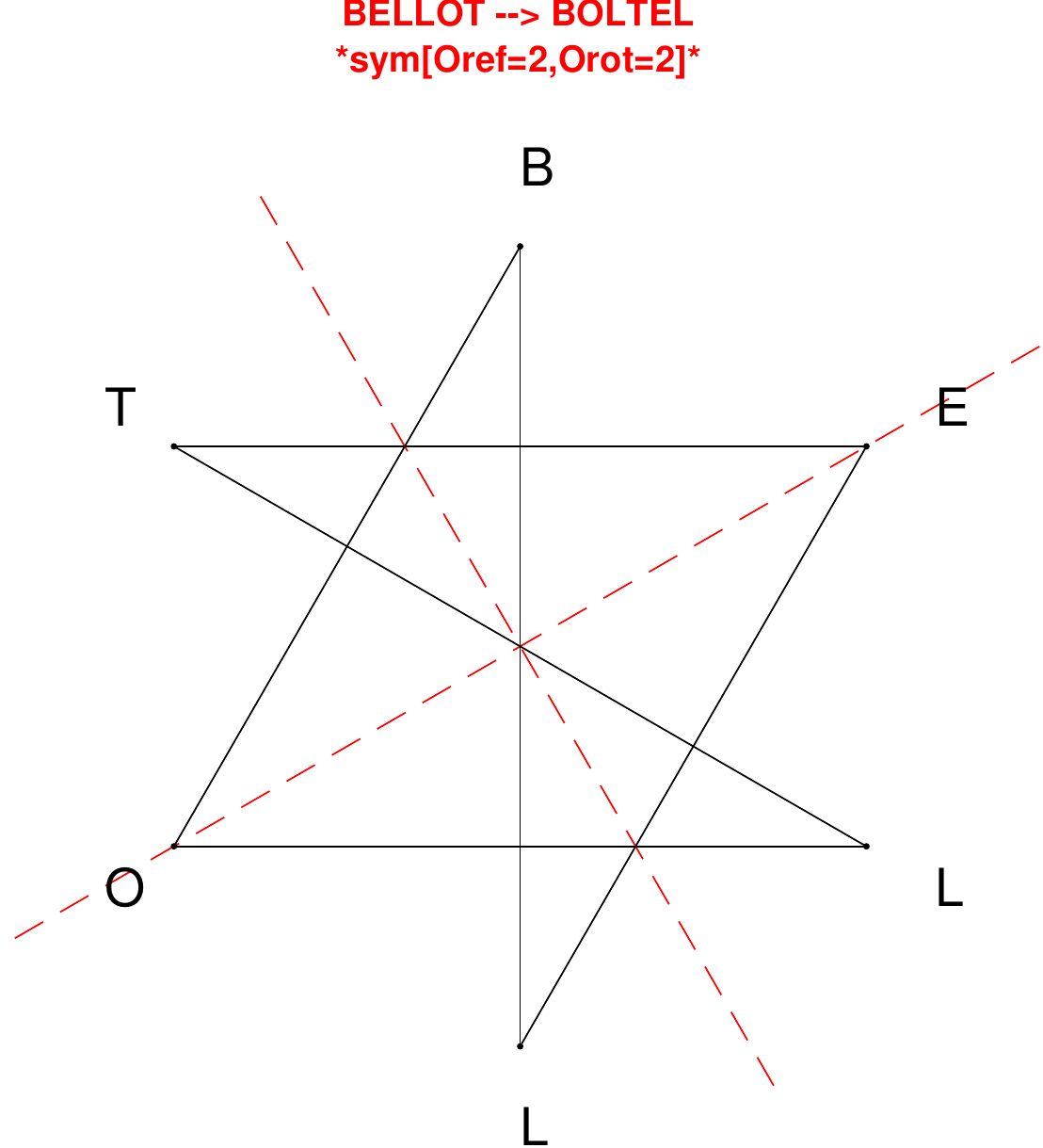}
\end{subfigure}
\hfill
\begin{subfigure}[T]{0.19\textwidth}
\centering
\includegraphics[width=\textwidth]{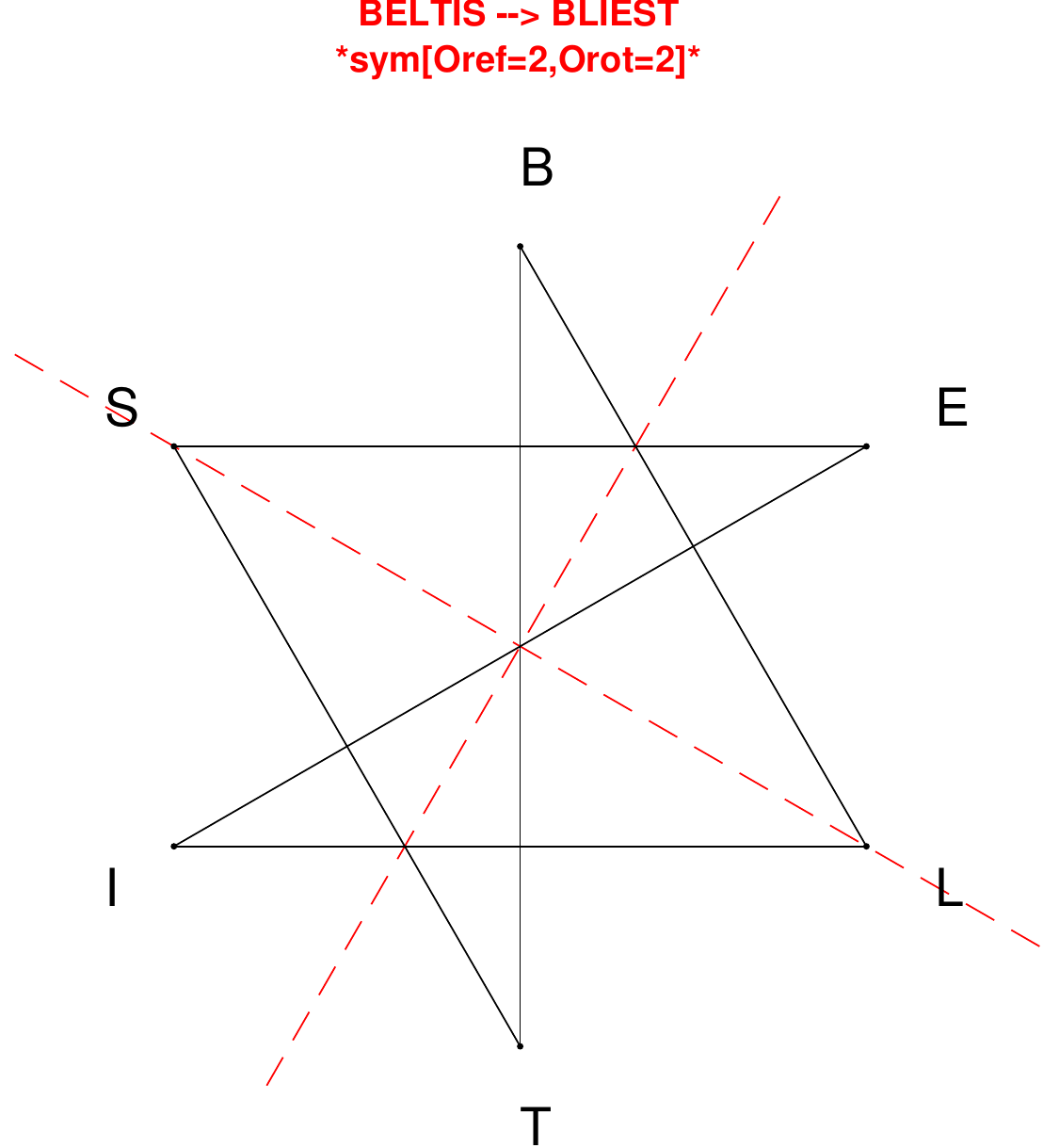}
\end{subfigure}
\end{figure}

\begin{figure}[H]
\centering
\begin{subfigure}[T]{0.19\textwidth}
\centering
\includegraphics[width=\textwidth]{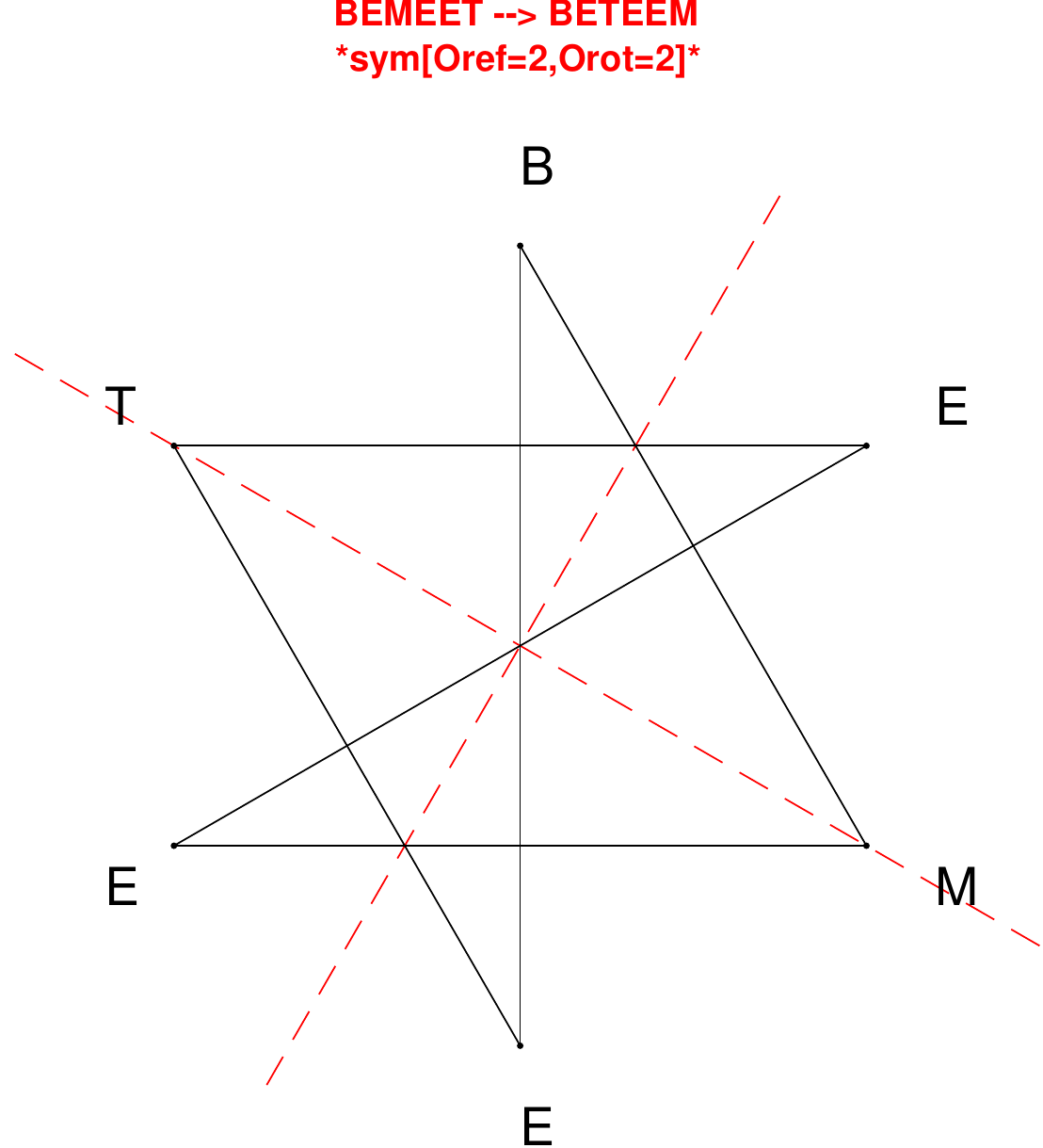}
\end{subfigure}
\hfill
\begin{subfigure}[T]{0.19\textwidth}
\centering
\includegraphics[width=\textwidth]{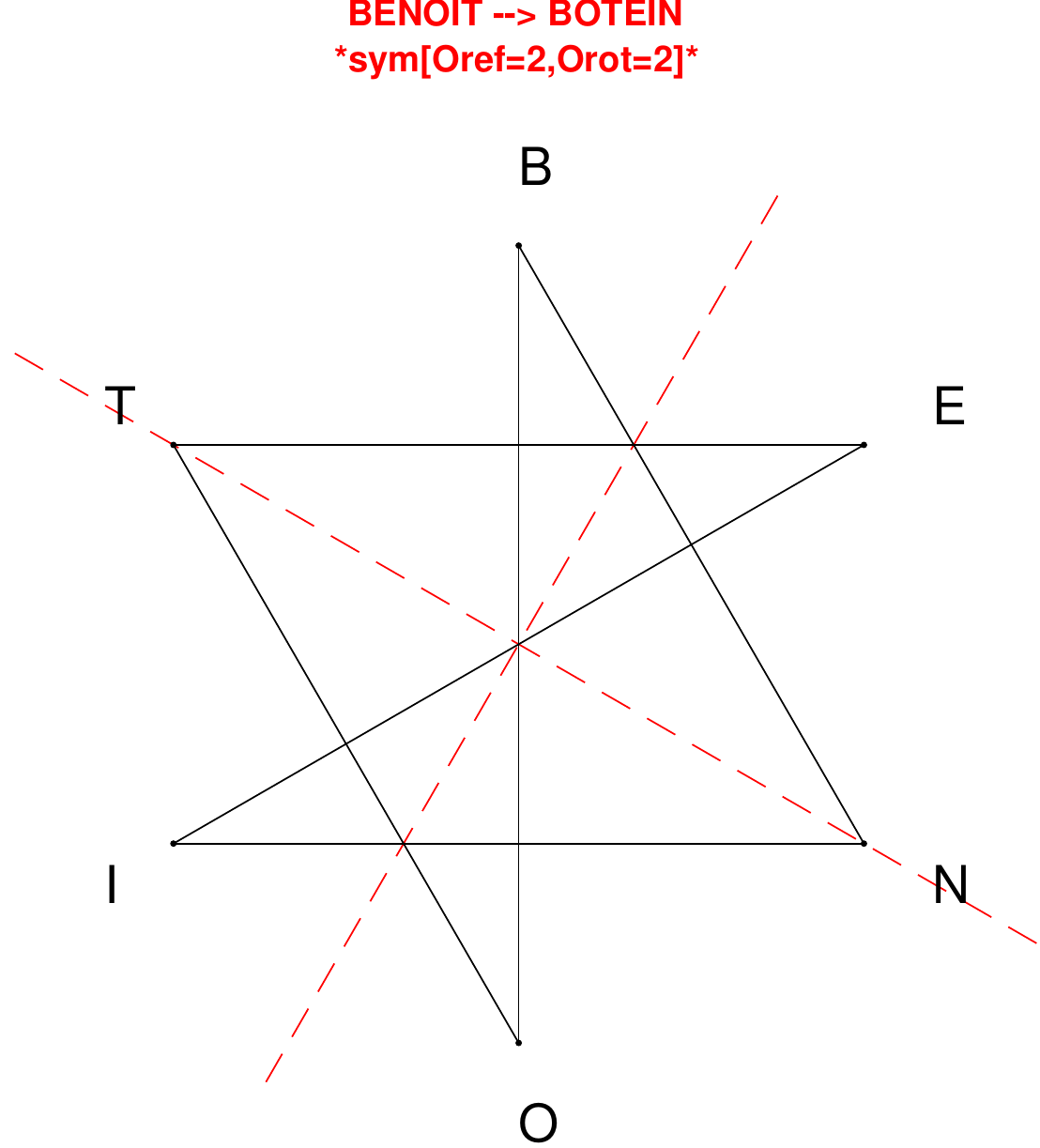}
\end{subfigure}
\hfill
\begin{subfigure}[T]{0.19\textwidth}
\centering
\includegraphics[width=\textwidth]{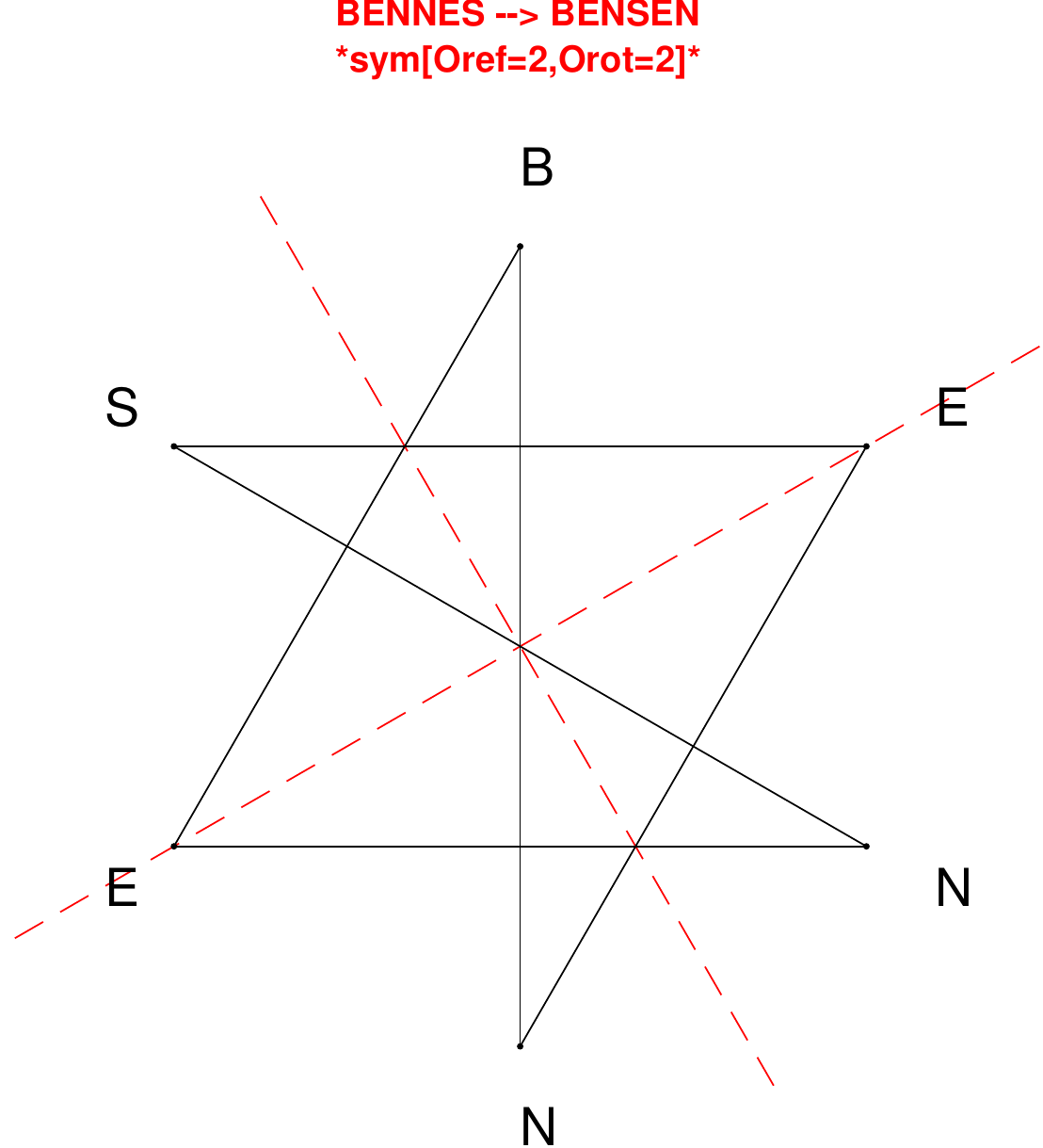}
\end{subfigure}
\hfill
\begin{subfigure}[T]{0.19\textwidth}
\centering
\includegraphics[width=\textwidth]{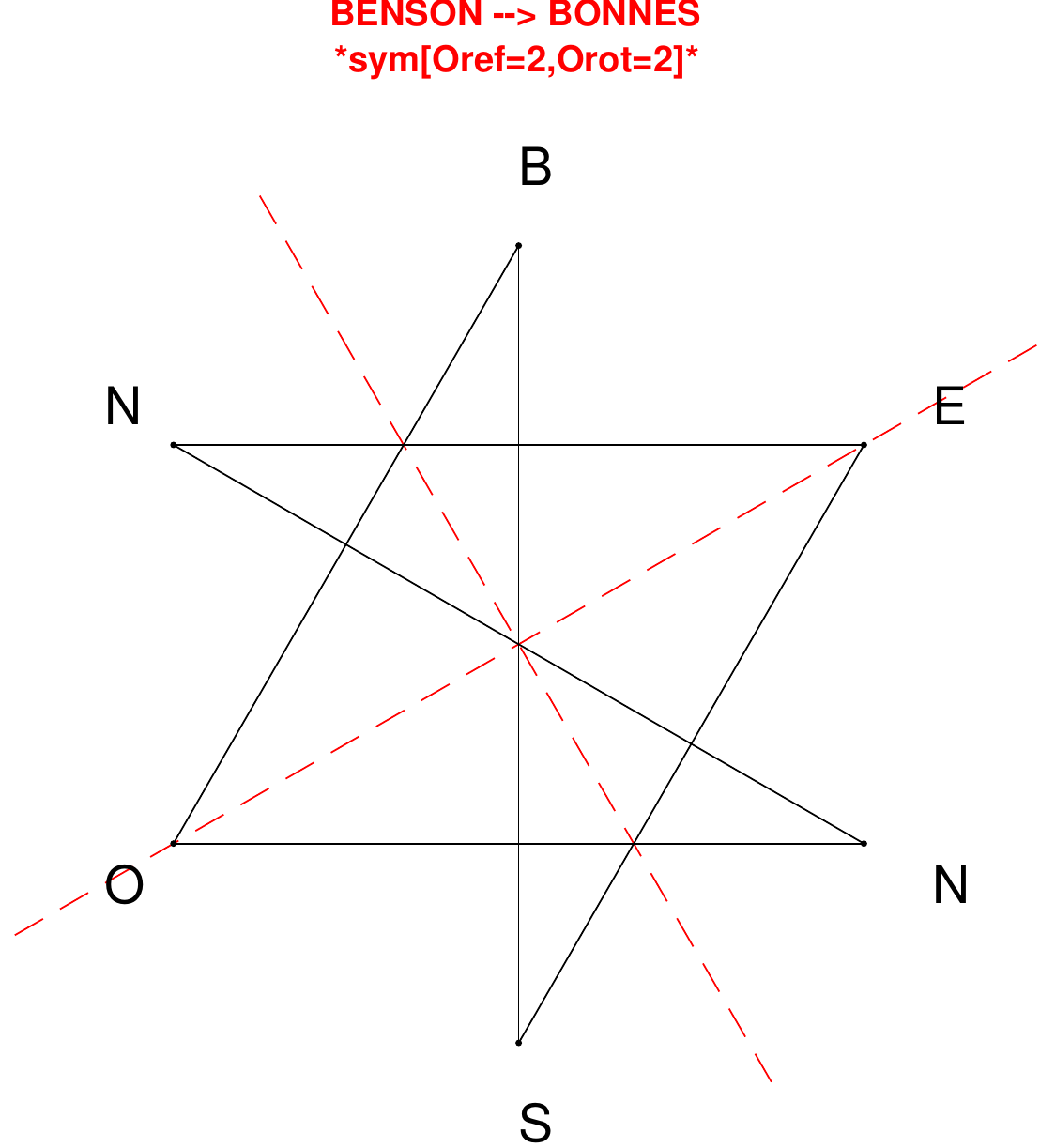}
\end{subfigure}
\hfill
\begin{subfigure}[T]{0.19\textwidth}
\centering
\includegraphics[width=\textwidth]{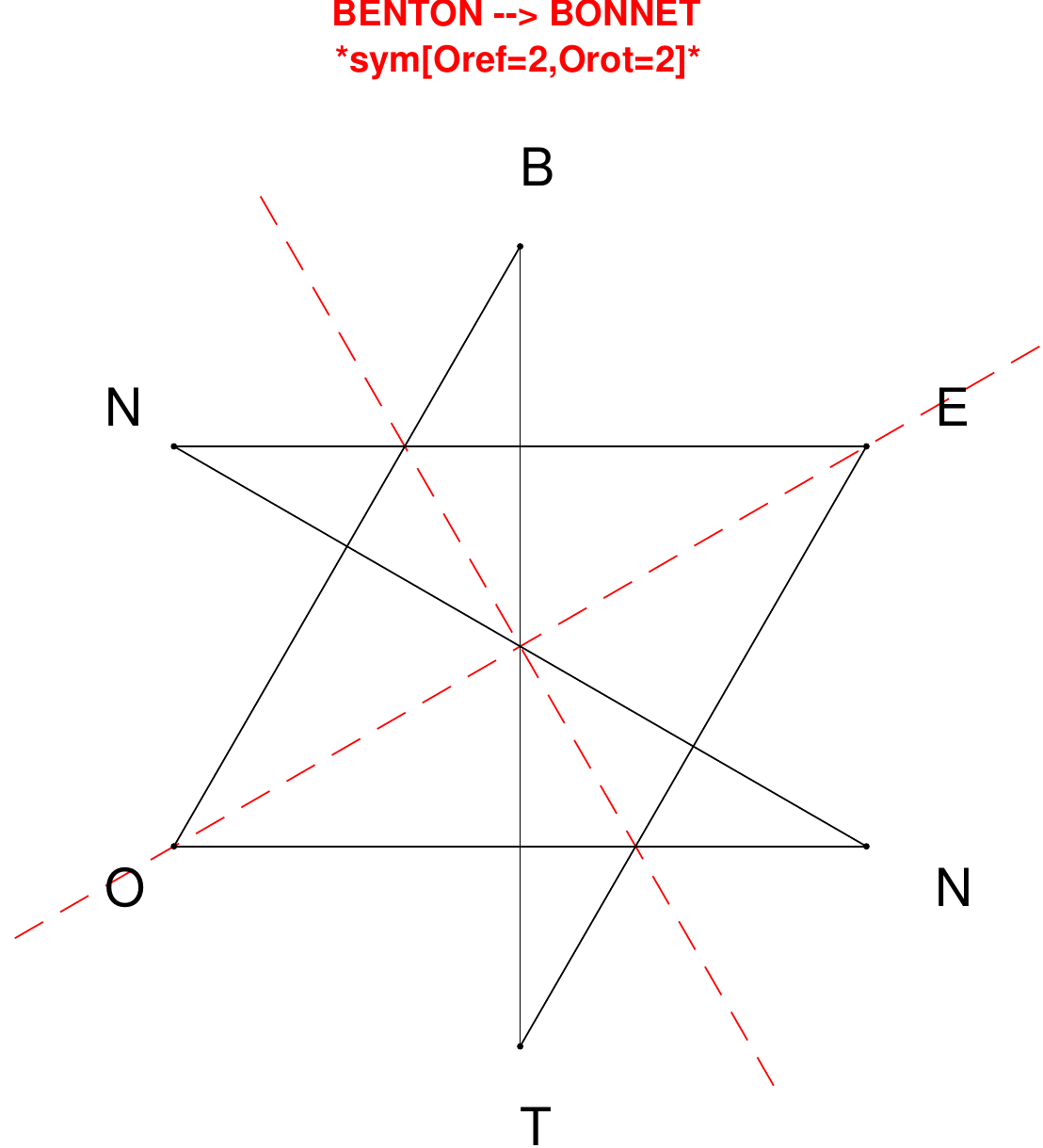}
\end{subfigure}
\end{figure}

\begin{figure}[H]
\centering
\begin{subfigure}[T]{0.19\textwidth}
\centering
\includegraphics[width=\textwidth]{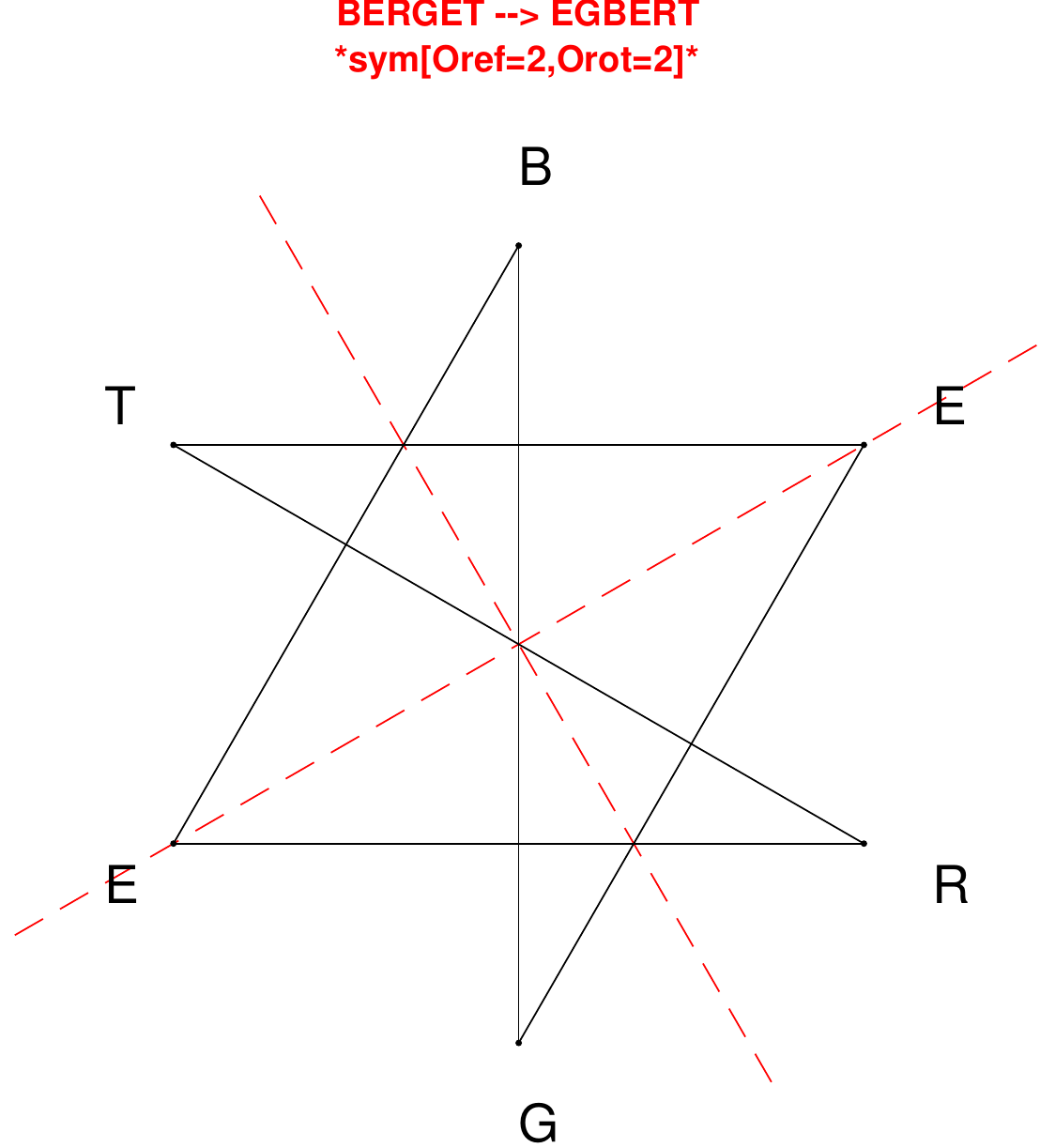}
\end{subfigure}
\hfill
\begin{subfigure}[T]{0.19\textwidth}
\centering
\includegraphics[width=\textwidth]{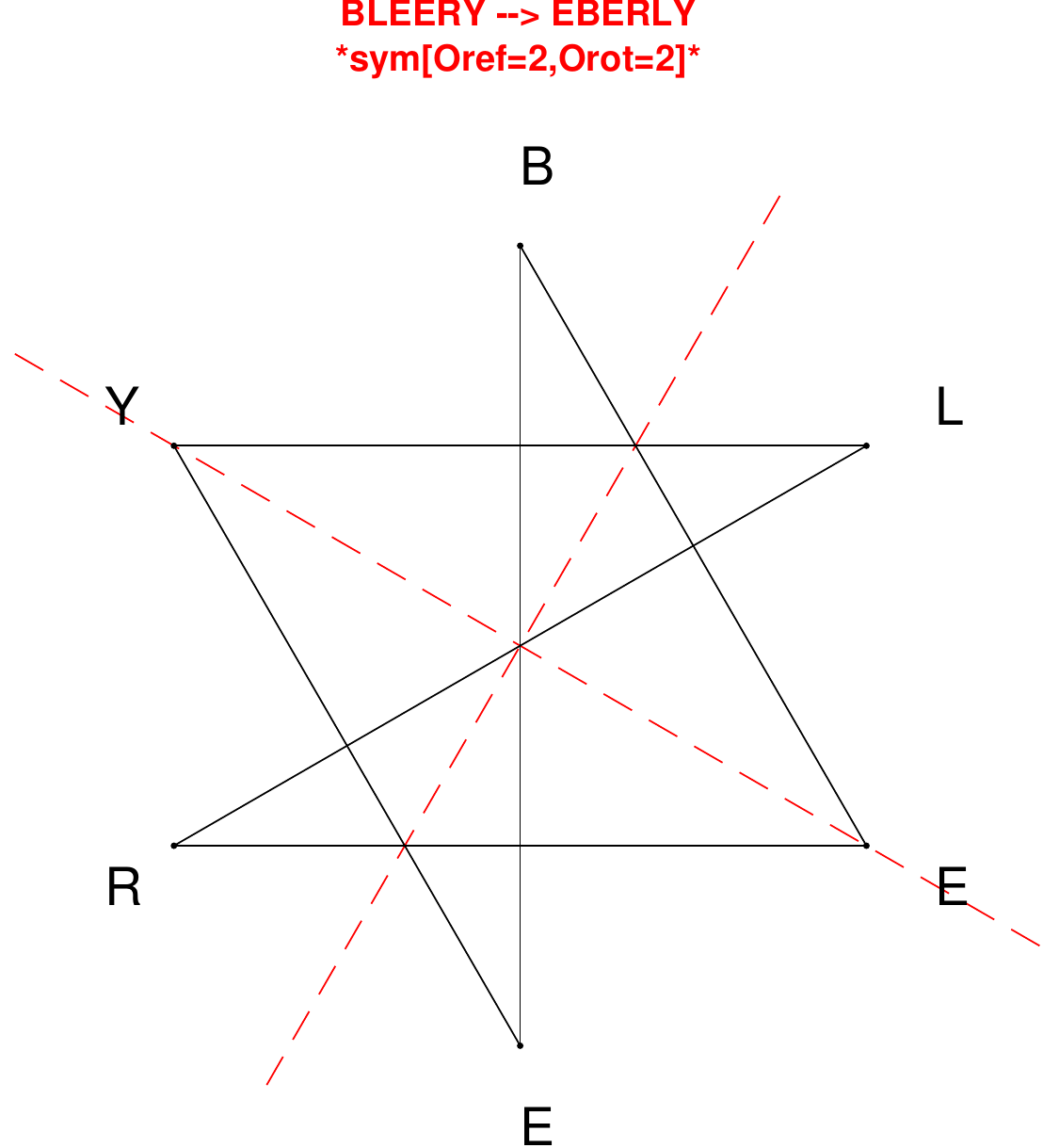}
\end{subfigure}
\hfill
\begin{subfigure}[T]{0.19\textwidth}
\centering
\includegraphics[width=\textwidth]{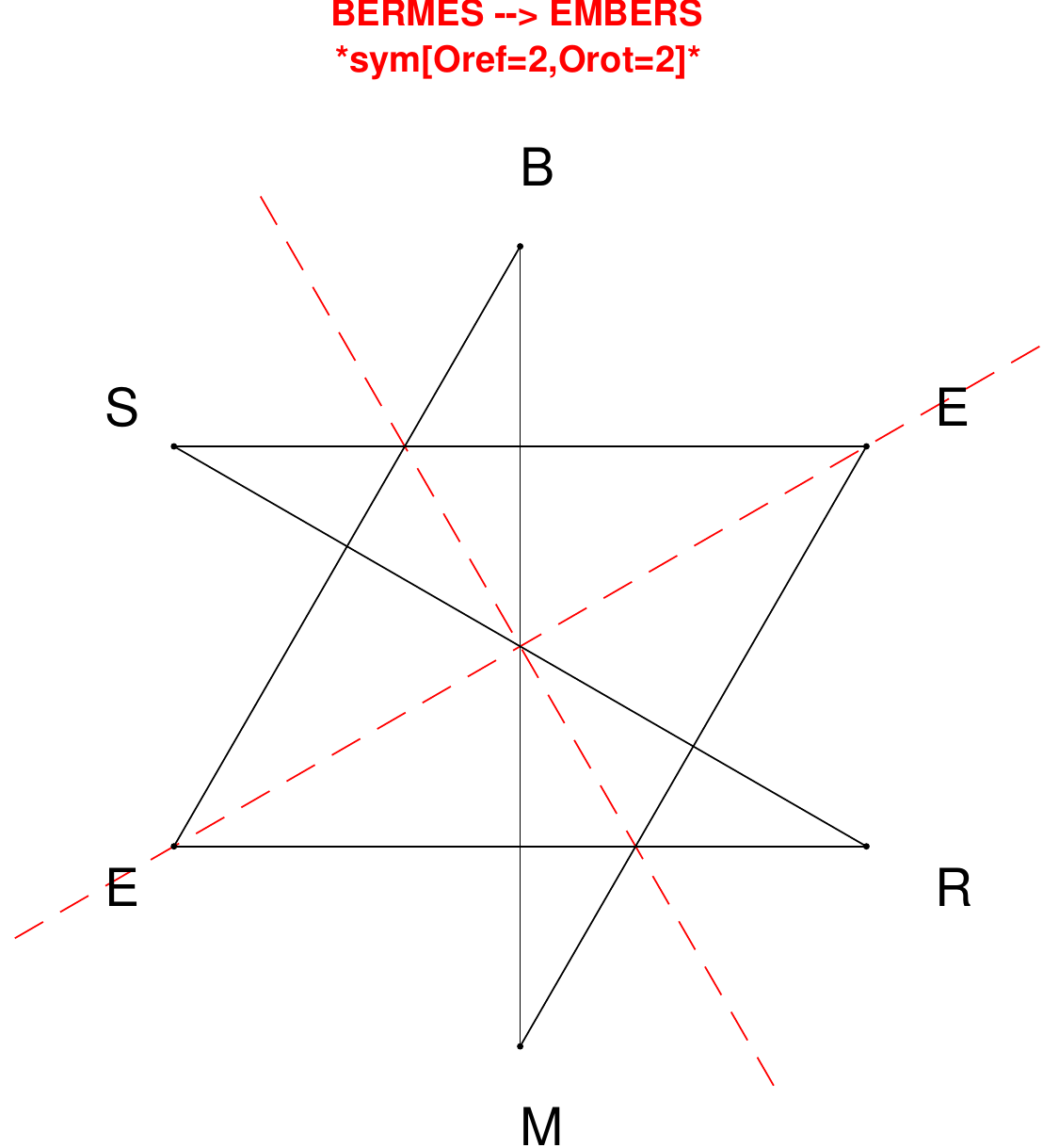}
\end{subfigure}
\hfill
\begin{subfigure}[T]{0.19\textwidth}
\centering
\includegraphics[width=\textwidth]{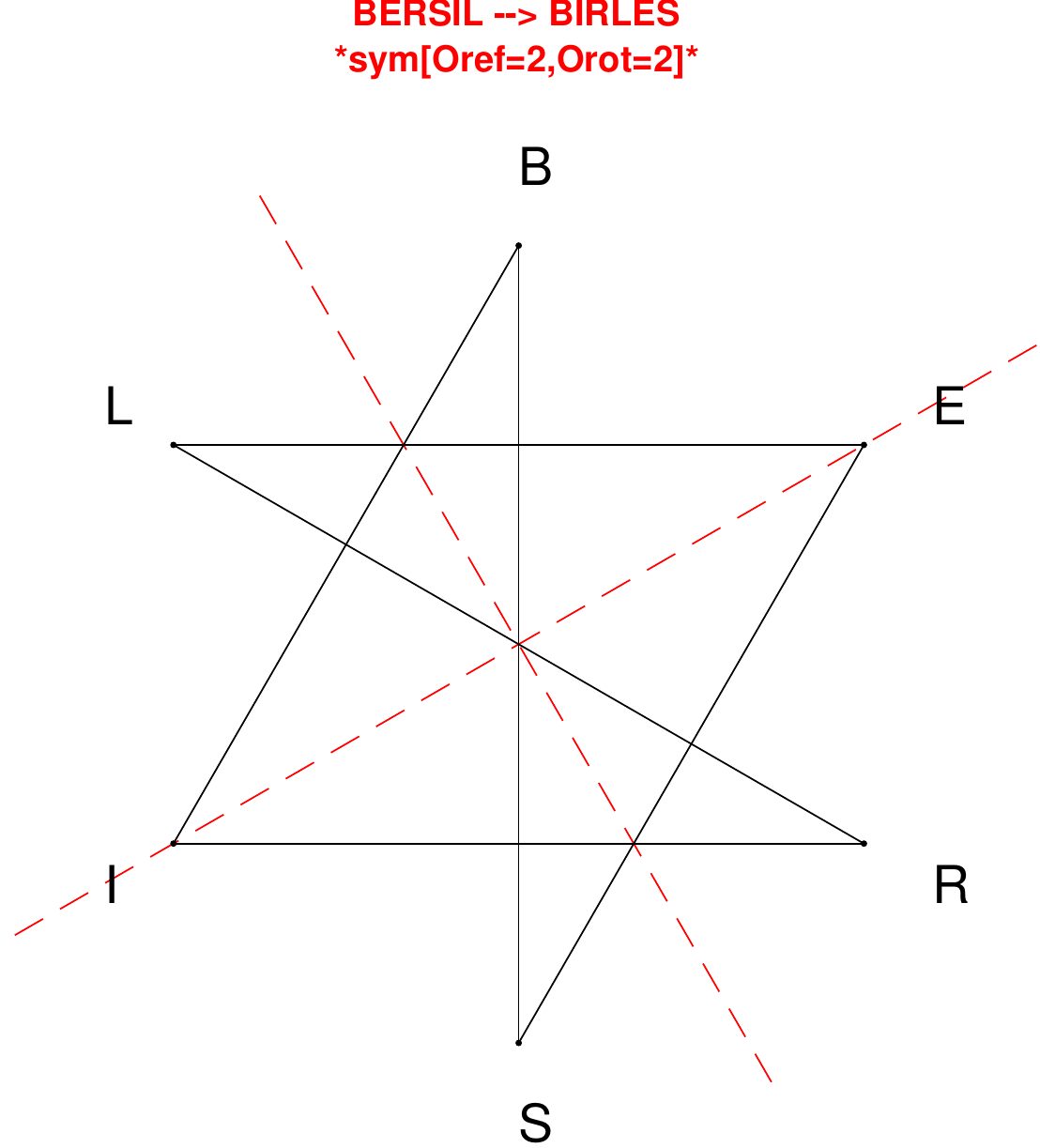}
\end{subfigure}
\hfill
\begin{subfigure}[T]{0.19\textwidth}
\centering
\includegraphics[width=\textwidth]{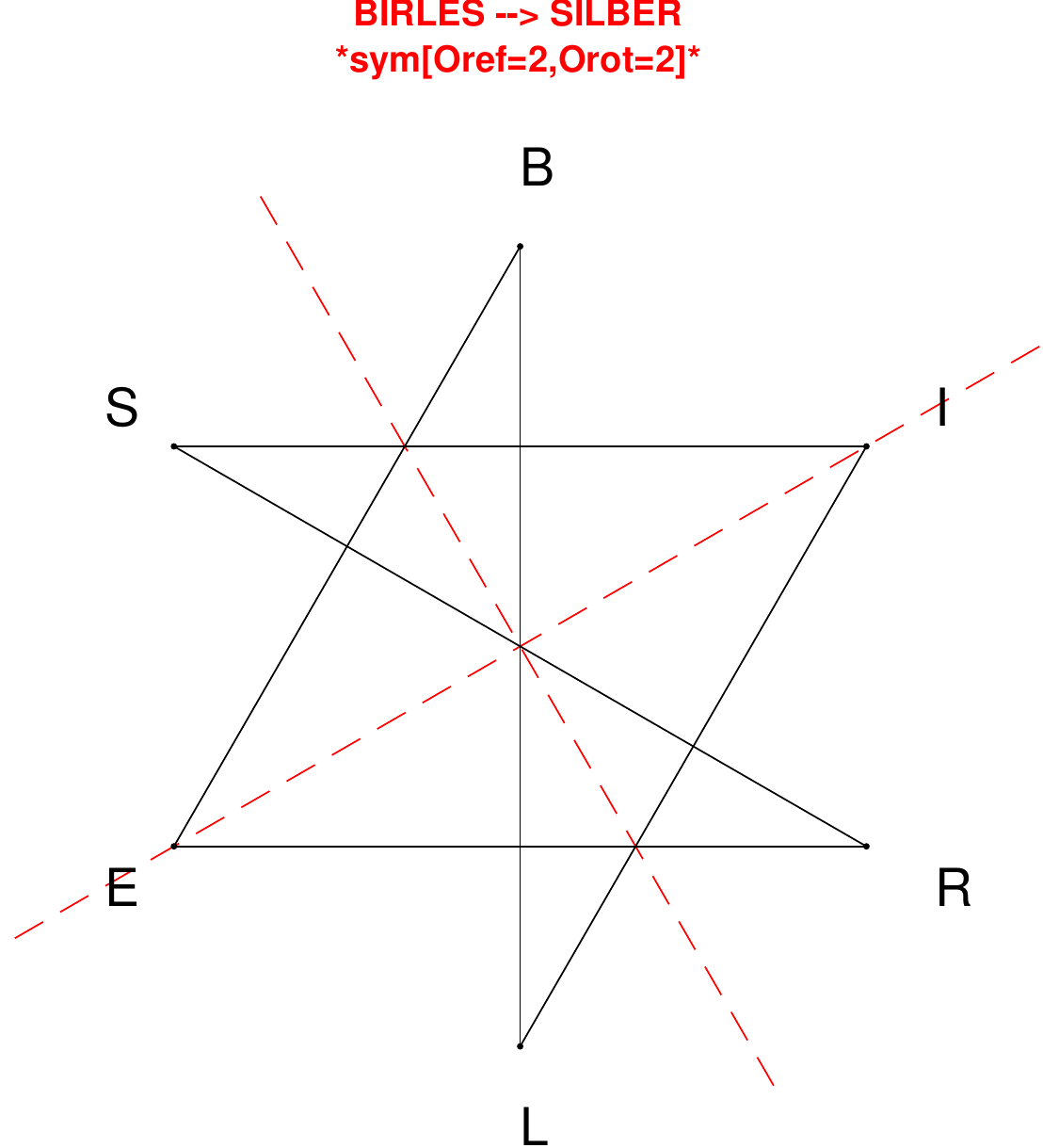}
\end{subfigure}
\end{figure}

\begin{figure}[H]
\centering
\begin{subfigure}[T]{0.19\textwidth}
\centering
\includegraphics[width=\textwidth]{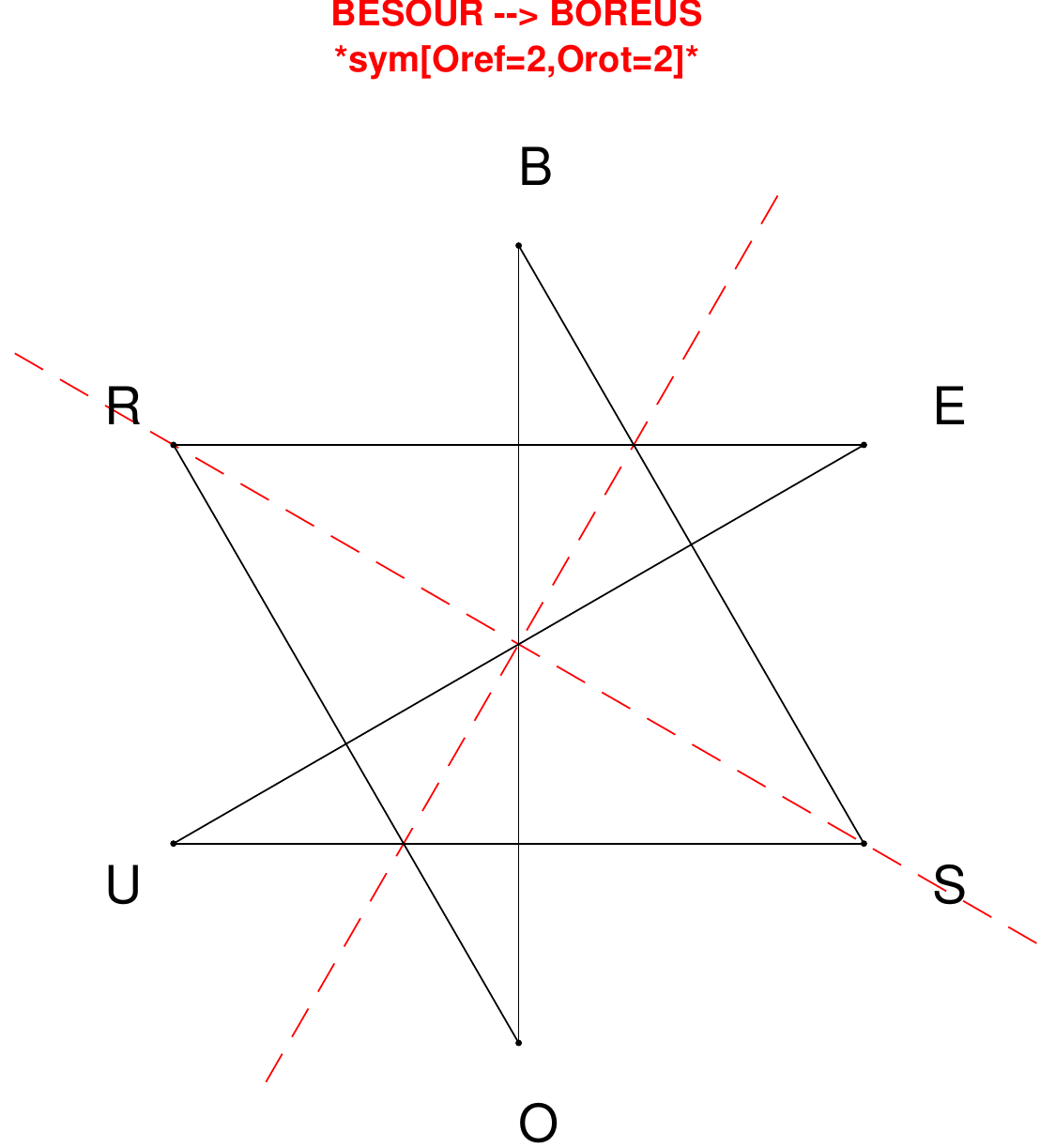}
\end{subfigure}
\hfill
\begin{subfigure}[T]{0.19\textwidth}
\centering
\includegraphics[width=\textwidth]{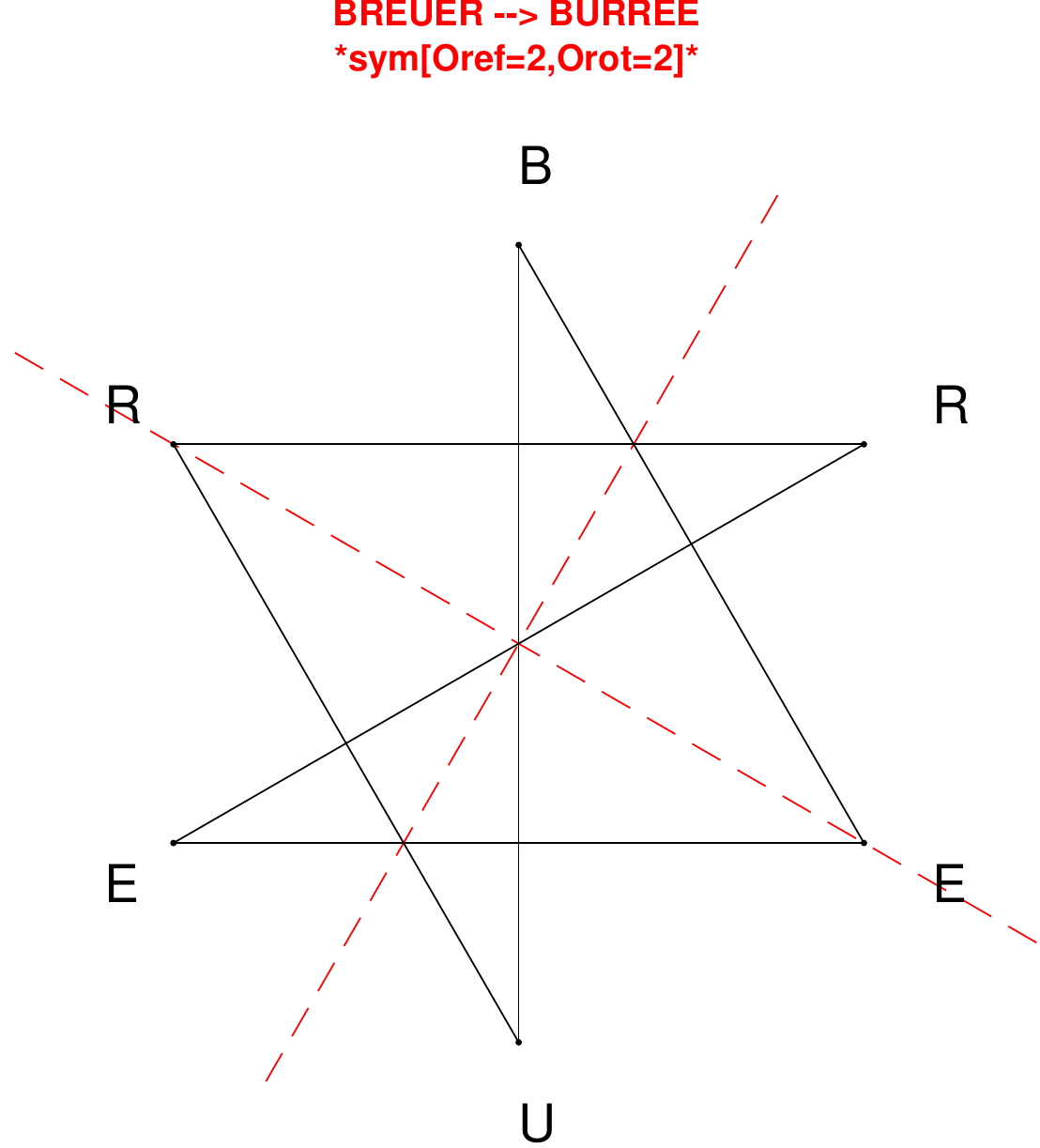}
\end{subfigure}
\hfill
\begin{subfigure}[T]{0.19\textwidth}
\centering
\includegraphics[width=\textwidth]{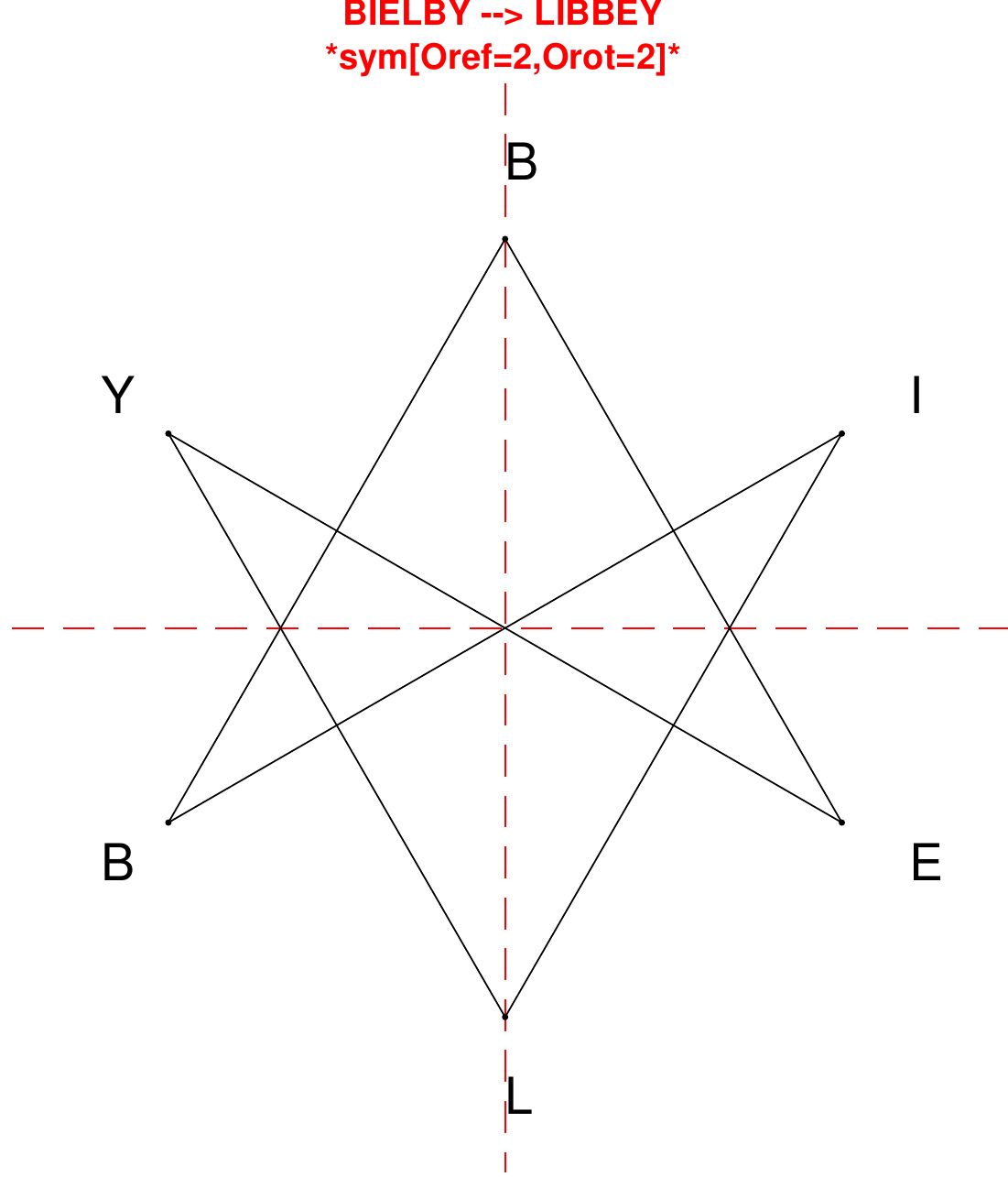}
\end{subfigure}
\hfill
\begin{subfigure}[T]{0.19\textwidth}
\centering
\includegraphics[width=\textwidth]{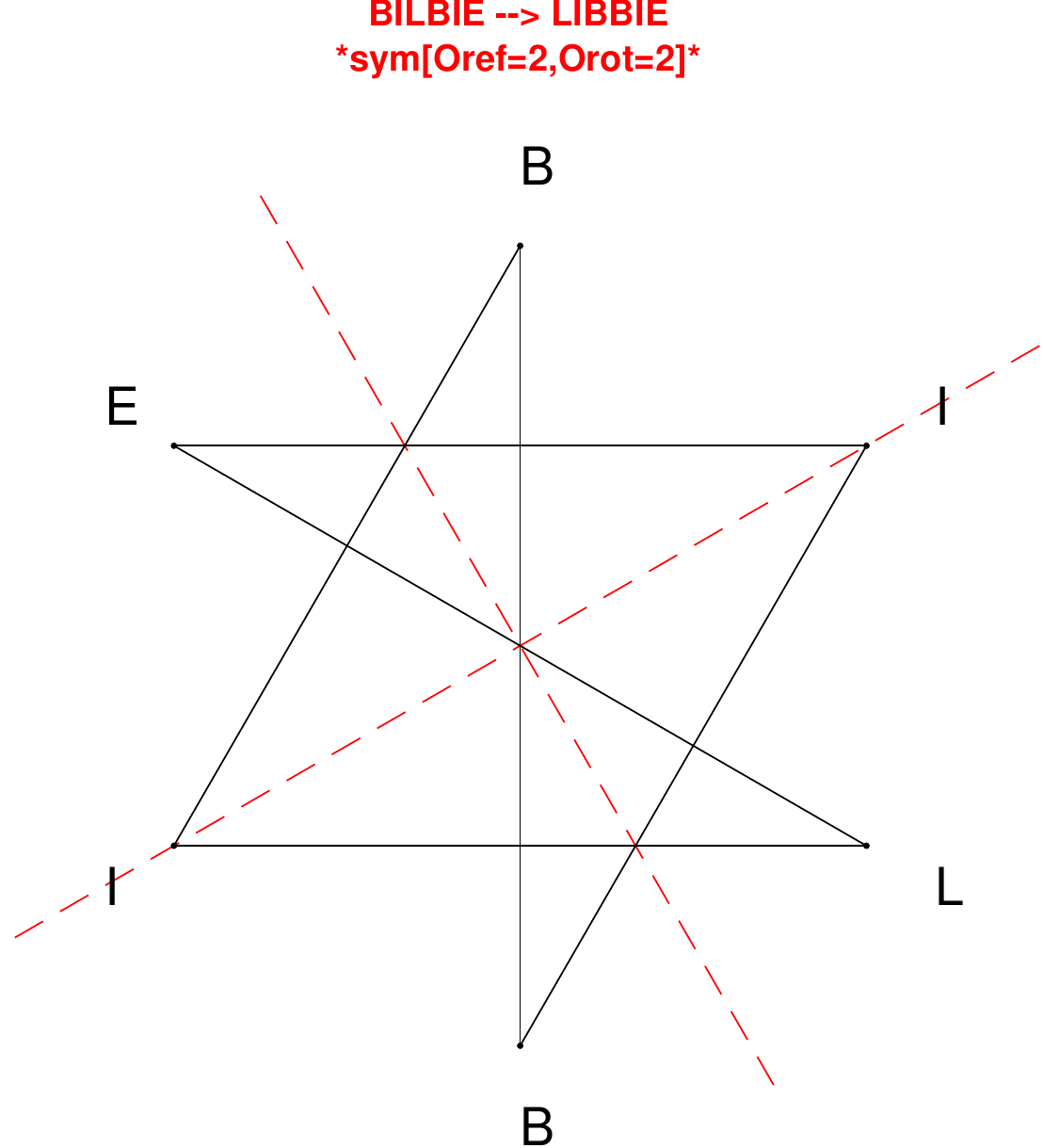}
\end{subfigure}
\hfill
\begin{subfigure}[T]{0.19\textwidth}
\centering
\includegraphics[width=\textwidth]{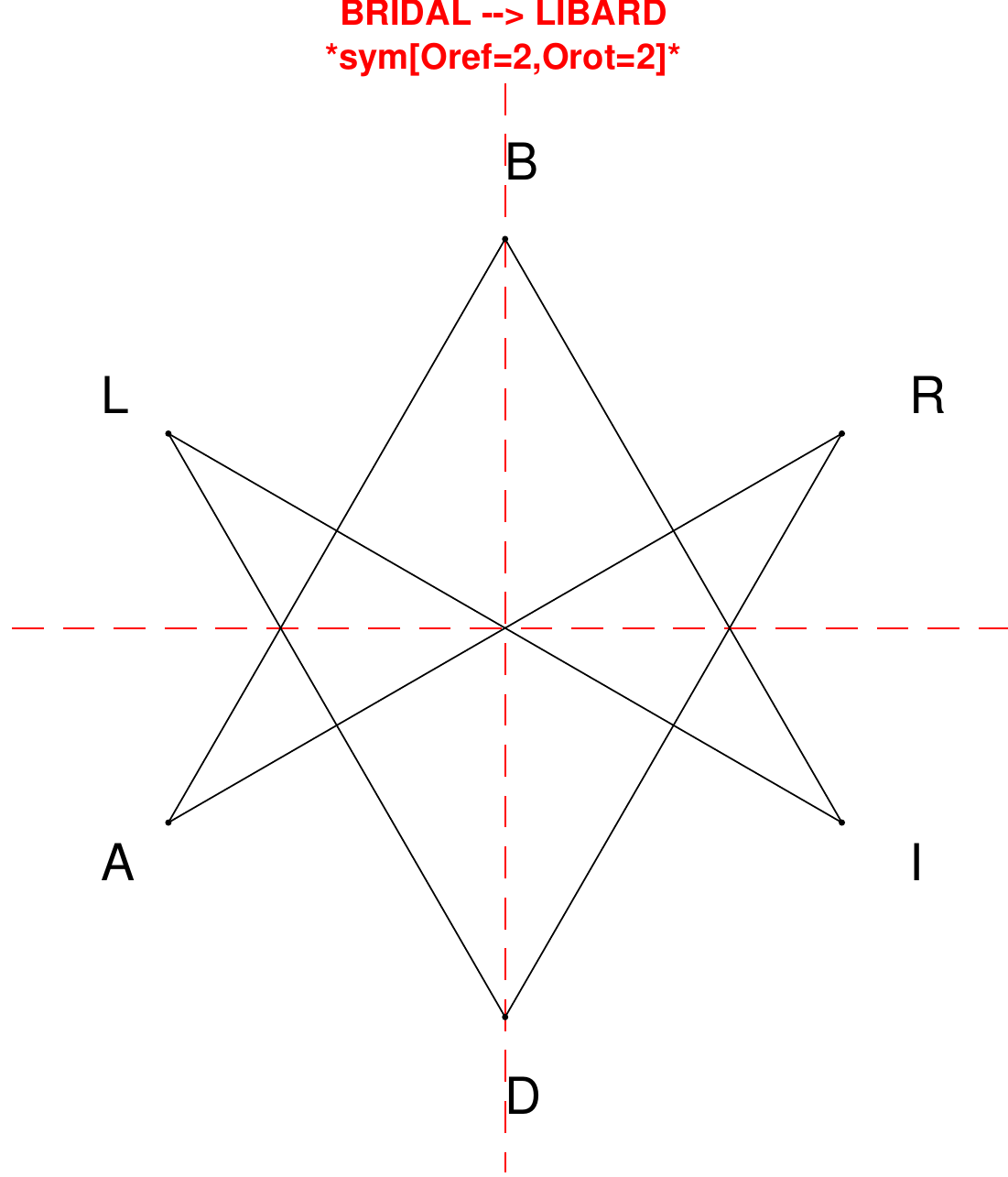}
\end{subfigure}
\end{figure}

\begin{figure}[H]
\centering
\begin{subfigure}[T]{0.19\textwidth}
\centering
\includegraphics[width=\textwidth]{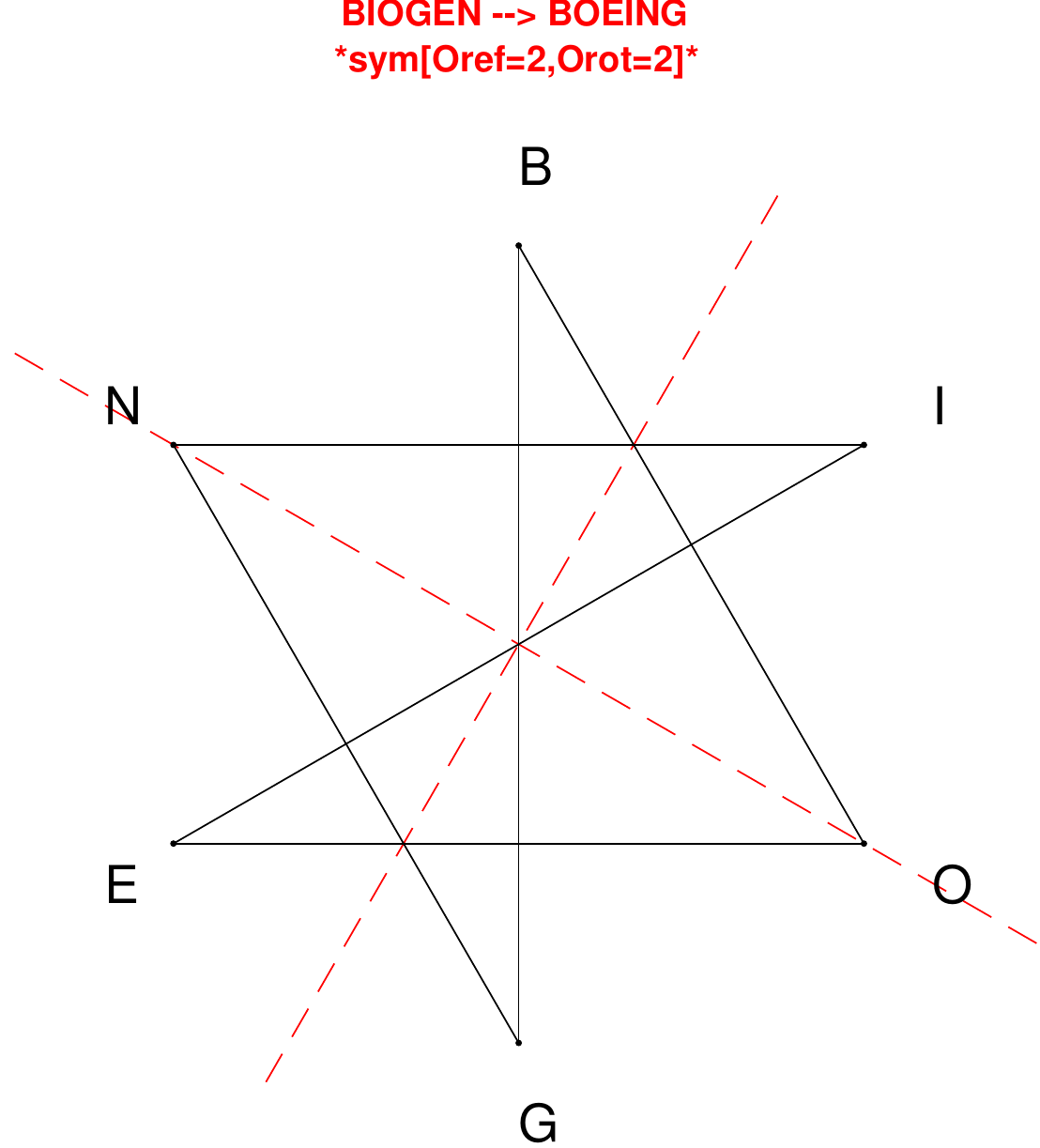}
\end{subfigure}
\hfill
\begin{subfigure}[T]{0.19\textwidth}
\centering
\includegraphics[width=\textwidth]{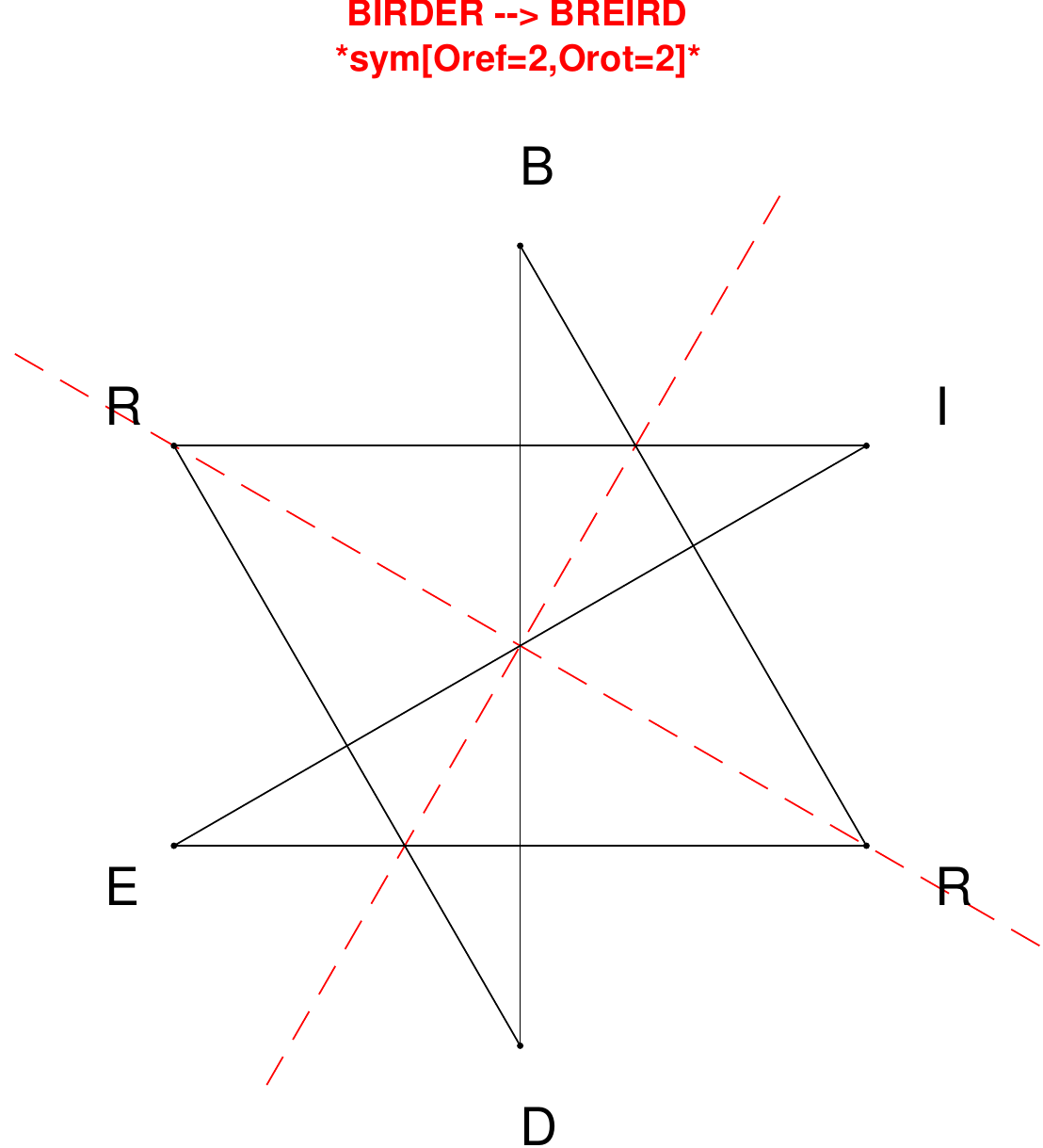}
\end{subfigure}
\hfill
\begin{subfigure}[T]{0.19\textwidth}
\centering
\includegraphics[width=\textwidth]{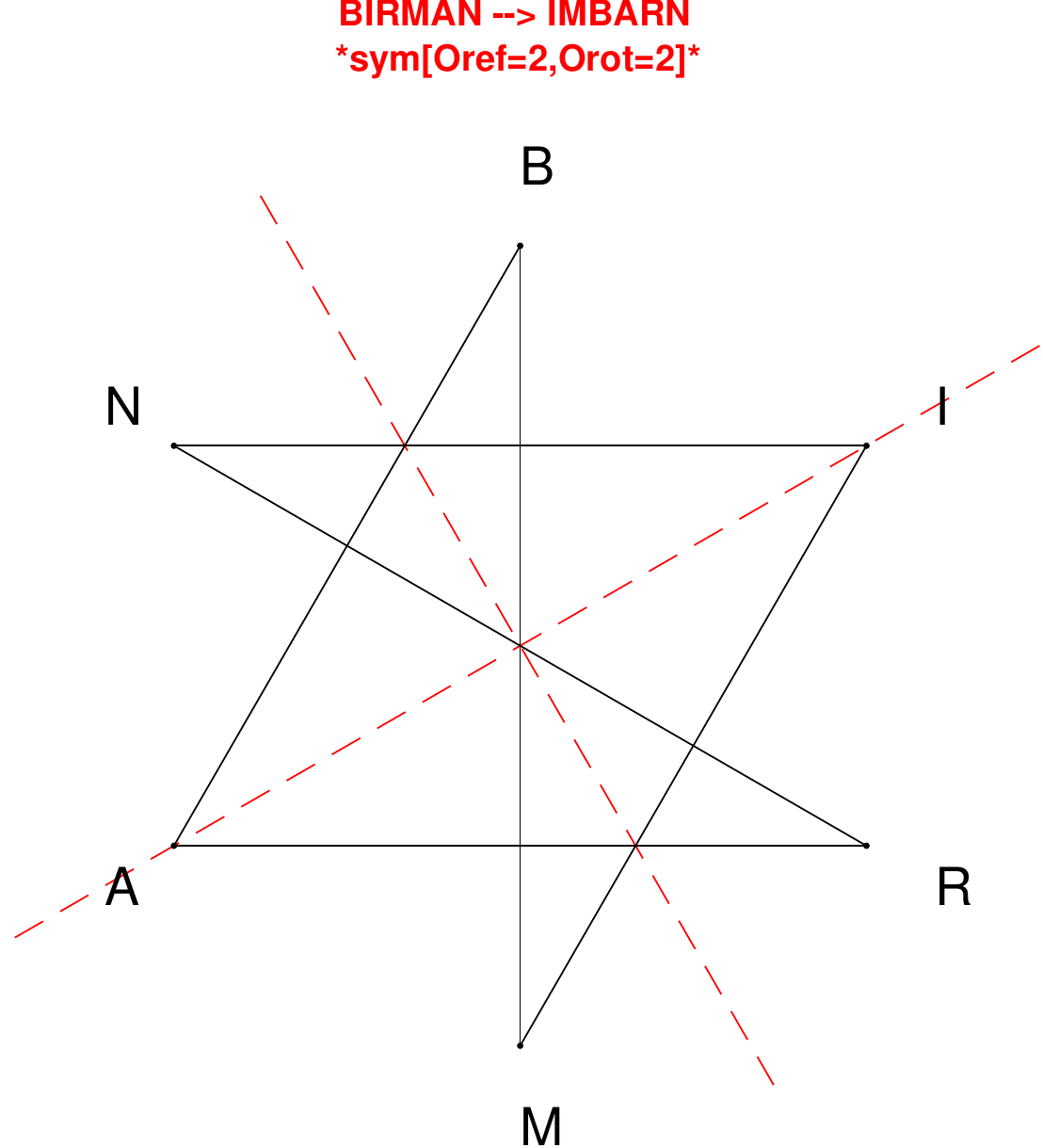}
\end{subfigure}
\hfill
\begin{subfigure}[T]{0.19\textwidth}
\centering
\includegraphics[width=\textwidth]{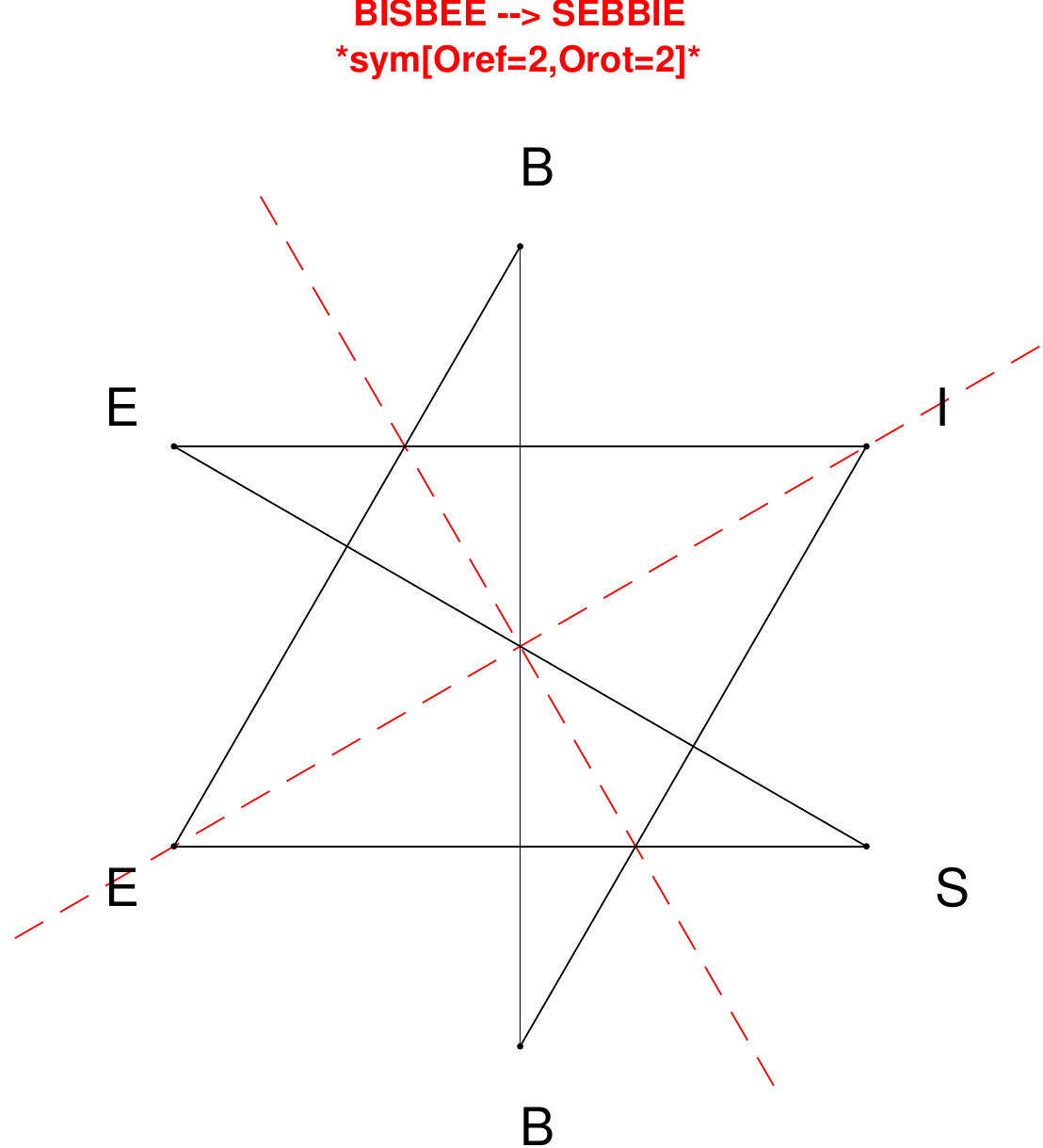}
\end{subfigure}
\hfill
\begin{subfigure}[T]{0.19\textwidth}
\centering
\includegraphics[width=\textwidth]{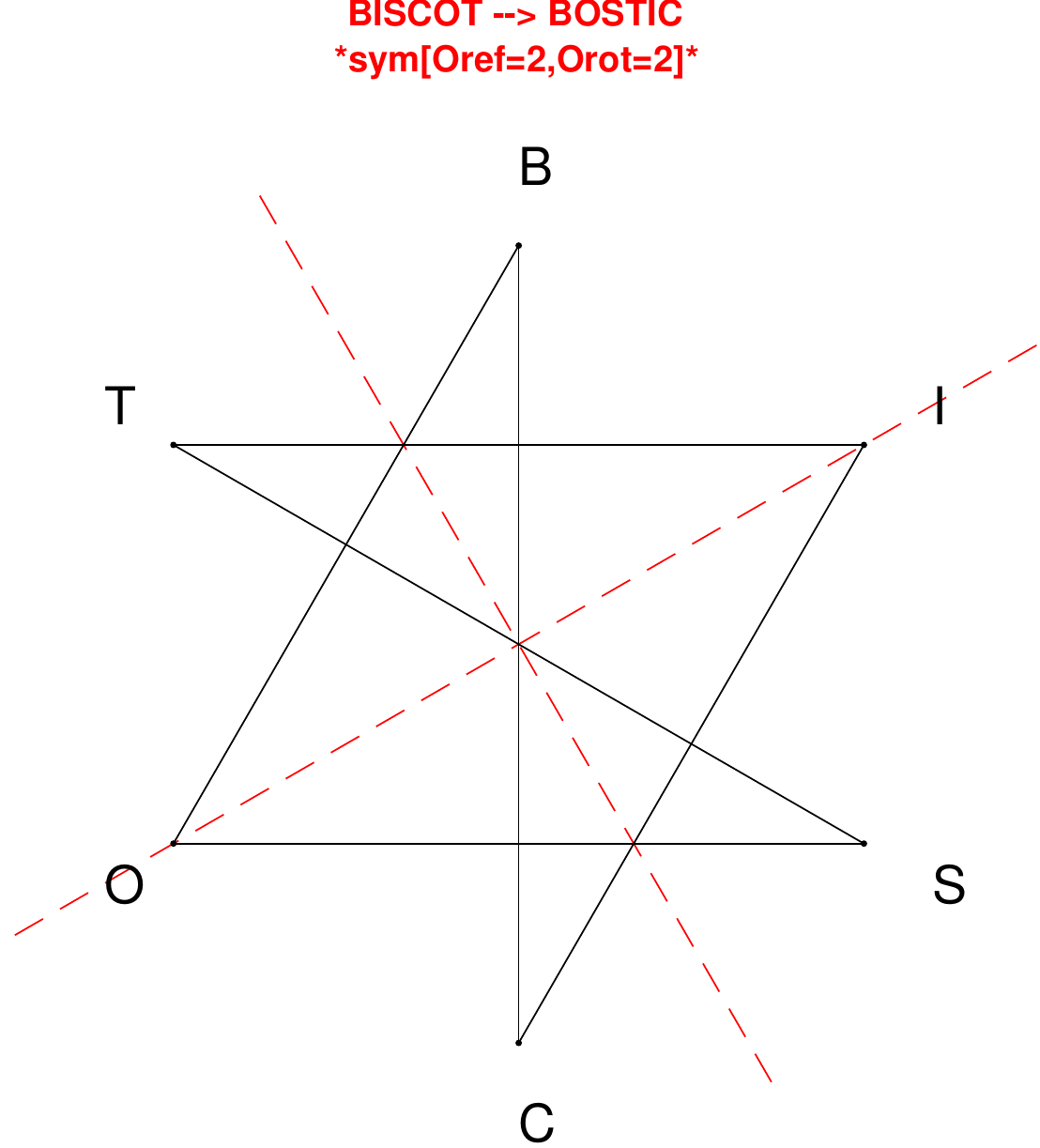}
\end{subfigure}
\end{figure}

\begin{figure}[H]
\centering
\begin{subfigure}[T]{0.19\textwidth}
\centering
\includegraphics[width=\textwidth]{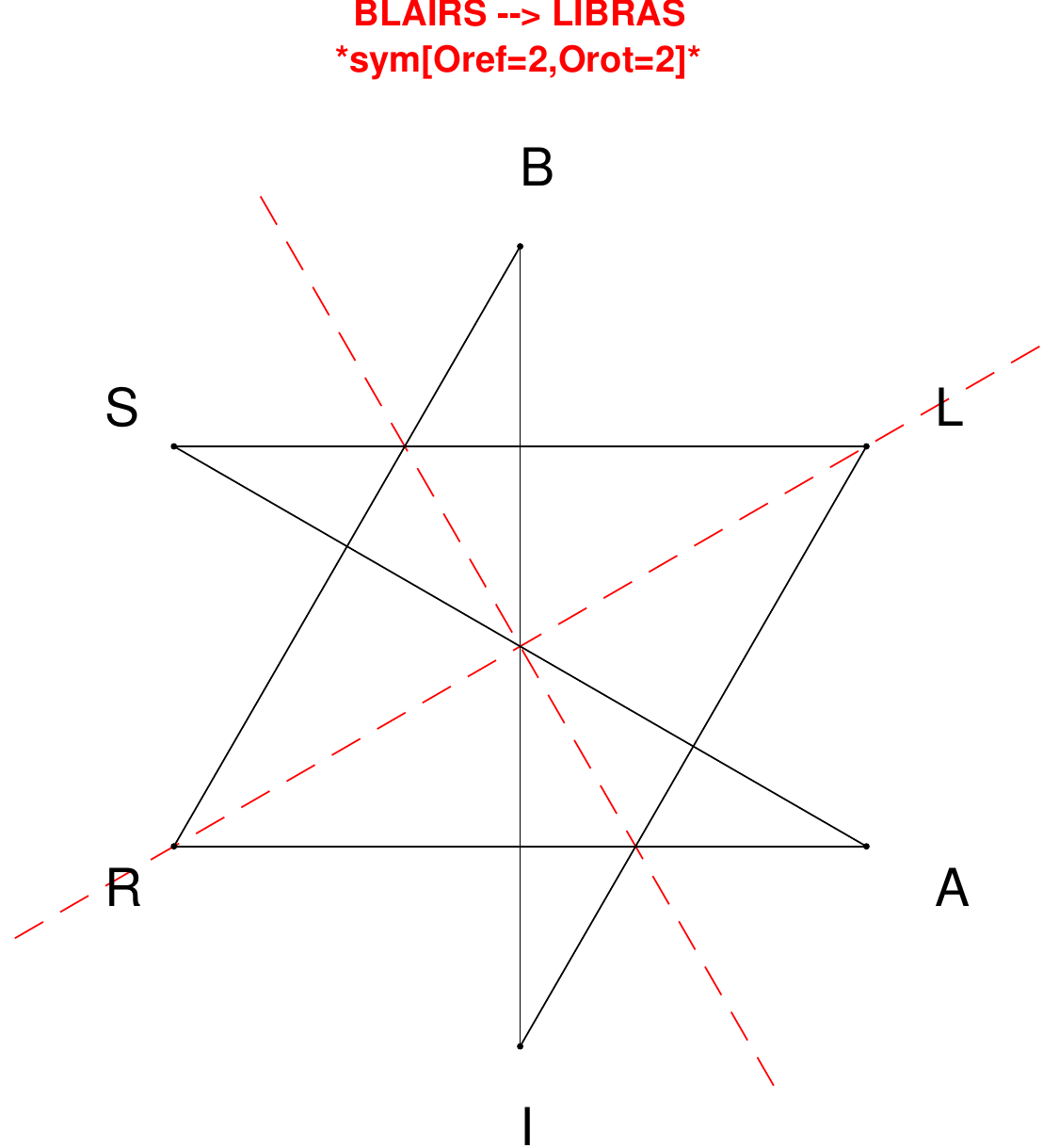}
\end{subfigure}
\hfill
\begin{subfigure}[T]{0.19\textwidth}
\centering
\includegraphics[width=\textwidth]{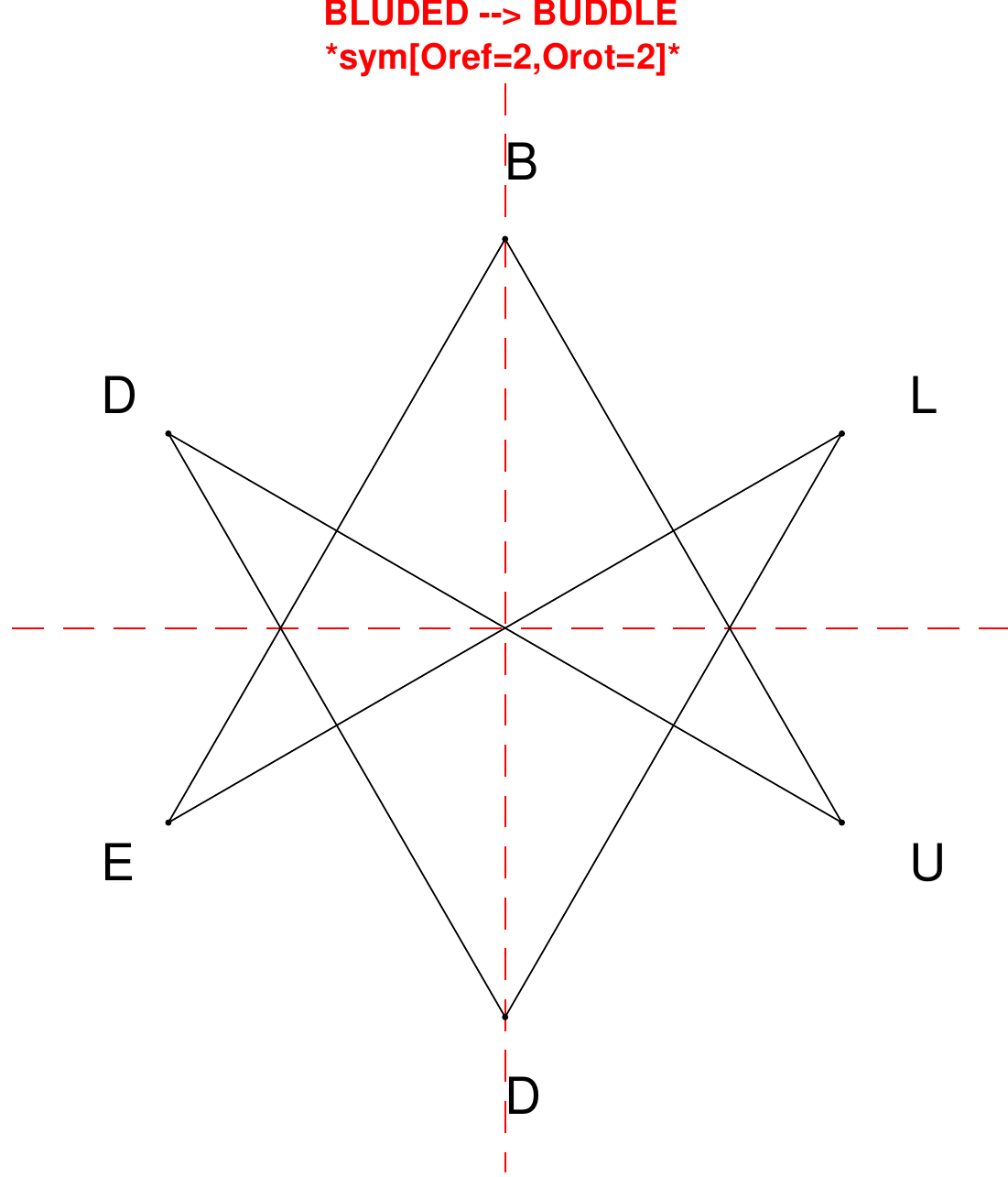}
\end{subfigure}
\hfill
\begin{subfigure}[T]{0.19\textwidth}
\centering
\includegraphics[width=\textwidth]{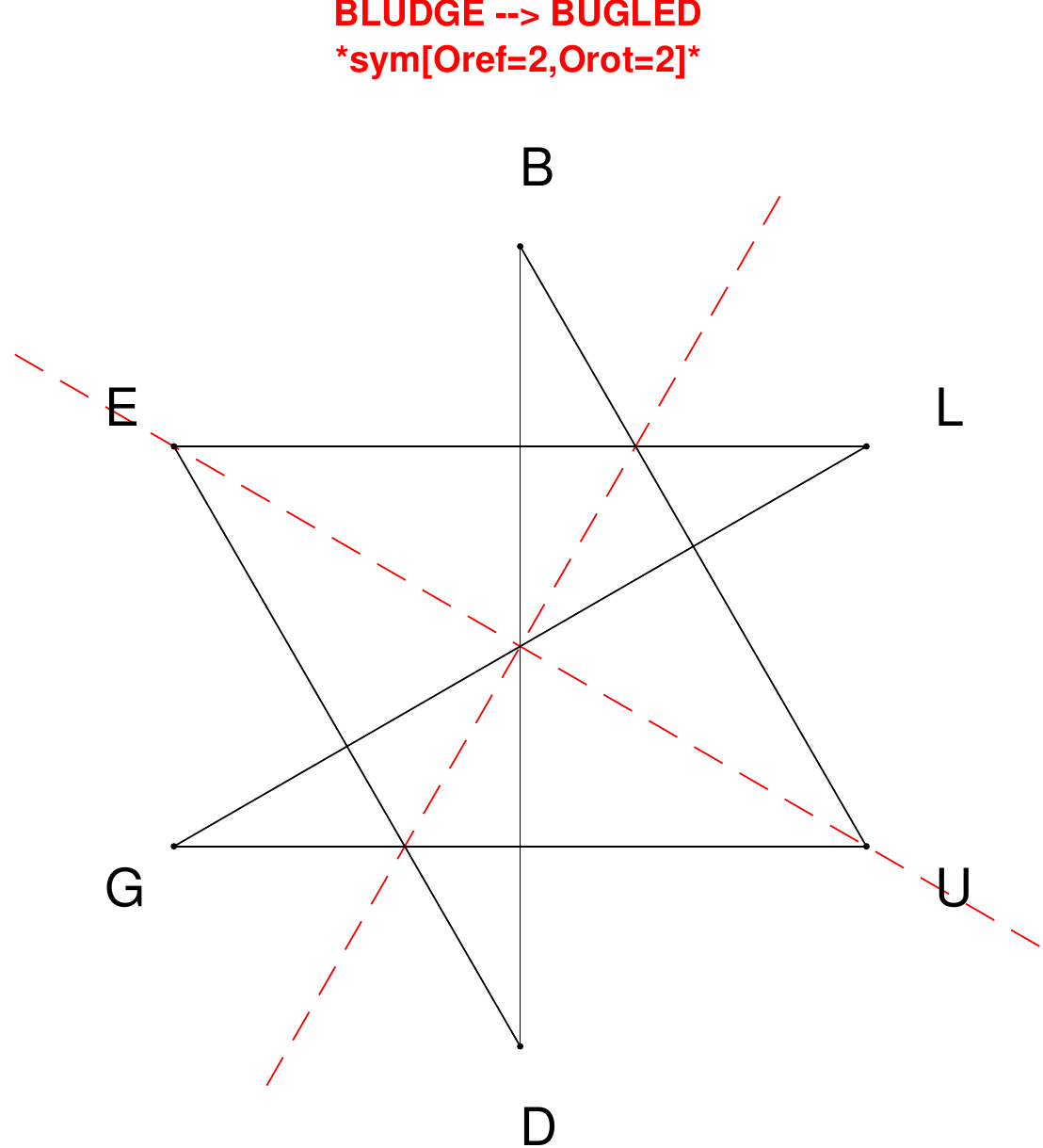}
\end{subfigure}
\hfill
\begin{subfigure}[T]{0.19\textwidth}
\centering
\includegraphics[width=\textwidth]{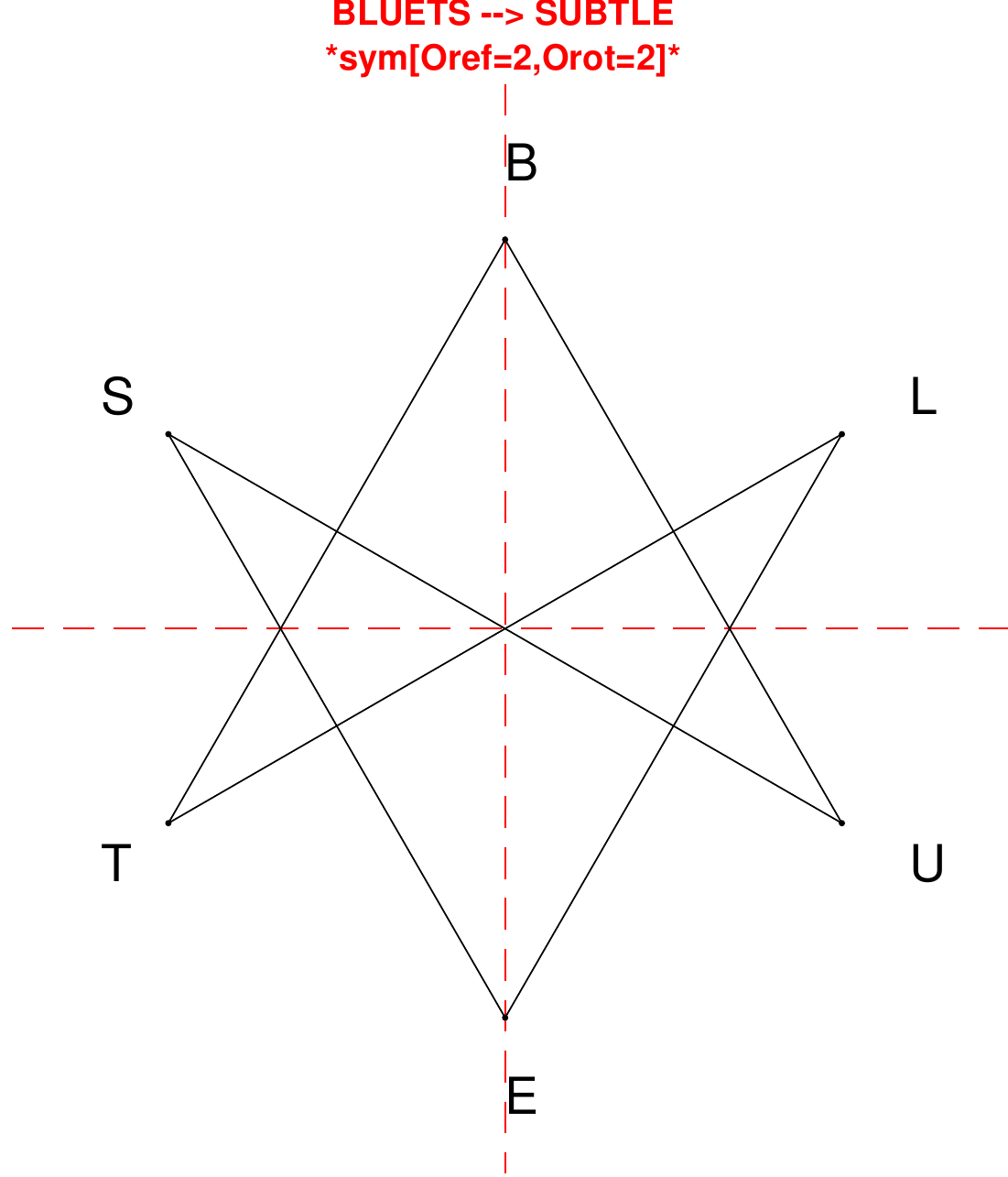}
\end{subfigure}
\hfill
\begin{subfigure}[T]{0.19\textwidth}
\centering
\includegraphics[width=\textwidth]{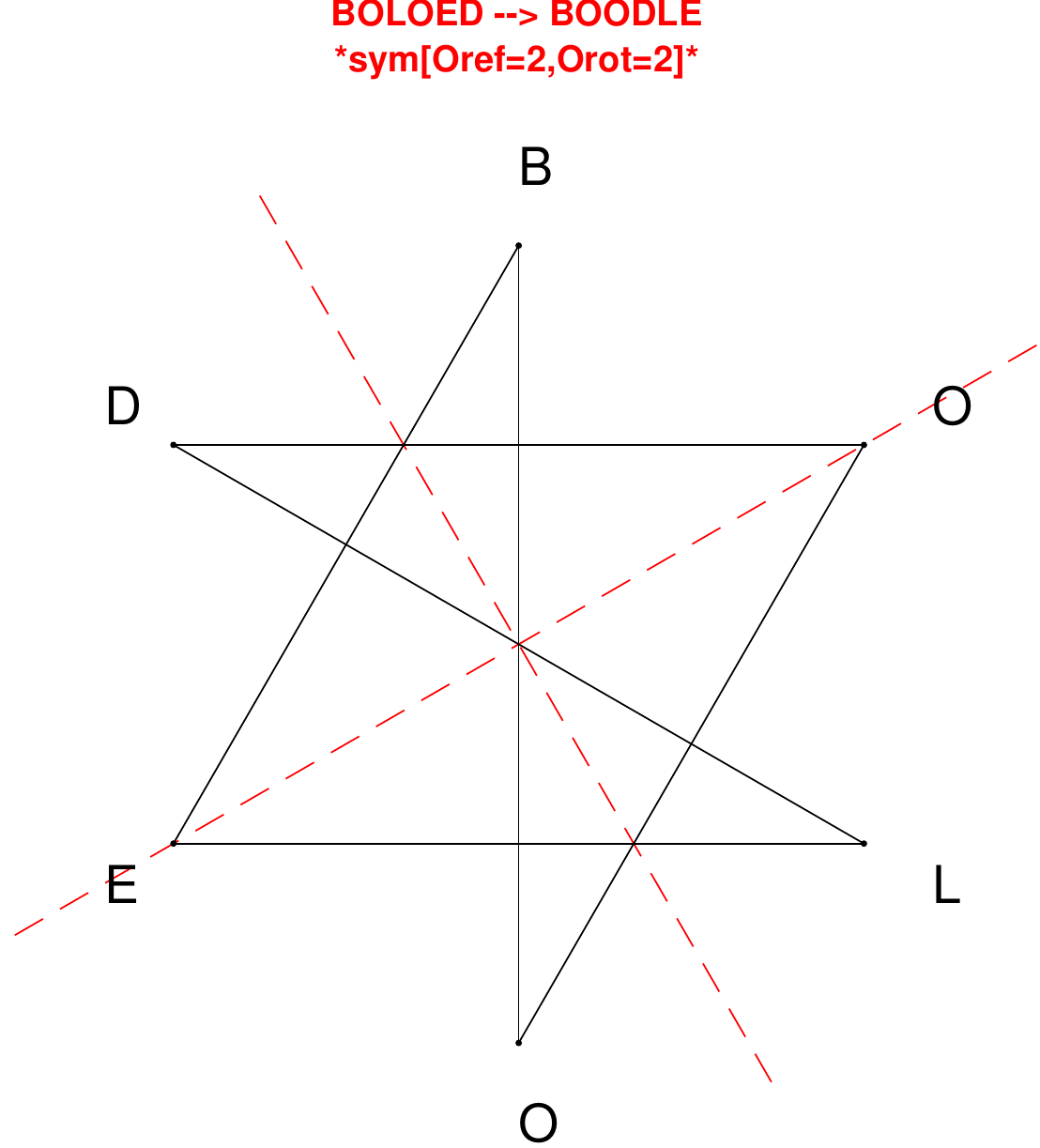}
\end{subfigure}
\end{figure}

\begin{figure}[H]
\centering
\begin{subfigure}[T]{0.19\textwidth}
\centering
\includegraphics[width=\textwidth]{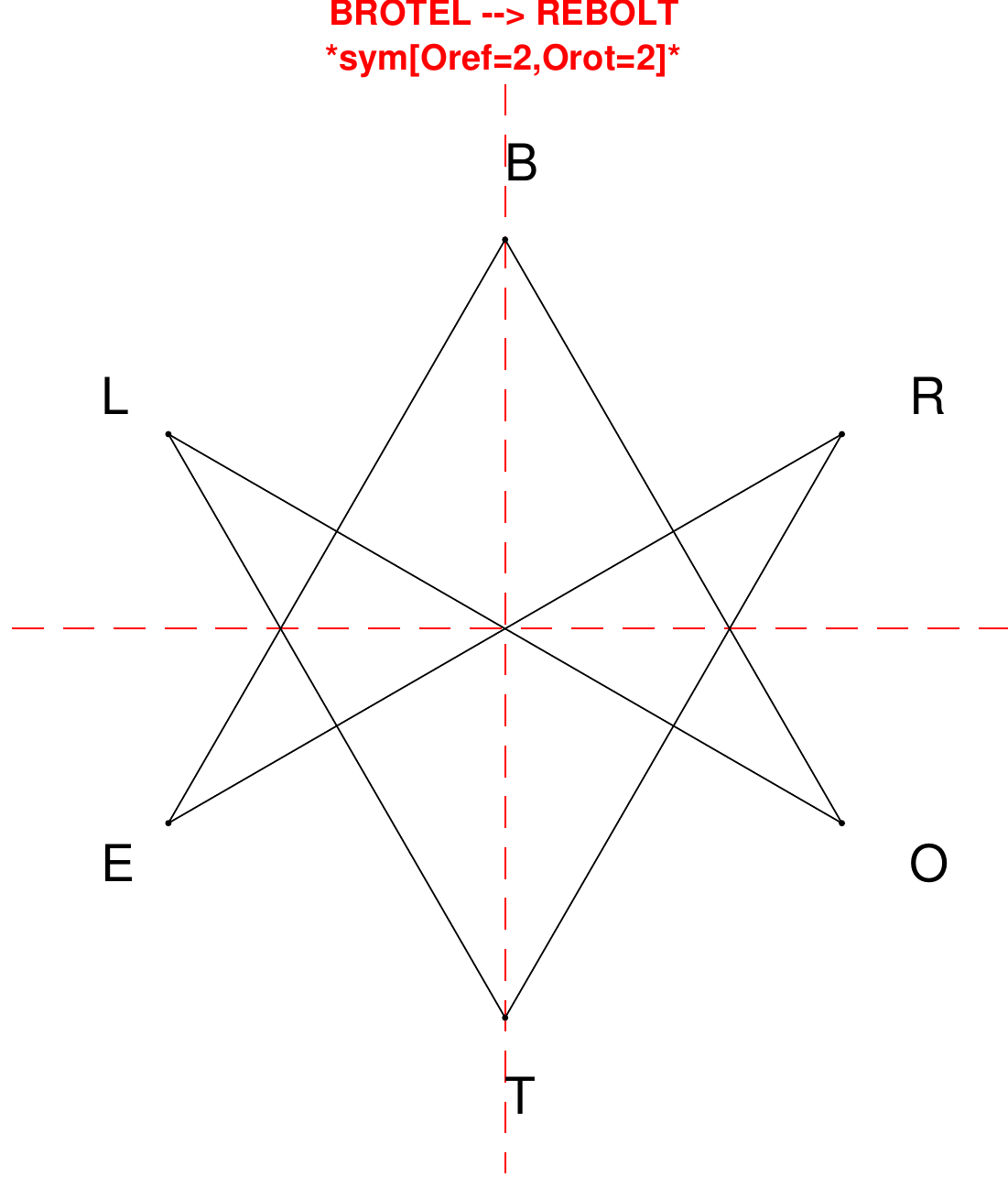}
\end{subfigure}
\hfill
\begin{subfigure}[T]{0.19\textwidth}
\centering
\includegraphics[width=\textwidth]{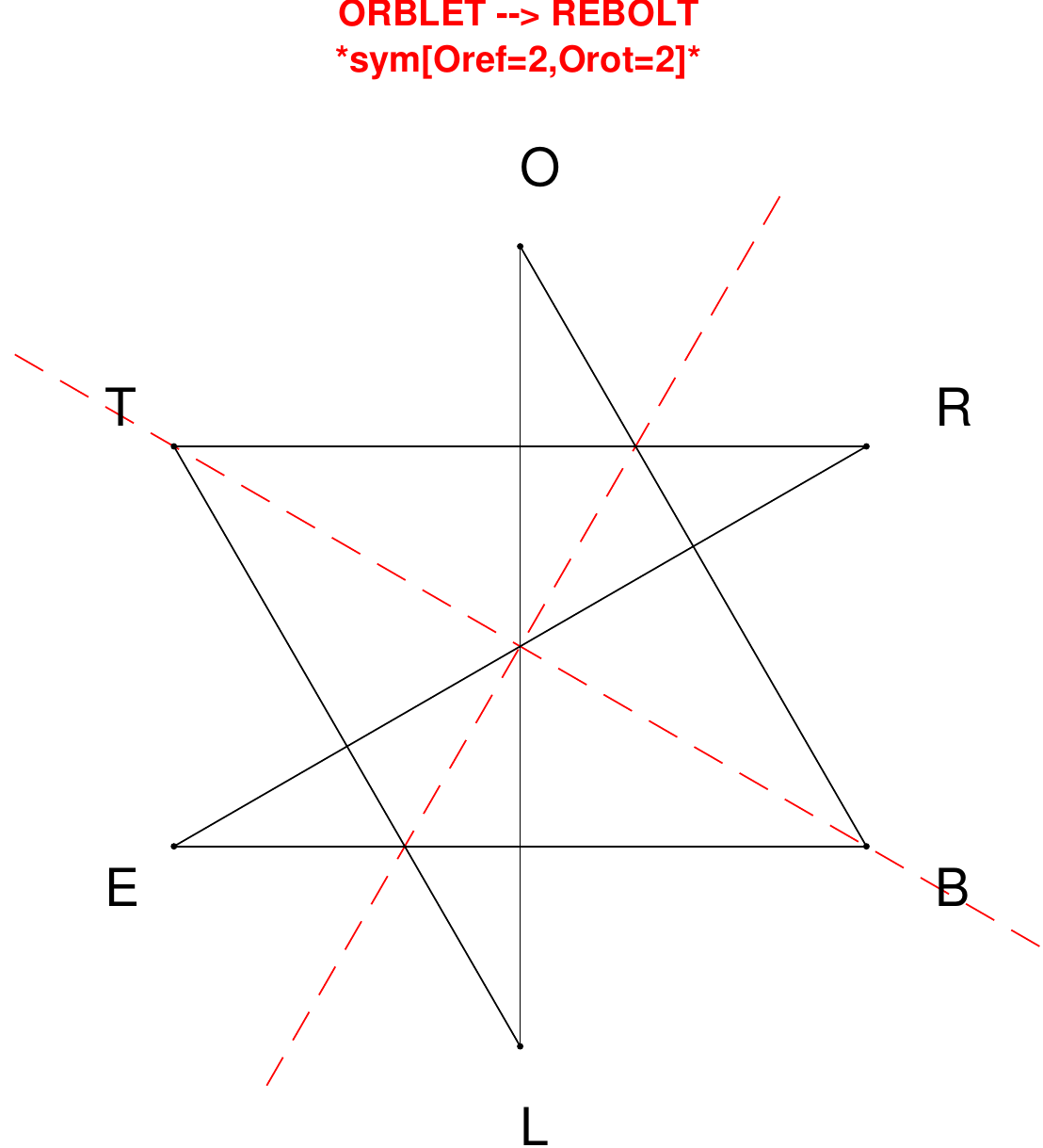}
\end{subfigure}
\hfill
\begin{subfigure}[T]{0.19\textwidth}
\centering
\includegraphics[width=\textwidth]{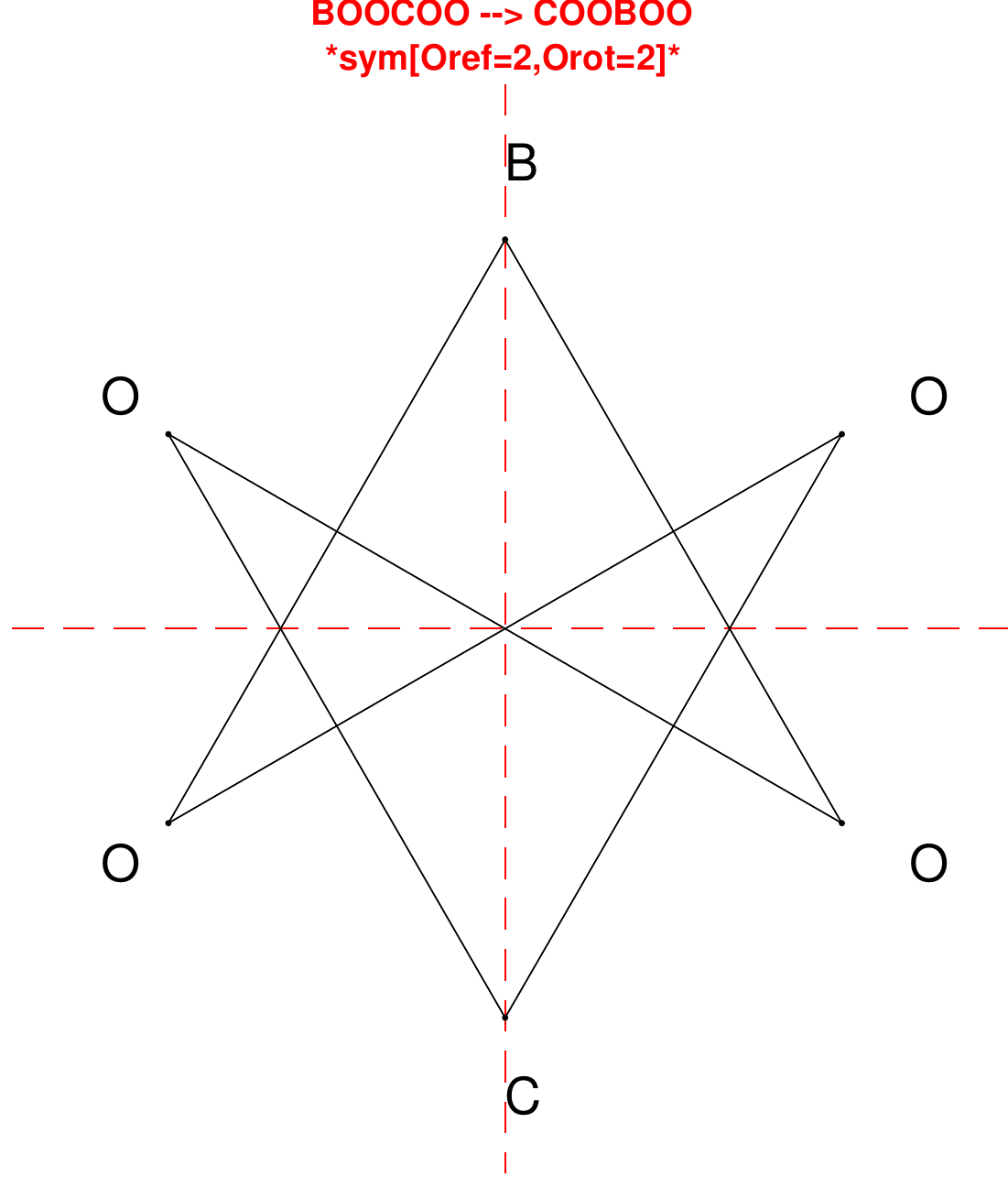}
\end{subfigure}
\hfill
\begin{subfigure}[T]{0.19\textwidth}
\centering
\includegraphics[width=\textwidth]{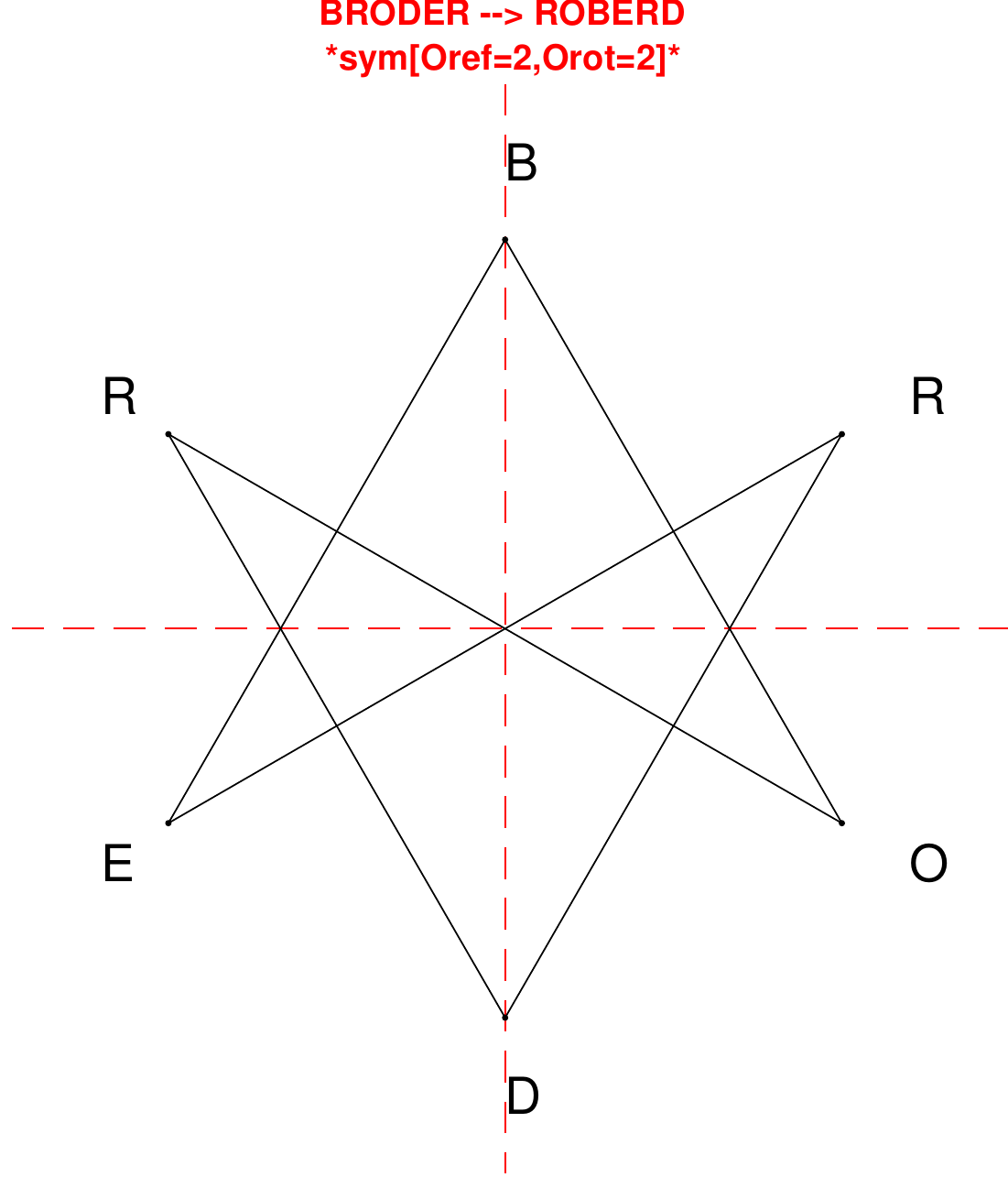}
\end{subfigure}
\hfill
\begin{subfigure}[T]{0.19\textwidth}
\centering
\includegraphics[width=\textwidth]{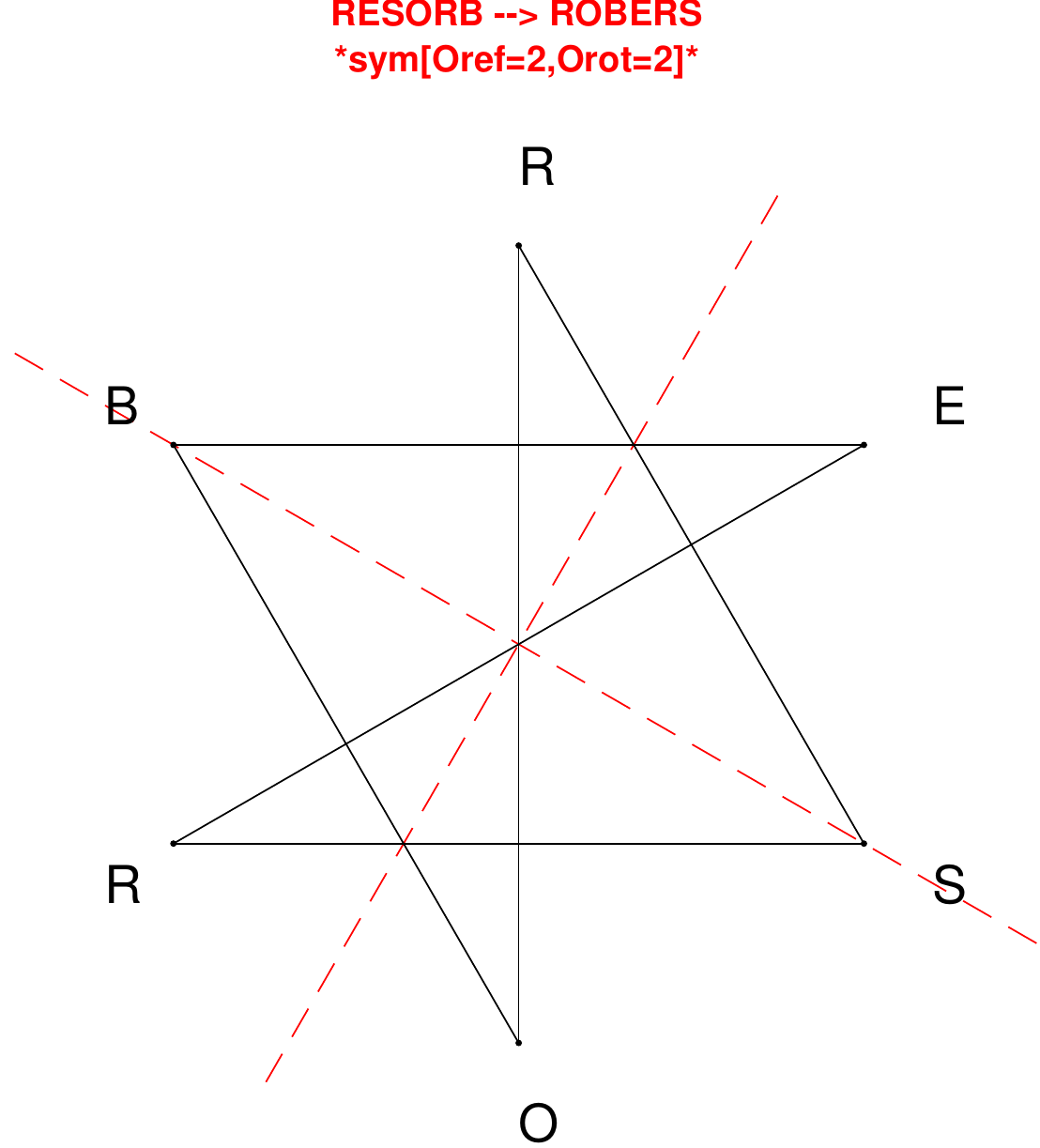}
\end{subfigure}
\end{figure}

\begin{figure}[H]
\centering
\begin{subfigure}[T]{0.19\textwidth}
\centering
\includegraphics[width=\textwidth]{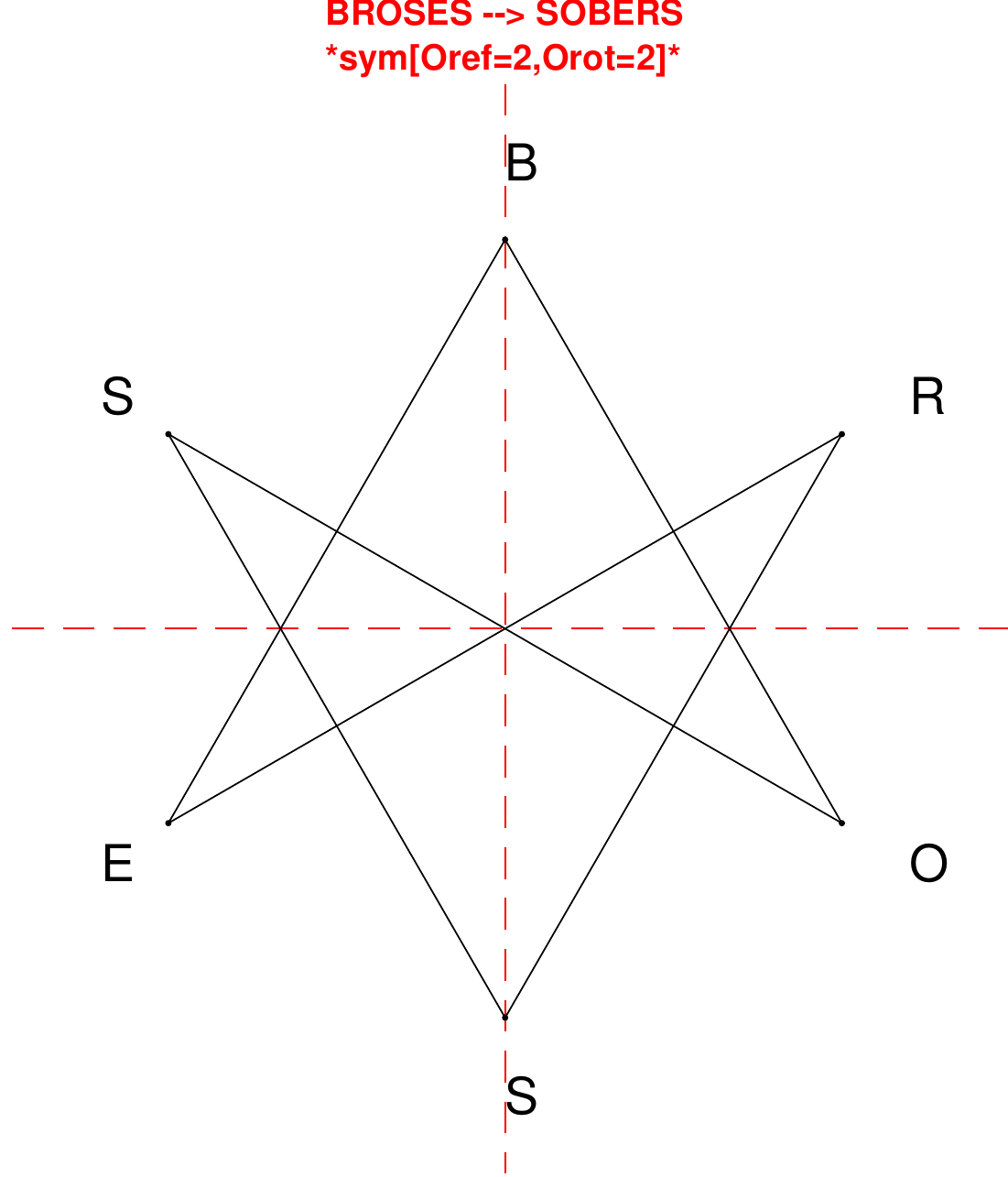}
\end{subfigure}
\hfill
\begin{subfigure}[T]{0.19\textwidth}
\centering
\includegraphics[width=\textwidth]{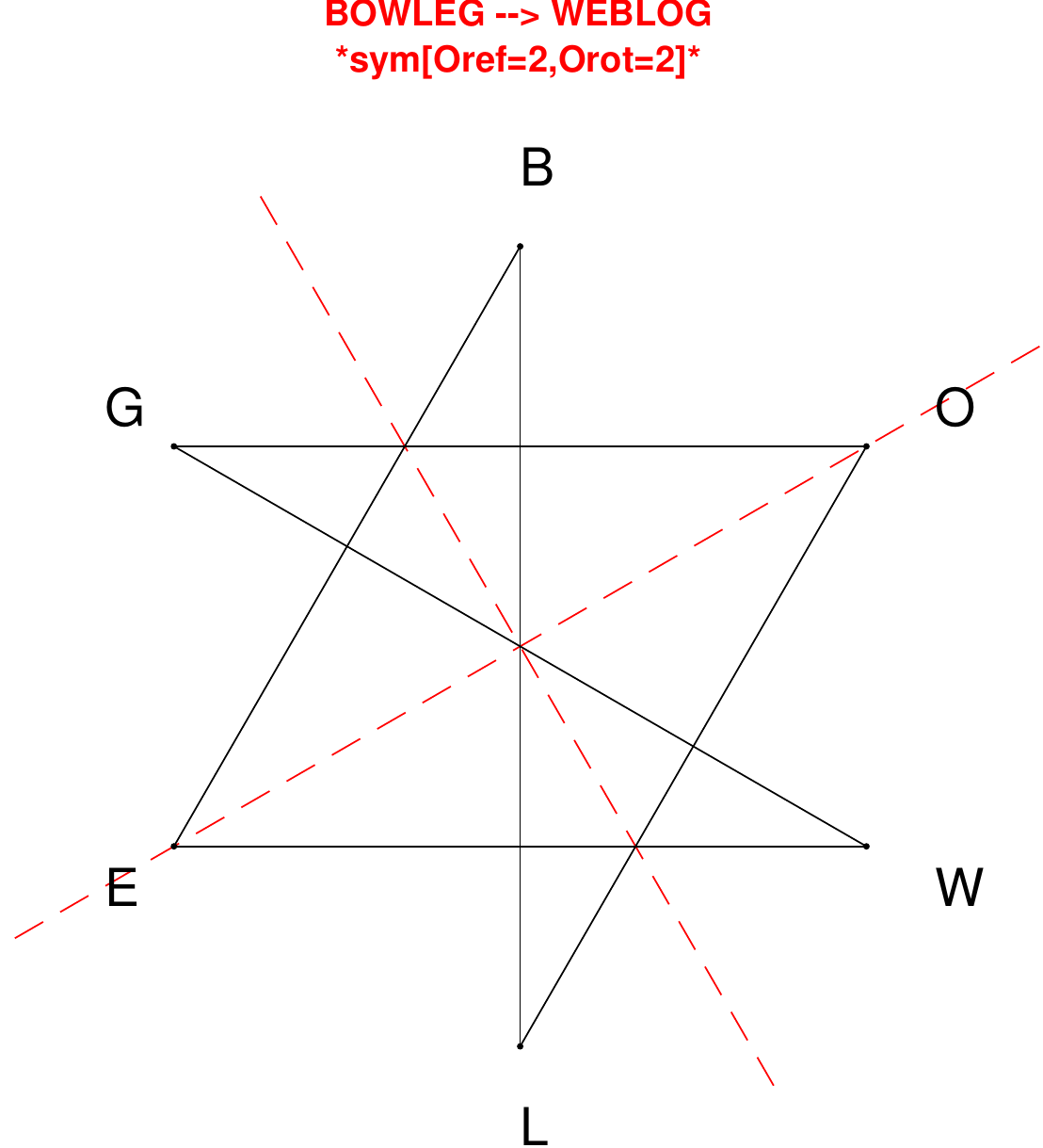}
\end{subfigure}
\hfill
\begin{subfigure}[T]{0.19\textwidth}
\centering
\includegraphics[width=\textwidth]{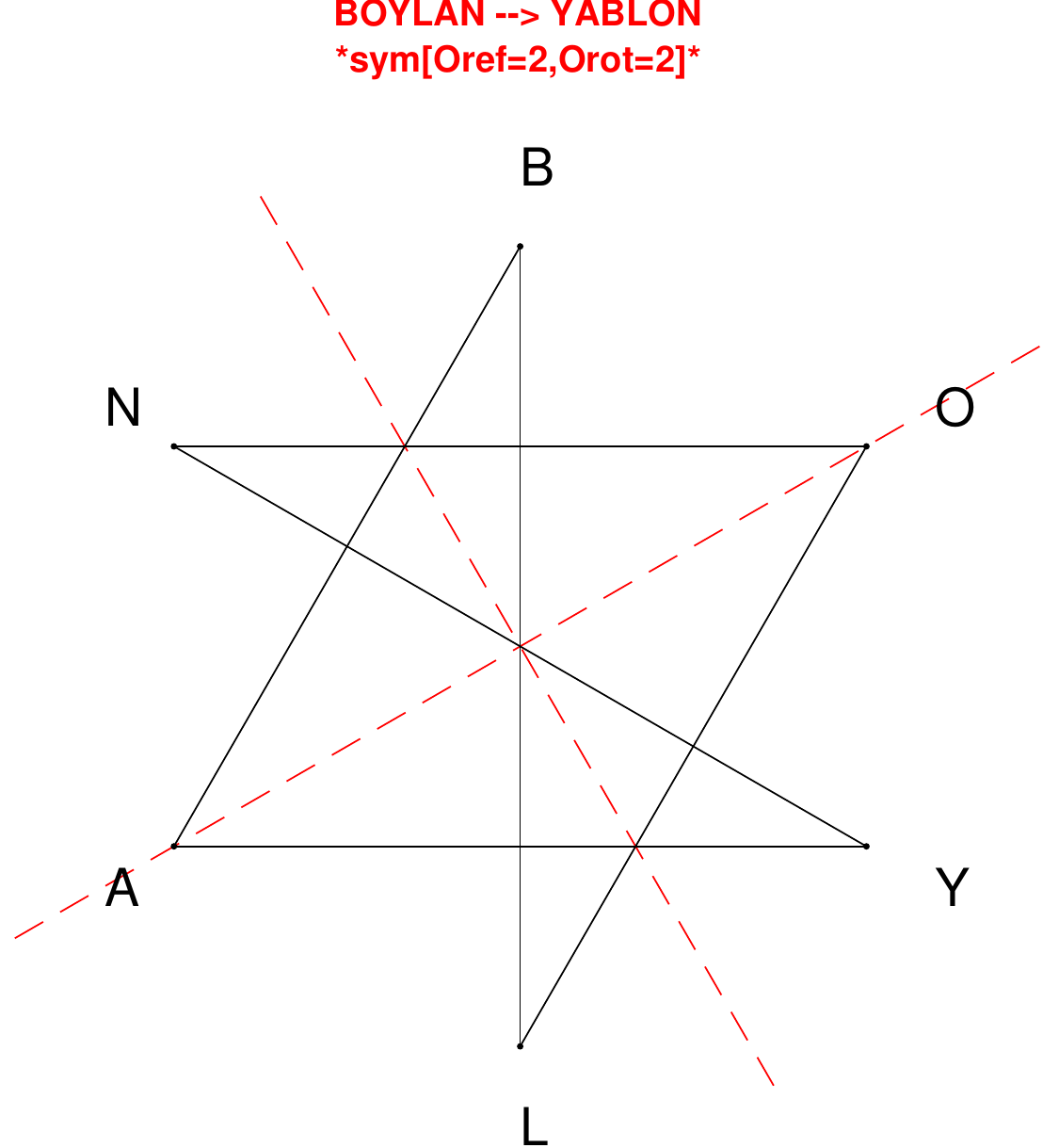}
\end{subfigure}
\hfill
\begin{subfigure}[T]{0.19\textwidth}
\centering
\includegraphics[width=\textwidth]{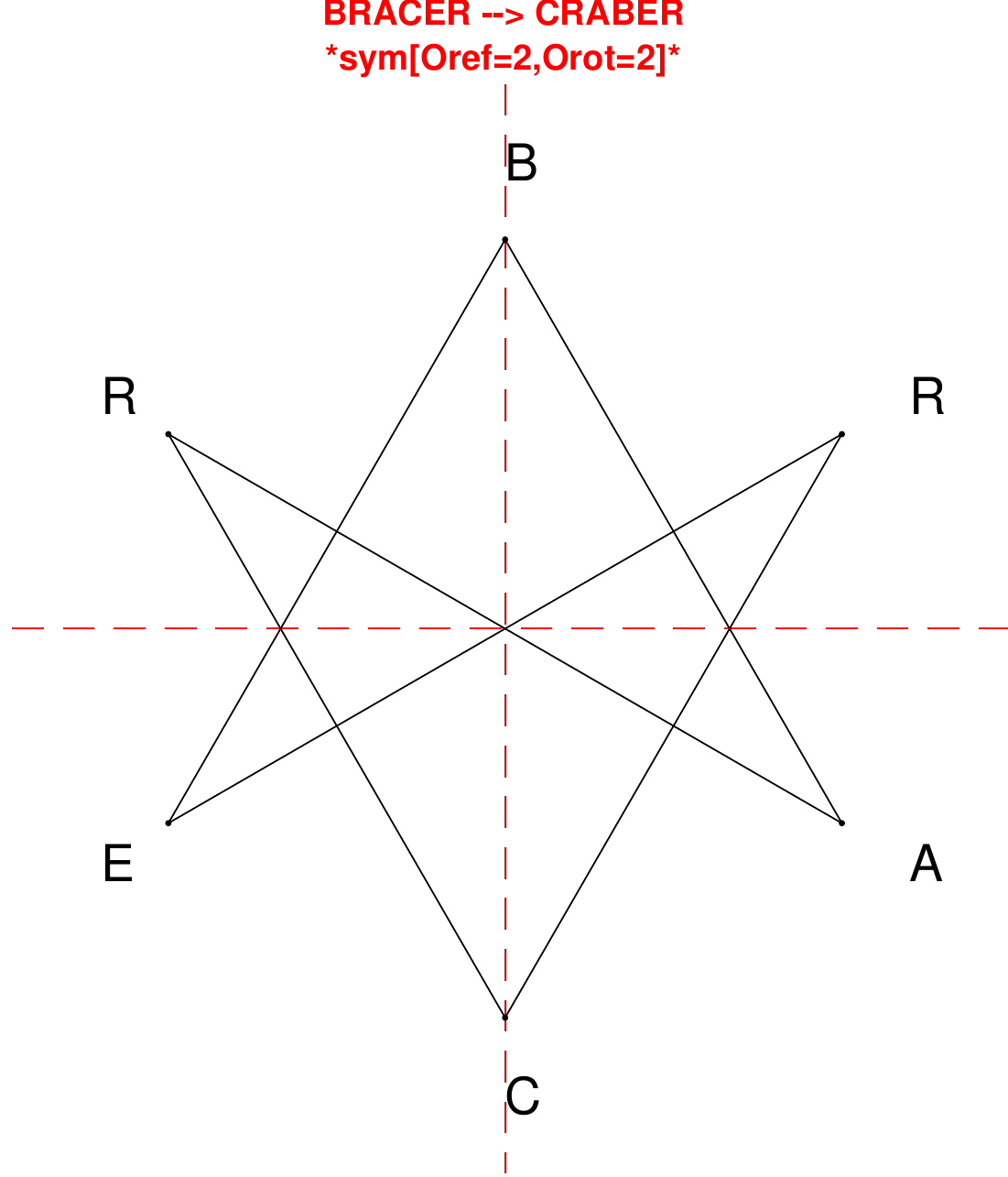}
\end{subfigure}
\hfill
\begin{subfigure}[T]{0.19\textwidth}
\centering
\includegraphics[width=\textwidth]{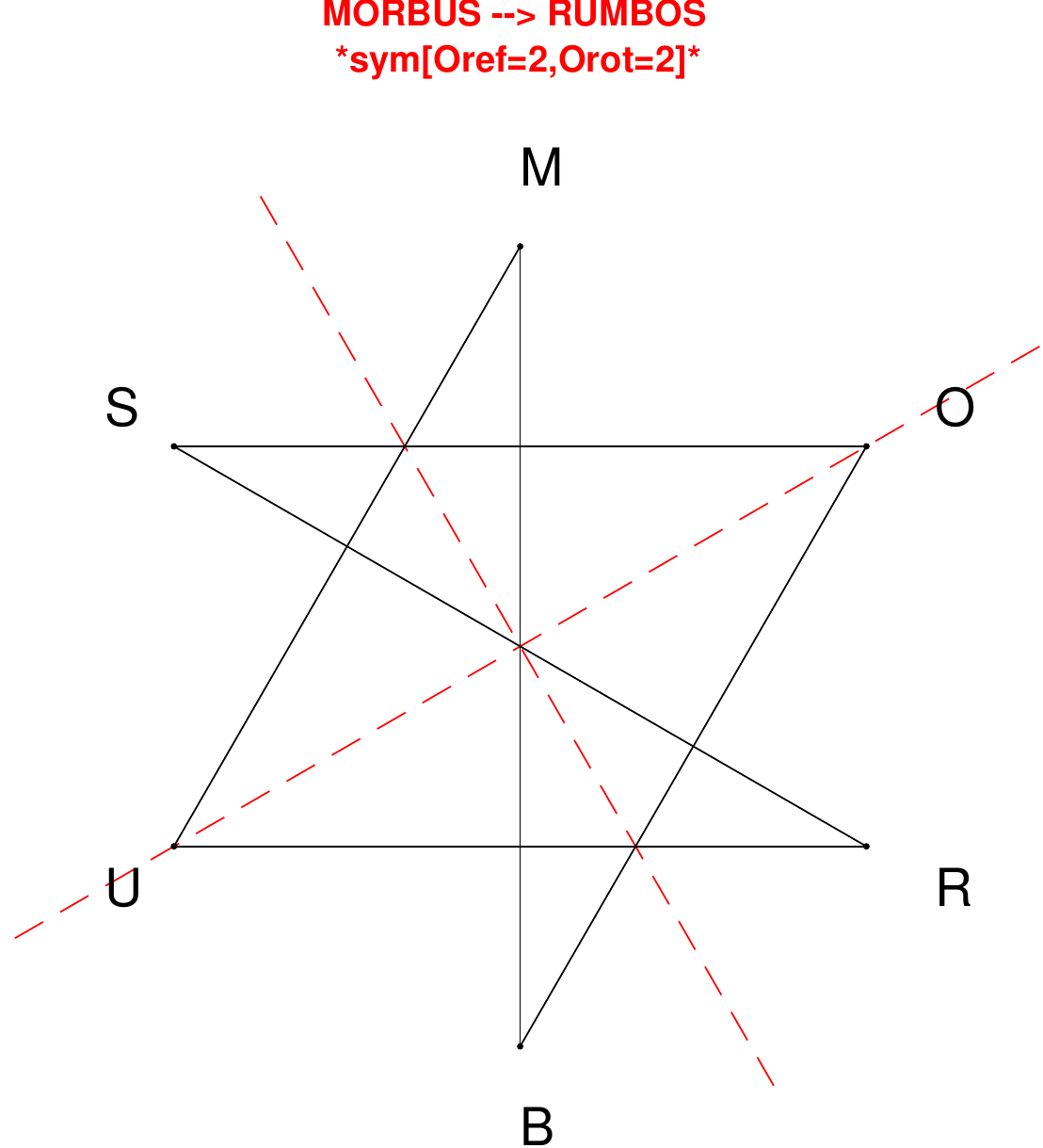}
\end{subfigure}
\end{figure}

\begin{figure}[H]
\centering
\begin{subfigure}[T]{0.19\textwidth}
\centering
\includegraphics[width=\textwidth]{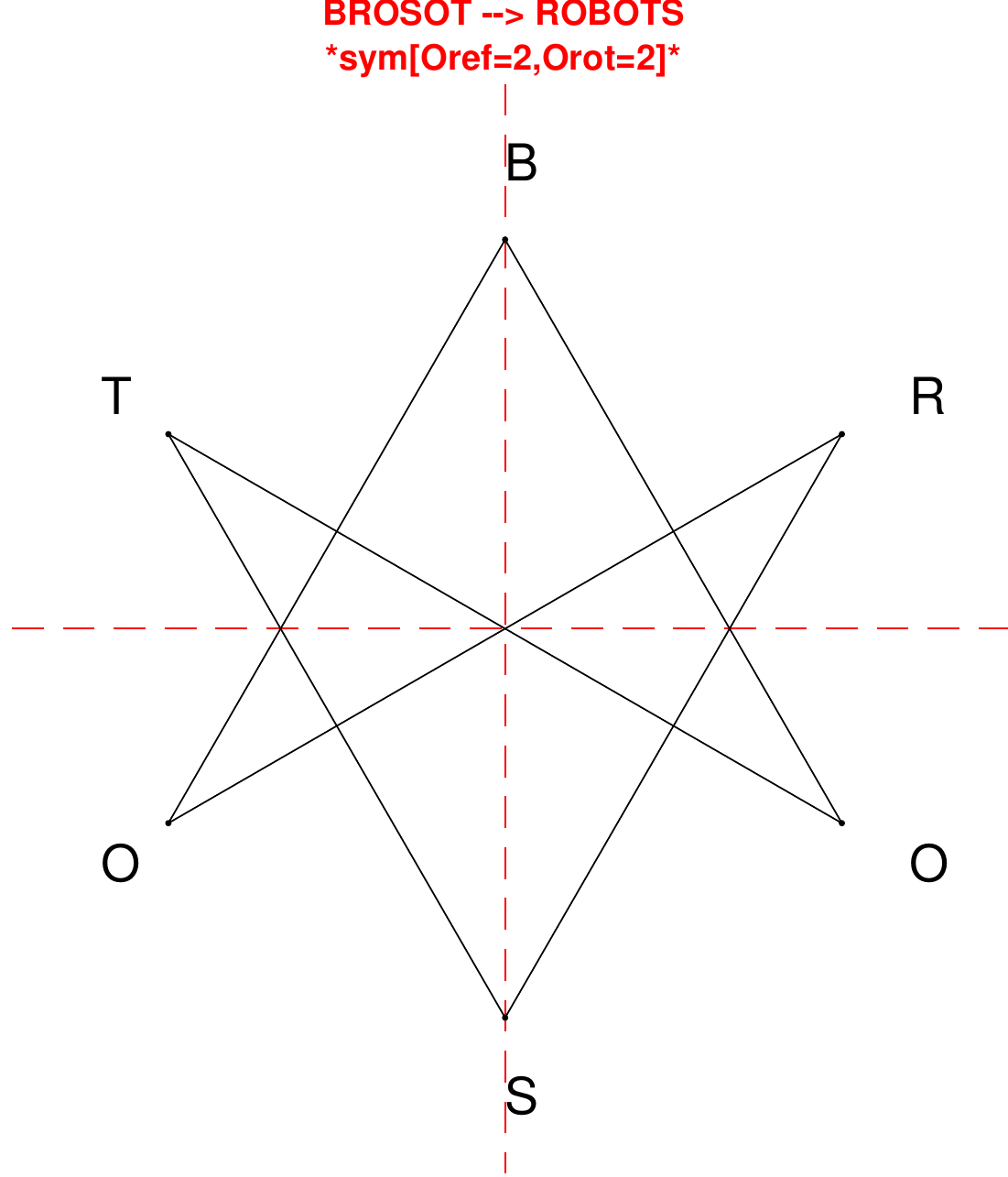}
\end{subfigure}
\hfill
\begin{subfigure}[T]{0.19\textwidth}
\centering
\includegraphics[width=\textwidth]{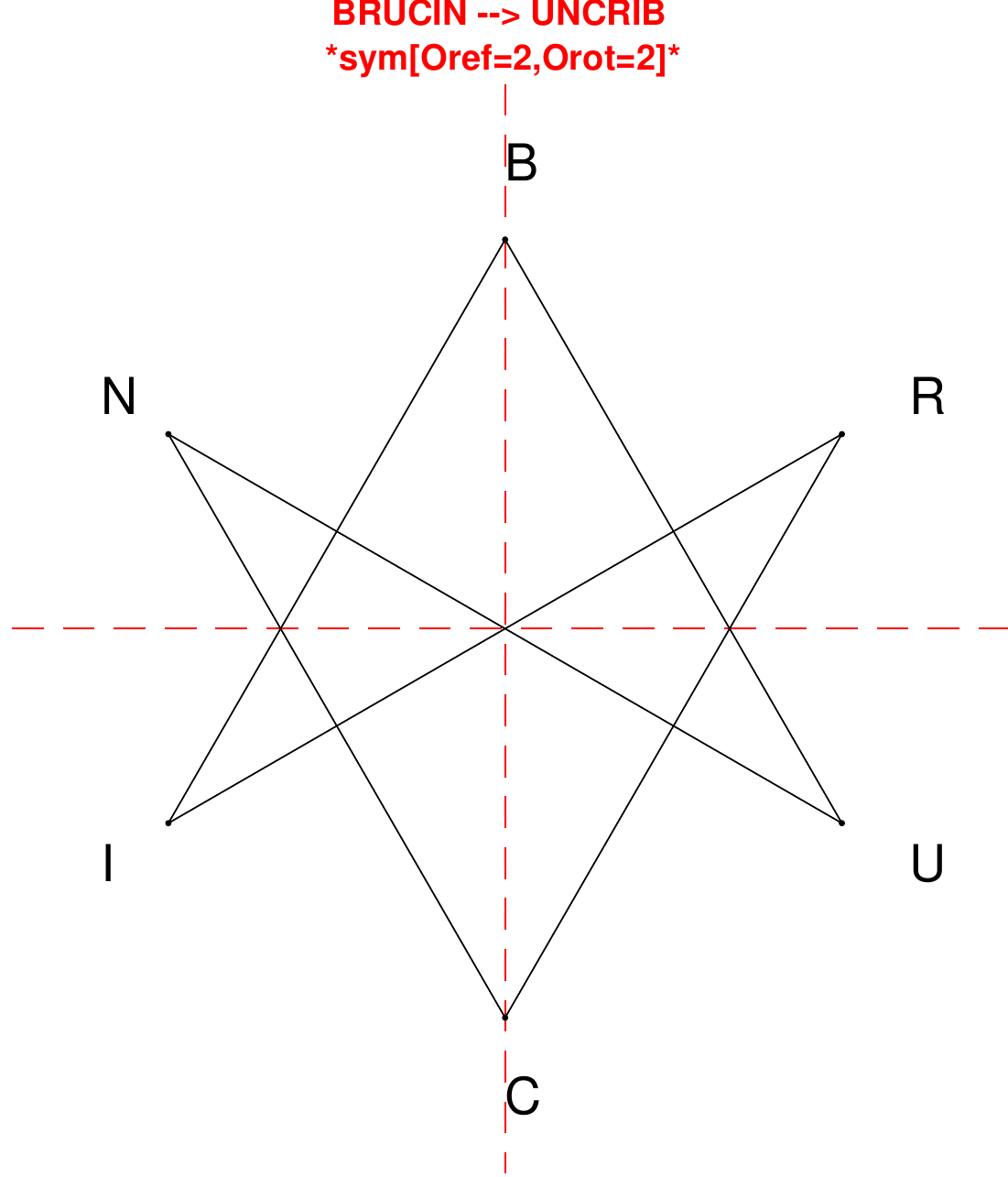}
\end{subfigure}
\hfill
\begin{subfigure}[T]{0.19\textwidth}
\centering
\includegraphics[width=\textwidth]{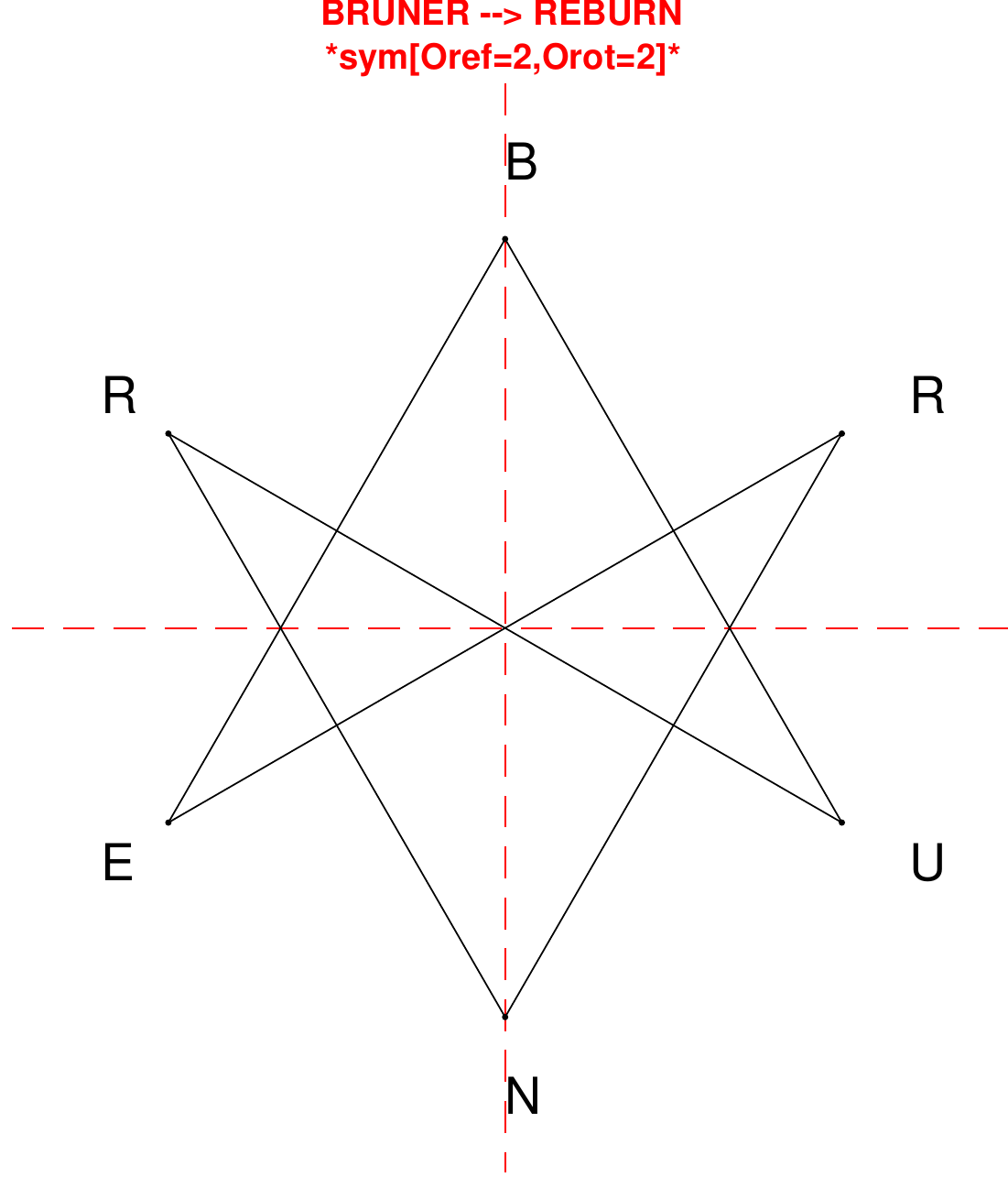}
\end{subfigure}
\hfill
\begin{subfigure}[T]{0.19\textwidth}
\centering
\includegraphics[width=\textwidth]{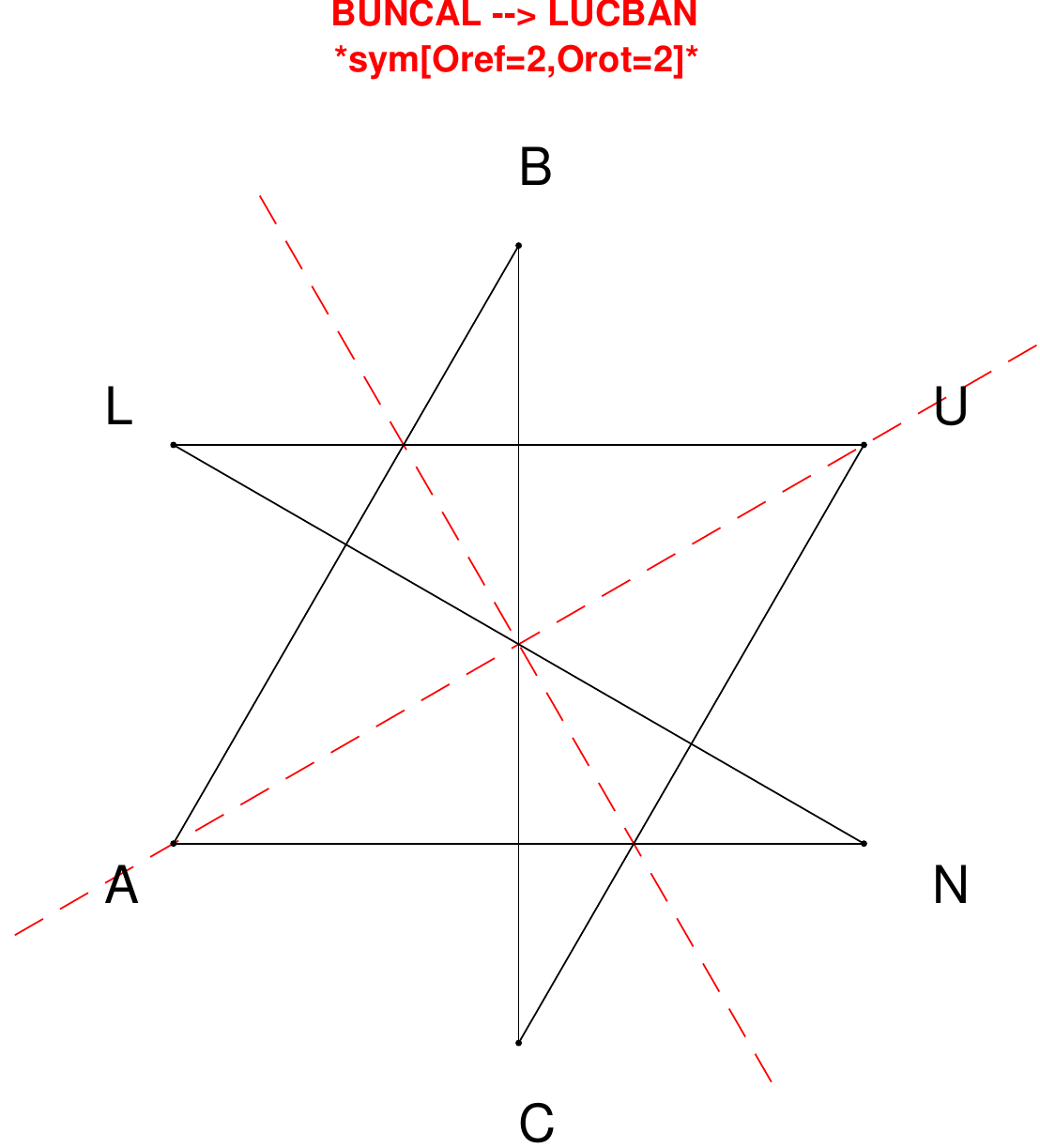}
\end{subfigure}
\hfill
\begin{subfigure}[T]{0.19\textwidth}
\centering
\includegraphics[width=\textwidth]{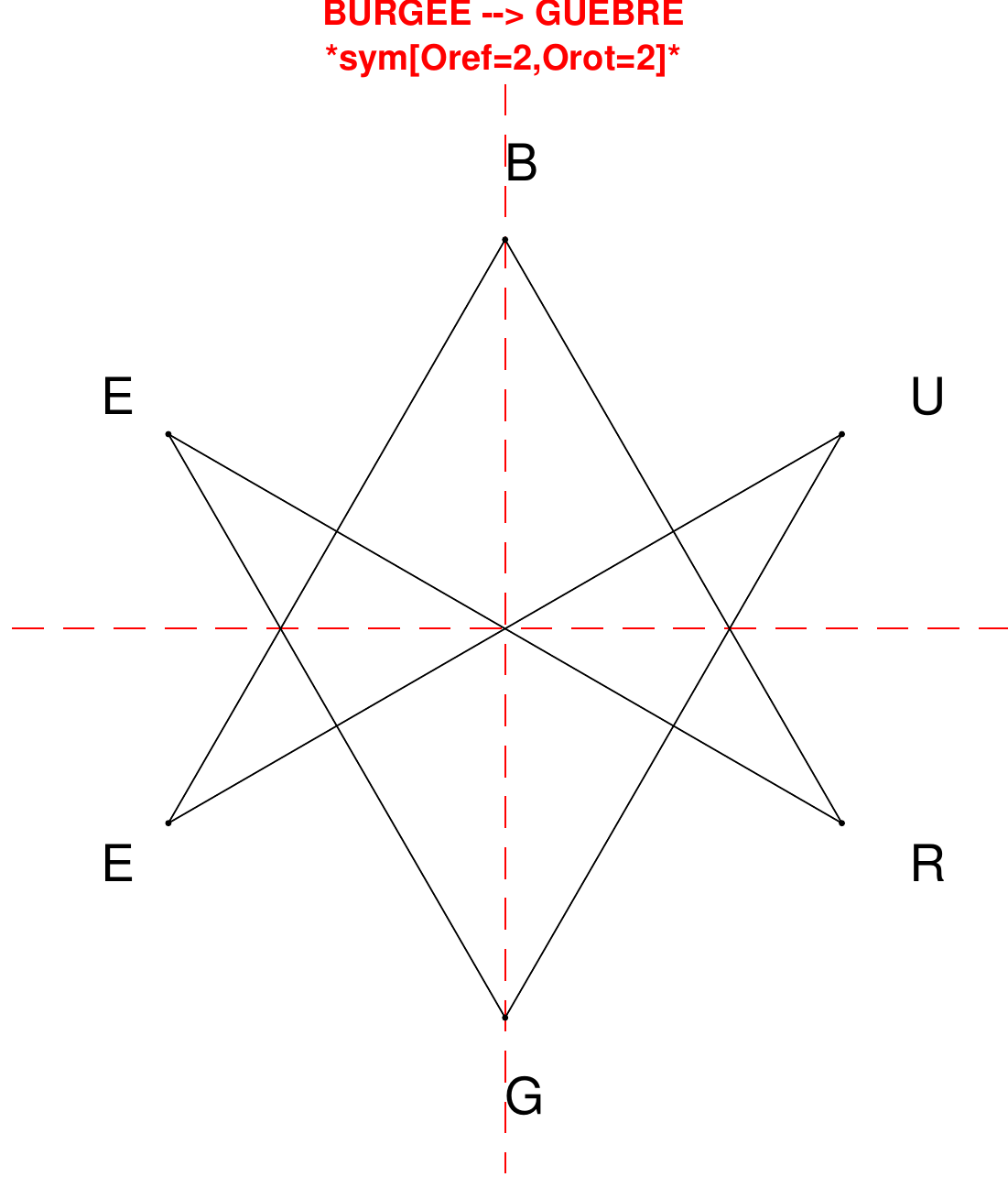}
\end{subfigure}
\end{figure}

\begin{figure}[H]
\centering
\begin{subfigure}[T]{0.19\textwidth}
\centering
\includegraphics[width=\textwidth]{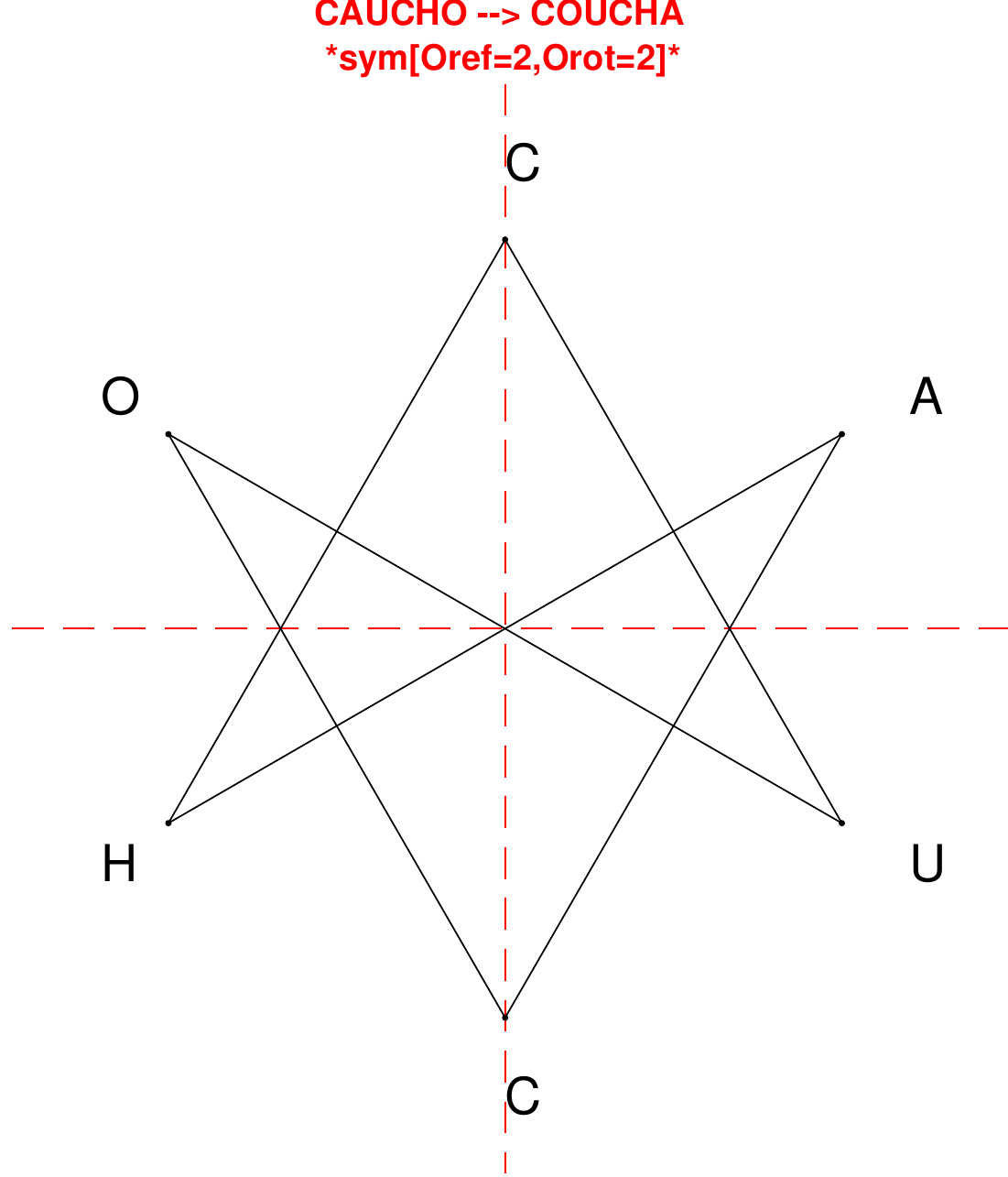}
\end{subfigure}
\hfill
\begin{subfigure}[T]{0.19\textwidth}
\centering
\includegraphics[width=\textwidth]{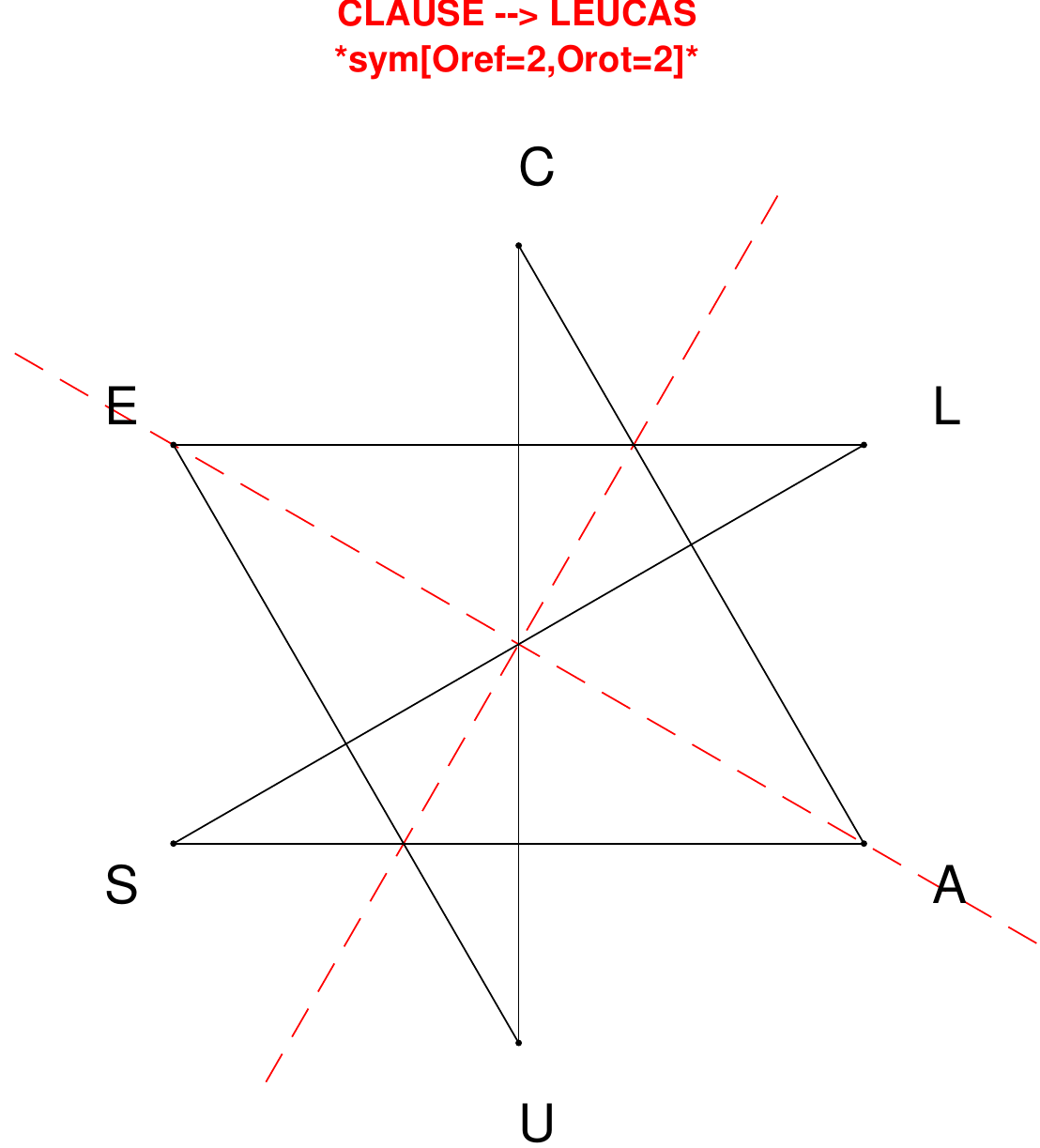}
\end{subfigure}
\hfill
\begin{subfigure}[T]{0.19\textwidth}
\centering
\includegraphics[width=\textwidth]{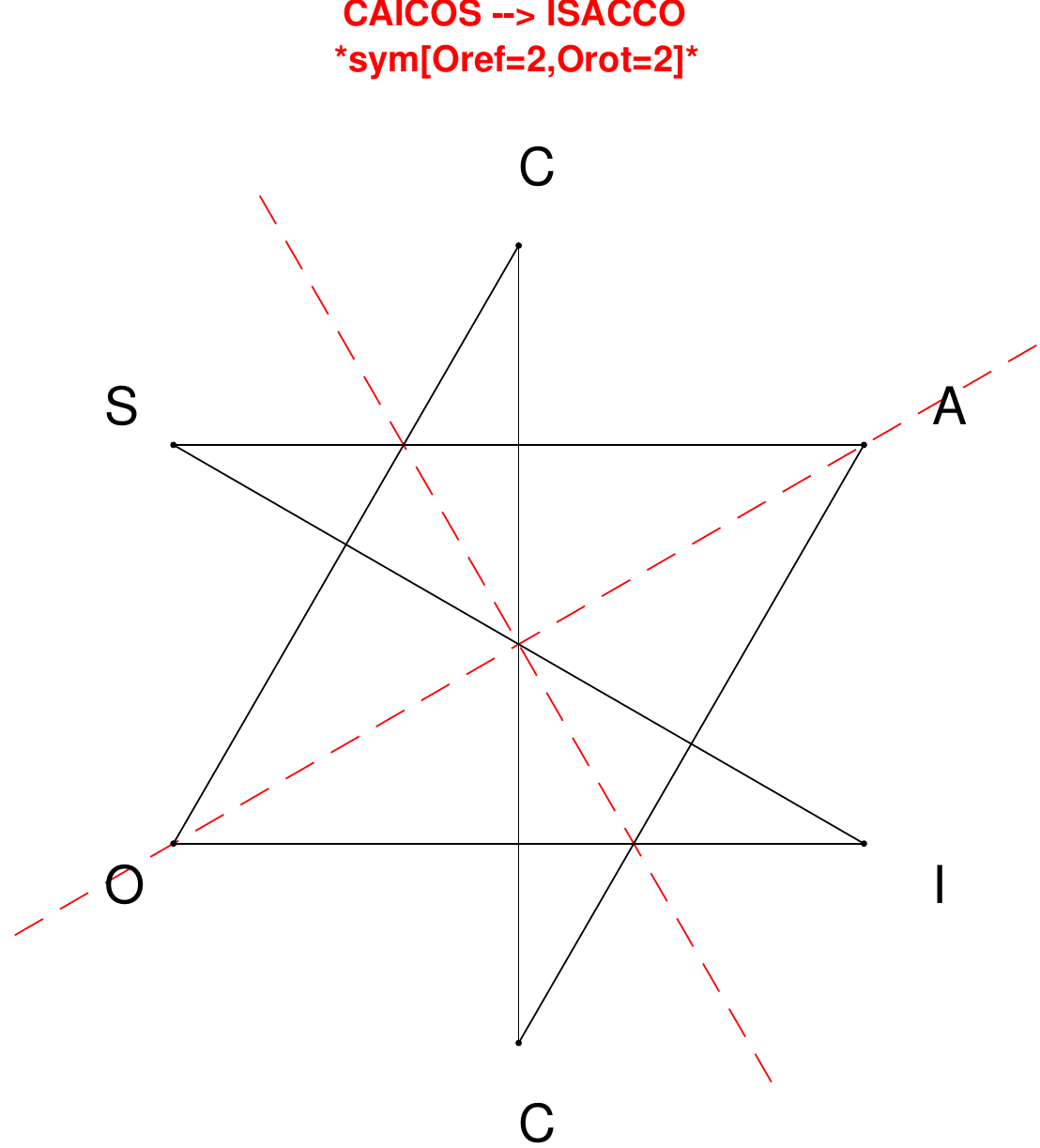}
\end{subfigure}
\hfill
\begin{subfigure}[T]{0.19\textwidth}
\centering
\includegraphics[width=\textwidth]{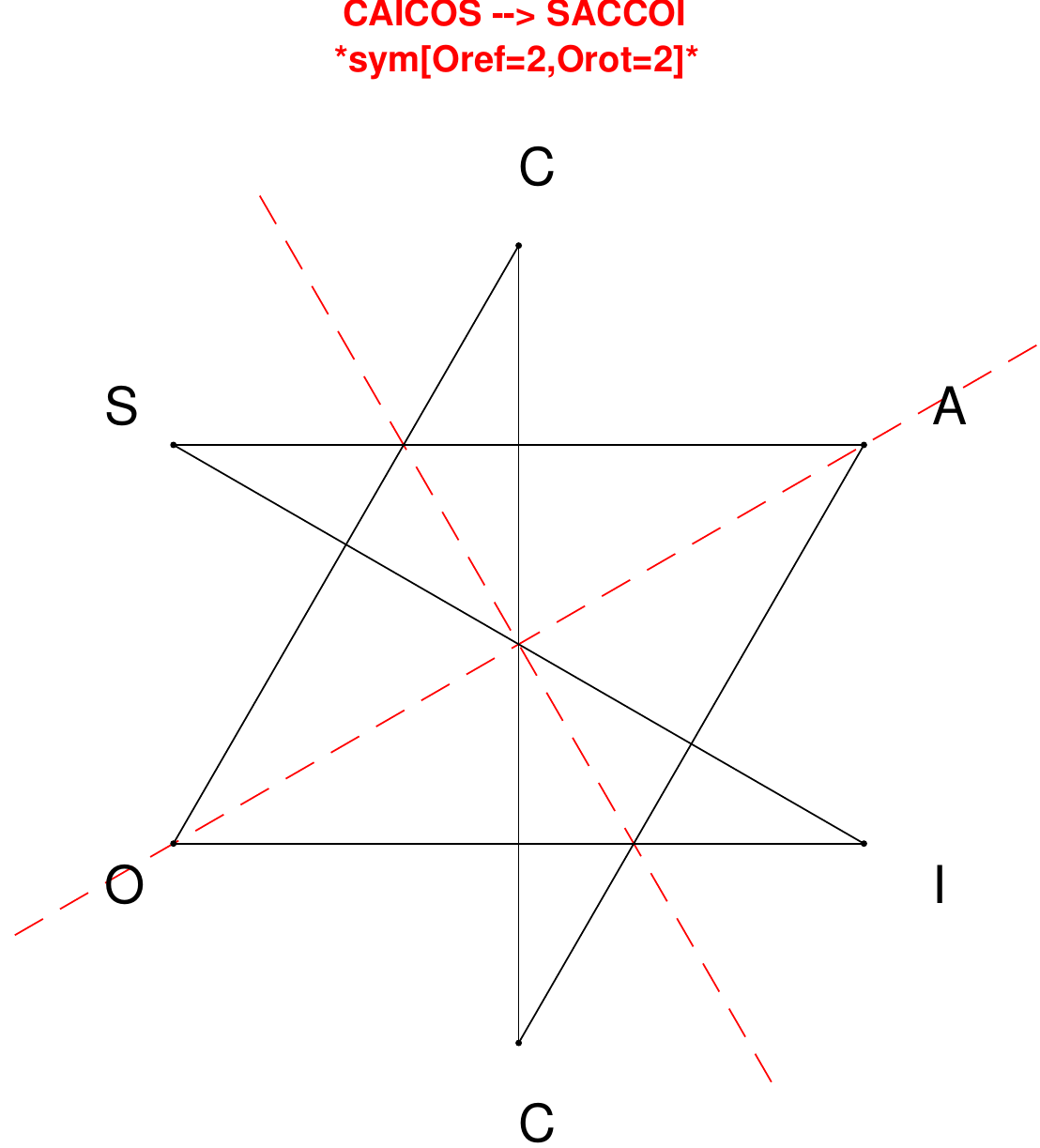}
\end{subfigure}
\hfill
\begin{subfigure}[T]{0.19\textwidth}
\centering
\includegraphics[width=\textwidth]{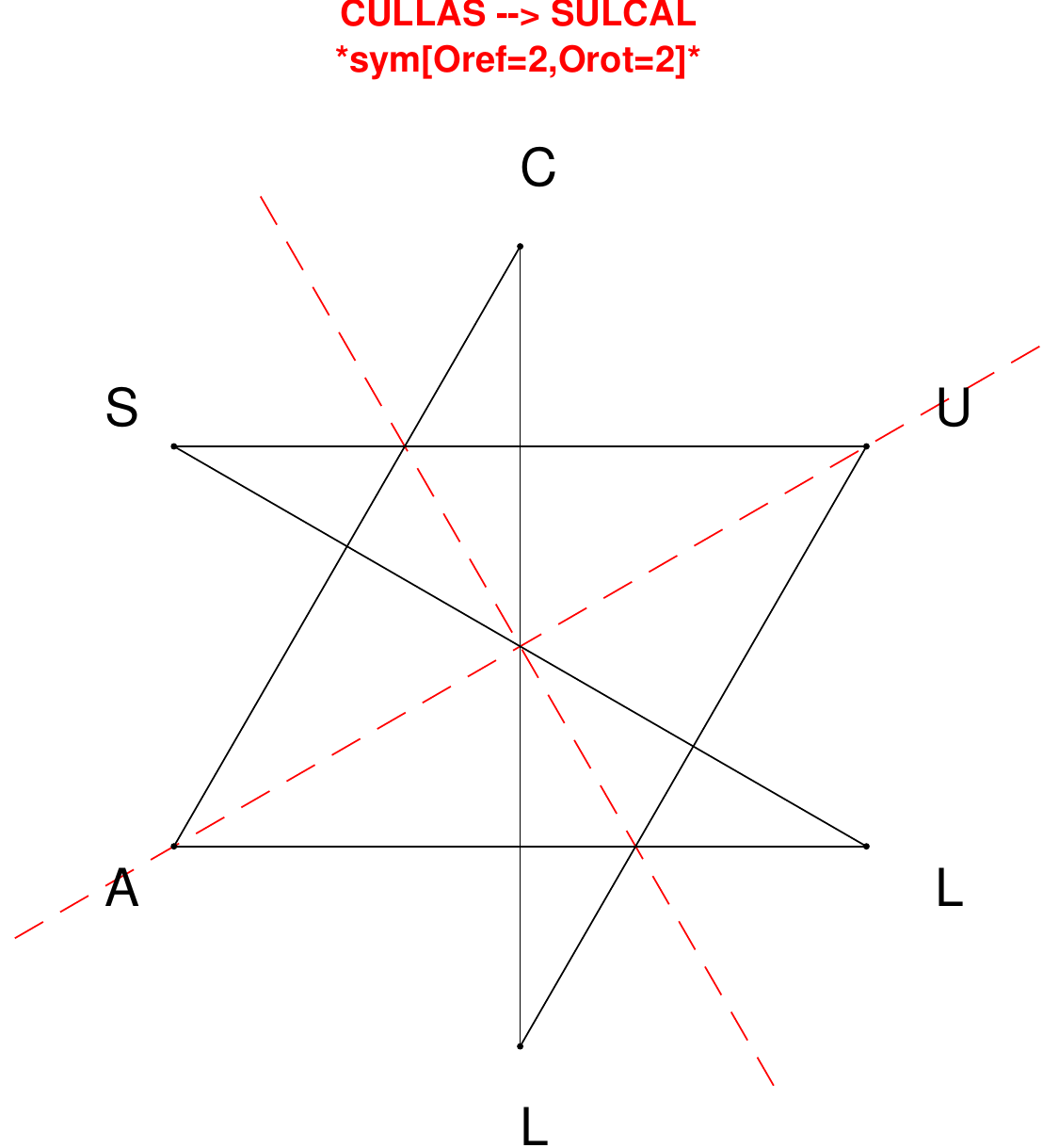}
\end{subfigure}
\end{figure}

\begin{figure}[H]
\centering
\begin{subfigure}[T]{0.19\textwidth}
\centering
\includegraphics[width=\textwidth]{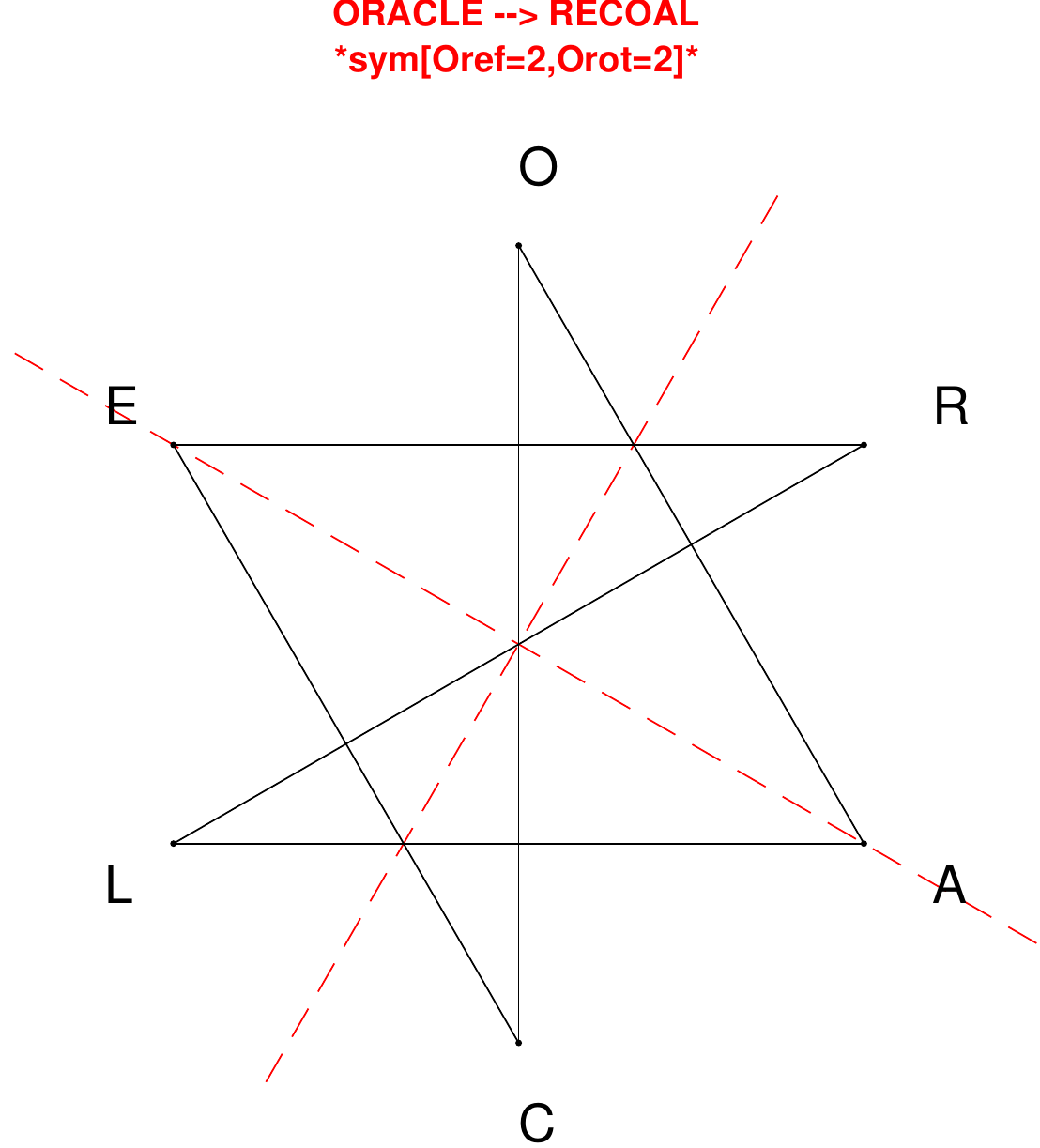}
\end{subfigure}
\hfill
\begin{subfigure}[T]{0.19\textwidth}
\centering
\includegraphics[width=\textwidth]{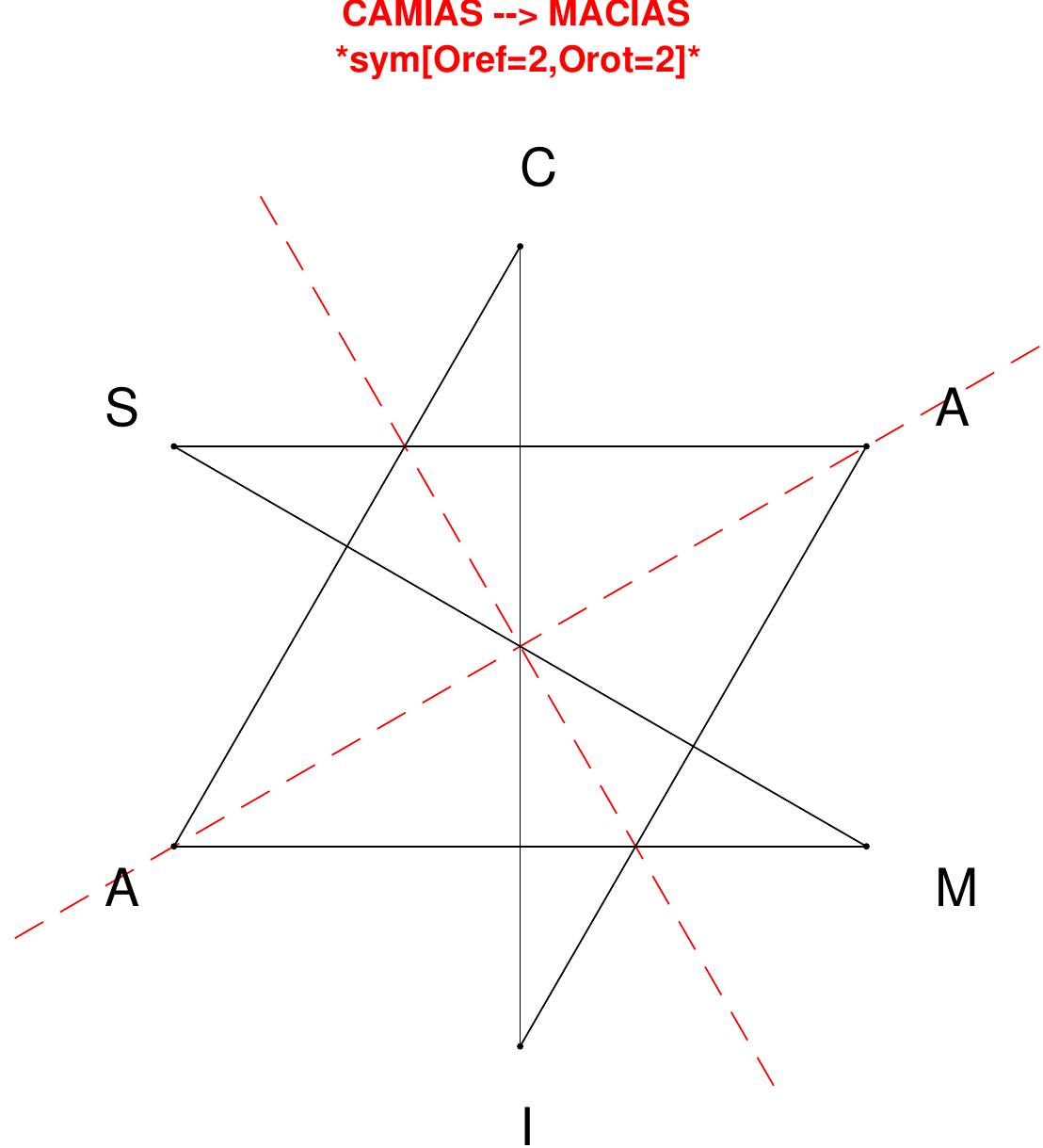}
\end{subfigure}
\hfill
\begin{subfigure}[T]{0.19\textwidth}
\centering
\includegraphics[width=\textwidth]{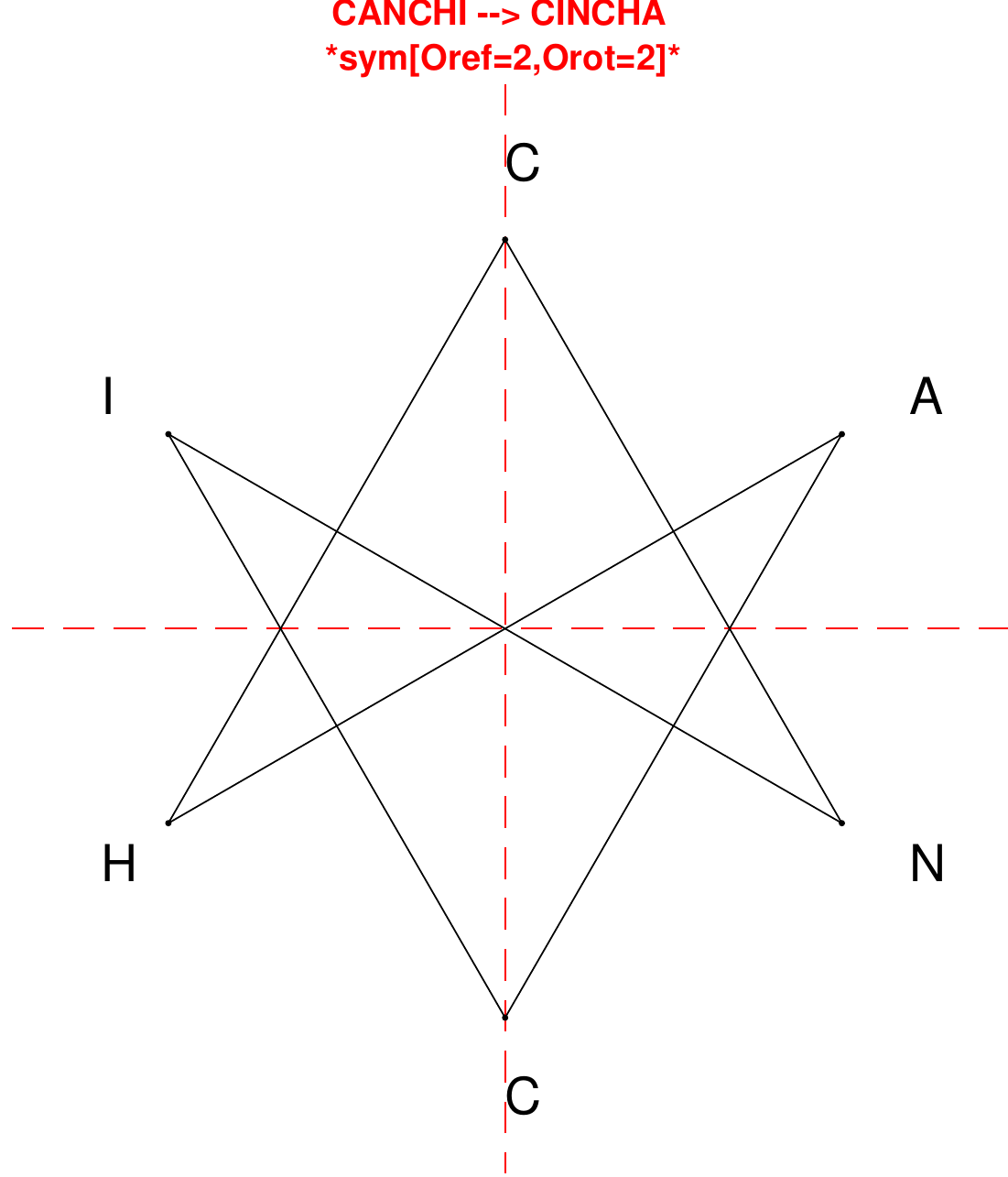}
\end{subfigure}
\hfill
\begin{subfigure}[T]{0.19\textwidth}
\centering
\includegraphics[width=\textwidth]{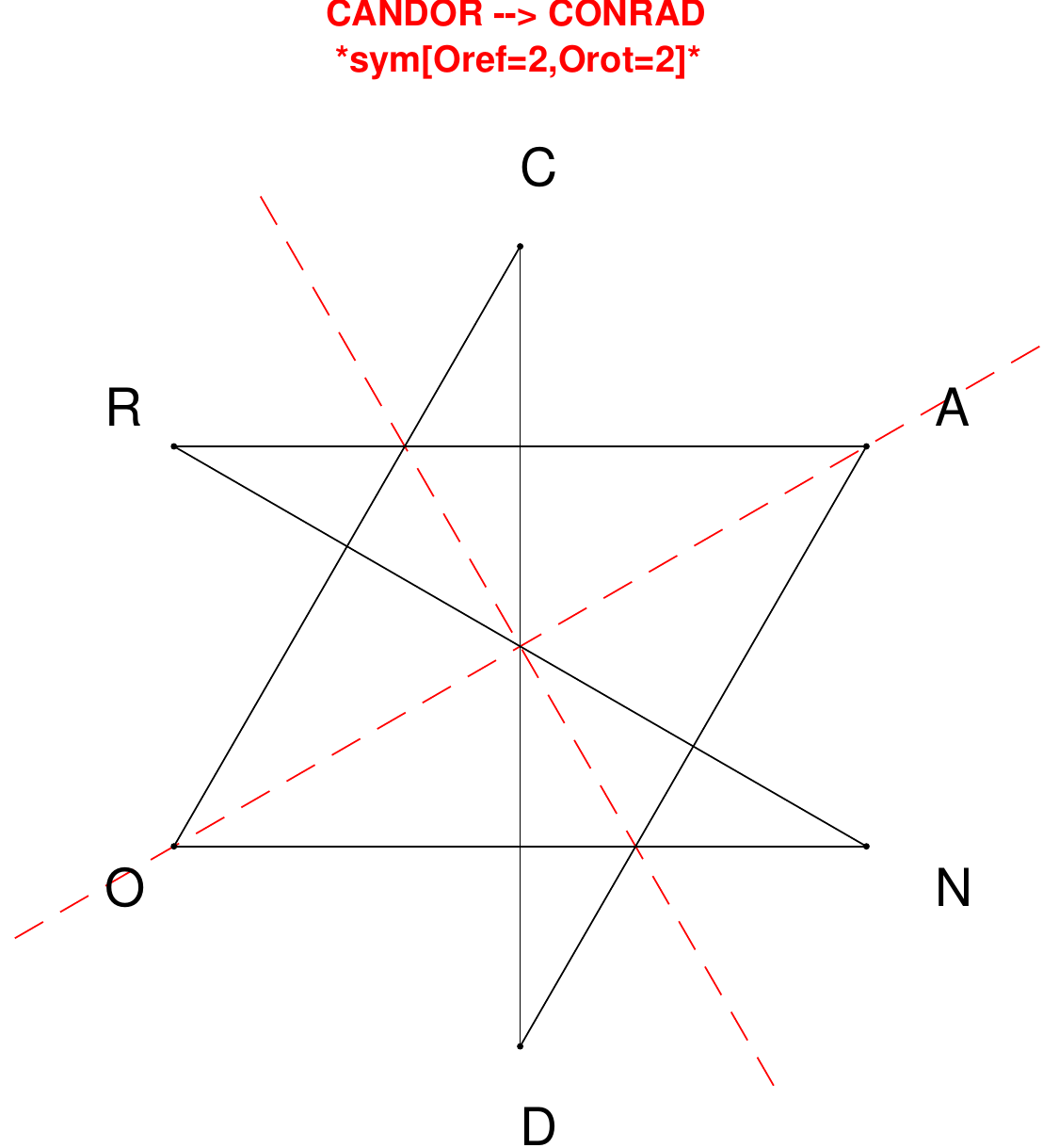}
\end{subfigure}
\hfill
\begin{subfigure}[T]{0.19\textwidth}
\centering
\includegraphics[width=\textwidth]{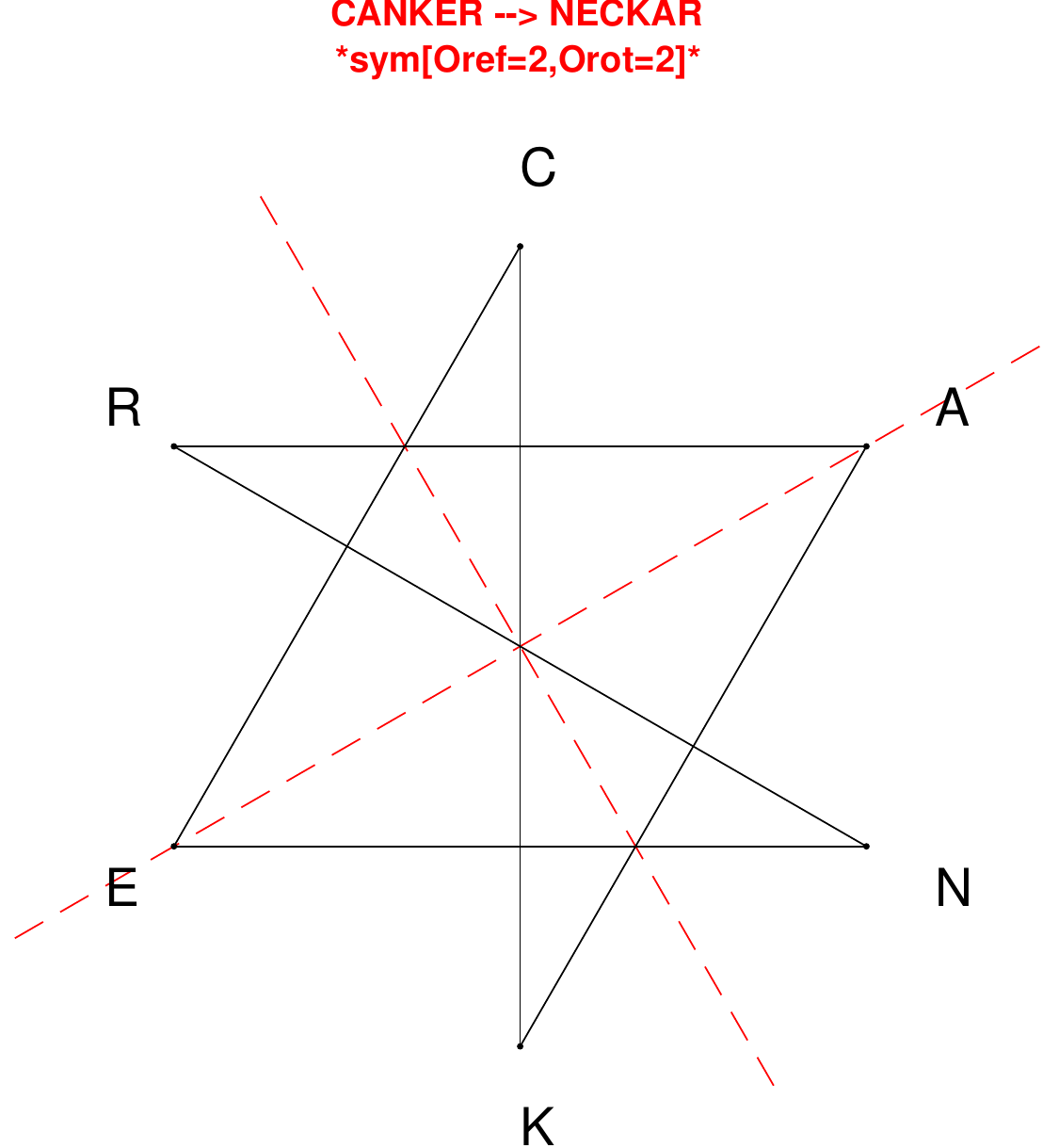}
\end{subfigure}
\end{figure}

\begin{figure}[H]
\centering
\begin{subfigure}[T]{0.19\textwidth}
\centering
\includegraphics[width=\textwidth]{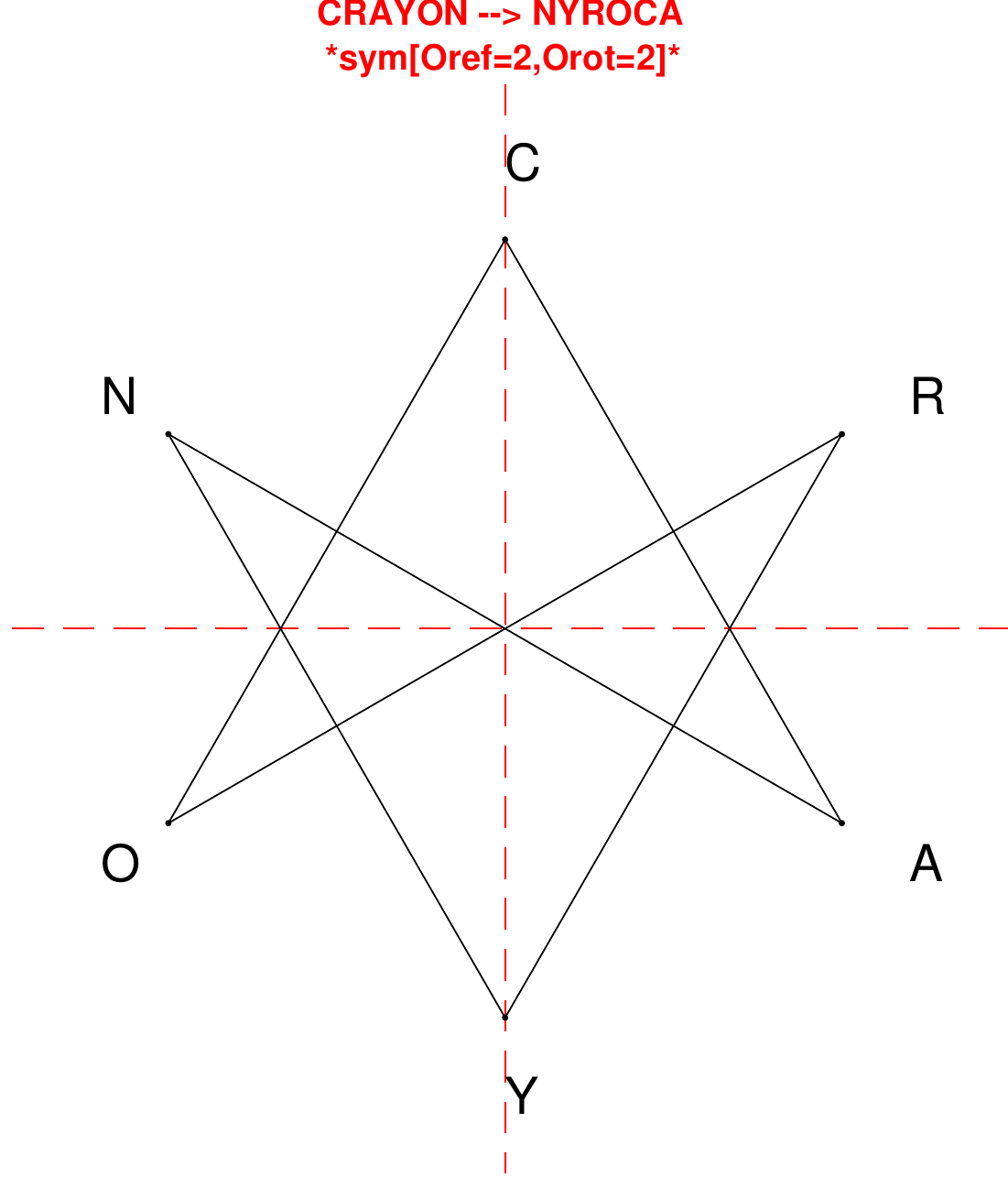}
\end{subfigure}
\hfill
\begin{subfigure}[T]{0.19\textwidth}
\centering
\includegraphics[width=\textwidth]{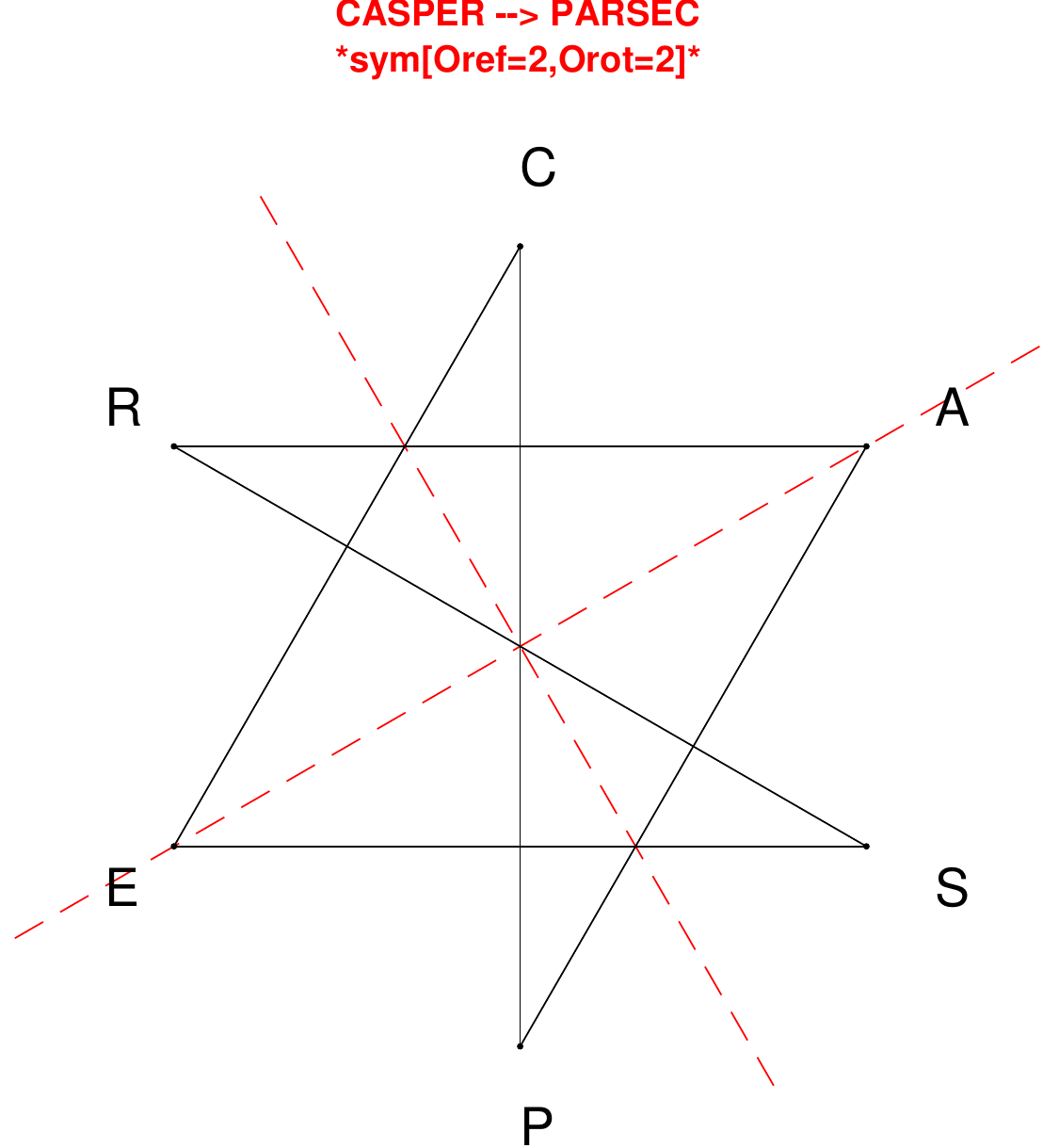}
\end{subfigure}
\hfill
\begin{subfigure}[T]{0.19\textwidth}
\centering
\includegraphics[width=\textwidth]{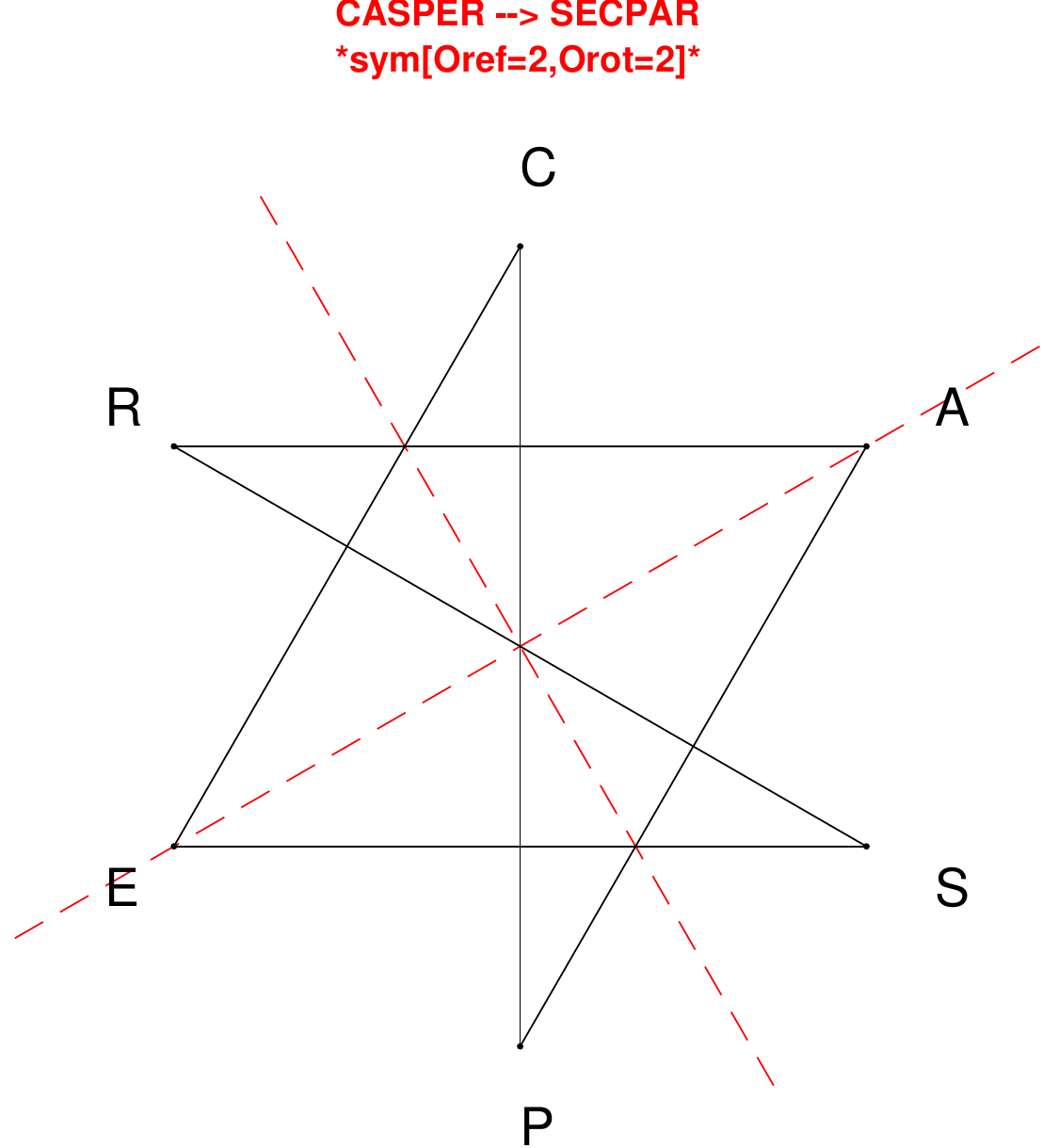}
\end{subfigure}
\hfill
\begin{subfigure}[T]{0.19\textwidth}
\centering
\includegraphics[width=\textwidth]{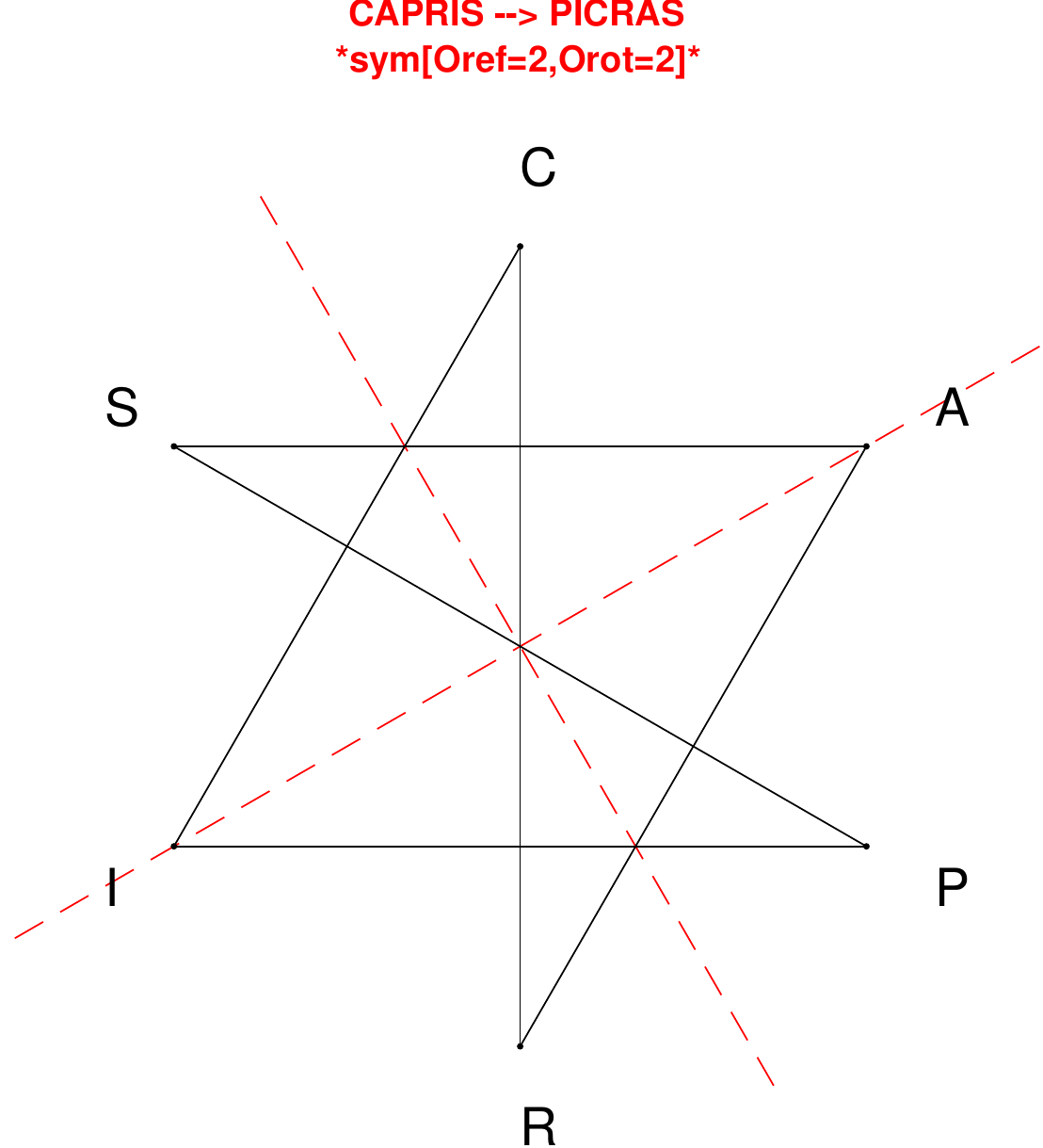}
\end{subfigure}
\hfill
\begin{subfigure}[T]{0.19\textwidth}
\centering
\includegraphics[width=\textwidth]{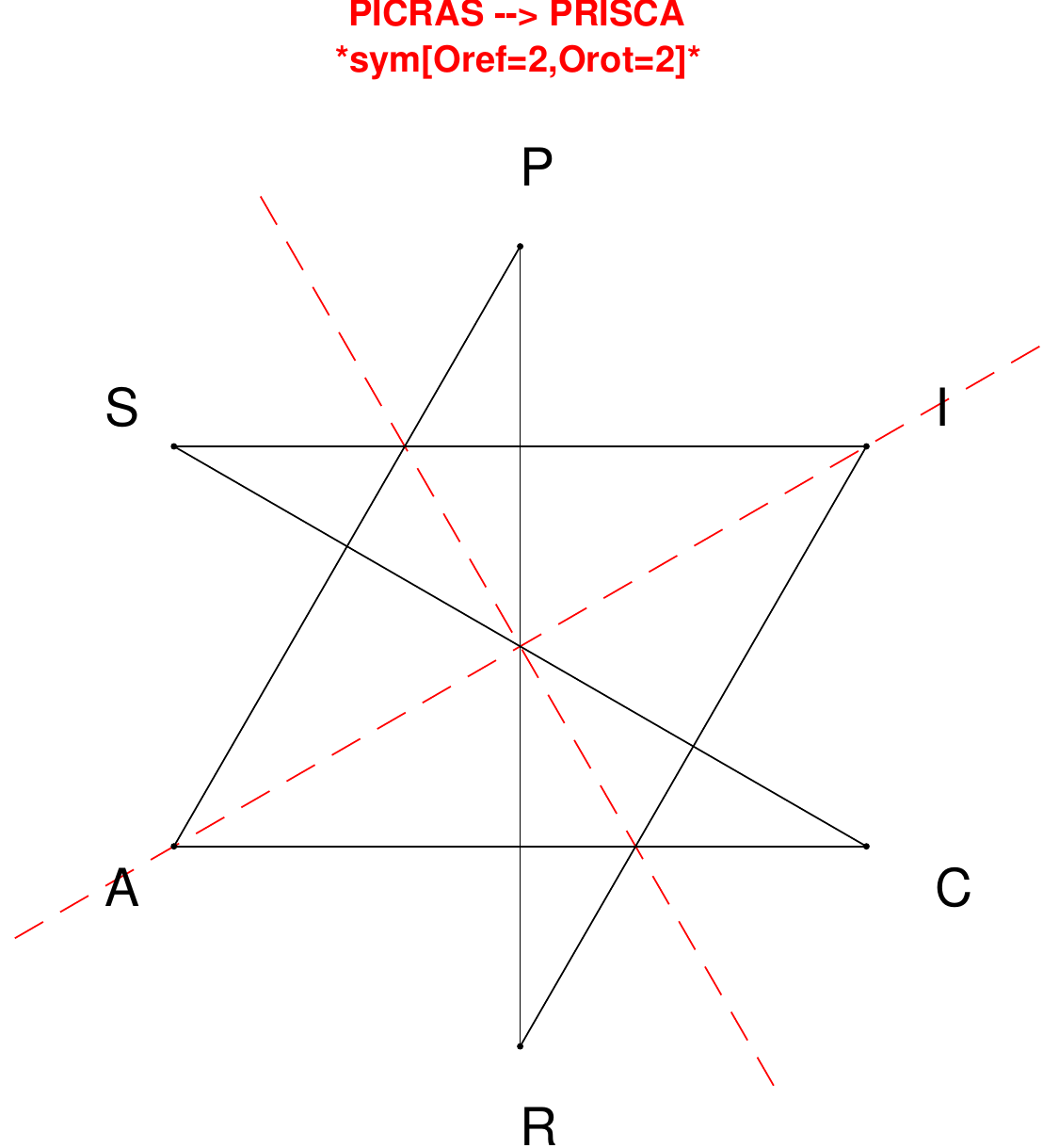}
\end{subfigure}
\end{figure}

\begin{figure}[H]
\centering
\begin{subfigure}[T]{0.19\textwidth}
\centering
\includegraphics[width=\textwidth]{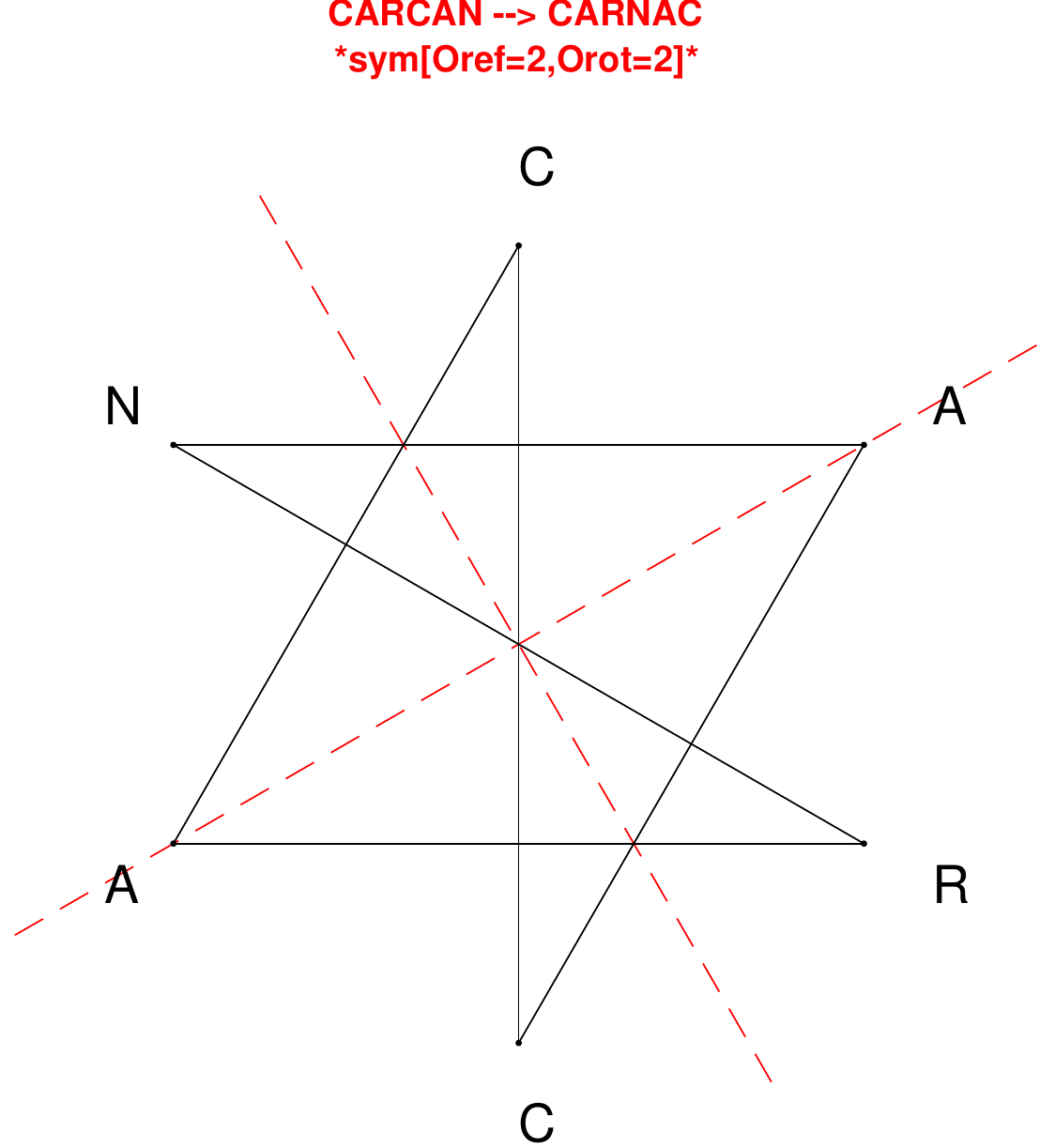}
\end{subfigure}
\hfill
\begin{subfigure}[T]{0.19\textwidth}
\centering
\includegraphics[width=\textwidth]{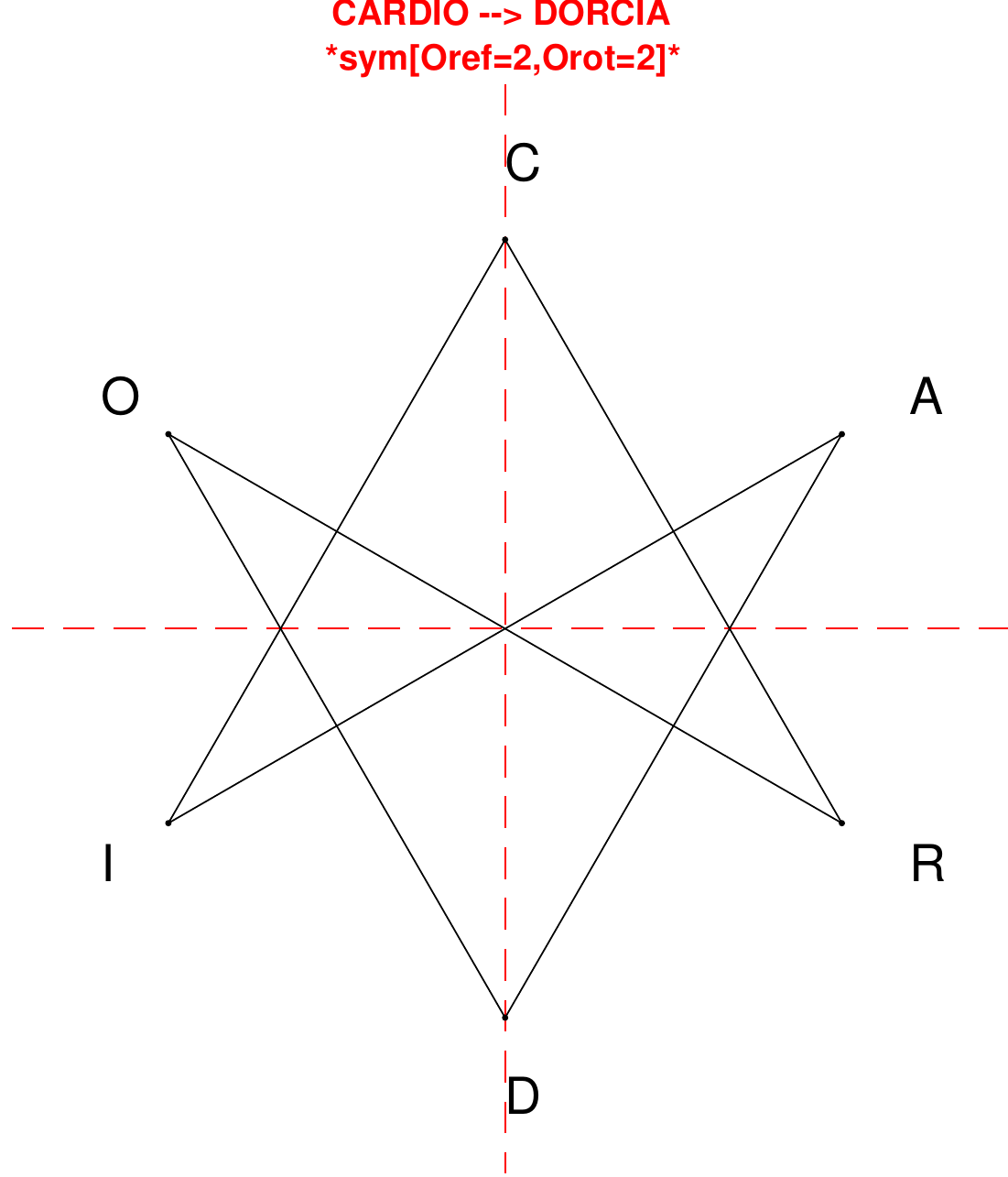}
\end{subfigure}
\hfill
\begin{subfigure}[T]{0.19\textwidth}
\centering
\includegraphics[width=\textwidth]{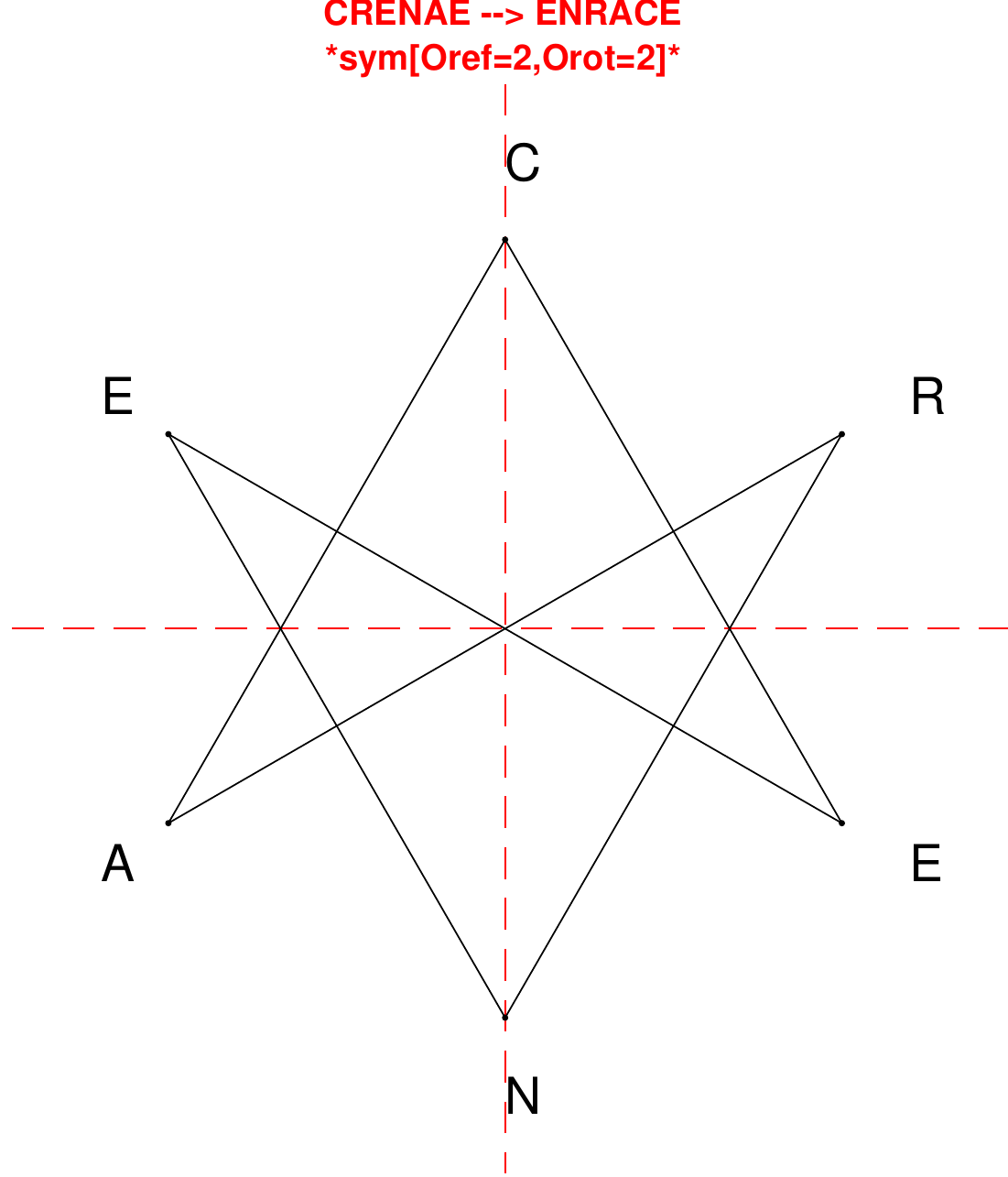}
\end{subfigure}
\hfill
\begin{subfigure}[T]{0.19\textwidth}
\centering
\includegraphics[width=\textwidth]{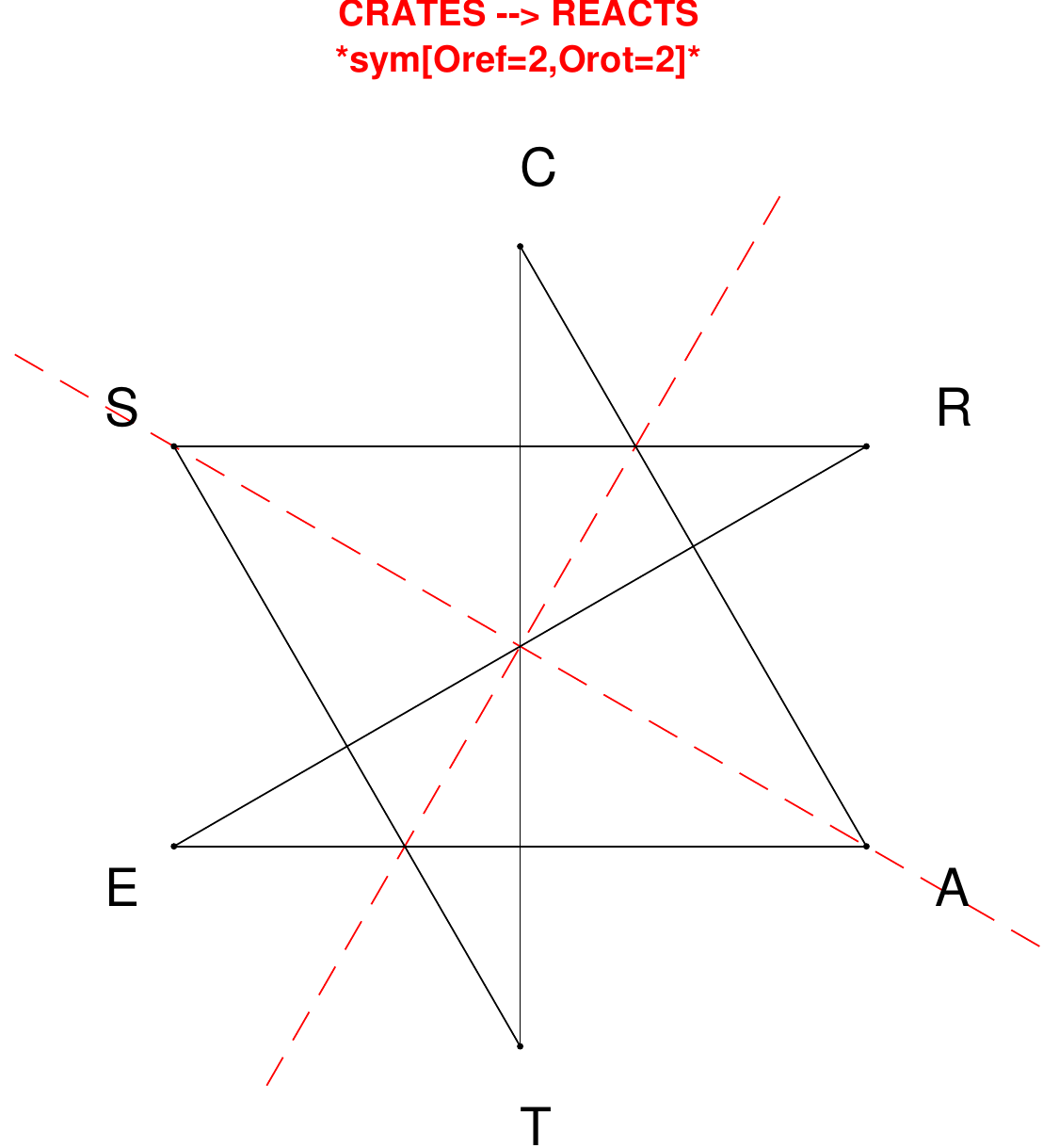}
\end{subfigure}
\hfill
\begin{subfigure}[T]{0.19\textwidth}
\centering
\includegraphics[width=\textwidth]{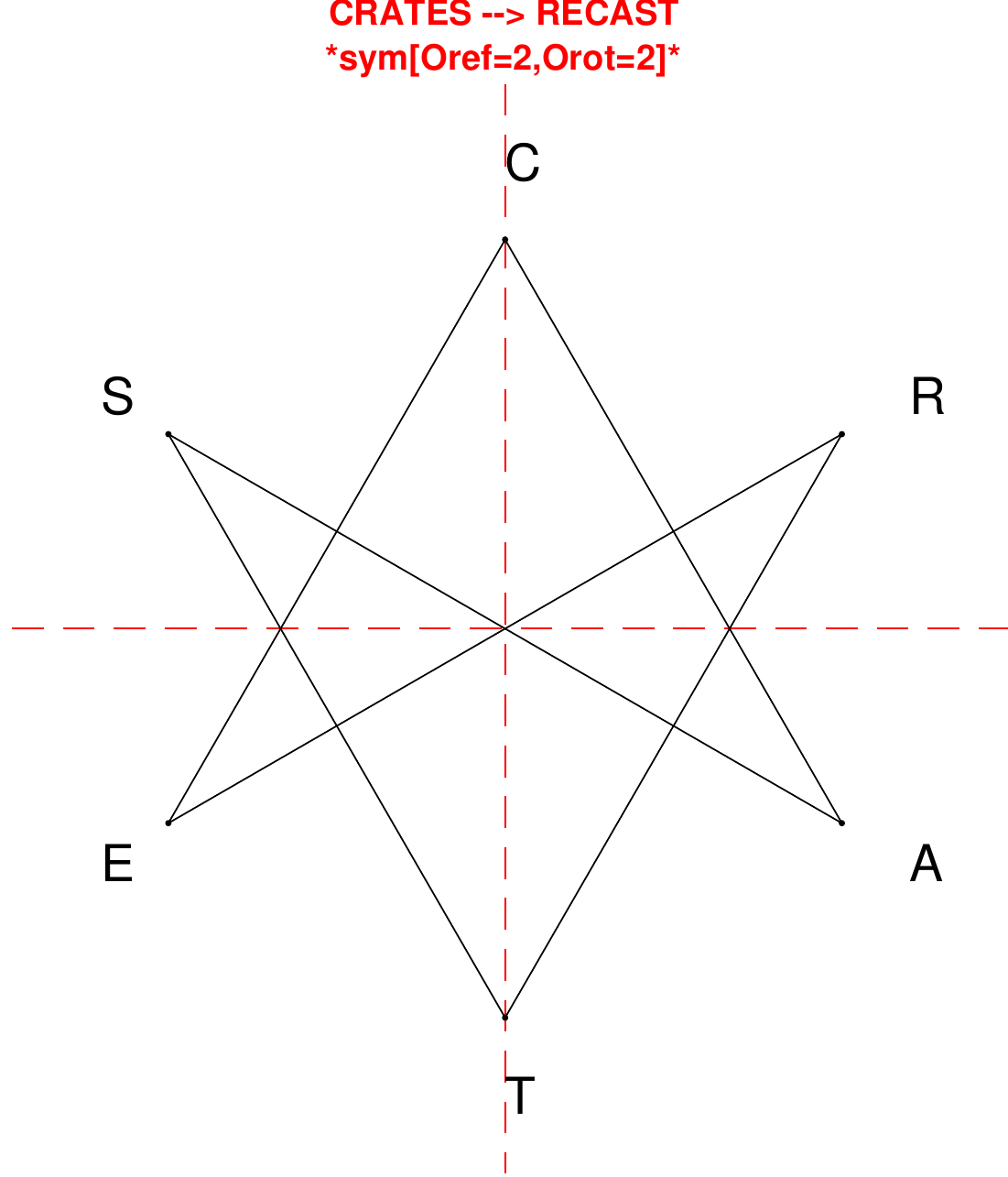}
\end{subfigure}
\end{figure}

\begin{figure}[H]
\centering
\begin{subfigure}[T]{0.19\textwidth}
\centering
\includegraphics[width=\textwidth]{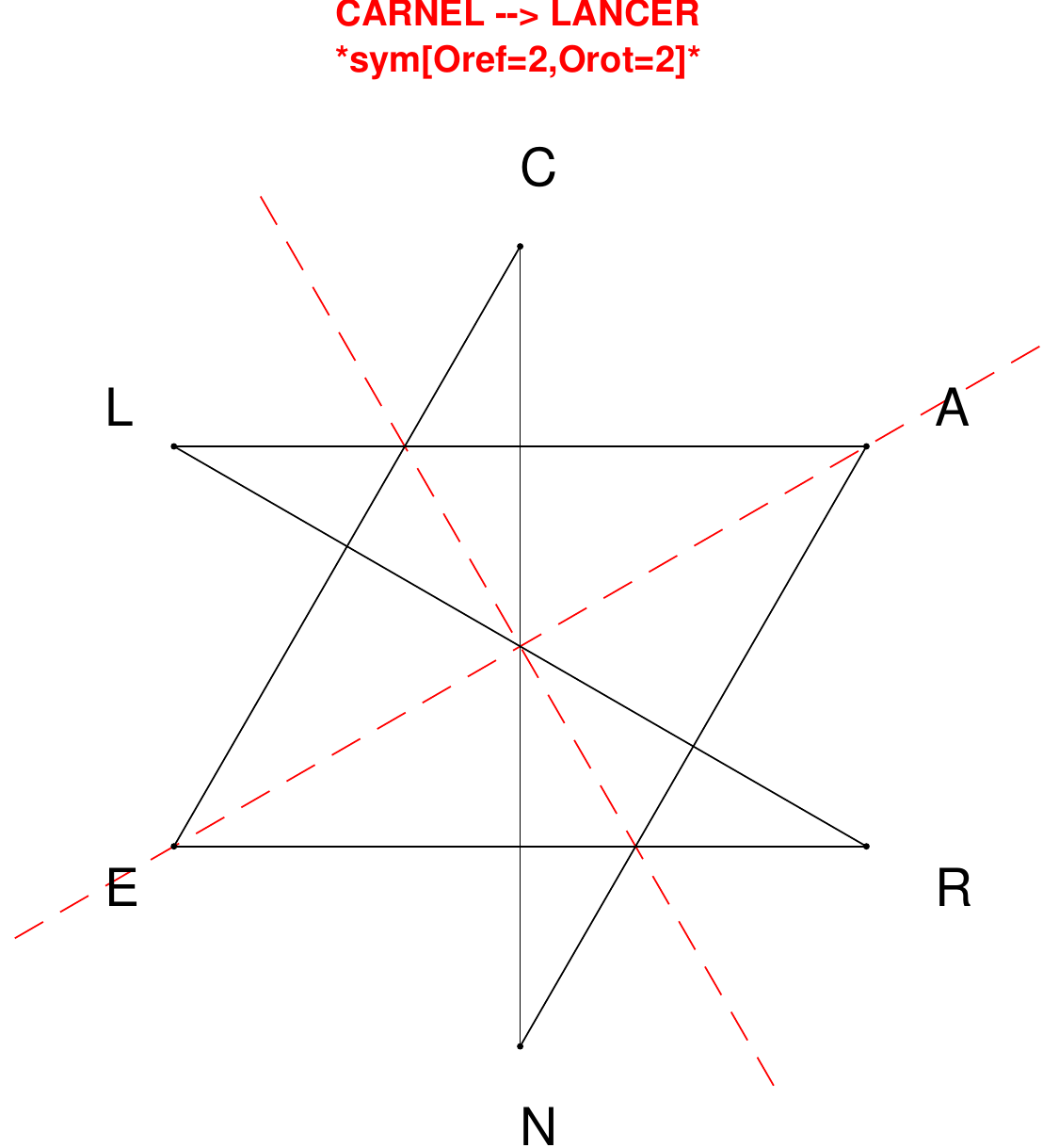}
\end{subfigure}
\hfill
\begin{subfigure}[T]{0.19\textwidth}
\centering
\includegraphics[width=\textwidth]{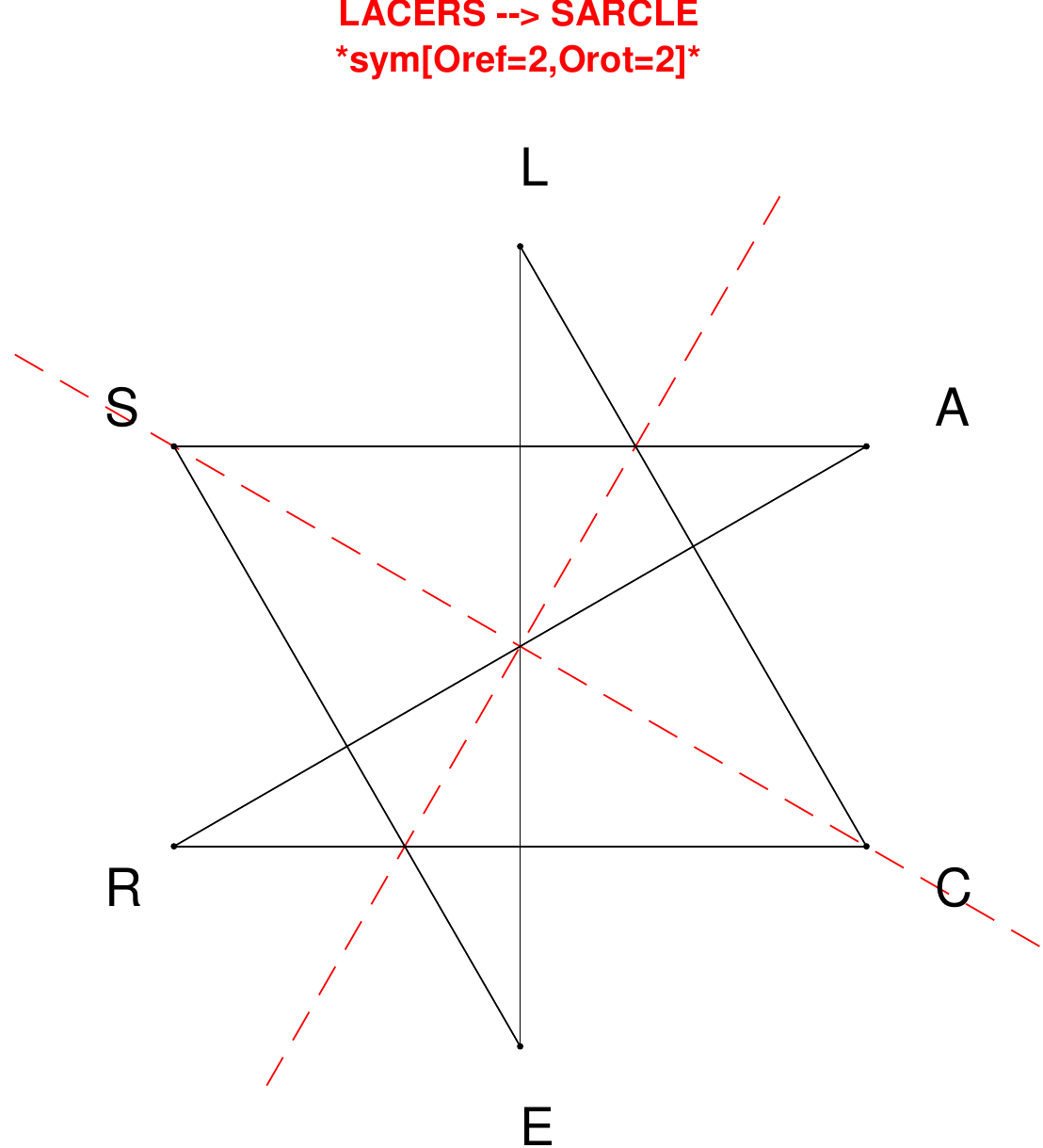}
\end{subfigure}
\hfill
\begin{subfigure}[T]{0.19\textwidth}
\centering
\includegraphics[width=\textwidth]{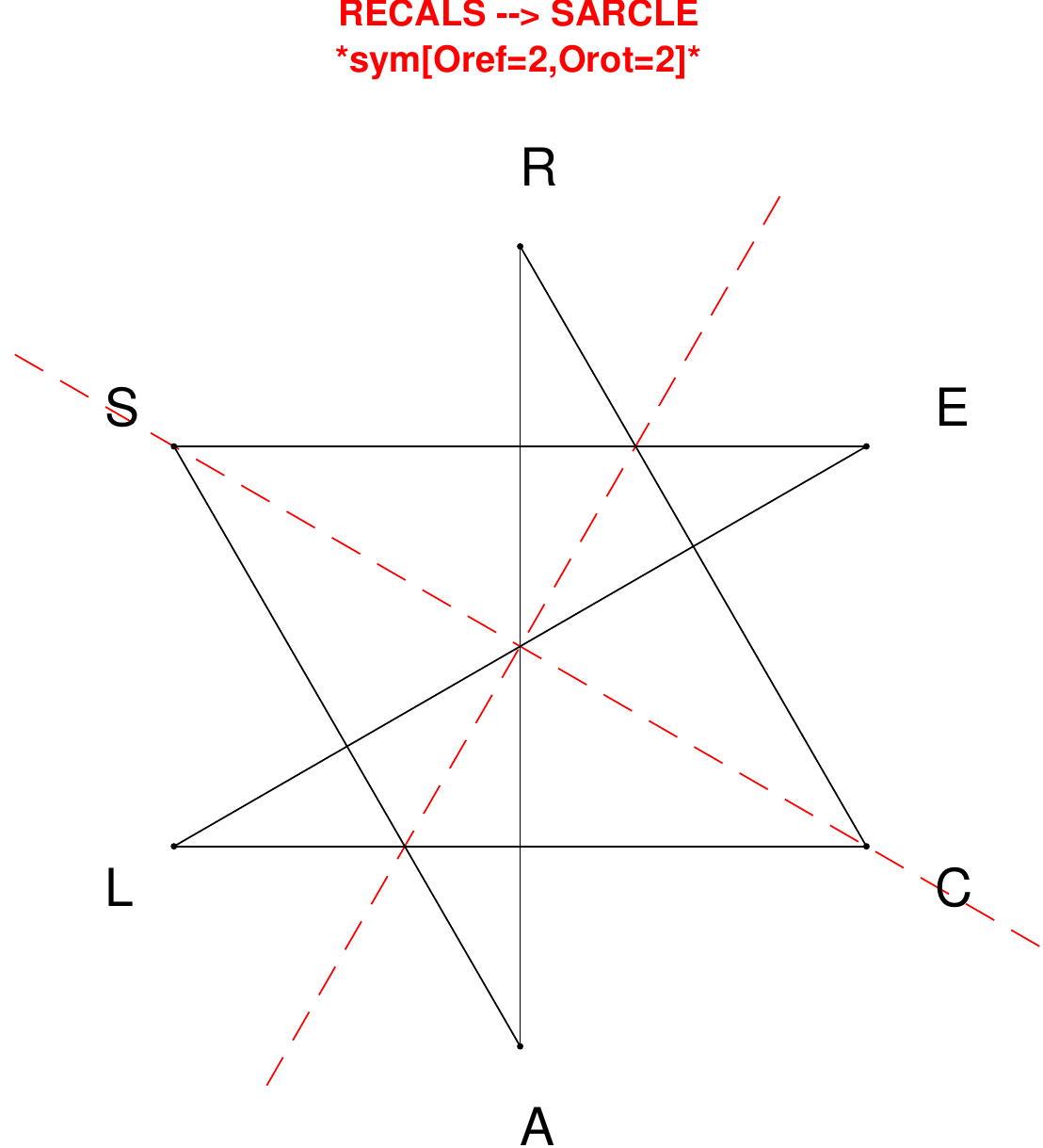}
\end{subfigure}
\hfill
\begin{subfigure}[T]{0.19\textwidth}
\centering
\includegraphics[width=\textwidth]{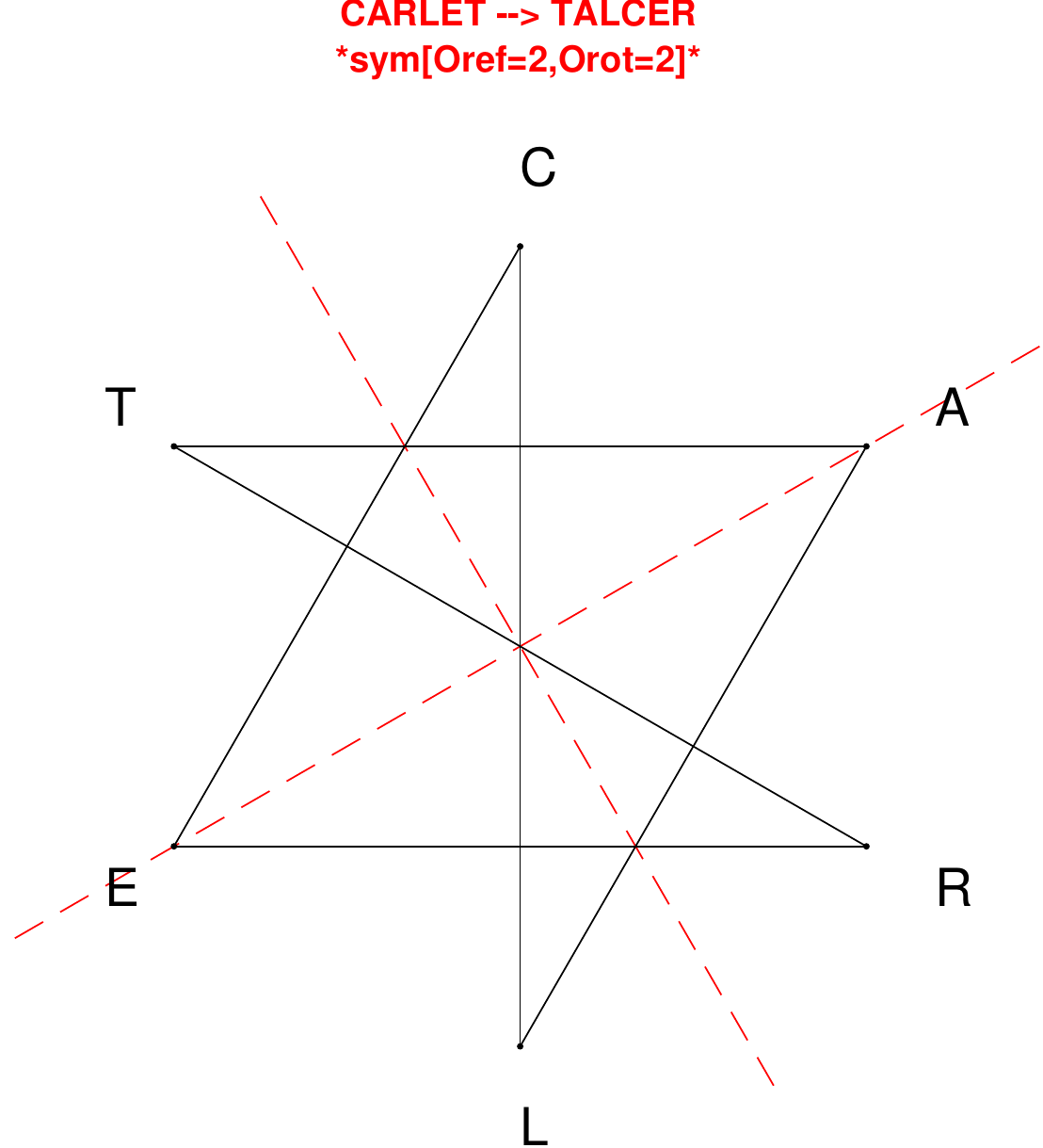}
\end{subfigure}
\hfill
\begin{subfigure}[T]{0.19\textwidth}
\centering
\includegraphics[width=\textwidth]{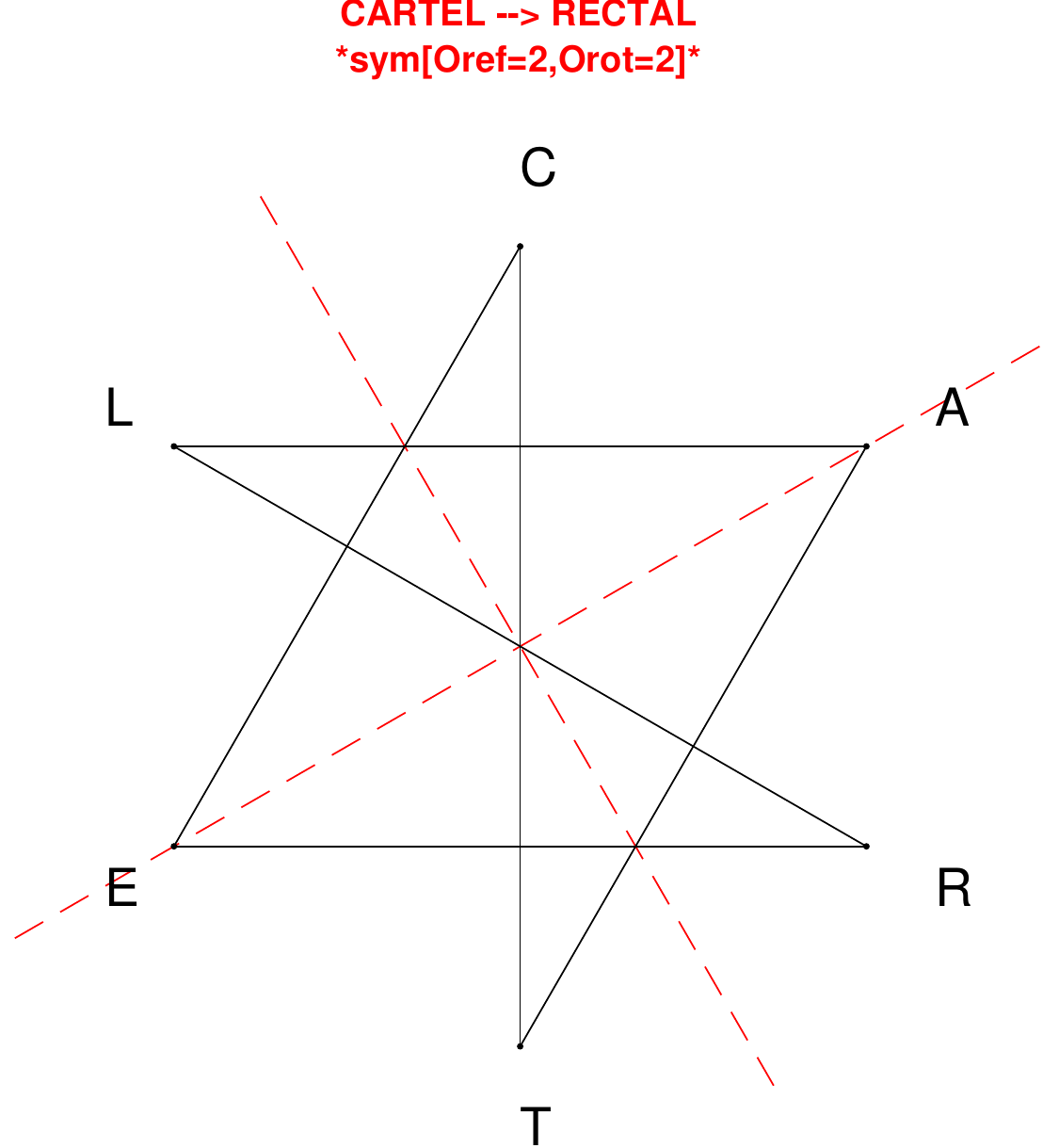}
\end{subfigure}
\end{figure}

\begin{figure}[H]
\centering
\begin{subfigure}[T]{0.19\textwidth}
\centering
\includegraphics[width=\textwidth]{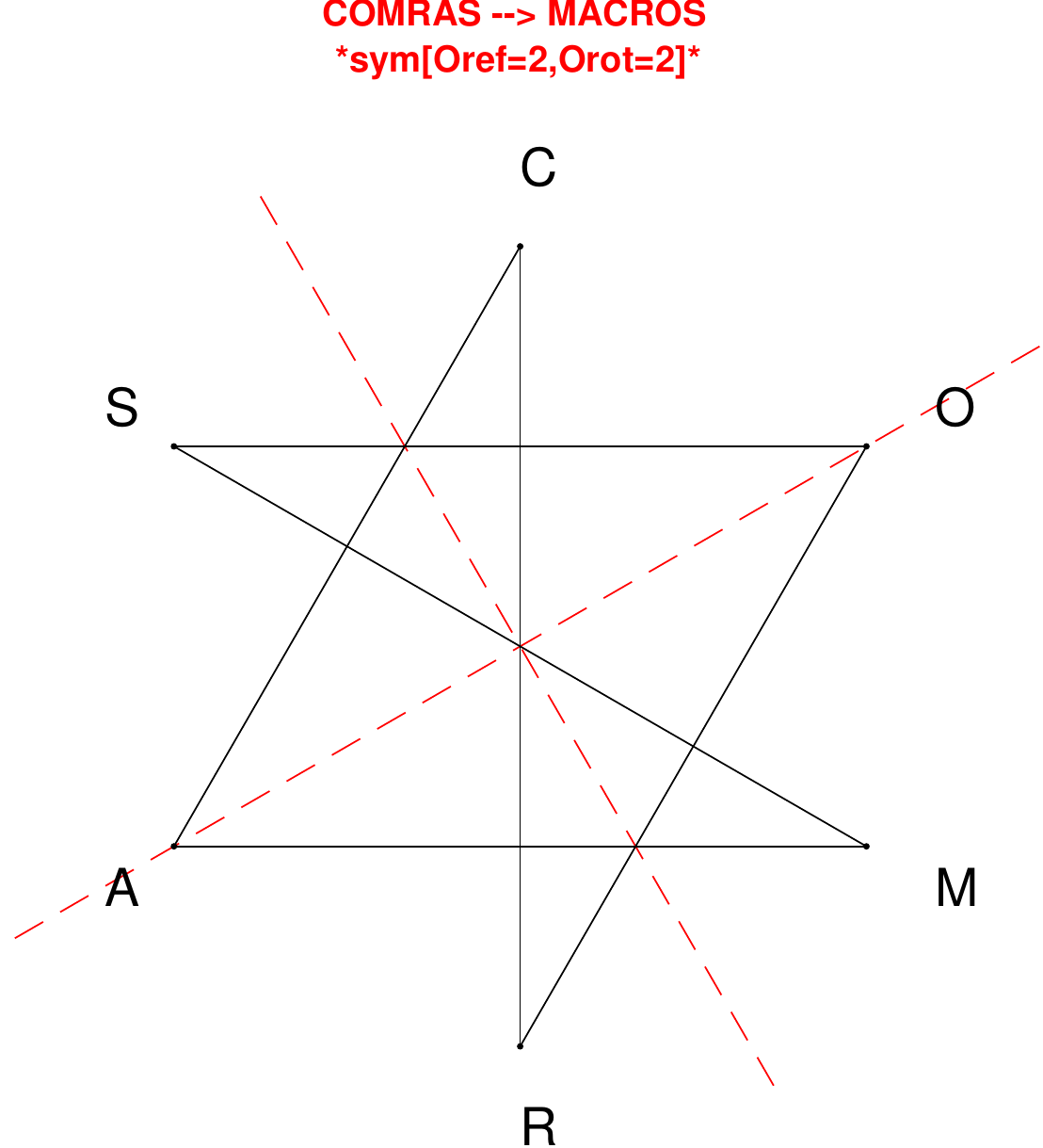}
\end{subfigure}
\hfill
\begin{subfigure}[T]{0.19\textwidth}
\centering
\includegraphics[width=\textwidth]{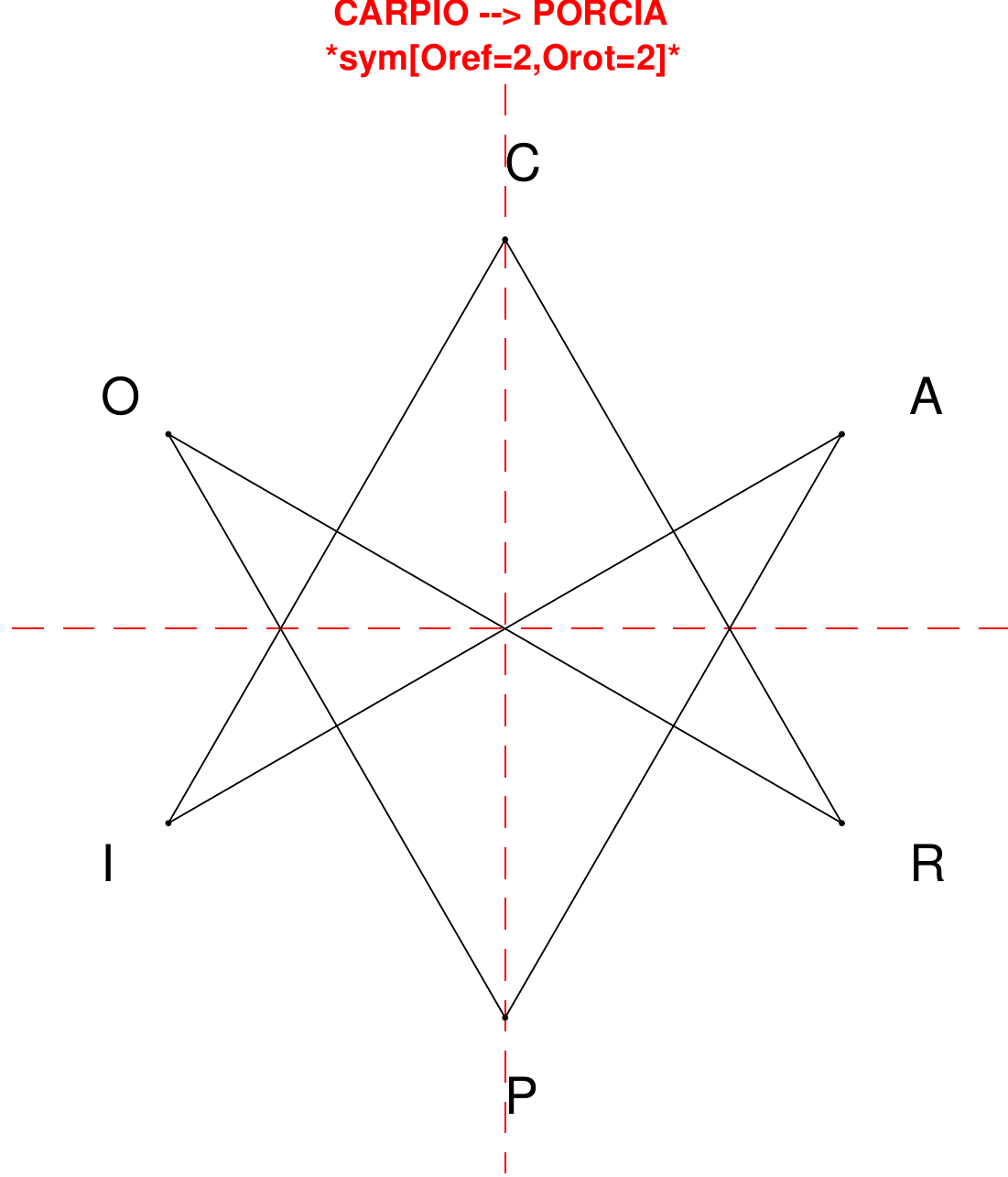}
\end{subfigure}
\hfill
\begin{subfigure}[T]{0.19\textwidth}
\centering
\includegraphics[width=\textwidth]{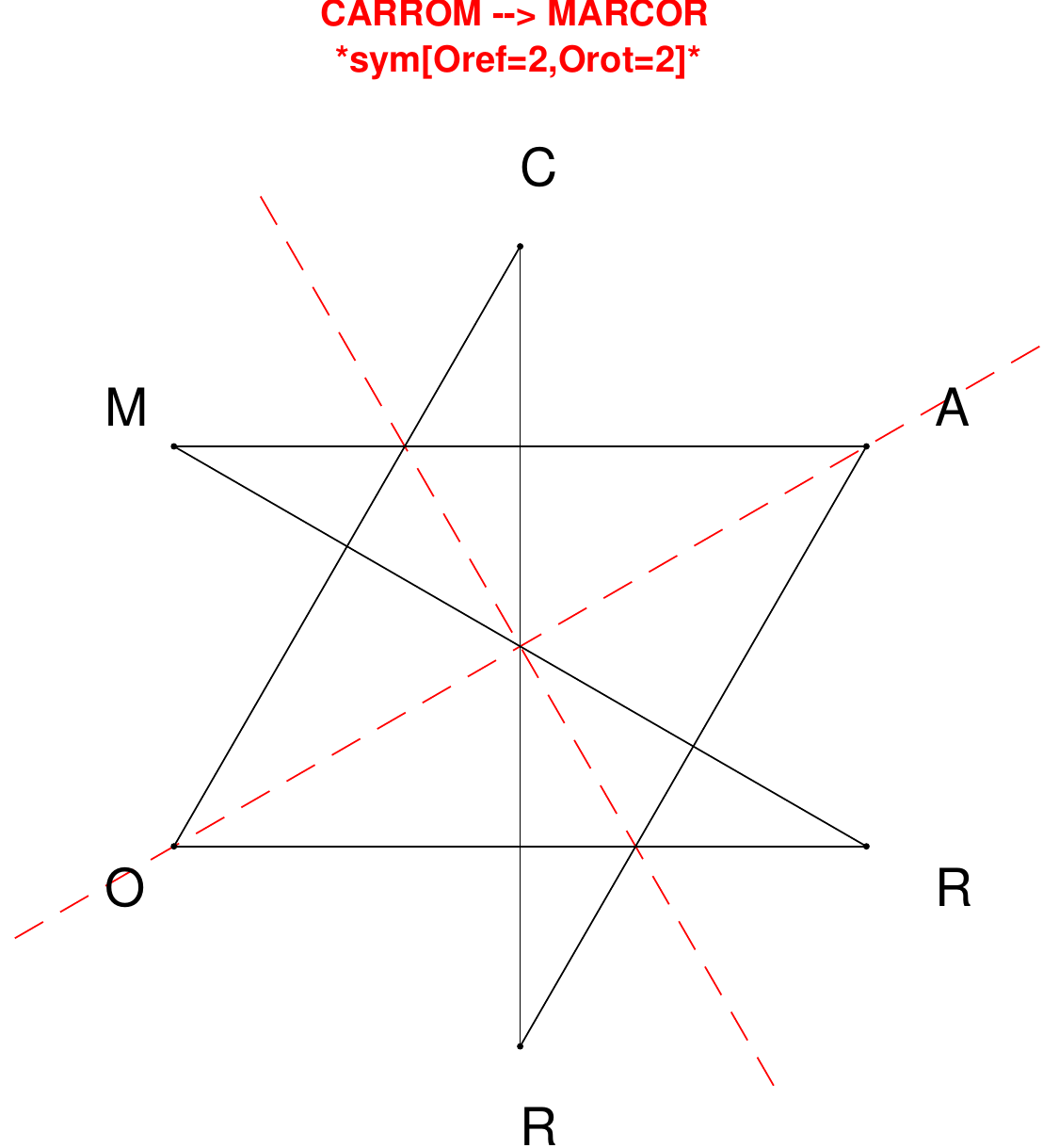}
\end{subfigure}
\hfill
\begin{subfigure}[T]{0.19\textwidth}
\centering
\includegraphics[width=\textwidth]{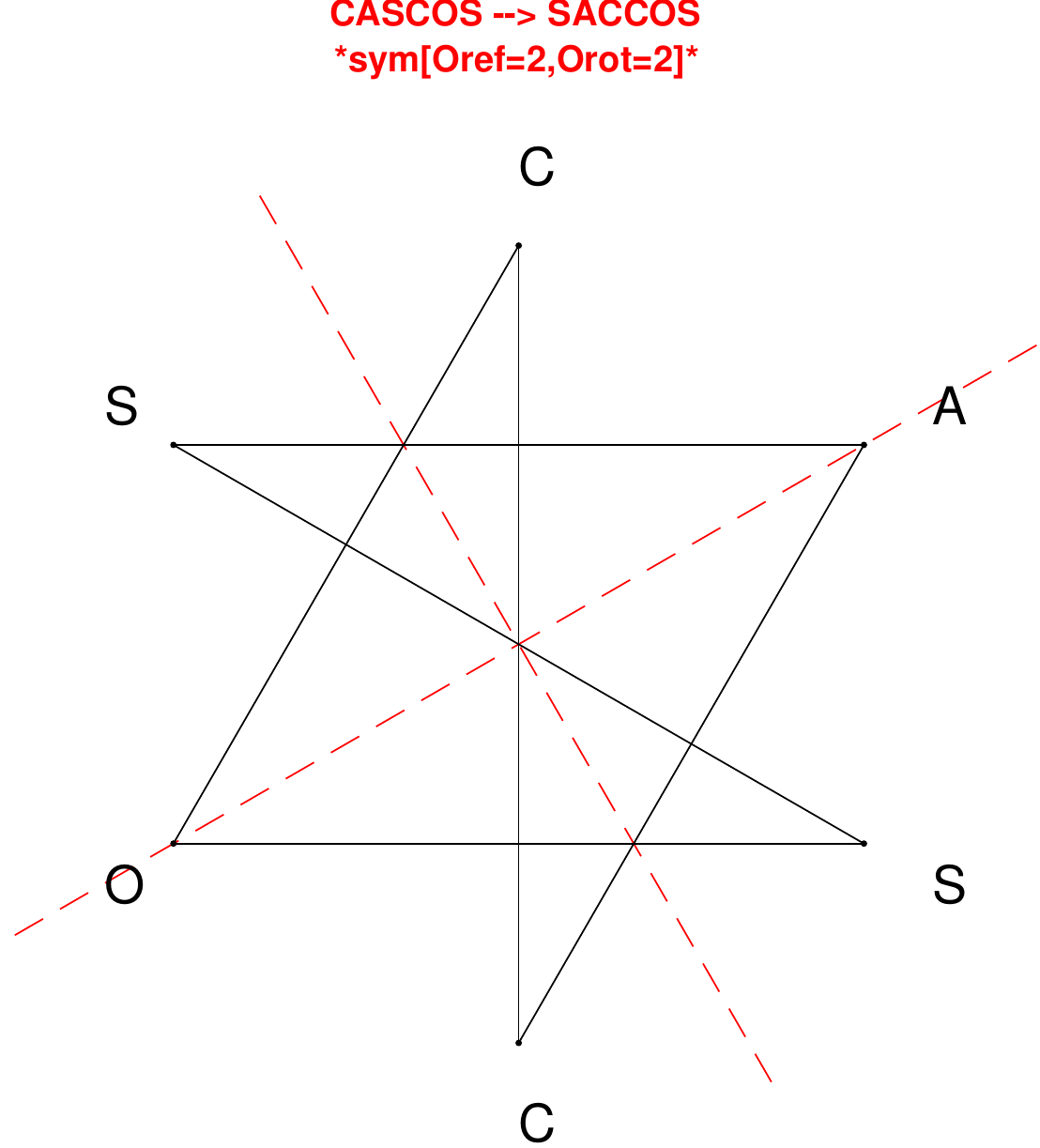}
\end{subfigure}
\hfill
\begin{subfigure}[T]{0.19\textwidth}
\centering
\includegraphics[width=\textwidth]{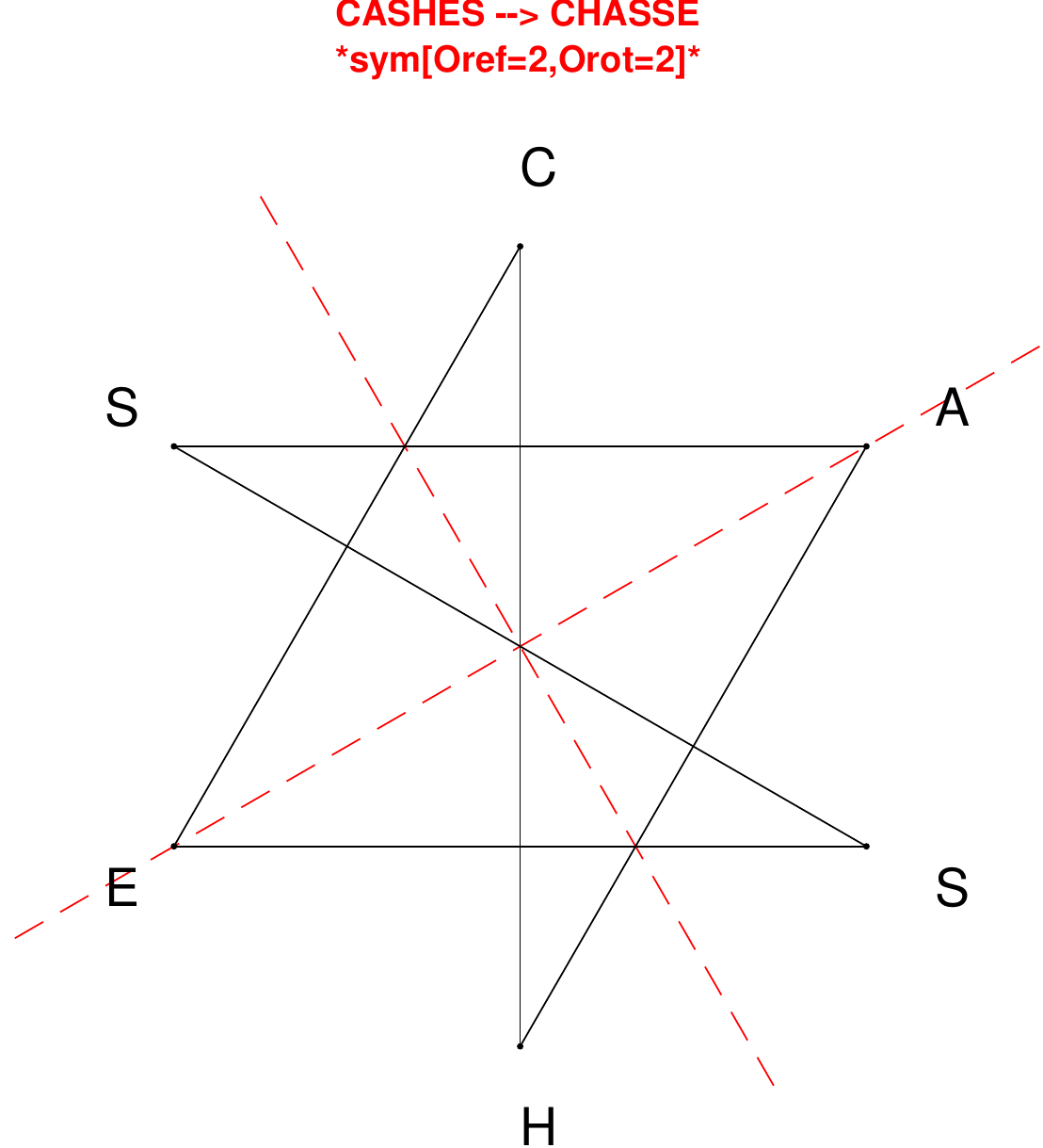}
\end{subfigure}
\end{figure}

\begin{figure}[H]
\centering
\begin{subfigure}[T]{0.19\textwidth}
\centering
\includegraphics[width=\textwidth]{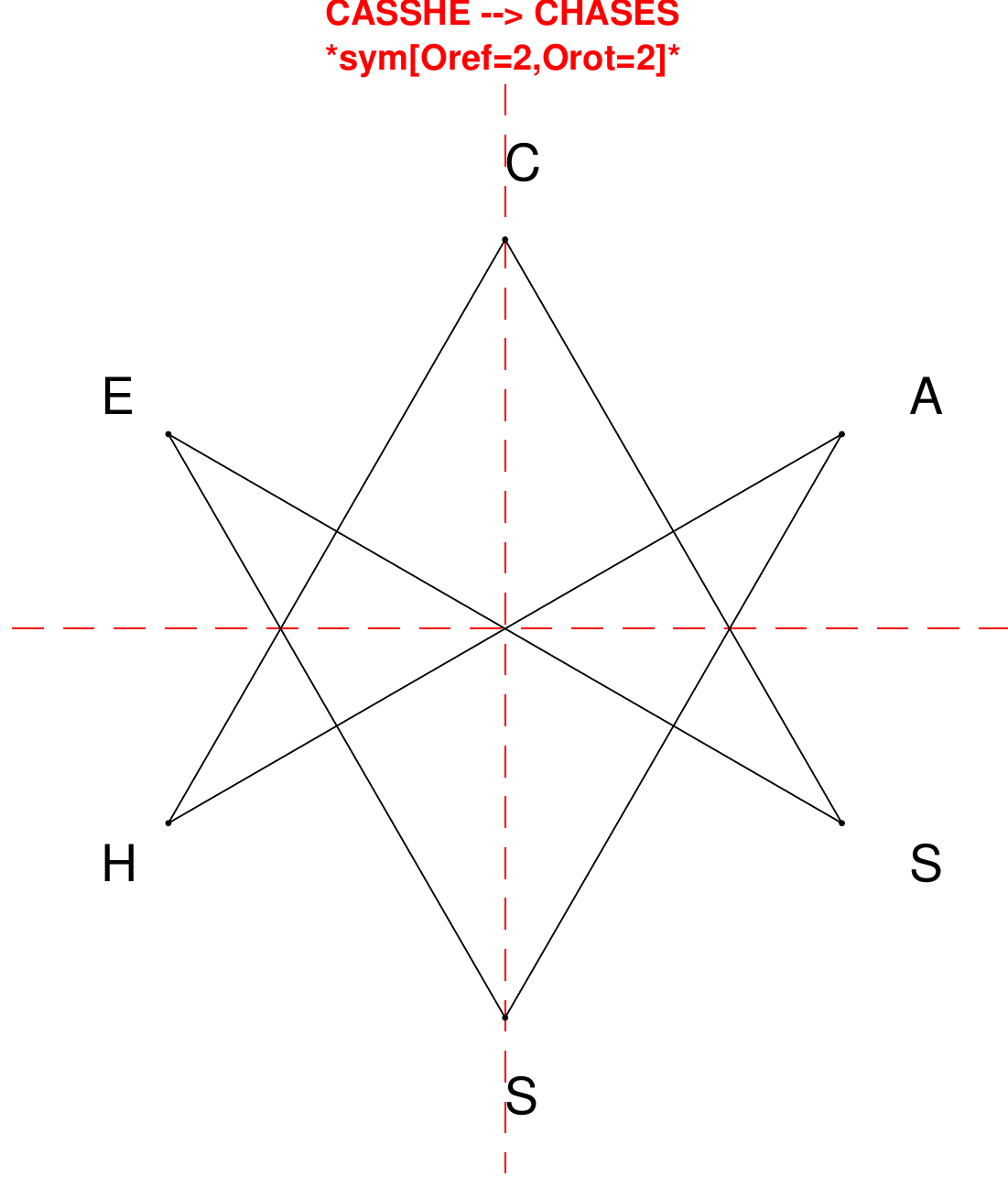}
\end{subfigure}
\hfill
\begin{subfigure}[T]{0.19\textwidth}
\centering
\includegraphics[width=\textwidth]{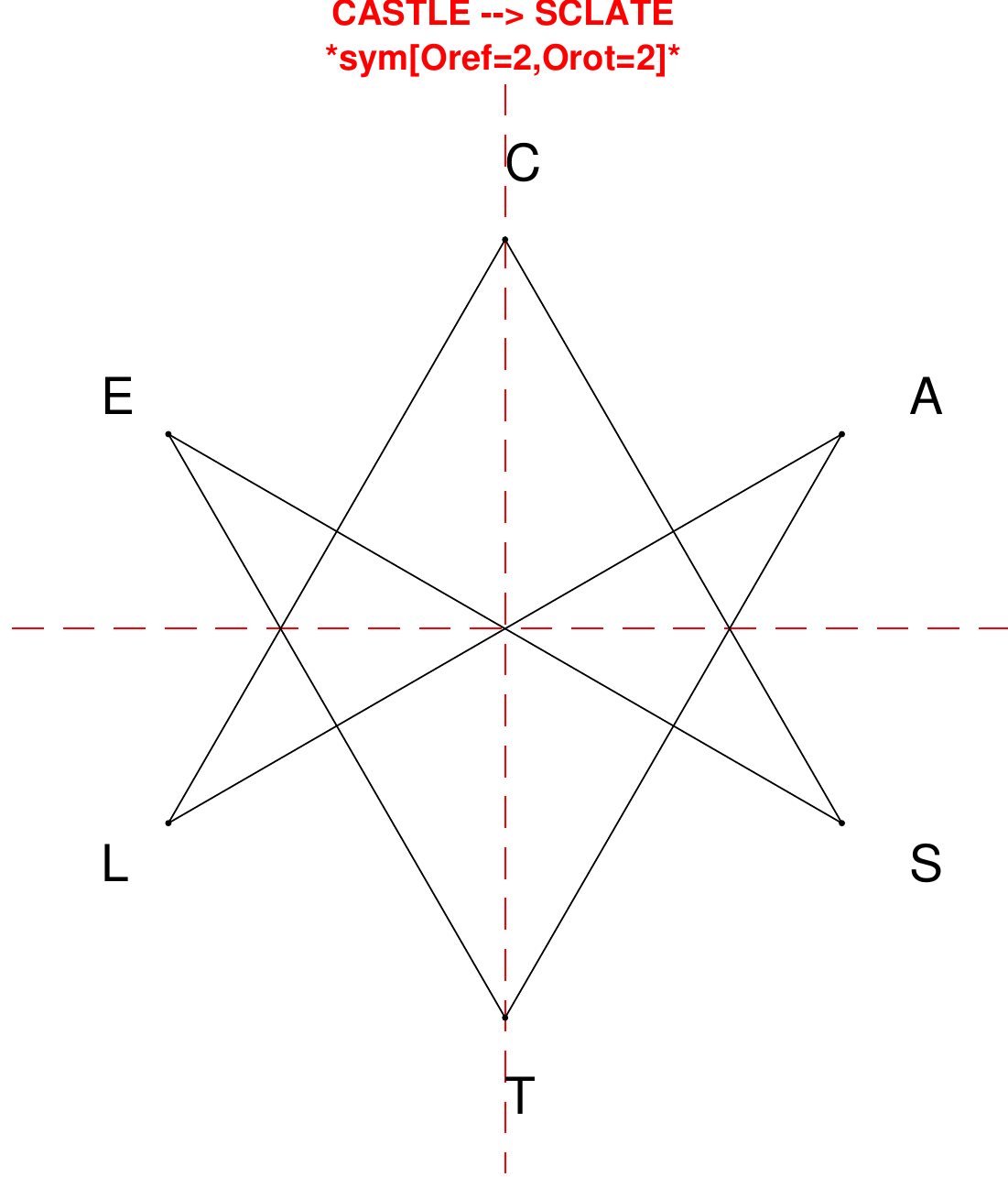}
\end{subfigure}
\hfill
\begin{subfigure}[T]{0.19\textwidth}
\centering
\includegraphics[width=\textwidth]{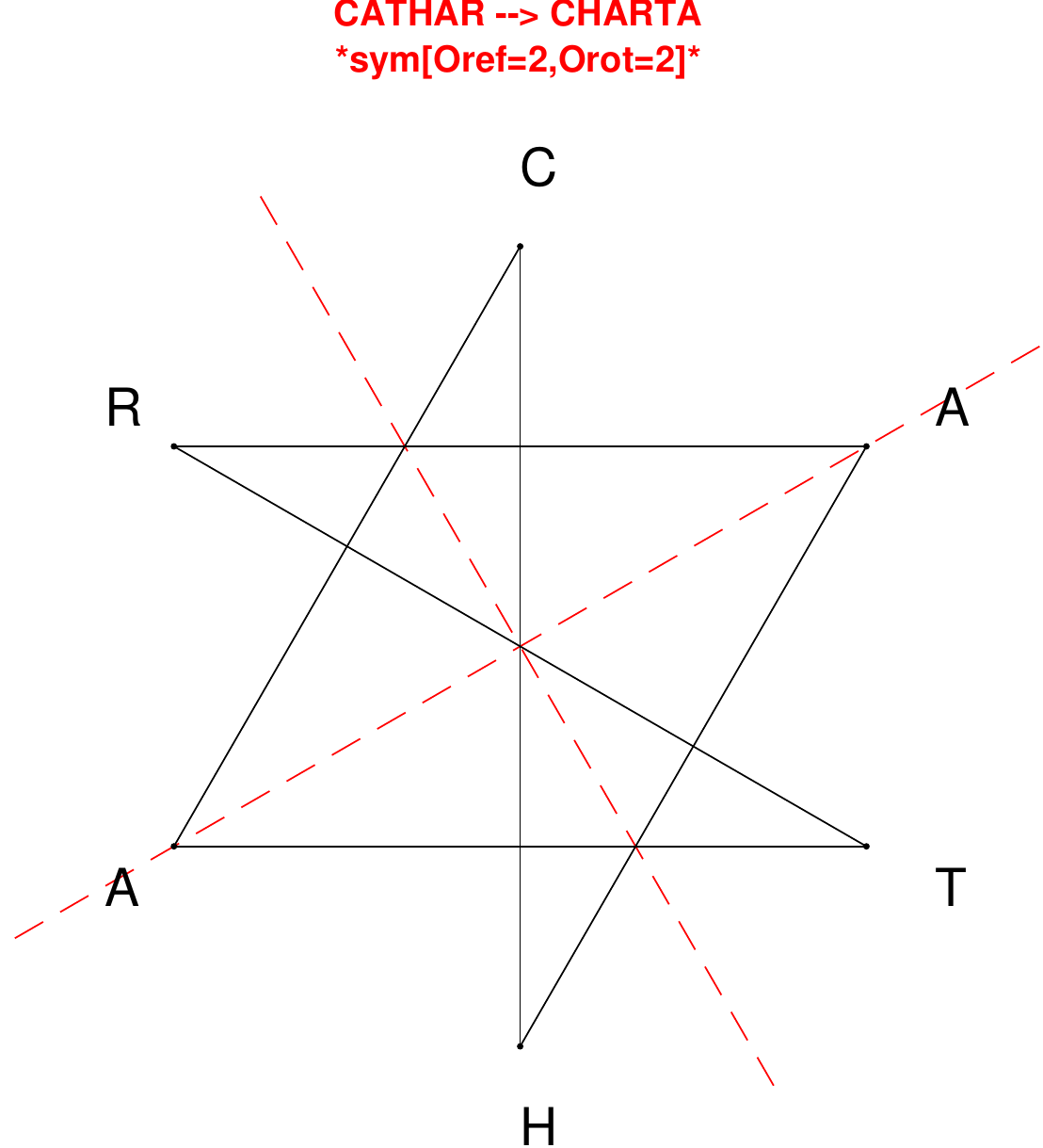}
\end{subfigure}
\hfill
\begin{subfigure}[T]{0.19\textwidth}
\centering
\includegraphics[width=\textwidth]{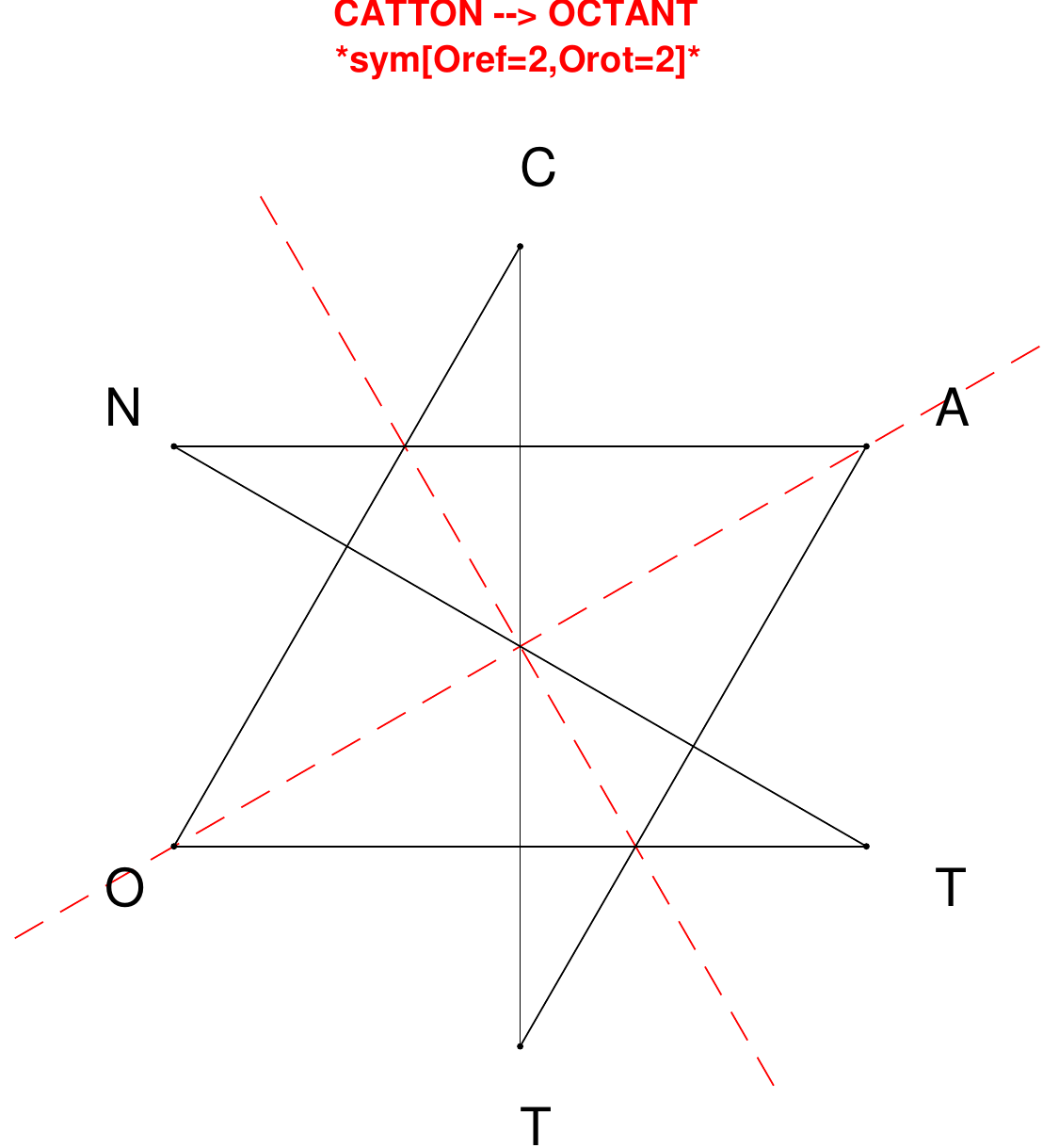}
\end{subfigure}
\hfill
\begin{subfigure}[T]{0.19\textwidth}
\centering
\includegraphics[width=\textwidth]{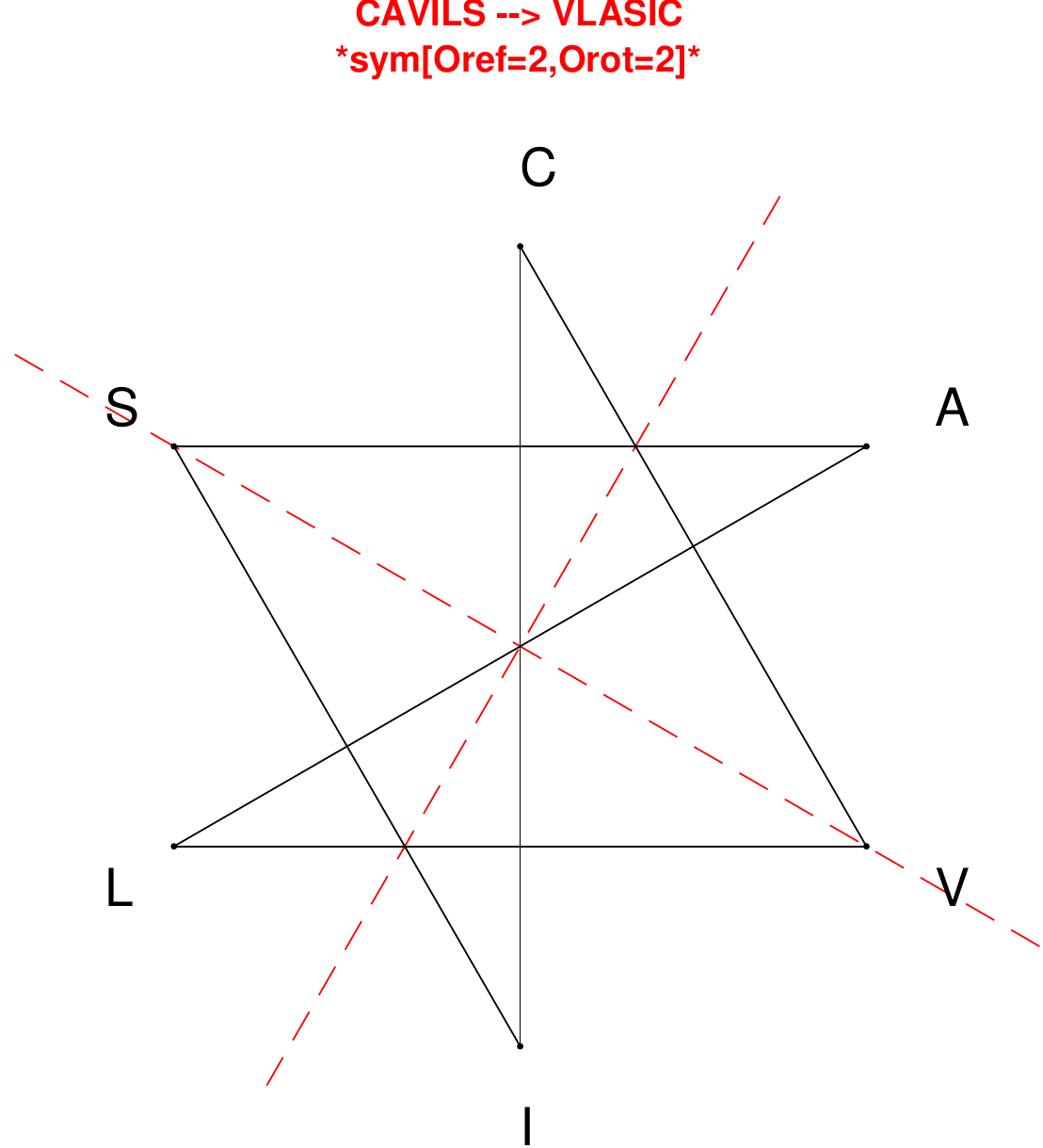}
\end{subfigure}
\end{figure}

\begin{figure}[H]
\centering
\begin{subfigure}[T]{0.19\textwidth}
\centering
\includegraphics[width=\textwidth]{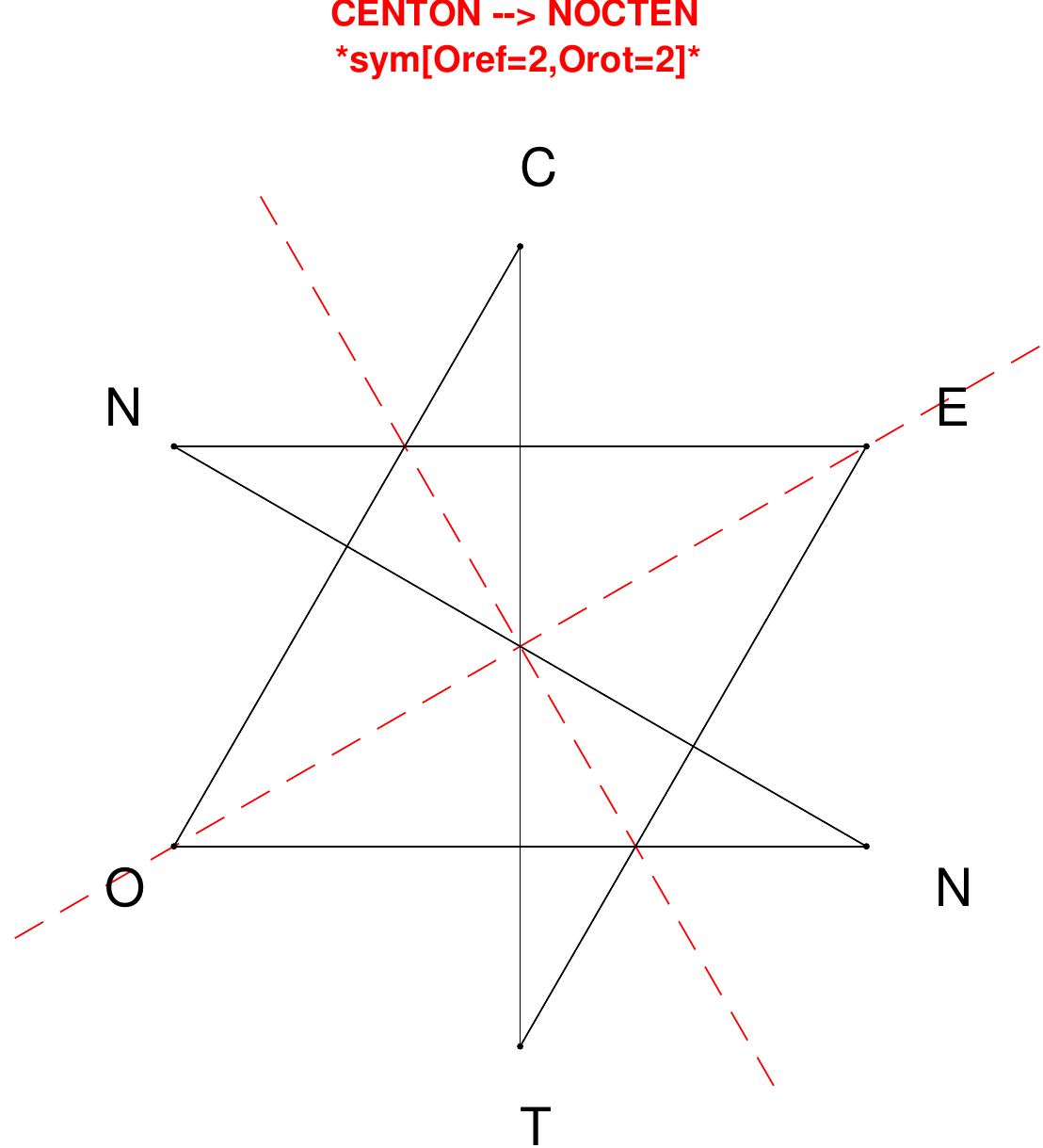}
\end{subfigure}
\hfill
\begin{subfigure}[T]{0.19\textwidth}
\centering
\includegraphics[width=\textwidth]{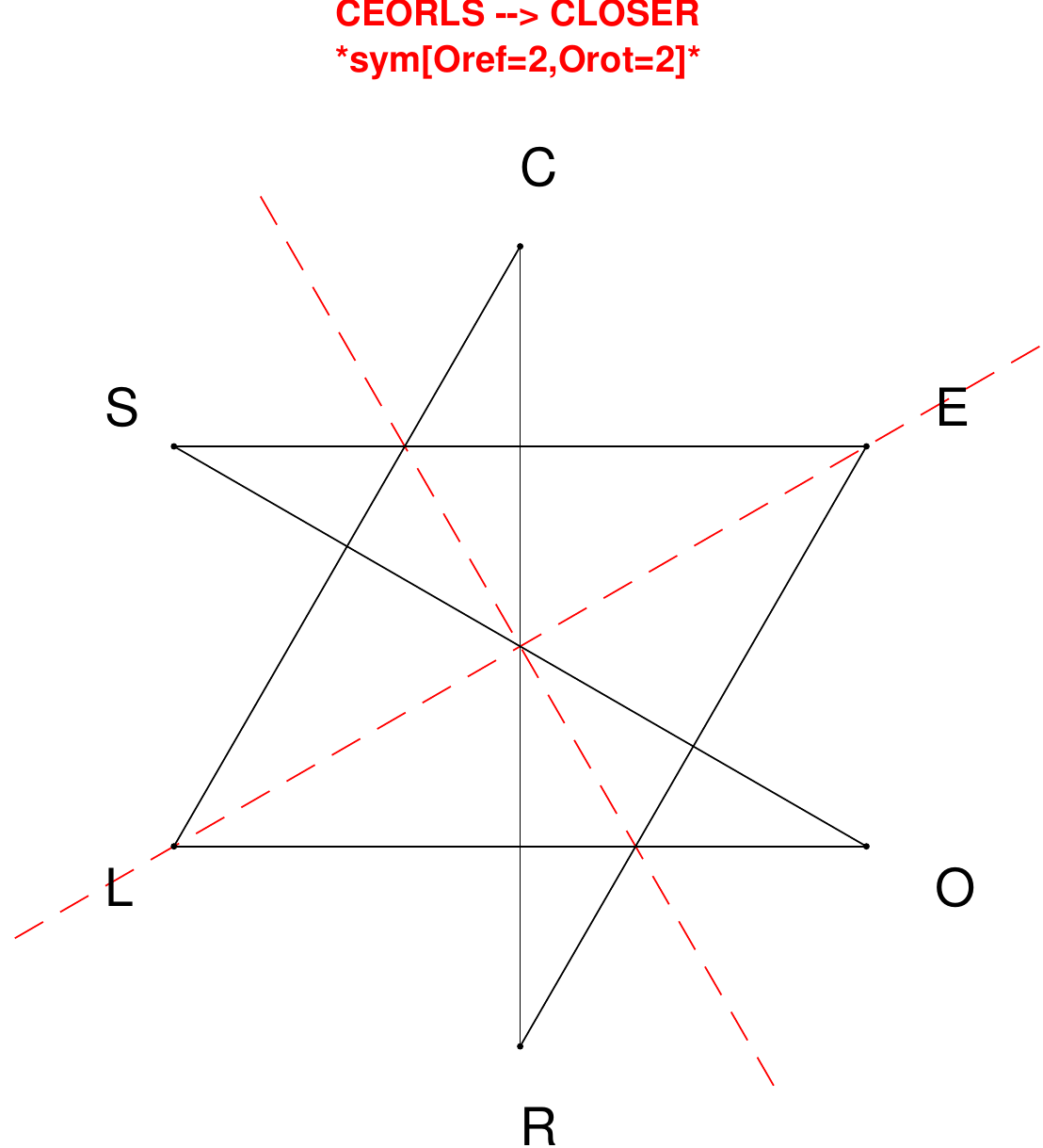}
\end{subfigure}
\hfill
\begin{subfigure}[T]{0.19\textwidth}
\centering
\includegraphics[width=\textwidth]{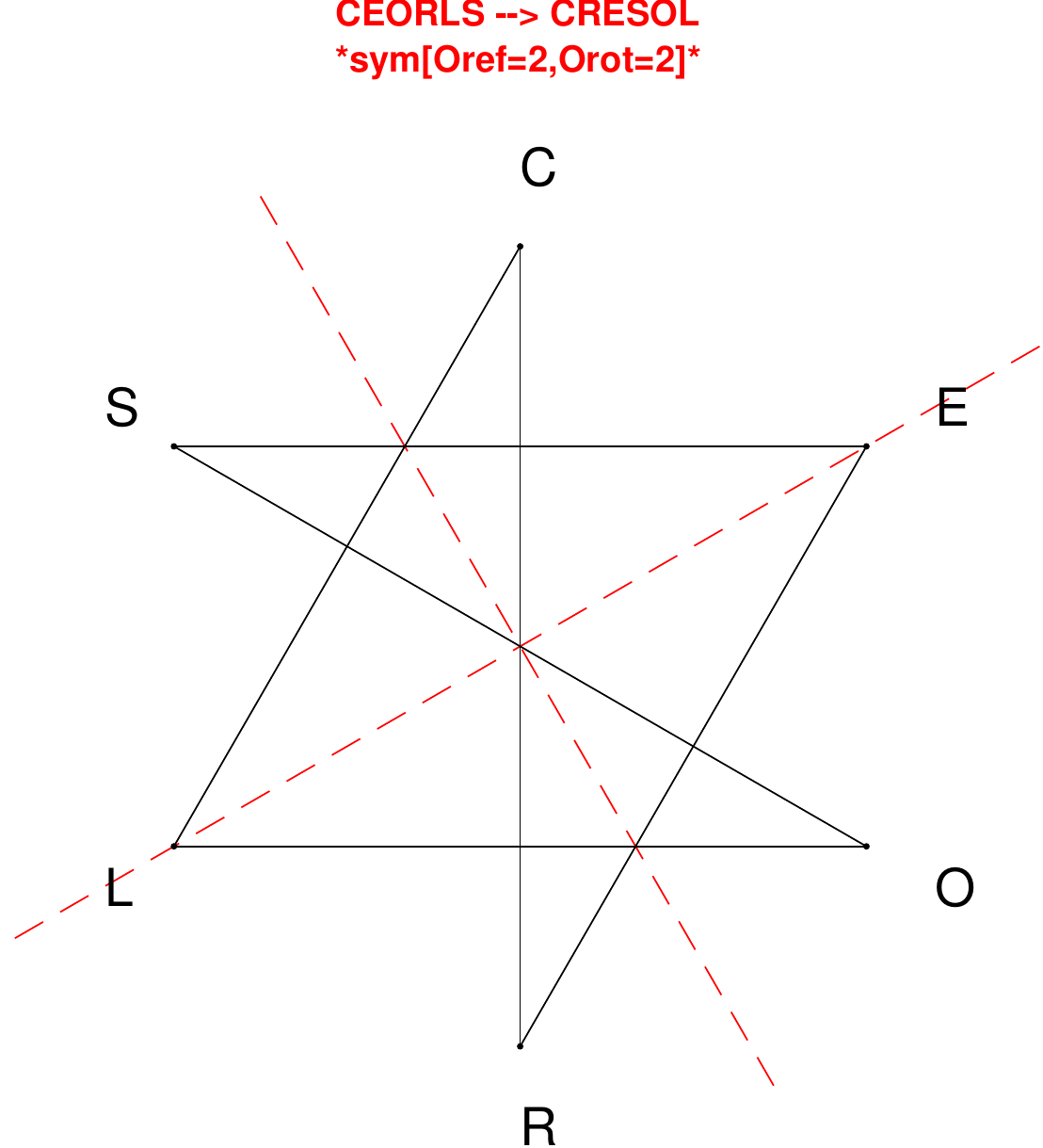}
\end{subfigure}
\hfill
\begin{subfigure}[T]{0.19\textwidth}
\centering
\includegraphics[width=\textwidth]{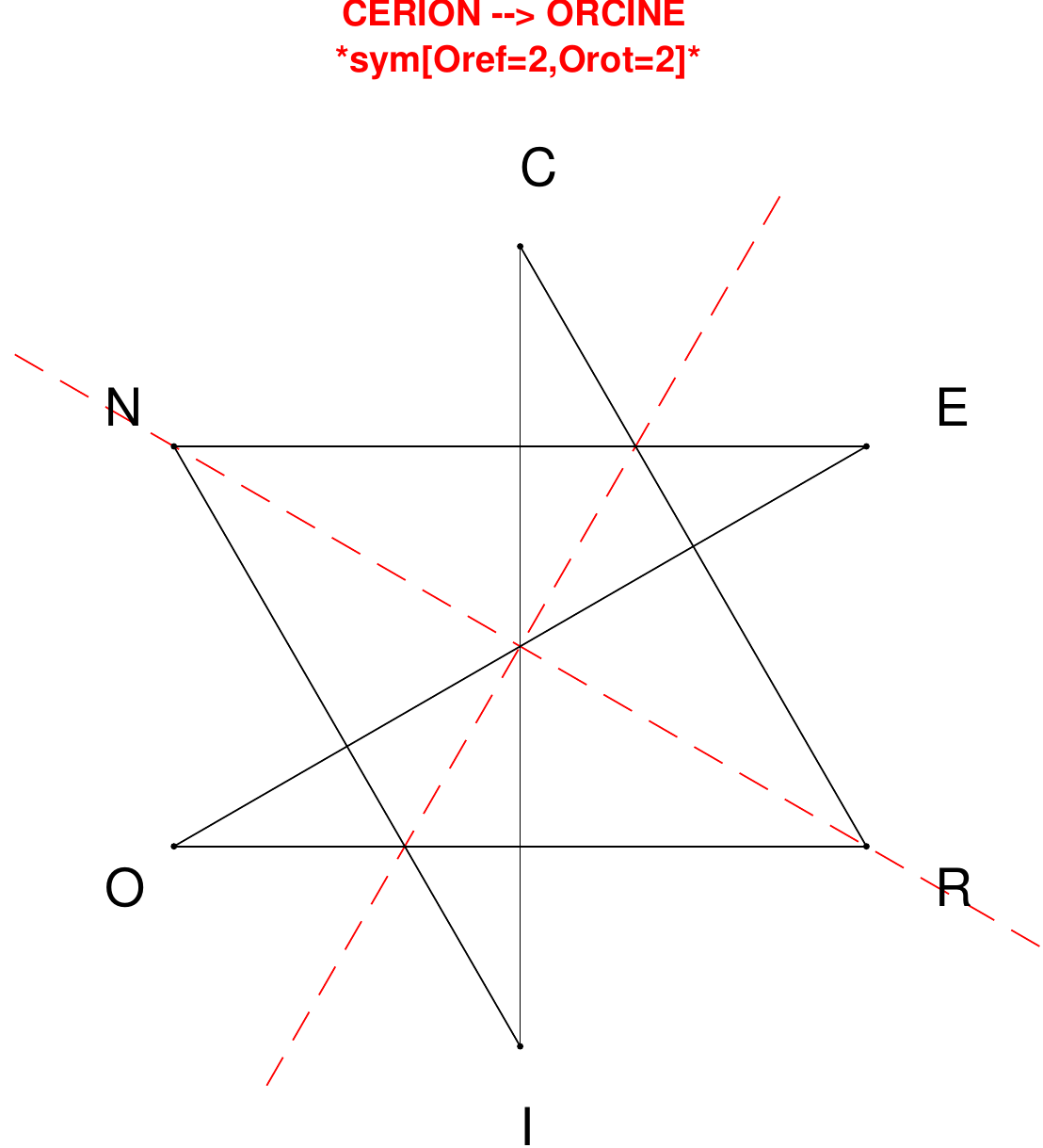}
\end{subfigure}
\hfill
\begin{subfigure}[T]{0.19\textwidth}
\centering
\includegraphics[width=\textwidth]{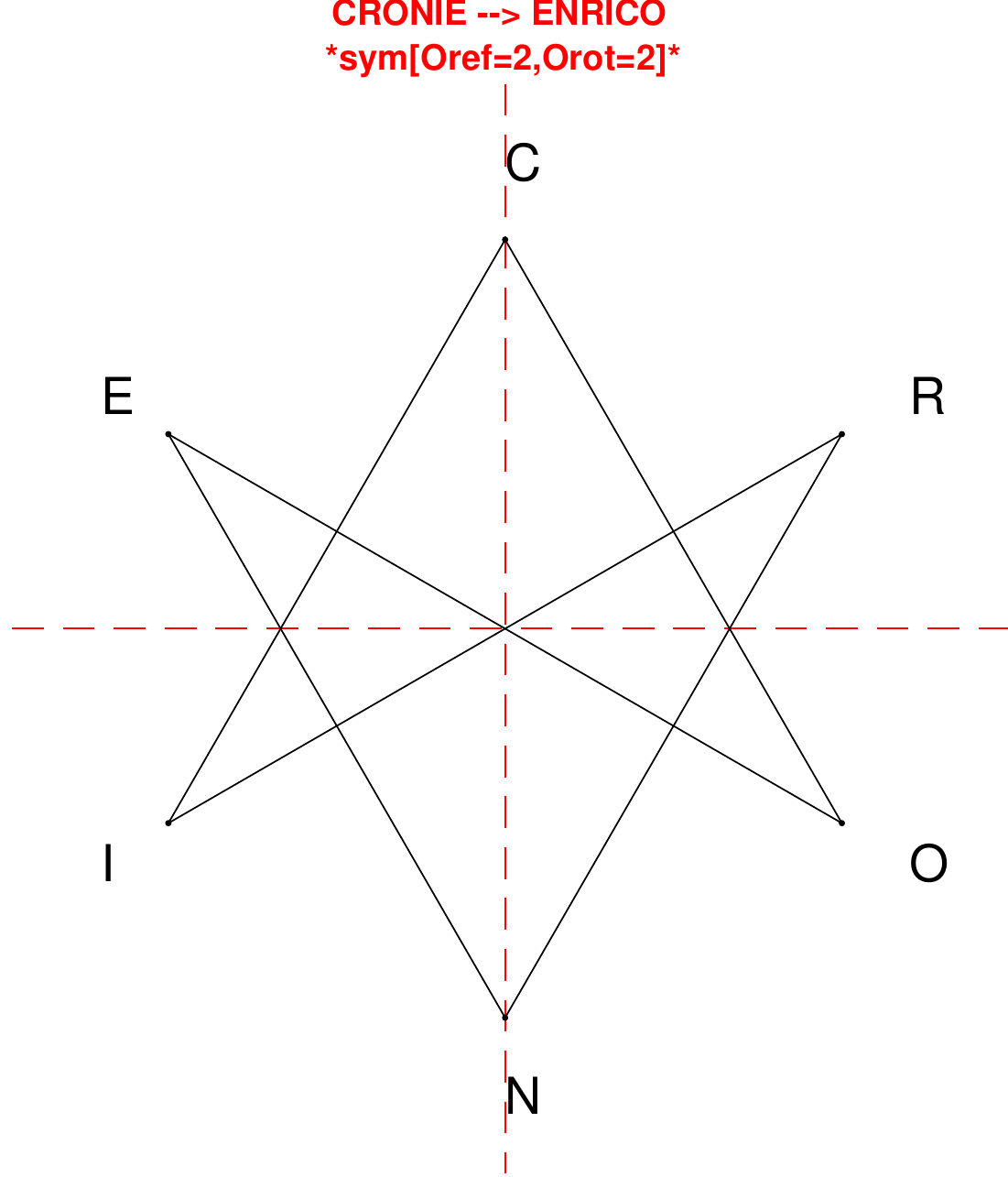}
\end{subfigure}
\end{figure}

\begin{figure}[H]
\centering
\begin{subfigure}[T]{0.19\textwidth}
\centering
\includegraphics[width=\textwidth]{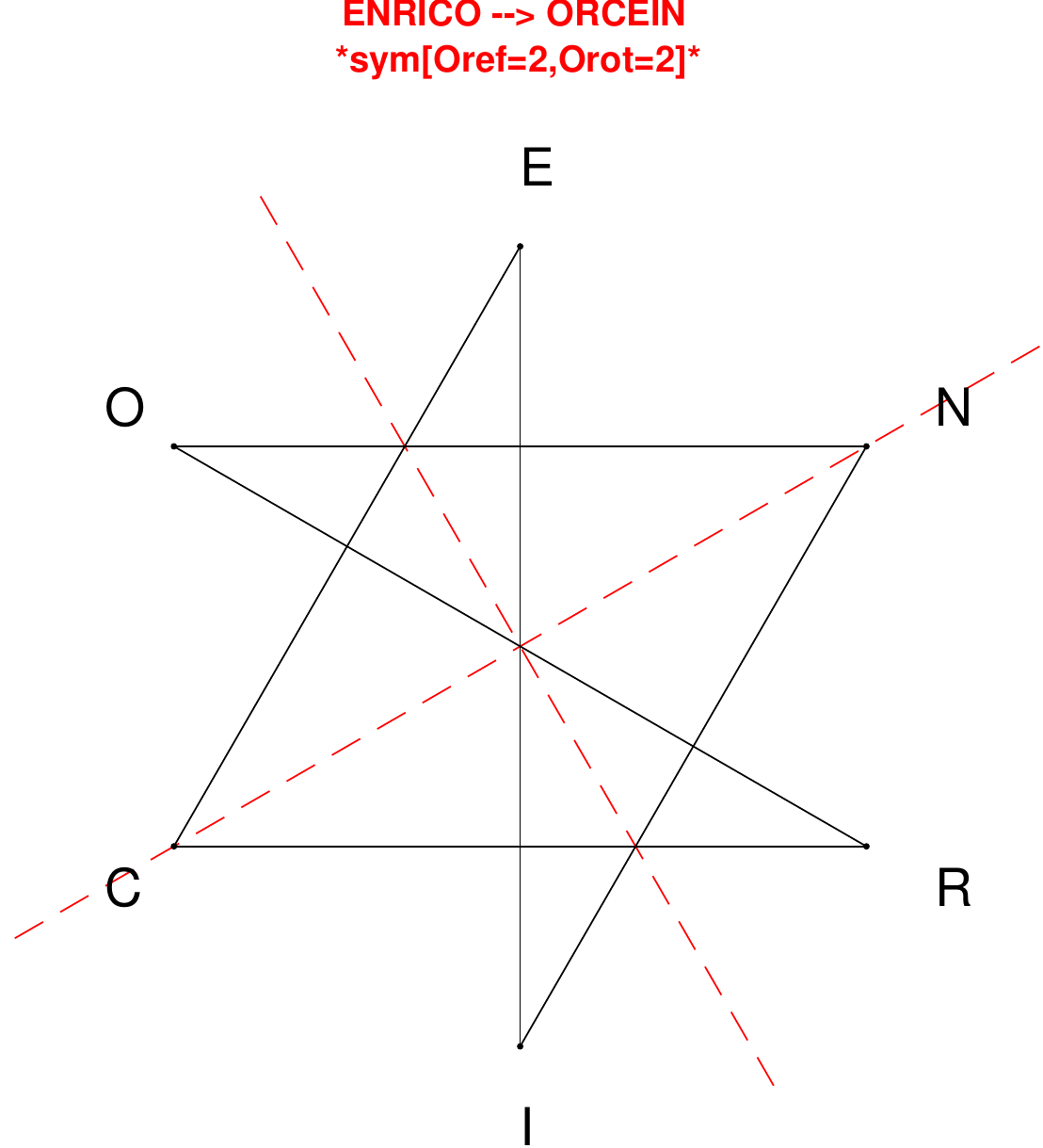}
\end{subfigure}
\hfill
\begin{subfigure}[T]{0.19\textwidth}
\centering
\includegraphics[width=\textwidth]{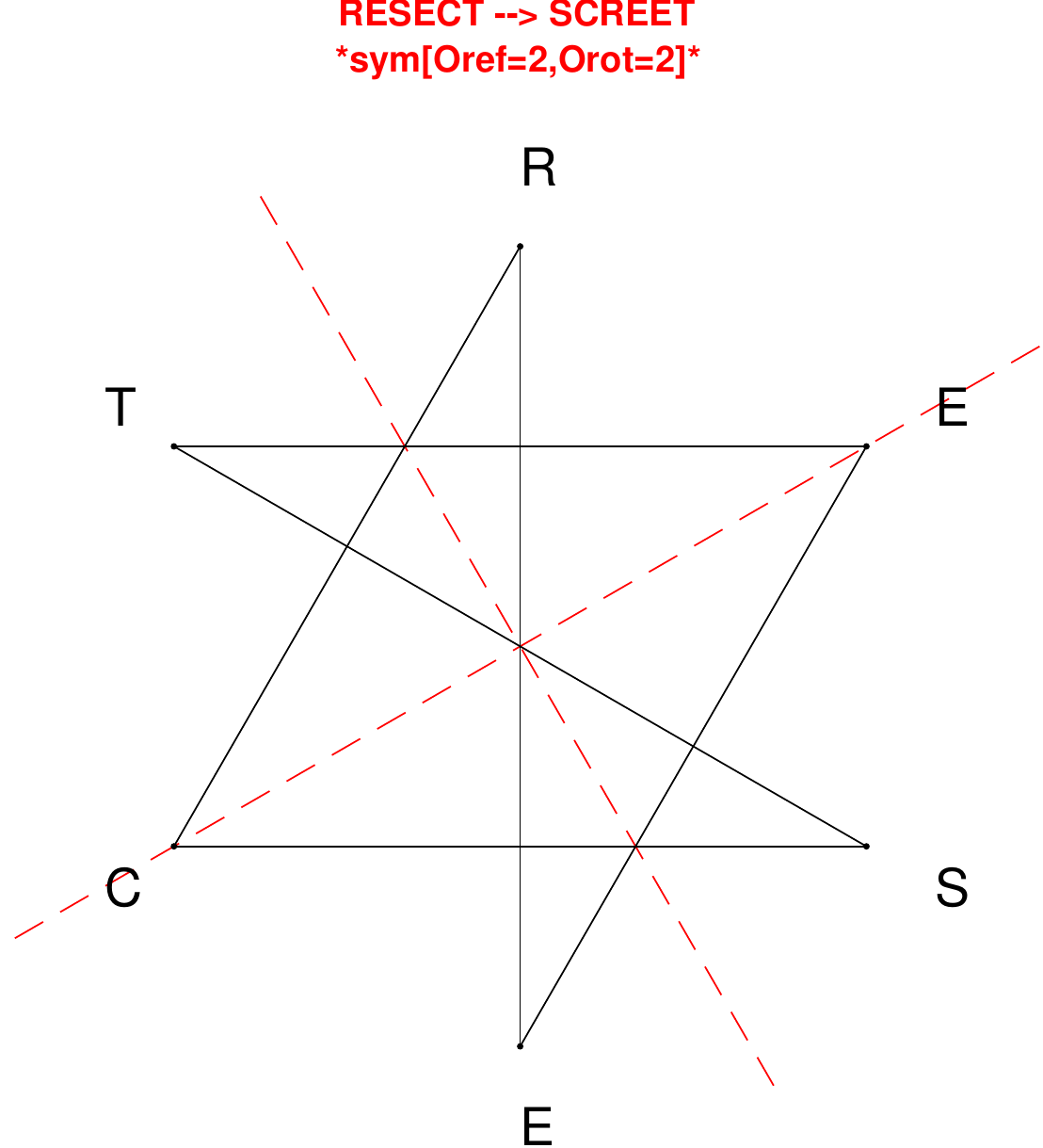}
\end{subfigure}
\hfill
\begin{subfigure}[T]{0.19\textwidth}
\centering
\includegraphics[width=\textwidth]{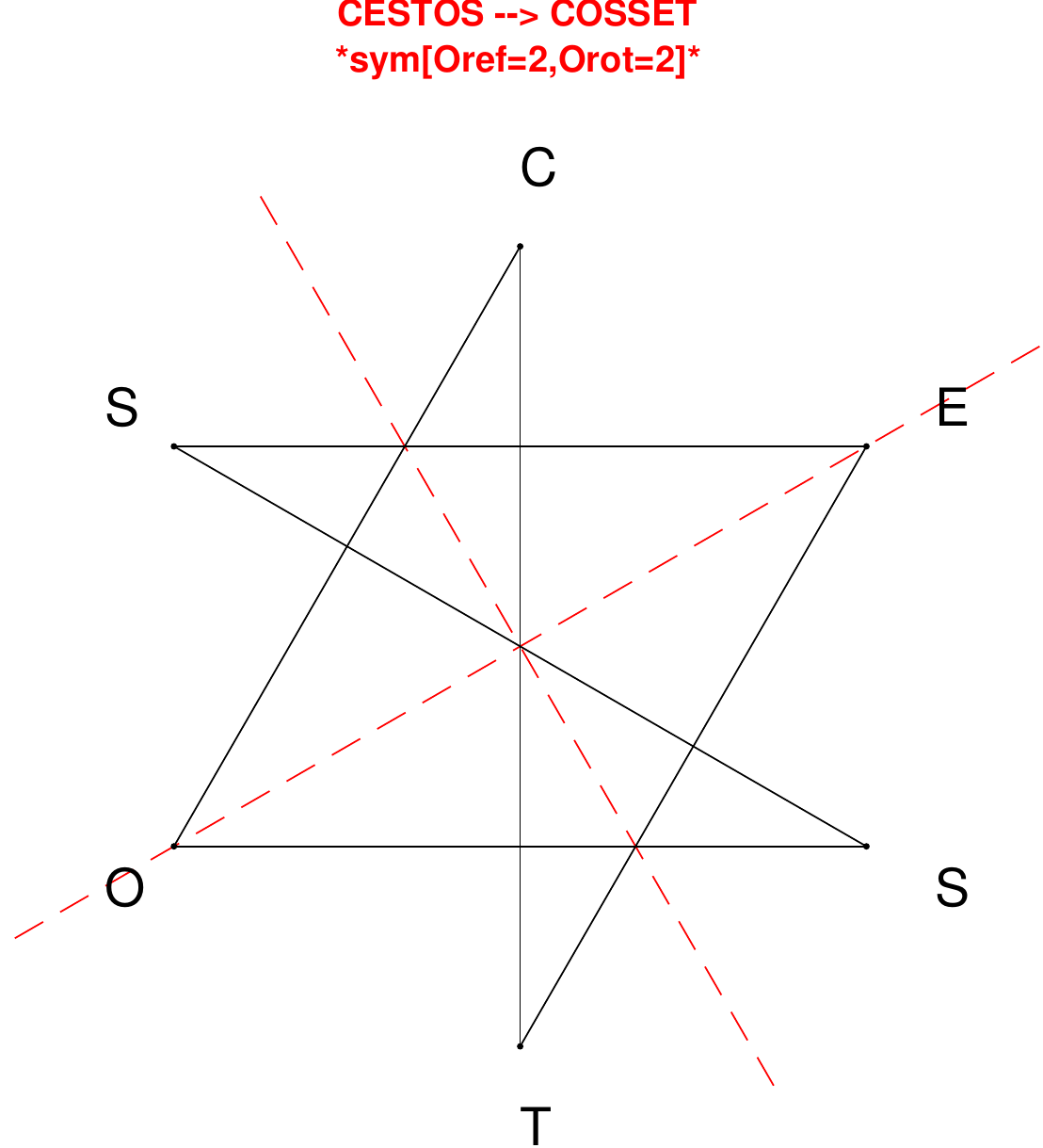}
\end{subfigure}
\hfill
\begin{subfigure}[T]{0.19\textwidth}
\centering
\includegraphics[width=\textwidth]{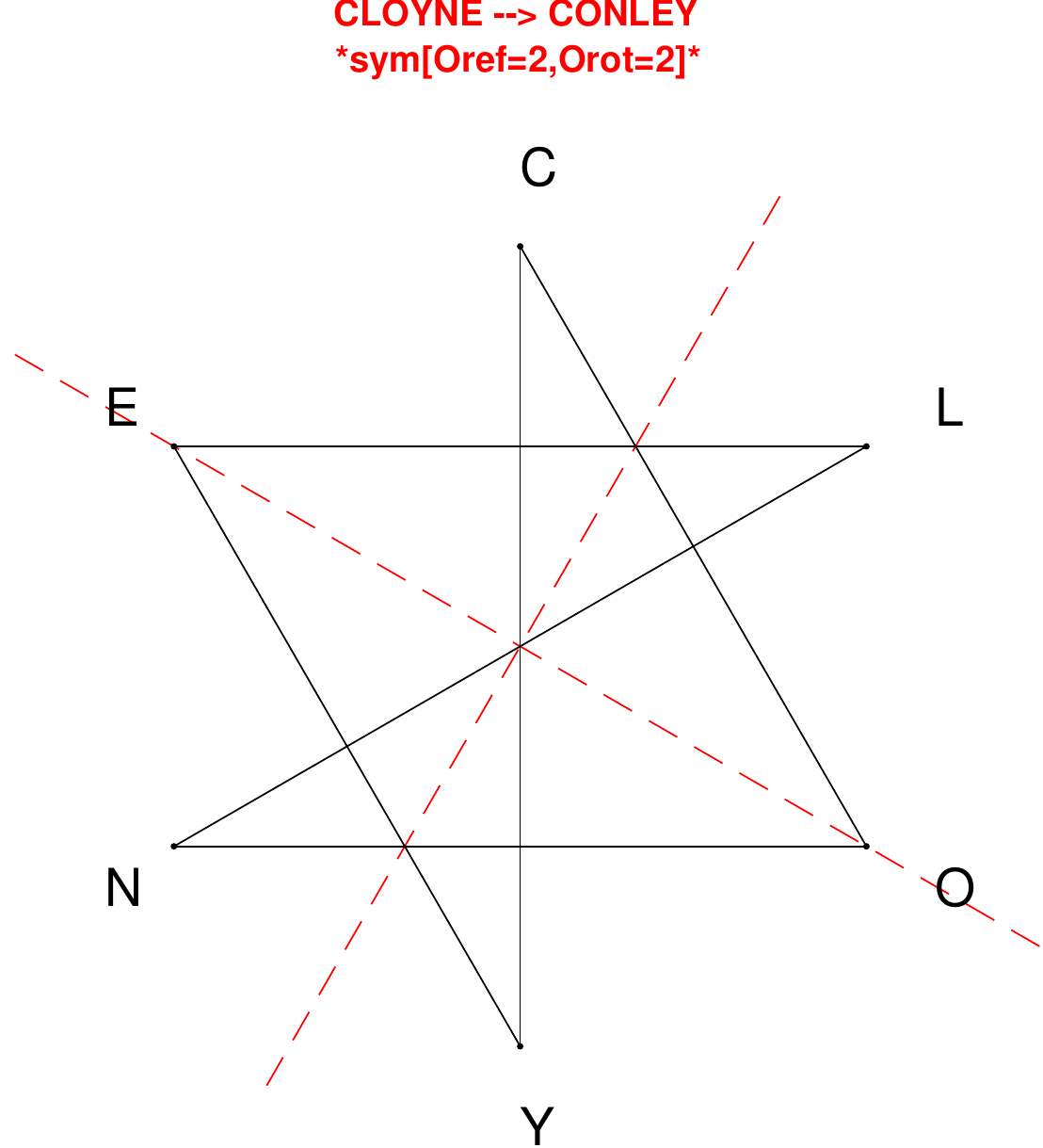}
\end{subfigure}
\hfill
\begin{subfigure}[T]{0.19\textwidth}
\centering
\includegraphics[width=\textwidth]{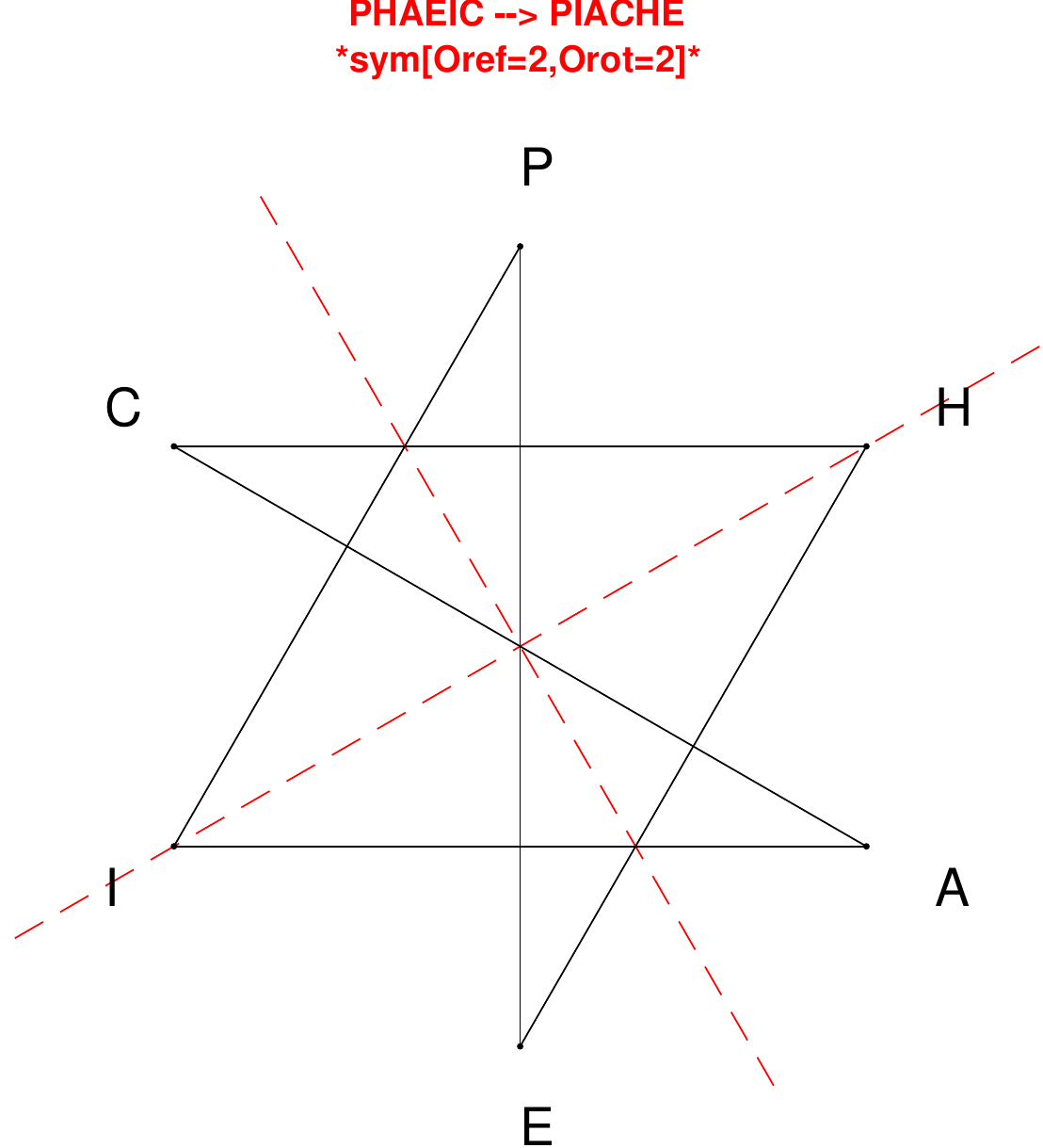}
\end{subfigure}
\end{figure}

\begin{figure}[H]
\centering
\begin{subfigure}[T]{0.19\textwidth}
\centering
\includegraphics[width=\textwidth]{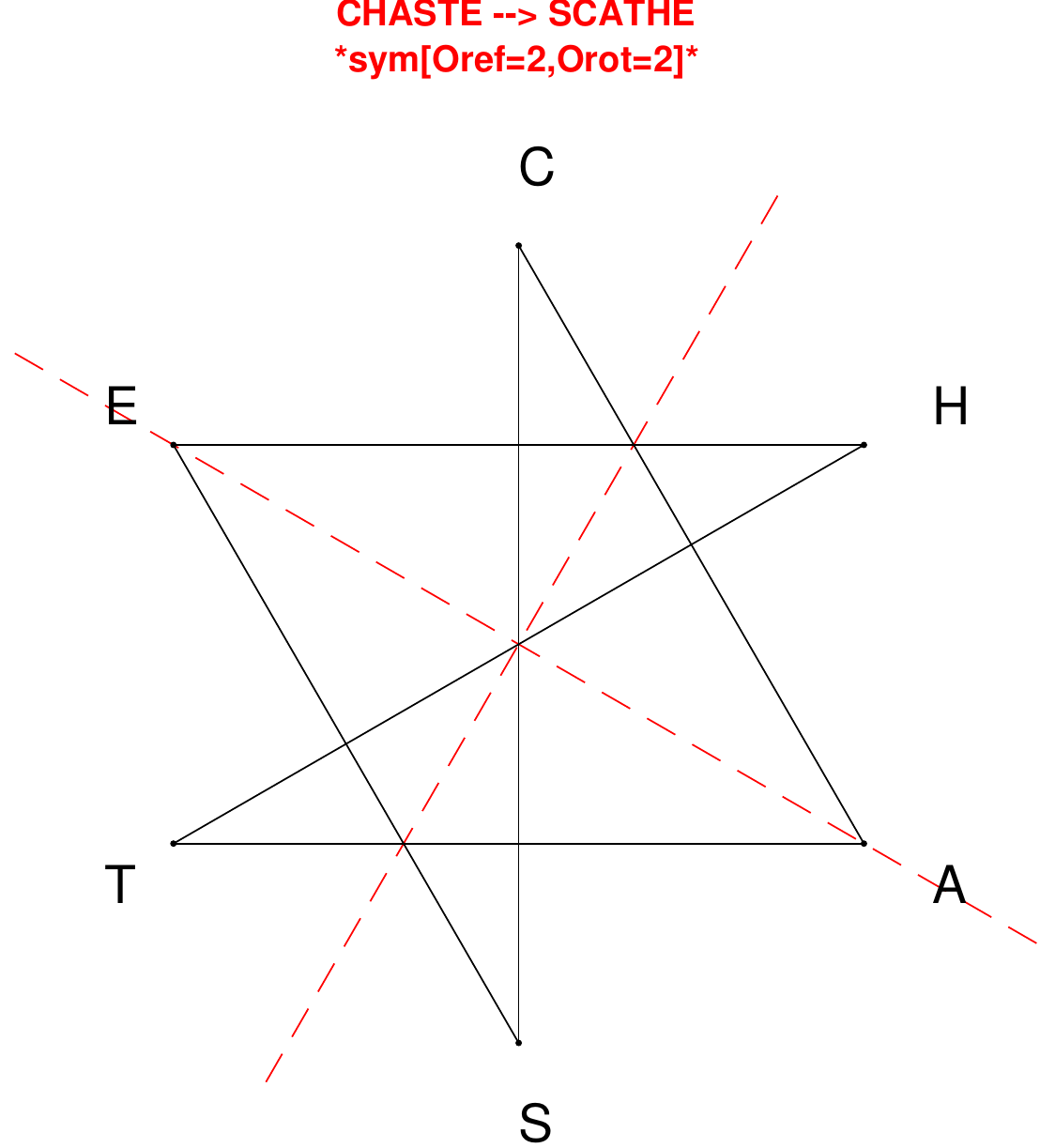}
\end{subfigure}
\hfill
\begin{subfigure}[T]{0.19\textwidth}
\centering
\includegraphics[width=\textwidth]{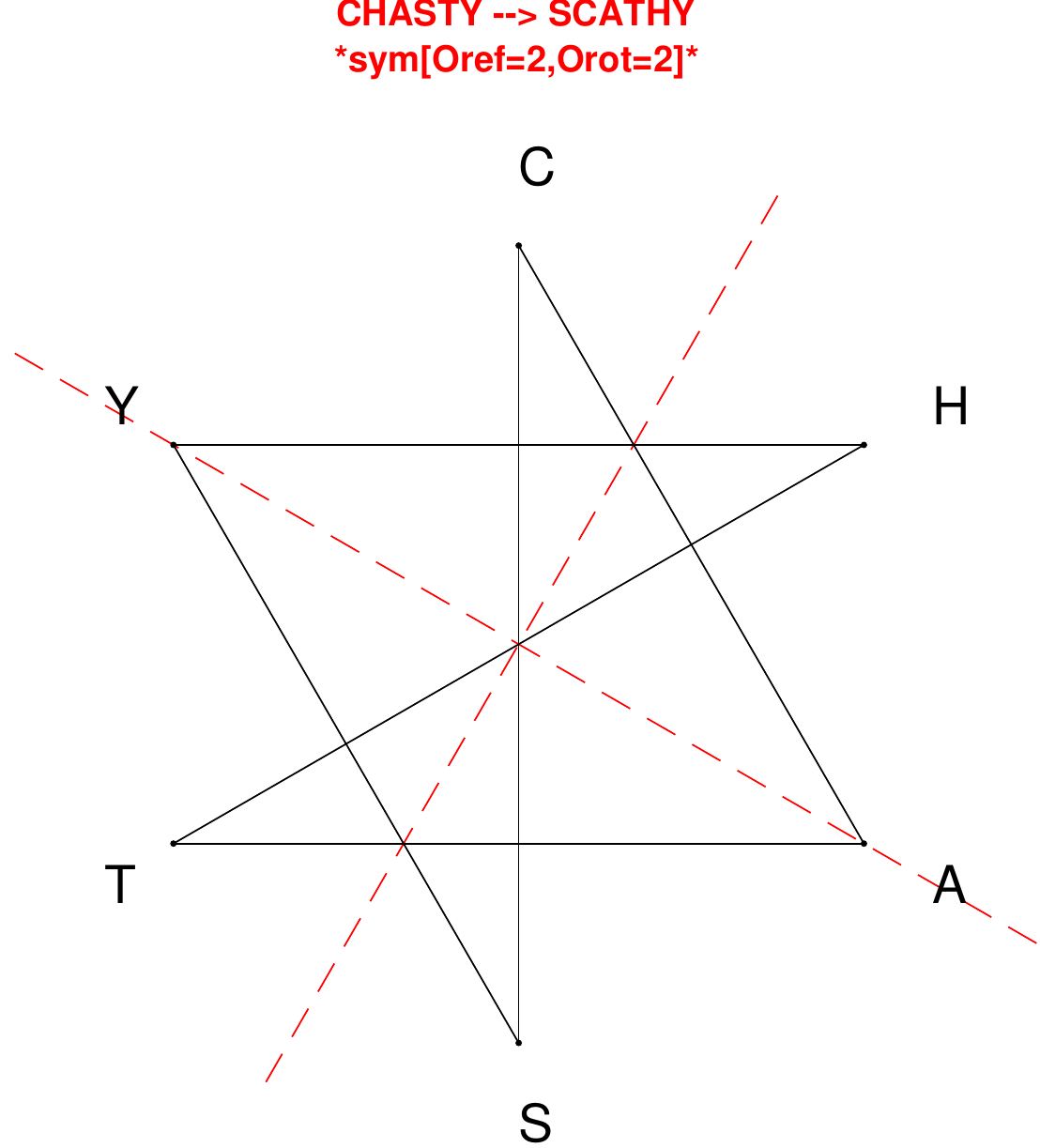}
\end{subfigure}
\hfill
\begin{subfigure}[T]{0.19\textwidth}
\centering
\includegraphics[width=\textwidth]{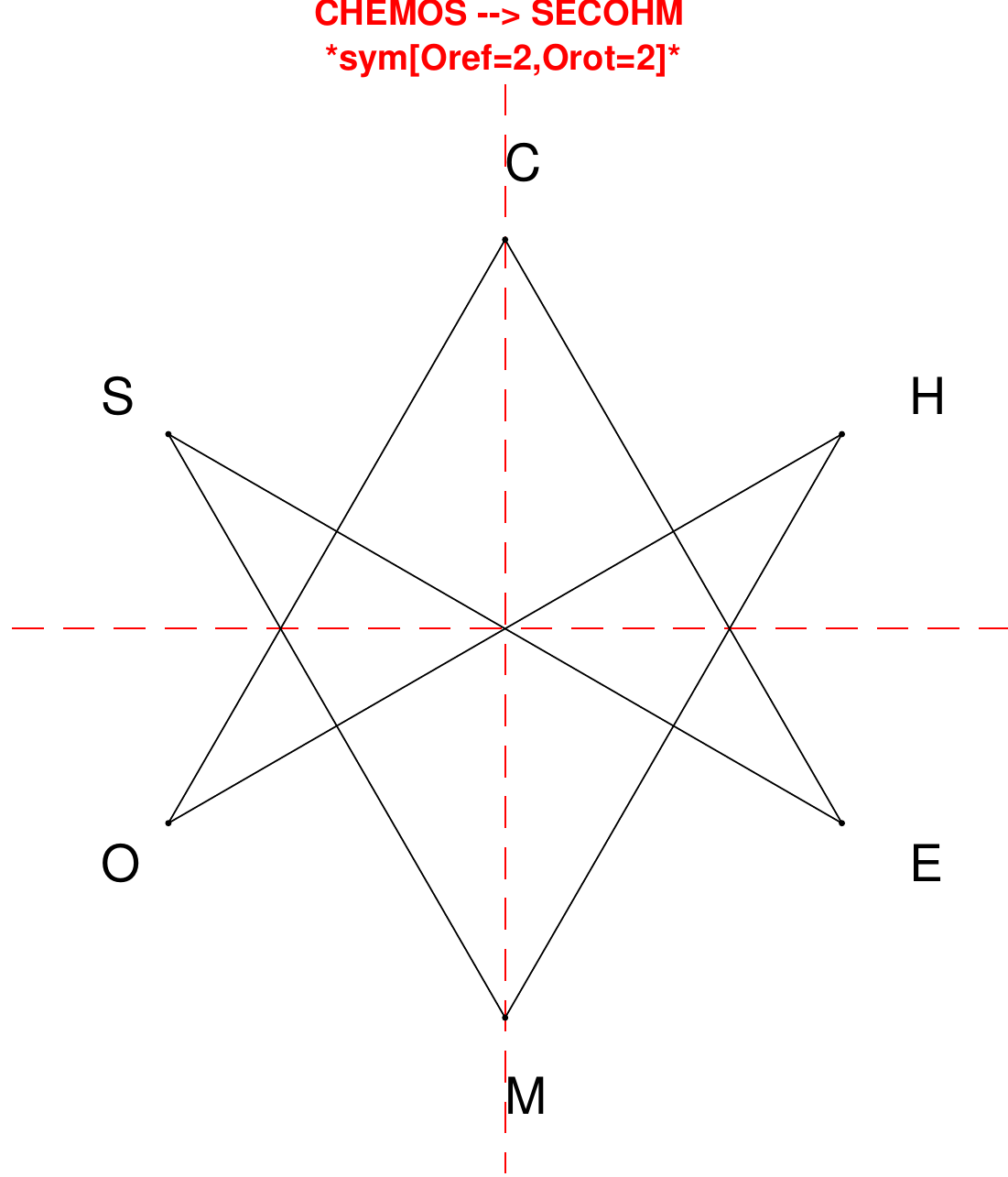}
\end{subfigure}
\hfill
\begin{subfigure}[T]{0.19\textwidth}
\centering
\includegraphics[width=\textwidth]{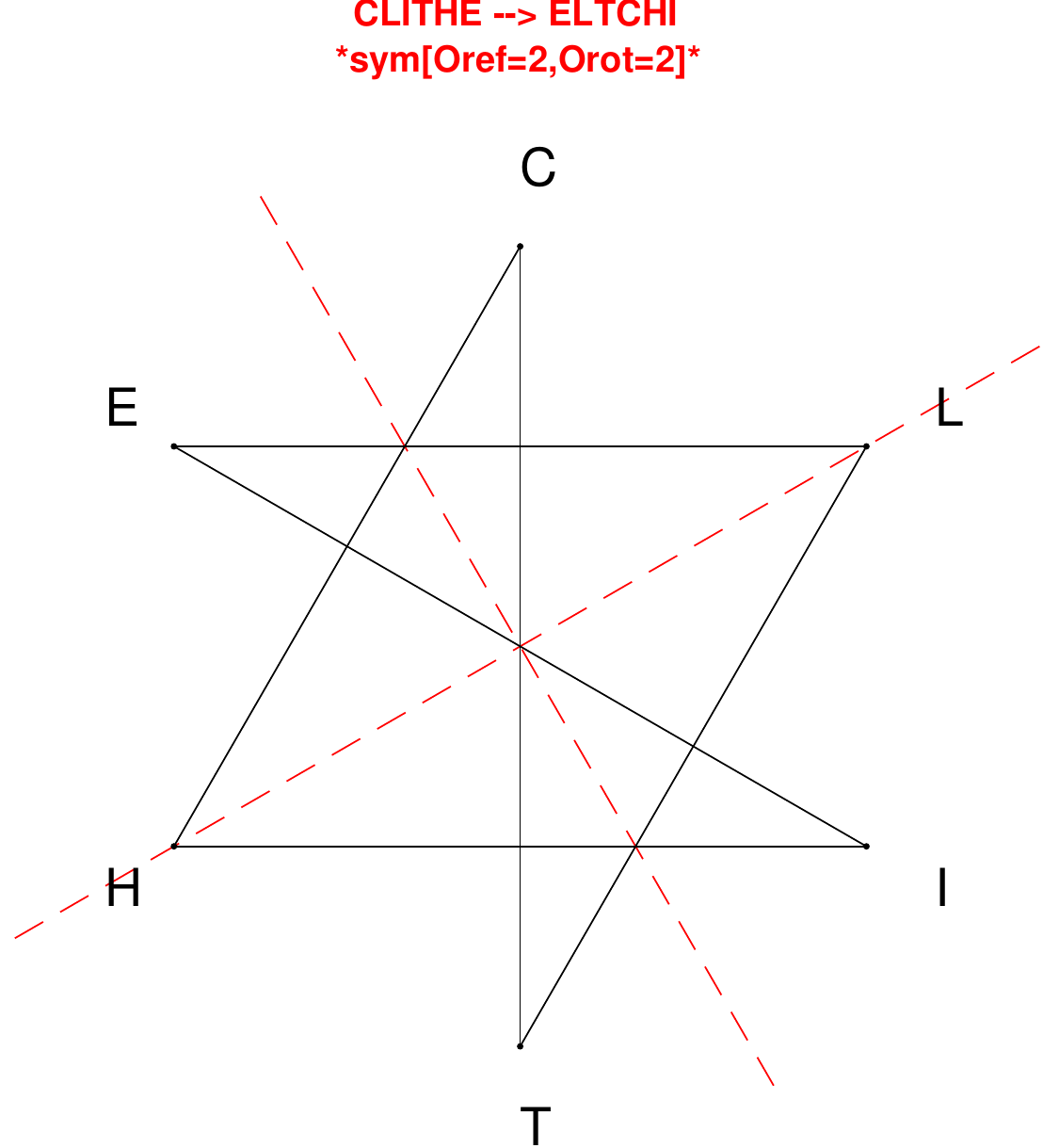}
\end{subfigure}
\hfill
\begin{subfigure}[T]{0.19\textwidth}
\centering
\includegraphics[width=\textwidth]{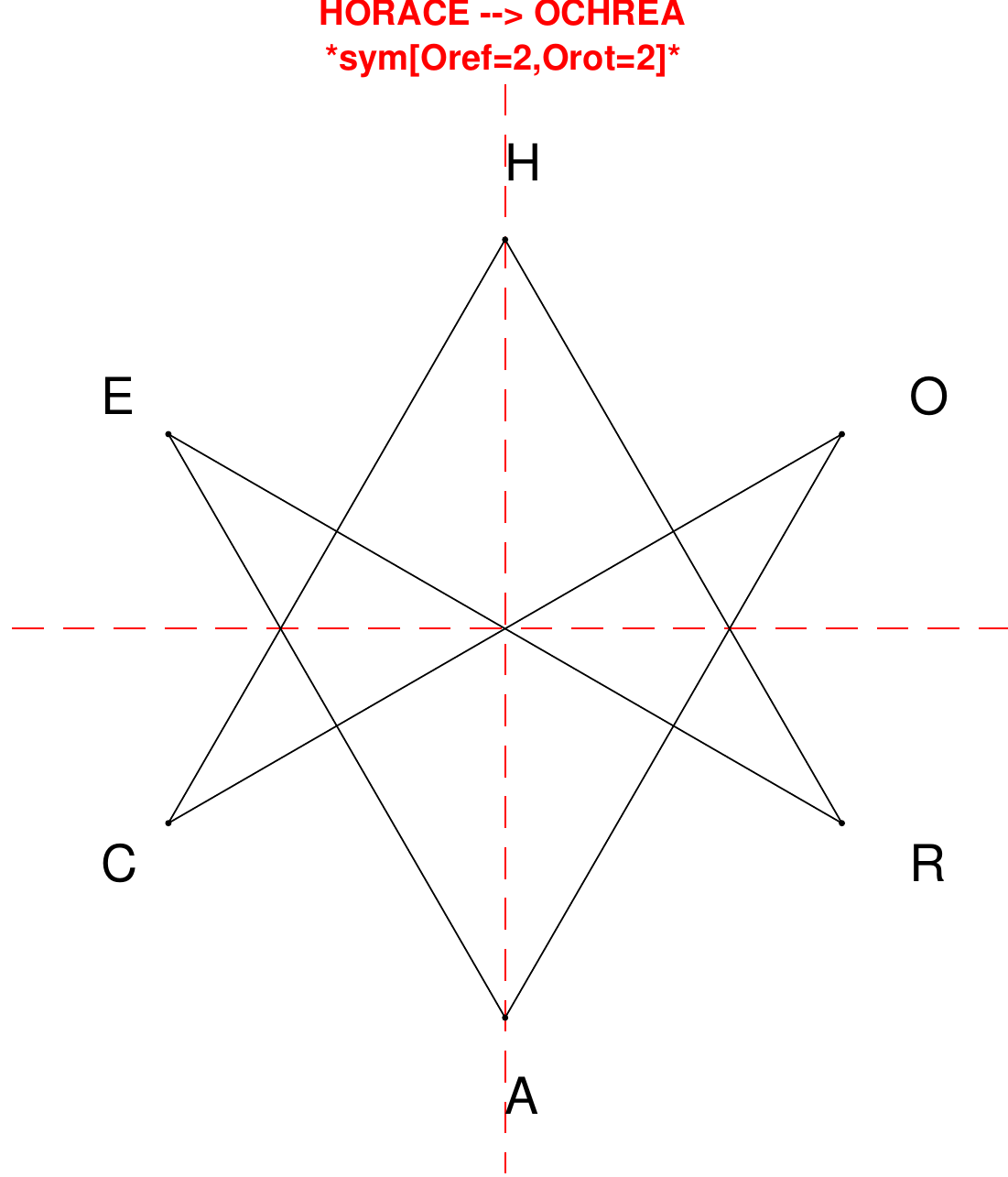}
\end{subfigure}
\end{figure}

\begin{figure}[H]
\centering
\begin{subfigure}[T]{0.19\textwidth}
\centering
\includegraphics[width=\textwidth]{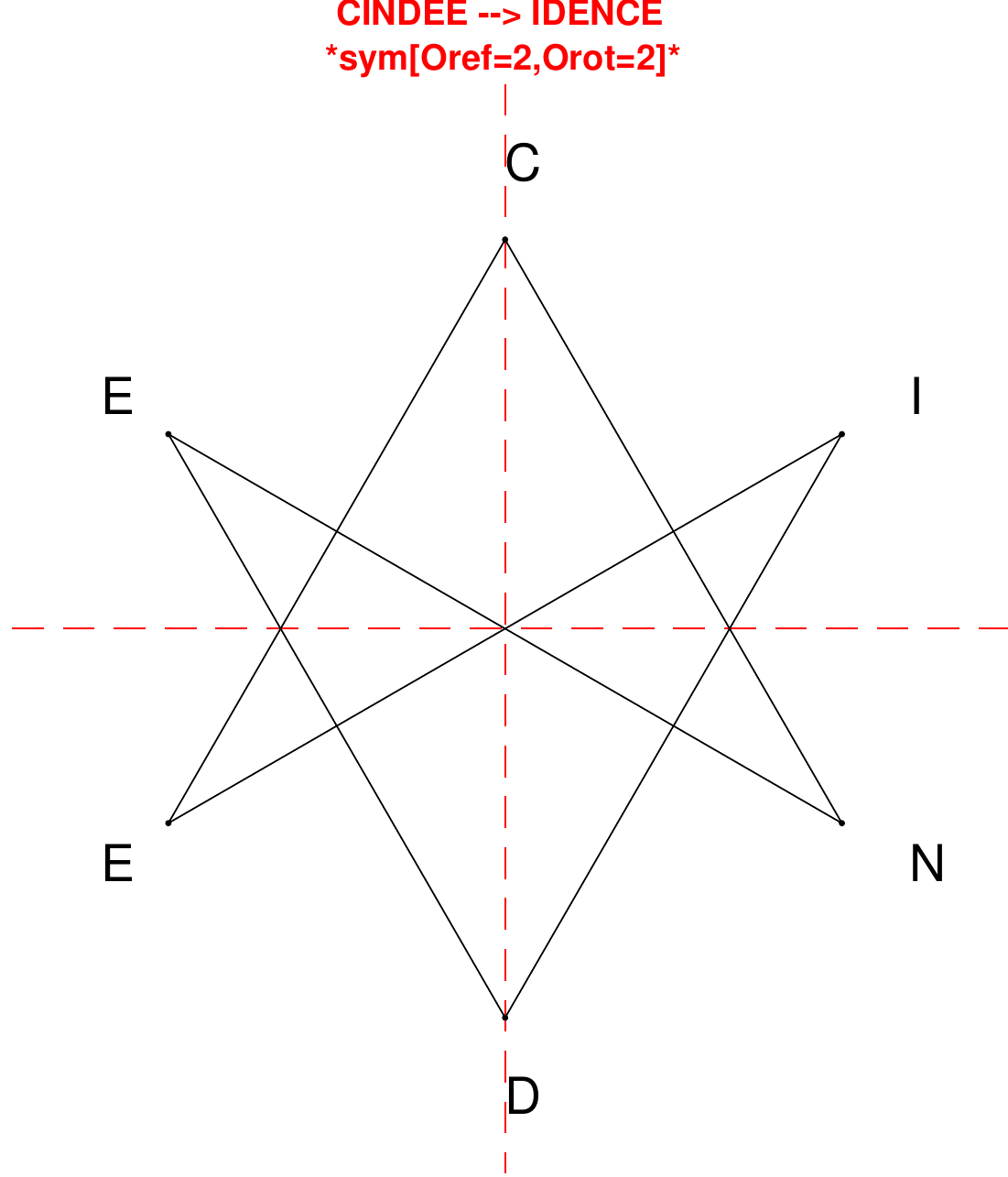}
\end{subfigure}
\hfill
\begin{subfigure}[T]{0.19\textwidth}
\centering
\includegraphics[width=\textwidth]{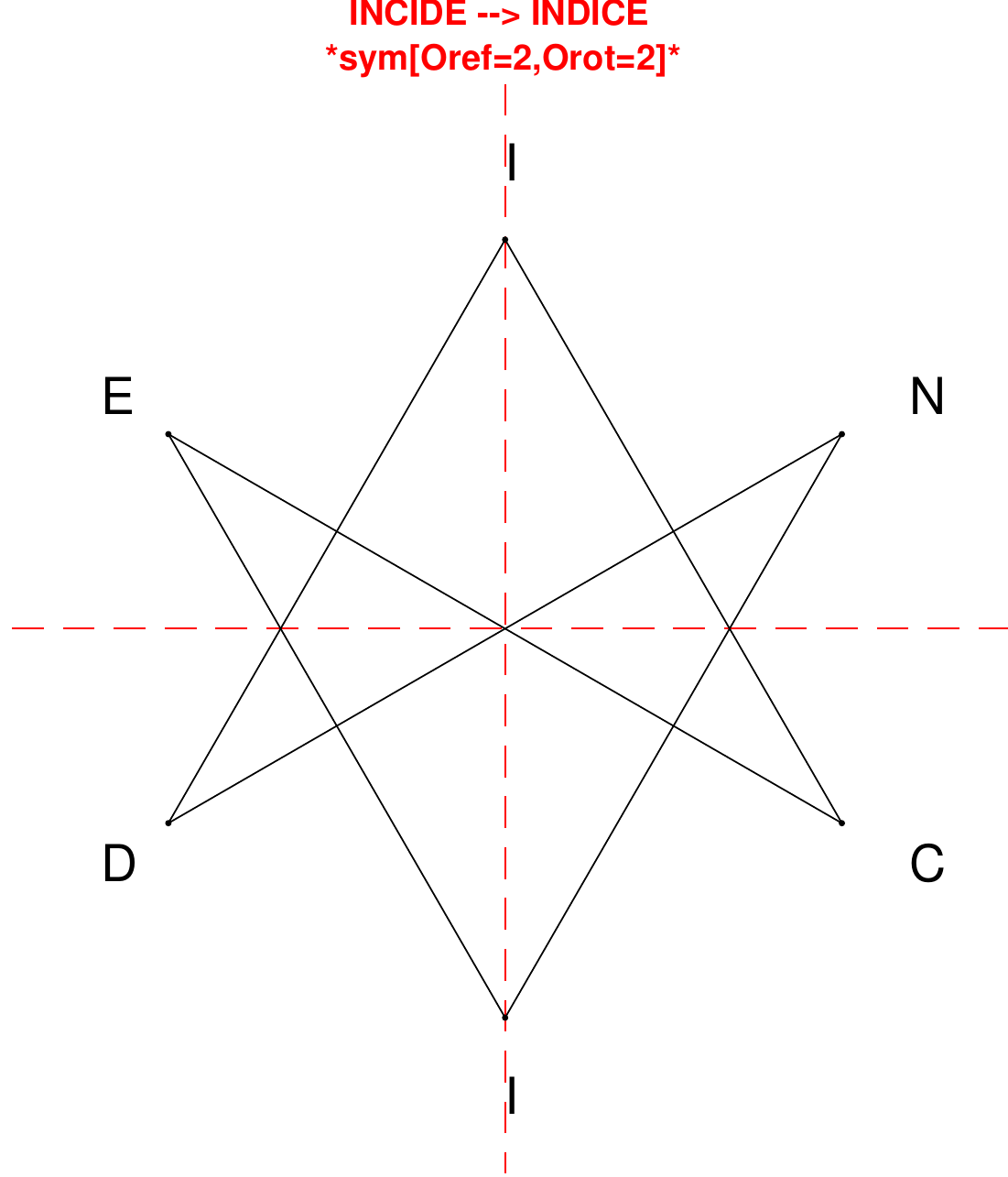}
\end{subfigure}
\hfill
\begin{subfigure}[T]{0.19\textwidth}
\centering
\includegraphics[width=\textwidth]{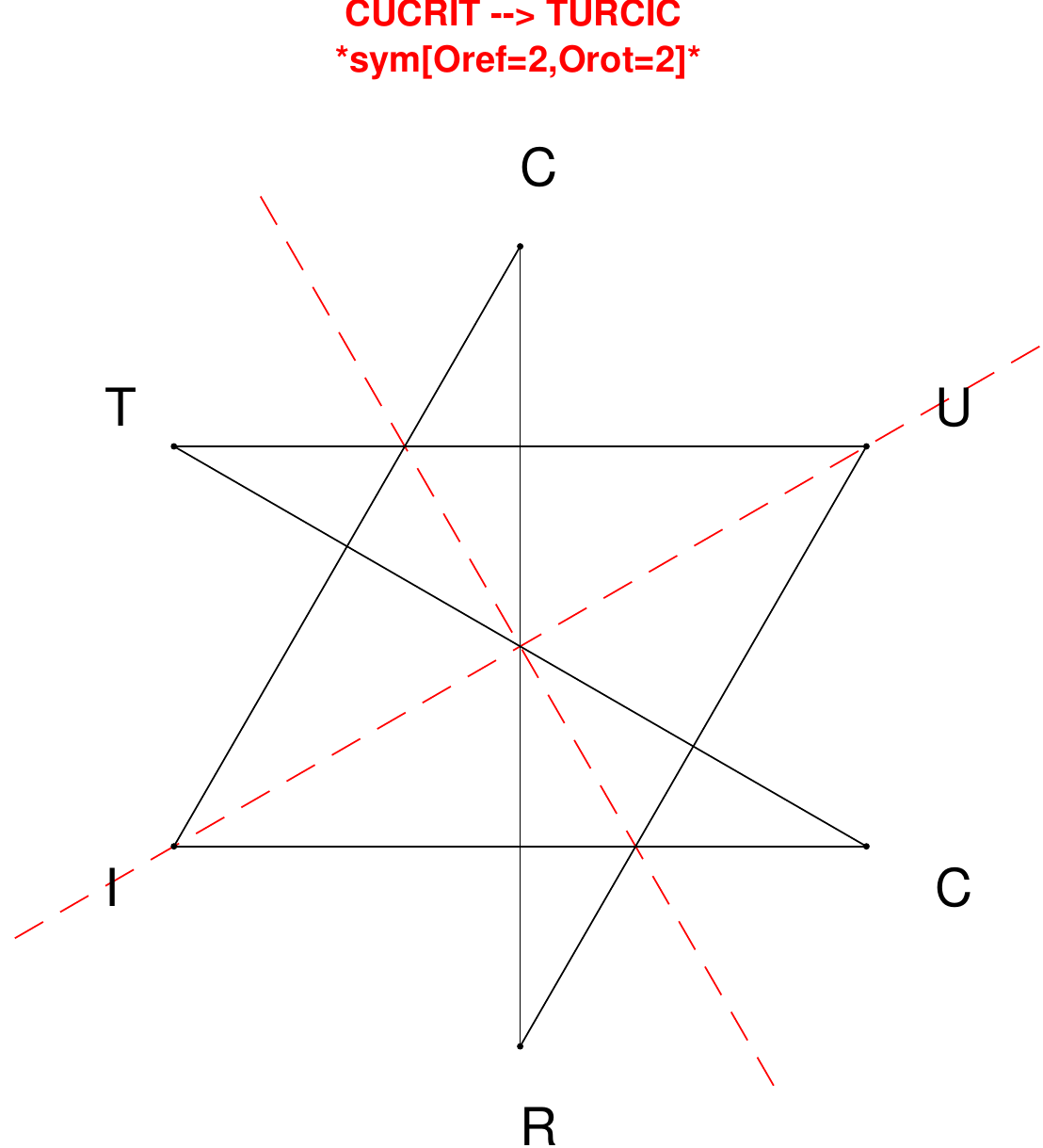}
\end{subfigure}
\hfill
\begin{subfigure}[T]{0.19\textwidth}
\centering
\includegraphics[width=\textwidth]{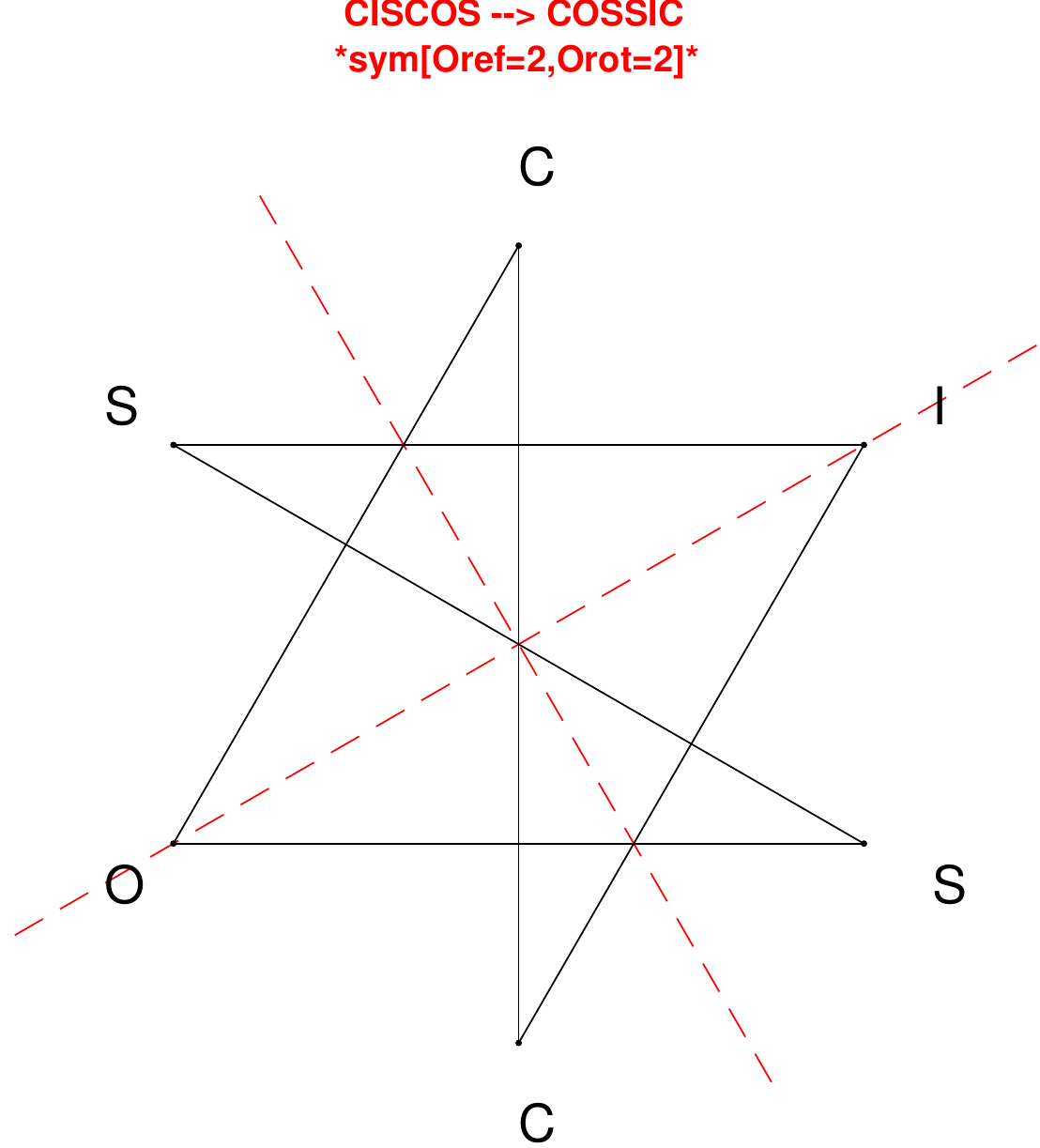}
\end{subfigure}
\hfill
\begin{subfigure}[T]{0.19\textwidth}
\centering
\includegraphics[width=\textwidth]{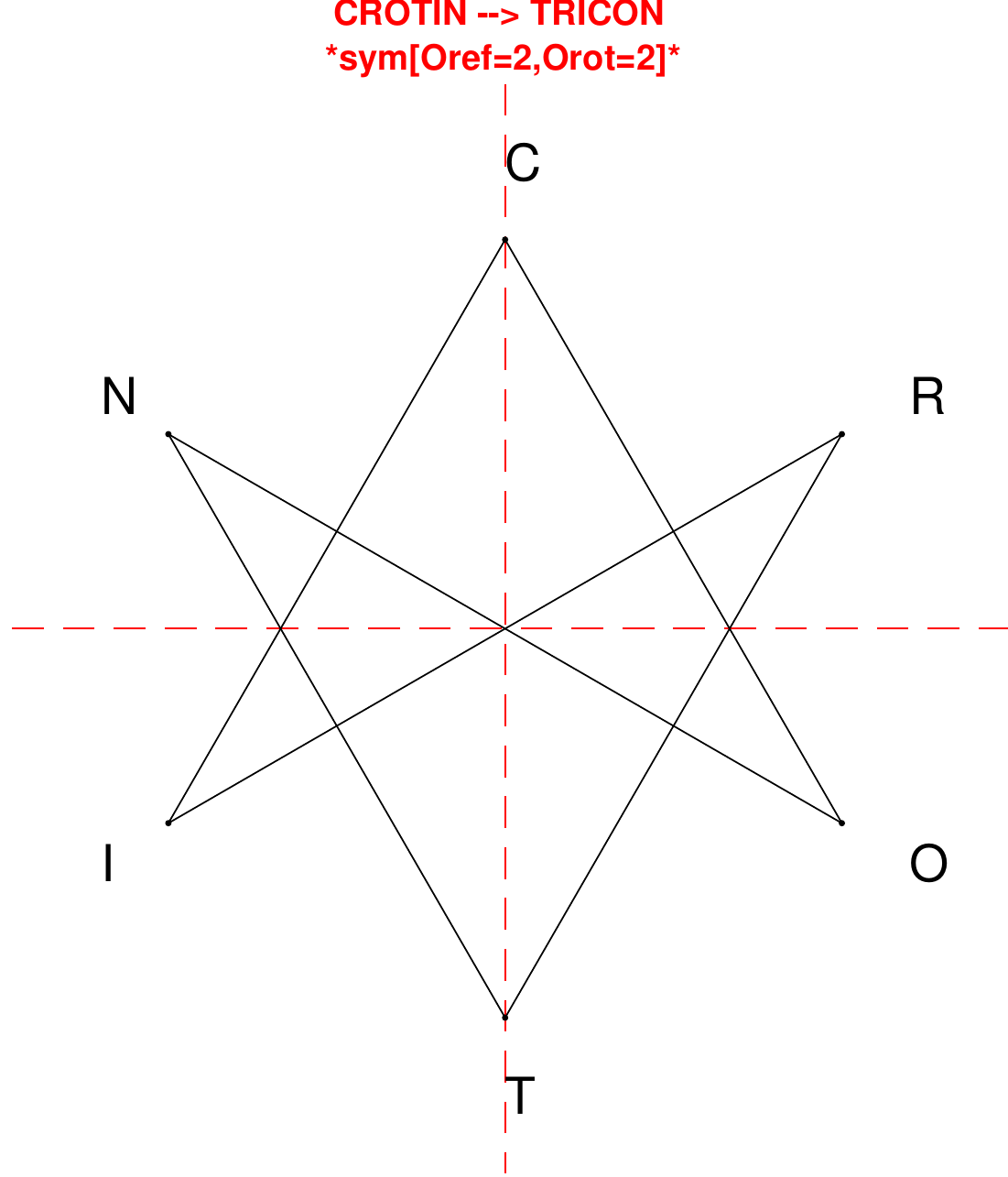}
\end{subfigure}
\end{figure}

\begin{figure}[H]
\centering
\begin{subfigure}[T]{0.19\textwidth}
\centering
\includegraphics[width=\textwidth]{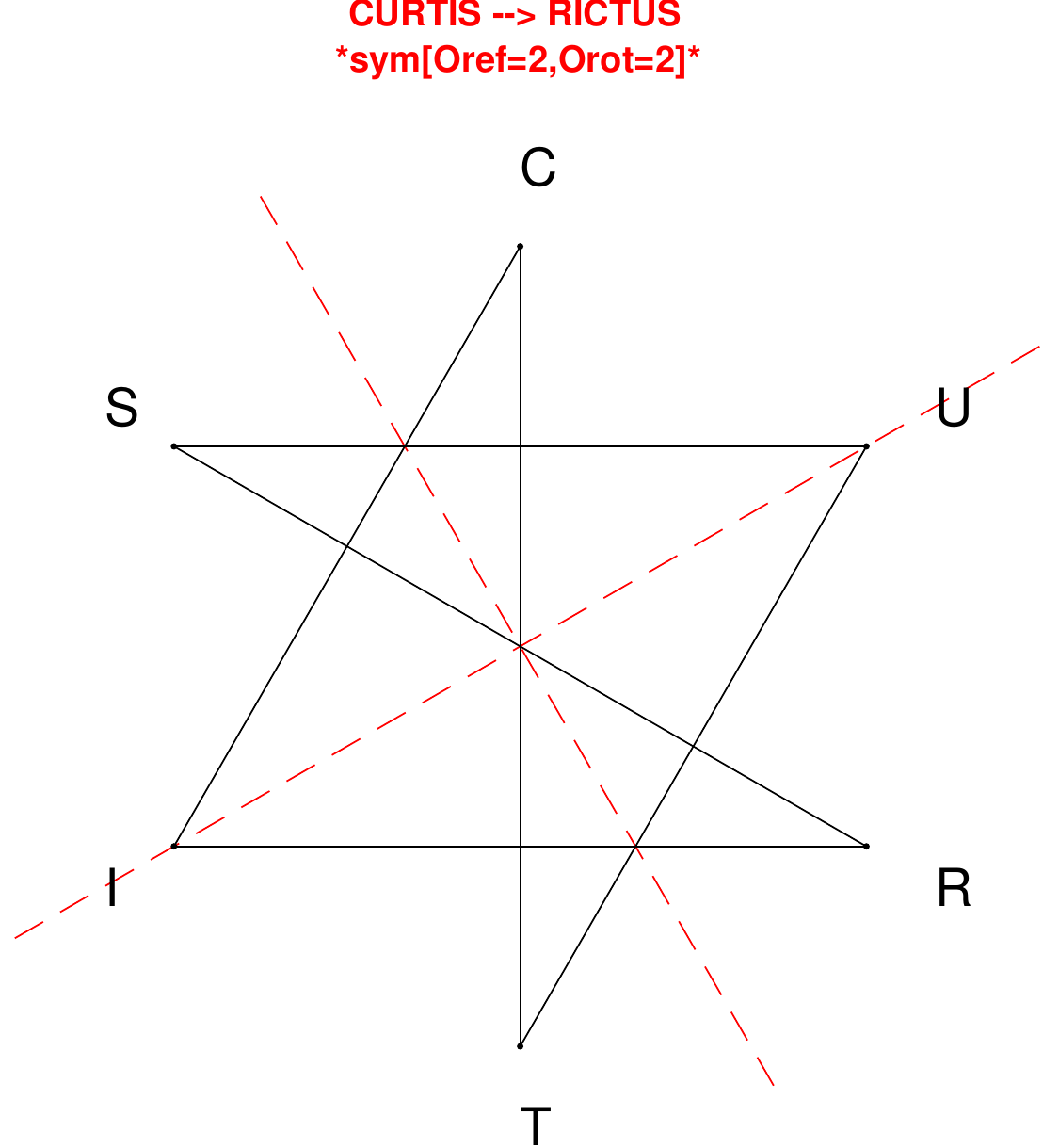}
\end{subfigure}
\hfill
\begin{subfigure}[T]{0.19\textwidth}
\centering
\includegraphics[width=\textwidth]{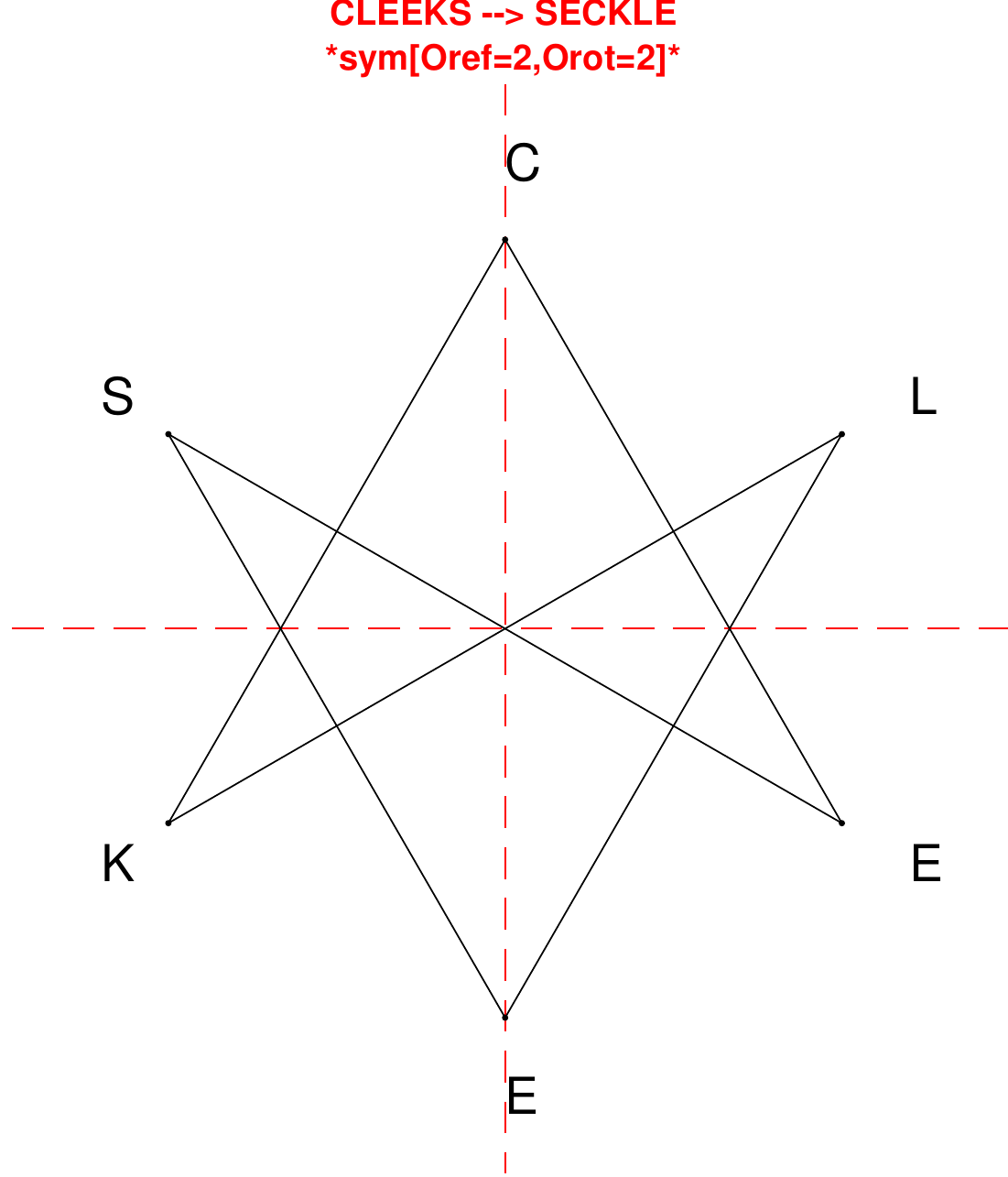}
\end{subfigure}
\hfill
\begin{subfigure}[T]{0.19\textwidth}
\centering
\includegraphics[width=\textwidth]{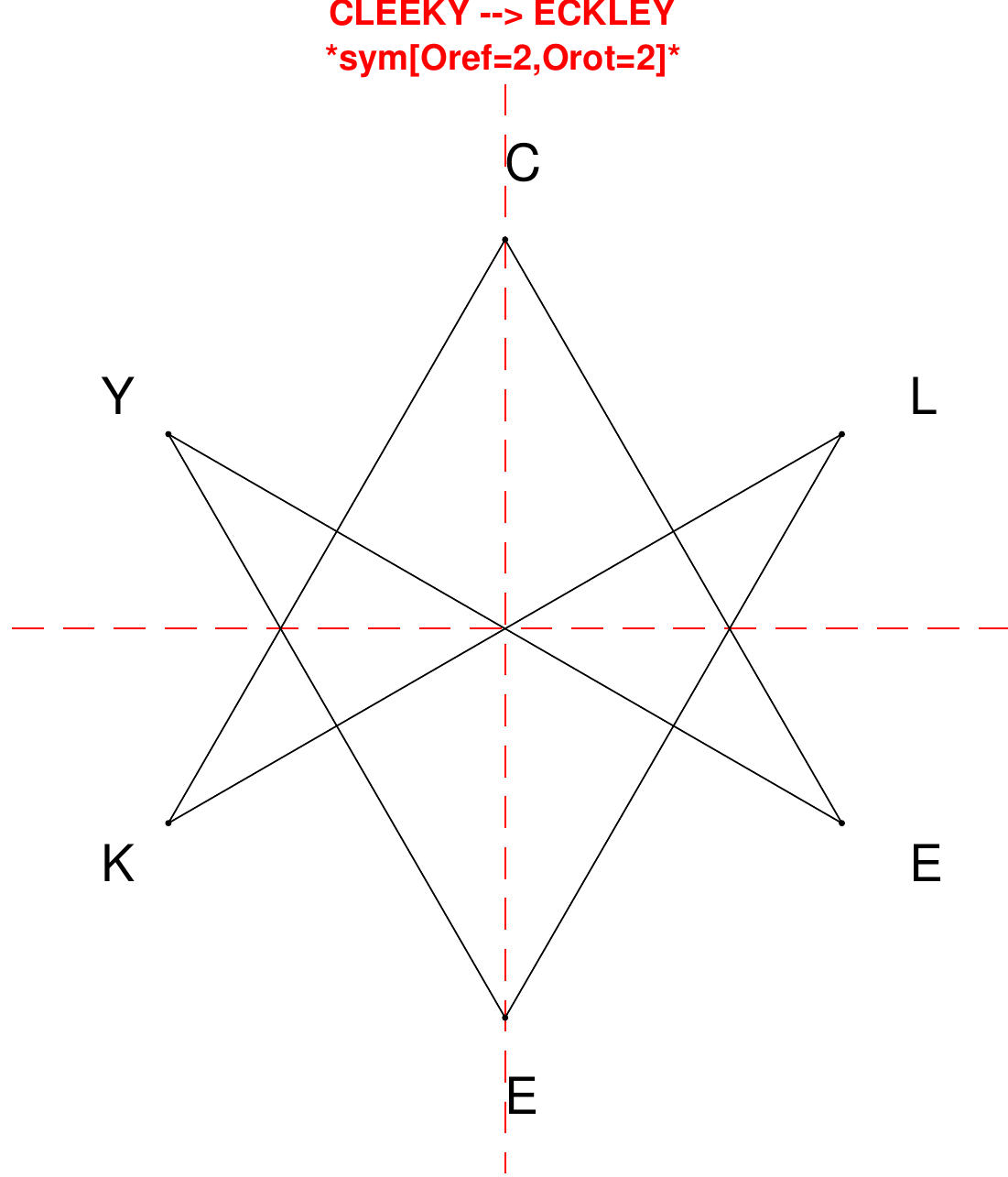}
\end{subfigure}
\hfill
\begin{subfigure}[T]{0.19\textwidth}
\centering
\includegraphics[width=\textwidth]{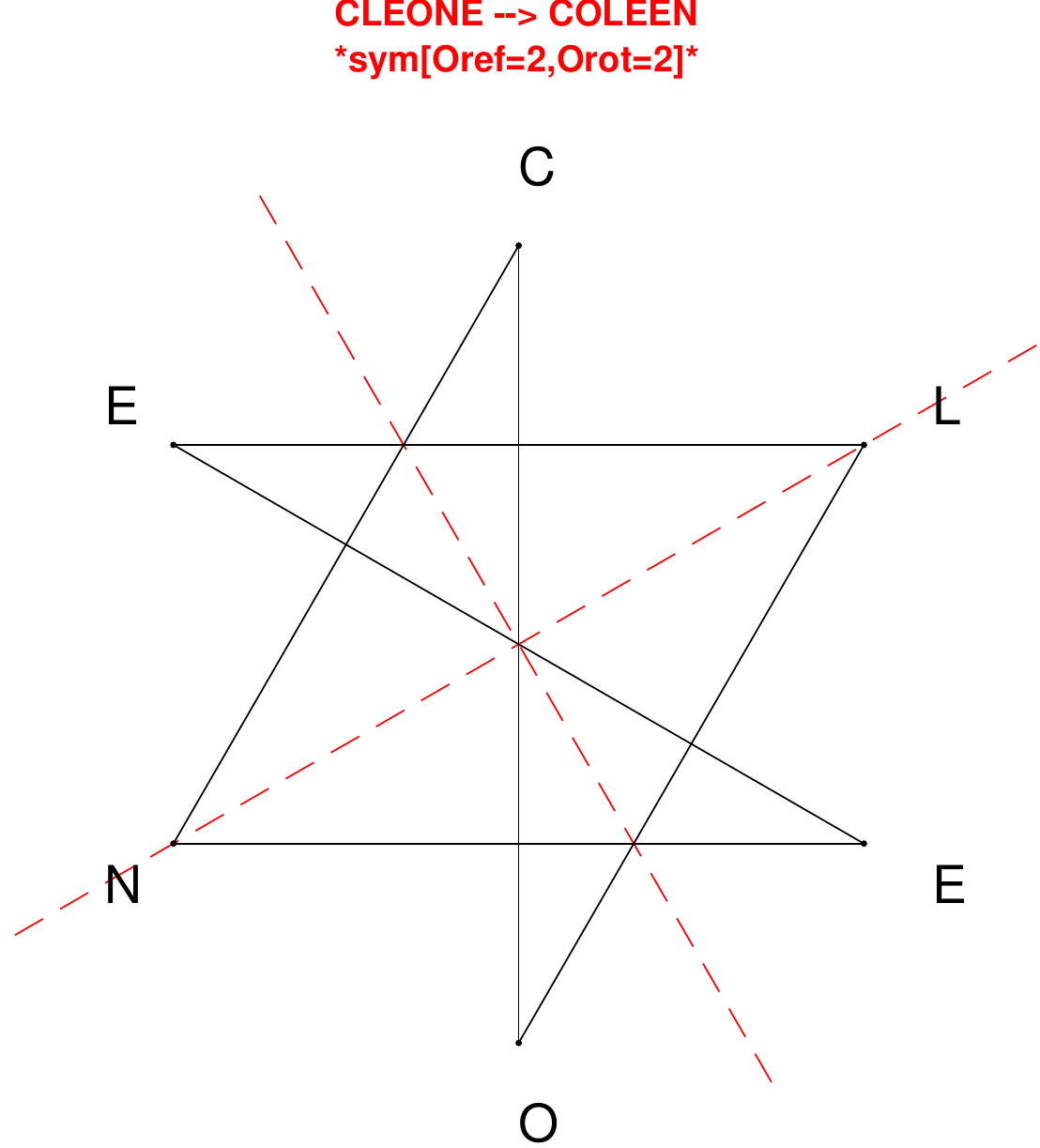}
\end{subfigure}
\hfill
\begin{subfigure}[T]{0.19\textwidth}
\centering
\includegraphics[width=\textwidth]{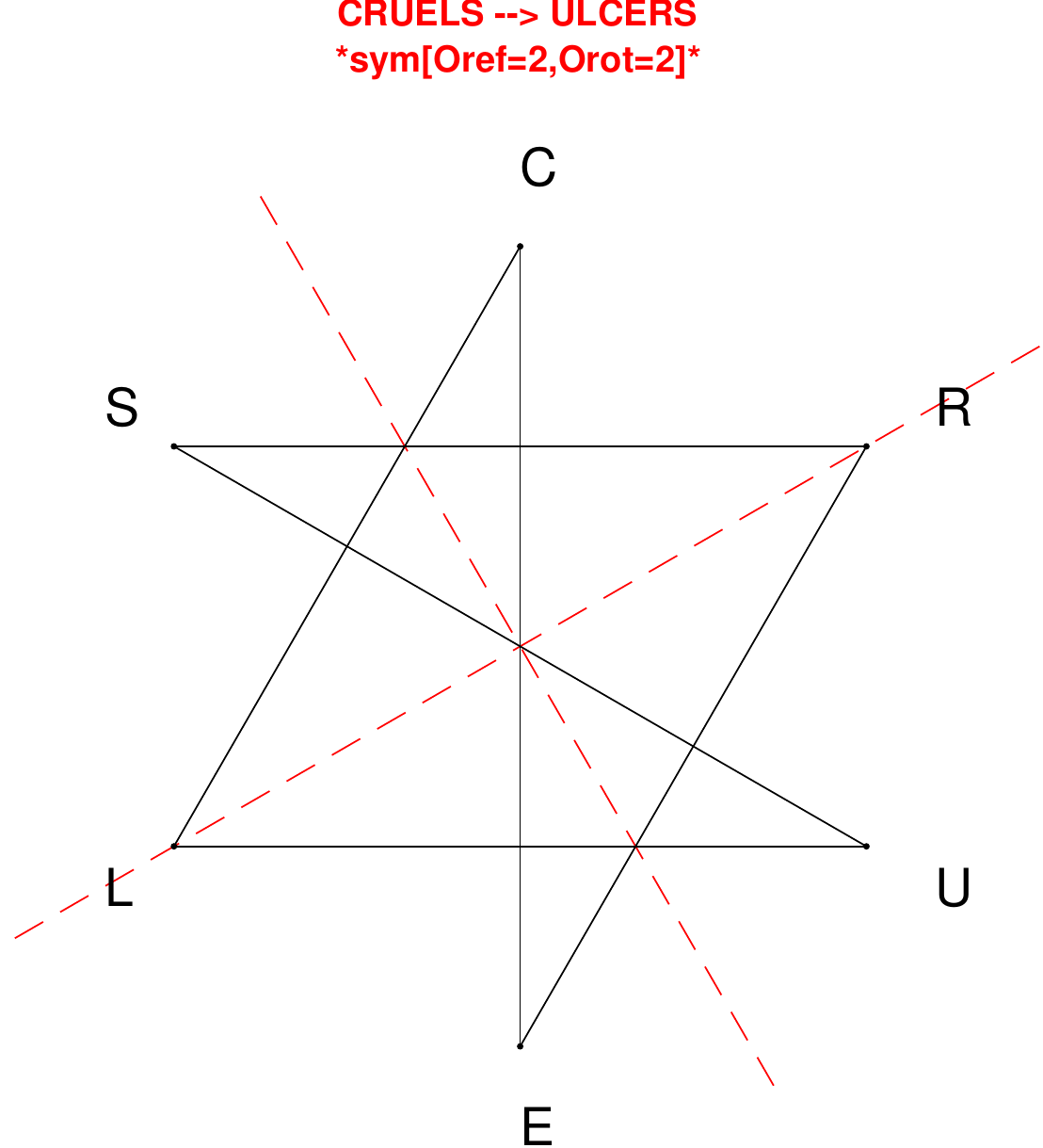}
\end{subfigure}
\end{figure}

\begin{figure}[H]
\centering
\begin{subfigure}[T]{0.19\textwidth}
\centering
\includegraphics[width=\textwidth]{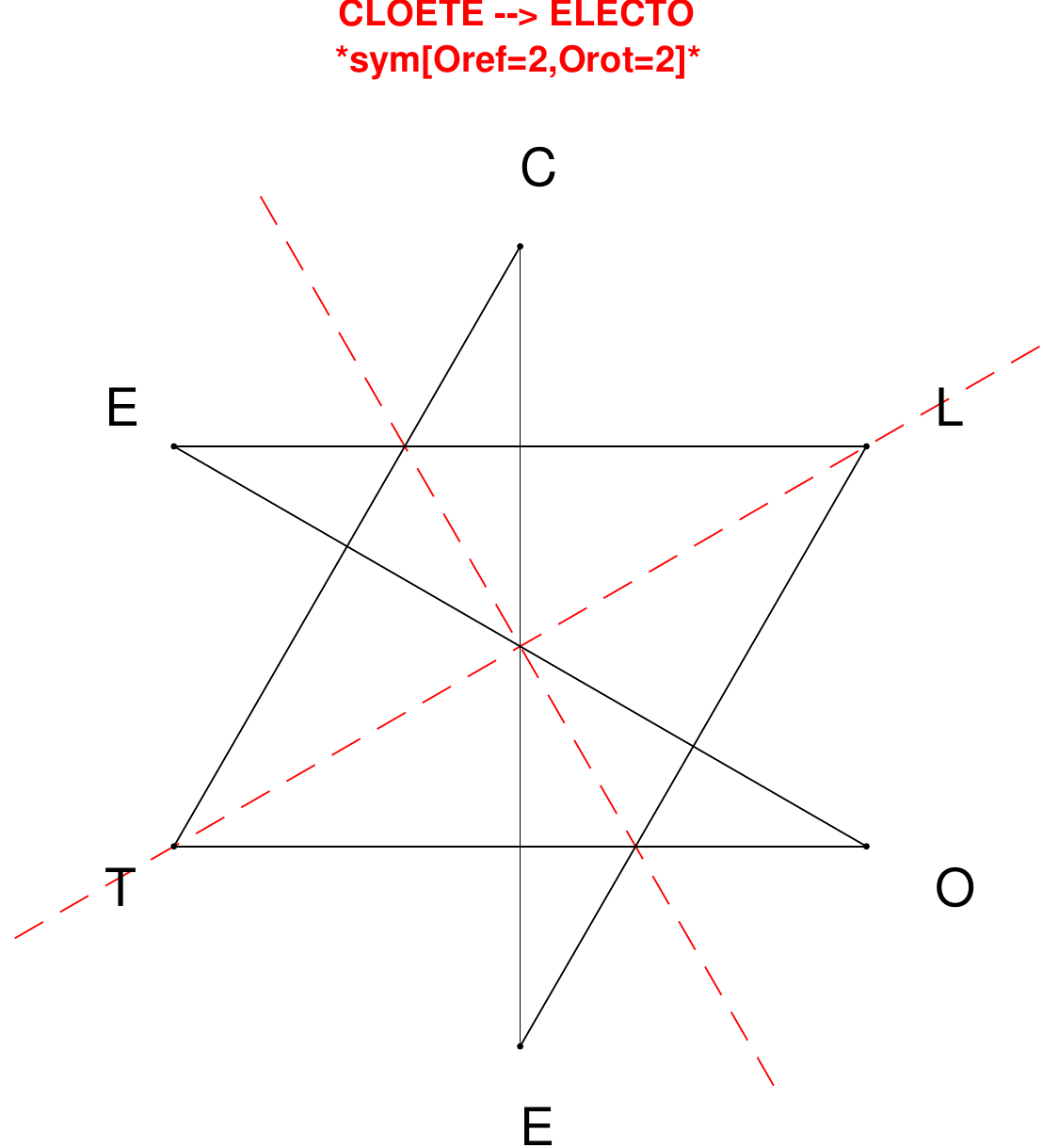}
\end{subfigure}
\hfill
\begin{subfigure}[T]{0.19\textwidth}
\centering
\includegraphics[width=\textwidth]{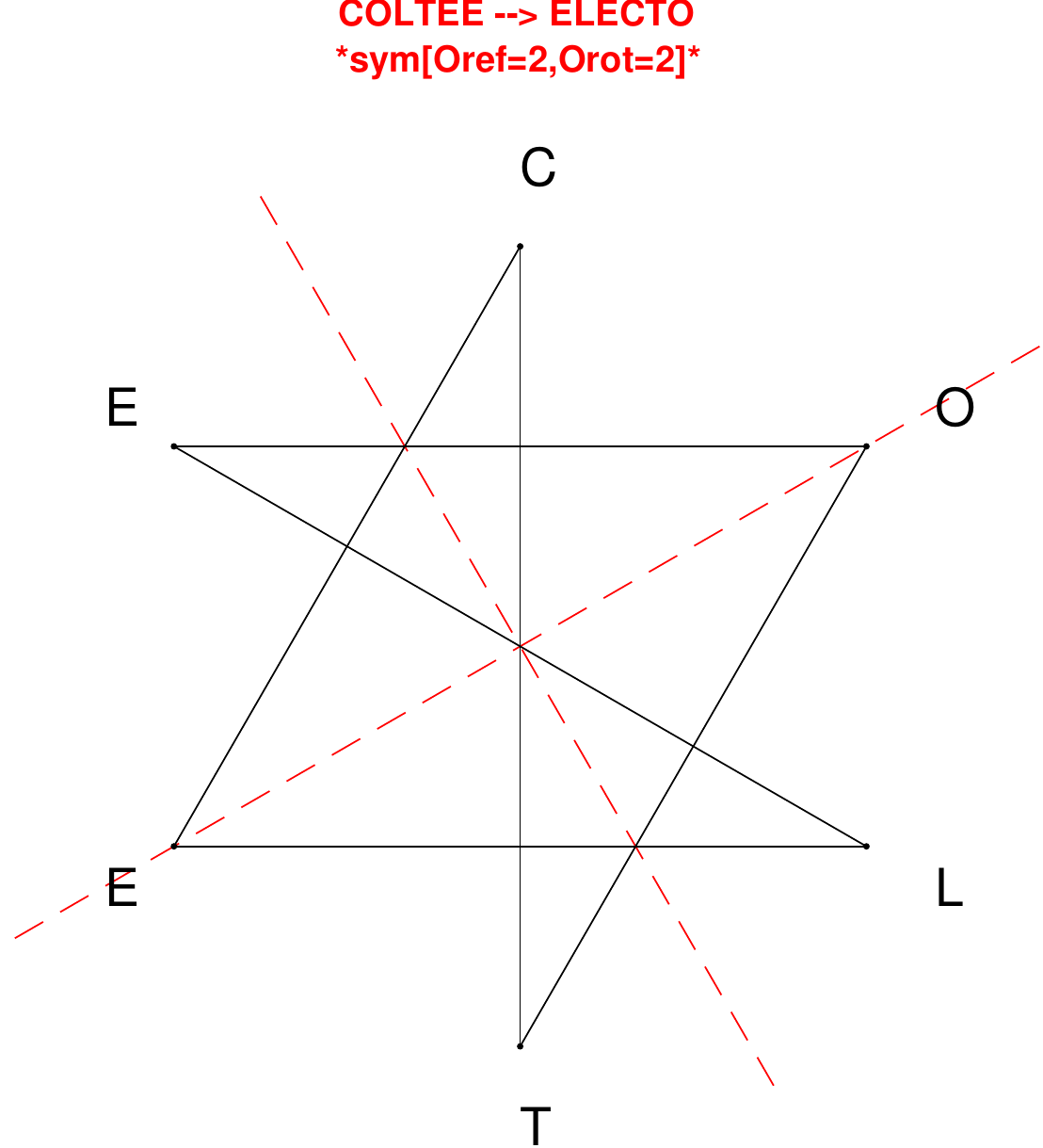}
\end{subfigure}
\hfill
\begin{subfigure}[T]{0.19\textwidth}
\centering
\includegraphics[width=\textwidth]{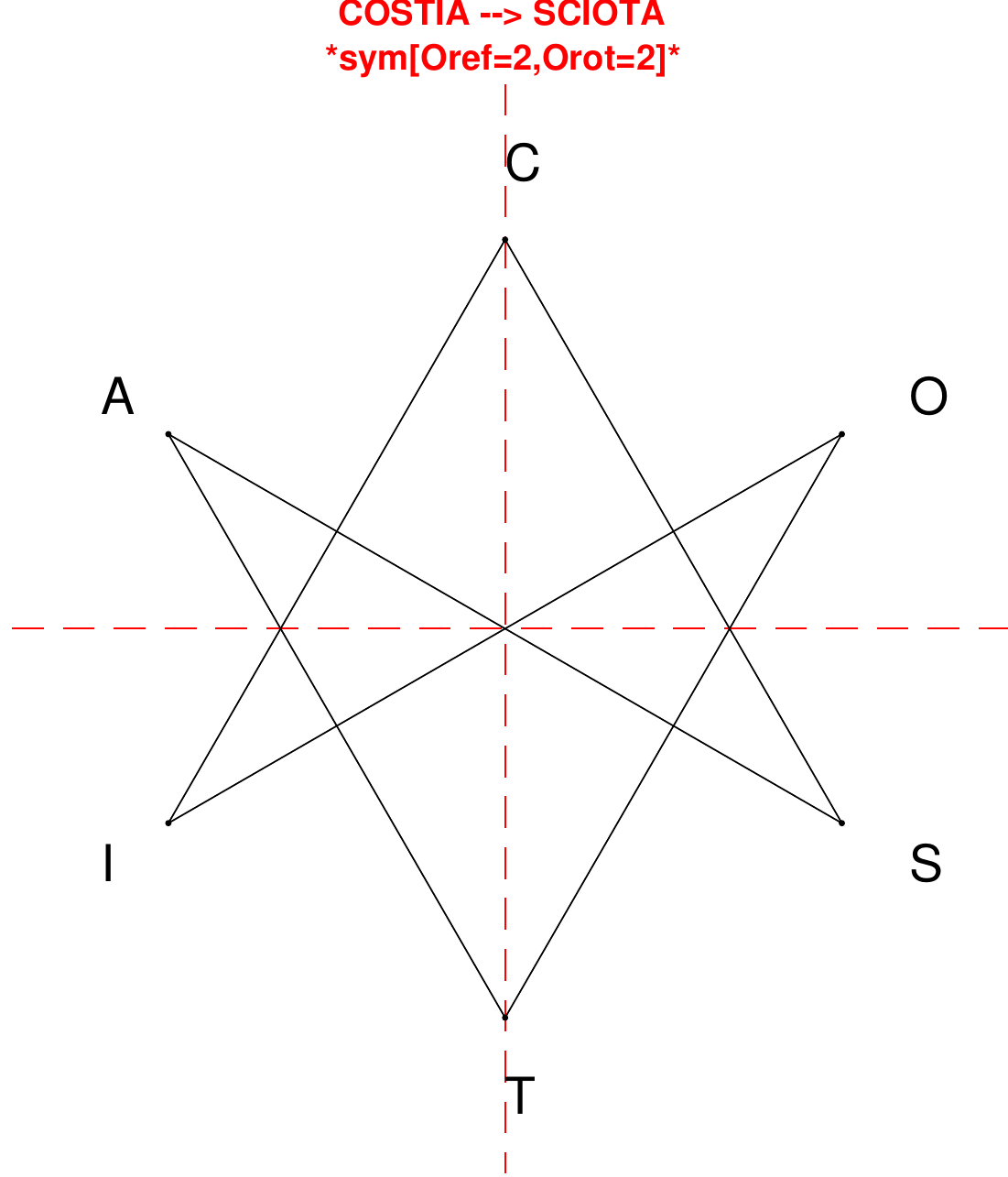}
\end{subfigure}
\hfill
\begin{subfigure}[T]{0.19\textwidth}
\centering
\includegraphics[width=\textwidth]{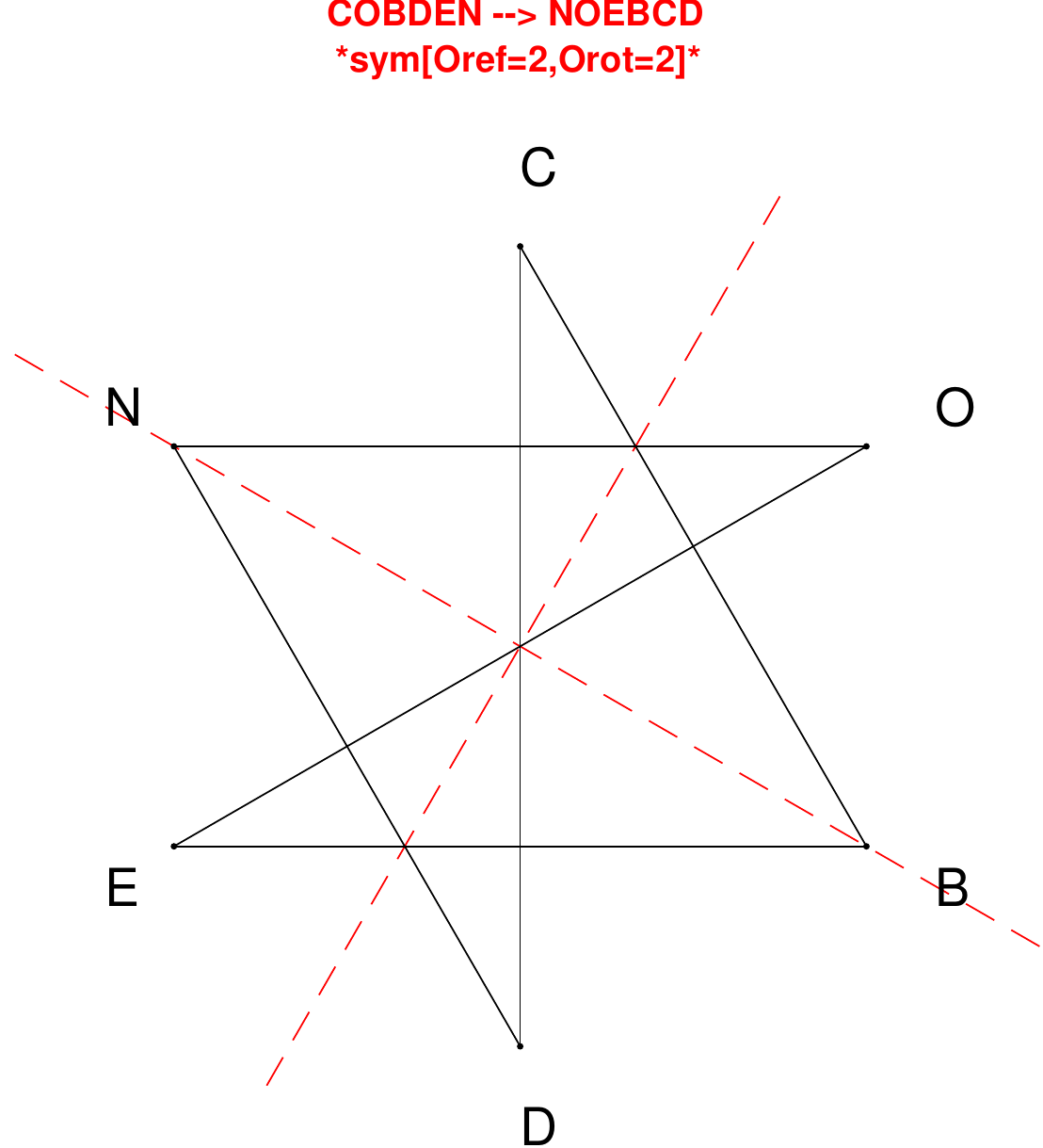}
\end{subfigure}
\hfill
\begin{subfigure}[T]{0.19\textwidth}
\centering
\includegraphics[width=\textwidth]{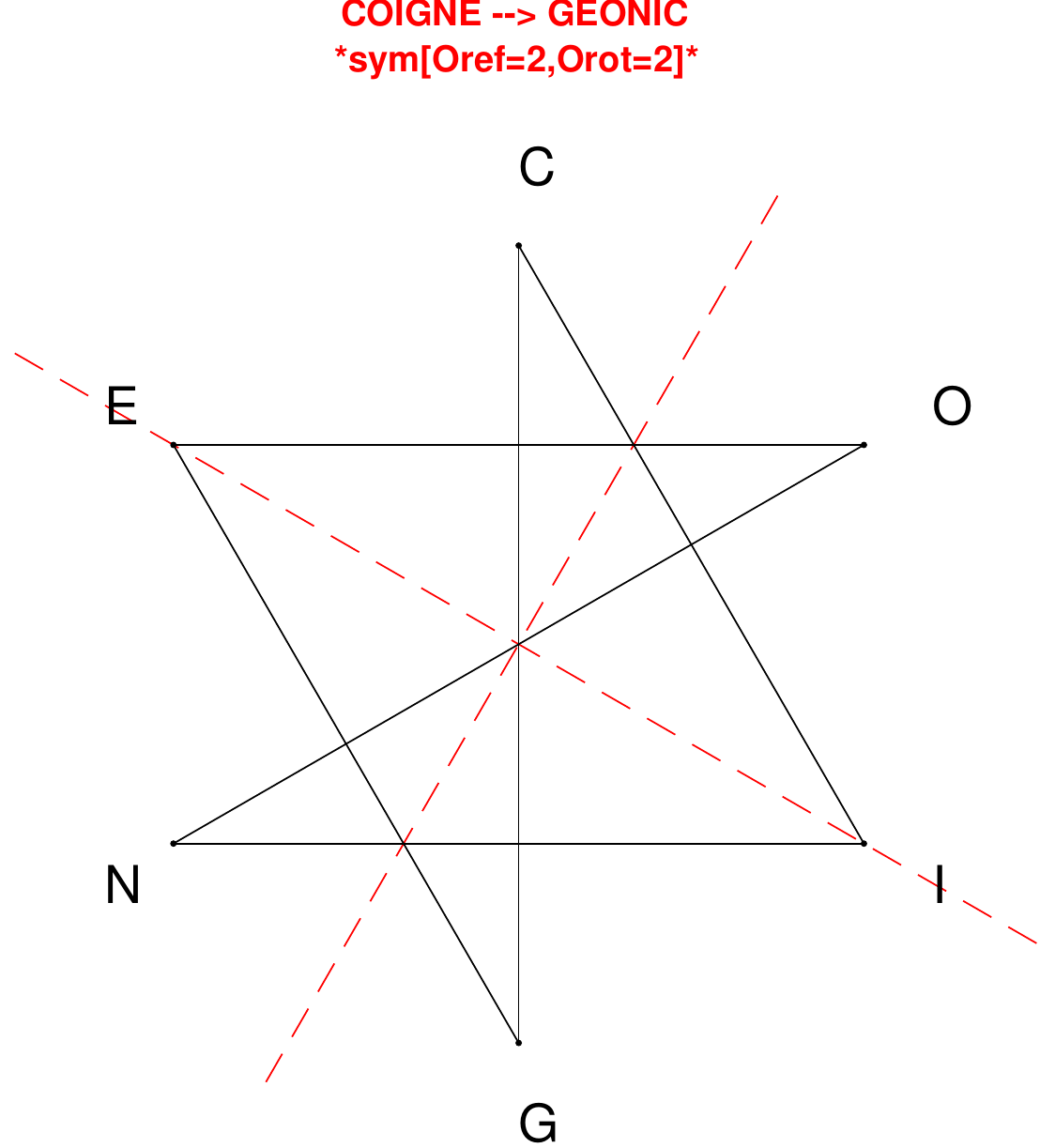}
\end{subfigure}
\end{figure}

\begin{figure}[H]
\centering
\begin{subfigure}[T]{0.19\textwidth}
\centering
\includegraphics[width=\textwidth]{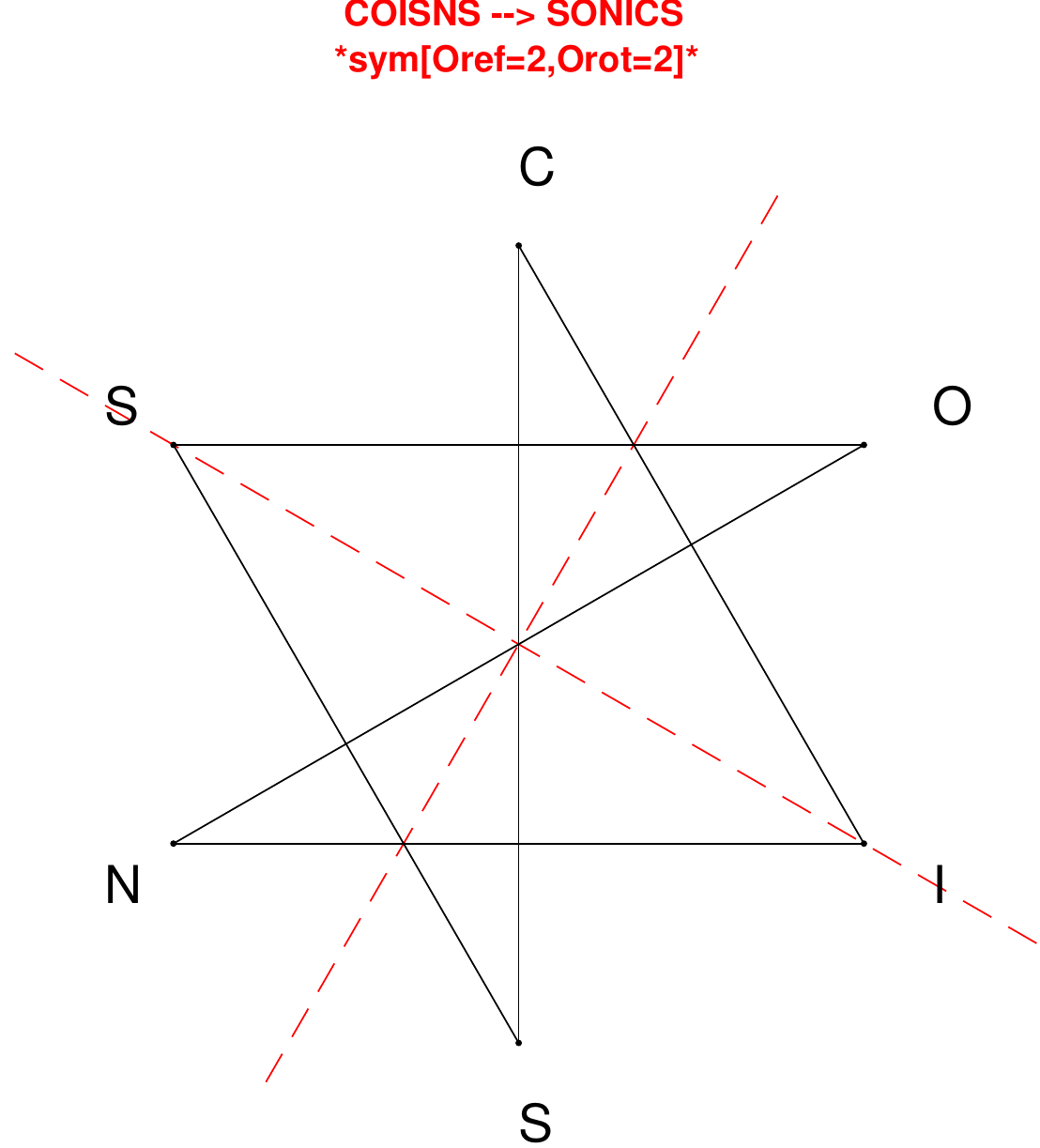}
\end{subfigure}
\hfill
\begin{subfigure}[T]{0.19\textwidth}
\centering
\includegraphics[width=\textwidth]{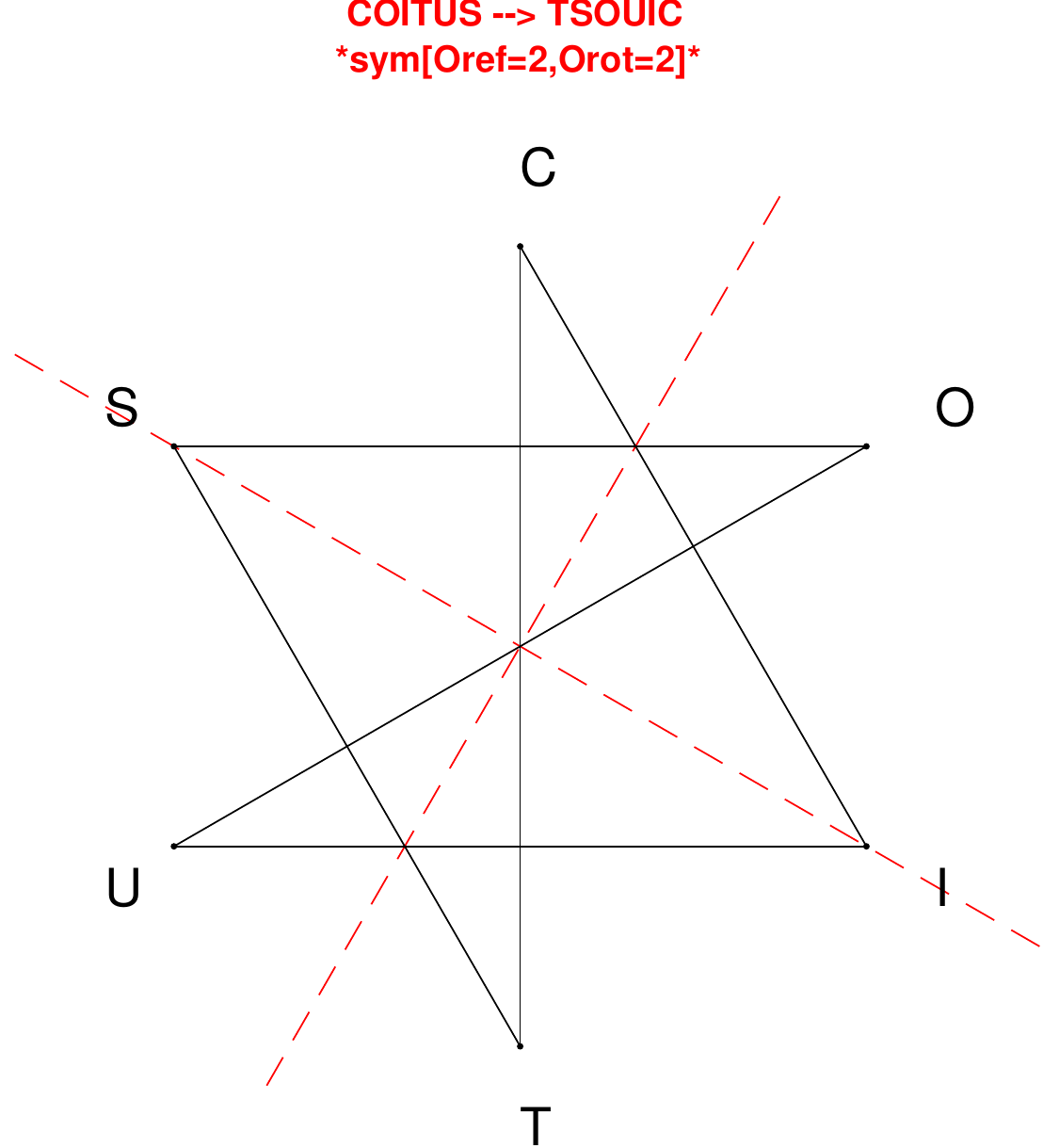}
\end{subfigure}
\hfill
\begin{subfigure}[T]{0.19\textwidth}
\centering
\includegraphics[width=\textwidth]{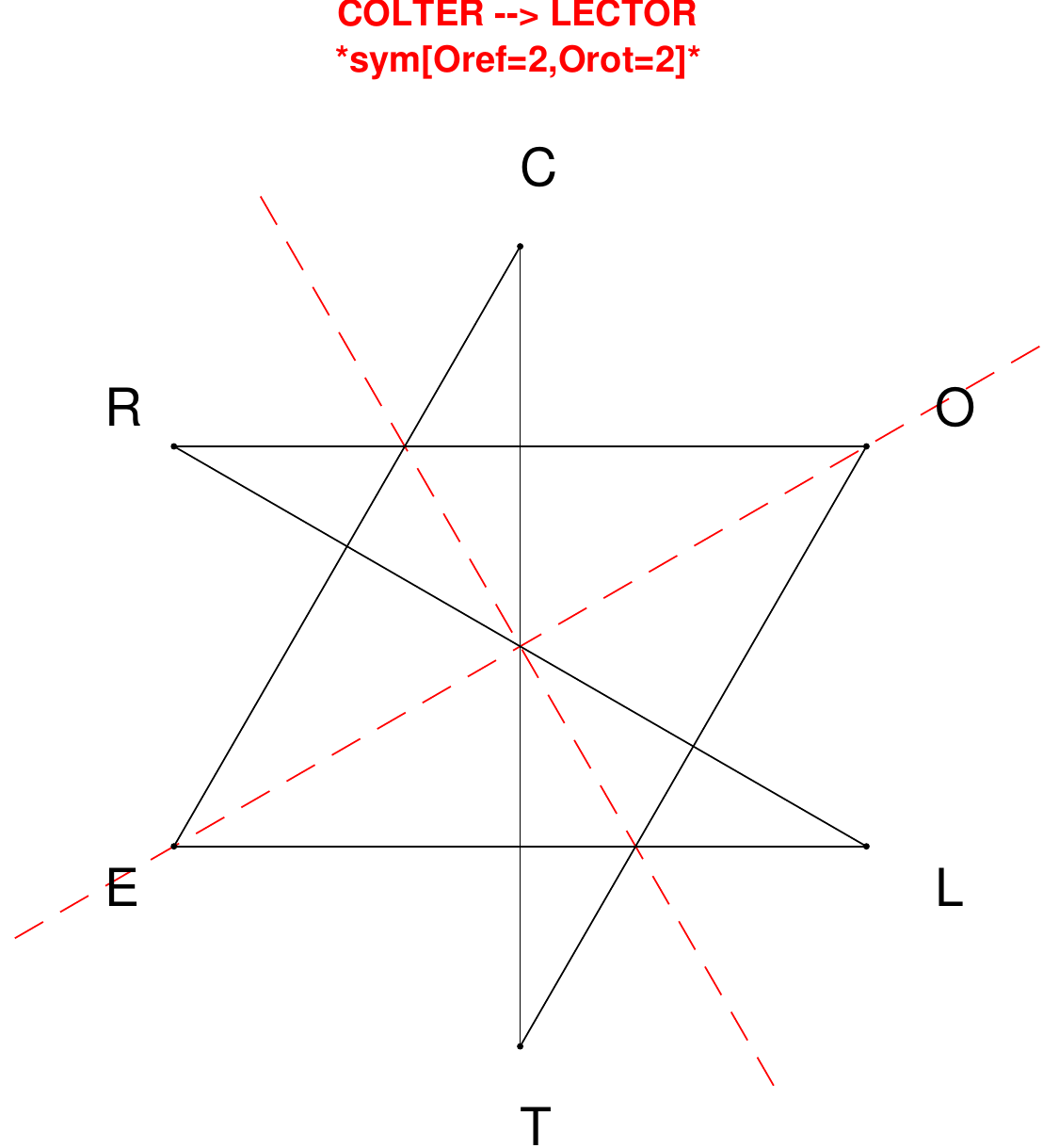}
\end{subfigure}
\hfill
\begin{subfigure}[T]{0.19\textwidth}
\centering
\includegraphics[width=\textwidth]{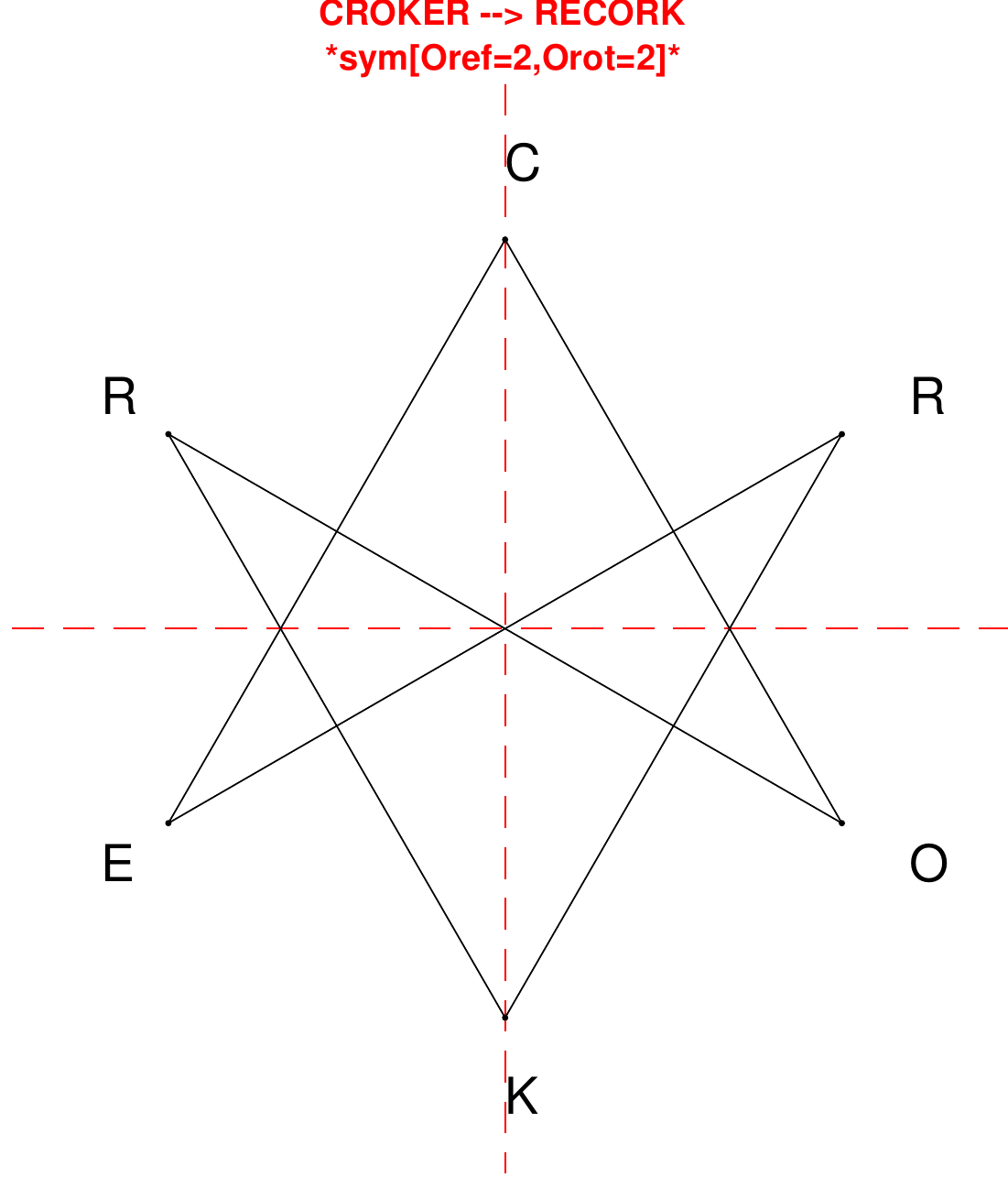}
\end{subfigure}
\hfill
\begin{subfigure}[T]{0.19\textwidth}
\centering
\includegraphics[width=\textwidth]{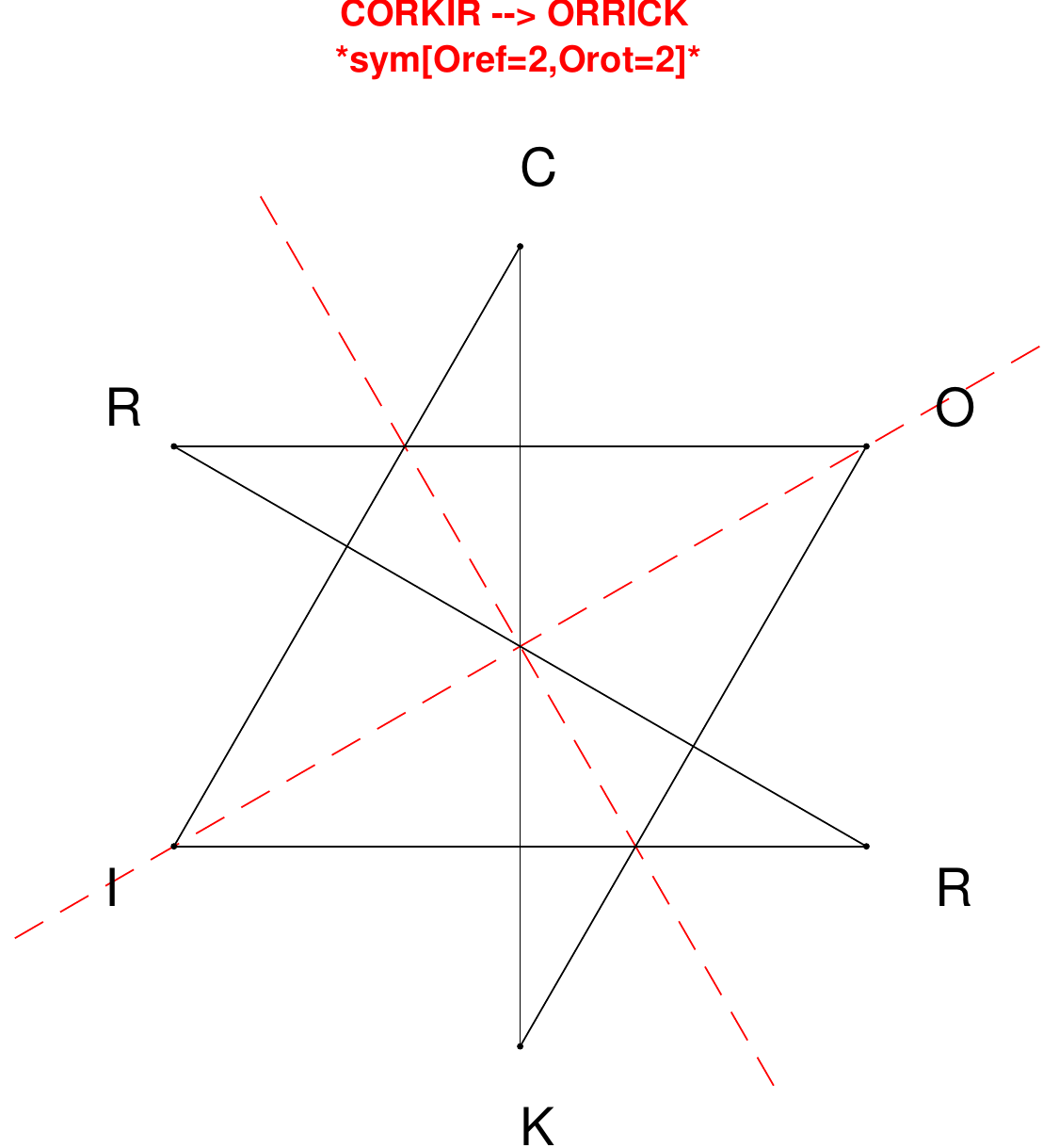}
\end{subfigure}
\end{figure}

\begin{figure}[H]
\centering
\begin{subfigure}[T]{0.19\textwidth}
\centering
\includegraphics[width=\textwidth]{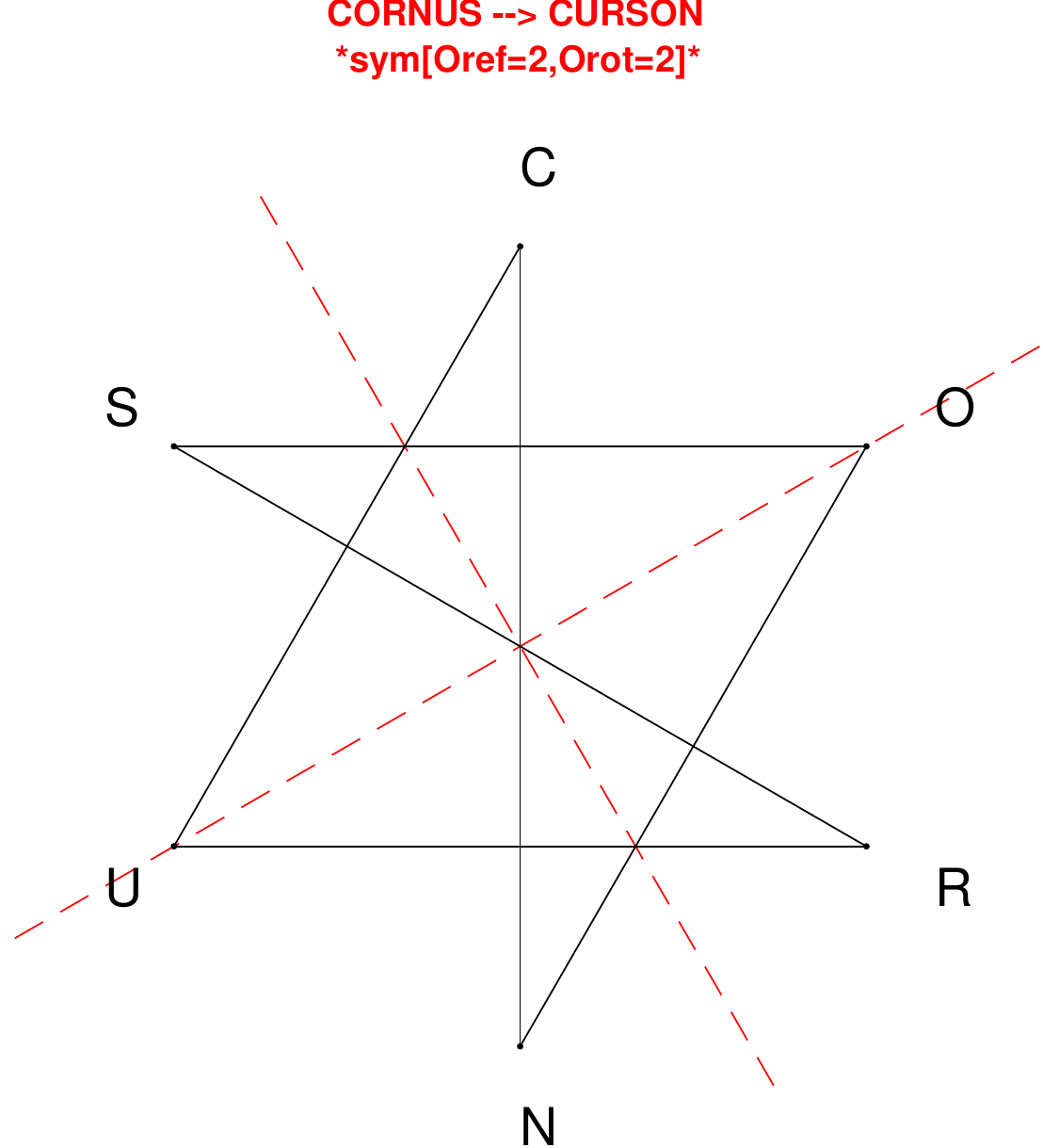}
\end{subfigure}
\hfill
\begin{subfigure}[T]{0.19\textwidth}
\centering
\includegraphics[width=\textwidth]{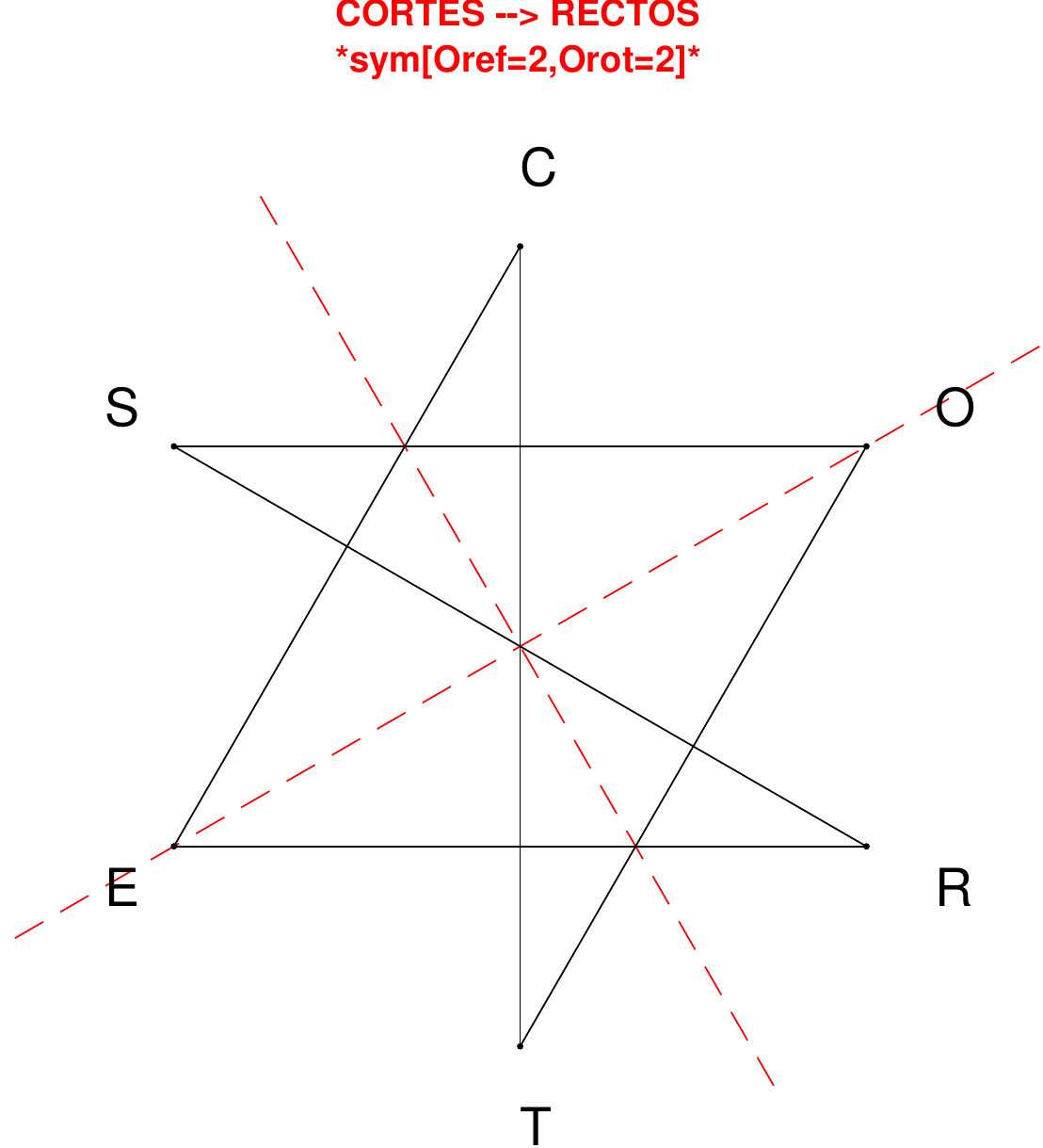}
\end{subfigure}
\hfill
\begin{subfigure}[T]{0.19\textwidth}
\centering
\includegraphics[width=\textwidth]{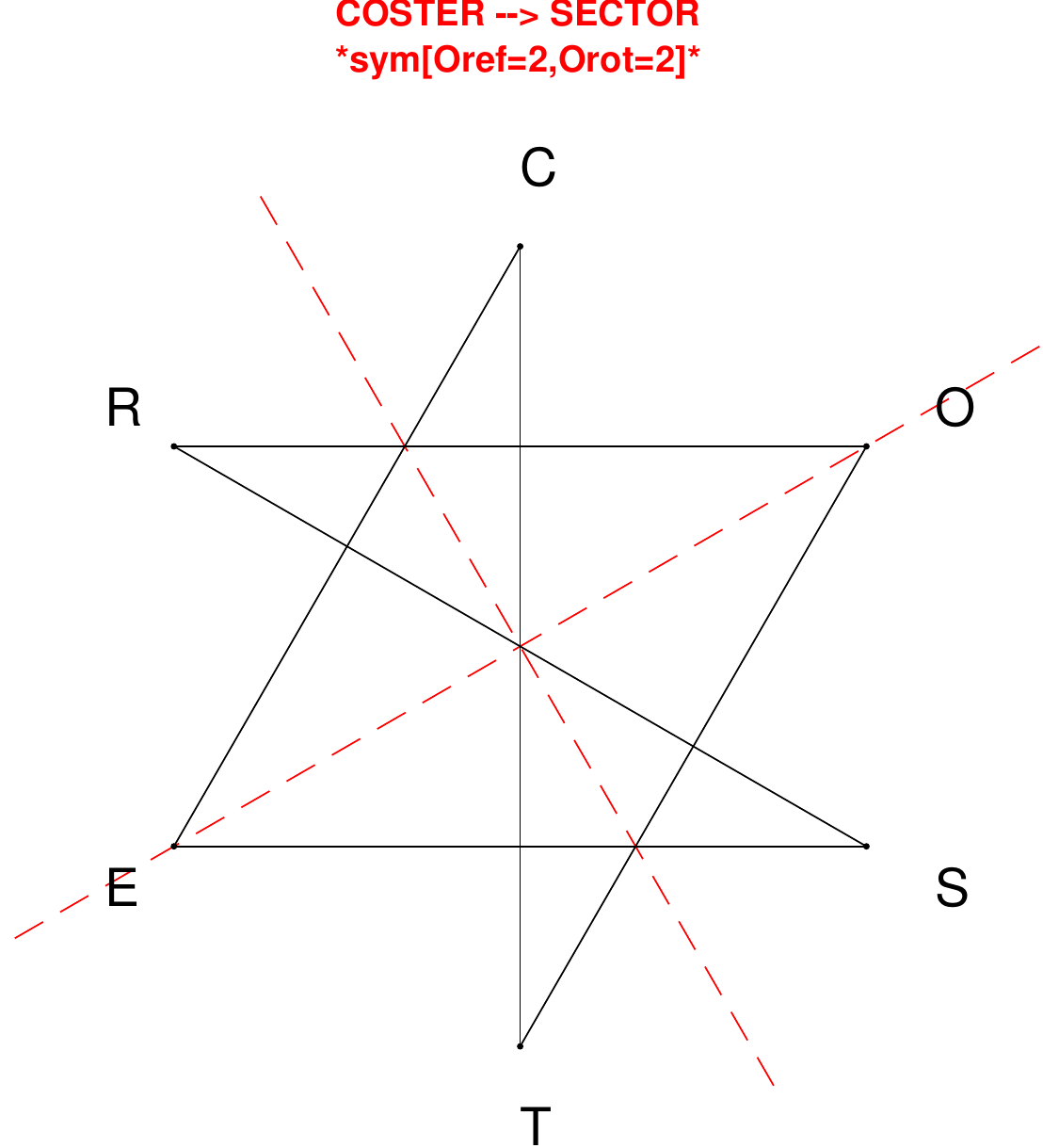}
\end{subfigure}
\hfill
\begin{subfigure}[T]{0.19\textwidth}
\centering
\includegraphics[width=\textwidth]{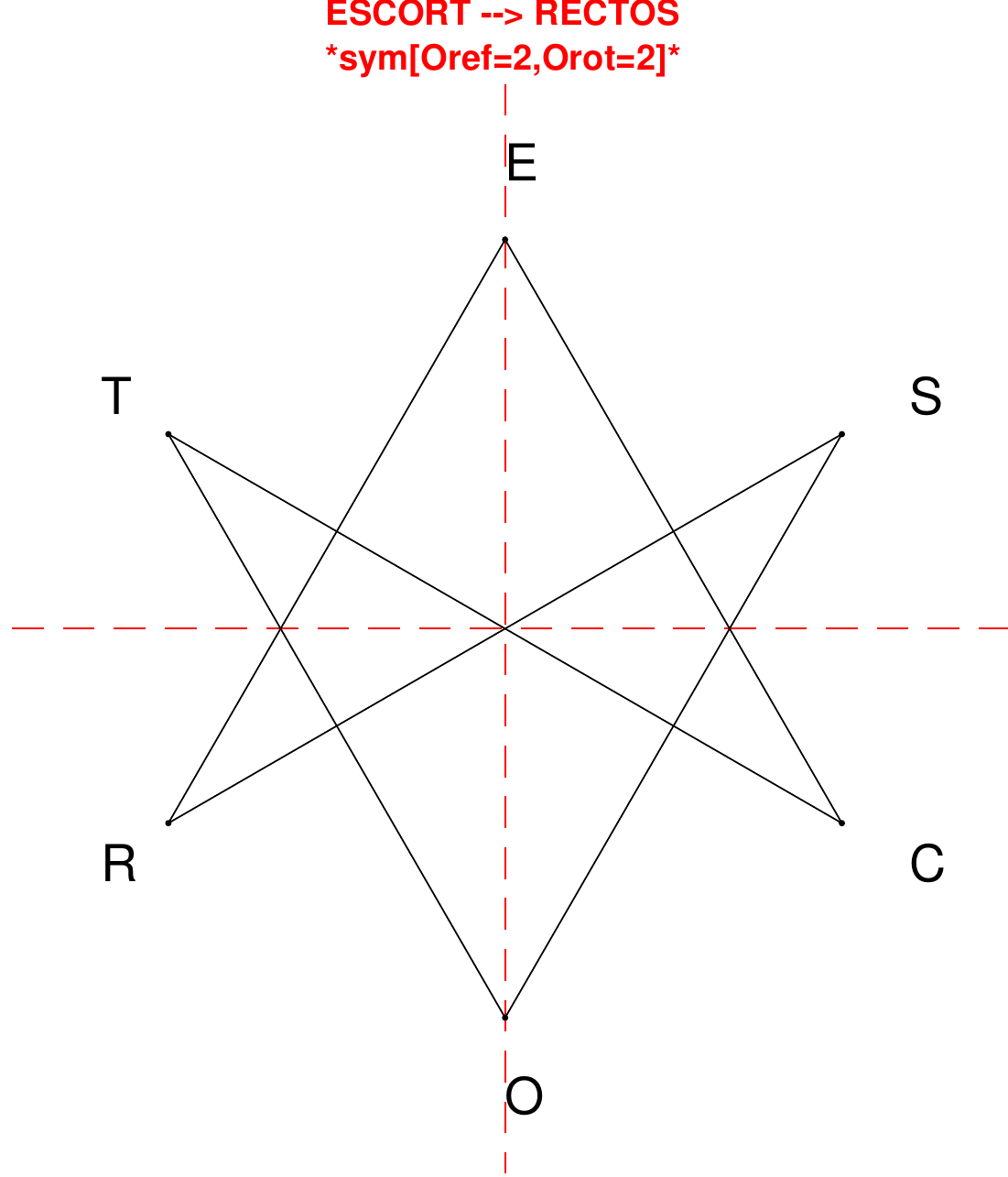}
\end{subfigure}
\hfill
\begin{subfigure}[T]{0.19\textwidth}
\centering
\includegraphics[width=\textwidth]{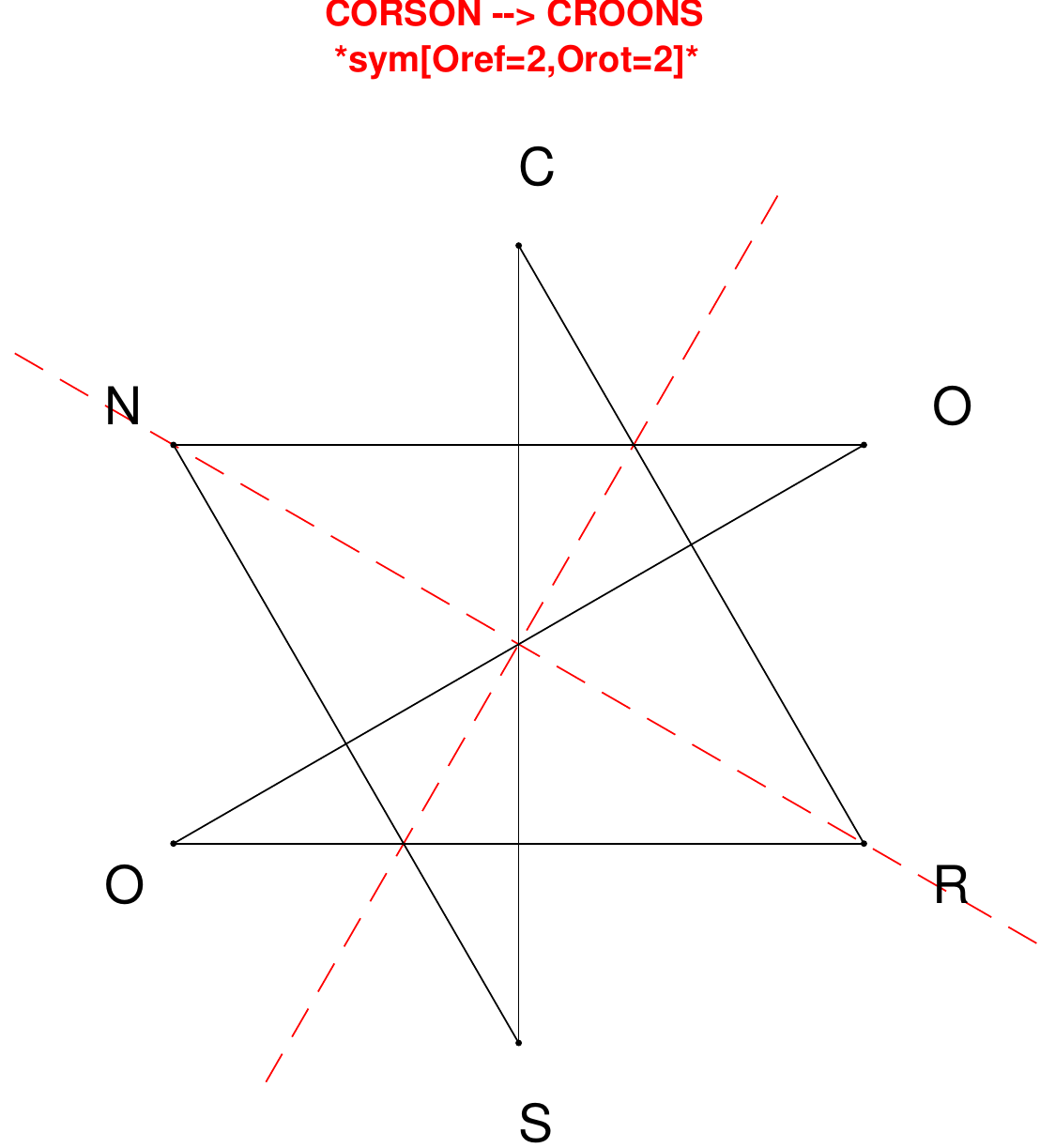}
\end{subfigure}
\end{figure}

\begin{figure}[H]
\centering
\begin{subfigure}[T]{0.19\textwidth}
\centering
\includegraphics[width=\textwidth]{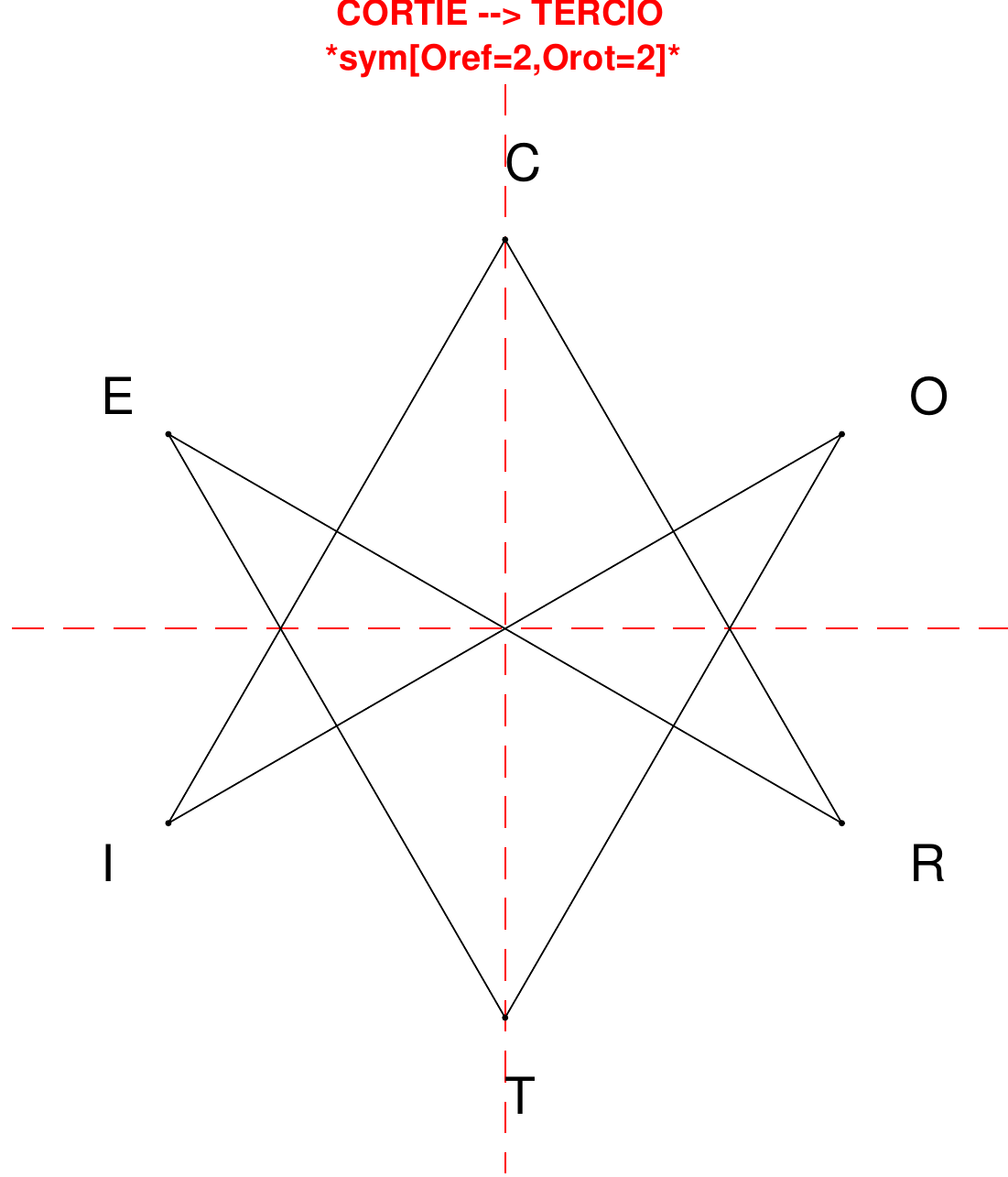}
\end{subfigure}
\hfill
\begin{subfigure}[T]{0.19\textwidth}
\centering
\includegraphics[width=\textwidth]{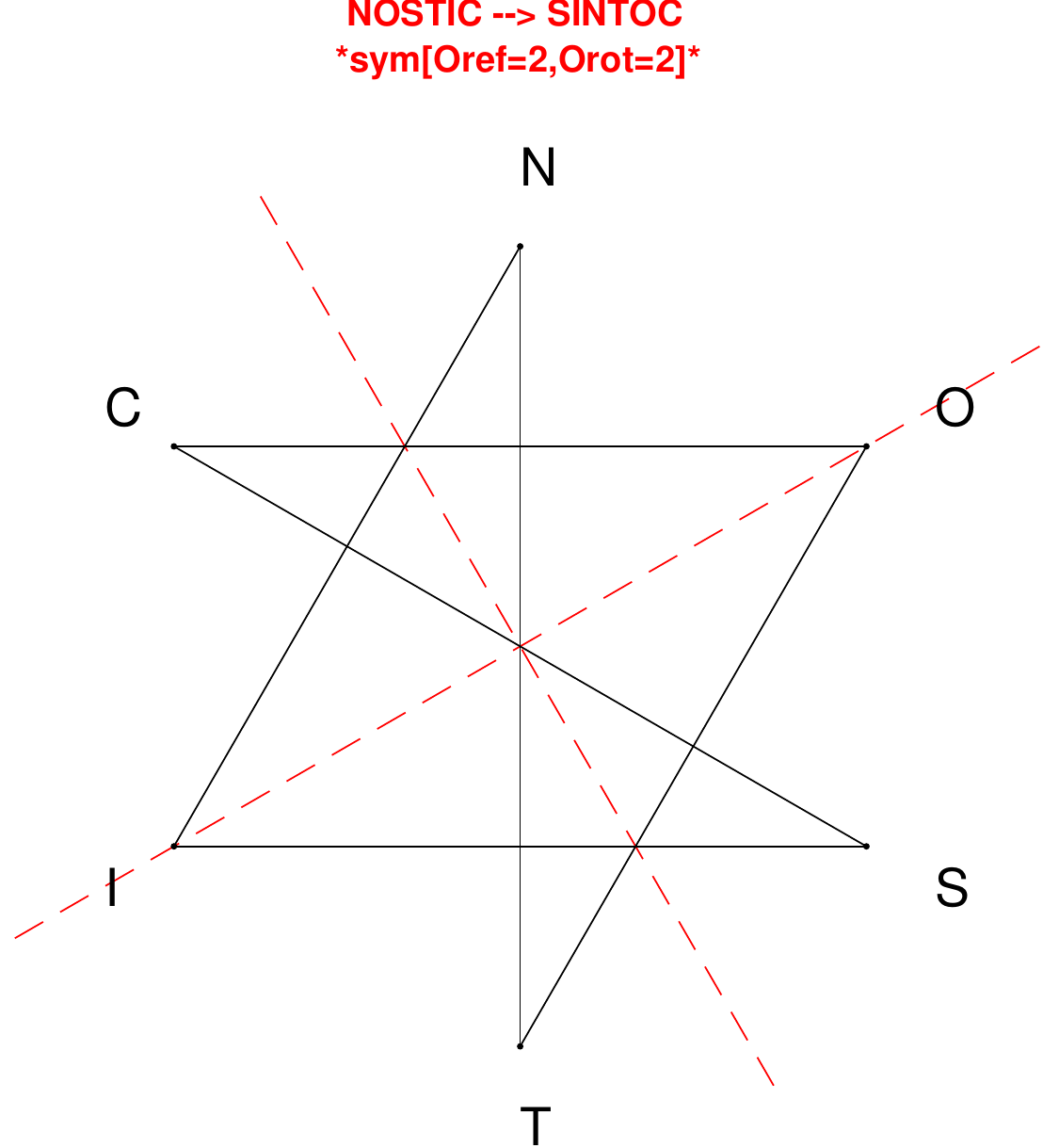}
\end{subfigure}
\hfill
\begin{subfigure}[T]{0.19\textwidth}
\centering
\includegraphics[width=\textwidth]{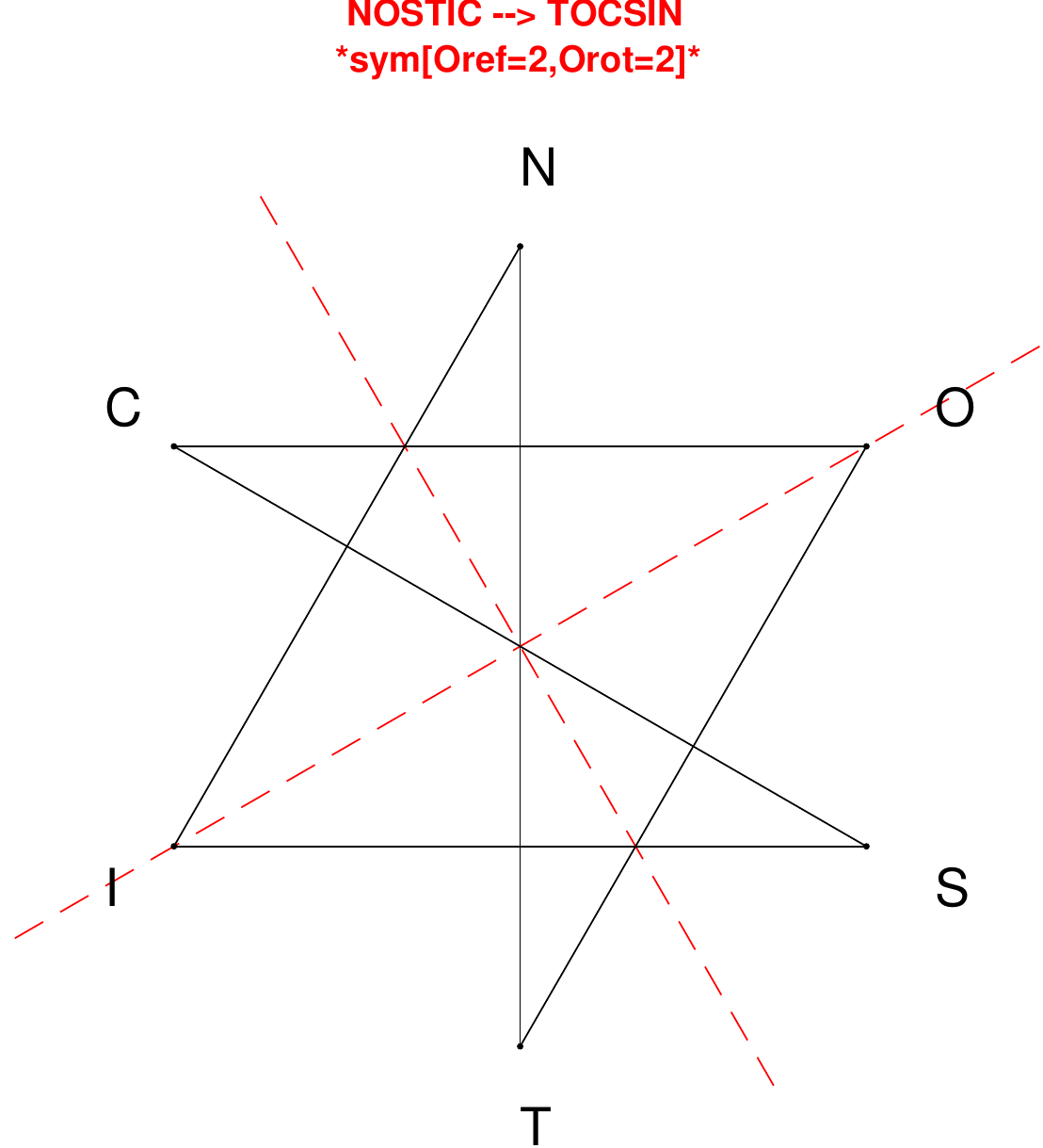}
\end{subfigure}
\hfill
\begin{subfigure}[T]{0.19\textwidth}
\centering
\includegraphics[width=\textwidth]{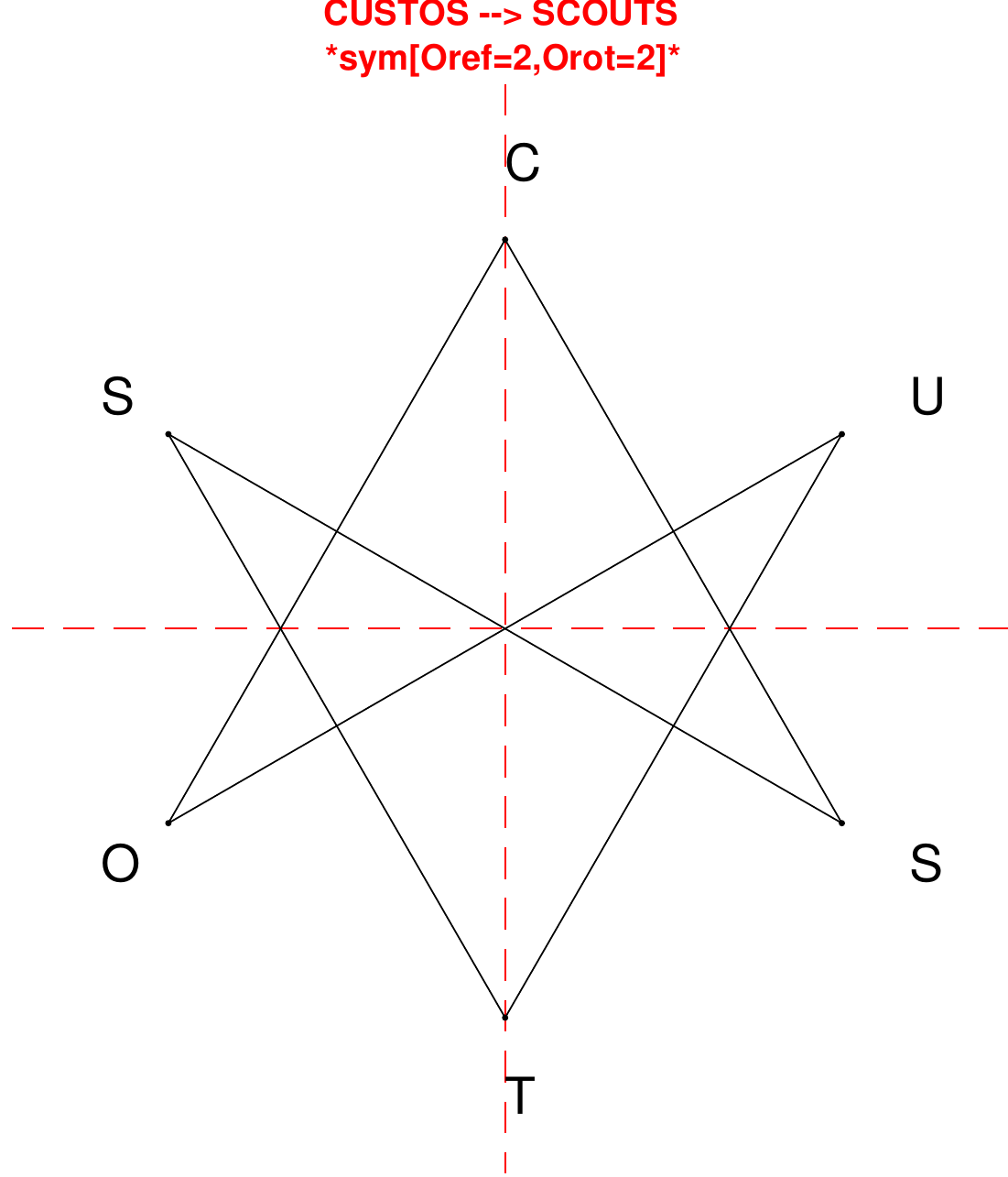}
\end{subfigure}
\hfill
\begin{subfigure}[T]{0.19\textwidth}
\centering
\includegraphics[width=\textwidth]{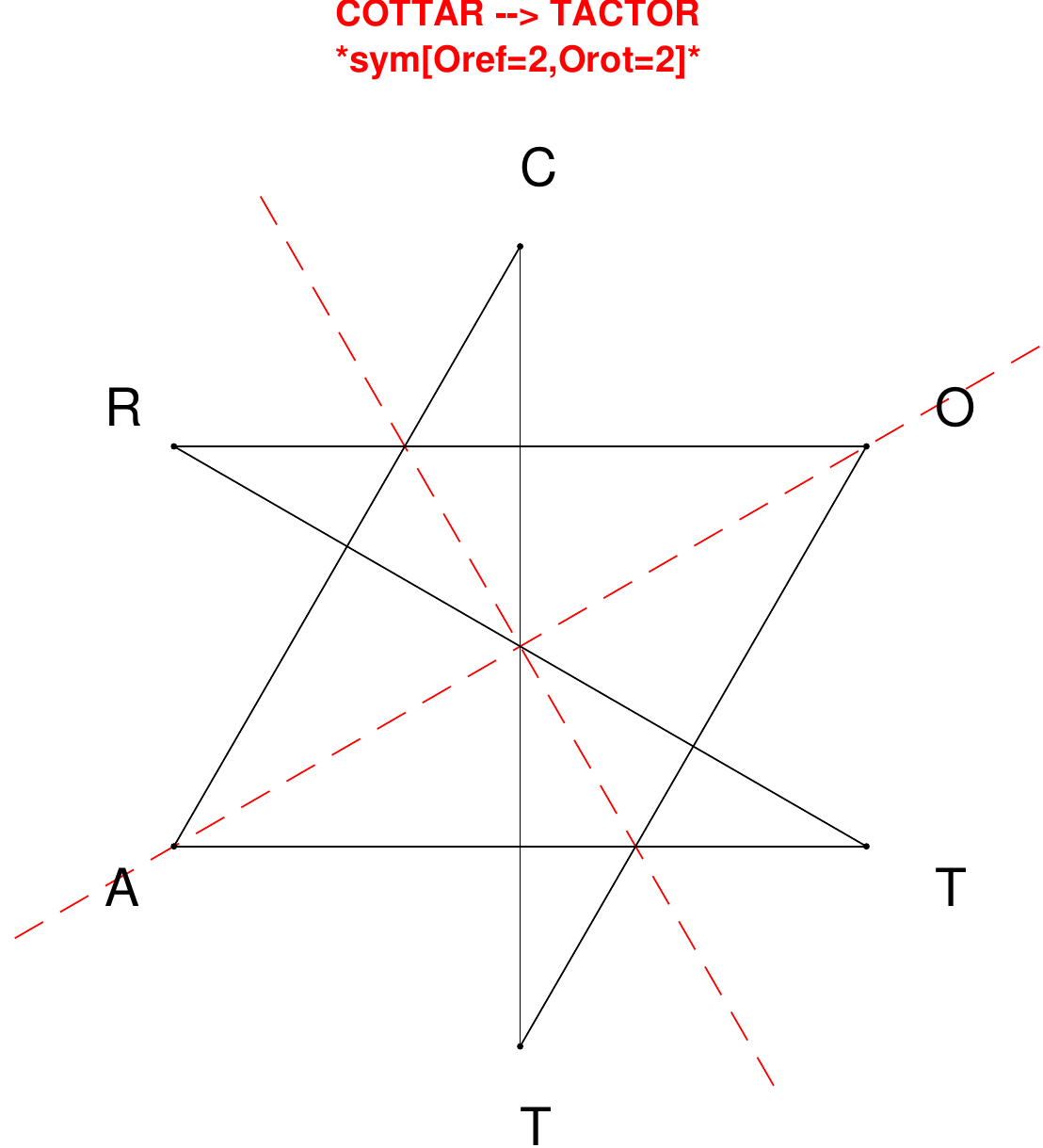}
\end{subfigure}
\end{figure}

\begin{figure}[H]
\centering
\begin{subfigure}[T]{0.19\textwidth}
\centering
\includegraphics[width=\textwidth]{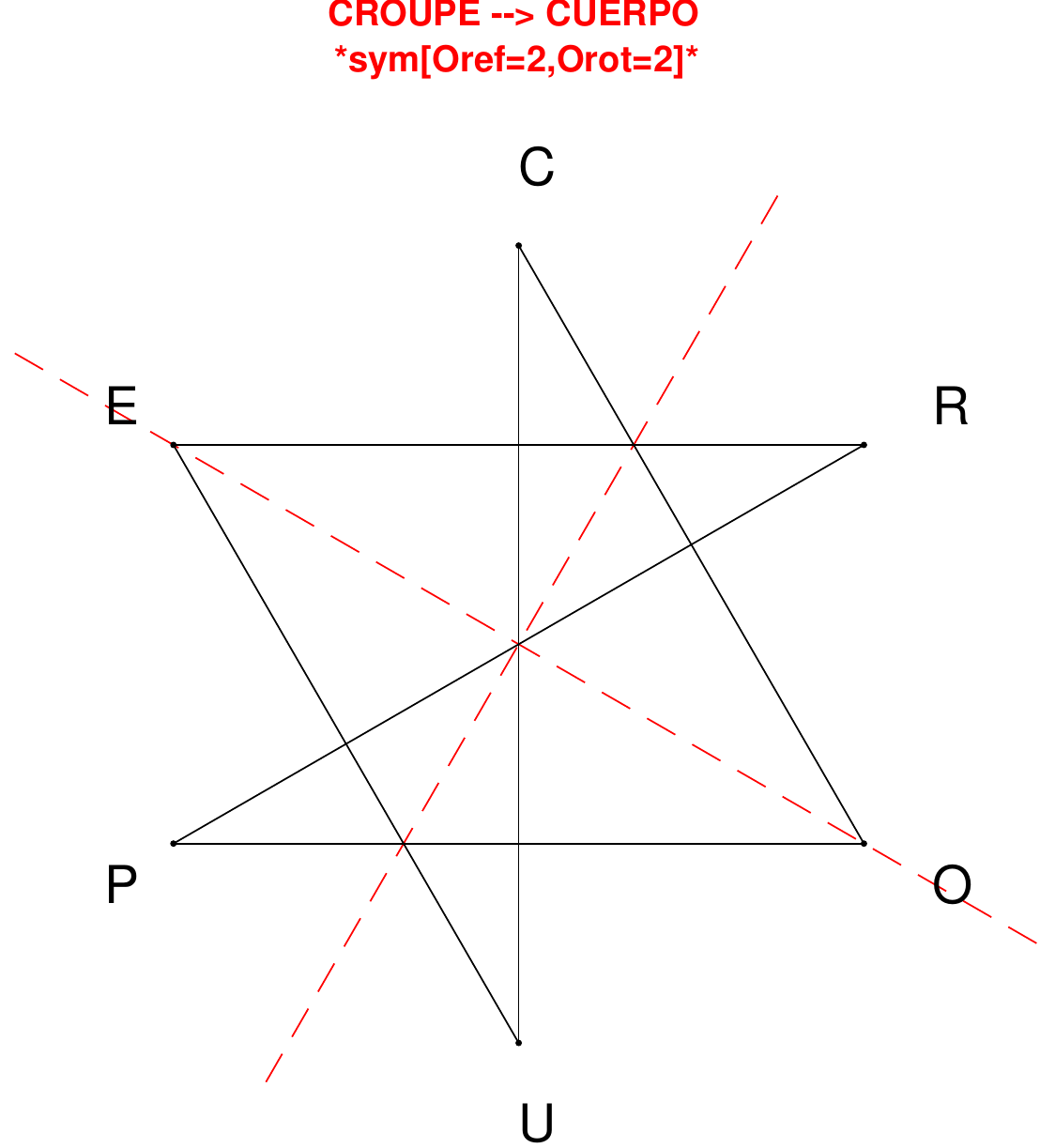}
\end{subfigure}
\hfill
\begin{subfigure}[T]{0.19\textwidth}
\centering
\includegraphics[width=\textwidth]{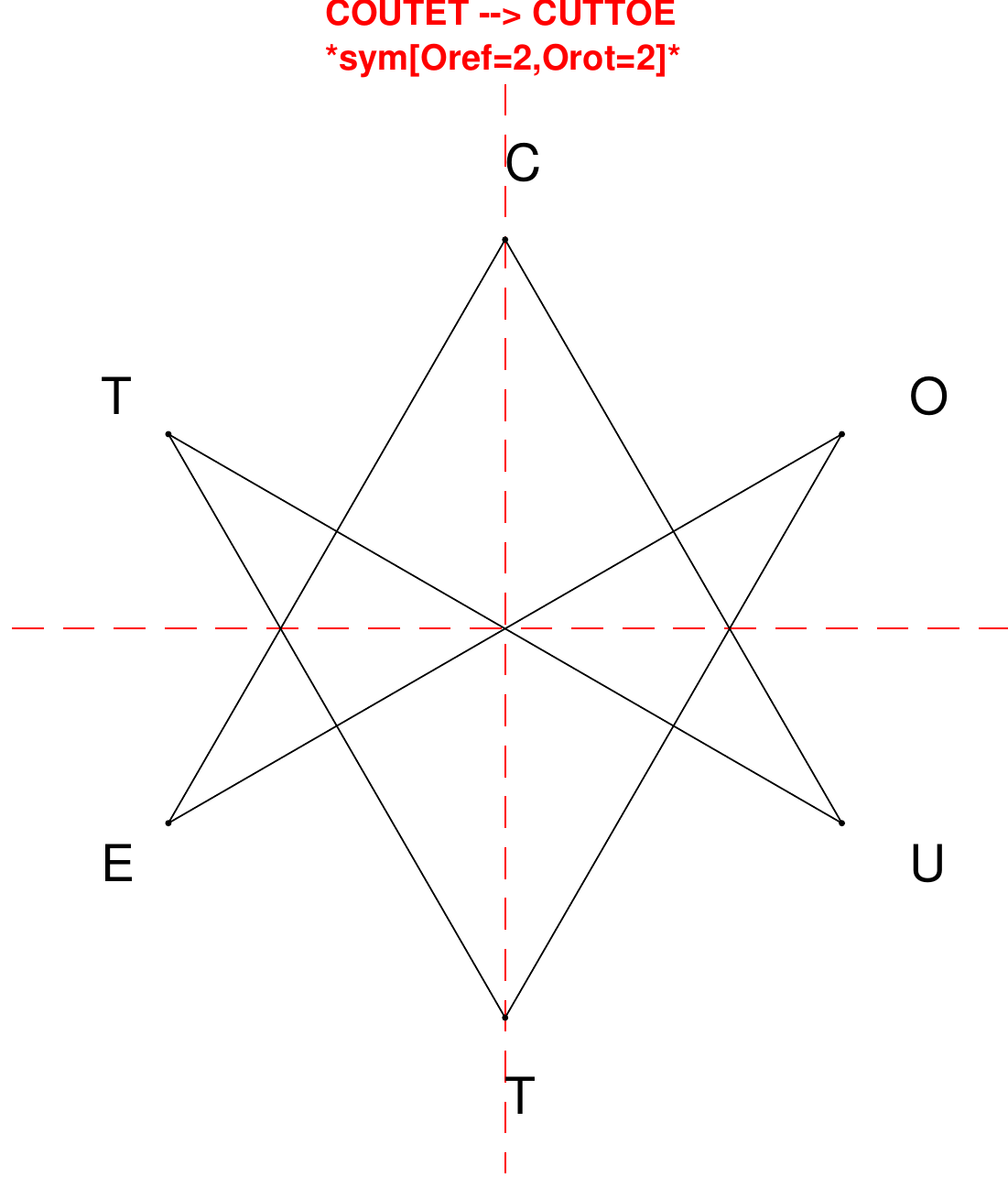}
\end{subfigure}
\hfill
\begin{subfigure}[T]{0.19\textwidth}
\centering
\includegraphics[width=\textwidth]{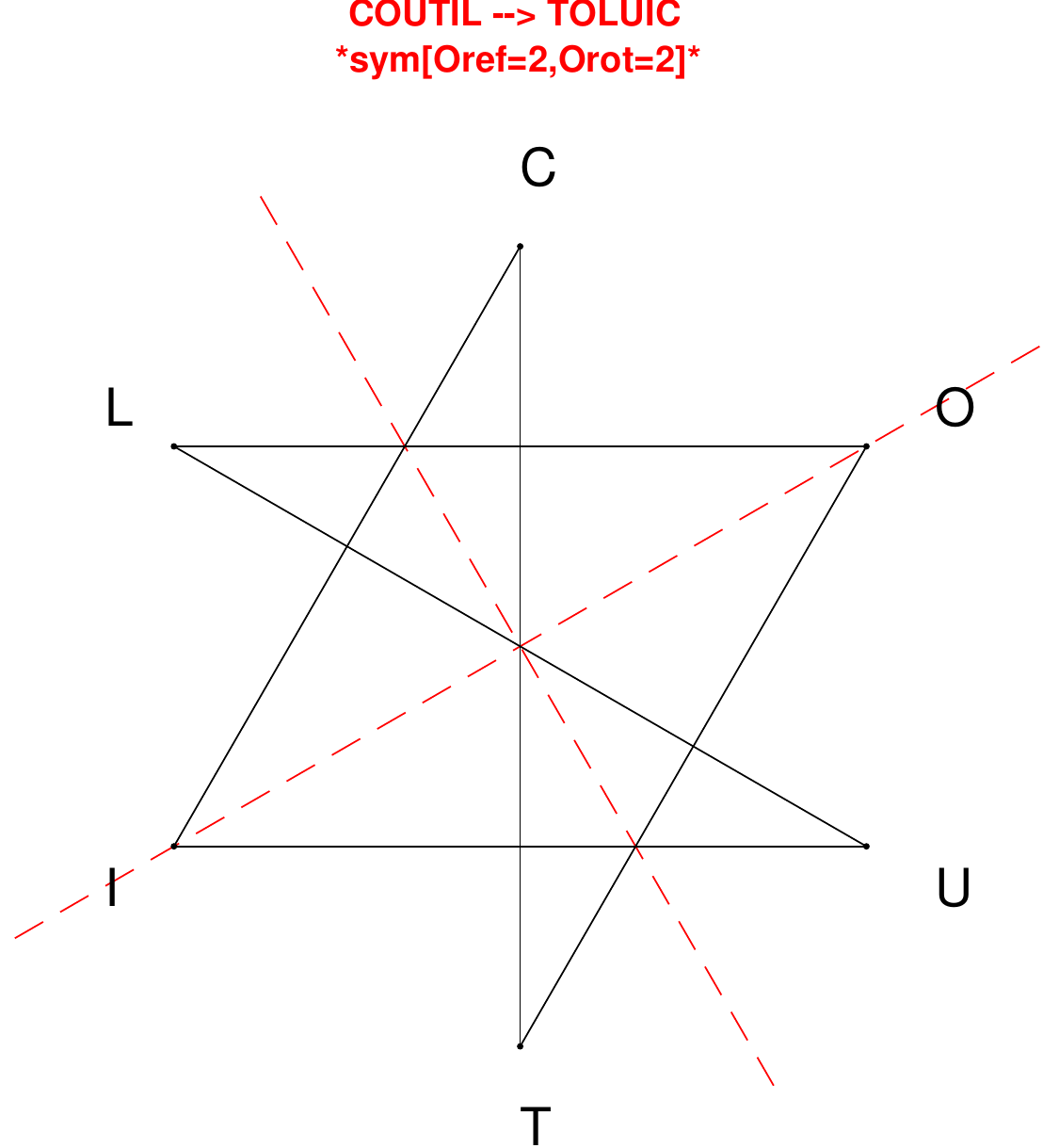}
\end{subfigure}
\hfill
\begin{subfigure}[T]{0.19\textwidth}
\centering
\includegraphics[width=\textwidth]{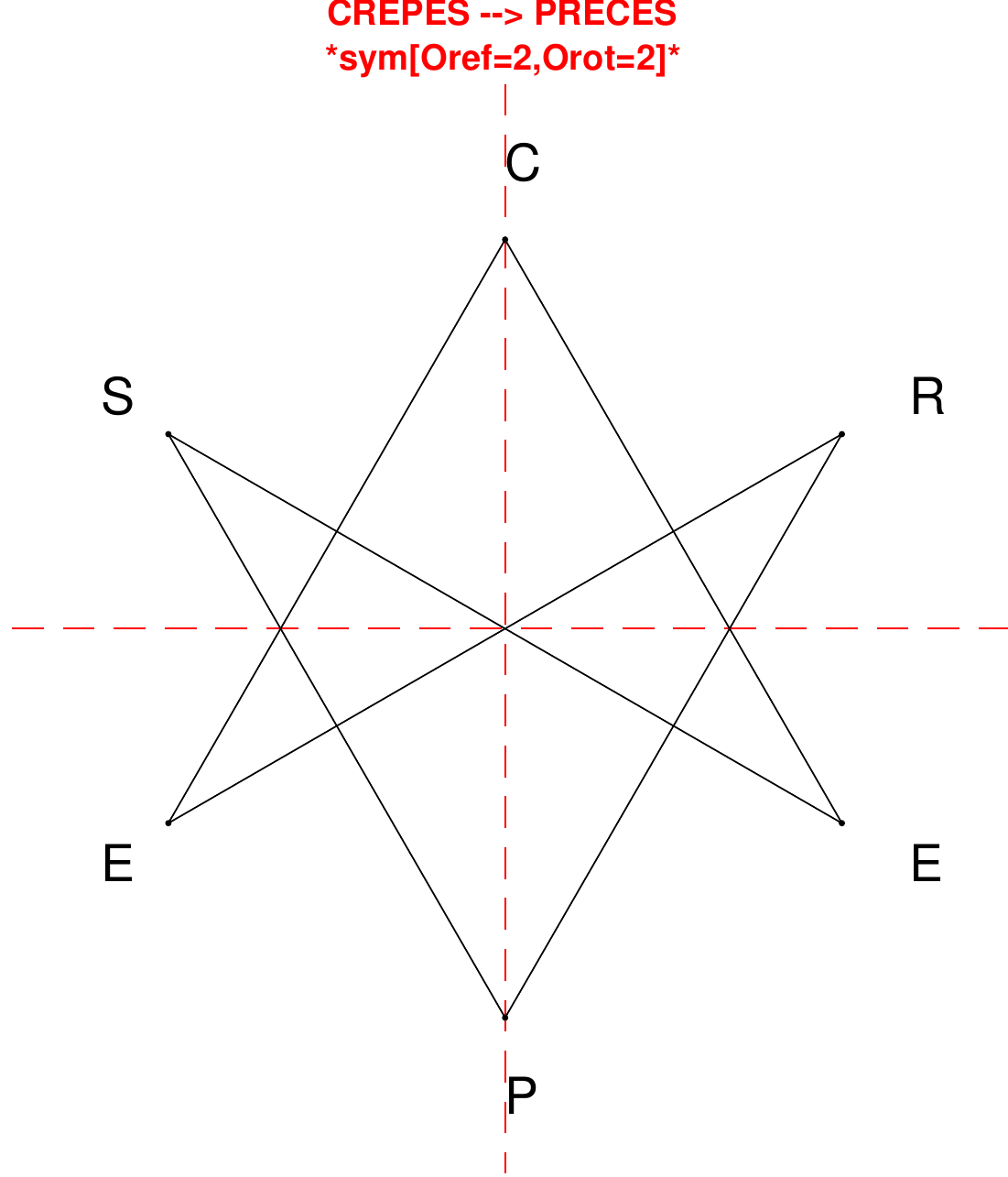}
\end{subfigure}
\hfill
\begin{subfigure}[T]{0.19\textwidth}
\centering
\includegraphics[width=\textwidth]{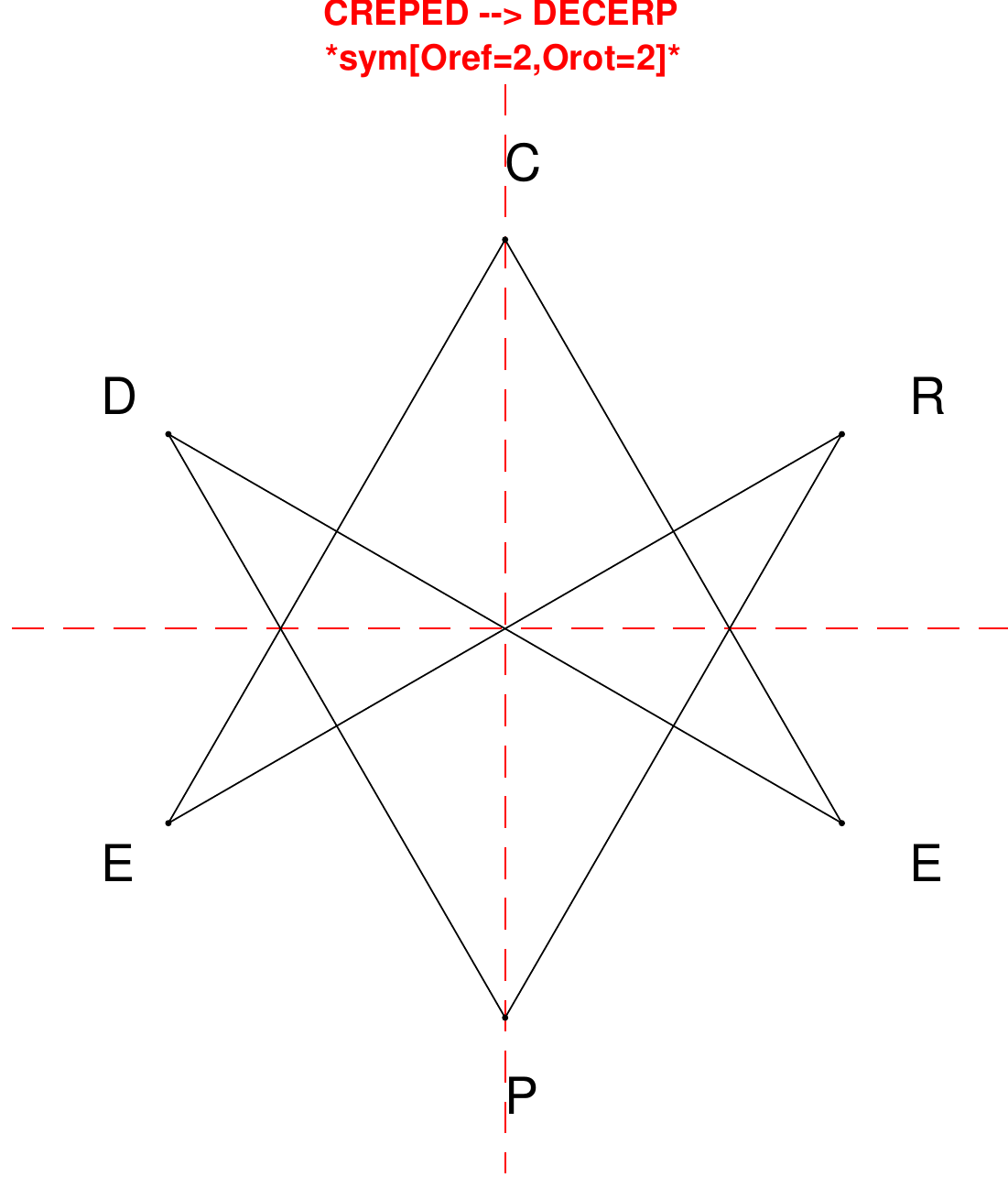}
\end{subfigure}
\end{figure}

\begin{figure}[H]
\centering
\begin{subfigure}[T]{0.19\textwidth}
\centering
\includegraphics[width=\textwidth]{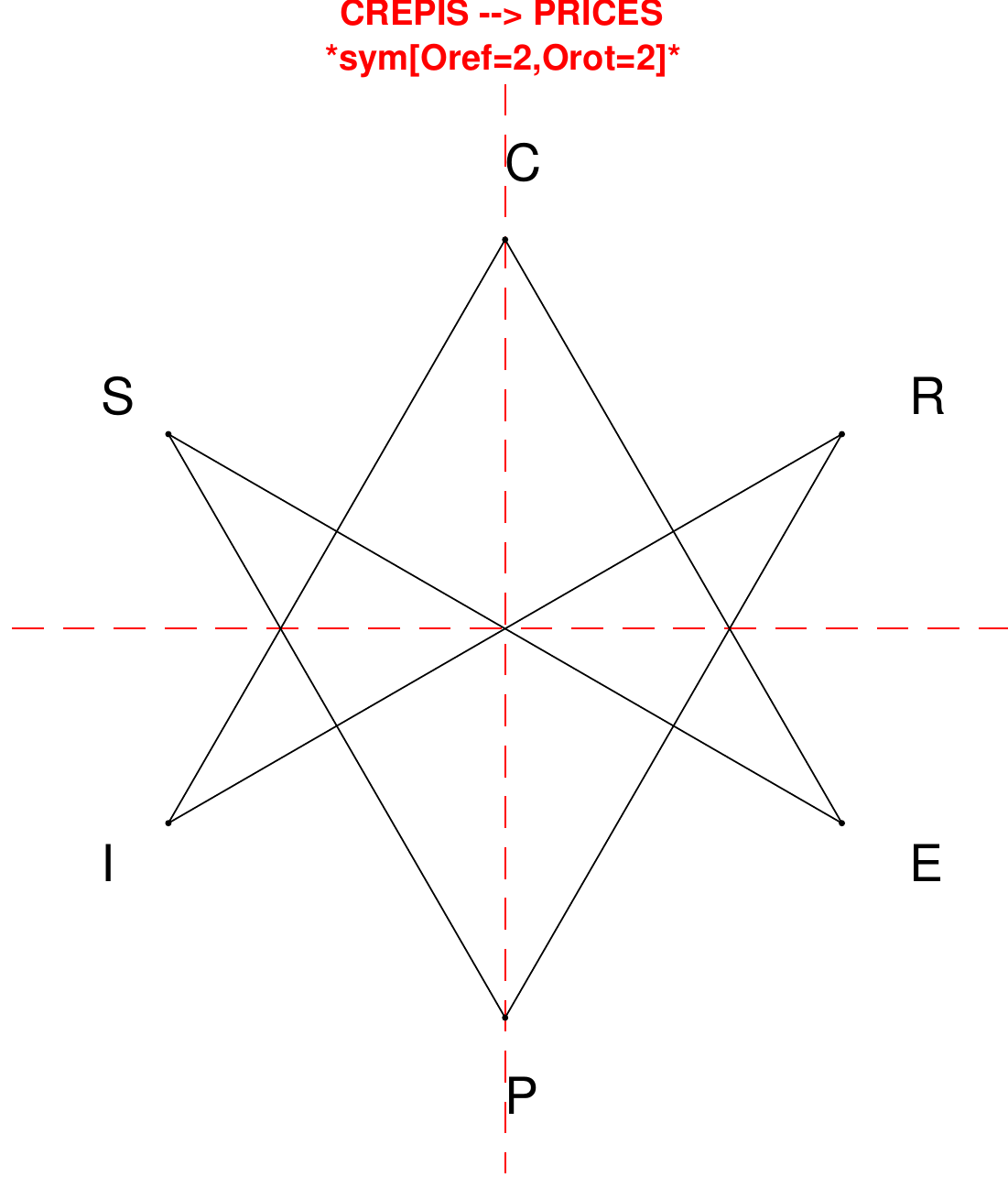}
\end{subfigure}
\hfill
\begin{subfigure}[T]{0.19\textwidth}
\centering
\includegraphics[width=\textwidth]{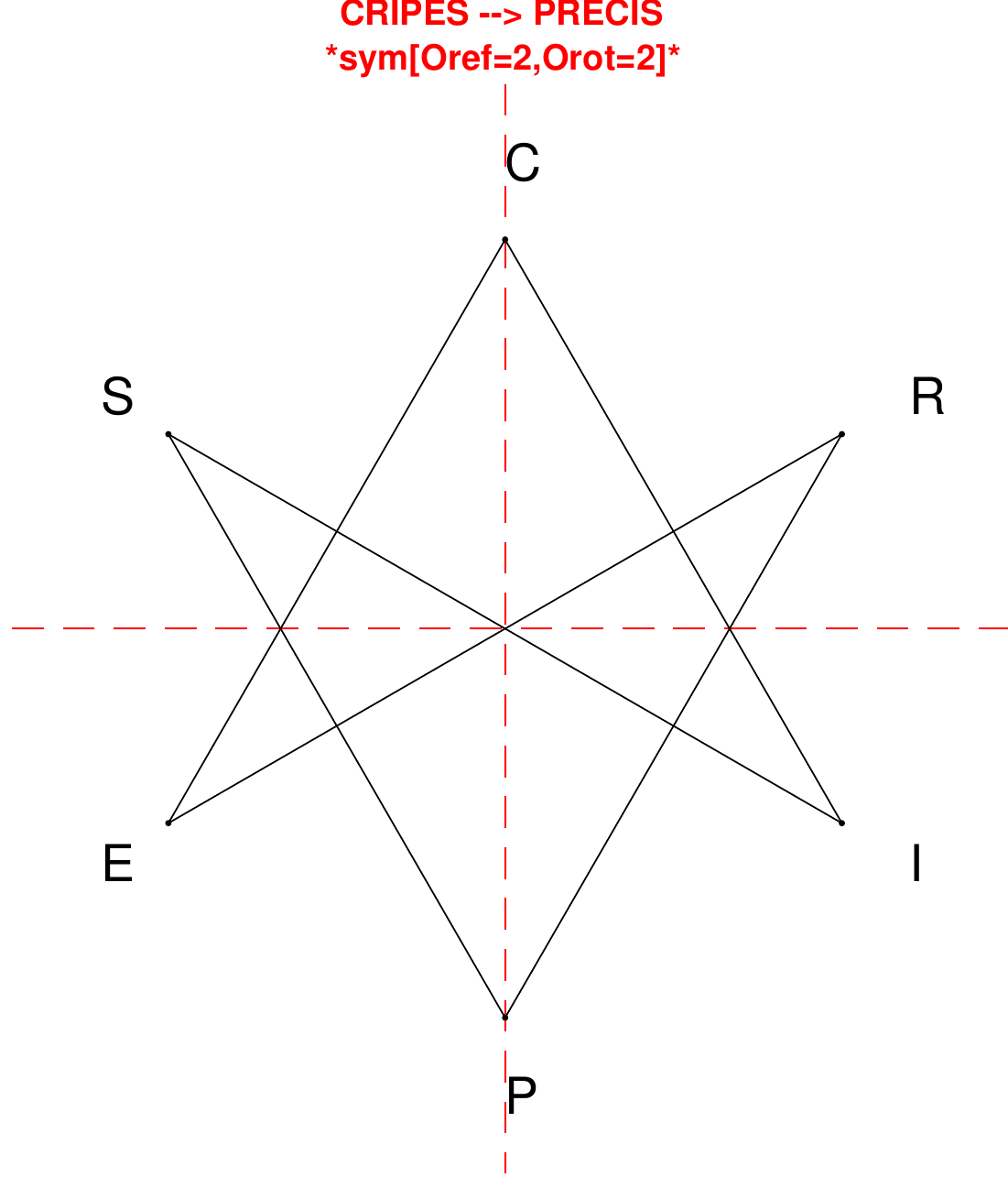}
\end{subfigure}
\hfill
\begin{subfigure}[T]{0.19\textwidth}
\centering
\includegraphics[width=\textwidth]{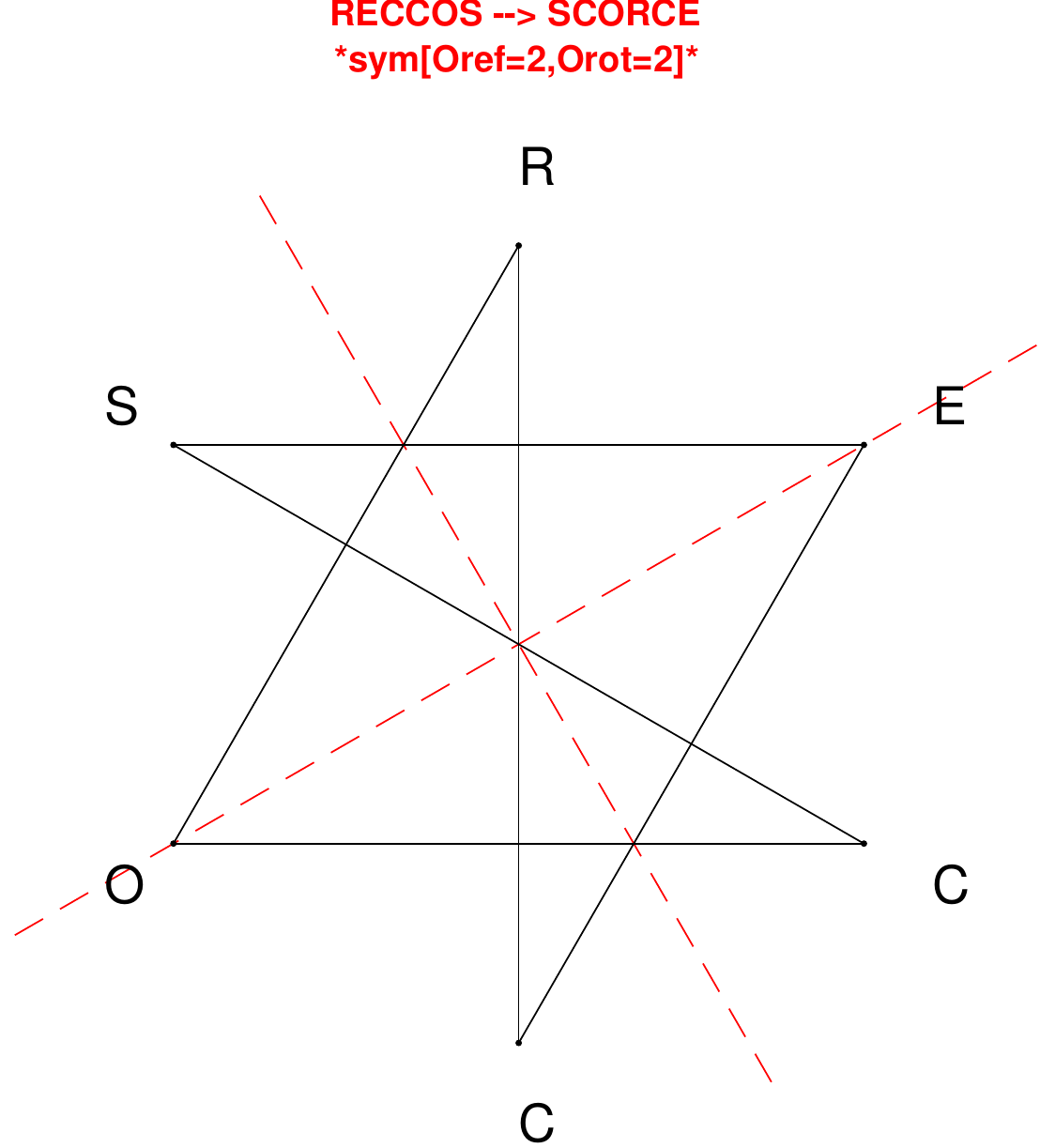}
\end{subfigure}
\hfill
\begin{subfigure}[T]{0.19\textwidth}
\centering
\includegraphics[width=\textwidth]{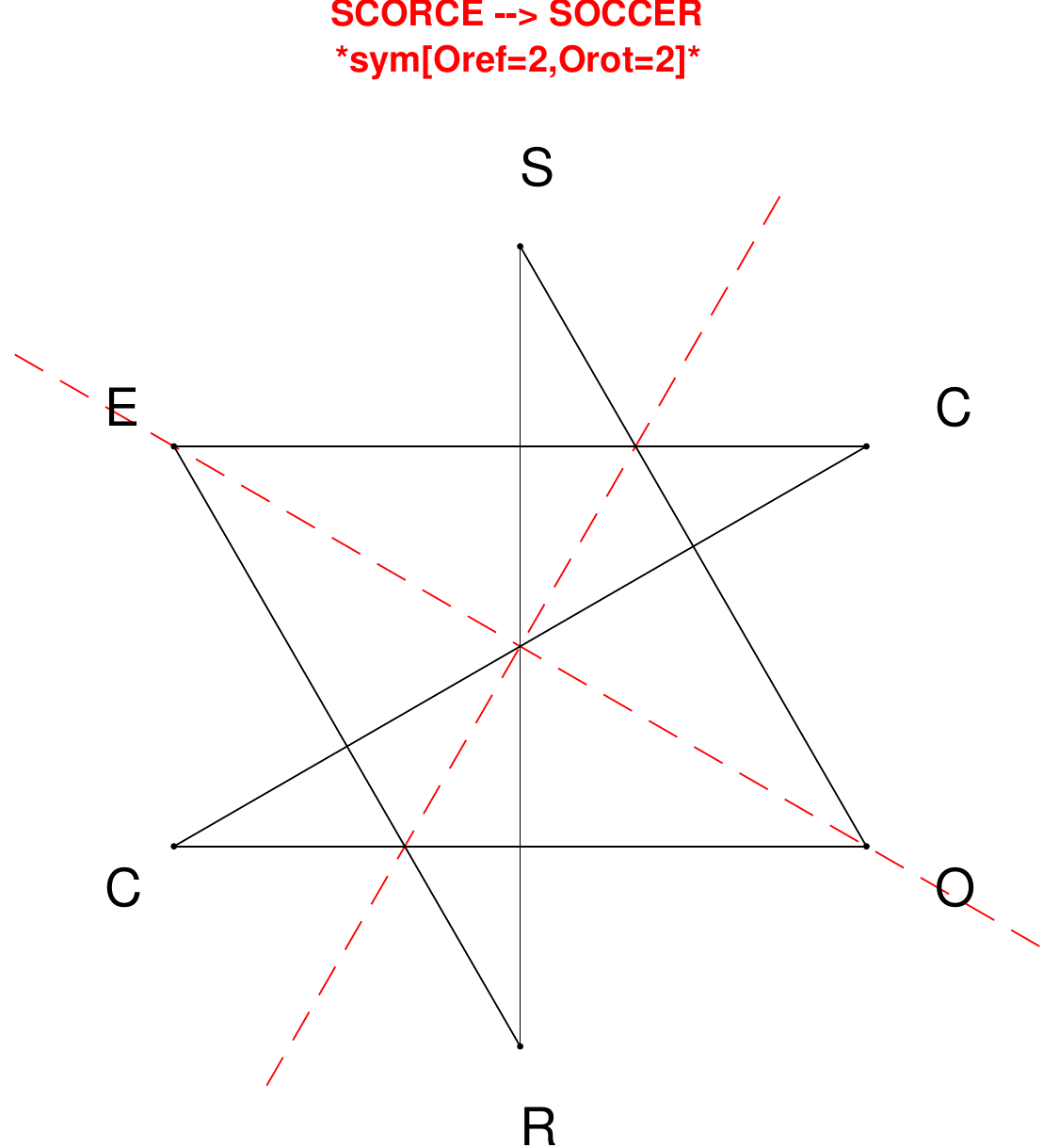}
\end{subfigure}
\hfill
\begin{subfigure}[T]{0.19\textwidth}
\centering
\includegraphics[width=\textwidth]{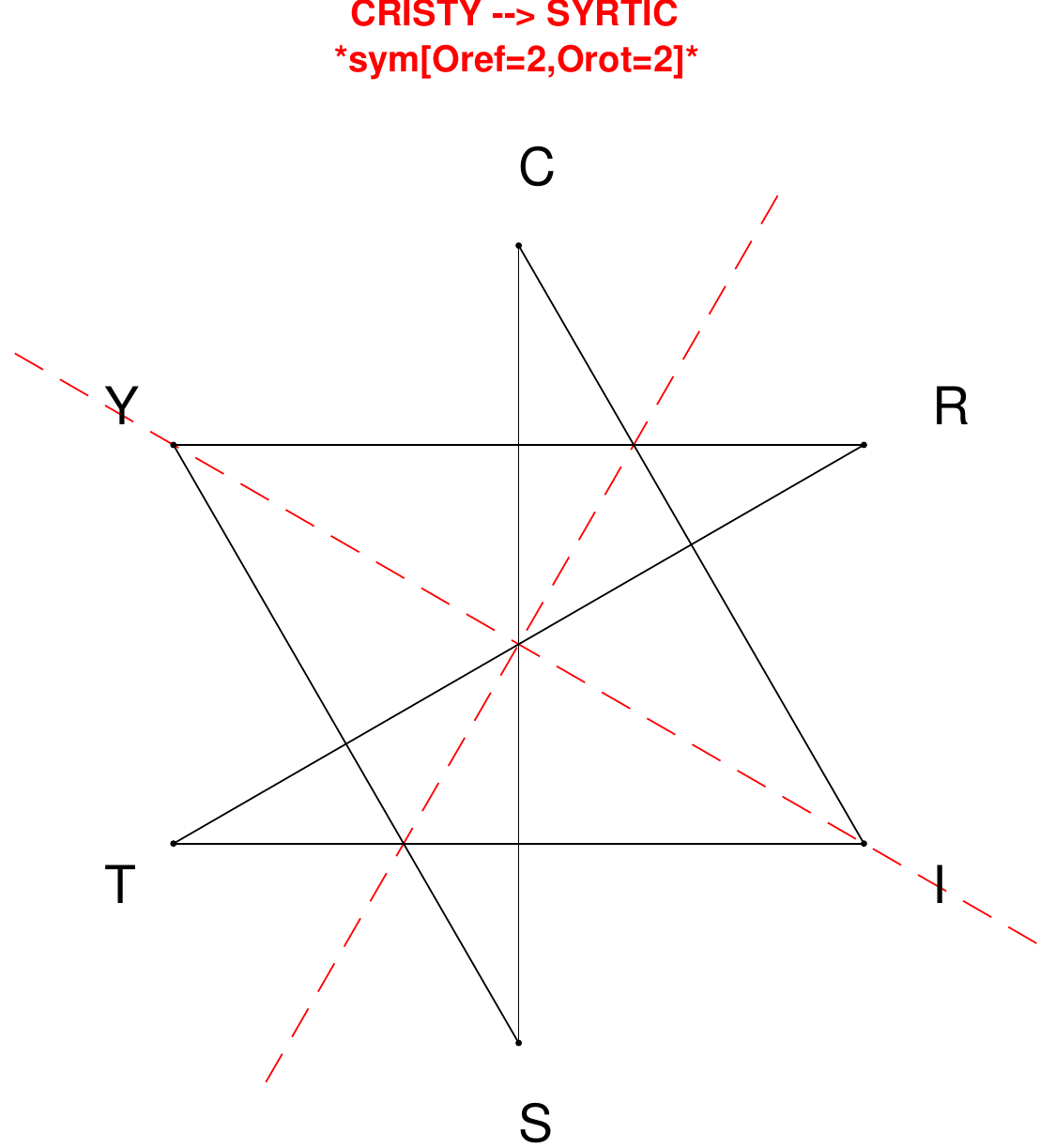}
\end{subfigure}
\end{figure}

\begin{figure}[H]
\centering
\begin{subfigure}[T]{0.19\textwidth}
\centering
\includegraphics[width=\textwidth]{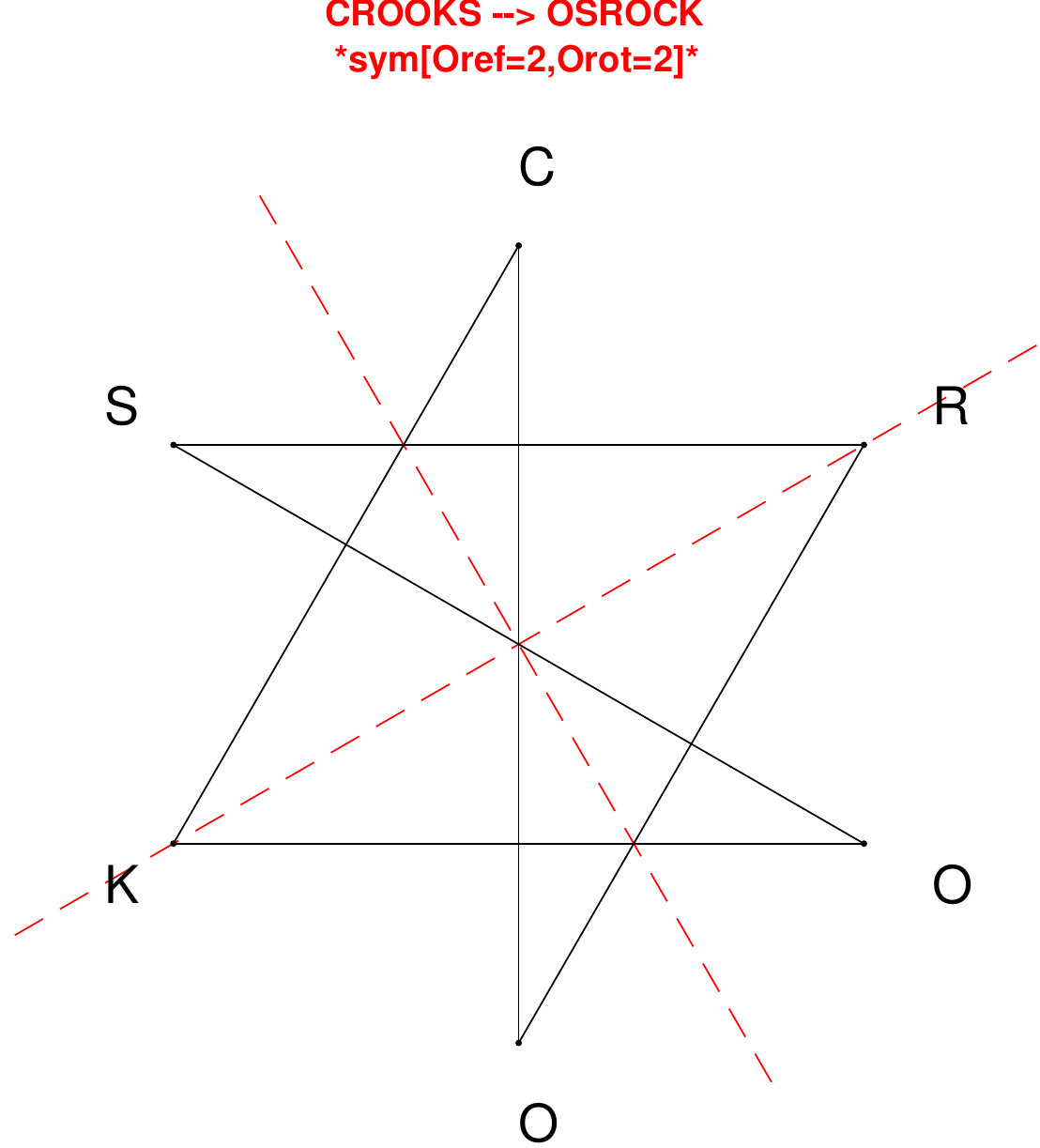}
\end{subfigure}
\hfill
\begin{subfigure}[T]{0.19\textwidth}
\centering
\includegraphics[width=\textwidth]{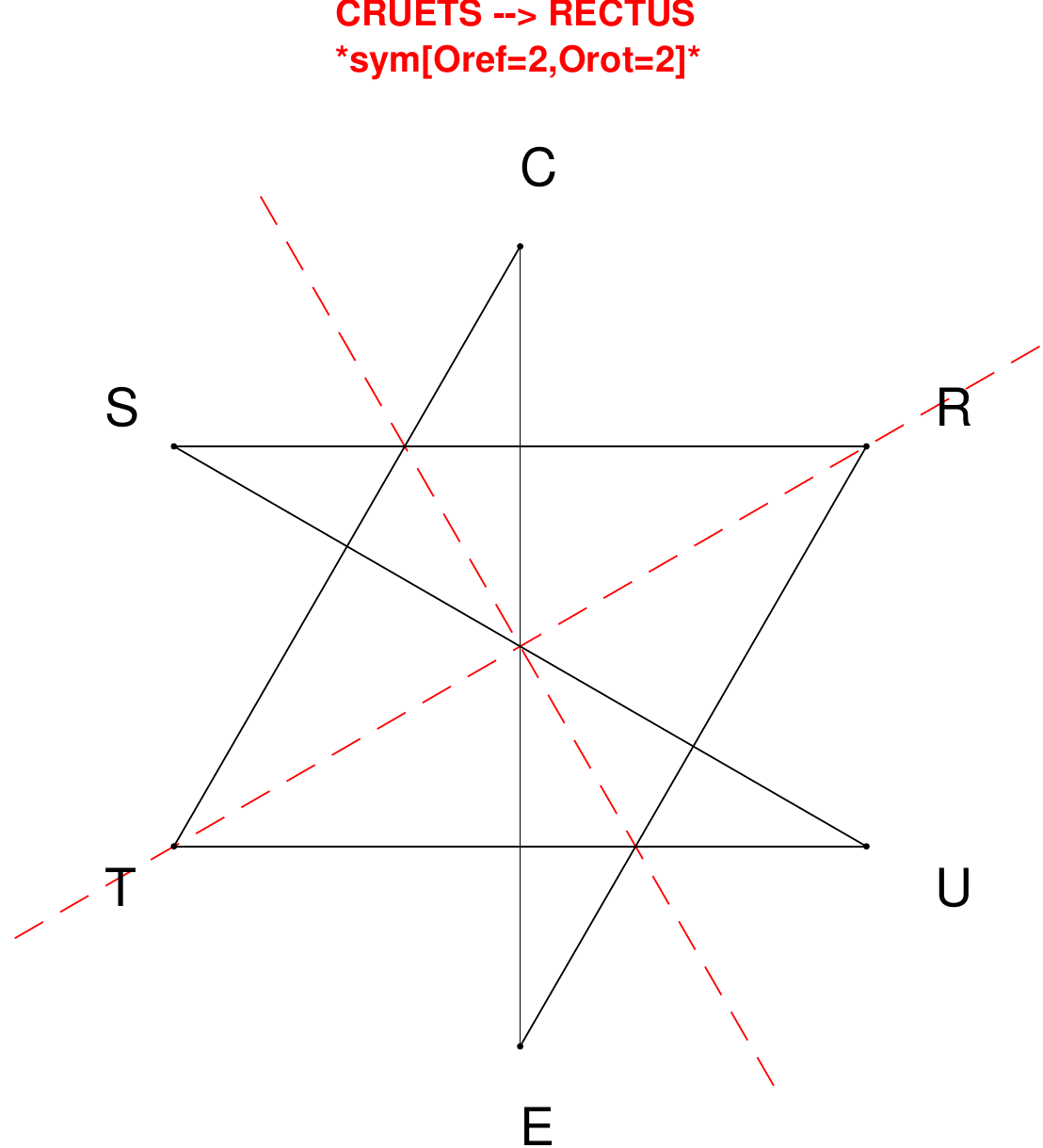}
\end{subfigure}
\hfill
\begin{subfigure}[T]{0.19\textwidth}
\centering
\includegraphics[width=\textwidth]{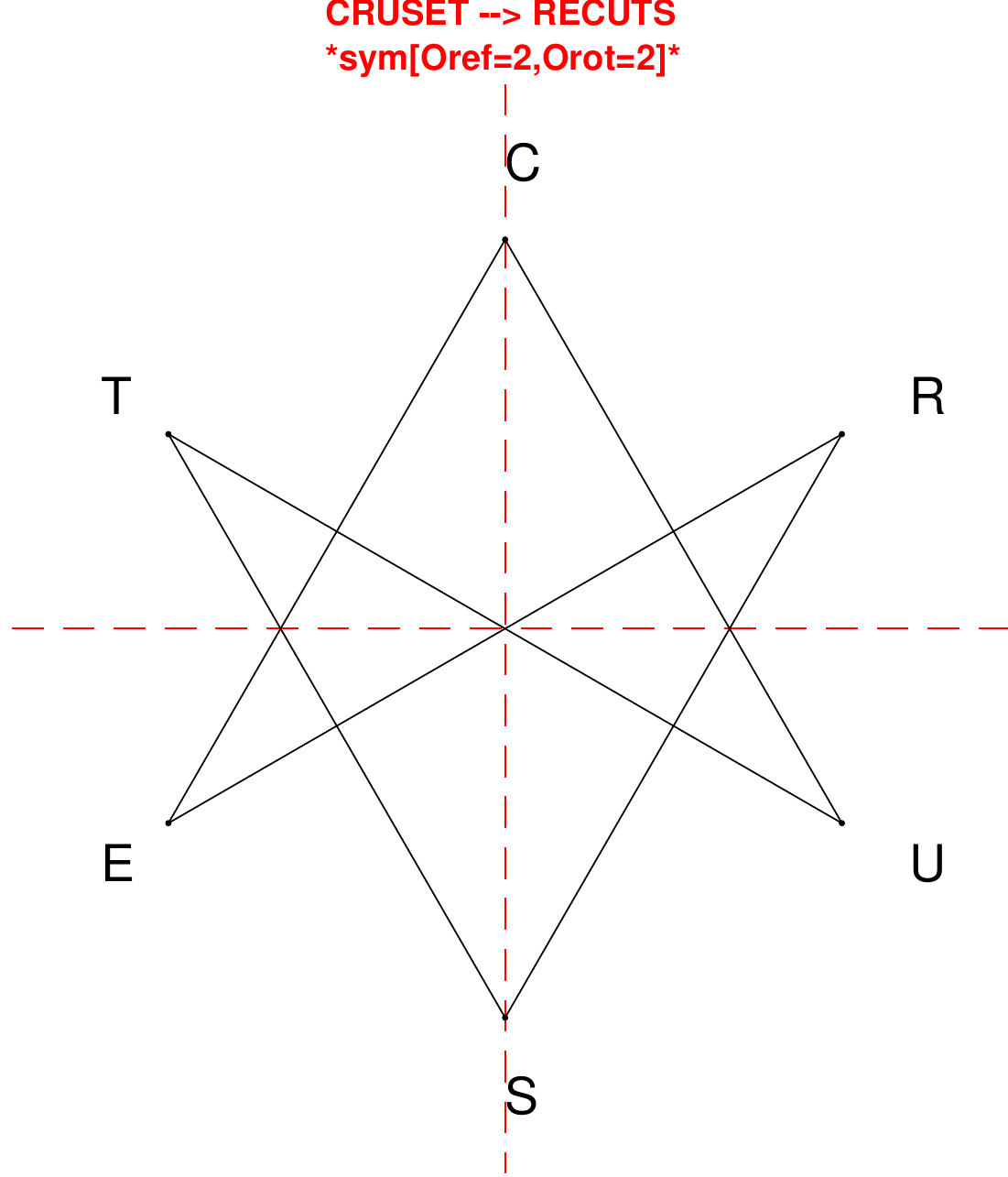}
\end{subfigure}
\hfill
\begin{subfigure}[T]{0.19\textwidth}
\centering
\includegraphics[width=\textwidth]{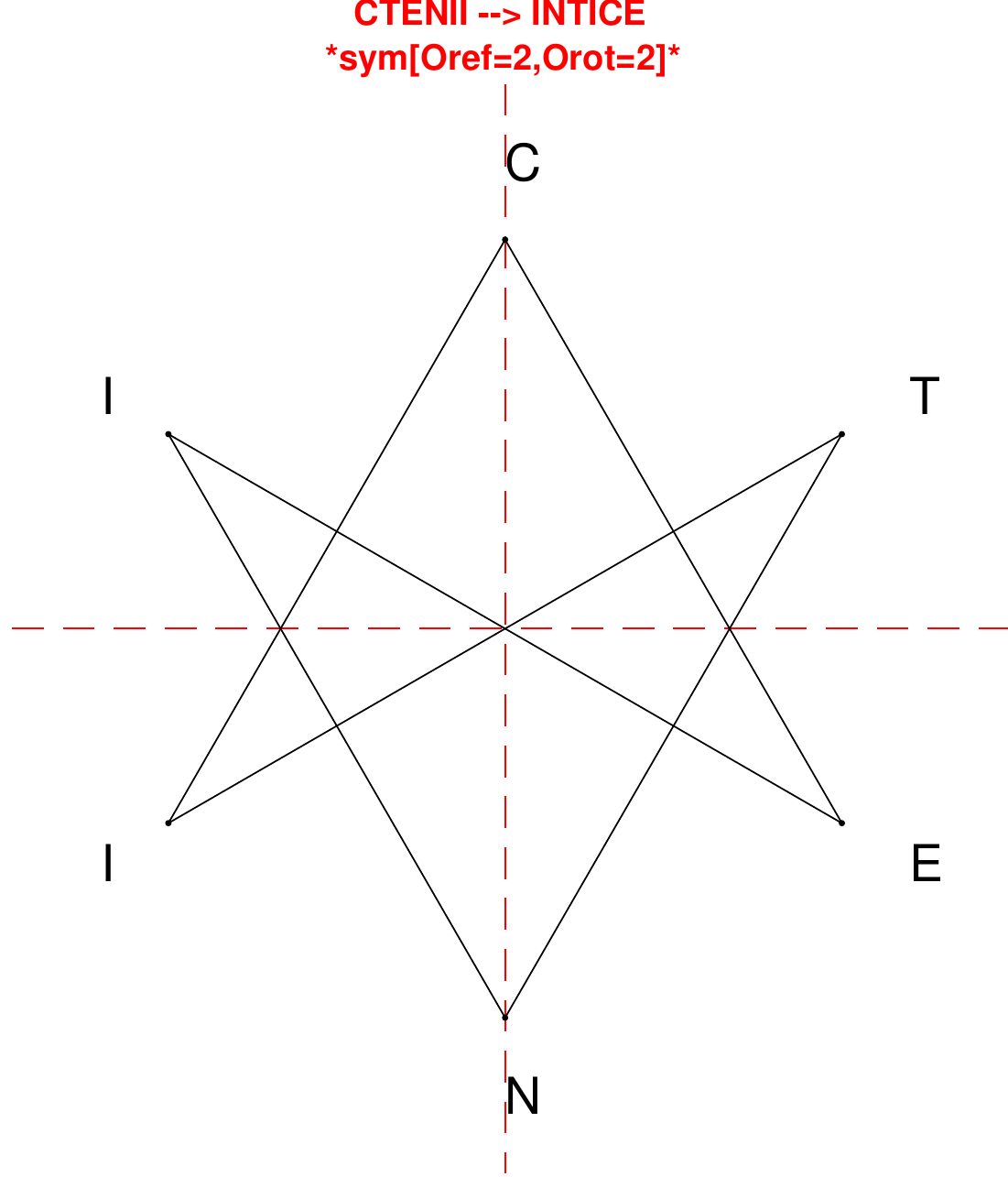}
\end{subfigure}
\hfill
\begin{subfigure}[T]{0.19\textwidth}
\centering
\includegraphics[width=\textwidth]{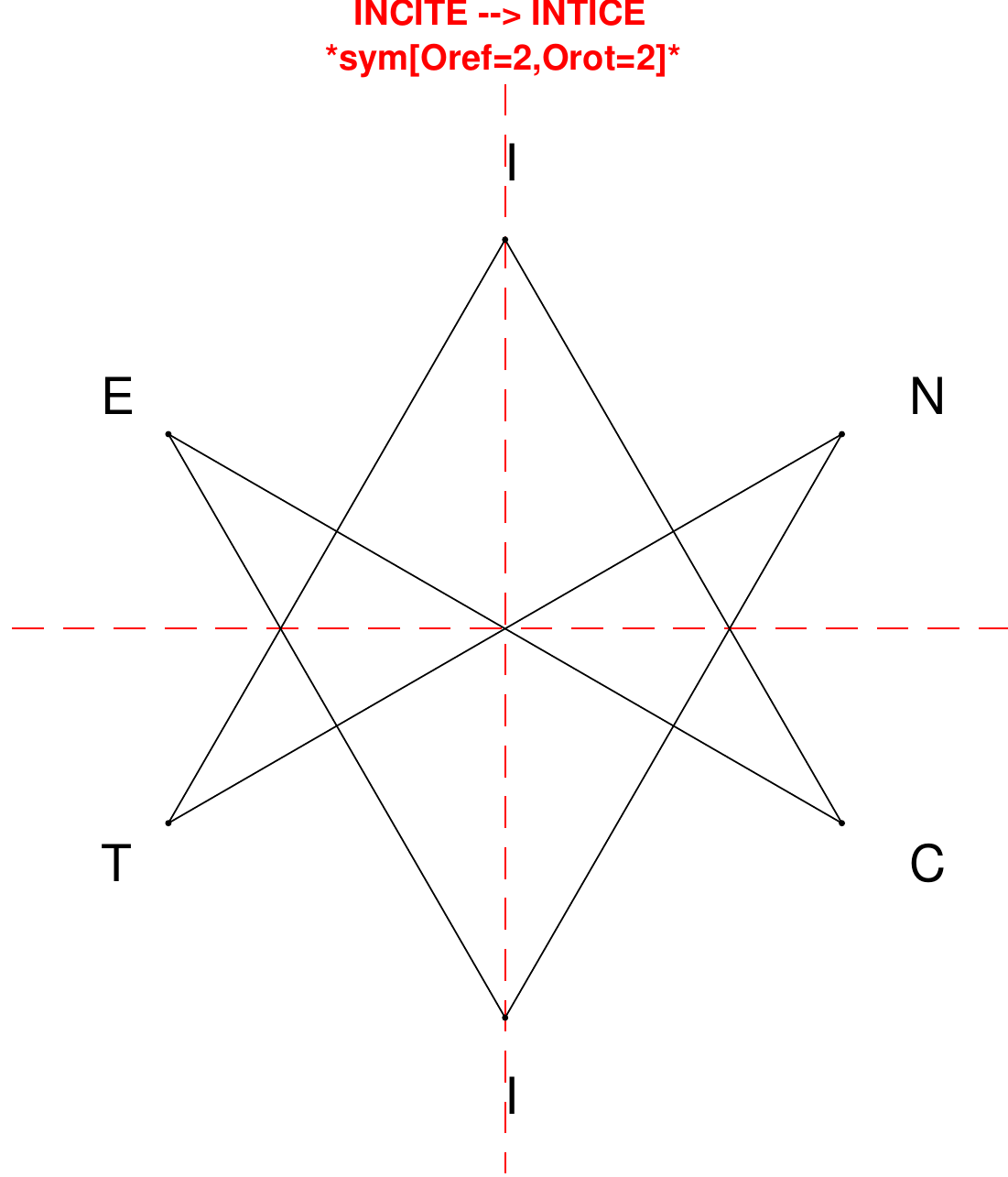}
\end{subfigure}
\end{figure}

\begin{figure}[H]
\centering
\begin{subfigure}[T]{0.19\textwidth}
\centering
\includegraphics[width=\textwidth]{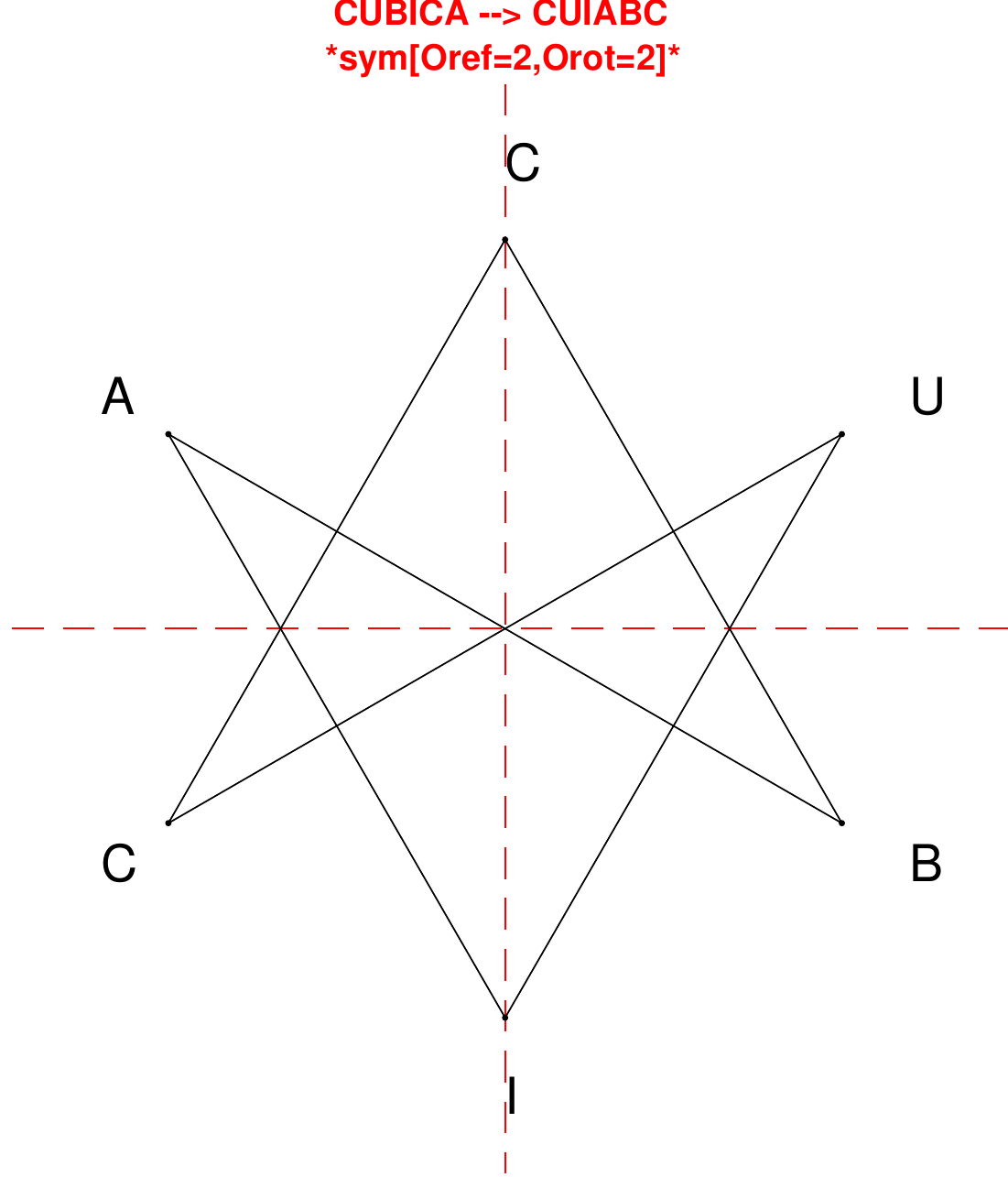}
\end{subfigure}
\hfill
\begin{subfigure}[T]{0.19\textwidth}
\centering
\includegraphics[width=\textwidth]{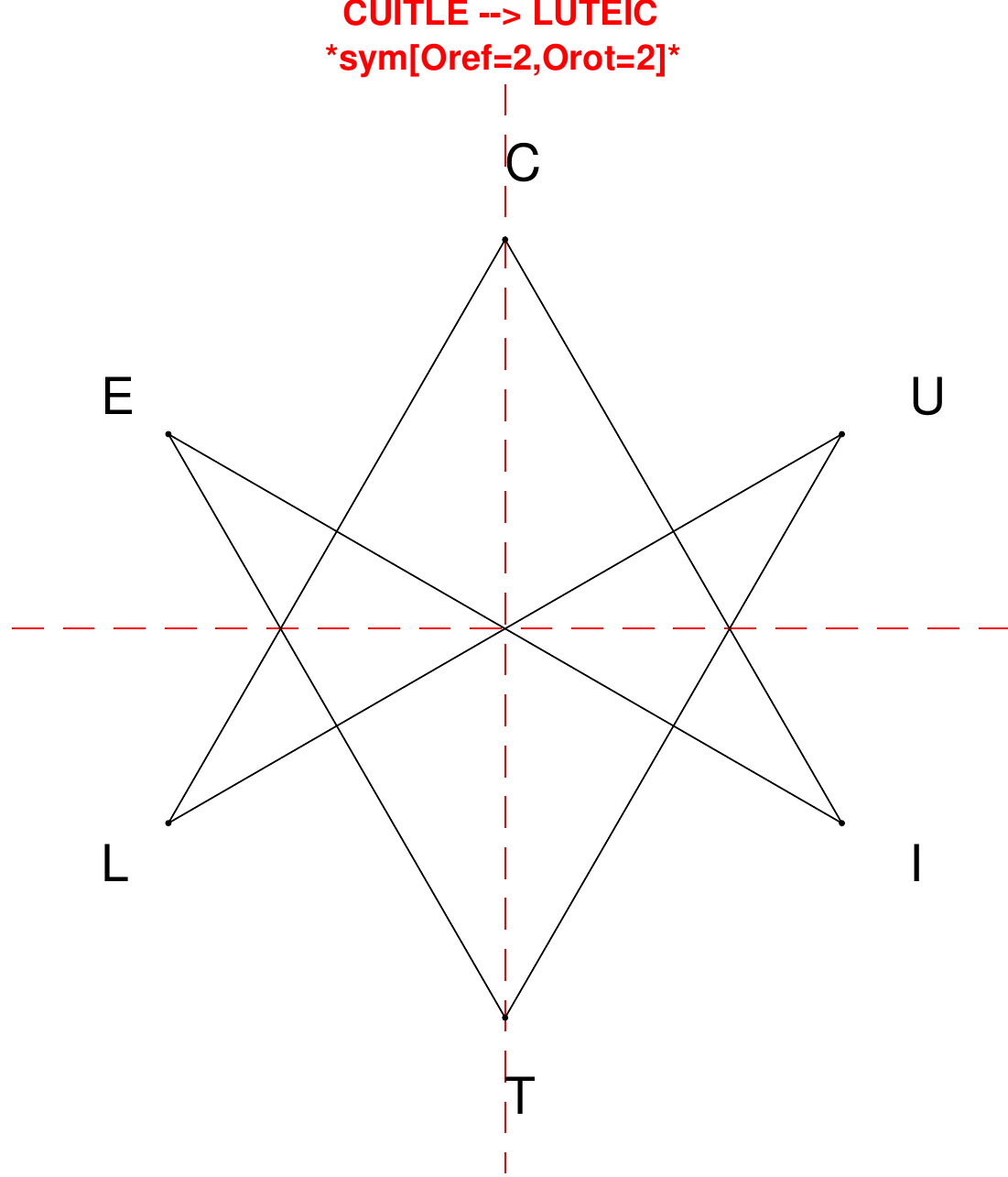}
\end{subfigure}
\hfill
\begin{subfigure}[T]{0.19\textwidth}
\centering
\includegraphics[width=\textwidth]{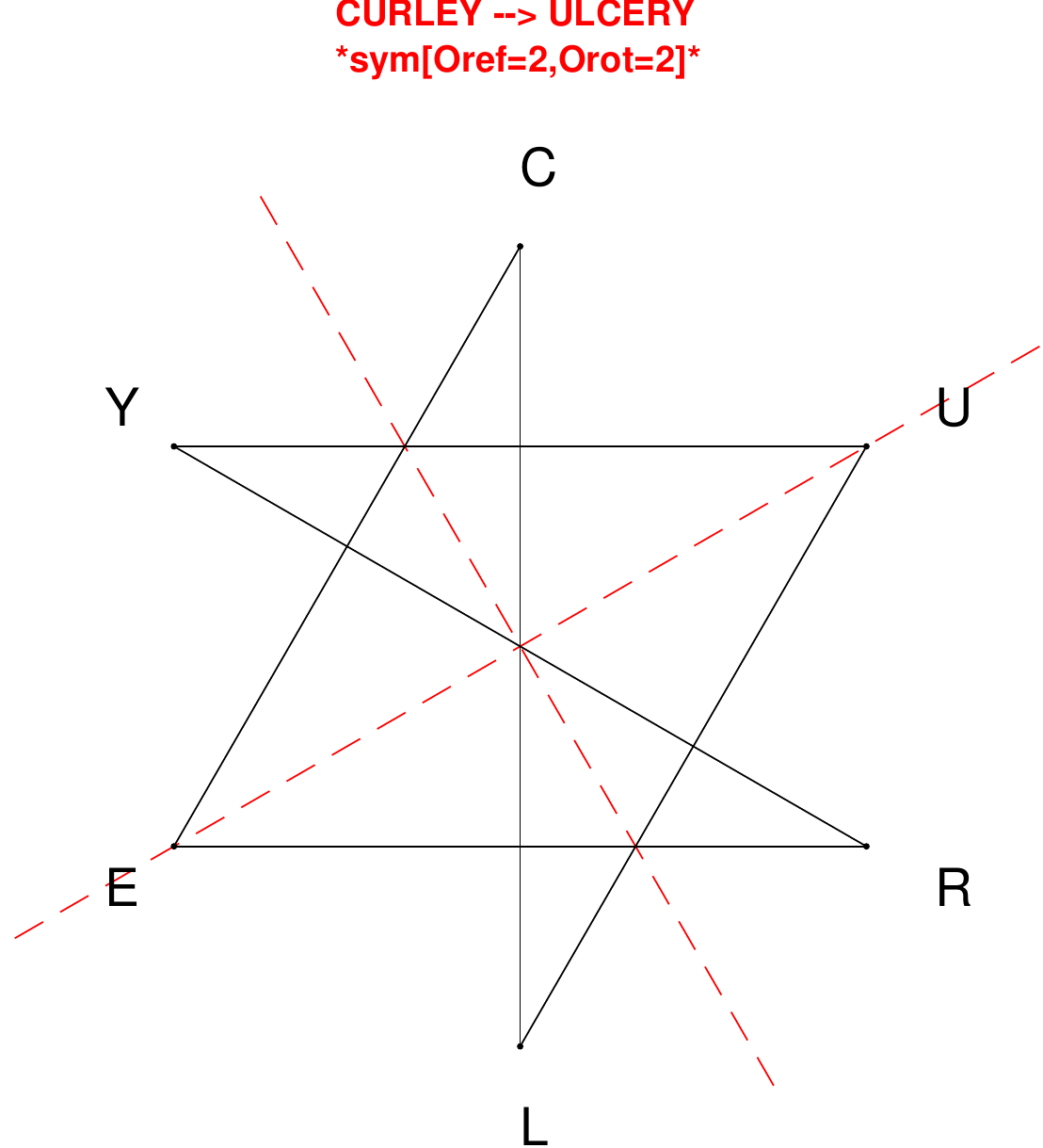}
\end{subfigure}
\hfill
\begin{subfigure}[T]{0.19\textwidth}
\centering
\includegraphics[width=\textwidth]{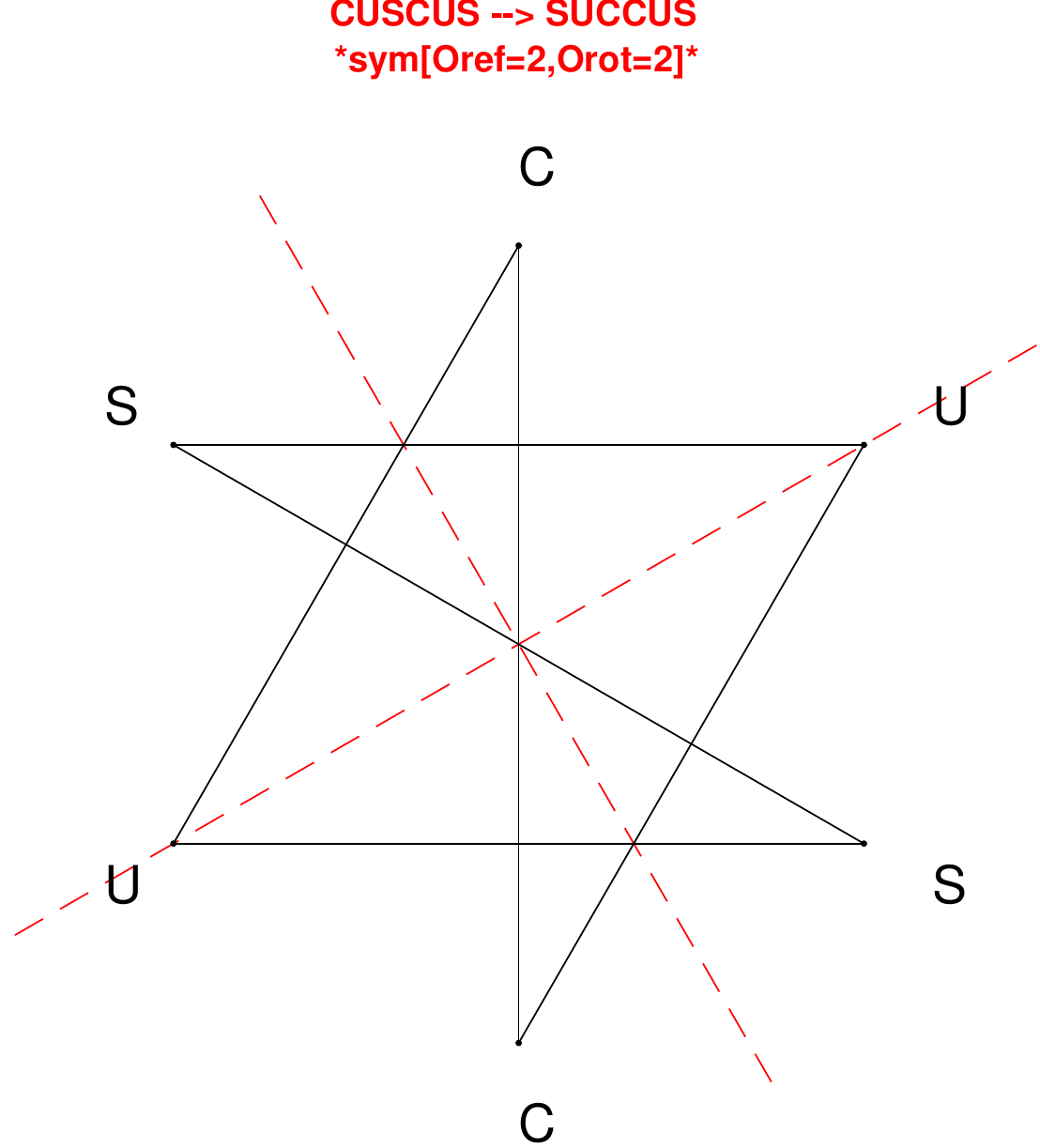}
\end{subfigure}
\hfill
\begin{subfigure}[T]{0.19\textwidth}
\centering
\includegraphics[width=\textwidth]{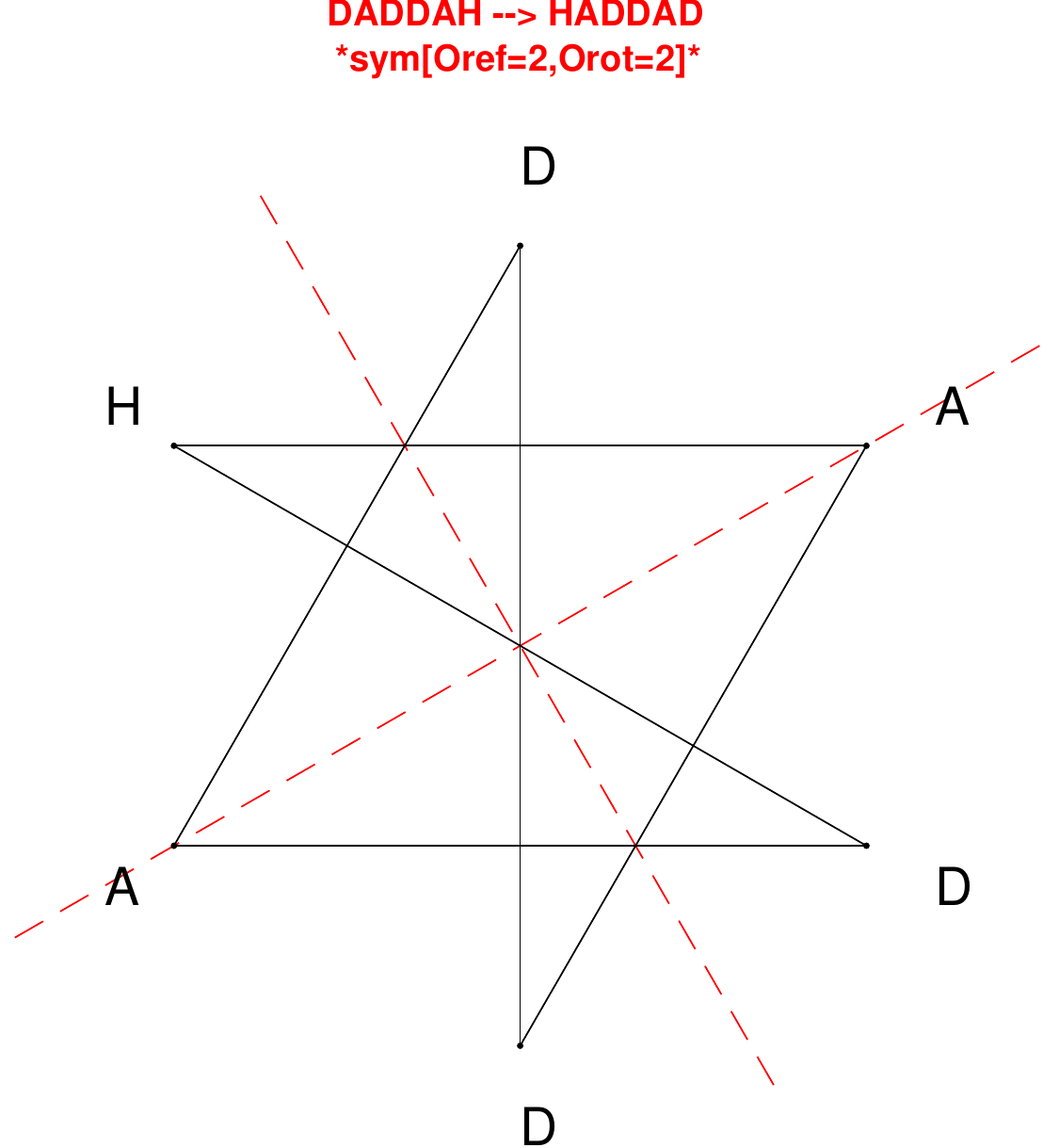}
\end{subfigure}
\end{figure}

\begin{figure}[H]
\centering
\begin{subfigure}[T]{0.19\textwidth}
\centering
\includegraphics[width=\textwidth]{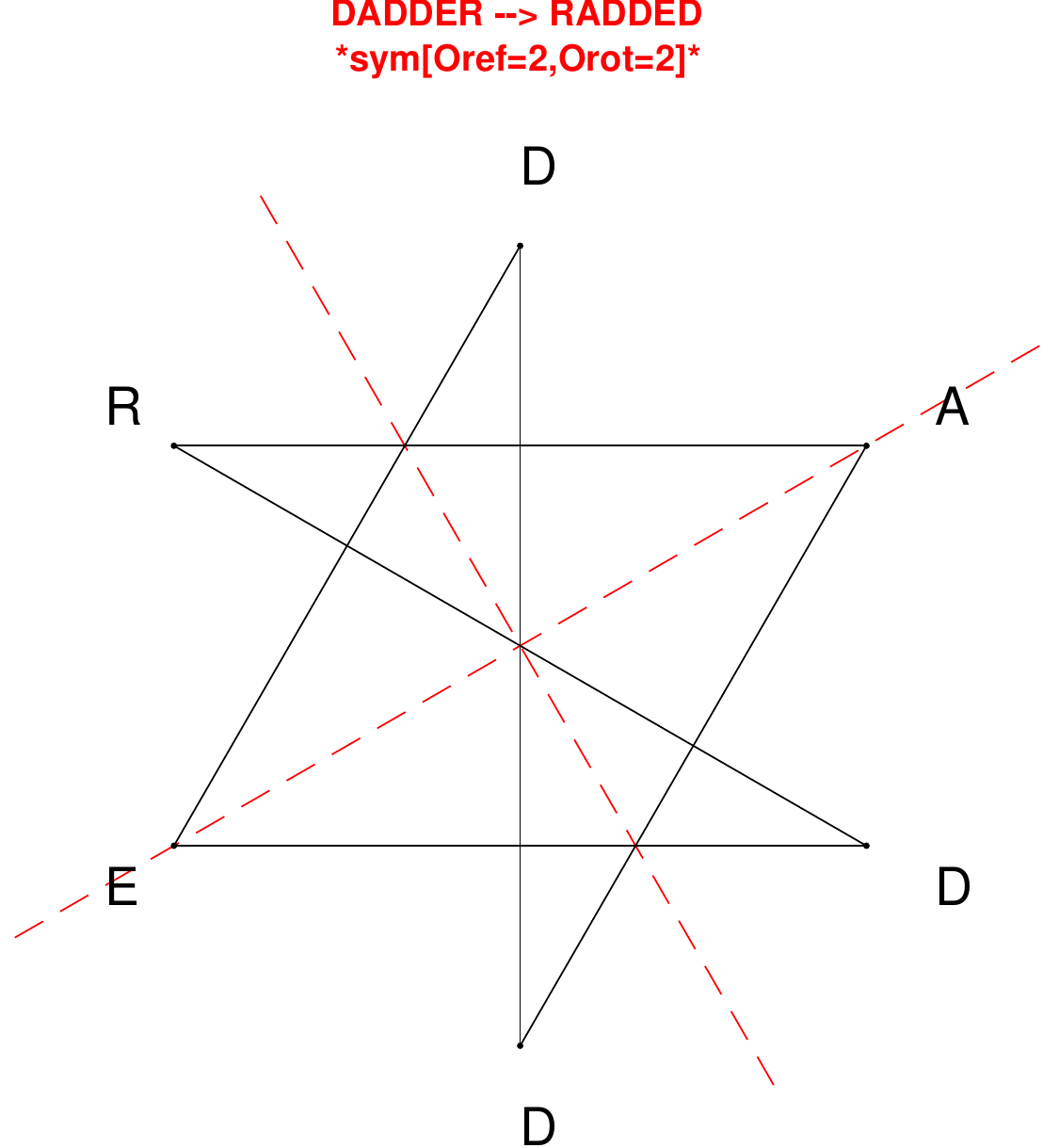}
\end{subfigure}
\hfill
\begin{subfigure}[T]{0.19\textwidth}
\centering
\includegraphics[width=\textwidth]{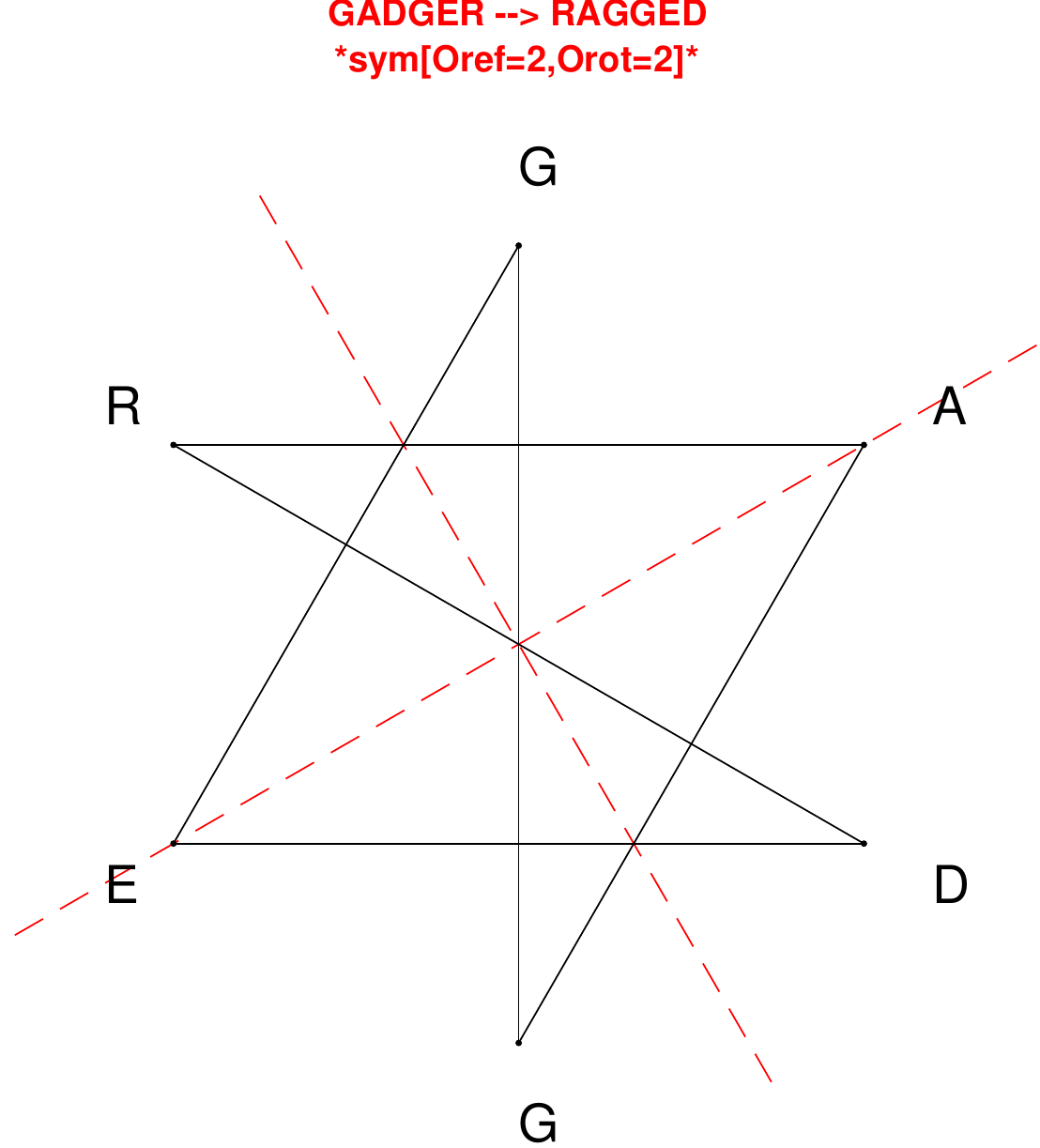}
\end{subfigure}
\hfill
\begin{subfigure}[T]{0.19\textwidth}
\centering
\includegraphics[width=\textwidth]{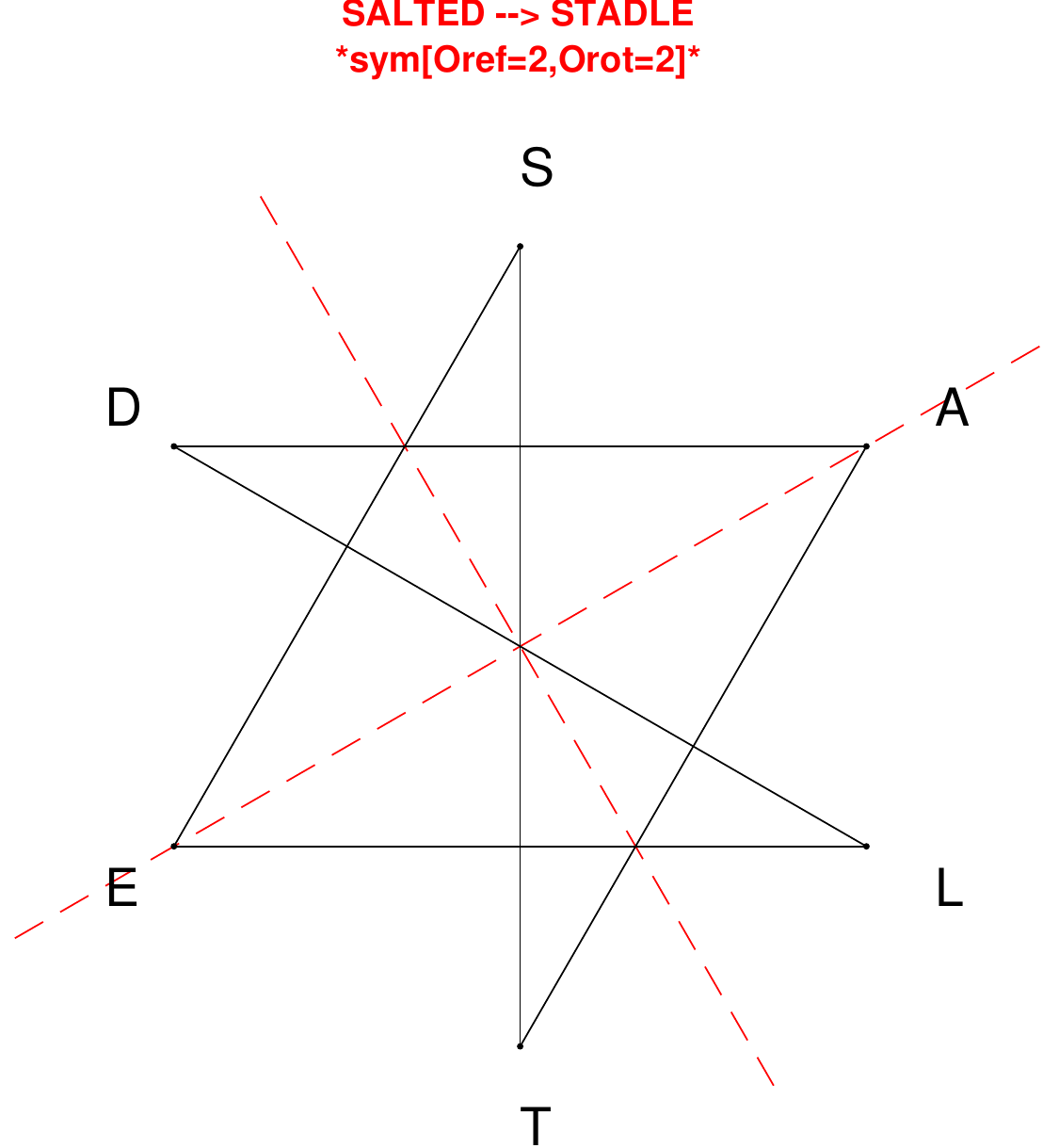}
\end{subfigure}
\hfill
\begin{subfigure}[T]{0.19\textwidth}
\centering
\includegraphics[width=\textwidth]{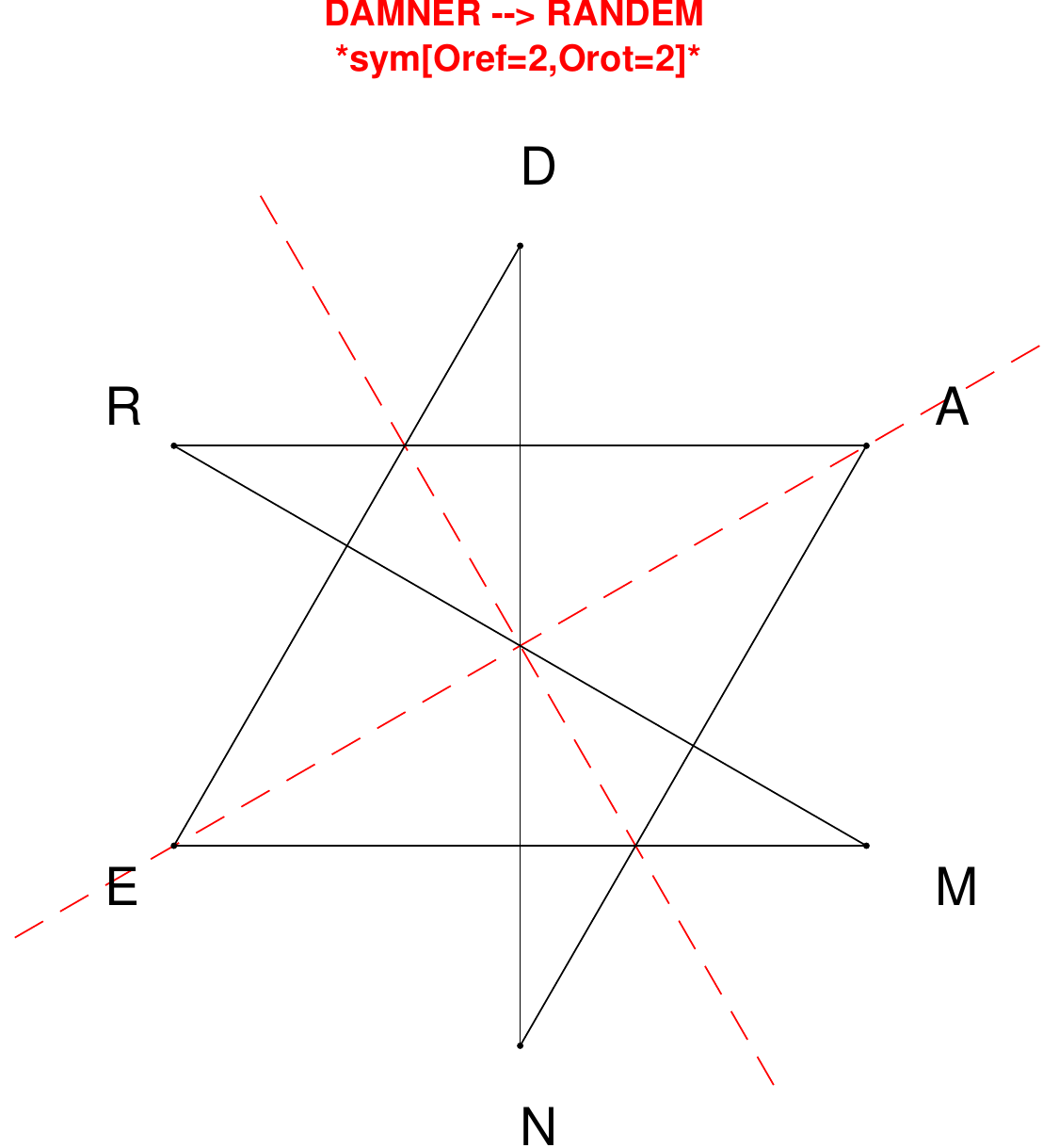}
\end{subfigure}
\hfill
\begin{subfigure}[T]{0.19\textwidth}
\centering
\includegraphics[width=\textwidth]{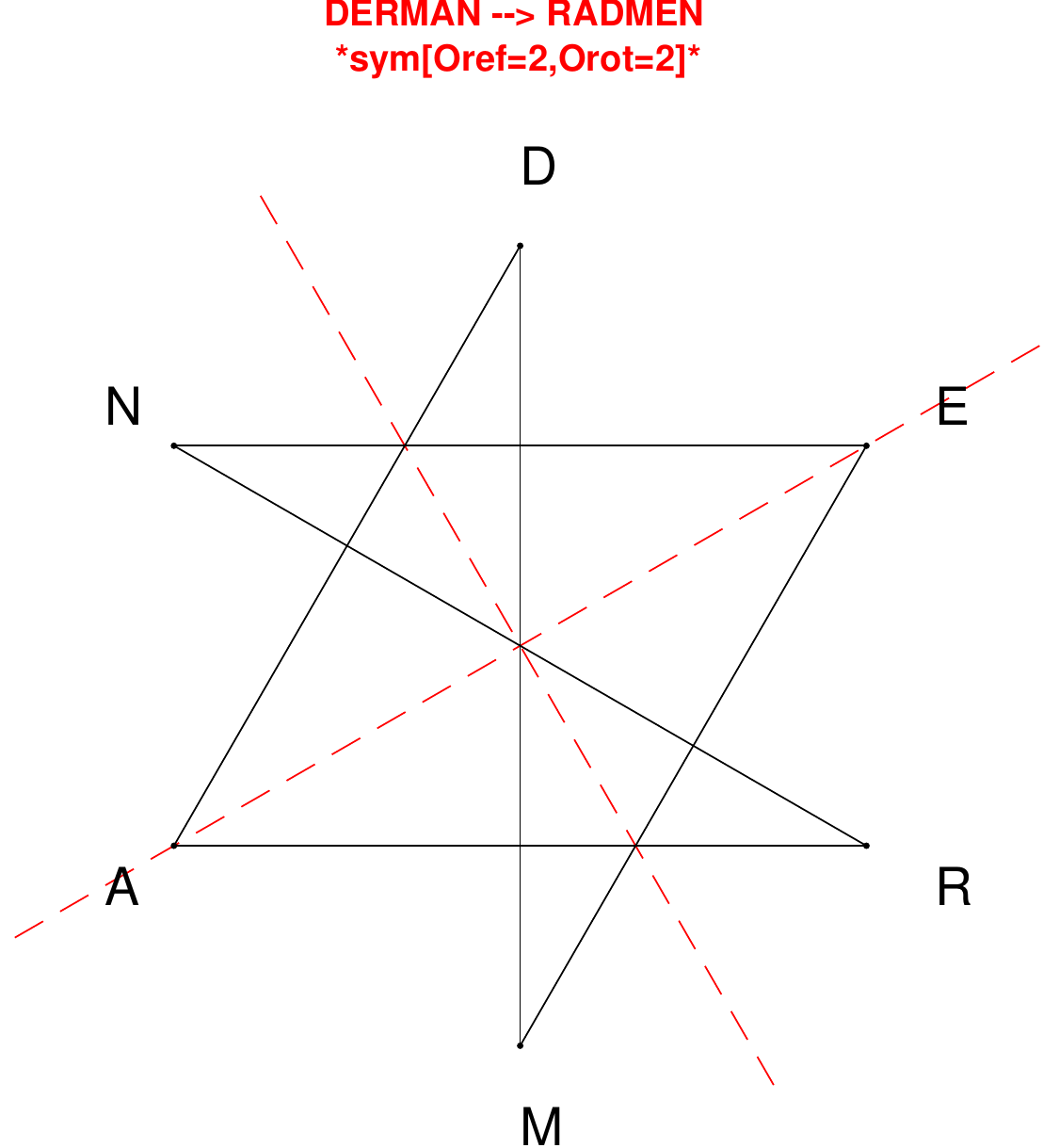}
\end{subfigure}
\end{figure}

\begin{figure}[H]
\centering
\begin{subfigure}[T]{0.19\textwidth}
\centering
\includegraphics[width=\textwidth]{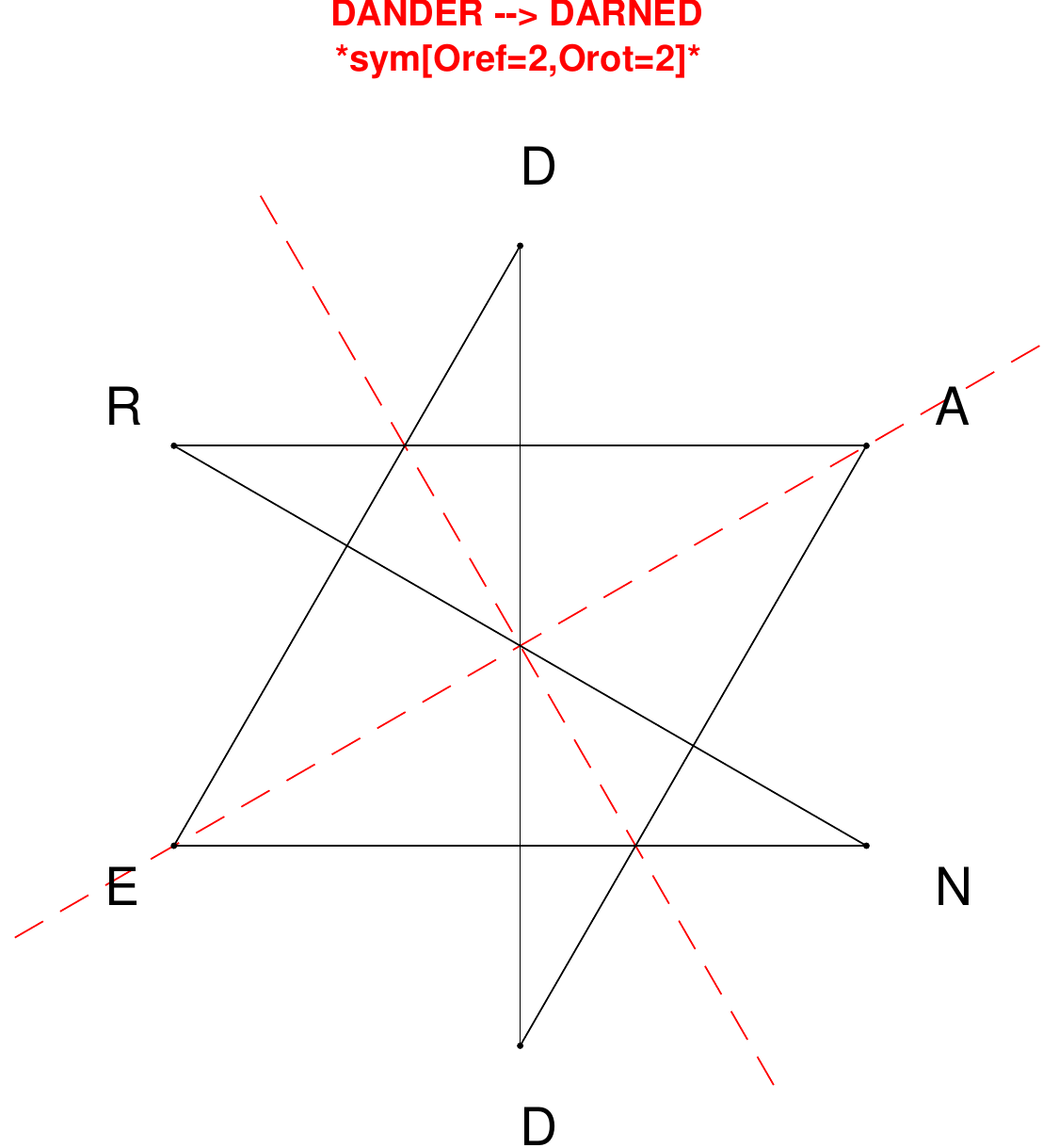}
\end{subfigure}
\hfill
\begin{subfigure}[T]{0.19\textwidth}
\centering
\includegraphics[width=\textwidth]{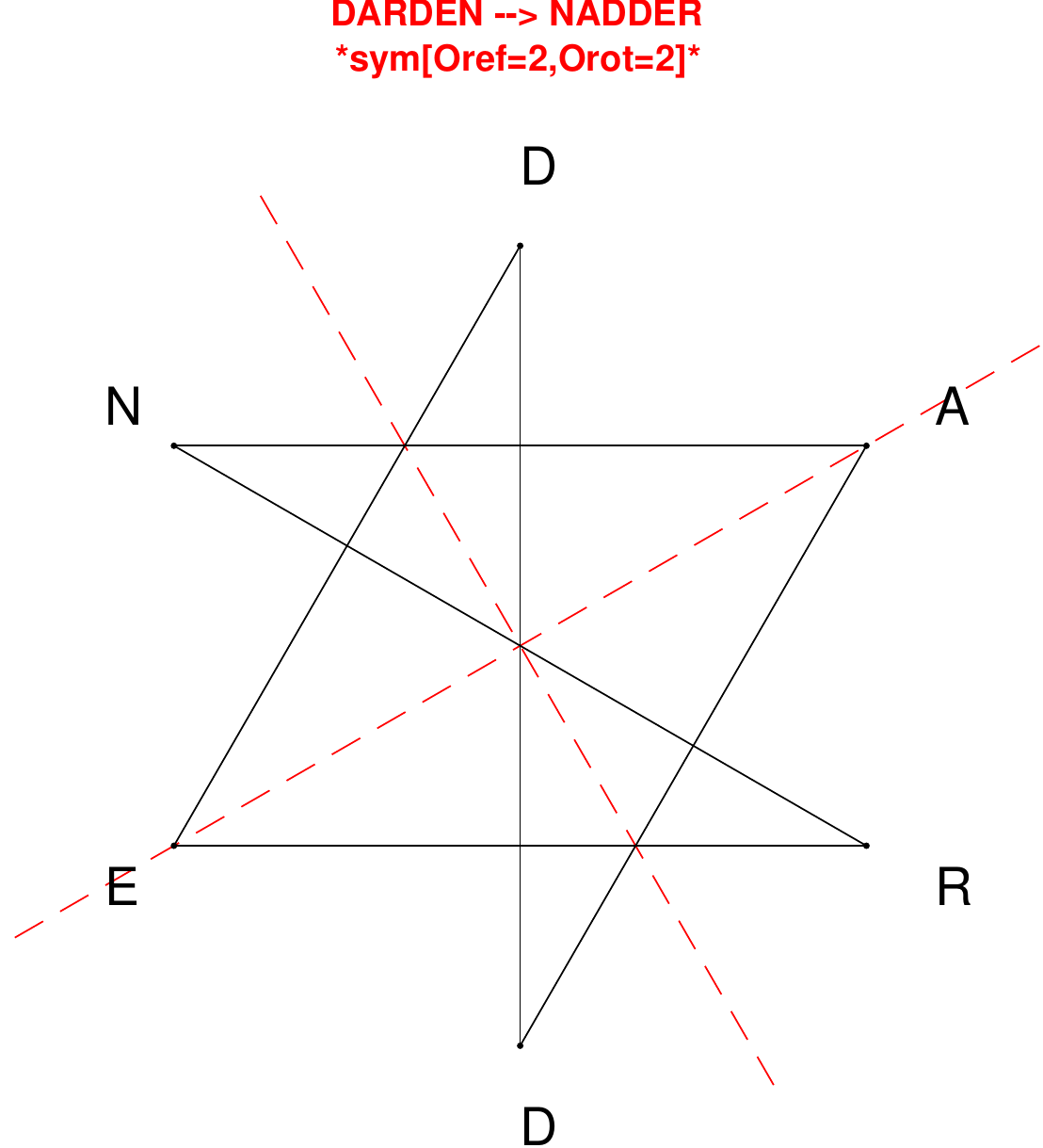}
\end{subfigure}
\hfill
\begin{subfigure}[T]{0.19\textwidth}
\centering
\includegraphics[width=\textwidth]{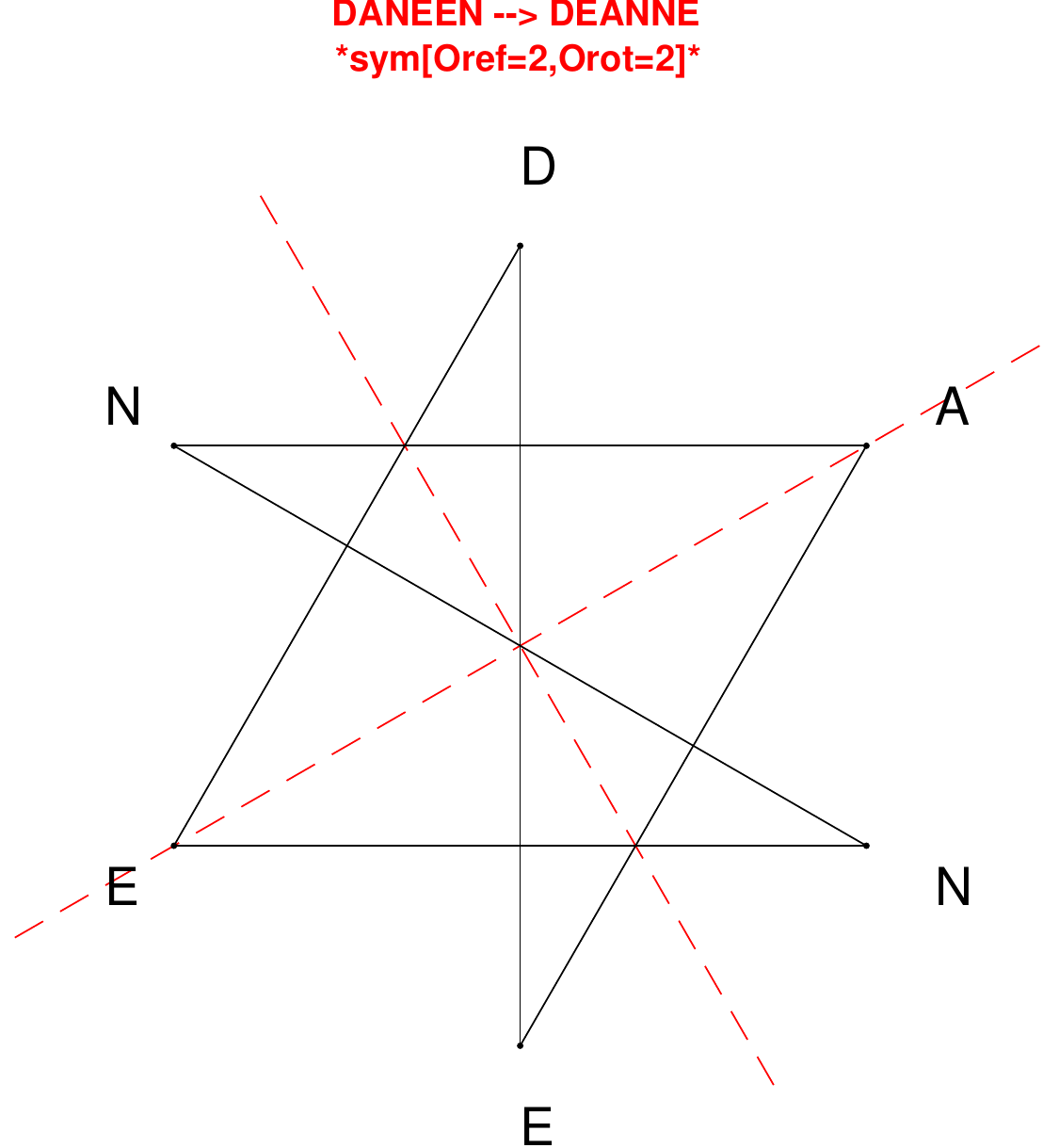}
\end{subfigure}
\hfill
\begin{subfigure}[T]{0.19\textwidth}
\centering
\includegraphics[width=\textwidth]{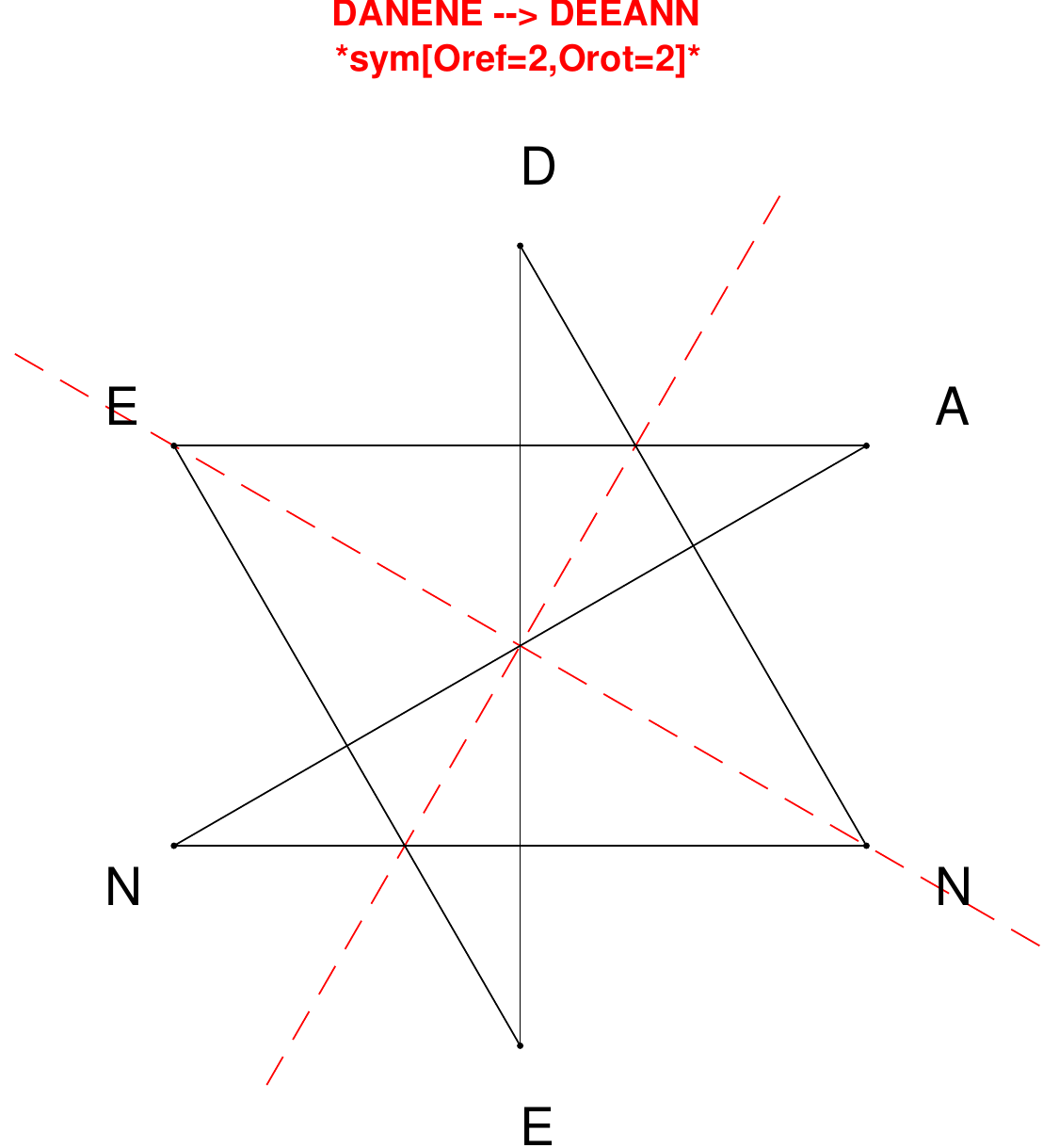}
\end{subfigure}
\hfill
\begin{subfigure}[T]{0.19\textwidth}
\centering
\includegraphics[width=\textwidth]{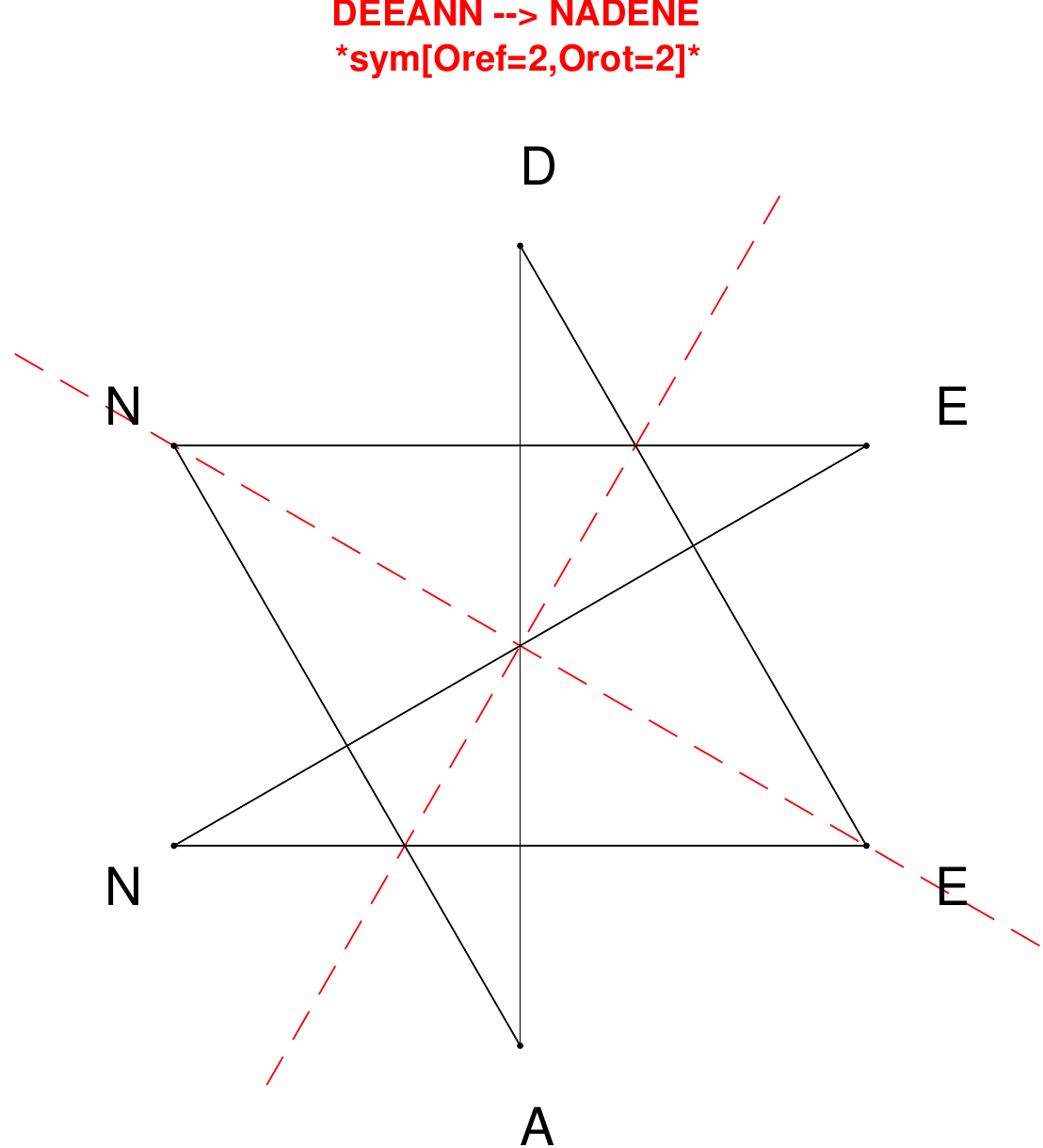}
\end{subfigure}
\end{figure}

\begin{figure}[H]
\centering
\begin{subfigure}[T]{0.19\textwidth}
\centering
\includegraphics[width=\textwidth]{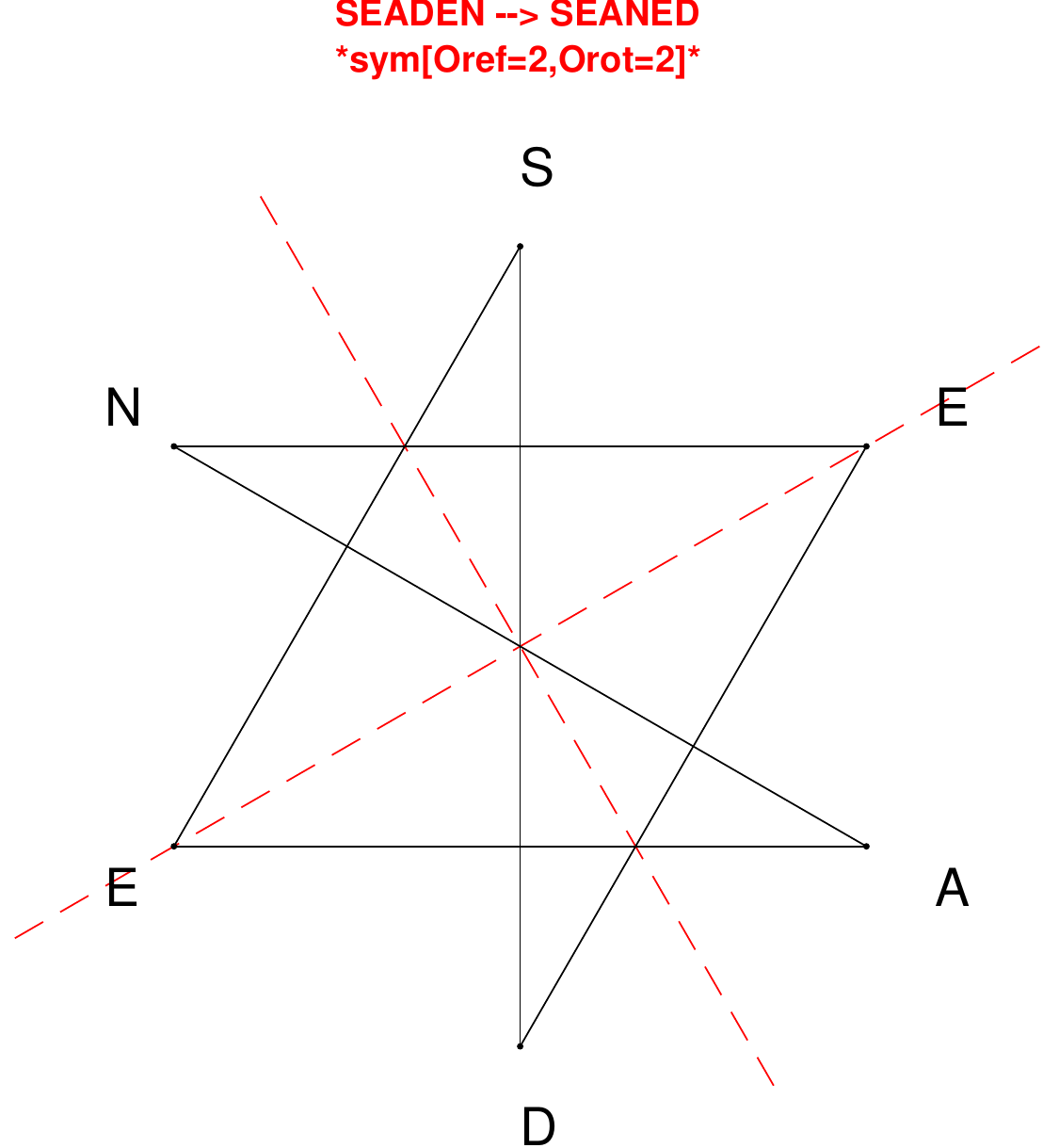}
\end{subfigure}
\hfill
\begin{subfigure}[T]{0.19\textwidth}
\centering
\includegraphics[width=\textwidth]{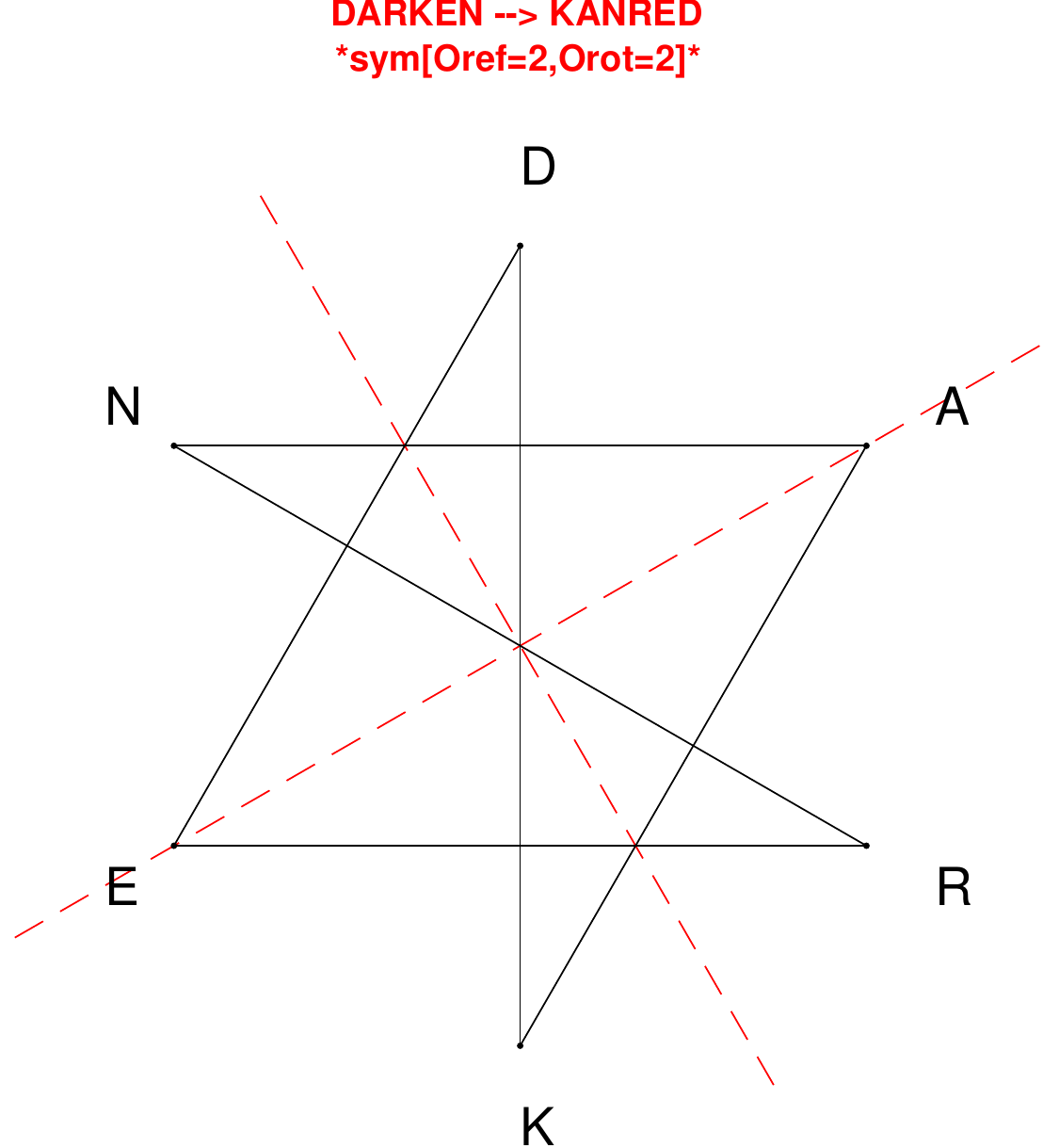}
\end{subfigure}
\hfill
\begin{subfigure}[T]{0.19\textwidth}
\centering
\includegraphics[width=\textwidth]{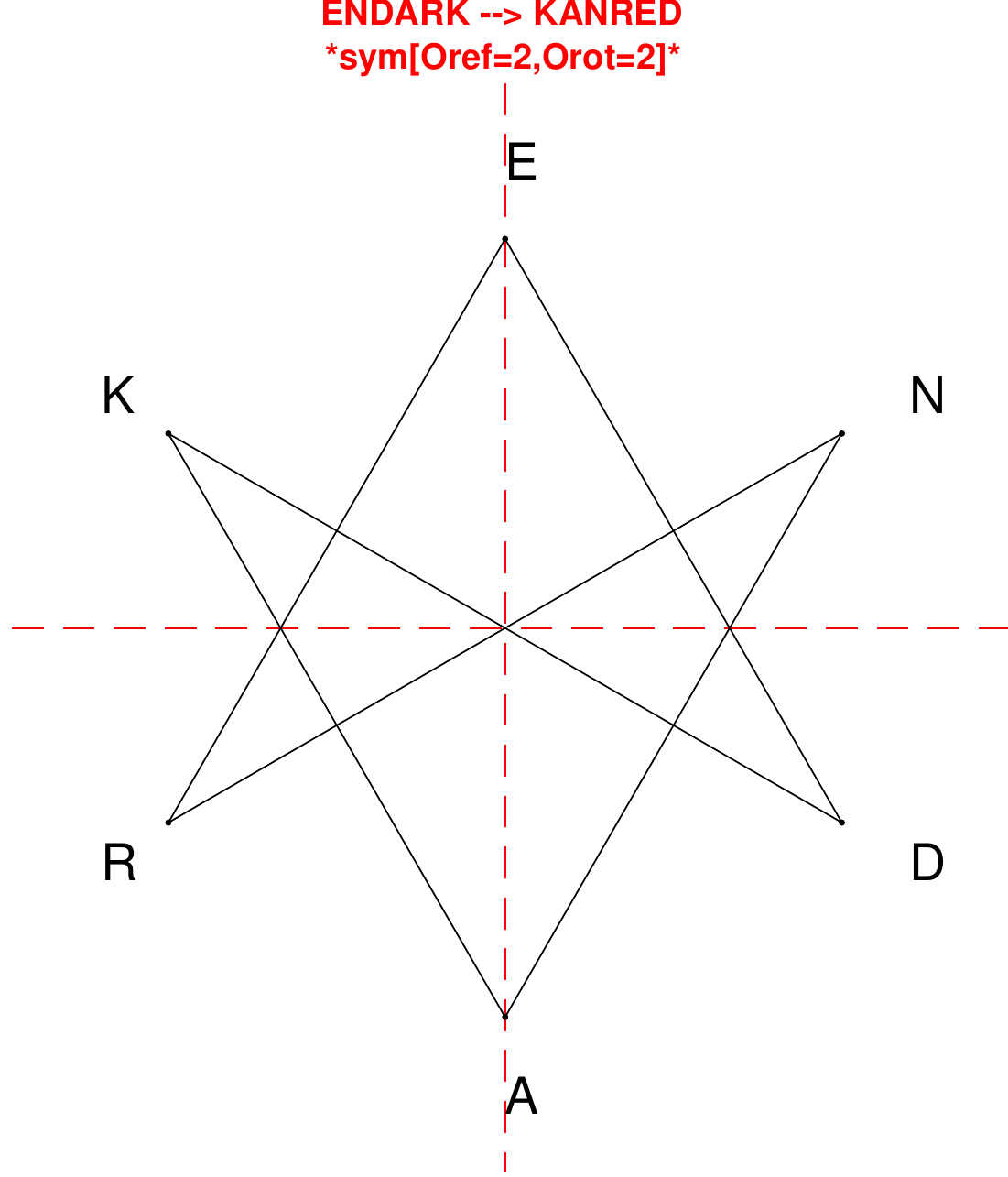}
\end{subfigure}
\hfill
\begin{subfigure}[T]{0.19\textwidth}
\centering
\includegraphics[width=\textwidth]{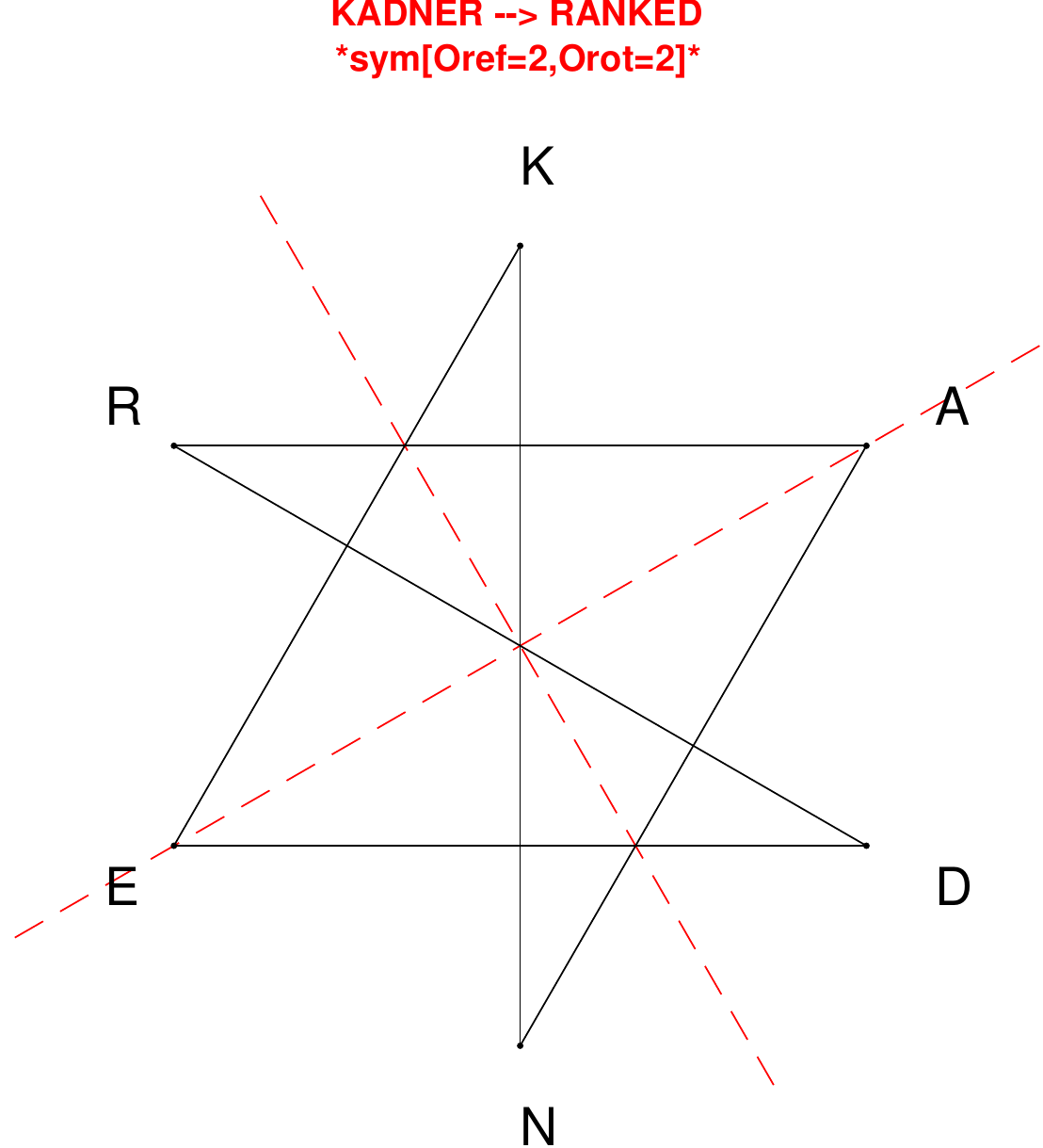}
\end{subfigure}
\hfill
\begin{subfigure}[T]{0.19\textwidth}
\centering
\includegraphics[width=\textwidth]{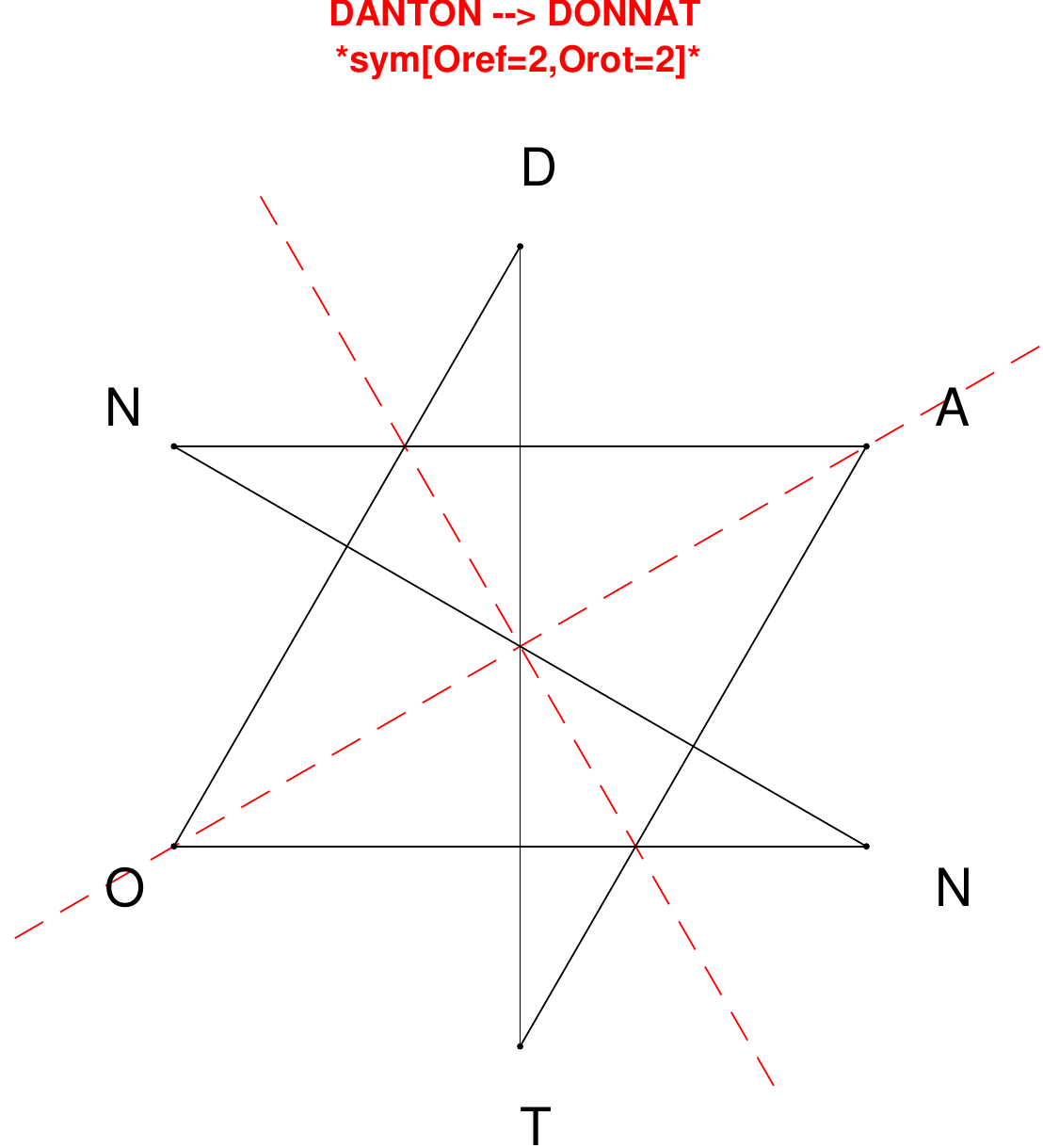}
\end{subfigure}
\end{figure}

\begin{figure}[H]
\centering
\begin{subfigure}[T]{0.19\textwidth}
\centering
\includegraphics[width=\textwidth]{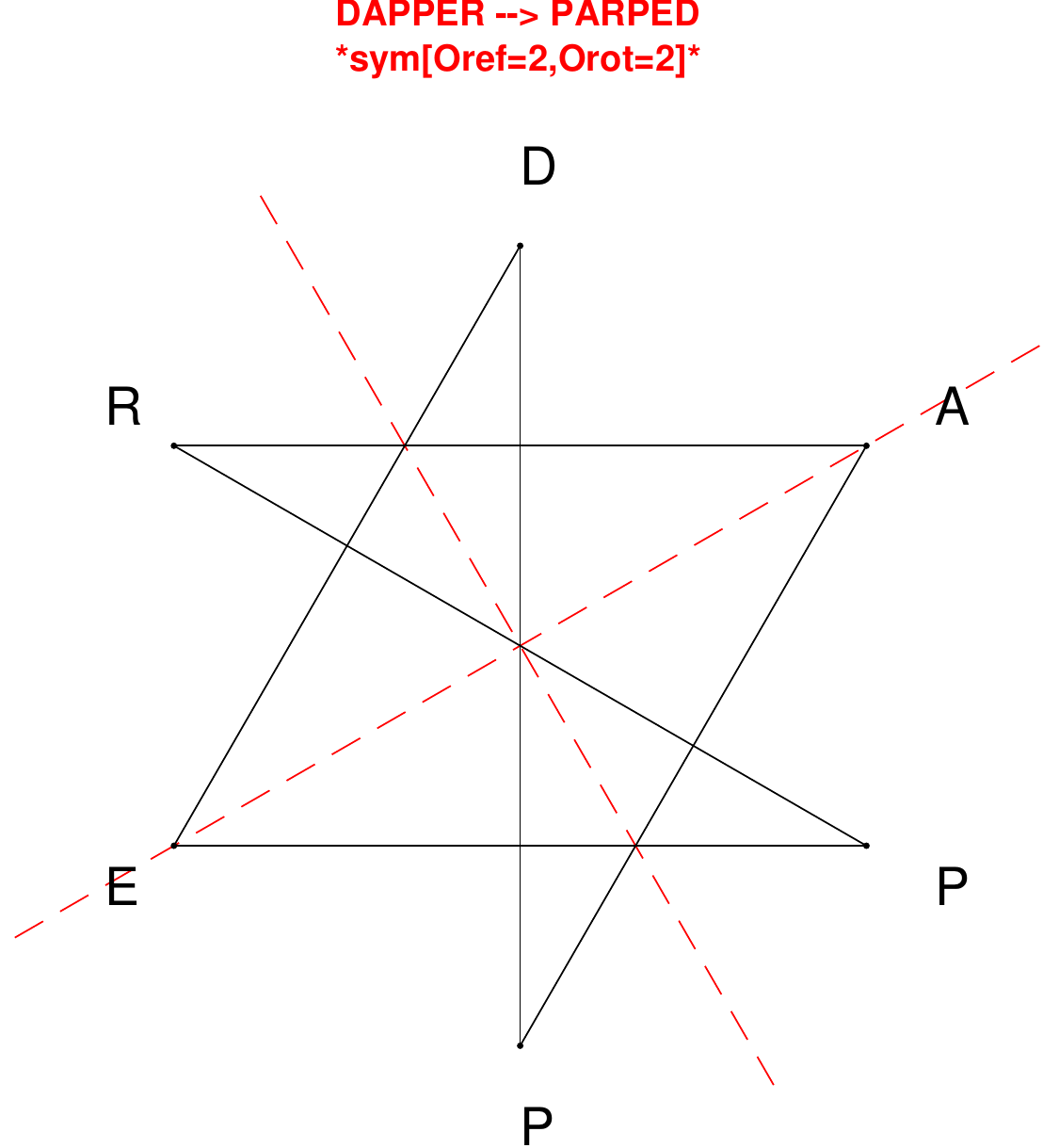}
\end{subfigure}
\hfill
\begin{subfigure}[T]{0.19\textwidth}
\centering
\includegraphics[width=\textwidth]{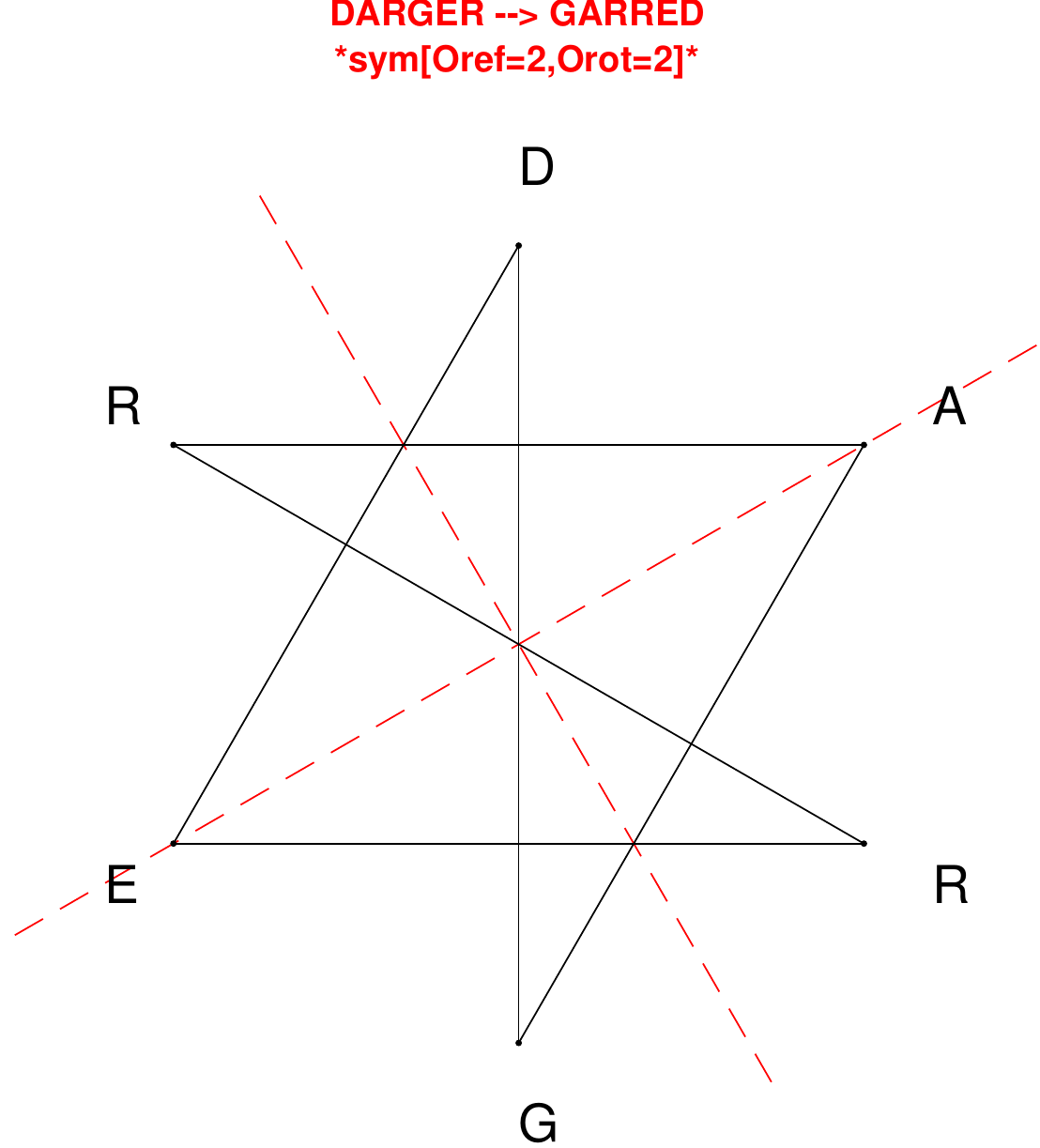}
\end{subfigure}
\hfill
\begin{subfigure}[T]{0.19\textwidth}
\centering
\includegraphics[width=\textwidth]{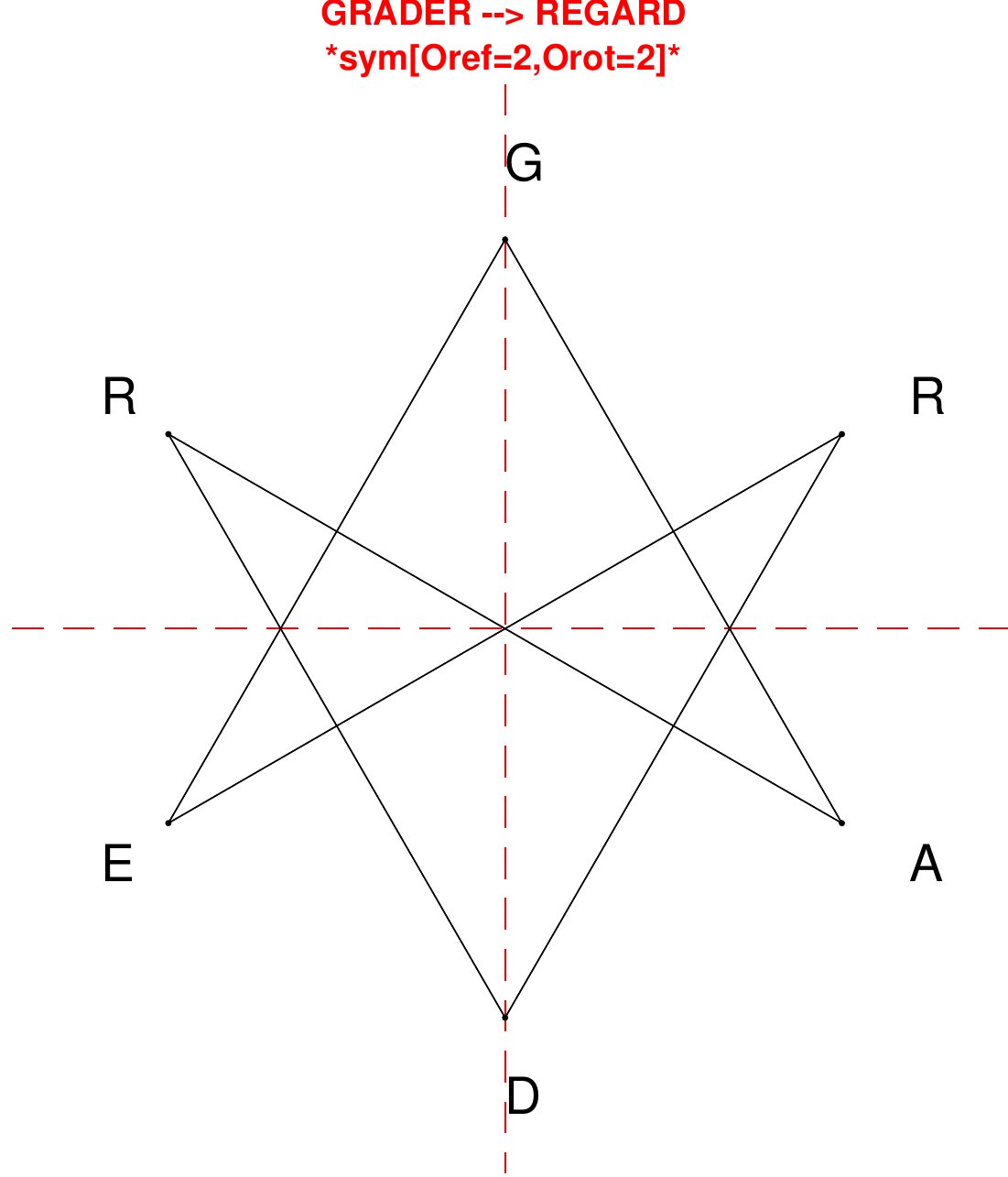}
\end{subfigure}
\hfill
\begin{subfigure}[T]{0.19\textwidth}
\centering
\includegraphics[width=\textwidth]{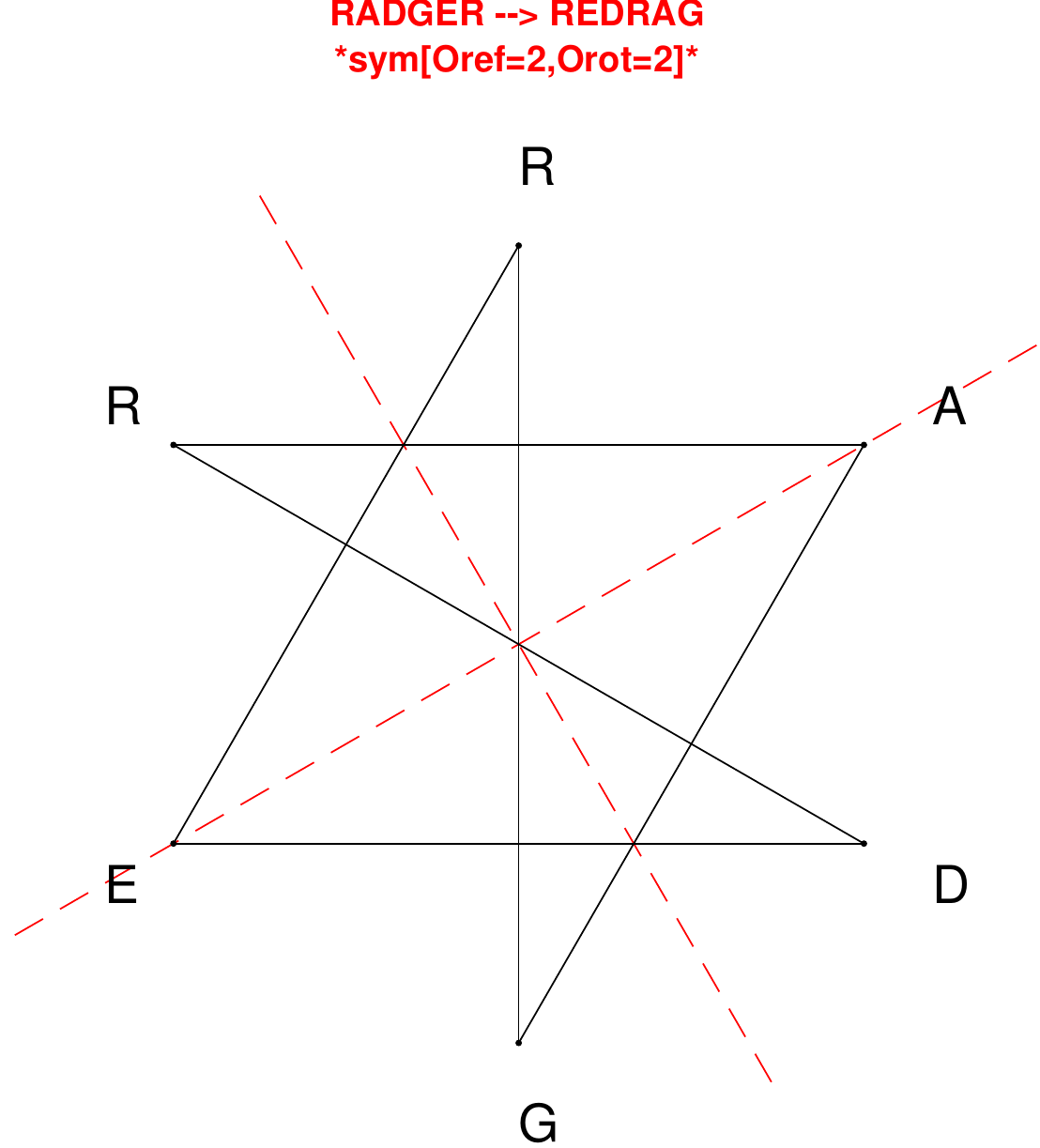}
\end{subfigure}
\hfill
\begin{subfigure}[T]{0.19\textwidth}
\centering
\includegraphics[width=\textwidth]{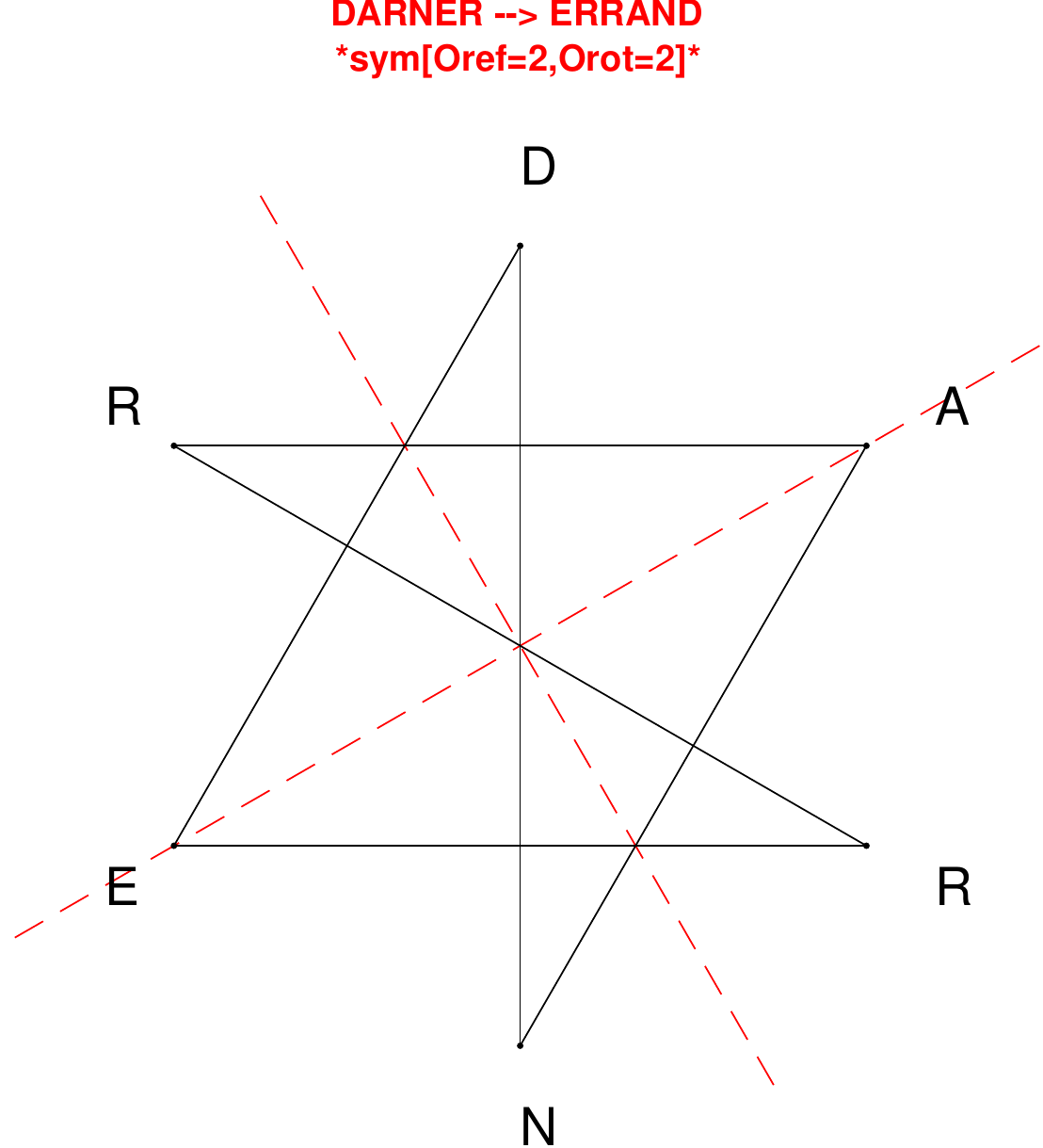}
\end{subfigure}
\end{figure}

\begin{figure}[H]
\centering
\begin{subfigure}[T]{0.19\textwidth}
\centering
\includegraphics[width=\textwidth]{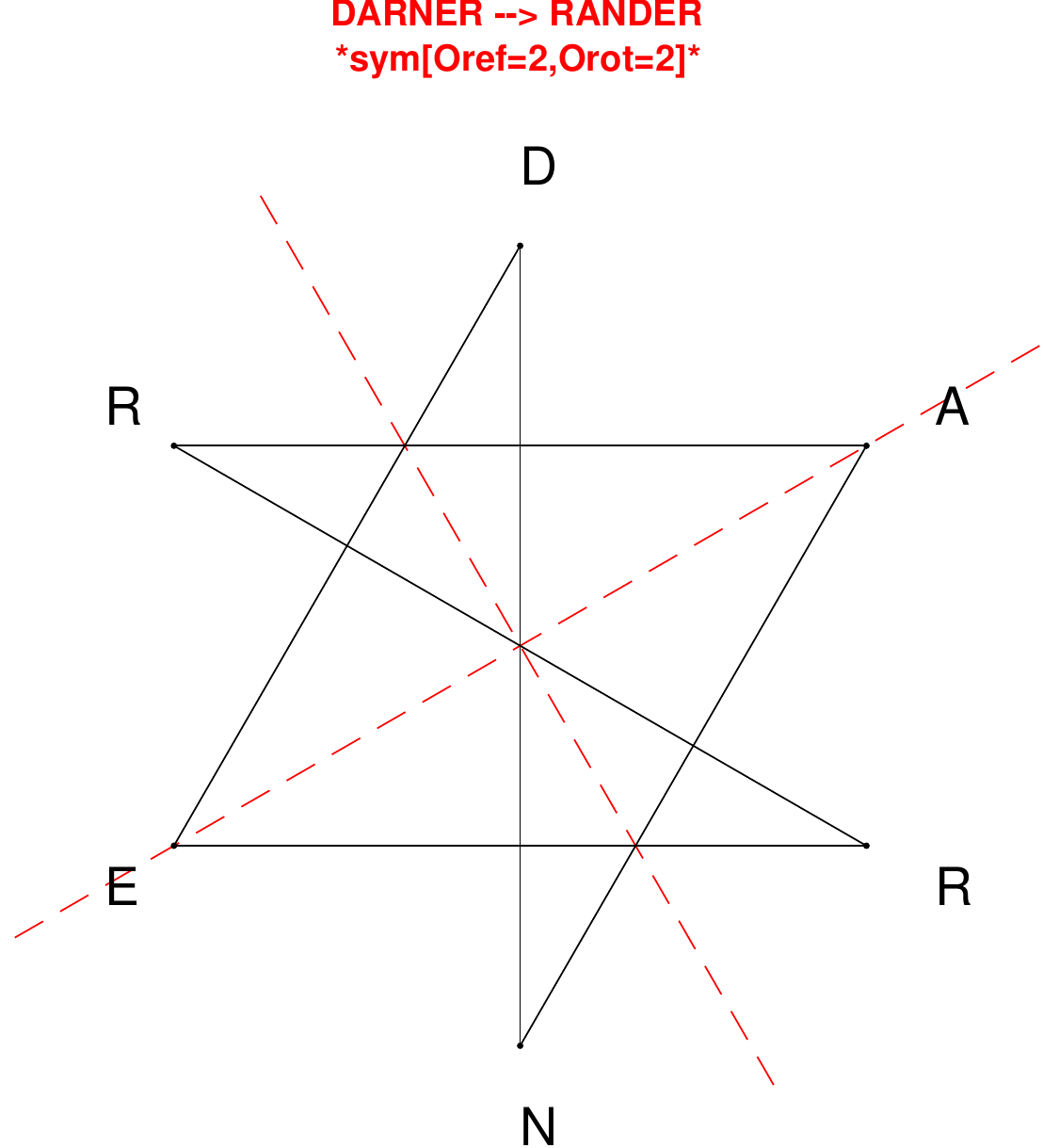}
\end{subfigure}
\hfill
\begin{subfigure}[T]{0.19\textwidth}
\centering
\includegraphics[width=\textwidth]{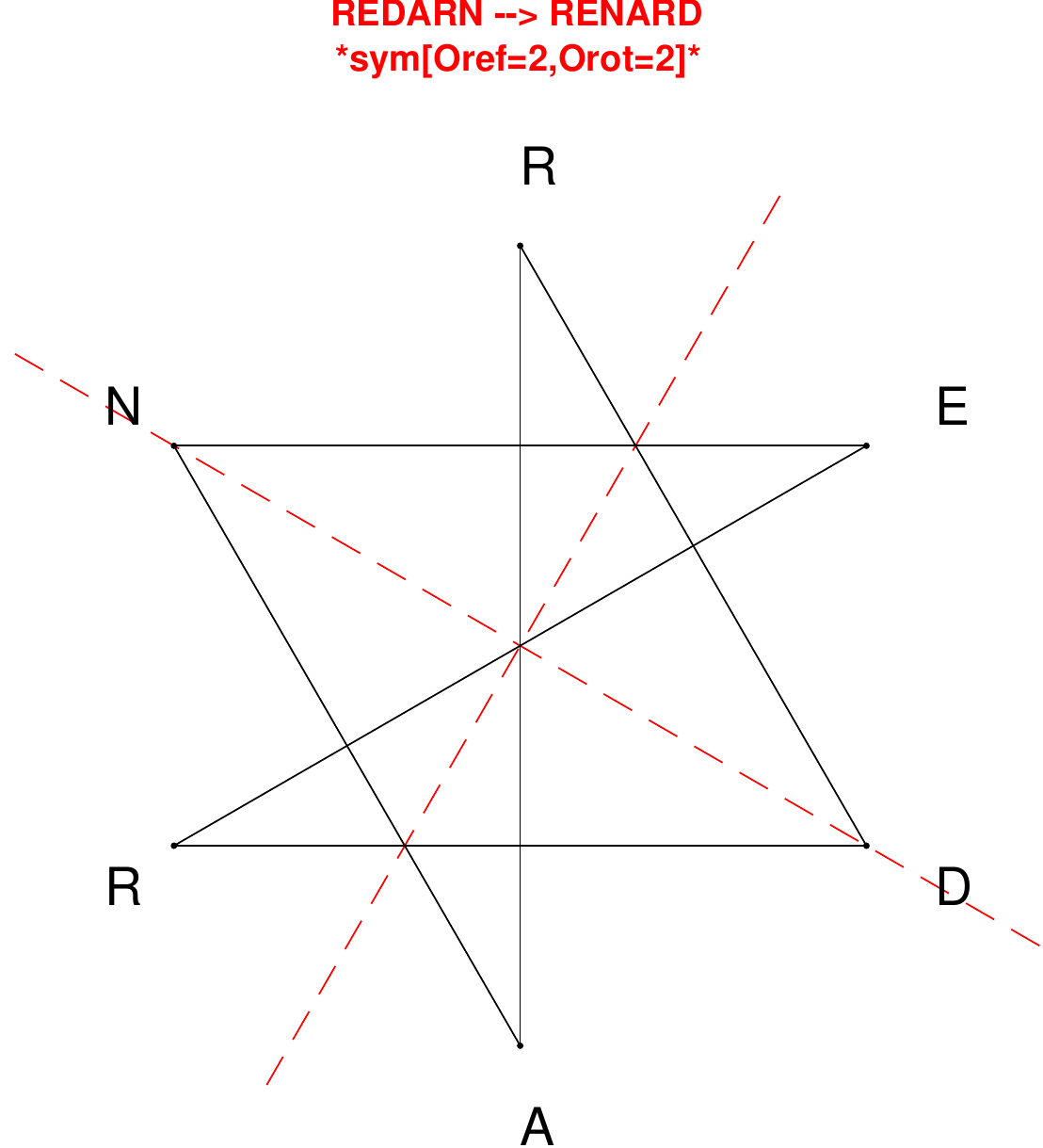}
\end{subfigure}
\hfill
\begin{subfigure}[T]{0.19\textwidth}
\centering
\includegraphics[width=\textwidth]{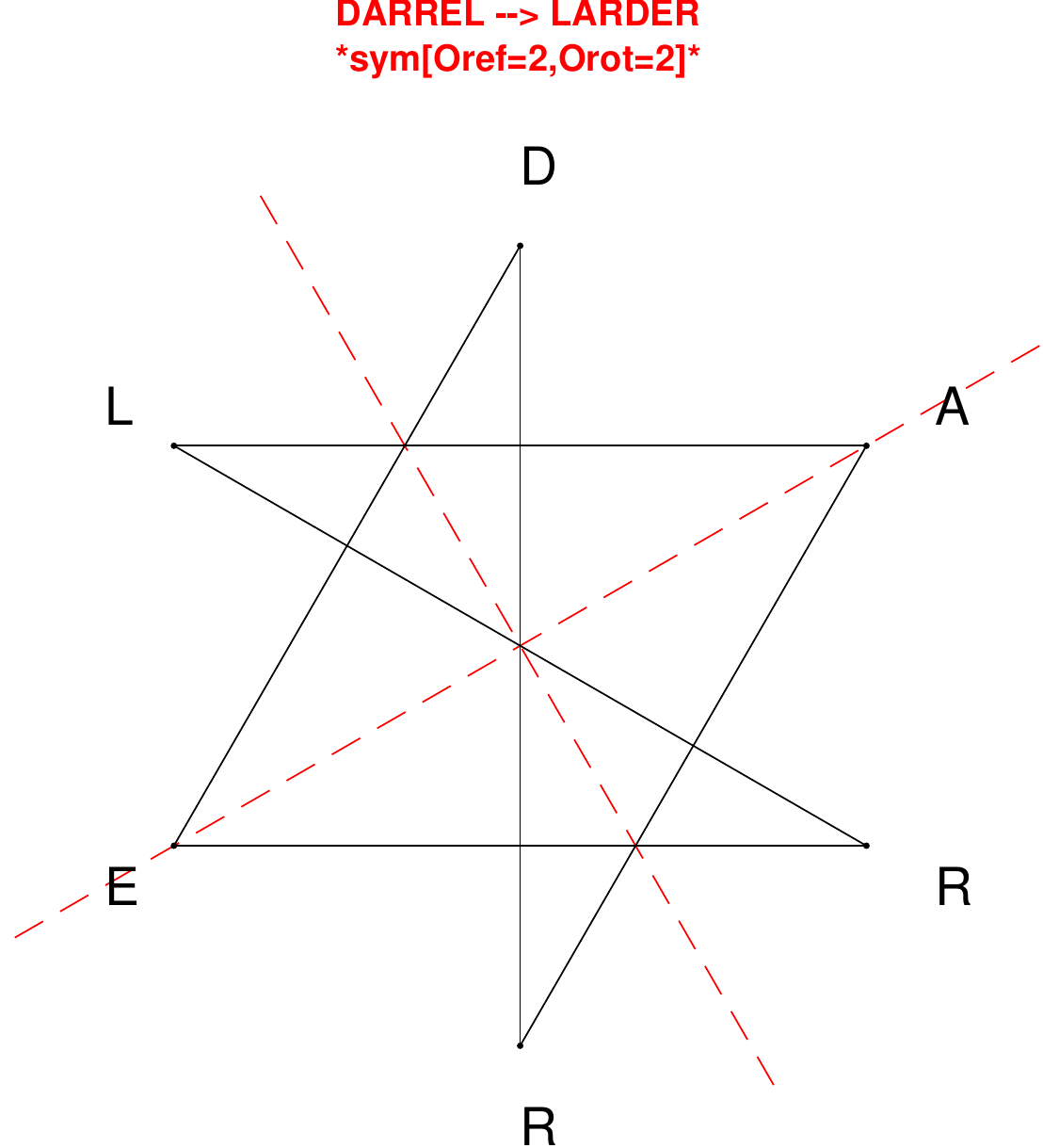}
\end{subfigure}
\hfill
\begin{subfigure}[T]{0.19\textwidth}
\centering
\includegraphics[width=\textwidth]{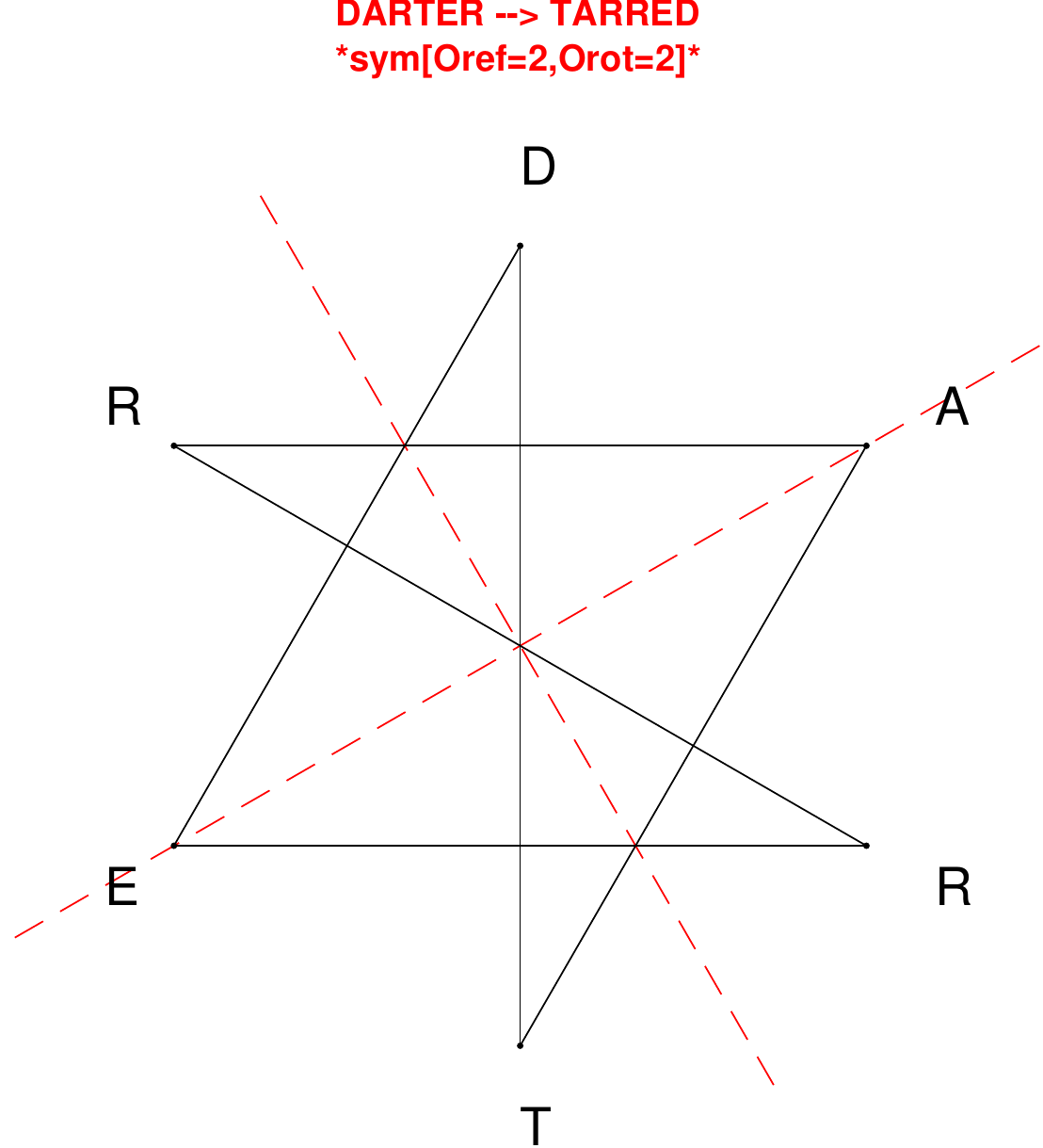}
\end{subfigure}
\hfill
\begin{subfigure}[T]{0.19\textwidth}
\centering
\includegraphics[width=\textwidth]{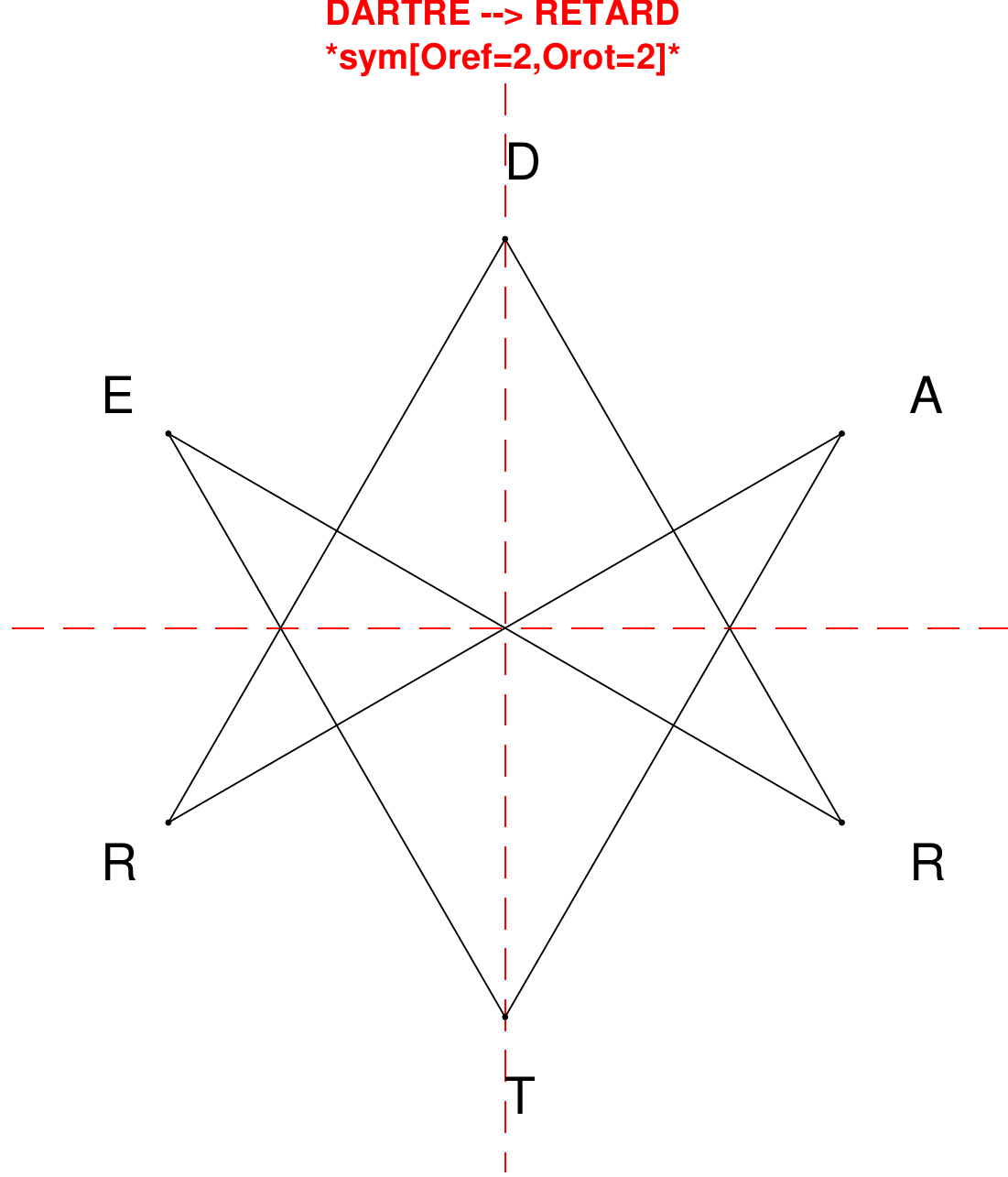}
\end{subfigure}
\end{figure}

\begin{figure}[H]
\centering
\begin{subfigure}[T]{0.19\textwidth}
\centering
\includegraphics[width=\textwidth]{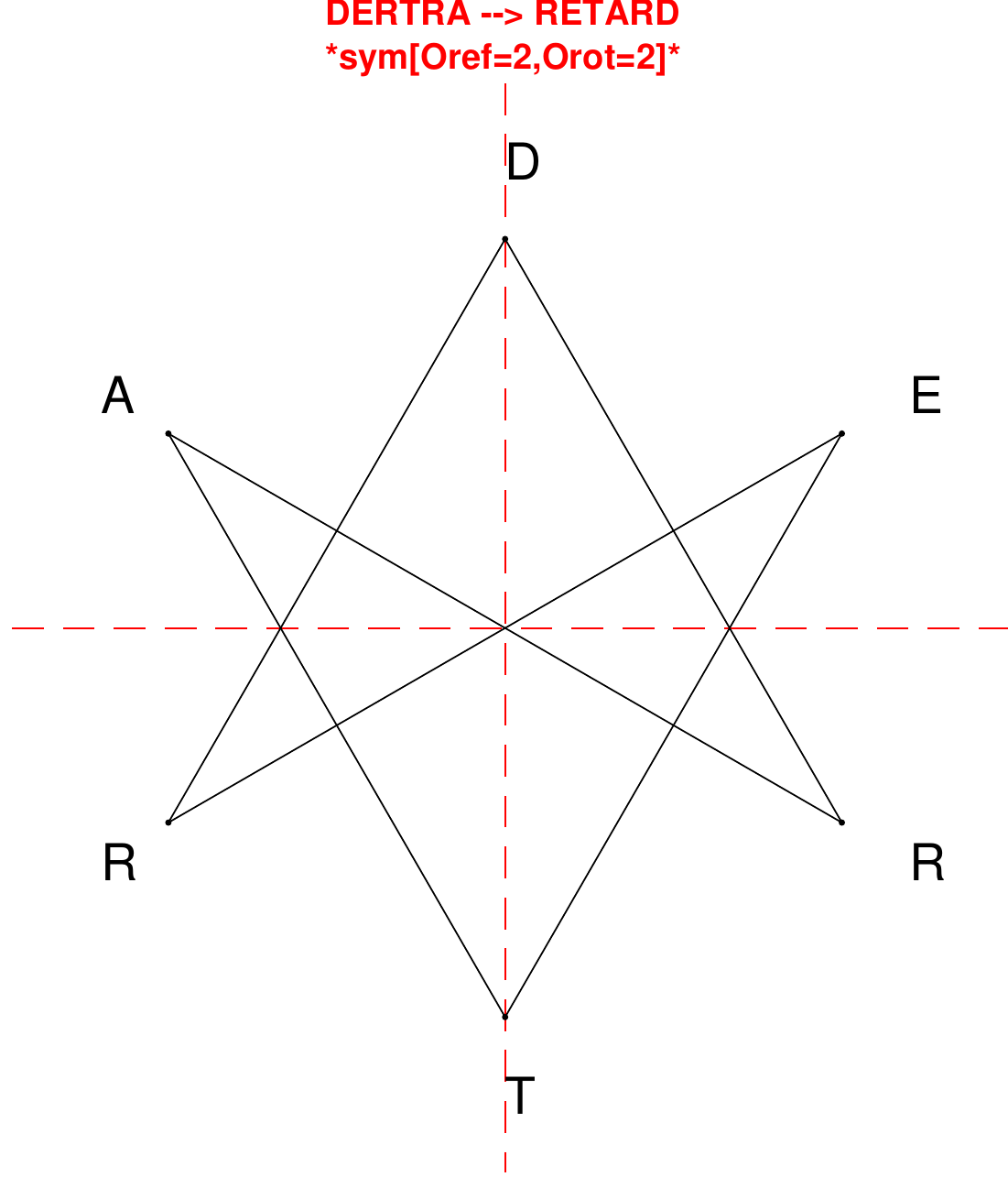}
\end{subfigure}
\hfill
\begin{subfigure}[T]{0.19\textwidth}
\centering
\includegraphics[width=\textwidth]{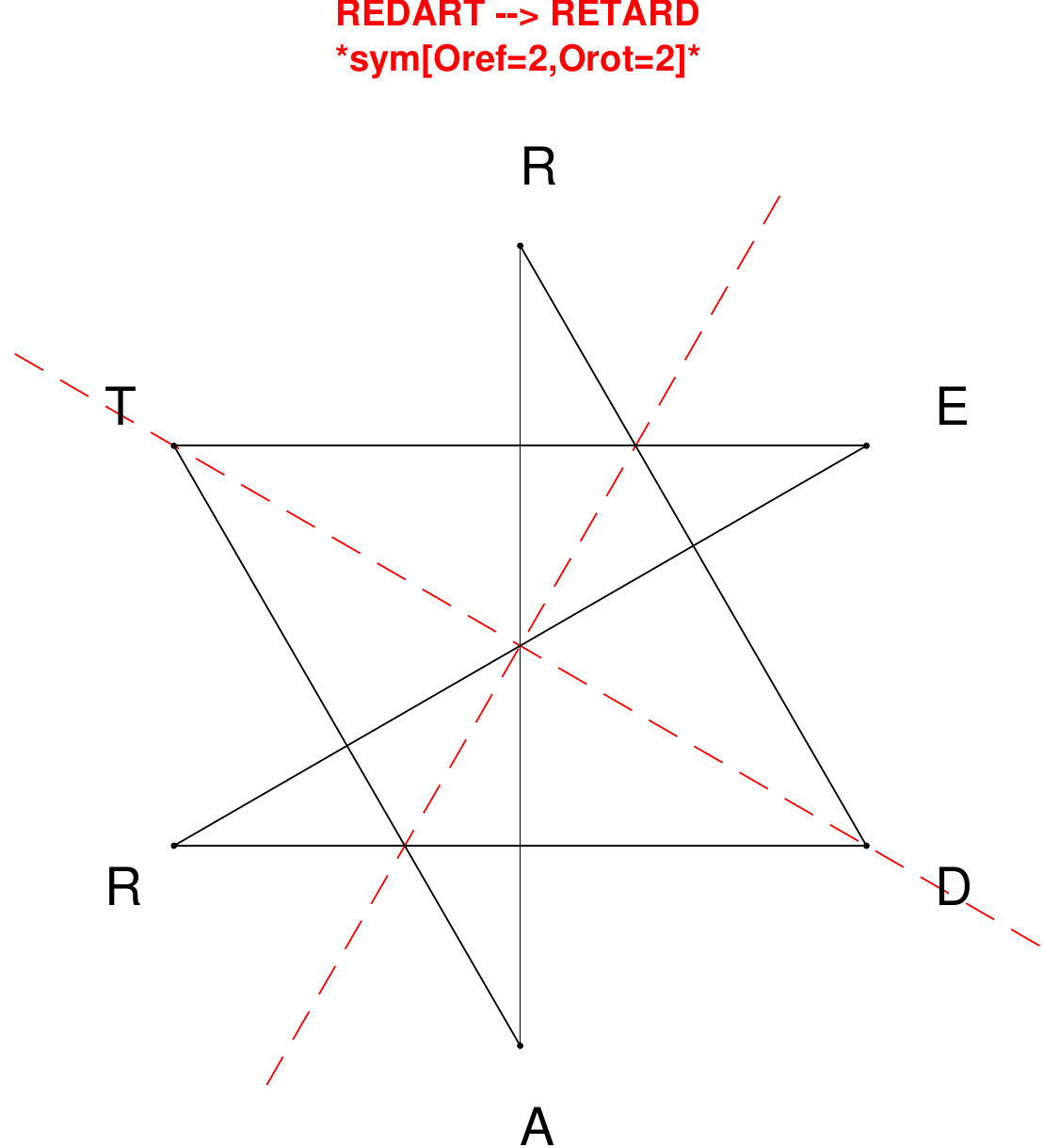}
\end{subfigure}
\hfill
\begin{subfigure}[T]{0.19\textwidth}
\centering
\includegraphics[width=\textwidth]{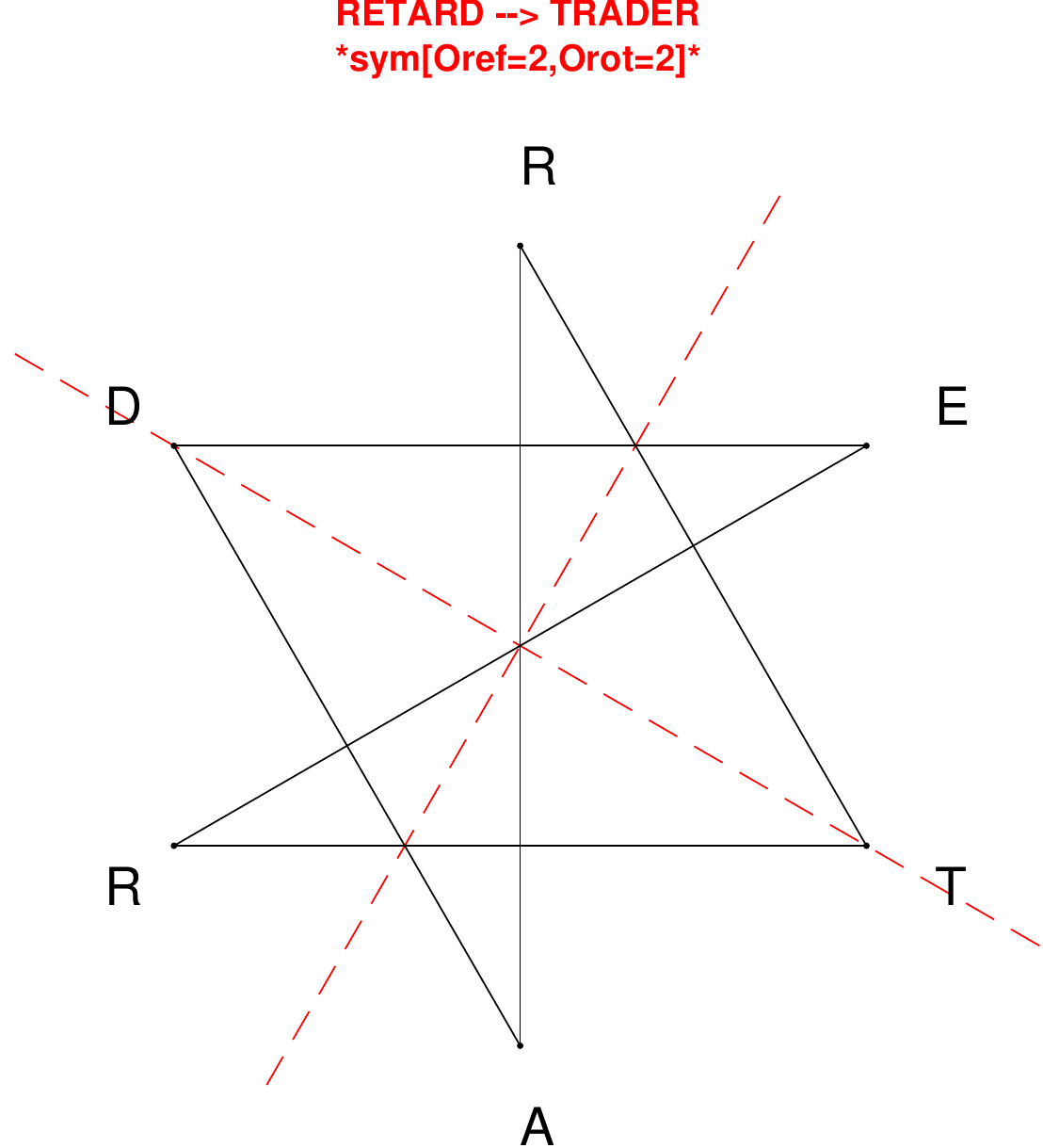}
\end{subfigure}
\hfill
\begin{subfigure}[T]{0.19\textwidth}
\centering
\includegraphics[width=\textwidth]{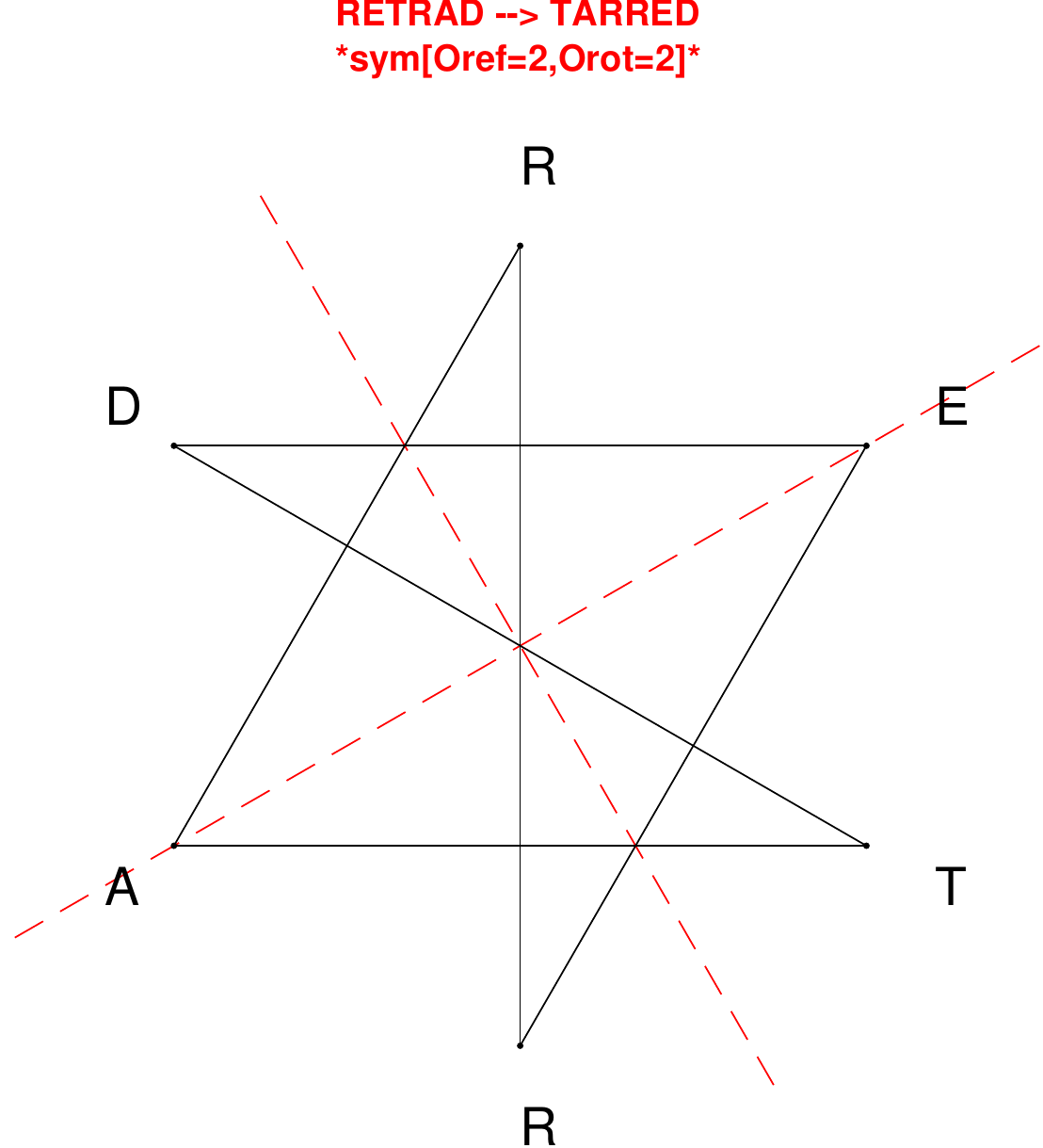}
\end{subfigure}
\hfill
\begin{subfigure}[T]{0.19\textwidth}
\centering
\includegraphics[width=\textwidth]{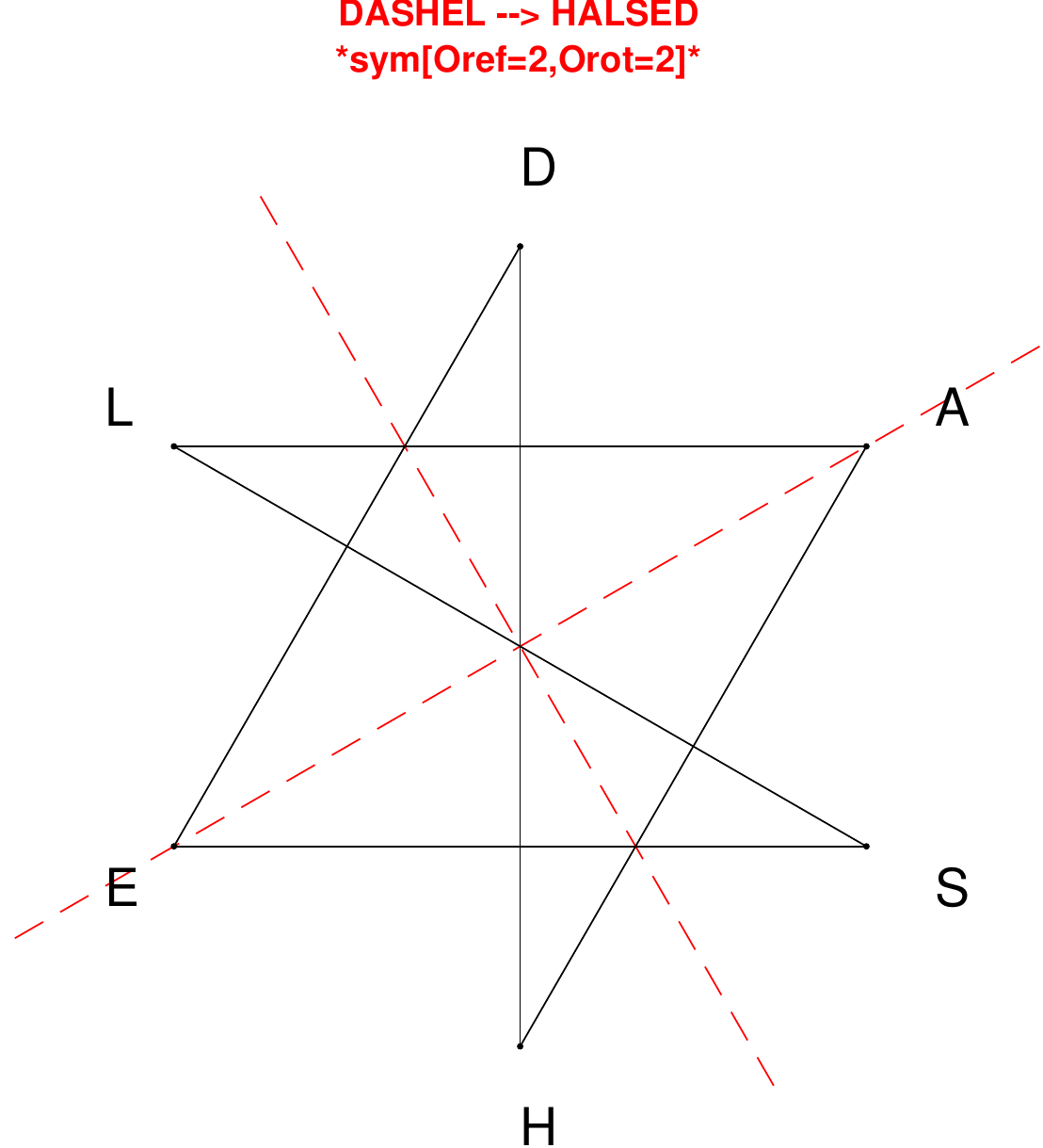}
\end{subfigure}
\end{figure}

\begin{figure}[H]
\centering
\begin{subfigure}[T]{0.19\textwidth}
\centering
\includegraphics[width=\textwidth]{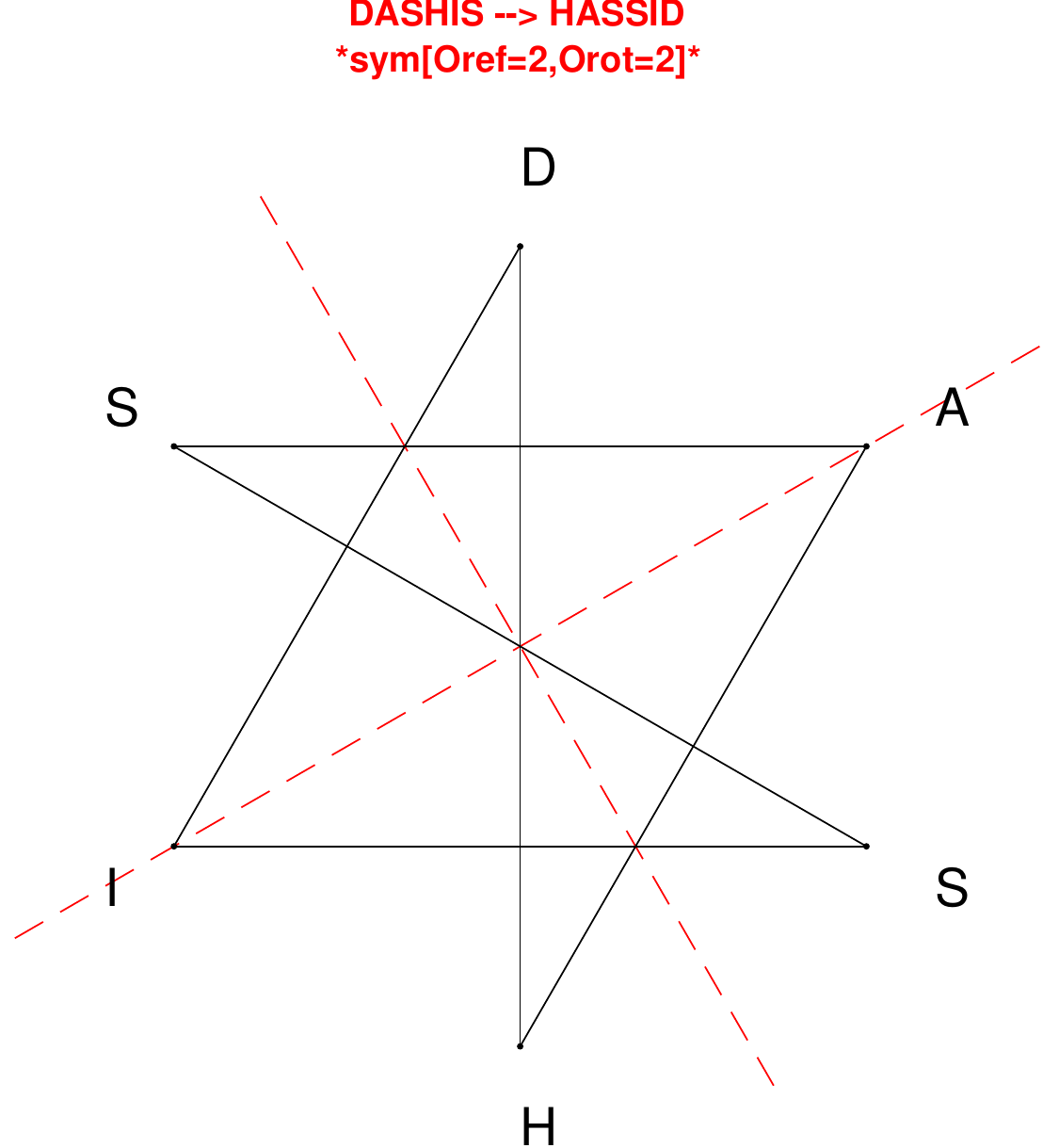}
\end{subfigure}
\hfill
\begin{subfigure}[T]{0.19\textwidth}
\centering
\includegraphics[width=\textwidth]{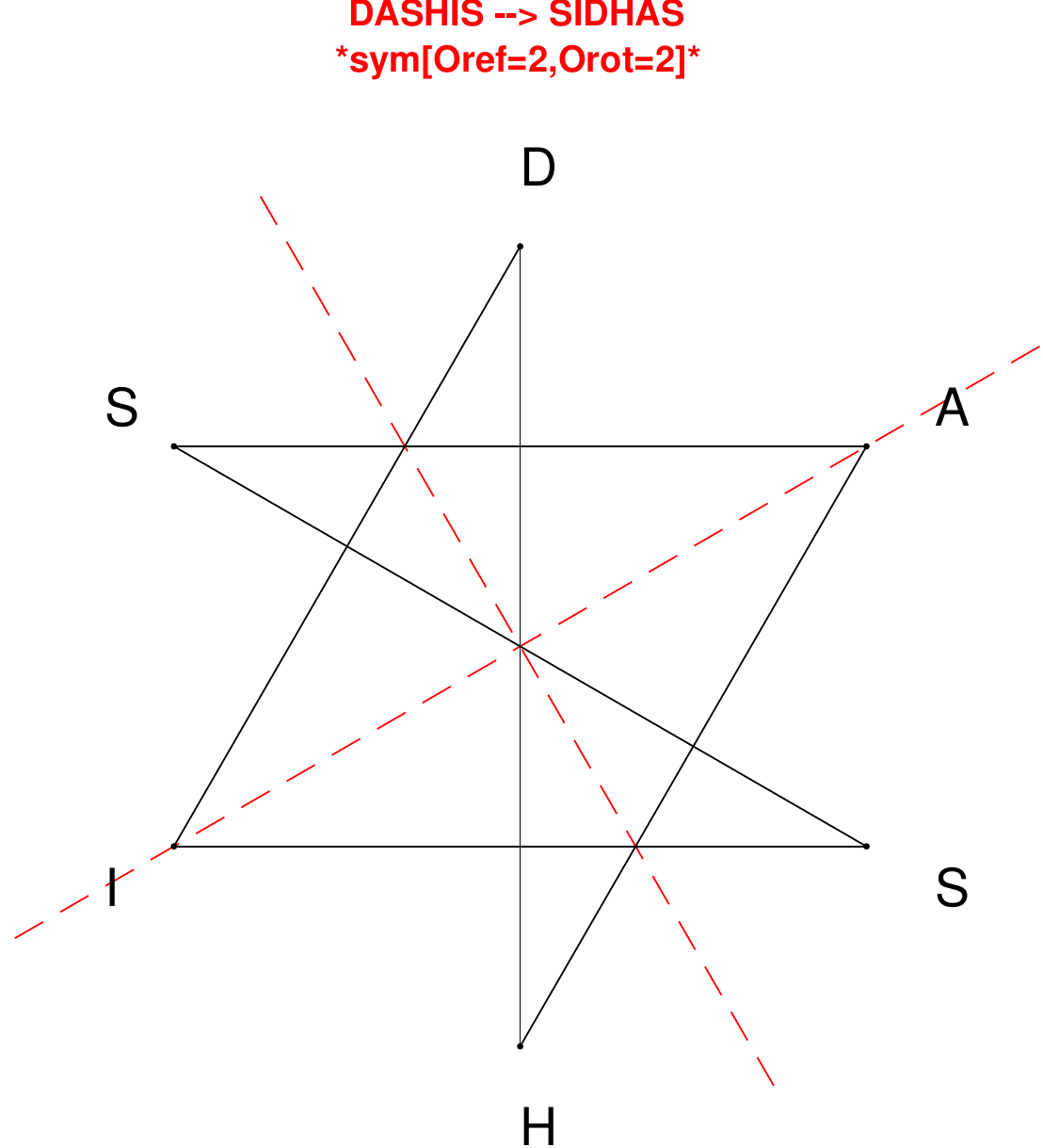}
\end{subfigure}
\hfill
\begin{subfigure}[T]{0.19\textwidth}
\centering
\includegraphics[width=\textwidth]{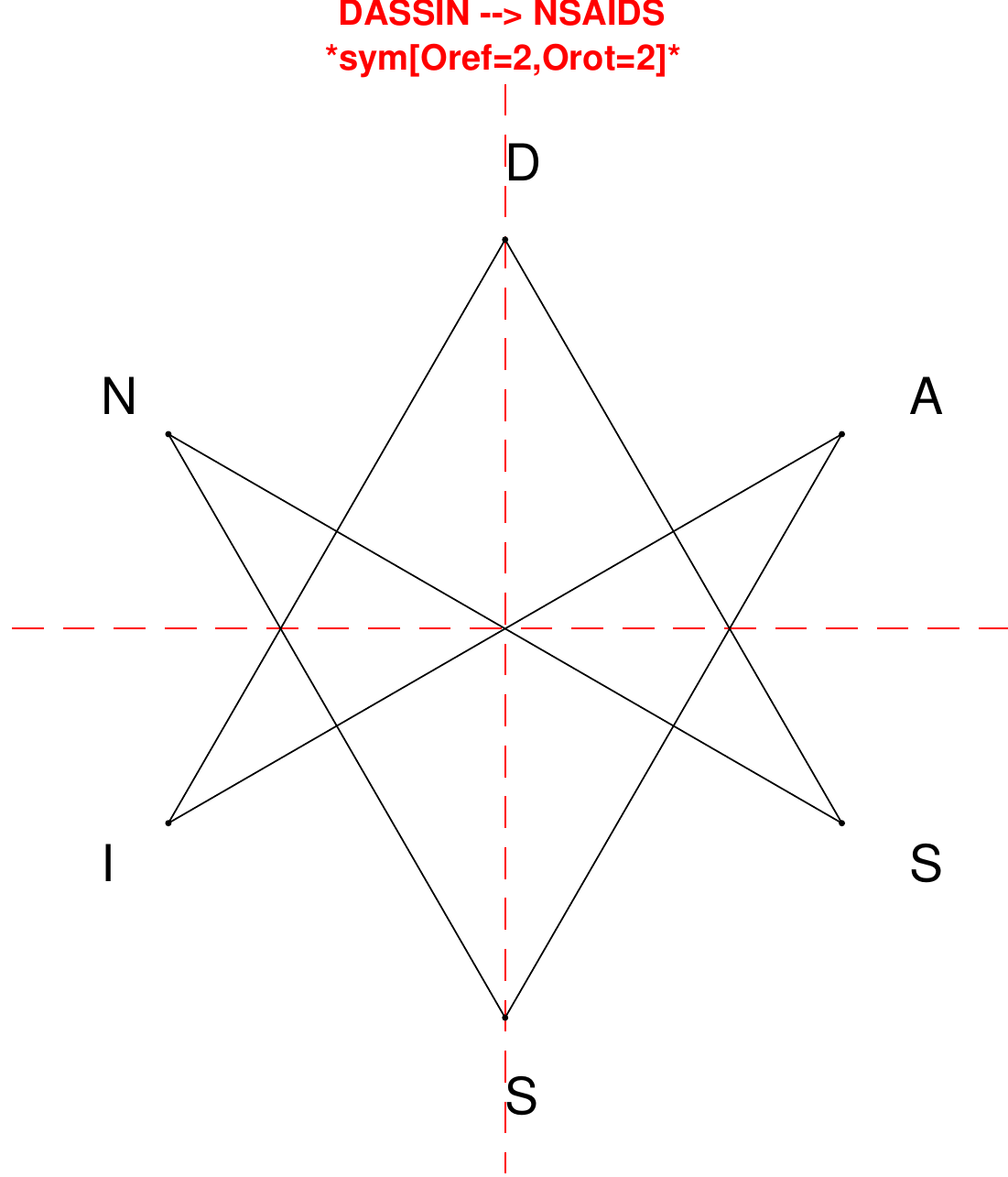}
\end{subfigure}
\hfill
\begin{subfigure}[T]{0.19\textwidth}
\centering
\includegraphics[width=\textwidth]{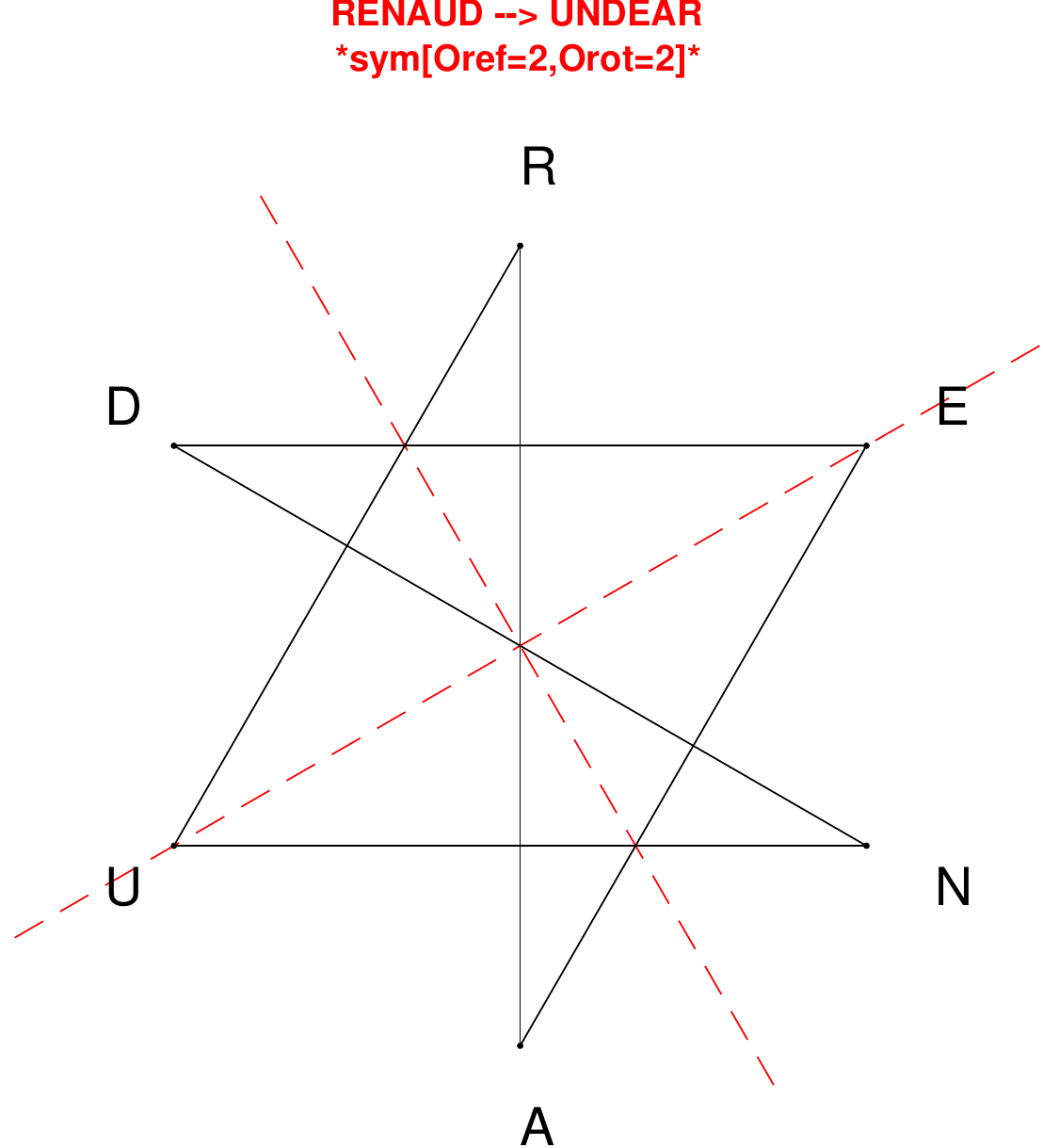}
\end{subfigure}
\hfill
\begin{subfigure}[T]{0.19\textwidth}
\centering
\includegraphics[width=\textwidth]{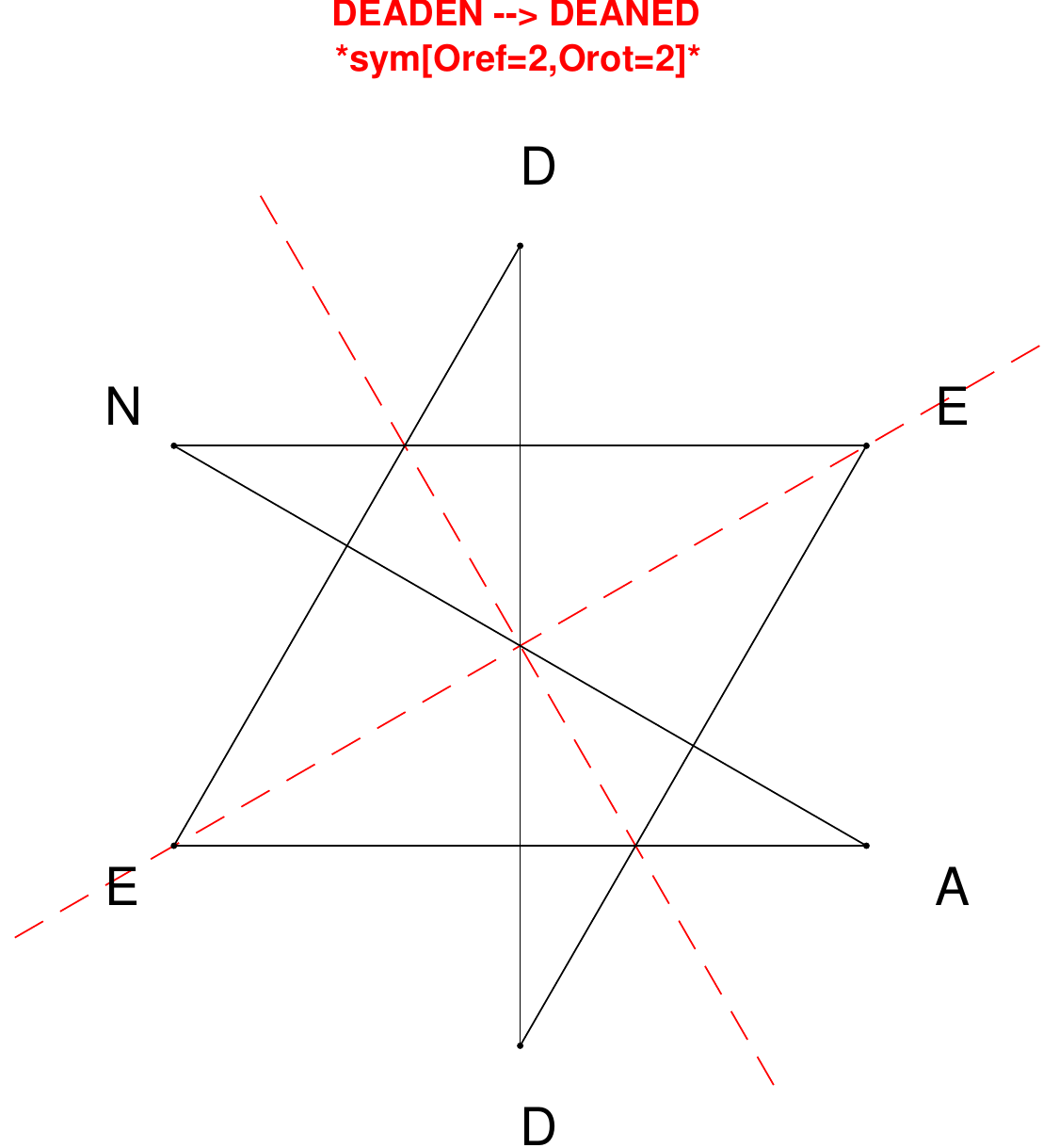}
\end{subfigure}
\end{figure}

\begin{figure}[H]
\centering
\begin{subfigure}[T]{0.19\textwidth}
\centering
\includegraphics[width=\textwidth]{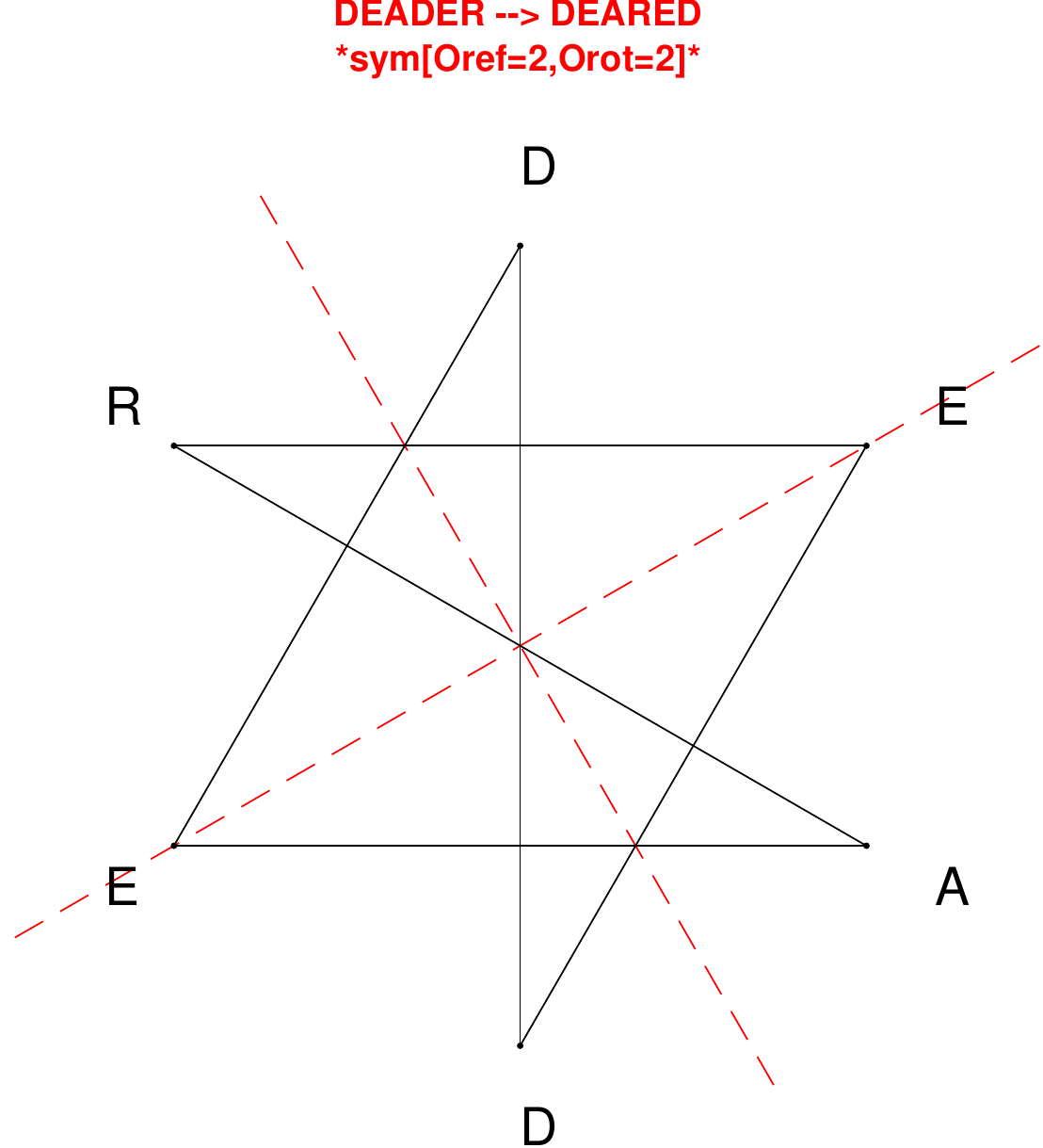}
\end{subfigure}
\hfill
\begin{subfigure}[T]{0.19\textwidth}
\centering
\includegraphics[width=\textwidth]{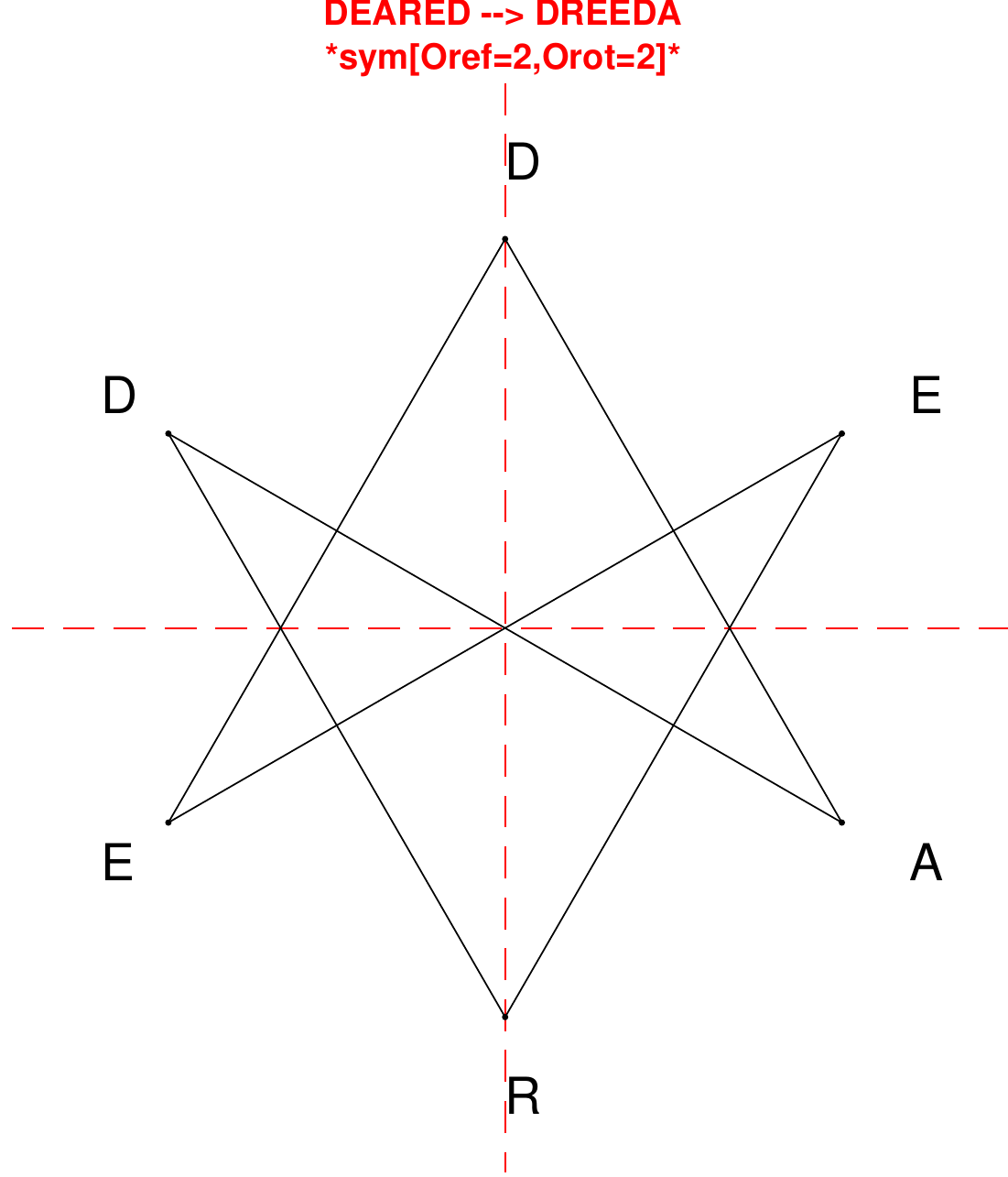}
\end{subfigure}
\hfill
\begin{subfigure}[T]{0.19\textwidth}
\centering
\includegraphics[width=\textwidth]{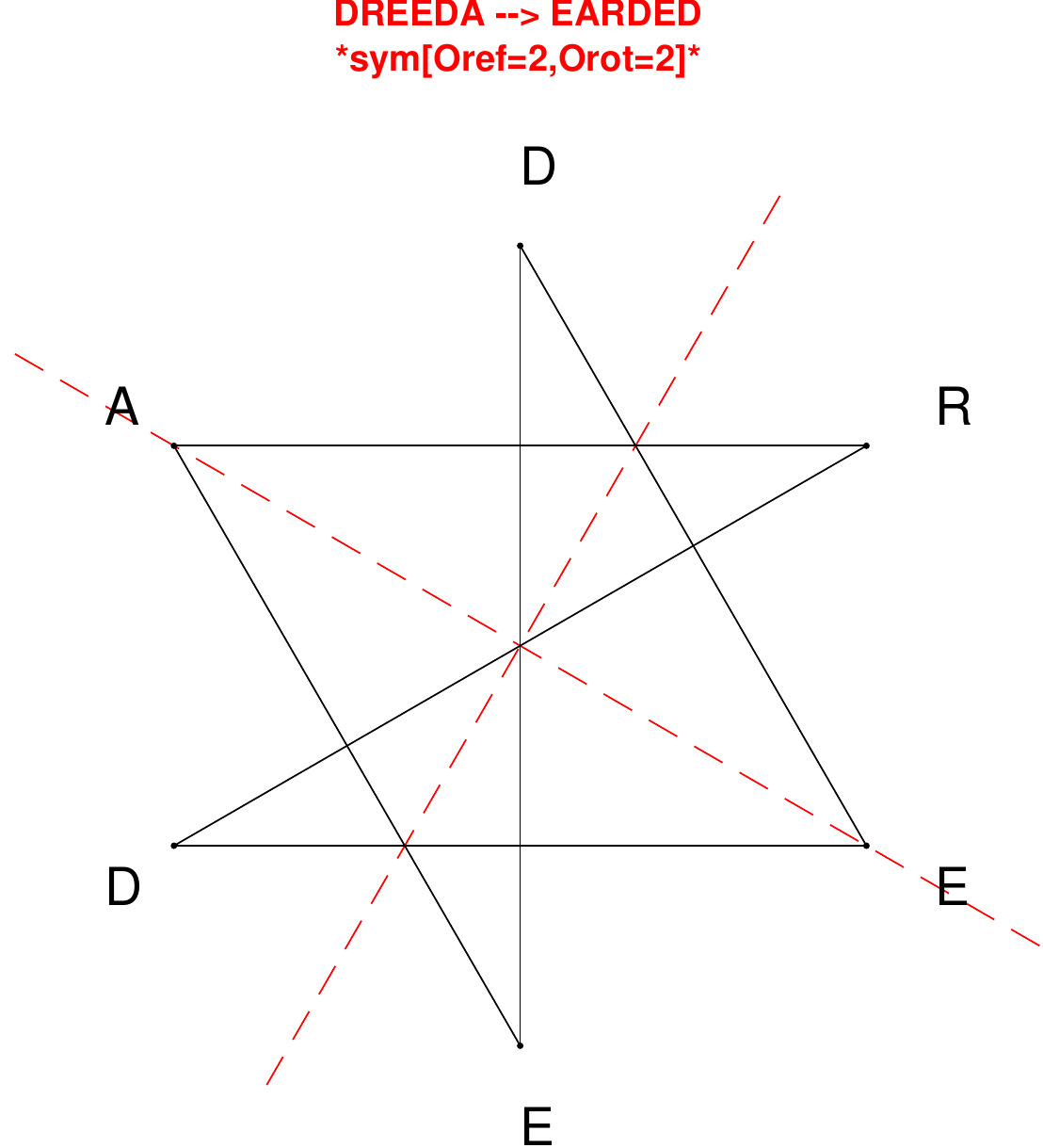}
\end{subfigure}
\hfill
\begin{subfigure}[T]{0.19\textwidth}
\centering
\includegraphics[width=\textwidth]{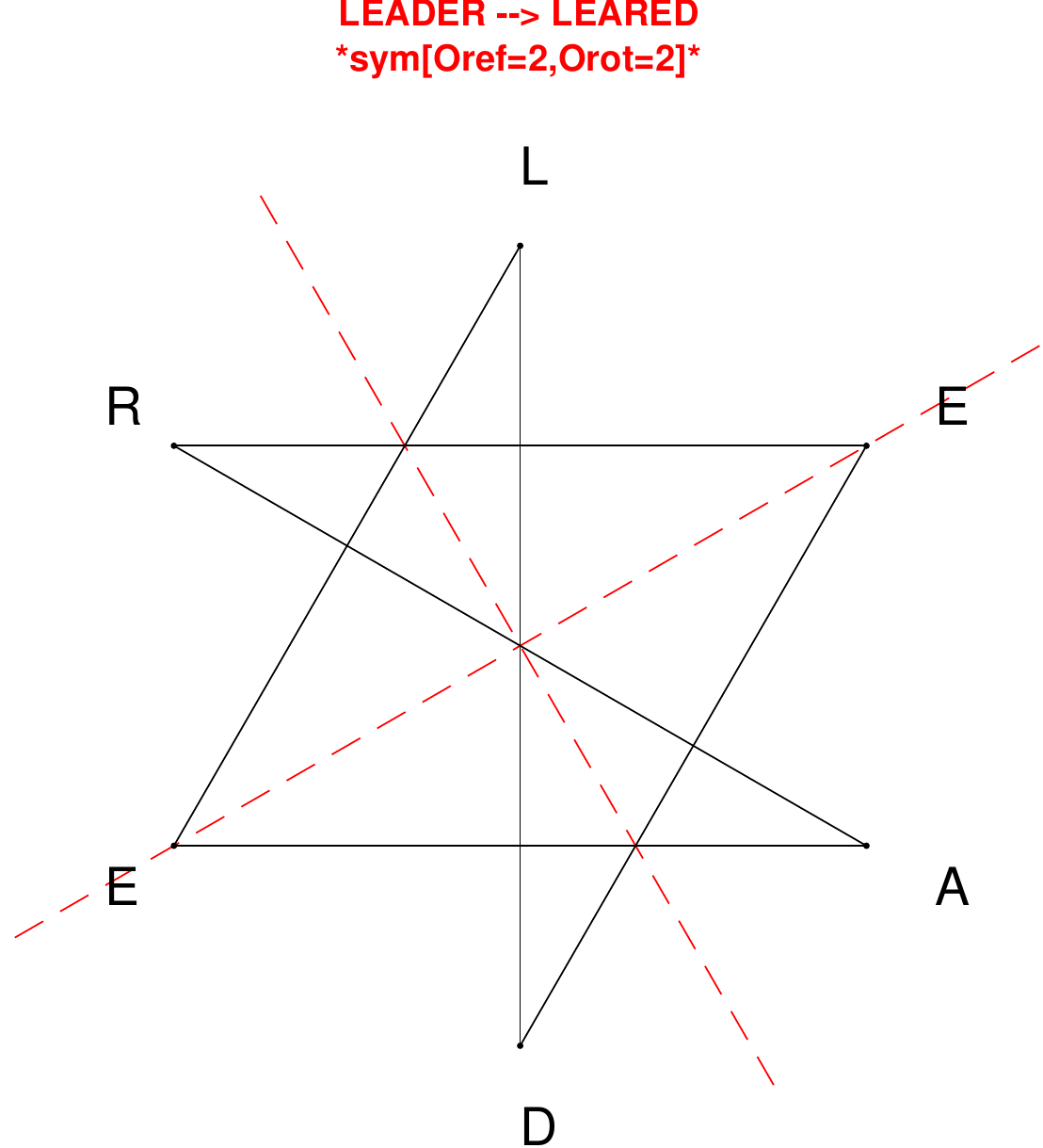}
\end{subfigure}
\hfill
\begin{subfigure}[T]{0.19\textwidth}
\centering
\includegraphics[width=\textwidth]{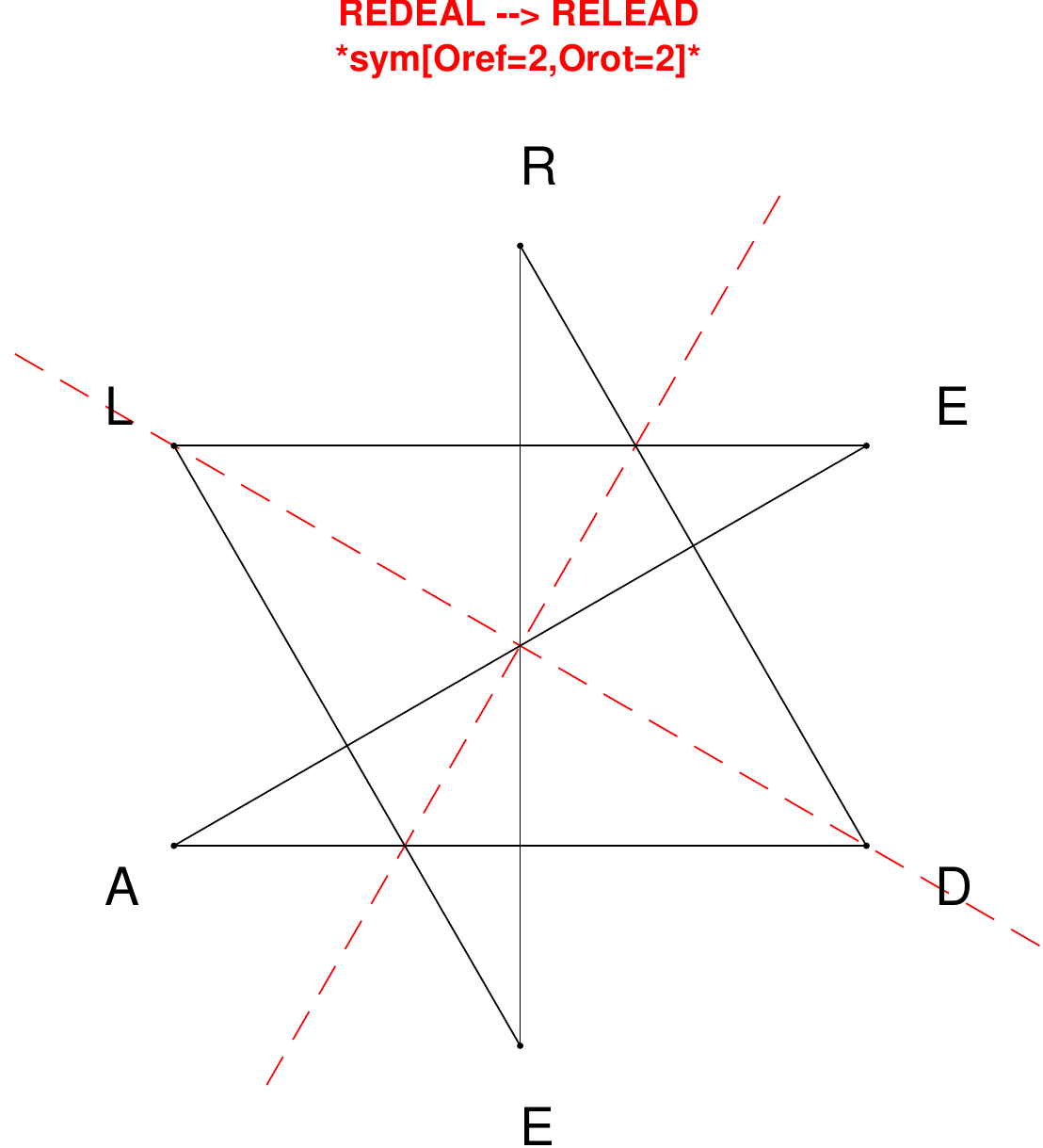}
\end{subfigure}
\end{figure}

\begin{figure}[H]
\centering
\begin{subfigure}[T]{0.19\textwidth}
\centering
\includegraphics[width=\textwidth]{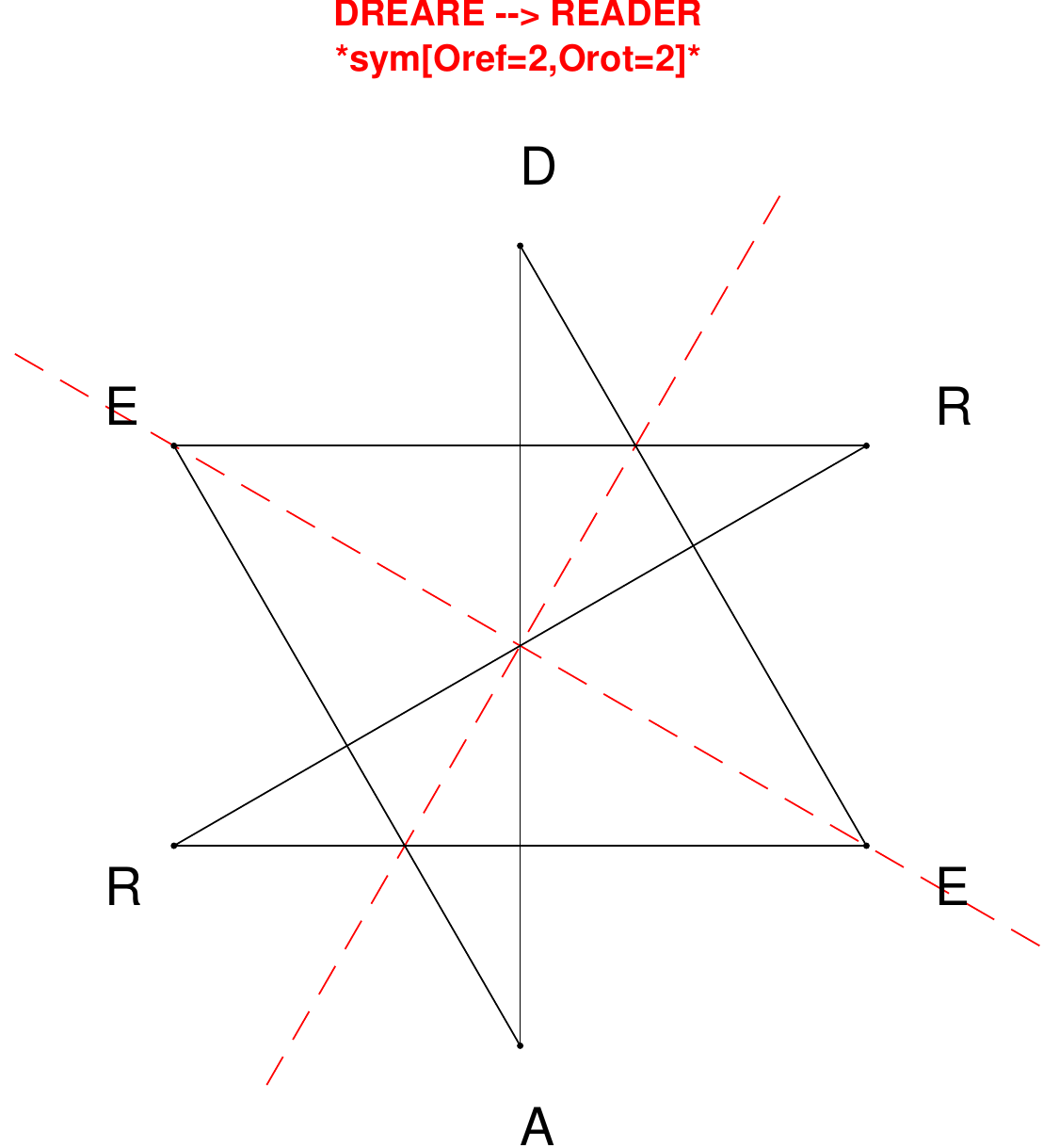}
\end{subfigure}
\hfill
\begin{subfigure}[T]{0.19\textwidth}
\centering
\includegraphics[width=\textwidth]{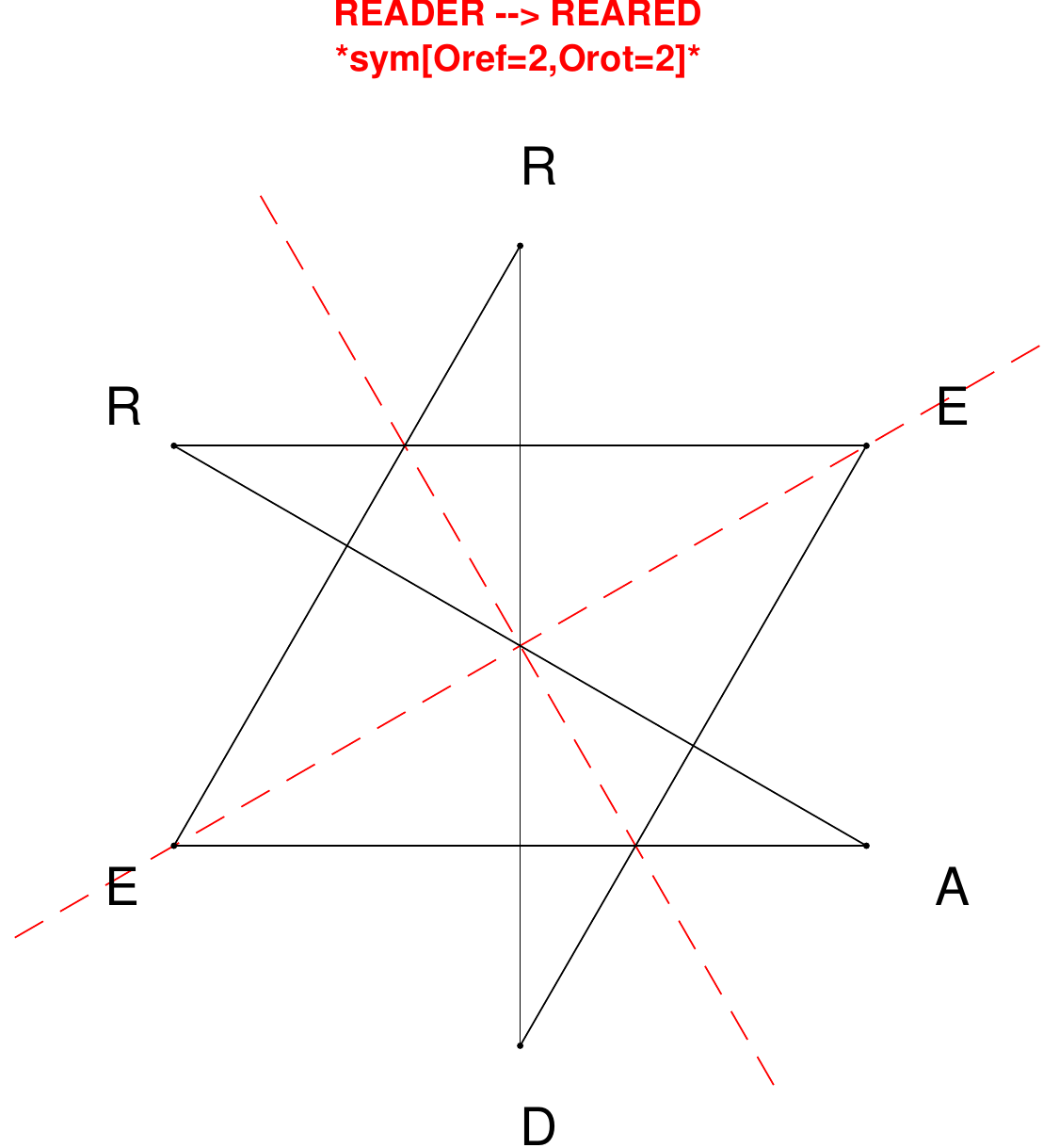}
\end{subfigure}
\hfill
\begin{subfigure}[T]{0.19\textwidth}
\centering
\includegraphics[width=\textwidth]{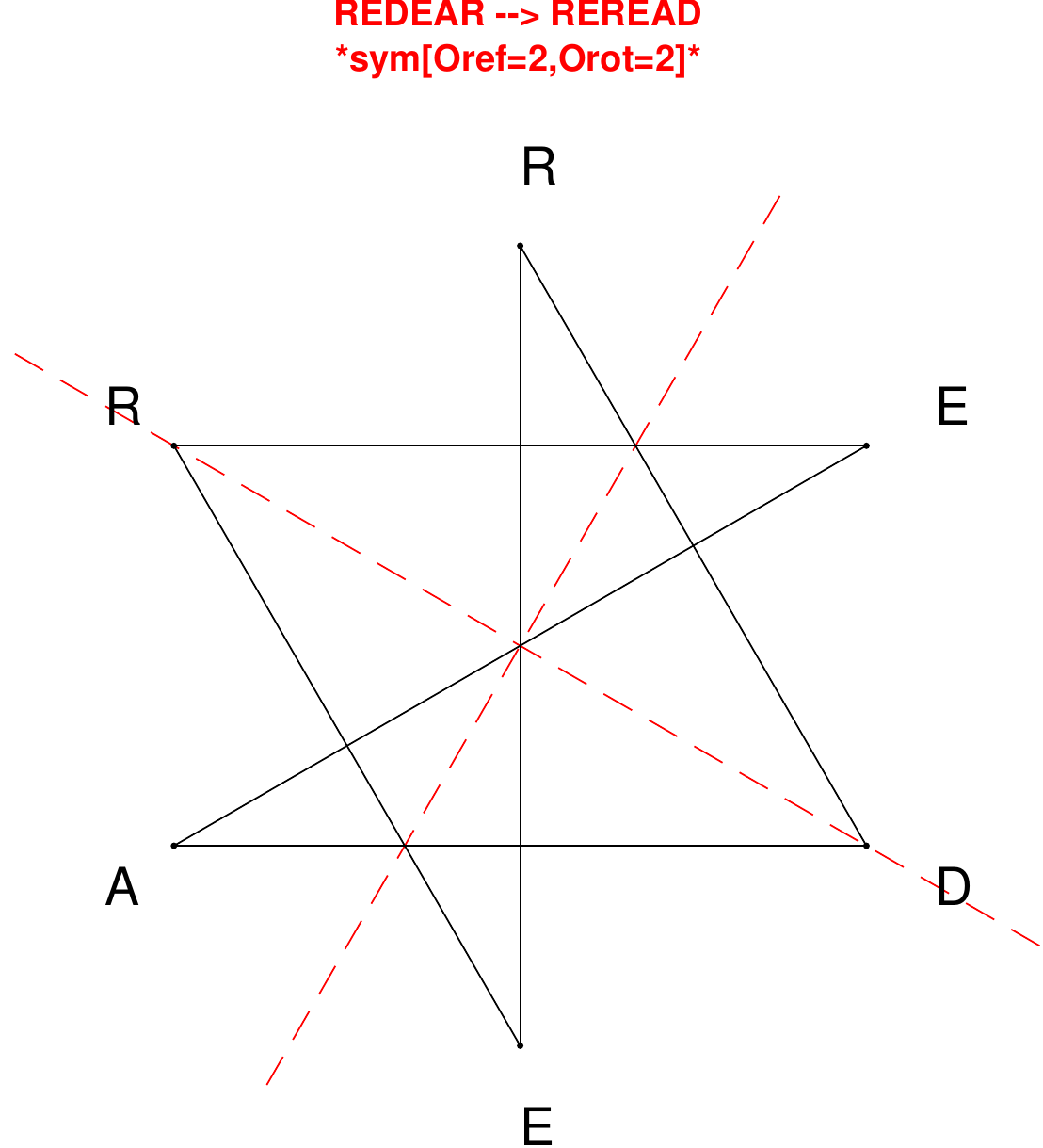}
\end{subfigure}
\hfill
\begin{subfigure}[T]{0.19\textwidth}
\centering
\includegraphics[width=\textwidth]{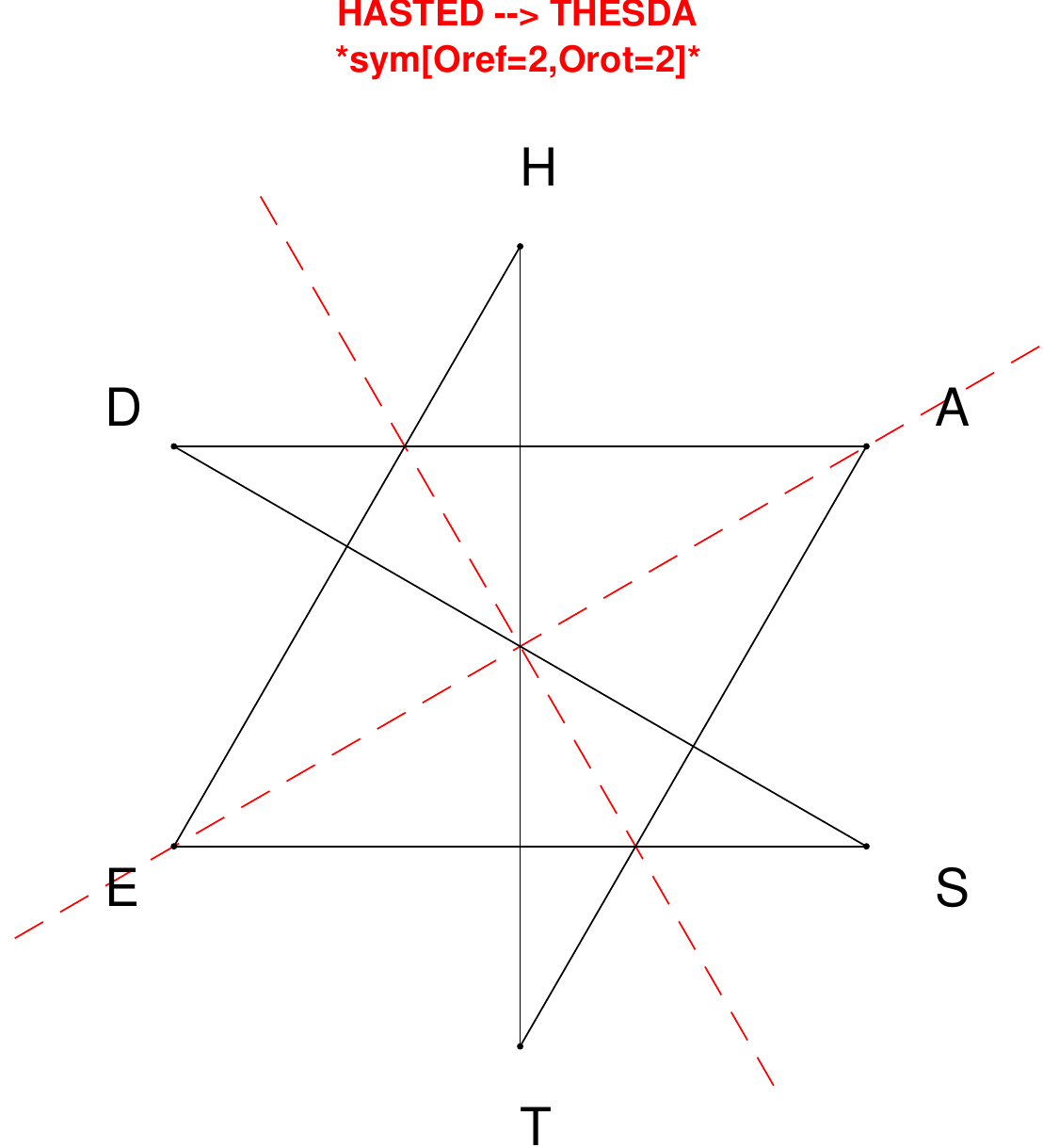}
\end{subfigure}
\hfill
\begin{subfigure}[T]{0.19\textwidth}
\centering
\includegraphics[width=\textwidth]{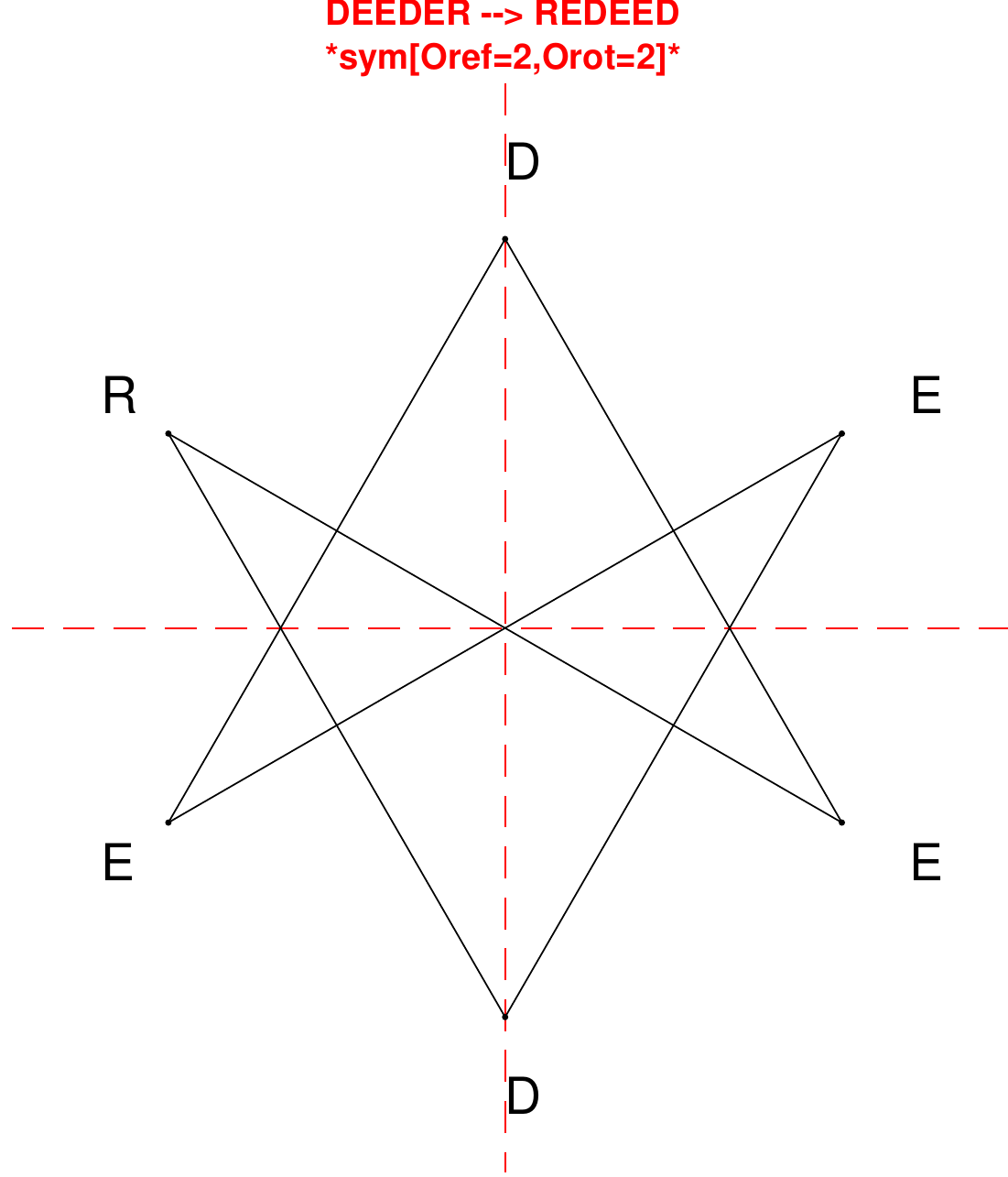}
\end{subfigure}
\end{figure}

\begin{figure}[H]
\centering
\begin{subfigure}[T]{0.19\textwidth}
\centering
\includegraphics[width=\textwidth]{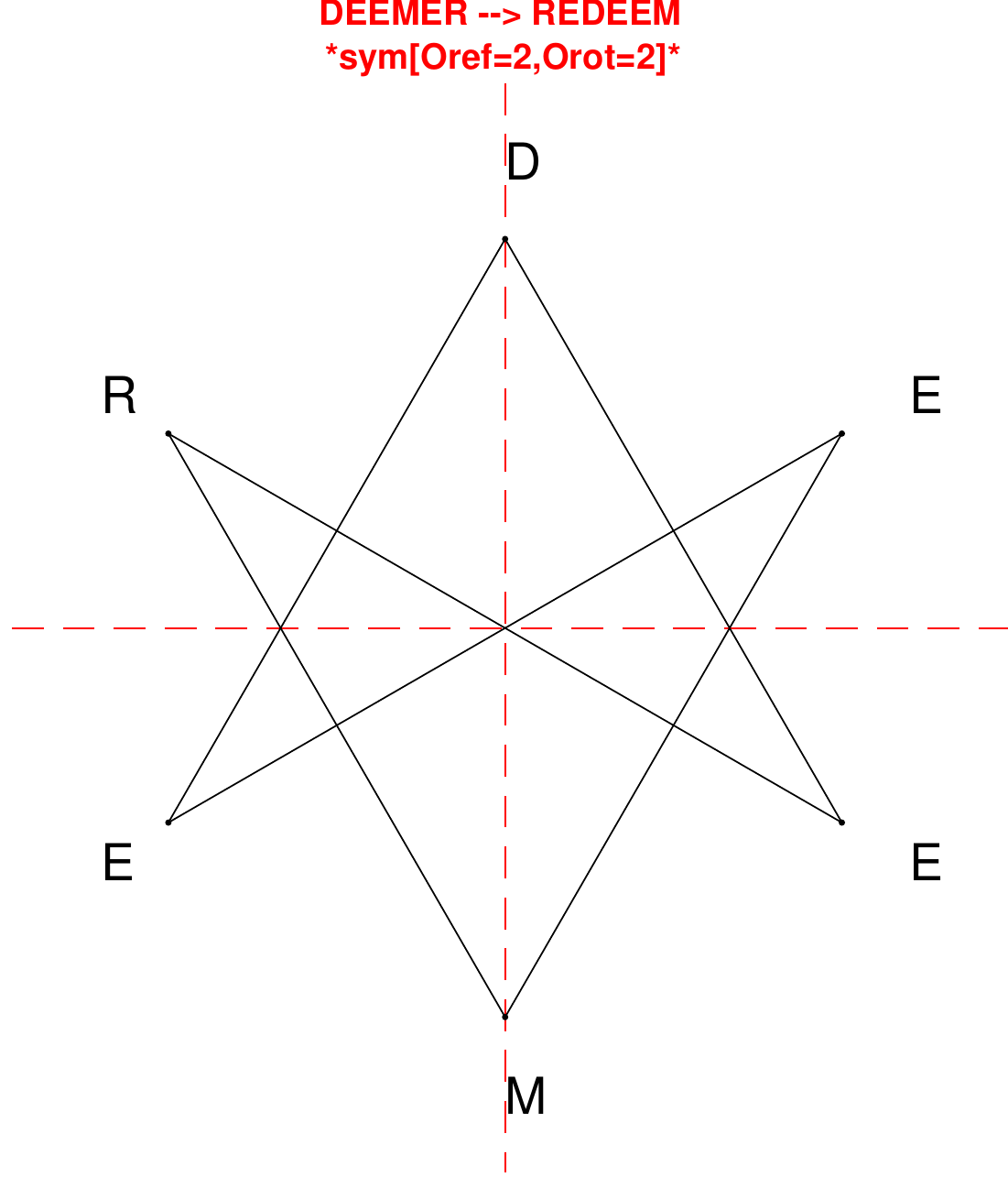}
\end{subfigure}
\hfill
\begin{subfigure}[T]{0.19\textwidth}
\centering
\includegraphics[width=\textwidth]{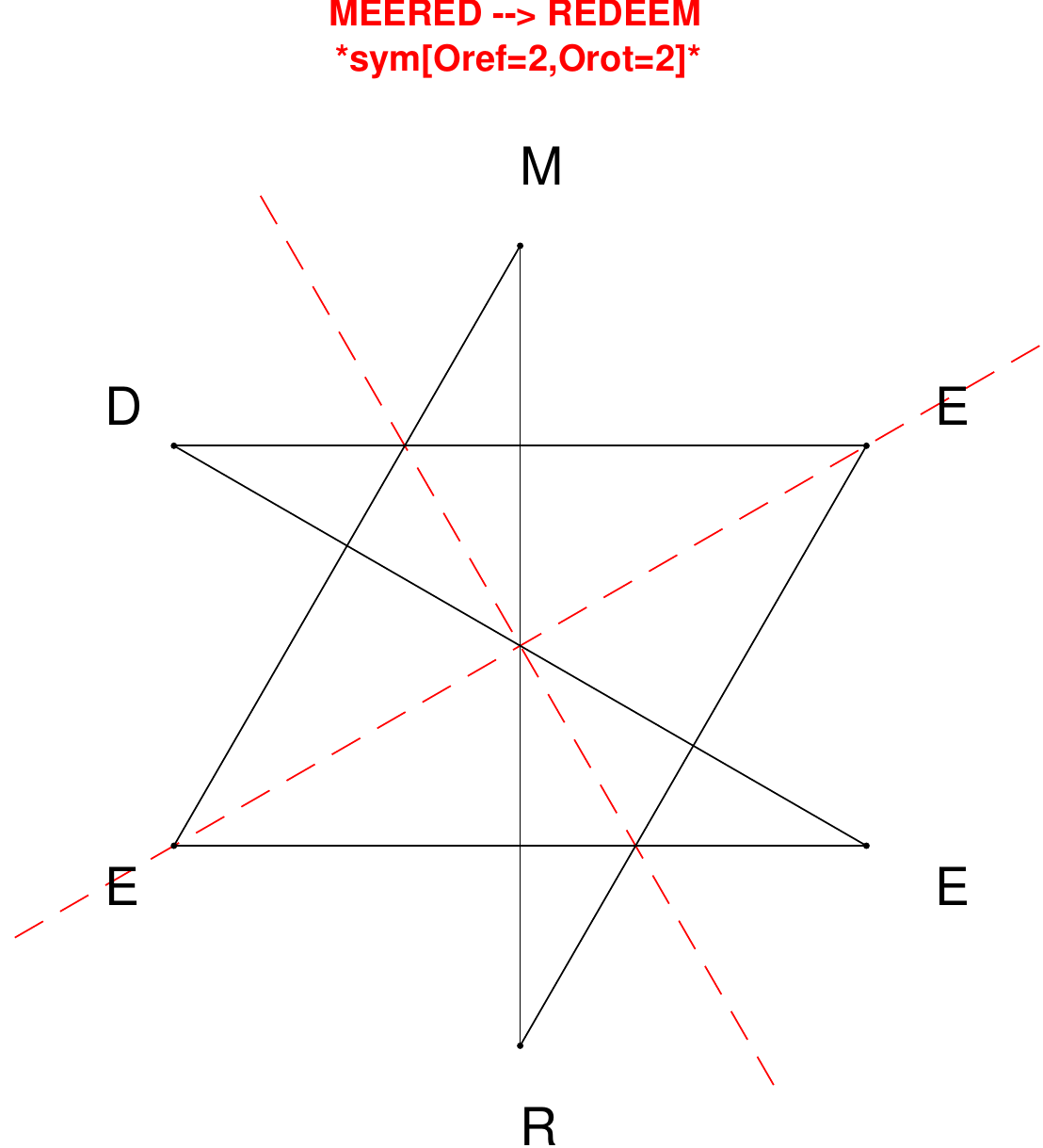}
\end{subfigure}
\hfill
\begin{subfigure}[T]{0.19\textwidth}
\centering
\includegraphics[width=\textwidth]{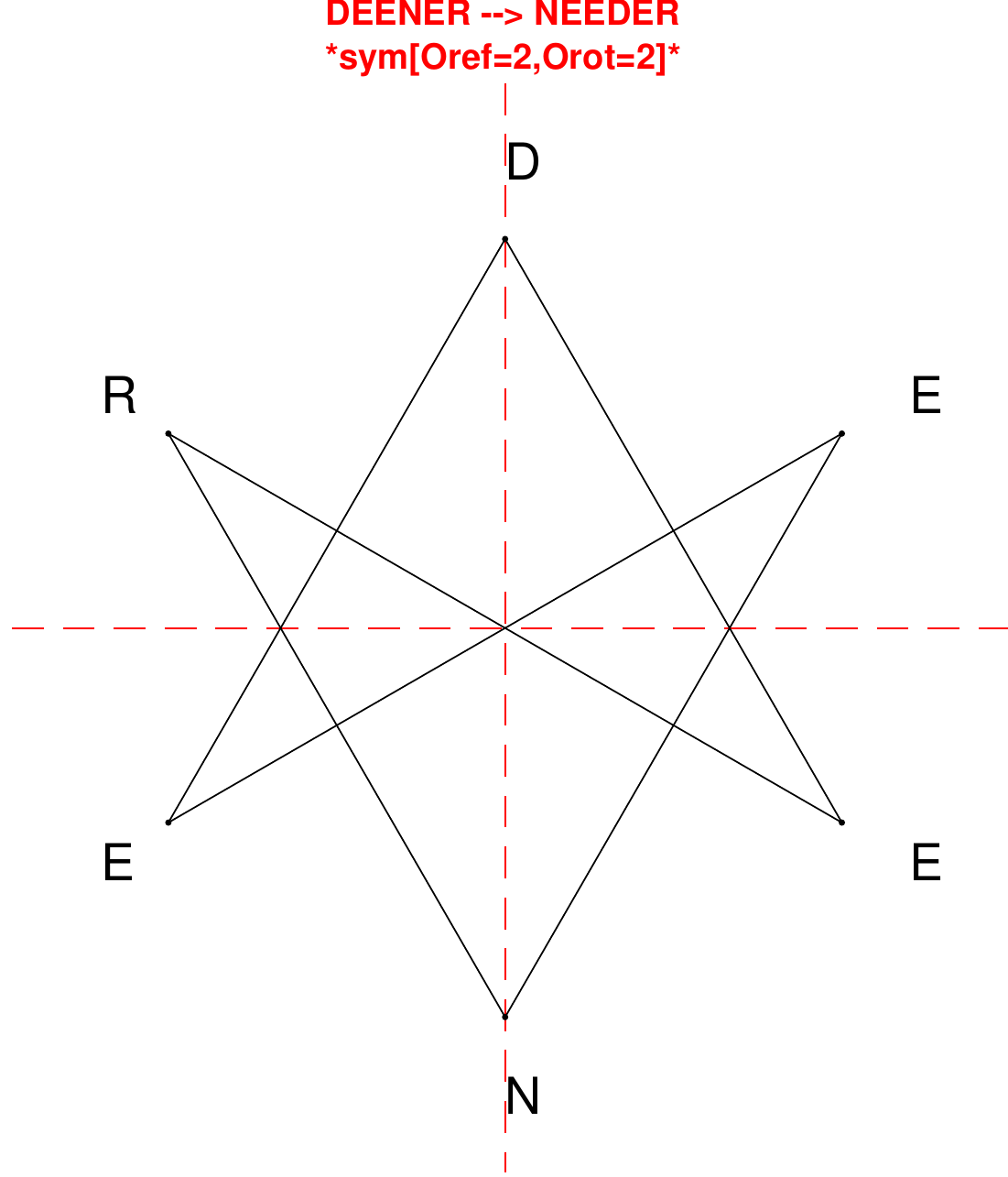}
\end{subfigure}
\hfill
\begin{subfigure}[T]{0.19\textwidth}
\centering
\includegraphics[width=\textwidth]{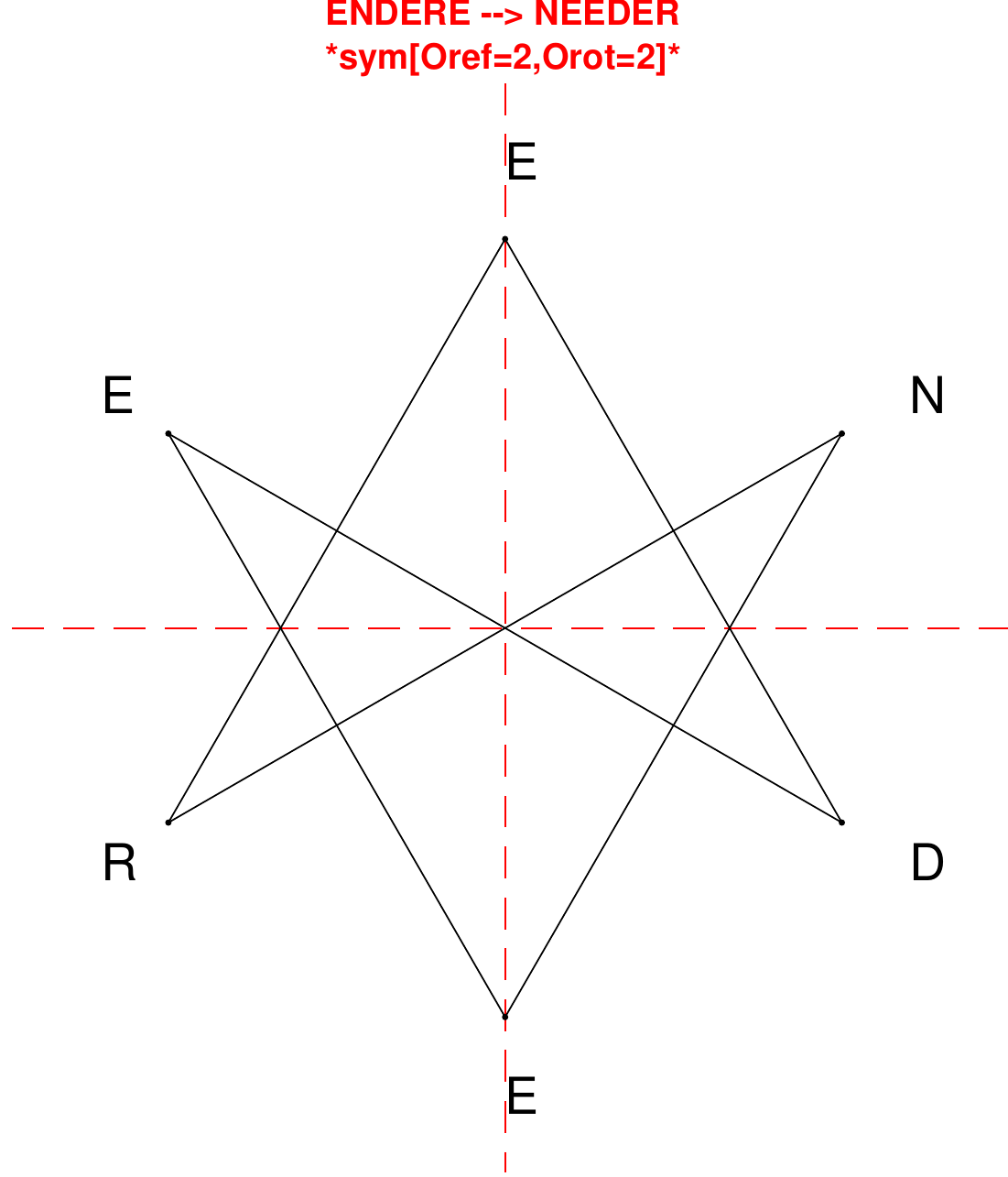}
\end{subfigure}
\hfill
\begin{subfigure}[T]{0.19\textwidth}
\centering
\includegraphics[width=\textwidth]{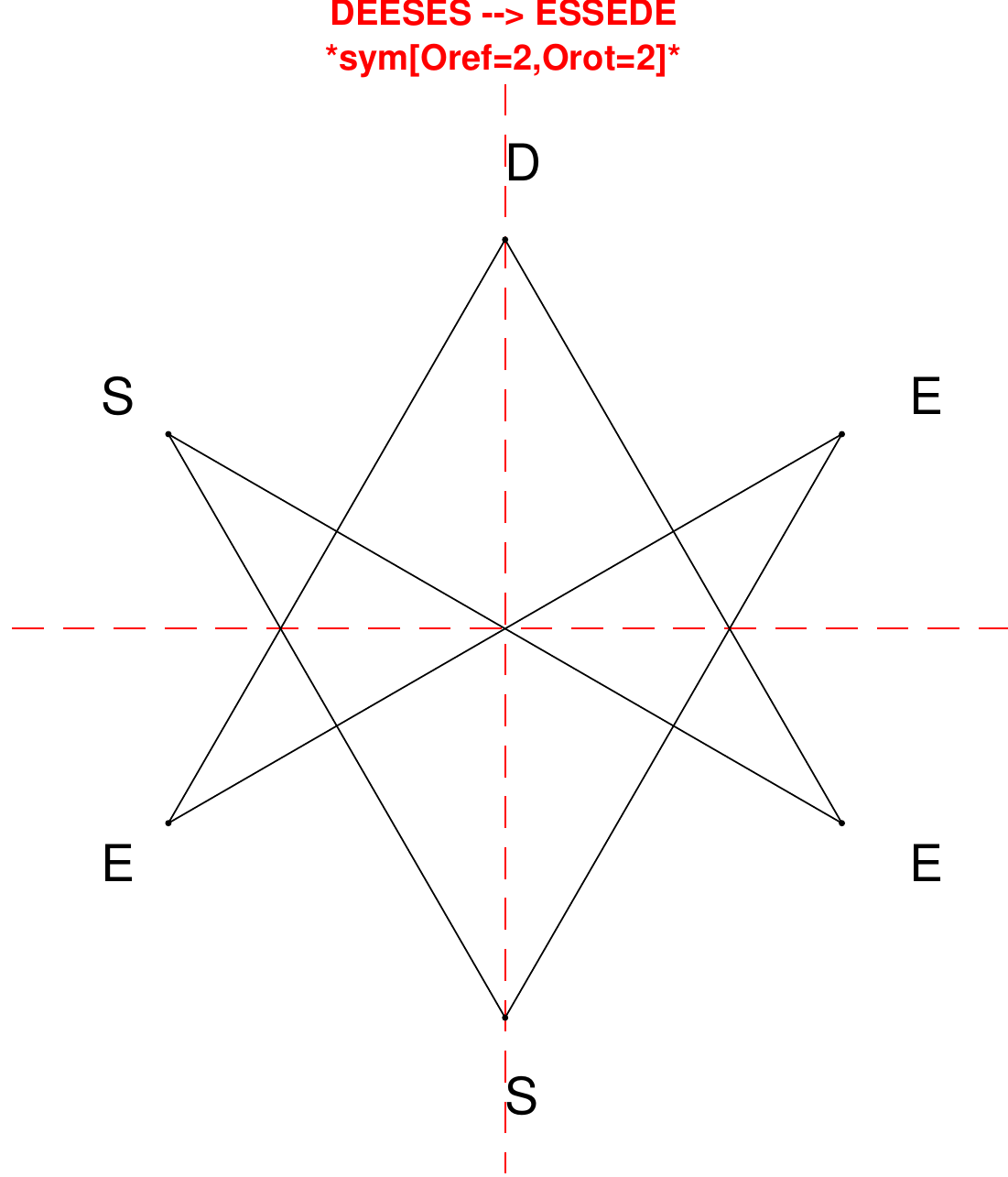}
\end{subfigure}
\end{figure}

\begin{figure}[H]
\centering
\begin{subfigure}[T]{0.19\textwidth}
\centering
\includegraphics[width=\textwidth]{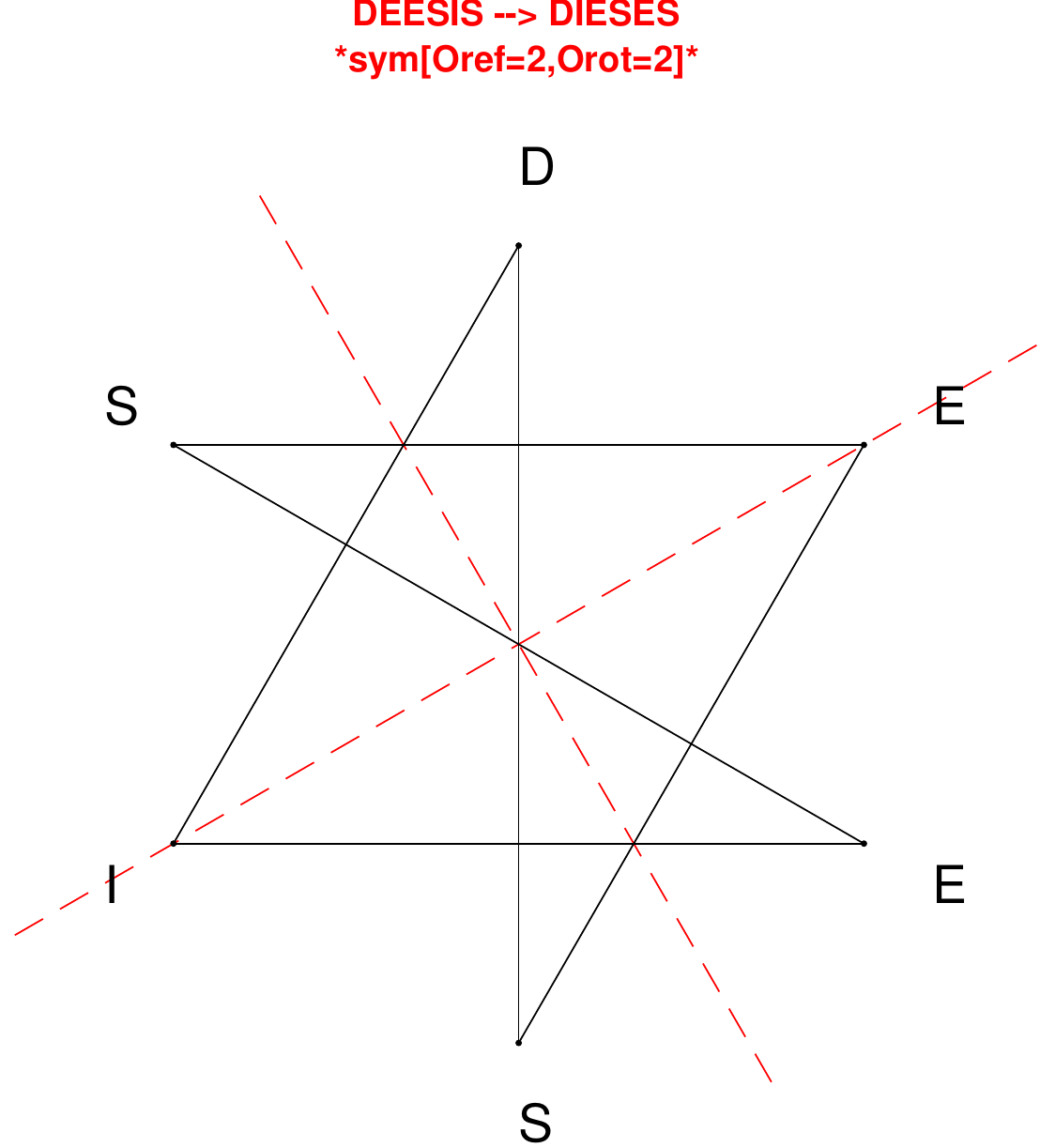}
\end{subfigure}
\hfill
\begin{subfigure}[T]{0.19\textwidth}
\centering
\includegraphics[width=\textwidth]{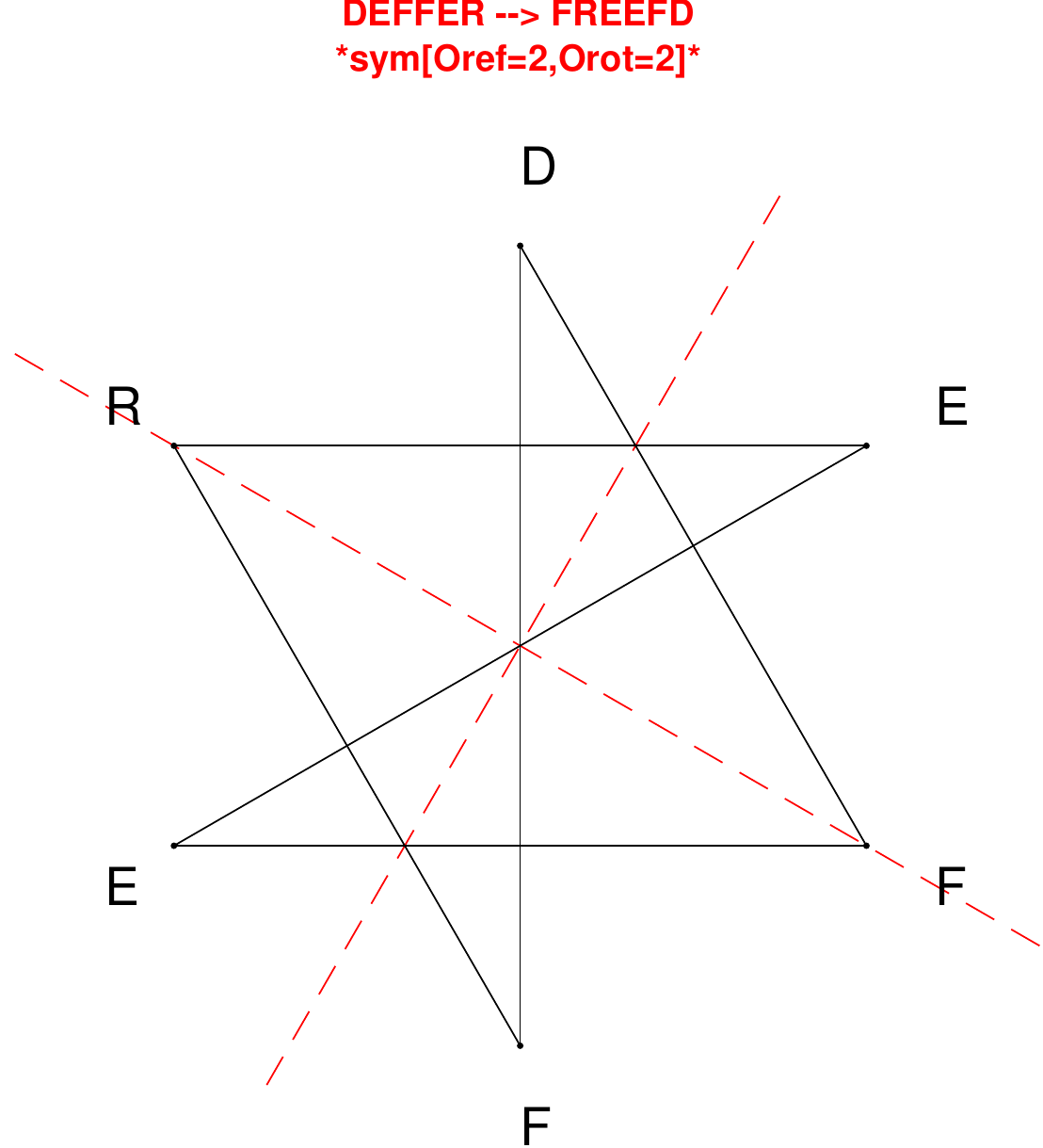}
\end{subfigure}
\hfill
\begin{subfigure}[T]{0.19\textwidth}
\centering
\includegraphics[width=\textwidth]{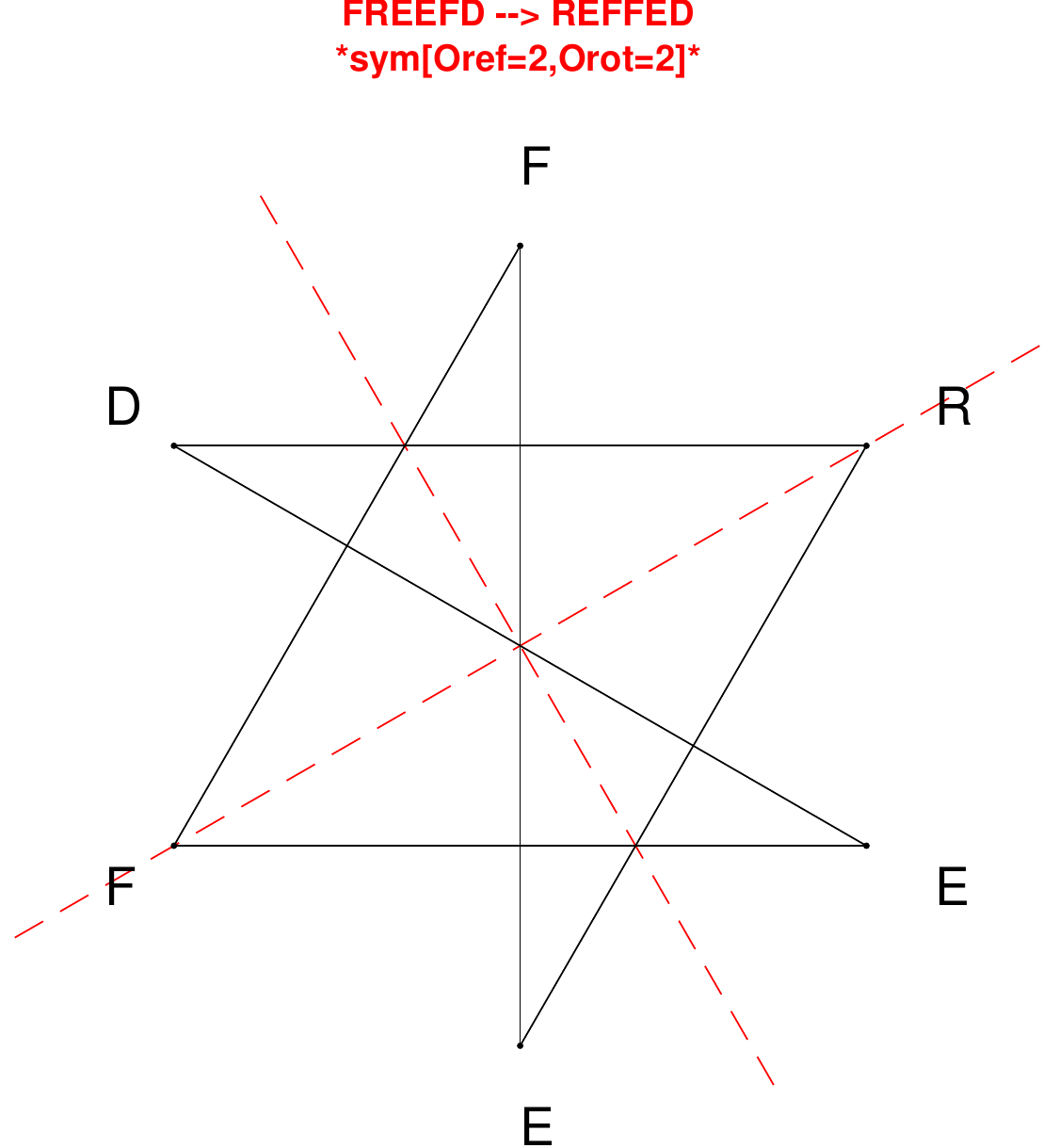}
\end{subfigure}
\hfill
\begin{subfigure}[T]{0.19\textwidth}
\centering
\includegraphics[width=\textwidth]{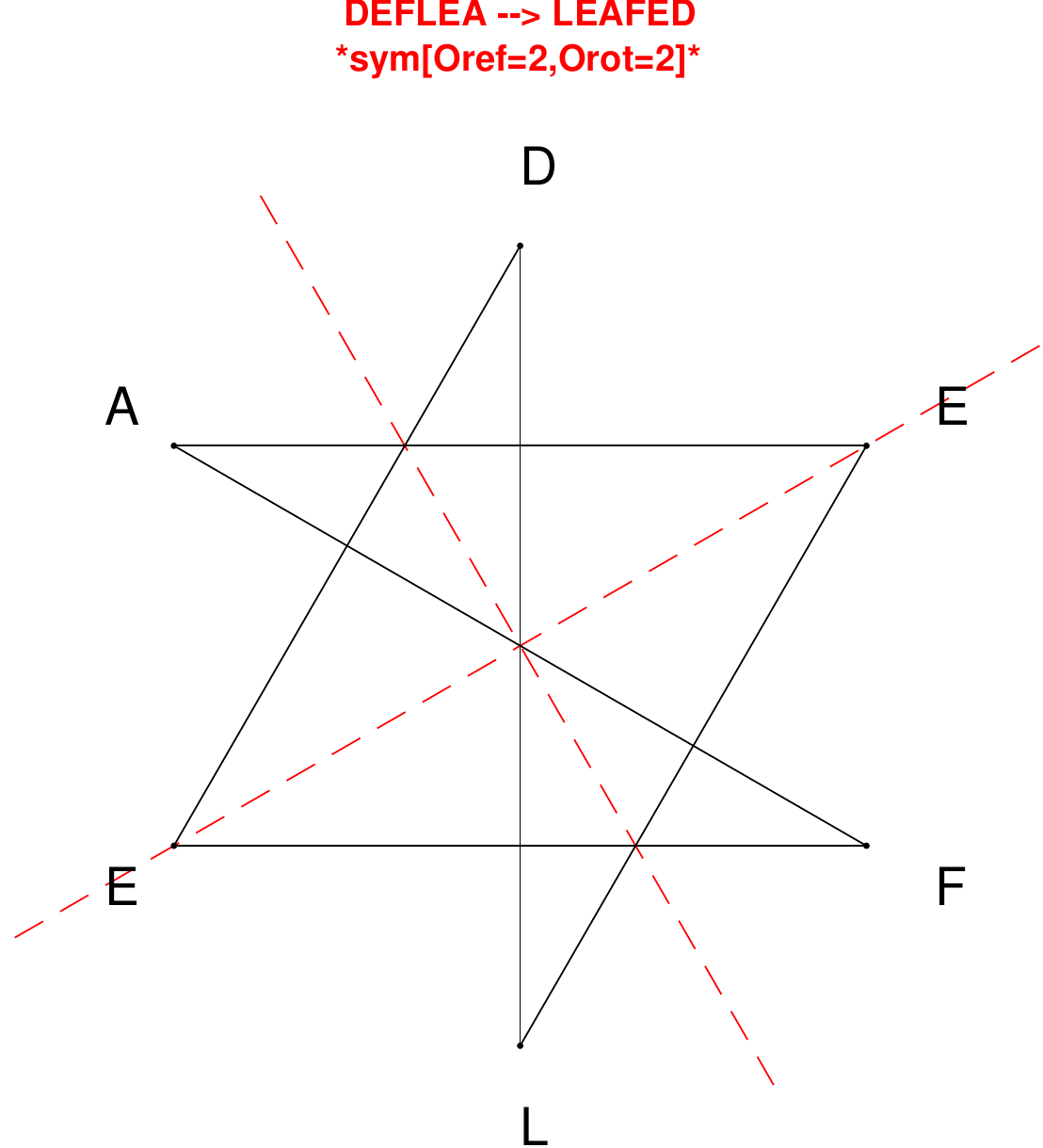}
\end{subfigure}
\hfill
\begin{subfigure}[T]{0.19\textwidth}
\centering
\includegraphics[width=\textwidth]{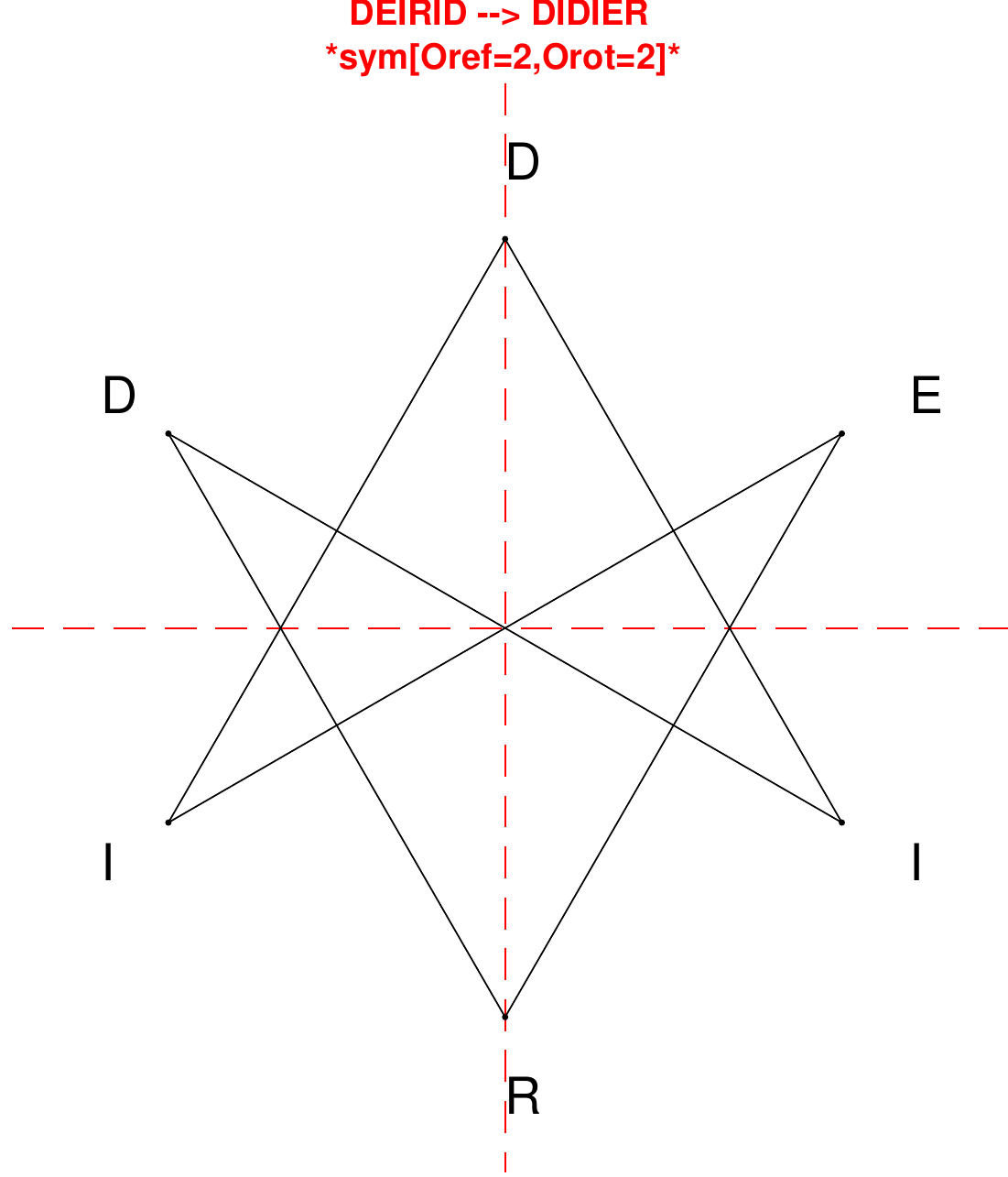}
\end{subfigure}
\end{figure}

\begin{figure}[H]
\centering
\begin{subfigure}[T]{0.19\textwidth}
\centering
\includegraphics[width=\textwidth]{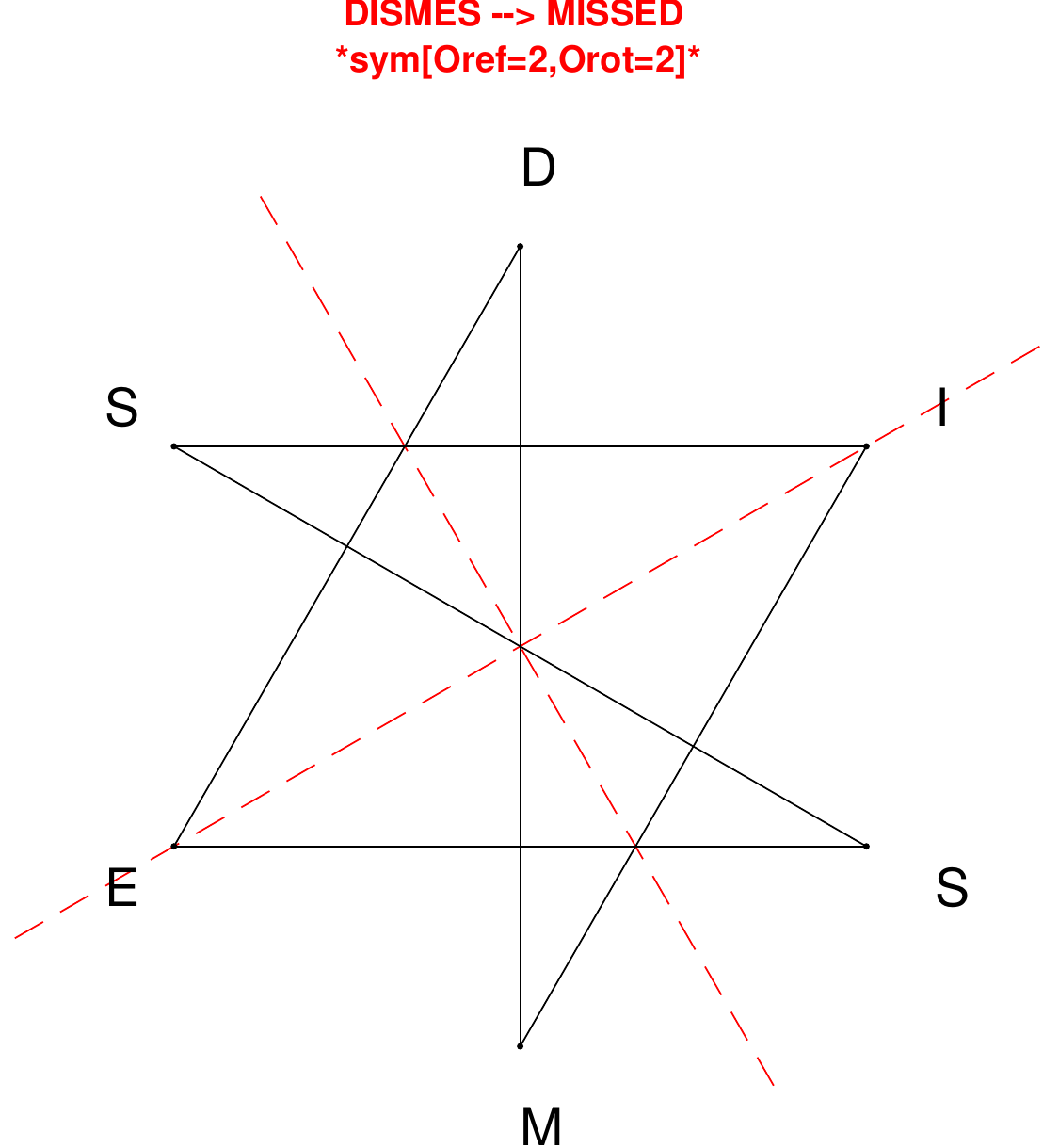}
\end{subfigure}
\hfill
\begin{subfigure}[T]{0.19\textwidth}
\centering
\includegraphics[width=\textwidth]{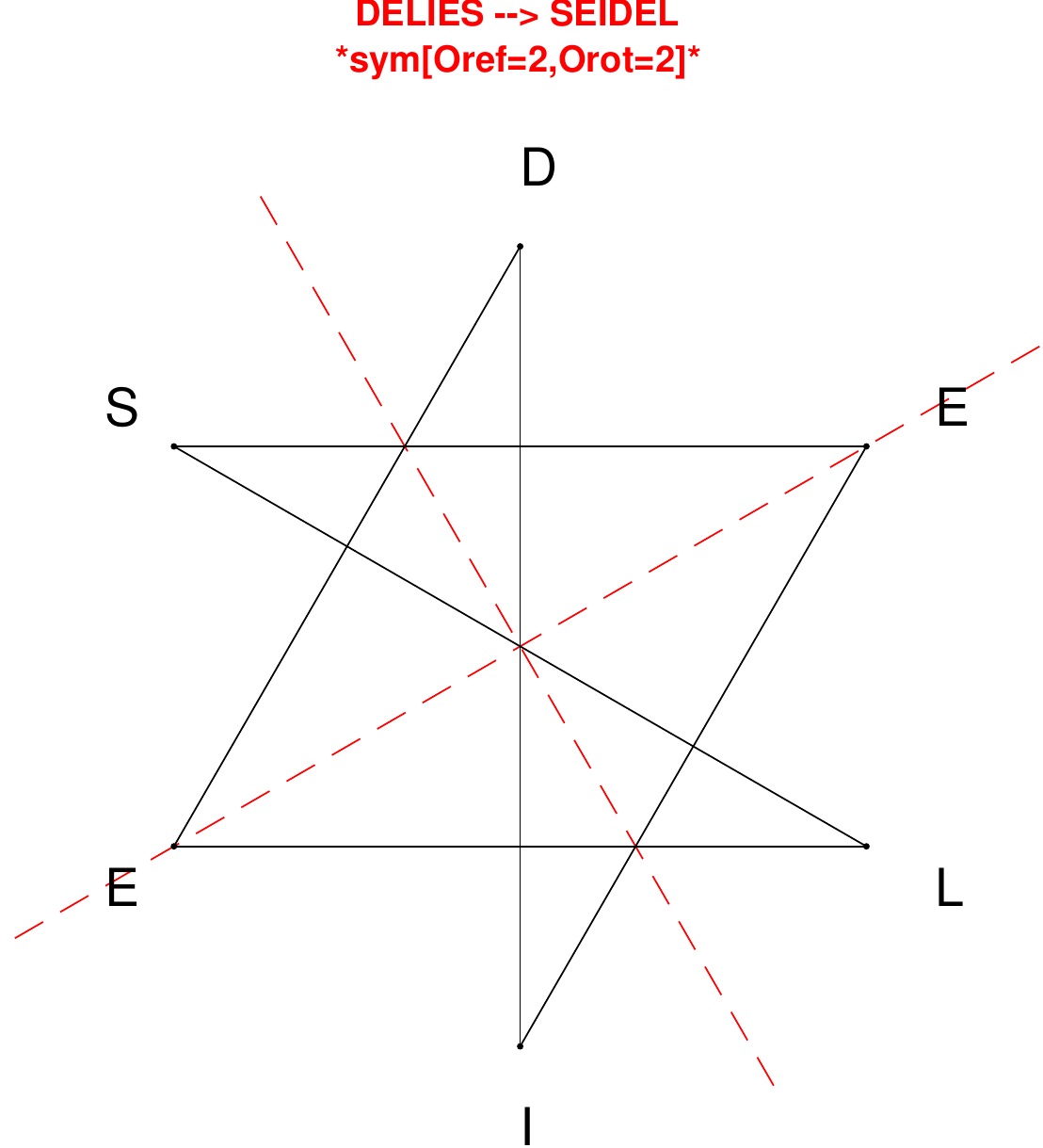}
\end{subfigure}
\hfill
\begin{subfigure}[T]{0.19\textwidth}
\centering
\includegraphics[width=\textwidth]{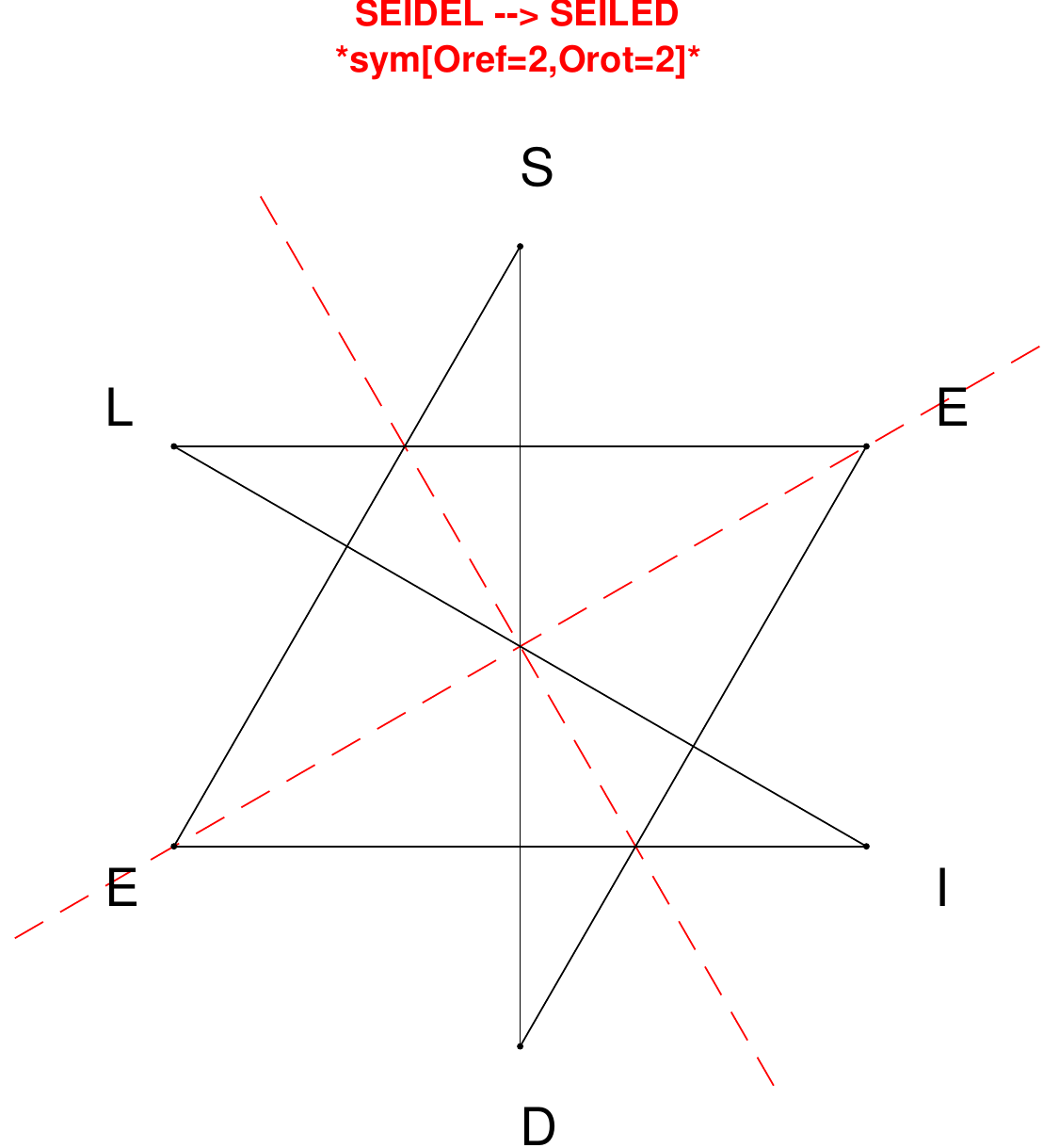}
\end{subfigure}
\hfill
\begin{subfigure}[T]{0.19\textwidth}
\centering
\includegraphics[width=\textwidth]{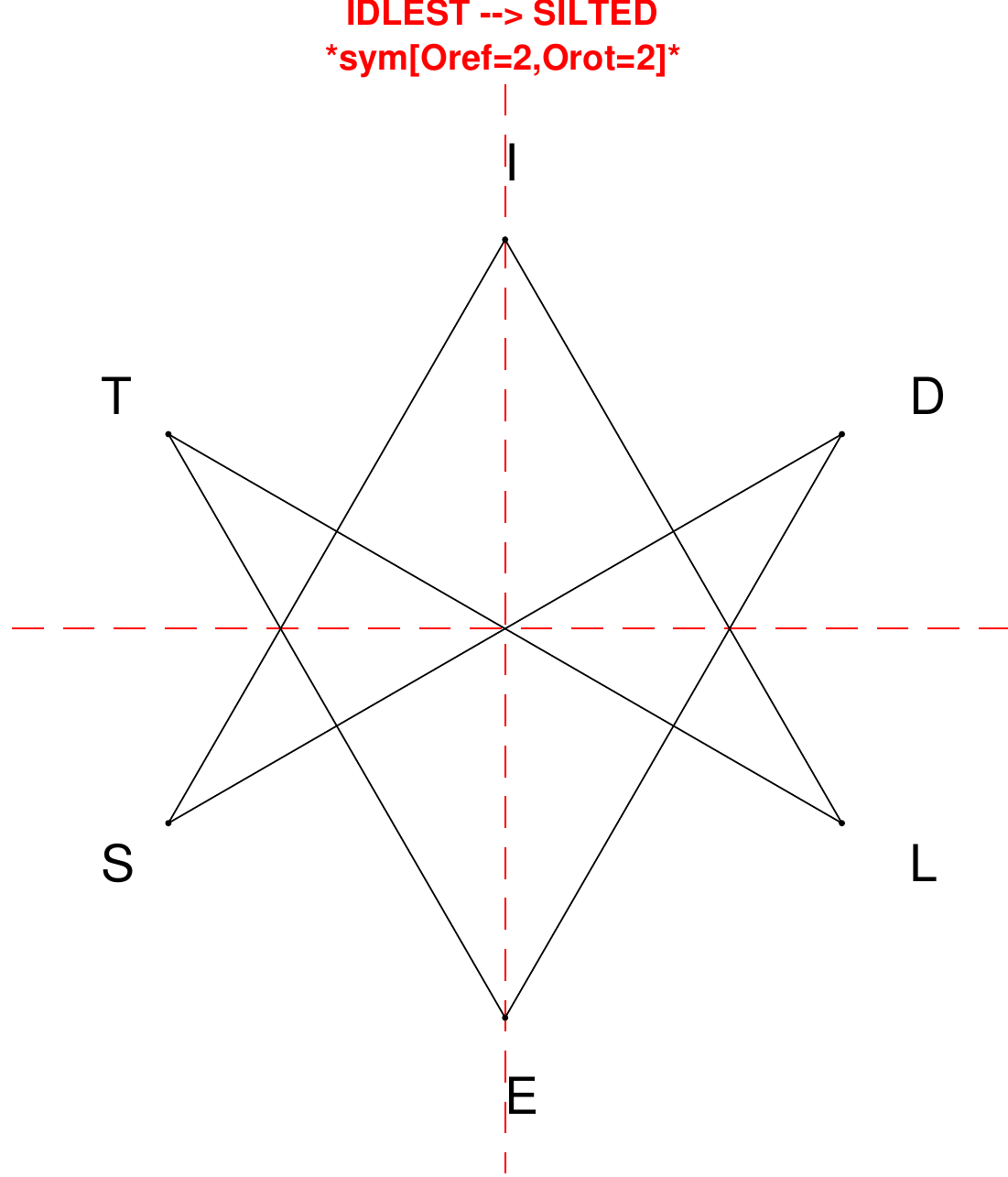}
\end{subfigure}
\hfill
\begin{subfigure}[T]{0.19\textwidth}
\centering
\includegraphics[width=\textwidth]{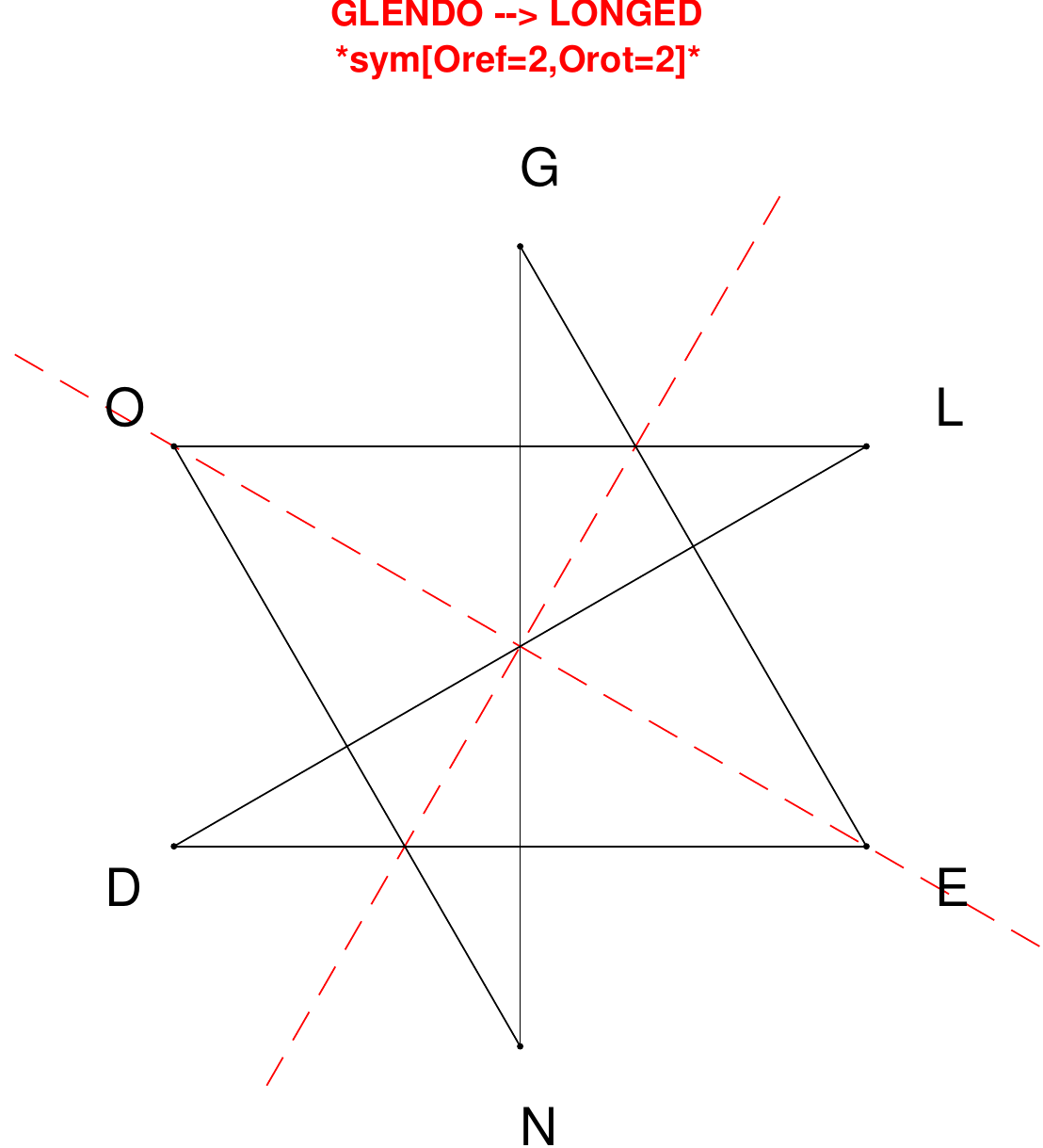}
\end{subfigure}
\end{figure}

\begin{figure}[H]
\centering
\begin{subfigure}[T]{0.19\textwidth}
\centering
\includegraphics[width=\textwidth]{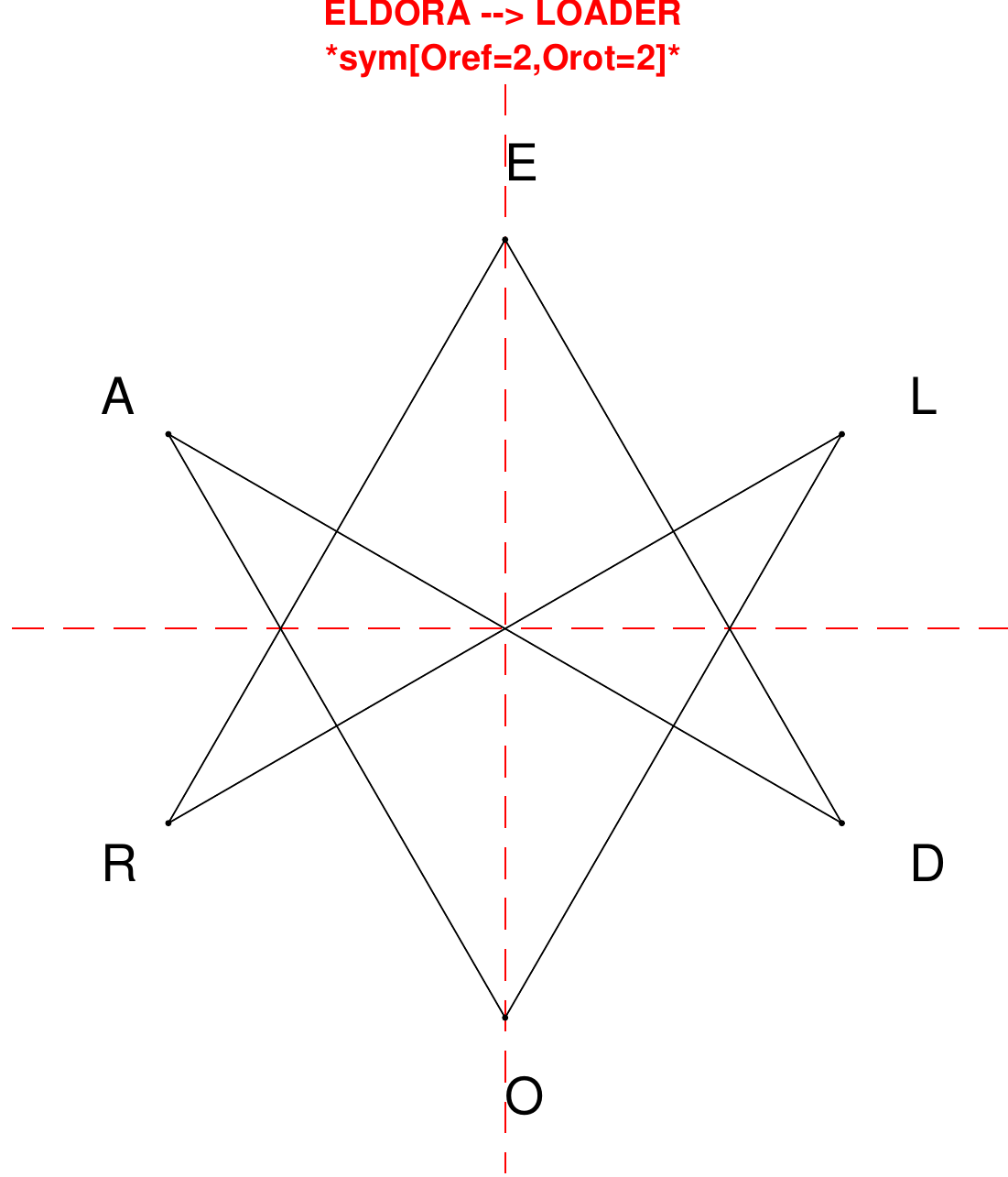}
\end{subfigure}
\hfill
\begin{subfigure}[T]{0.19\textwidth}
\centering
\includegraphics[width=\textwidth]{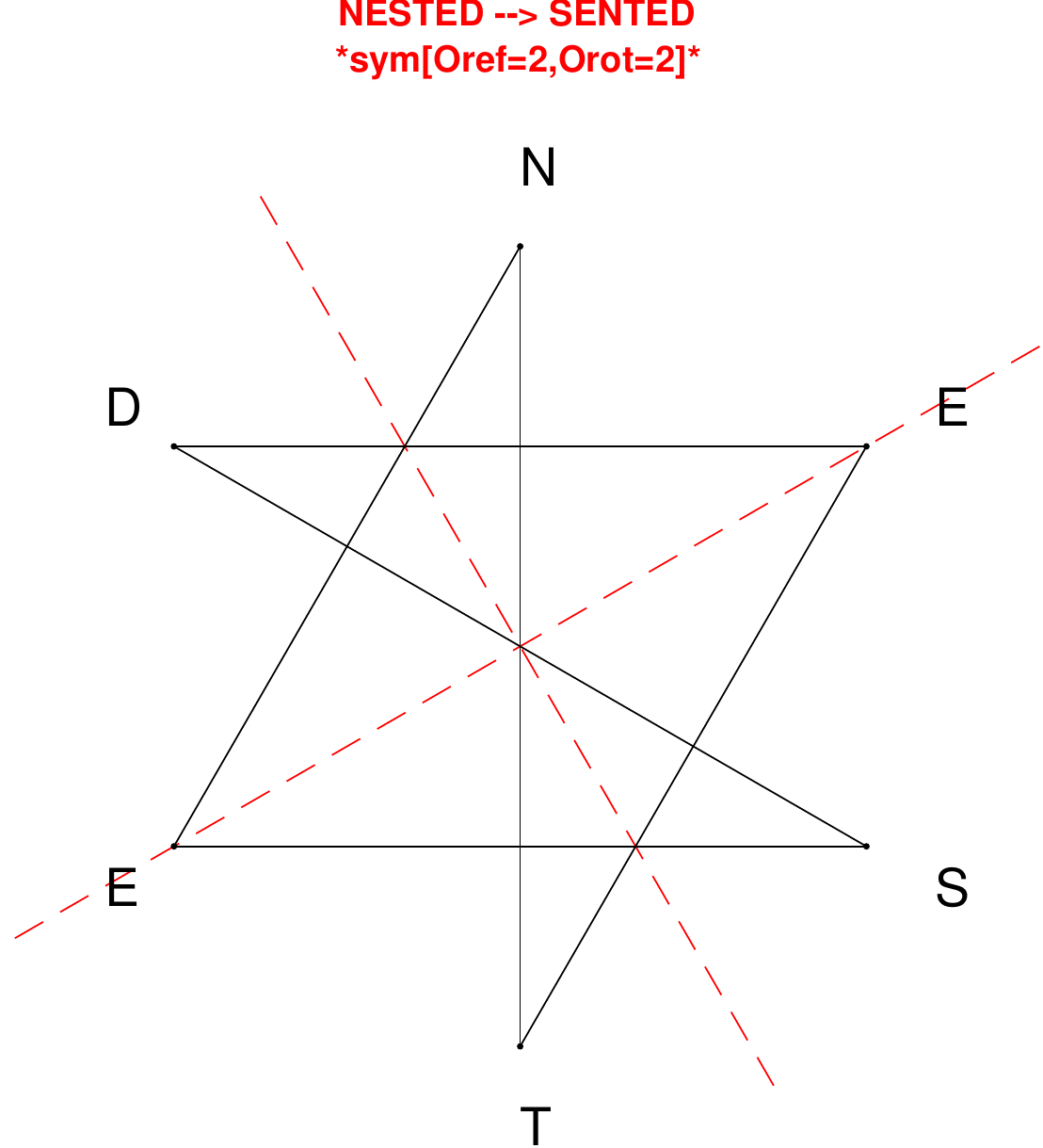}
\end{subfigure}
\hfill
\begin{subfigure}[T]{0.19\textwidth}
\centering
\includegraphics[width=\textwidth]{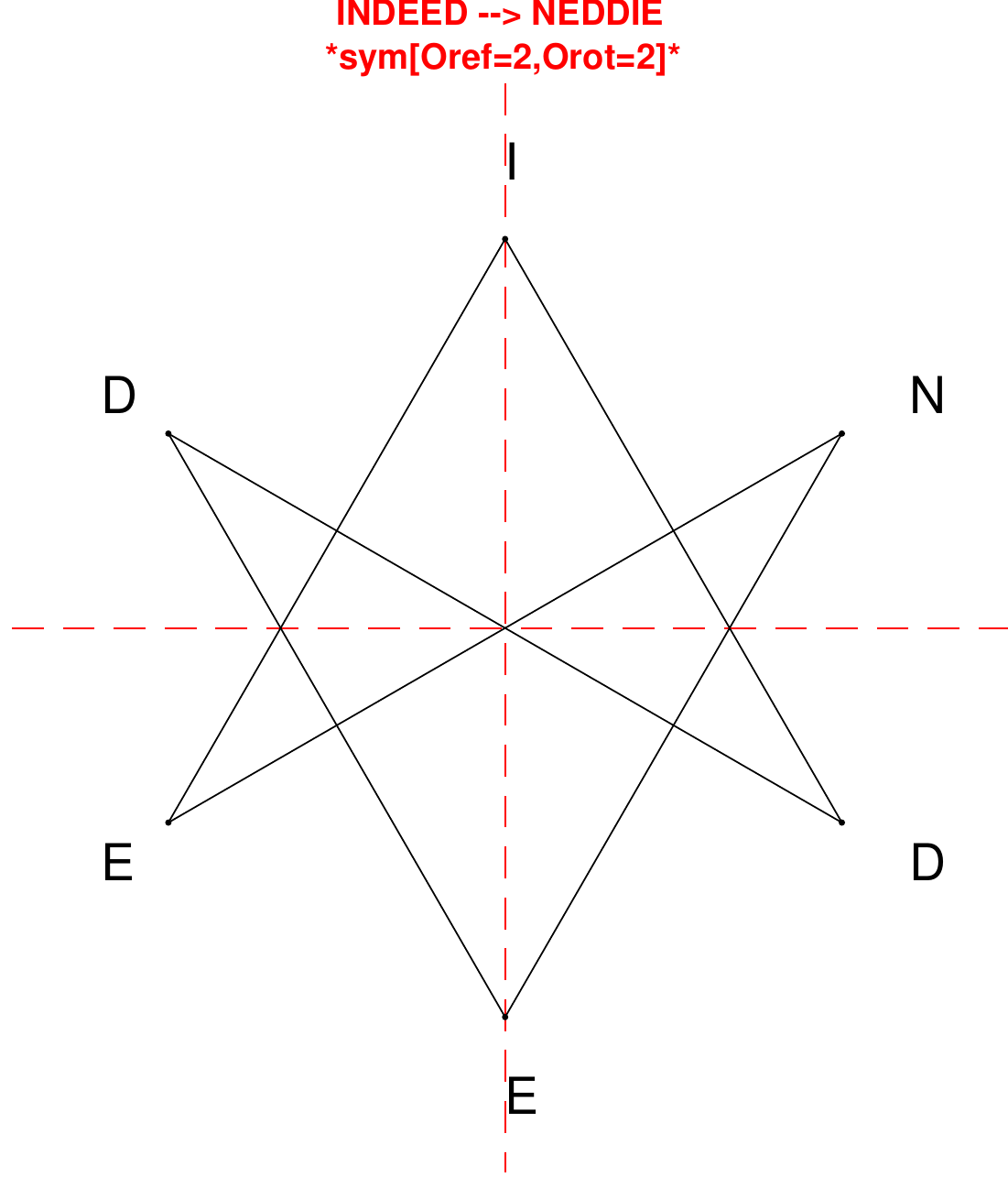}
\end{subfigure}
\hfill
\begin{subfigure}[T]{0.19\textwidth}
\centering
\includegraphics[width=\textwidth]{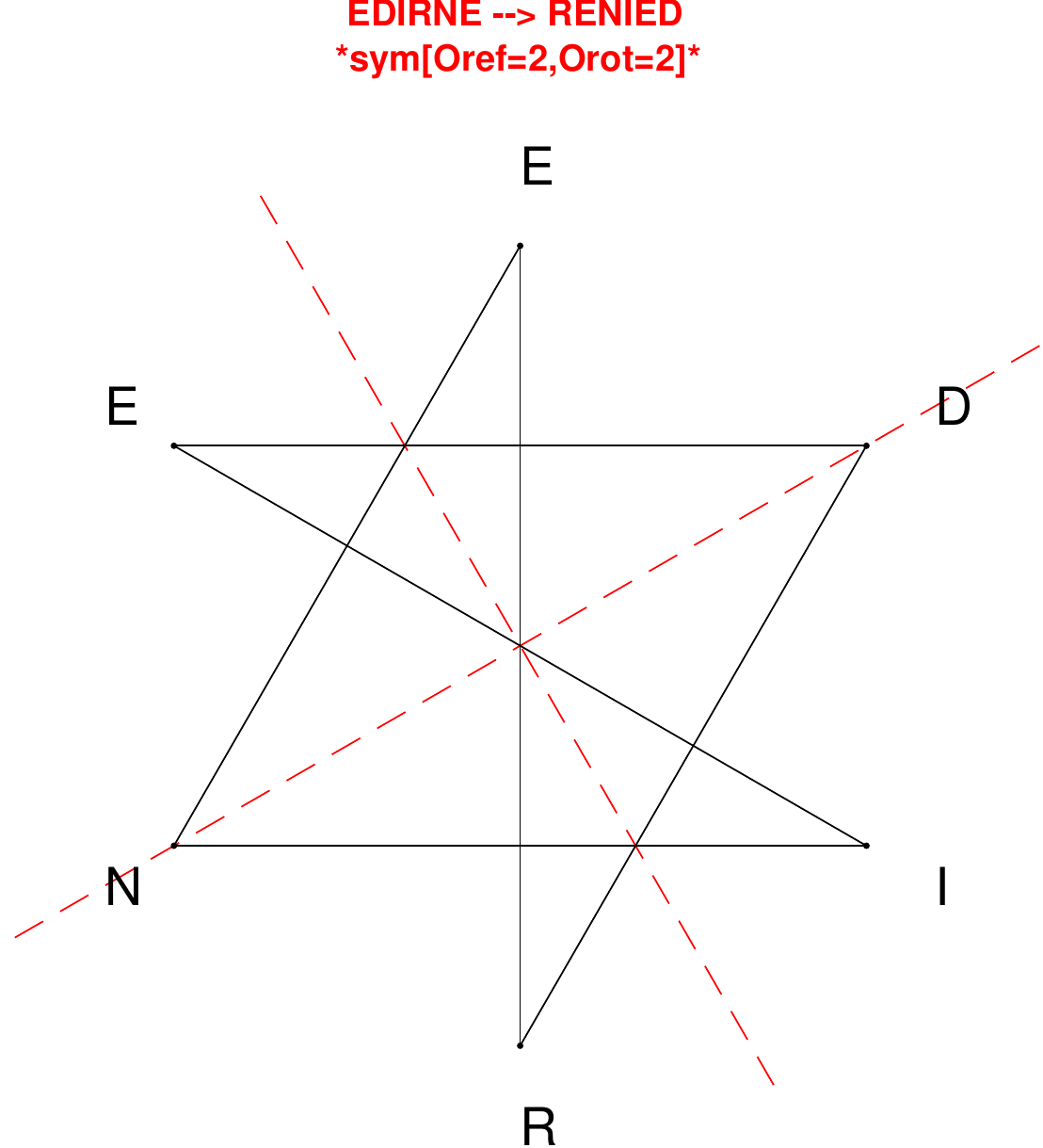}
\end{subfigure}
\hfill
\begin{subfigure}[T]{0.19\textwidth}
\centering
\includegraphics[width=\textwidth]{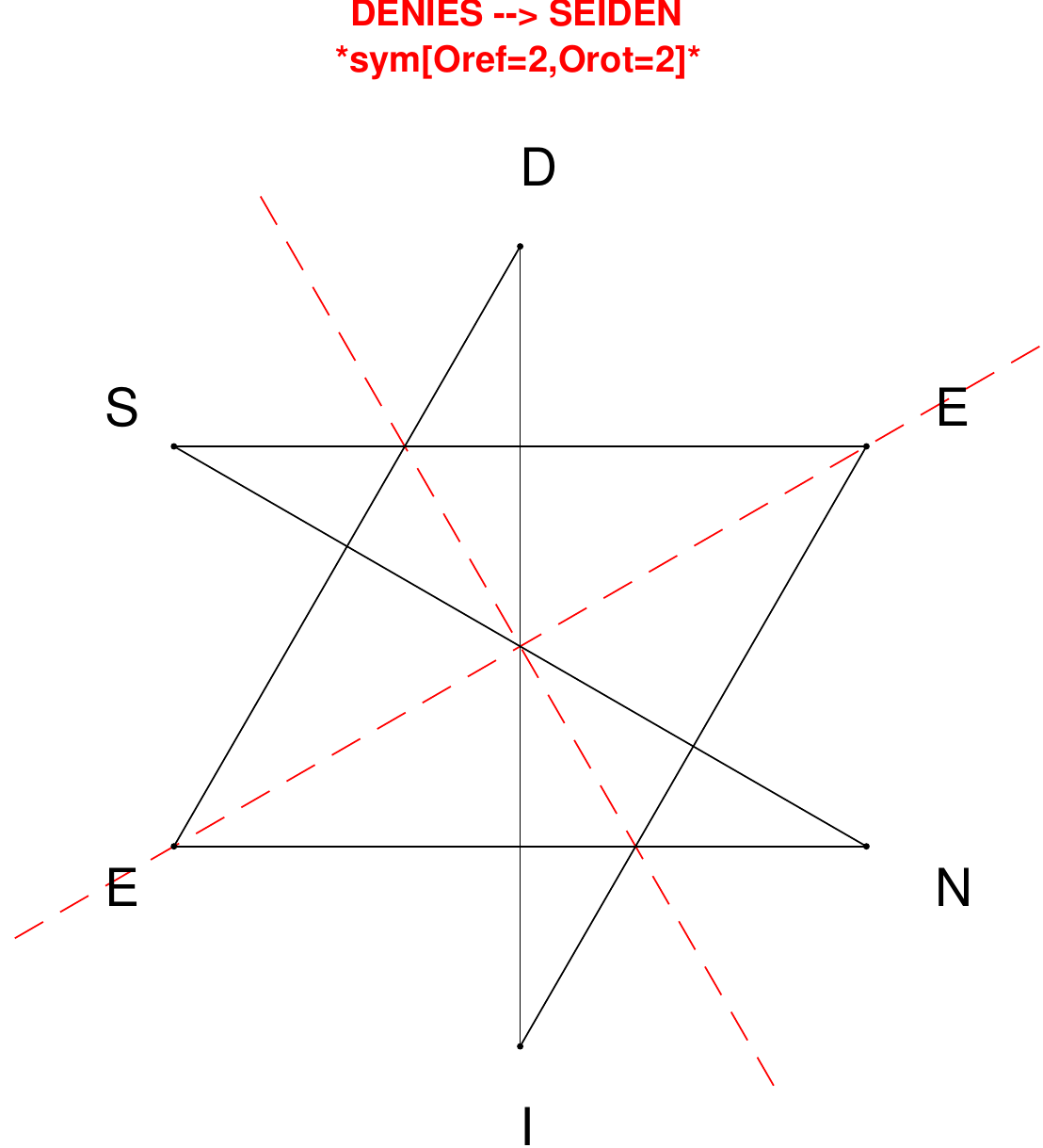}
\end{subfigure}
\end{figure}

\begin{figure}[H]
\centering
\begin{subfigure}[T]{0.19\textwidth}
\centering
\includegraphics[width=\textwidth]{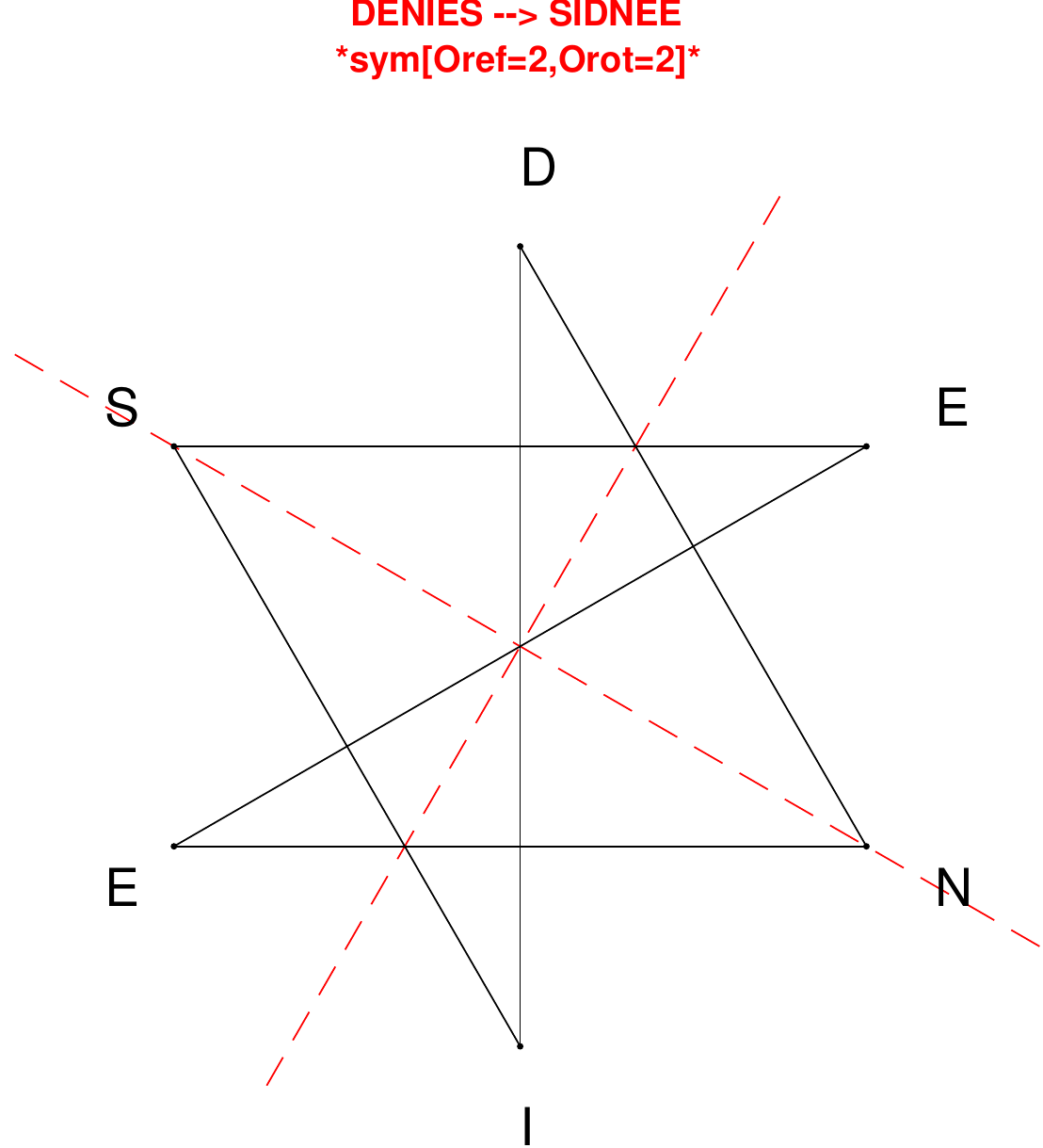}
\end{subfigure}
\hfill
\begin{subfigure}[T]{0.19\textwidth}
\centering
\includegraphics[width=\textwidth]{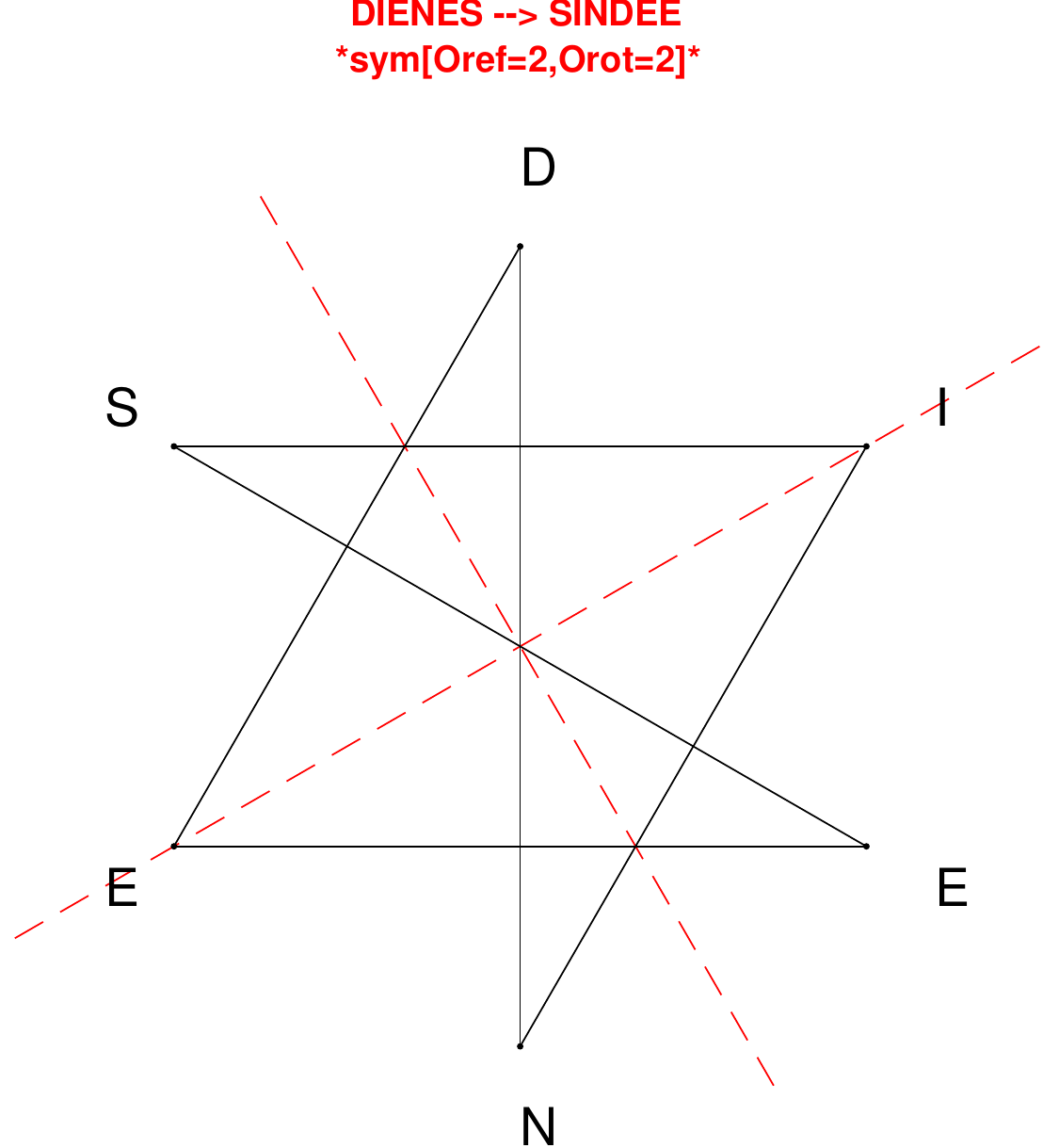}
\end{subfigure}
\hfill
\begin{subfigure}[T]{0.19\textwidth}
\centering
\includegraphics[width=\textwidth]{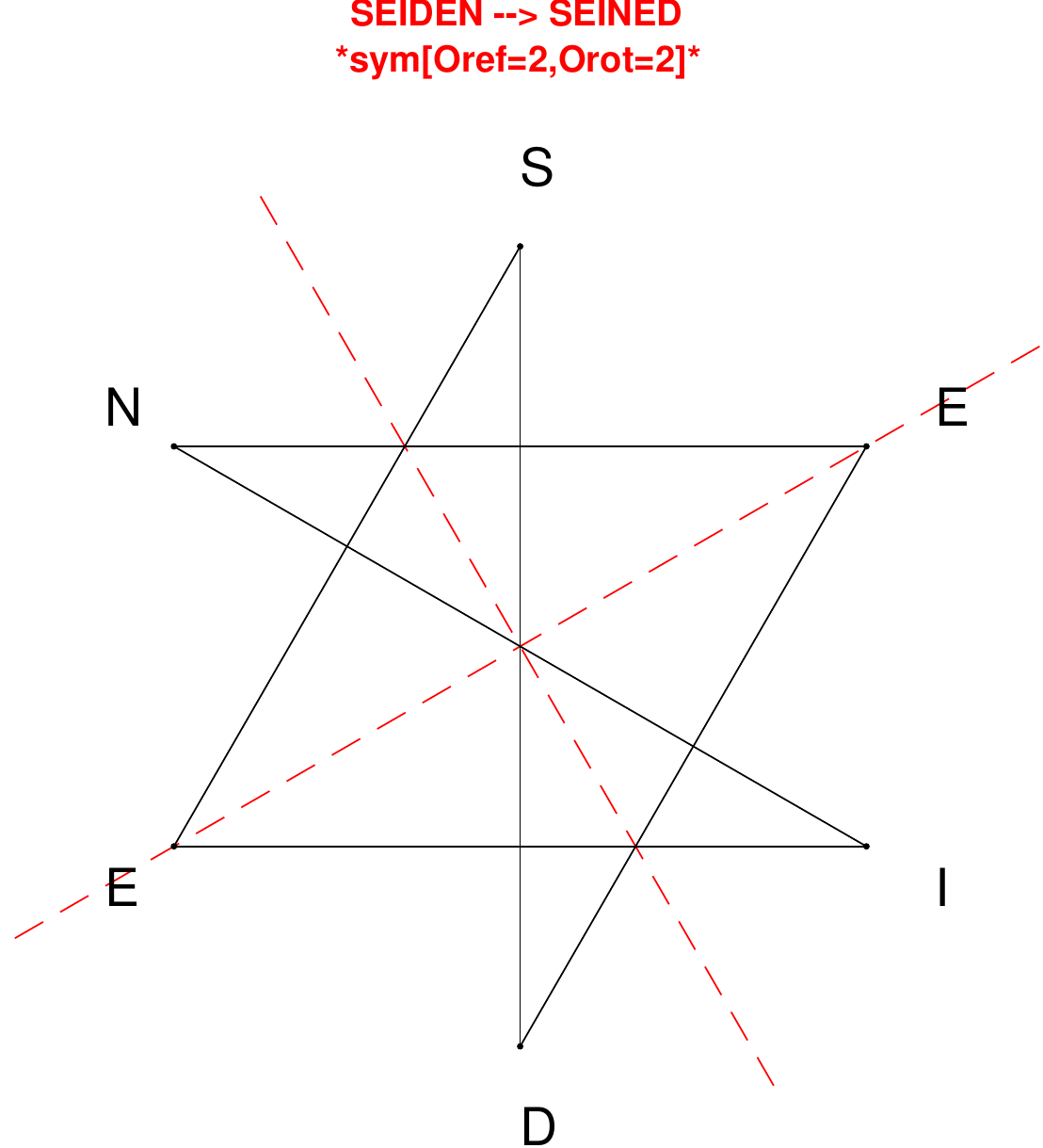}
\end{subfigure}
\hfill
\begin{subfigure}[T]{0.19\textwidth}
\centering
\includegraphics[width=\textwidth]{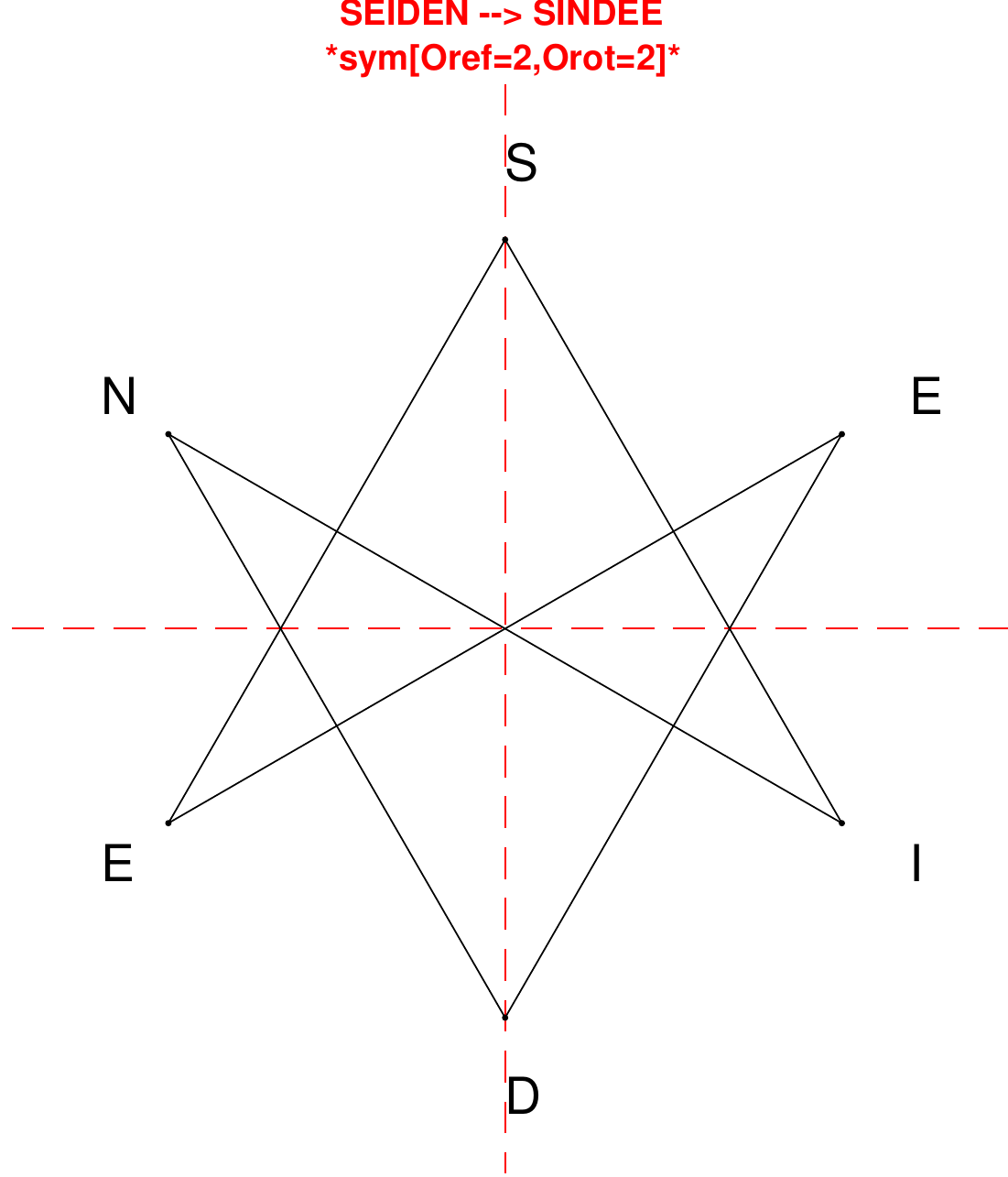}
\end{subfigure}
\hfill
\begin{subfigure}[T]{0.19\textwidth}
\centering
\includegraphics[width=\textwidth]{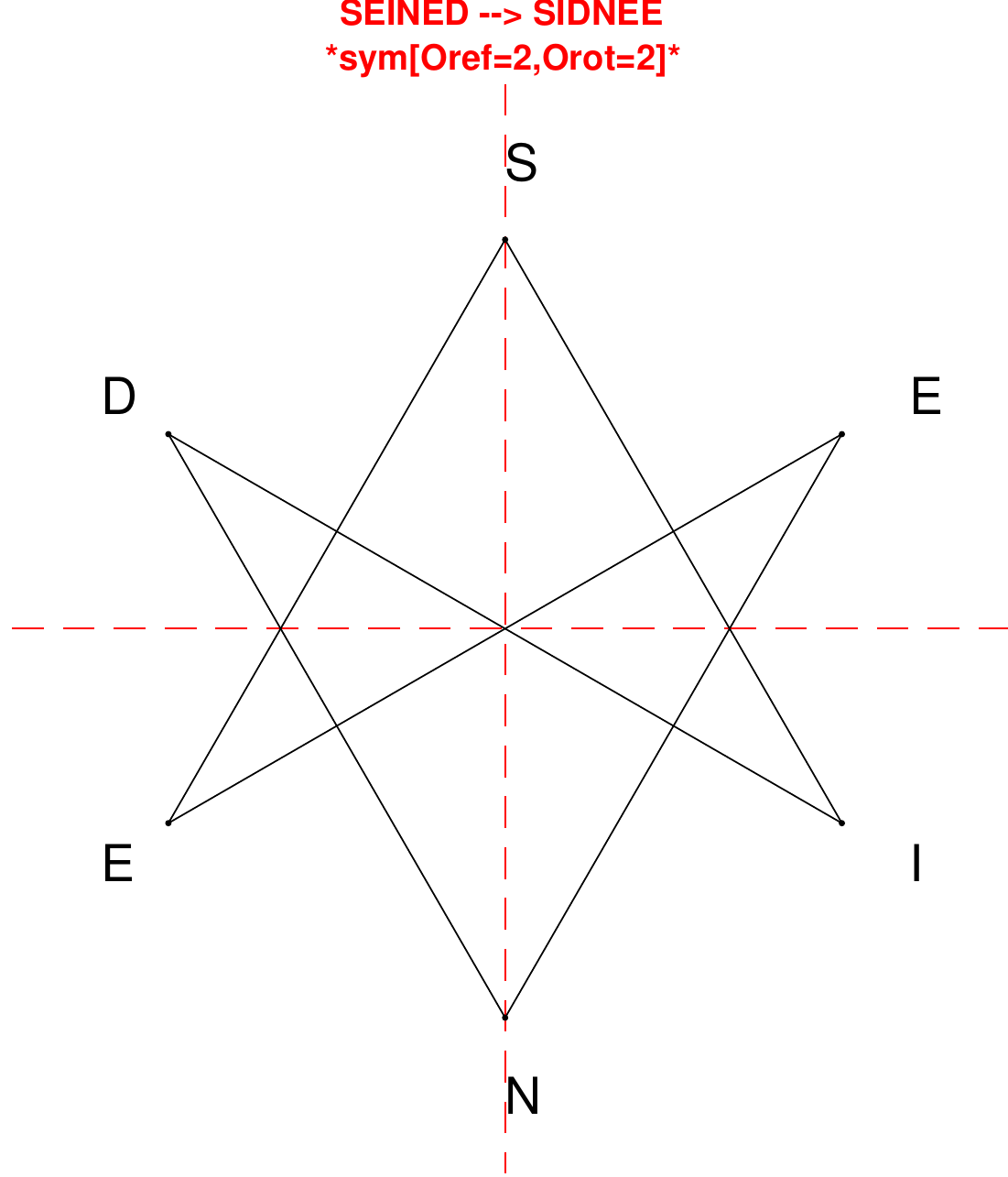}
\end{subfigure}
\end{figure}

\begin{figure}[H]
\centering
\begin{subfigure}[T]{0.19\textwidth}
\centering
\includegraphics[width=\textwidth]{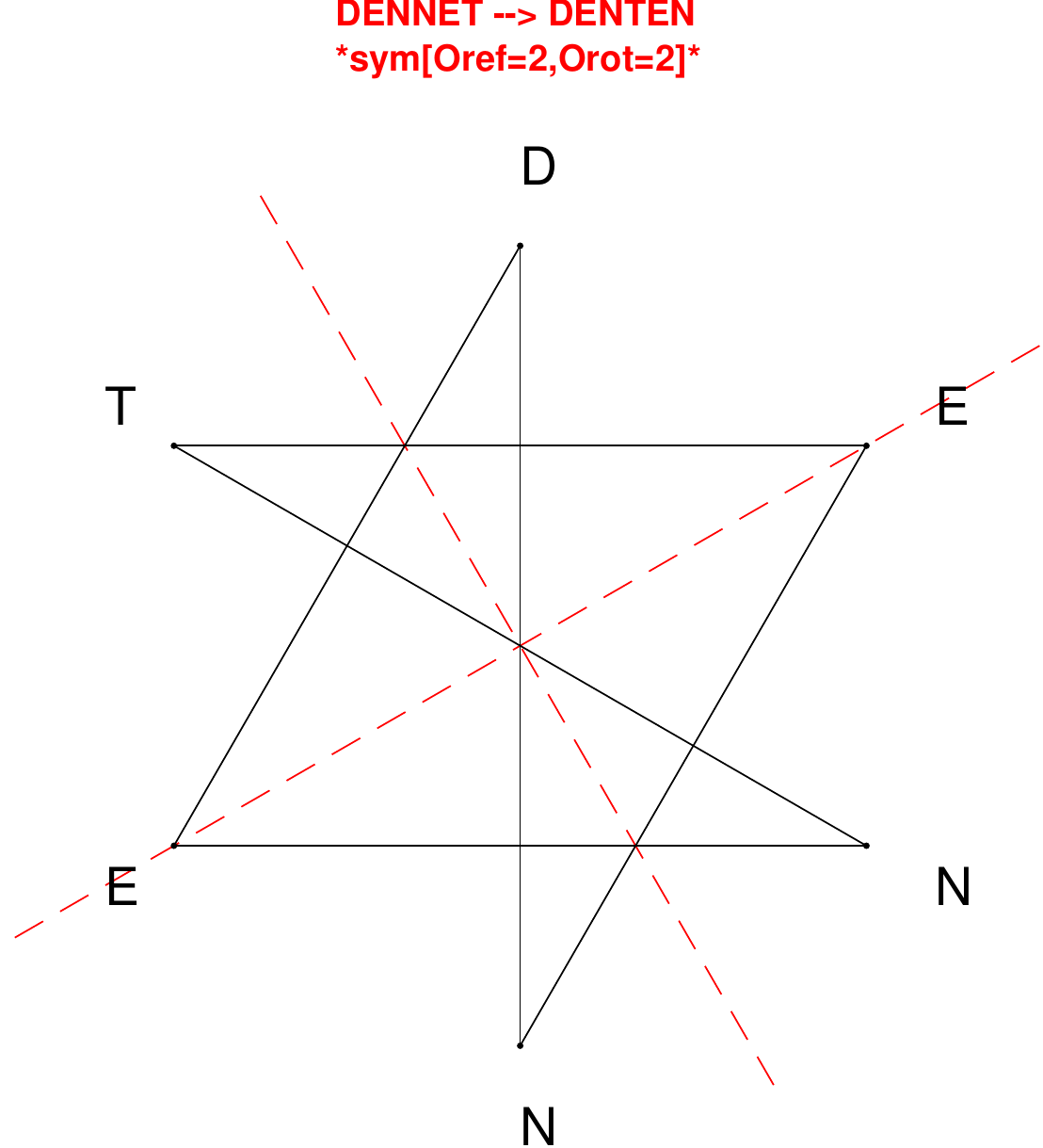}
\end{subfigure}
\hfill
\begin{subfigure}[T]{0.19\textwidth}
\centering
\includegraphics[width=\textwidth]{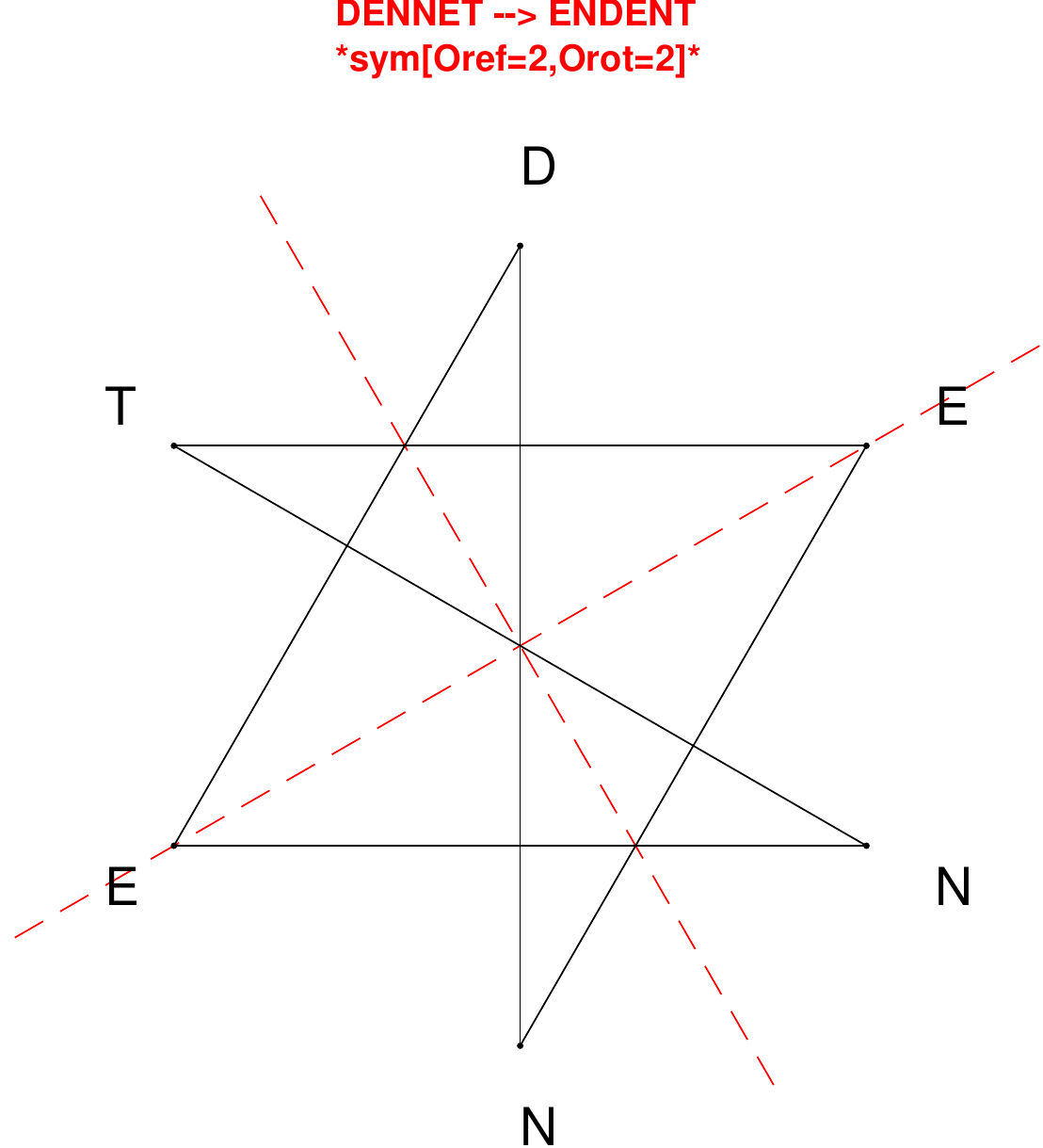}
\end{subfigure}
\hfill
\begin{subfigure}[T]{0.19\textwidth}
\centering
\includegraphics[width=\textwidth]{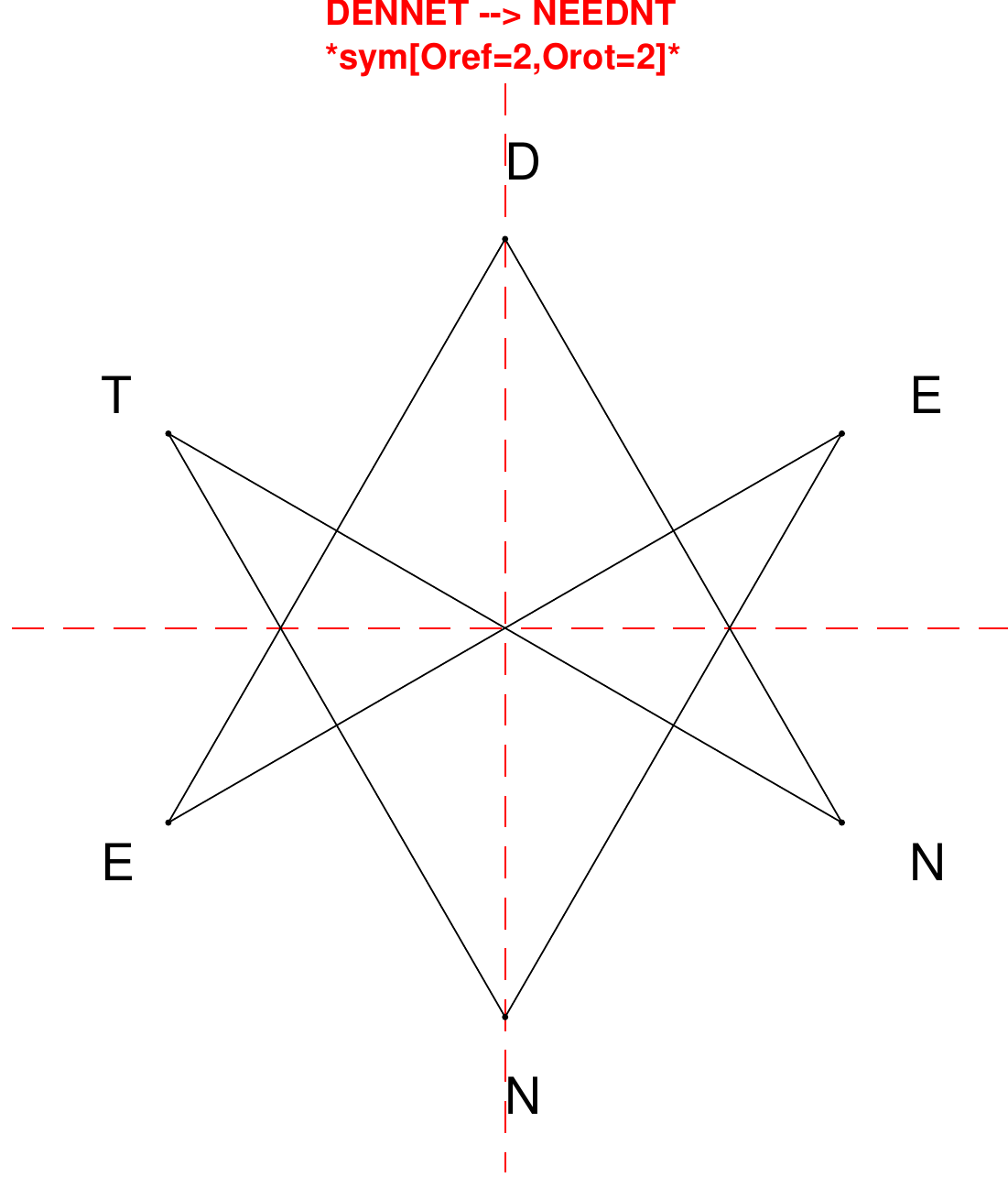}
\end{subfigure}
\hfill
\begin{subfigure}[T]{0.19\textwidth}
\centering
\includegraphics[width=\textwidth]{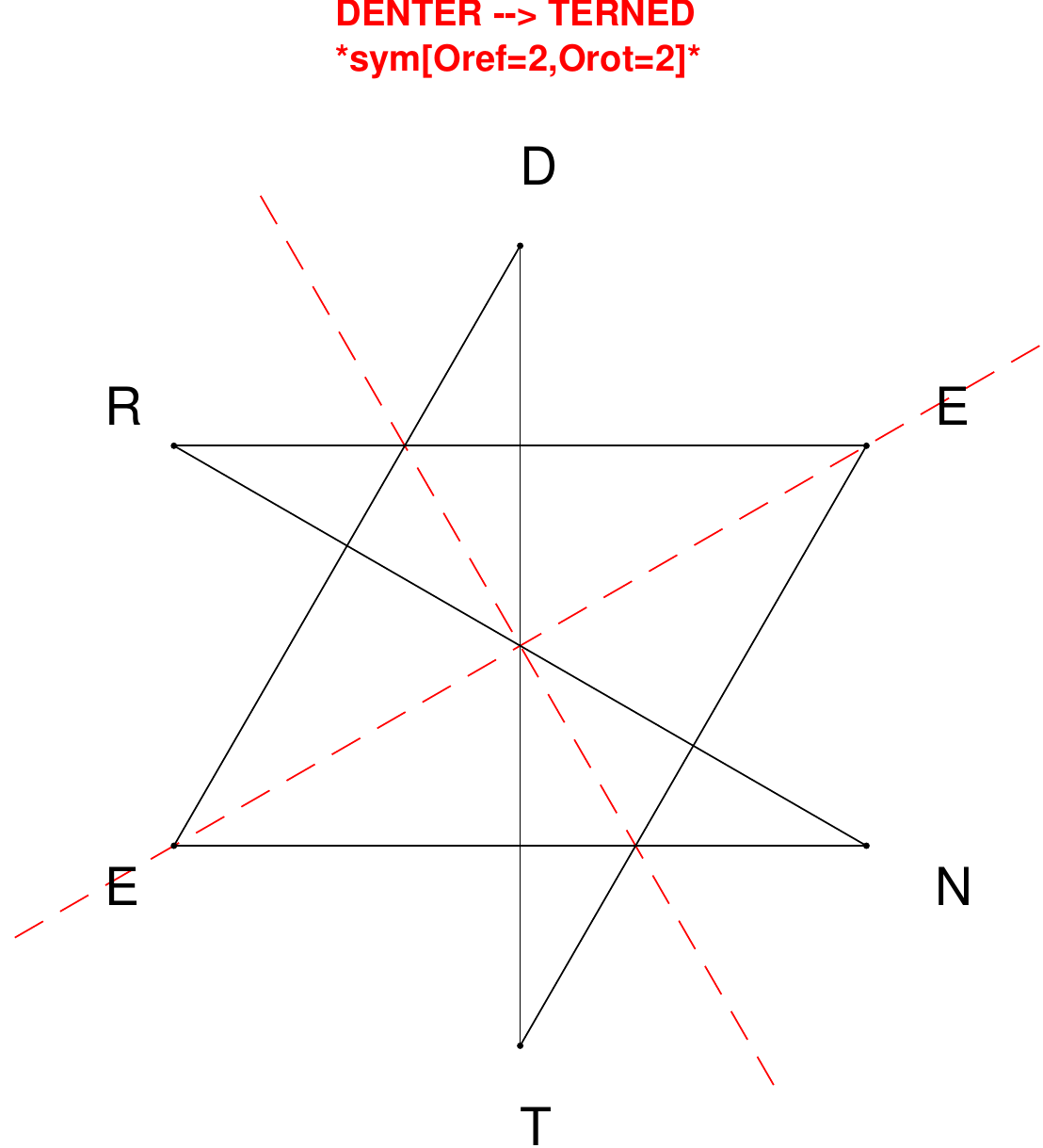}
\end{subfigure}
\hfill
\begin{subfigure}[T]{0.19\textwidth}
\centering
\includegraphics[width=\textwidth]{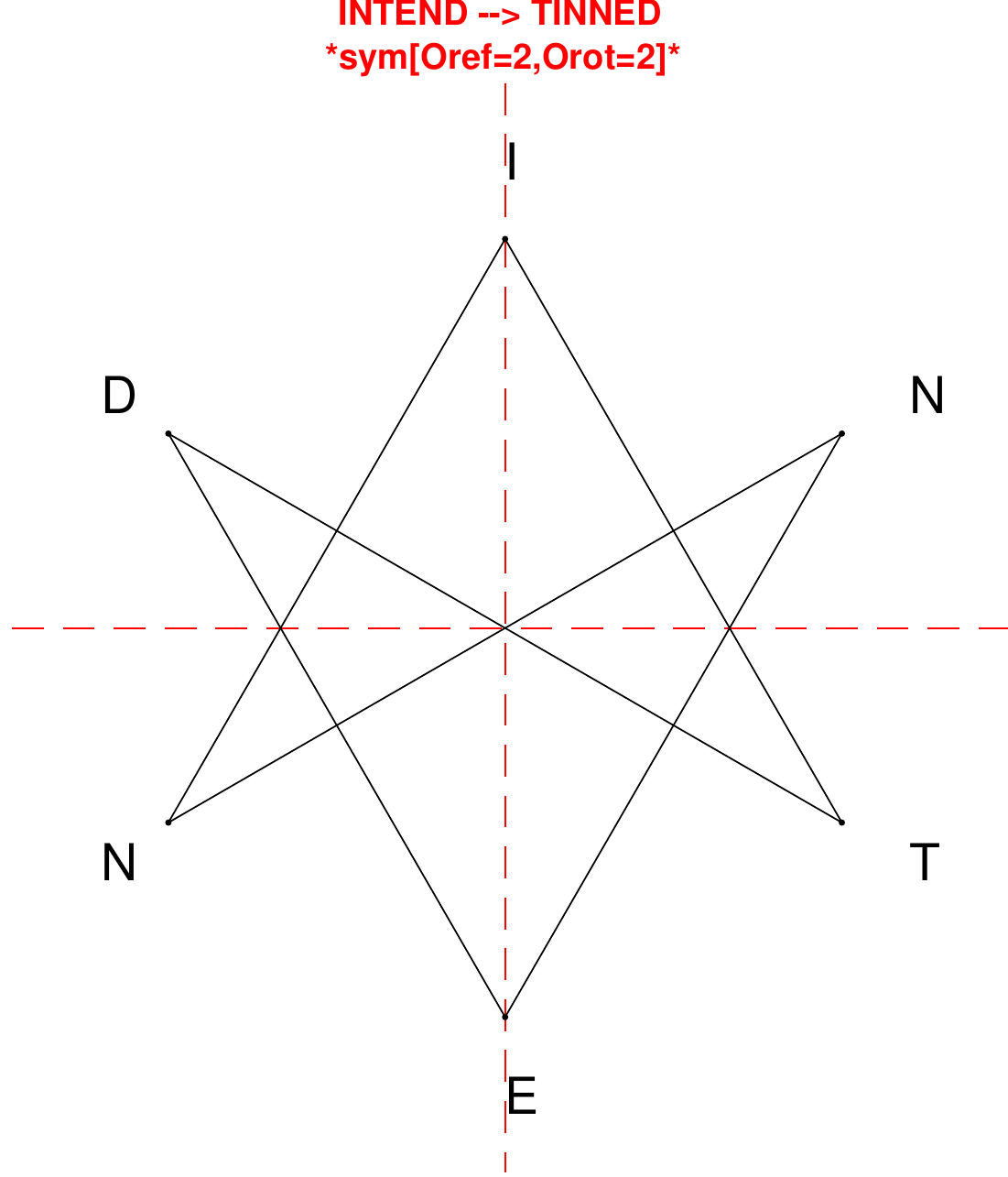}
\end{subfigure}
\end{figure}

\begin{figure}[H]
\centering
\begin{subfigure}[T]{0.19\textwidth}
\centering
\includegraphics[width=\textwidth]{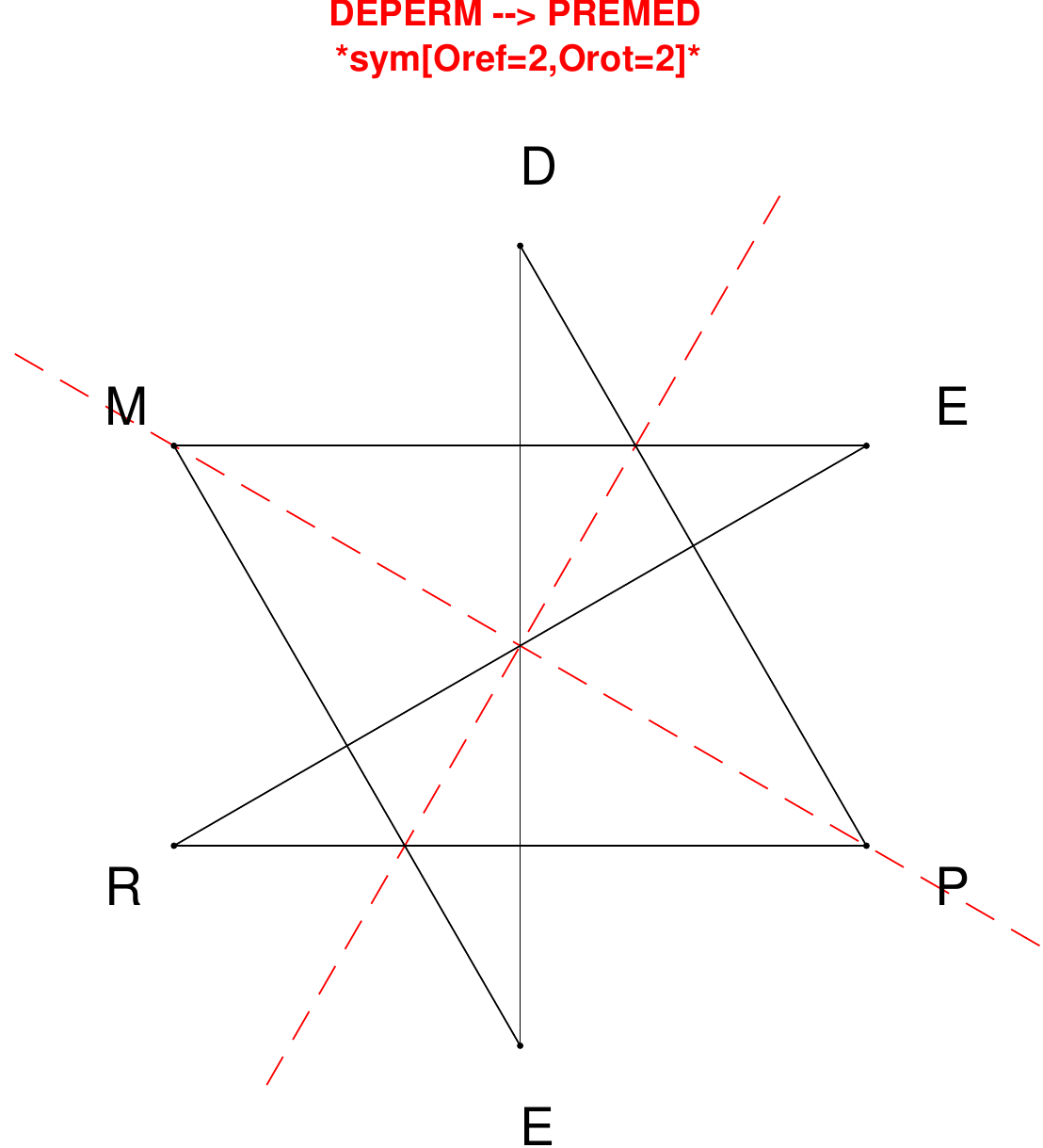}
\end{subfigure}
\hfill
\begin{subfigure}[T]{0.19\textwidth}
\centering
\includegraphics[width=\textwidth]{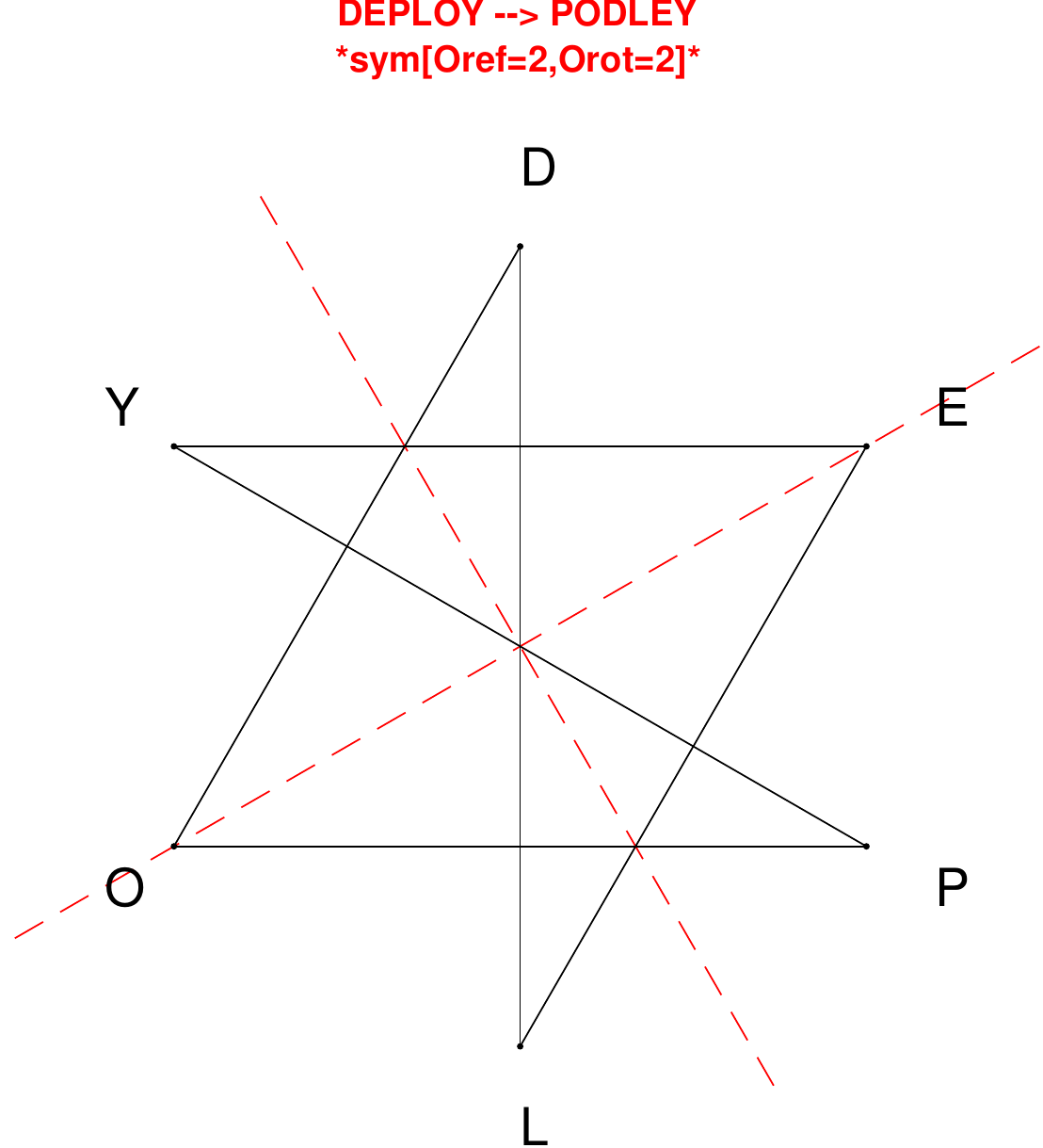}
\end{subfigure}
\hfill
\begin{subfigure}[T]{0.19\textwidth}
\centering
\includegraphics[width=\textwidth]{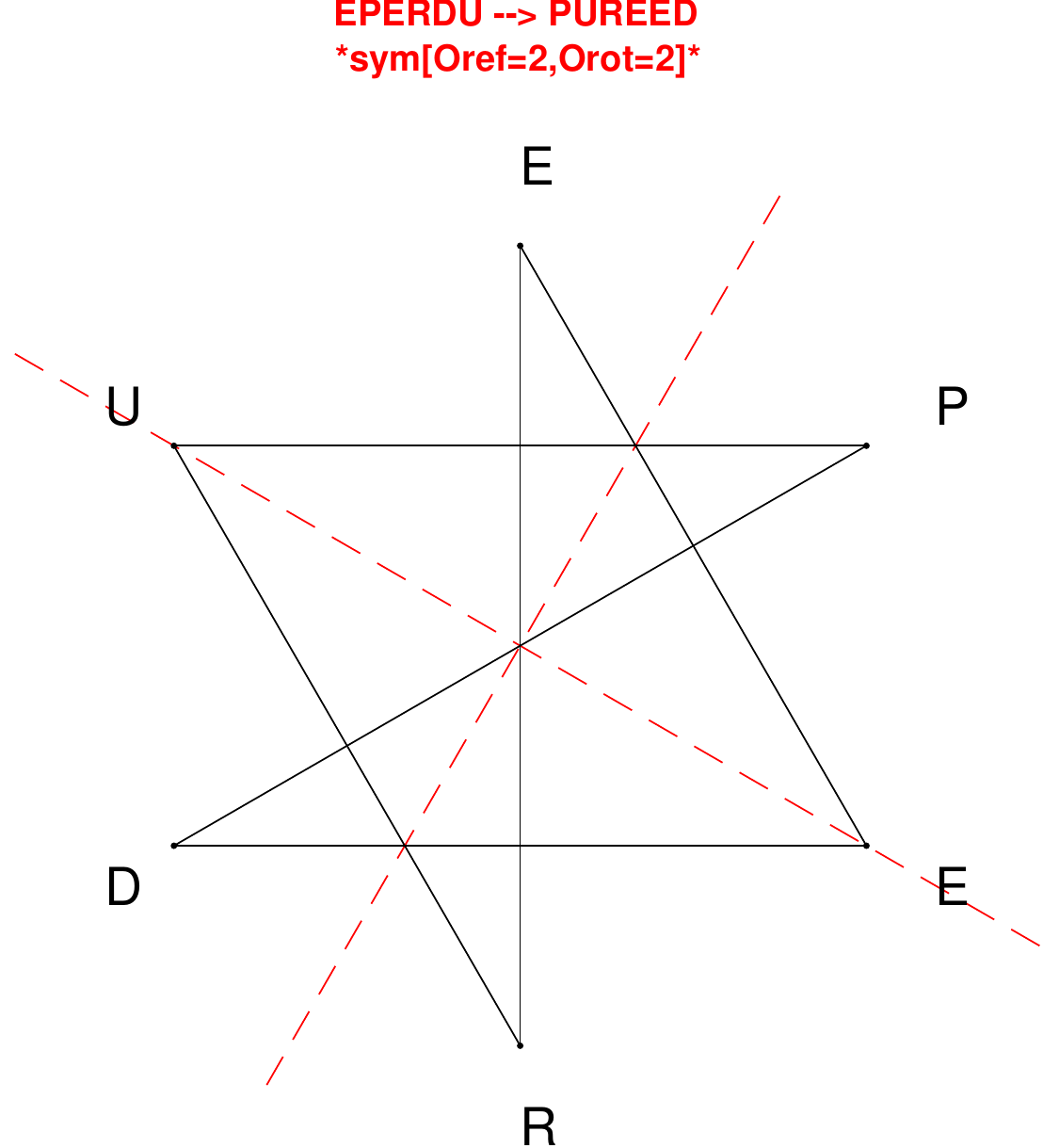}
\end{subfigure}
\hfill
\begin{subfigure}[T]{0.19\textwidth}
\centering
\includegraphics[width=\textwidth]{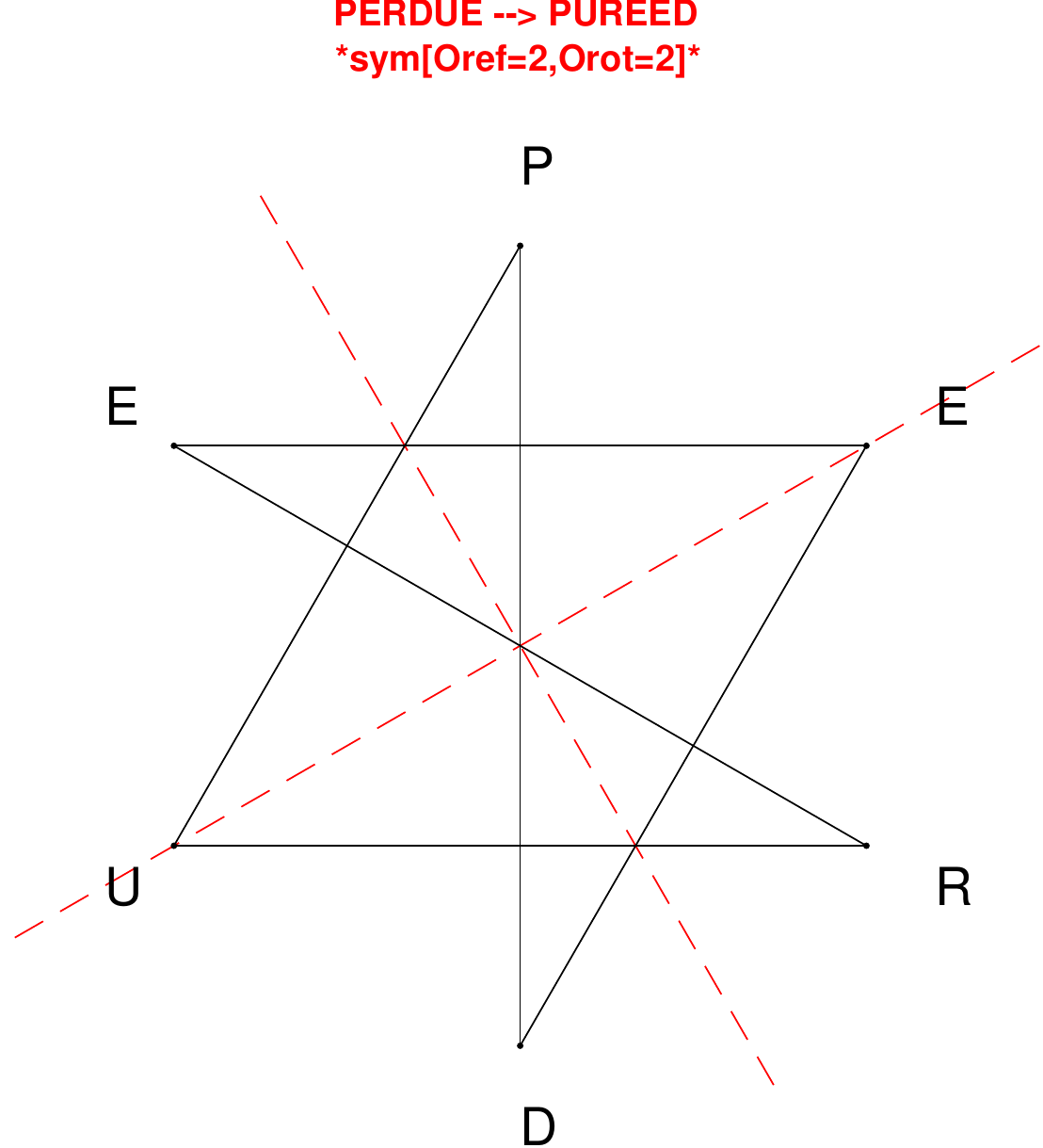}
\end{subfigure}
\hfill
\begin{subfigure}[T]{0.19\textwidth}
\centering
\includegraphics[width=\textwidth]{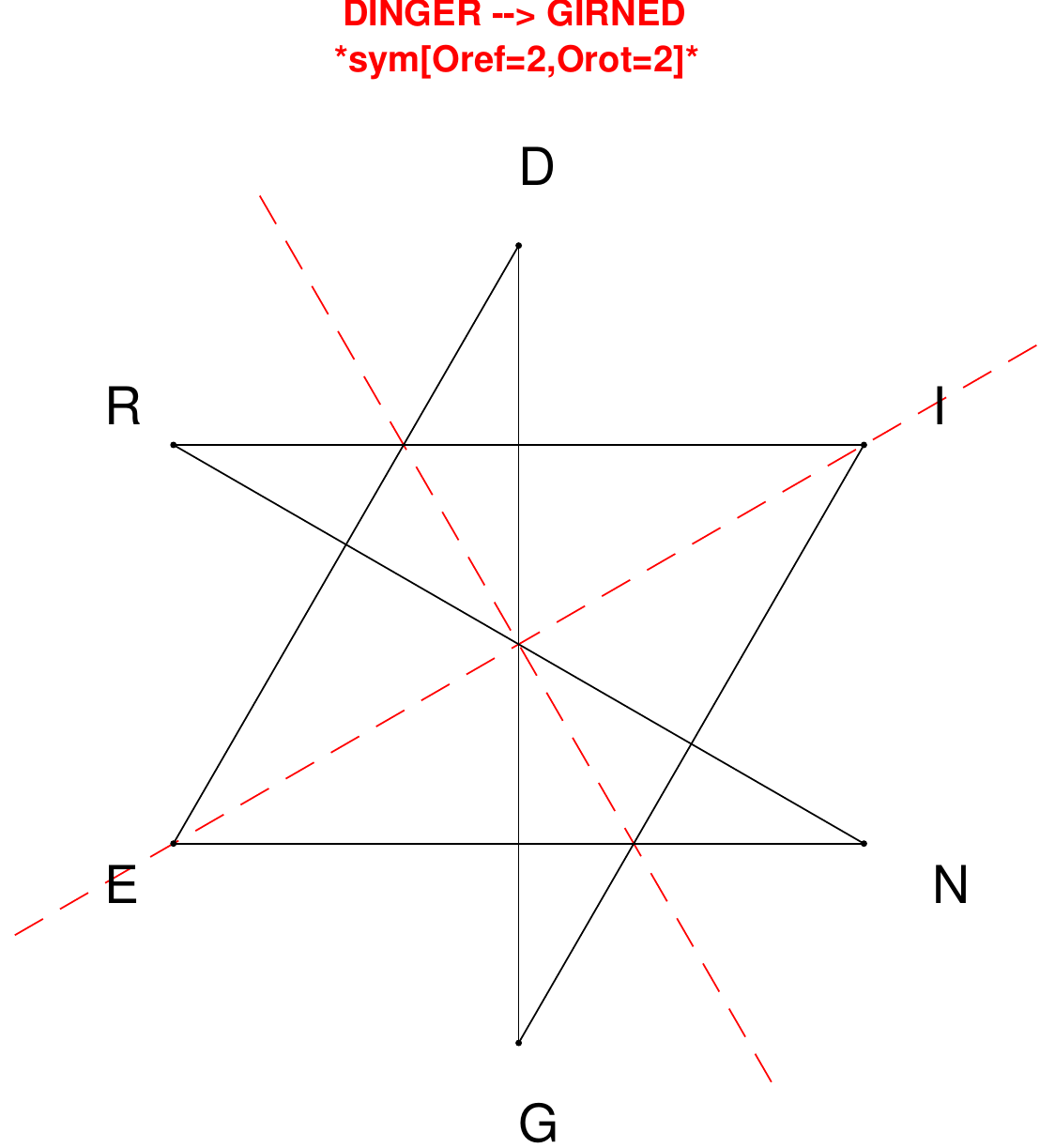}
\end{subfigure}
\end{figure}

\begin{figure}[H]
\centering
\begin{subfigure}[T]{0.19\textwidth}
\centering
\includegraphics[width=\textwidth]{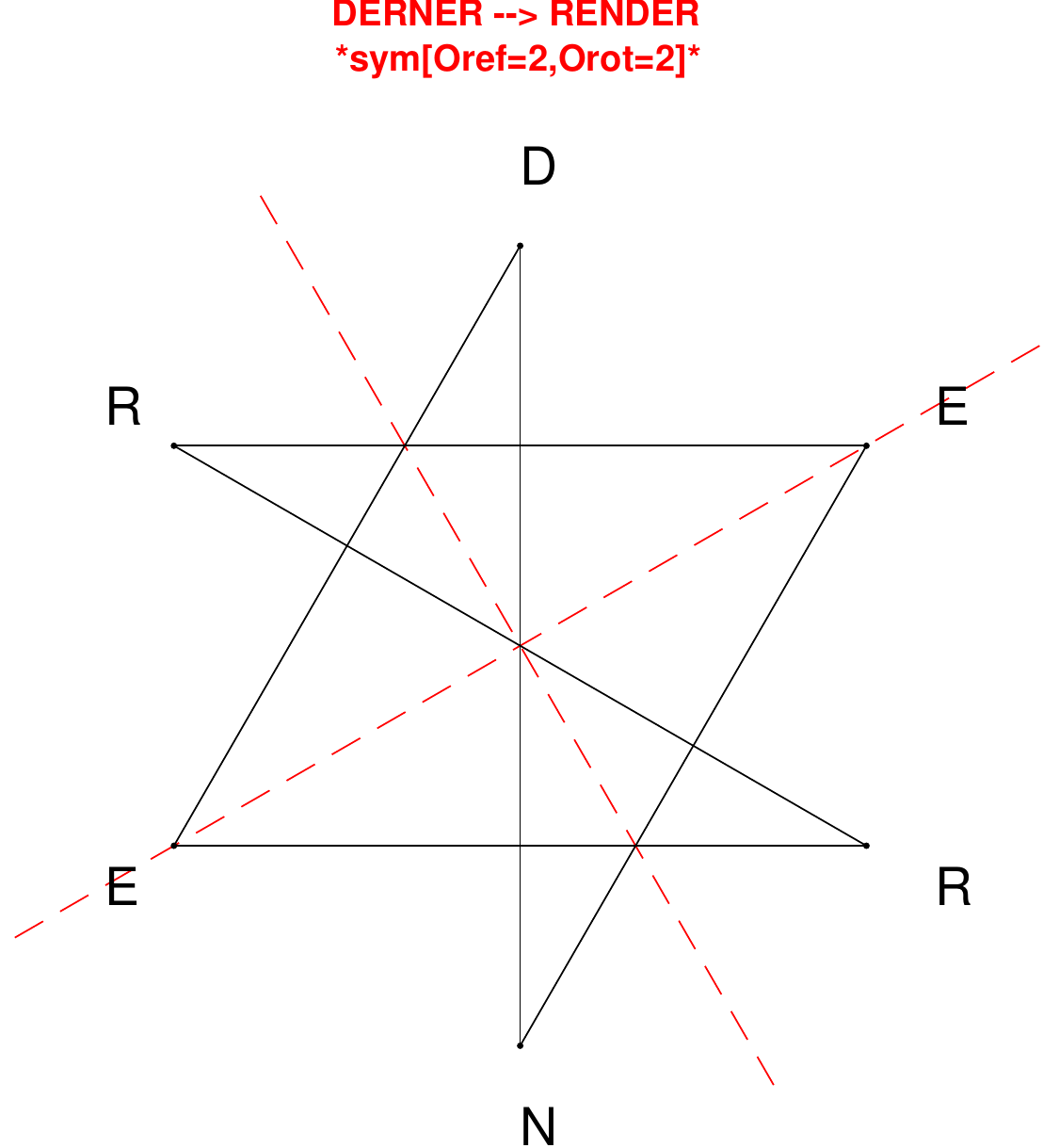}
\end{subfigure}
\hfill
\begin{subfigure}[T]{0.19\textwidth}
\centering
\includegraphics[width=\textwidth]{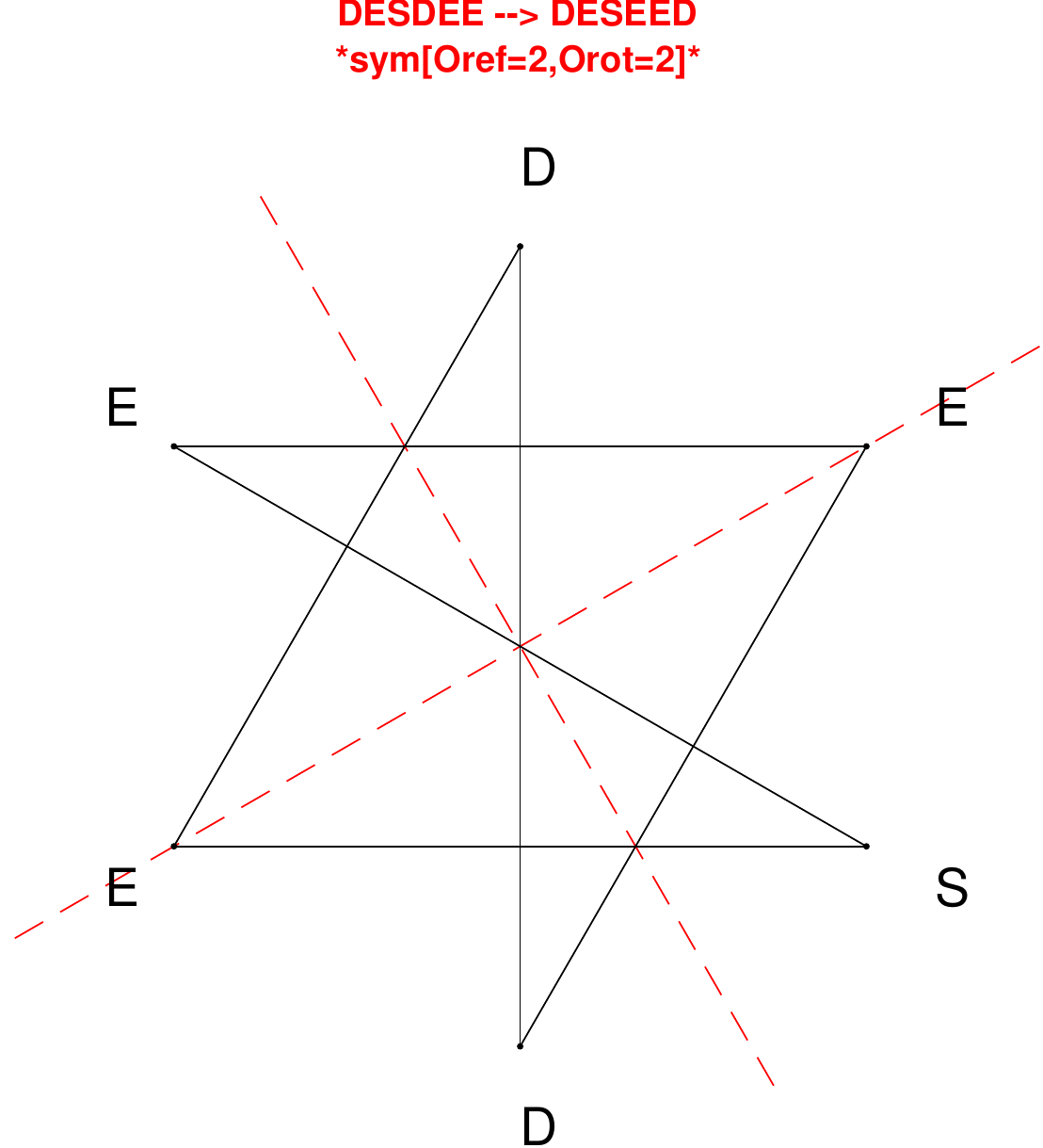}
\end{subfigure}
\hfill
\begin{subfigure}[T]{0.19\textwidth}
\centering
\includegraphics[width=\textwidth]{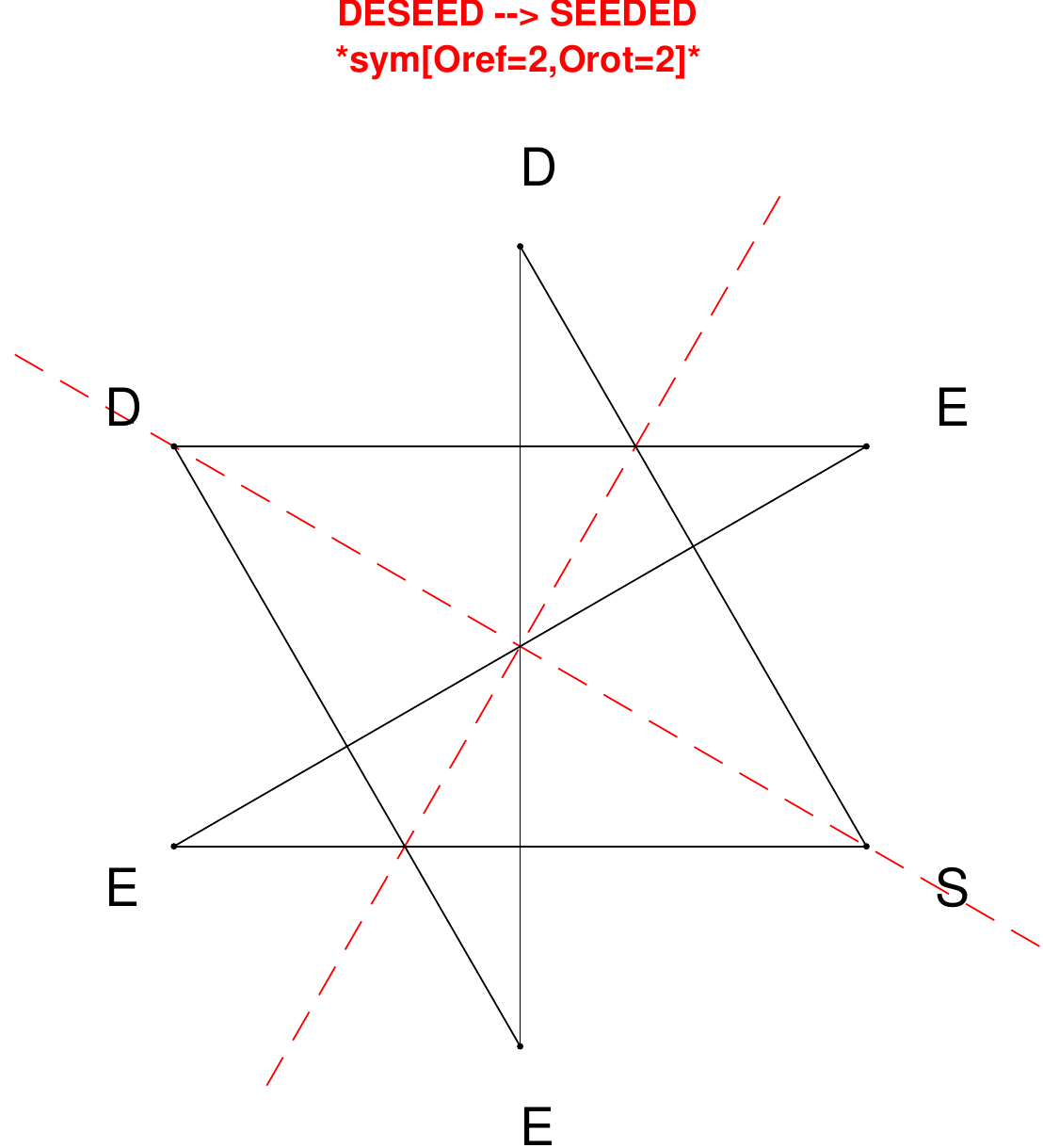}
\end{subfigure}
\hfill
\begin{subfigure}[T]{0.19\textwidth}
\centering
\includegraphics[width=\textwidth]{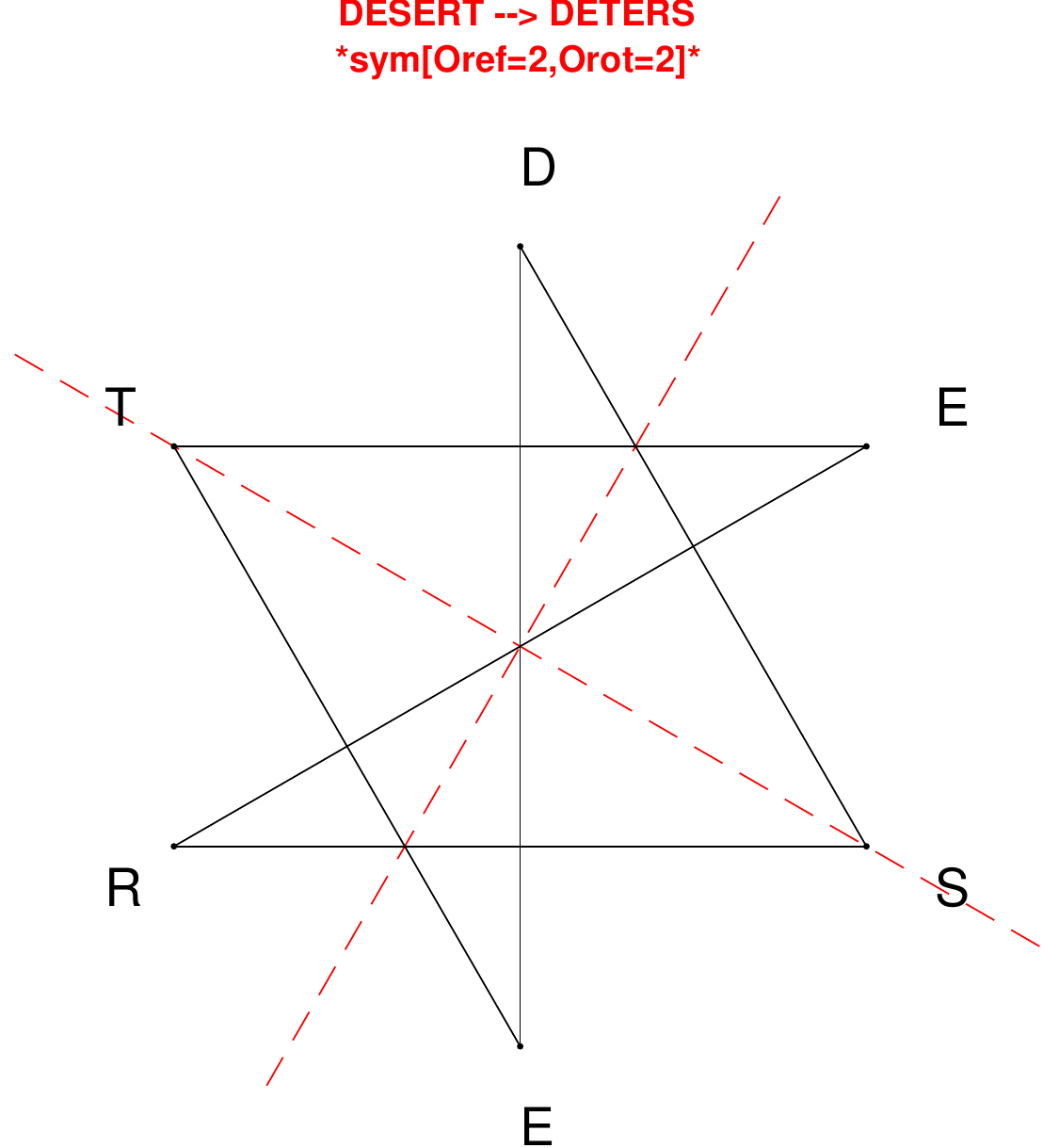}
\end{subfigure}
\hfill
\begin{subfigure}[T]{0.19\textwidth}
\centering
\includegraphics[width=\textwidth]{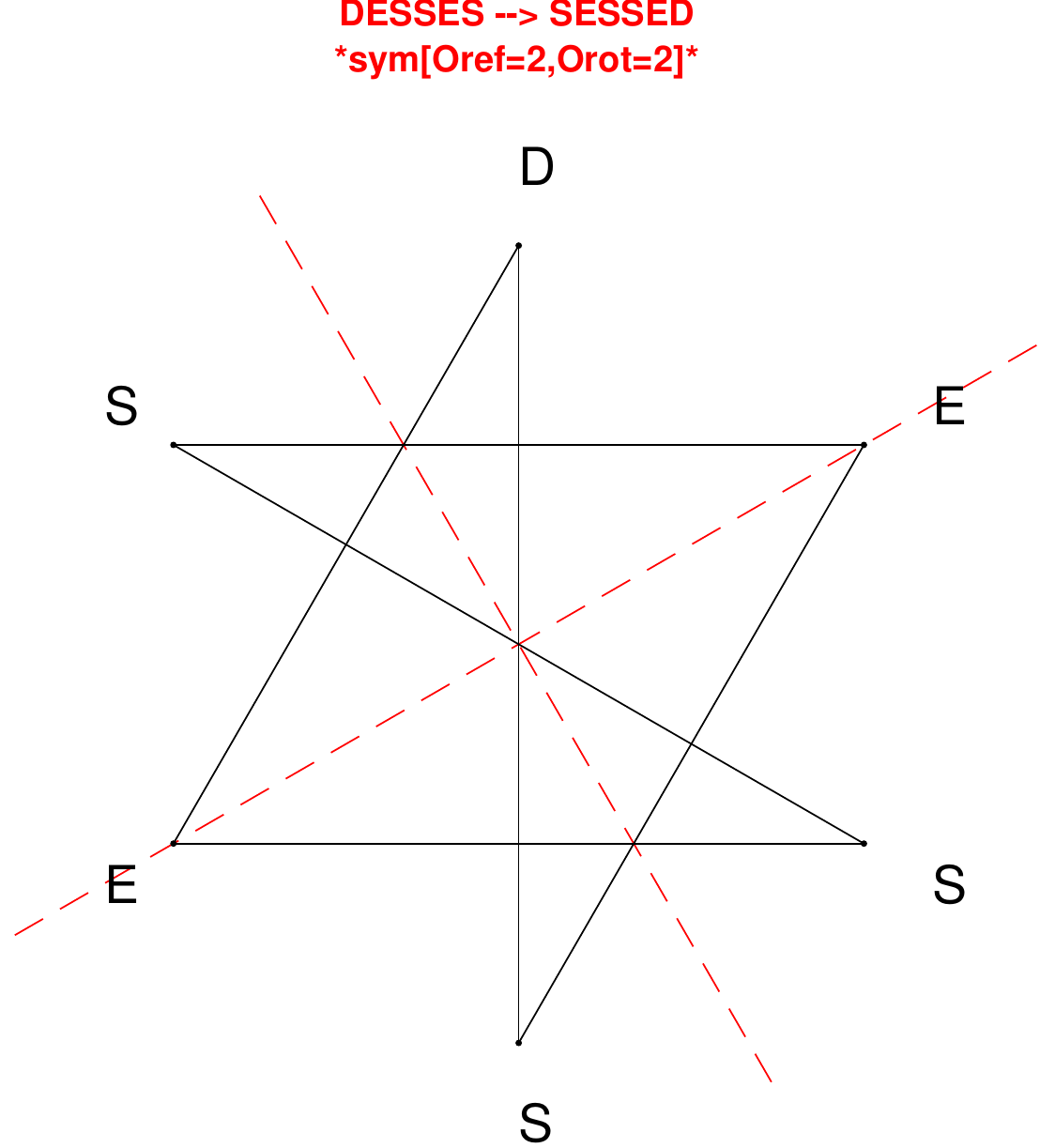}
\end{subfigure}
\end{figure}

\begin{figure}[H]
\centering
\begin{subfigure}[T]{0.19\textwidth}
\centering
\includegraphics[width=\textwidth]{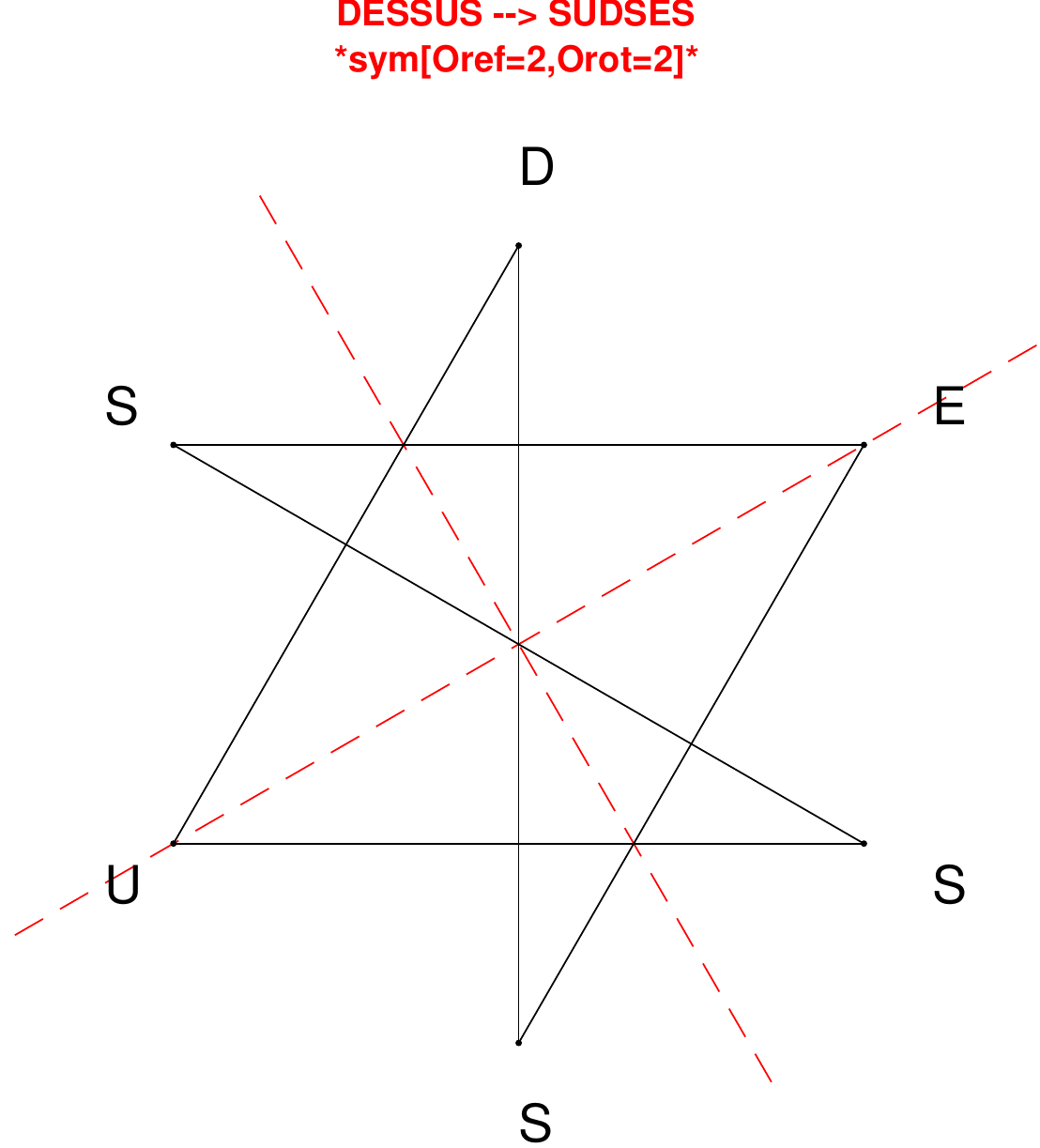}
\end{subfigure}
\hfill
\begin{subfigure}[T]{0.19\textwidth}
\centering
\includegraphics[width=\textwidth]{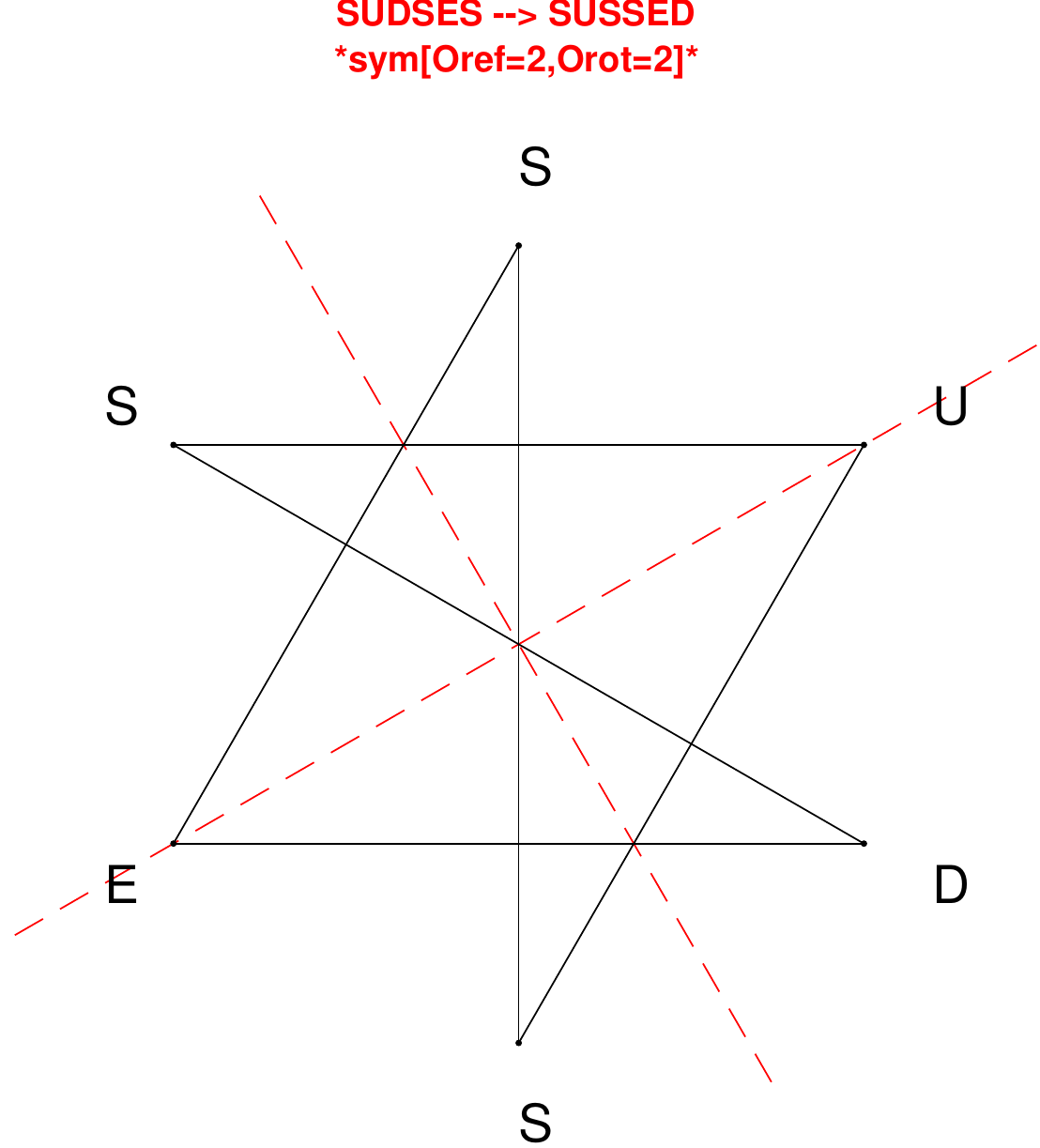}
\end{subfigure}
\hfill
\begin{subfigure}[T]{0.19\textwidth}
\centering
\includegraphics[width=\textwidth]{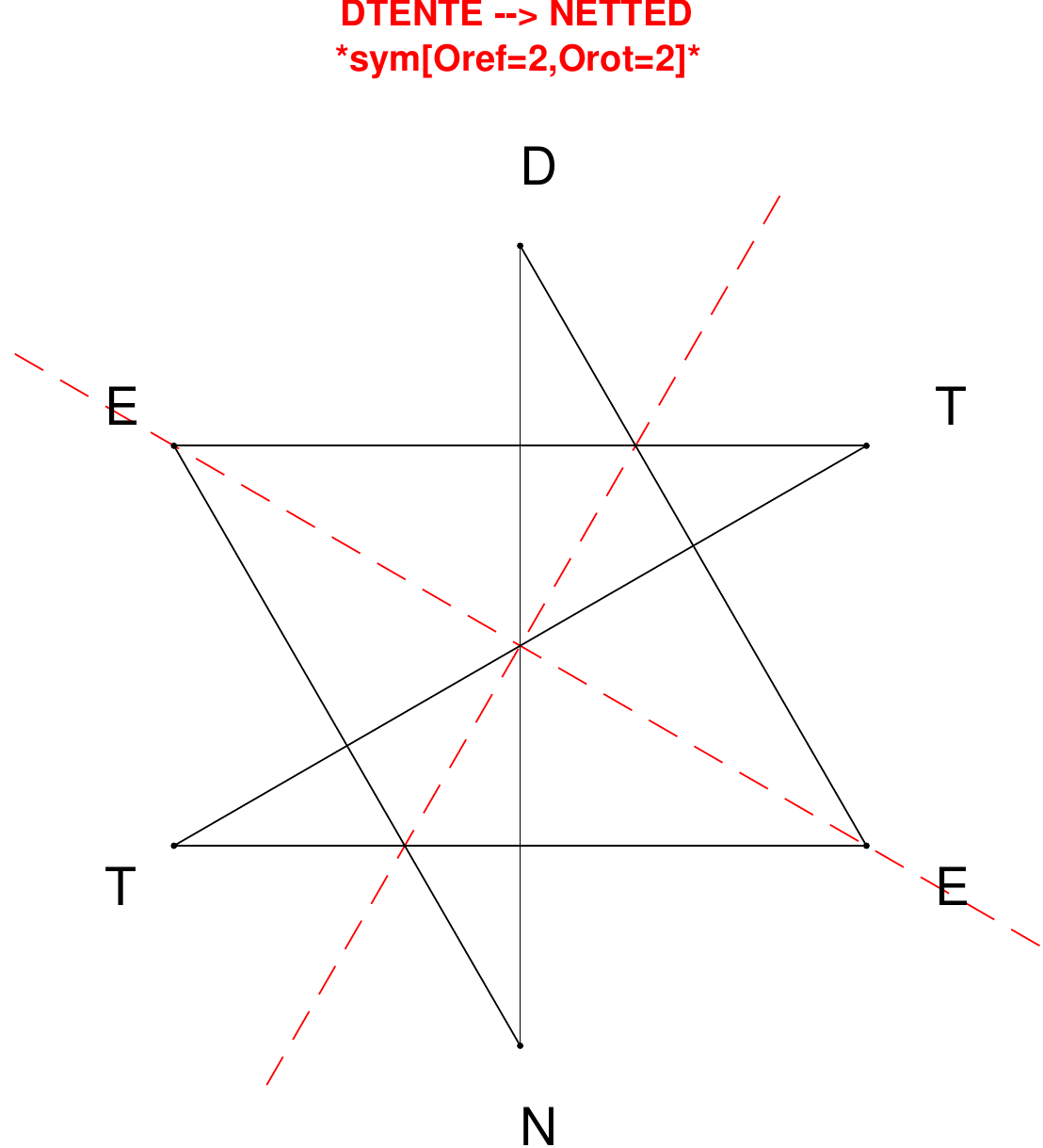}
\end{subfigure}
\hfill
\begin{subfigure}[T]{0.19\textwidth}
\centering
\includegraphics[width=\textwidth]{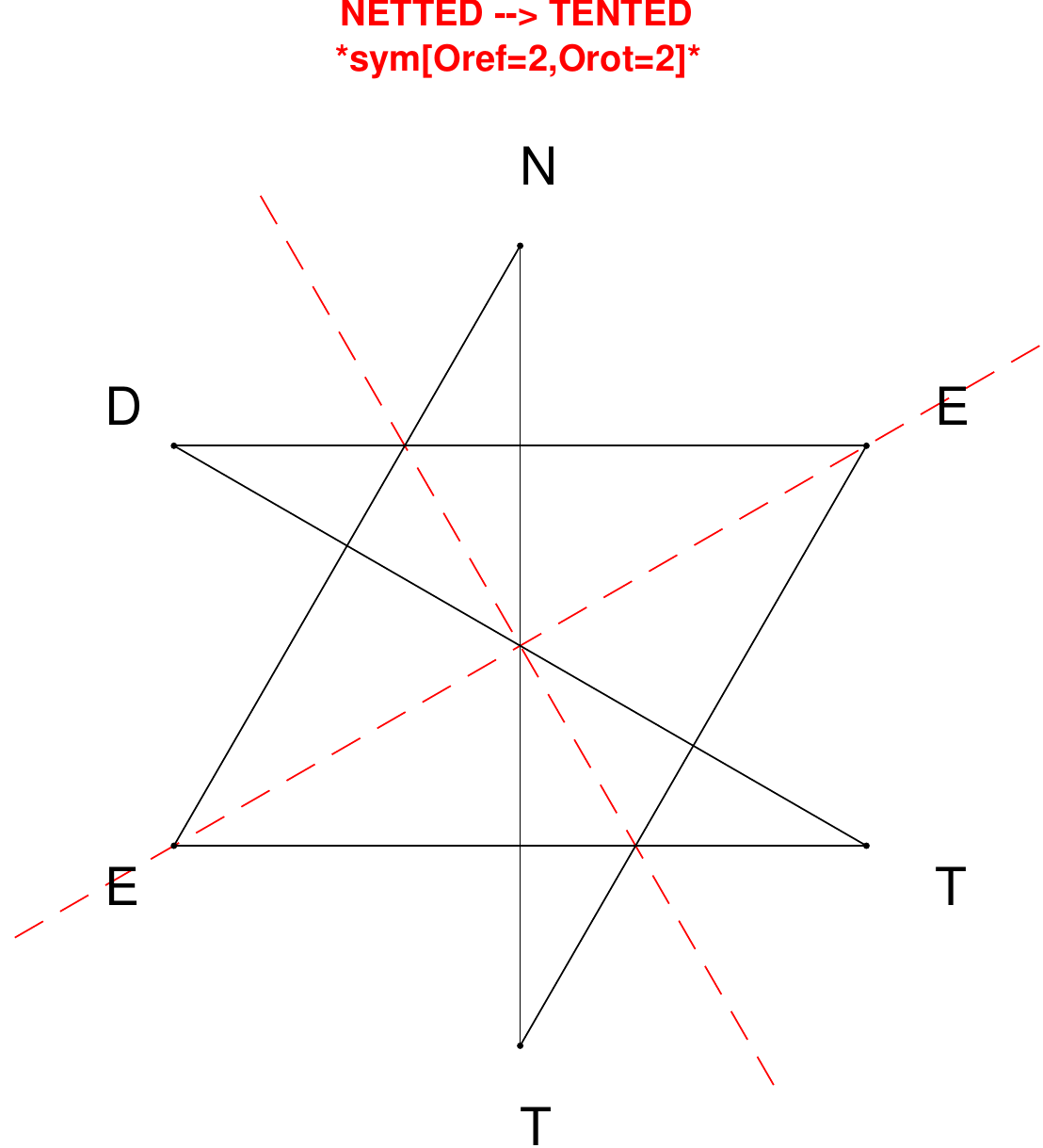}
\end{subfigure}
\hfill
\begin{subfigure}[T]{0.19\textwidth}
\centering
\includegraphics[width=\textwidth]{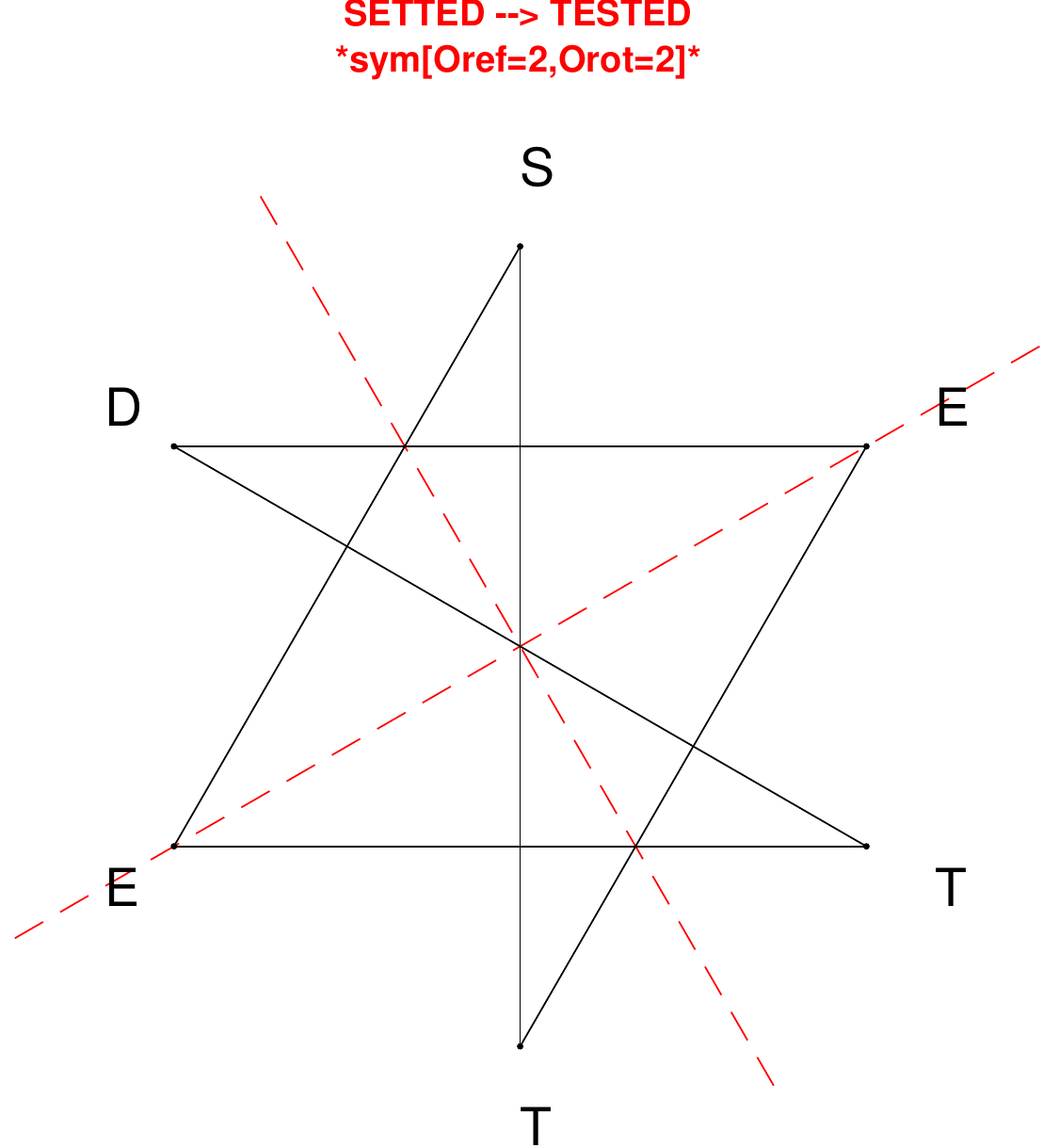}
\end{subfigure}
\end{figure}

\begin{figure}[H]
\centering
\begin{subfigure}[T]{0.19\textwidth}
\centering
\includegraphics[width=\textwidth]{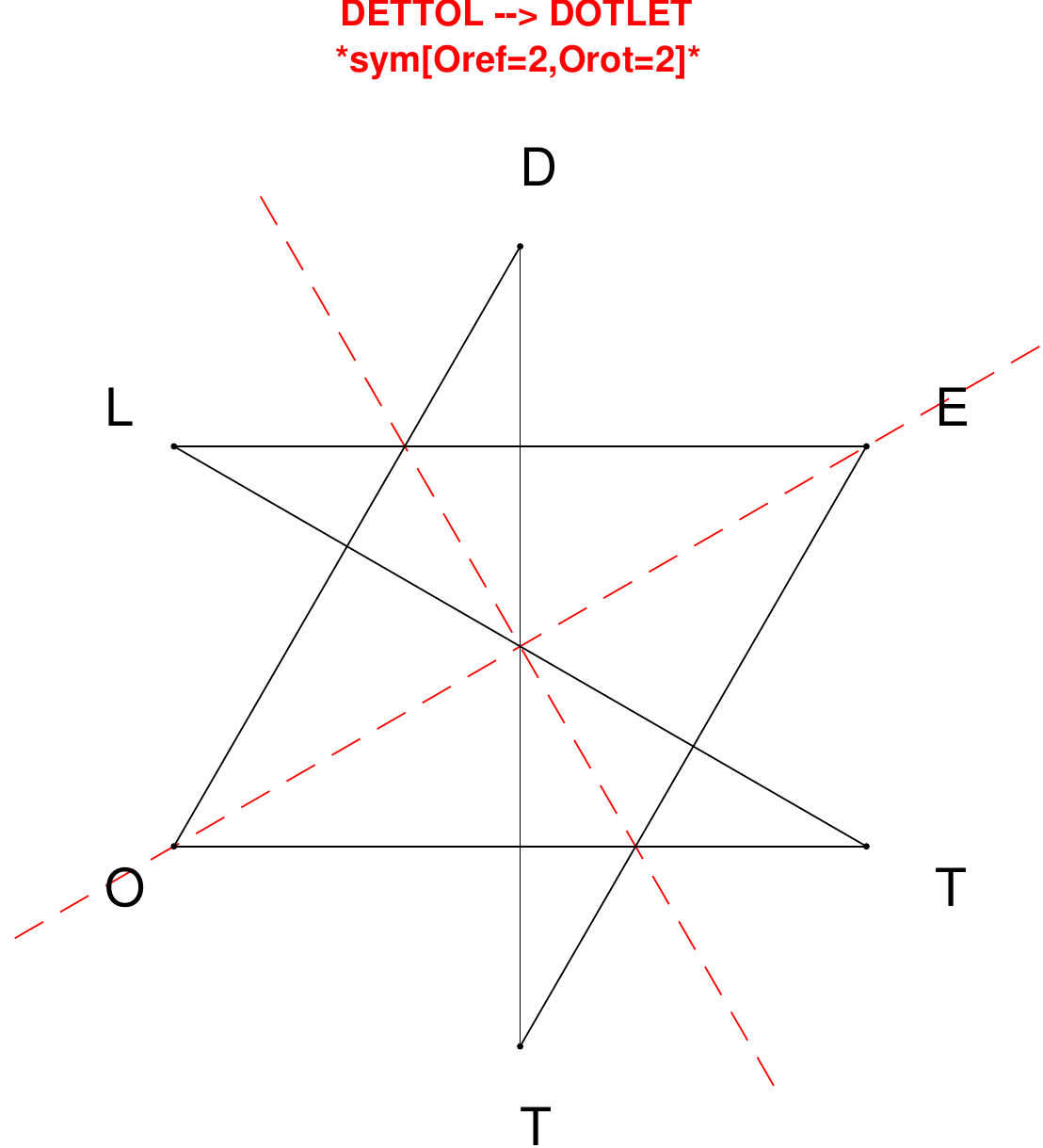}
\end{subfigure}
\hfill
\begin{subfigure}[T]{0.19\textwidth}
\centering
\includegraphics[width=\textwidth]{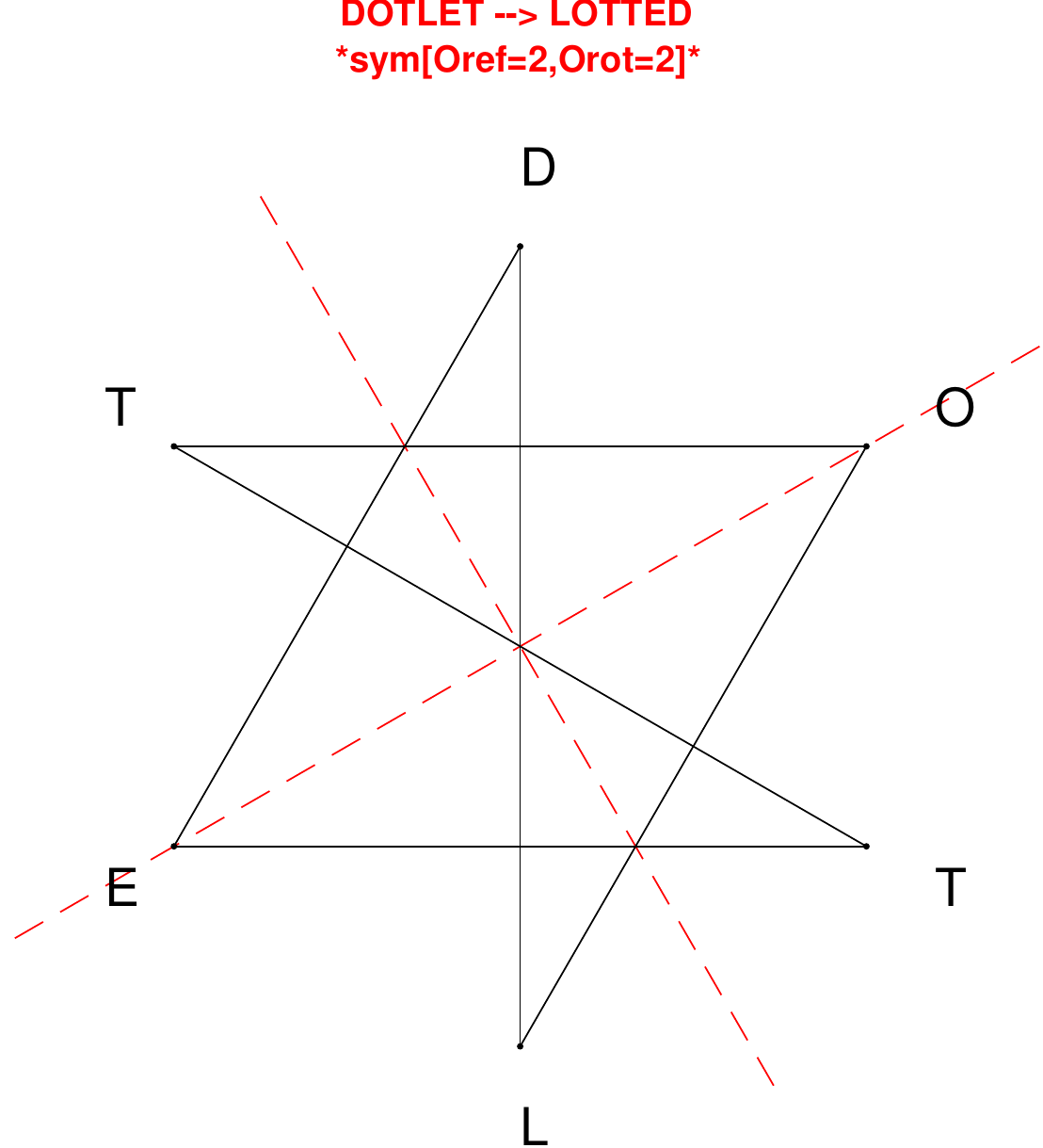}
\end{subfigure}
\hfill
\begin{subfigure}[T]{0.19\textwidth}
\centering
\includegraphics[width=\textwidth]{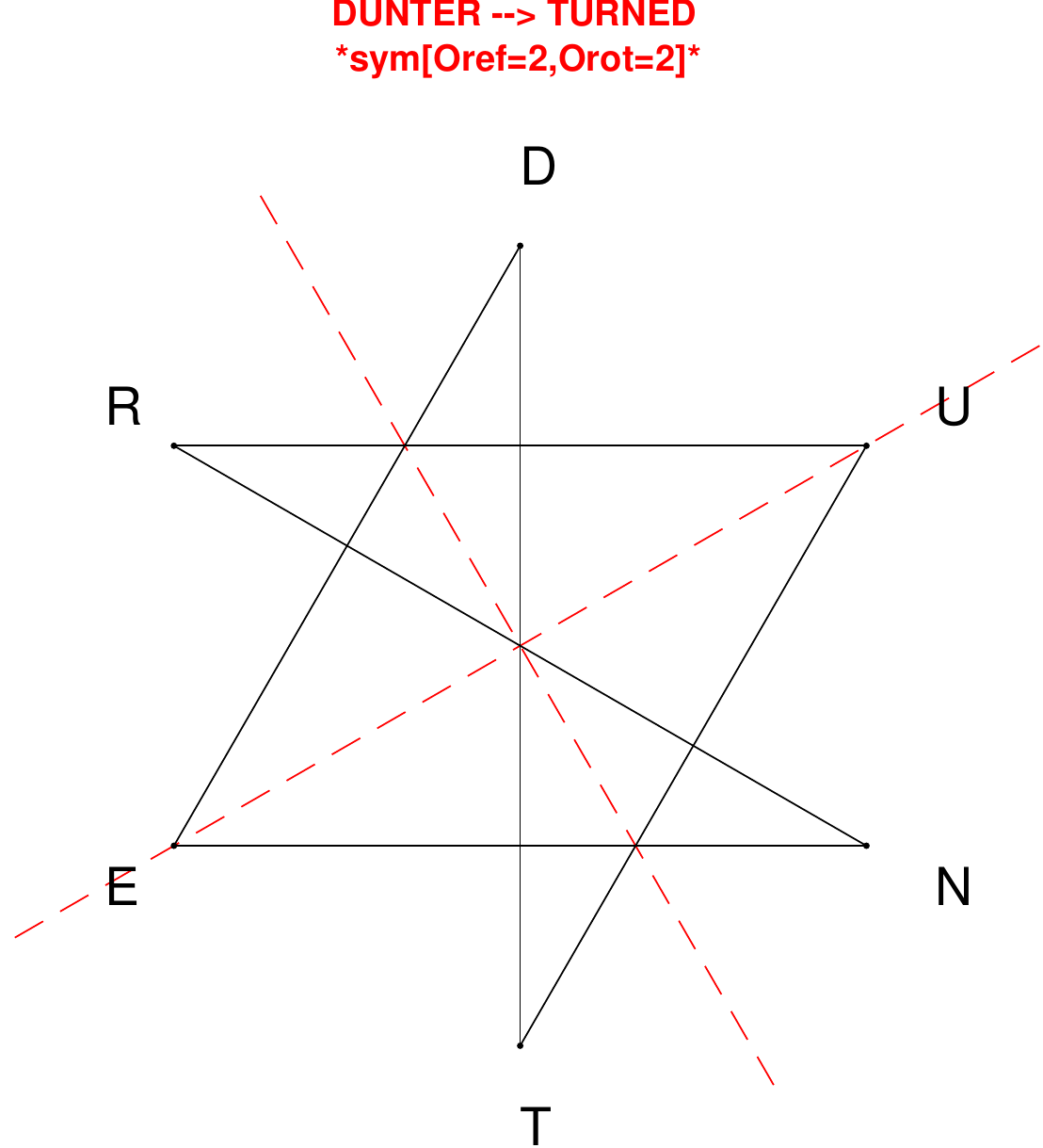}
\end{subfigure}
\hfill
\begin{subfigure}[T]{0.19\textwidth}
\centering
\includegraphics[width=\textwidth]{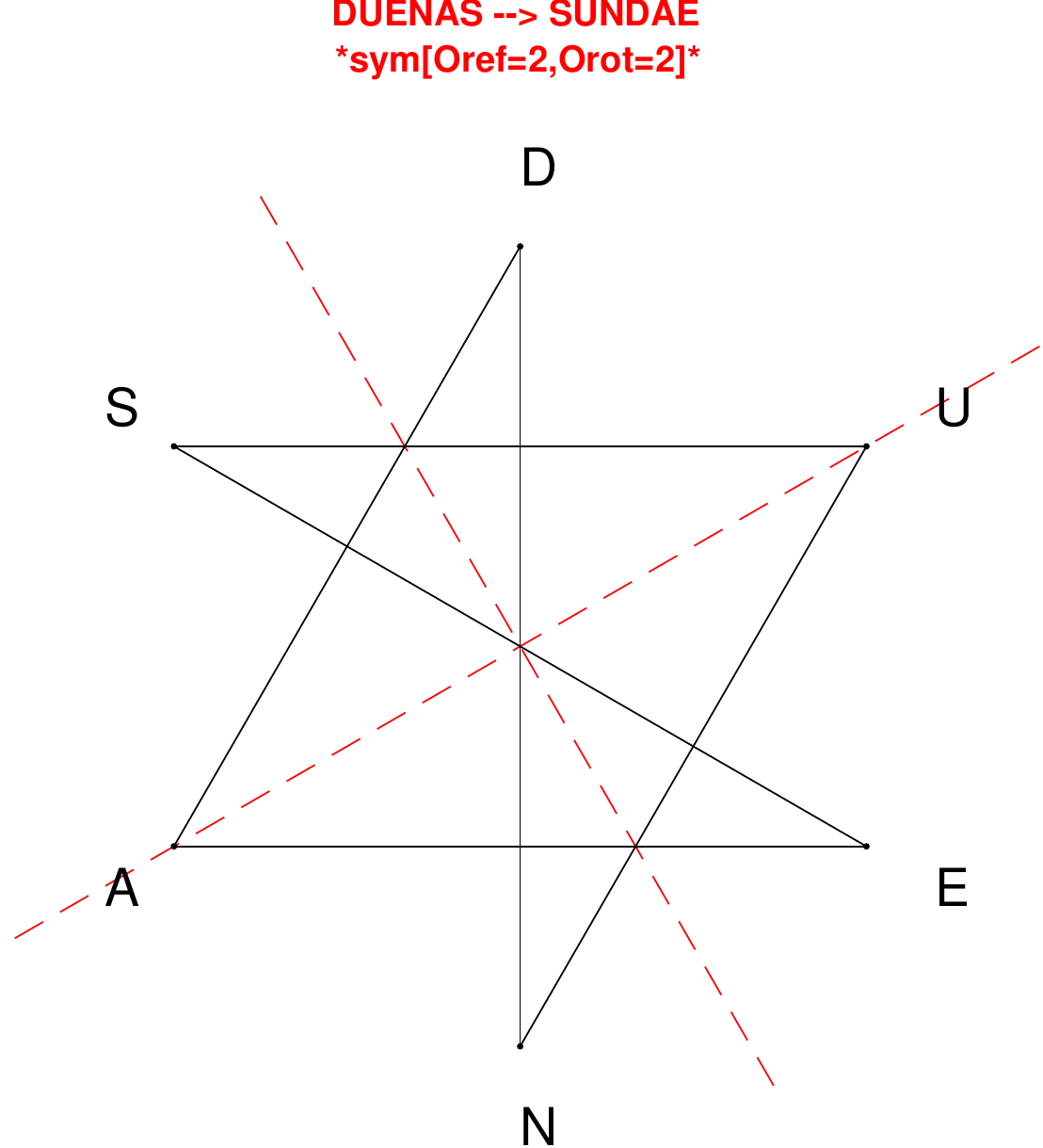}
\end{subfigure}
\hfill
\begin{subfigure}[T]{0.19\textwidth}
\centering
\includegraphics[width=\textwidth]{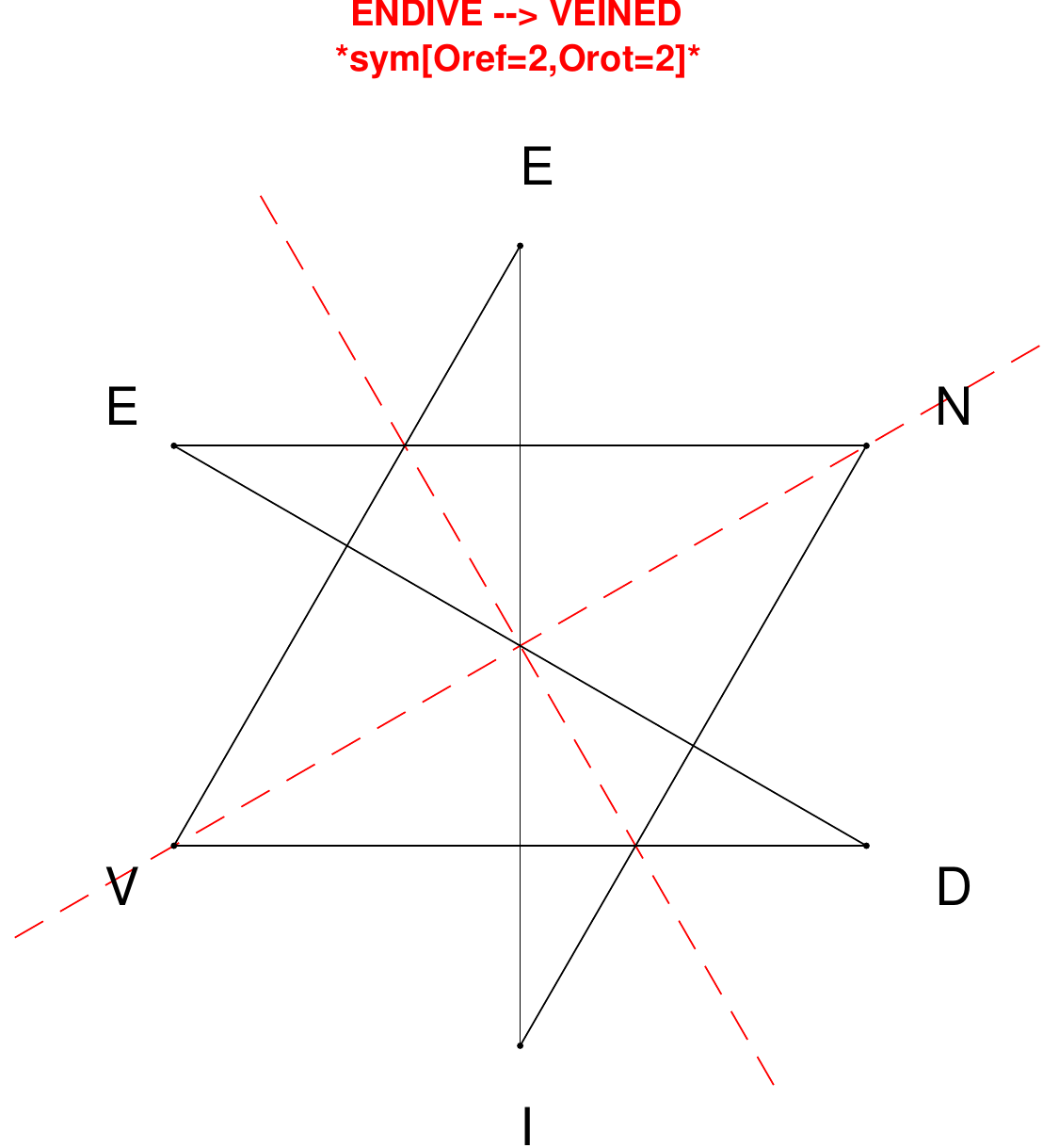}
\end{subfigure}
\end{figure}

\begin{figure}[H]
\centering
\begin{subfigure}[T]{0.19\textwidth}
\centering
\includegraphics[width=\textwidth]{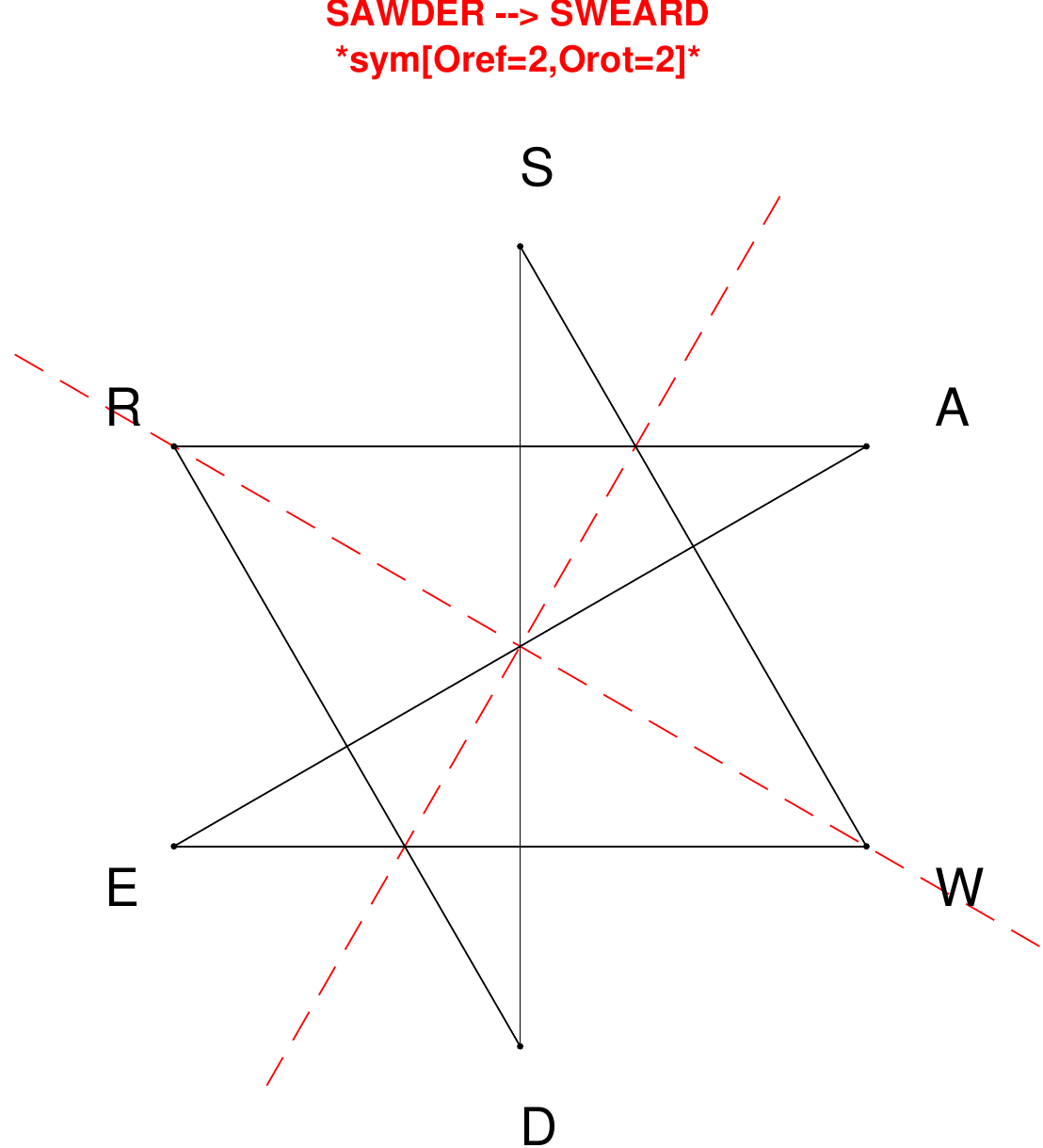}
\end{subfigure}
\hfill
\begin{subfigure}[T]{0.19\textwidth}
\centering
\includegraphics[width=\textwidth]{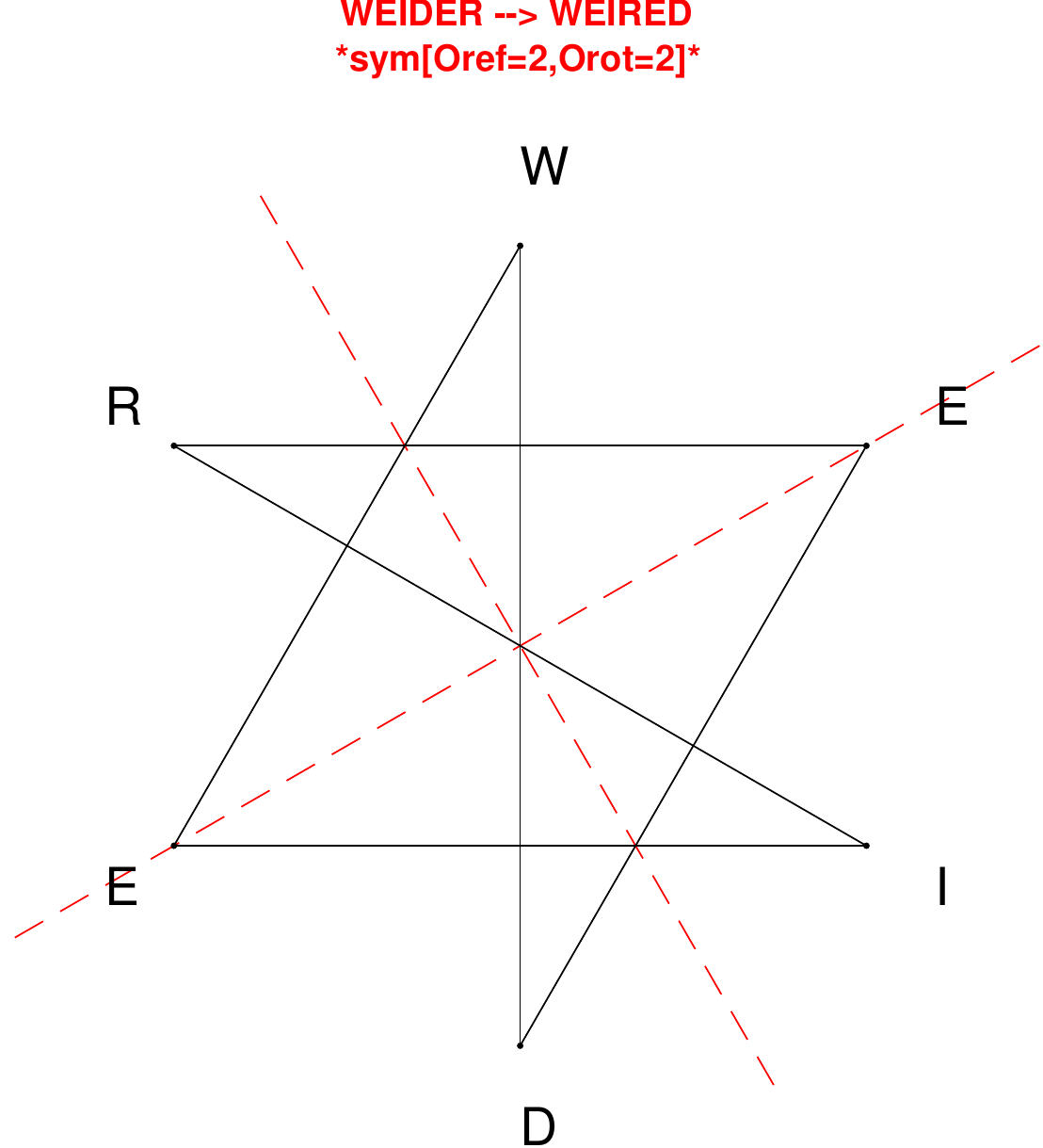}
\end{subfigure}
\hfill
\begin{subfigure}[T]{0.19\textwidth}
\centering
\includegraphics[width=\textwidth]{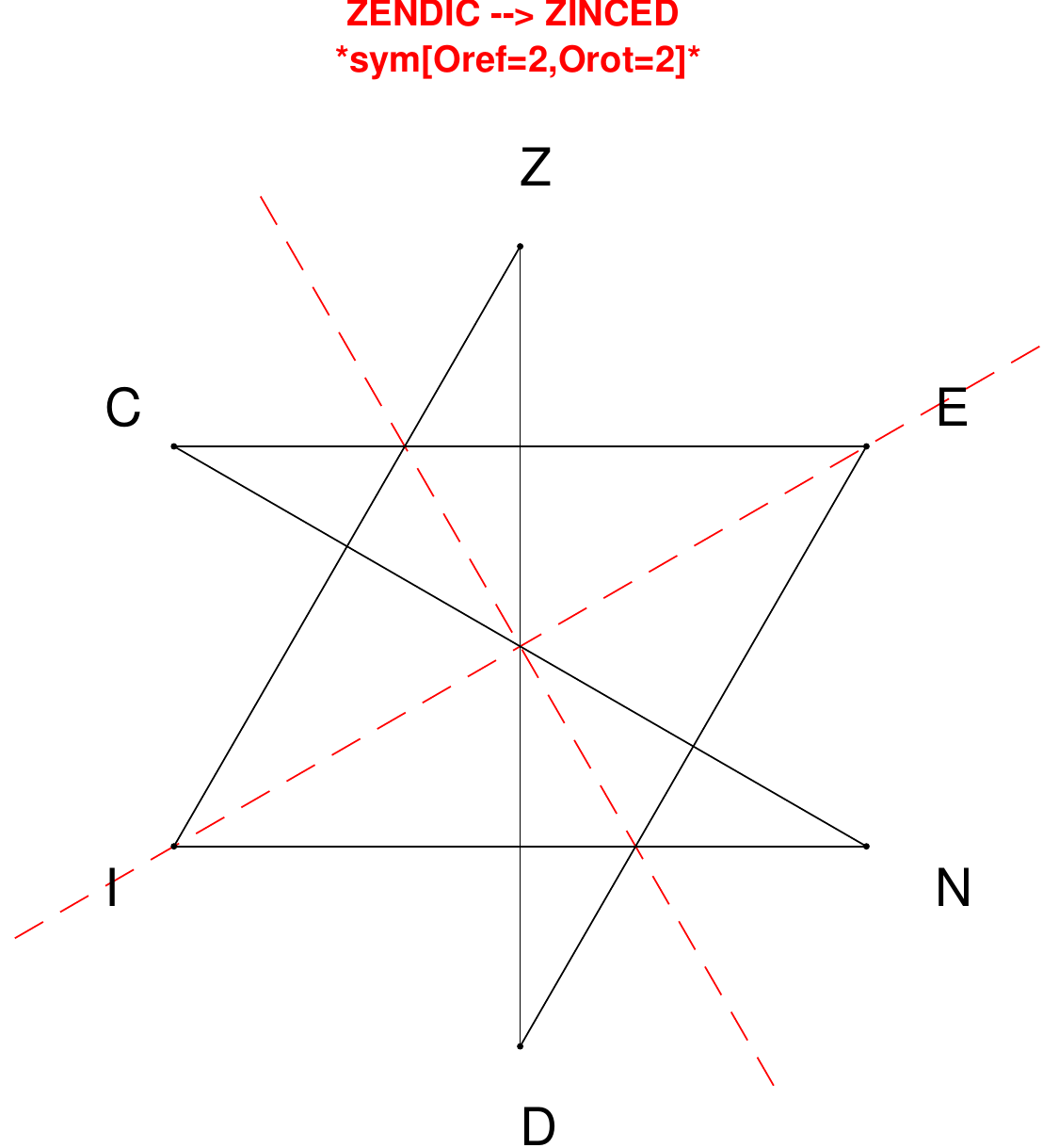}
\end{subfigure}
\hfill
\begin{subfigure}[T]{0.19\textwidth}
\centering
\includegraphics[width=\textwidth]{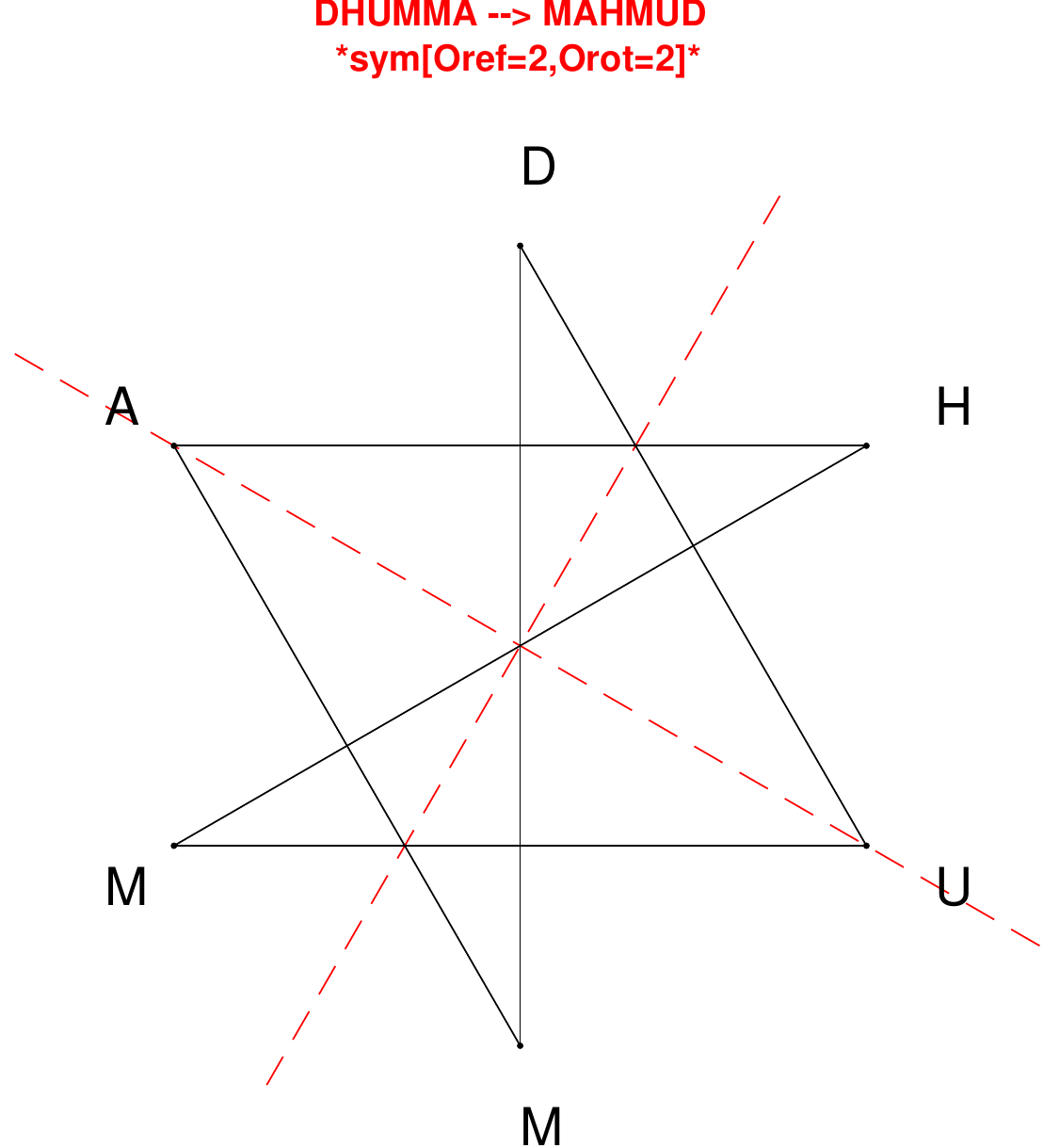}
\end{subfigure}
\hfill
\begin{subfigure}[T]{0.19\textwidth}
\centering
\includegraphics[width=\textwidth]{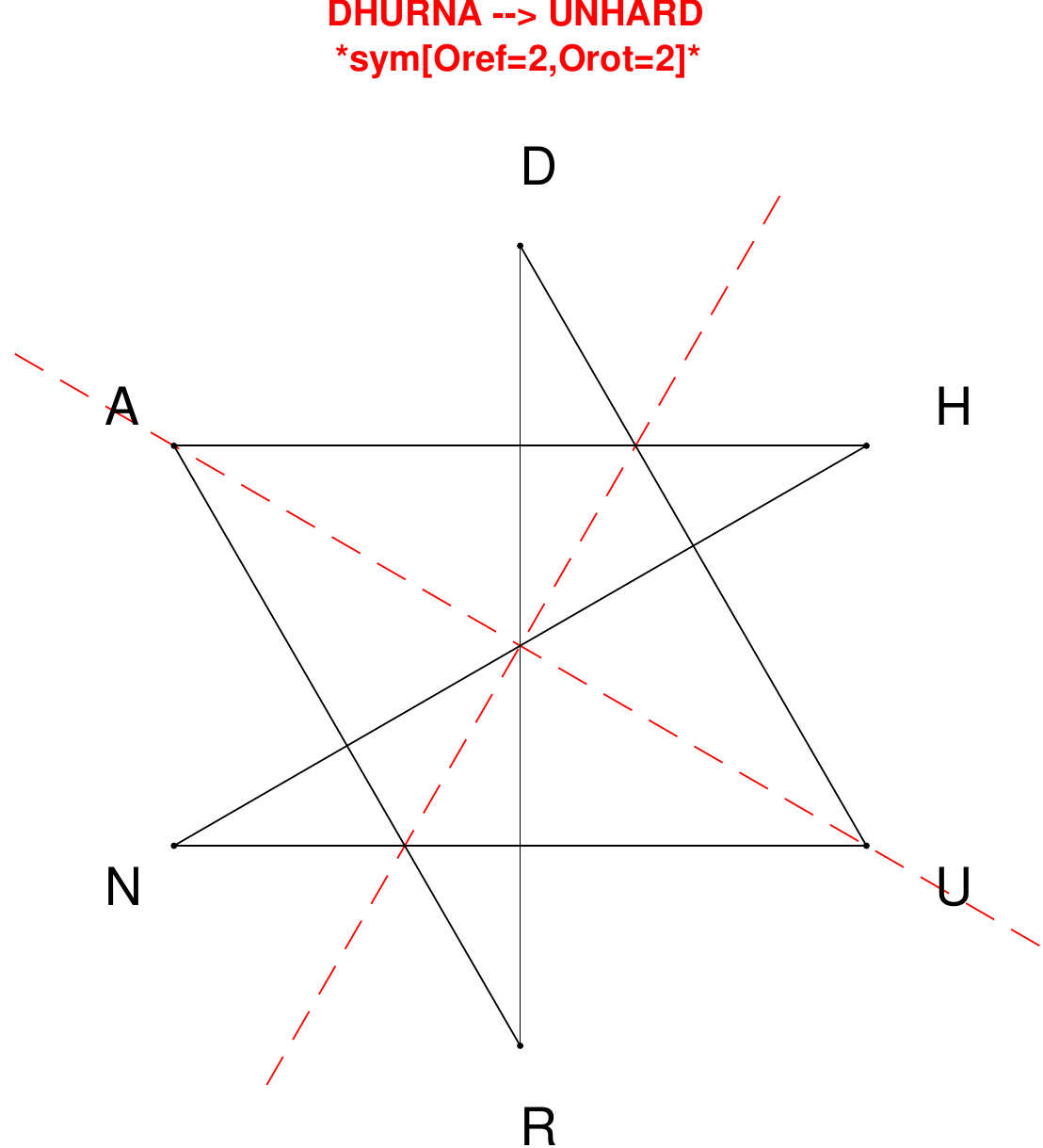}
\end{subfigure}
\end{figure}

\begin{figure}[H]
\centering
\begin{subfigure}[T]{0.19\textwidth}
\centering
\includegraphics[width=\textwidth]{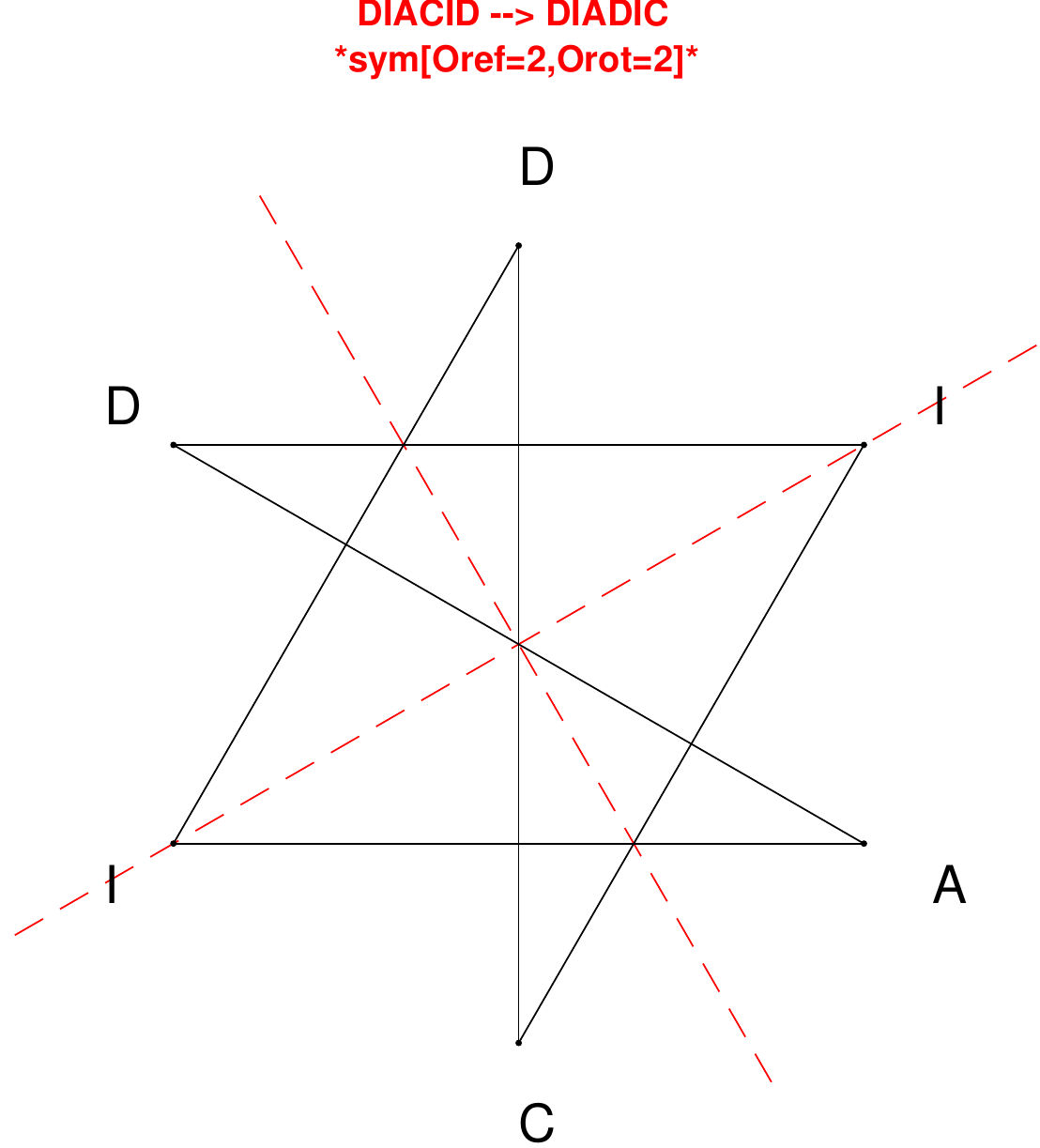}
\end{subfigure}
\hfill
\begin{subfigure}[T]{0.19\textwidth}
\centering
\includegraphics[width=\textwidth]{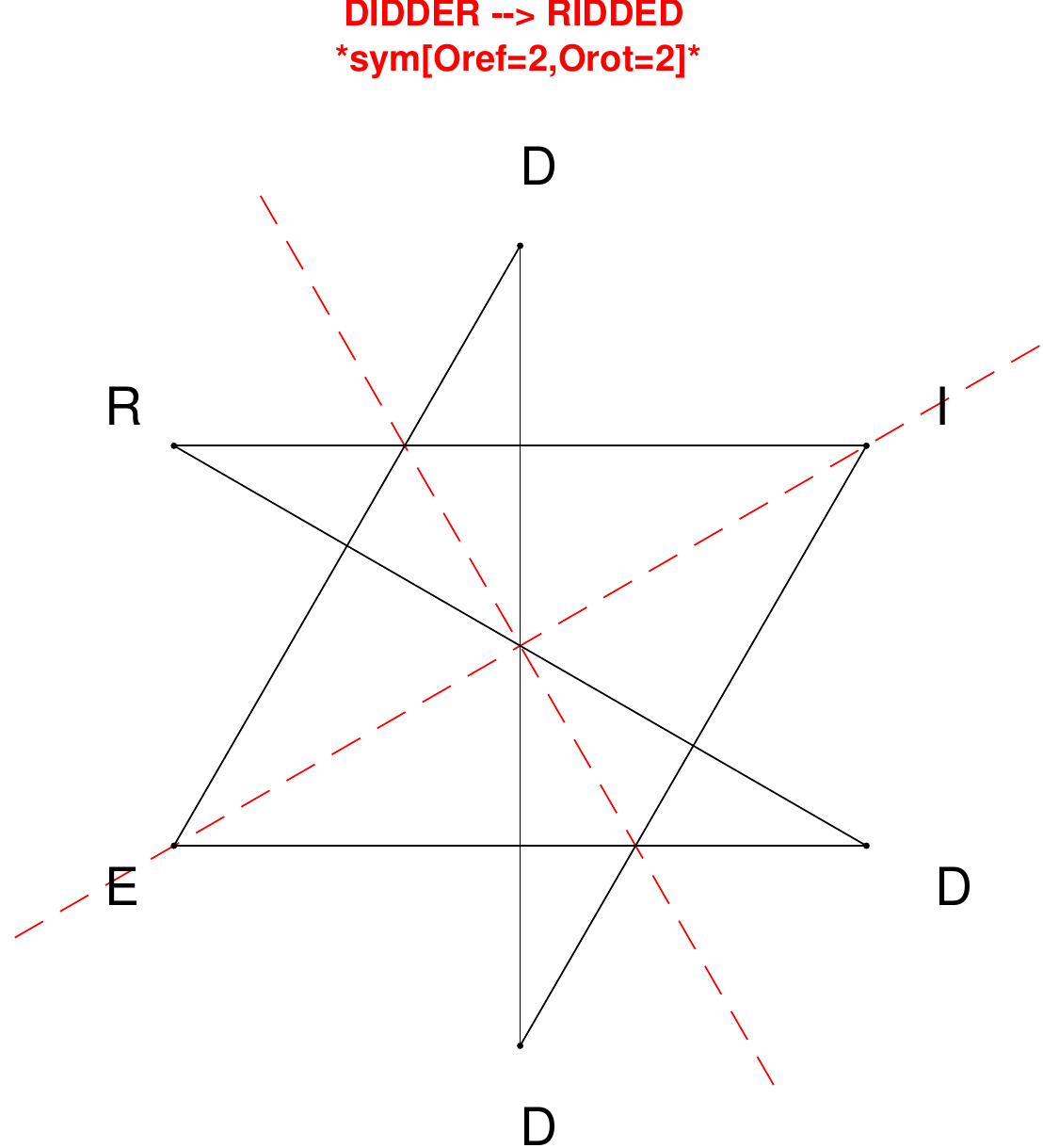}
\end{subfigure}
\hfill
\begin{subfigure}[T]{0.19\textwidth}
\centering
\includegraphics[width=\textwidth]{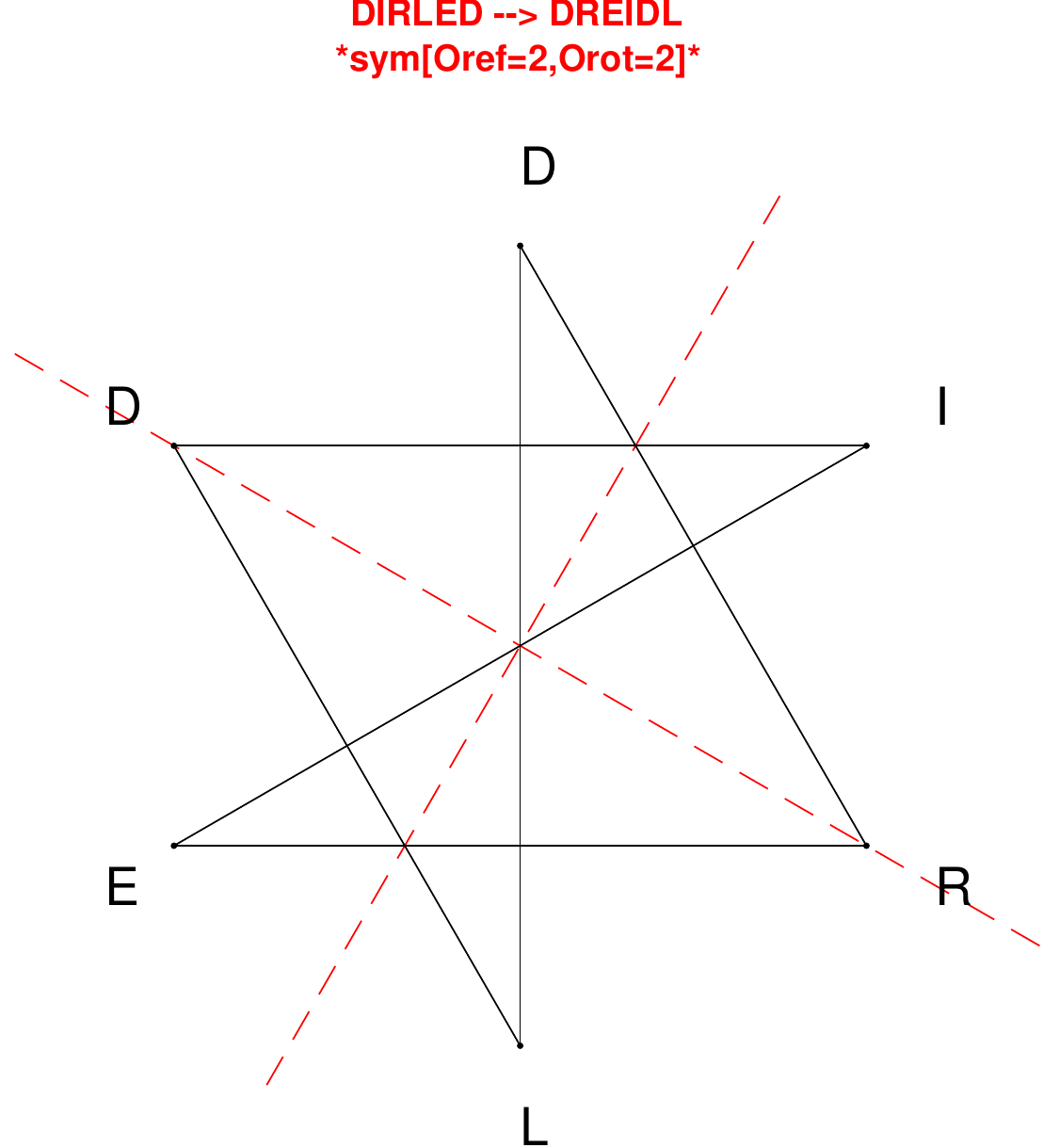}
\end{subfigure}
\hfill
\begin{subfigure}[T]{0.19\textwidth}
\centering
\includegraphics[width=\textwidth]{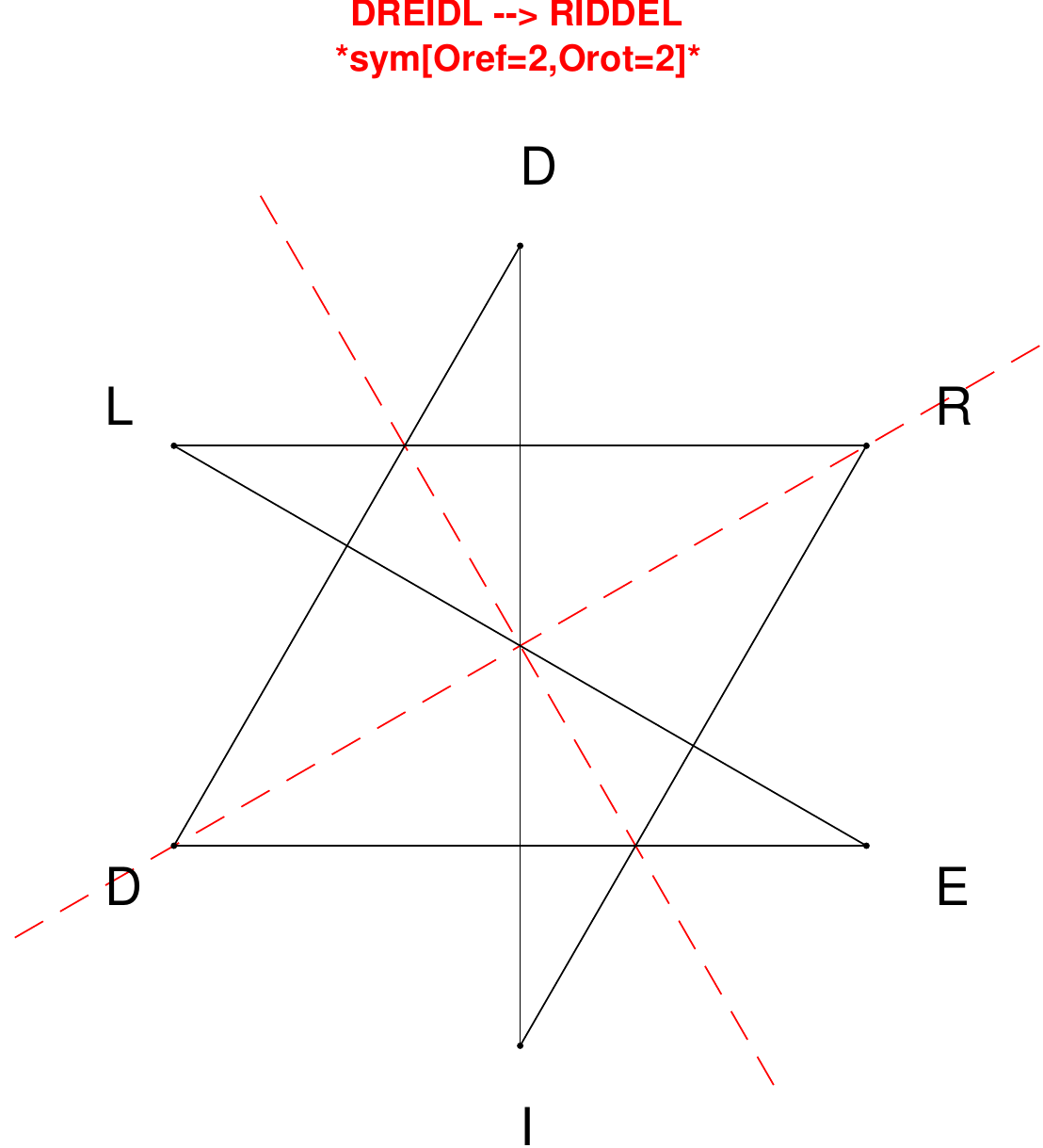}
\end{subfigure}
\hfill
\begin{subfigure}[T]{0.19\textwidth}
\centering
\includegraphics[width=\textwidth]{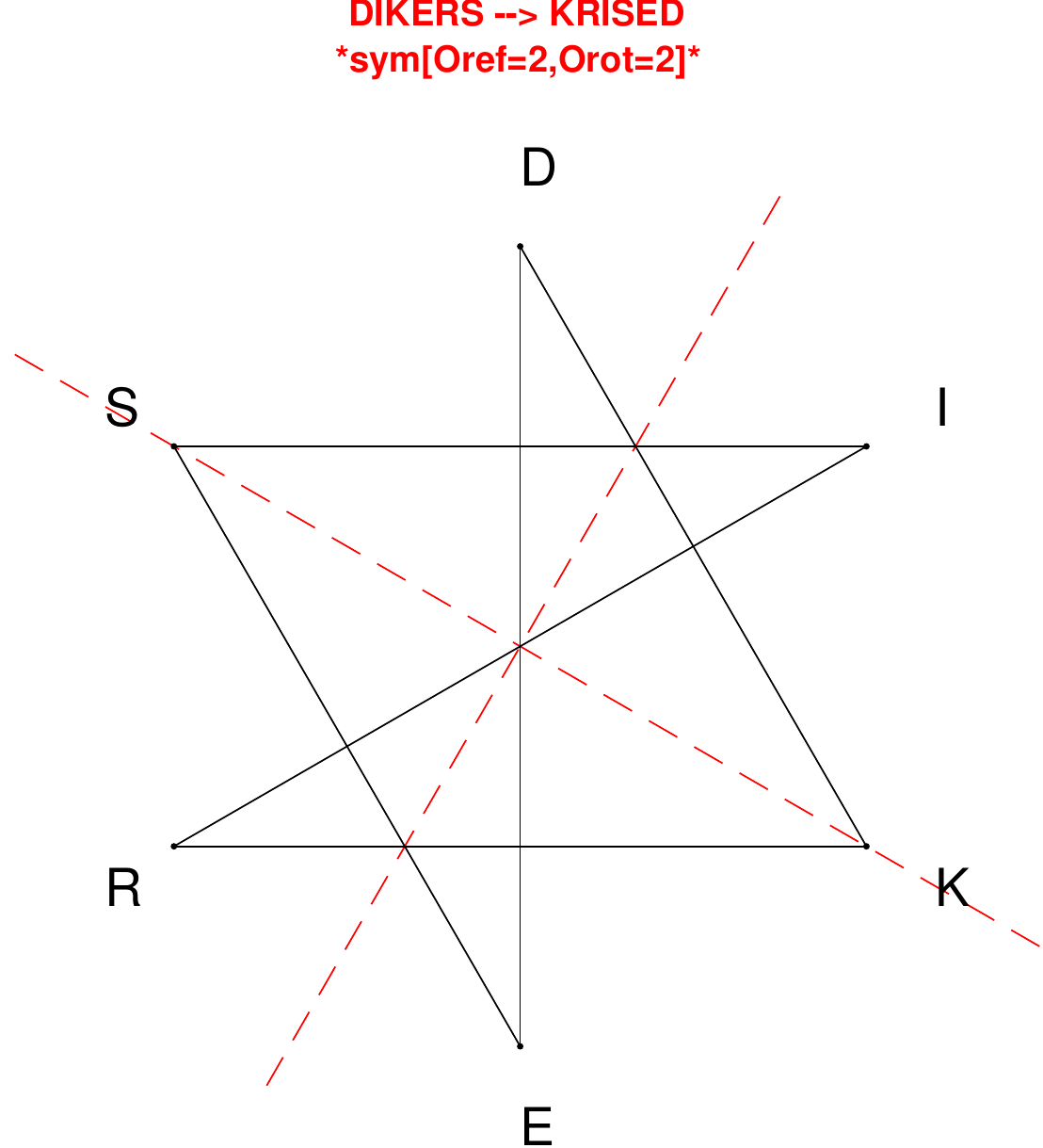}
\end{subfigure}
\end{figure}

\begin{figure}[H]
\centering
\begin{subfigure}[T]{0.19\textwidth}
\centering
\includegraphics[width=\textwidth]{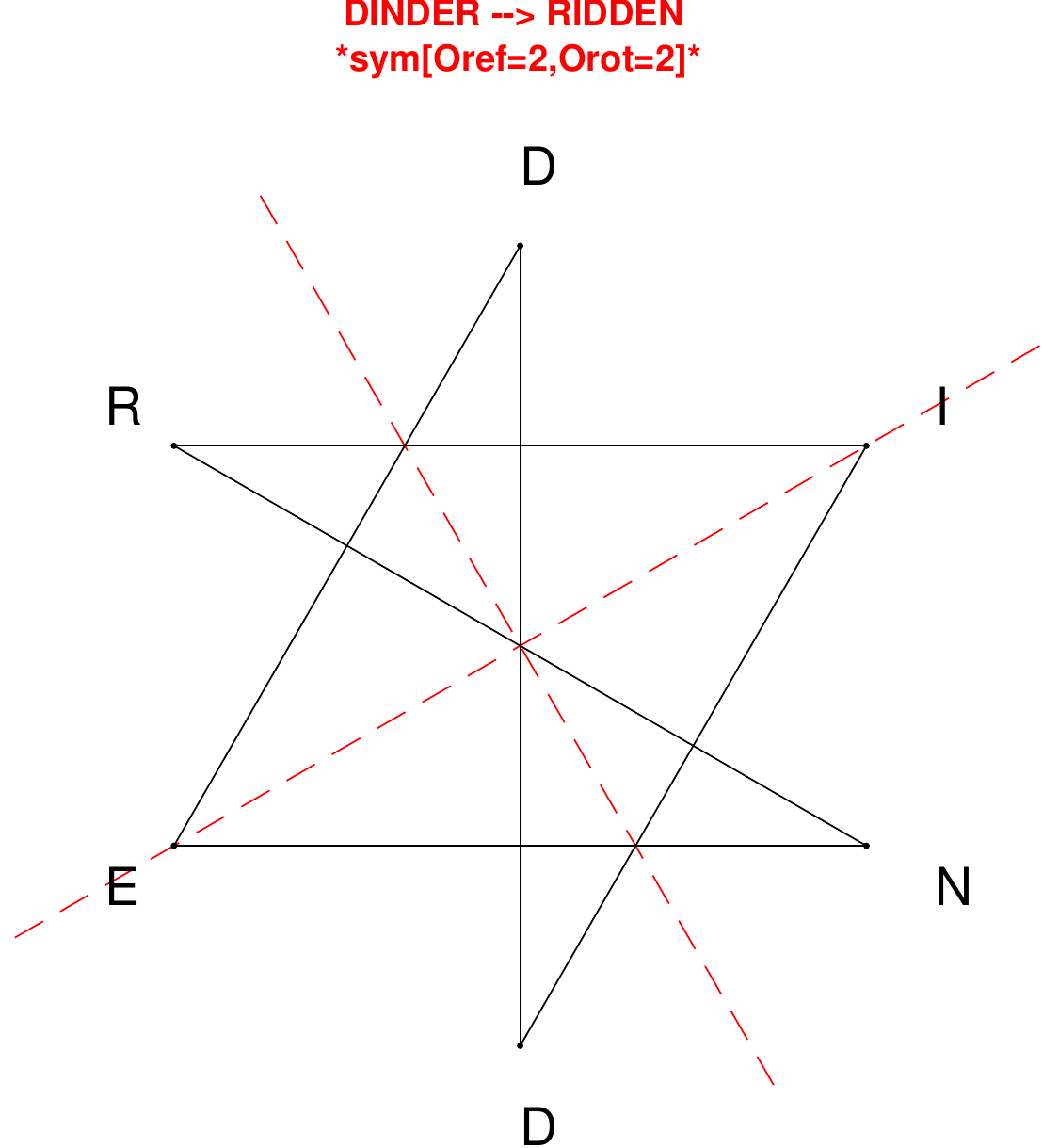}
\end{subfigure}
\hfill
\begin{subfigure}[T]{0.19\textwidth}
\centering
\includegraphics[width=\textwidth]{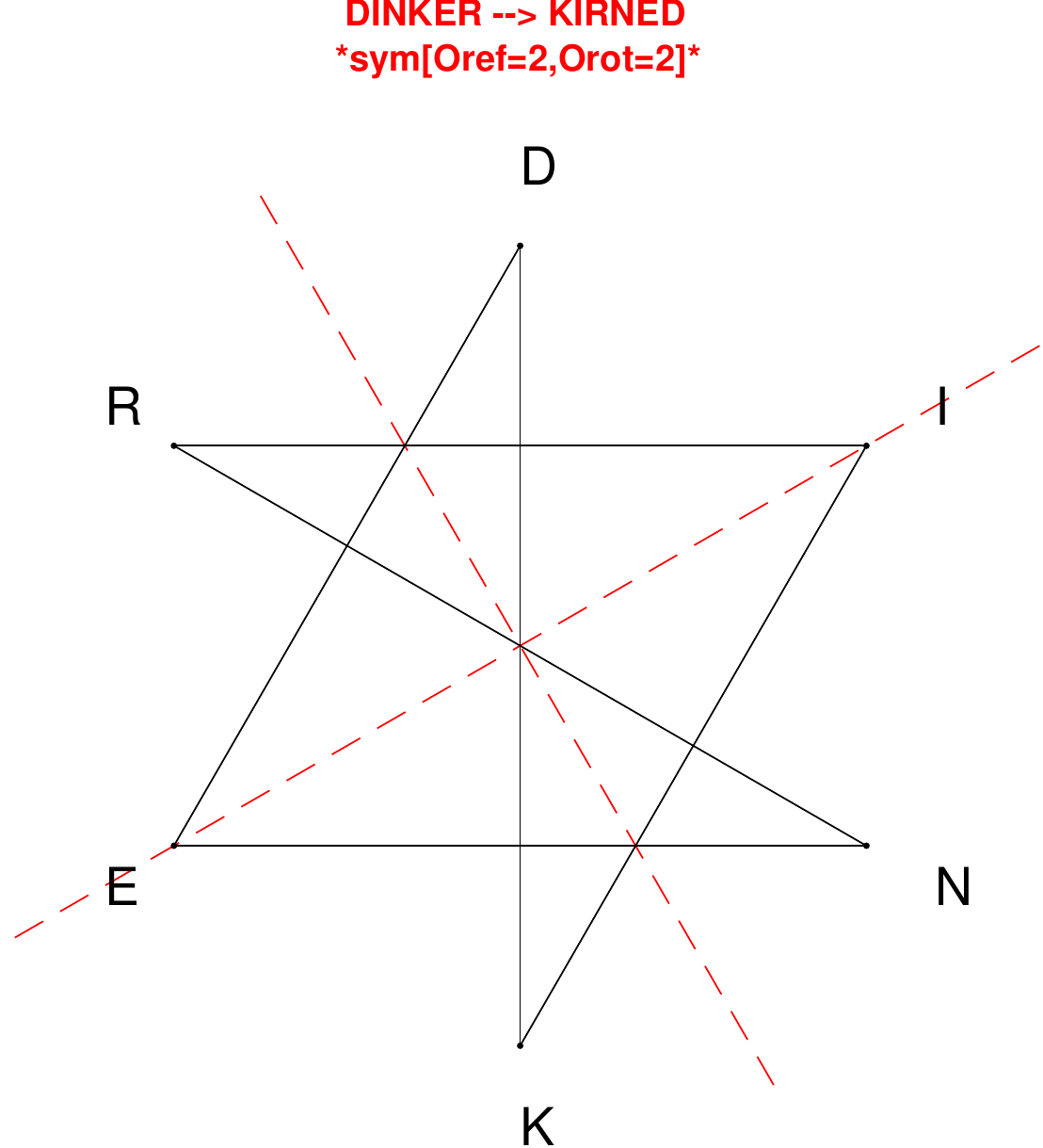}
\end{subfigure}
\hfill
\begin{subfigure}[T]{0.19\textwidth}
\centering
\includegraphics[width=\textwidth]{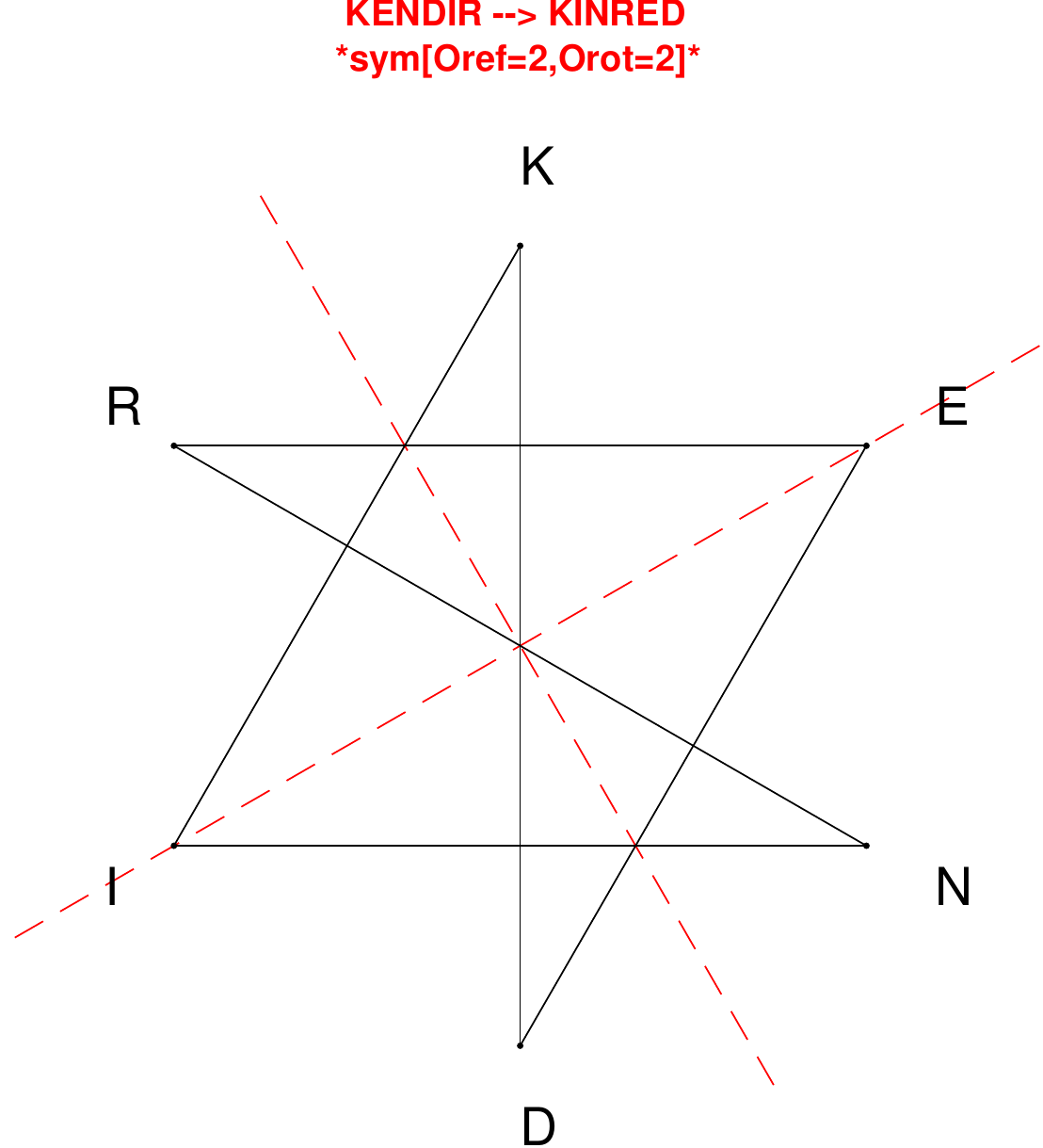}
\end{subfigure}
\hfill
\begin{subfigure}[T]{0.19\textwidth}
\centering
\includegraphics[width=\textwidth]{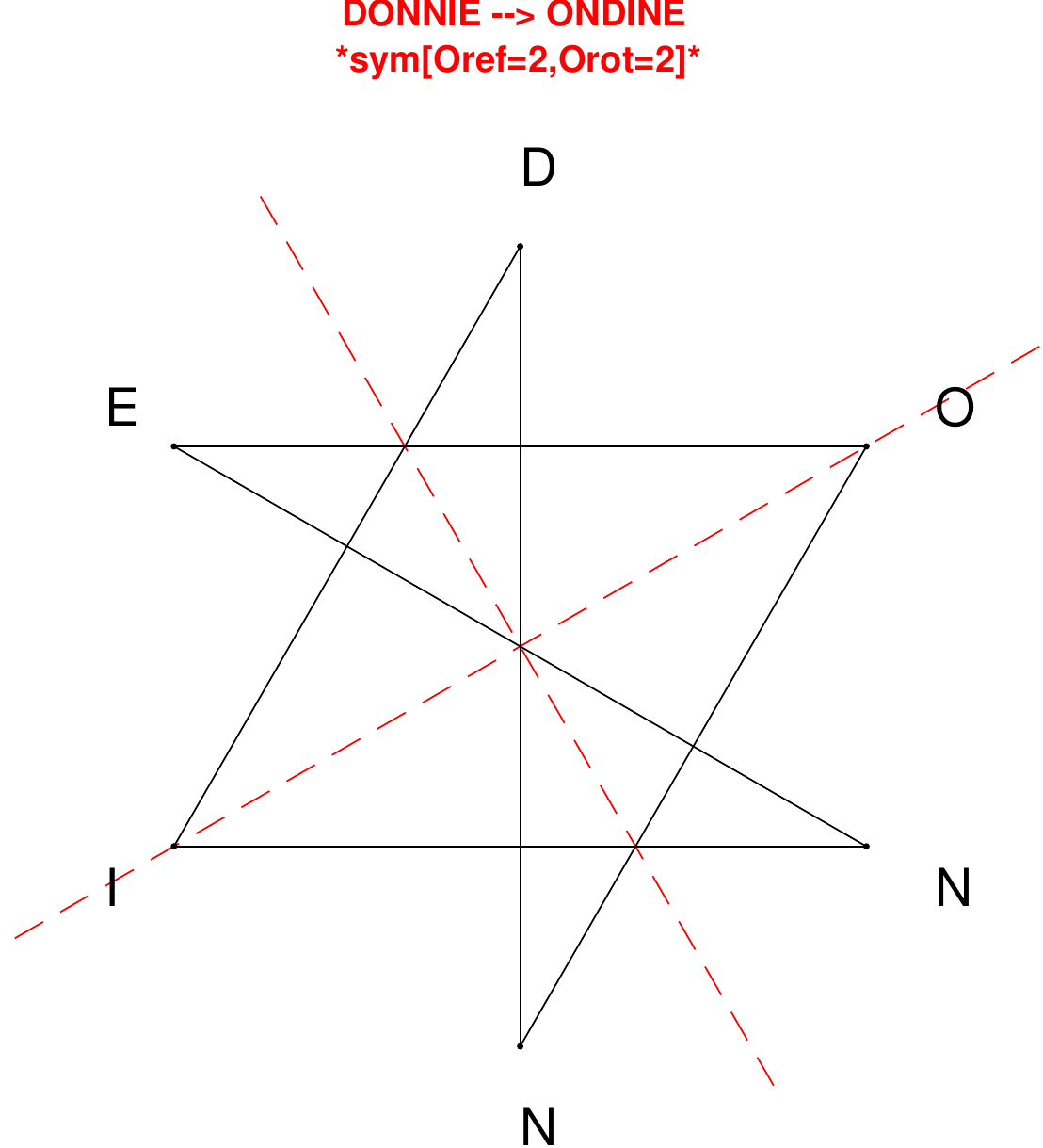}
\end{subfigure}
\hfill
\begin{subfigure}[T]{0.19\textwidth}
\centering
\includegraphics[width=\textwidth]{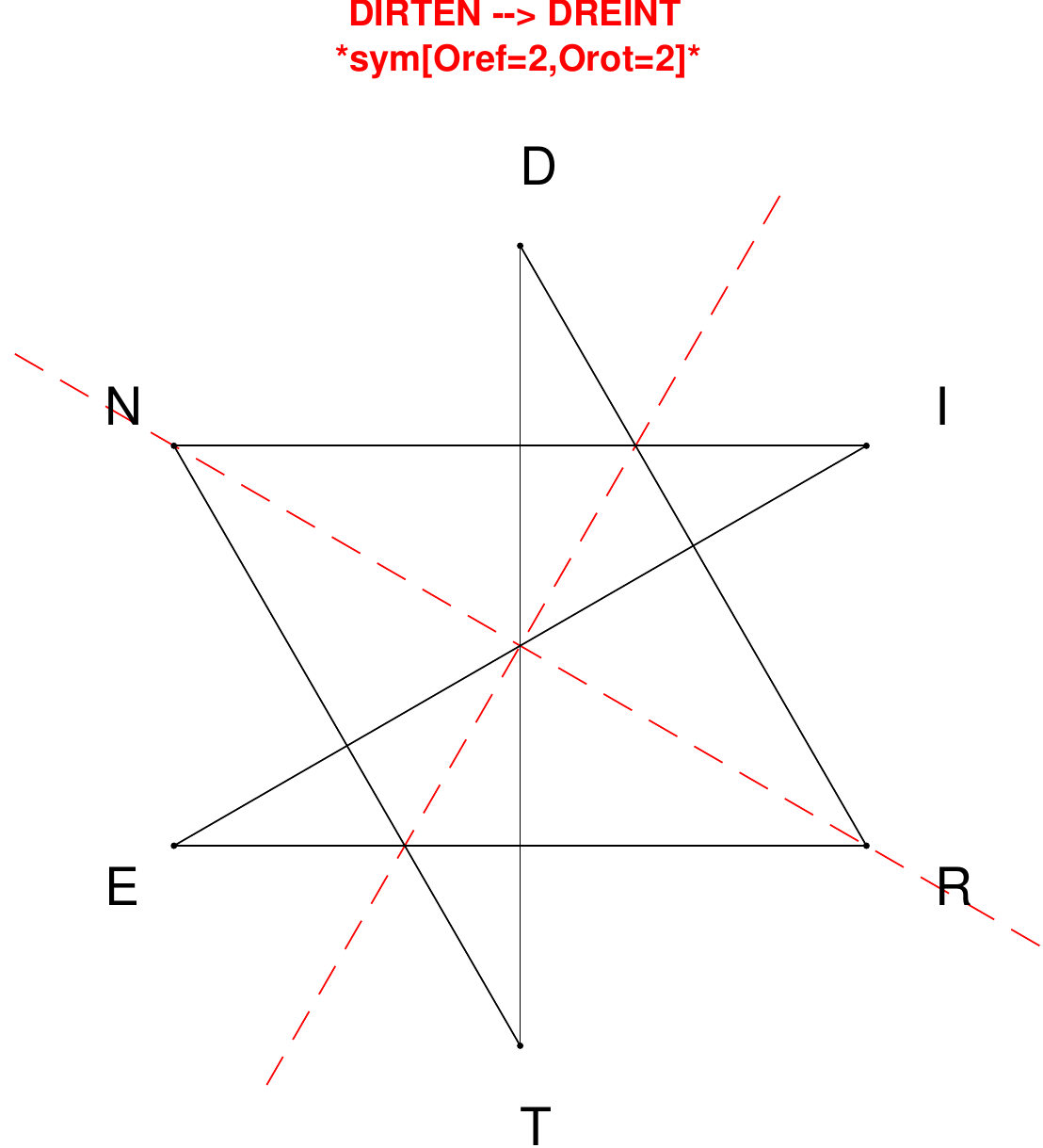}
\end{subfigure}
\end{figure}

\begin{figure}[H]
\centering
\begin{subfigure}[T]{0.19\textwidth}
\centering
\includegraphics[width=\textwidth]{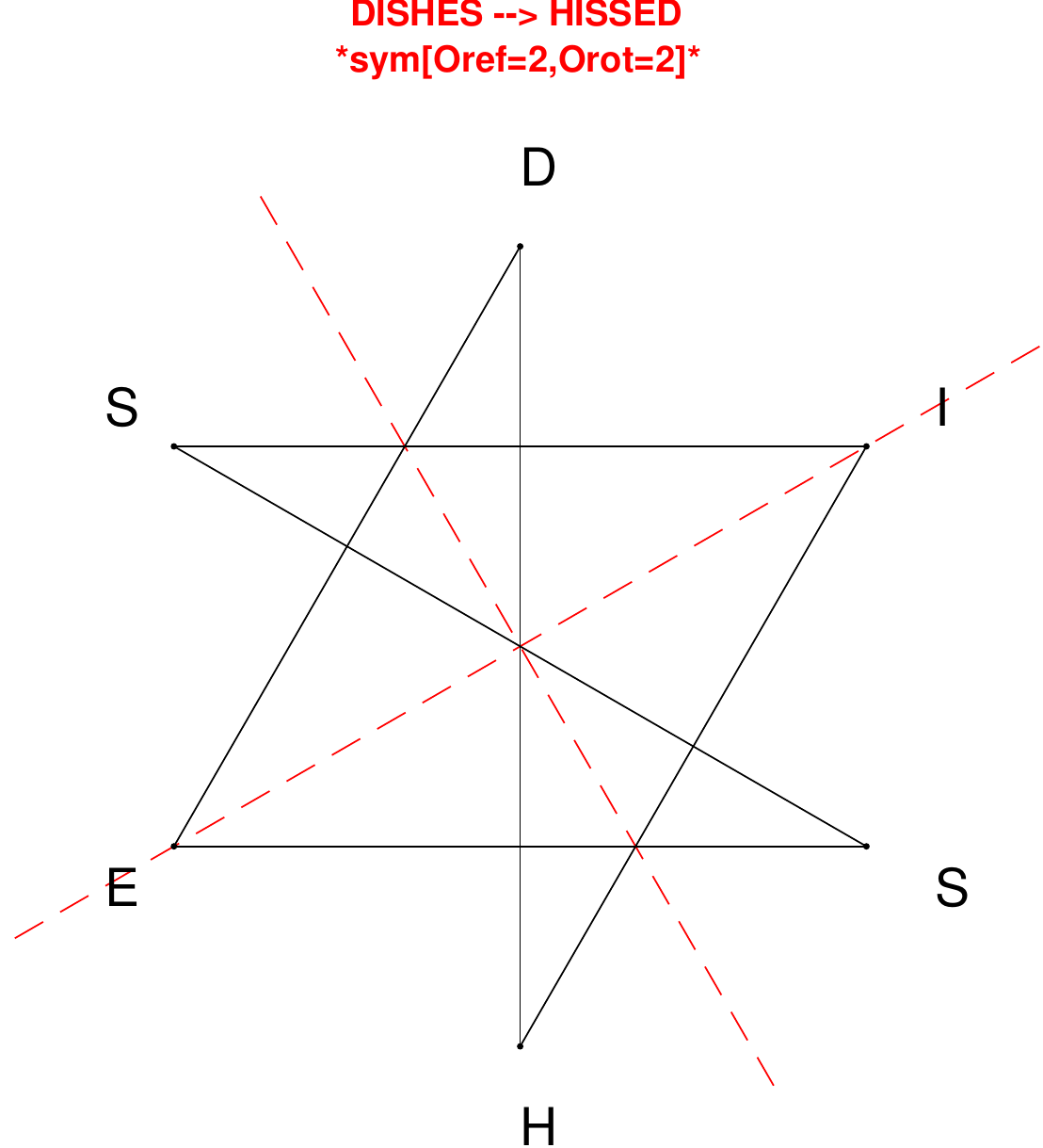}
\end{subfigure}
\hfill
\begin{subfigure}[T]{0.19\textwidth}
\centering
\includegraphics[width=\textwidth]{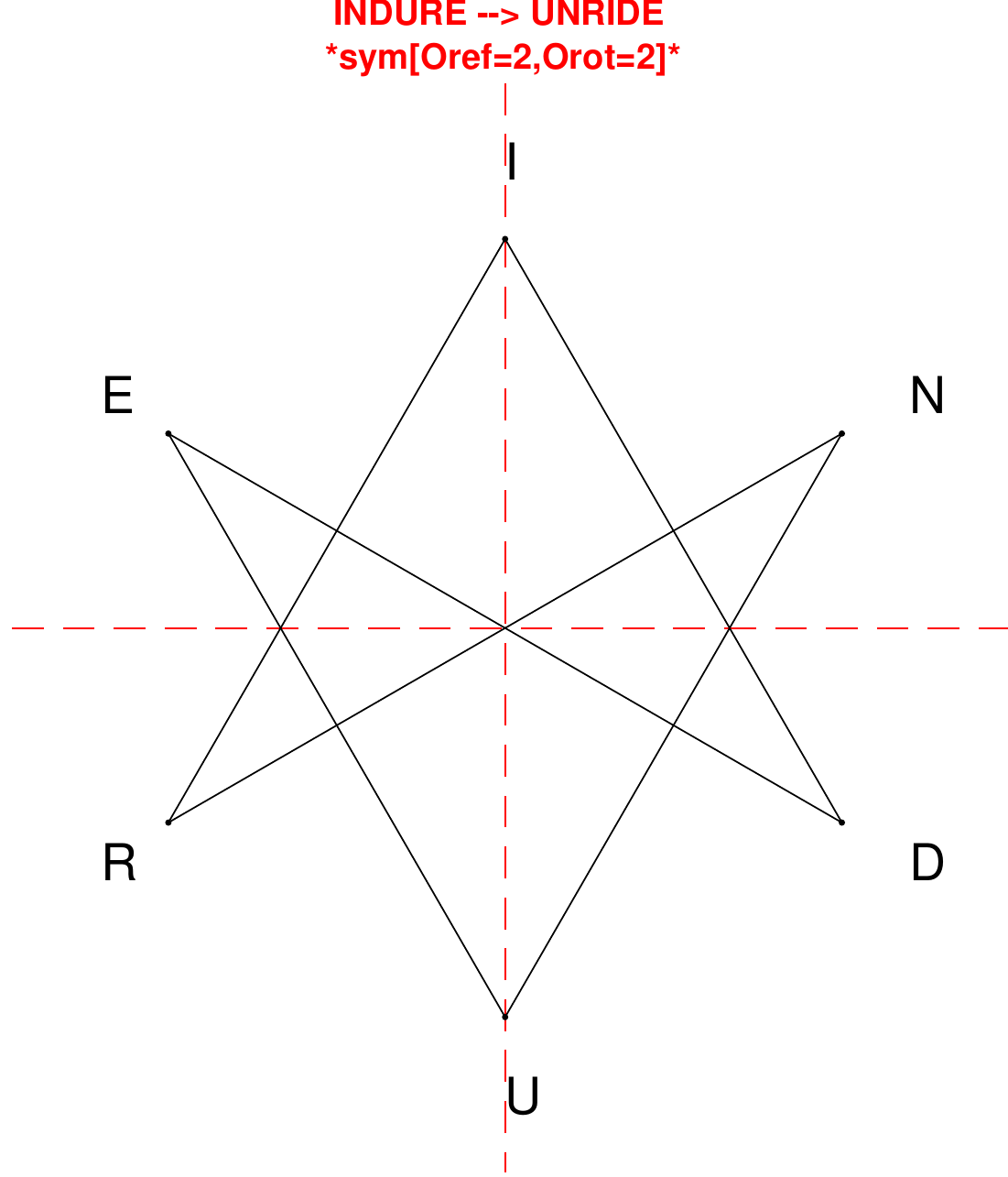}
\end{subfigure}
\hfill
\begin{subfigure}[T]{0.19\textwidth}
\centering
\includegraphics[width=\textwidth]{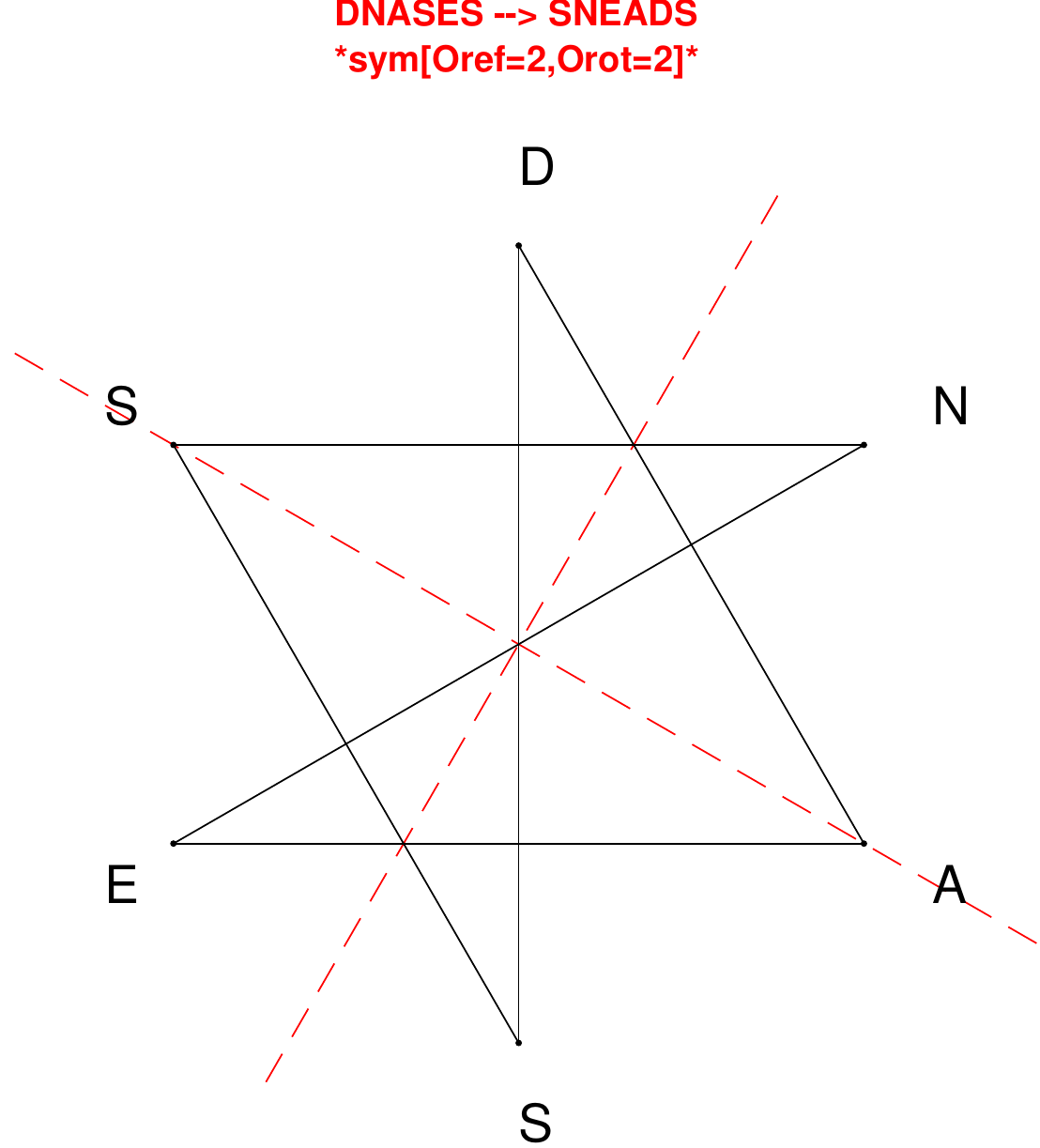}
\end{subfigure}
\hfill
\begin{subfigure}[T]{0.19\textwidth}
\centering
\includegraphics[width=\textwidth]{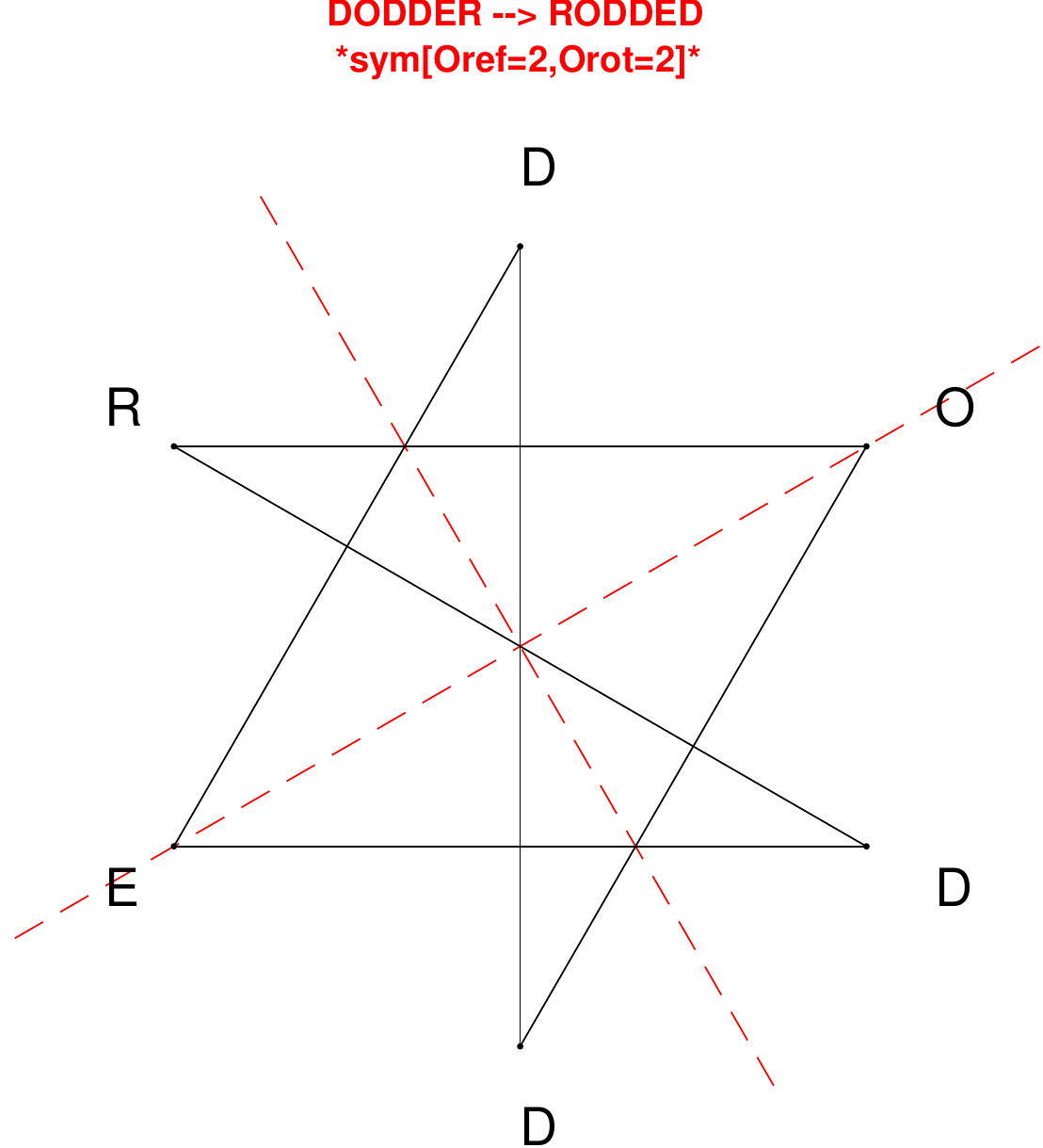}
\end{subfigure}
\hfill
\begin{subfigure}[T]{0.19\textwidth}
\centering
\includegraphics[width=\textwidth]{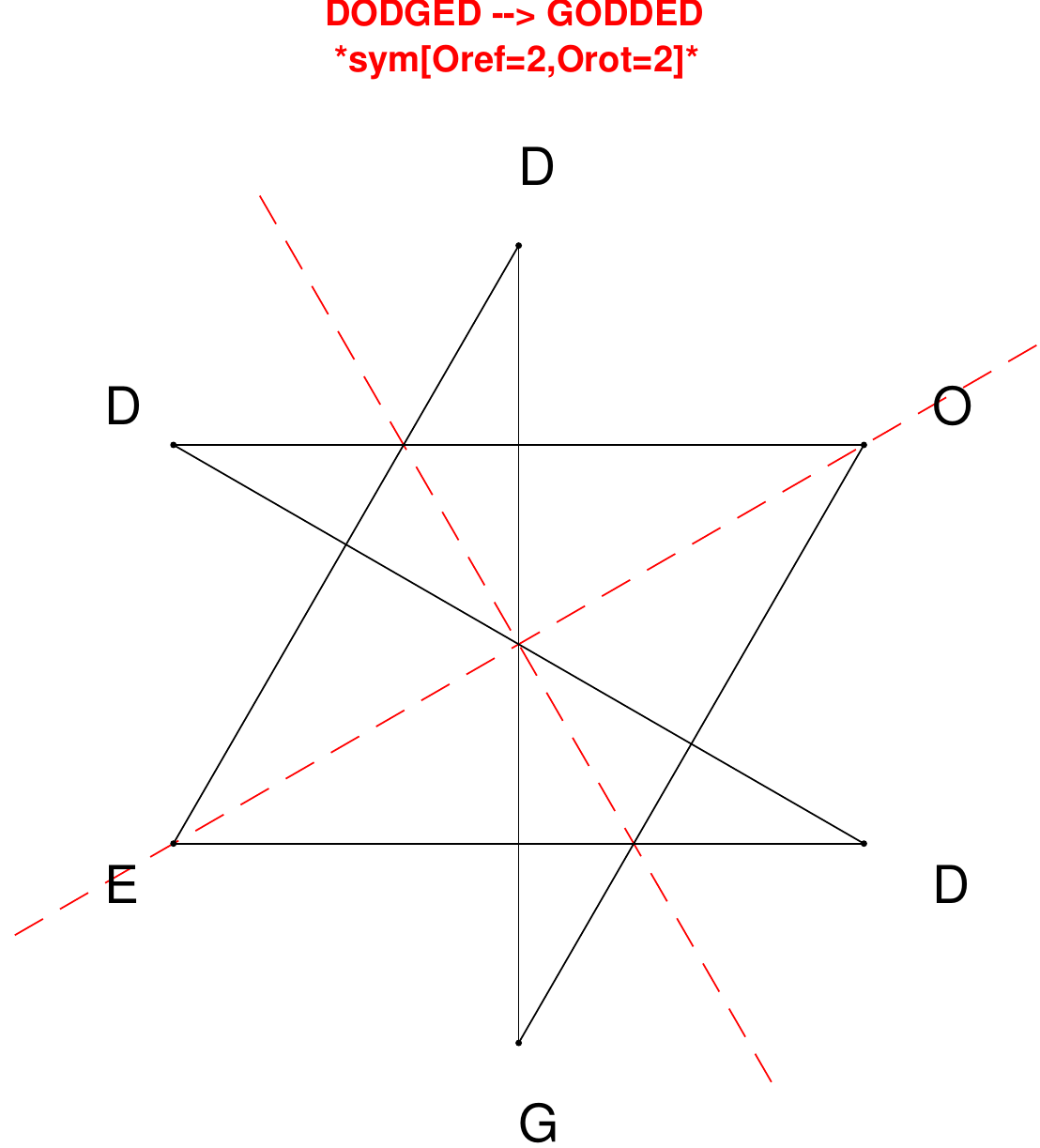}
\end{subfigure}
\end{figure}

\begin{figure}[H]
\centering
\begin{subfigure}[T]{0.19\textwidth}
\centering
\includegraphics[width=\textwidth]{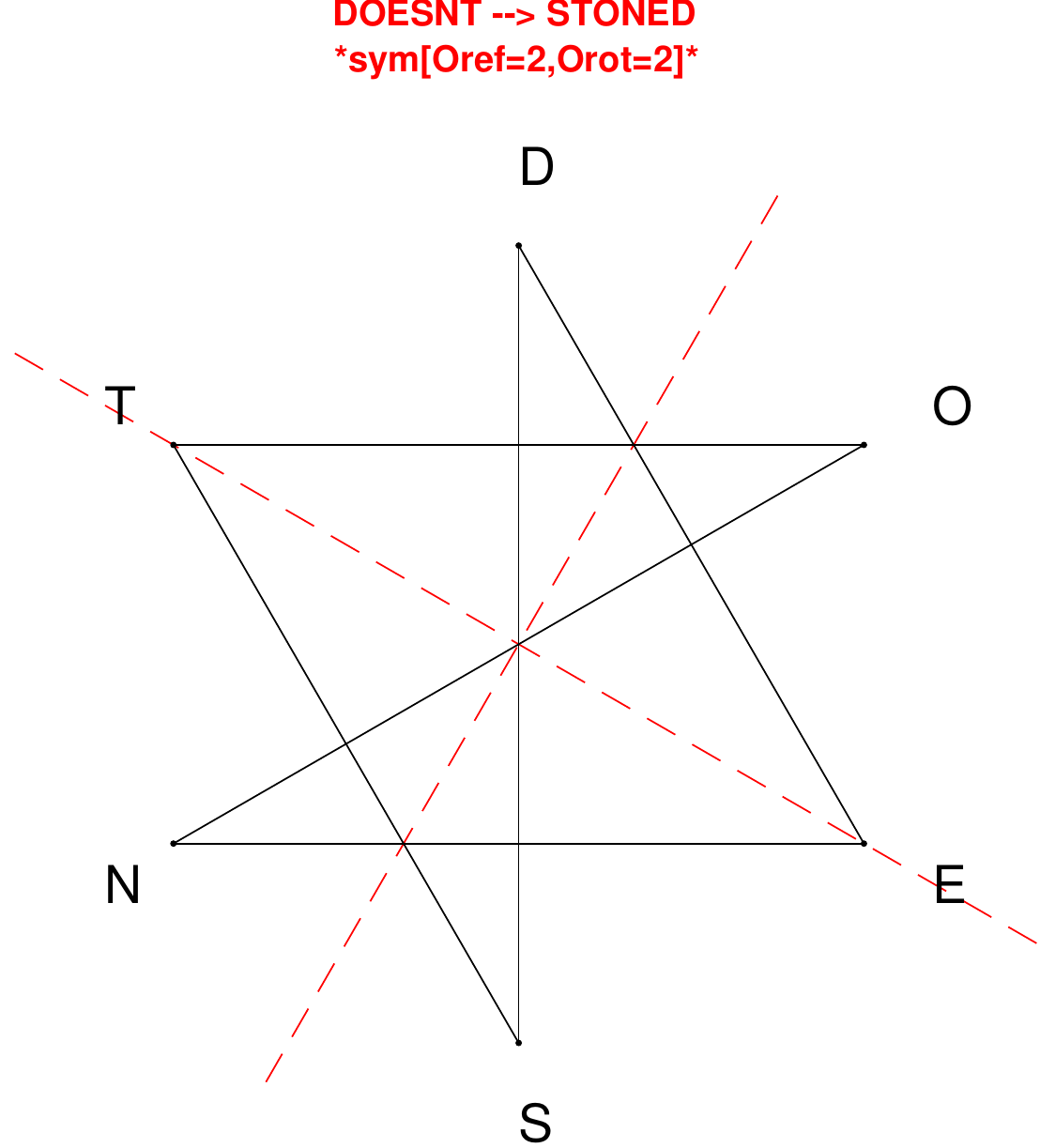}
\end{subfigure}
\hfill
\begin{subfigure}[T]{0.19\textwidth}
\centering
\includegraphics[width=\textwidth]{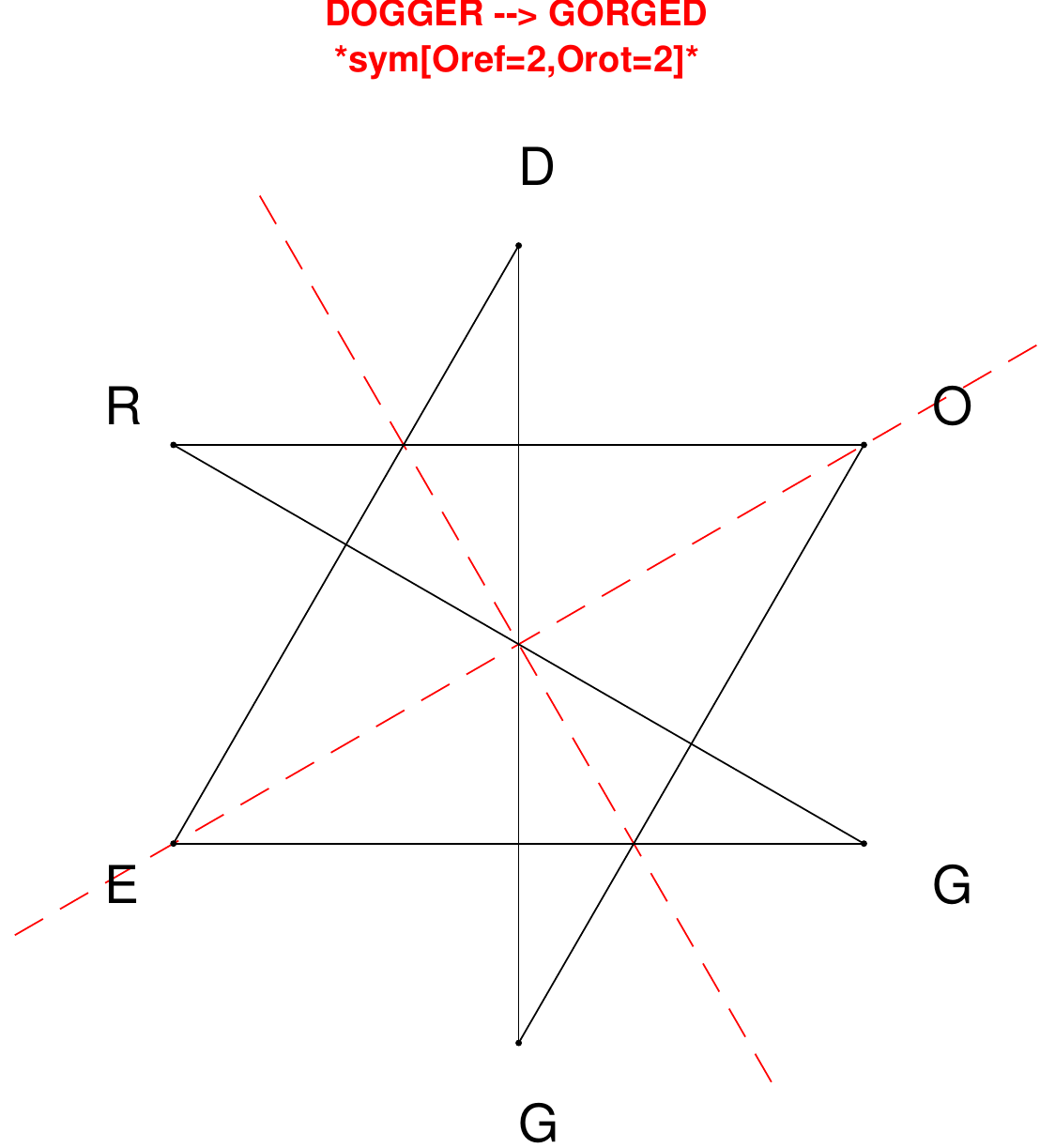}
\end{subfigure}
\hfill
\begin{subfigure}[T]{0.19\textwidth}
\centering
\includegraphics[width=\textwidth]{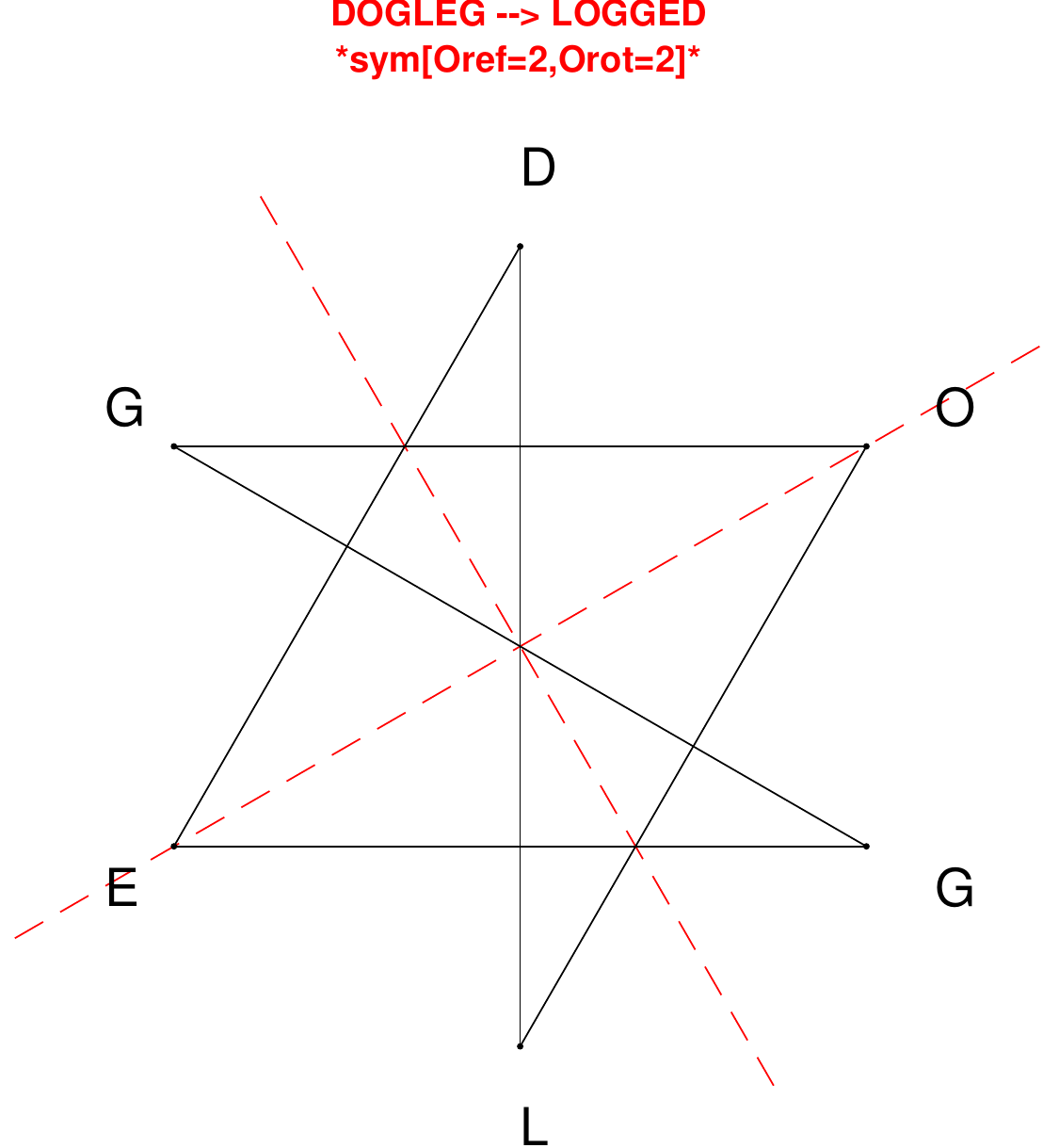}
\end{subfigure}
\hfill
\begin{subfigure}[T]{0.19\textwidth}
\centering
\includegraphics[width=\textwidth]{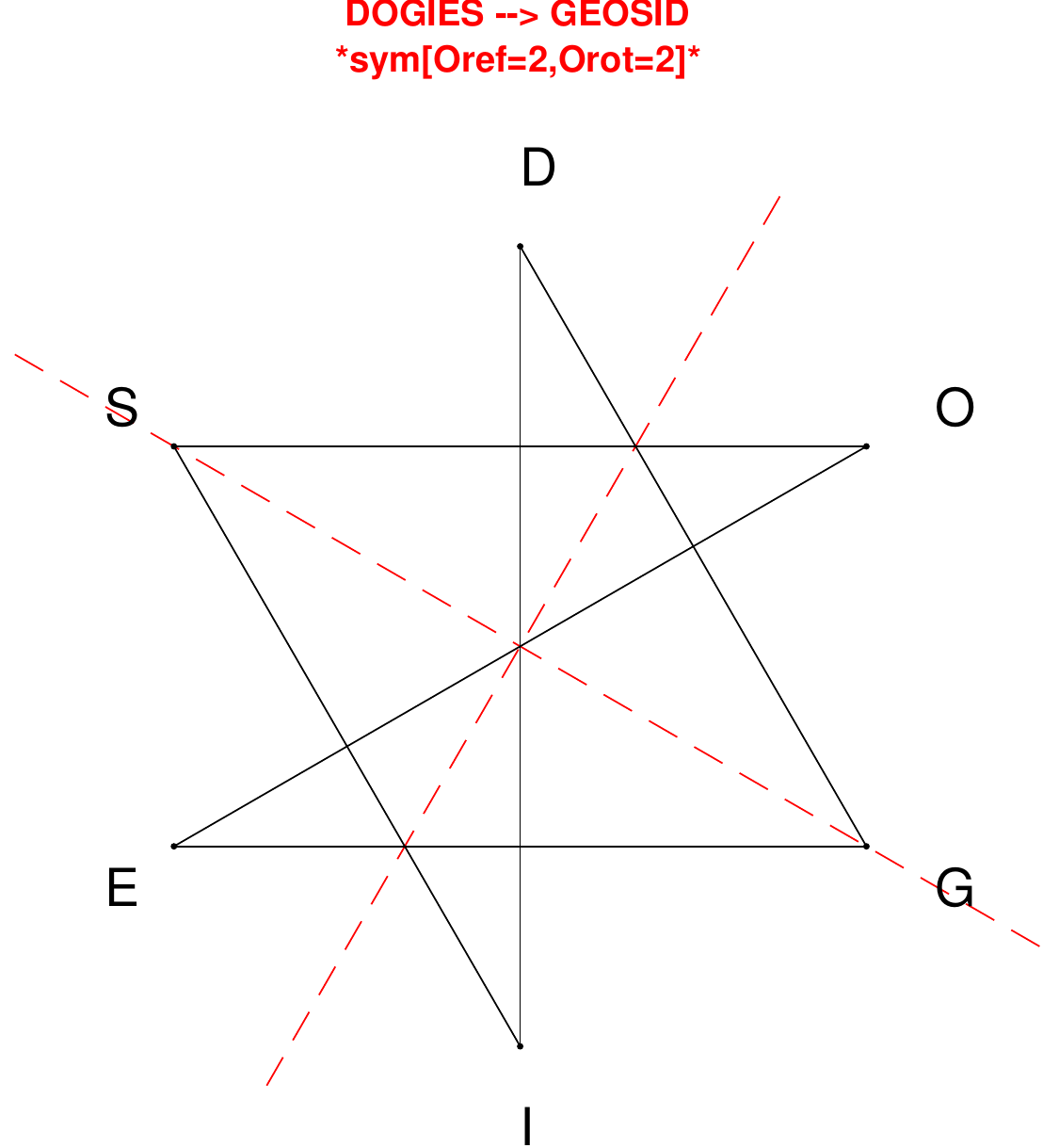}
\end{subfigure}
\hfill
\begin{subfigure}[T]{0.19\textwidth}
\centering
\includegraphics[width=\textwidth]{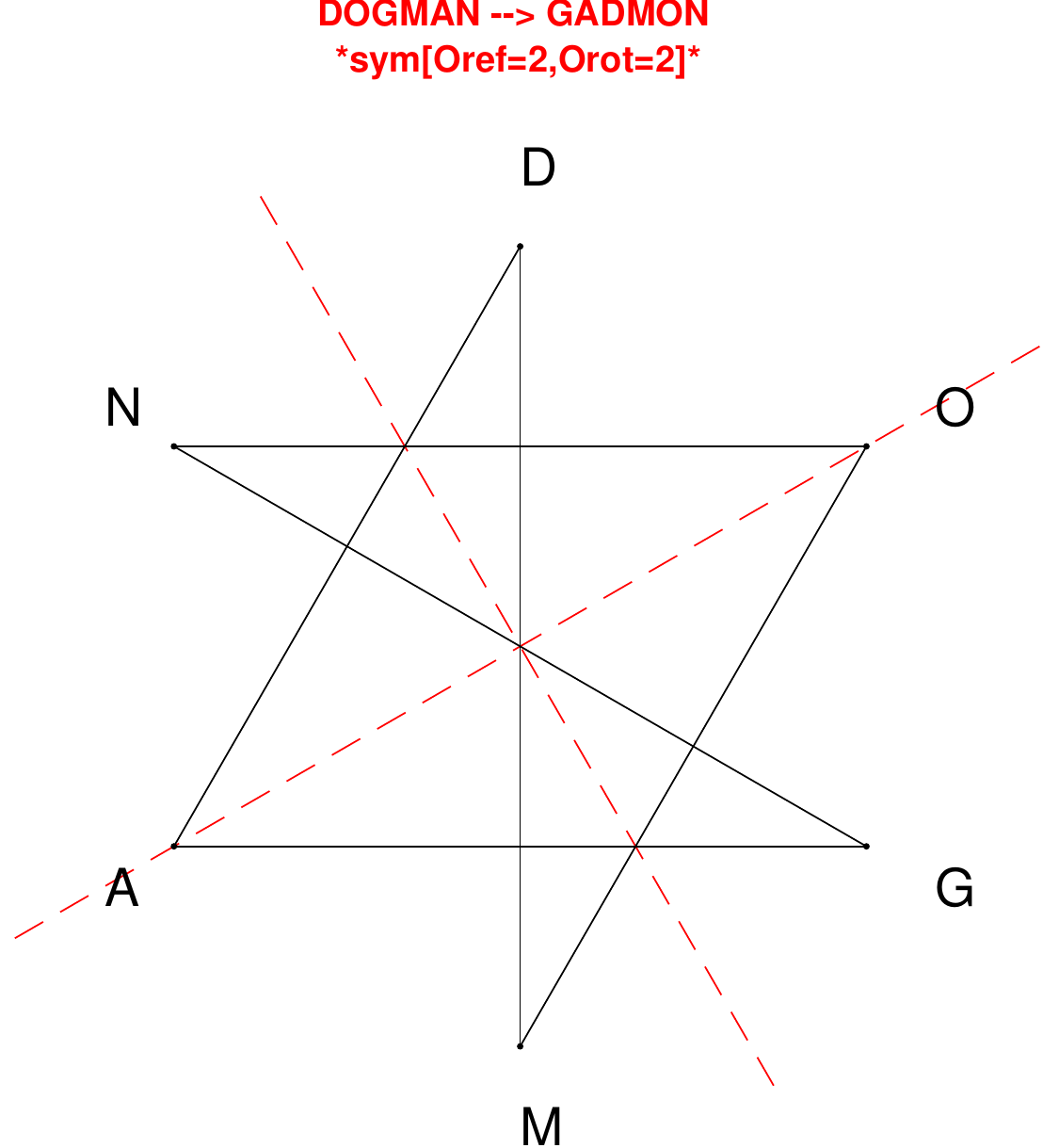}
\end{subfigure}
\end{figure}

\begin{figure}[H]
\centering
\begin{subfigure}[T]{0.19\textwidth}
\centering
\includegraphics[width=\textwidth]{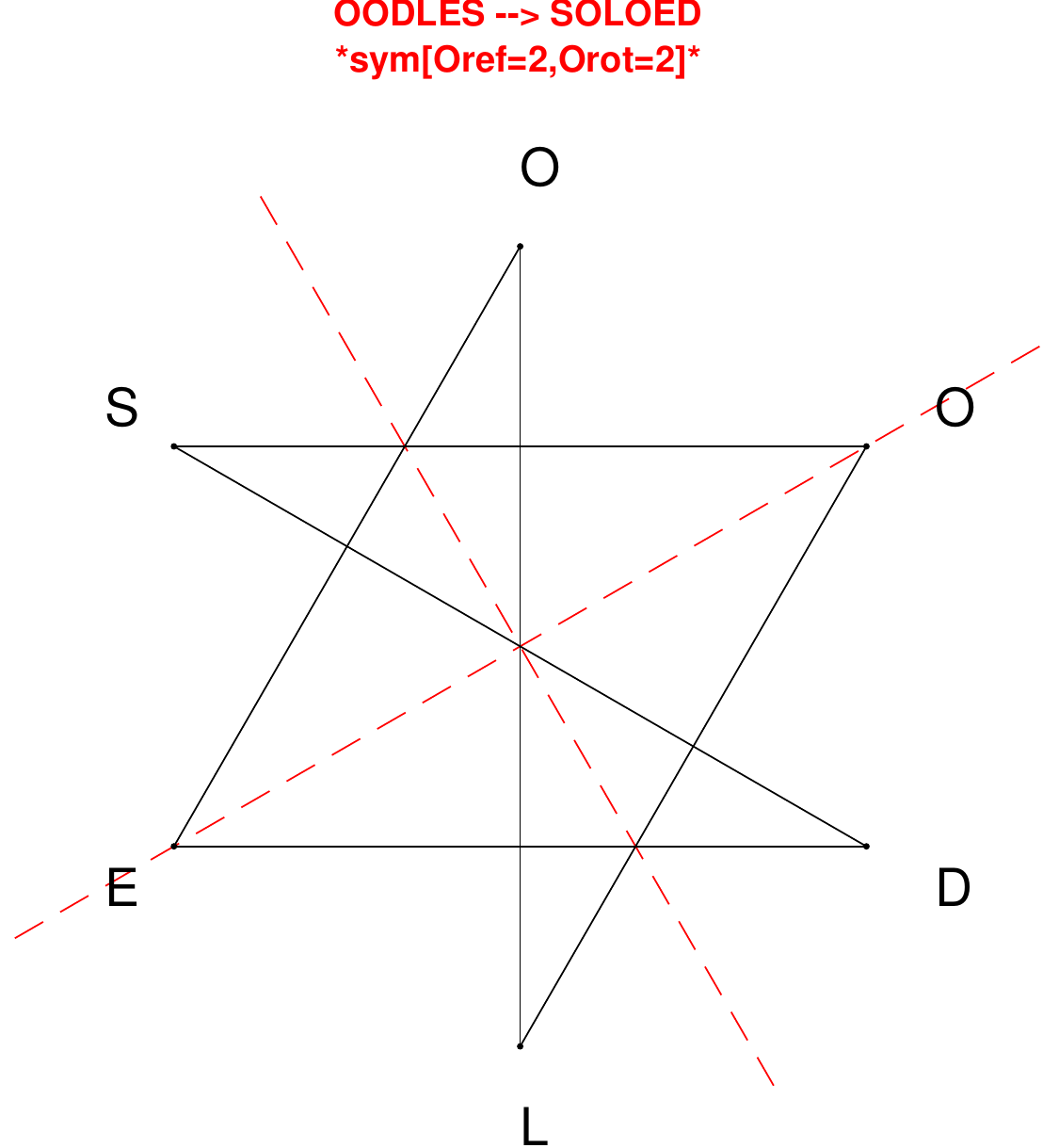}
\end{subfigure}
\hfill
\begin{subfigure}[T]{0.19\textwidth}
\centering
\includegraphics[width=\textwidth]{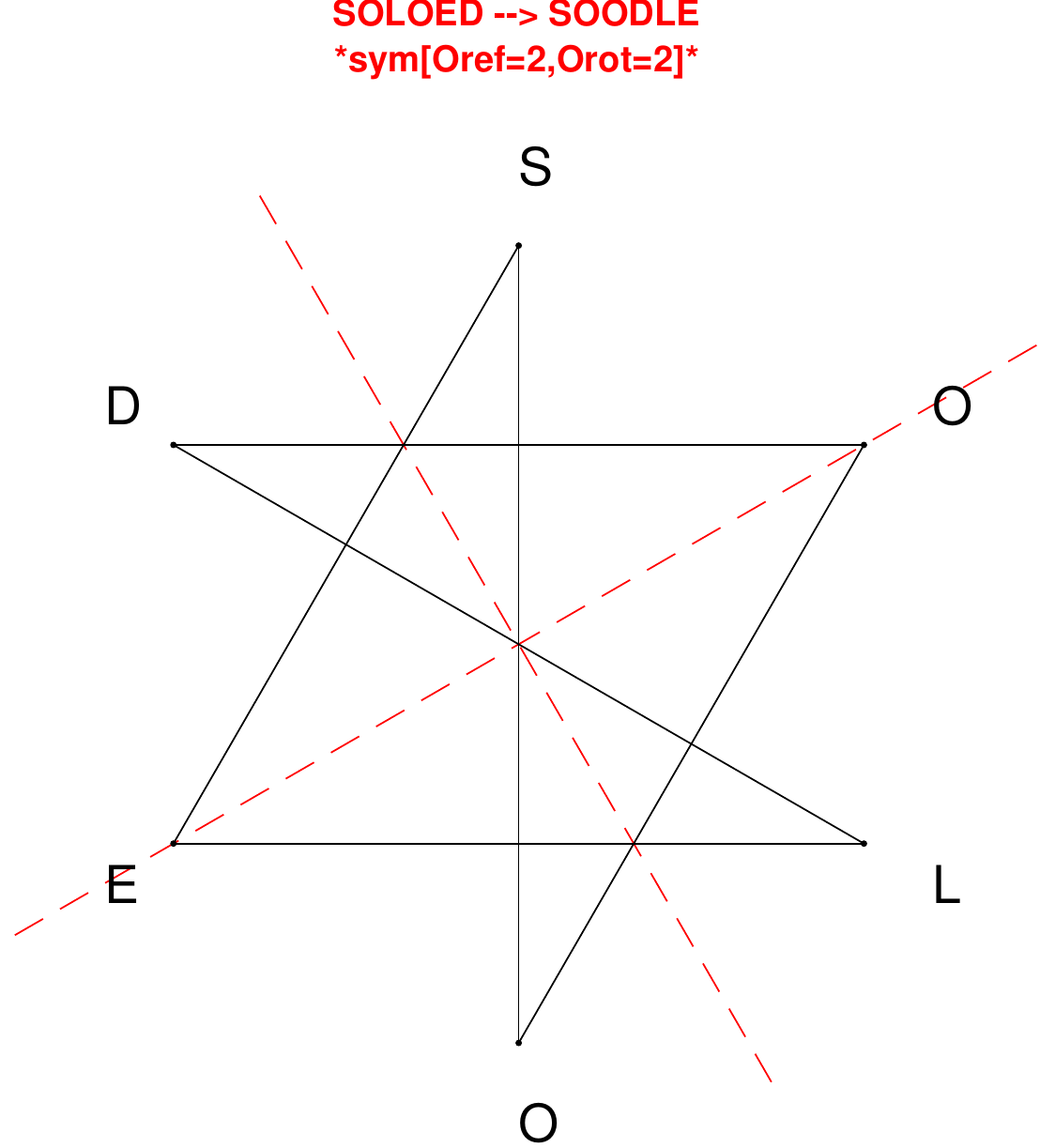}
\end{subfigure}
\hfill
\begin{subfigure}[T]{0.19\textwidth}
\centering
\includegraphics[width=\textwidth]{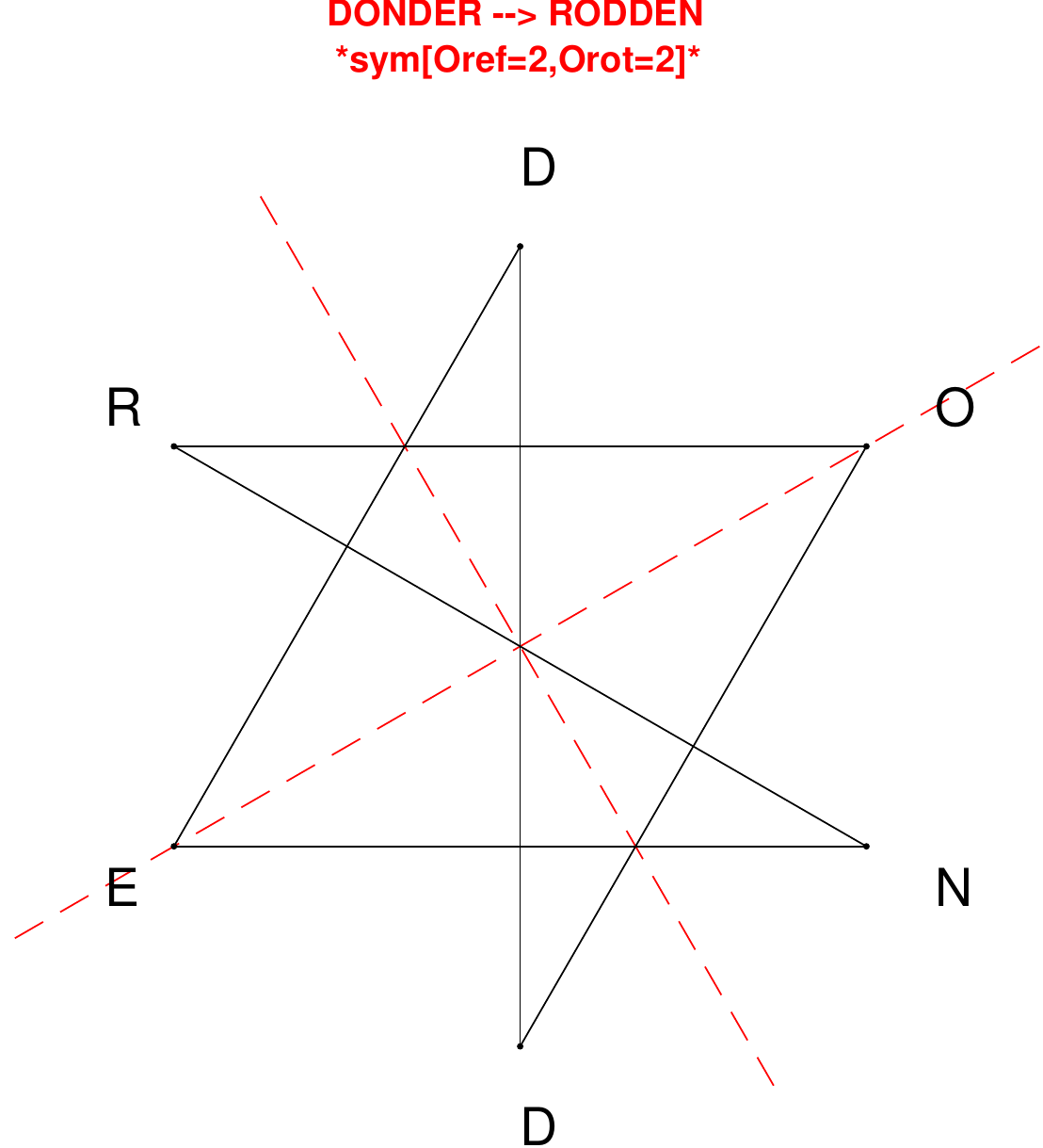}
\end{subfigure}
\hfill
\begin{subfigure}[T]{0.19\textwidth}
\centering
\includegraphics[width=\textwidth]{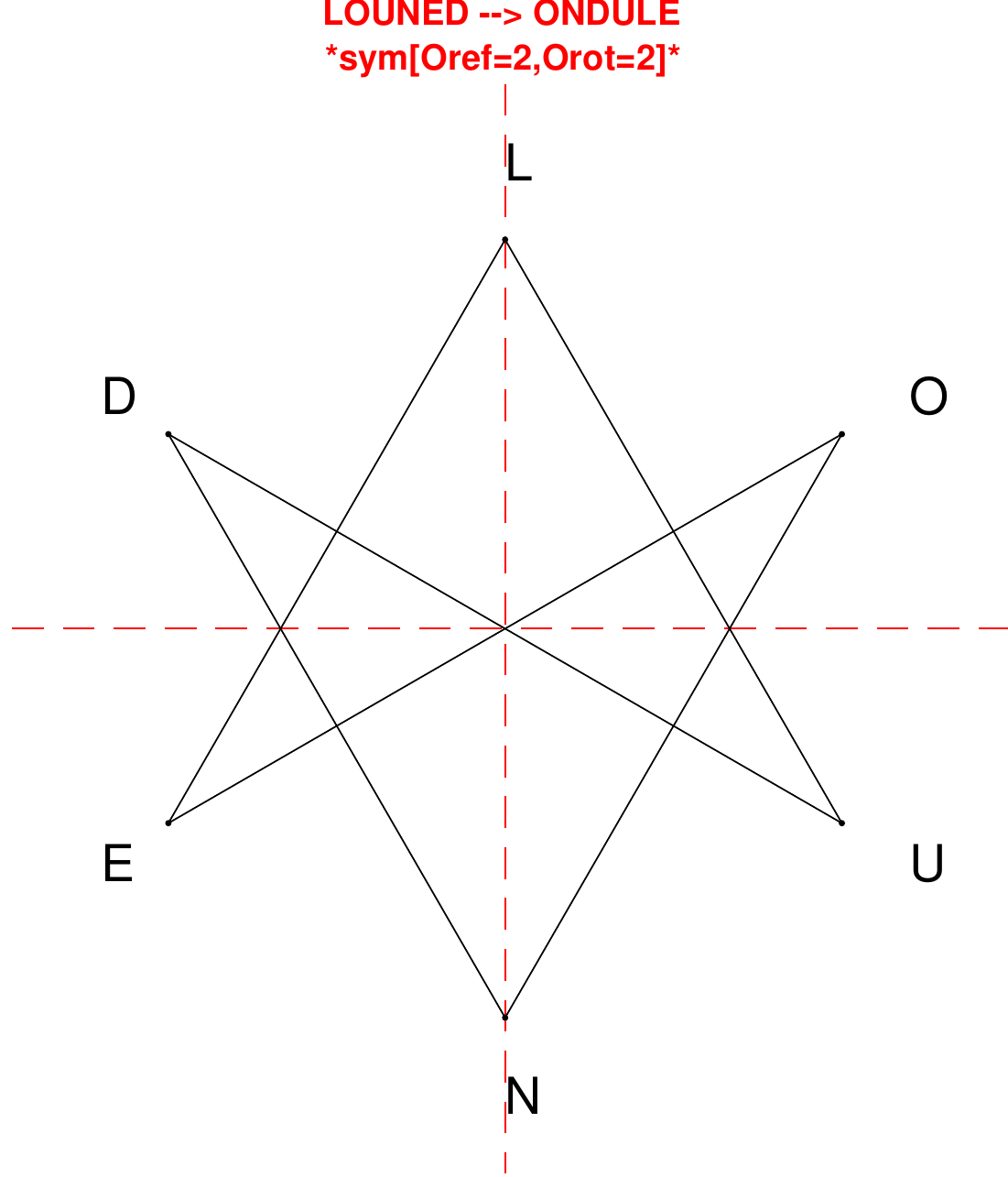}
\end{subfigure}
\hfill
\begin{subfigure}[T]{0.19\textwidth}
\centering
\includegraphics[width=\textwidth]{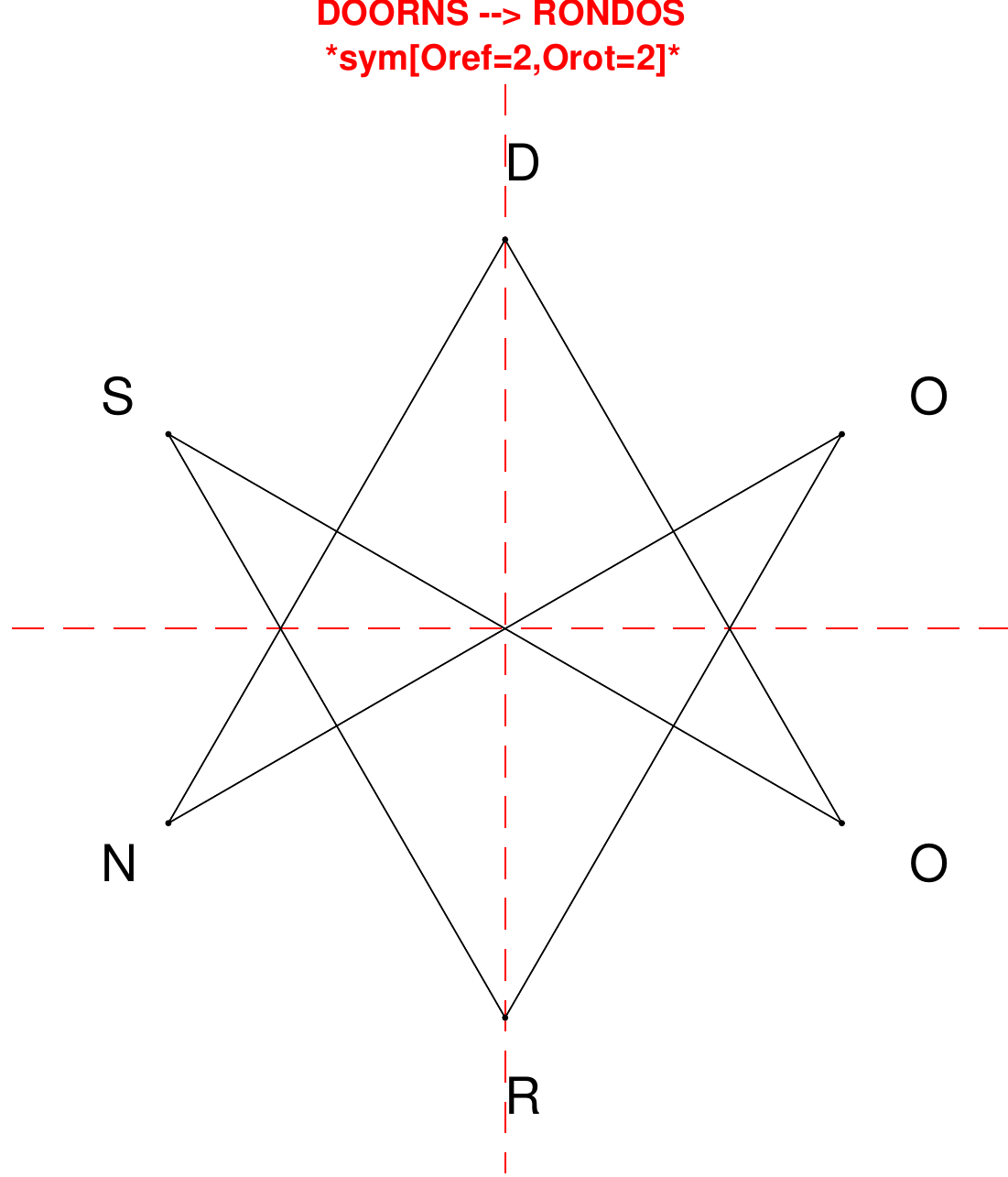}
\end{subfigure}
\end{figure}

\begin{figure}[H]
\centering
\begin{subfigure}[T]{0.19\textwidth}
\centering
\includegraphics[width=\textwidth]{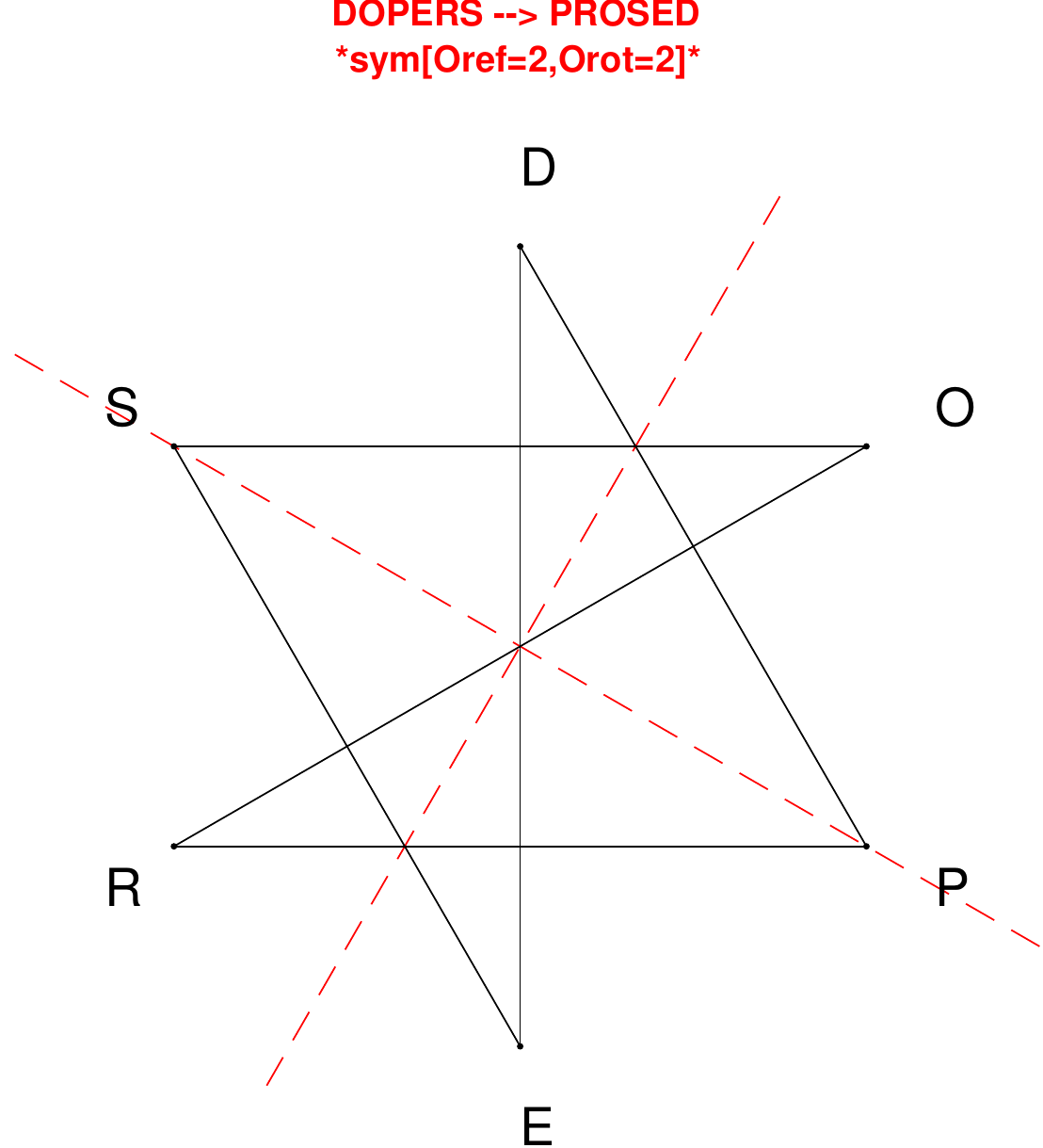}
\end{subfigure}
\hfill
\begin{subfigure}[T]{0.19\textwidth}
\centering
\includegraphics[width=\textwidth]{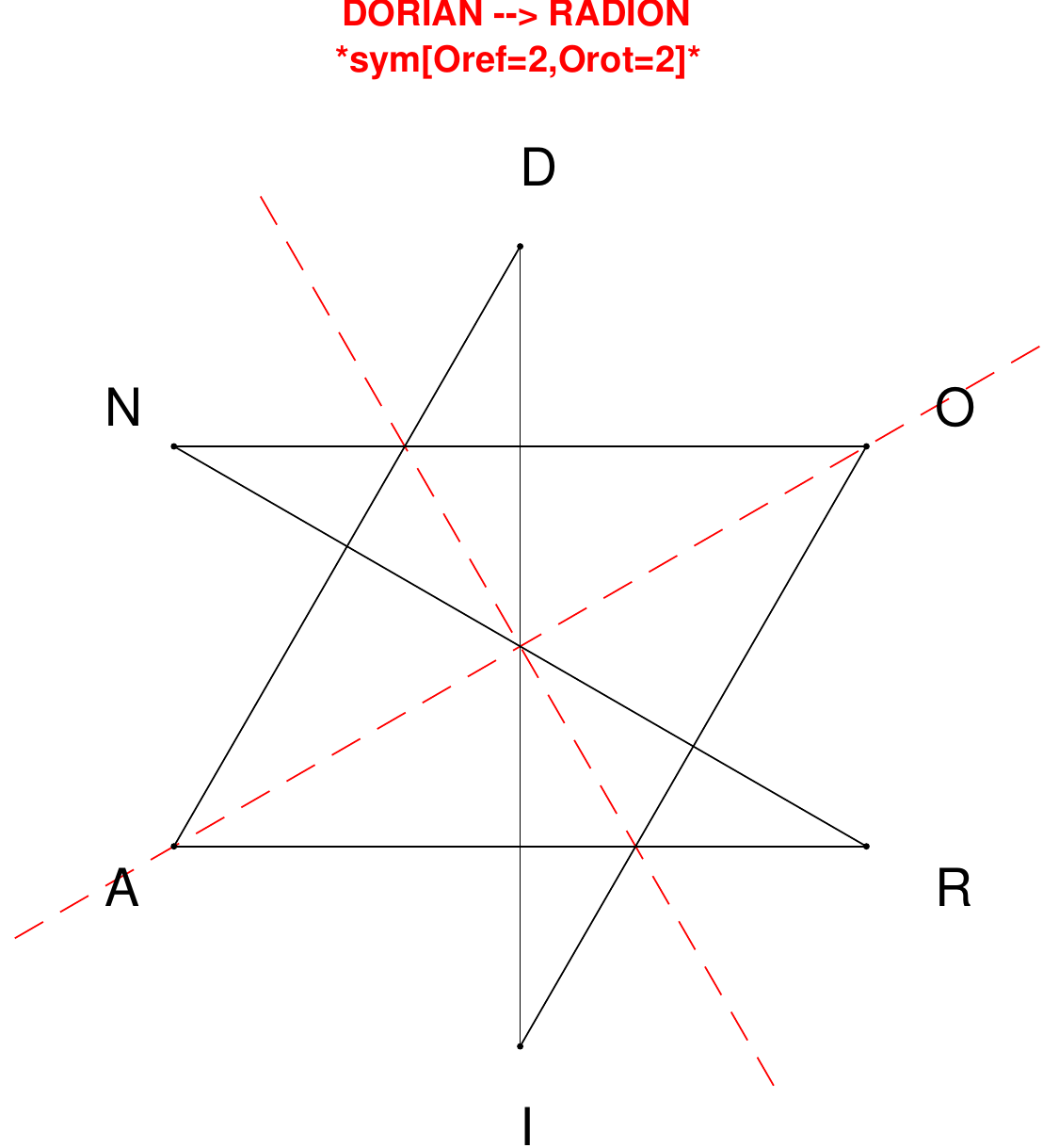}
\end{subfigure}
\hfill
\begin{subfigure}[T]{0.19\textwidth}
\centering
\includegraphics[width=\textwidth]{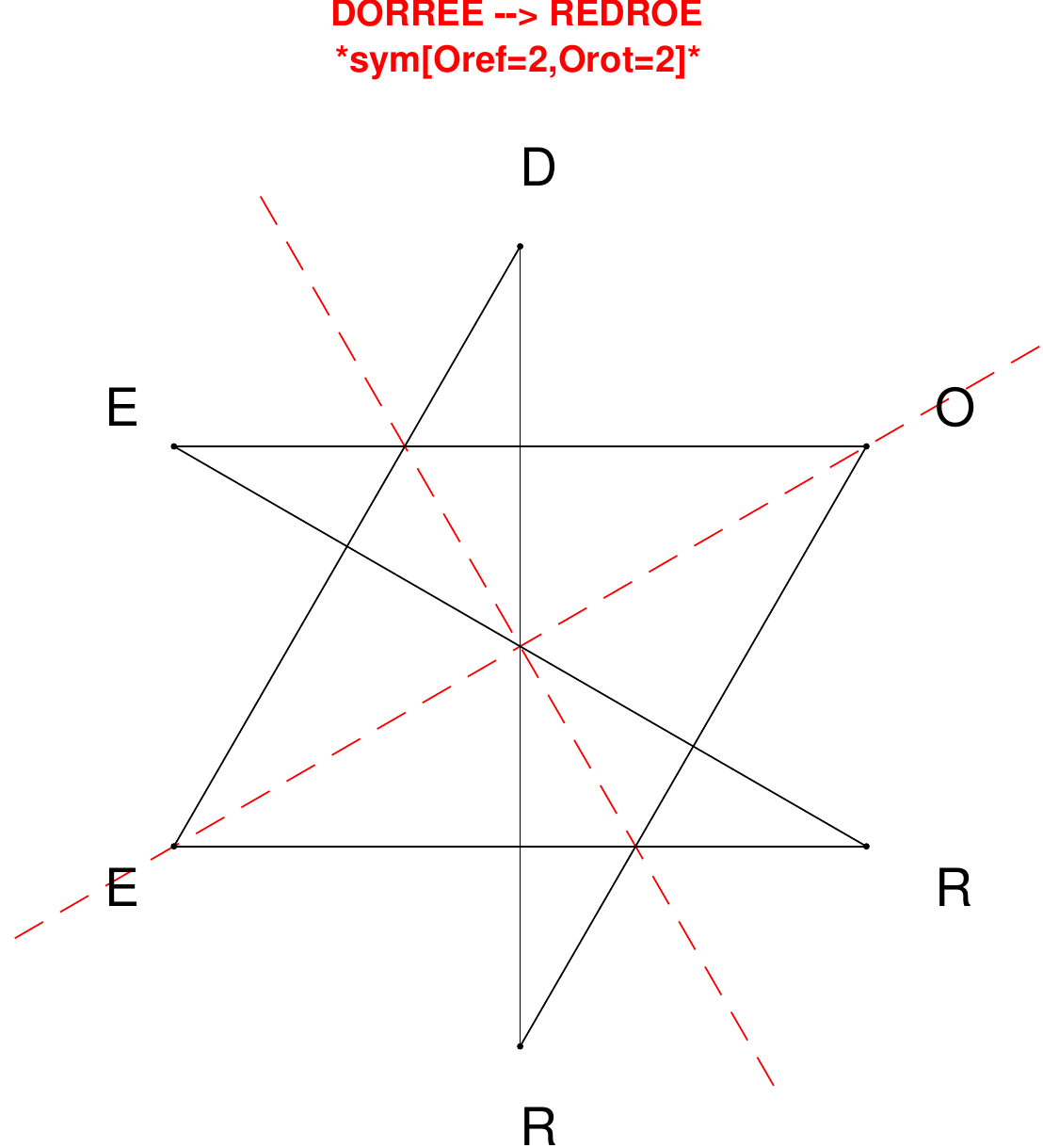}
\end{subfigure}
\hfill
\begin{subfigure}[T]{0.19\textwidth}
\centering
\includegraphics[width=\textwidth]{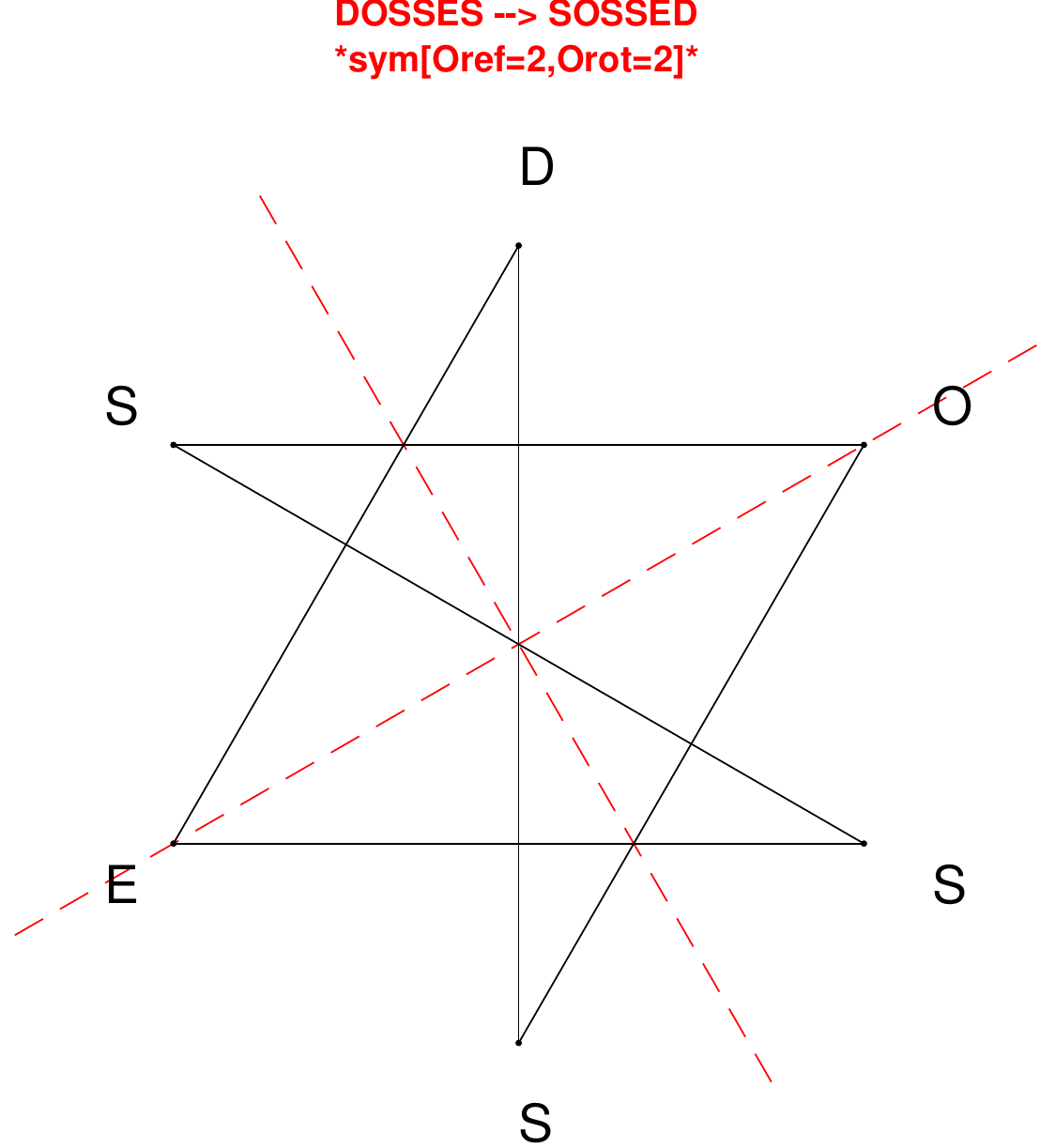}
\end{subfigure}
\hfill
\begin{subfigure}[T]{0.19\textwidth}
\centering
\includegraphics[width=\textwidth]{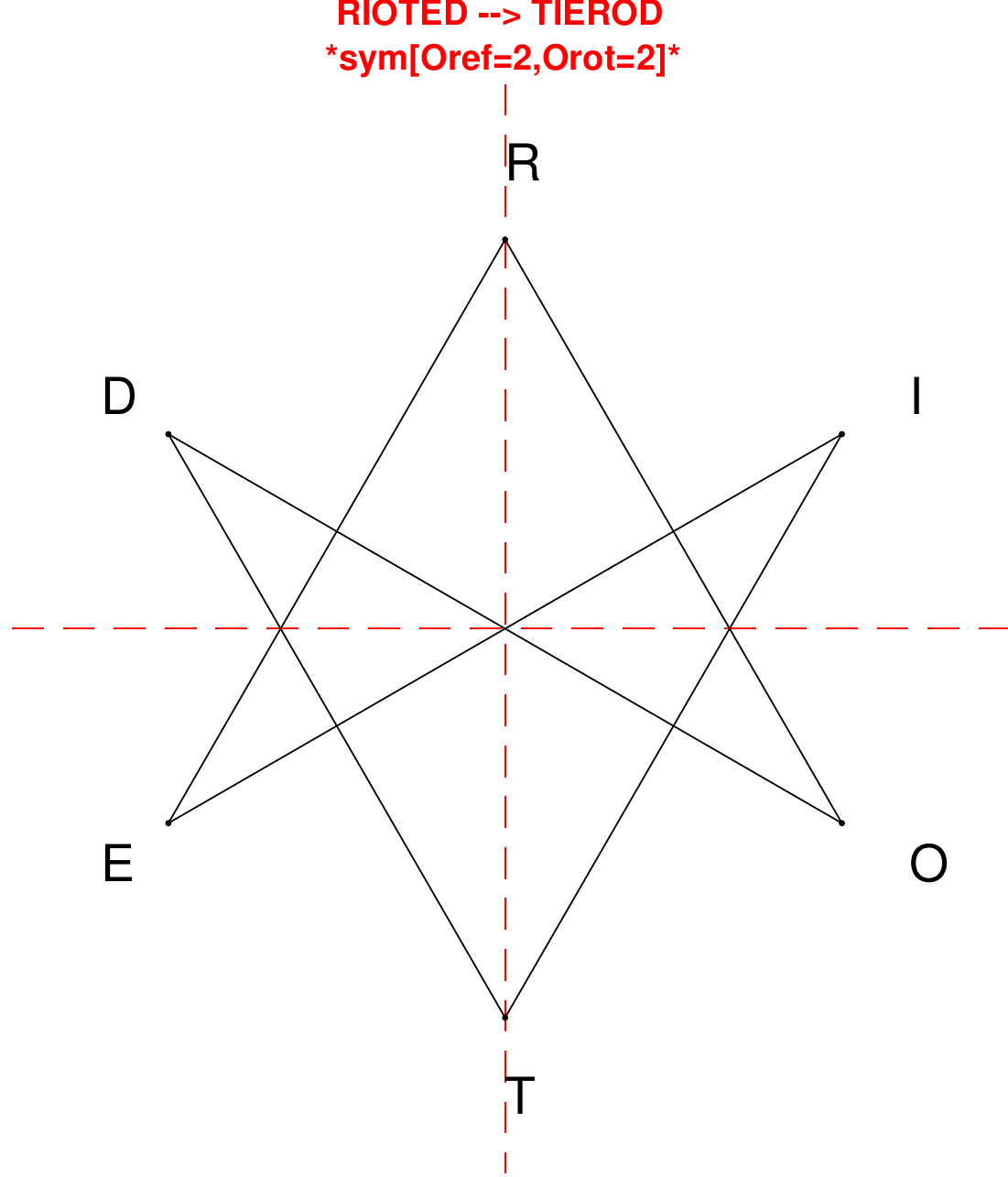}
\end{subfigure}
\end{figure}

\begin{figure}[H]
\centering
\begin{subfigure}[T]{0.19\textwidth}
\centering
\includegraphics[width=\textwidth]{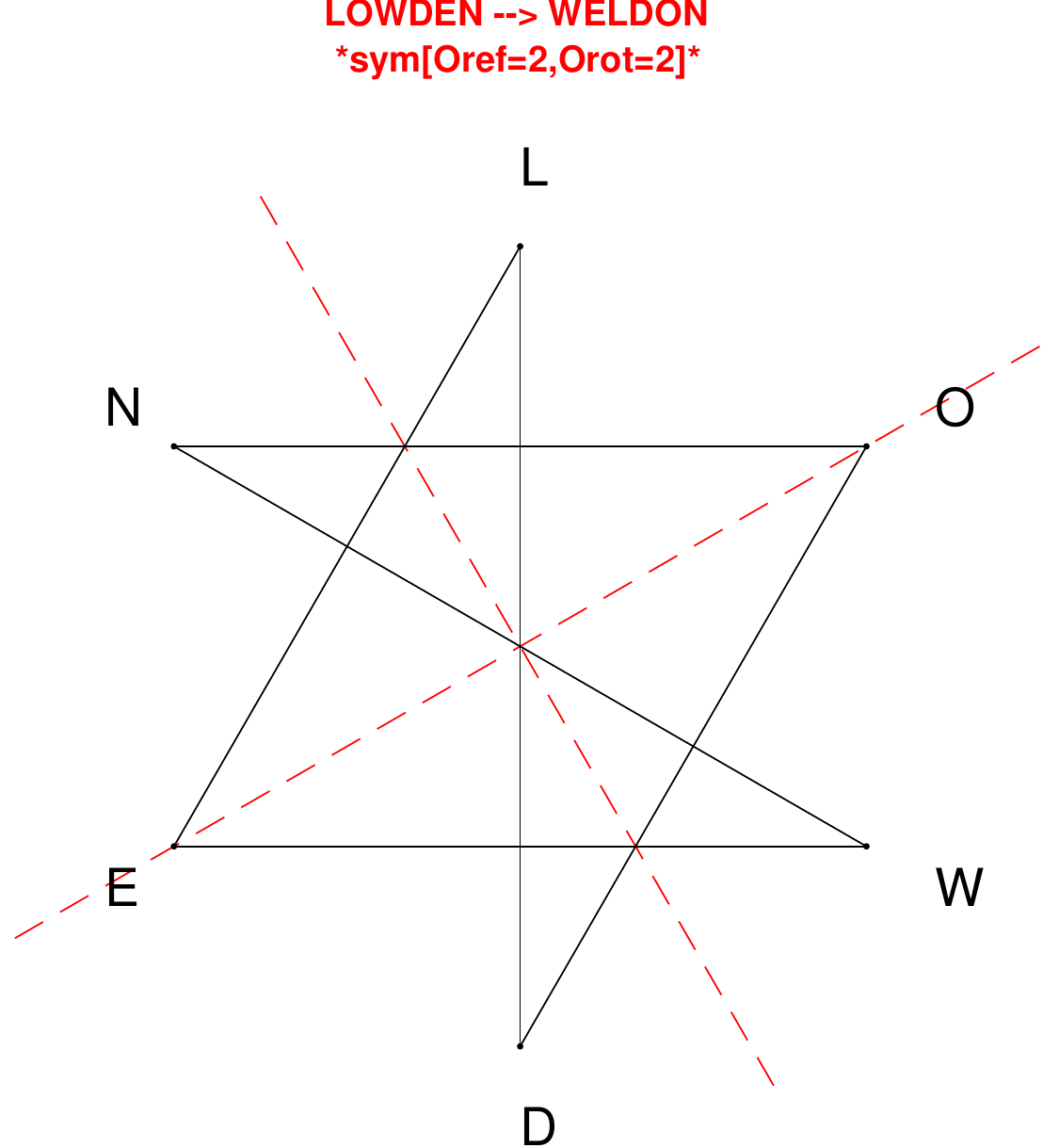}
\end{subfigure}
\hfill
\begin{subfigure}[T]{0.19\textwidth}
\centering
\includegraphics[width=\textwidth]{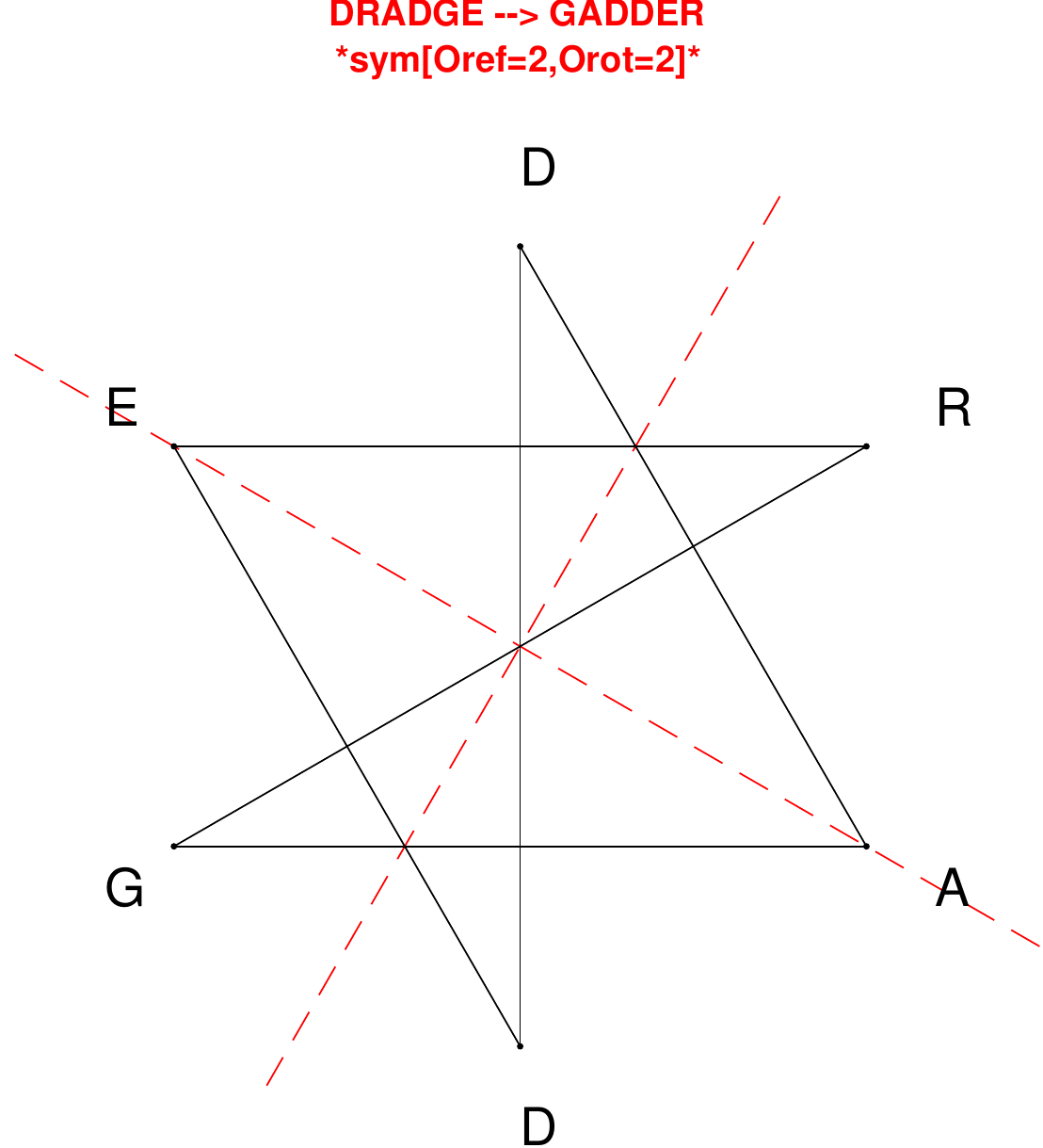}
\end{subfigure}
\hfill
\begin{subfigure}[T]{0.19\textwidth}
\centering
\includegraphics[width=\textwidth]{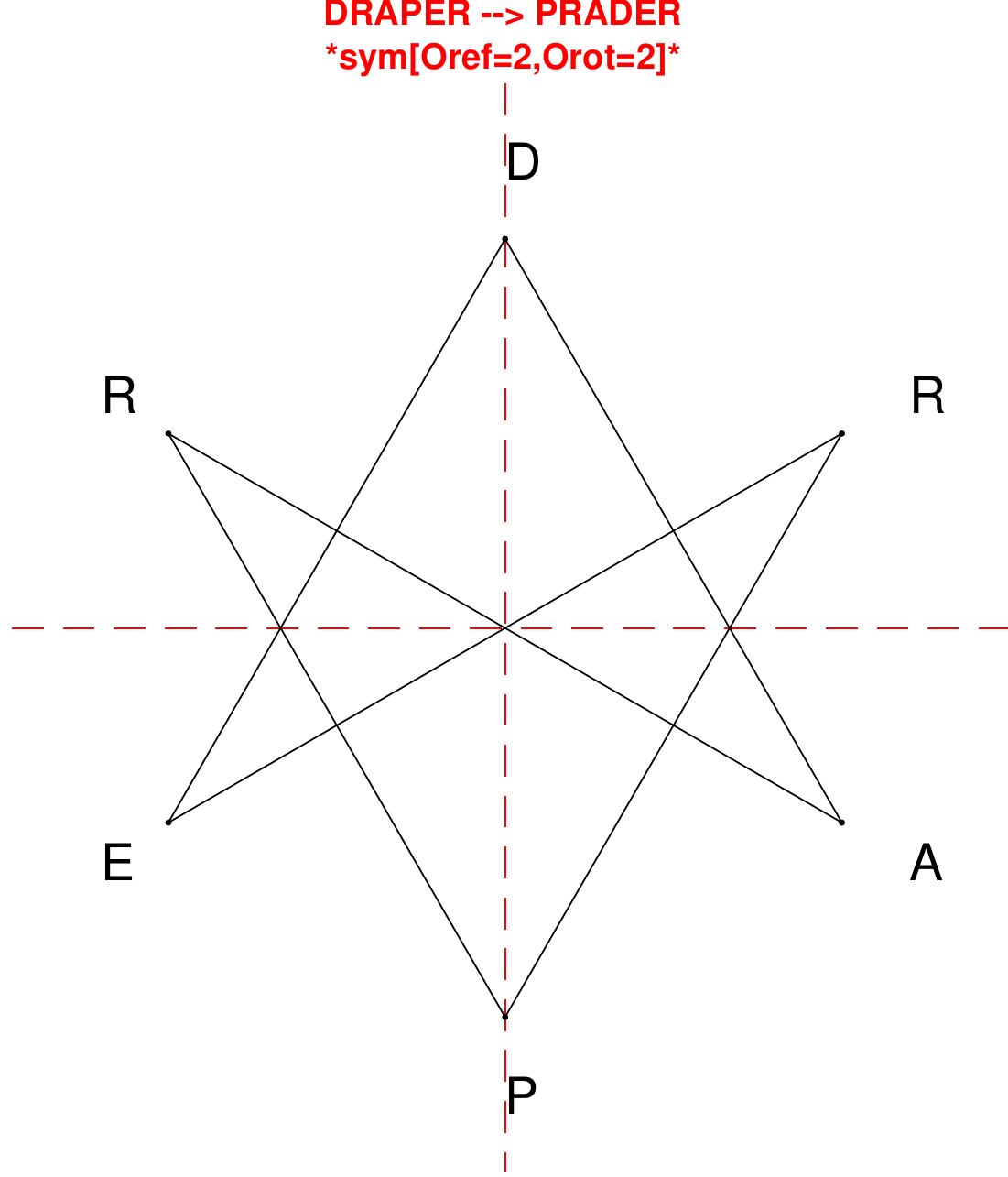}
\end{subfigure}
\hfill
\begin{subfigure}[T]{0.19\textwidth}
\centering
\includegraphics[width=\textwidth]{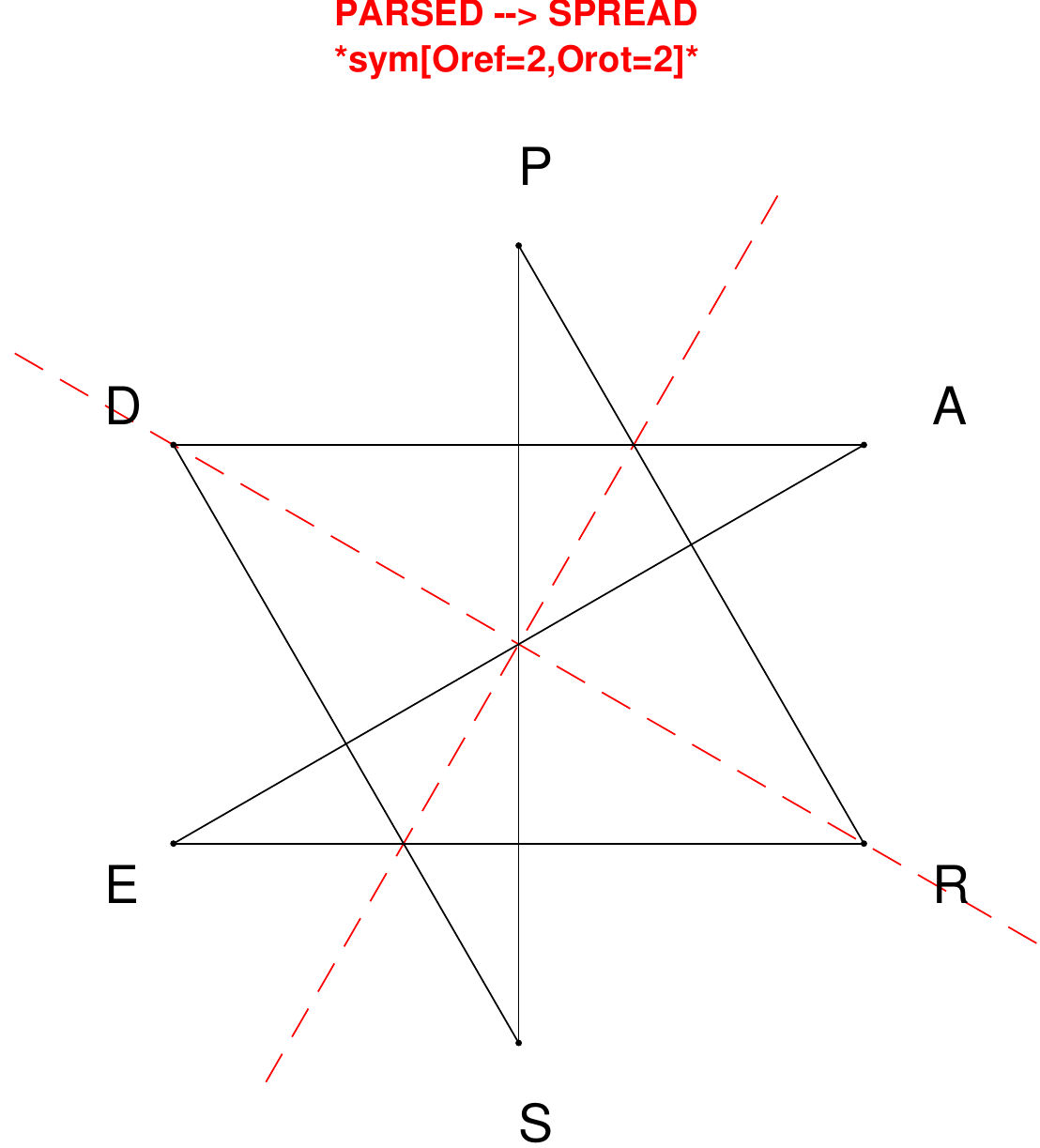}
\end{subfigure}
\hfill
\begin{subfigure}[T]{0.19\textwidth}
\centering
\includegraphics[width=\textwidth]{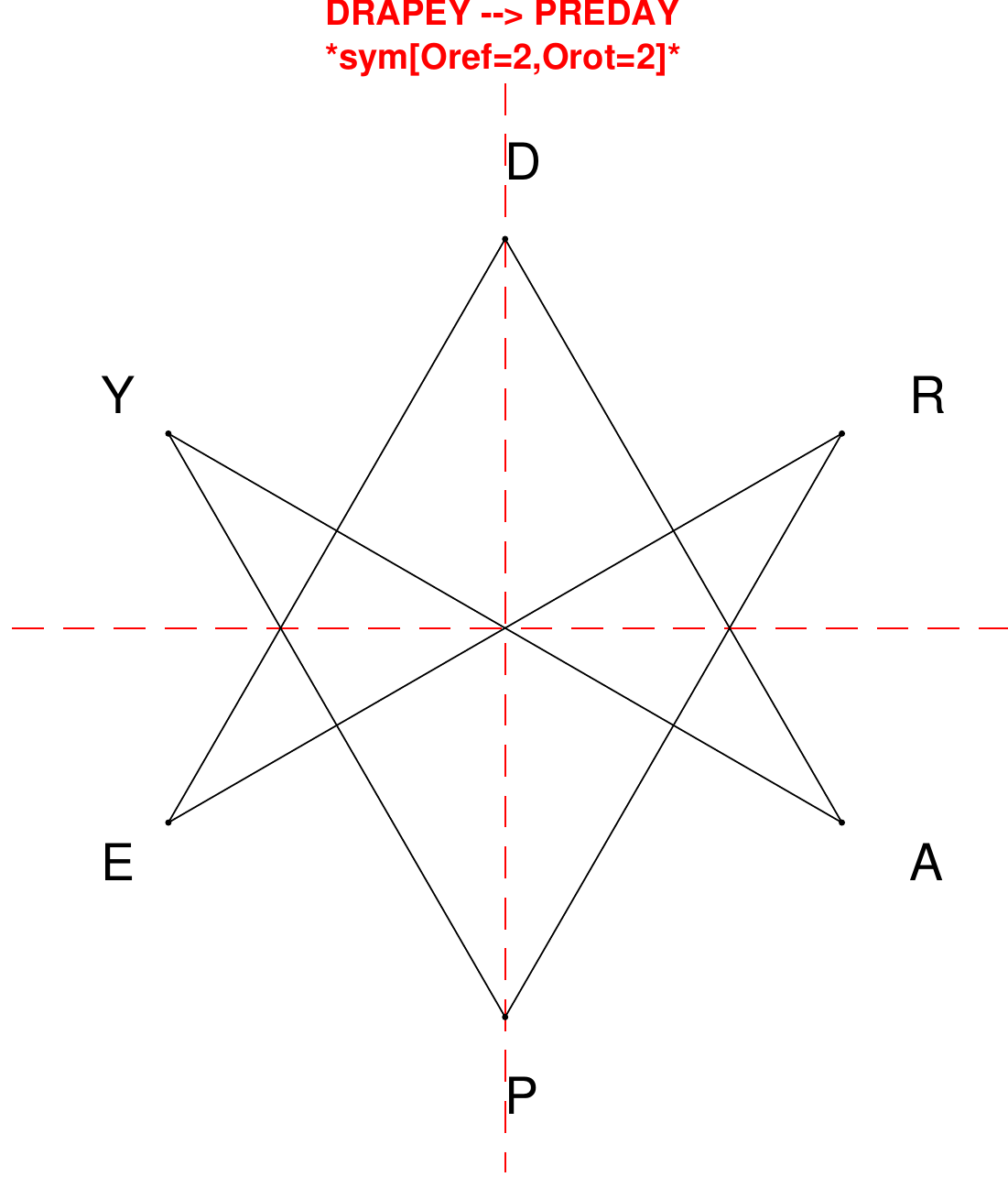}
\end{subfigure}
\end{figure}

\begin{figure}[H]
\centering
\begin{subfigure}[T]{0.19\textwidth}
\centering
\includegraphics[width=\textwidth]{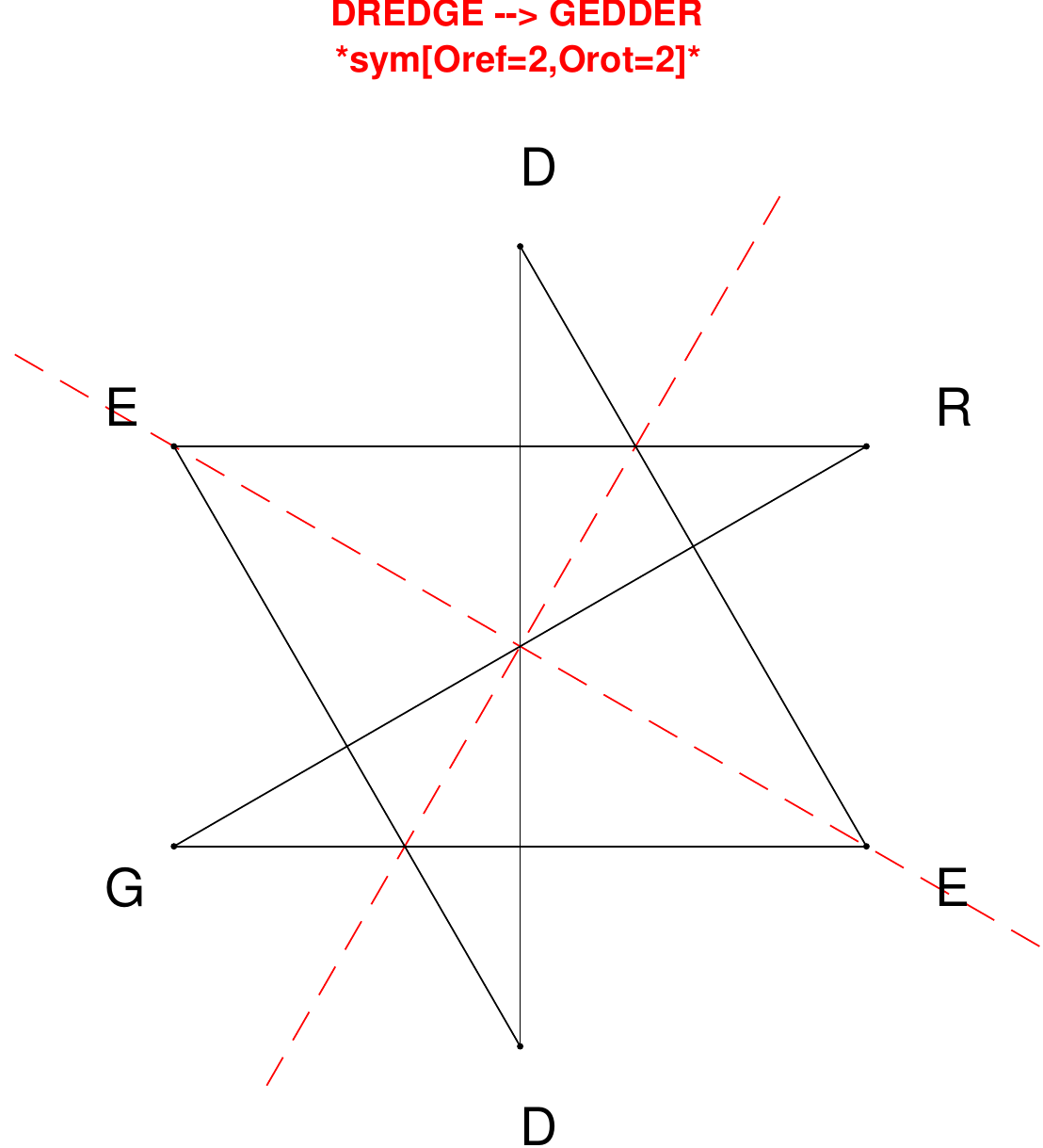}
\end{subfigure}
\hfill
\begin{subfigure}[T]{0.19\textwidth}
\centering
\includegraphics[width=\textwidth]{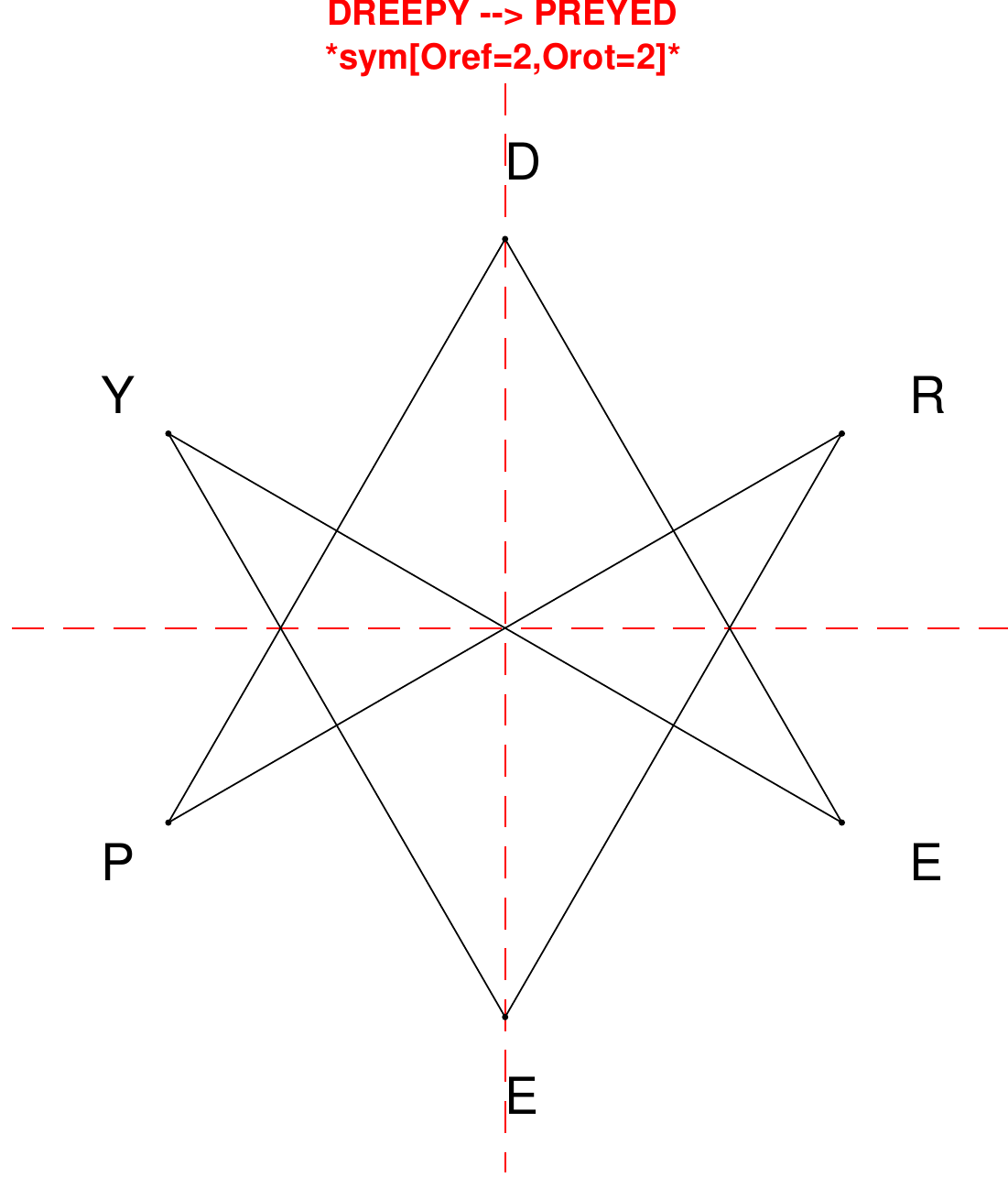}
\end{subfigure}
\hfill
\begin{subfigure}[T]{0.19\textwidth}
\centering
\includegraphics[width=\textwidth]{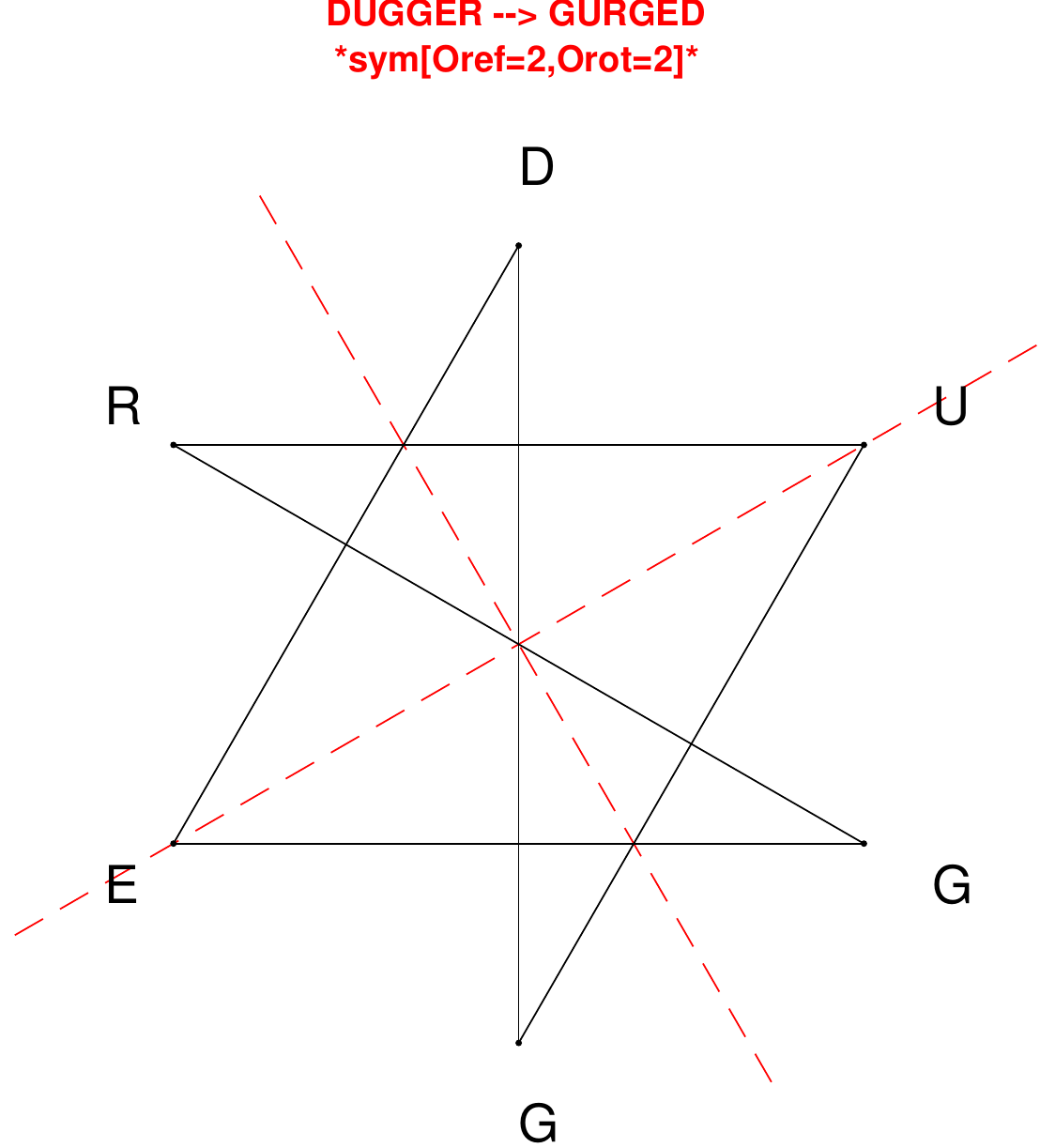}
\end{subfigure}
\hfill
\begin{subfigure}[T]{0.19\textwidth}
\centering
\includegraphics[width=\textwidth]{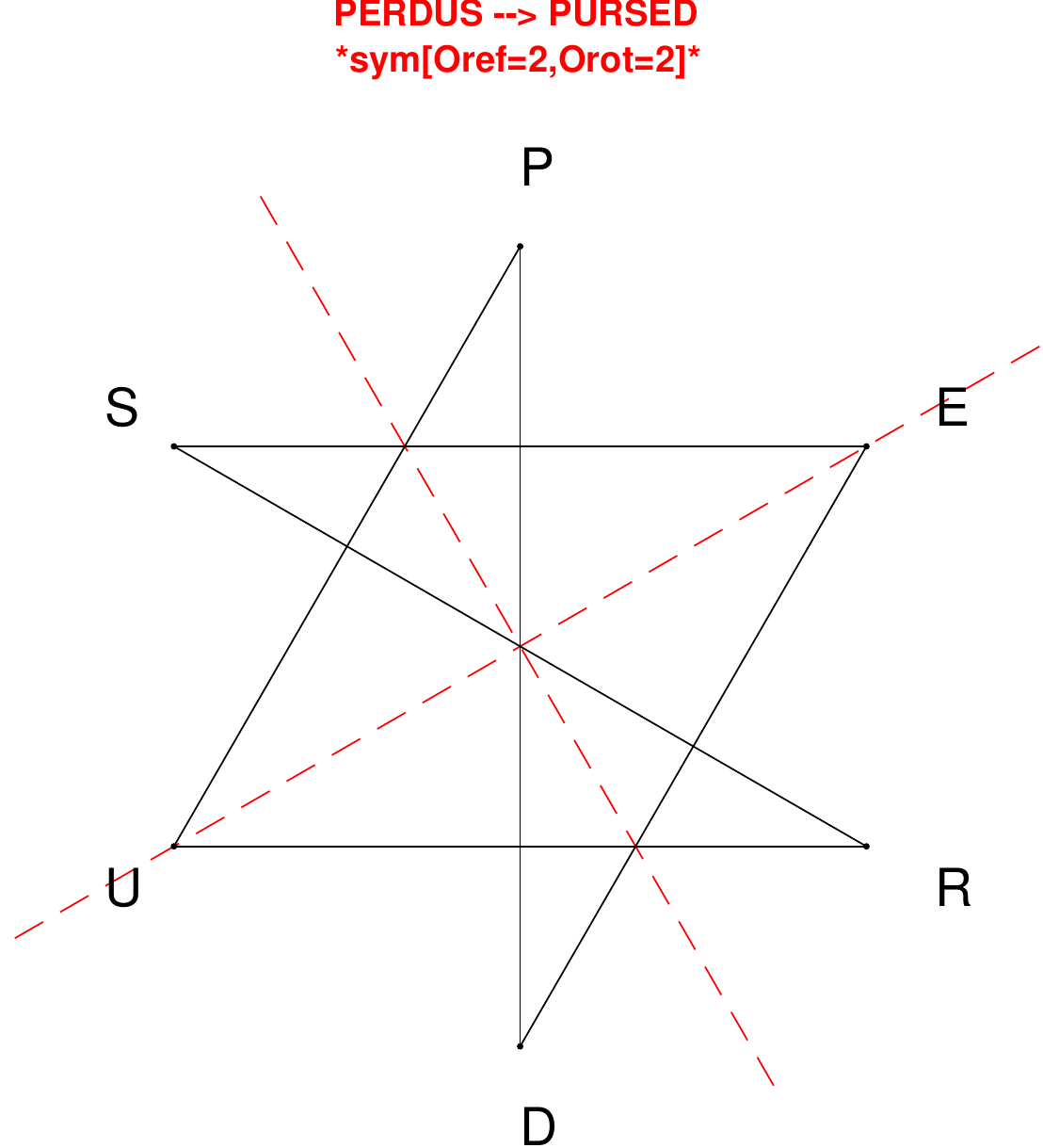}
\end{subfigure}
\hfill
\begin{subfigure}[T]{0.19\textwidth}
\centering
\includegraphics[width=\textwidth]{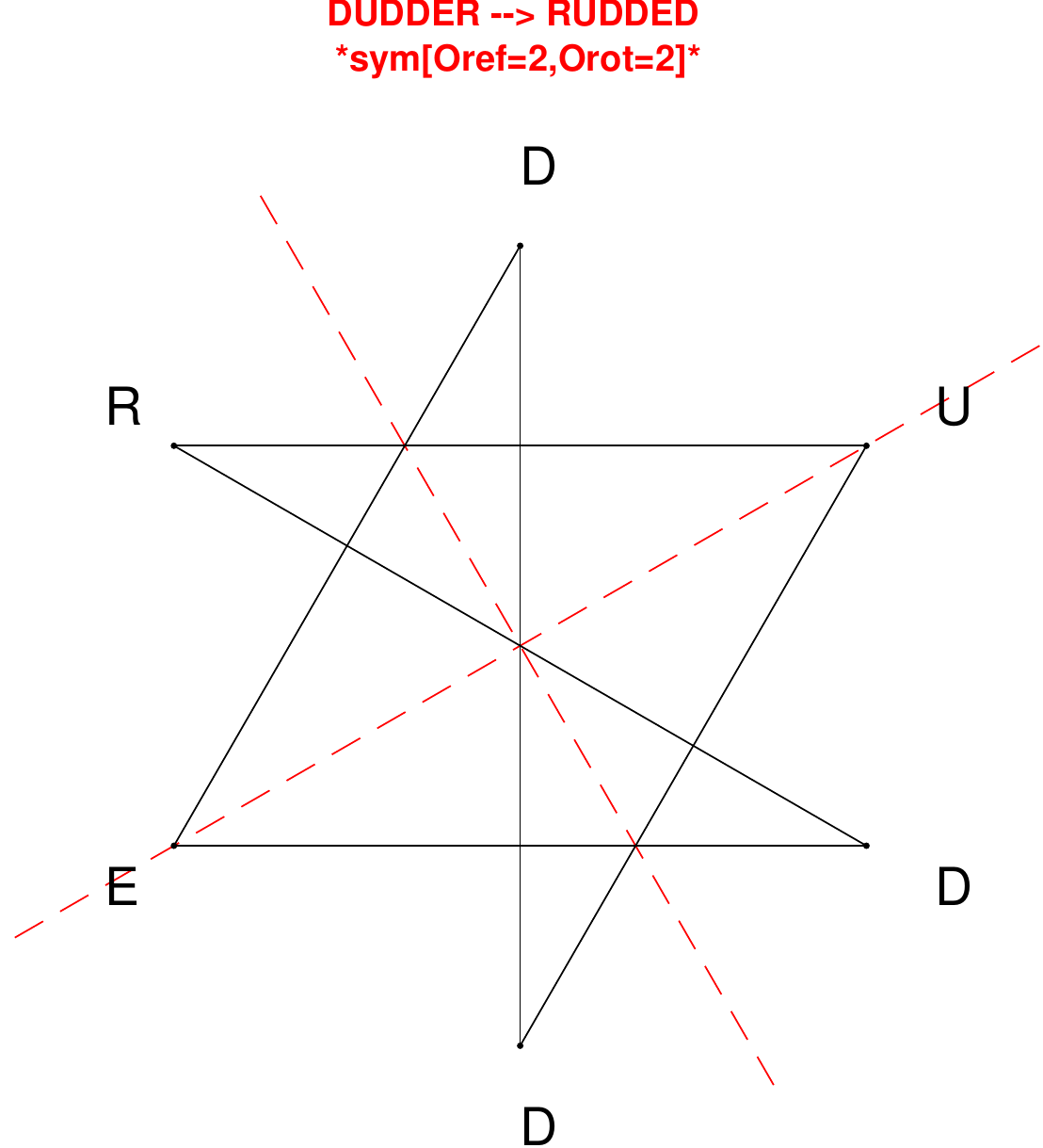}
\end{subfigure}
\end{figure}

\begin{figure}[H]
\centering
\begin{subfigure}[T]{0.19\textwidth}
\centering
\includegraphics[width=\textwidth]{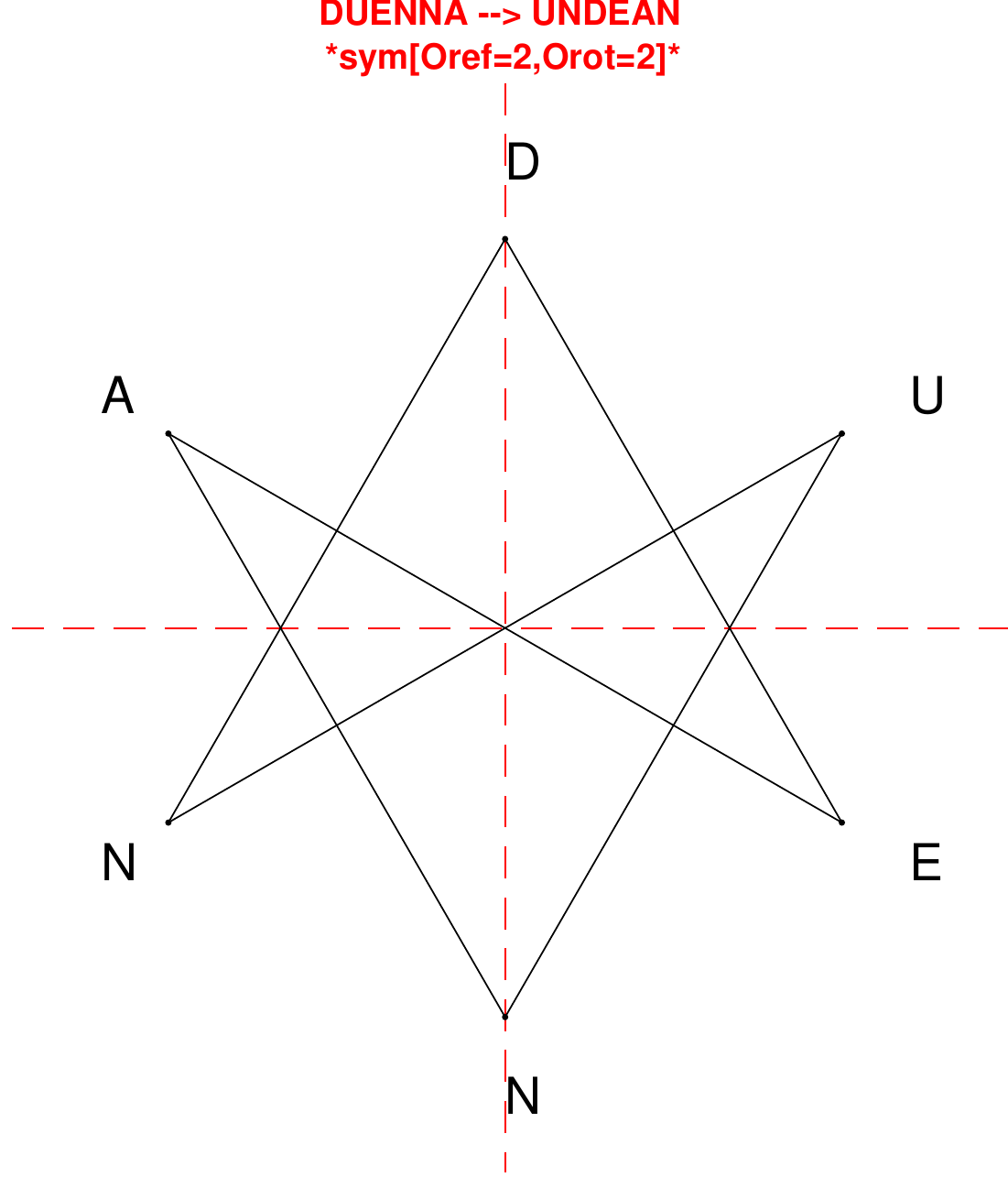}
\end{subfigure}
\hfill
\begin{subfigure}[T]{0.19\textwidth}
\centering
\includegraphics[width=\textwidth]{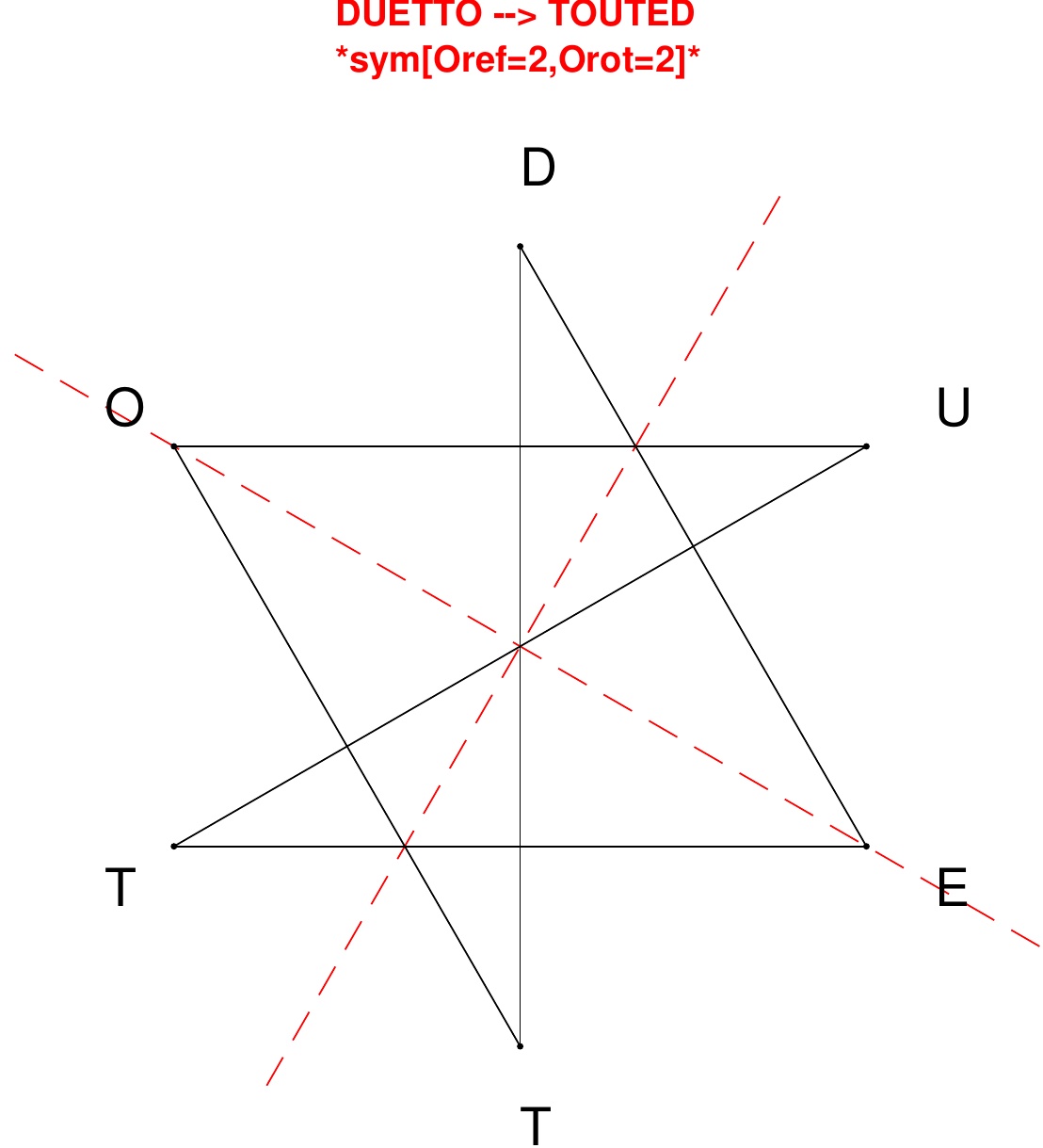}
\end{subfigure}
\hfill
\begin{subfigure}[T]{0.19\textwidth}
\centering
\includegraphics[width=\textwidth]{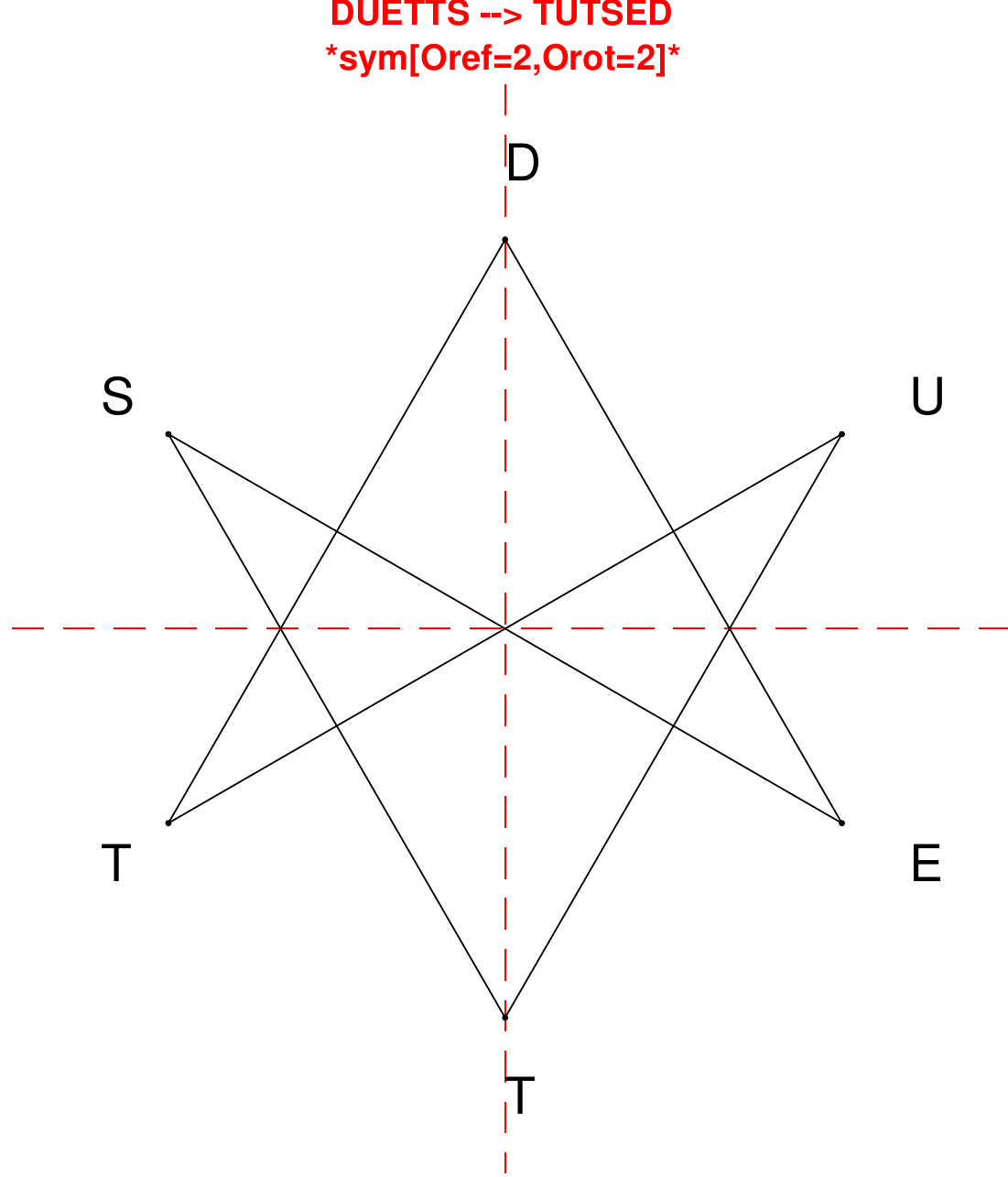}
\end{subfigure}
\hfill
\begin{subfigure}[T]{0.19\textwidth}
\centering
\includegraphics[width=\textwidth]{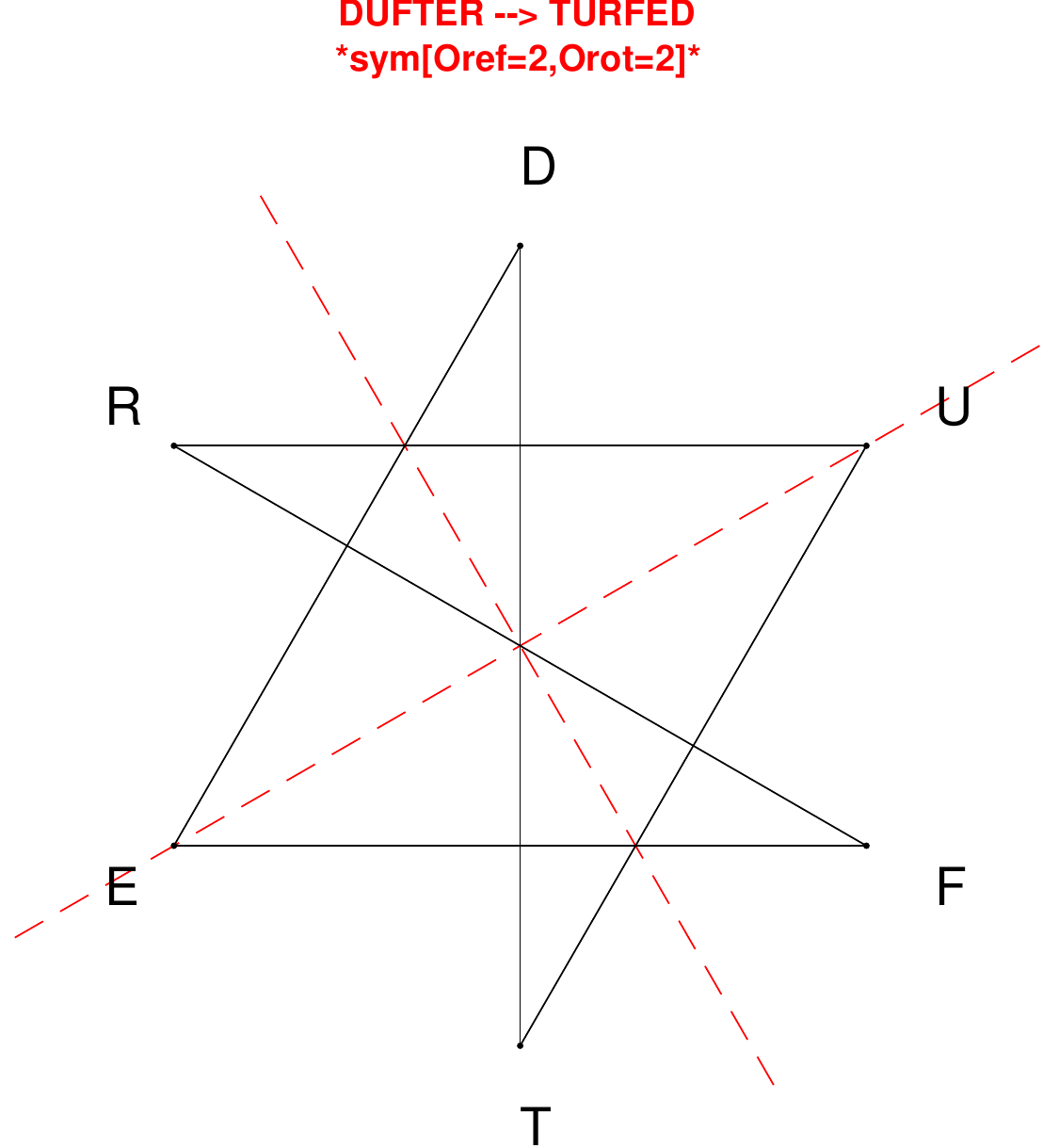}
\end{subfigure}
\hfill
\begin{subfigure}[T]{0.19\textwidth}
\centering
\includegraphics[width=\textwidth]{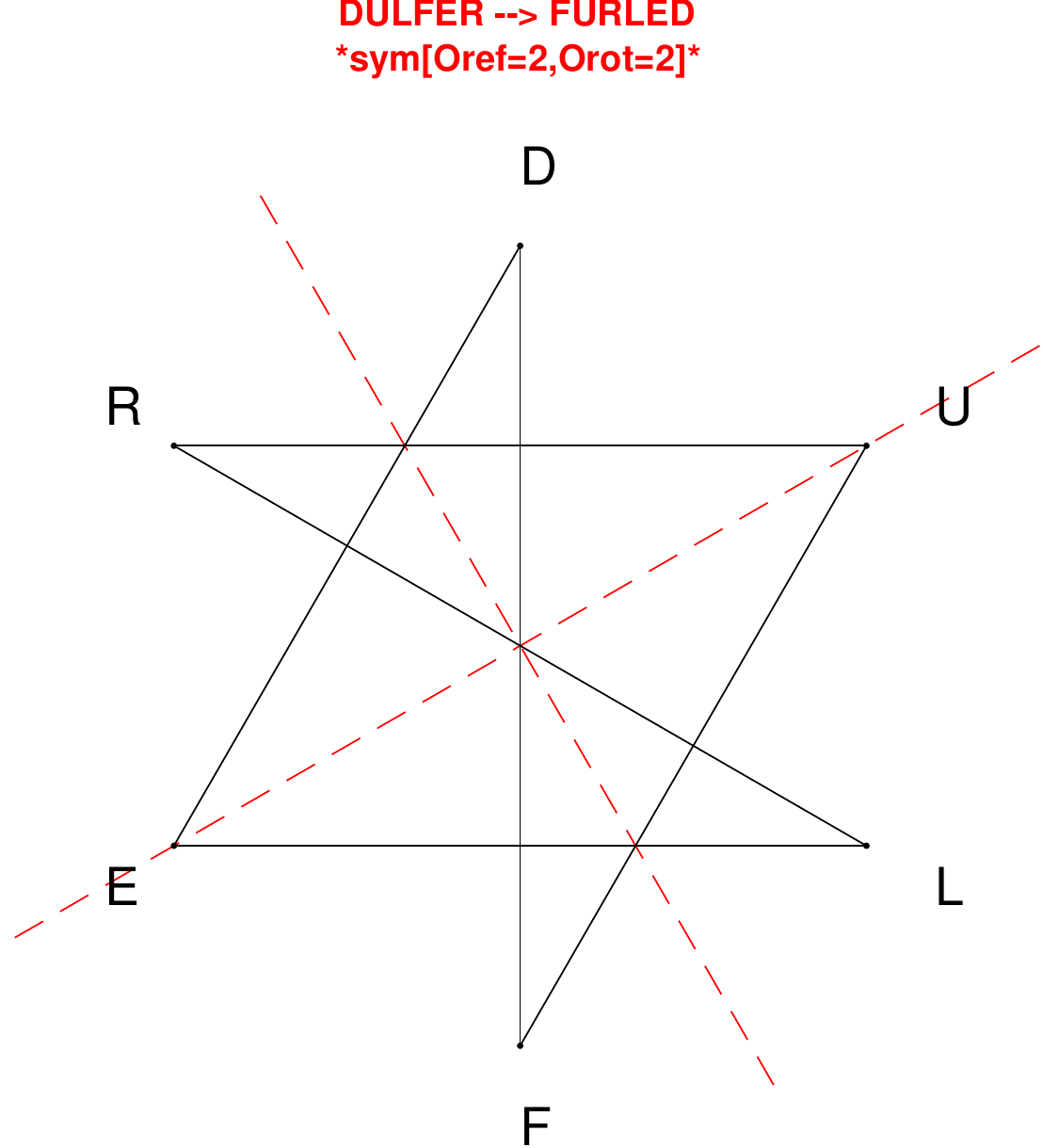}
\end{subfigure}
\end{figure}

\begin{figure}[H]
\centering
\begin{subfigure}[T]{0.19\textwidth}
\centering
\includegraphics[width=\textwidth]{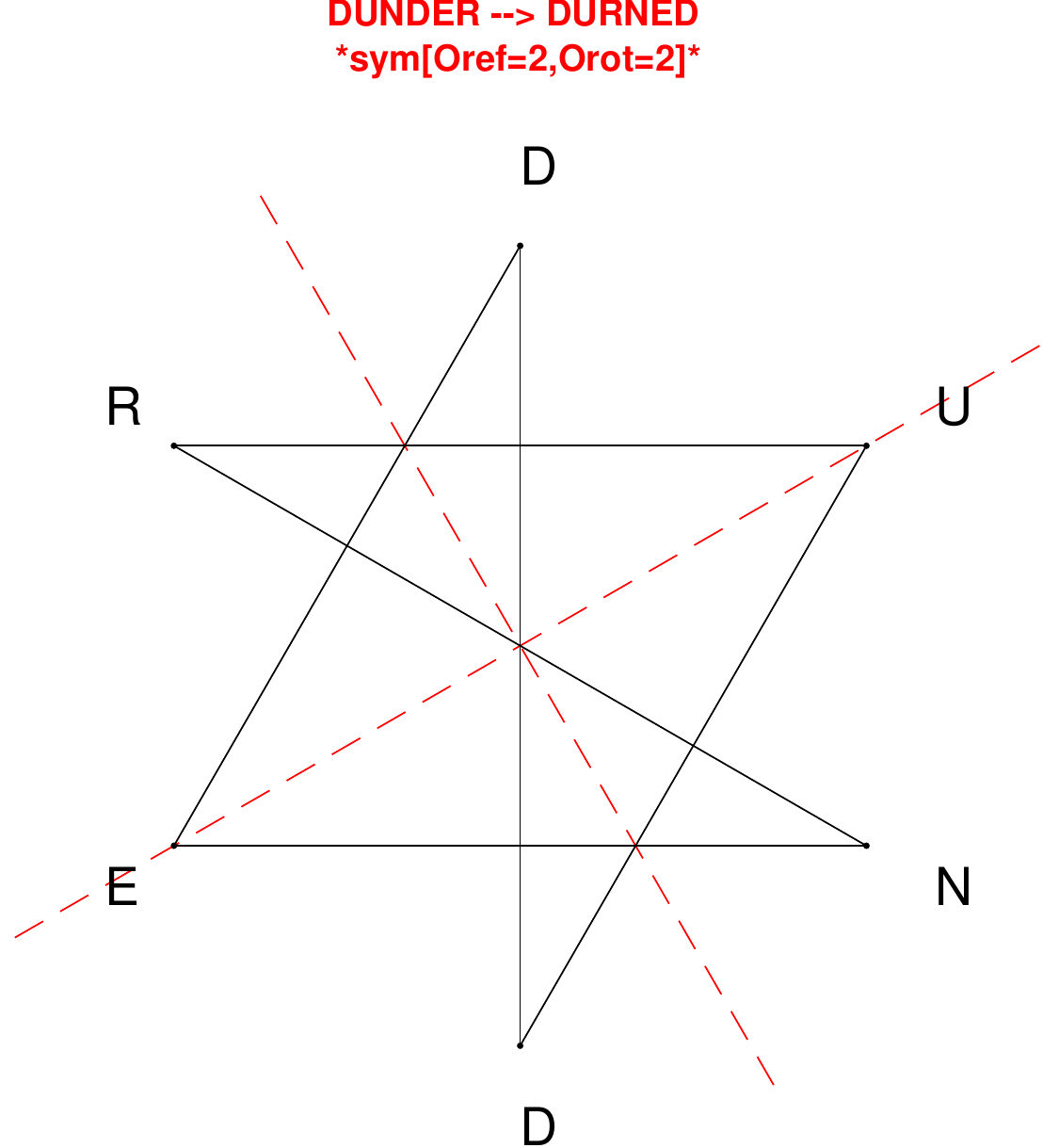}
\end{subfigure}
\hfill
\begin{subfigure}[T]{0.19\textwidth}
\centering
\includegraphics[width=\textwidth]{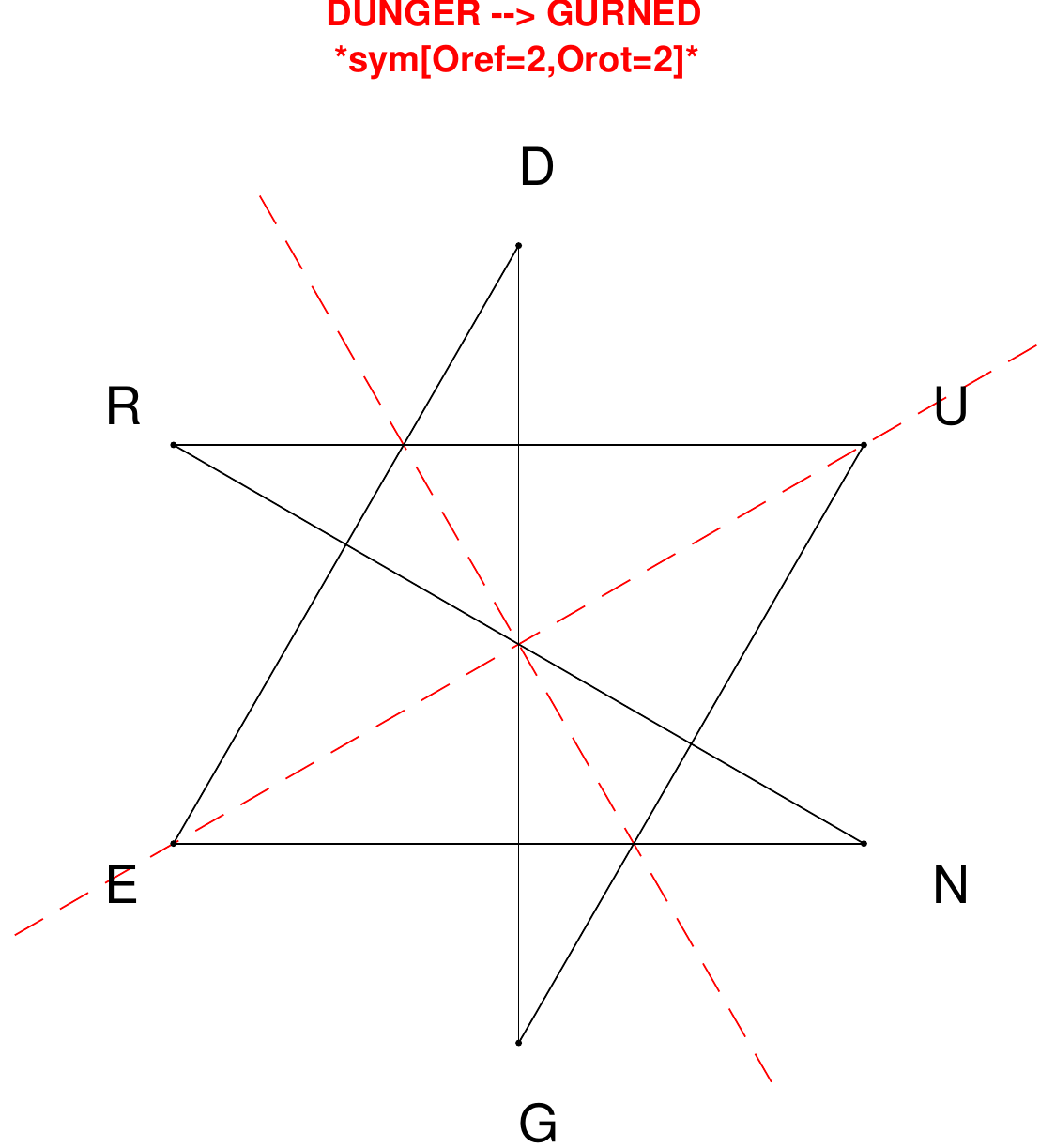}
\end{subfigure}
\hfill
\begin{subfigure}[T]{0.19\textwidth}
\centering
\includegraphics[width=\textwidth]{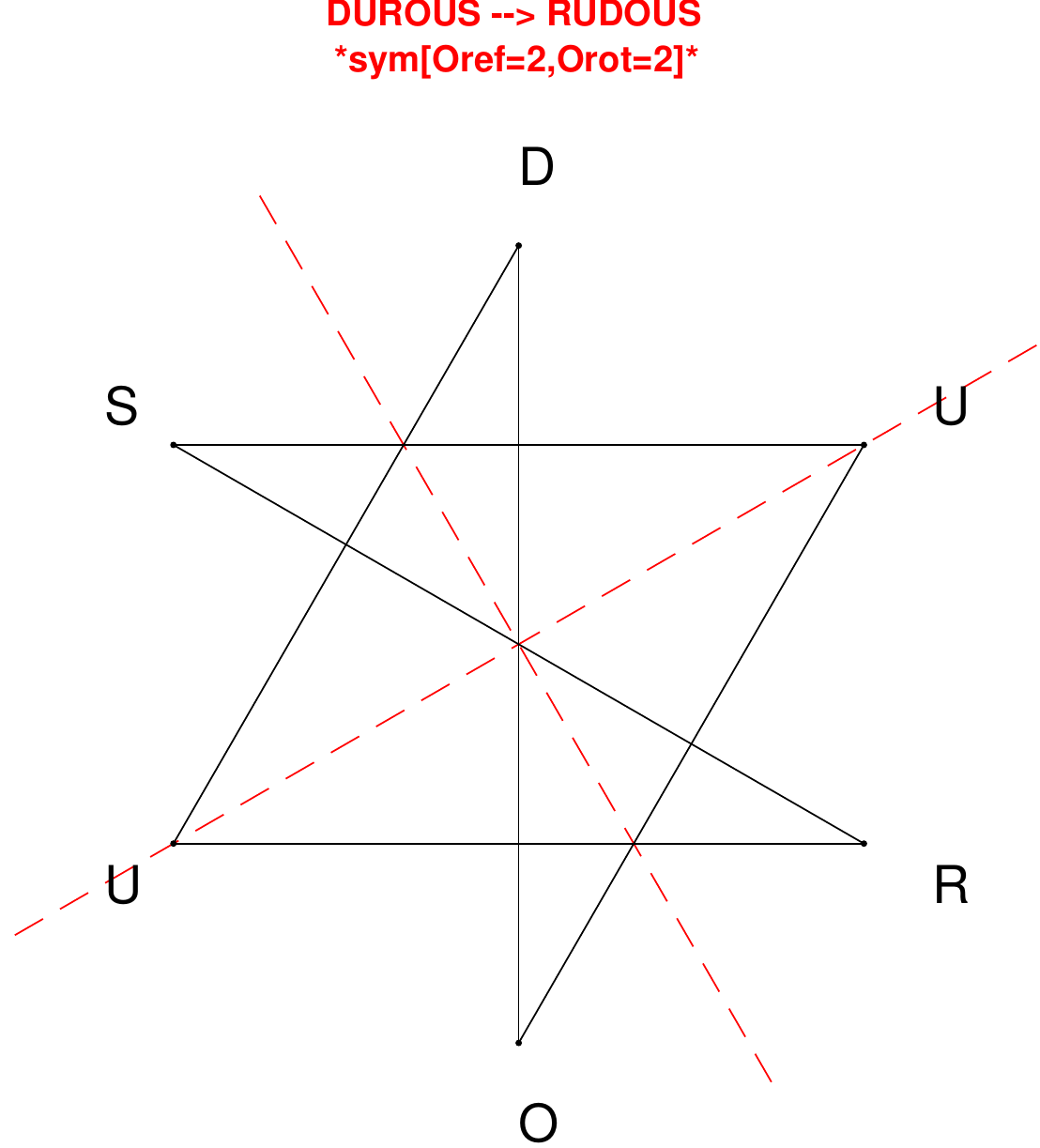}
\end{subfigure}
\hfill
\begin{subfigure}[T]{0.19\textwidth}
\centering
\includegraphics[width=\textwidth]{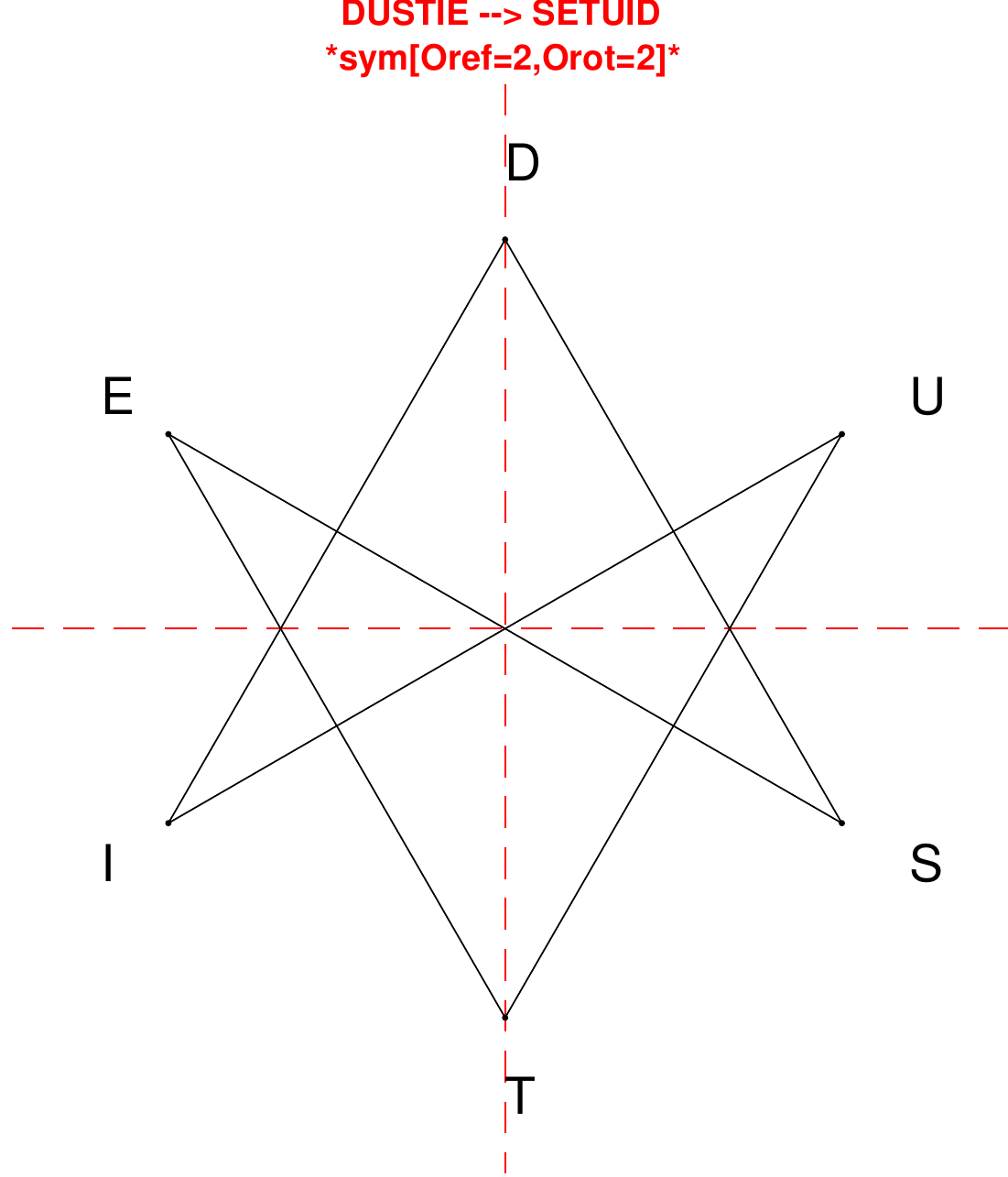}
\end{subfigure}
\hfill
\begin{subfigure}[T]{0.19\textwidth}
\centering
\includegraphics[width=\textwidth]{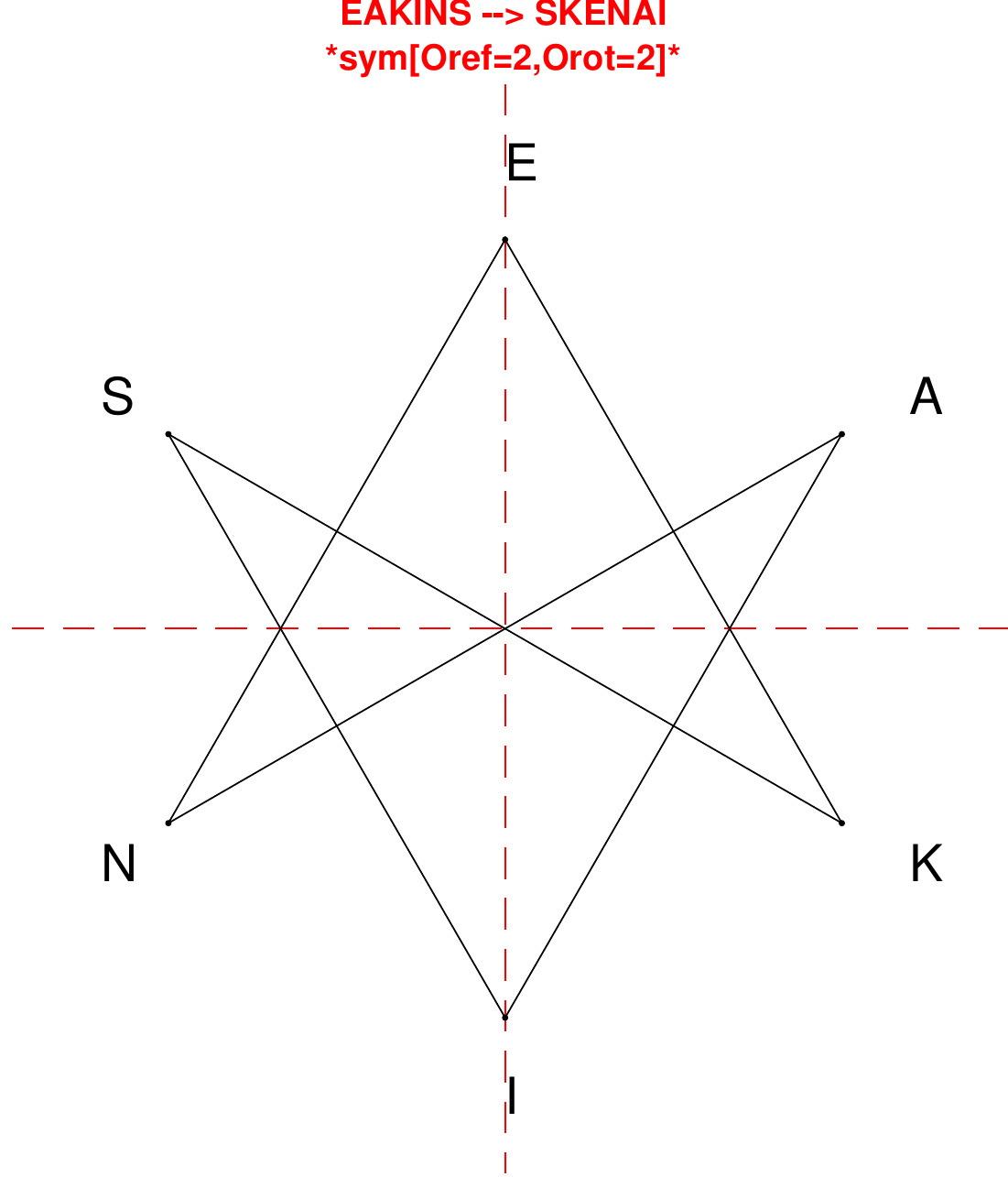}
\end{subfigure}
\end{figure}

\begin{figure}[H]
\centering
\begin{subfigure}[T]{0.19\textwidth}
\centering
\includegraphics[width=\textwidth]{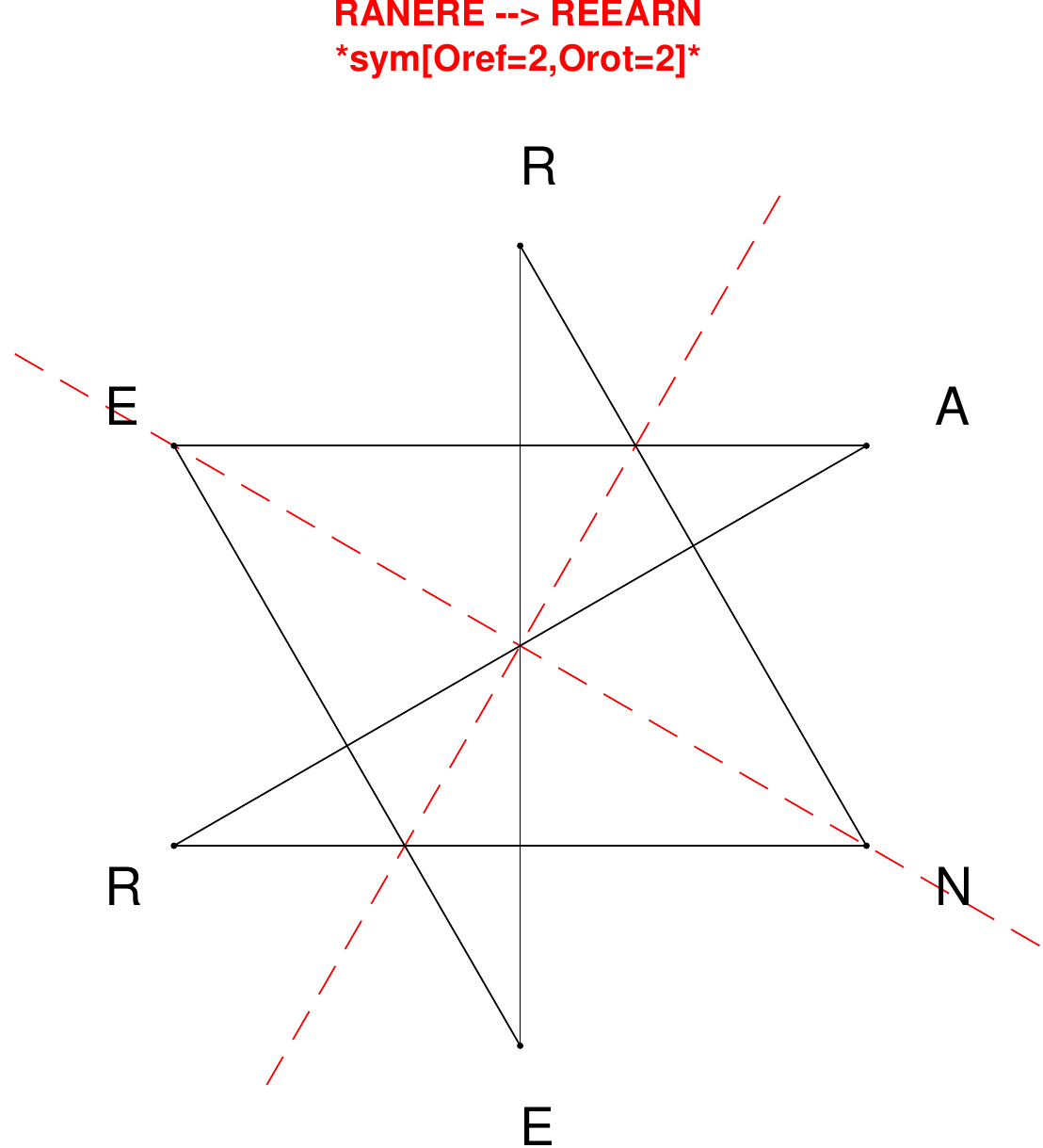}
\end{subfigure}
\hfill
\begin{subfigure}[T]{0.19\textwidth}
\centering
\includegraphics[width=\textwidth]{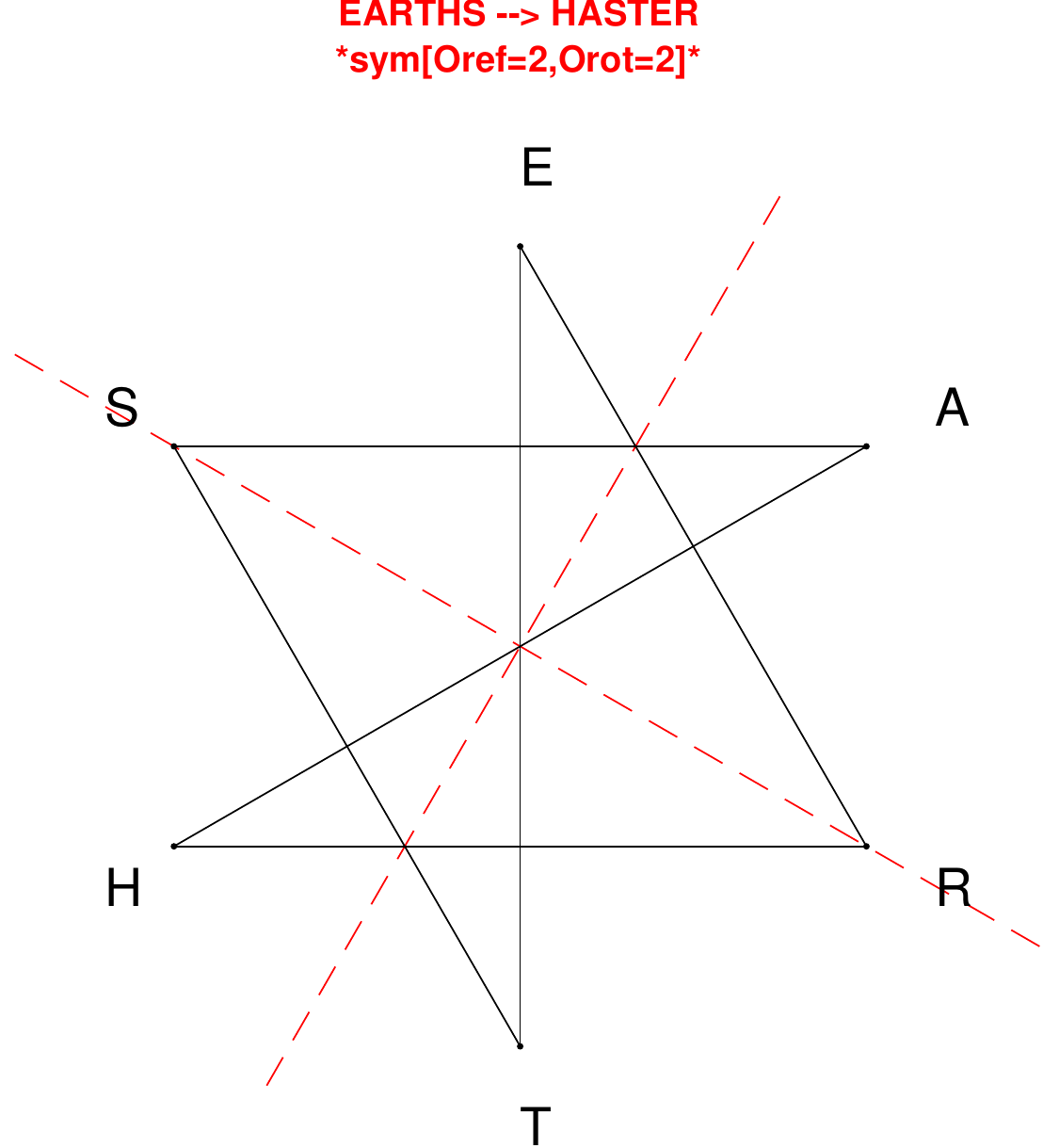}
\end{subfigure}
\hfill
\begin{subfigure}[T]{0.19\textwidth}
\centering
\includegraphics[width=\textwidth]{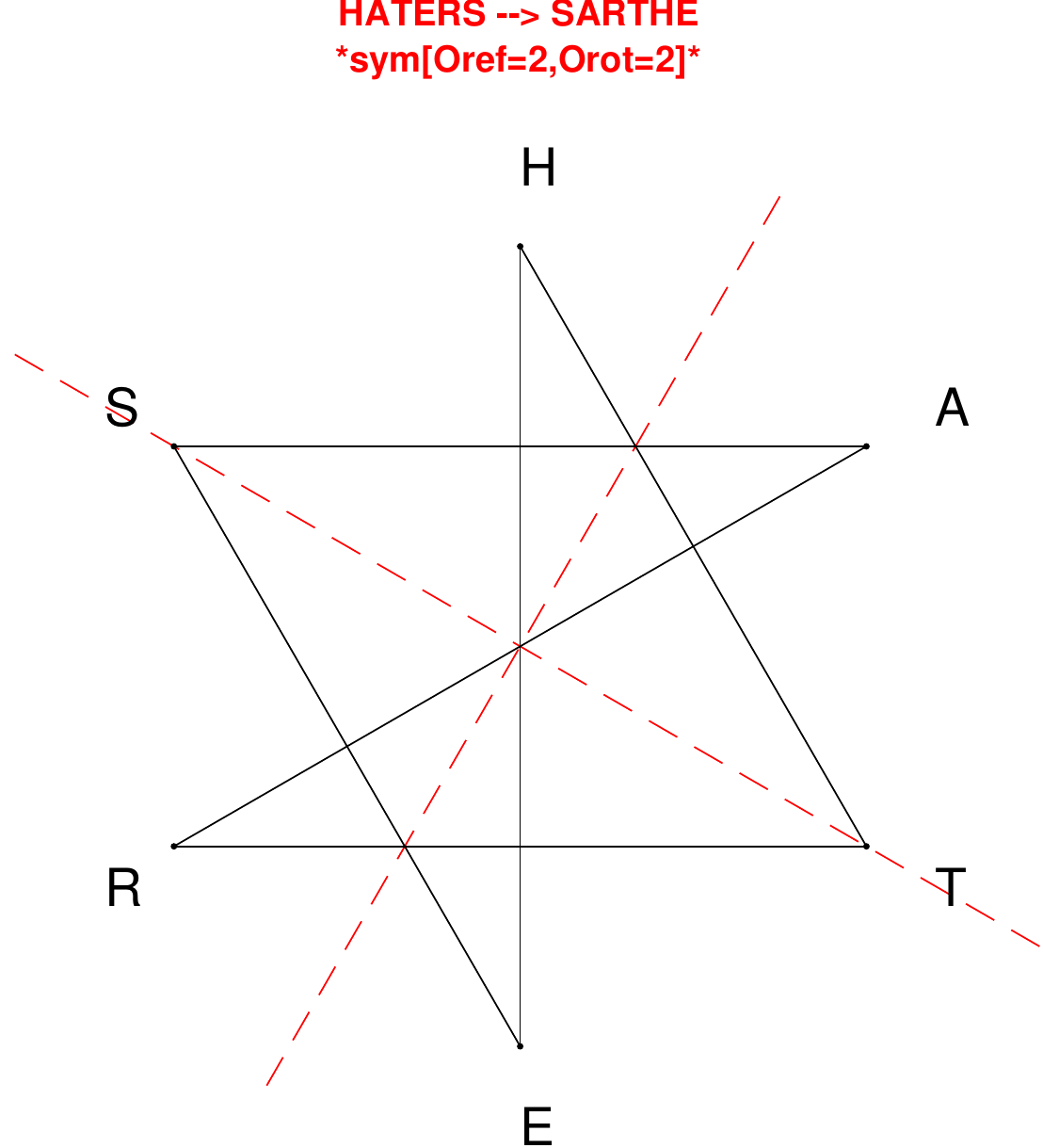}
\end{subfigure}
\hfill
\begin{subfigure}[T]{0.19\textwidth}
\centering
\includegraphics[width=\textwidth]{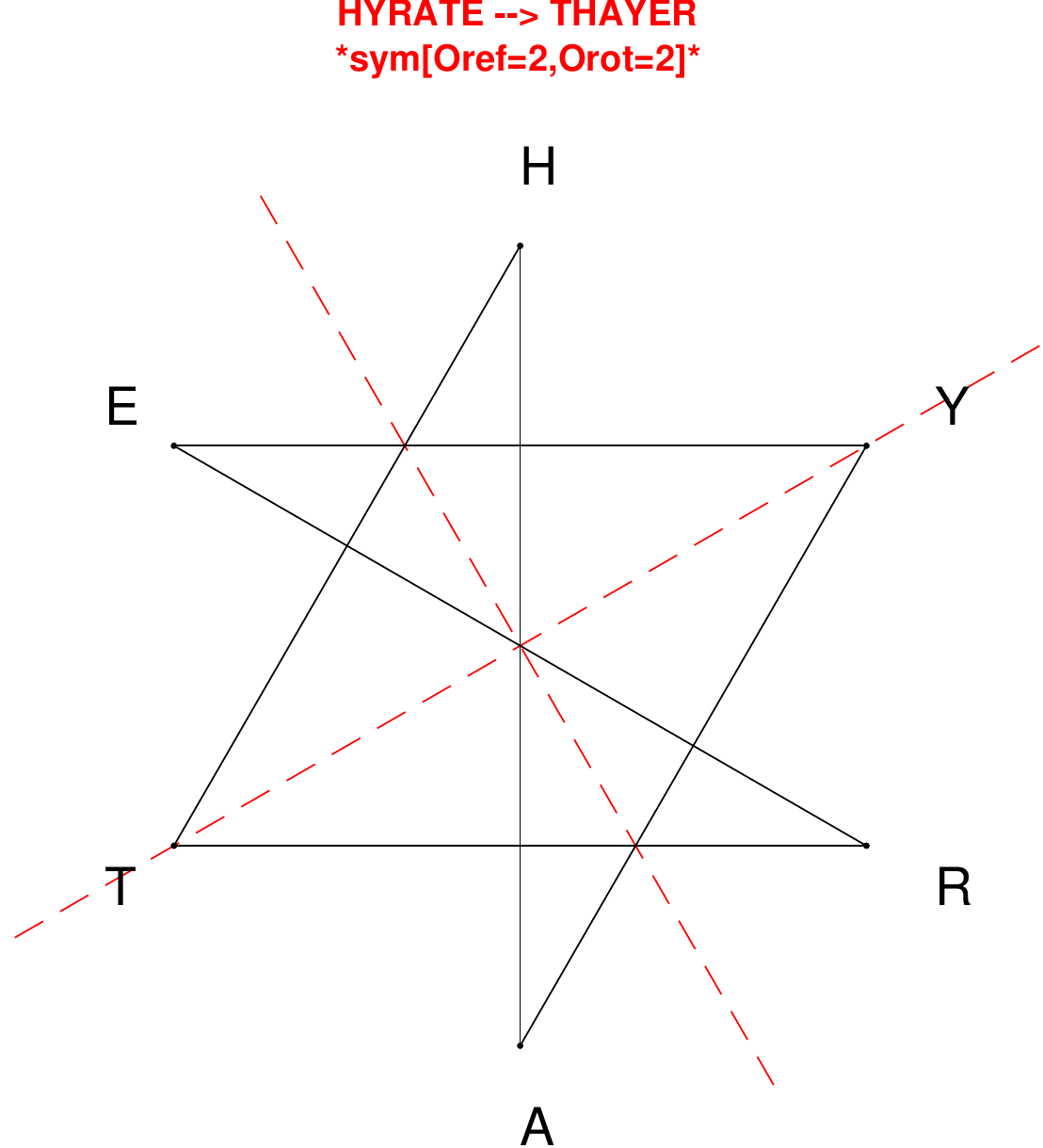}
\end{subfigure}
\hfill
\begin{subfigure}[T]{0.19\textwidth}
\centering
\includegraphics[width=\textwidth]{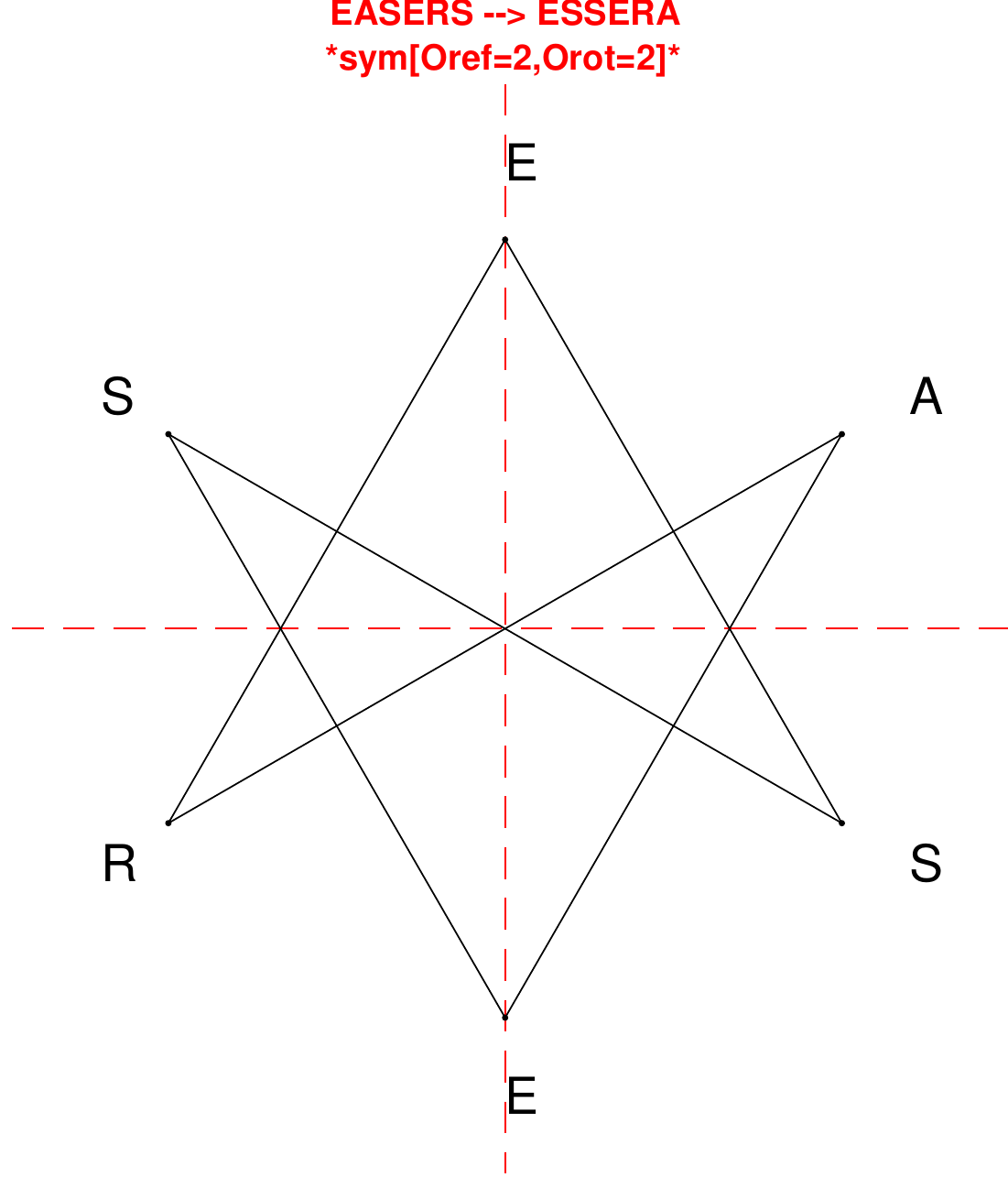}
\end{subfigure}
\end{figure}

\begin{figure}[H]
\centering
\begin{subfigure}[T]{0.19\textwidth}
\centering
\includegraphics[width=\textwidth]{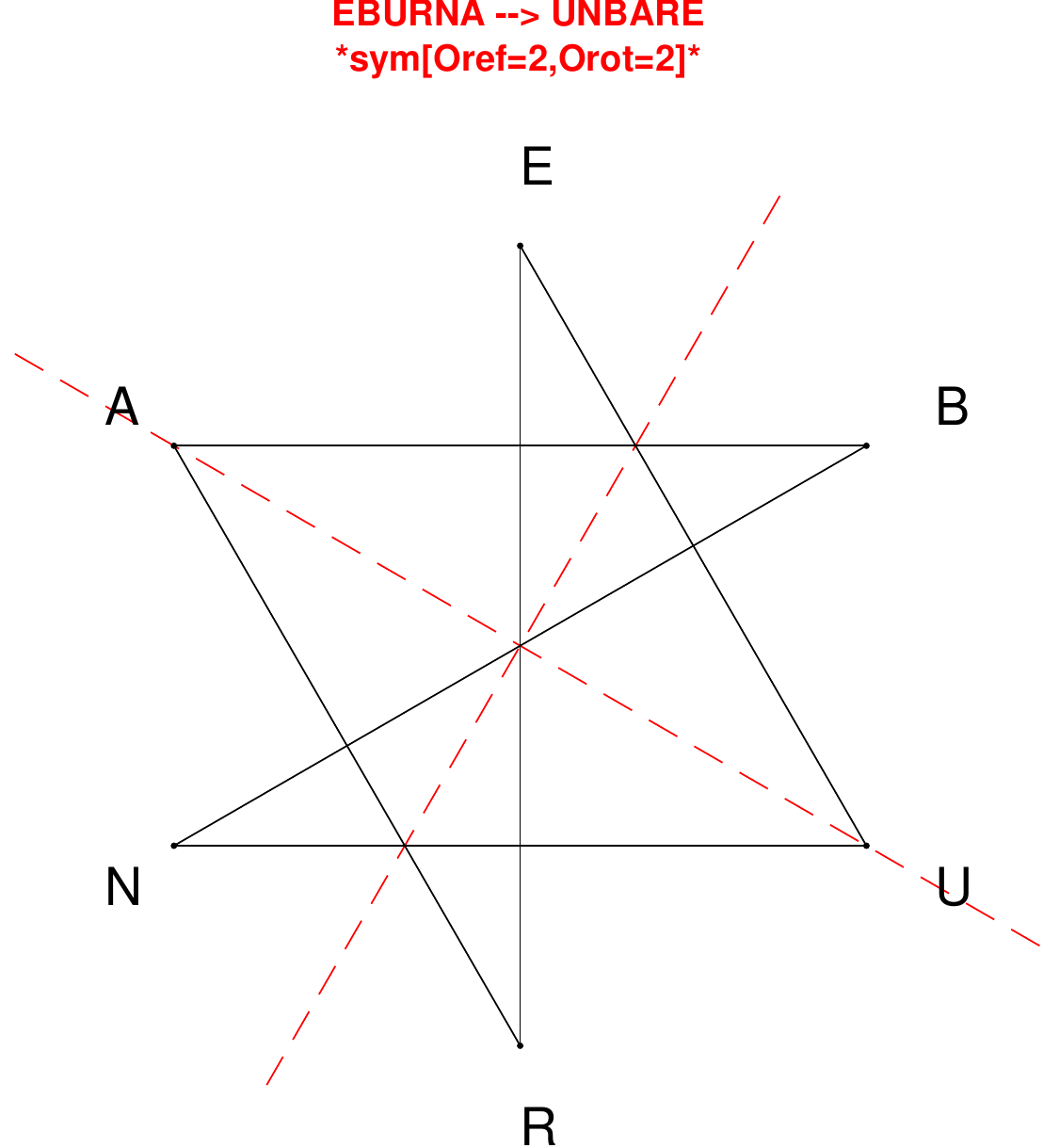}
\end{subfigure}
\hfill
\begin{subfigure}[T]{0.19\textwidth}
\centering
\includegraphics[width=\textwidth]{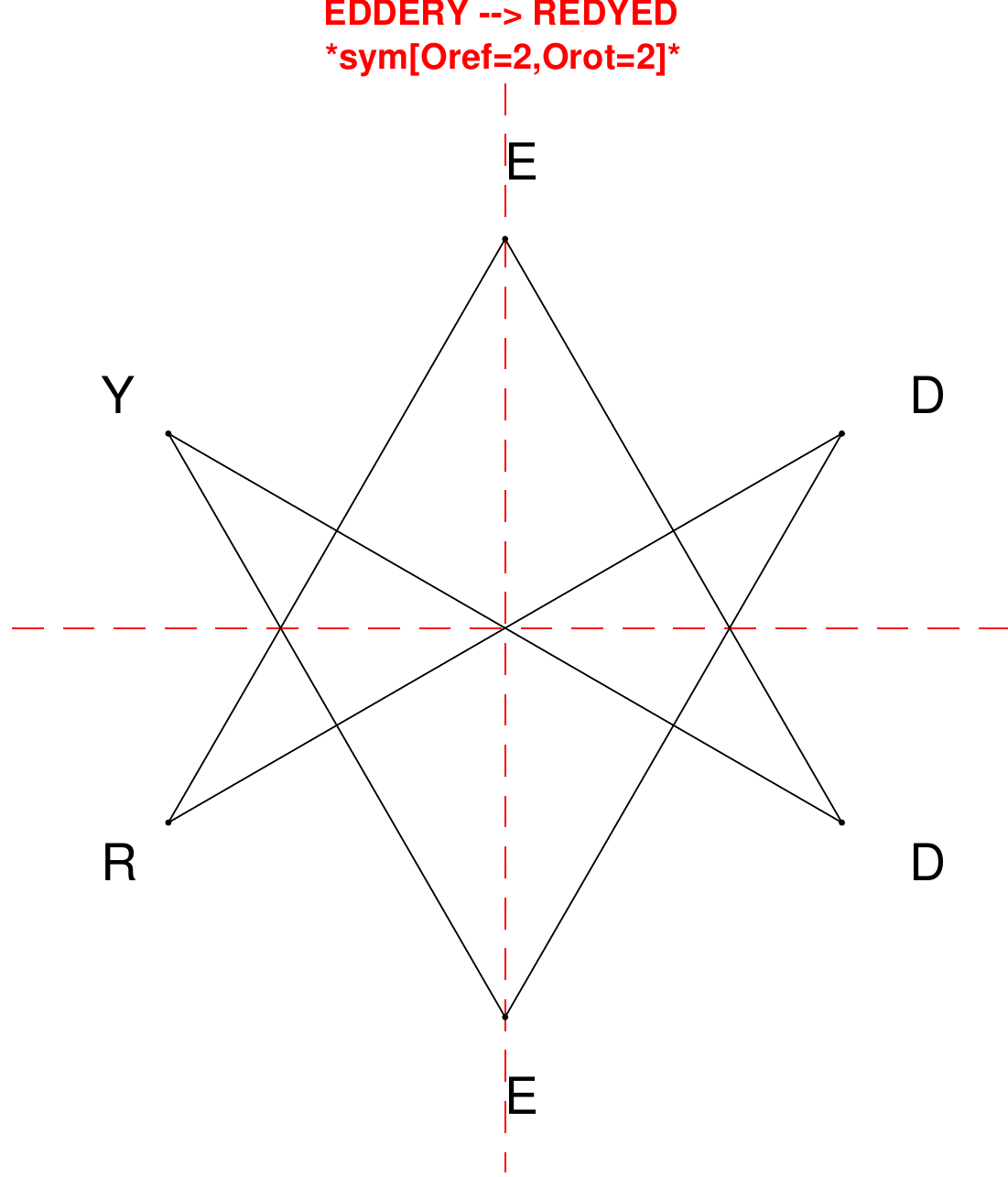}
\end{subfigure}
\hfill
\begin{subfigure}[T]{0.19\textwidth}
\centering
\includegraphics[width=\textwidth]{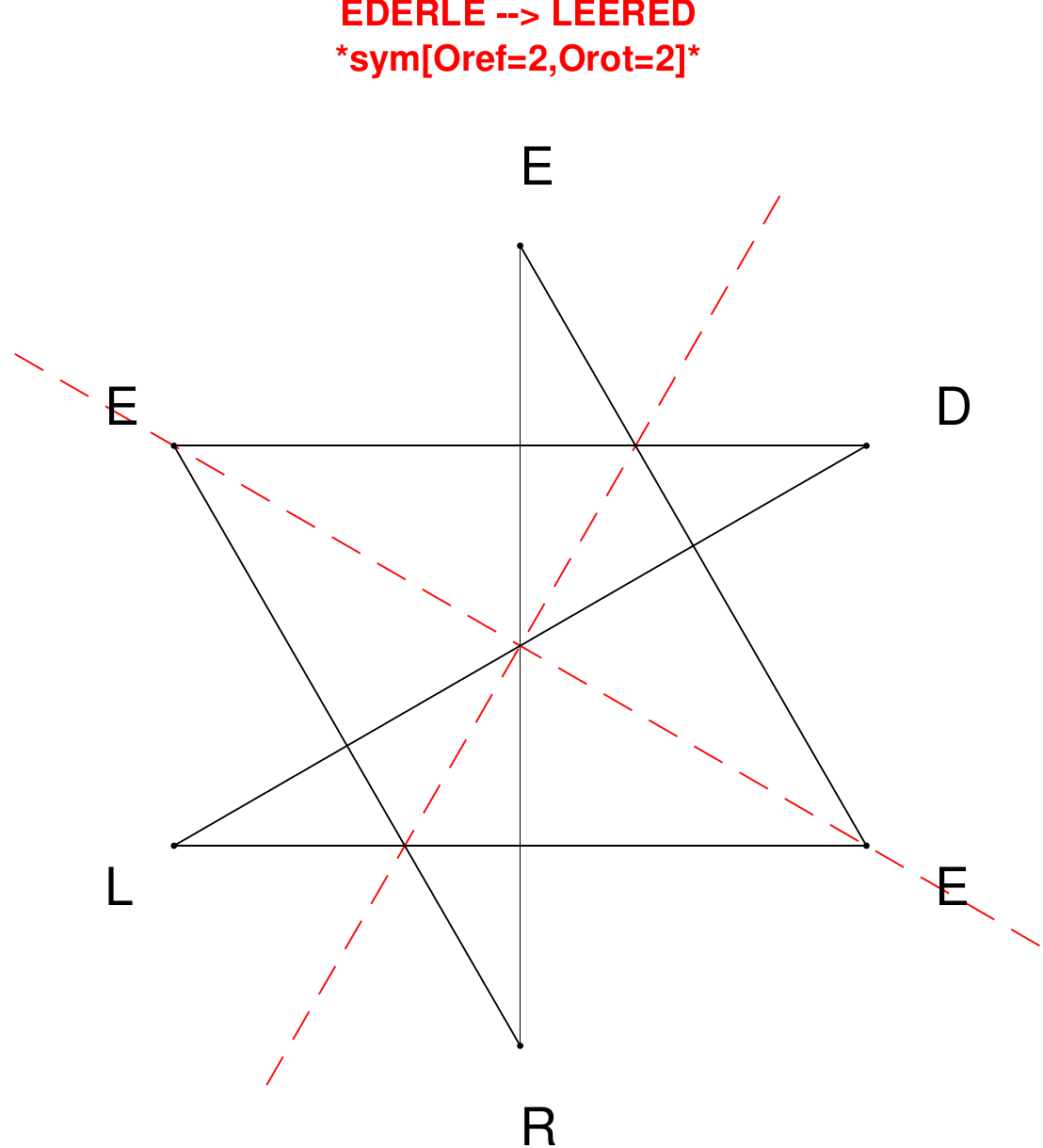}
\end{subfigure}
\hfill
\begin{subfigure}[T]{0.19\textwidth}
\centering
\includegraphics[width=\textwidth]{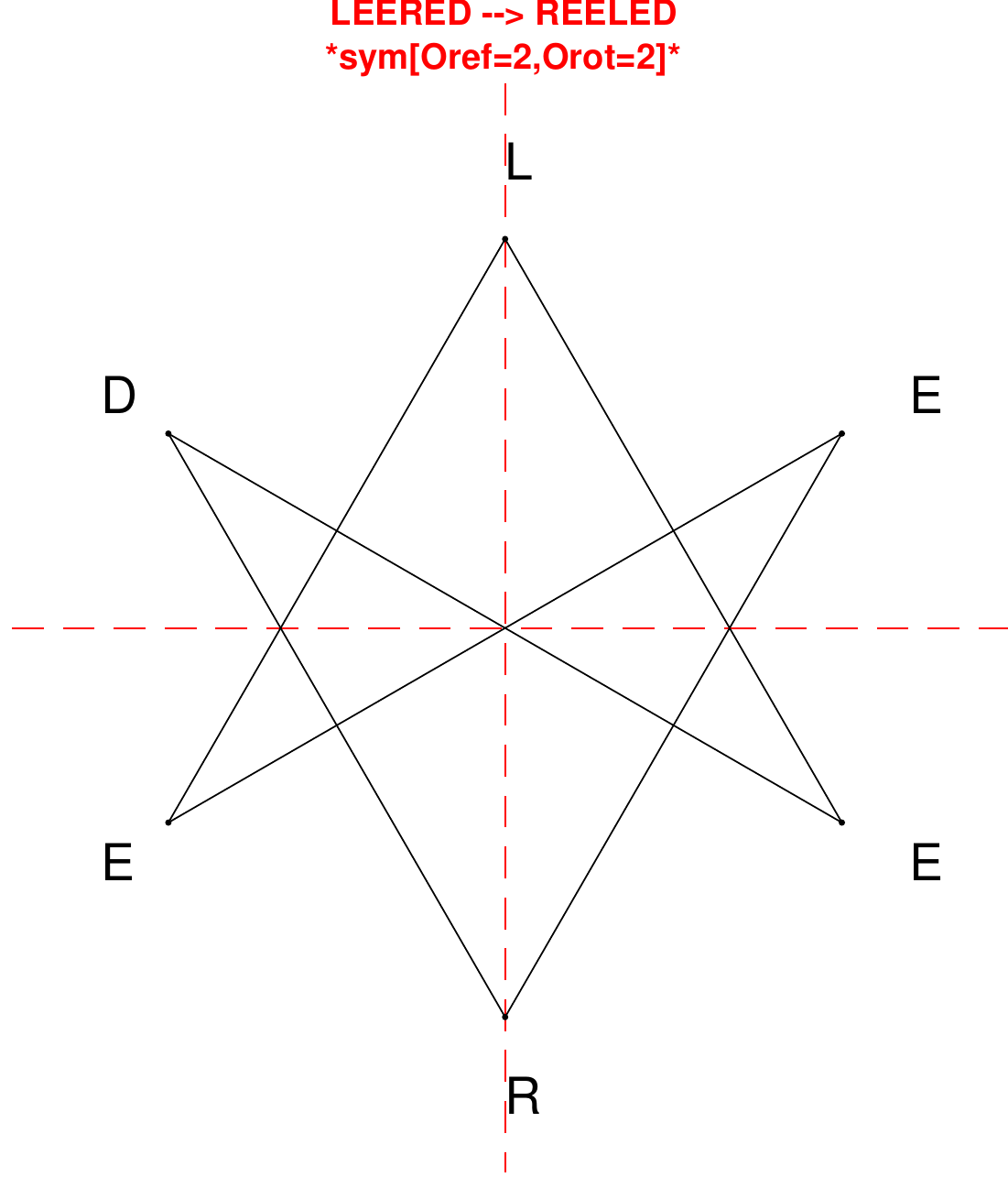}
\end{subfigure}
\hfill
\begin{subfigure}[T]{0.19\textwidth}
\centering
\includegraphics[width=\textwidth]{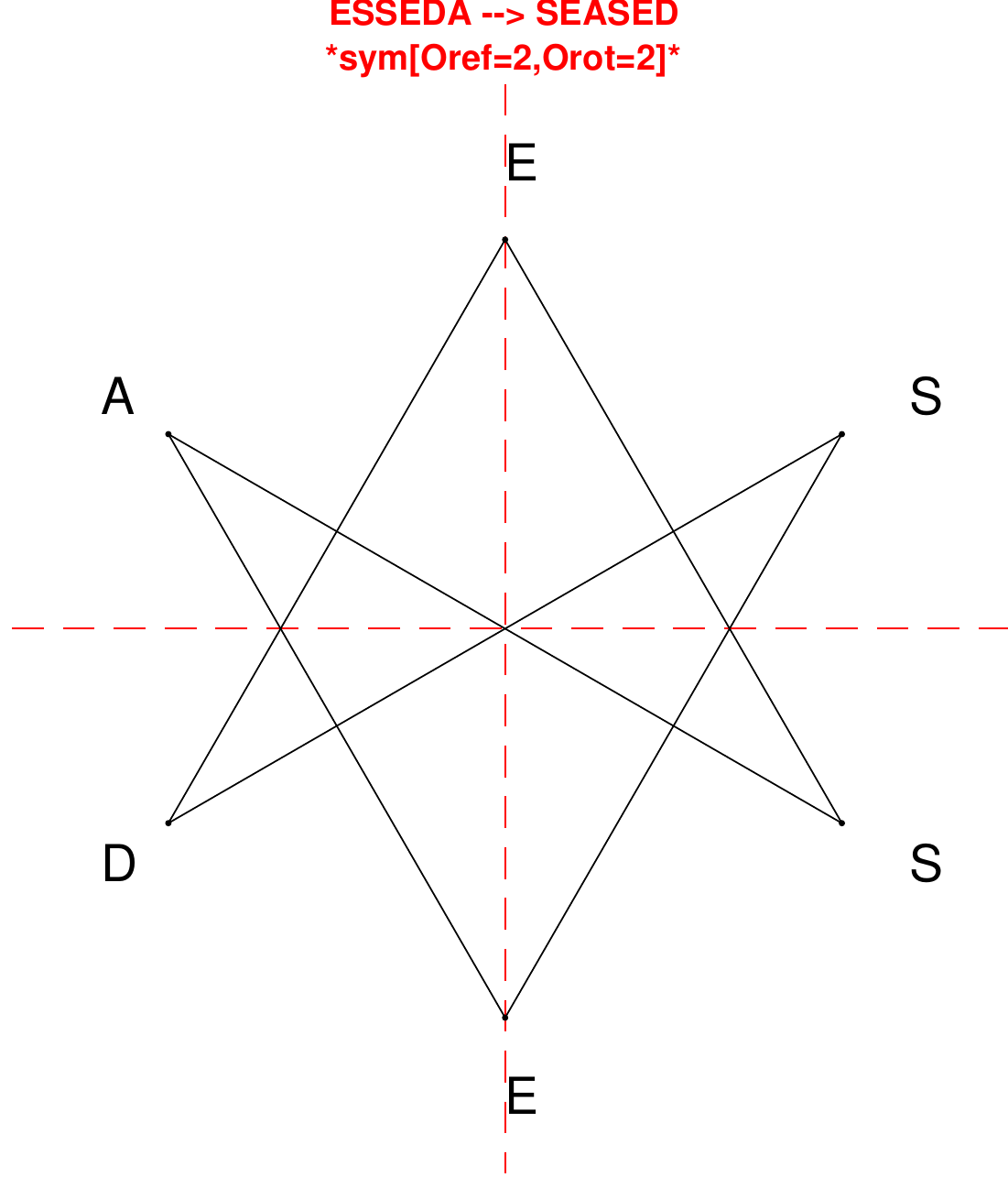}
\end{subfigure}
\end{figure}

\begin{figure}[H]
\centering
\begin{subfigure}[T]{0.19\textwidth}
\centering
\includegraphics[width=\textwidth]{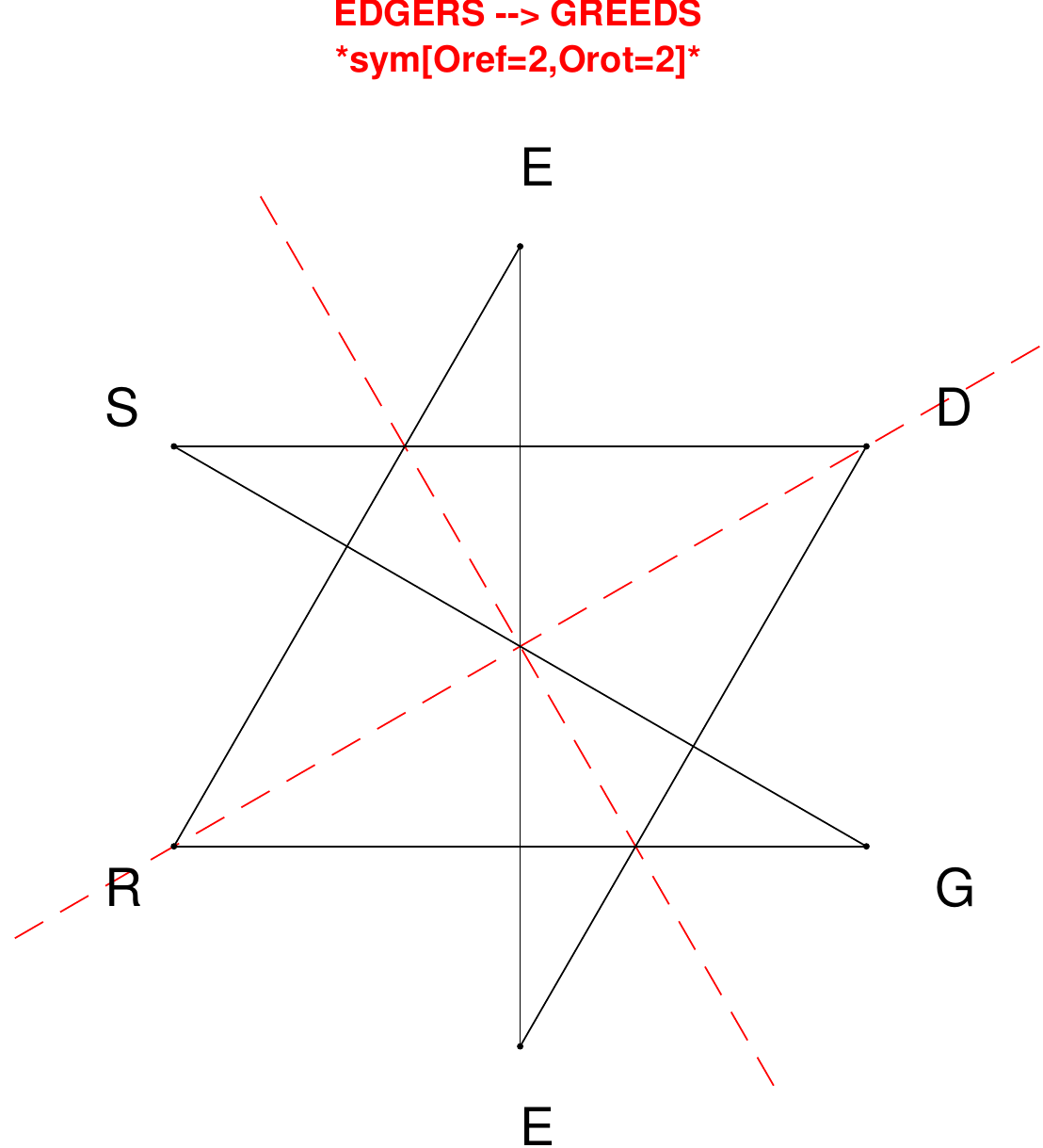}
\end{subfigure}
\hfill
\begin{subfigure}[T]{0.19\textwidth}
\centering
\includegraphics[width=\textwidth]{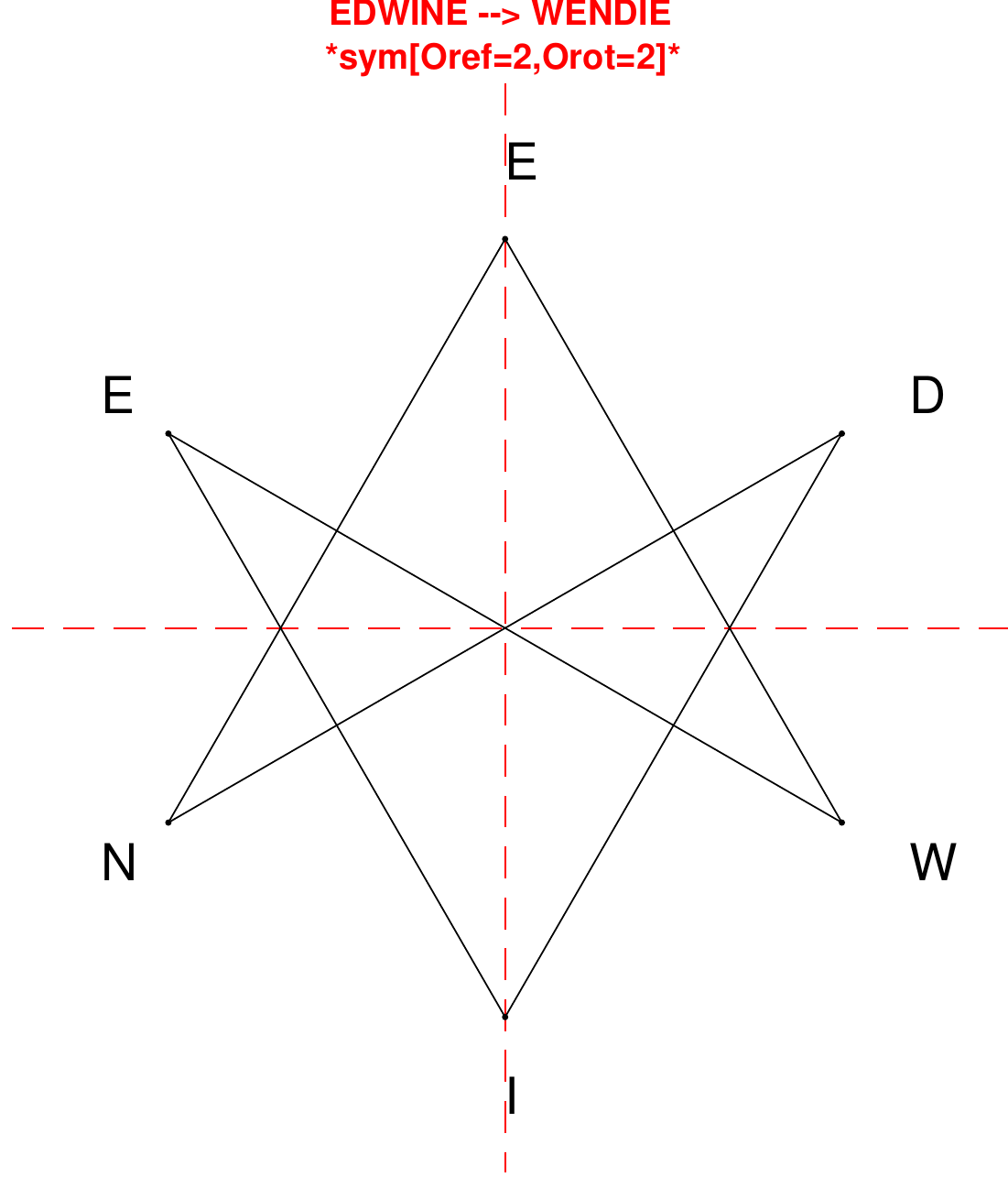}
\end{subfigure}
\hfill
\begin{subfigure}[T]{0.19\textwidth}
\centering
\includegraphics[width=\textwidth]{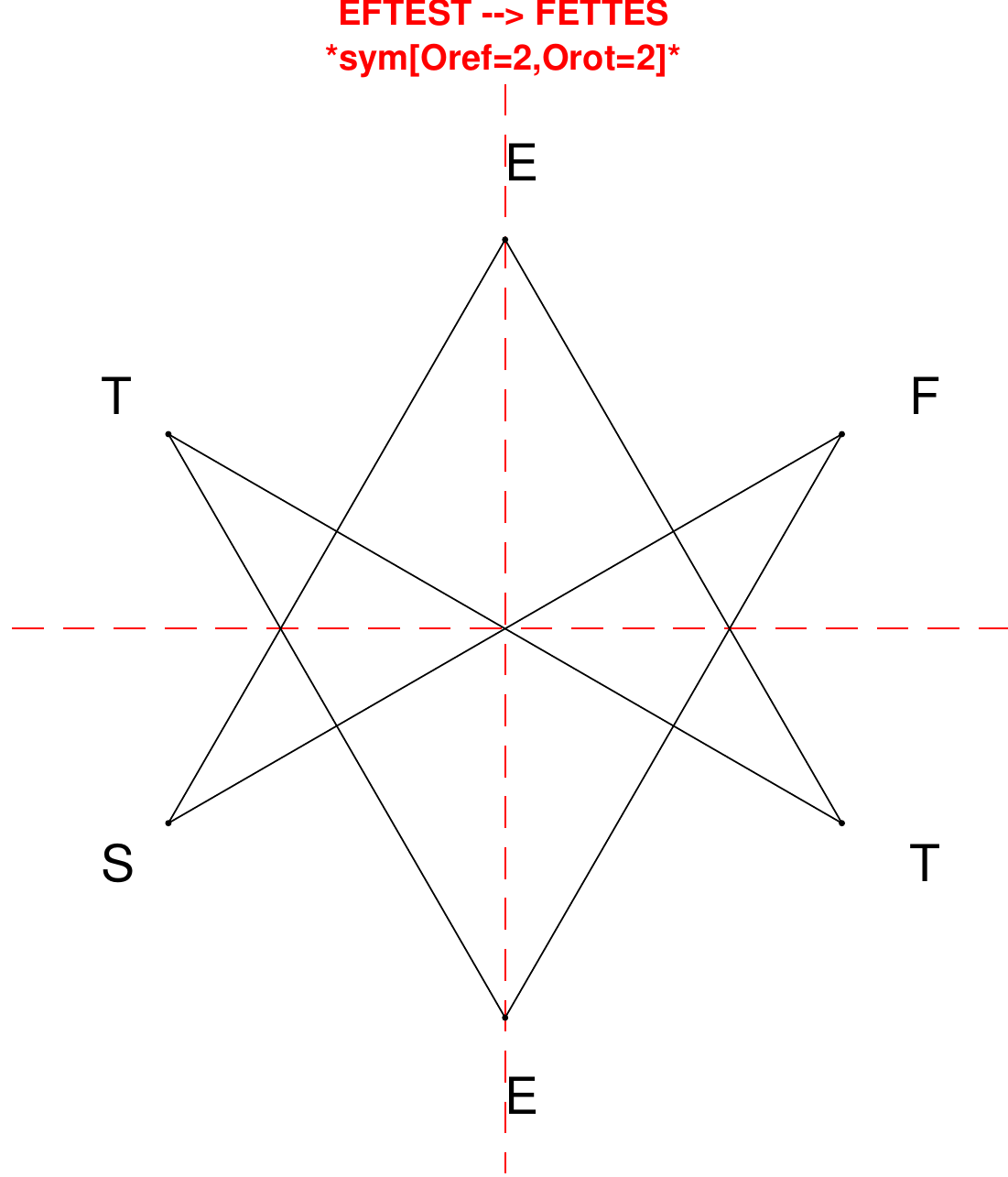}
\end{subfigure}
\hfill
\begin{subfigure}[T]{0.19\textwidth}
\centering
\includegraphics[width=\textwidth]{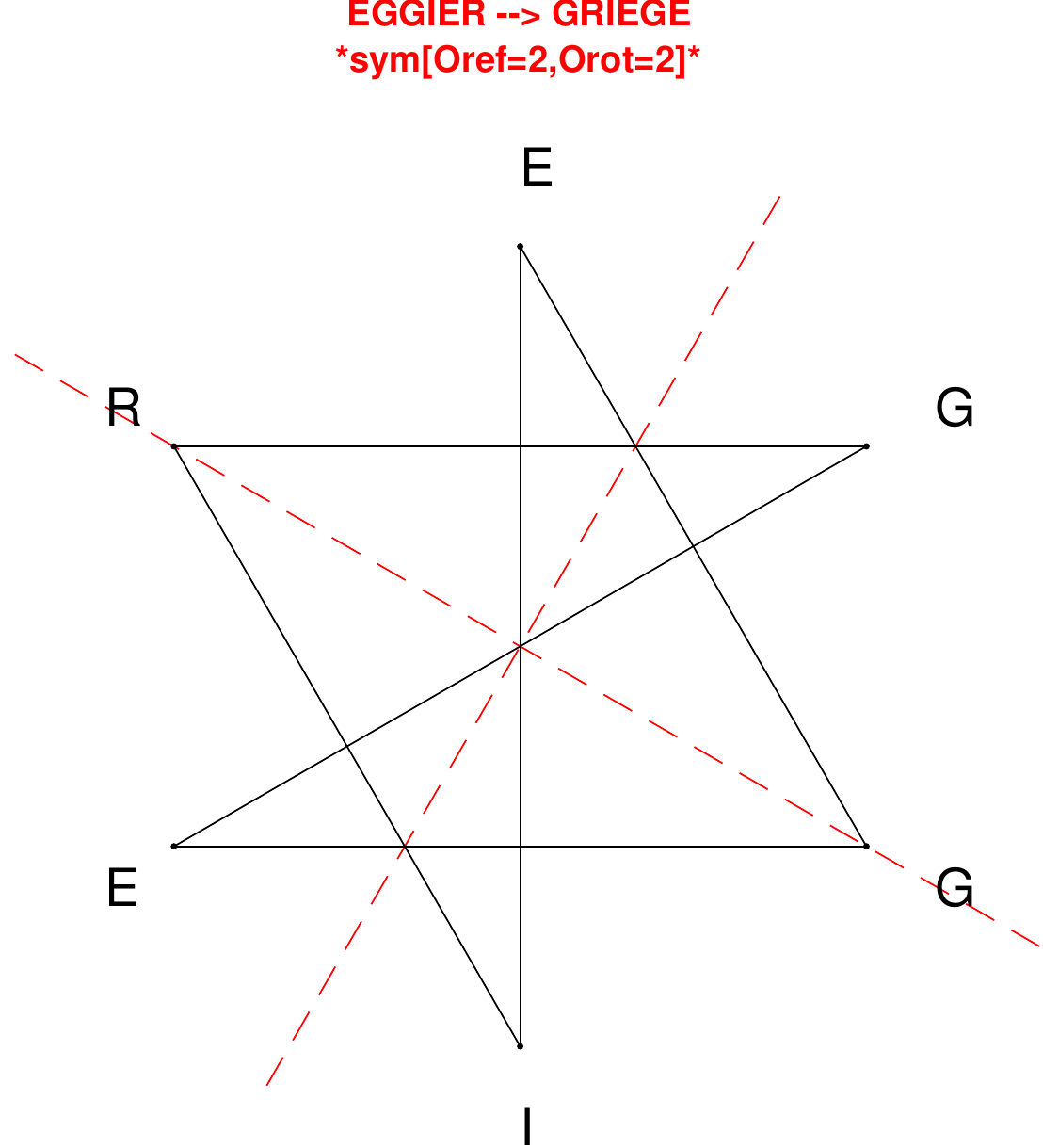}
\end{subfigure}
\hfill
\begin{subfigure}[T]{0.19\textwidth}
\centering
\includegraphics[width=\textwidth]{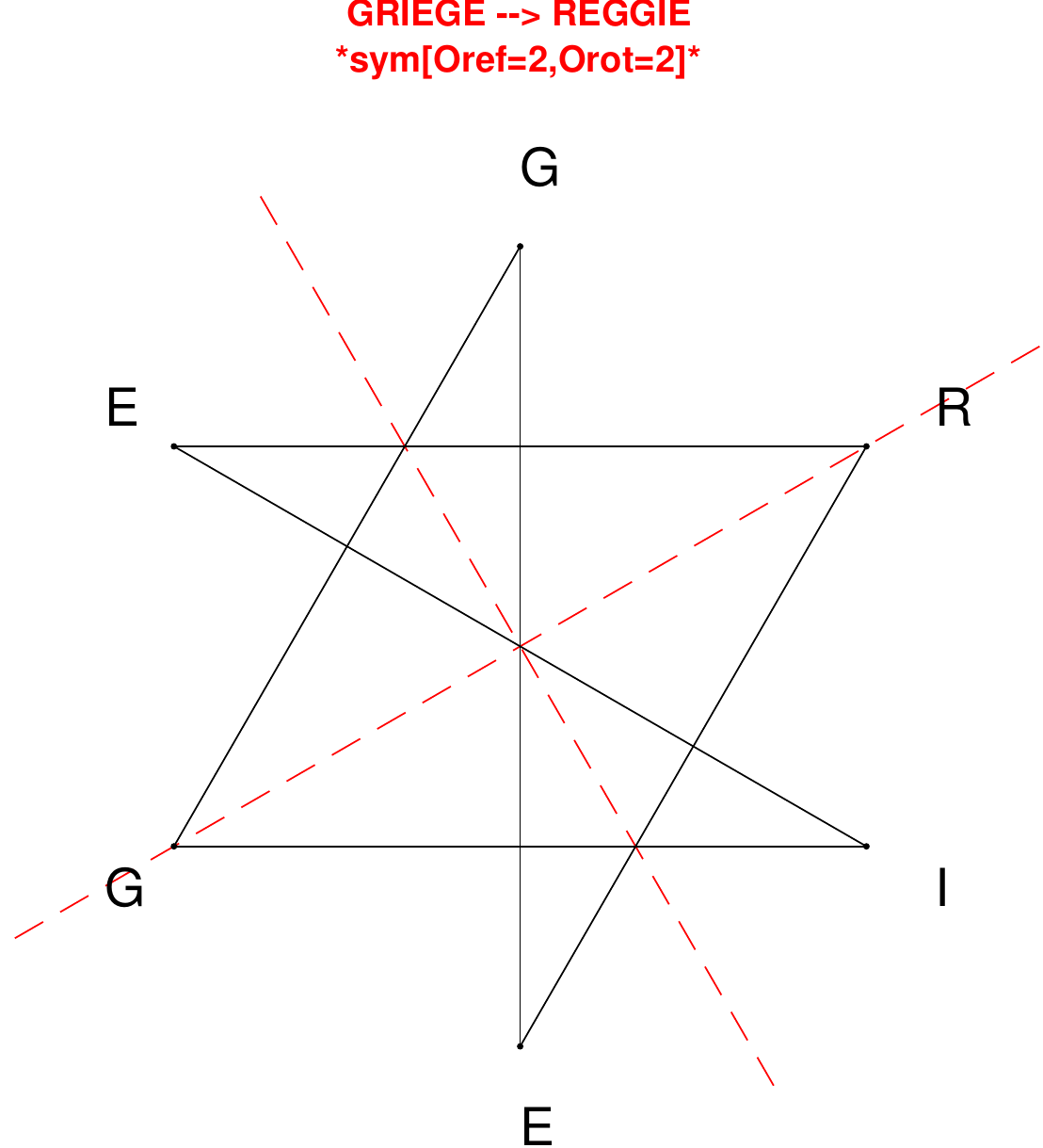}
\end{subfigure}
\end{figure}

\begin{figure}[H]
\centering
\begin{subfigure}[T]{0.19\textwidth}
\centering
\includegraphics[width=\textwidth]{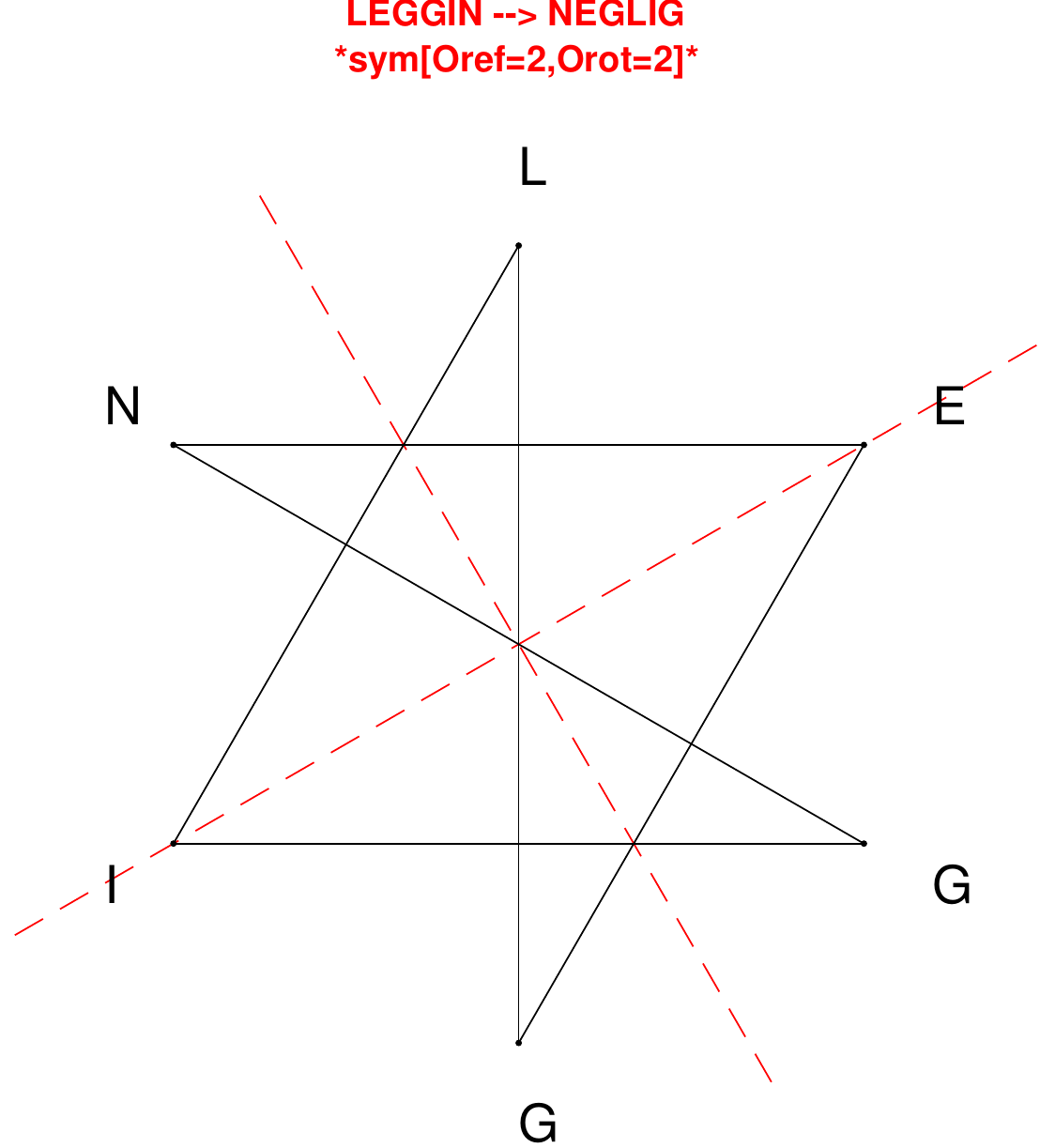}
\end{subfigure}
\hfill
\begin{subfigure}[T]{0.19\textwidth}
\centering
\includegraphics[width=\textwidth]{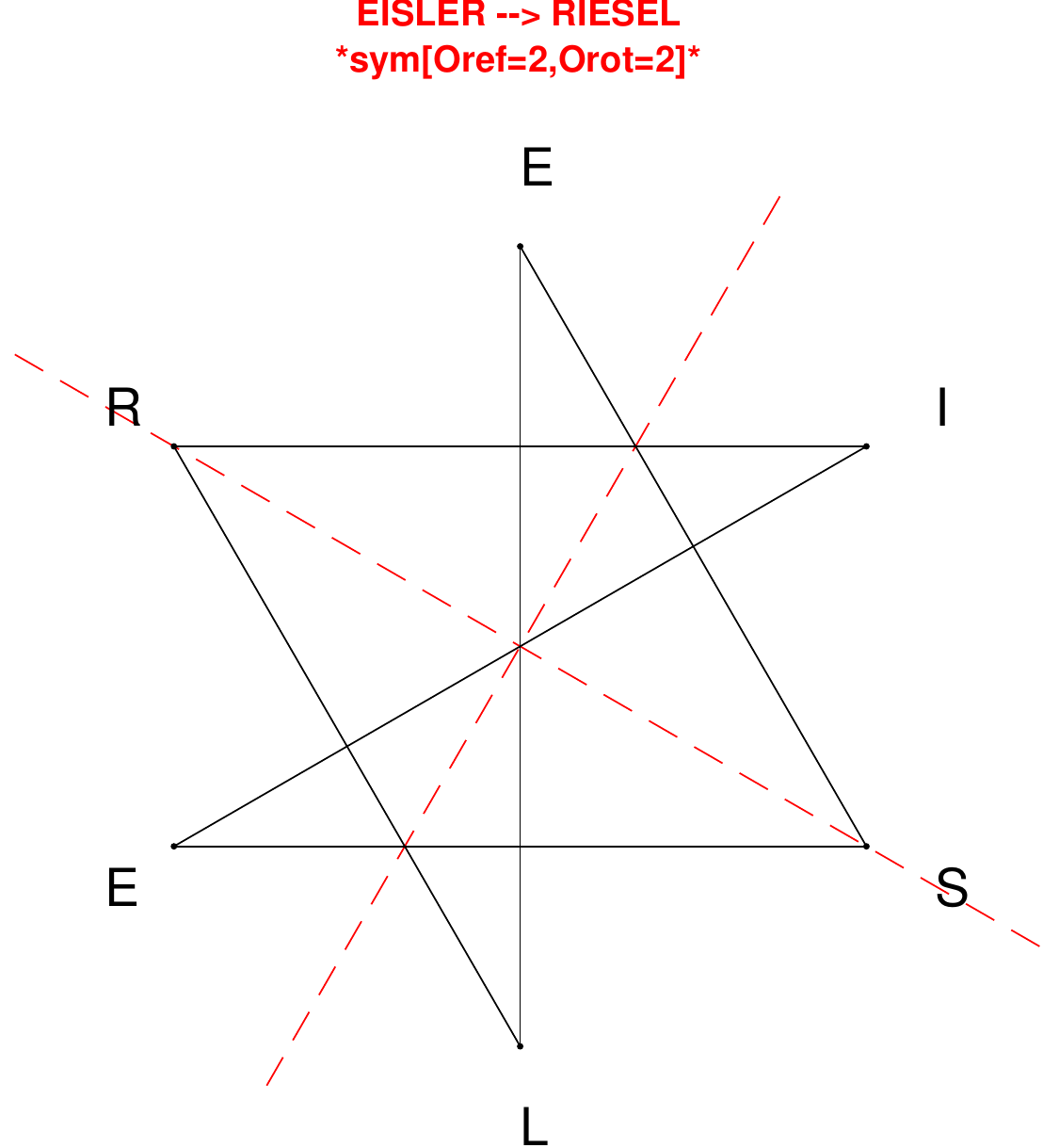}
\end{subfigure}
\hfill
\begin{subfigure}[T]{0.19\textwidth}
\centering
\includegraphics[width=\textwidth]{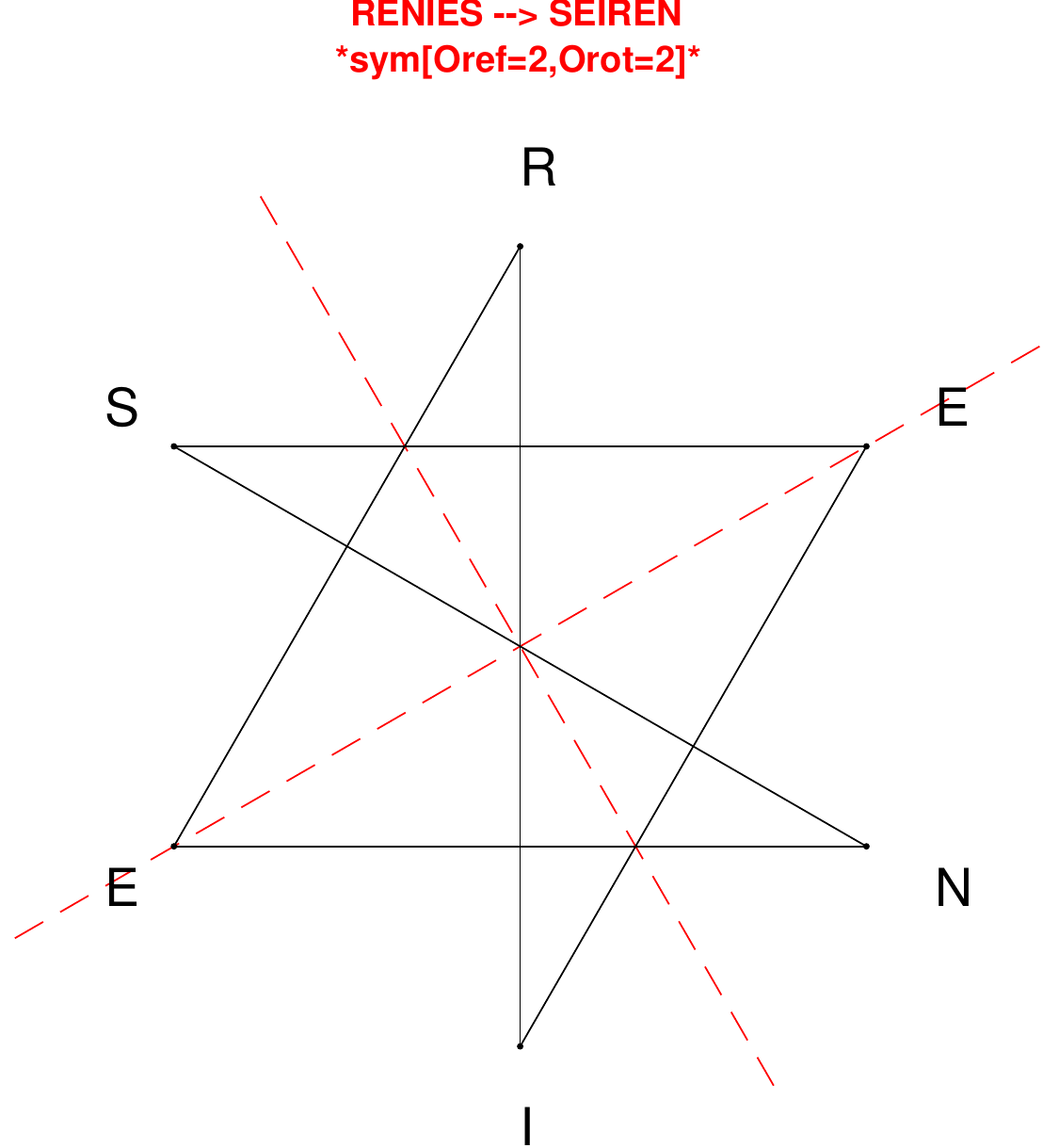}
\end{subfigure}
\hfill
\begin{subfigure}[T]{0.19\textwidth}
\centering
\includegraphics[width=\textwidth]{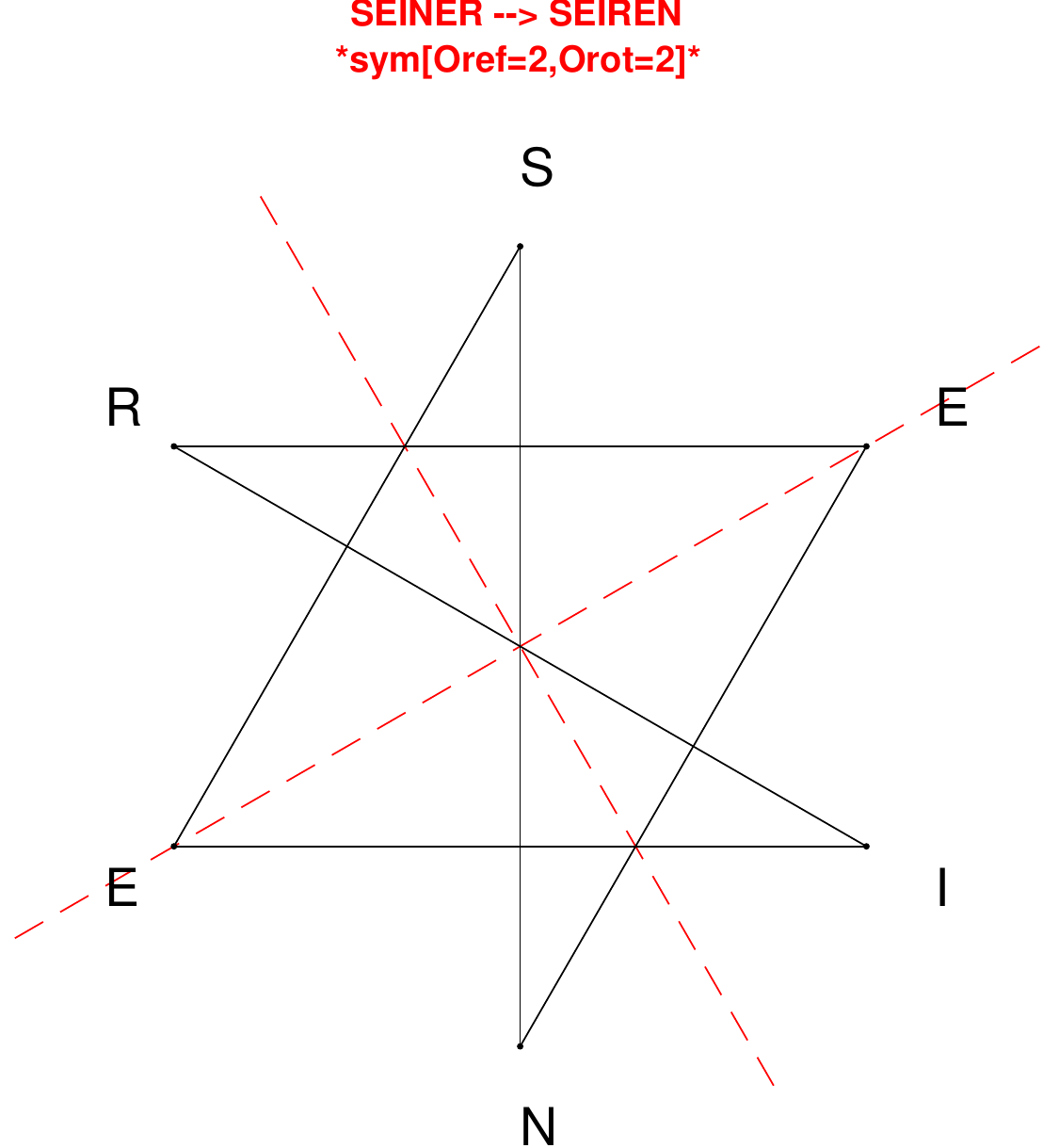}
\end{subfigure}
\hfill
\begin{subfigure}[T]{0.19\textwidth}
\centering
\includegraphics[width=\textwidth]{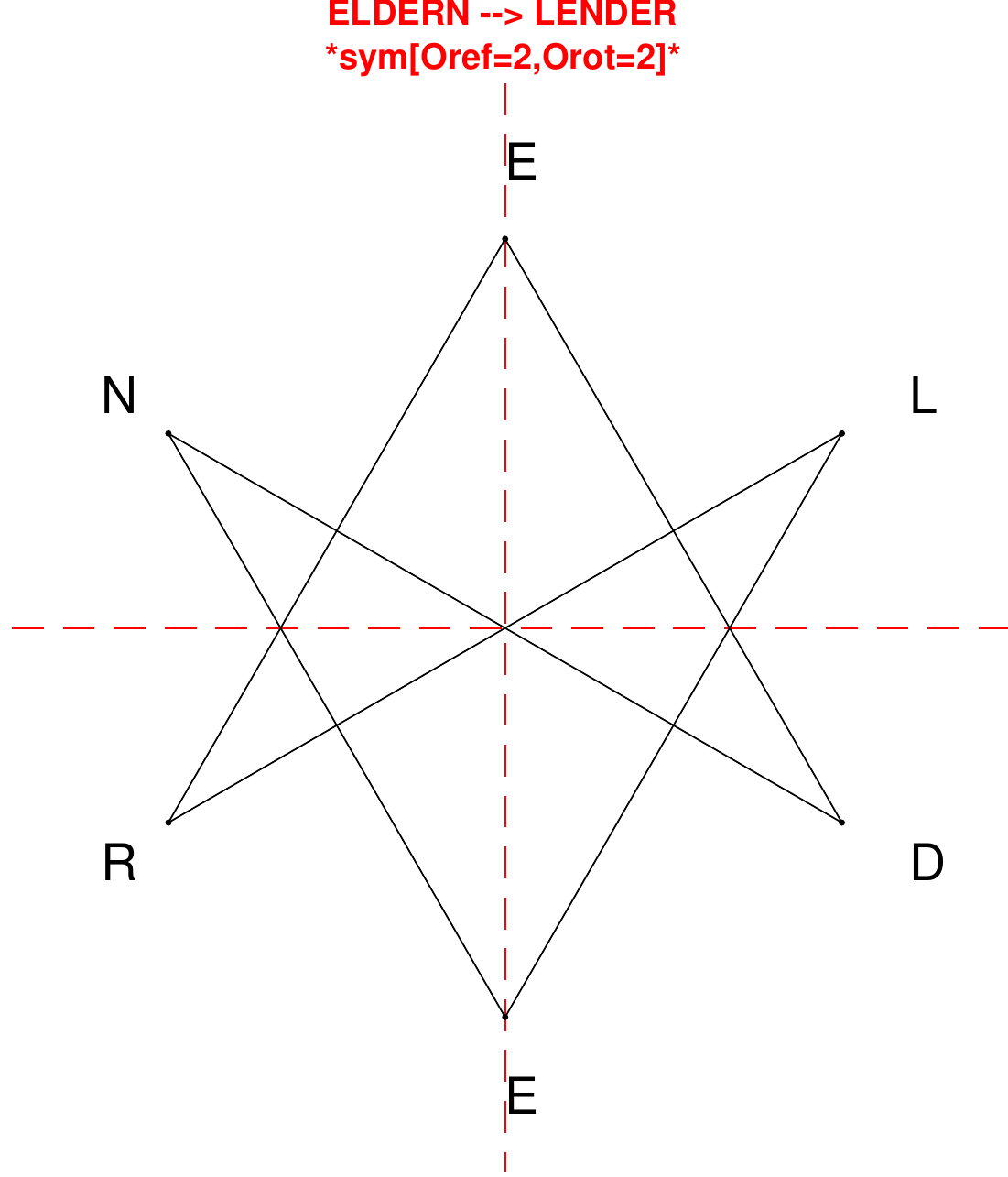}
\end{subfigure}
\end{figure}

\begin{figure}[H]
\centering
\begin{subfigure}[T]{0.19\textwidth}
\centering
\includegraphics[width=\textwidth]{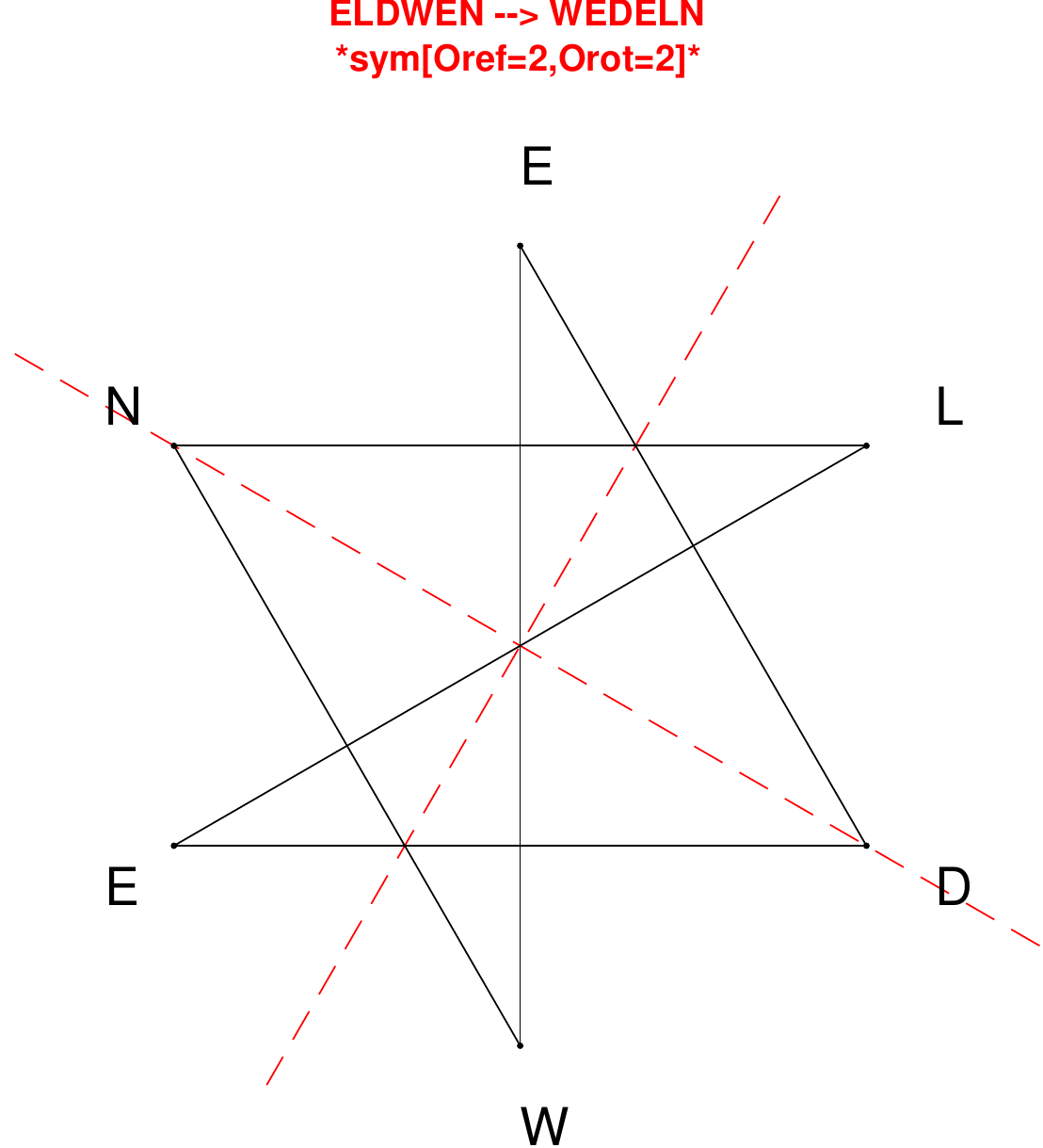}
\end{subfigure}
\hfill
\begin{subfigure}[T]{0.19\textwidth}
\centering
\includegraphics[width=\textwidth]{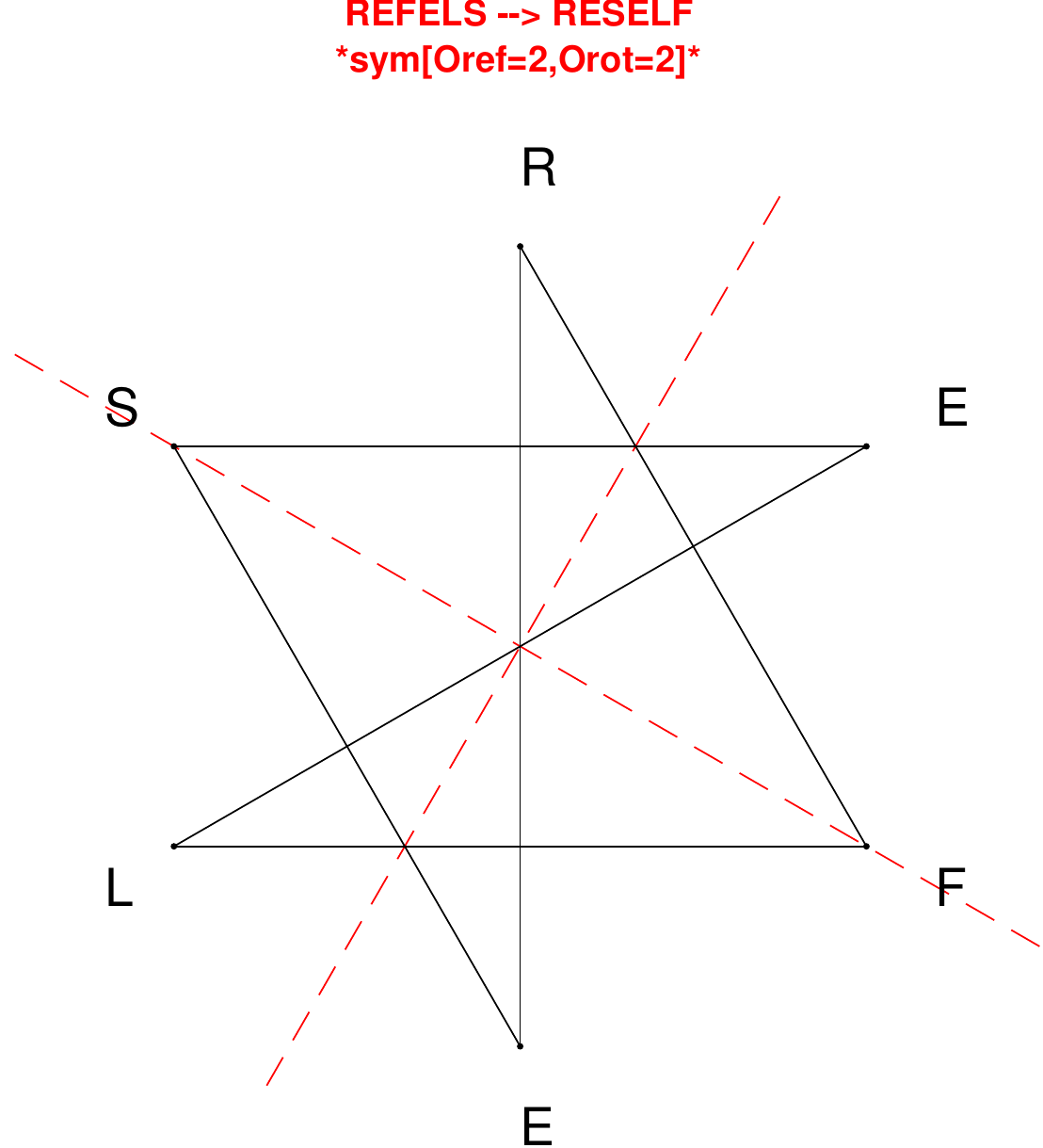}
\end{subfigure}
\hfill
\begin{subfigure}[T]{0.19\textwidth}
\centering
\includegraphics[width=\textwidth]{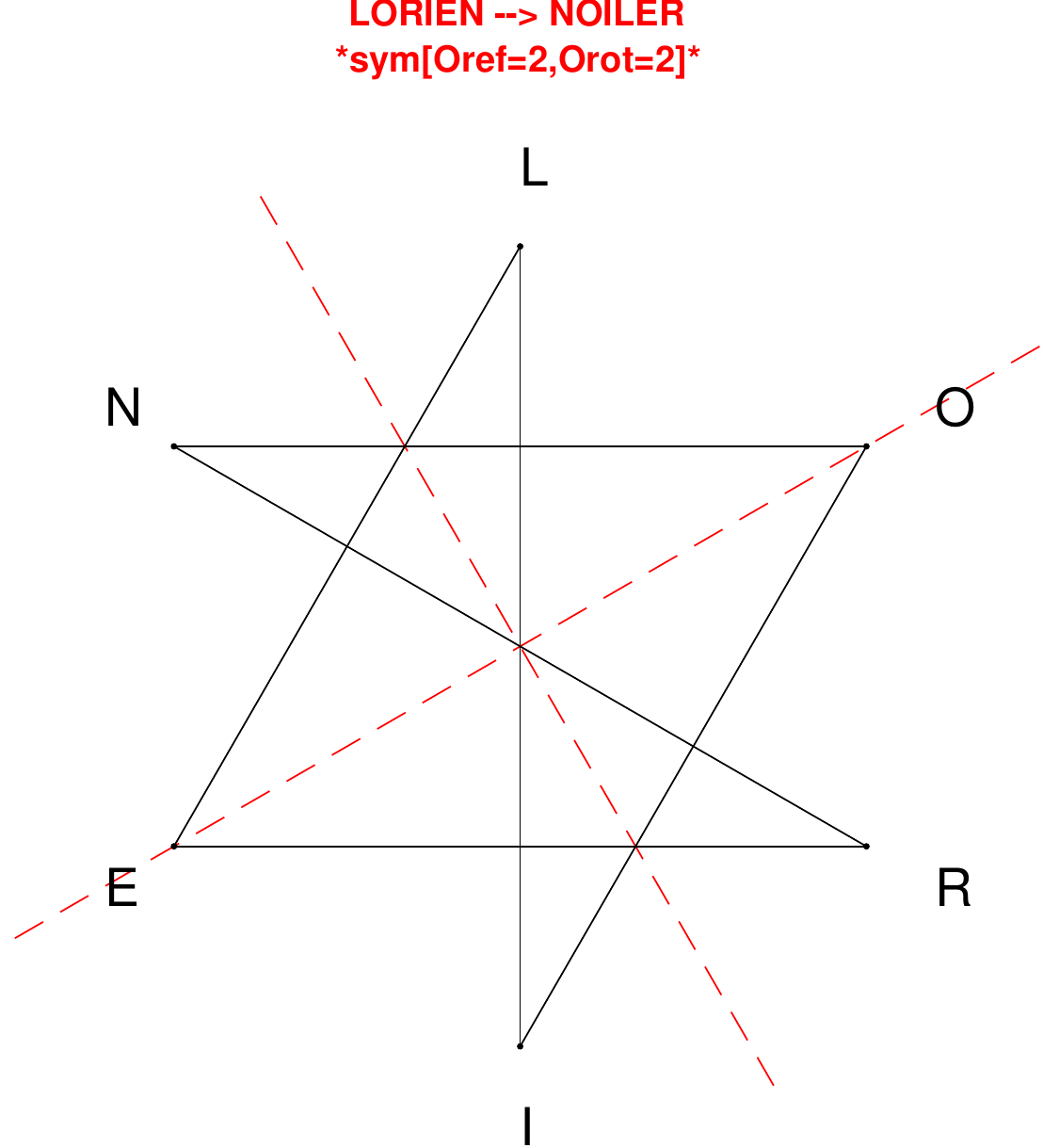}
\end{subfigure}
\hfill
\begin{subfigure}[T]{0.19\textwidth}
\centering
\includegraphics[width=\textwidth]{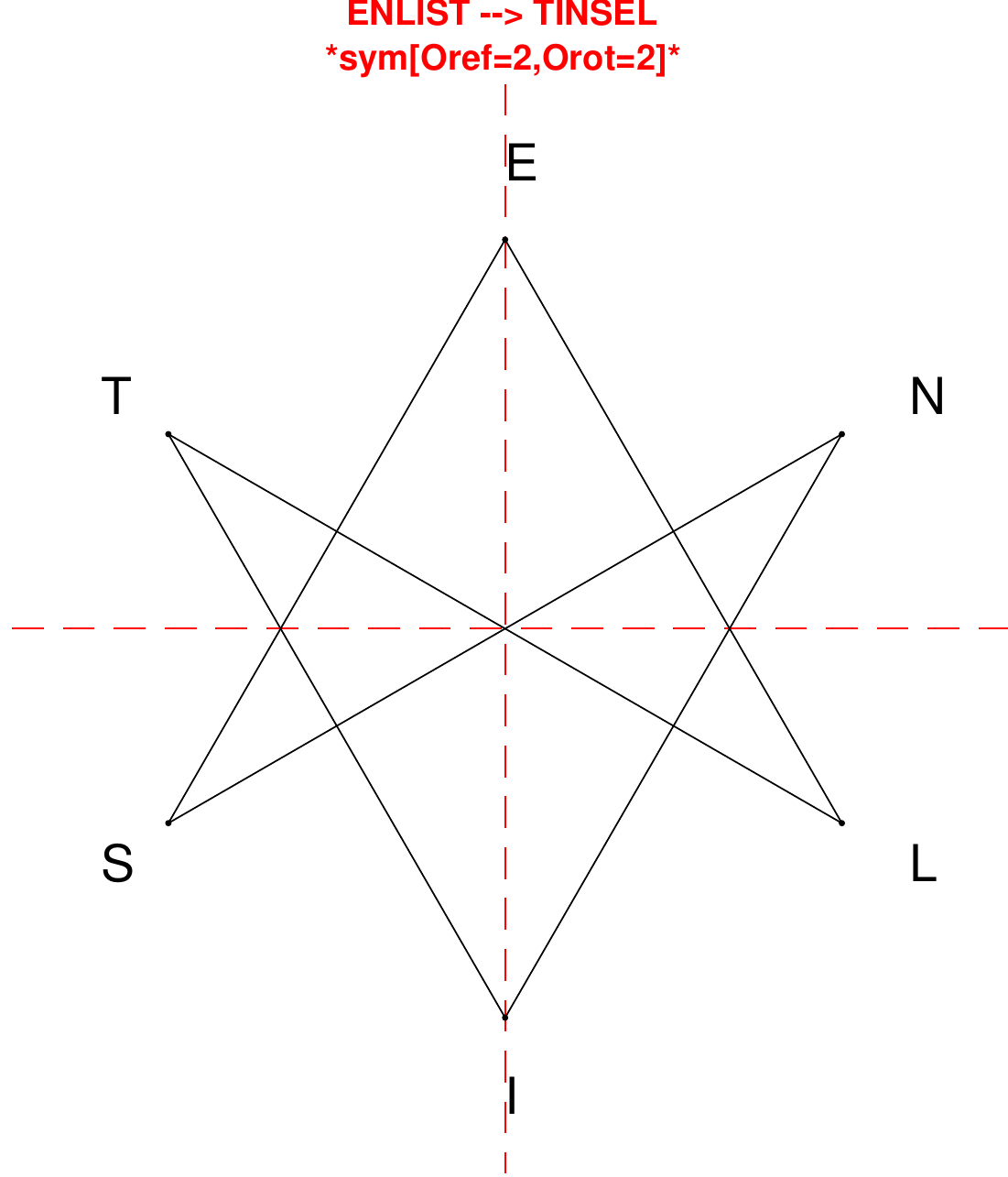}
\end{subfigure}
\hfill
\begin{subfigure}[T]{0.19\textwidth}
\centering
\includegraphics[width=\textwidth]{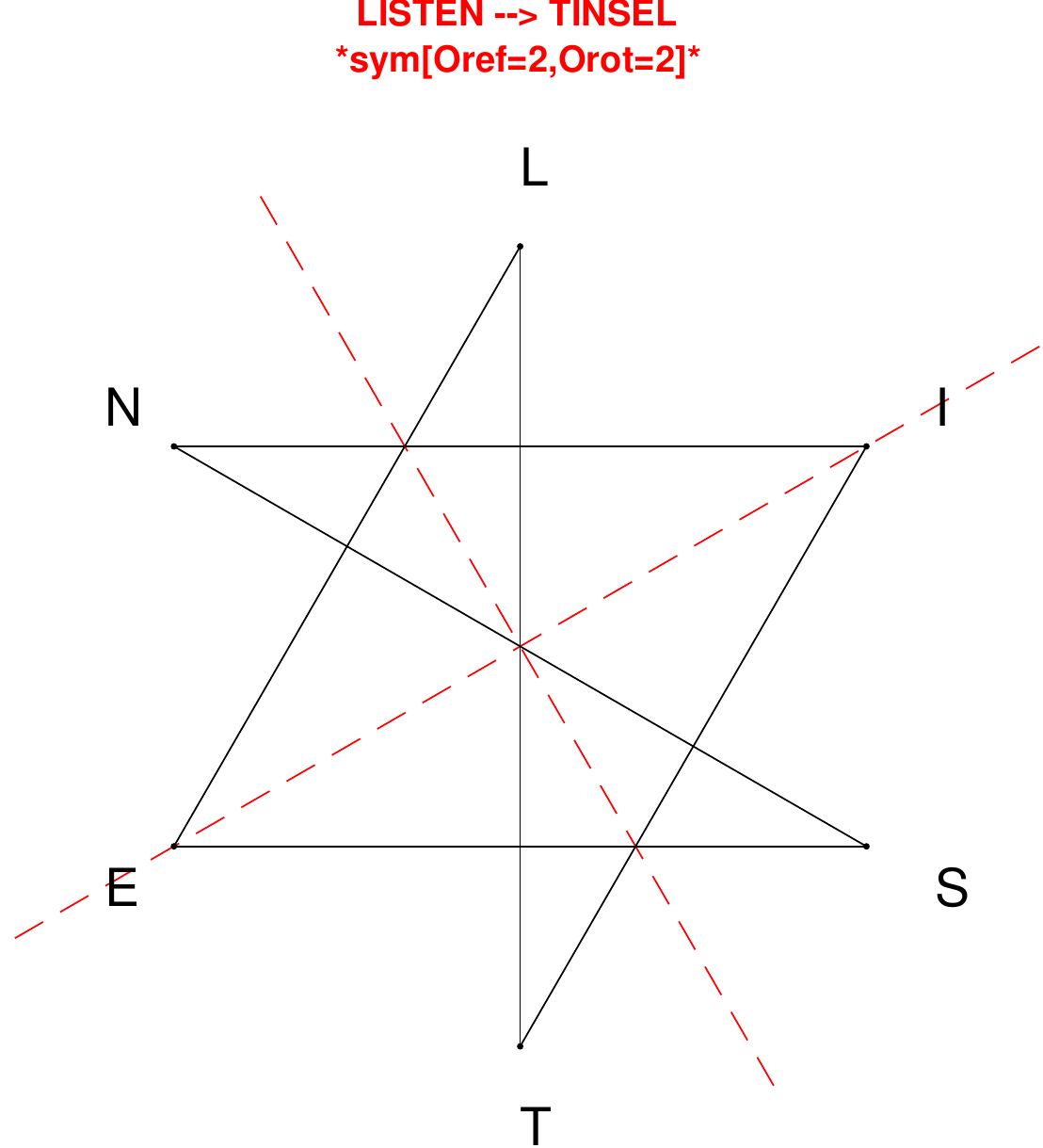}
\end{subfigure}
\end{figure}

\begin{figure}[H]
\centering
\begin{subfigure}[T]{0.19\textwidth}
\centering
\includegraphics[width=\textwidth]{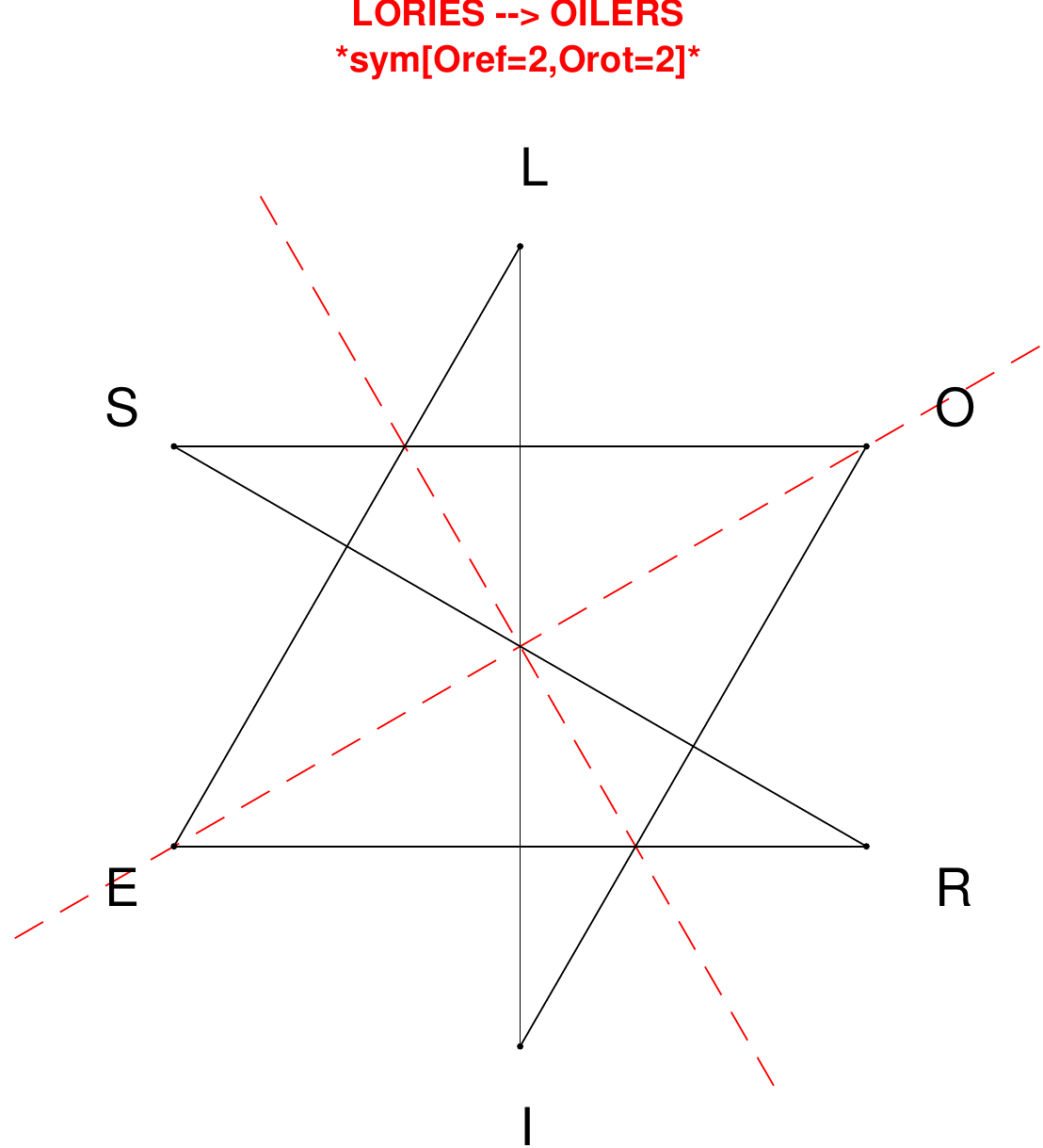}
\end{subfigure}
\hfill
\begin{subfigure}[T]{0.19\textwidth}
\centering
\includegraphics[width=\textwidth]{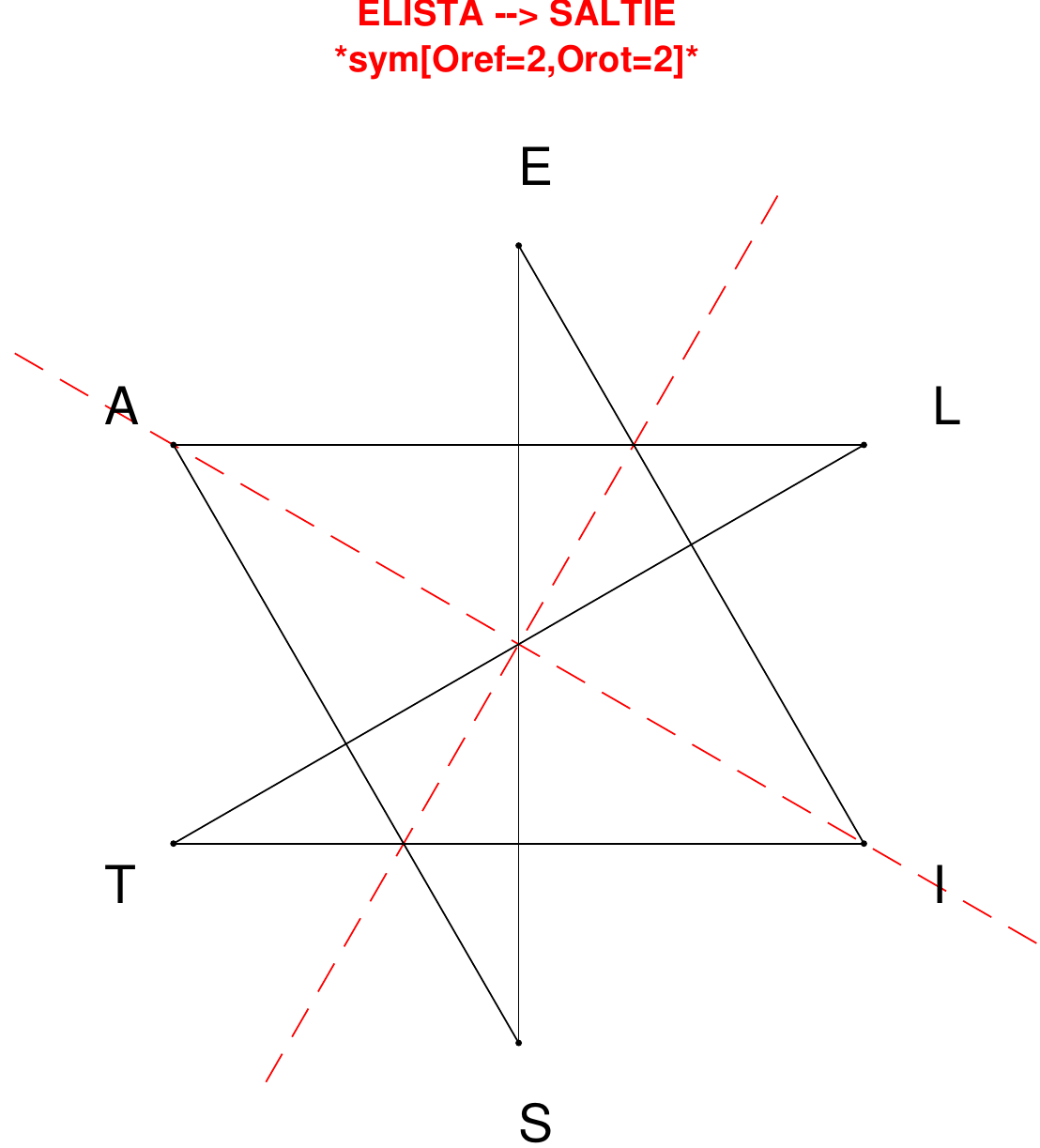}
\end{subfigure}
\hfill
\begin{subfigure}[T]{0.19\textwidth}
\centering
\includegraphics[width=\textwidth]{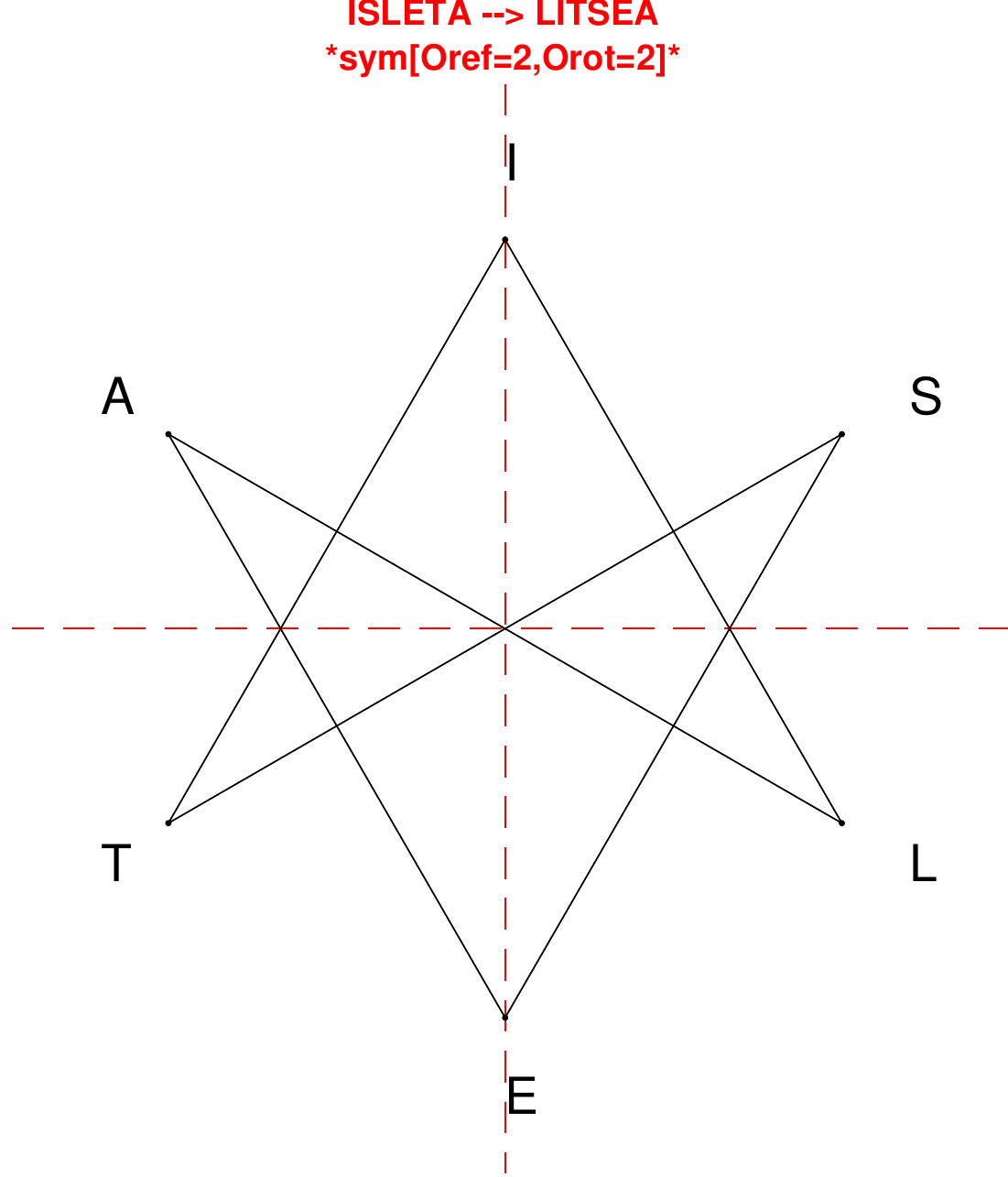}
\end{subfigure}
\hfill
\begin{subfigure}[T]{0.19\textwidth}
\centering
\includegraphics[width=\textwidth]{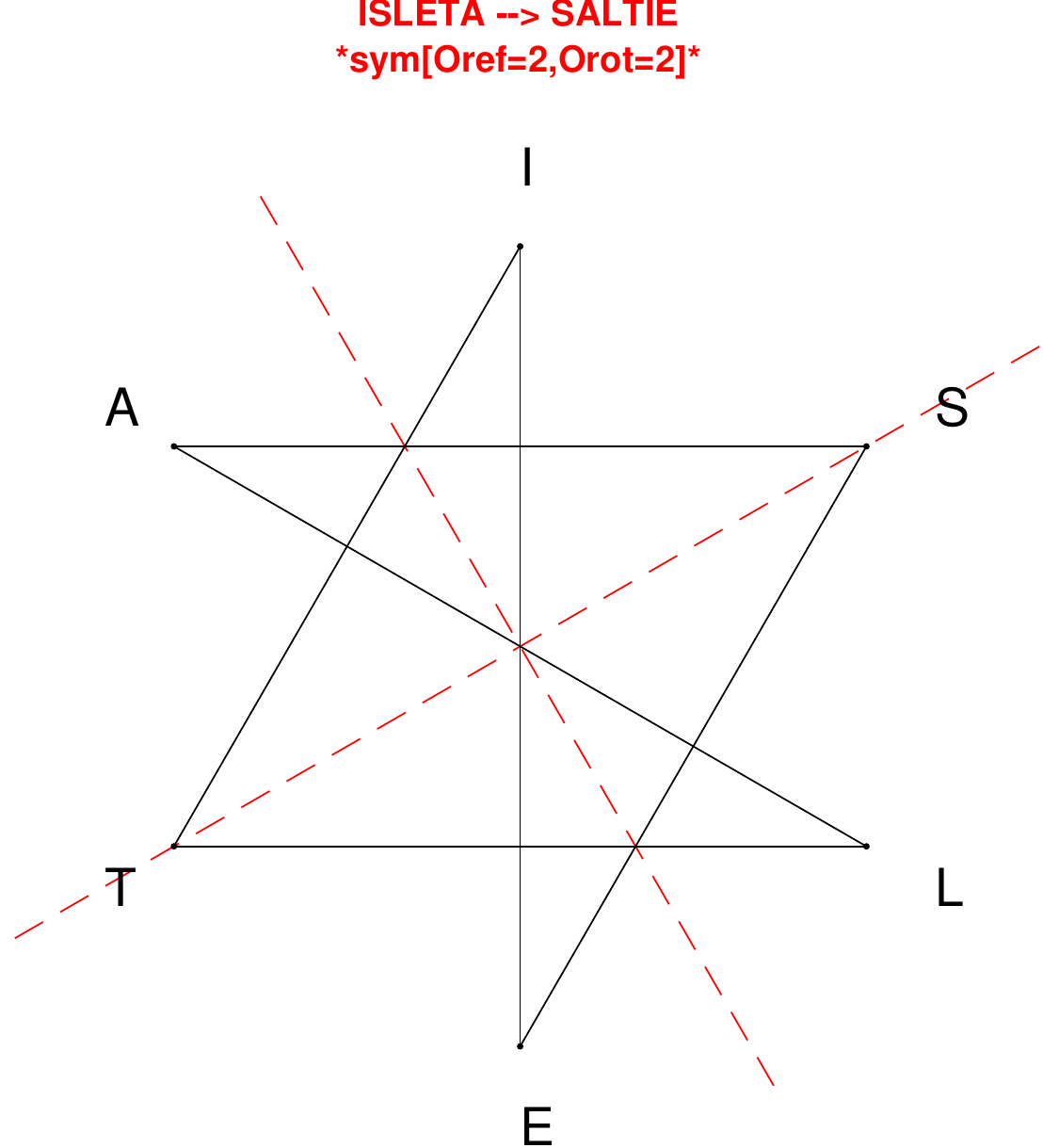}
\end{subfigure}
\hfill
\begin{subfigure}[T]{0.19\textwidth}
\centering
\includegraphics[width=\textwidth]{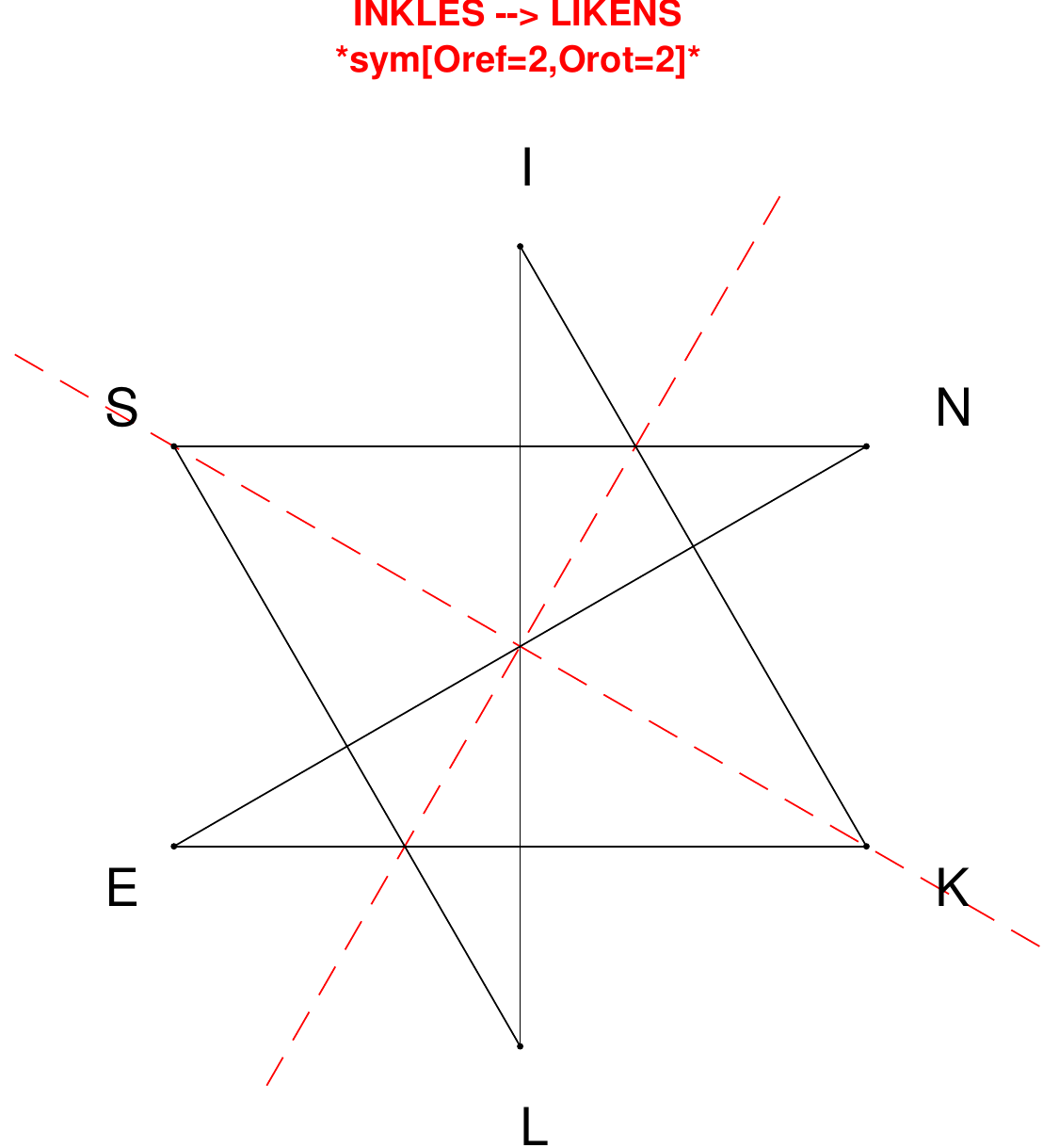}
\end{subfigure}
\end{figure}

\begin{figure}[H]
\centering
\begin{subfigure}[T]{0.19\textwidth}
\centering
\includegraphics[width=\textwidth]{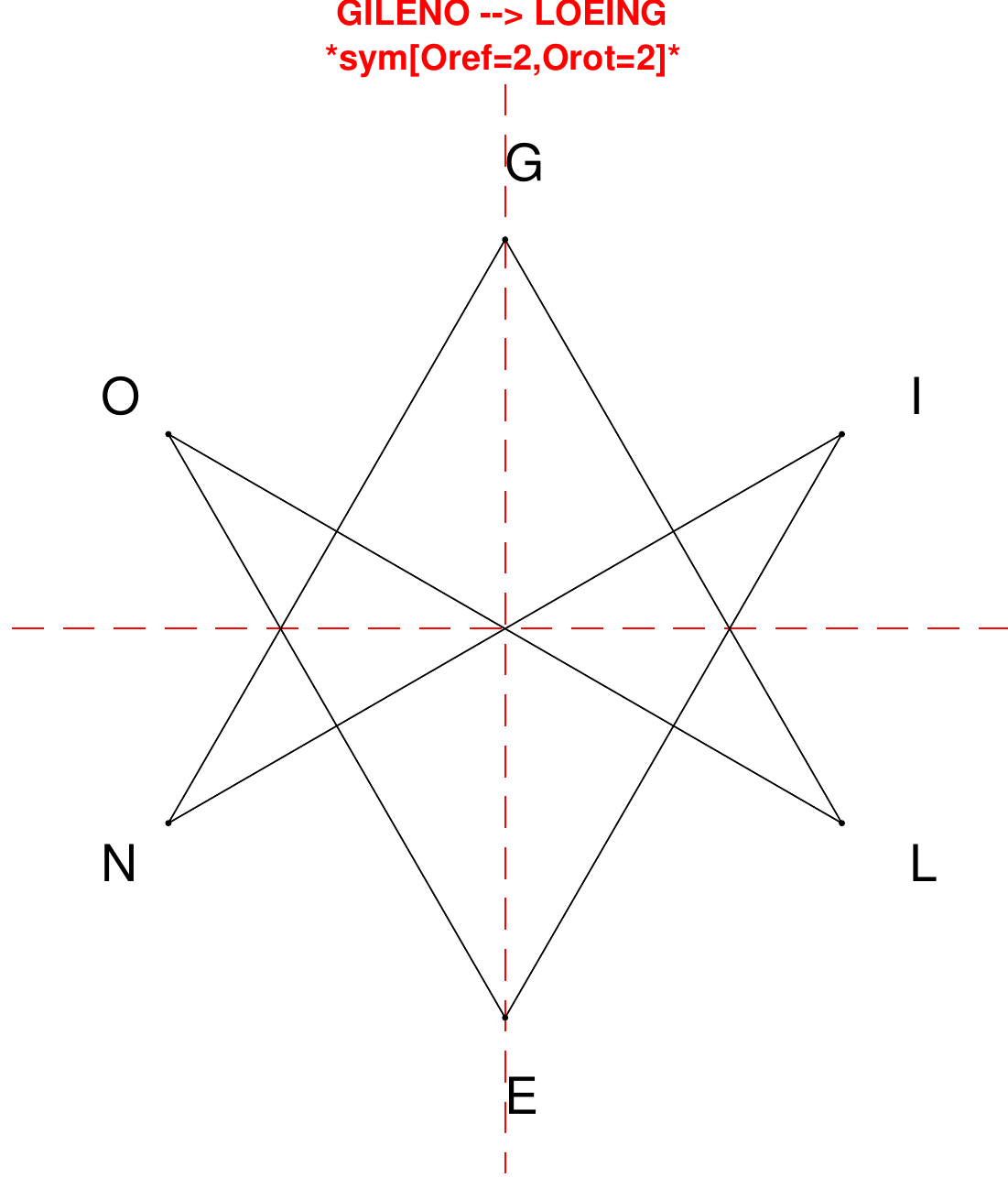}
\end{subfigure}
\hfill
\begin{subfigure}[T]{0.19\textwidth}
\centering
\includegraphics[width=\textwidth]{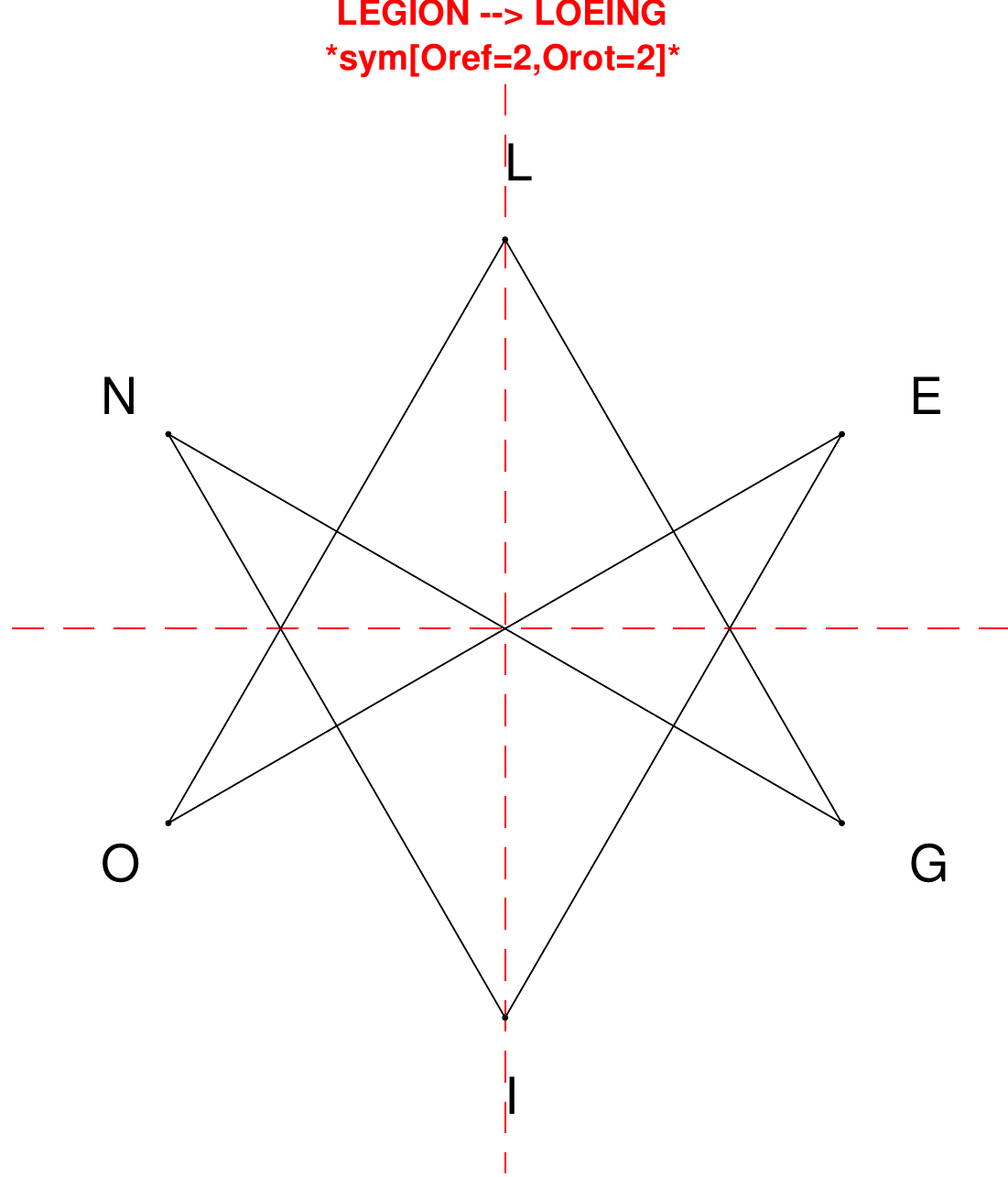}
\end{subfigure}
\hfill
\begin{subfigure}[T]{0.19\textwidth}
\centering
\includegraphics[width=\textwidth]{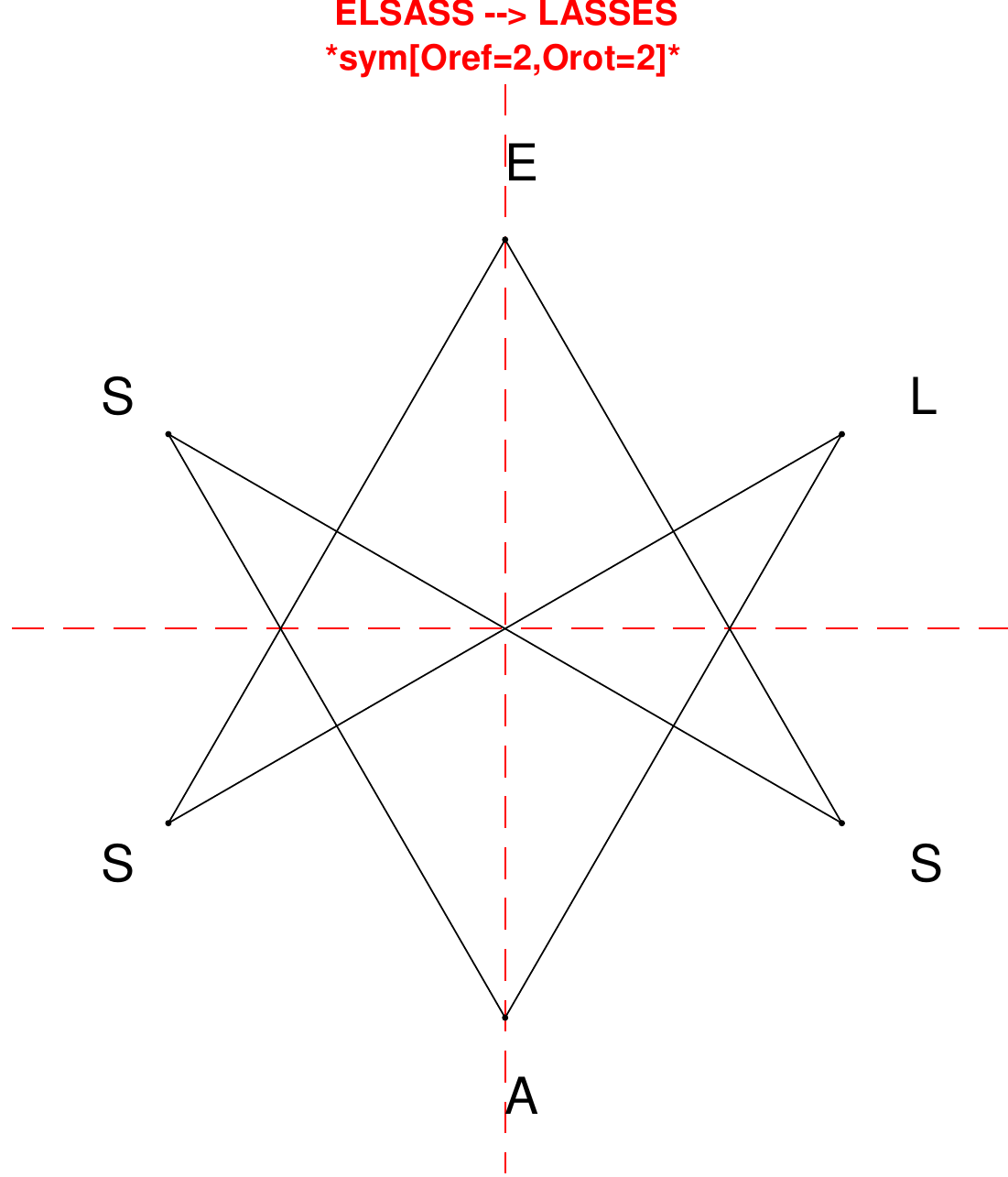}
\end{subfigure}
\hfill
\begin{subfigure}[T]{0.19\textwidth}
\centering
\includegraphics[width=\textwidth]{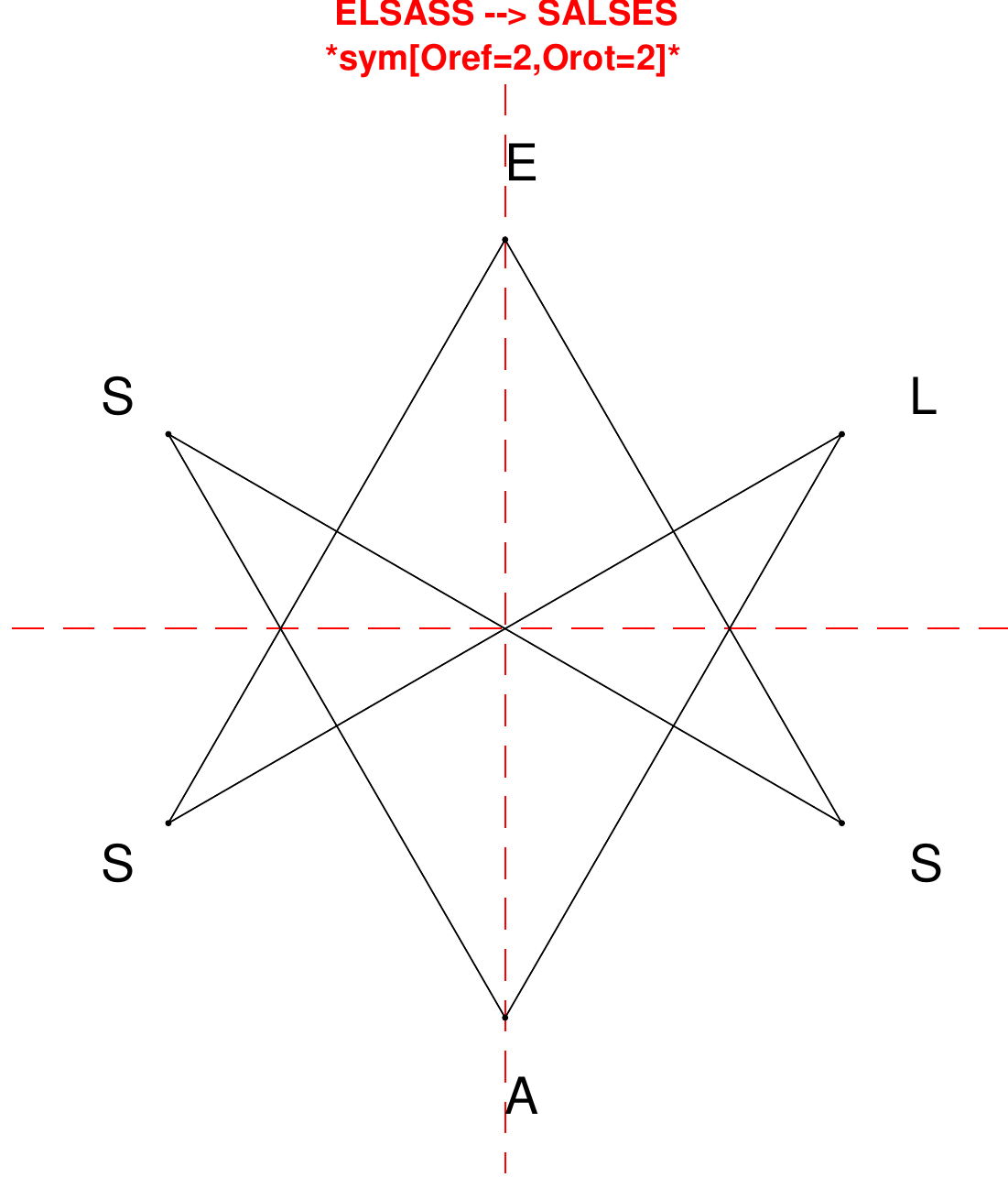}
\end{subfigure}
\hfill
\begin{subfigure}[T]{0.19\textwidth}
\centering
\includegraphics[width=\textwidth]{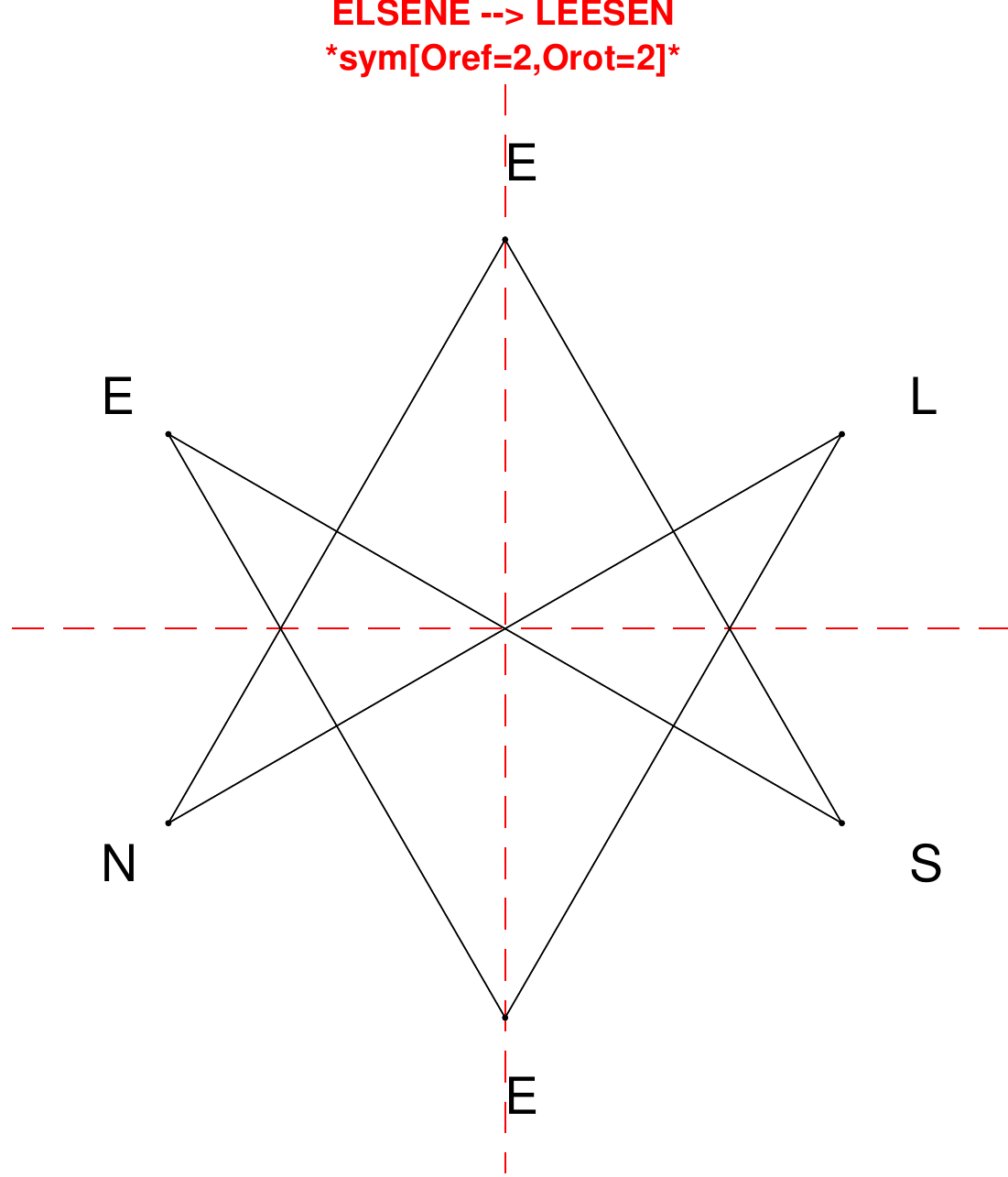}
\end{subfigure}
\end{figure}

\begin{figure}[H]
\centering
\begin{subfigure}[T]{0.19\textwidth}
\centering
\includegraphics[width=\textwidth]{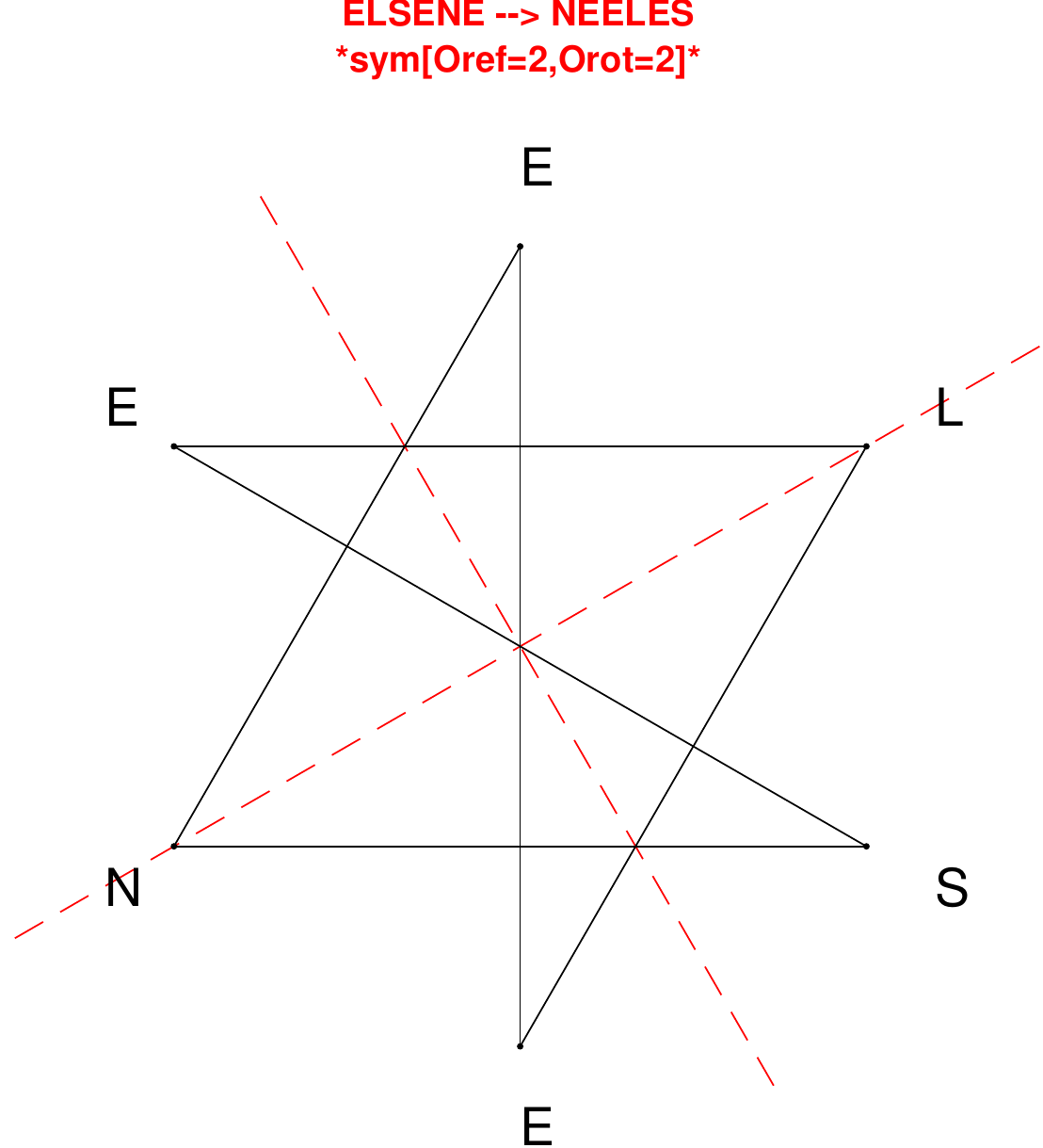}
\end{subfigure}
\hfill
\begin{subfigure}[T]{0.19\textwidth}
\centering
\includegraphics[width=\textwidth]{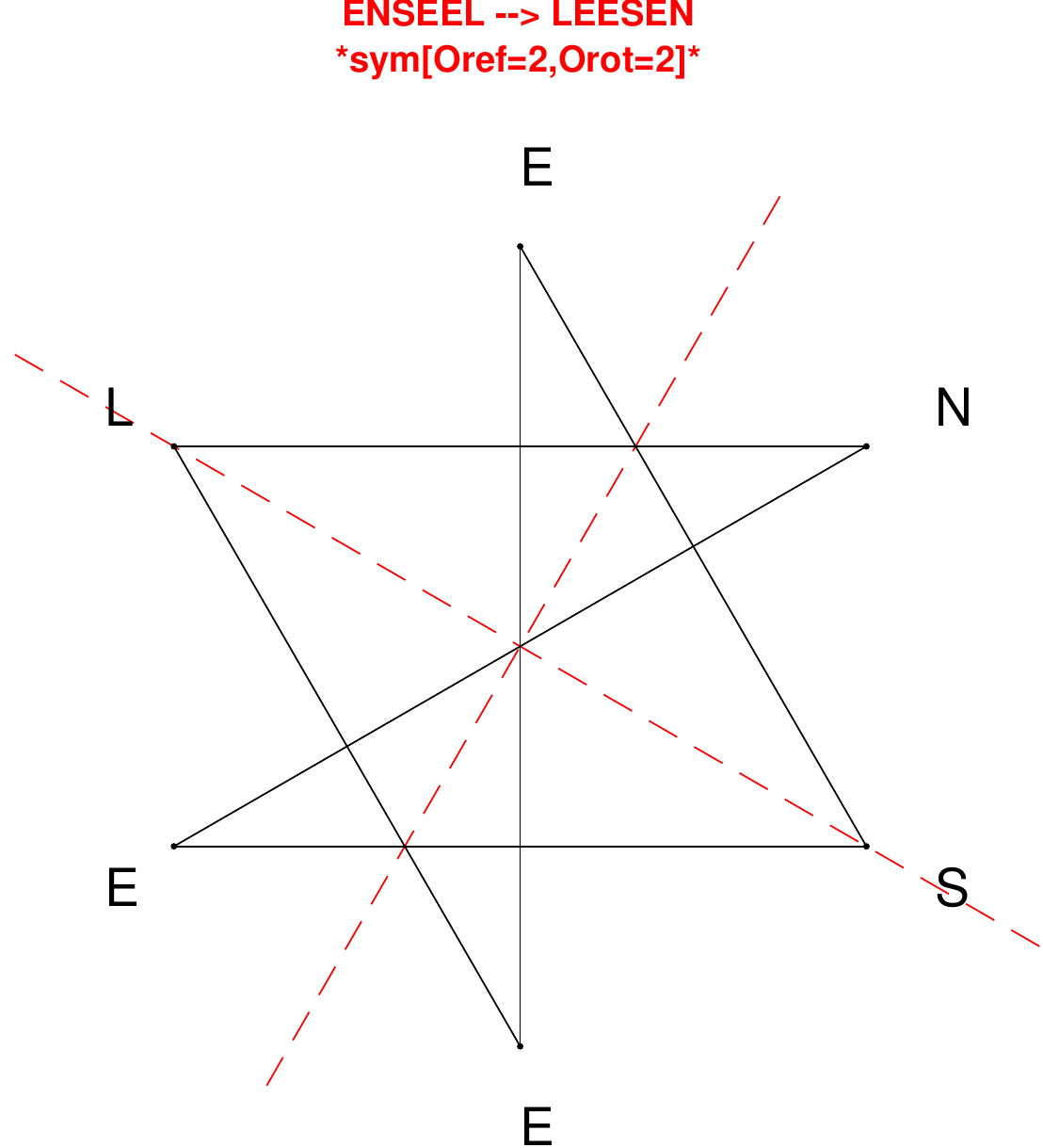}
\end{subfigure}
\hfill
\begin{subfigure}[T]{0.19\textwidth}
\centering
\includegraphics[width=\textwidth]{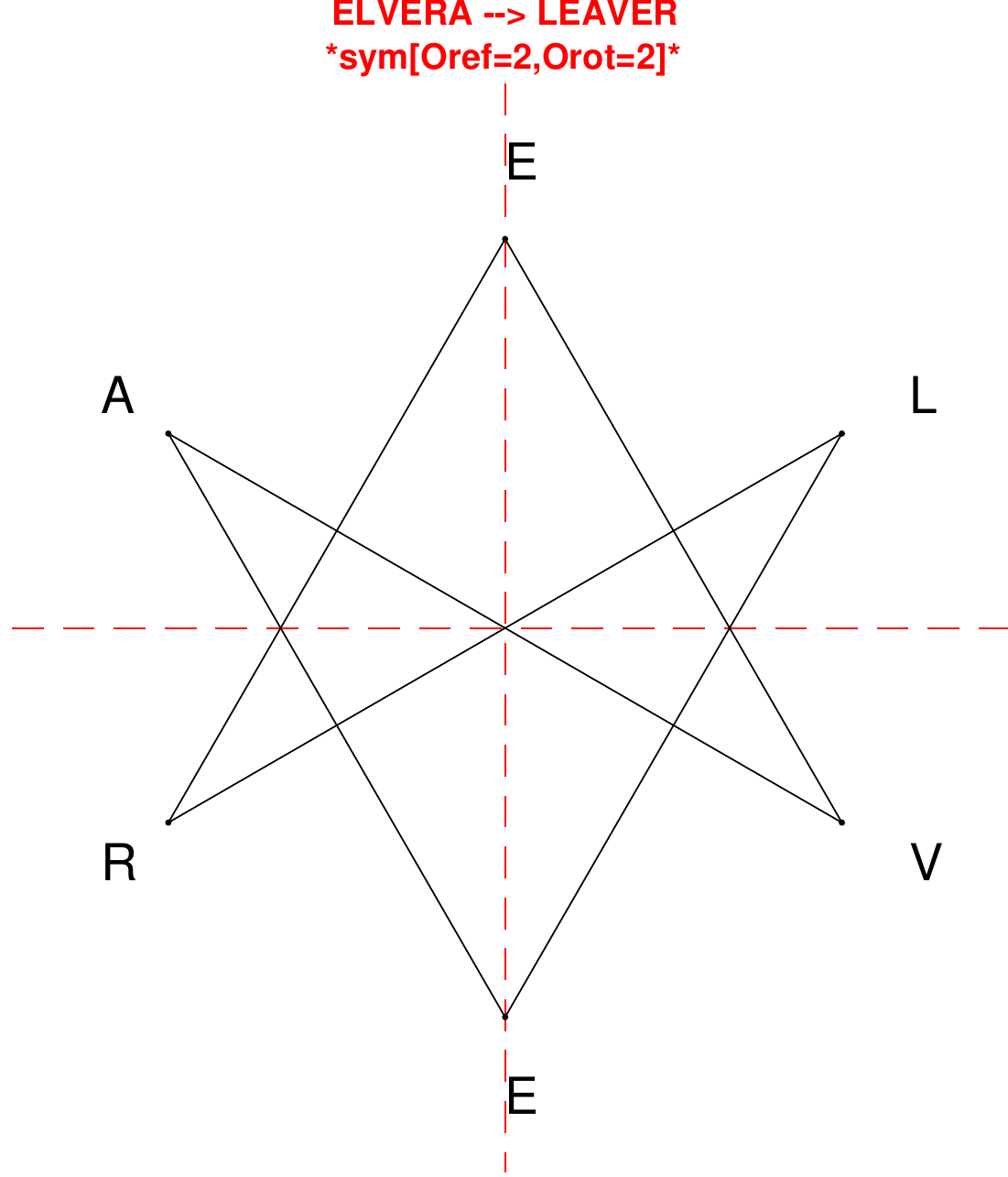}
\end{subfigure}
\hfill
\begin{subfigure}[T]{0.19\textwidth}
\centering
\includegraphics[width=\textwidth]{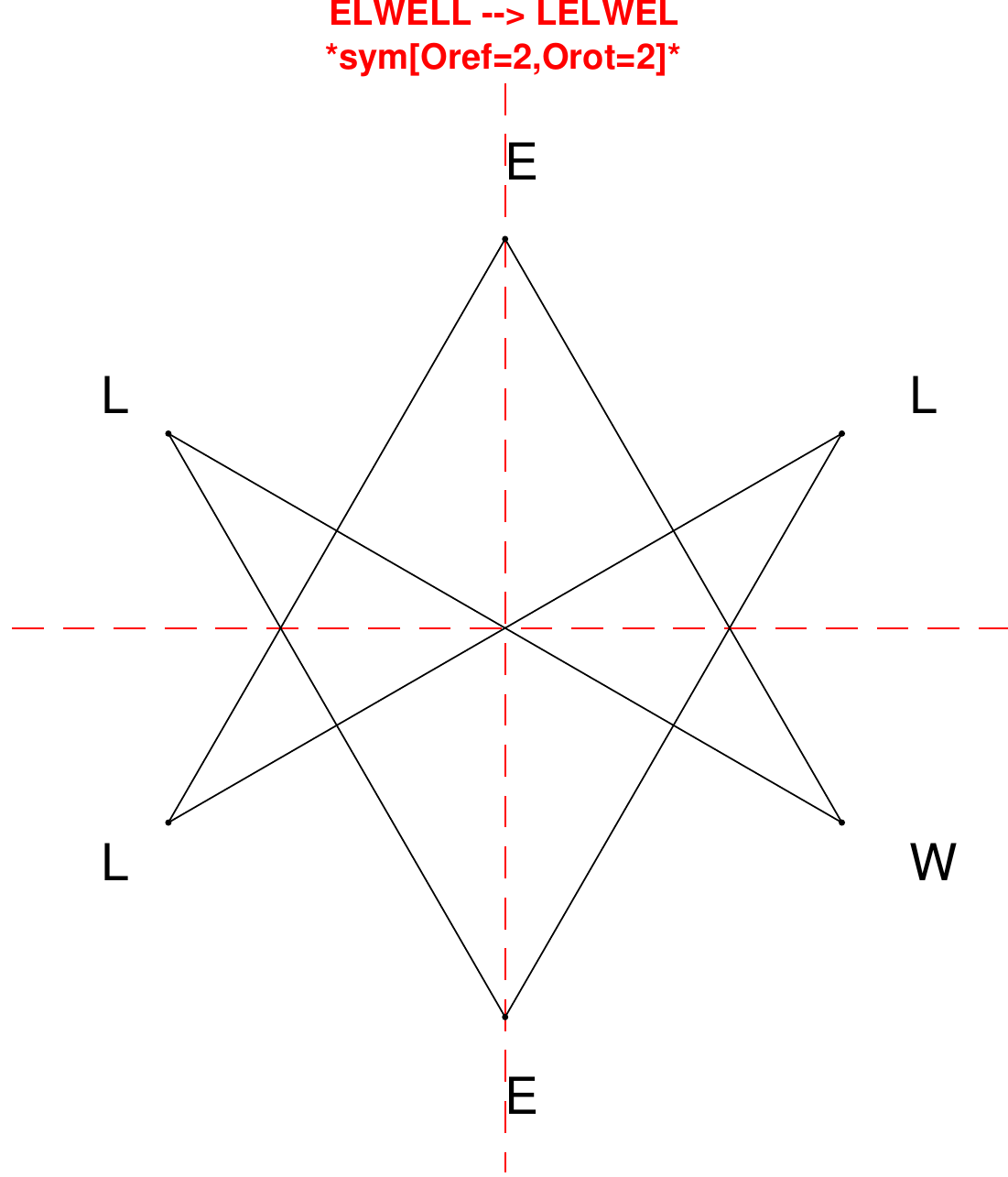}
\end{subfigure}
\hfill
\begin{subfigure}[T]{0.19\textwidth}
\centering
\includegraphics[width=\textwidth]{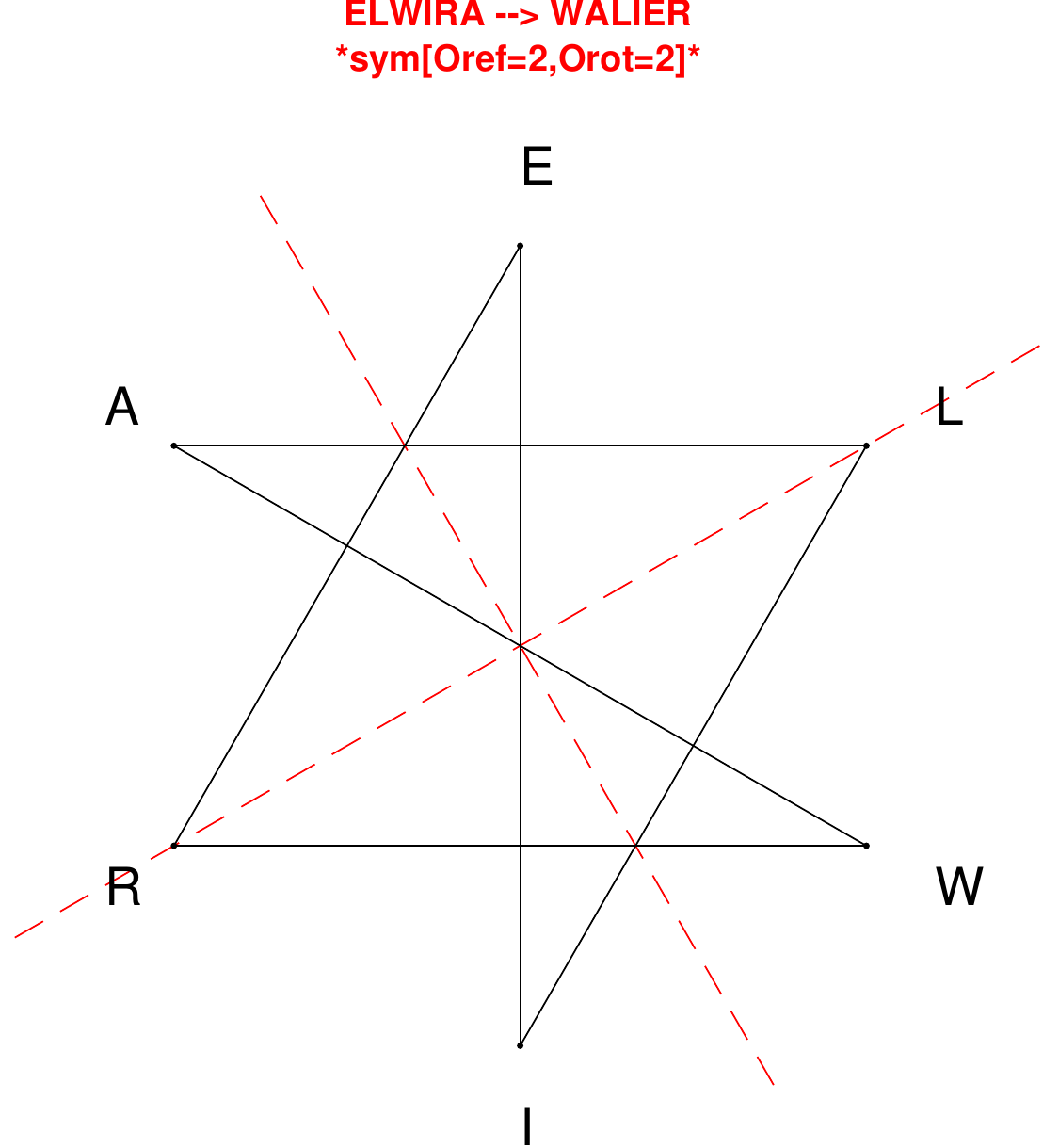}
\end{subfigure}
\end{figure}

\begin{figure}[H]
\centering
\begin{subfigure}[T]{0.19\textwidth}
\centering
\includegraphics[width=\textwidth]{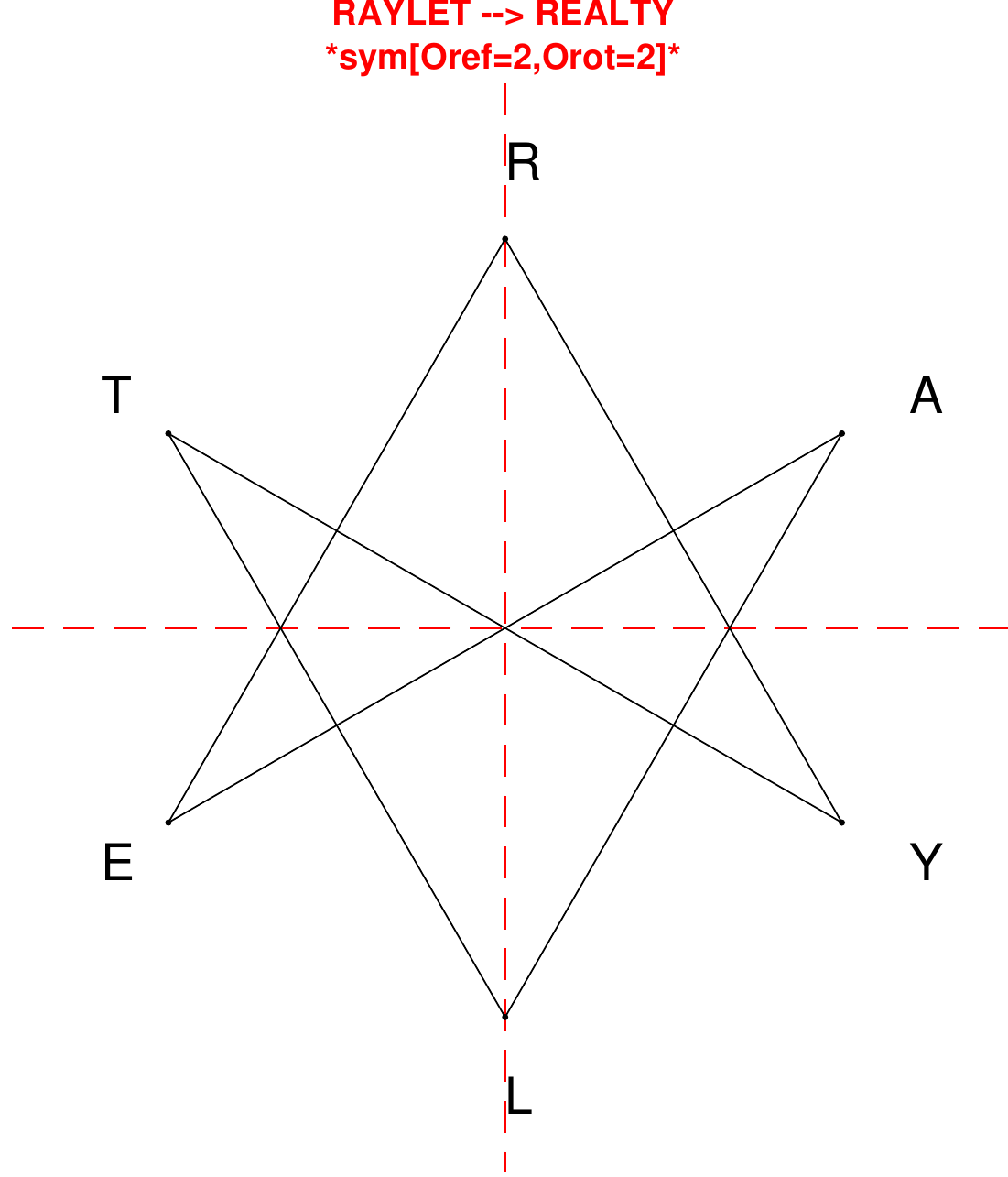}
\end{subfigure}
\hfill
\begin{subfigure}[T]{0.19\textwidth}
\centering
\includegraphics[width=\textwidth]{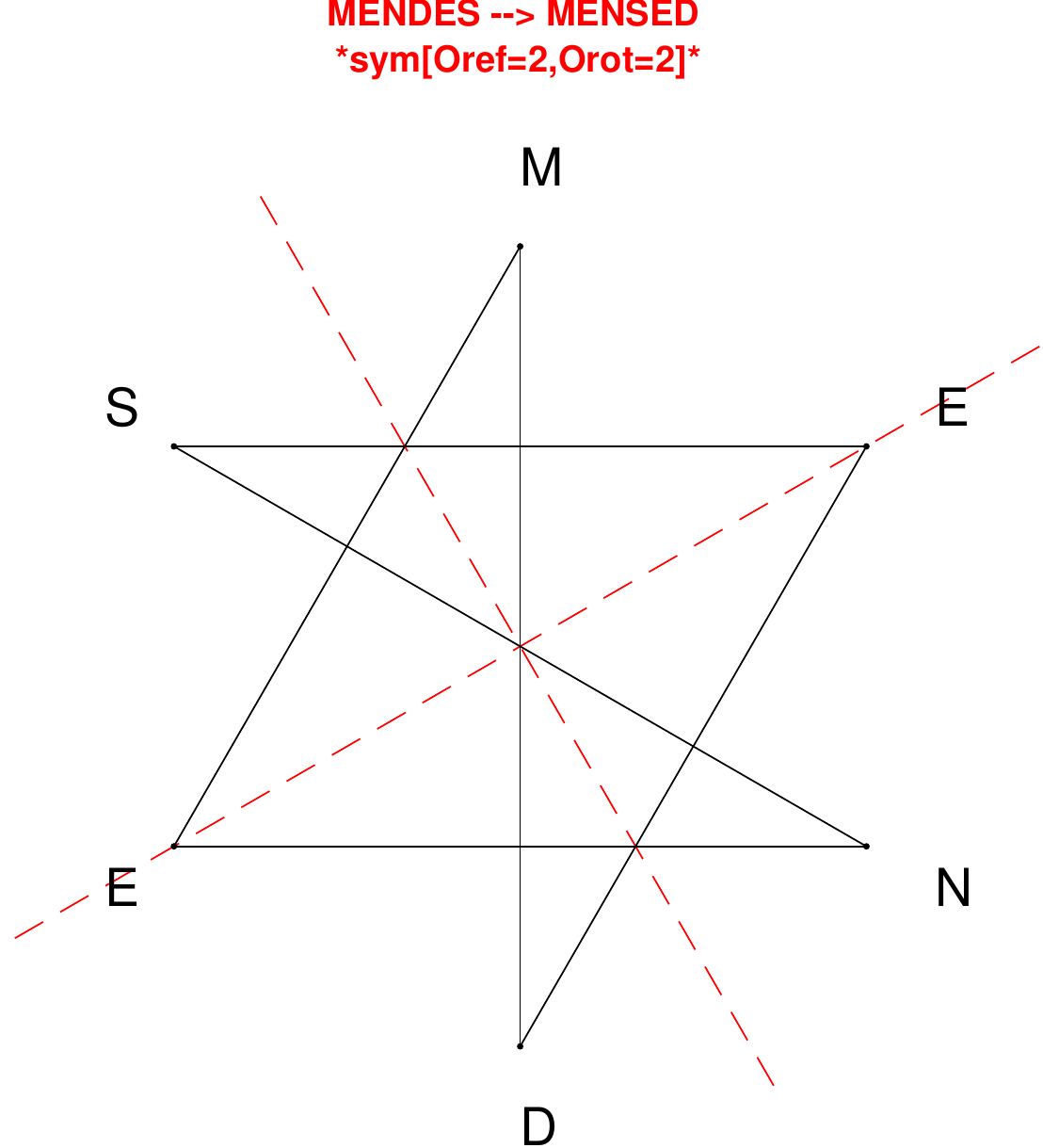}
\end{subfigure}
\hfill
\begin{subfigure}[T]{0.19\textwidth}
\centering
\includegraphics[width=\textwidth]{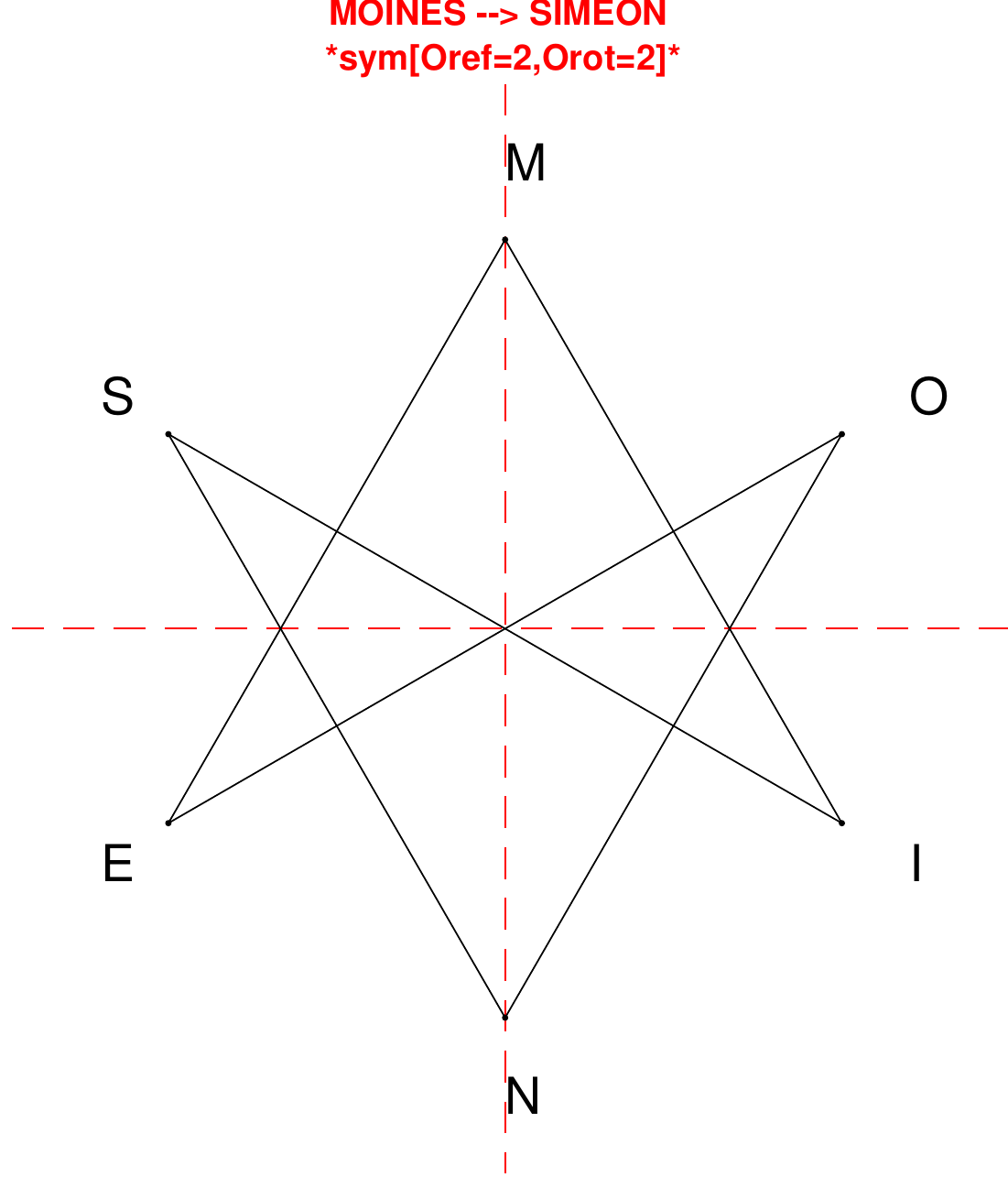}
\end{subfigure}
\hfill
\begin{subfigure}[T]{0.19\textwidth}
\centering
\includegraphics[width=\textwidth]{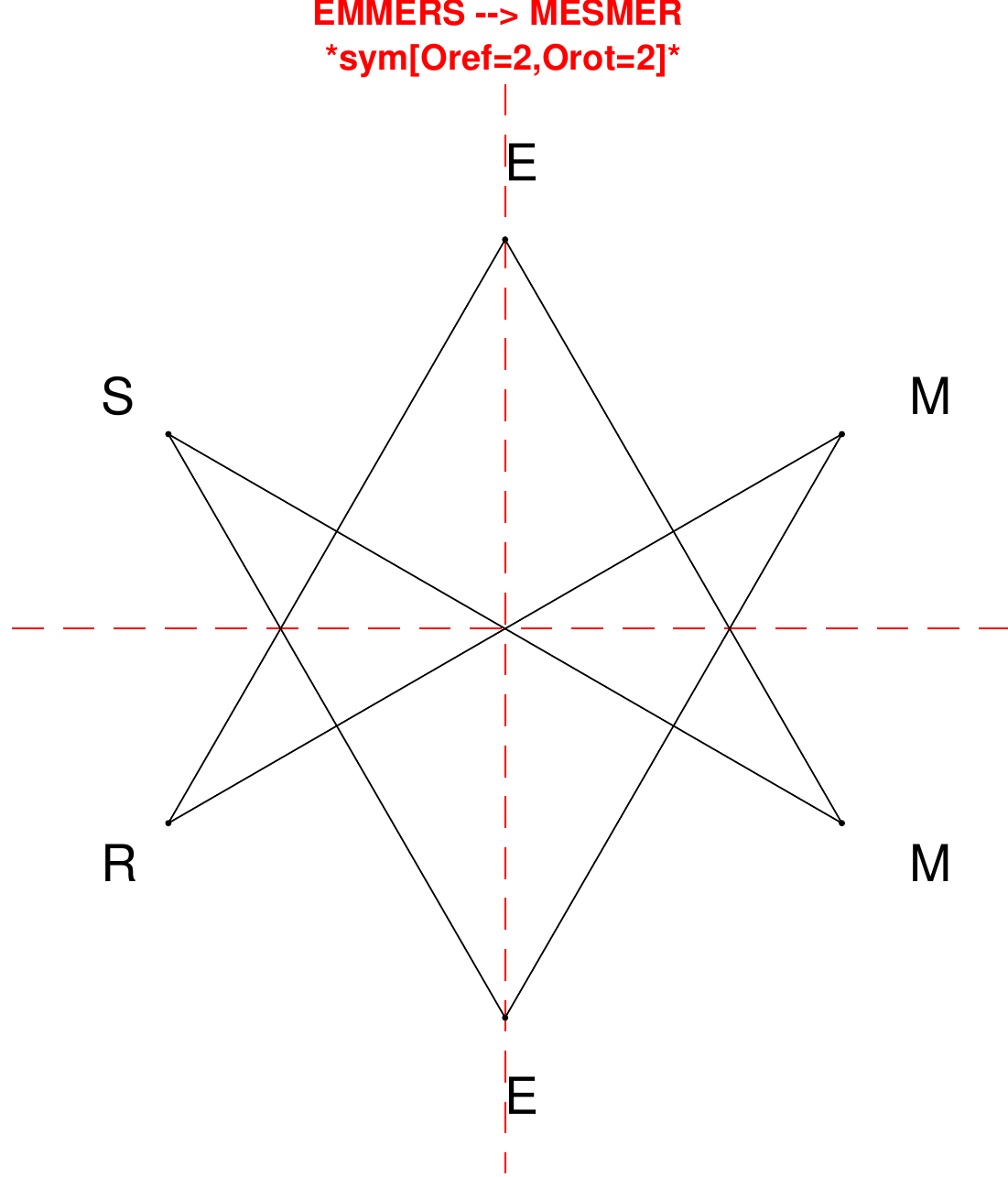}
\end{subfigure}
\hfill
\begin{subfigure}[T]{0.19\textwidth}
\centering
\includegraphics[width=\textwidth]{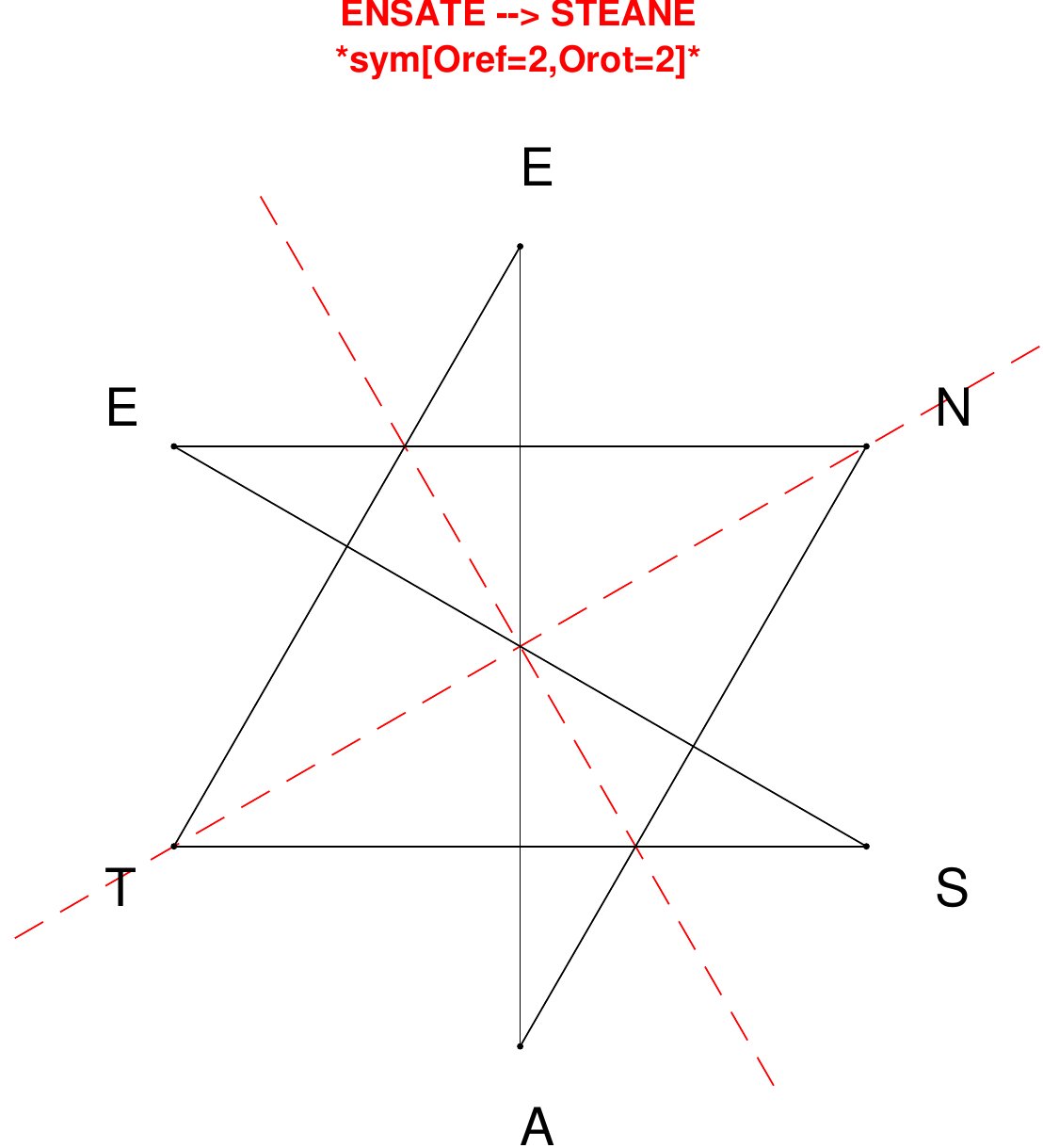}
\end{subfigure}
\end{figure}

\begin{figure}[H]
\centering
\begin{subfigure}[T]{0.19\textwidth}
\centering
\includegraphics[width=\textwidth]{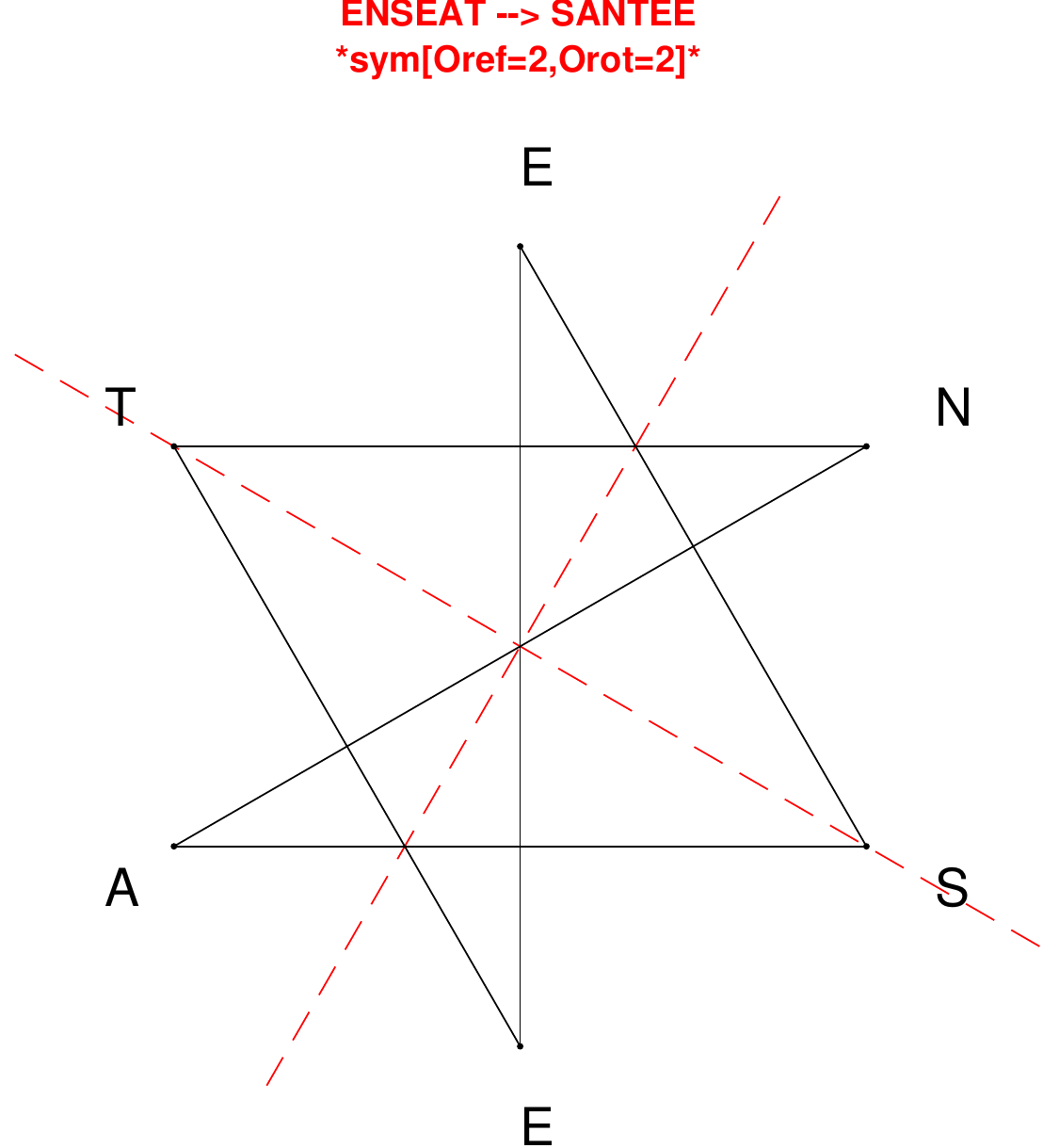}
\end{subfigure}
\hfill
\begin{subfigure}[T]{0.19\textwidth}
\centering
\includegraphics[width=\textwidth]{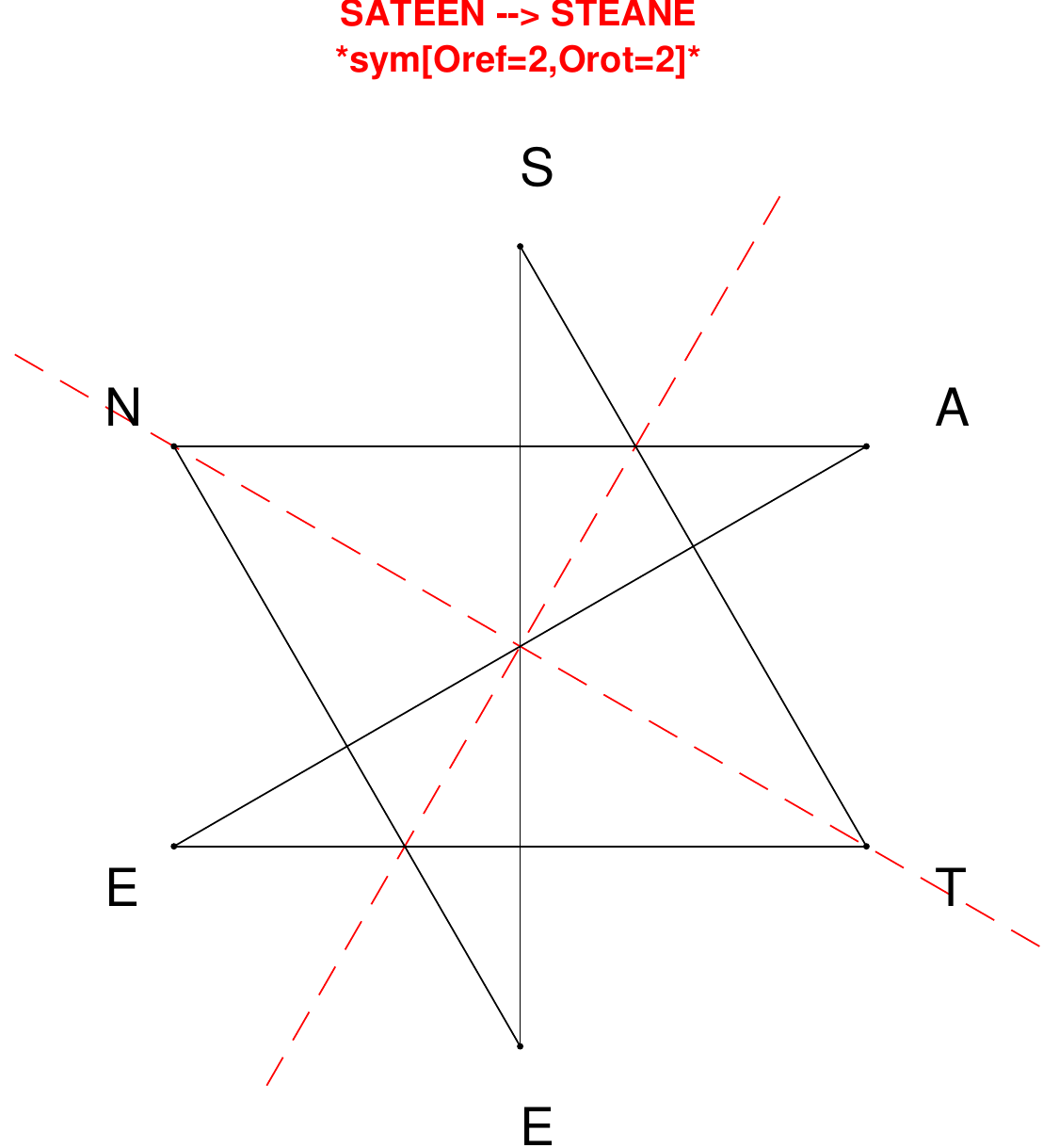}
\end{subfigure}
\hfill
\begin{subfigure}[T]{0.19\textwidth}
\centering
\includegraphics[width=\textwidth]{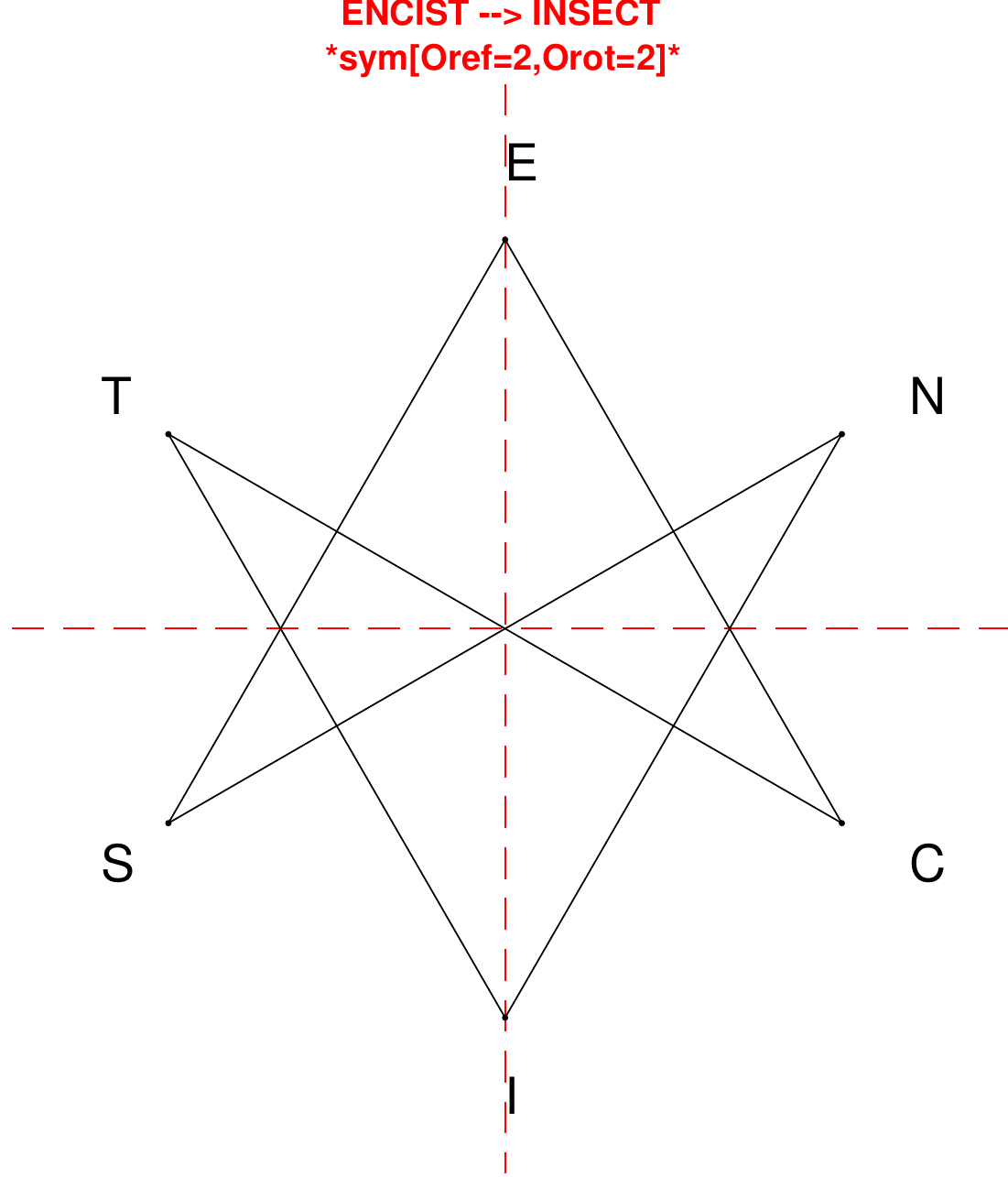}
\end{subfigure}
\hfill
\begin{subfigure}[T]{0.19\textwidth}
\centering
\includegraphics[width=\textwidth]{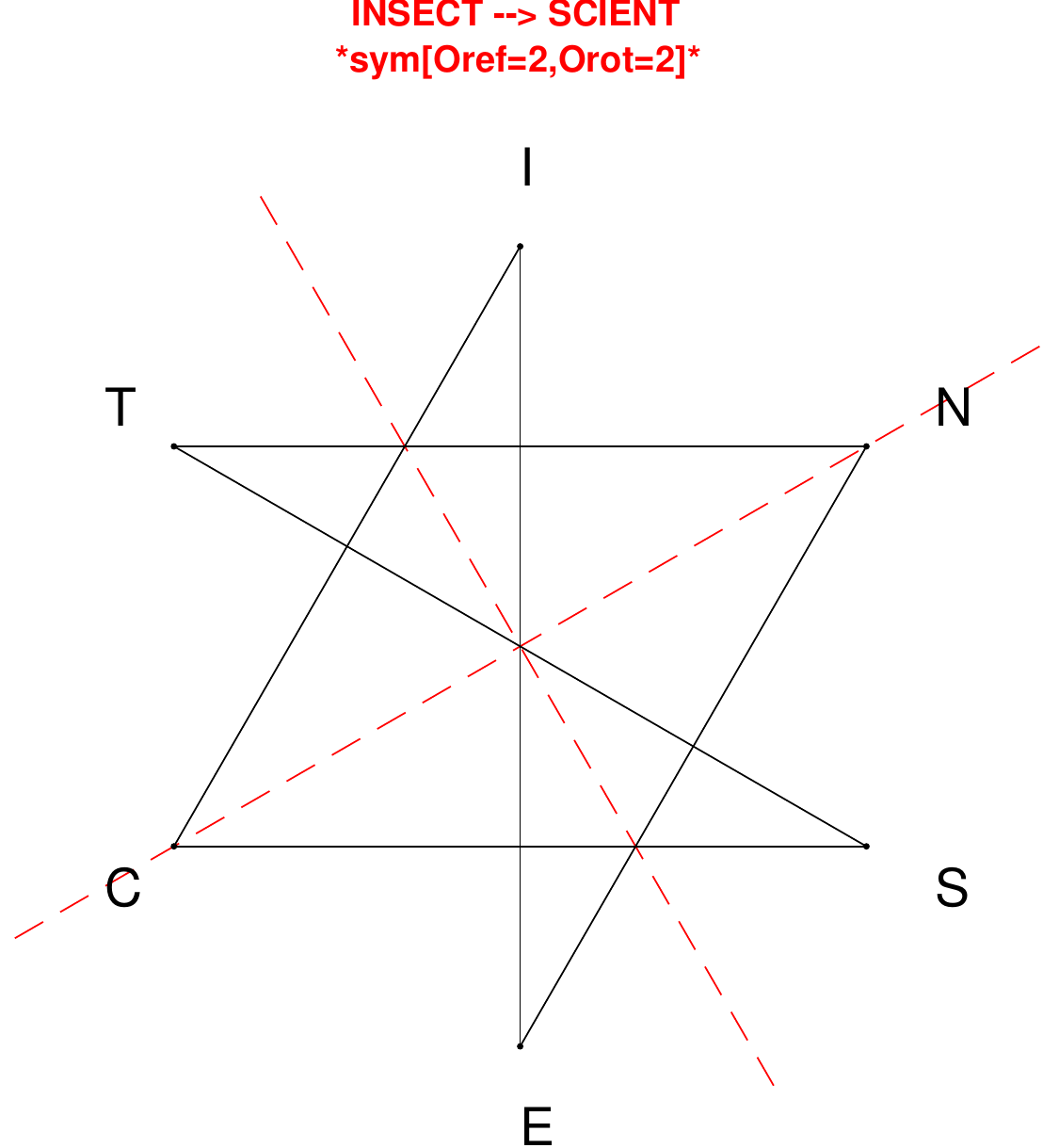}
\end{subfigure}
\hfill
\begin{subfigure}[T]{0.19\textwidth}
\centering
\includegraphics[width=\textwidth]{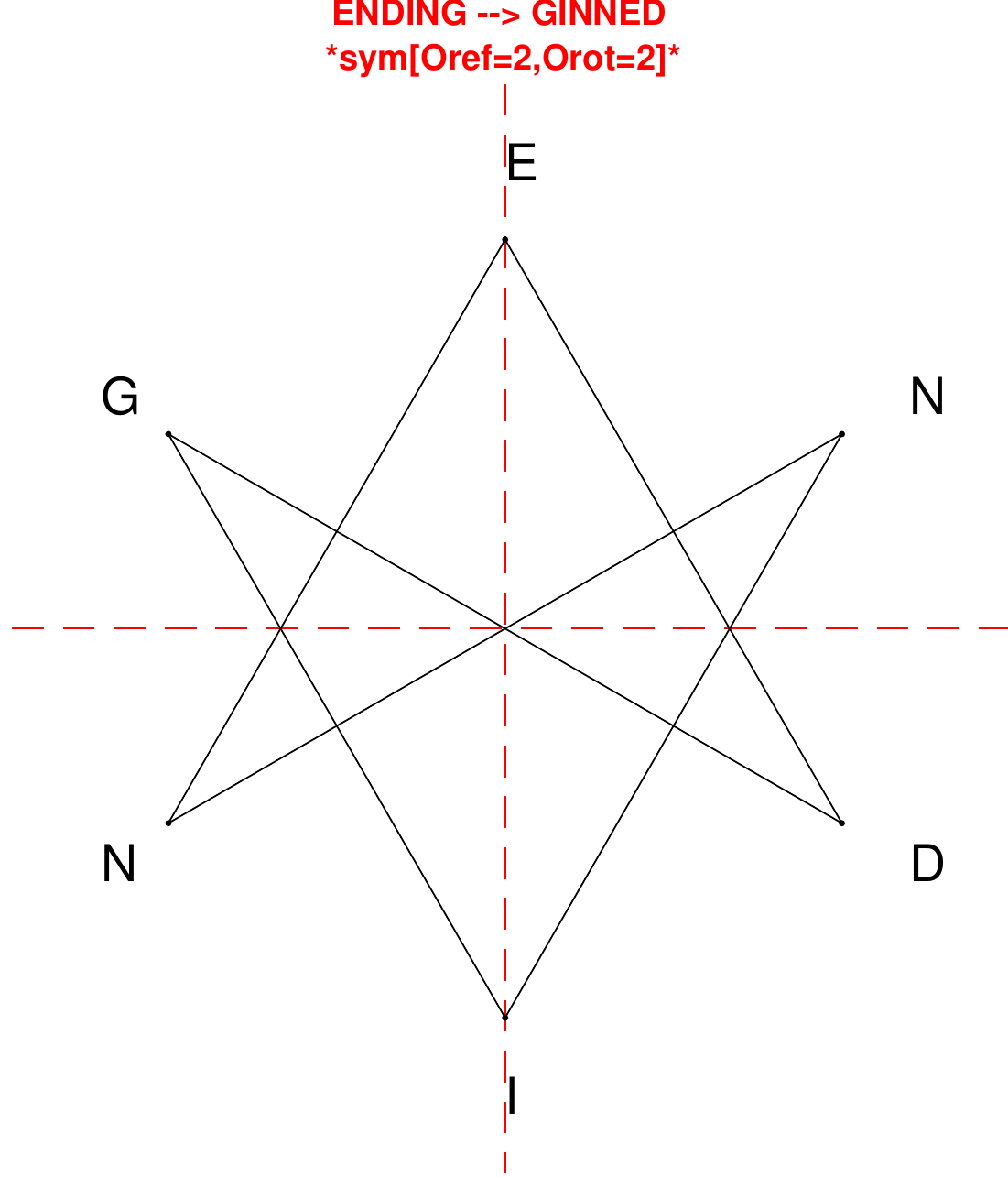}
\end{subfigure}
\end{figure}

\begin{figure}[H]
\centering
\begin{subfigure}[T]{0.19\textwidth}
\centering
\includegraphics[width=\textwidth]{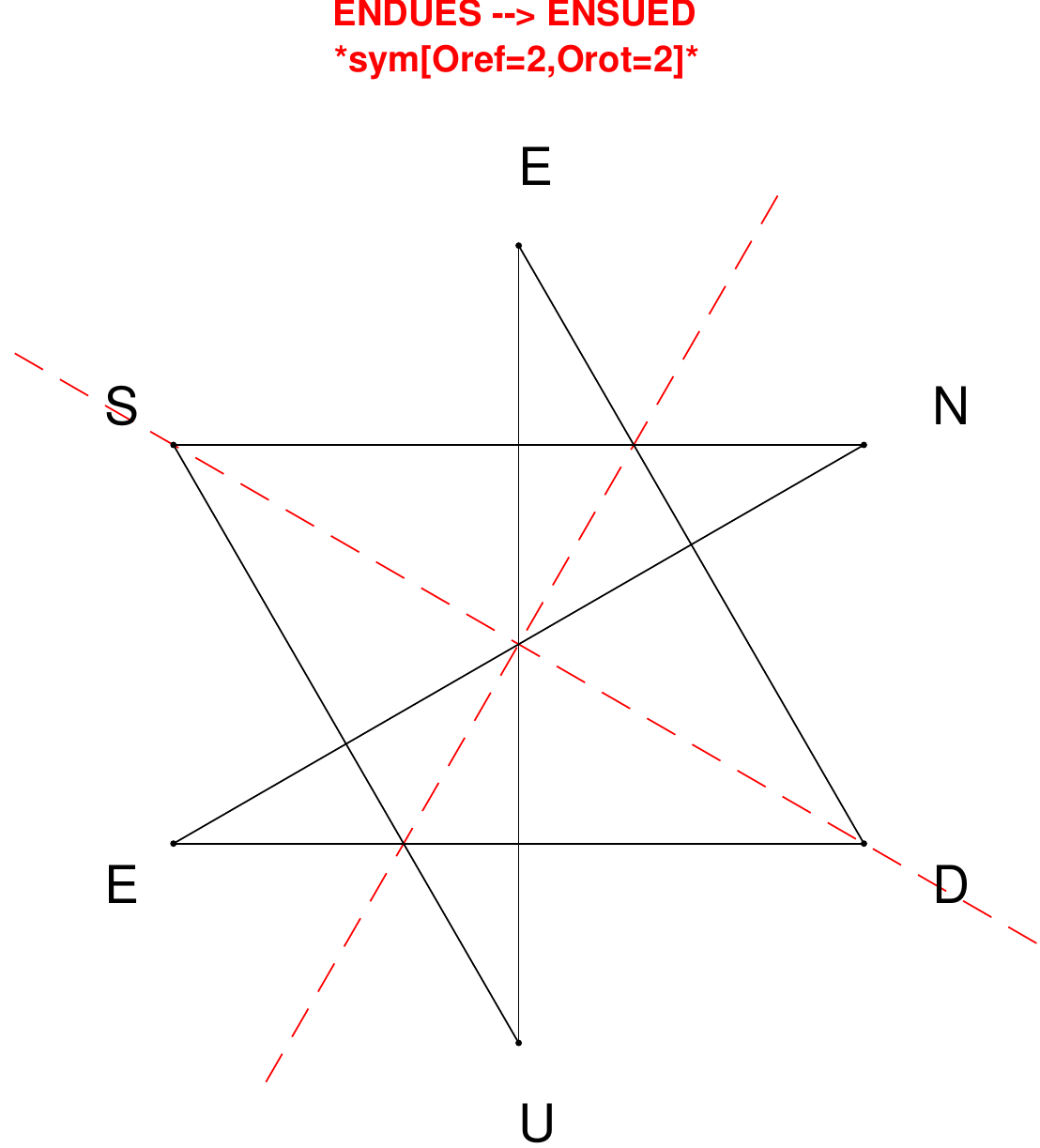}
\end{subfigure}
\hfill
\begin{subfigure}[T]{0.19\textwidth}
\centering
\includegraphics[width=\textwidth]{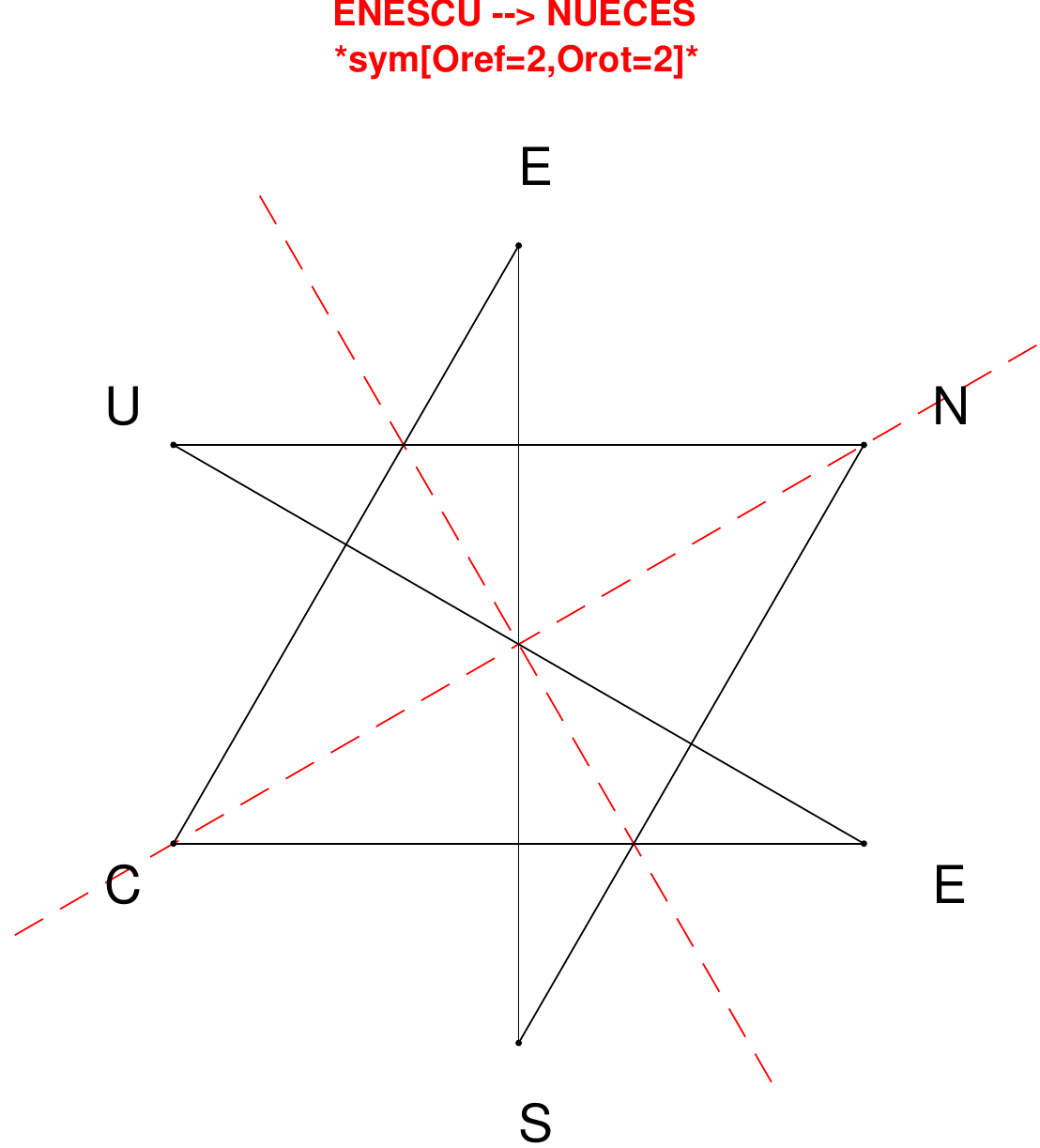}
\end{subfigure}
\hfill
\begin{subfigure}[T]{0.19\textwidth}
\centering
\includegraphics[width=\textwidth]{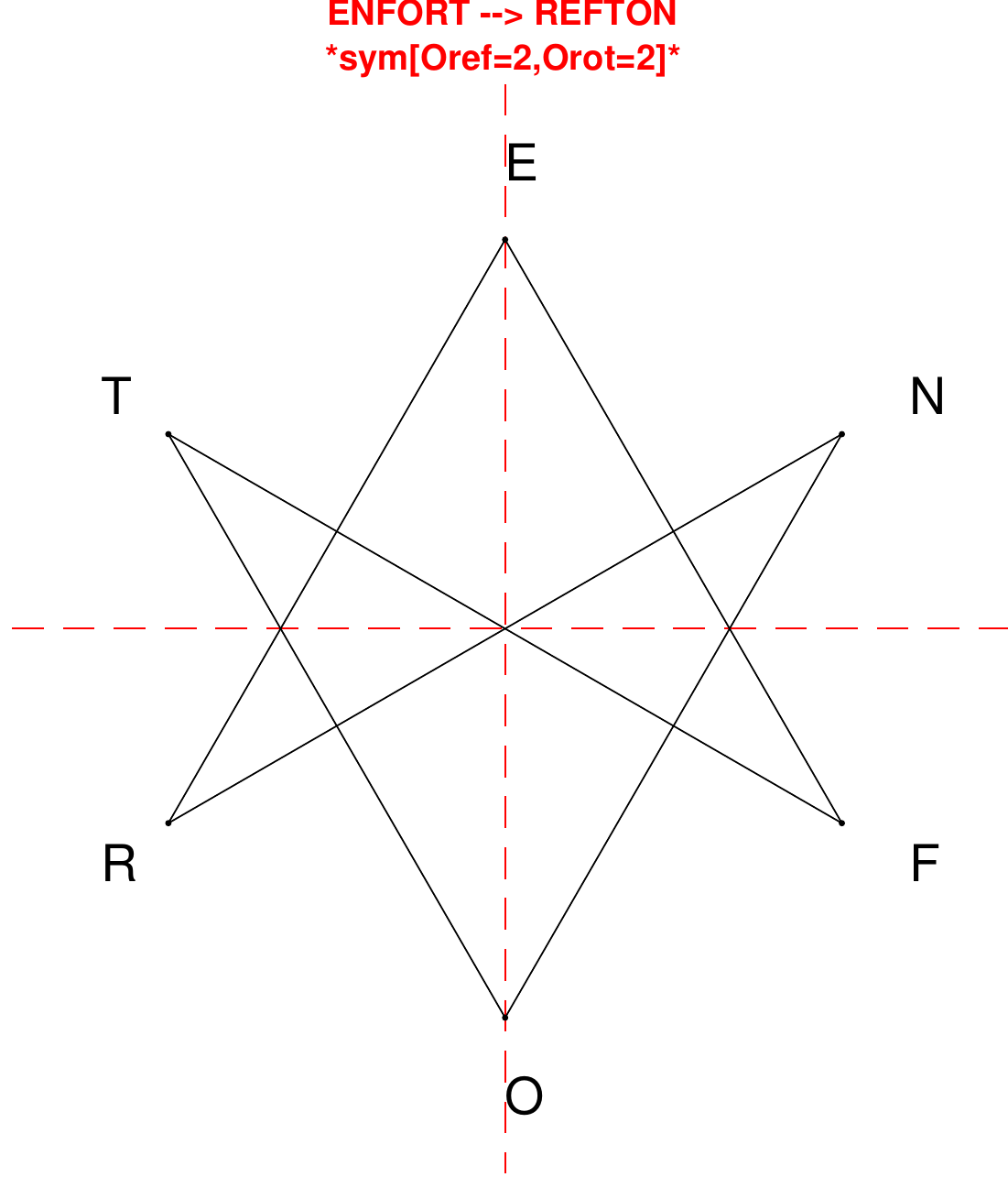}
\end{subfigure}
\hfill
\begin{subfigure}[T]{0.19\textwidth}
\centering
\includegraphics[width=\textwidth]{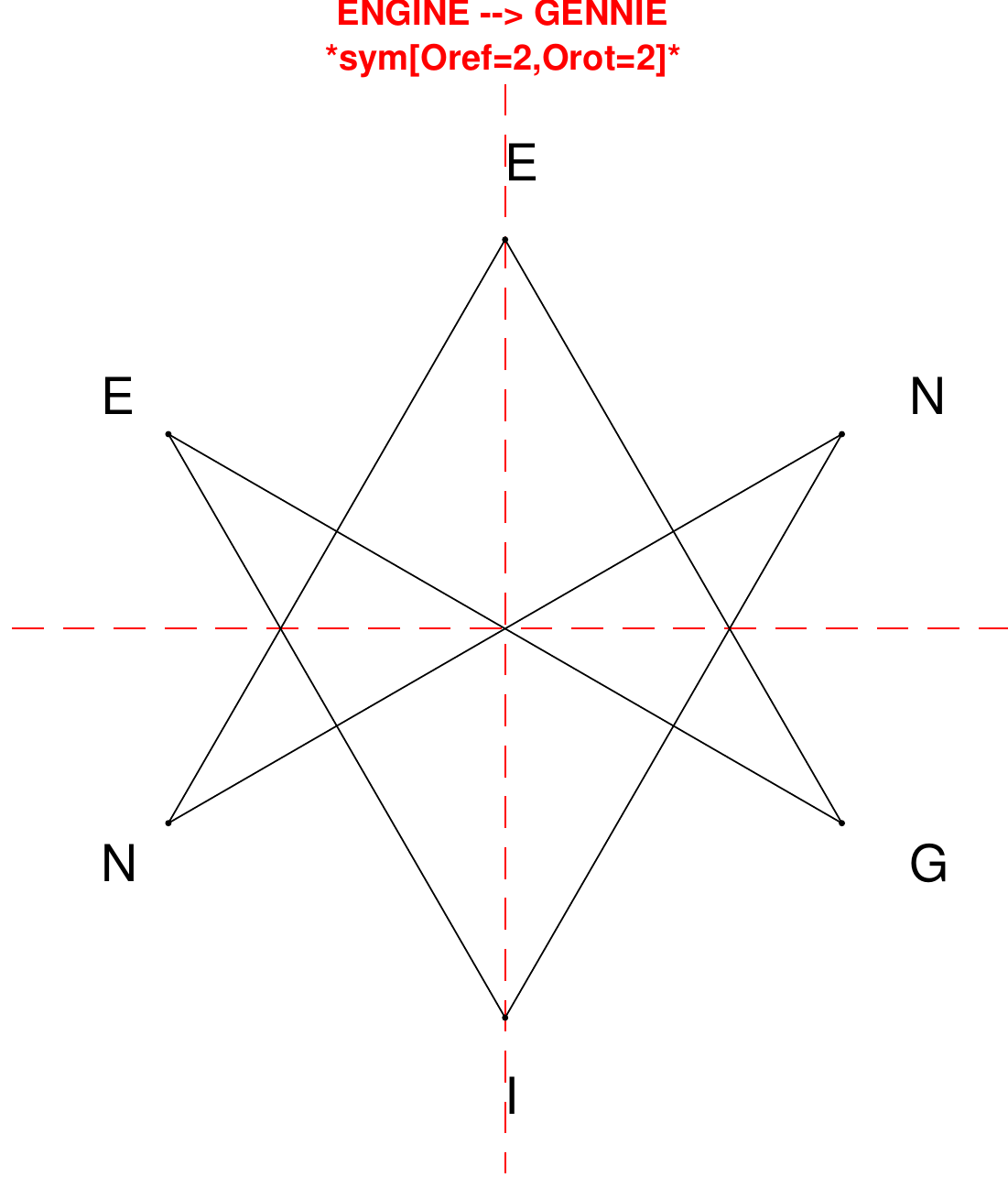}
\end{subfigure}
\hfill
\begin{subfigure}[T]{0.19\textwidth}
\centering
\includegraphics[width=\textwidth]{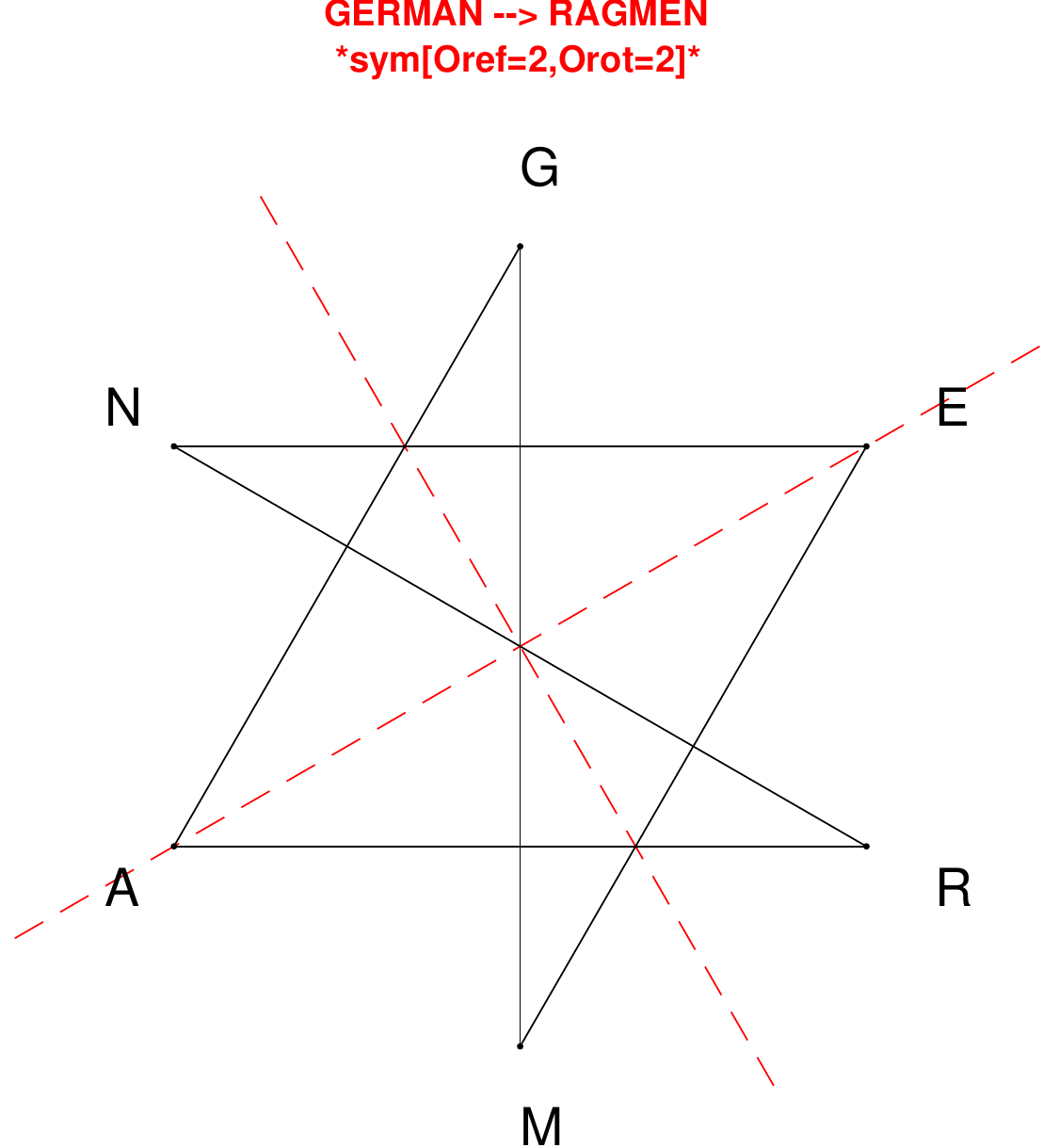}
\end{subfigure}
\end{figure}

\begin{figure}[H]
\centering
\begin{subfigure}[T]{0.19\textwidth}
\centering
\includegraphics[width=\textwidth]{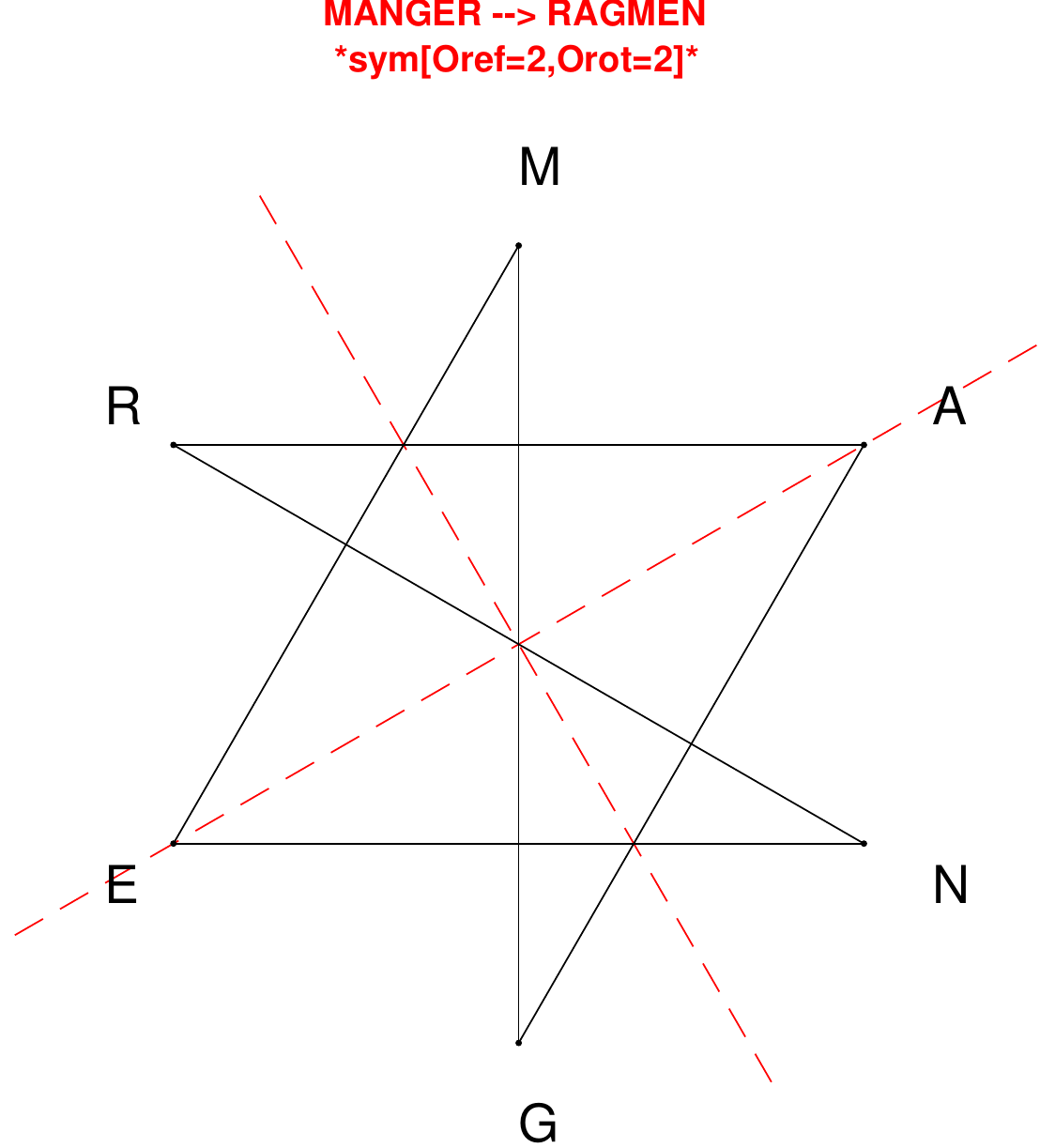}
\end{subfigure}
\hfill
\begin{subfigure}[T]{0.19\textwidth}
\centering
\includegraphics[width=\textwidth]{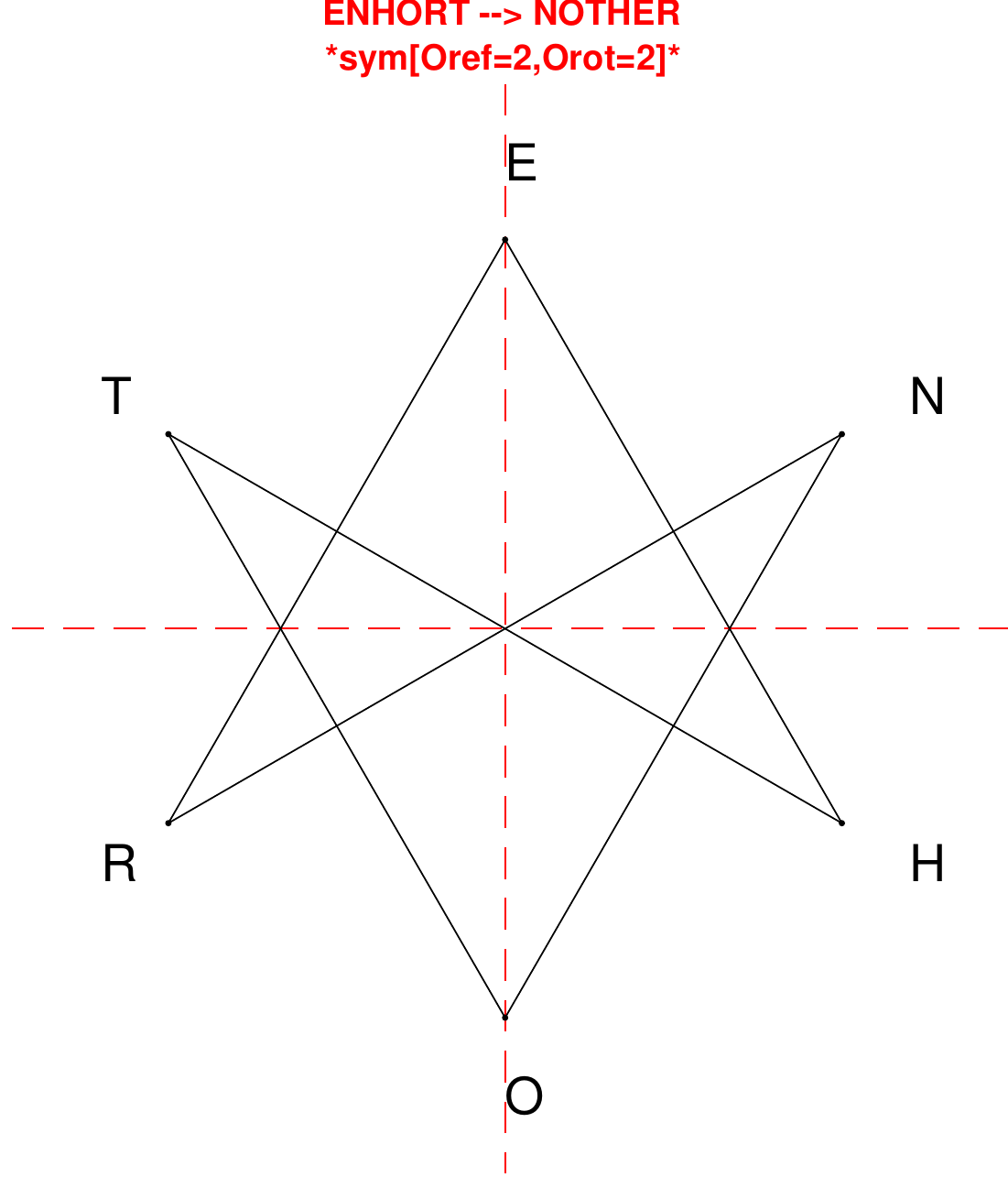}
\end{subfigure}
\hfill
\begin{subfigure}[T]{0.19\textwidth}
\centering
\includegraphics[width=\textwidth]{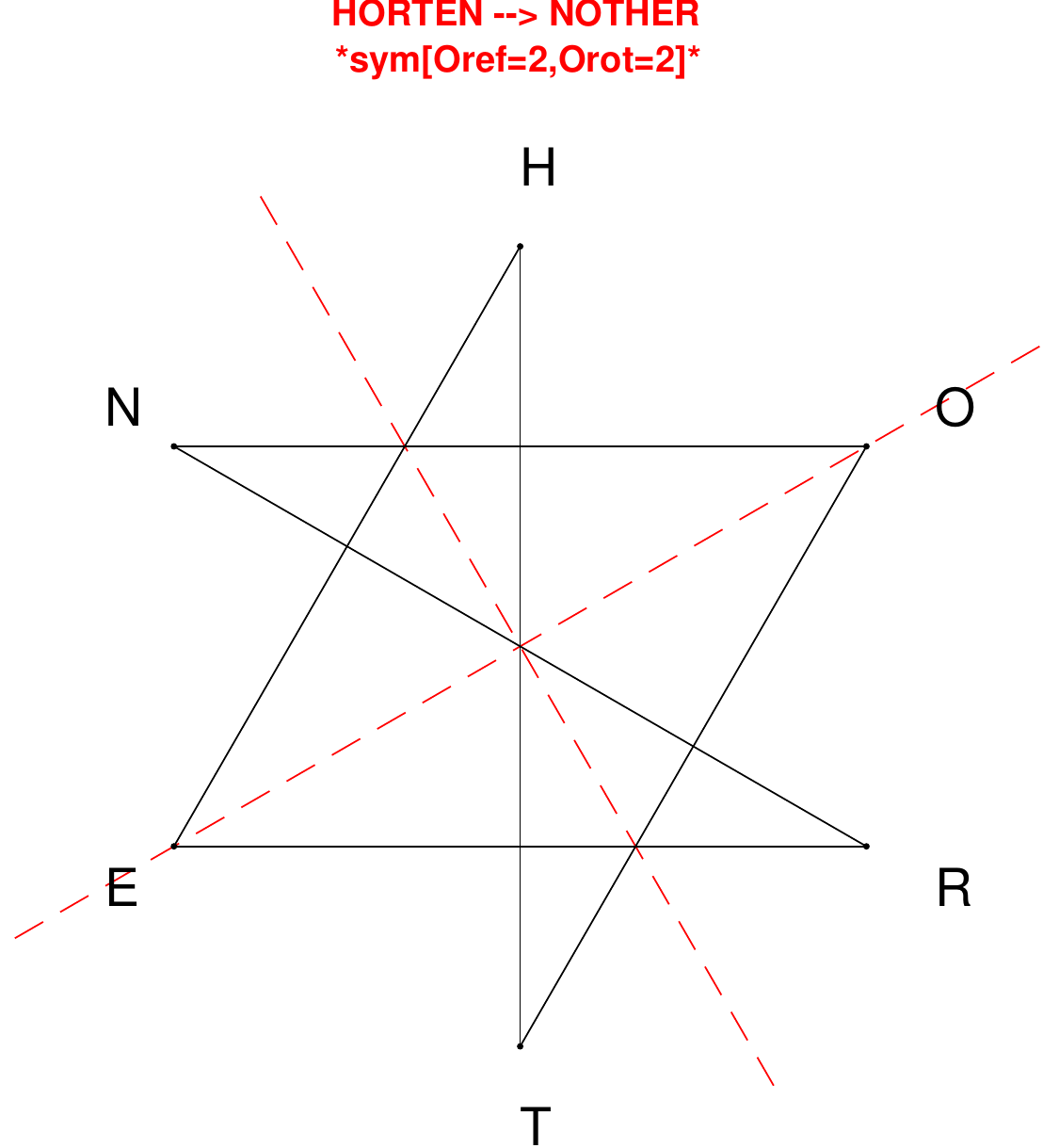}
\end{subfigure}
\hfill
\begin{subfigure}[T]{0.19\textwidth}
\centering
\includegraphics[width=\textwidth]{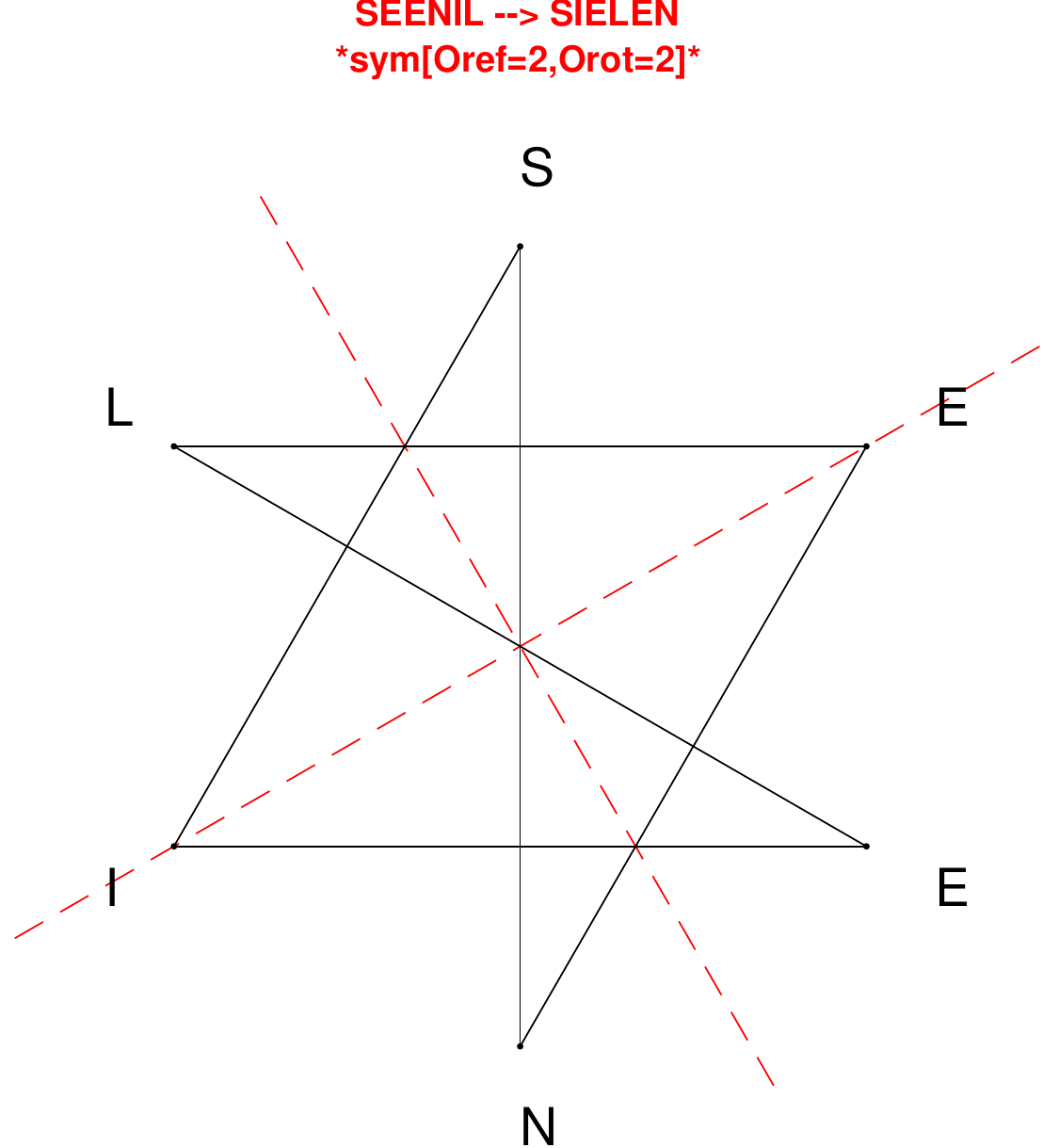}
\end{subfigure}
\hfill
\begin{subfigure}[T]{0.19\textwidth}
\centering
\includegraphics[width=\textwidth]{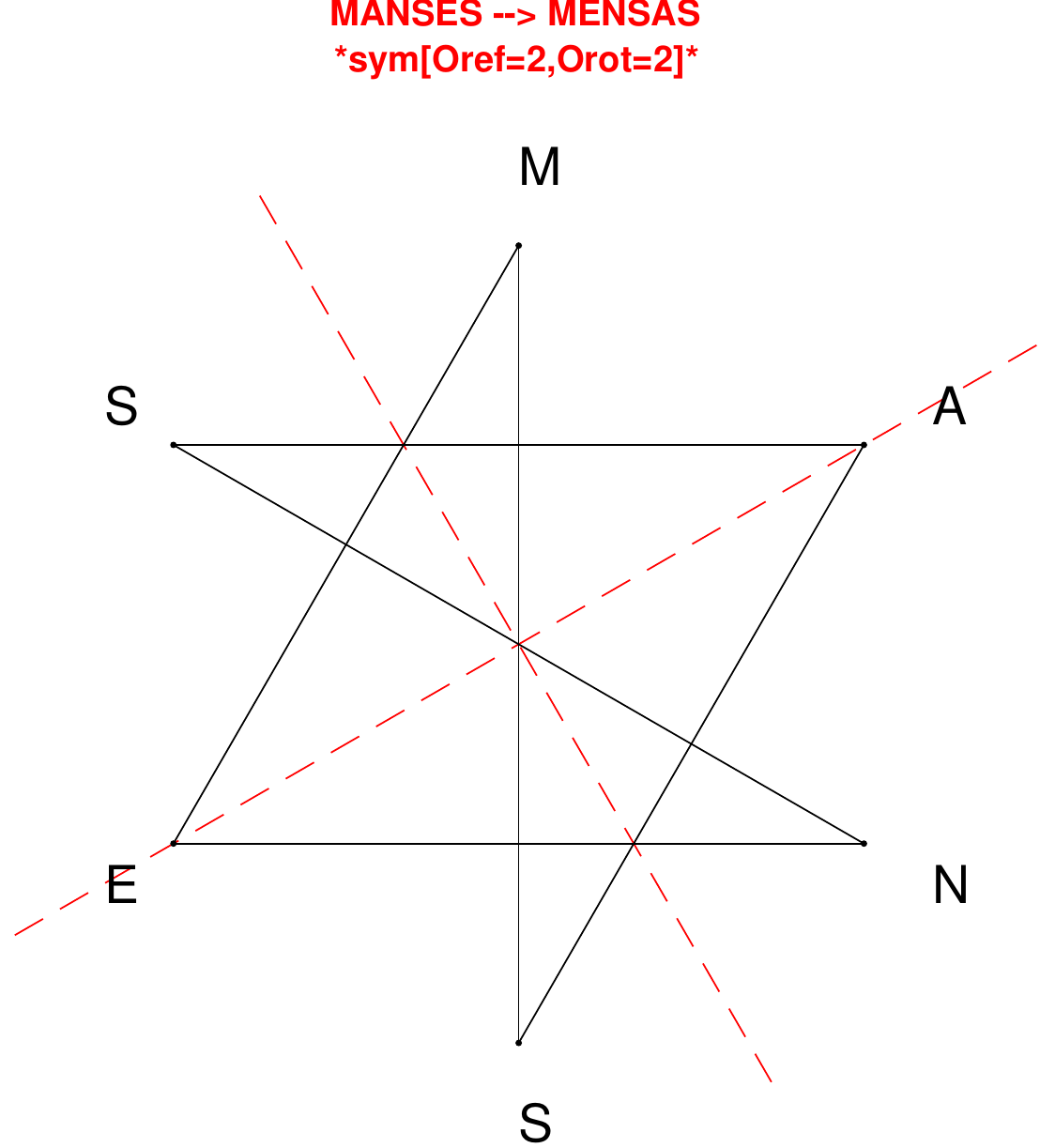}
\end{subfigure}
\end{figure}

\begin{figure}[H]
\centering
\begin{subfigure}[T]{0.19\textwidth}
\centering
\includegraphics[width=\textwidth]{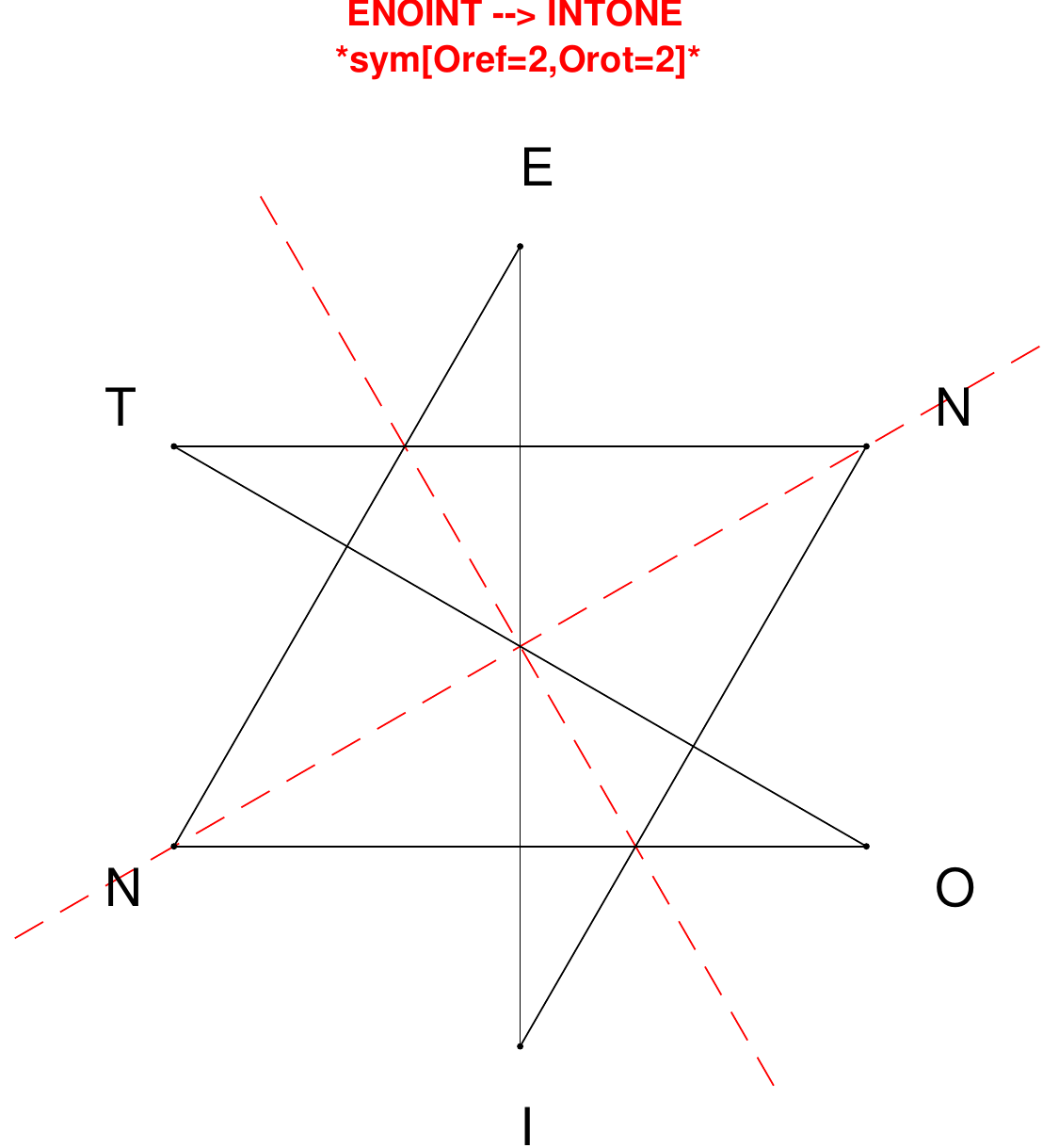}
\end{subfigure}
\hfill
\begin{subfigure}[T]{0.19\textwidth}
\centering
\includegraphics[width=\textwidth]{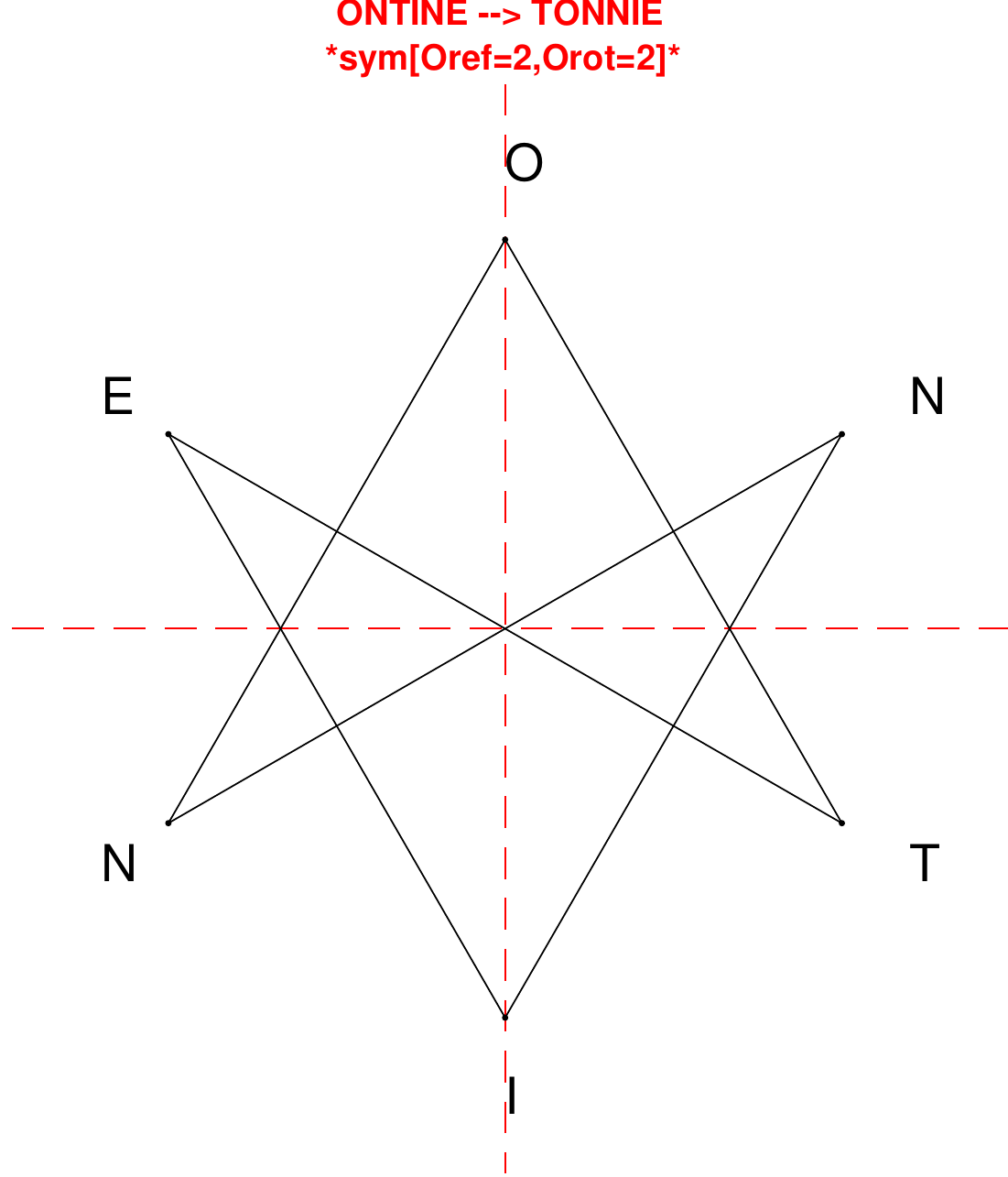}
\end{subfigure}
\hfill
\begin{subfigure}[T]{0.19\textwidth}
\centering
\includegraphics[width=\textwidth]{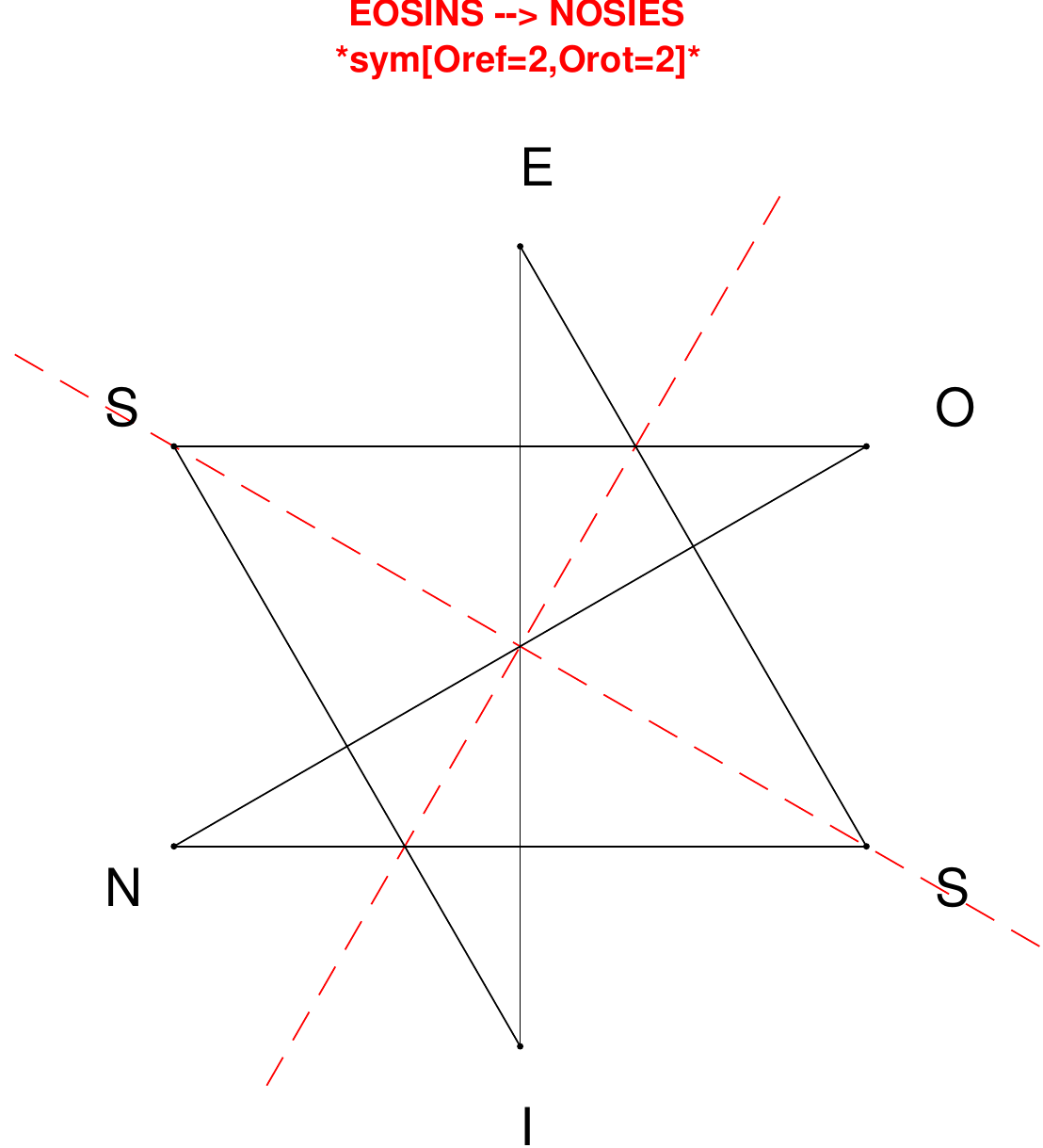}
\end{subfigure}
\hfill
\begin{subfigure}[T]{0.19\textwidth}
\centering
\includegraphics[width=\textwidth]{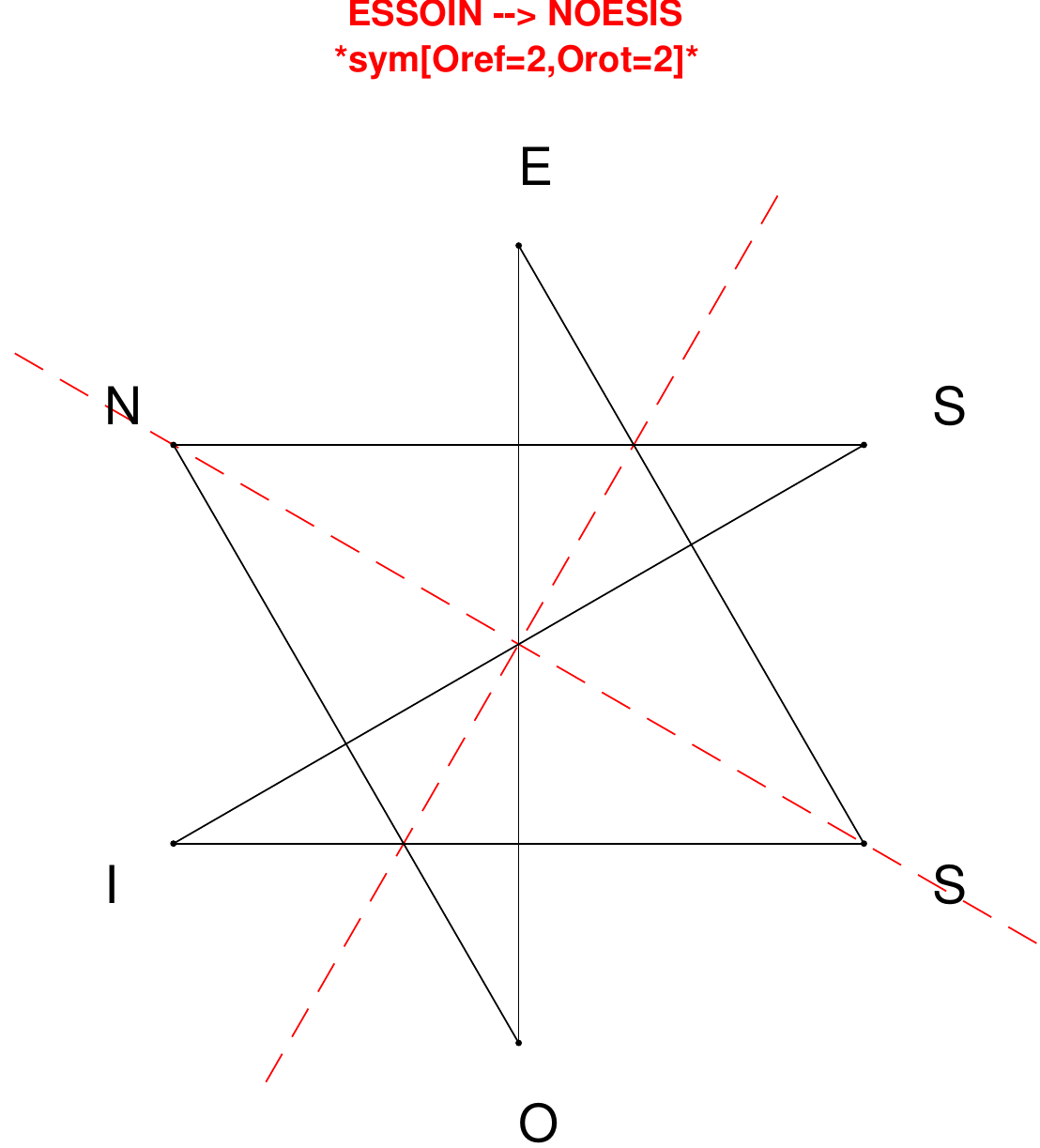}
\end{subfigure}
\hfill
\begin{subfigure}[T]{0.19\textwidth}
\centering
\includegraphics[width=\textwidth]{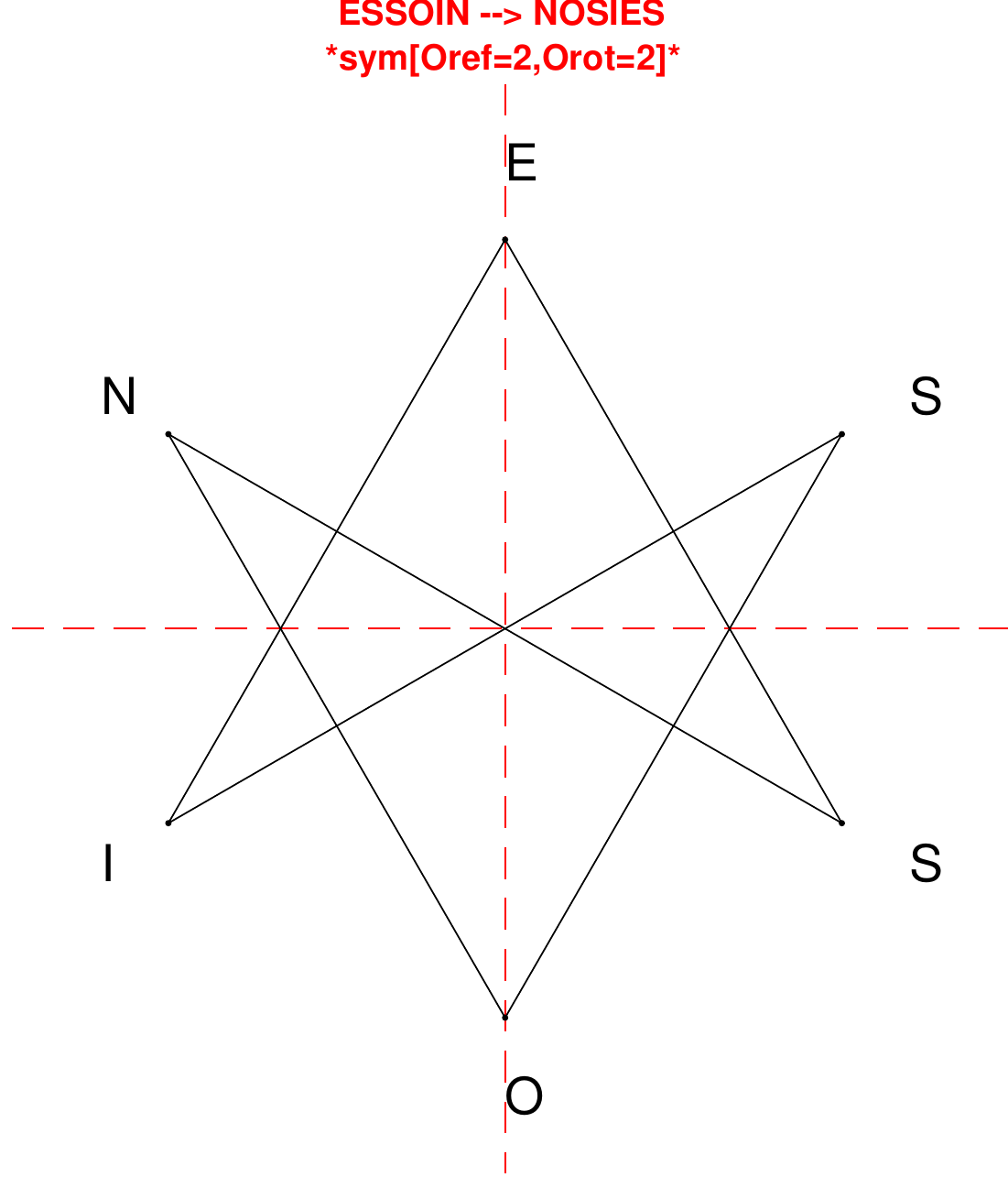}
\end{subfigure}
\end{figure}

\begin{figure}[H]
\centering
\begin{subfigure}[T]{0.19\textwidth}
\centering
\includegraphics[width=\textwidth]{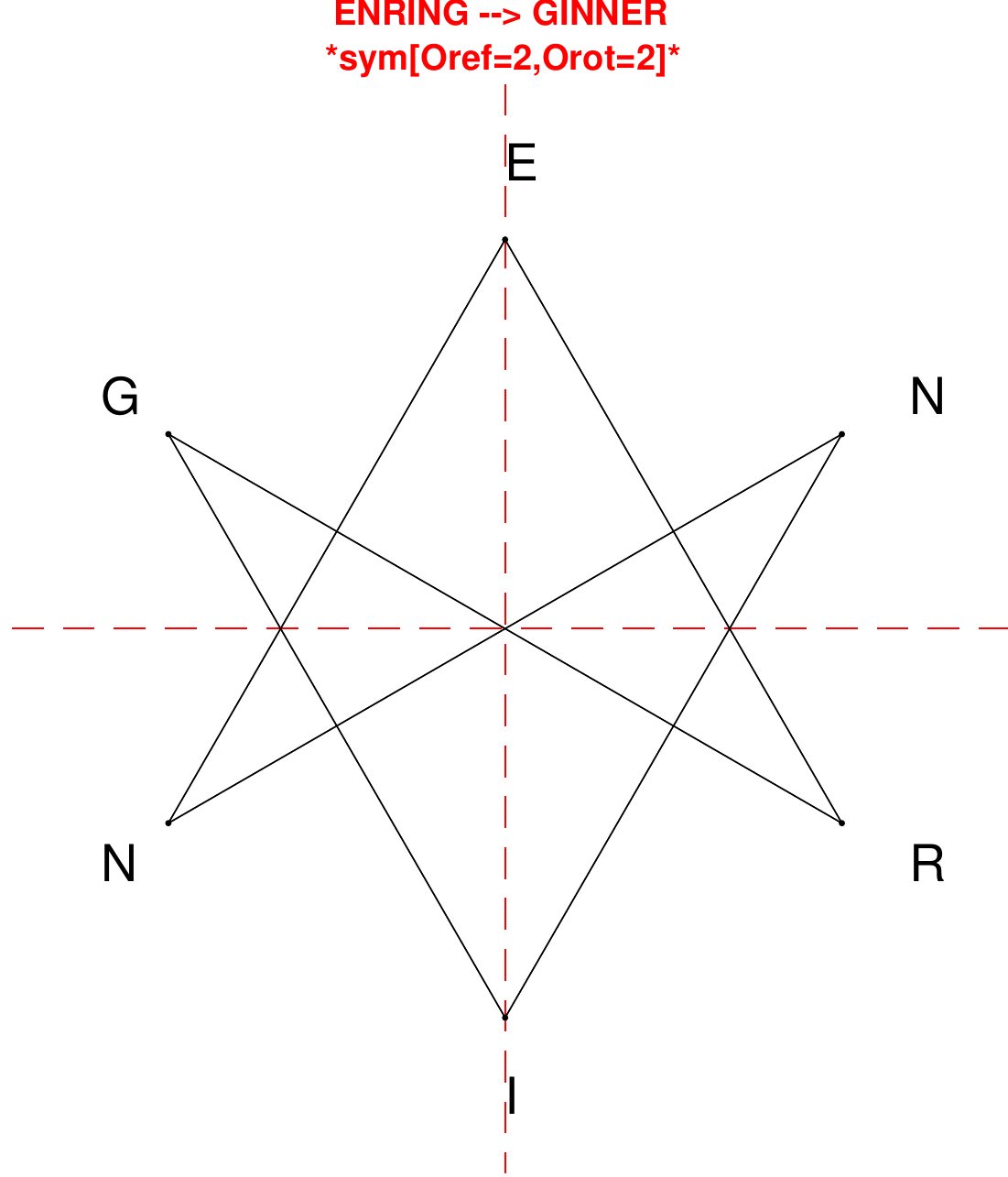}
\end{subfigure}
\hfill
\begin{subfigure}[T]{0.19\textwidth}
\centering
\includegraphics[width=\textwidth]{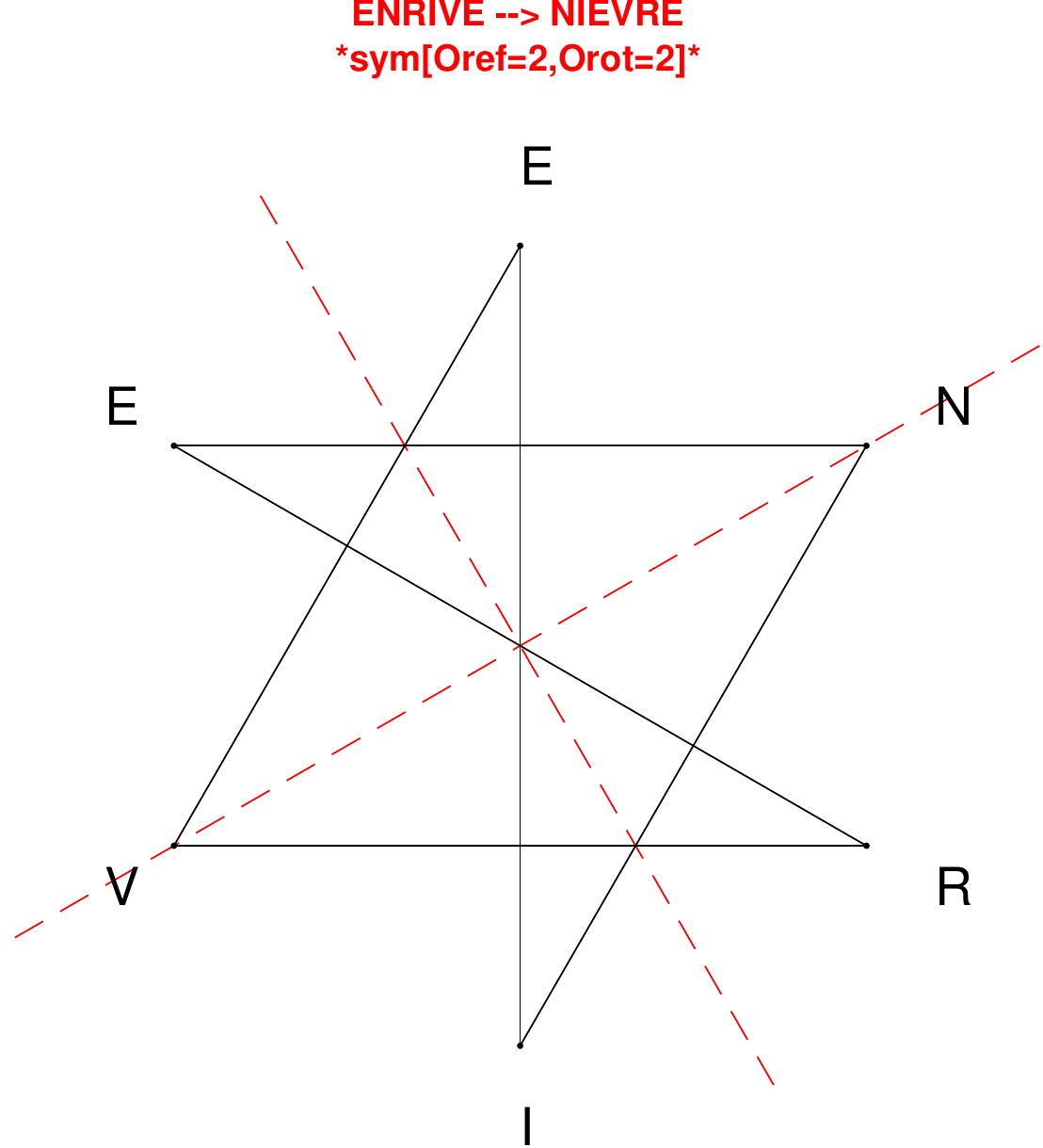}
\end{subfigure}
\hfill
\begin{subfigure}[T]{0.19\textwidth}
\centering
\includegraphics[width=\textwidth]{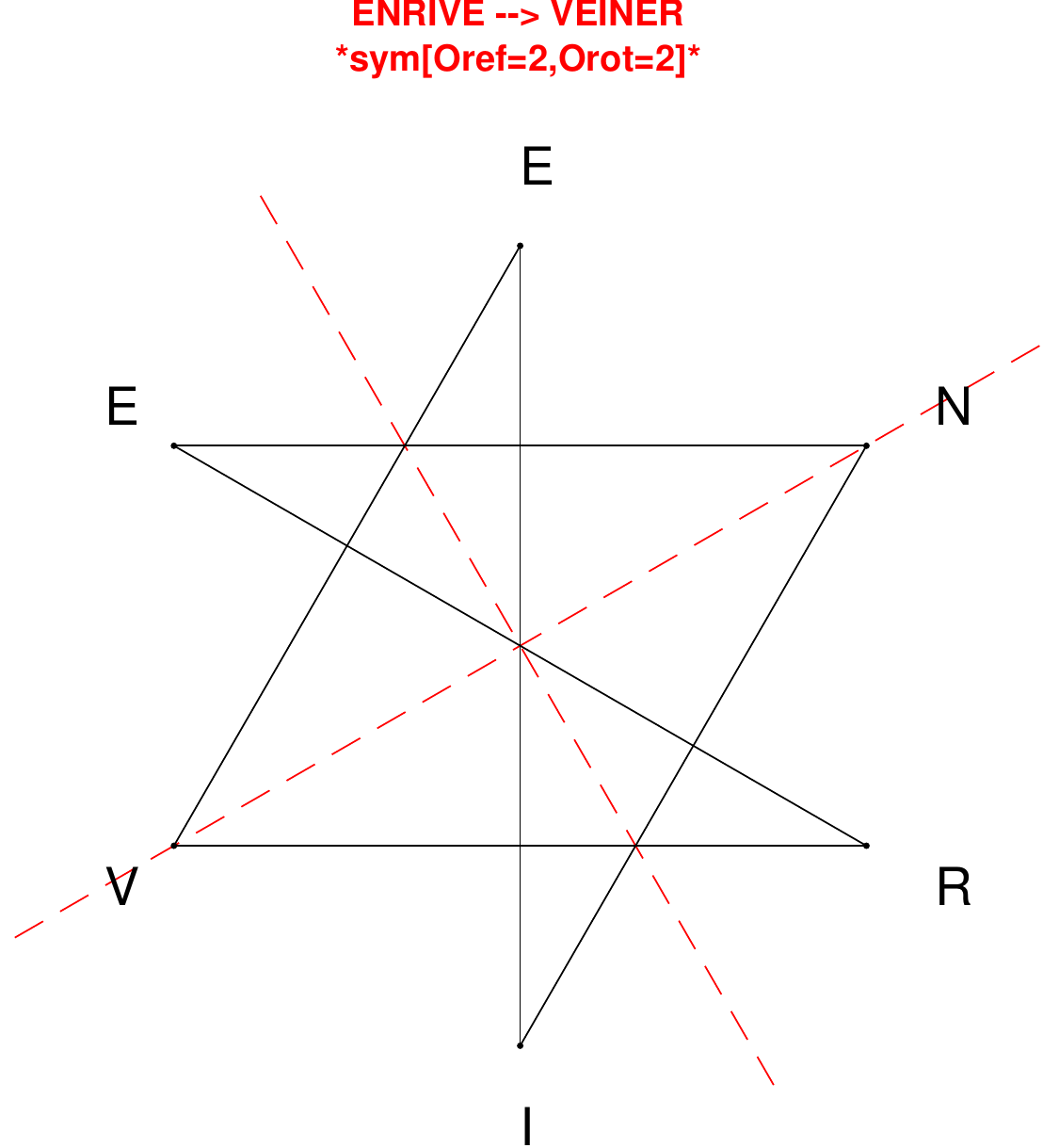}
\end{subfigure}
\hfill
\begin{subfigure}[T]{0.19\textwidth}
\centering
\includegraphics[width=\textwidth]{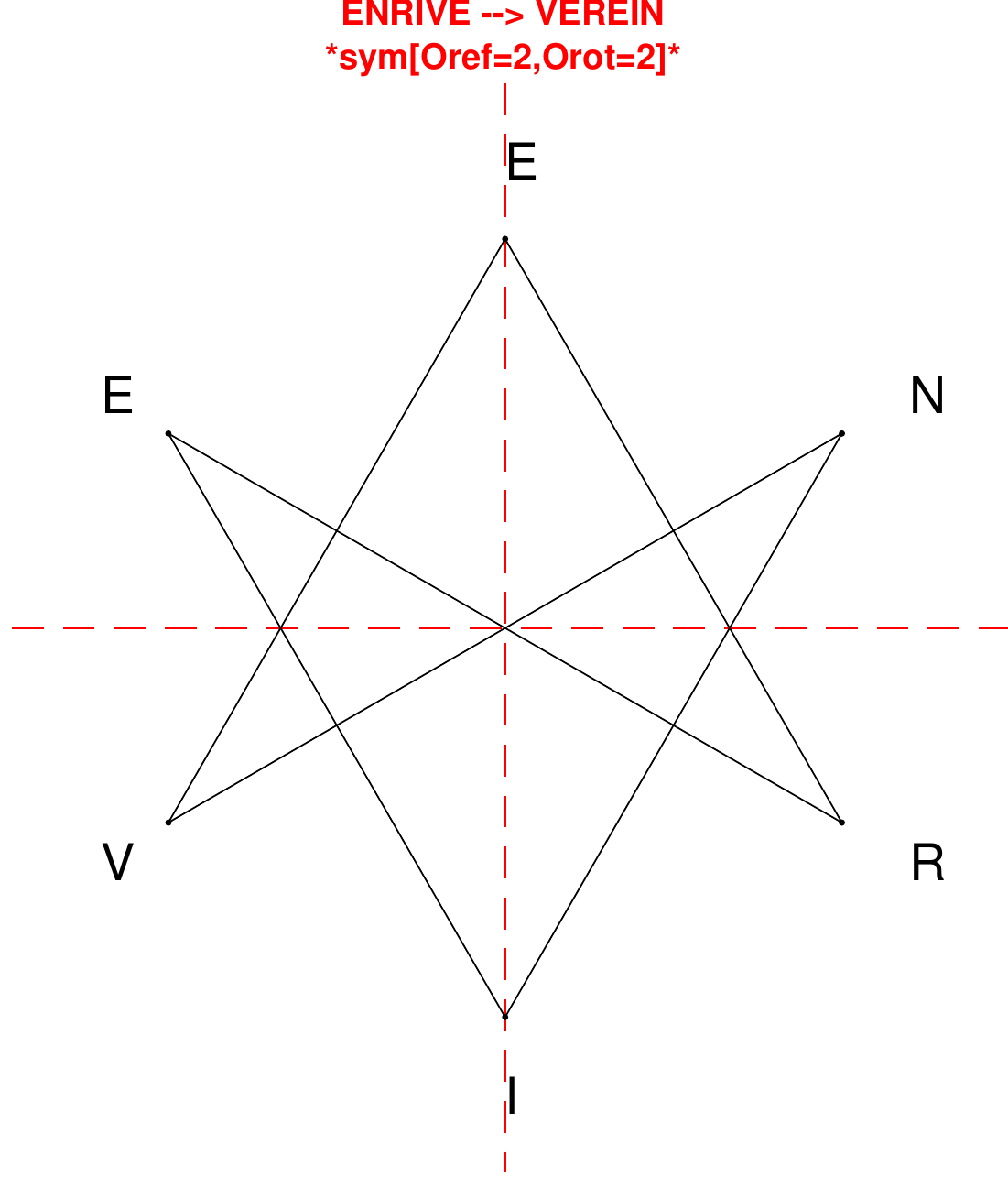}
\end{subfigure}
\hfill
\begin{subfigure}[T]{0.19\textwidth}
\centering
\includegraphics[width=\textwidth]{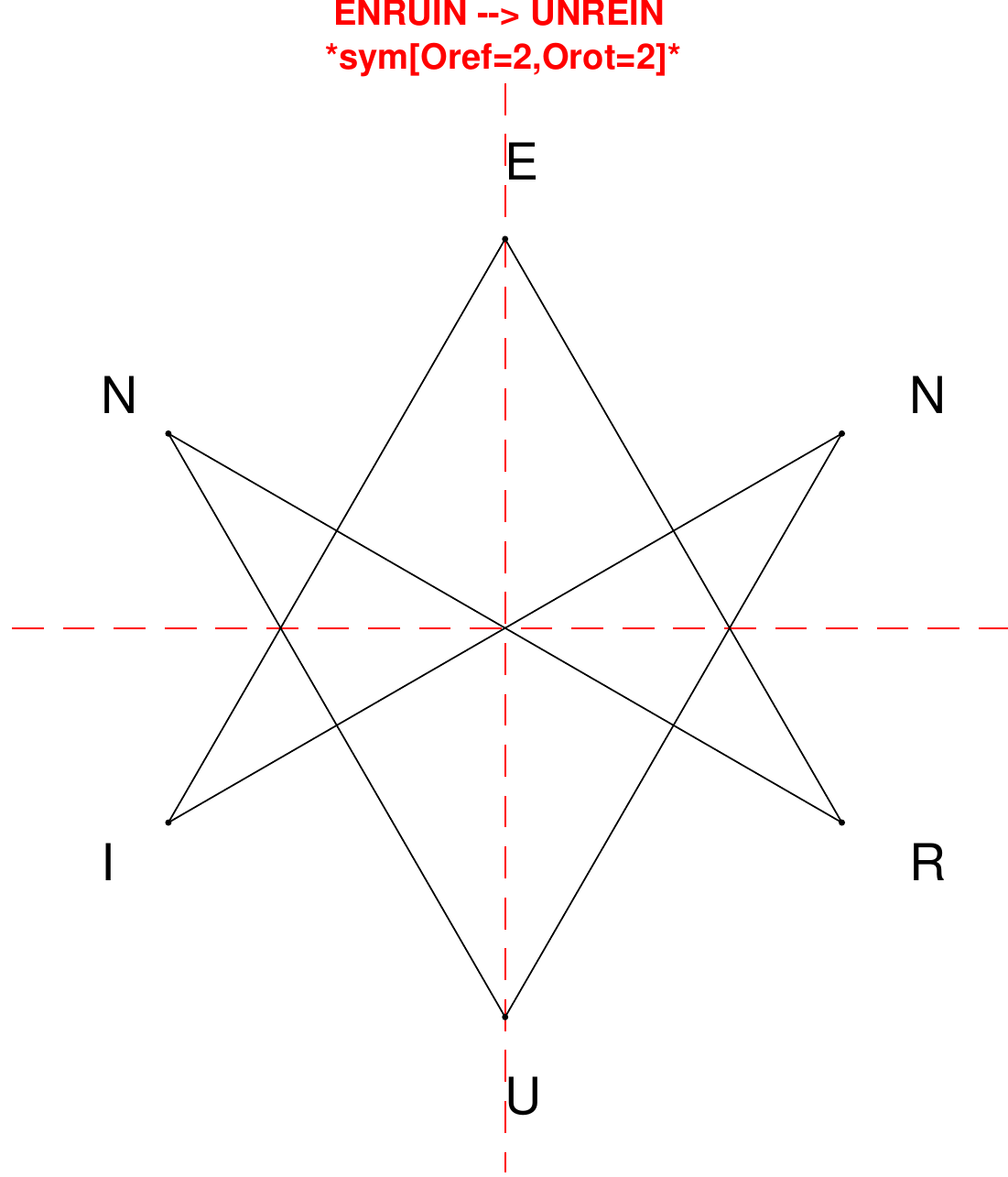}
\end{subfigure}
\end{figure}

\begin{figure}[H]
\centering
\begin{subfigure}[T]{0.19\textwidth}
\centering
\includegraphics[width=\textwidth]{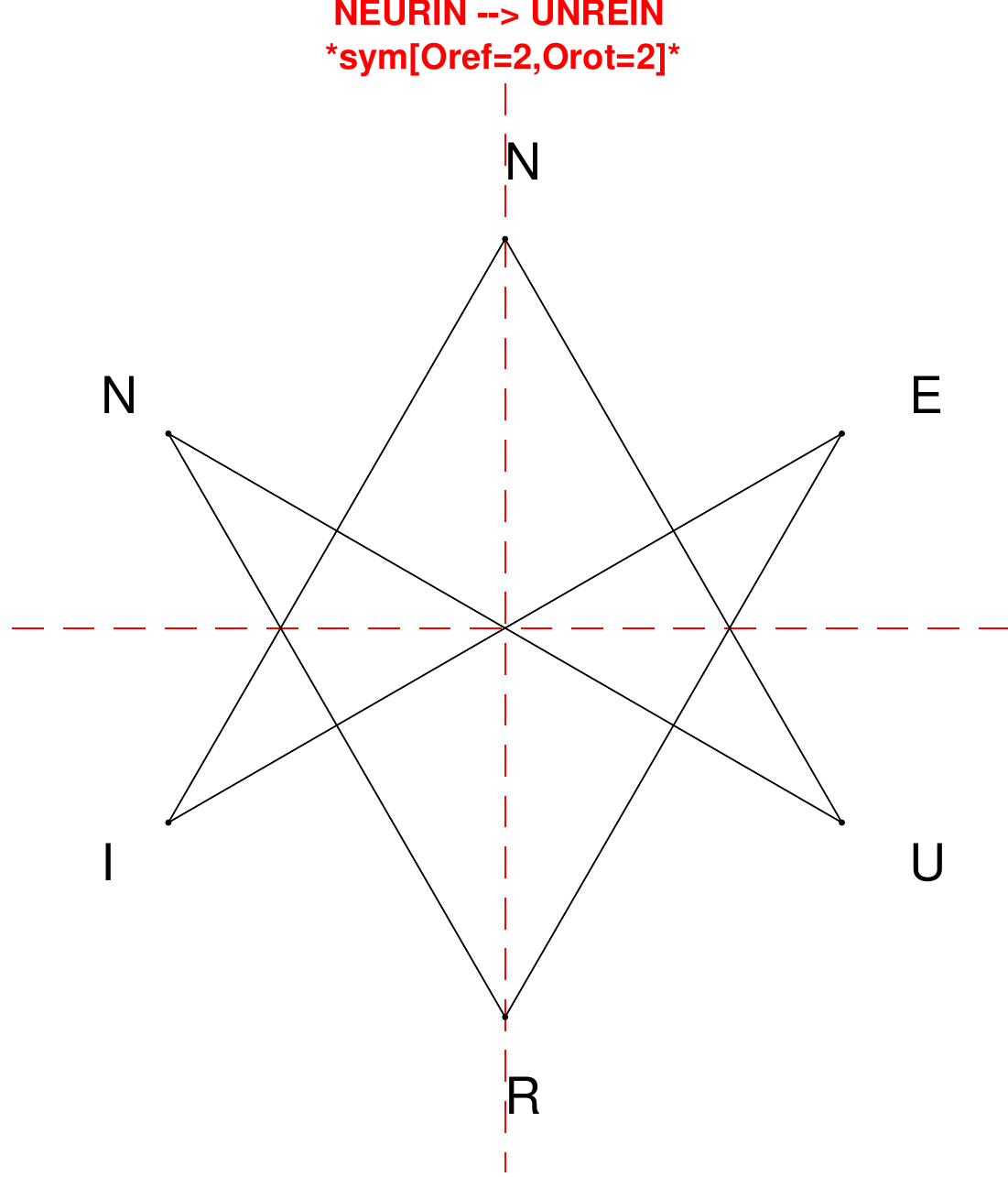}
\end{subfigure}
\hfill
\begin{subfigure}[T]{0.19\textwidth}
\centering
\includegraphics[width=\textwidth]{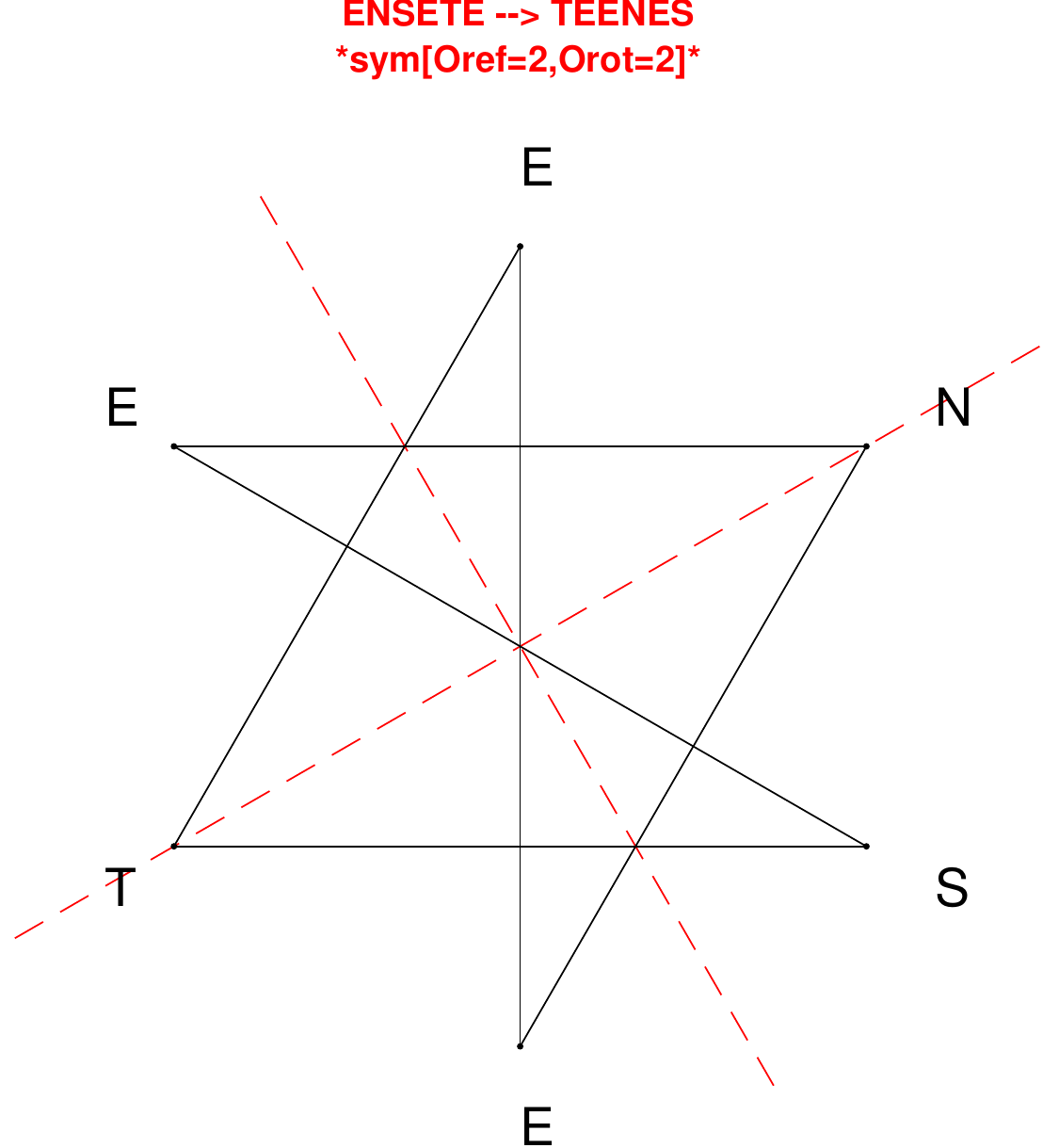}
\end{subfigure}
\hfill
\begin{subfigure}[T]{0.19\textwidth}
\centering
\includegraphics[width=\textwidth]{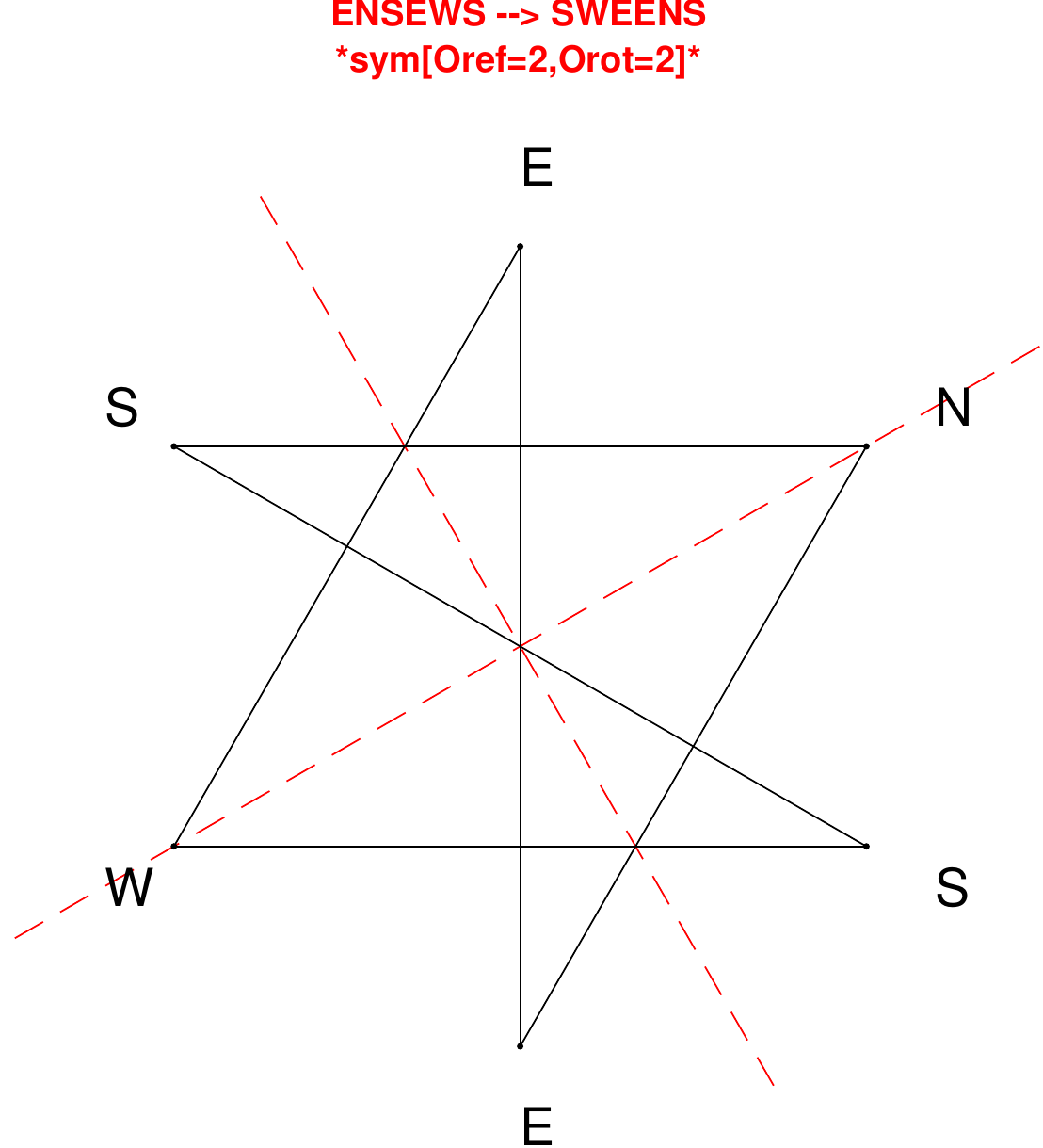}
\end{subfigure}
\hfill
\begin{subfigure}[T]{0.19\textwidth}
\centering
\includegraphics[width=\textwidth]{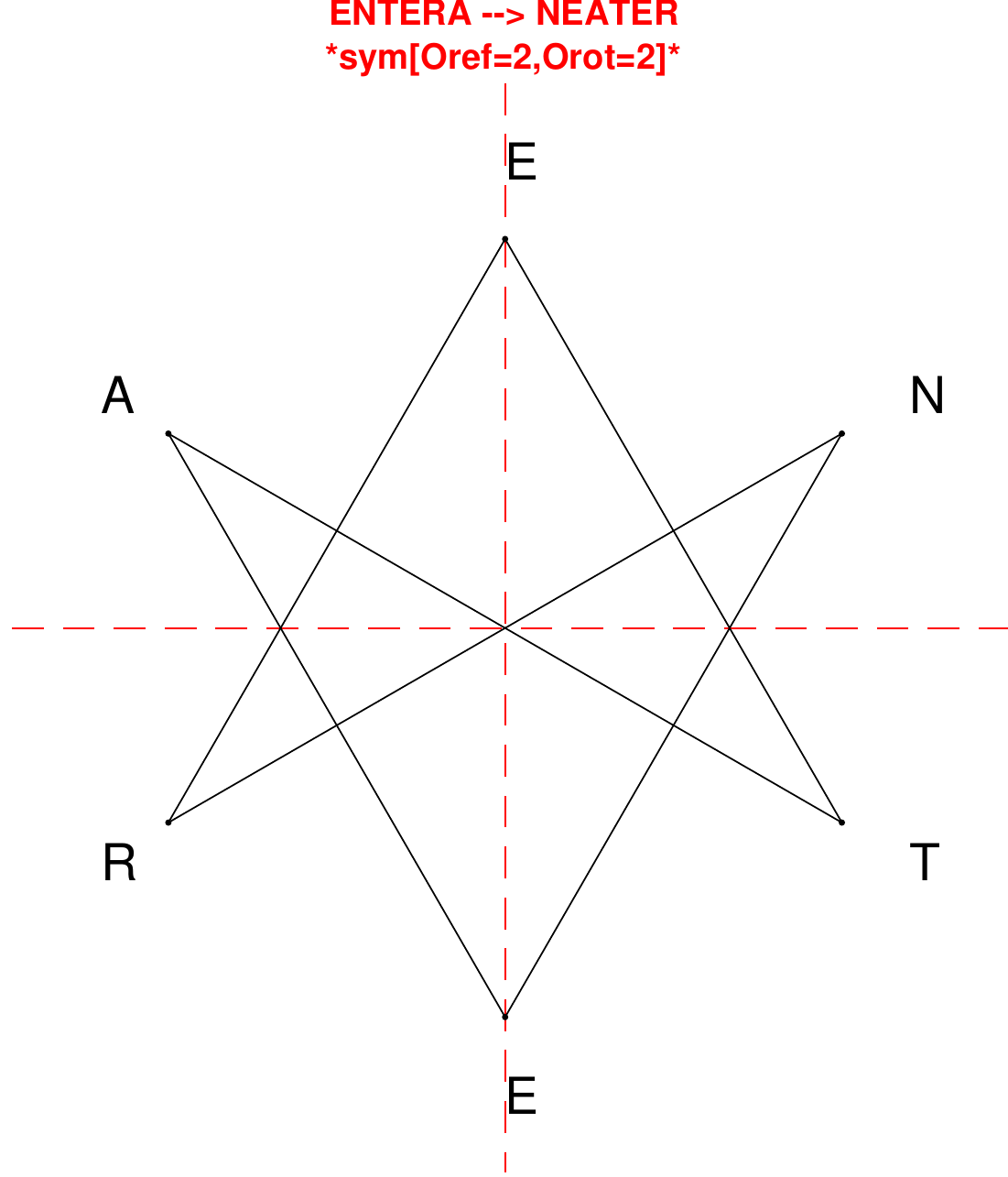}
\end{subfigure}
\hfill
\begin{subfigure}[T]{0.19\textwidth}
\centering
\includegraphics[width=\textwidth]{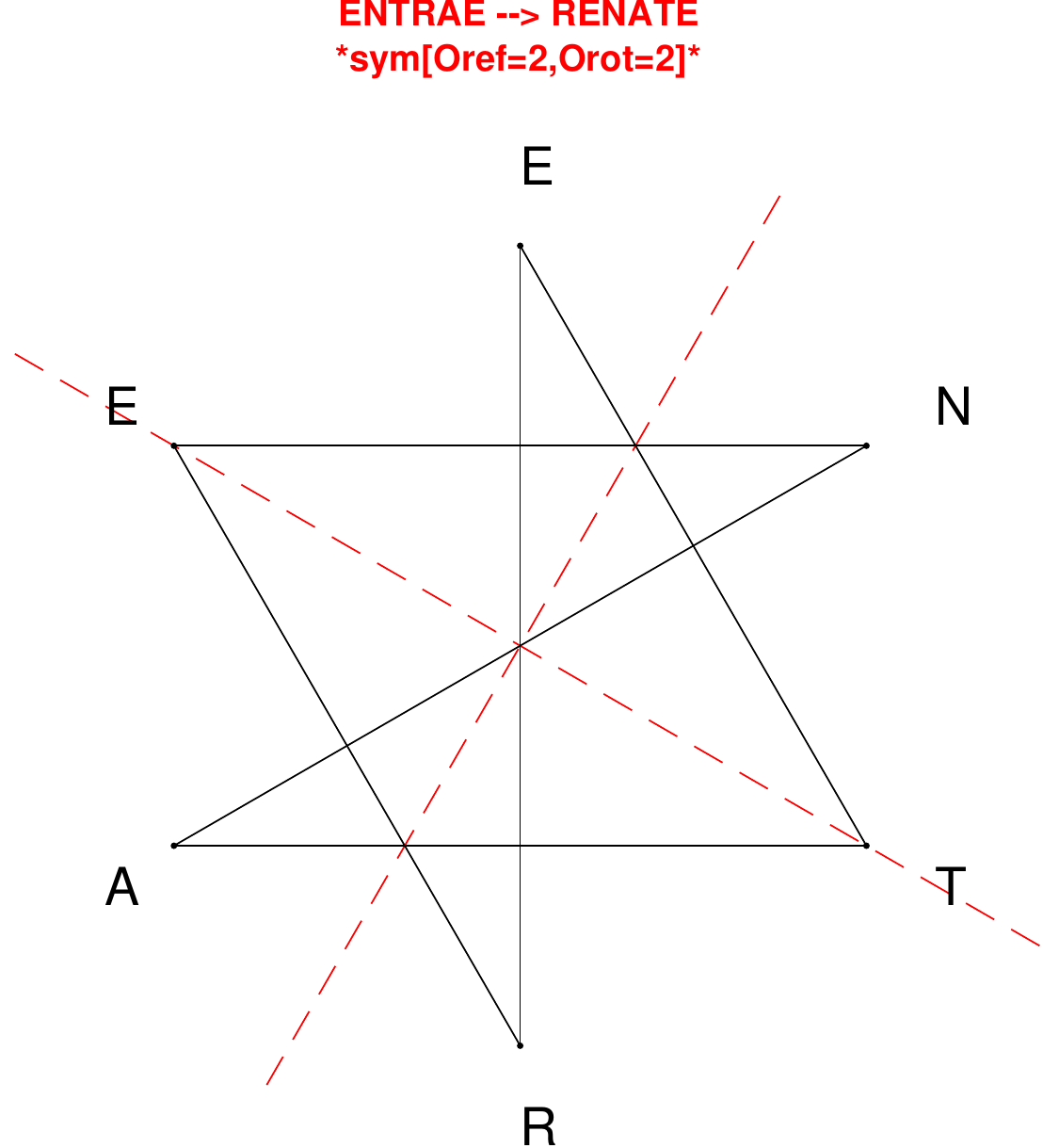}
\end{subfigure}
\end{figure}

\begin{figure}[H]
\centering
\begin{subfigure}[T]{0.19\textwidth}
\centering
\includegraphics[width=\textwidth]{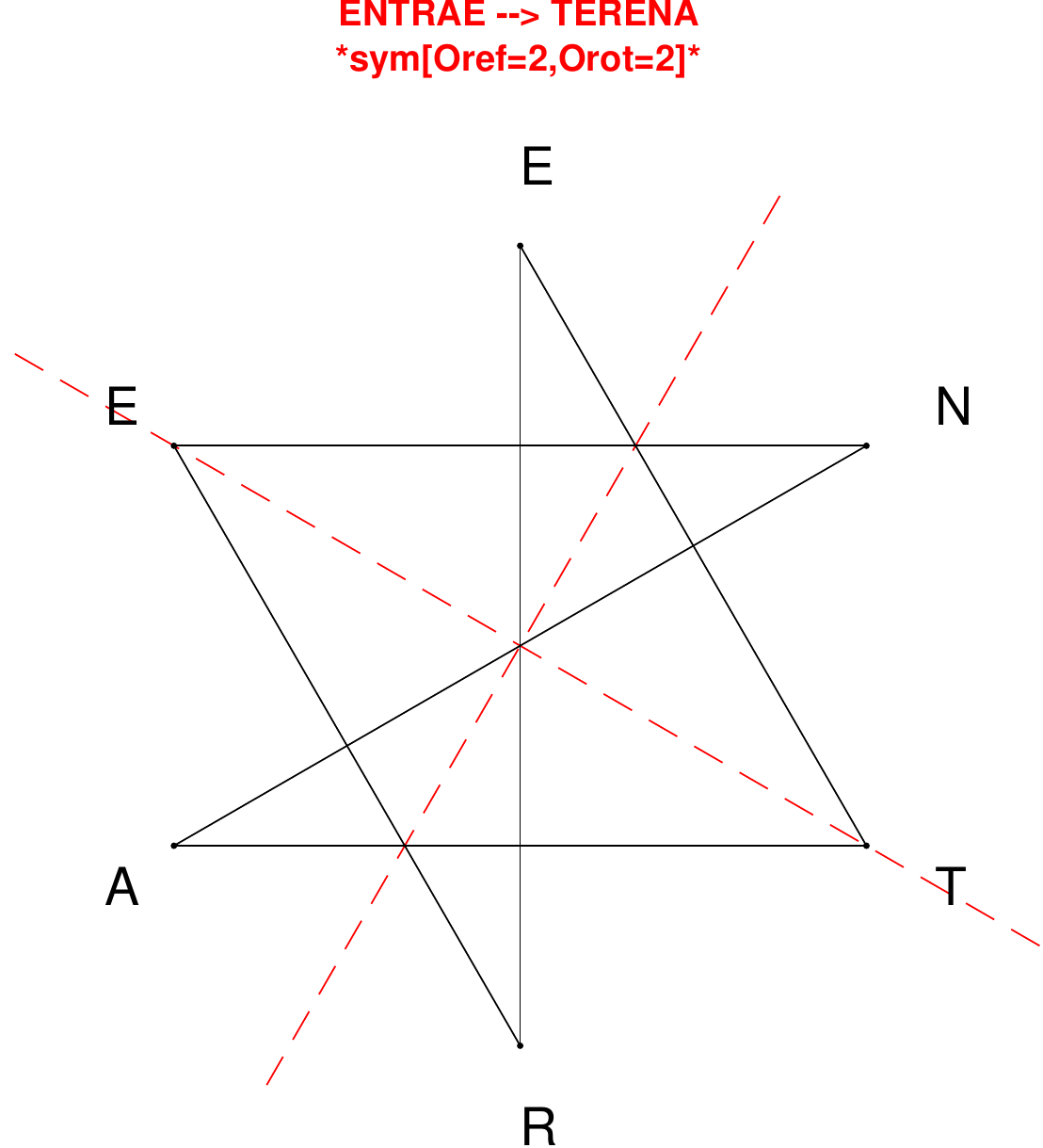}
\end{subfigure}
\hfill
\begin{subfigure}[T]{0.19\textwidth}
\centering
\includegraphics[width=\textwidth]{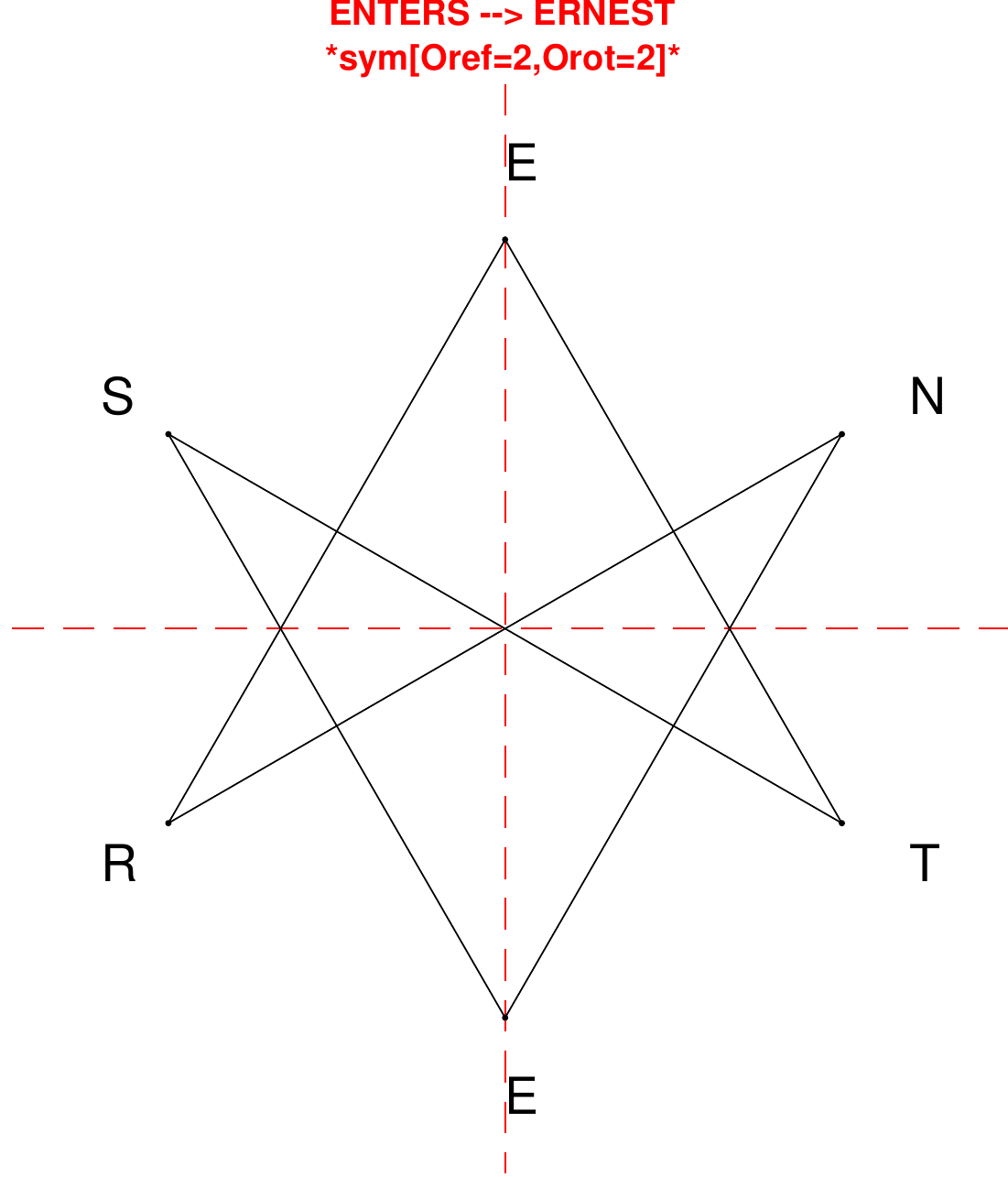}
\end{subfigure}
\hfill
\begin{subfigure}[T]{0.19\textwidth}
\centering
\includegraphics[width=\textwidth]{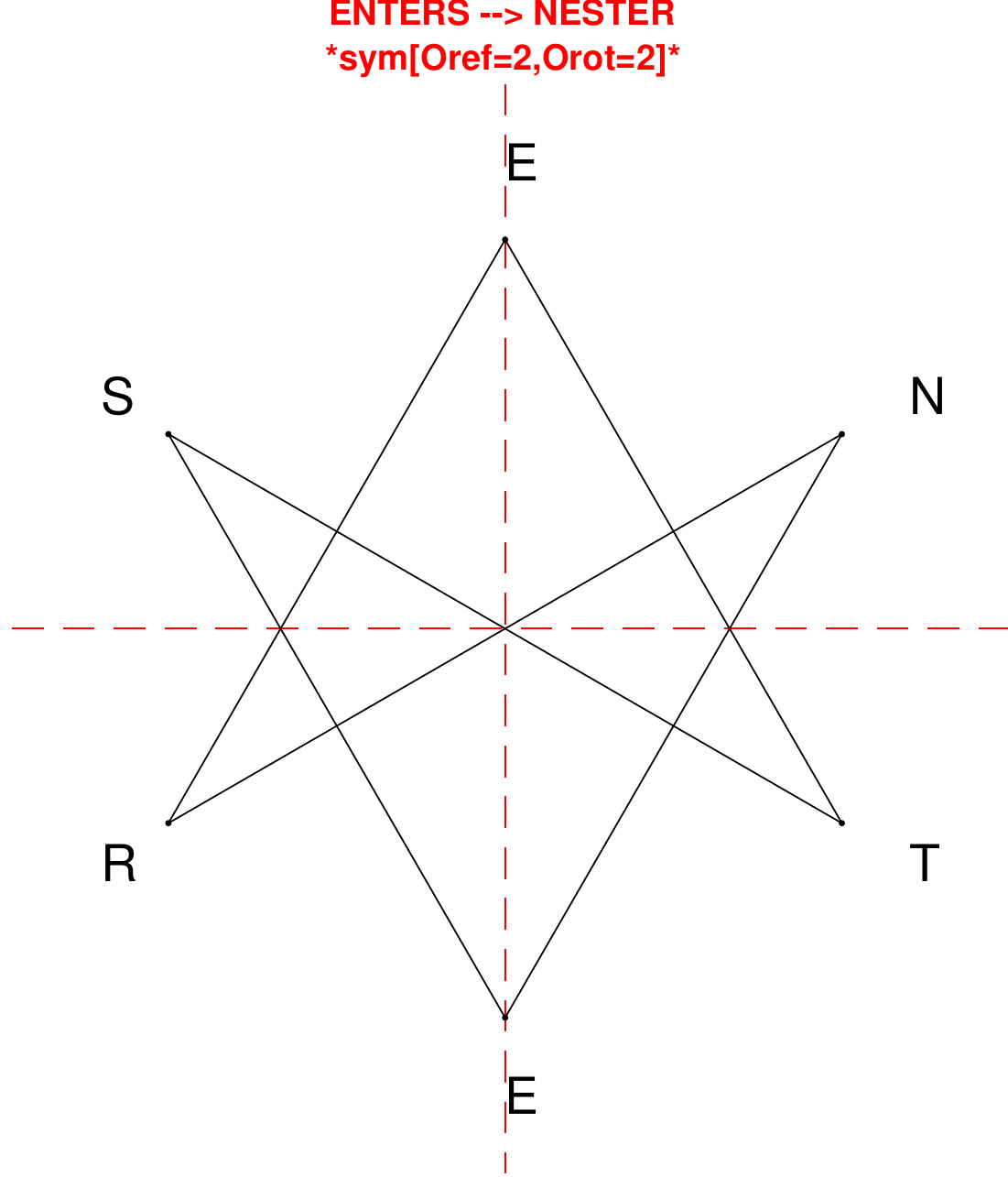}
\end{subfigure}
\hfill
\begin{subfigure}[T]{0.19\textwidth}
\centering
\includegraphics[width=\textwidth]{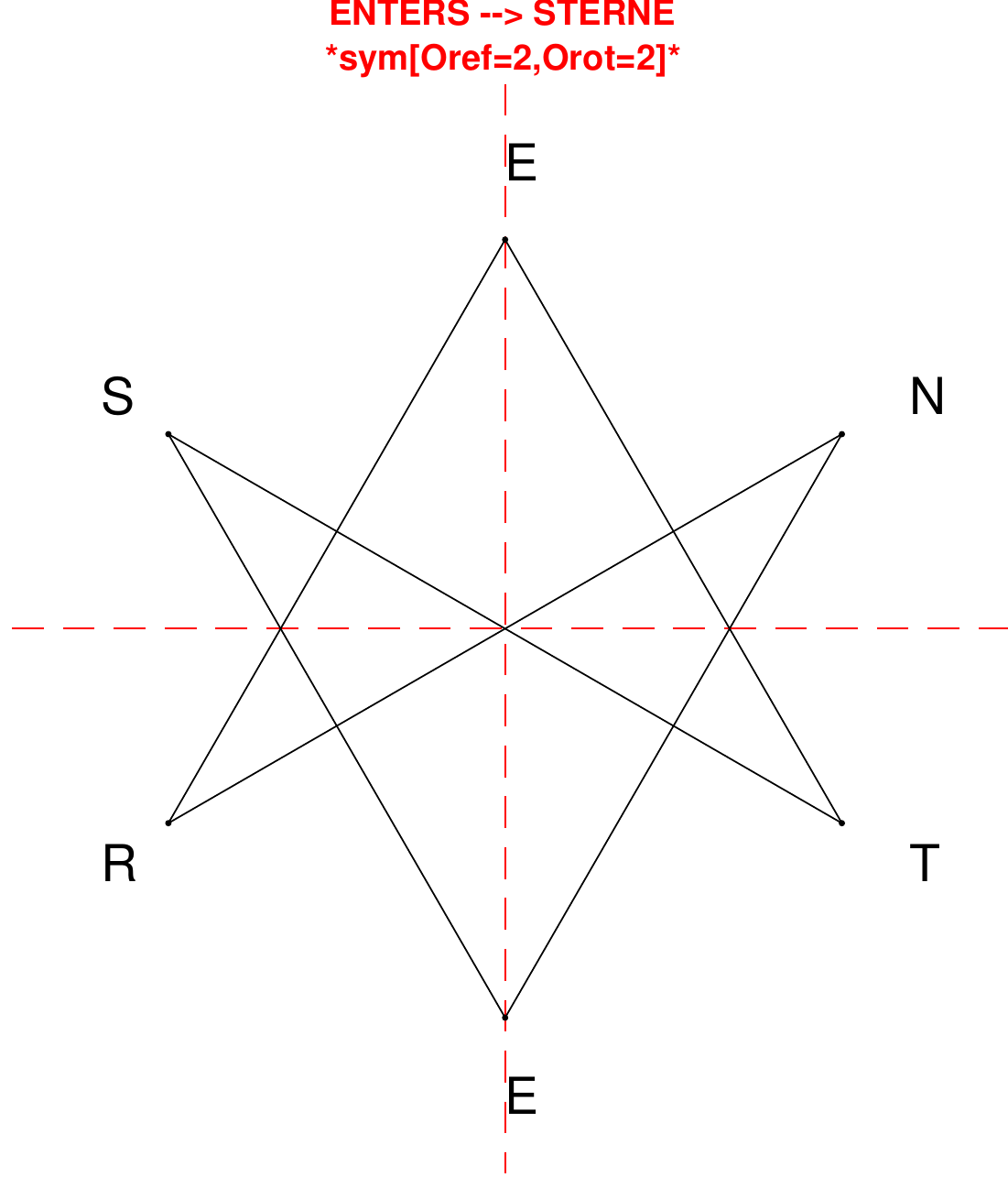}
\end{subfigure}
\hfill
\begin{subfigure}[T]{0.19\textwidth}
\centering
\includegraphics[width=\textwidth]{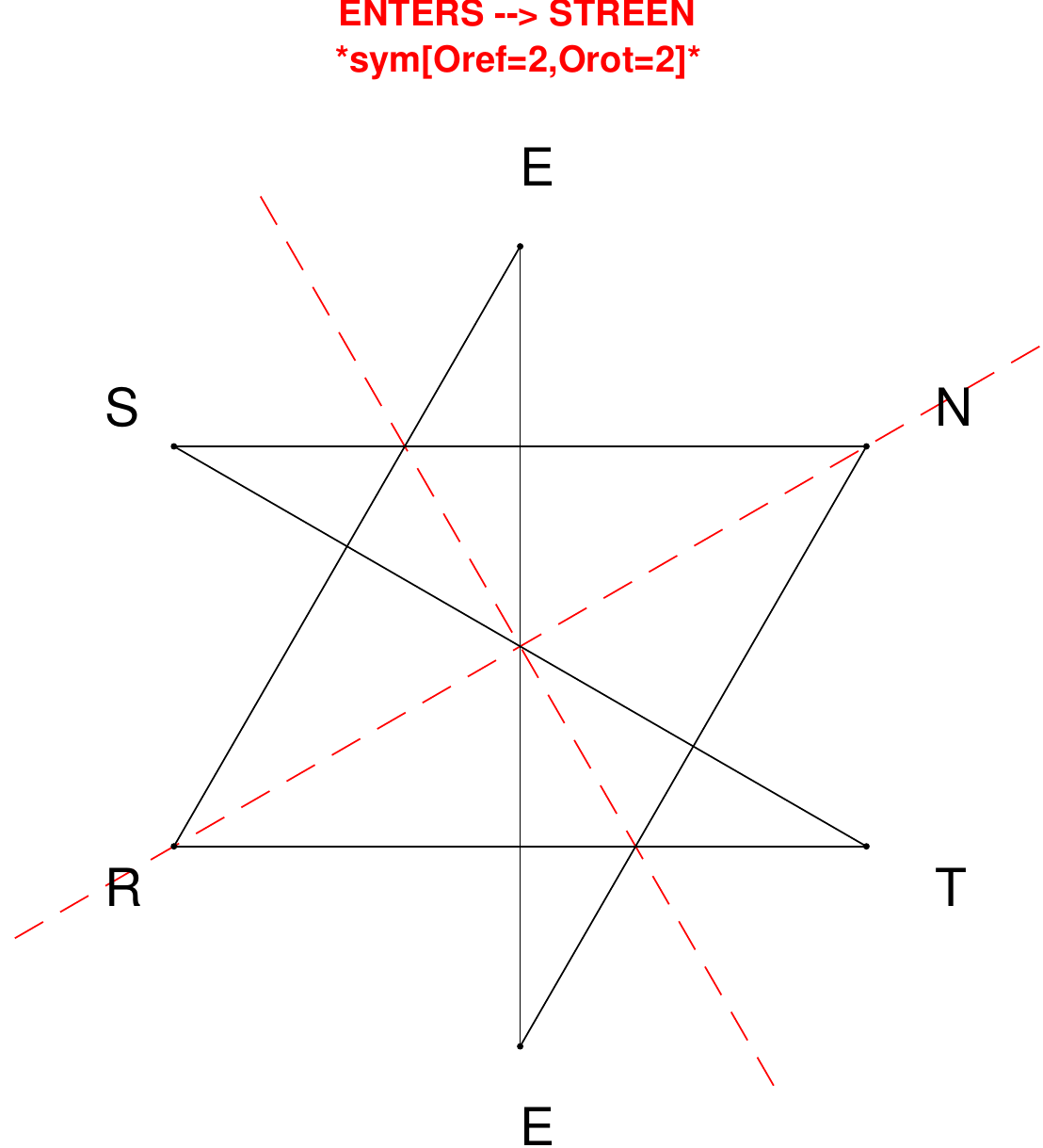}
\end{subfigure}
\end{figure}

\begin{figure}[H]
\centering
\begin{subfigure}[T]{0.19\textwidth}
\centering
\includegraphics[width=\textwidth]{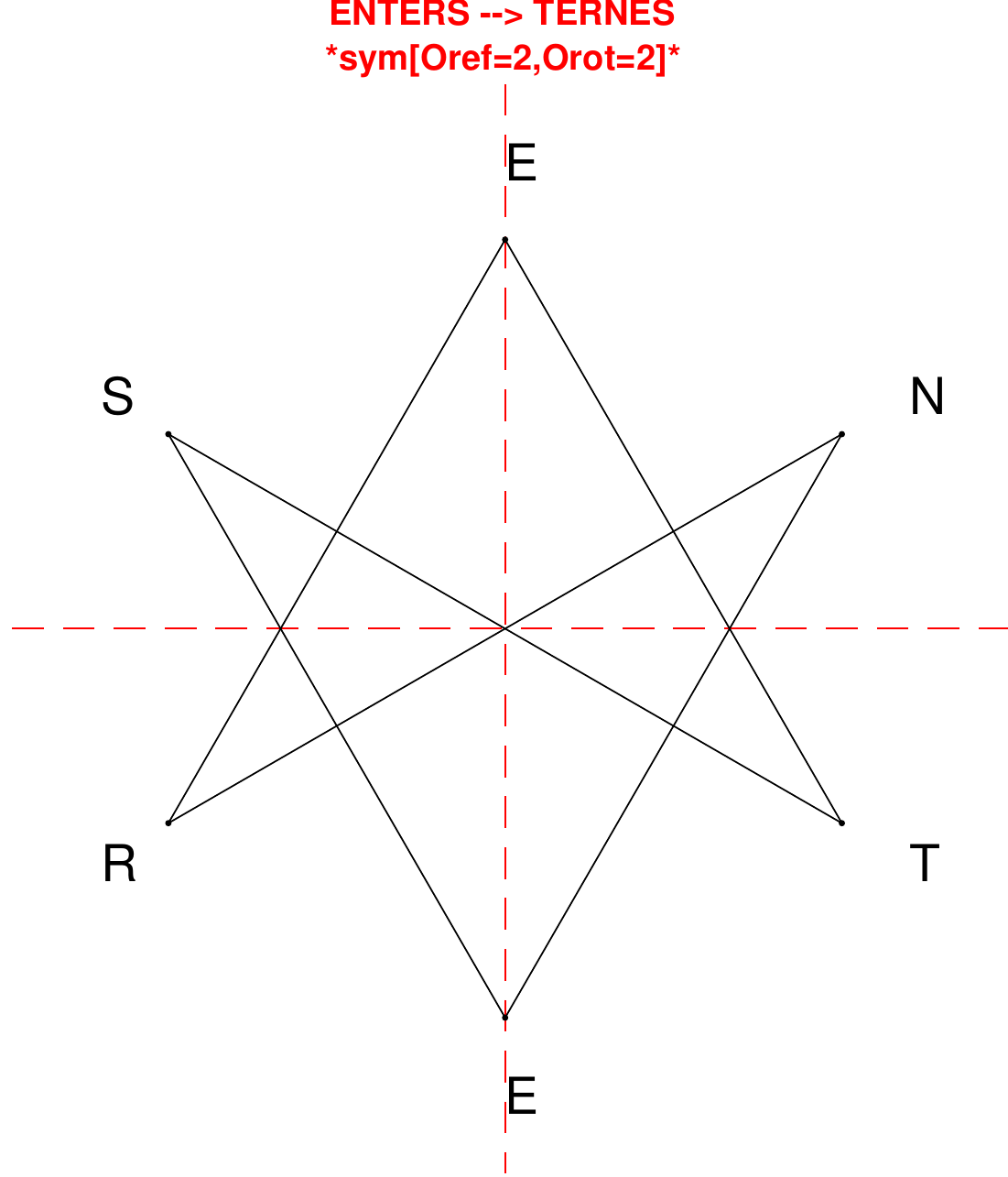}
\end{subfigure}
\hfill
\begin{subfigure}[T]{0.19\textwidth}
\centering
\includegraphics[width=\textwidth]{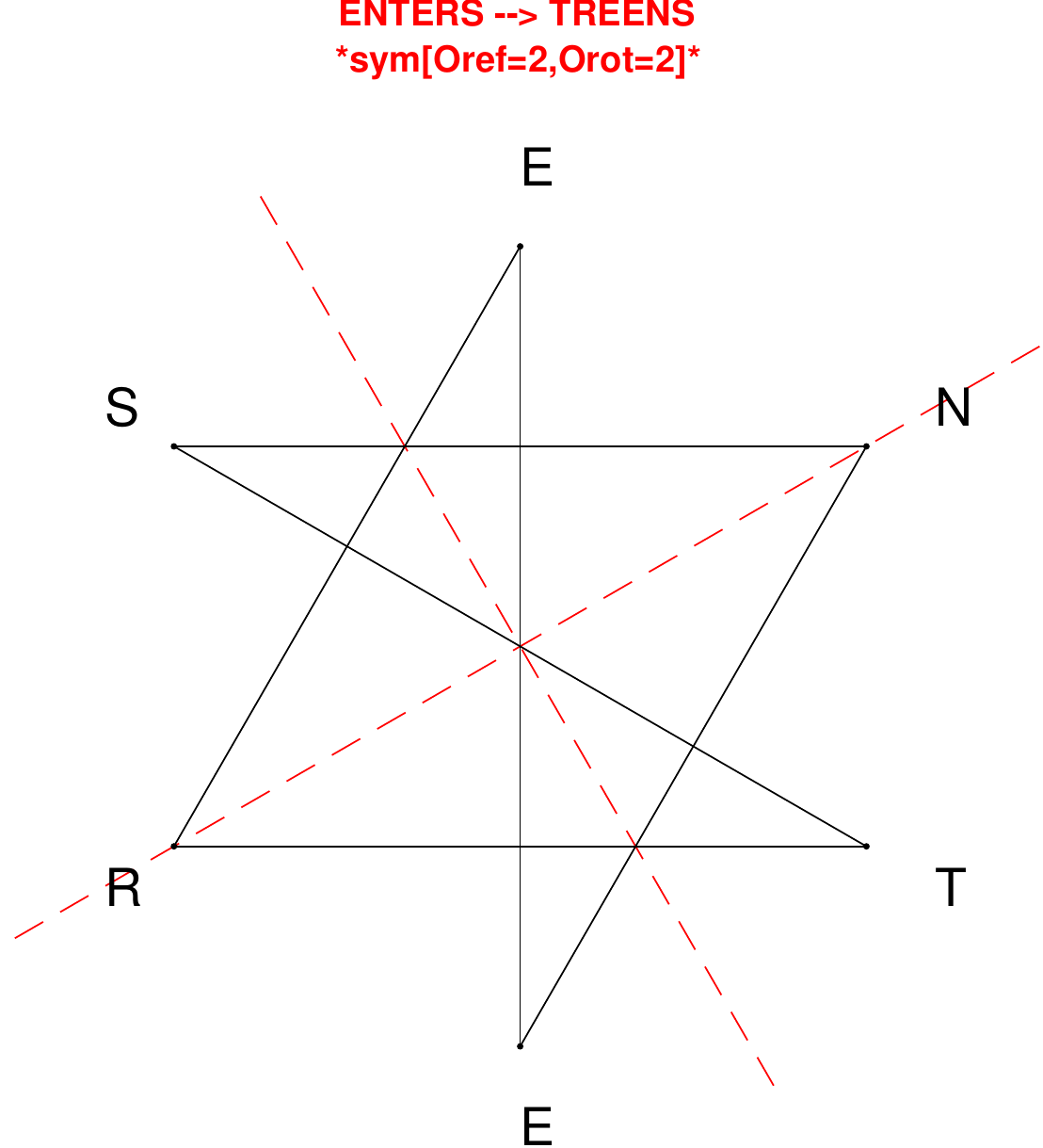}
\end{subfigure}
\hfill
\begin{subfigure}[T]{0.19\textwidth}
\centering
\includegraphics[width=\textwidth]{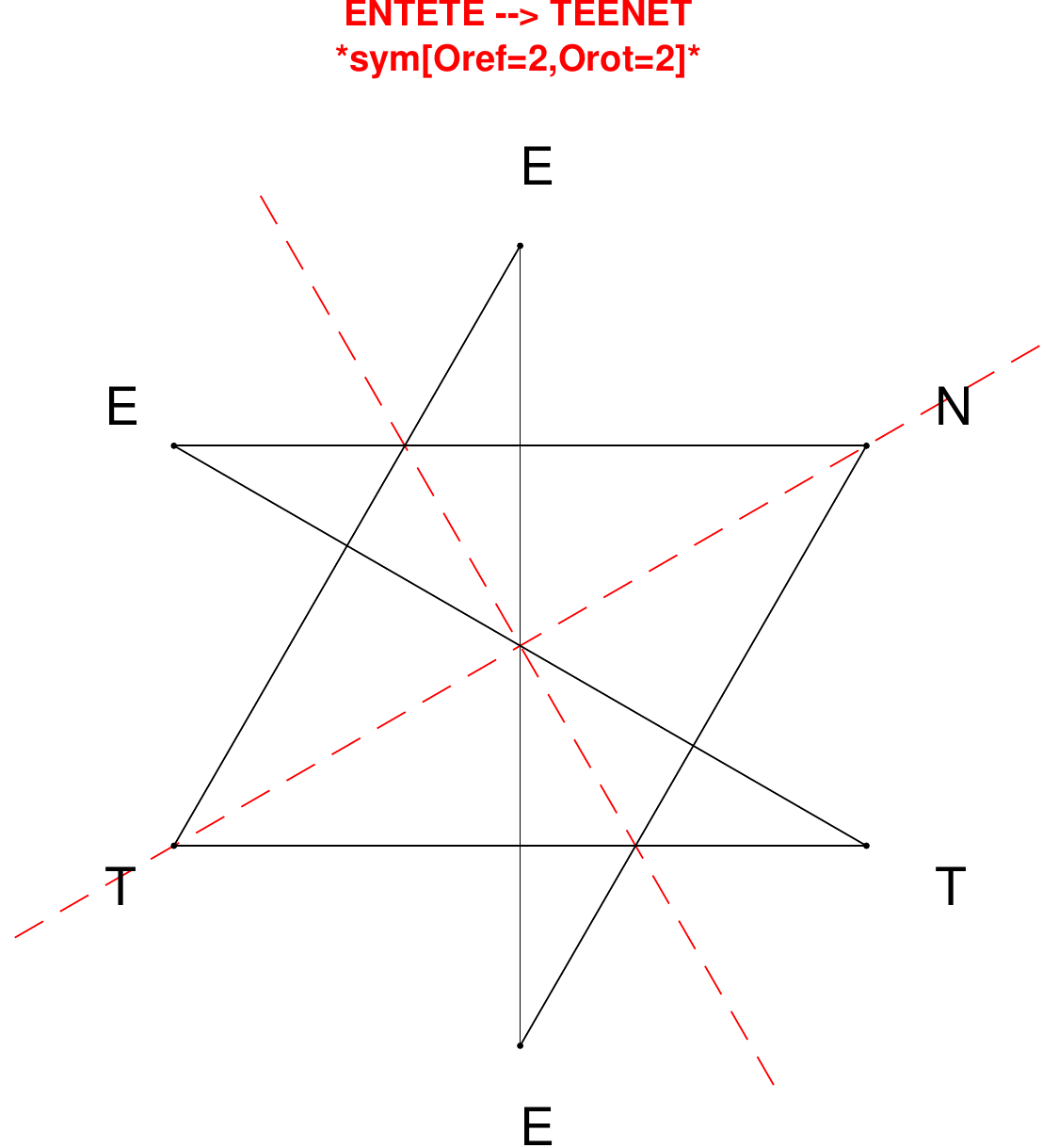}
\end{subfigure}
\hfill
\begin{subfigure}[T]{0.19\textwidth}
\centering
\includegraphics[width=\textwidth]{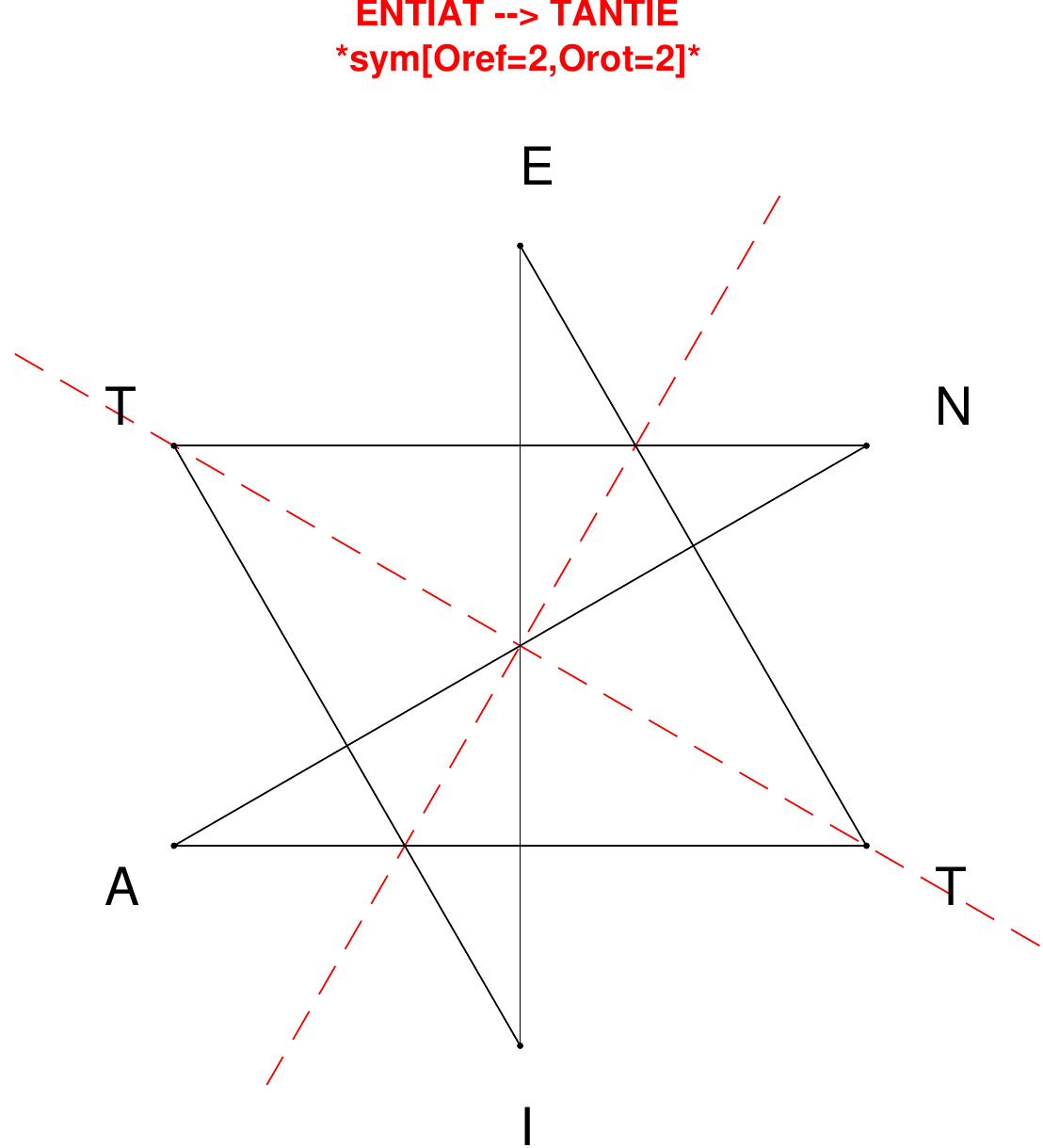}
\end{subfigure}
\hfill
\begin{subfigure}[T]{0.19\textwidth}
\centering
\includegraphics[width=\textwidth]{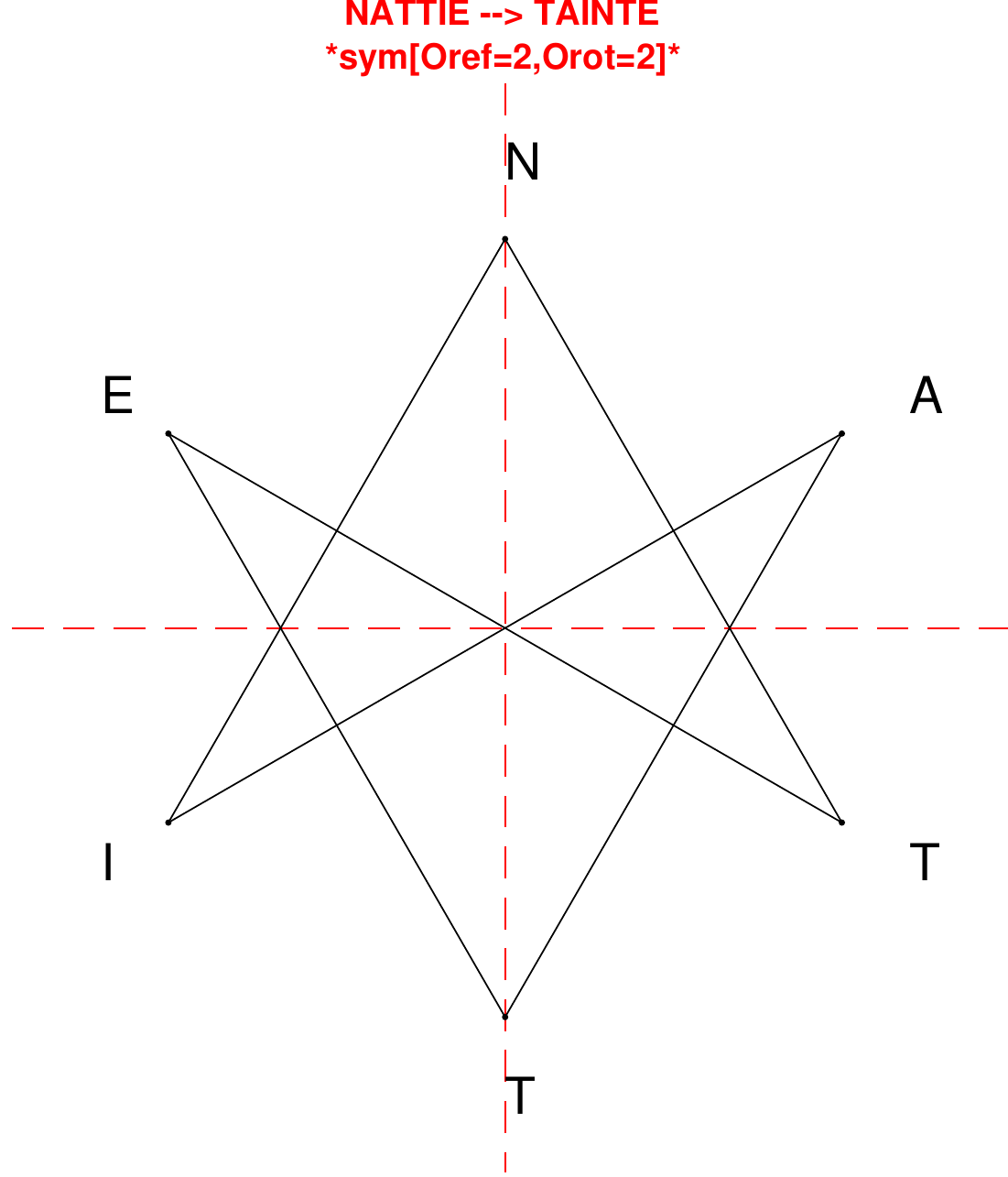}
\end{subfigure}
\end{figure}

\begin{figure}[H]
\centering
\begin{subfigure}[T]{0.19\textwidth}
\centering
\includegraphics[width=\textwidth]{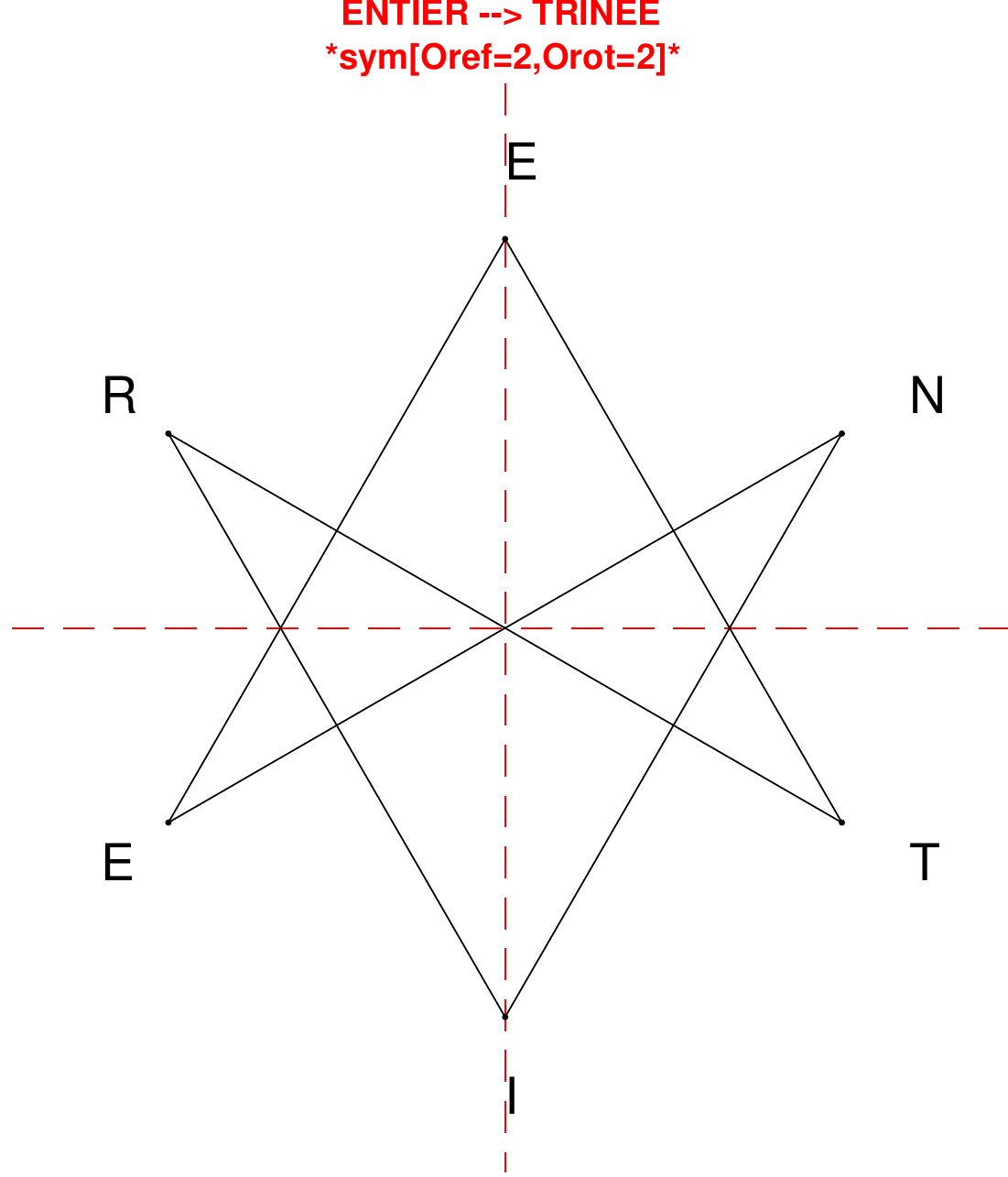}
\end{subfigure}
\hfill
\begin{subfigure}[T]{0.19\textwidth}
\centering
\includegraphics[width=\textwidth]{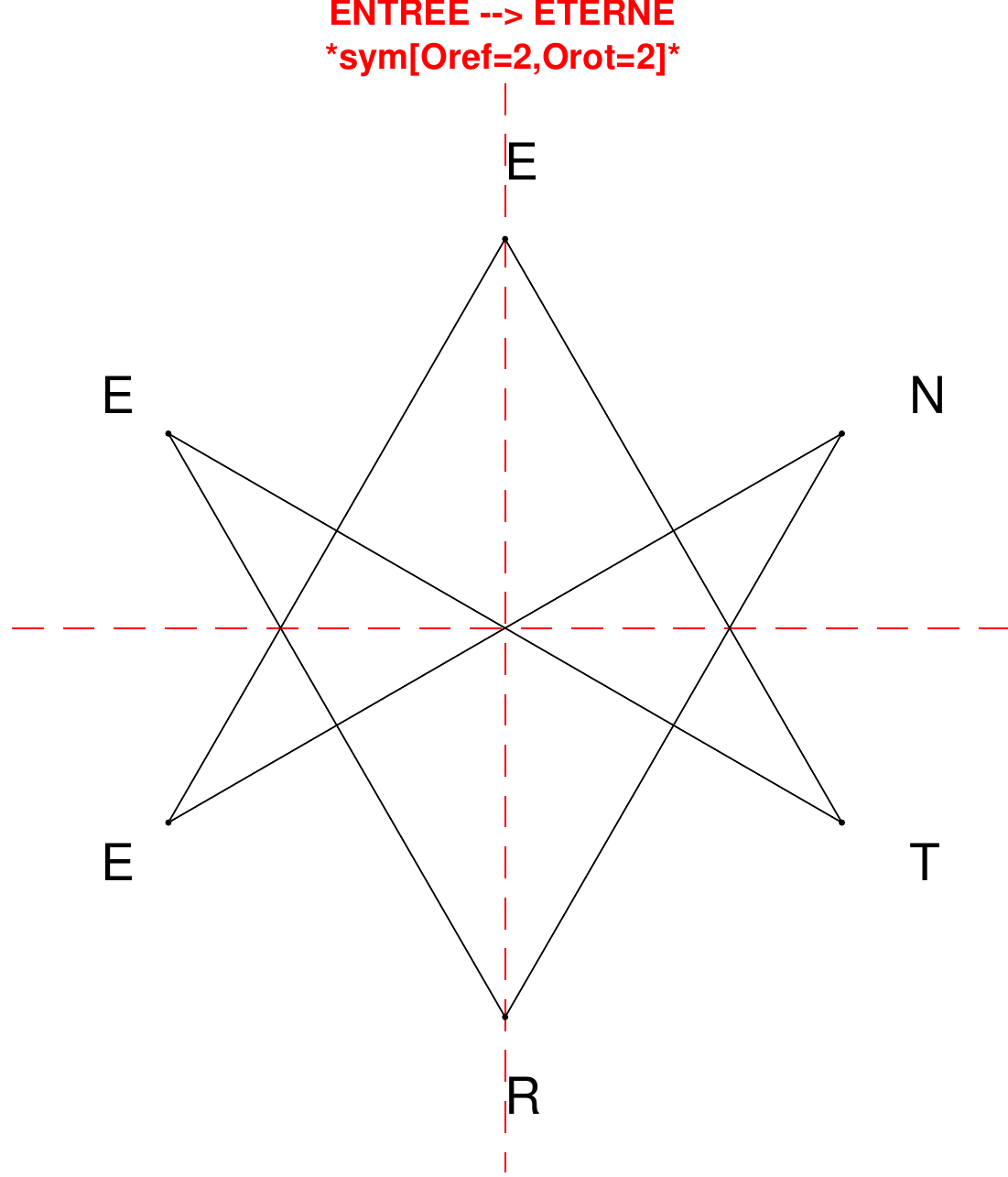}
\end{subfigure}
\hfill
\begin{subfigure}[T]{0.19\textwidth}
\centering
\includegraphics[width=\textwidth]{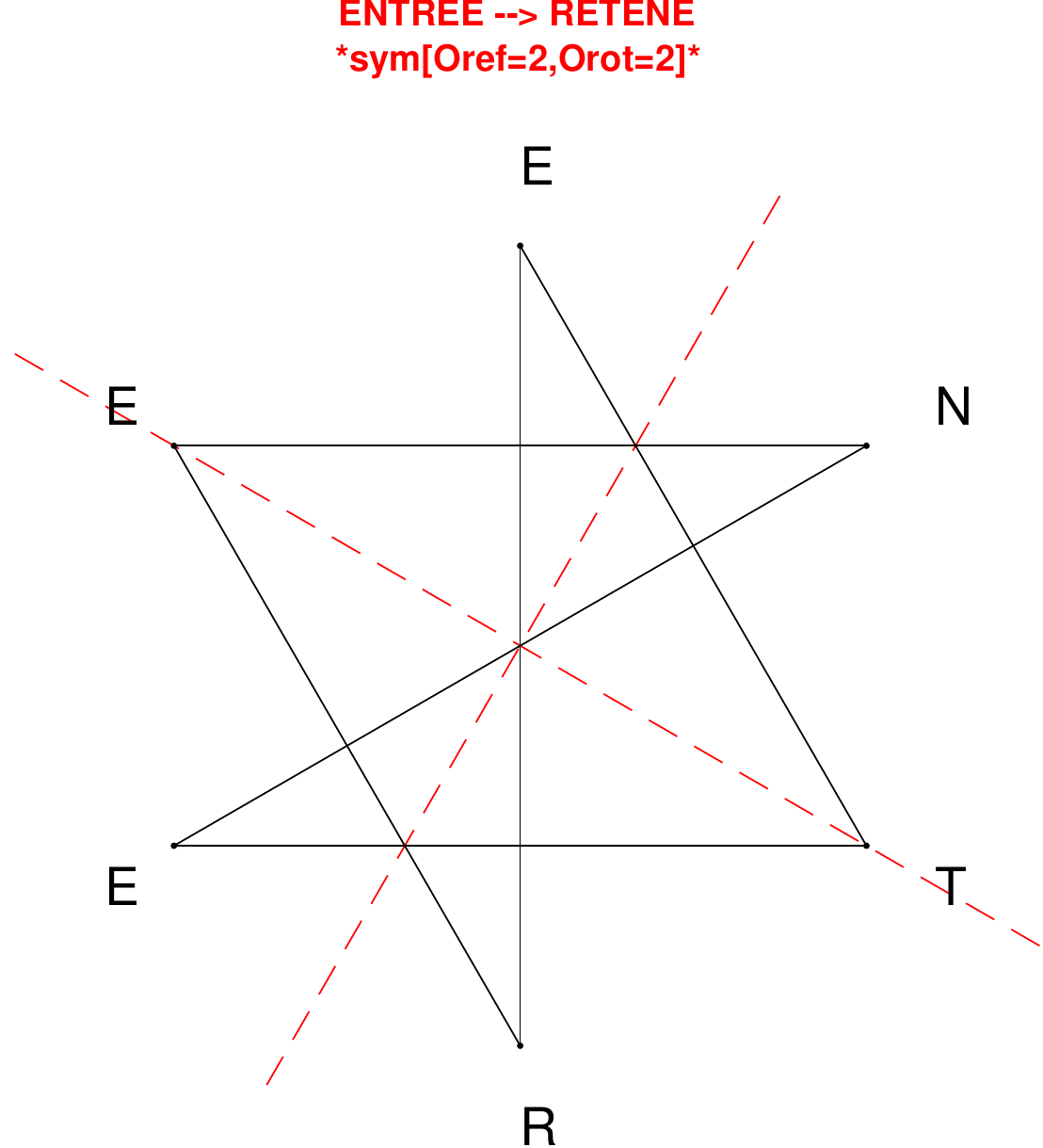}
\end{subfigure}
\hfill
\begin{subfigure}[T]{0.19\textwidth}
\centering
\includegraphics[width=\textwidth]{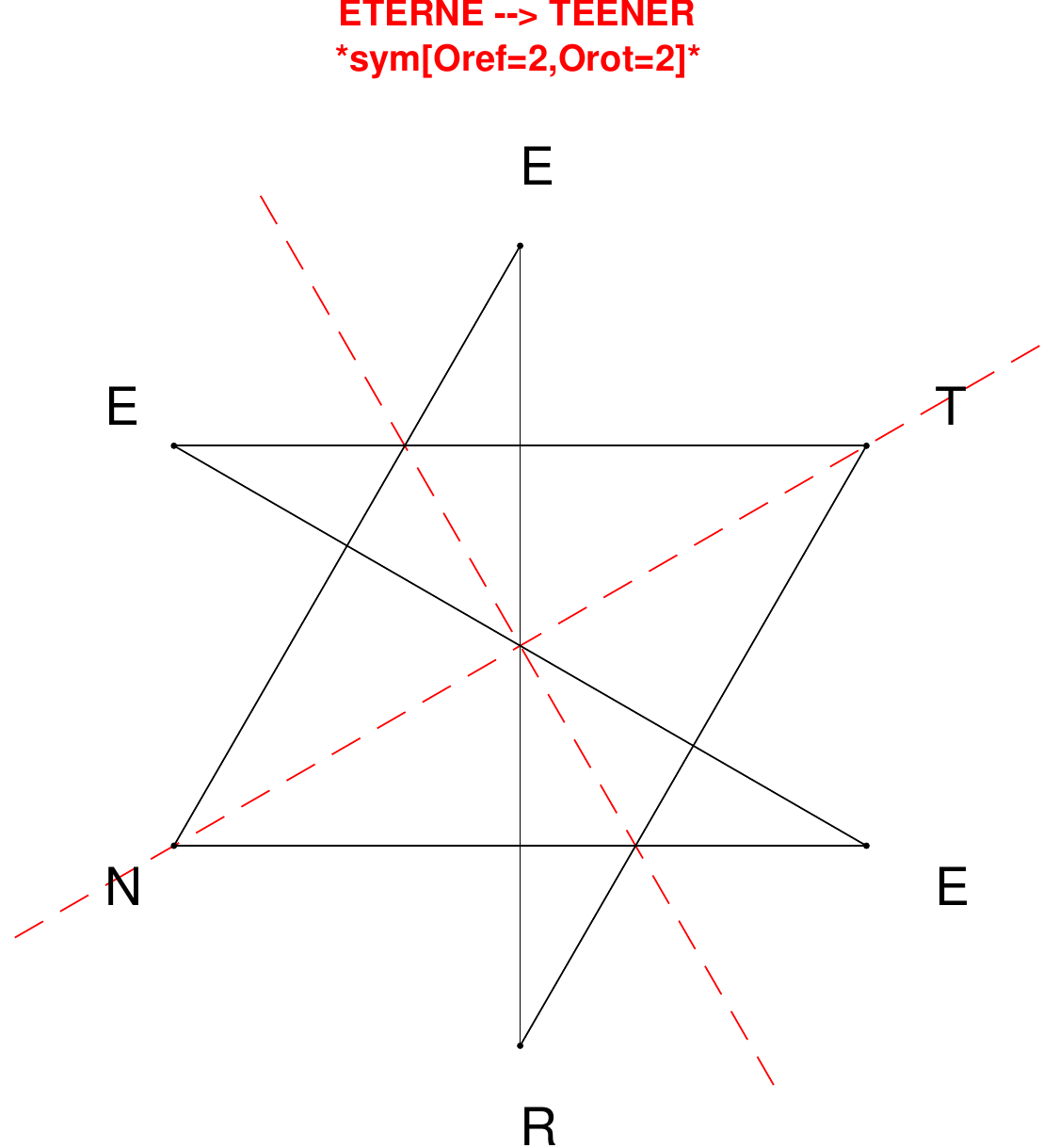}
\end{subfigure}
\hfill
\begin{subfigure}[T]{0.19\textwidth}
\centering
\includegraphics[width=\textwidth]{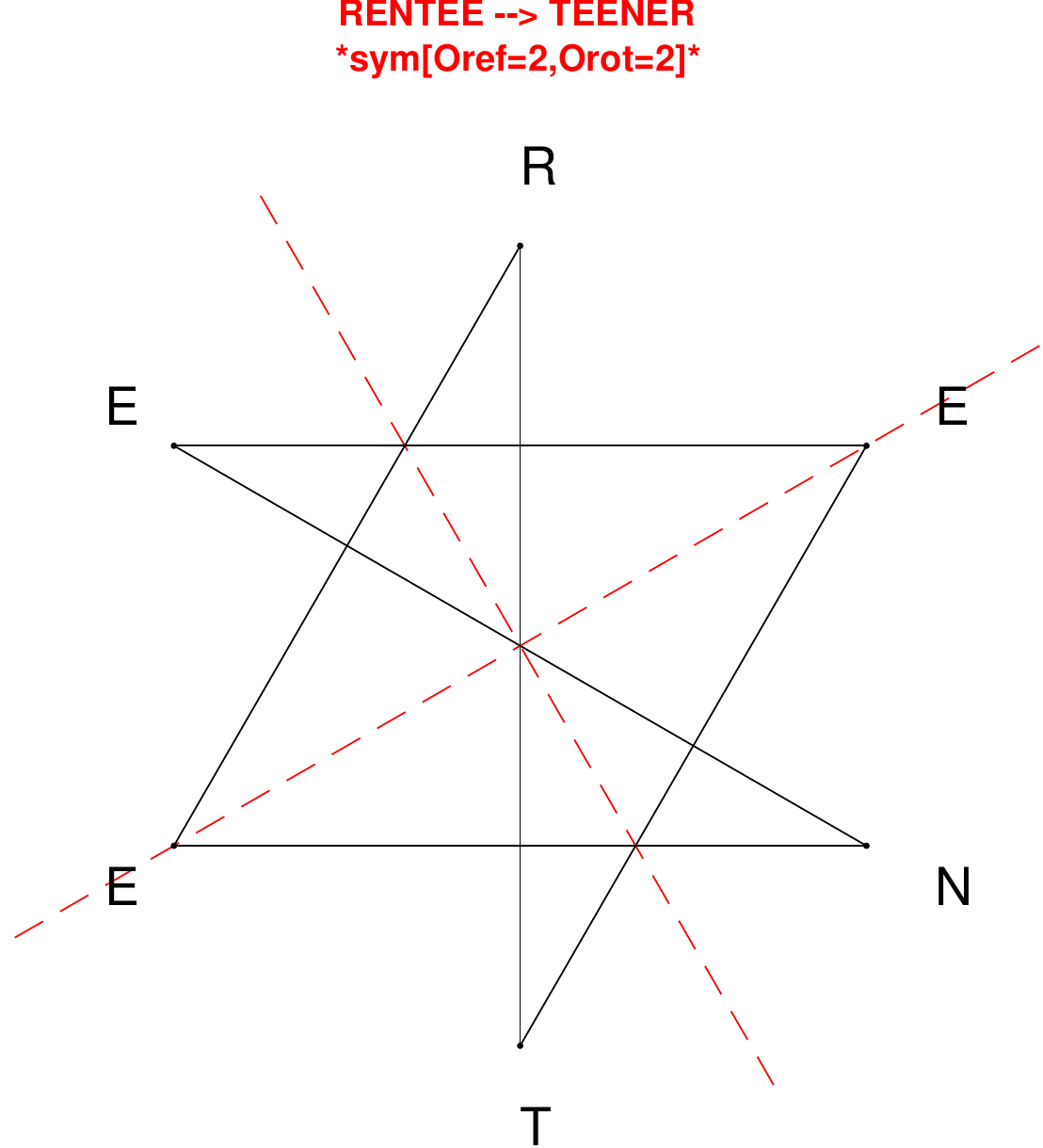}
\end{subfigure}
\end{figure}

\begin{figure}[H]
\centering
\begin{subfigure}[T]{0.19\textwidth}
\centering
\includegraphics[width=\textwidth]{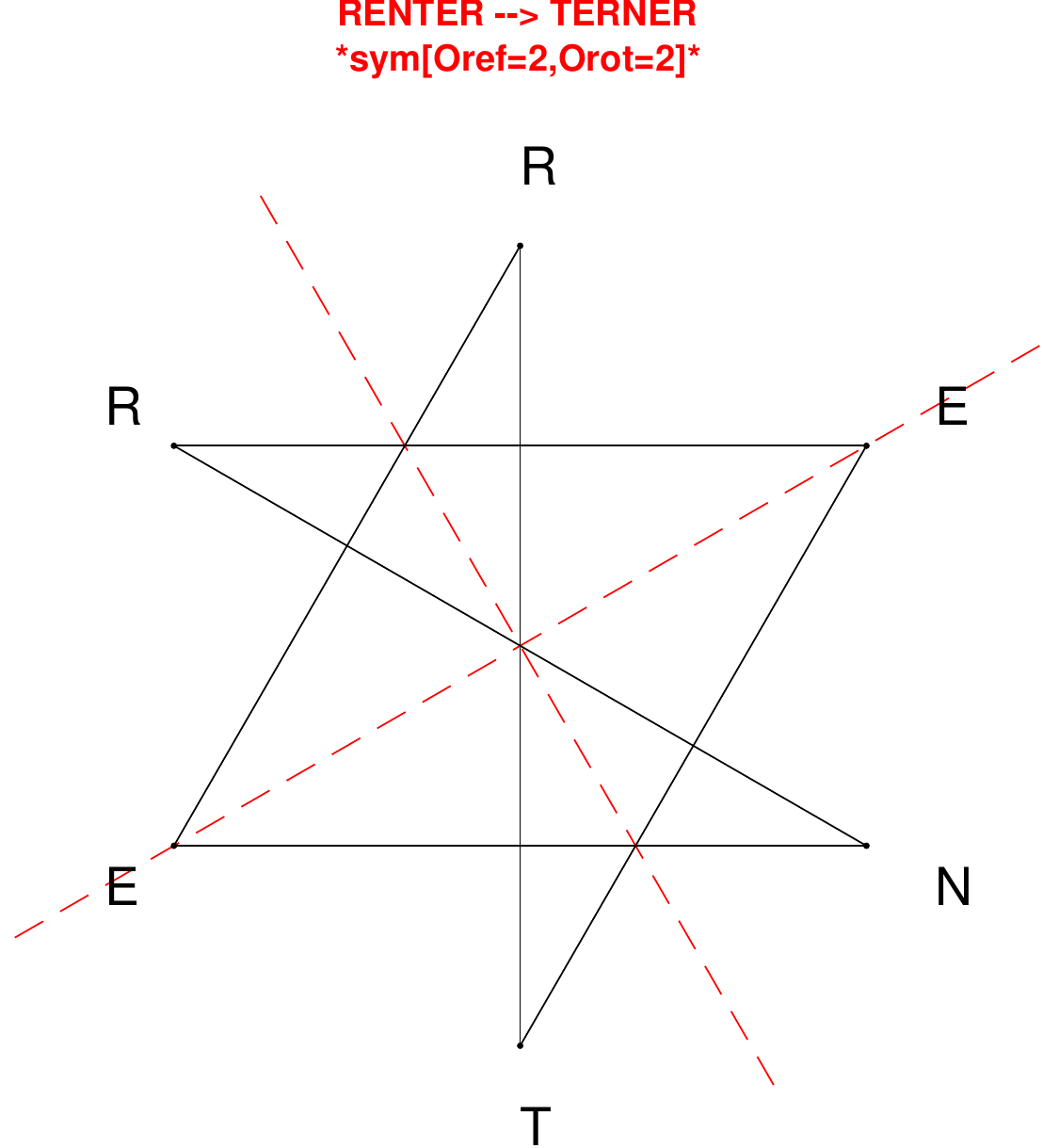}
\end{subfigure}
\hfill
\begin{subfigure}[T]{0.19\textwidth}
\centering
\includegraphics[width=\textwidth]{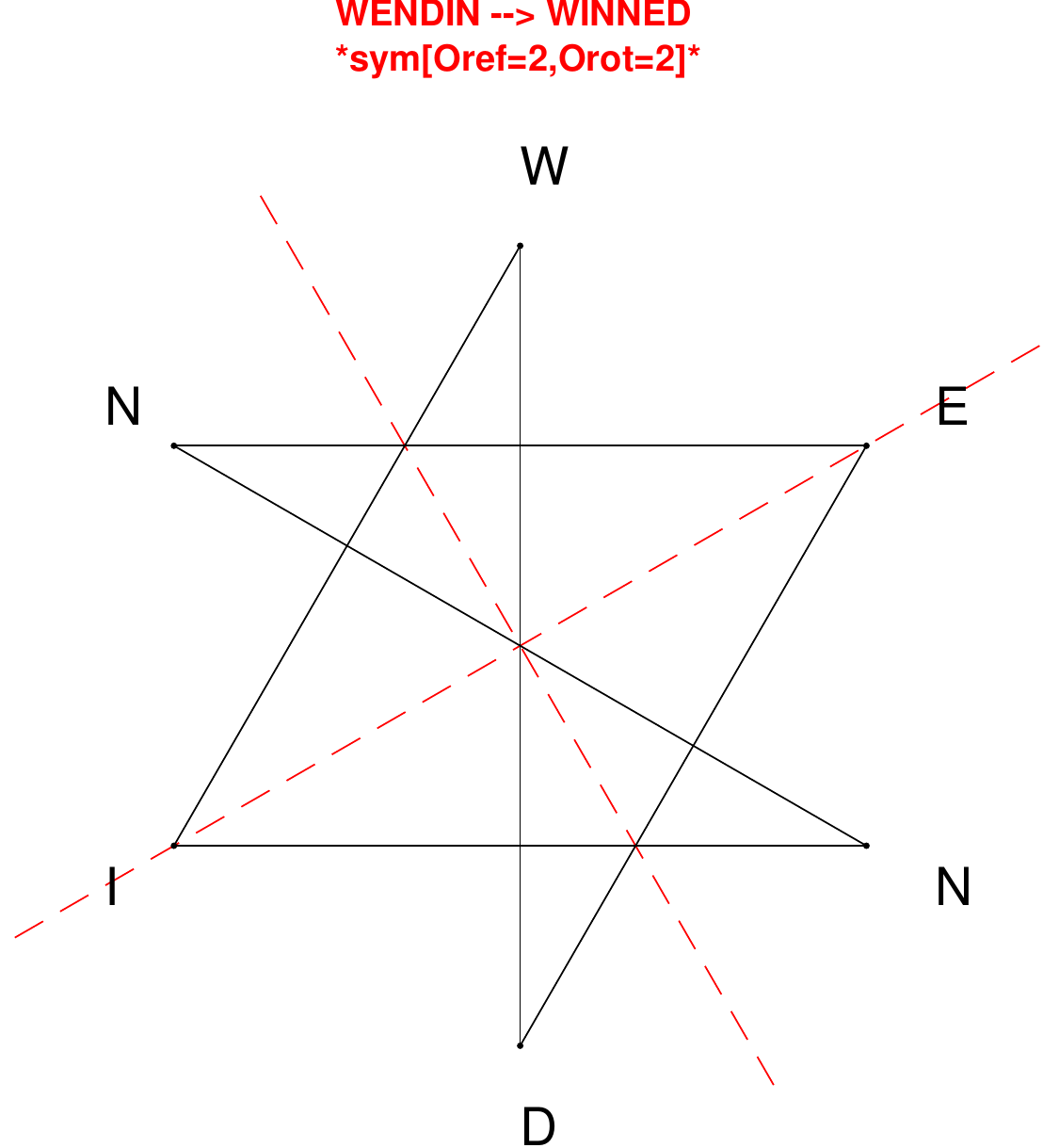}
\end{subfigure}
\hfill
\begin{subfigure}[T]{0.19\textwidth}
\centering
\includegraphics[width=\textwidth]{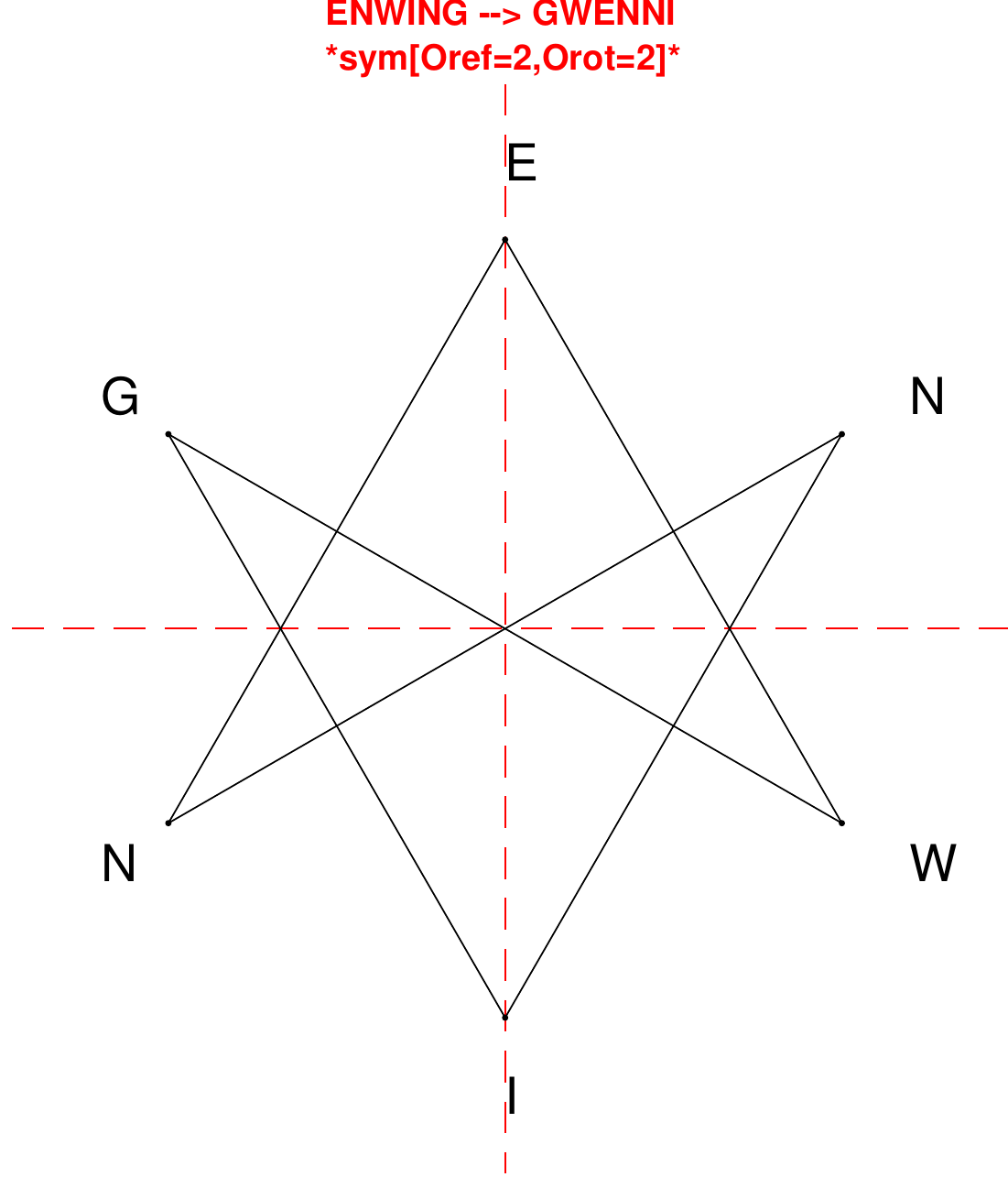}
\end{subfigure}
\hfill
\begin{subfigure}[T]{0.19\textwidth}
\centering
\includegraphics[width=\textwidth]{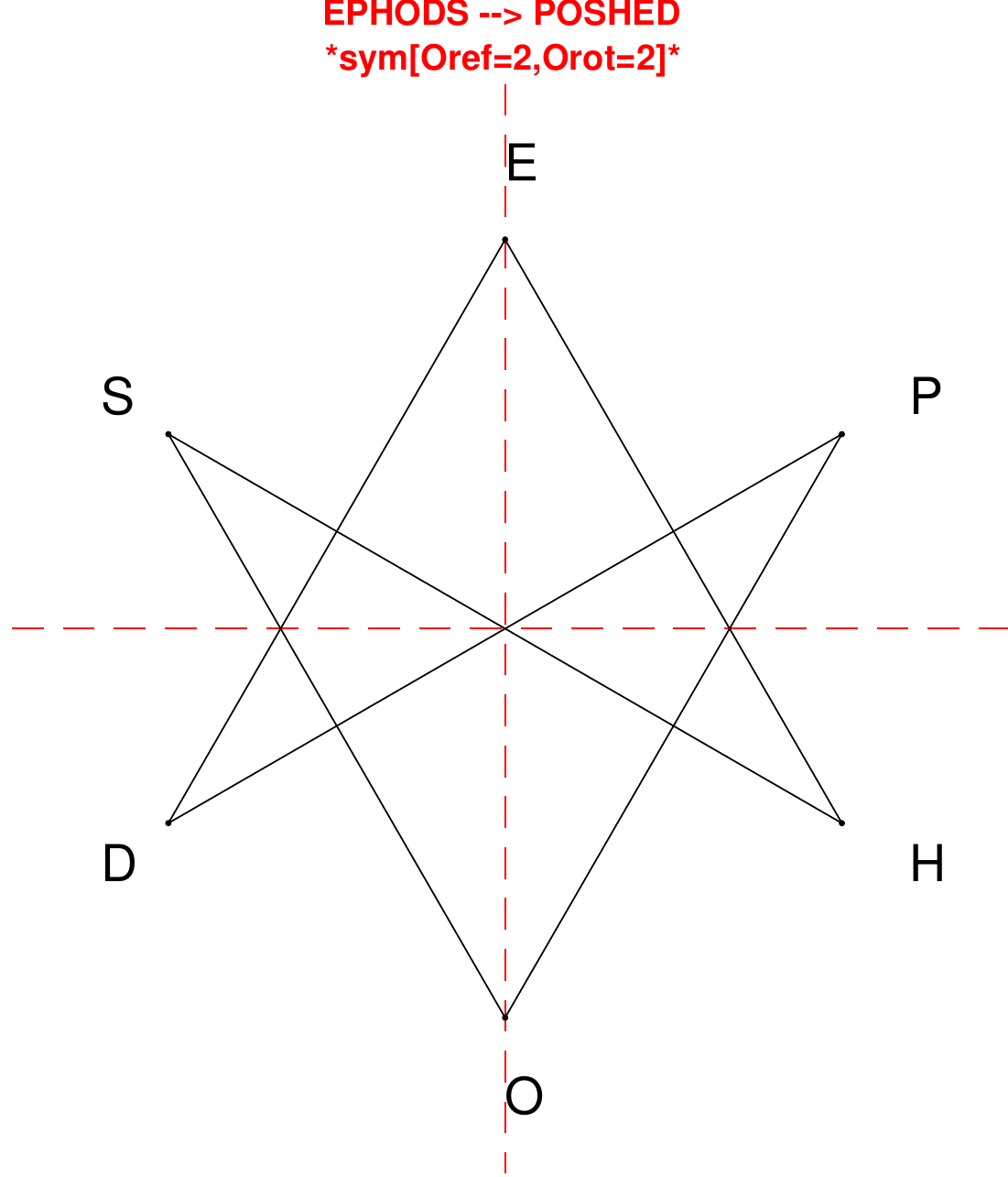}
\end{subfigure}
\hfill
\begin{subfigure}[T]{0.19\textwidth}
\centering
\includegraphics[width=\textwidth]{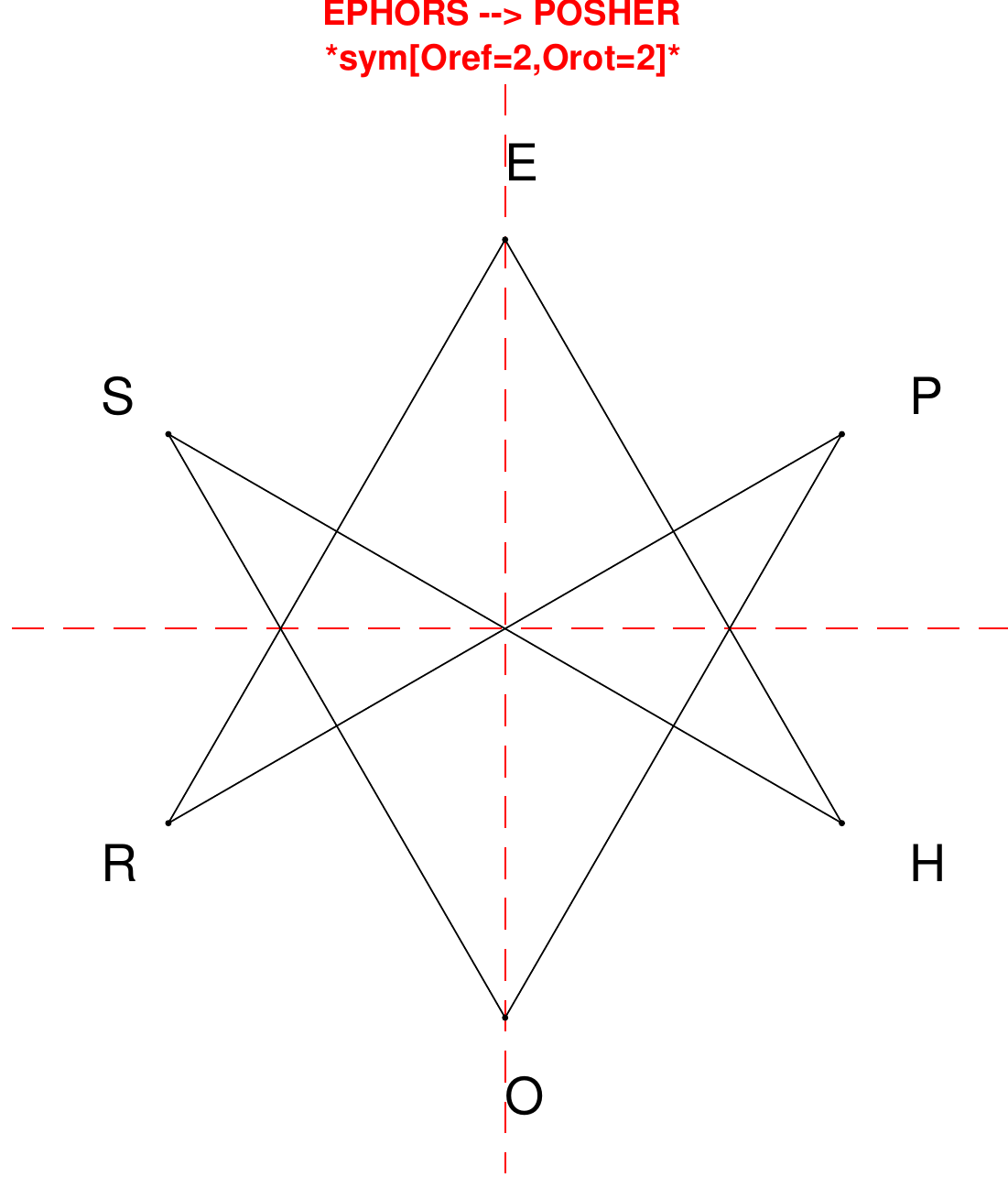}
\end{subfigure}
\end{figure}

\begin{figure}[H]
\centering
\begin{subfigure}[T]{0.19\textwidth}
\centering
\includegraphics[width=\textwidth]{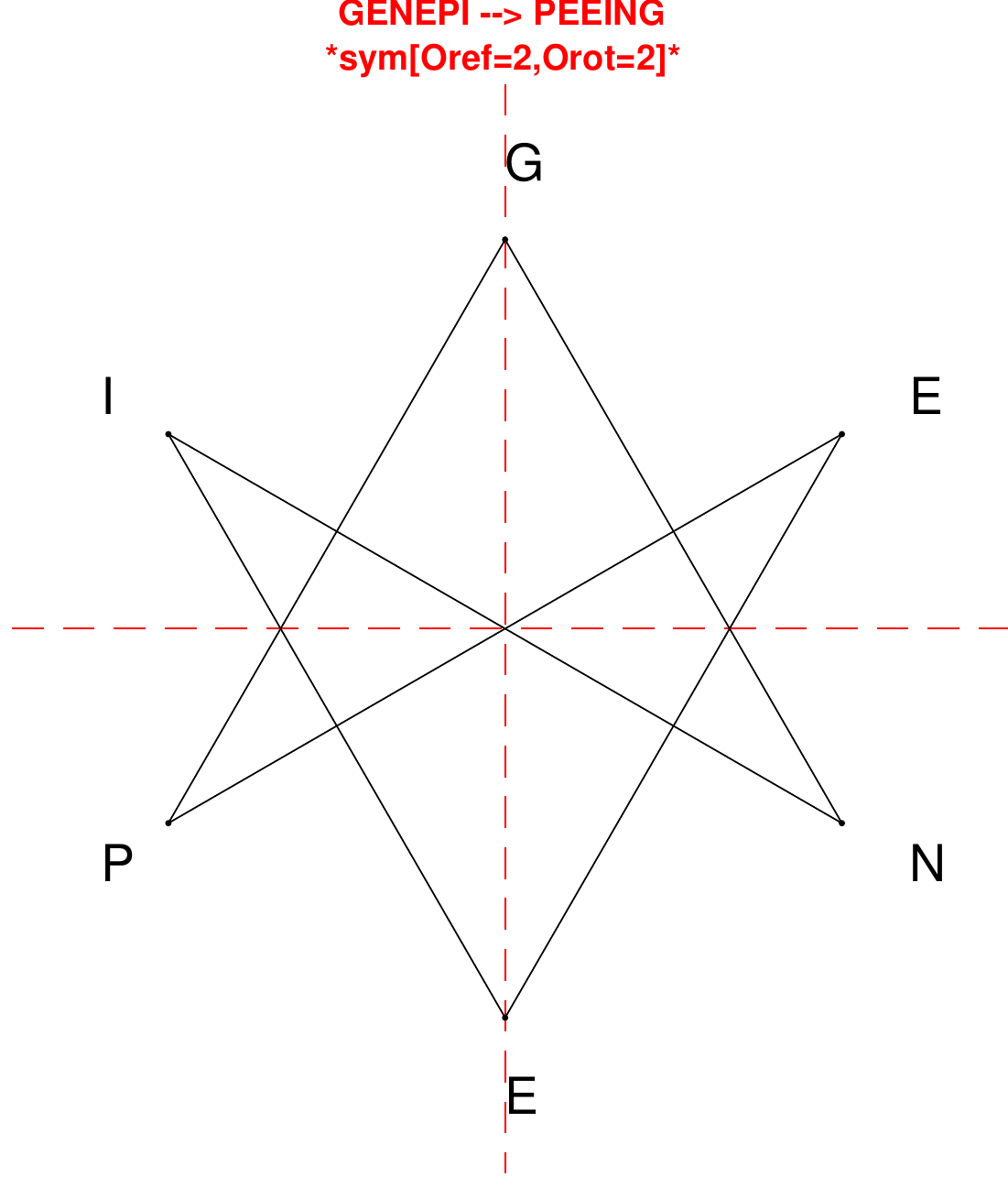}
\end{subfigure}
\hfill
\begin{subfigure}[T]{0.19\textwidth}
\centering
\includegraphics[width=\textwidth]{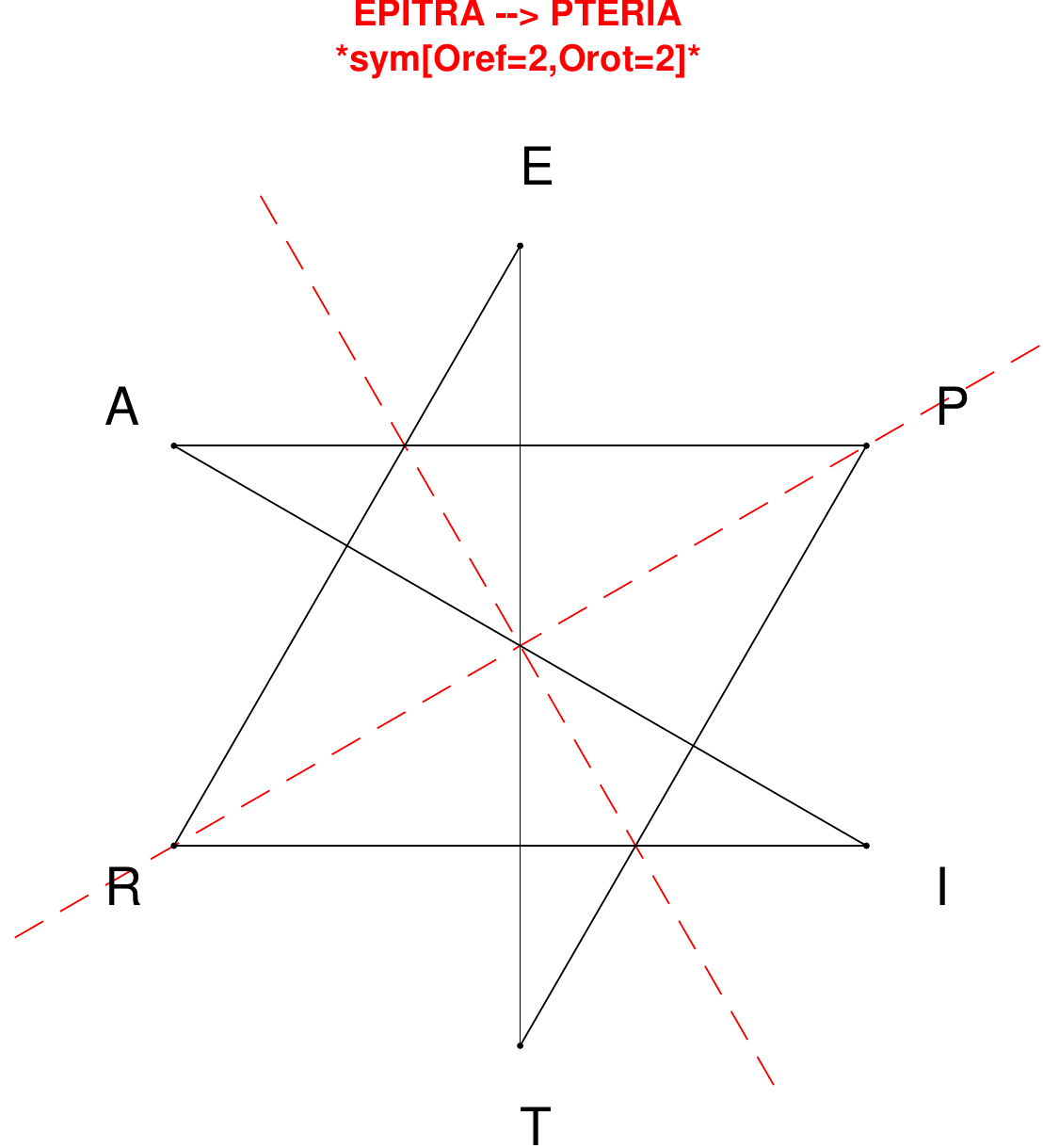}
\end{subfigure}
\hfill
\begin{subfigure}[T]{0.19\textwidth}
\centering
\includegraphics[width=\textwidth]{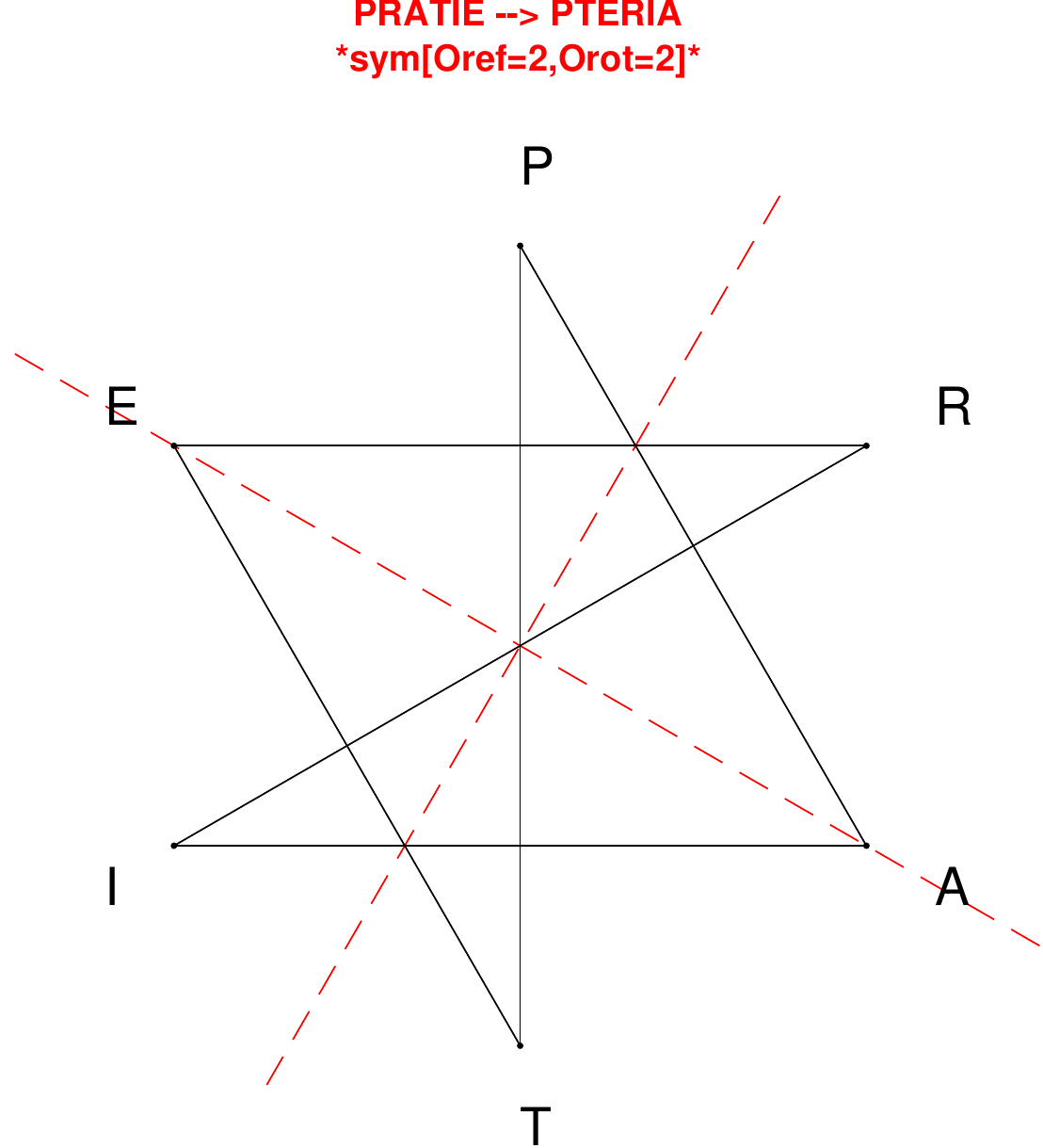}
\end{subfigure}
\hfill
\begin{subfigure}[T]{0.19\textwidth}
\centering
\includegraphics[width=\textwidth]{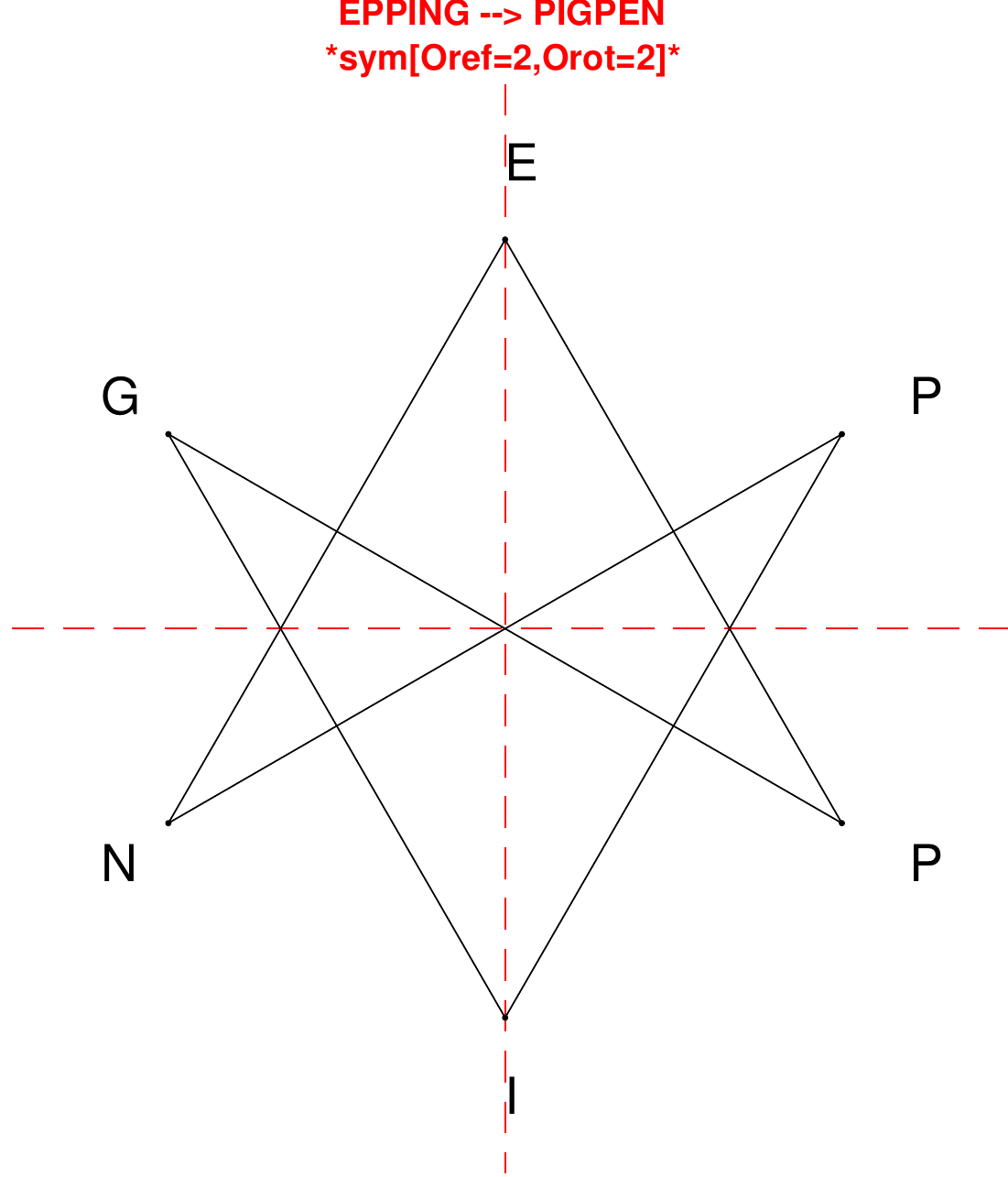}
\end{subfigure}
\hfill
\begin{subfigure}[T]{0.19\textwidth}
\centering
\includegraphics[width=\textwidth]{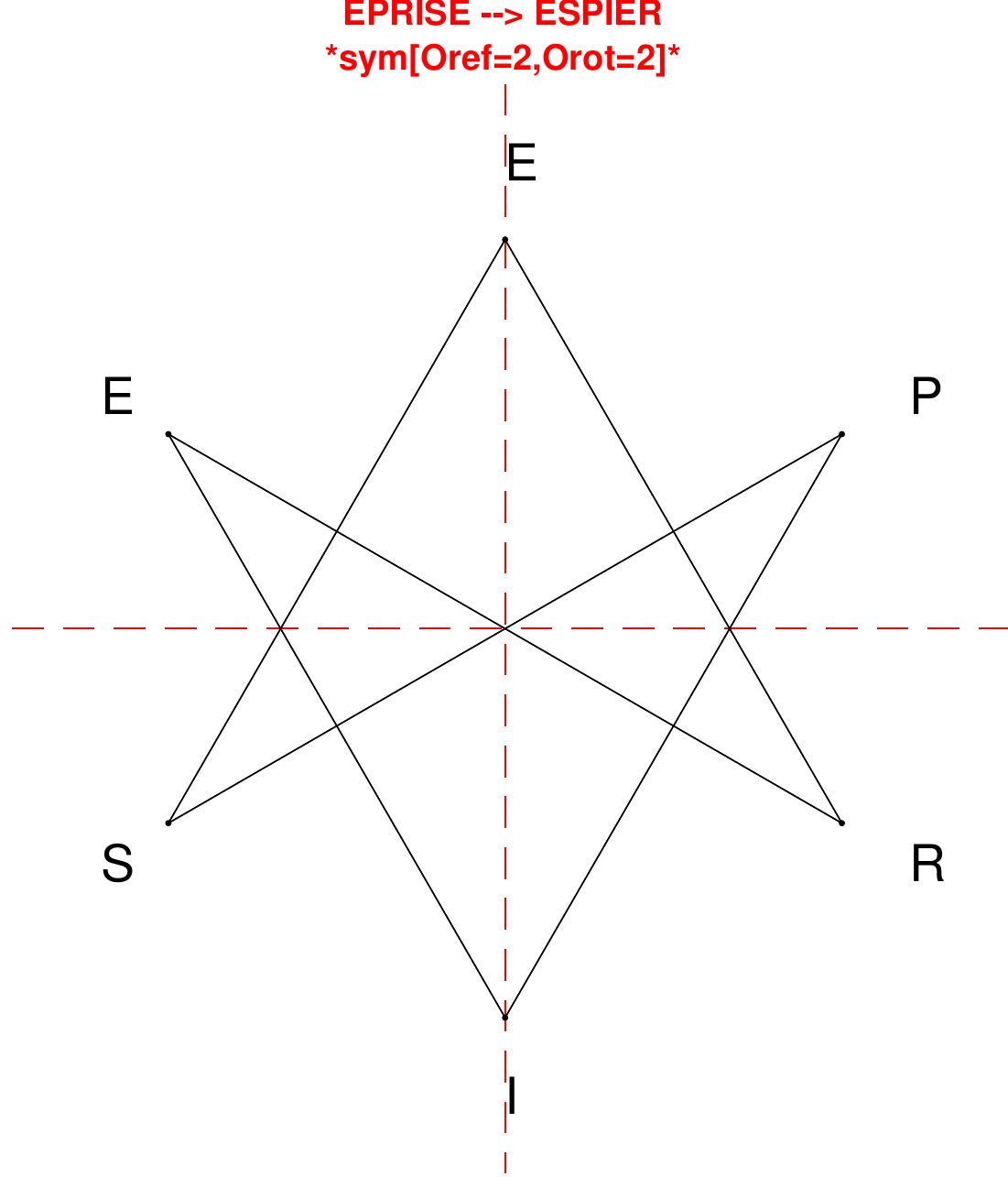}
\end{subfigure}
\end{figure}

\begin{figure}[H]
\centering
\begin{subfigure}[T]{0.19\textwidth}
\centering
\includegraphics[width=\textwidth]{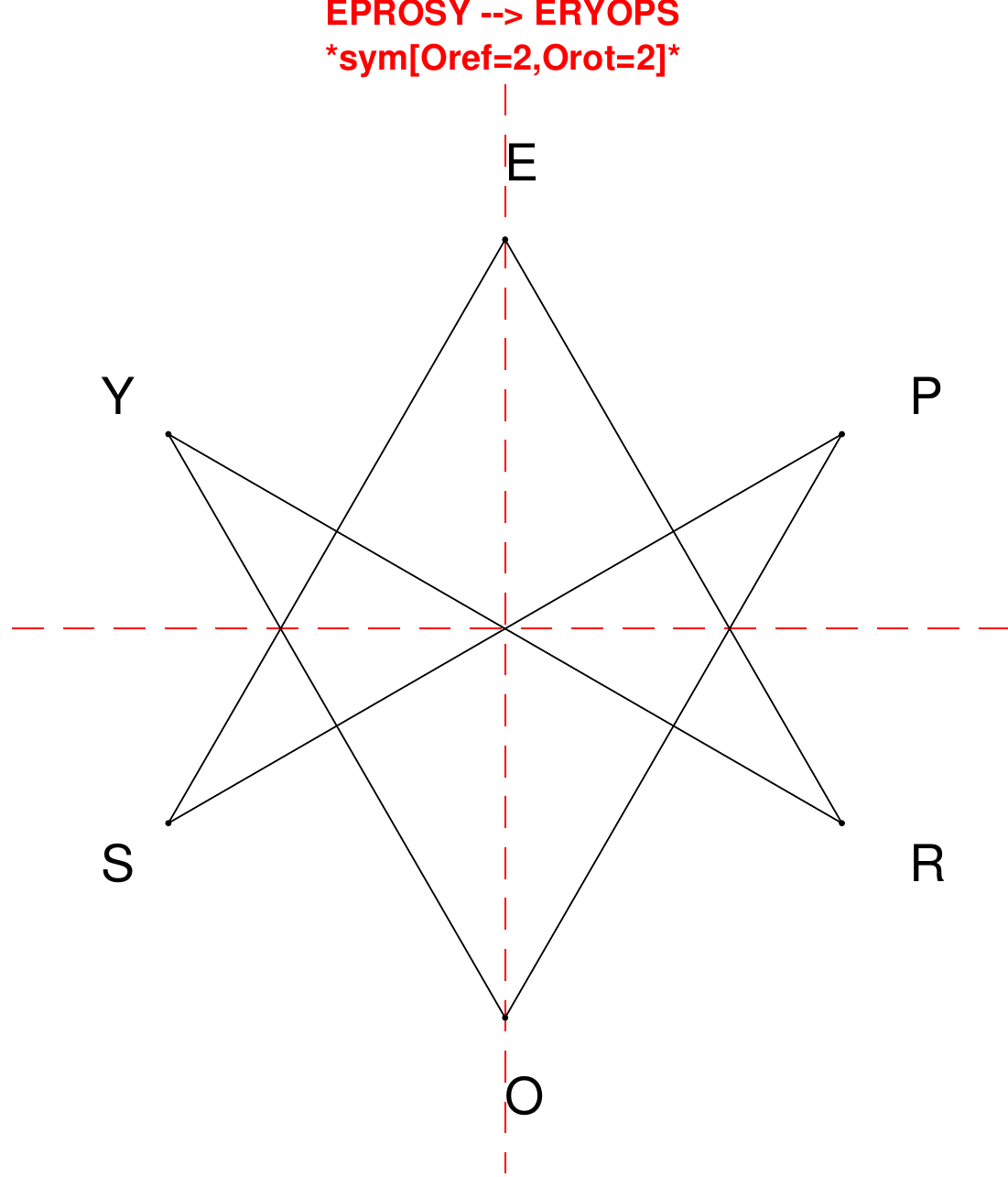}
\end{subfigure}
\hfill
\begin{subfigure}[T]{0.19\textwidth}
\centering
\includegraphics[width=\textwidth]{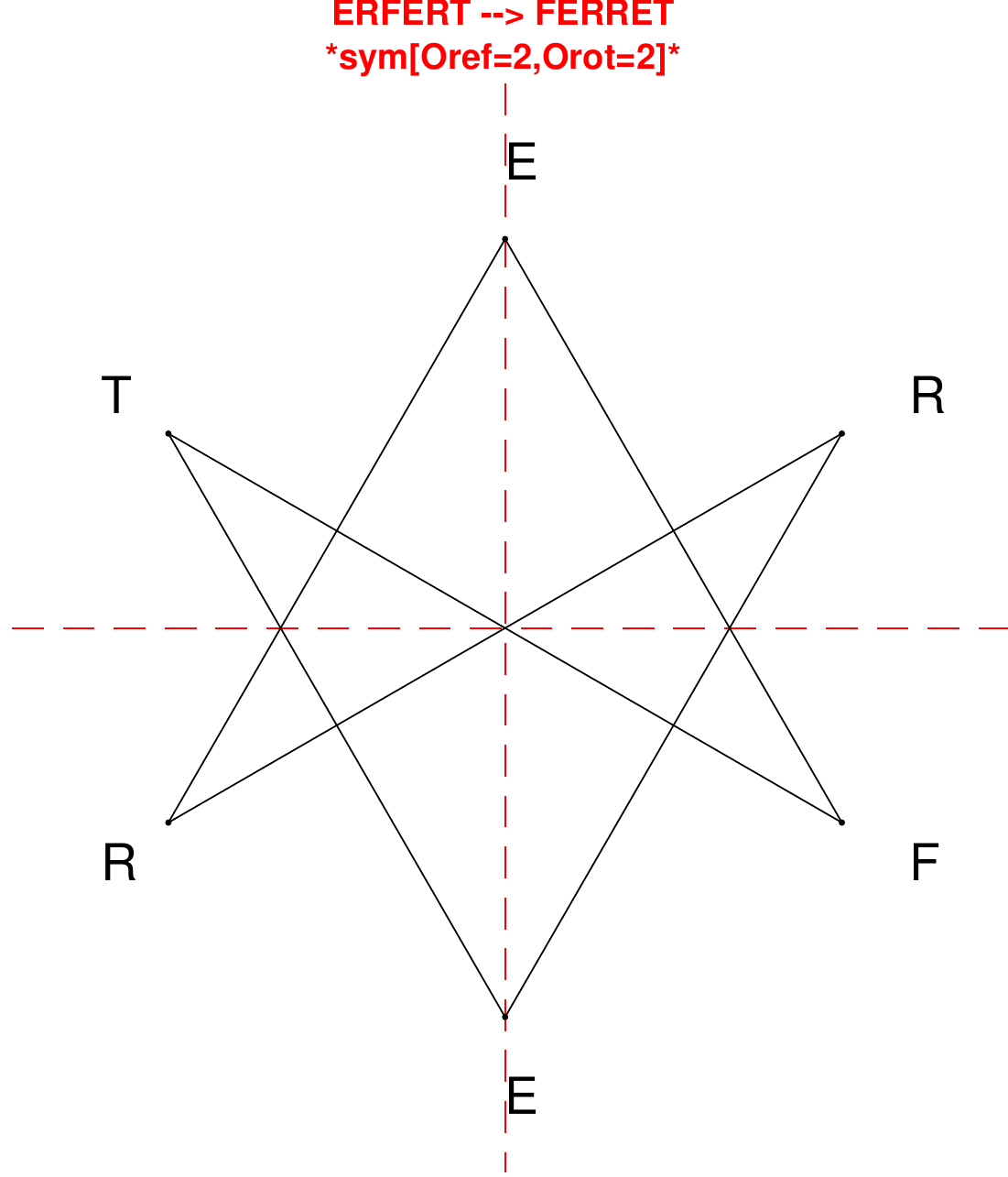}
\end{subfigure}
\hfill
\begin{subfigure}[T]{0.19\textwidth}
\centering
\includegraphics[width=\textwidth]{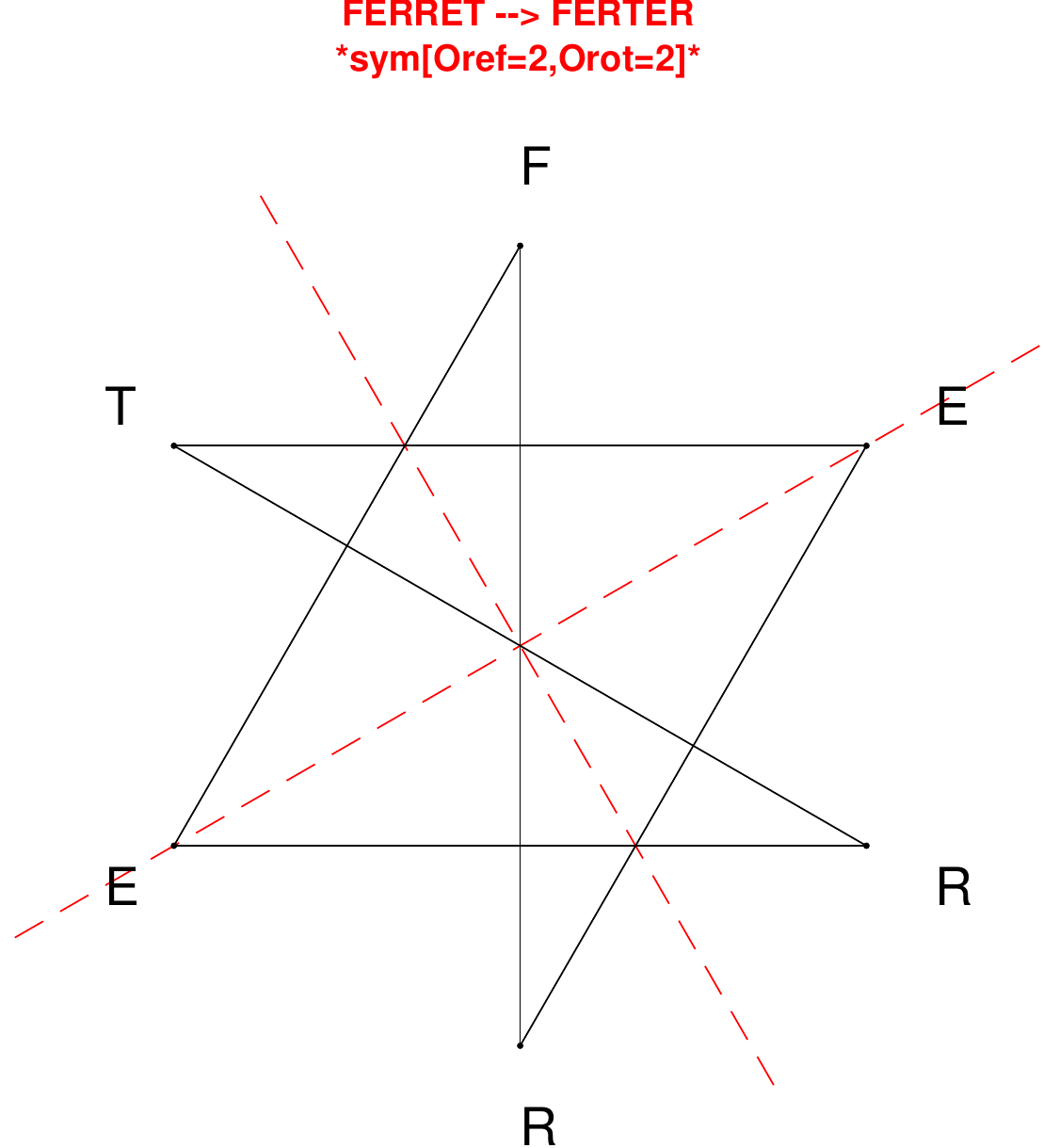}
\end{subfigure}
\hfill
\begin{subfigure}[T]{0.19\textwidth}
\centering
\includegraphics[width=\textwidth]{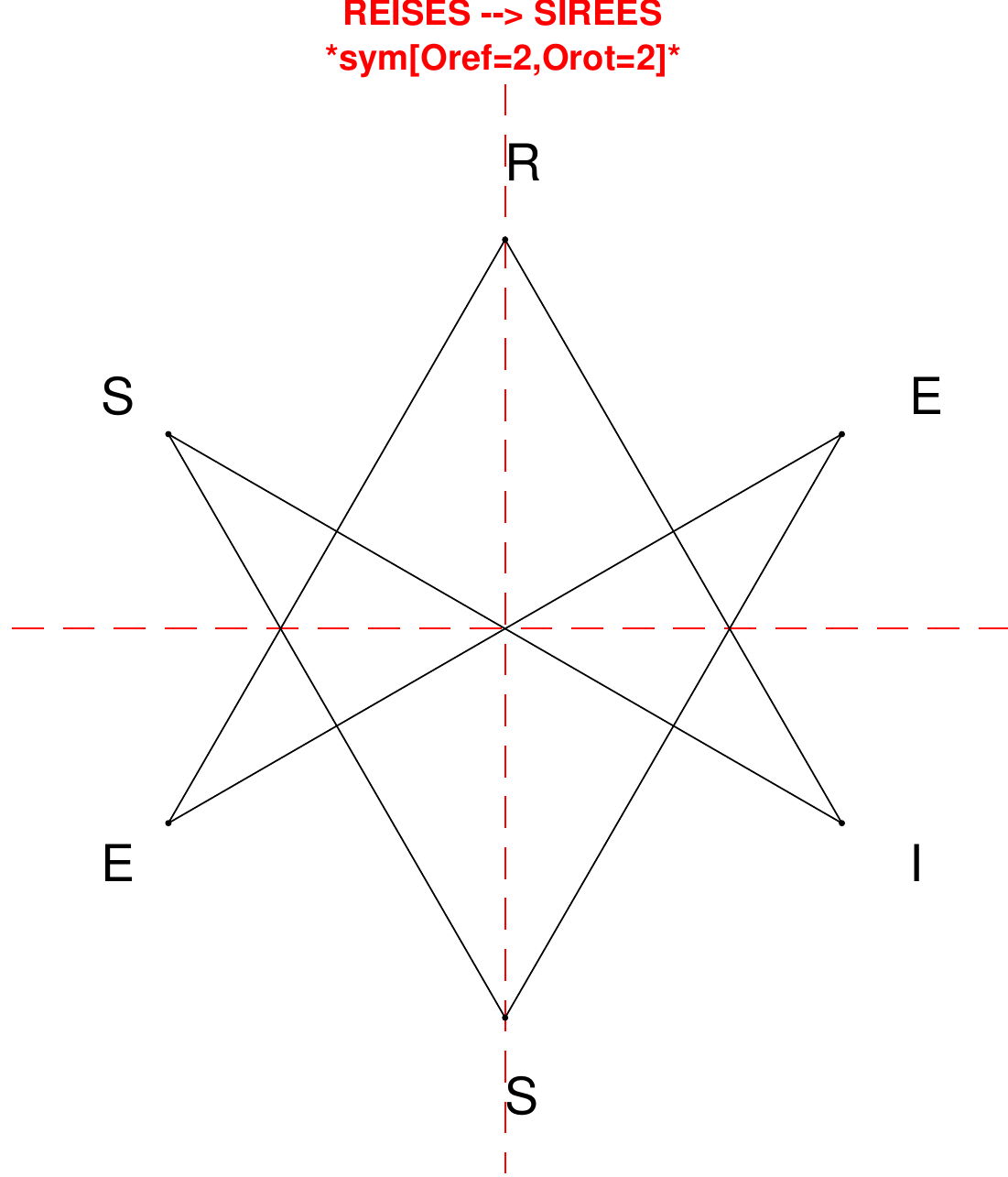}
\end{subfigure}
\hfill
\begin{subfigure}[T]{0.19\textwidth}
\centering
\includegraphics[width=\textwidth]{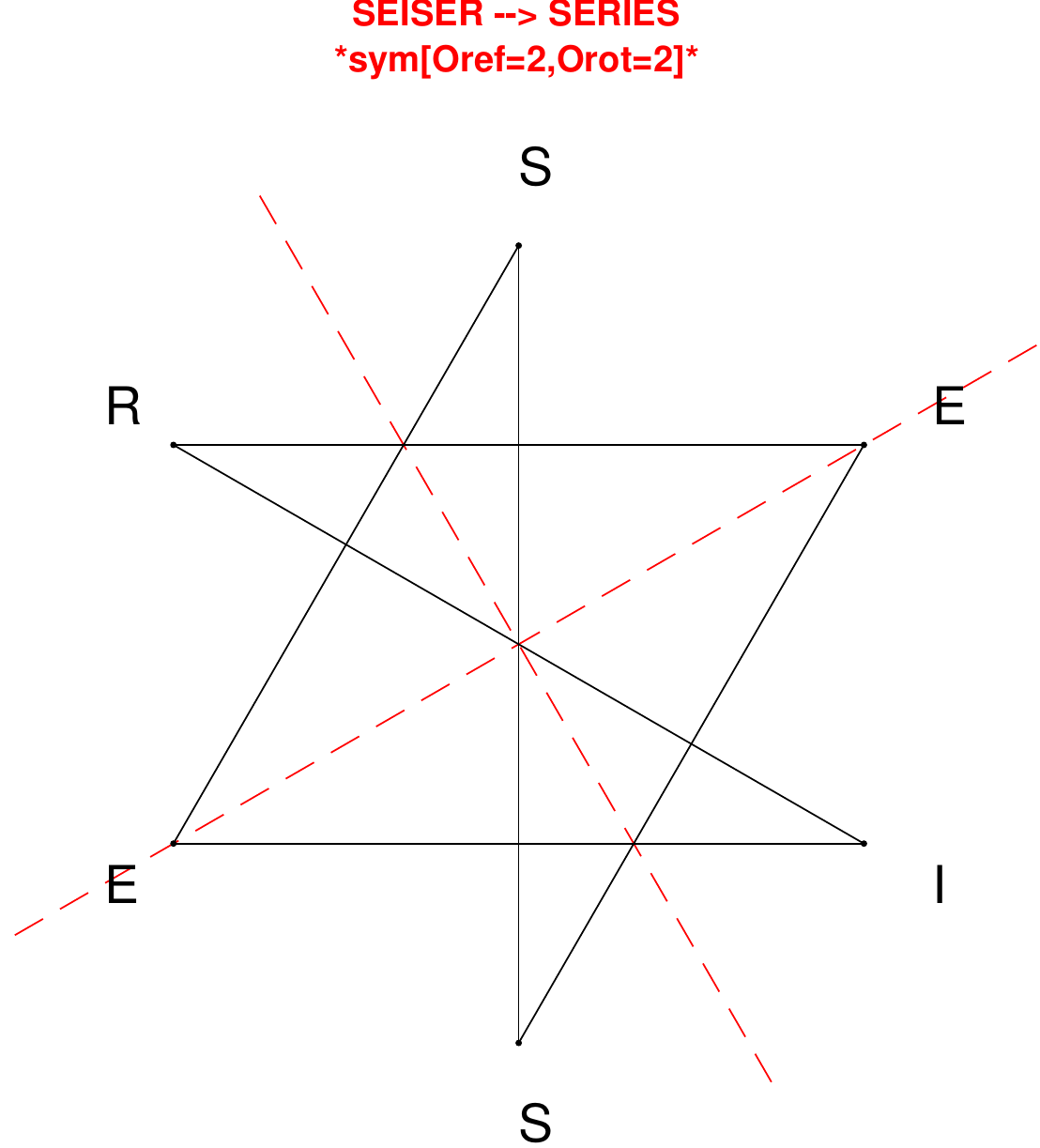}
\end{subfigure}
\end{figure}

\begin{figure}[H]
\centering
\begin{subfigure}[T]{0.19\textwidth}
\centering
\includegraphics[width=\textwidth]{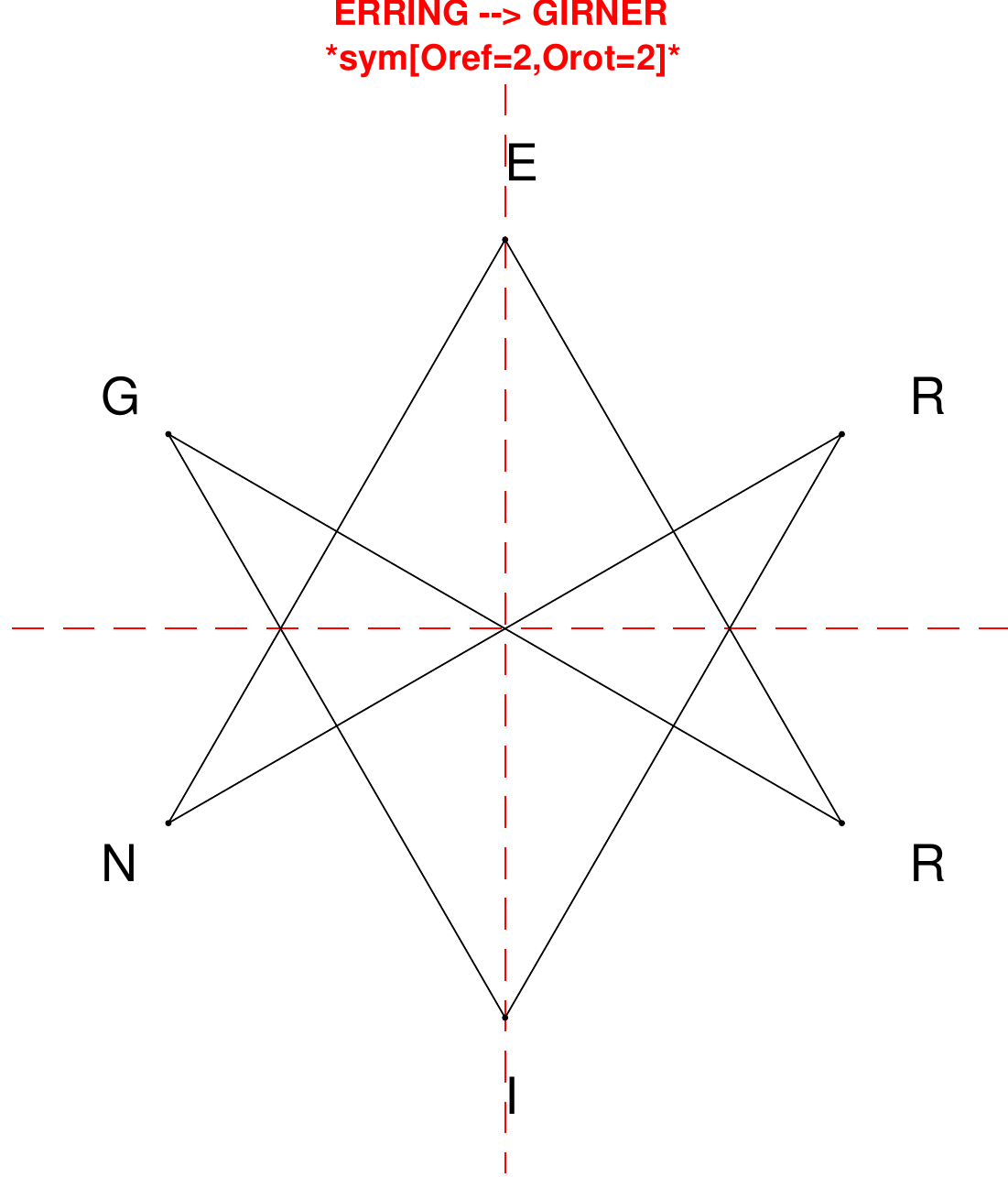}
\end{subfigure}
\hfill
\begin{subfigure}[T]{0.19\textwidth}
\centering
\includegraphics[width=\textwidth]{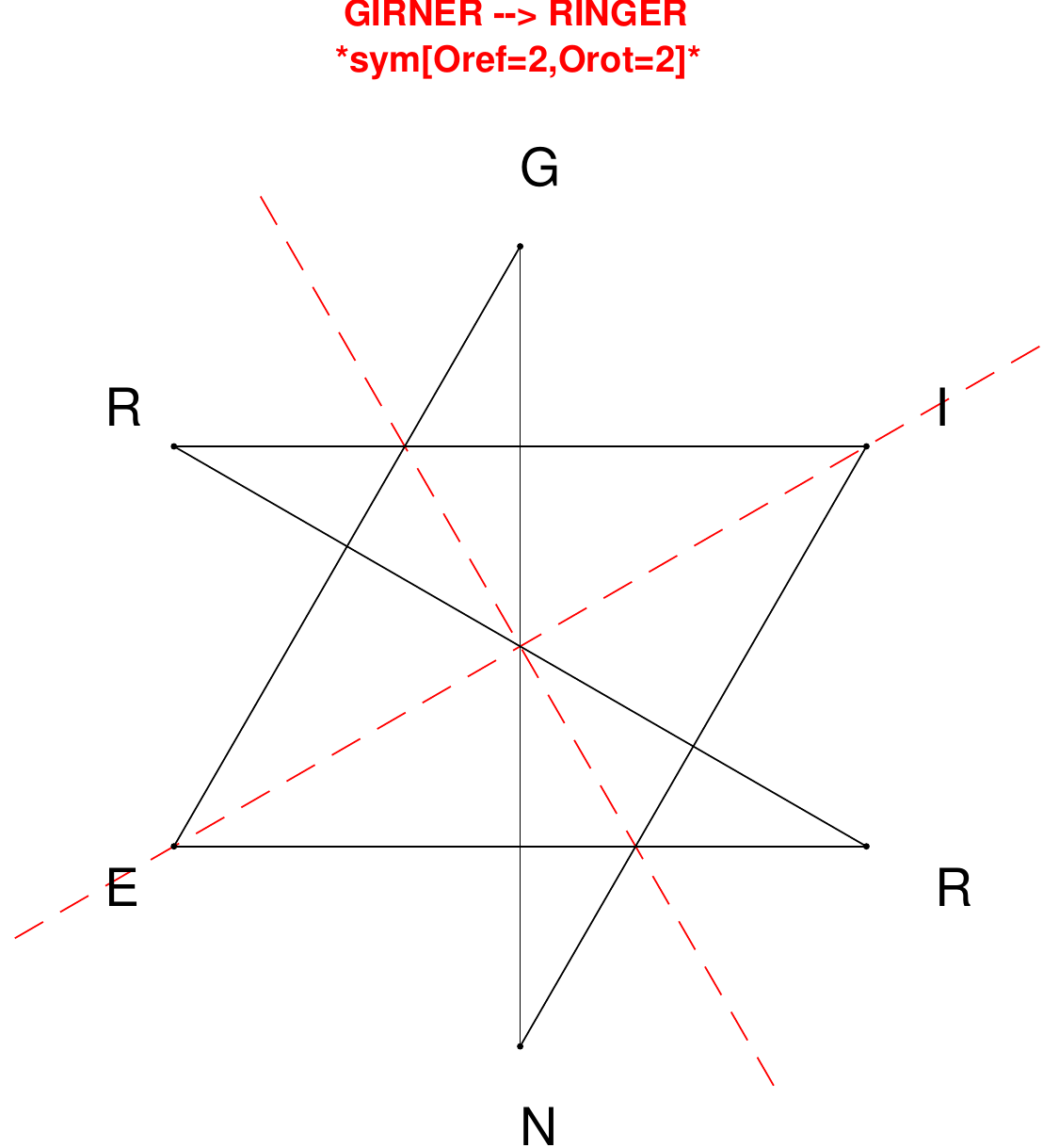}
\end{subfigure}
\hfill
\begin{subfigure}[T]{0.19\textwidth}
\centering
\includegraphics[width=\textwidth]{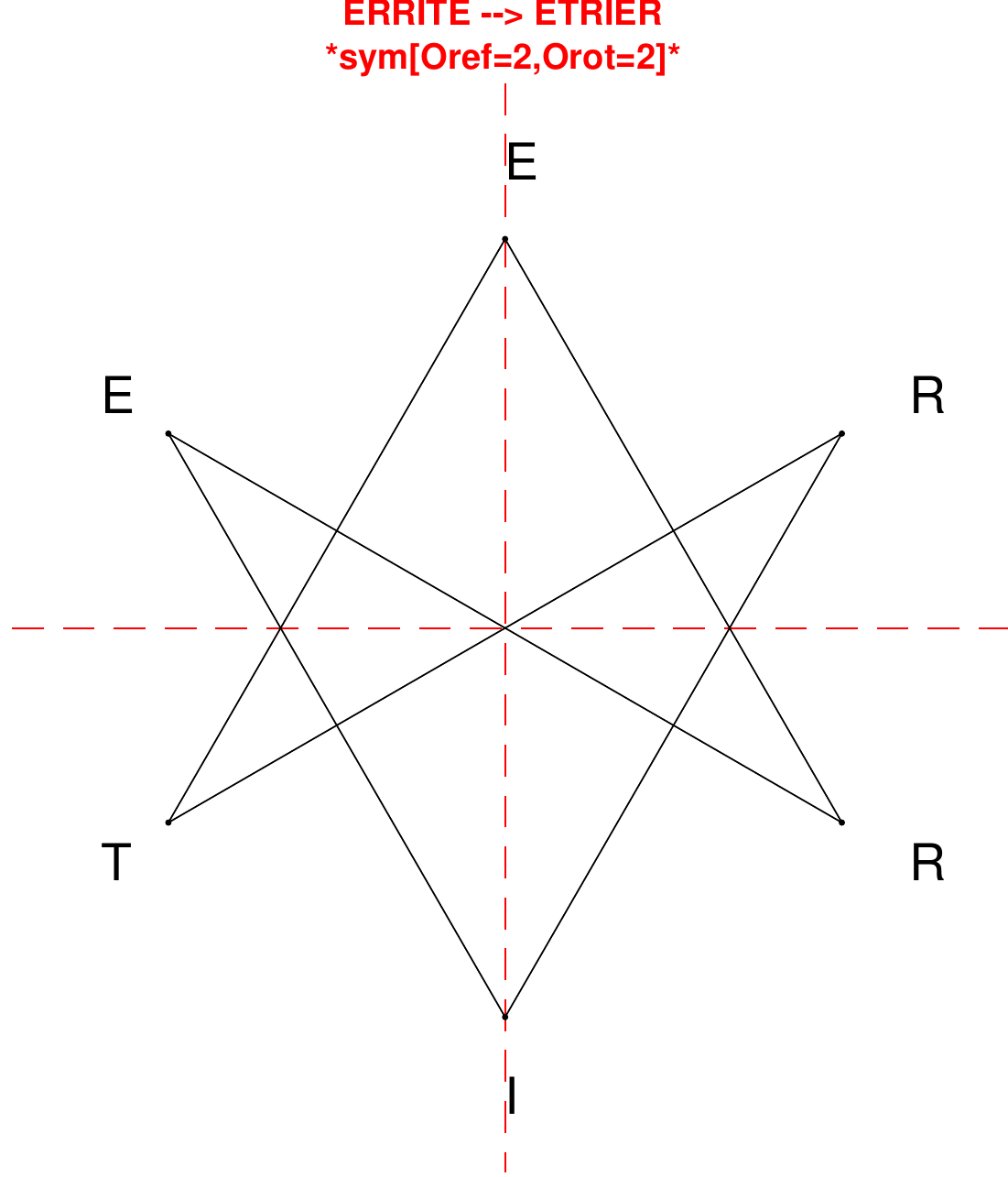}
\end{subfigure}
\hfill
\begin{subfigure}[T]{0.19\textwidth}
\centering
\includegraphics[width=\textwidth]{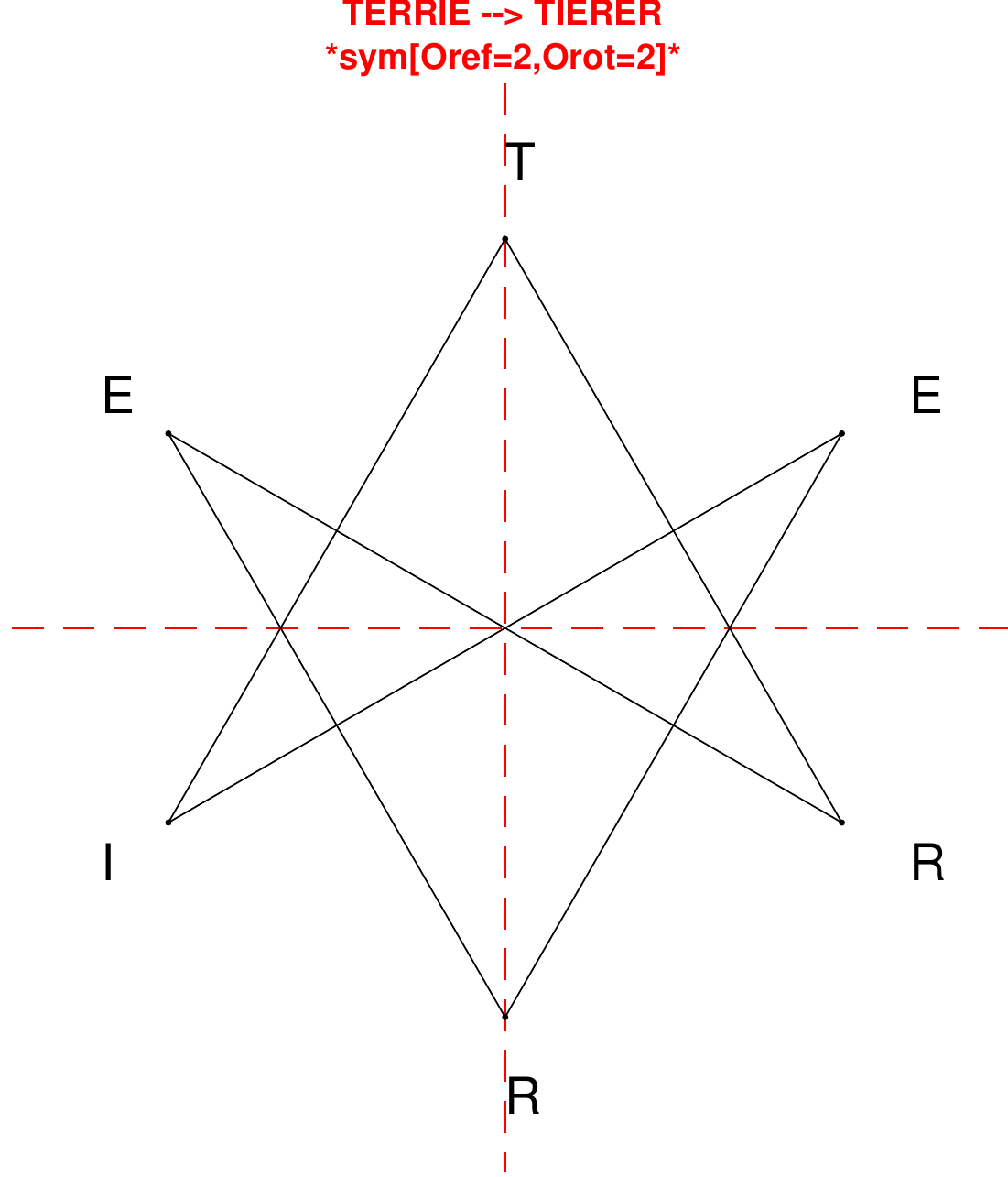}
\end{subfigure}
\hfill
\begin{subfigure}[T]{0.19\textwidth}
\centering
\includegraphics[width=\textwidth]{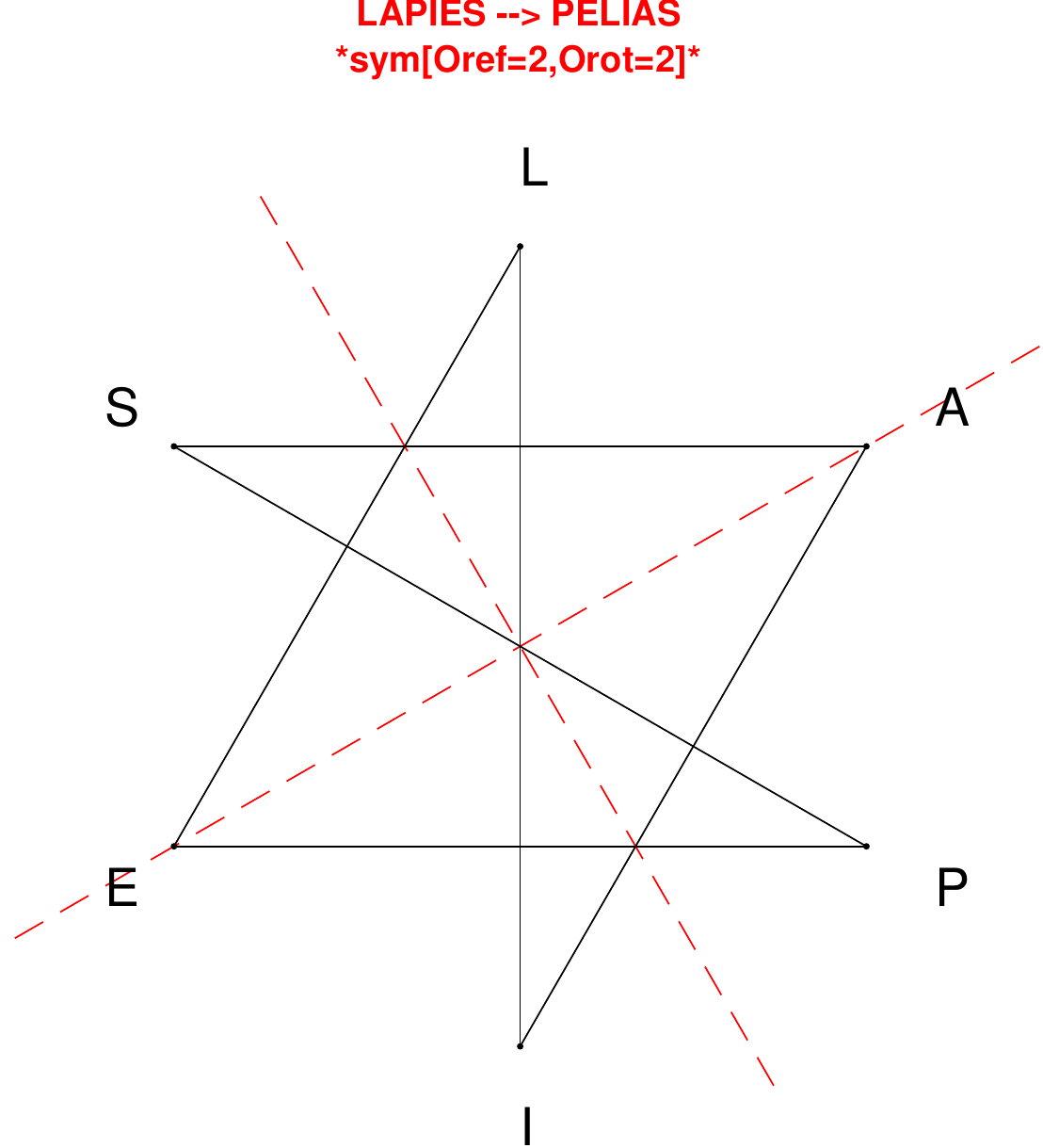}
\end{subfigure}
\end{figure}

\begin{figure}[H]
\centering
\begin{subfigure}[T]{0.19\textwidth}
\centering
\includegraphics[width=\textwidth]{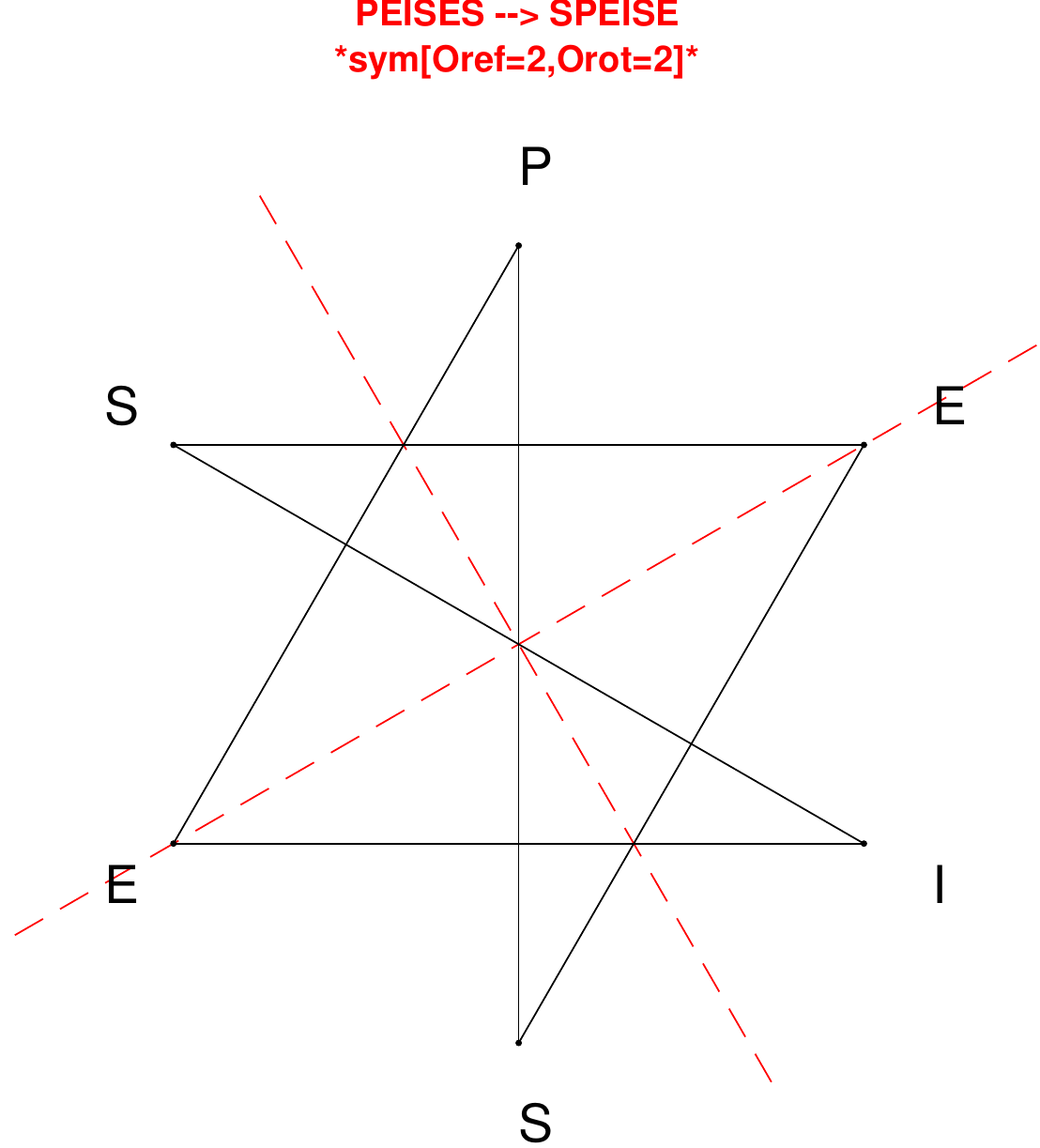}
\end{subfigure}
\hfill
\begin{subfigure}[T]{0.19\textwidth}
\centering
\includegraphics[width=\textwidth]{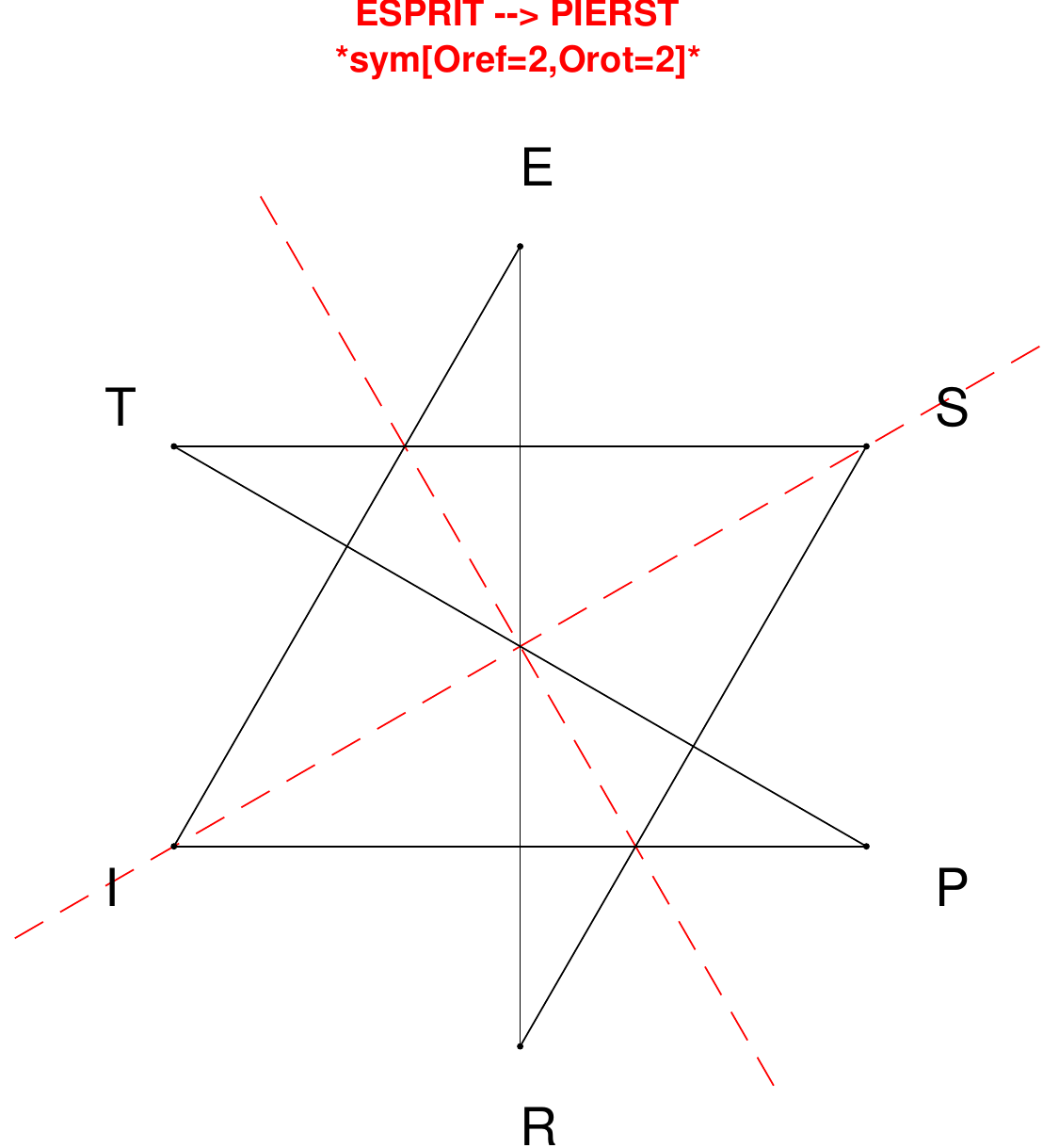}
\end{subfigure}
\hfill
\begin{subfigure}[T]{0.19\textwidth}
\centering
\includegraphics[width=\textwidth]{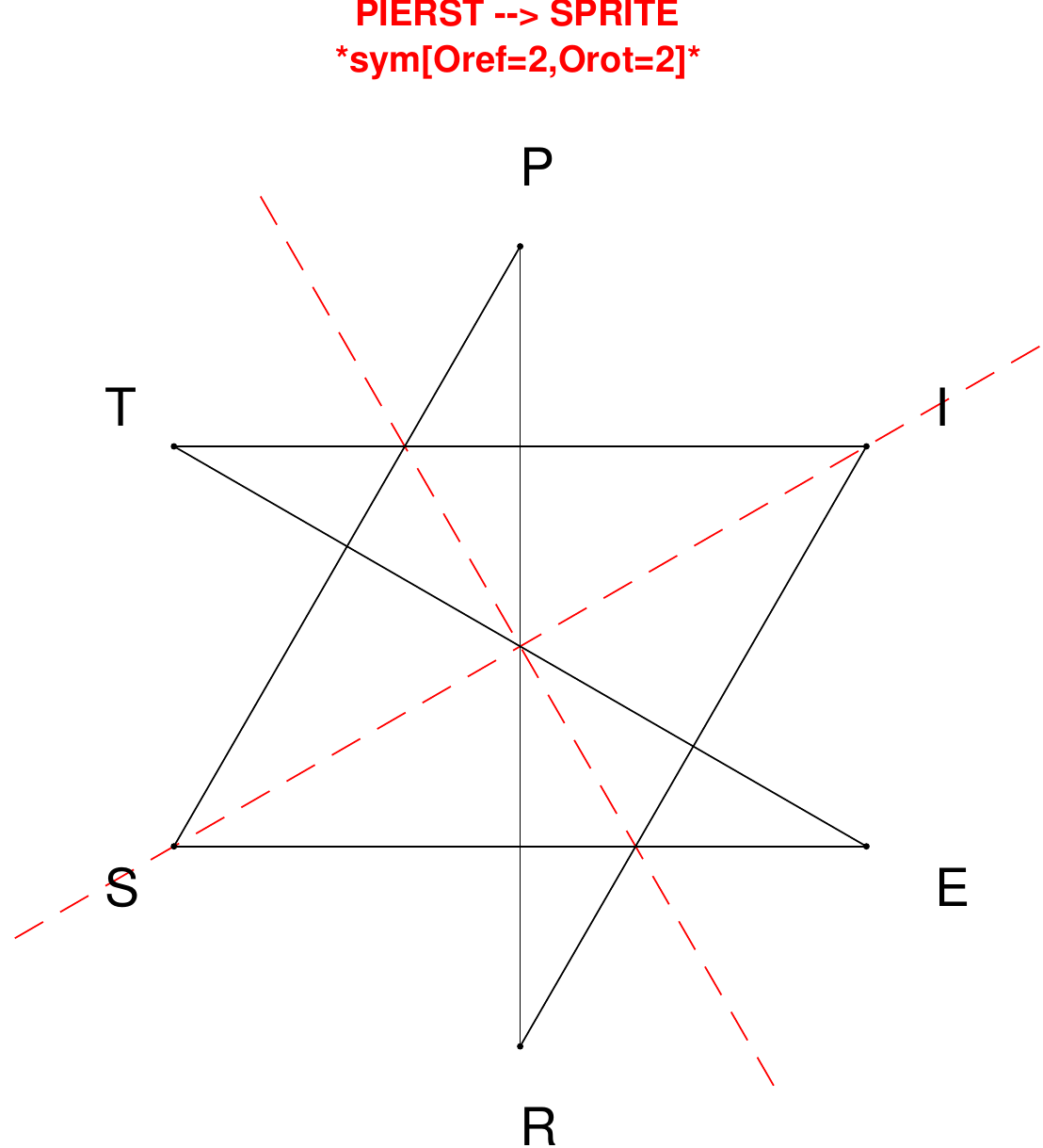}
\end{subfigure}
\hfill
\begin{subfigure}[T]{0.19\textwidth}
\centering
\includegraphics[width=\textwidth]{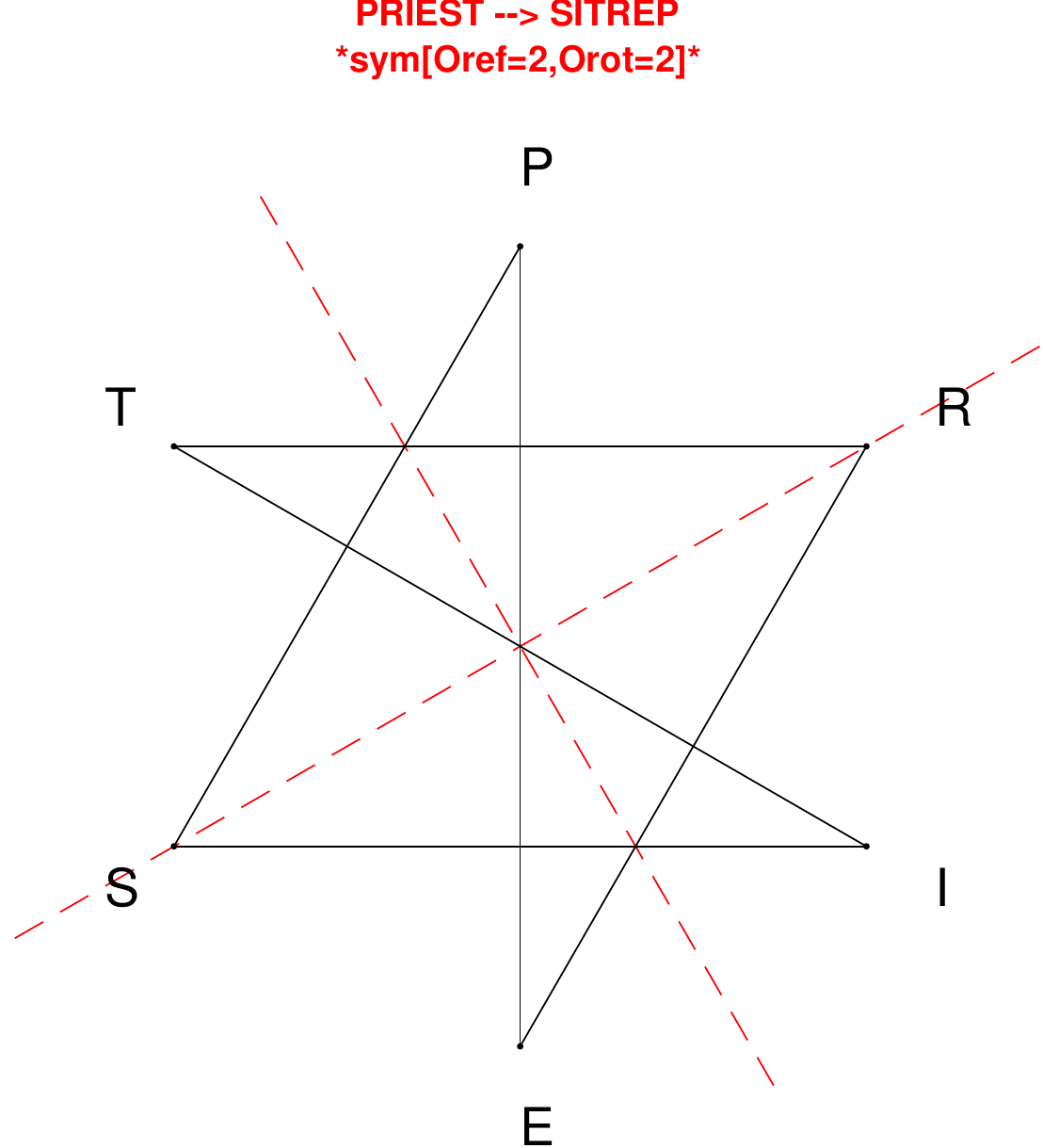}
\end{subfigure}
\hfill
\begin{subfigure}[T]{0.19\textwidth}
\centering
\includegraphics[width=\textwidth]{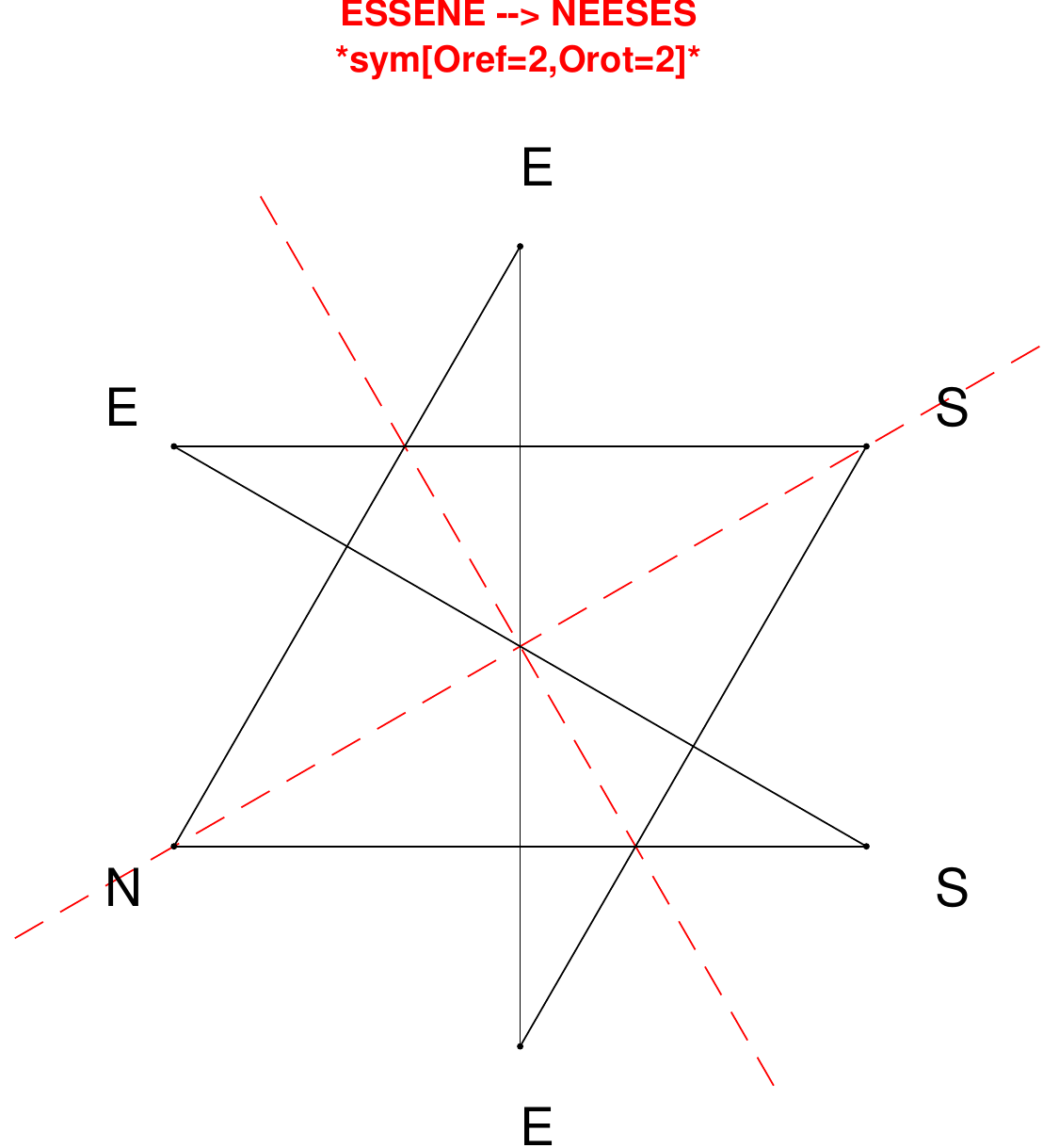}
\end{subfigure}
\end{figure}

\begin{figure}[H]
\centering
\begin{subfigure}[T]{0.19\textwidth}
\centering
\includegraphics[width=\textwidth]{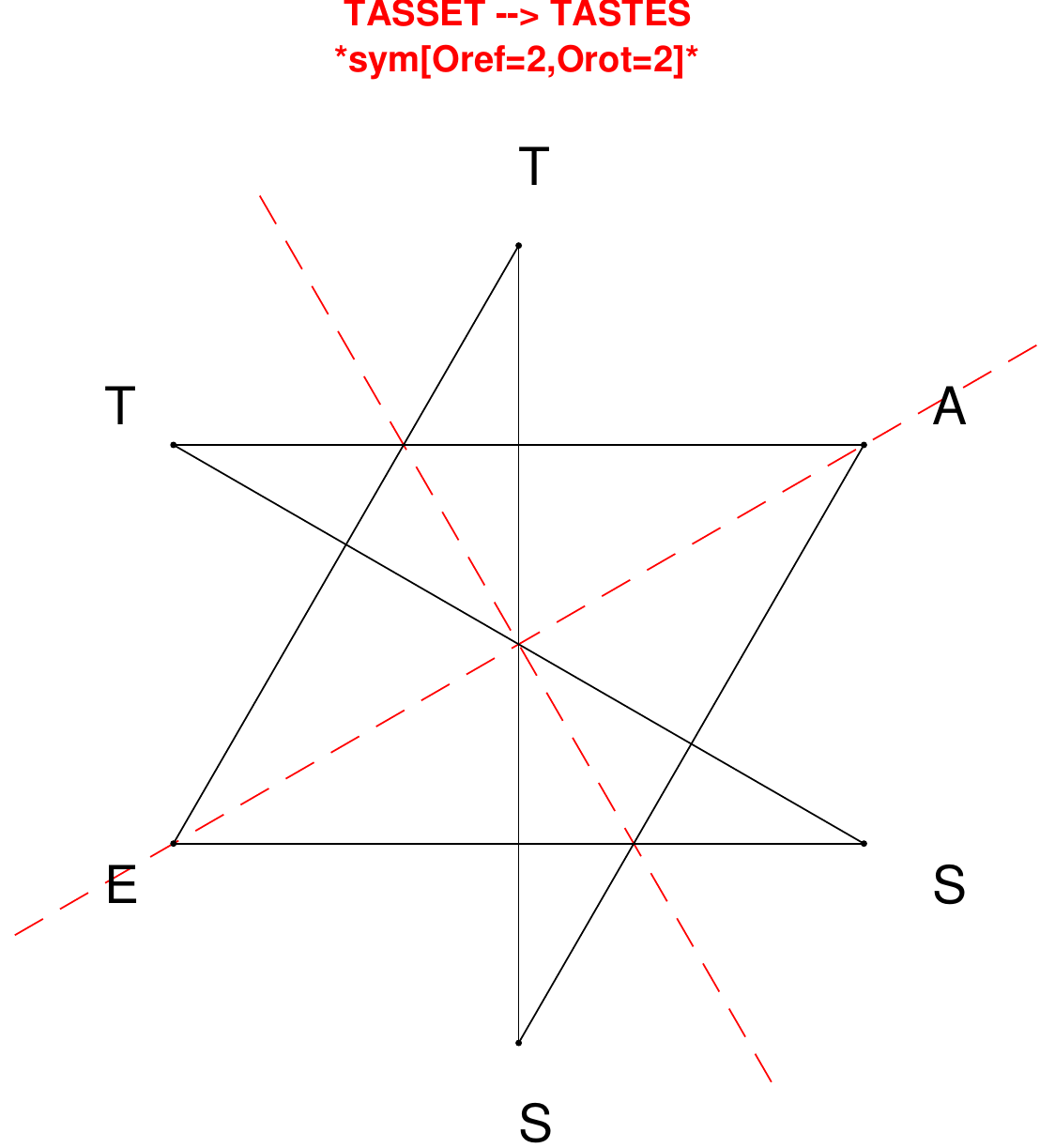}
\end{subfigure}
\hfill
\begin{subfigure}[T]{0.19\textwidth}
\centering
\includegraphics[width=\textwidth]{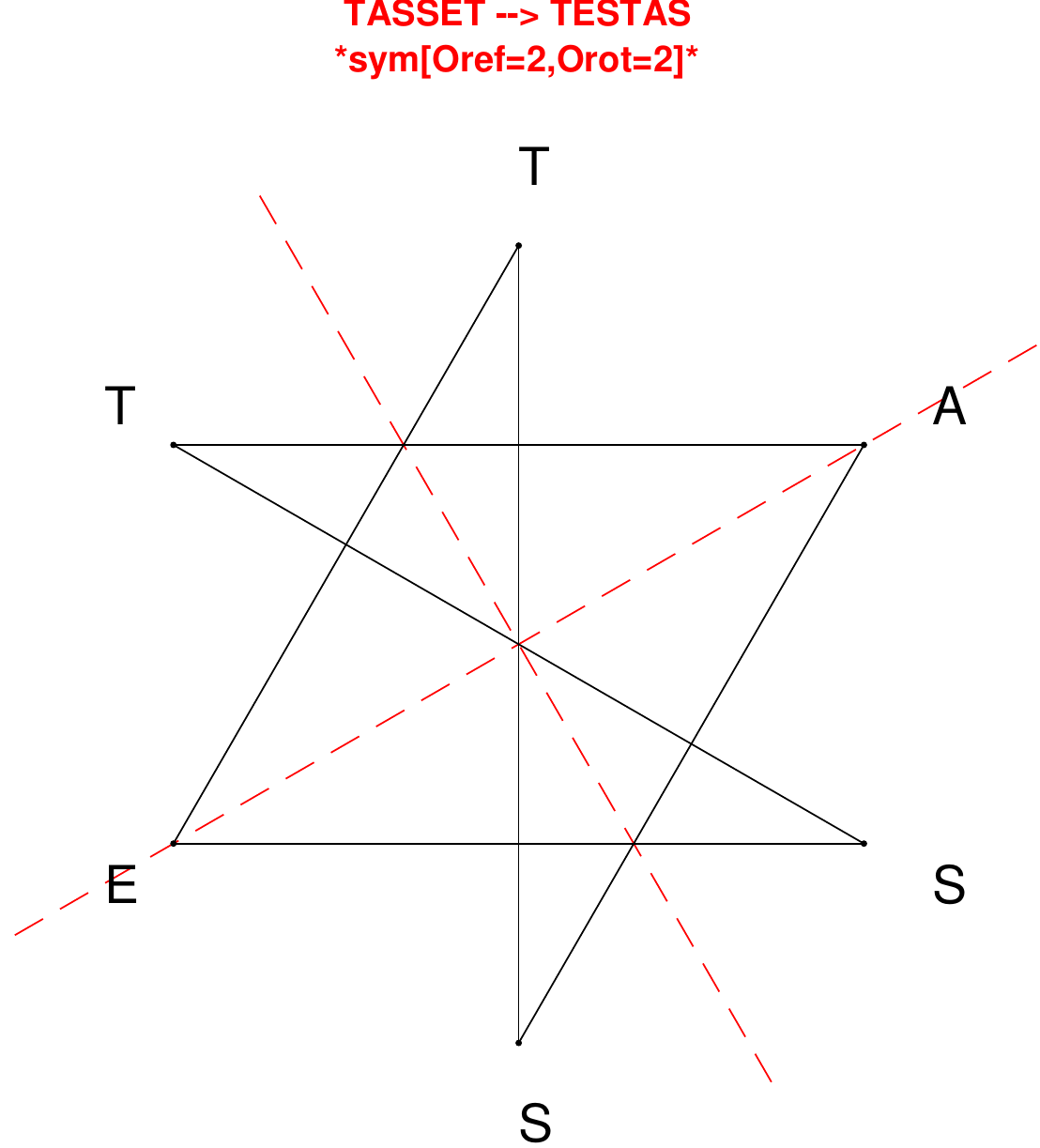}
\end{subfigure}
\hfill
\begin{subfigure}[T]{0.19\textwidth}
\centering
\includegraphics[width=\textwidth]{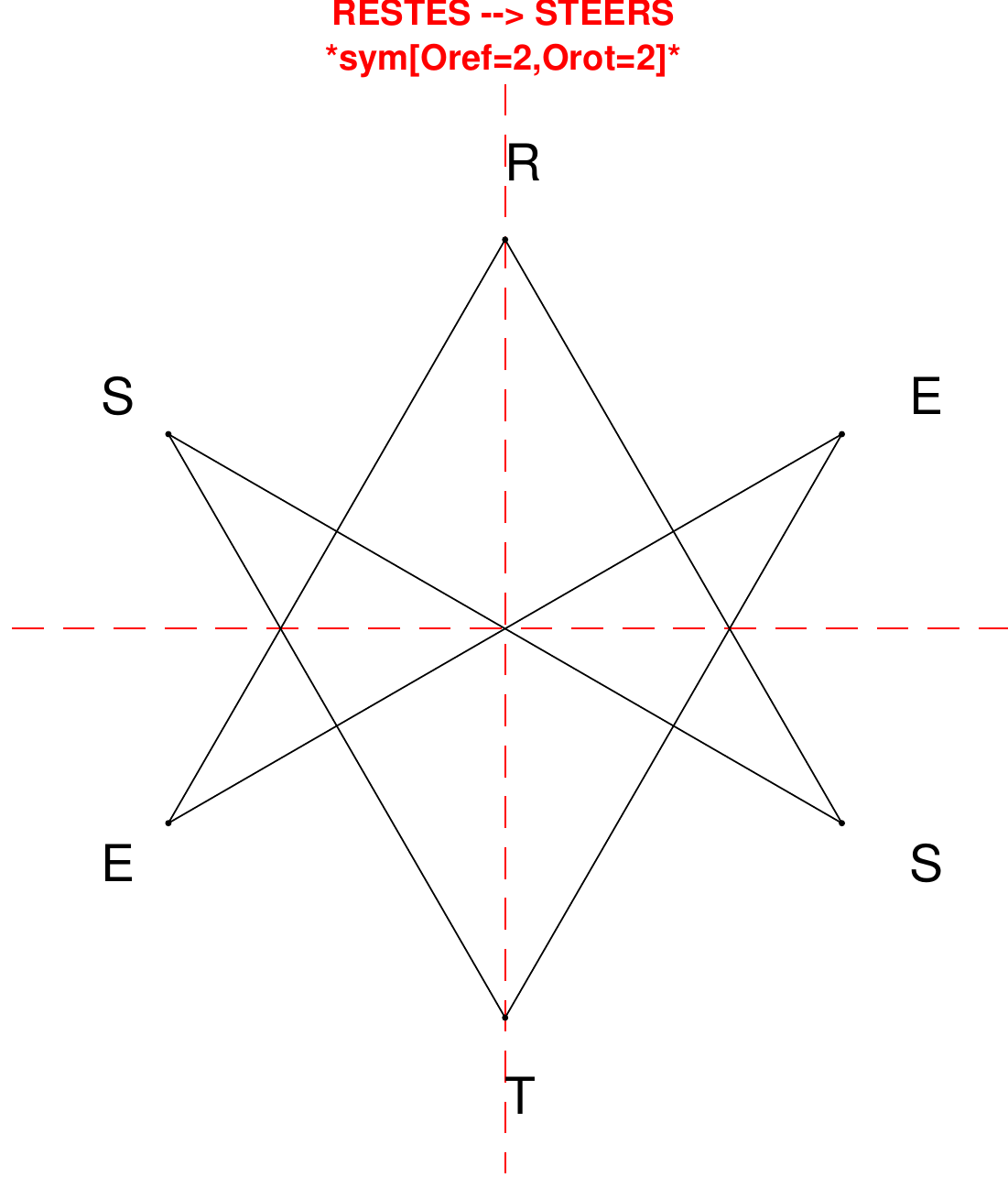}
\end{subfigure}
\hfill
\begin{subfigure}[T]{0.19\textwidth}
\centering
\includegraphics[width=\textwidth]{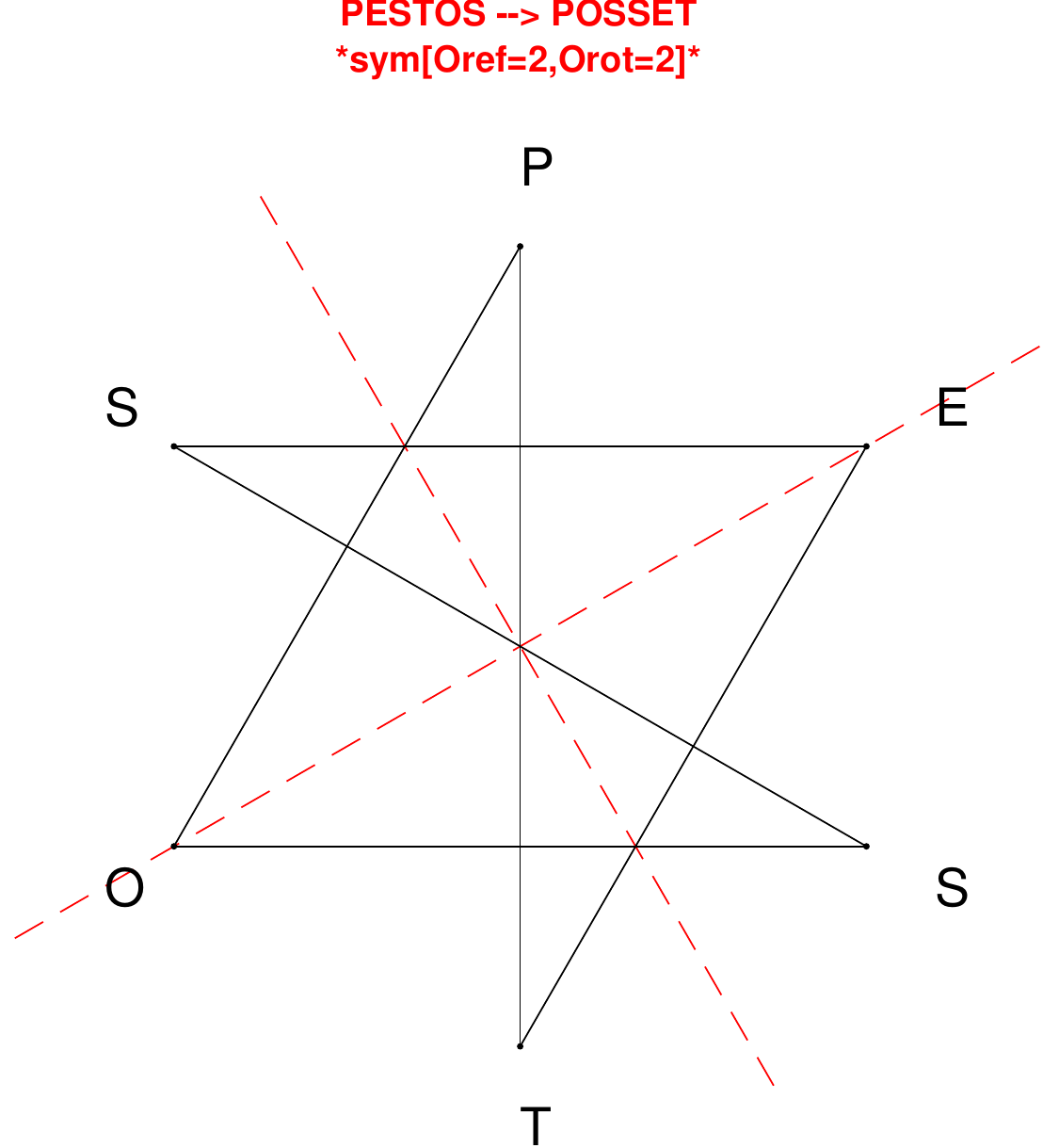}
\end{subfigure}
\hfill
\begin{subfigure}[T]{0.19\textwidth}
\centering
\includegraphics[width=\textwidth]{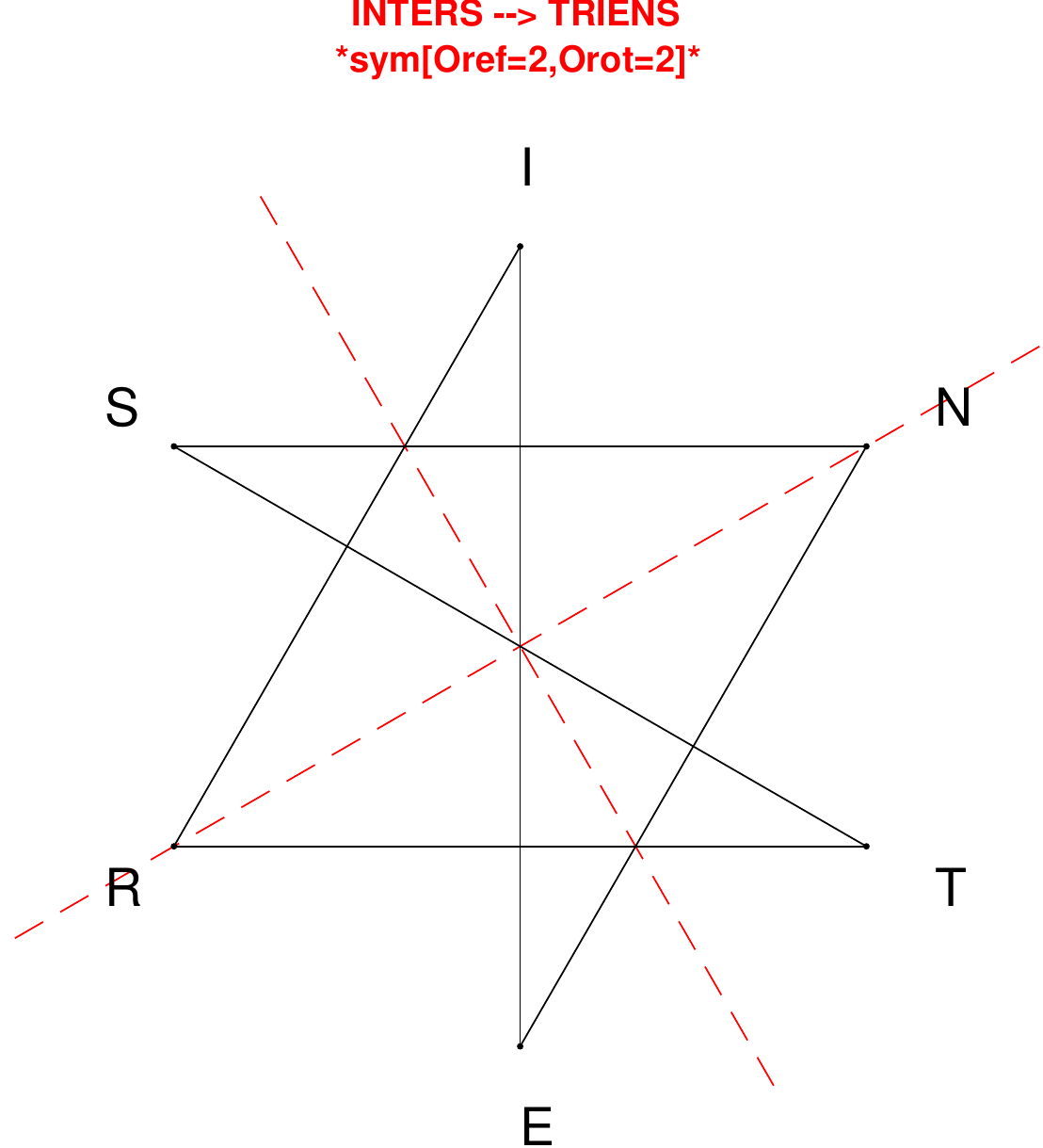}
\end{subfigure}
\end{figure}

\begin{figure}[H]
\centering
\begin{subfigure}[T]{0.19\textwidth}
\centering
\includegraphics[width=\textwidth]{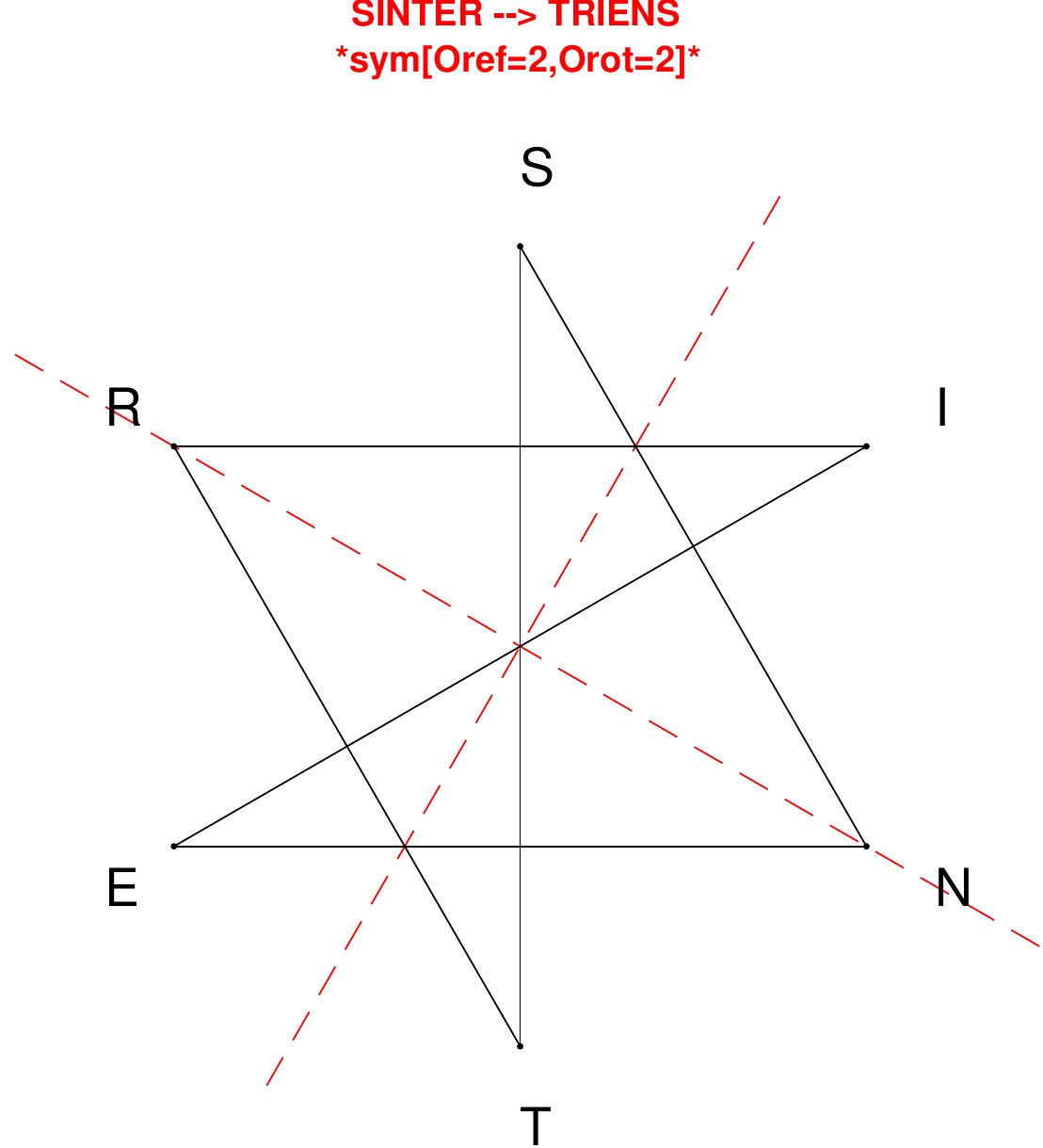}
\end{subfigure}
\hfill
\begin{subfigure}[T]{0.19\textwidth}
\centering
\includegraphics[width=\textwidth]{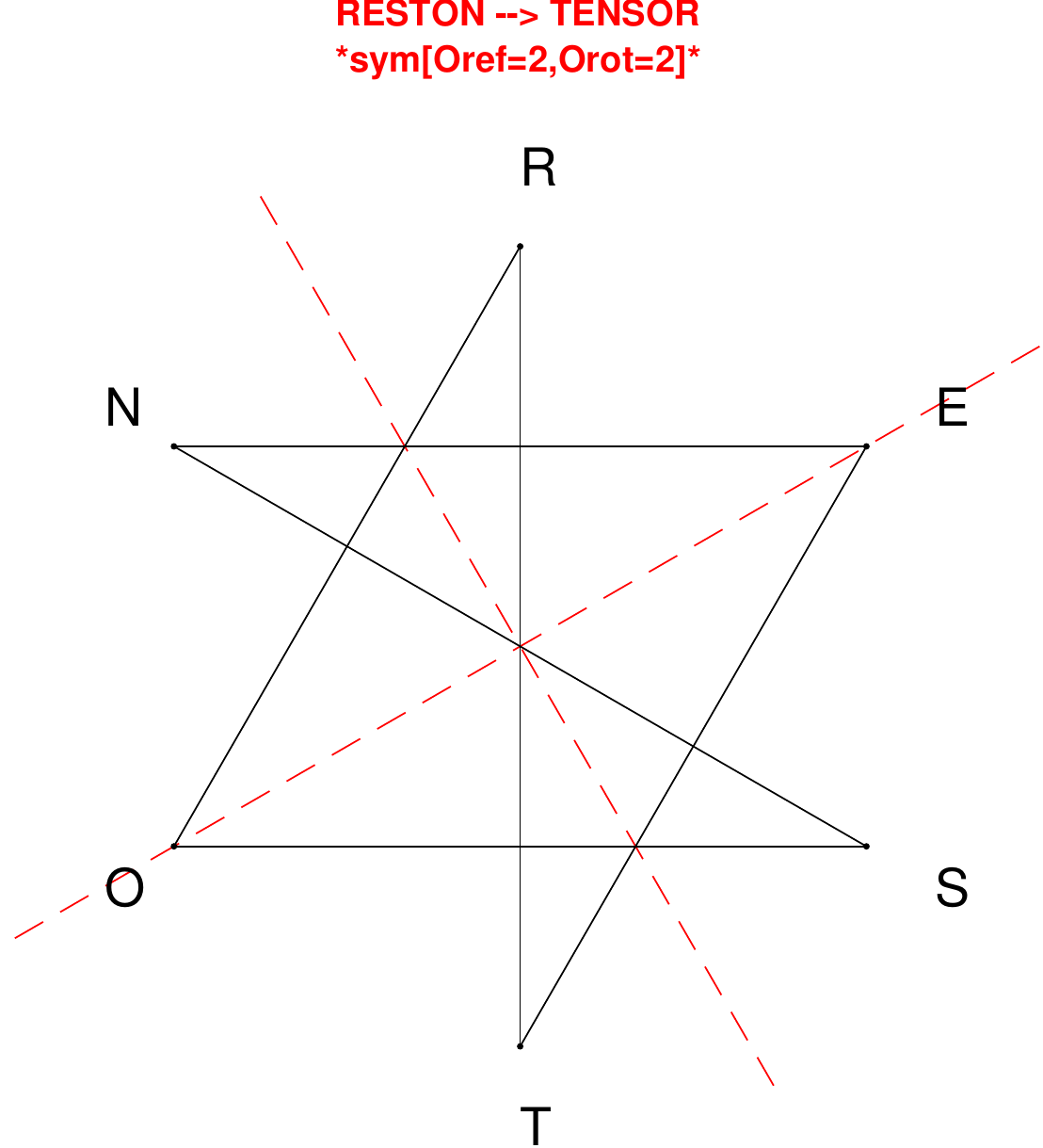}
\end{subfigure}
\hfill
\begin{subfigure}[T]{0.19\textwidth}
\centering
\includegraphics[width=\textwidth]{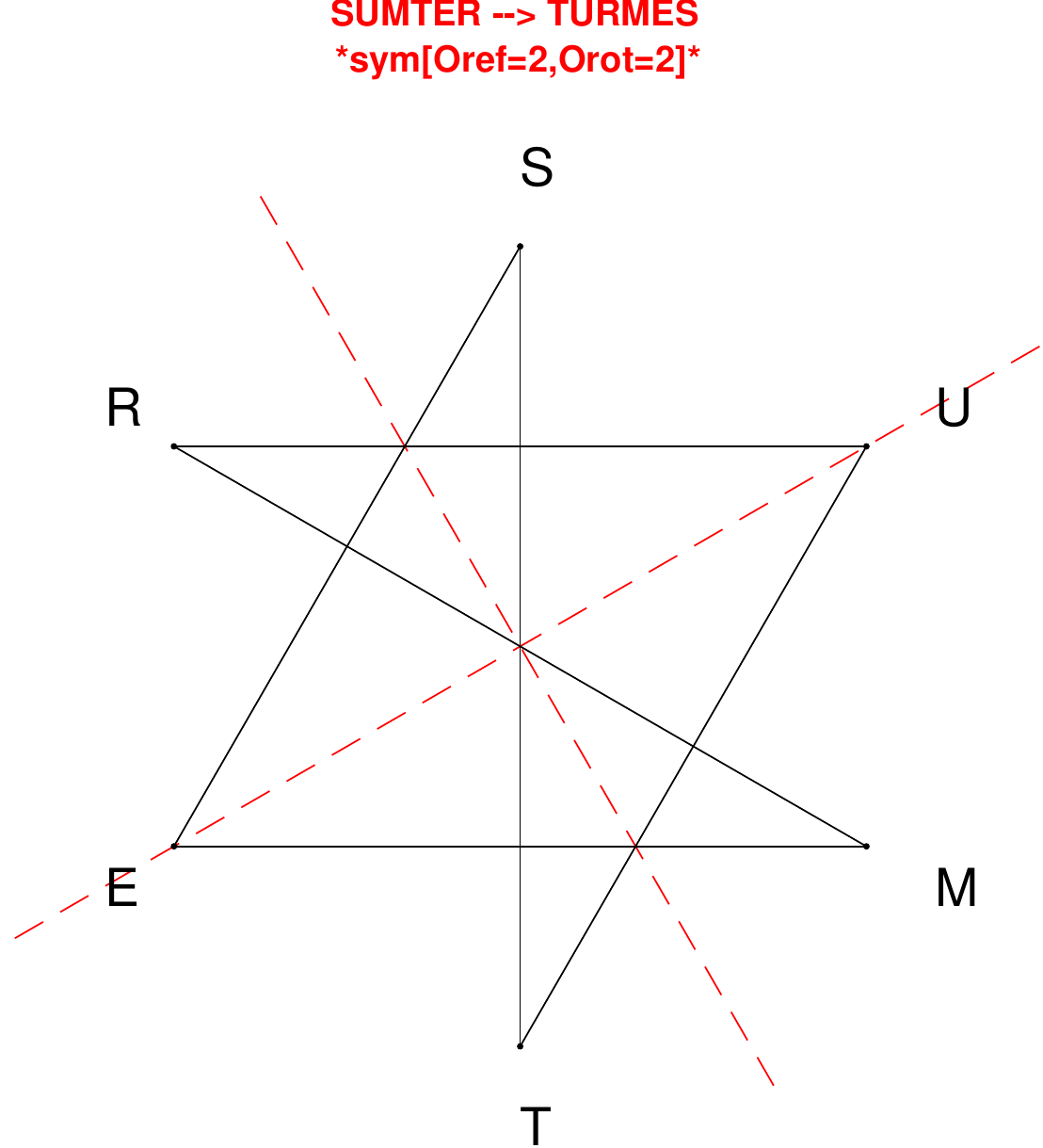}
\end{subfigure}
\hfill
\begin{subfigure}[T]{0.19\textwidth}
\centering
\includegraphics[width=\textwidth]{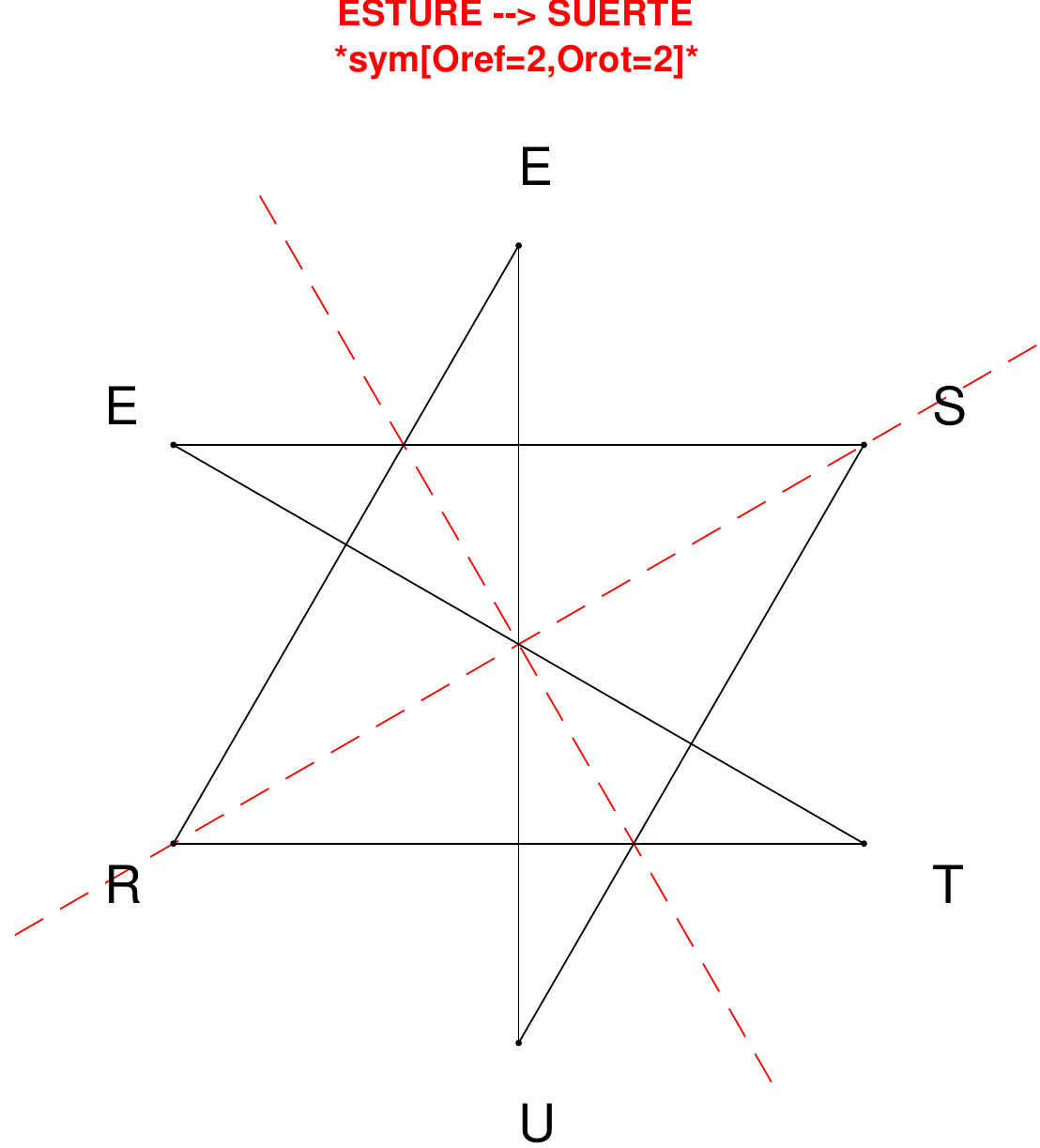}
\end{subfigure}
\hfill
\begin{subfigure}[T]{0.19\textwidth}
\centering
\includegraphics[width=\textwidth]{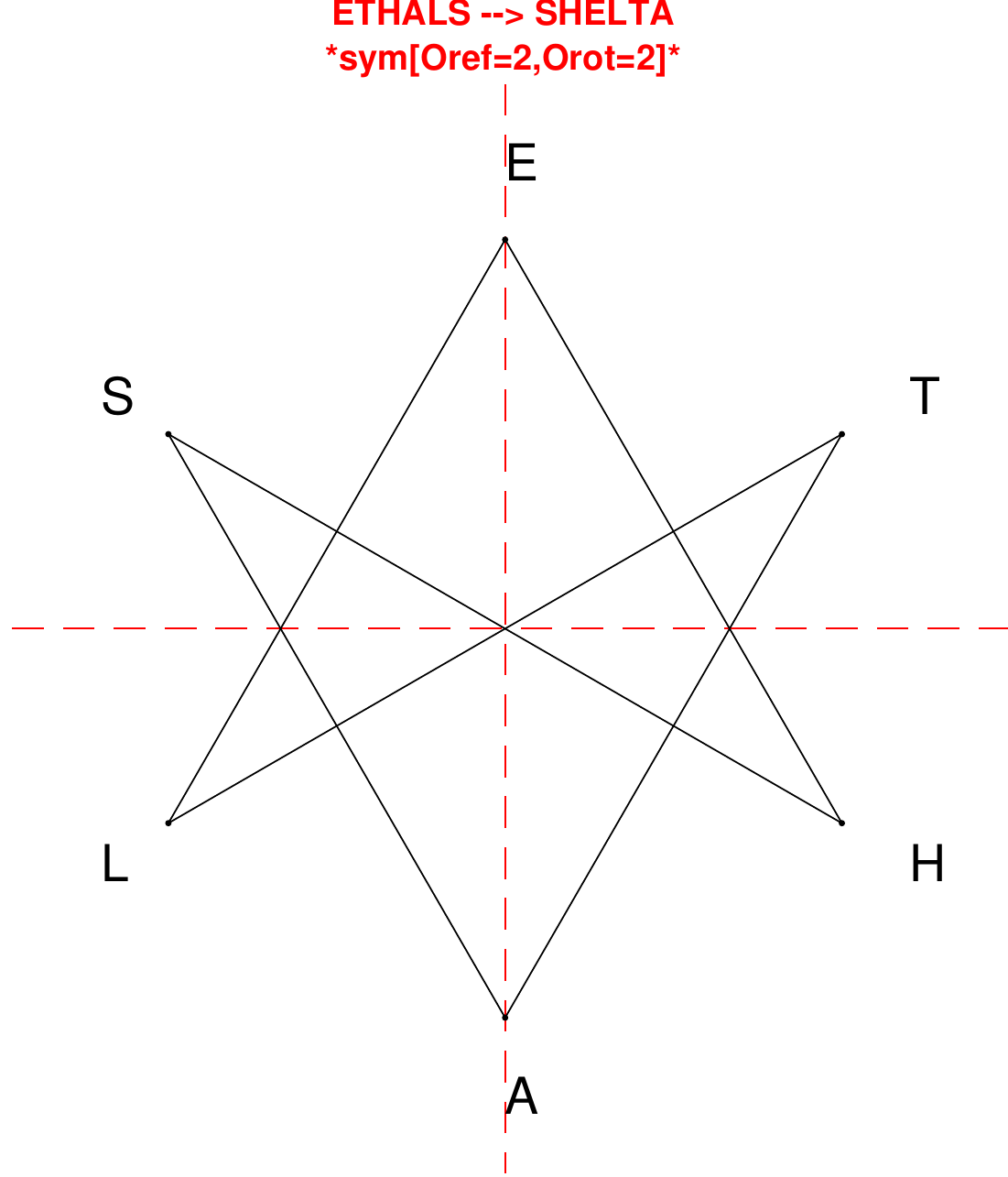}
\end{subfigure}
\end{figure}

\begin{figure}[H]
\centering
\begin{subfigure}[T]{0.19\textwidth}
\centering
\includegraphics[width=\textwidth]{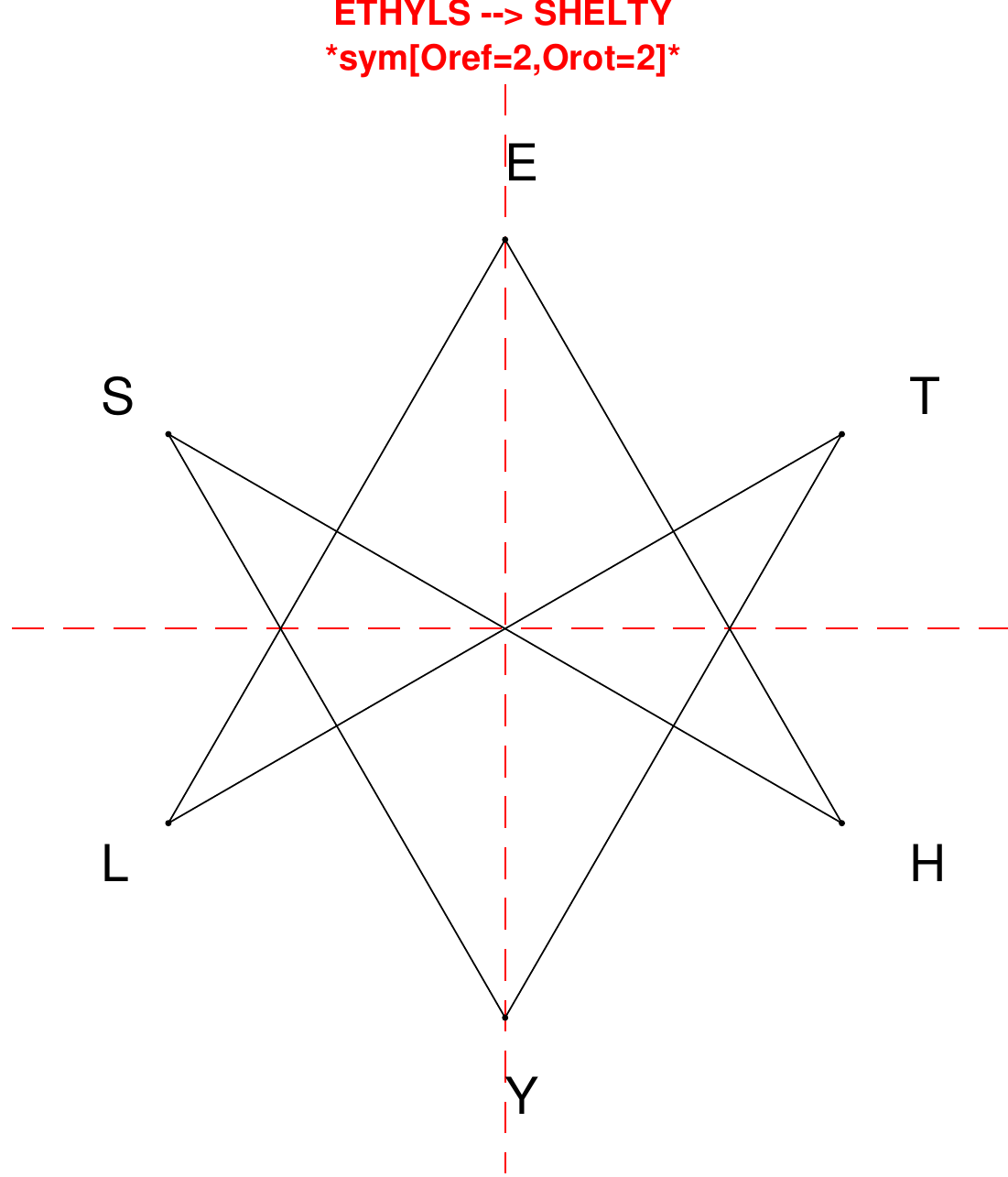}
\end{subfigure}
\hfill
\begin{subfigure}[T]{0.19\textwidth}
\centering
\includegraphics[width=\textwidth]{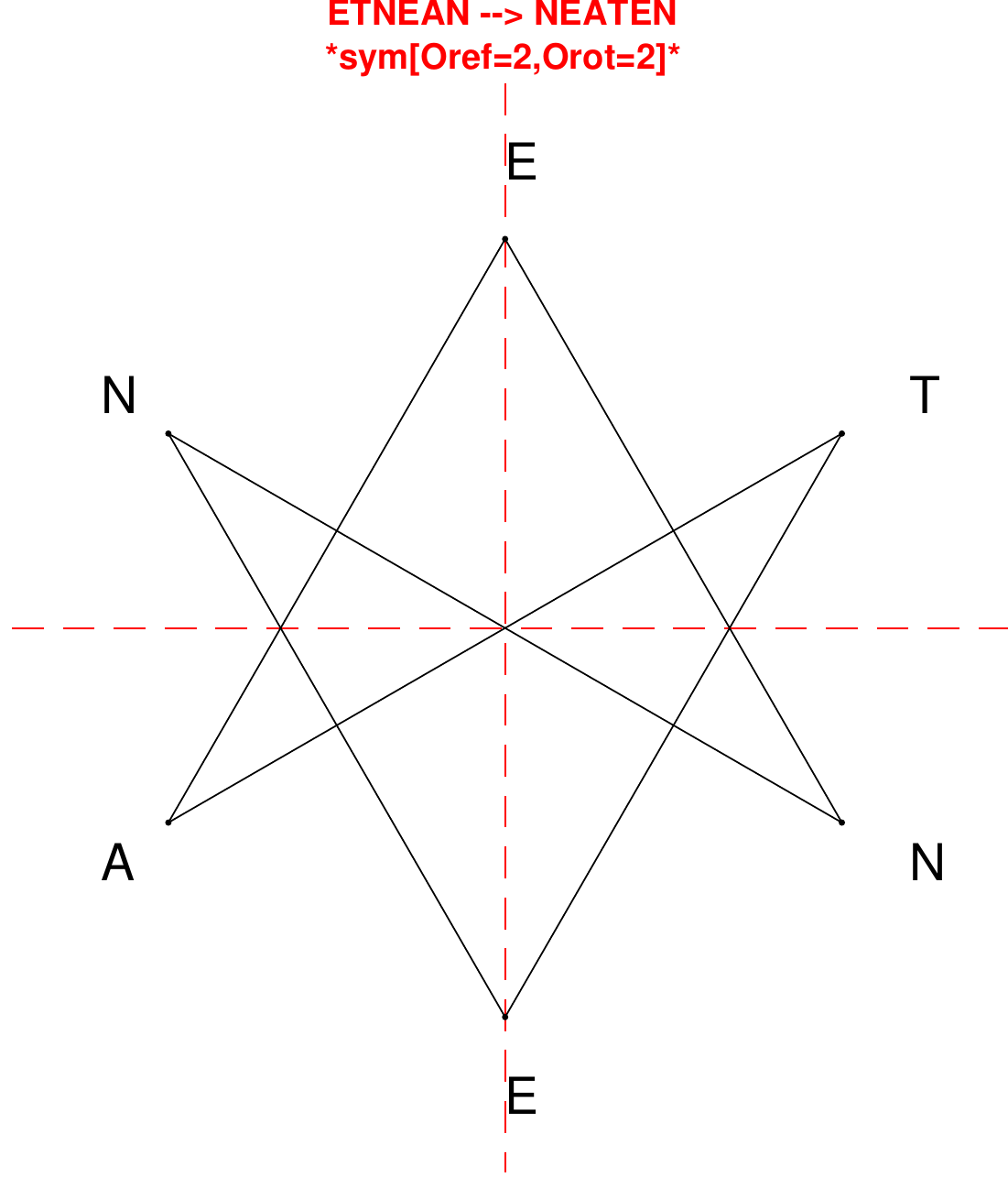}
\end{subfigure}
\hfill
\begin{subfigure}[T]{0.19\textwidth}
\centering
\includegraphics[width=\textwidth]{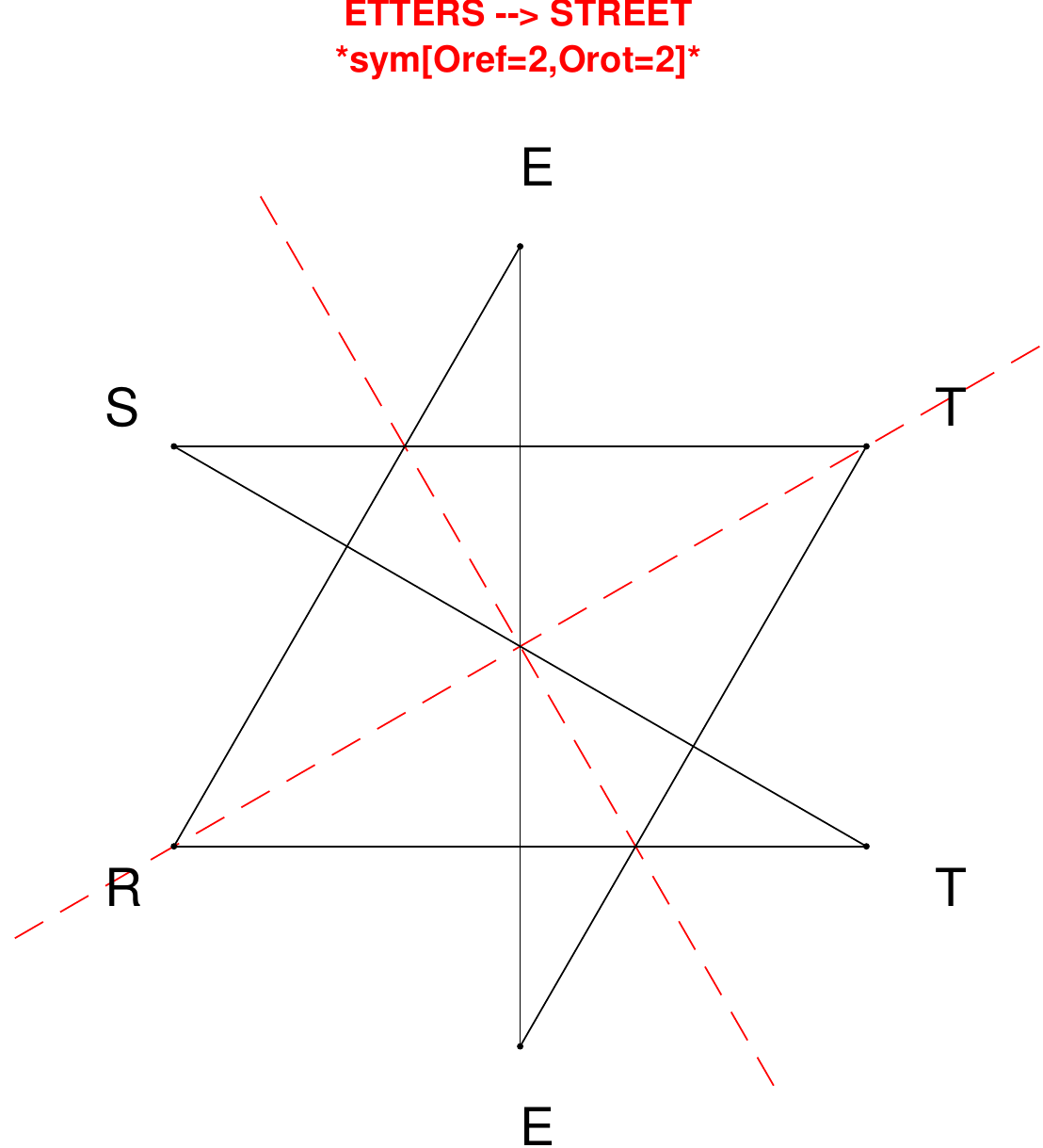}
\end{subfigure}
\hfill
\begin{subfigure}[T]{0.19\textwidth}
\centering
\includegraphics[width=\textwidth]{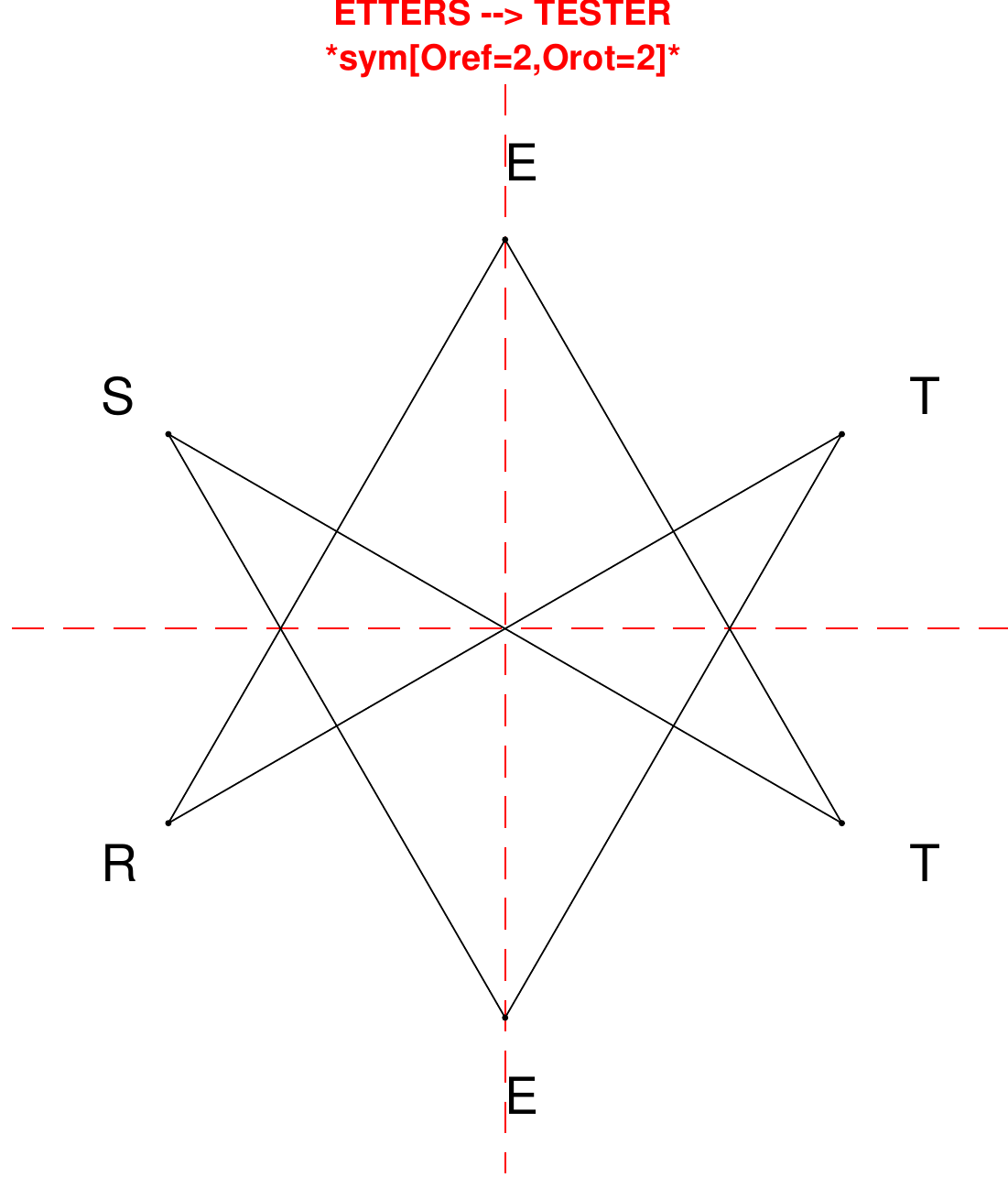}
\end{subfigure}
\hfill
\begin{subfigure}[T]{0.19\textwidth}
\centering
\includegraphics[width=\textwidth]{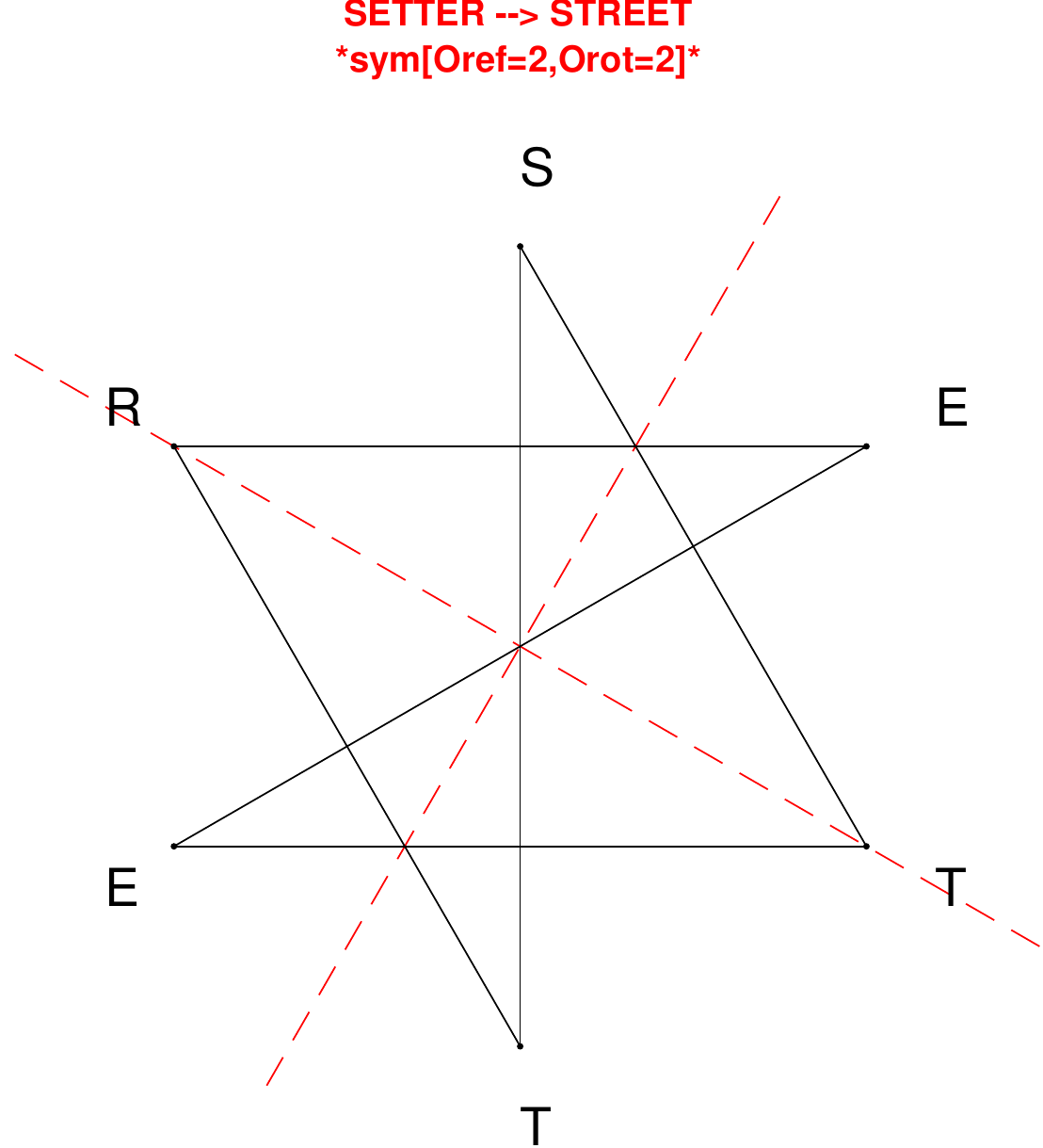}
\end{subfigure}
\end{figure}

\begin{figure}[H]
\centering
\begin{subfigure}[T]{0.19\textwidth}
\centering
\includegraphics[width=\textwidth]{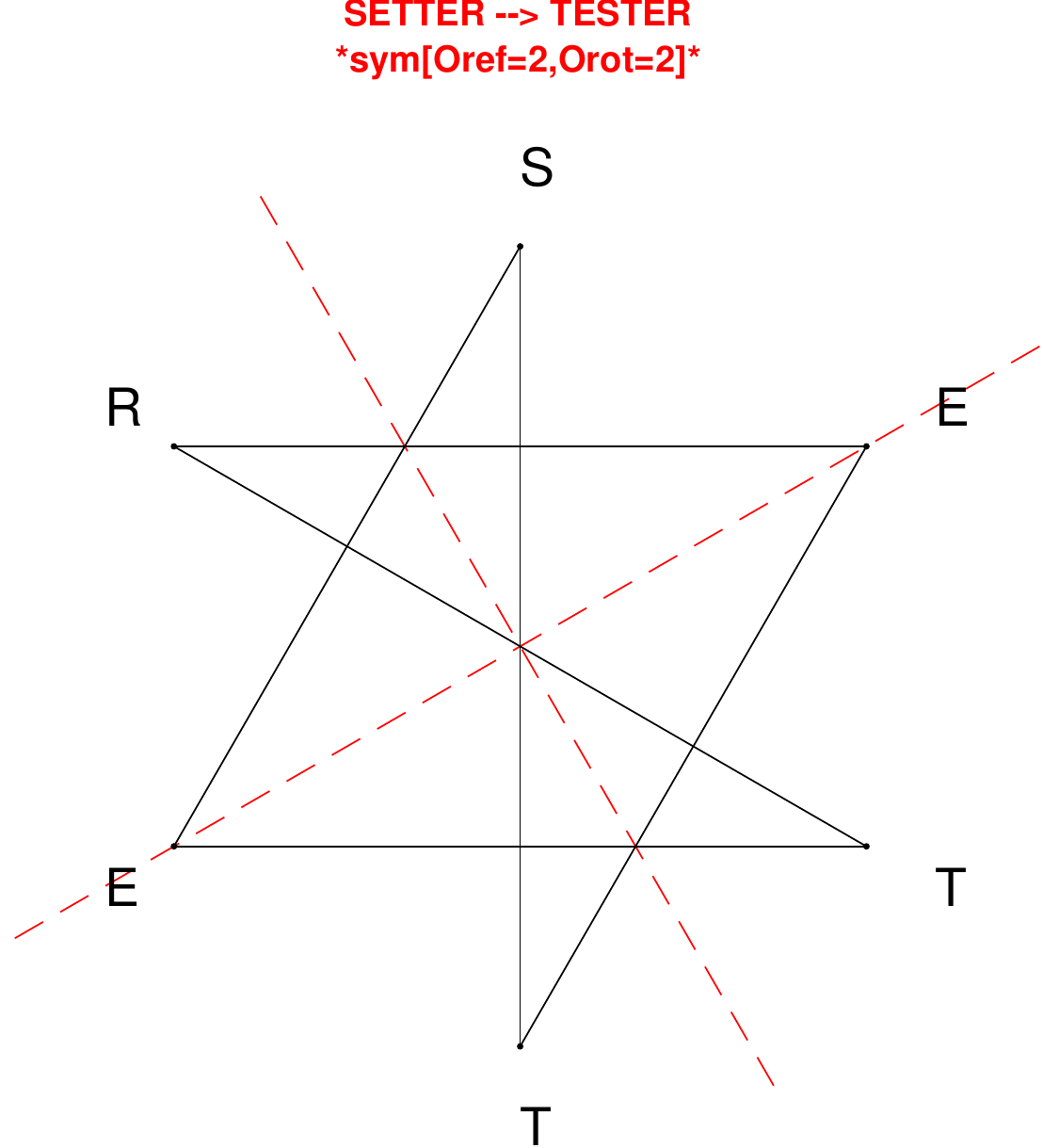}
\end{subfigure}
\hfill
\begin{subfigure}[T]{0.19\textwidth}
\centering
\includegraphics[width=\textwidth]{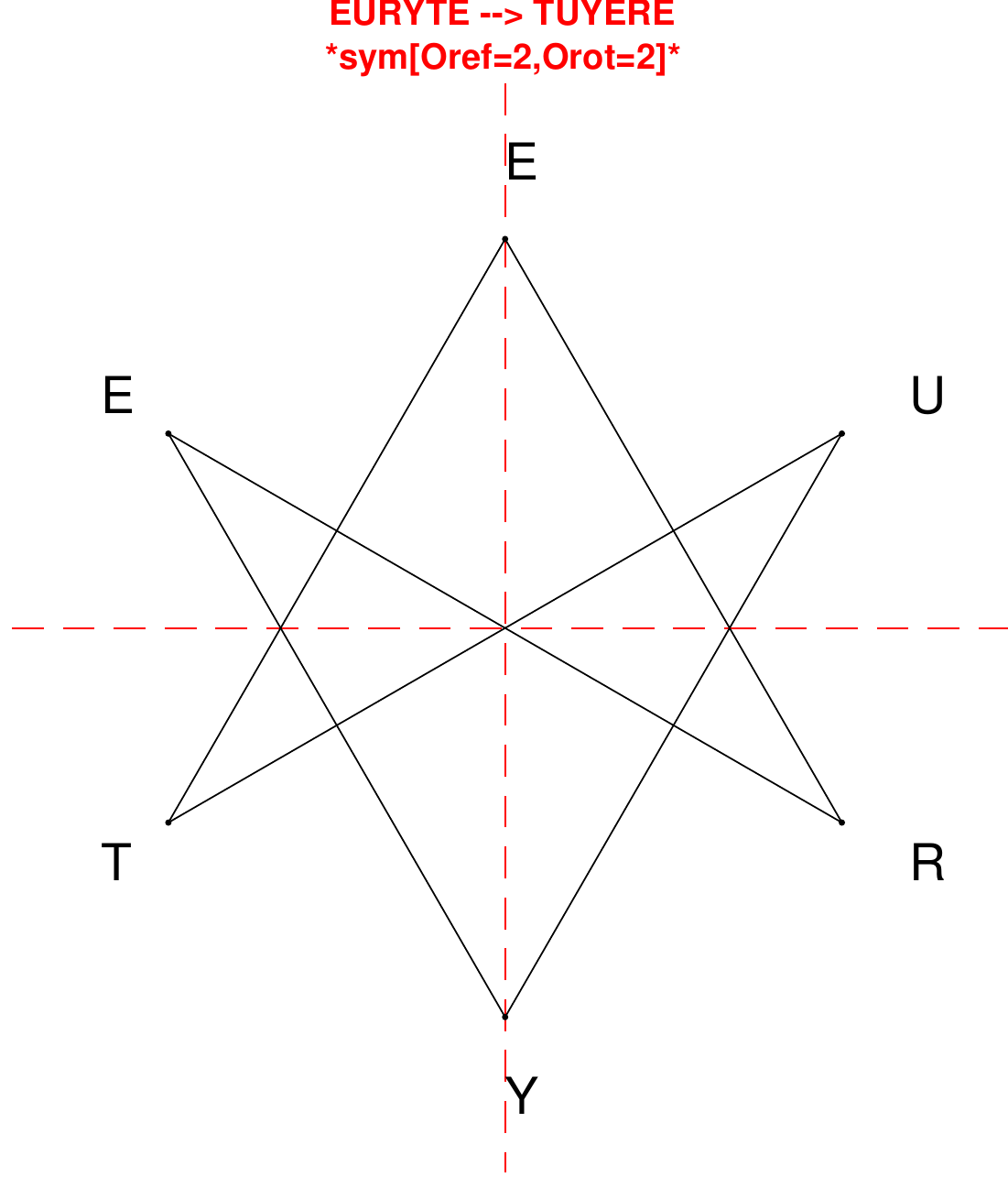}
\end{subfigure}
\hfill
\begin{subfigure}[T]{0.19\textwidth}
\centering
\includegraphics[width=\textwidth]{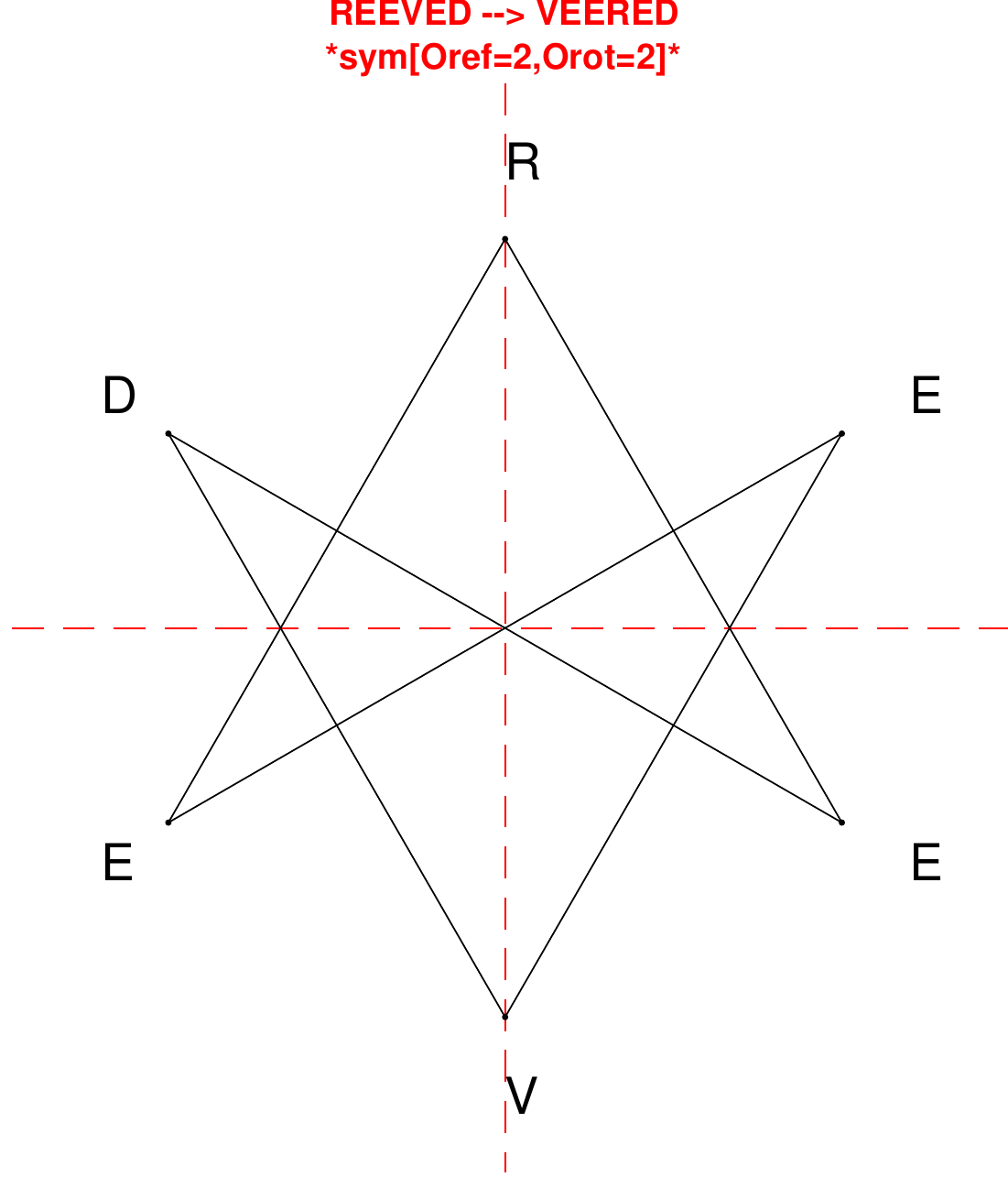}
\end{subfigure}
\hfill
\begin{subfigure}[T]{0.19\textwidth}
\centering
\includegraphics[width=\textwidth]{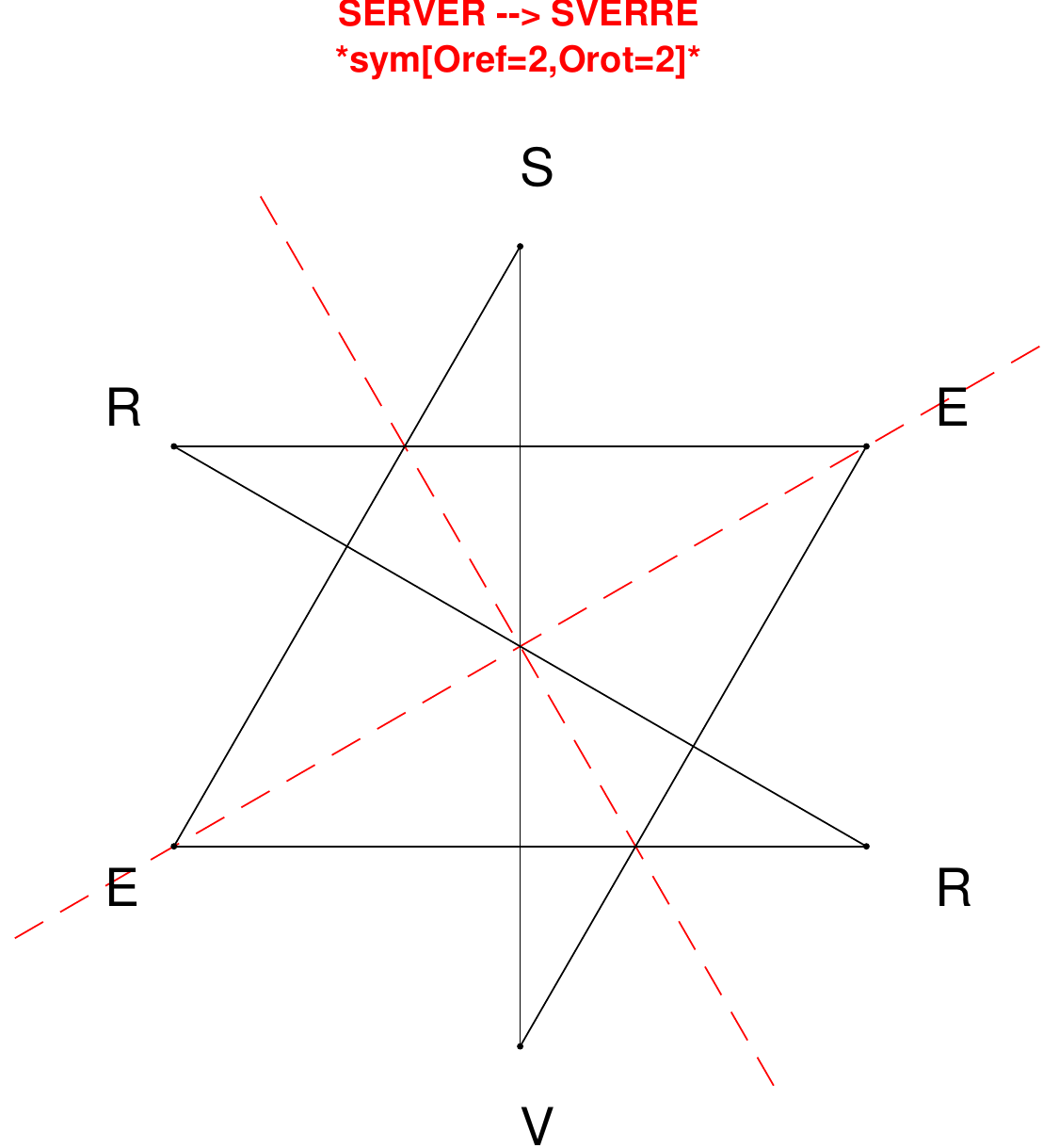}
\end{subfigure}
\hfill
\begin{subfigure}[T]{0.19\textwidth}
\centering
\includegraphics[width=\textwidth]{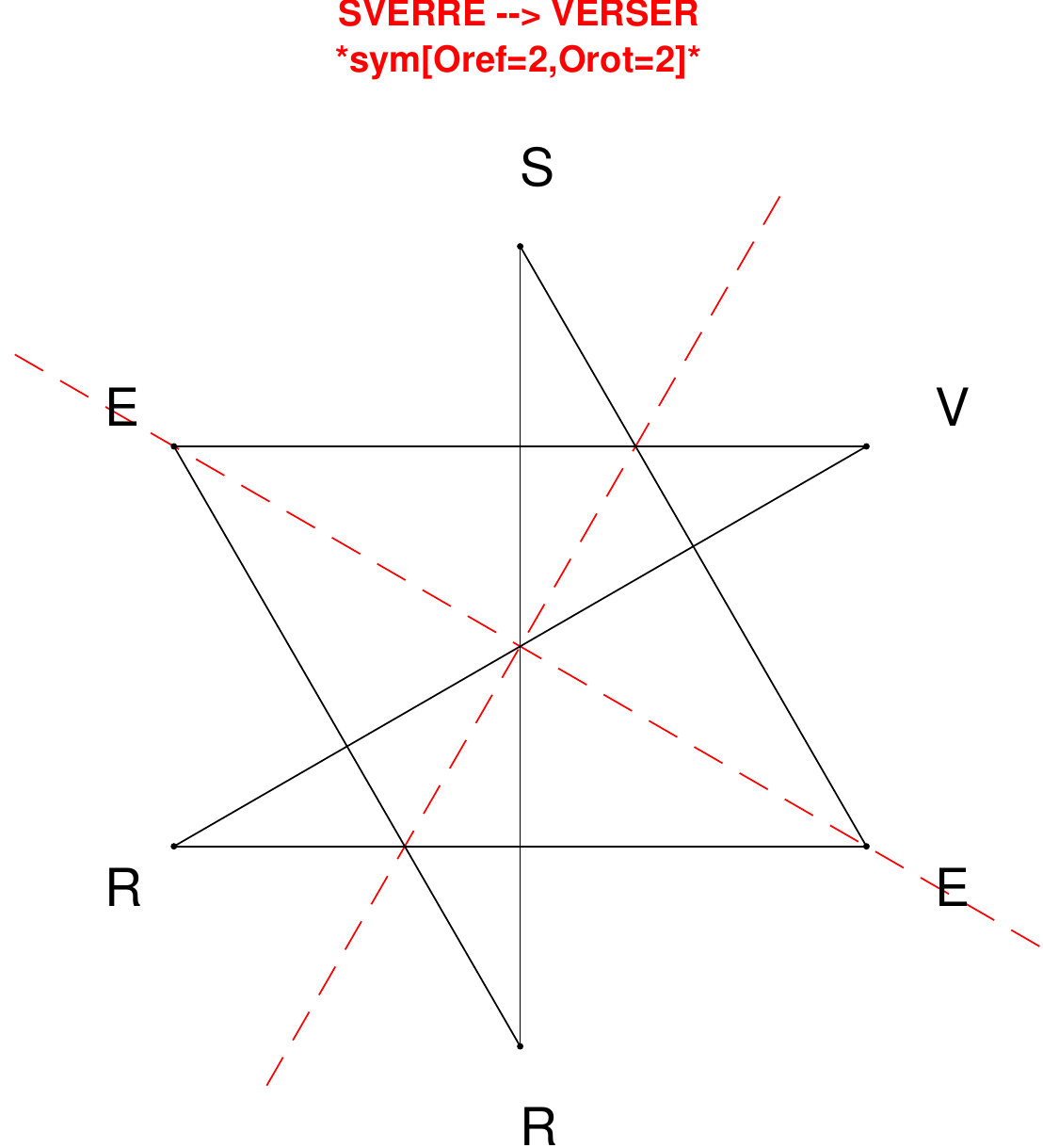}
\end{subfigure}
\end{figure}

\begin{figure}[H]
\centering
\begin{subfigure}[T]{0.19\textwidth}
\centering
\includegraphics[width=\textwidth]{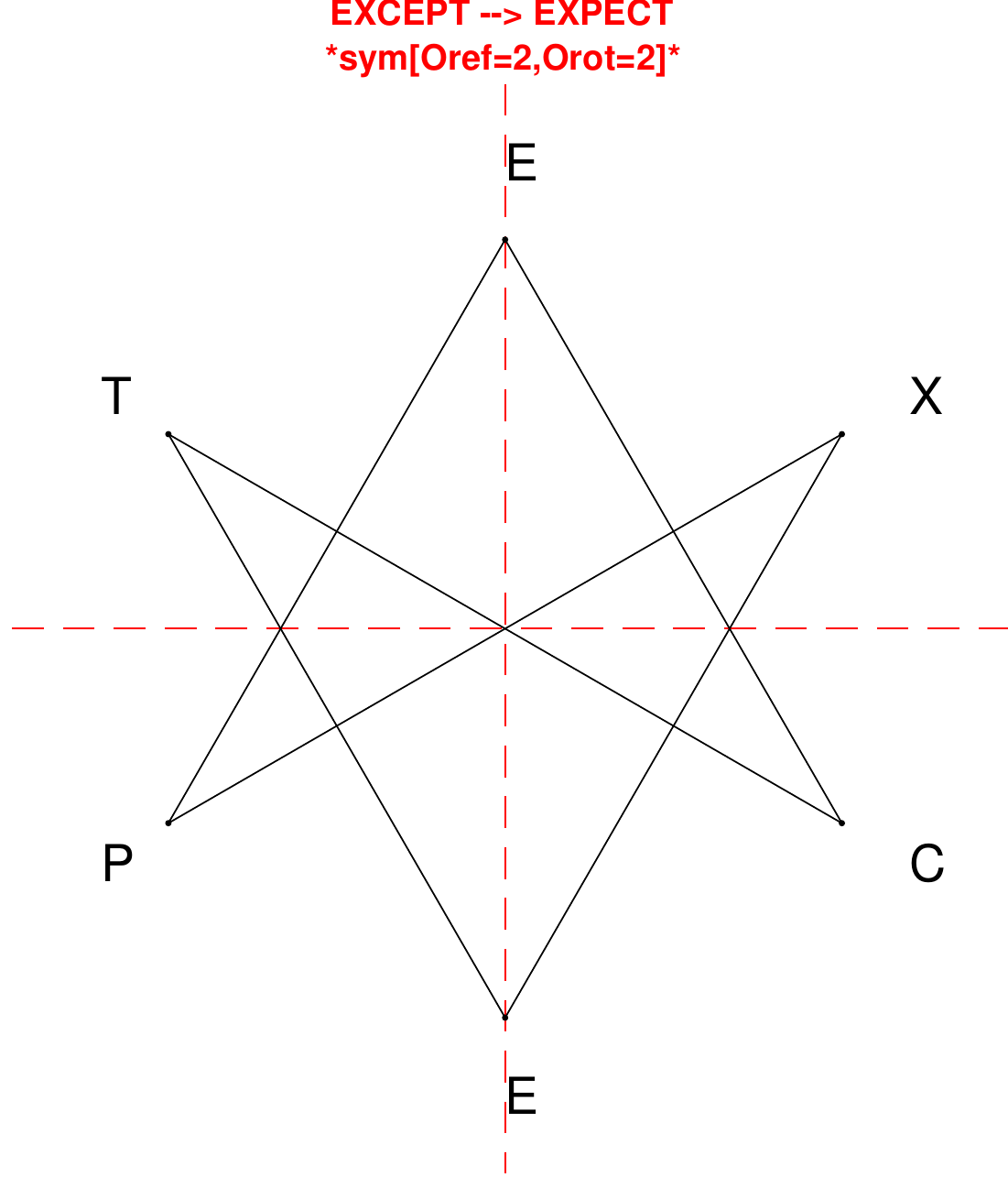}
\end{subfigure}
\hfill
\begin{subfigure}[T]{0.19\textwidth}
\centering
\includegraphics[width=\textwidth]{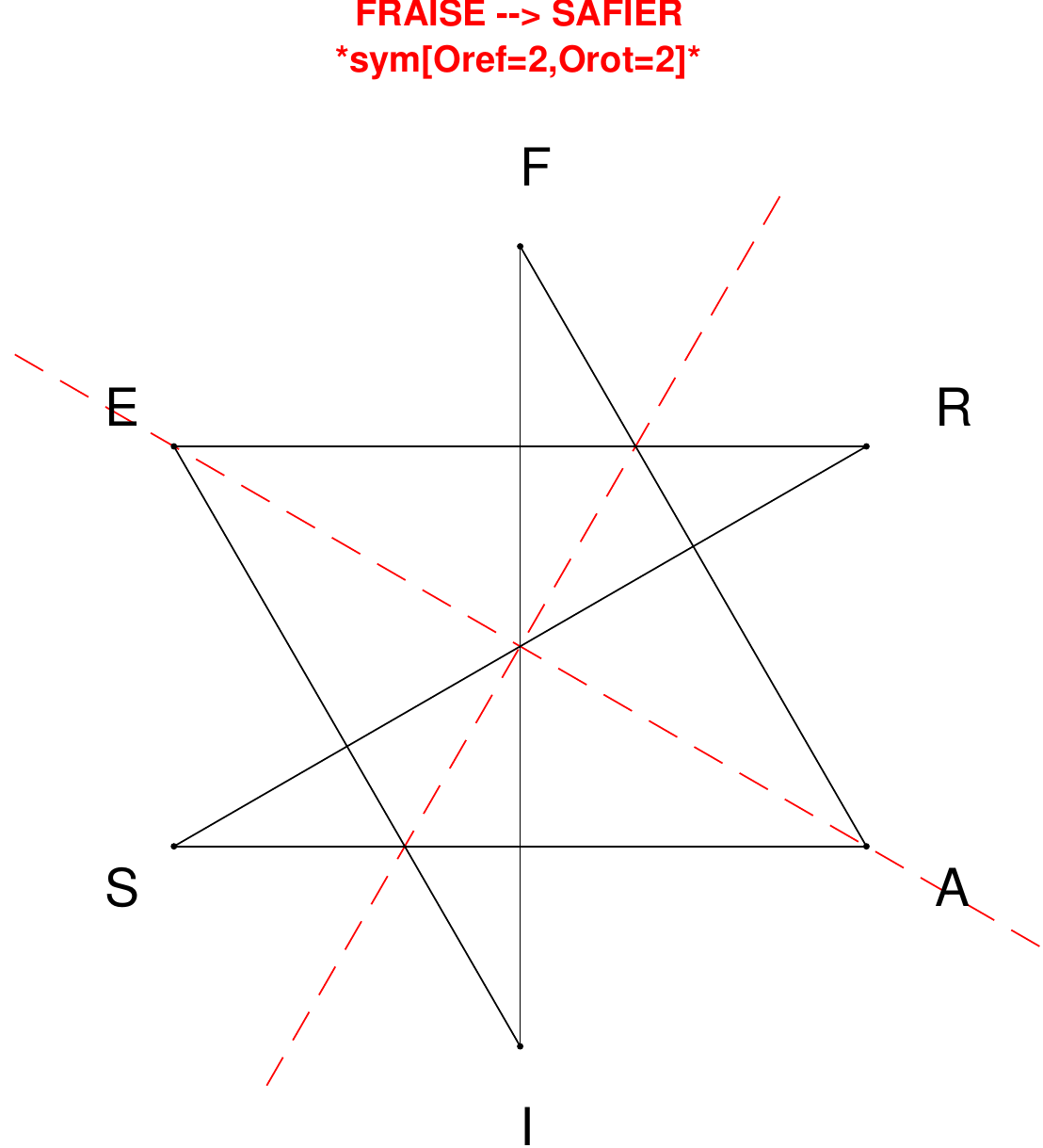}
\end{subfigure}
\hfill
\begin{subfigure}[T]{0.19\textwidth}
\centering
\includegraphics[width=\textwidth]{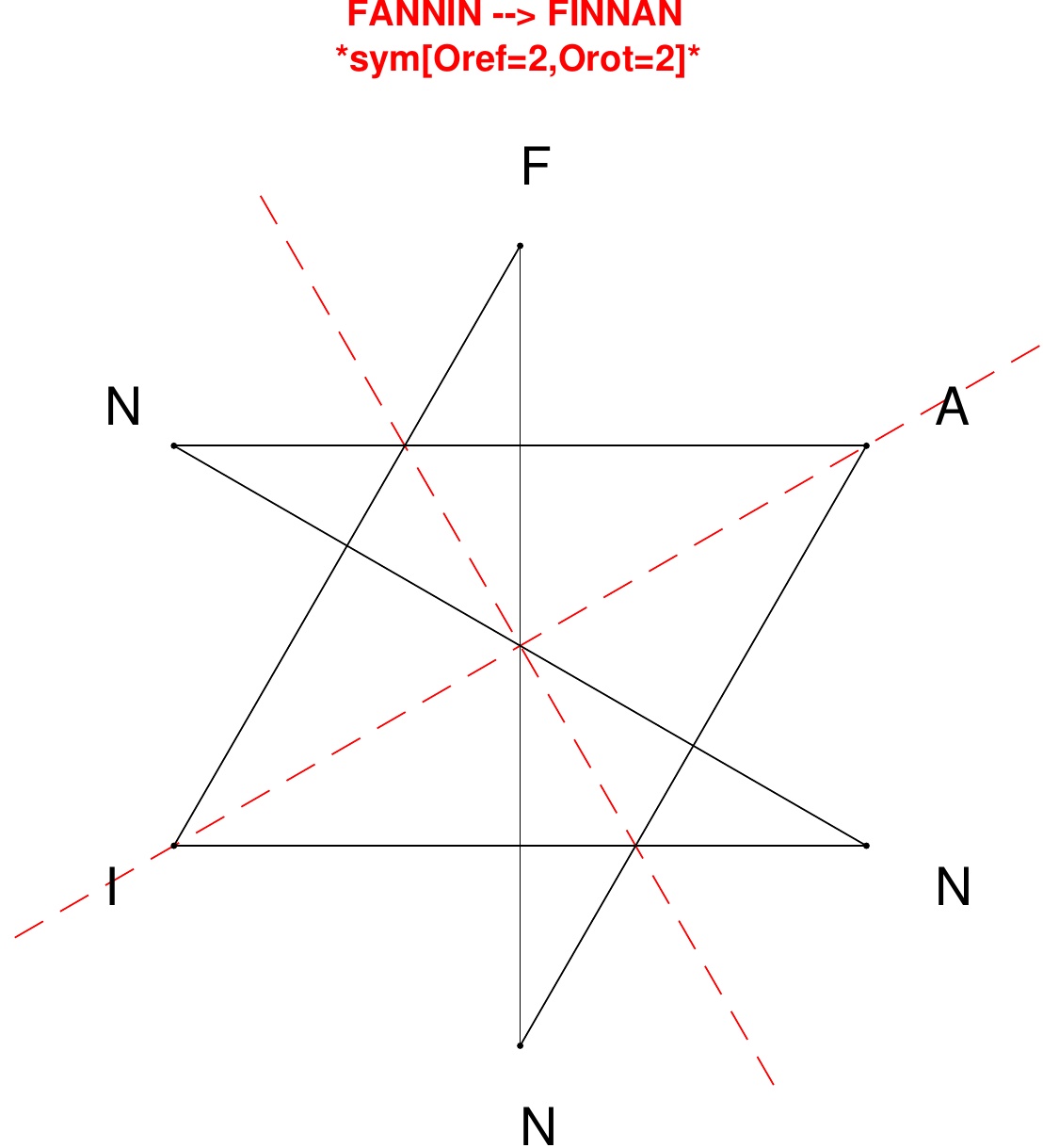}
\end{subfigure}
\hfill
\begin{subfigure}[T]{0.19\textwidth}
\centering
\includegraphics[width=\textwidth]{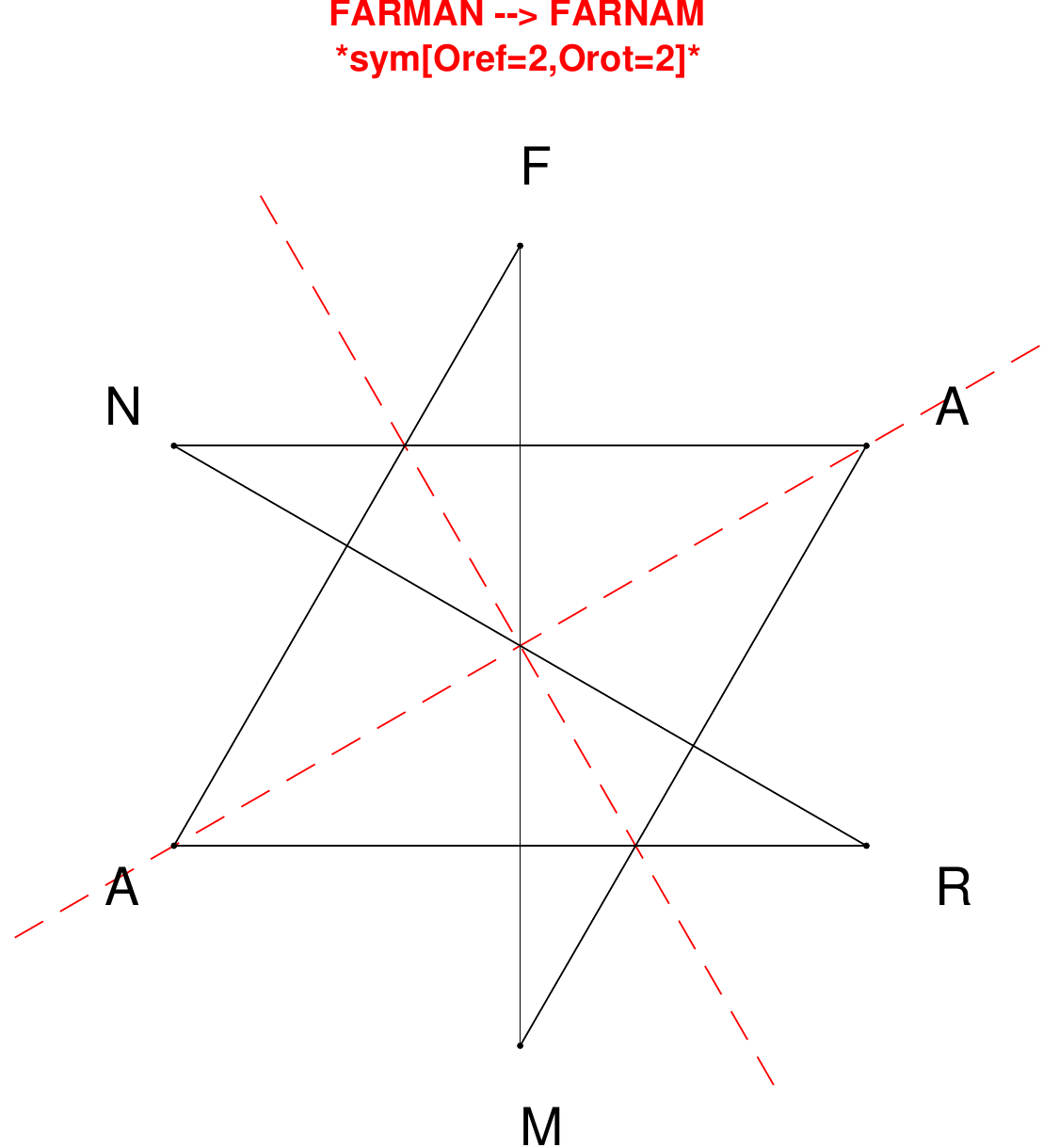}
\end{subfigure}
\hfill
\begin{subfigure}[T]{0.19\textwidth}
\centering
\includegraphics[width=\textwidth]{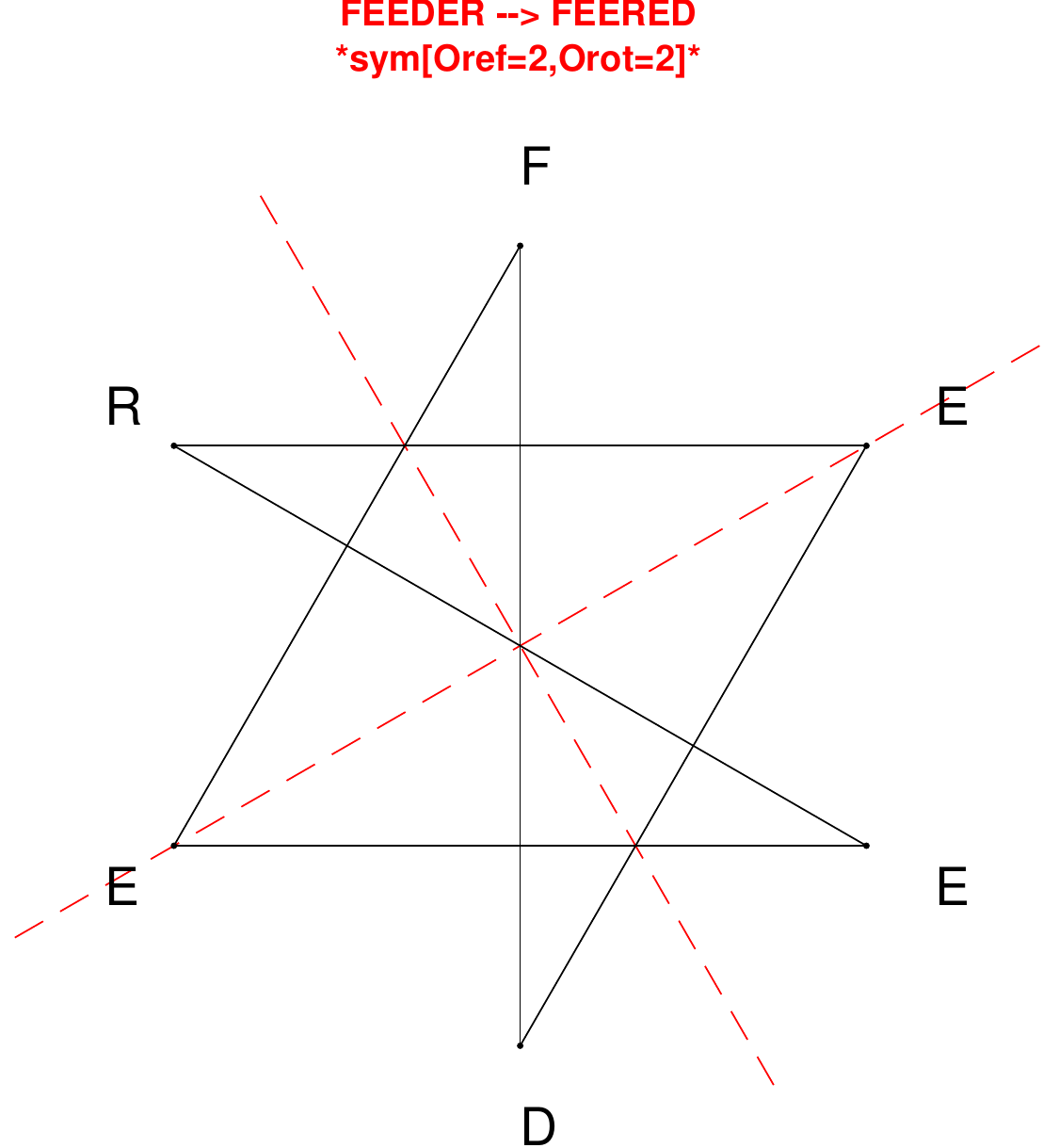}
\end{subfigure}
\end{figure}

\begin{figure}[H]
\centering
\begin{subfigure}[T]{0.19\textwidth}
\centering
\includegraphics[width=\textwidth]{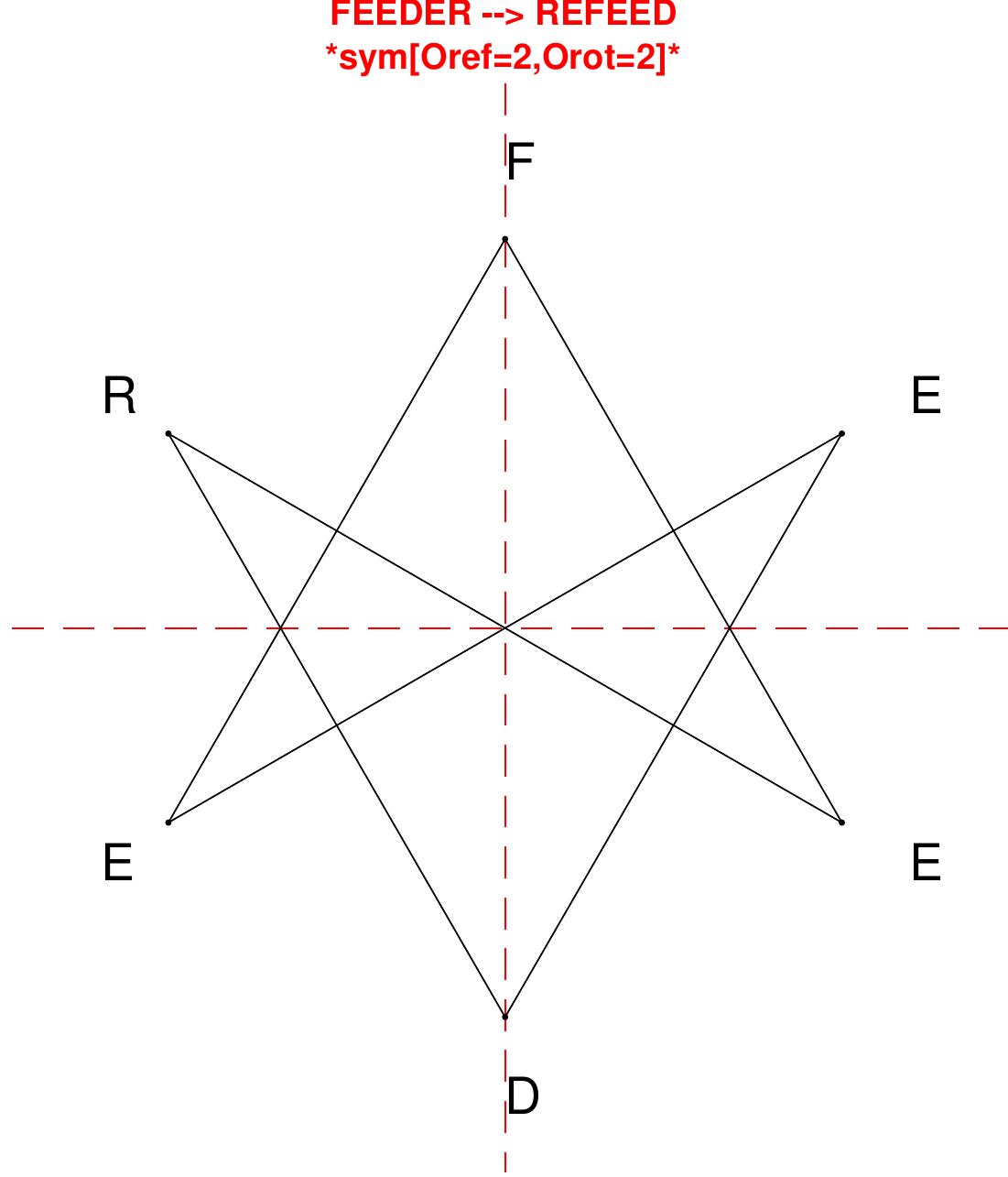}
\end{subfigure}
\hfill
\begin{subfigure}[T]{0.19\textwidth}
\centering
\includegraphics[width=\textwidth]{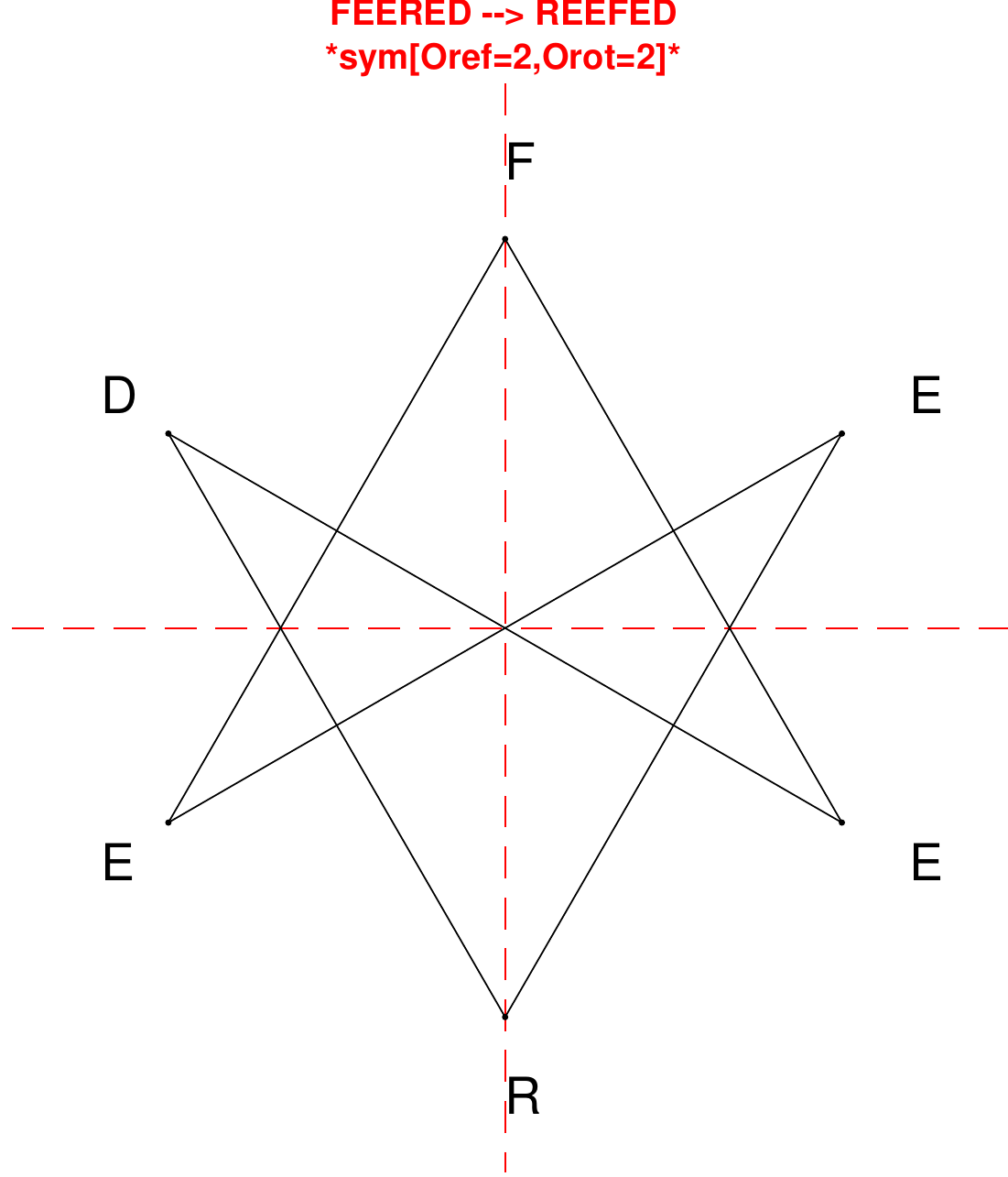}
\end{subfigure}
\hfill
\begin{subfigure}[T]{0.19\textwidth}
\centering
\includegraphics[width=\textwidth]{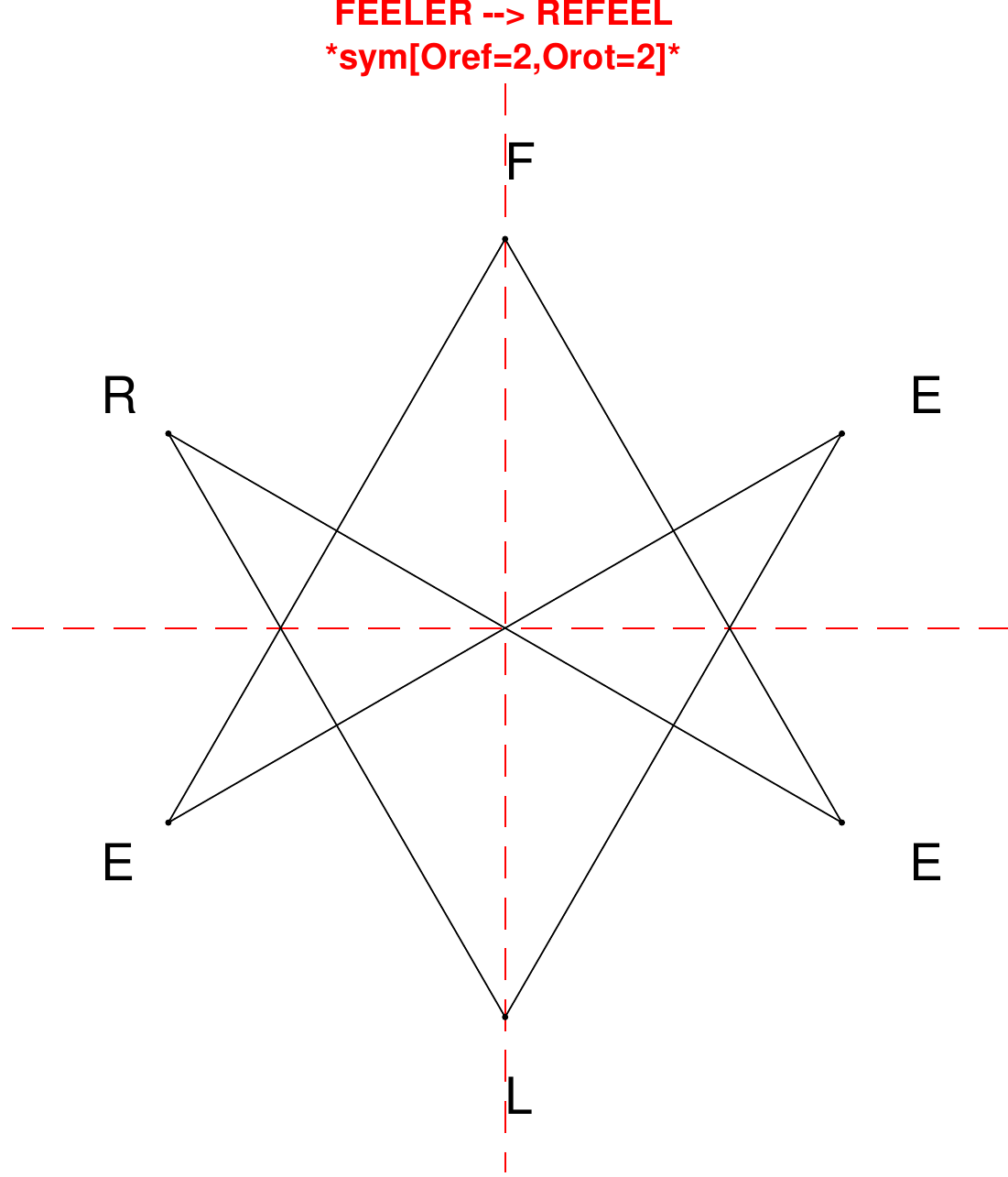}
\end{subfigure}
\hfill
\begin{subfigure}[T]{0.19\textwidth}
\centering
\includegraphics[width=\textwidth]{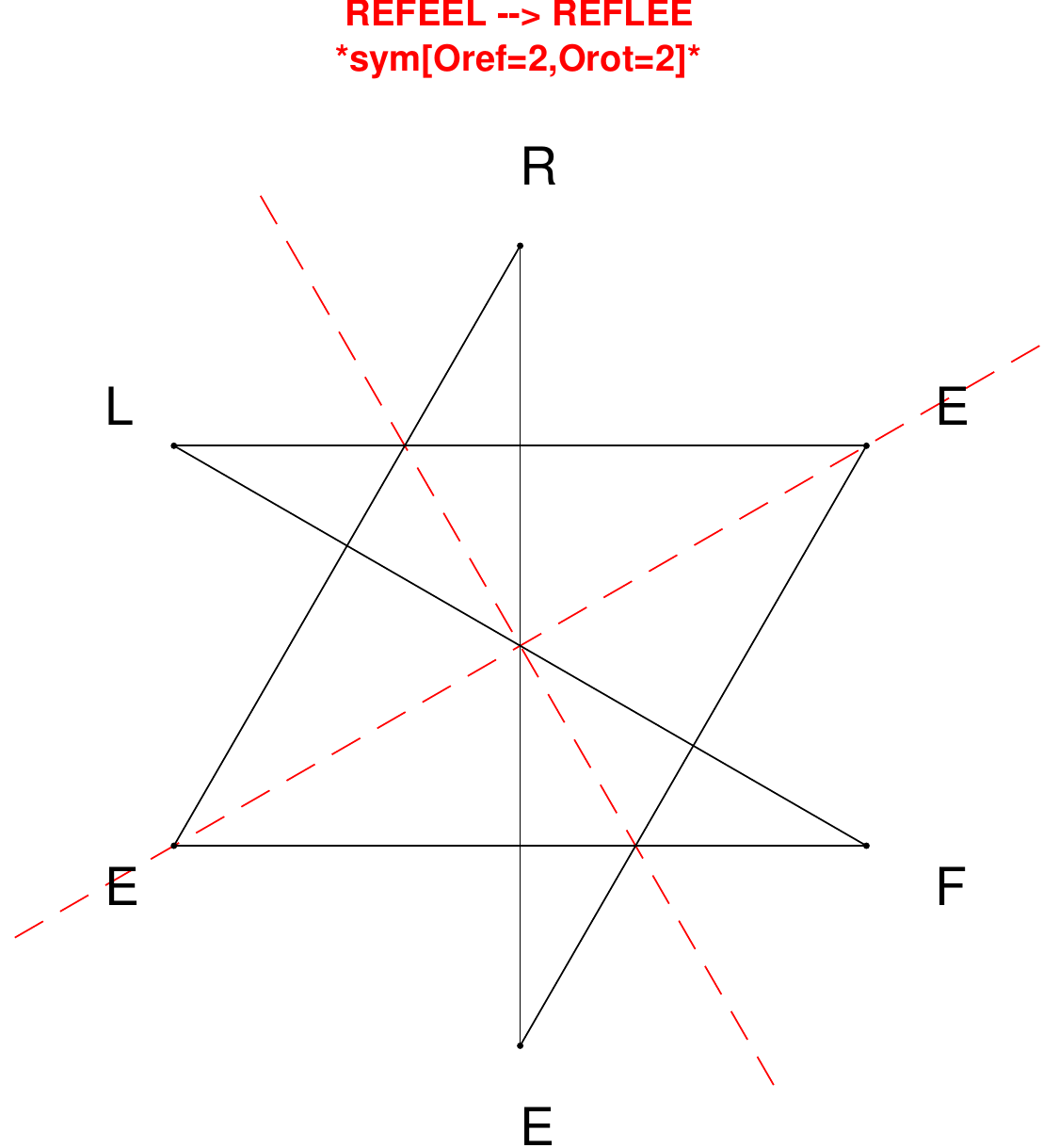}
\end{subfigure}
\hfill
\begin{subfigure}[T]{0.19\textwidth}
\centering
\includegraphics[width=\textwidth]{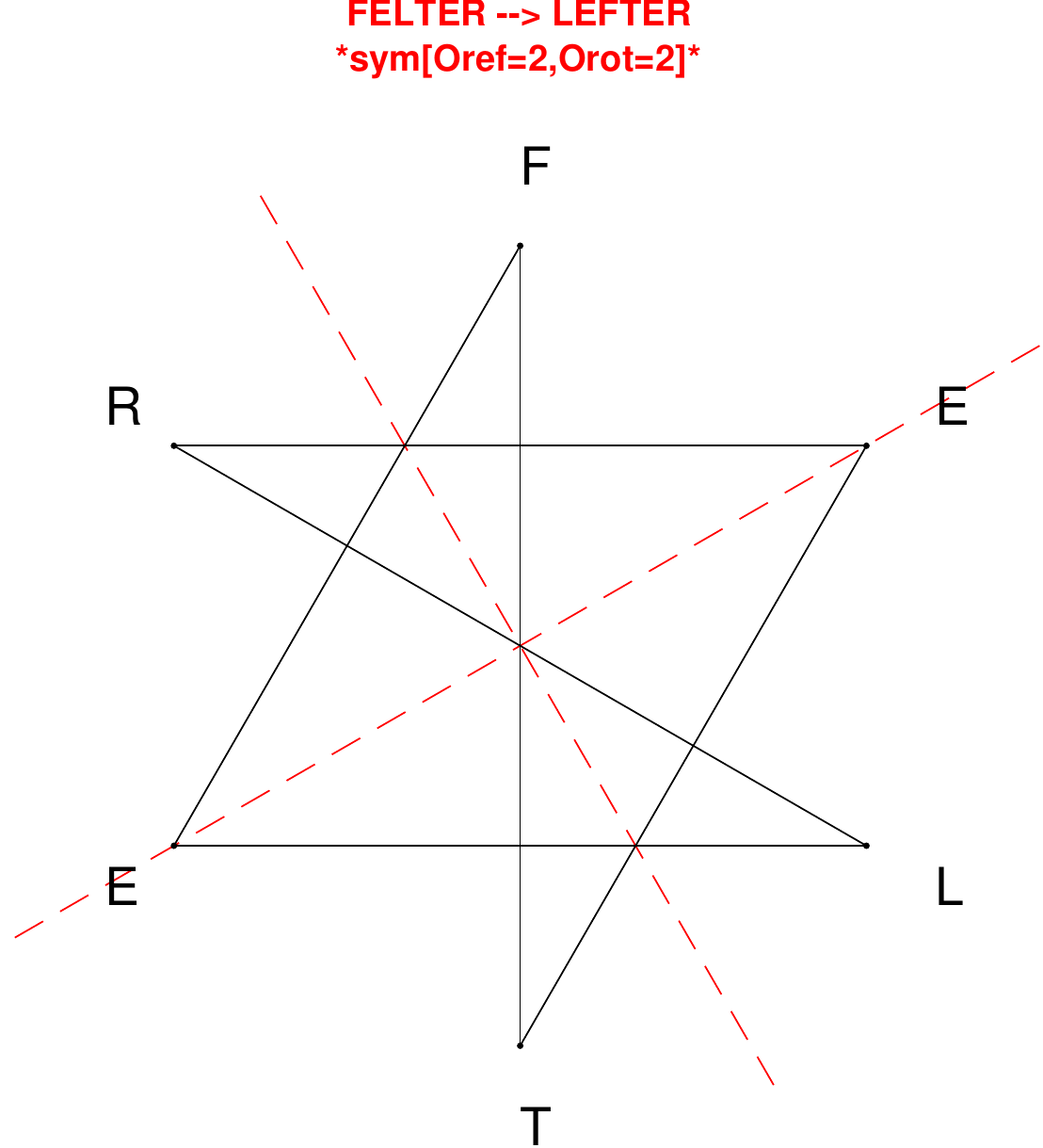}
\end{subfigure}
\end{figure}

\begin{figure}[H]
\centering
\begin{subfigure}[T]{0.19\textwidth}
\centering
\includegraphics[width=\textwidth]{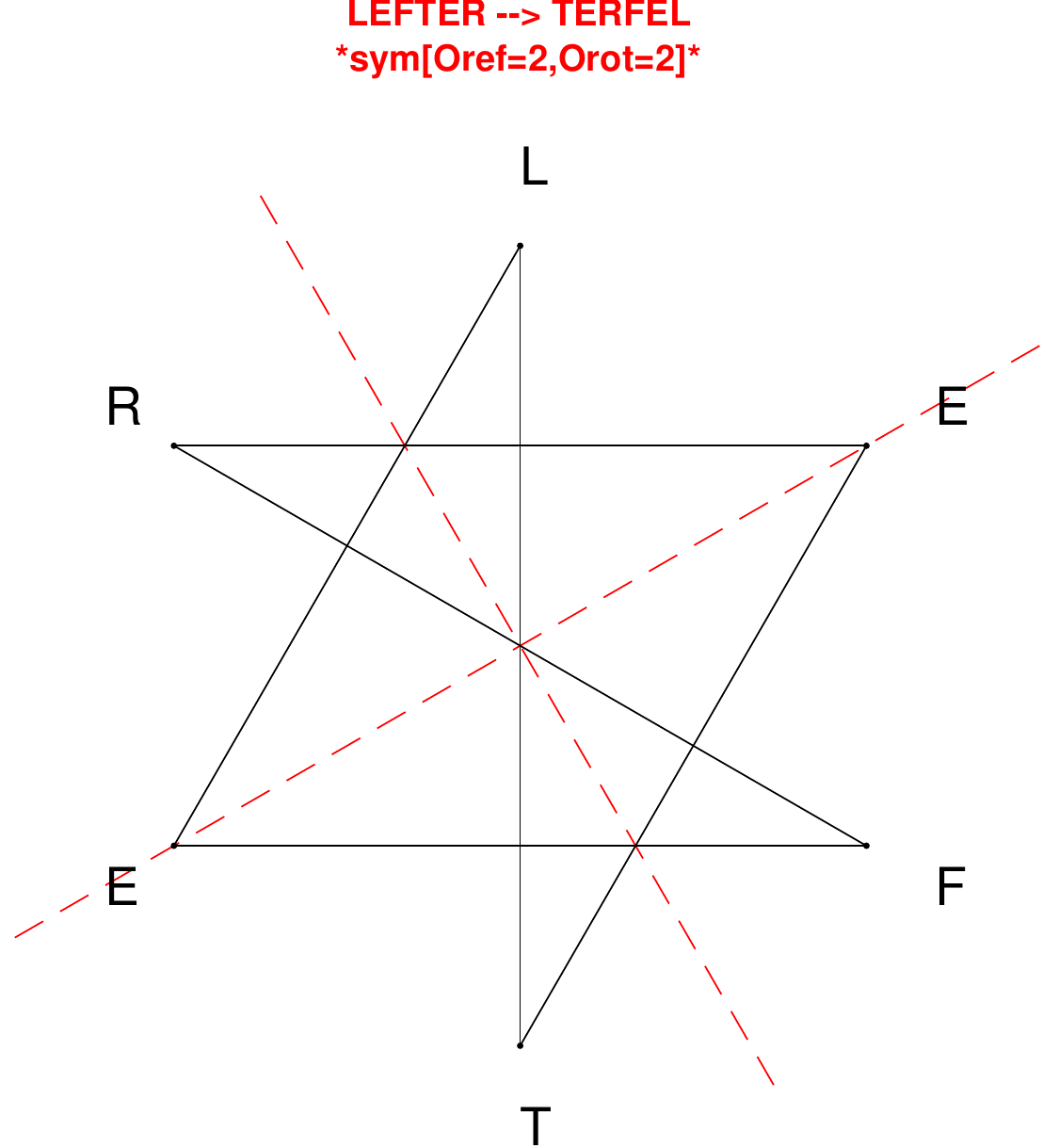}
\end{subfigure}
\hfill
\begin{subfigure}[T]{0.19\textwidth}
\centering
\includegraphics[width=\textwidth]{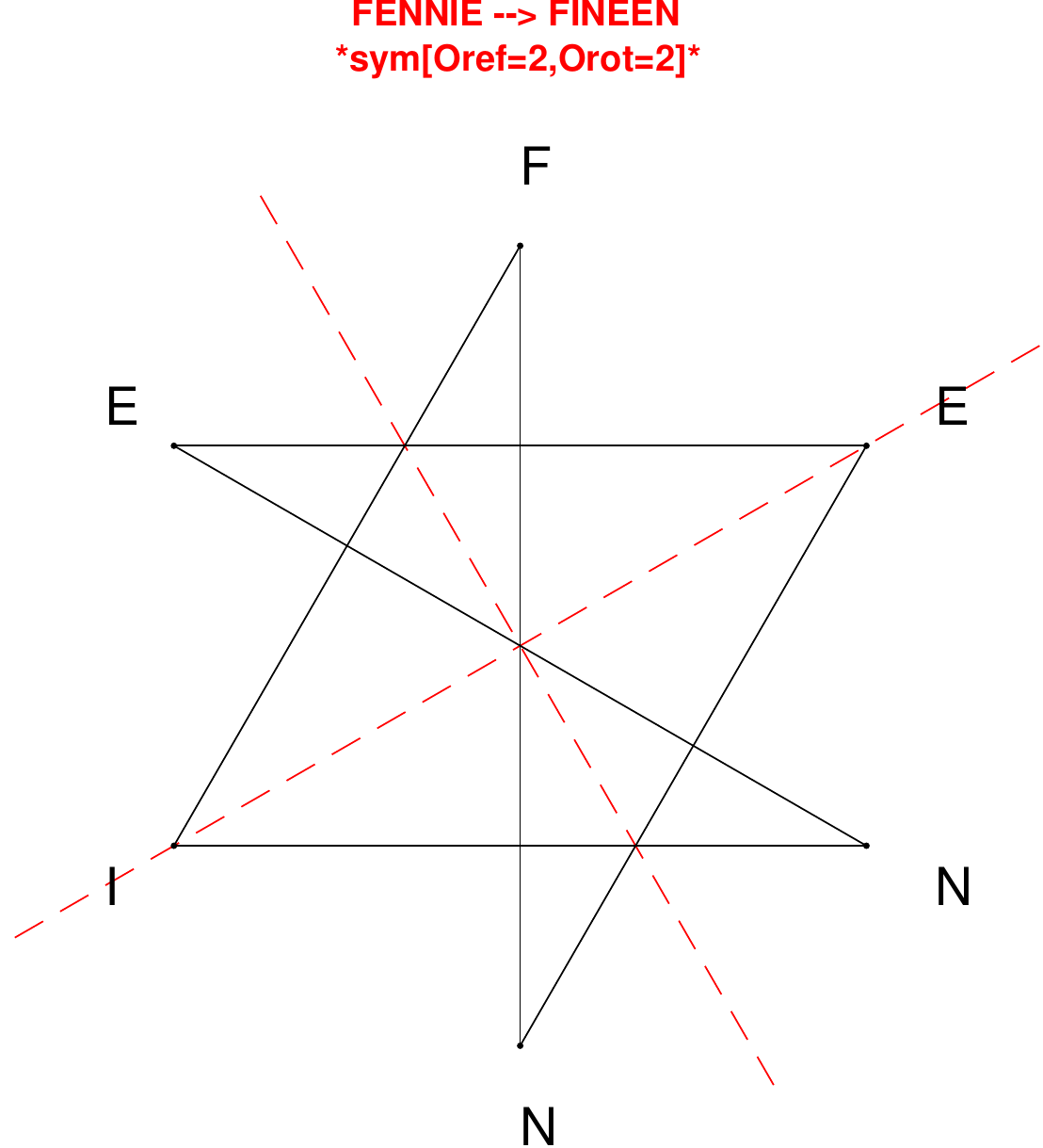}
\end{subfigure}
\hfill
\begin{subfigure}[T]{0.19\textwidth}
\centering
\includegraphics[width=\textwidth]{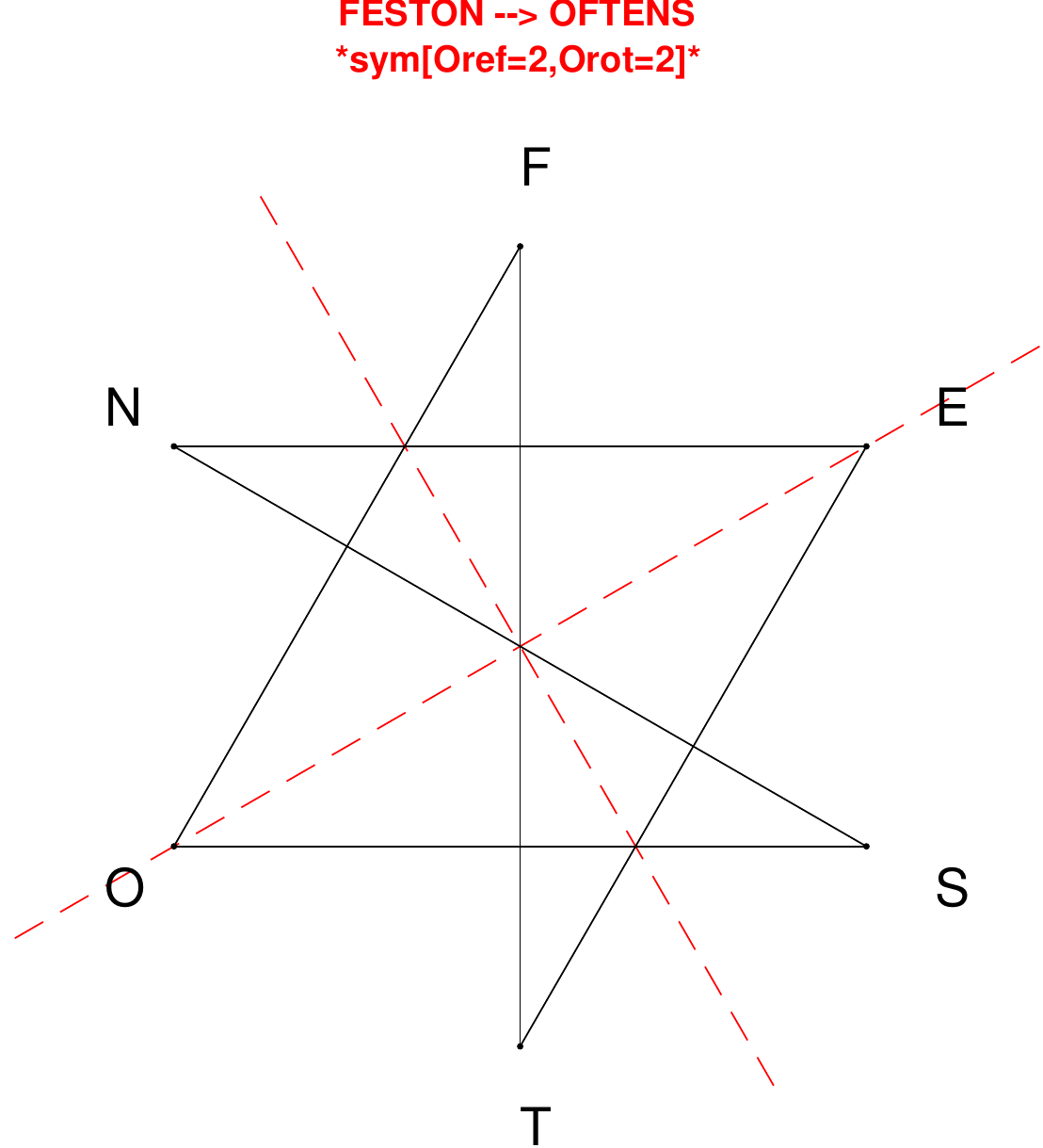}
\end{subfigure}
\hfill
\begin{subfigure}[T]{0.19\textwidth}
\centering
\includegraphics[width=\textwidth]{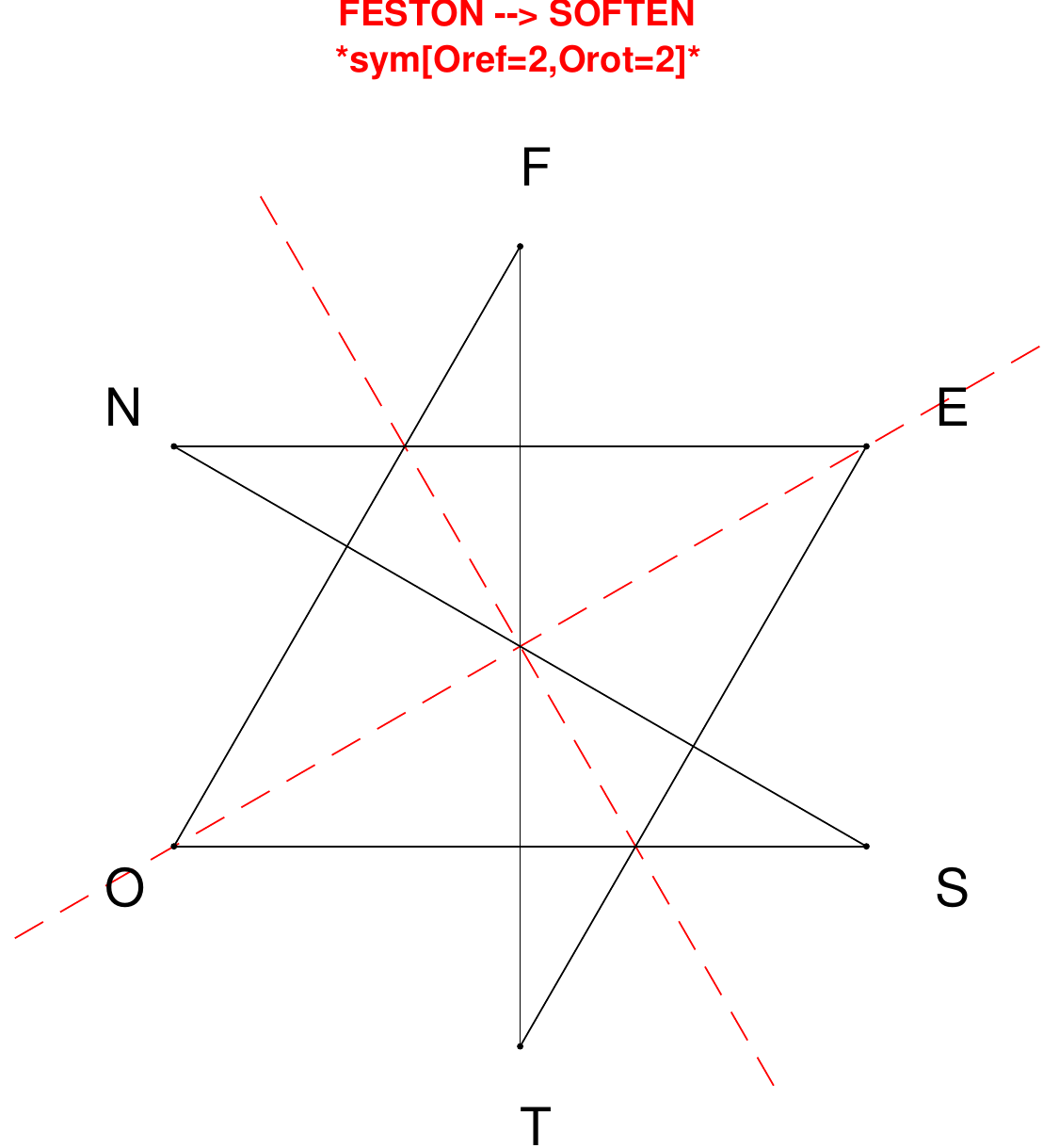}
\end{subfigure}
\hfill
\begin{subfigure}[T]{0.19\textwidth}
\centering
\includegraphics[width=\textwidth]{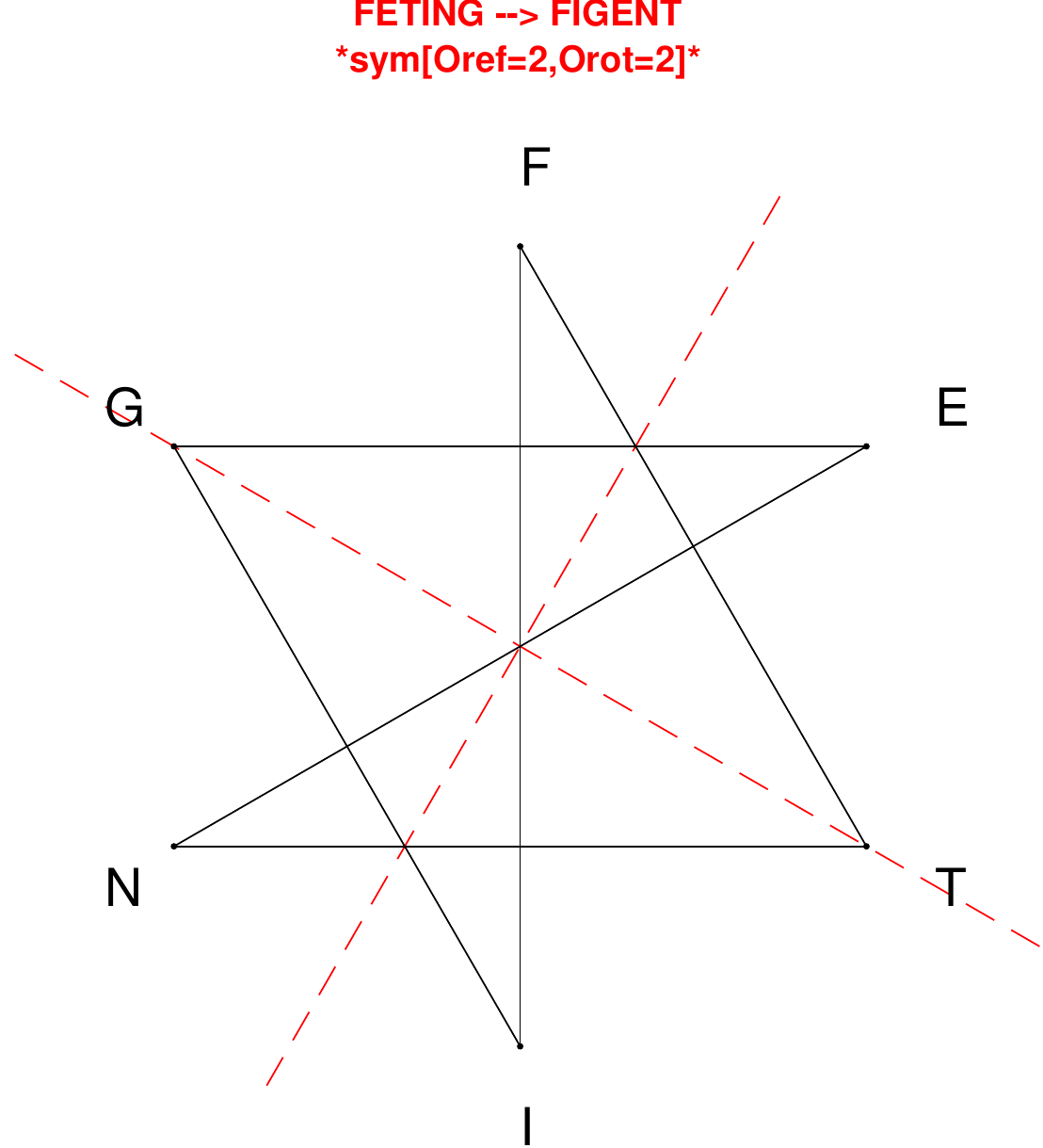}
\end{subfigure}
\end{figure}

\begin{figure}[H]
\centering
\begin{subfigure}[T]{0.19\textwidth}
\centering
\includegraphics[width=\textwidth]{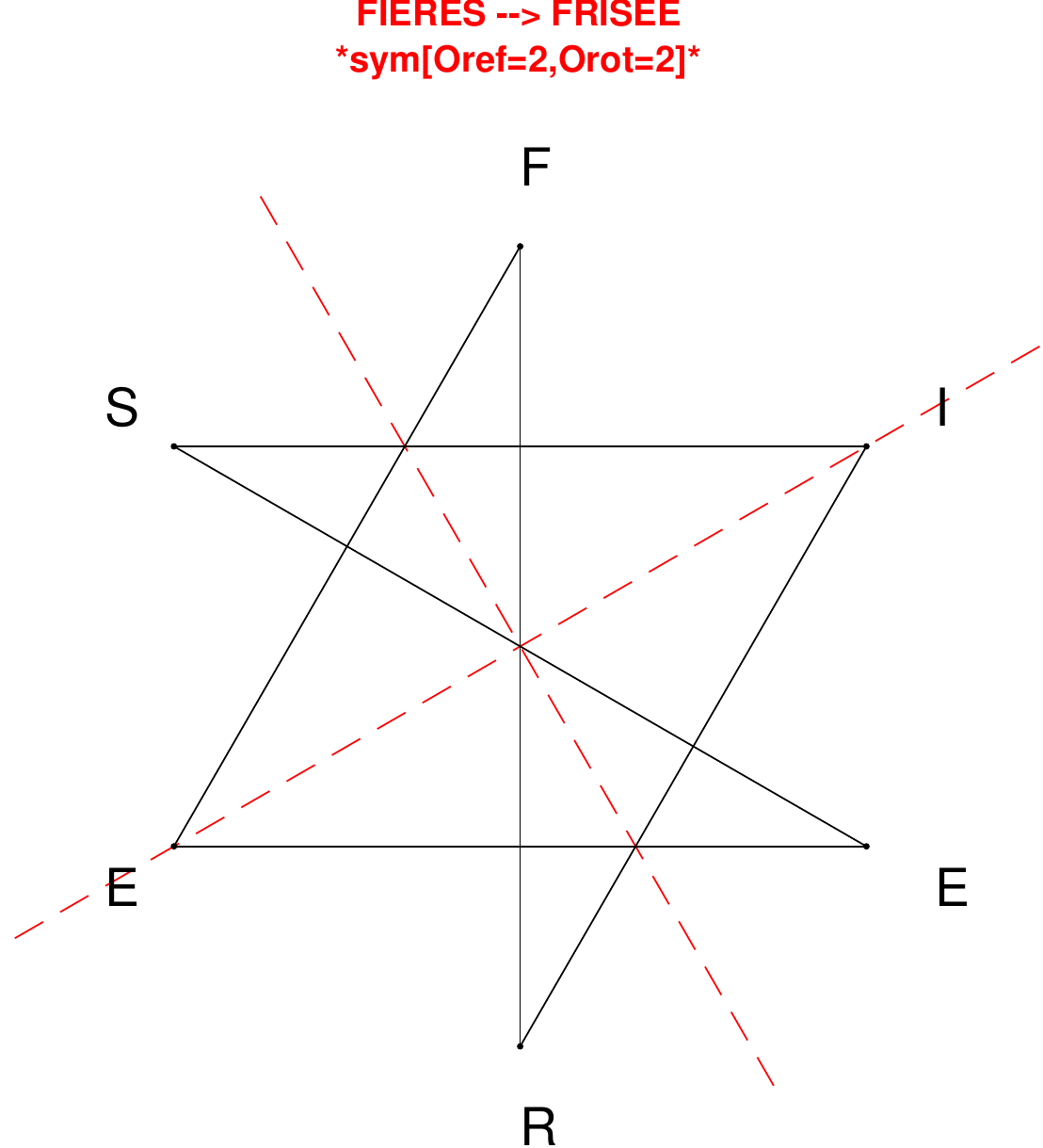}
\end{subfigure}
\hfill
\begin{subfigure}[T]{0.19\textwidth}
\centering
\includegraphics[width=\textwidth]{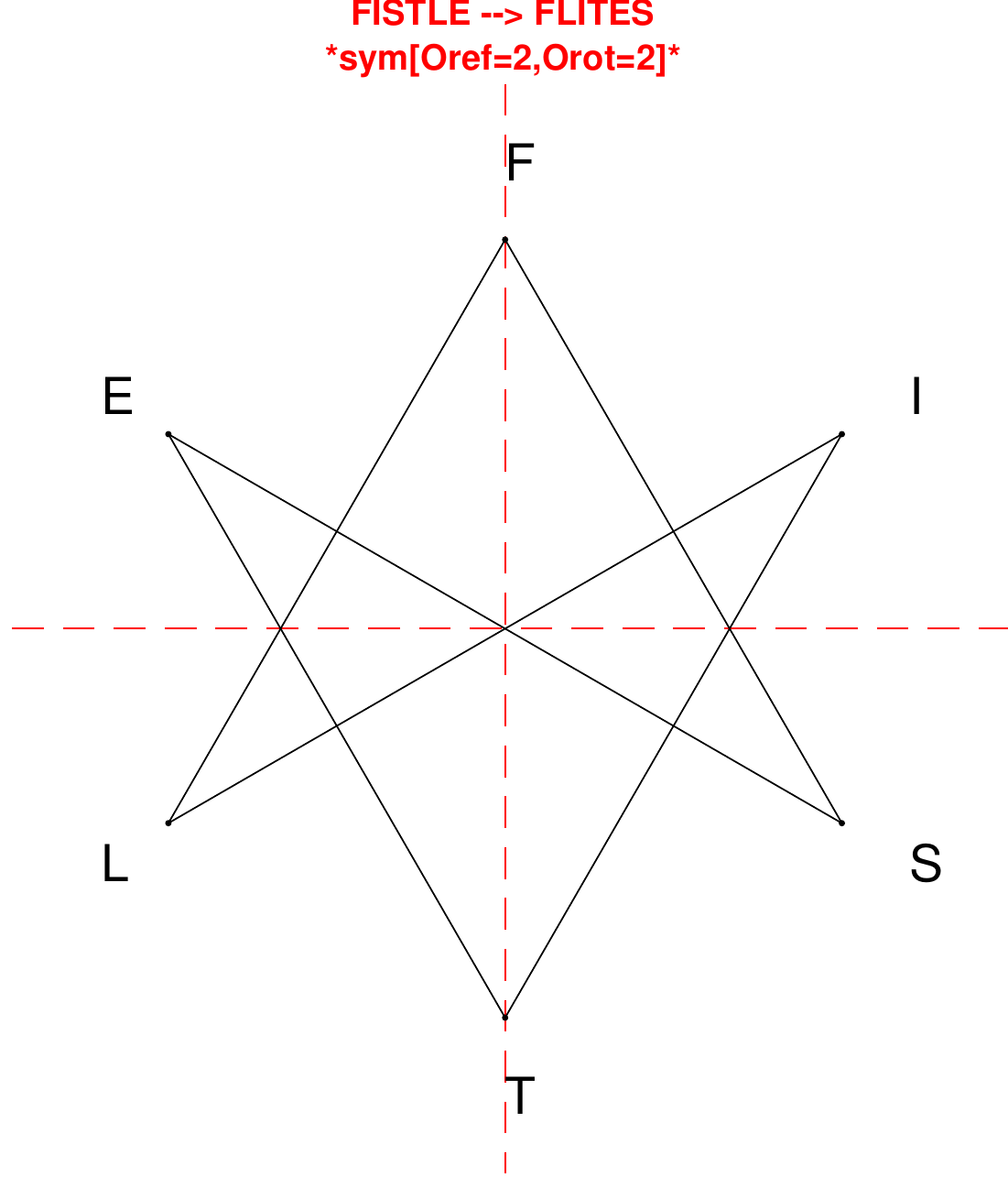}
\end{subfigure}
\hfill
\begin{subfigure}[T]{0.19\textwidth}
\centering
\includegraphics[width=\textwidth]{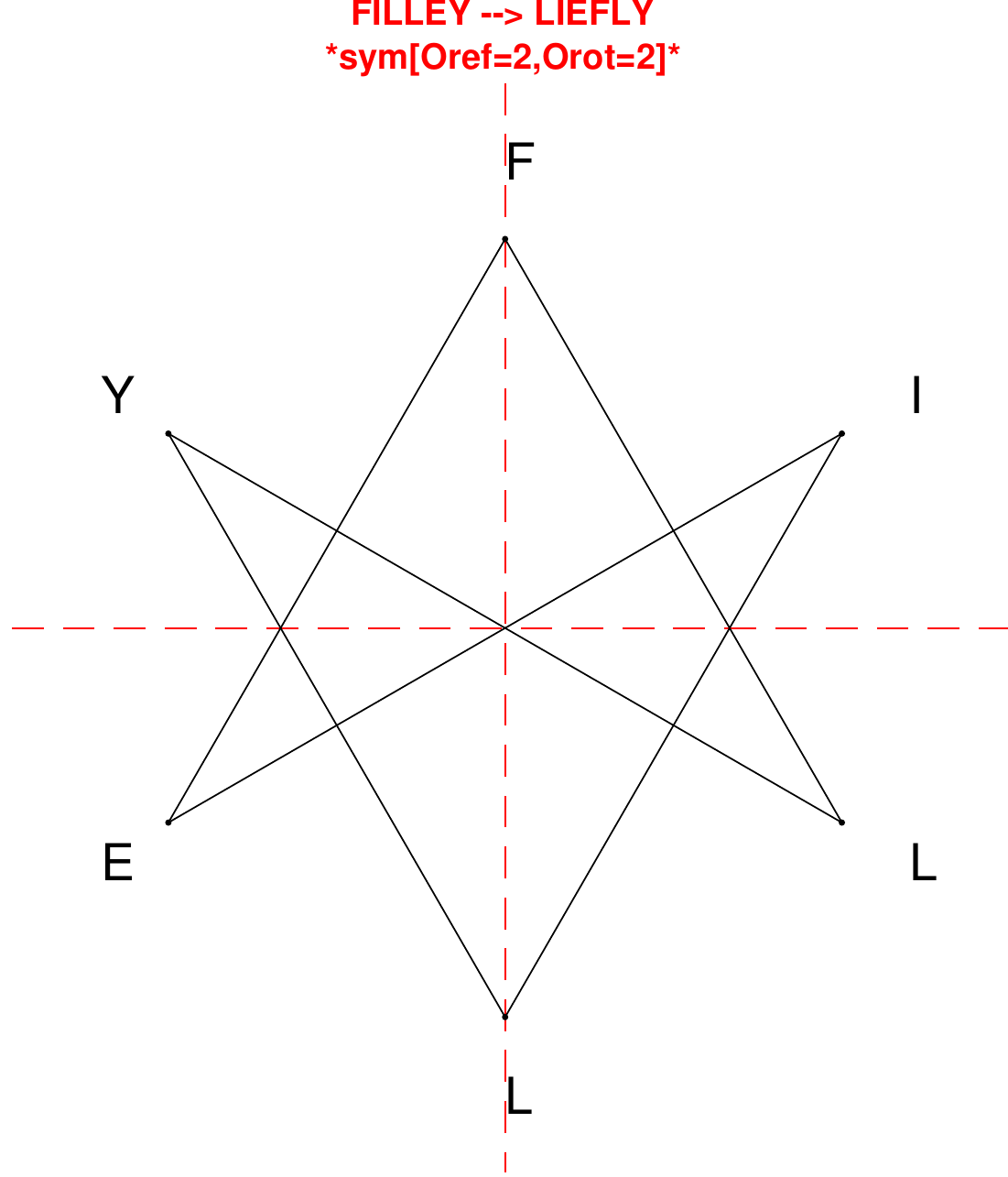}
\end{subfigure}
\hfill
\begin{subfigure}[T]{0.19\textwidth}
\centering
\includegraphics[width=\textwidth]{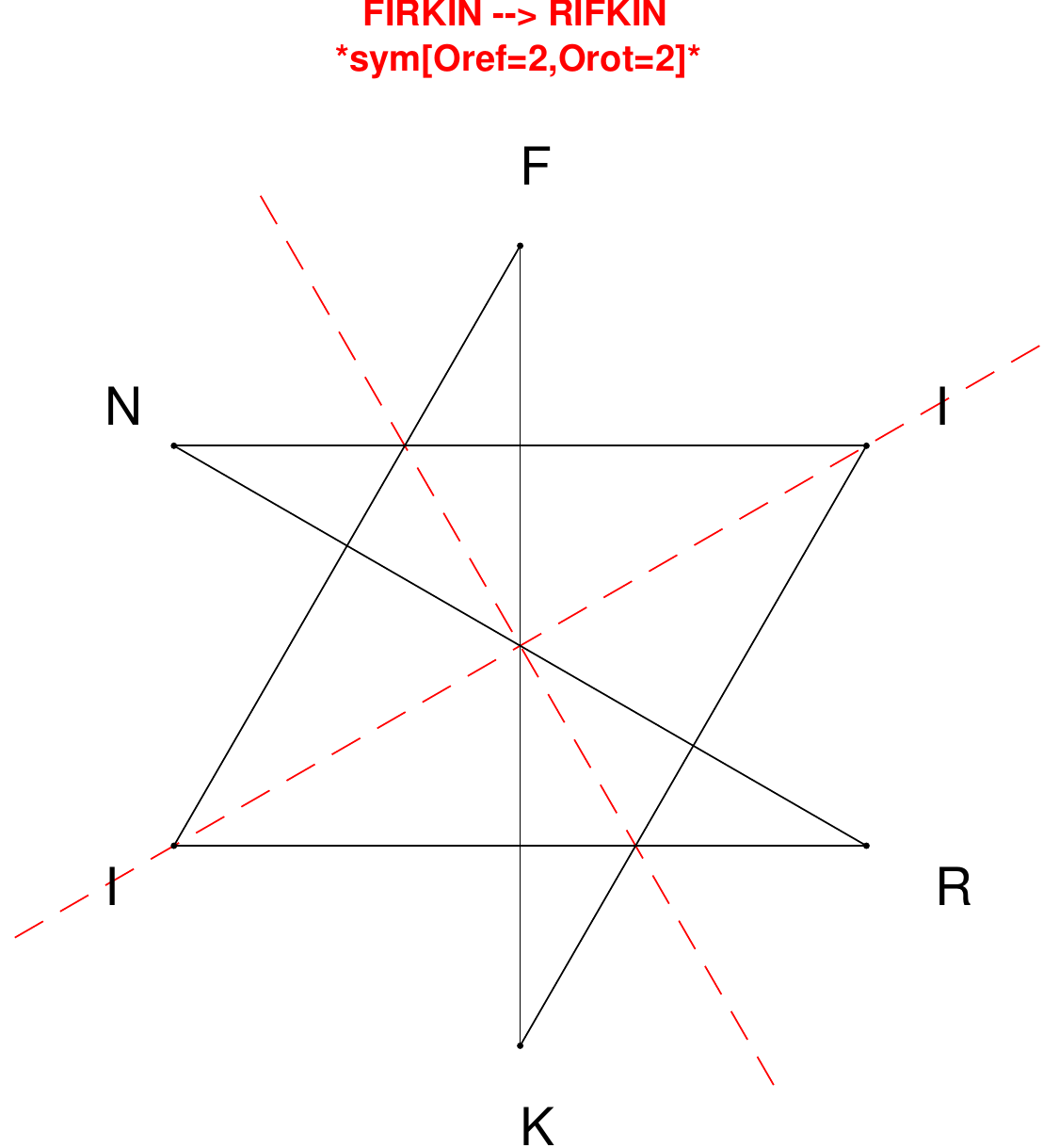}
\end{subfigure}
\hfill
\begin{subfigure}[T]{0.19\textwidth}
\centering
\includegraphics[width=\textwidth]{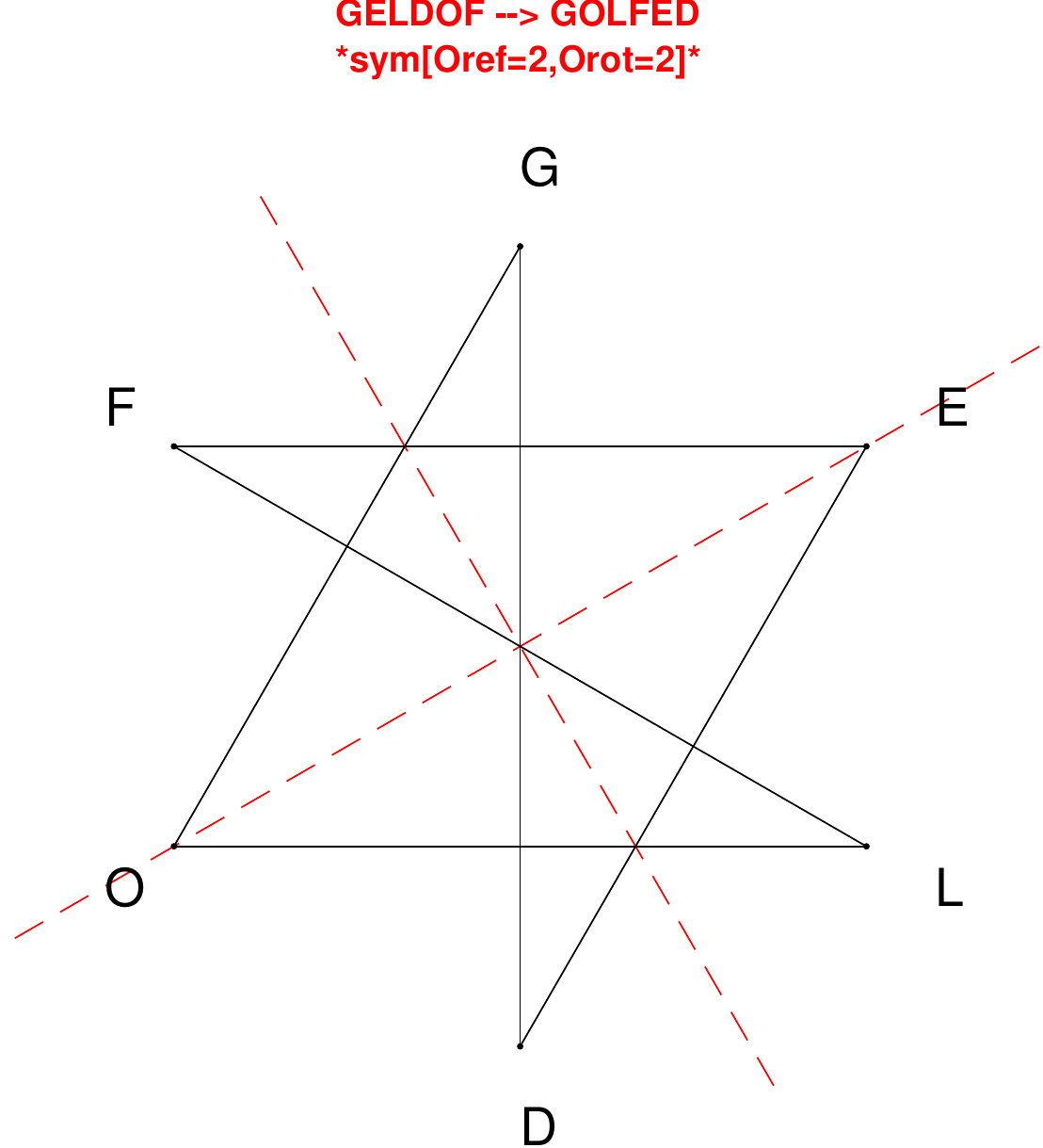}
\end{subfigure}
\end{figure}

\begin{figure}[H]
\centering
\begin{subfigure}[T]{0.19\textwidth}
\centering
\includegraphics[width=\textwidth]{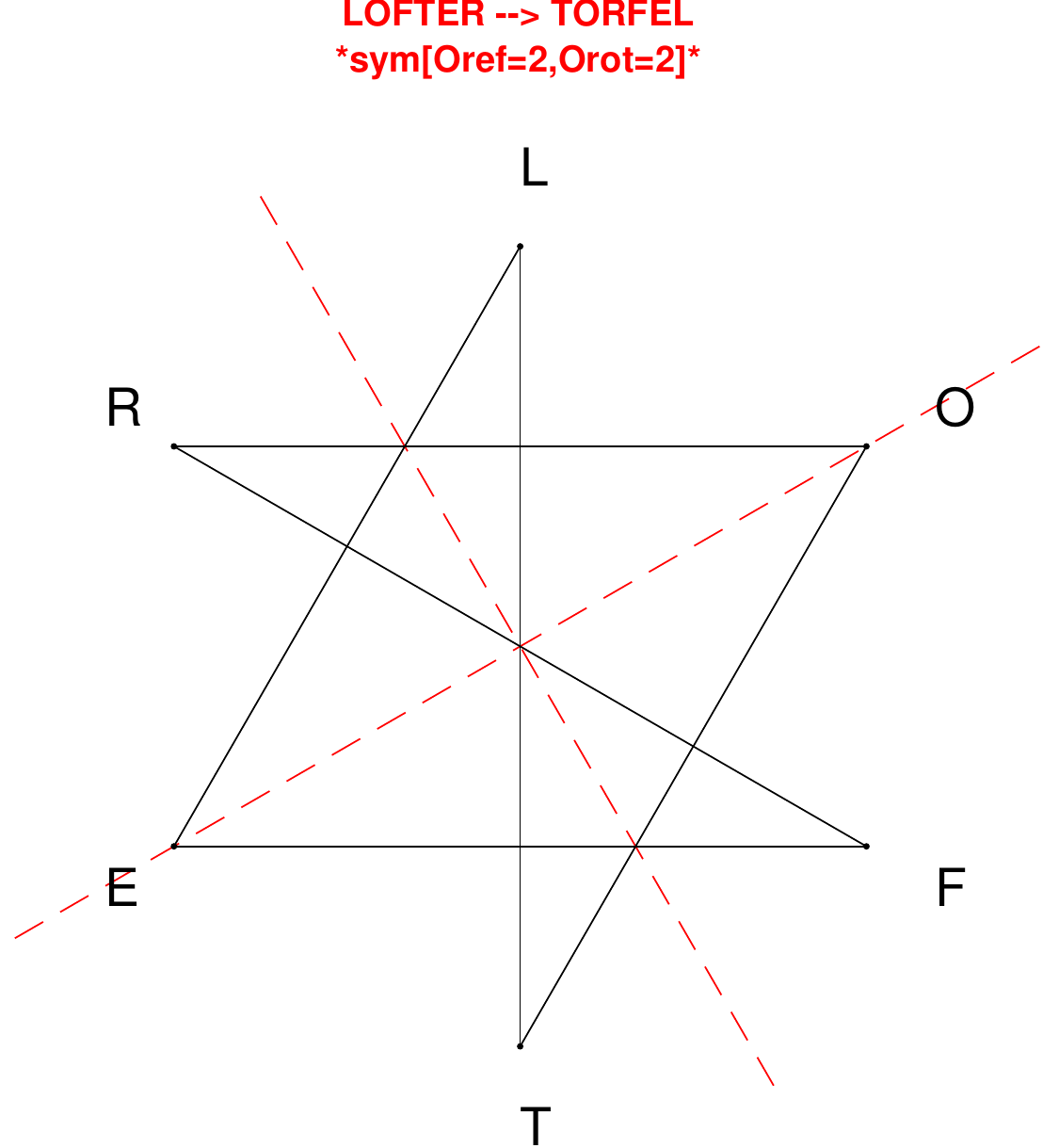}
\end{subfigure}
\hfill
\begin{subfigure}[T]{0.19\textwidth}
\centering
\includegraphics[width=\textwidth]{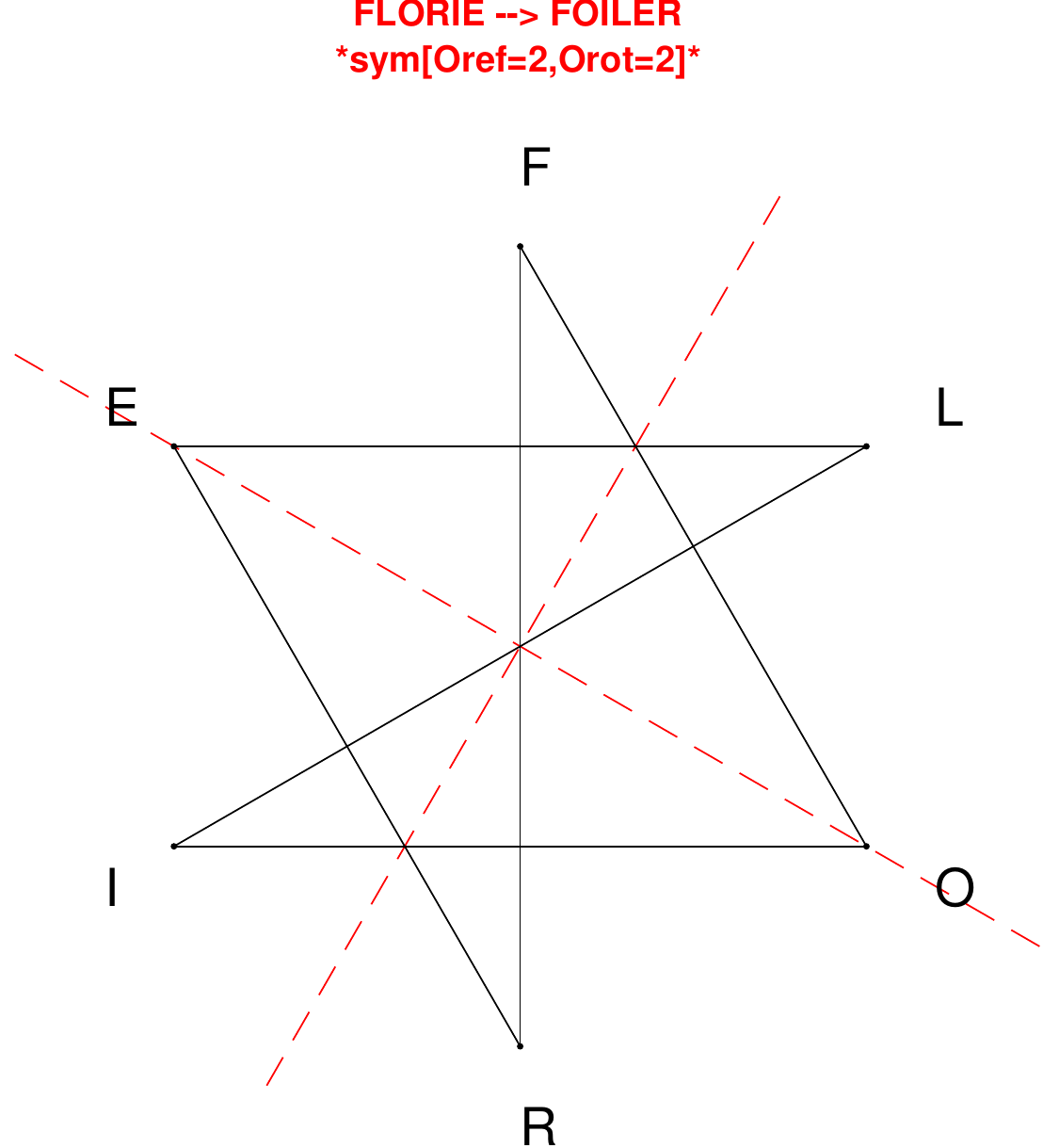}
\end{subfigure}
\hfill
\begin{subfigure}[T]{0.19\textwidth}
\centering
\includegraphics[width=\textwidth]{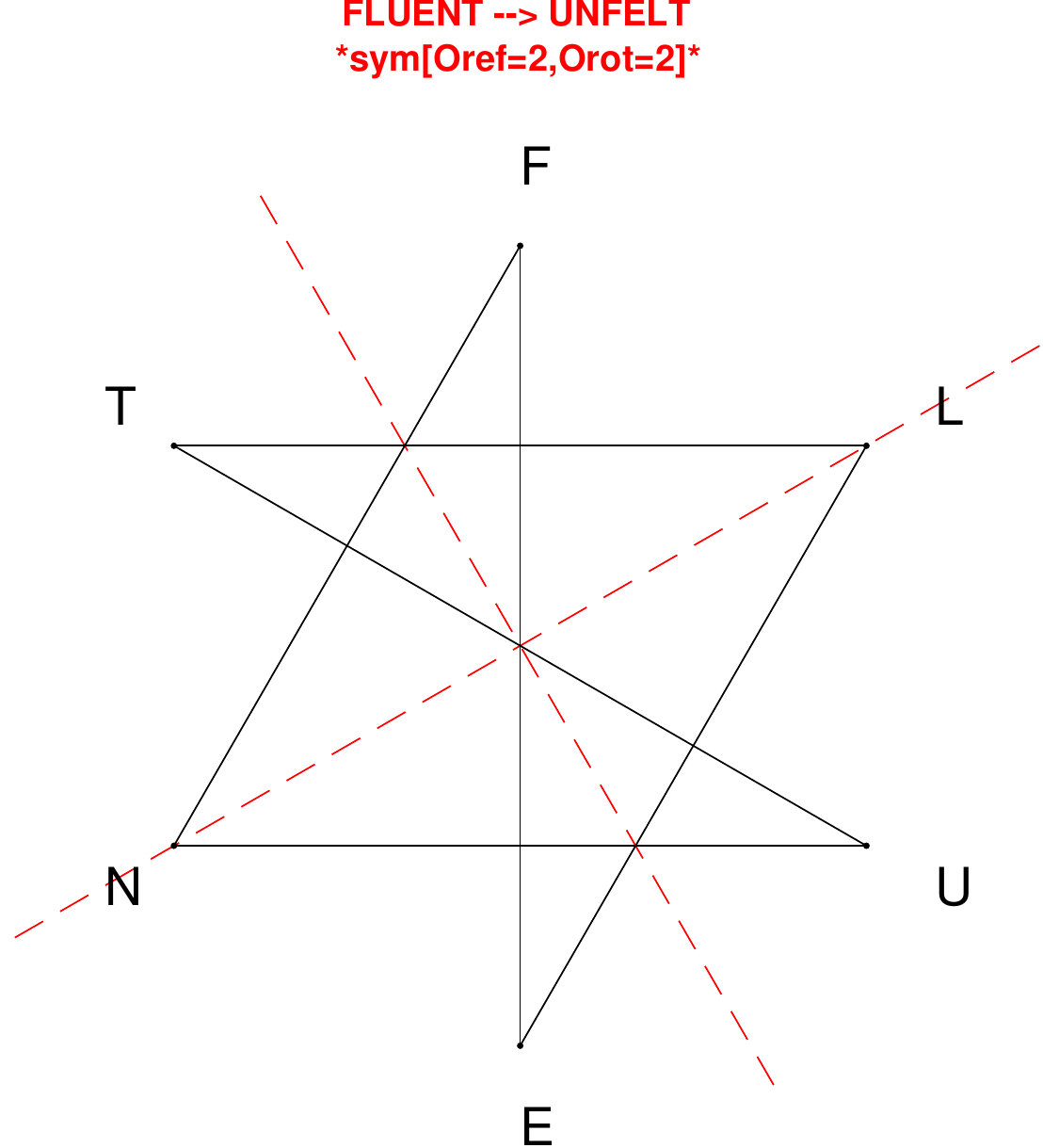}
\end{subfigure}
\hfill
\begin{subfigure}[T]{0.19\textwidth}
\centering
\includegraphics[width=\textwidth]{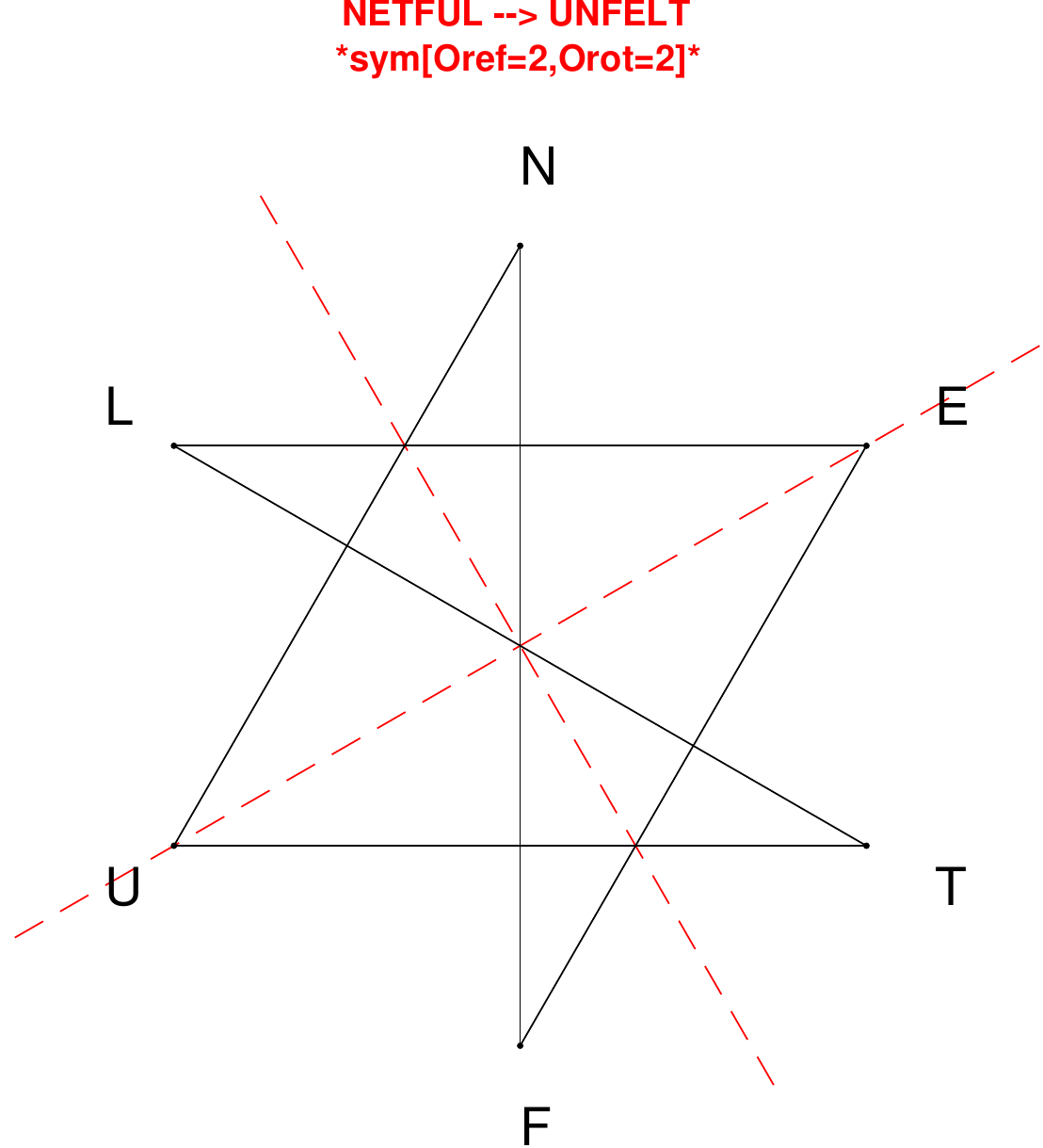}
\end{subfigure}
\hfill
\begin{subfigure}[T]{0.19\textwidth}
\centering
\includegraphics[width=\textwidth]{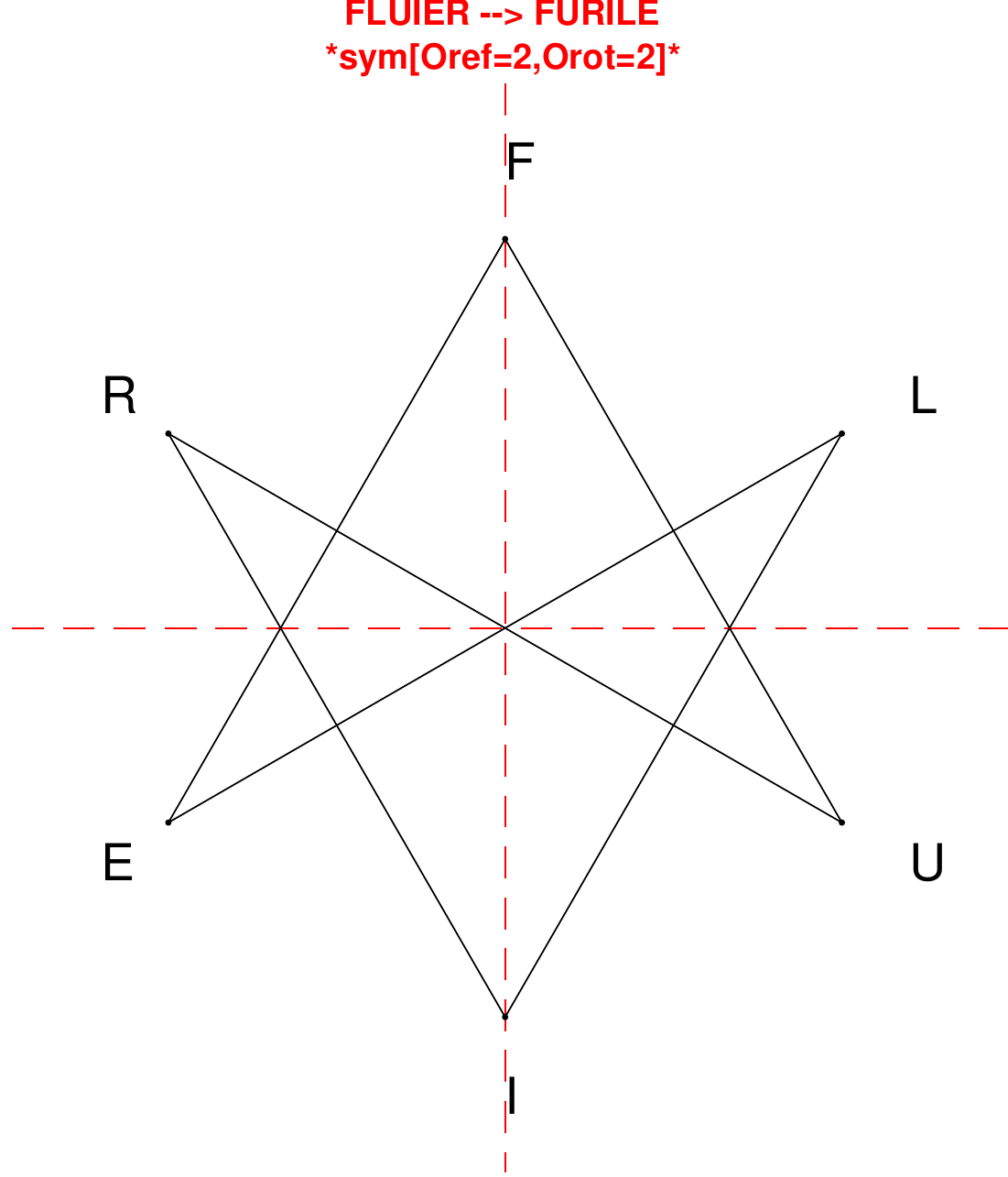}
\end{subfigure}
\end{figure}

\begin{figure}[H]
\centering
\begin{subfigure}[T]{0.19\textwidth}
\centering
\includegraphics[width=\textwidth]{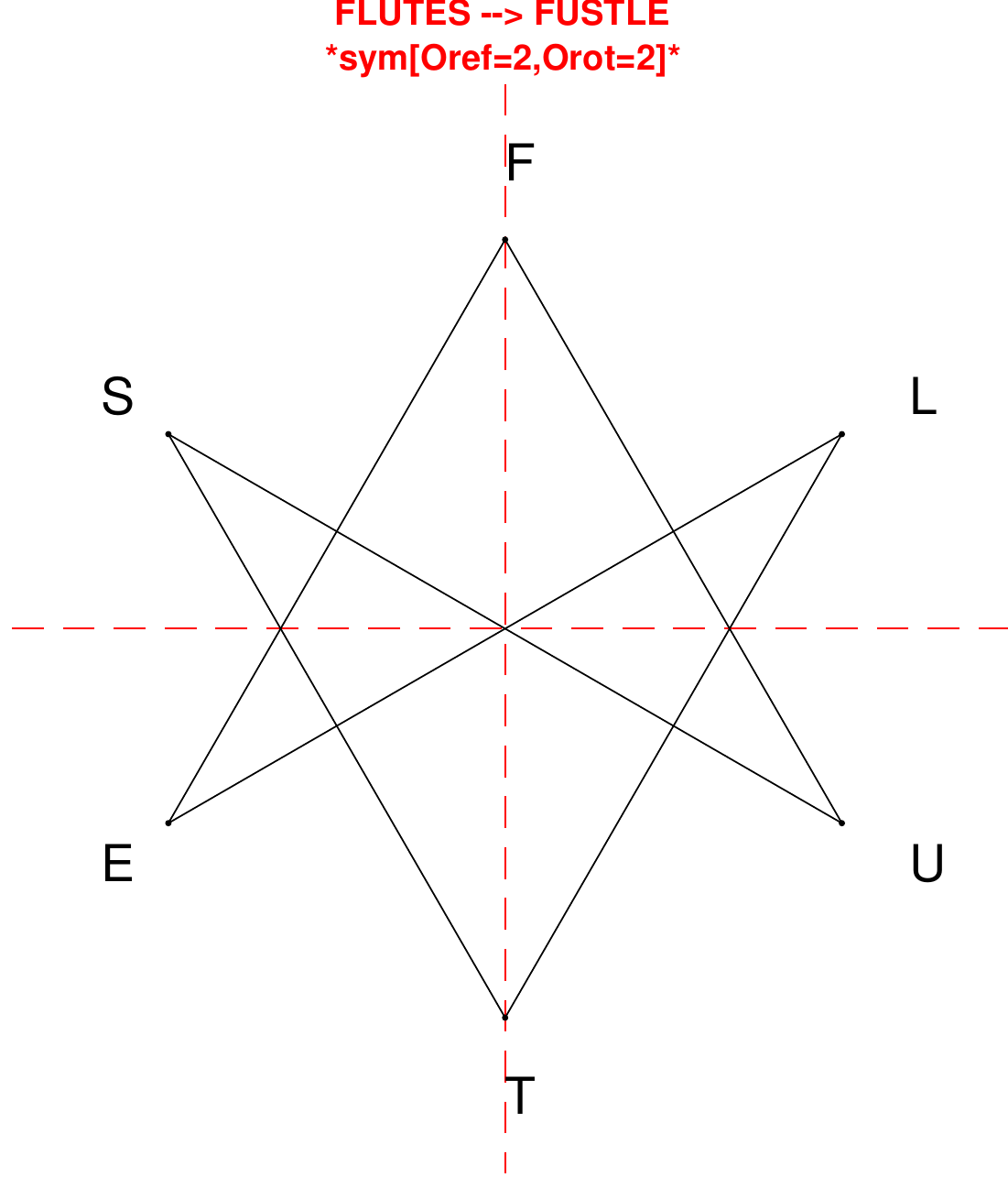}
\end{subfigure}
\hfill
\begin{subfigure}[T]{0.19\textwidth}
\centering
\includegraphics[width=\textwidth]{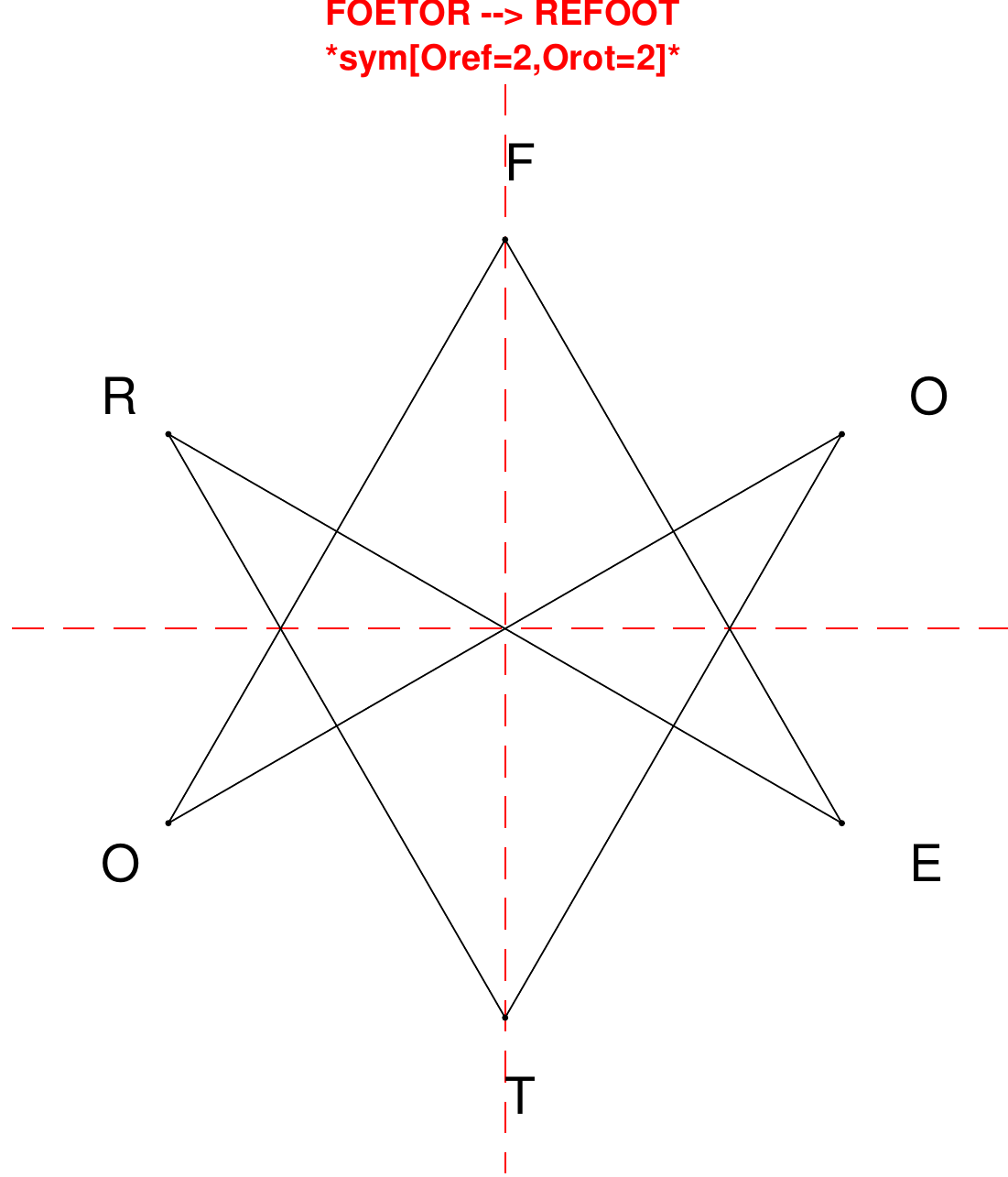}
\end{subfigure}
\hfill
\begin{subfigure}[T]{0.19\textwidth}
\centering
\includegraphics[width=\textwidth]{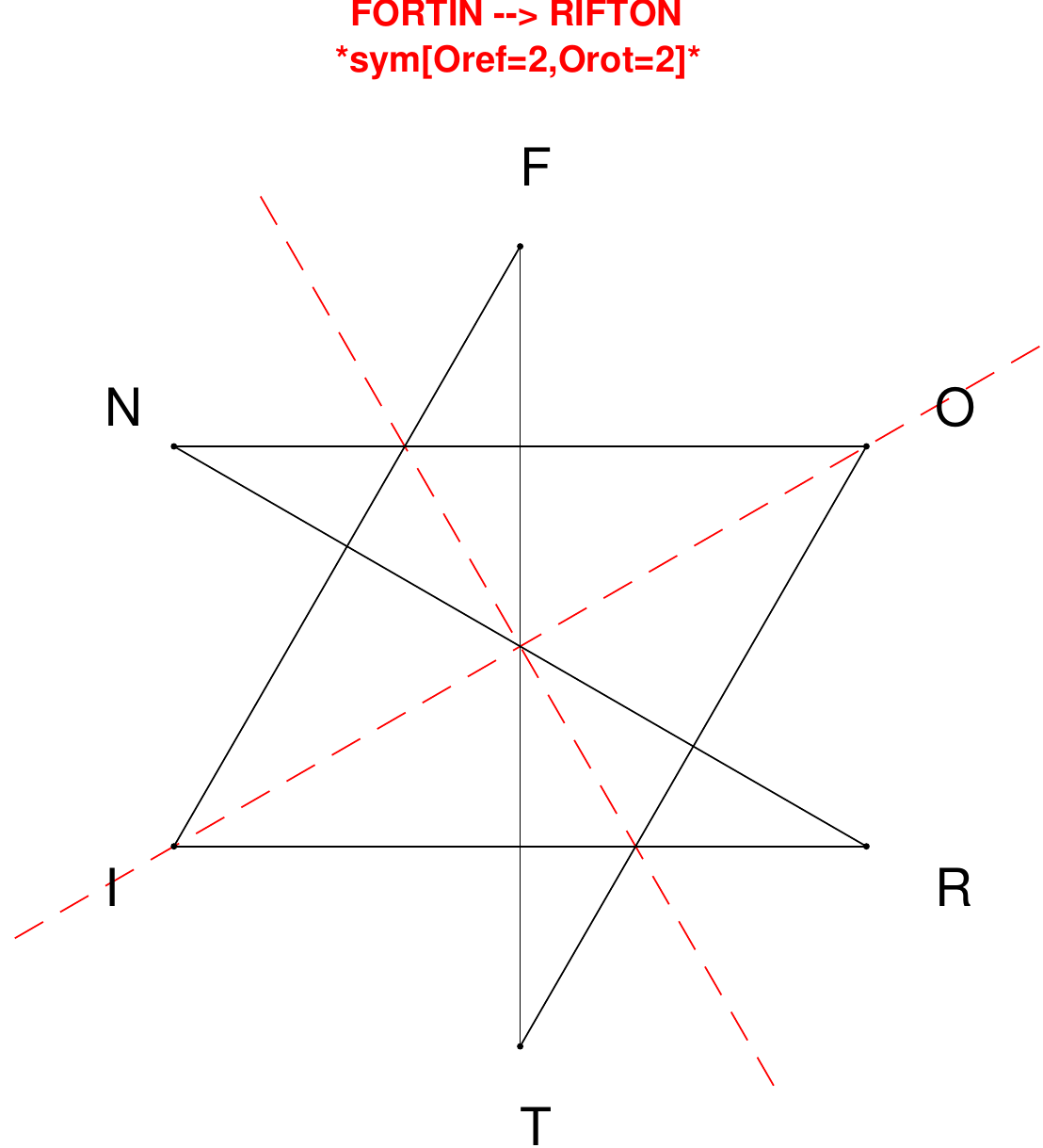}
\end{subfigure}
\hfill
\begin{subfigure}[T]{0.19\textwidth}
\centering
\includegraphics[width=\textwidth]{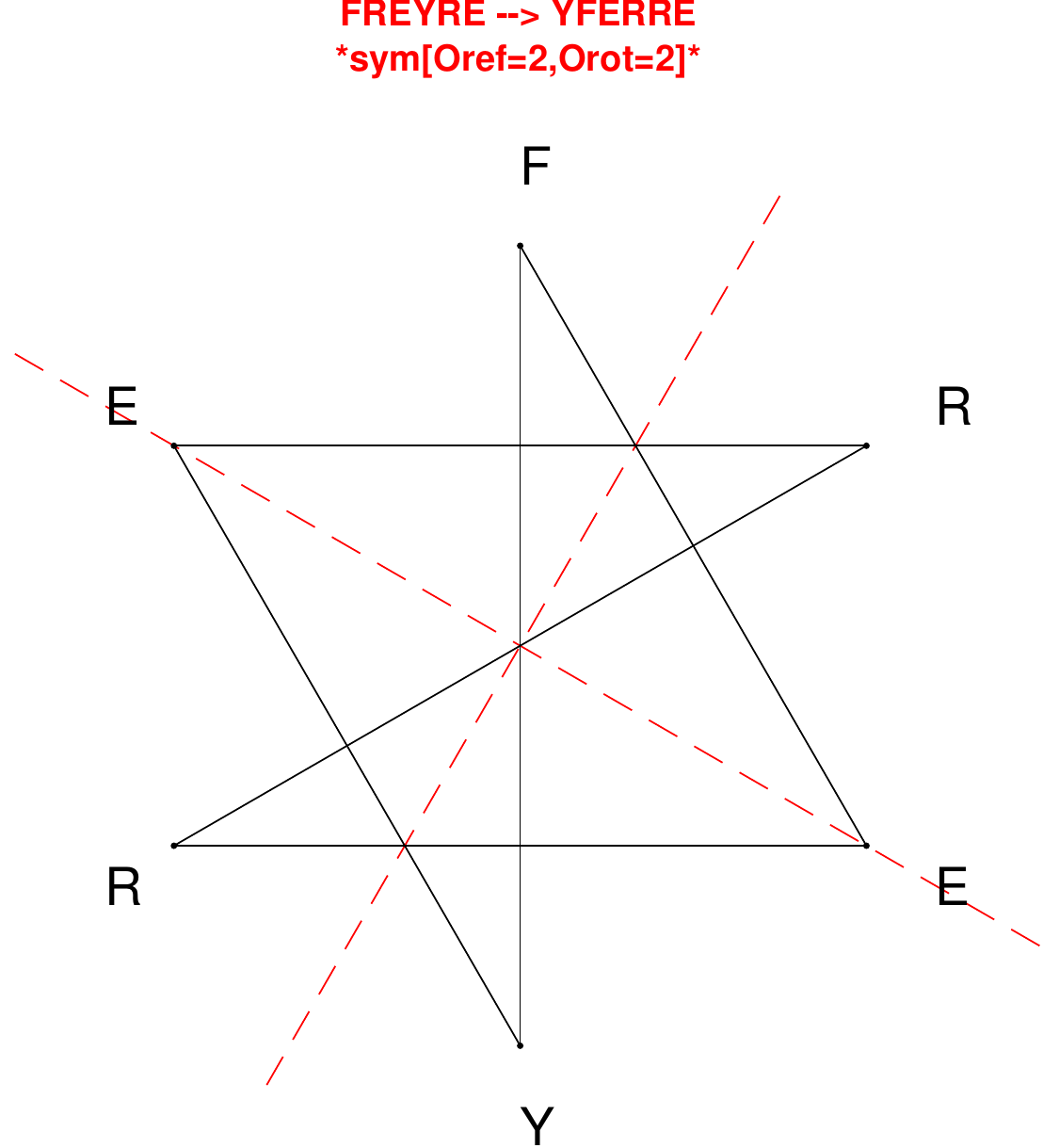}
\end{subfigure}
\hfill
\begin{subfigure}[T]{0.19\textwidth}
\centering
\includegraphics[width=\textwidth]{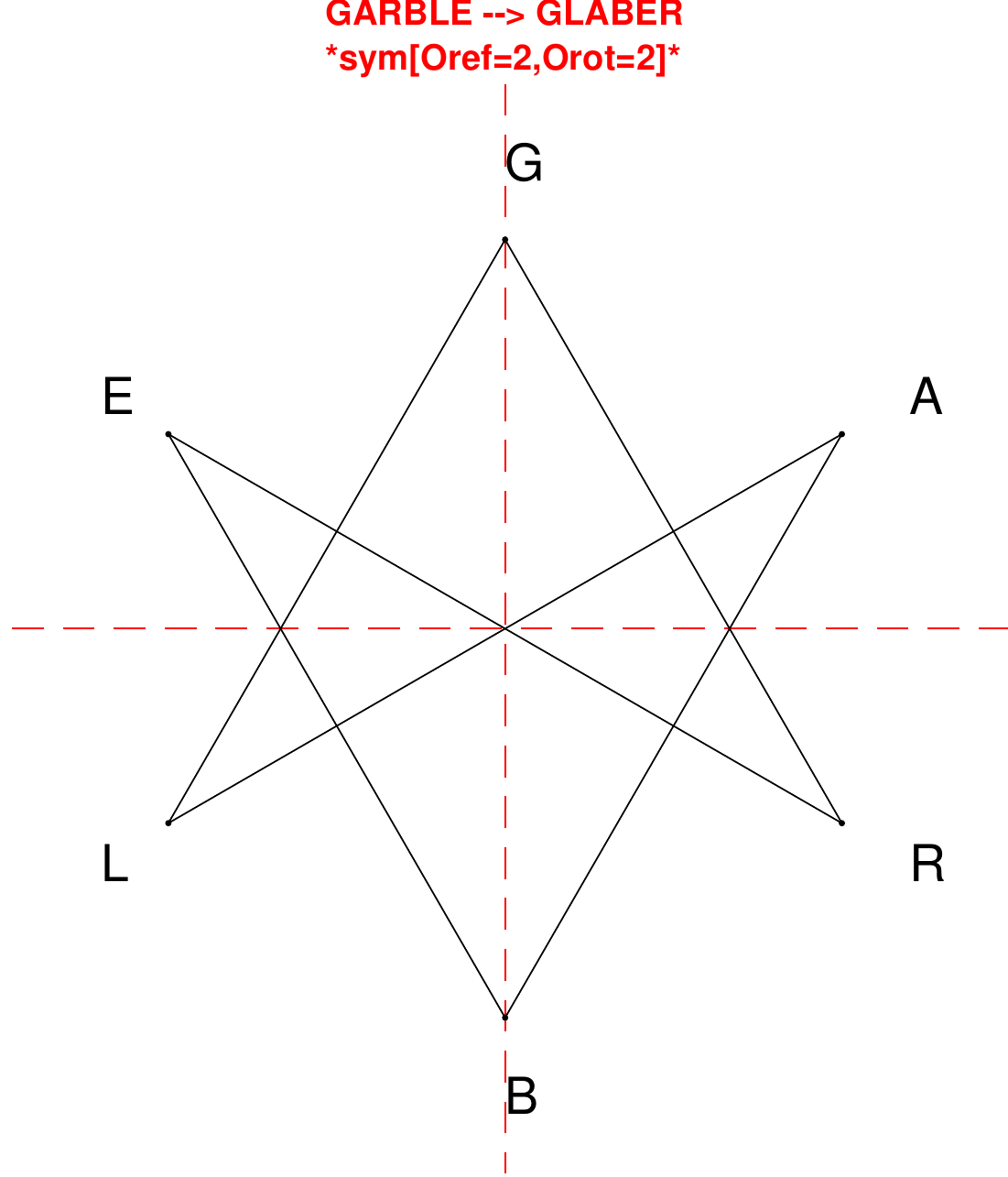}
\end{subfigure}
\end{figure}

\begin{figure}[H]
\centering
\begin{subfigure}[T]{0.19\textwidth}
\centering
\includegraphics[width=\textwidth]{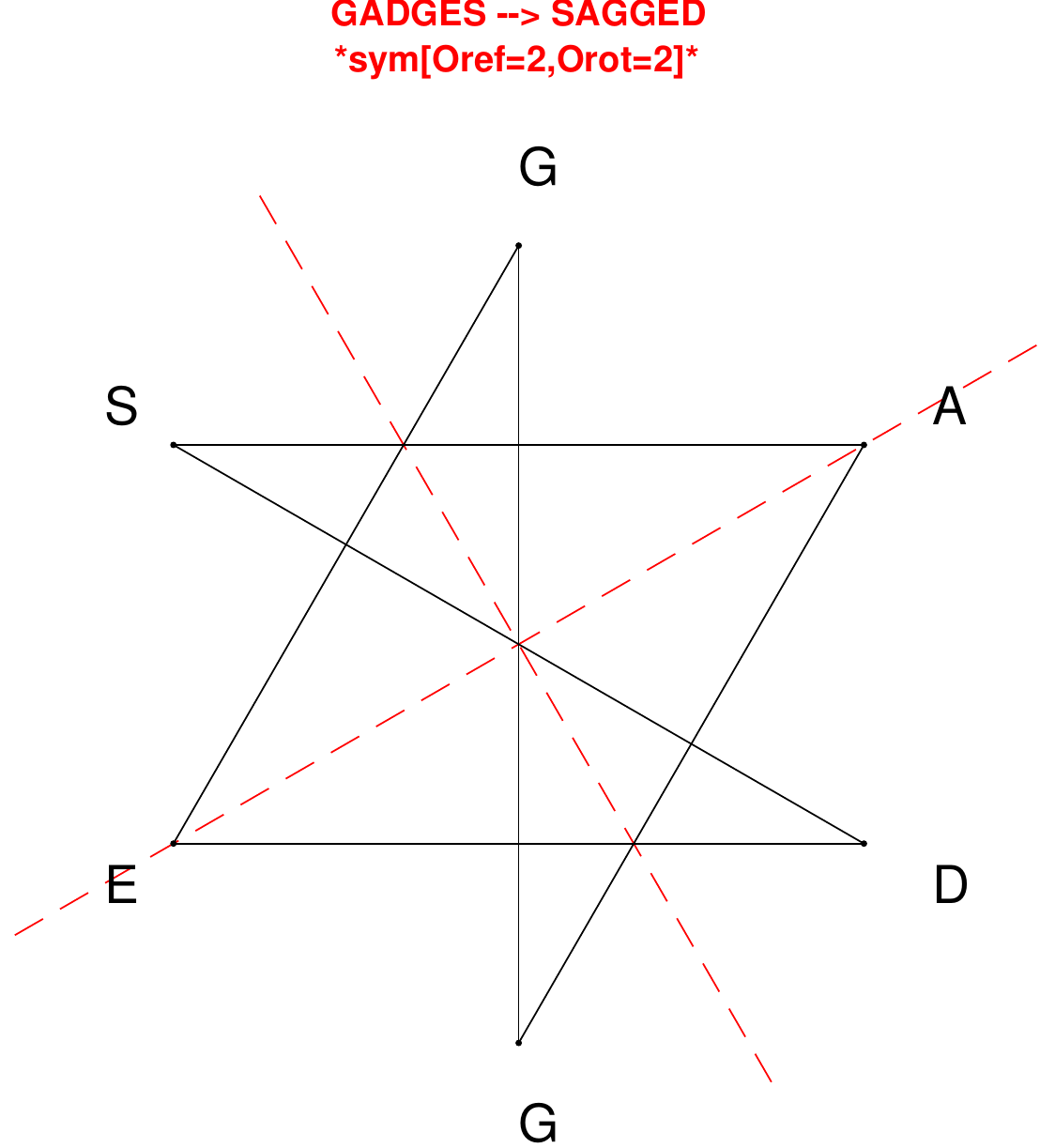}
\end{subfigure}
\hfill
\begin{subfigure}[T]{0.19\textwidth}
\centering
\includegraphics[width=\textwidth]{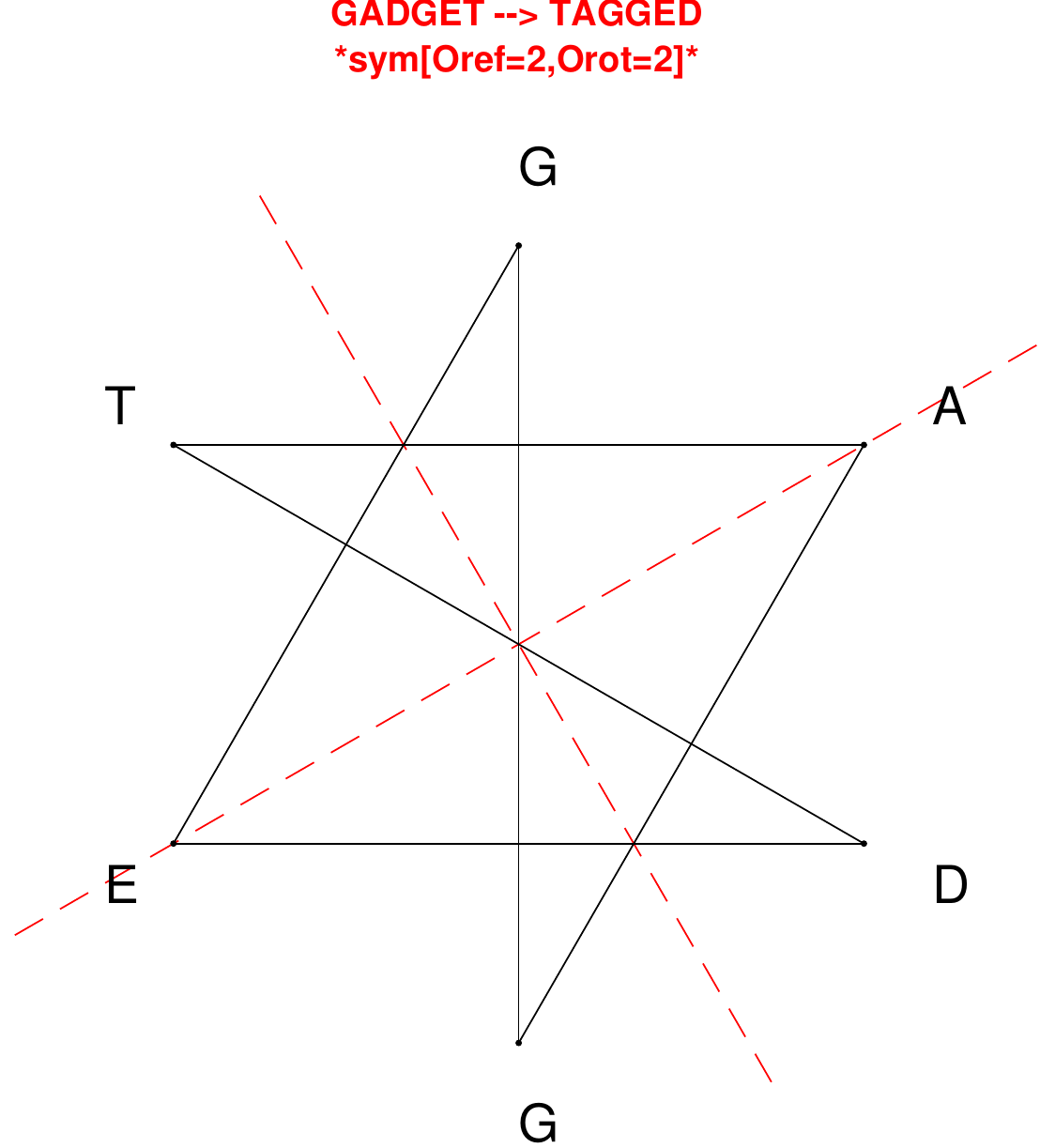}
\end{subfigure}
\hfill
\begin{subfigure}[T]{0.19\textwidth}
\centering
\includegraphics[width=\textwidth]{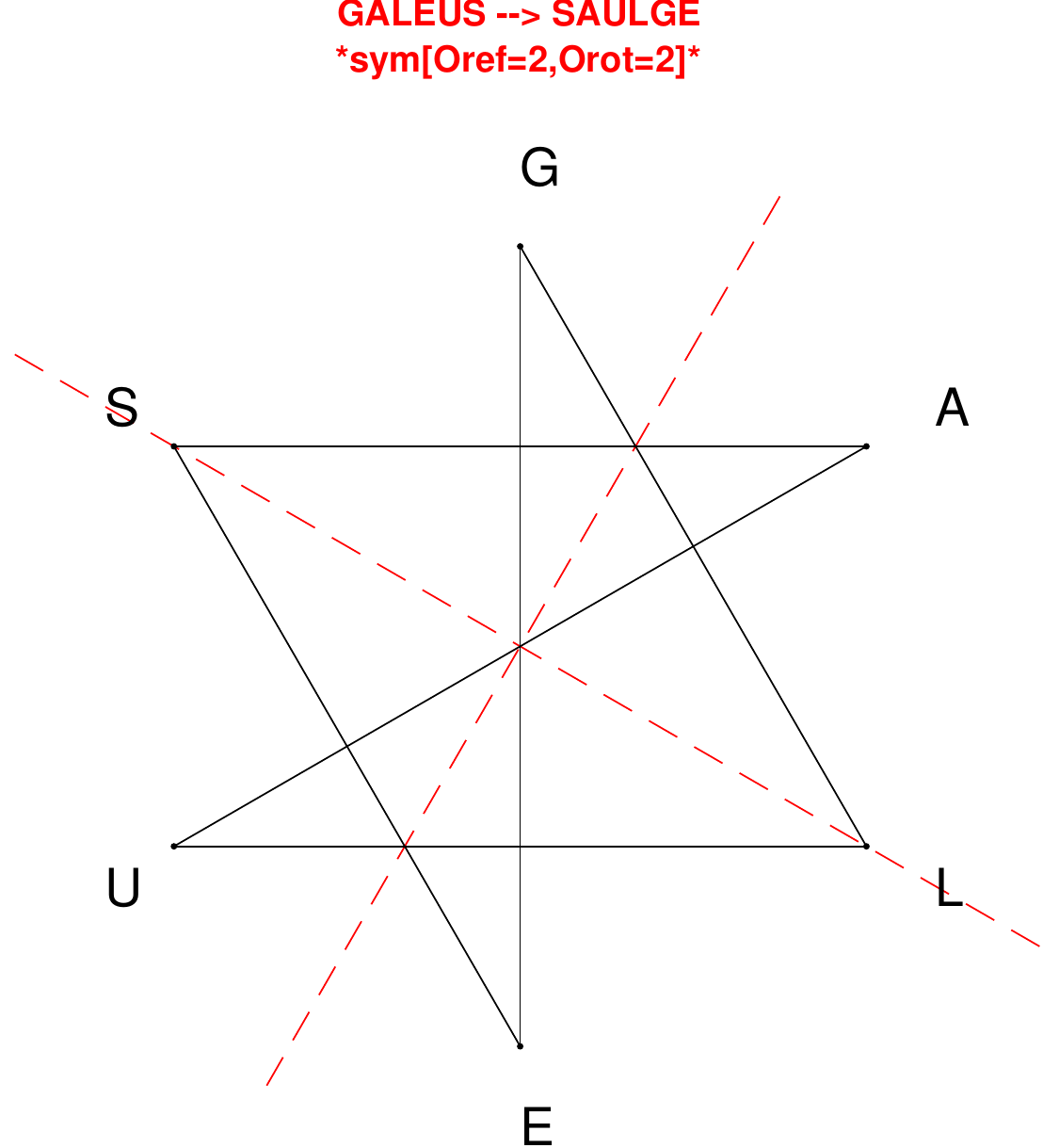}
\end{subfigure}
\hfill
\begin{subfigure}[T]{0.19\textwidth}
\centering
\includegraphics[width=\textwidth]{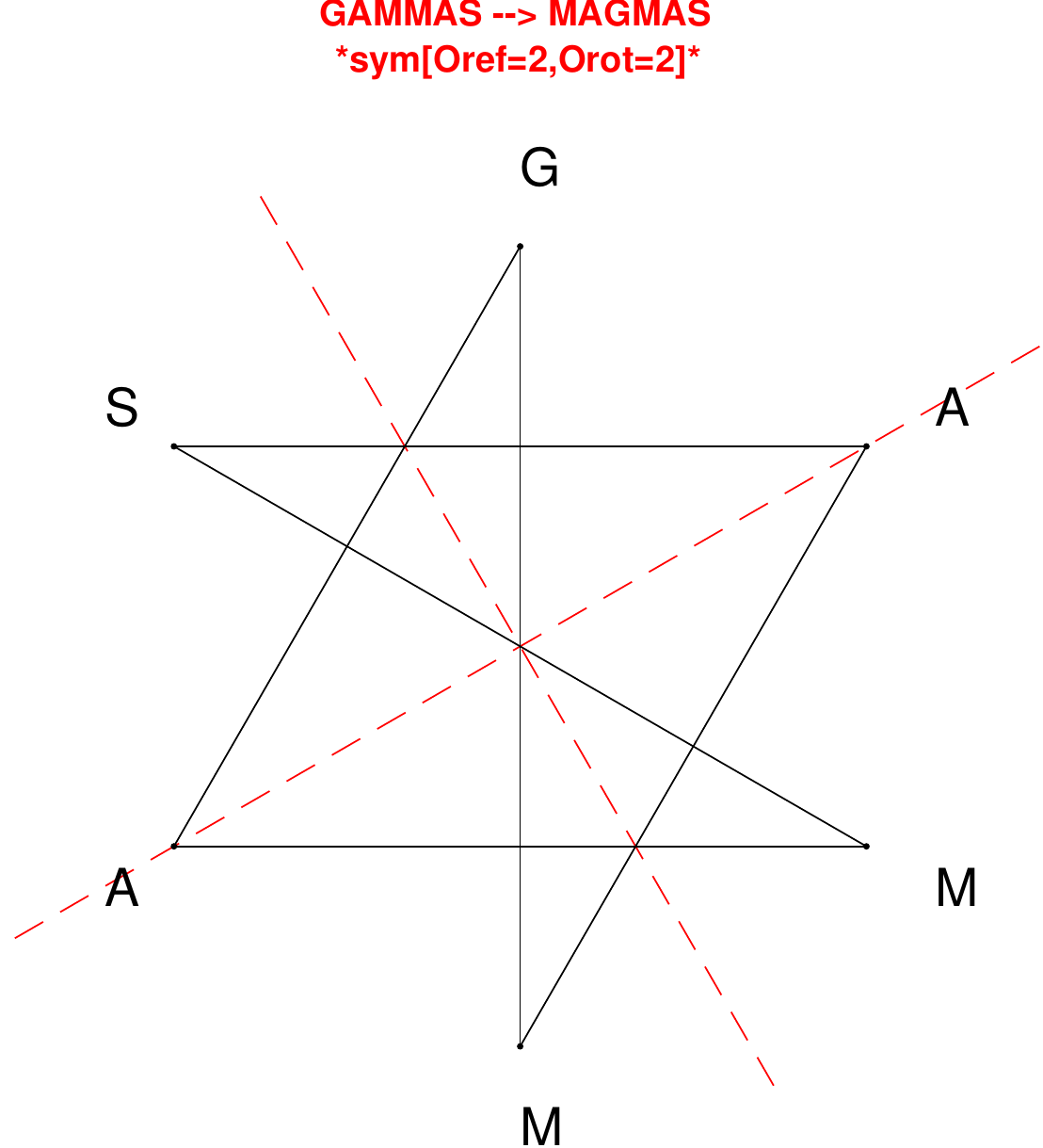}
\end{subfigure}
\hfill
\begin{subfigure}[T]{0.19\textwidth}
\centering
\includegraphics[width=\textwidth]{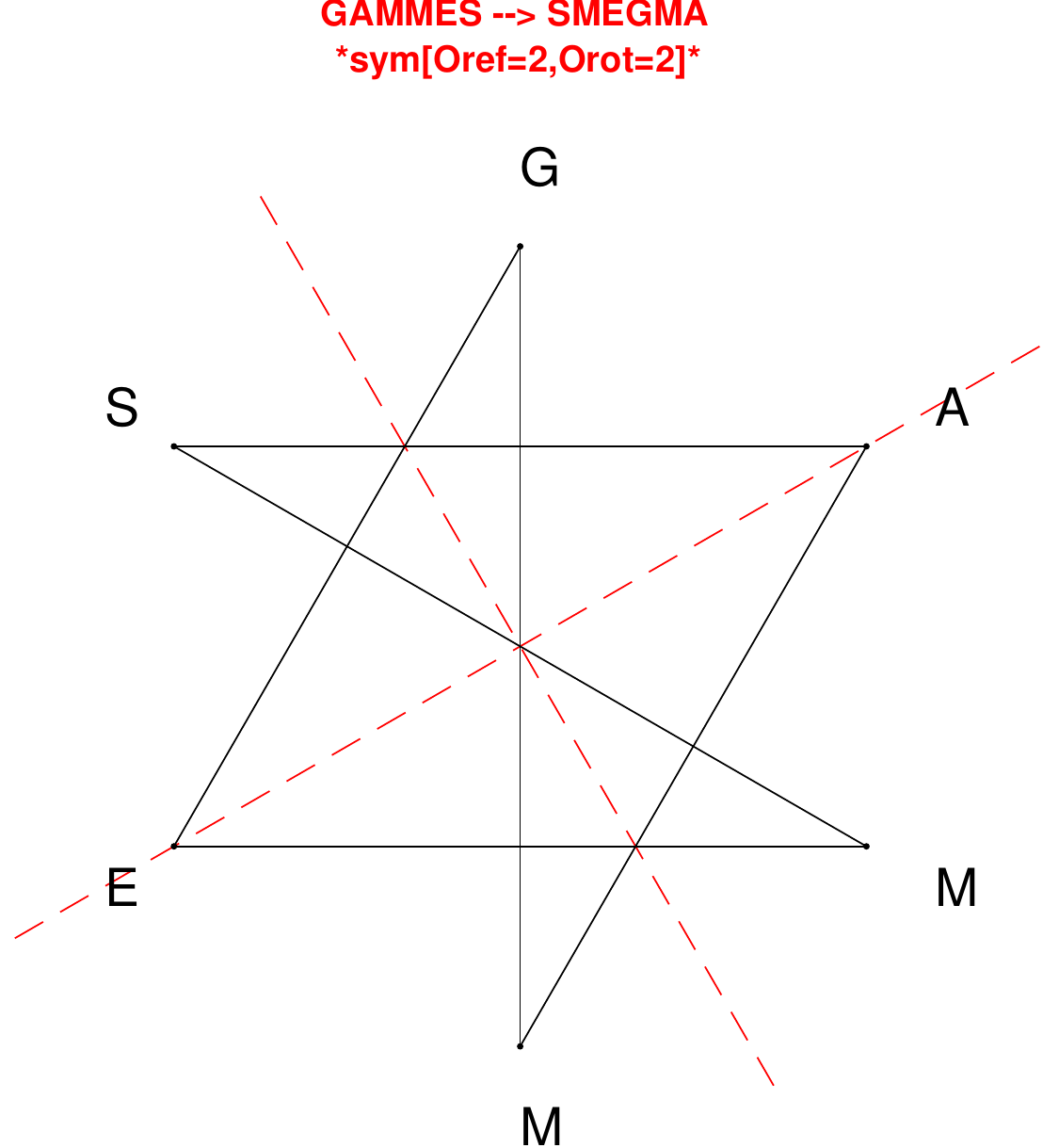}
\end{subfigure}
\end{figure}

\begin{figure}[H]
\centering
\begin{subfigure}[T]{0.19\textwidth}
\centering
\includegraphics[width=\textwidth]{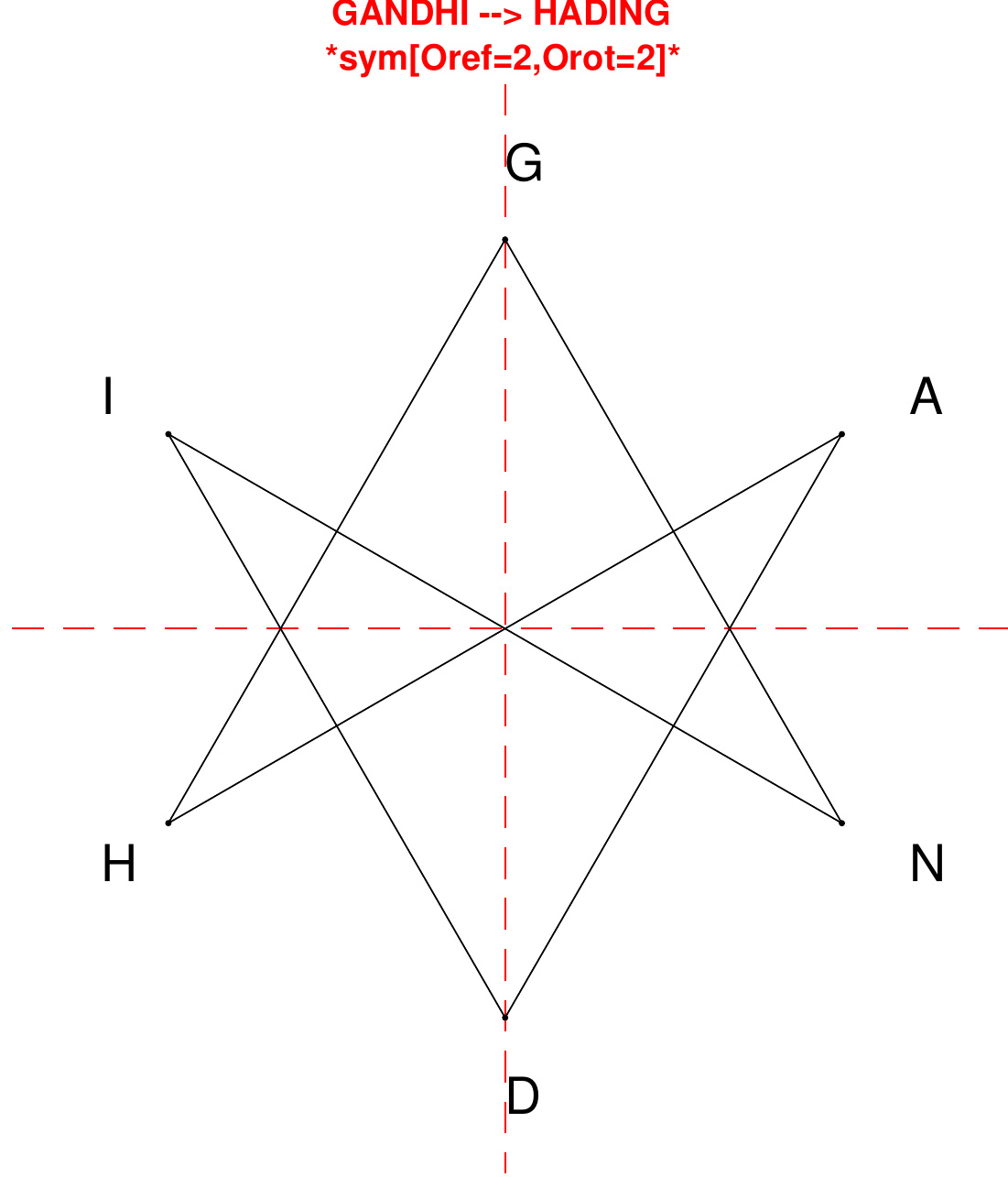}
\end{subfigure}
\hfill
\begin{subfigure}[T]{0.19\textwidth}
\centering
\includegraphics[width=\textwidth]{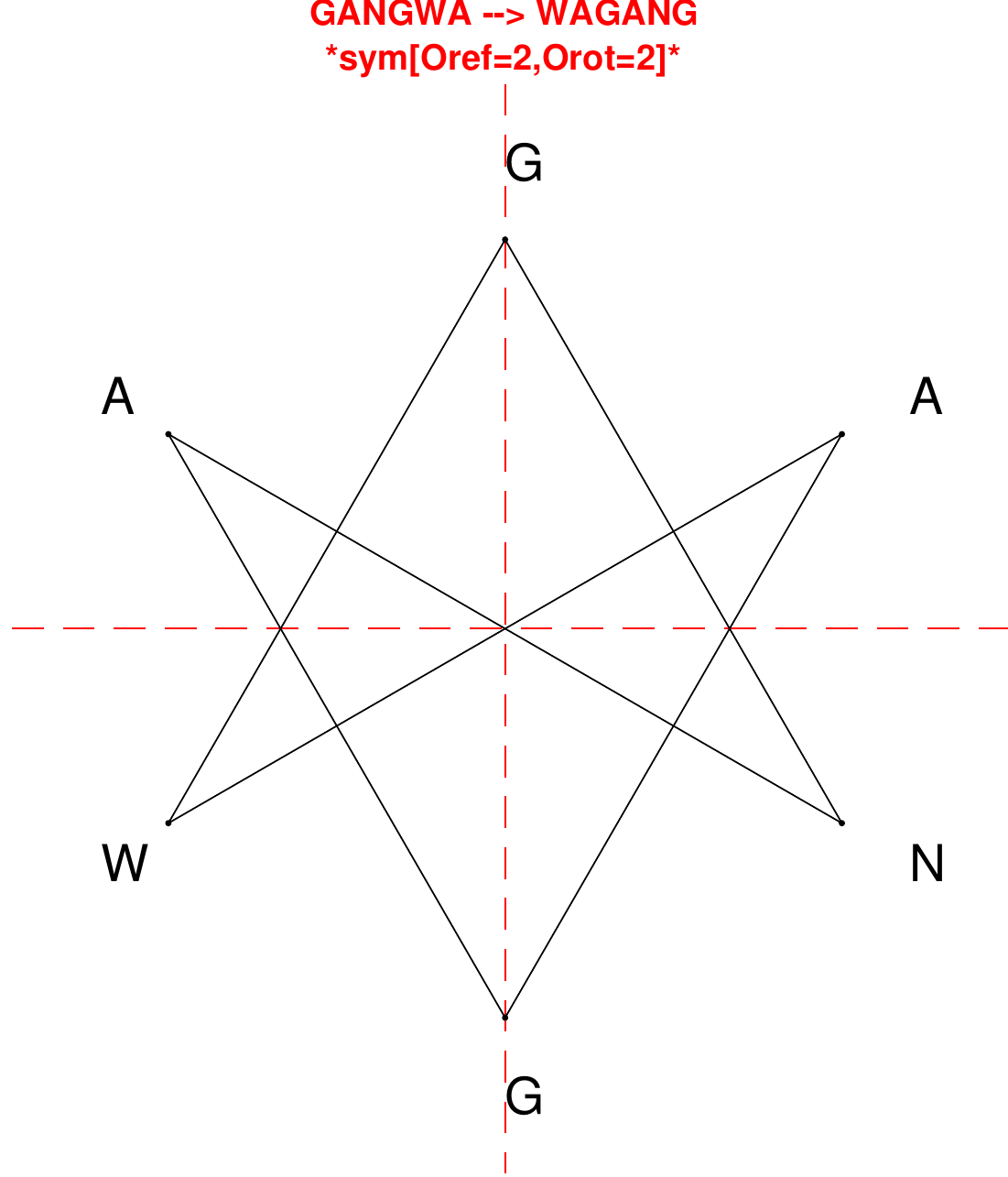}
\end{subfigure}
\hfill
\begin{subfigure}[T]{0.19\textwidth}
\centering
\includegraphics[width=\textwidth]{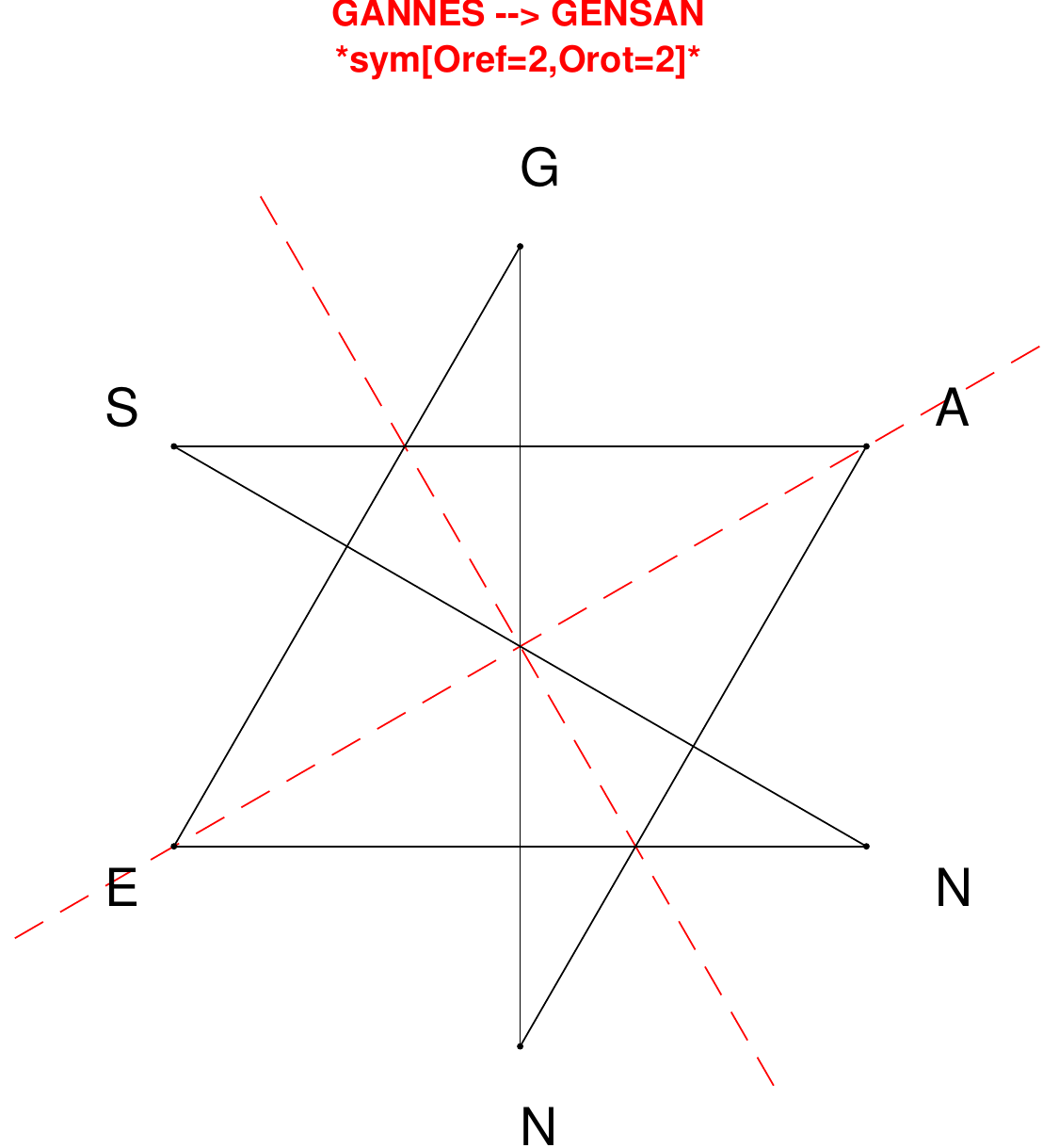}
\end{subfigure}
\hfill
\begin{subfigure}[T]{0.19\textwidth}
\centering
\includegraphics[width=\textwidth]{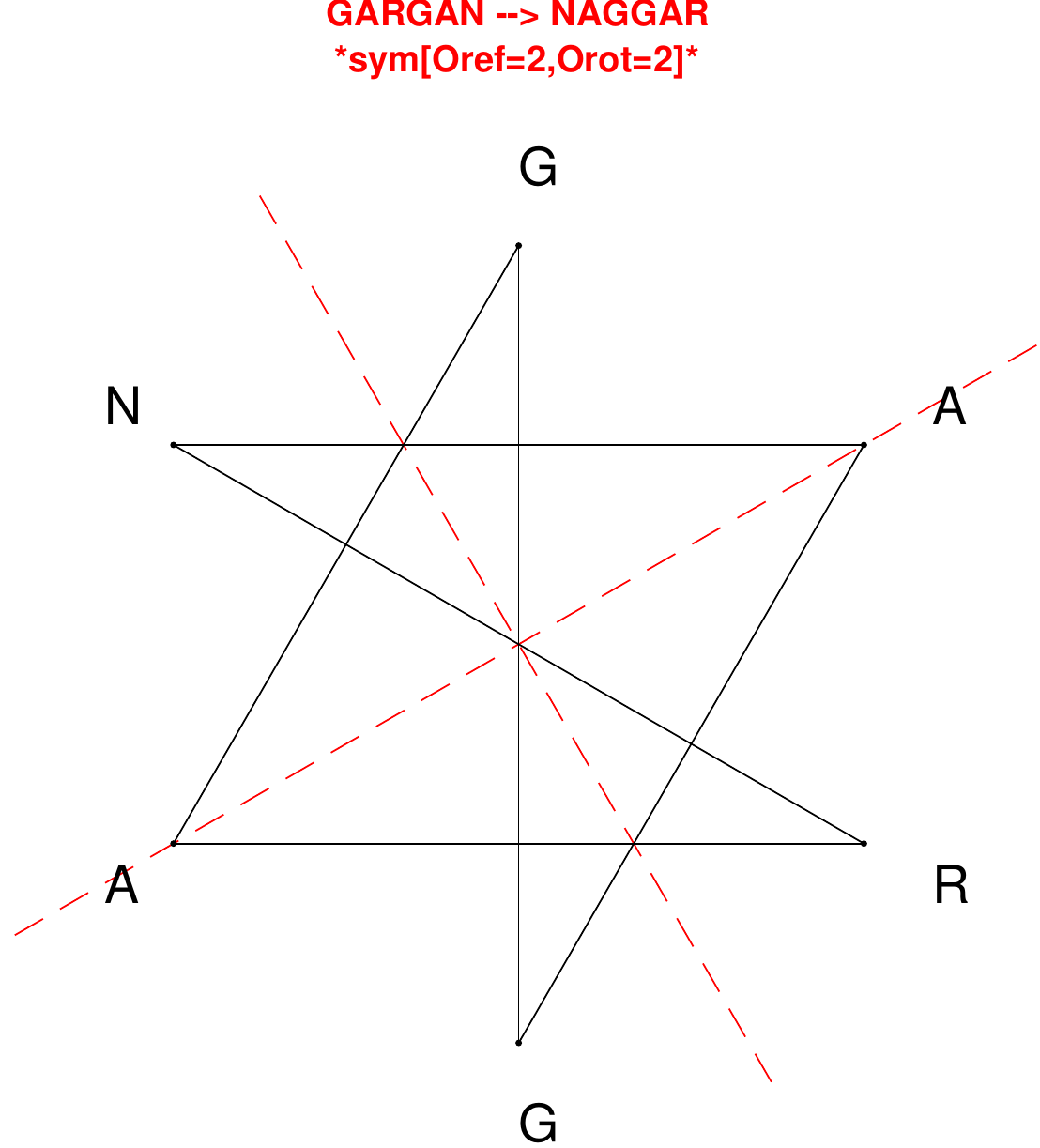}
\end{subfigure}
\hfill
\begin{subfigure}[T]{0.19\textwidth}
\centering
\includegraphics[width=\textwidth]{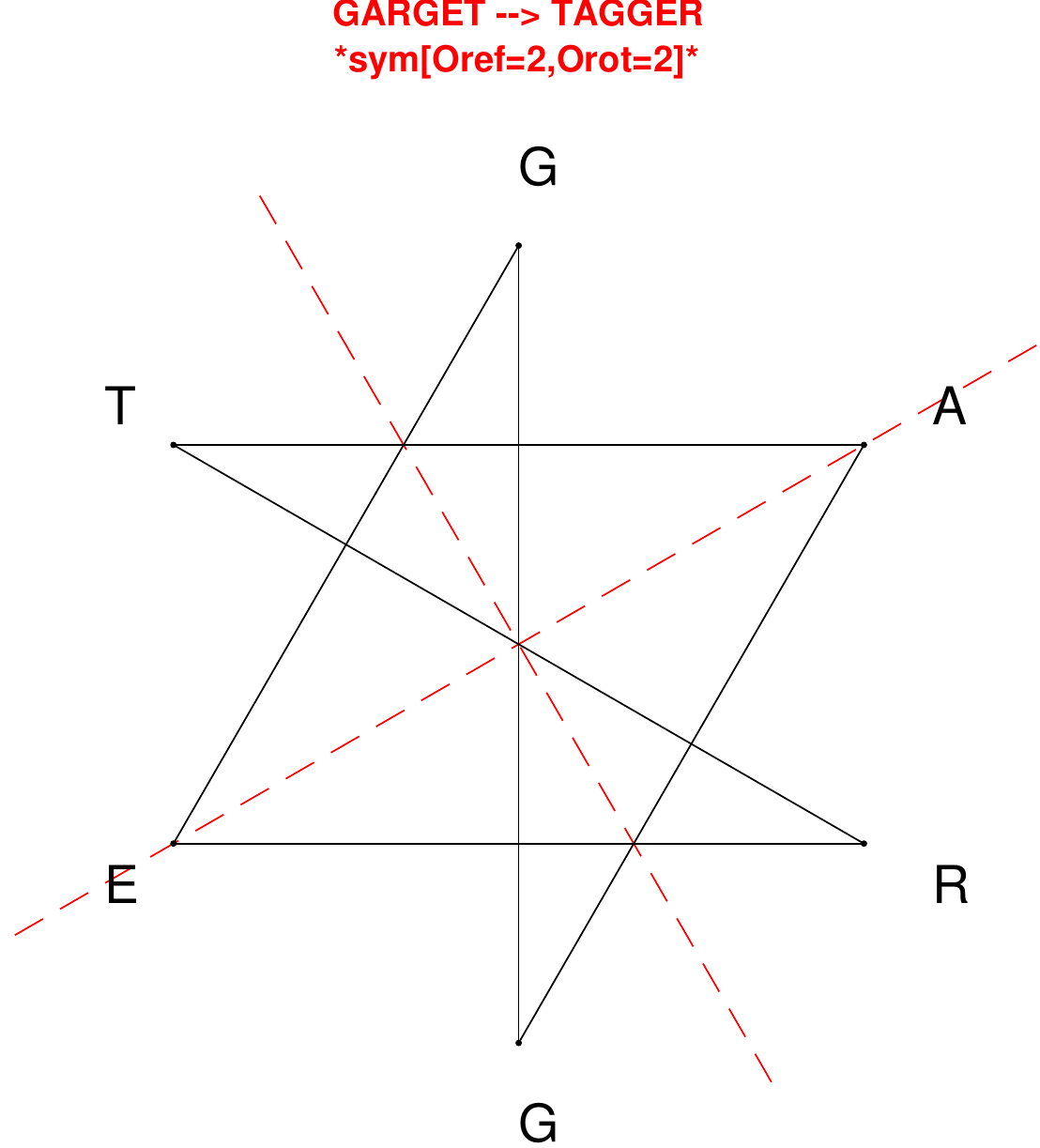}
\end{subfigure}
\end{figure}

\begin{figure}[H]
\centering
\begin{subfigure}[T]{0.19\textwidth}
\centering
\includegraphics[width=\textwidth]{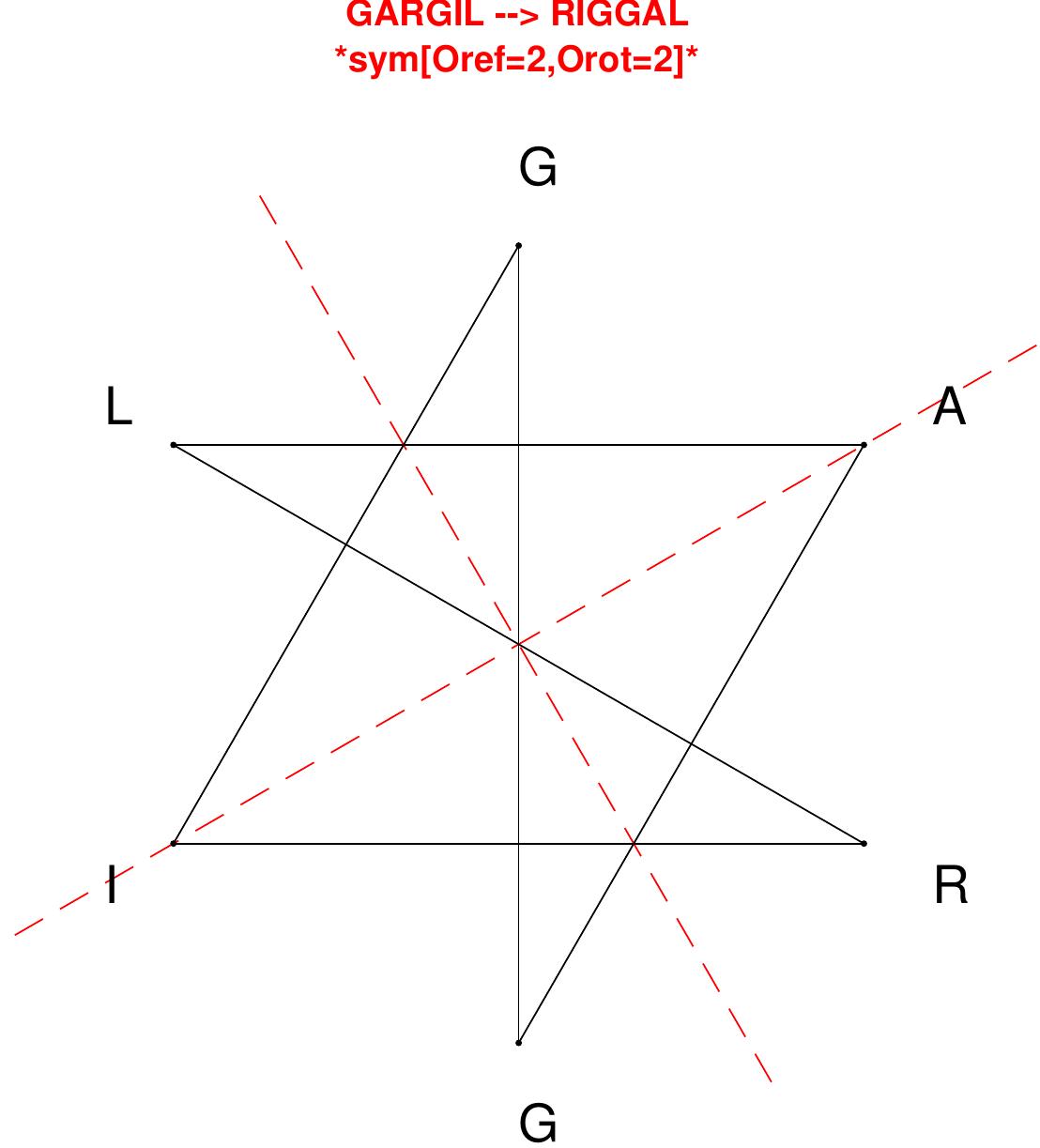}
\end{subfigure}
\hfill
\begin{subfigure}[T]{0.19\textwidth}
\centering
\includegraphics[width=\textwidth]{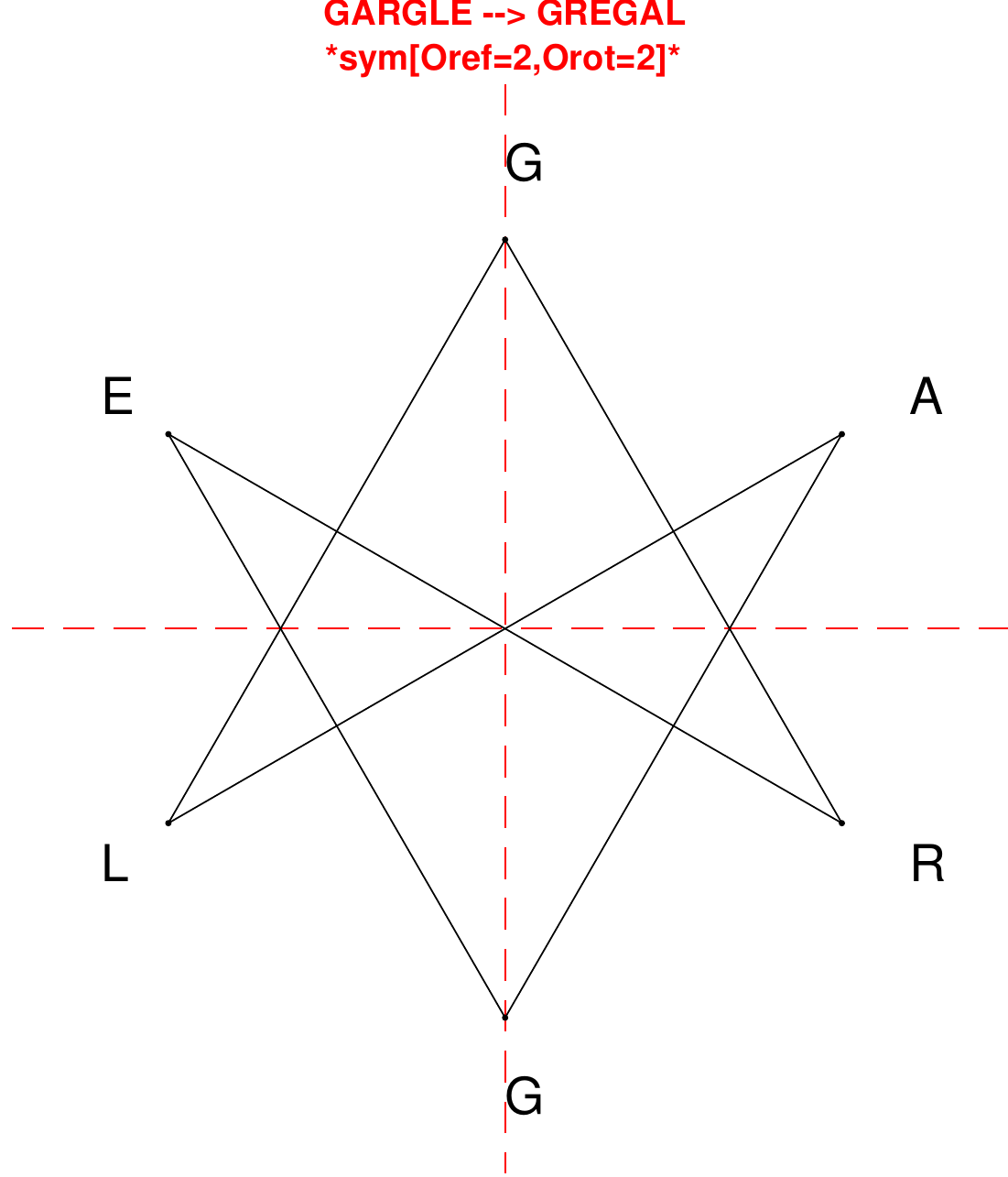}
\end{subfigure}
\hfill
\begin{subfigure}[T]{0.19\textwidth}
\centering
\includegraphics[width=\textwidth]{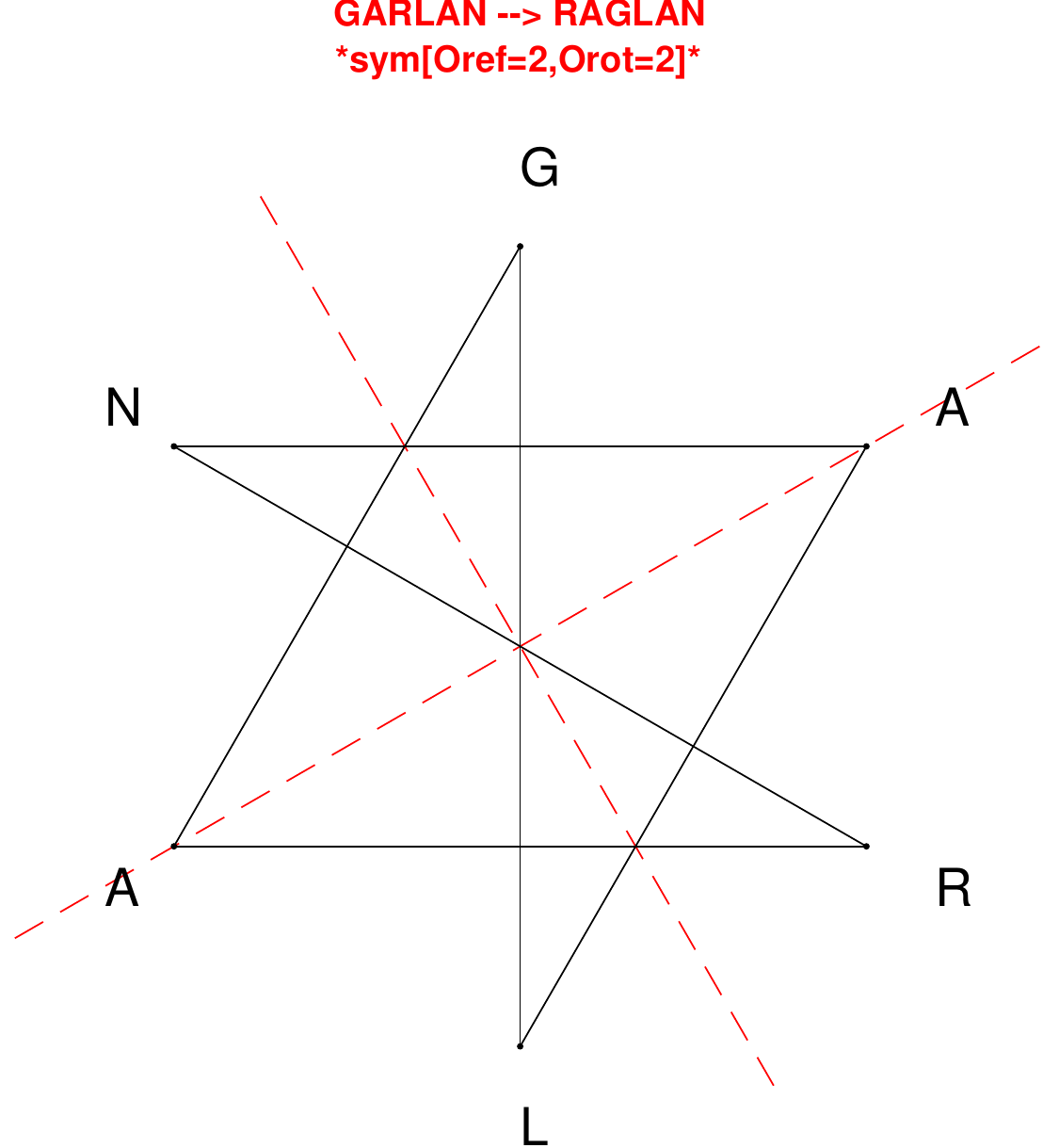}
\end{subfigure}
\hfill
\begin{subfigure}[T]{0.19\textwidth}
\centering
\includegraphics[width=\textwidth]{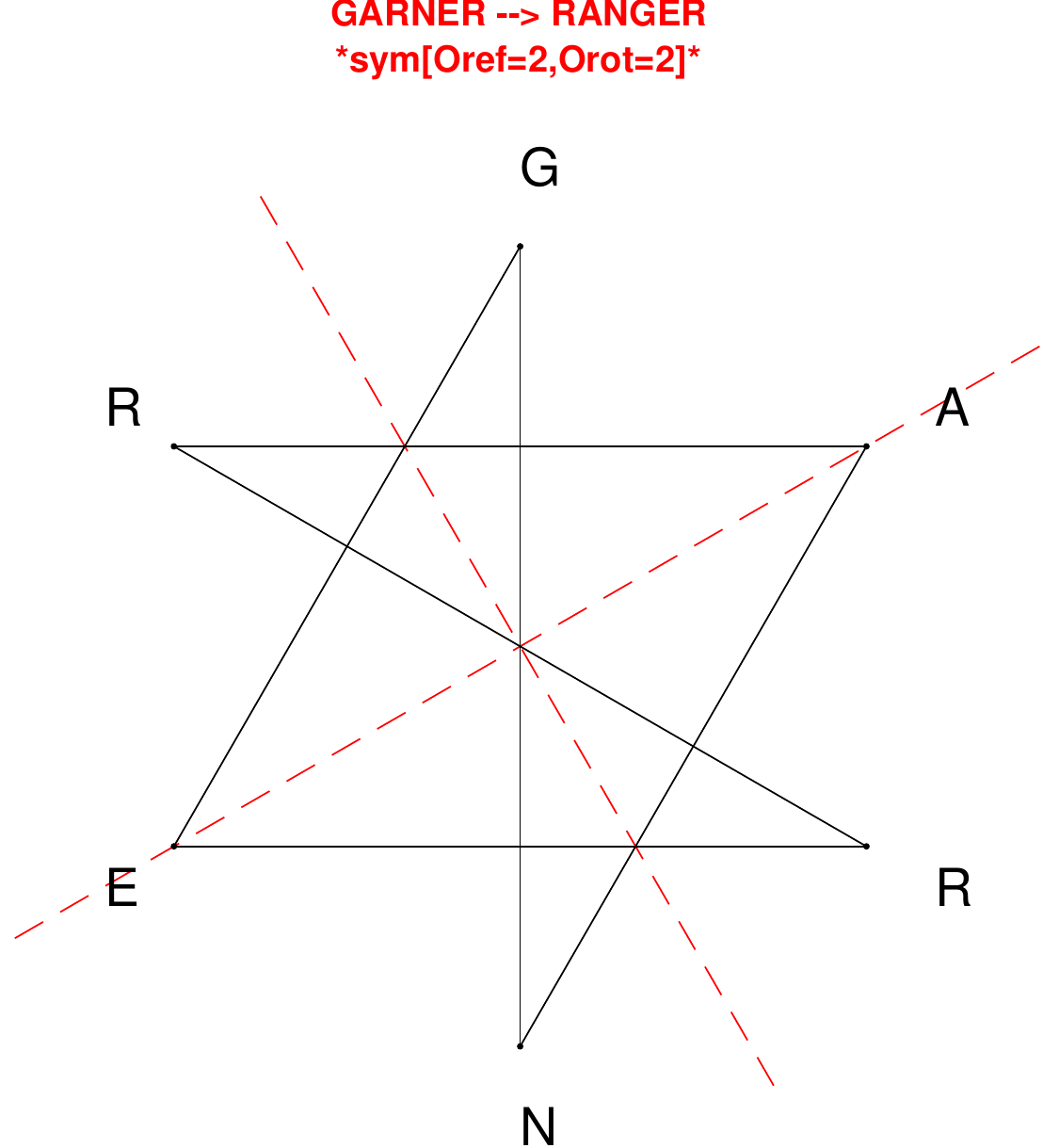}
\end{subfigure}
\hfill
\begin{subfigure}[T]{0.19\textwidth}
\centering
\includegraphics[width=\textwidth]{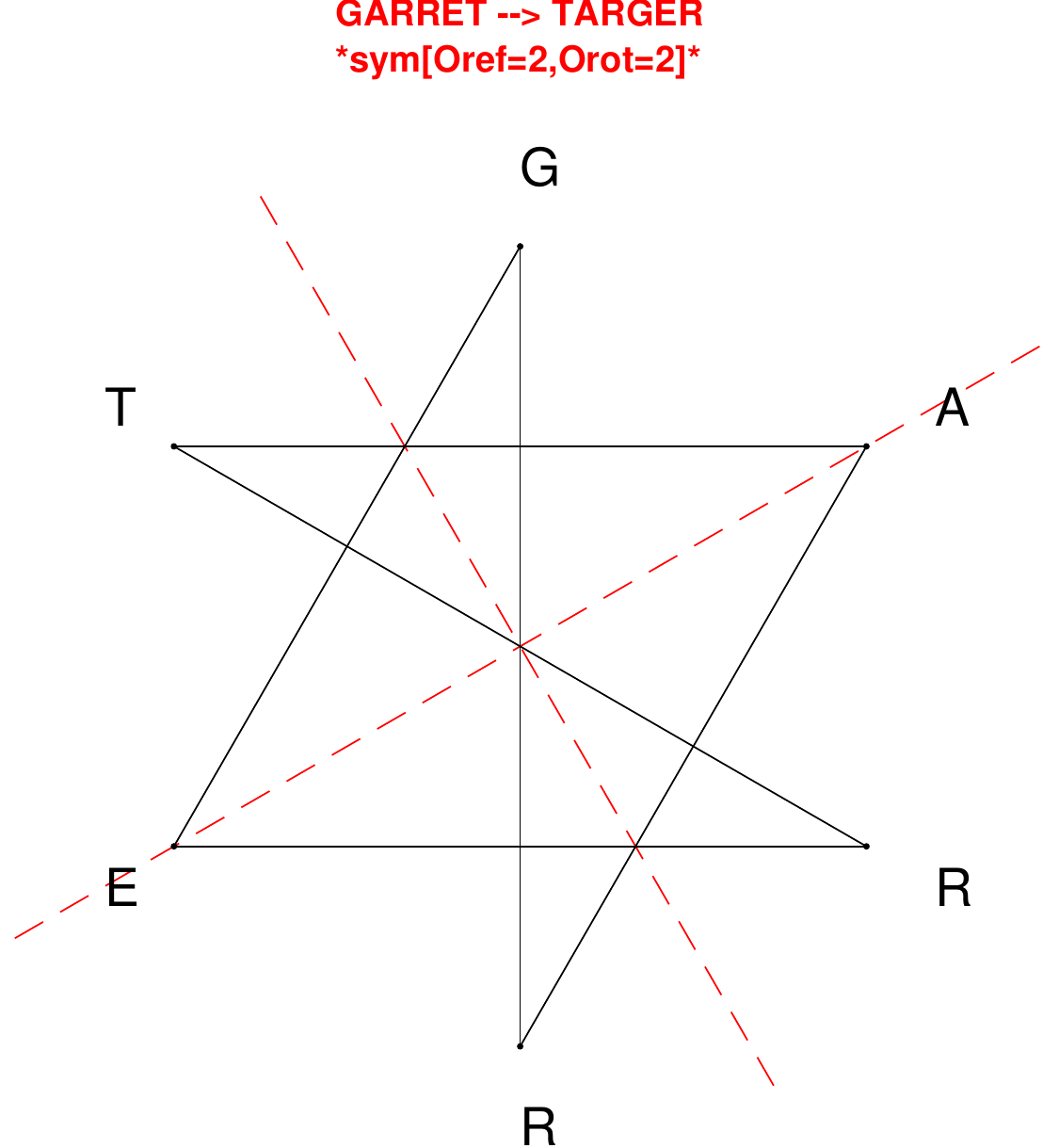}
\end{subfigure}
\end{figure}

\begin{figure}[H]
\centering
\begin{subfigure}[T]{0.19\textwidth}
\centering
\includegraphics[width=\textwidth]{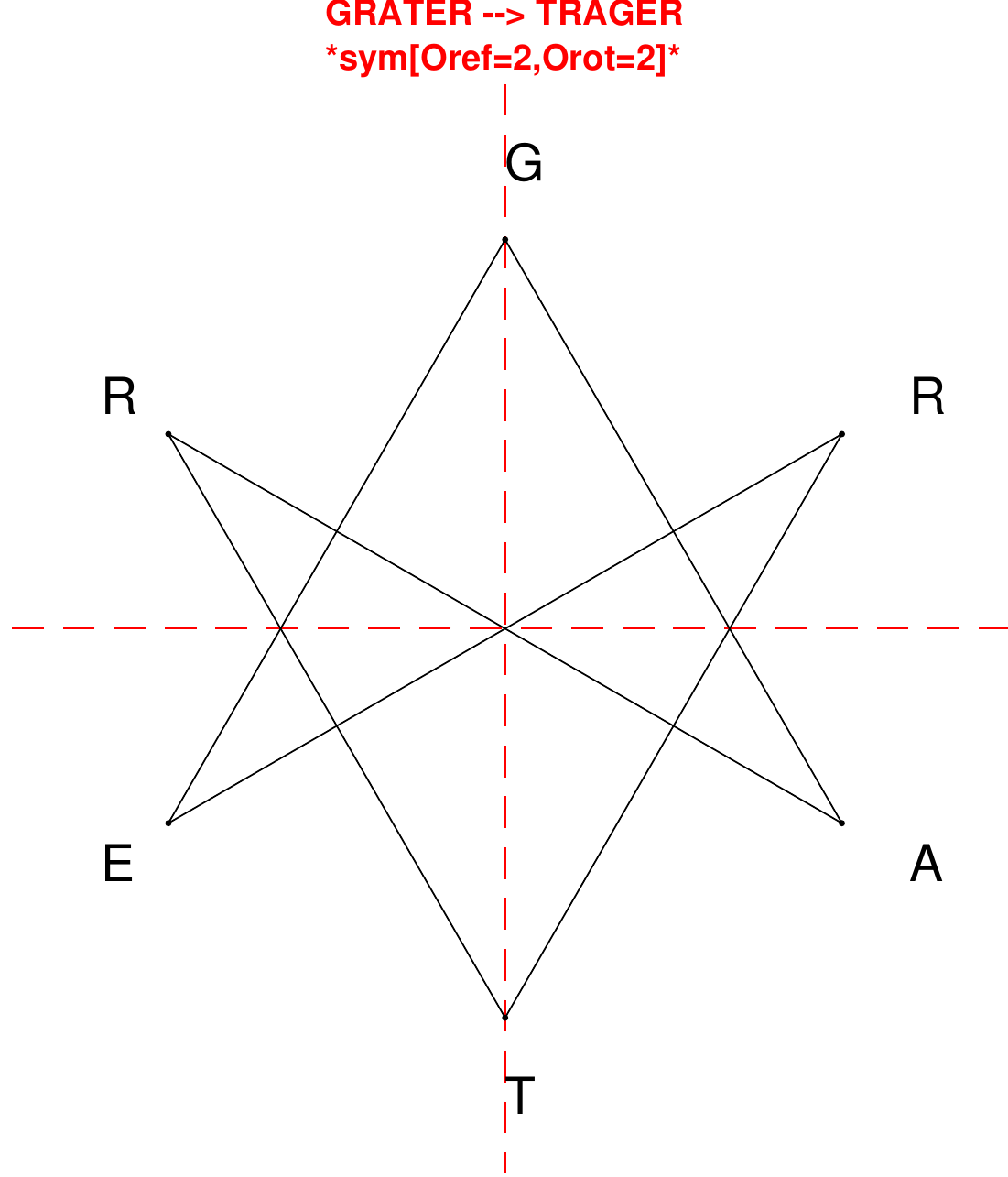}
\end{subfigure}
\hfill
\begin{subfigure}[T]{0.19\textwidth}
\centering
\includegraphics[width=\textwidth]{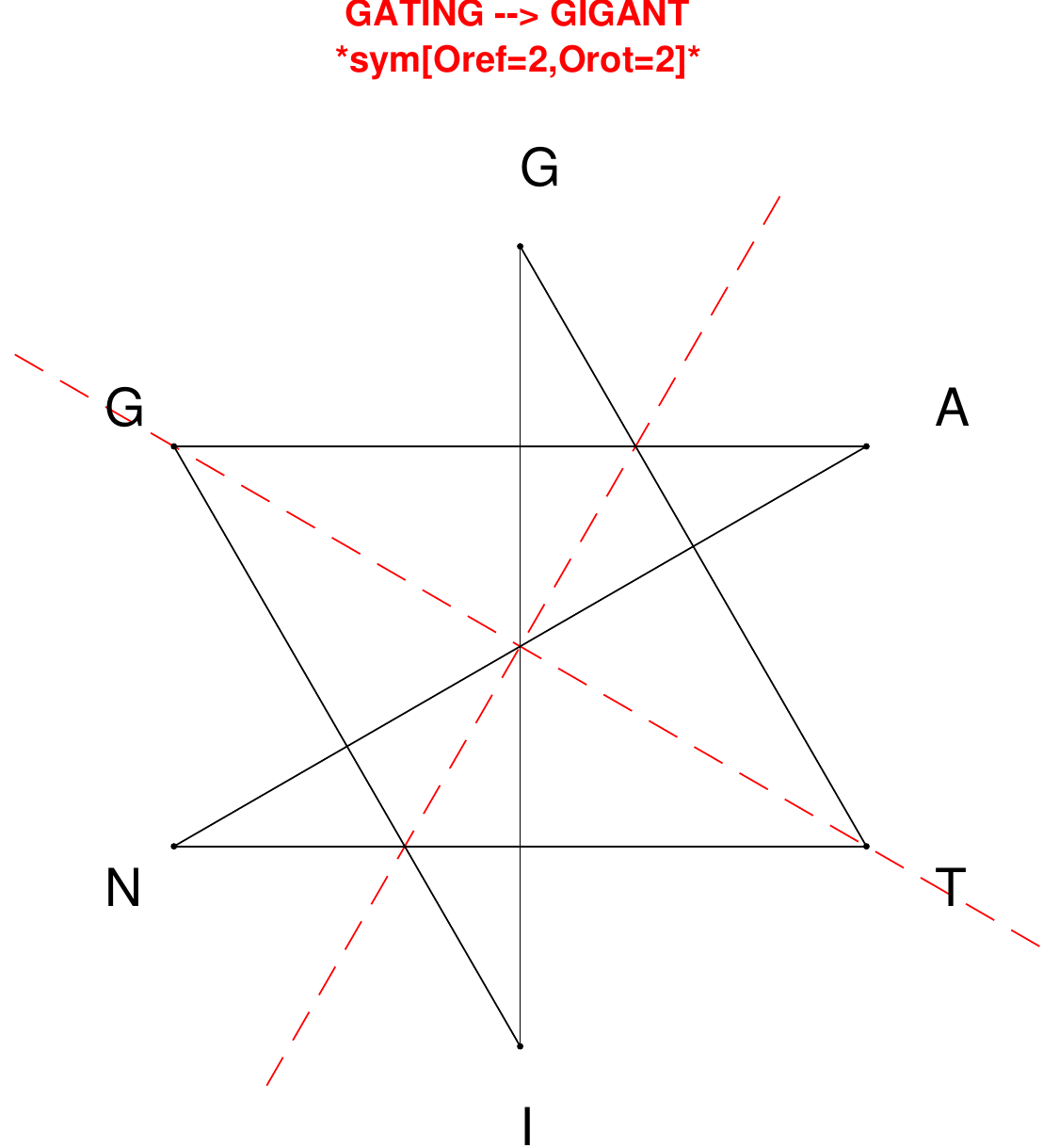}
\end{subfigure}
\hfill
\begin{subfigure}[T]{0.19\textwidth}
\centering
\includegraphics[width=\textwidth]{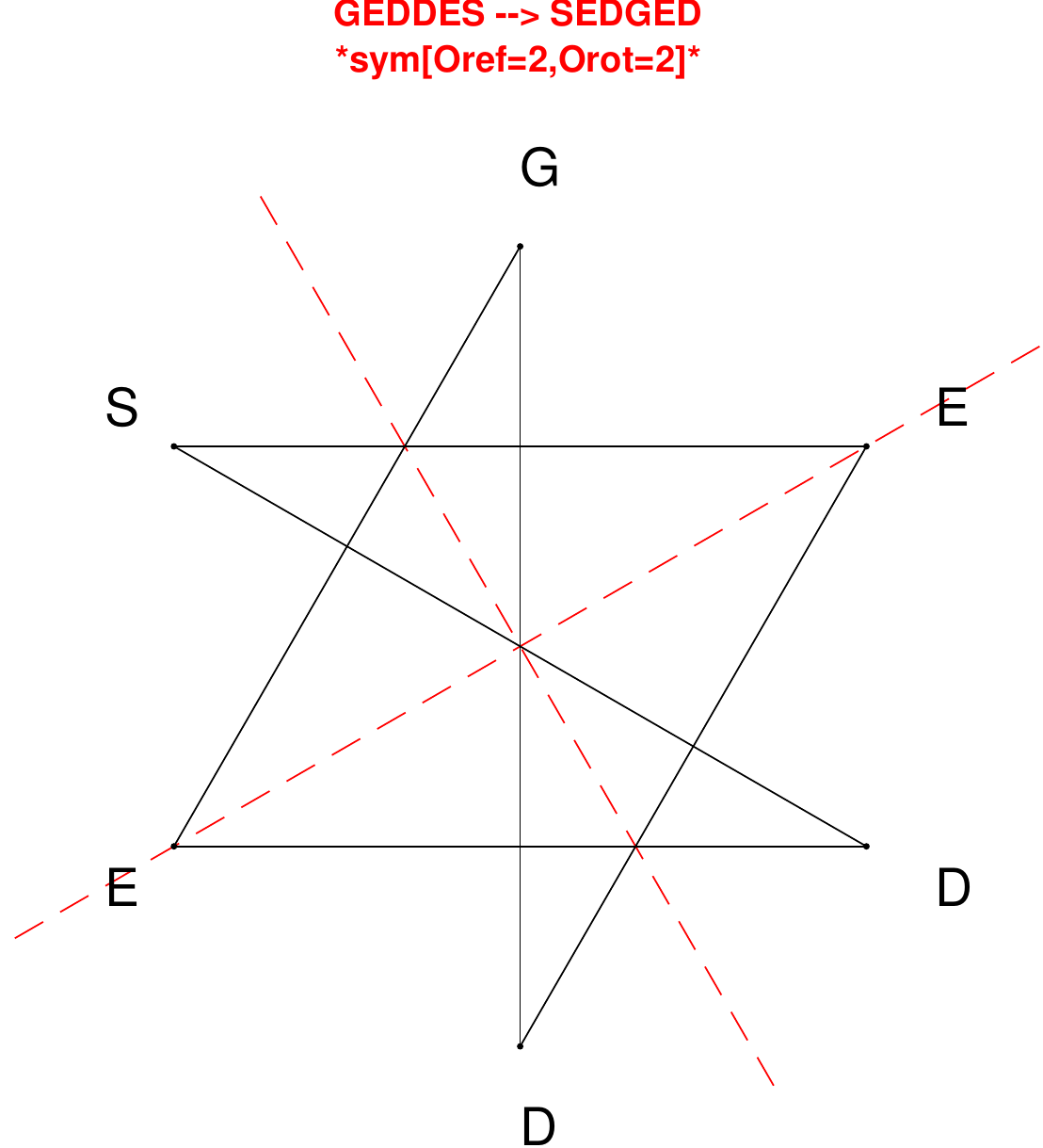}
\end{subfigure}
\hfill
\begin{subfigure}[T]{0.19\textwidth}
\centering
\includegraphics[width=\textwidth]{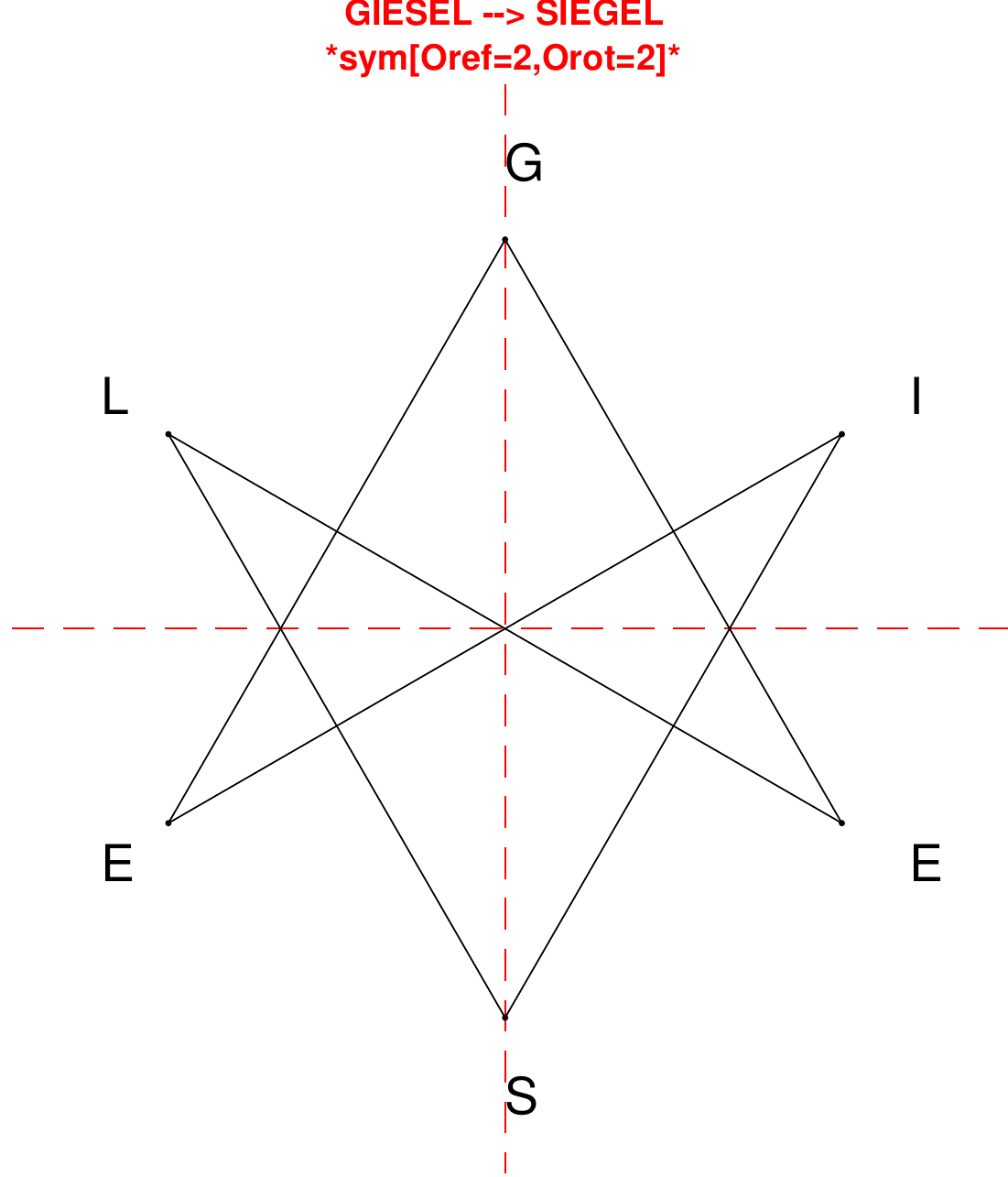}
\end{subfigure}
\hfill
\begin{subfigure}[T]{0.19\textwidth}
\centering
\includegraphics[width=\textwidth]{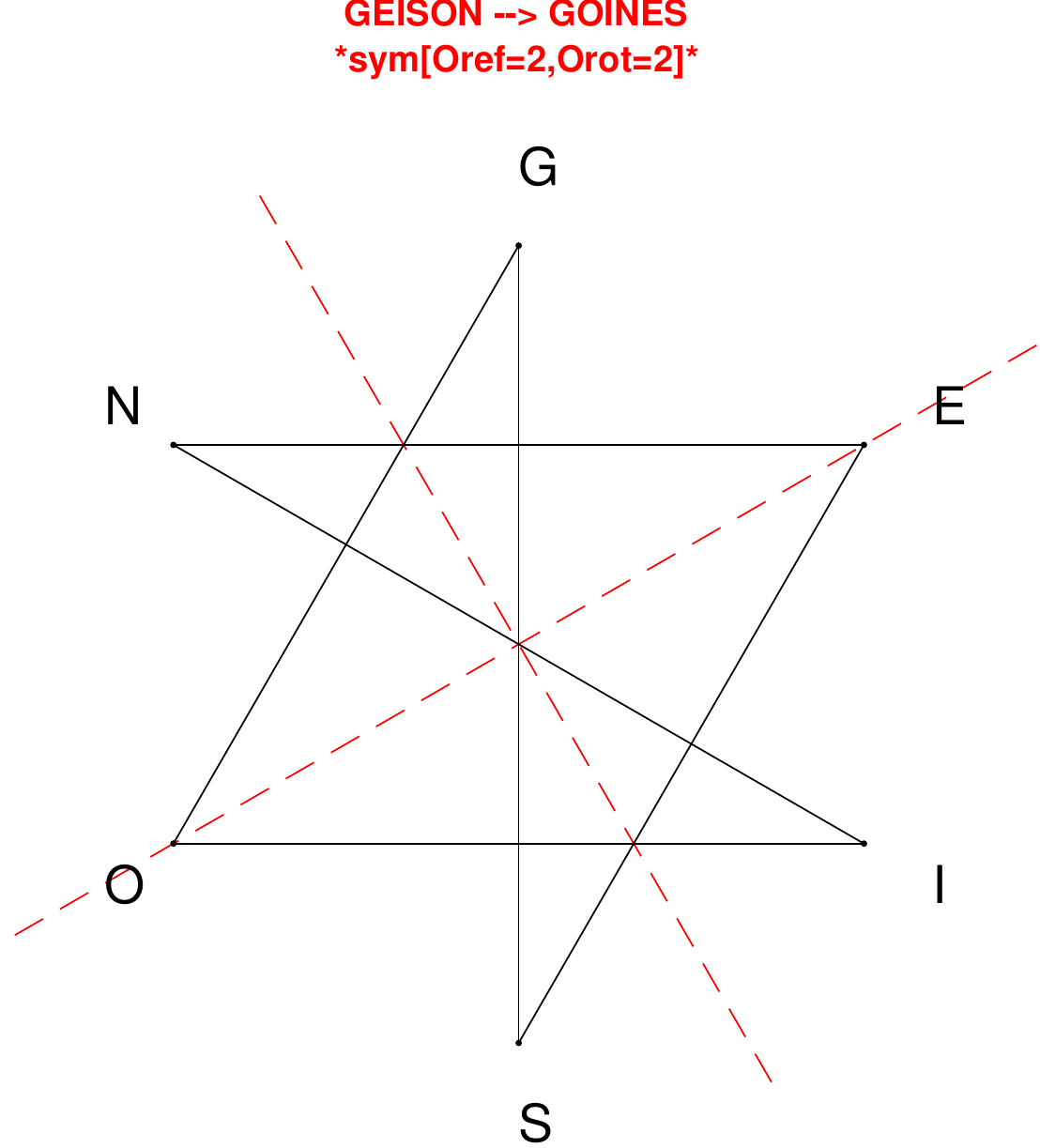}
\end{subfigure}
\end{figure}

\begin{figure}[H]
\centering
\begin{subfigure}[T]{0.19\textwidth}
\centering
\includegraphics[width=\textwidth]{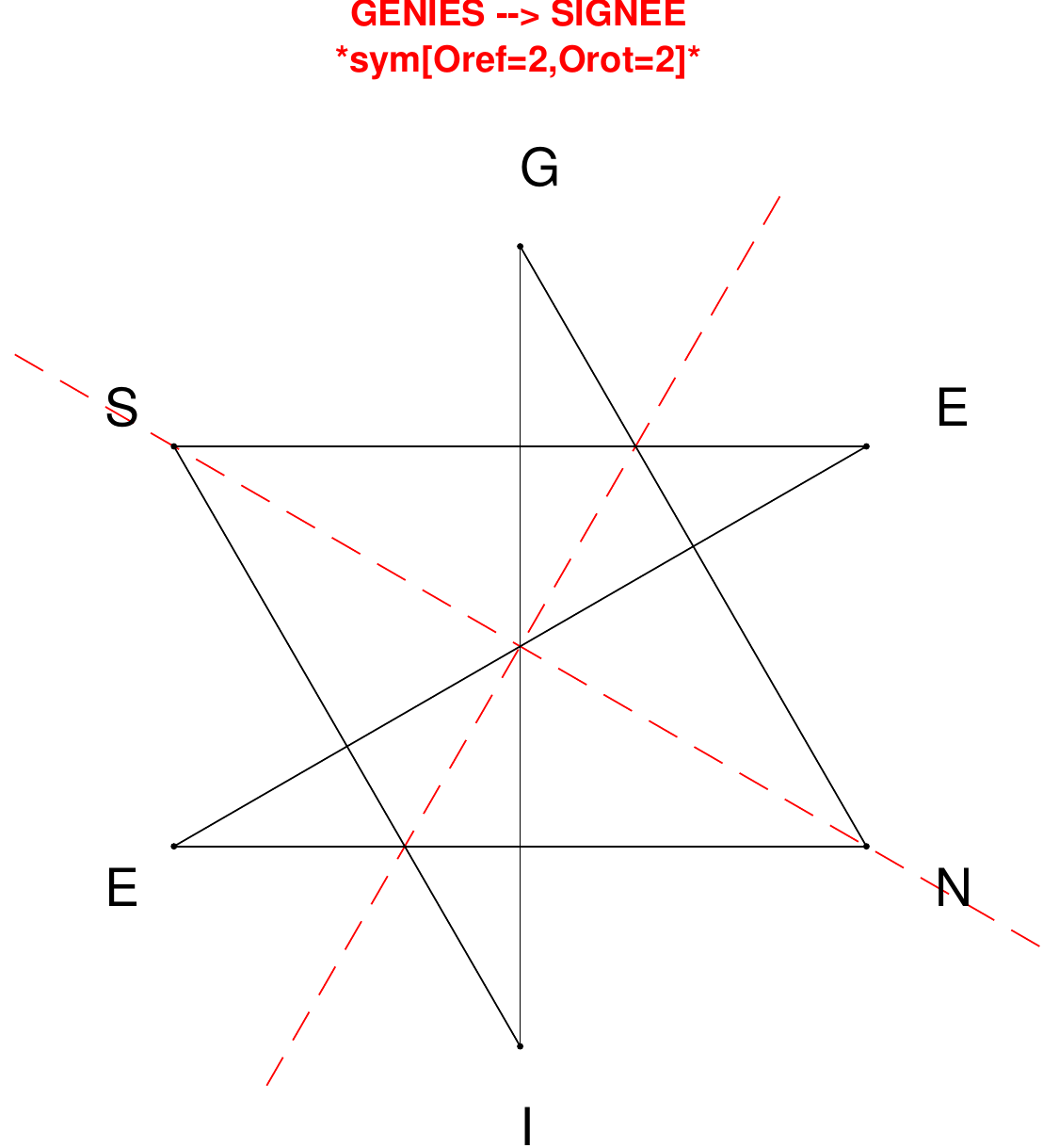}
\end{subfigure}
\hfill
\begin{subfigure}[T]{0.19\textwidth}
\centering
\includegraphics[width=\textwidth]{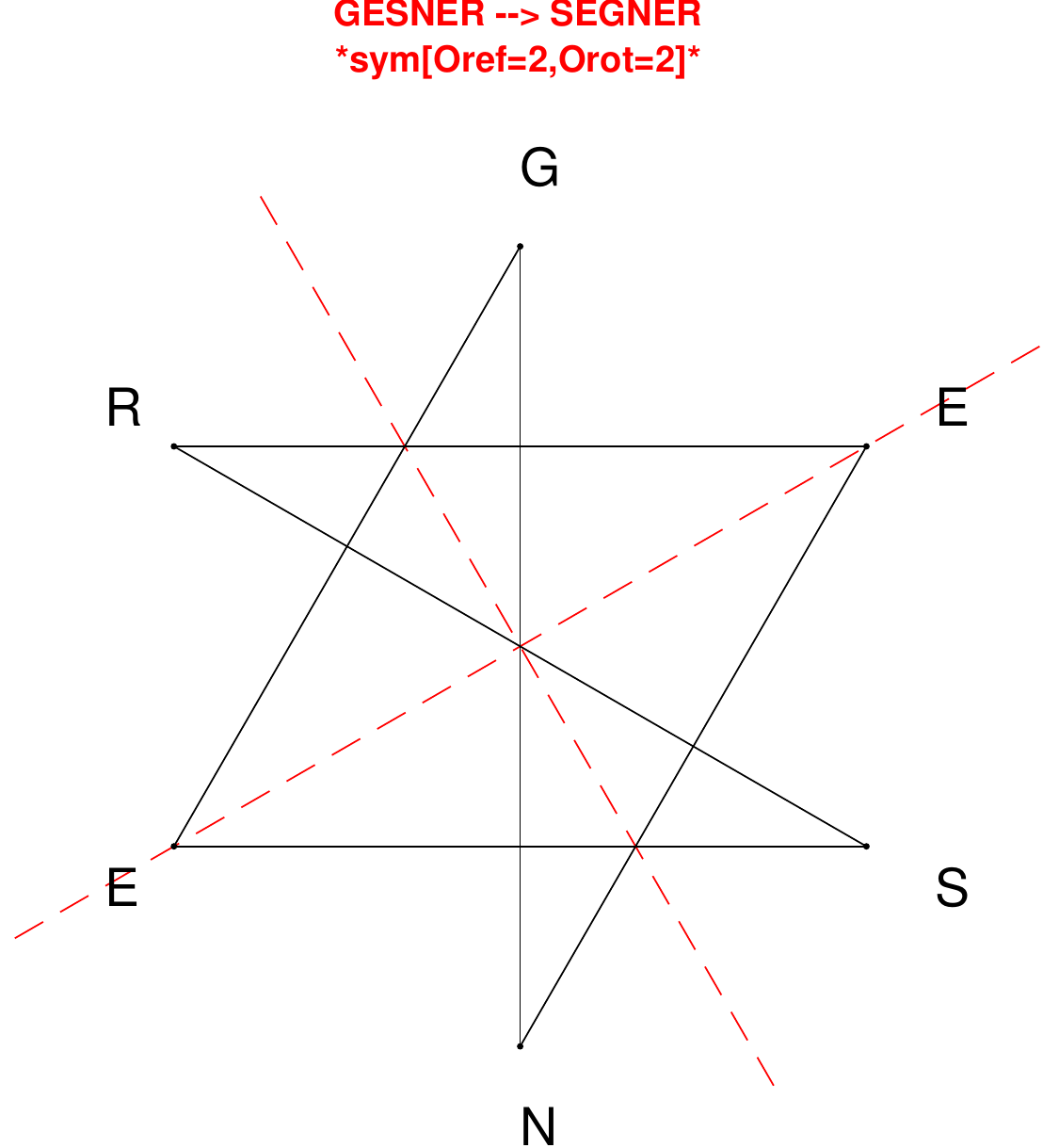}
\end{subfigure}
\hfill
\begin{subfigure}[T]{0.19\textwidth}
\centering
\includegraphics[width=\textwidth]{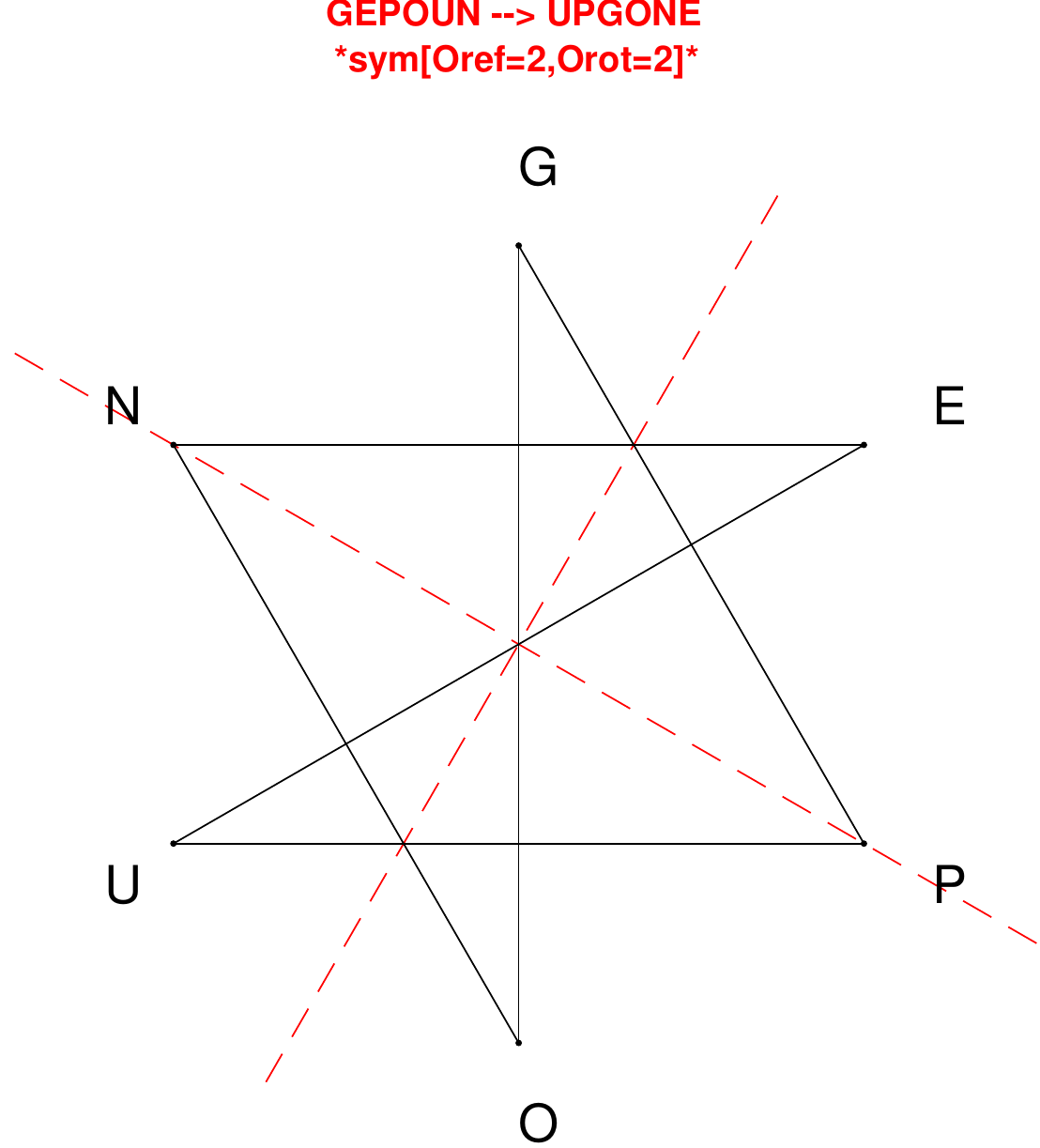}
\end{subfigure}
\hfill
\begin{subfigure}[T]{0.19\textwidth}
\centering
\includegraphics[width=\textwidth]{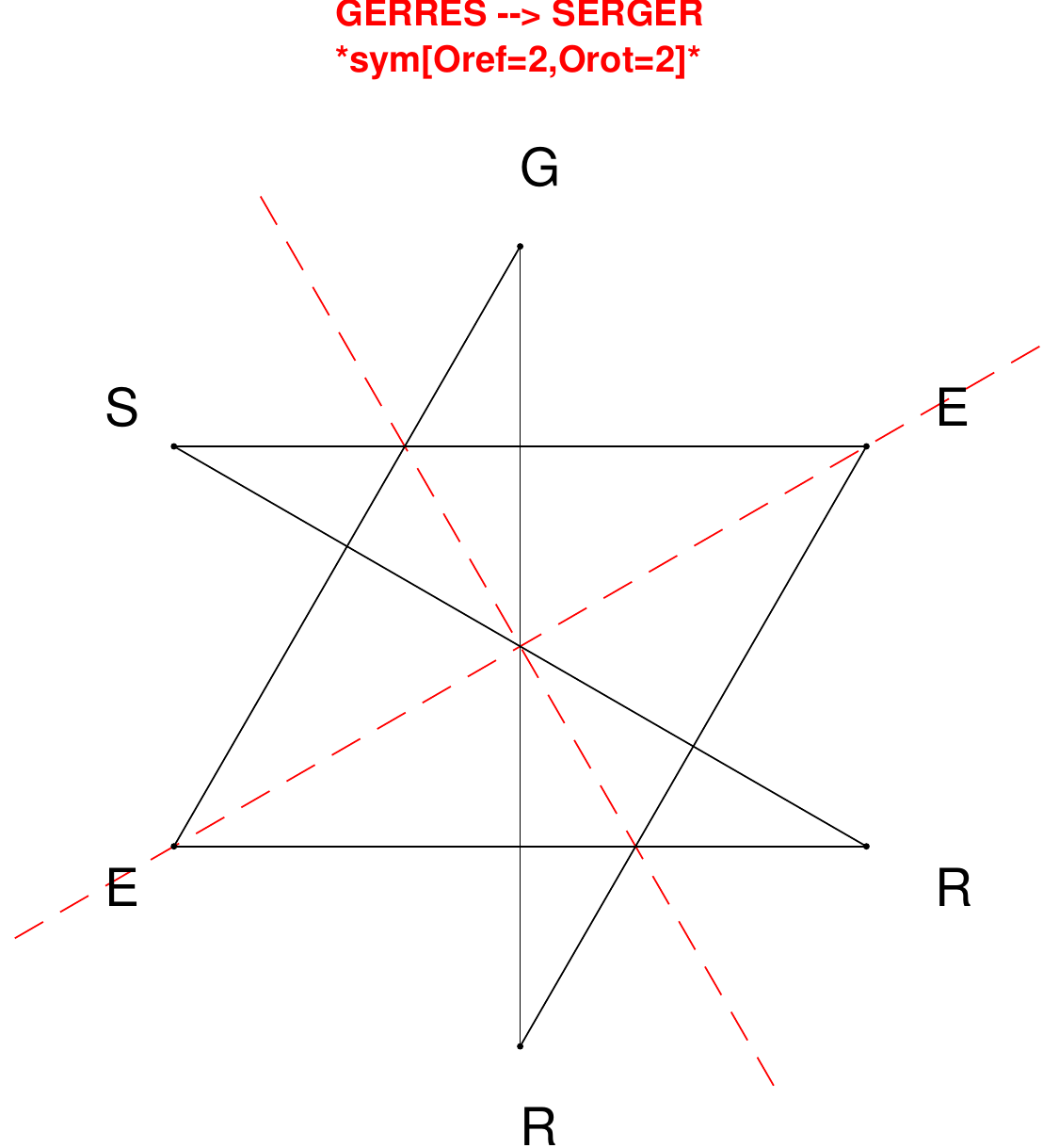}
\end{subfigure}
\hfill
\begin{subfigure}[T]{0.19\textwidth}
\centering
\includegraphics[width=\textwidth]{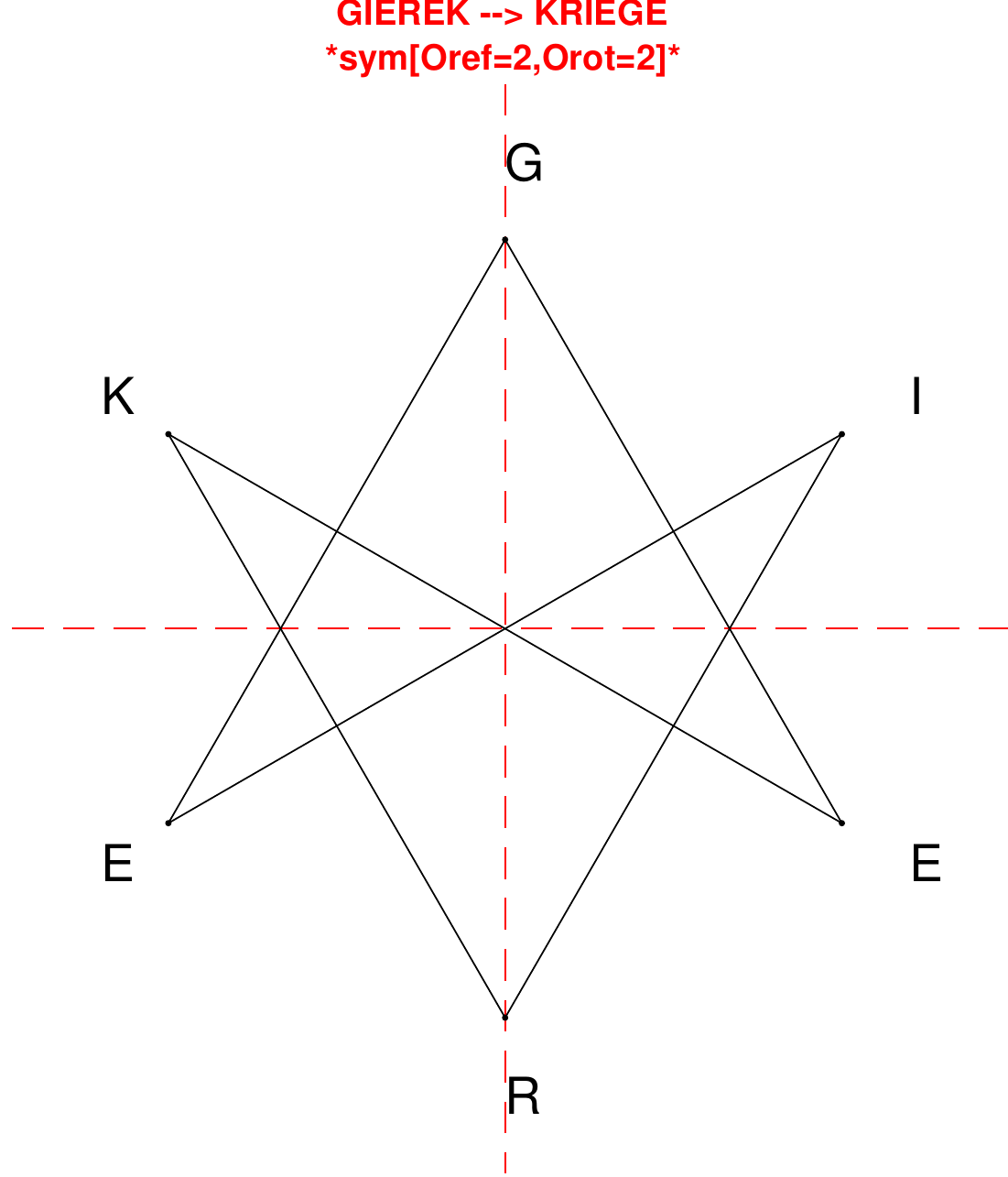}
\end{subfigure}
\end{figure}

\begin{figure}[H]
\centering
\begin{subfigure}[T]{0.19\textwidth}
\centering
\includegraphics[width=\textwidth]{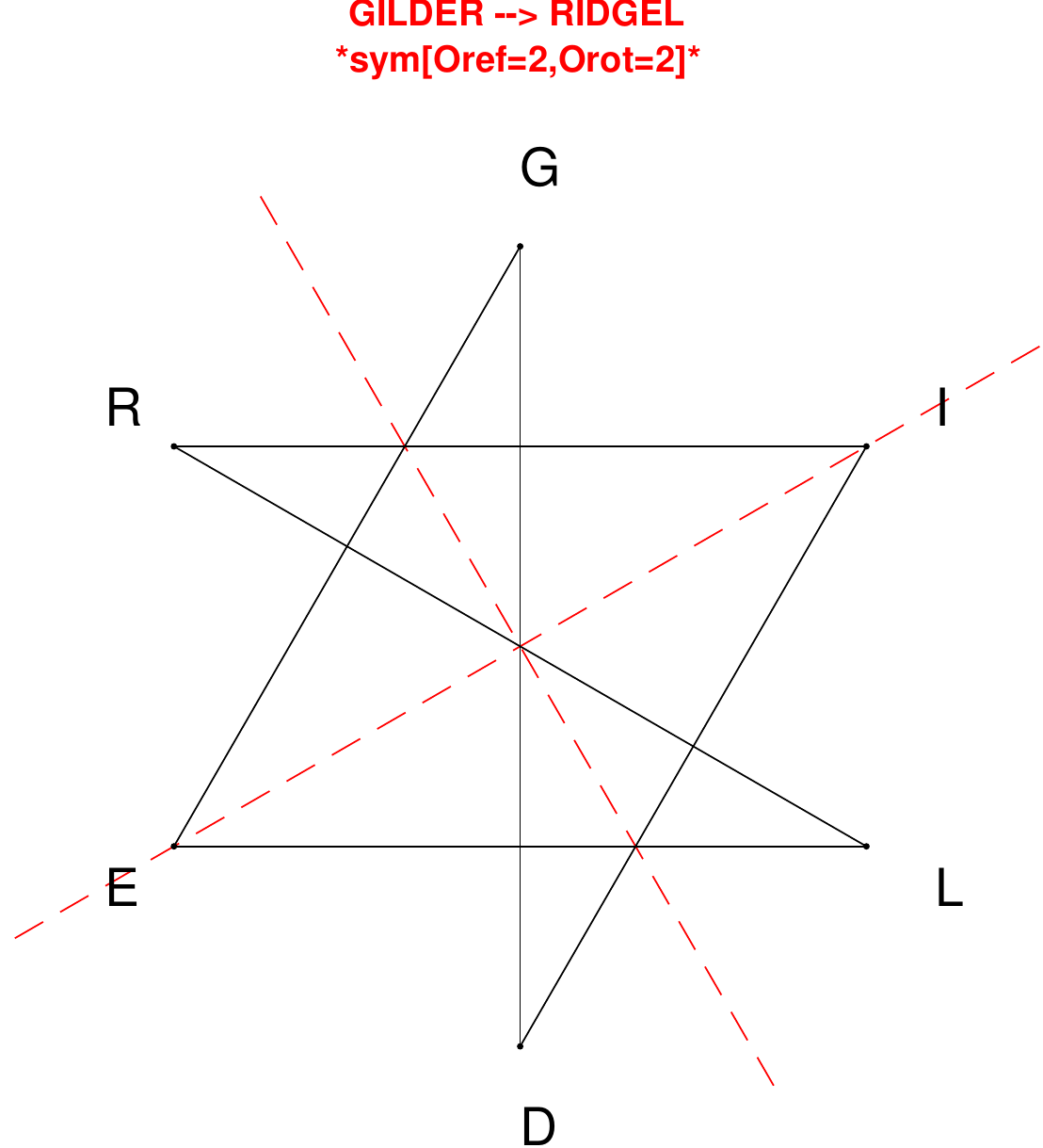}
\end{subfigure}
\hfill
\begin{subfigure}[T]{0.19\textwidth}
\centering
\includegraphics[width=\textwidth]{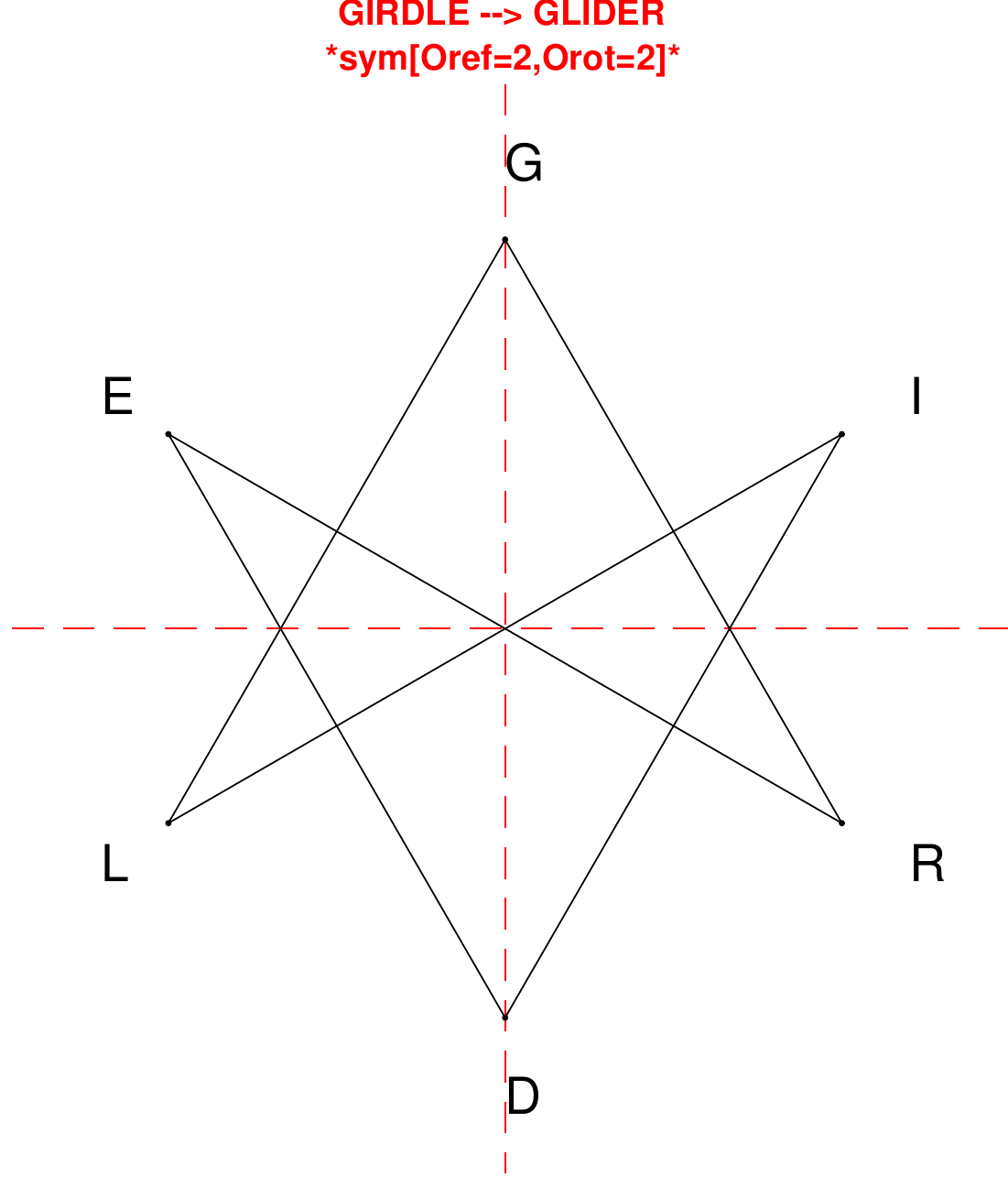}
\end{subfigure}
\hfill
\begin{subfigure}[T]{0.19\textwidth}
\centering
\includegraphics[width=\textwidth]{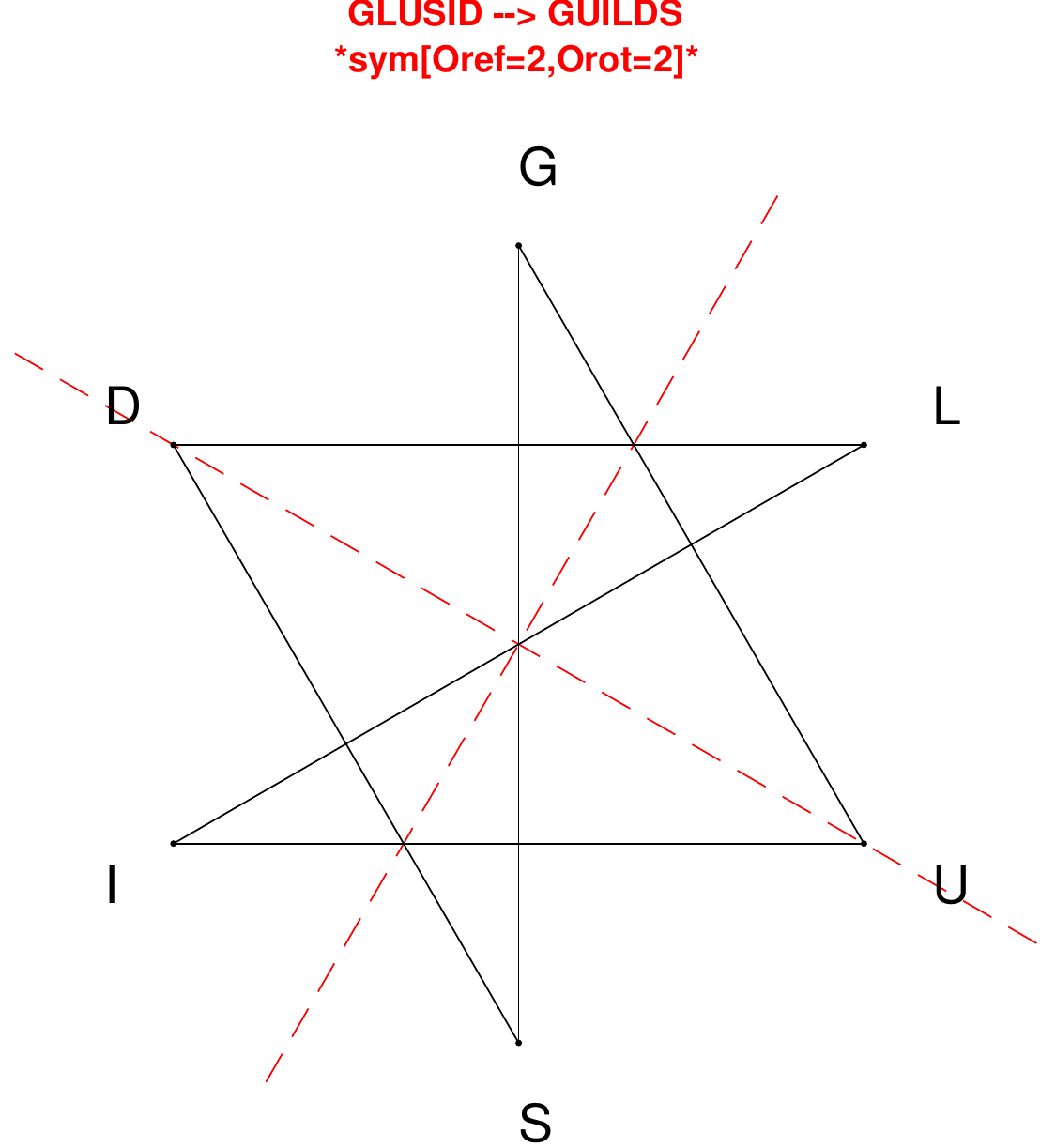}
\end{subfigure}
\hfill
\begin{subfigure}[T]{0.19\textwidth}
\centering
\includegraphics[width=\textwidth]{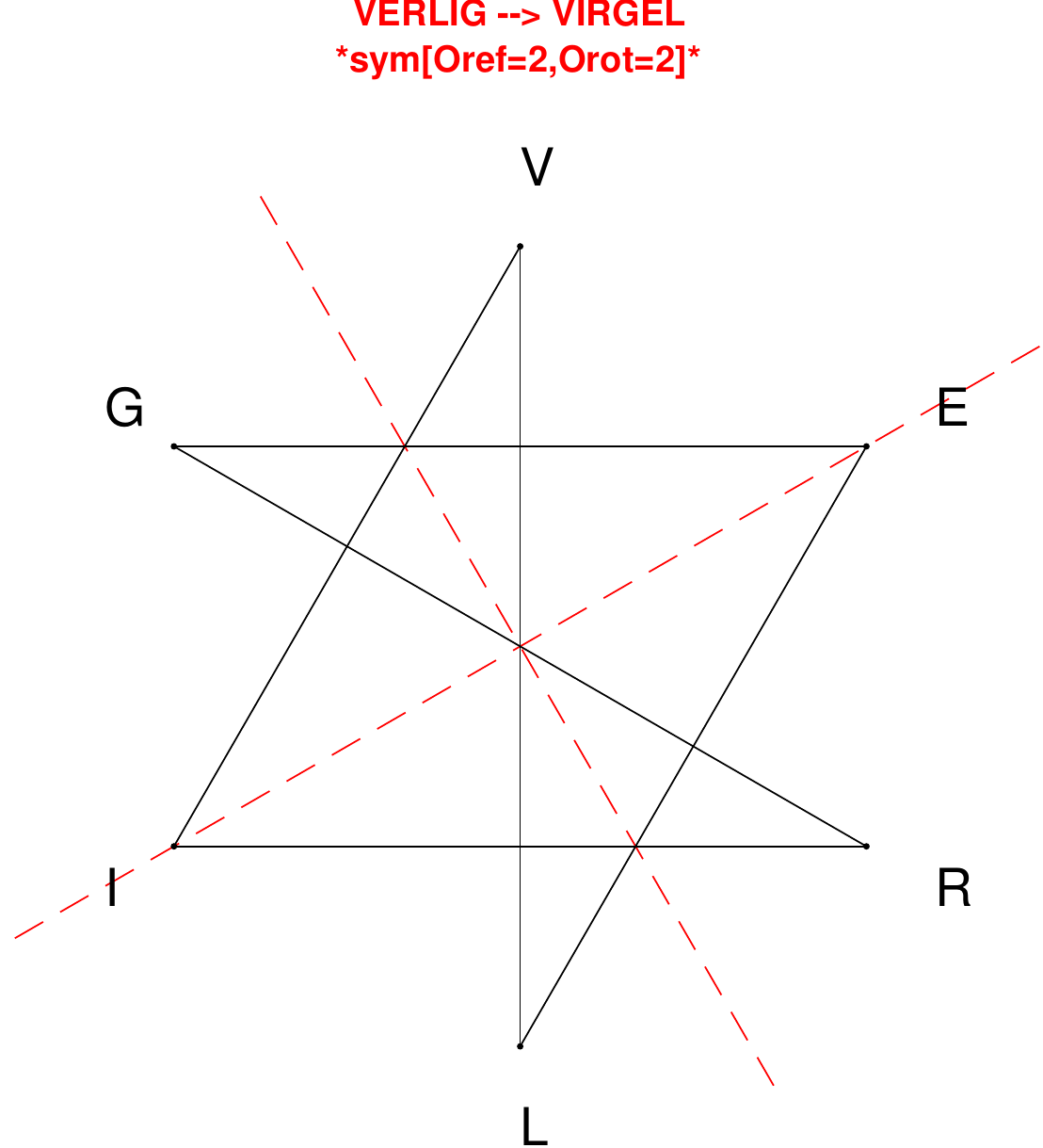}
\end{subfigure}
\hfill
\begin{subfigure}[T]{0.19\textwidth}
\centering
\includegraphics[width=\textwidth]{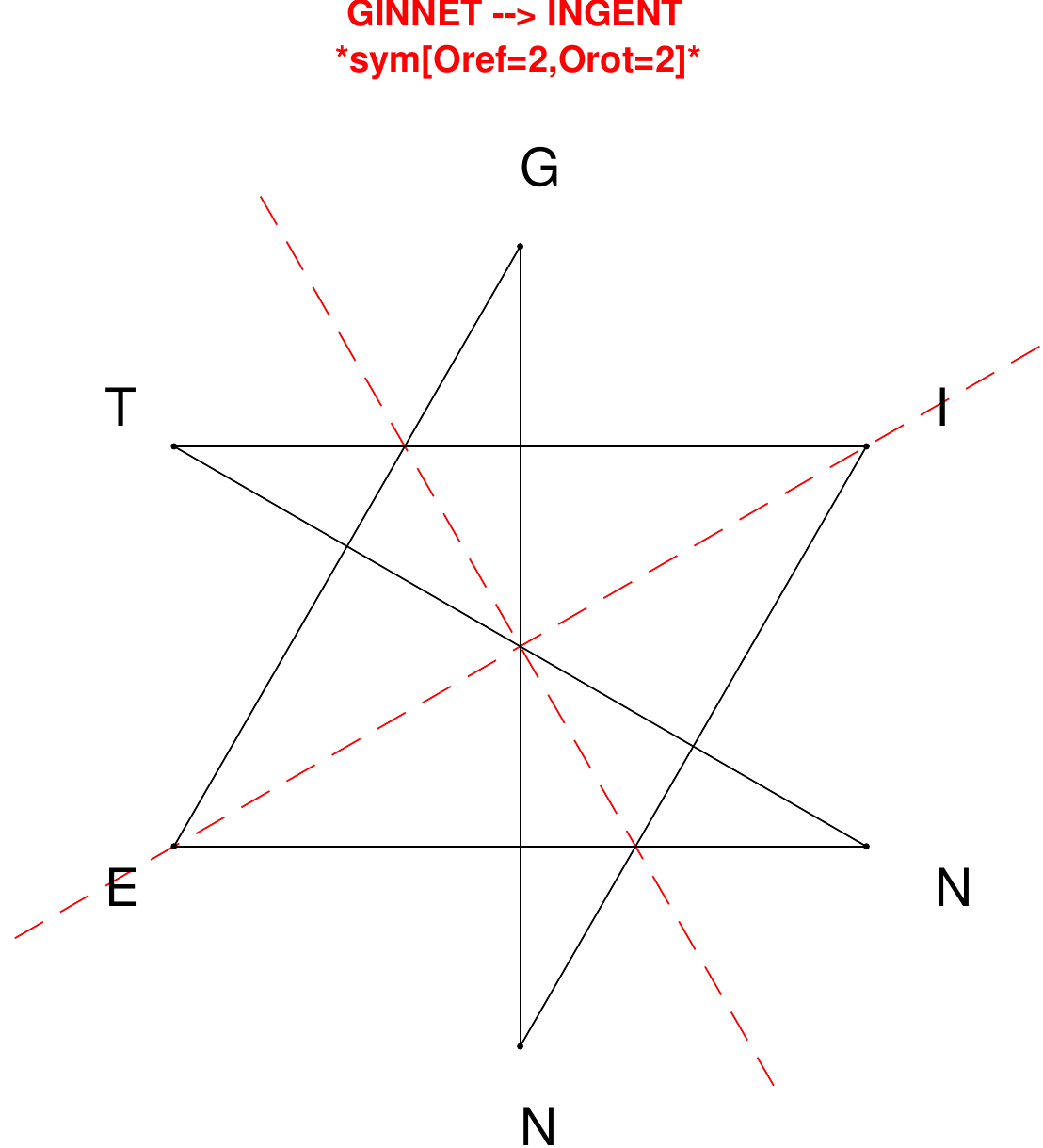}
\end{subfigure}
\end{figure}

\begin{figure}[H]
\centering
\begin{subfigure}[T]{0.19\textwidth}
\centering
\includegraphics[width=\textwidth]{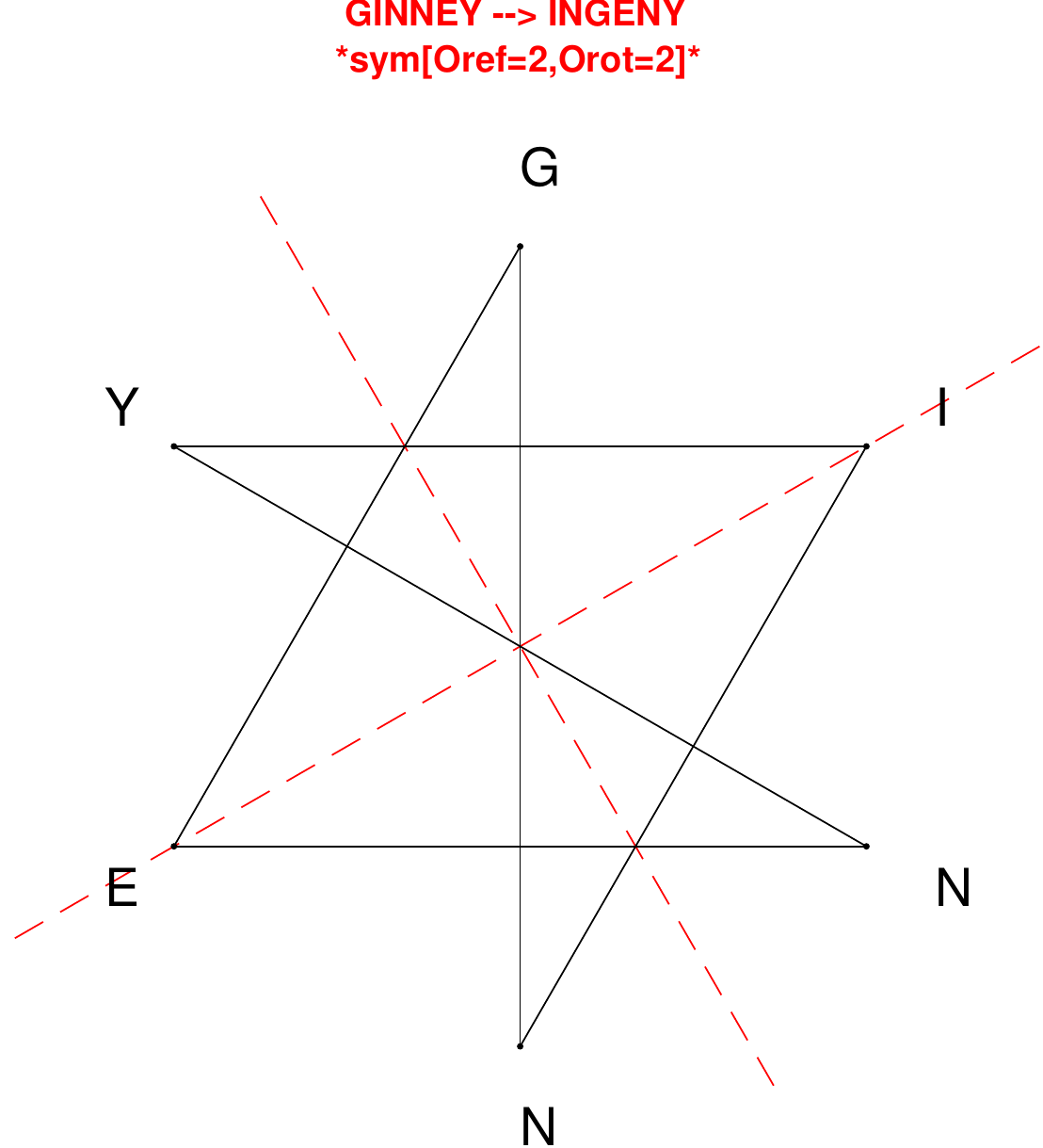}
\end{subfigure}
\hfill
\begin{subfigure}[T]{0.19\textwidth}
\centering
\includegraphics[width=\textwidth]{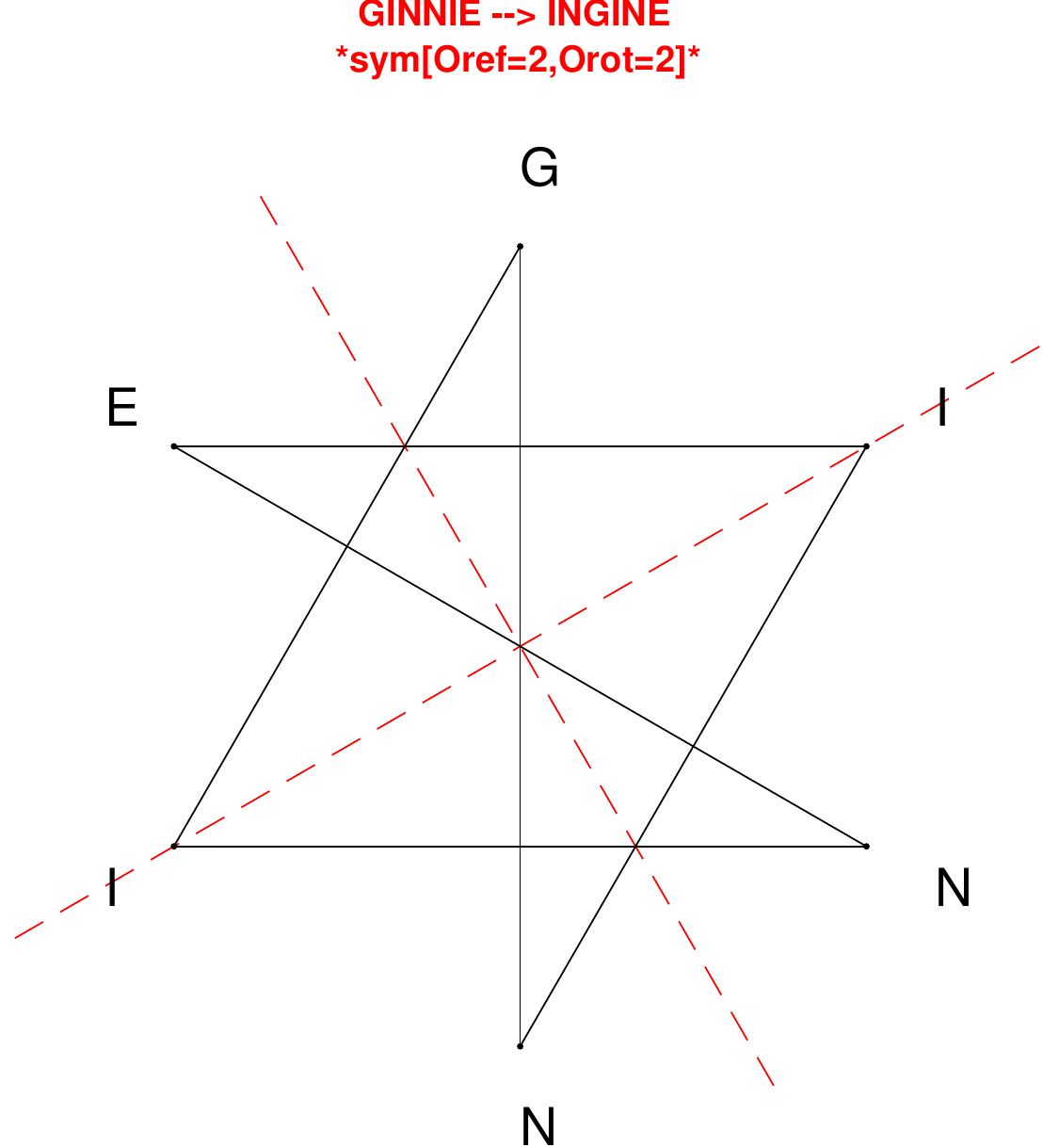}
\end{subfigure}
\hfill
\begin{subfigure}[T]{0.19\textwidth}
\centering
\includegraphics[width=\textwidth]{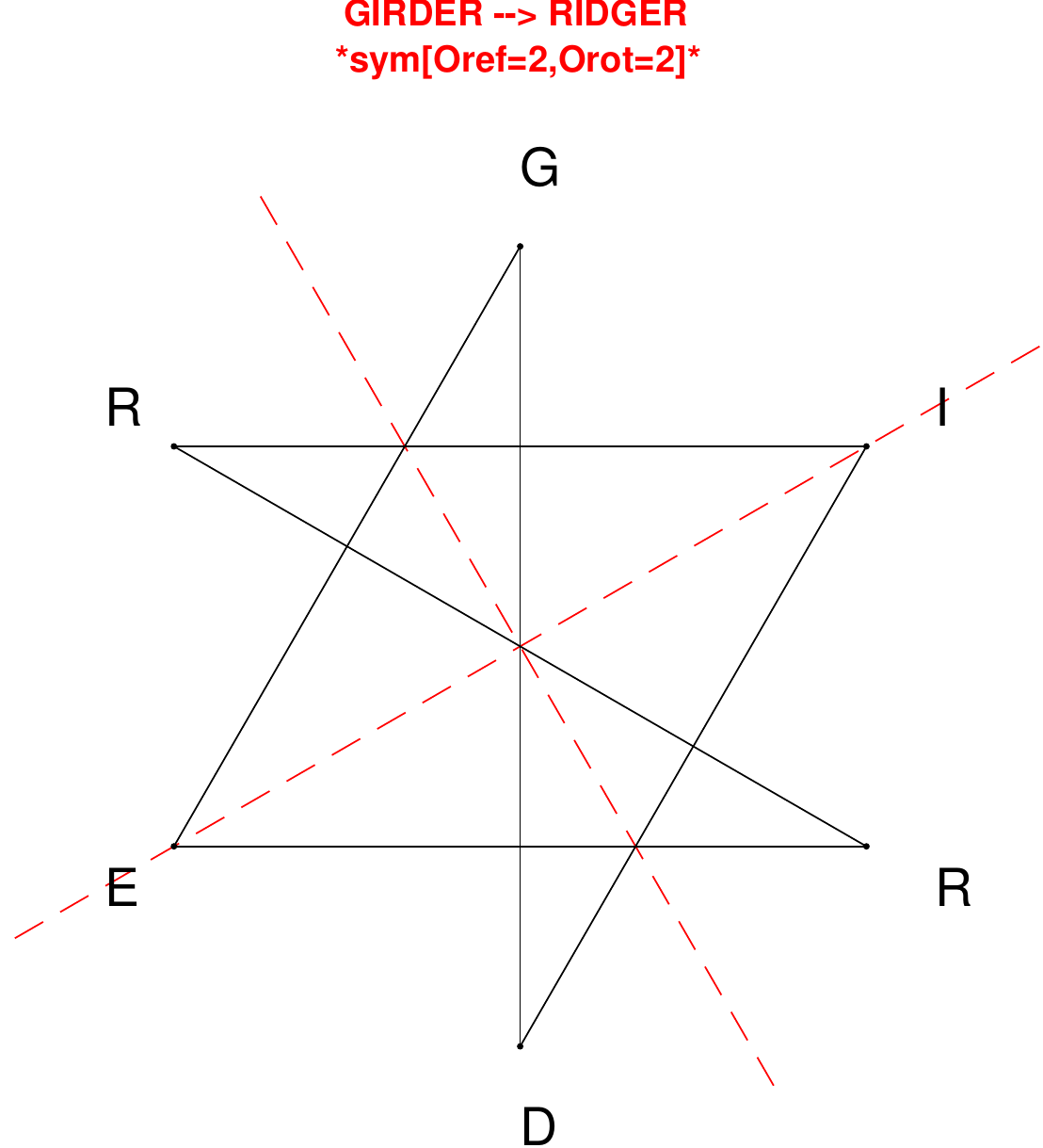}
\end{subfigure}
\hfill
\begin{subfigure}[T]{0.19\textwidth}
\centering
\includegraphics[width=\textwidth]{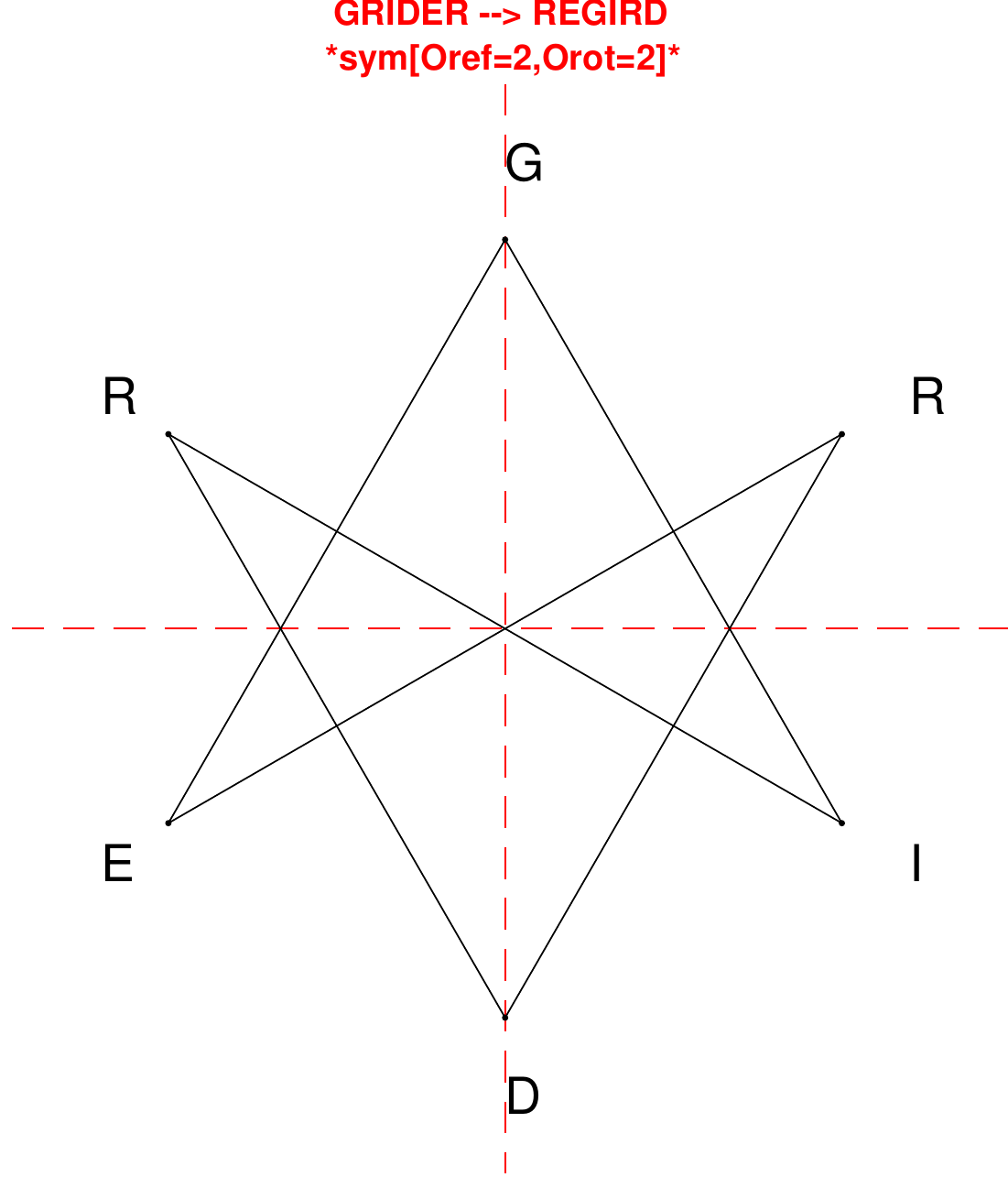}
\end{subfigure}
\hfill
\begin{subfigure}[T]{0.19\textwidth}
\centering
\includegraphics[width=\textwidth]{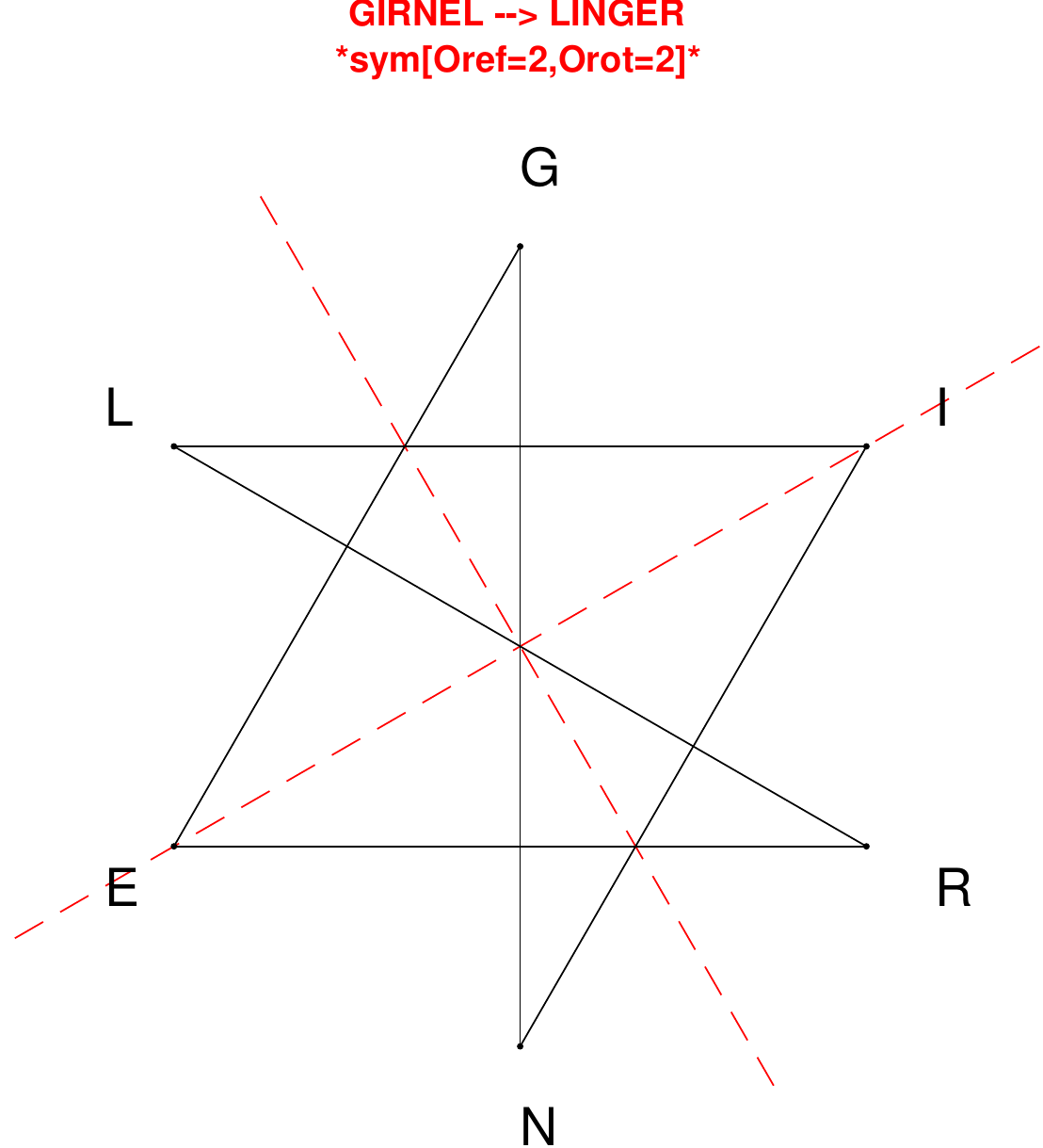}
\end{subfigure}
\end{figure}

\begin{figure}[H]
\centering
\begin{subfigure}[T]{0.19\textwidth}
\centering
\includegraphics[width=\textwidth]{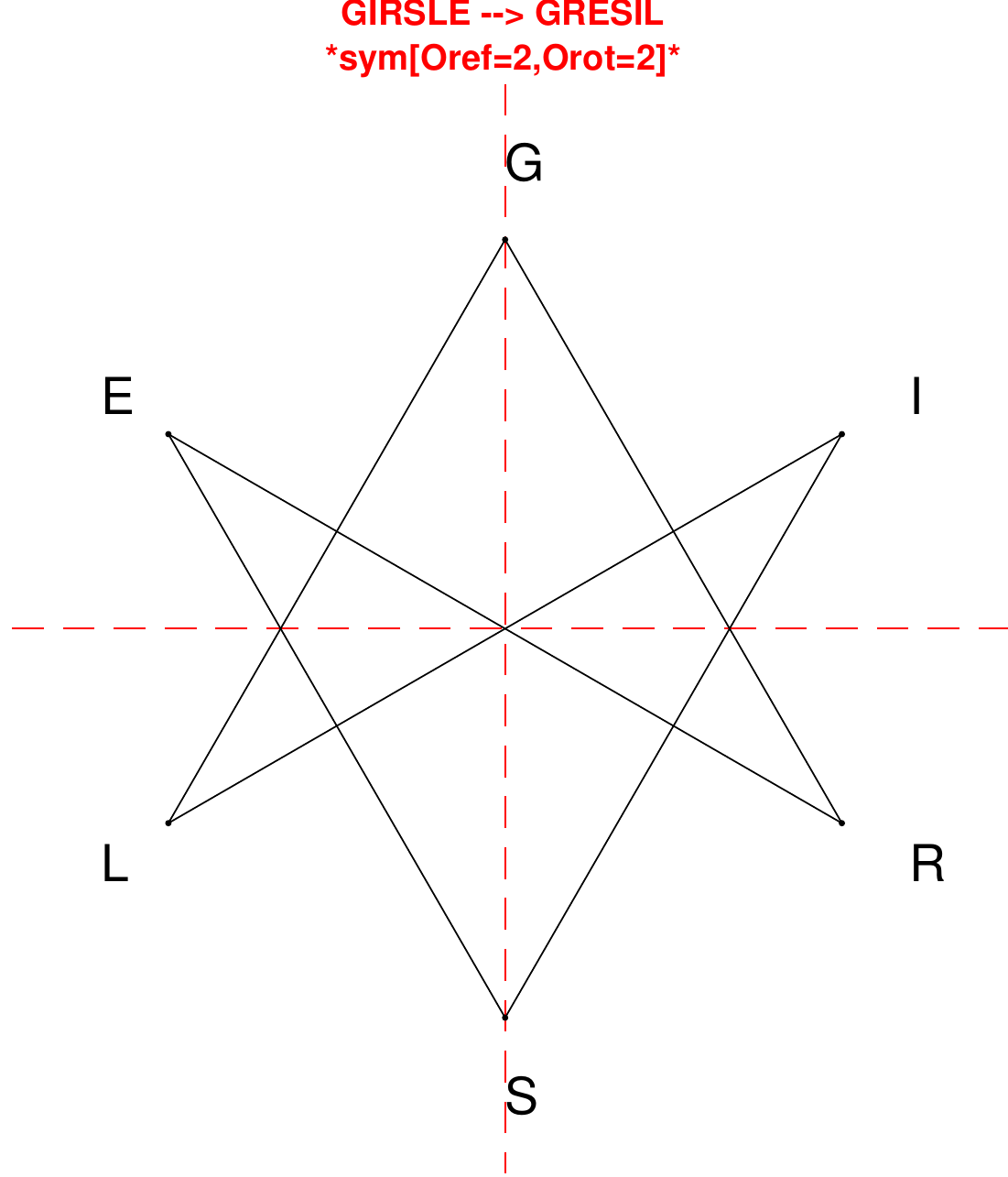}
\end{subfigure}
\hfill
\begin{subfigure}[T]{0.19\textwidth}
\centering
\includegraphics[width=\textwidth]{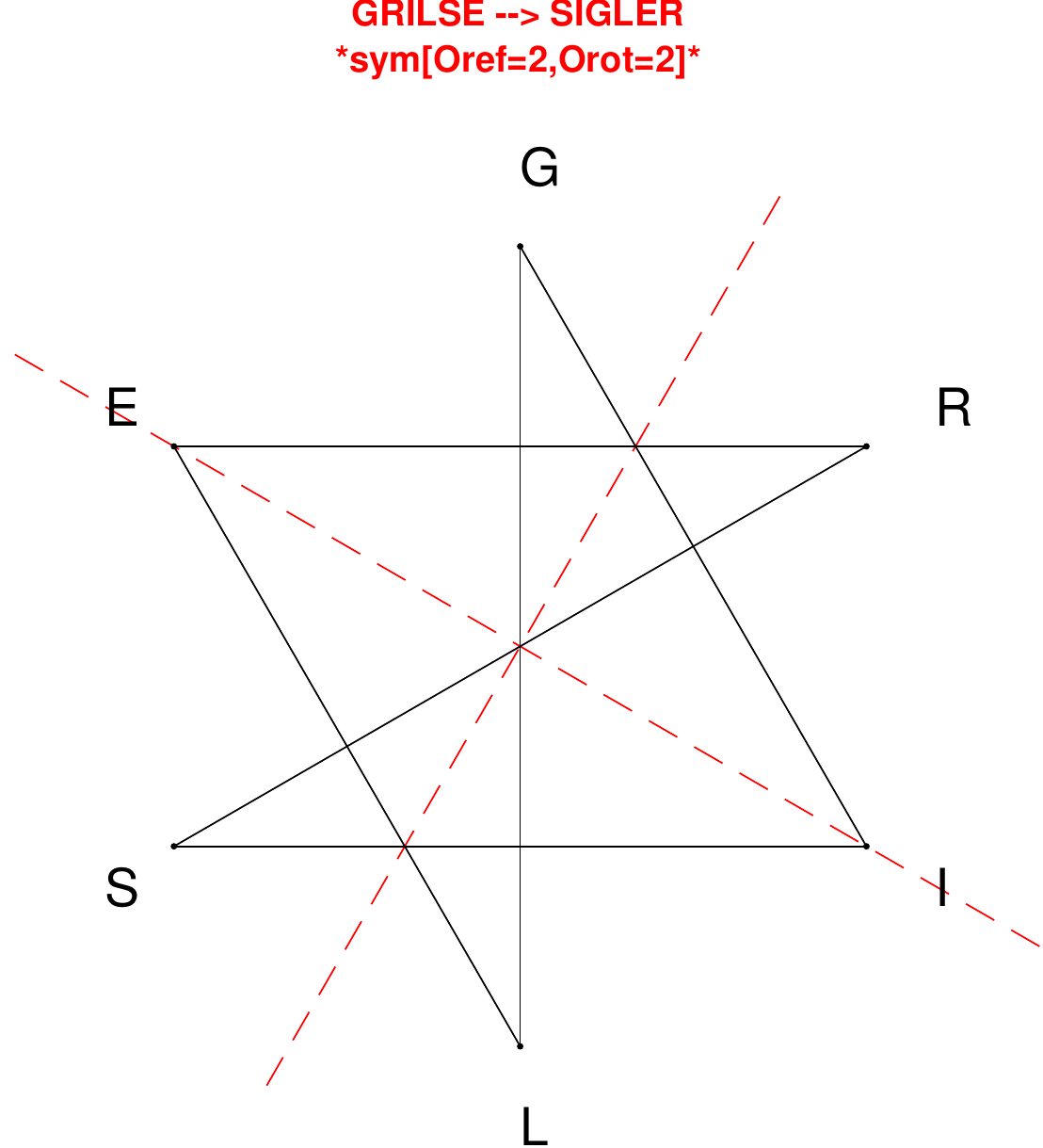}
\end{subfigure}
\hfill
\begin{subfigure}[T]{0.19\textwidth}
\centering
\includegraphics[width=\textwidth]{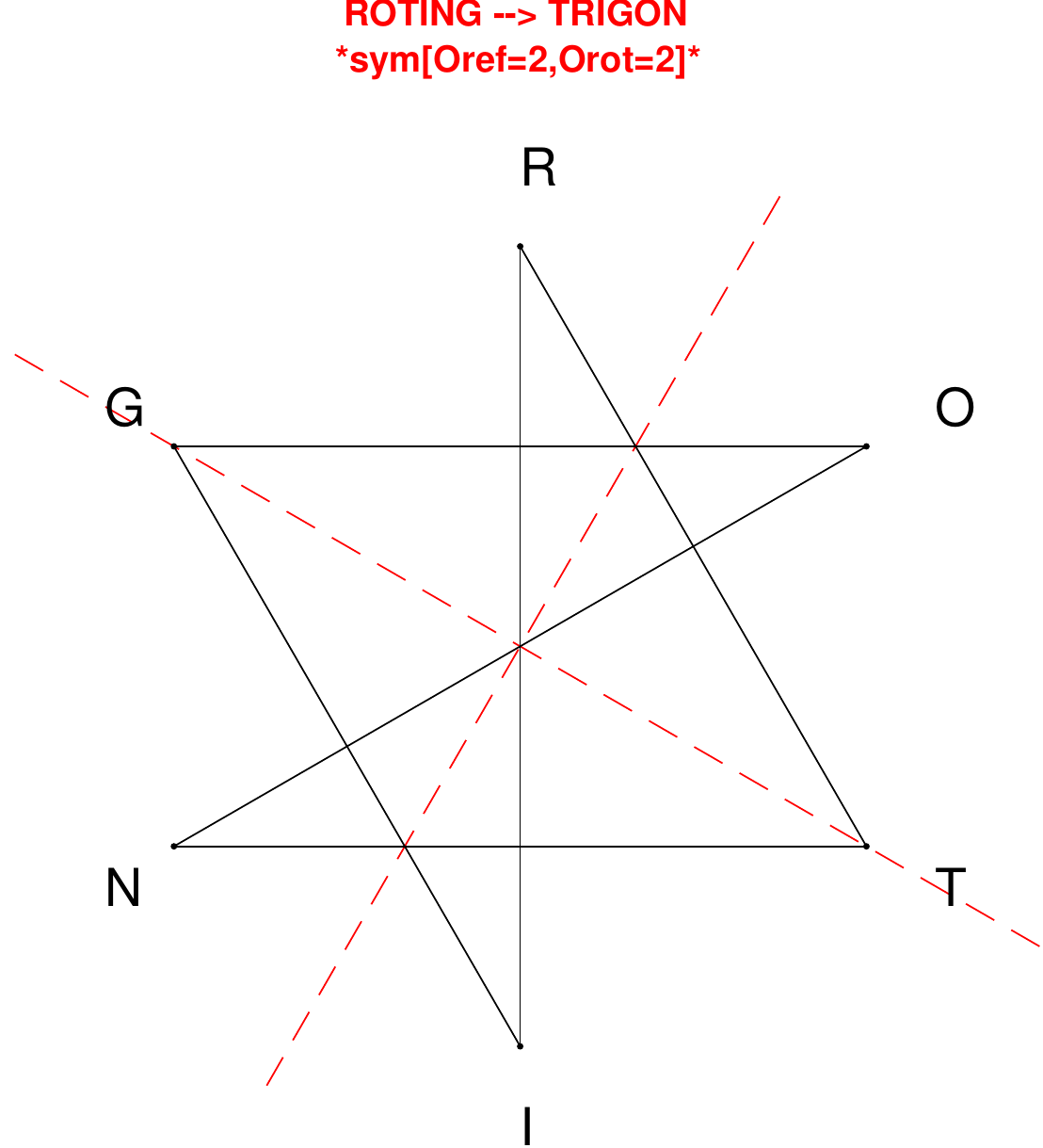}
\end{subfigure}
\hfill
\begin{subfigure}[T]{0.19\textwidth}
\centering
\includegraphics[width=\textwidth]{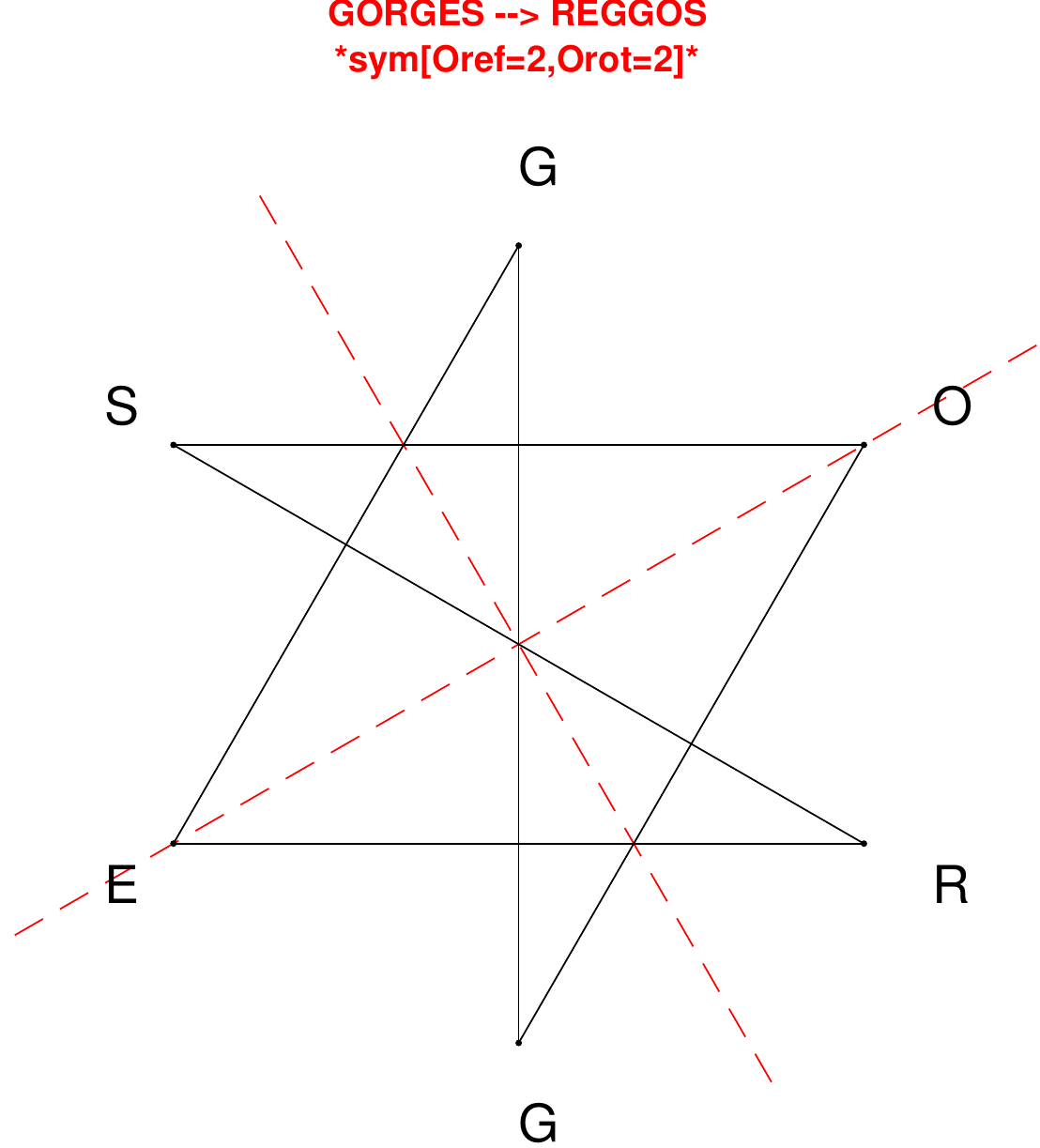}
\end{subfigure}
\hfill
\begin{subfigure}[T]{0.19\textwidth}
\centering
\includegraphics[width=\textwidth]{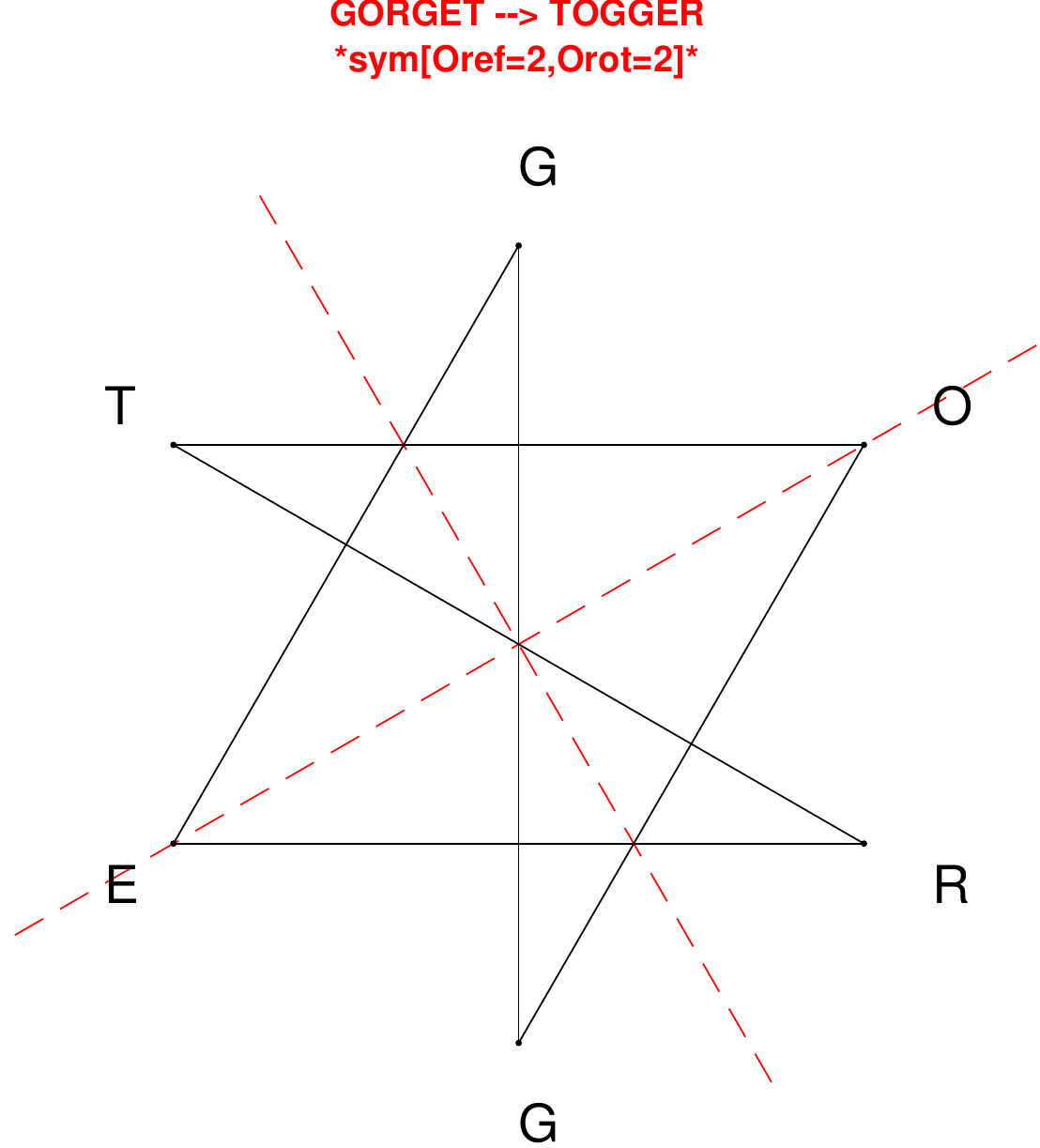}
\end{subfigure}
\end{figure}

\begin{figure}[H]
\centering
\begin{subfigure}[T]{0.19\textwidth}
\centering
\includegraphics[width=\textwidth]{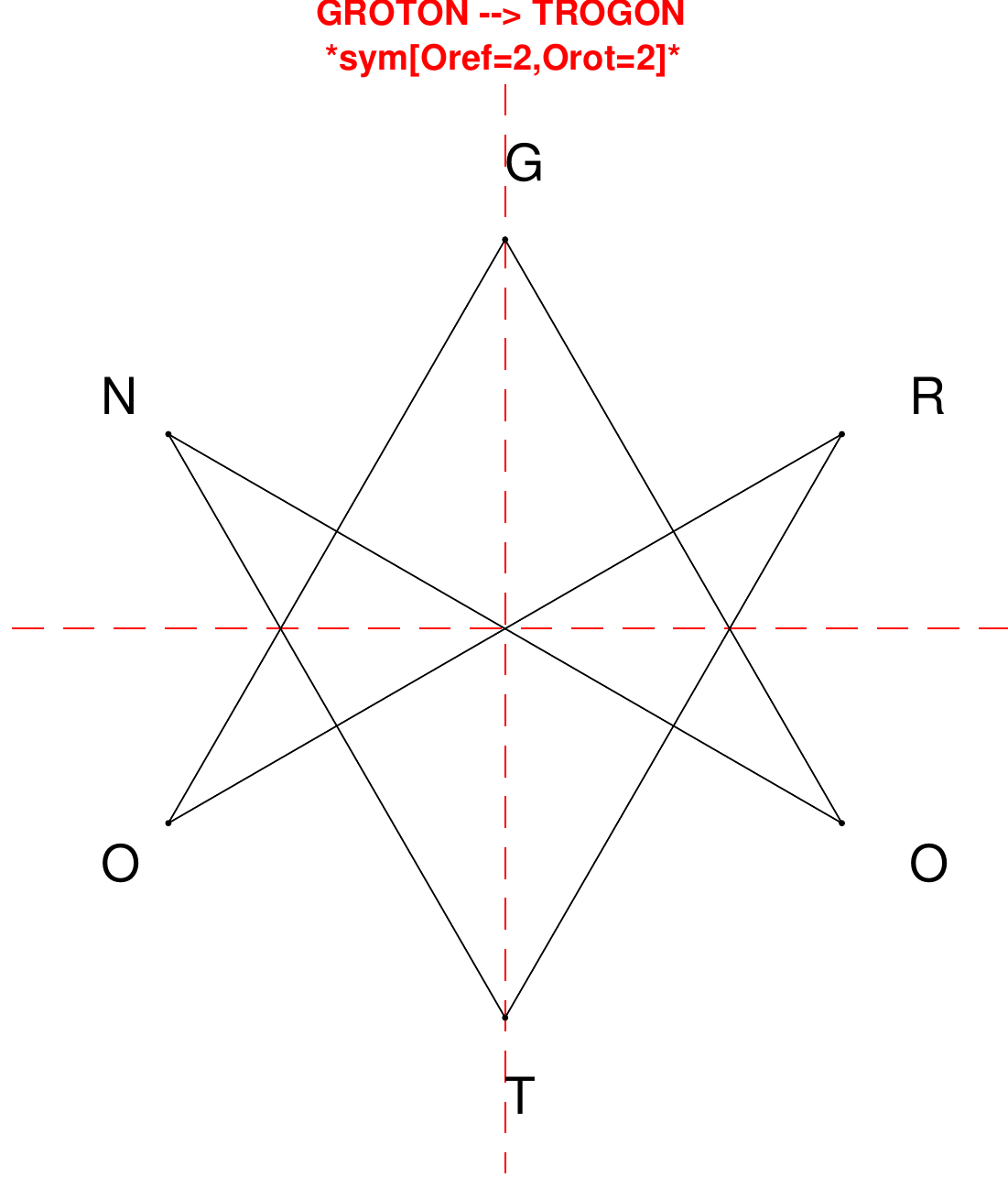}
\end{subfigure}
\hfill
\begin{subfigure}[T]{0.19\textwidth}
\centering
\includegraphics[width=\textwidth]{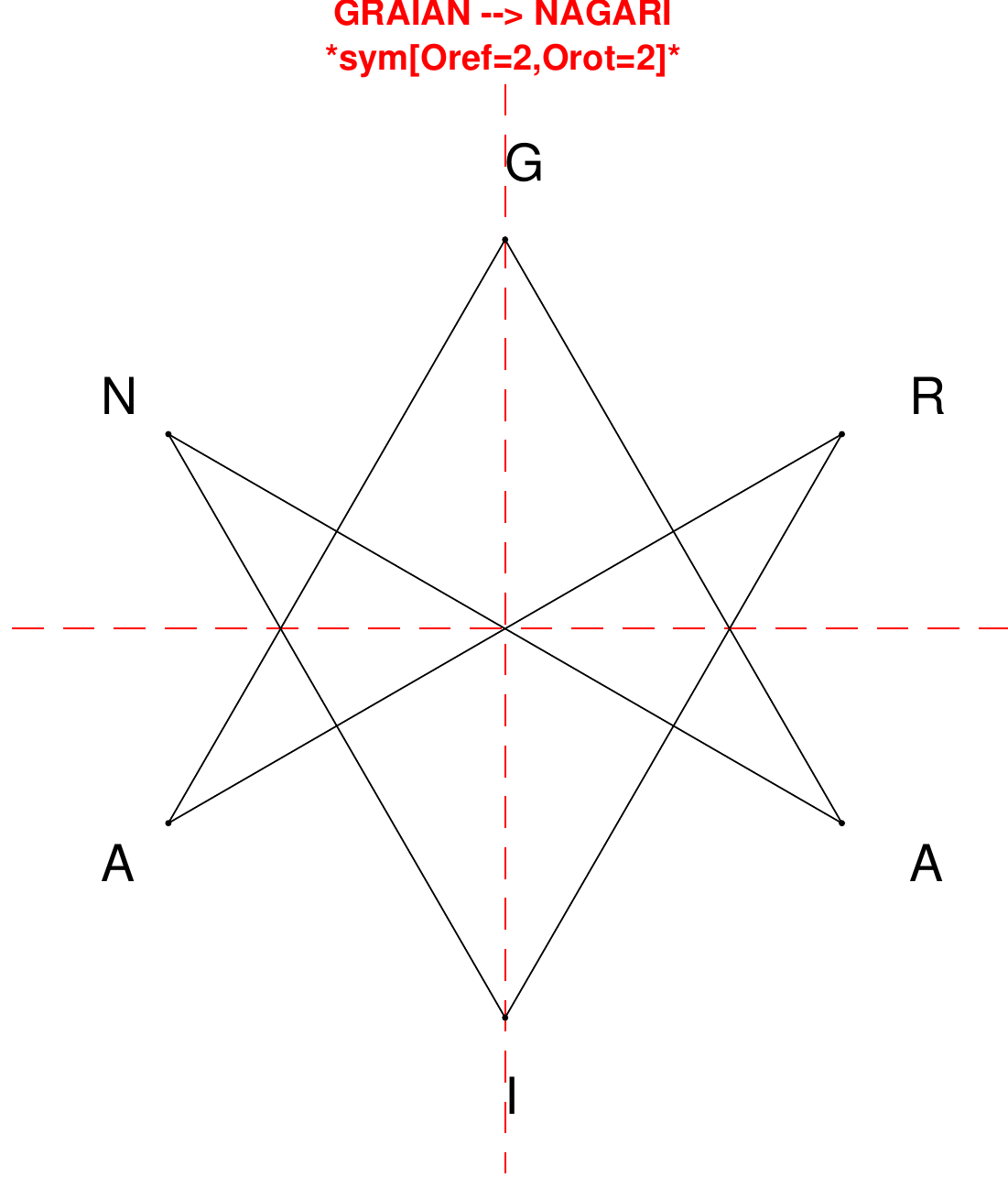}
\end{subfigure}
\hfill
\begin{subfigure}[T]{0.19\textwidth}
\centering
\includegraphics[width=\textwidth]{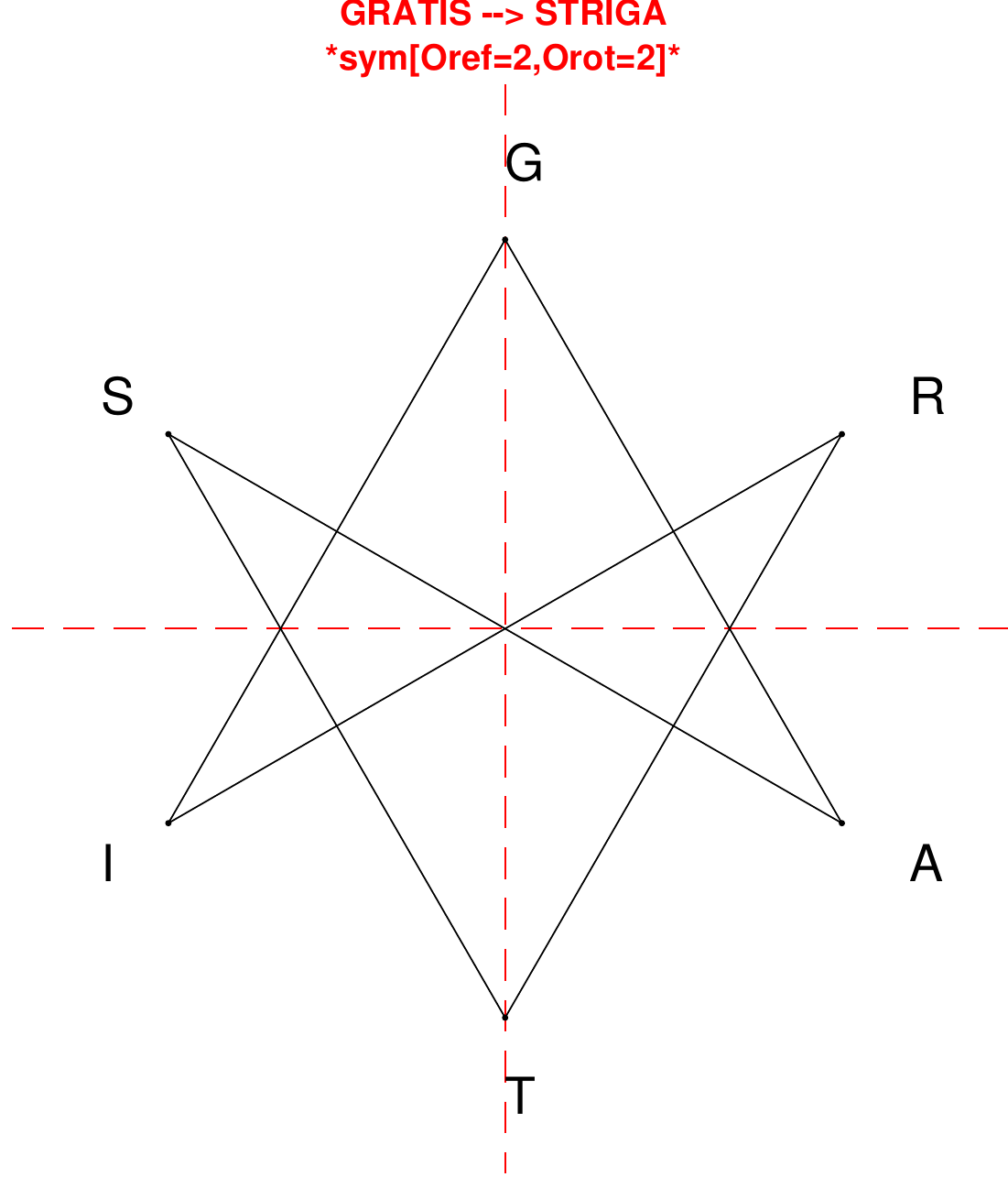}
\end{subfigure}
\hfill
\begin{subfigure}[T]{0.19\textwidth}
\centering
\includegraphics[width=\textwidth]{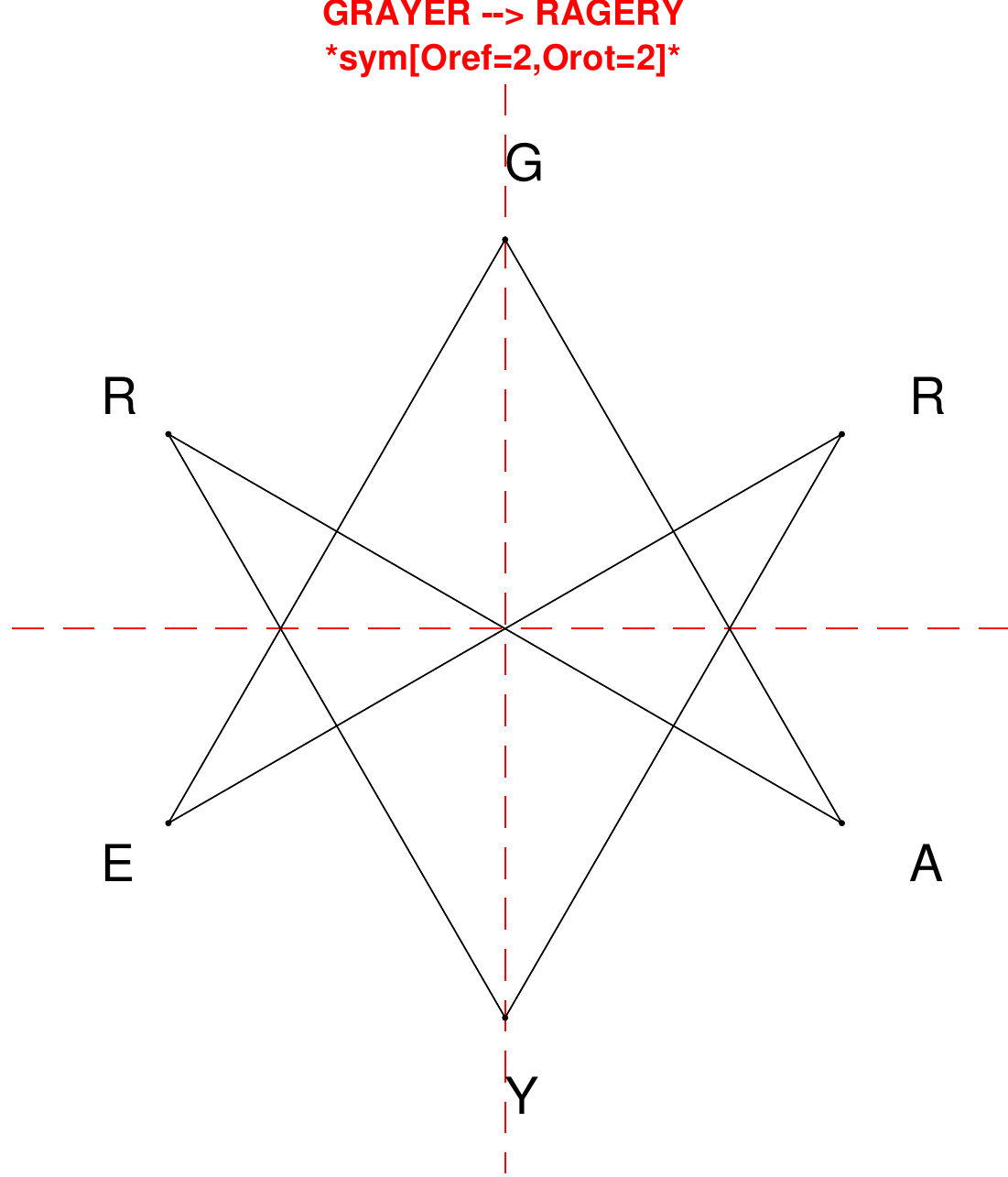}
\end{subfigure}
\hfill
\begin{subfigure}[T]{0.19\textwidth}
\centering
\includegraphics[width=\textwidth]{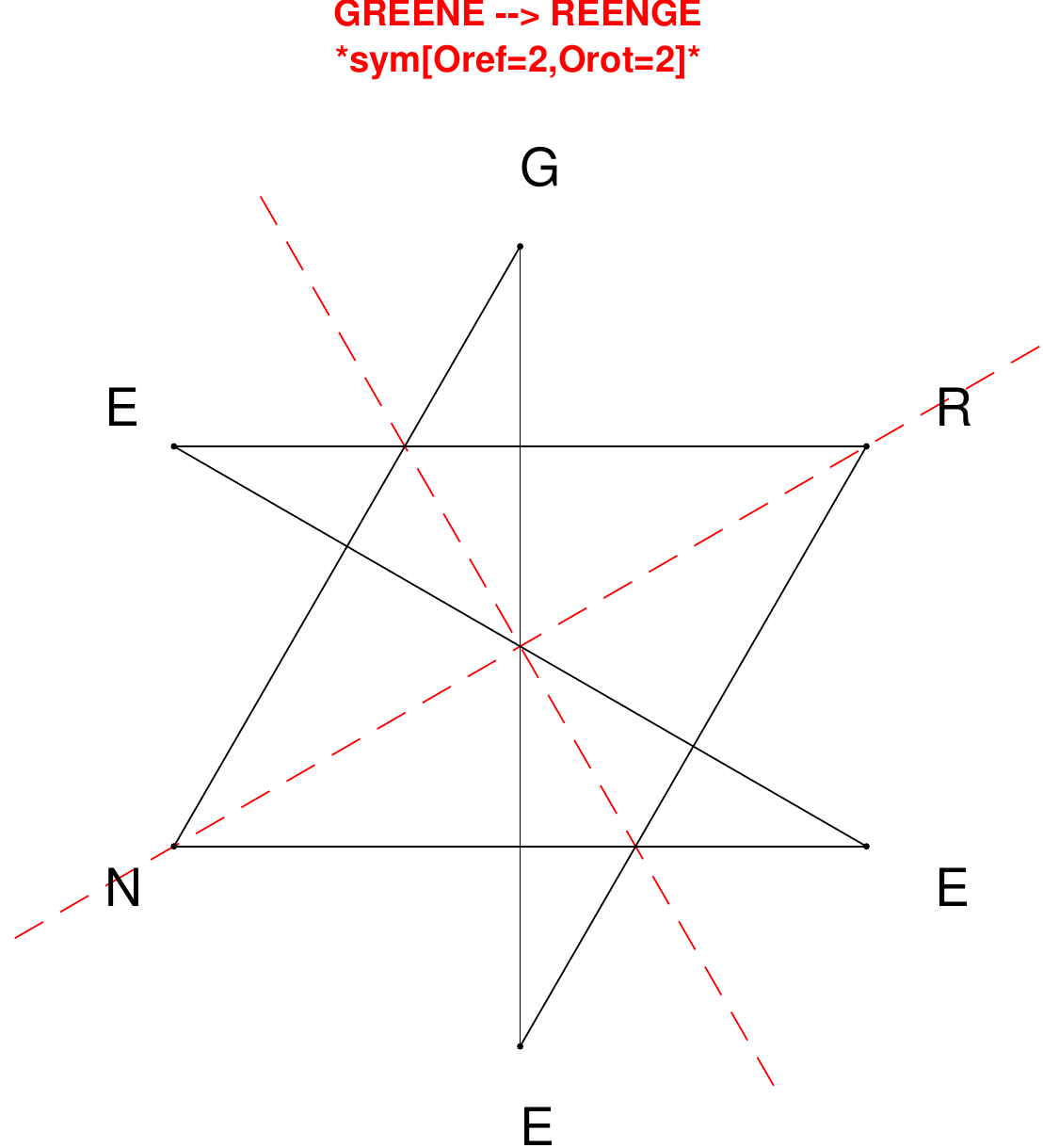}
\end{subfigure}
\end{figure}

\begin{figure}[H]
\centering
\begin{subfigure}[T]{0.19\textwidth}
\centering
\includegraphics[width=\textwidth]{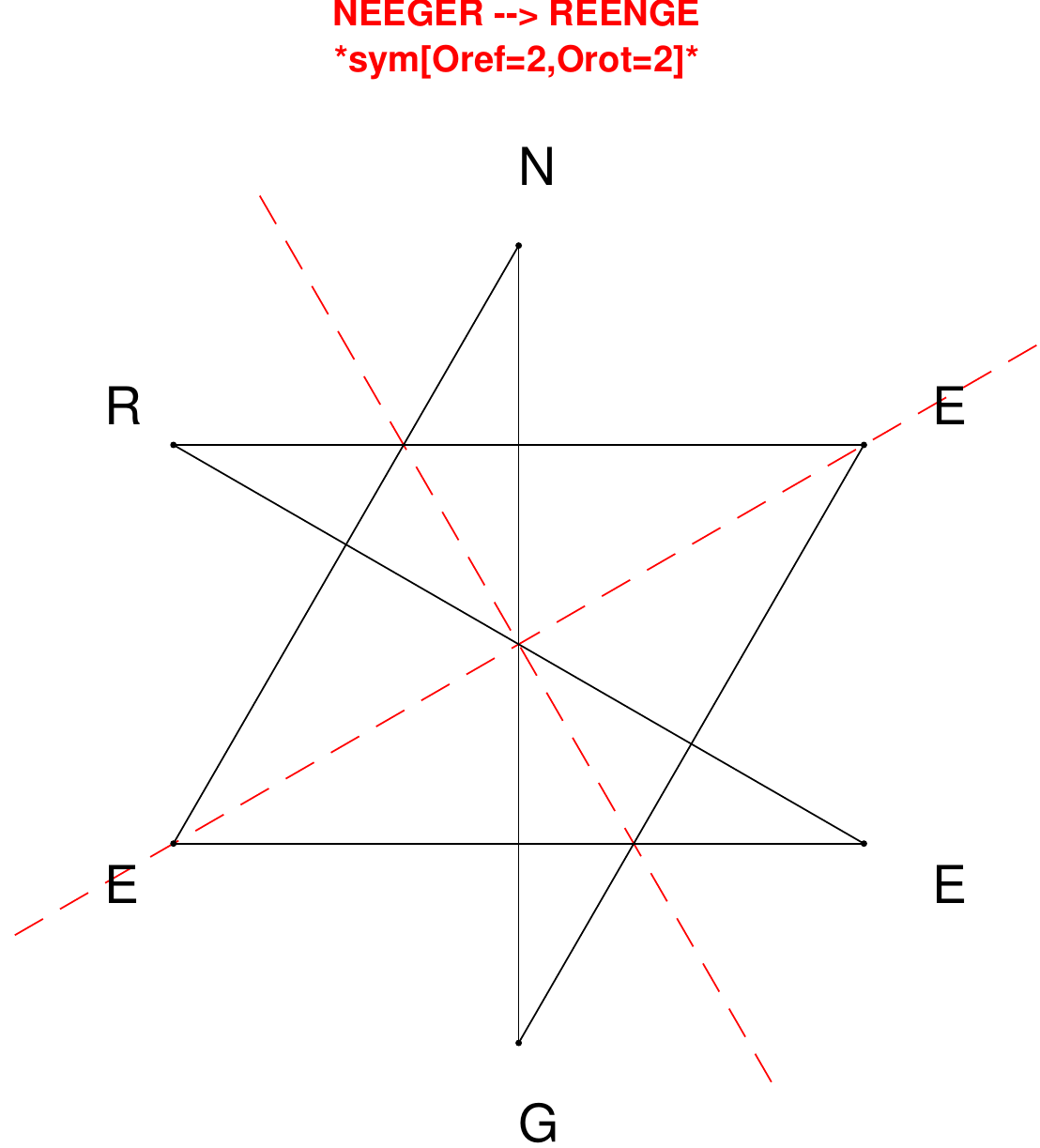}
\end{subfigure}
\hfill
\begin{subfigure}[T]{0.19\textwidth}
\centering
\includegraphics[width=\textwidth]{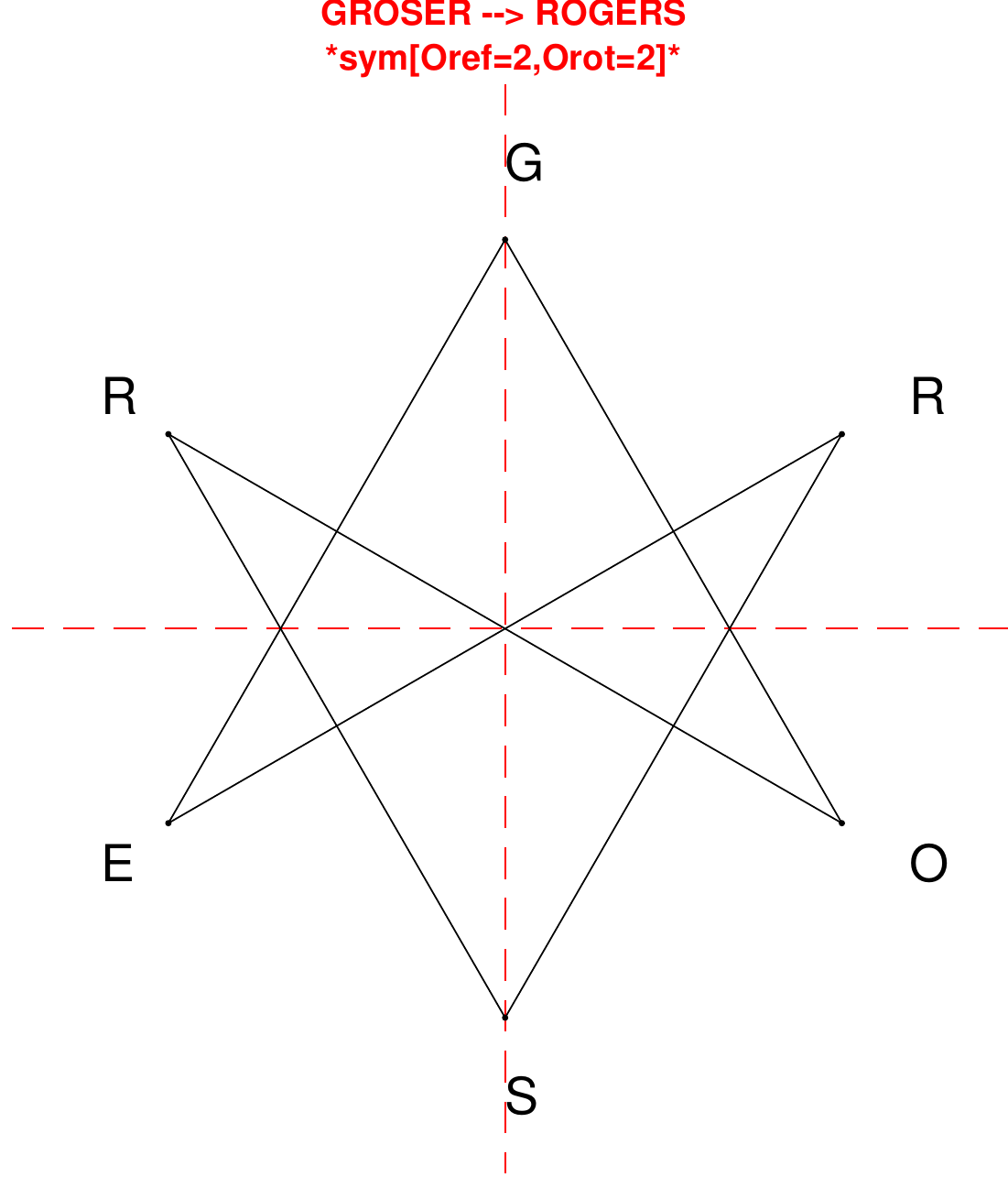}
\end{subfigure}
\hfill
\begin{subfigure}[T]{0.19\textwidth}
\centering
\includegraphics[width=\textwidth]{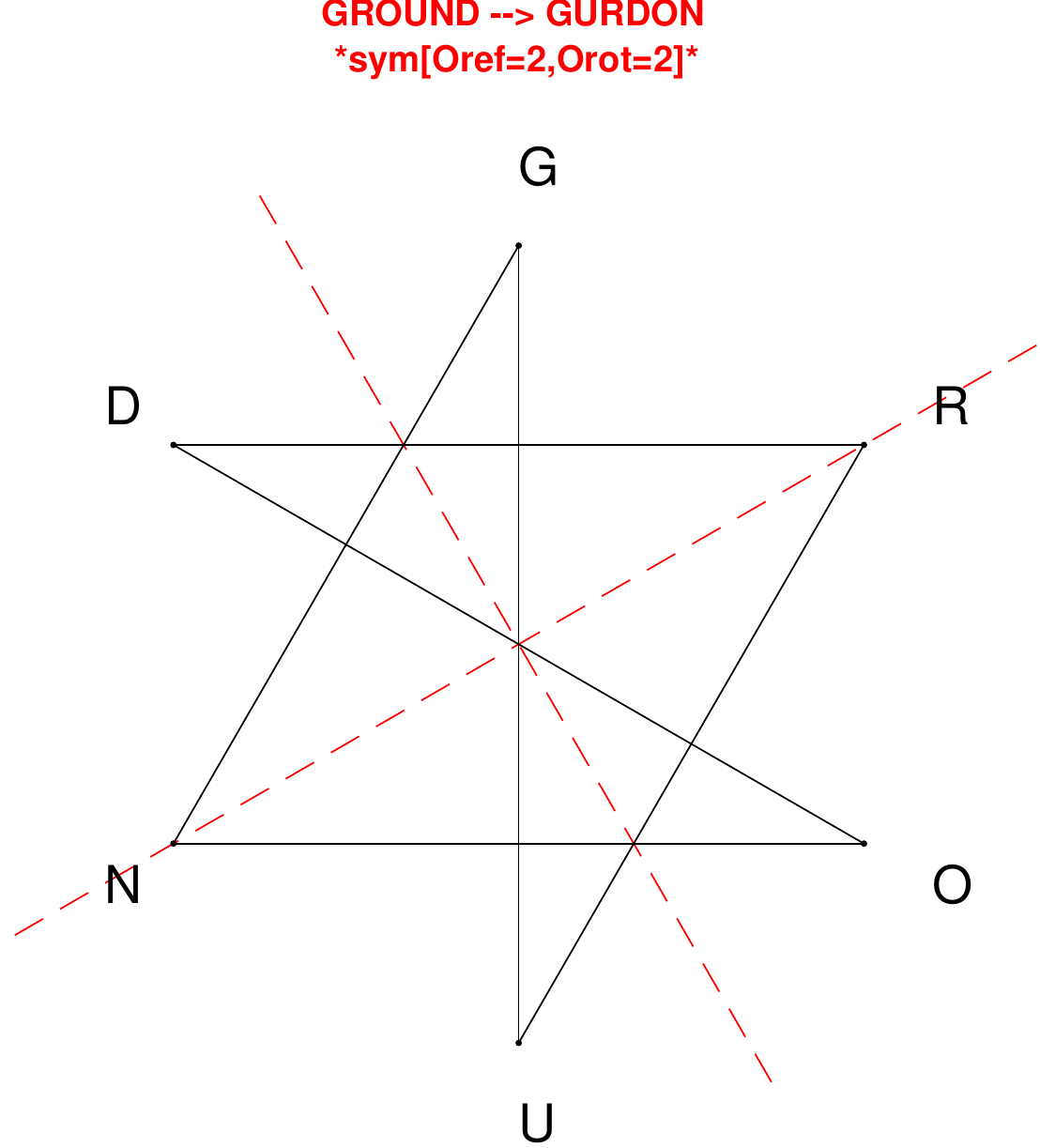}
\end{subfigure}
\hfill
\begin{subfigure}[T]{0.19\textwidth}
\centering
\includegraphics[width=\textwidth]{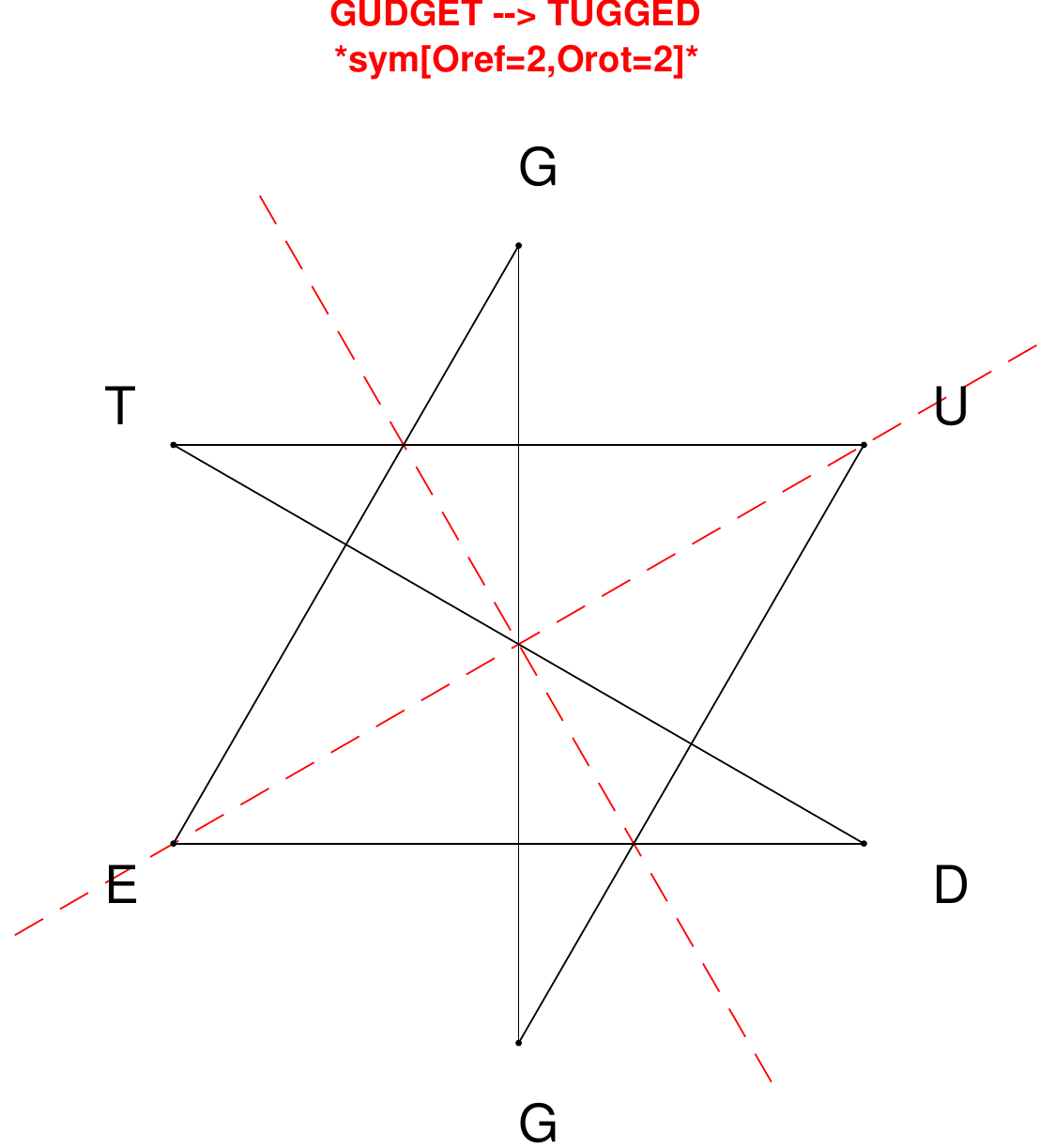}
\end{subfigure}
\hfill
\begin{subfigure}[T]{0.19\textwidth}
\centering
\includegraphics[width=\textwidth]{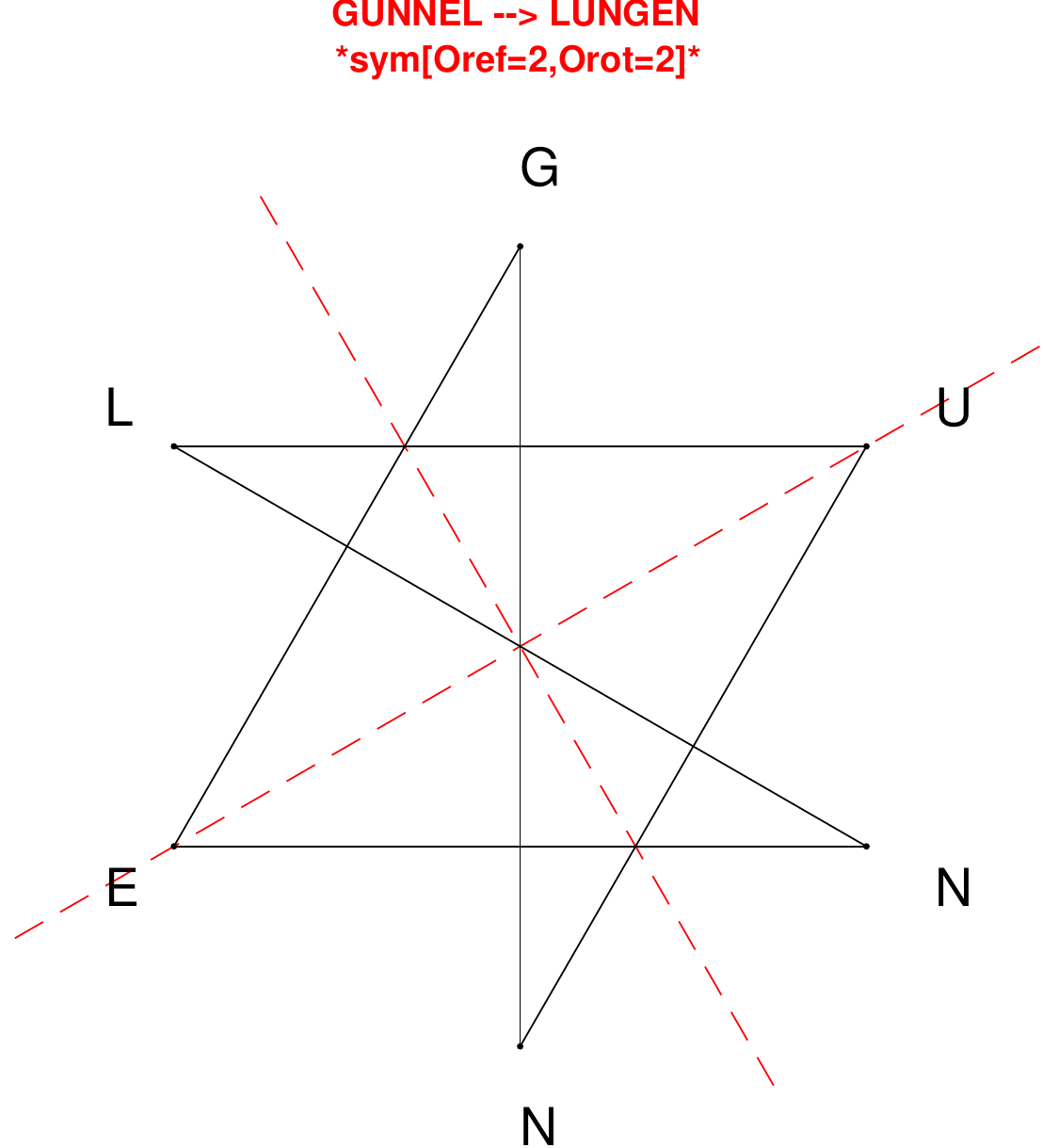}
\end{subfigure}
\end{figure}

\begin{figure}[H]
\centering
\begin{subfigure}[T]{0.19\textwidth}
\centering
\includegraphics[width=\textwidth]{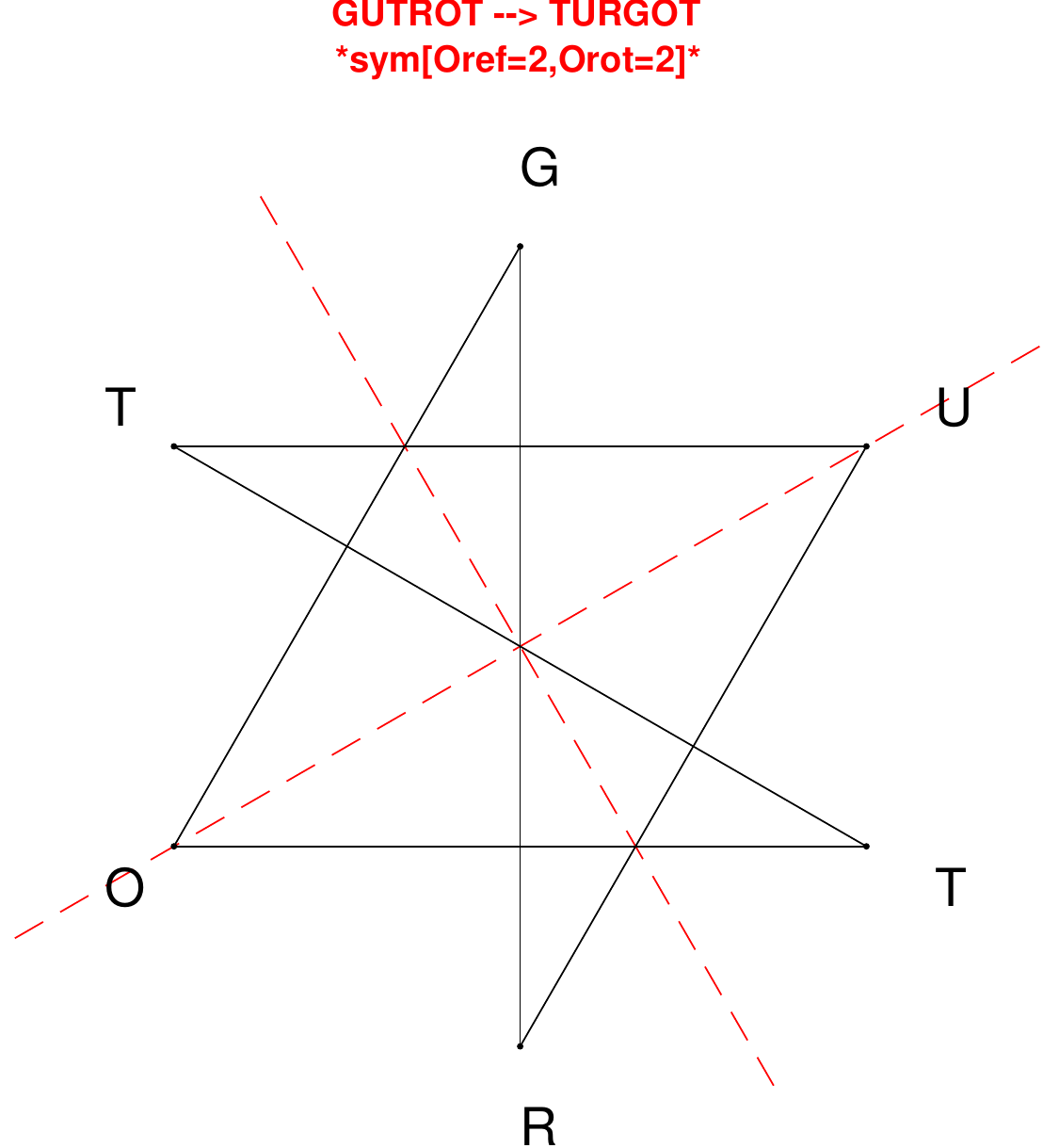}
\end{subfigure}
\hfill
\begin{subfigure}[T]{0.19\textwidth}
\centering
\includegraphics[width=\textwidth]{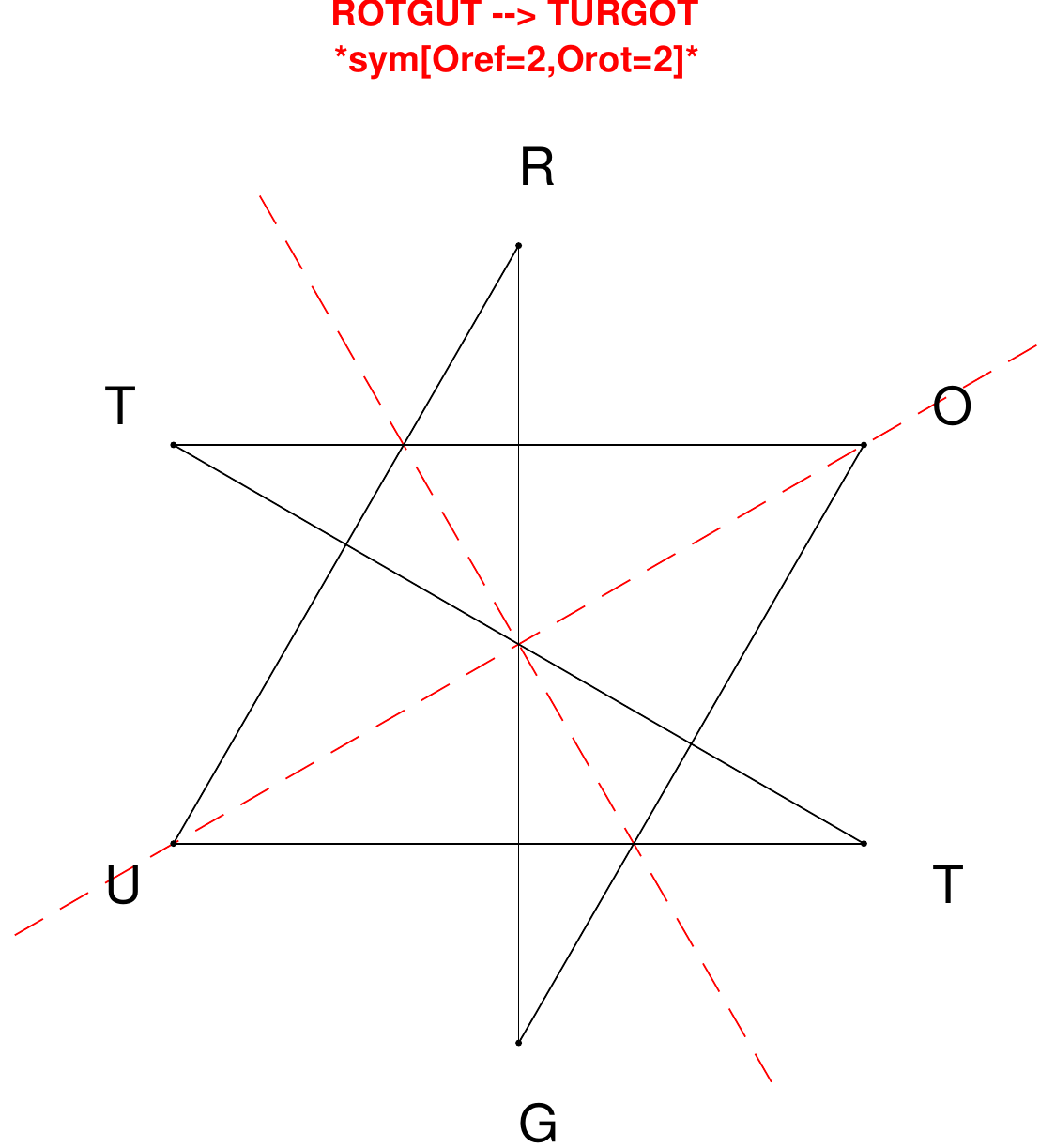}
\end{subfigure}
\hfill
\begin{subfigure}[T]{0.19\textwidth}
\centering
\includegraphics[width=\textwidth]{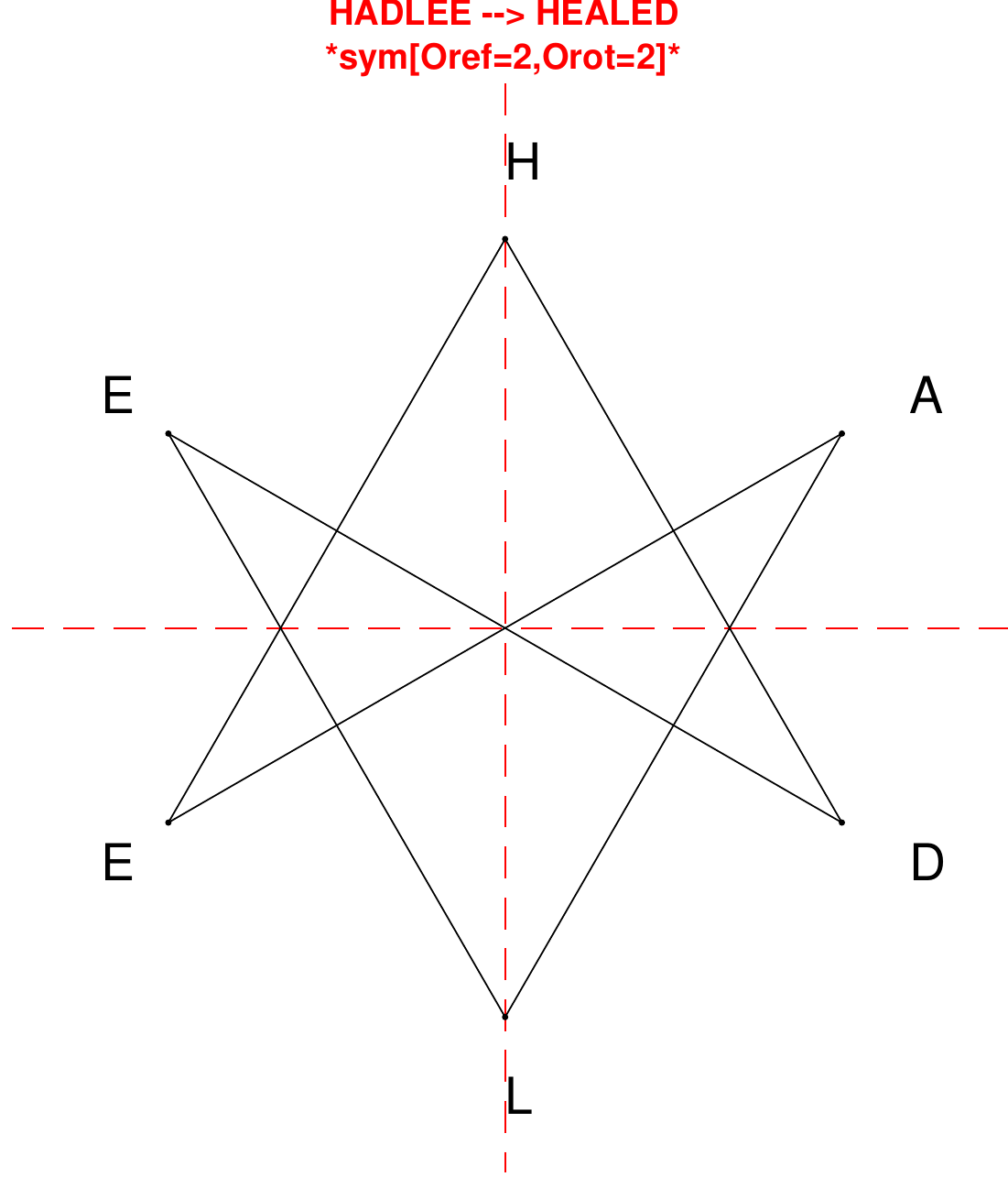}
\end{subfigure}
\hfill
\begin{subfigure}[T]{0.19\textwidth}
\centering
\includegraphics[width=\textwidth]{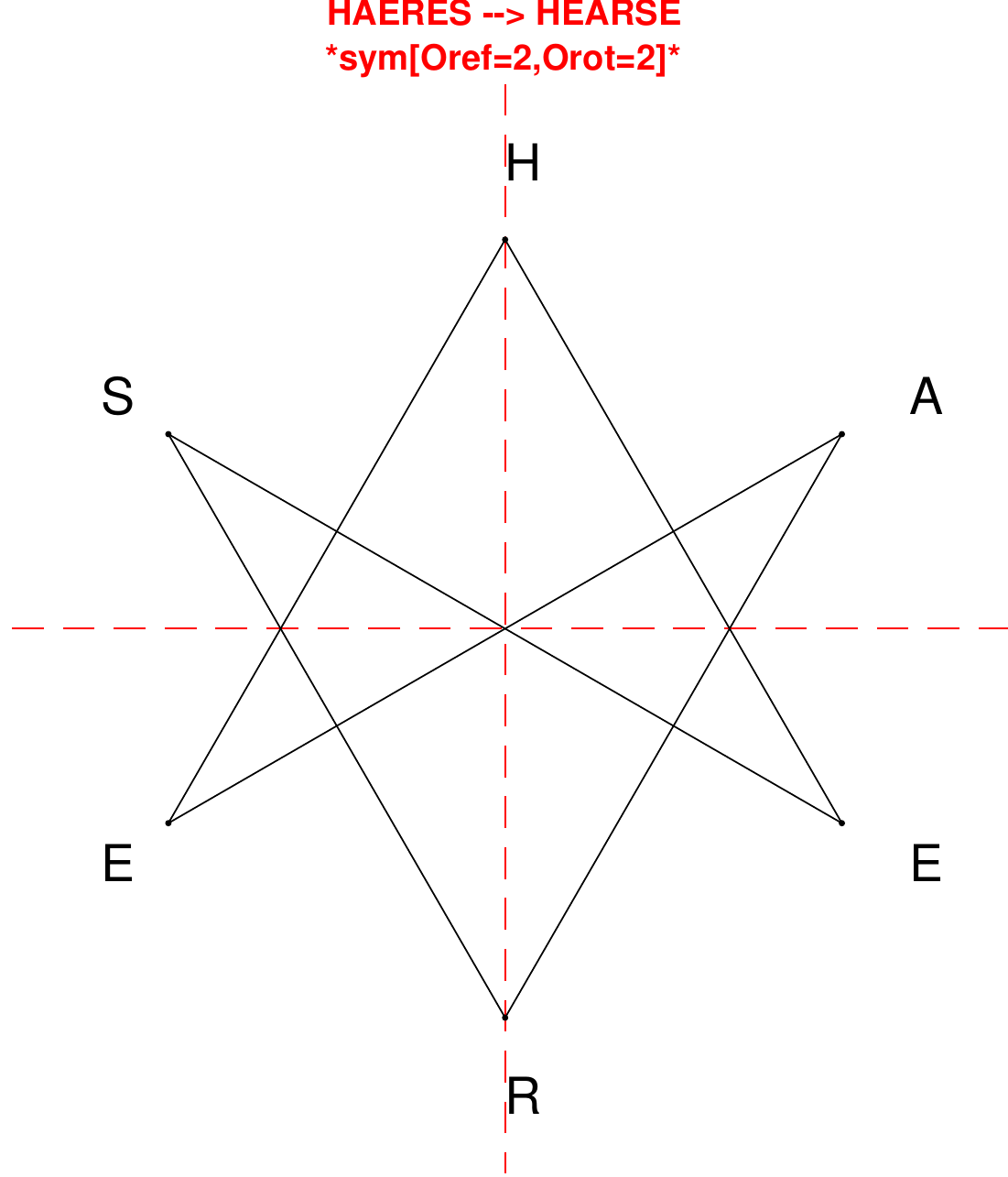}
\end{subfigure}
\hfill
\begin{subfigure}[T]{0.19\textwidth}
\centering
\includegraphics[width=\textwidth]{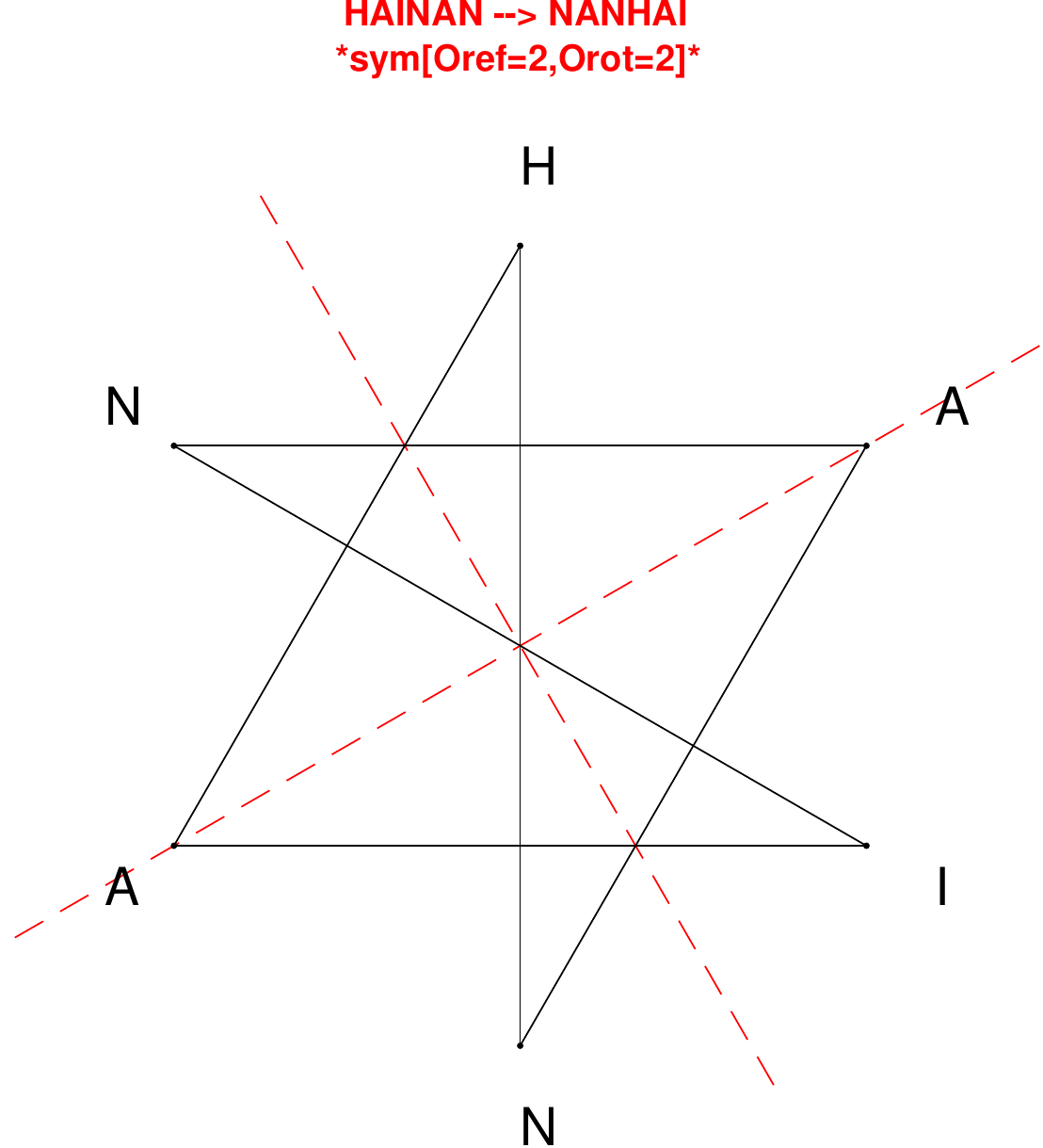}
\end{subfigure}
\end{figure}

\begin{figure}[H]
\centering
\begin{subfigure}[T]{0.19\textwidth}
\centering
\includegraphics[width=\textwidth]{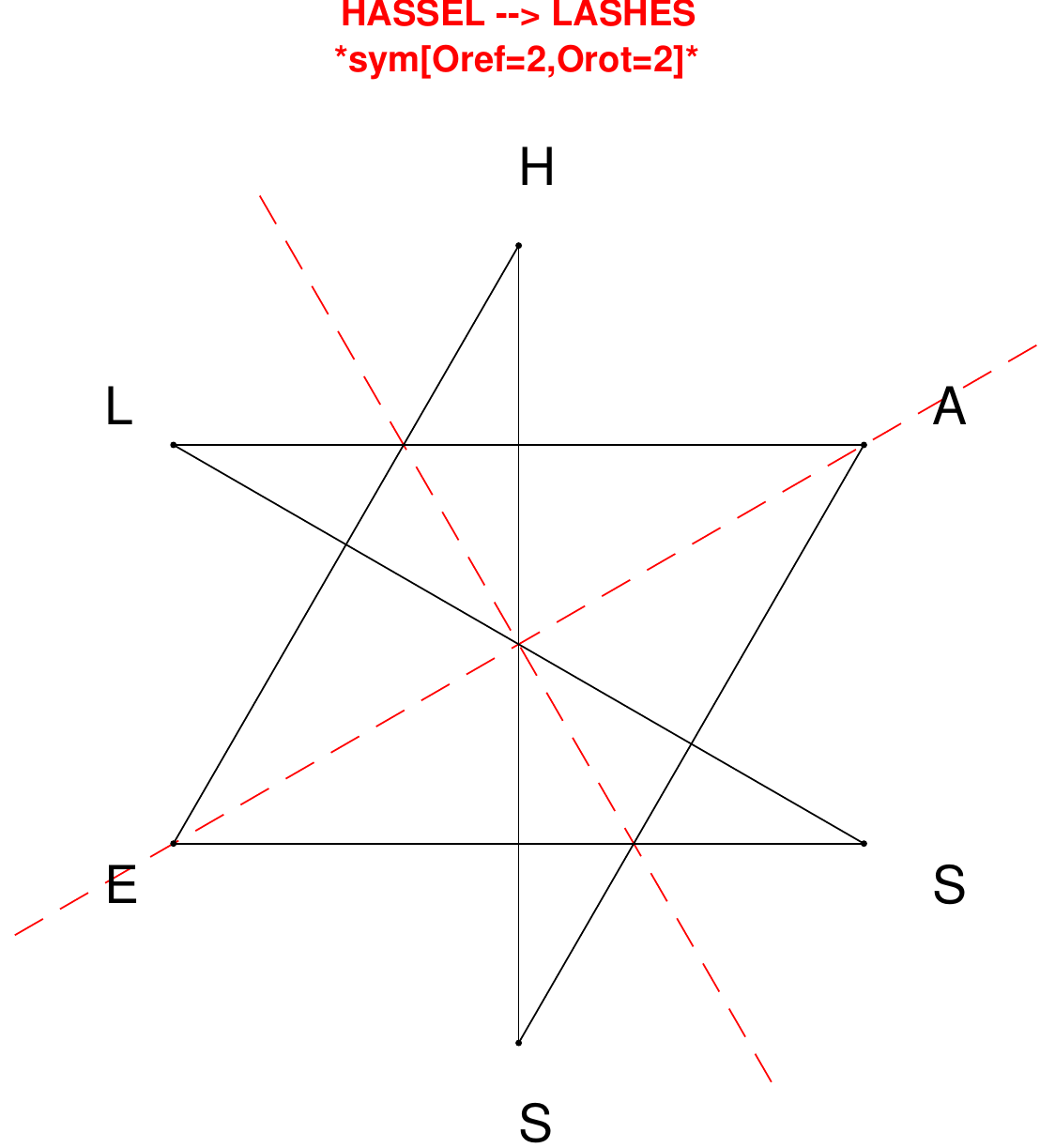}
\end{subfigure}
\hfill
\begin{subfigure}[T]{0.19\textwidth}
\centering
\includegraphics[width=\textwidth]{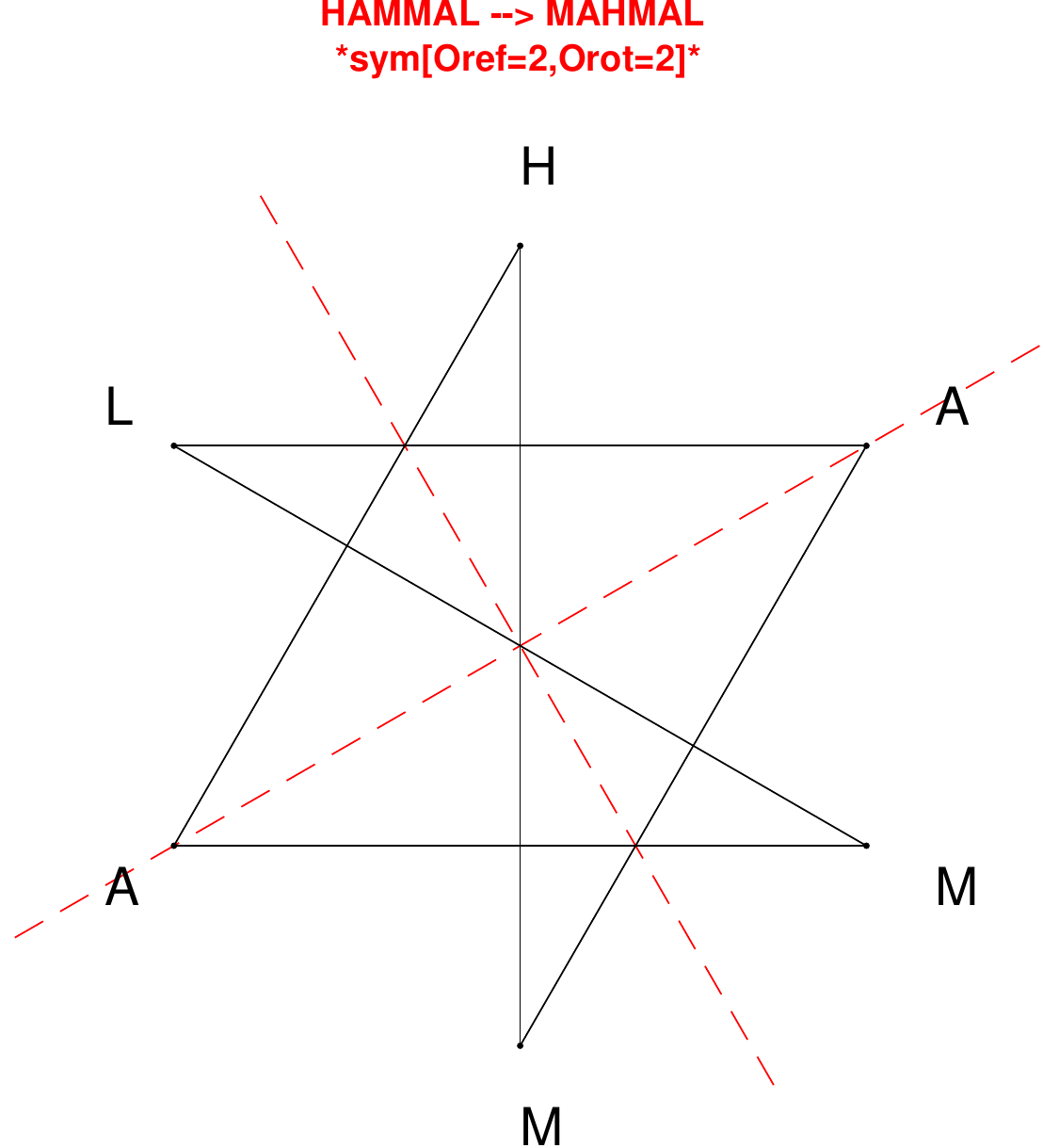}
\end{subfigure}
\hfill
\begin{subfigure}[T]{0.19\textwidth}
\centering
\includegraphics[width=\textwidth]{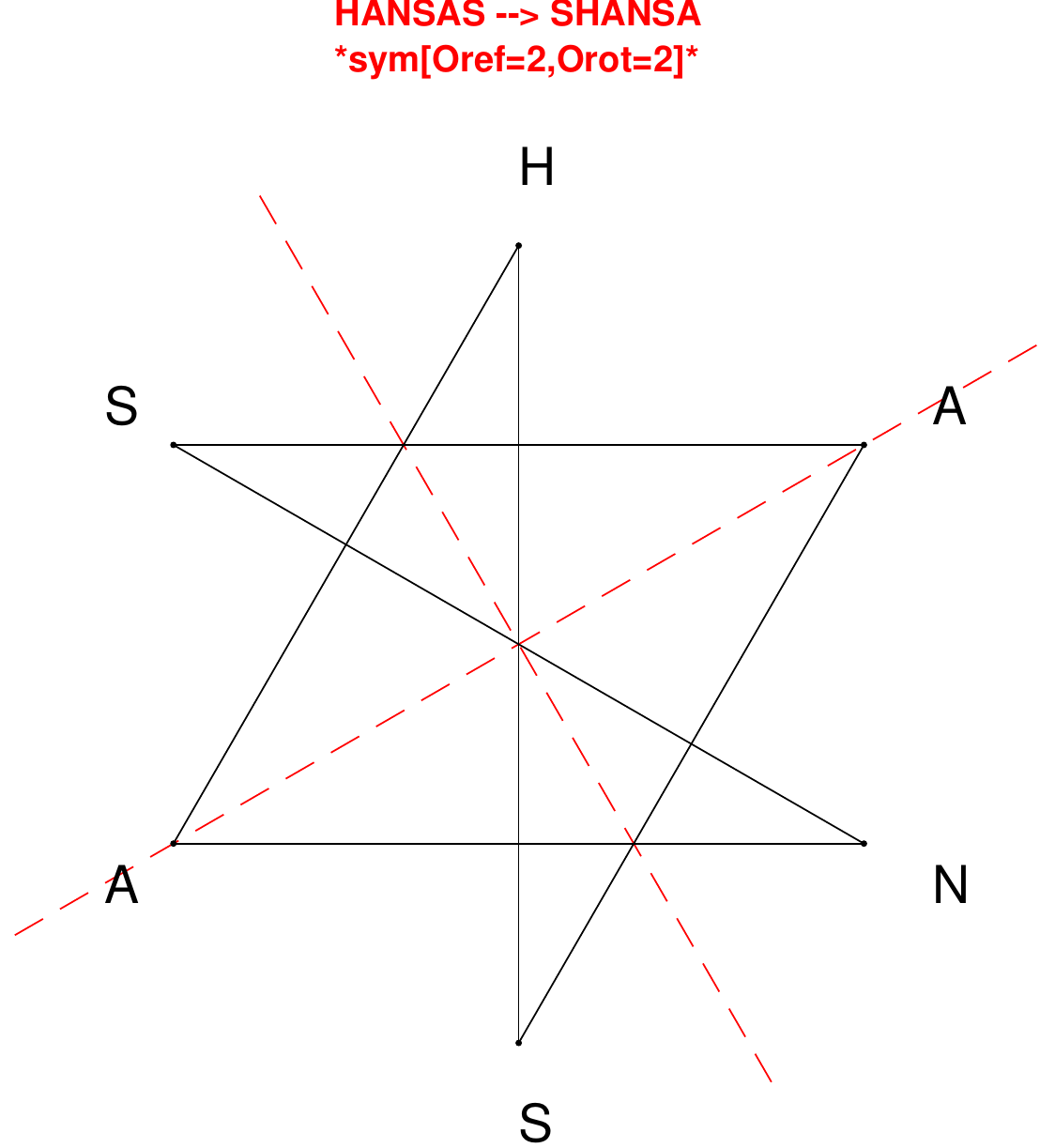}
\end{subfigure}
\hfill
\begin{subfigure}[T]{0.19\textwidth}
\centering
\includegraphics[width=\textwidth]{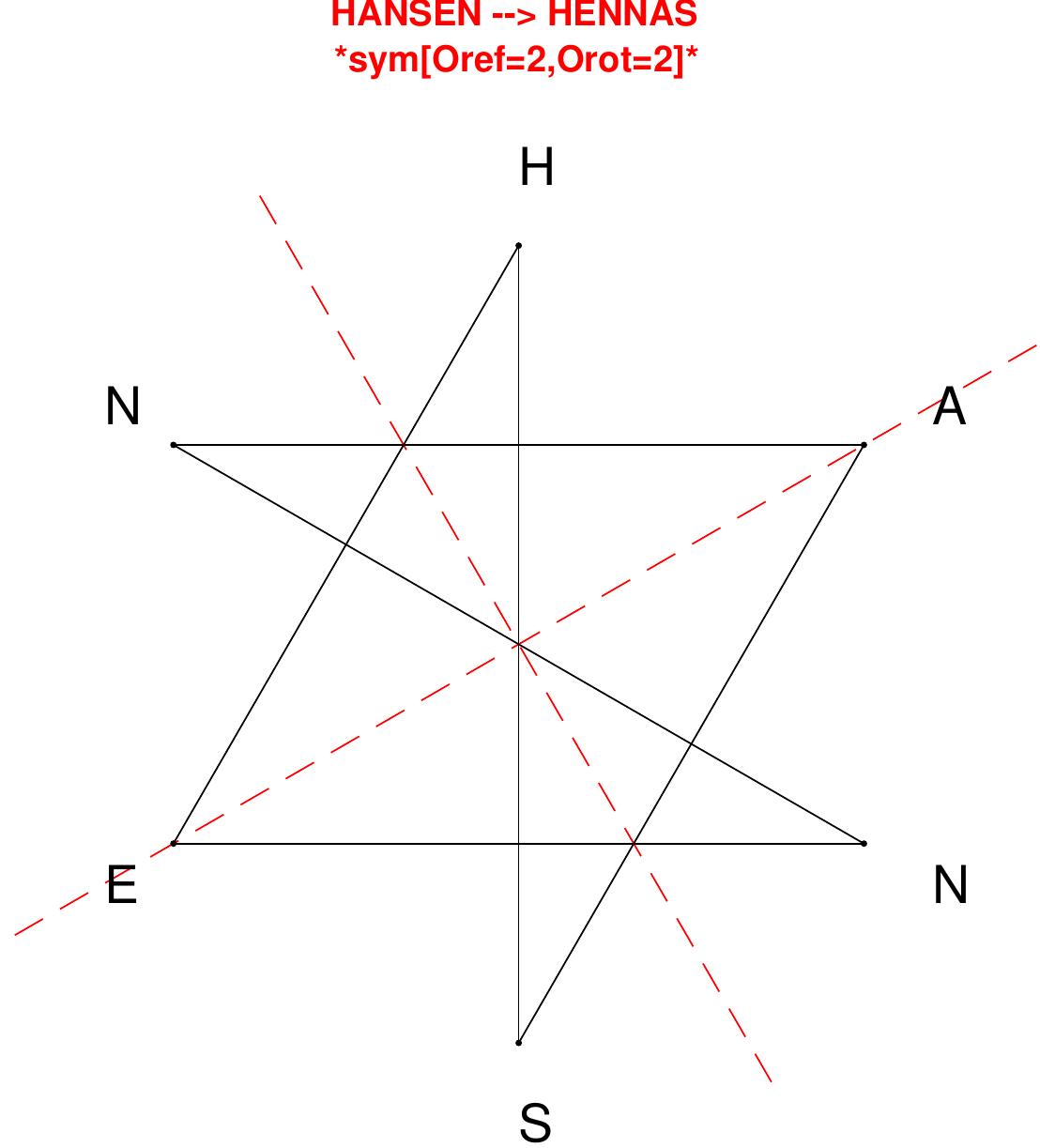}
\end{subfigure}
\hfill
\begin{subfigure}[T]{0.19\textwidth}
\centering
\includegraphics[width=\textwidth]{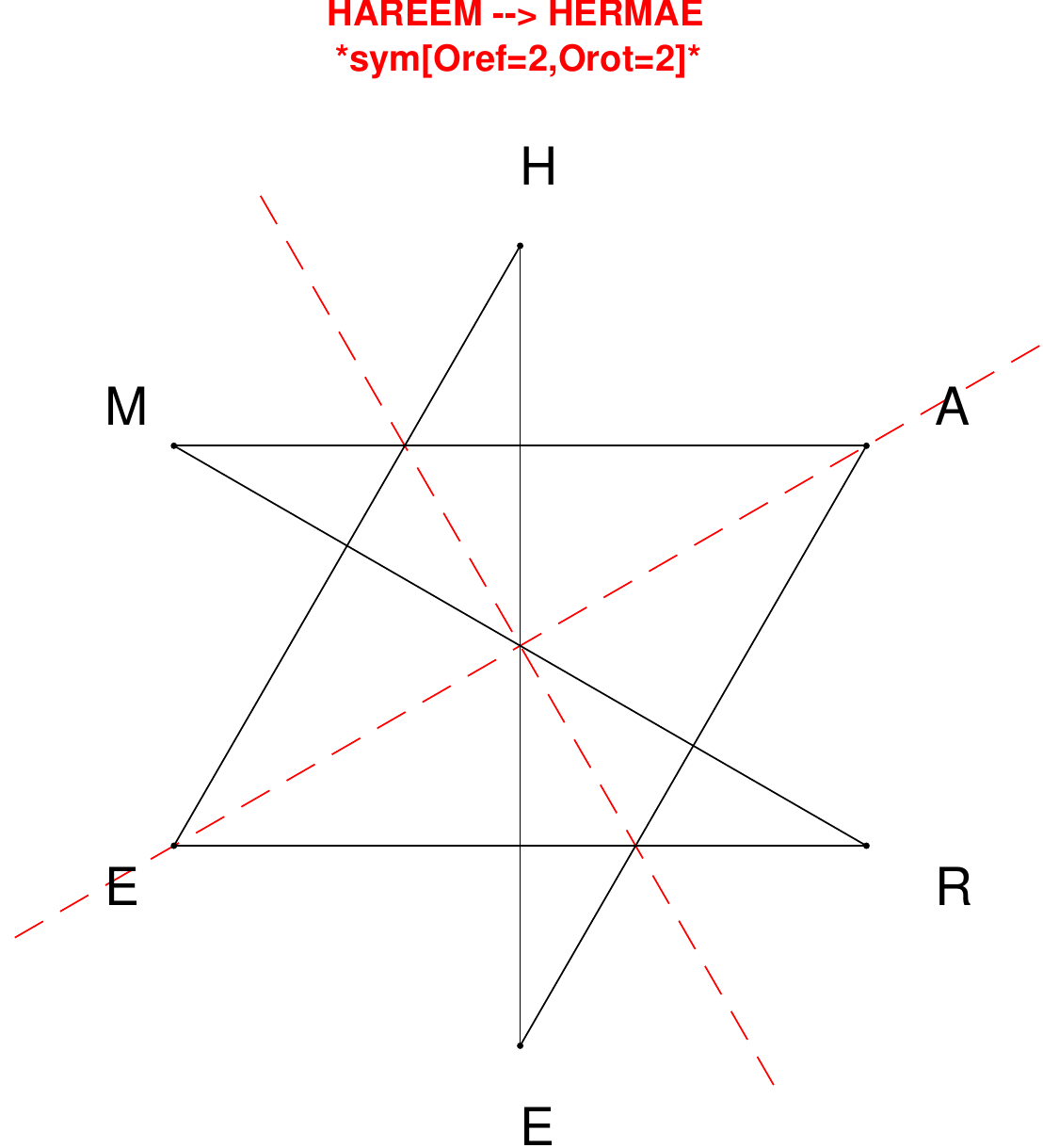}
\end{subfigure}
\end{figure}

\begin{figure}[H]
\centering
\begin{subfigure}[T]{0.19\textwidth}
\centering
\includegraphics[width=\textwidth]{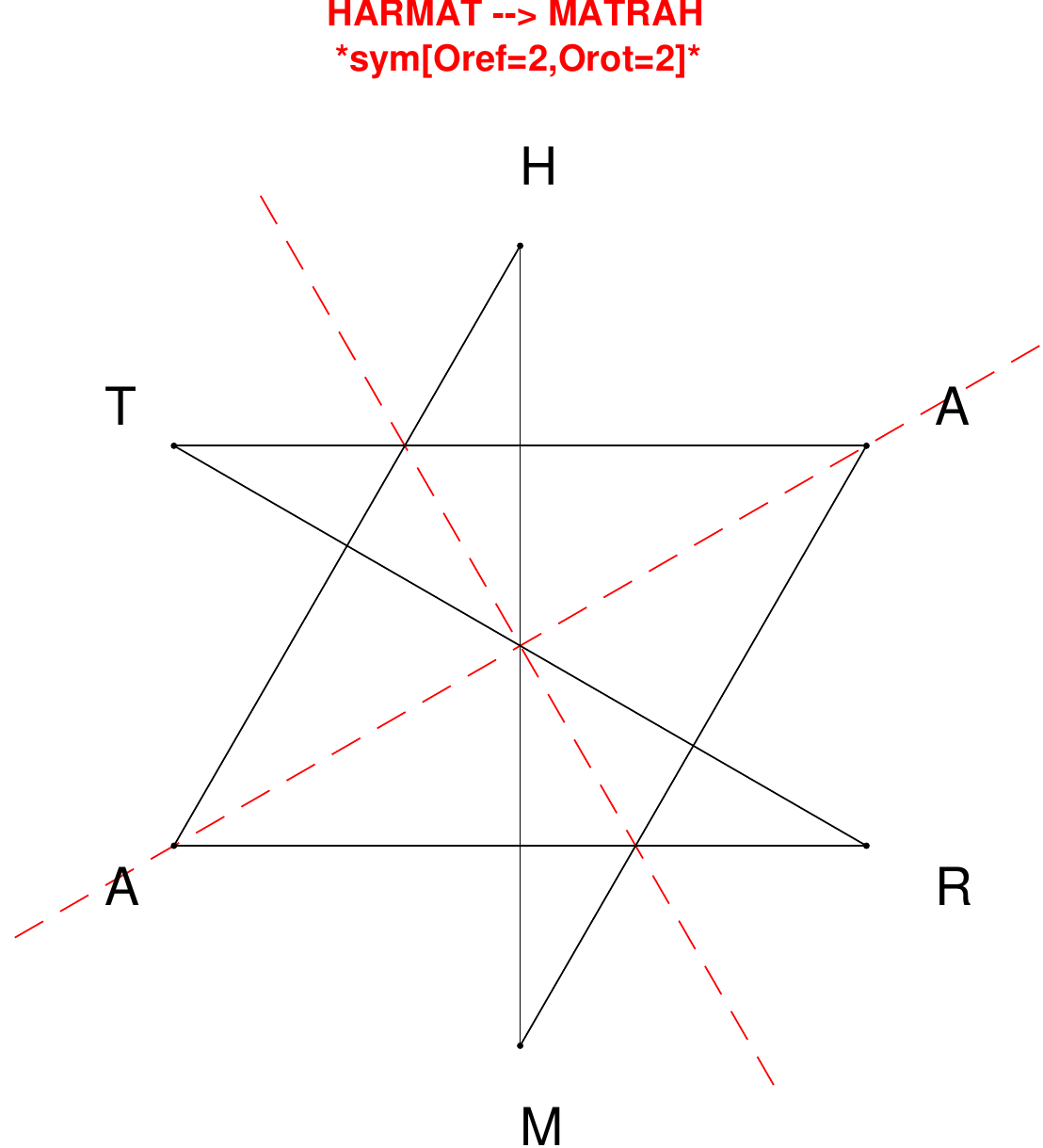}
\end{subfigure}
\hfill
\begin{subfigure}[T]{0.19\textwidth}
\centering
\includegraphics[width=\textwidth]{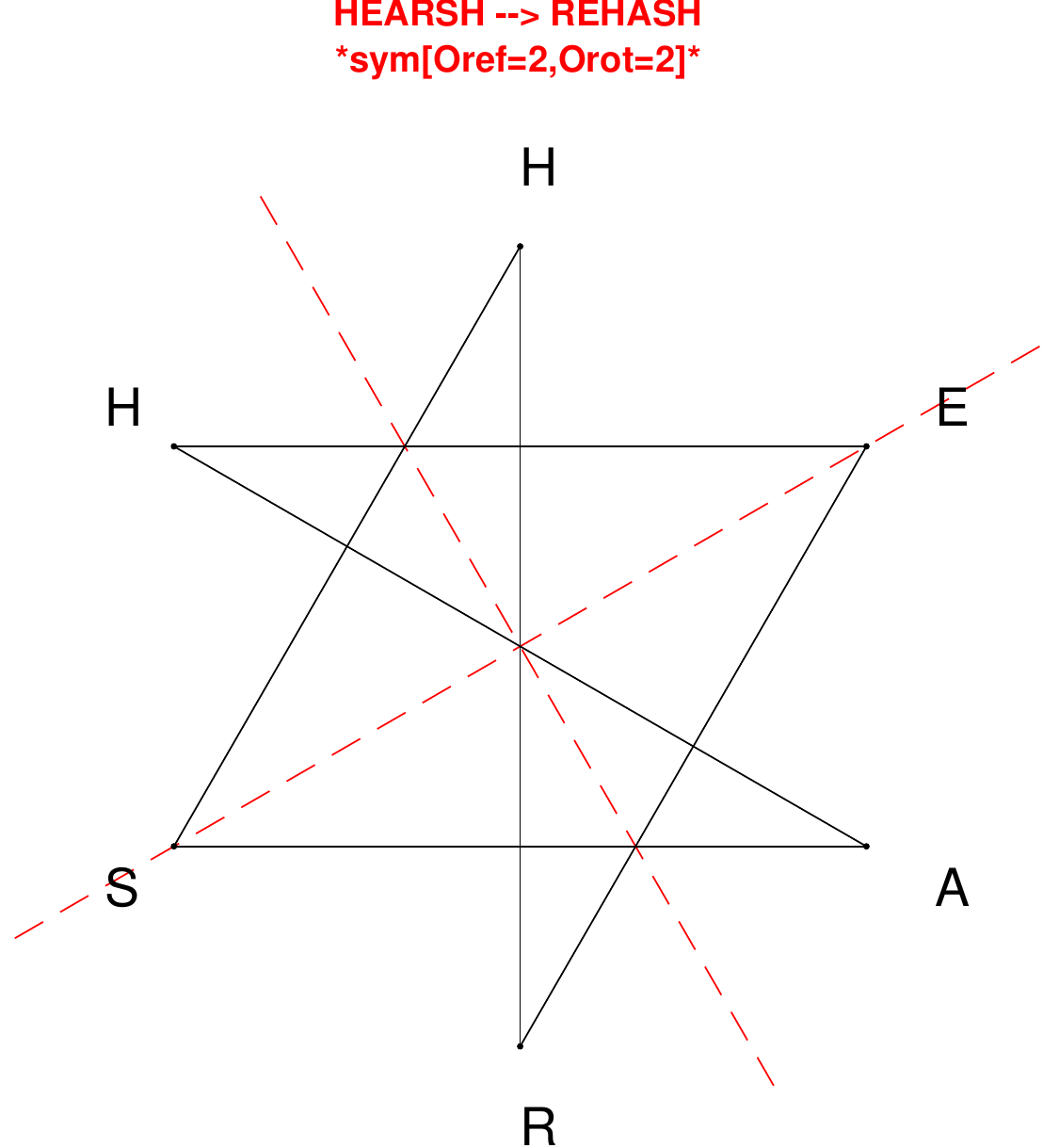}
\end{subfigure}
\hfill
\begin{subfigure}[T]{0.19\textwidth}
\centering
\includegraphics[width=\textwidth]{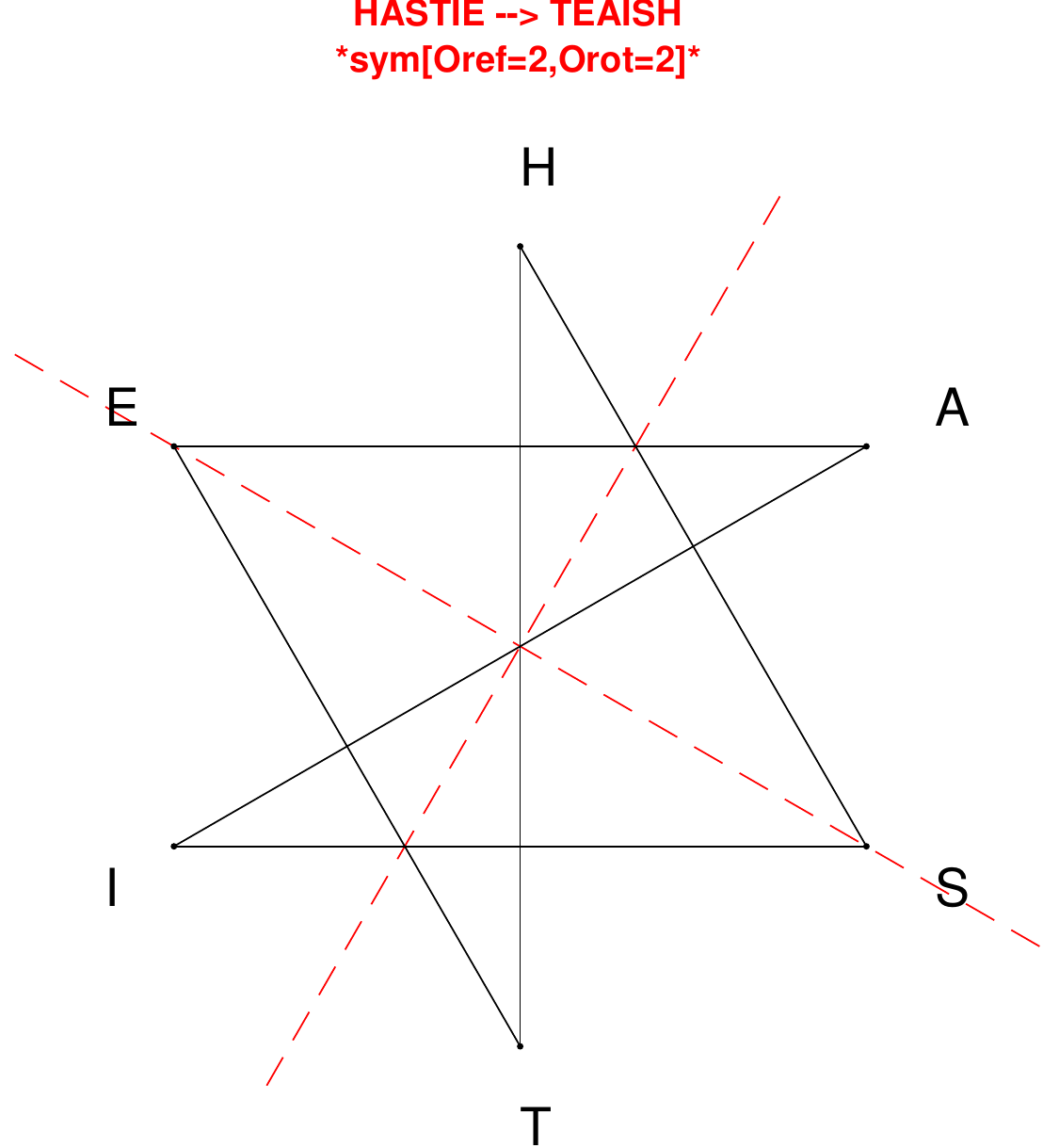}
\end{subfigure}
\hfill
\begin{subfigure}[T]{0.19\textwidth}
\centering
\includegraphics[width=\textwidth]{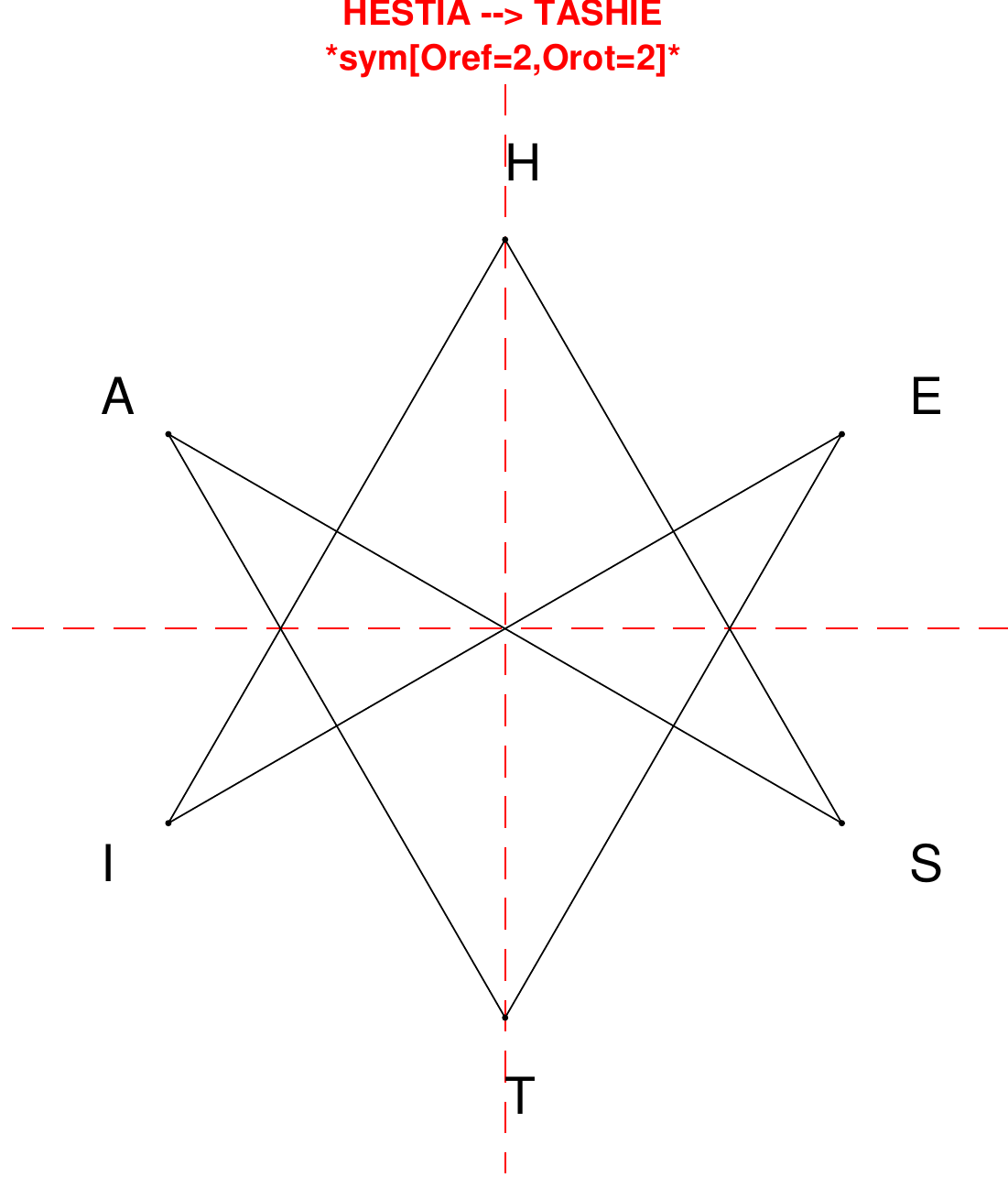}
\end{subfigure}
\hfill
\begin{subfigure}[T]{0.19\textwidth}
\centering
\includegraphics[width=\textwidth]{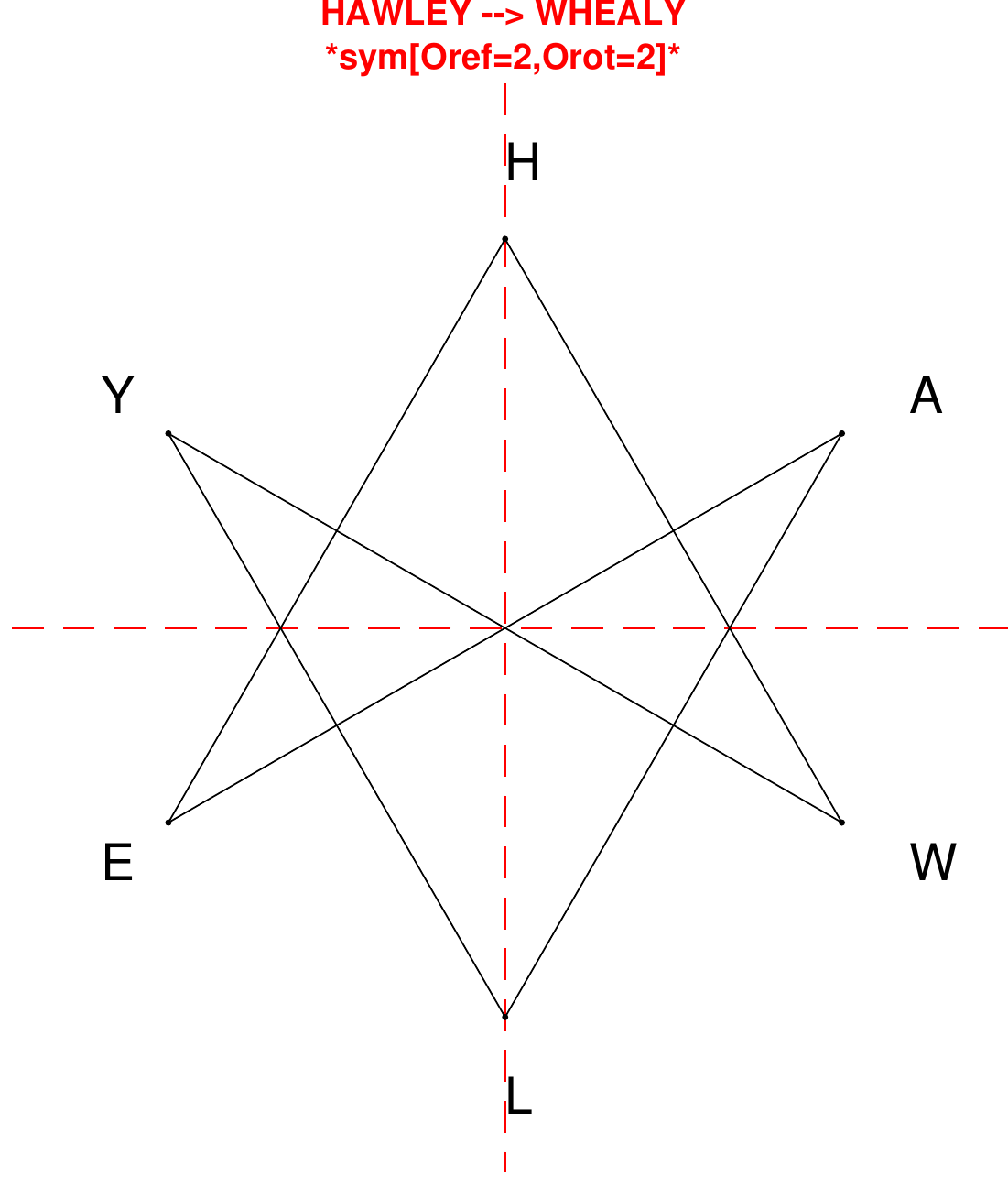}
\end{subfigure}
\end{figure}

\begin{figure}[H]
\centering
\begin{subfigure}[T]{0.19\textwidth}
\centering
\includegraphics[width=\textwidth]{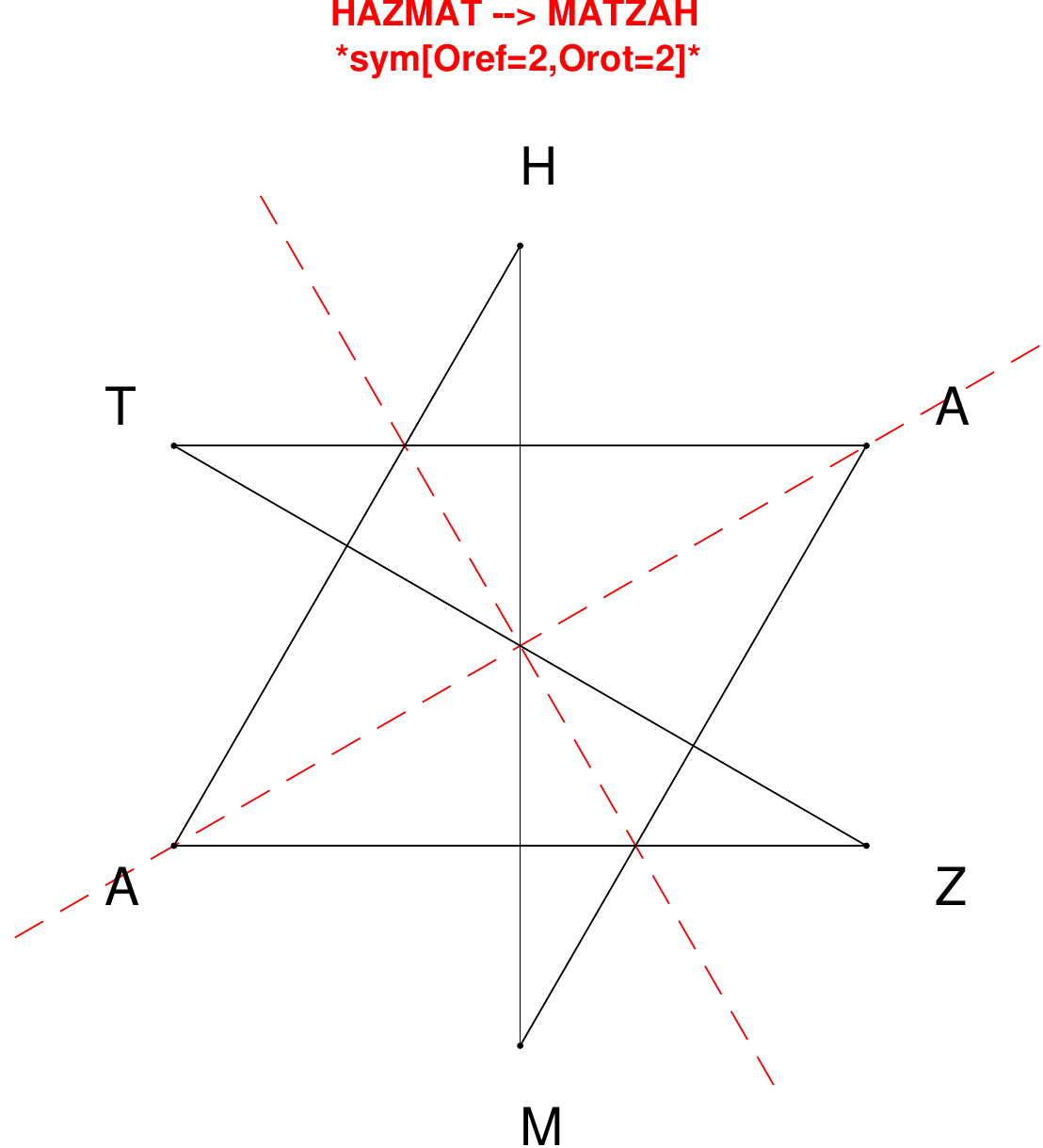}
\end{subfigure}
\hfill
\begin{subfigure}[T]{0.19\textwidth}
\centering
\includegraphics[width=\textwidth]{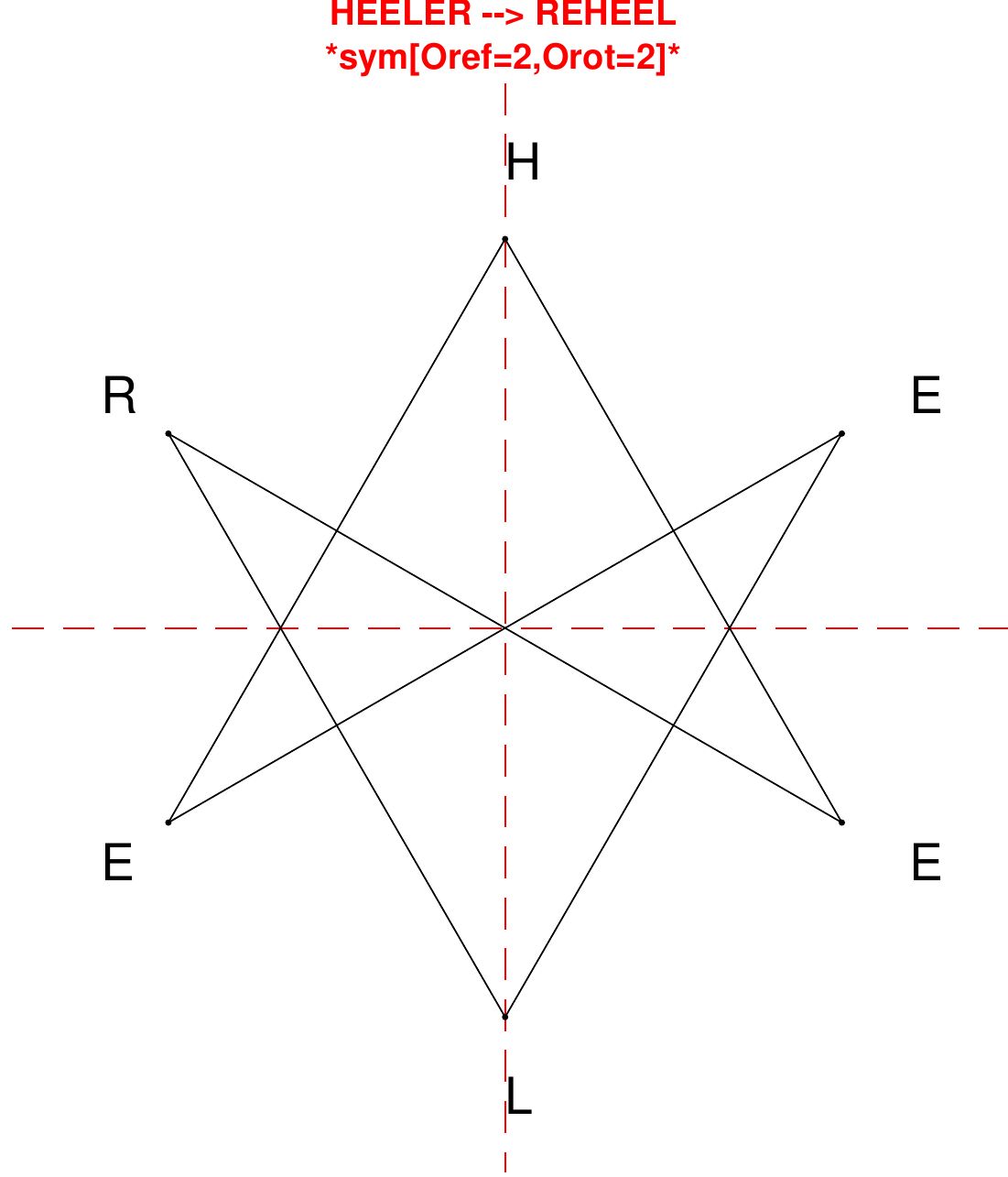}
\end{subfigure}
\hfill
\begin{subfigure}[T]{0.19\textwidth}
\centering
\includegraphics[width=\textwidth]{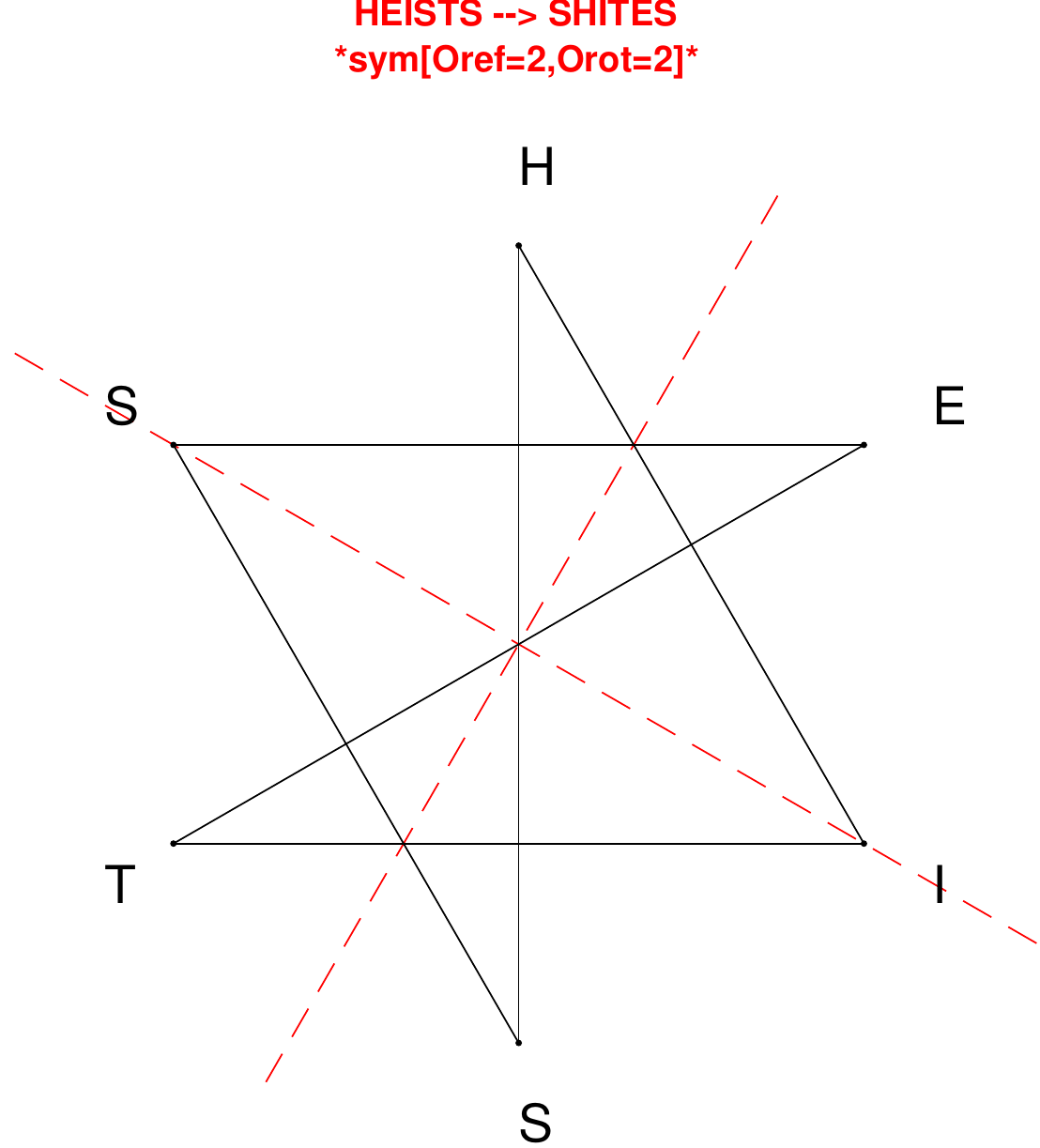}
\end{subfigure}
\hfill
\begin{subfigure}[T]{0.19\textwidth}
\centering
\includegraphics[width=\textwidth]{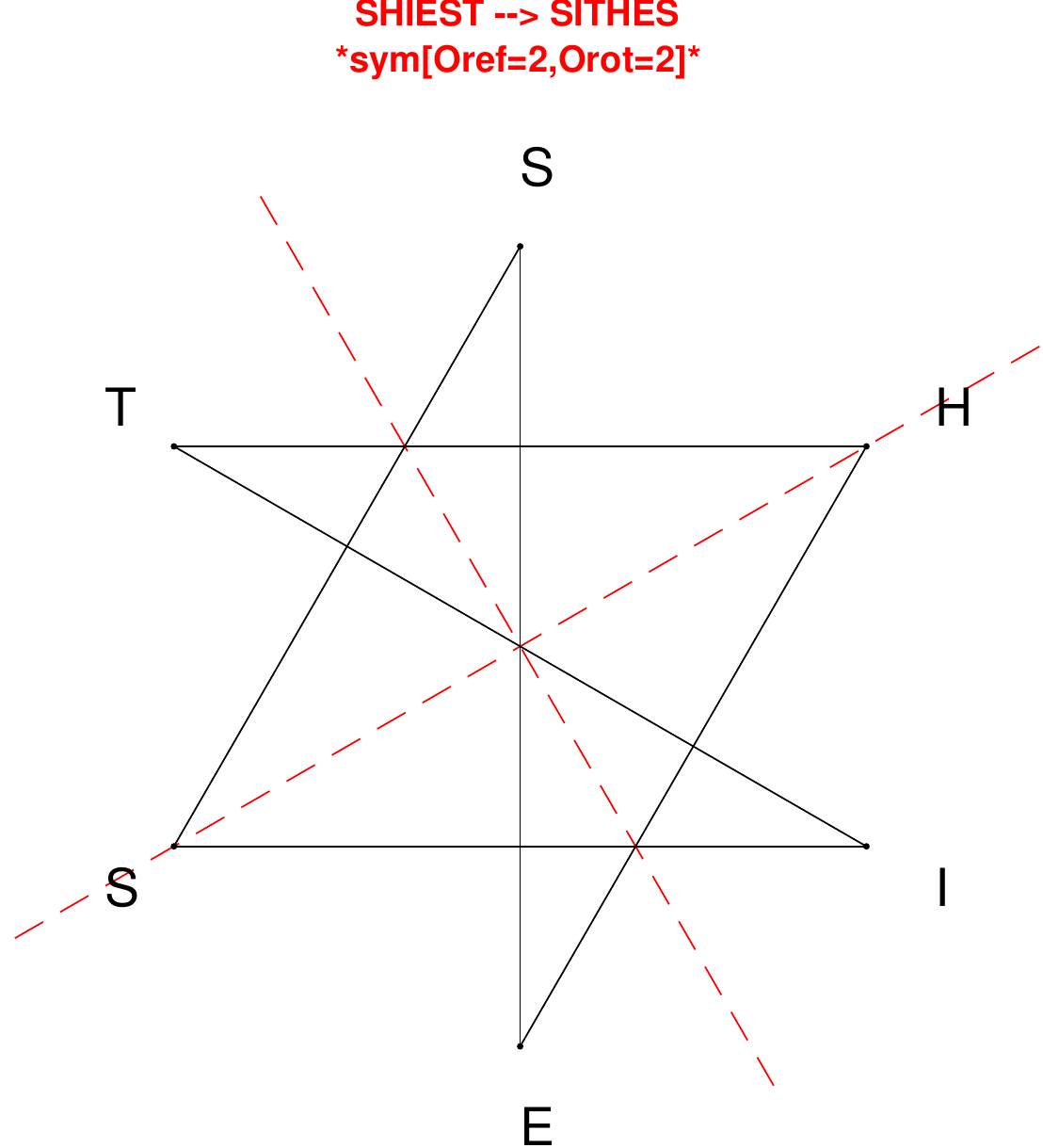}
\end{subfigure}
\hfill
\begin{subfigure}[T]{0.19\textwidth}
\centering
\includegraphics[width=\textwidth]{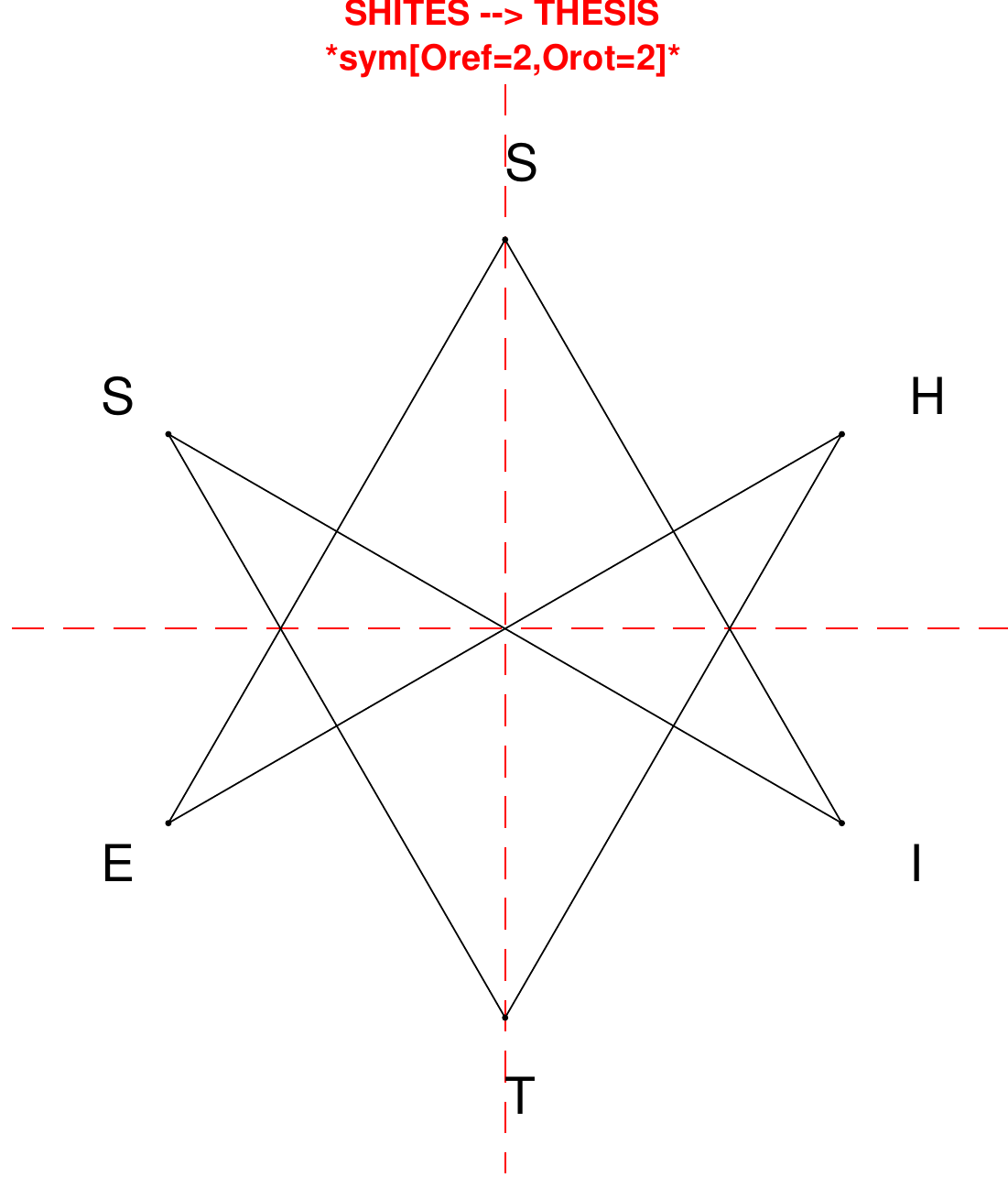}
\end{subfigure}
\end{figure}

\begin{figure}[H]
\centering
\begin{subfigure}[T]{0.19\textwidth}
\centering
\includegraphics[width=\textwidth]{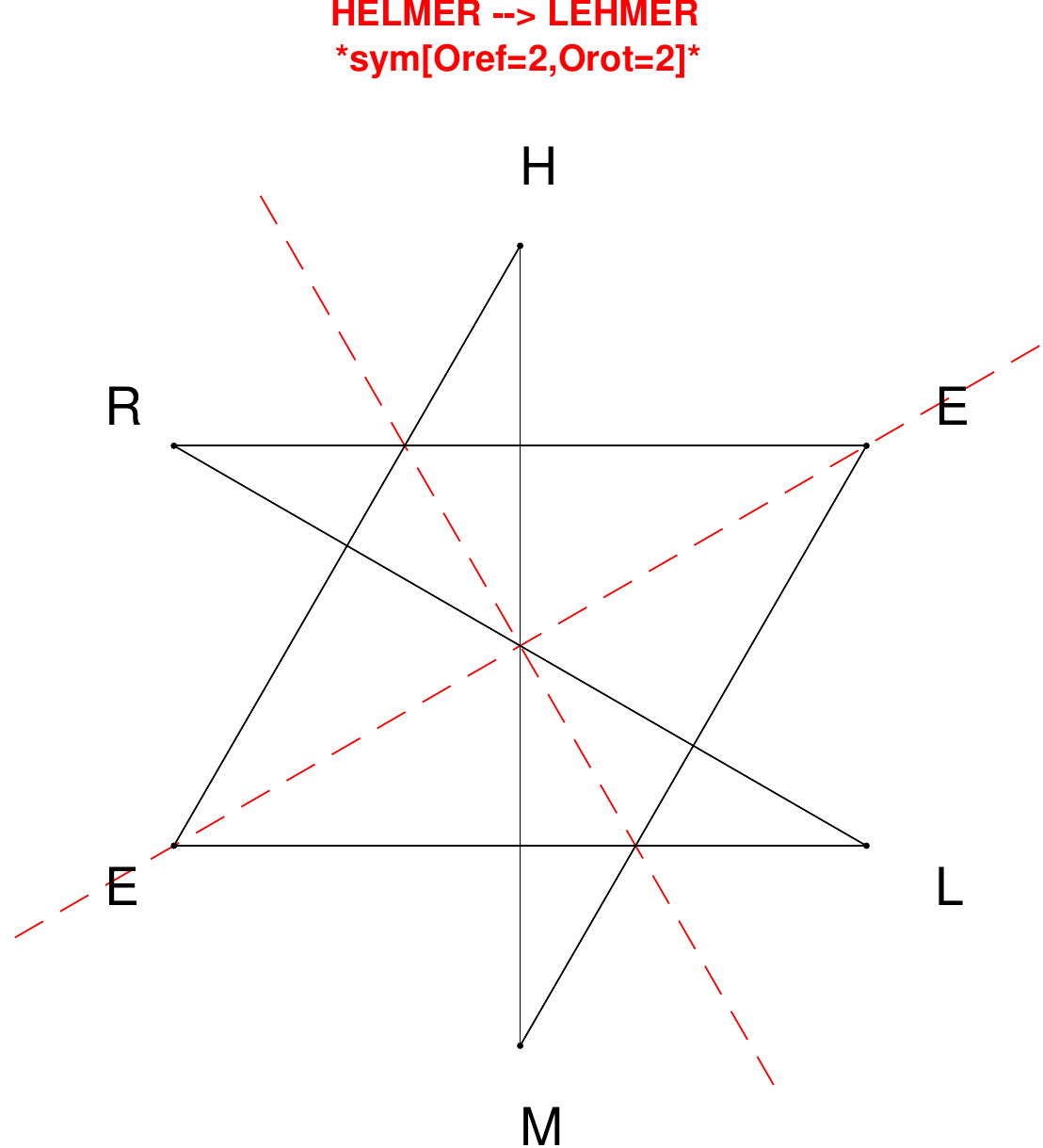}
\end{subfigure}
\hfill
\begin{subfigure}[T]{0.19\textwidth}
\centering
\includegraphics[width=\textwidth]{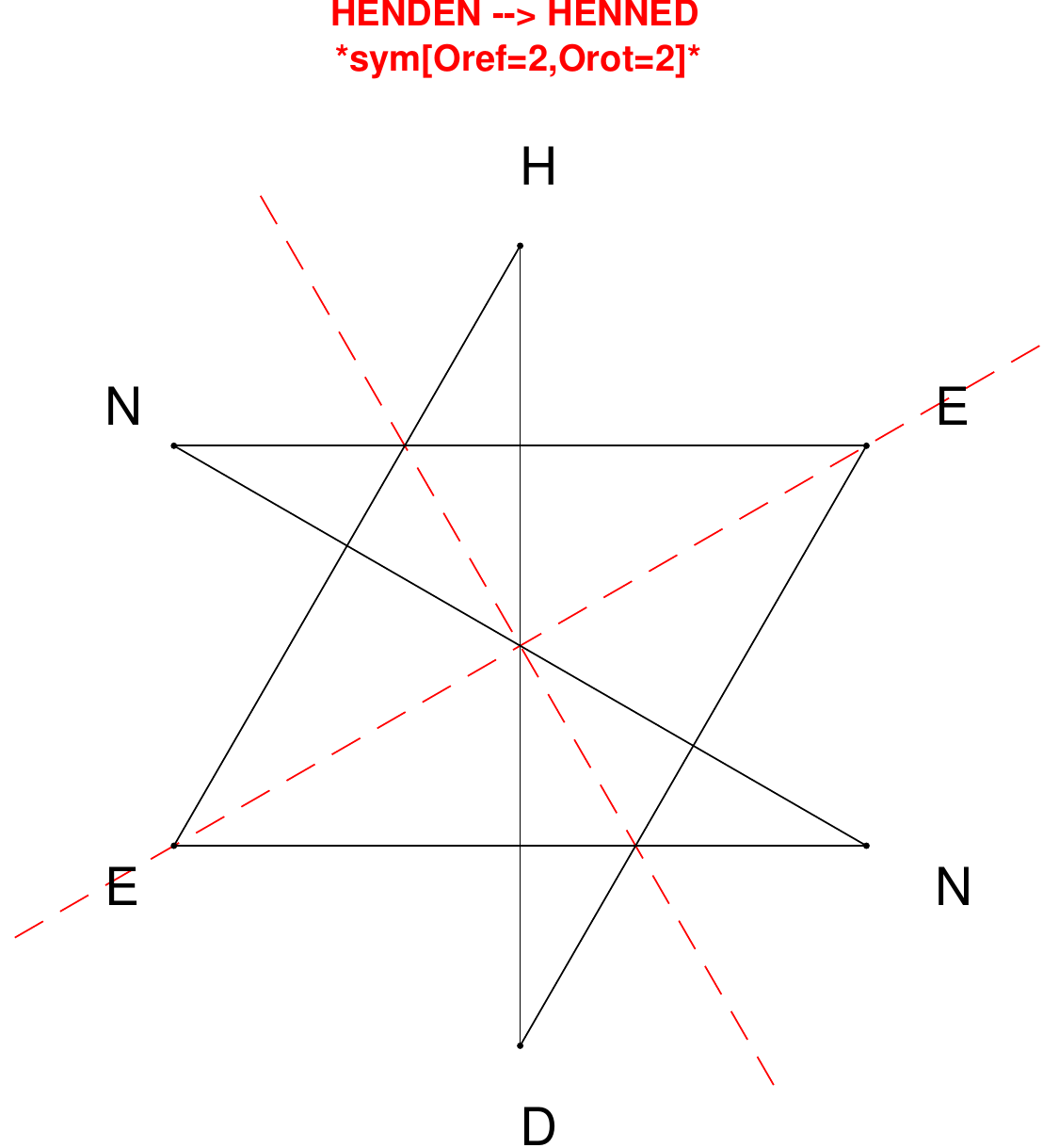}
\end{subfigure}
\hfill
\begin{subfigure}[T]{0.19\textwidth}
\centering
\includegraphics[width=\textwidth]{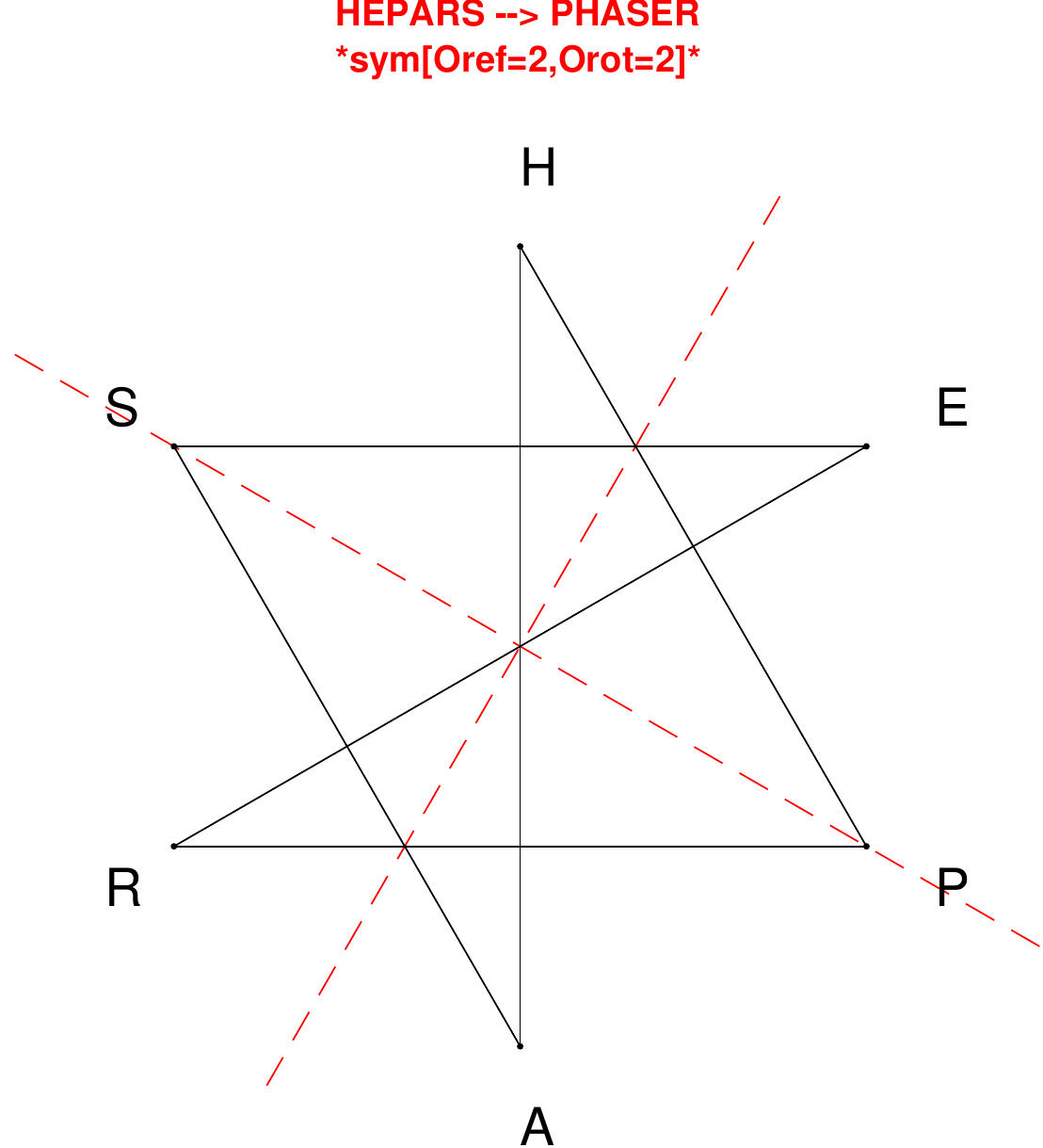}
\end{subfigure}
\hfill
\begin{subfigure}[T]{0.19\textwidth}
\centering
\includegraphics[width=\textwidth]{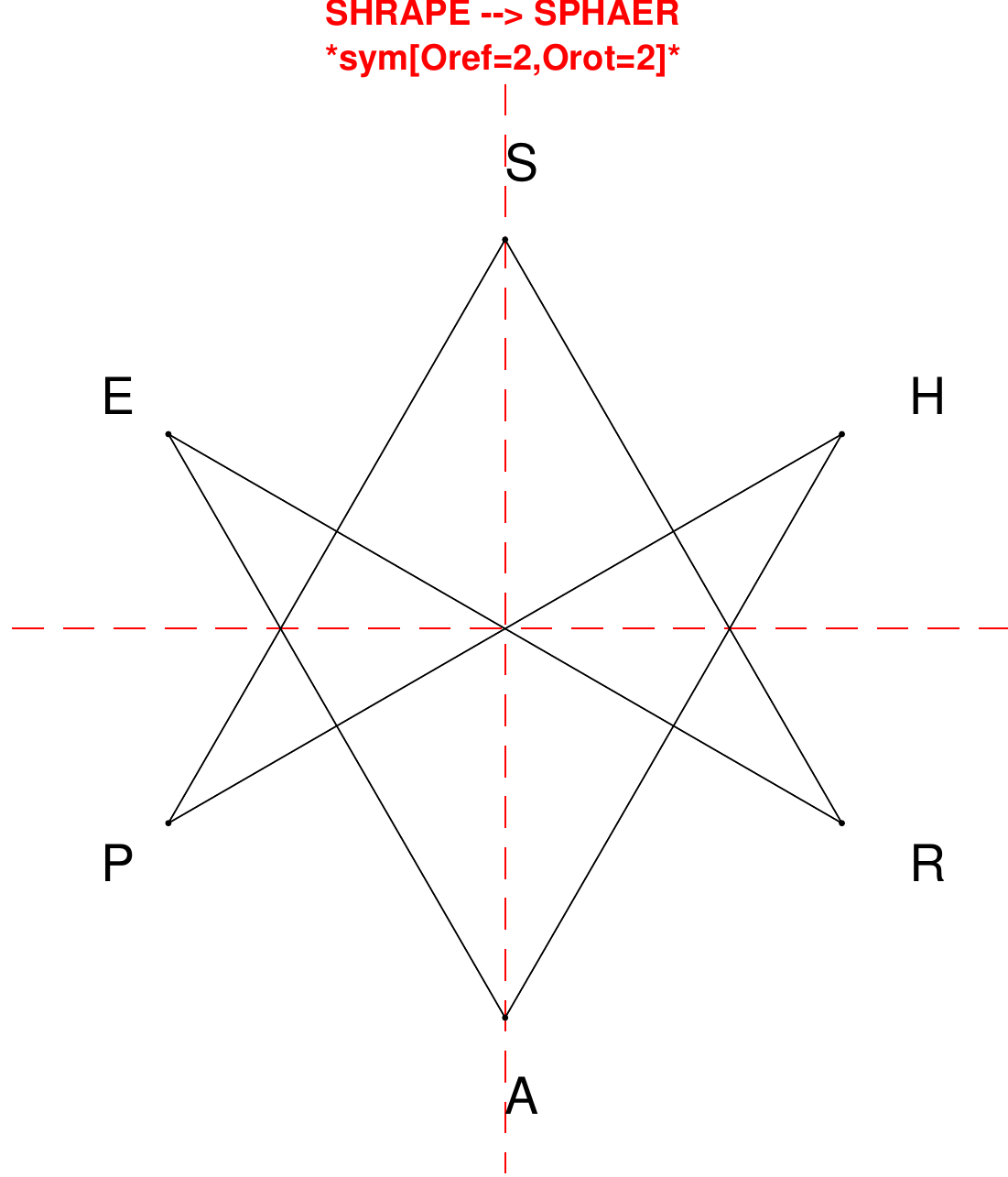}
\end{subfigure}
\hfill
\begin{subfigure}[T]{0.19\textwidth}
\centering
\includegraphics[width=\textwidth]{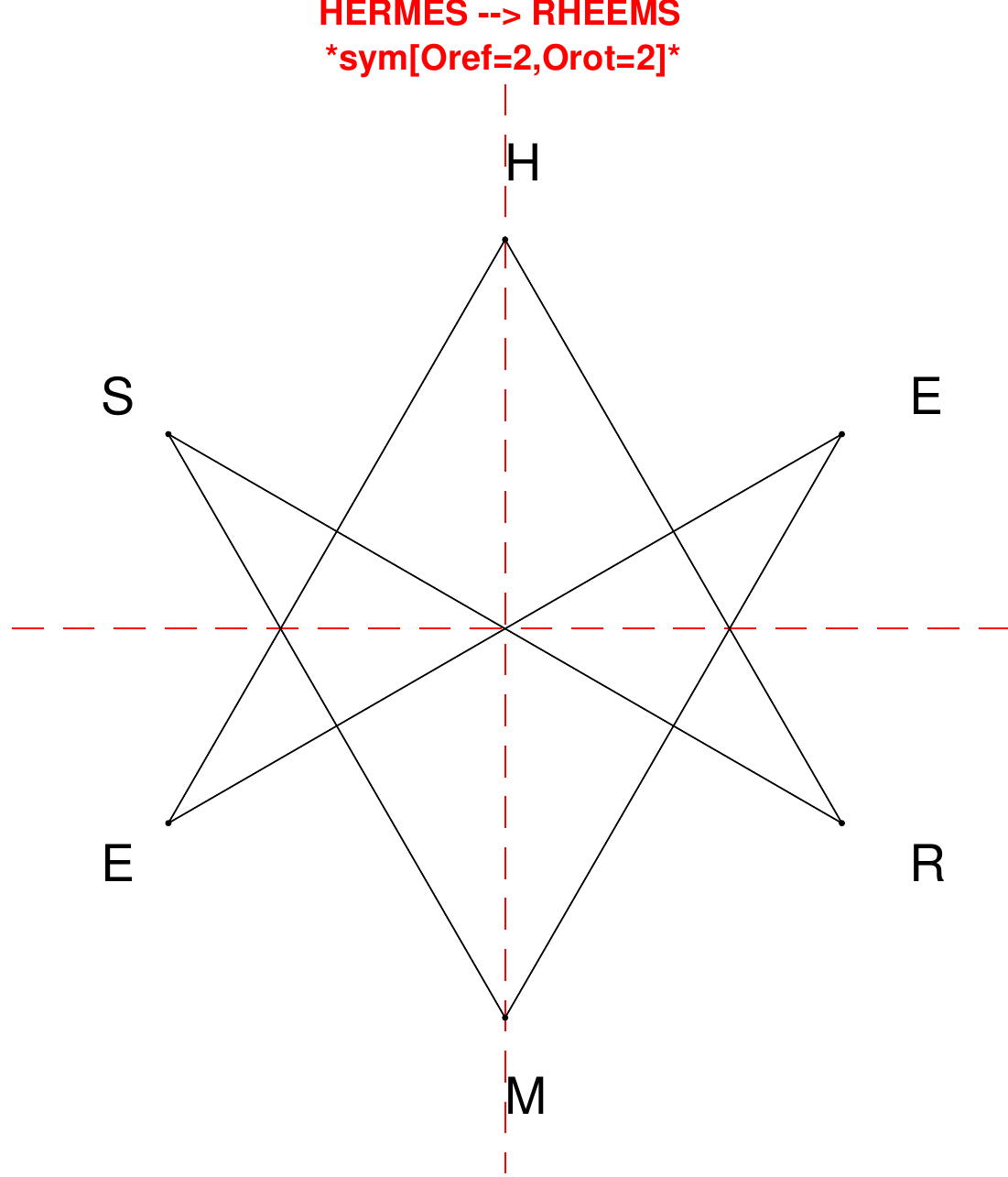}
\end{subfigure}
\end{figure}

\begin{figure}[H]
\centering
\begin{subfigure}[T]{0.19\textwidth}
\centering
\includegraphics[width=\textwidth]{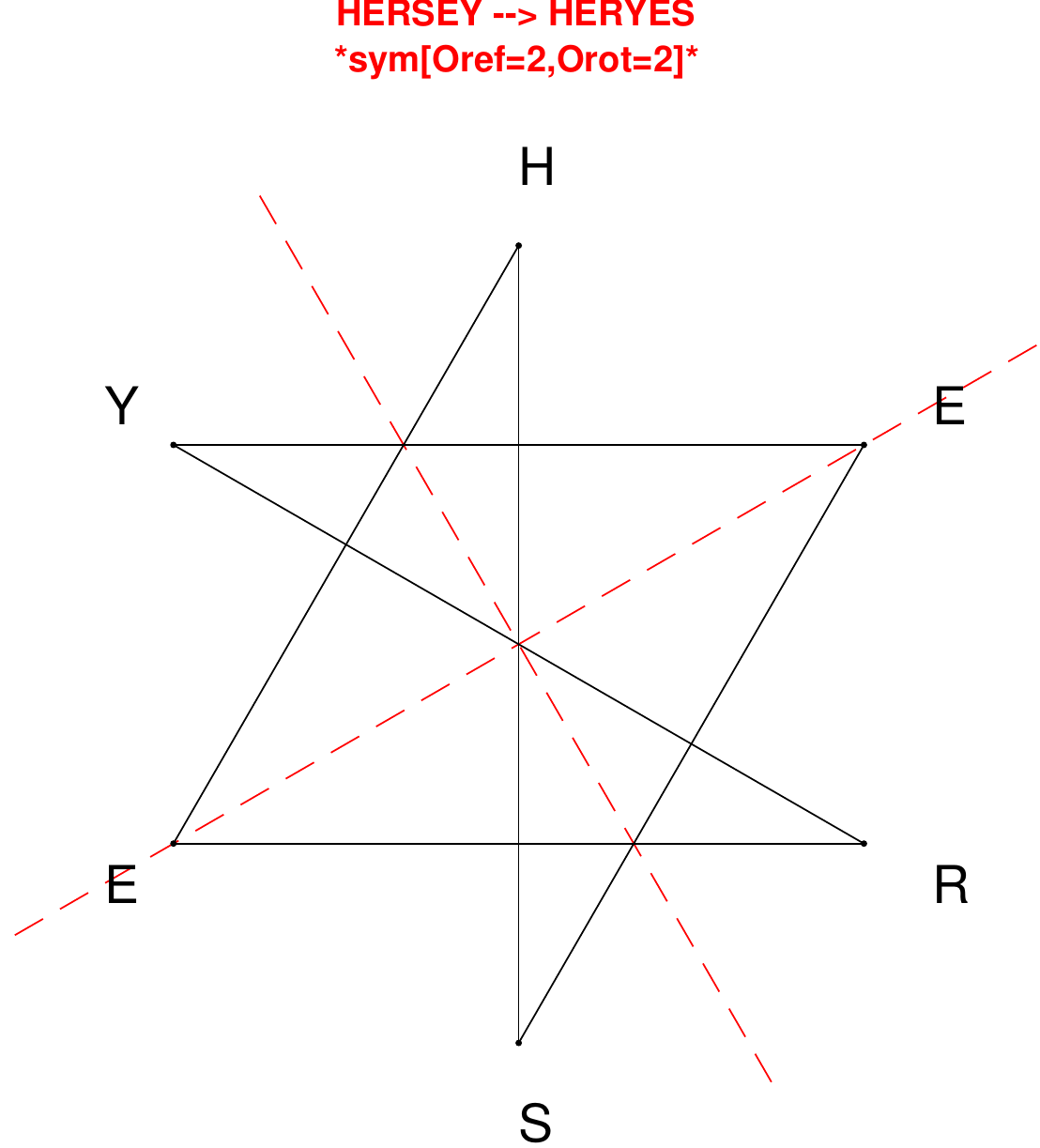}
\end{subfigure}
\hfill
\begin{subfigure}[T]{0.19\textwidth}
\centering
\includegraphics[width=\textwidth]{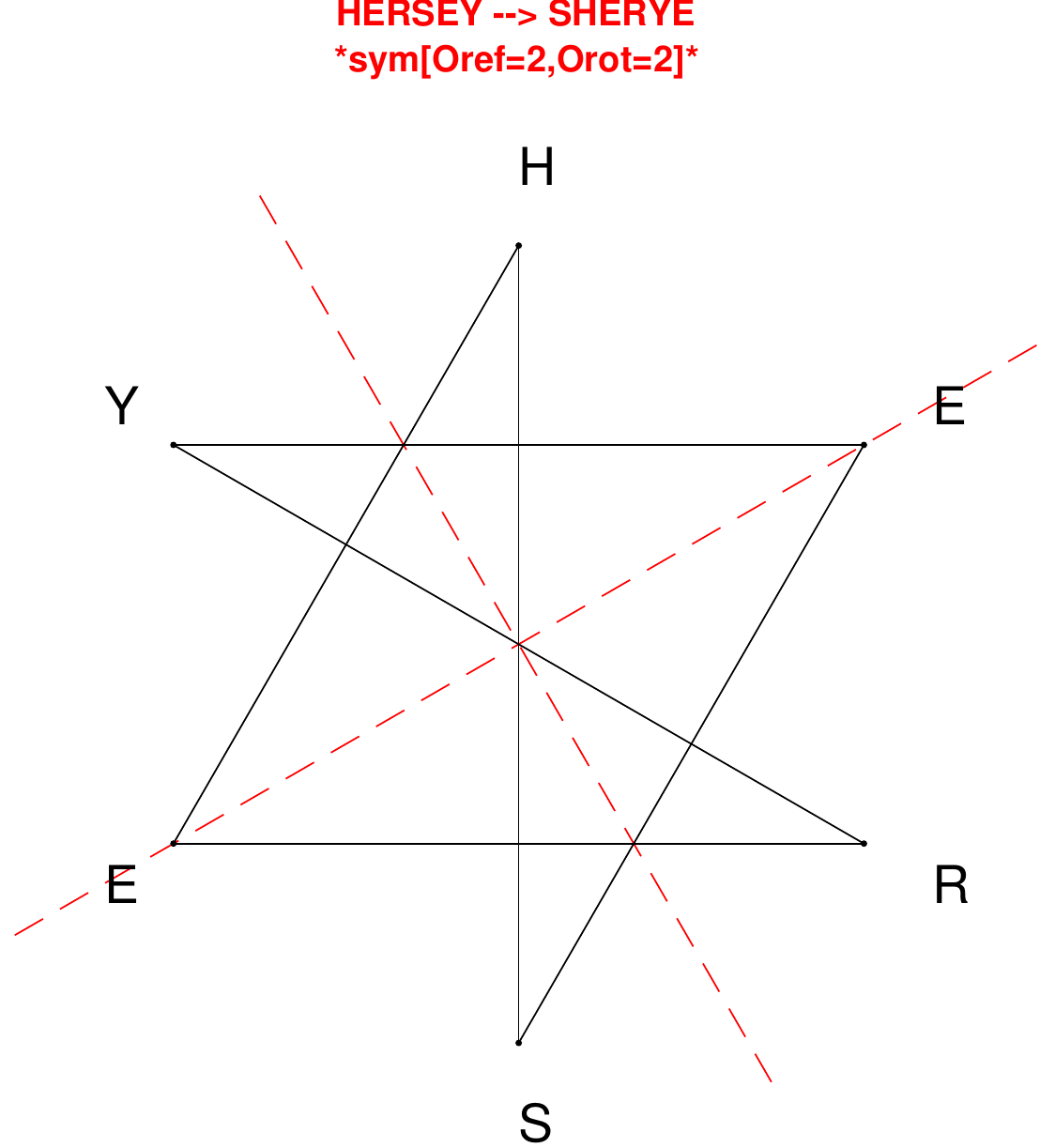}
\end{subfigure}
\hfill
\begin{subfigure}[T]{0.19\textwidth}
\centering
\includegraphics[width=\textwidth]{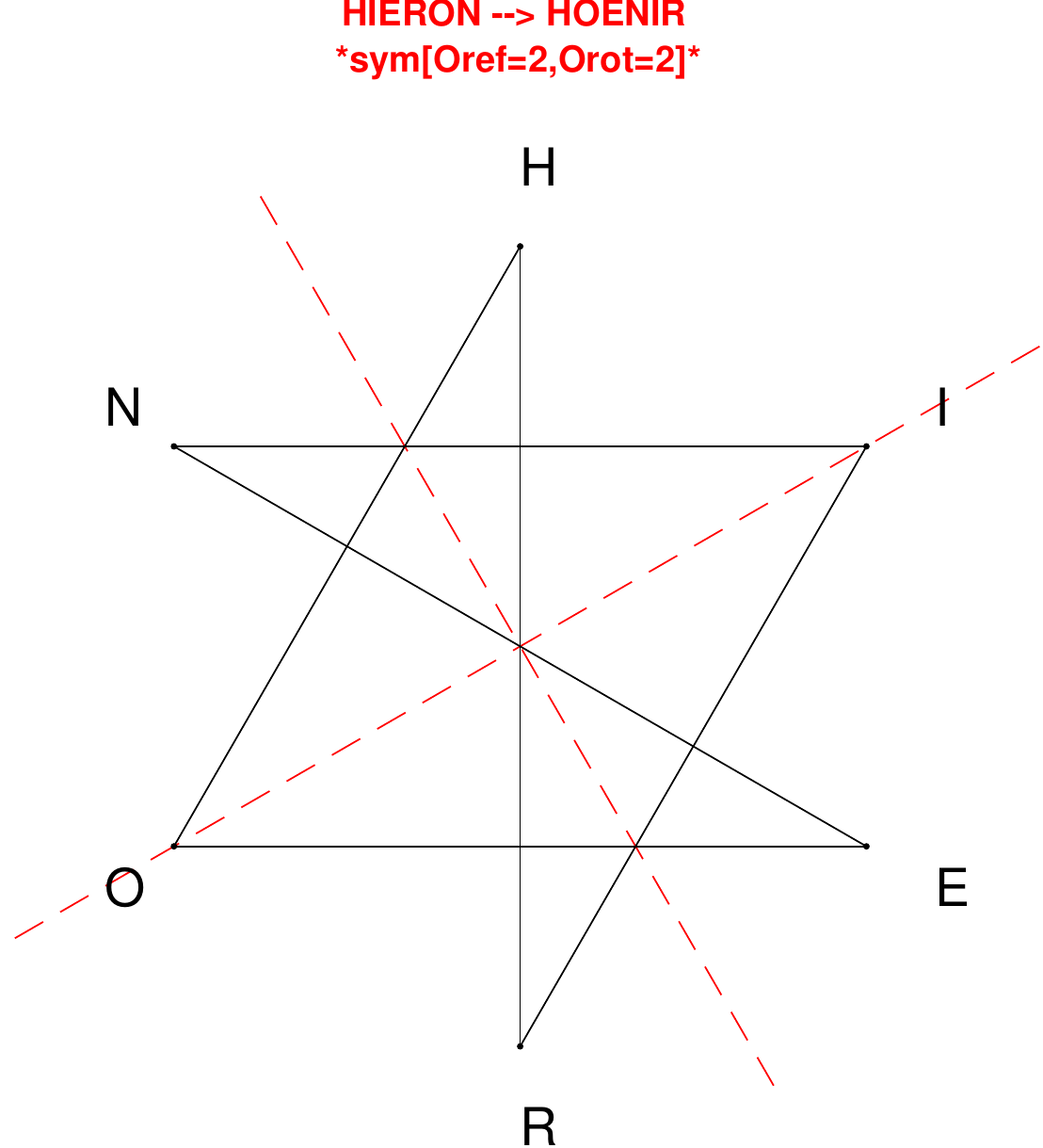}
\end{subfigure}
\hfill
\begin{subfigure}[T]{0.19\textwidth}
\centering
\includegraphics[width=\textwidth]{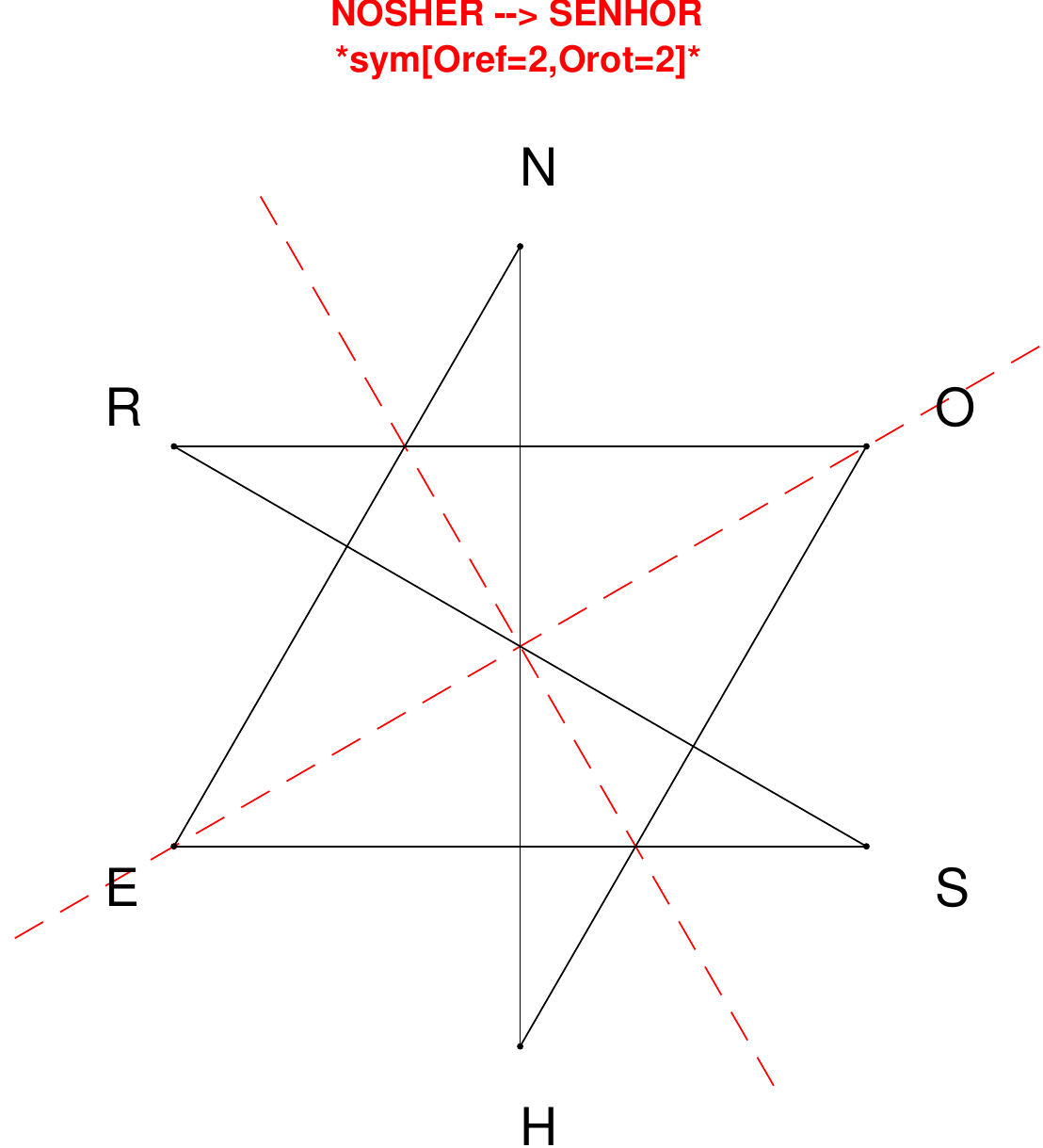}
\end{subfigure}
\hfill
\begin{subfigure}[T]{0.19\textwidth}
\centering
\includegraphics[width=\textwidth]{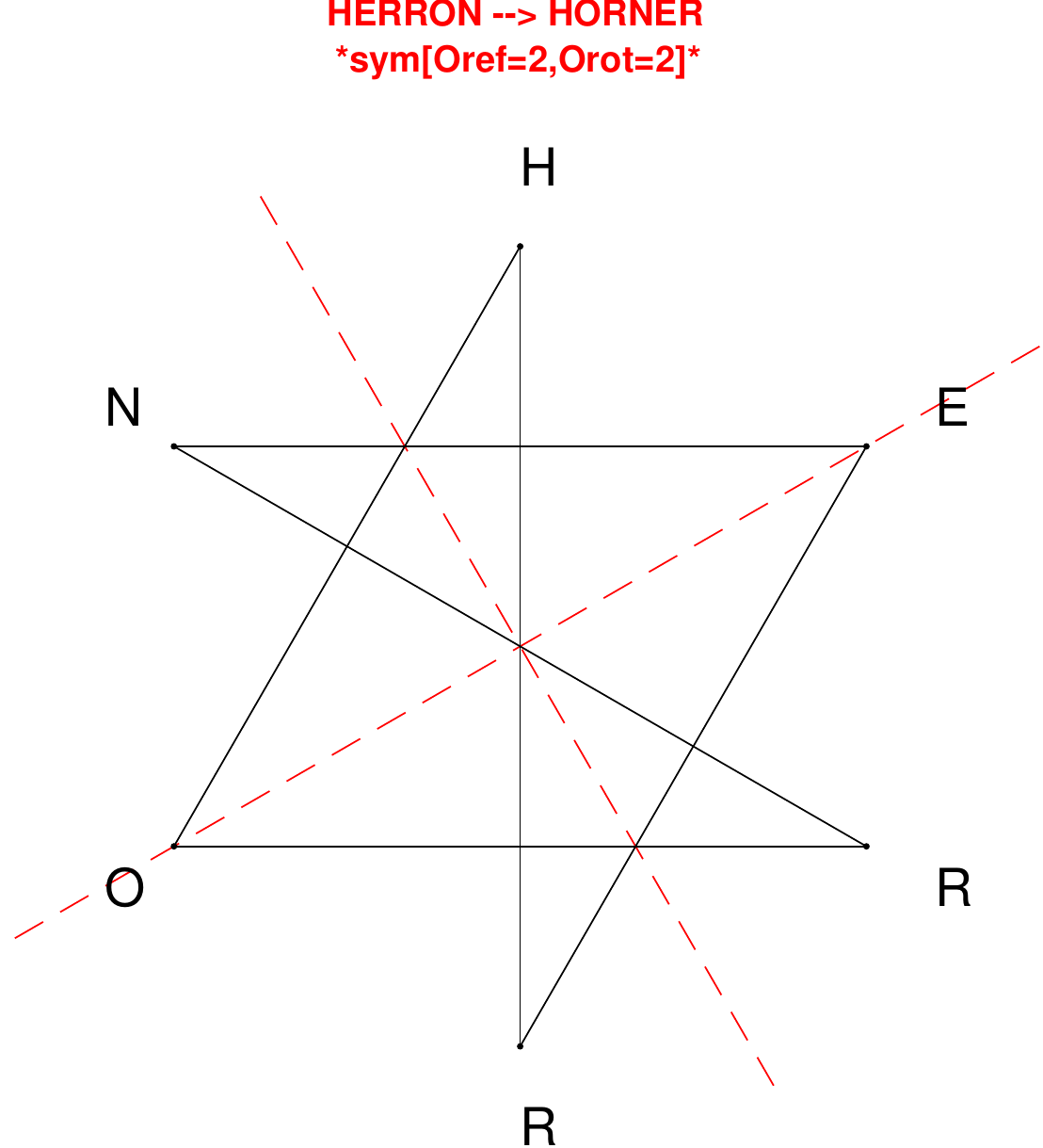}
\end{subfigure}
\end{figure}

\begin{figure}[H]
\centering
\begin{subfigure}[T]{0.19\textwidth}
\centering
\includegraphics[width=\textwidth]{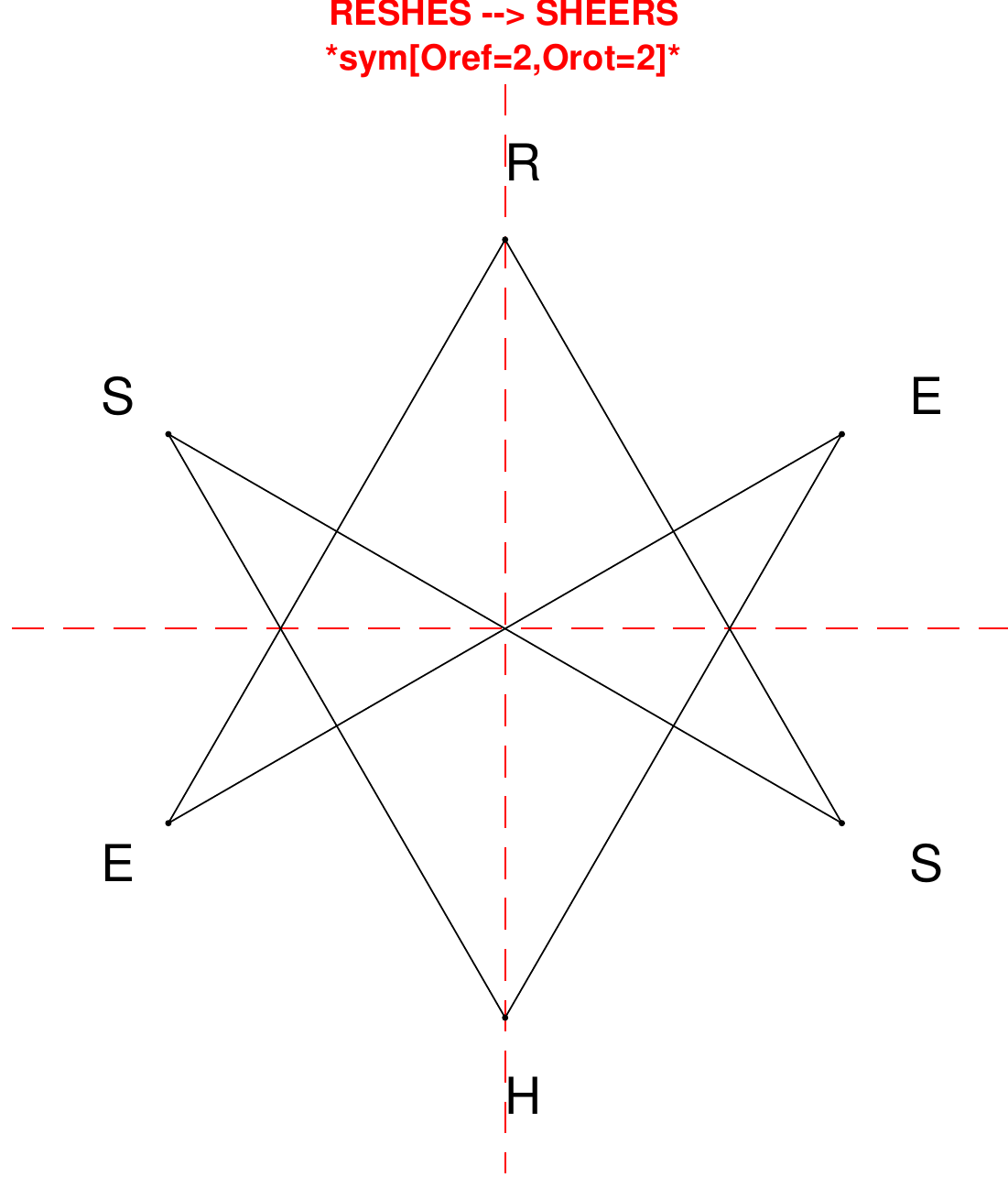}
\end{subfigure}
\hfill
\begin{subfigure}[T]{0.19\textwidth}
\centering
\includegraphics[width=\textwidth]{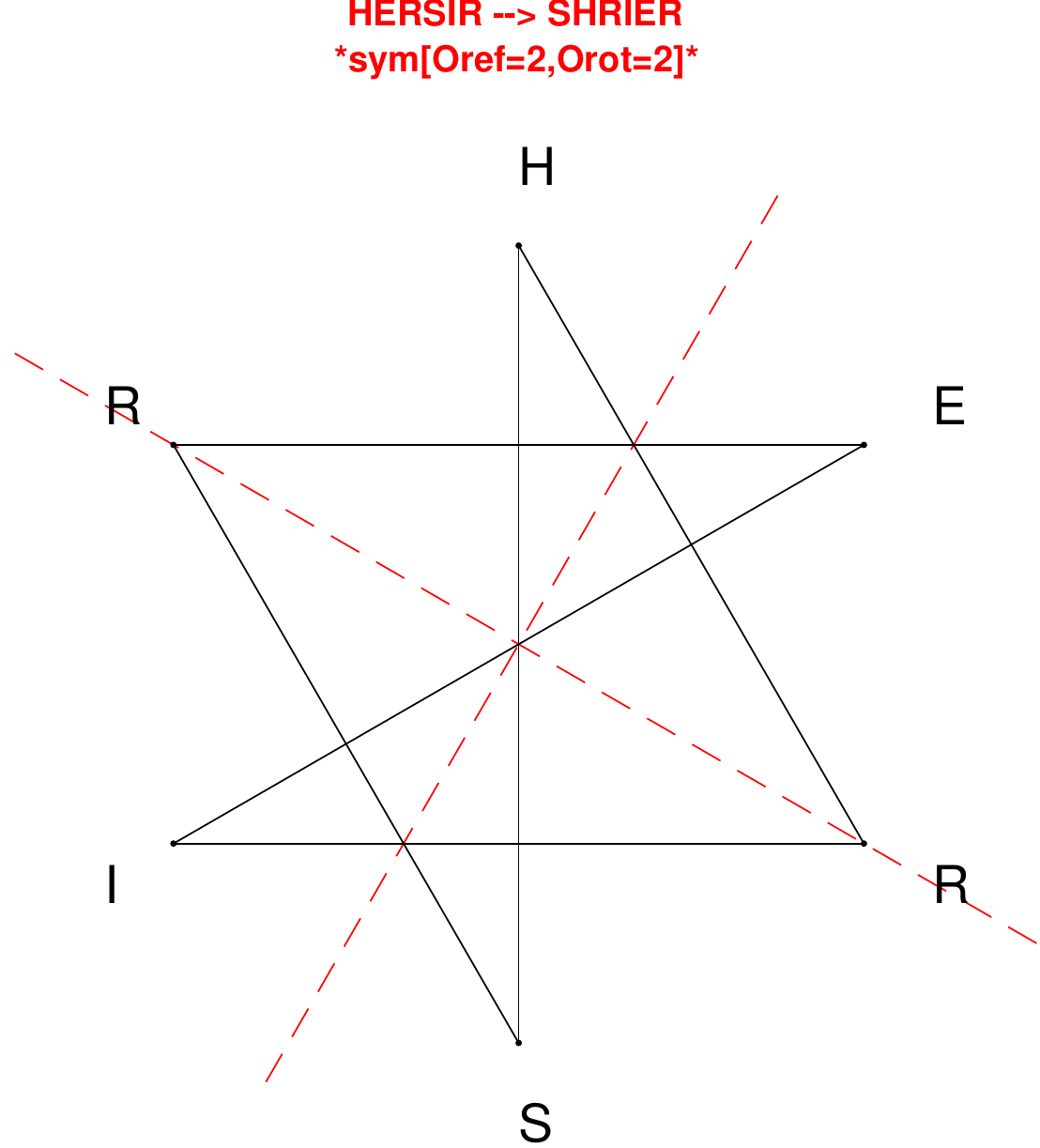}
\end{subfigure}
\hfill
\begin{subfigure}[T]{0.19\textwidth}
\centering
\includegraphics[width=\textwidth]{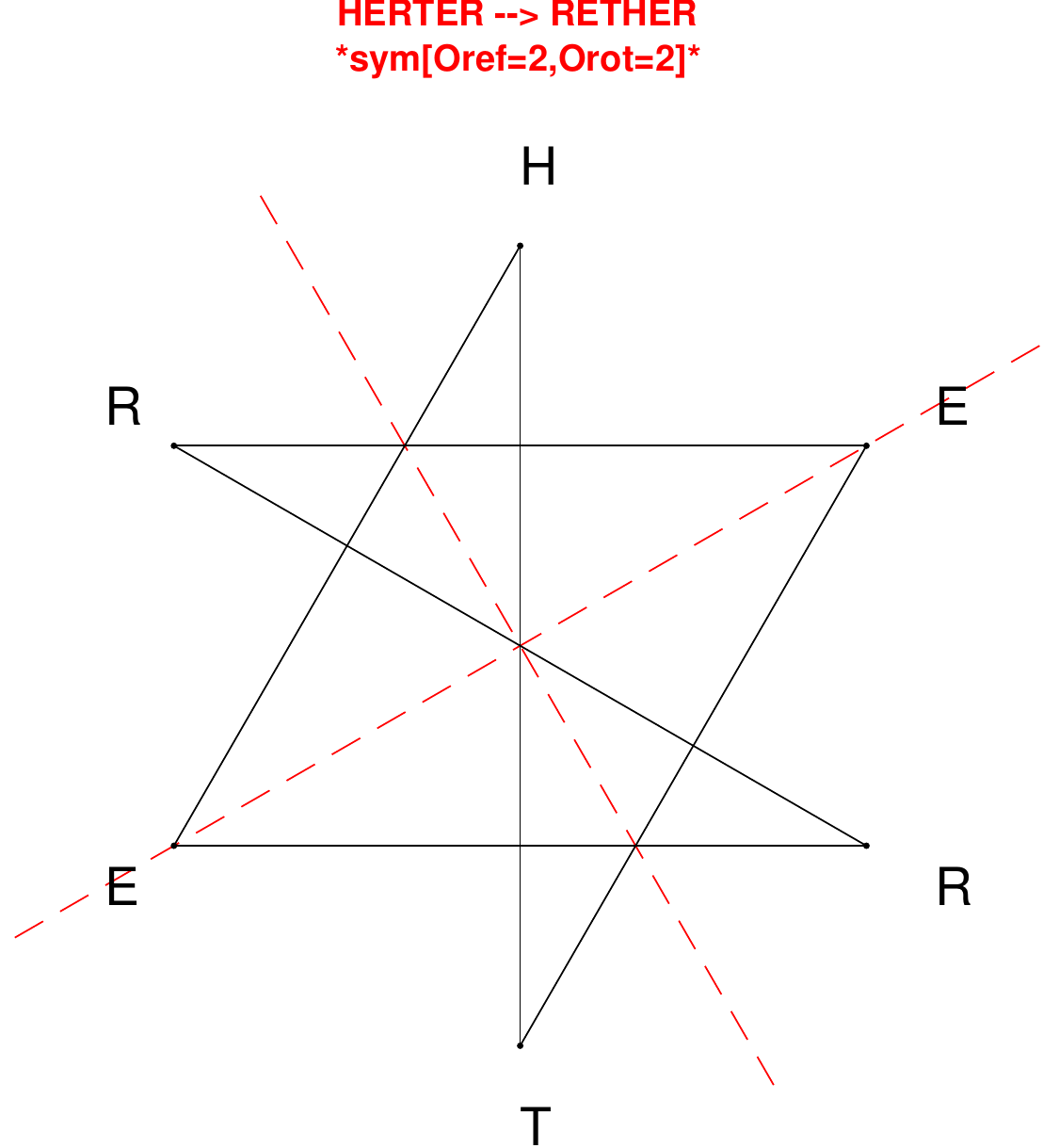}
\end{subfigure}
\hfill
\begin{subfigure}[T]{0.19\textwidth}
\centering
\includegraphics[width=\textwidth]{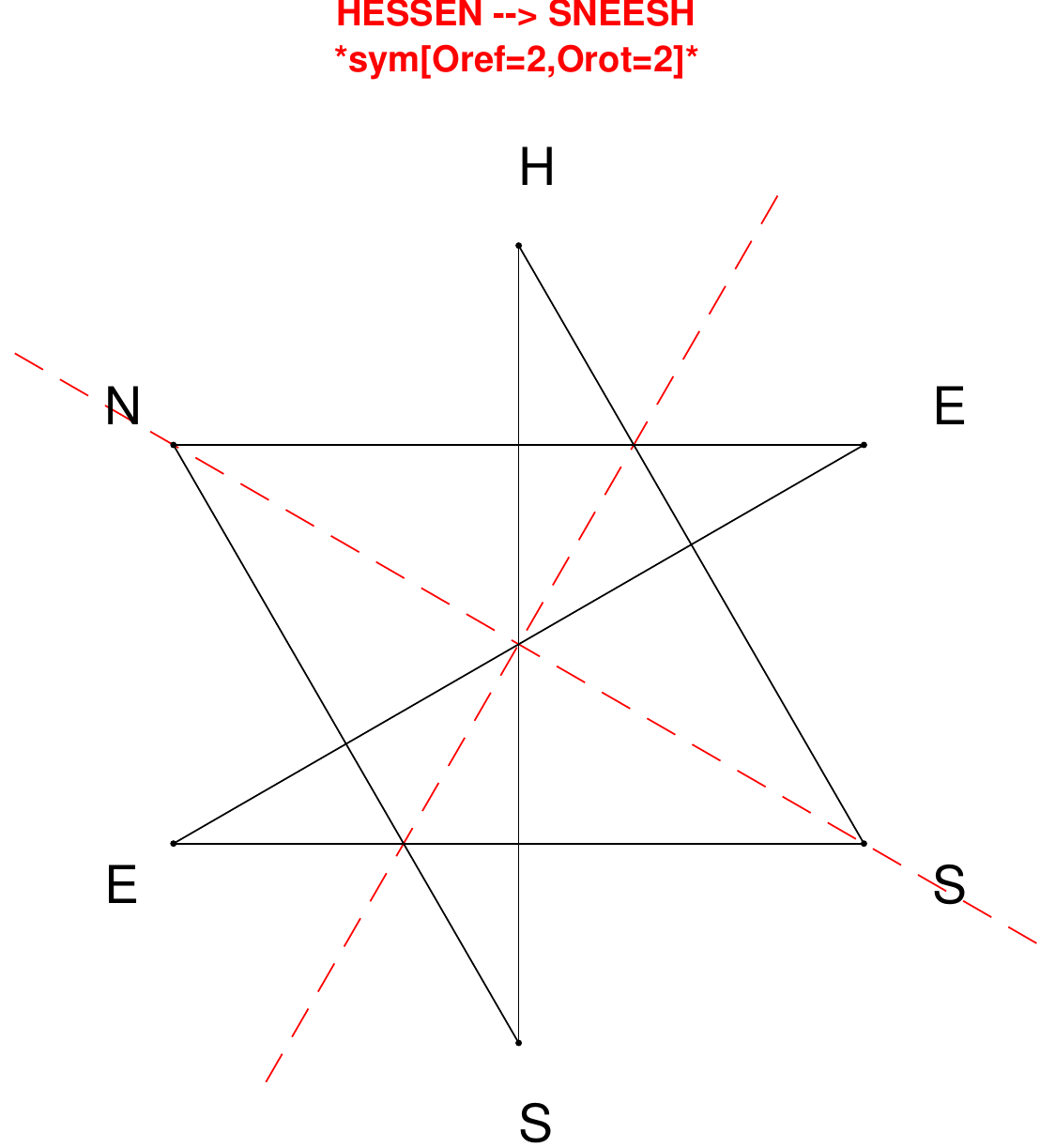}
\end{subfigure}
\hfill
\begin{subfigure}[T]{0.19\textwidth}
\centering
\includegraphics[width=\textwidth]{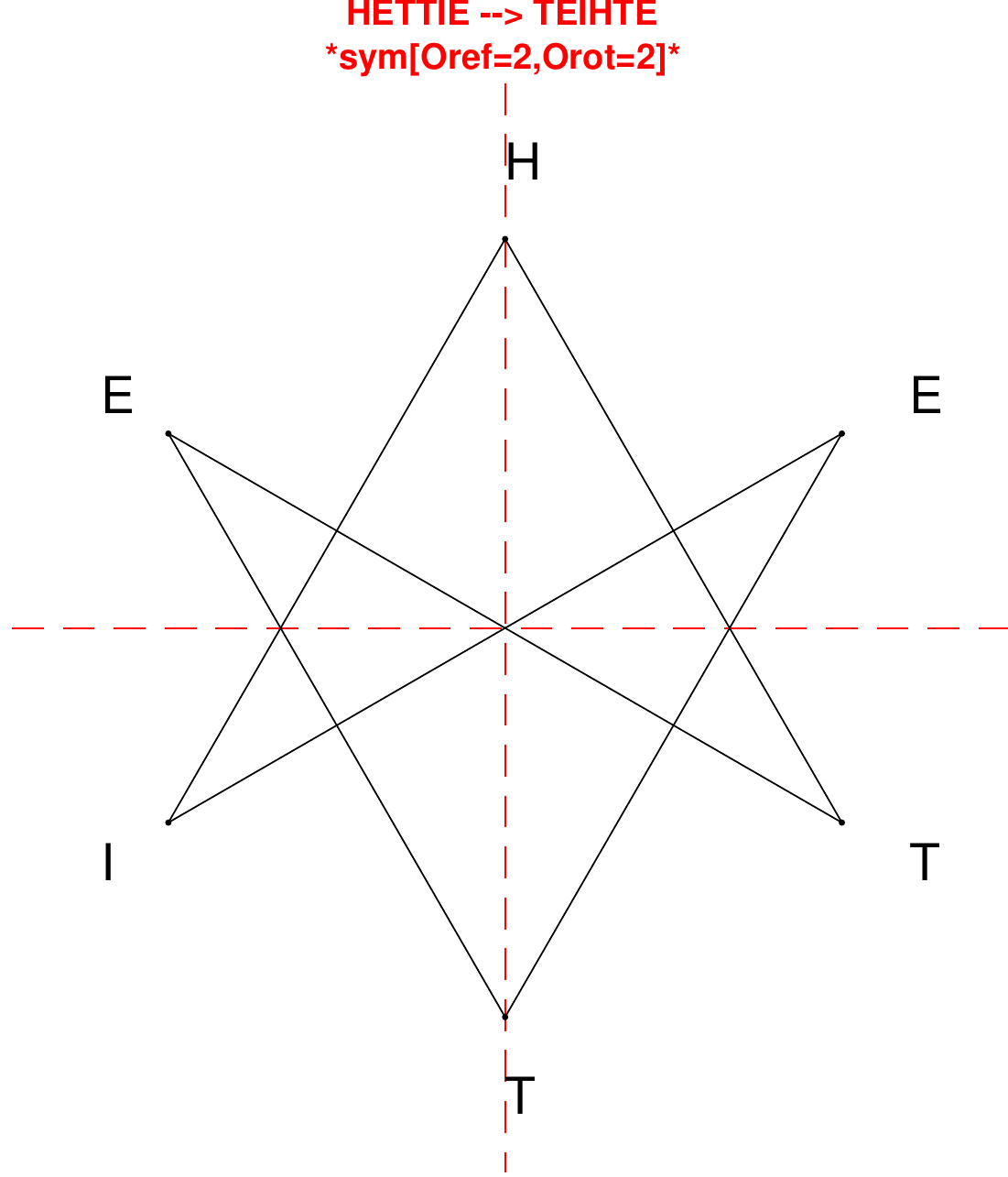}
\end{subfigure}
\end{figure}

\begin{figure}[H]
\centering
\begin{subfigure}[T]{0.19\textwidth}
\centering
\includegraphics[width=\textwidth]{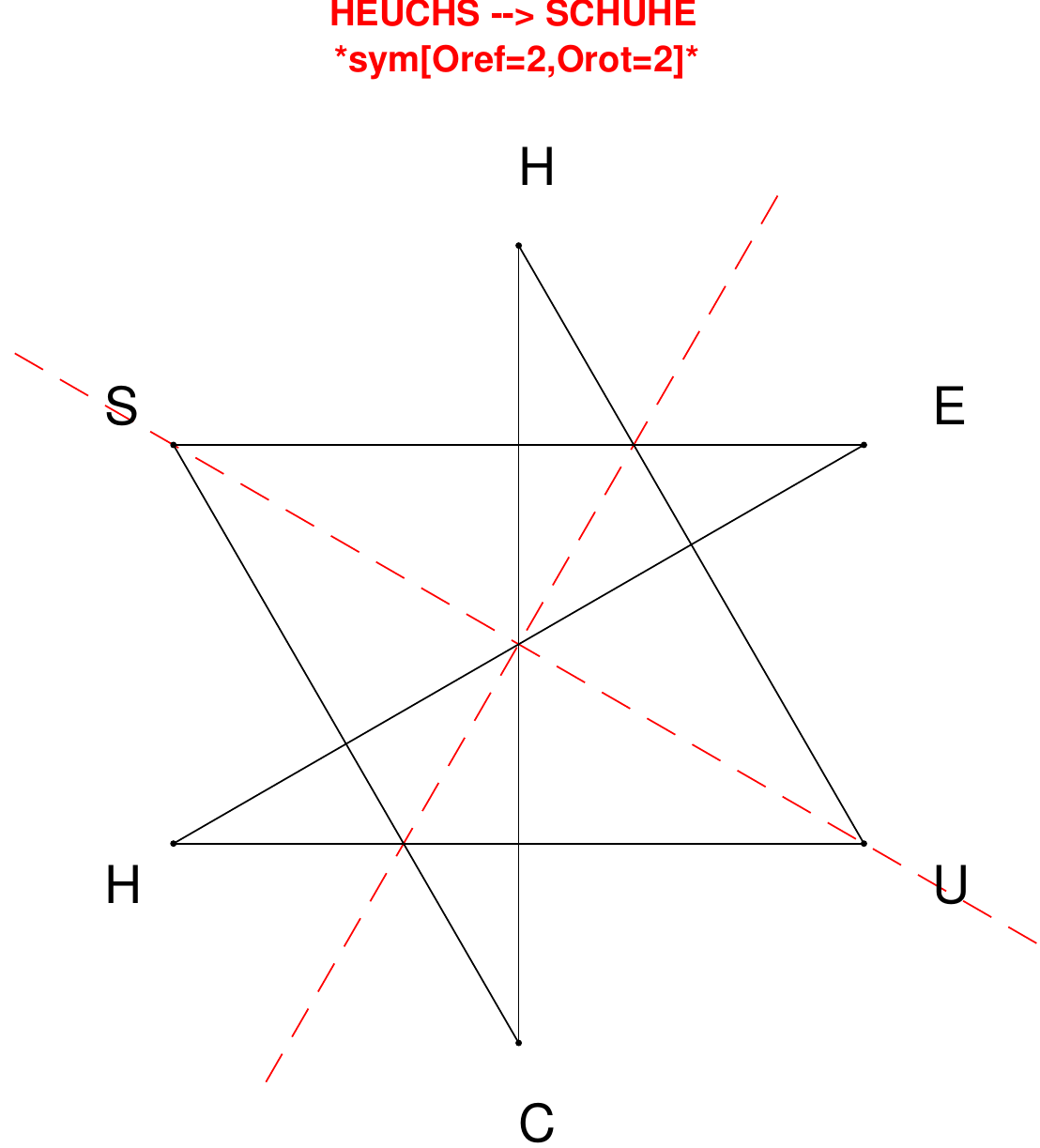}
\end{subfigure}
\hfill
\begin{subfigure}[T]{0.19\textwidth}
\centering
\includegraphics[width=\textwidth]{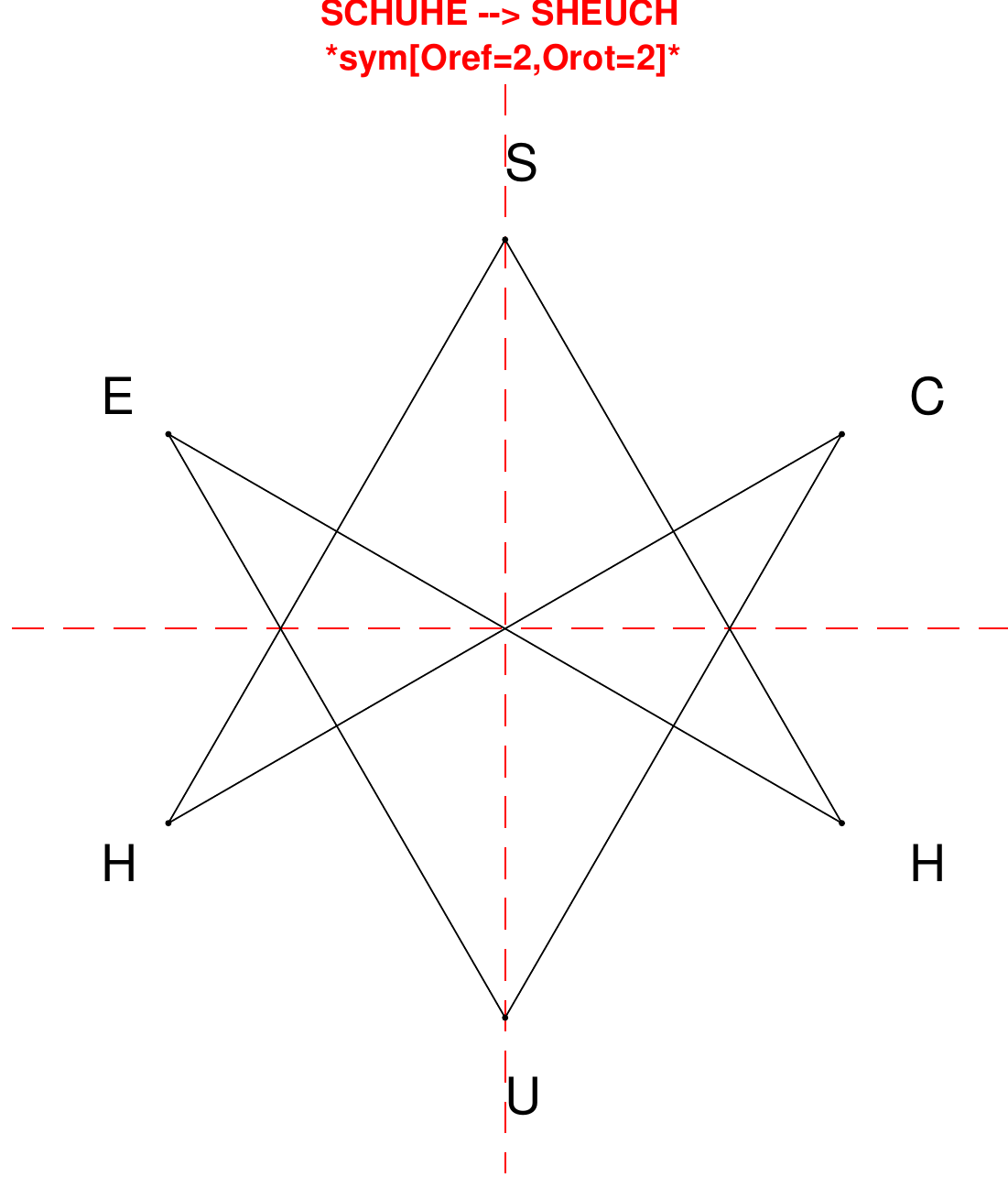}
\end{subfigure}
\hfill
\begin{subfigure}[T]{0.19\textwidth}
\centering
\includegraphics[width=\textwidth]{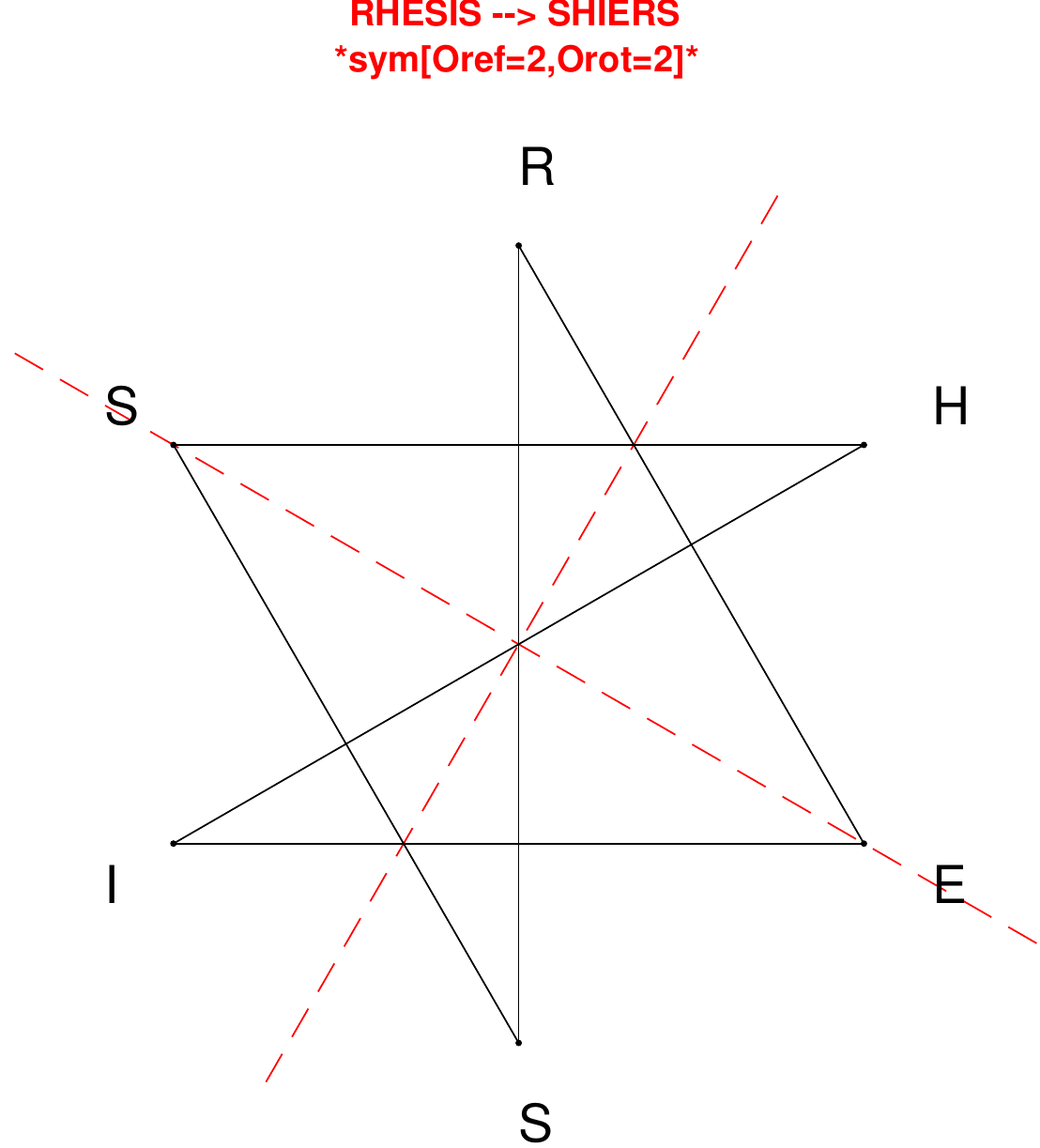}
\end{subfigure}
\hfill
\begin{subfigure}[T]{0.19\textwidth}
\centering
\includegraphics[width=\textwidth]{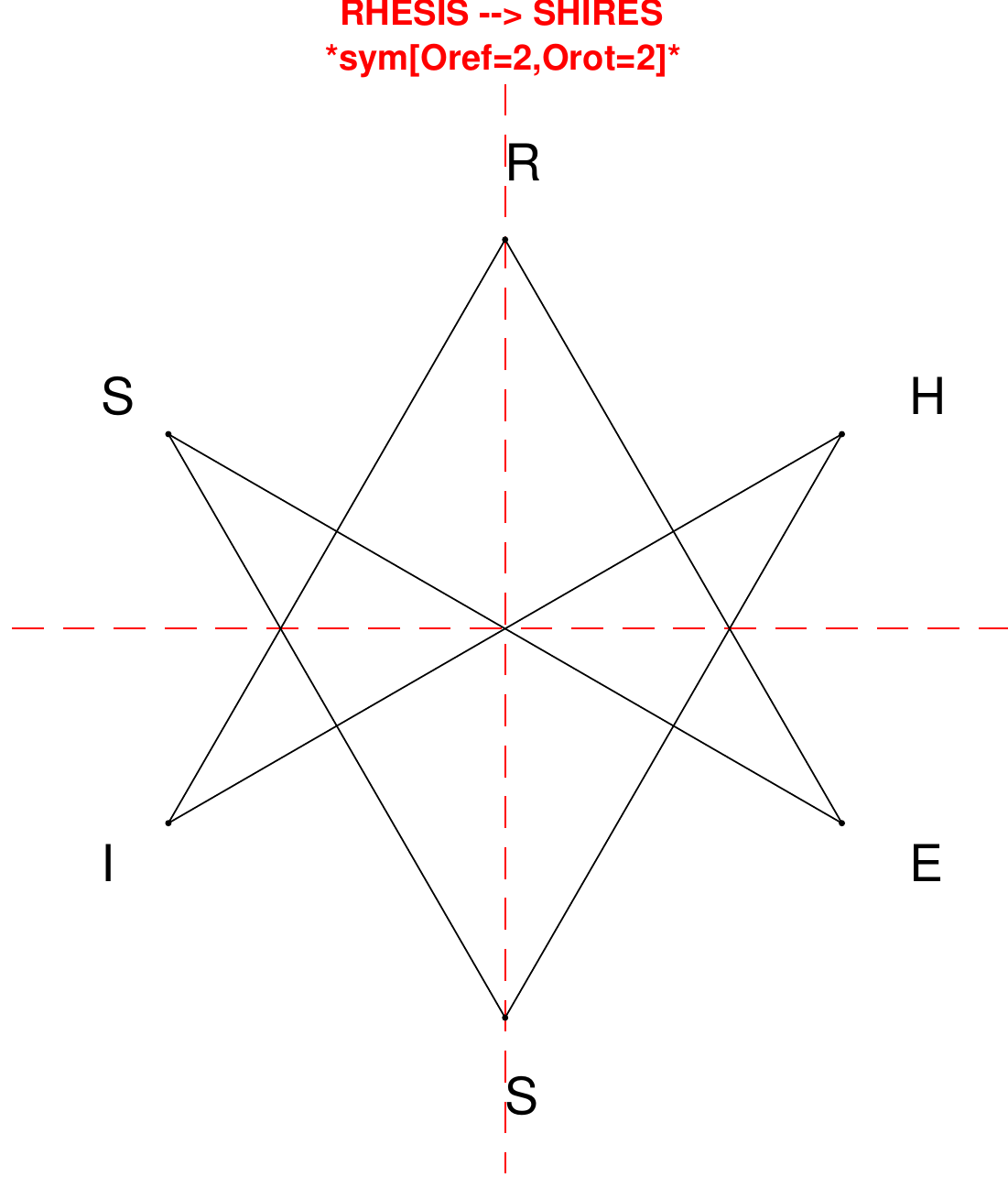}
\end{subfigure}
\hfill
\begin{subfigure}[T]{0.19\textwidth}
\centering
\includegraphics[width=\textwidth]{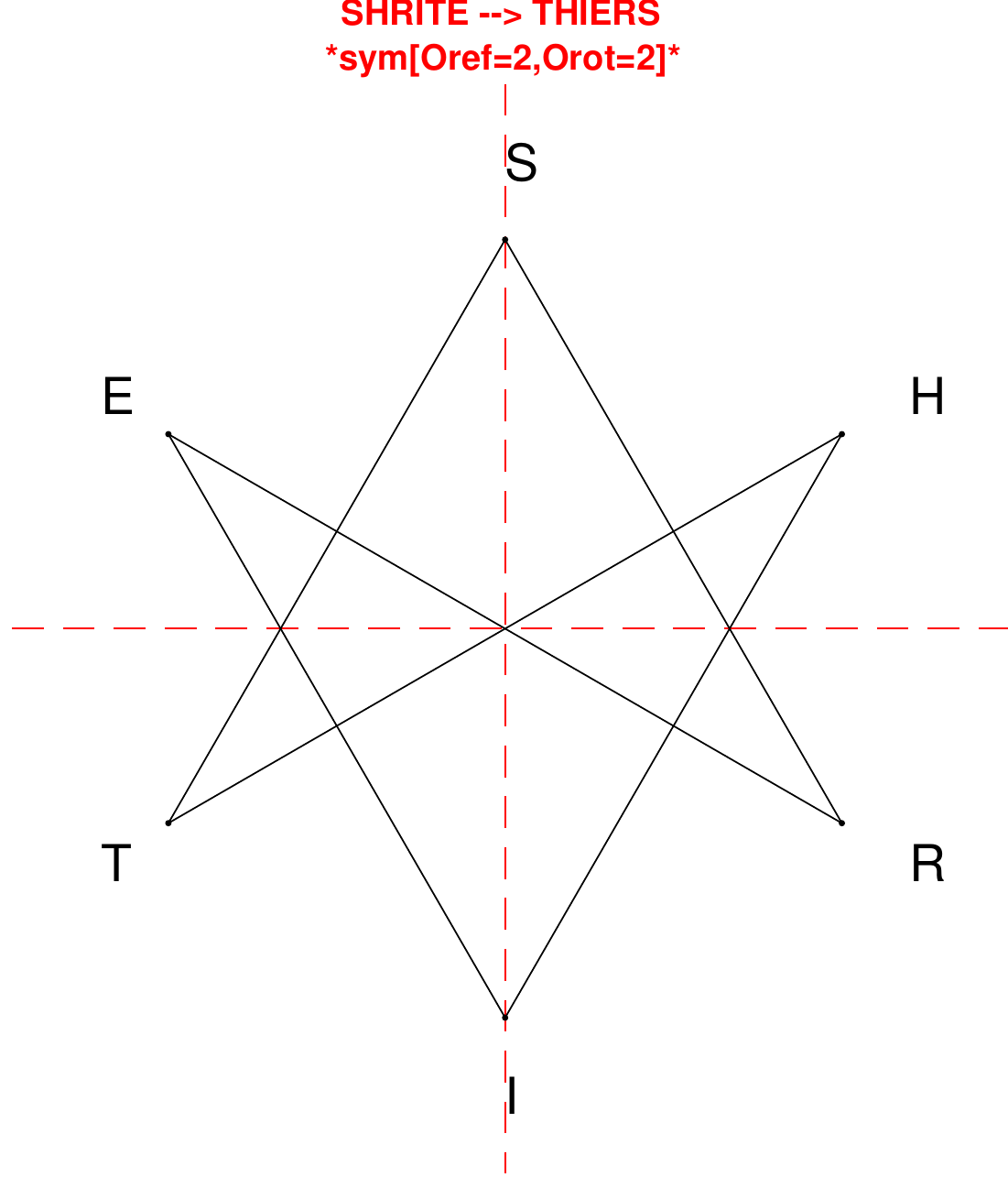}
\end{subfigure}
\end{figure}

\begin{figure}[H]
\centering
\begin{subfigure}[T]{0.19\textwidth}
\centering
\includegraphics[width=\textwidth]{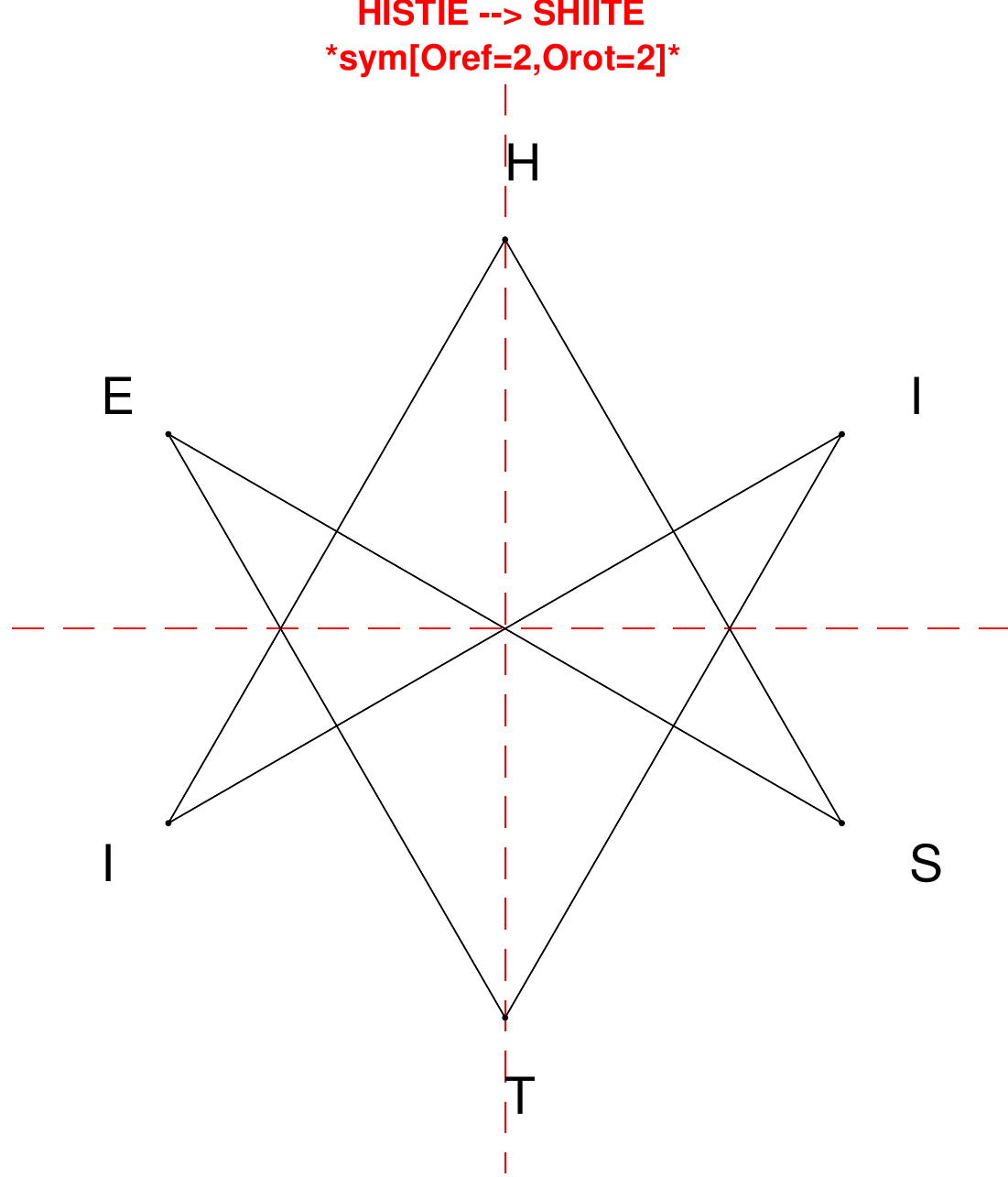}
\end{subfigure}
\hfill
\begin{subfigure}[T]{0.19\textwidth}
\centering
\includegraphics[width=\textwidth]{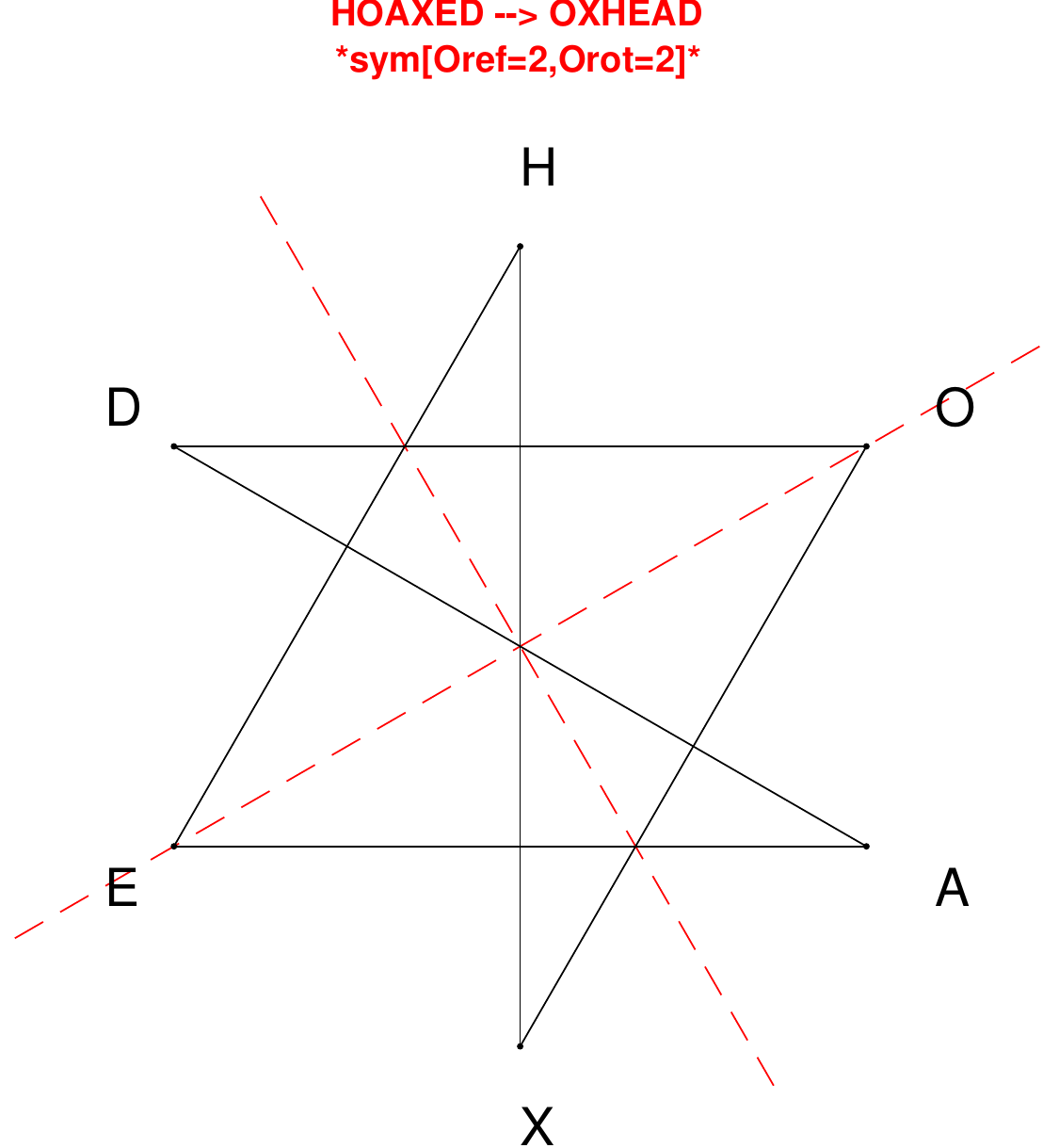}
\end{subfigure}
\hfill
\begin{subfigure}[T]{0.19\textwidth}
\centering
\includegraphics[width=\textwidth]{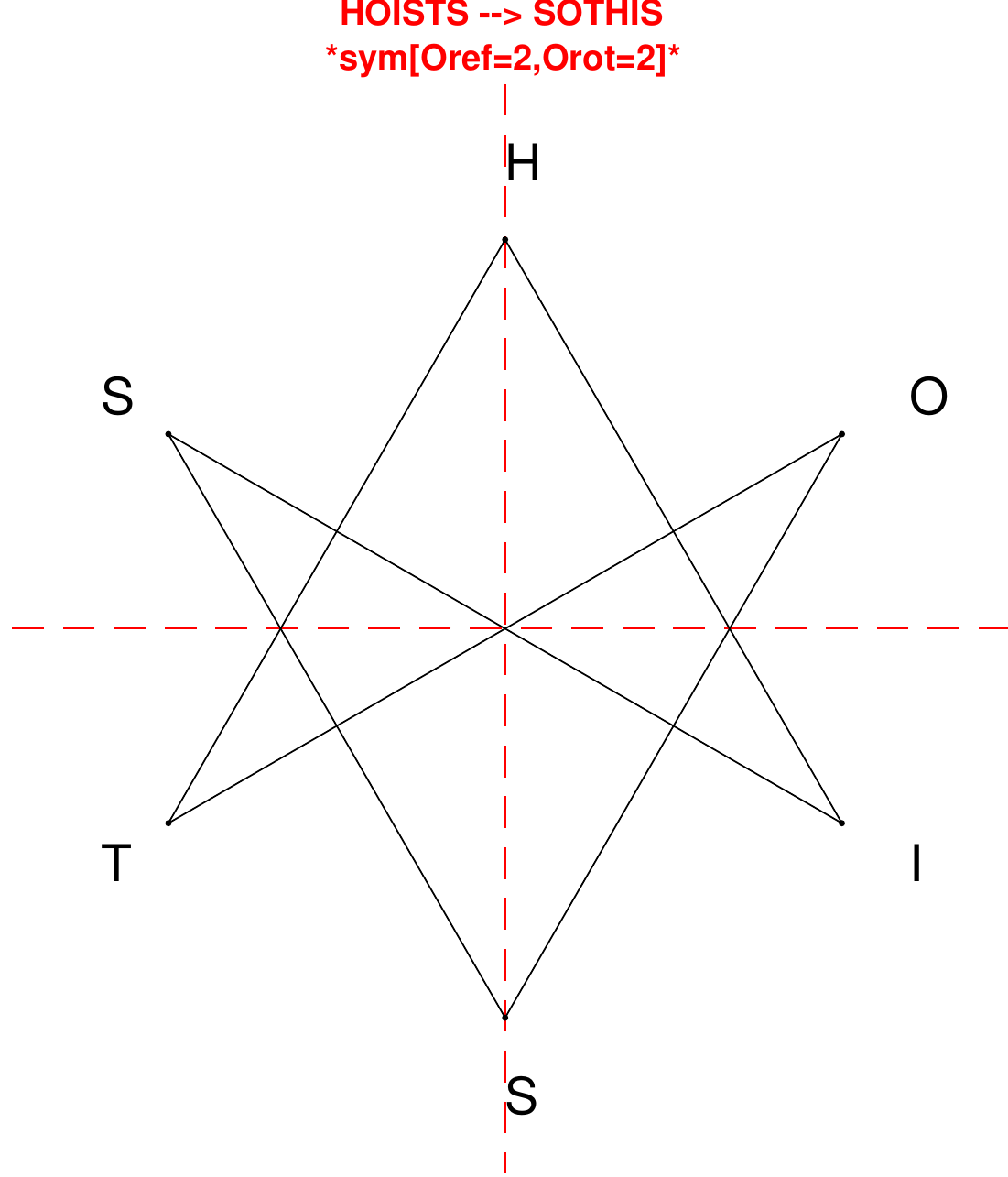}
\end{subfigure}
\hfill
\begin{subfigure}[T]{0.19\textwidth}
\centering
\includegraphics[width=\textwidth]{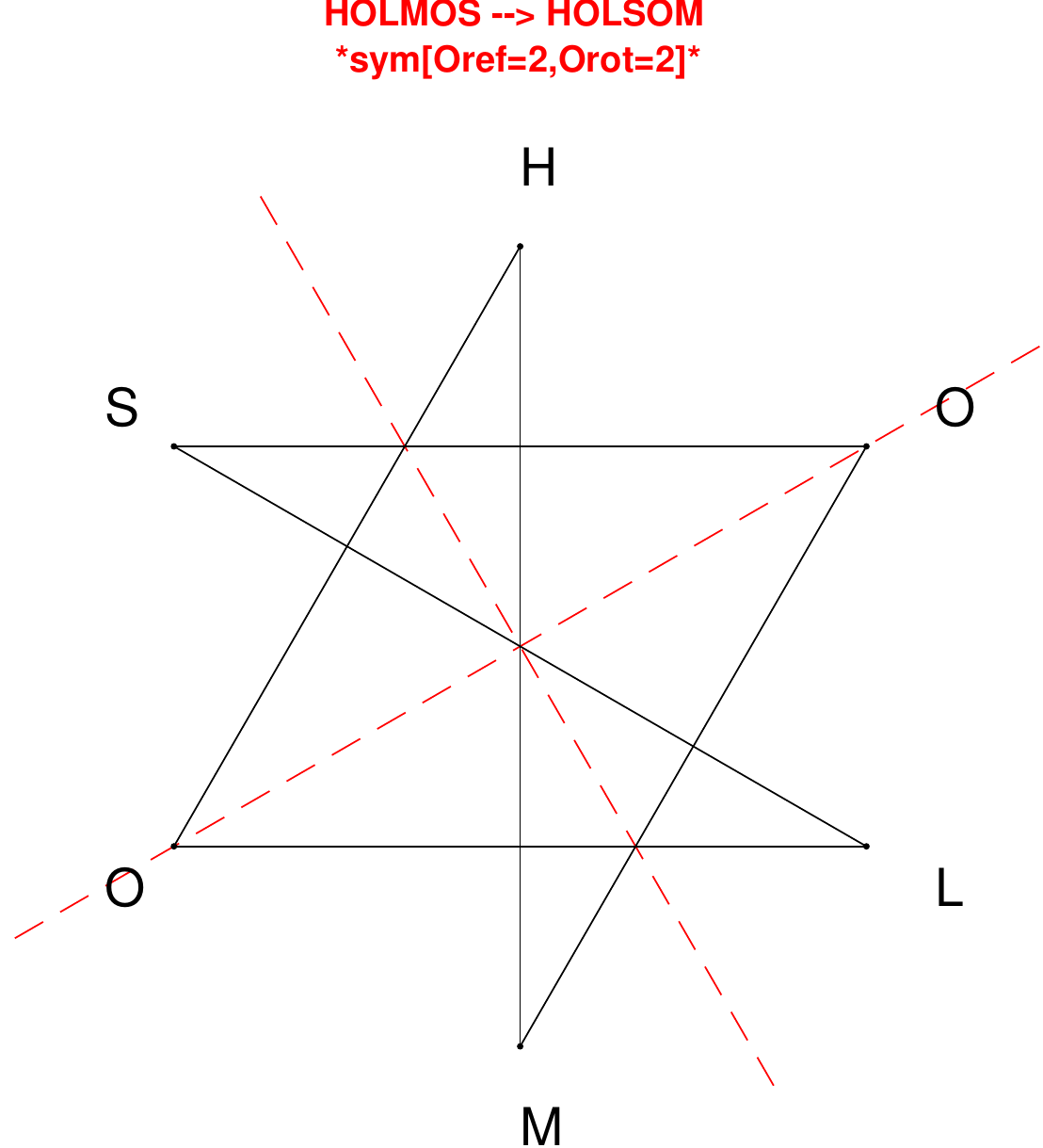}
\end{subfigure}
\hfill
\begin{subfigure}[T]{0.19\textwidth}
\centering
\includegraphics[width=\textwidth]{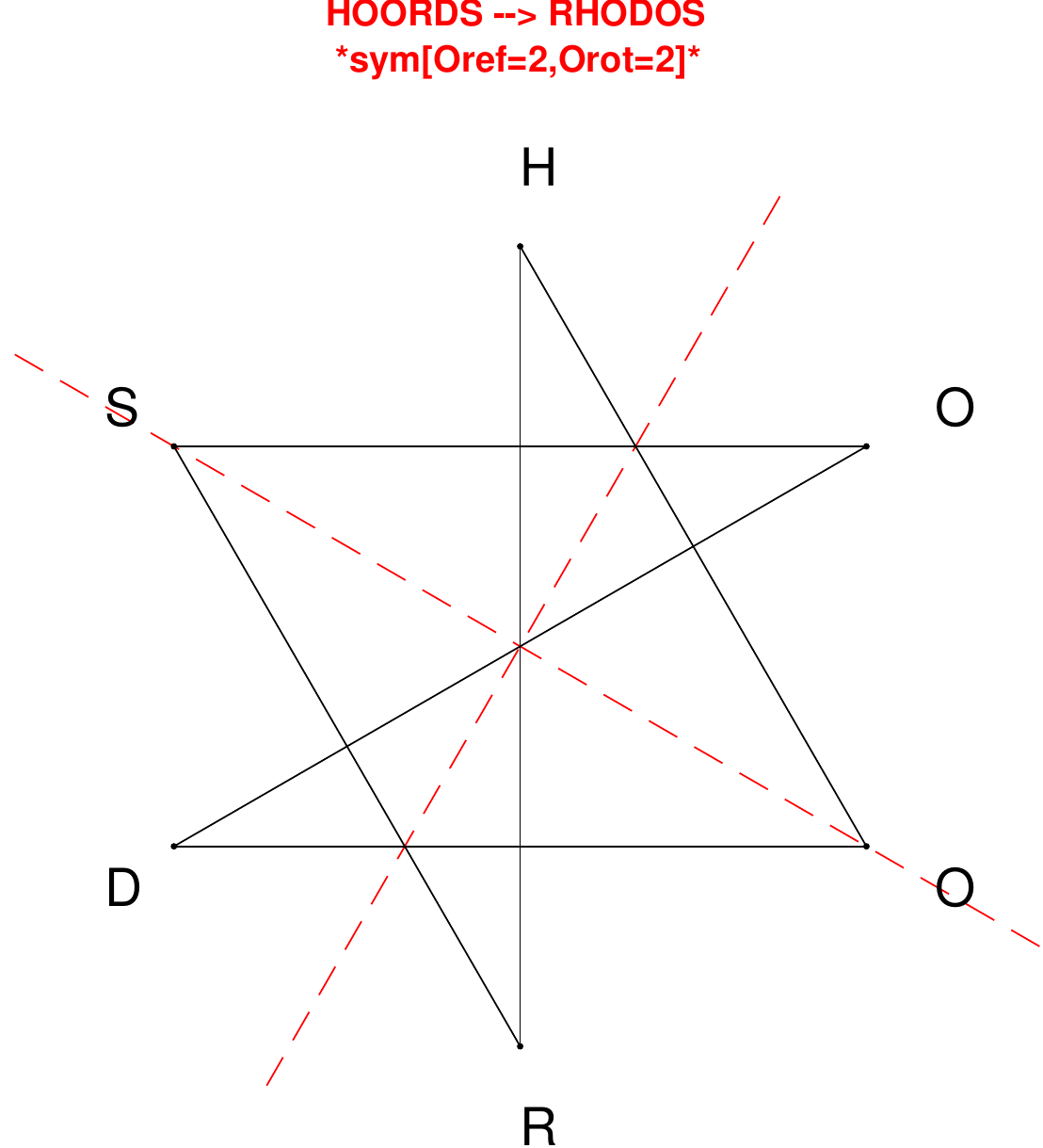}
\end{subfigure}
\end{figure}

\begin{figure}[H]
\centering
\begin{subfigure}[T]{0.19\textwidth}
\centering
\includegraphics[width=\textwidth]{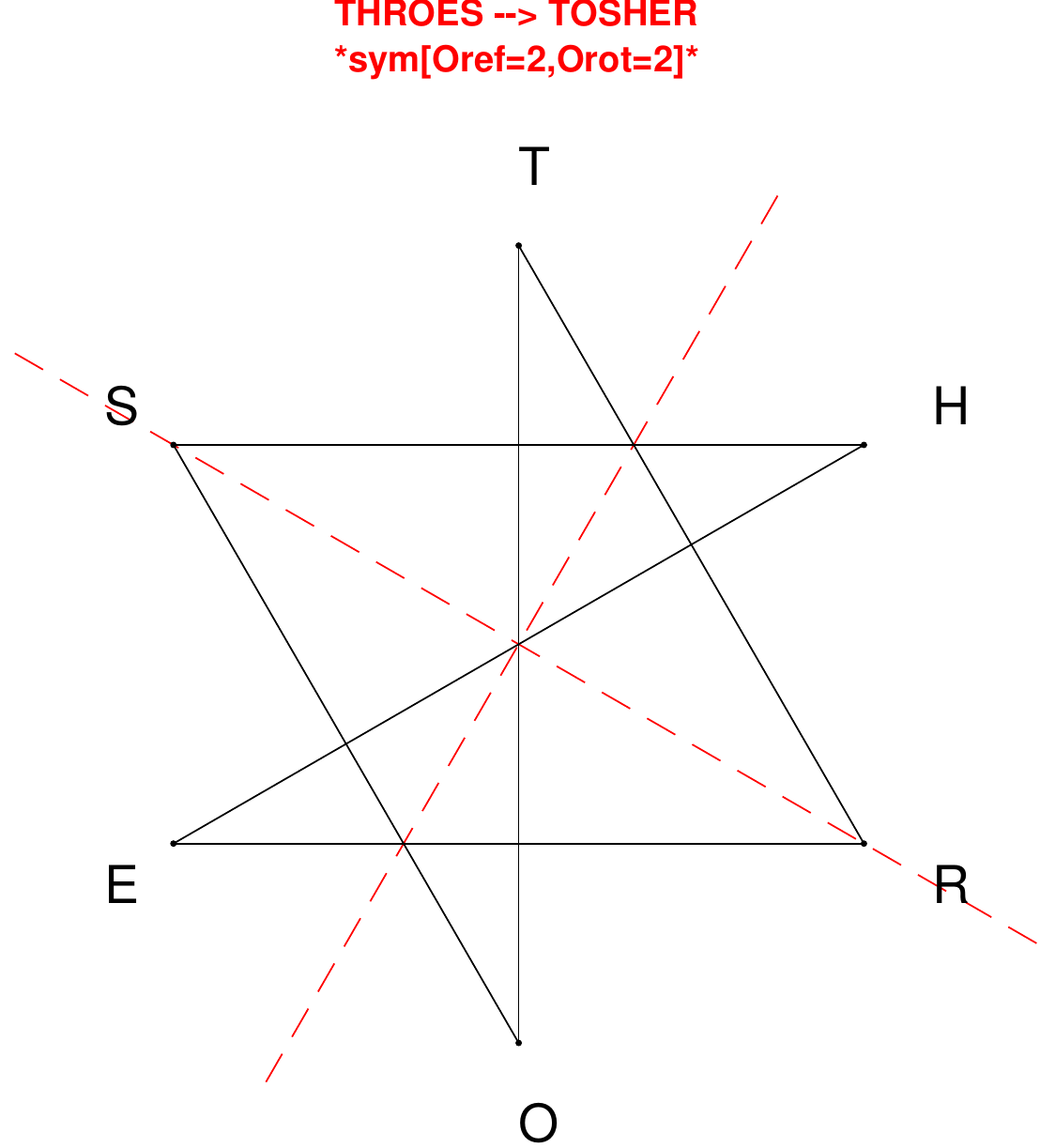}
\end{subfigure}
\hfill
\begin{subfigure}[T]{0.19\textwidth}
\centering
\includegraphics[width=\textwidth]{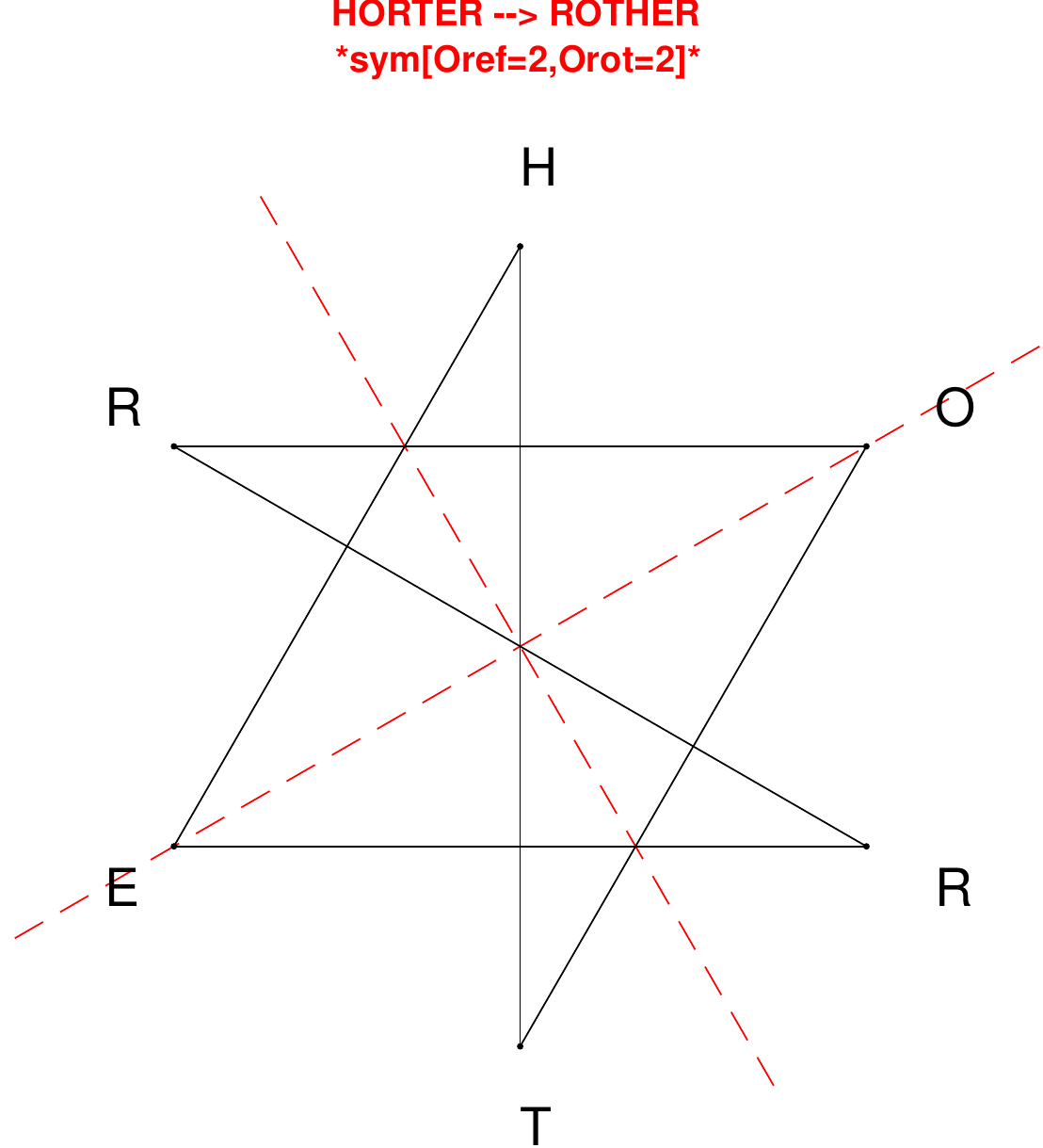}
\end{subfigure}
\hfill
\begin{subfigure}[T]{0.19\textwidth}
\centering
\includegraphics[width=\textwidth]{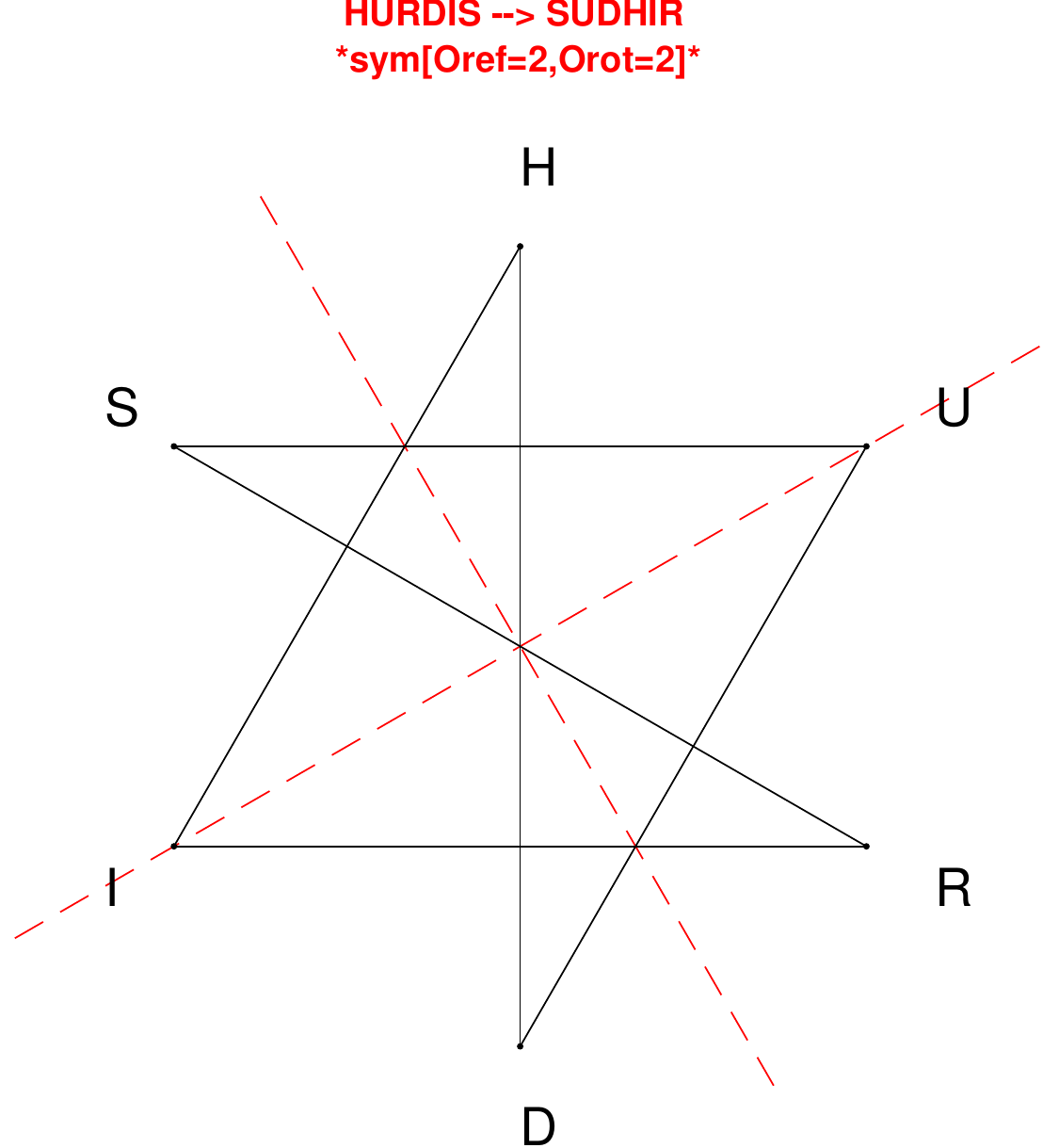}
\end{subfigure}
\hfill
\begin{subfigure}[T]{0.19\textwidth}
\centering
\includegraphics[width=\textwidth]{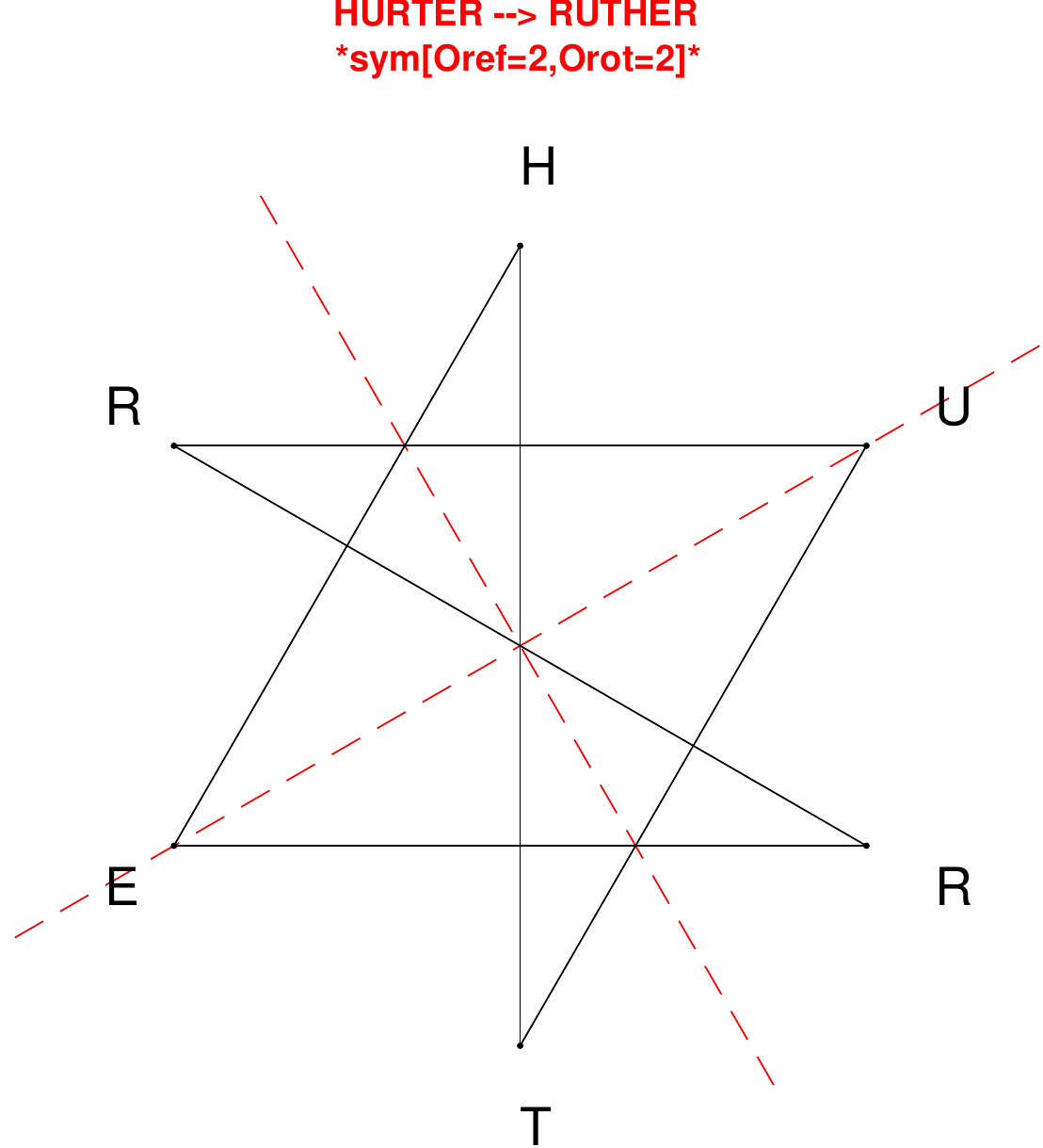}
\end{subfigure}
\hfill
\begin{subfigure}[T]{0.19\textwidth}
\centering
\includegraphics[width=\textwidth]{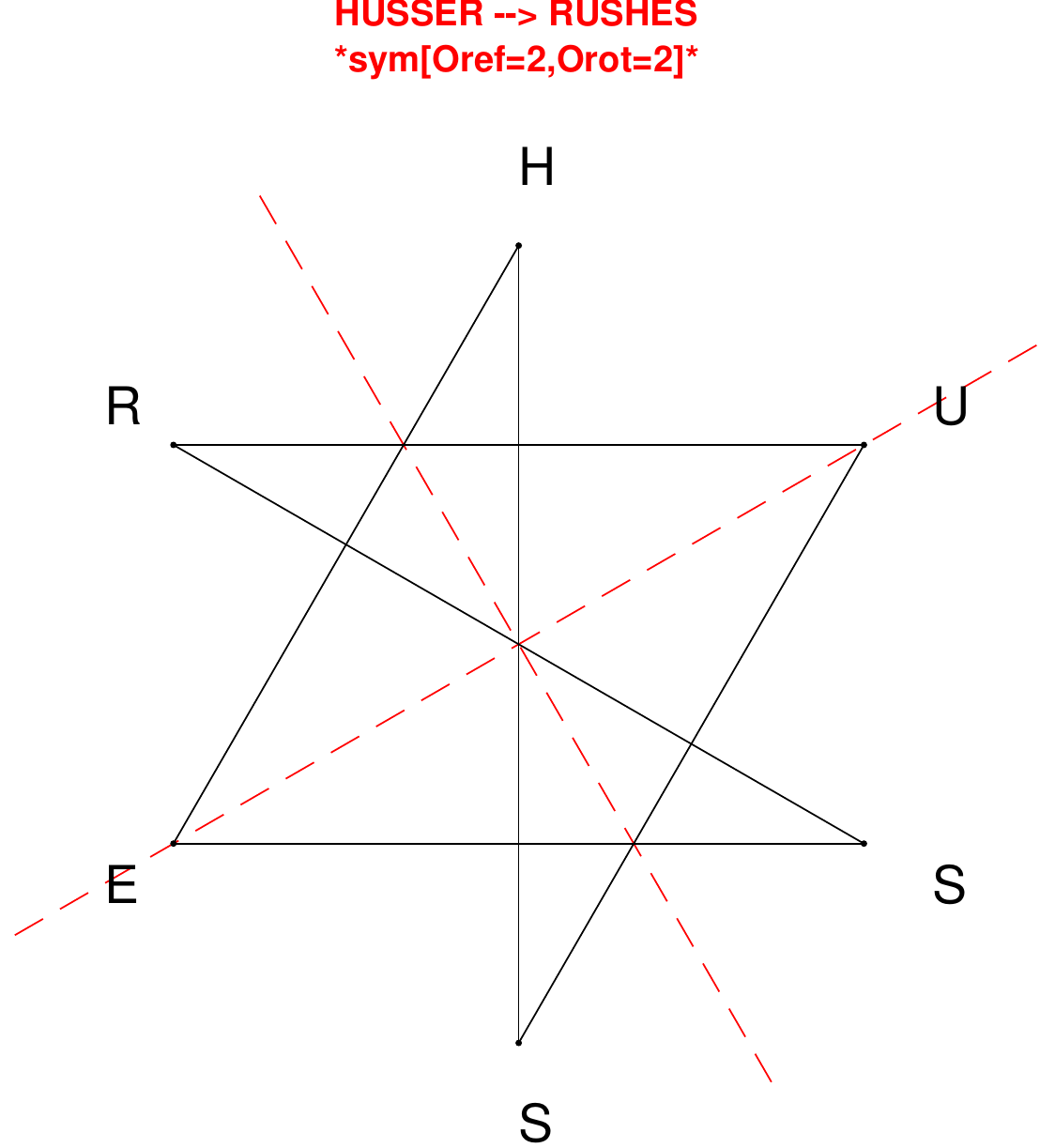}
\end{subfigure}
\end{figure}

\begin{figure}[H]
\centering
\begin{subfigure}[T]{0.19\textwidth}
\centering
\includegraphics[width=\textwidth]{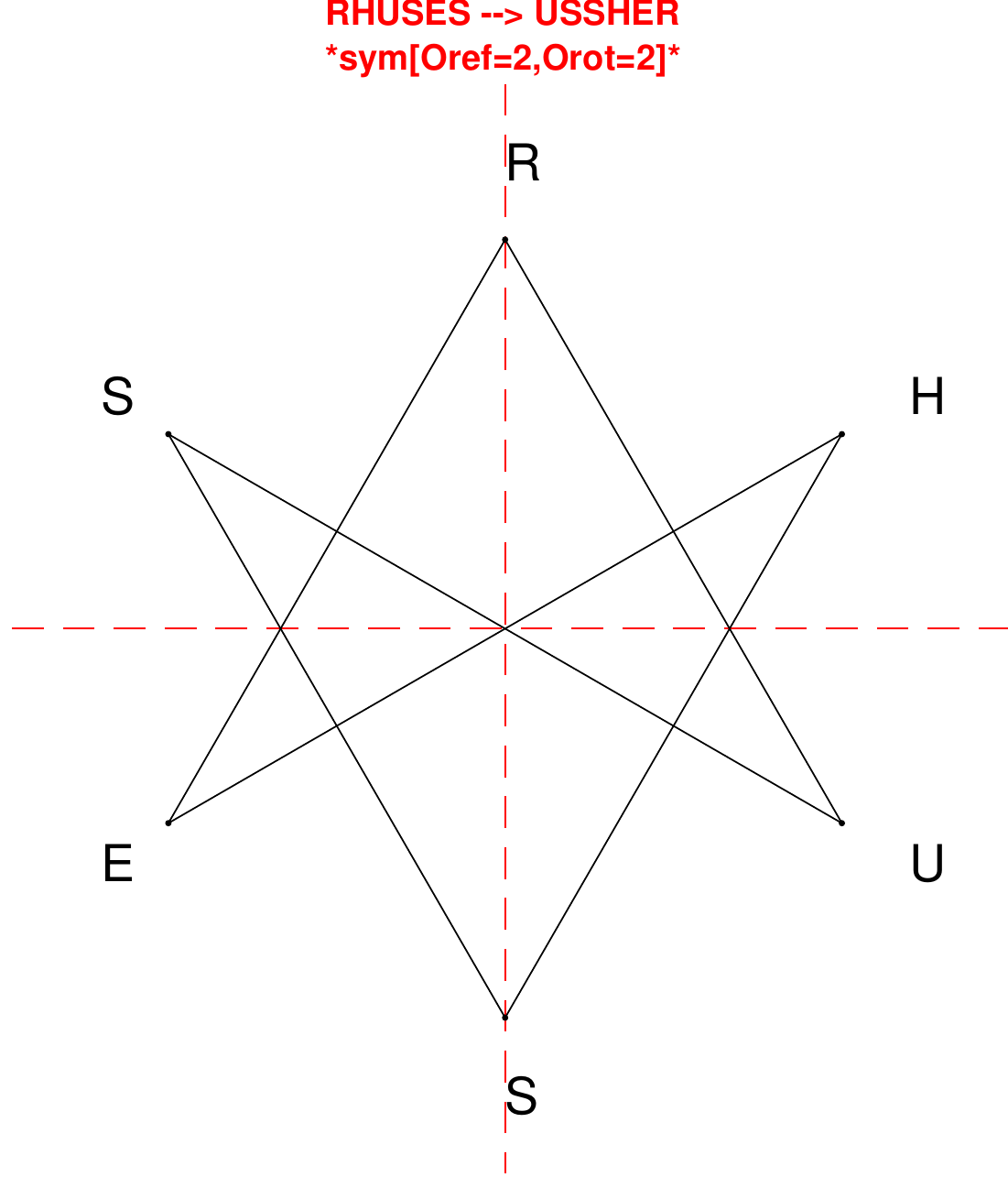}
\end{subfigure}
\hfill
\begin{subfigure}[T]{0.19\textwidth}
\centering
\includegraphics[width=\textwidth]{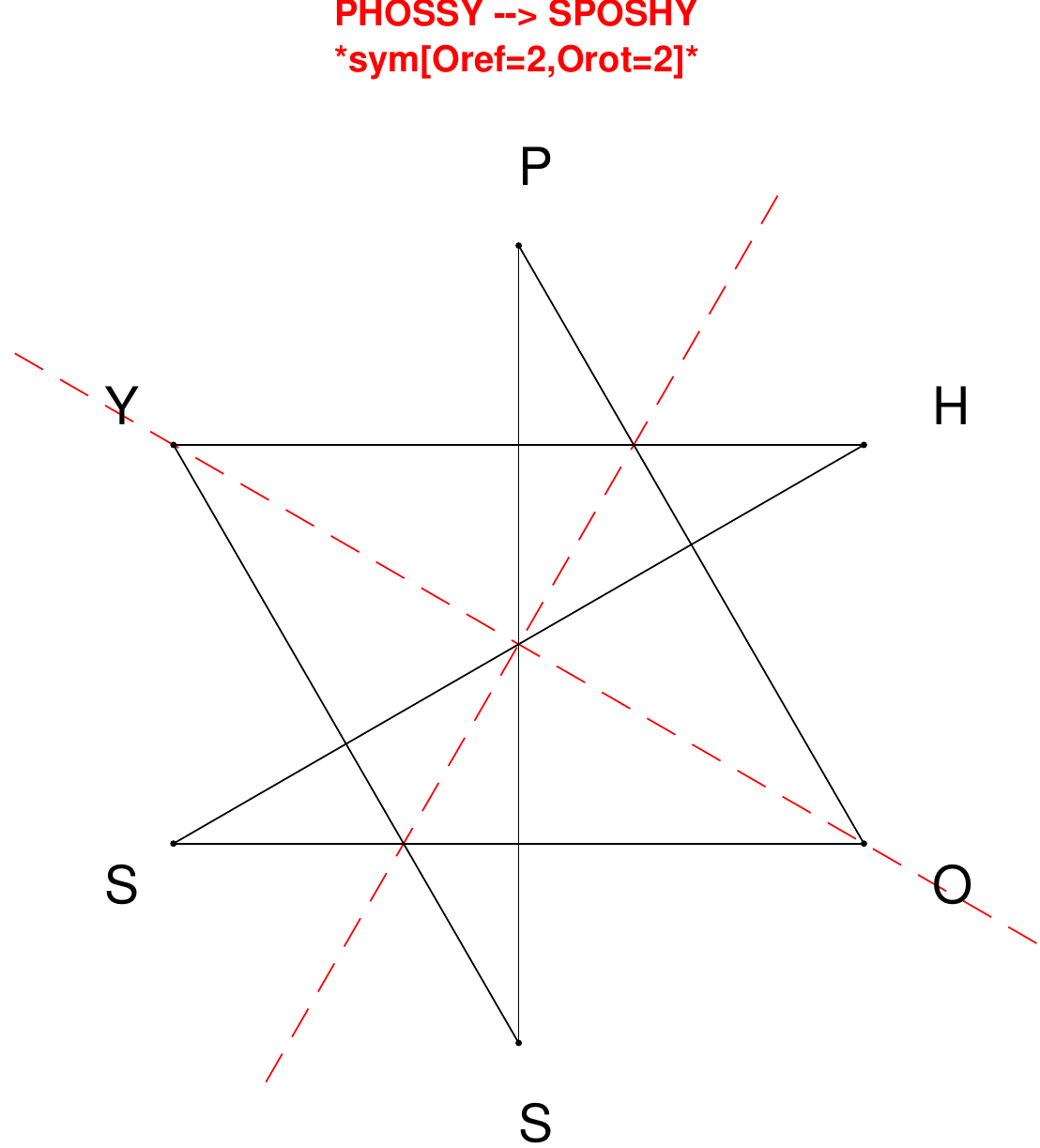}
\end{subfigure}
\hfill
\begin{subfigure}[T]{0.19\textwidth}
\centering
\includegraphics[width=\textwidth]{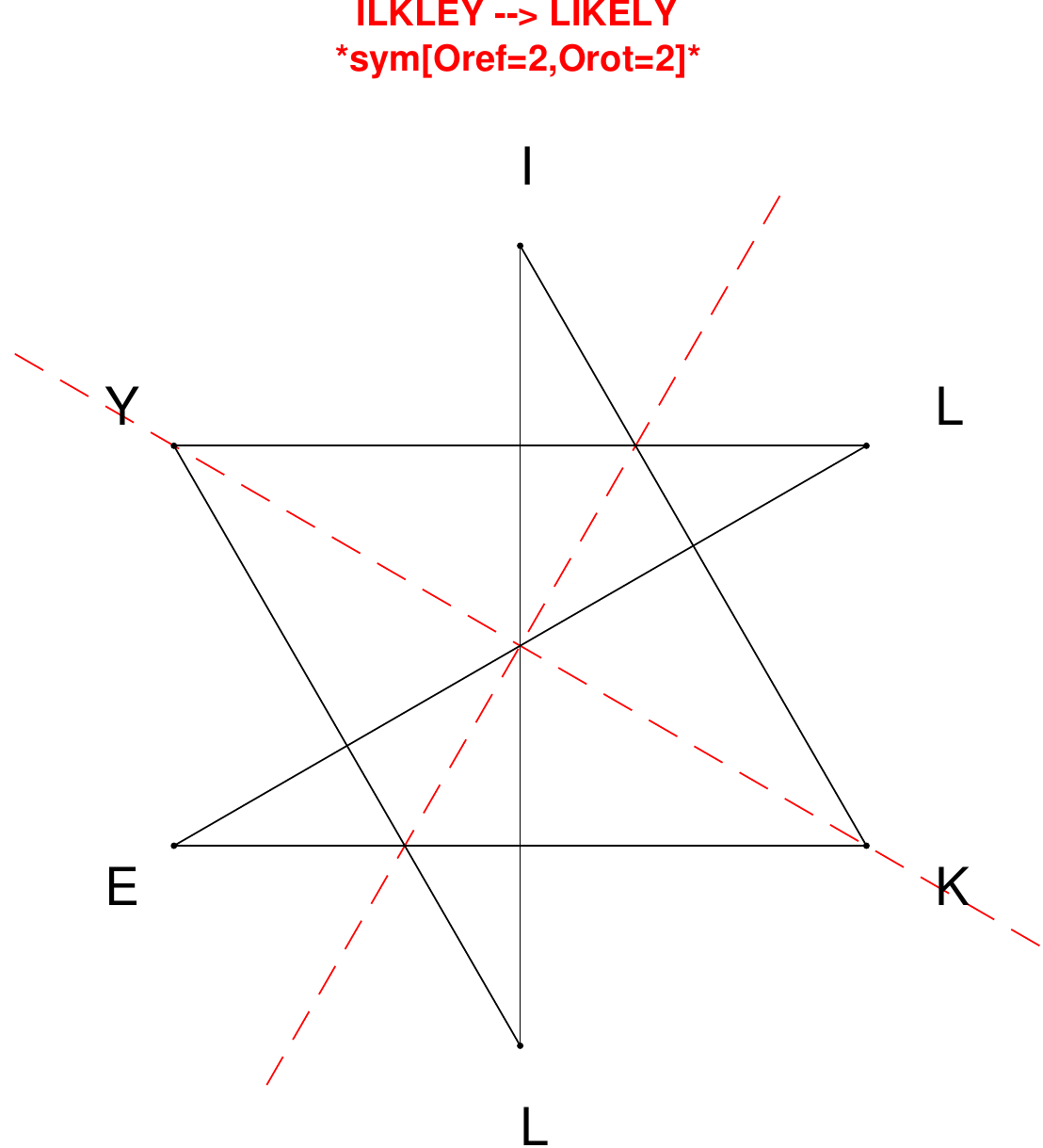}
\end{subfigure}
\hfill
\begin{subfigure}[T]{0.19\textwidth}
\centering
\includegraphics[width=\textwidth]{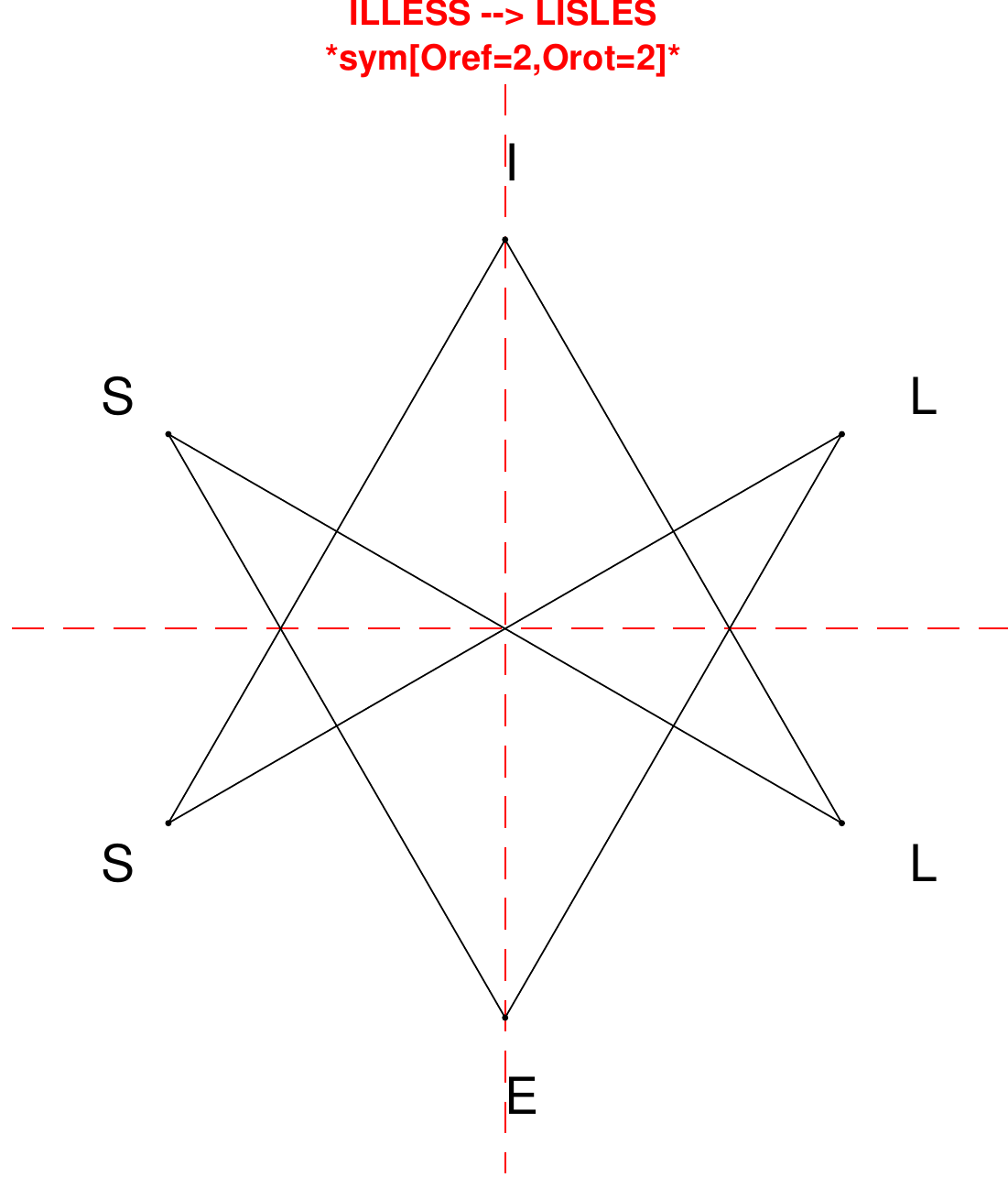}
\end{subfigure}
\hfill
\begin{subfigure}[T]{0.19\textwidth}
\centering
\includegraphics[width=\textwidth]{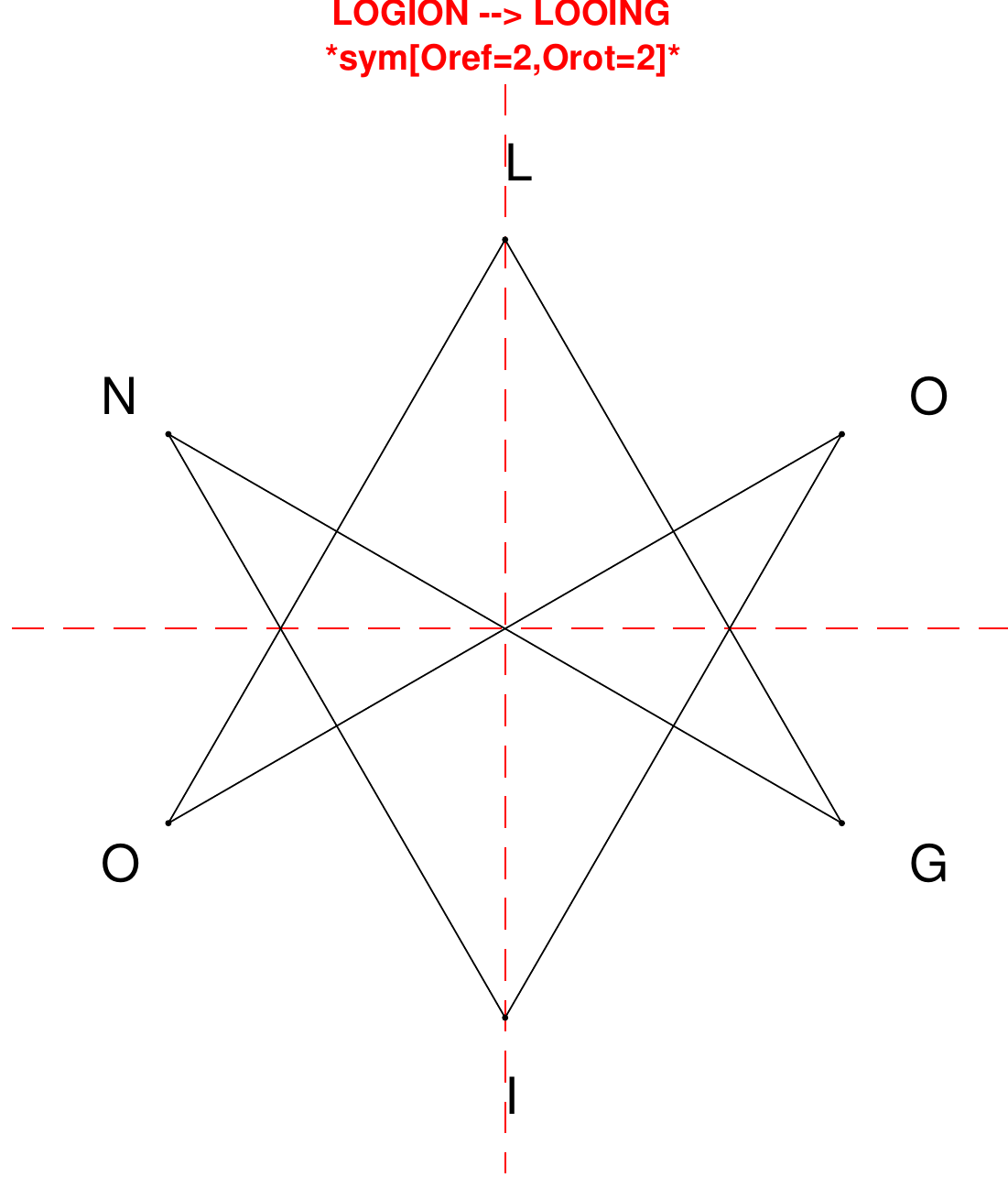}
\end{subfigure}
\end{figure}

\begin{figure}[H]
\centering
\begin{subfigure}[T]{0.19\textwidth}
\centering
\includegraphics[width=\textwidth]{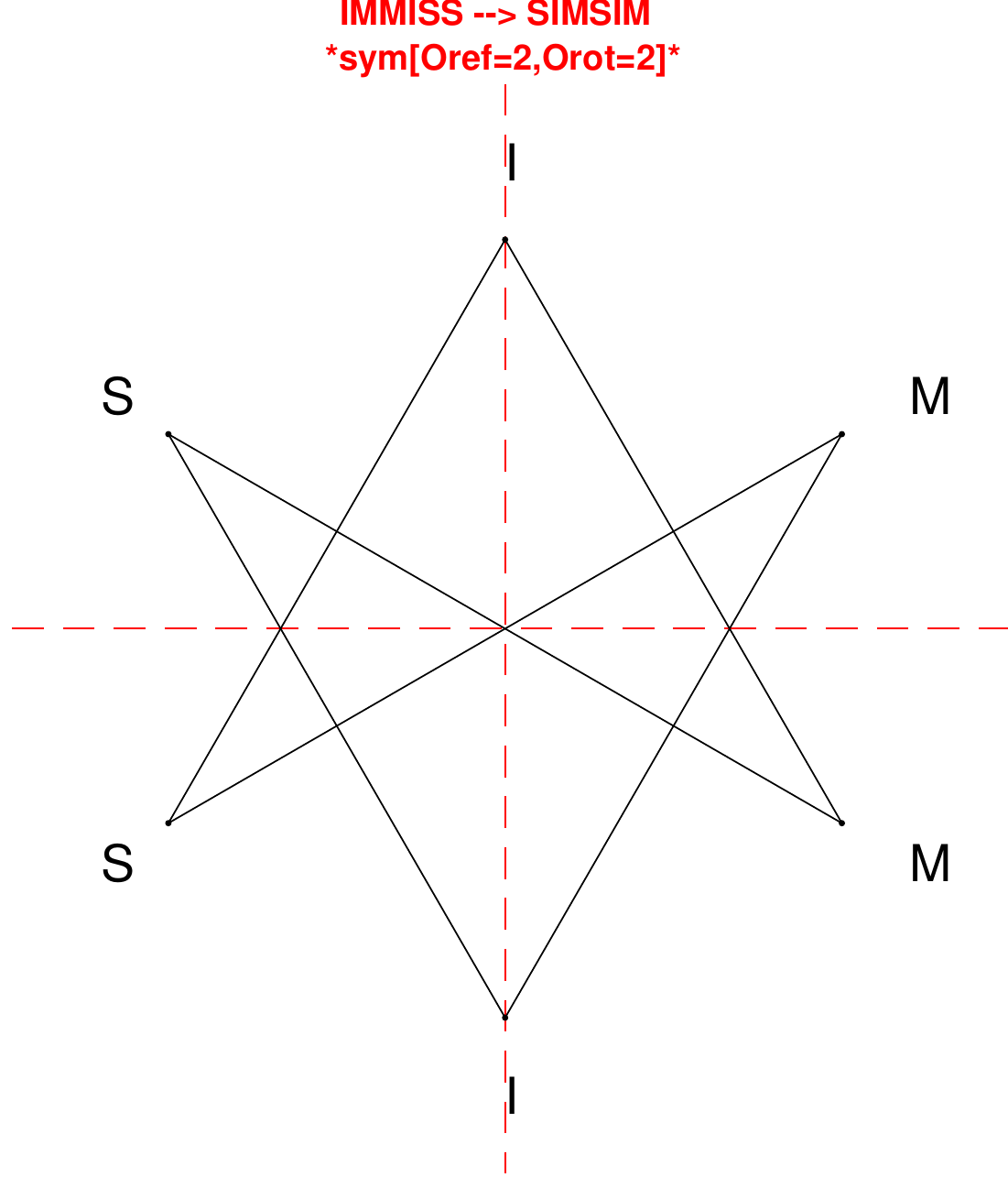}
\end{subfigure}
\hfill
\begin{subfigure}[T]{0.19\textwidth}
\centering
\includegraphics[width=\textwidth]{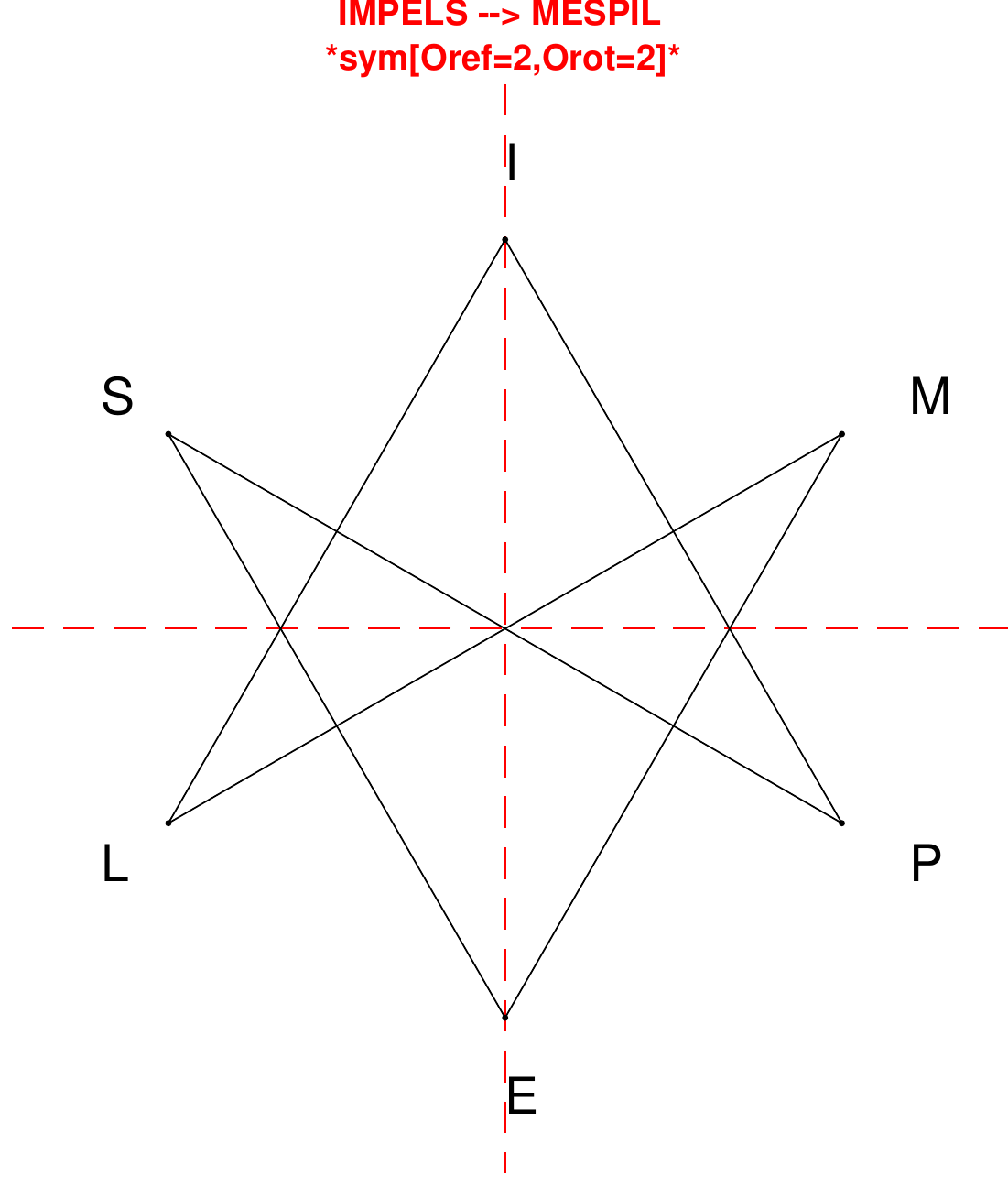}
\end{subfigure}
\hfill
\begin{subfigure}[T]{0.19\textwidth}
\centering
\includegraphics[width=\textwidth]{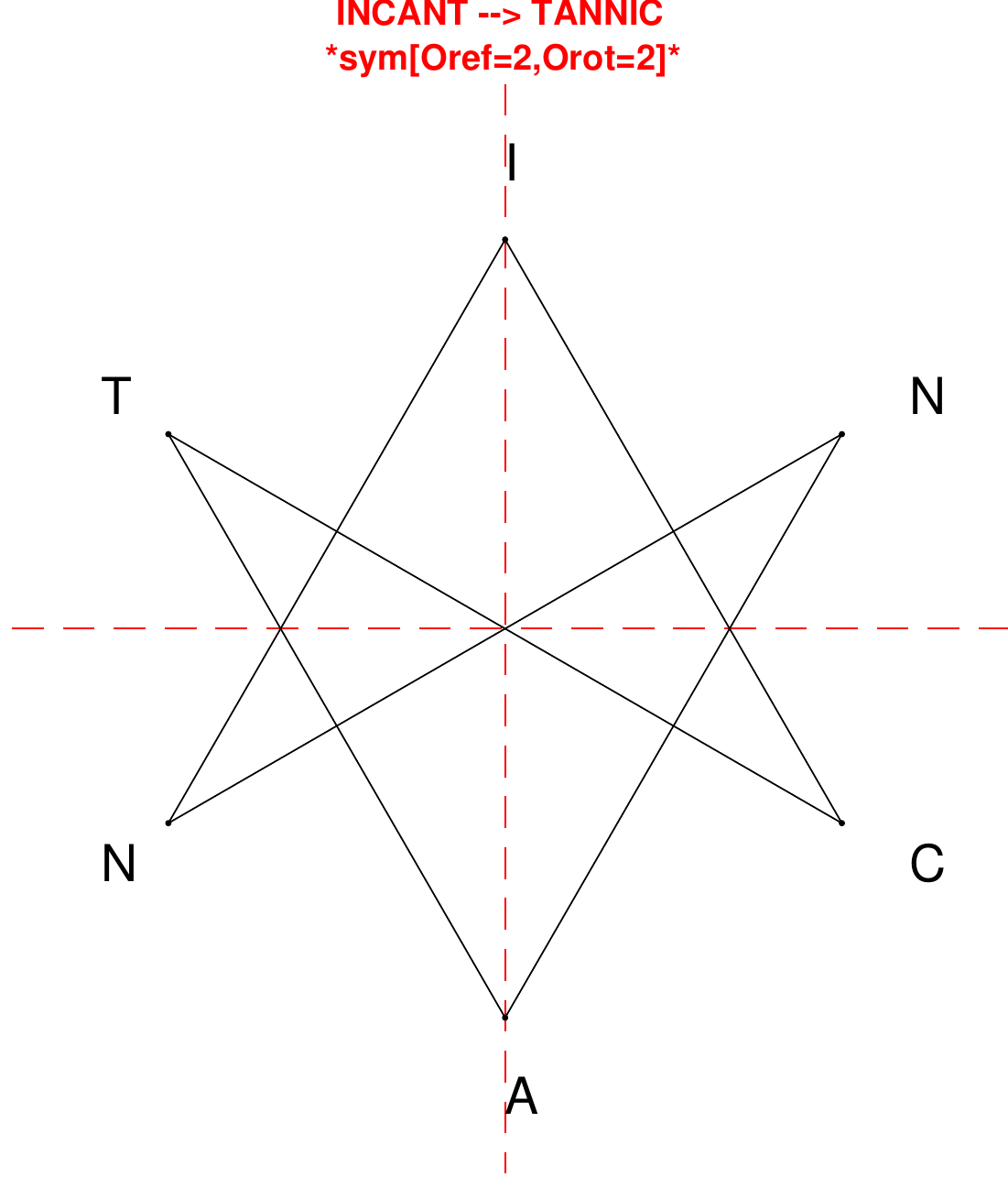}
\end{subfigure}
\hfill
\begin{subfigure}[T]{0.19\textwidth}
\centering
\includegraphics[width=\textwidth]{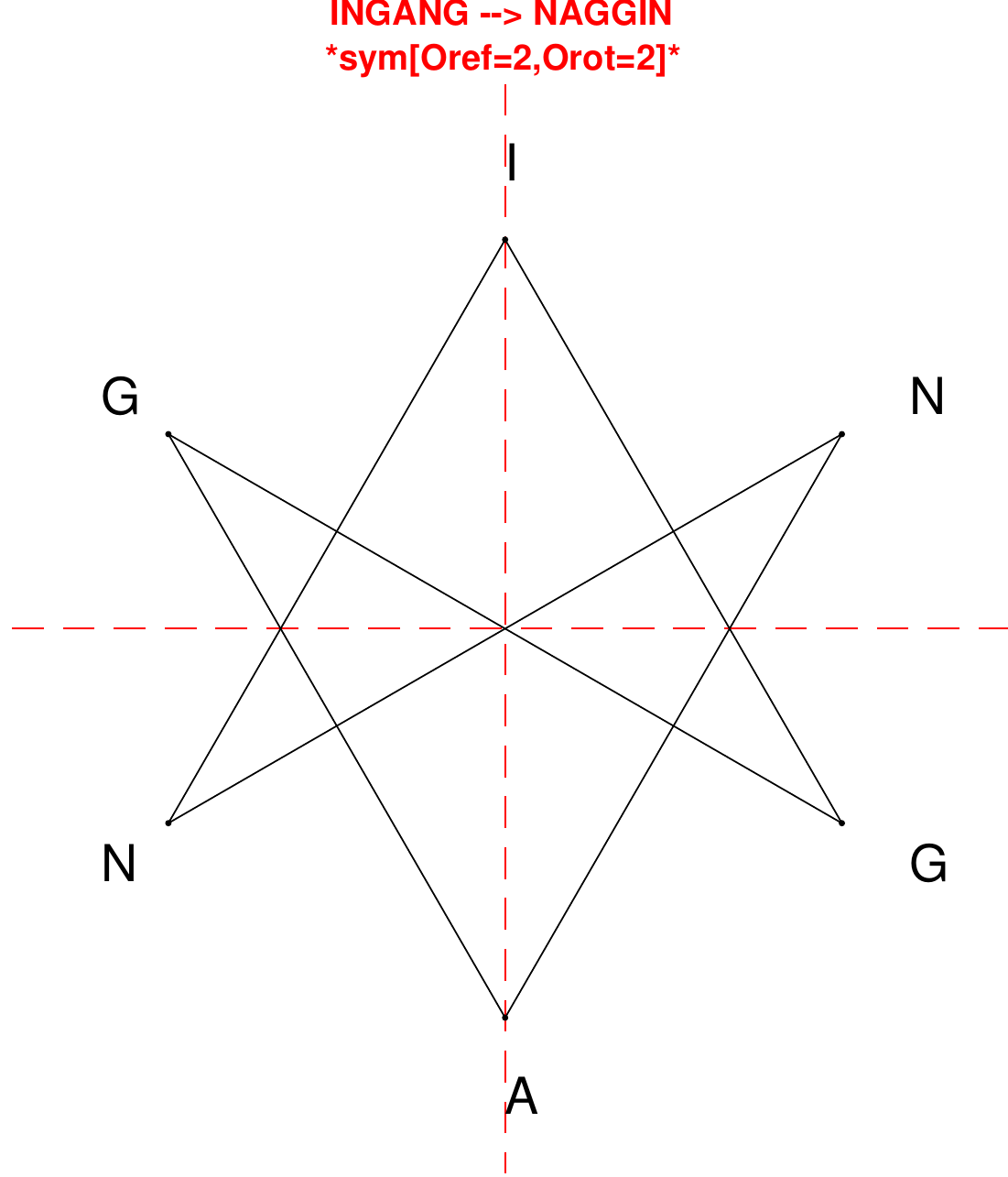}
\end{subfigure}
\hfill
\begin{subfigure}[T]{0.19\textwidth}
\centering
\includegraphics[width=\textwidth]{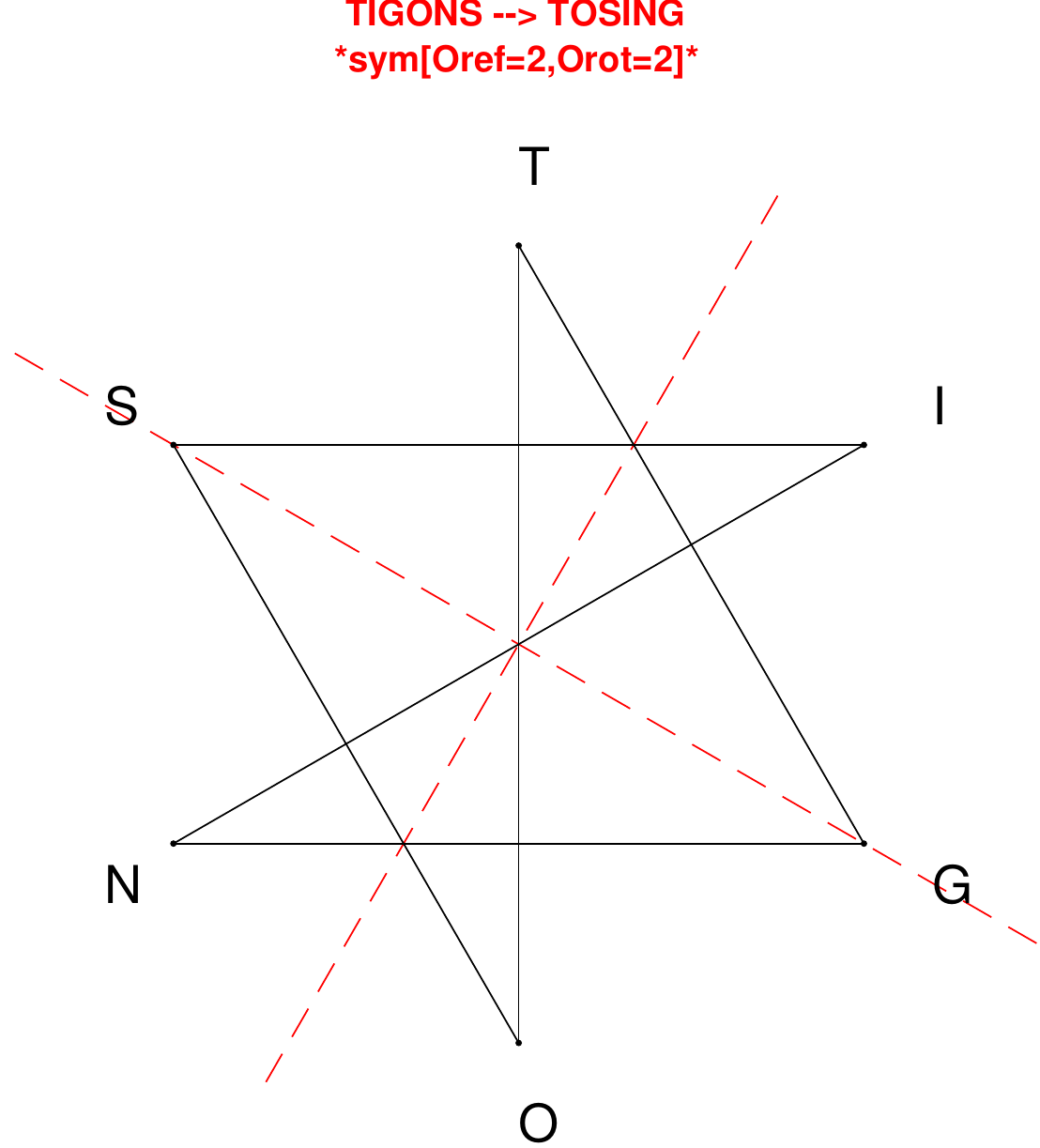}
\end{subfigure}
\end{figure}

\begin{figure}[H]
\centering
\begin{subfigure}[T]{0.19\textwidth}
\centering
\includegraphics[width=\textwidth]{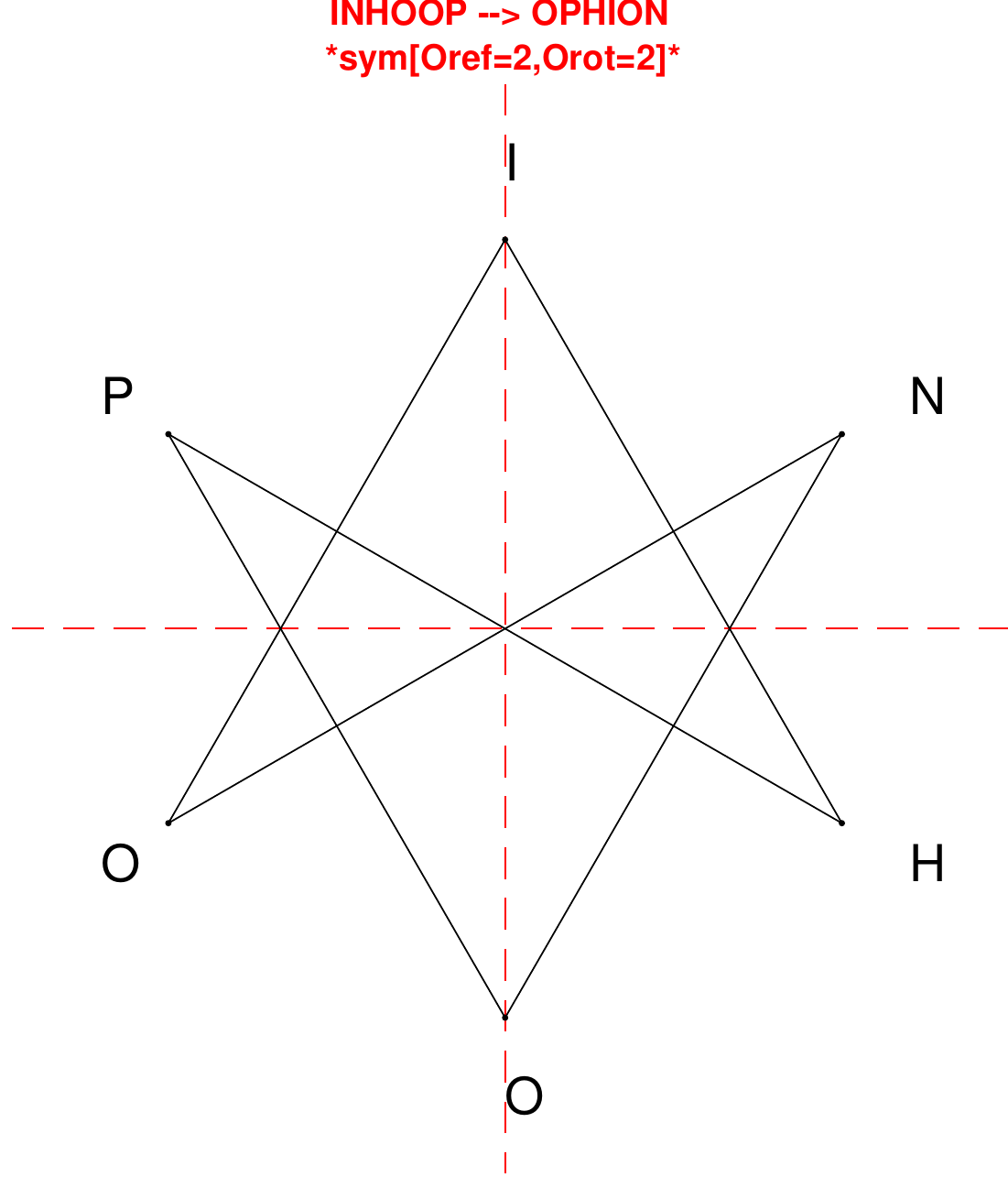}
\end{subfigure}
\hfill
\begin{subfigure}[T]{0.19\textwidth}
\centering
\includegraphics[width=\textwidth]{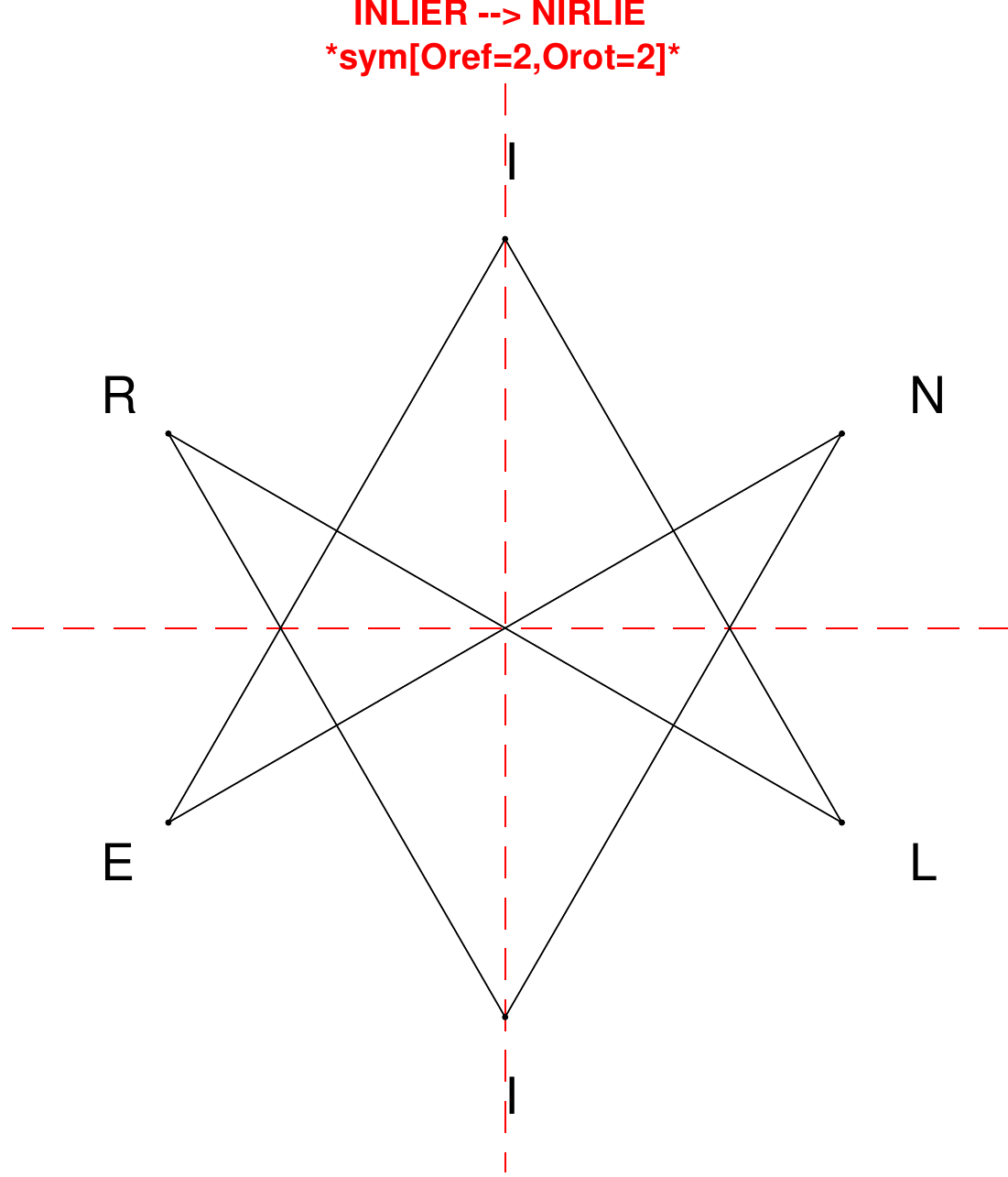}
\end{subfigure}
\hfill
\begin{subfigure}[T]{0.19\textwidth}
\centering
\includegraphics[width=\textwidth]{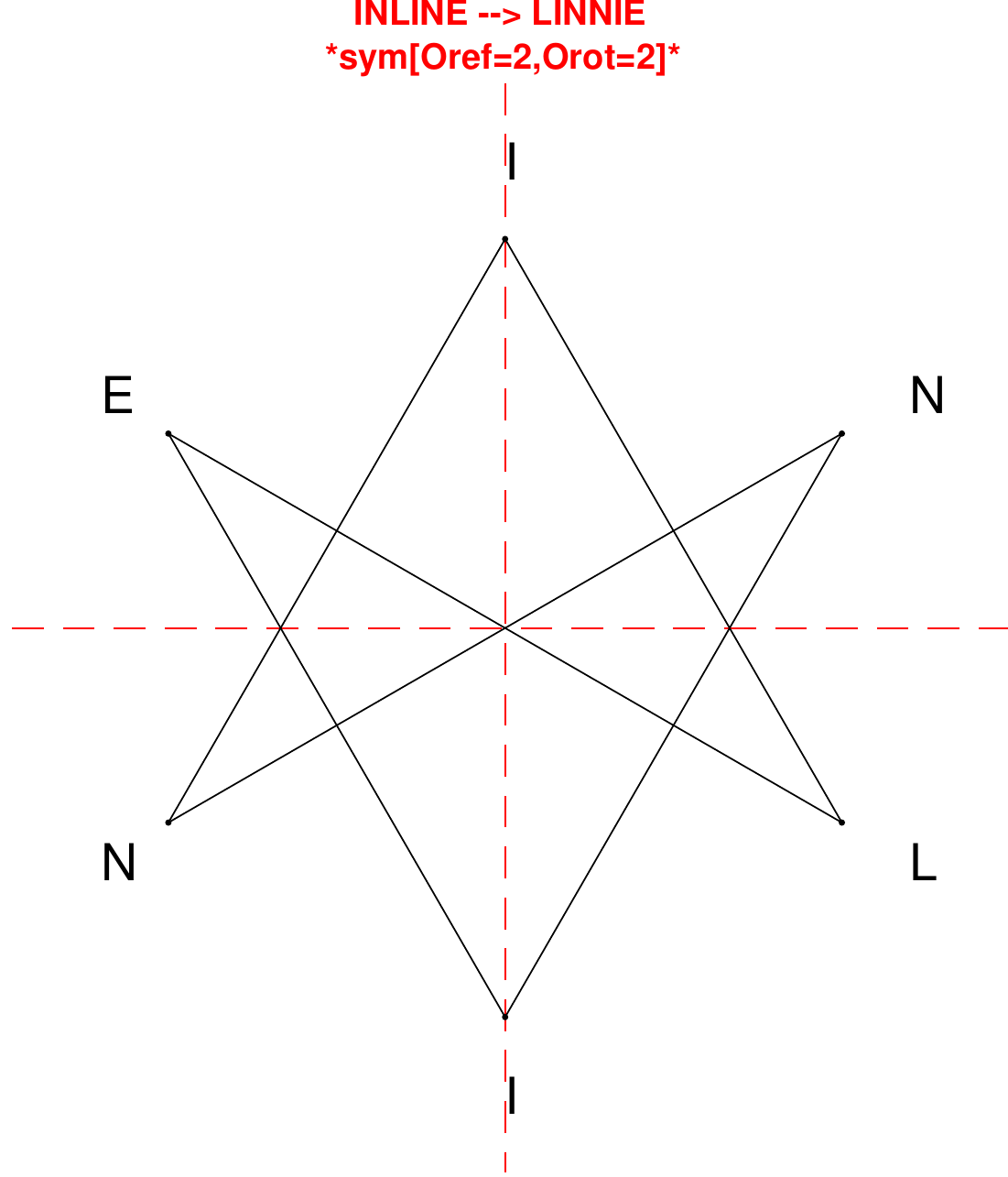}
\end{subfigure}
\hfill
\begin{subfigure}[T]{0.19\textwidth}
\centering
\includegraphics[width=\textwidth]{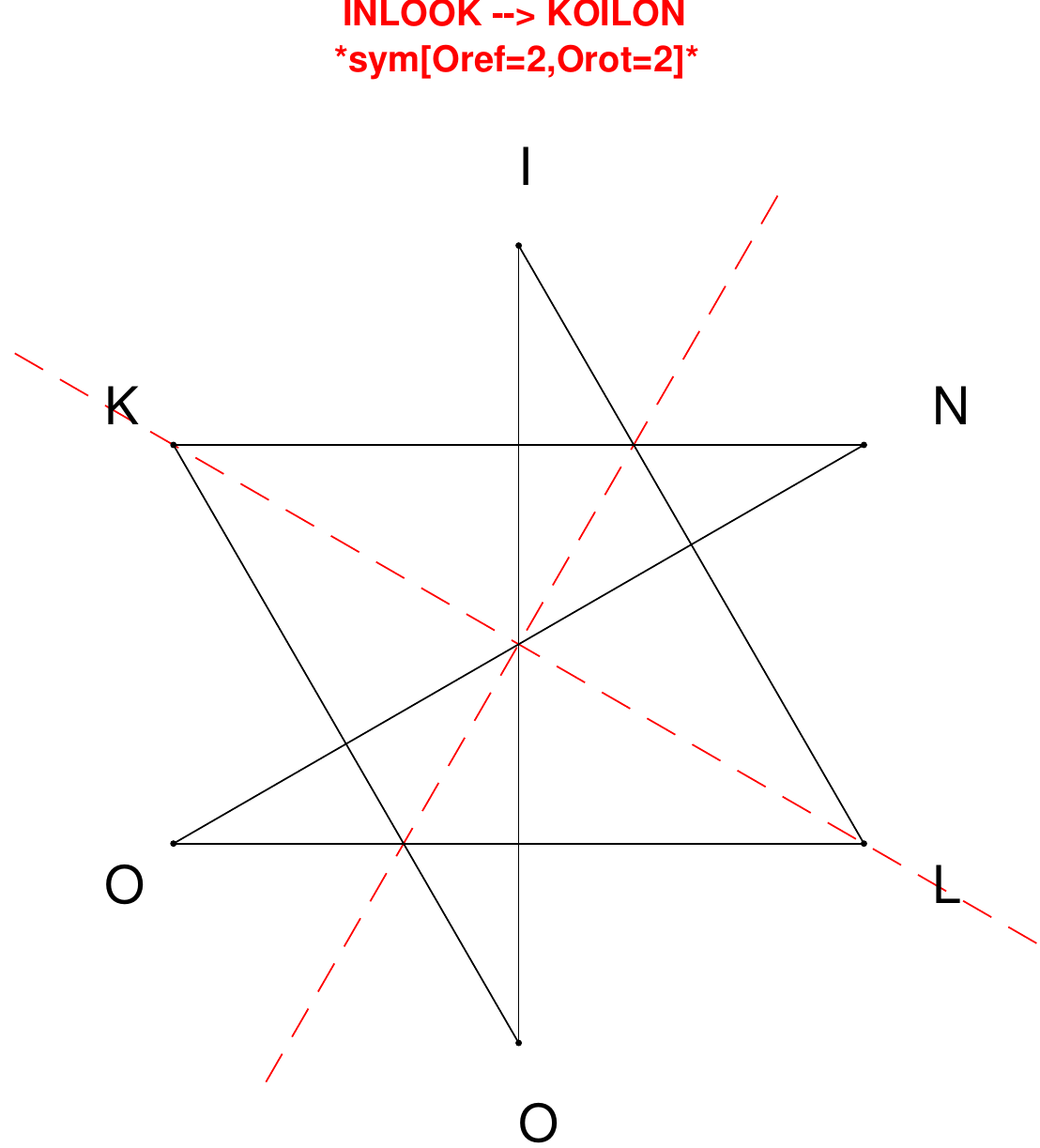}
\end{subfigure}
\hfill
\begin{subfigure}[T]{0.19\textwidth}
\centering
\includegraphics[width=\textwidth]{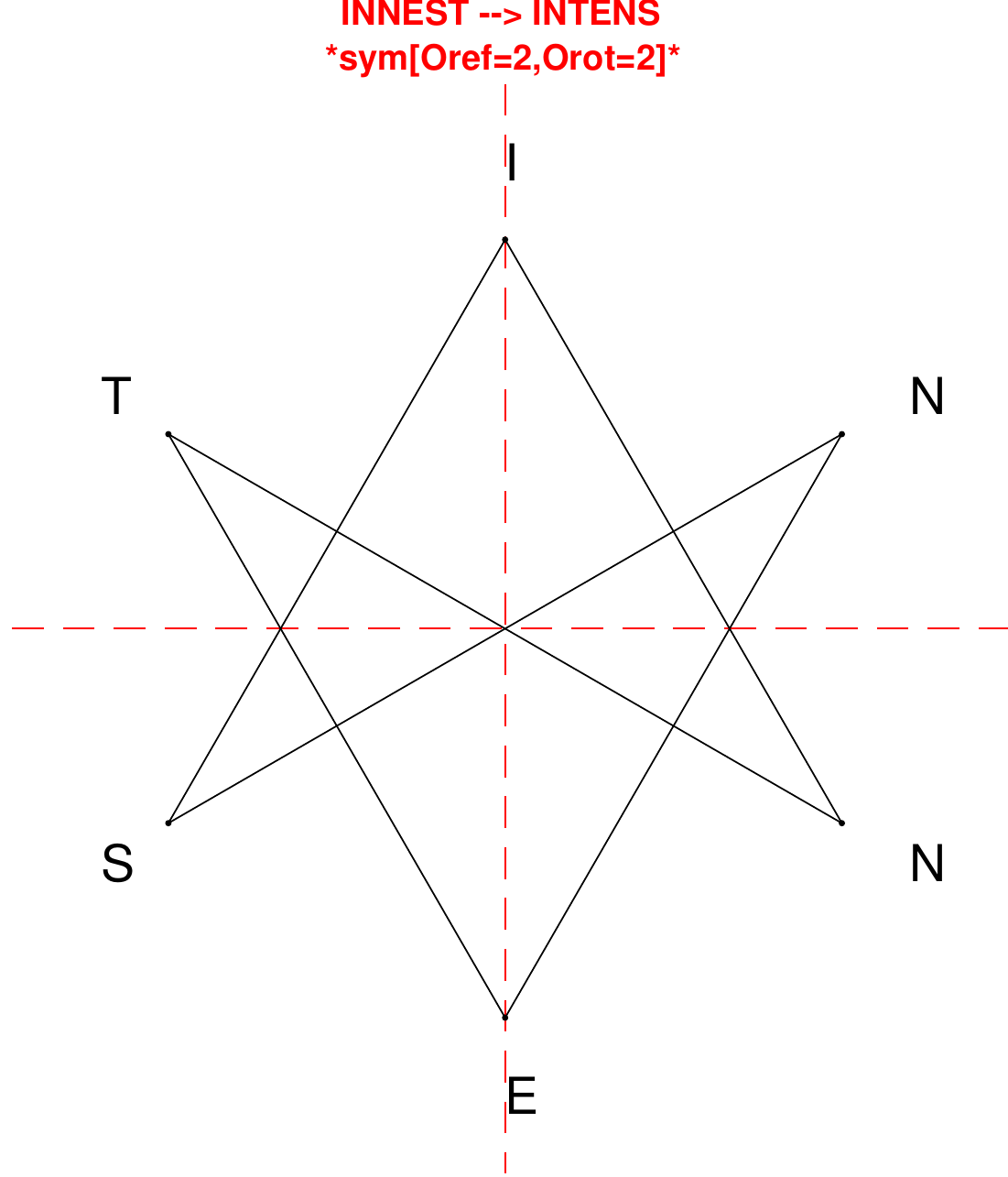}
\end{subfigure}
\end{figure}

\begin{figure}[H]
\centering
\begin{subfigure}[T]{0.19\textwidth}
\centering
\includegraphics[width=\textwidth]{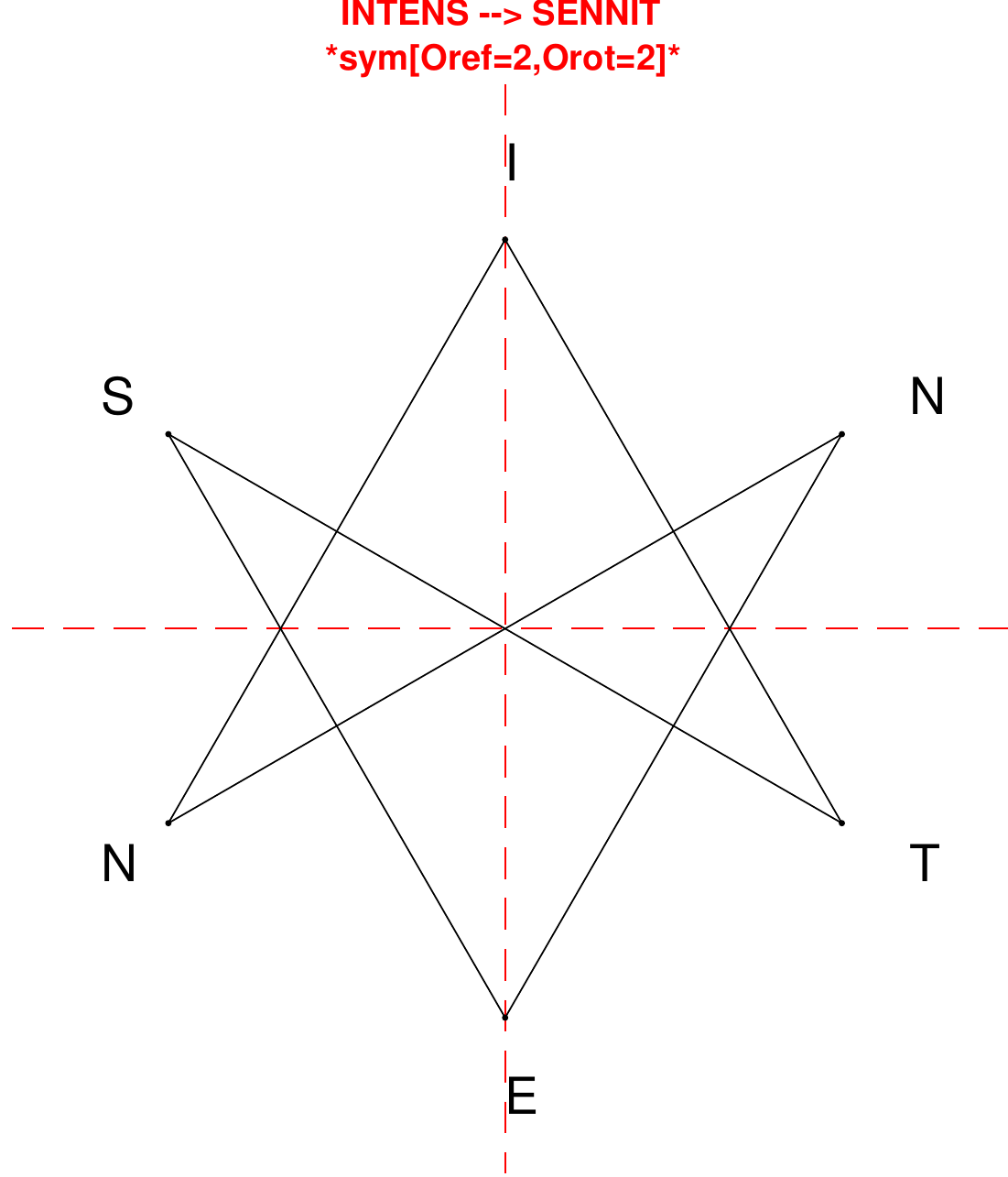}
\end{subfigure}
\hfill
\begin{subfigure}[T]{0.19\textwidth}
\centering
\includegraphics[width=\textwidth]{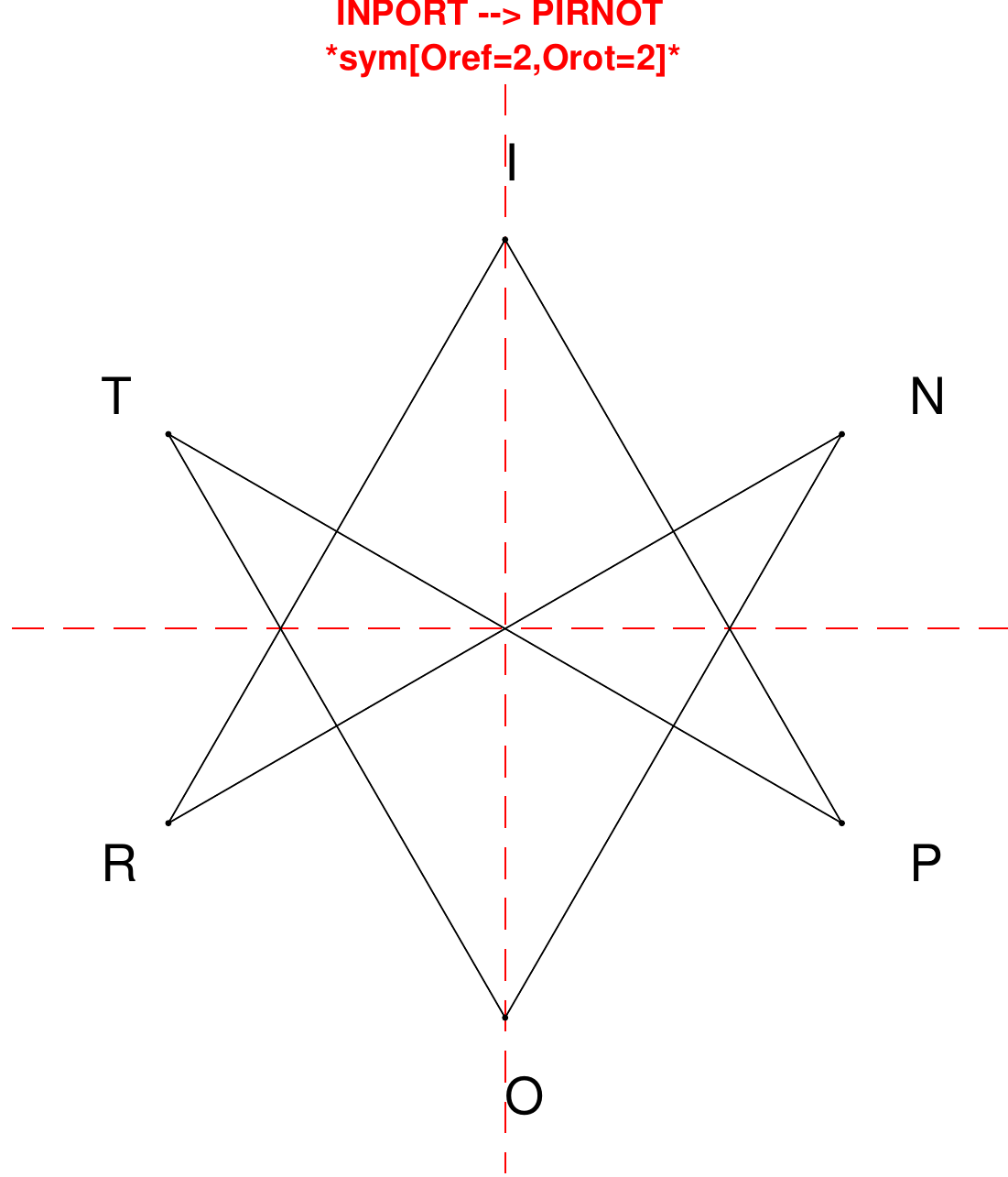}
\end{subfigure}
\hfill
\begin{subfigure}[T]{0.19\textwidth}
\centering
\includegraphics[width=\textwidth]{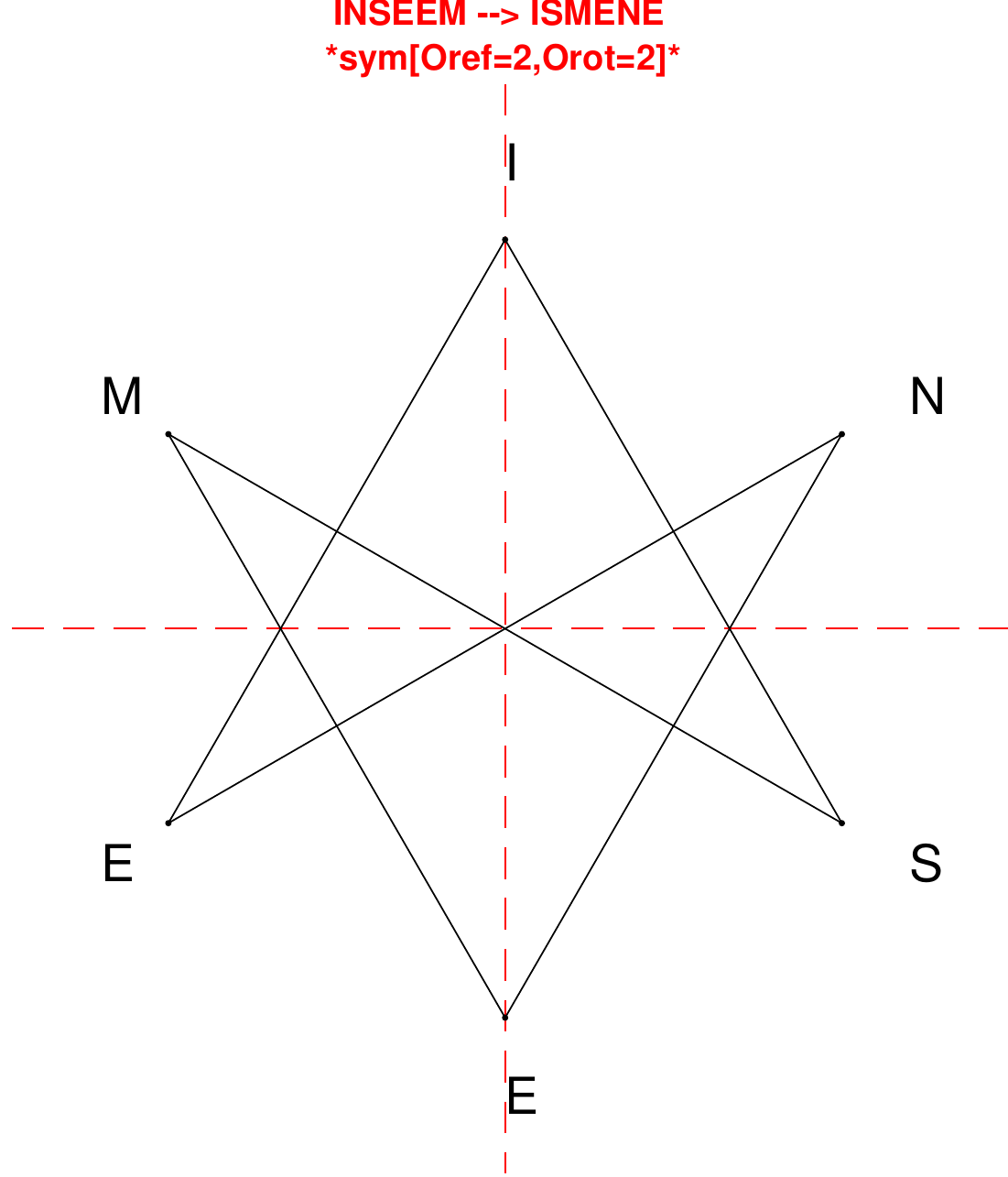}
\end{subfigure}
\hfill
\begin{subfigure}[T]{0.19\textwidth}
\centering
\includegraphics[width=\textwidth]{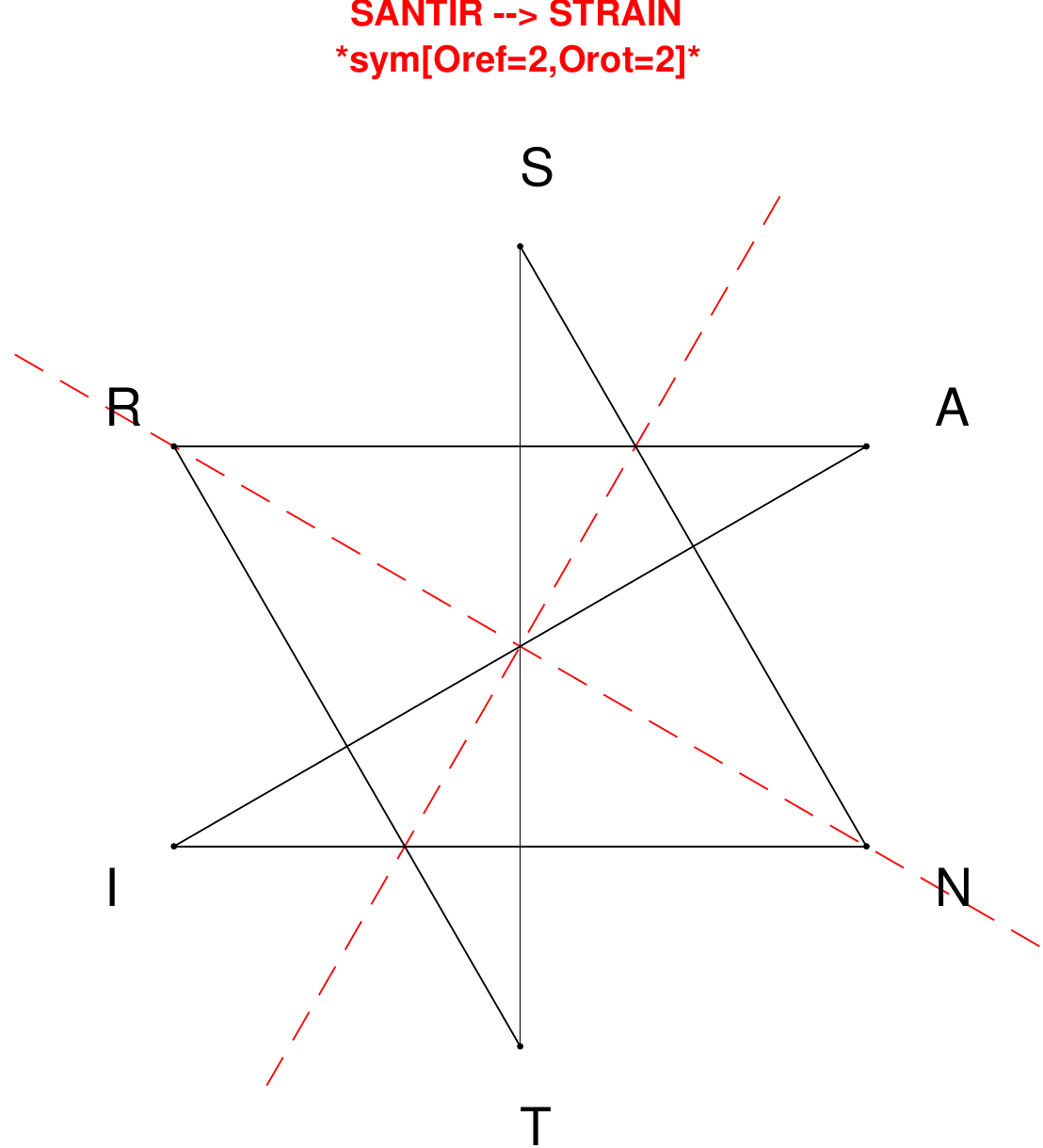}
\end{subfigure}
\hfill
\begin{subfigure}[T]{0.19\textwidth}
\centering
\includegraphics[width=\textwidth]{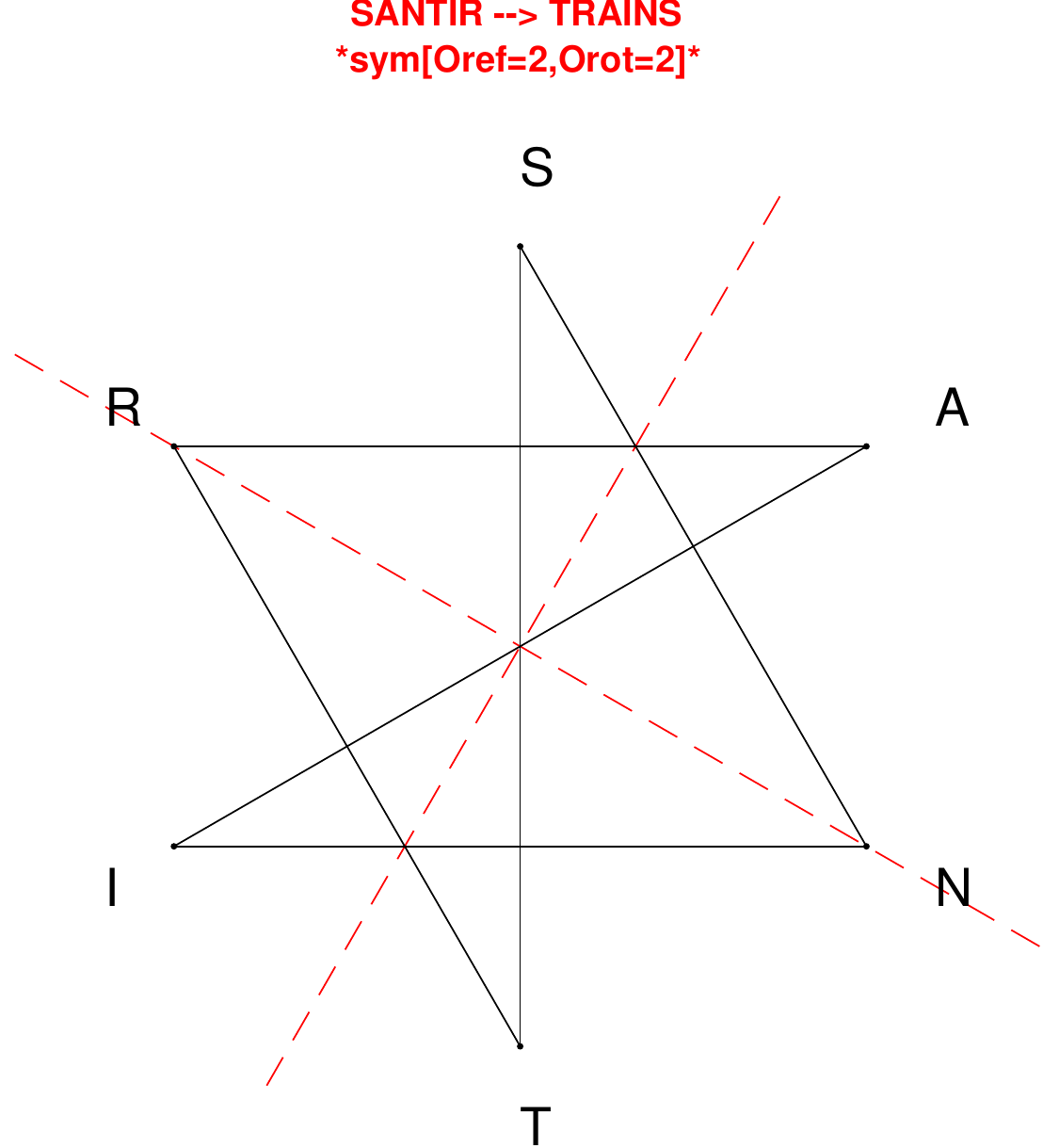}
\end{subfigure}
\end{figure}

\begin{figure}[H]
\centering
\begin{subfigure}[T]{0.19\textwidth}
\centering
\includegraphics[width=\textwidth]{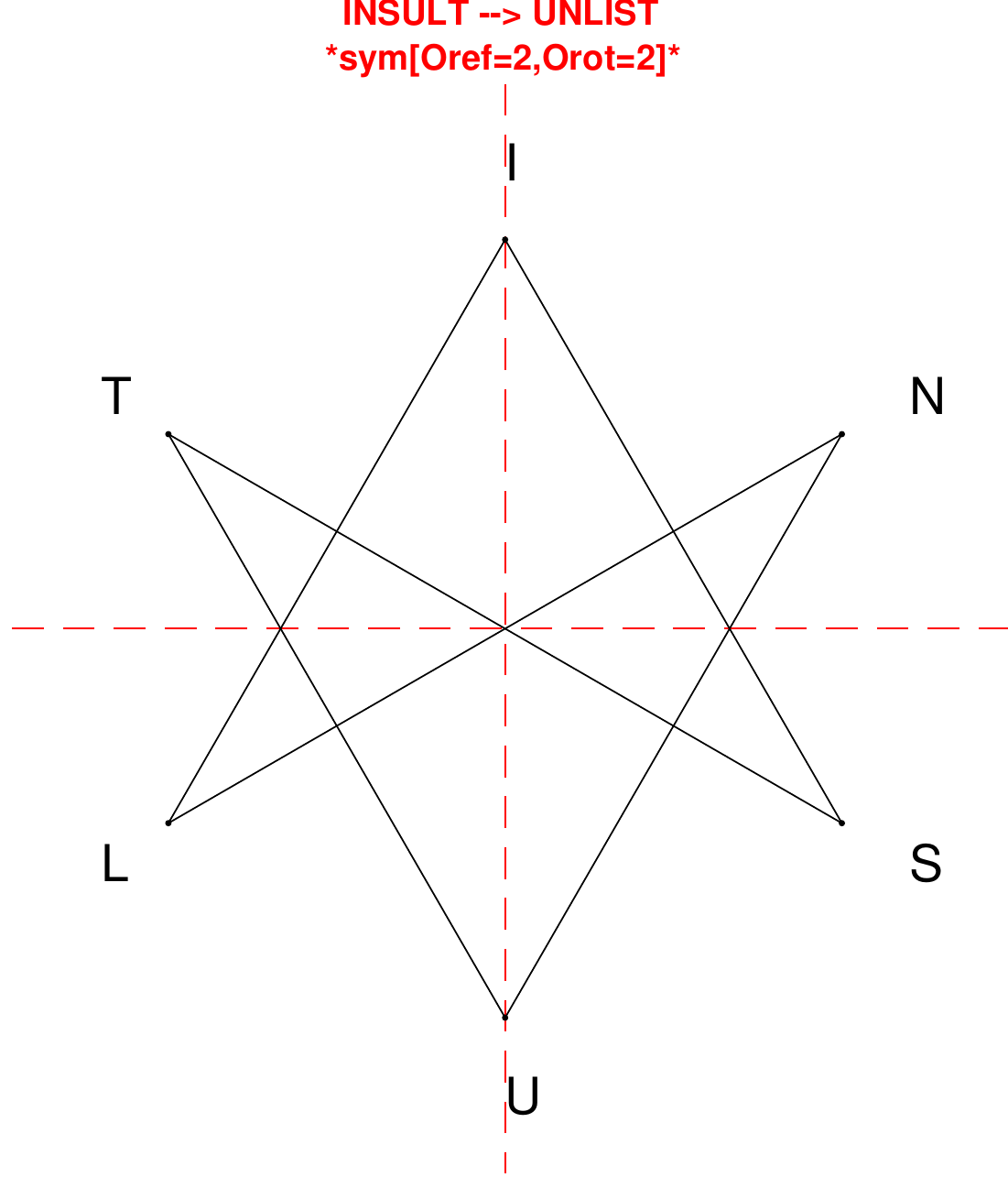}
\end{subfigure}
\hfill
\begin{subfigure}[T]{0.19\textwidth}
\centering
\includegraphics[width=\textwidth]{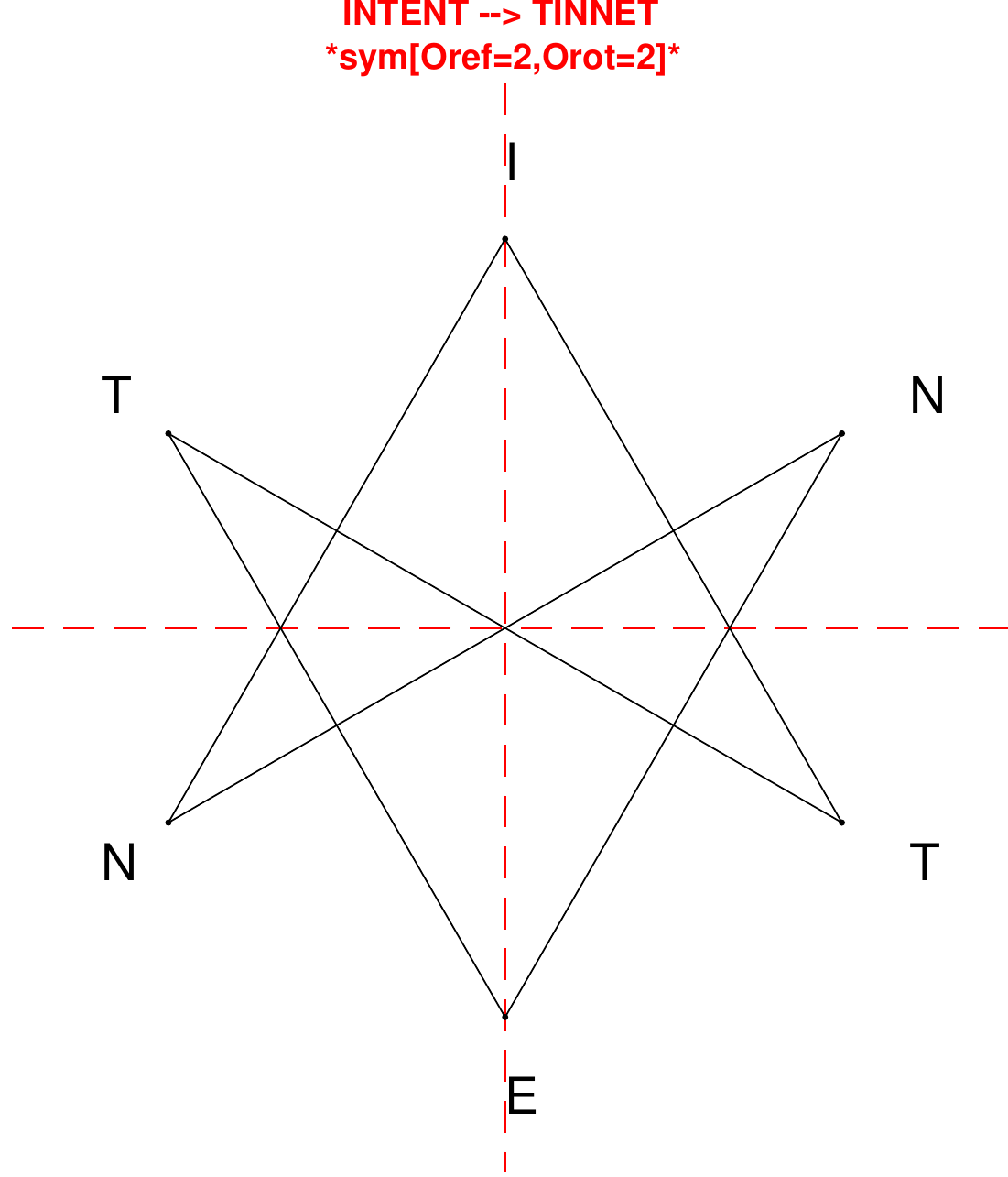}
\end{subfigure}
\hfill
\begin{subfigure}[T]{0.19\textwidth}
\centering
\includegraphics[width=\textwidth]{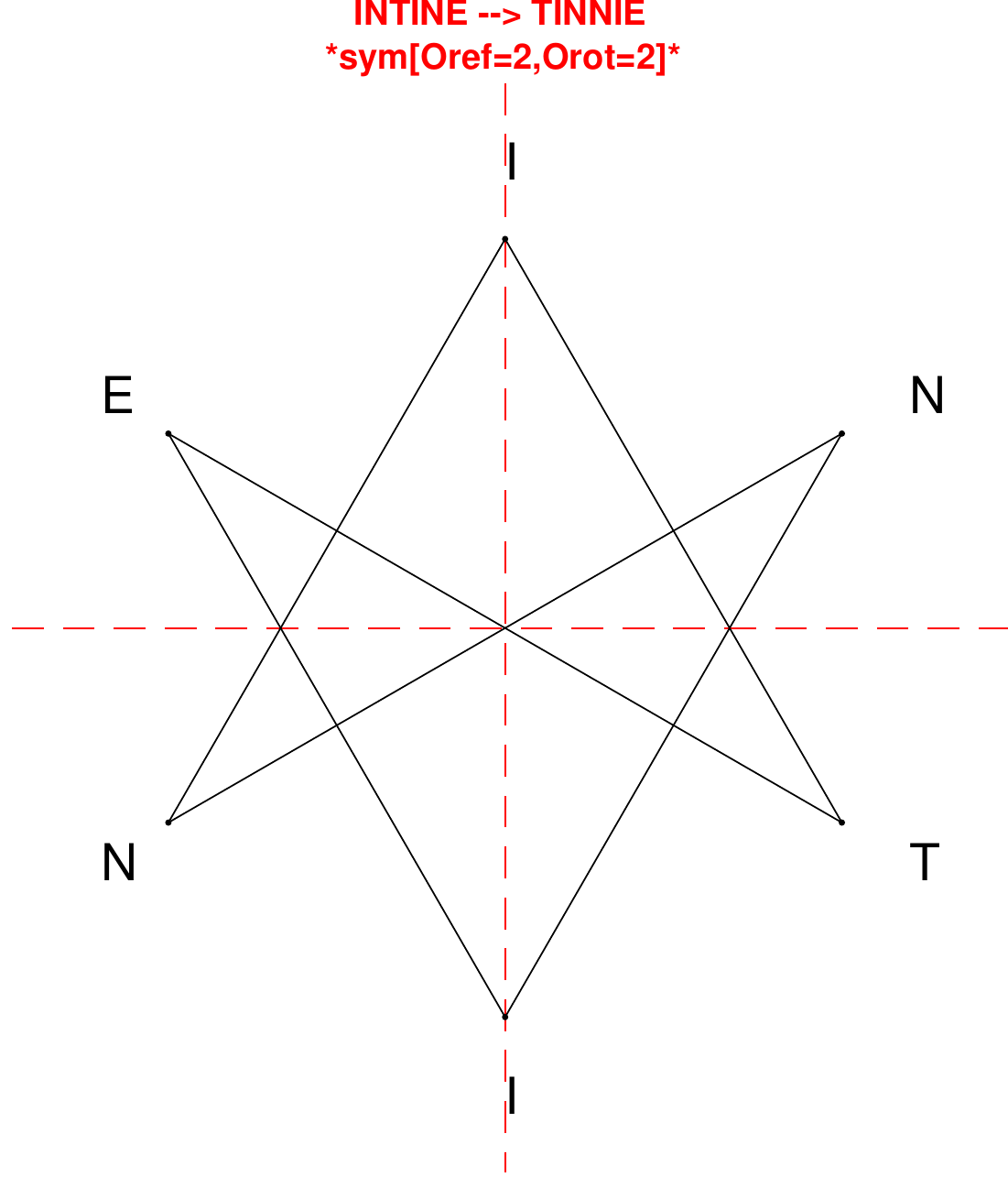}
\end{subfigure}
\hfill
\begin{subfigure}[T]{0.19\textwidth}
\centering
\includegraphics[width=\textwidth]{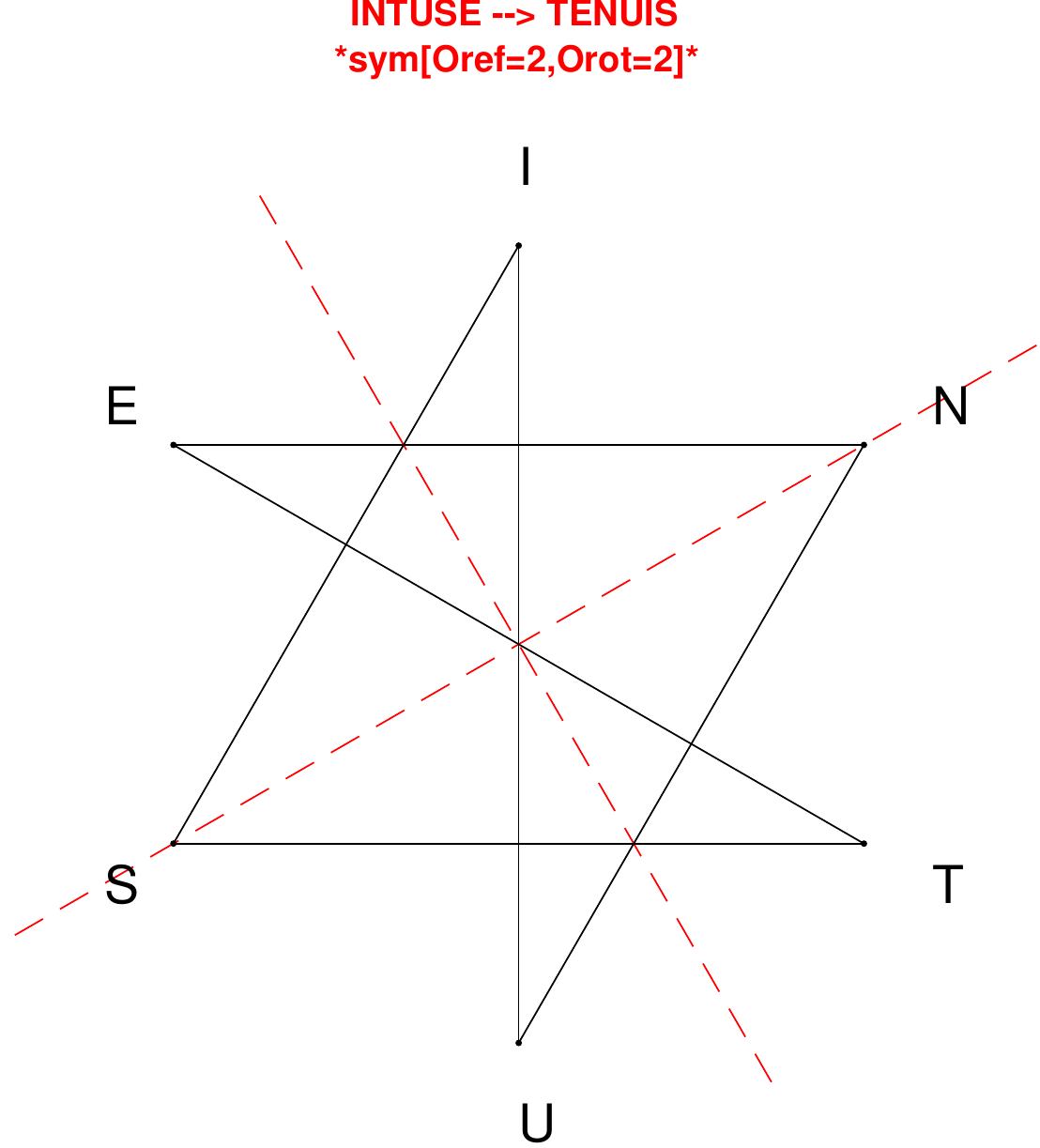}
\end{subfigure}
\hfill
\begin{subfigure}[T]{0.19\textwidth}
\centering
\includegraphics[width=\textwidth]{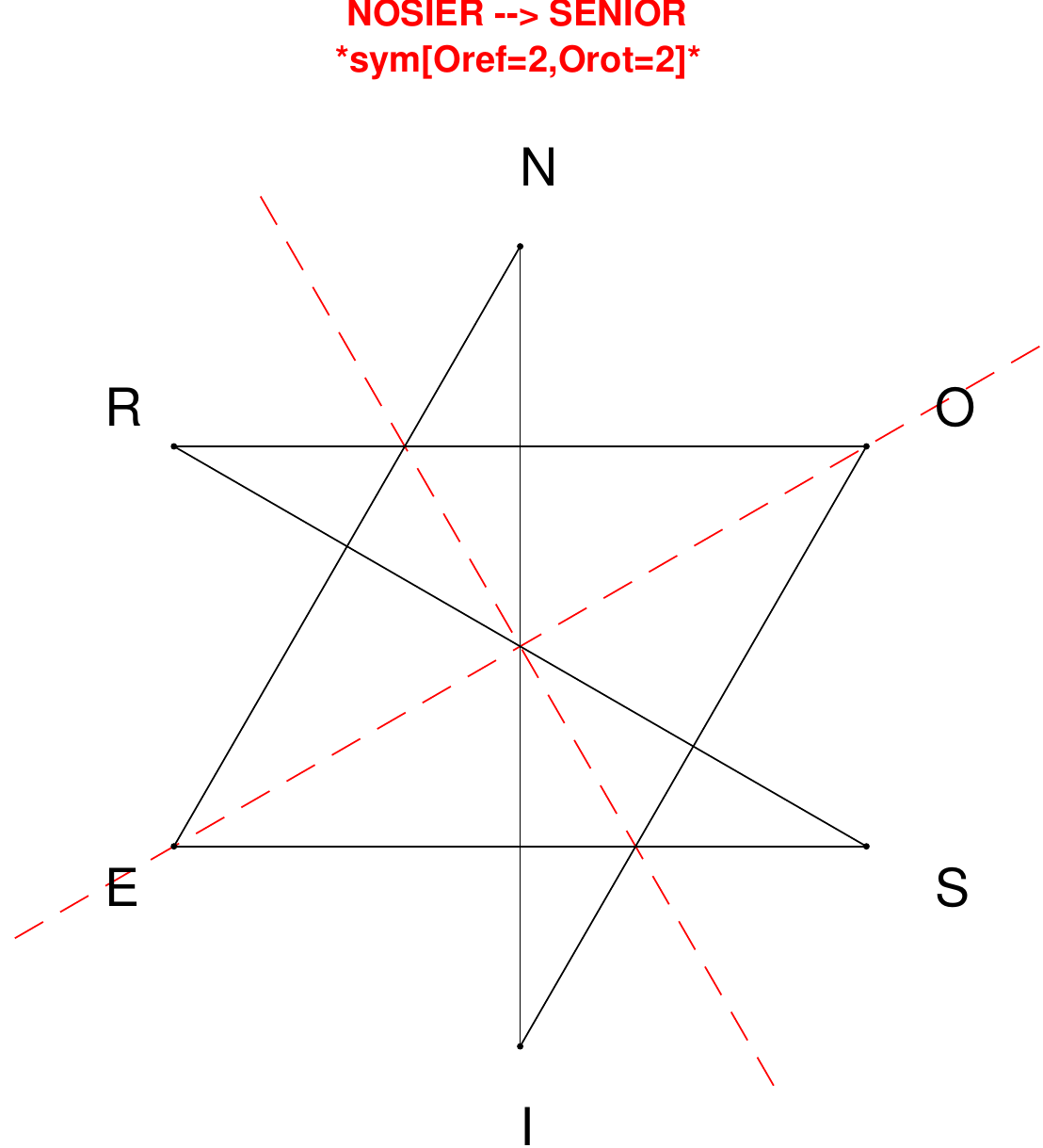}
\end{subfigure}
\end{figure}

\begin{figure}[H]
\centering
\begin{subfigure}[T]{0.19\textwidth}
\centering
\includegraphics[width=\textwidth]{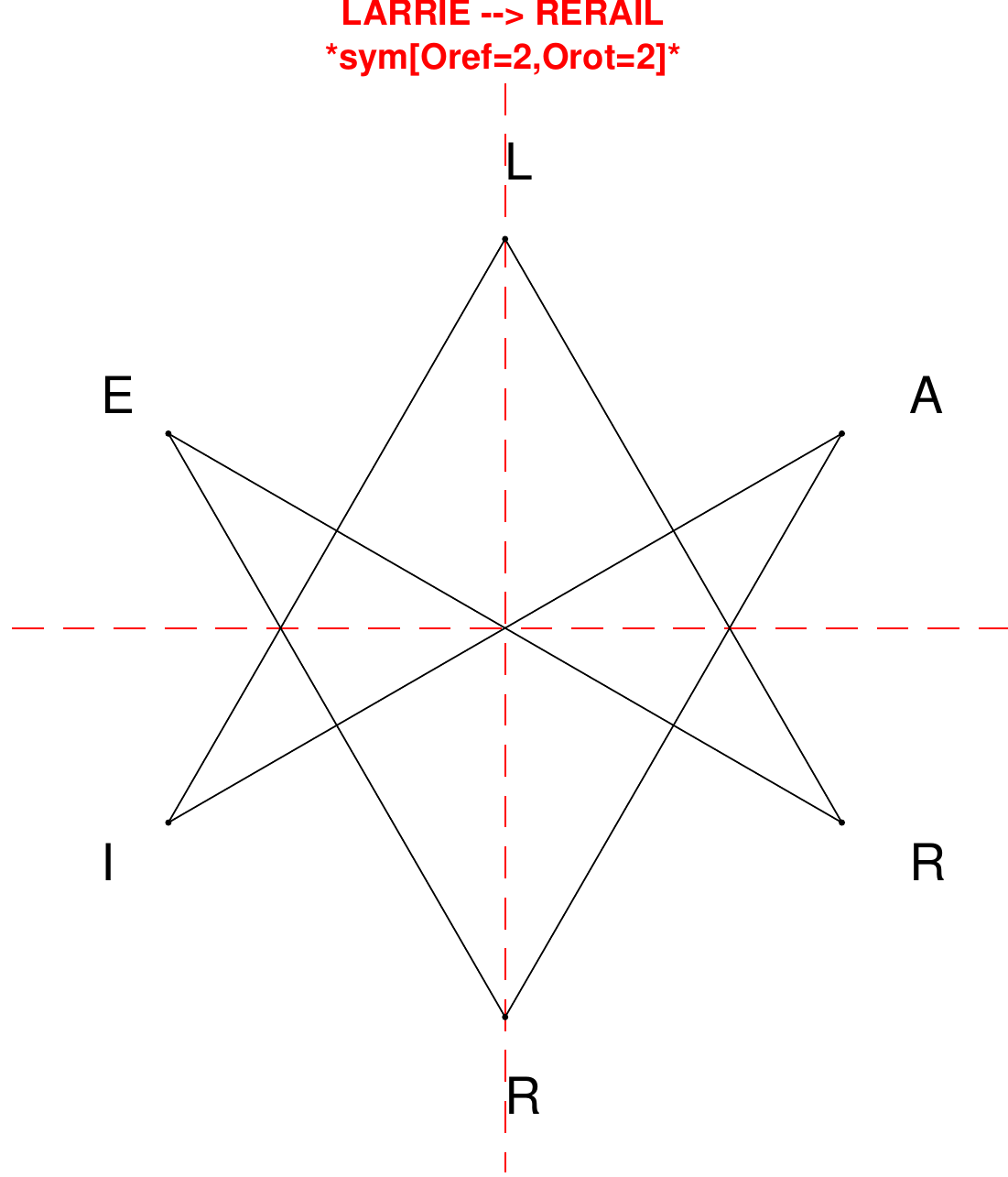}
\end{subfigure}
\hfill
\begin{subfigure}[T]{0.19\textwidth}
\centering
\includegraphics[width=\textwidth]{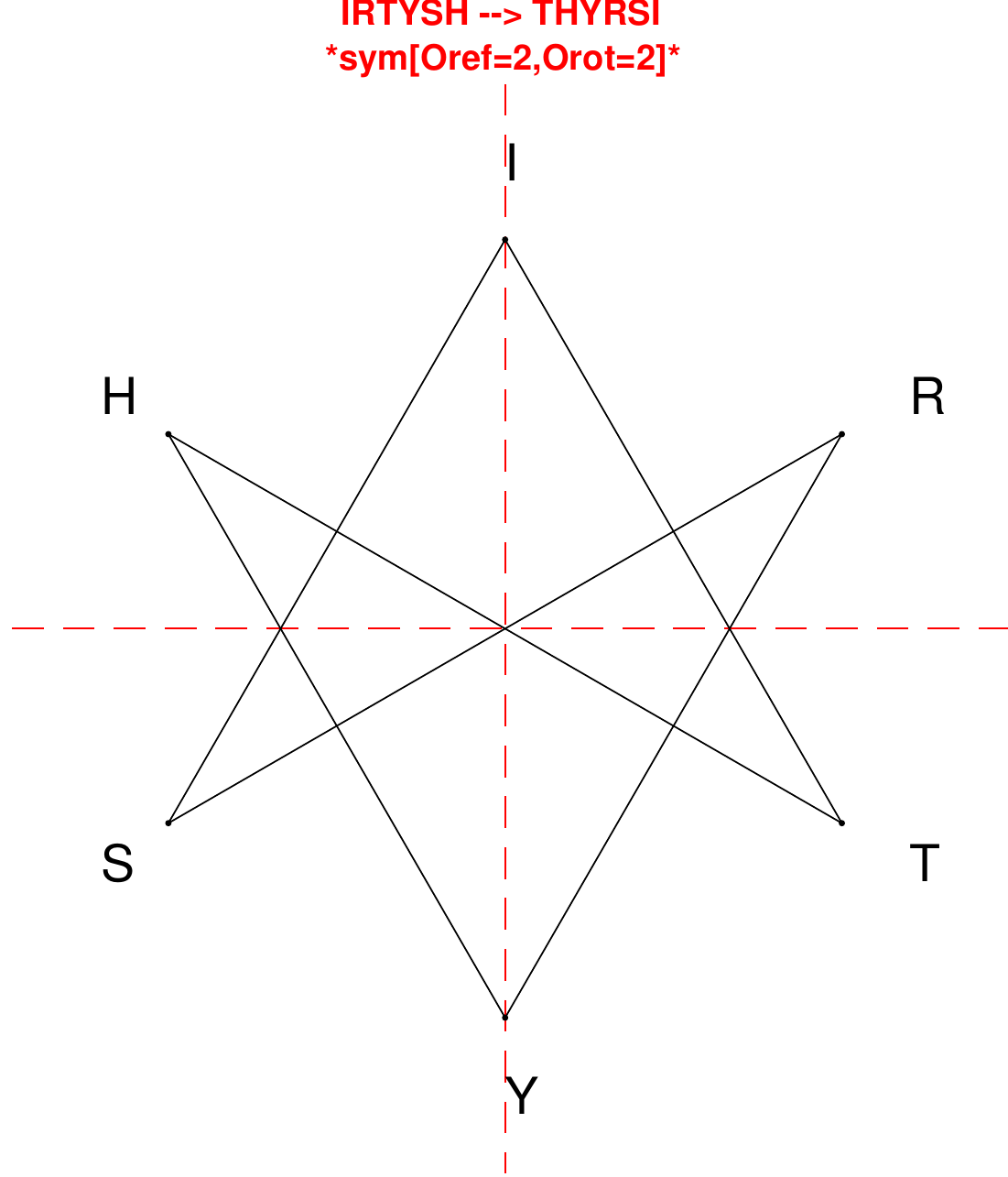}
\end{subfigure}
\hfill
\begin{subfigure}[T]{0.19\textwidth}
\centering
\includegraphics[width=\textwidth]{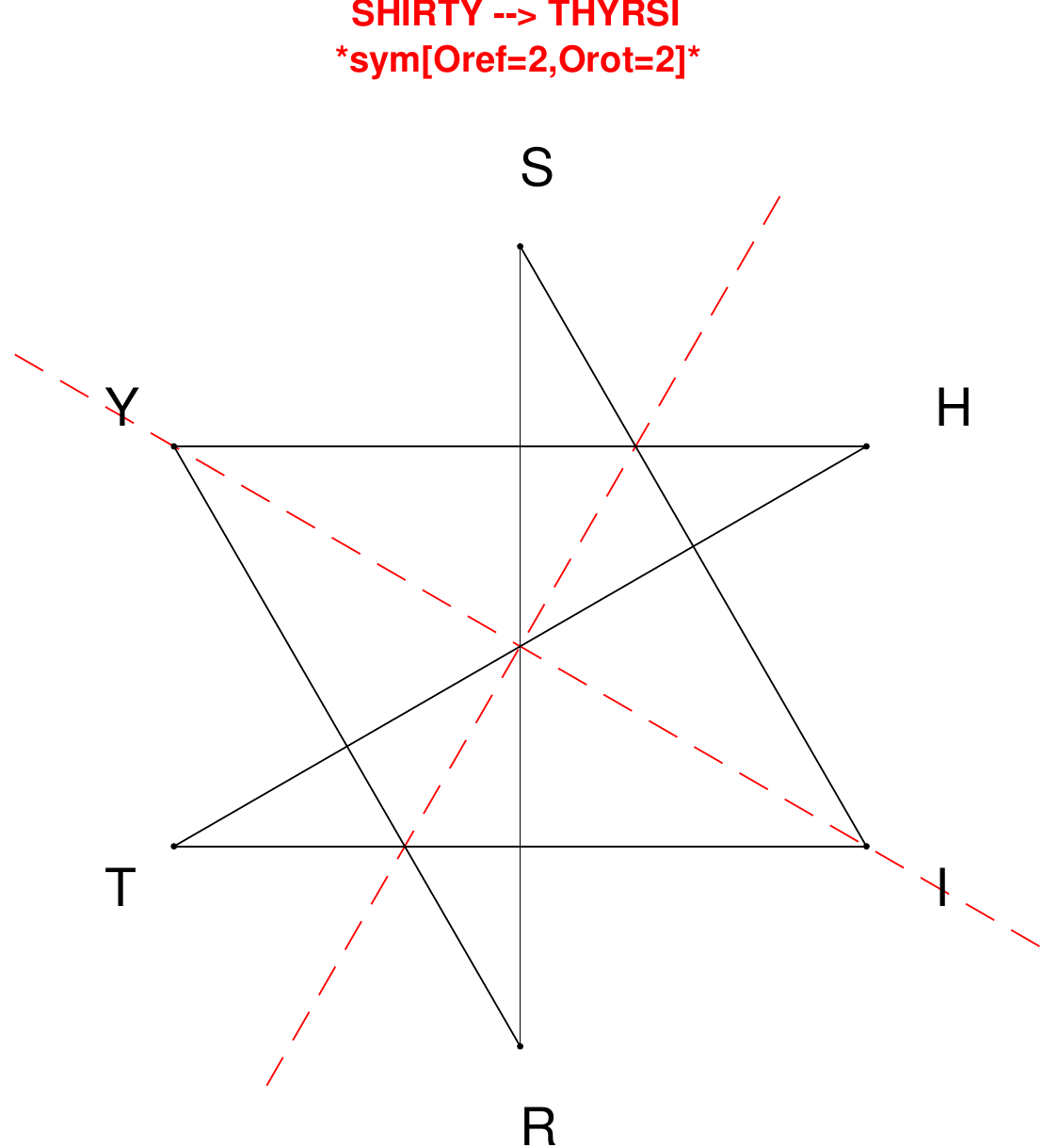}
\end{subfigure}
\hfill
\begin{subfigure}[T]{0.19\textwidth}
\centering
\includegraphics[width=\textwidth]{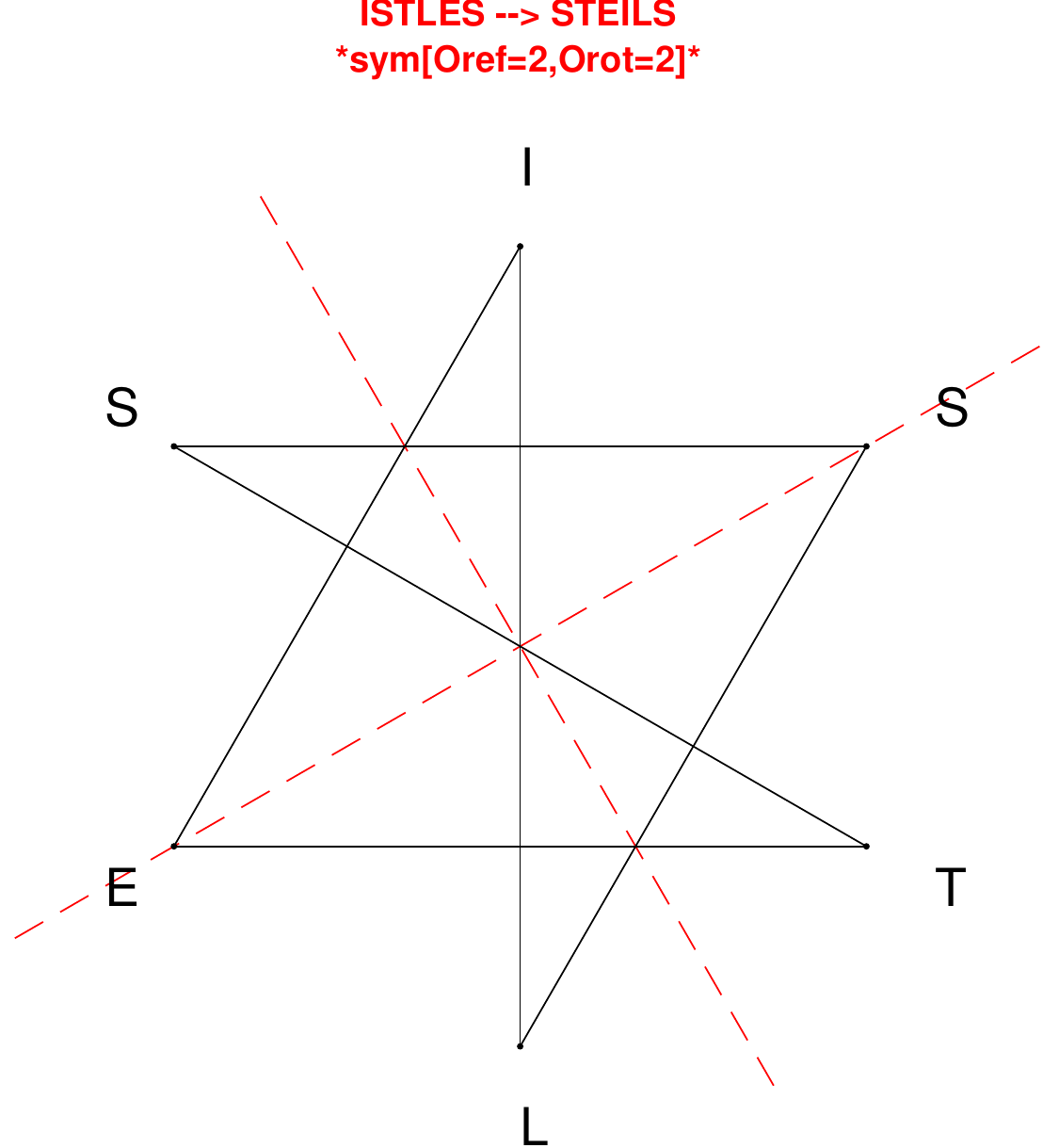}
\end{subfigure}
\hfill
\begin{subfigure}[T]{0.19\textwidth}
\centering
\includegraphics[width=\textwidth]{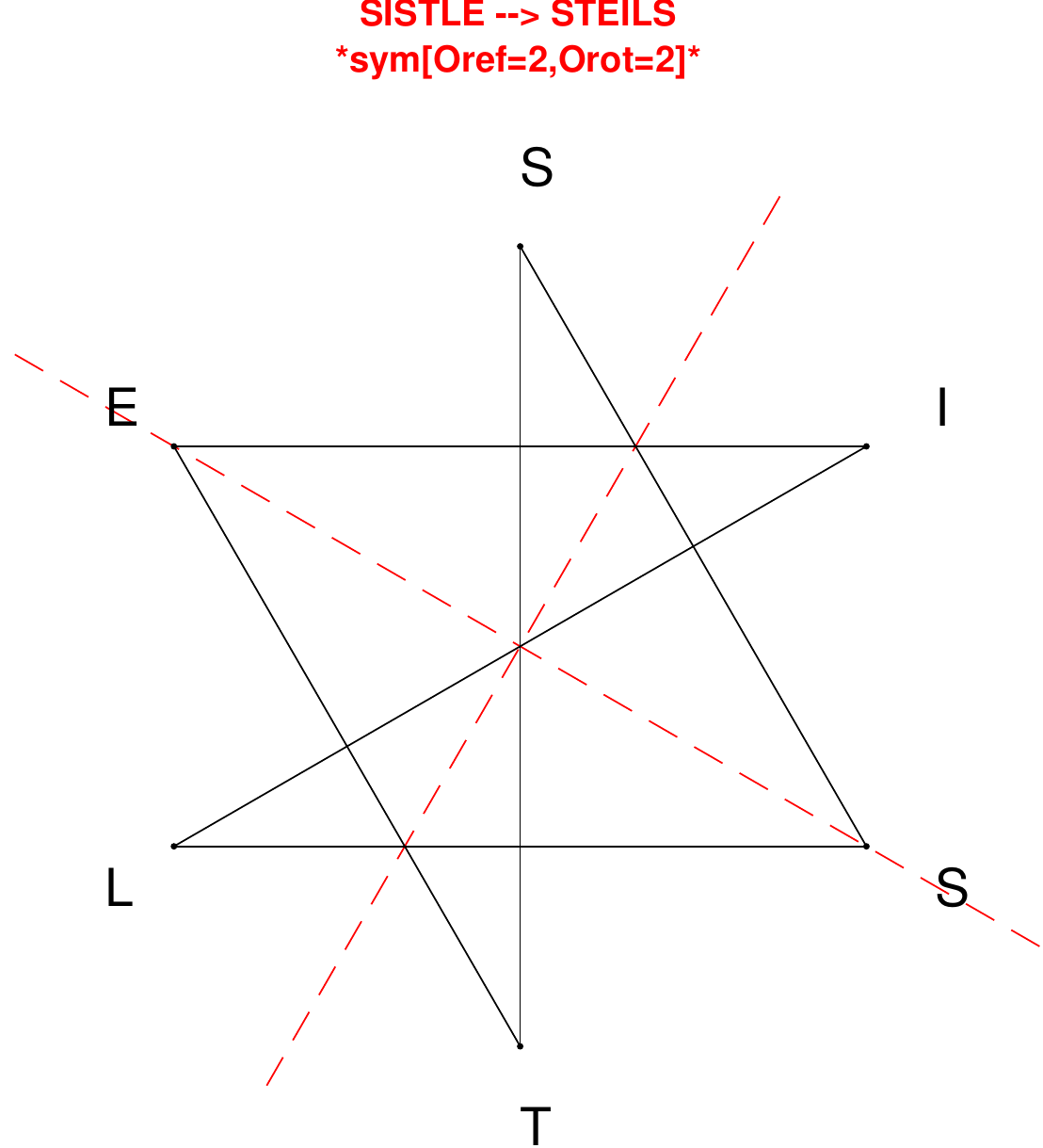}
\end{subfigure}
\end{figure}

\begin{figure}[H]
\centering
\begin{subfigure}[T]{0.19\textwidth}
\centering
\includegraphics[width=\textwidth]{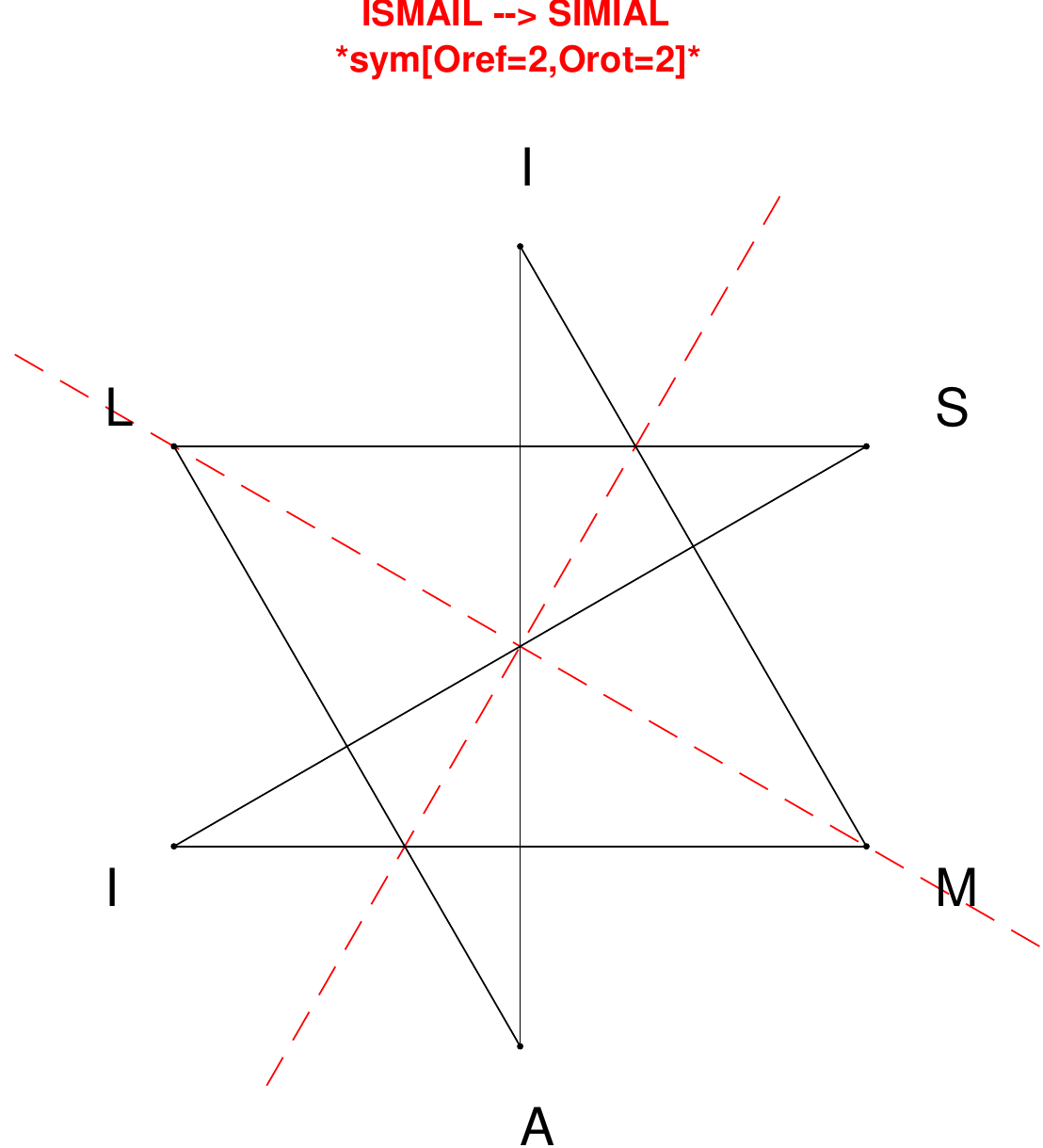}
\end{subfigure}
\hfill
\begin{subfigure}[T]{0.19\textwidth}
\centering
\includegraphics[width=\textwidth]{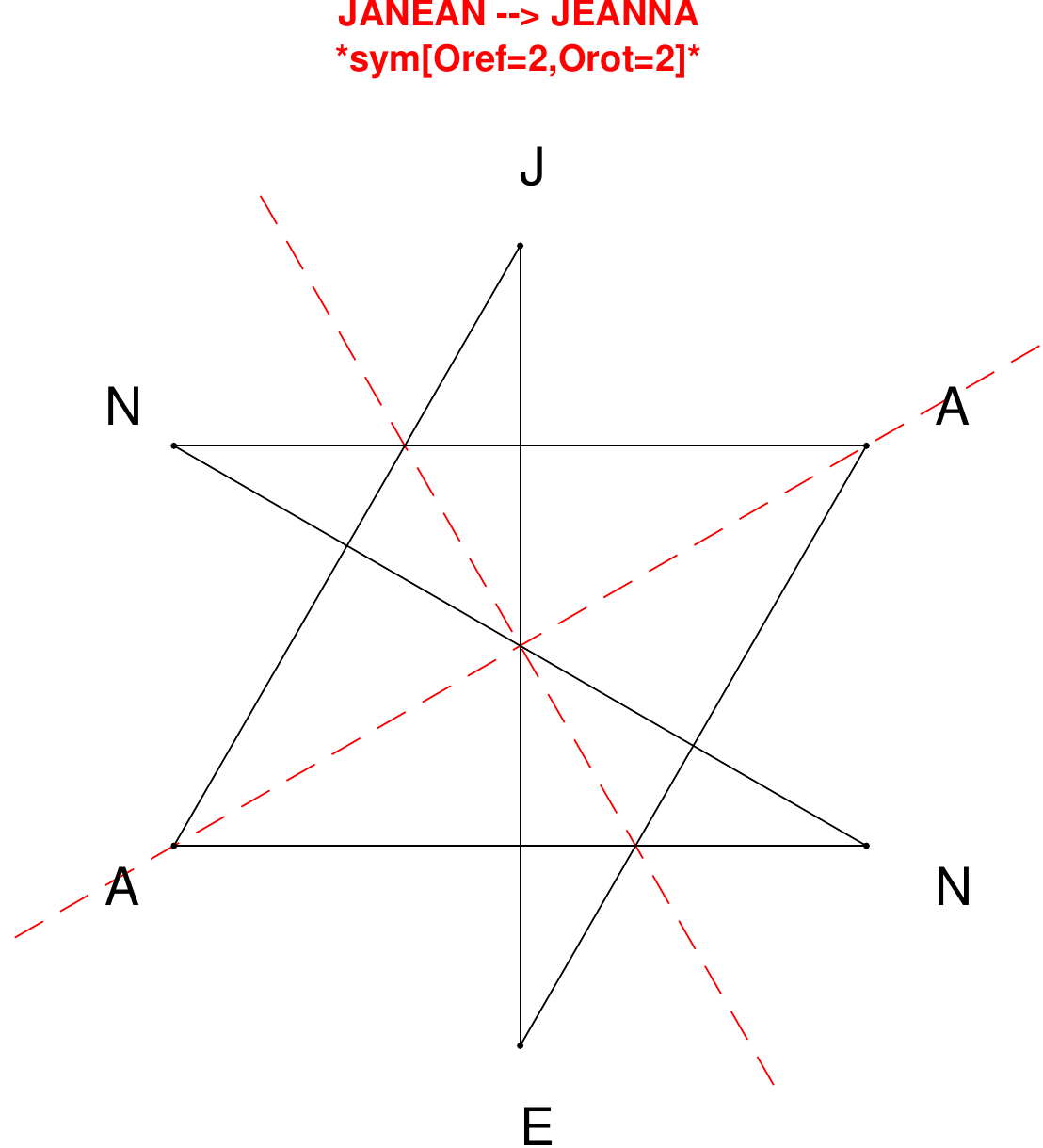}
\end{subfigure}
\hfill
\begin{subfigure}[T]{0.19\textwidth}
\centering
\includegraphics[width=\textwidth]{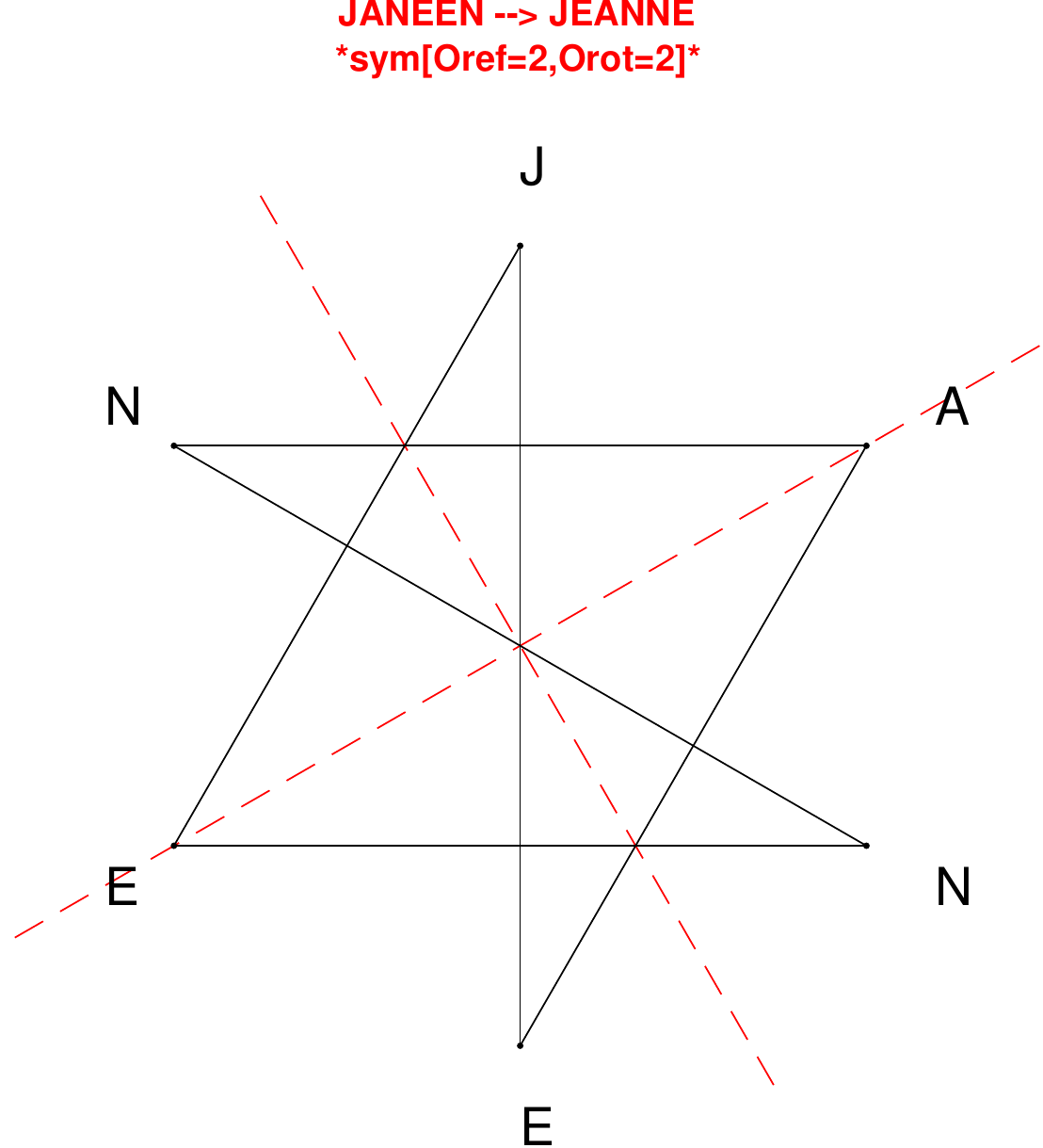}
\end{subfigure}
\hfill
\begin{subfigure}[T]{0.19\textwidth}
\centering
\includegraphics[width=\textwidth]{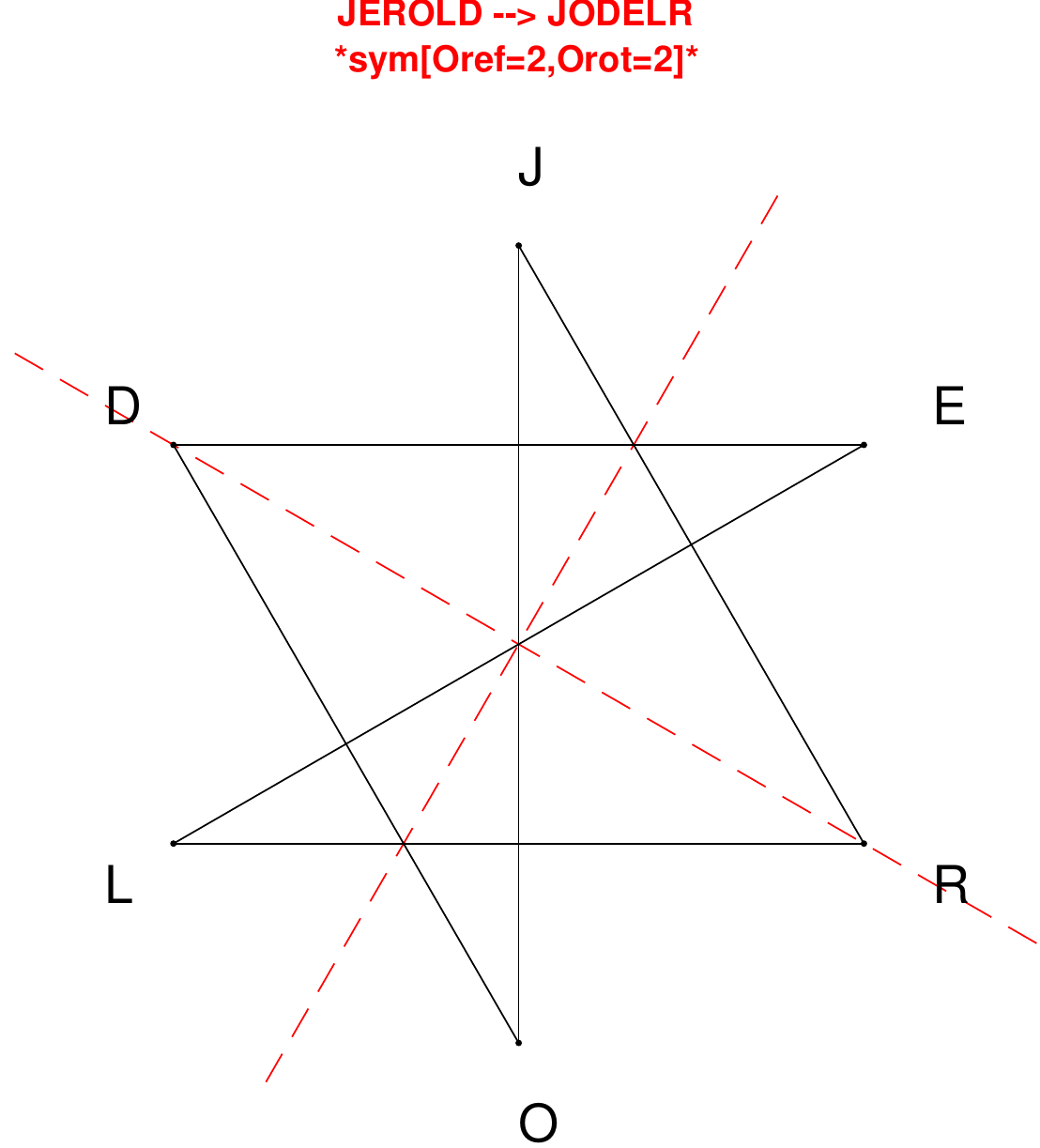}
\end{subfigure}
\hfill
\begin{subfigure}[T]{0.19\textwidth}
\centering
\includegraphics[width=\textwidth]{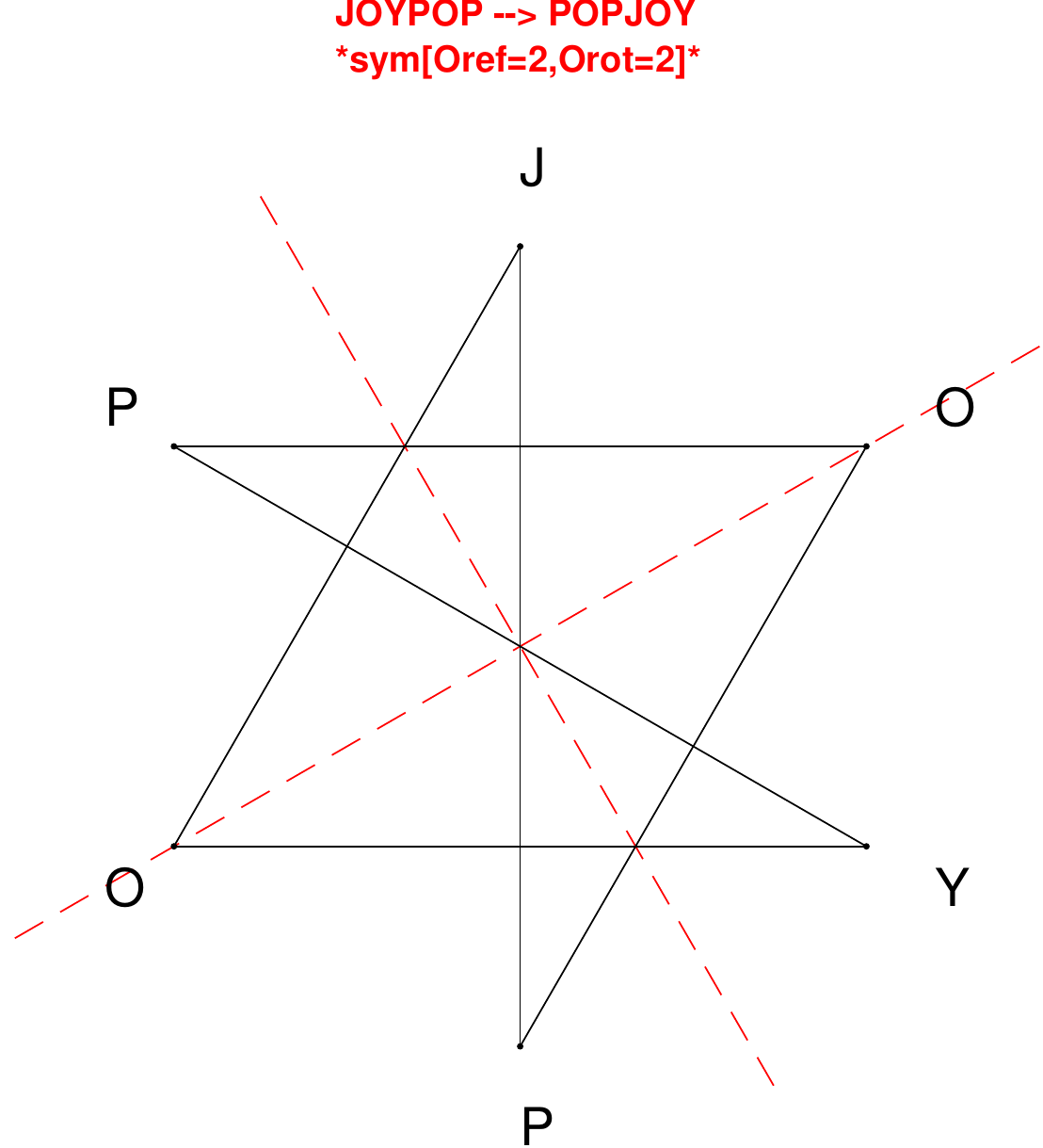}
\end{subfigure}
\end{figure}

\begin{figure}[H]
\centering
\begin{subfigure}[T]{0.19\textwidth}
\centering
\includegraphics[width=\textwidth]{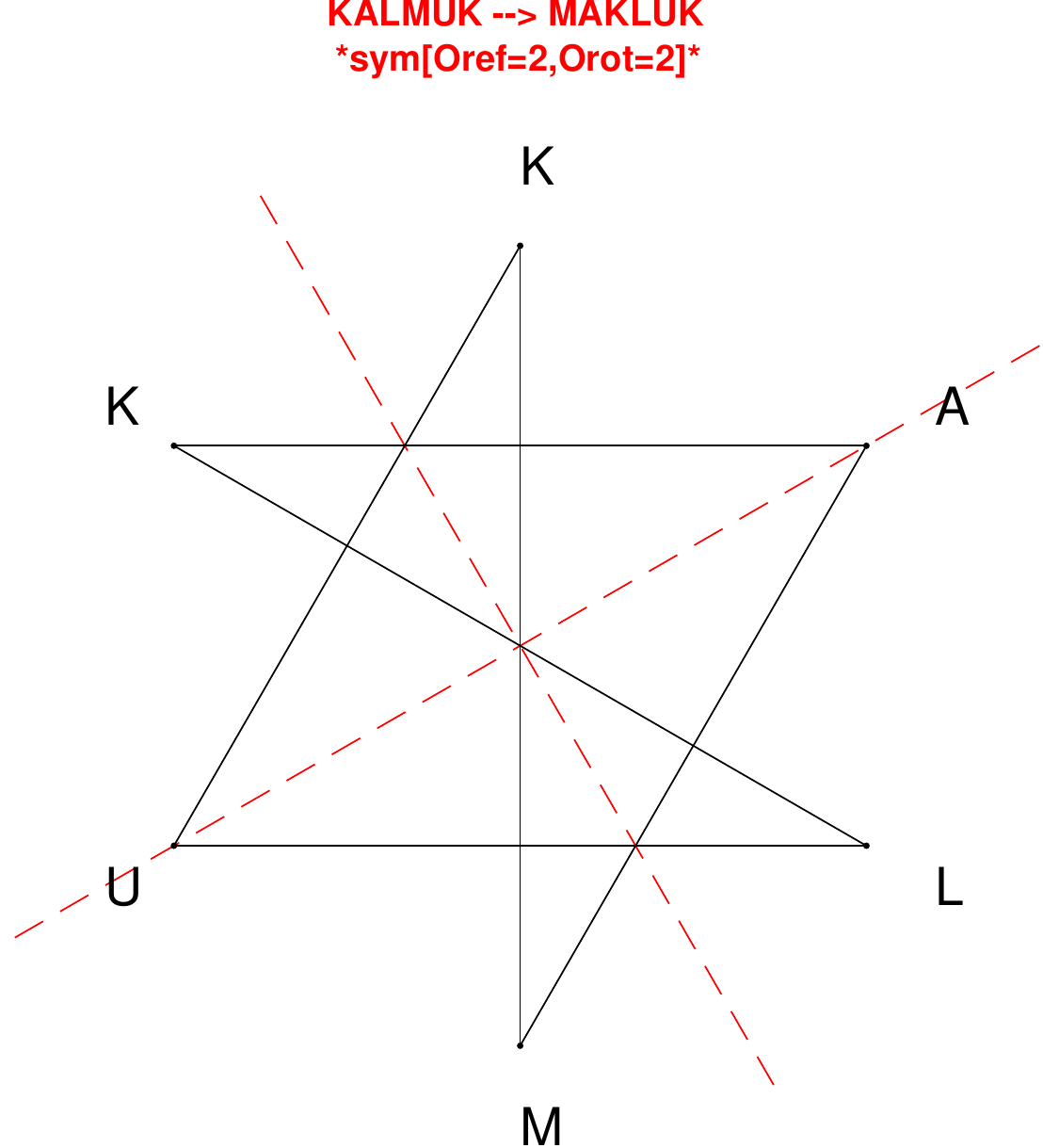}
\end{subfigure}
\hfill
\begin{subfigure}[T]{0.19\textwidth}
\centering
\includegraphics[width=\textwidth]{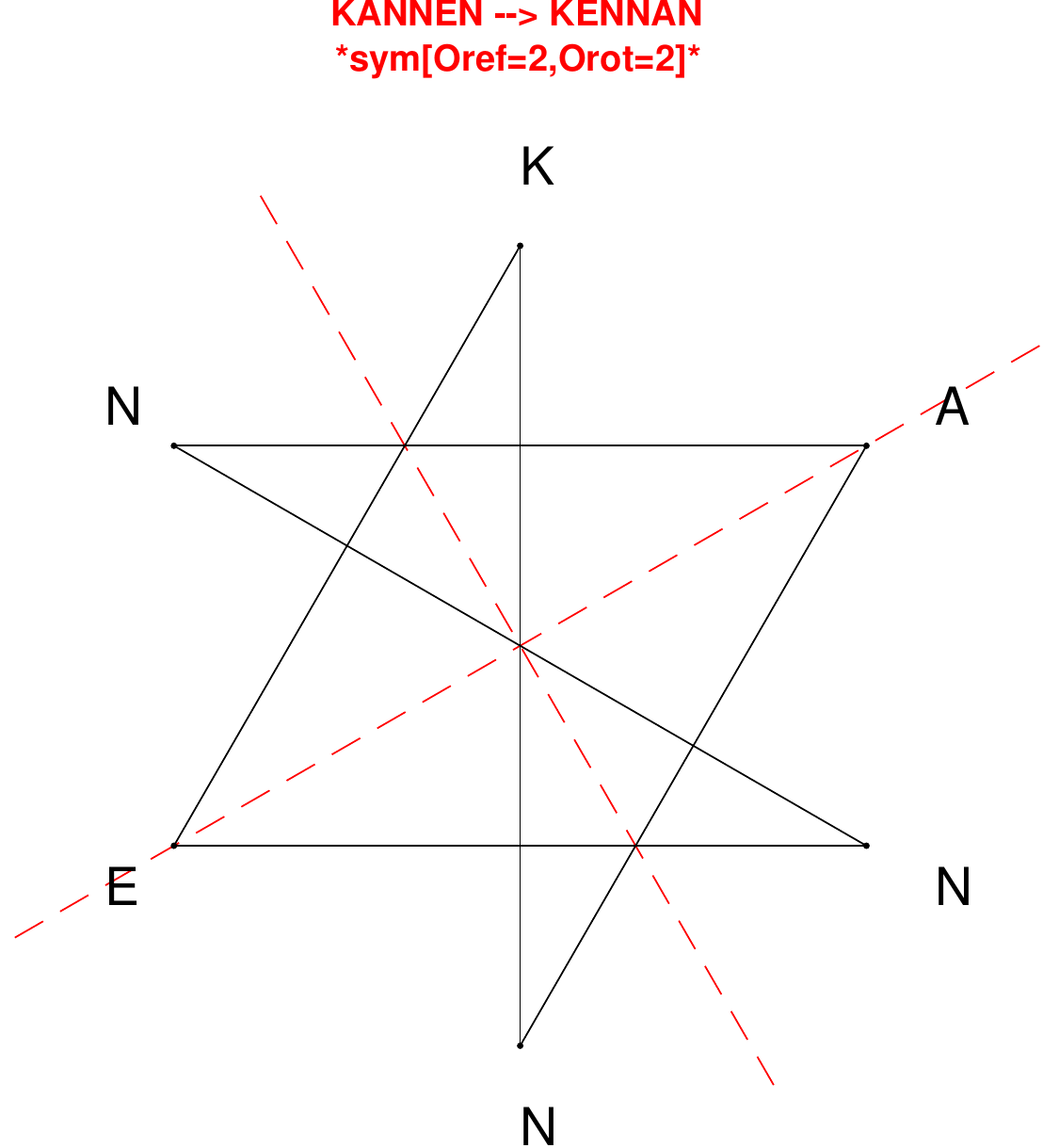}
\end{subfigure}
\hfill
\begin{subfigure}[T]{0.19\textwidth}
\centering
\includegraphics[width=\textwidth]{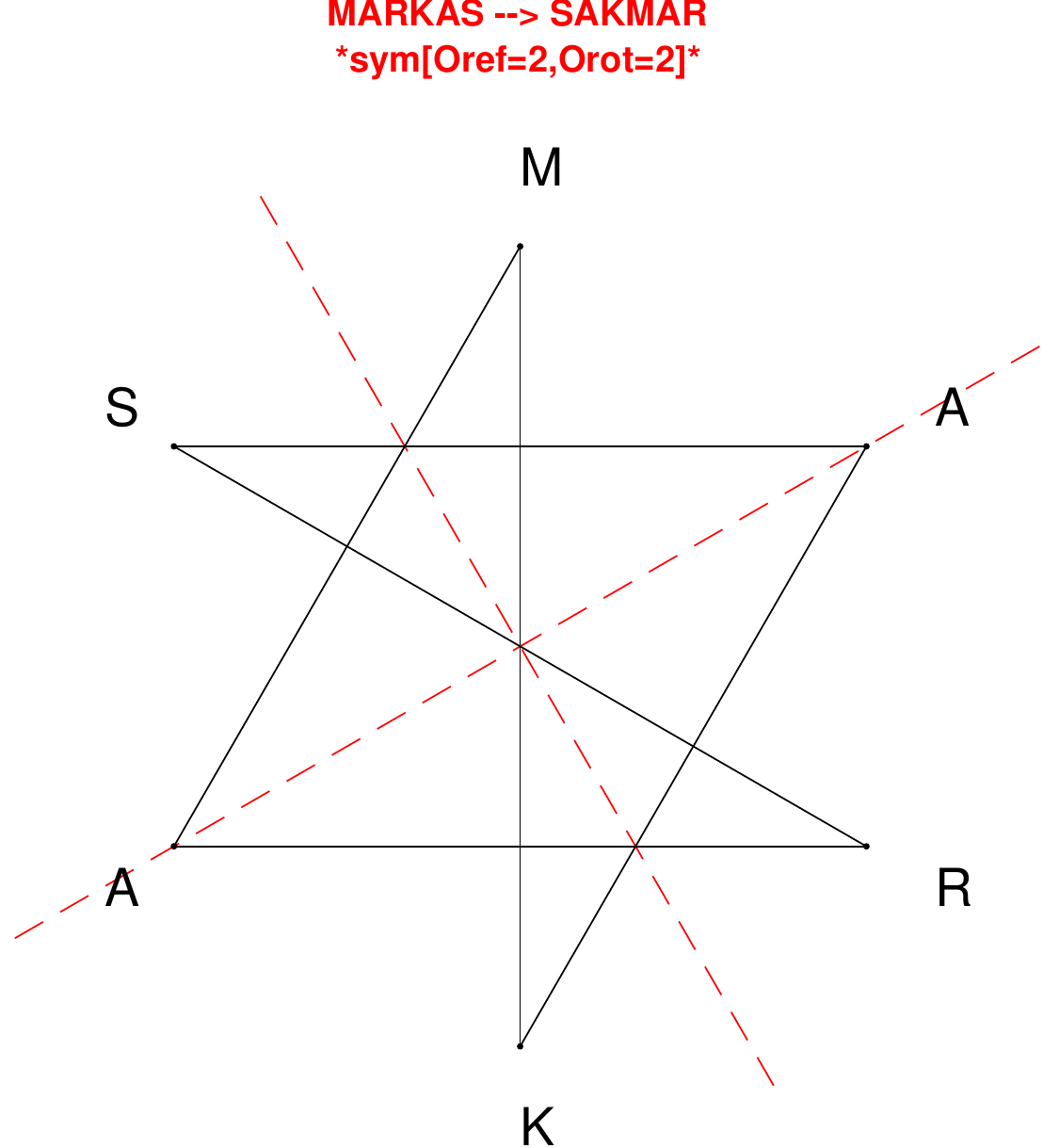}
\end{subfigure}
\hfill
\begin{subfigure}[T]{0.19\textwidth}
\centering
\includegraphics[width=\textwidth]{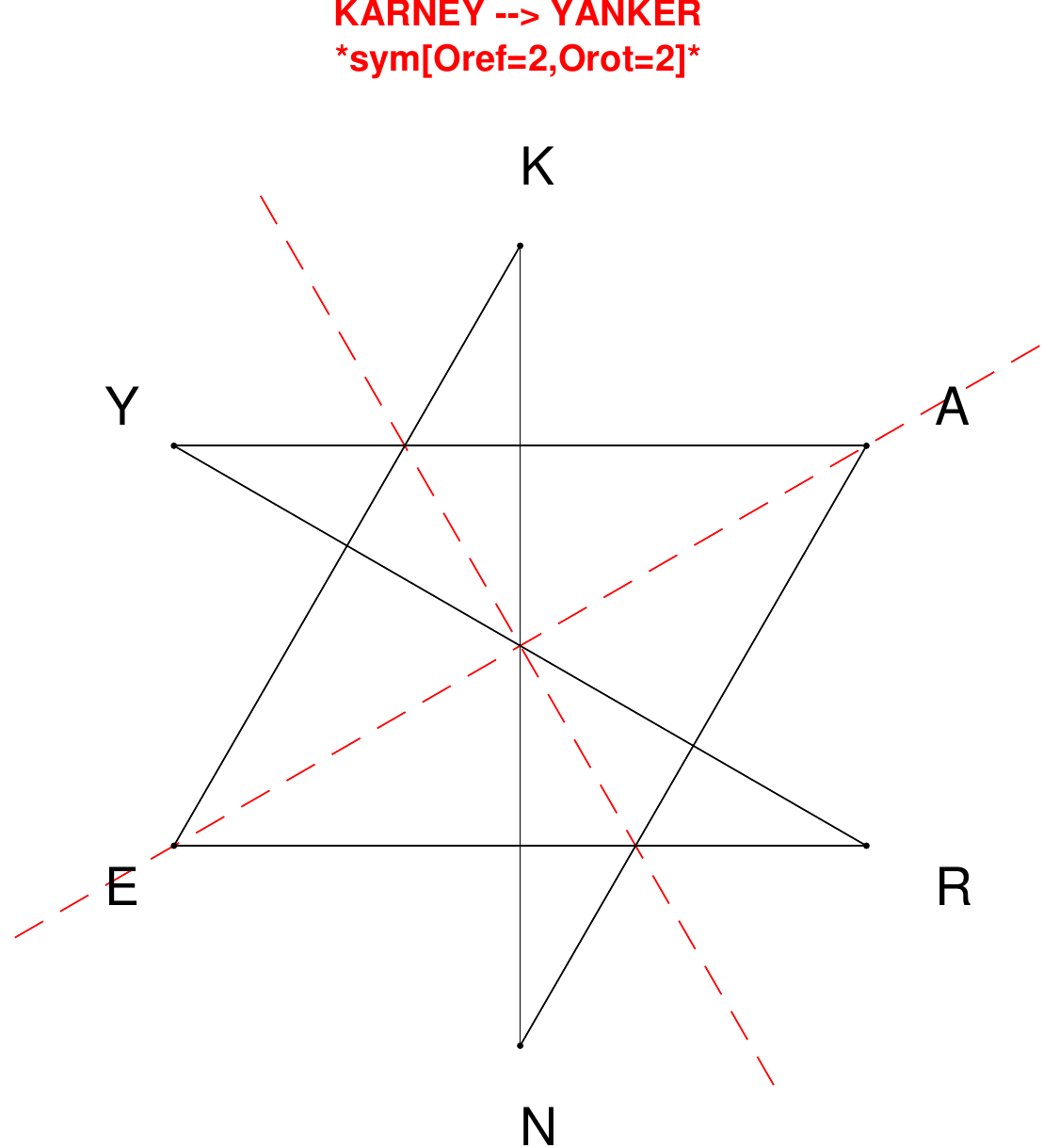}
\end{subfigure}
\hfill
\begin{subfigure}[T]{0.19\textwidth}
\centering
\includegraphics[width=\textwidth]{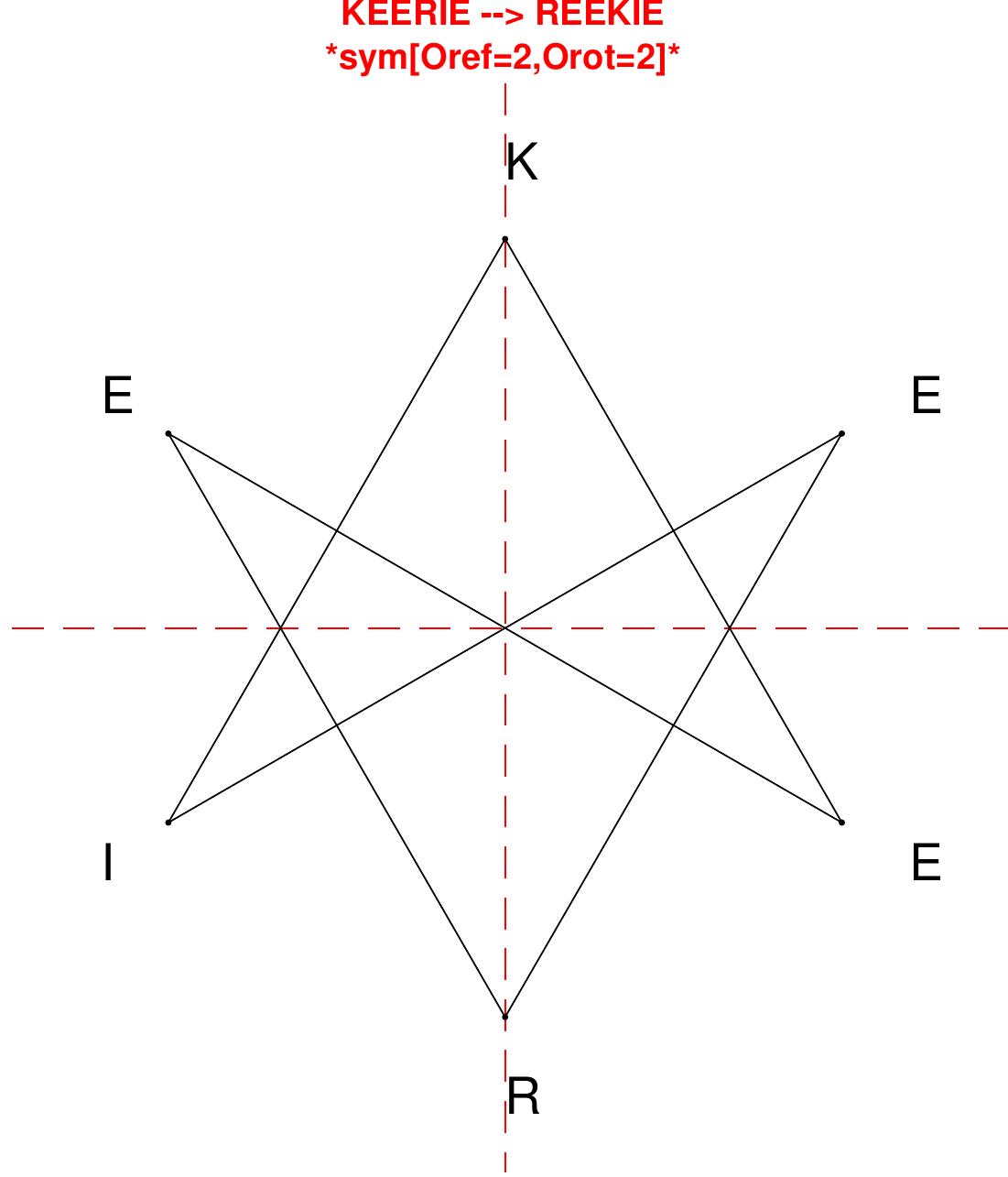}
\end{subfigure}
\end{figure}

\begin{figure}[H]
\centering
\begin{subfigure}[T]{0.19\textwidth}
\centering
\includegraphics[width=\textwidth]{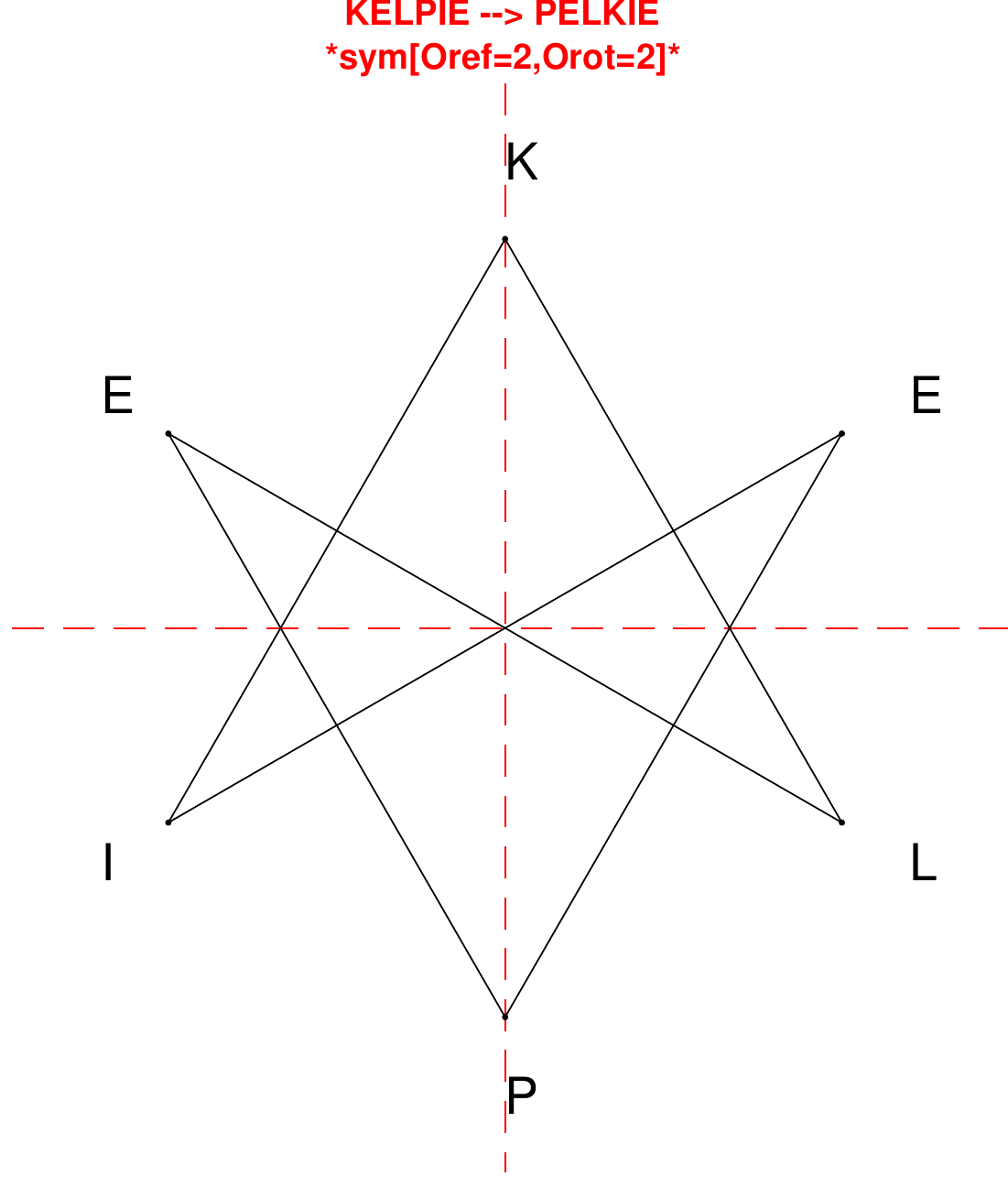}
\end{subfigure}
\hfill
\begin{subfigure}[T]{0.19\textwidth}
\centering
\includegraphics[width=\textwidth]{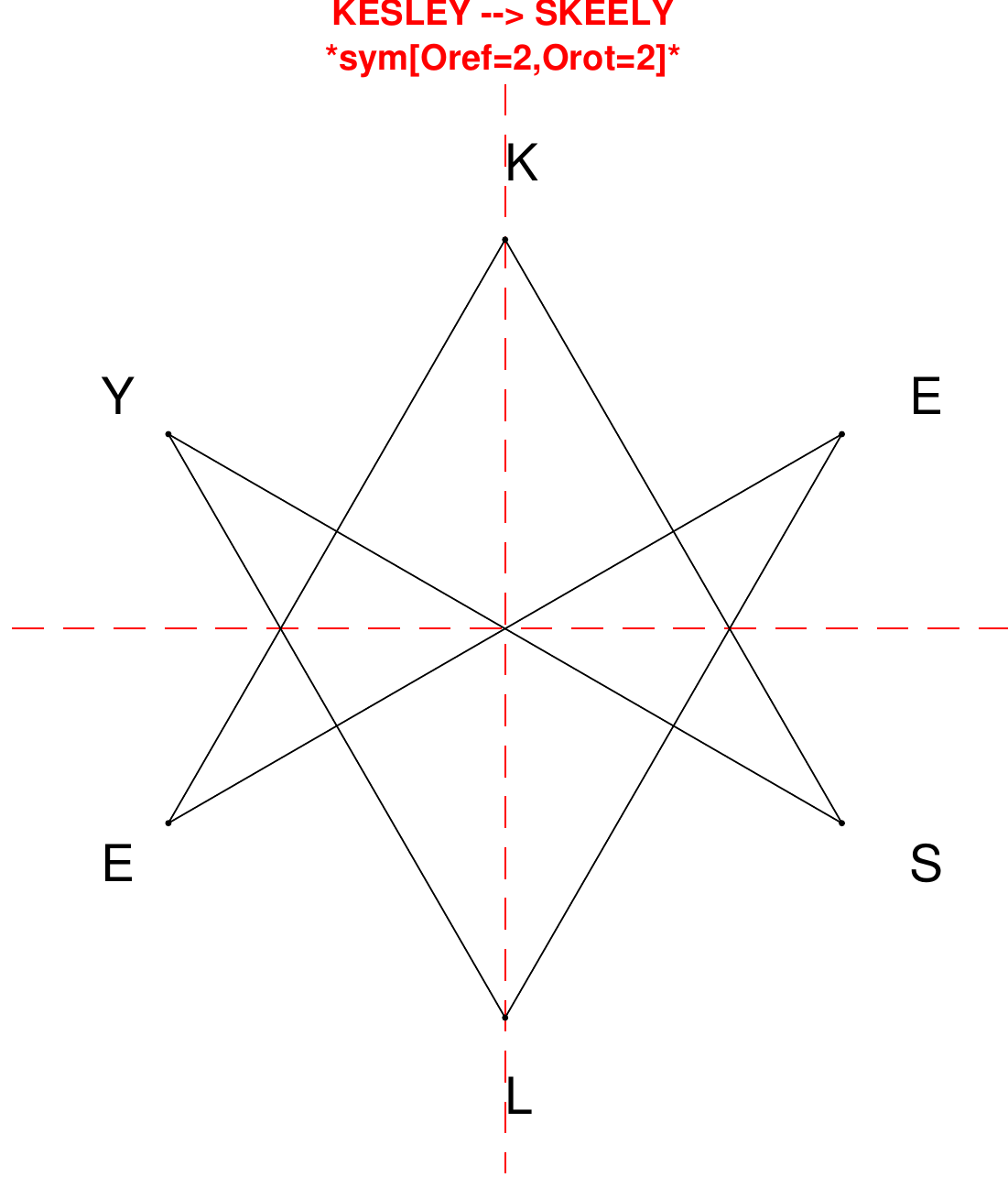}
\end{subfigure}
\hfill
\begin{subfigure}[T]{0.19\textwidth}
\centering
\includegraphics[width=\textwidth]{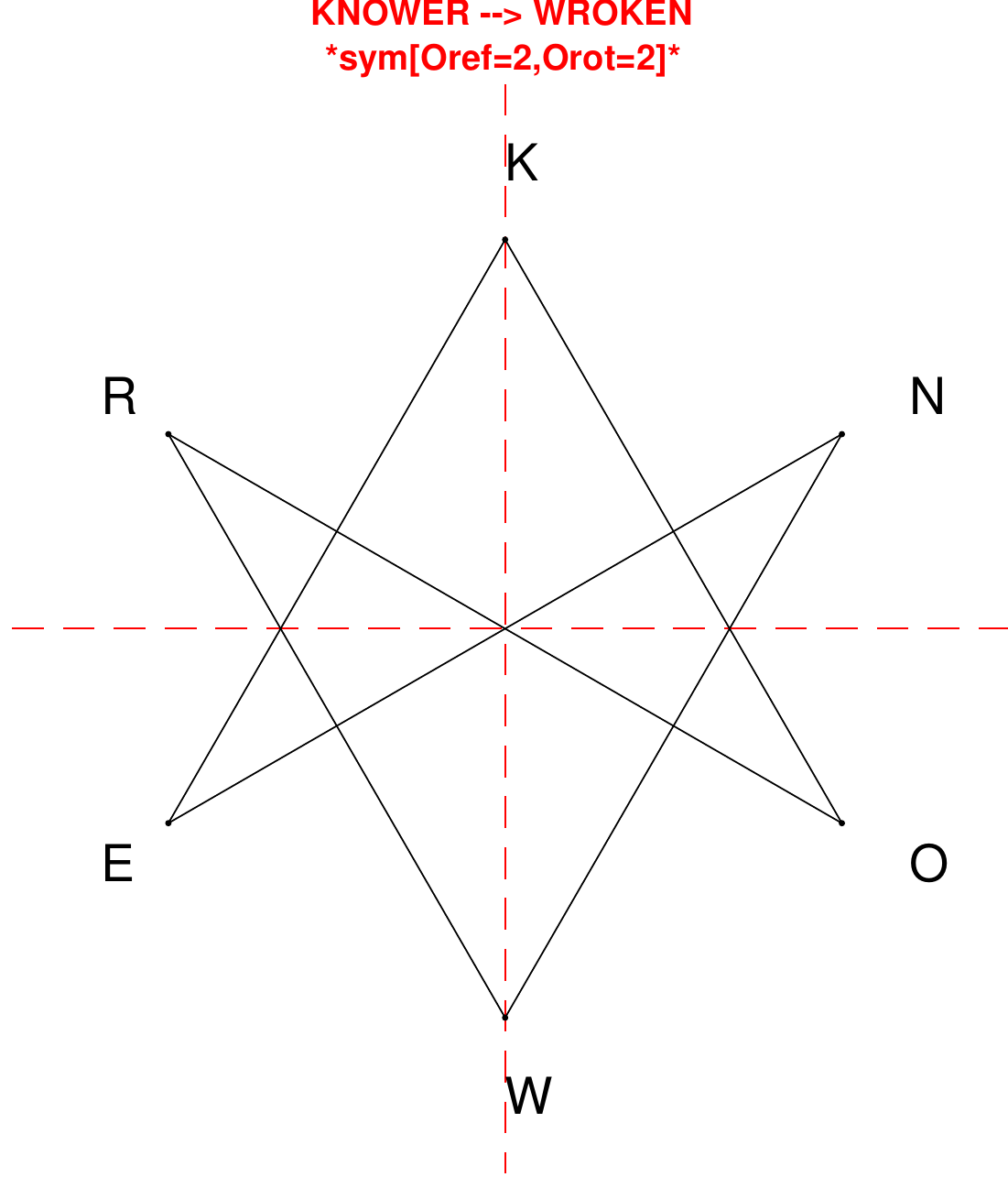}
\end{subfigure}
\hfill
\begin{subfigure}[T]{0.19\textwidth}
\centering
\includegraphics[width=\textwidth]{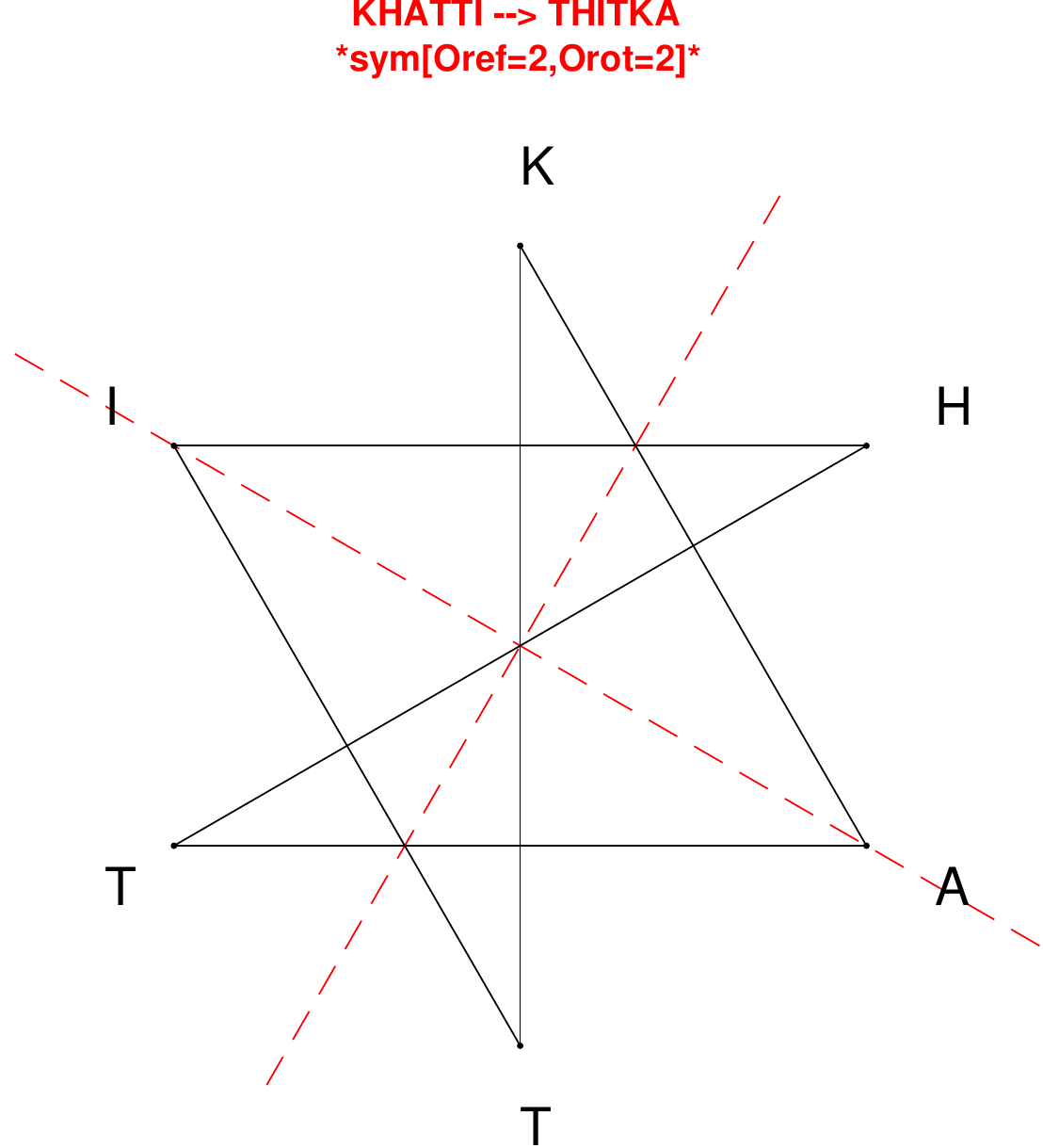}
\end{subfigure}
\hfill
\begin{subfigure}[T]{0.19\textwidth}
\centering
\includegraphics[width=\textwidth]{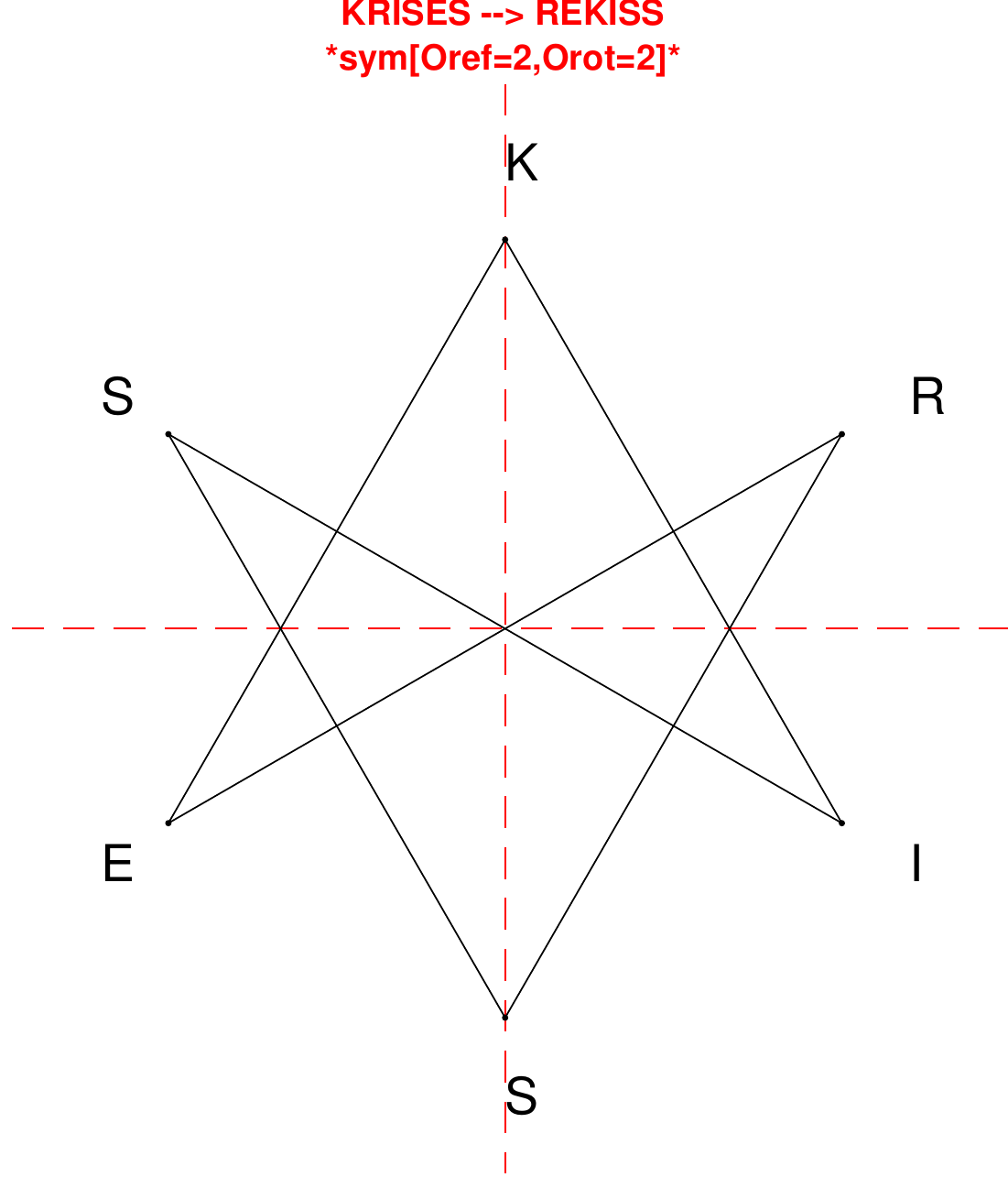}
\end{subfigure}
\end{figure}

\begin{figure}[H]
\centering
\begin{subfigure}[T]{0.19\textwidth}
\centering
\includegraphics[width=\textwidth]{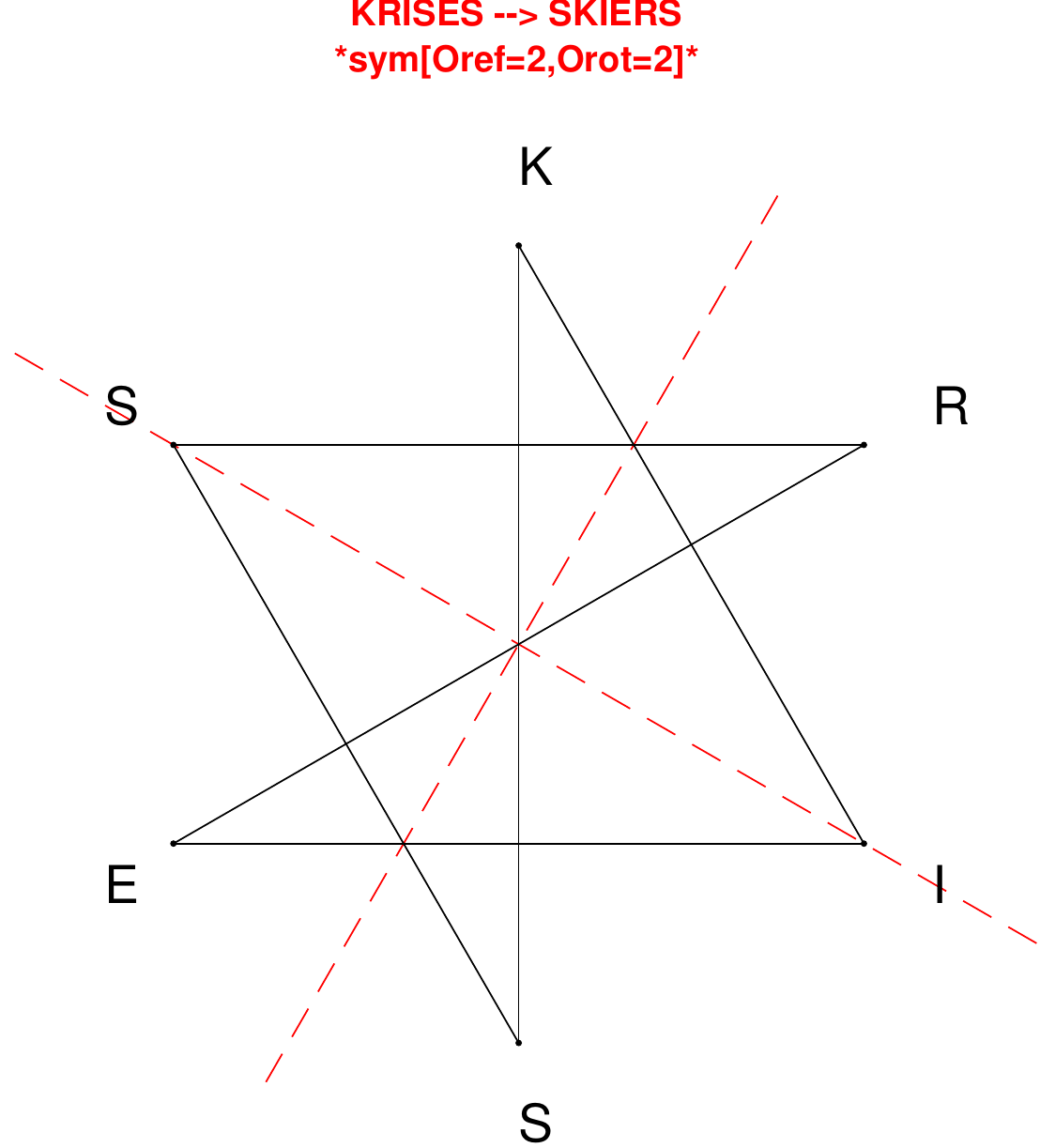}
\end{subfigure}
\hfill
\begin{subfigure}[T]{0.19\textwidth}
\centering
\includegraphics[width=\textwidth]{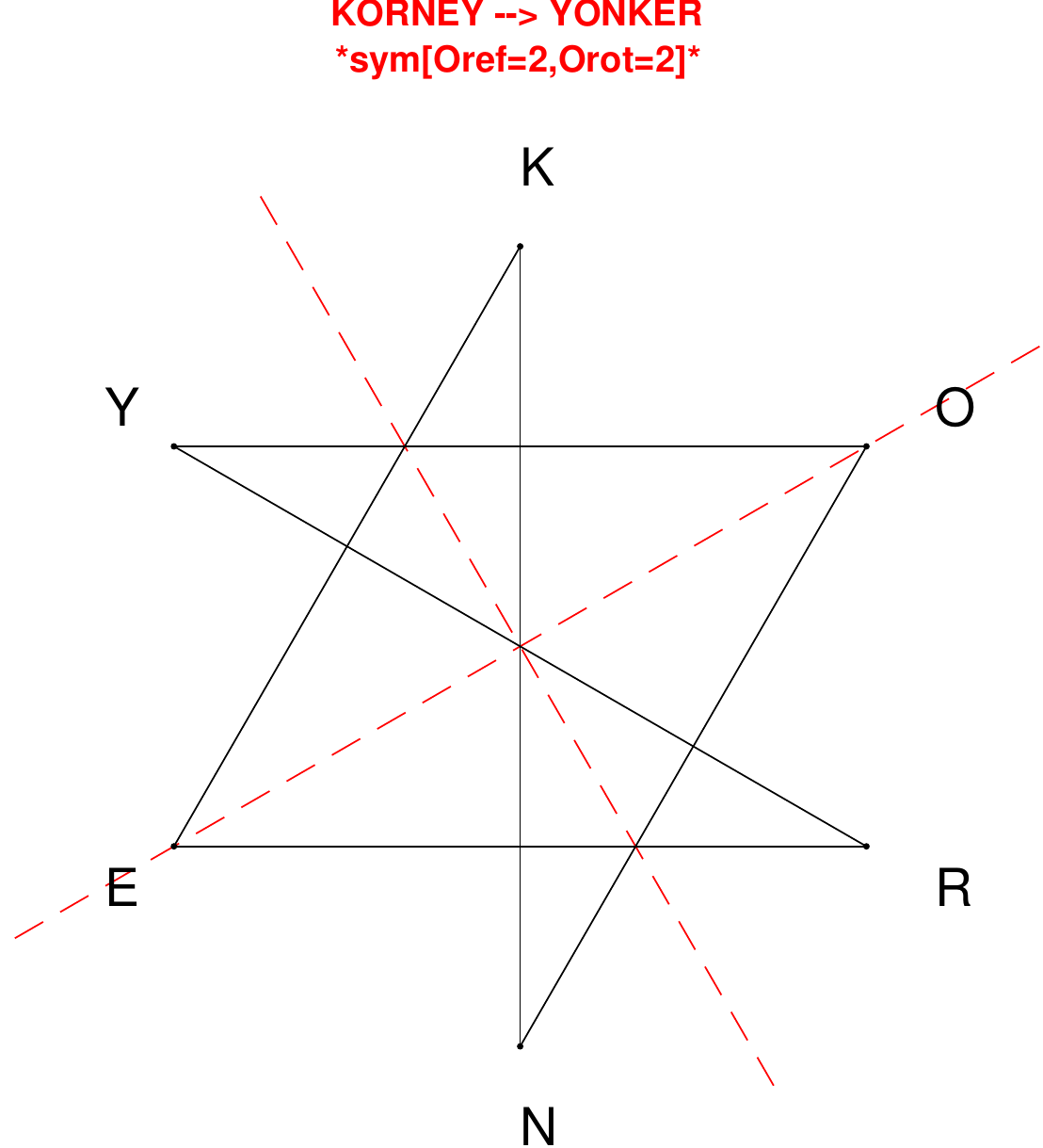}
\end{subfigure}
\hfill
\begin{subfigure}[T]{0.19\textwidth}
\centering
\includegraphics[width=\textwidth]{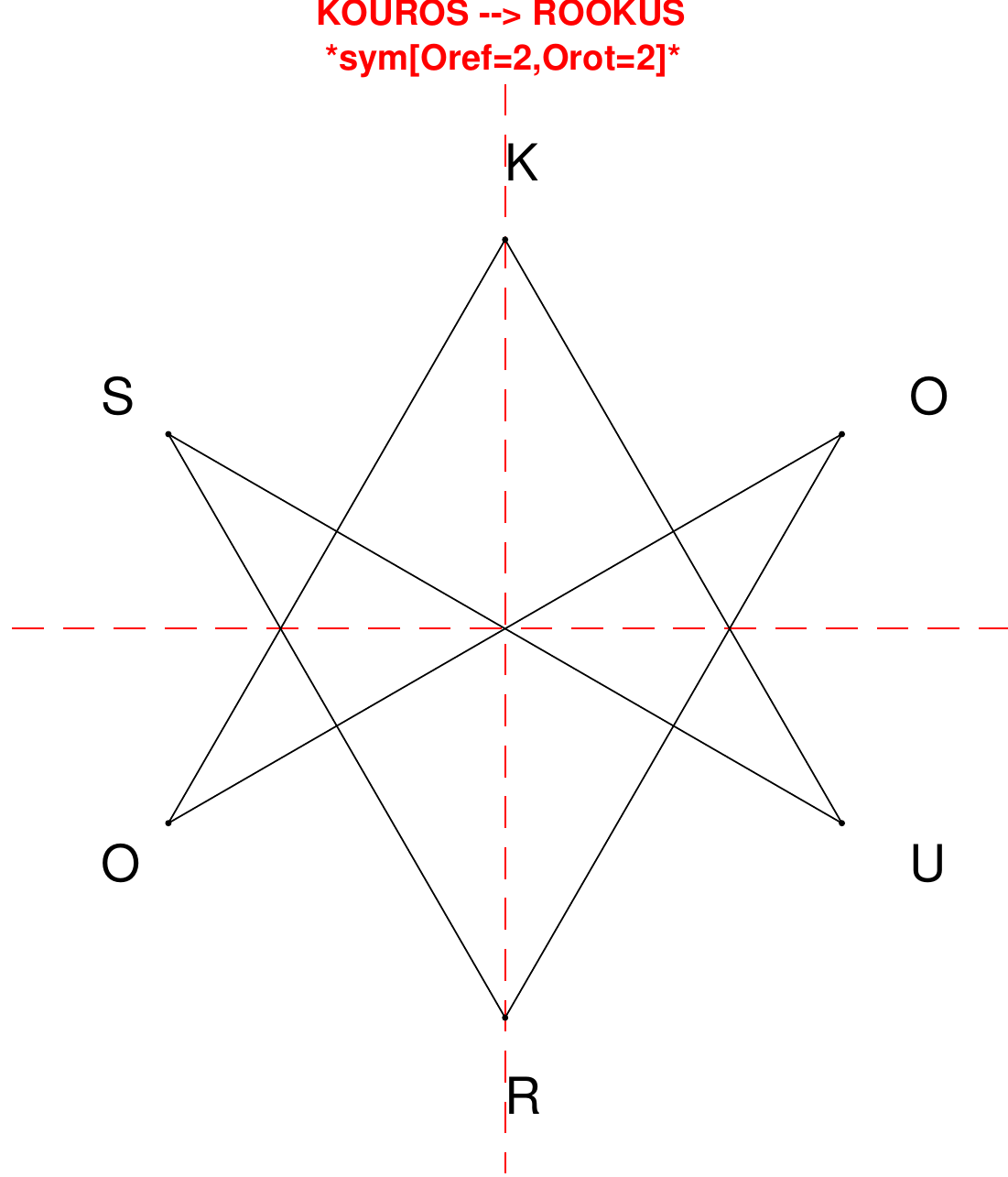}
\end{subfigure}
\hfill
\begin{subfigure}[T]{0.19\textwidth}
\centering
\includegraphics[width=\textwidth]{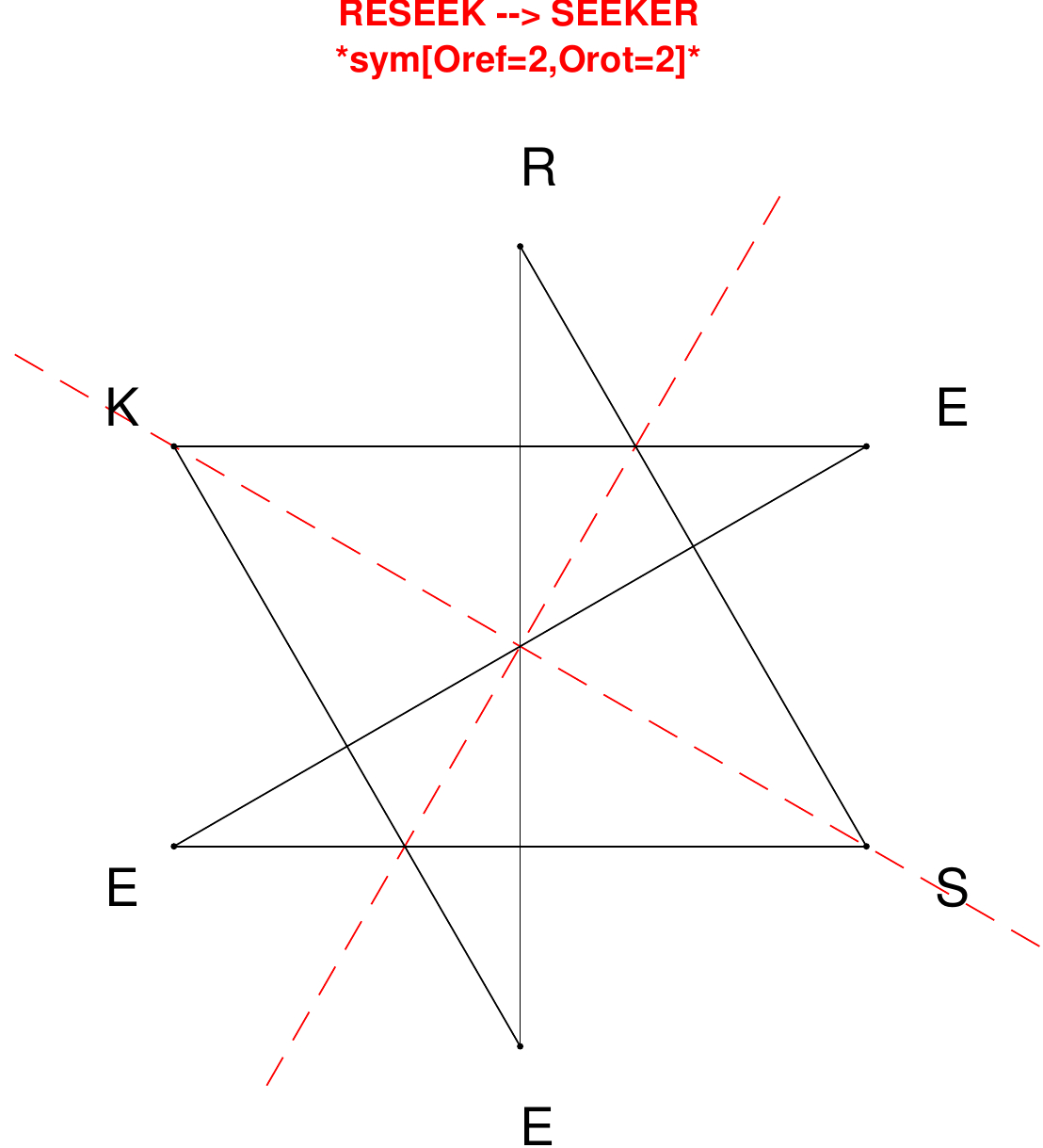}
\end{subfigure}
\hfill
\begin{subfigure}[T]{0.19\textwidth}
\centering
\includegraphics[width=\textwidth]{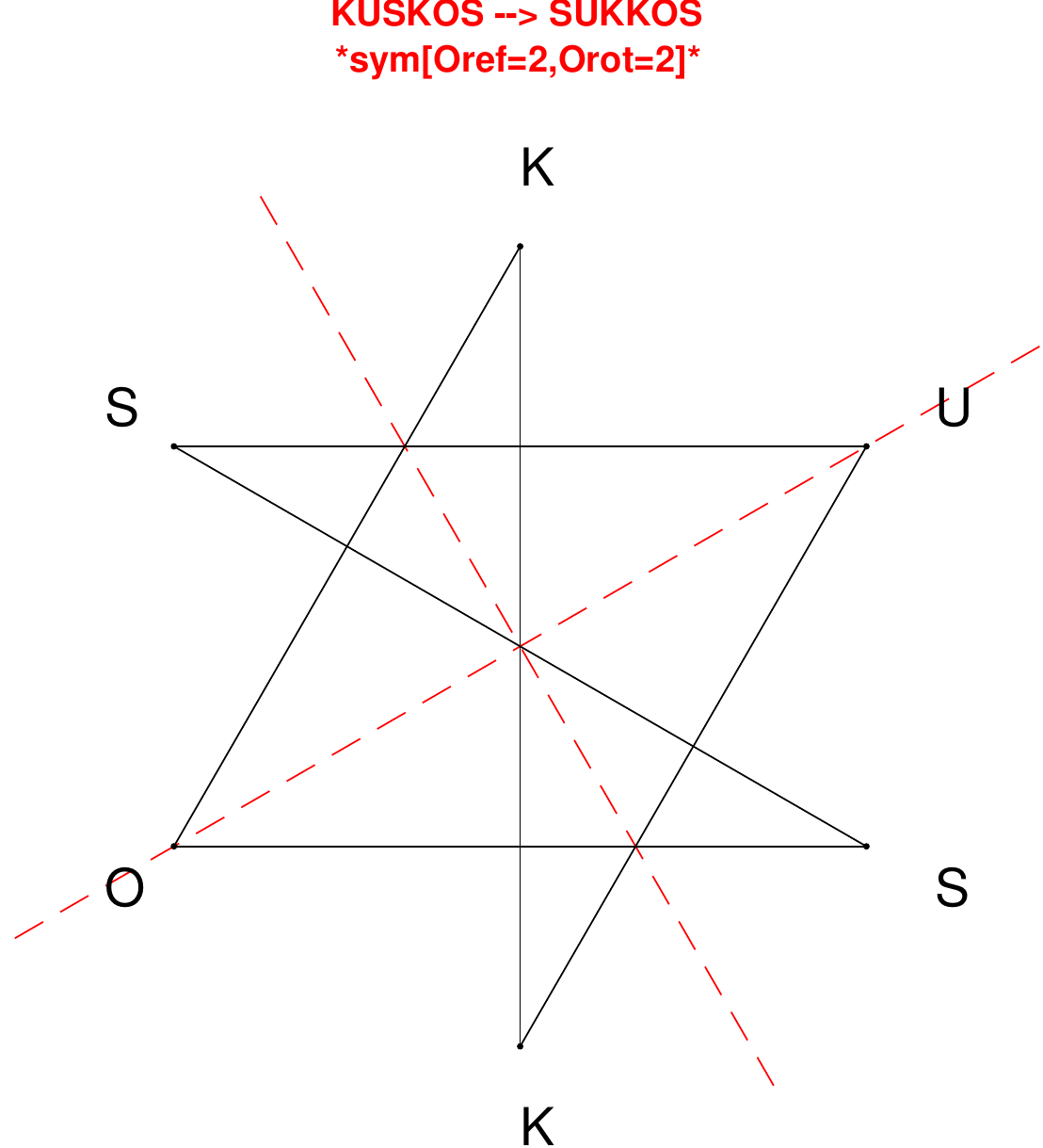}
\end{subfigure}
\end{figure}

\begin{figure}[H]
\centering
\begin{subfigure}[T]{0.19\textwidth}
\centering
\includegraphics[width=\textwidth]{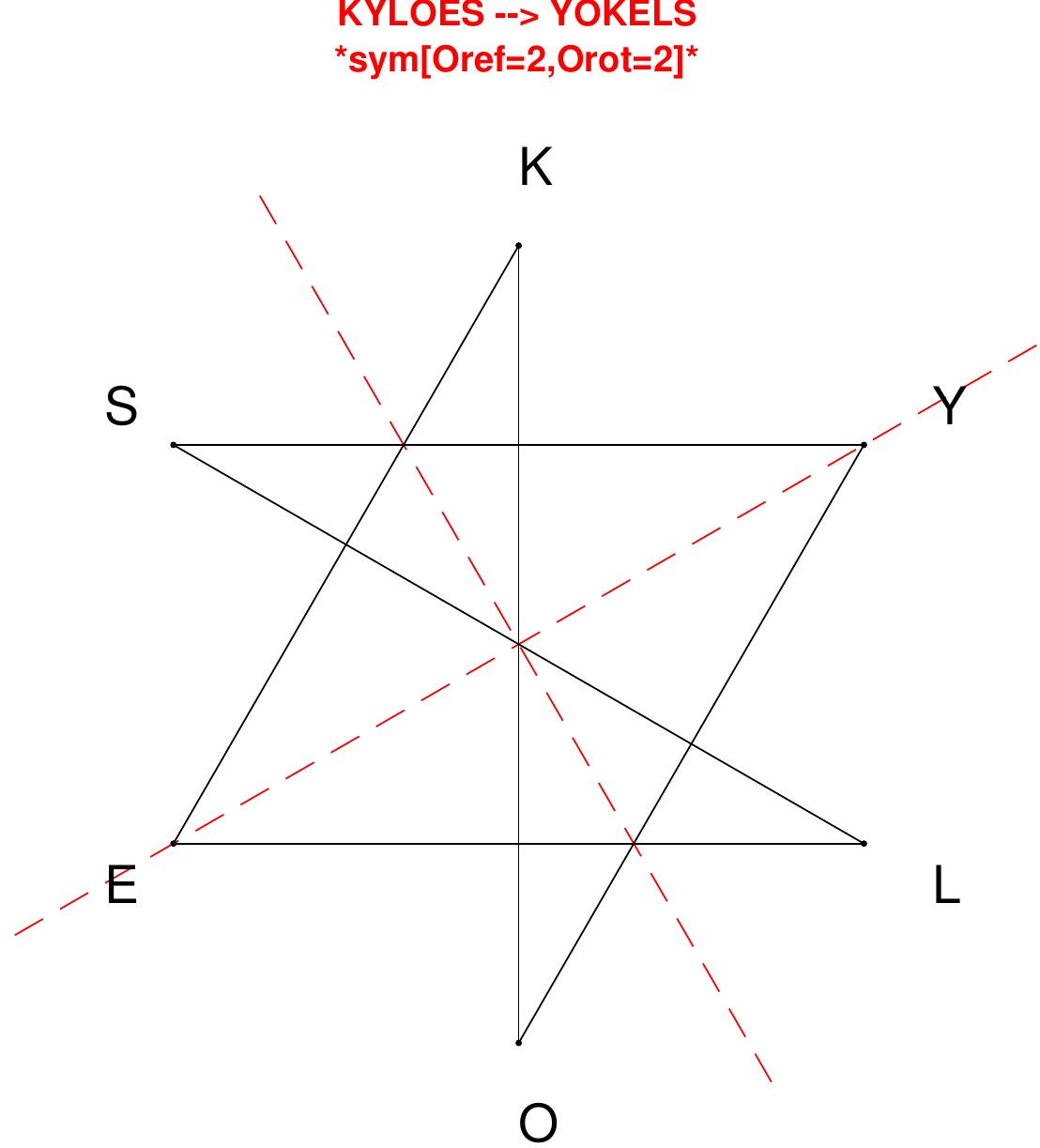}
\end{subfigure}
\hfill
\begin{subfigure}[T]{0.19\textwidth}
\centering
\includegraphics[width=\textwidth]{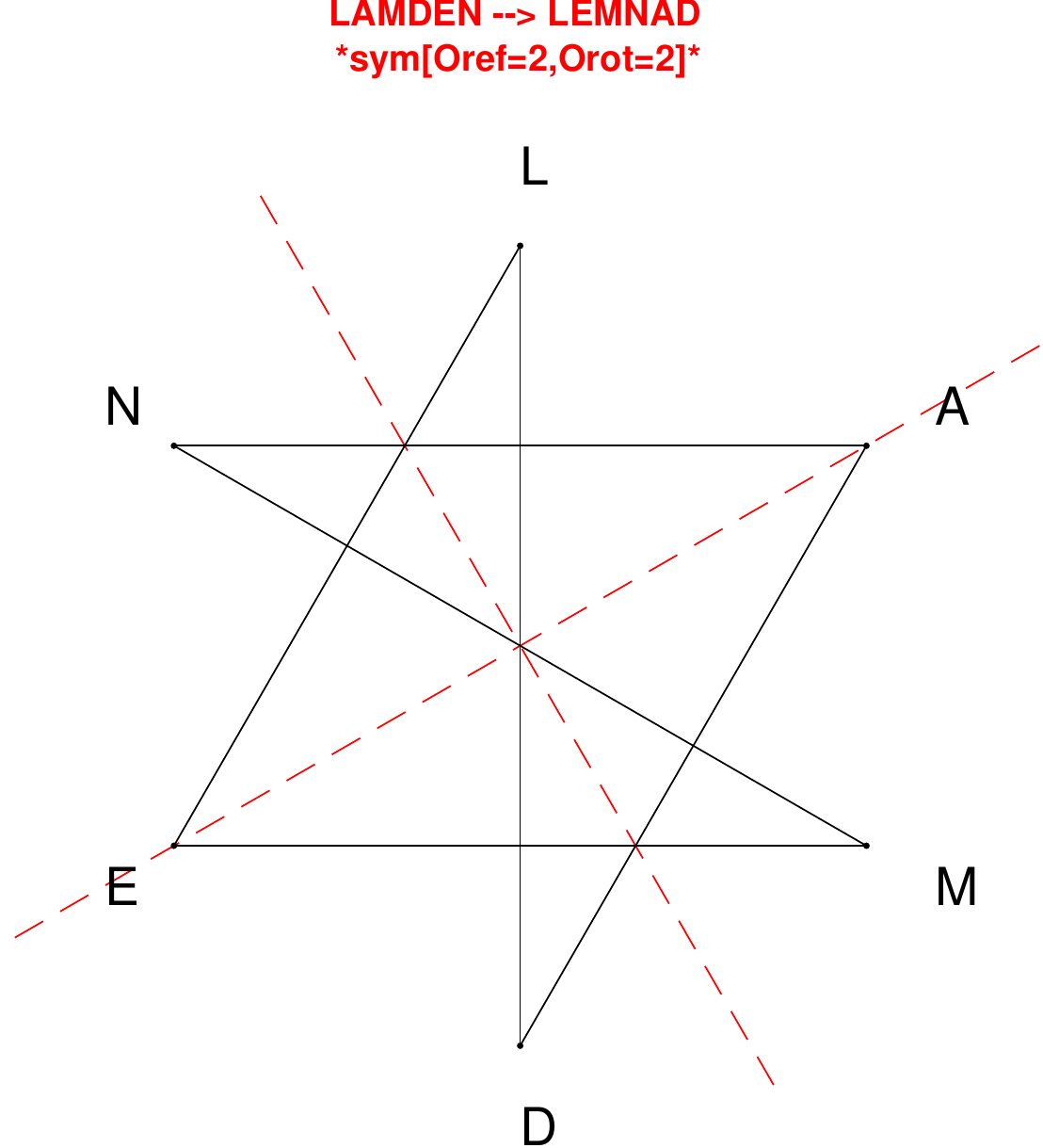}
\end{subfigure}
\hfill
\begin{subfigure}[T]{0.19\textwidth}
\centering
\includegraphics[width=\textwidth]{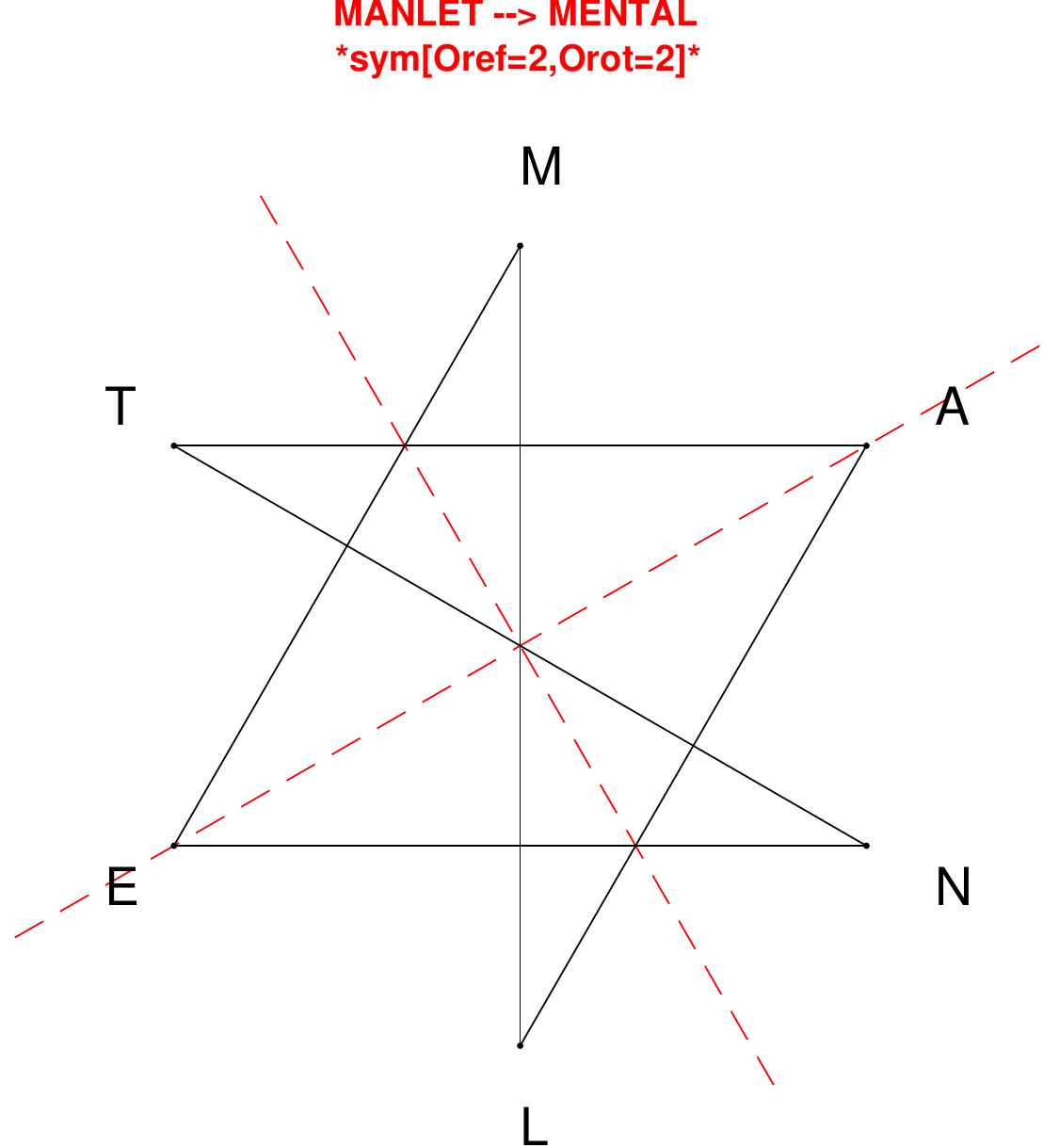}
\end{subfigure}
\hfill
\begin{subfigure}[T]{0.19\textwidth}
\centering
\includegraphics[width=\textwidth]{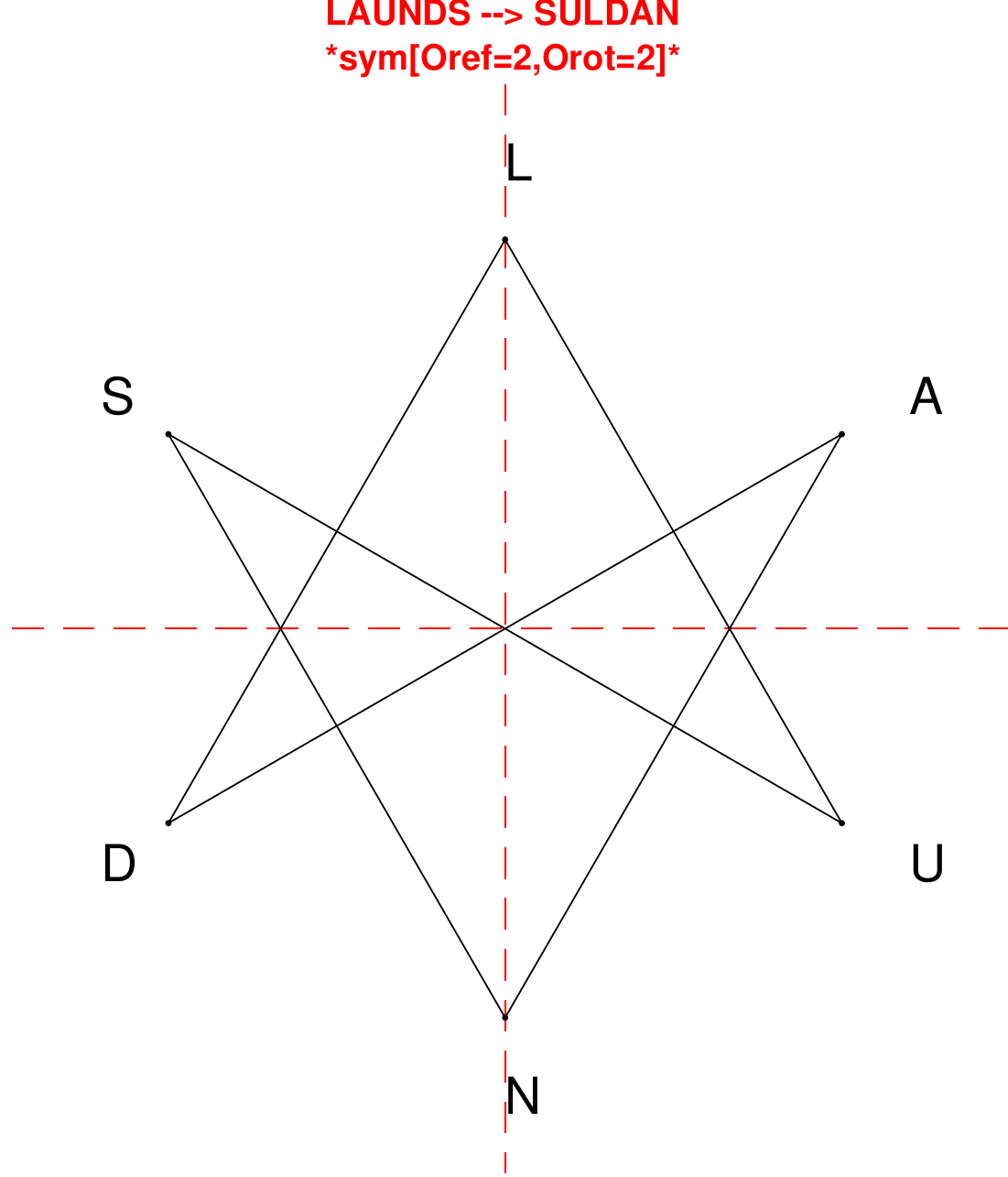}
\end{subfigure}
\hfill
\begin{subfigure}[T]{0.19\textwidth}
\centering
\includegraphics[width=\textwidth]{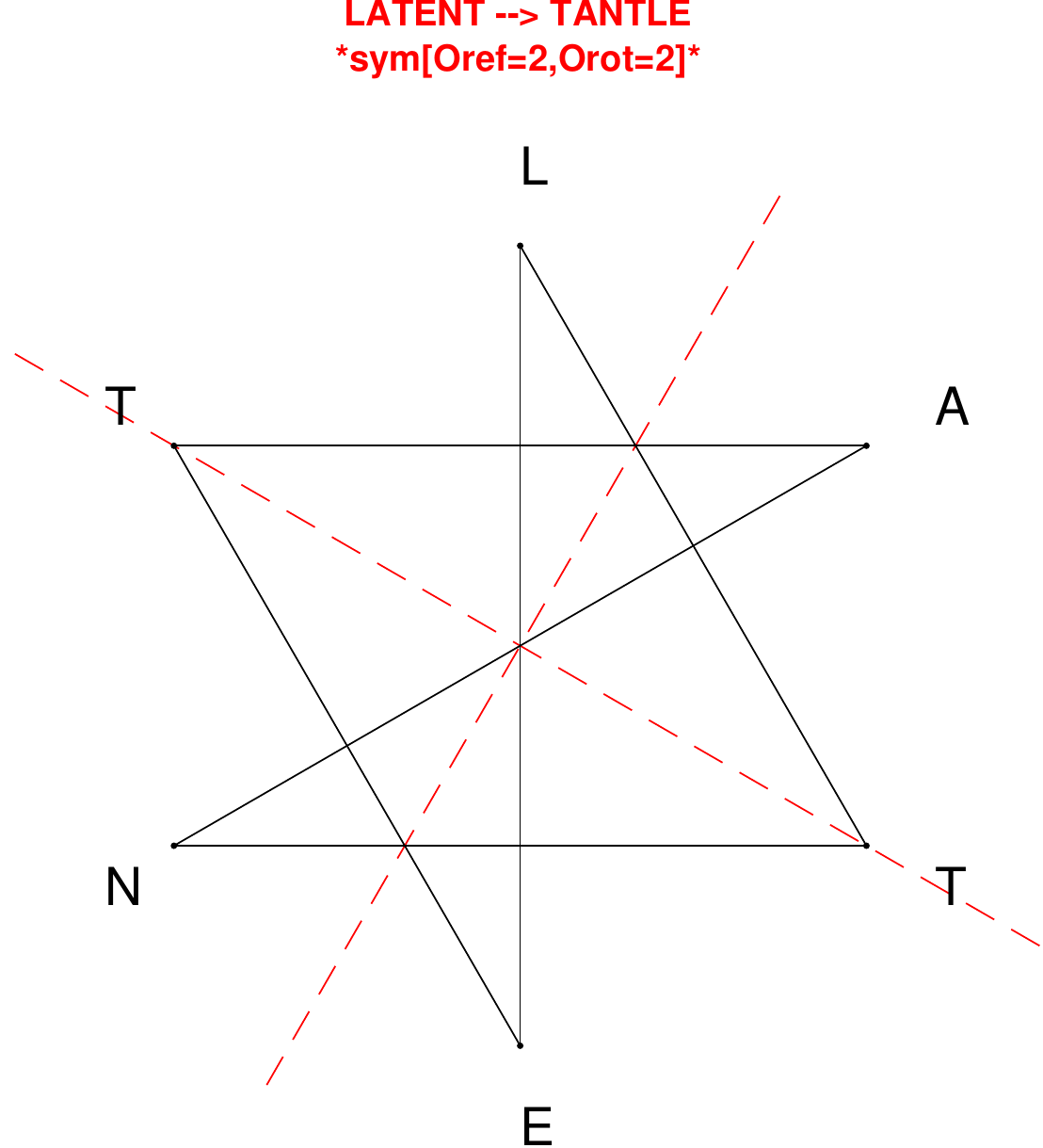}
\end{subfigure}
\end{figure}

\begin{figure}[H]
\centering
\begin{subfigure}[T]{0.19\textwidth}
\centering
\includegraphics[width=\textwidth]{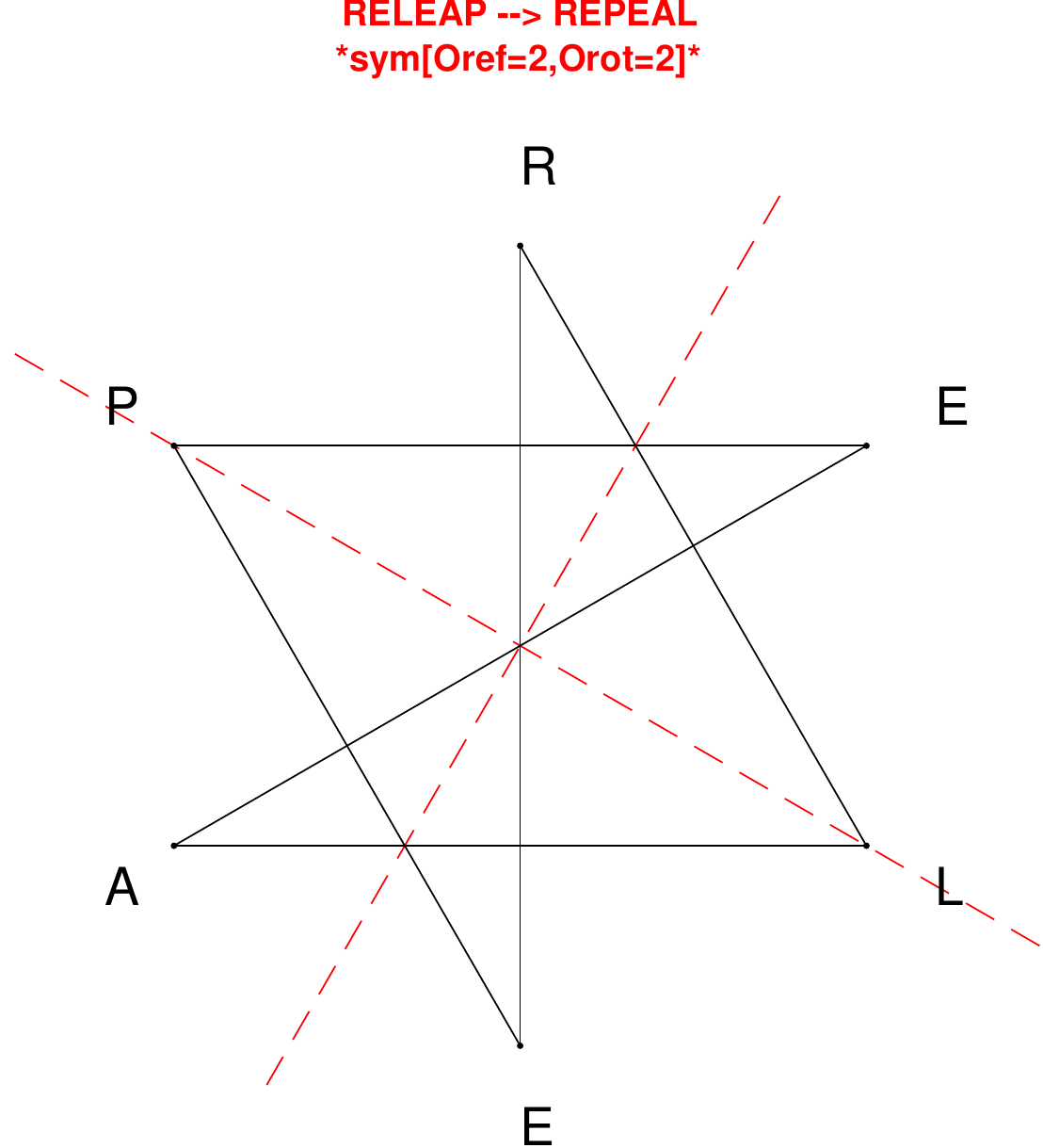}
\end{subfigure}
\hfill
\begin{subfigure}[T]{0.19\textwidth}
\centering
\includegraphics[width=\textwidth]{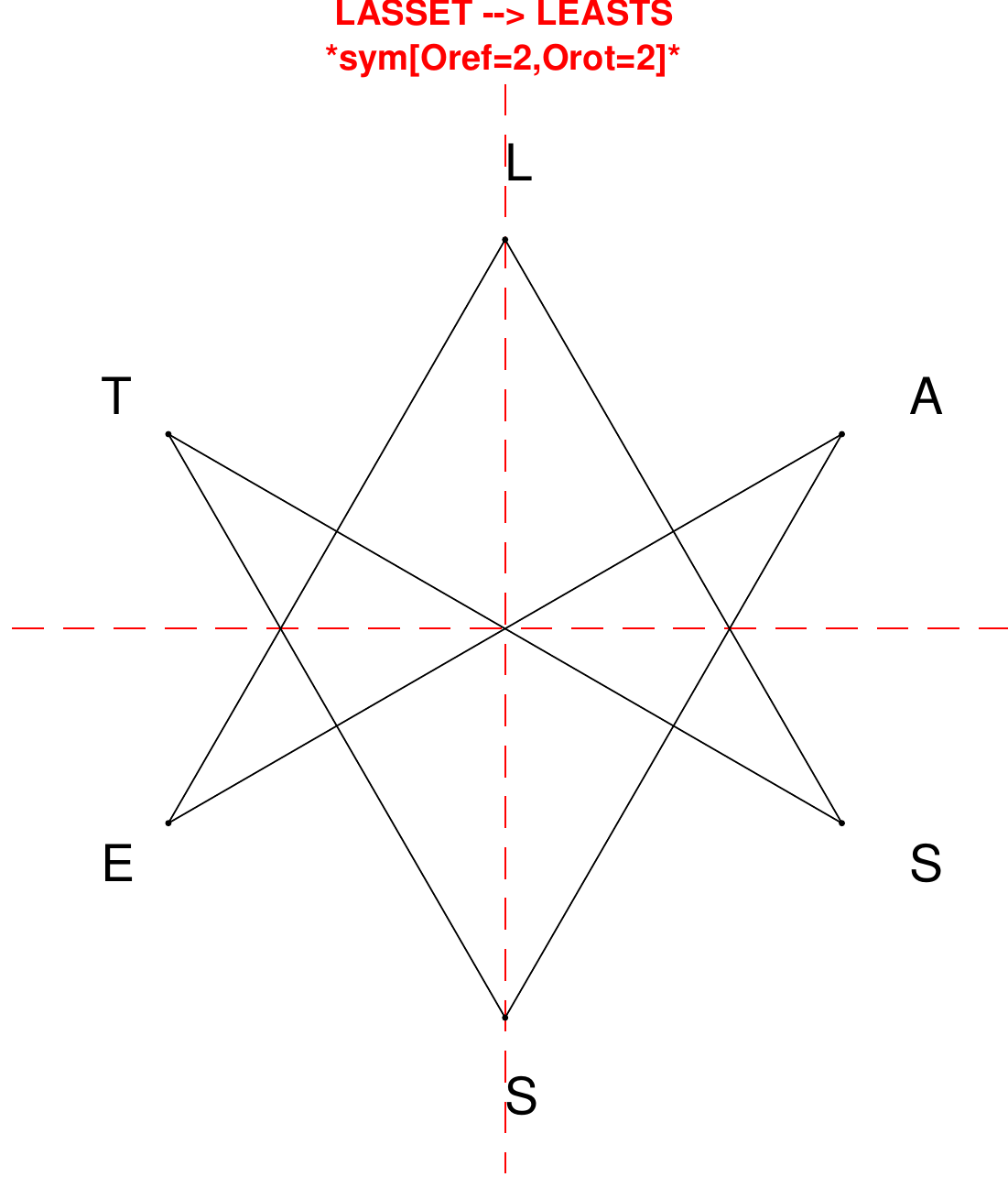}
\end{subfigure}
\hfill
\begin{subfigure}[T]{0.19\textwidth}
\centering
\includegraphics[width=\textwidth]{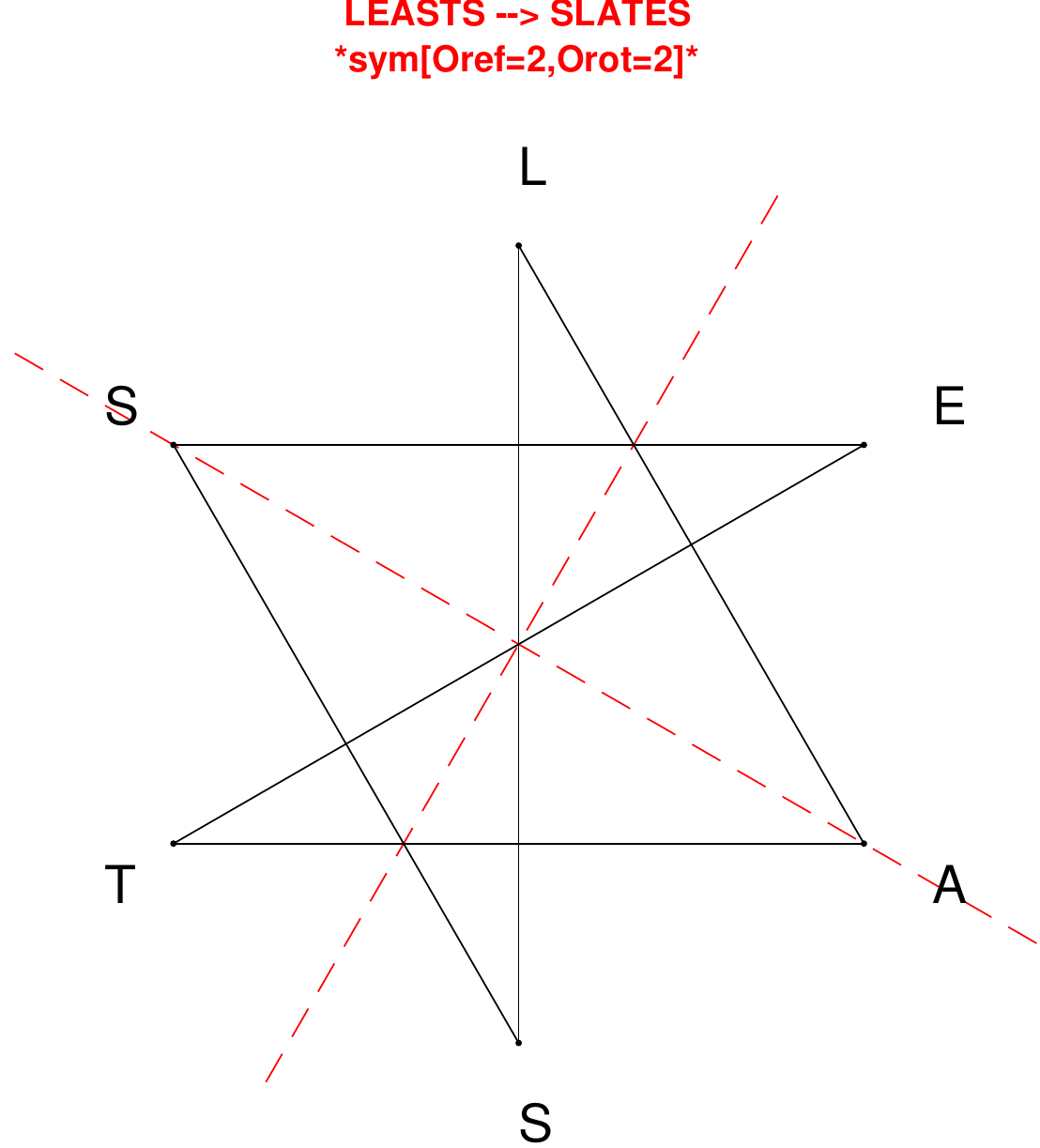}
\end{subfigure}
\hfill
\begin{subfigure}[T]{0.19\textwidth}
\centering
\includegraphics[width=\textwidth]{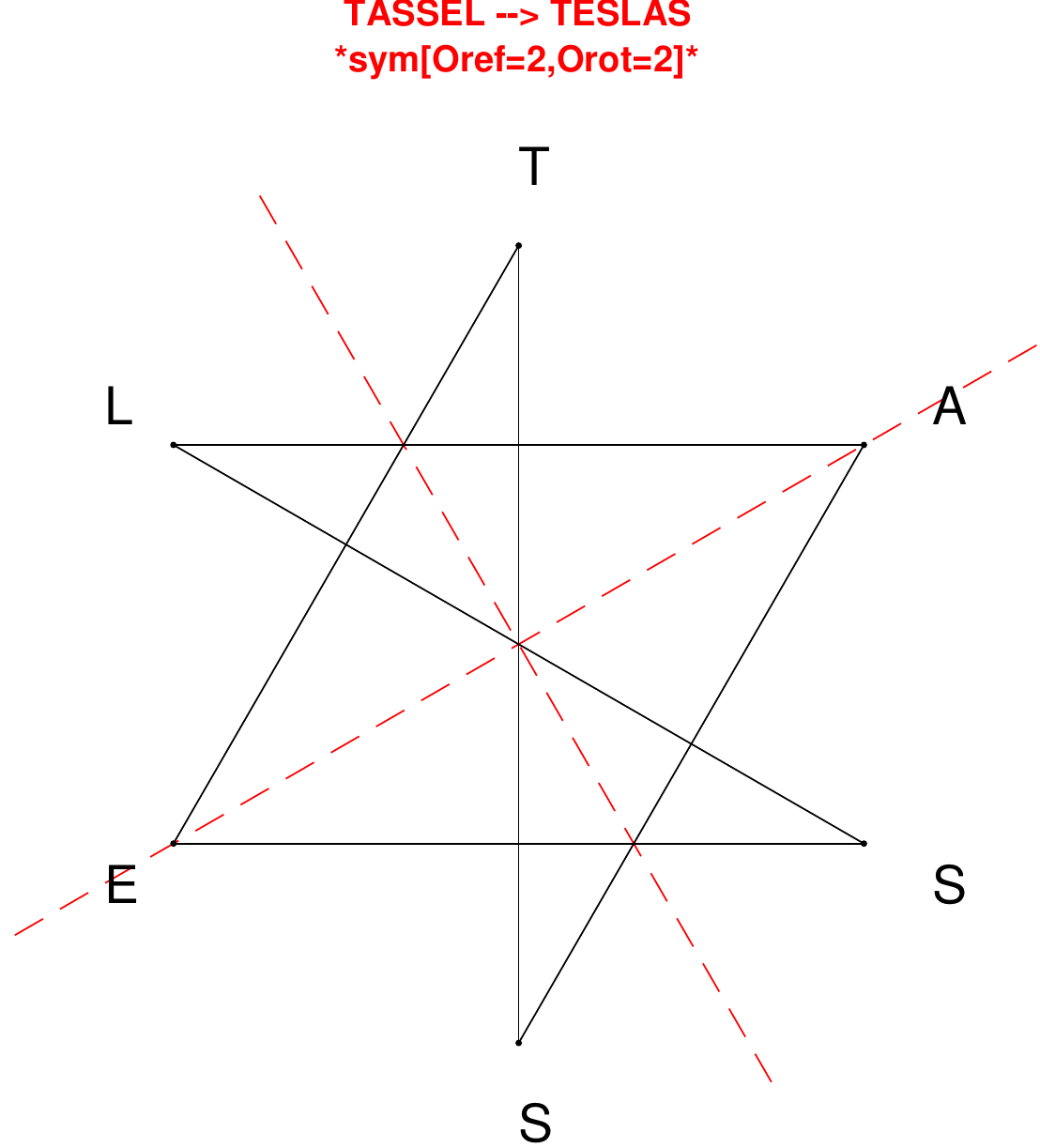}
\end{subfigure}
\hfill
\begin{subfigure}[T]{0.19\textwidth}
\centering
\includegraphics[width=\textwidth]{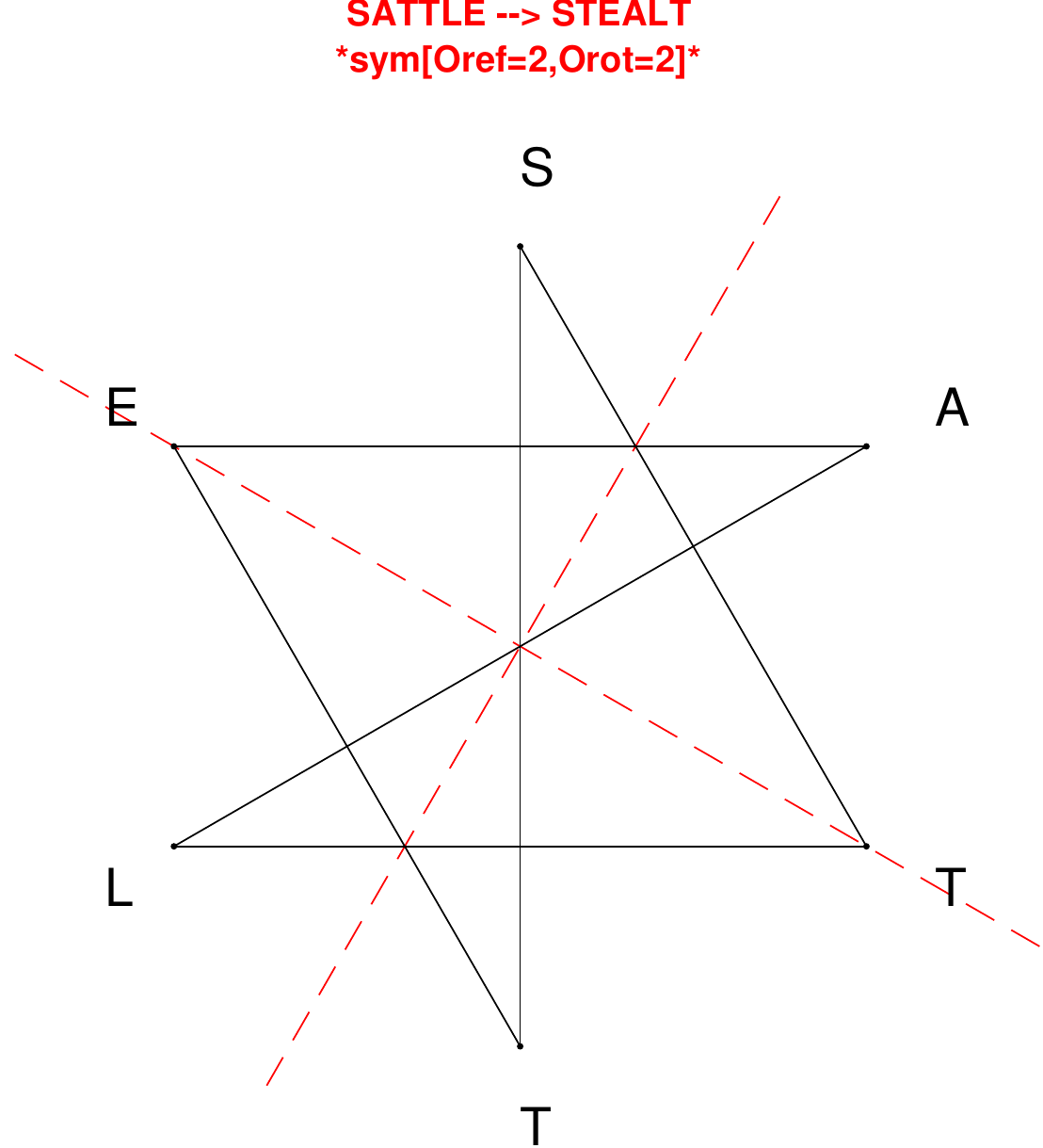}
\end{subfigure}
\end{figure}

\begin{figure}[H]
\centering
\begin{subfigure}[T]{0.19\textwidth}
\centering
\includegraphics[width=\textwidth]{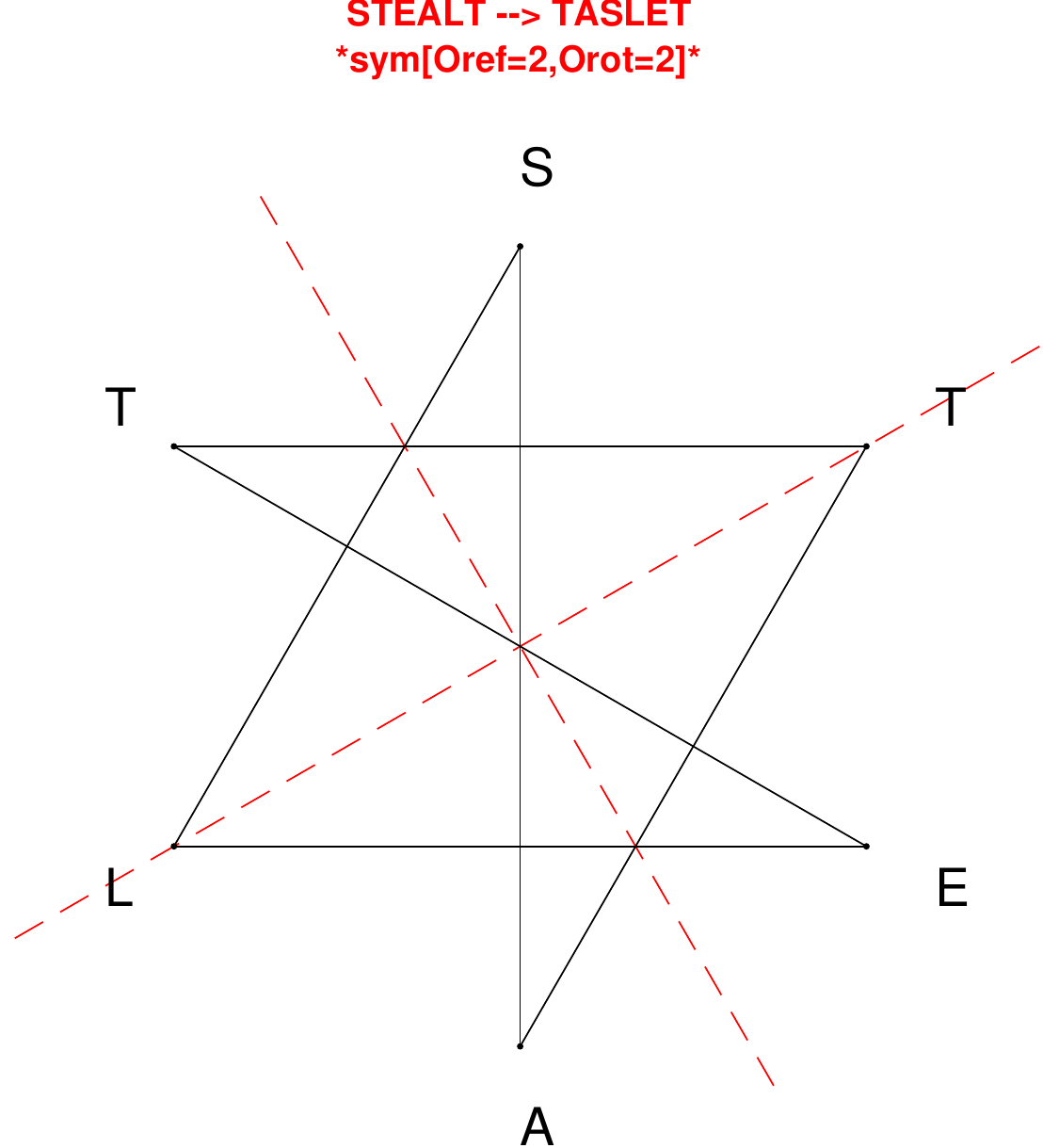}
\end{subfigure}
\hfill
\begin{subfigure}[T]{0.19\textwidth}
\centering
\includegraphics[width=\textwidth]{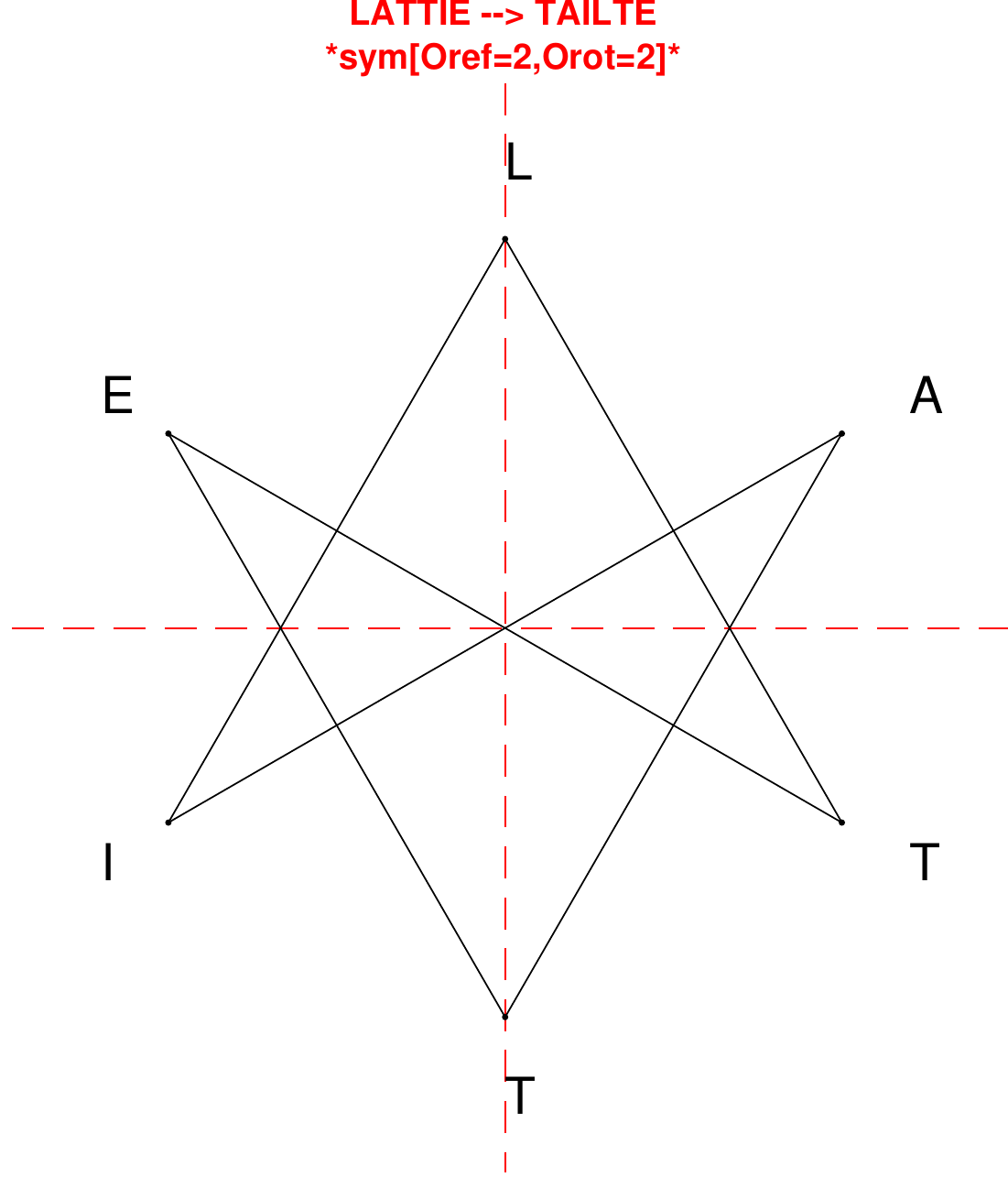}
\end{subfigure}
\hfill
\begin{subfigure}[T]{0.19\textwidth}
\centering
\includegraphics[width=\textwidth]{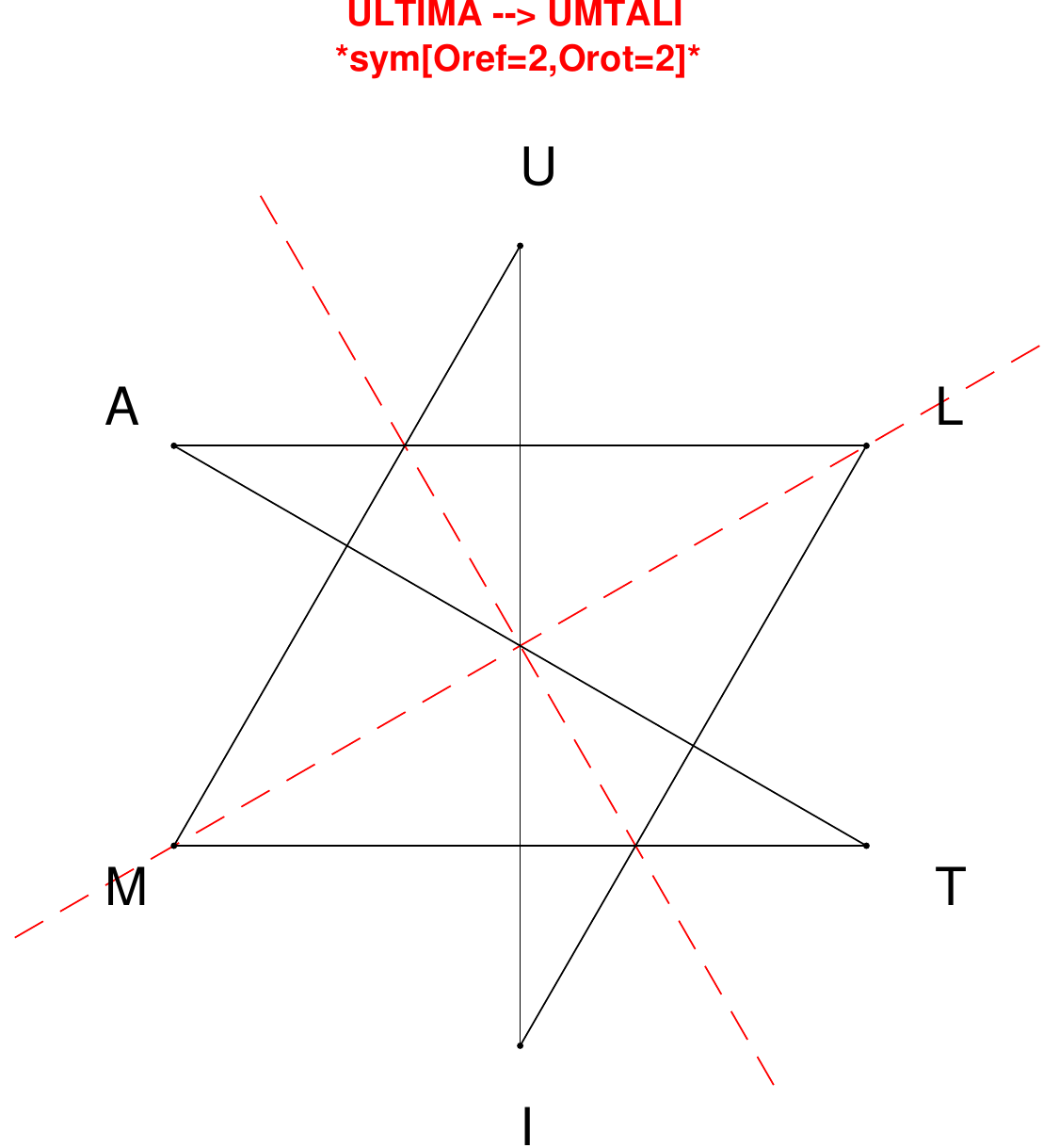}
\end{subfigure}
\hfill
\begin{subfigure}[T]{0.19\textwidth}
\centering
\includegraphics[width=\textwidth]{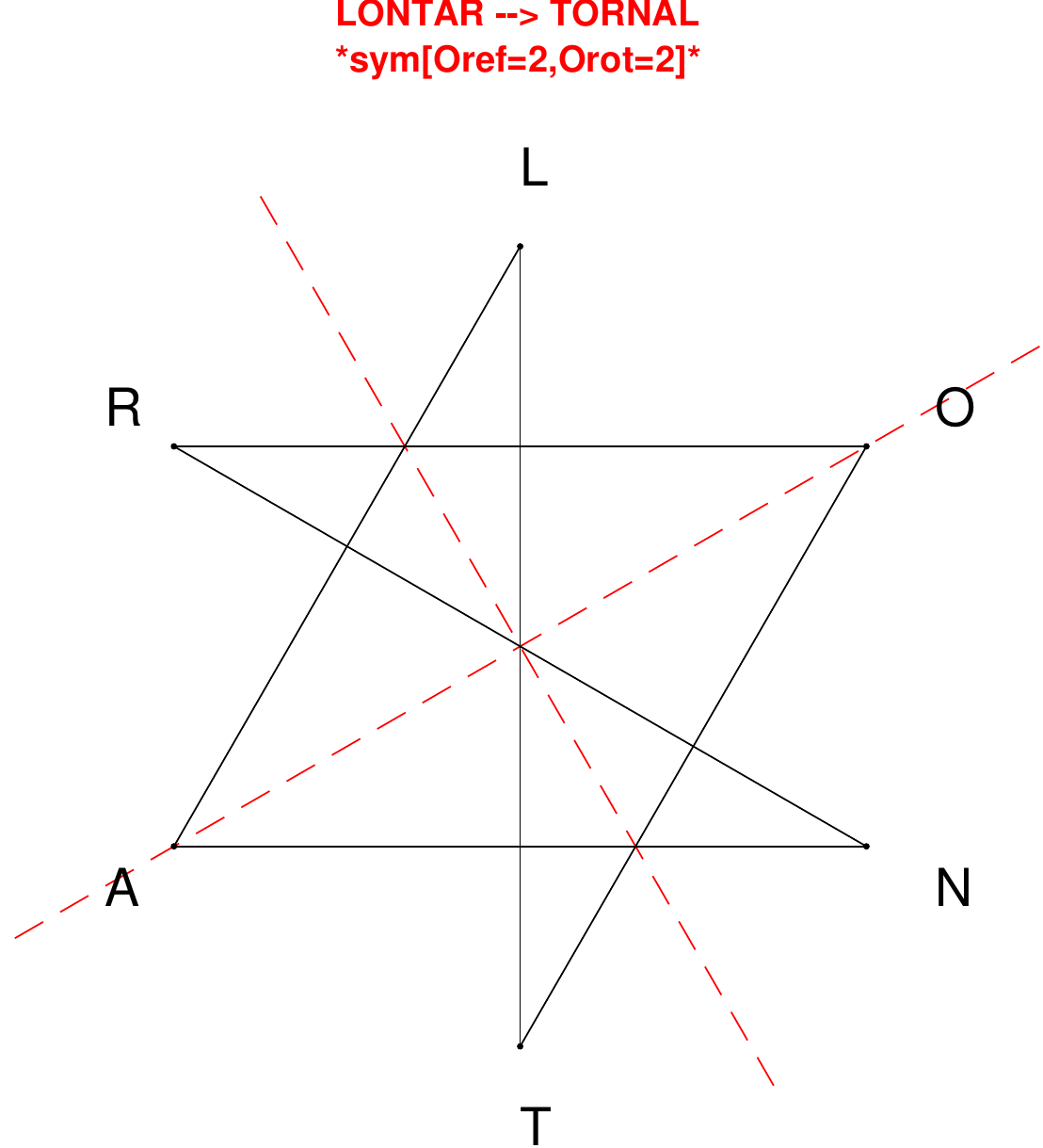}
\end{subfigure}
\hfill
\begin{subfigure}[T]{0.19\textwidth}
\centering
\includegraphics[width=\textwidth]{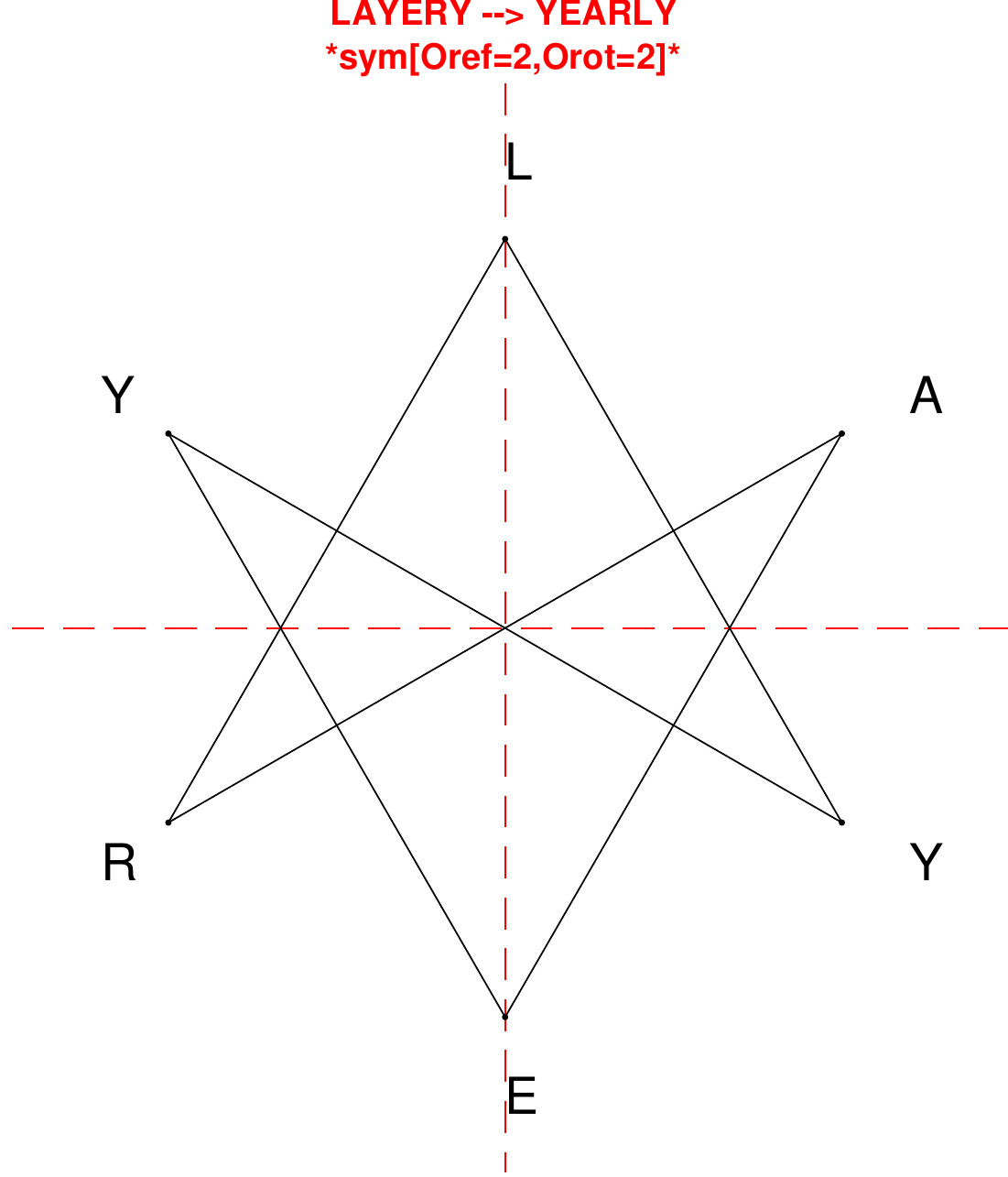}
\end{subfigure}
\end{figure}

\begin{figure}[H]
\centering
\begin{subfigure}[T]{0.19\textwidth}
\centering
\includegraphics[width=\textwidth]{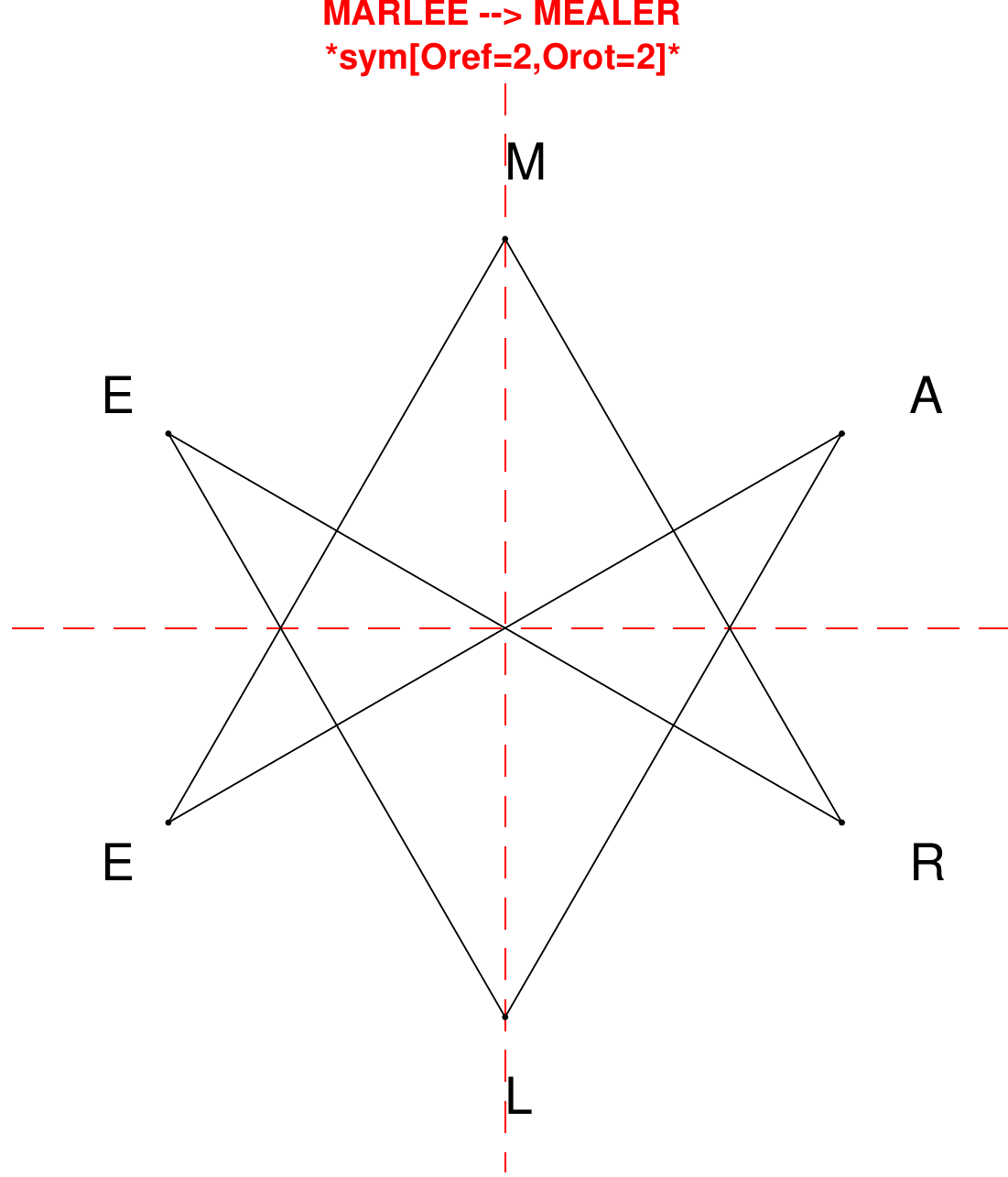}
\end{subfigure}
\hfill
\begin{subfigure}[T]{0.19\textwidth}
\centering
\includegraphics[width=\textwidth]{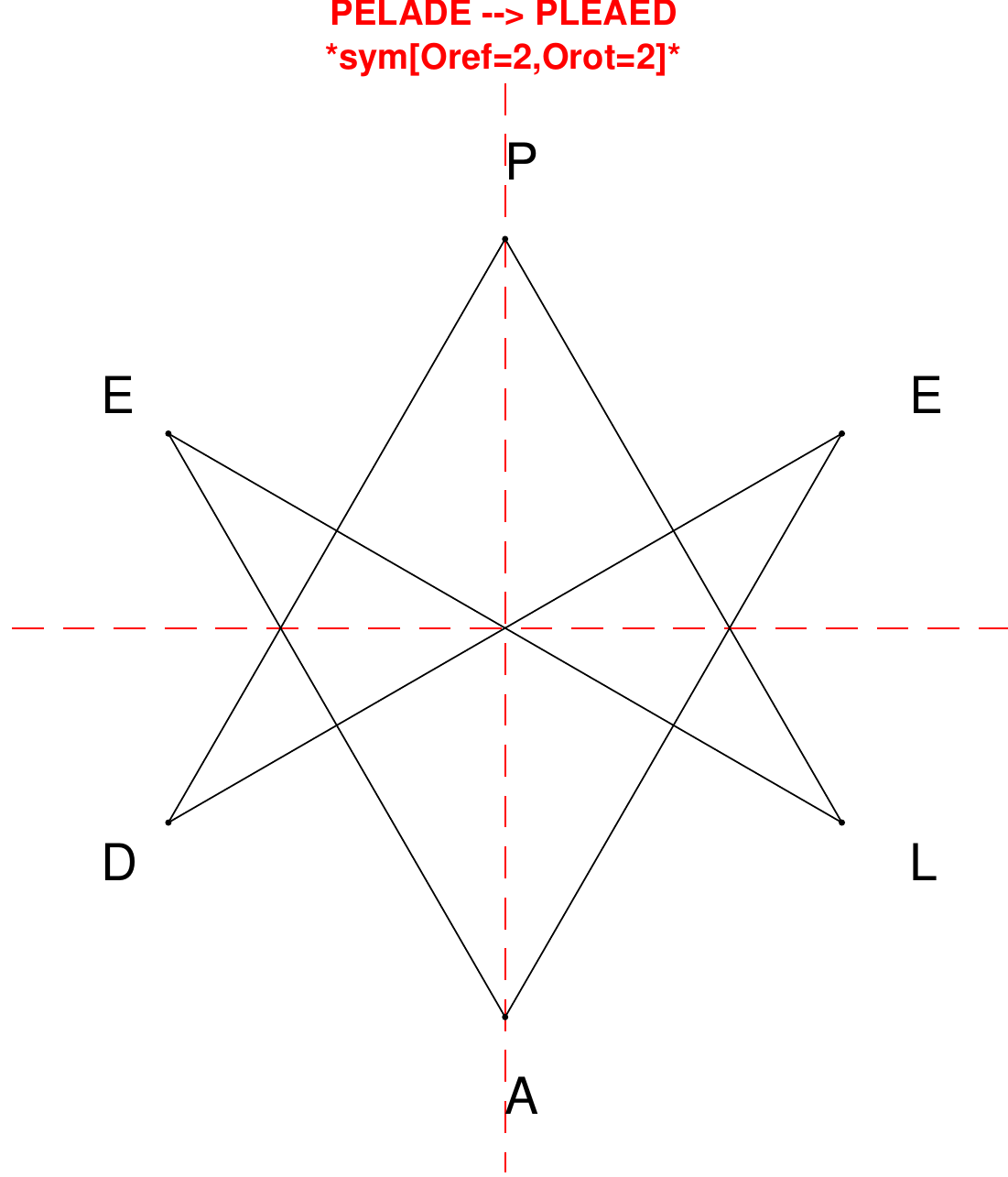}
\end{subfigure}
\hfill
\begin{subfigure}[T]{0.19\textwidth}
\centering
\includegraphics[width=\textwidth]{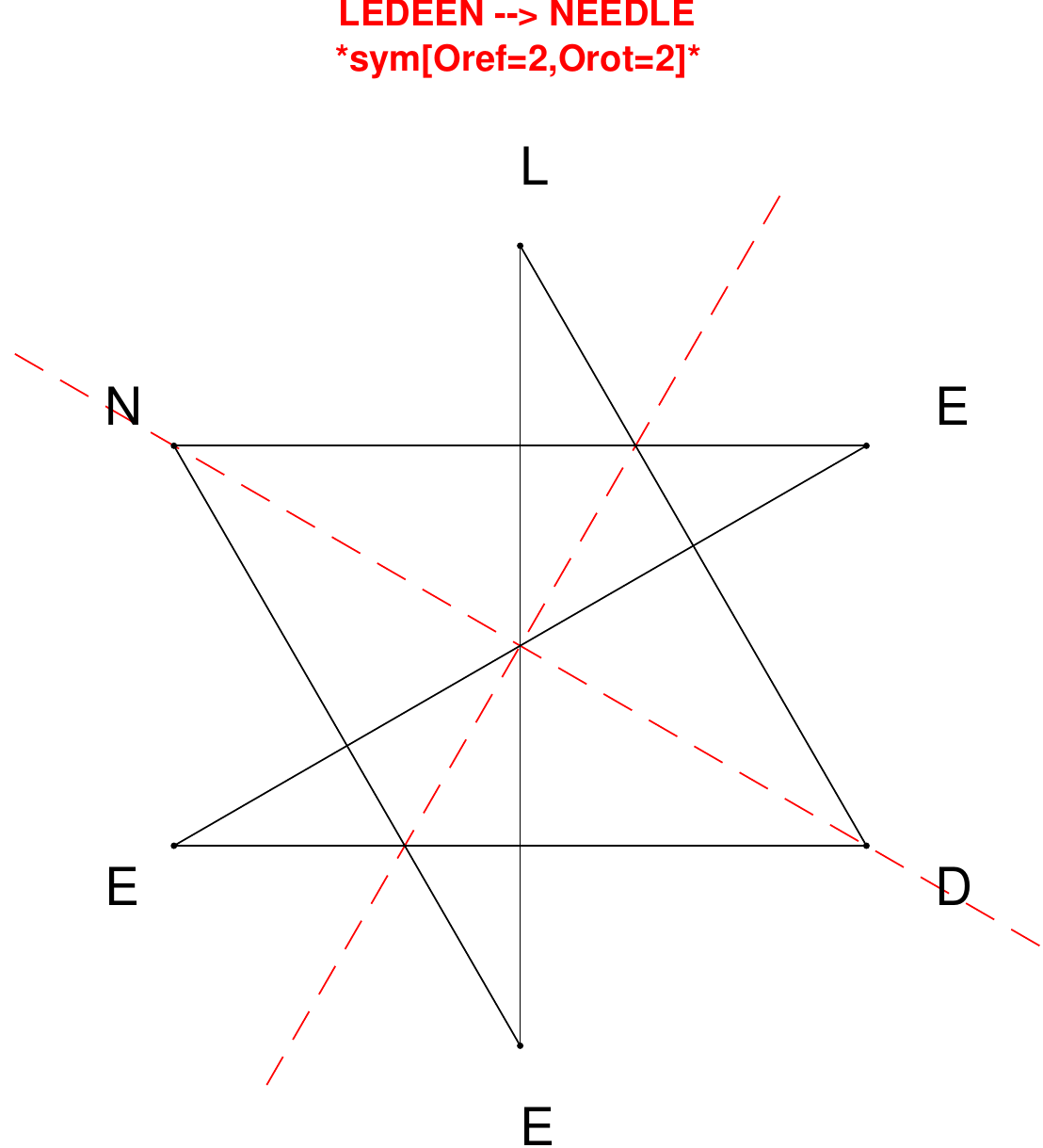}
\end{subfigure}
\hfill
\begin{subfigure}[T]{0.19\textwidth}
\centering
\includegraphics[width=\textwidth]{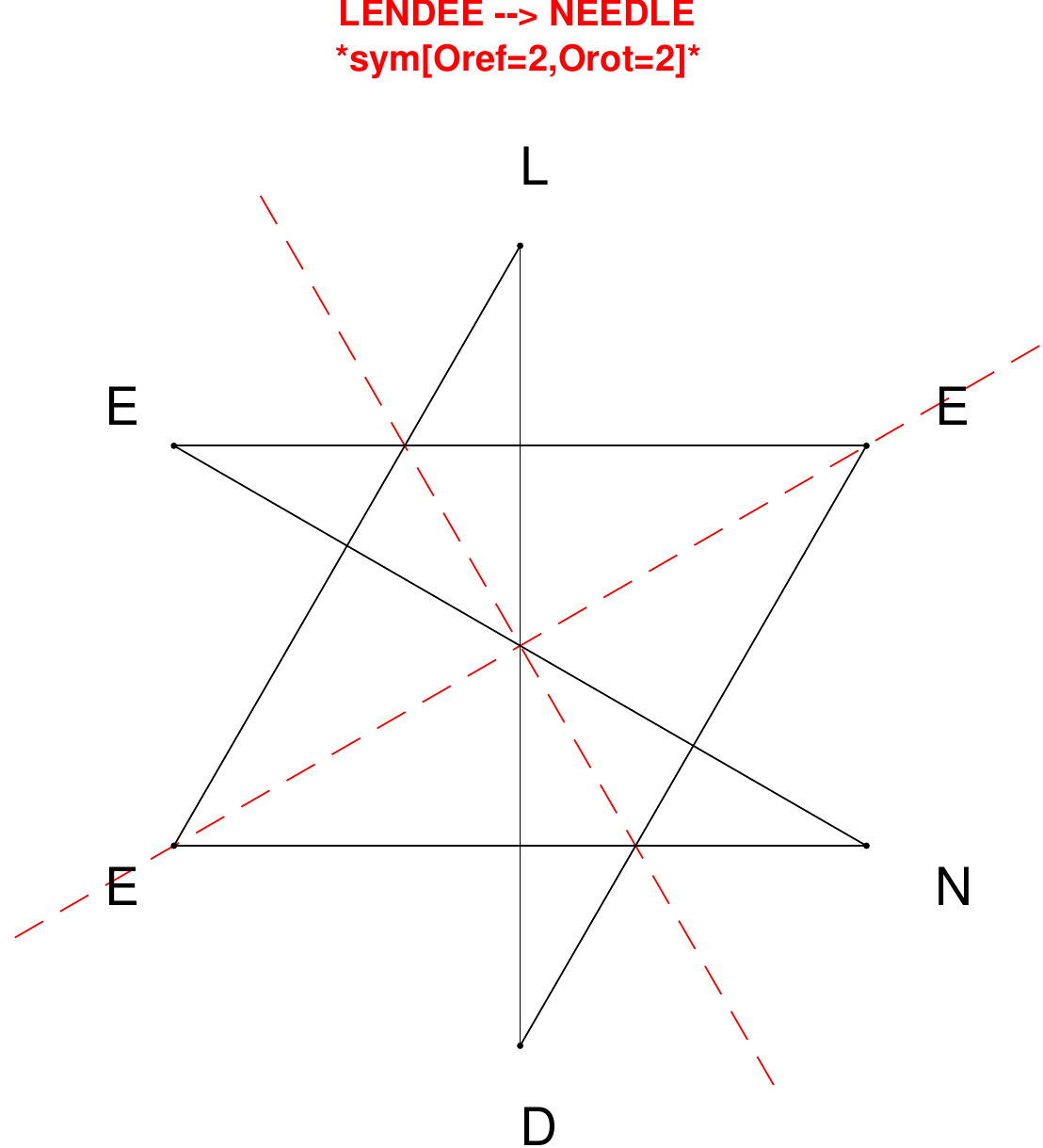}
\end{subfigure}
\hfill
\begin{subfigure}[T]{0.19\textwidth}
\centering
\includegraphics[width=\textwidth]{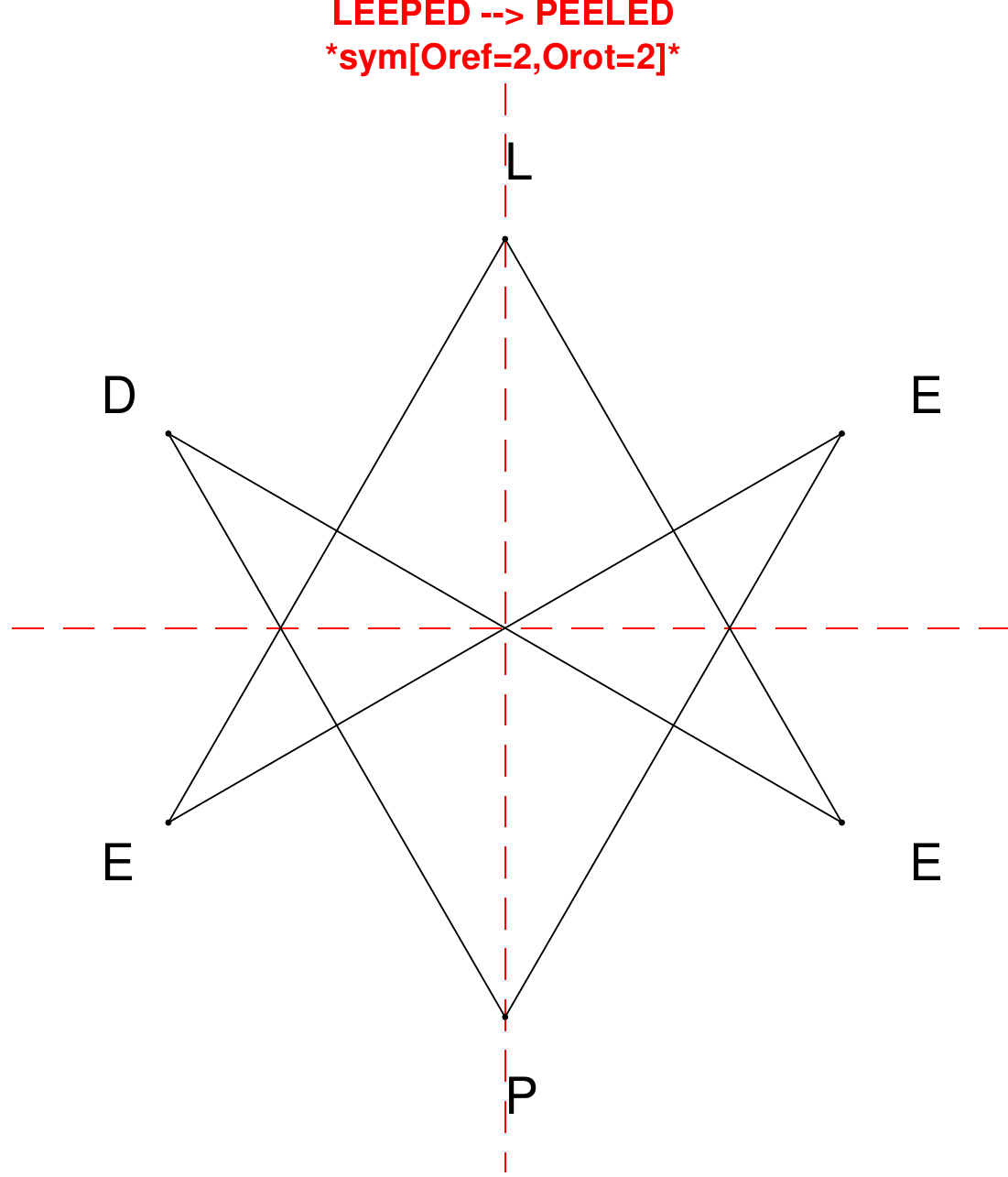}
\end{subfigure}
\end{figure}

\begin{figure}[H]
\centering
\begin{subfigure}[T]{0.19\textwidth}
\centering
\includegraphics[width=\textwidth]{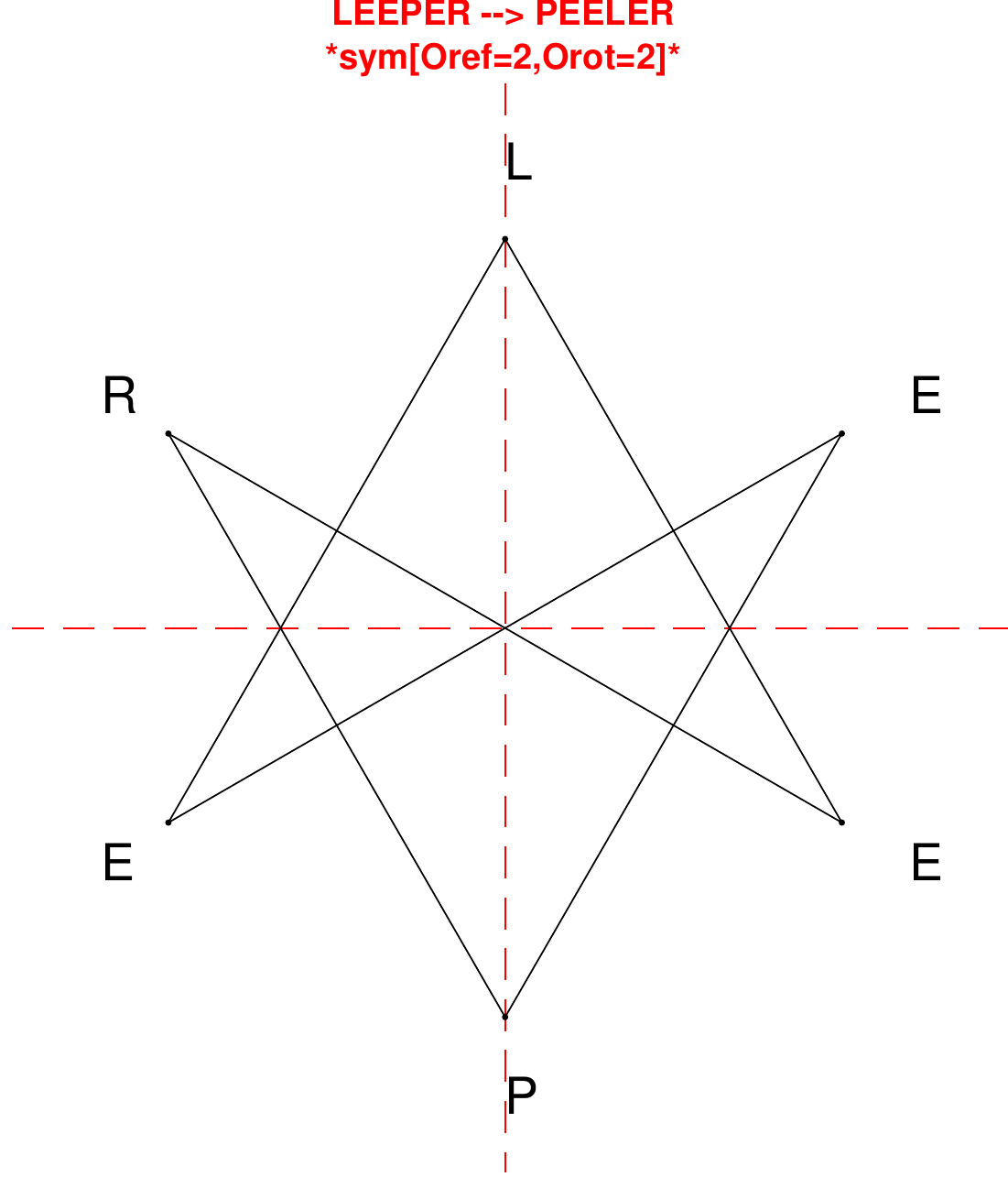}
\end{subfigure}
\hfill
\begin{subfigure}[T]{0.19\textwidth}
\centering
\includegraphics[width=\textwidth]{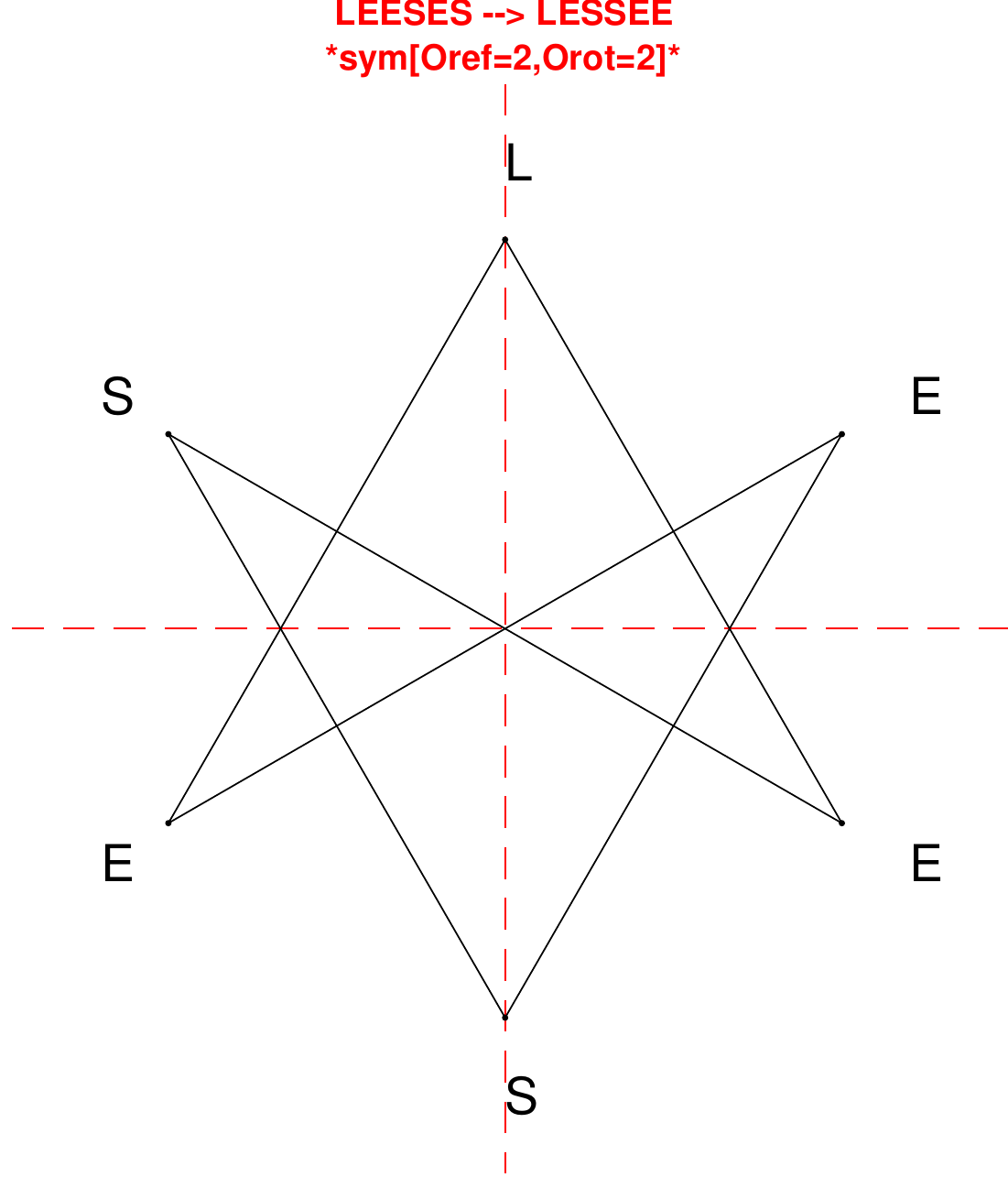}
\end{subfigure}
\hfill
\begin{subfigure}[T]{0.19\textwidth}
\centering
\includegraphics[width=\textwidth]{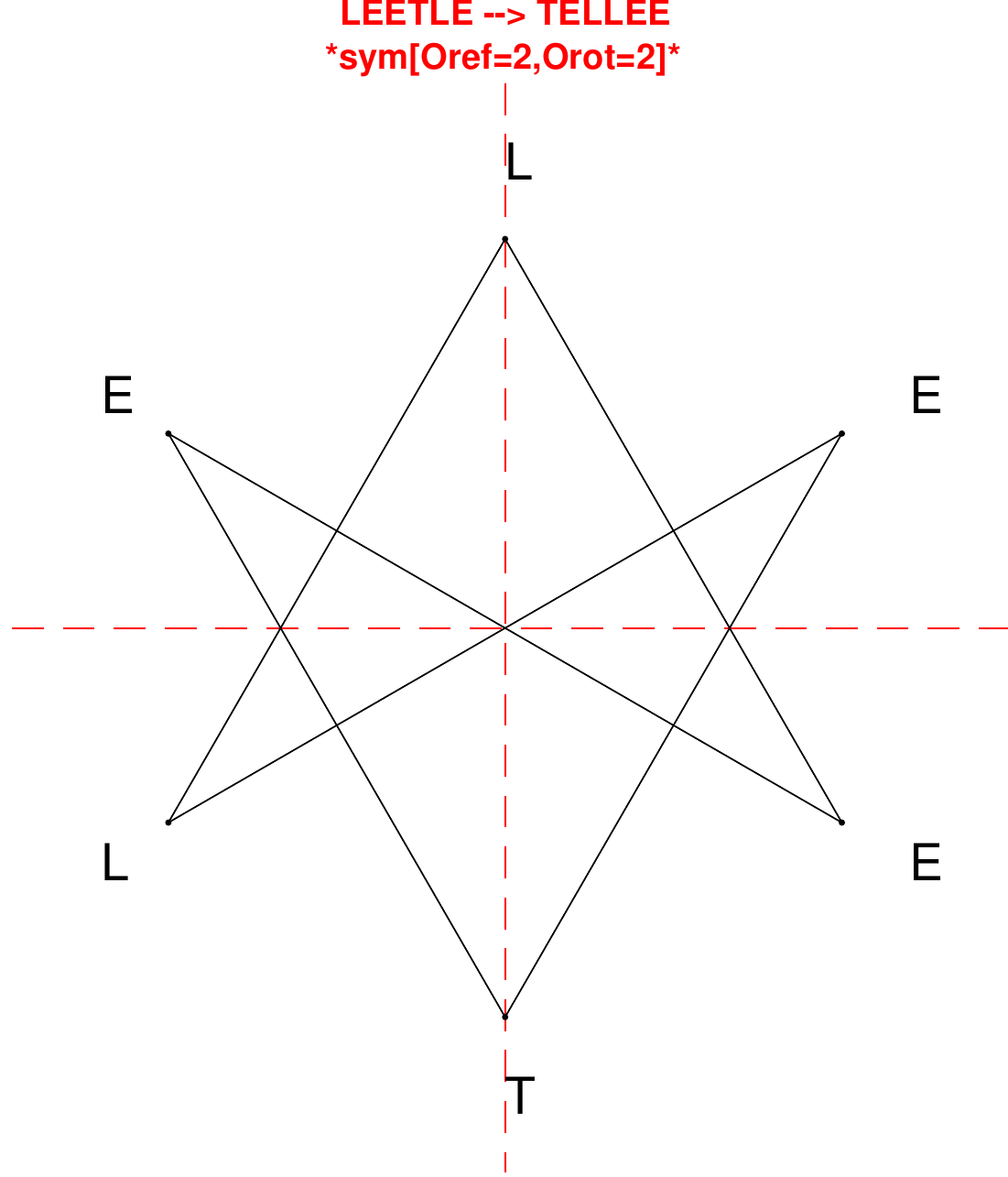}
\end{subfigure}
\hfill
\begin{subfigure}[T]{0.19\textwidth}
\centering
\includegraphics[width=\textwidth]{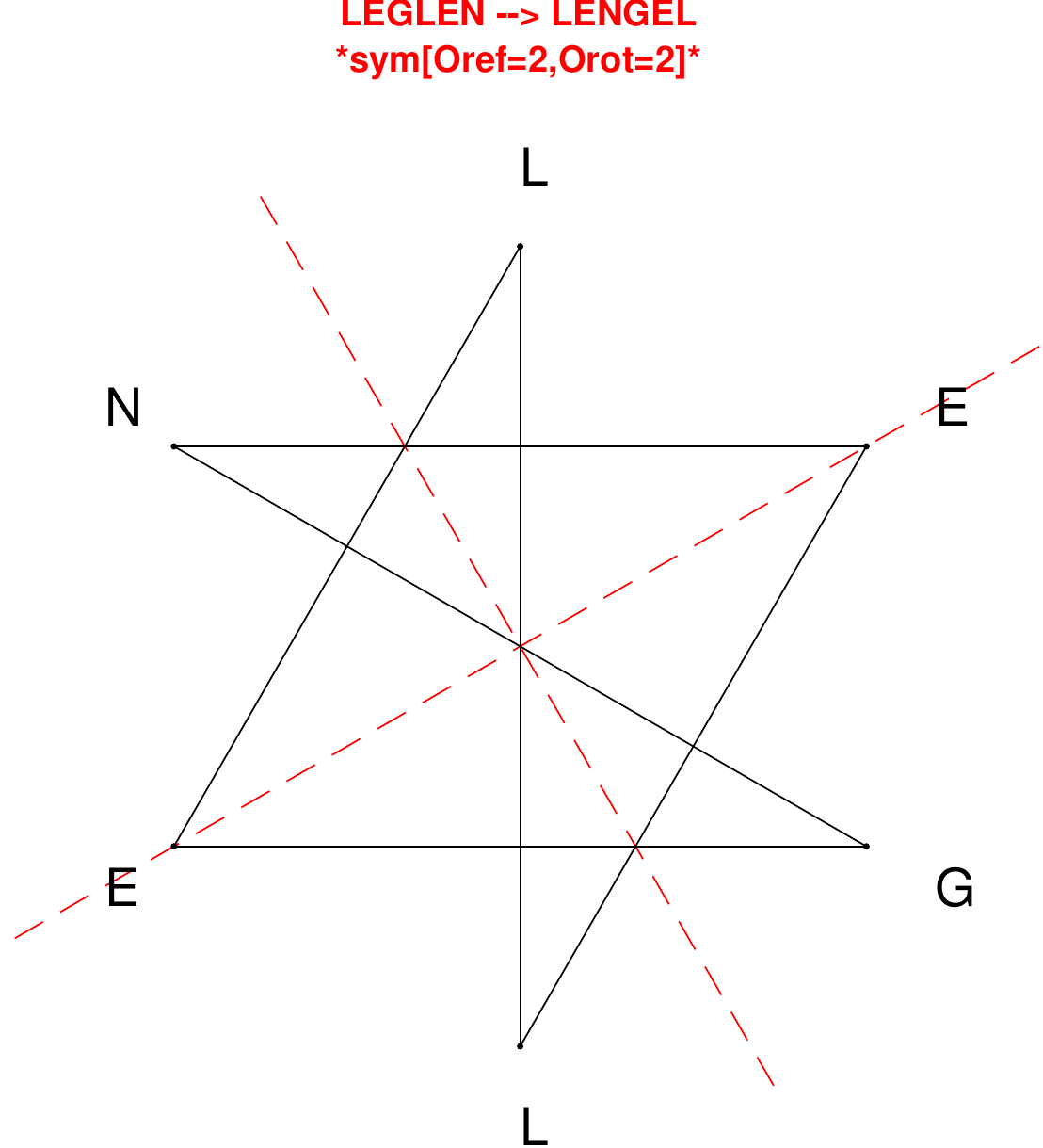}
\end{subfigure}
\hfill
\begin{subfigure}[T]{0.19\textwidth}
\centering
\includegraphics[width=\textwidth]{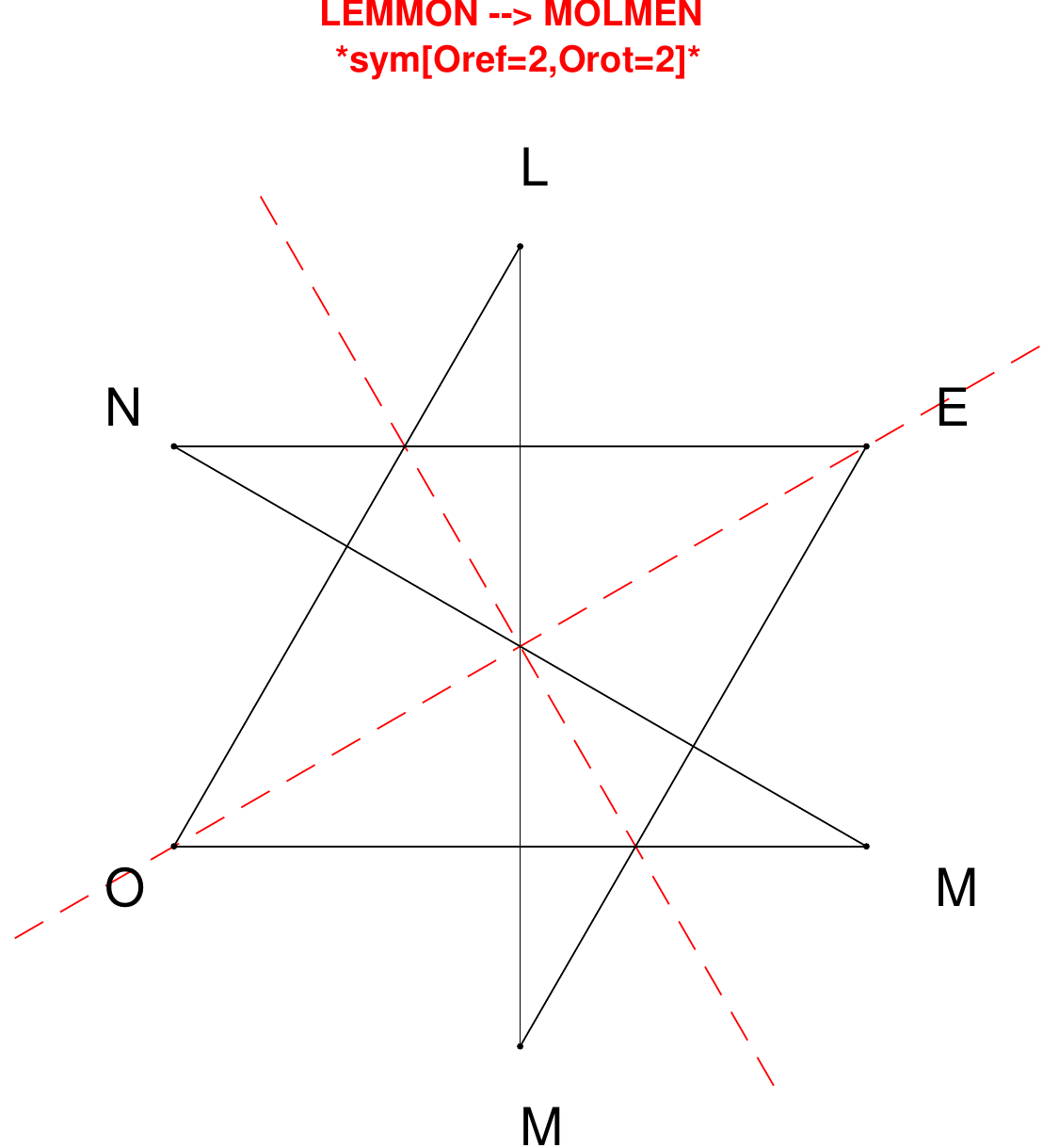}
\end{subfigure}
\end{figure}

\begin{figure}[H]
\centering
\begin{subfigure}[T]{0.19\textwidth}
\centering
\includegraphics[width=\textwidth]{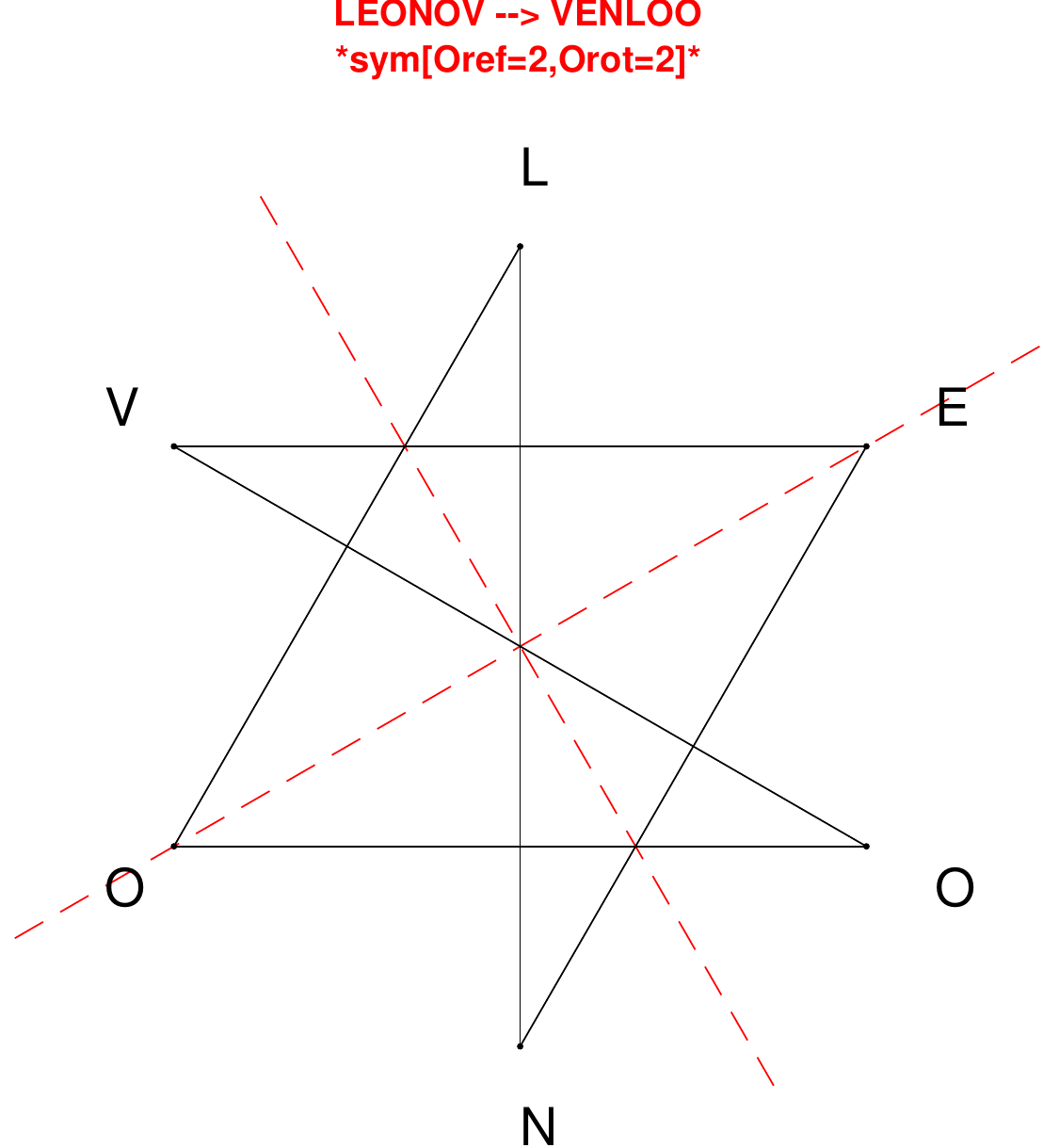}
\end{subfigure}
\hfill
\begin{subfigure}[T]{0.19\textwidth}
\centering
\includegraphics[width=\textwidth]{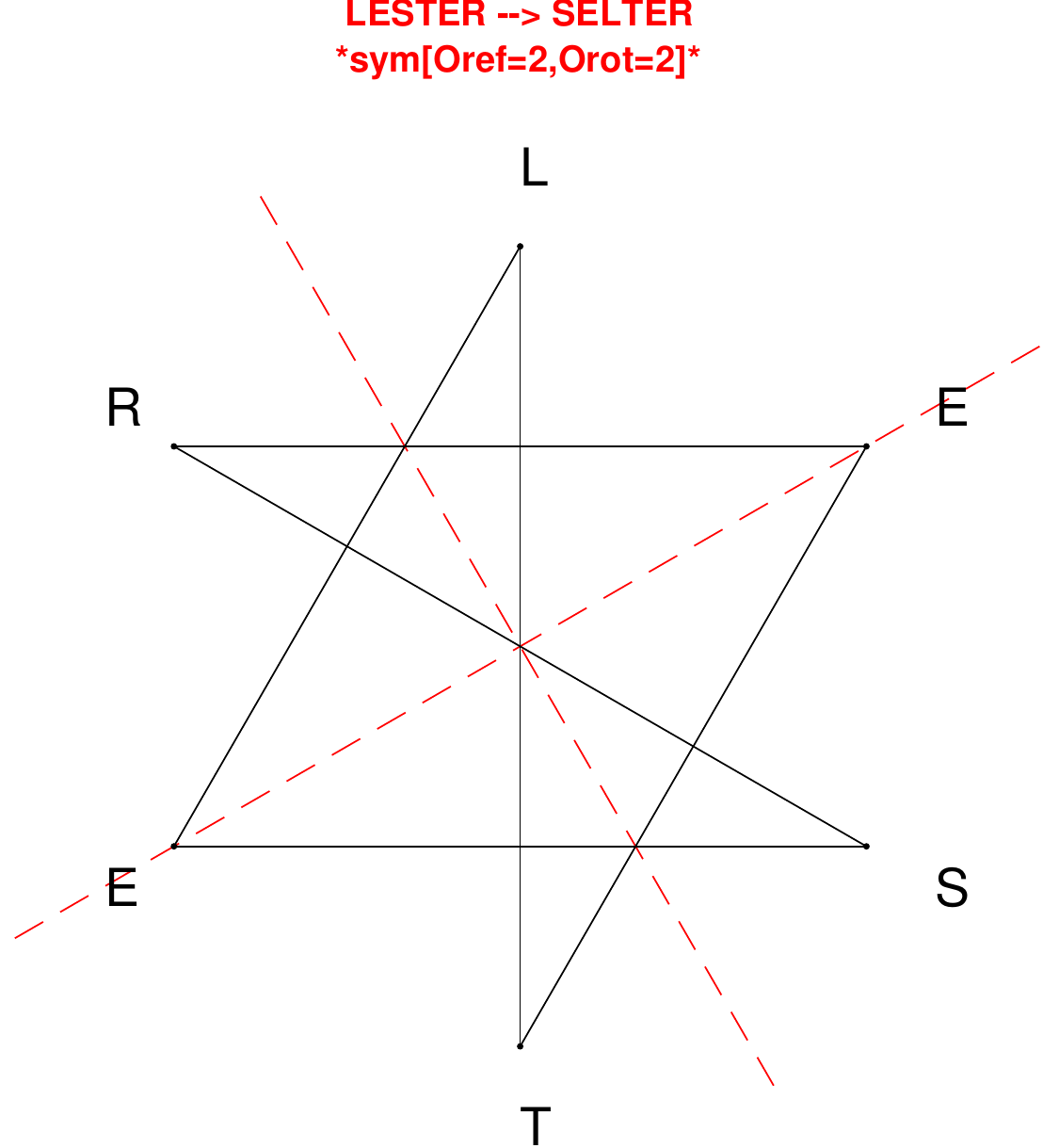}
\end{subfigure}
\hfill
\begin{subfigure}[T]{0.19\textwidth}
\centering
\includegraphics[width=\textwidth]{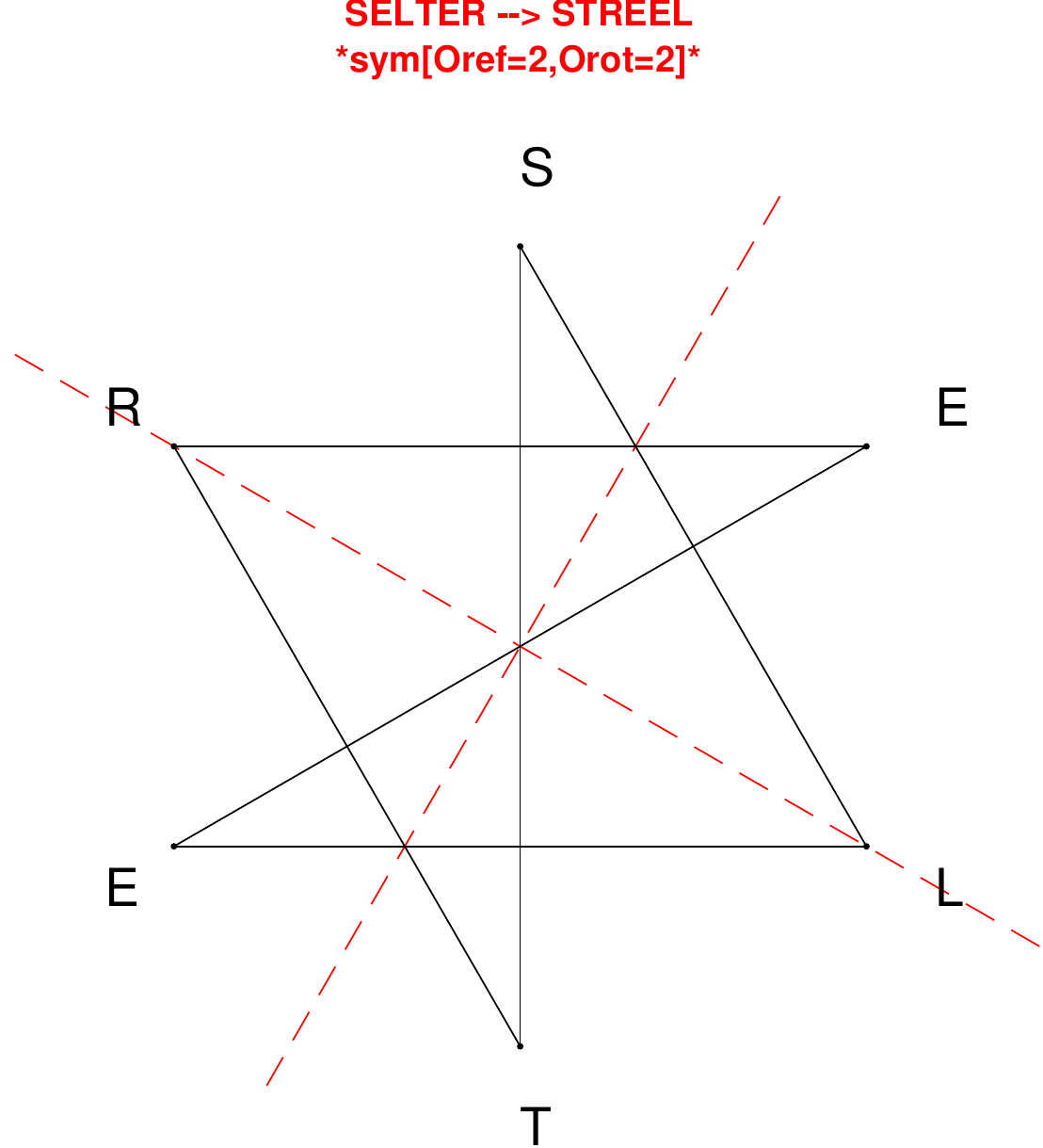}
\end{subfigure}
\hfill
\begin{subfigure}[T]{0.19\textwidth}
\centering
\includegraphics[width=\textwidth]{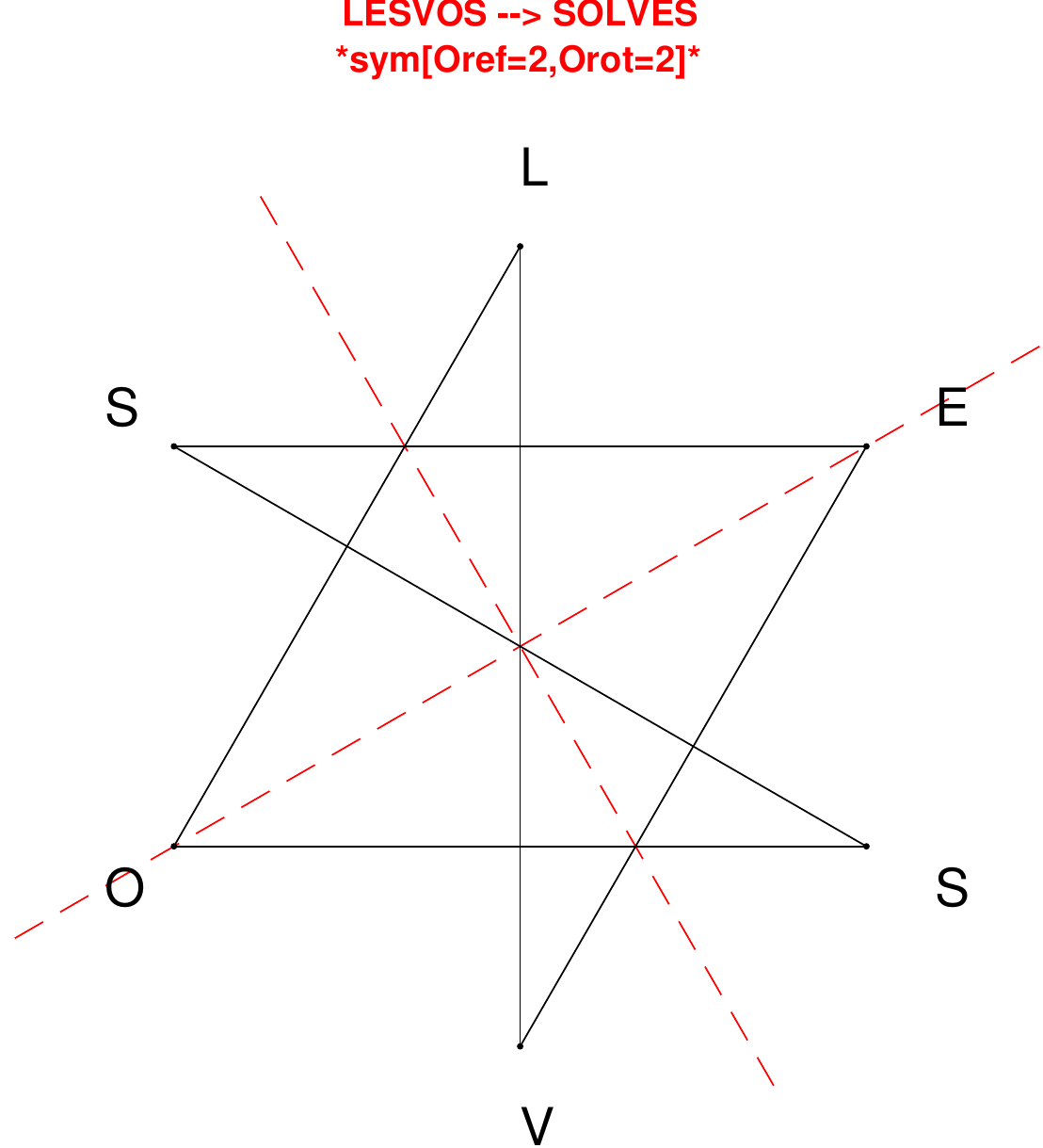}
\end{subfigure}
\hfill
\begin{subfigure}[T]{0.19\textwidth}
\centering
\includegraphics[width=\textwidth]{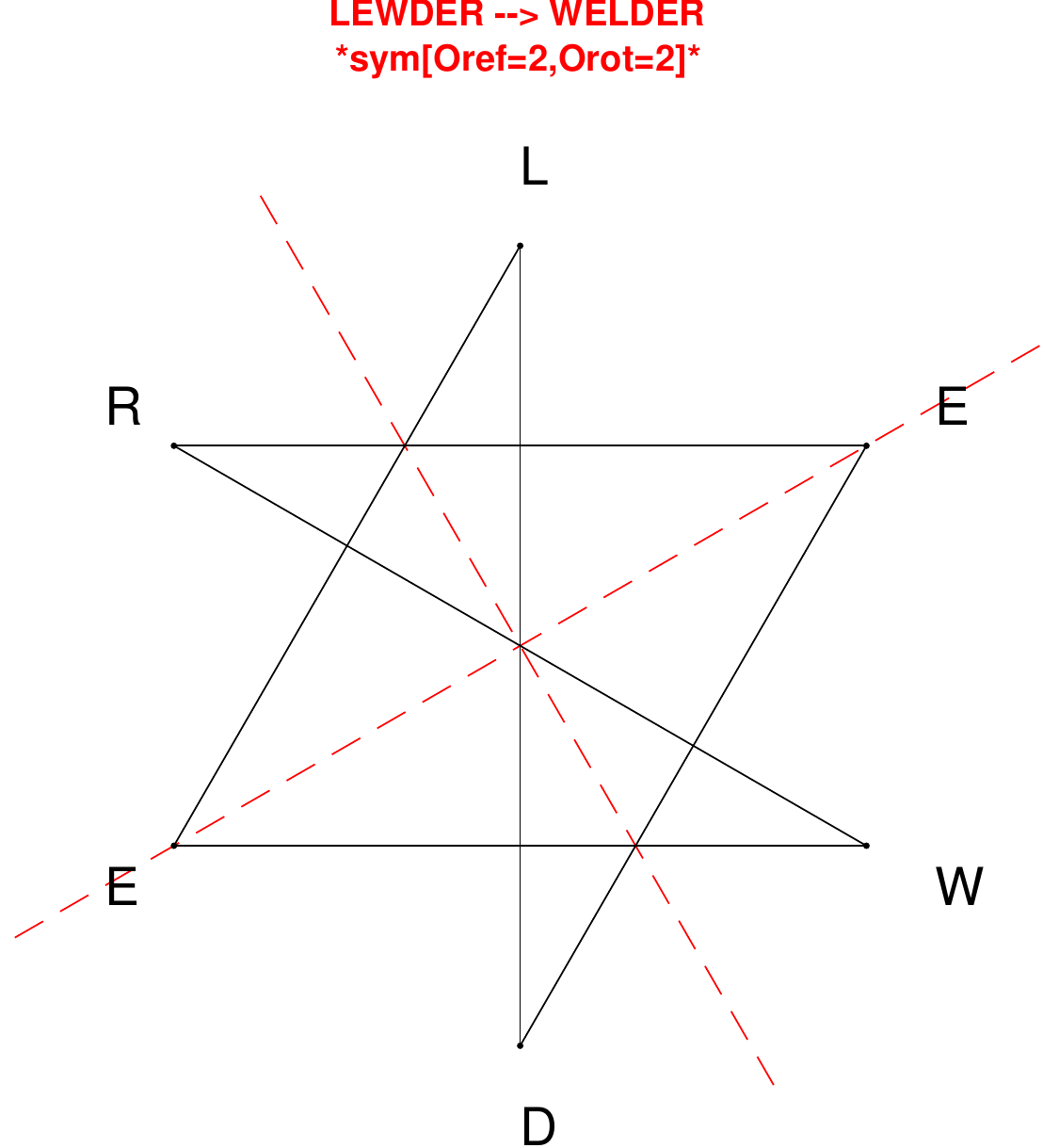}
\end{subfigure}
\end{figure}

\begin{figure}[H]
\centering
\begin{subfigure}[T]{0.19\textwidth}
\centering
\includegraphics[width=\textwidth]{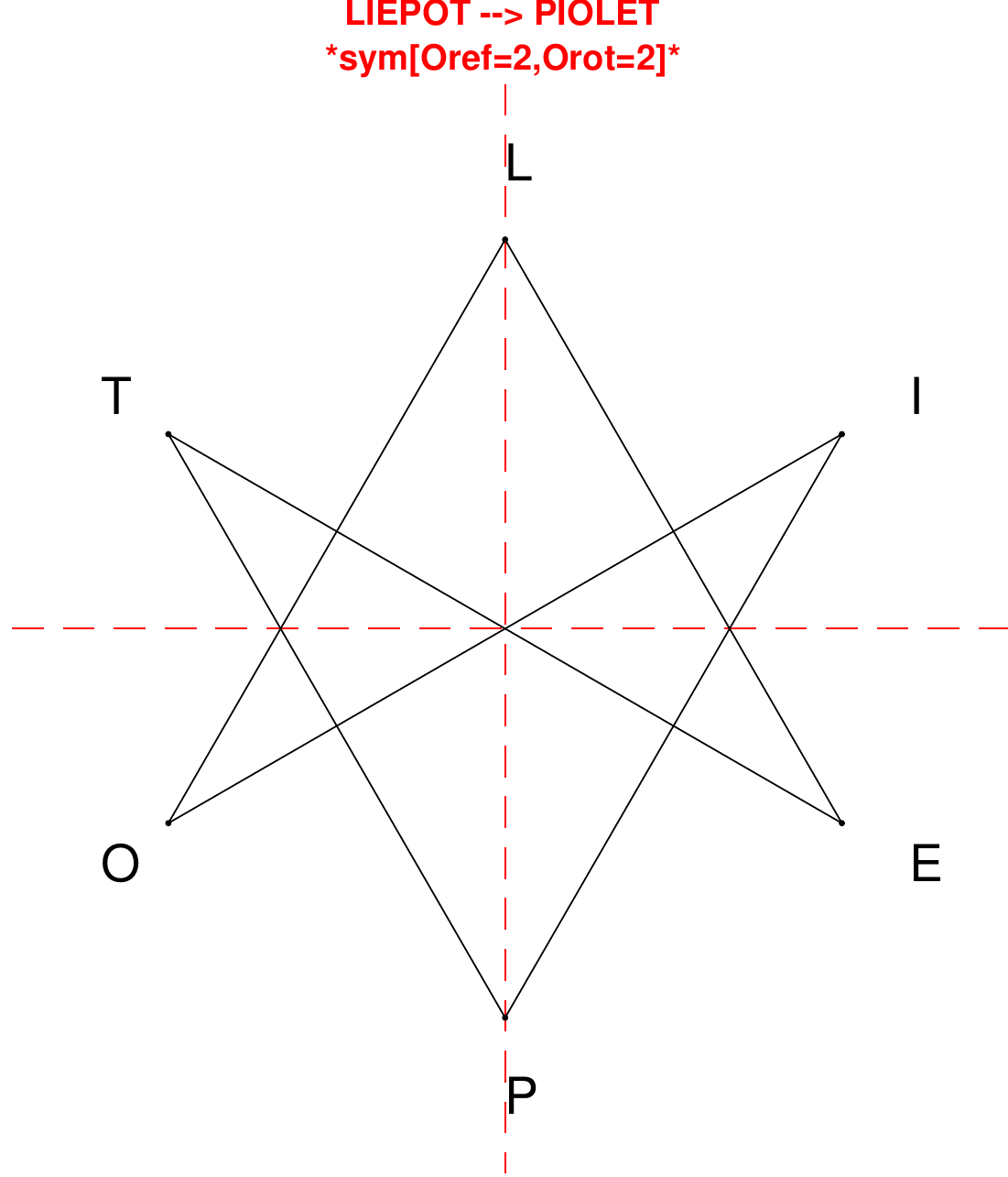}
\end{subfigure}
\hfill
\begin{subfigure}[T]{0.19\textwidth}
\centering
\includegraphics[width=\textwidth]{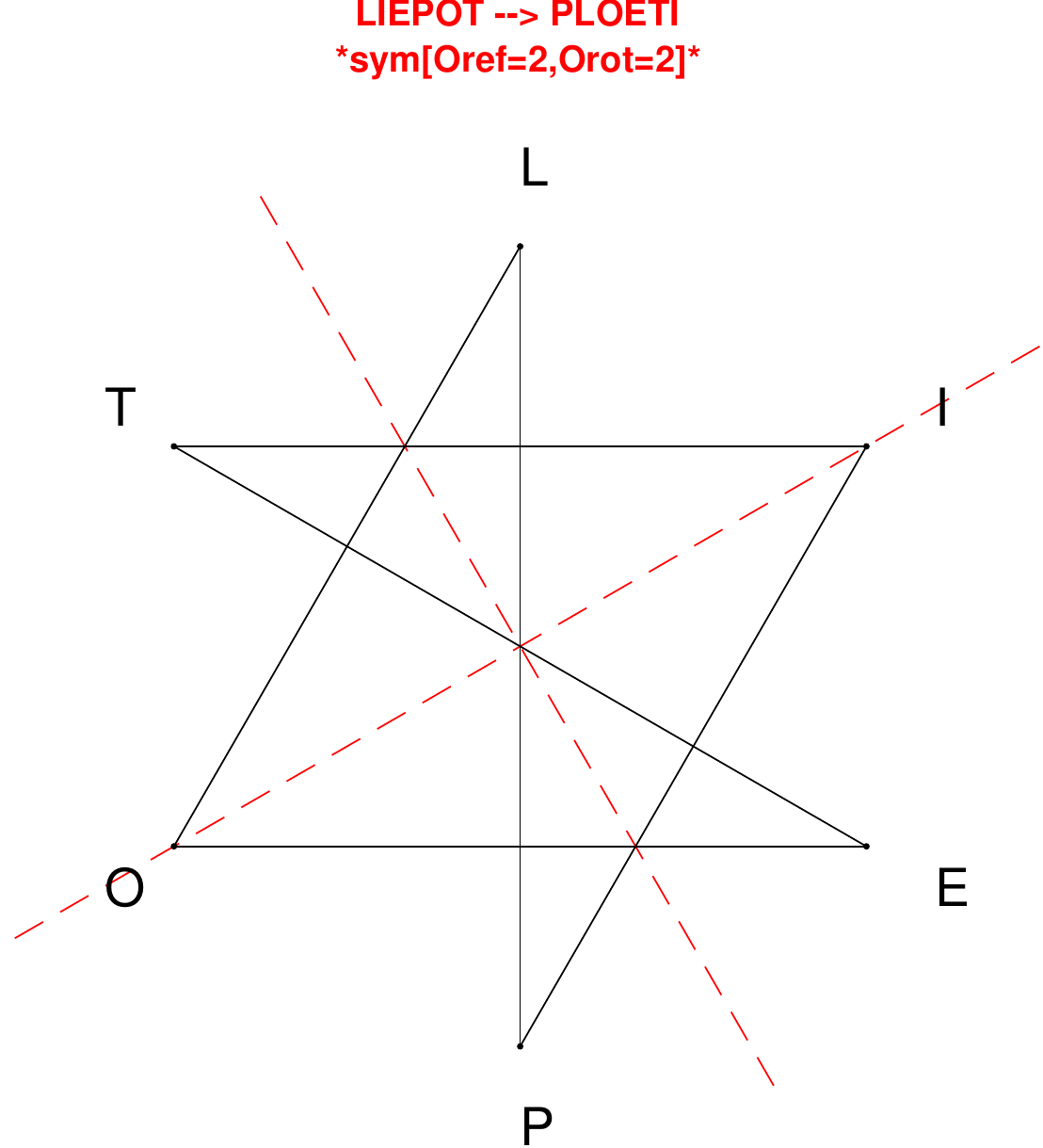}
\end{subfigure}
\hfill
\begin{subfigure}[T]{0.19\textwidth}
\centering
\includegraphics[width=\textwidth]{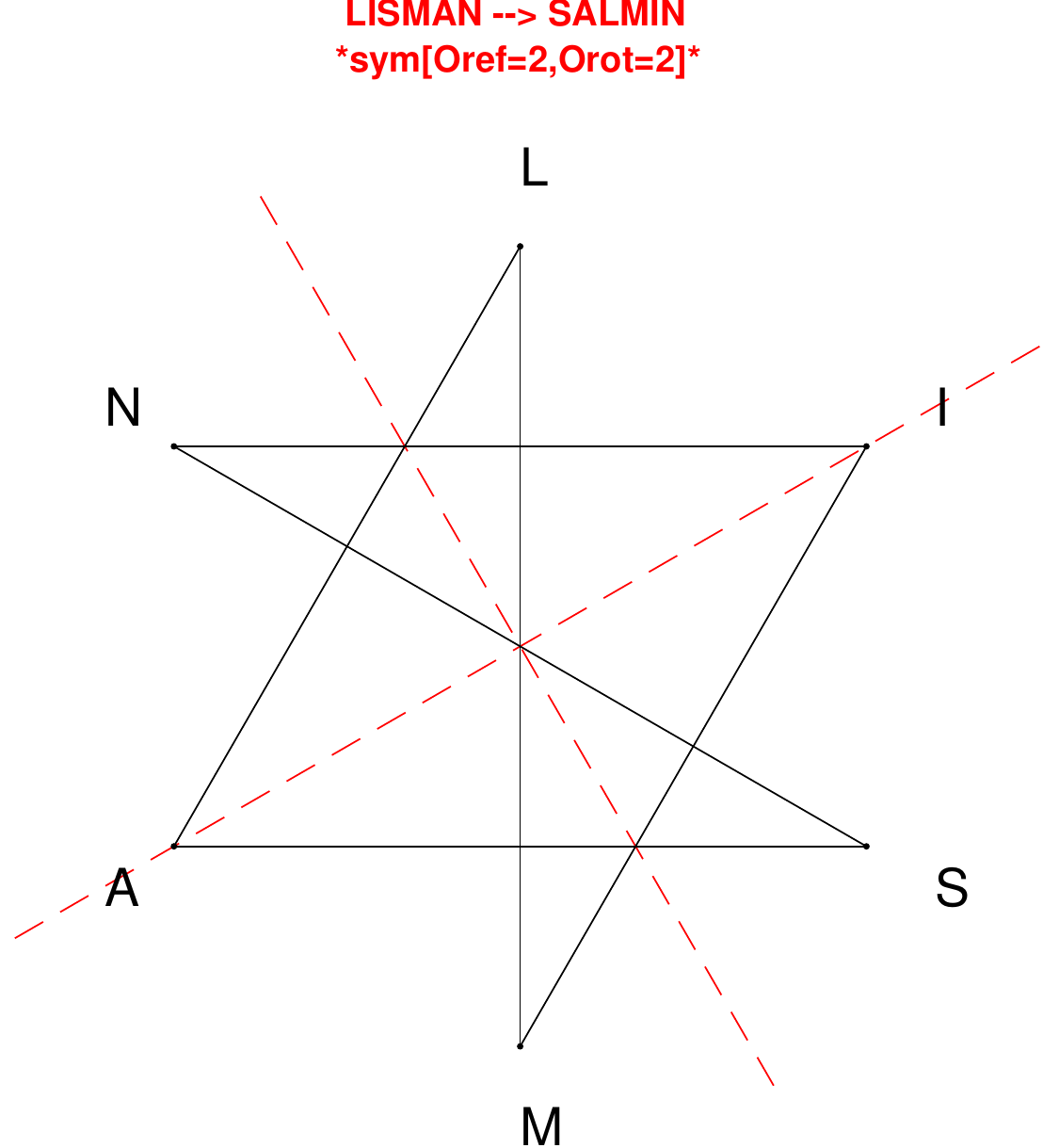}
\end{subfigure}
\hfill
\begin{subfigure}[T]{0.19\textwidth}
\centering
\includegraphics[width=\textwidth]{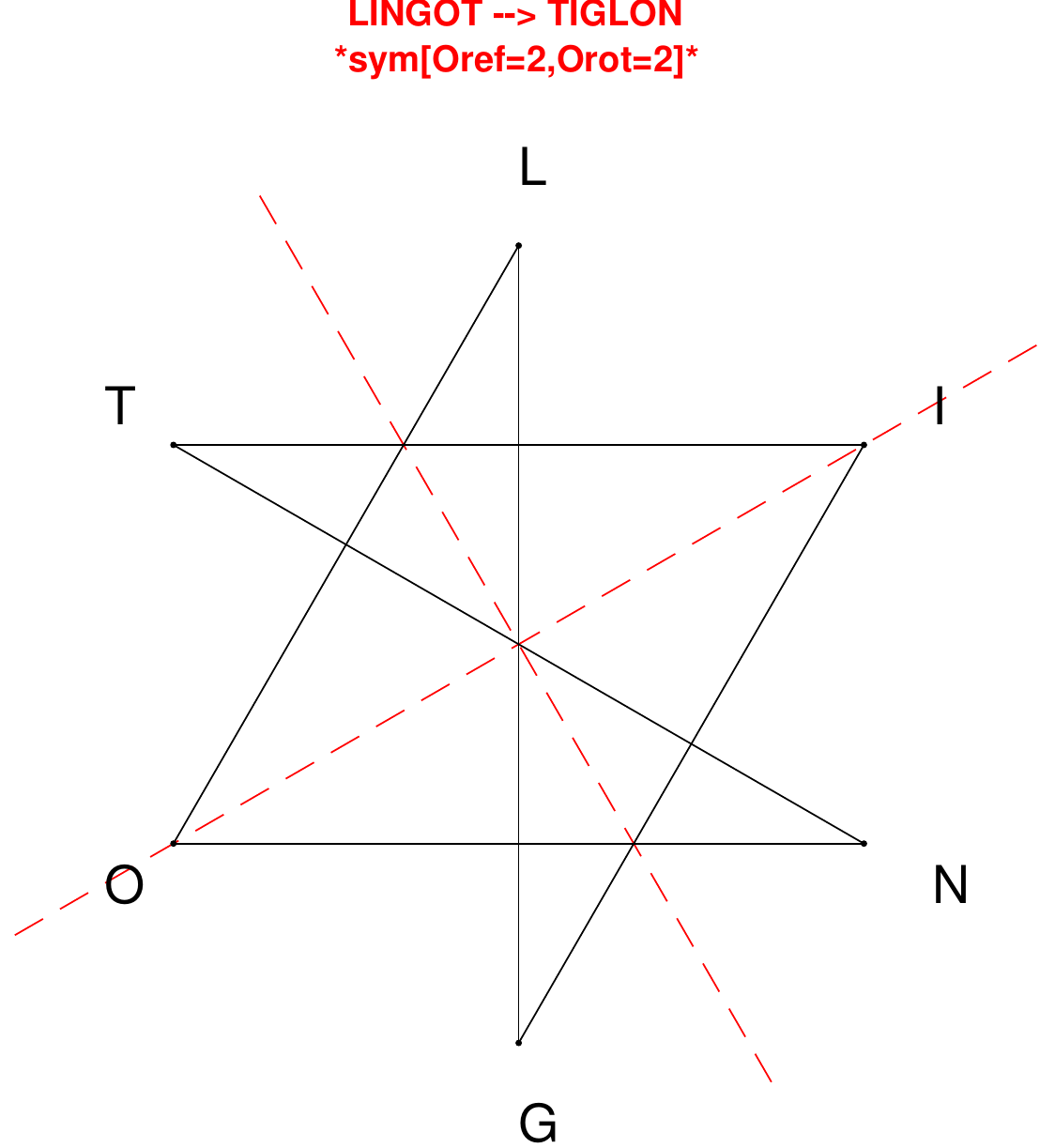}
\end{subfigure}
\hfill
\begin{subfigure}[T]{0.19\textwidth}
\centering
\includegraphics[width=\textwidth]{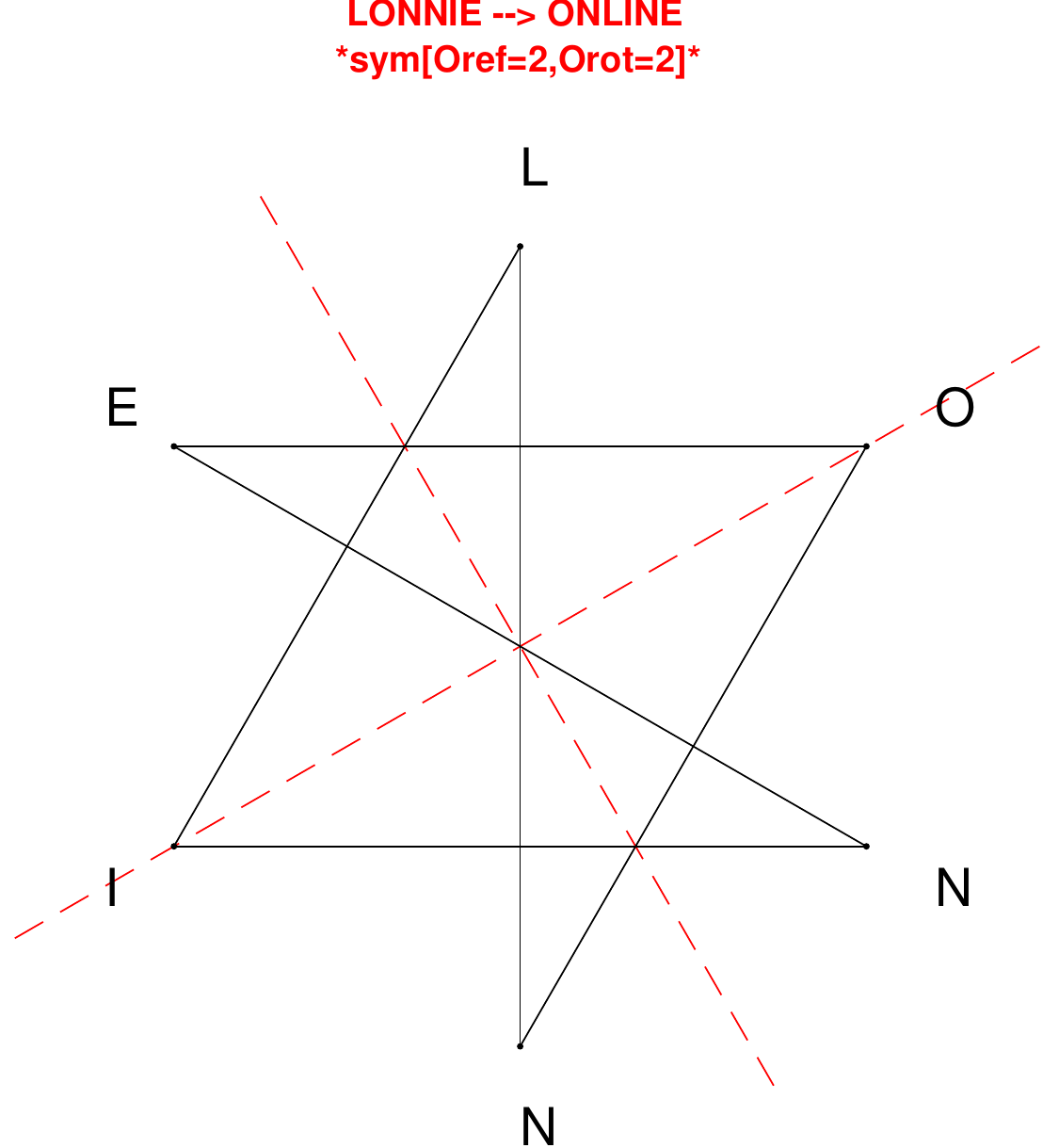}
\end{subfigure}
\end{figure}

\begin{figure}[H]
\centering
\begin{subfigure}[T]{0.19\textwidth}
\centering
\includegraphics[width=\textwidth]{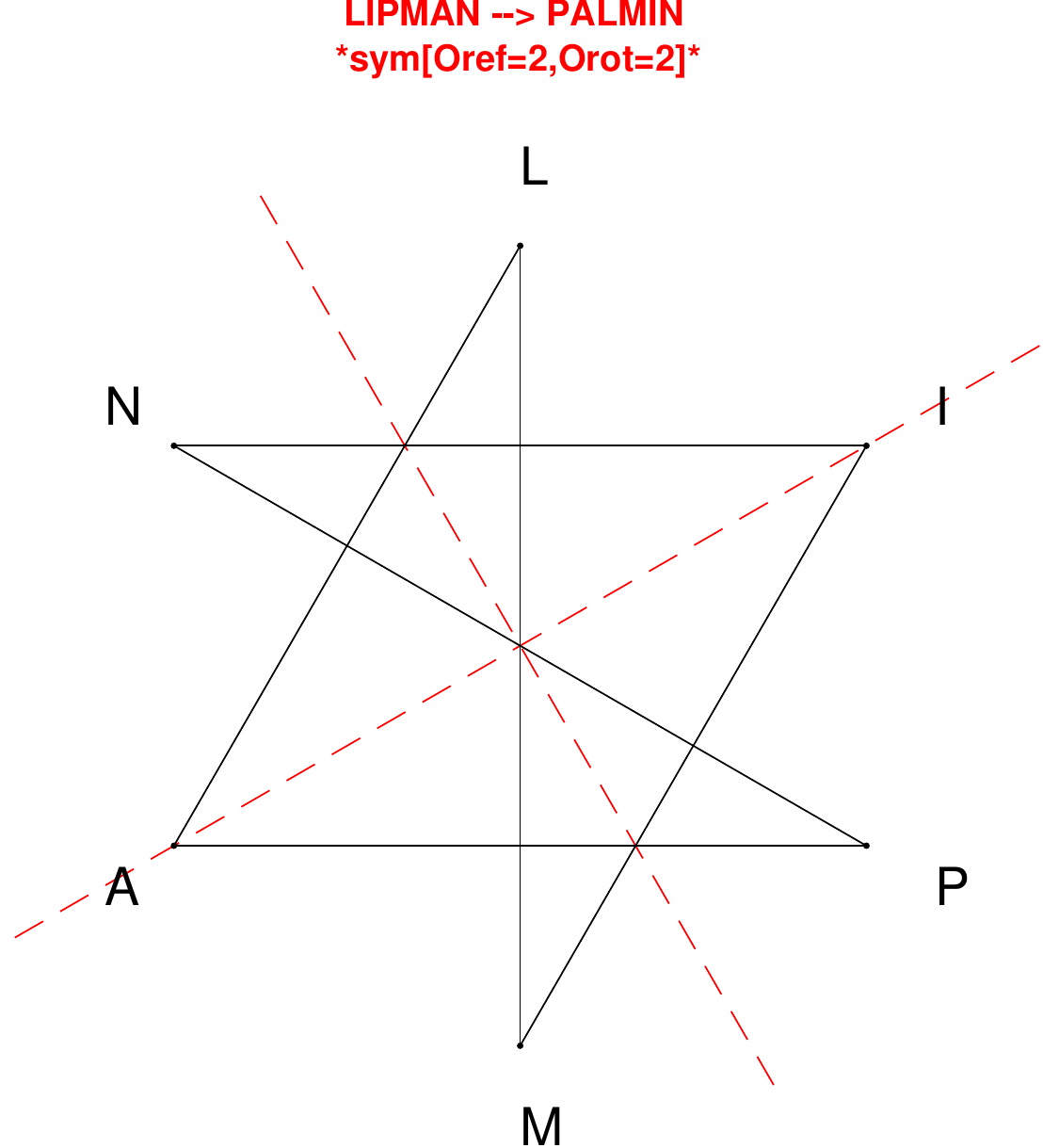}
\end{subfigure}
\hfill
\begin{subfigure}[T]{0.19\textwidth}
\centering
\includegraphics[width=\textwidth]{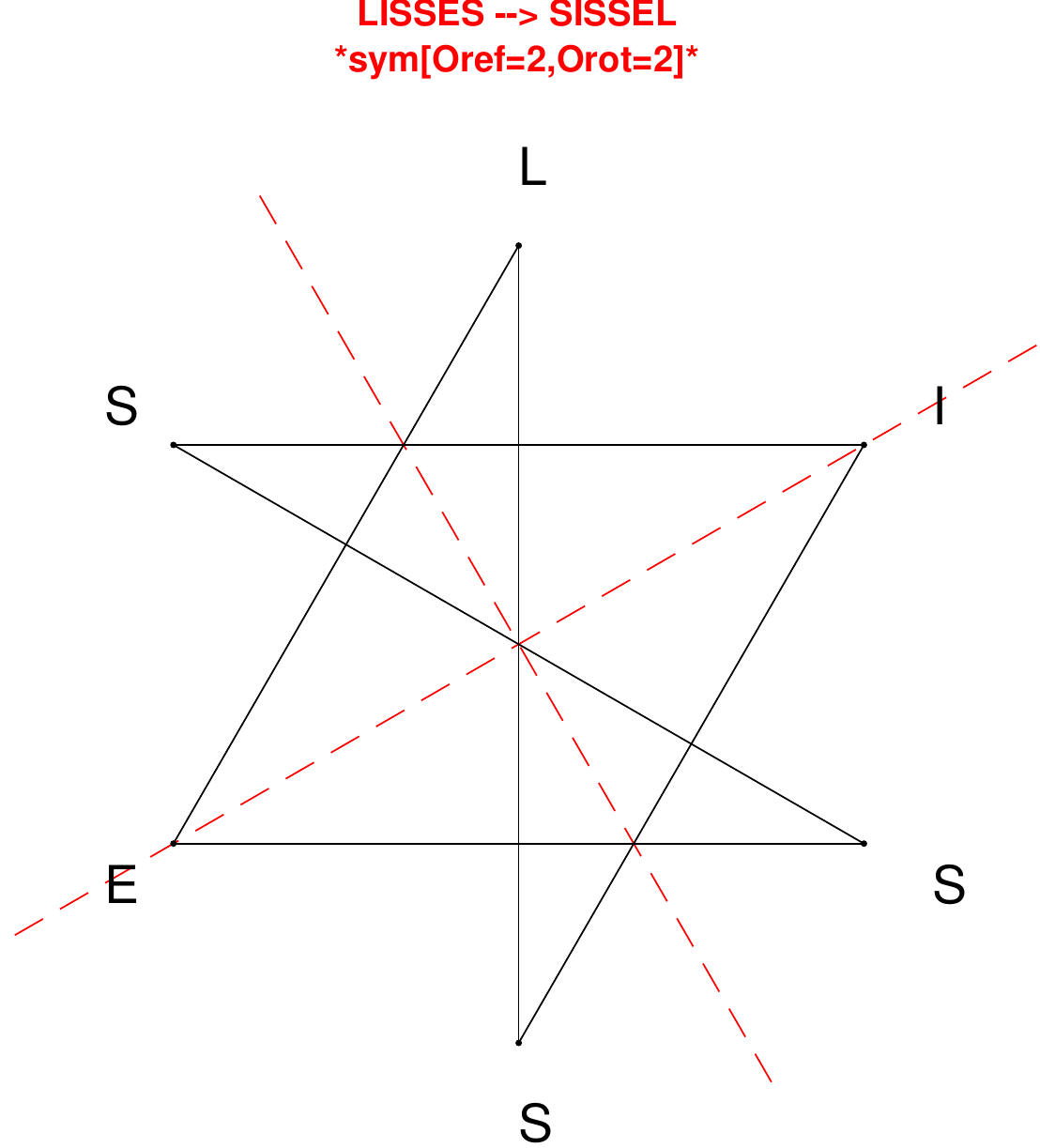}
\end{subfigure}
\hfill
\begin{subfigure}[T]{0.19\textwidth}
\centering
\includegraphics[width=\textwidth]{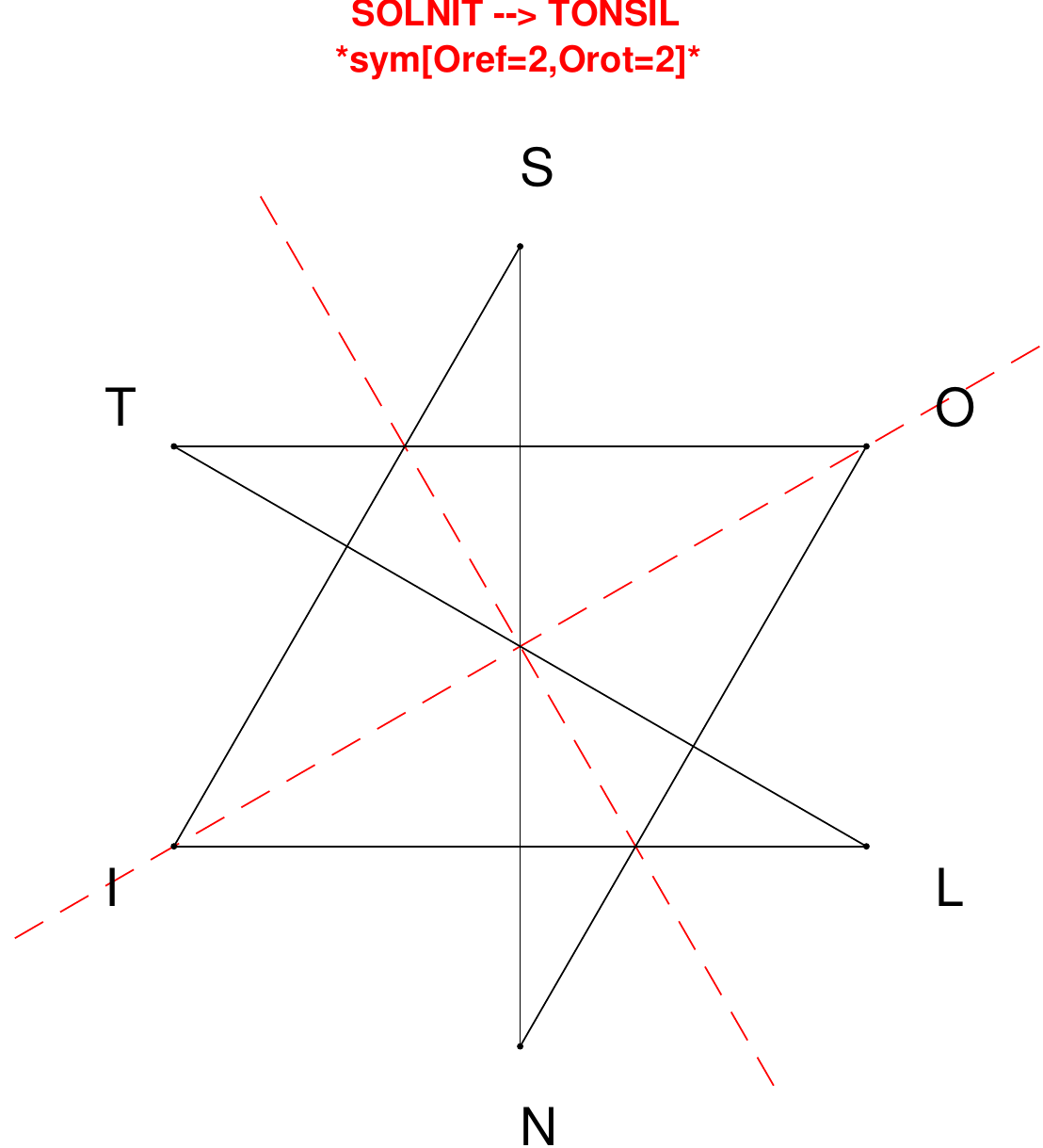}
\end{subfigure}
\hfill
\begin{subfigure}[T]{0.19\textwidth}
\centering
\includegraphics[width=\textwidth]{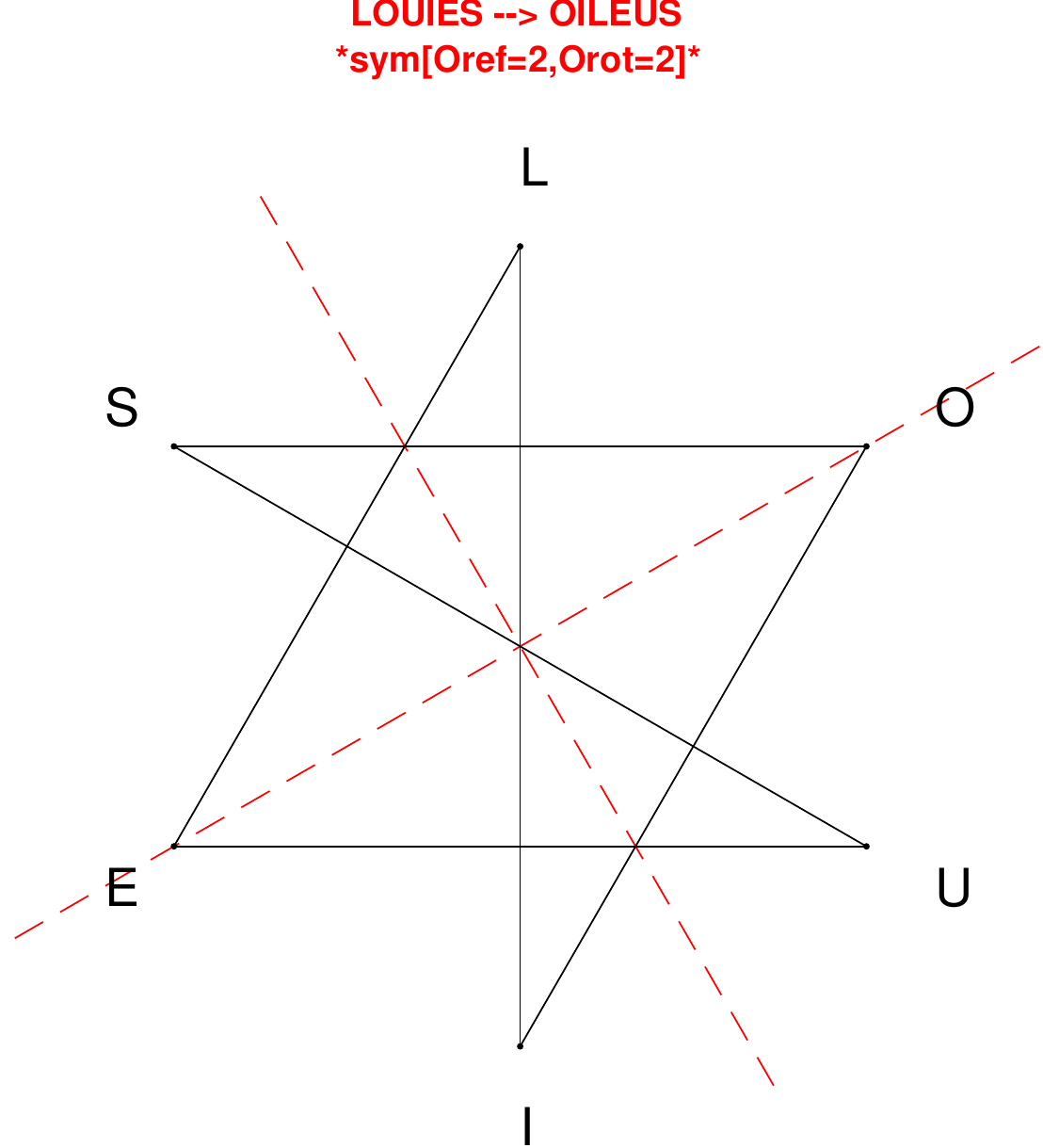}
\end{subfigure}
\hfill
\begin{subfigure}[T]{0.19\textwidth}
\centering
\includegraphics[width=\textwidth]{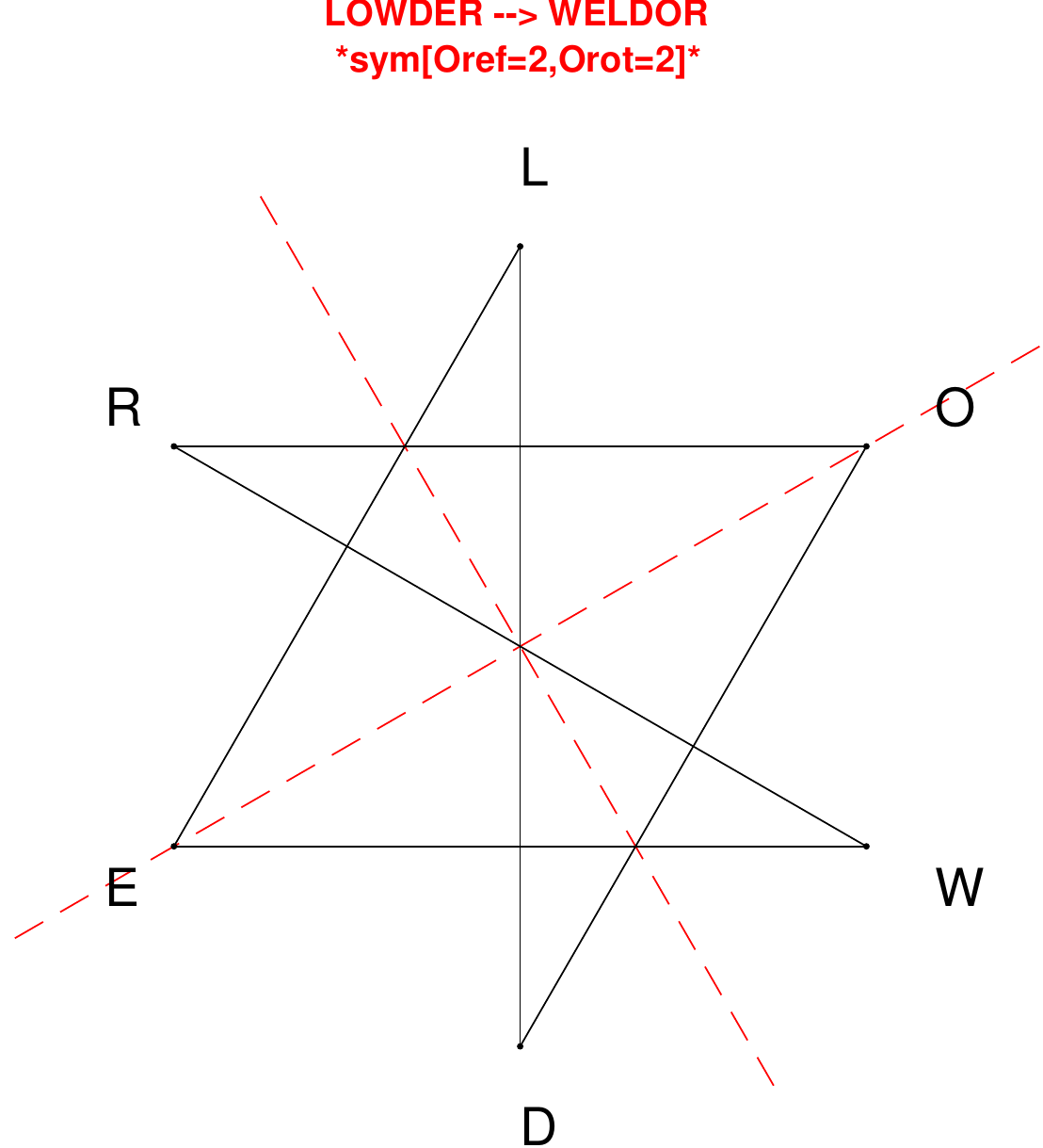}
\end{subfigure}
\end{figure}

\begin{figure}[H]
\centering
\begin{subfigure}[T]{0.19\textwidth}
\centering
\includegraphics[width=\textwidth]{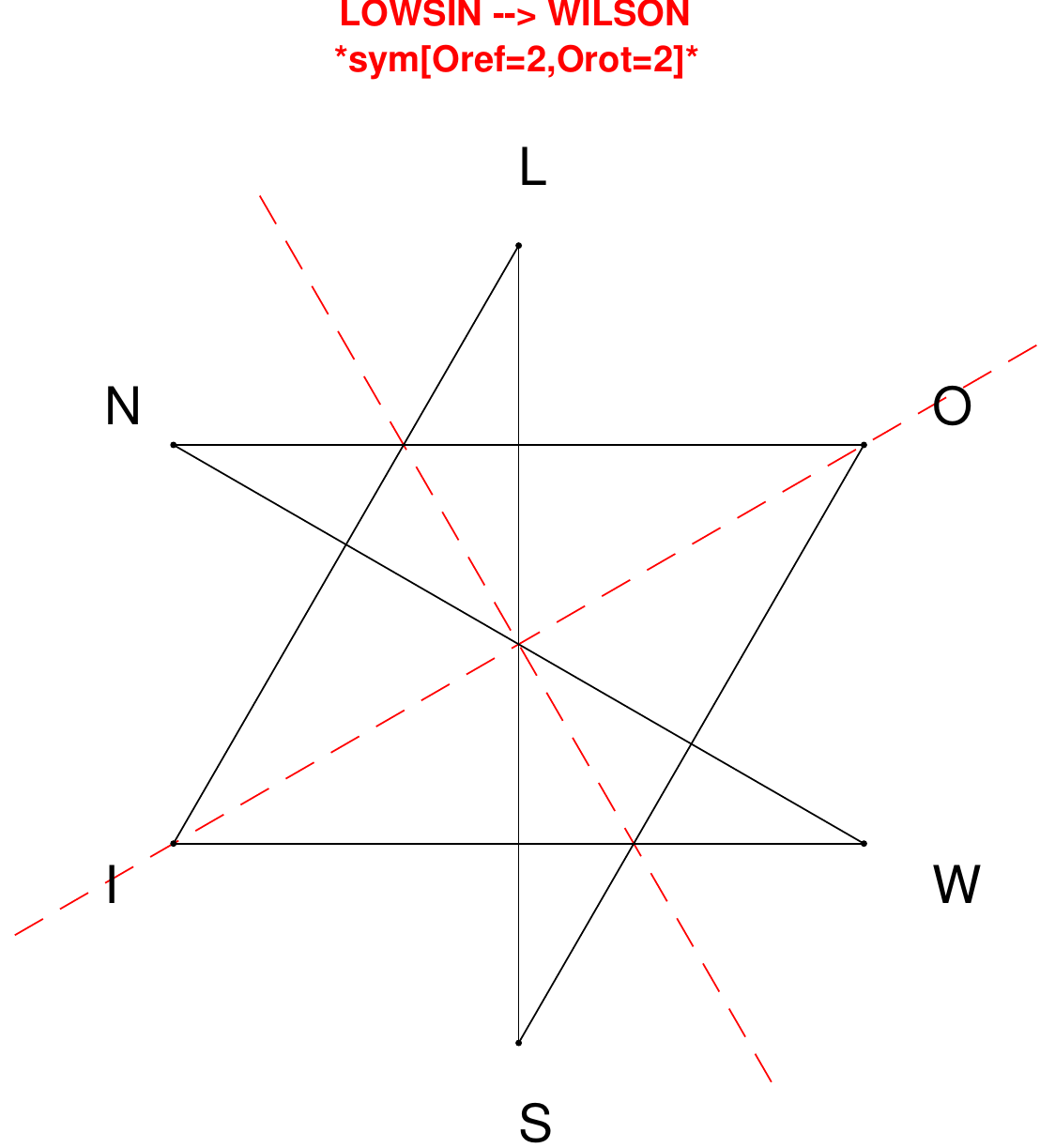}
\end{subfigure}
\hfill
\begin{subfigure}[T]{0.19\textwidth}
\centering
\includegraphics[width=\textwidth]{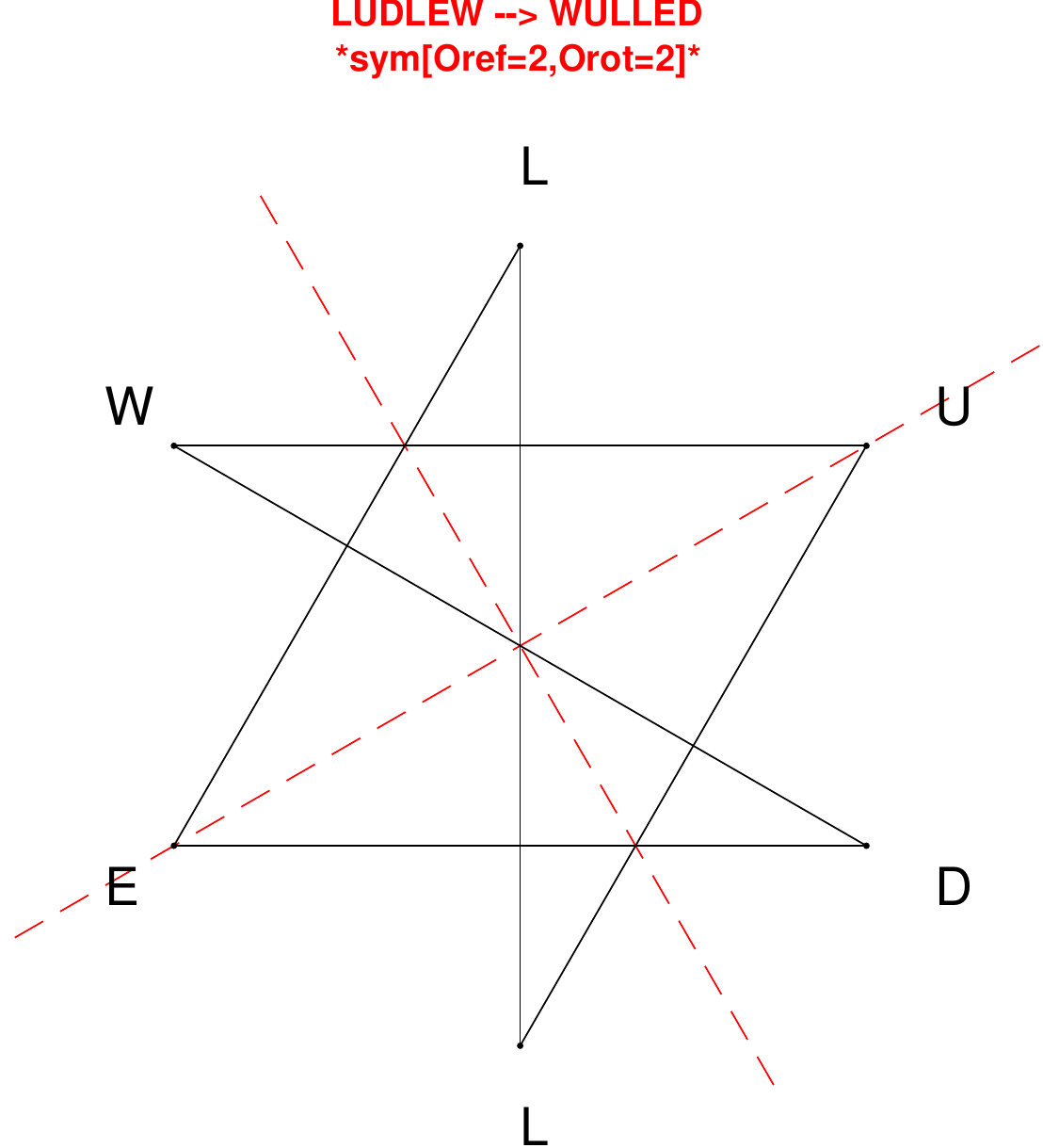}
\end{subfigure}
\hfill
\begin{subfigure}[T]{0.19\textwidth}
\centering
\includegraphics[width=\textwidth]{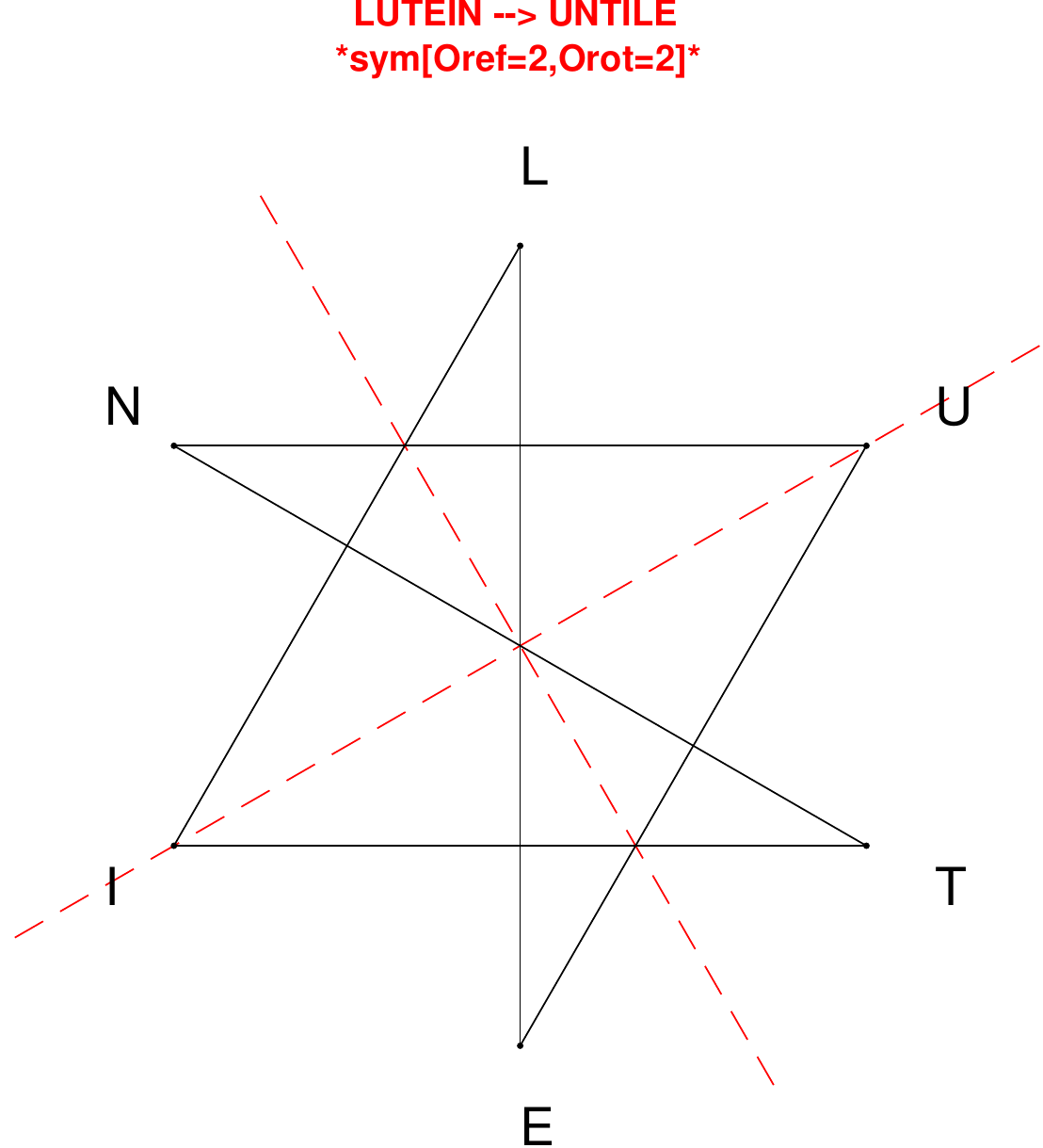}
\end{subfigure}
\hfill
\begin{subfigure}[T]{0.19\textwidth}
\centering
\includegraphics[width=\textwidth]{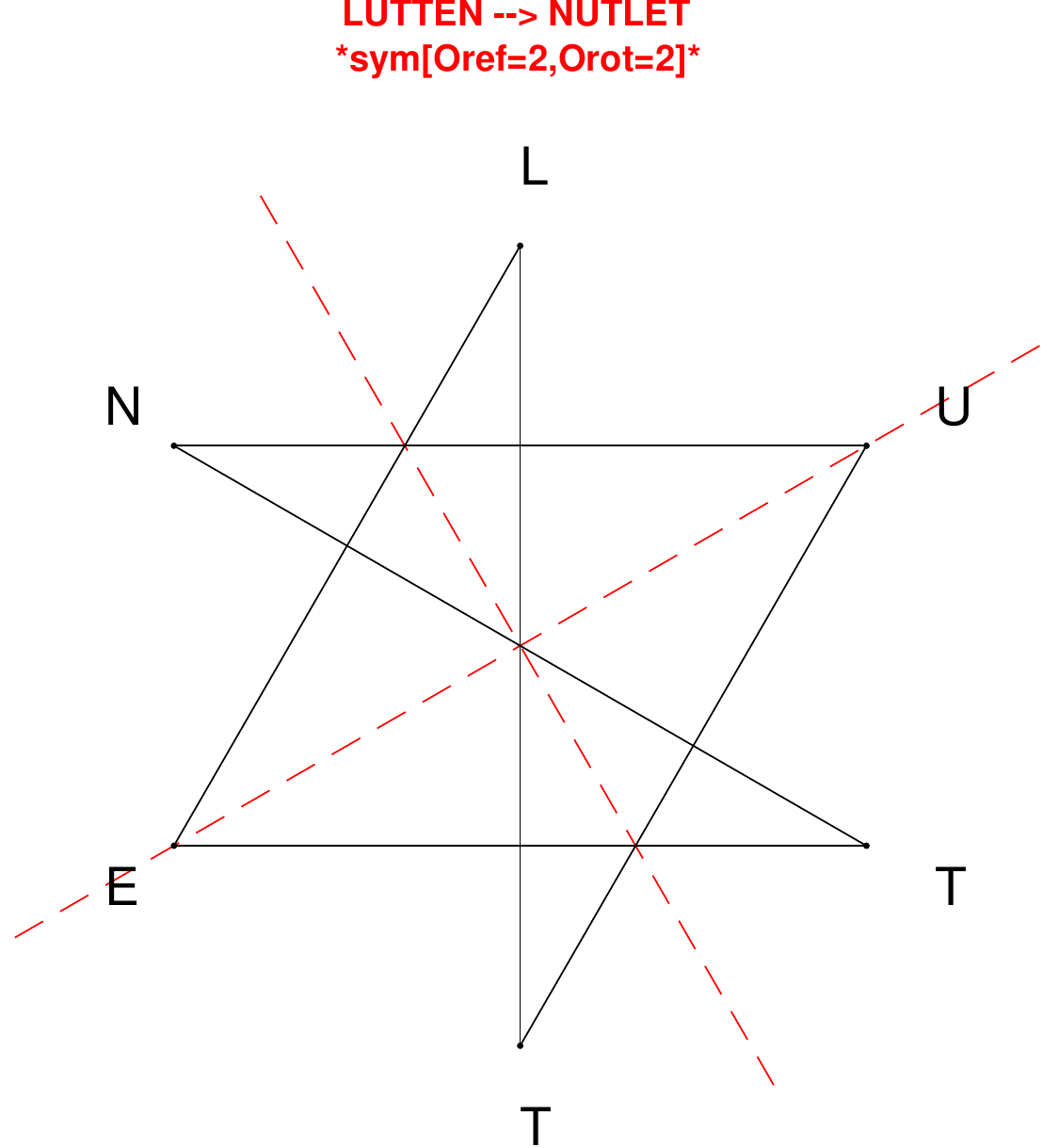}
\end{subfigure}
\hfill
\begin{subfigure}[T]{0.19\textwidth}
\centering
\includegraphics[width=\textwidth]{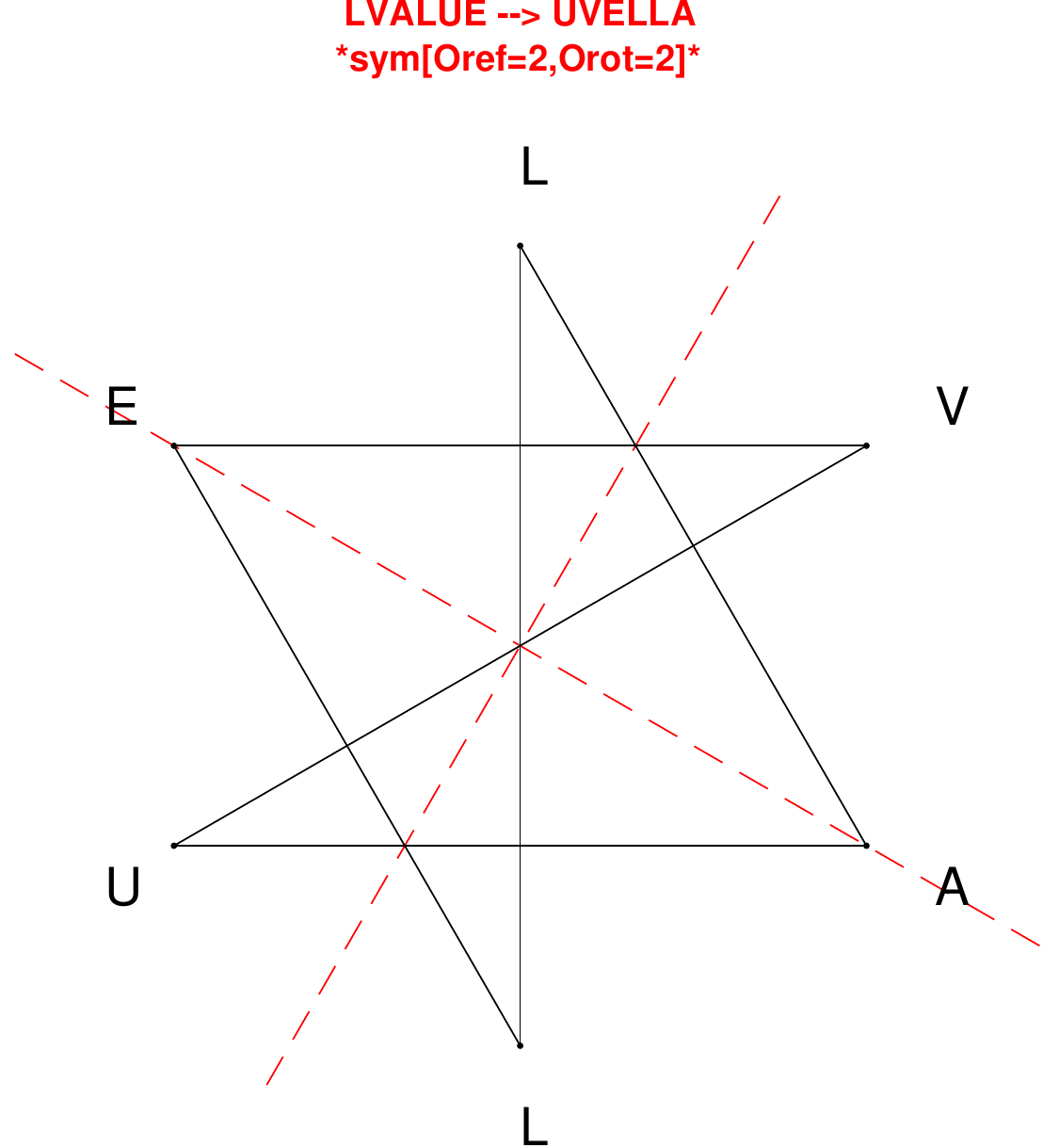}
\end{subfigure}
\end{figure}

\begin{figure}[H]
\centering
\begin{subfigure}[T]{0.19\textwidth}
\centering
\includegraphics[width=\textwidth]{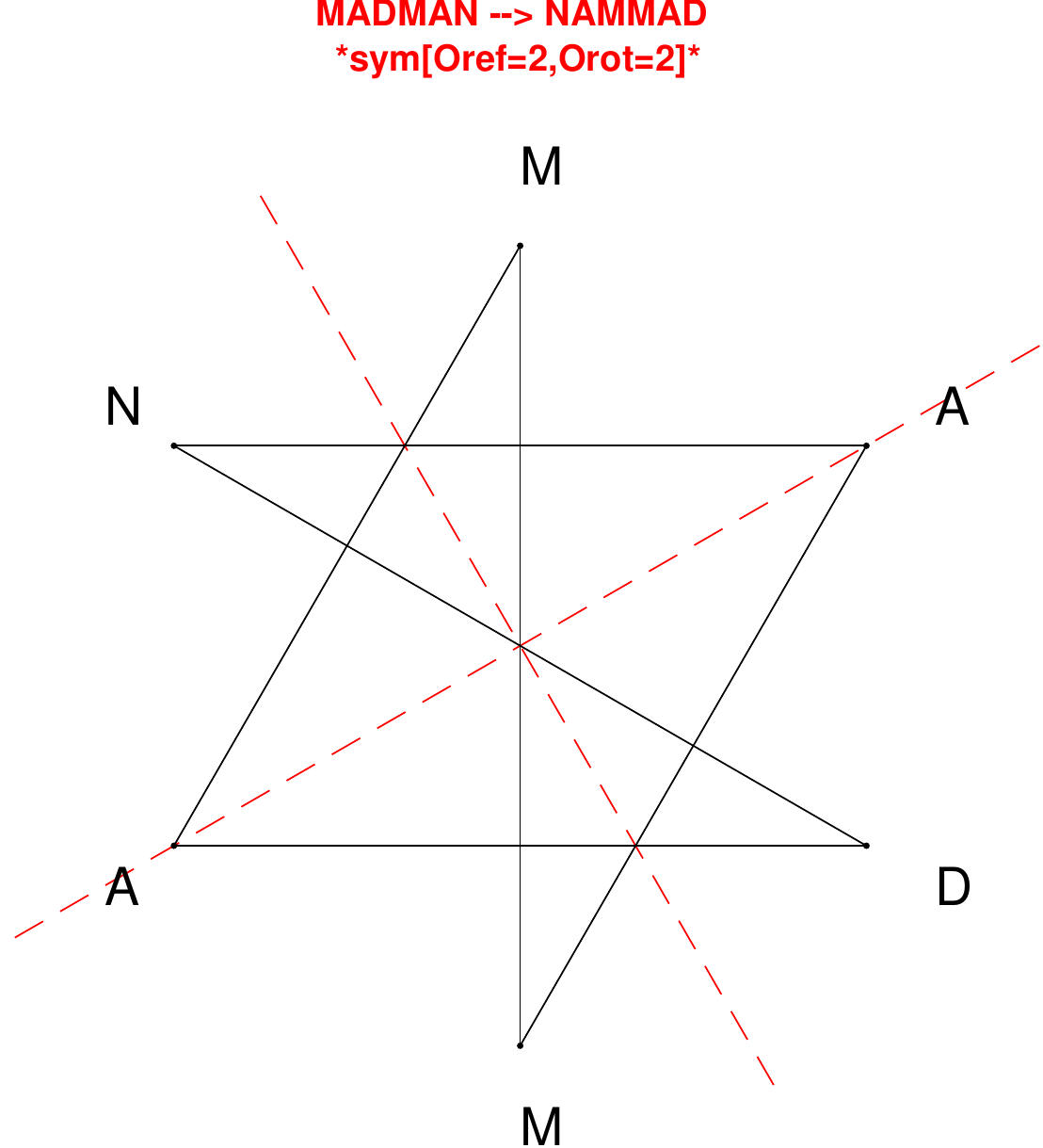}
\end{subfigure}
\hfill
\begin{subfigure}[T]{0.19\textwidth}
\centering
\includegraphics[width=\textwidth]{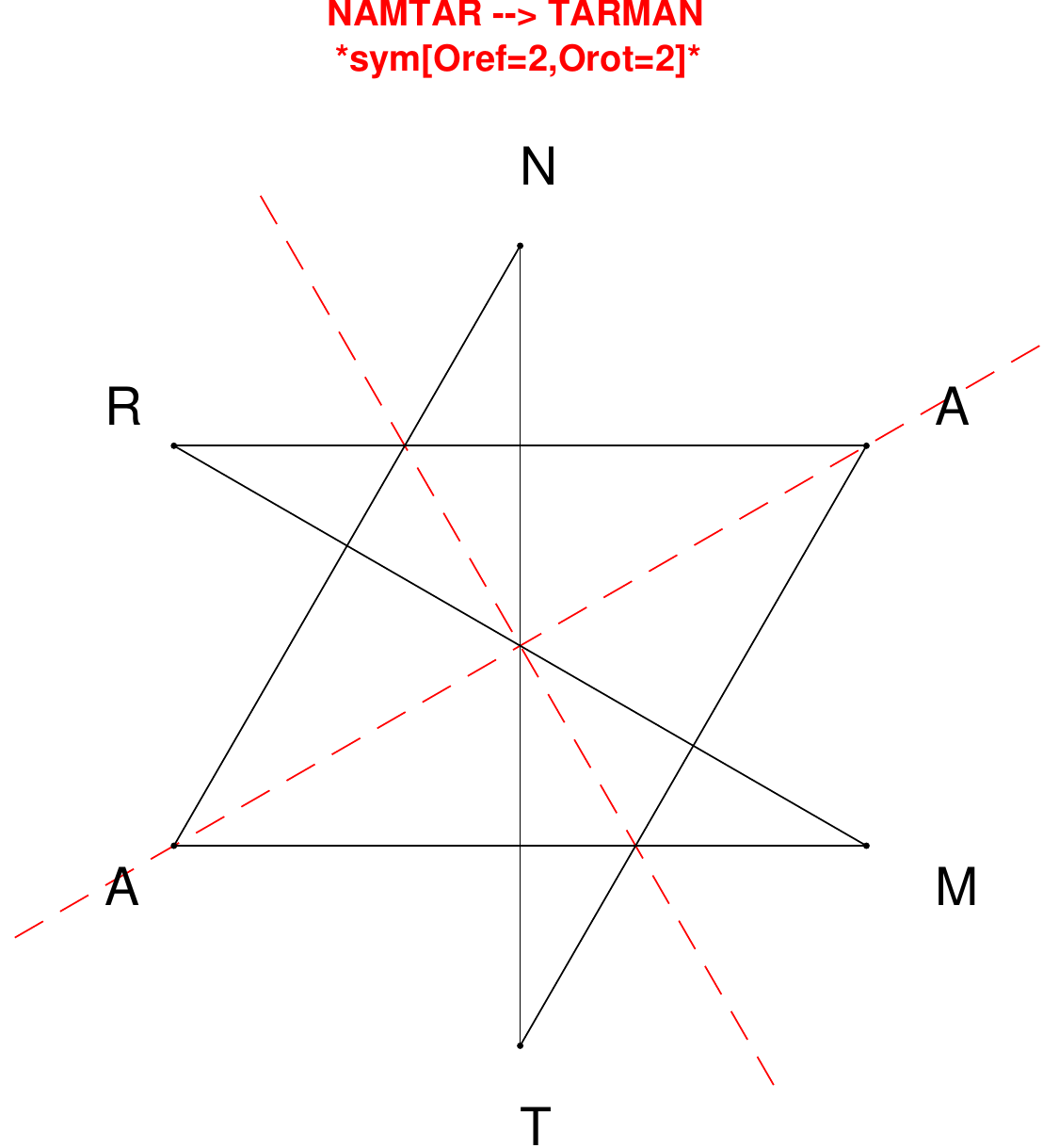}
\end{subfigure}
\hfill
\begin{subfigure}[T]{0.19\textwidth}
\centering
\includegraphics[width=\textwidth]{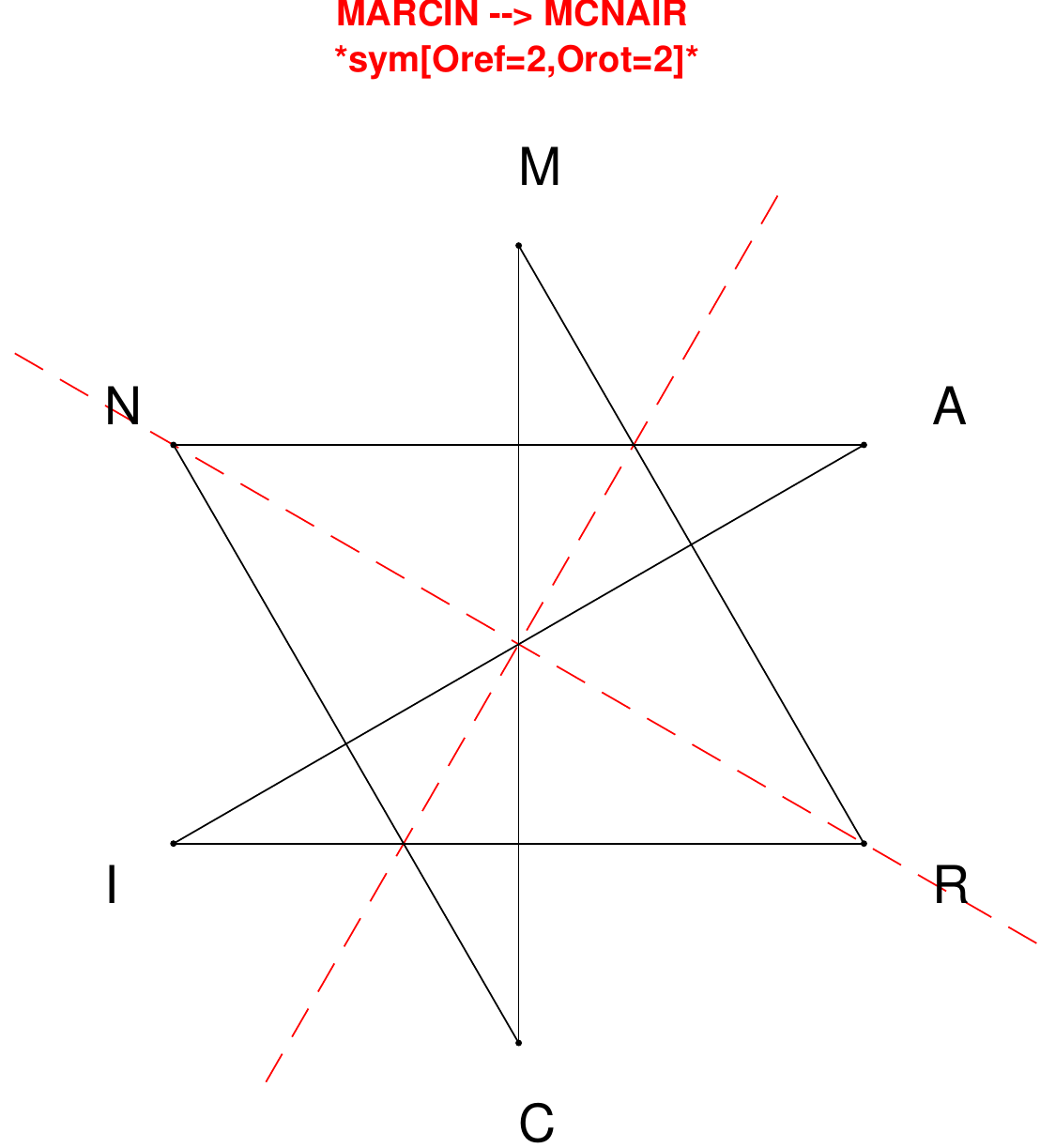}
\end{subfigure}
\hfill
\begin{subfigure}[T]{0.19\textwidth}
\centering
\includegraphics[width=\textwidth]{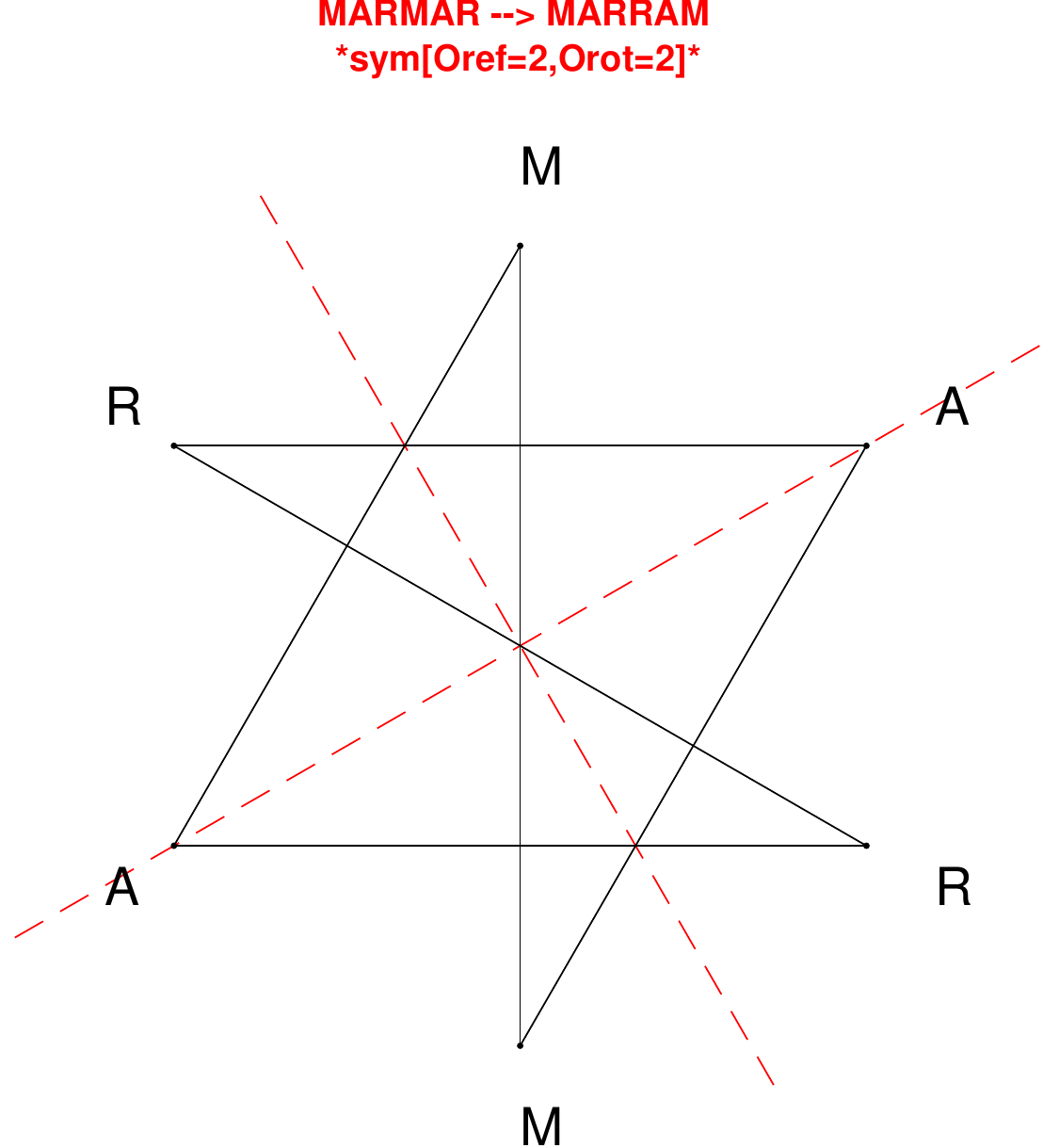}
\end{subfigure}
\hfill
\begin{subfigure}[T]{0.19\textwidth}
\centering
\includegraphics[width=\textwidth]{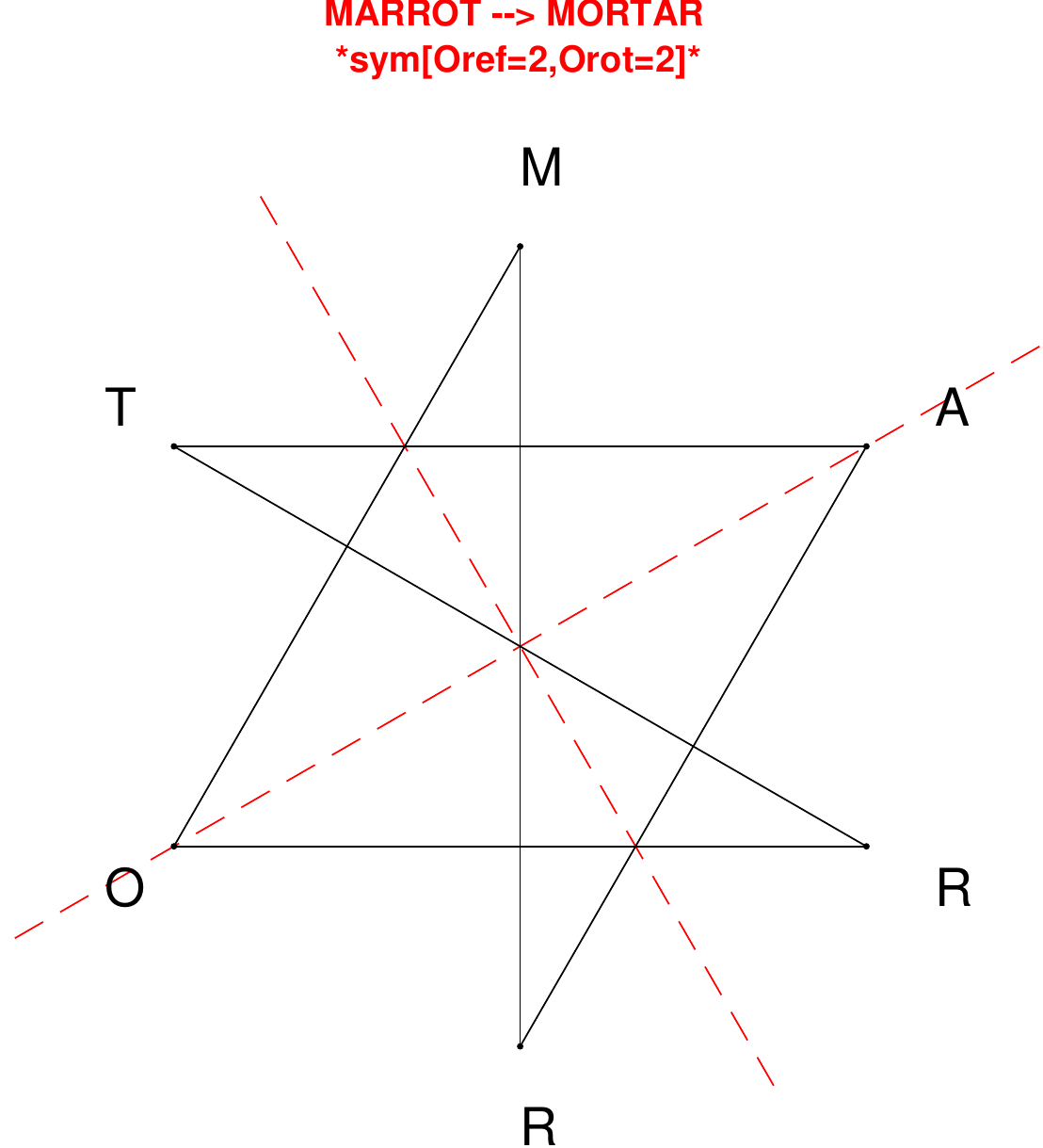}
\end{subfigure}
\end{figure}

\begin{figure}[H]
\centering
\begin{subfigure}[T]{0.19\textwidth}
\centering
\includegraphics[width=\textwidth]{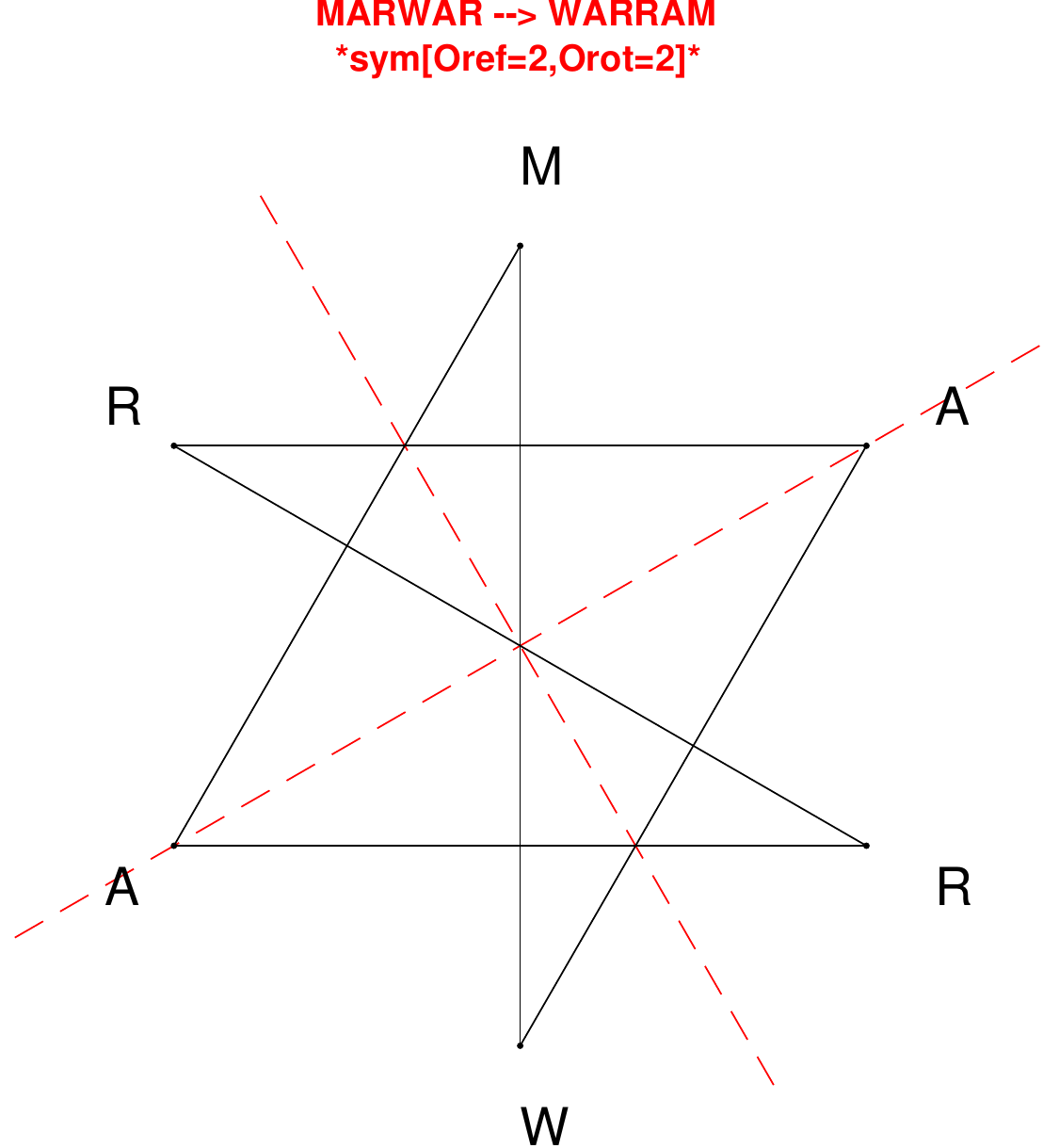}
\end{subfigure}
\hfill
\begin{subfigure}[T]{0.19\textwidth}
\centering
\includegraphics[width=\textwidth]{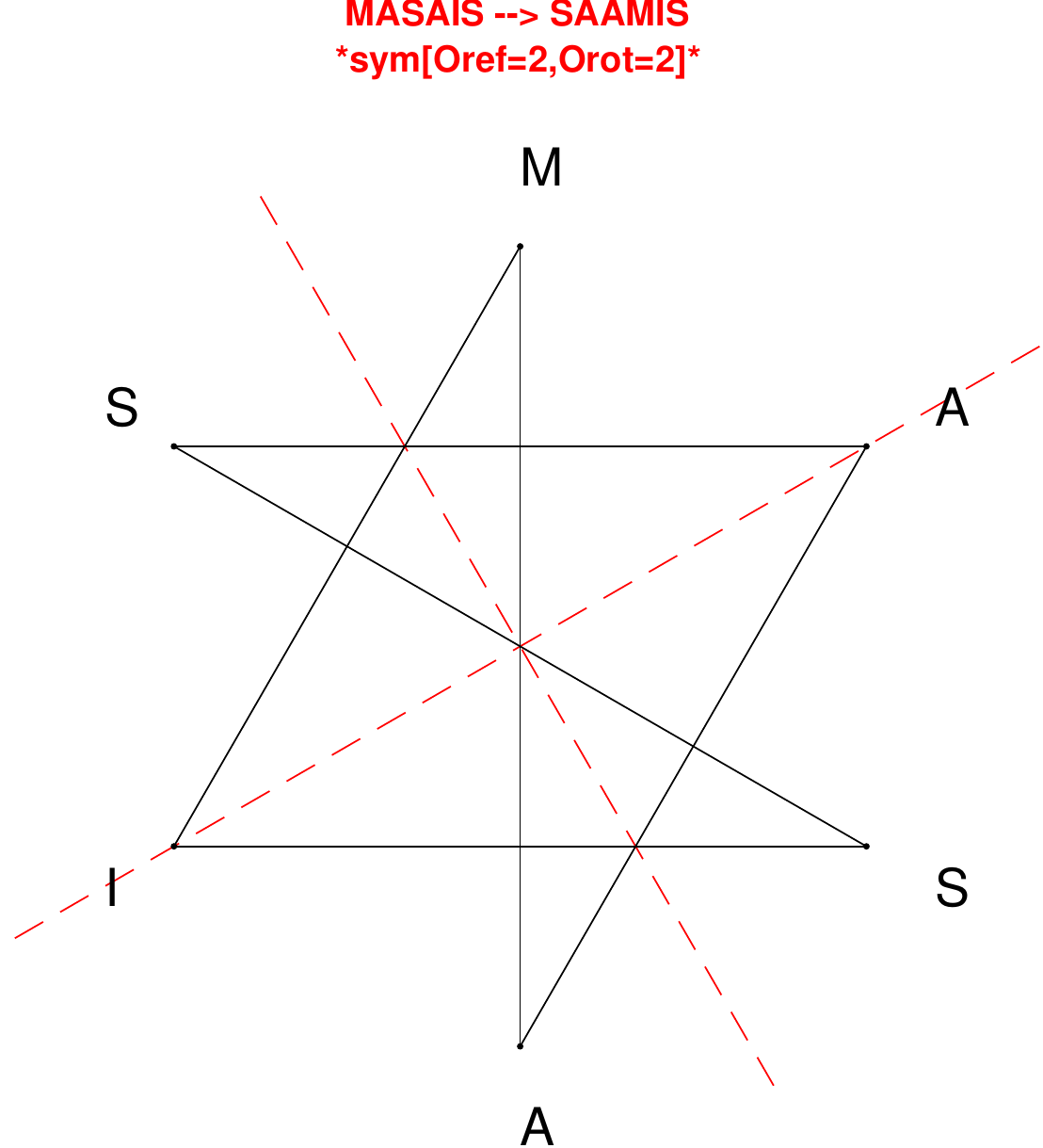}
\end{subfigure}
\hfill
\begin{subfigure}[T]{0.19\textwidth}
\centering
\includegraphics[width=\textwidth]{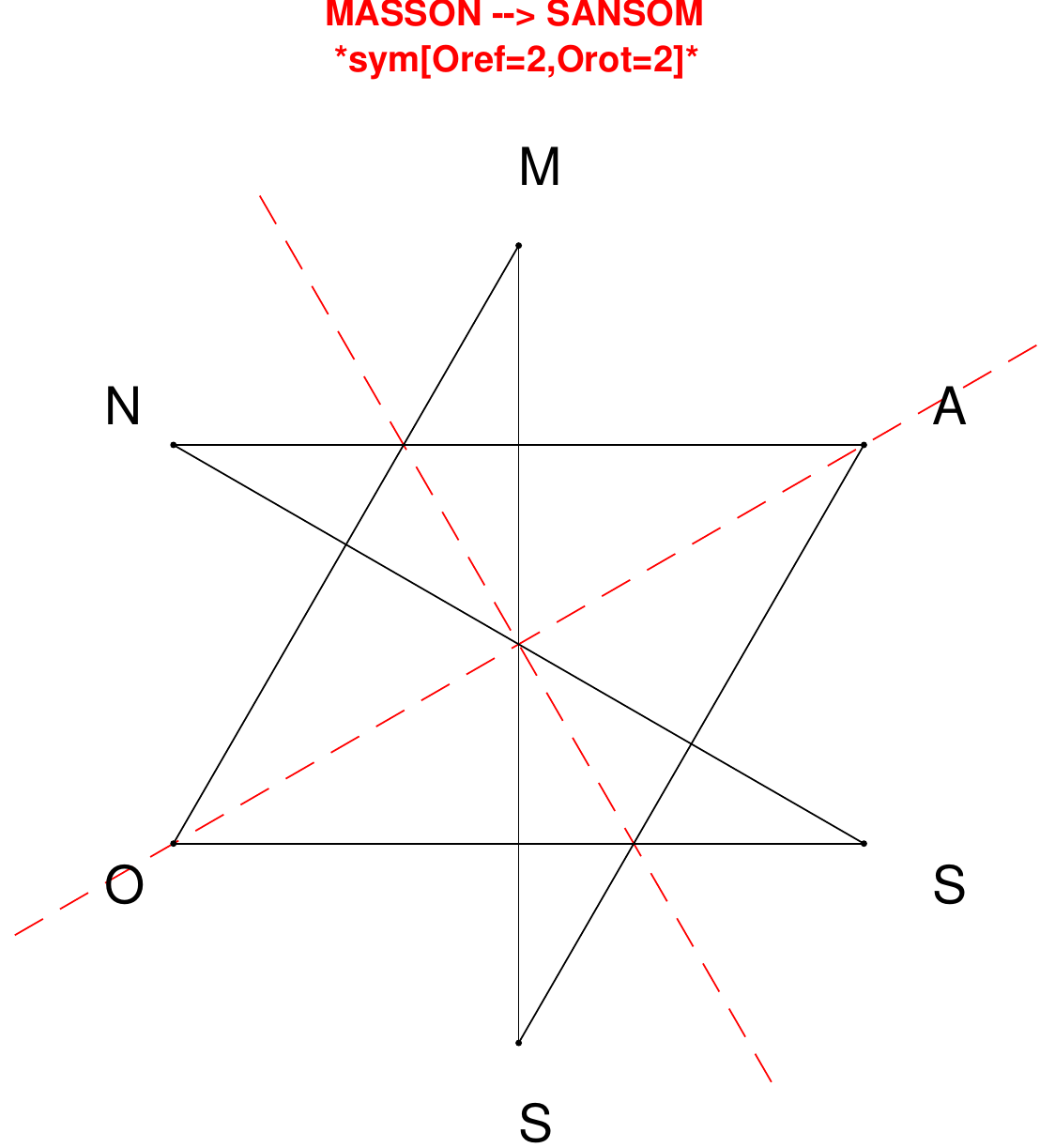}
\end{subfigure}
\hfill
\begin{subfigure}[T]{0.19\textwidth}
\centering
\includegraphics[width=\textwidth]{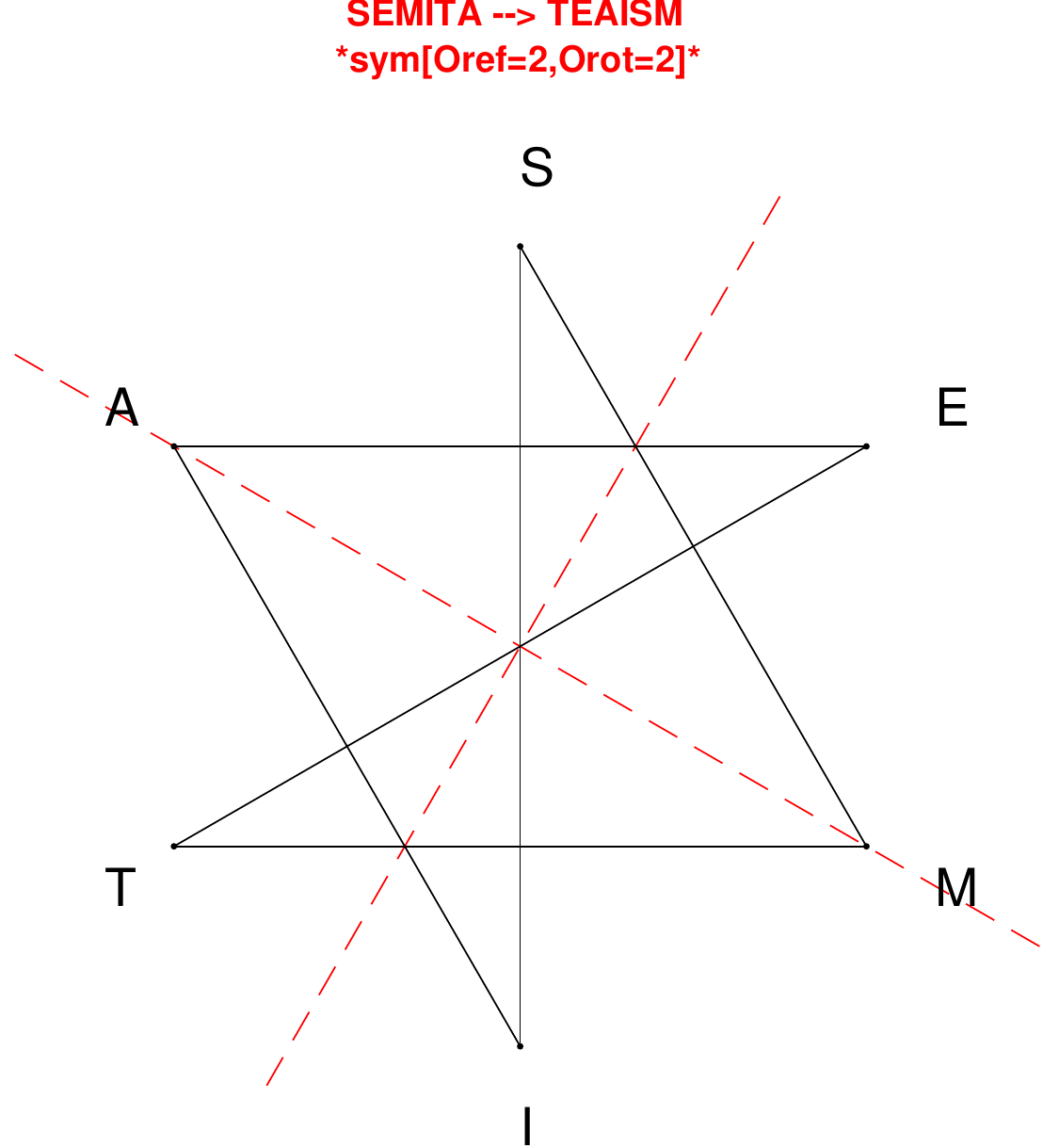}
\end{subfigure}
\hfill
\begin{subfigure}[T]{0.19\textwidth}
\centering
\includegraphics[width=\textwidth]{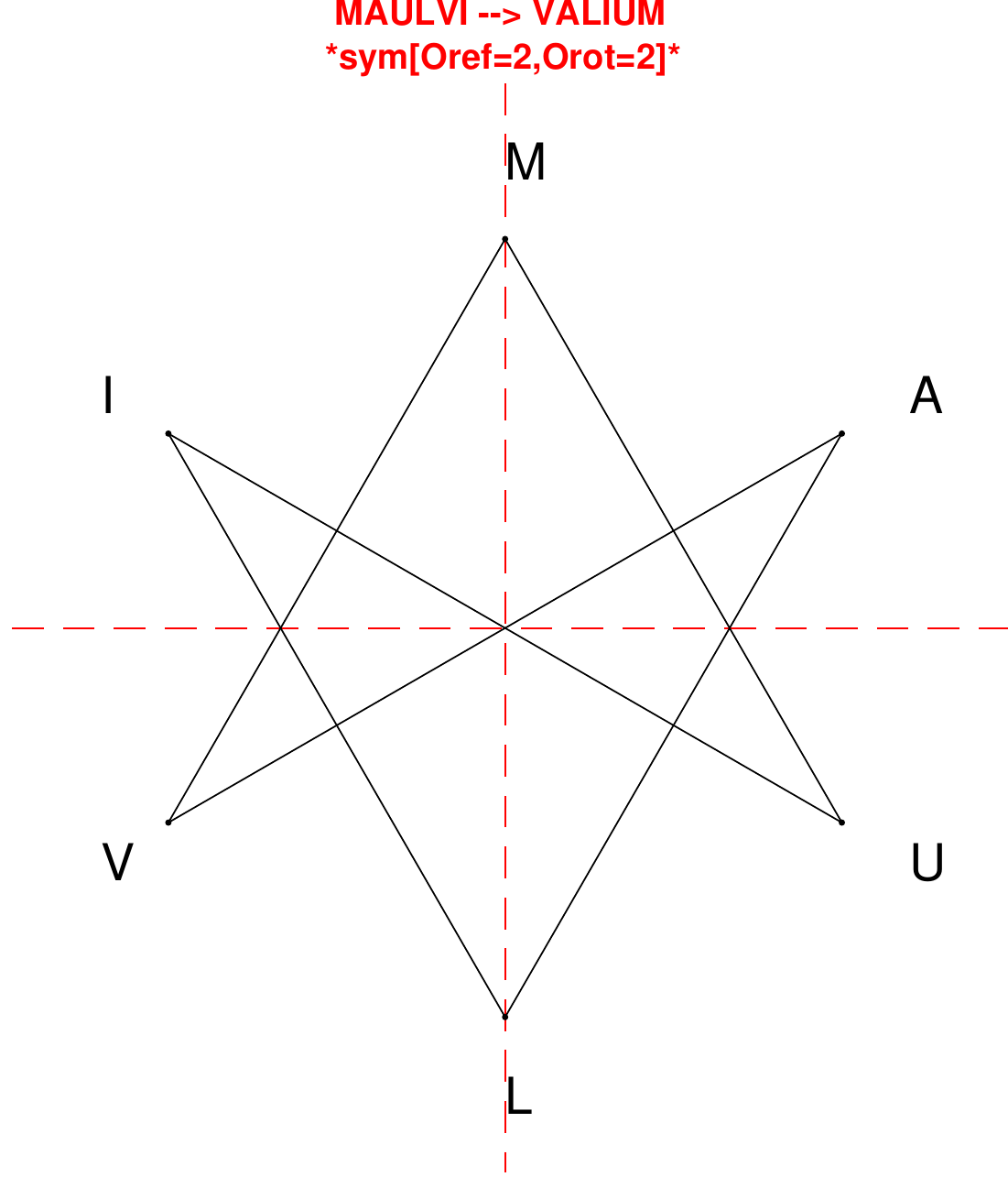}
\end{subfigure}
\end{figure}

\begin{figure}[H]
\centering
\begin{subfigure}[T]{0.19\textwidth}
\centering
\includegraphics[width=\textwidth]{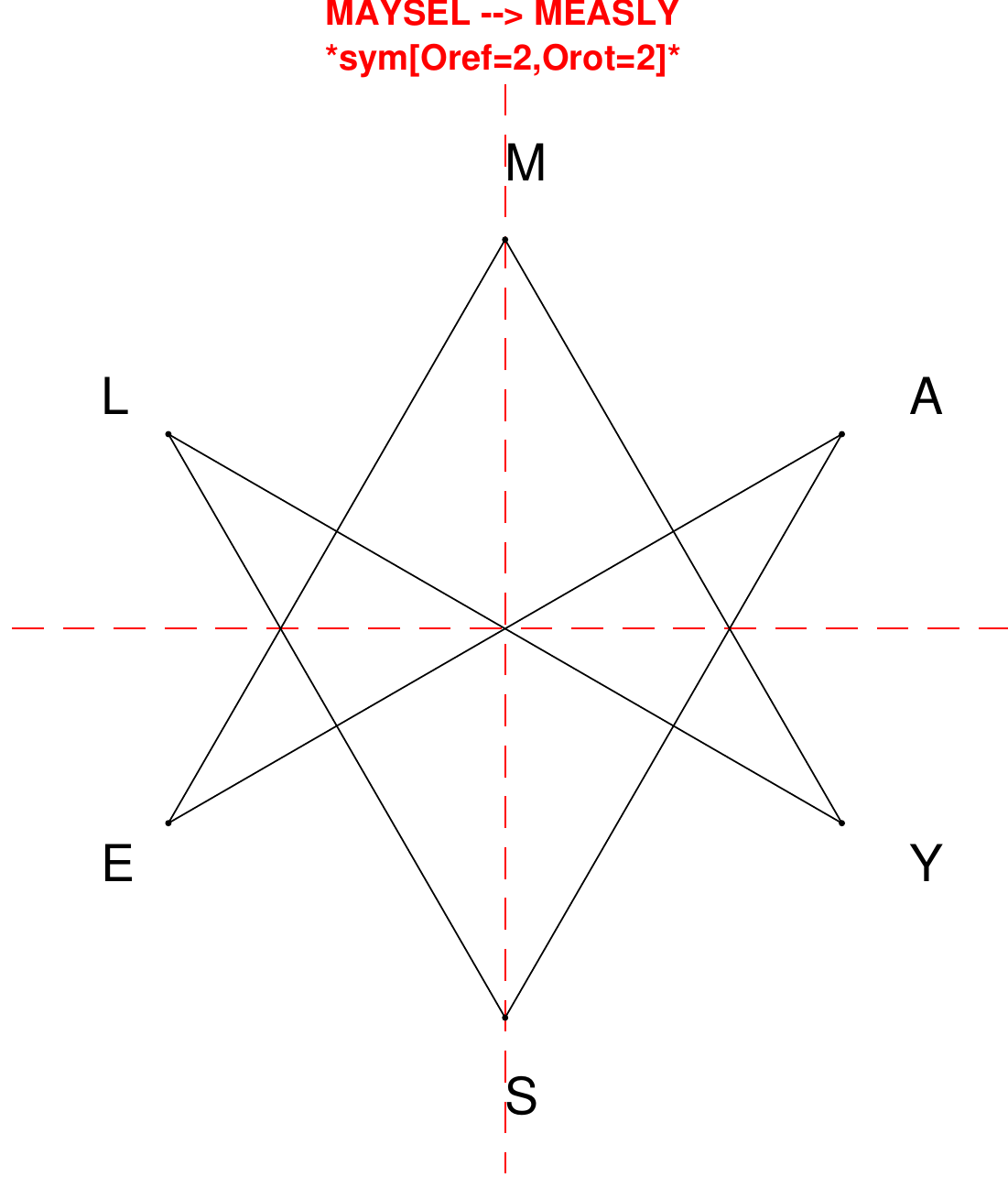}
\end{subfigure}
\hfill
\begin{subfigure}[T]{0.19\textwidth}
\centering
\includegraphics[width=\textwidth]{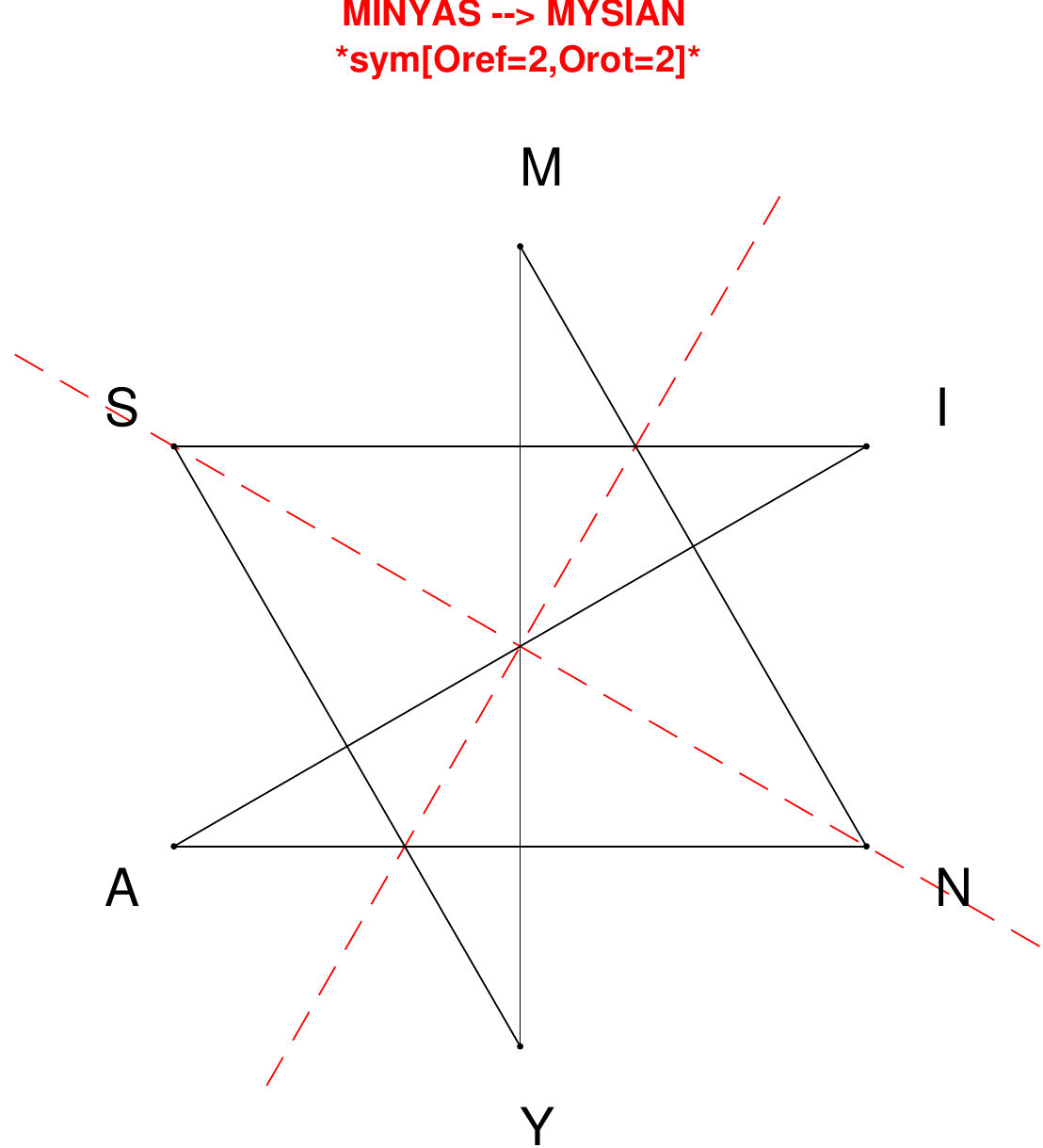}
\end{subfigure}
\hfill
\begin{subfigure}[T]{0.19\textwidth}
\centering
\includegraphics[width=\textwidth]{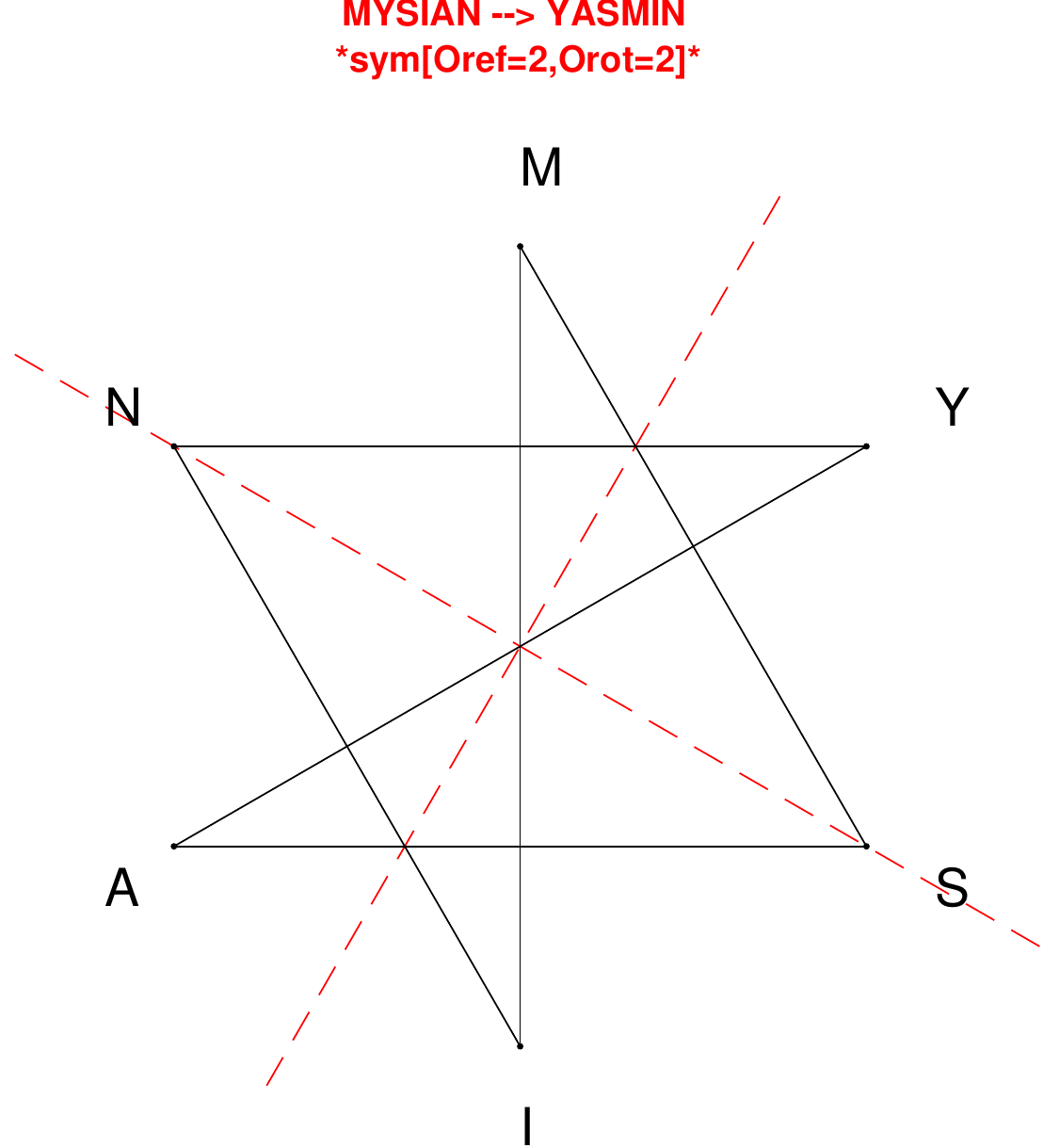}
\end{subfigure}
\hfill
\begin{subfigure}[T]{0.19\textwidth}
\centering
\includegraphics[width=\textwidth]{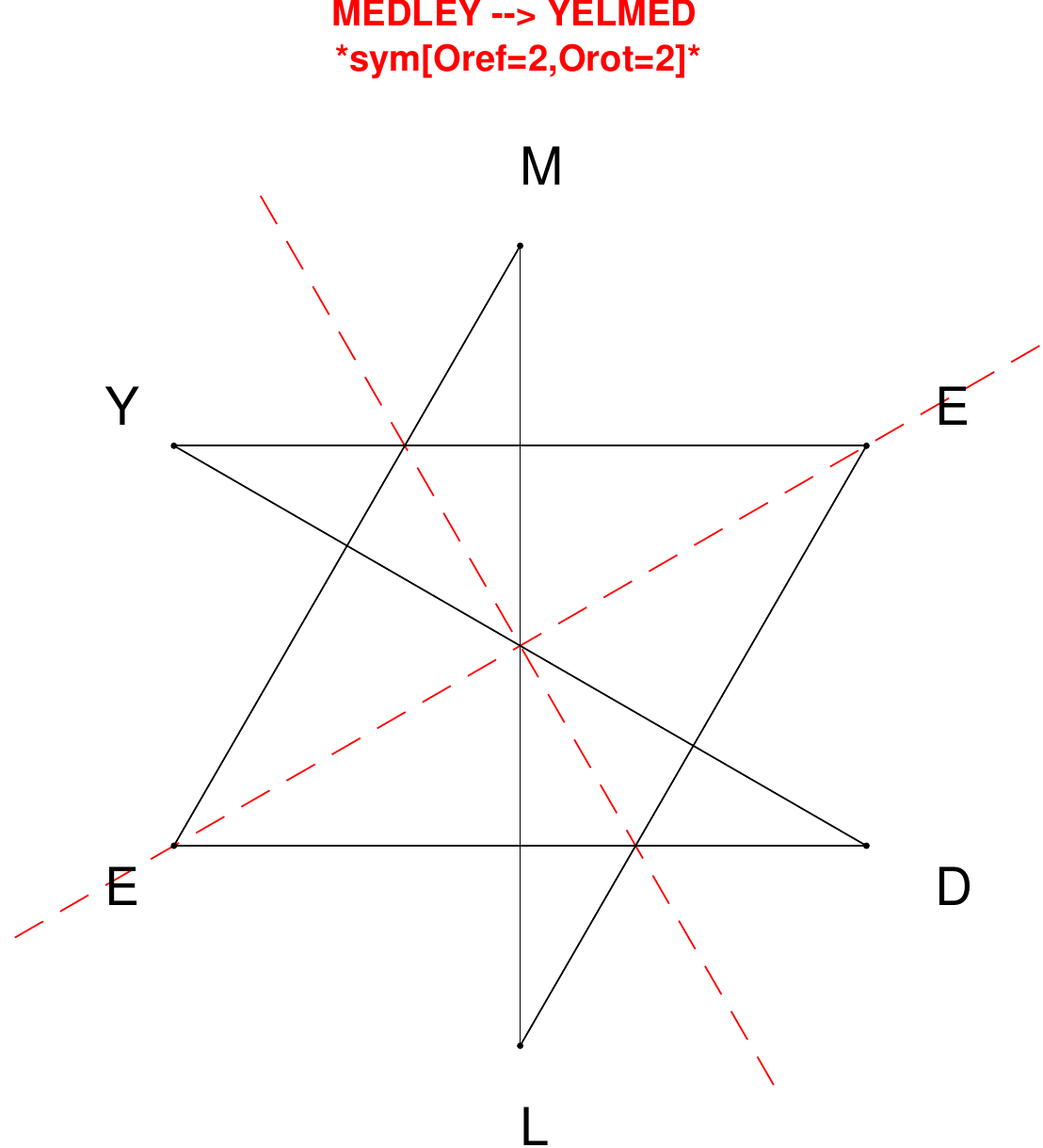}
\end{subfigure}
\hfill
\begin{subfigure}[T]{0.19\textwidth}
\centering
\includegraphics[width=\textwidth]{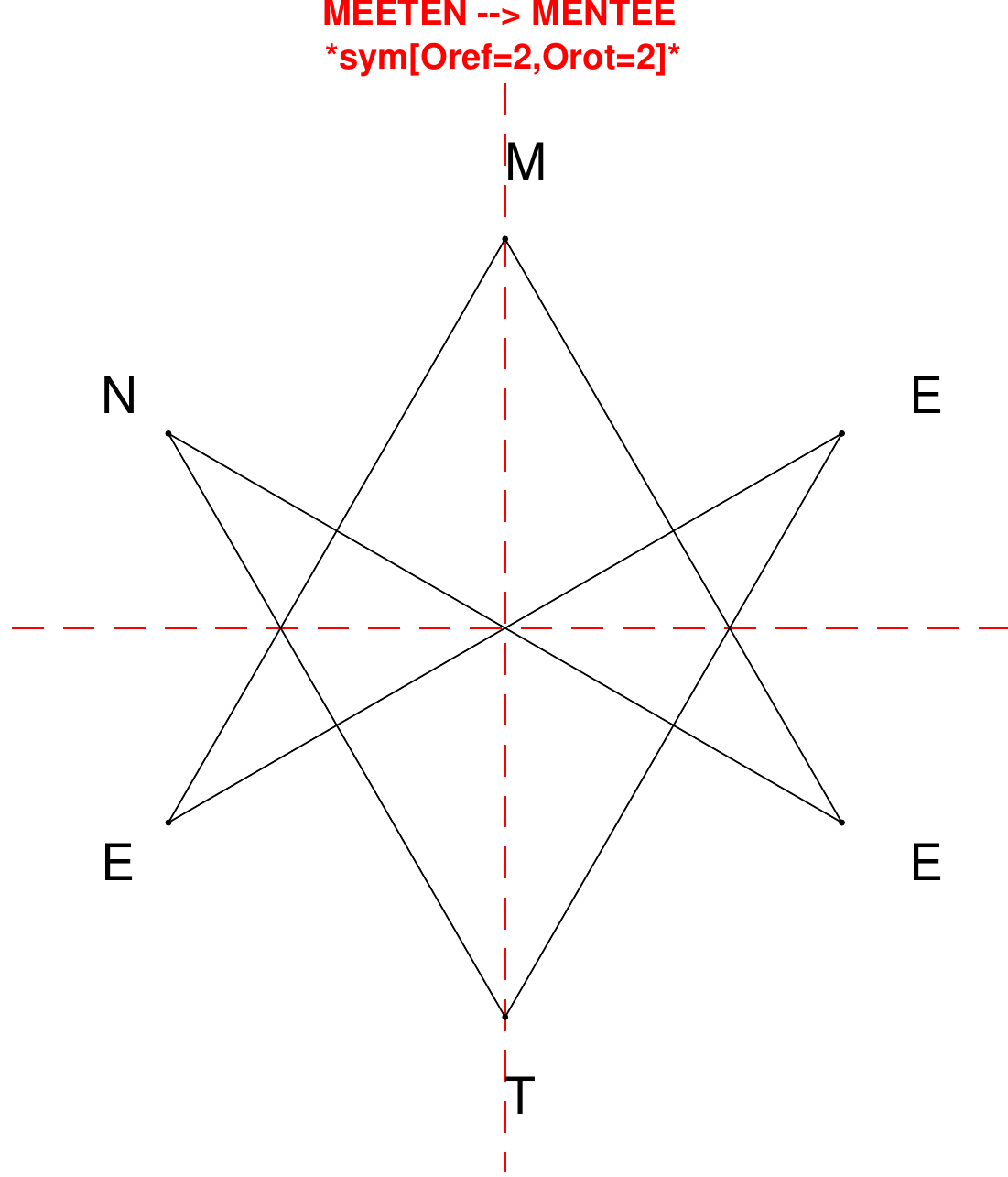}
\end{subfigure}
\end{figure}

\begin{figure}[H]
\centering
\begin{subfigure}[T]{0.19\textwidth}
\centering
\includegraphics[width=\textwidth]{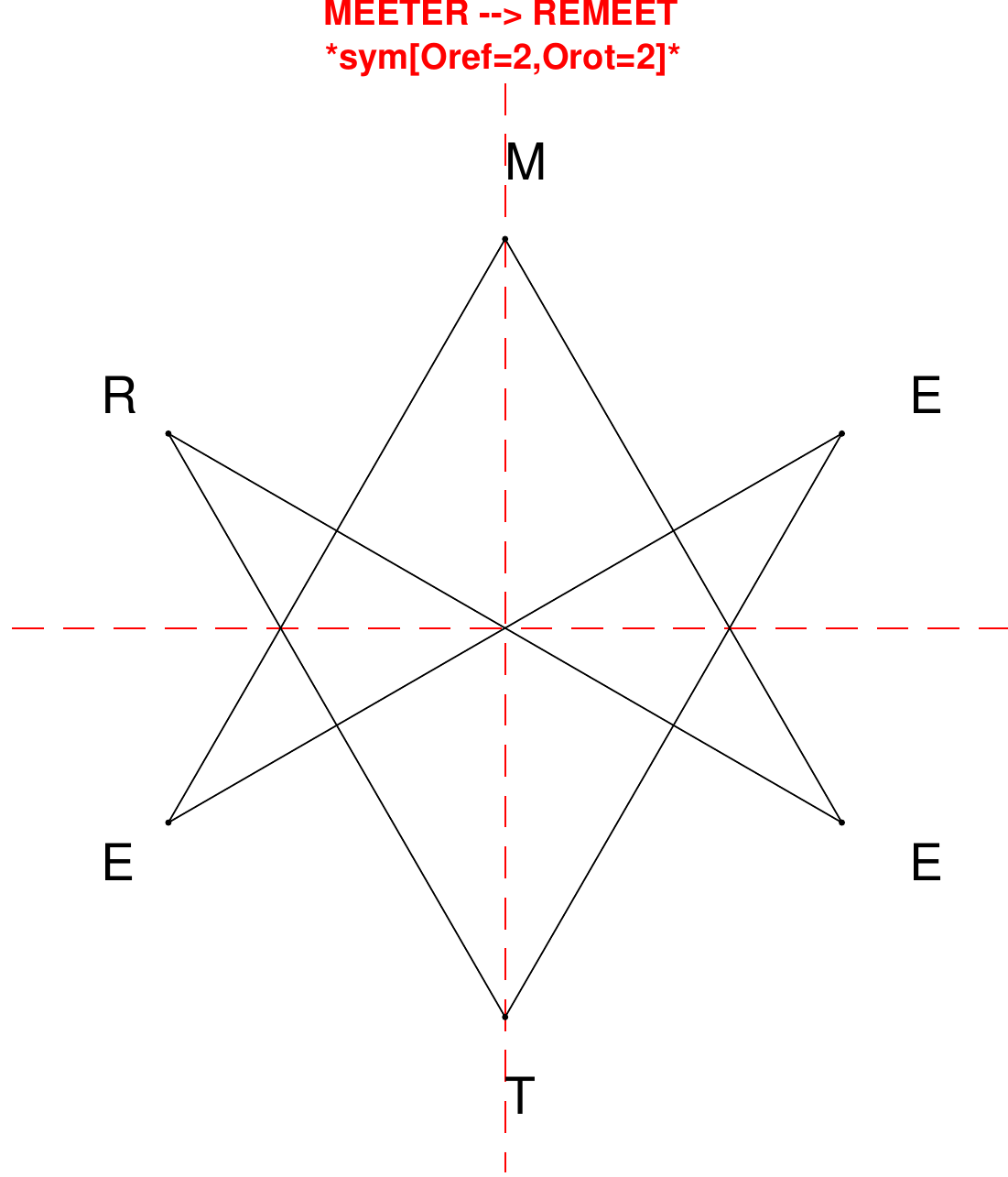}
\end{subfigure}
\hfill
\begin{subfigure}[T]{0.19\textwidth}
\centering
\includegraphics[width=\textwidth]{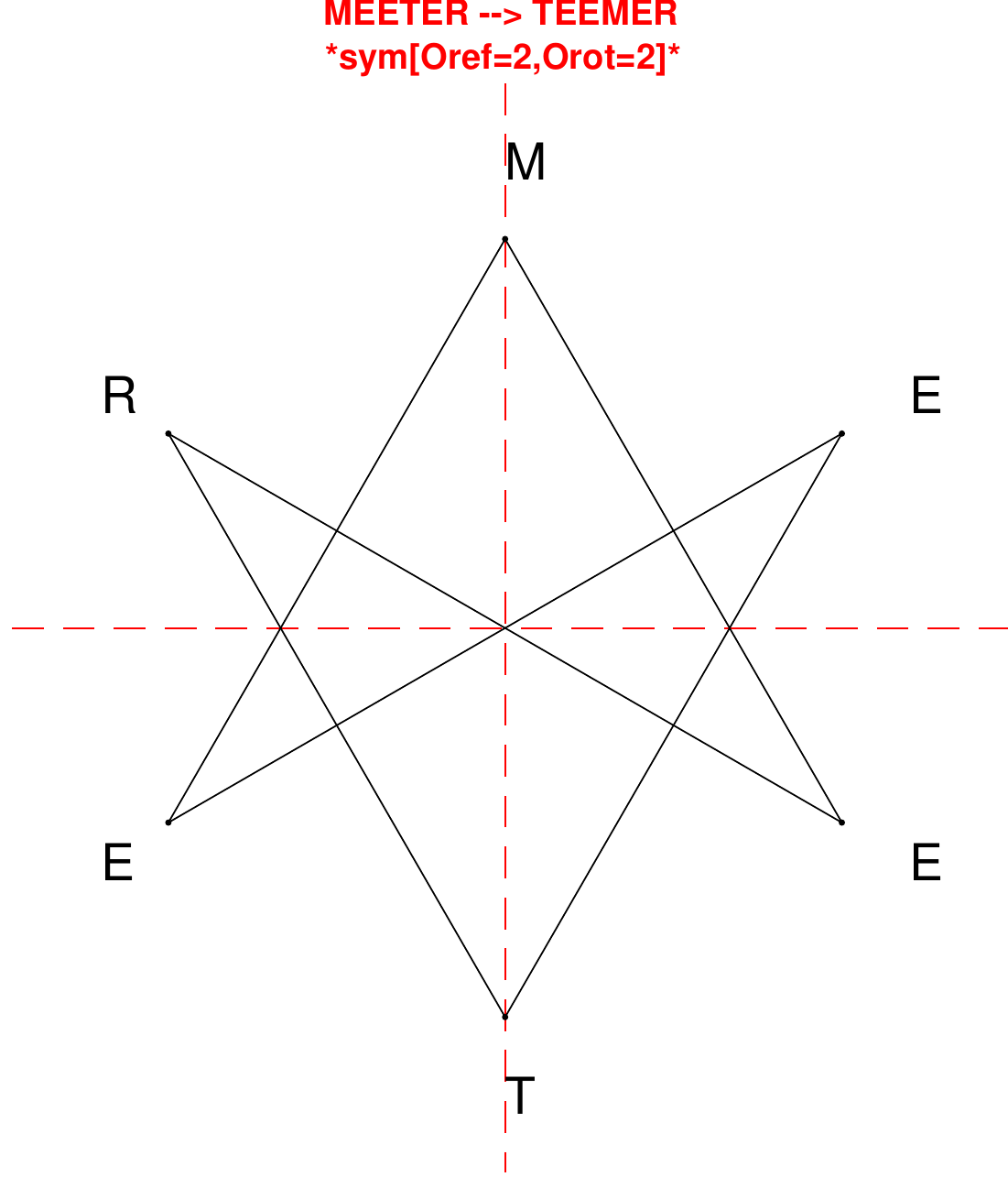}
\end{subfigure}
\hfill
\begin{subfigure}[T]{0.19\textwidth}
\centering
\includegraphics[width=\textwidth]{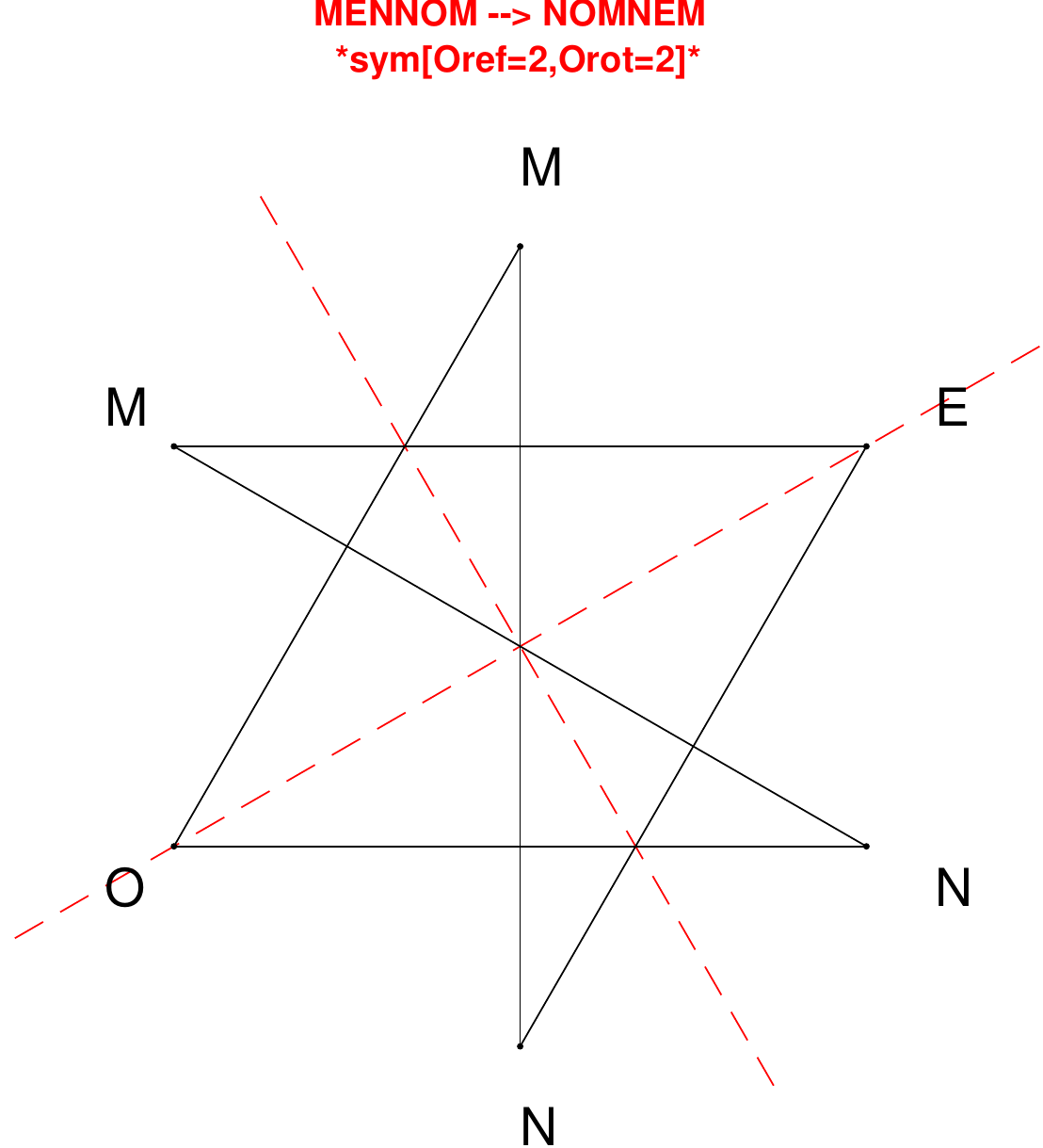}
\end{subfigure}
\hfill
\begin{subfigure}[T]{0.19\textwidth}
\centering
\includegraphics[width=\textwidth]{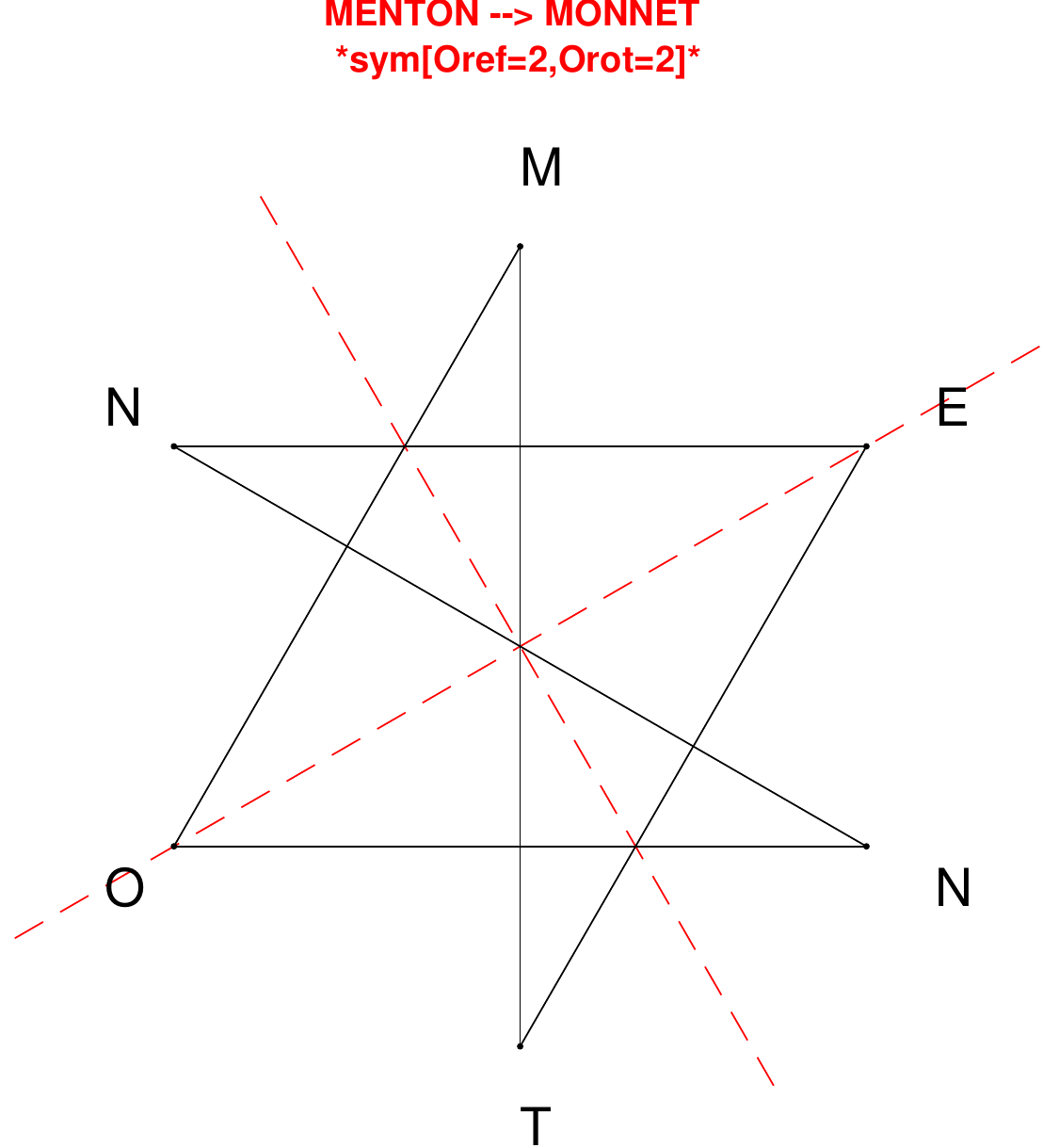}
\end{subfigure}
\hfill
\begin{subfigure}[T]{0.19\textwidth}
\centering
\includegraphics[width=\textwidth]{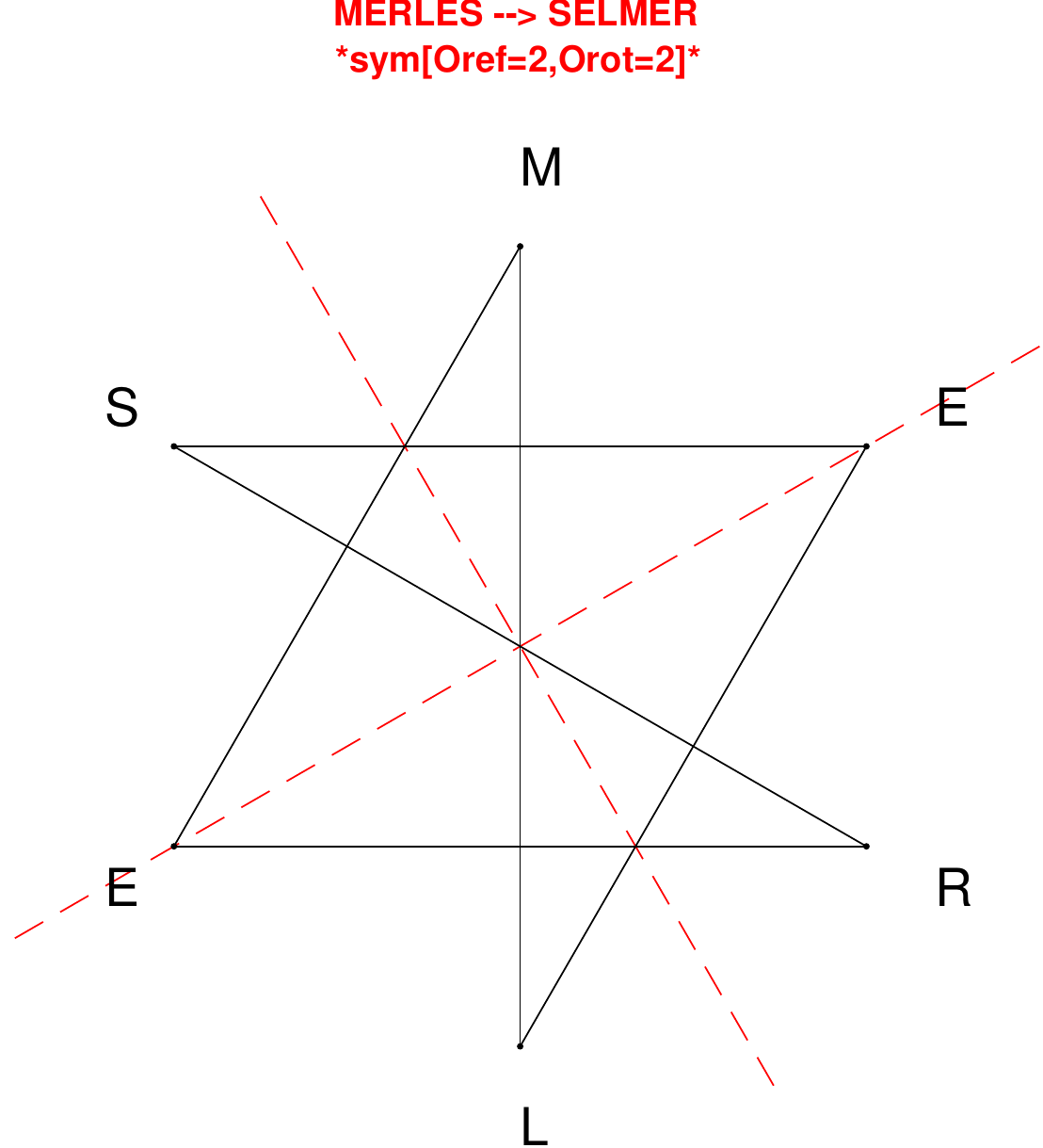}
\end{subfigure}
\end{figure}

\begin{figure}[H]
\centering
\begin{subfigure}[T]{0.19\textwidth}
\centering
\includegraphics[width=\textwidth]{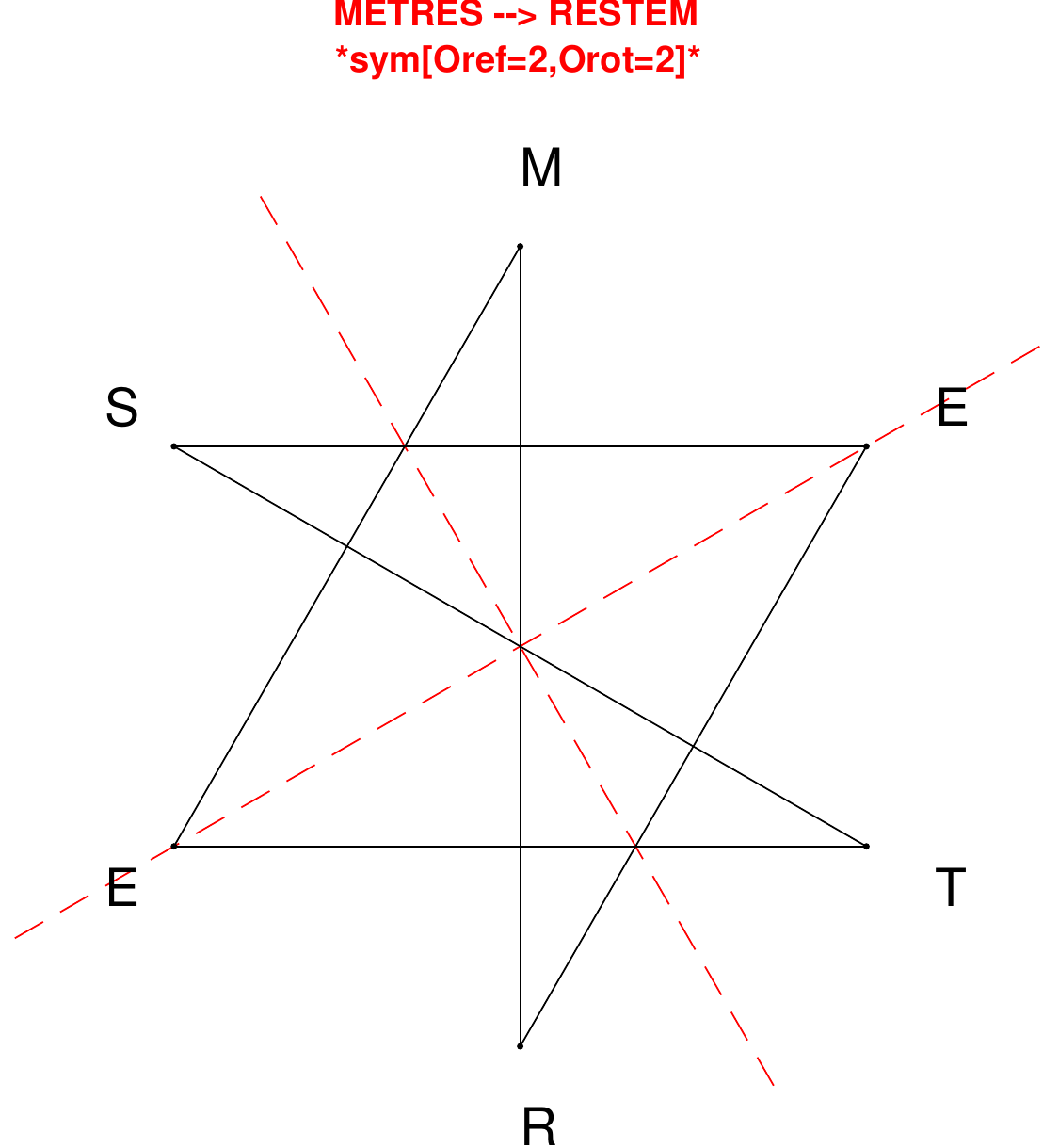}
\end{subfigure}
\hfill
\begin{subfigure}[T]{0.19\textwidth}
\centering
\includegraphics[width=\textwidth]{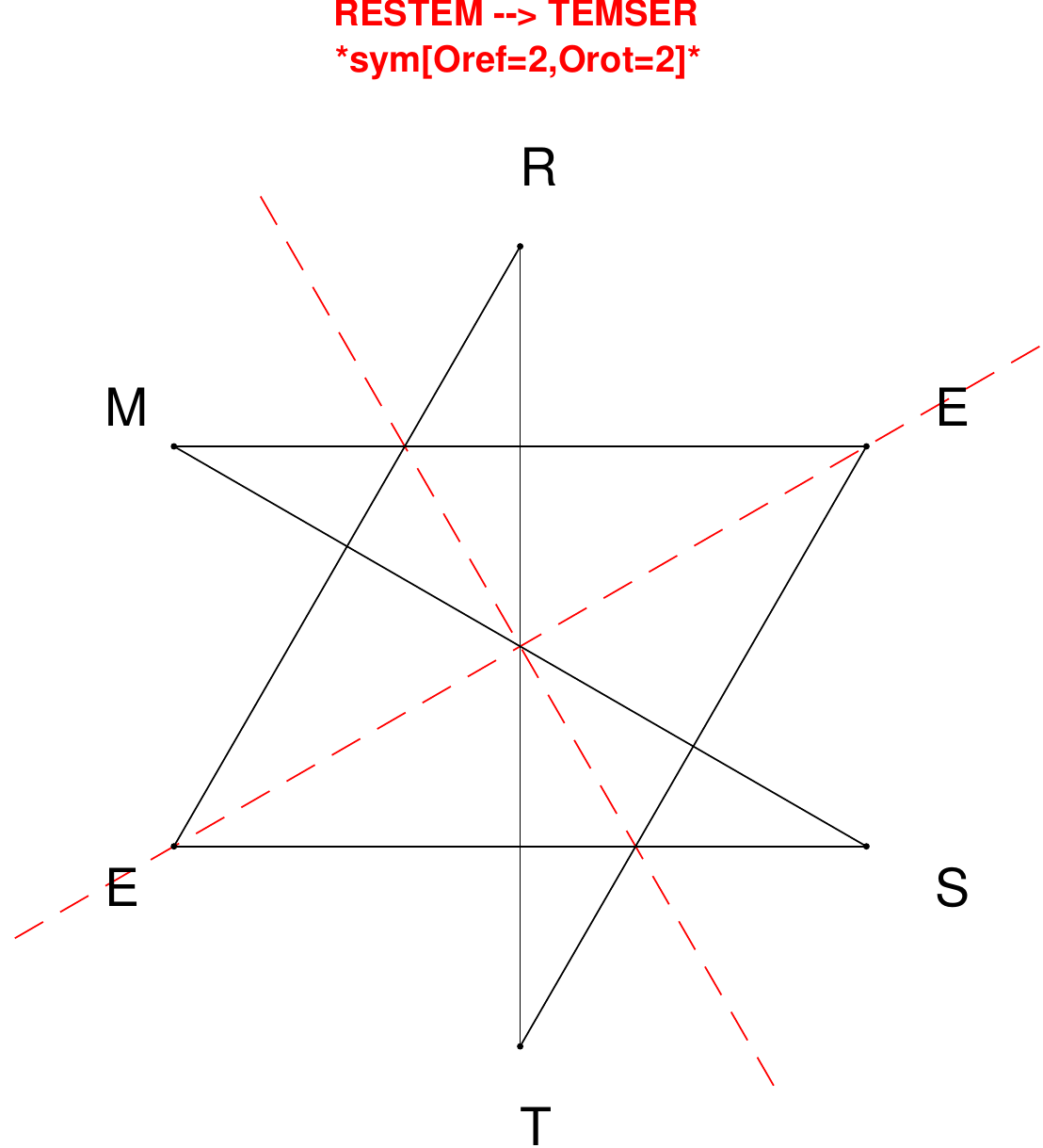}
\end{subfigure}
\hfill
\begin{subfigure}[T]{0.19\textwidth}
\centering
\includegraphics[width=\textwidth]{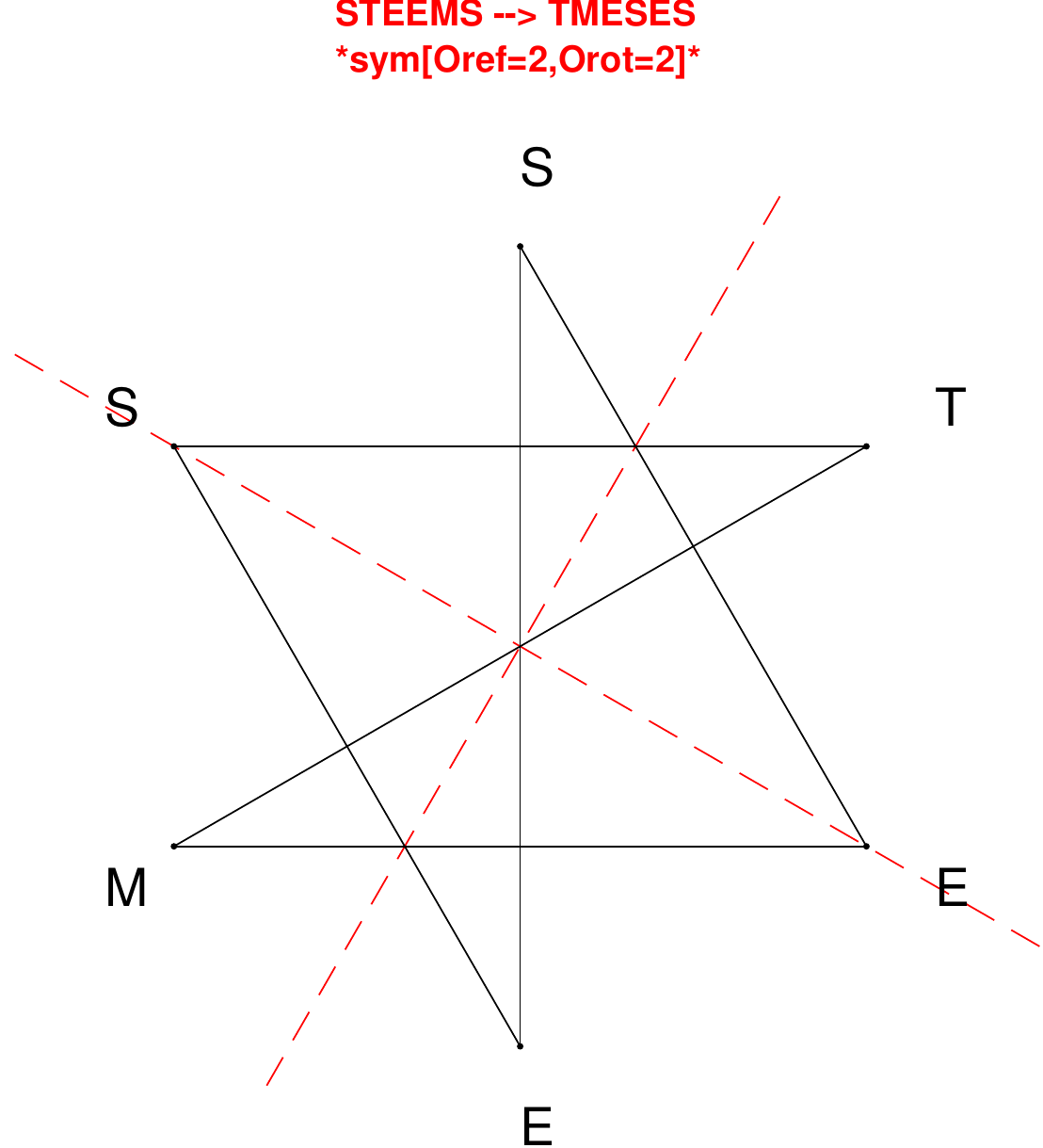}
\end{subfigure}
\hfill
\begin{subfigure}[T]{0.19\textwidth}
\centering
\includegraphics[width=\textwidth]{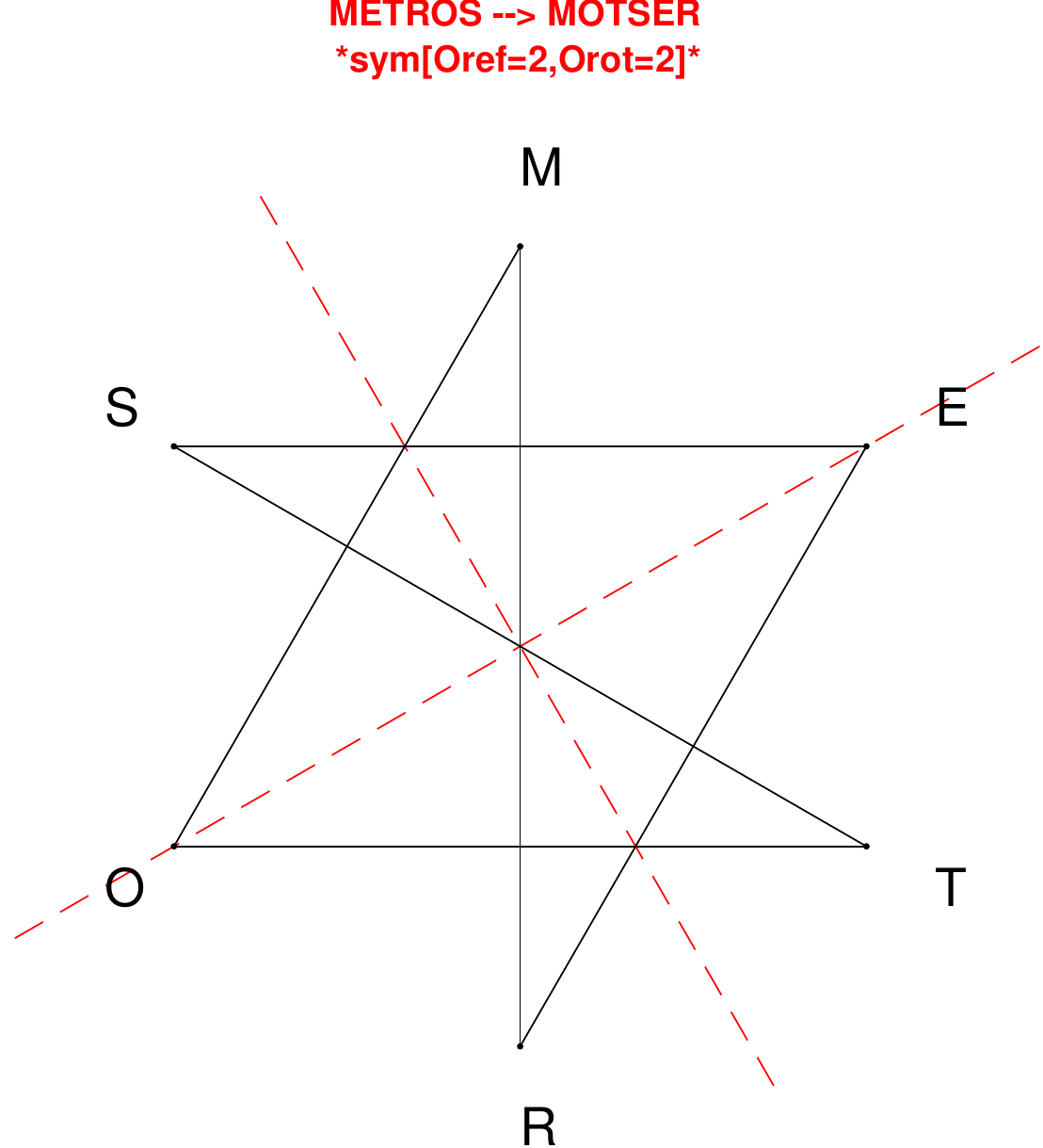}
\end{subfigure}
\hfill
\begin{subfigure}[T]{0.19\textwidth}
\centering
\includegraphics[width=\textwidth]{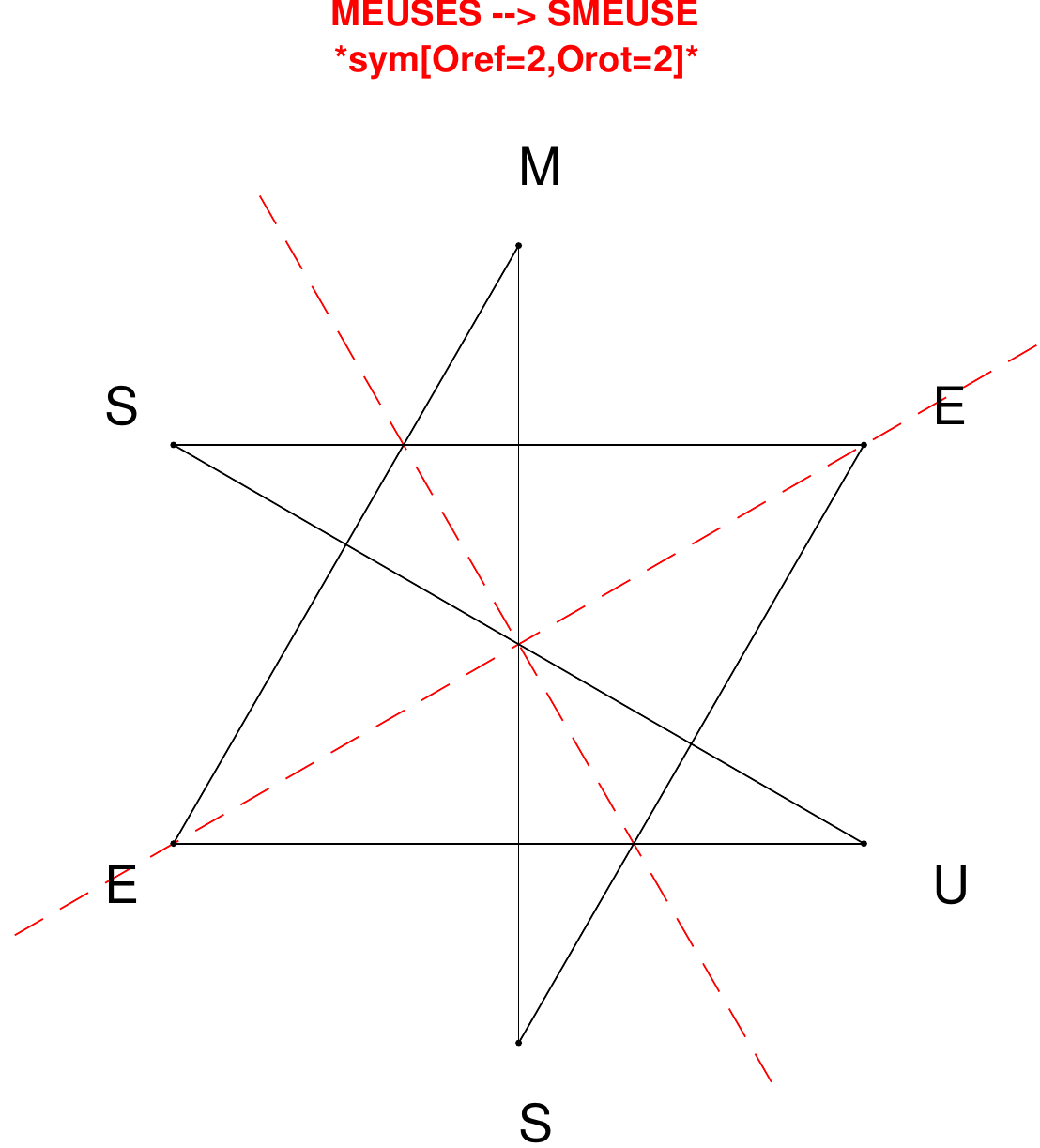}
\end{subfigure}
\end{figure}

\begin{figure}[H]
\centering
\begin{subfigure}[T]{0.19\textwidth}
\centering
\includegraphics[width=\textwidth]{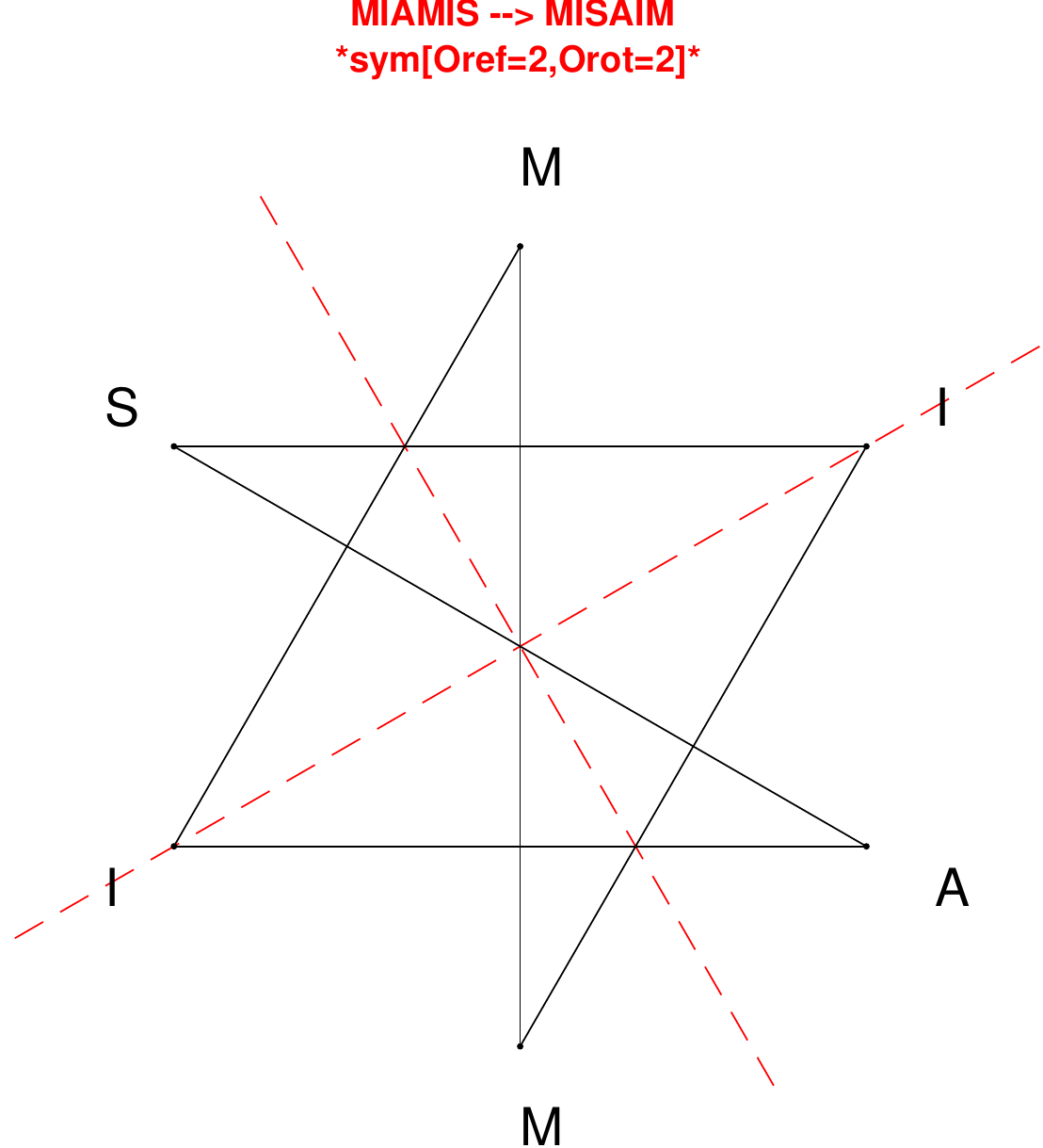}
\end{subfigure}
\hfill
\begin{subfigure}[T]{0.19\textwidth}
\centering
\includegraphics[width=\textwidth]{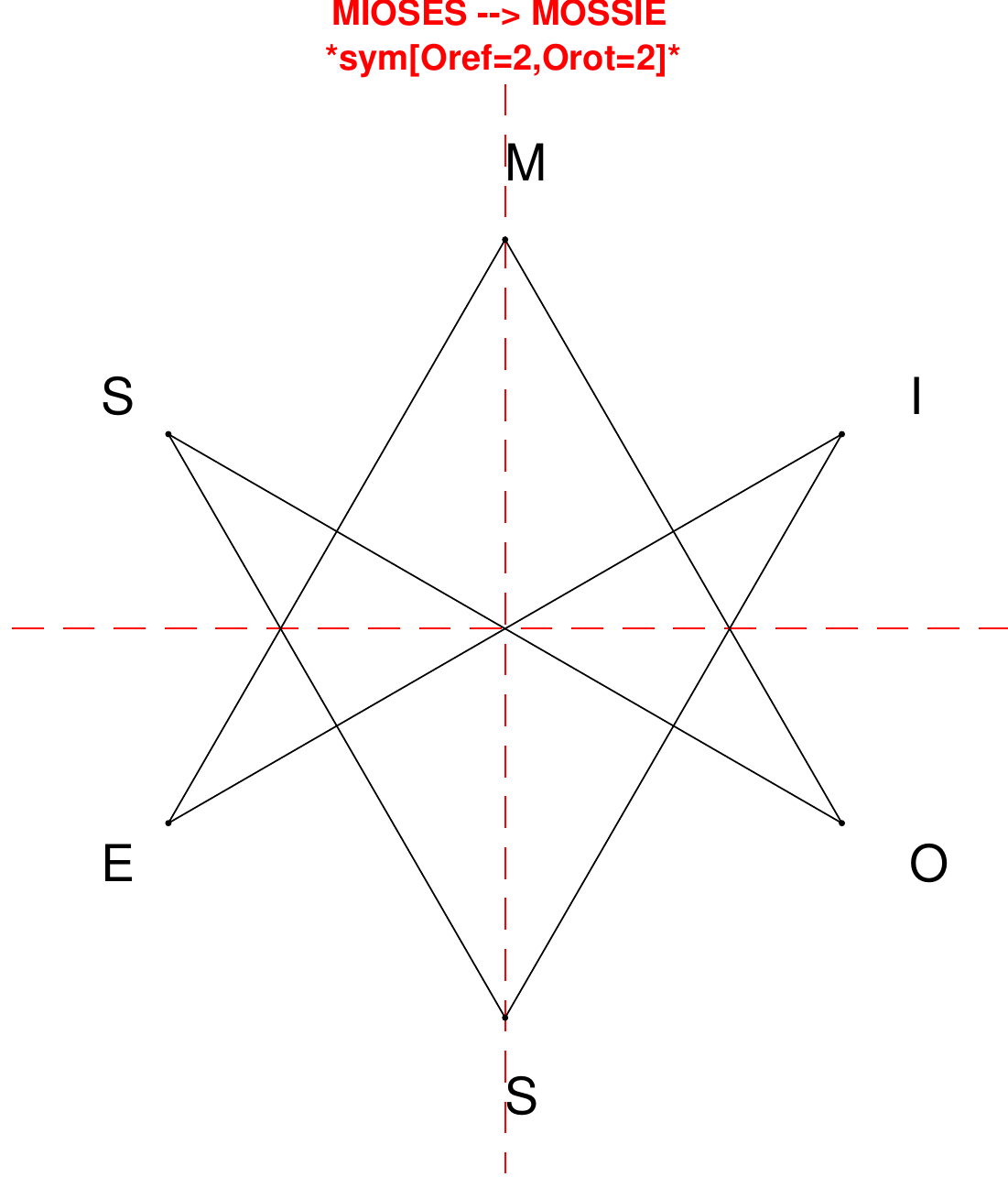}
\end{subfigure}
\hfill
\begin{subfigure}[T]{0.19\textwidth}
\centering
\includegraphics[width=\textwidth]{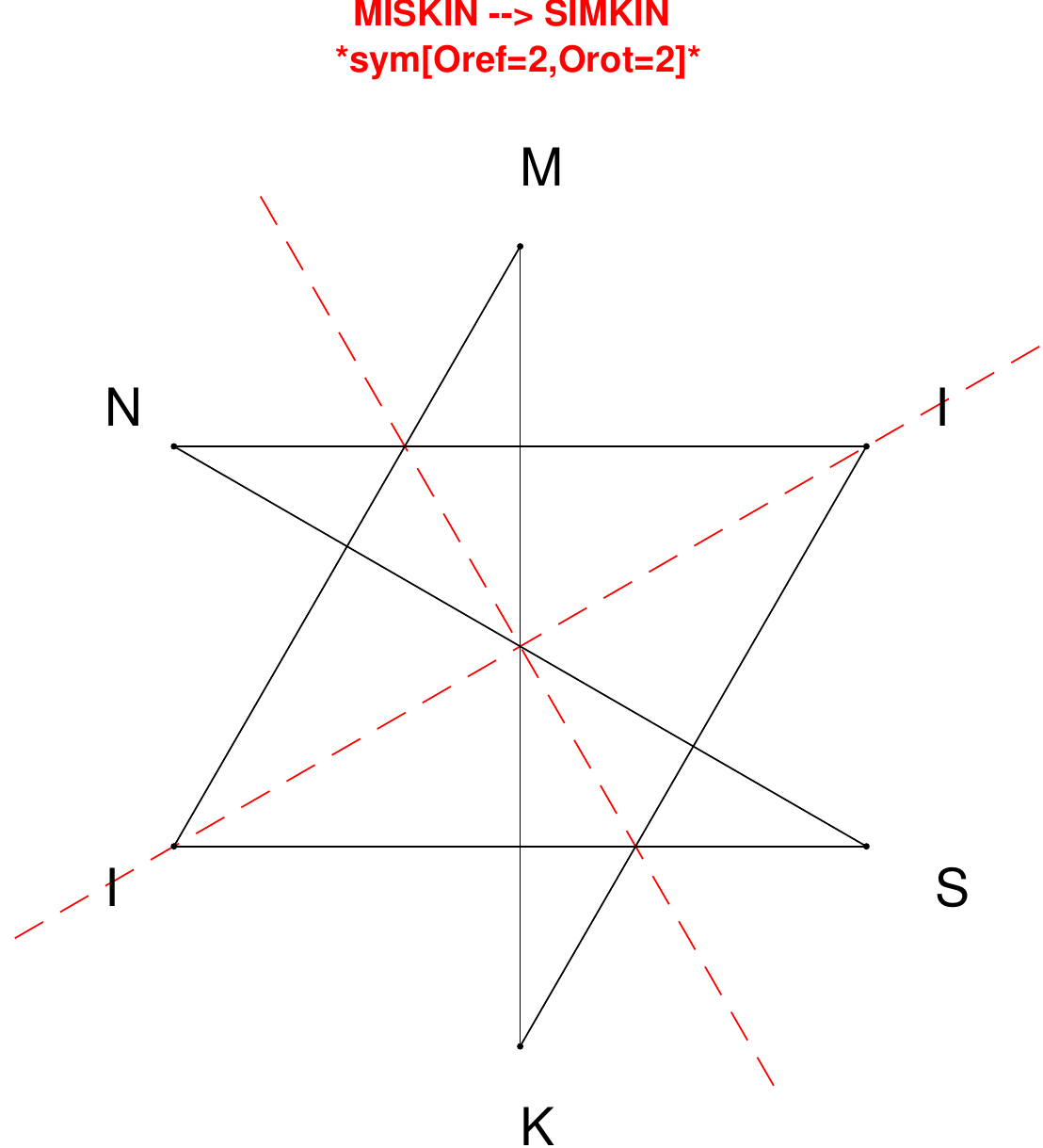}
\end{subfigure}
\hfill
\begin{subfigure}[T]{0.19\textwidth}
\centering
\includegraphics[width=\textwidth]{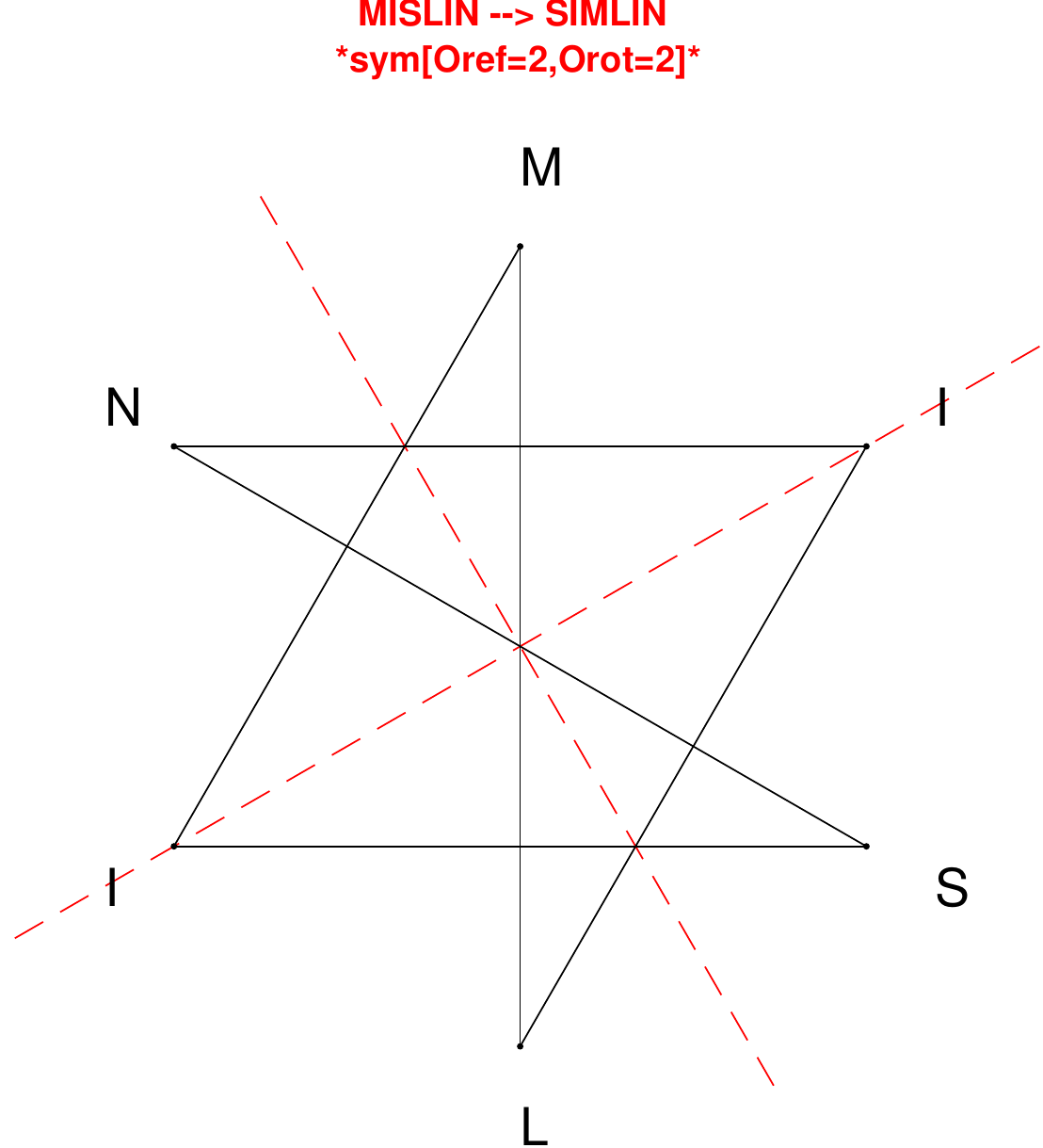}
\end{subfigure}
\hfill
\begin{subfigure}[T]{0.19\textwidth}
\centering
\includegraphics[width=\textwidth]{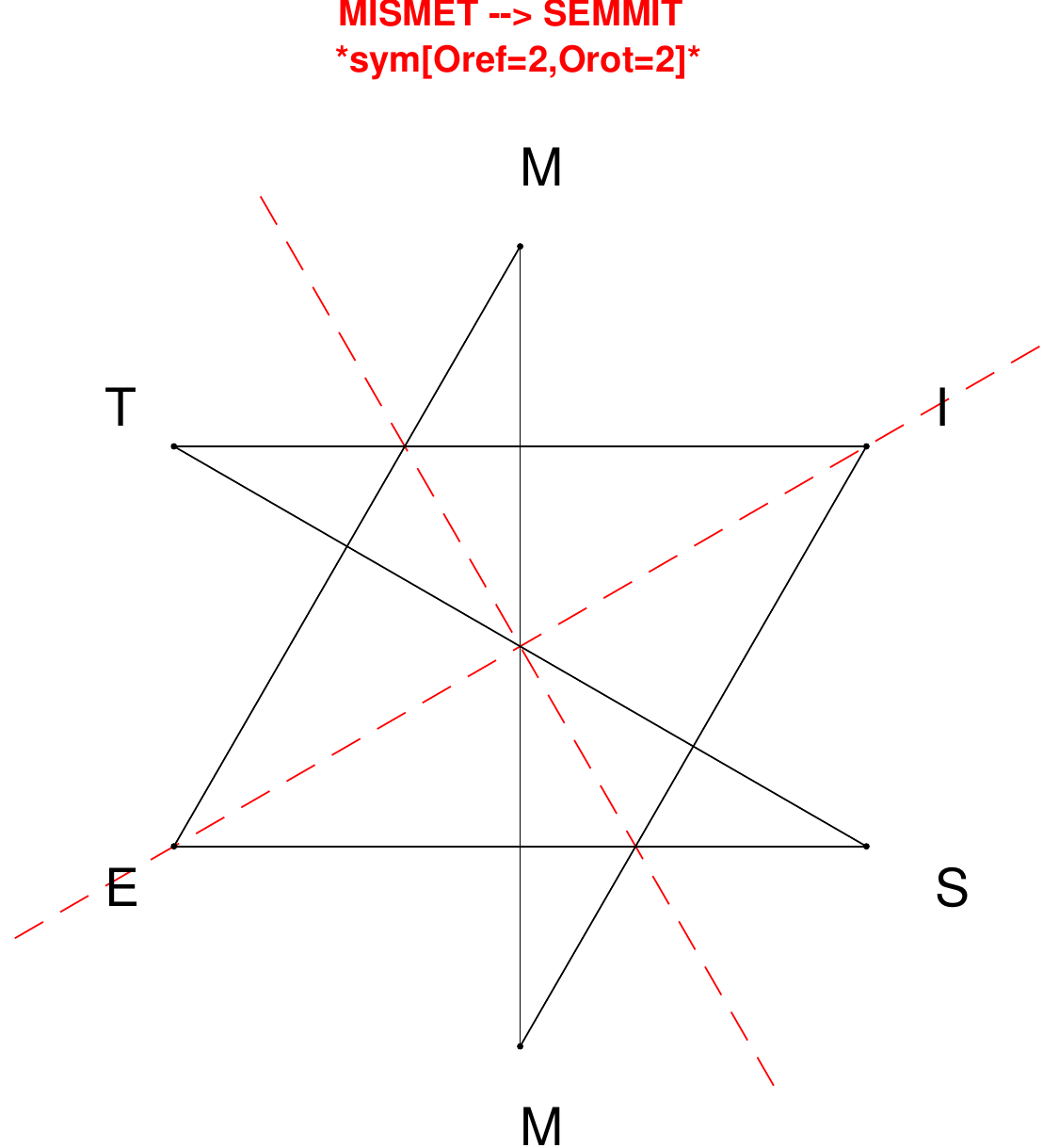}
\end{subfigure}
\end{figure}

\begin{figure}[H]
\centering
\begin{subfigure}[T]{0.19\textwidth}
\centering
\includegraphics[width=\textwidth]{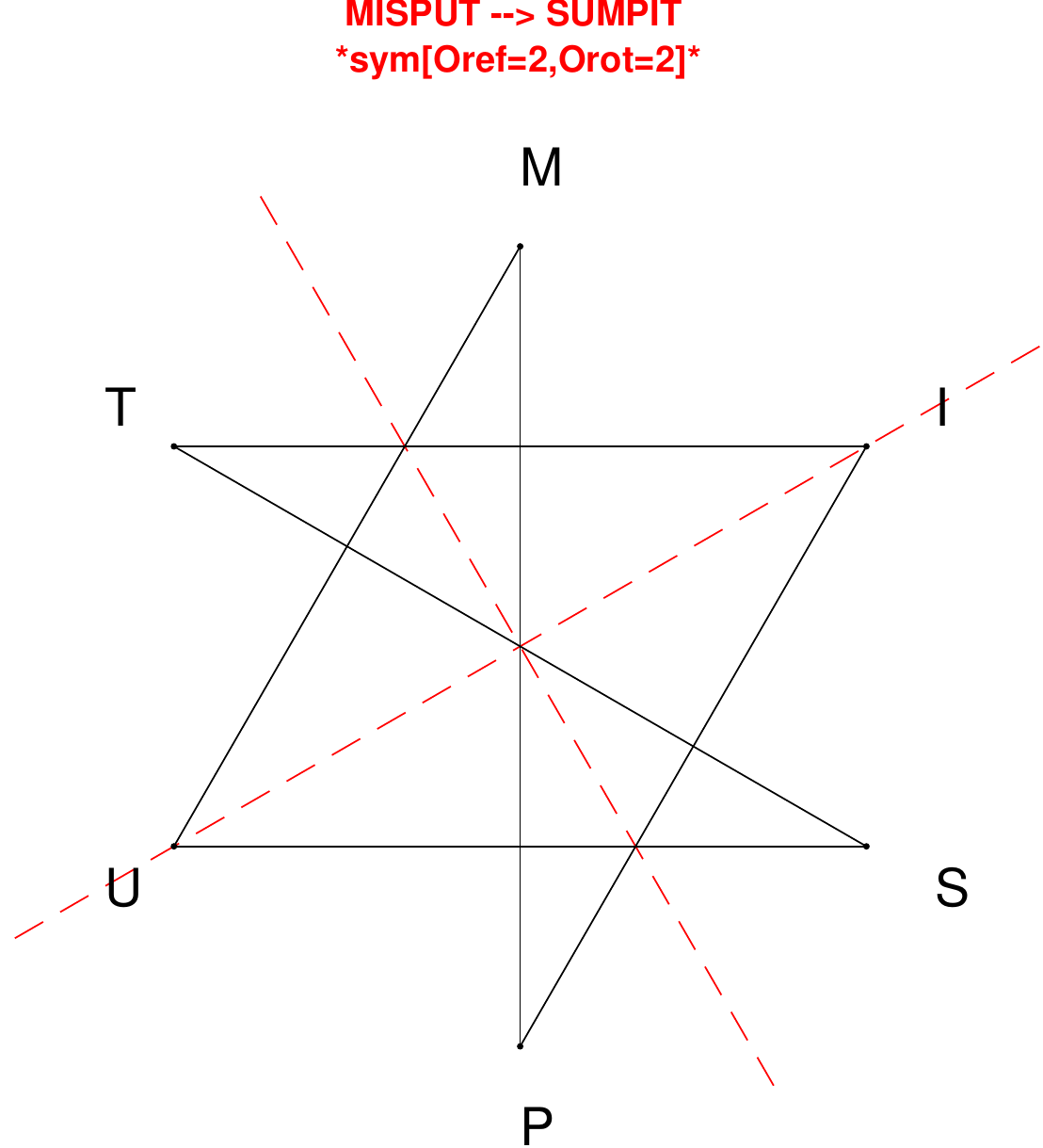}
\end{subfigure}
\hfill
\begin{subfigure}[T]{0.19\textwidth}
\centering
\includegraphics[width=\textwidth]{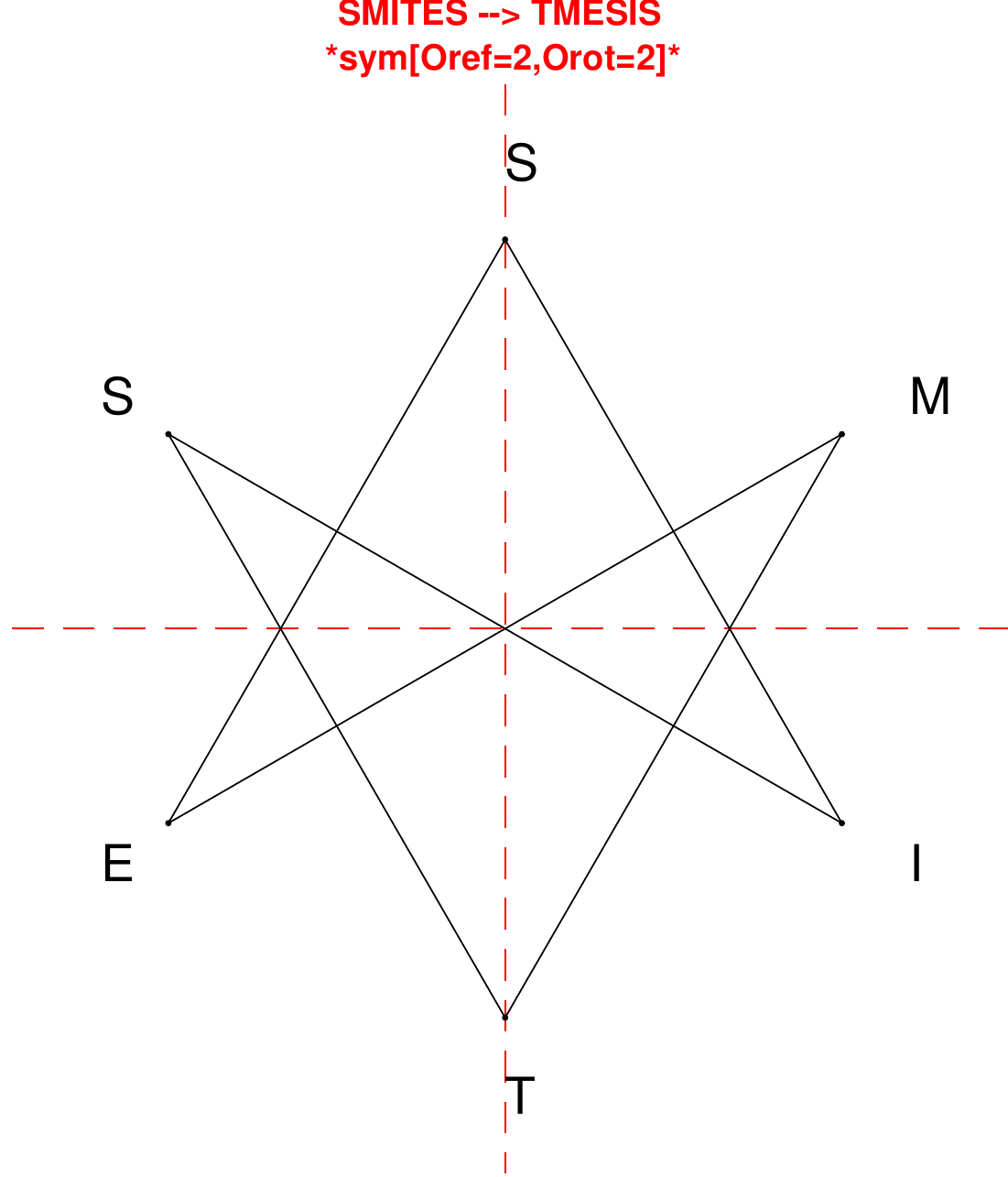}
\end{subfigure}
\hfill
\begin{subfigure}[T]{0.19\textwidth}
\centering
\includegraphics[width=\textwidth]{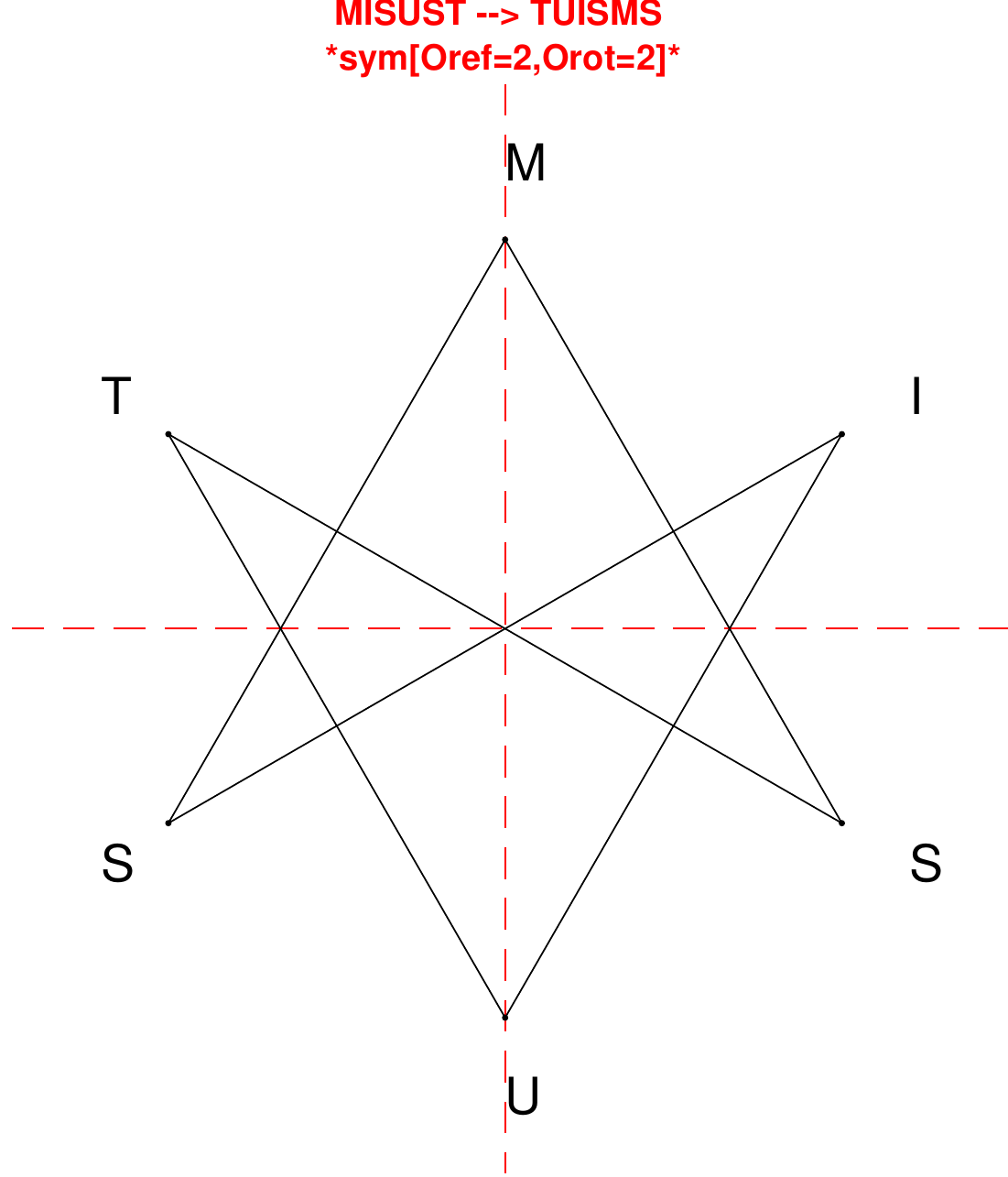}
\end{subfigure}
\hfill
\begin{subfigure}[T]{0.19\textwidth}
\centering
\includegraphics[width=\textwidth]{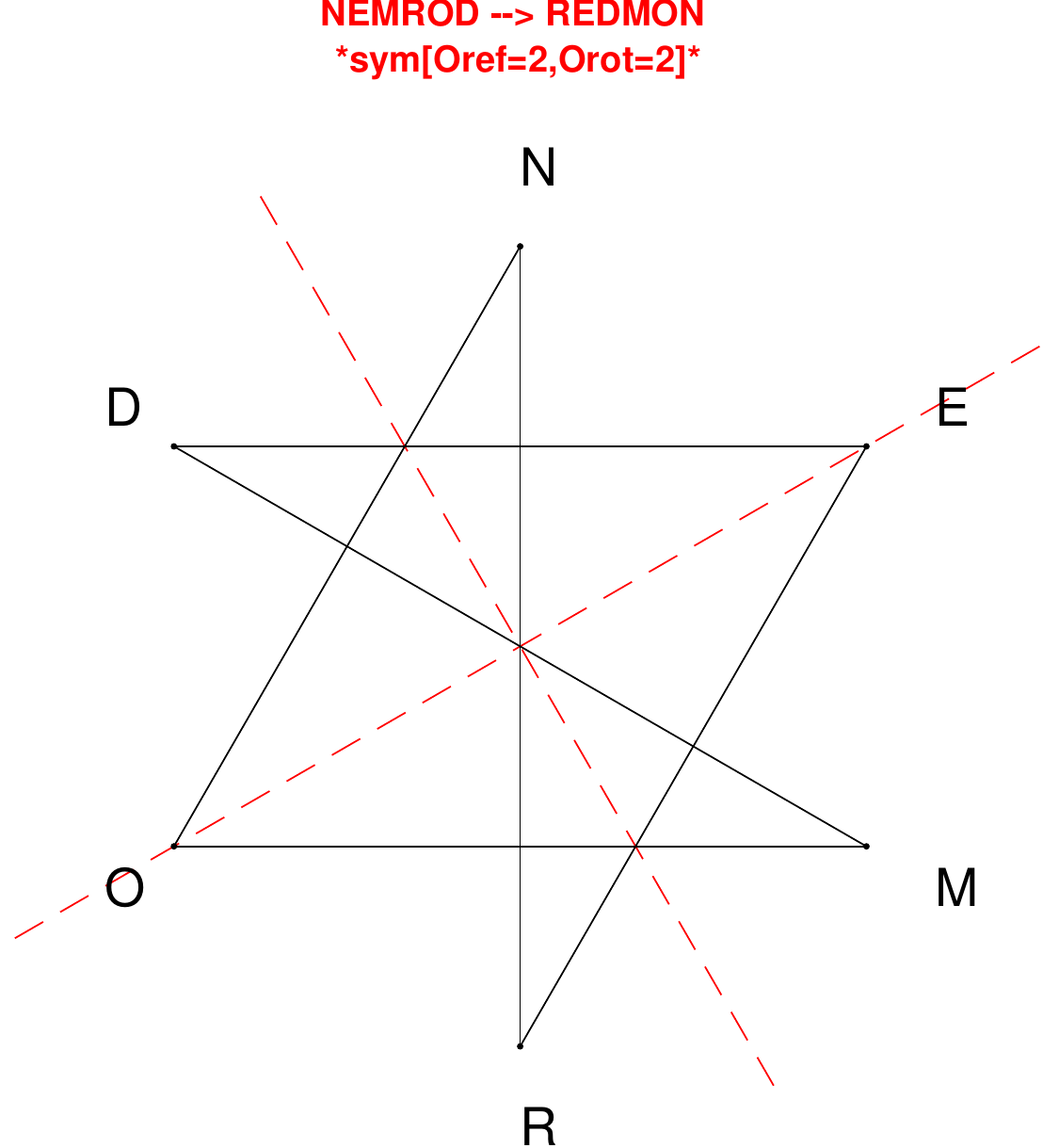}
\end{subfigure}
\hfill
\begin{subfigure}[T]{0.19\textwidth}
\centering
\includegraphics[width=\textwidth]{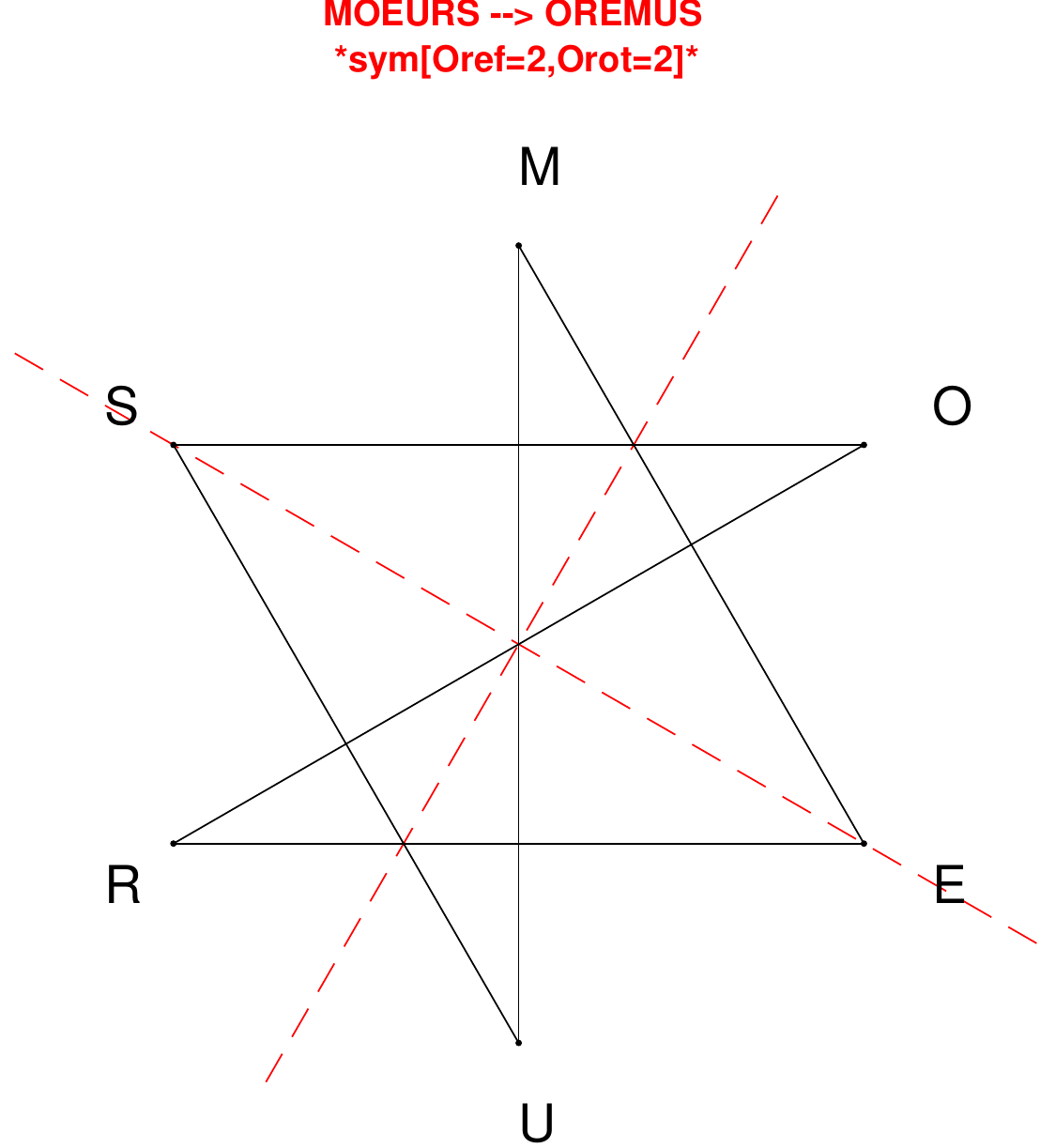}
\end{subfigure}
\end{figure}

\begin{figure}[H]
\centering
\begin{subfigure}[T]{0.19\textwidth}
\centering
\includegraphics[width=\textwidth]{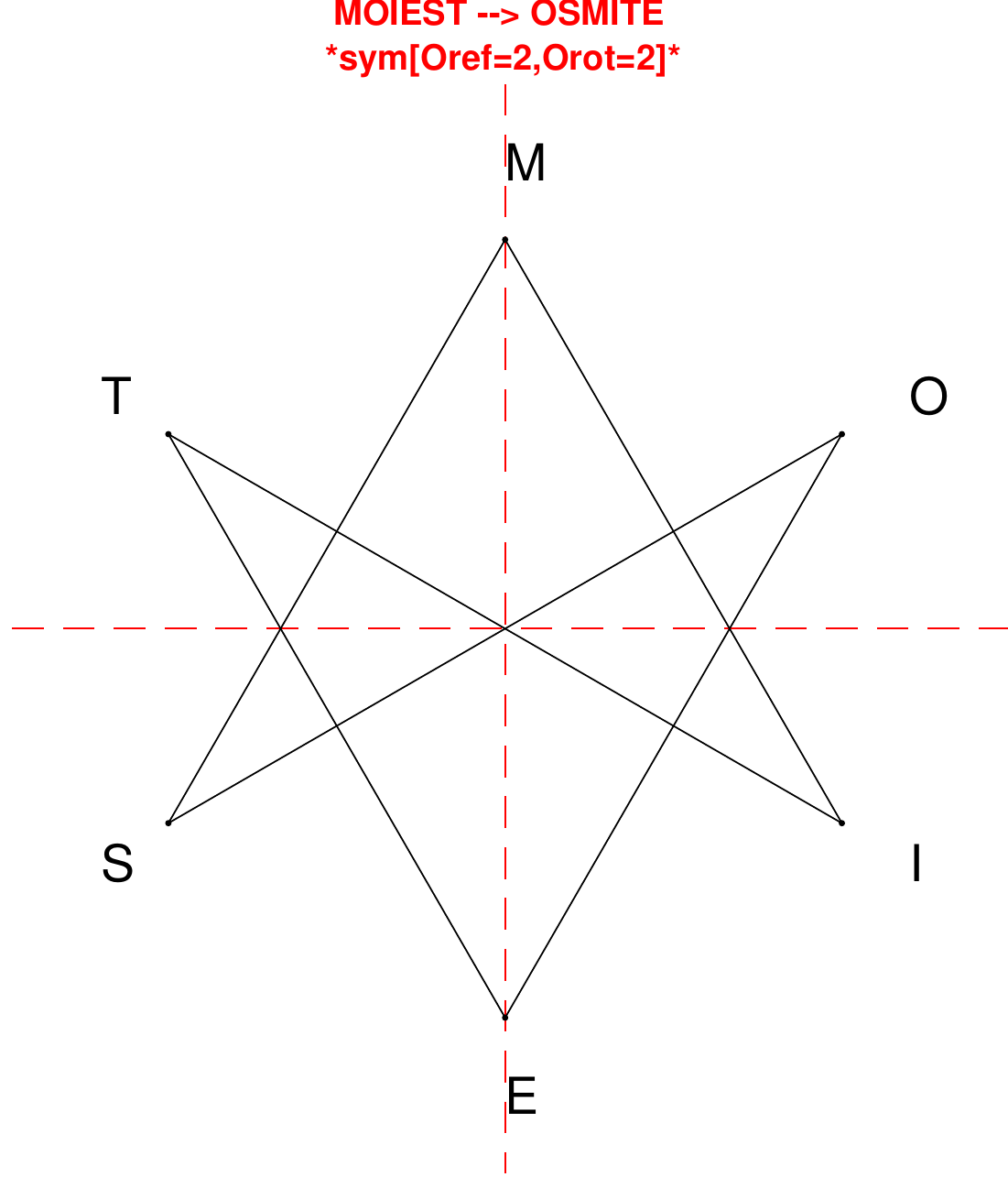}
\end{subfigure}
\hfill
\begin{subfigure}[T]{0.19\textwidth}
\centering
\includegraphics[width=\textwidth]{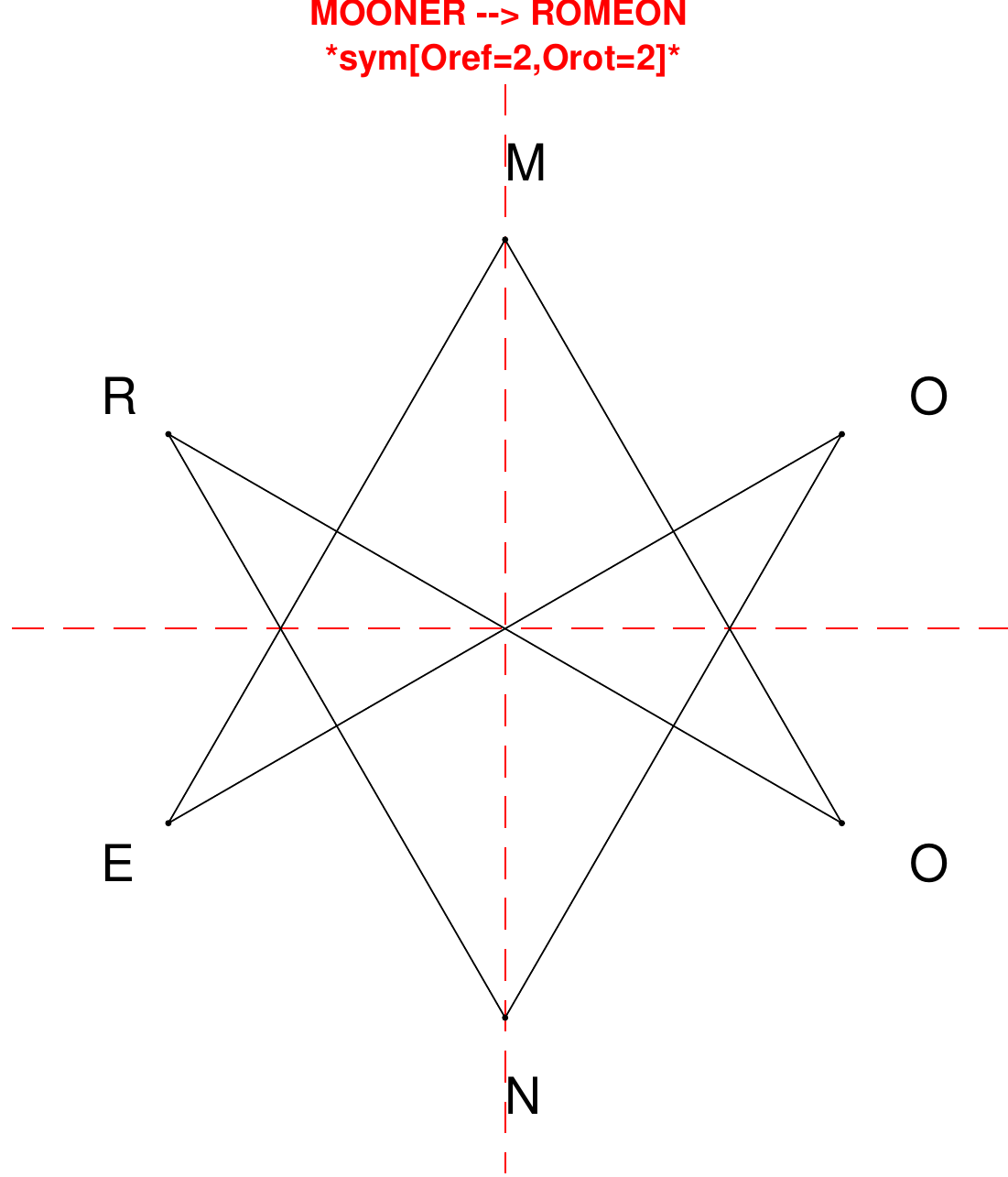}
\end{subfigure}
\hfill
\begin{subfigure}[T]{0.19\textwidth}
\centering
\includegraphics[width=\textwidth]{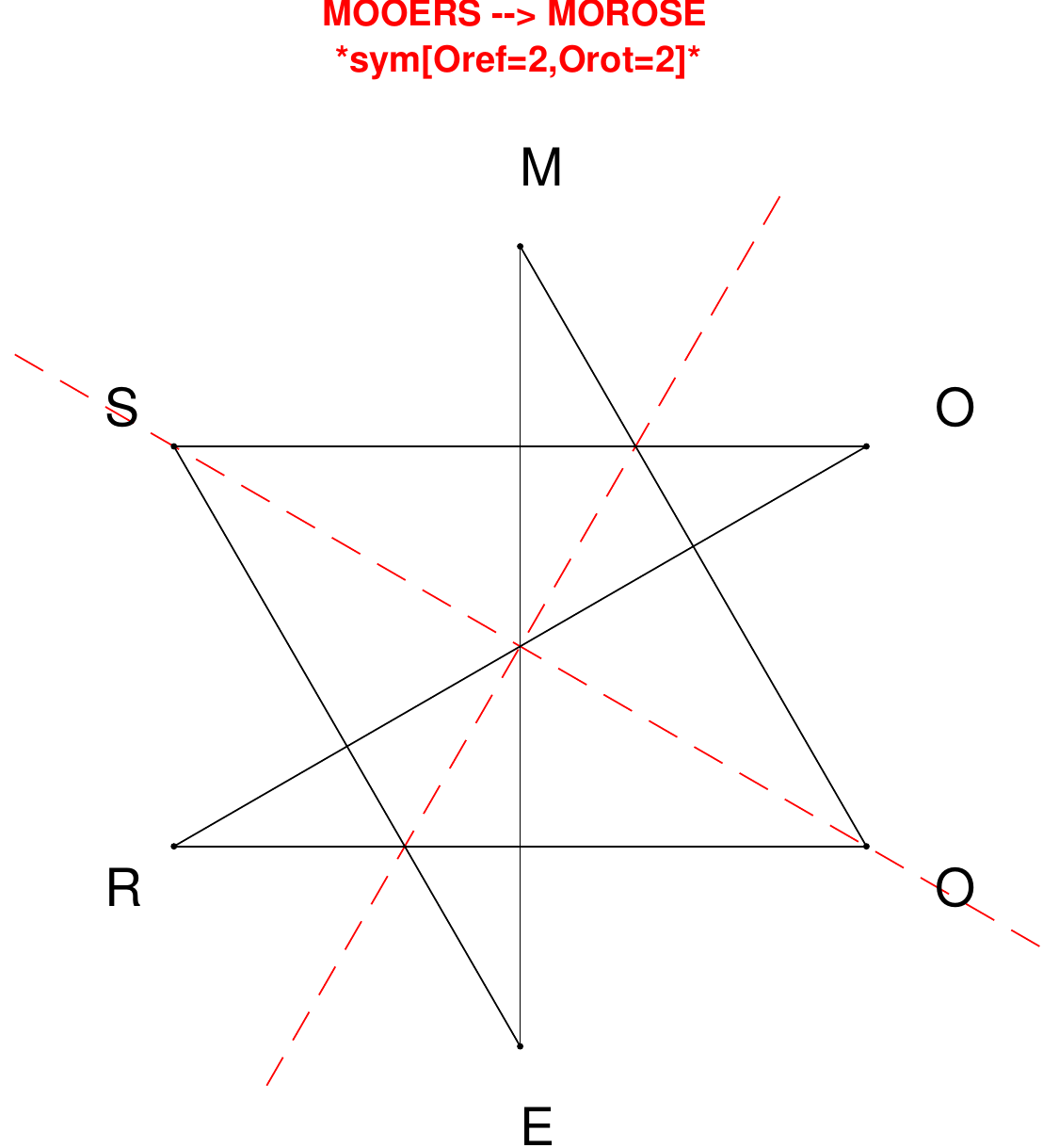}
\end{subfigure}
\hfill
\begin{subfigure}[T]{0.19\textwidth}
\centering
\includegraphics[width=\textwidth]{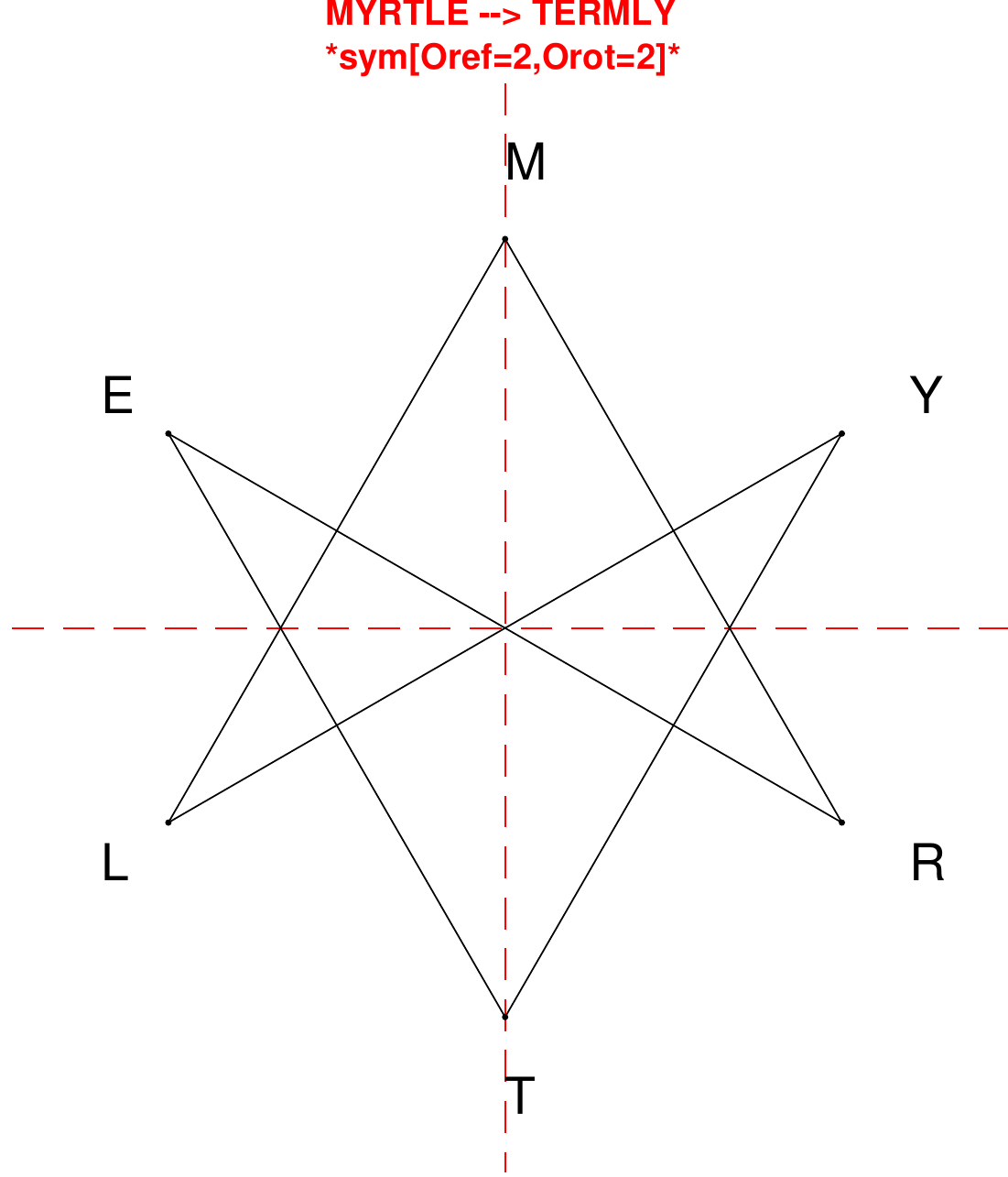}
\end{subfigure}
\hfill
\begin{subfigure}[T]{0.19\textwidth}
\centering
\includegraphics[width=\textwidth]{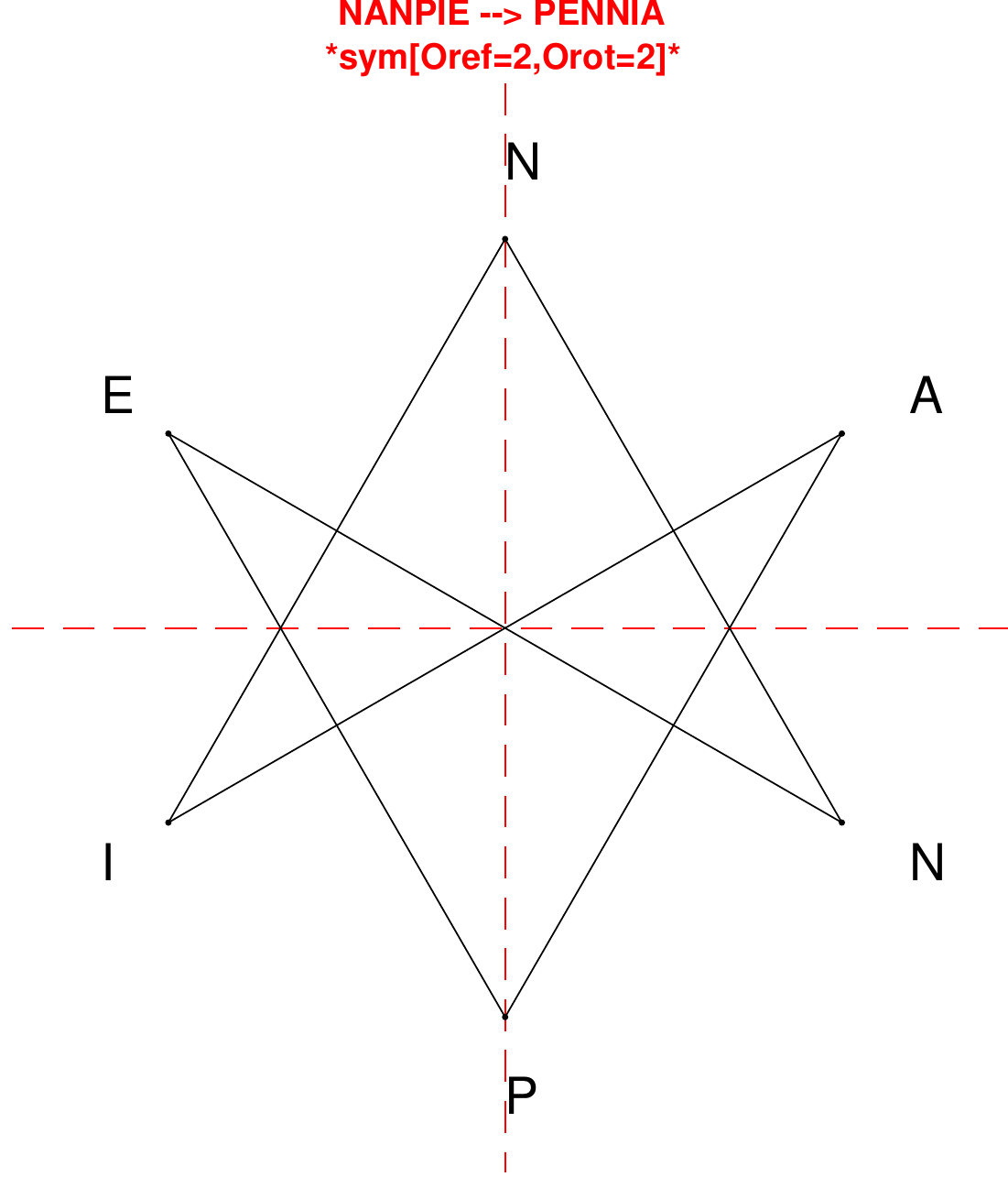}
\end{subfigure}
\end{figure}

\begin{figure}[H]
\centering
\begin{subfigure}[T]{0.19\textwidth}
\centering
\includegraphics[width=\textwidth]{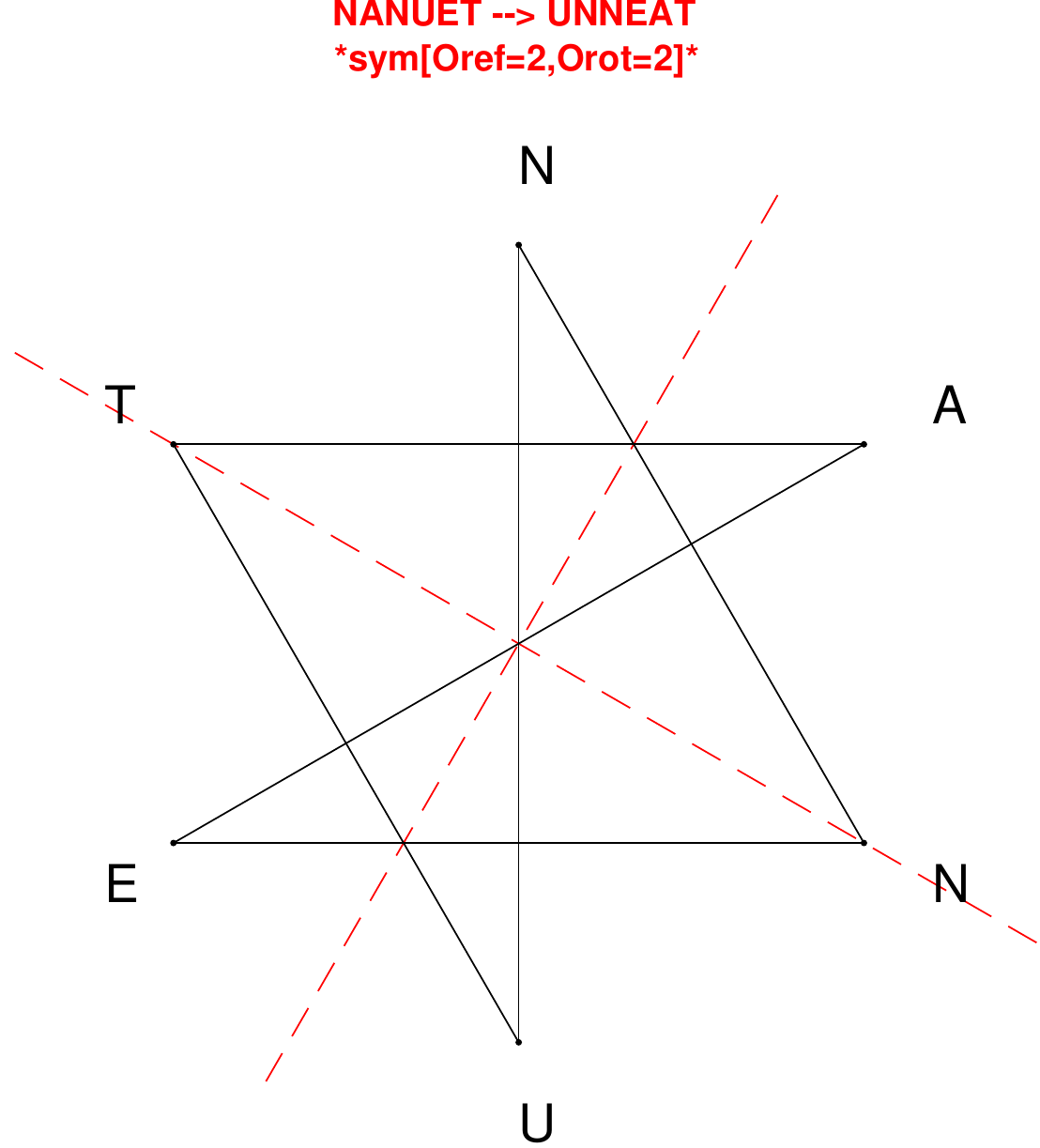}
\end{subfigure}
\hfill
\begin{subfigure}[T]{0.19\textwidth}
\centering
\includegraphics[width=\textwidth]{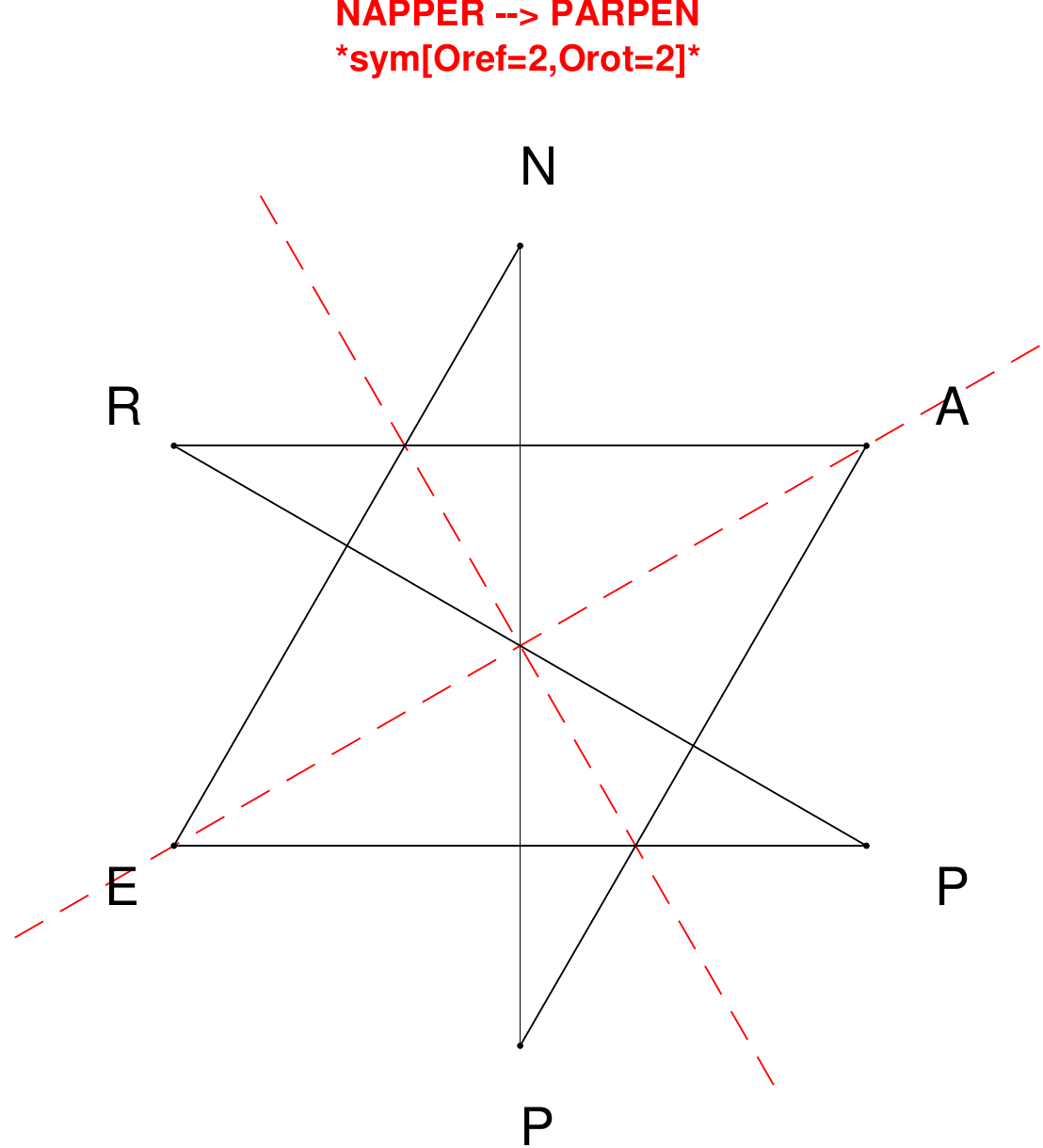}
\end{subfigure}
\hfill
\begin{subfigure}[T]{0.19\textwidth}
\centering
\includegraphics[width=\textwidth]{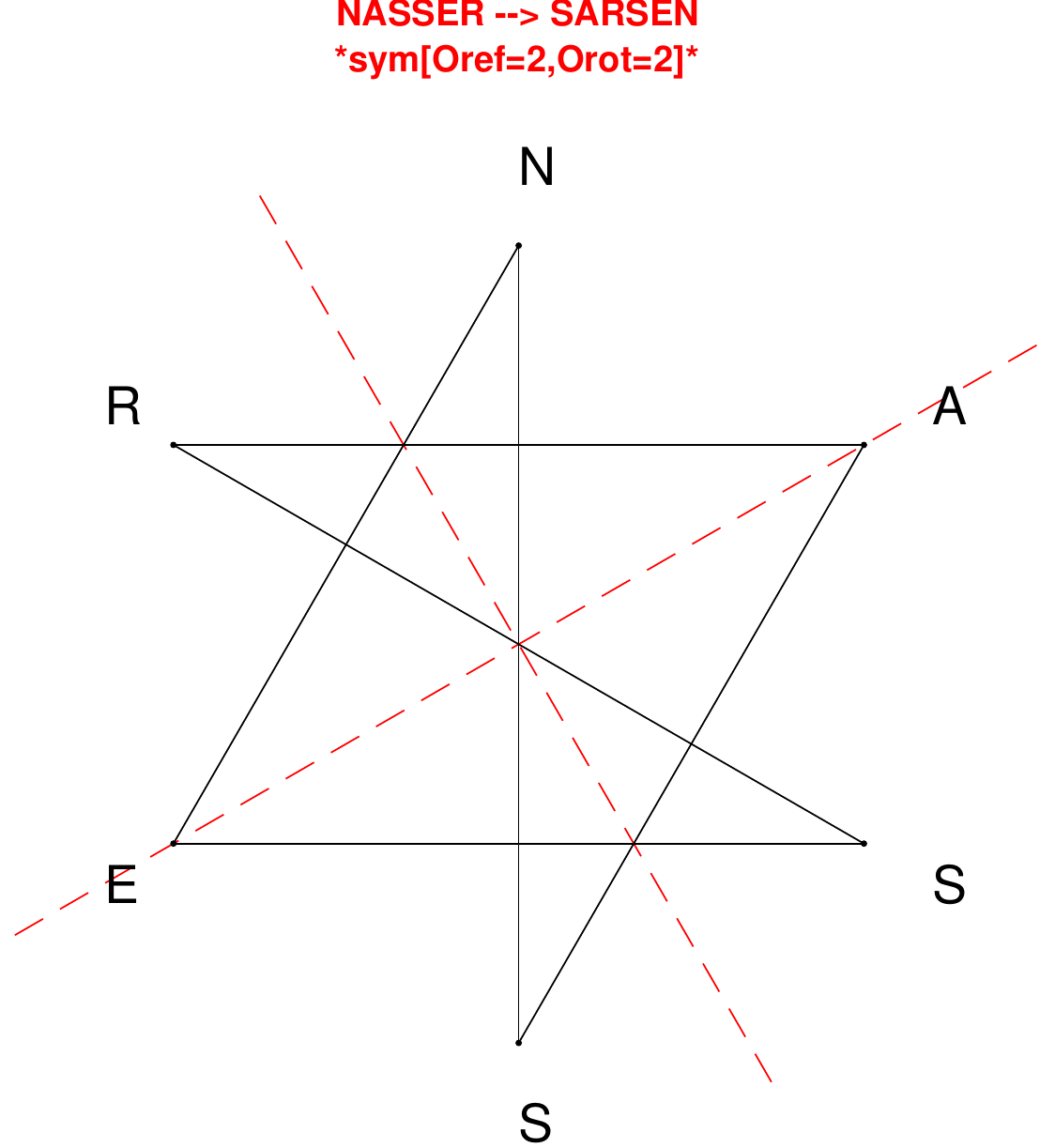}
\end{subfigure}
\hfill
\begin{subfigure}[T]{0.19\textwidth}
\centering
\includegraphics[width=\textwidth]{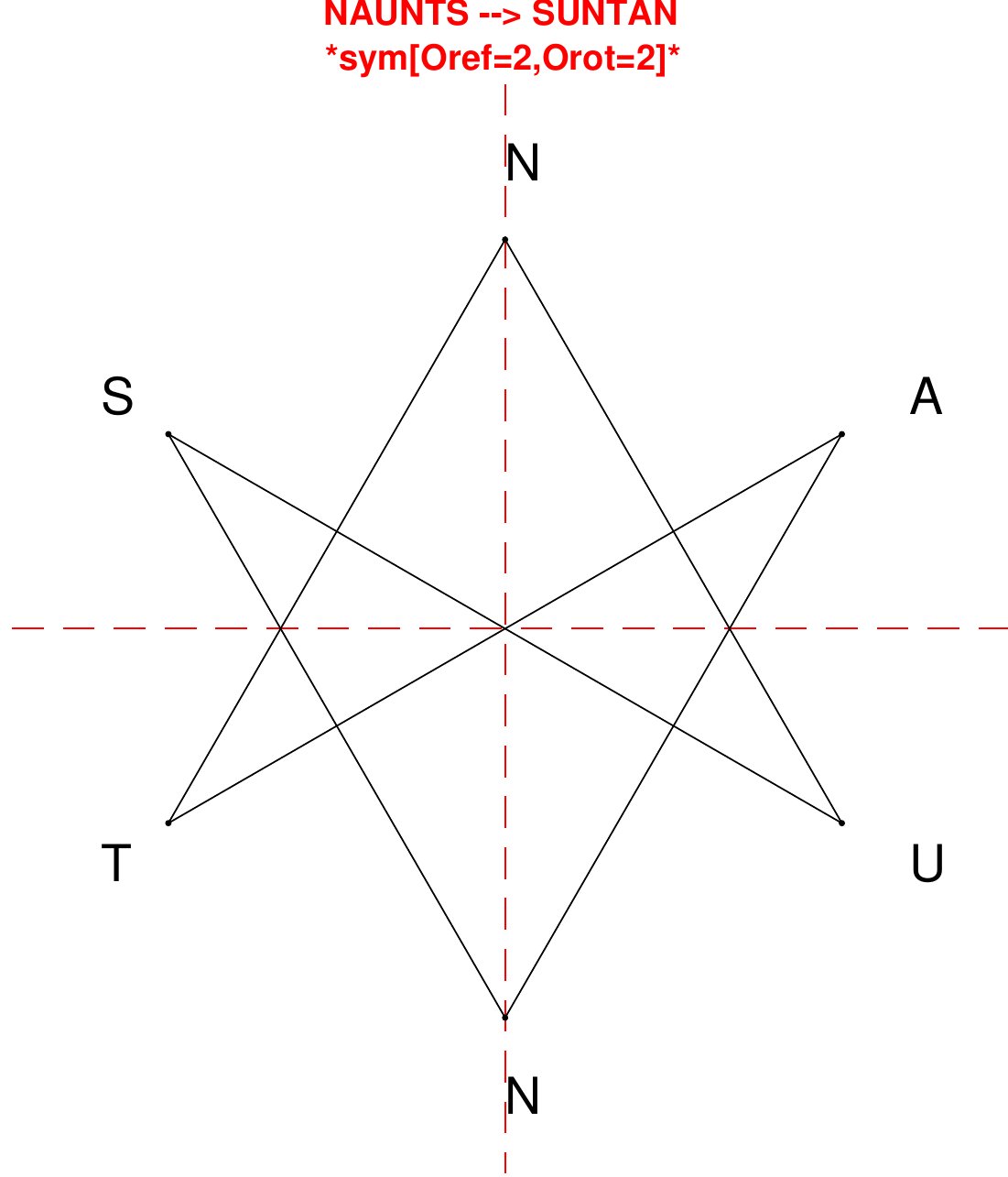}
\end{subfigure}
\hfill
\begin{subfigure}[T]{0.19\textwidth}
\centering
\includegraphics[width=\textwidth]{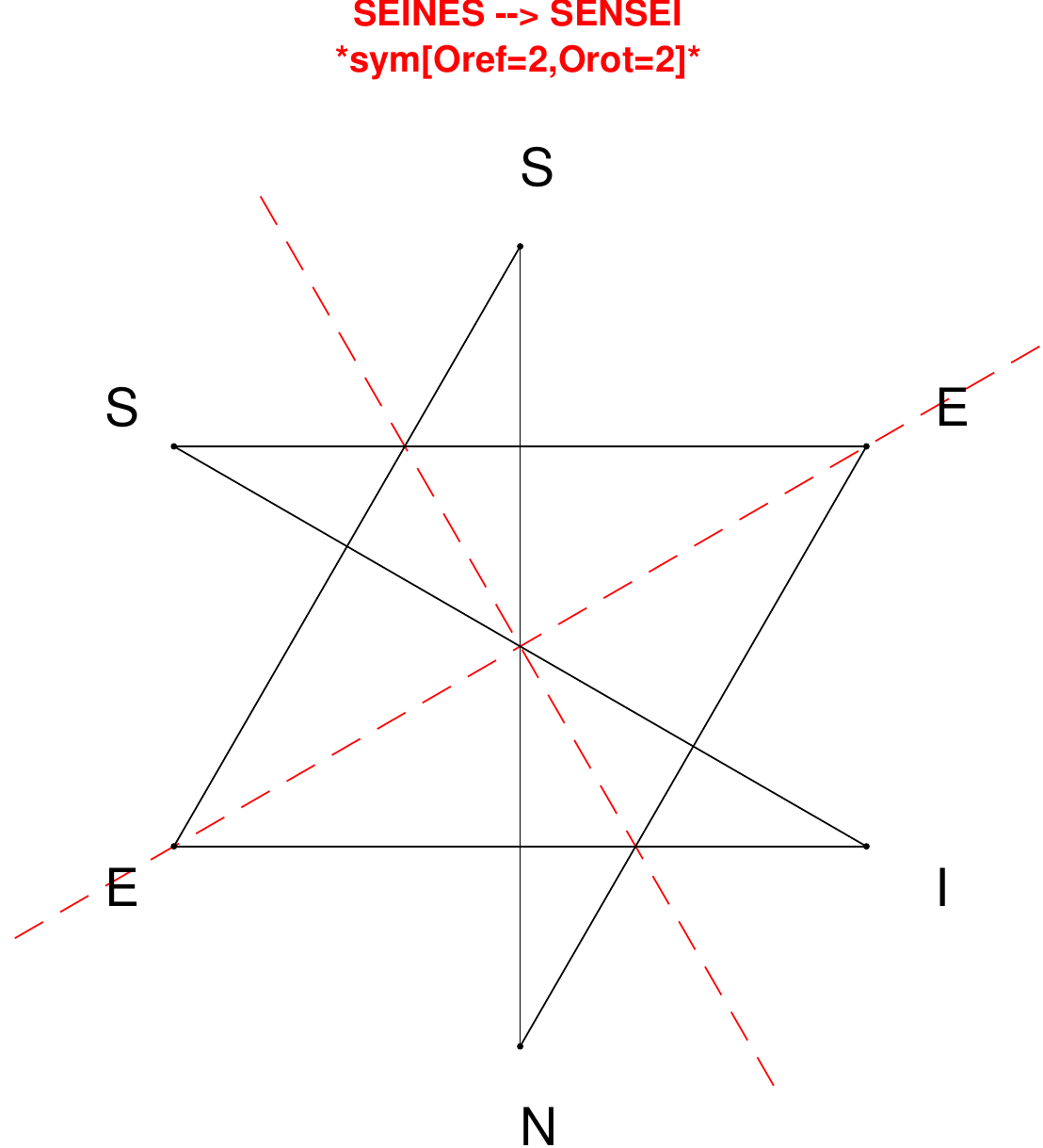}
\end{subfigure}
\end{figure}

\begin{figure}[H]
\centering
\begin{subfigure}[T]{0.19\textwidth}
\centering
\includegraphics[width=\textwidth]{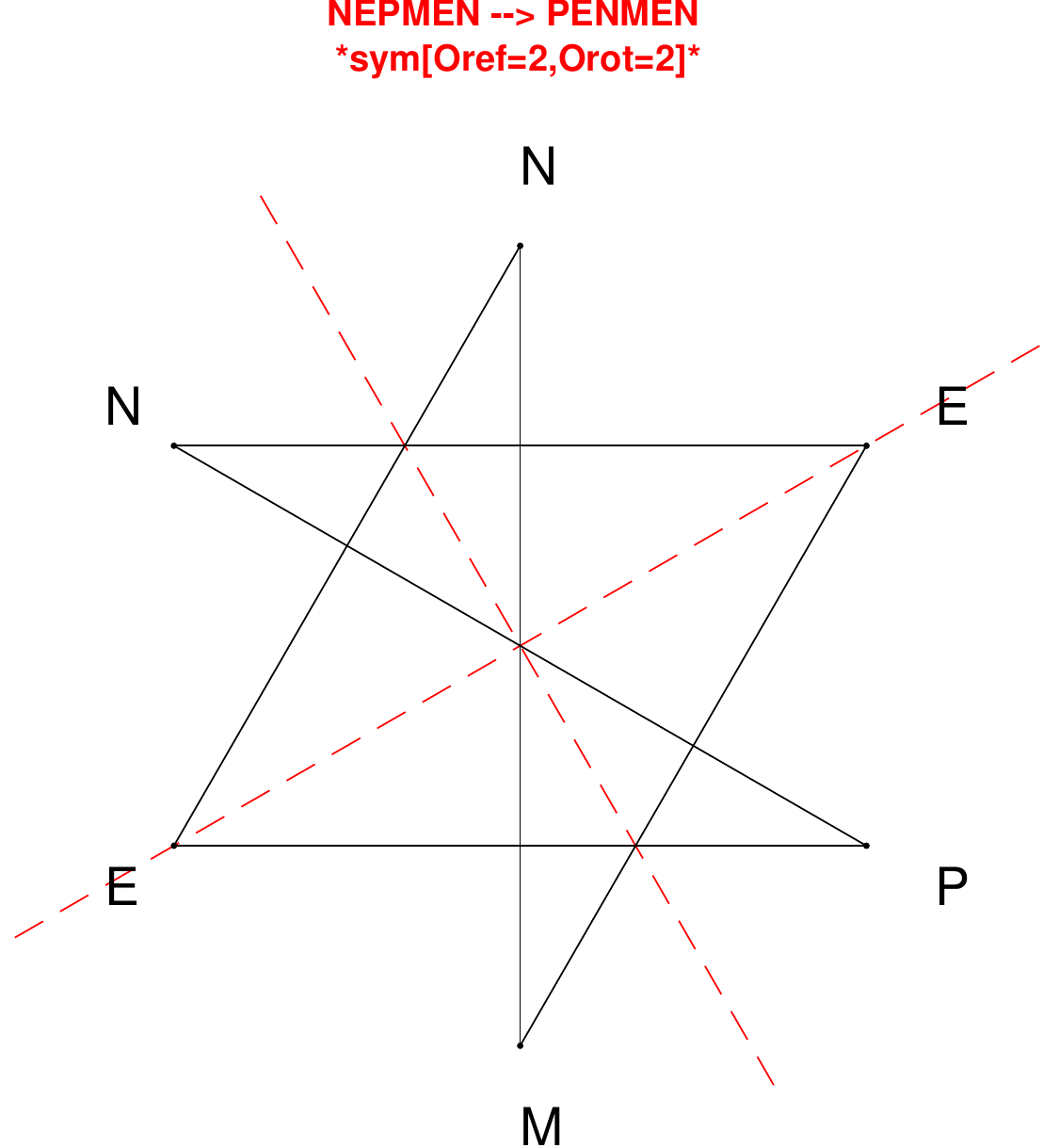}
\end{subfigure}
\hfill
\begin{subfigure}[T]{0.19\textwidth}
\centering
\includegraphics[width=\textwidth]{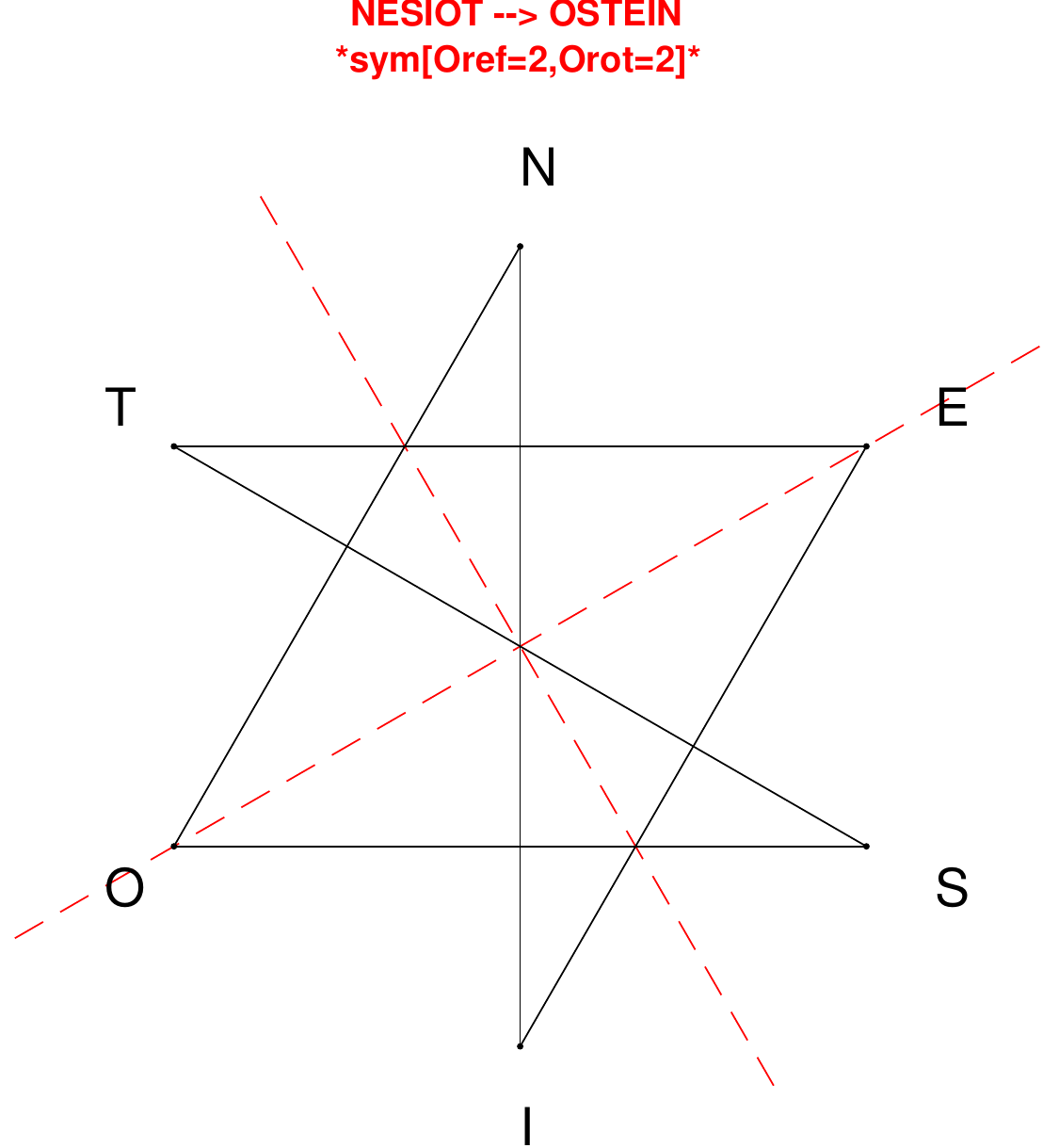}
\end{subfigure}
\hfill
\begin{subfigure}[T]{0.19\textwidth}
\centering
\includegraphics[width=\textwidth]{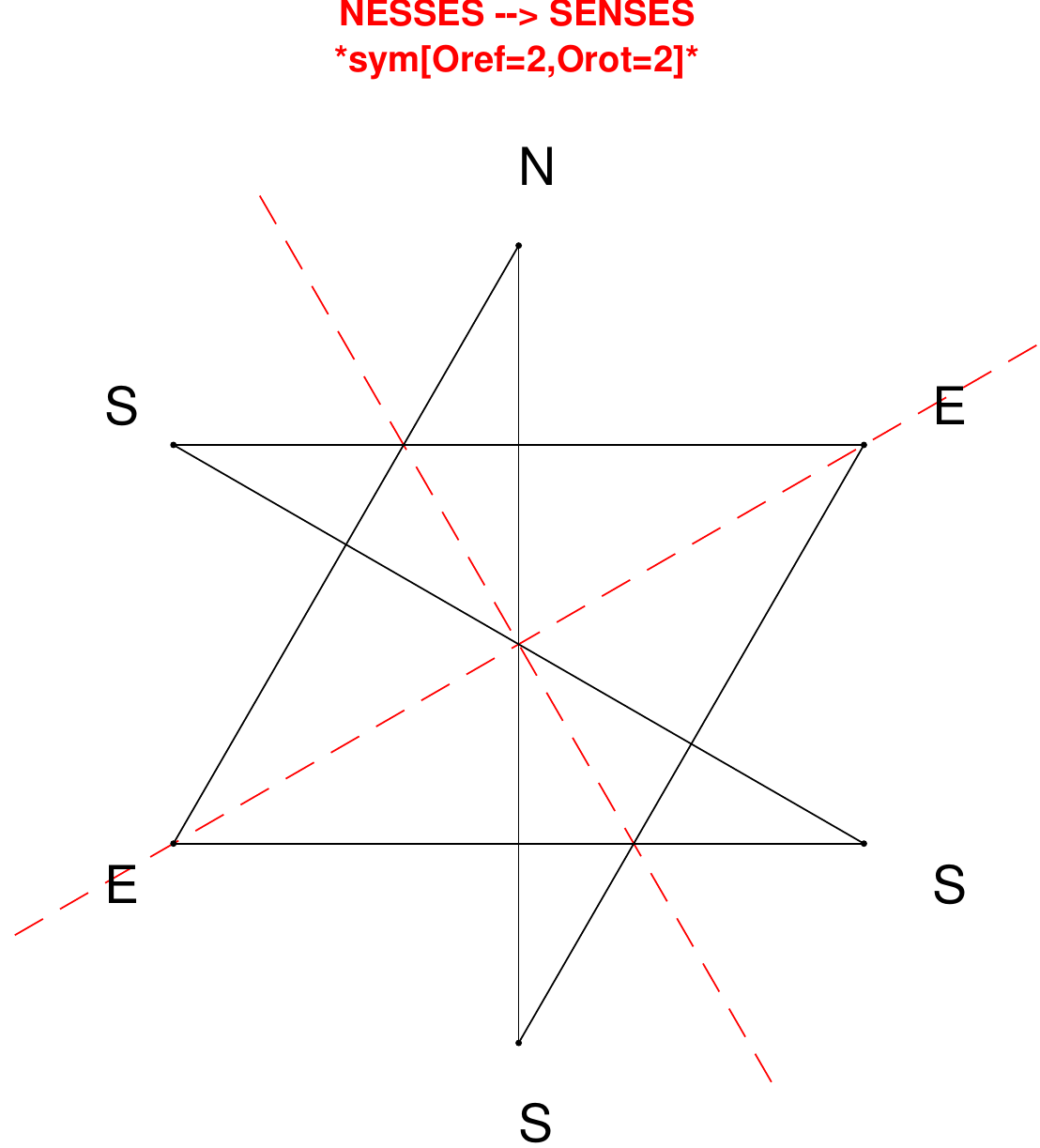}
\end{subfigure}
\hfill
\begin{subfigure}[T]{0.19\textwidth}
\centering
\includegraphics[width=\textwidth]{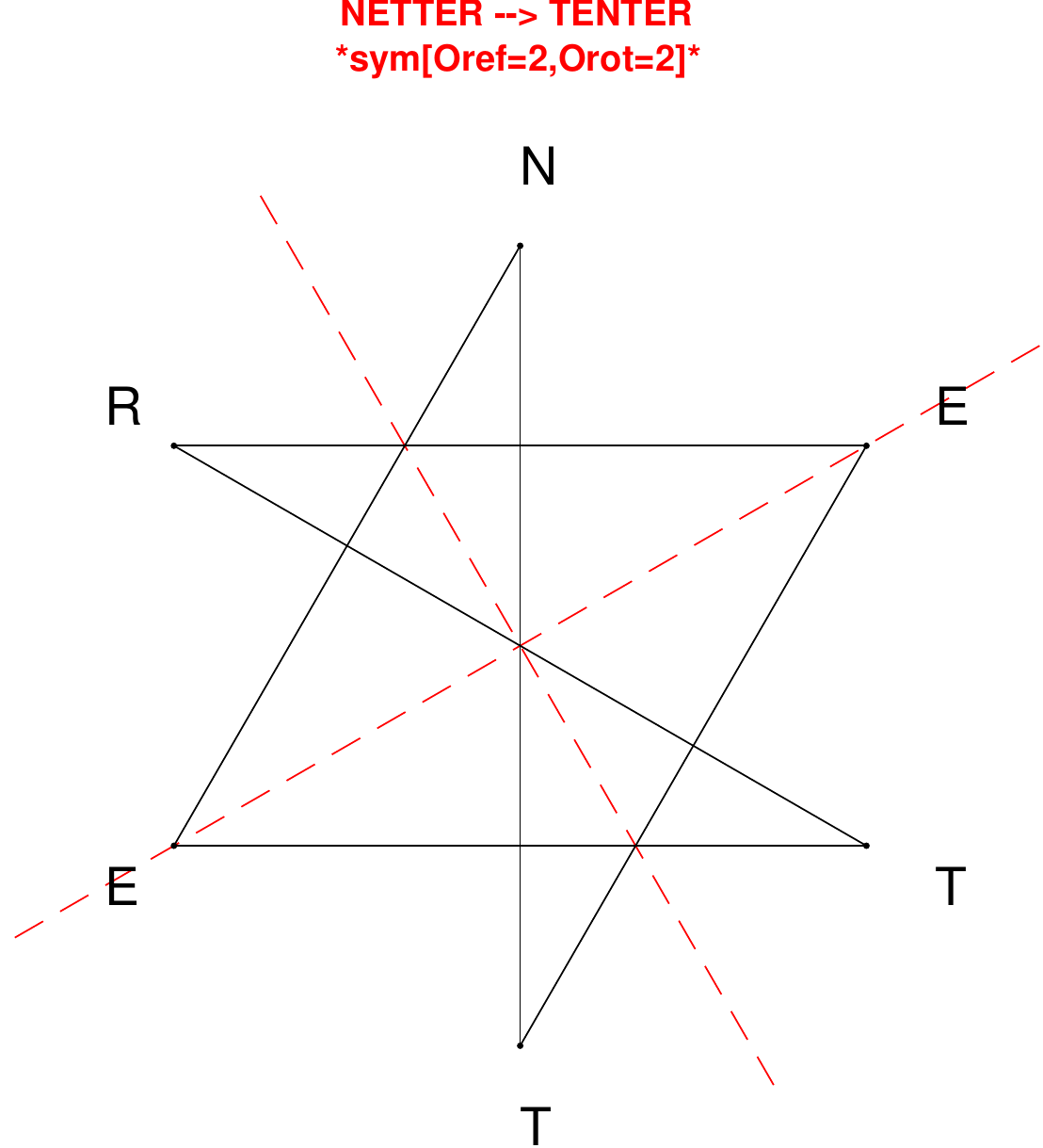}
\end{subfigure}
\hfill
\begin{subfigure}[T]{0.19\textwidth}
\centering
\includegraphics[width=\textwidth]{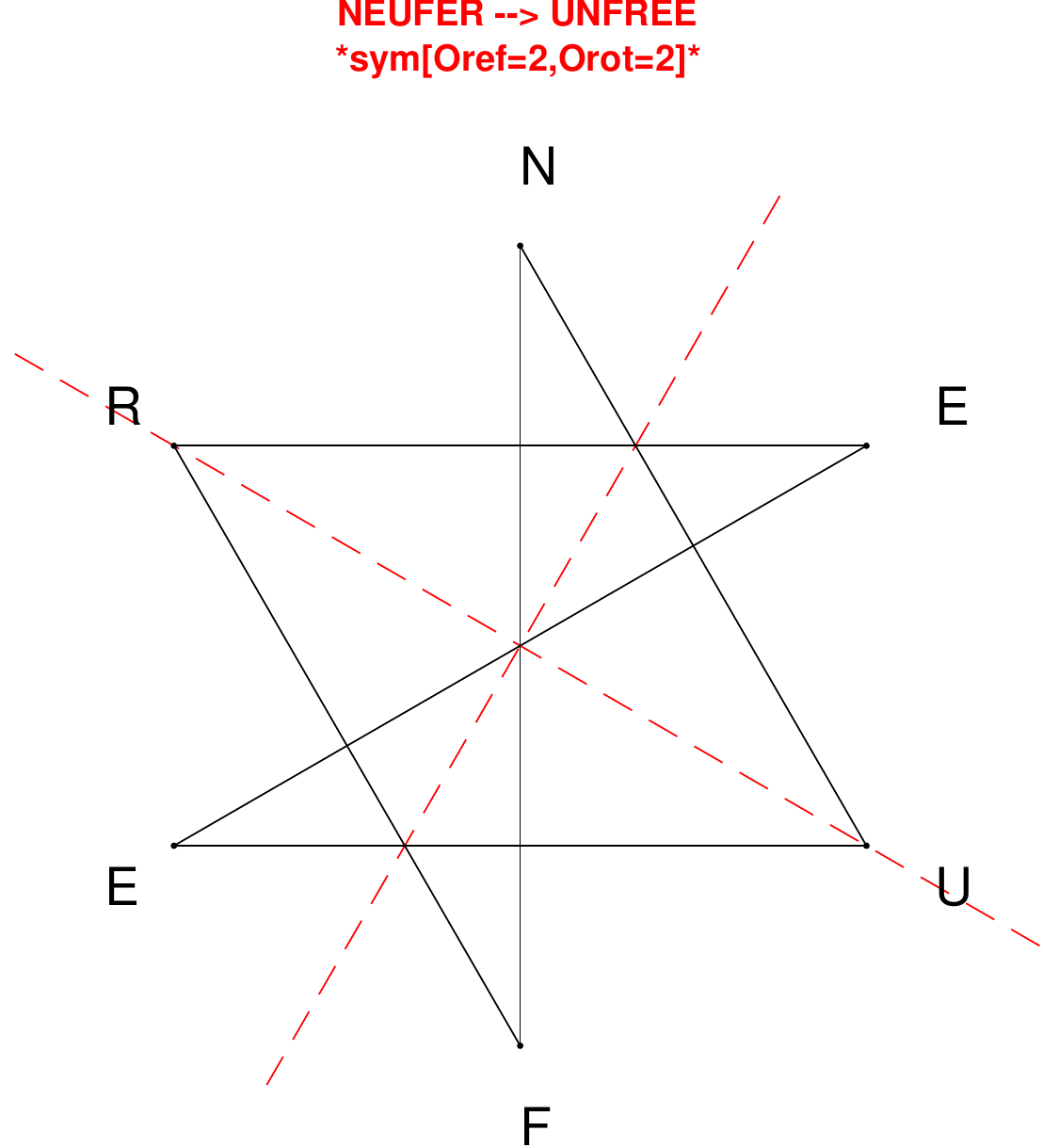}
\end{subfigure}
\end{figure}

\begin{figure}[H]
\centering
\begin{subfigure}[T]{0.19\textwidth}
\centering
\includegraphics[width=\textwidth]{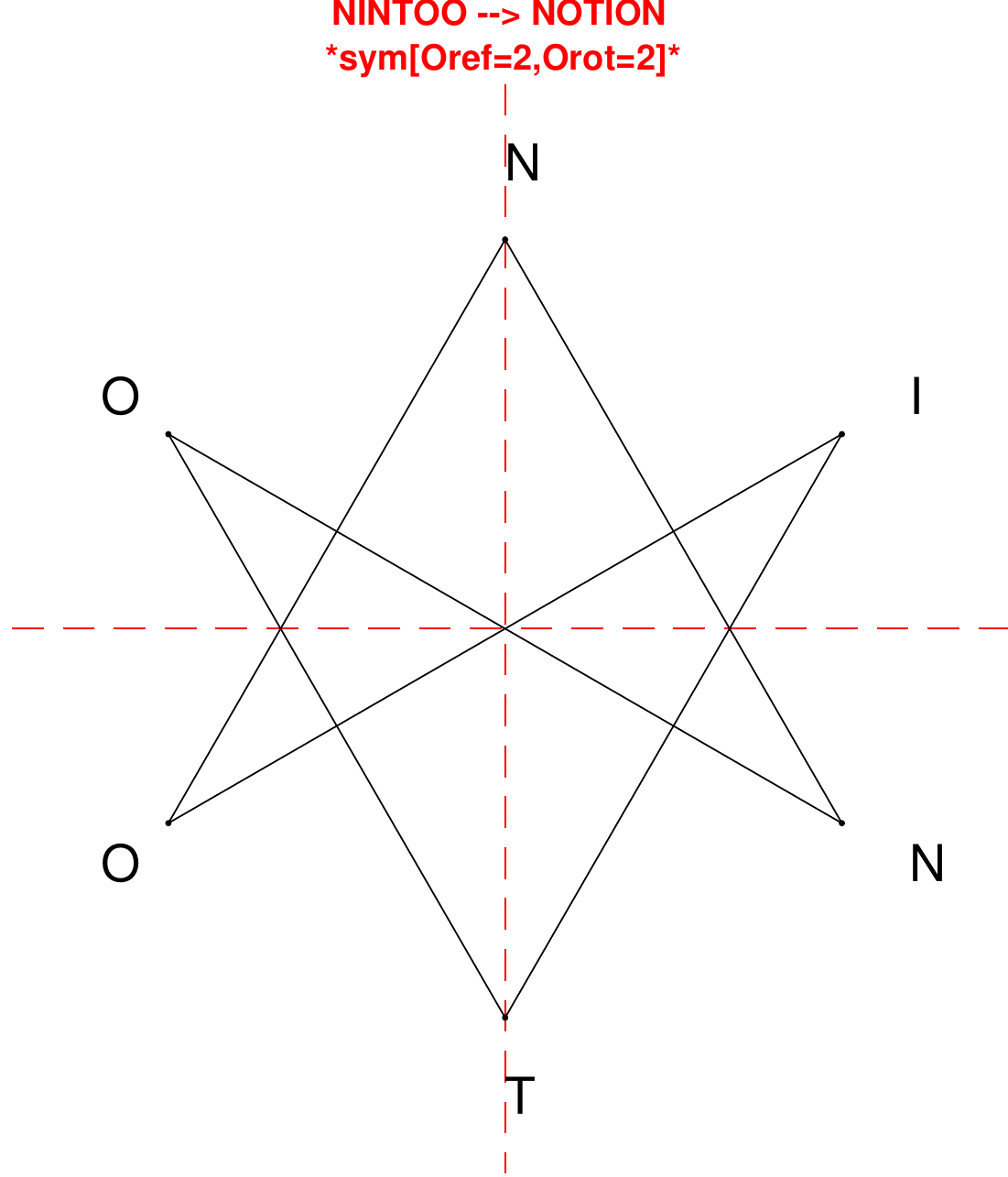}
\end{subfigure}
\hfill
\begin{subfigure}[T]{0.19\textwidth}
\centering
\includegraphics[width=\textwidth]{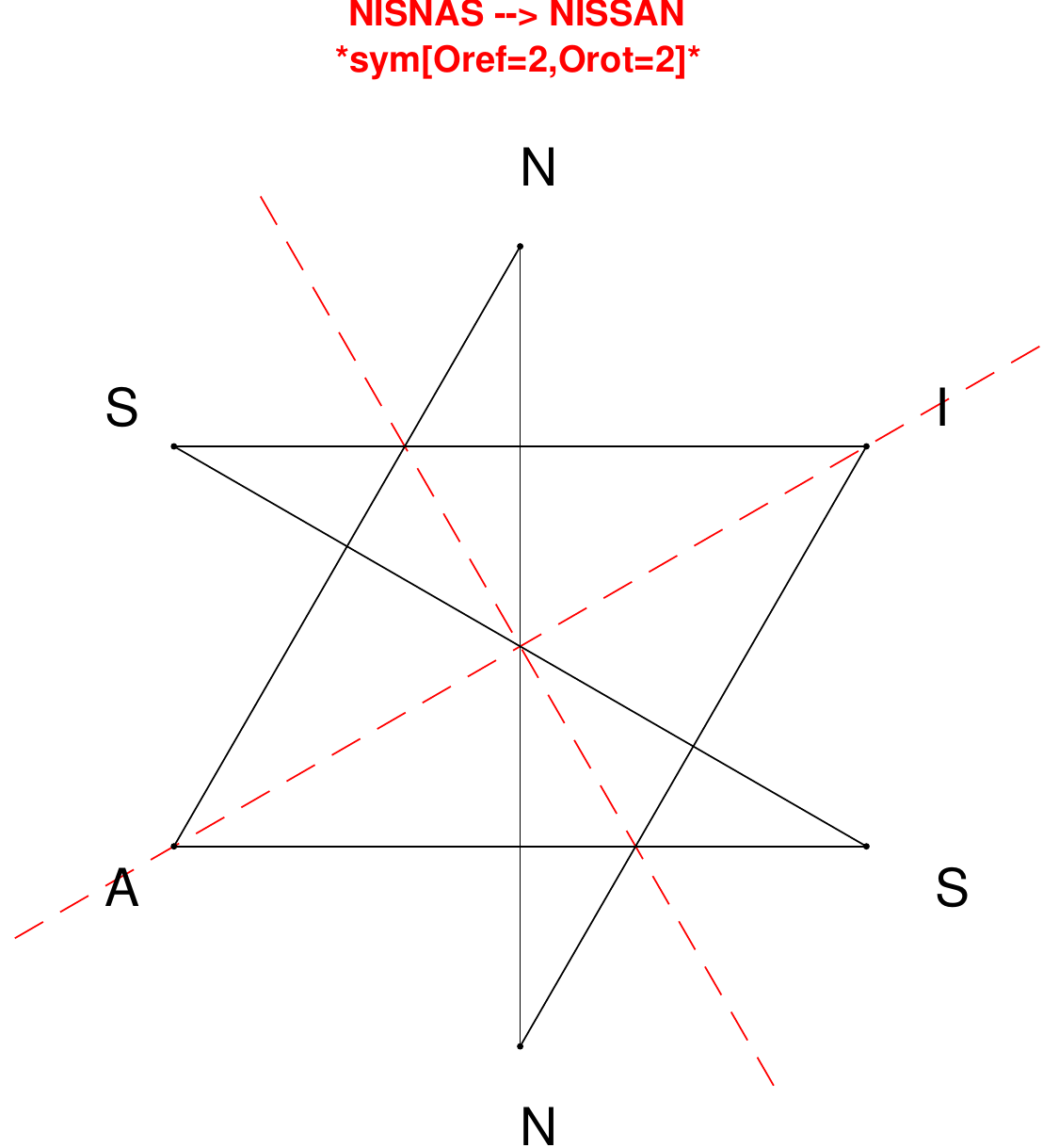}
\end{subfigure}
\hfill
\begin{subfigure}[T]{0.19\textwidth}
\centering
\includegraphics[width=\textwidth]{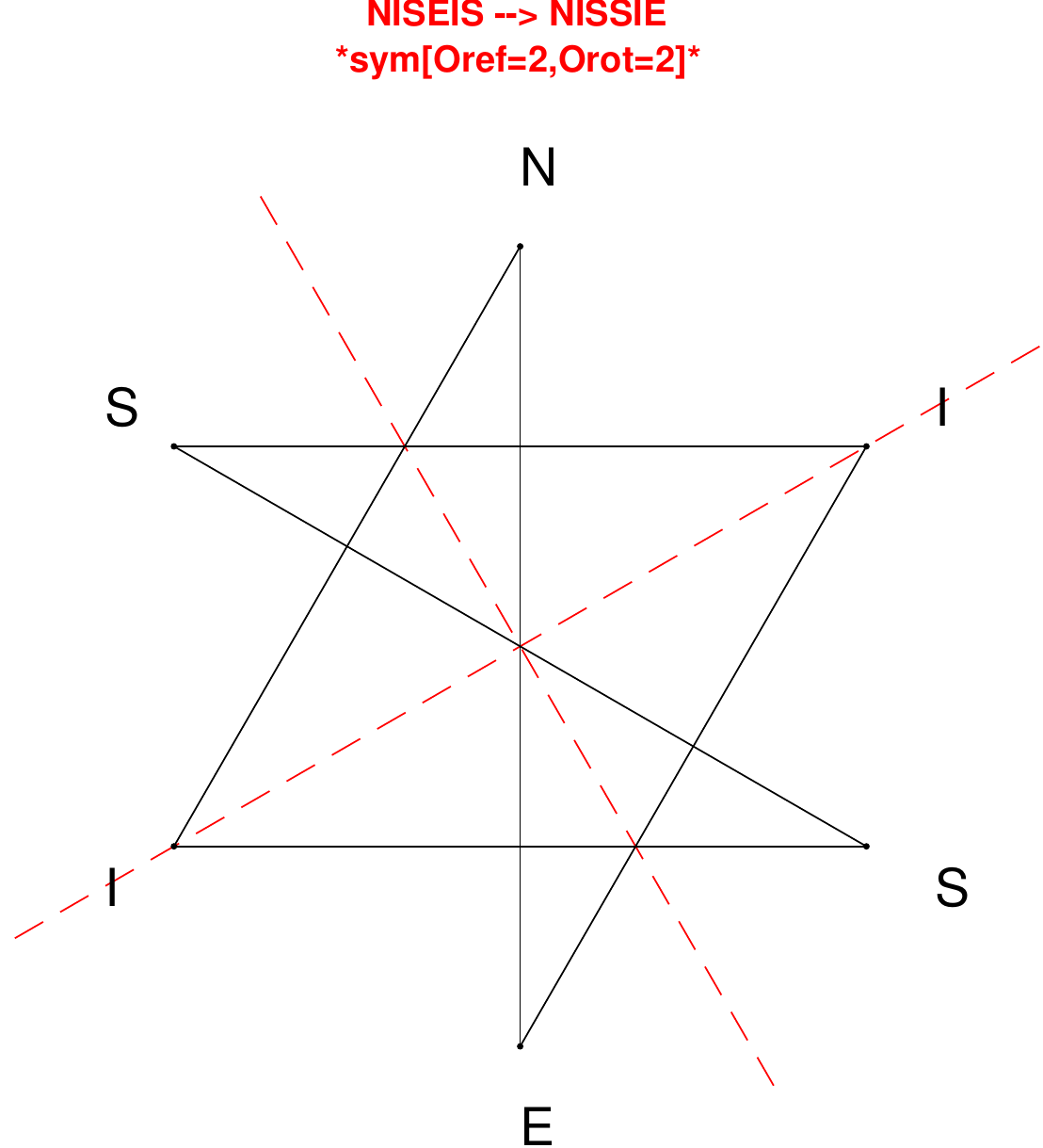}
\end{subfigure}
\hfill
\begin{subfigure}[T]{0.19\textwidth}
\centering
\includegraphics[width=\textwidth]{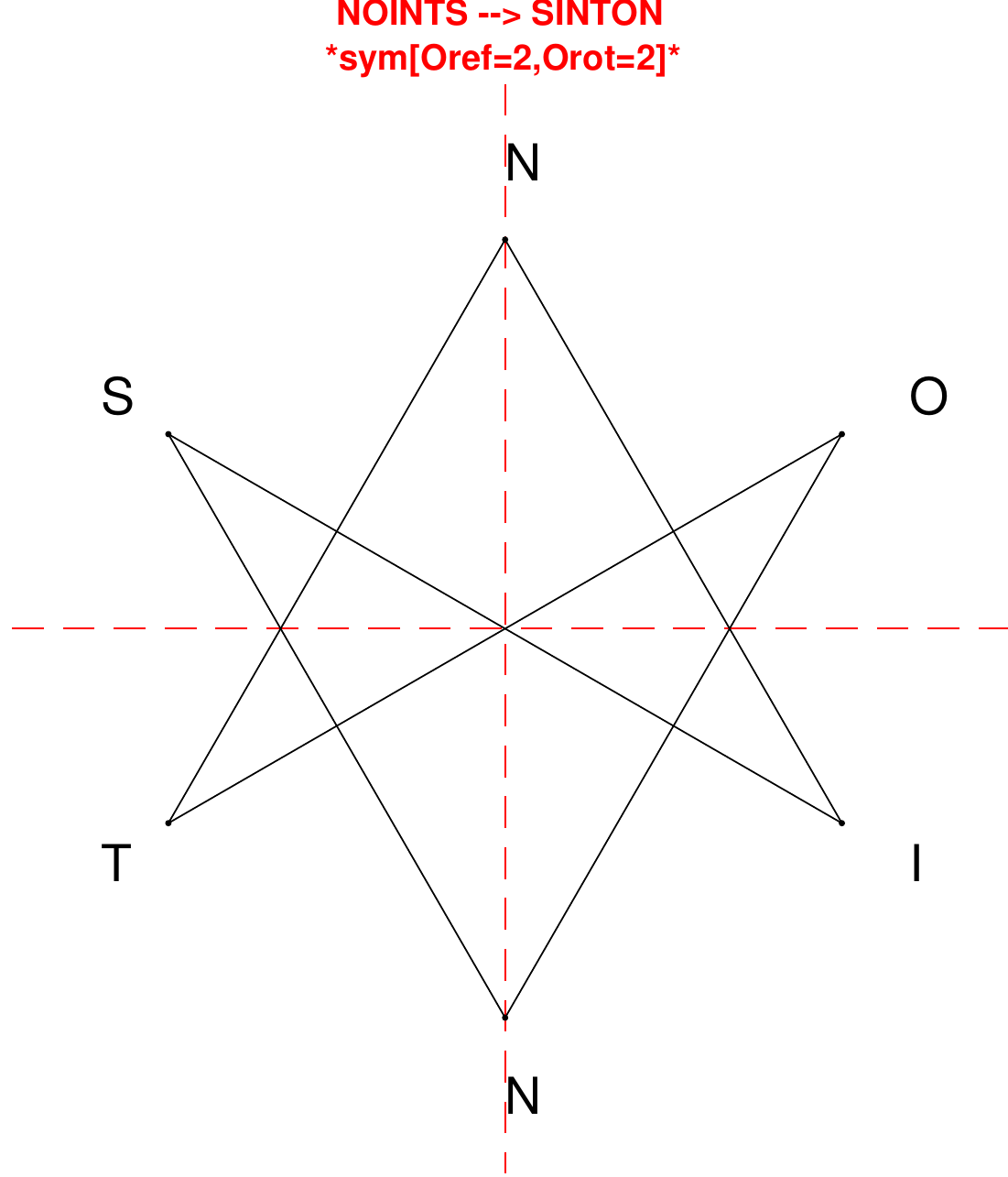}
\end{subfigure}
\hfill
\begin{subfigure}[T]{0.19\textwidth}
\centering
\includegraphics[width=\textwidth]{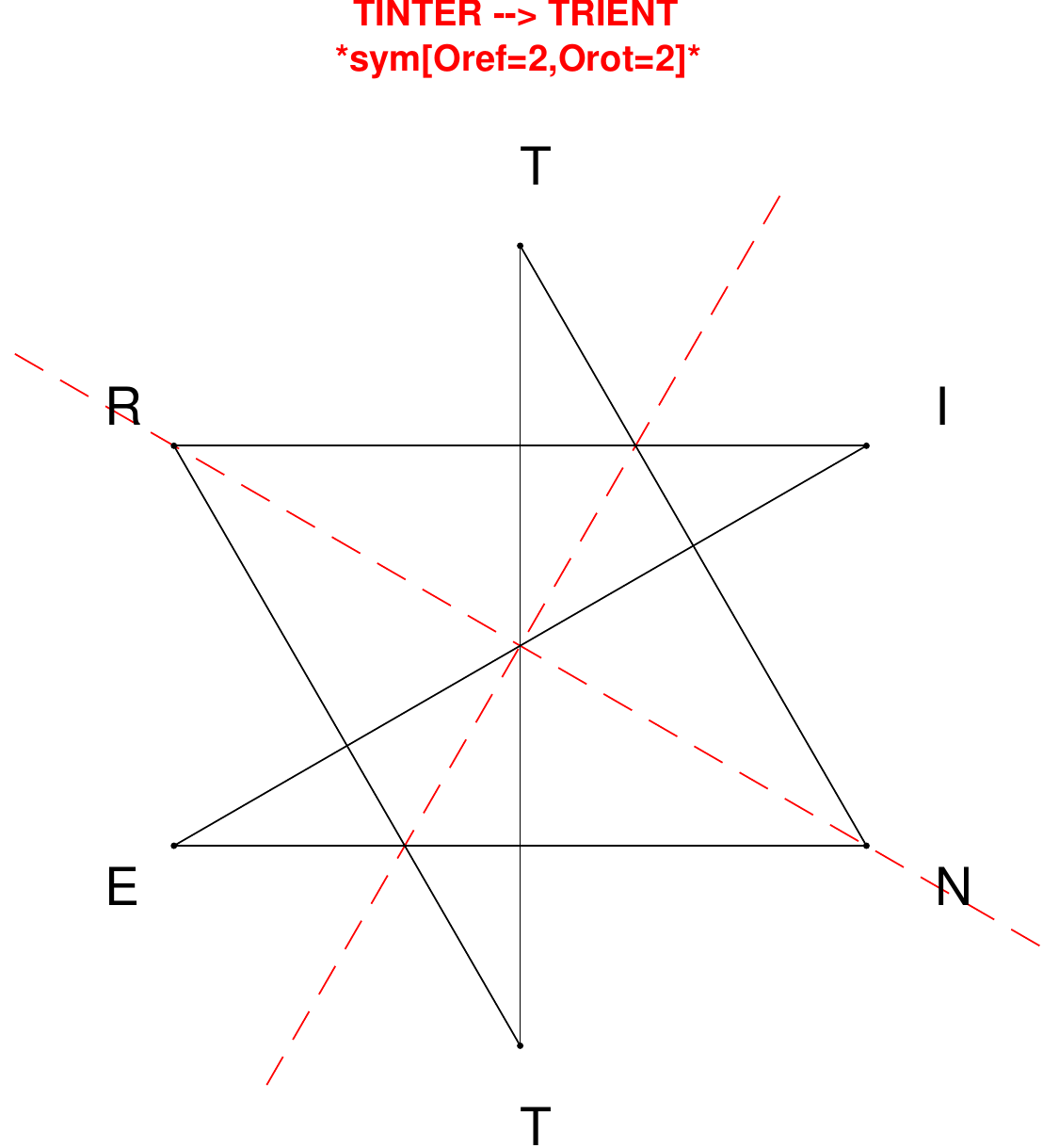}
\end{subfigure}
\end{figure}

\begin{figure}[H]
\centering
\begin{subfigure}[T]{0.19\textwidth}
\centering
\includegraphics[width=\textwidth]{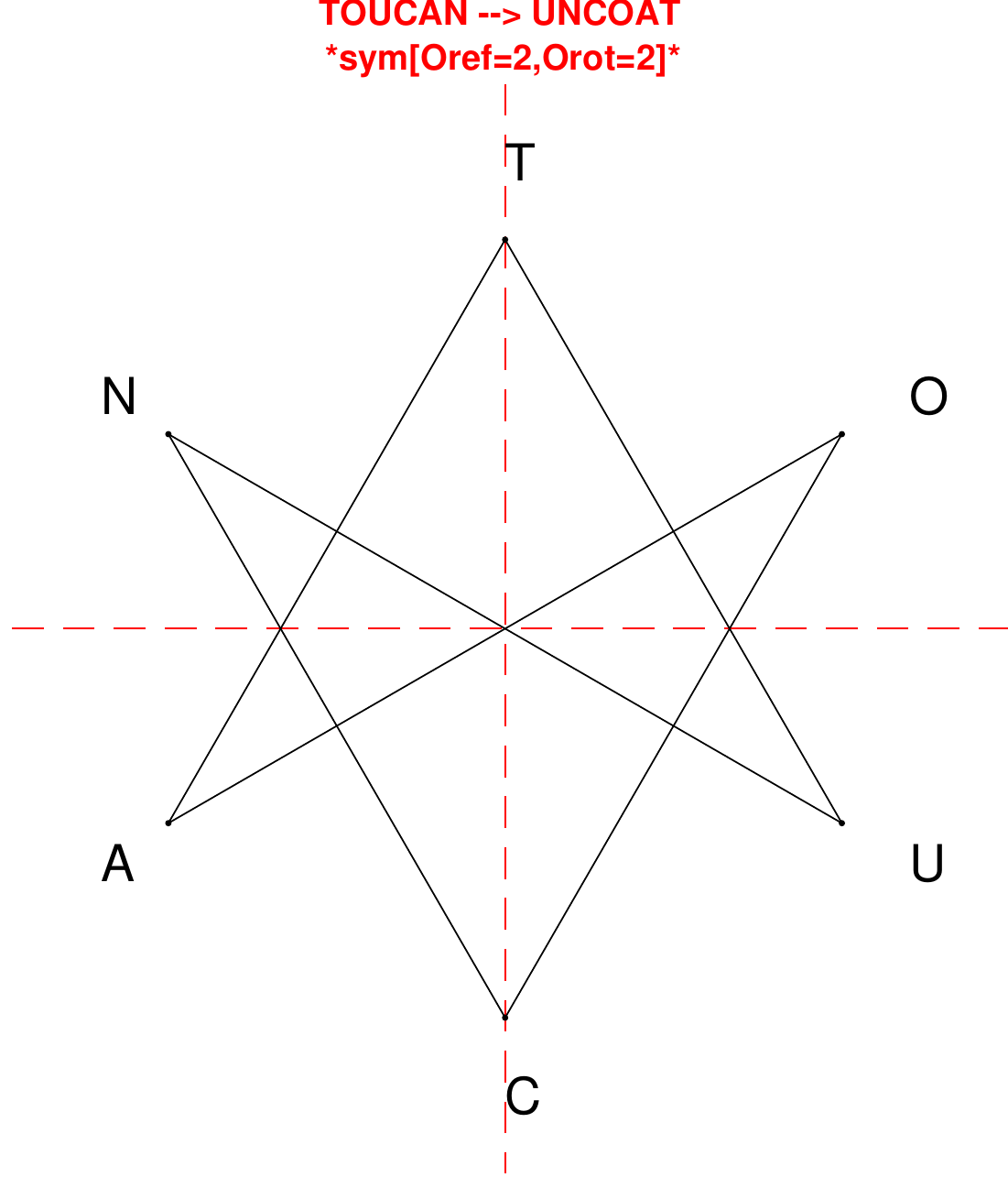}
\end{subfigure}
\hfill
\begin{subfigure}[T]{0.19\textwidth}
\centering
\includegraphics[width=\textwidth]{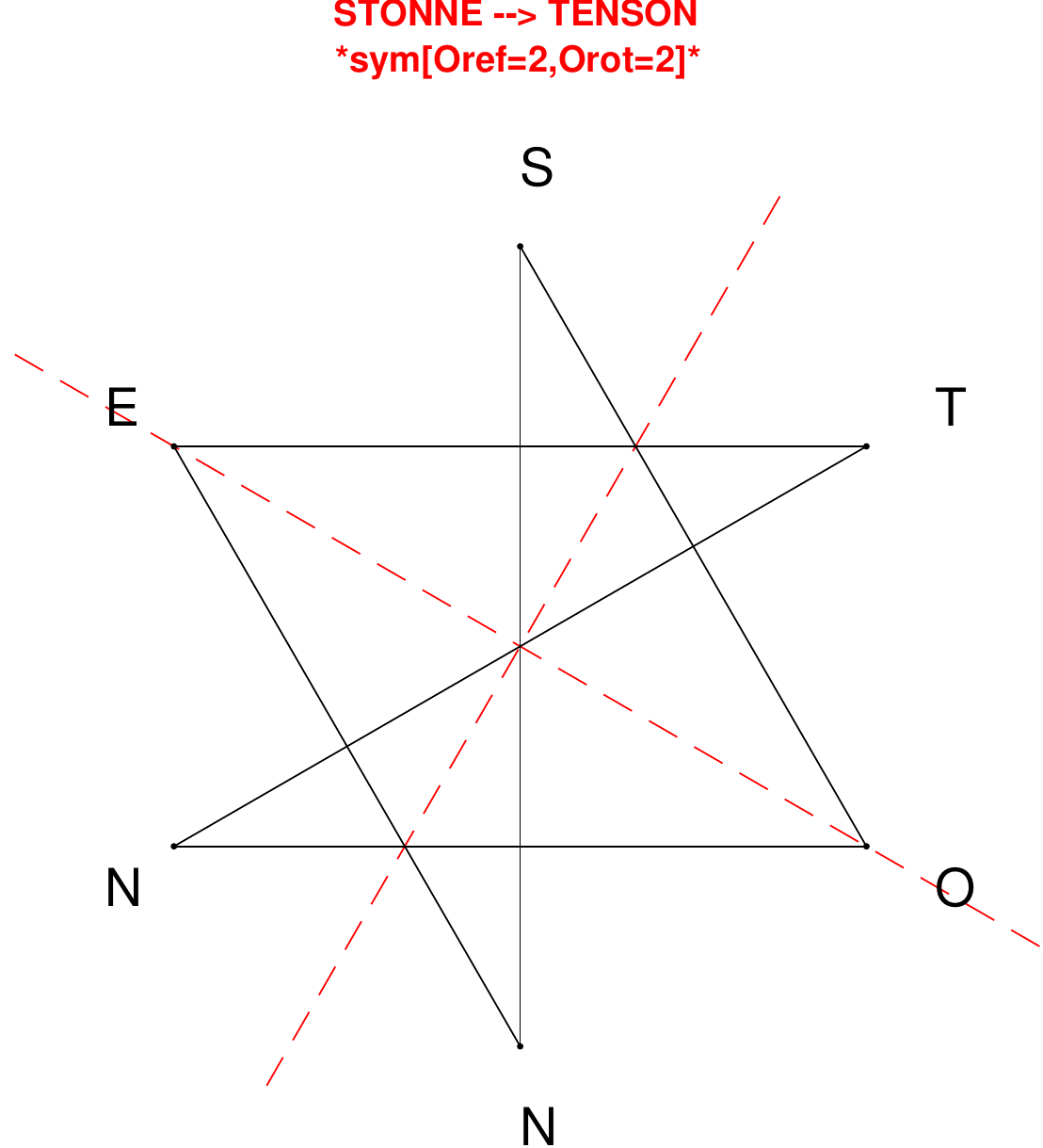}
\end{subfigure}
\hfill
\begin{subfigure}[T]{0.19\textwidth}
\centering
\includegraphics[width=\textwidth]{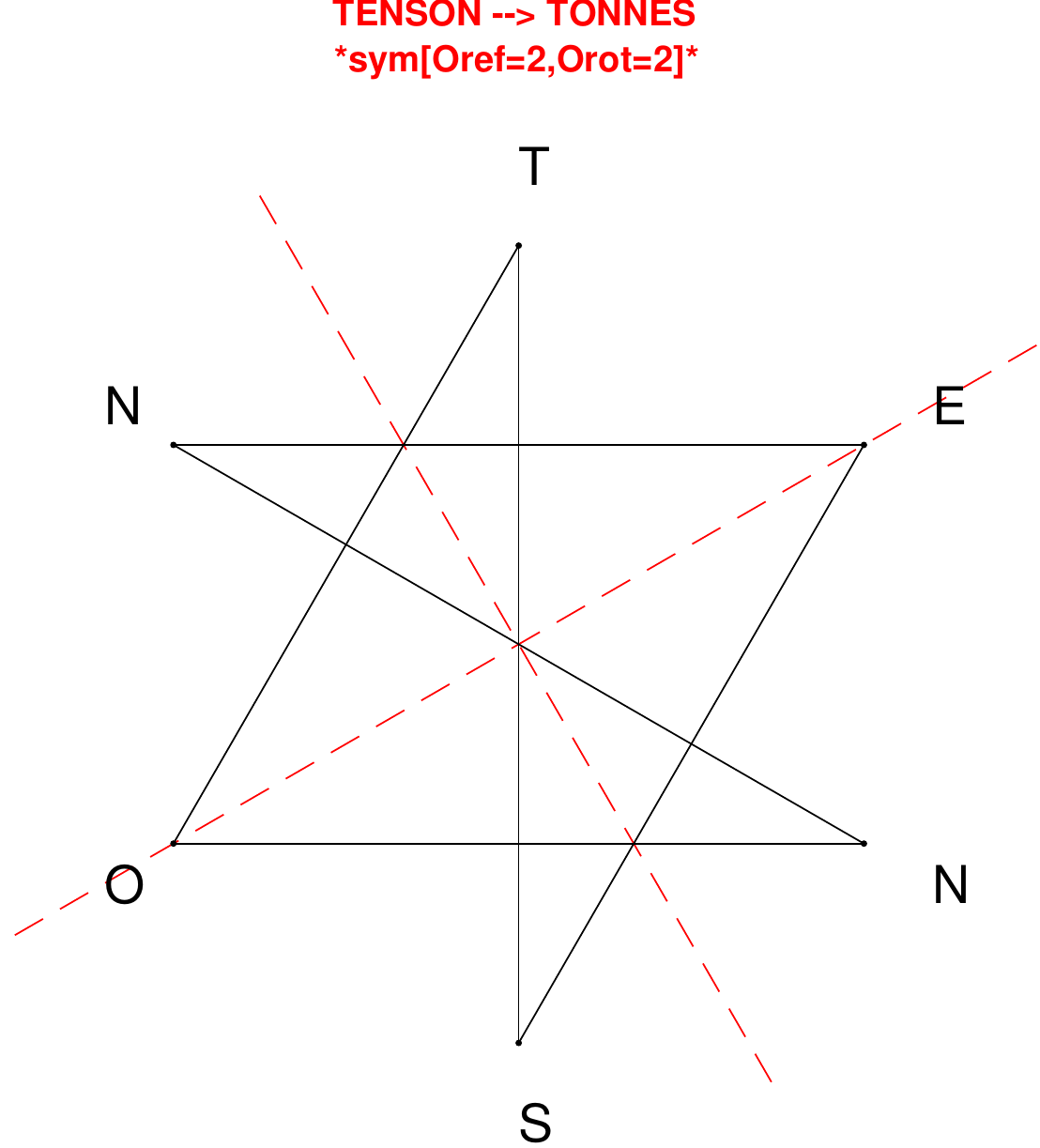}
\end{subfigure}
\hfill
\begin{subfigure}[T]{0.19\textwidth}
\centering
\includegraphics[width=\textwidth]{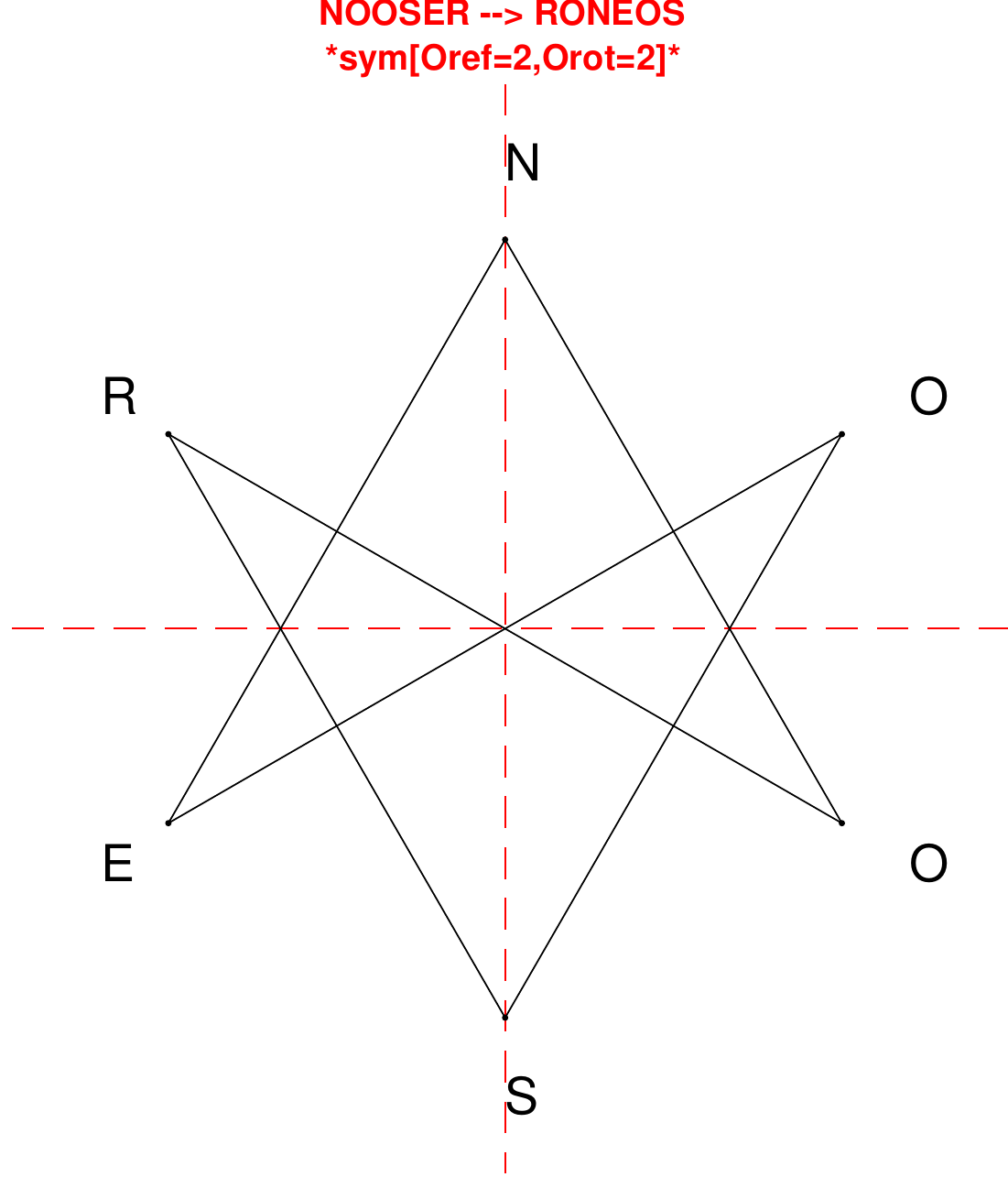}
\end{subfigure}
\hfill
\begin{subfigure}[T]{0.19\textwidth}
\centering
\includegraphics[width=\textwidth]{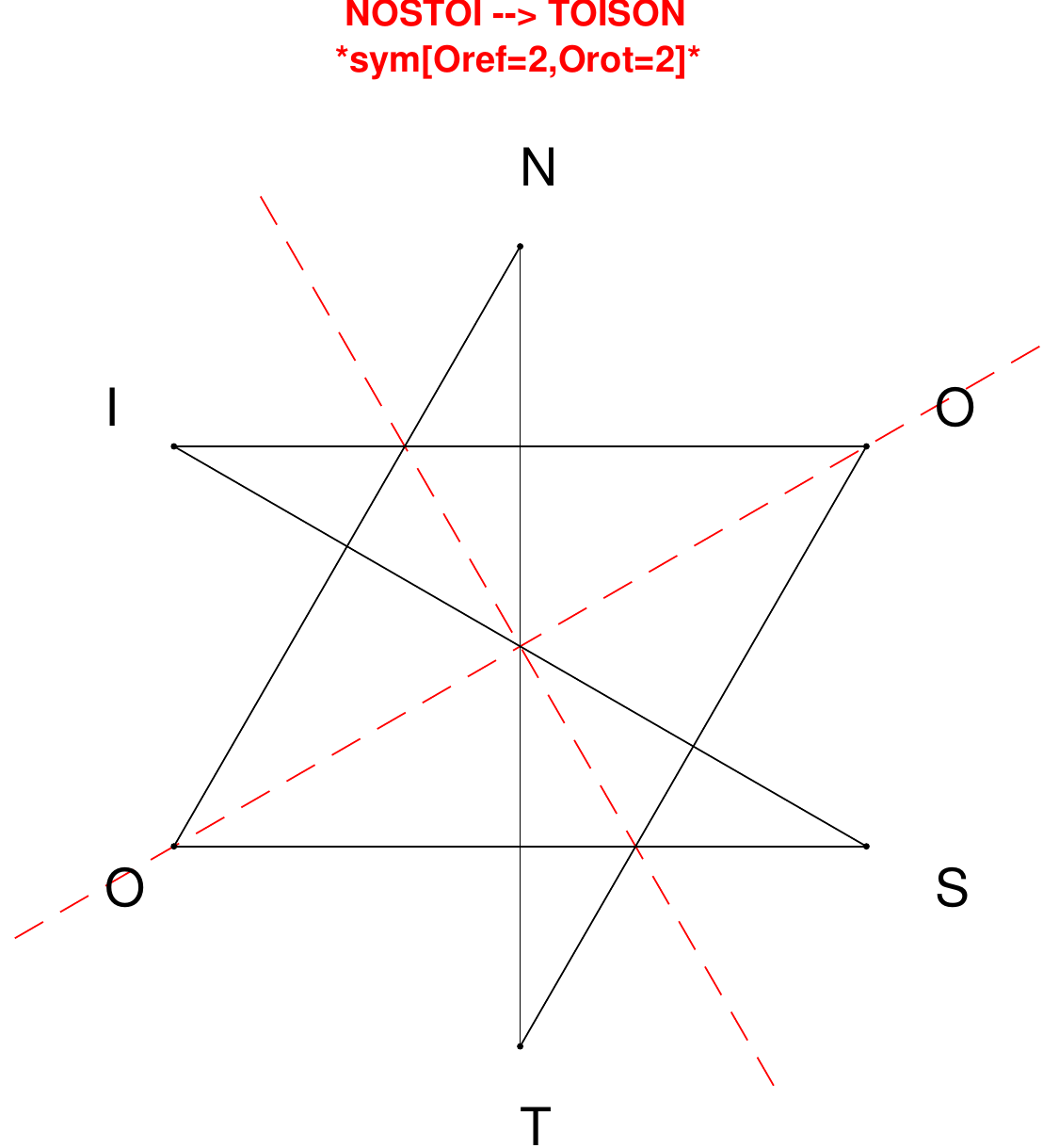}
\end{subfigure}
\end{figure}

\begin{figure}[H]
\centering
\begin{subfigure}[T]{0.19\textwidth}
\centering
\includegraphics[width=\textwidth]{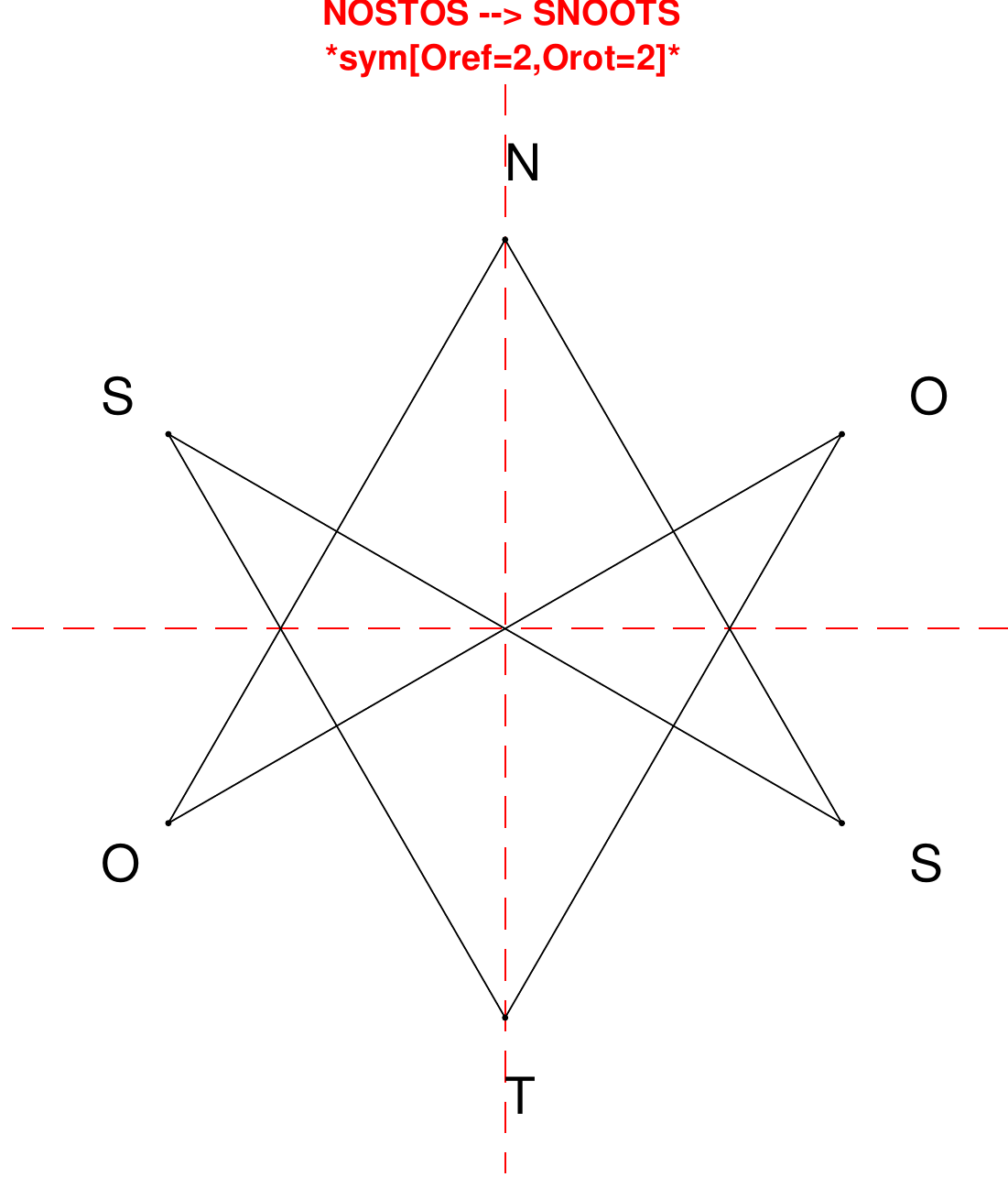}
\end{subfigure}
\hfill
\begin{subfigure}[T]{0.19\textwidth}
\centering
\includegraphics[width=\textwidth]{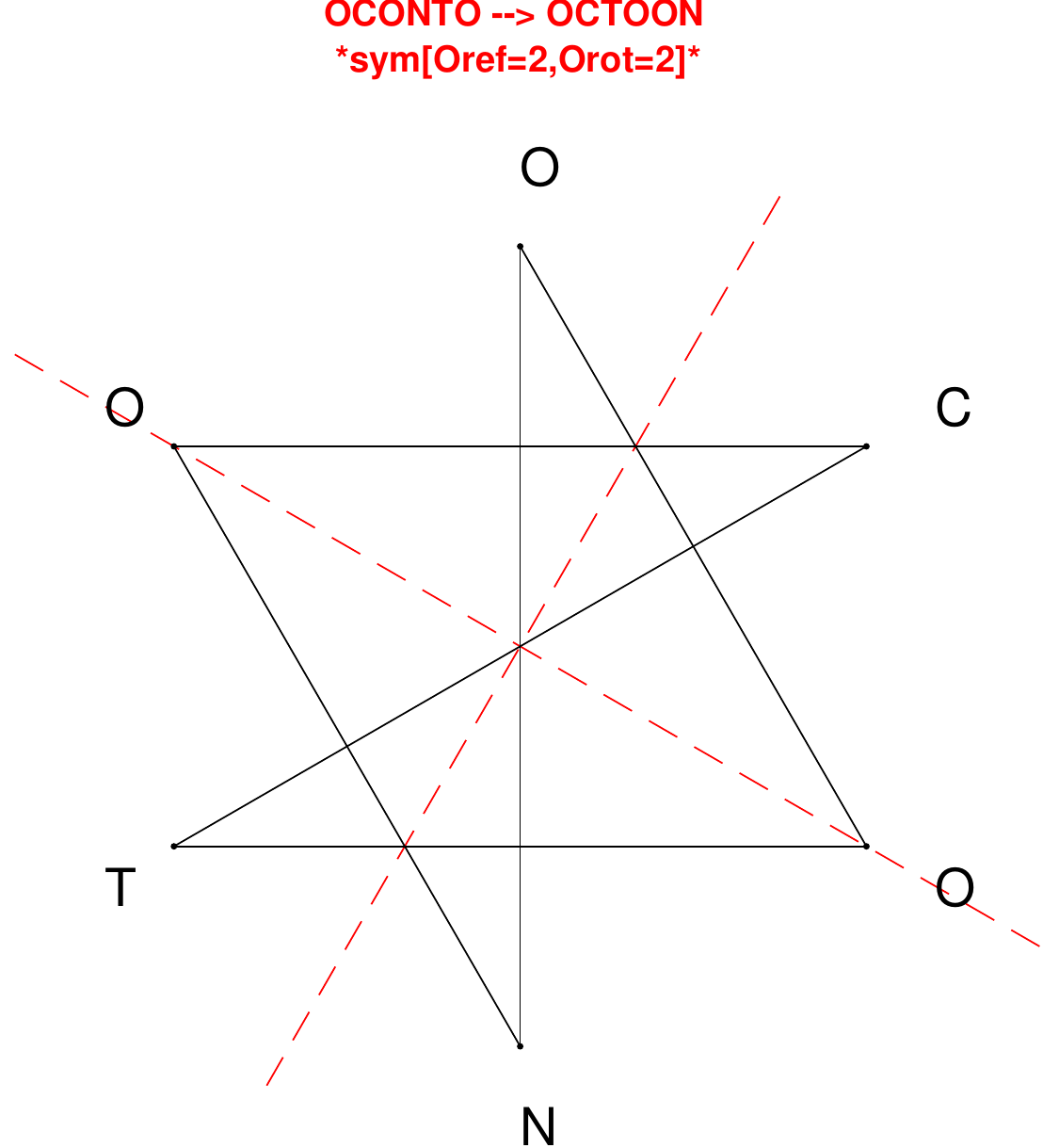}
\end{subfigure}
\hfill
\begin{subfigure}[T]{0.19\textwidth}
\centering
\includegraphics[width=\textwidth]{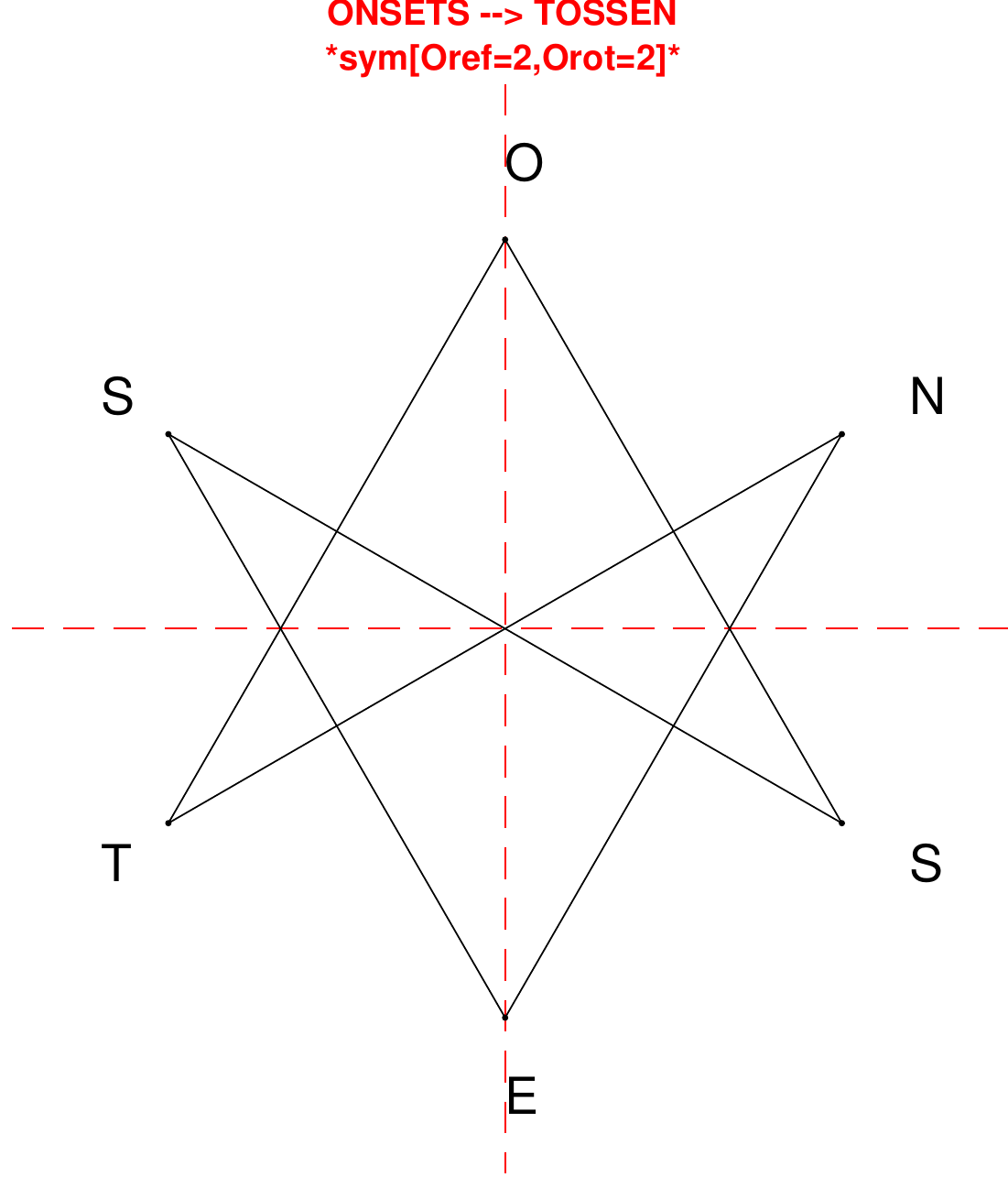}
\end{subfigure}
\hfill
\begin{subfigure}[T]{0.19\textwidth}
\centering
\includegraphics[width=\textwidth]{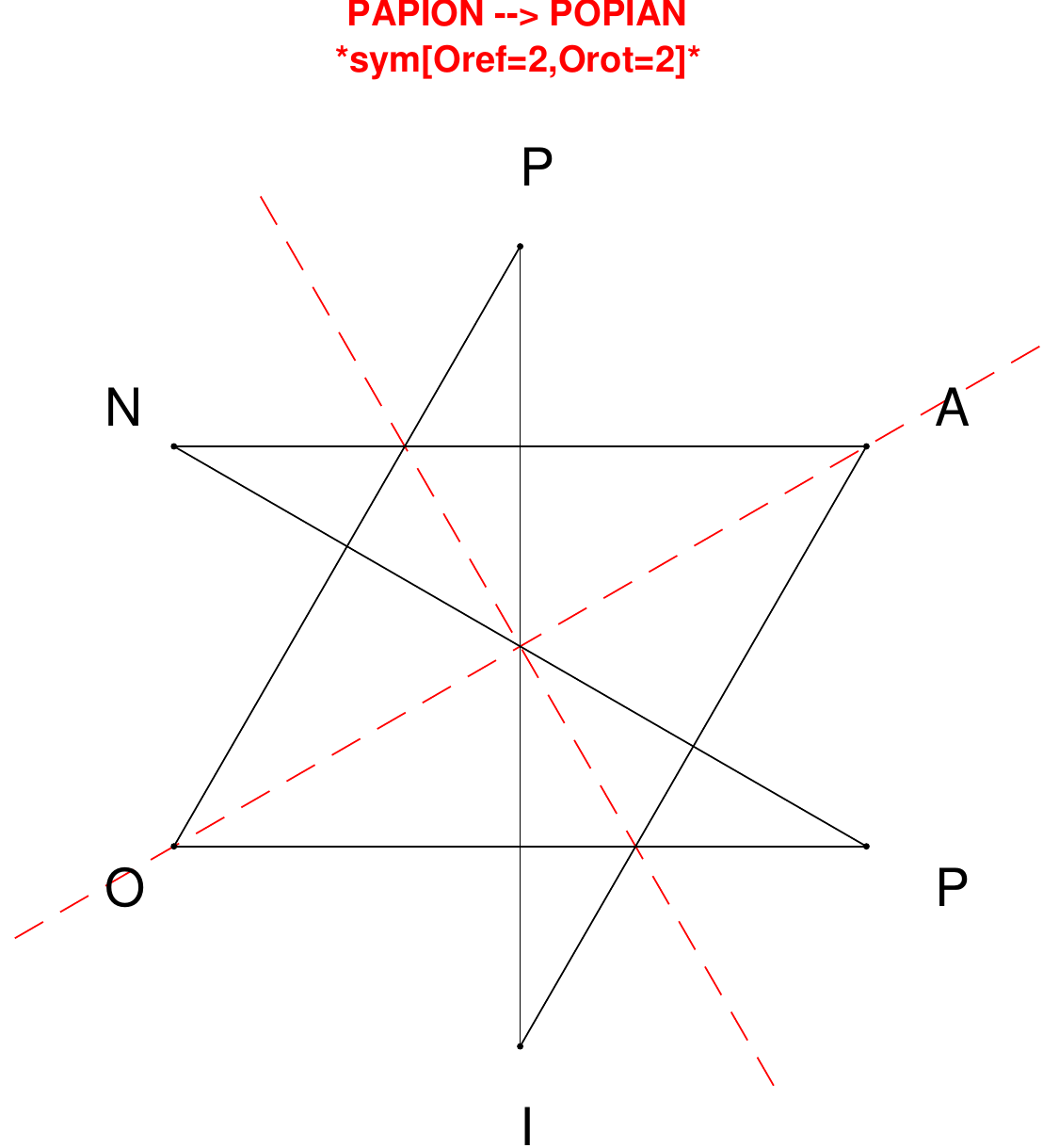}
\end{subfigure}
\hfill
\begin{subfigure}[T]{0.19\textwidth}
\centering
\includegraphics[width=\textwidth]{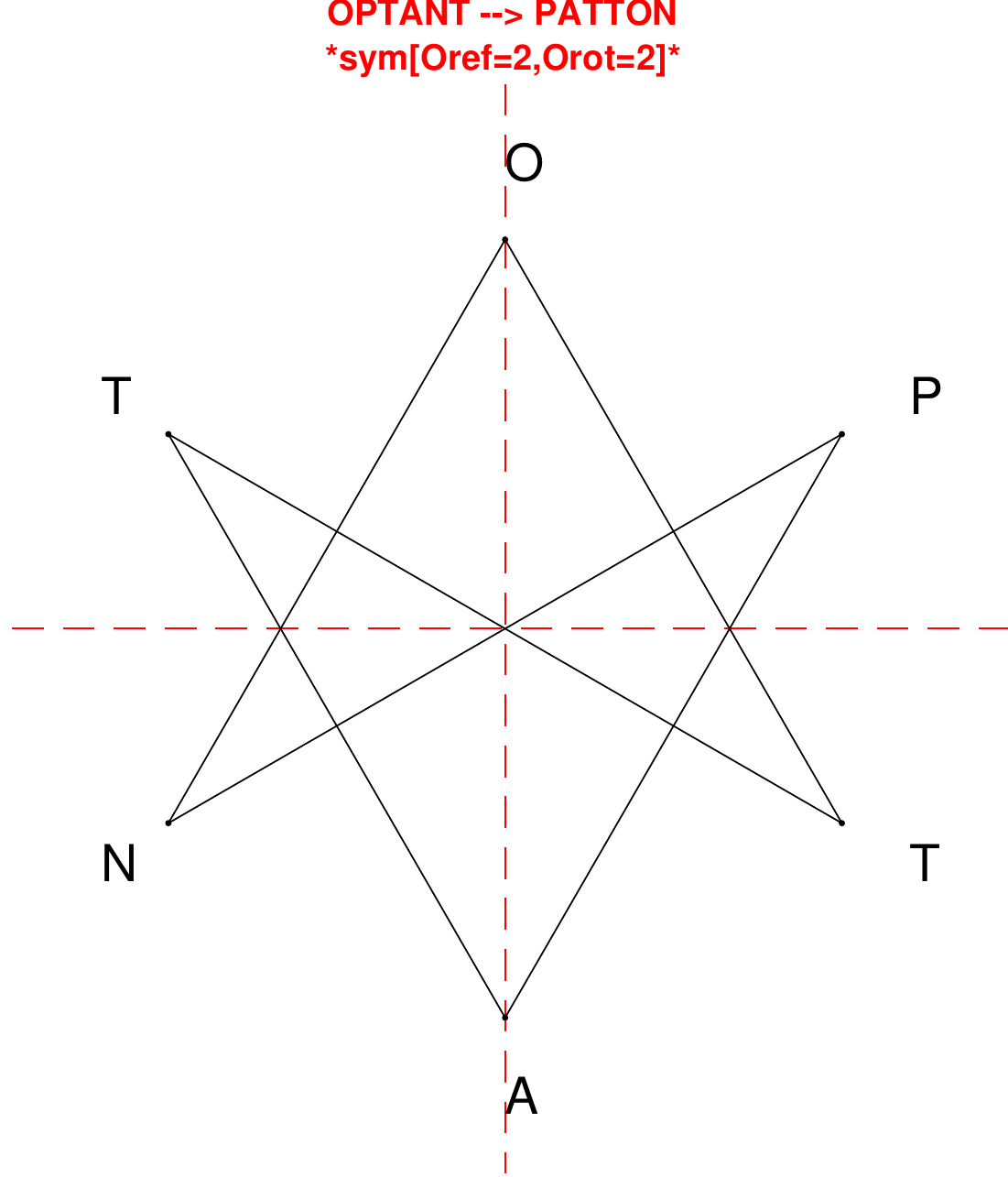}
\end{subfigure}
\end{figure}

\begin{figure}[H]
\centering
\begin{subfigure}[T]{0.19\textwidth}
\centering
\includegraphics[width=\textwidth]{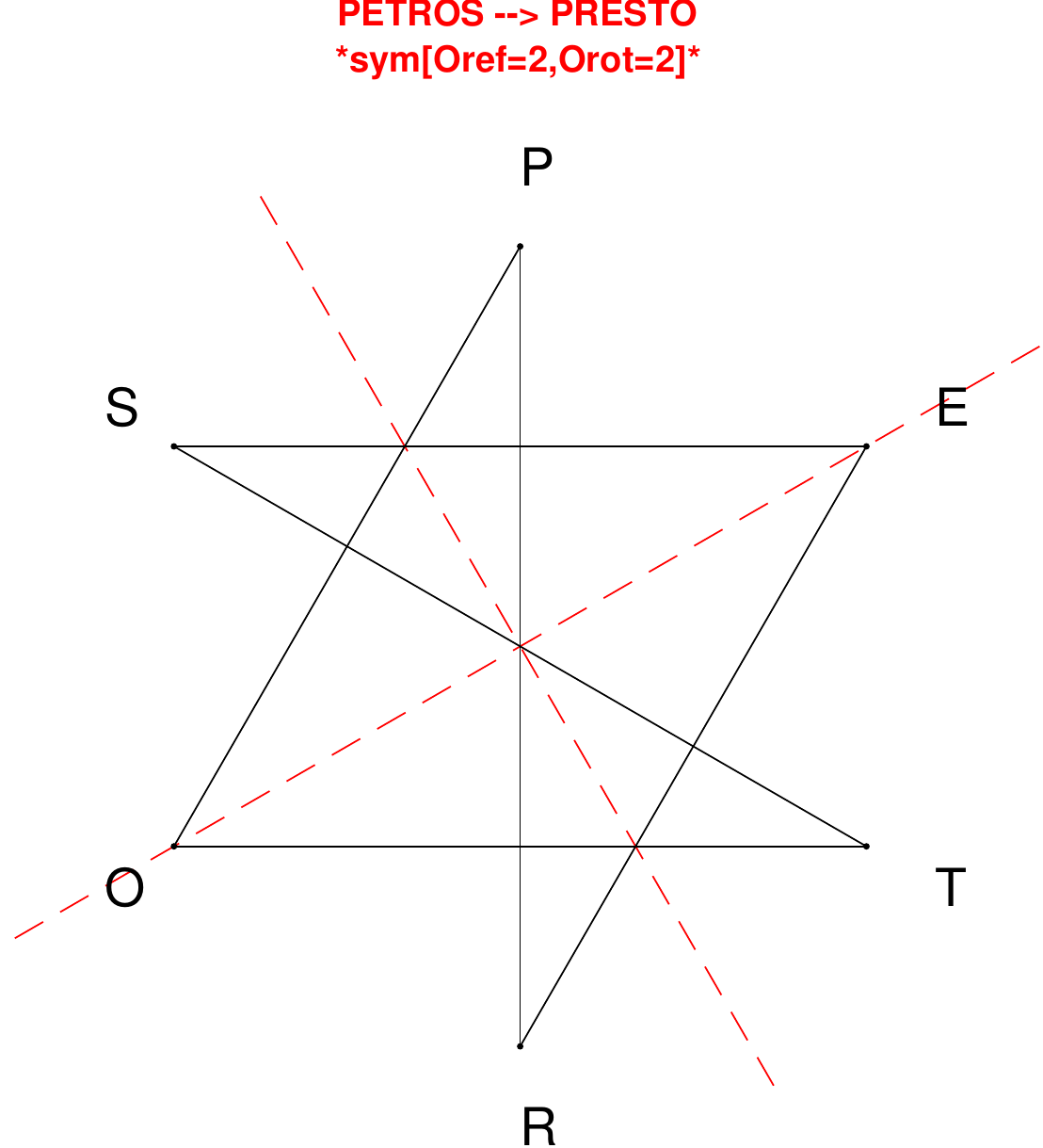}
\end{subfigure}
\hfill
\begin{subfigure}[T]{0.19\textwidth}
\centering
\includegraphics[width=\textwidth]{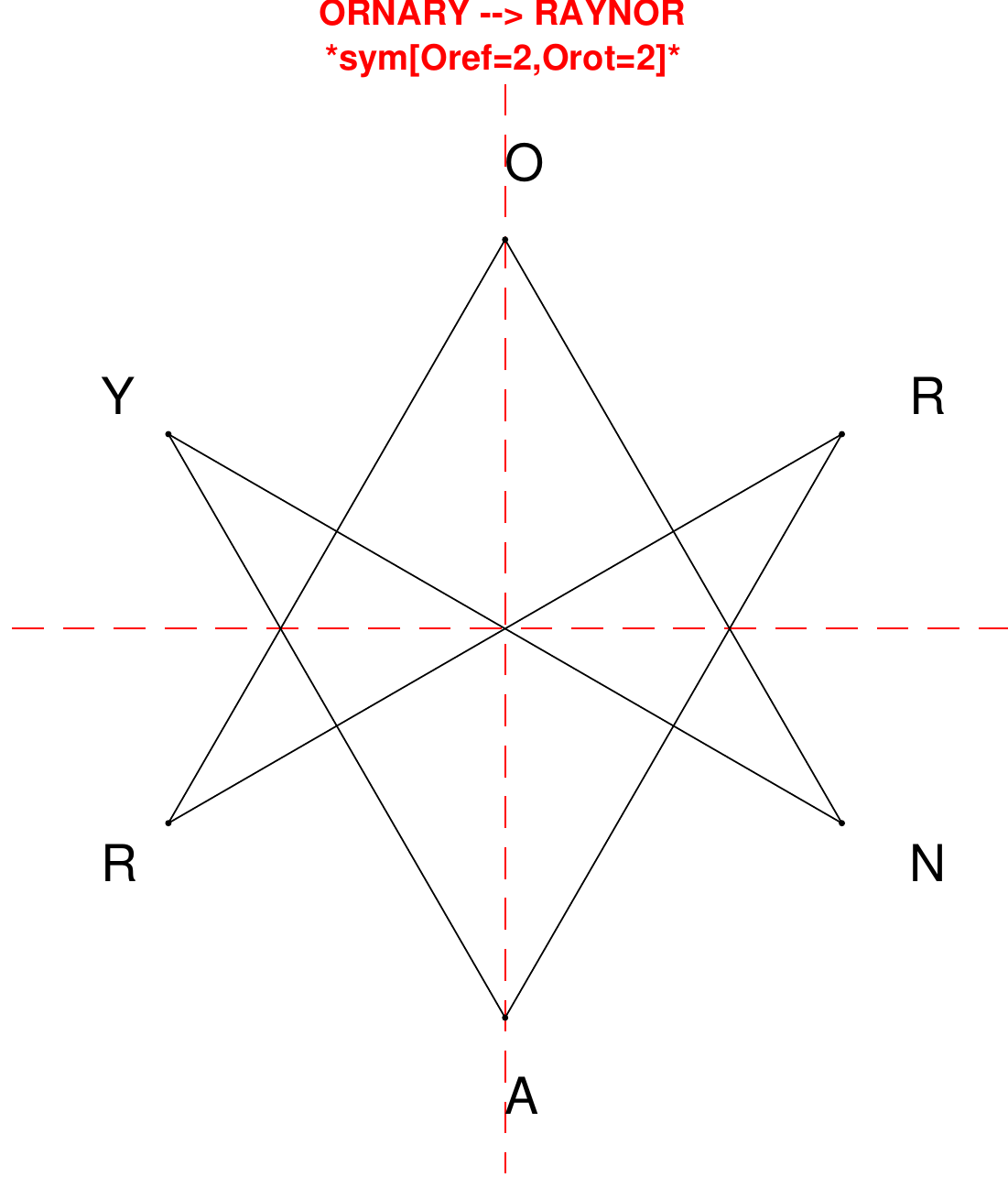}
\end{subfigure}
\hfill
\begin{subfigure}[T]{0.19\textwidth}
\centering
\includegraphics[width=\textwidth]{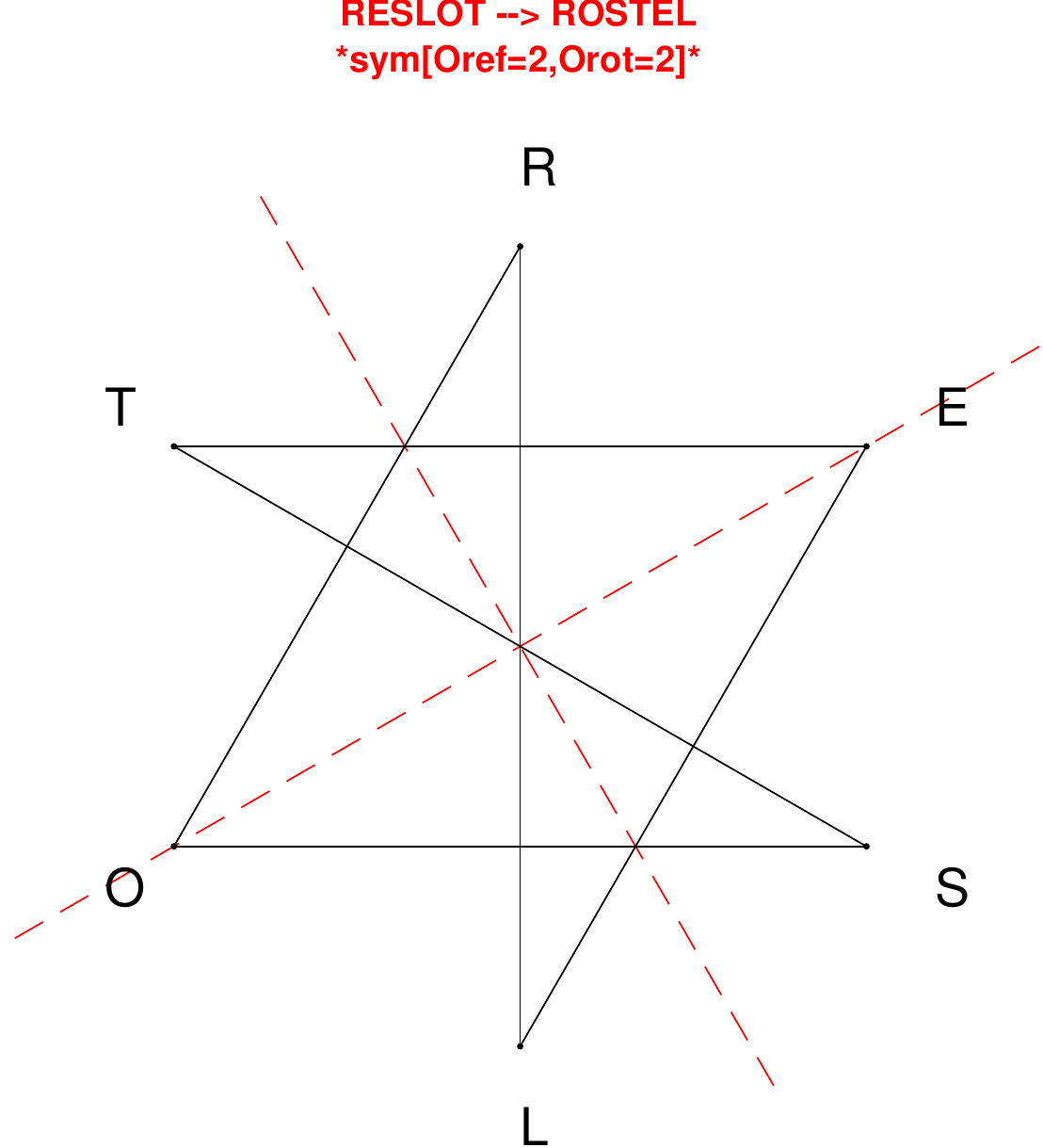}
\end{subfigure}
\hfill
\begin{subfigure}[T]{0.19\textwidth}
\centering
\includegraphics[width=\textwidth]{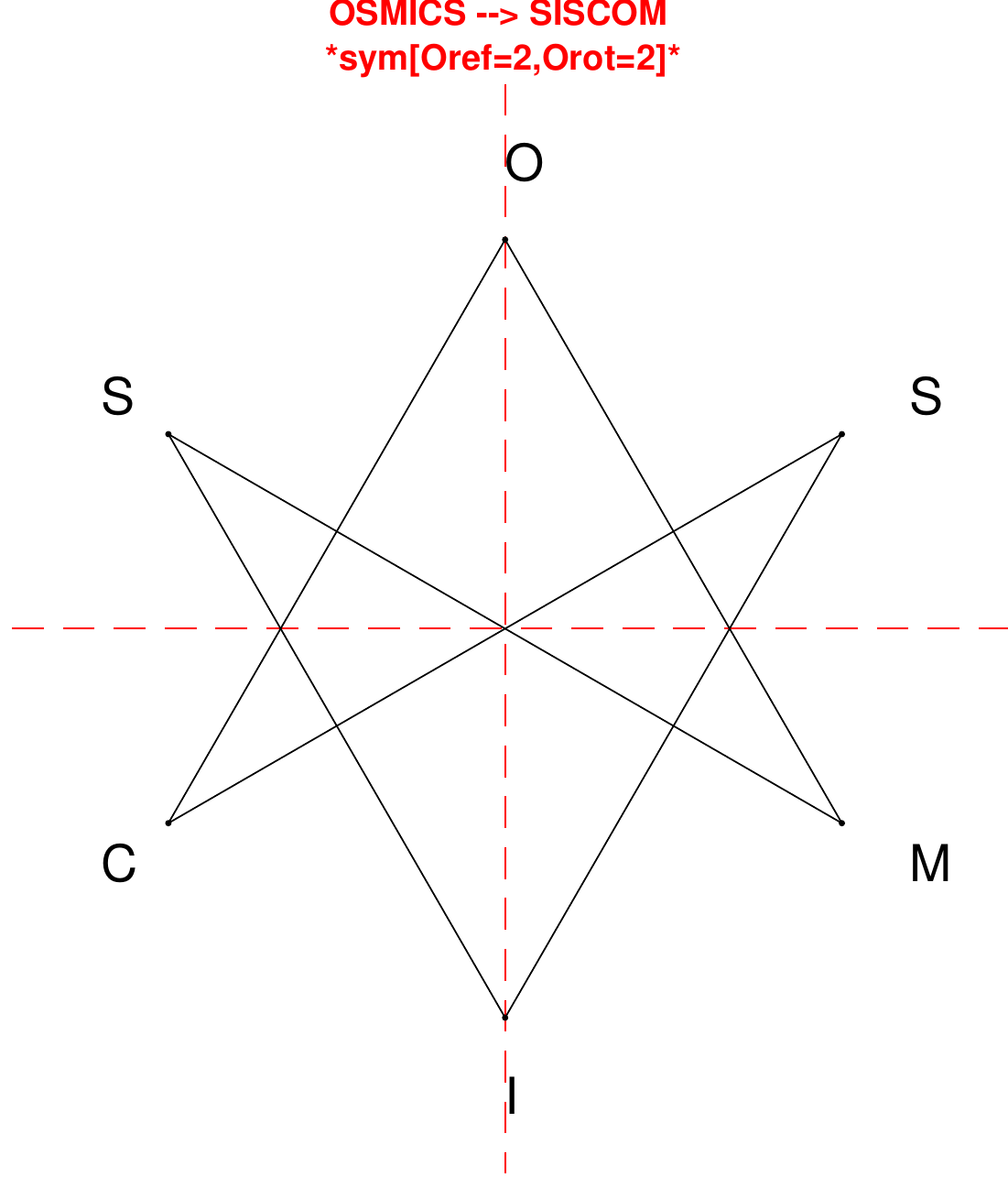}
\end{subfigure}
\hfill
\begin{subfigure}[T]{0.19\textwidth}
\centering
\includegraphics[width=\textwidth]{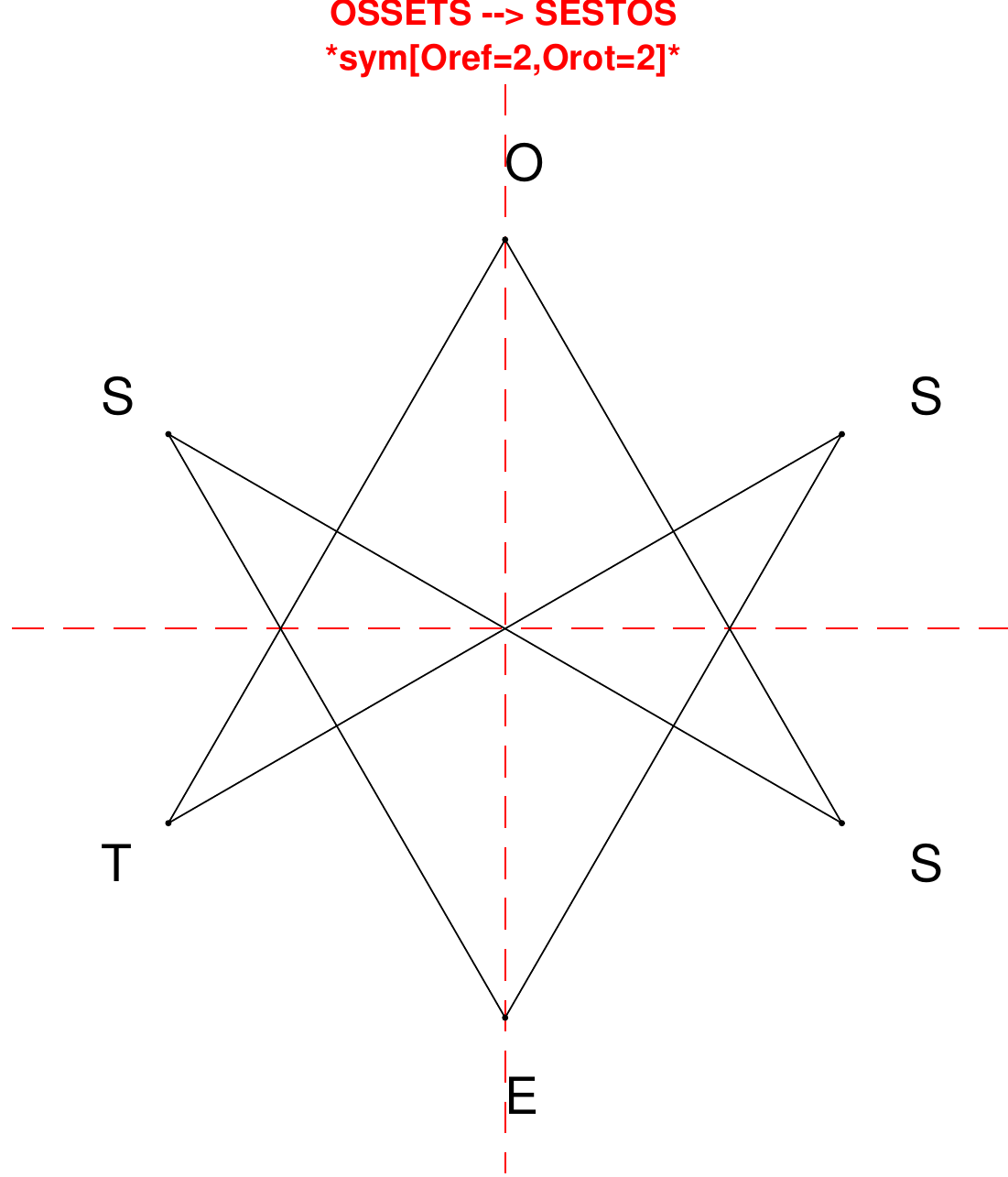}
\end{subfigure}
\end{figure}

\begin{figure}[H]
\centering
\begin{subfigure}[T]{0.19\textwidth}
\centering
\includegraphics[width=\textwidth]{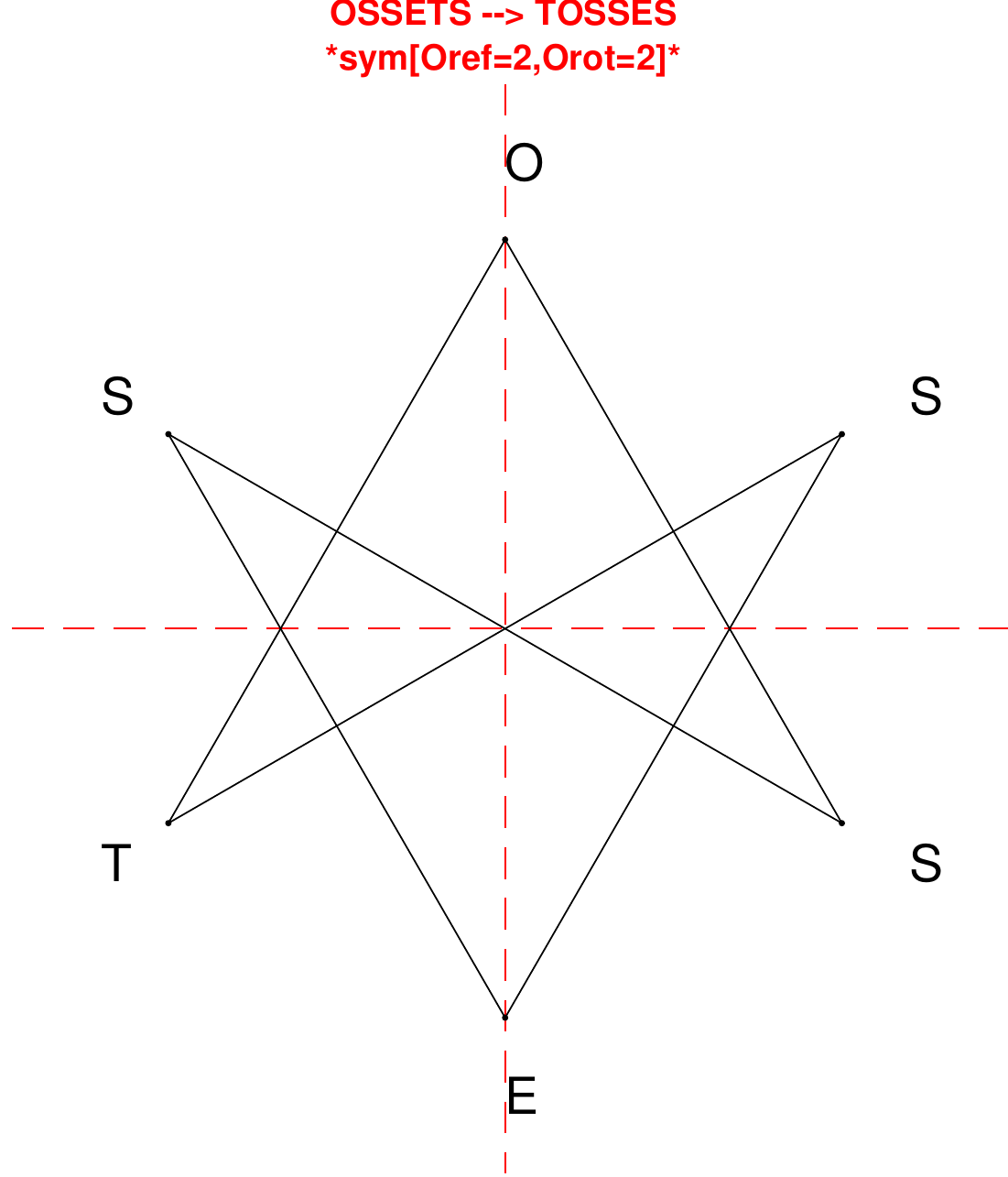}
\end{subfigure}
\hfill
\begin{subfigure}[T]{0.19\textwidth}
\centering
\includegraphics[width=\textwidth]{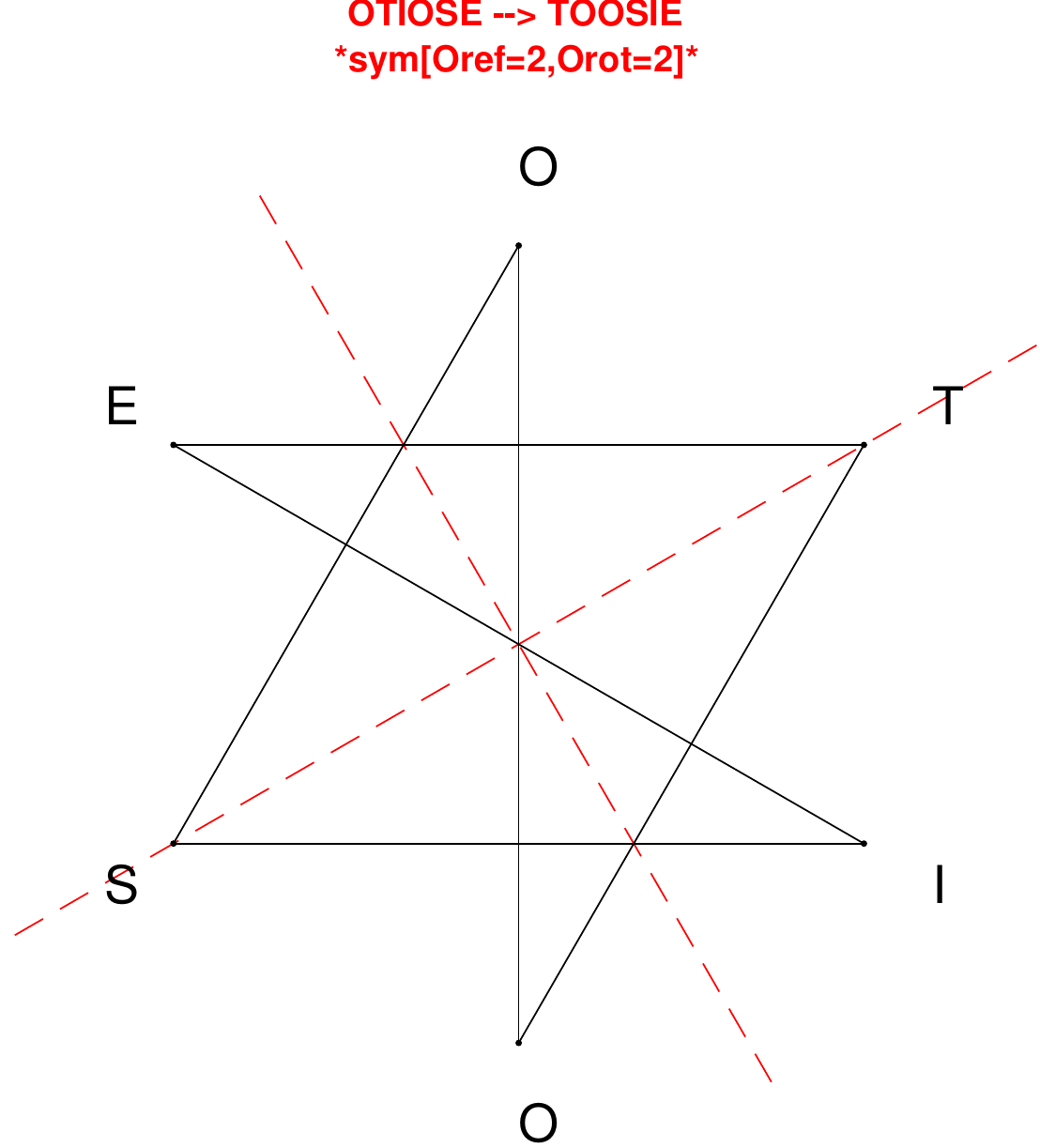}
\end{subfigure}
\hfill
\begin{subfigure}[T]{0.19\textwidth}
\centering
\includegraphics[width=\textwidth]{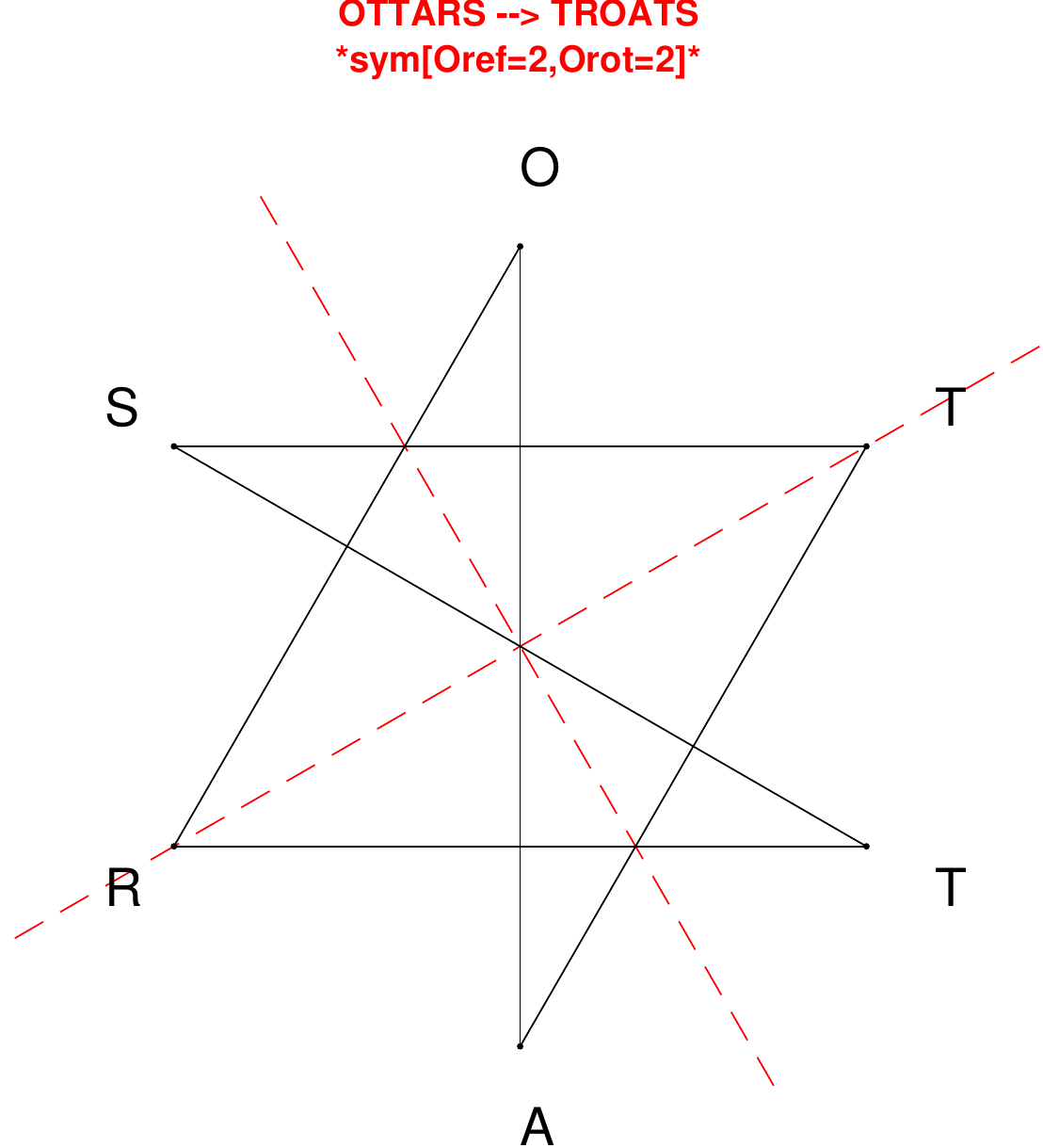}
\end{subfigure}
\hfill
\begin{subfigure}[T]{0.19\textwidth}
\centering
\includegraphics[width=\textwidth]{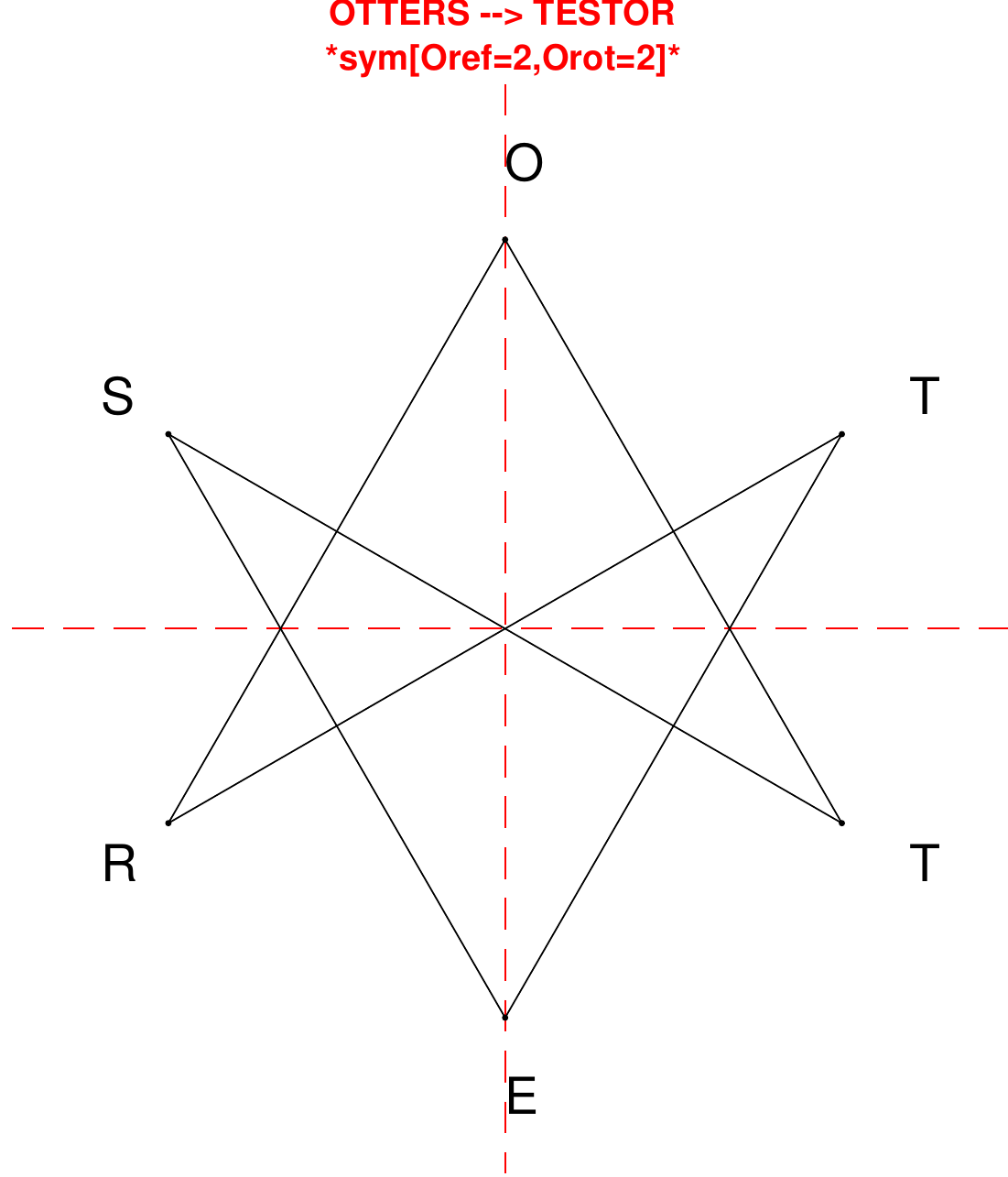}
\end{subfigure}
\hfill
\begin{subfigure}[T]{0.19\textwidth}
\centering
\includegraphics[width=\textwidth]{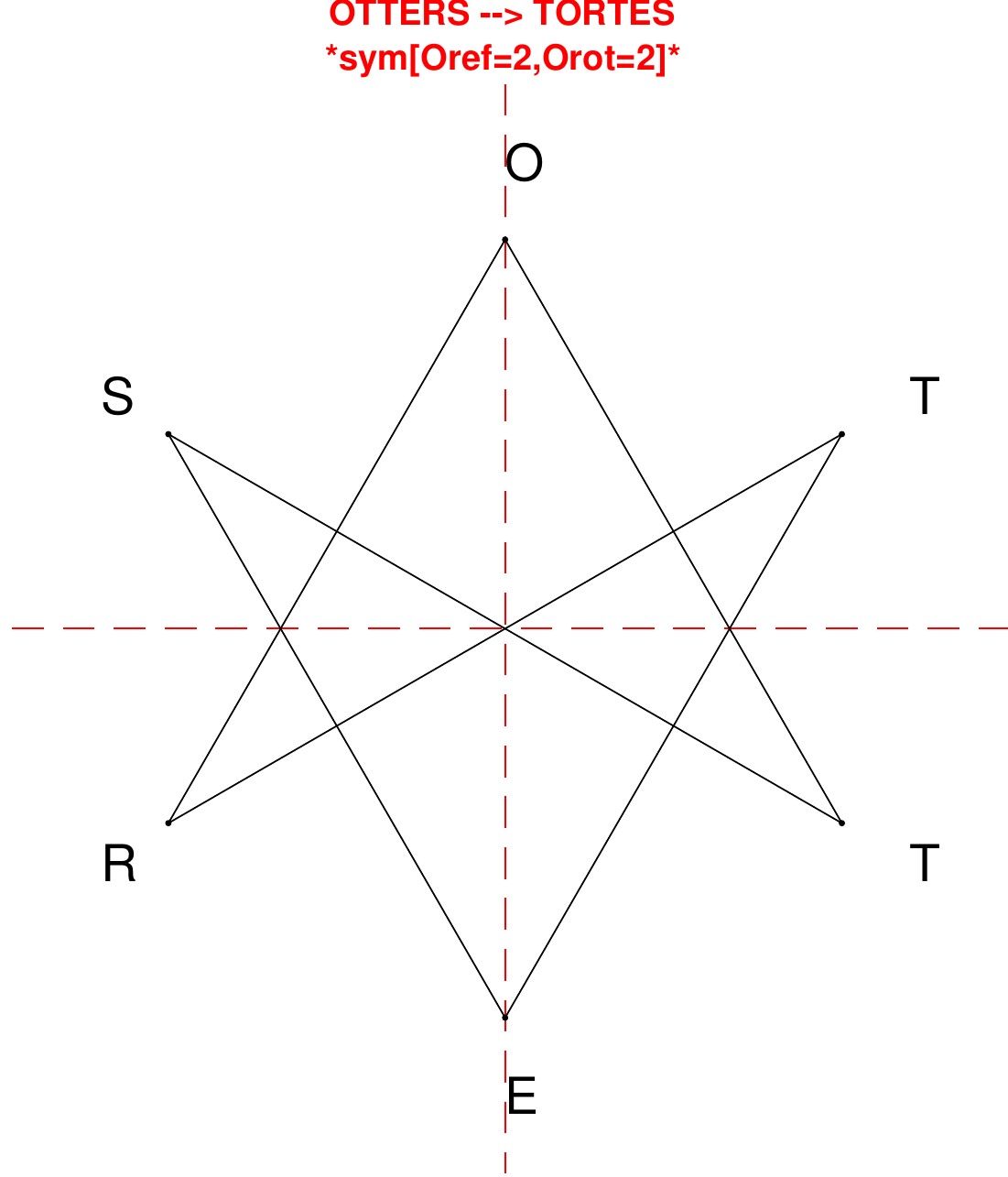}
\end{subfigure}
\end{figure}

\begin{figure}[H]
\centering
\begin{subfigure}[T]{0.19\textwidth}
\centering
\includegraphics[width=\textwidth]{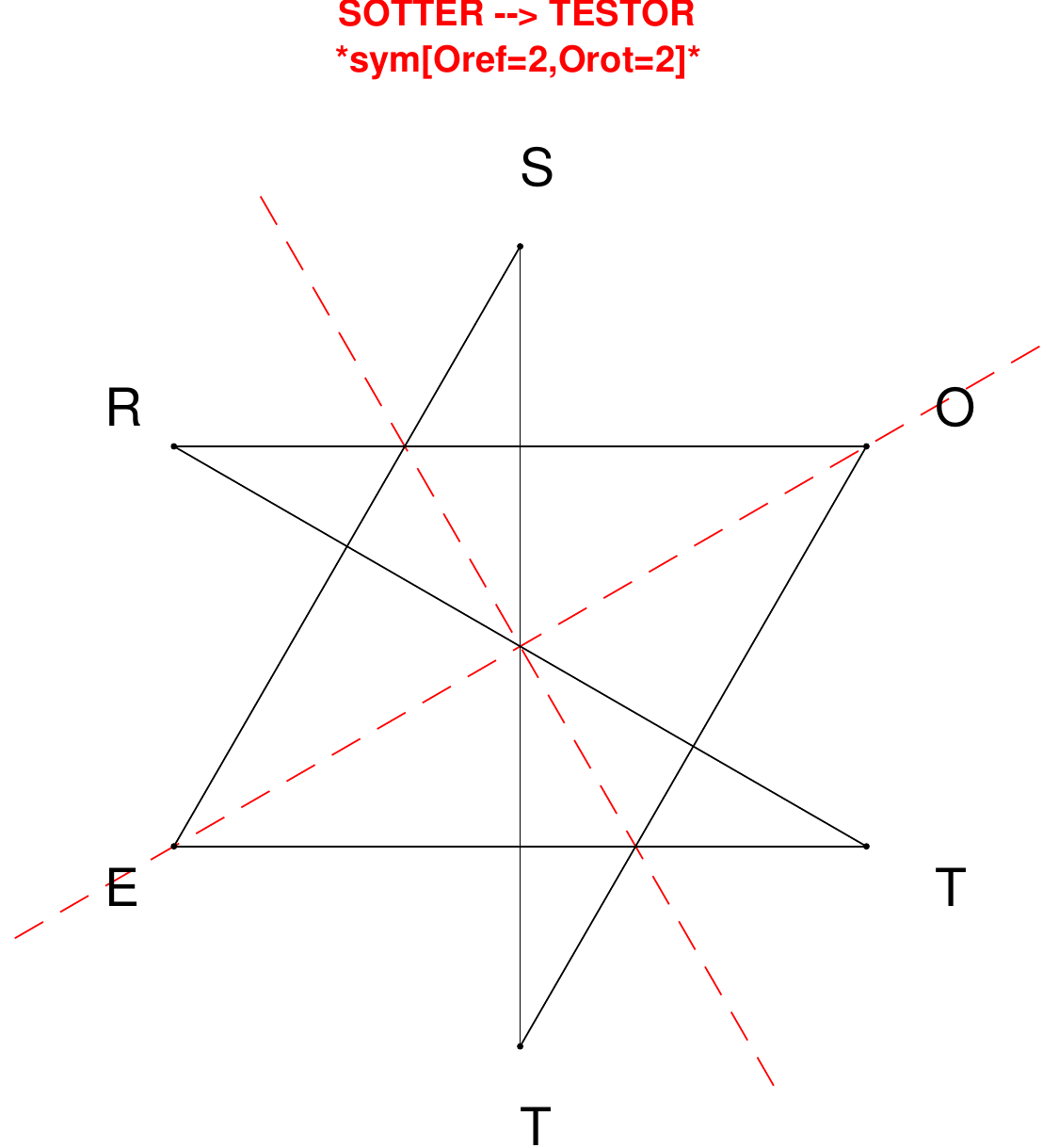}
\end{subfigure}
\hfill
\begin{subfigure}[T]{0.19\textwidth}
\centering
\includegraphics[width=\textwidth]{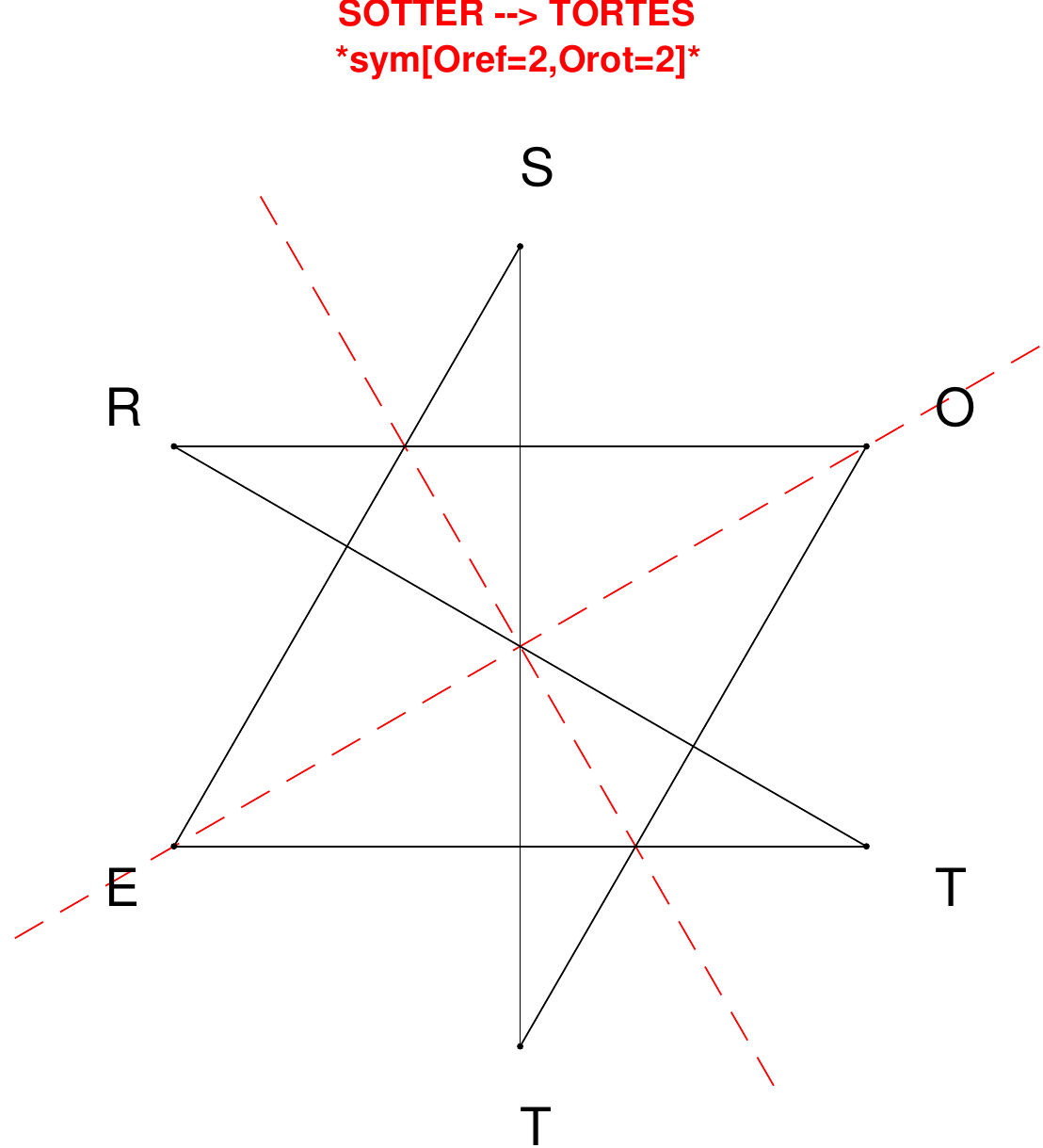}
\end{subfigure}
\hfill
\begin{subfigure}[T]{0.19\textwidth}
\centering
\includegraphics[width=\textwidth]{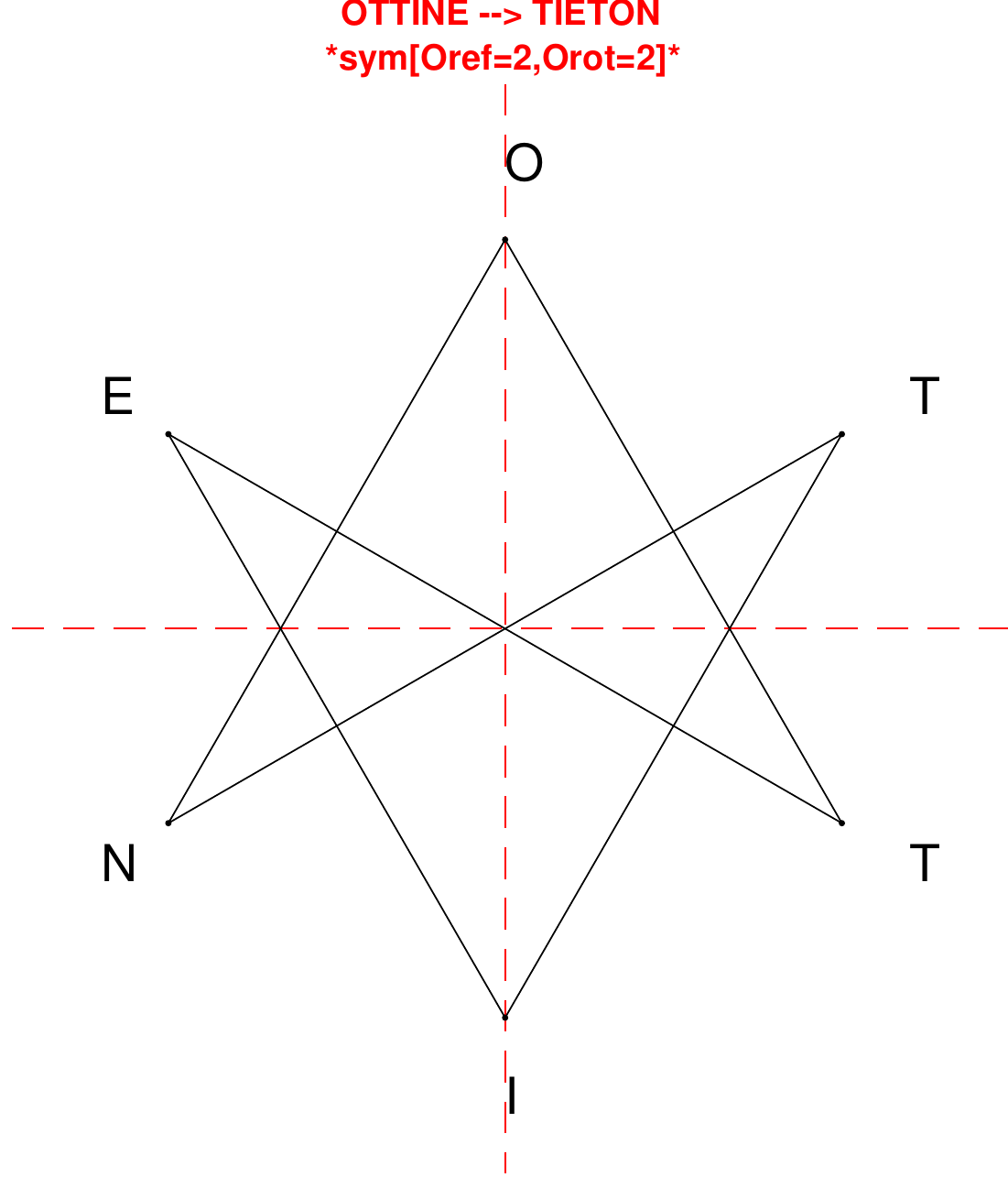}
\end{subfigure}
\hfill
\begin{subfigure}[T]{0.19\textwidth}
\centering
\includegraphics[width=\textwidth]{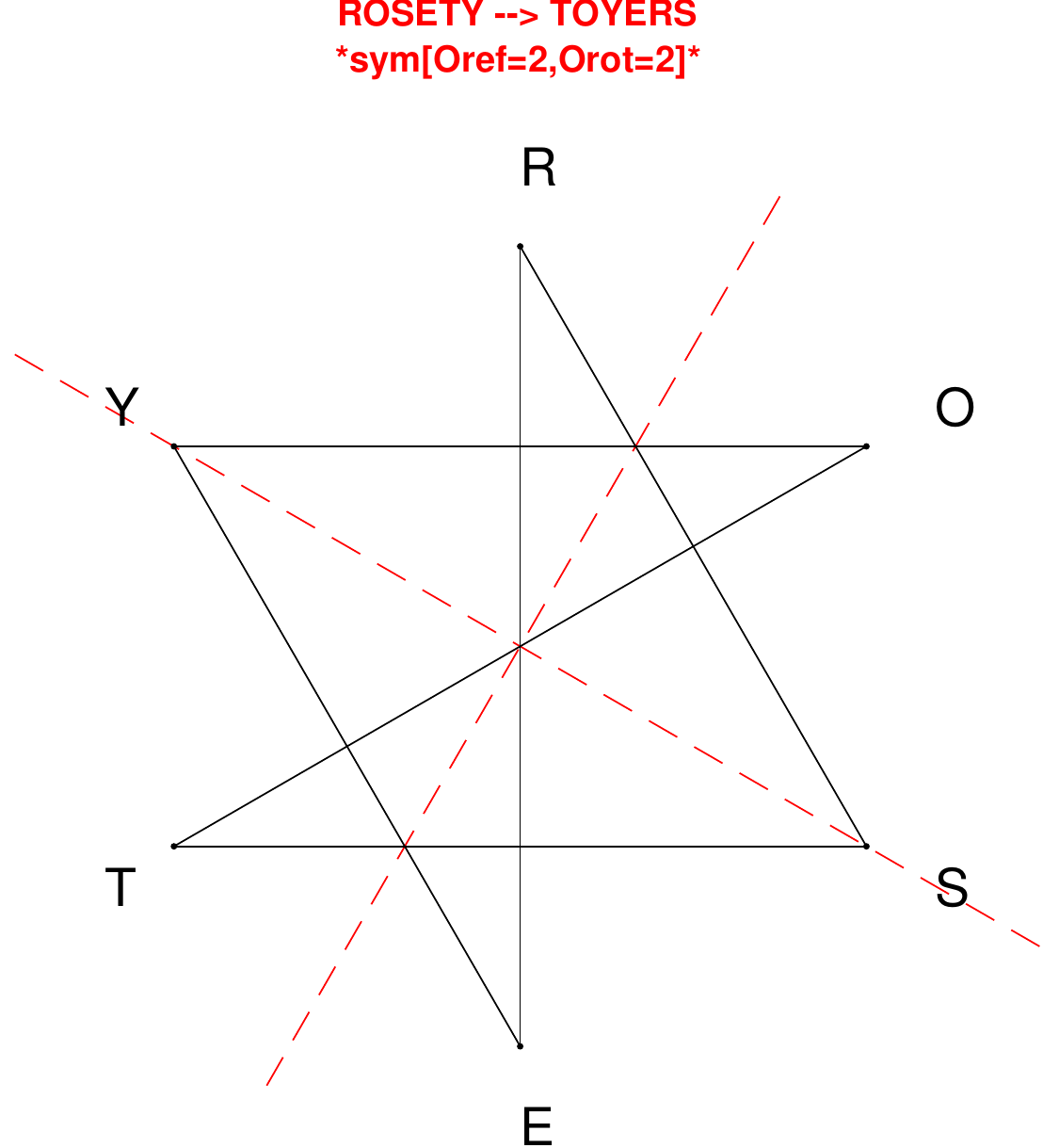}
\end{subfigure}
\hfill
\begin{subfigure}[T]{0.19\textwidth}
\centering
\includegraphics[width=\textwidth]{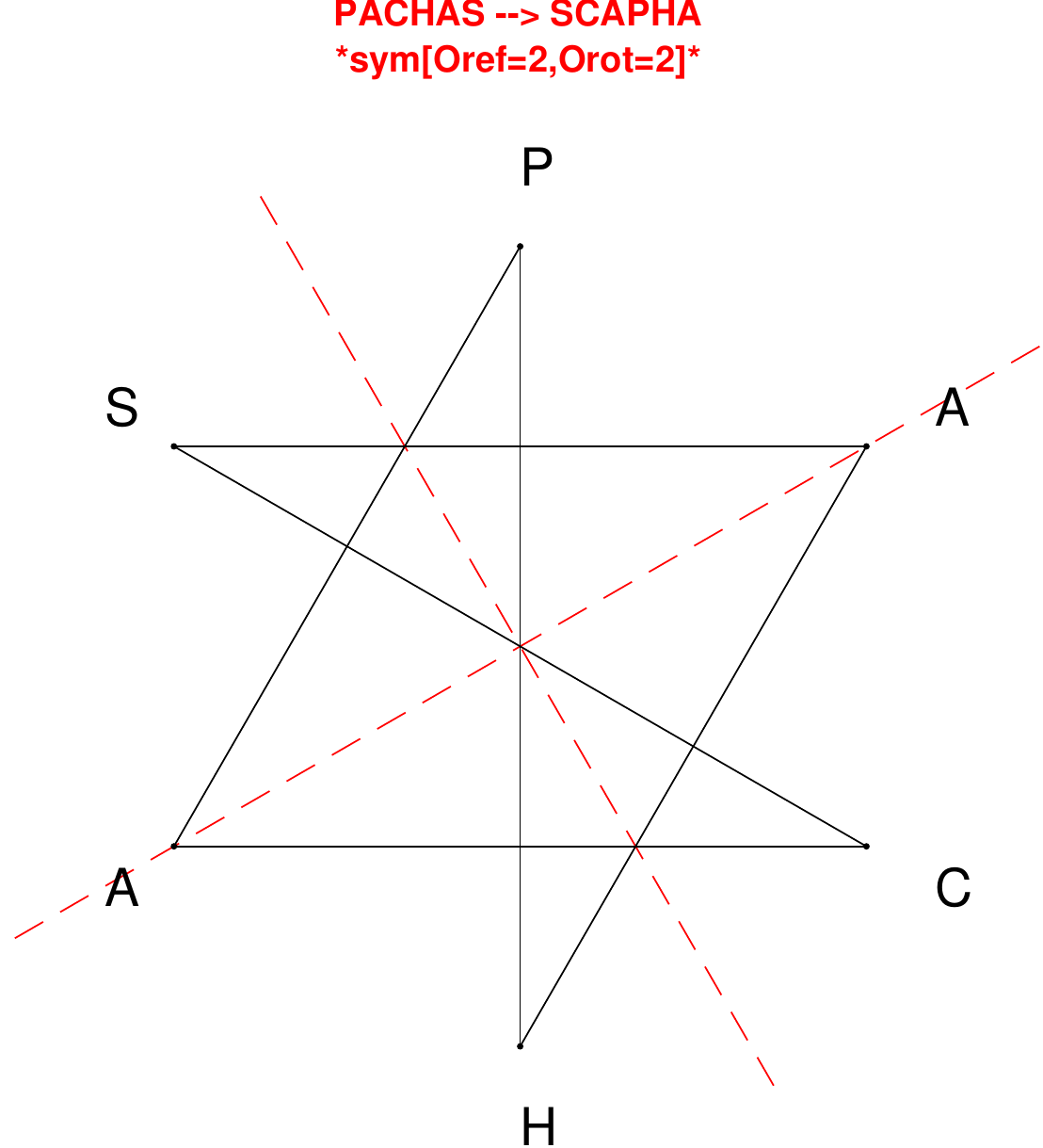}
\end{subfigure}
\end{figure}

\begin{figure}[H]
\centering
\begin{subfigure}[T]{0.19\textwidth}
\centering
\includegraphics[width=\textwidth]{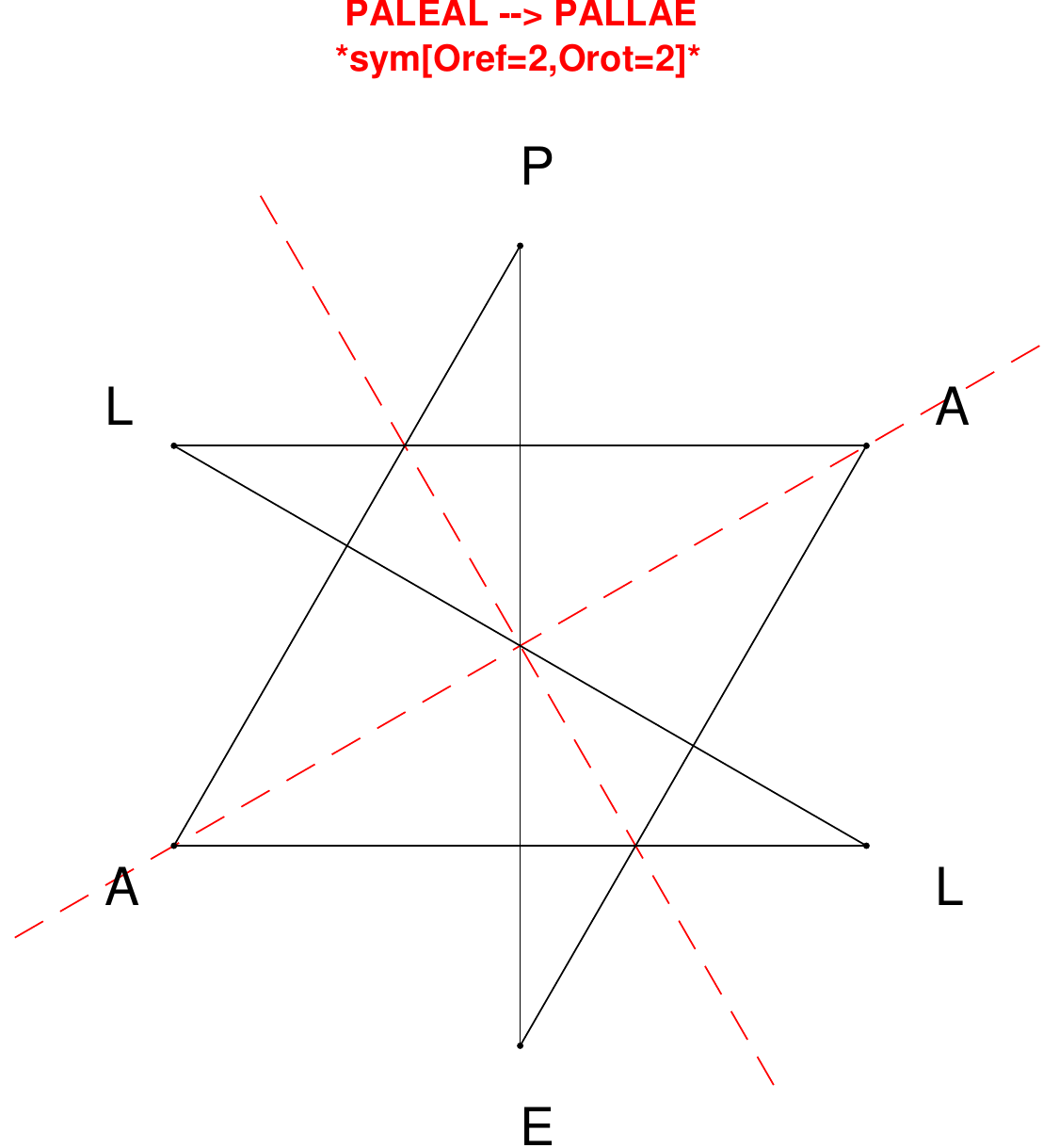}
\end{subfigure}
\hfill
\begin{subfigure}[T]{0.19\textwidth}
\centering
\includegraphics[width=\textwidth]{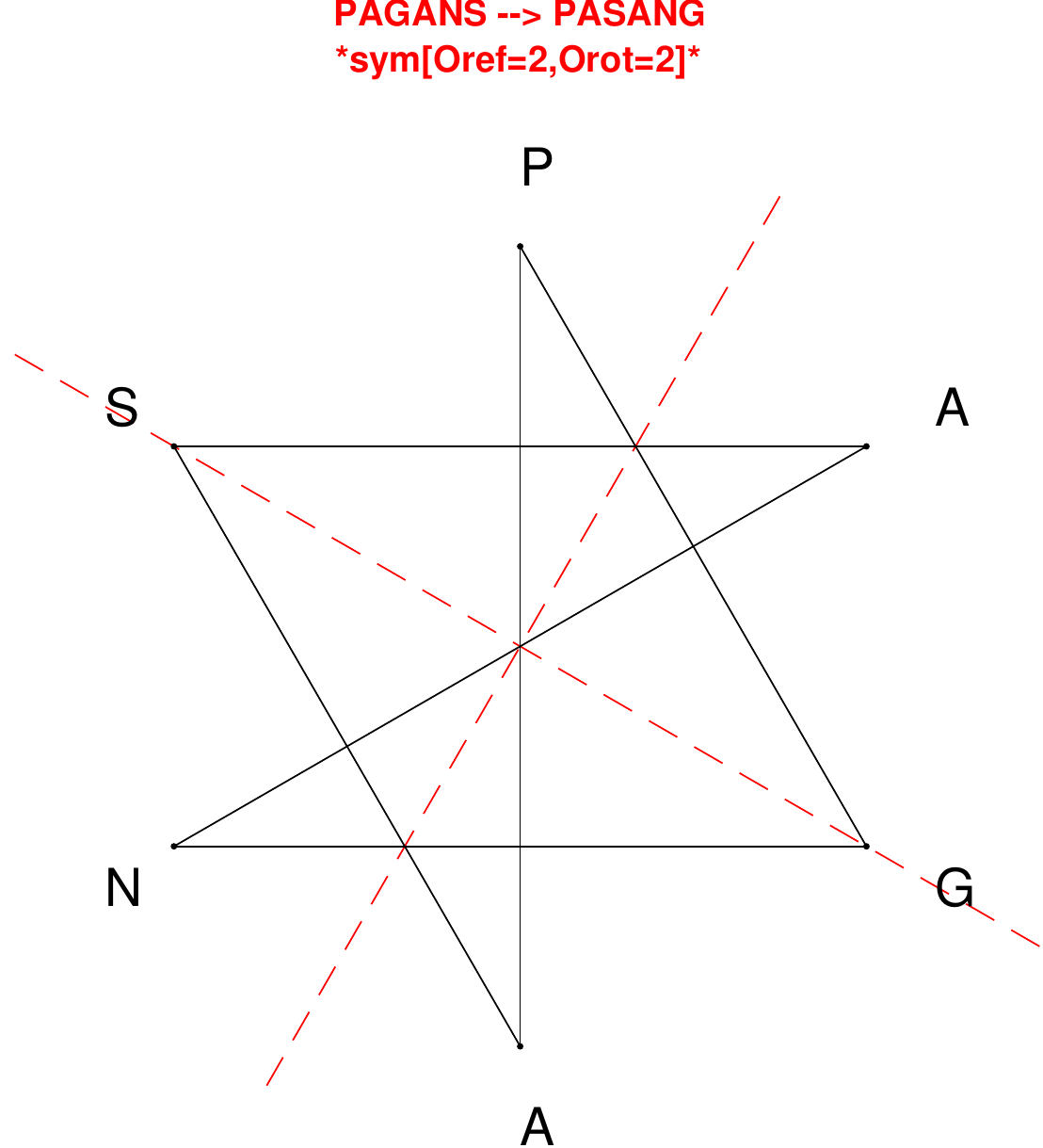}
\end{subfigure}
\hfill
\begin{subfigure}[T]{0.19\textwidth}
\centering
\includegraphics[width=\textwidth]{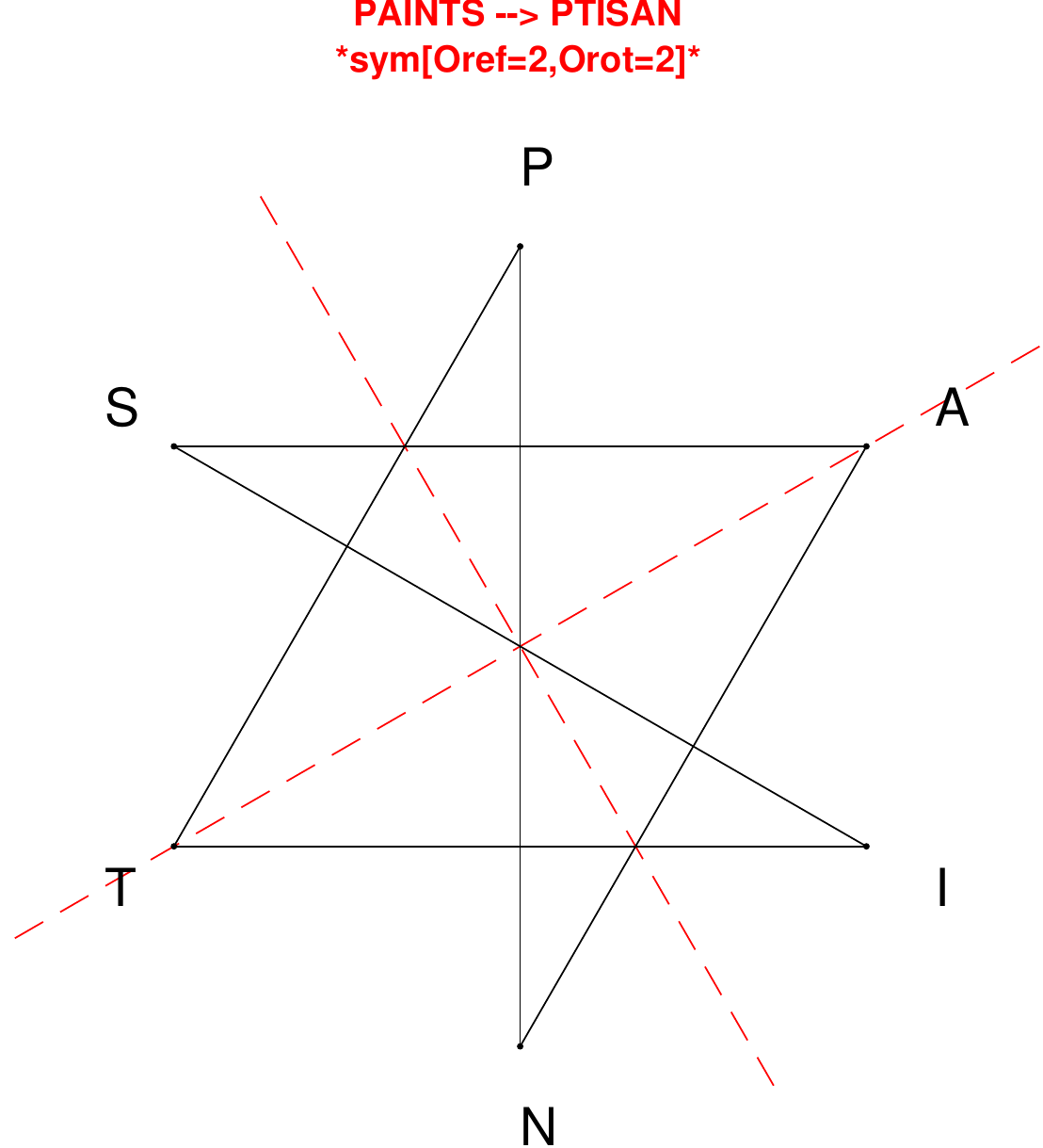}
\end{subfigure}
\hfill
\begin{subfigure}[T]{0.19\textwidth}
\centering
\includegraphics[width=\textwidth]{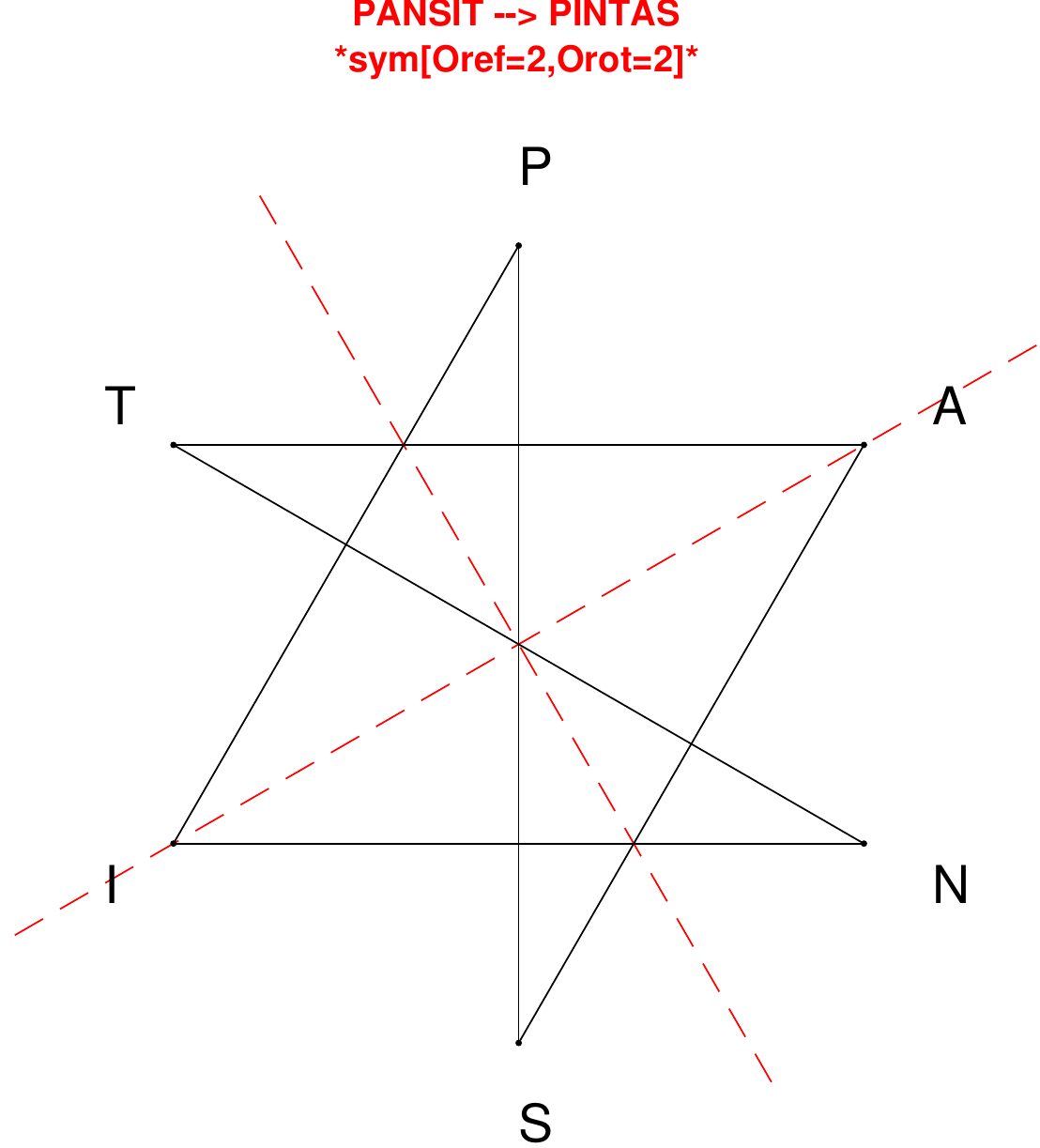}
\end{subfigure}
\hfill
\begin{subfigure}[T]{0.19\textwidth}
\centering
\includegraphics[width=\textwidth]{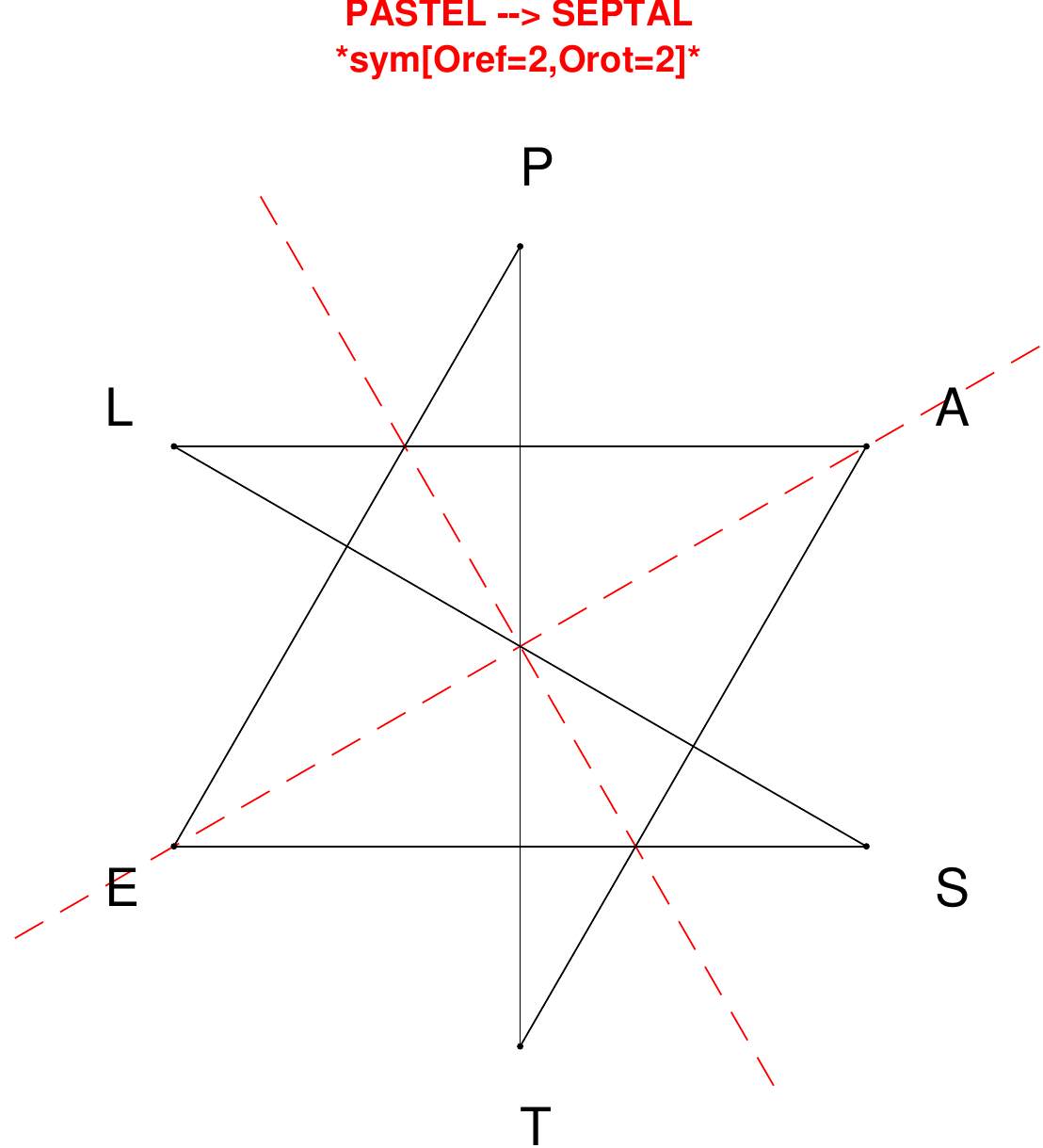}
\end{subfigure}
\end{figure}

\begin{figure}[H]
\centering
\begin{subfigure}[T]{0.19\textwidth}
\centering
\includegraphics[width=\textwidth]{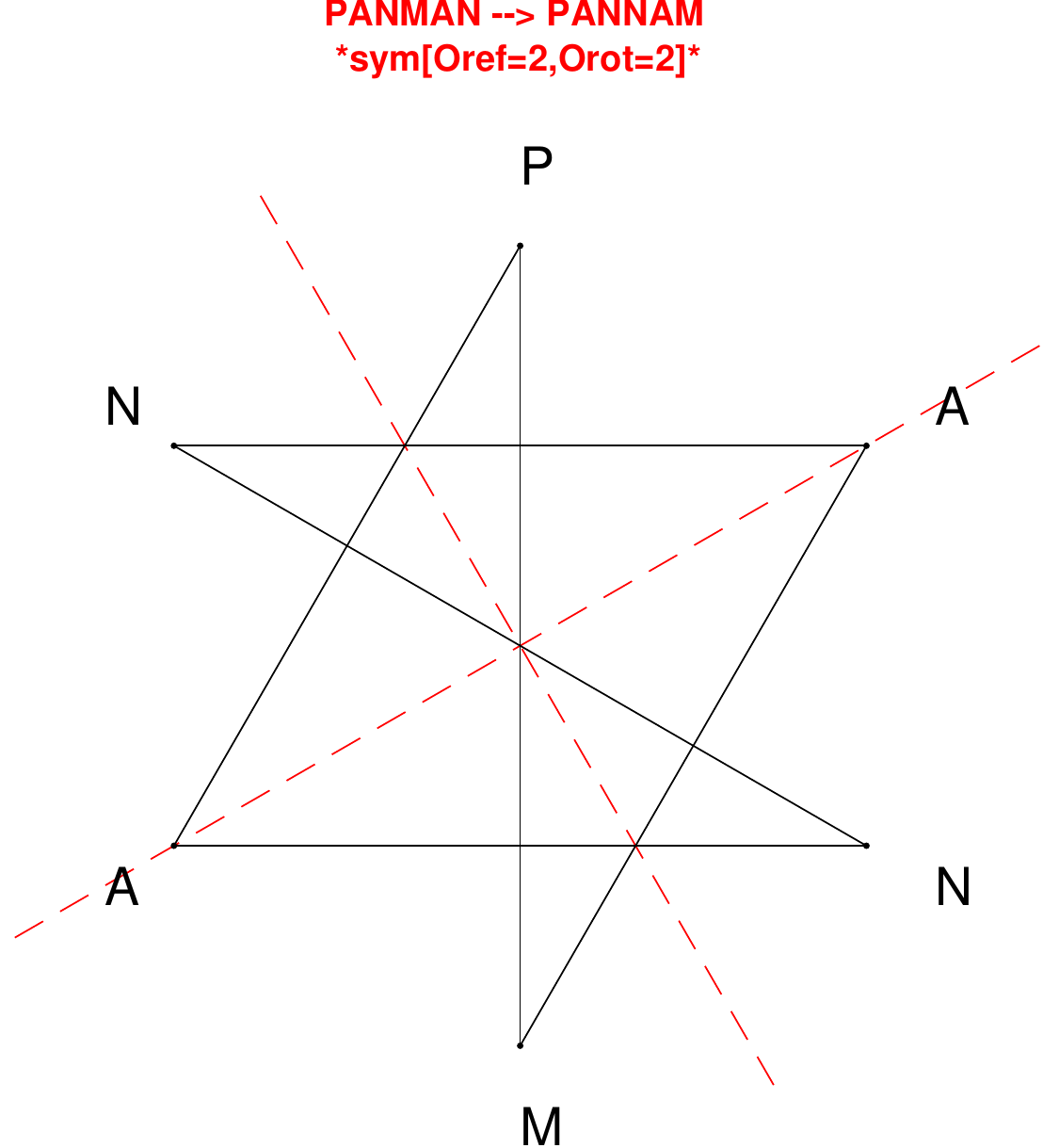}
\end{subfigure}
\hfill
\begin{subfigure}[T]{0.19\textwidth}
\centering
\includegraphics[width=\textwidth]{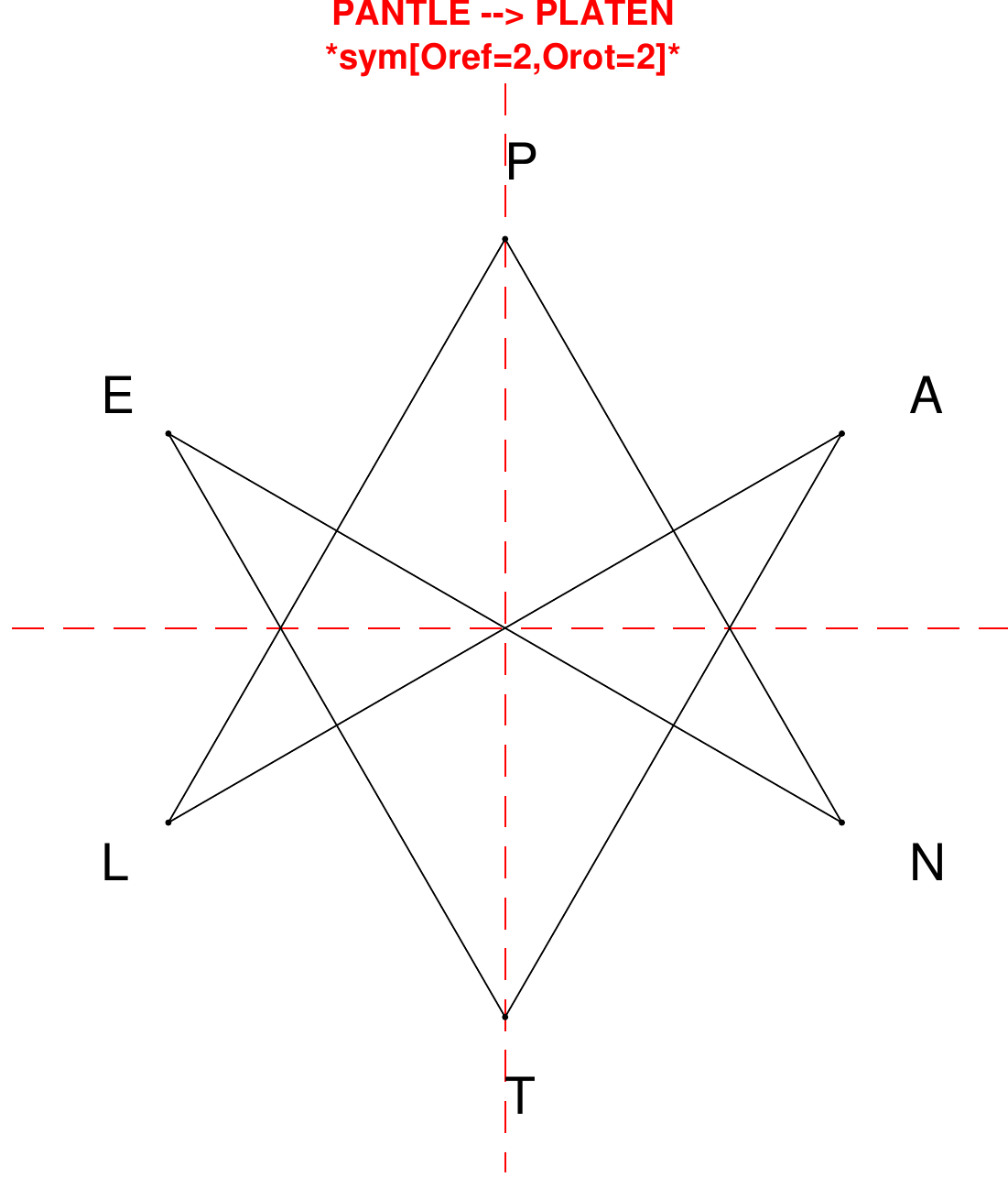}
\end{subfigure}
\hfill
\begin{subfigure}[T]{0.19\textwidth}
\centering
\includegraphics[width=\textwidth]{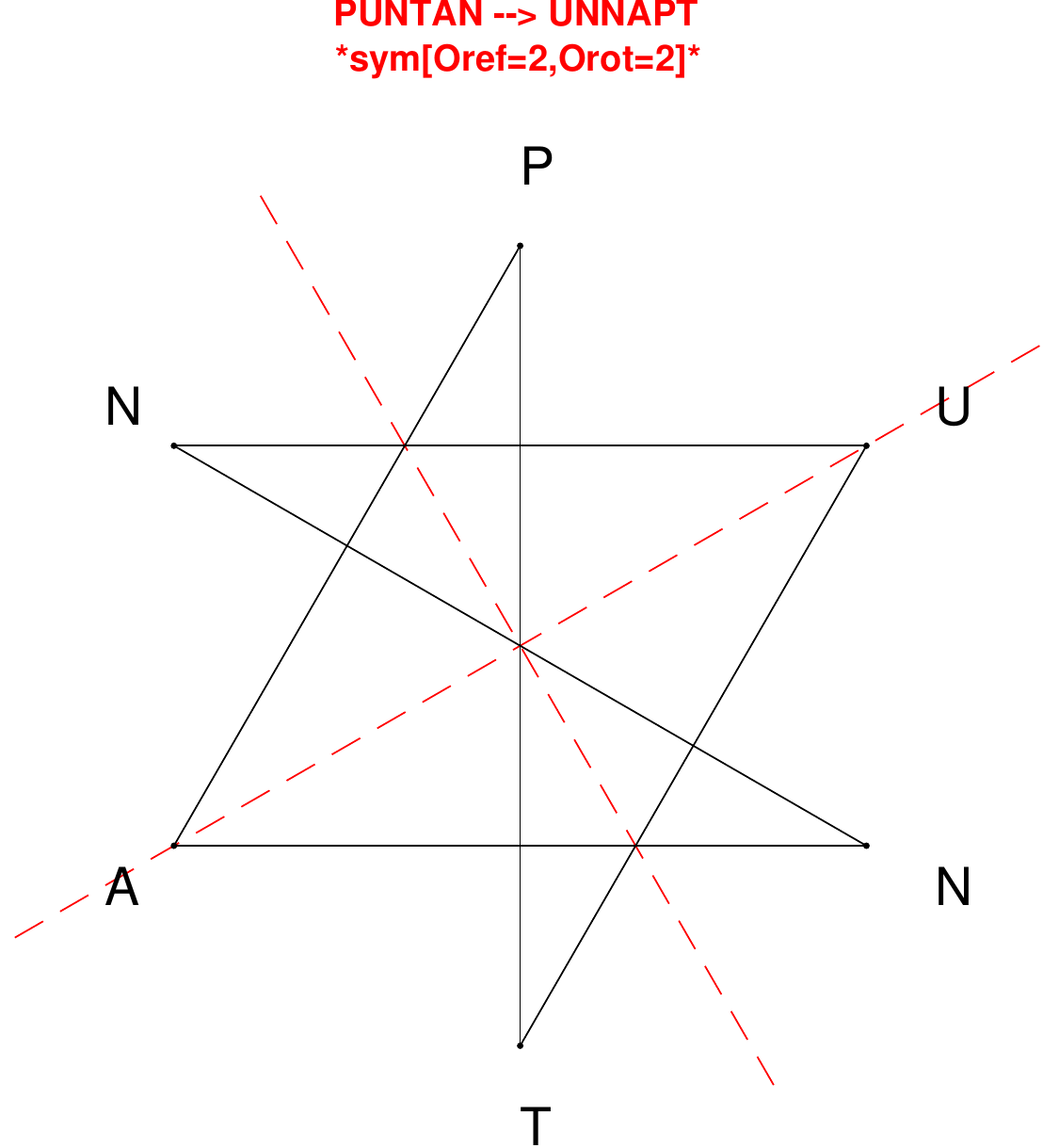}
\end{subfigure}
\hfill
\begin{subfigure}[T]{0.19\textwidth}
\centering
\includegraphics[width=\textwidth]{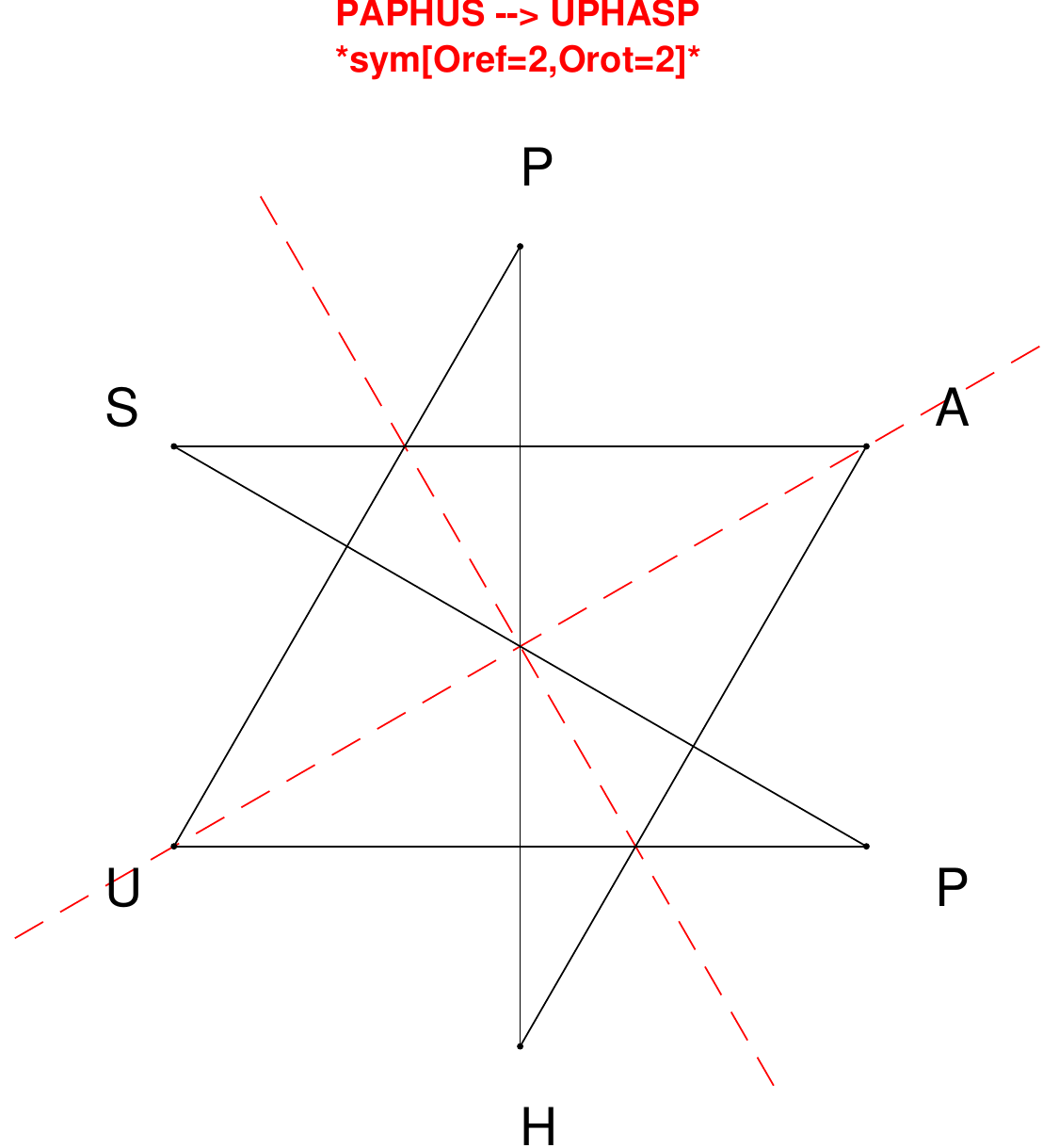}
\end{subfigure}
\hfill
\begin{subfigure}[T]{0.19\textwidth}
\centering
\includegraphics[width=\textwidth]{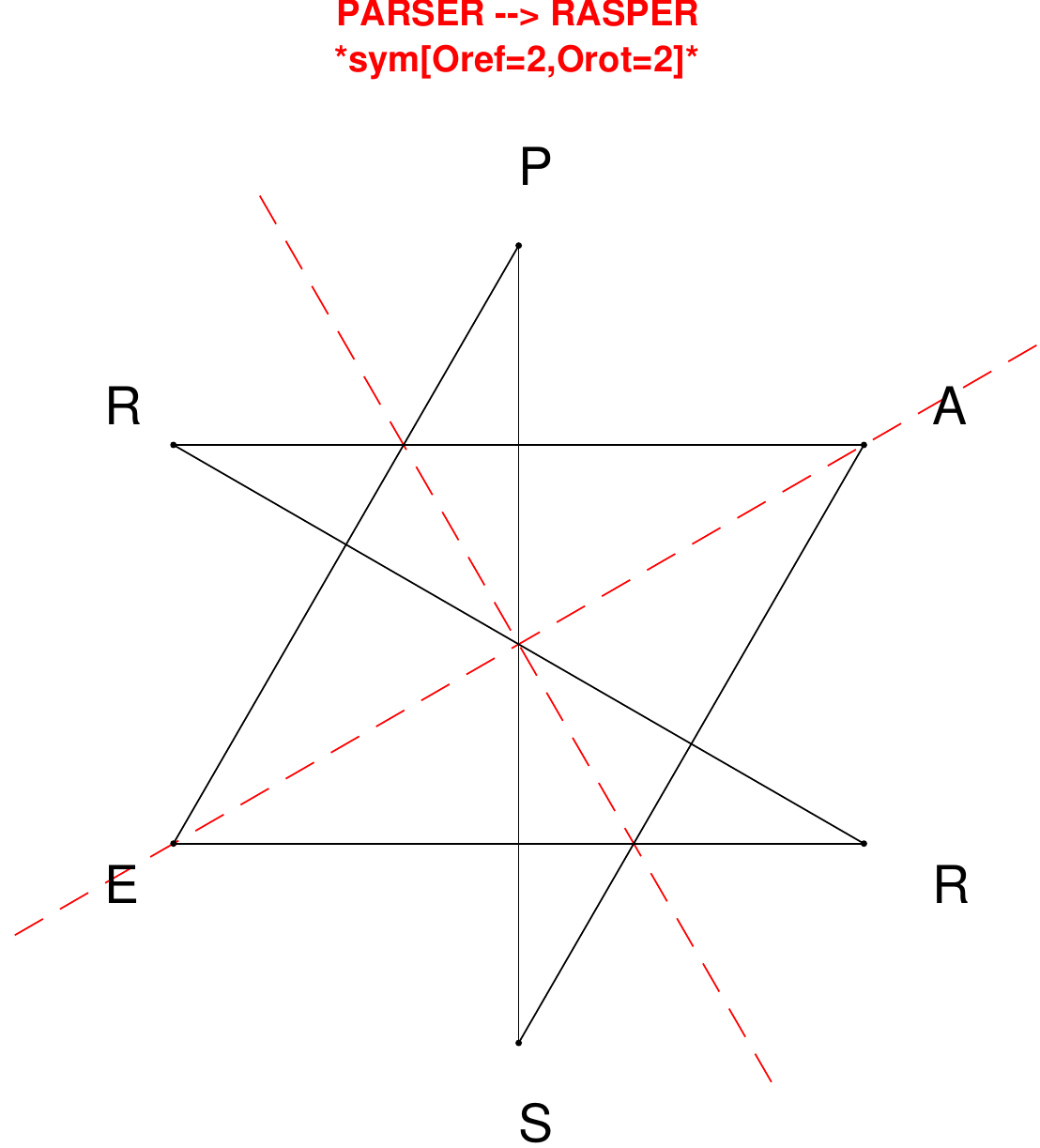}
\end{subfigure}
\end{figure}

\begin{figure}[H]
\centering
\begin{subfigure}[T]{0.19\textwidth}
\centering
\includegraphics[width=\textwidth]{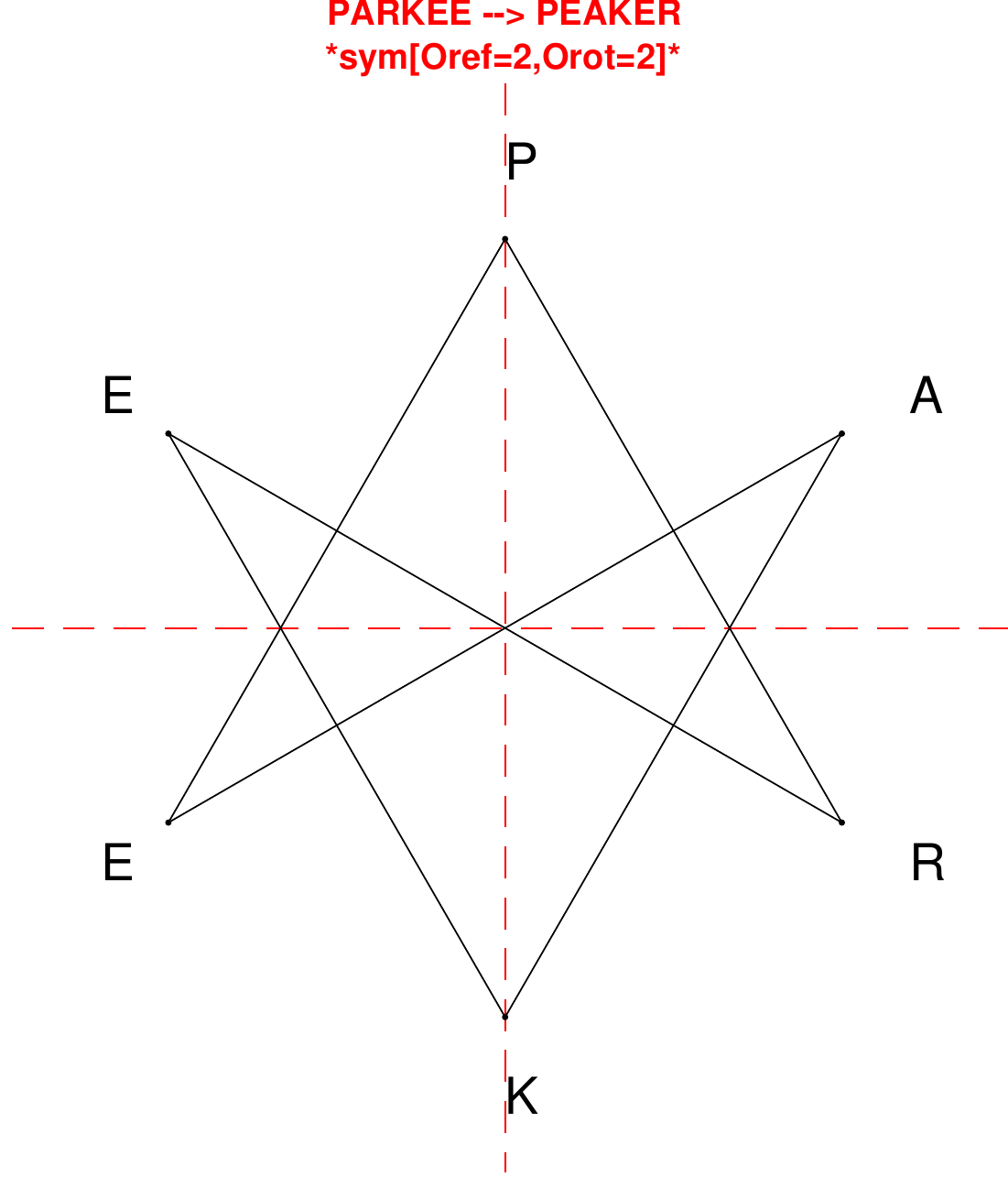}
\end{subfigure}
\hfill
\begin{subfigure}[T]{0.19\textwidth}
\centering
\includegraphics[width=\textwidth]{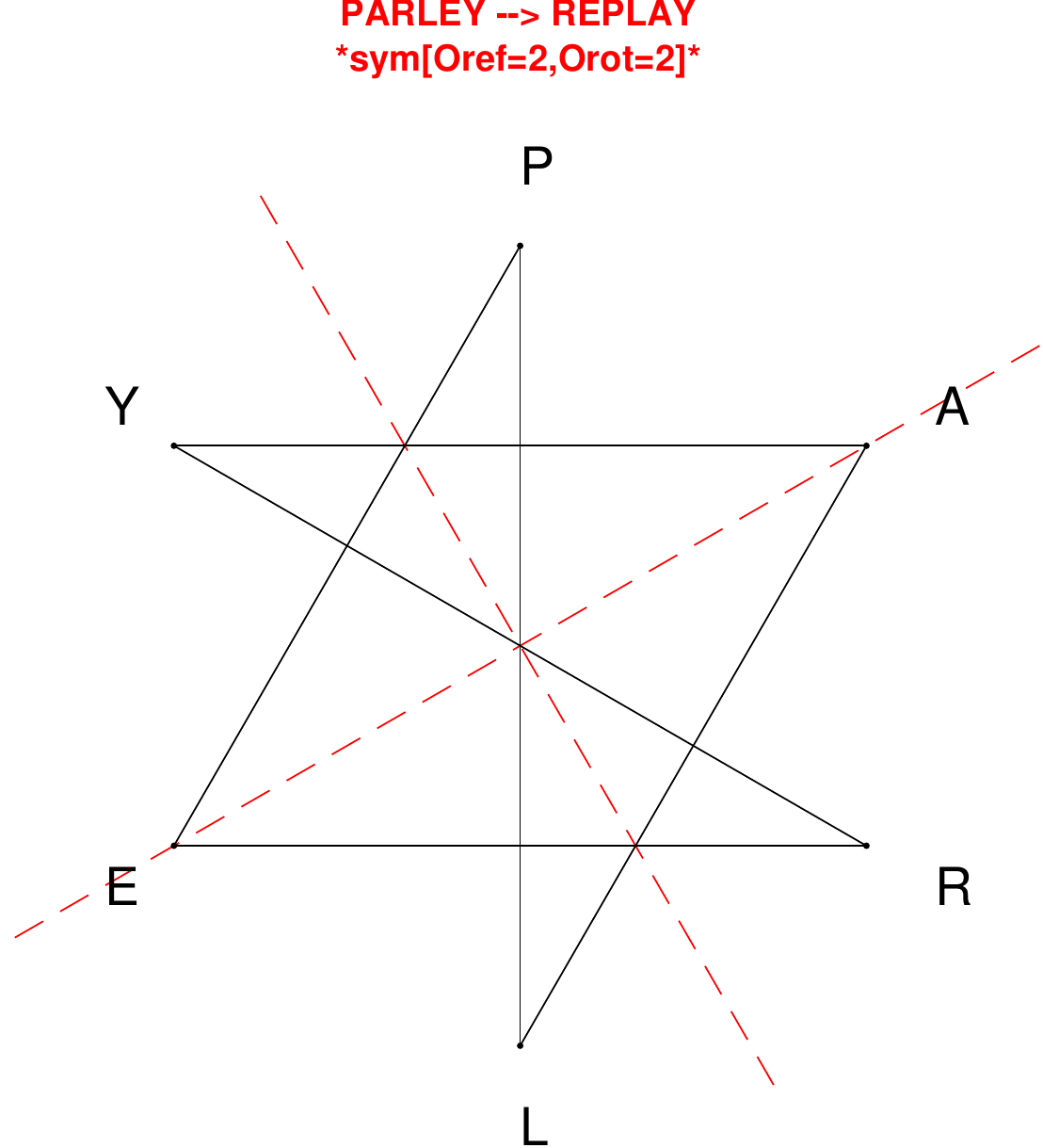}
\end{subfigure}
\hfill
\begin{subfigure}[T]{0.19\textwidth}
\centering
\includegraphics[width=\textwidth]{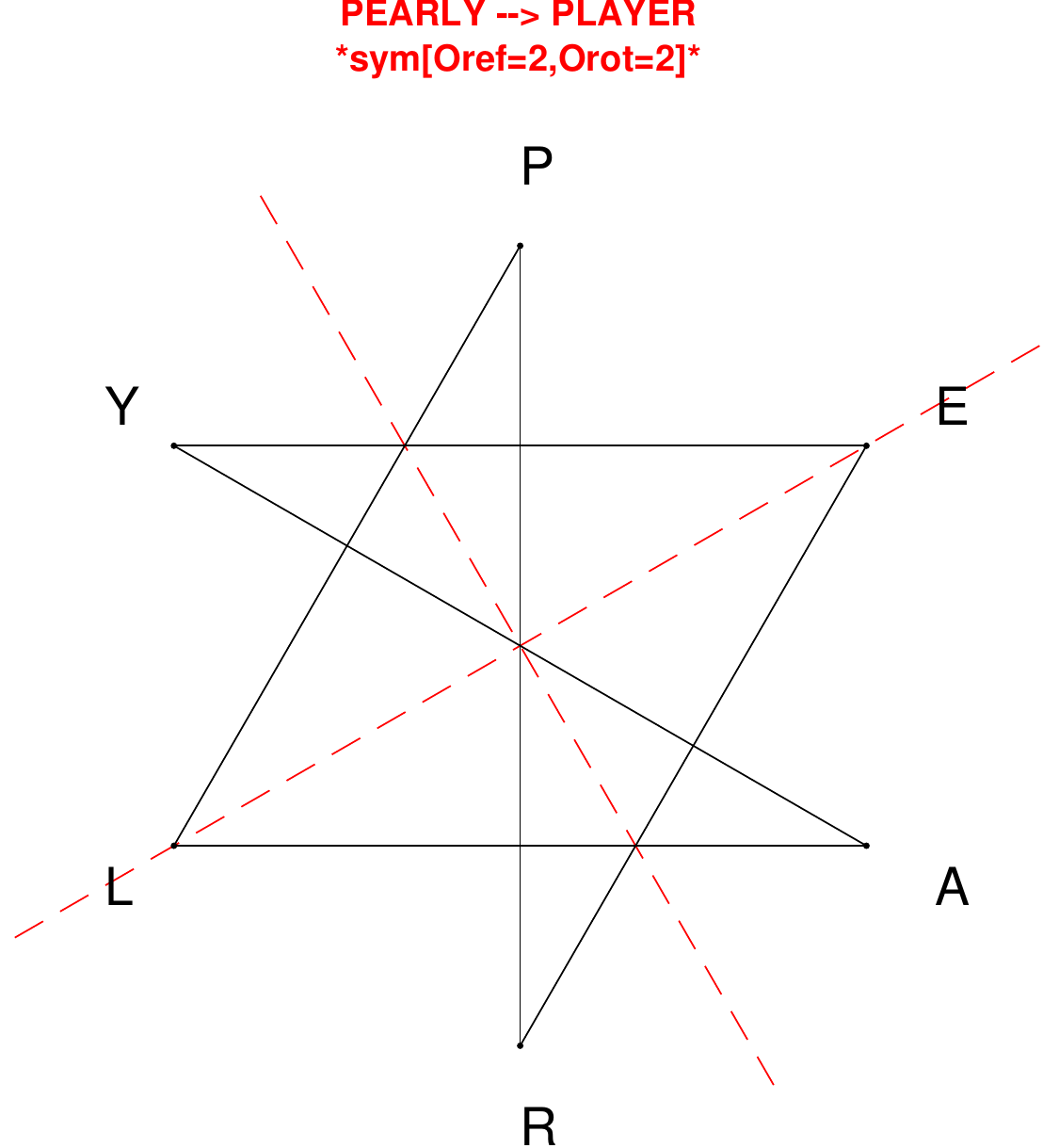}
\end{subfigure}
\hfill
\begin{subfigure}[T]{0.19\textwidth}
\centering
\includegraphics[width=\textwidth]{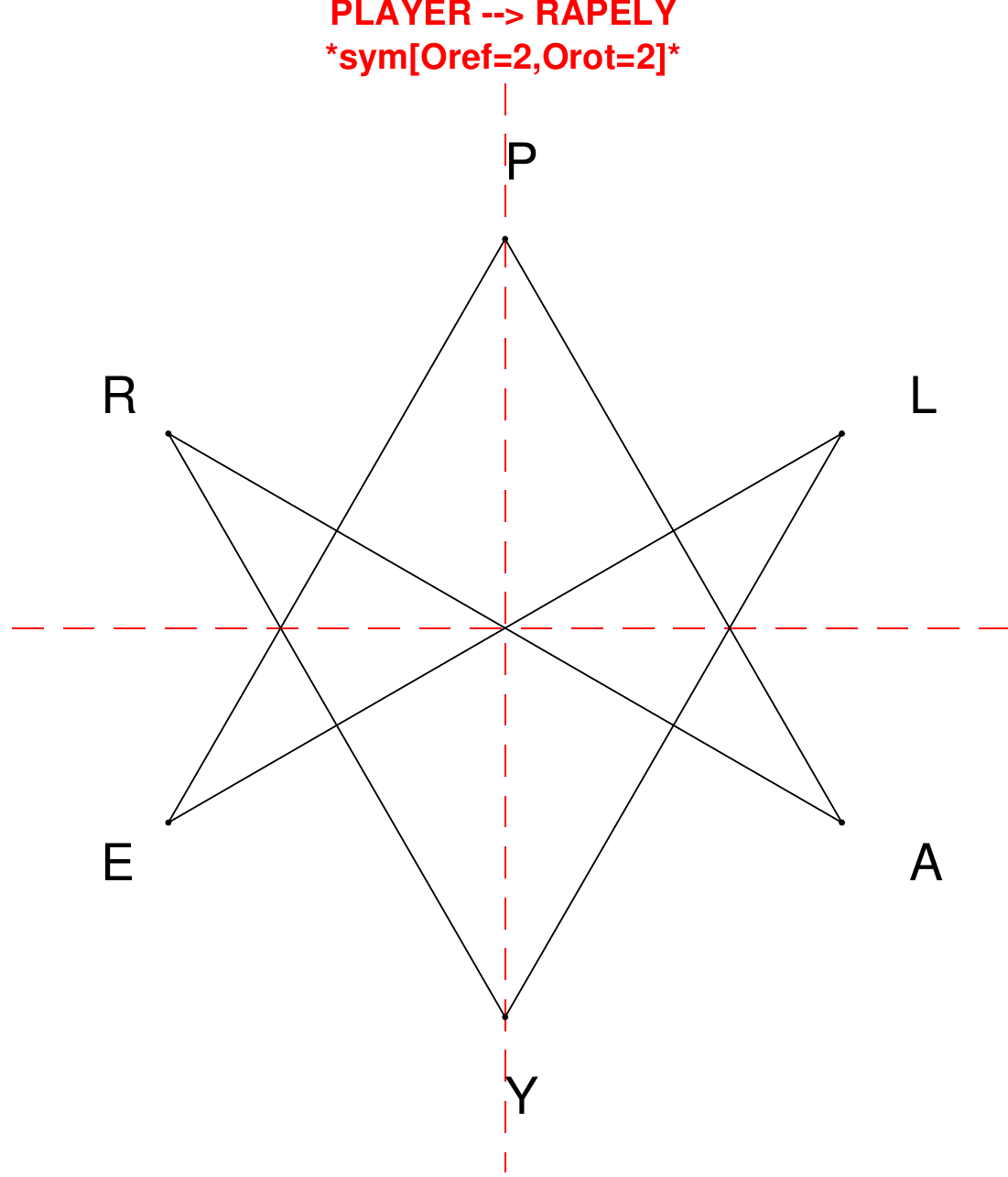}
\end{subfigure}
\hfill
\begin{subfigure}[T]{0.19\textwidth}
\centering
\includegraphics[width=\textwidth]{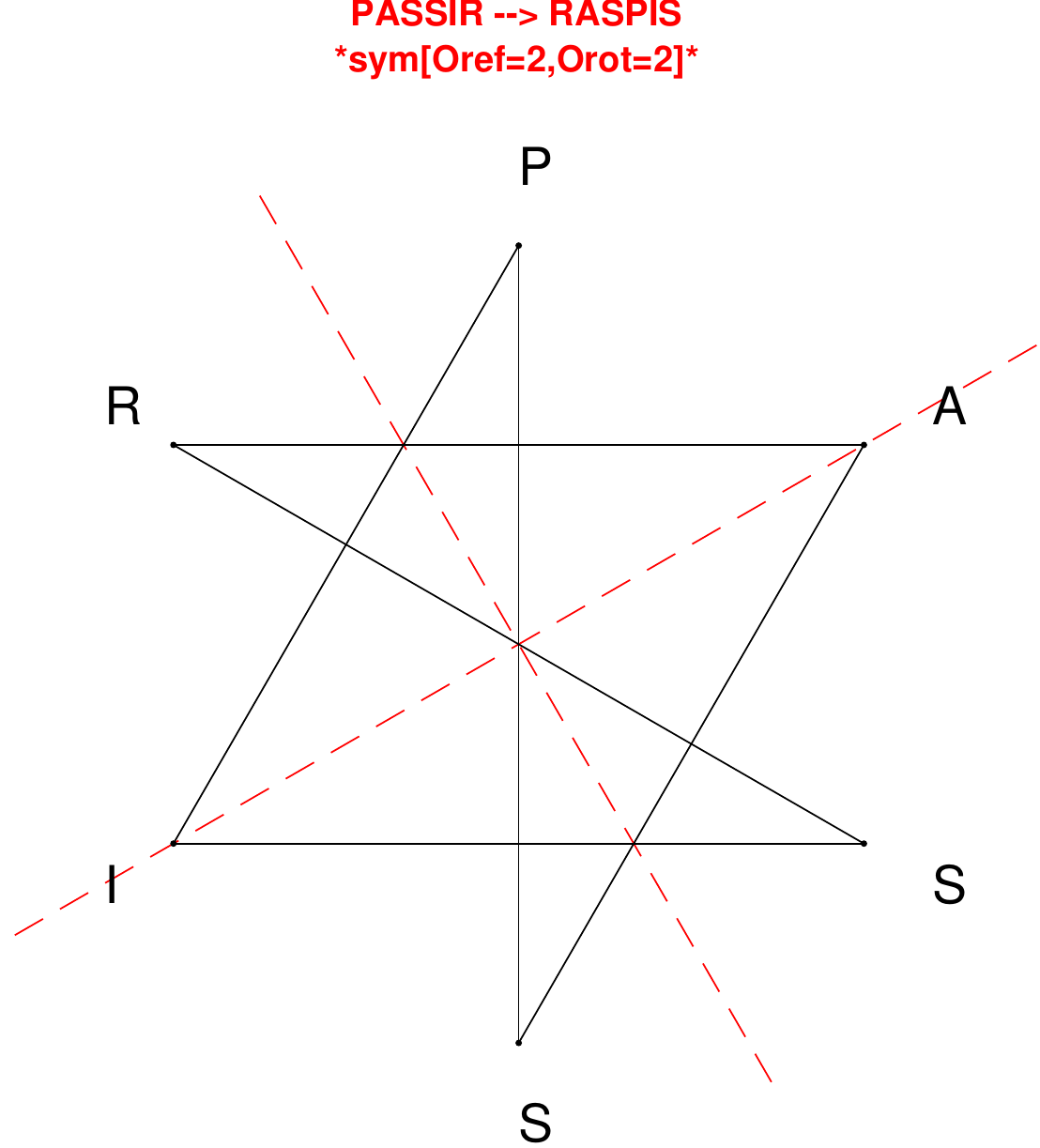}
\end{subfigure}
\end{figure}

\begin{figure}[H]
\centering
\begin{subfigure}[T]{0.19\textwidth}
\centering
\includegraphics[width=\textwidth]{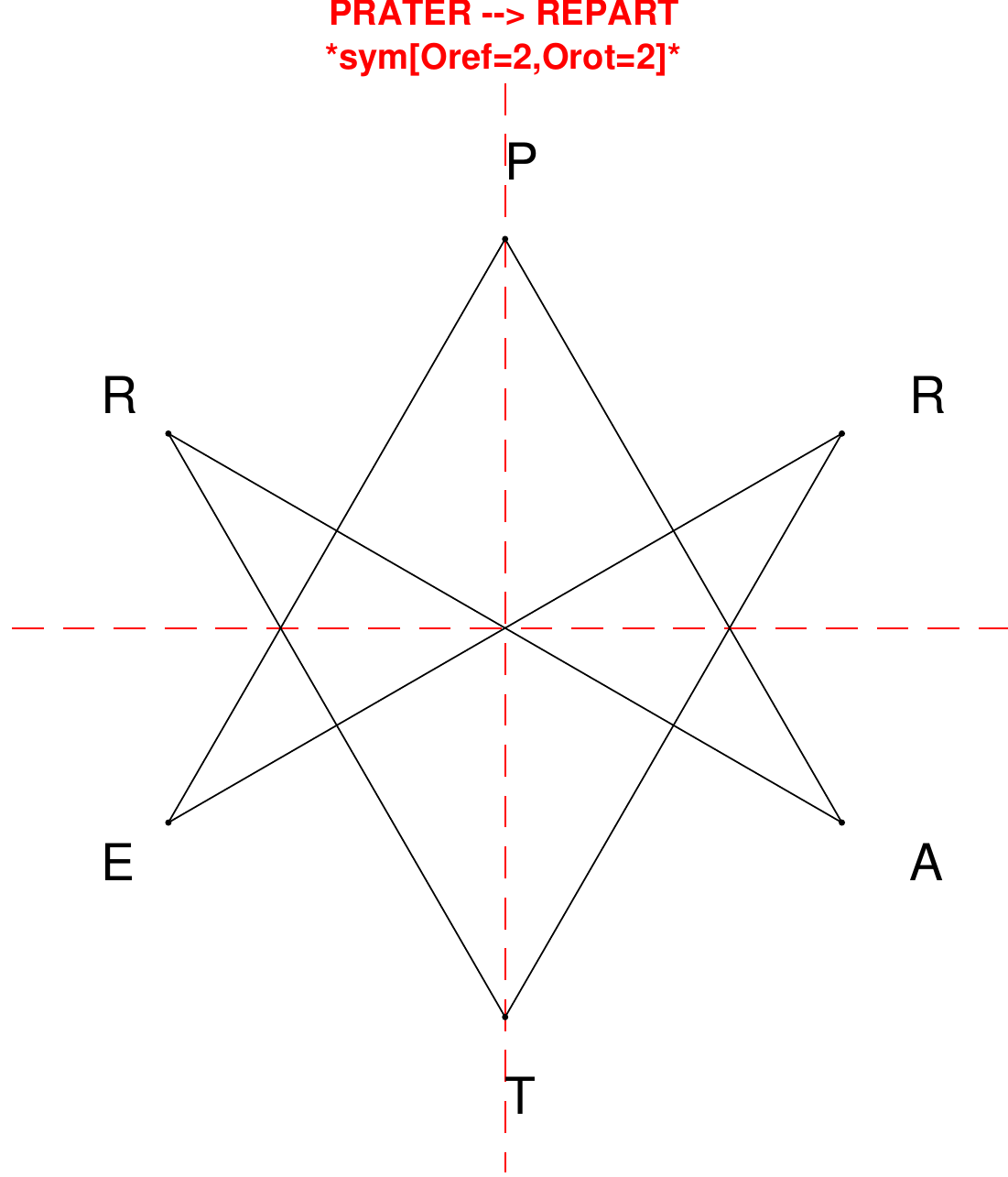}
\end{subfigure}
\hfill
\begin{subfigure}[T]{0.19\textwidth}
\centering
\includegraphics[width=\textwidth]{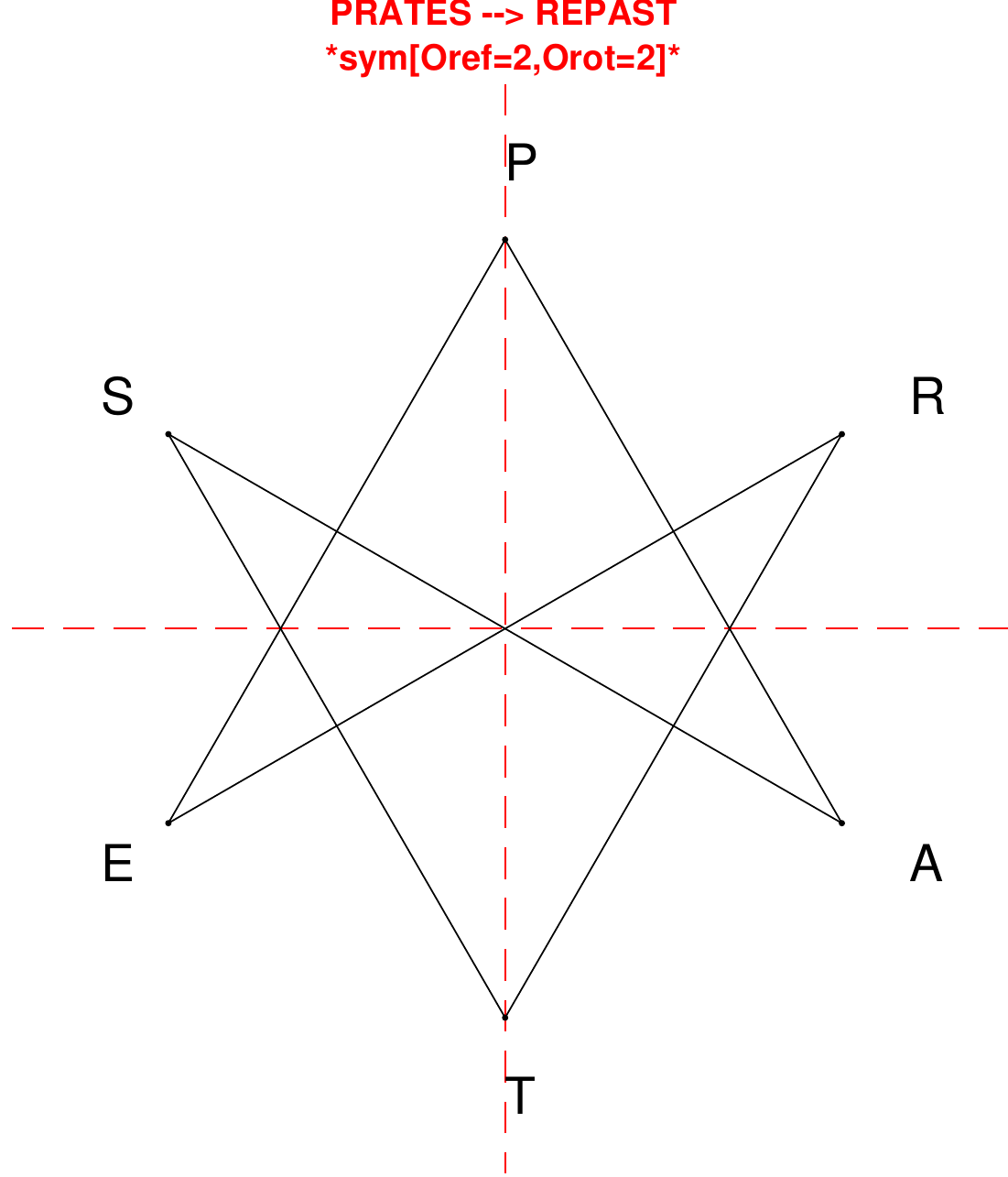}
\end{subfigure}
\hfill
\begin{subfigure}[T]{0.19\textwidth}
\centering
\includegraphics[width=\textwidth]{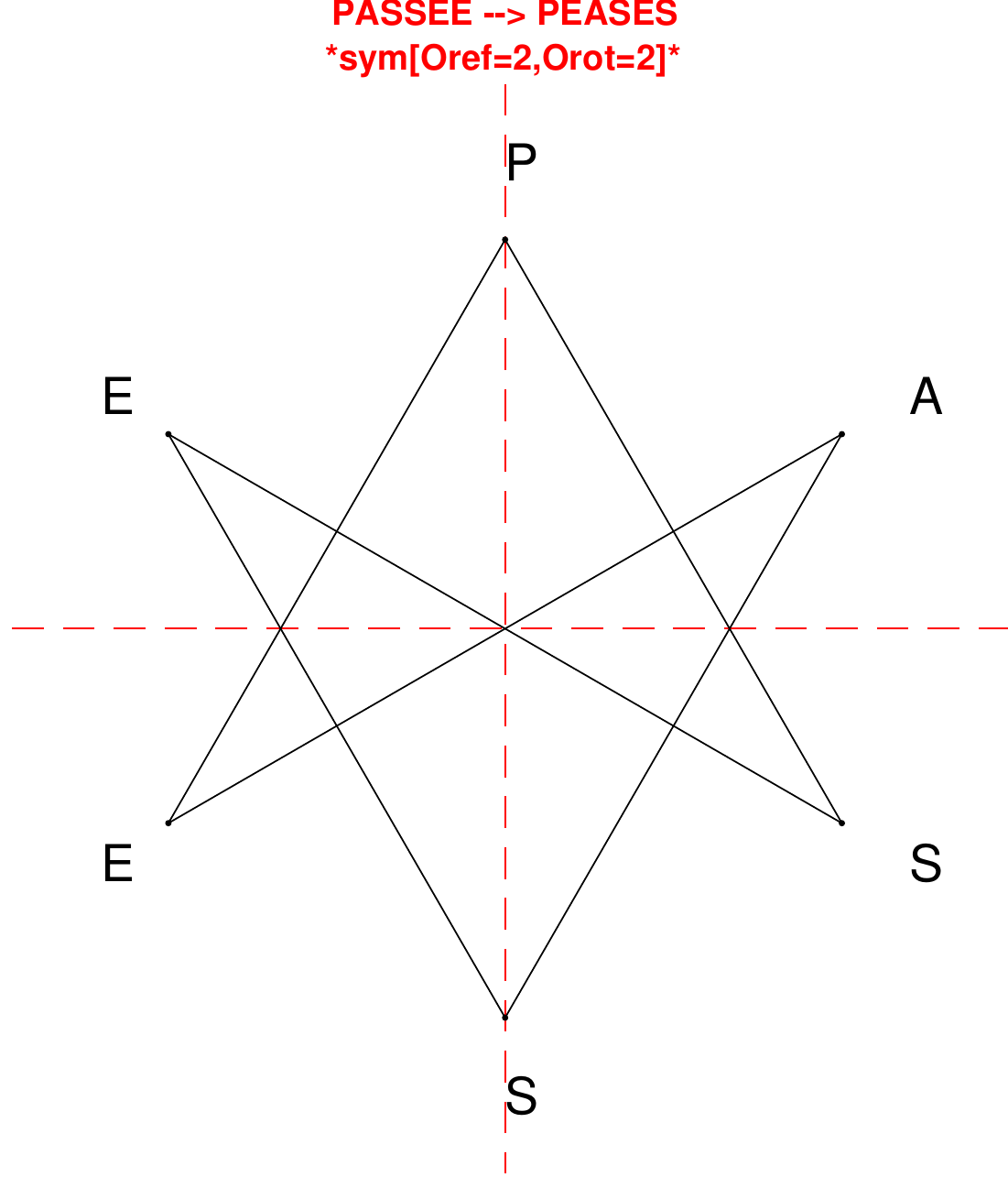}
\end{subfigure}
\hfill
\begin{subfigure}[T]{0.19\textwidth}
\centering
\includegraphics[width=\textwidth]{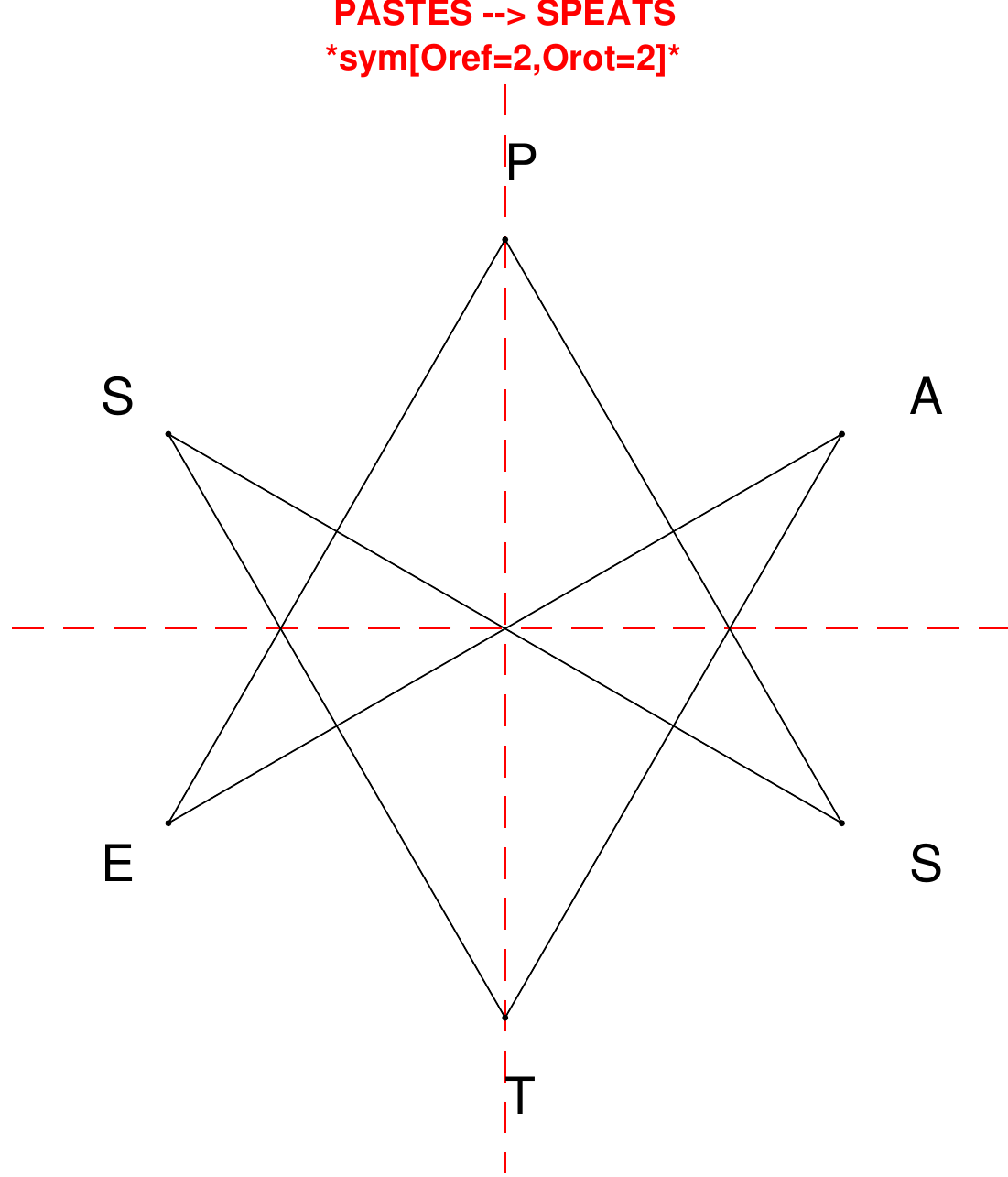}
\end{subfigure}
\hfill
\begin{subfigure}[T]{0.19\textwidth}
\centering
\includegraphics[width=\textwidth]{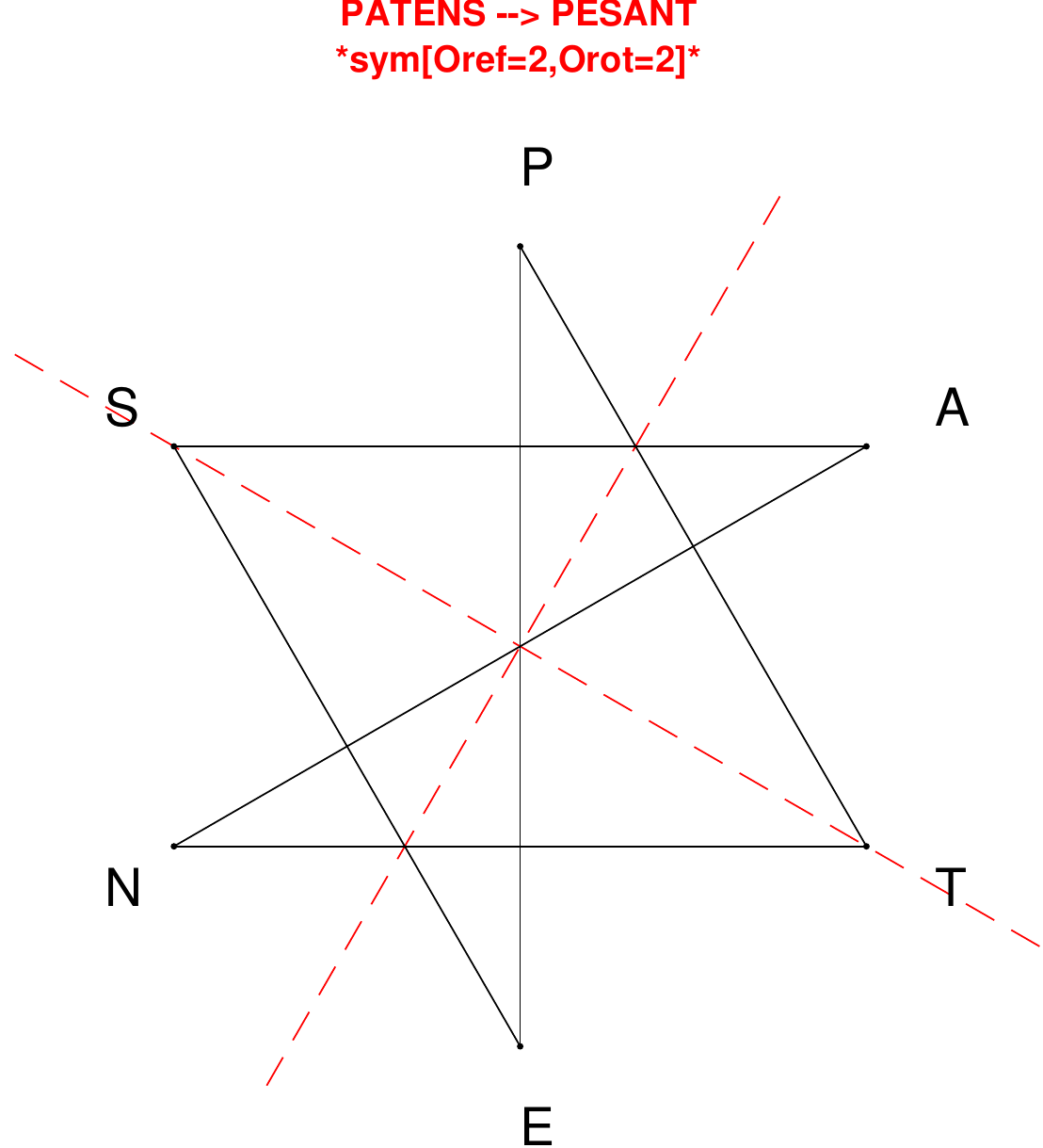}
\end{subfigure}
\end{figure}

\begin{figure}[H]
\centering
\begin{subfigure}[T]{0.19\textwidth}
\centering
\includegraphics[width=\textwidth]{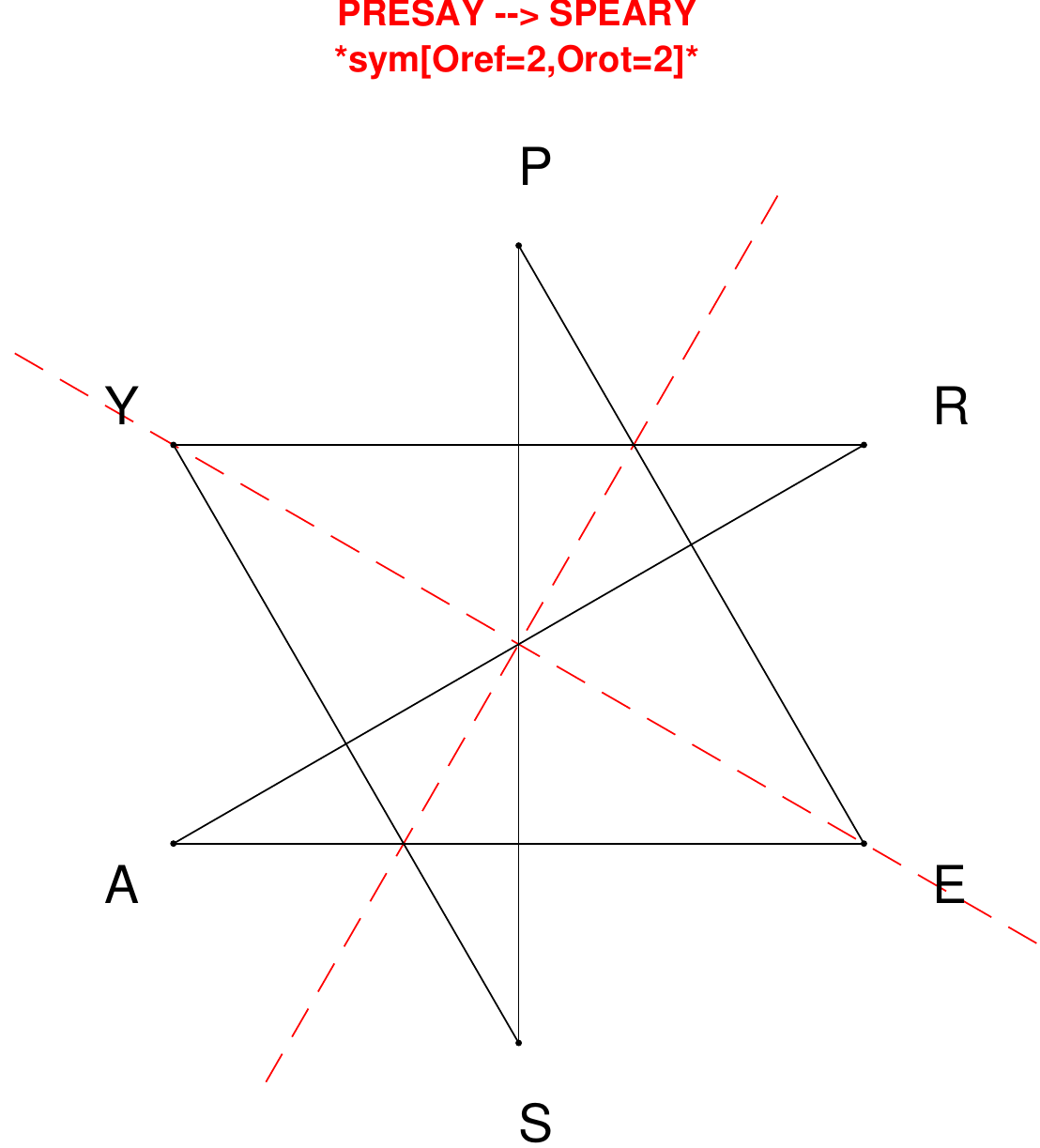}
\end{subfigure}
\hfill
\begin{subfigure}[T]{0.19\textwidth}
\centering
\includegraphics[width=\textwidth]{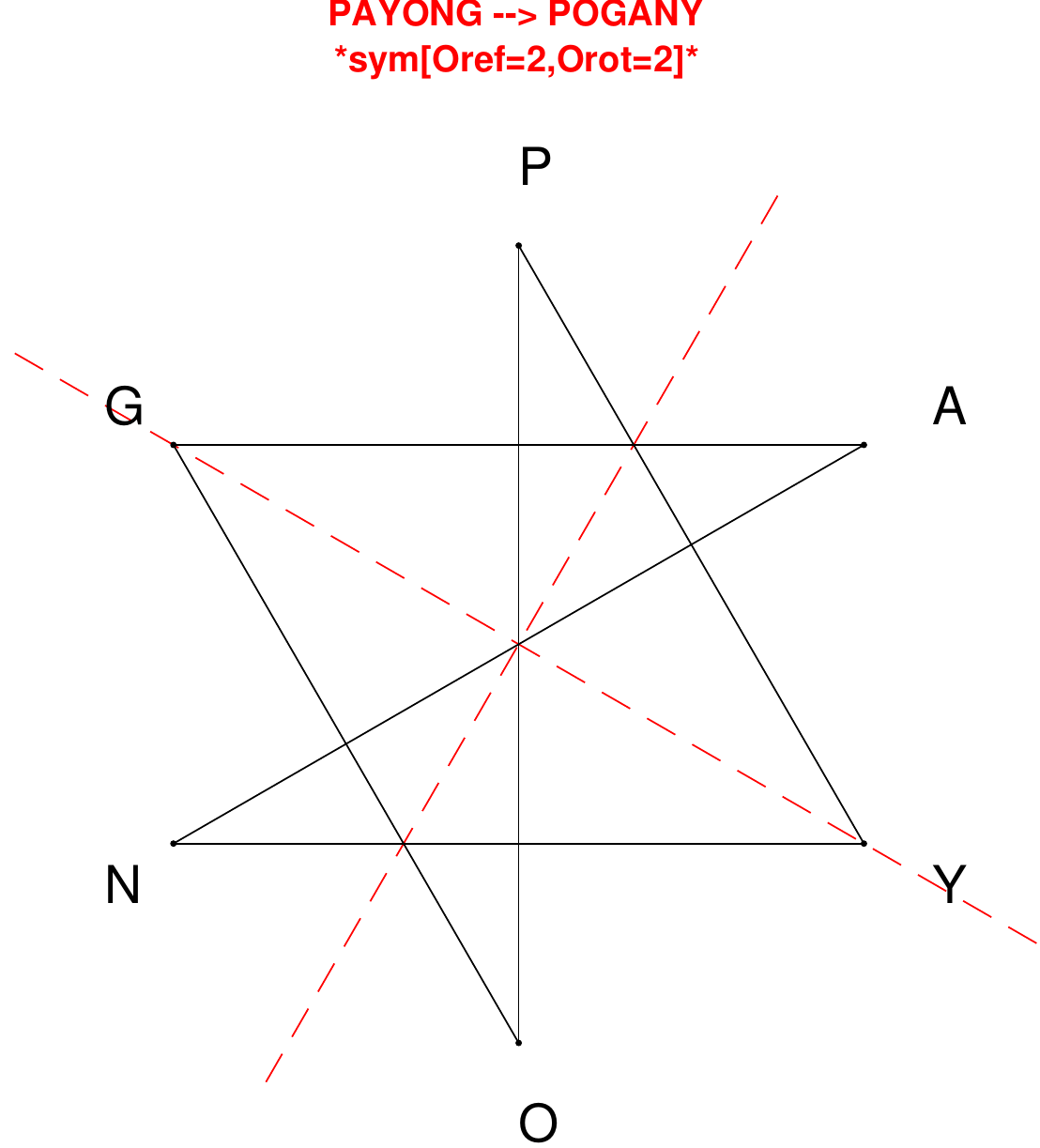}
\end{subfigure}
\hfill
\begin{subfigure}[T]{0.19\textwidth}
\centering
\includegraphics[width=\textwidth]{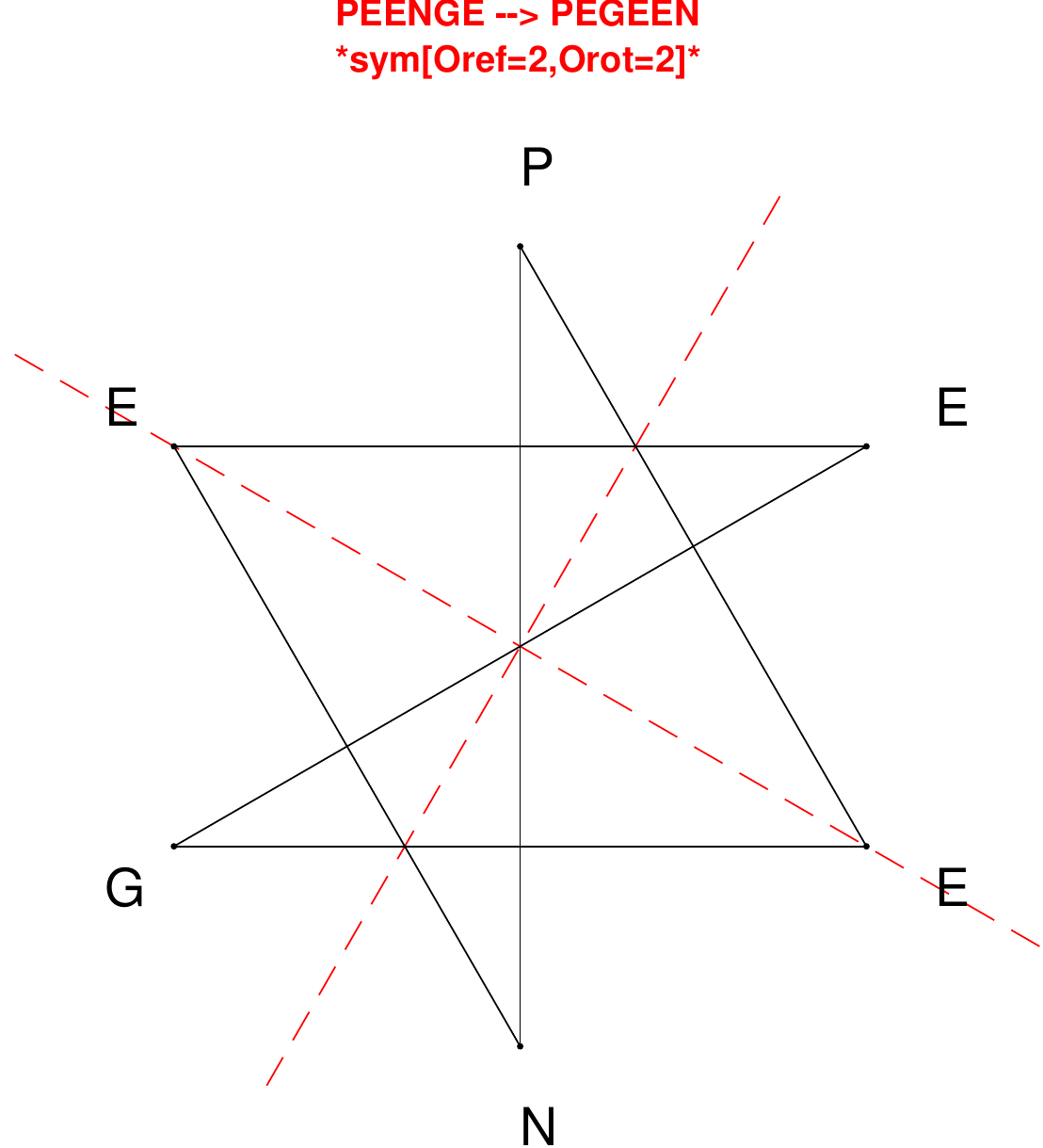}
\end{subfigure}
\hfill
\begin{subfigure}[T]{0.19\textwidth}
\centering
\includegraphics[width=\textwidth]{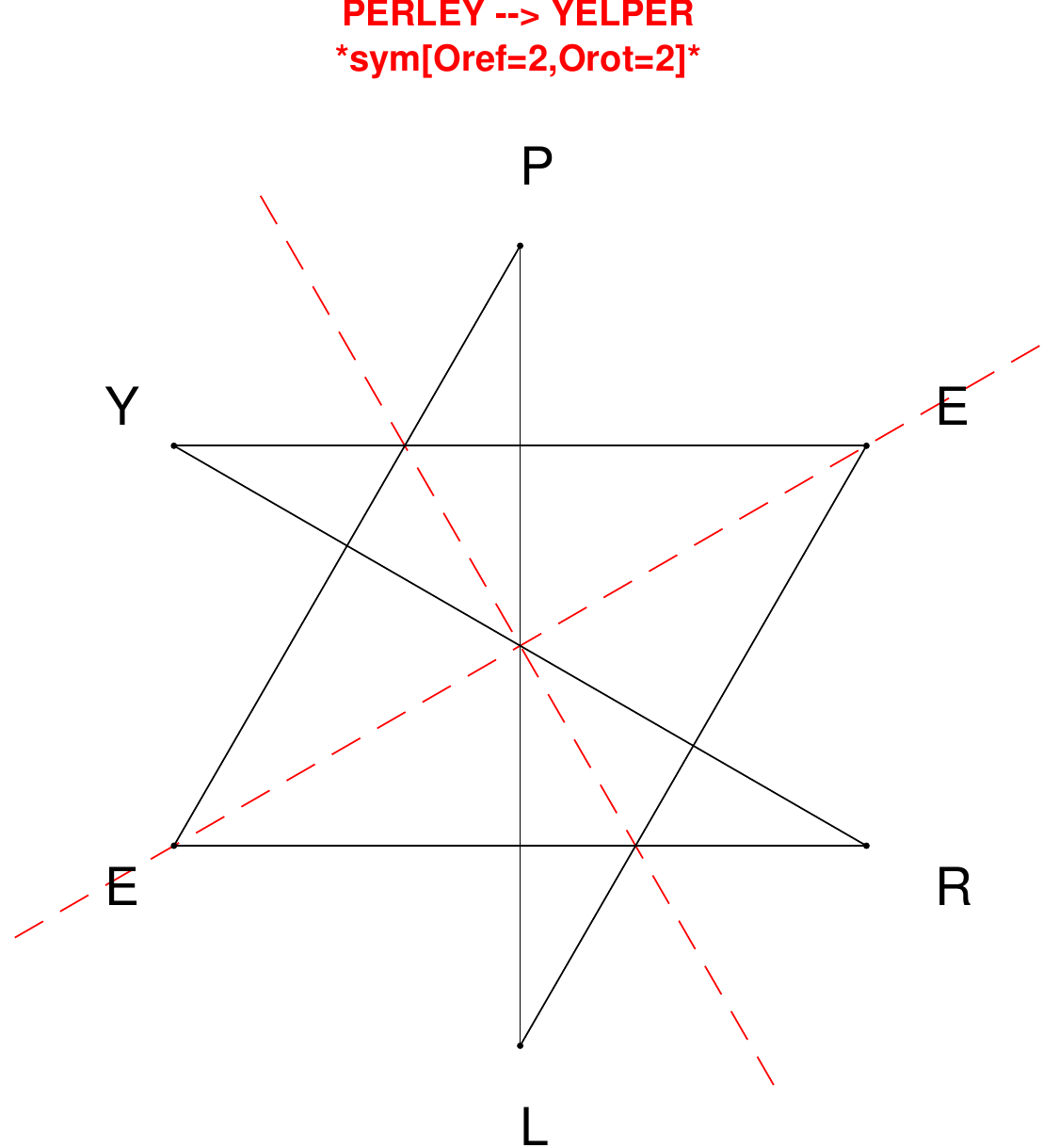}
\end{subfigure}
\hfill
\begin{subfigure}[T]{0.19\textwidth}
\centering
\includegraphics[width=\textwidth]{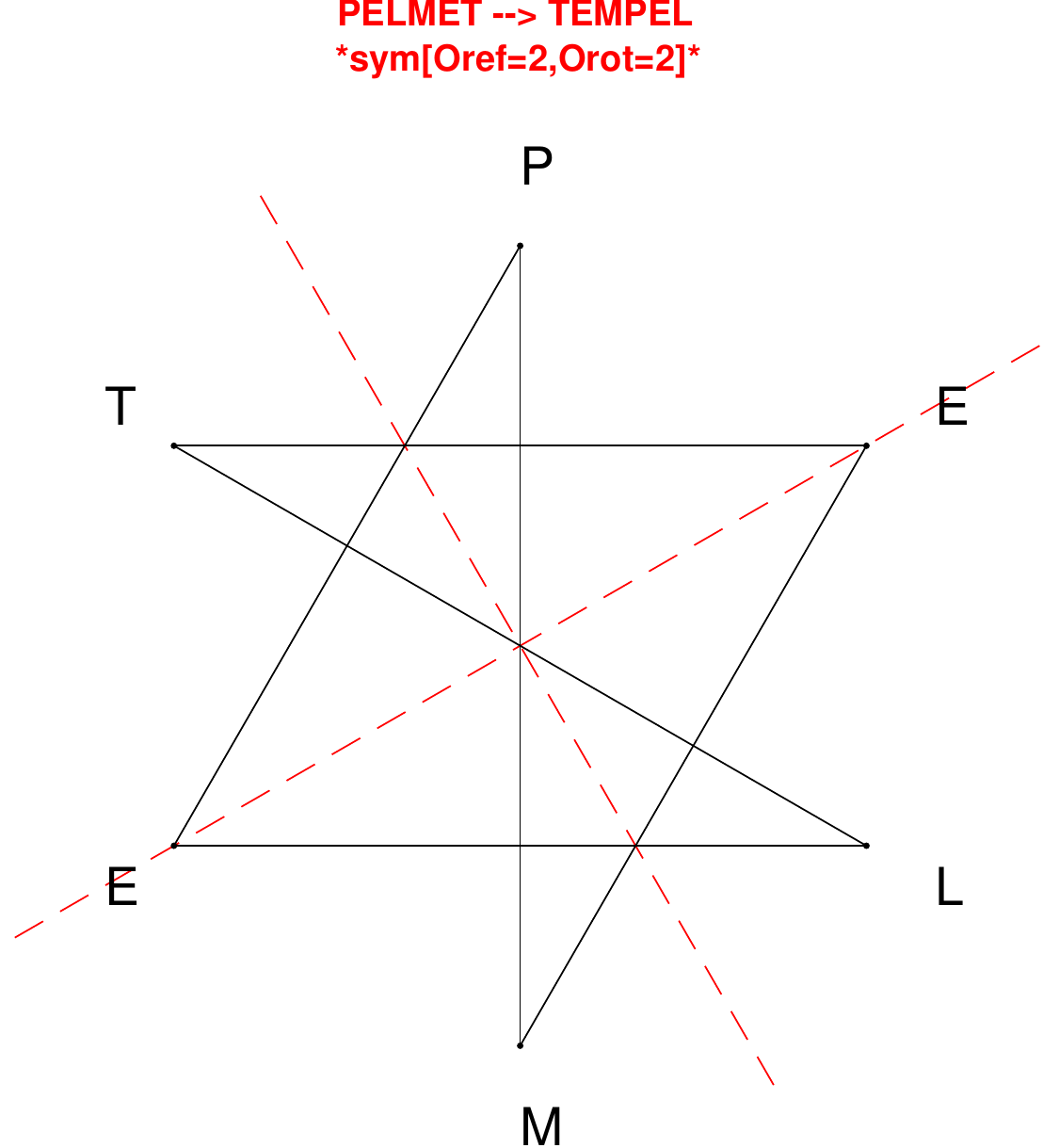}
\end{subfigure}
\end{figure}

\begin{figure}[H]
\centering
\begin{subfigure}[T]{0.19\textwidth}
\centering
\includegraphics[width=\textwidth]{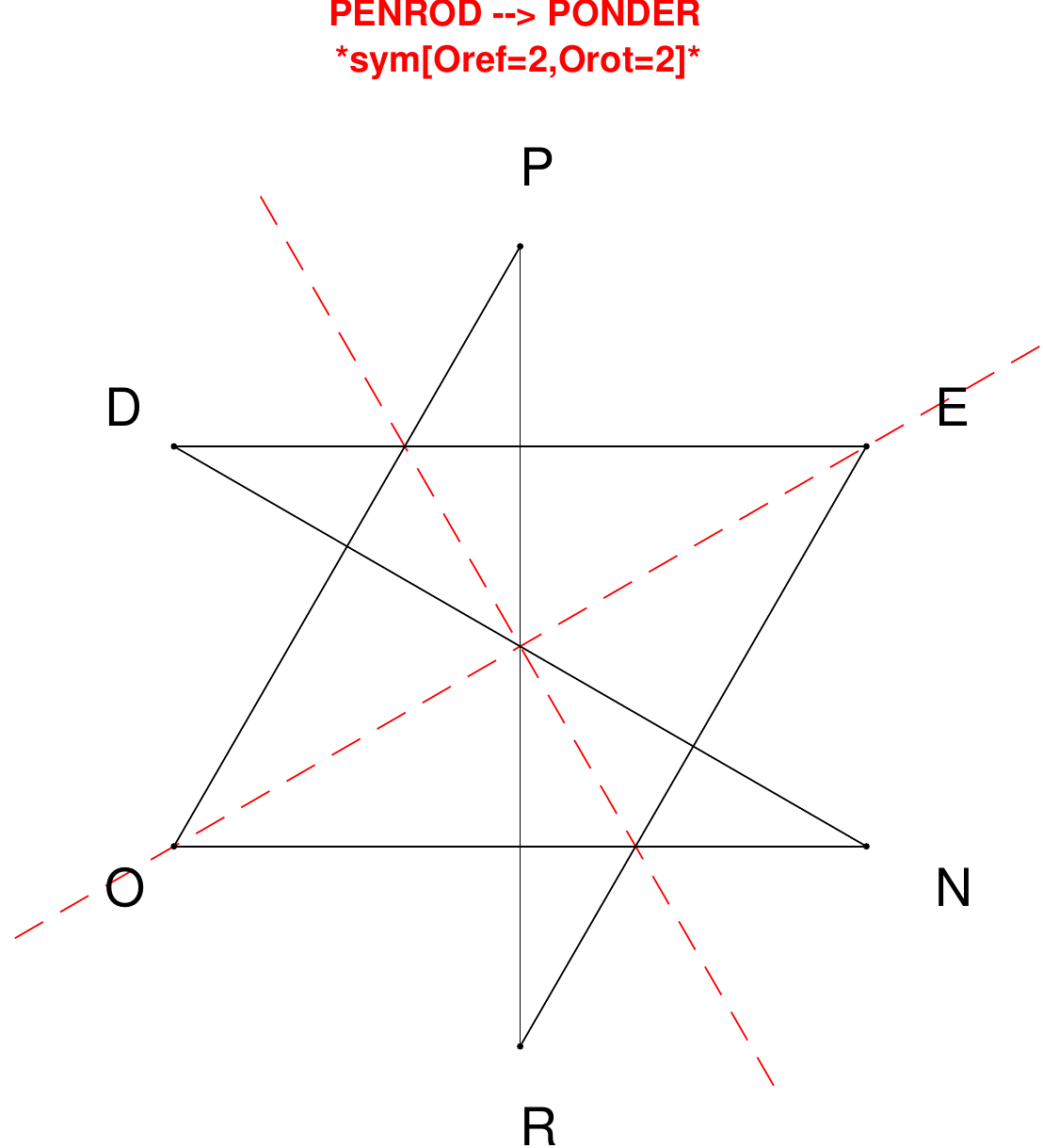}
\end{subfigure}
\hfill
\begin{subfigure}[T]{0.19\textwidth}
\centering
\includegraphics[width=\textwidth]{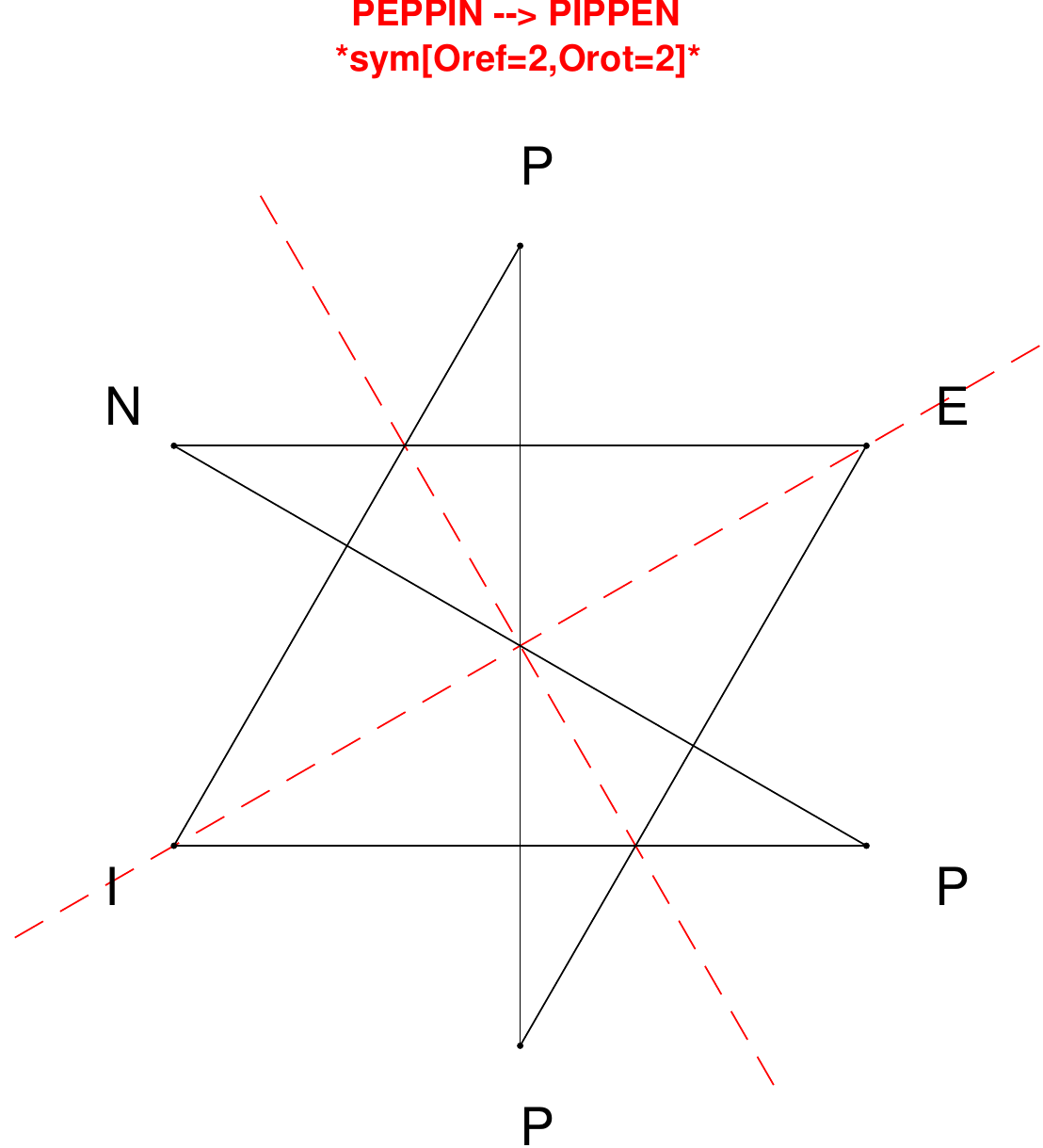}
\end{subfigure}
\hfill
\begin{subfigure}[T]{0.19\textwidth}
\centering
\includegraphics[width=\textwidth]{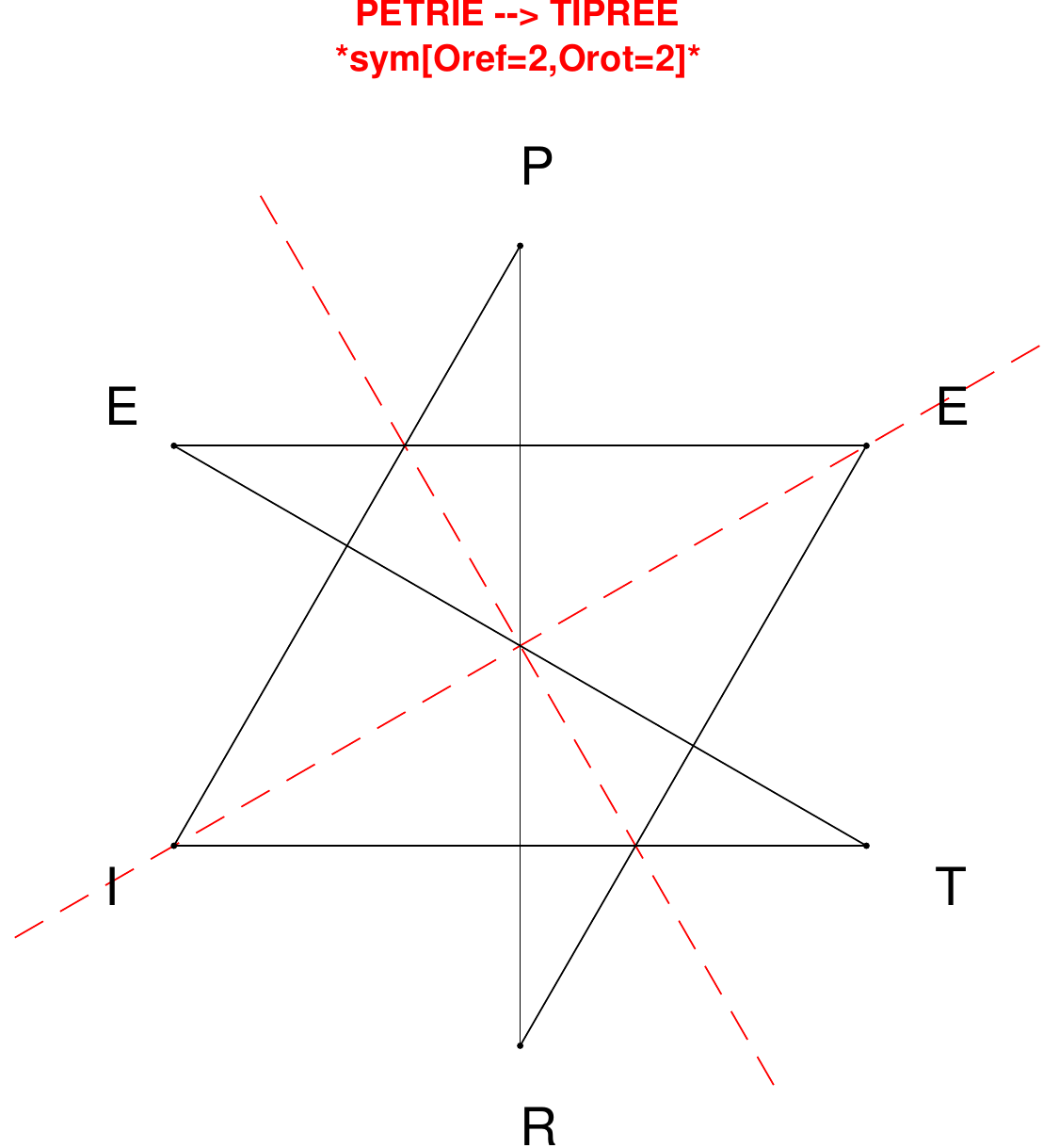}
\end{subfigure}
\hfill
\begin{subfigure}[T]{0.19\textwidth}
\centering
\includegraphics[width=\textwidth]{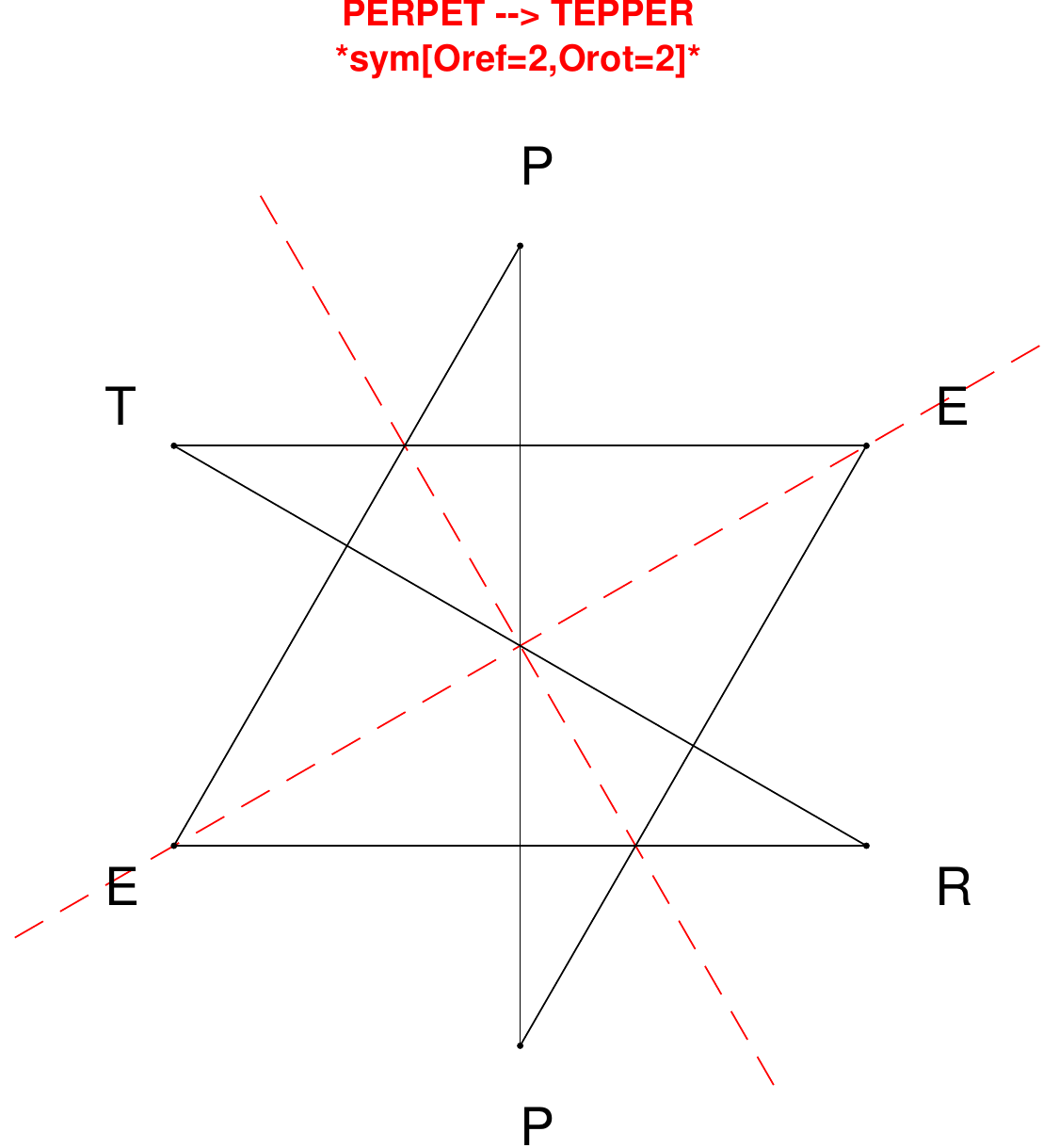}
\end{subfigure}
\hfill
\begin{subfigure}[T]{0.19\textwidth}
\centering
\includegraphics[width=\textwidth]{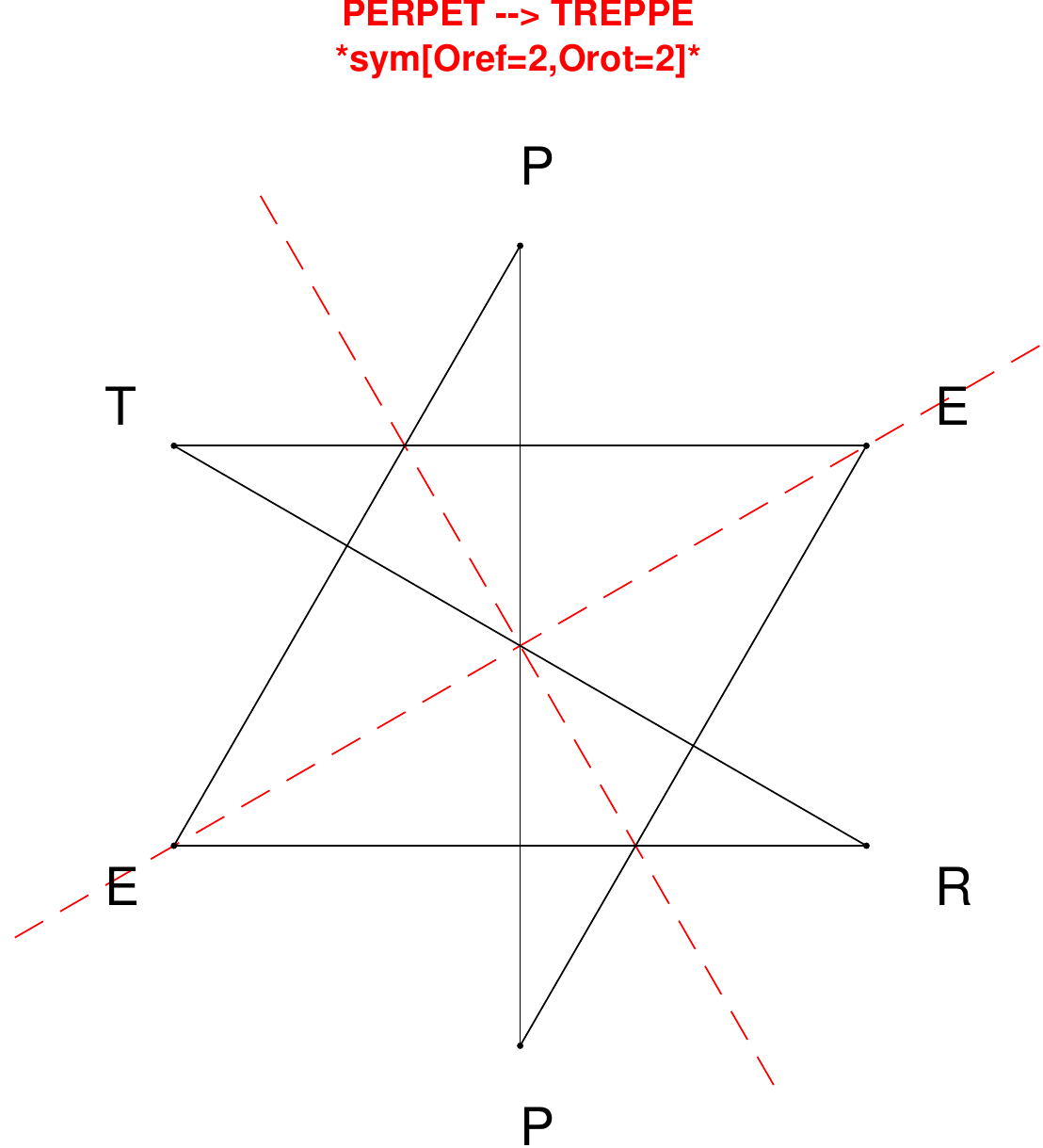}
\end{subfigure}
\end{figure}

\begin{figure}[H]
\centering
\begin{subfigure}[T]{0.19\textwidth}
\centering
\includegraphics[width=\textwidth]{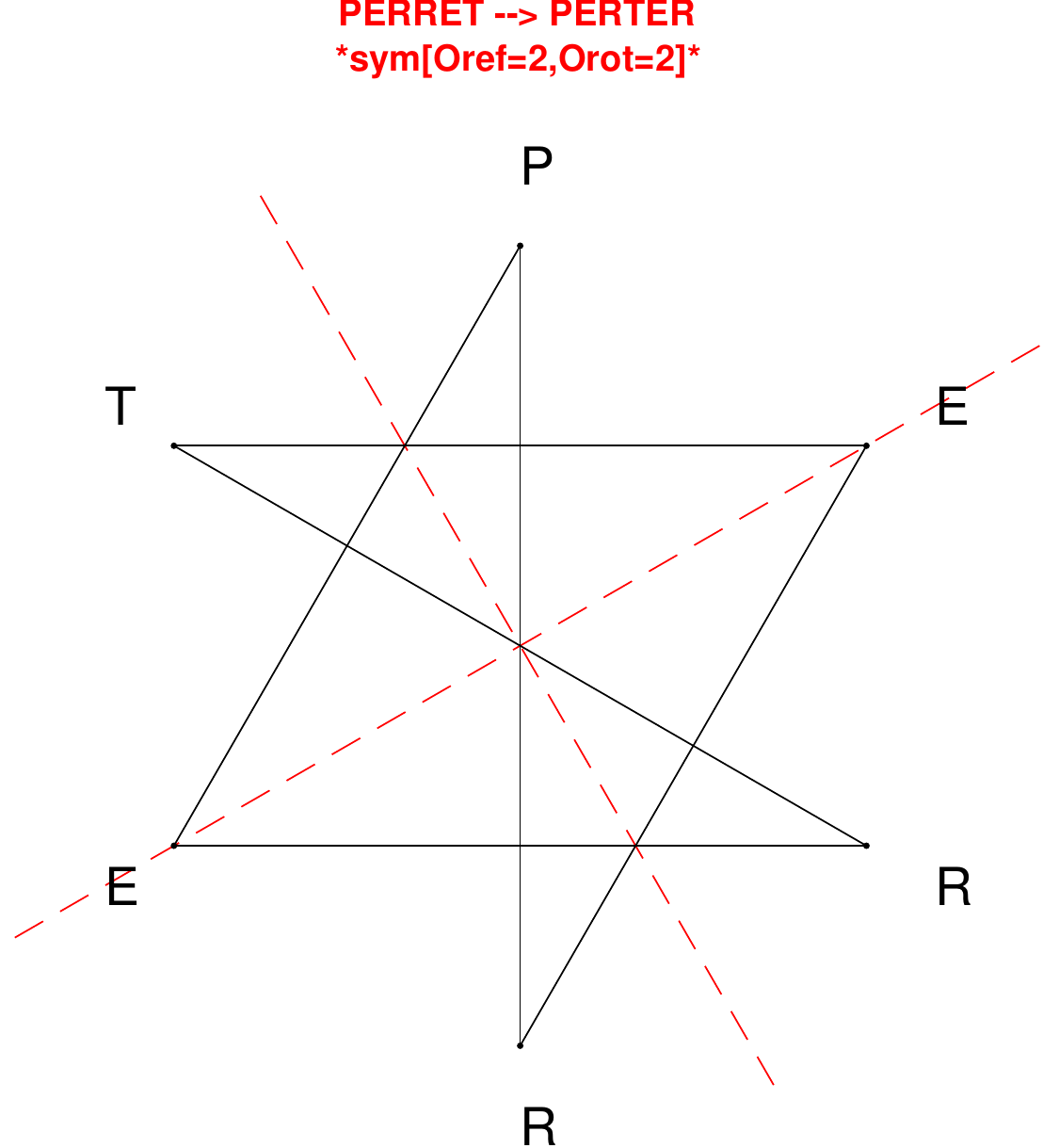}
\end{subfigure}
\hfill
\begin{subfigure}[T]{0.19\textwidth}
\centering
\includegraphics[width=\textwidth]{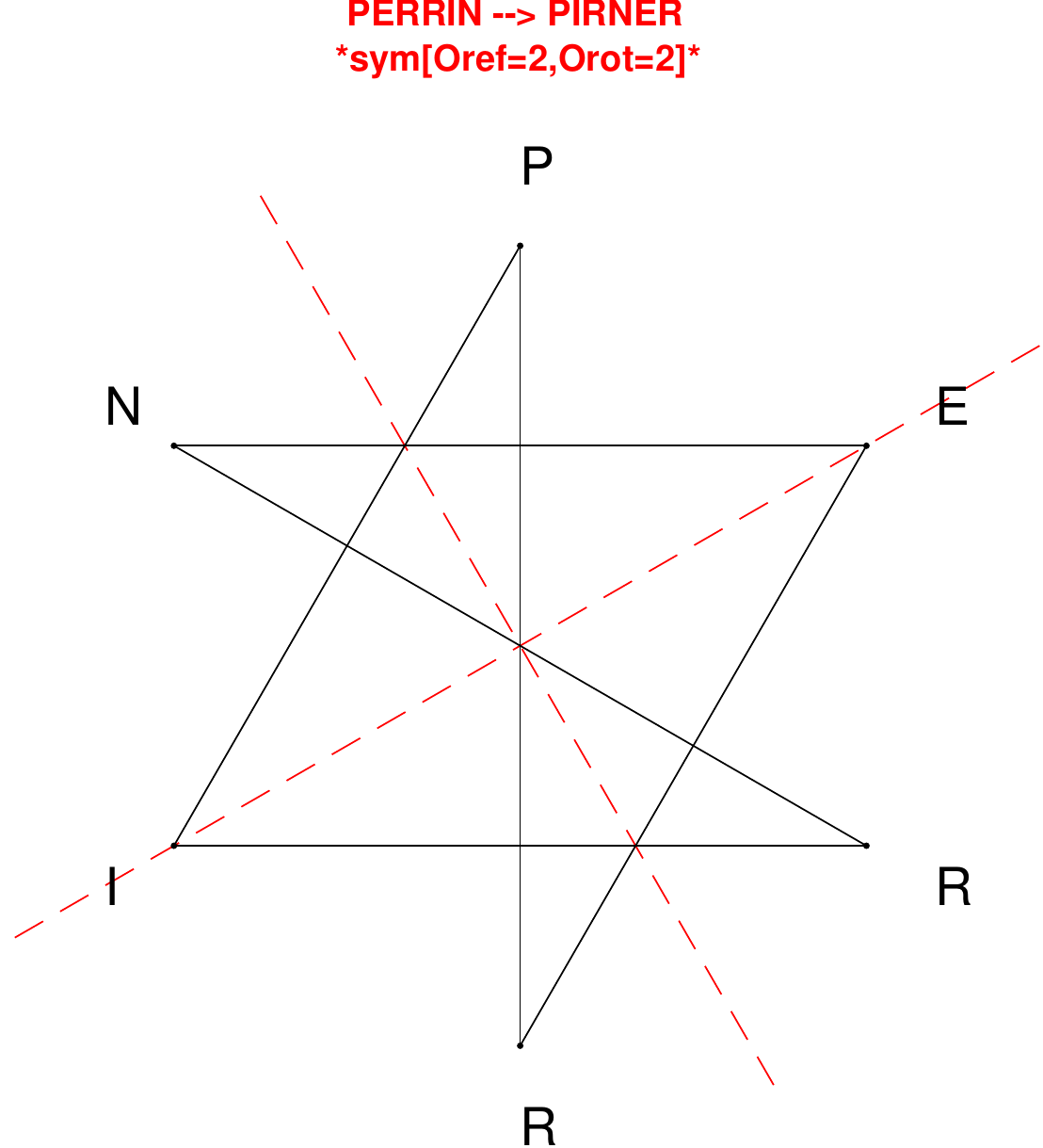}
\end{subfigure}
\hfill
\begin{subfigure}[T]{0.19\textwidth}
\centering
\includegraphics[width=\textwidth]{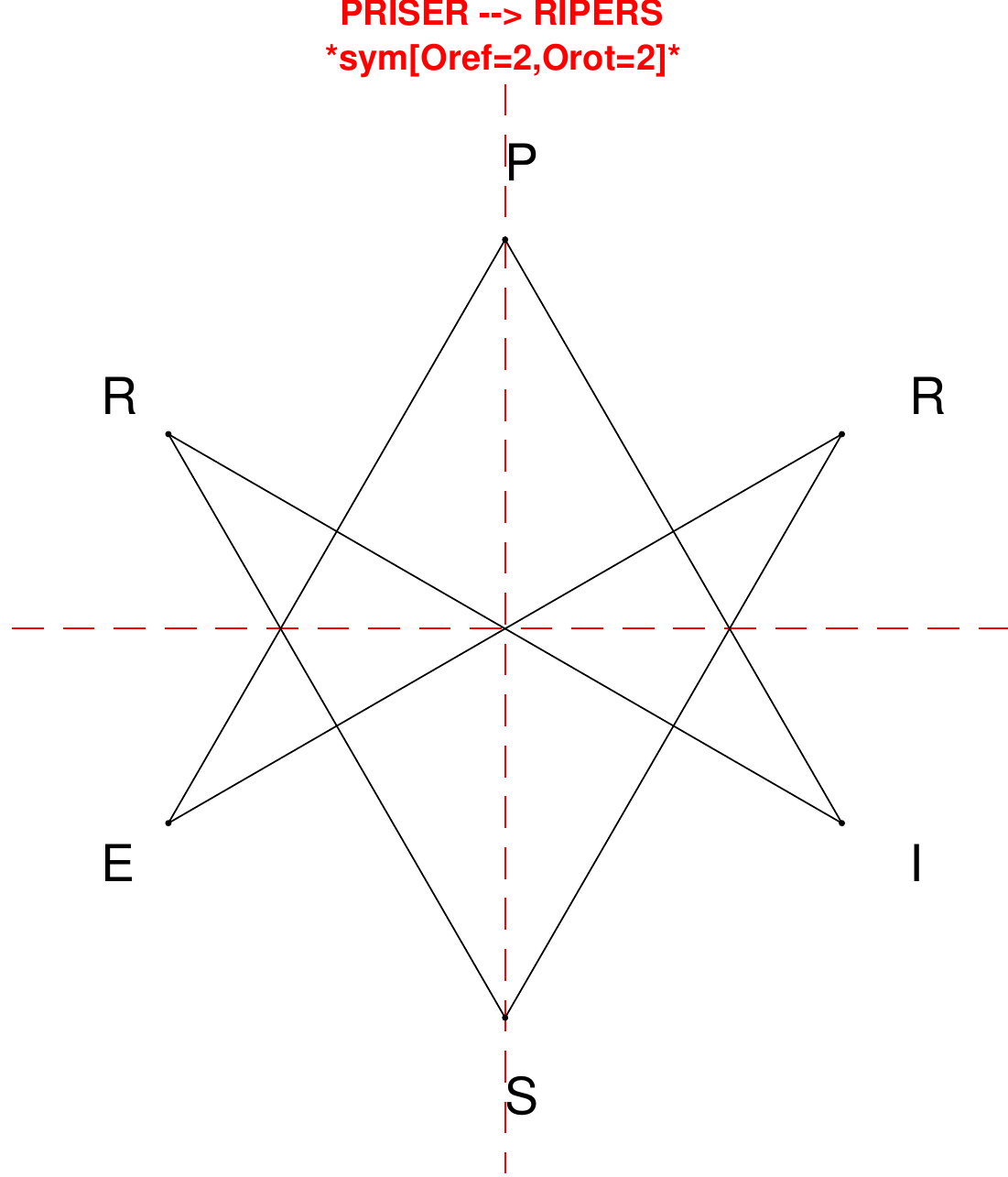}
\end{subfigure}
\hfill
\begin{subfigure}[T]{0.19\textwidth}
\centering
\includegraphics[width=\textwidth]{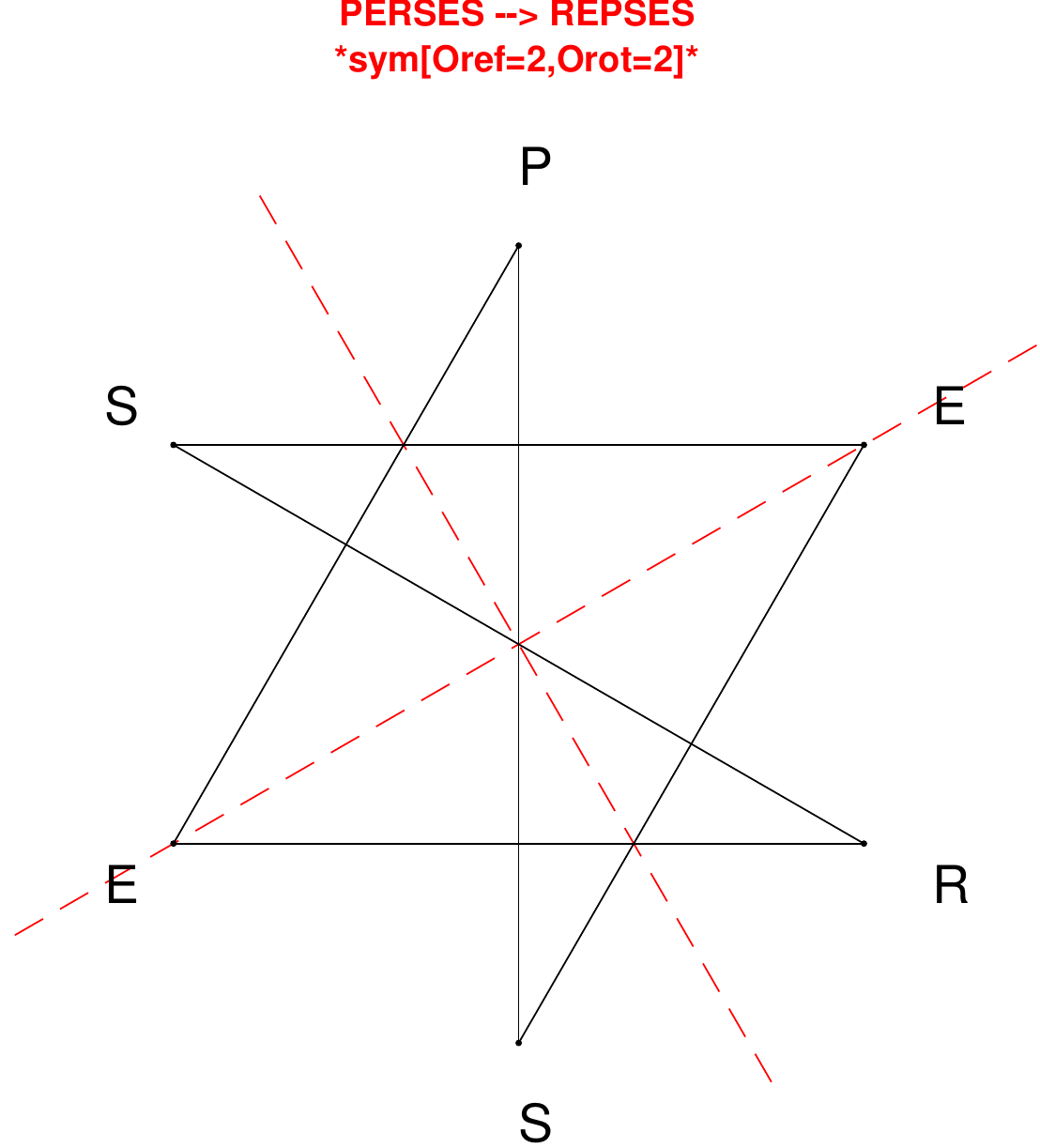}
\end{subfigure}
\hfill
\begin{subfigure}[T]{0.19\textwidth}
\centering
\includegraphics[width=\textwidth]{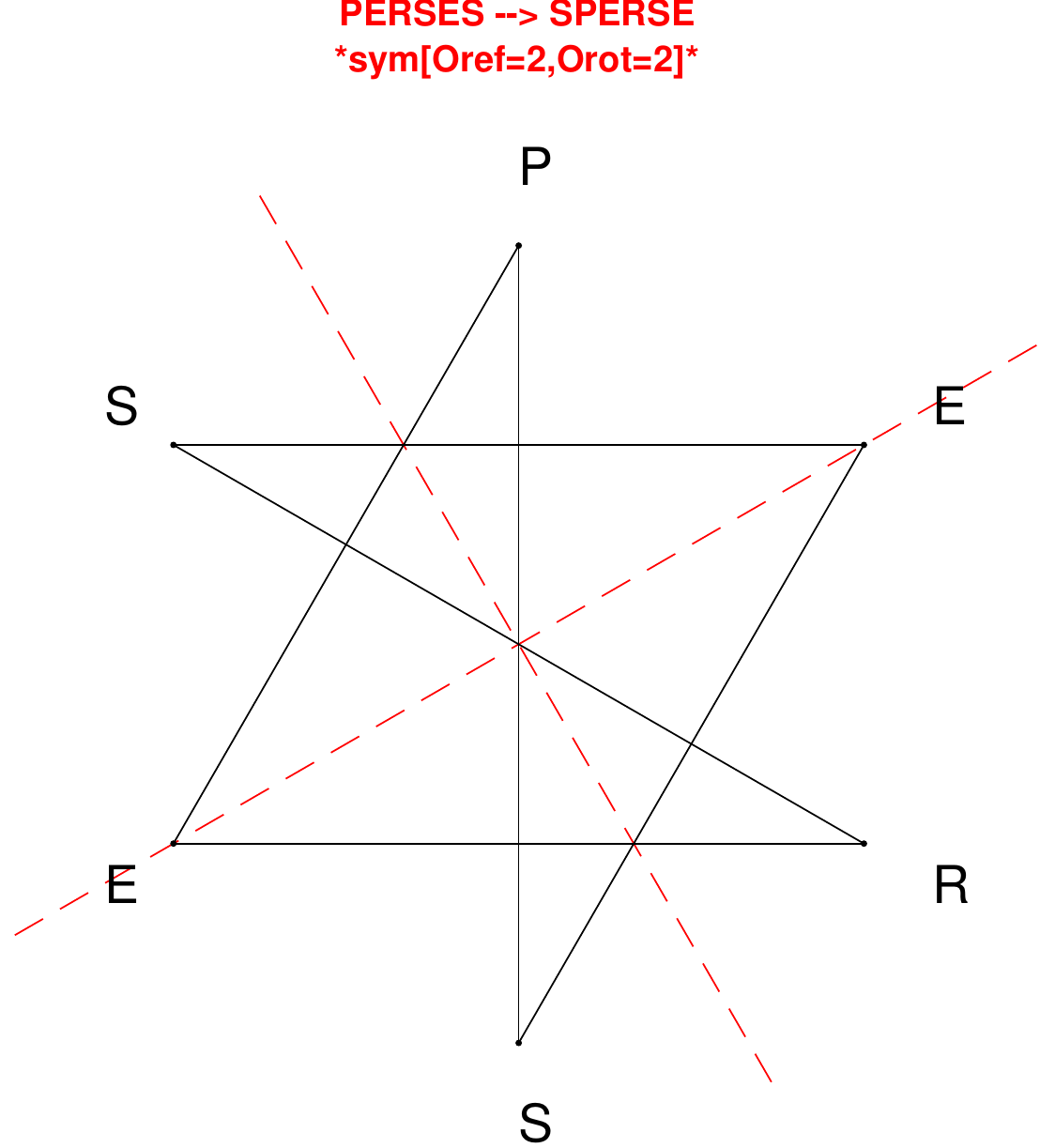}
\end{subfigure}
\end{figure}

\begin{figure}[H]
\centering
\begin{subfigure}[T]{0.19\textwidth}
\centering
\includegraphics[width=\textwidth]{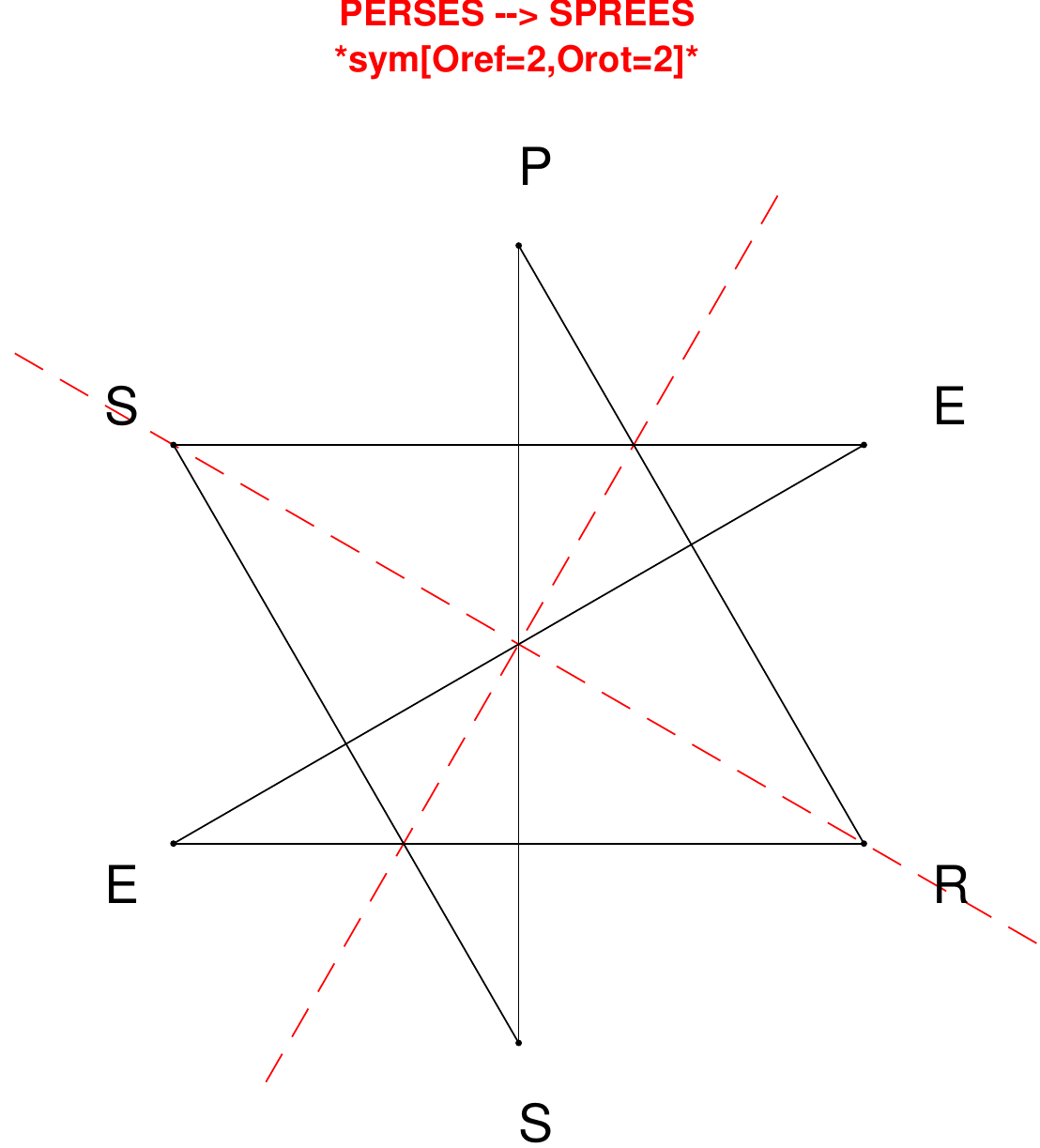}
\end{subfigure}
\hfill
\begin{subfigure}[T]{0.19\textwidth}
\centering
\includegraphics[width=\textwidth]{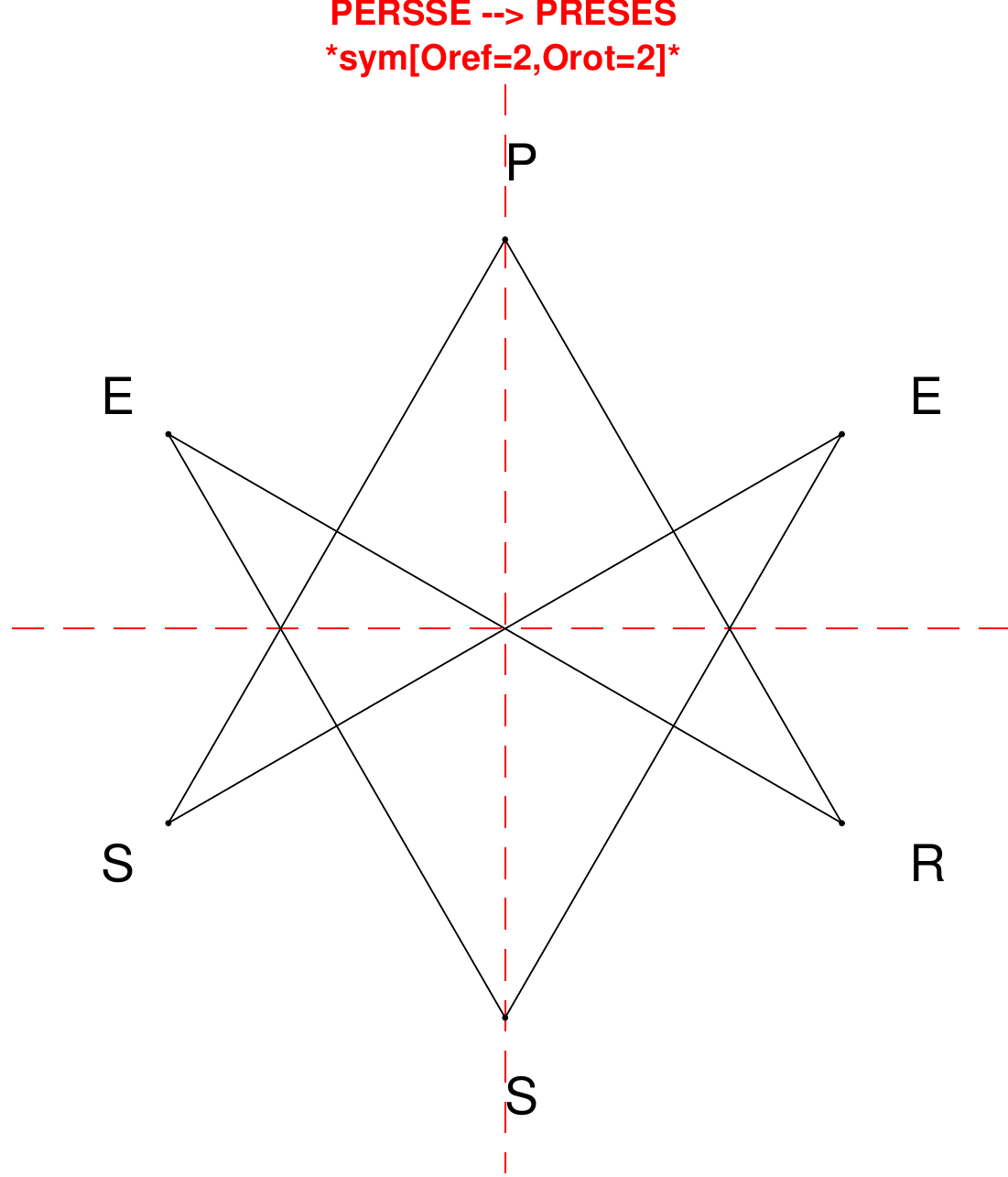}
\end{subfigure}
\hfill
\begin{subfigure}[T]{0.19\textwidth}
\centering
\includegraphics[width=\textwidth]{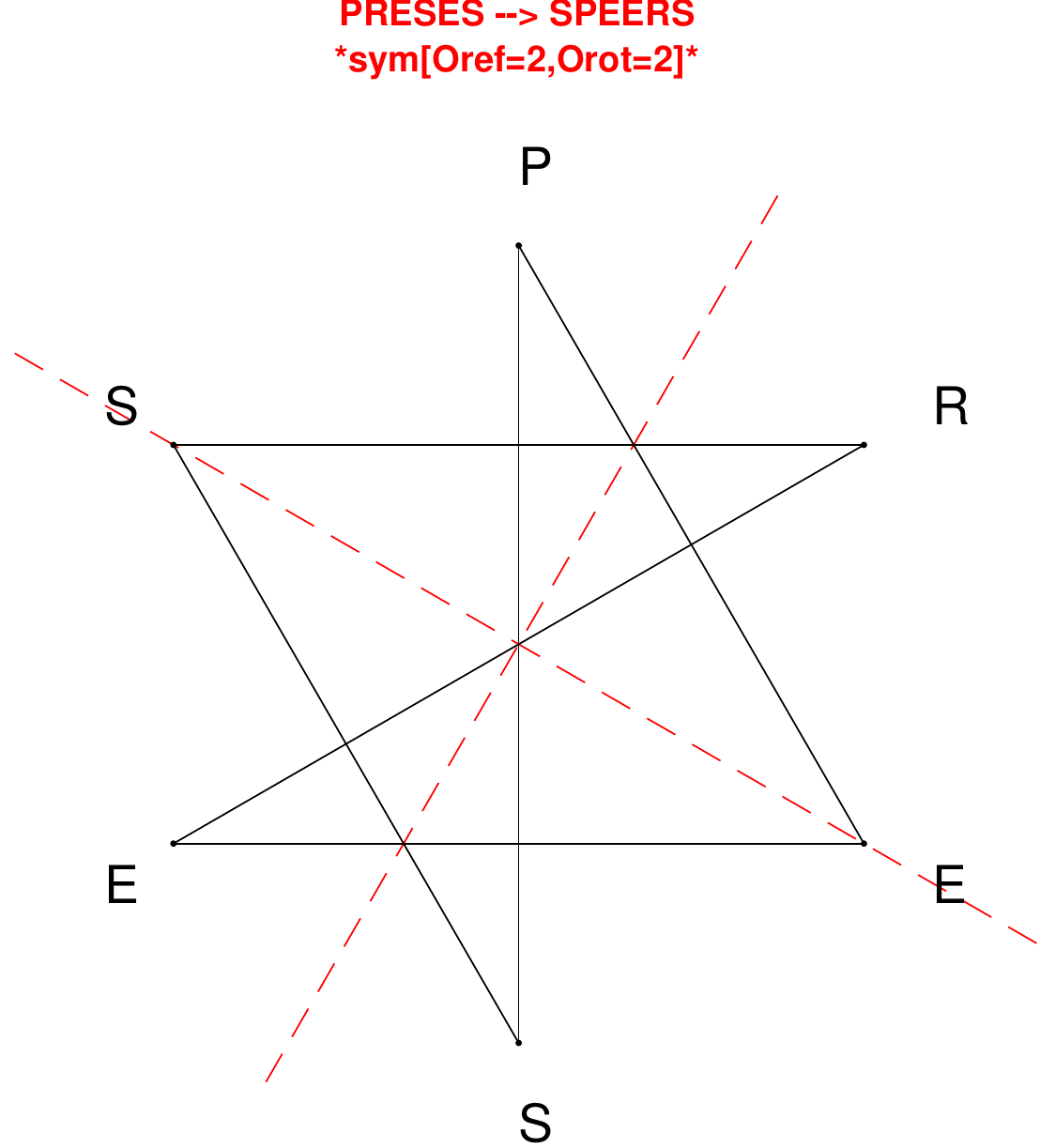}
\end{subfigure}
\hfill
\begin{subfigure}[T]{0.19\textwidth}
\centering
\includegraphics[width=\textwidth]{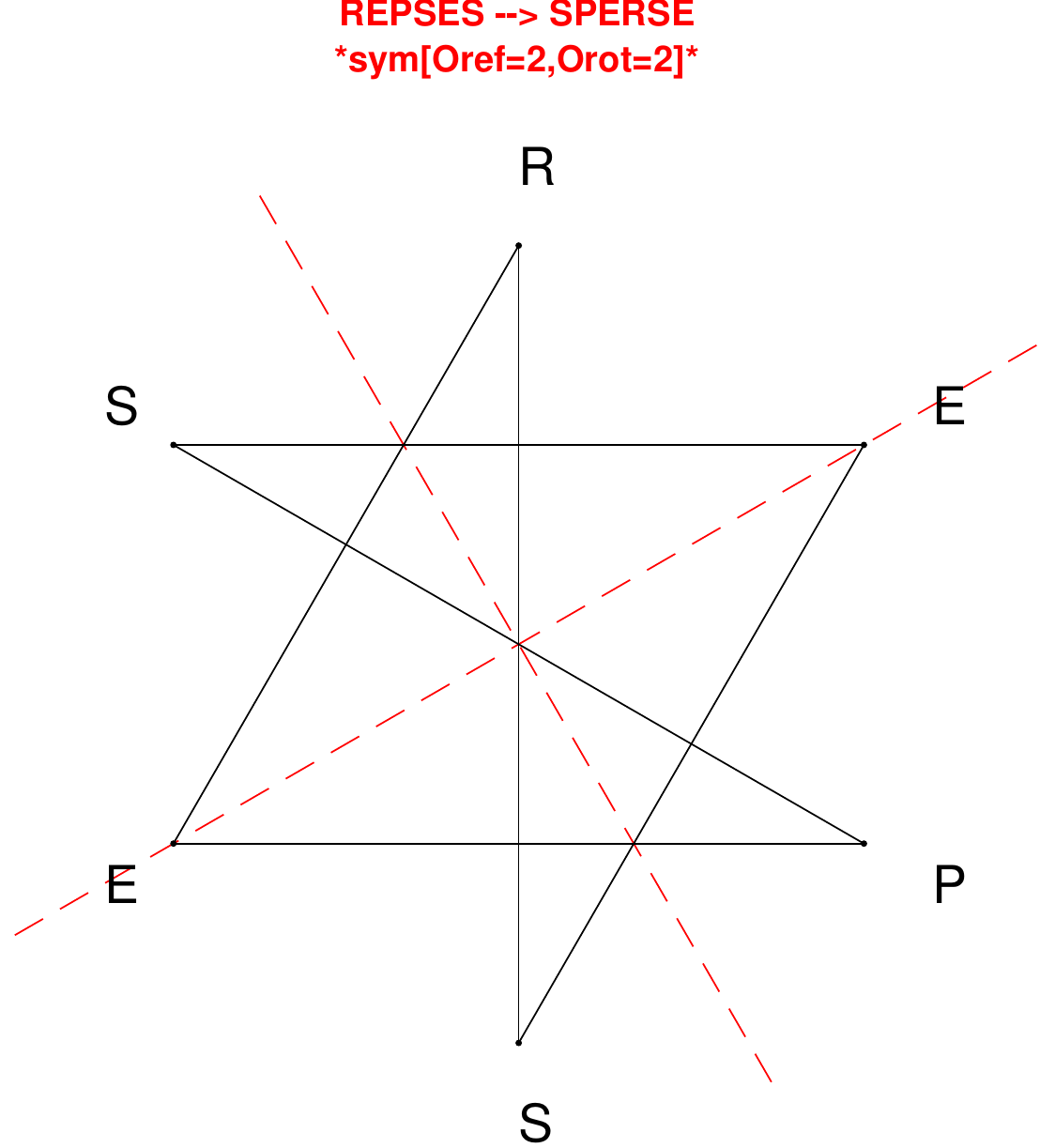}
\end{subfigure}
\hfill
\begin{subfigure}[T]{0.19\textwidth}
\centering
\includegraphics[width=\textwidth]{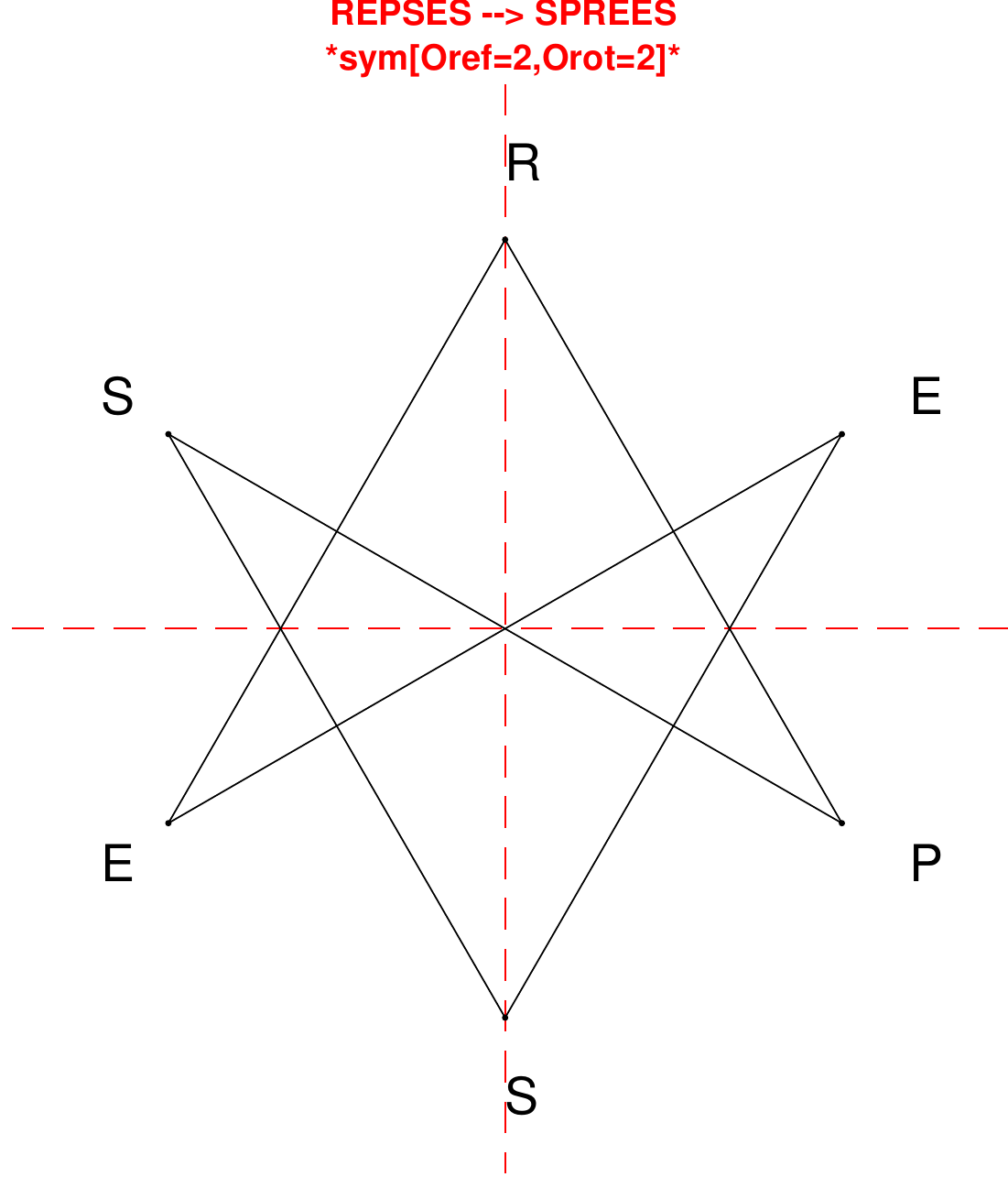}
\end{subfigure}
\end{figure}

\begin{figure}[H]
\centering
\begin{subfigure}[T]{0.19\textwidth}
\centering
\includegraphics[width=\textwidth]{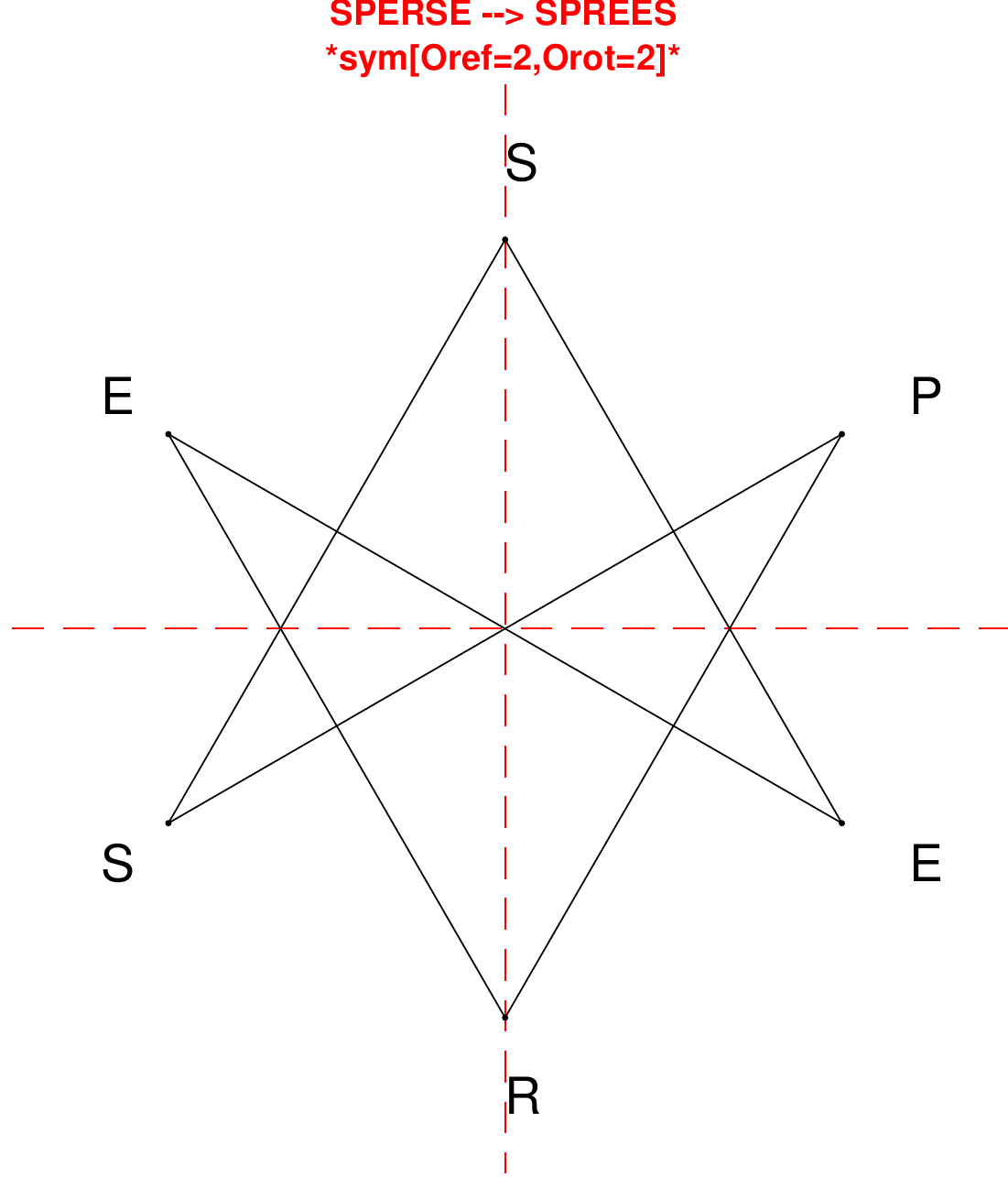}
\end{subfigure}
\hfill
\begin{subfigure}[T]{0.19\textwidth}
\centering
\includegraphics[width=\textwidth]{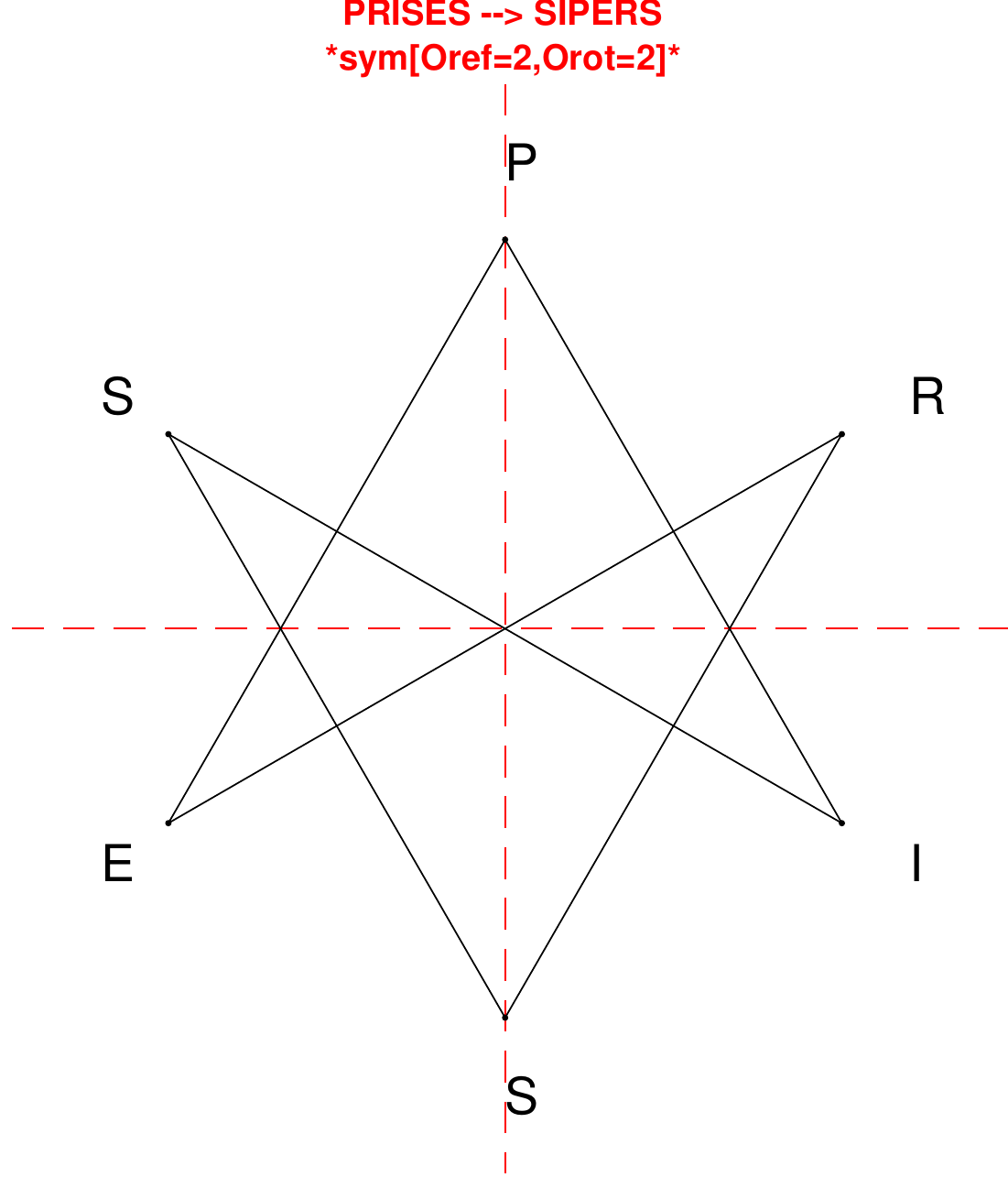}
\end{subfigure}
\hfill
\begin{subfigure}[T]{0.19\textwidth}
\centering
\includegraphics[width=\textwidth]{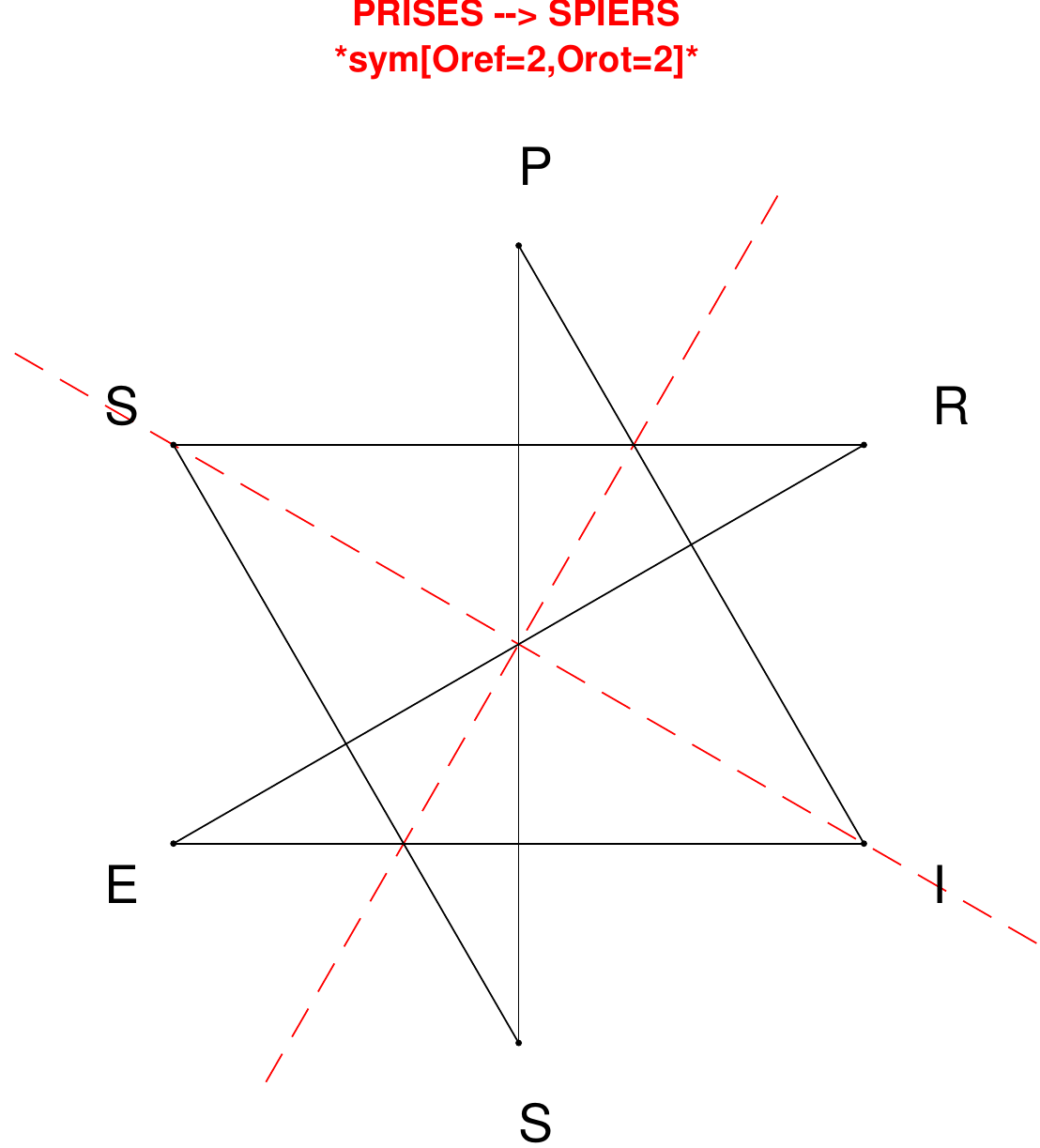}
\end{subfigure}
\hfill
\begin{subfigure}[T]{0.19\textwidth}
\centering
\includegraphics[width=\textwidth]{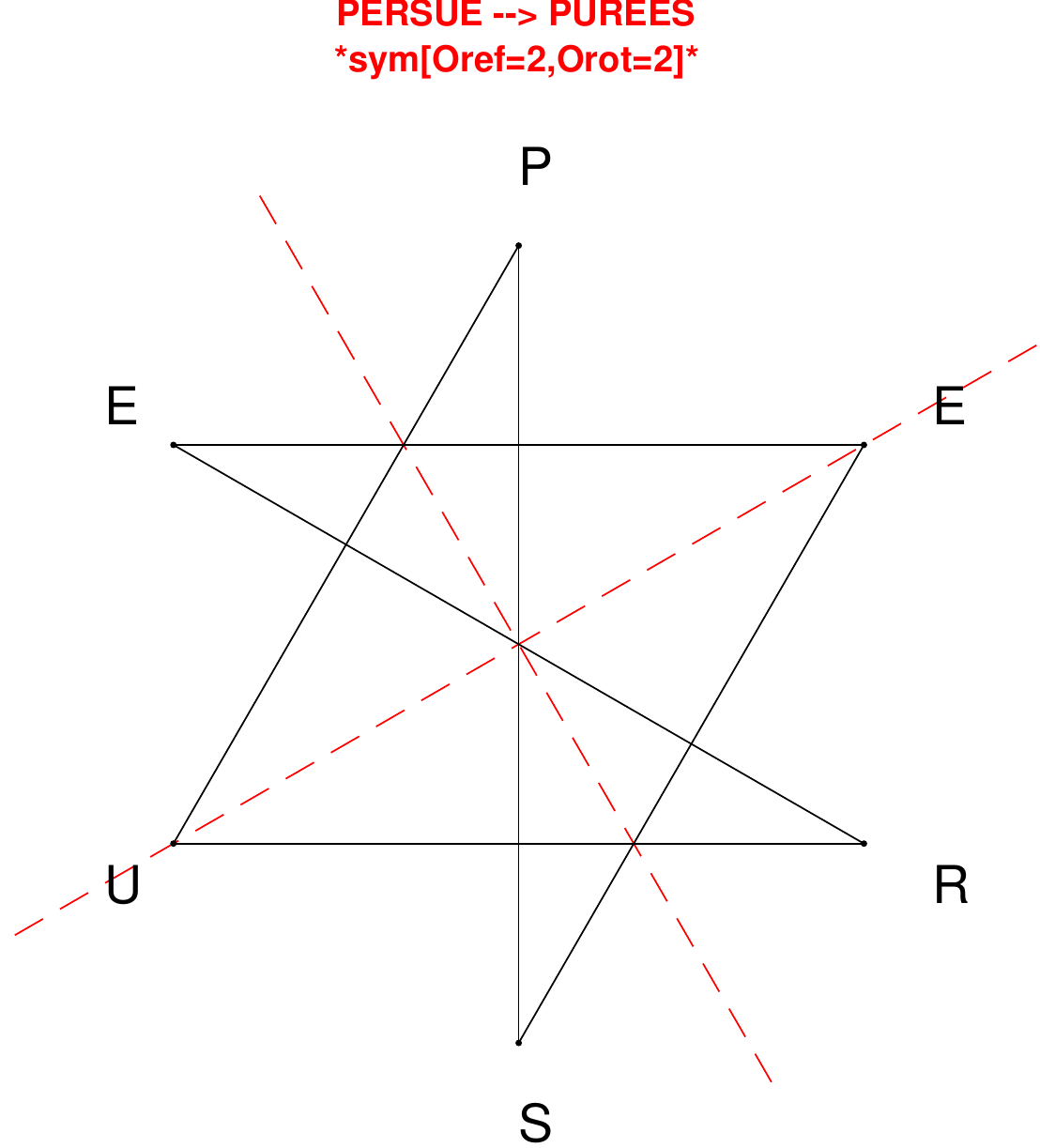}
\end{subfigure}
\hfill
\begin{subfigure}[T]{0.19\textwidth}
\centering
\includegraphics[width=\textwidth]{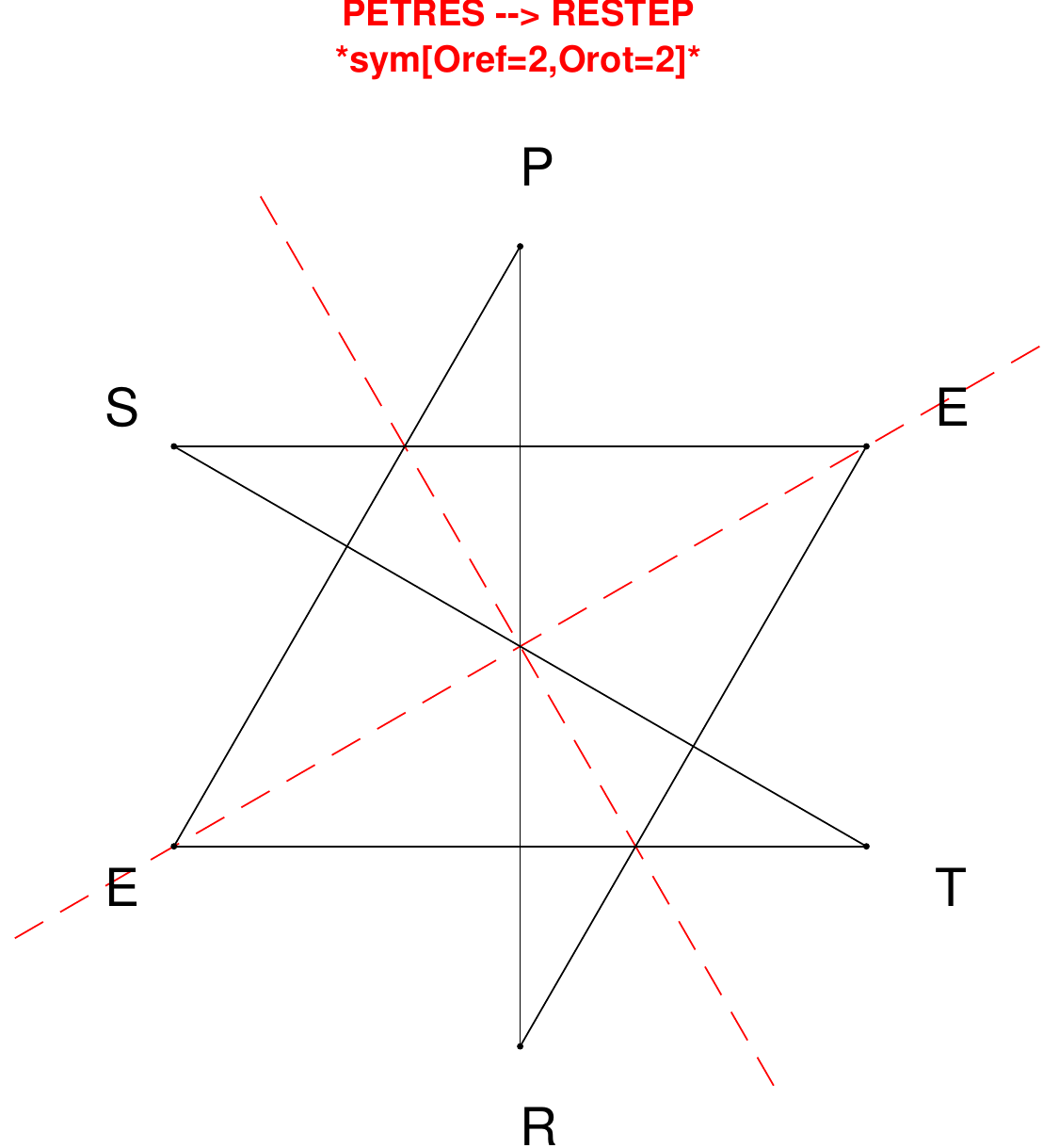}
\end{subfigure}
\end{figure}

\begin{figure}[H]
\centering
\begin{subfigure}[T]{0.19\textwidth}
\centering
\includegraphics[width=\textwidth]{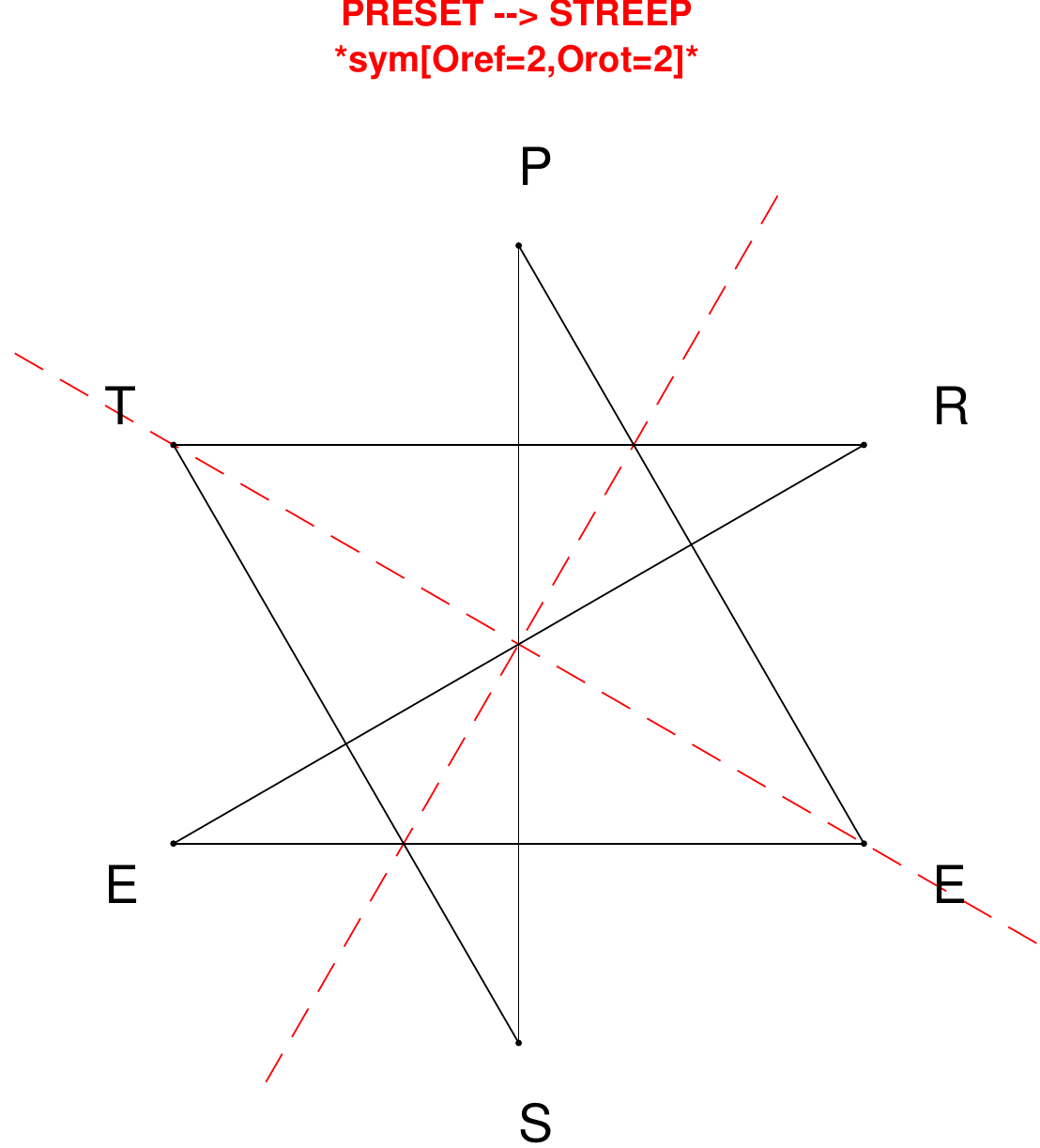}
\end{subfigure}
\hfill
\begin{subfigure}[T]{0.19\textwidth}
\centering
\includegraphics[width=\textwidth]{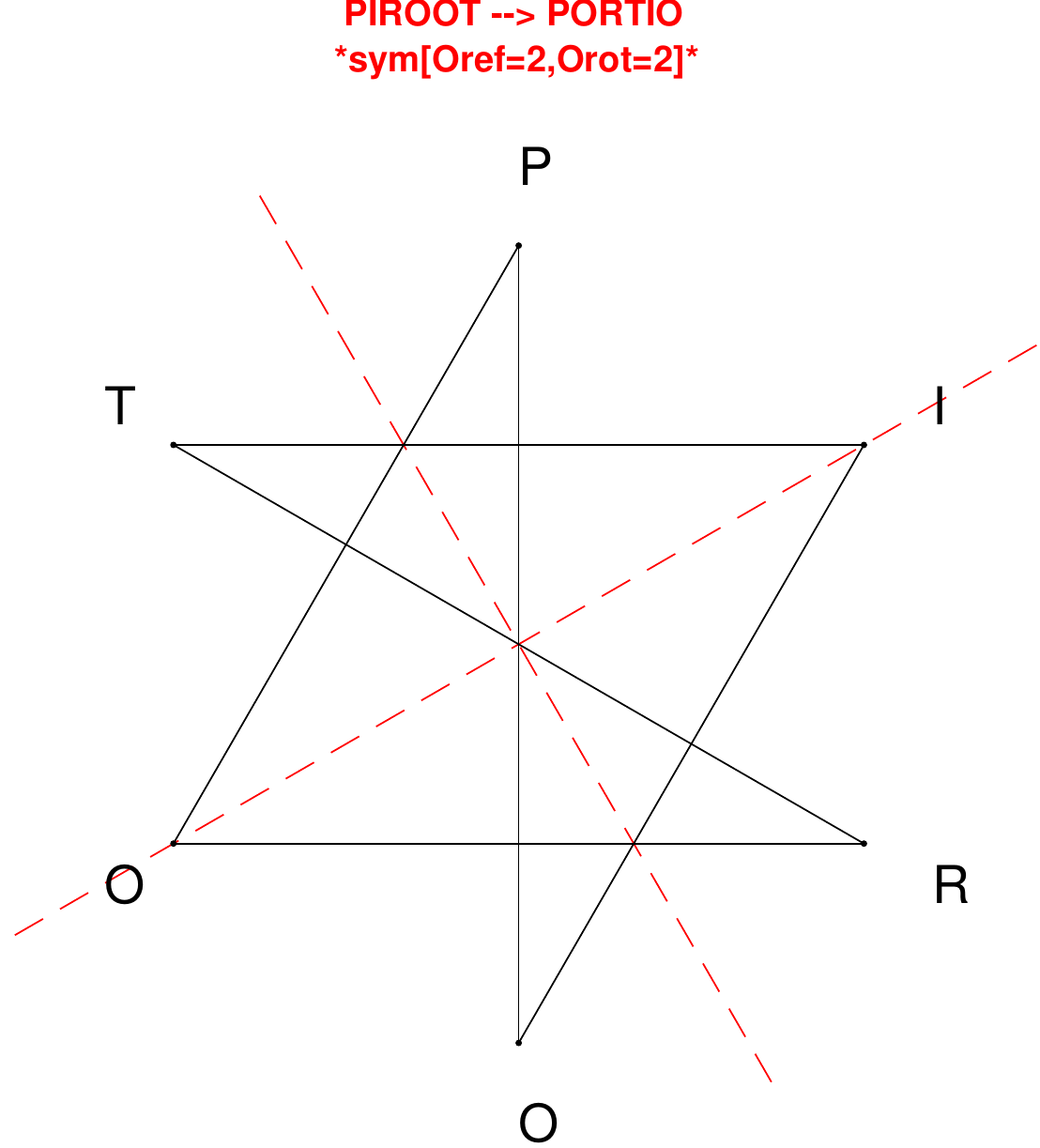}
\end{subfigure}
\hfill
\begin{subfigure}[T]{0.19\textwidth}
\centering
\includegraphics[width=\textwidth]{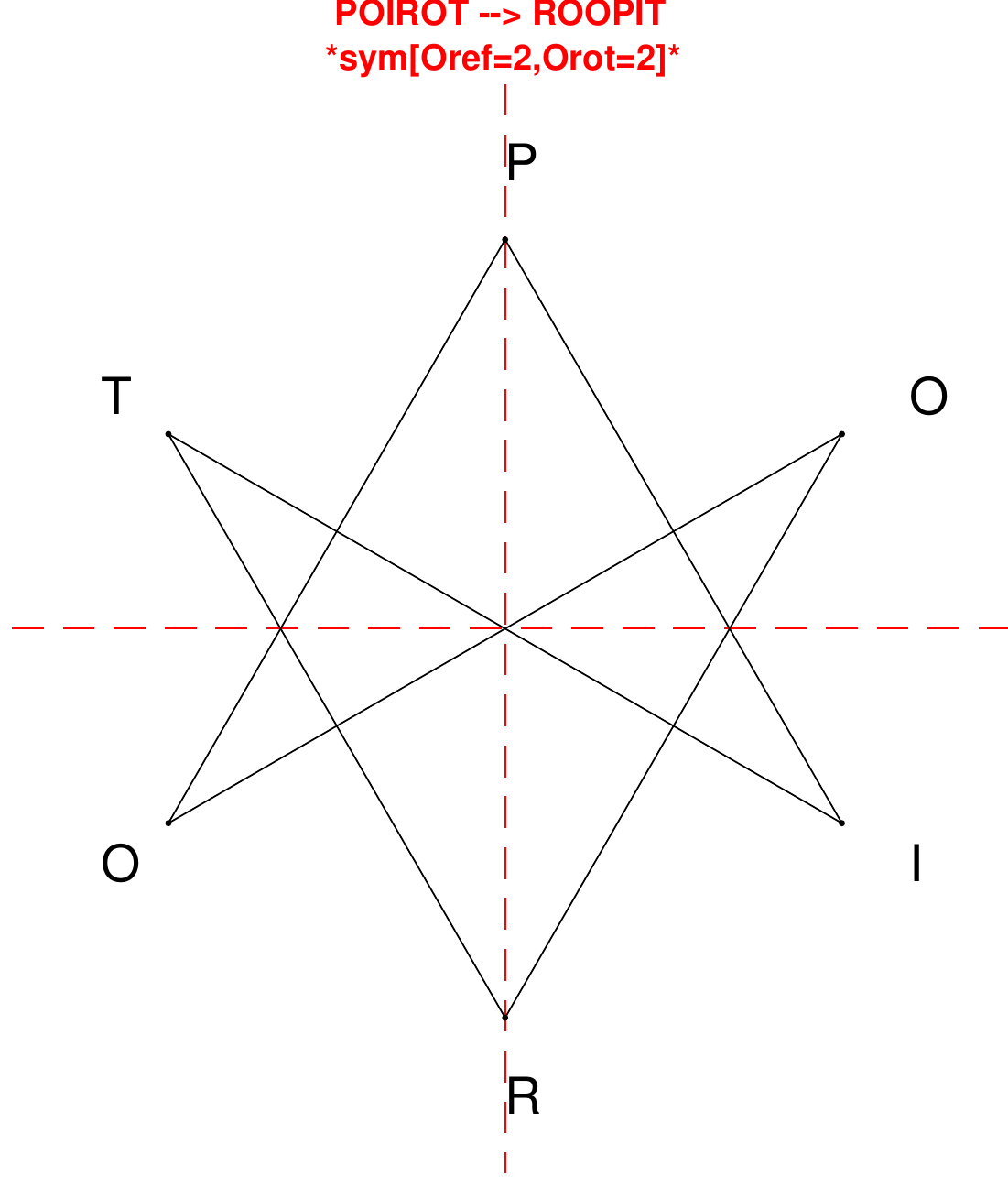}
\end{subfigure}
\hfill
\begin{subfigure}[T]{0.19\textwidth}
\centering
\includegraphics[width=\textwidth]{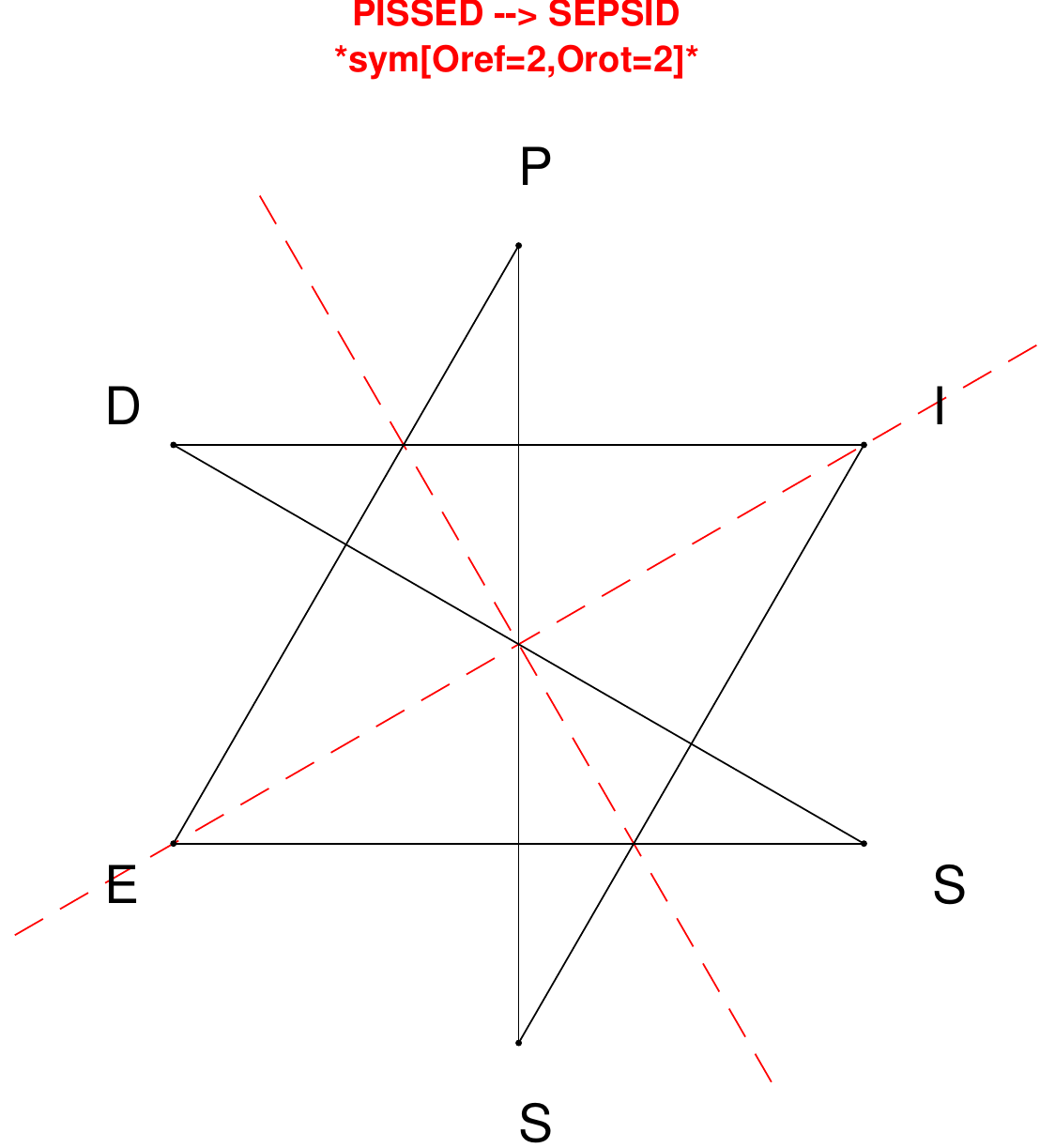}
\end{subfigure}
\hfill
\begin{subfigure}[T]{0.19\textwidth}
\centering
\includegraphics[width=\textwidth]{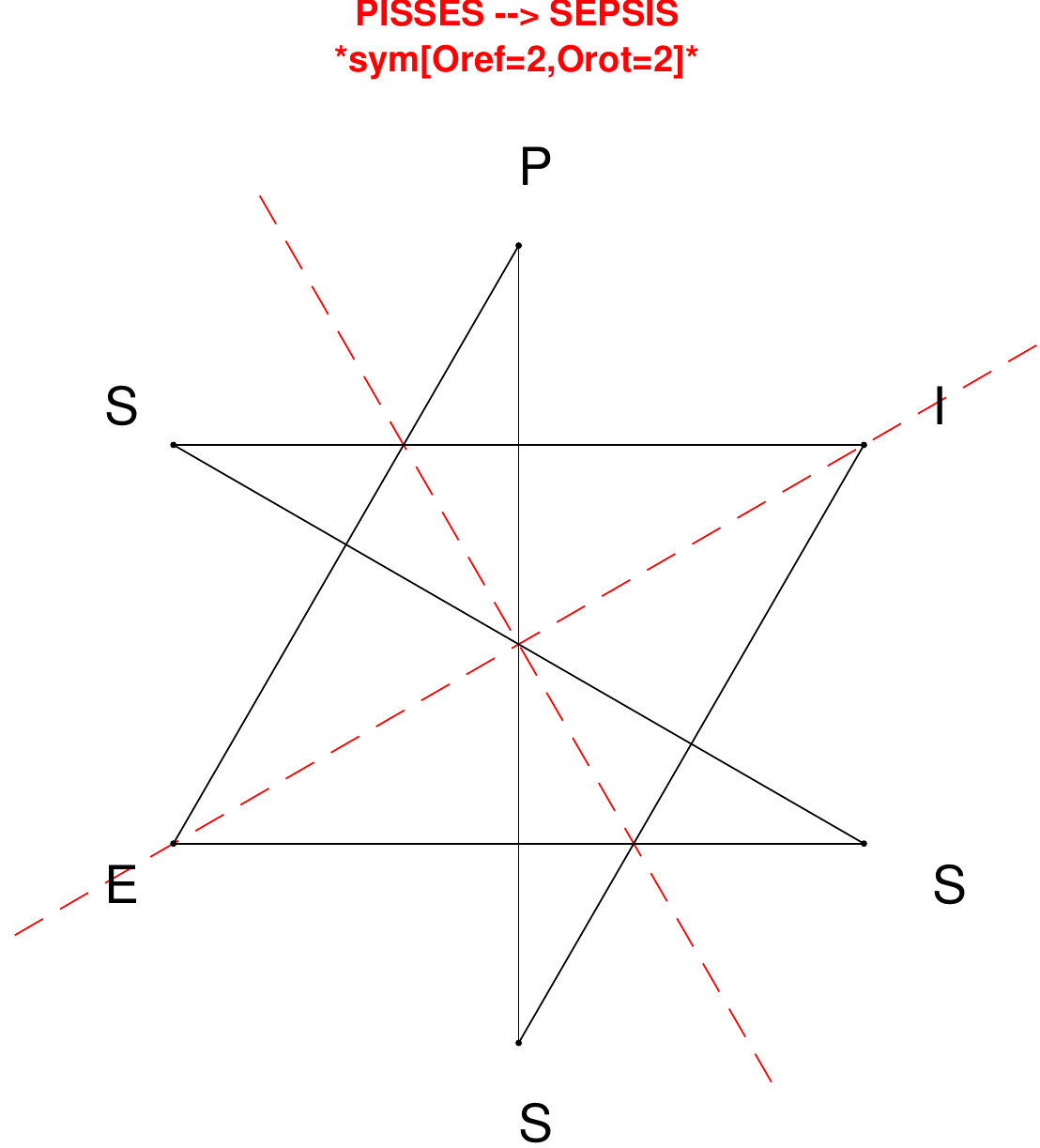}
\end{subfigure}
\end{figure}

\begin{figure}[H]
\centering
\begin{subfigure}[T]{0.19\textwidth}
\centering
\includegraphics[width=\textwidth]{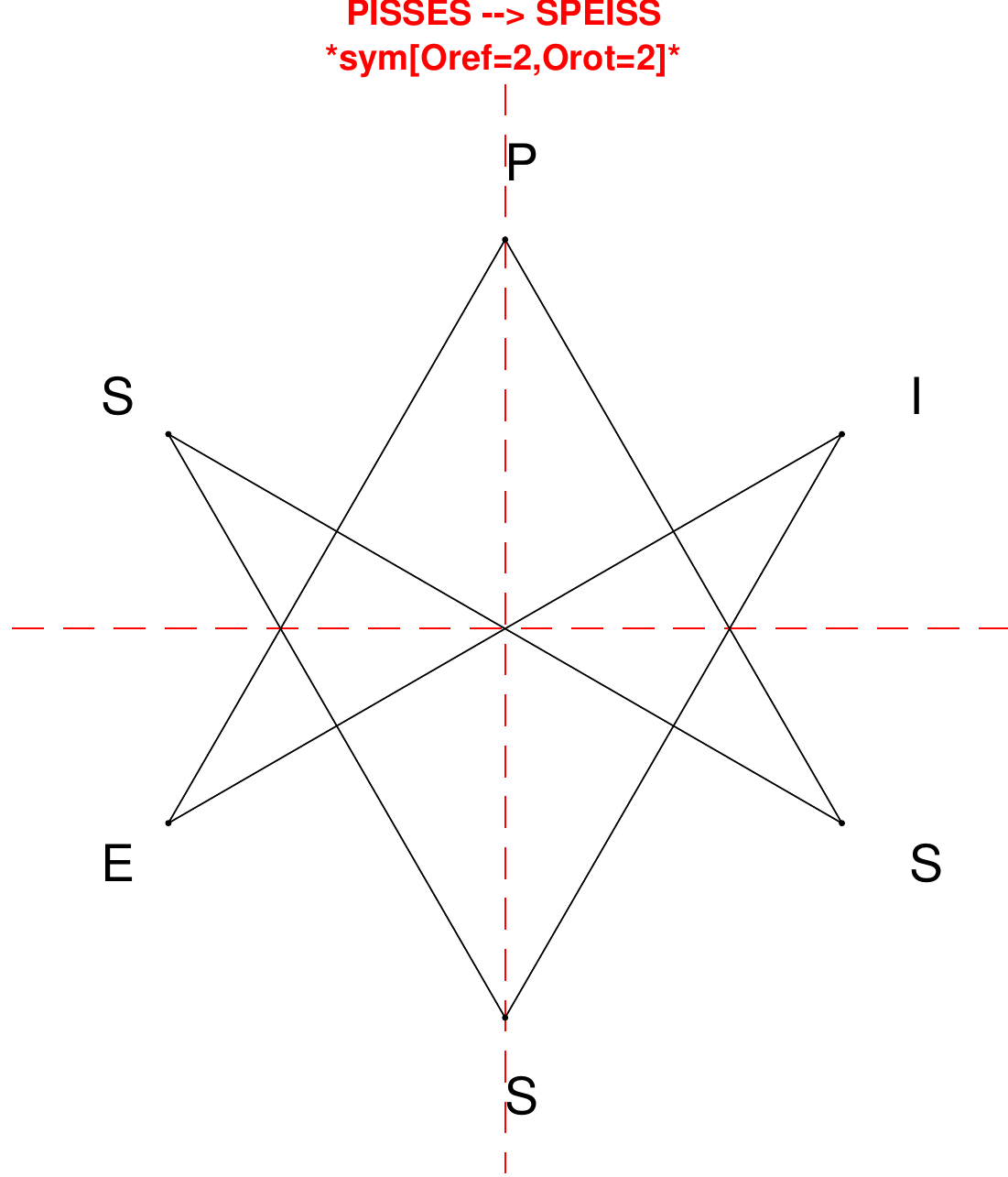}
\end{subfigure}
\hfill
\begin{subfigure}[T]{0.19\textwidth}
\centering
\includegraphics[width=\textwidth]{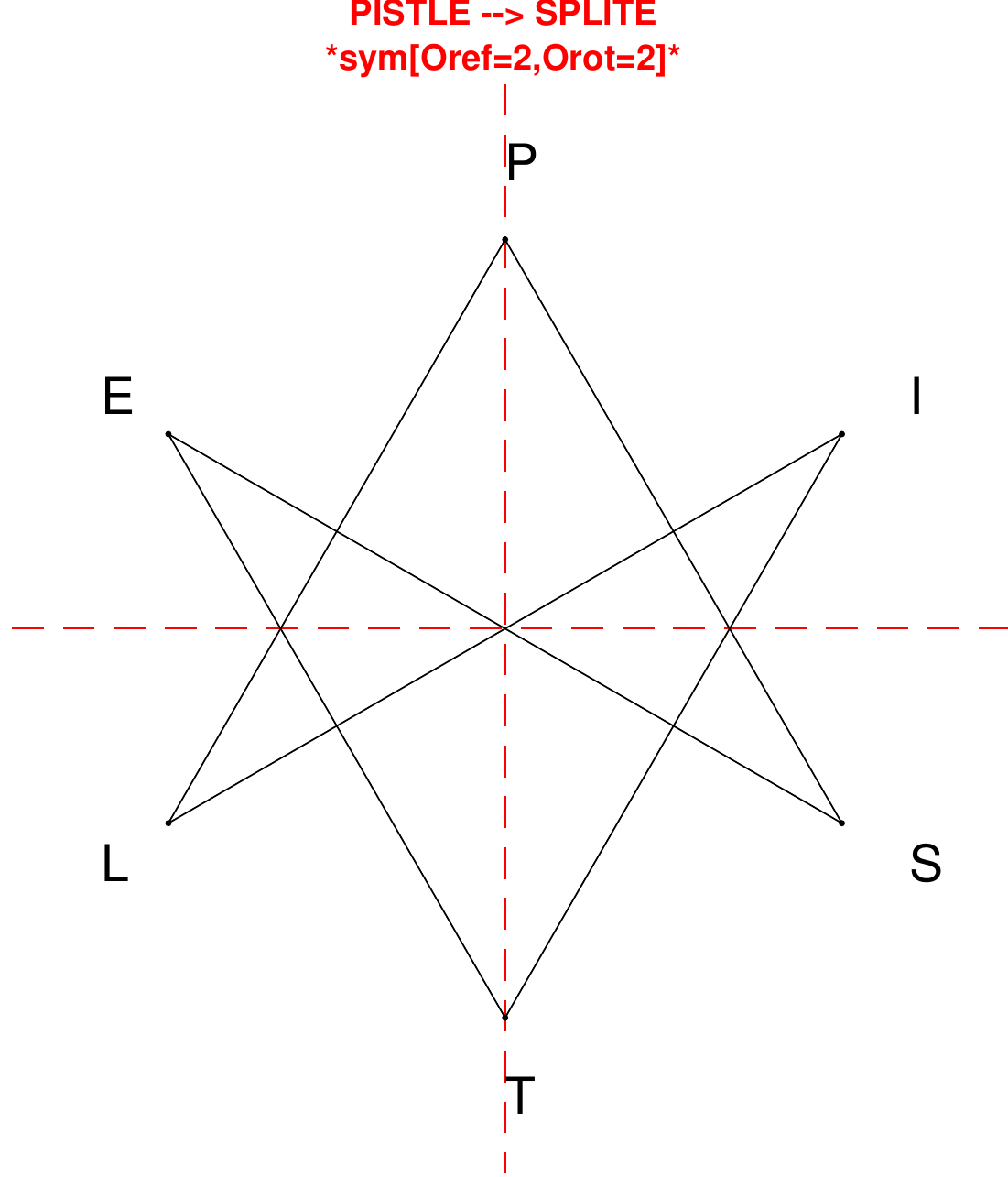}
\end{subfigure}
\hfill
\begin{subfigure}[T]{0.19\textwidth}
\centering
\includegraphics[width=\textwidth]{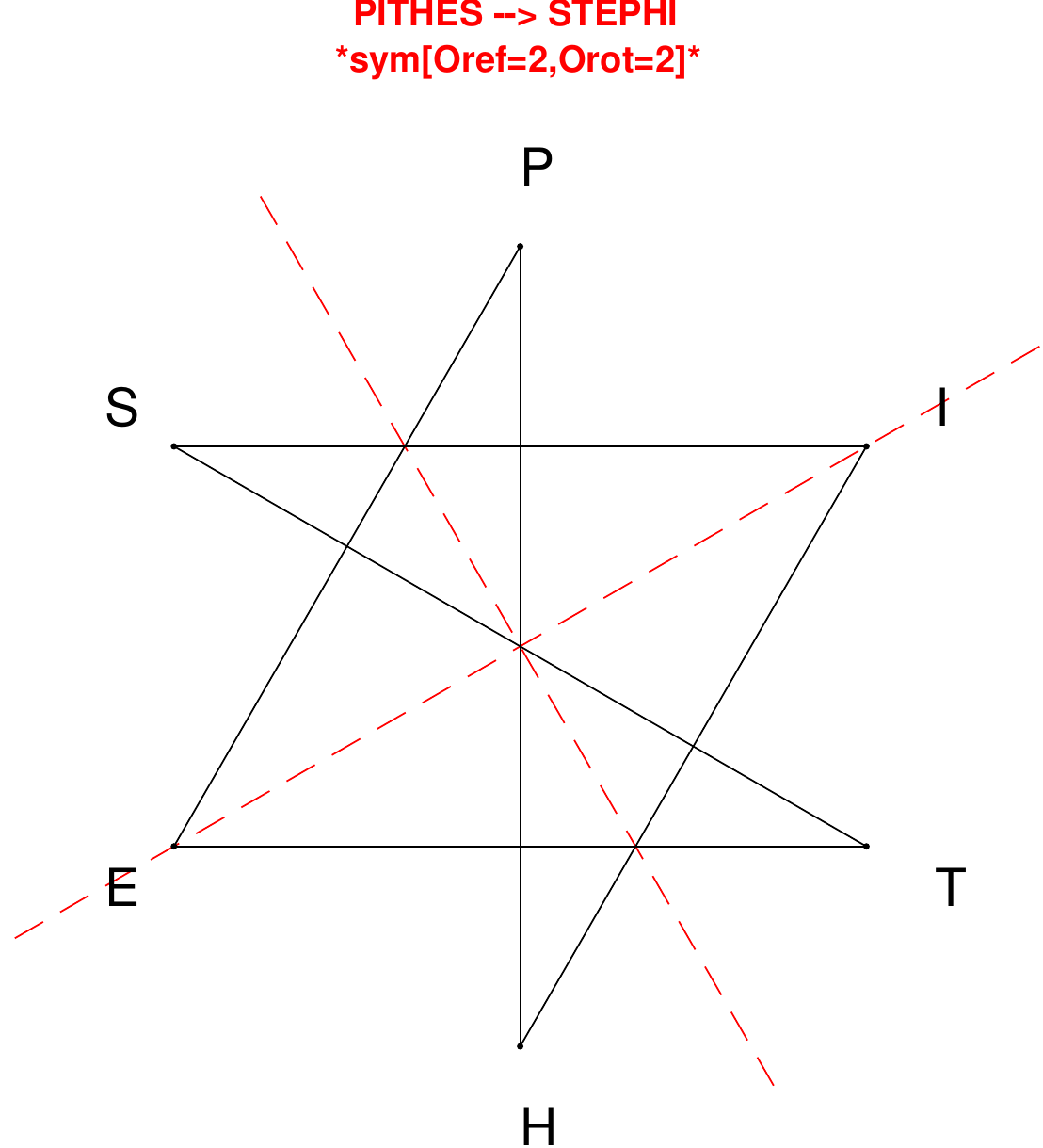}
\end{subfigure}
\hfill
\begin{subfigure}[T]{0.19\textwidth}
\centering
\includegraphics[width=\textwidth]{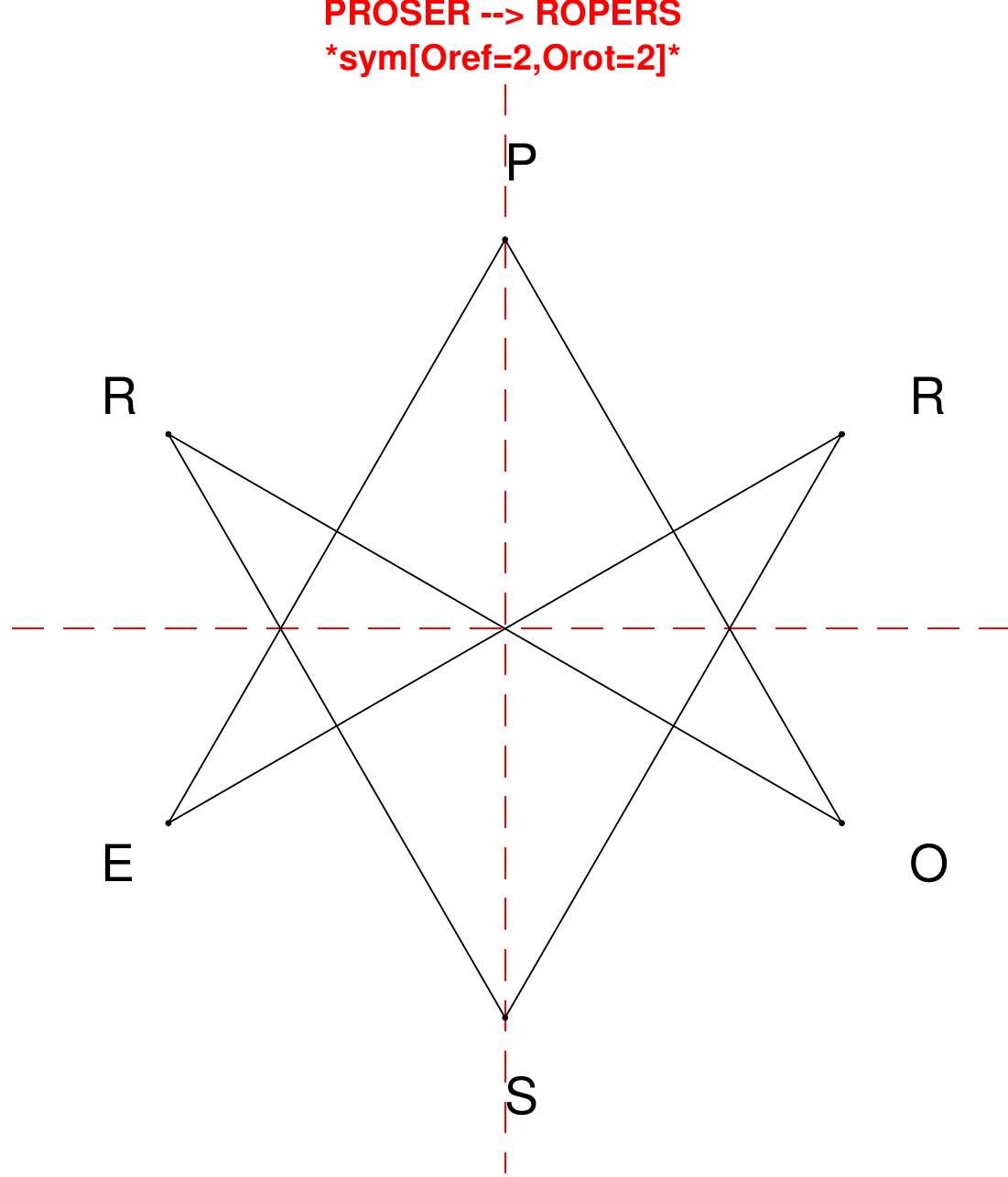}
\end{subfigure}
\hfill
\begin{subfigure}[T]{0.19\textwidth}
\centering
\includegraphics[width=\textwidth]{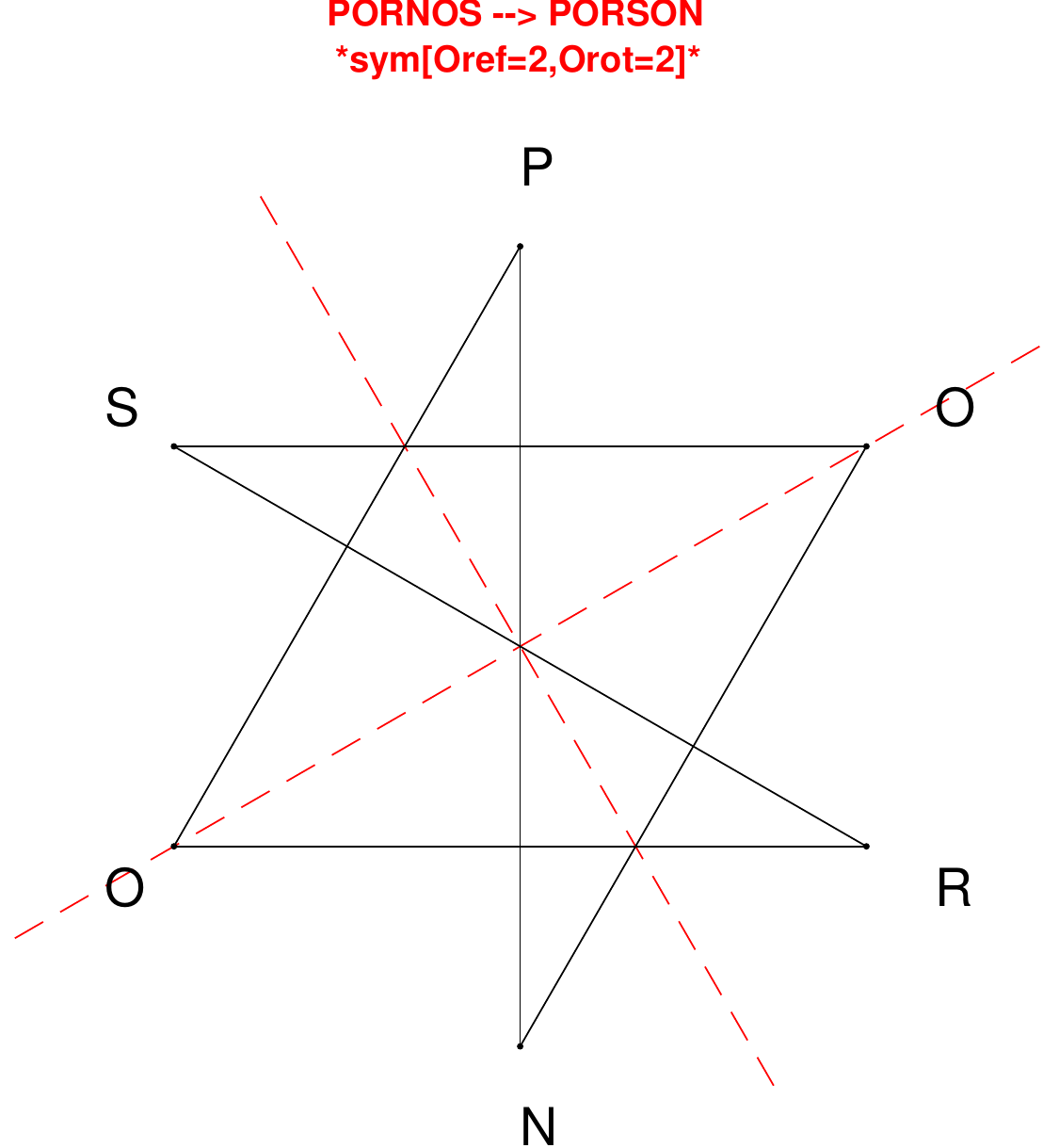}
\end{subfigure}
\end{figure}

\begin{figure}[H]
\centering
\begin{subfigure}[T]{0.19\textwidth}
\centering
\includegraphics[width=\textwidth]{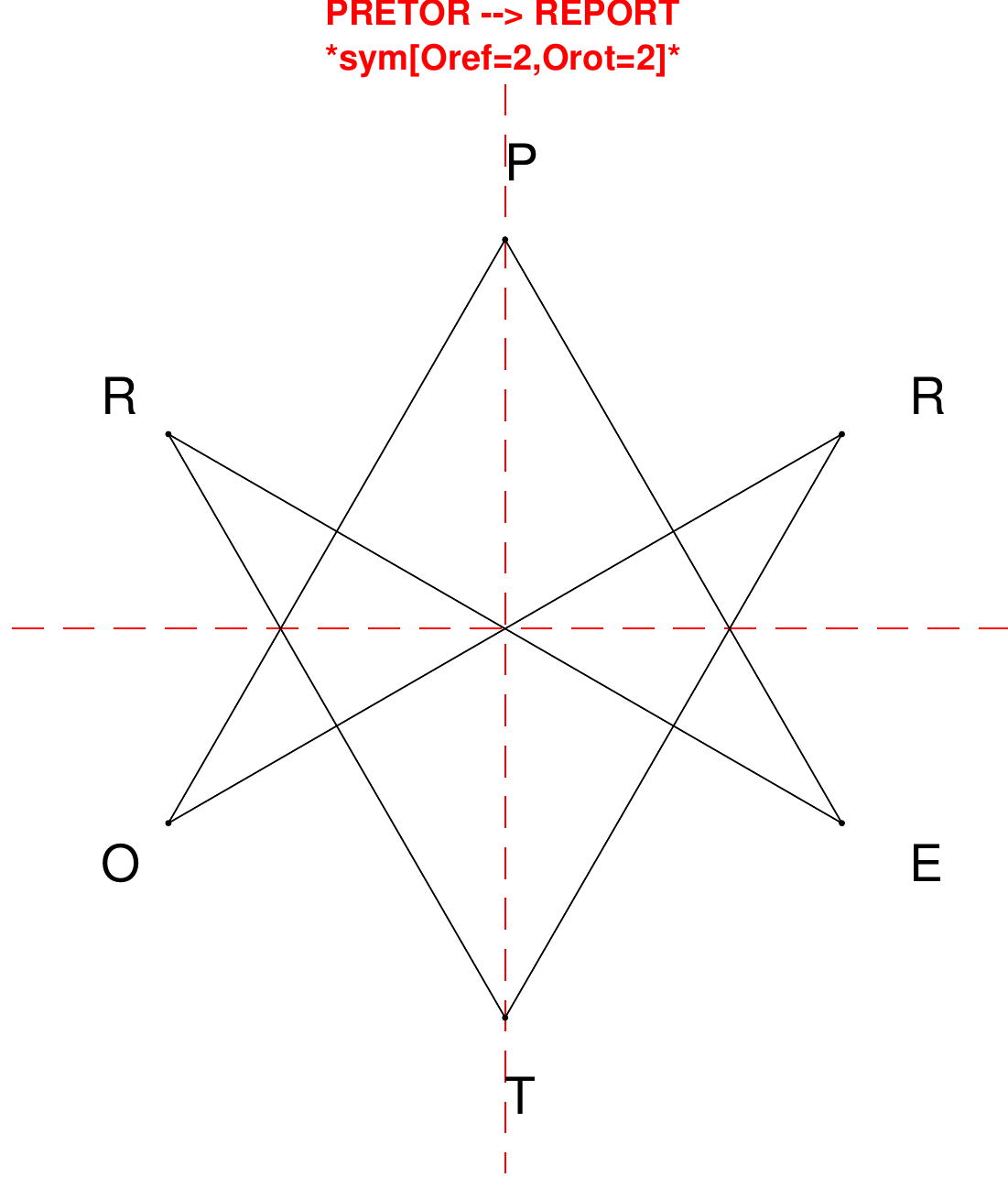}
\end{subfigure}
\hfill
\begin{subfigure}[T]{0.19\textwidth}
\centering
\includegraphics[width=\textwidth]{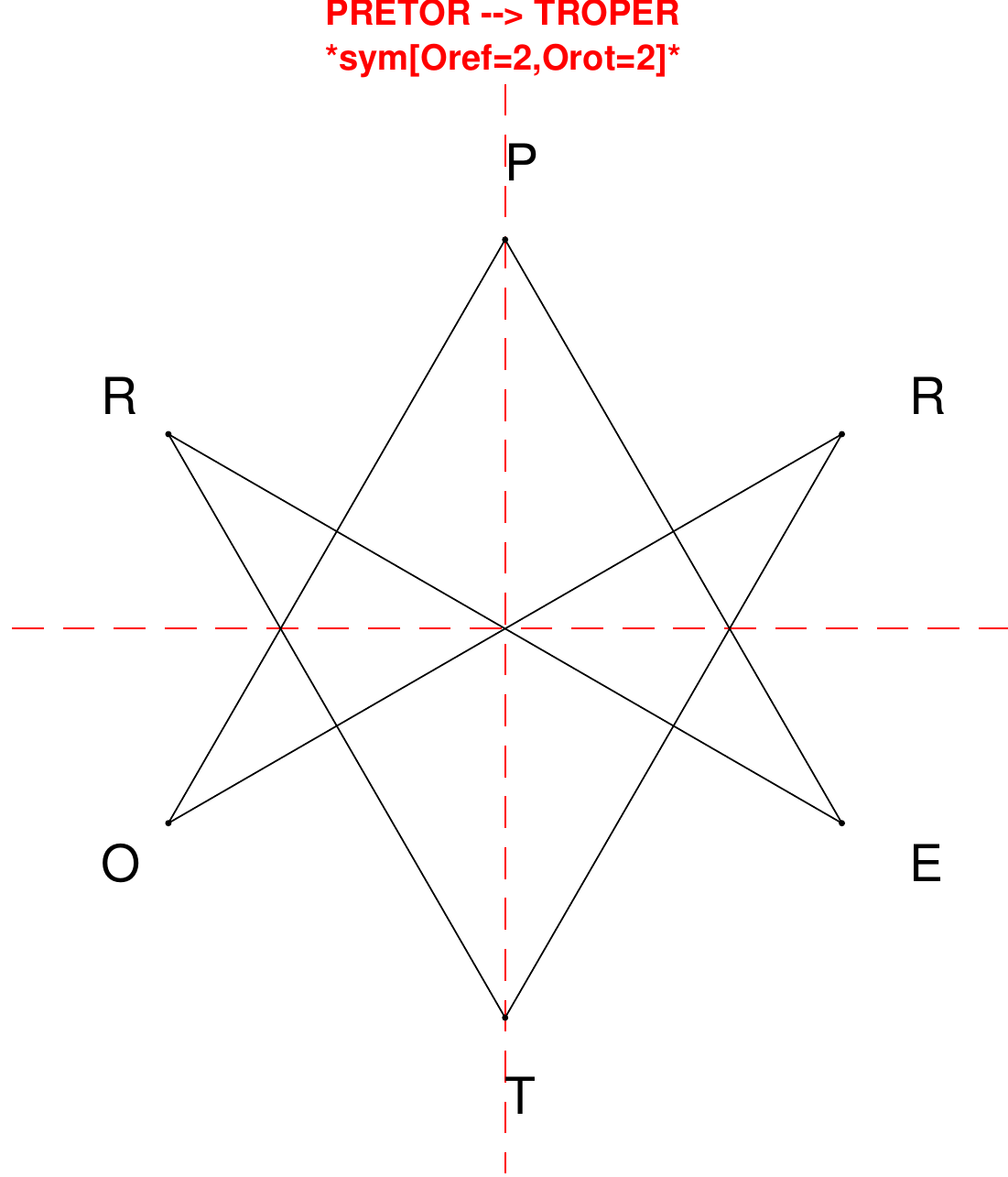}
\end{subfigure}
\hfill
\begin{subfigure}[T]{0.19\textwidth}
\centering
\includegraphics[width=\textwidth]{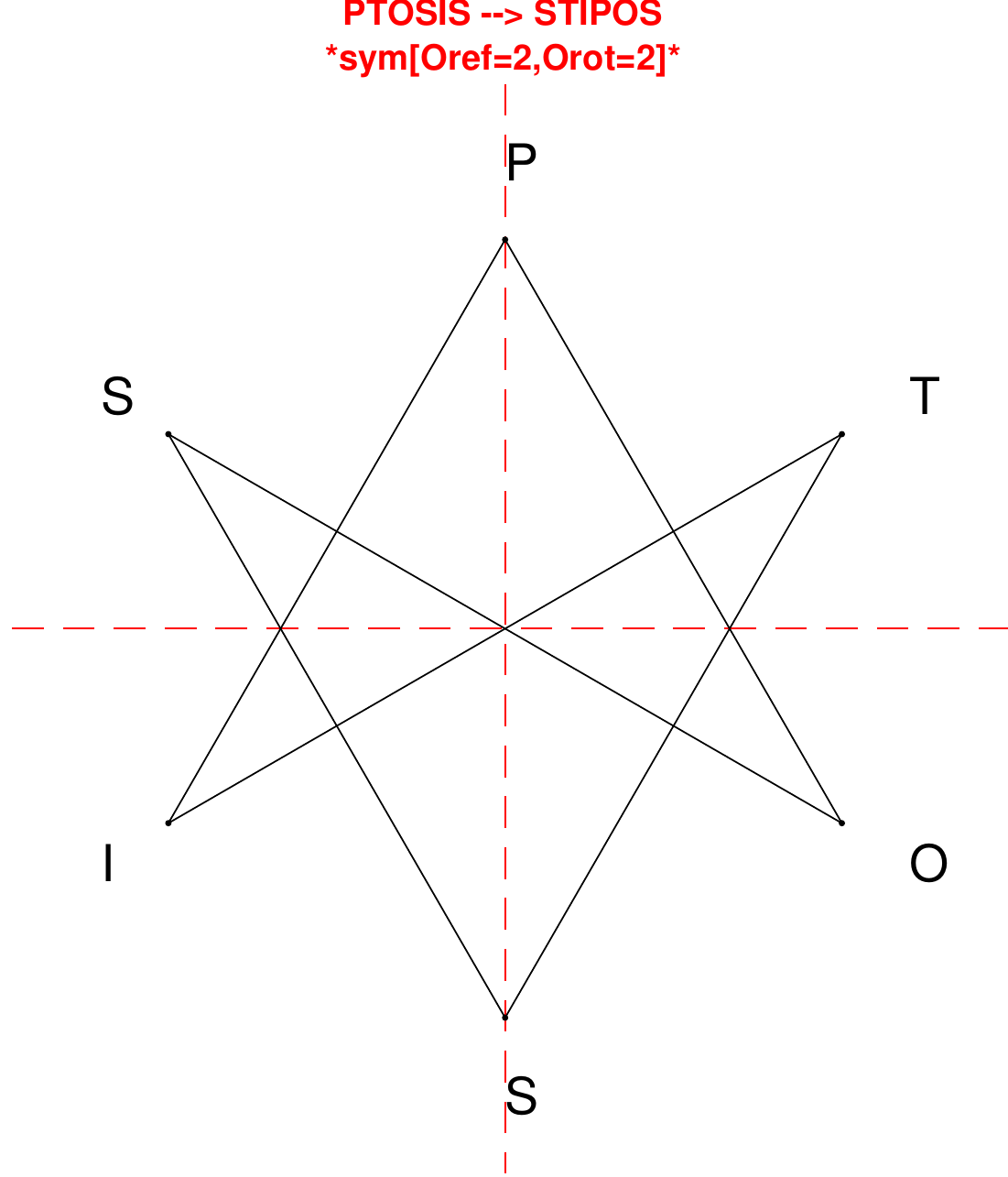}
\end{subfigure}
\hfill
\begin{subfigure}[T]{0.19\textwidth}
\centering
\includegraphics[width=\textwidth]{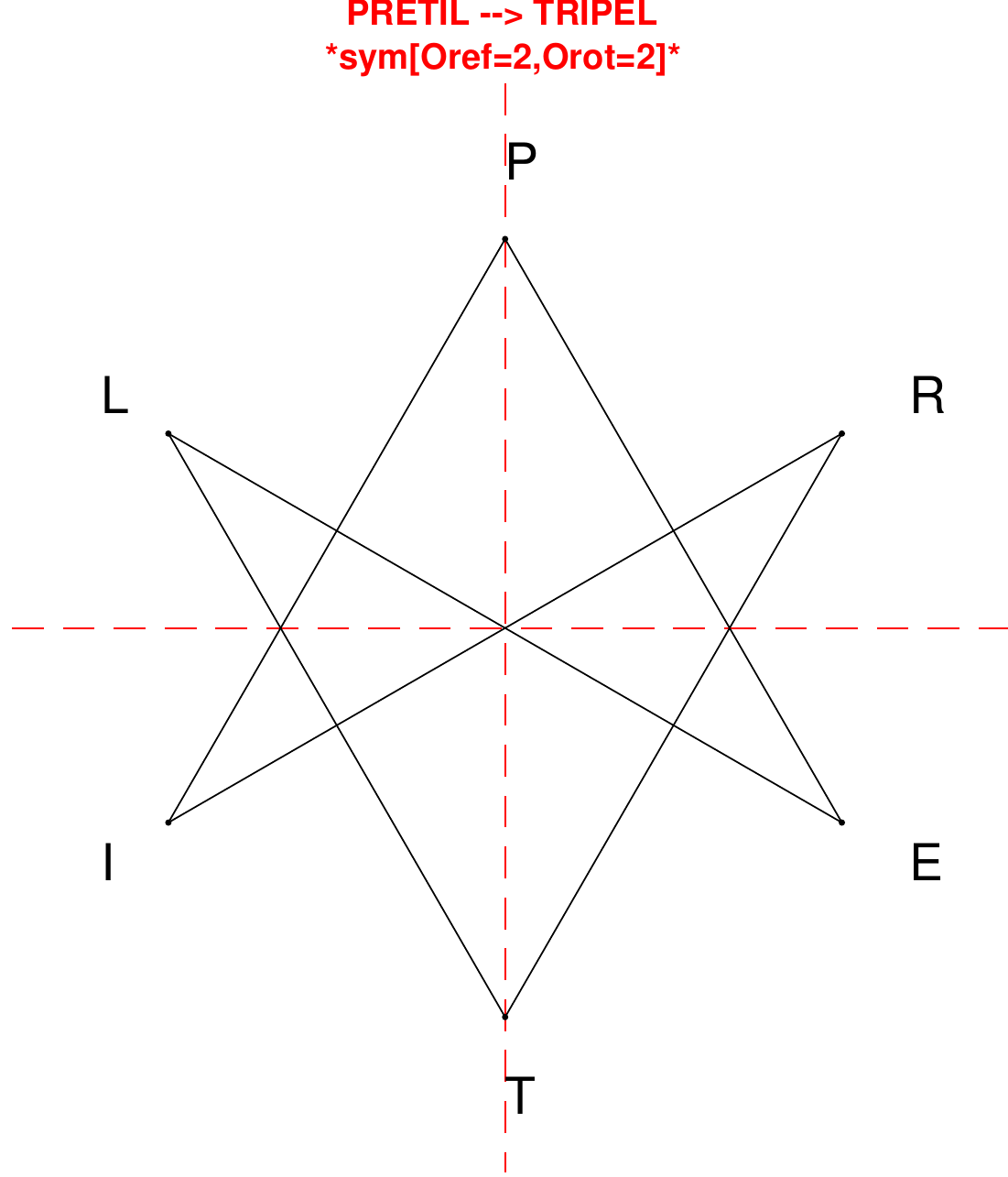}
\end{subfigure}
\hfill
\begin{subfigure}[T]{0.19\textwidth}
\centering
\includegraphics[width=\textwidth]{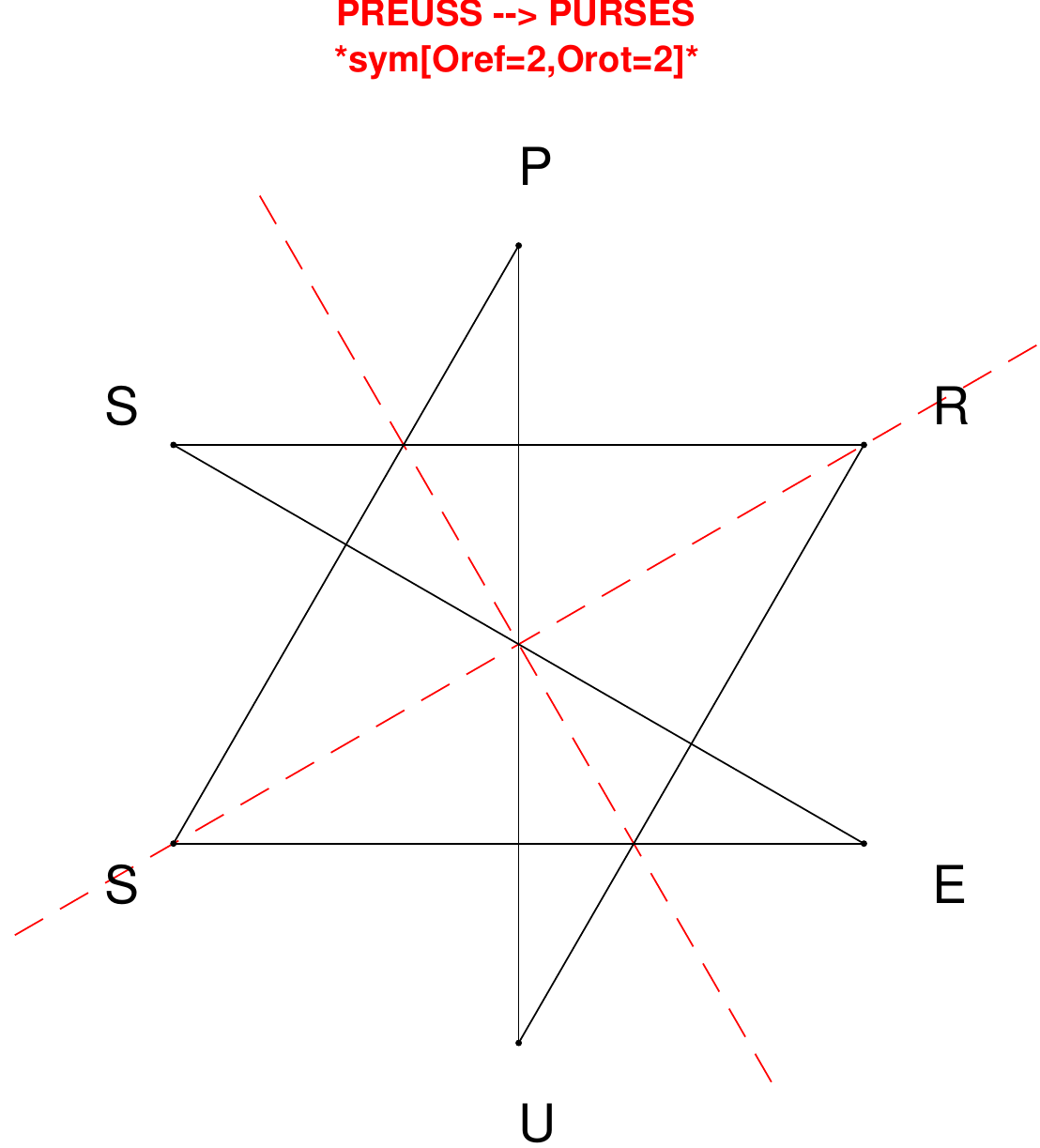}
\end{subfigure}
\end{figure}

\begin{figure}[H]
\centering
\begin{subfigure}[T]{0.19\textwidth}
\centering
\includegraphics[width=\textwidth]{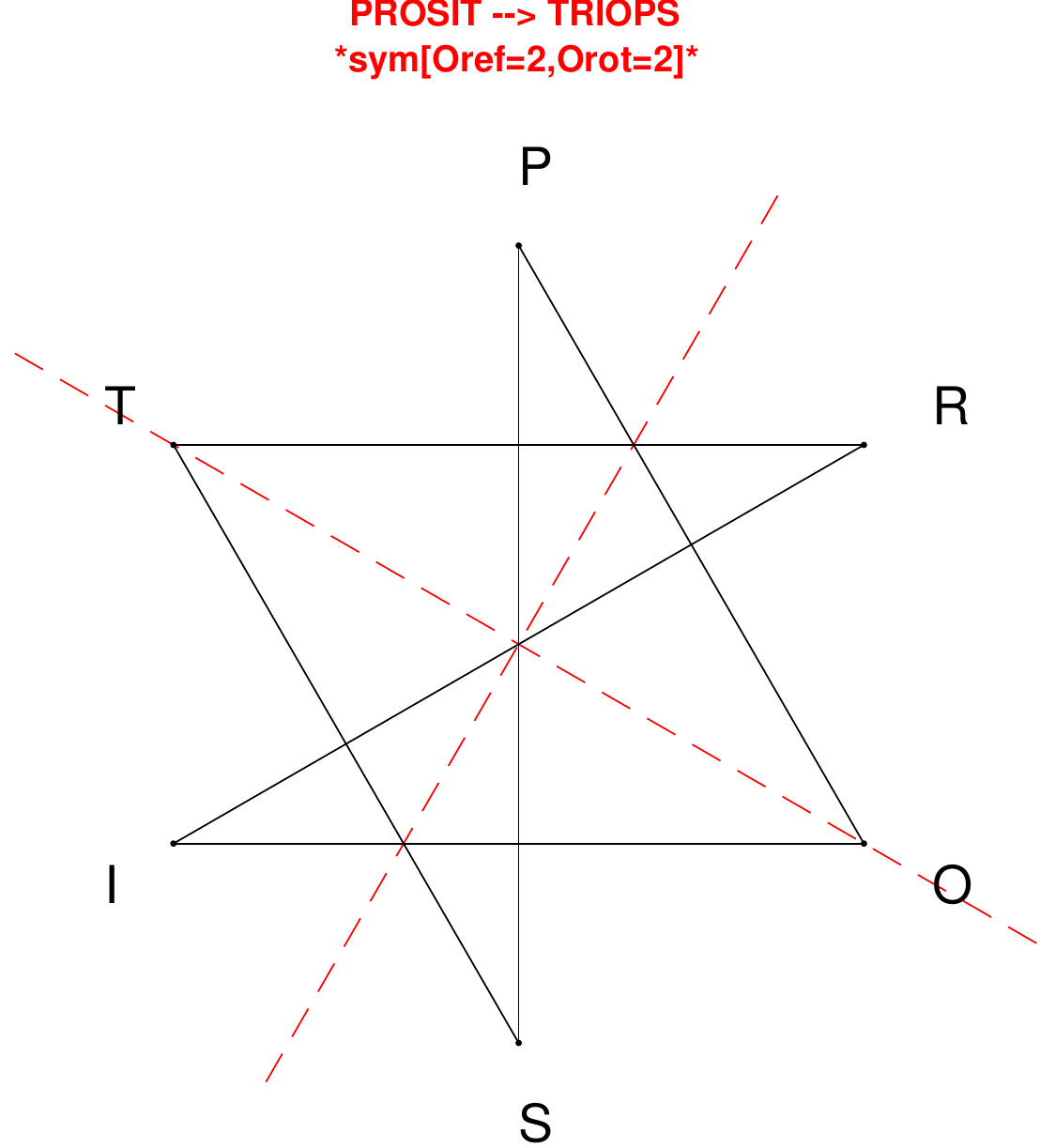}
\end{subfigure}
\hfill
\begin{subfigure}[T]{0.19\textwidth}
\centering
\includegraphics[width=\textwidth]{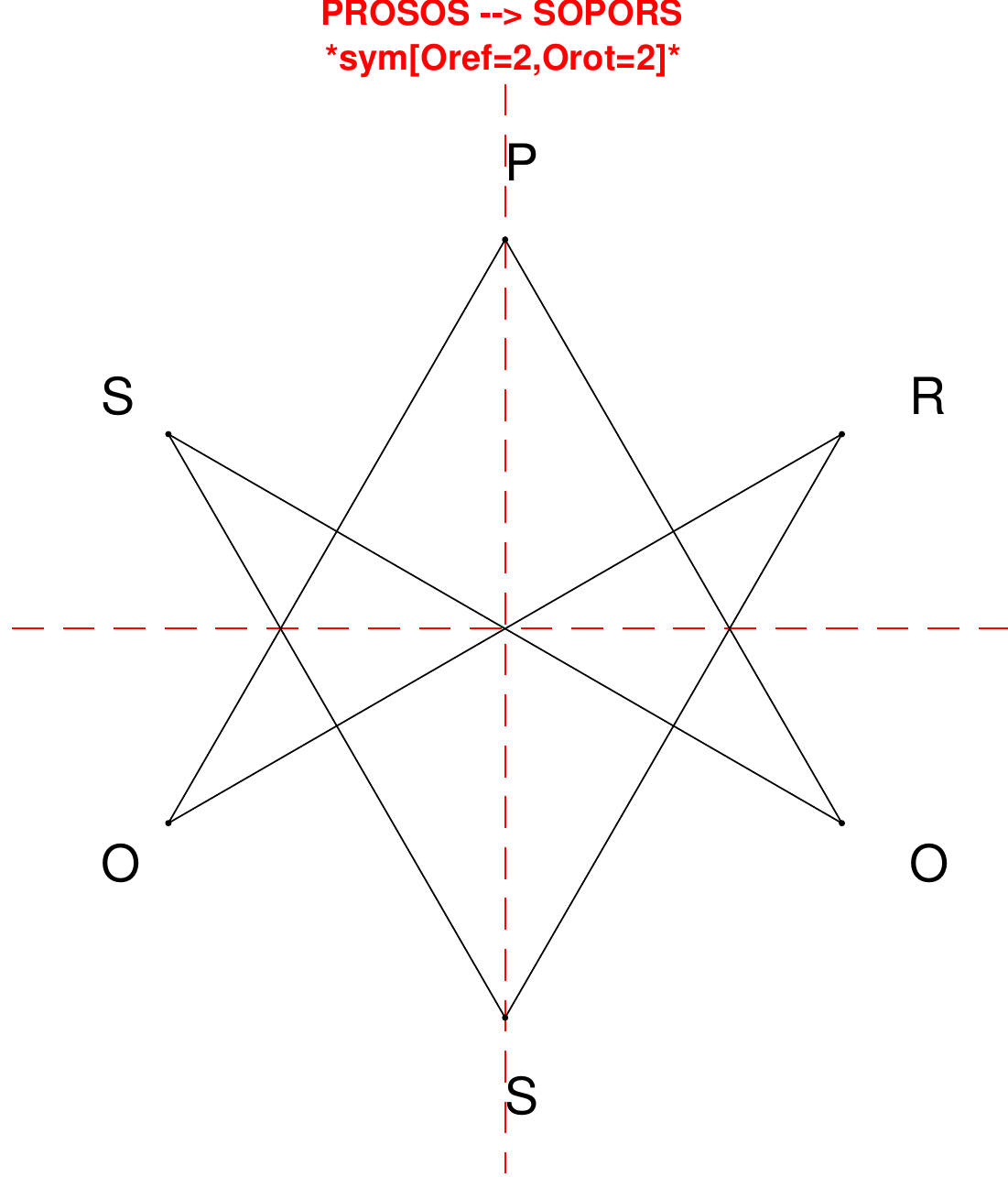}
\end{subfigure}
\hfill
\begin{subfigure}[T]{0.19\textwidth}
\centering
\includegraphics[width=\textwidth]{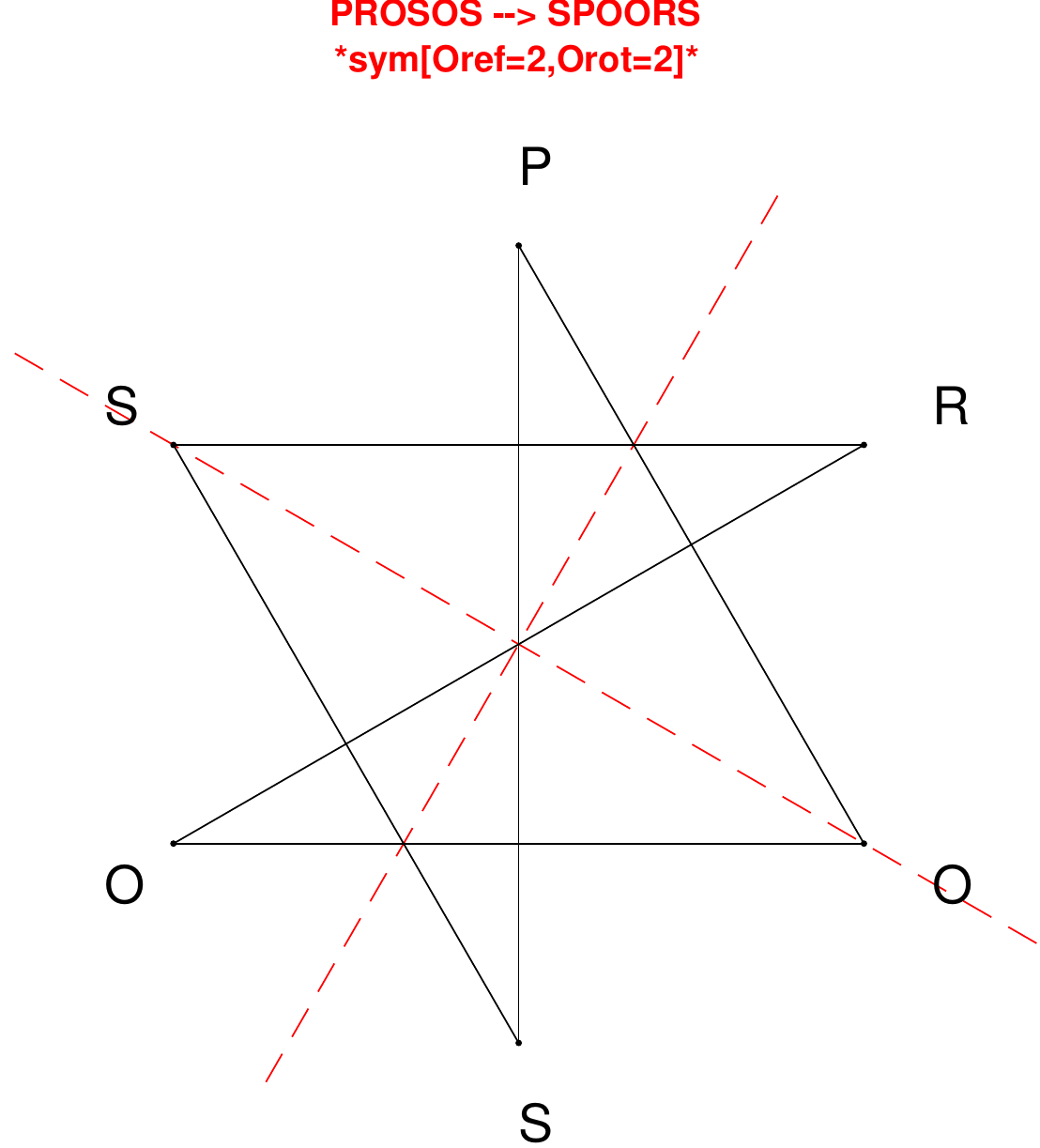}
\end{subfigure}
\hfill
\begin{subfigure}[T]{0.19\textwidth}
\centering
\includegraphics[width=\textwidth]{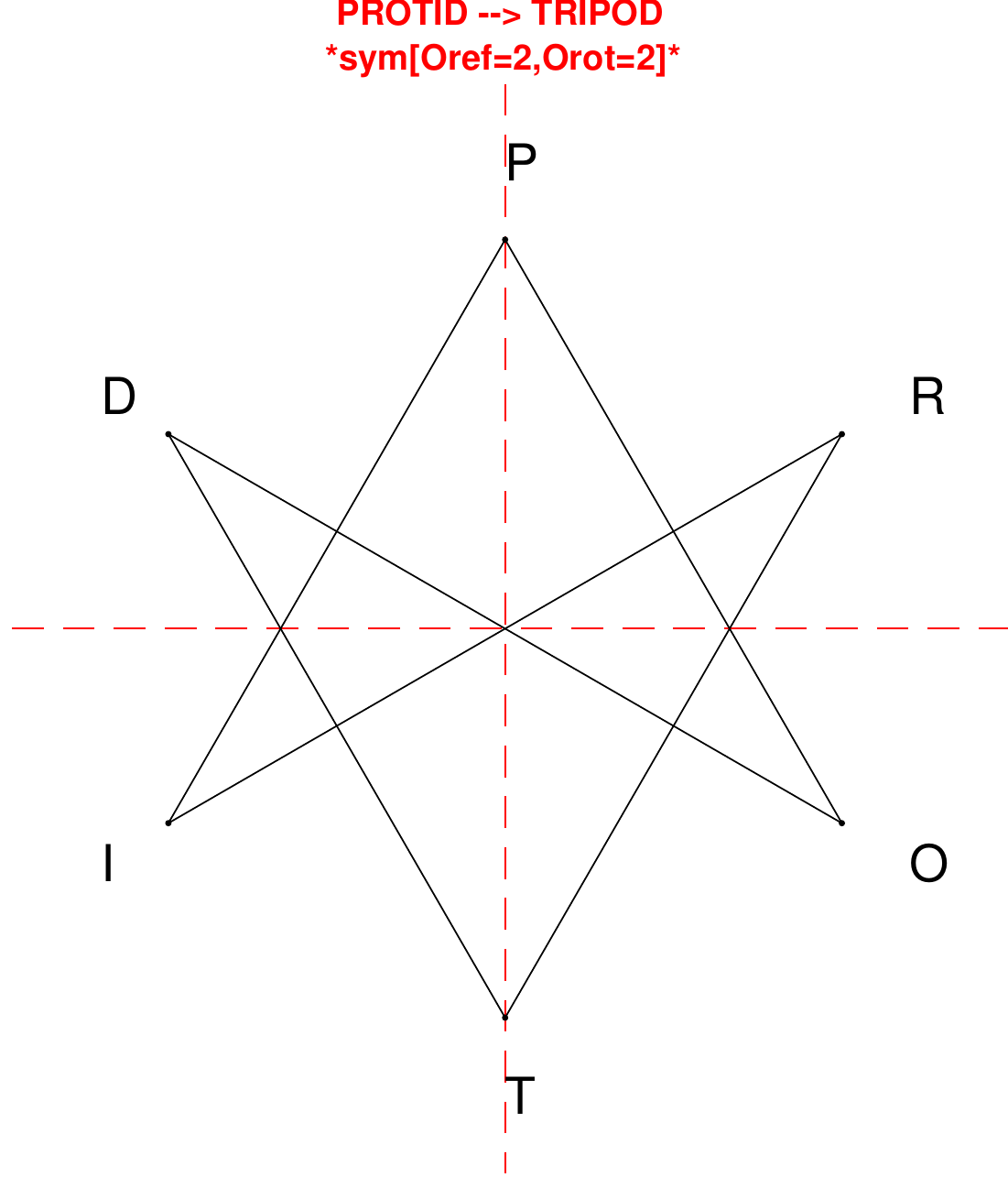}
\end{subfigure}
\hfill
\begin{subfigure}[T]{0.19\textwidth}
\centering
\includegraphics[width=\textwidth]{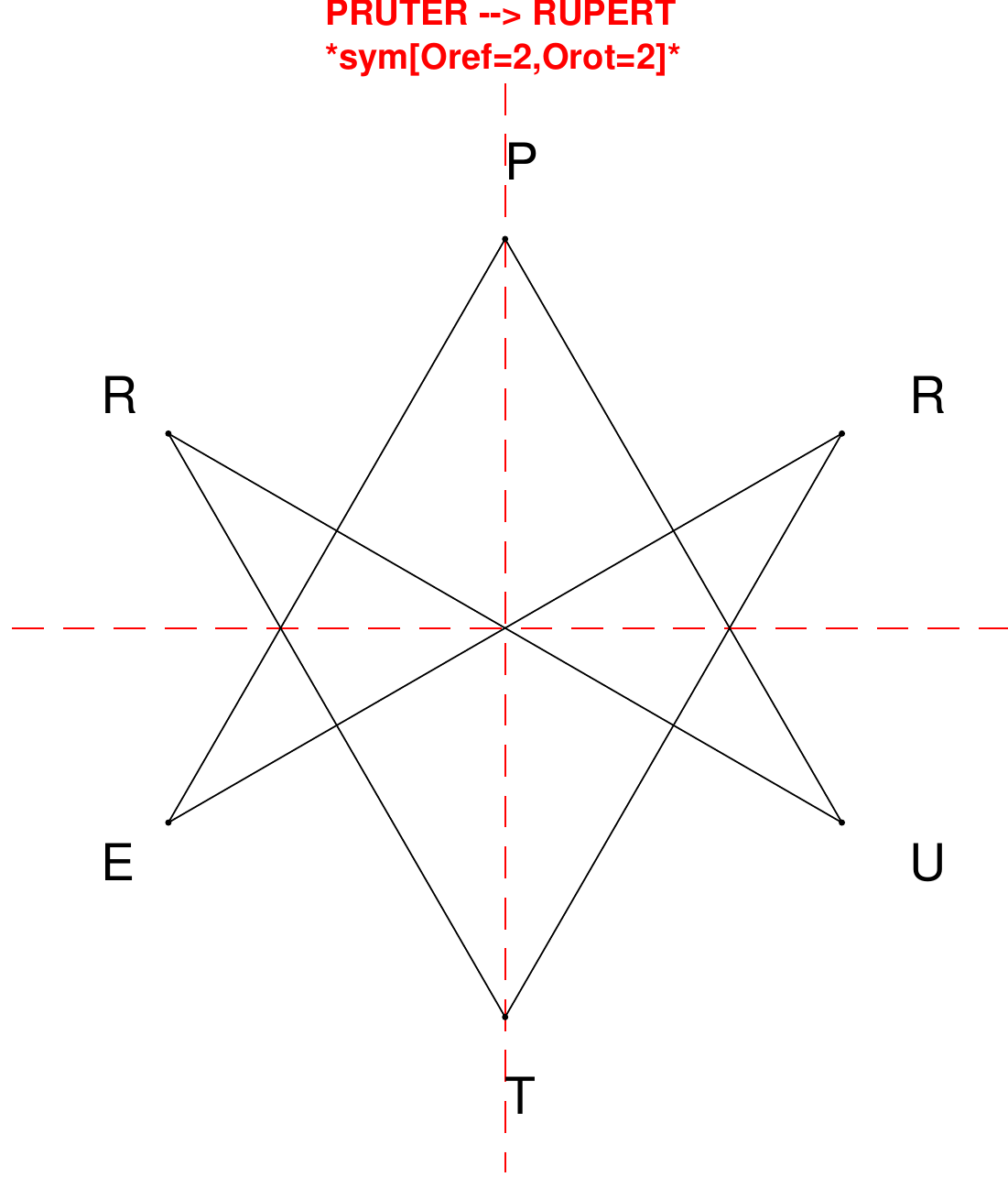}
\end{subfigure}
\end{figure}

\begin{figure}[H]
\centering
\begin{subfigure}[T]{0.19\textwidth}
\centering
\includegraphics[width=\textwidth]{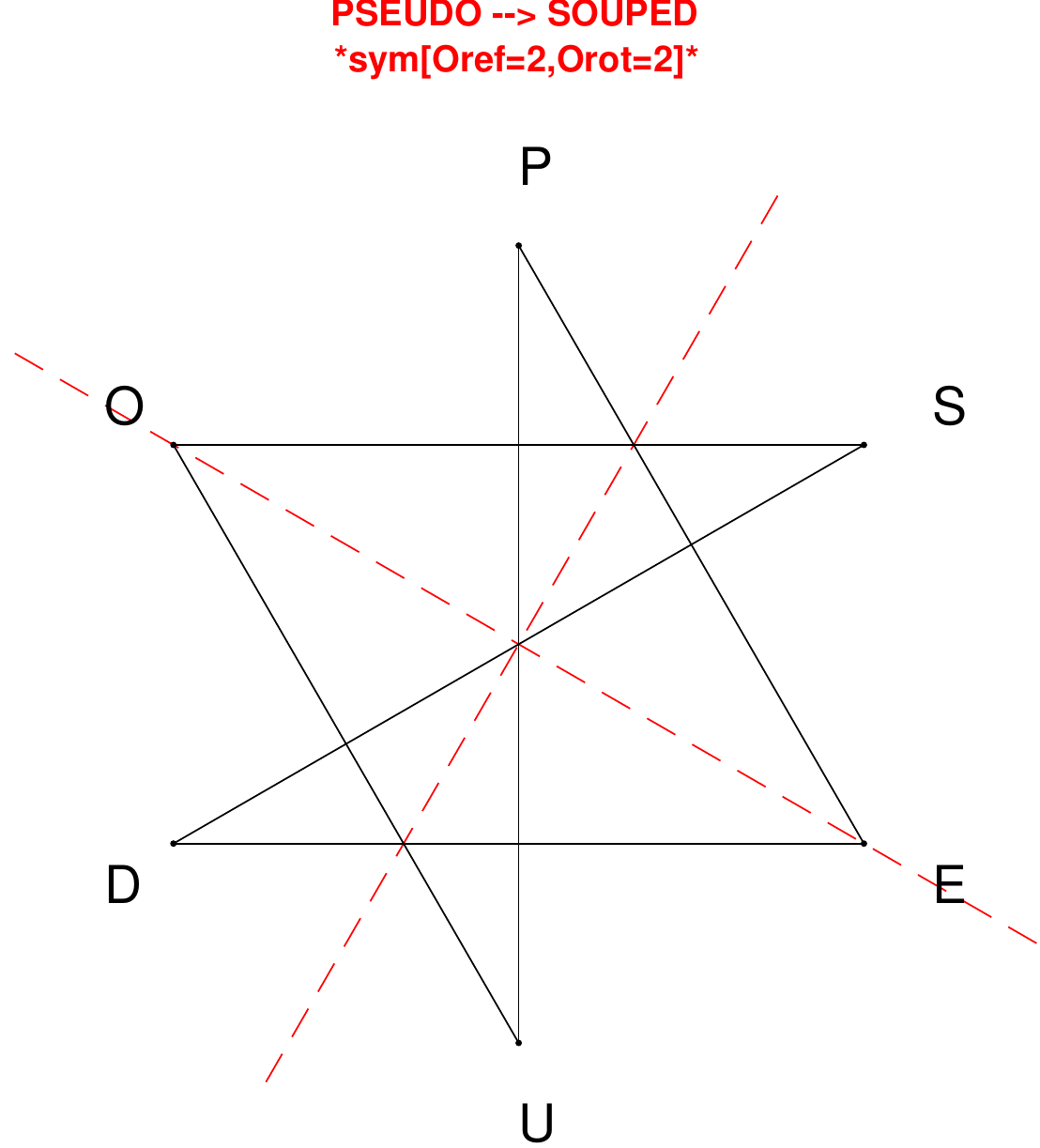}
\end{subfigure}
\hfill
\begin{subfigure}[T]{0.19\textwidth}
\centering
\includegraphics[width=\textwidth]{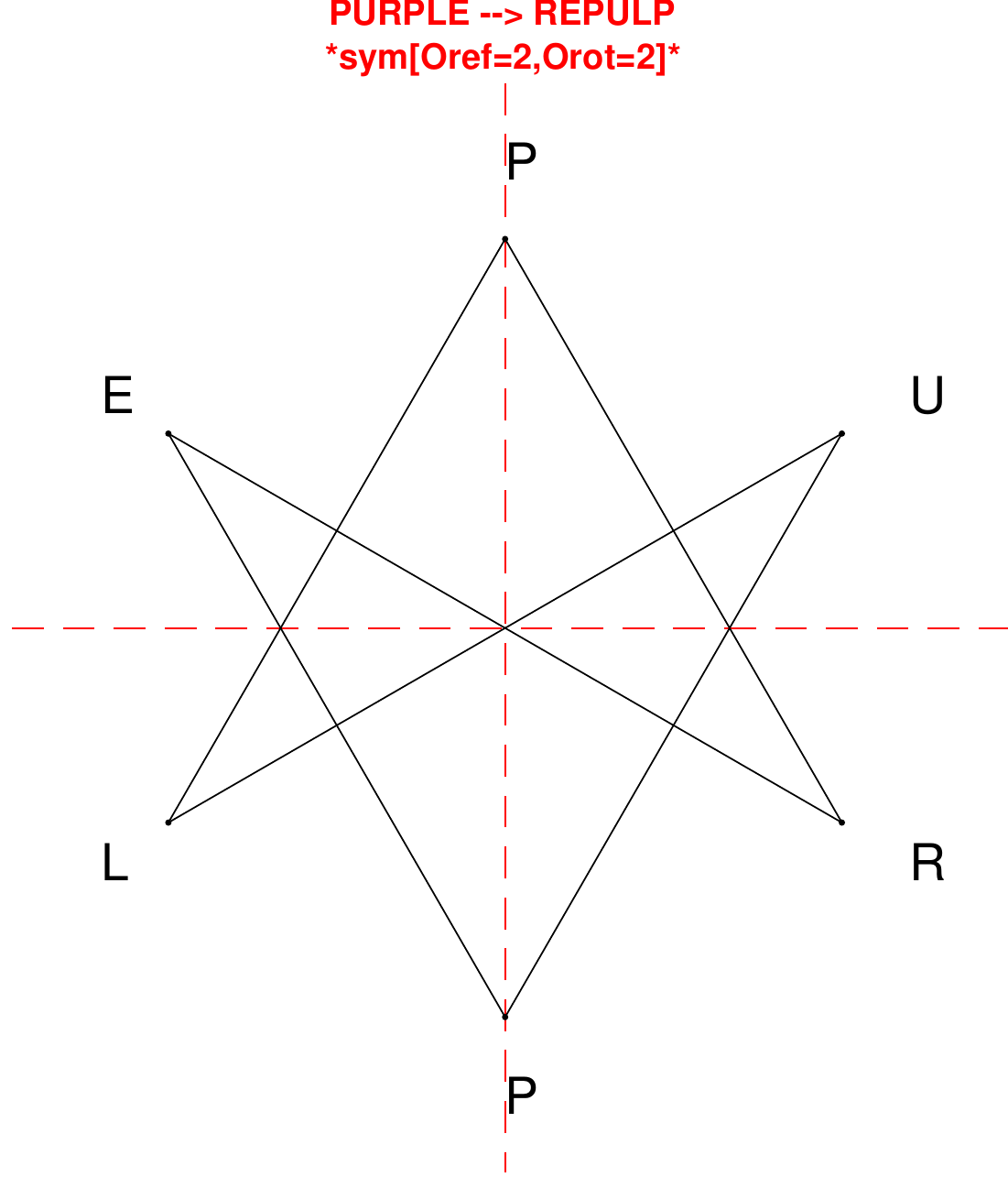}
\end{subfigure}
\hfill
\begin{subfigure}[T]{0.19\textwidth}
\centering
\includegraphics[width=\textwidth]{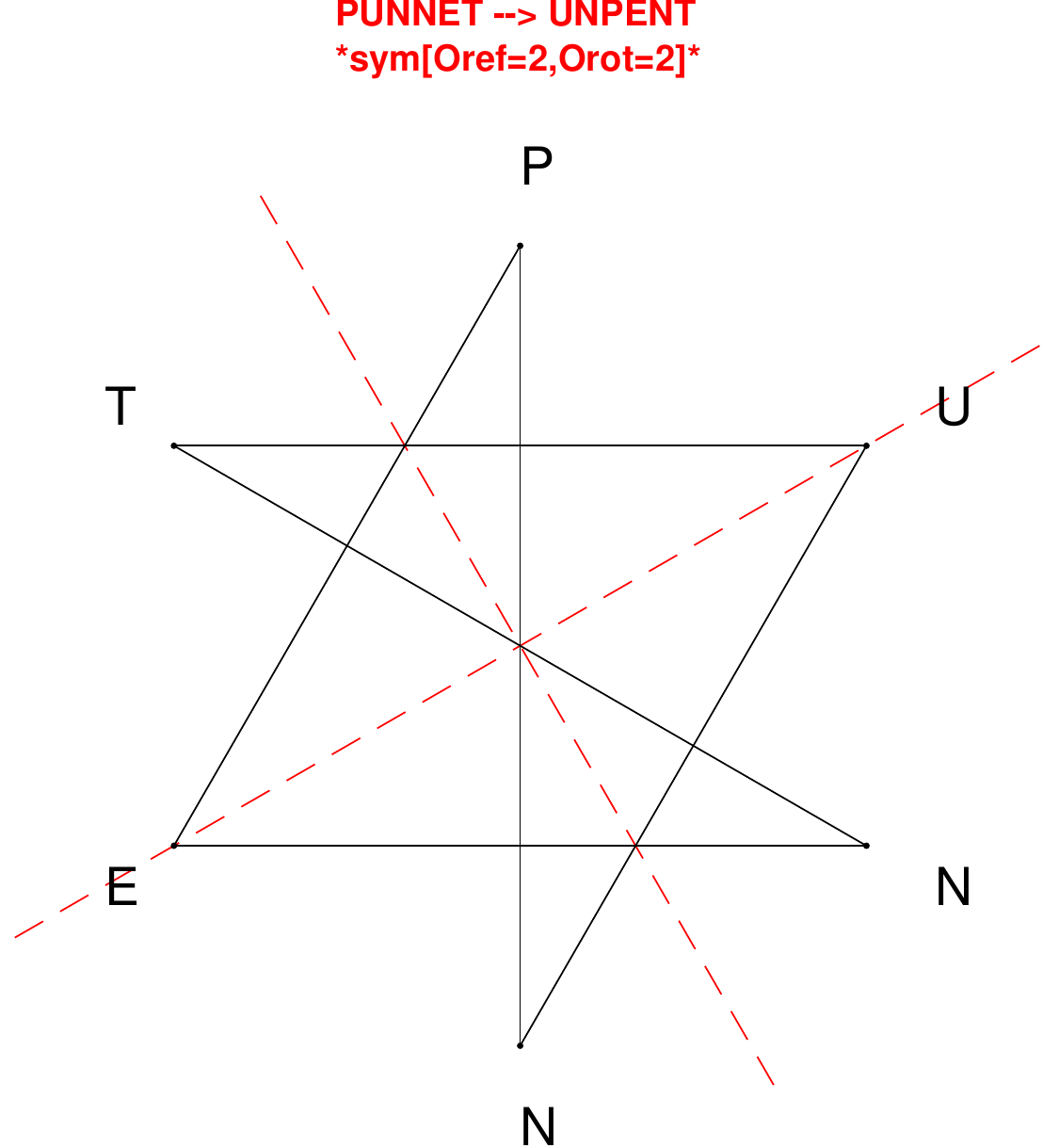}
\end{subfigure}
\hfill
\begin{subfigure}[T]{0.19\textwidth}
\centering
\includegraphics[width=\textwidth]{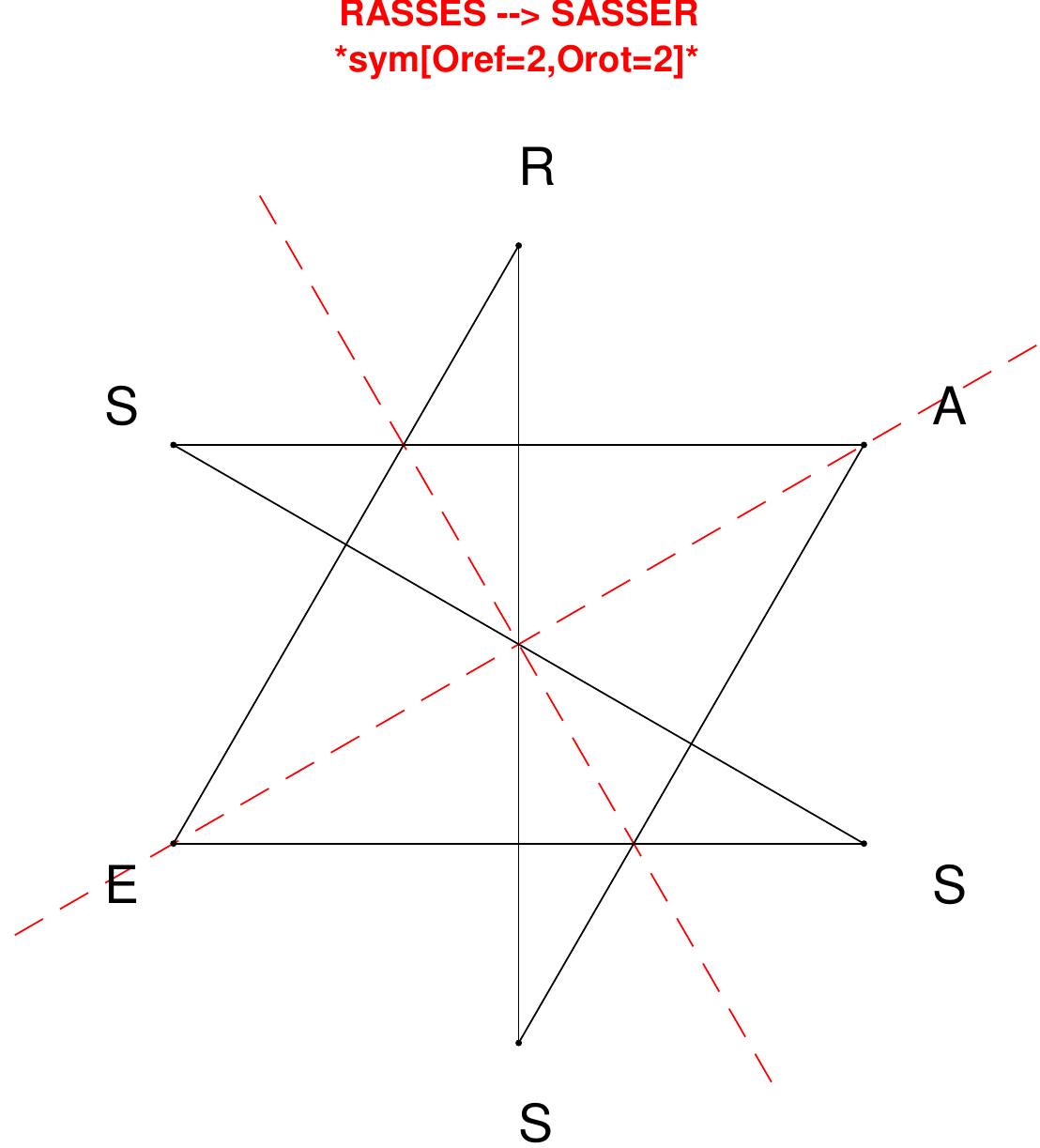}
\end{subfigure}
\hfill
\begin{subfigure}[T]{0.19\textwidth}
\centering
\includegraphics[width=\textwidth]{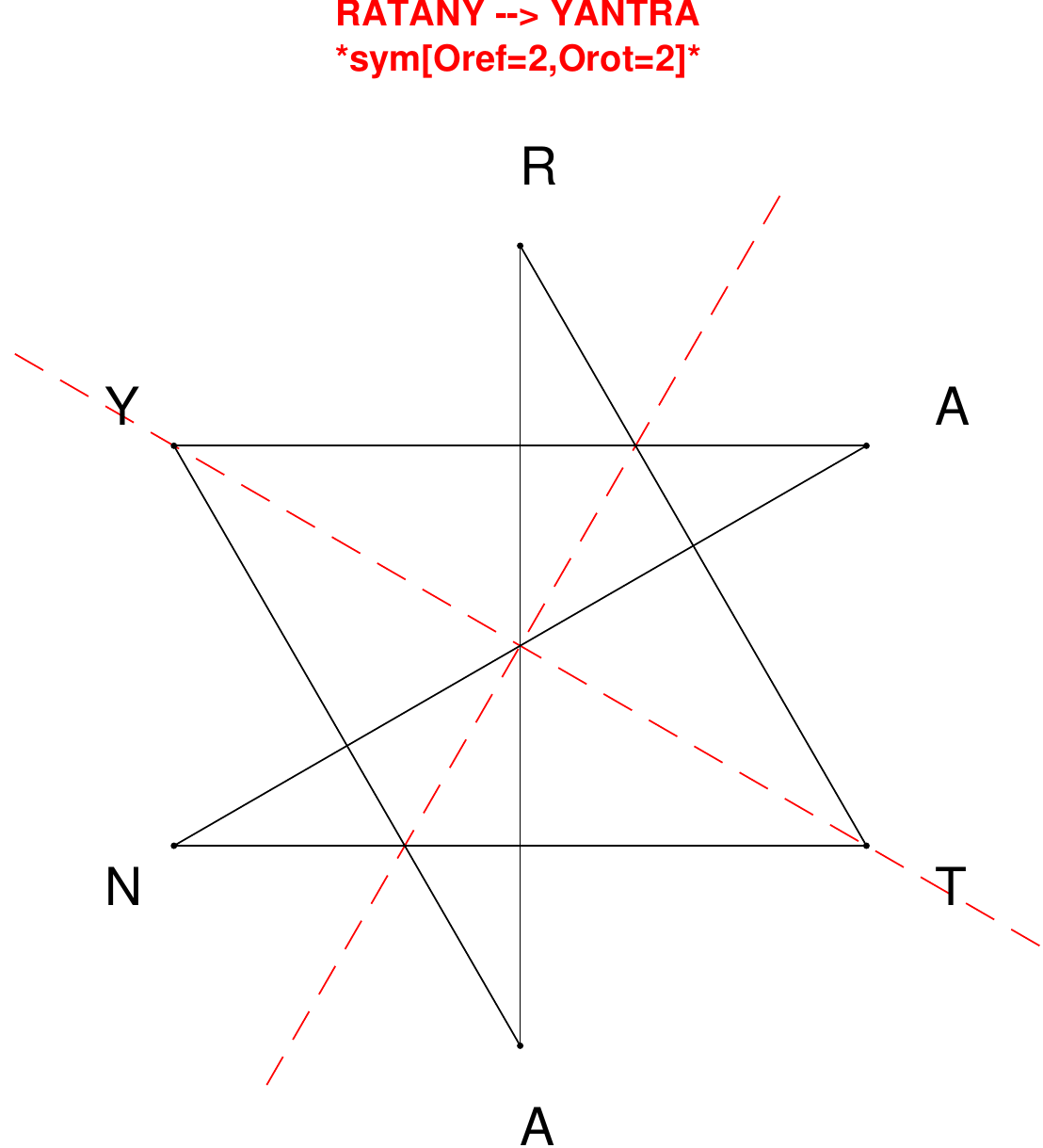}
\end{subfigure}
\end{figure}

\begin{figure}[H]
\centering
\begin{subfigure}[T]{0.19\textwidth}
\centering
\includegraphics[width=\textwidth]{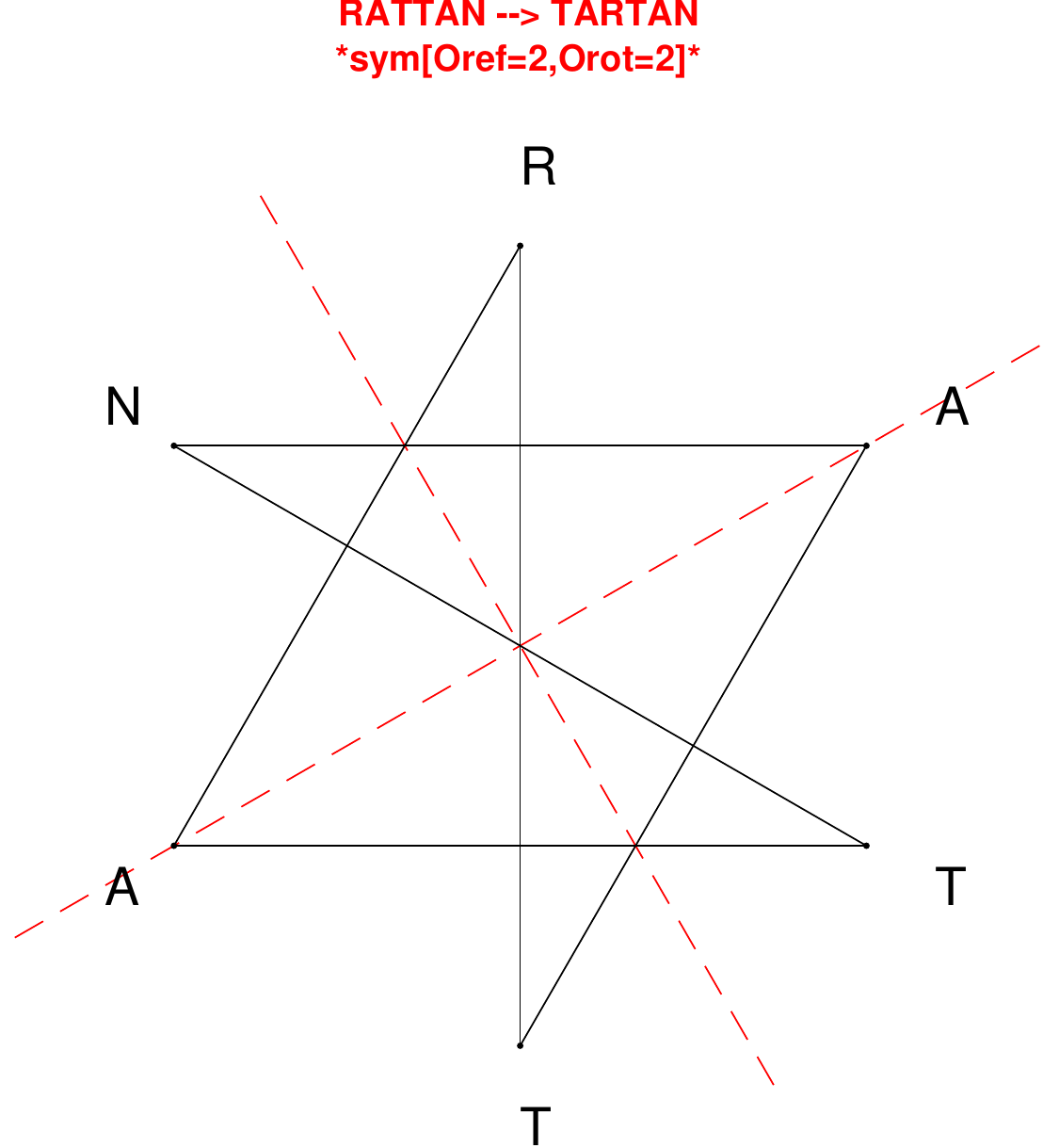}
\end{subfigure}
\hfill
\begin{subfigure}[T]{0.19\textwidth}
\centering
\includegraphics[width=\textwidth]{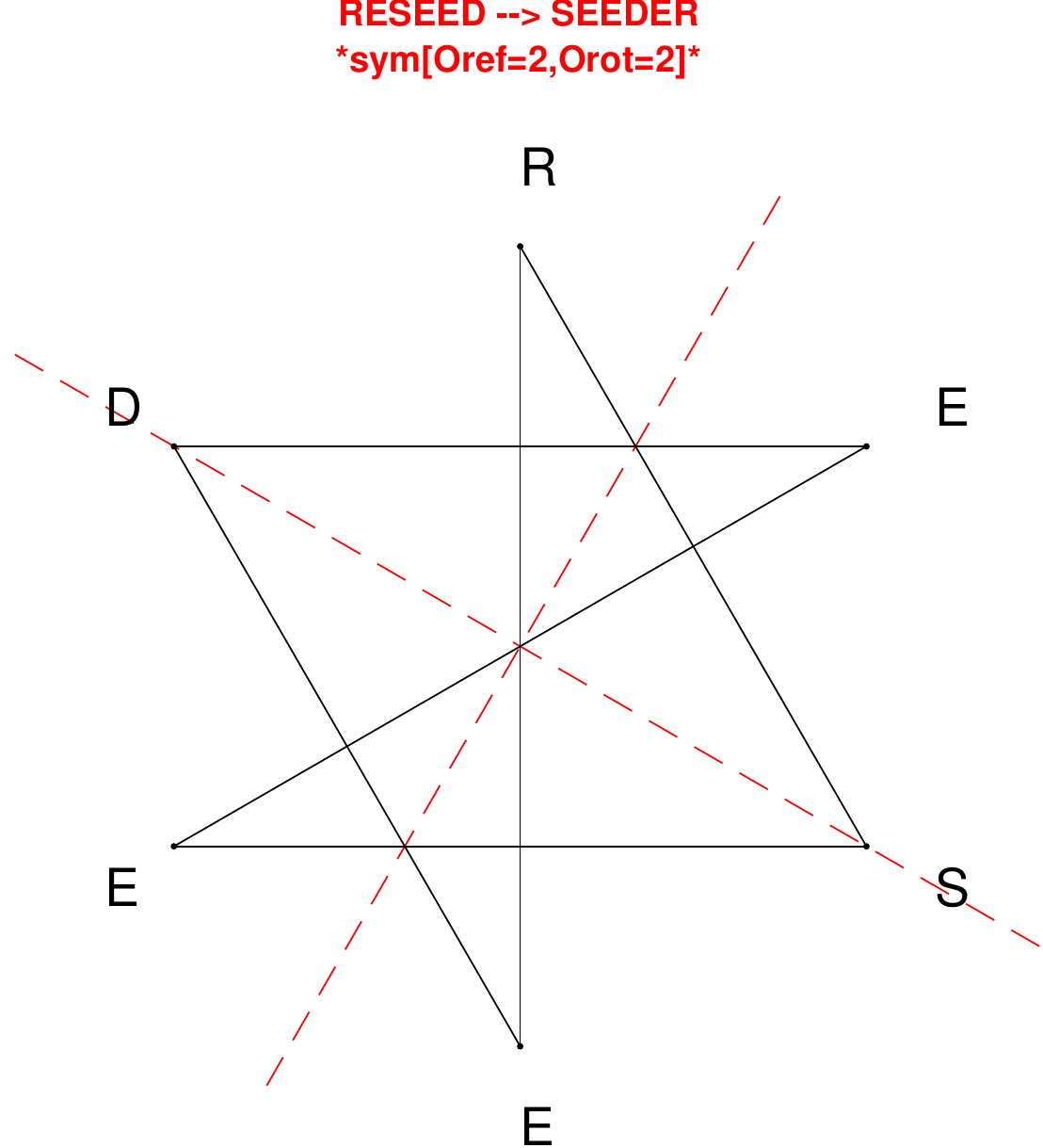}
\end{subfigure}
\hfill
\begin{subfigure}[T]{0.19\textwidth}
\centering
\includegraphics[width=\textwidth]{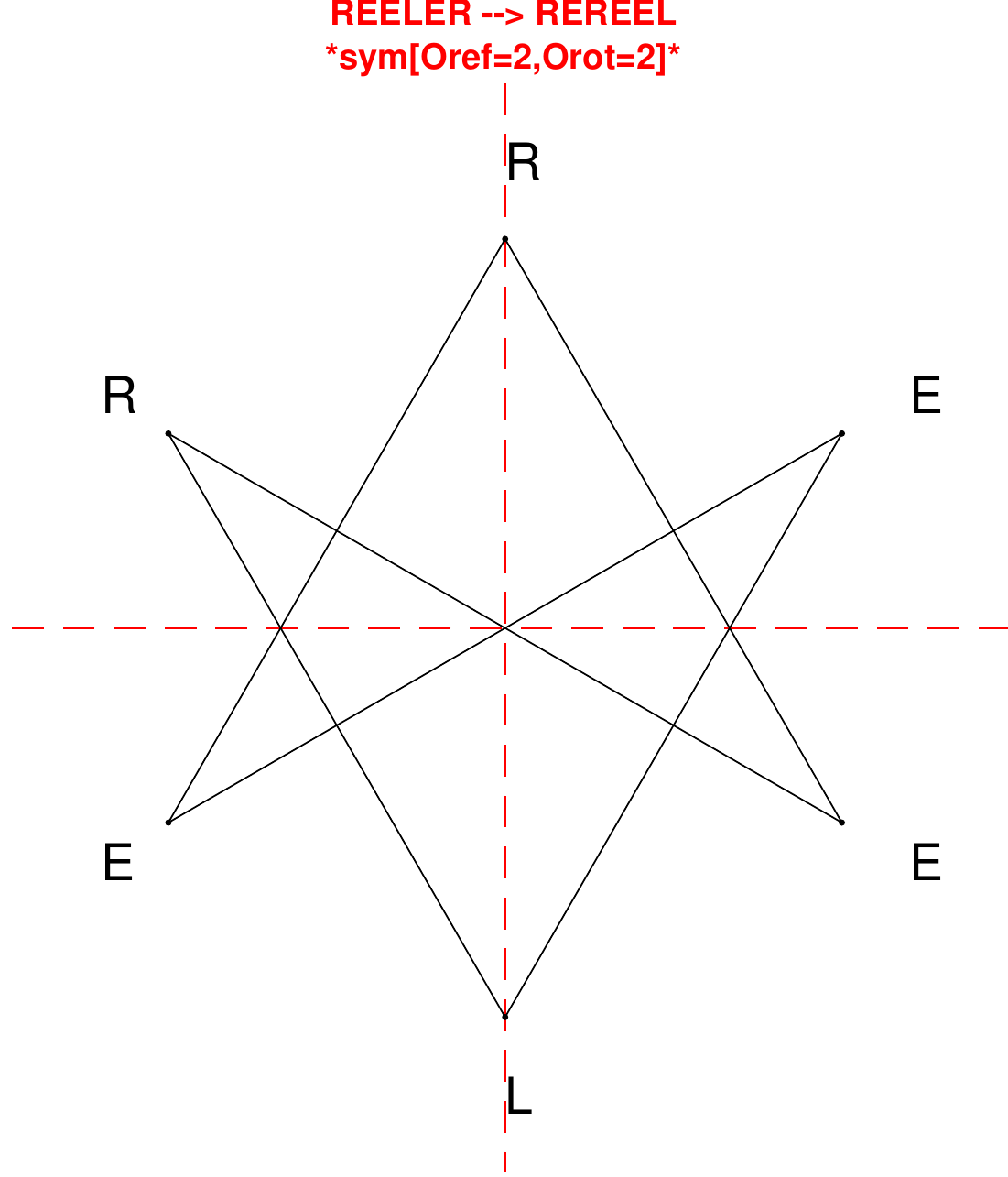}
\end{subfigure}
\hfill
\begin{subfigure}[T]{0.19\textwidth}
\centering
\includegraphics[width=\textwidth]{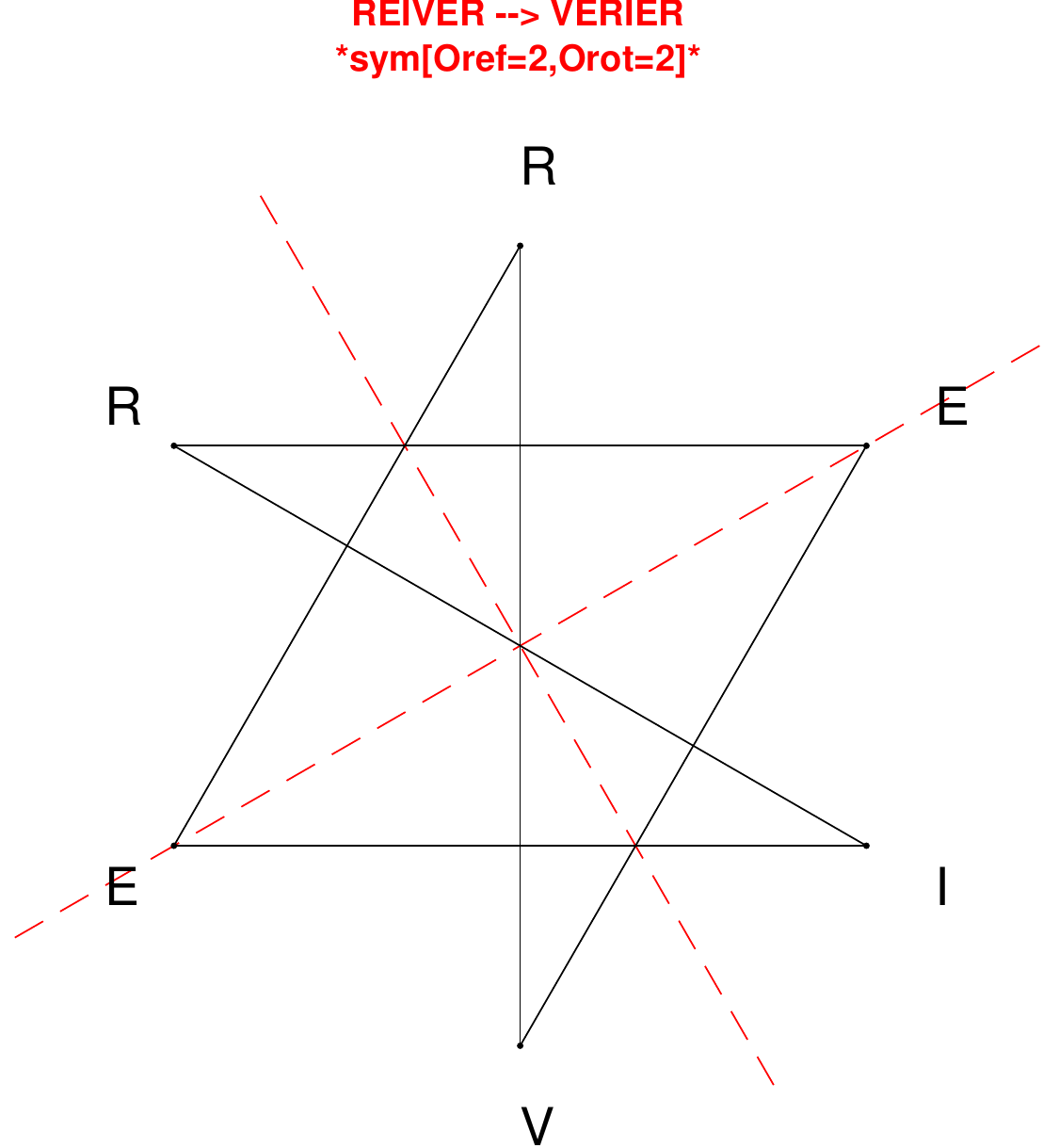}
\end{subfigure}
\hfill
\begin{subfigure}[T]{0.19\textwidth}
\centering
\includegraphics[width=\textwidth]{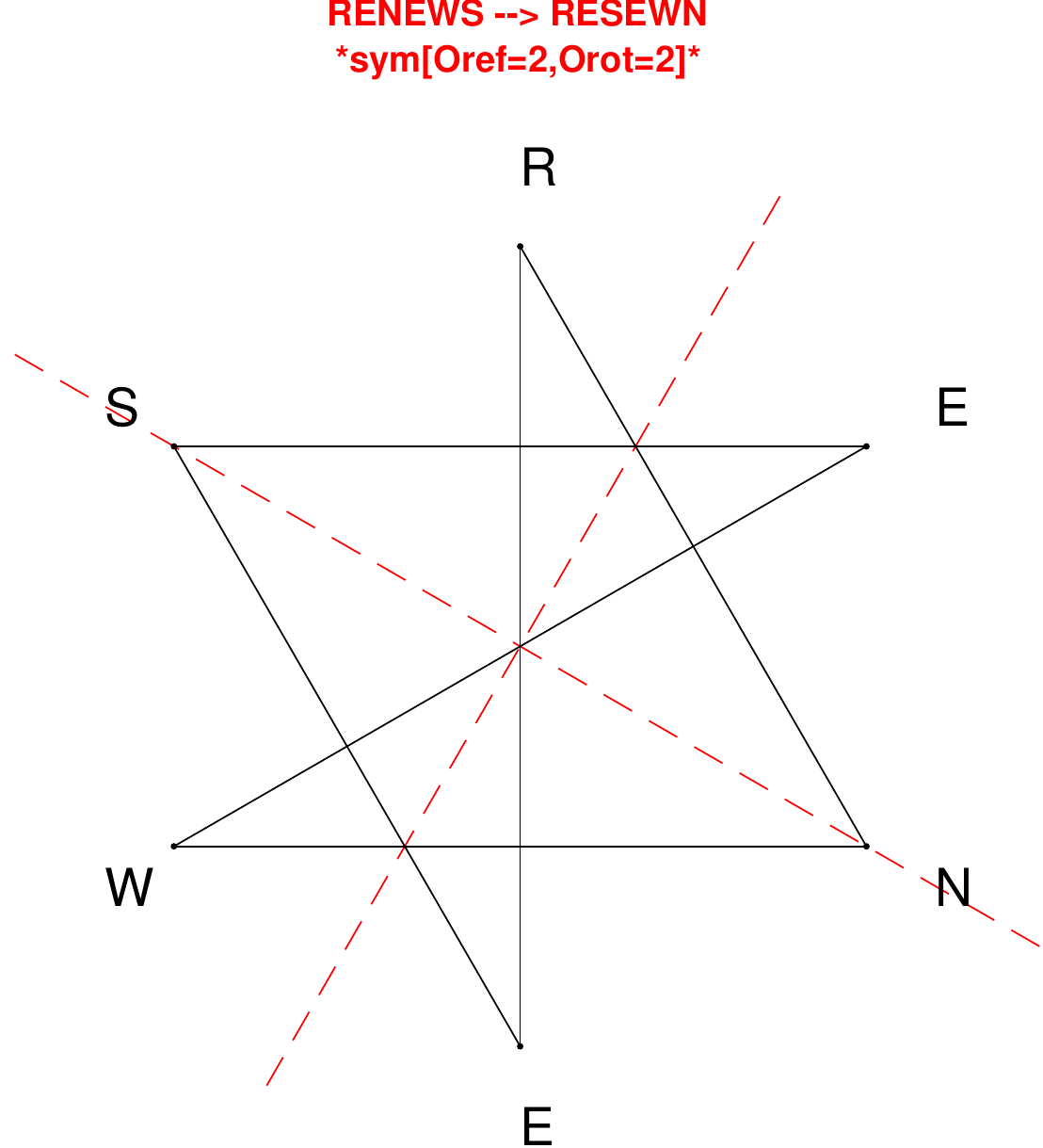}
\end{subfigure}
\end{figure}

\begin{figure}[H]
\centering
\begin{subfigure}[T]{0.19\textwidth}
\centering
\includegraphics[width=\textwidth]{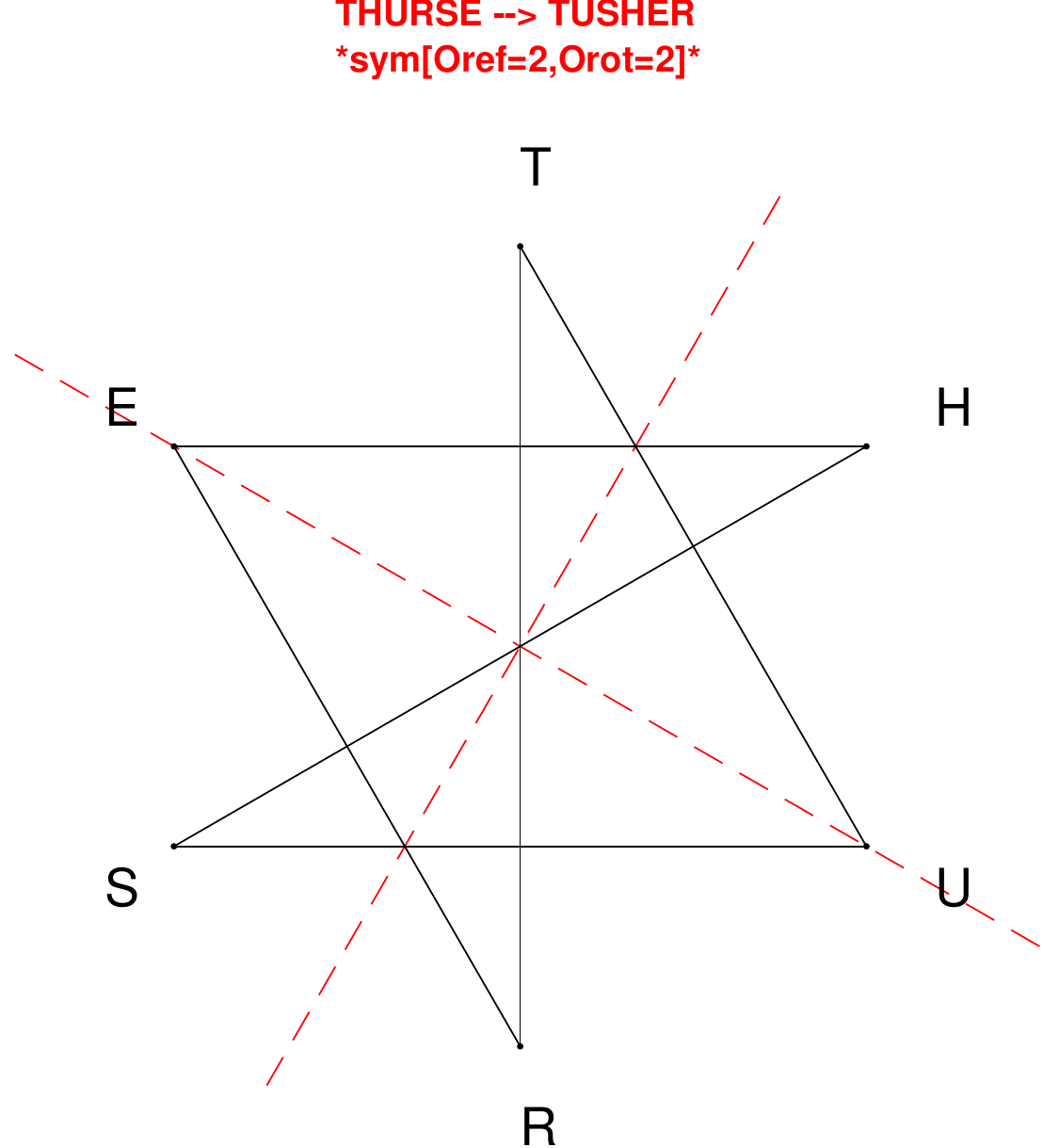}
\end{subfigure}
\hfill
\begin{subfigure}[T]{0.19\textwidth}
\centering
\includegraphics[width=\textwidth]{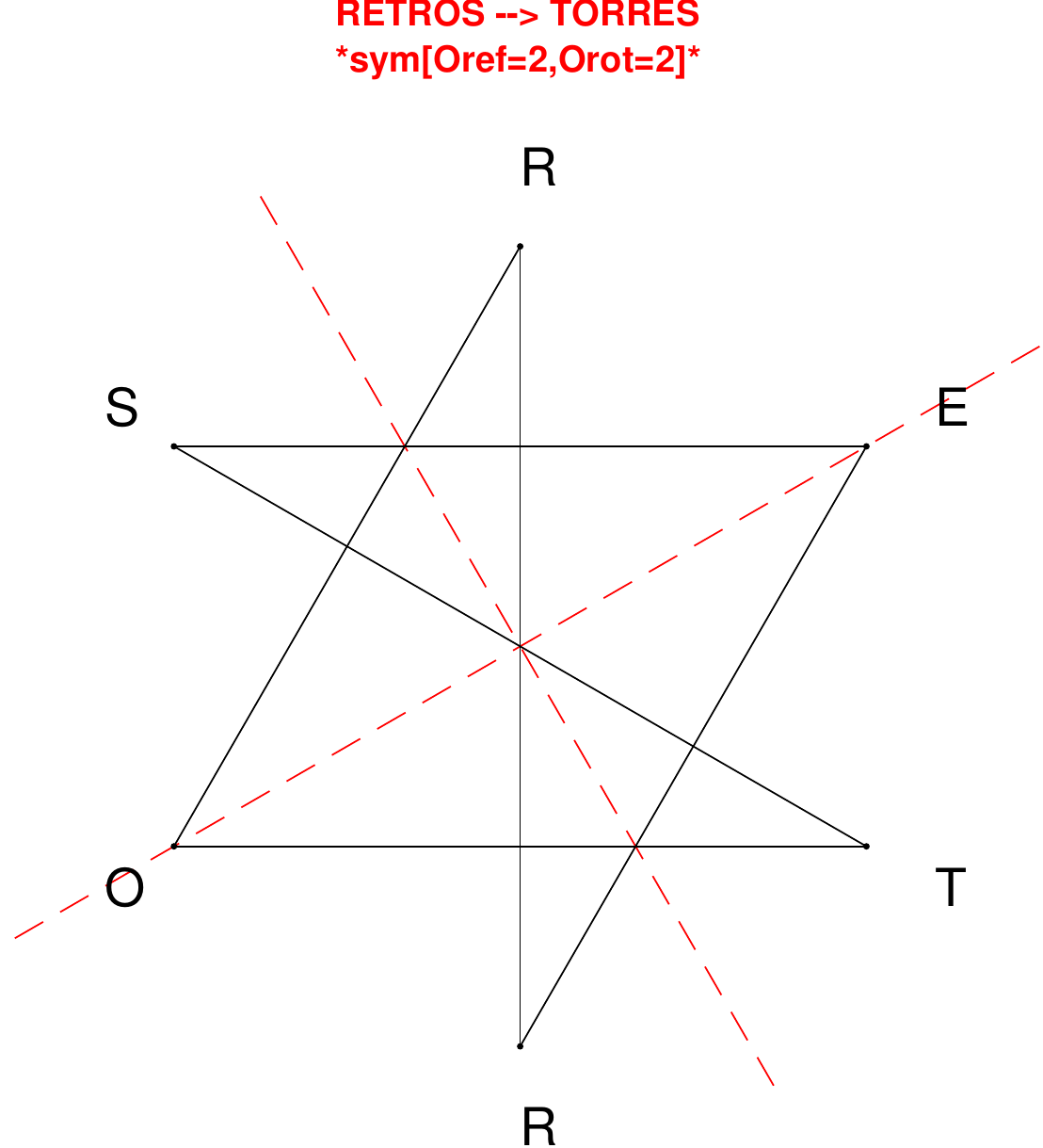}
\end{subfigure}
\hfill
\begin{subfigure}[T]{0.19\textwidth}
\centering
\includegraphics[width=\textwidth]{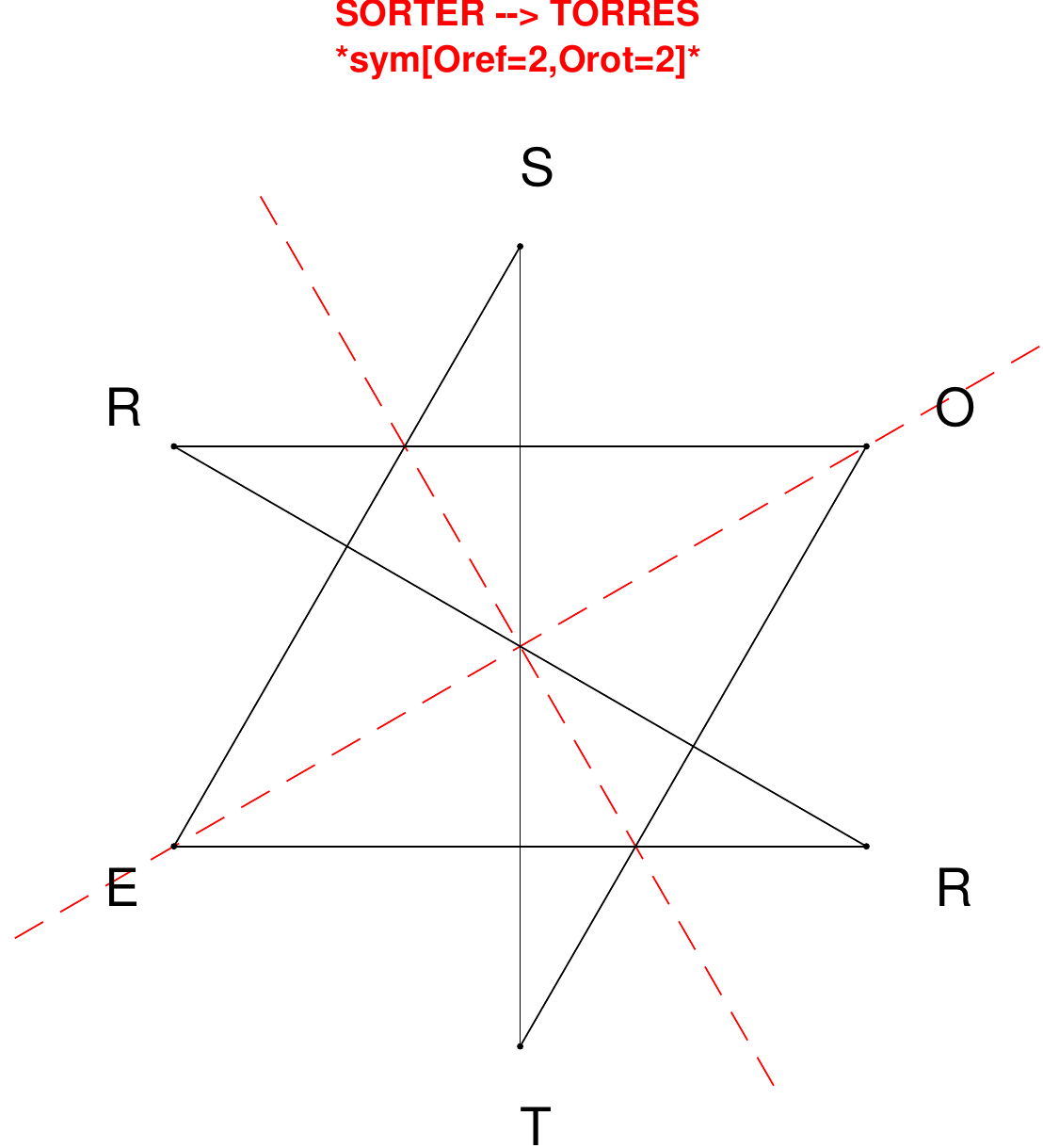}
\end{subfigure}
\hfill
\begin{subfigure}[T]{0.19\textwidth}
\centering
\includegraphics[width=\textwidth]{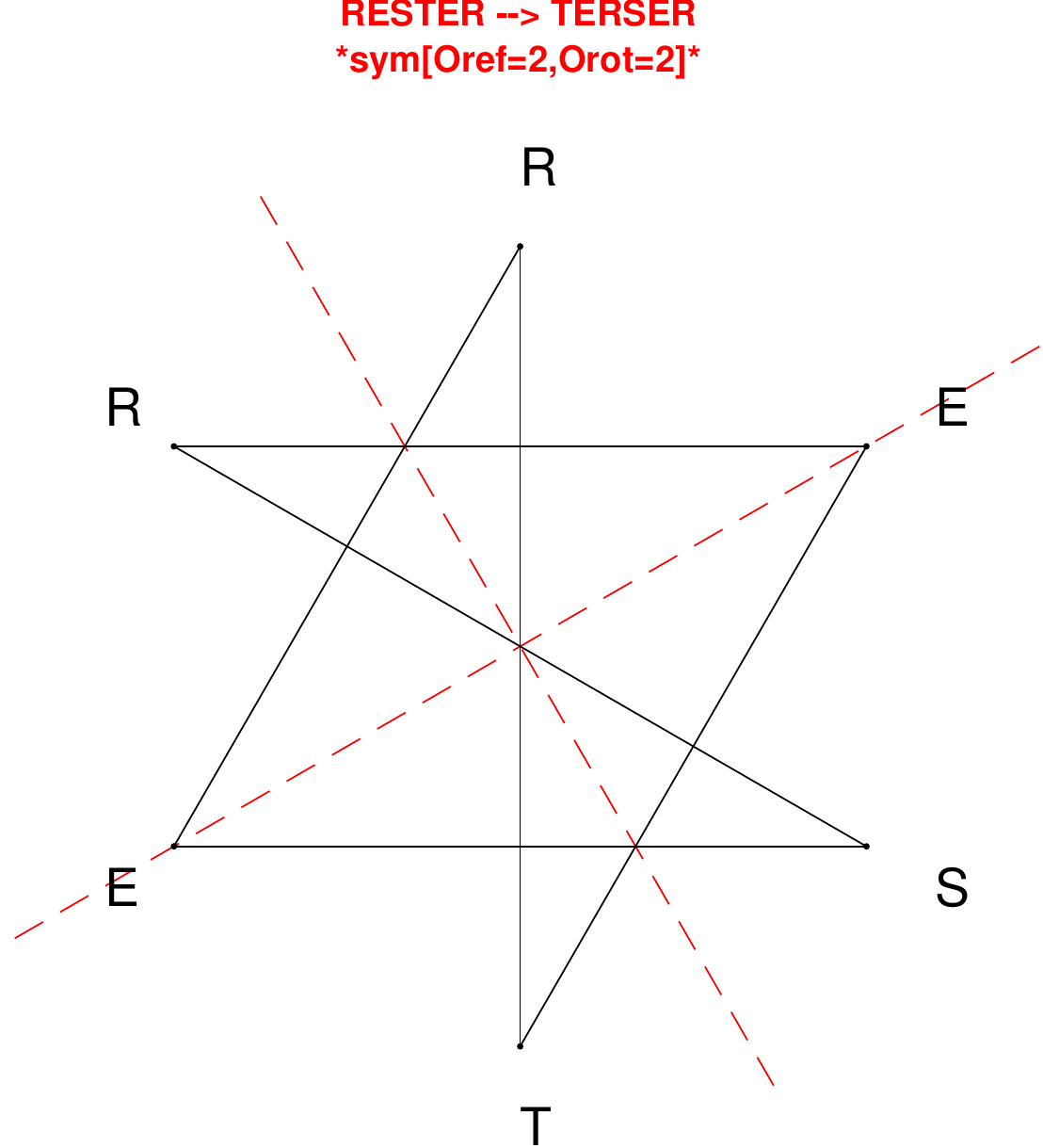}
\end{subfigure}
\hfill
\begin{subfigure}[T]{0.19\textwidth}
\centering
\includegraphics[width=\textwidth]{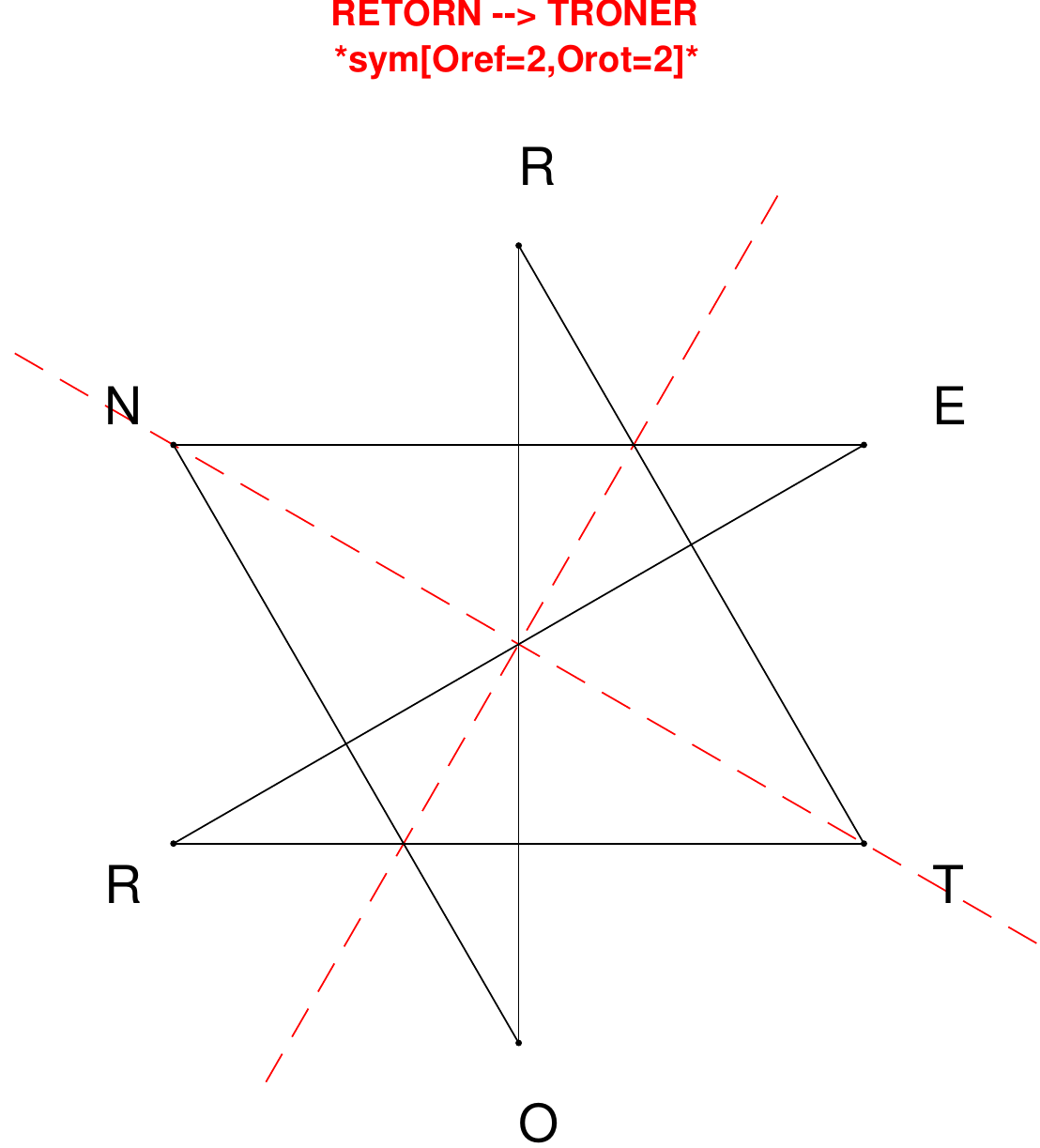}
\end{subfigure}
\end{figure}

\begin{figure}[H]
\centering
\begin{subfigure}[T]{0.19\textwidth}
\centering
\includegraphics[width=\textwidth]{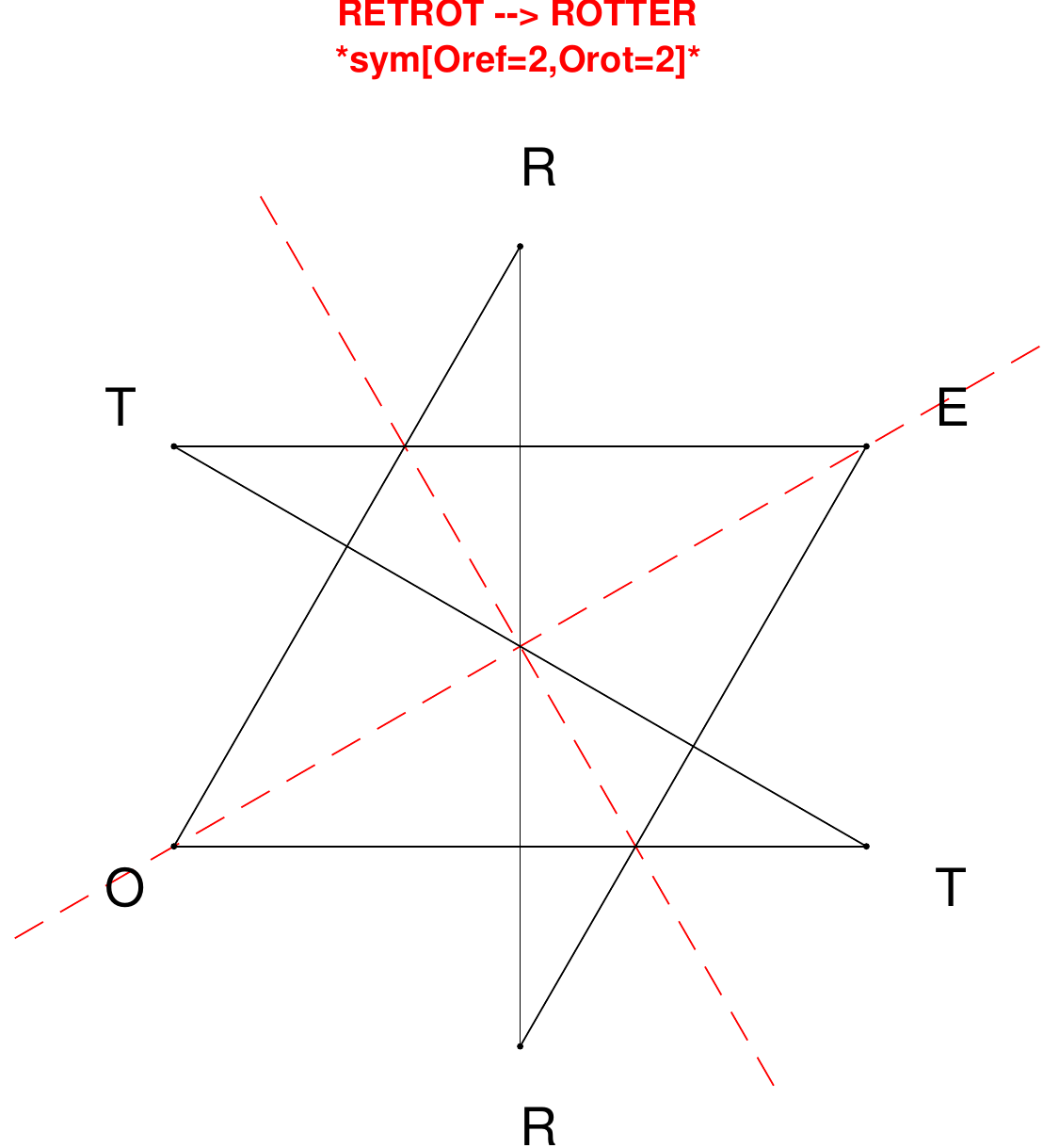}
\end{subfigure}
\hfill
\begin{subfigure}[T]{0.19\textwidth}
\centering
\includegraphics[width=\textwidth]{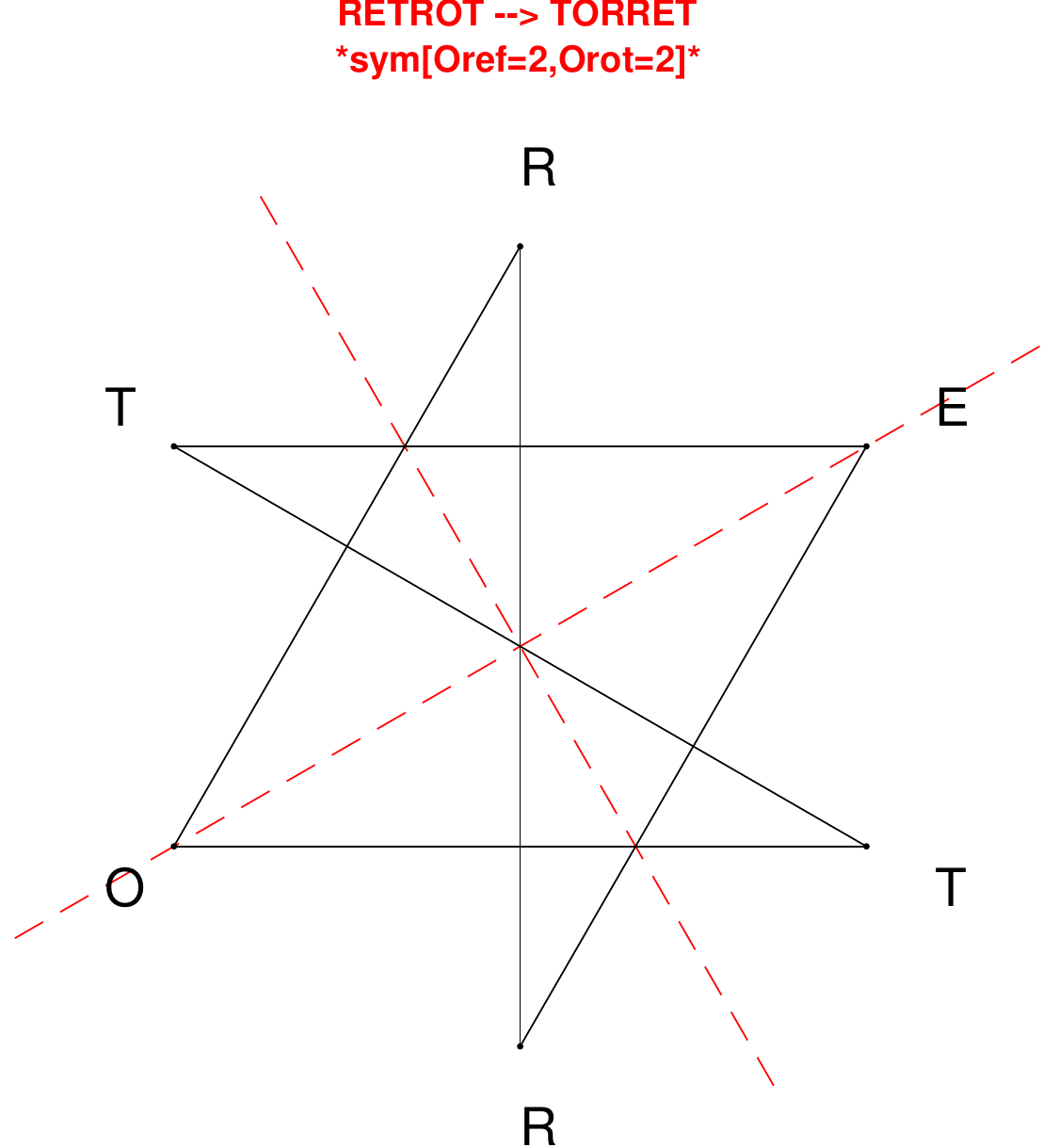}
\end{subfigure}
\hfill
\begin{subfigure}[T]{0.19\textwidth}
\centering
\includegraphics[width=\textwidth]{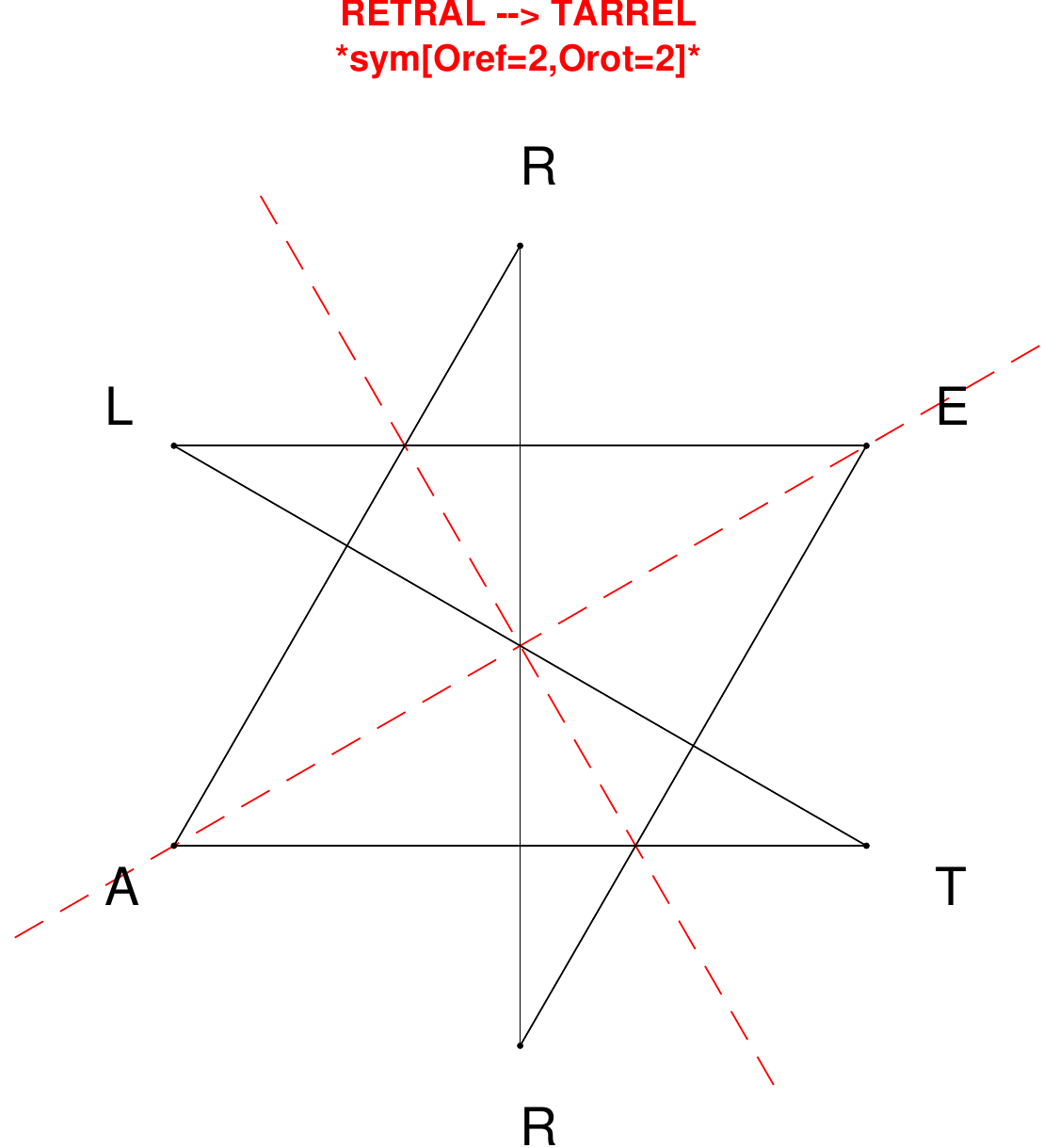}
\end{subfigure}
\hfill
\begin{subfigure}[T]{0.19\textwidth}
\centering
\includegraphics[width=\textwidth]{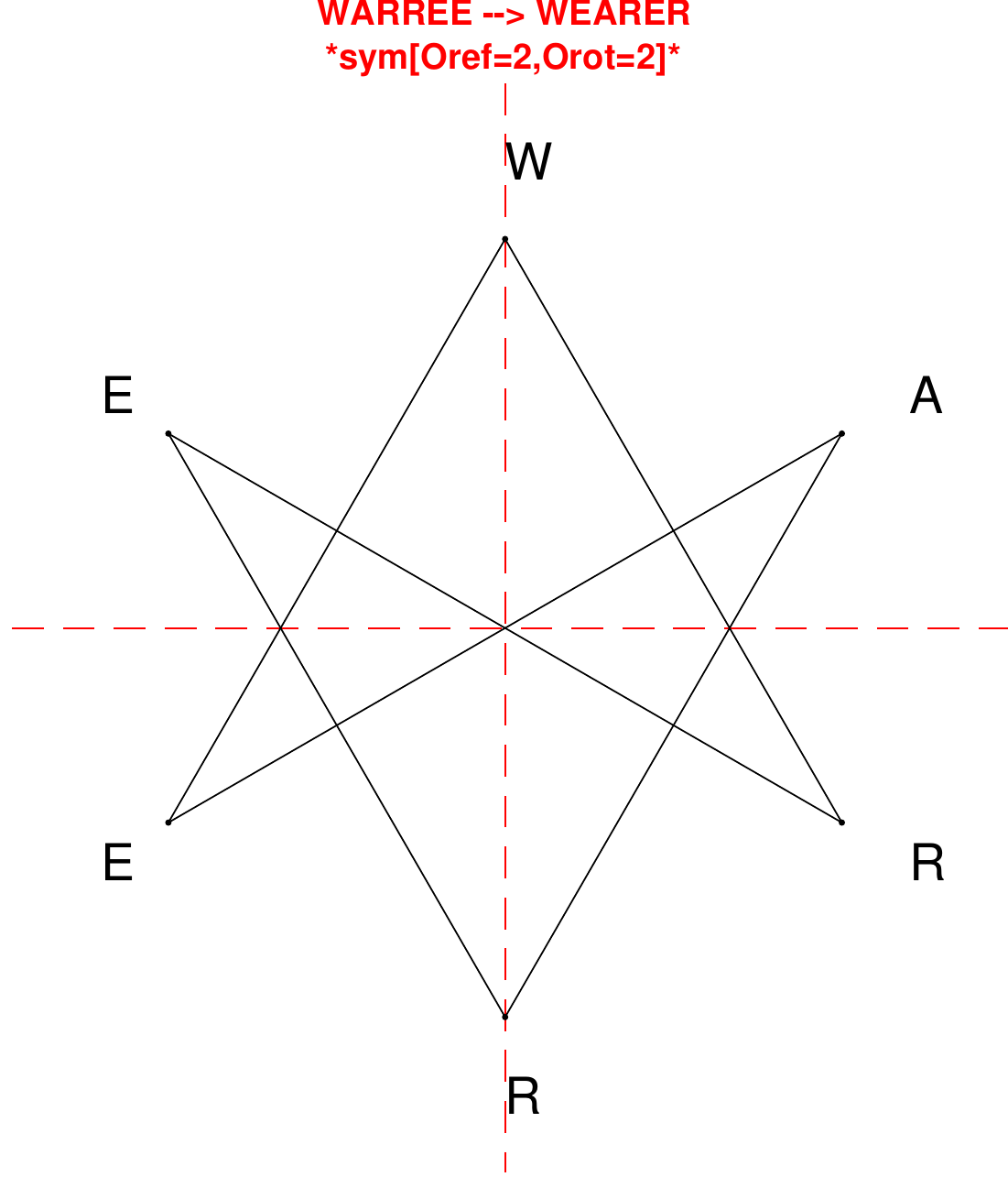}
\end{subfigure}
\hfill
\begin{subfigure}[T]{0.19\textwidth}
\centering
\includegraphics[width=\textwidth]{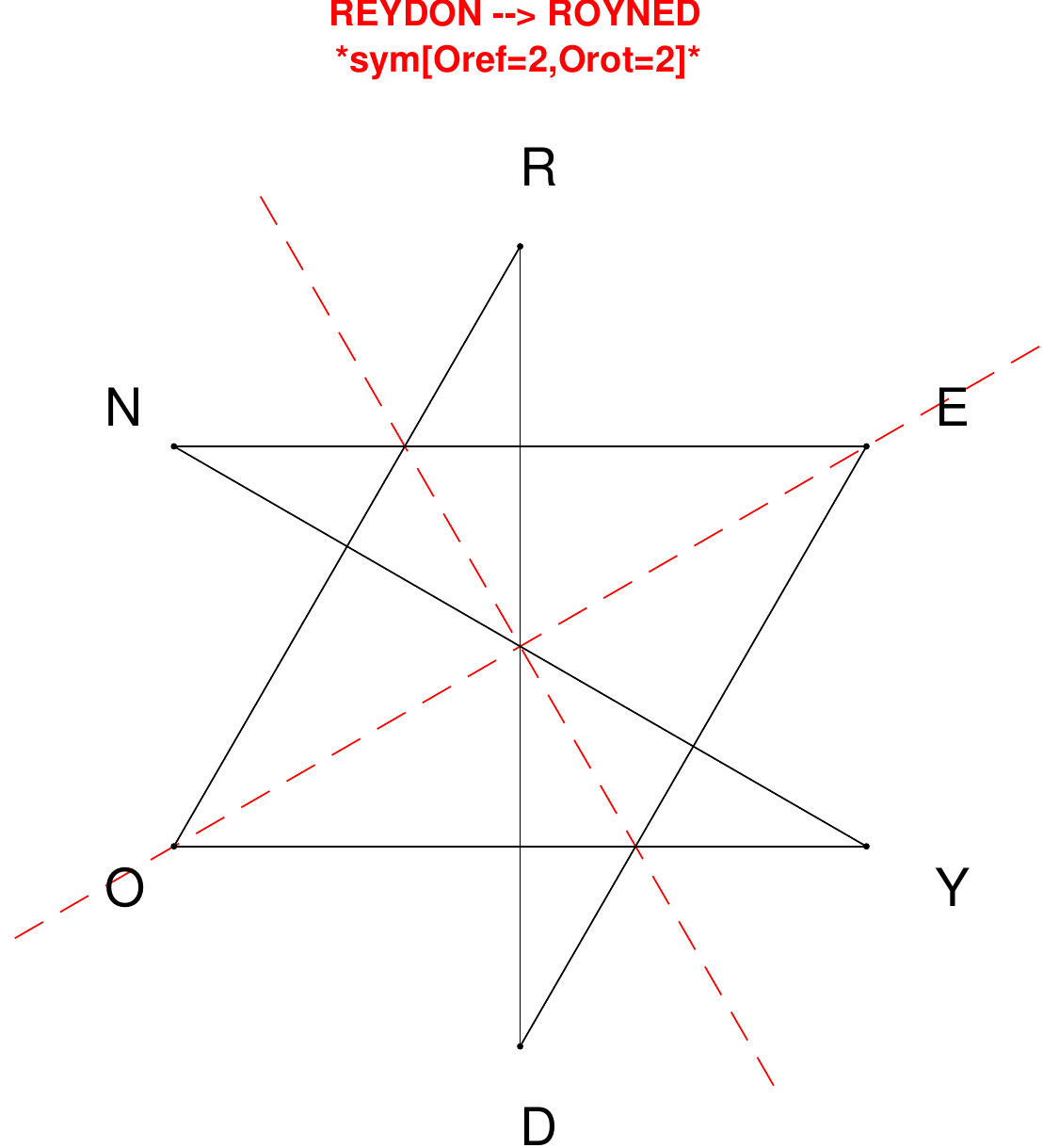}
\end{subfigure}
\end{figure}

\begin{figure}[H]
\centering
\begin{subfigure}[T]{0.19\textwidth}
\centering
\includegraphics[width=\textwidth]{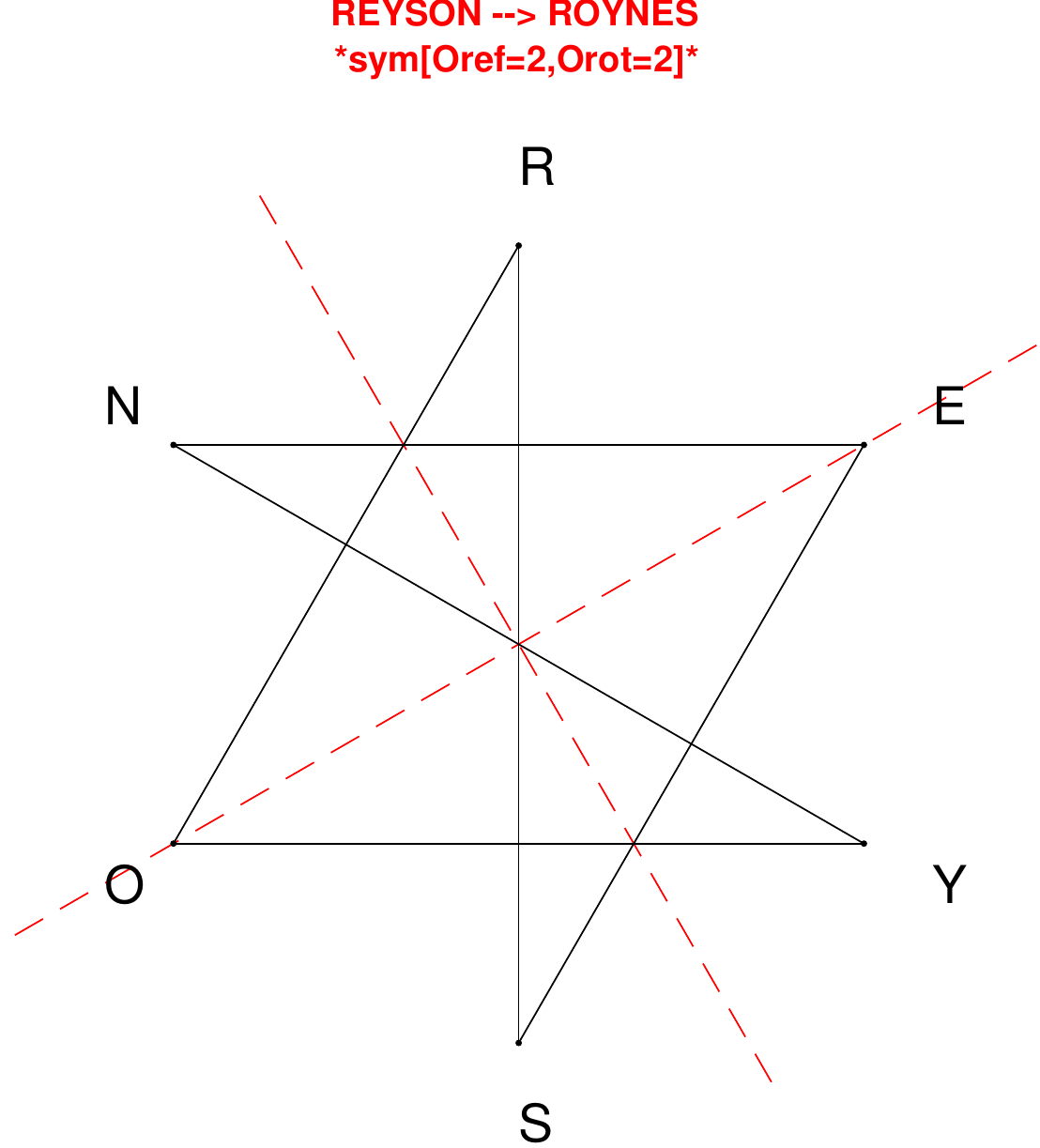}
\end{subfigure}
\hfill
\begin{subfigure}[T]{0.19\textwidth}
\centering
\includegraphics[width=\textwidth]{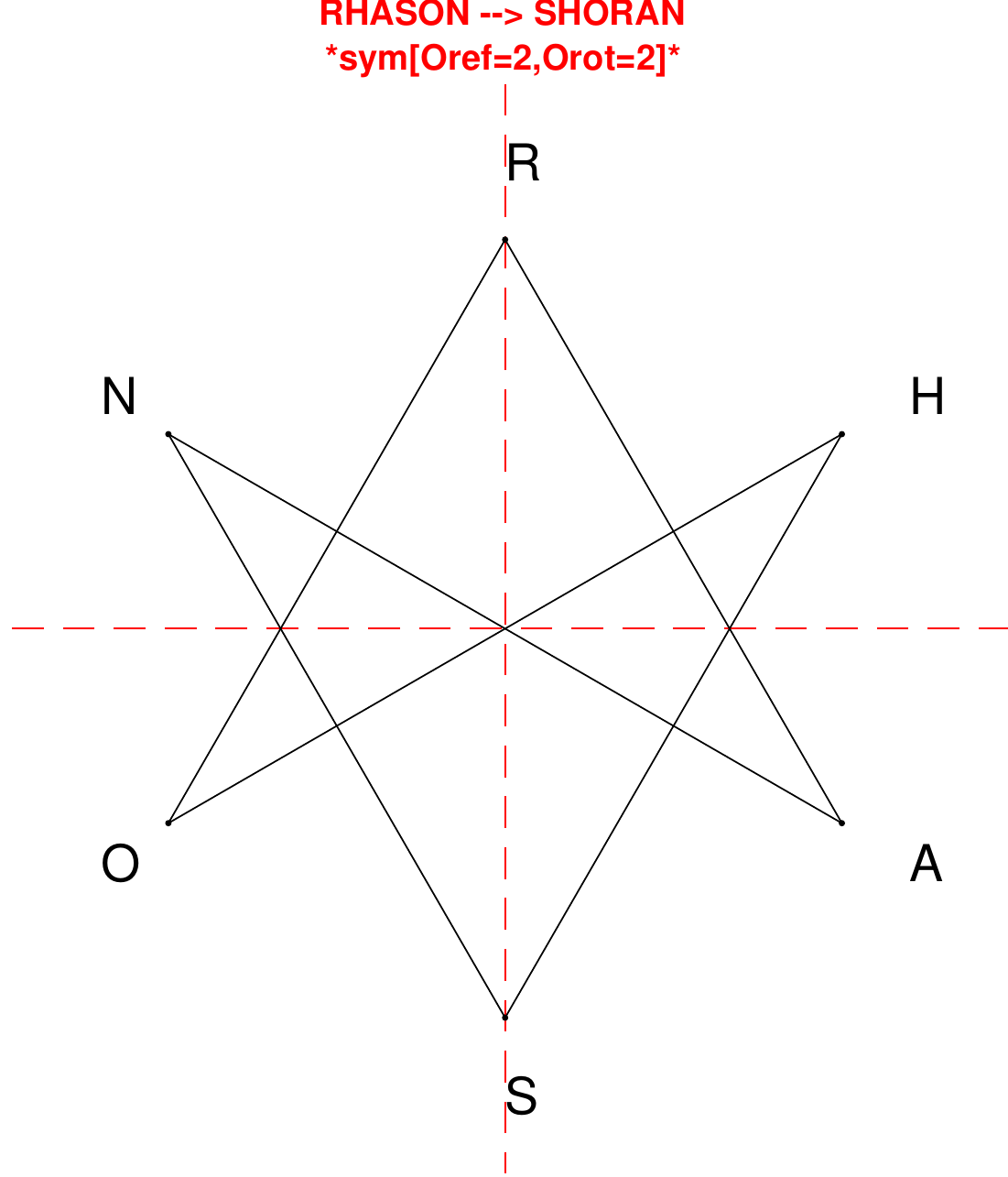}
\end{subfigure}
\hfill
\begin{subfigure}[T]{0.19\textwidth}
\centering
\includegraphics[width=\textwidth]{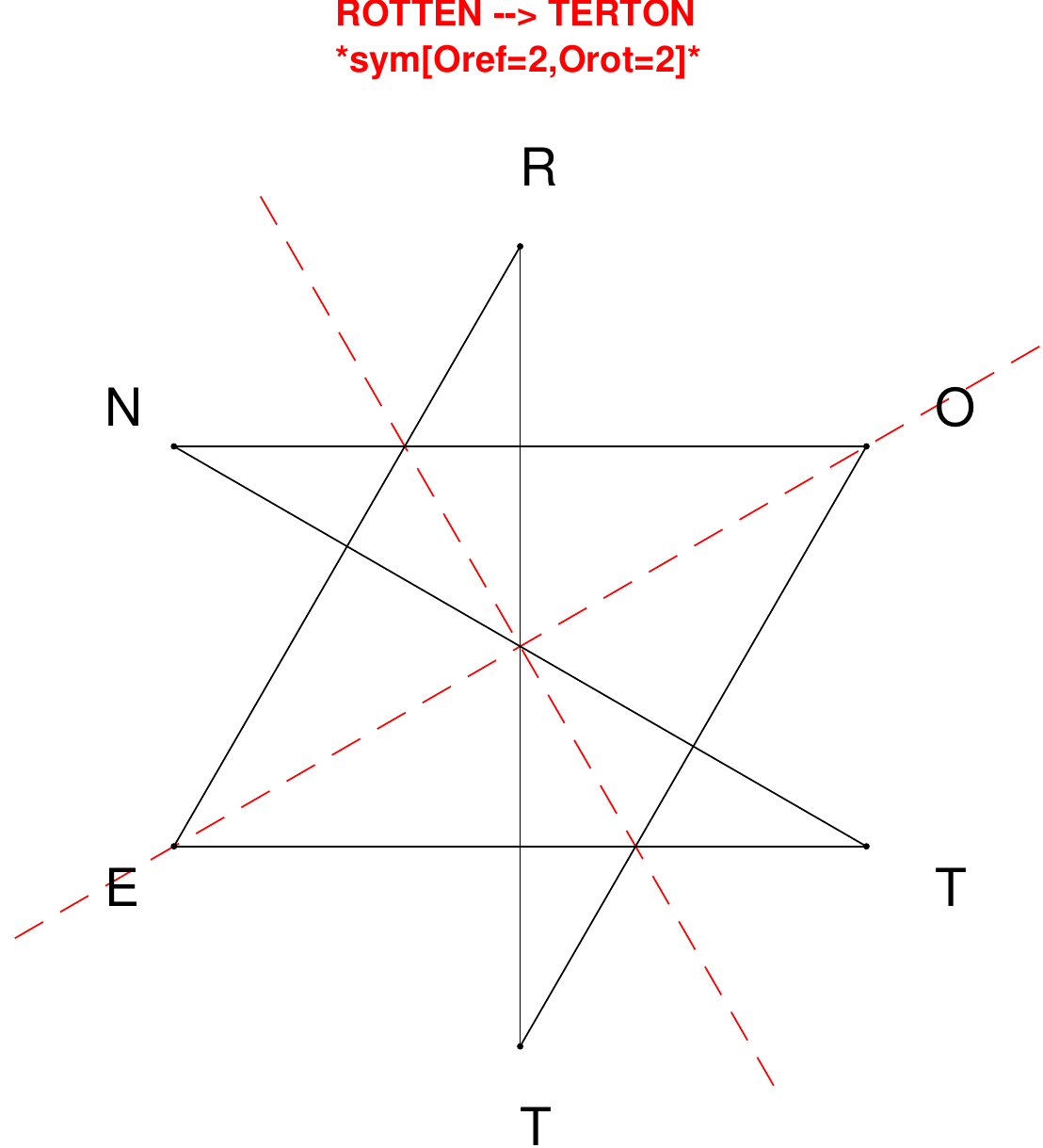}
\end{subfigure}
\hfill
\begin{subfigure}[T]{0.19\textwidth}
\centering
\includegraphics[width=\textwidth]{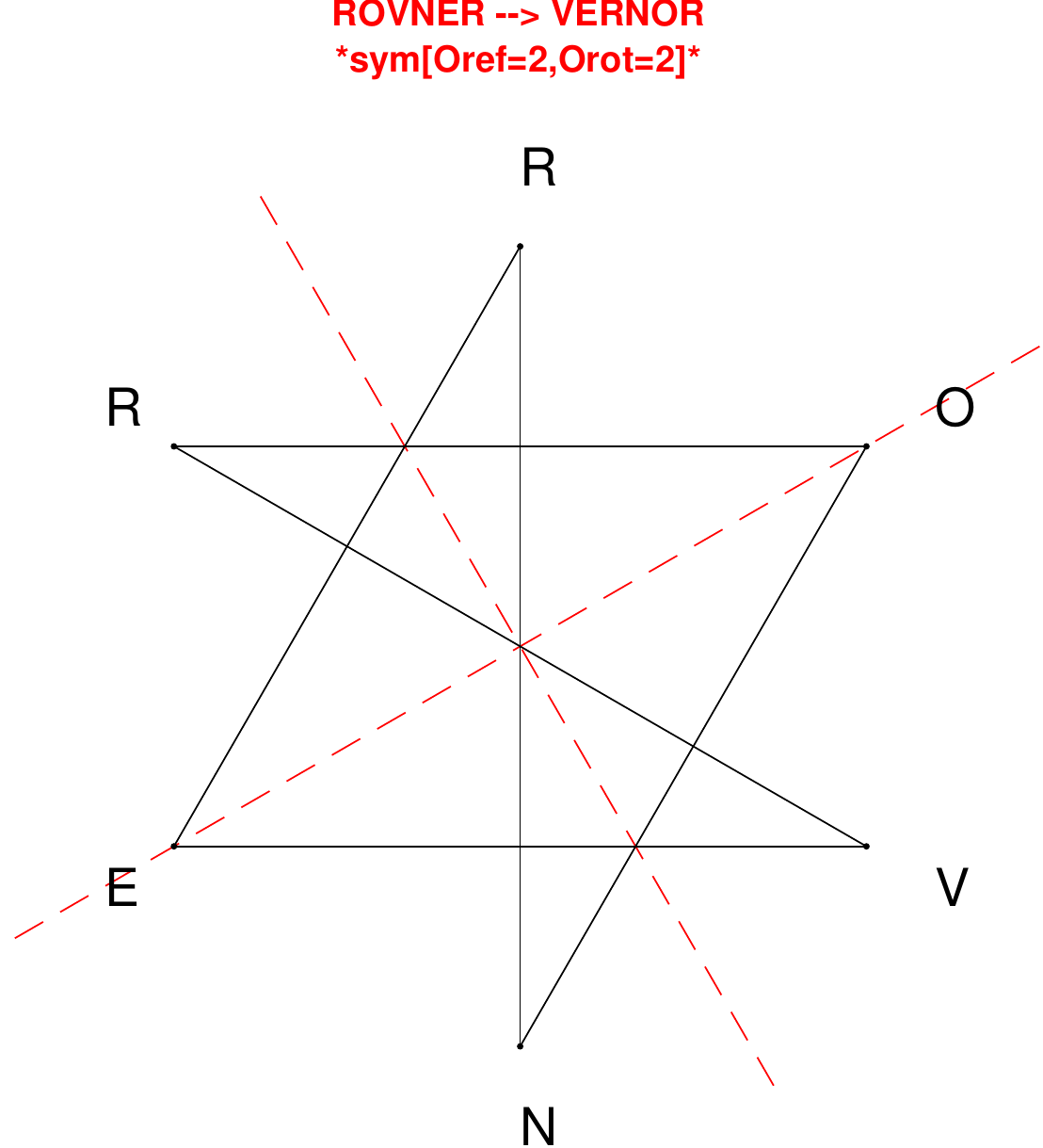}
\end{subfigure}
\hfill
\begin{subfigure}[T]{0.19\textwidth}
\centering
\includegraphics[width=\textwidth]{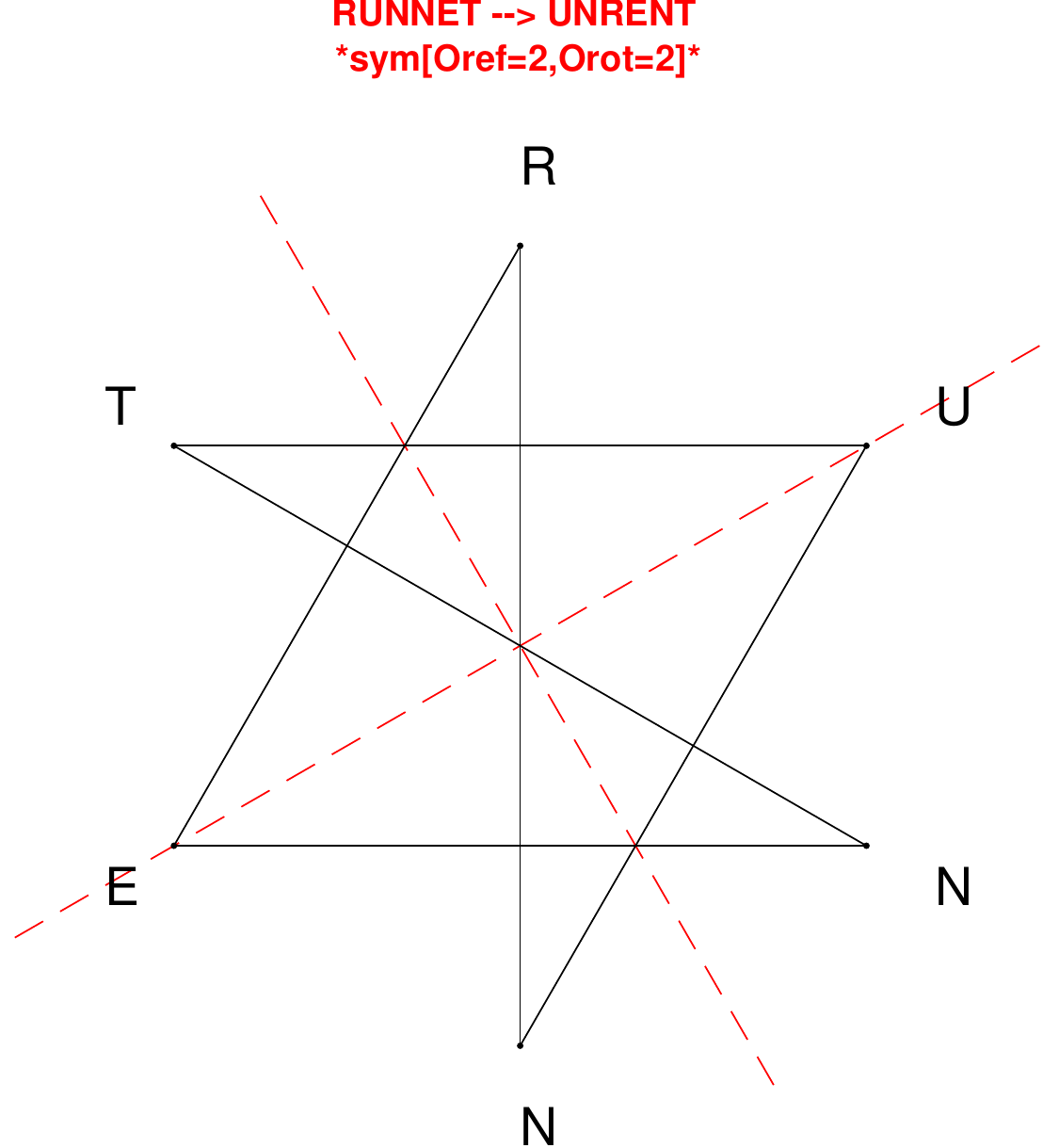}
\end{subfigure}
\end{figure}

\begin{figure}[H]
\centering
\begin{subfigure}[T]{0.19\textwidth}
\centering
\includegraphics[width=\textwidth]{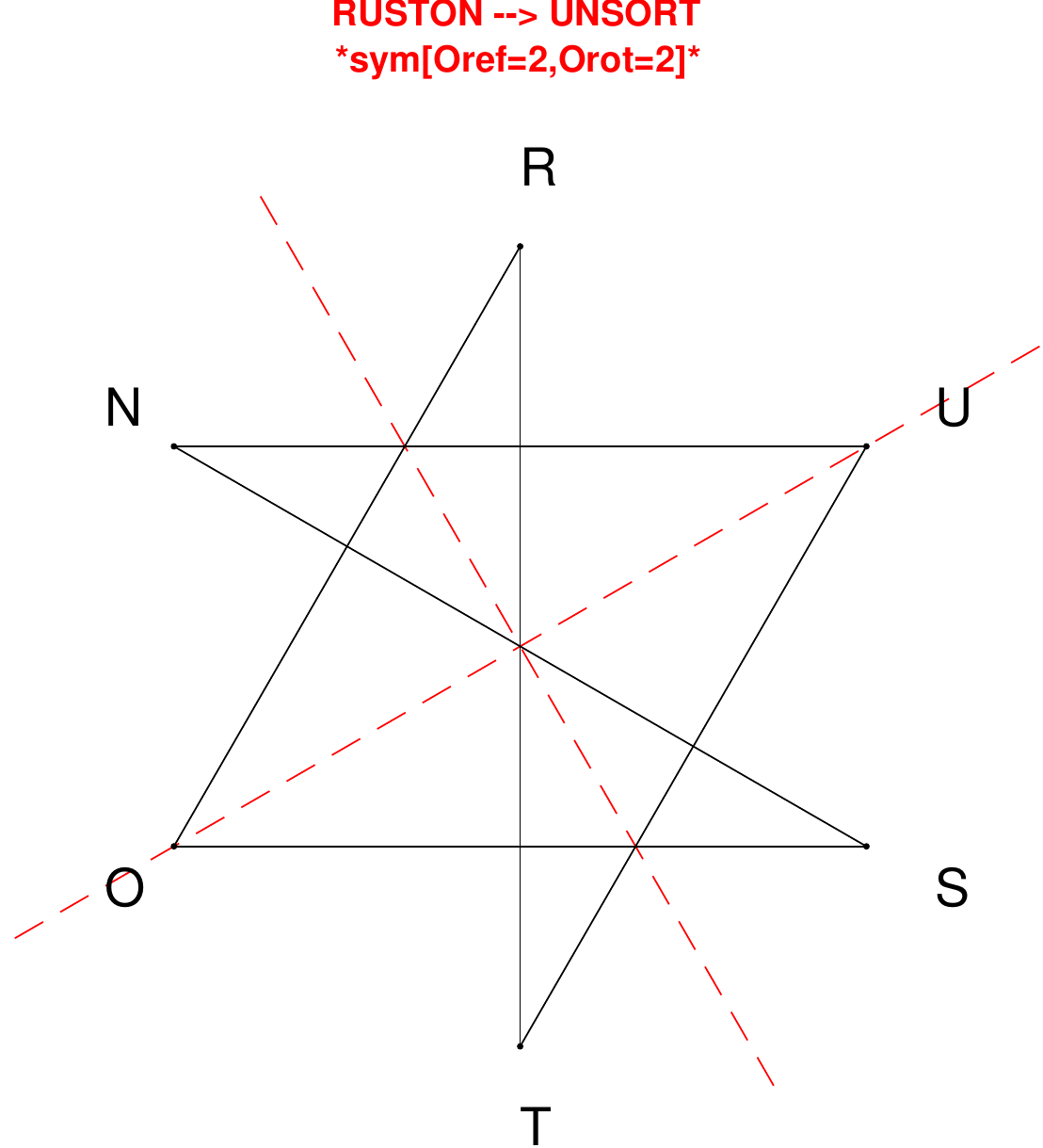}
\end{subfigure}
\hfill
\begin{subfigure}[T]{0.19\textwidth}
\centering
\includegraphics[width=\textwidth]{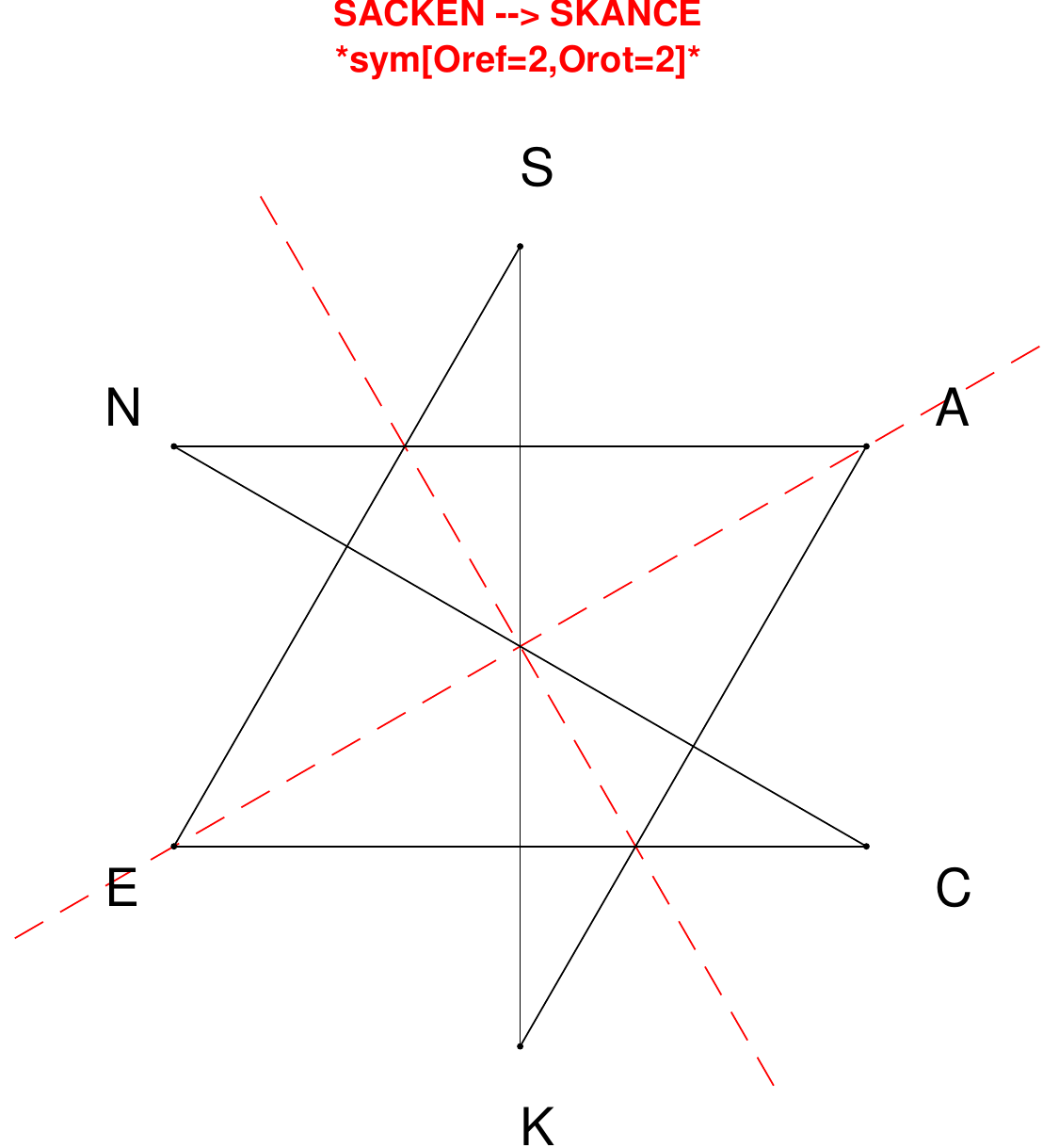}
\end{subfigure}
\hfill
\begin{subfigure}[T]{0.19\textwidth}
\centering
\includegraphics[width=\textwidth]{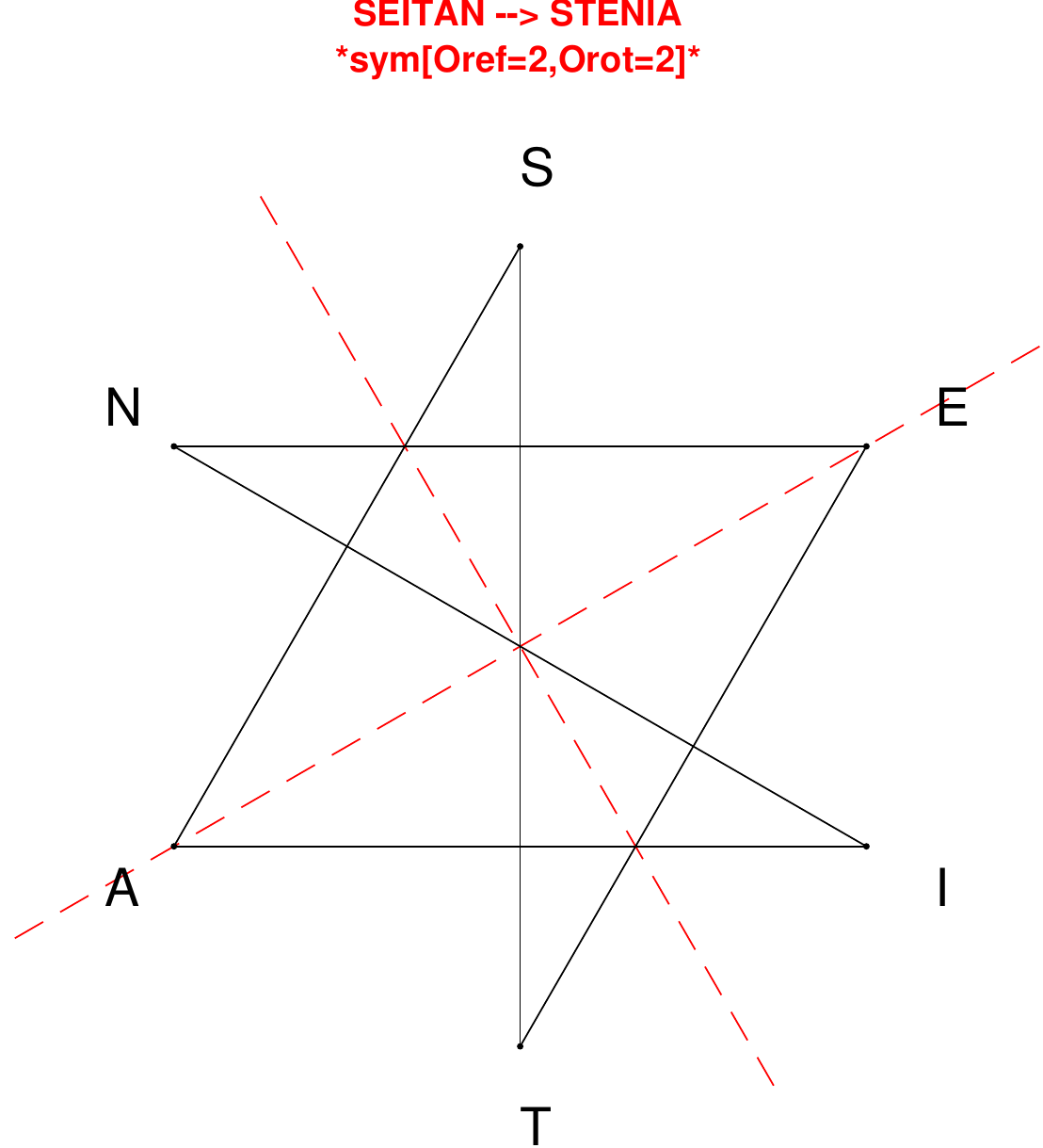}
\end{subfigure}
\hfill
\begin{subfigure}[T]{0.19\textwidth}
\centering
\includegraphics[width=\textwidth]{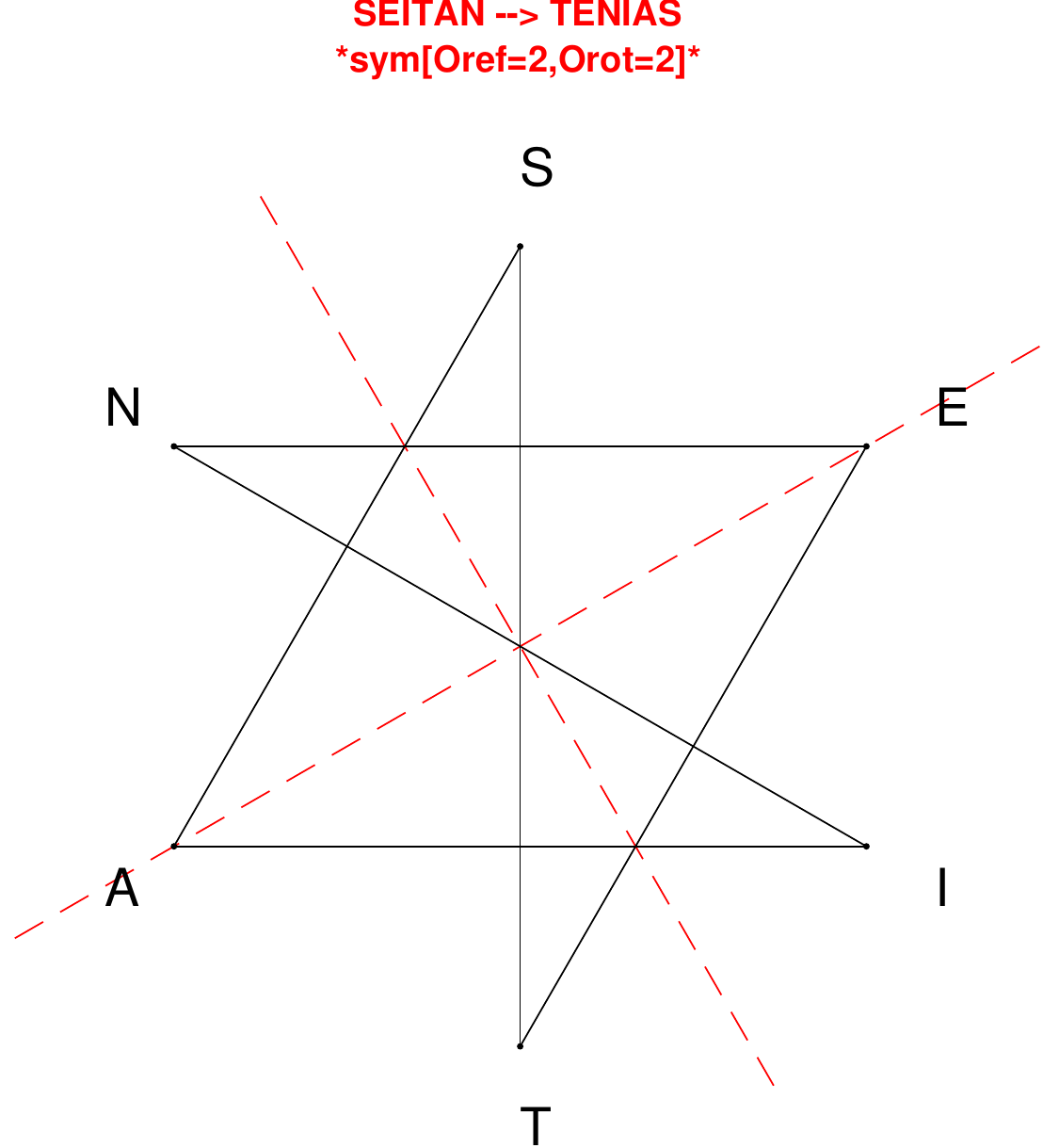}
\end{subfigure}
\hfill
\begin{subfigure}[T]{0.19\textwidth}
\centering
\includegraphics[width=\textwidth]{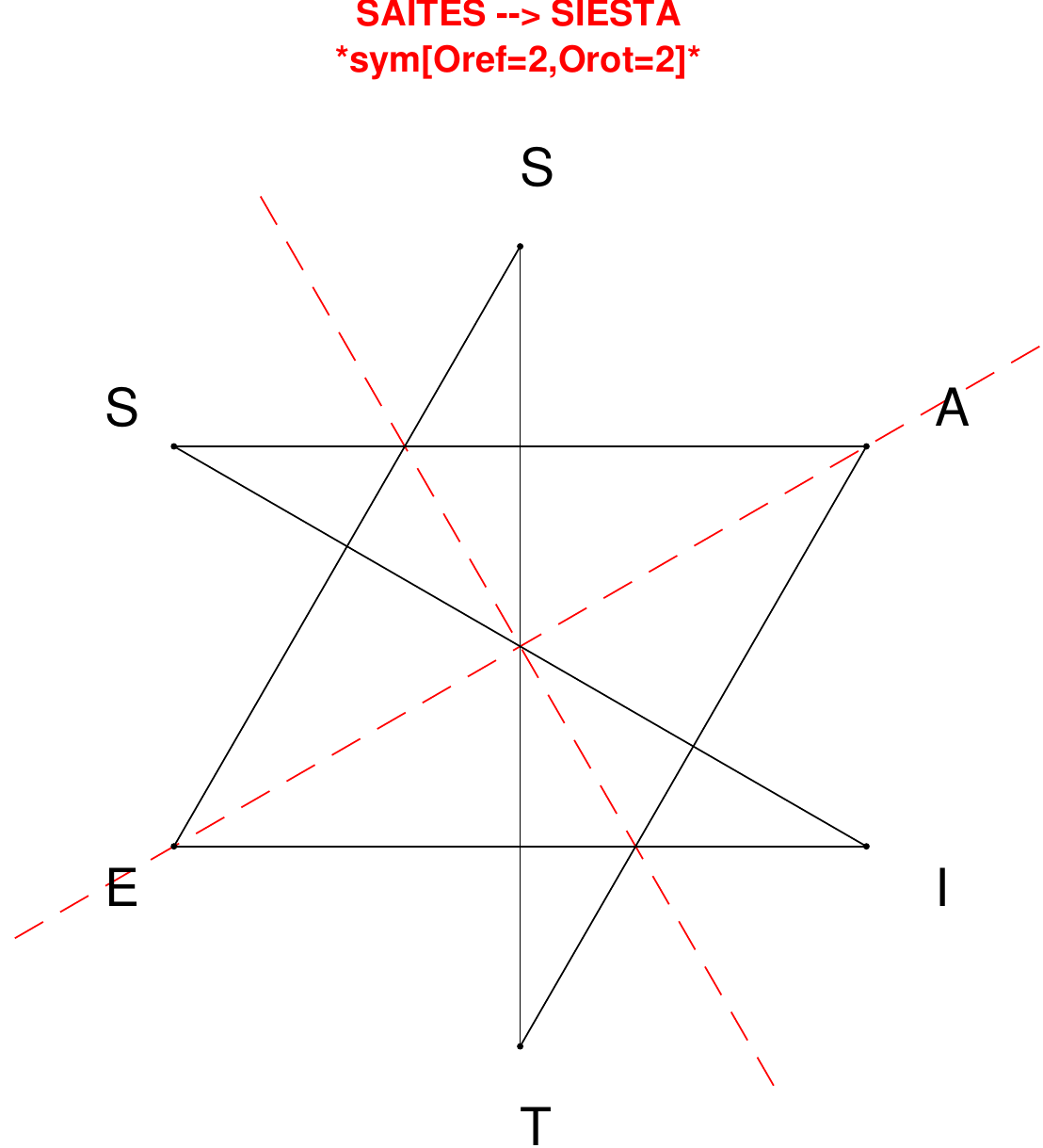}
\end{subfigure}
\end{figure}

\begin{figure}[H]
\centering
\begin{subfigure}[T]{0.19\textwidth}
\centering
\includegraphics[width=\textwidth]{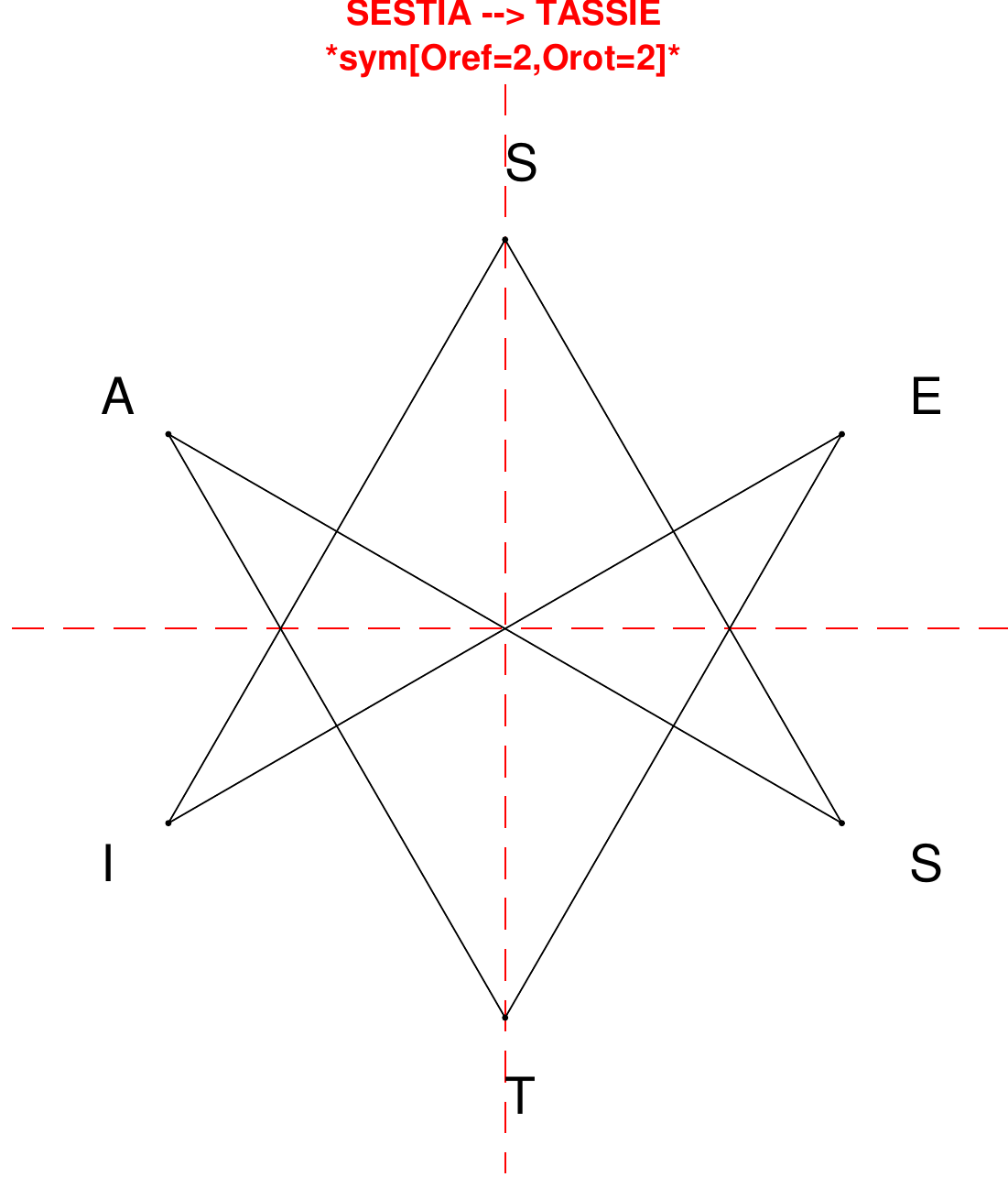}
\end{subfigure}
\hfill
\begin{subfigure}[T]{0.19\textwidth}
\centering
\includegraphics[width=\textwidth]{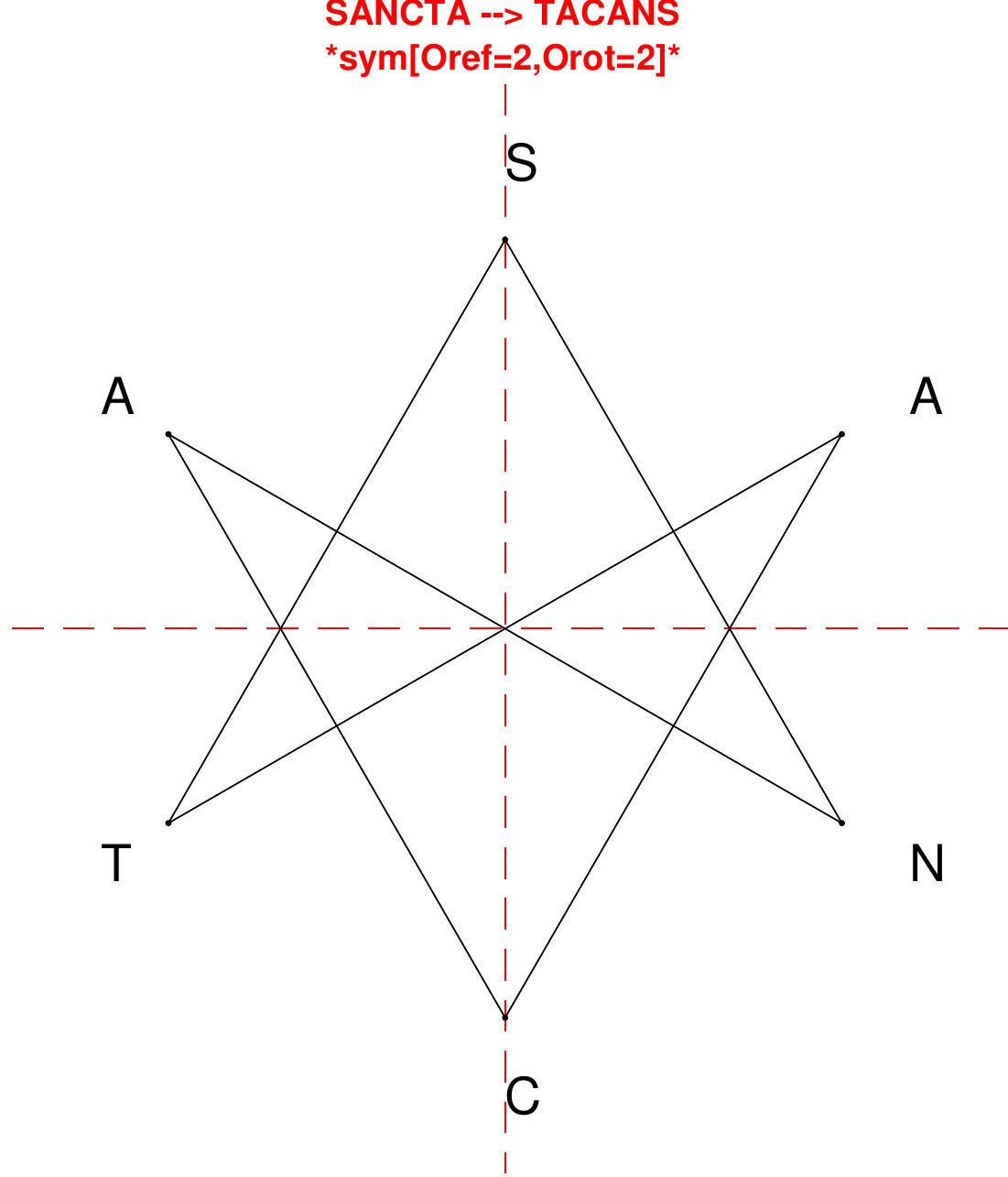}
\end{subfigure}
\hfill
\begin{subfigure}[T]{0.19\textwidth}
\centering
\includegraphics[width=\textwidth]{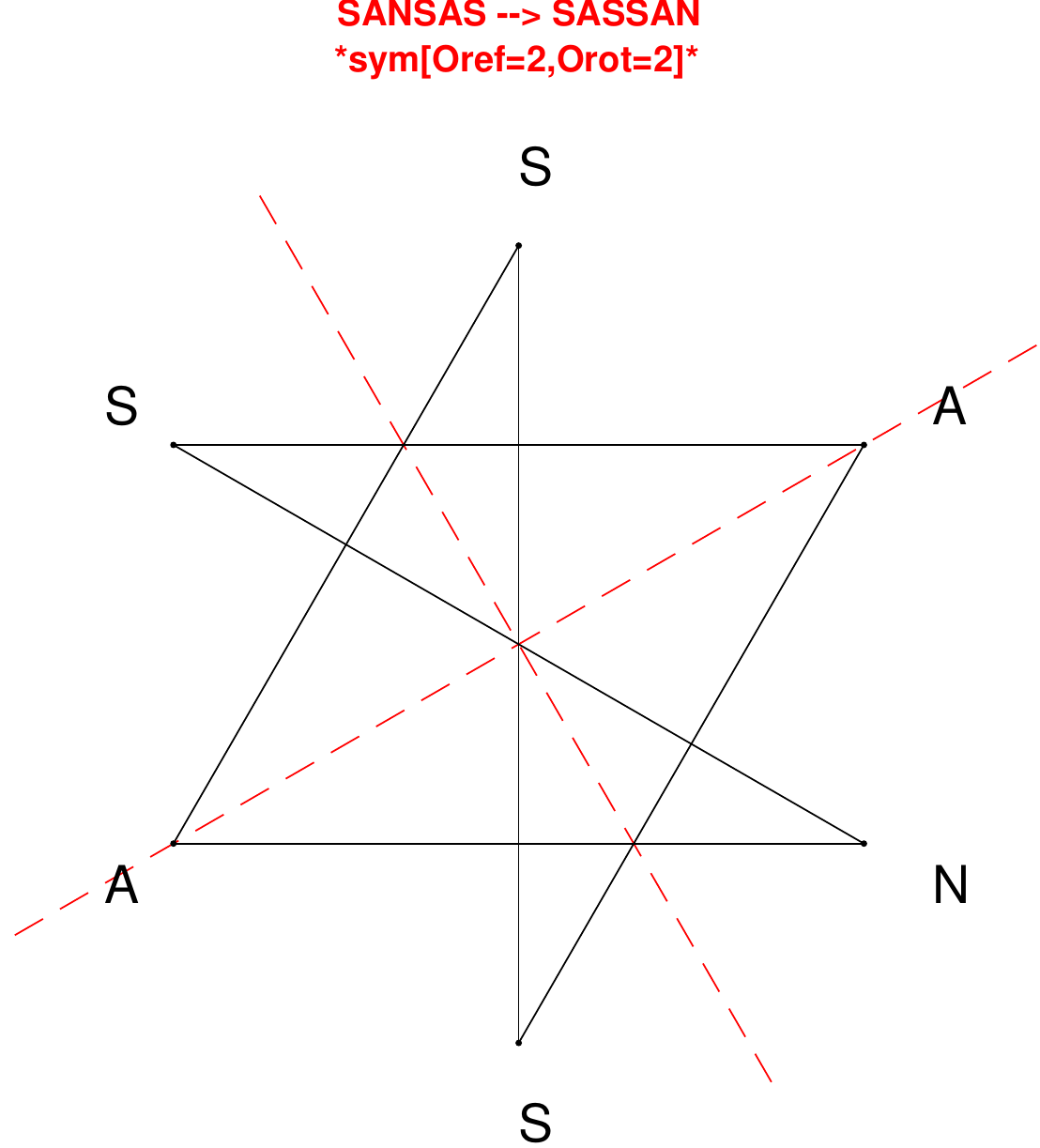}
\end{subfigure}
\hfill
\begin{subfigure}[T]{0.19\textwidth}
\centering
\includegraphics[width=\textwidth]{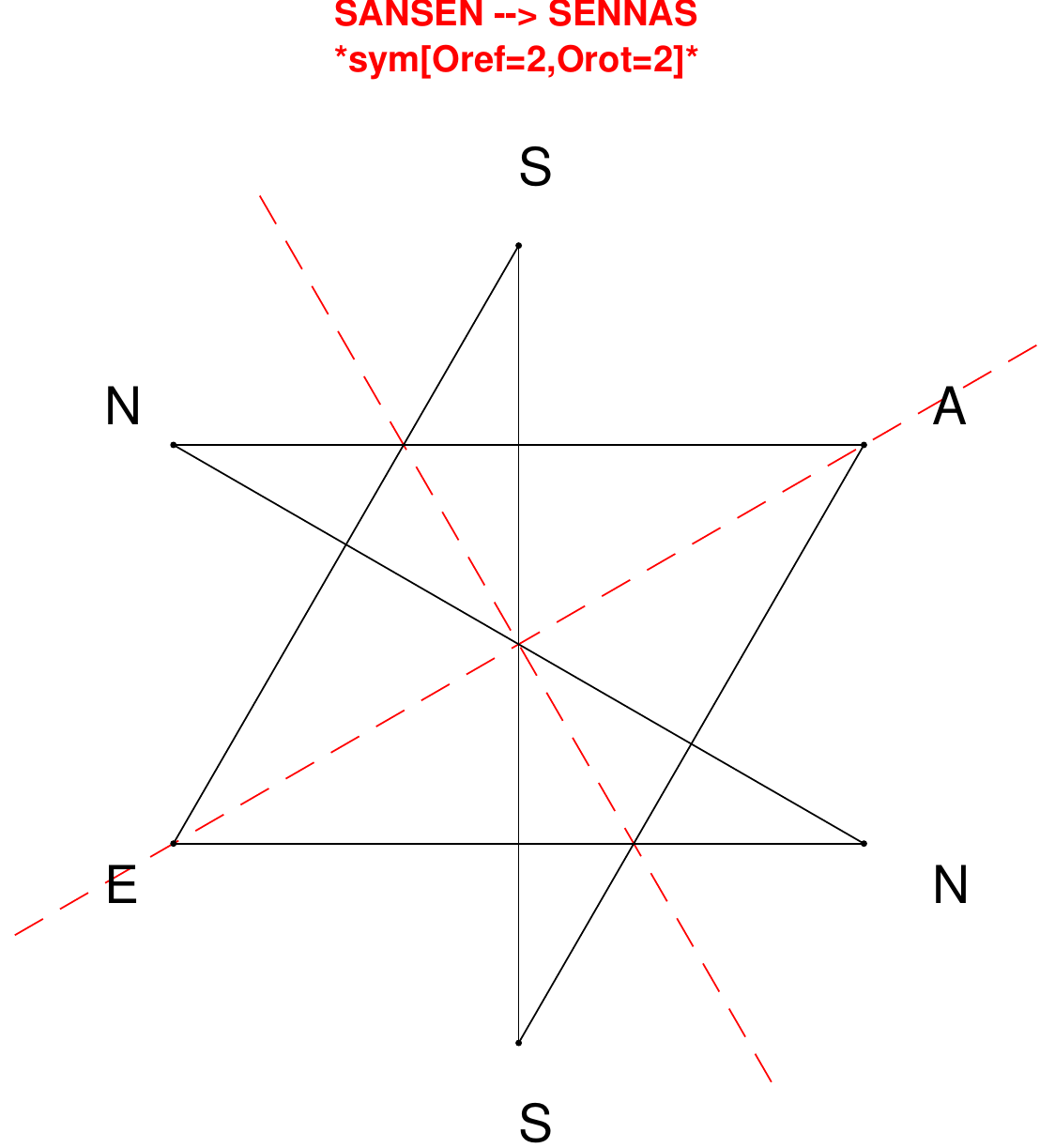}
\end{subfigure}
\hfill
\begin{subfigure}[T]{0.19\textwidth}
\centering
\includegraphics[width=\textwidth]{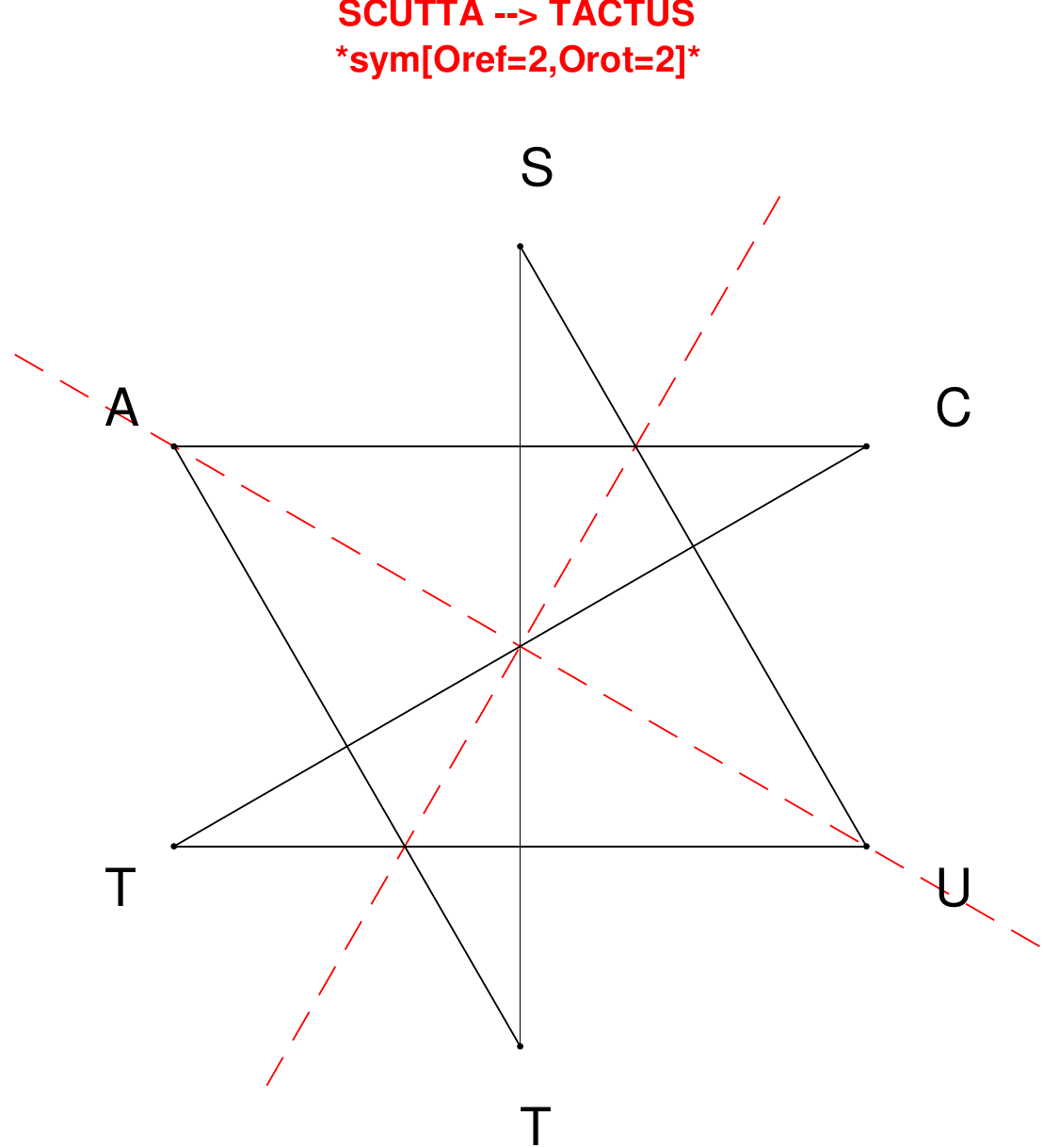}
\end{subfigure}
\end{figure}

\begin{figure}[H]
\centering
\begin{subfigure}[T]{0.19\textwidth}
\centering
\includegraphics[width=\textwidth]{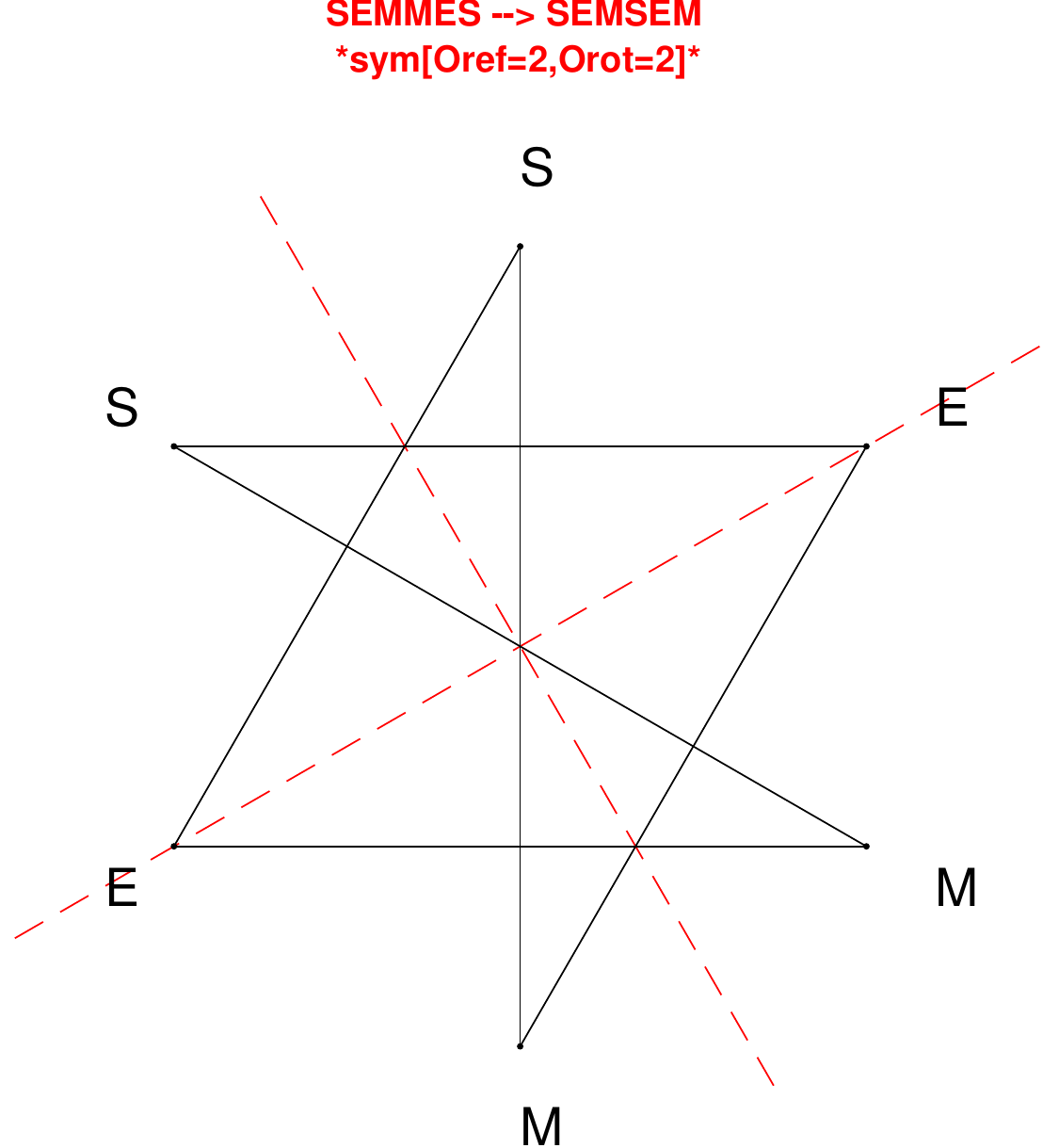}
\end{subfigure}
\hfill
\begin{subfigure}[T]{0.19\textwidth}
\centering
\includegraphics[width=\textwidth]{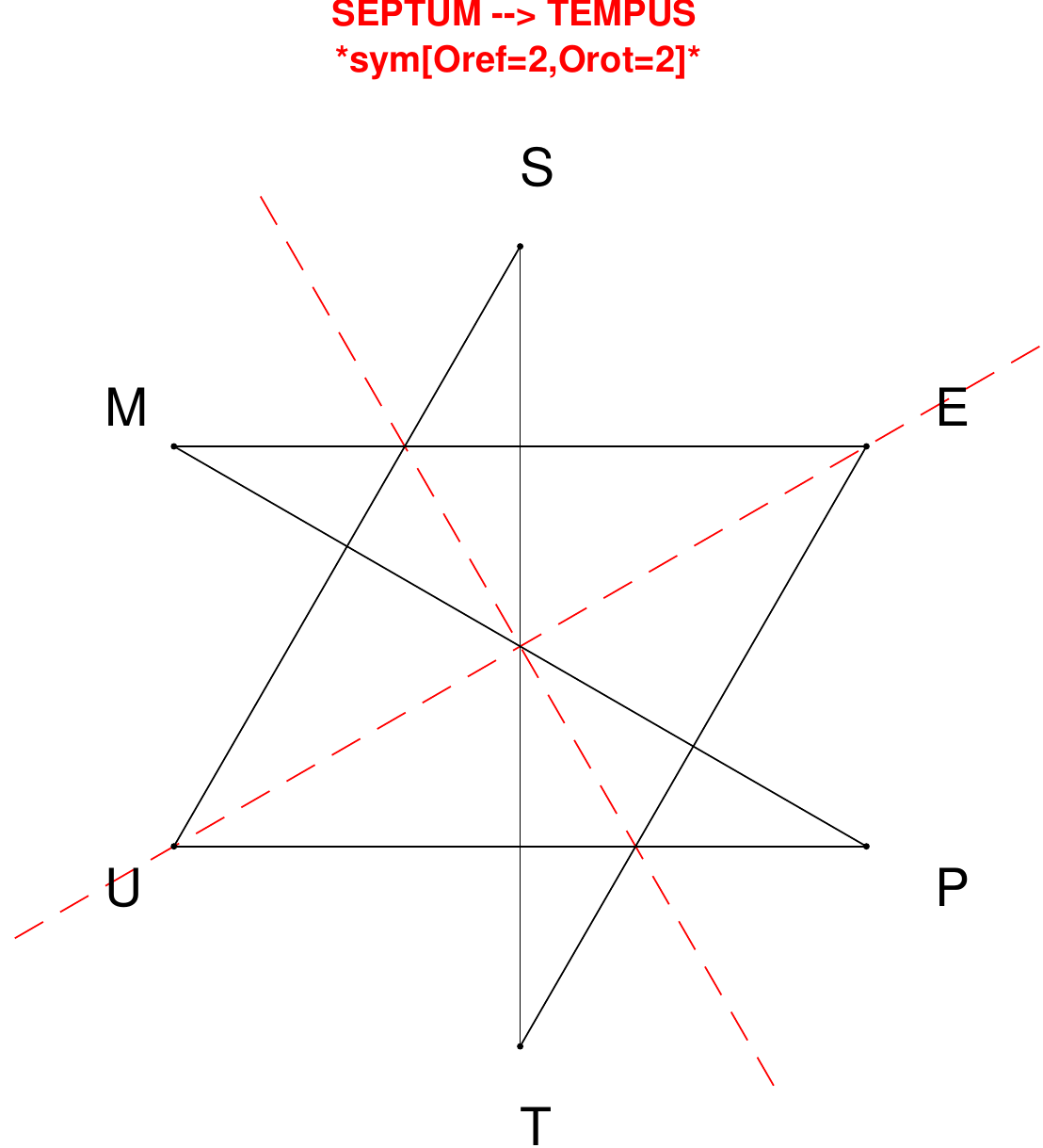}
\end{subfigure}
\hfill
\begin{subfigure}[T]{0.19\textwidth}
\centering
\includegraphics[width=\textwidth]{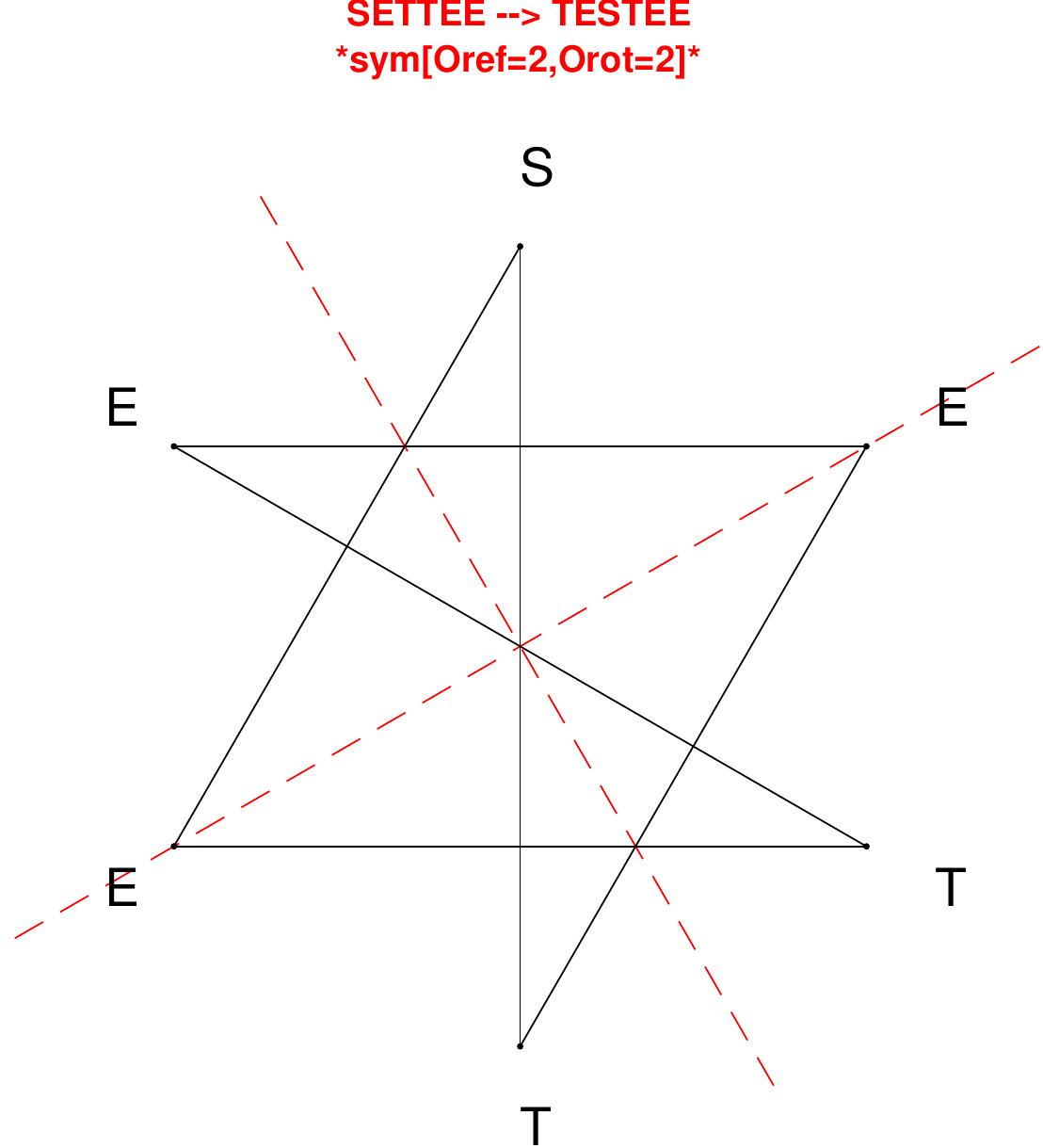}
\end{subfigure}
\hfill
\begin{subfigure}[T]{0.19\textwidth}
\centering
\includegraphics[width=\textwidth]{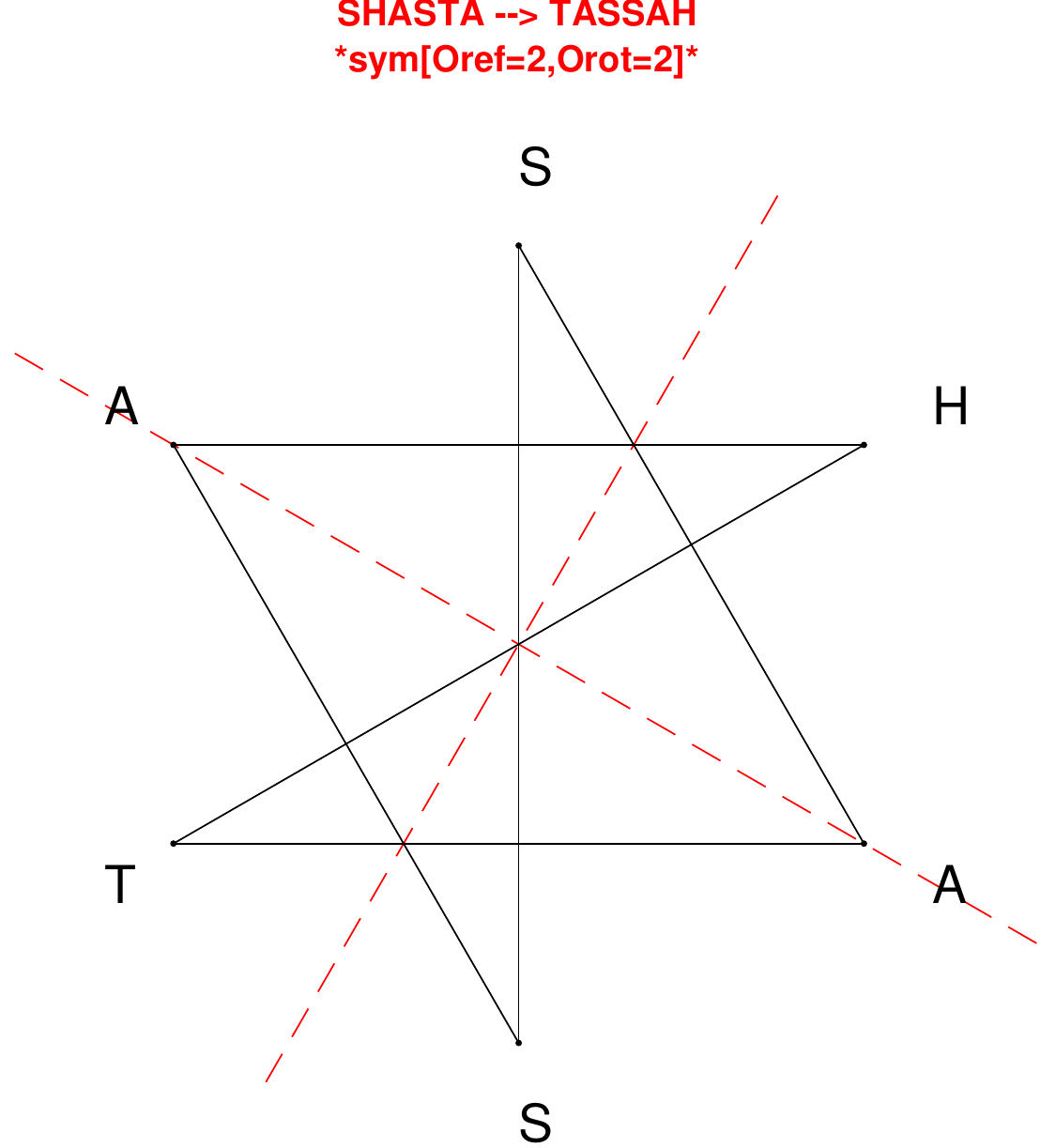}
\end{subfigure}
\hfill
\begin{subfigure}[T]{0.19\textwidth}
\centering
\includegraphics[width=\textwidth]{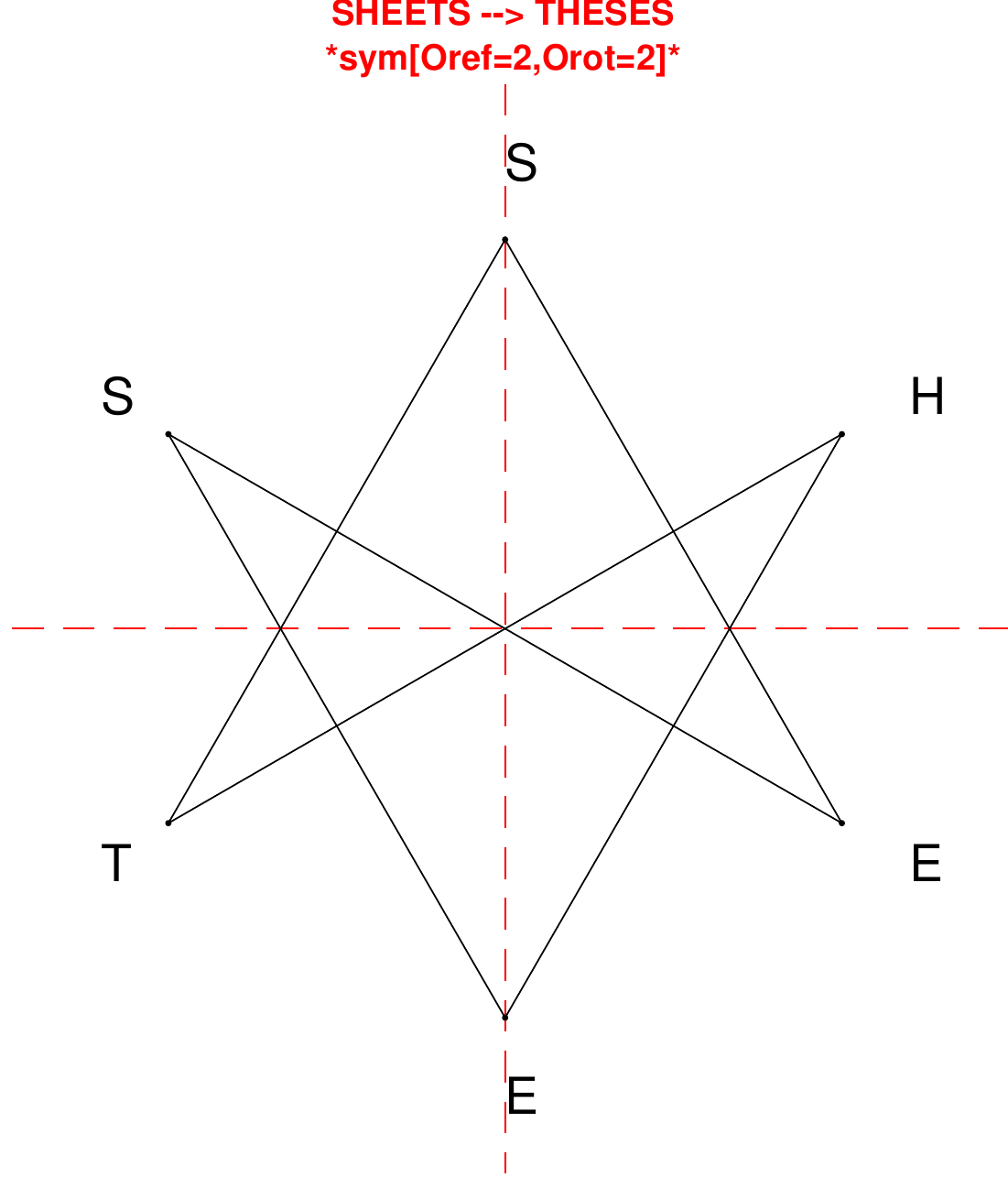}
\end{subfigure}
\end{figure}

\begin{figure}[H]
\centering
\begin{subfigure}[T]{0.19\textwidth}
\centering
\includegraphics[width=\textwidth]{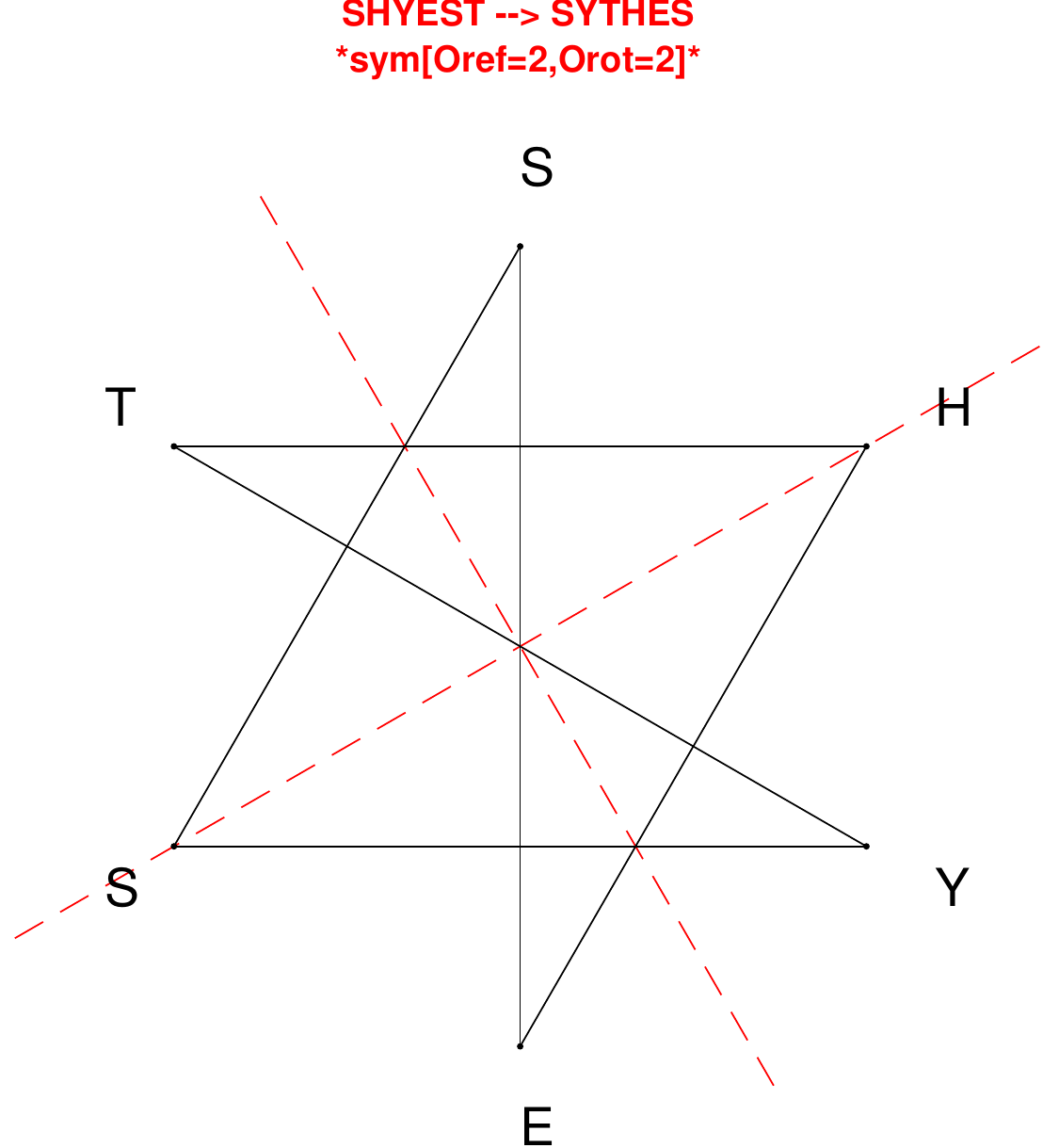}
\end{subfigure}
\hfill
\begin{subfigure}[T]{0.19\textwidth}
\centering
\includegraphics[width=\textwidth]{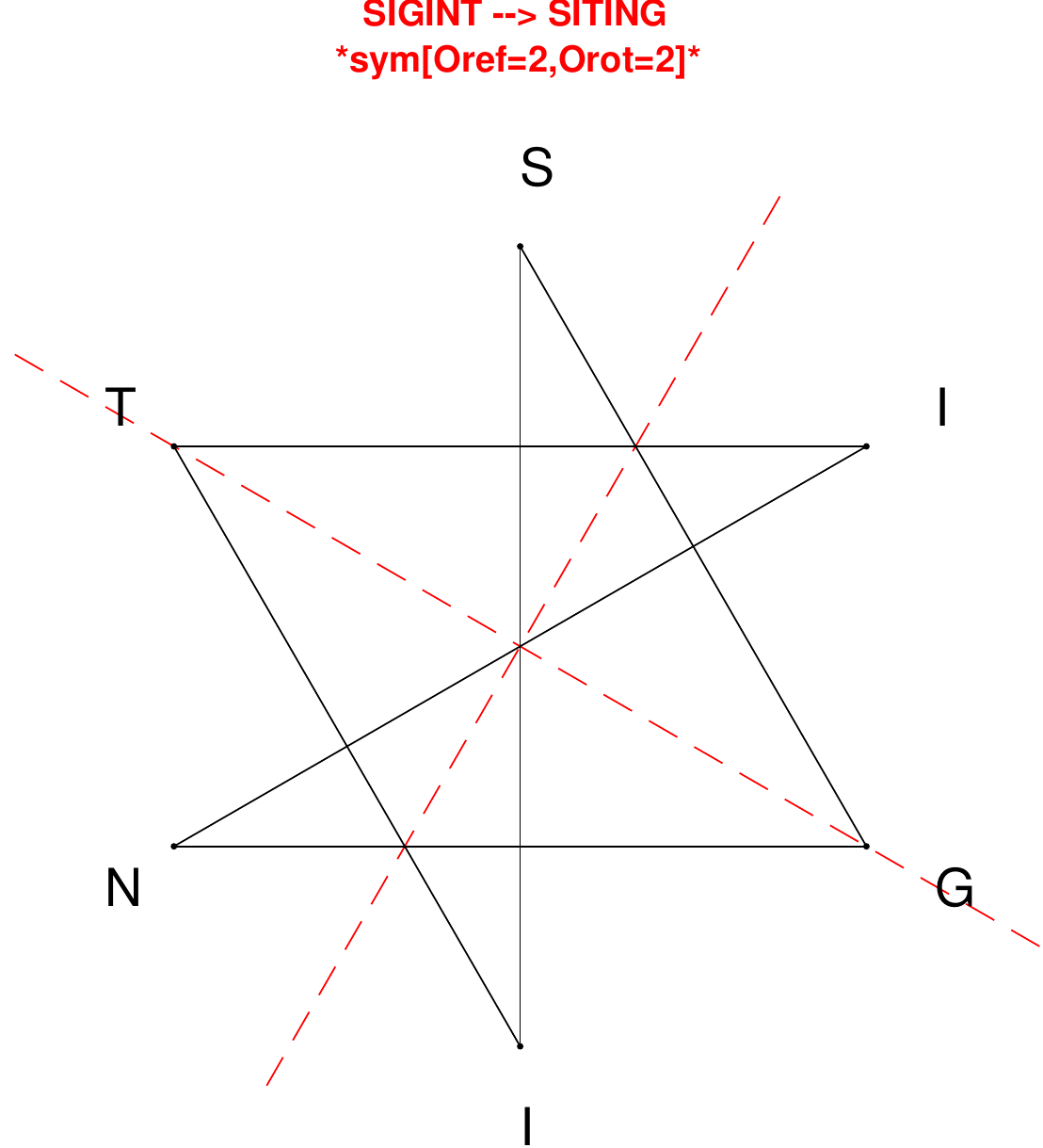}
\end{subfigure}
\hfill
\begin{subfigure}[T]{0.19\textwidth}
\centering
\includegraphics[width=\textwidth]{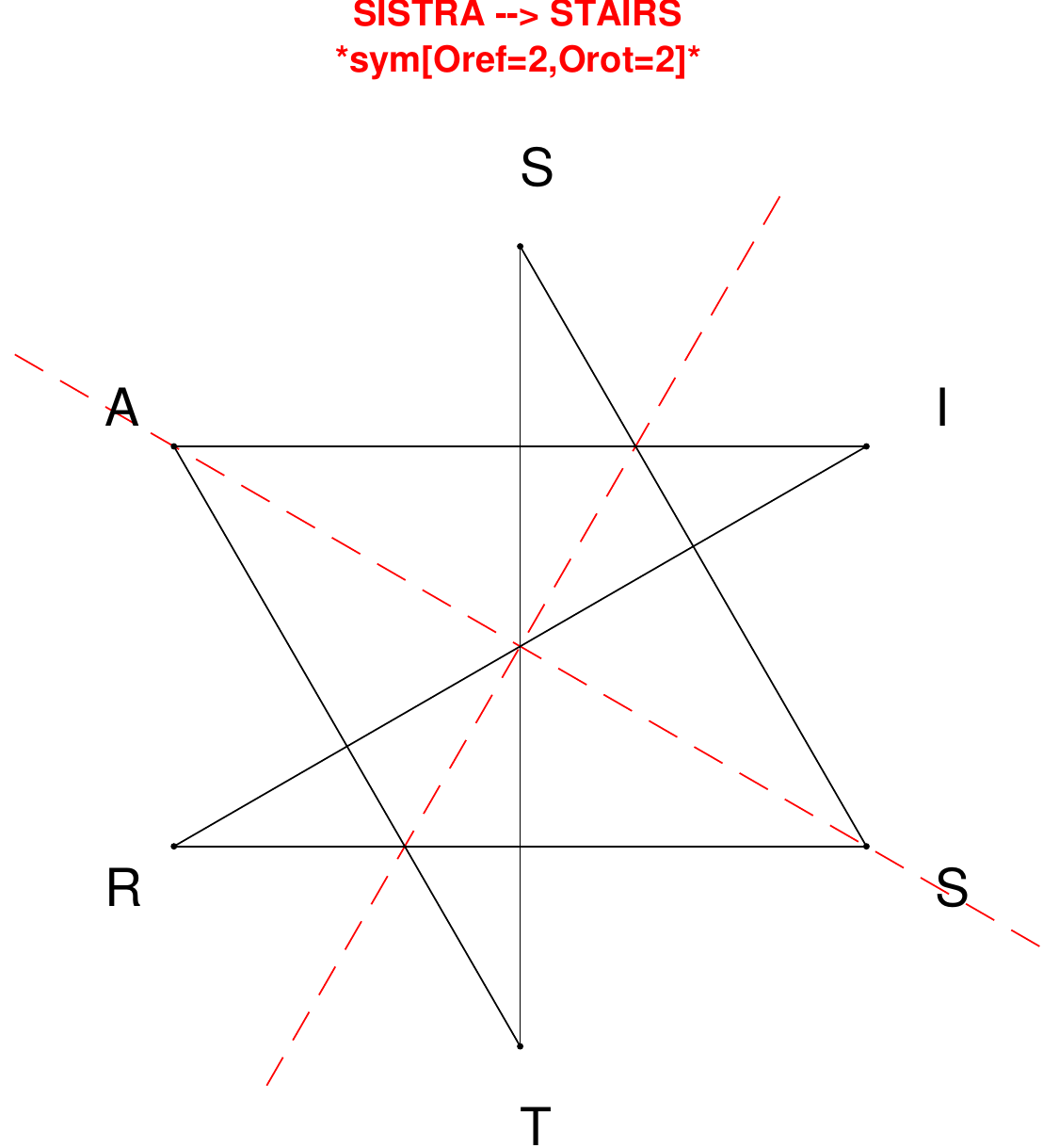}
\end{subfigure}
\hfill
\begin{subfigure}[T]{0.19\textwidth}
\centering
\includegraphics[width=\textwidth]{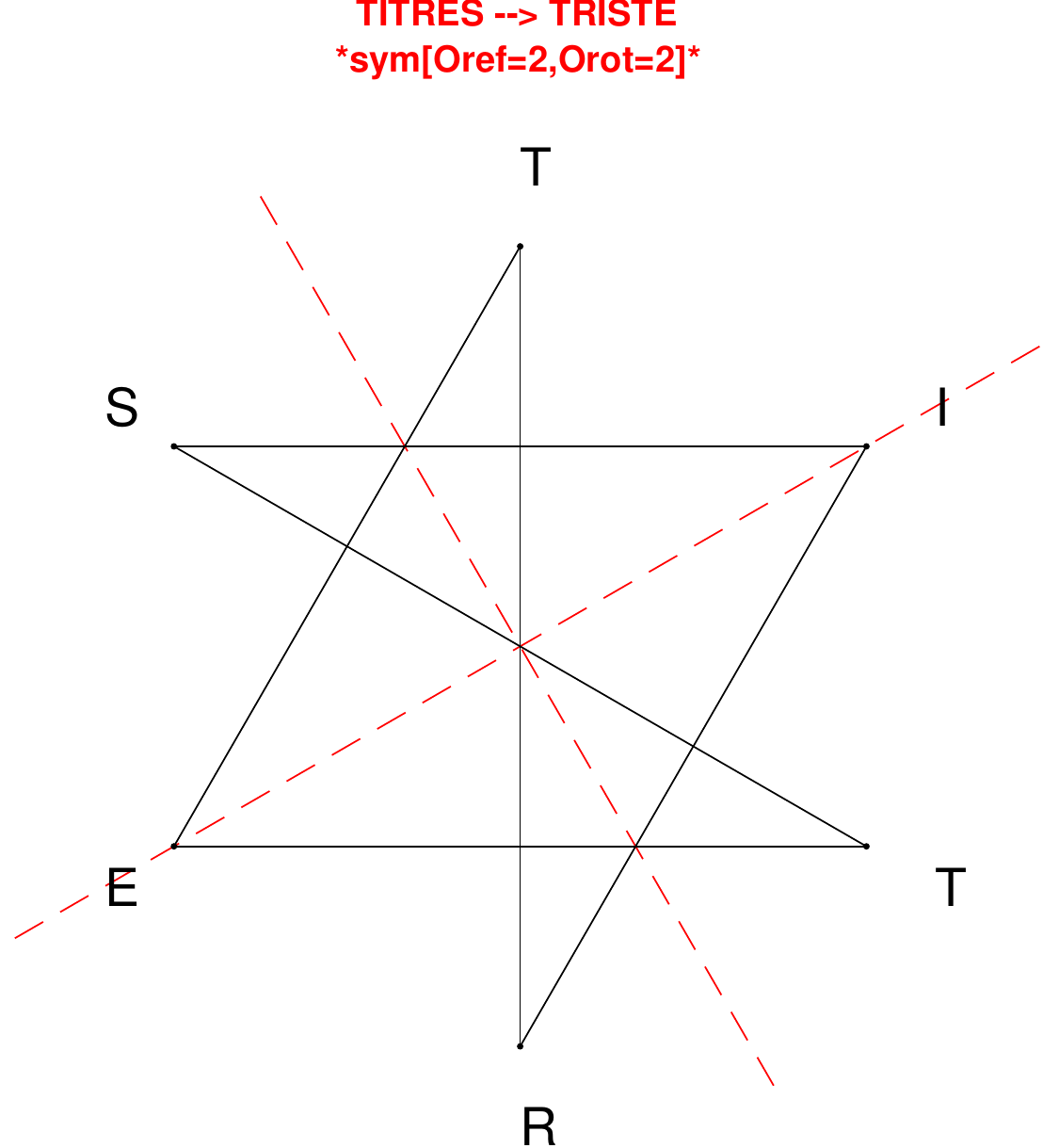}
\end{subfigure}
\hfill
\begin{subfigure}[T]{0.19\textwidth}
\centering
\includegraphics[width=\textwidth]{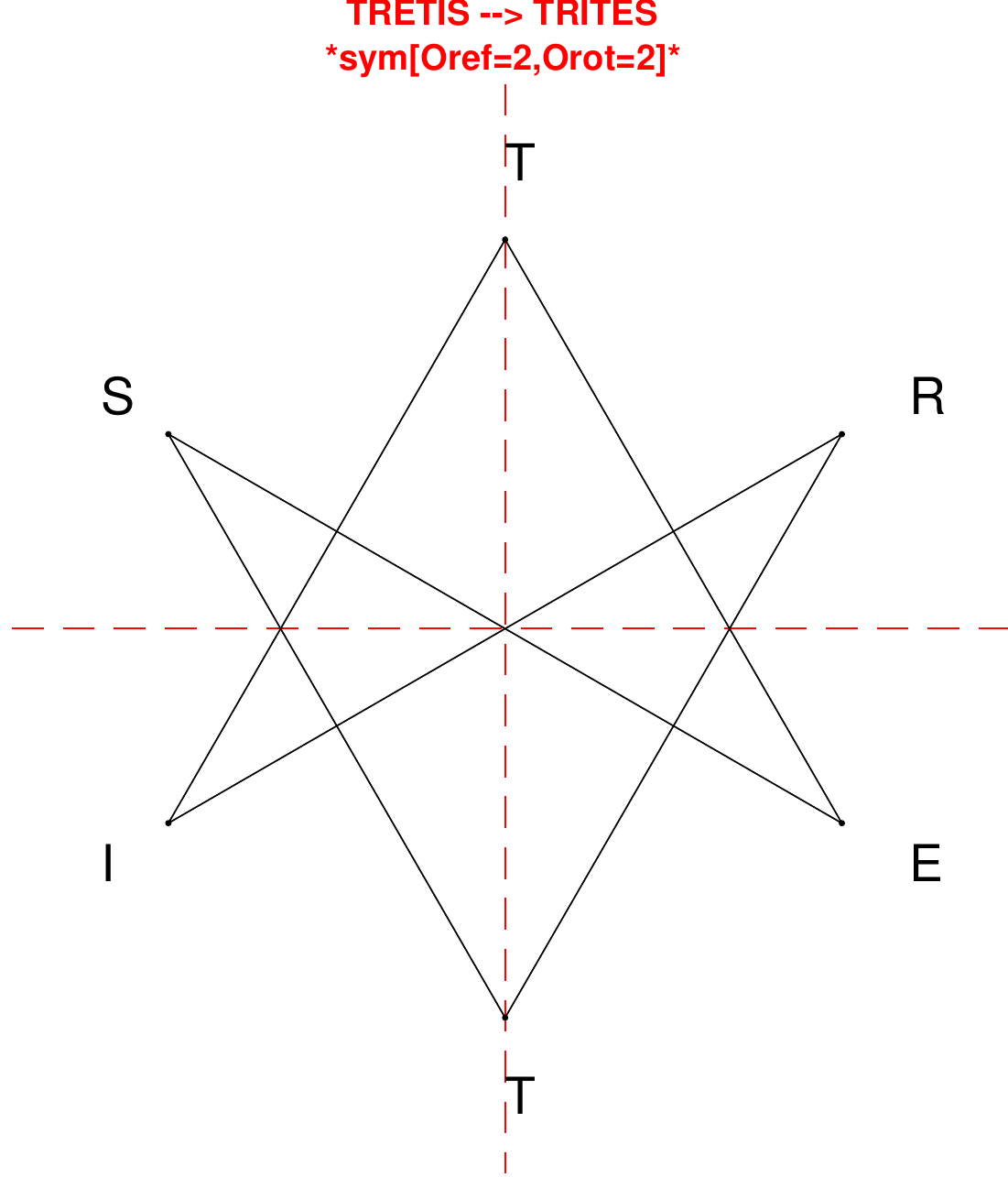}
\end{subfigure}
\end{figure}

\begin{figure}[H]
\centering
\begin{subfigure}[T]{0.19\textwidth}
\centering
\includegraphics[width=\textwidth]{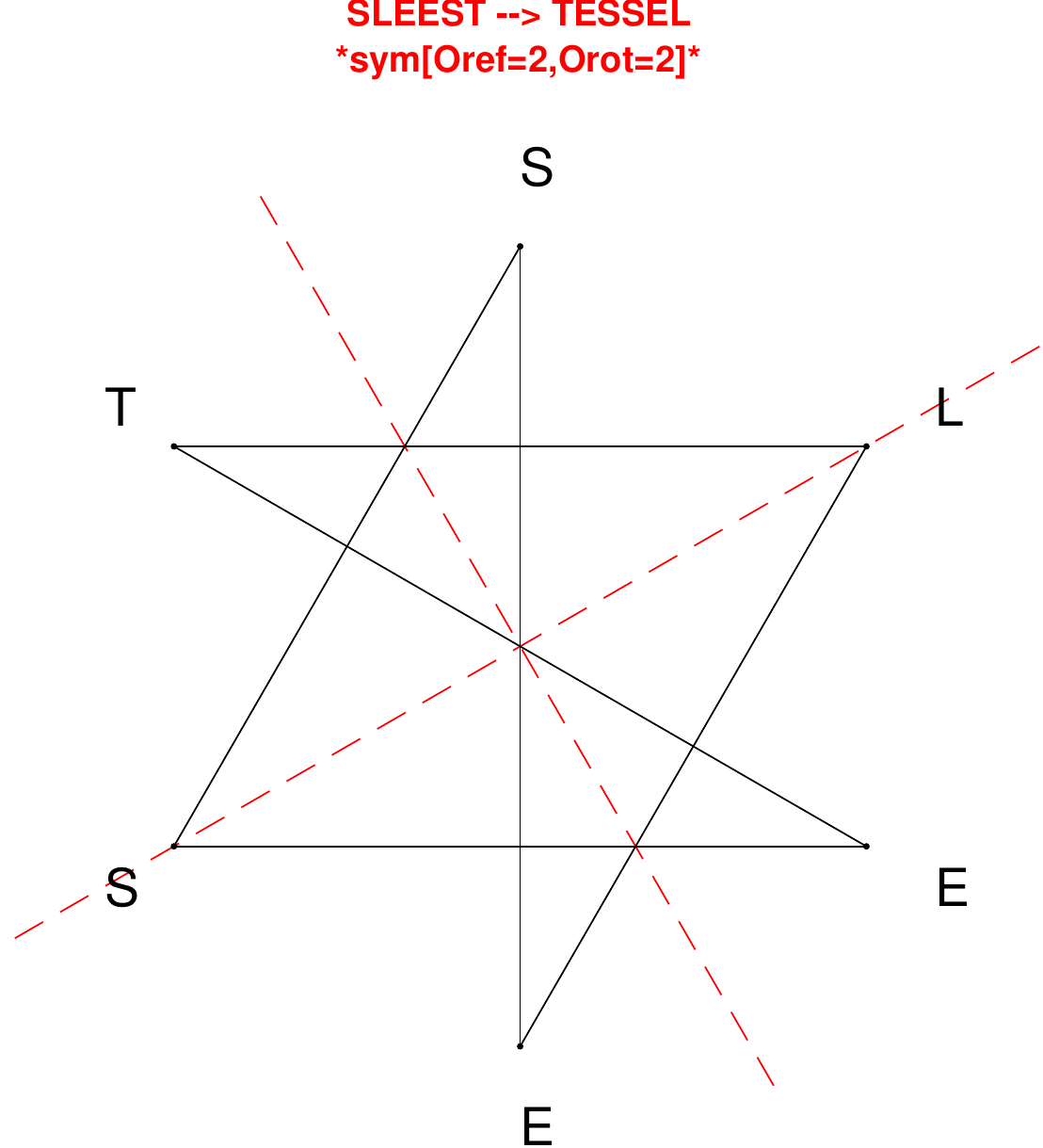}
\end{subfigure}
\hfill
\begin{subfigure}[T]{0.19\textwidth}
\centering
\includegraphics[width=\textwidth]{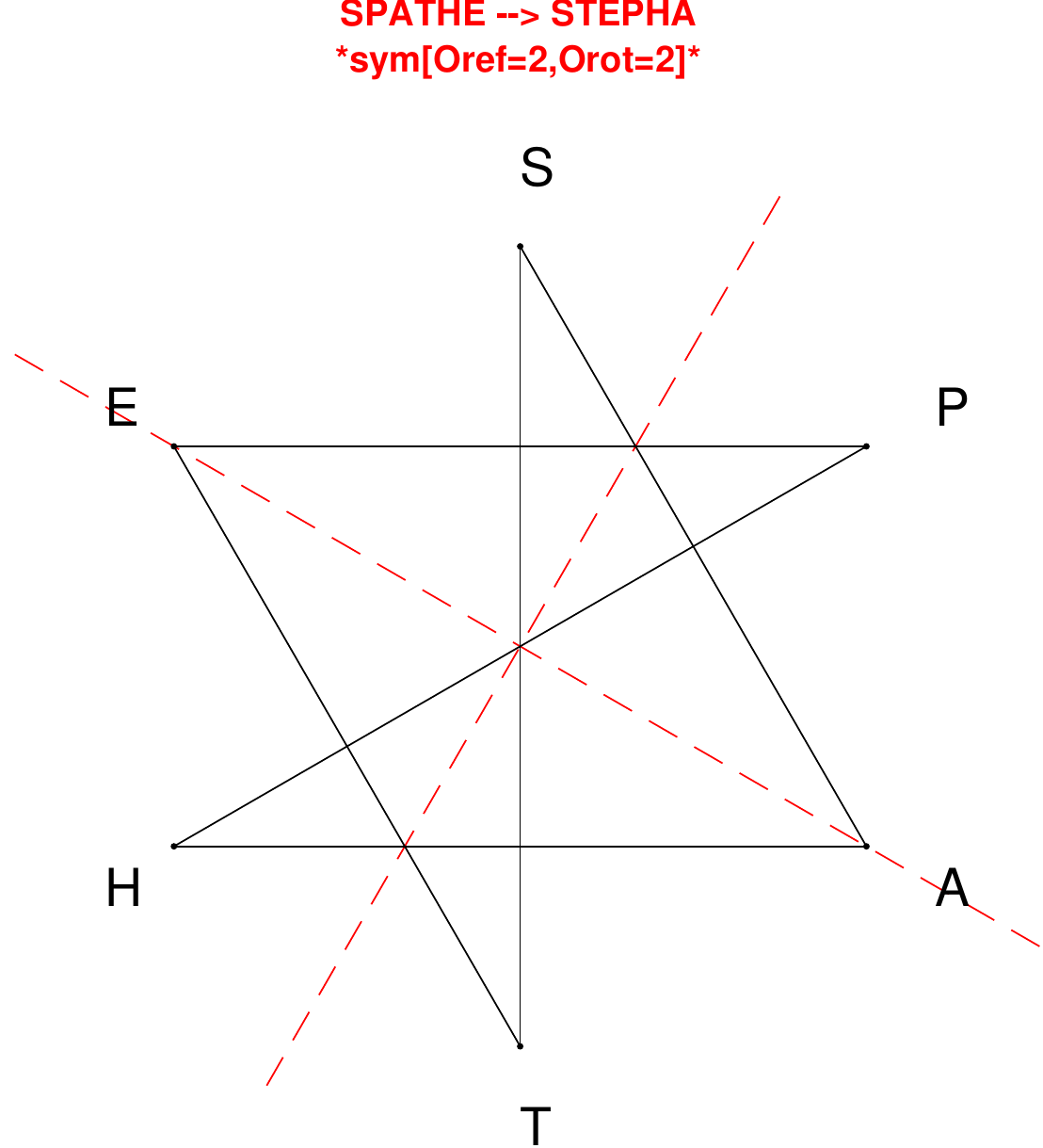}
\end{subfigure}
\hfill
\begin{subfigure}[T]{0.19\textwidth}
\centering
\includegraphics[width=\textwidth]{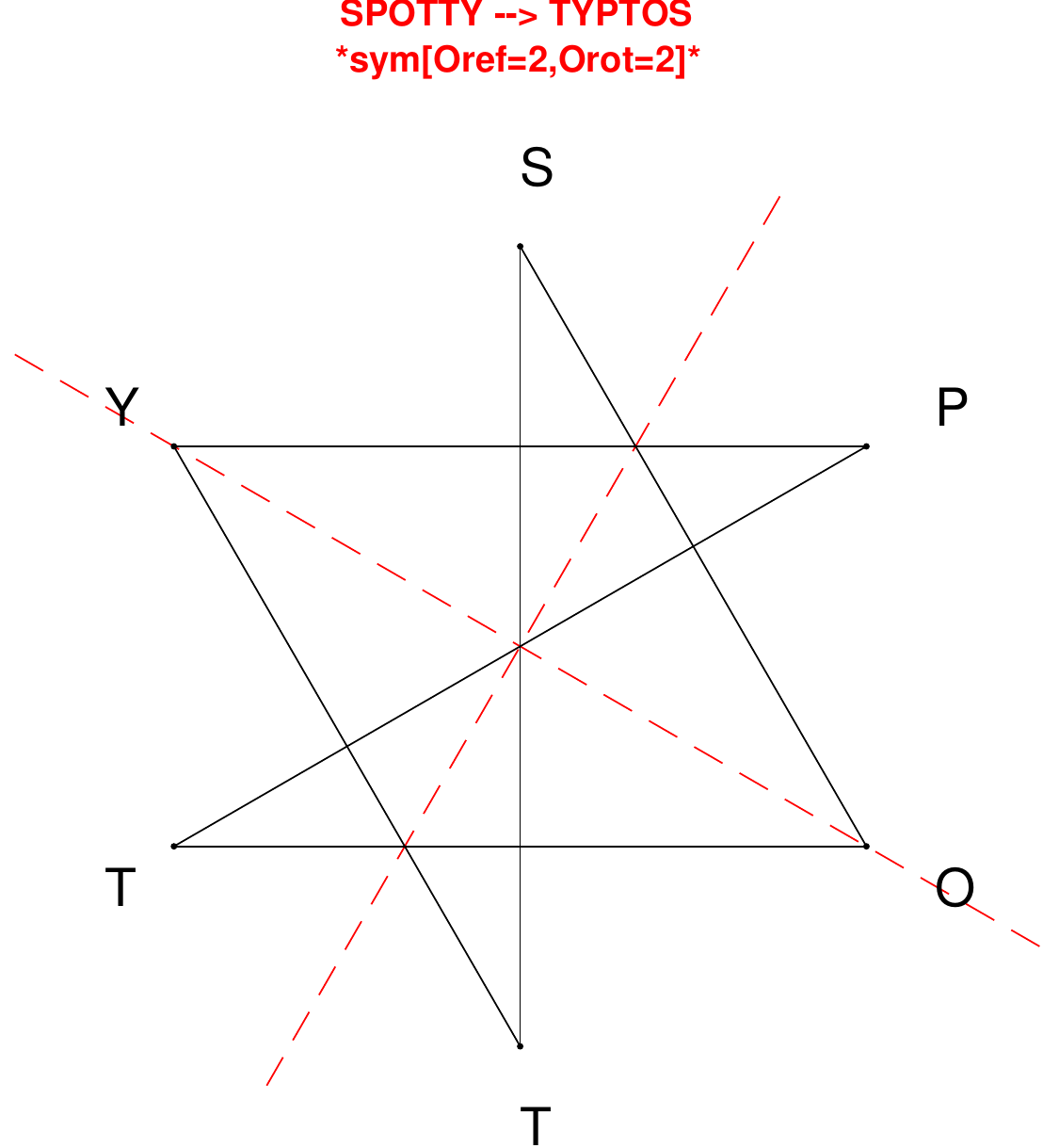}
\end{subfigure}
\hfill
\begin{subfigure}[T]{0.19\textwidth}
\centering
\includegraphics[width=\textwidth]{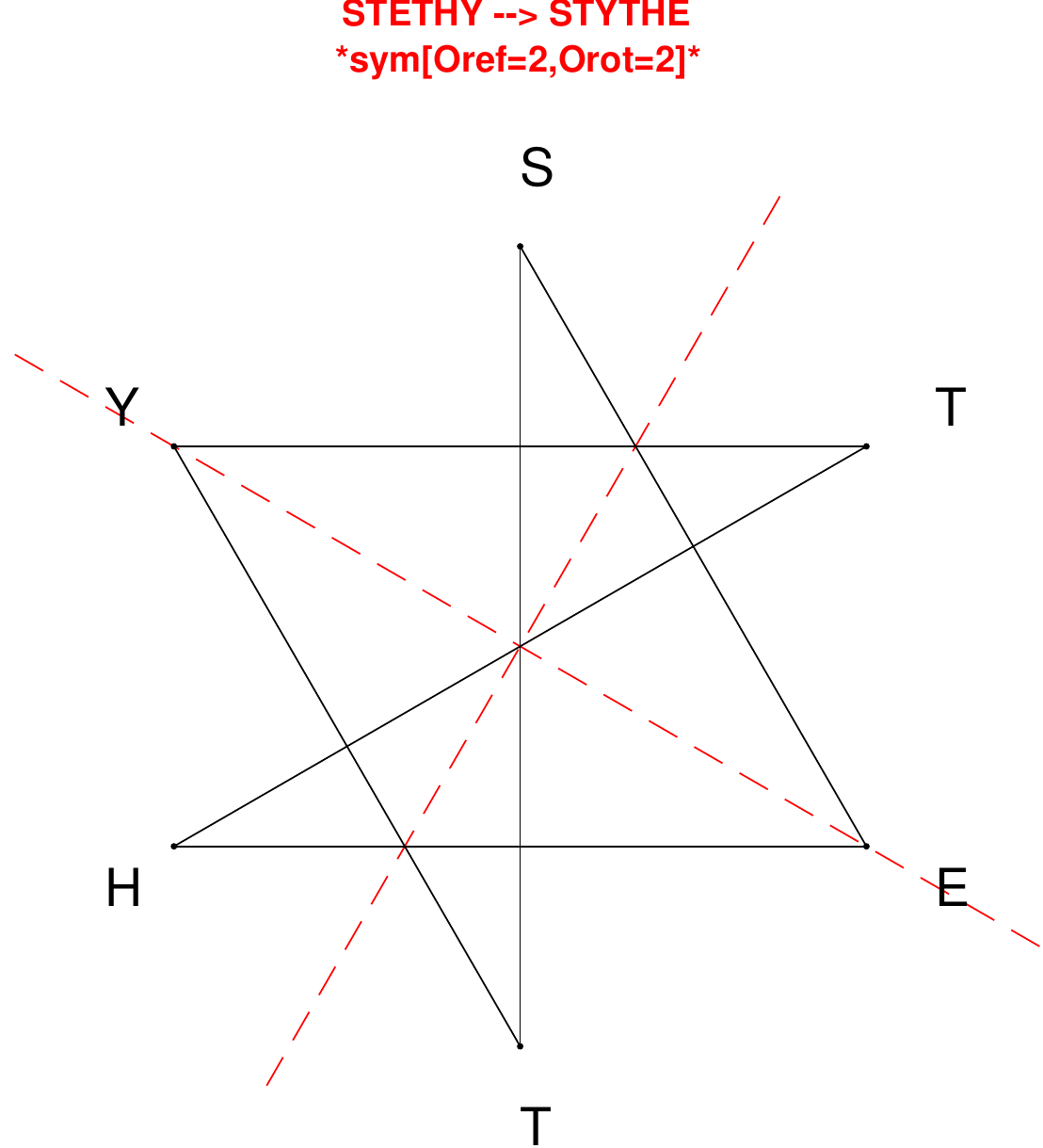}
\end{subfigure}
\hfill
\begin{subfigure}[T]{0.19\textwidth}
\centering
\includegraphics[width=\textwidth]{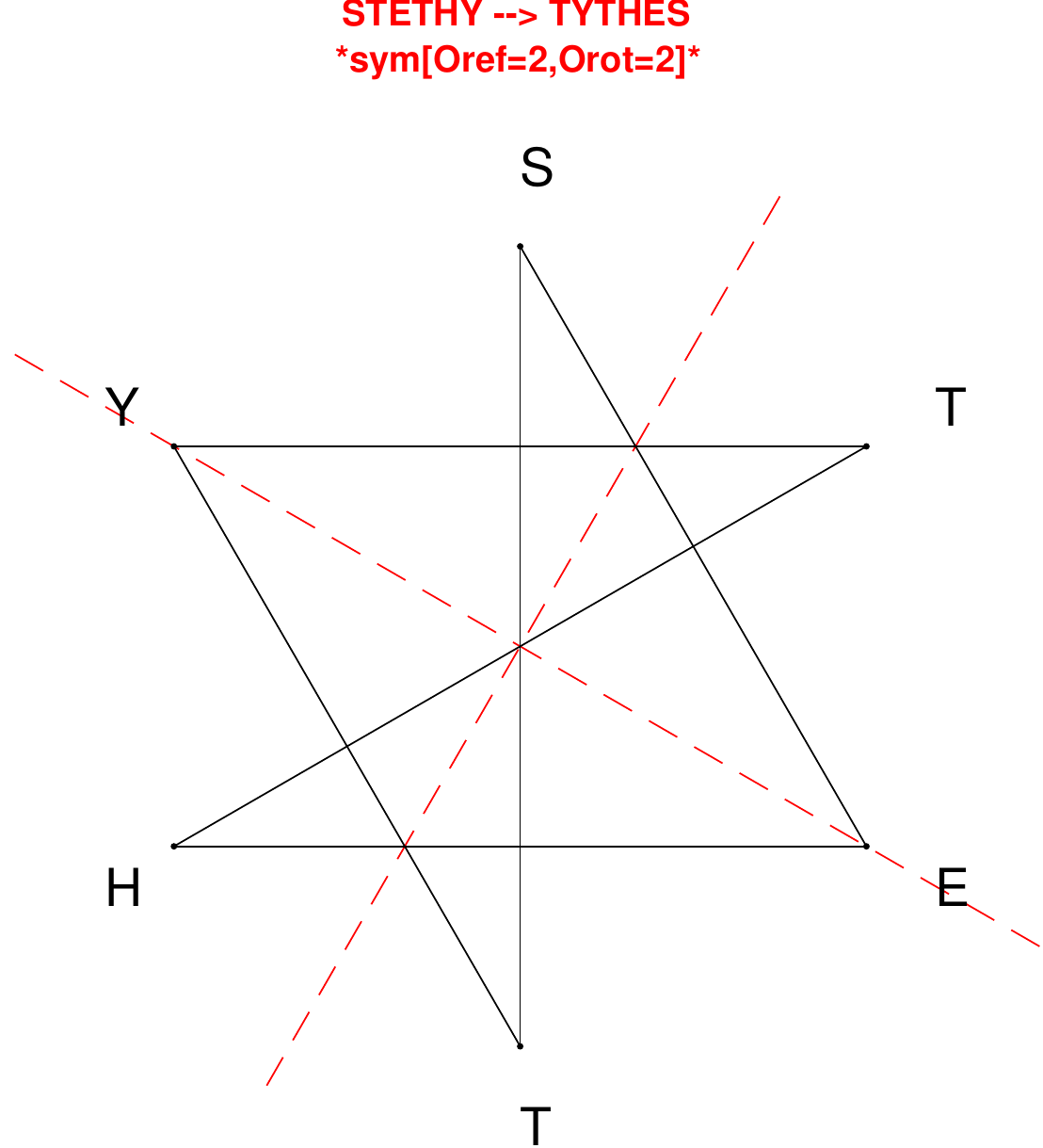}
\end{subfigure}
\end{figure}

\begin{figure}[H]
\centering
\begin{subfigure}[T]{0.19\textwidth}
\centering
\includegraphics[width=\textwidth]{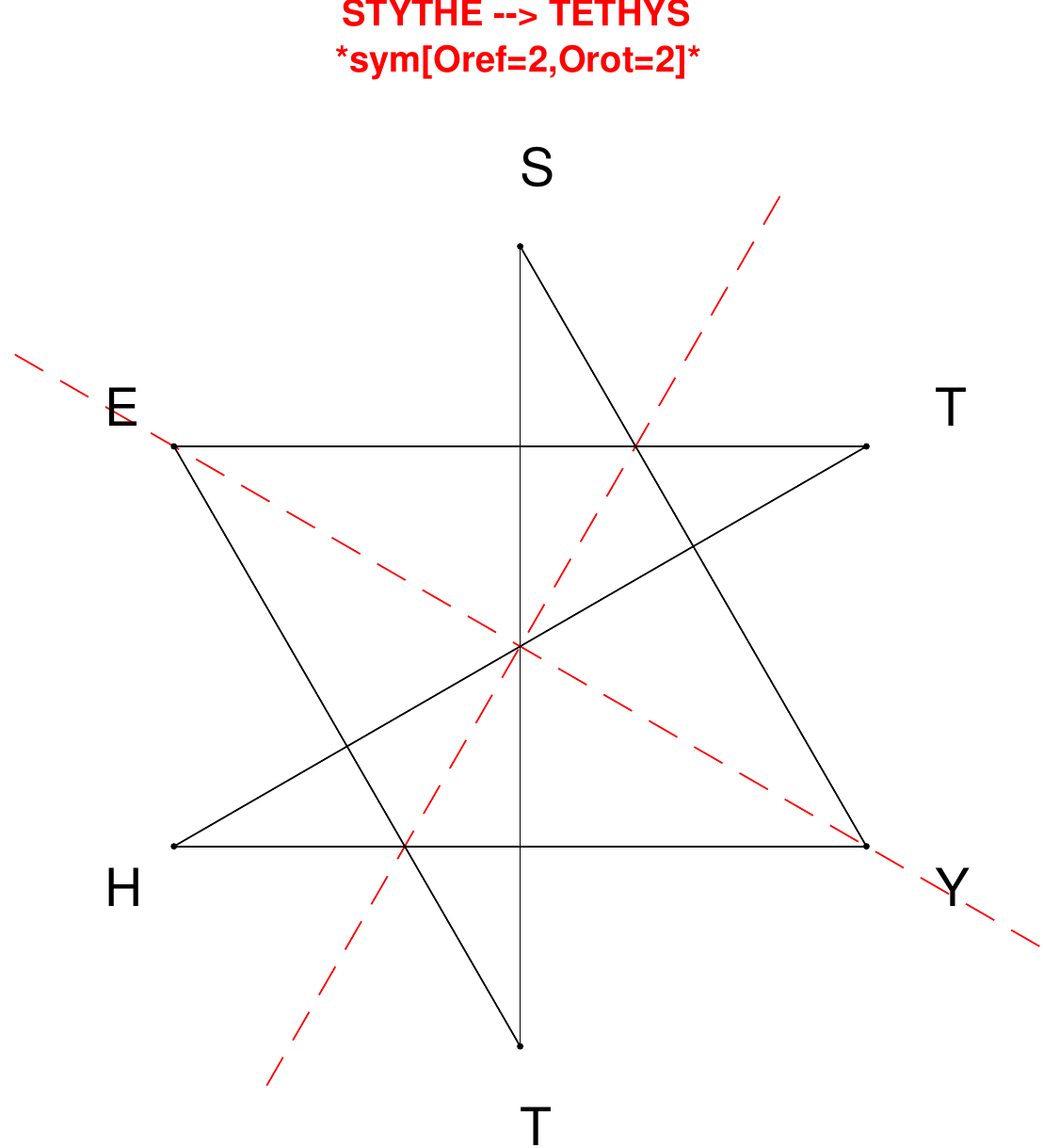}
\end{subfigure}
\hfill
\begin{subfigure}[T]{0.19\textwidth}
\centering
\includegraphics[width=\textwidth]{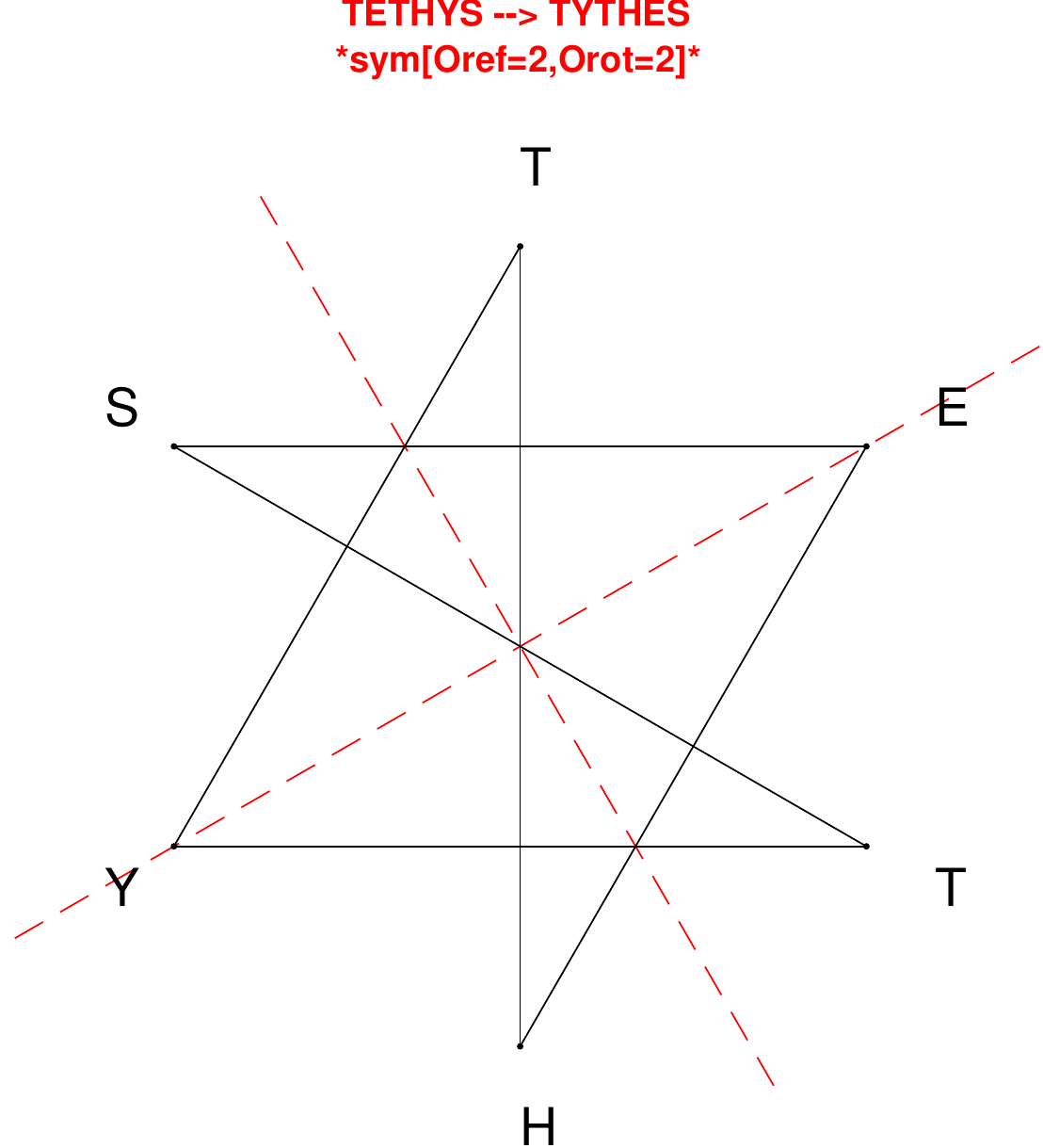}
\end{subfigure}
\hfill
\begin{subfigure}[T]{0.19\textwidth}
\centering
\includegraphics[width=\textwidth]{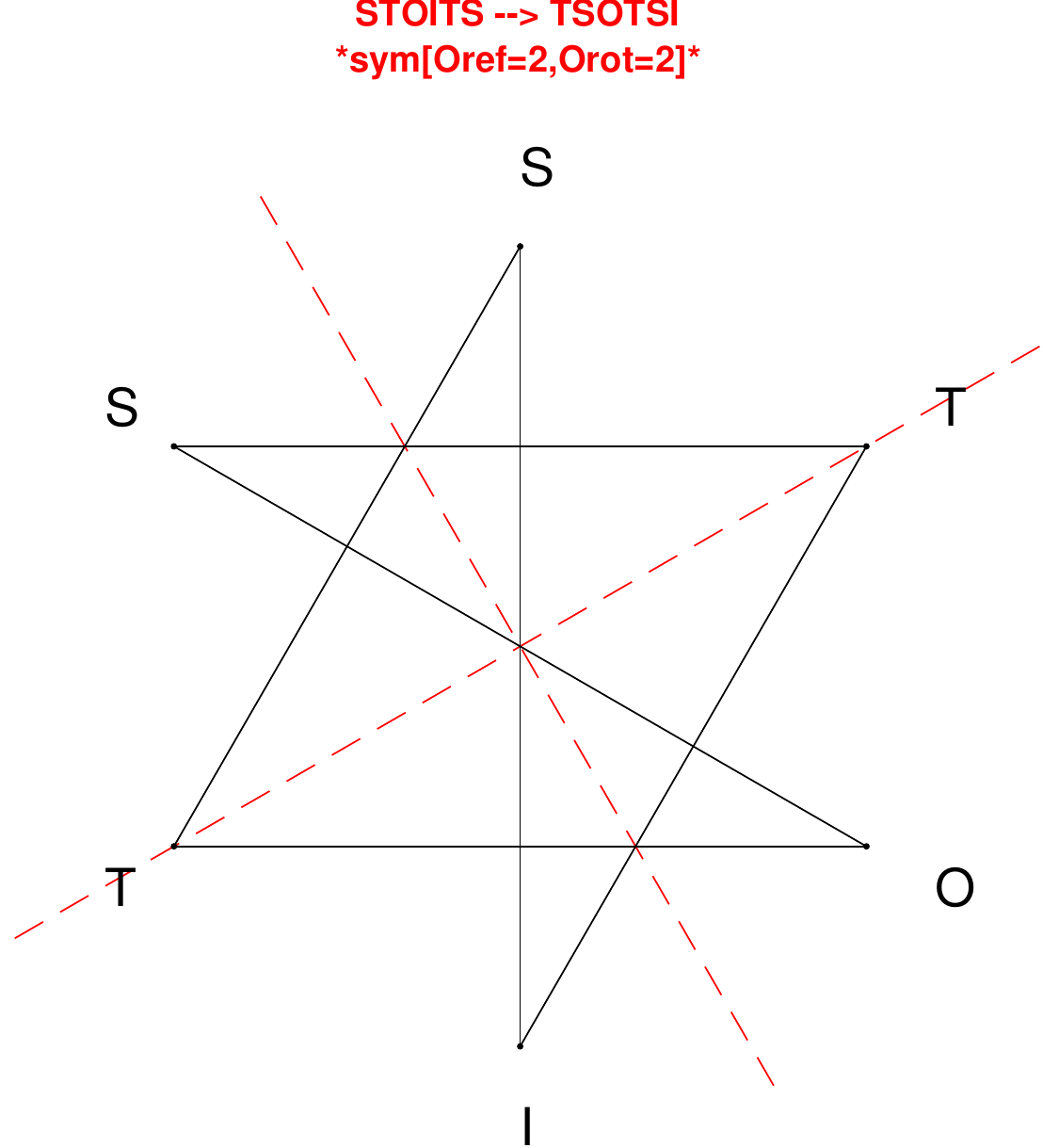}
\end{subfigure}
\hfill
\begin{subfigure}[T]{0.19\textwidth}
\centering
\includegraphics[width=\textwidth]{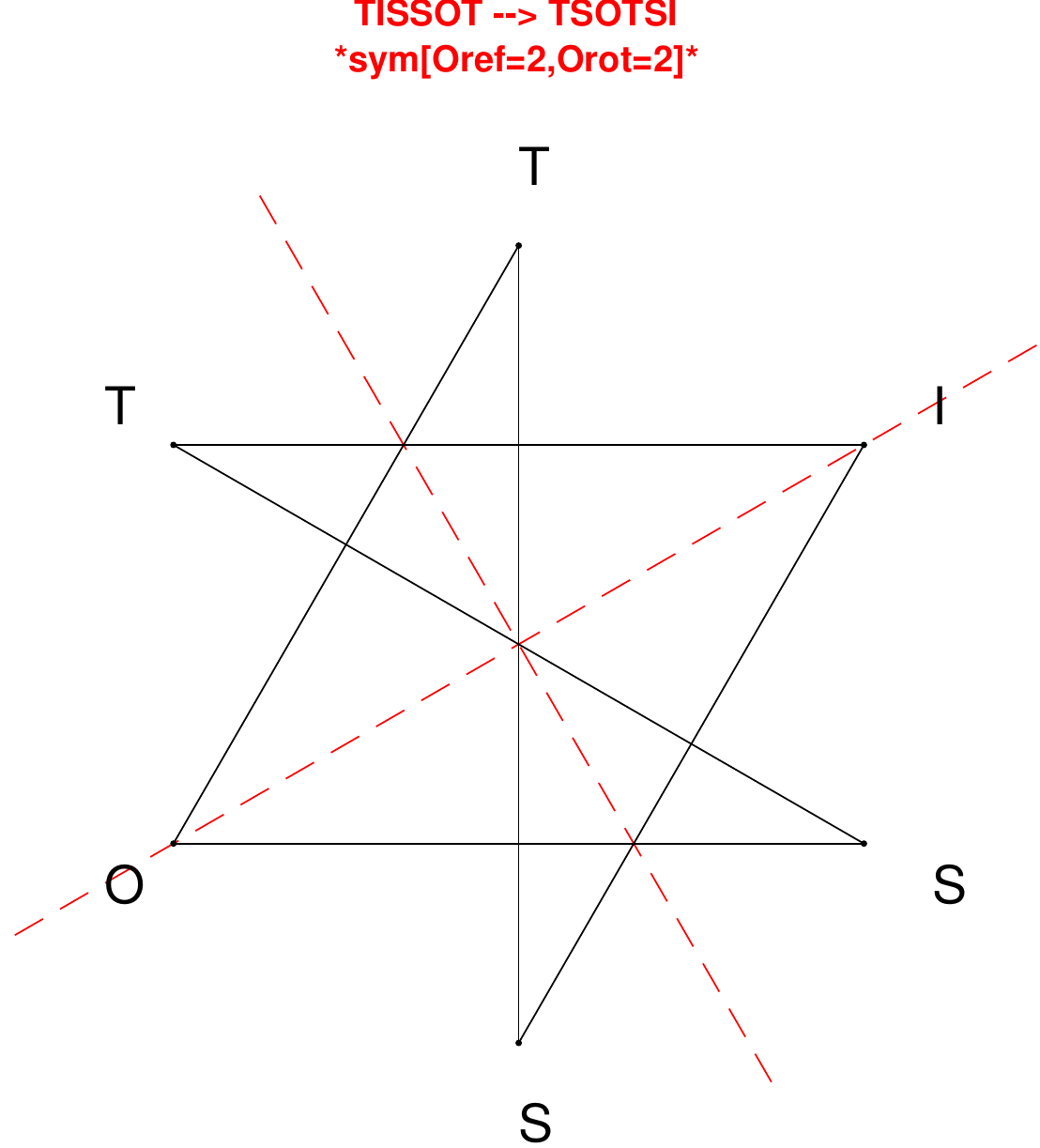}
\end{subfigure}
\hfill
\begin{subfigure}[T]{0.19\textwidth}
\centering
\includegraphics[width=\textwidth]{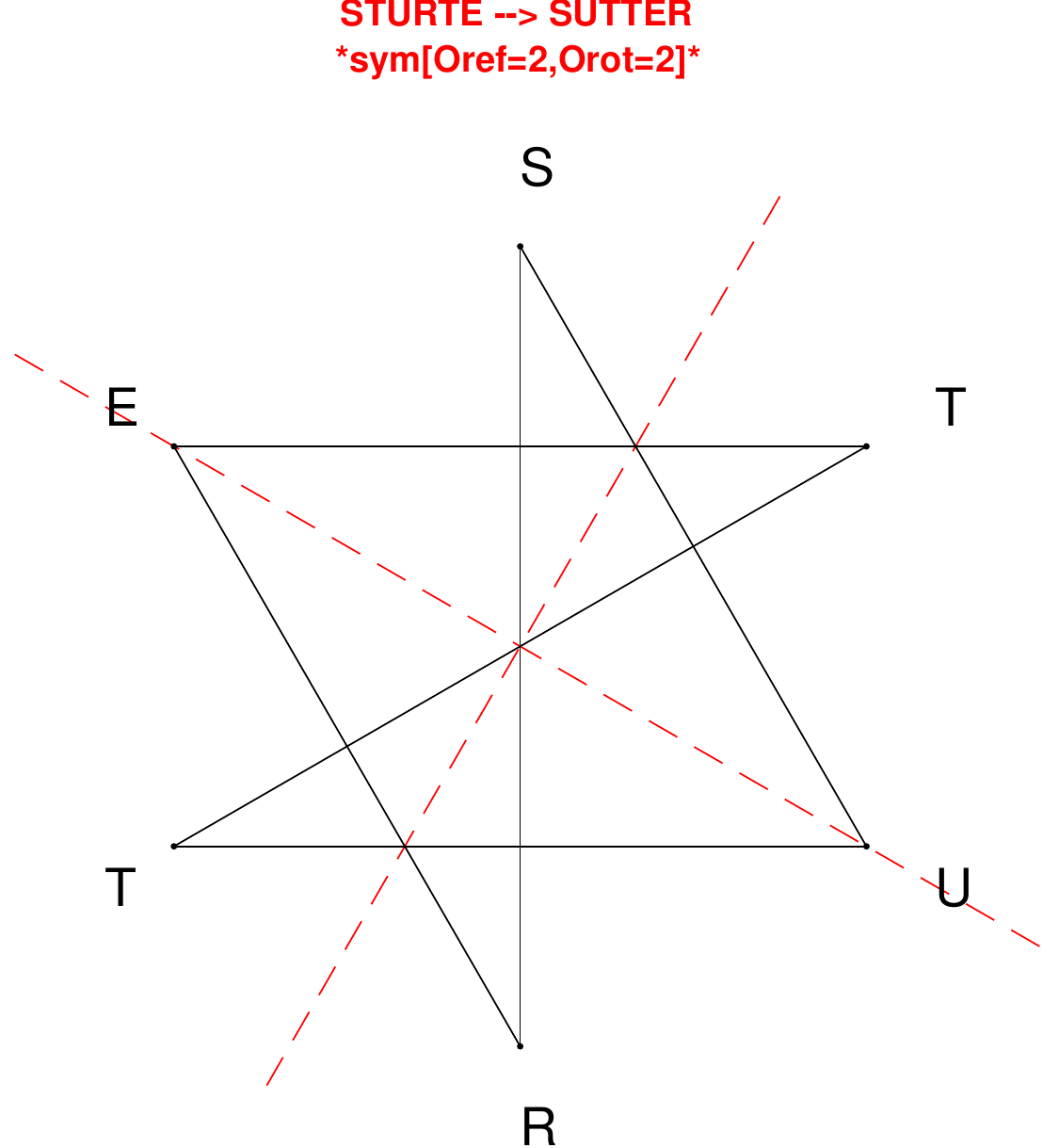}
\end{subfigure}
\end{figure}

\begin{figure}[H]
\centering
\begin{subfigure}[T]{0.19\textwidth}
\centering
\includegraphics[width=\textwidth]{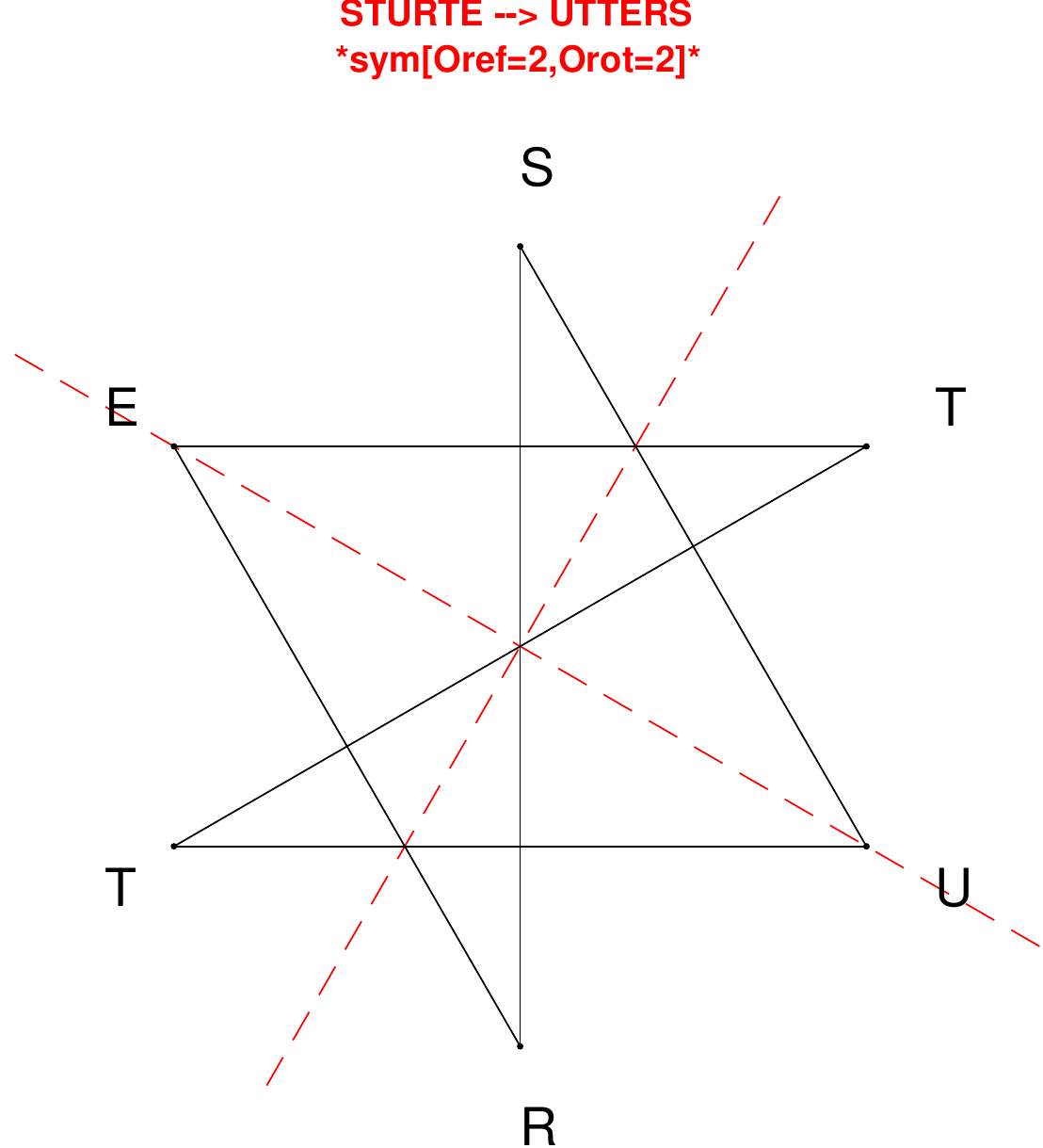}
\end{subfigure}
\hfill
\begin{subfigure}[T]{0.19\textwidth}
\centering
\includegraphics[width=\textwidth]{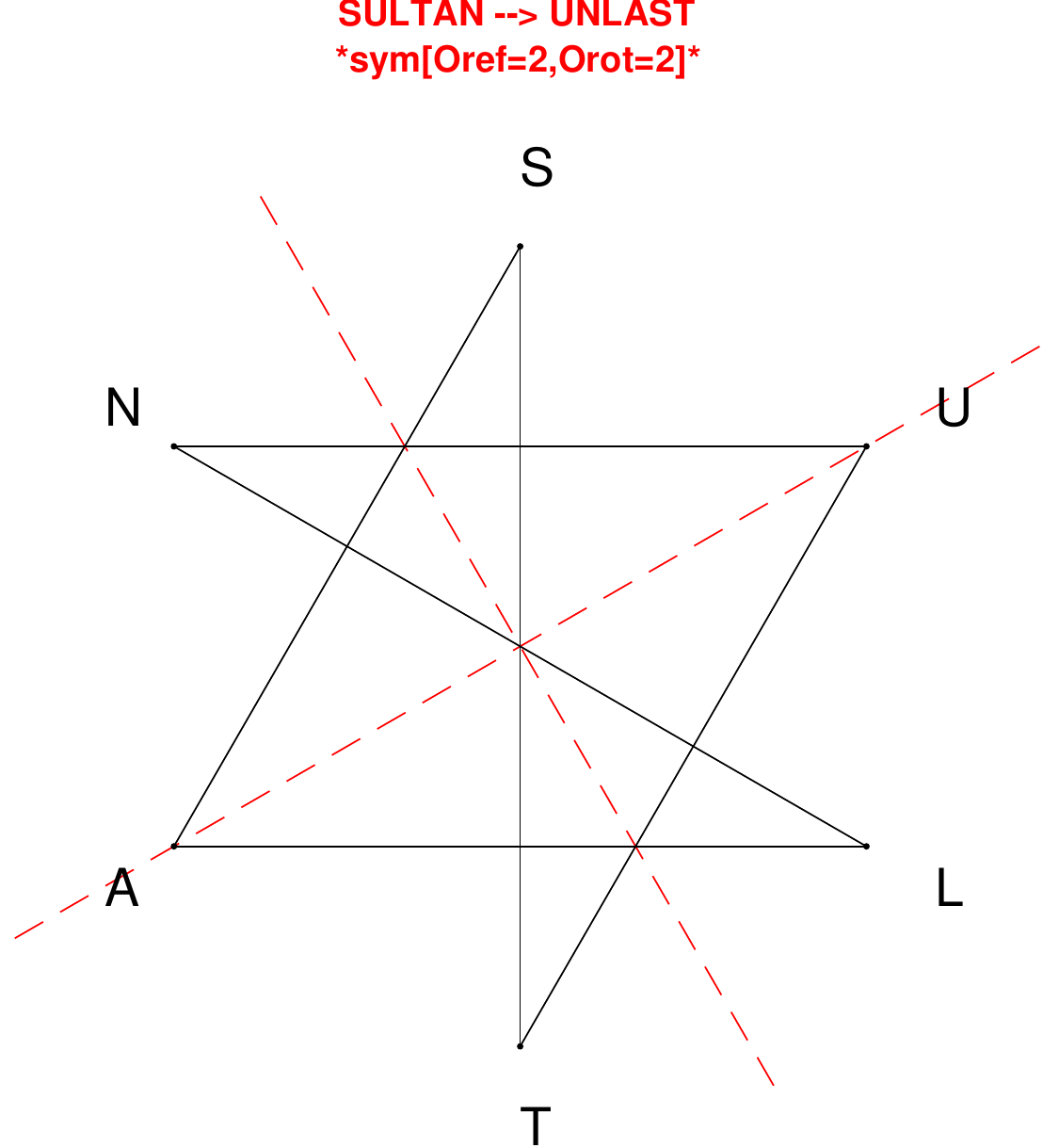}
\end{subfigure}
\hfill
\begin{subfigure}[T]{0.19\textwidth}
\centering
\includegraphics[width=\textwidth]{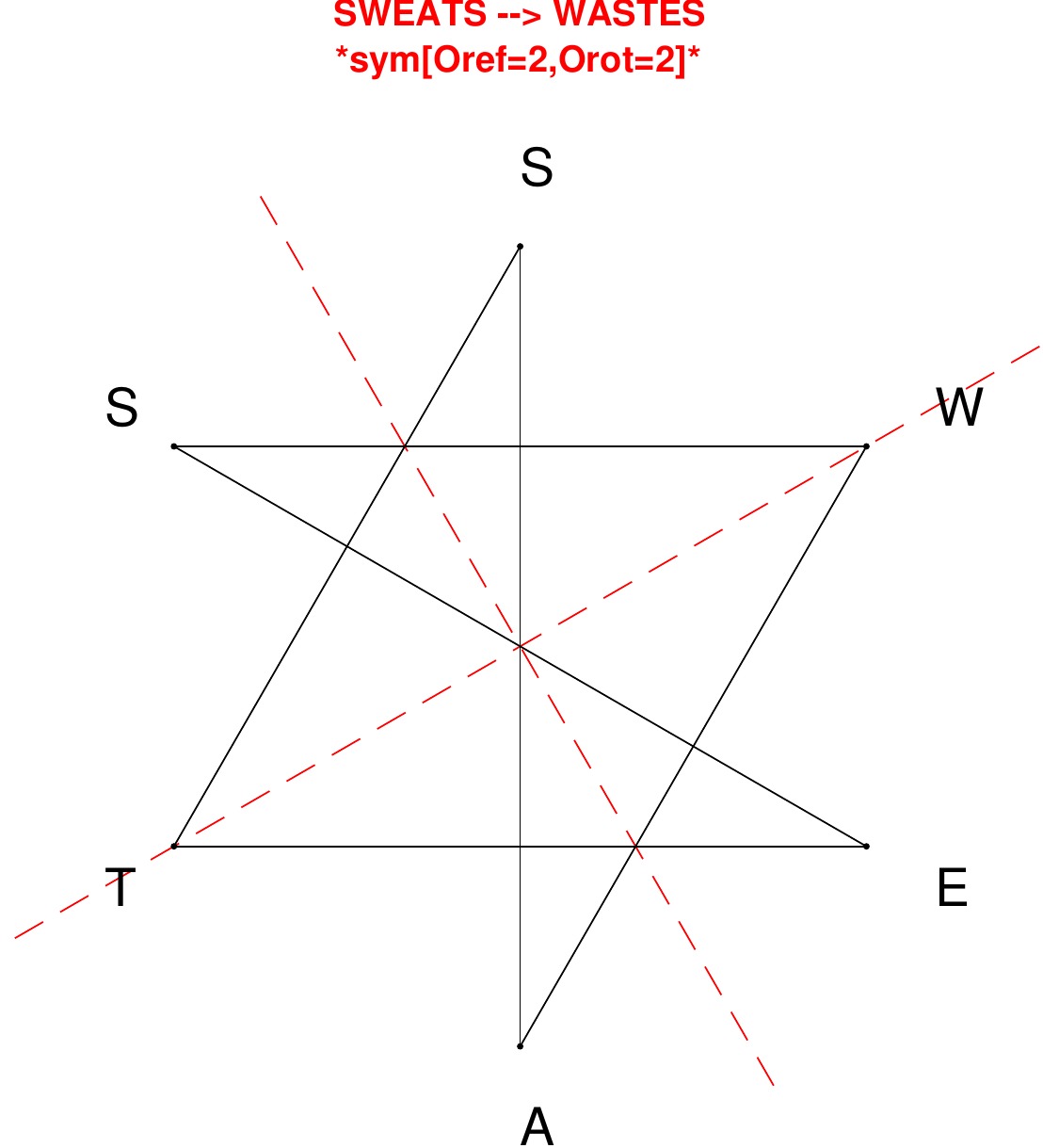}
\end{subfigure}
\hfill
\begin{subfigure}[T]{0.19\textwidth}
\centering
\includegraphics[width=\textwidth]{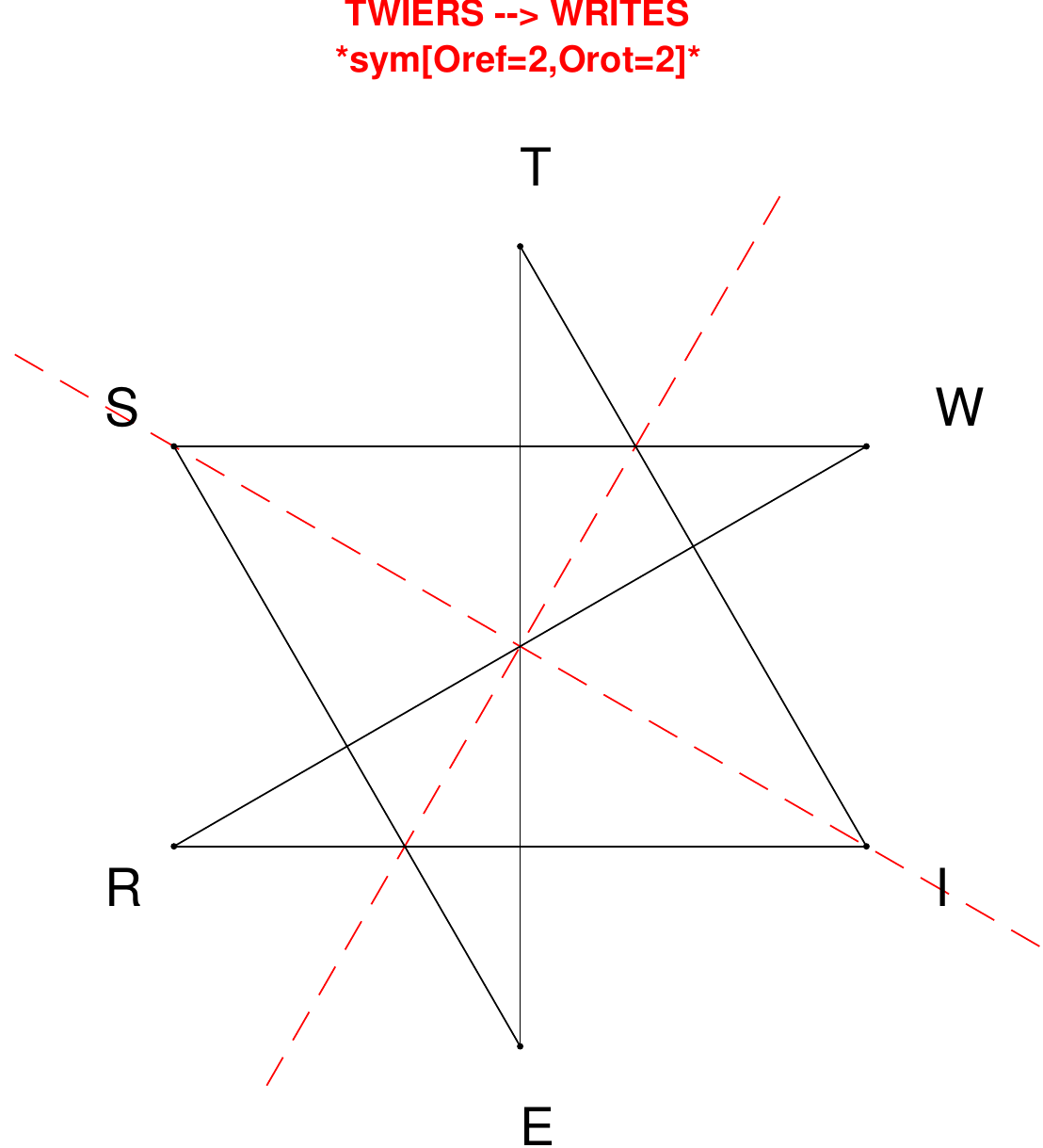}
\end{subfigure}
\hfill
\begin{subfigure}[T]{0.19\textwidth}
\centering
\includegraphics[width=\textwidth]{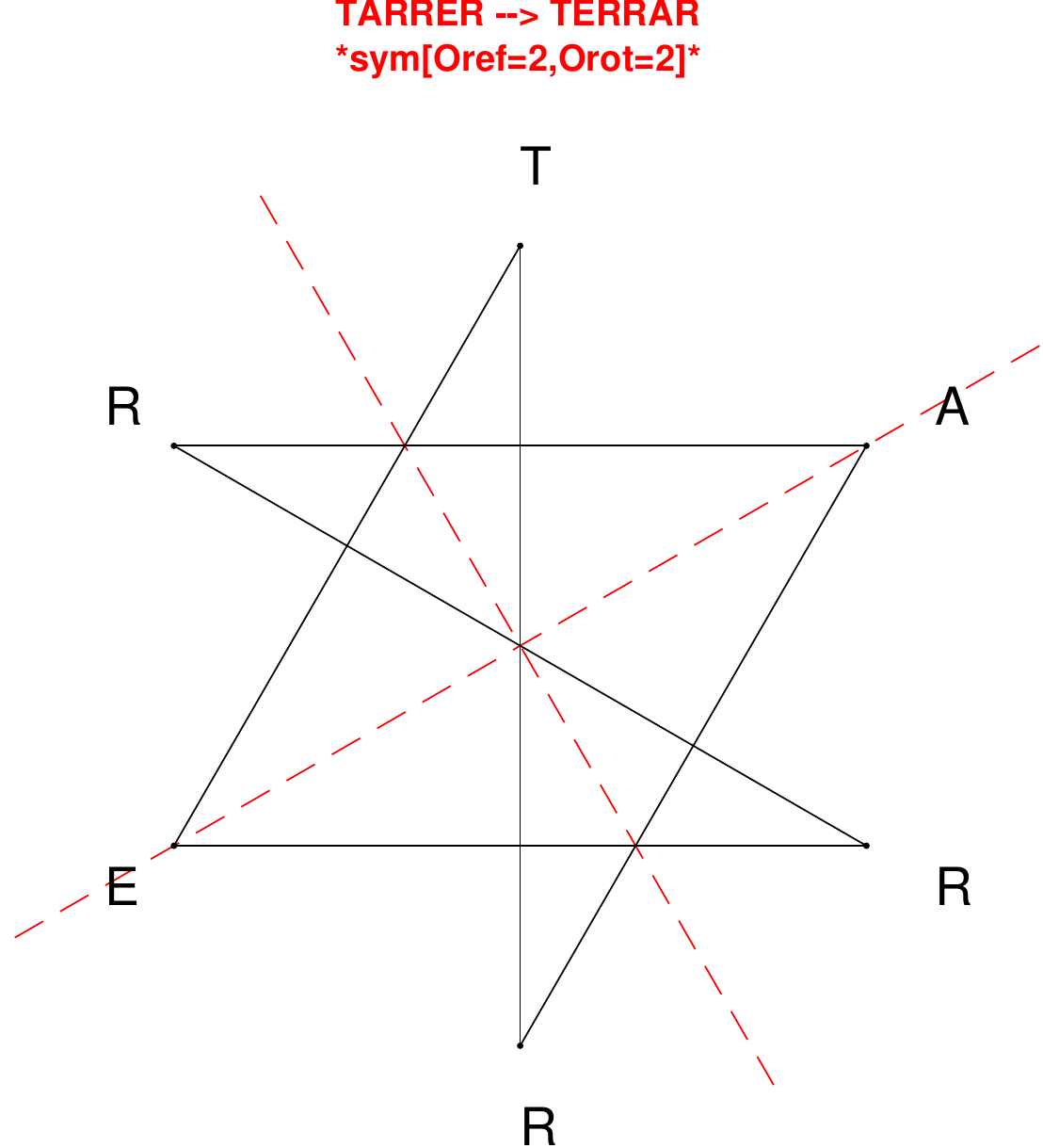}
\end{subfigure}
\end{figure}

\begin{figure}[H]
\centering
\begin{subfigure}[T]{0.19\textwidth}
\centering
\includegraphics[width=\textwidth]{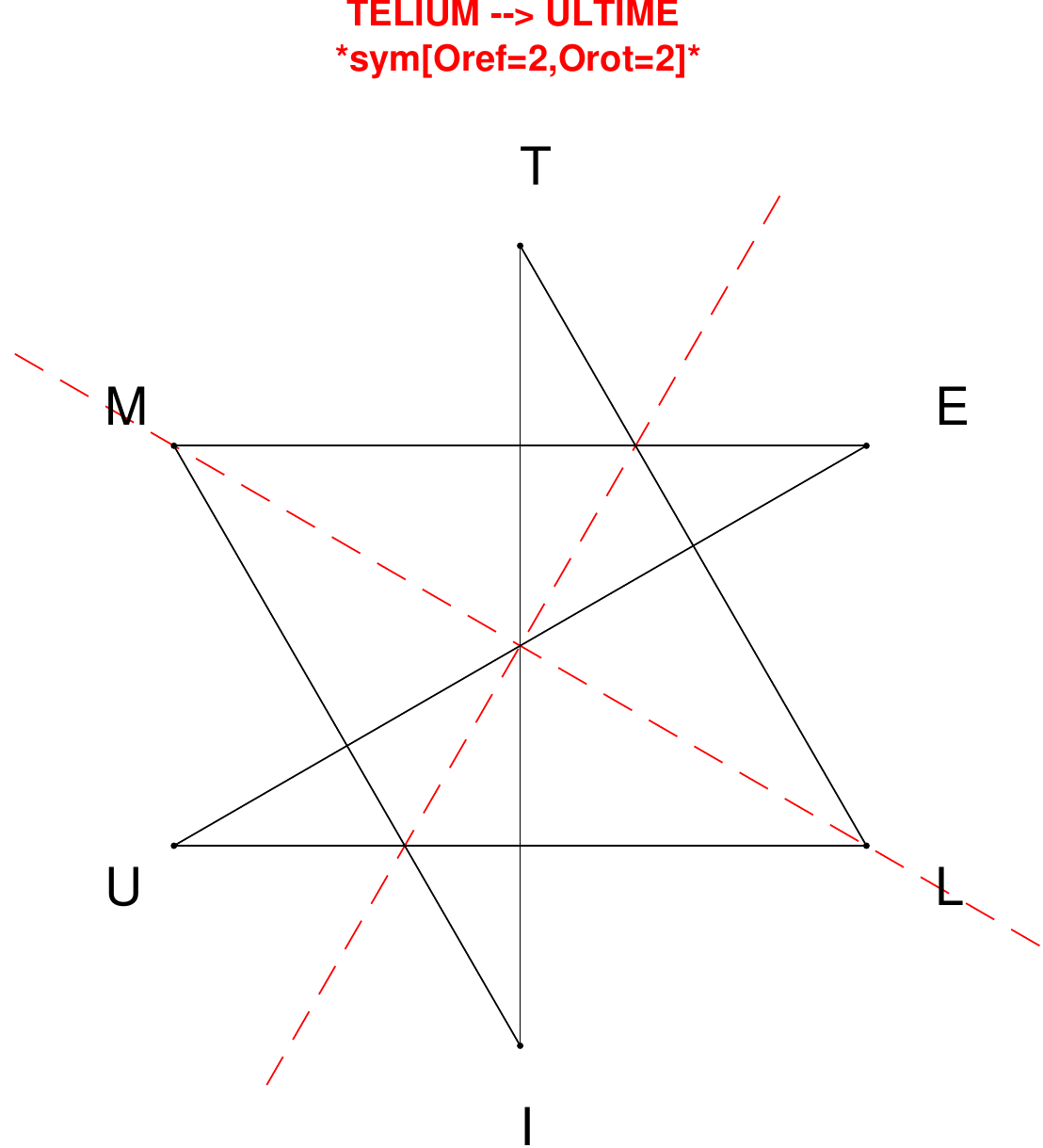}
\end{subfigure}
\hfill
\begin{subfigure}[T]{0.19\textwidth}
\centering
\includegraphics[width=\textwidth]{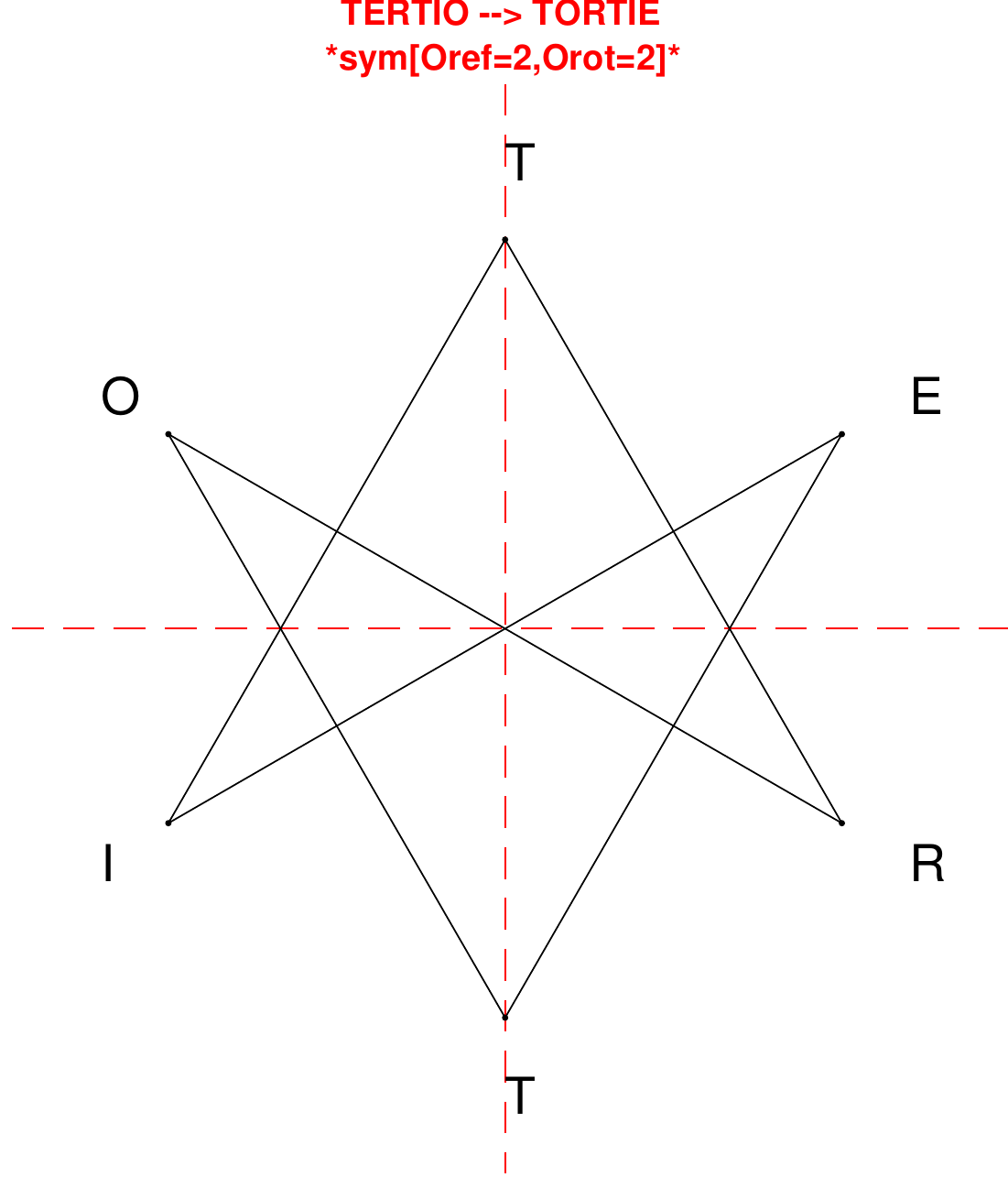}
\end{subfigure}
\hfill
\begin{subfigure}[T]{0.19\textwidth}
\centering
\includegraphics[width=\textwidth]{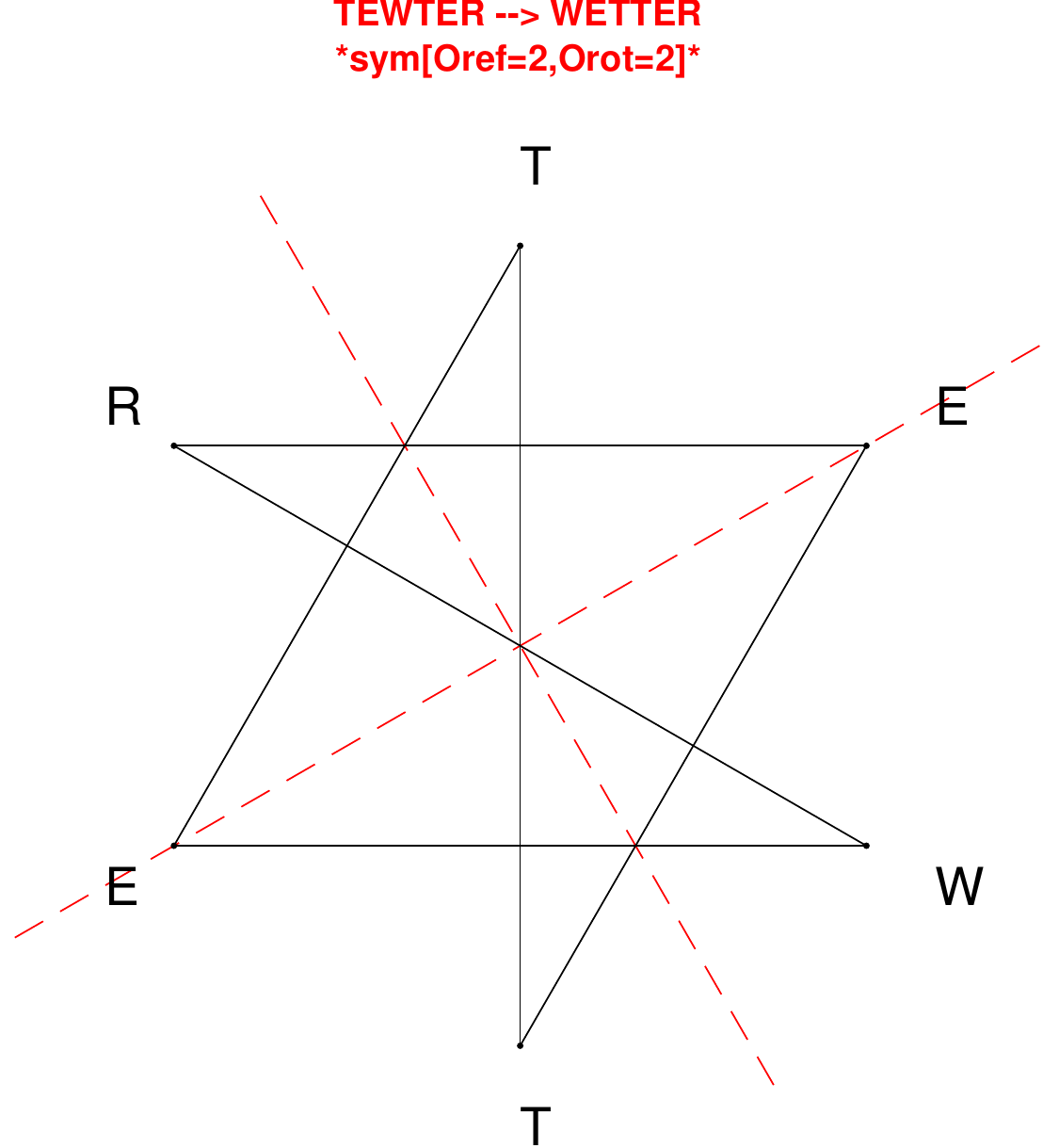}
\end{subfigure}
\hfill
\begin{subfigure}[T]{0.19\textwidth}
\centering
\includegraphics[width=\textwidth]{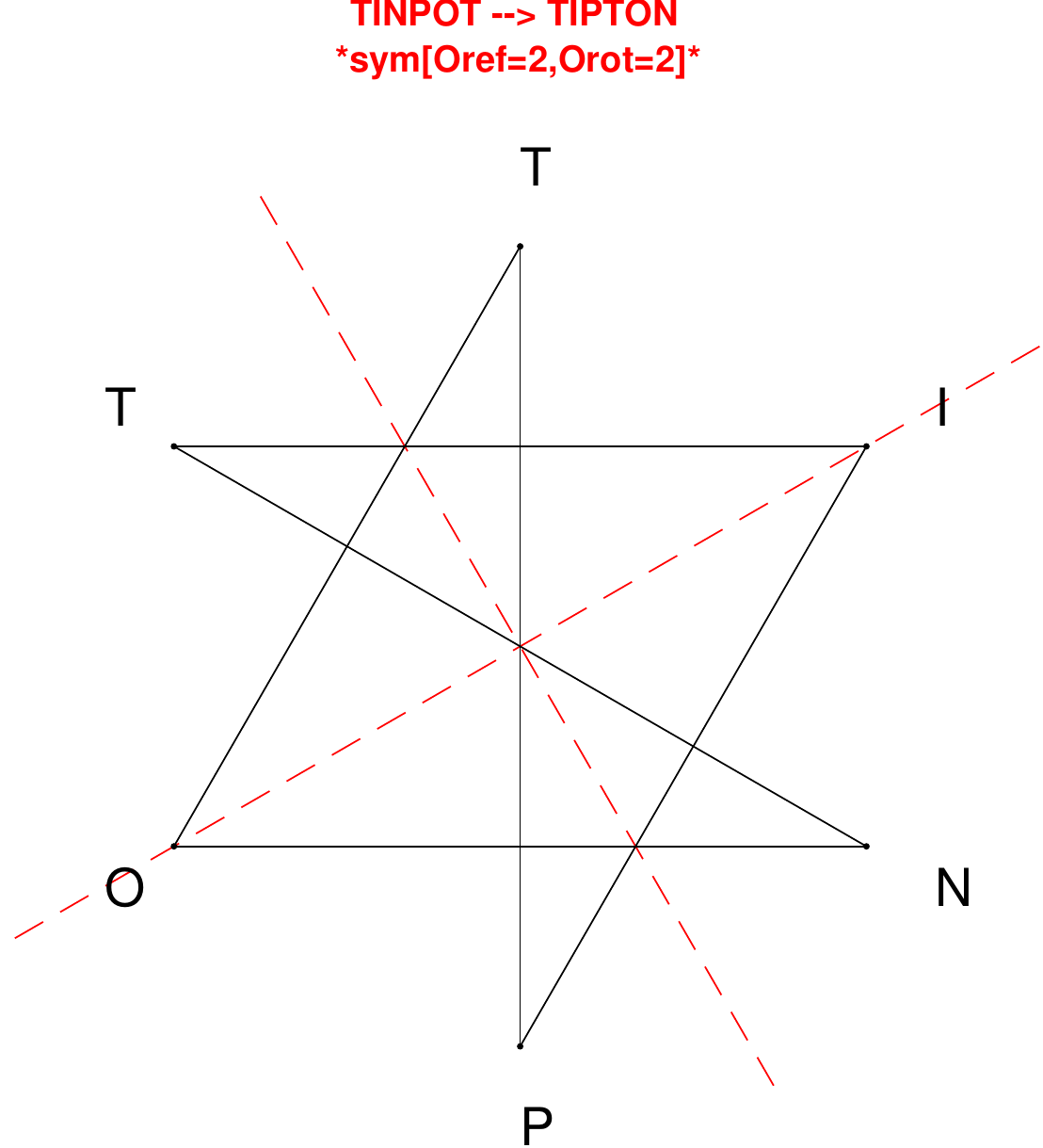}
\end{subfigure}
\hfill
\begin{subfigure}[T]{0.19\textwidth}
\centering
\includegraphics[width=\textwidth]{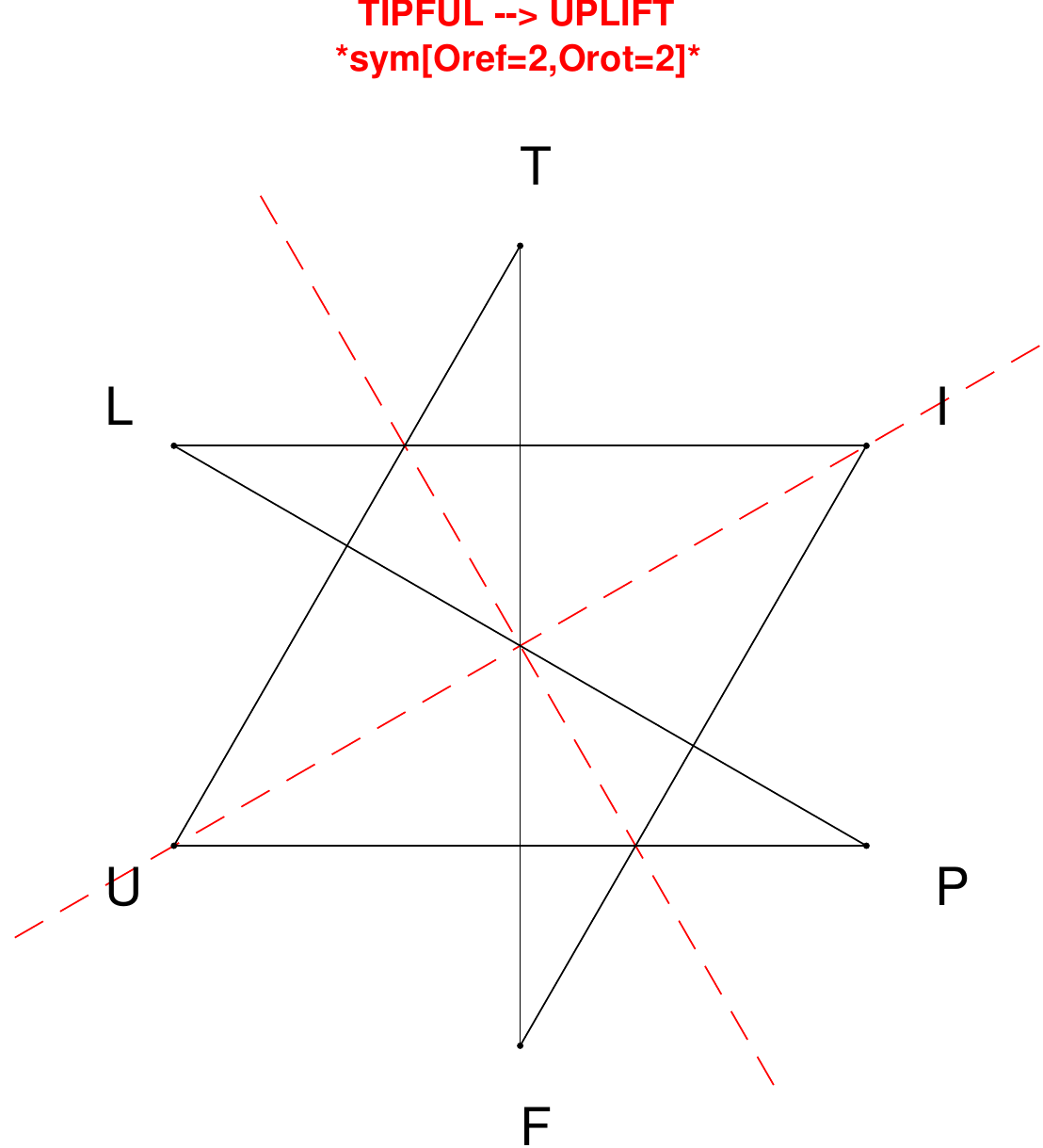}
\end{subfigure}
\end{figure}

\begin{figure}[H]
\centering
\begin{subfigure}[T]{0.19\textwidth}
\centering
\includegraphics[width=\textwidth]{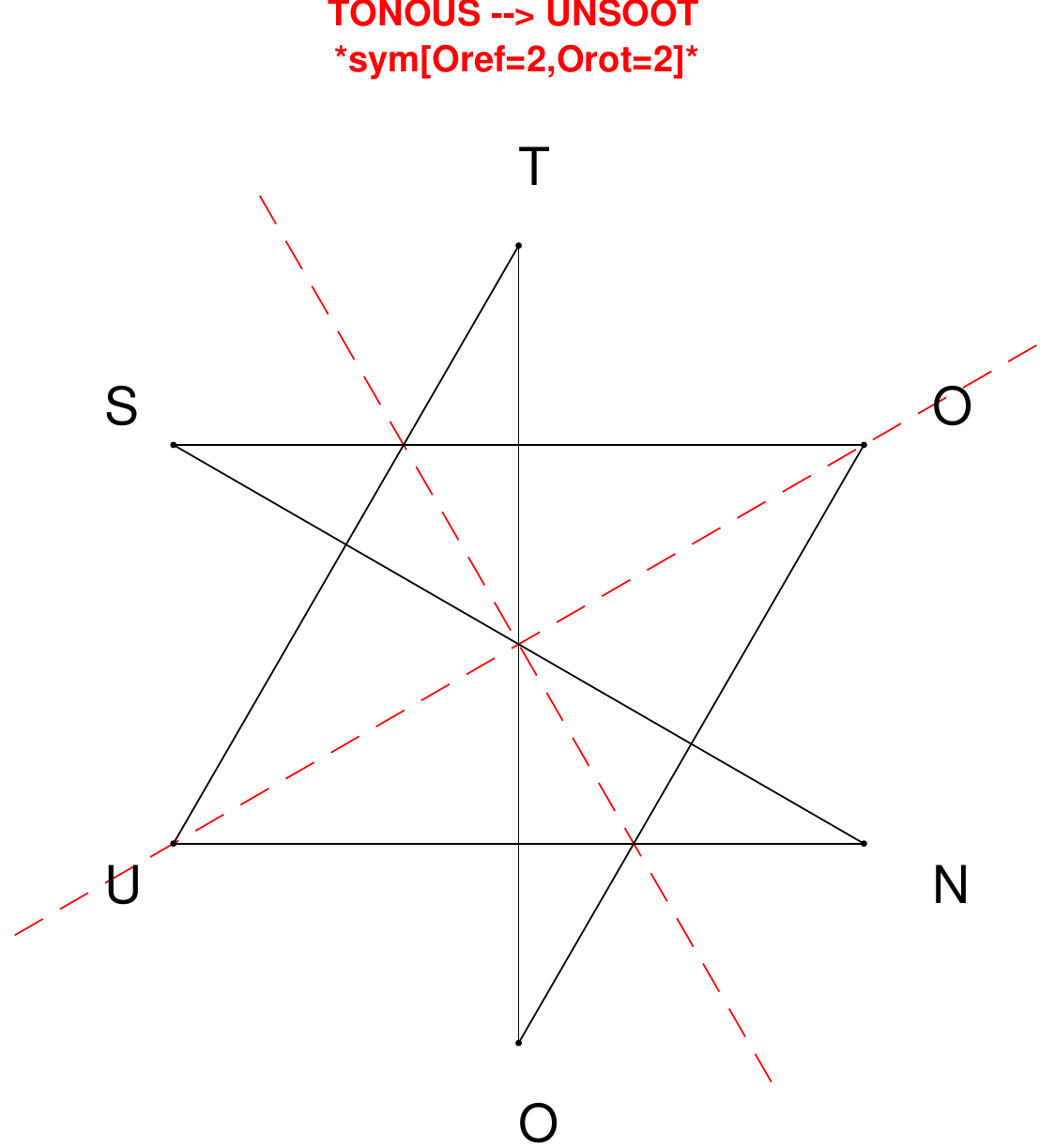}
\end{subfigure}
\hfill
\begin{subfigure}[T]{0.19\textwidth}
\centering
\includegraphics[width=\textwidth]{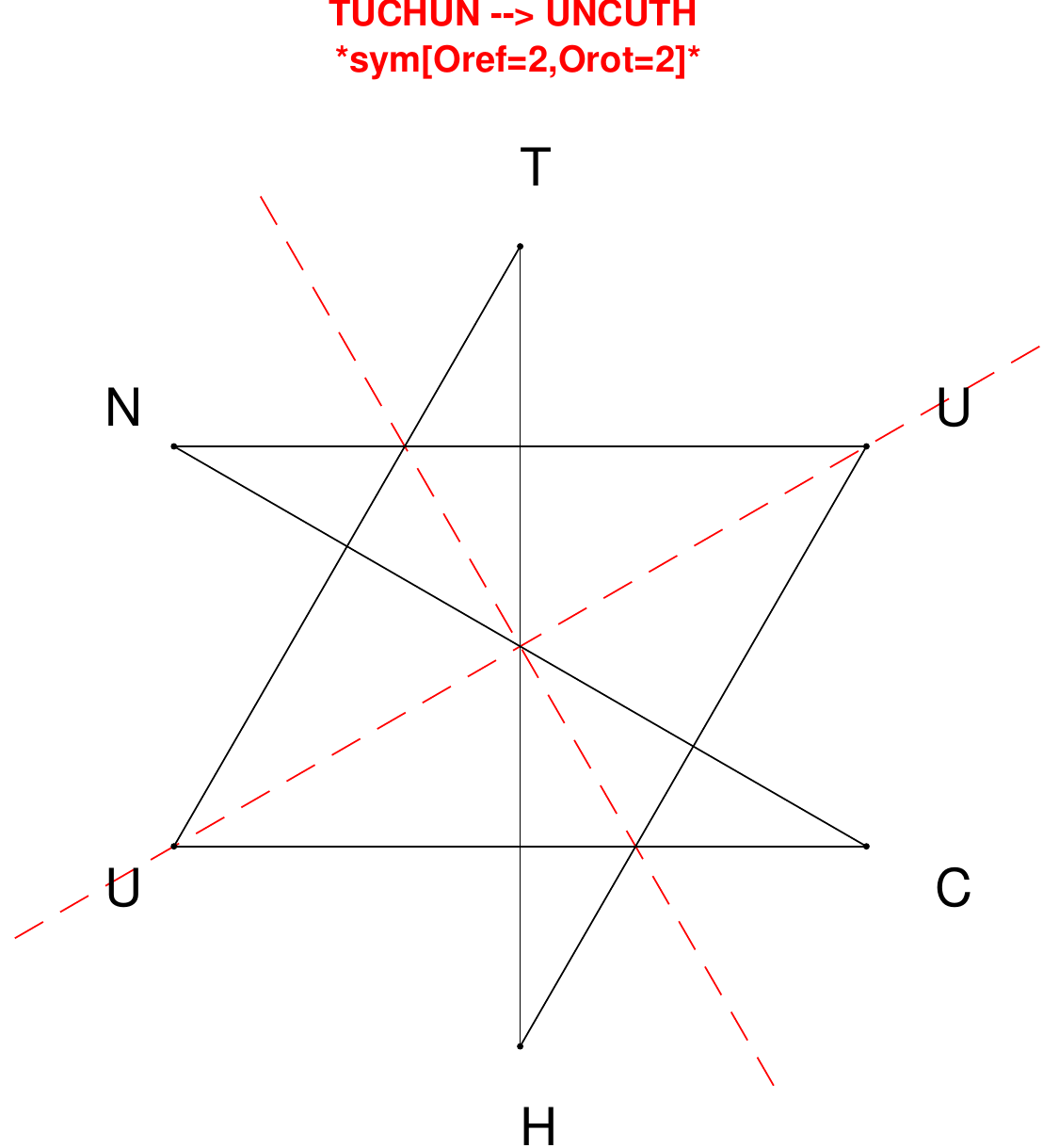}
\end{subfigure}
\hfill
\begin{subfigure}[T]{0.19\textwidth}
\centering
\includegraphics[width=\textwidth]{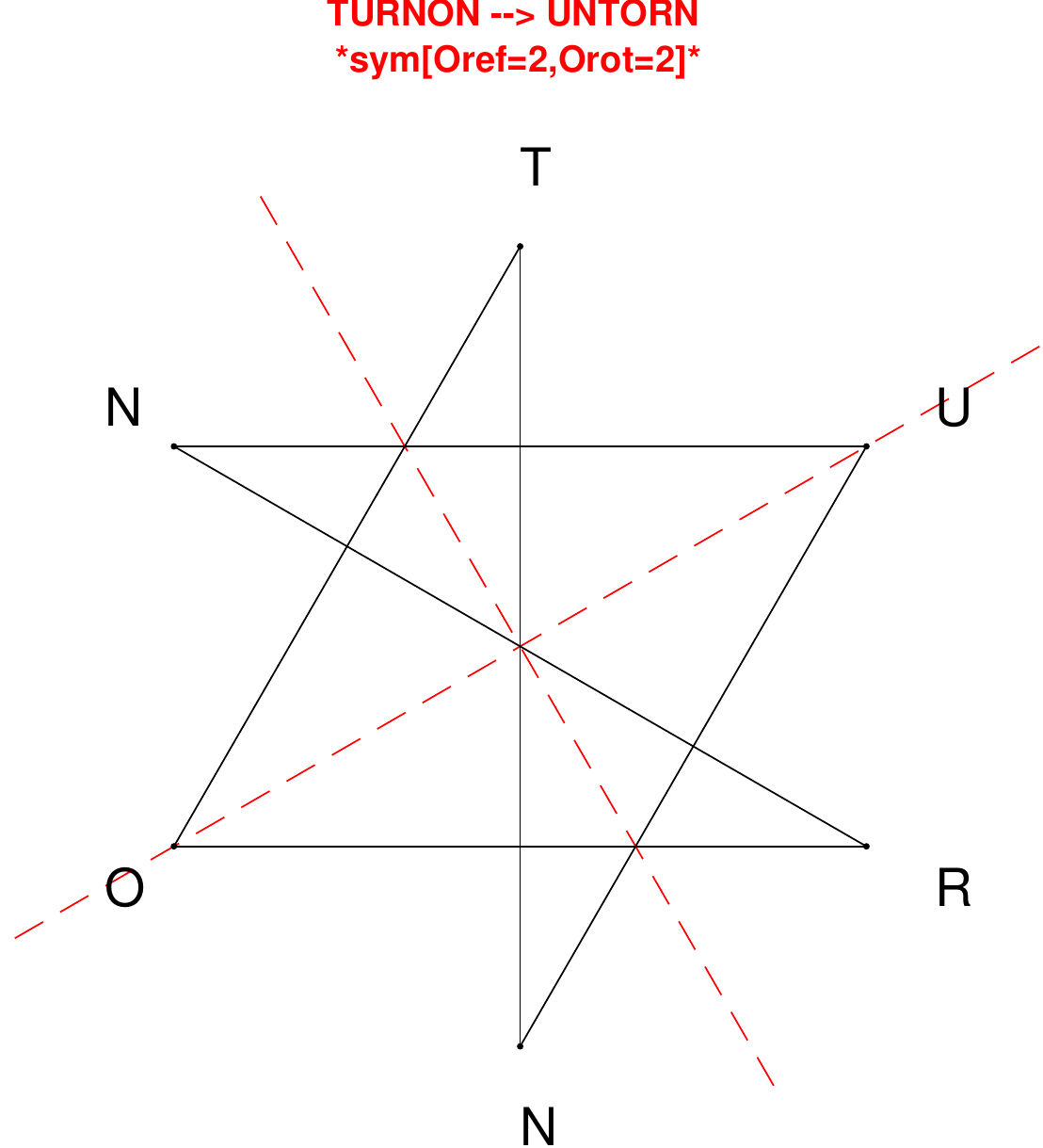}
\end{subfigure}
\hfill
\begin{subfigure}[T]{0.19\textwidth}
\centering
\includegraphics[width=\textwidth]{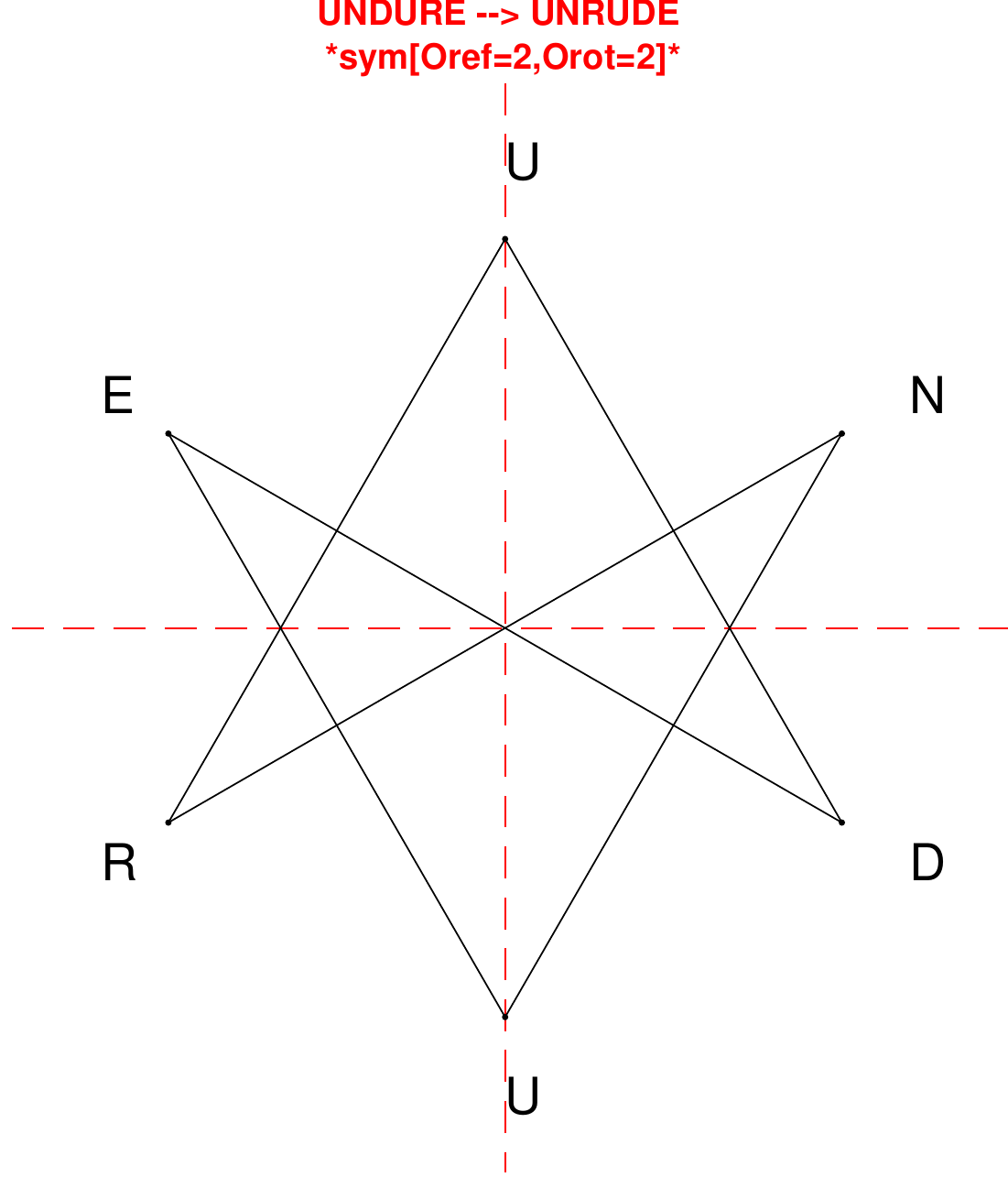}
\end{subfigure}
\hfill
\begin{subfigure}[T]{0.19\textwidth}
\centering
\includegraphics[width=\textwidth]{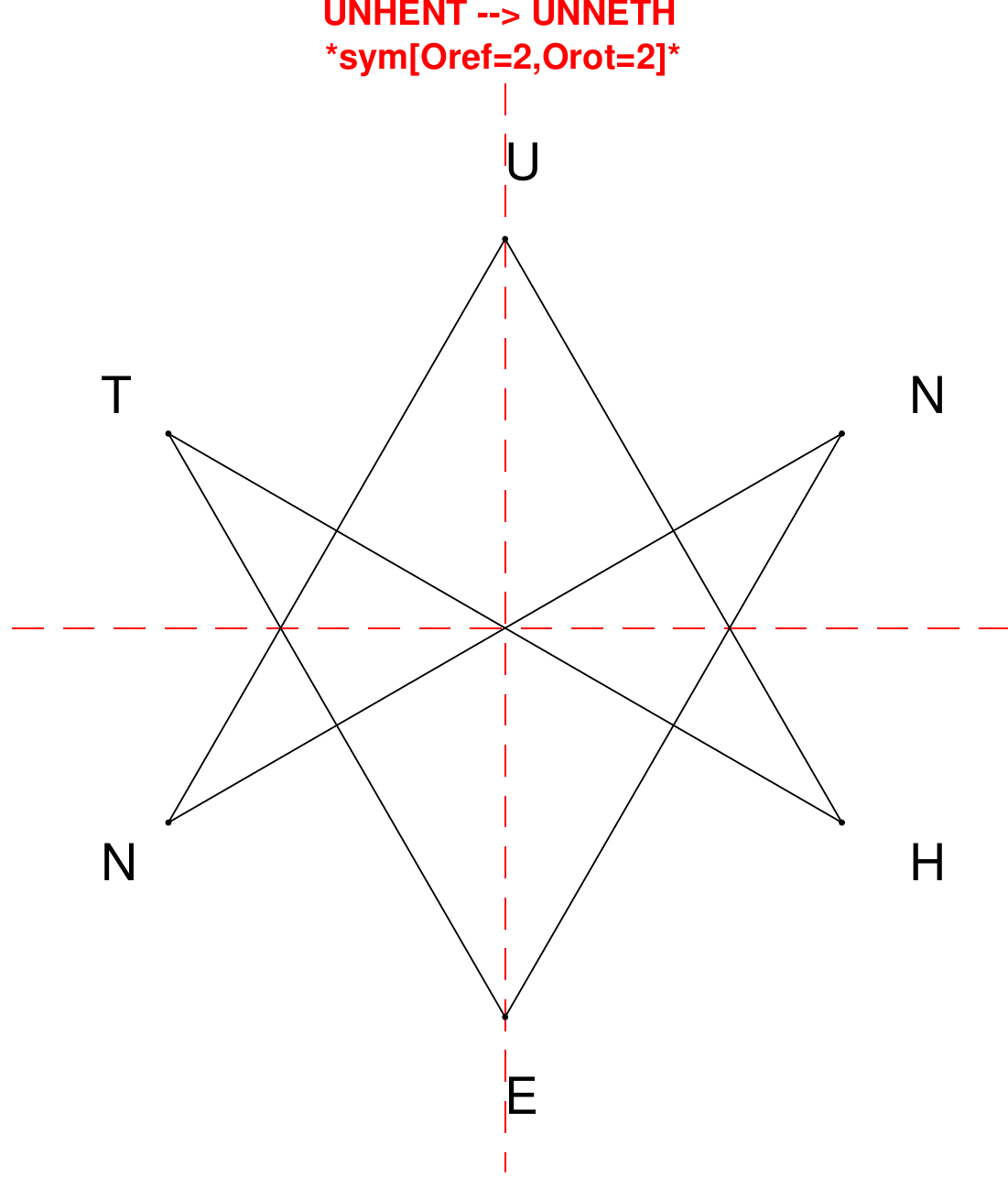}
\end{subfigure}
\end{figure}

\begin{figure}[H]
\centering
\begin{subfigure}[T]{0.19\textwidth}
\centering
\includegraphics[width=\textwidth]{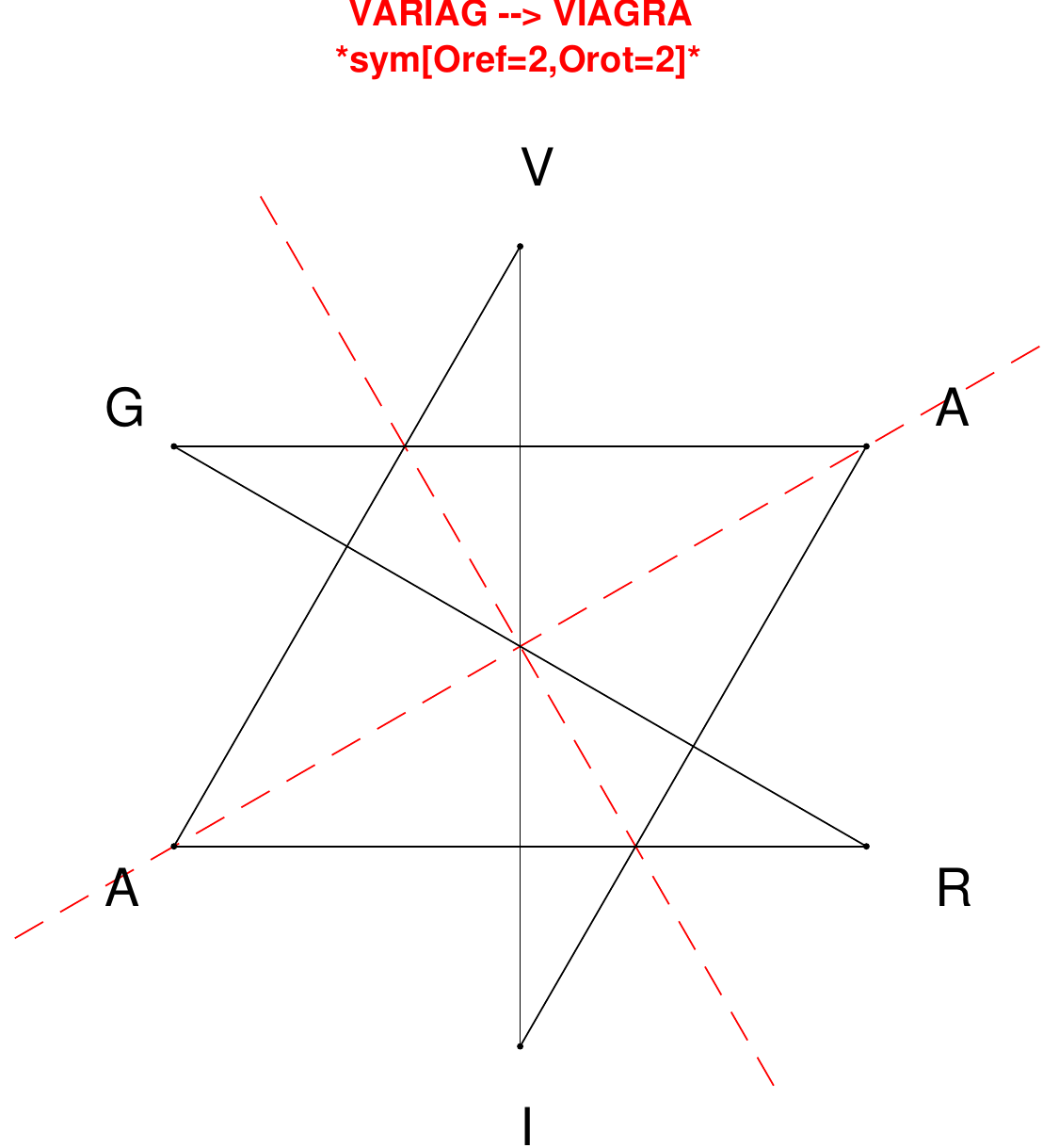}
\end{subfigure}
\hfill
\begin{subfigure}[T]{0.19\textwidth}
\centering
\includegraphics[width=\textwidth]{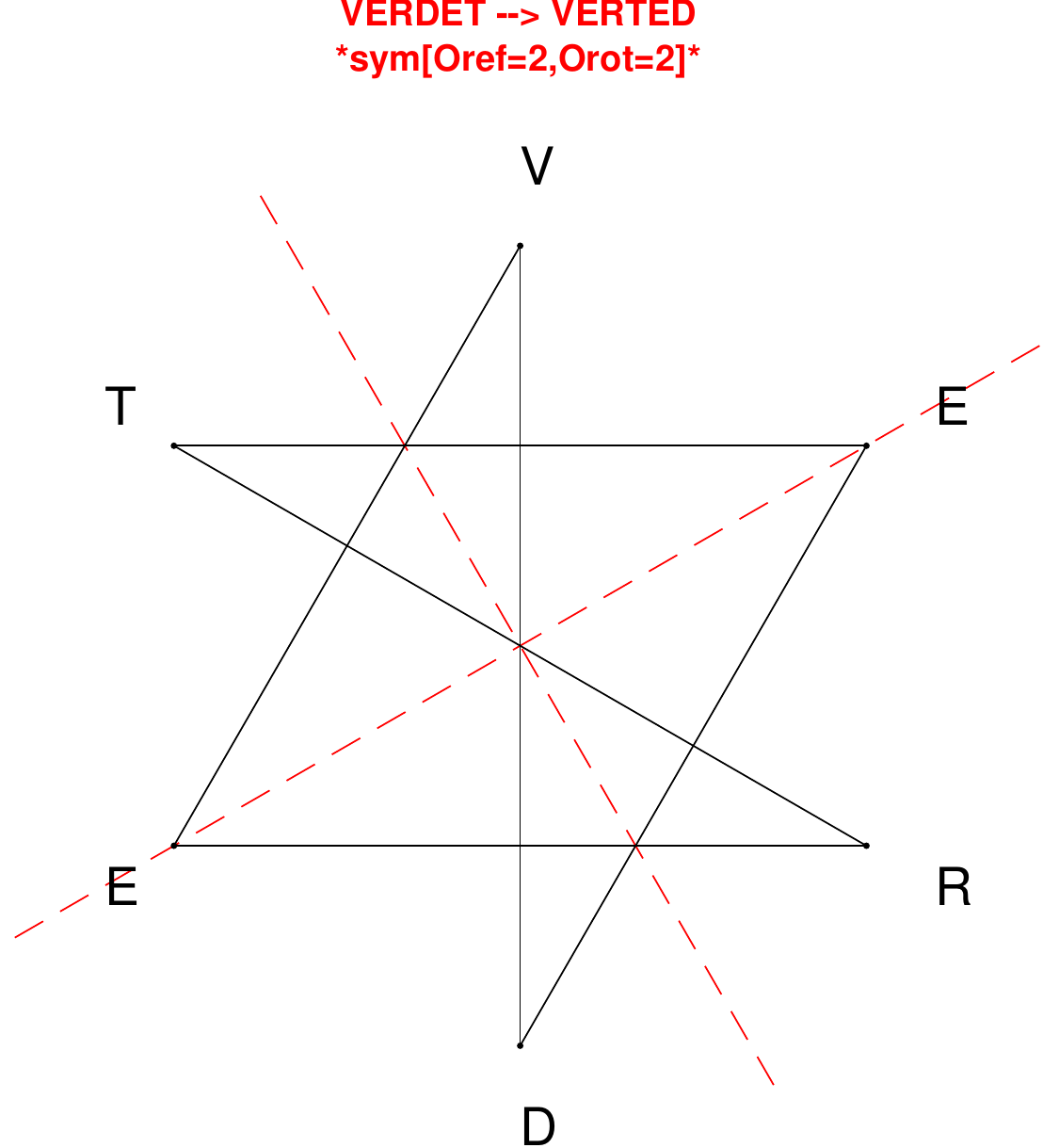}
\end{subfigure}
\hfill
\begin{subfigure}[T]{0.19\textwidth}
\centering
\includegraphics[width=\textwidth]{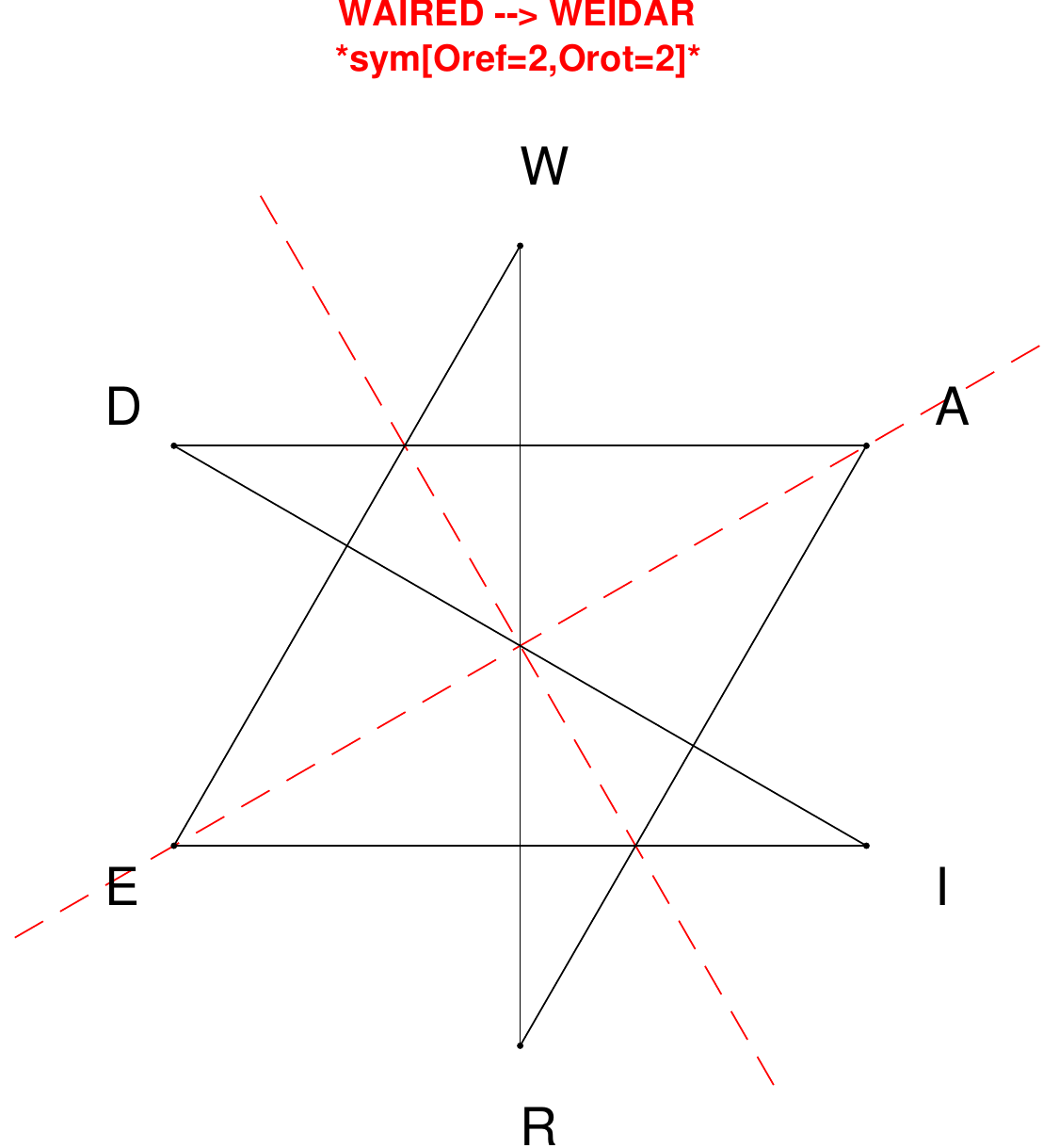}
\end{subfigure}
\hfill
\begin{subfigure}[T]{0.19\textwidth}
\centering
\includegraphics[width=\textwidth]{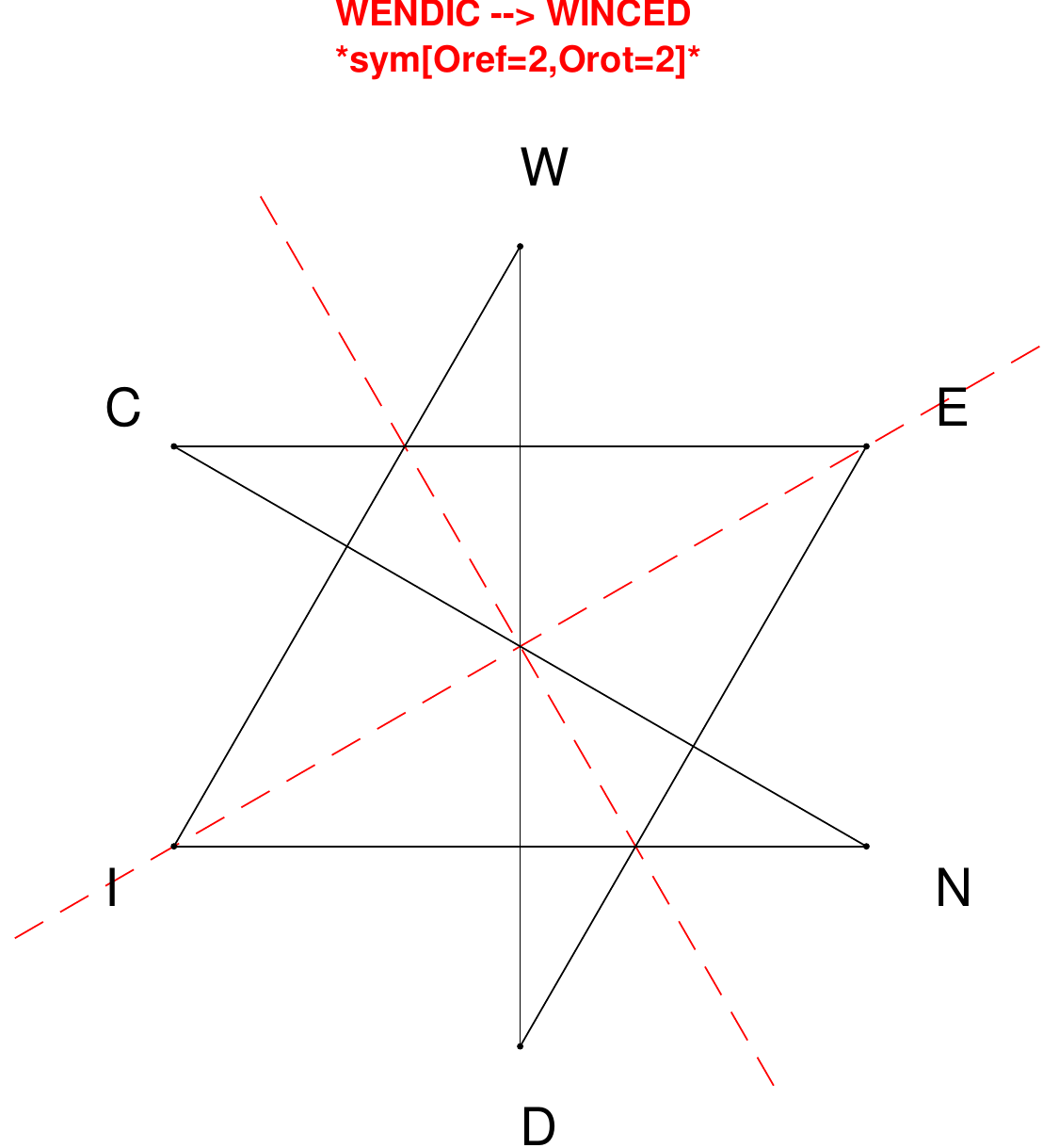}
\end{subfigure}
\hfill
\begin{subfigure}[T]{0.19\textwidth}
\centering
\includegraphics[width=\textwidth]{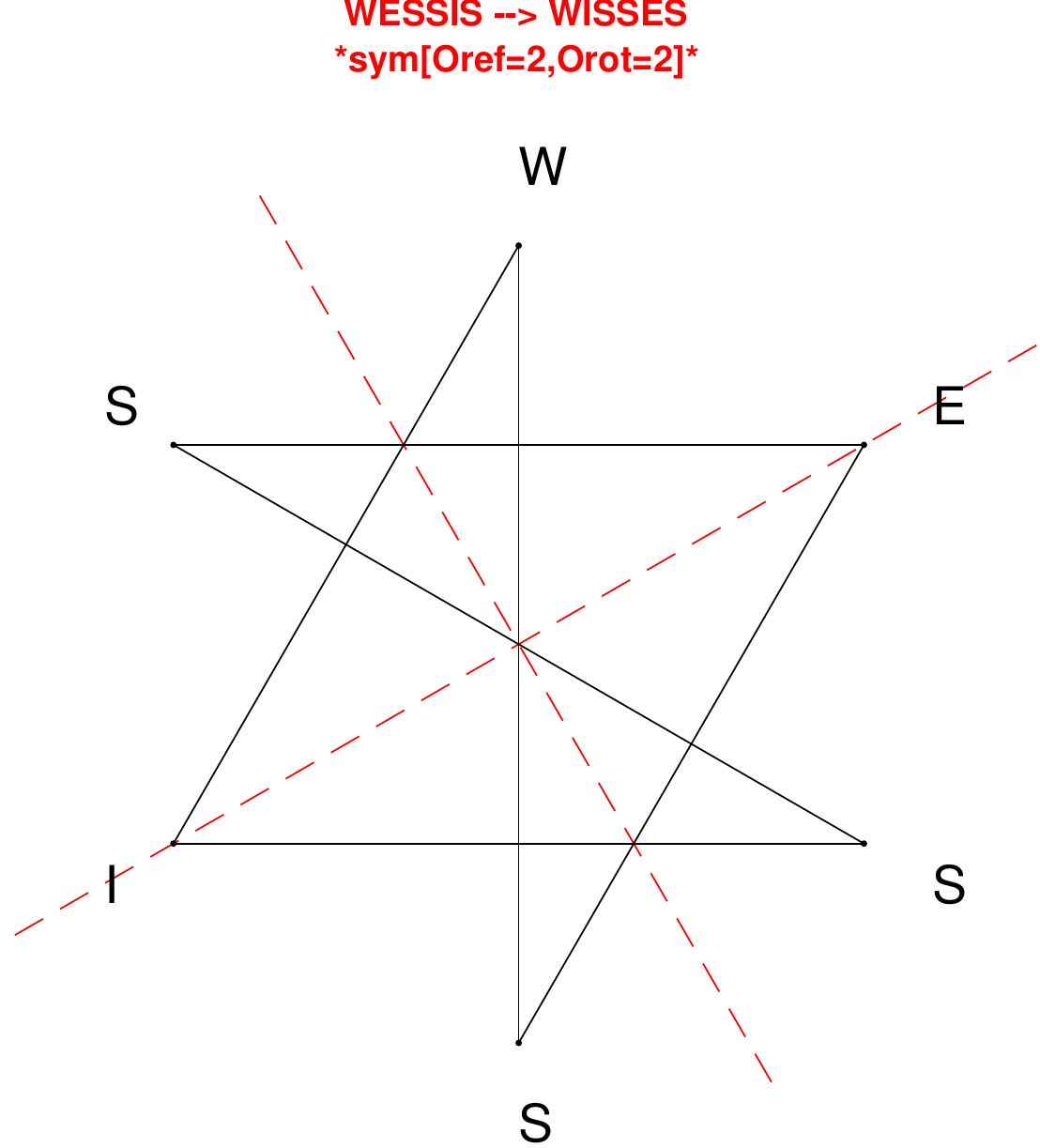}
\end{subfigure}
\end{figure}

\begin{figure}[H]
\centering
\begin{subfigure}[T]{0.19\textwidth}
\centering
\includegraphics[width=\textwidth]{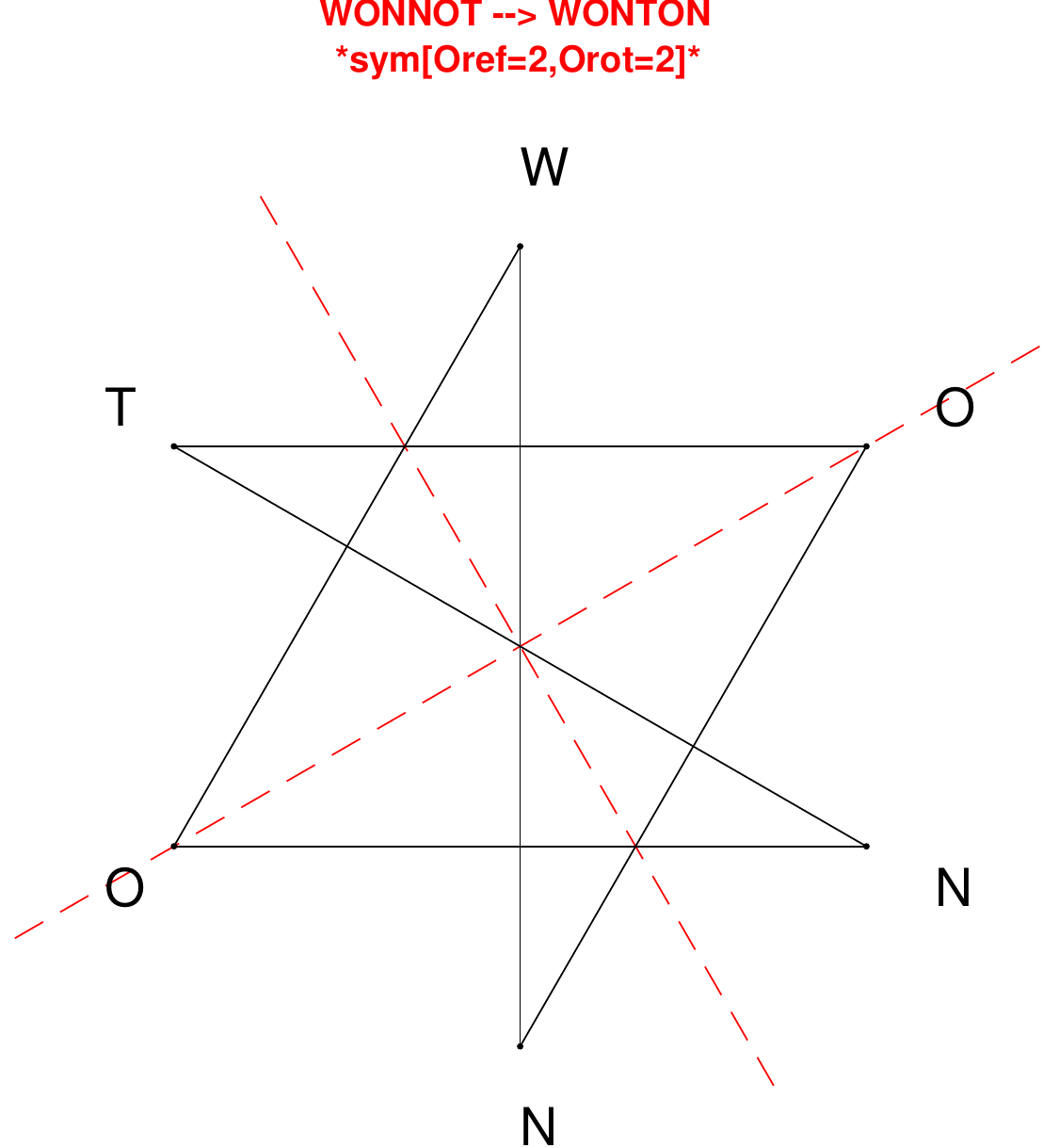}
\end{subfigure}
\hfill
\begin{subfigure}[T]{0.19\textwidth}
\centering
\includegraphics[width=\textwidth]{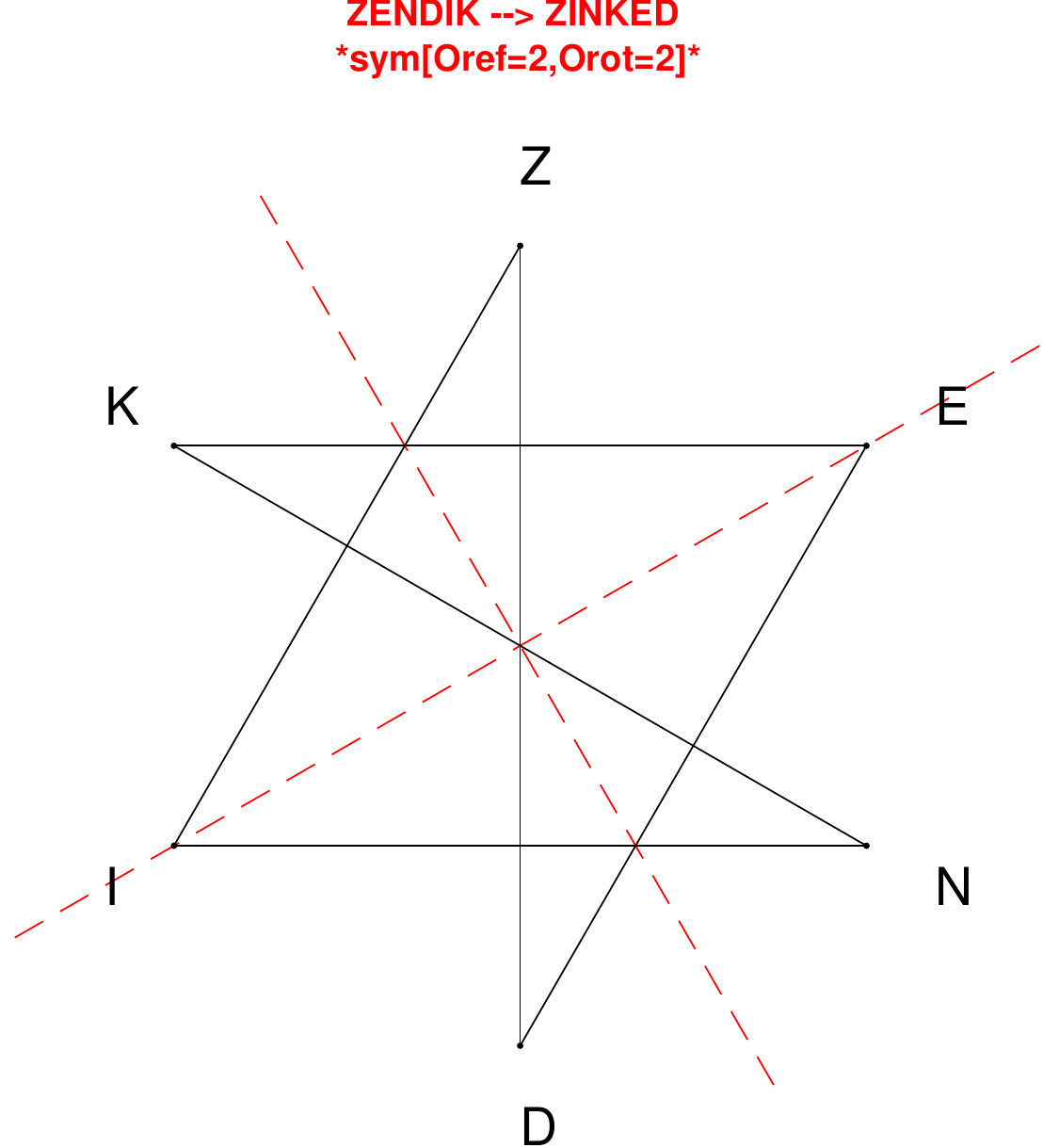}
\end{subfigure}
\hfill
\end{figure}

%%%%%%%%%%%%%%%%%%
\clearpage
\subsection{Star Anagrams $N = 5$}
Finally, all of the stars for $N=5$ are perfect.

\begin{figure}[H]
\centering
\begin{subfigure}[T]{0.19\textwidth}
\centering
\includegraphics[width=\textwidth]{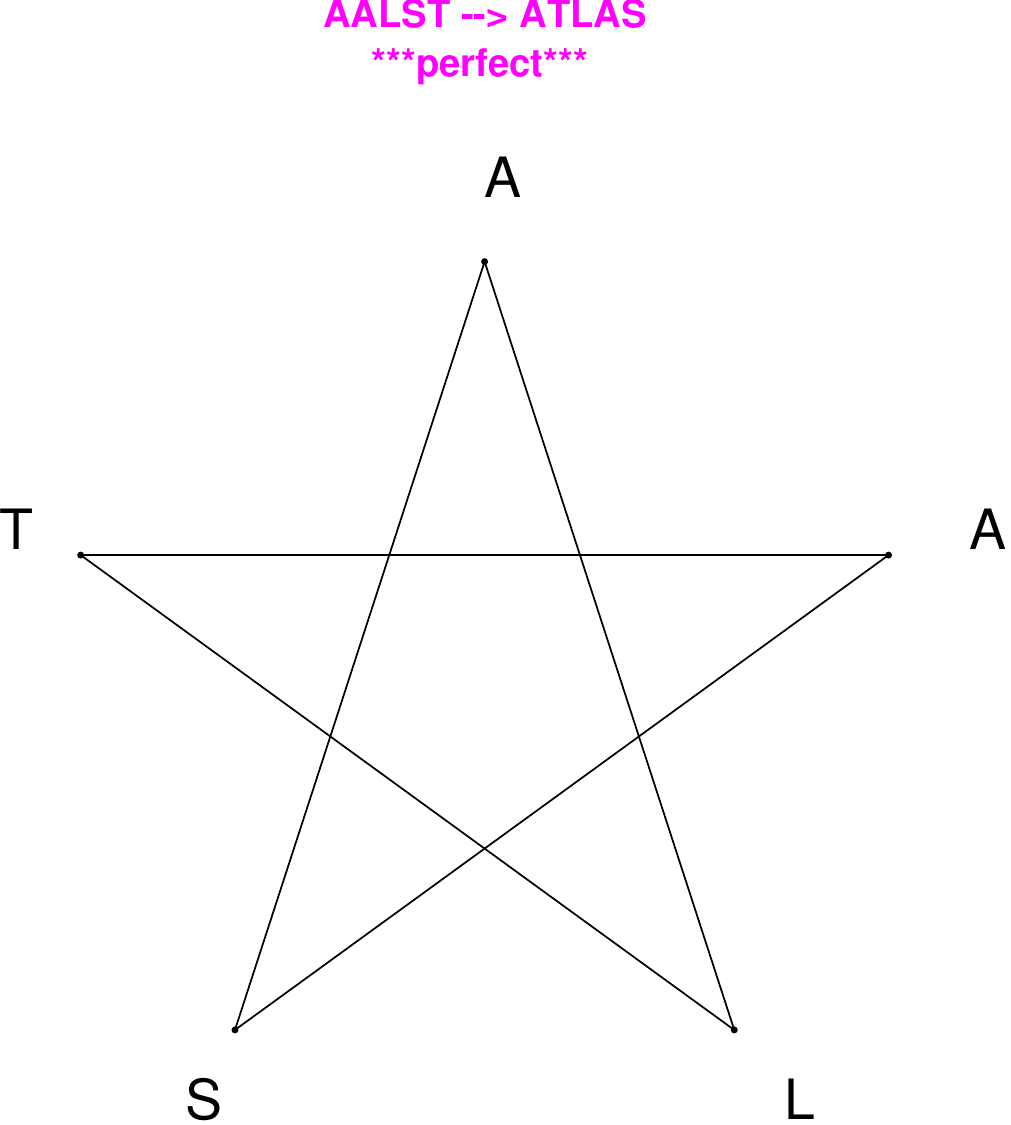}
\end{subfigure}
\hfill
\begin{subfigure}[T]{0.19\textwidth}
\centering
\includegraphics[width=\textwidth]{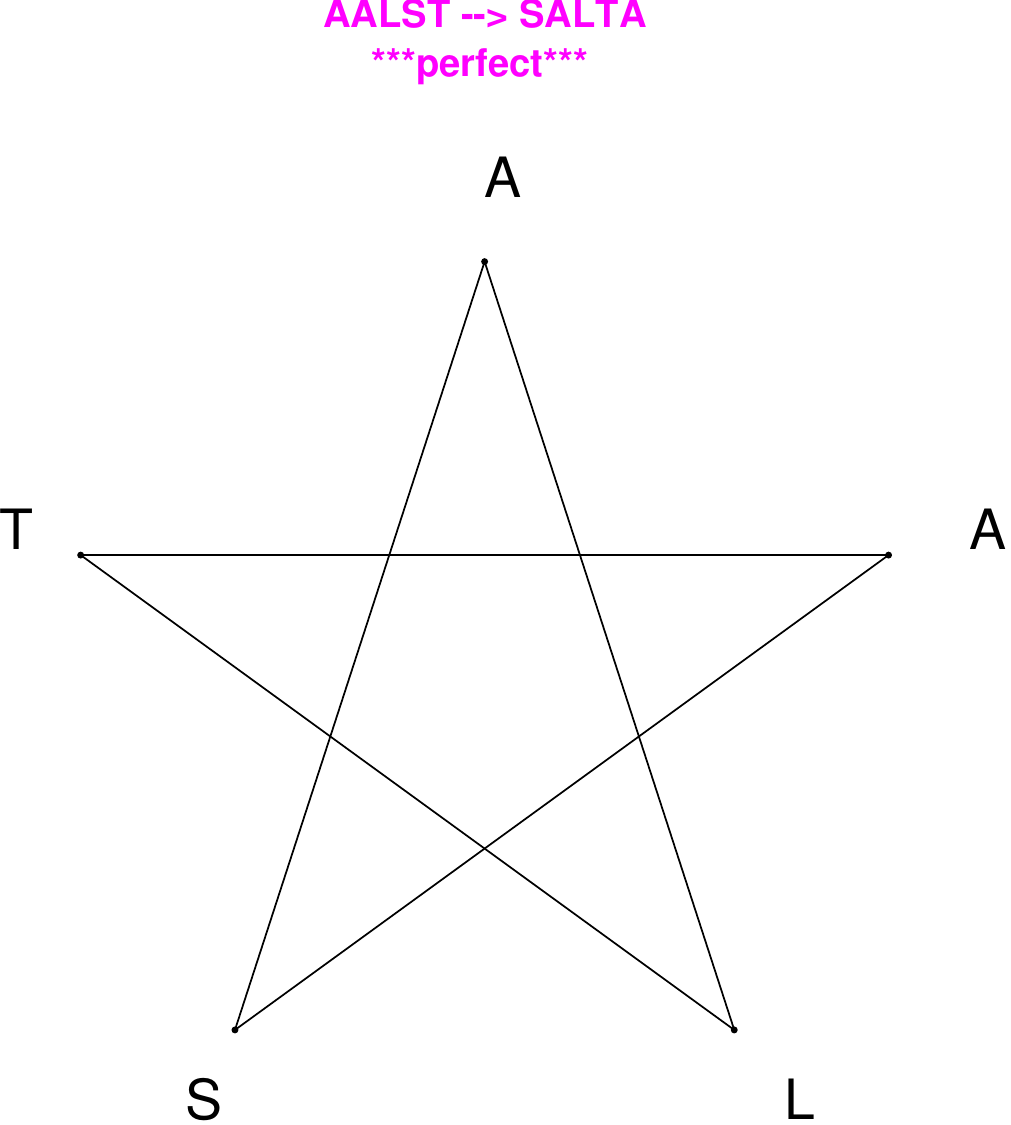}
\end{subfigure}
\hfill
\begin{subfigure}[T]{0.19\textwidth}
\centering
\includegraphics[width=\textwidth]{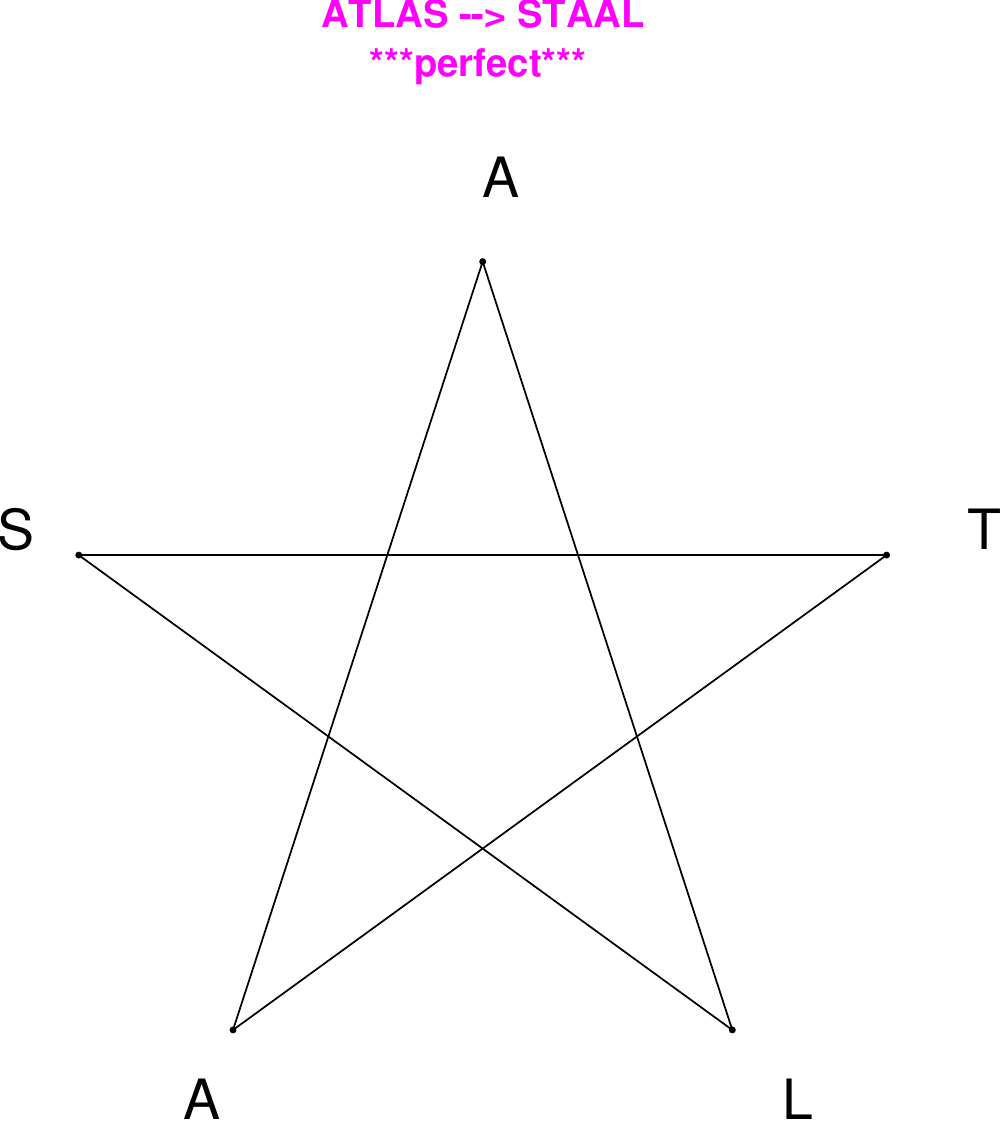}
\end{subfigure}
\hfill
\begin{subfigure}[T]{0.19\textwidth}
\centering
\includegraphics[width=\textwidth]{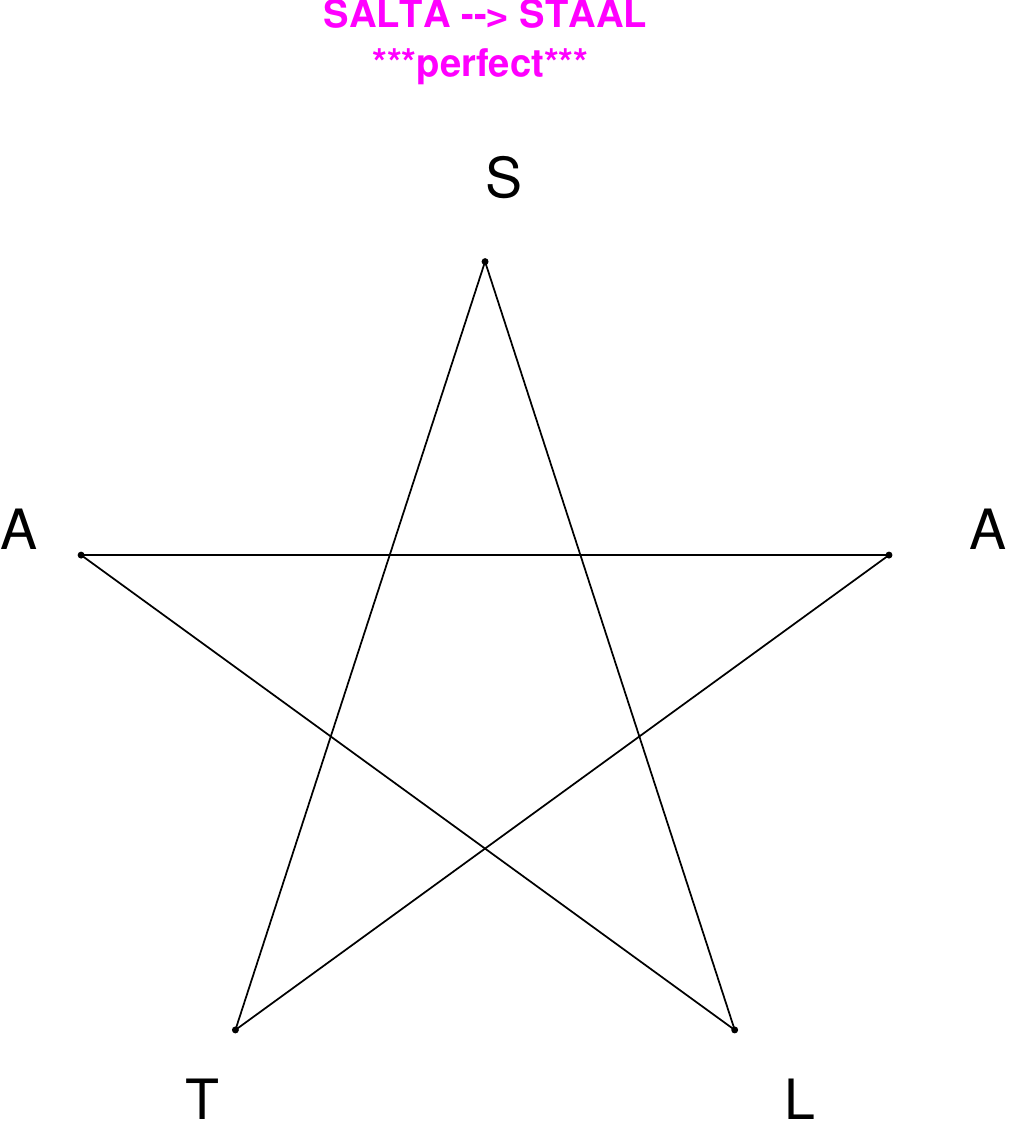}
\end{subfigure}
\hfill
\begin{subfigure}[T]{0.19\textwidth}
\centering
\includegraphics[width=\textwidth]{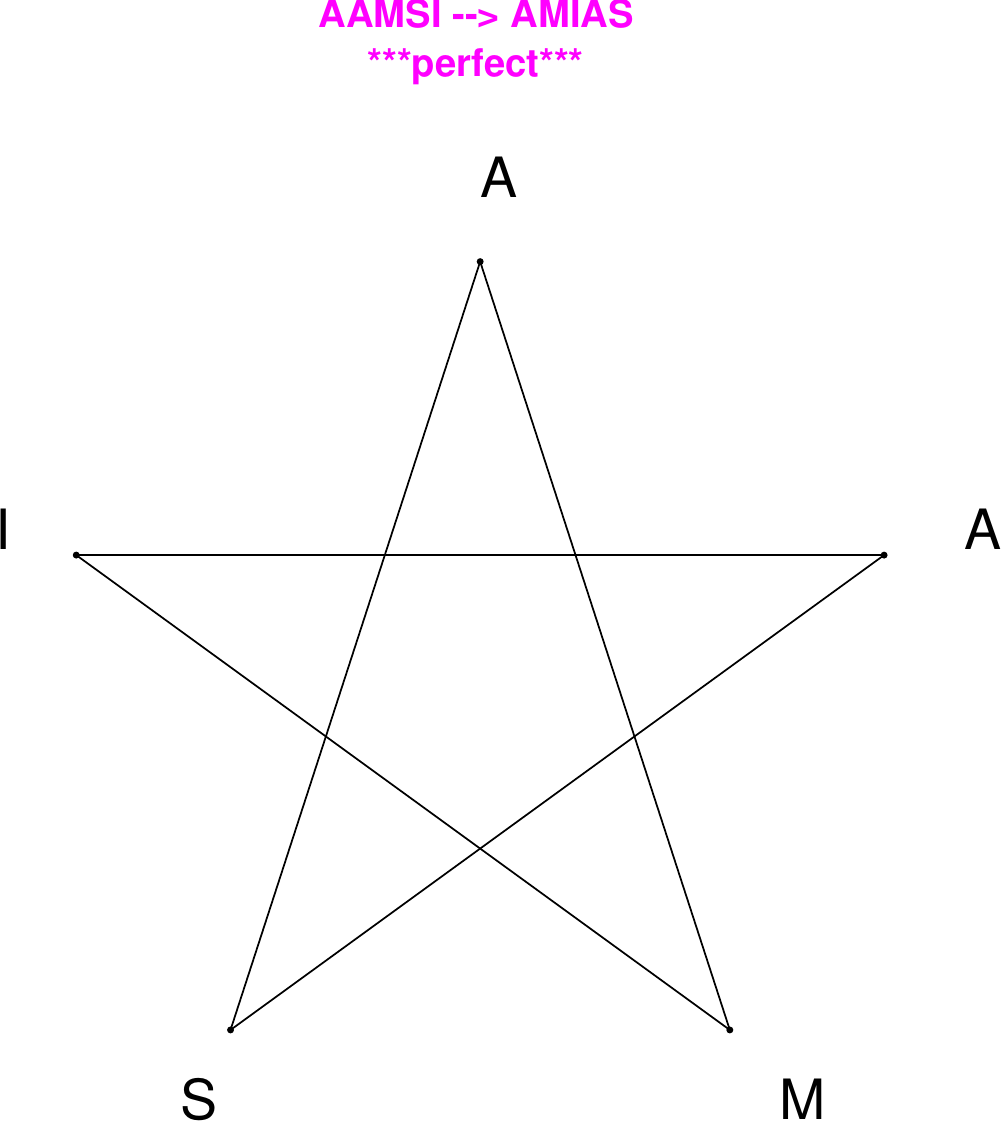}
\end{subfigure}
\end{figure}

\begin{figure}[H]
\centering
\begin{subfigure}[T]{0.19\textwidth}
\centering
\includegraphics[width=\textwidth]{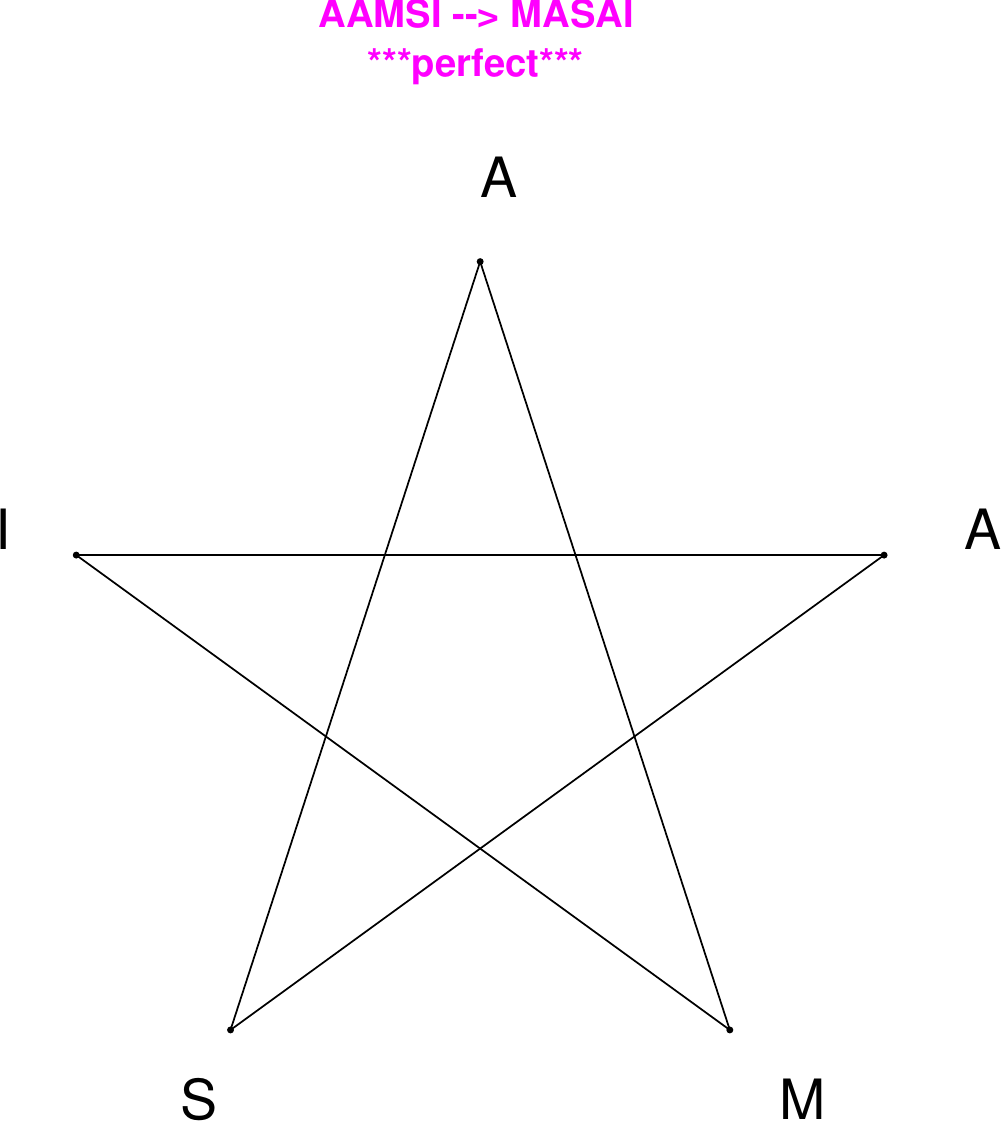}
\end{subfigure}
\hfill
\begin{subfigure}[T]{0.19\textwidth}
\centering
\includegraphics[width=\textwidth]{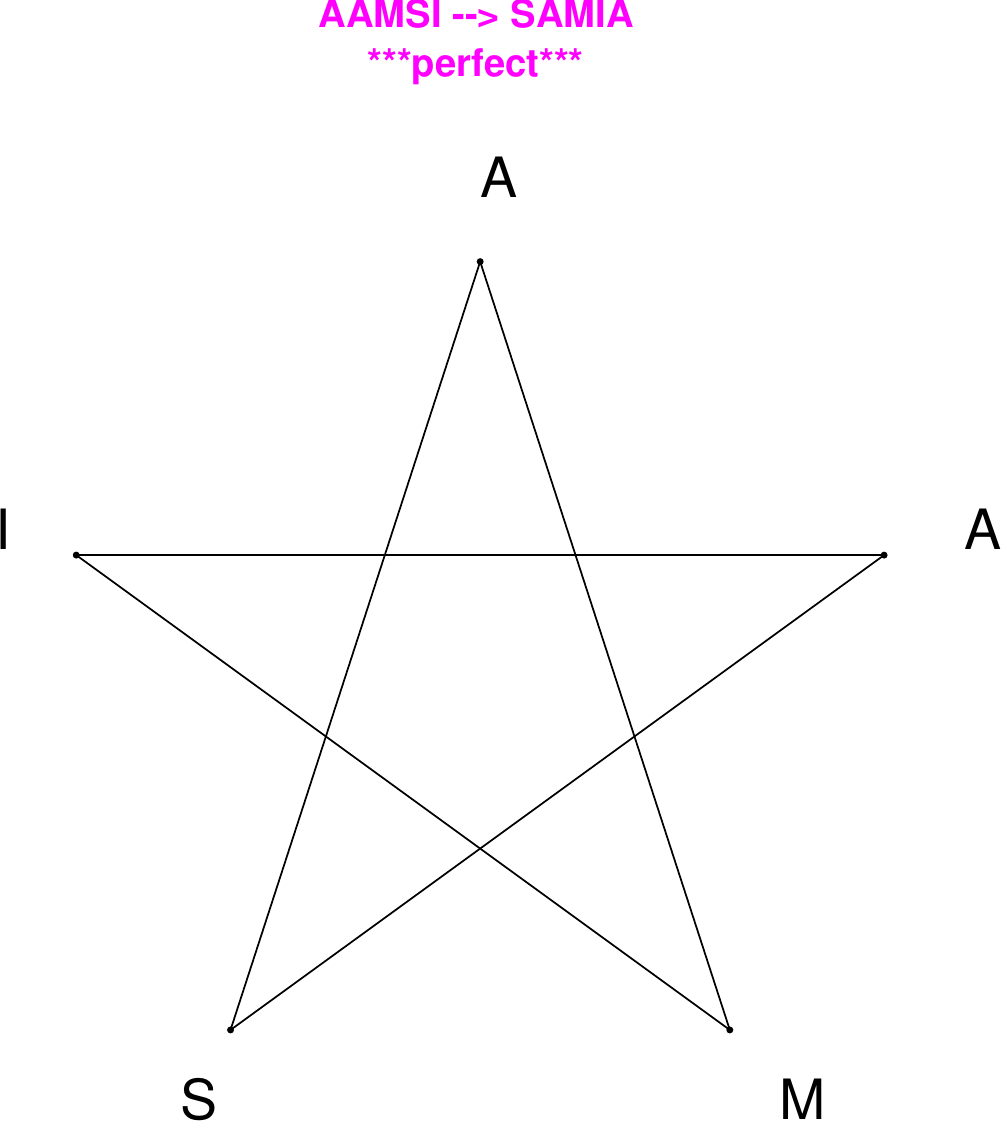}
\end{subfigure}
\hfill
\begin{subfigure}[T]{0.19\textwidth}
\centering
\includegraphics[width=\textwidth]{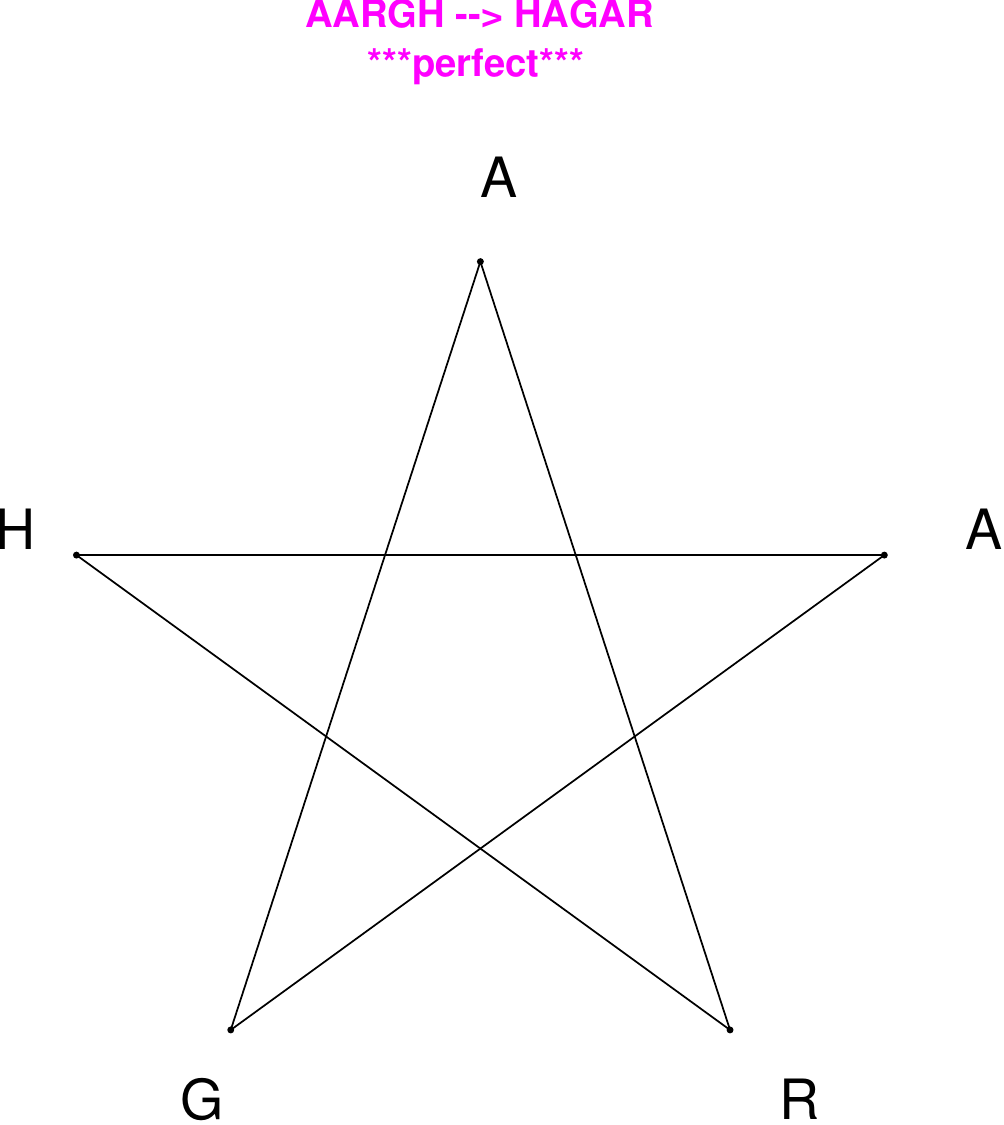}
\end{subfigure}
\hfill
\begin{subfigure}[T]{0.19\textwidth}
\centering
\includegraphics[width=\textwidth]{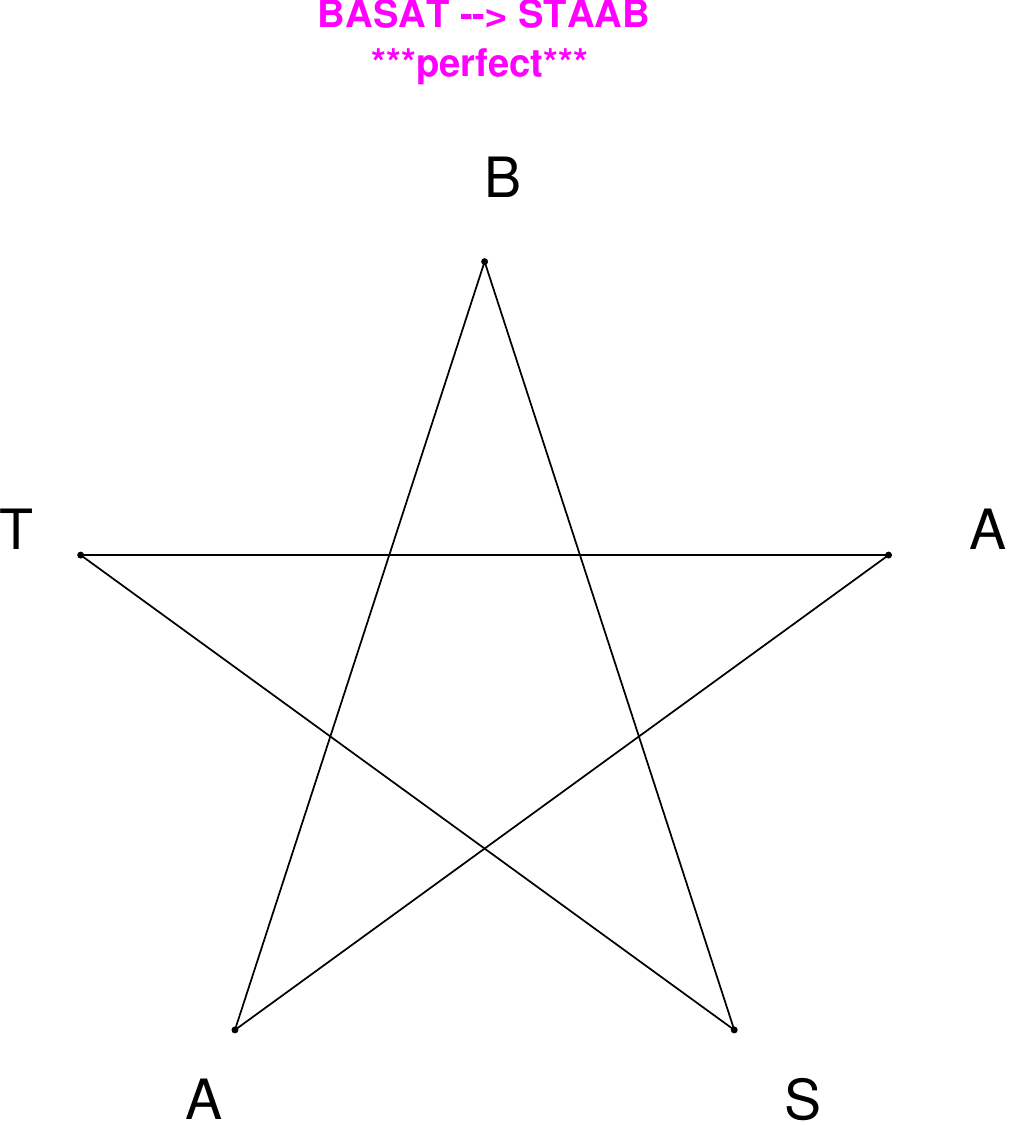}
\end{subfigure}
\hfill
\begin{subfigure}[T]{0.19\textwidth}
\centering
\includegraphics[width=\textwidth]{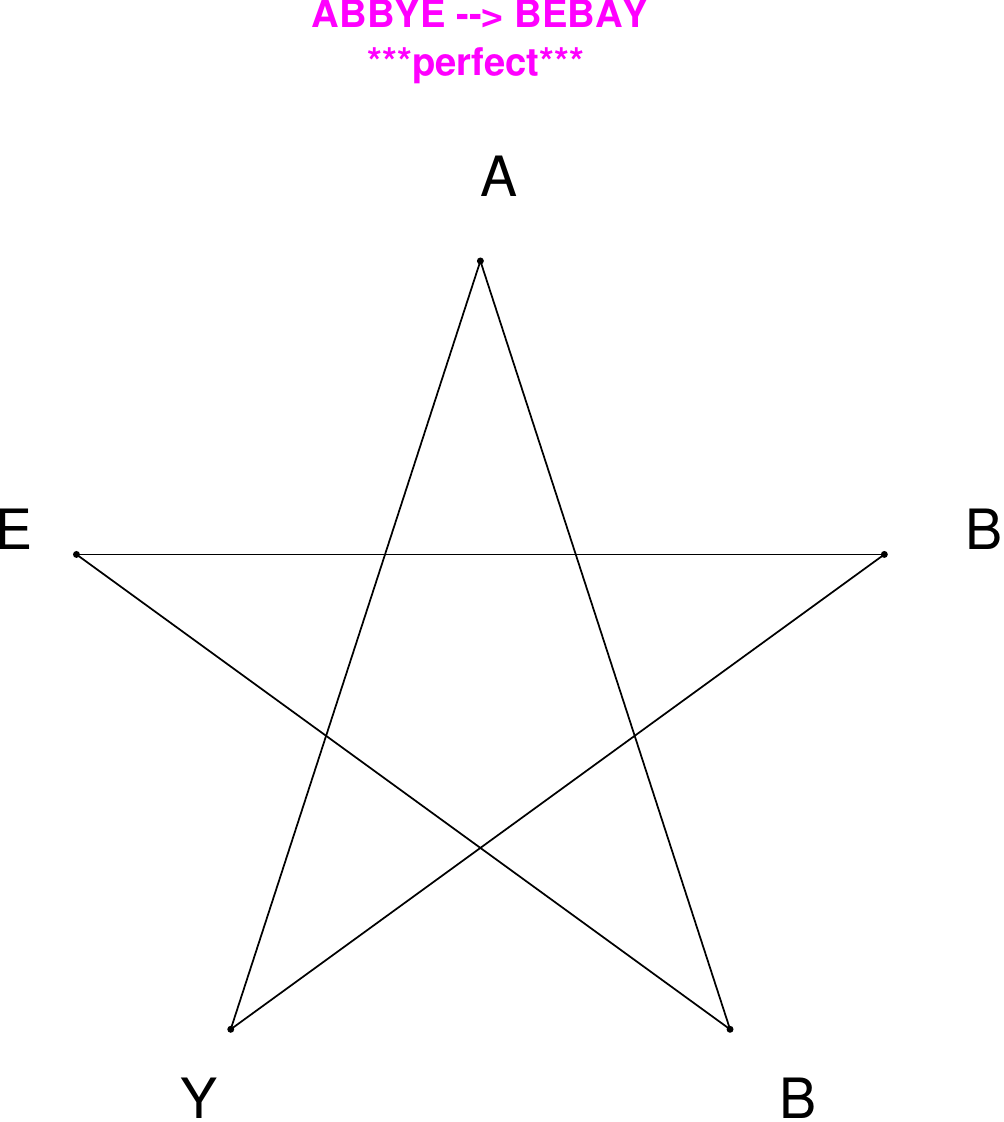}
\end{subfigure}
\end{figure}

\begin{figure}[H]
\centering
\begin{subfigure}[T]{0.19\textwidth}
\centering
\includegraphics[width=\textwidth]{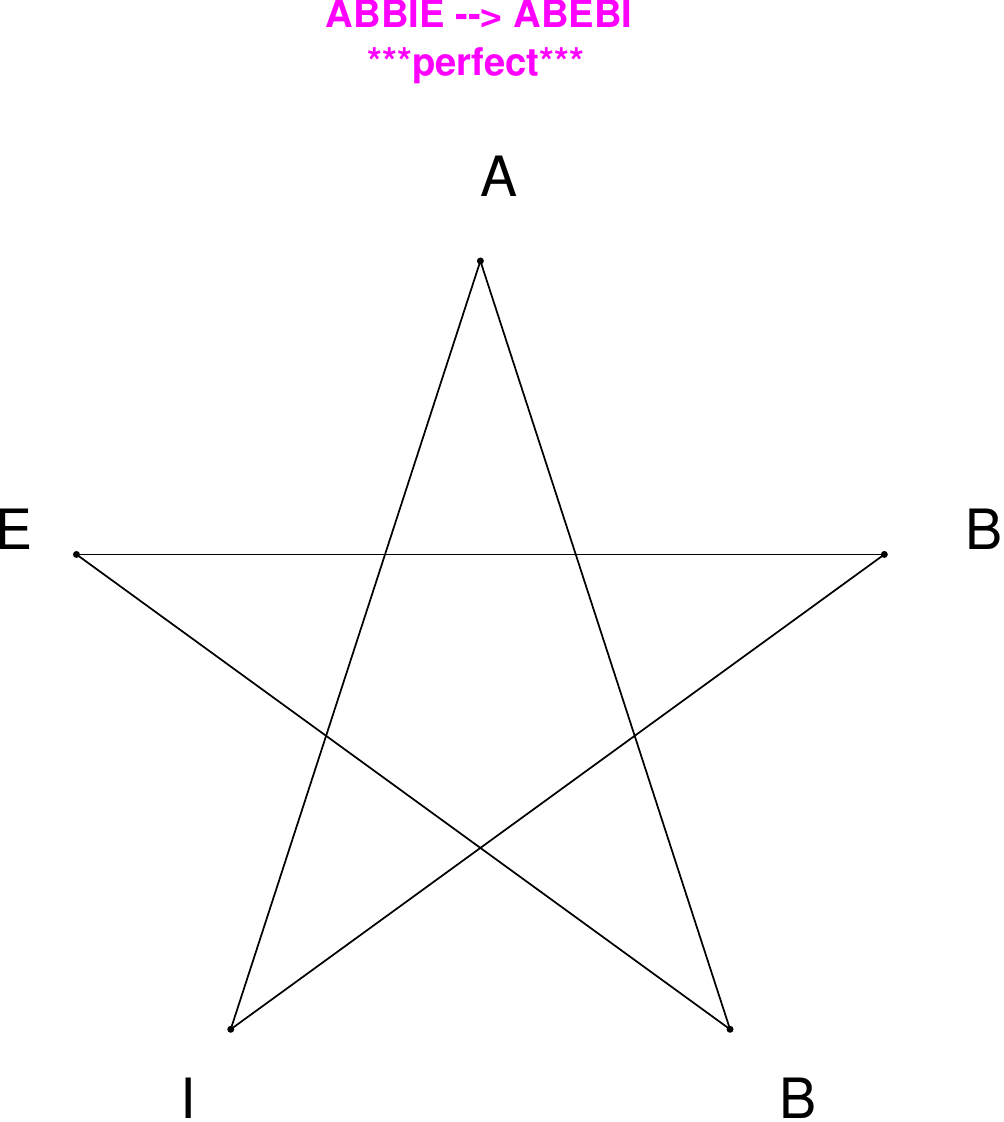}
\end{subfigure}
\hfill
\begin{subfigure}[T]{0.19\textwidth}
\centering
\includegraphics[width=\textwidth]{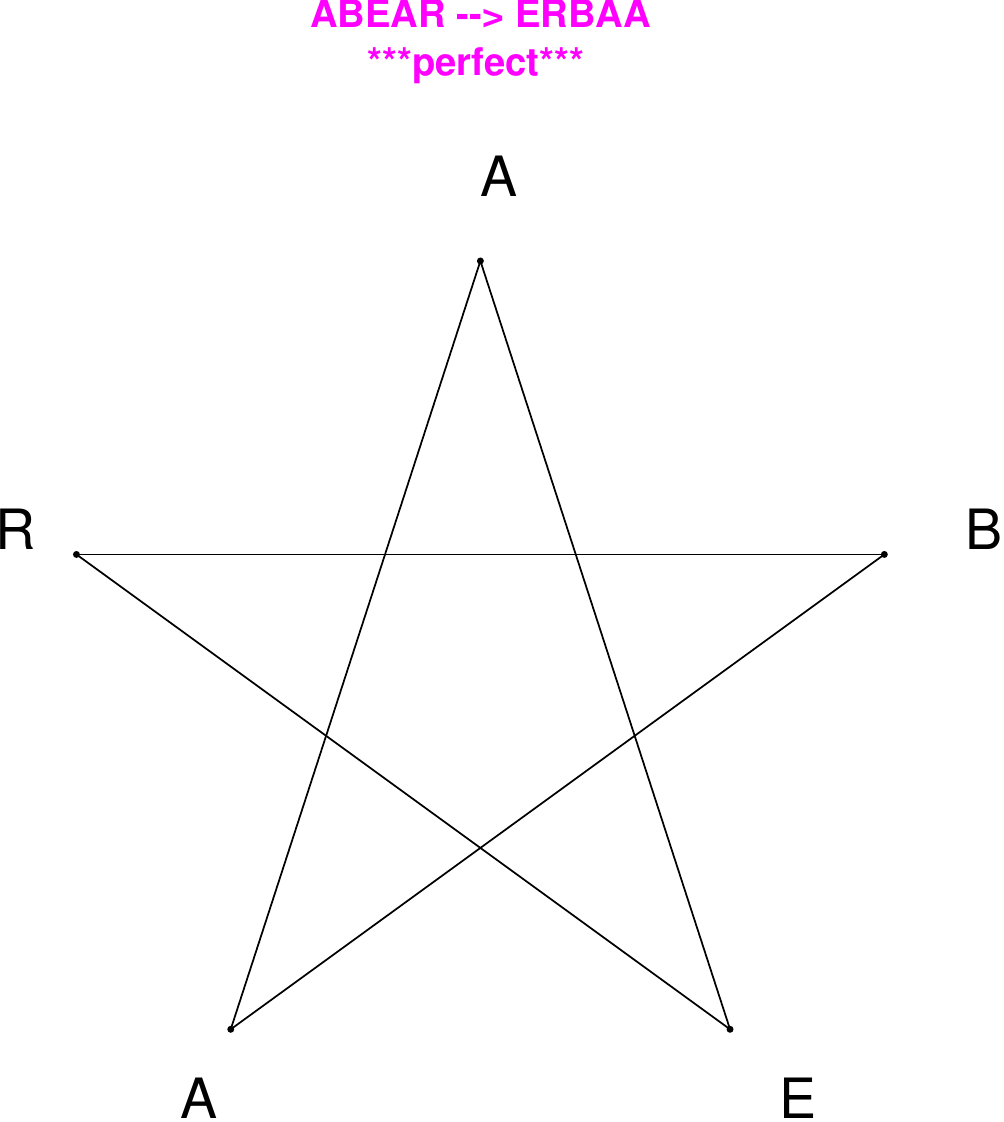}
\end{subfigure}
\hfill
\begin{subfigure}[T]{0.19\textwidth}
\centering
\includegraphics[width=\textwidth]{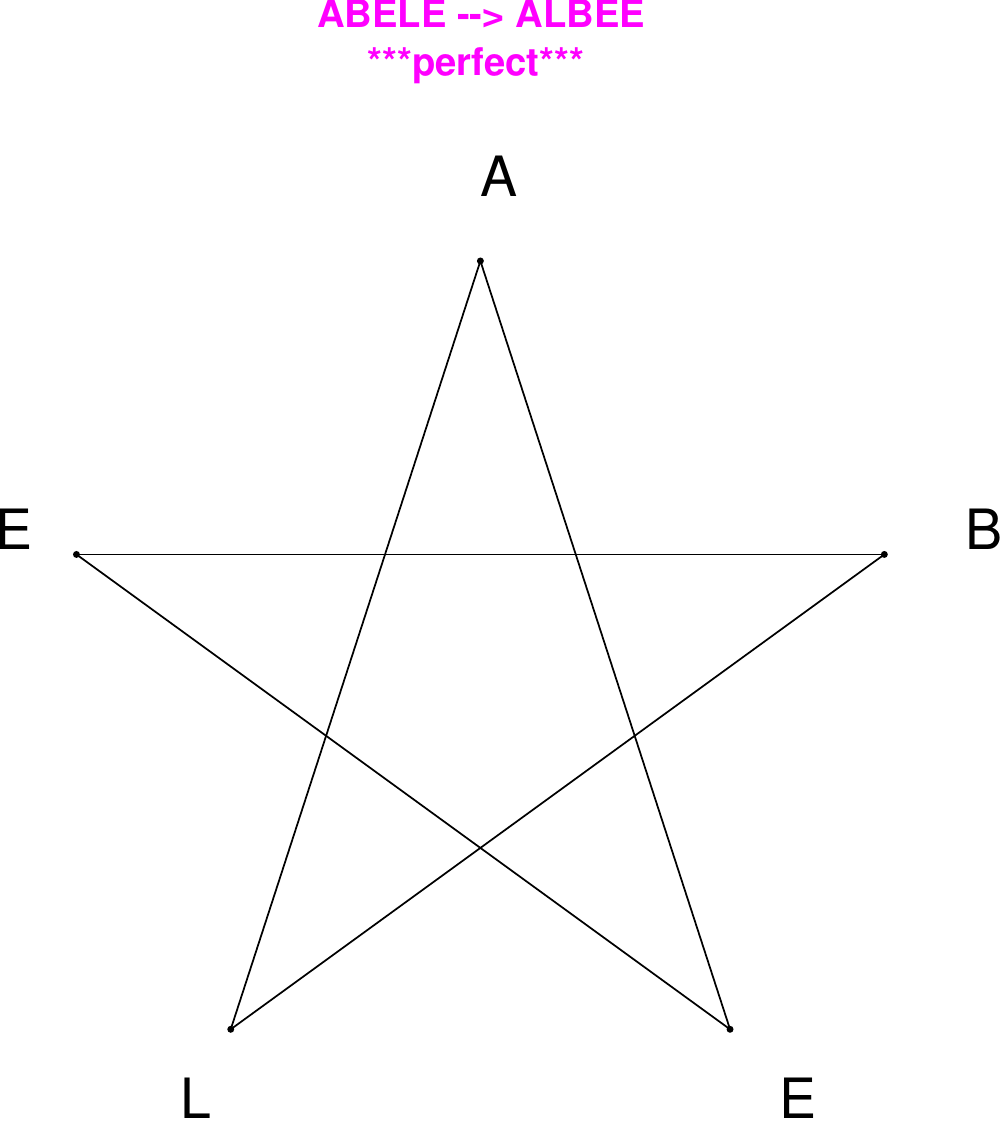}
\end{subfigure}
\hfill
\begin{subfigure}[T]{0.19\textwidth}
\centering
\includegraphics[width=\textwidth]{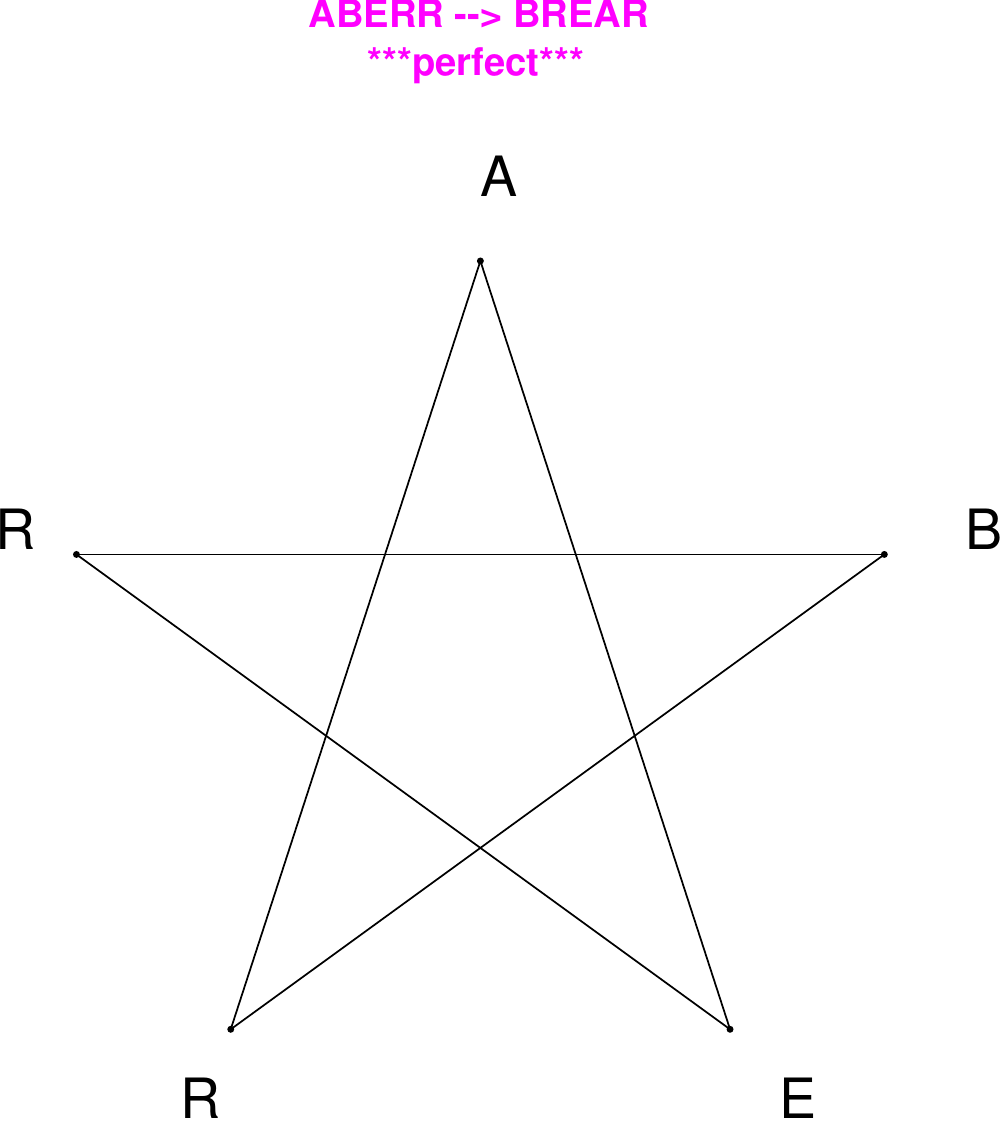}
\end{subfigure}
\hfill
\begin{subfigure}[T]{0.19\textwidth}
\centering
\includegraphics[width=\textwidth]{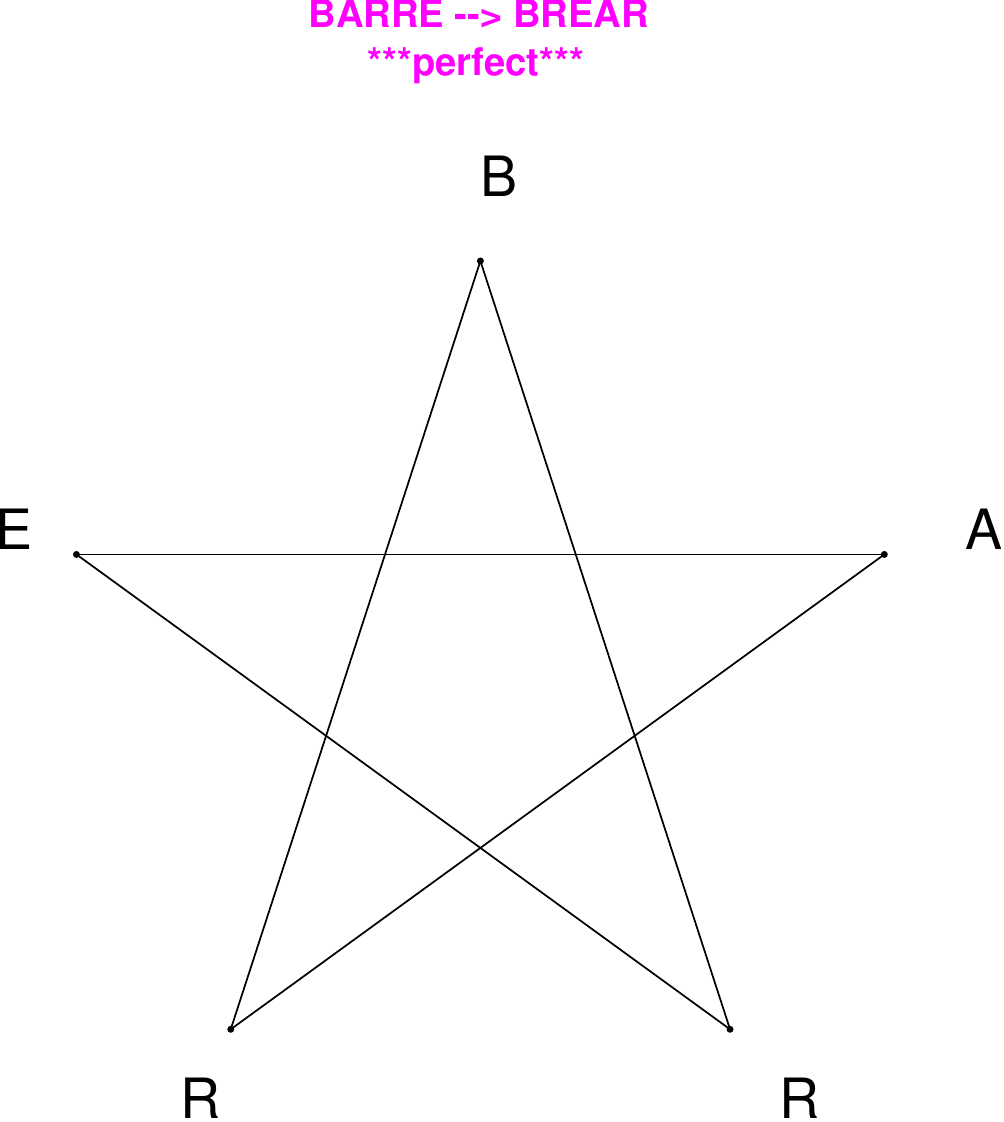}
\end{subfigure}
\end{figure}

\begin{figure}[H]
\centering
\begin{subfigure}[T]{0.19\textwidth}
\centering
\includegraphics[width=\textwidth]{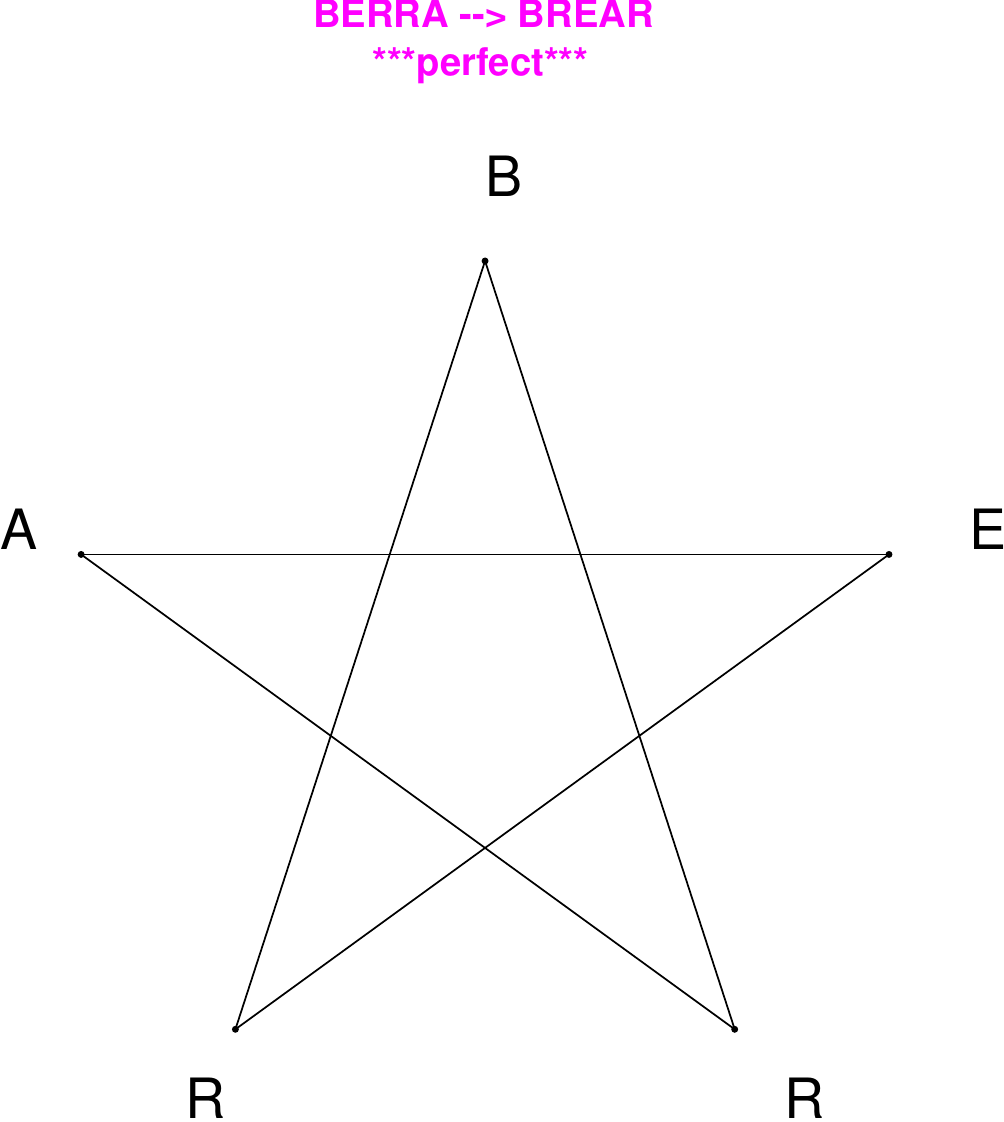}
\end{subfigure}
\hfill
\begin{subfigure}[T]{0.19\textwidth}
\centering
\includegraphics[width=\textwidth]{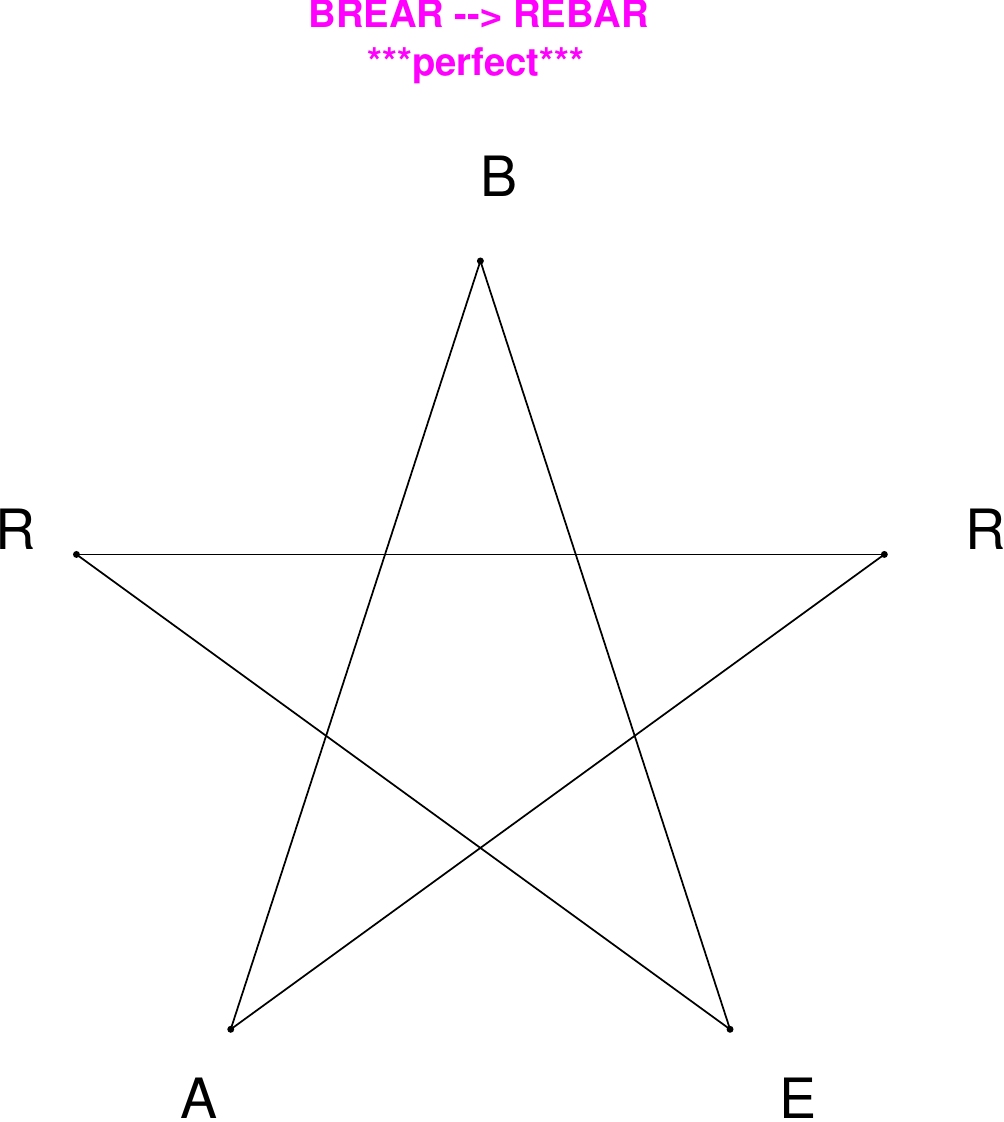}
\end{subfigure}
\hfill
\begin{subfigure}[T]{0.19\textwidth}
\centering
\includegraphics[width=\textwidth]{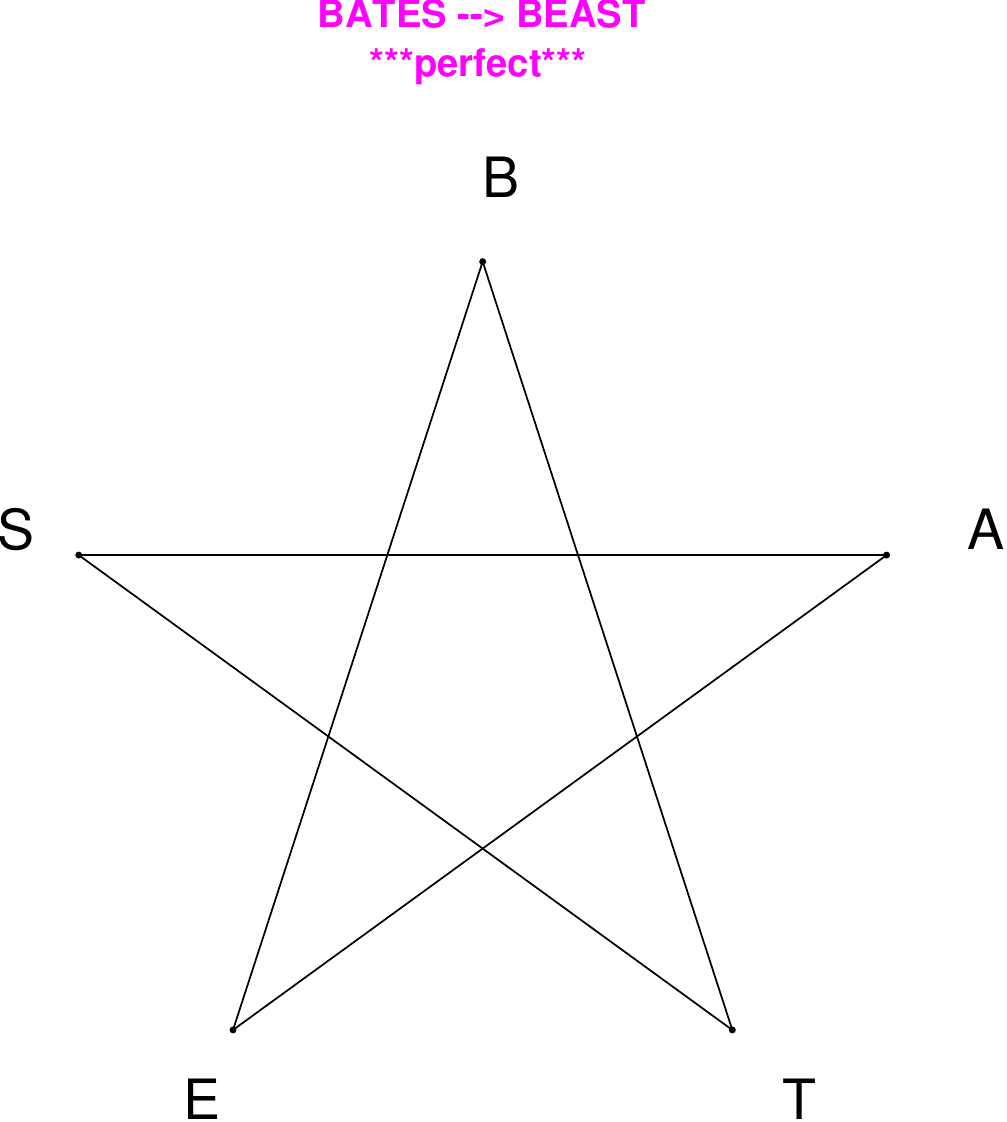}
\end{subfigure}
\hfill
\begin{subfigure}[T]{0.19\textwidth}
\centering
\includegraphics[width=\textwidth]{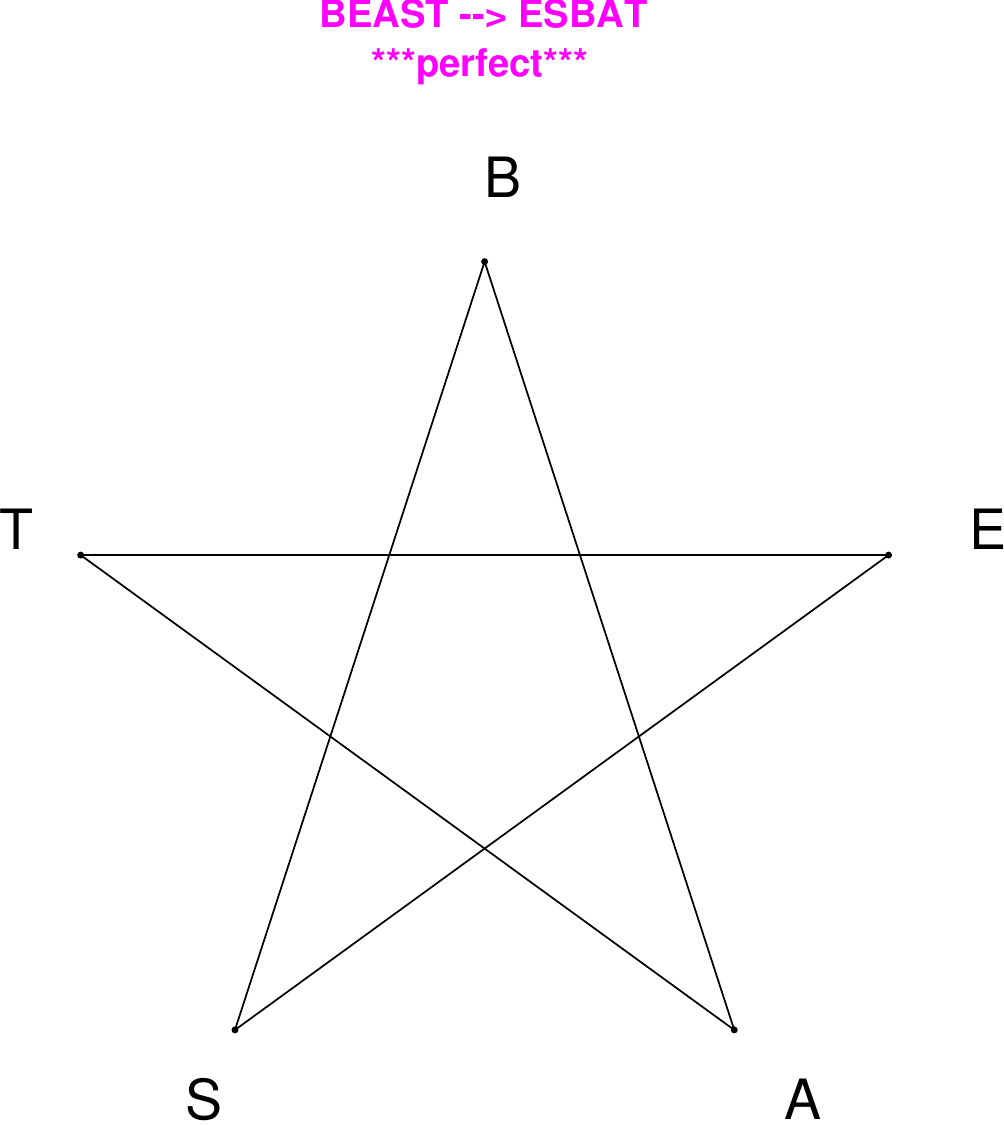}
\end{subfigure}
\hfill
\begin{subfigure}[T]{0.19\textwidth}
\centering
\includegraphics[width=\textwidth]{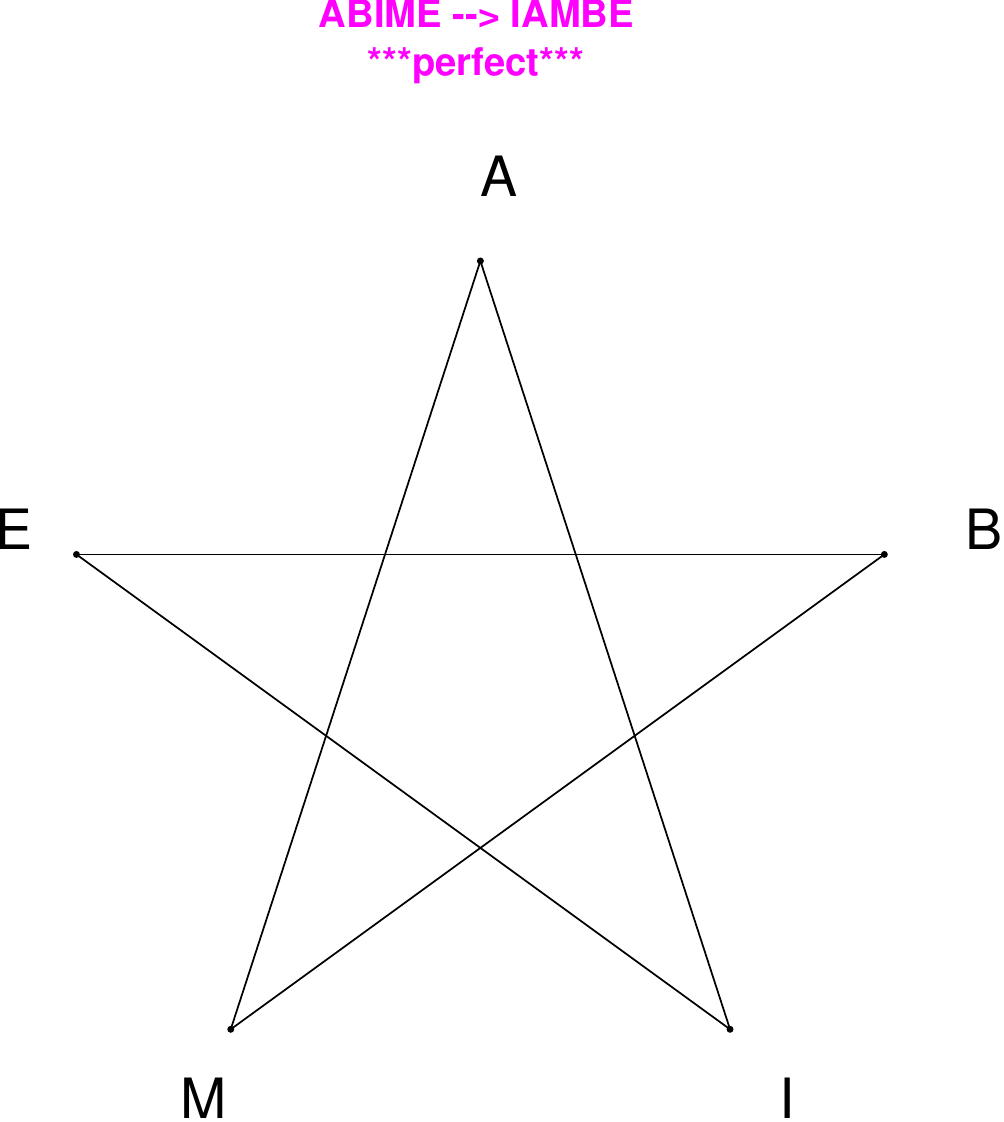}
\end{subfigure}
\end{figure}

\begin{figure}[H]
\centering
\begin{subfigure}[T]{0.19\textwidth}
\centering
\includegraphics[width=\textwidth]{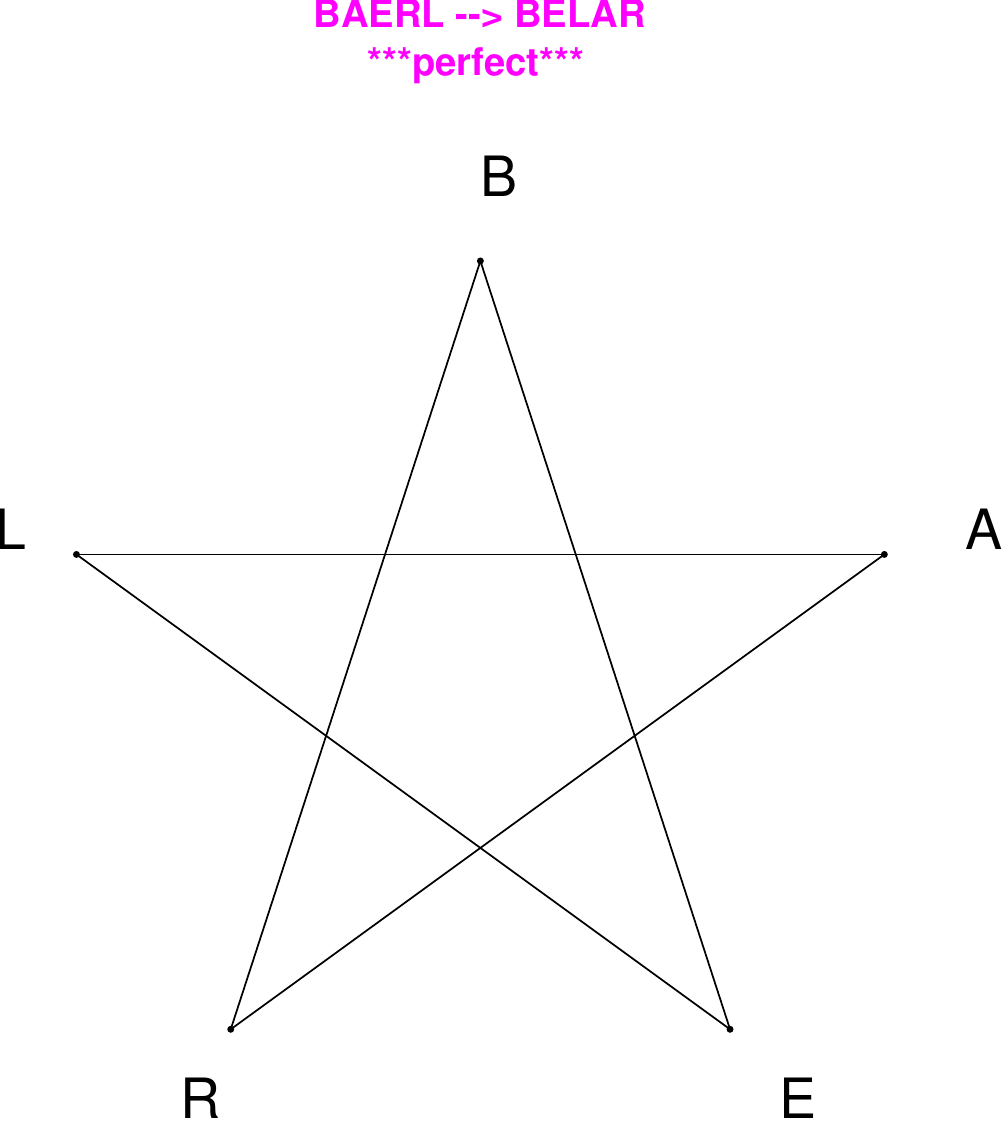}
\end{subfigure}
\hfill
\begin{subfigure}[T]{0.19\textwidth}
\centering
\includegraphics[width=\textwidth]{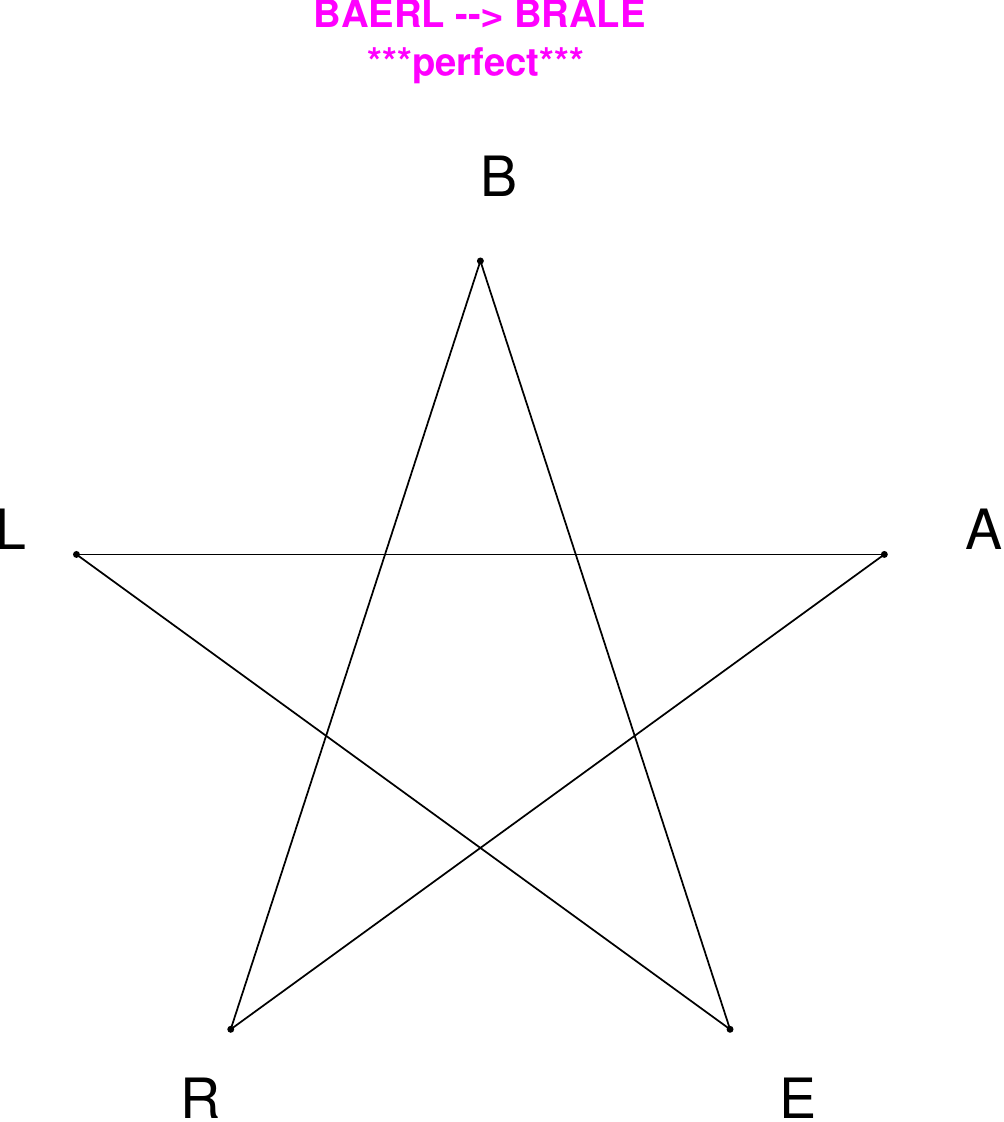}
\end{subfigure}
\hfill
\begin{subfigure}[T]{0.19\textwidth}
\centering
\includegraphics[width=\textwidth]{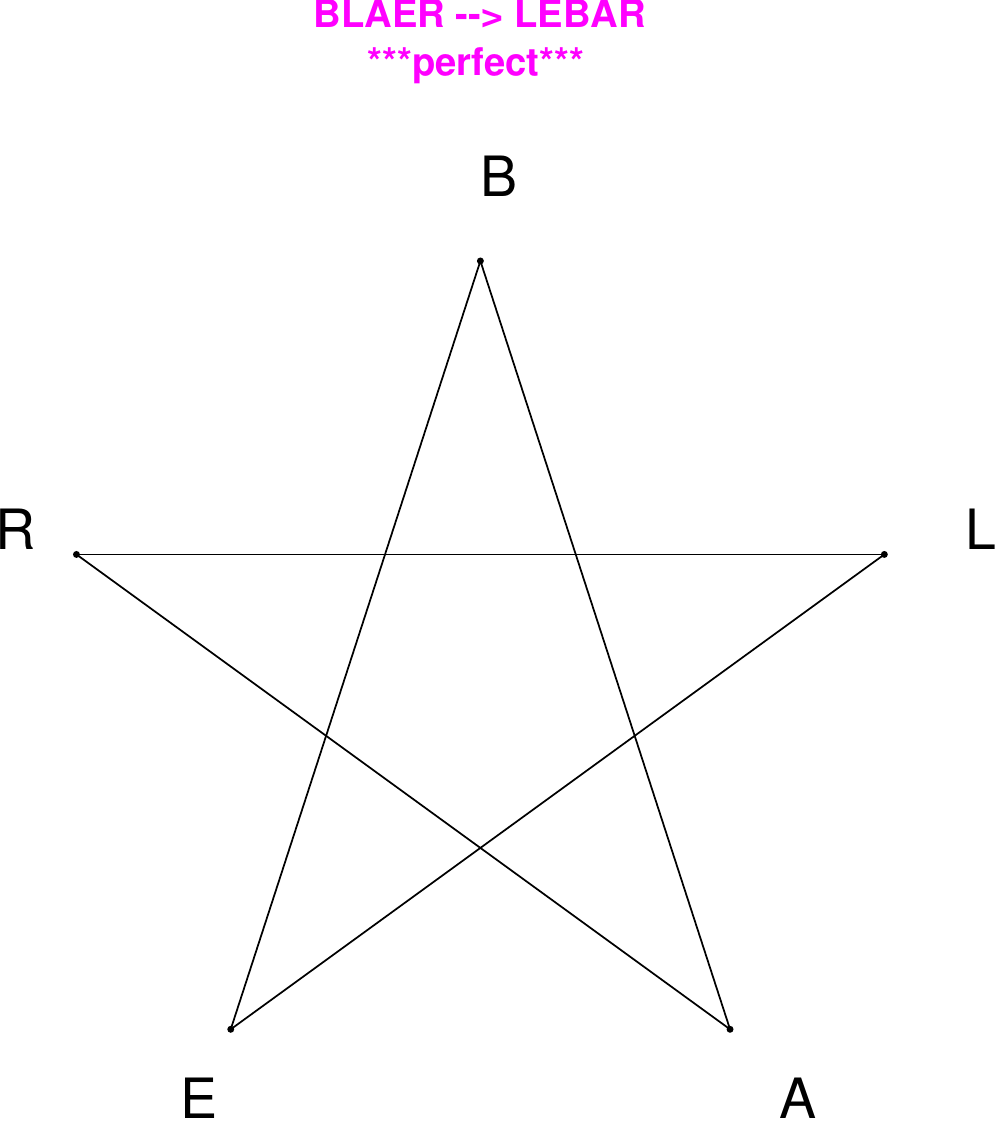}
\end{subfigure}
\hfill
\begin{subfigure}[T]{0.19\textwidth}
\centering
\includegraphics[width=\textwidth]{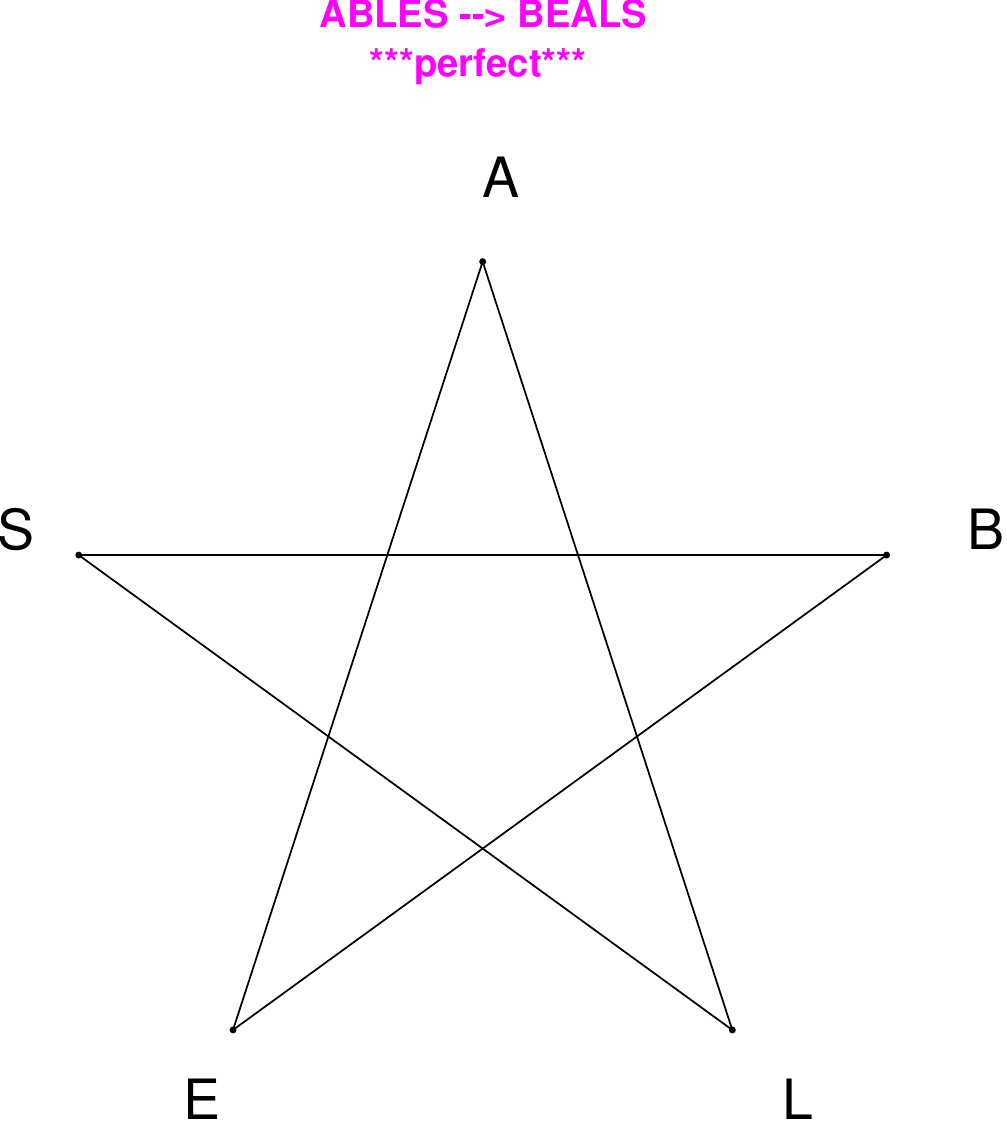}
\end{subfigure}
\hfill
\begin{subfigure}[T]{0.19\textwidth}
\centering
\includegraphics[width=\textwidth]{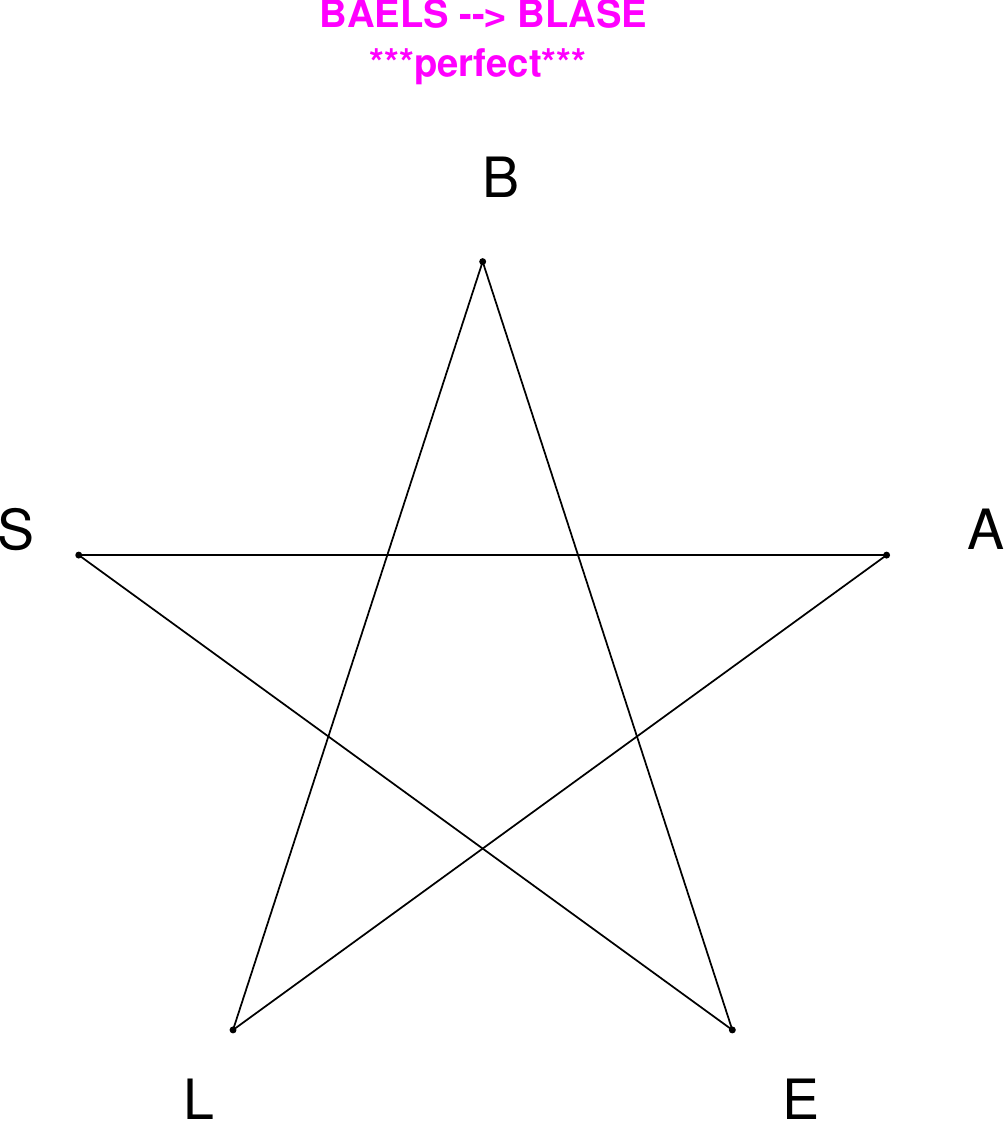}
\end{subfigure}
\end{figure}

\begin{figure}[H]
\centering
\begin{subfigure}[T]{0.19\textwidth}
\centering
\includegraphics[width=\textwidth]{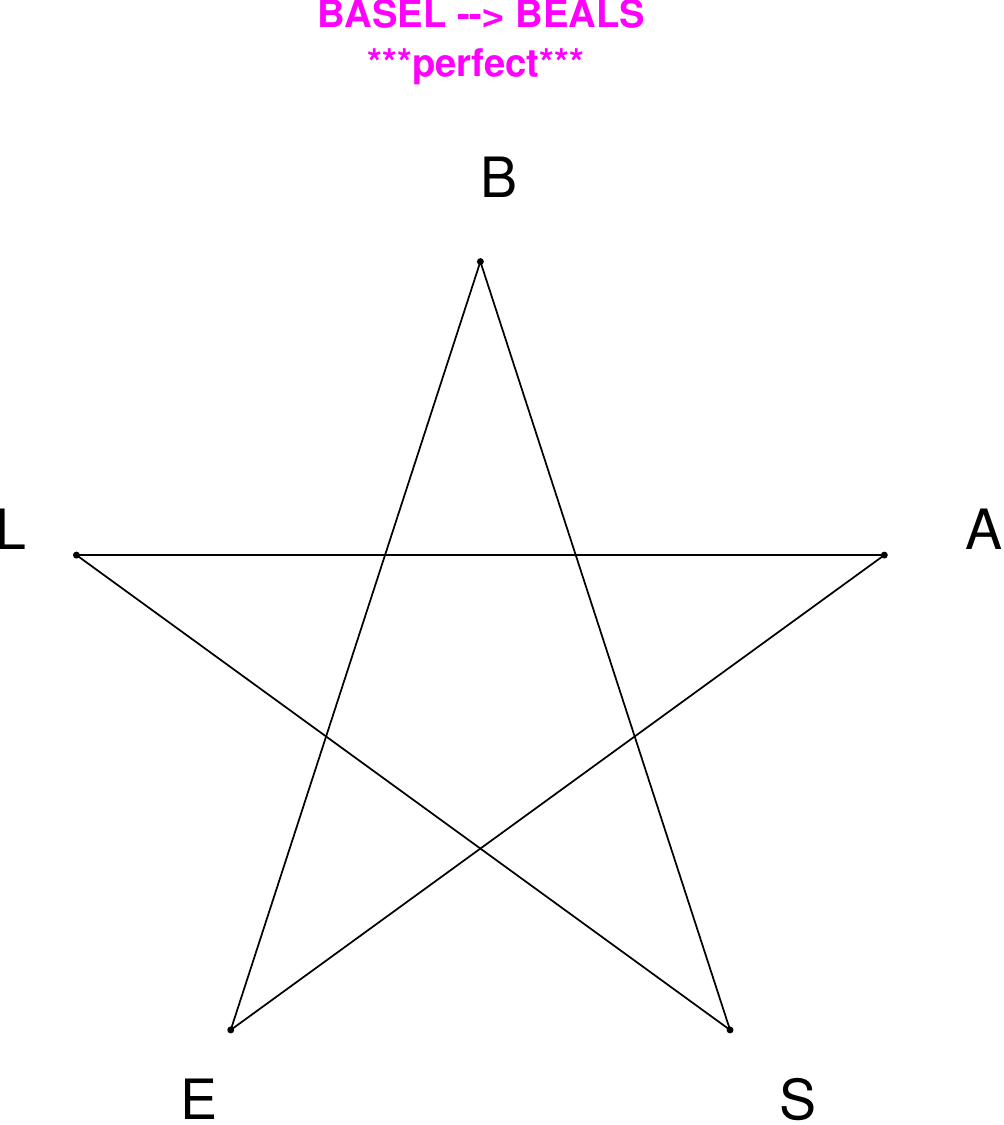}
\end{subfigure}
\hfill
\begin{subfigure}[T]{0.19\textwidth}
\centering
\includegraphics[width=\textwidth]{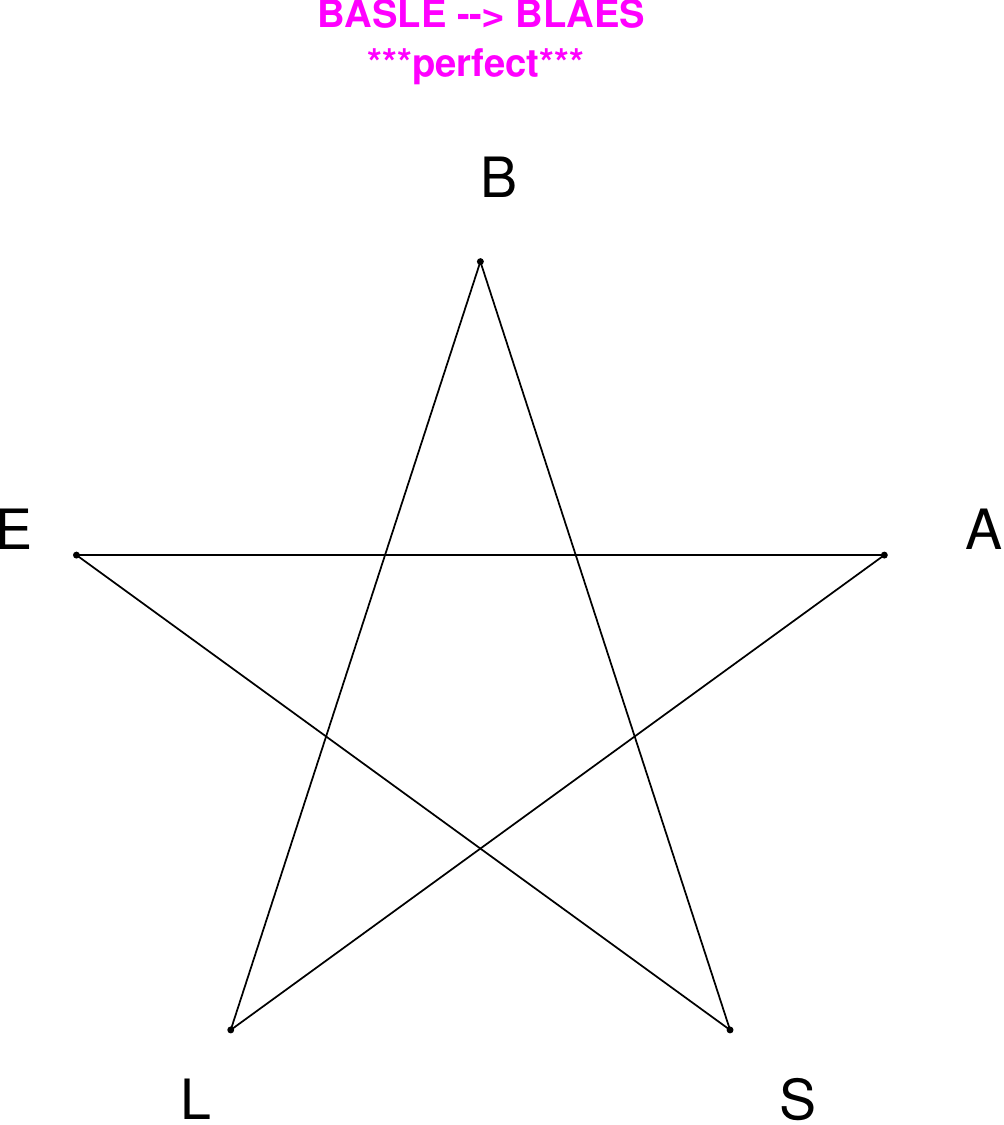}
\end{subfigure}
\hfill
\begin{subfigure}[T]{0.19\textwidth}
\centering
\includegraphics[width=\textwidth]{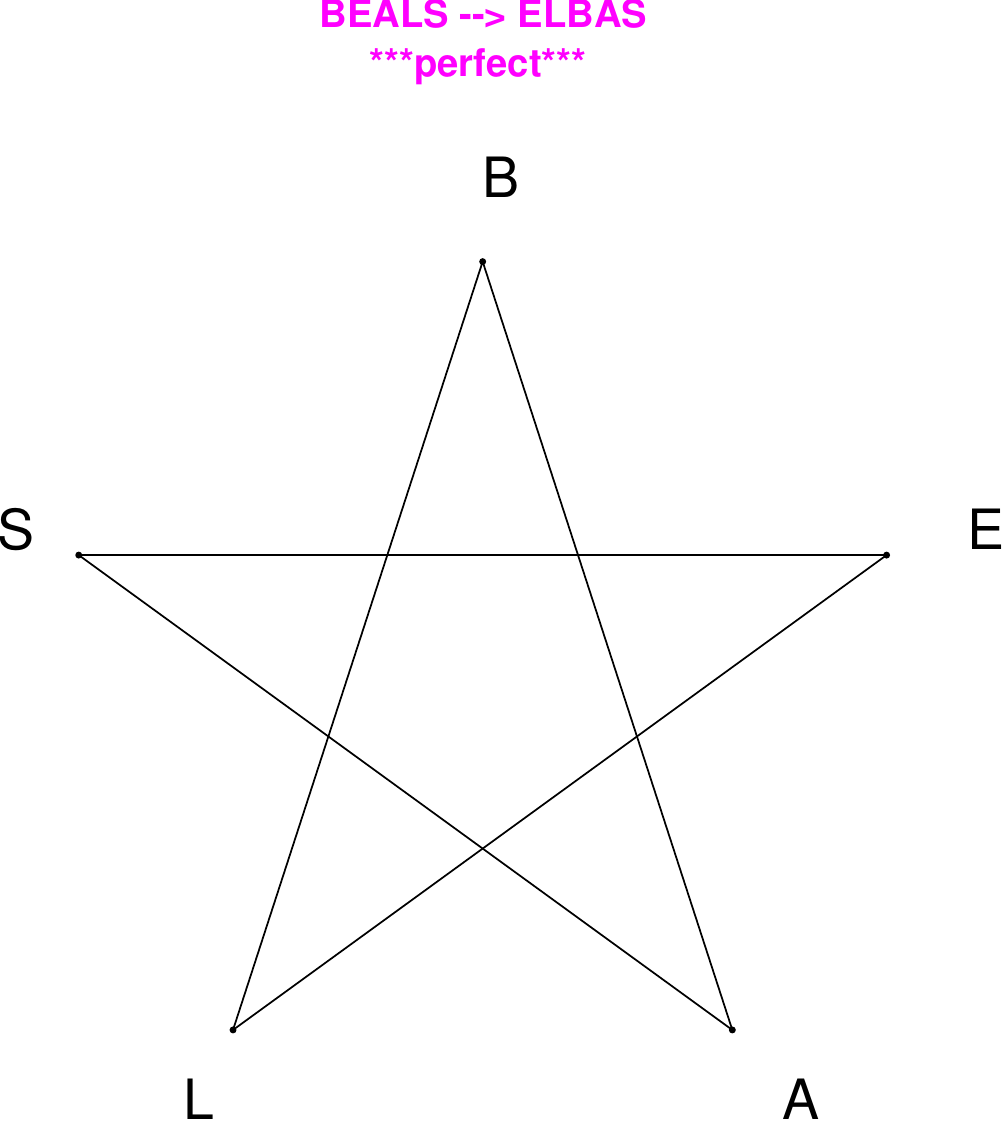}
\end{subfigure}
\hfill
\begin{subfigure}[T]{0.19\textwidth}
\centering
\includegraphics[width=\textwidth]{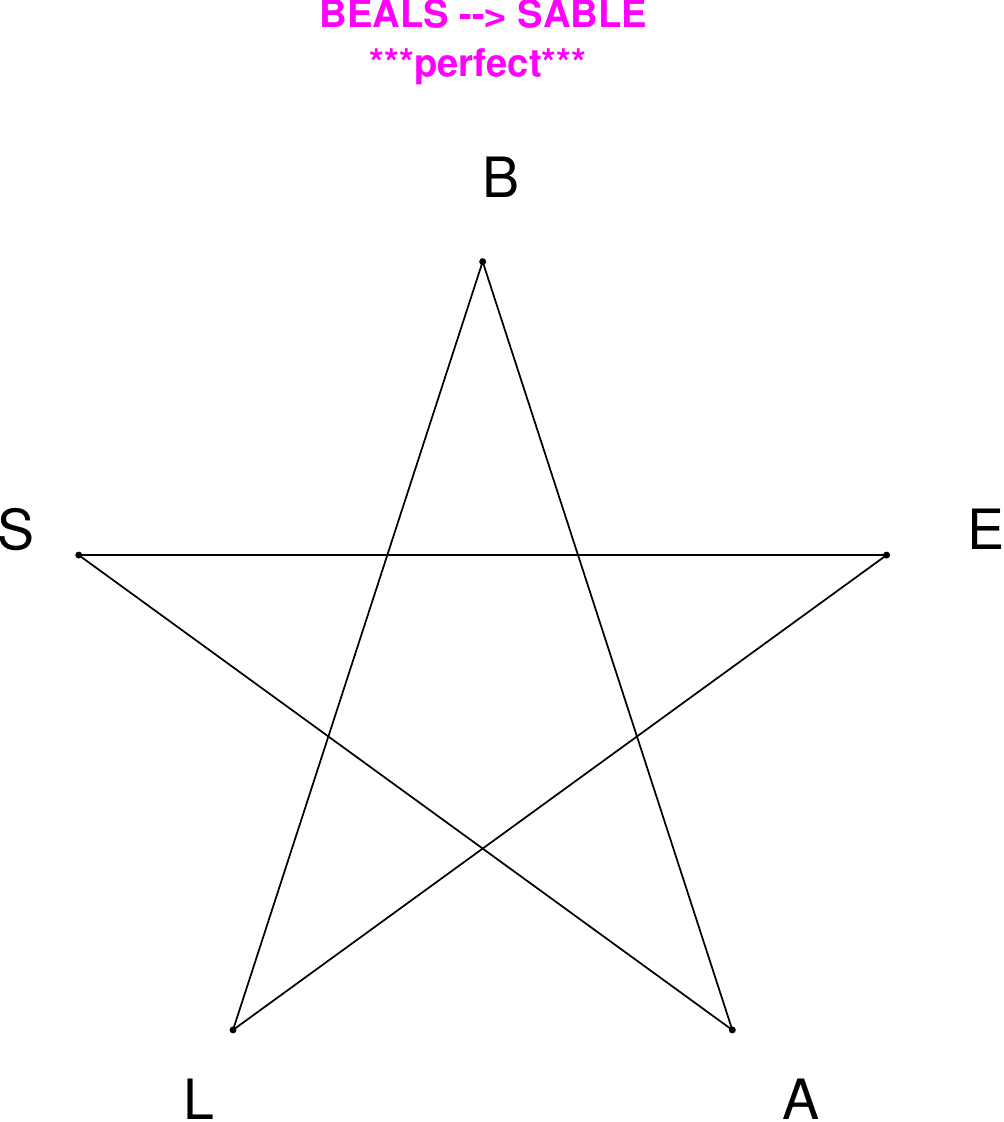}
\end{subfigure}
\hfill
\begin{subfigure}[T]{0.19\textwidth}
\centering
\includegraphics[width=\textwidth]{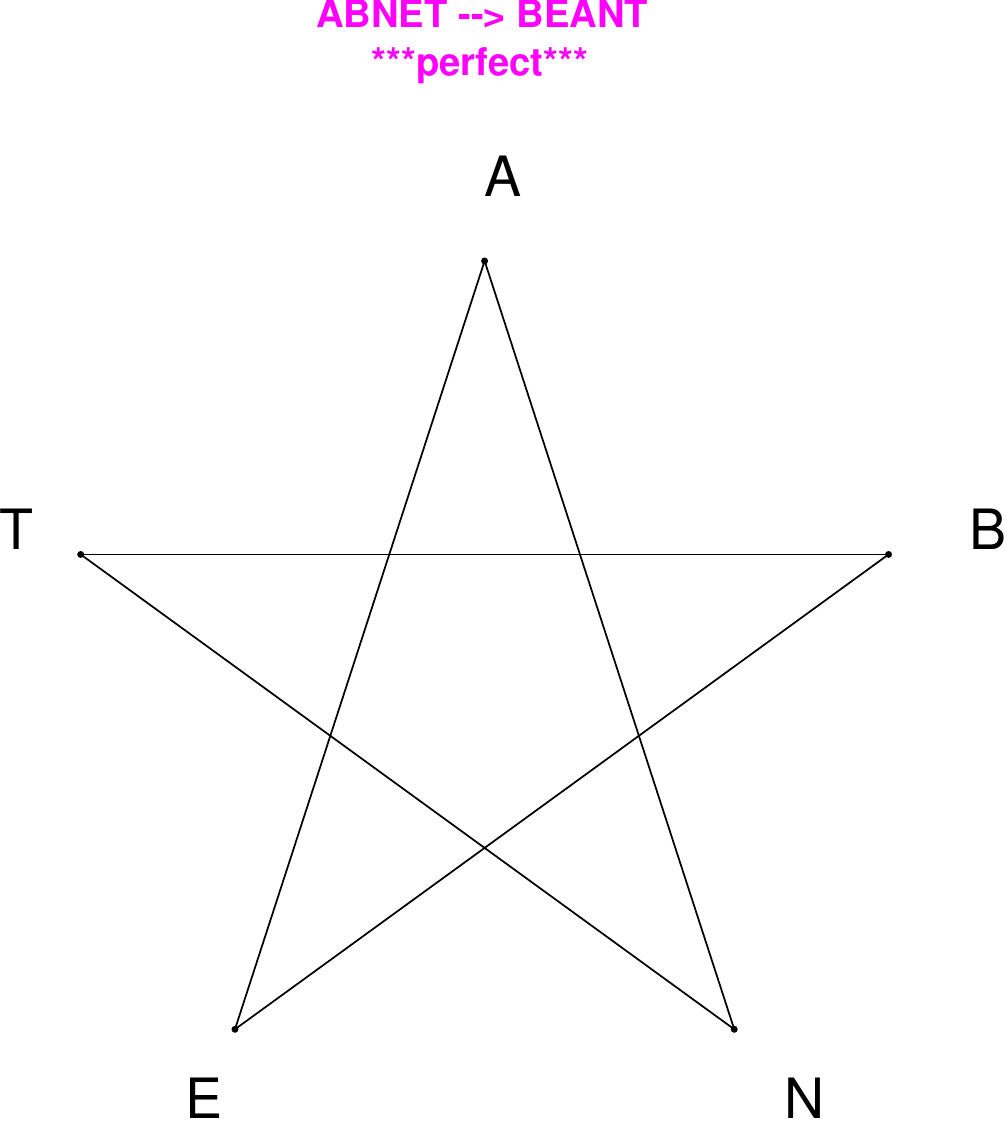}
\end{subfigure}
\end{figure}

\begin{figure}[H]
\centering
\begin{subfigure}[T]{0.19\textwidth}
\centering
\includegraphics[width=\textwidth]{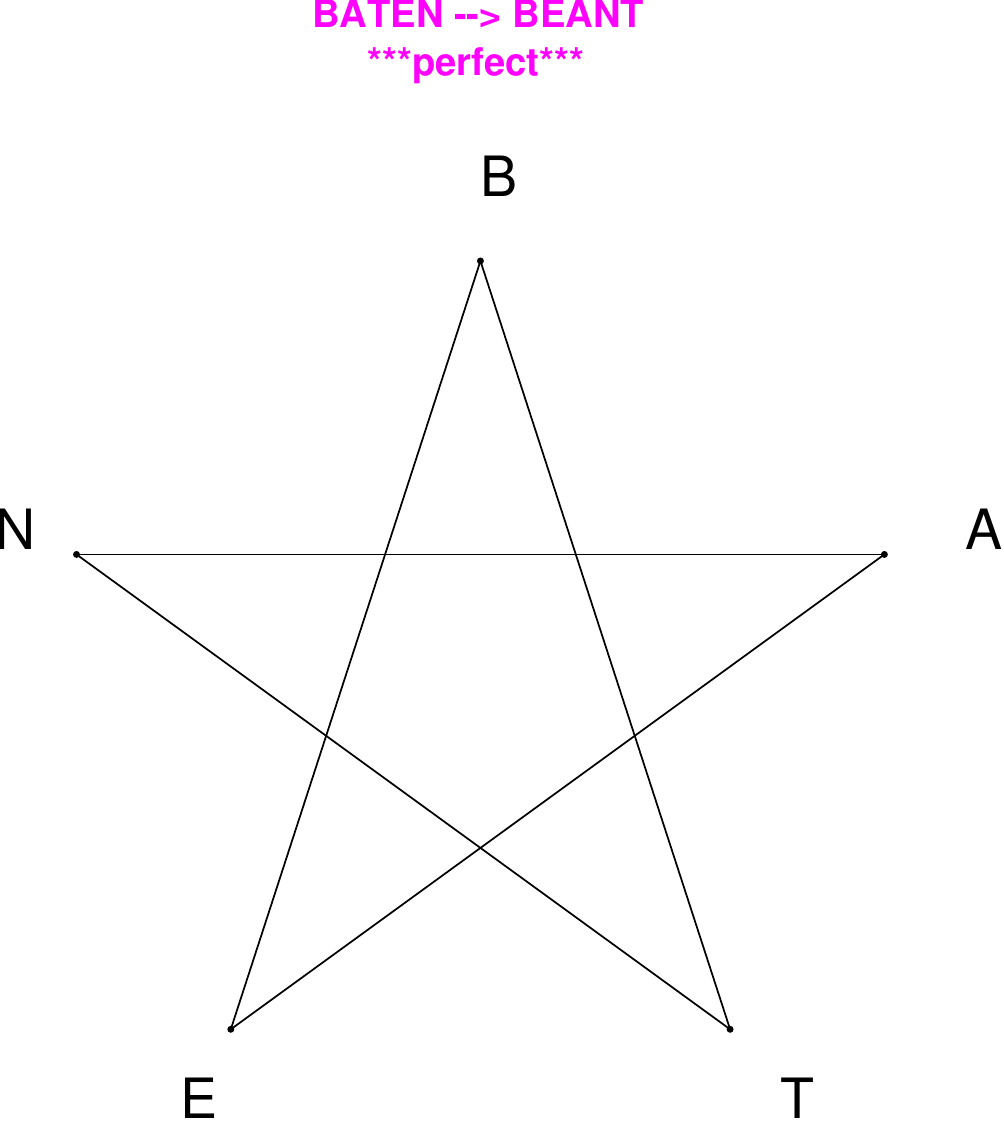}
\end{subfigure}
\hfill
\begin{subfigure}[T]{0.19\textwidth}
\centering
\includegraphics[width=\textwidth]{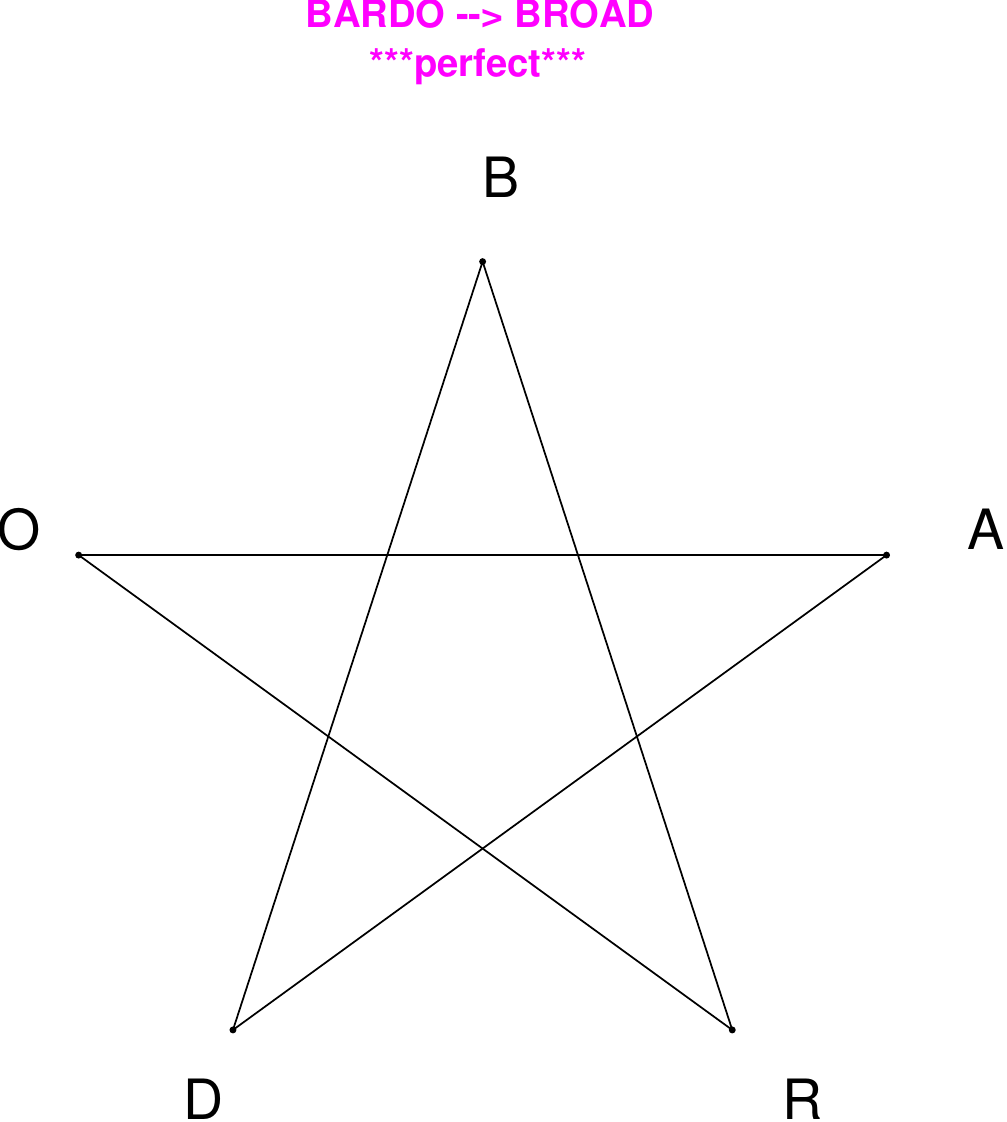}
\end{subfigure}
\hfill
\begin{subfigure}[T]{0.19\textwidth}
\centering
\includegraphics[width=\textwidth]{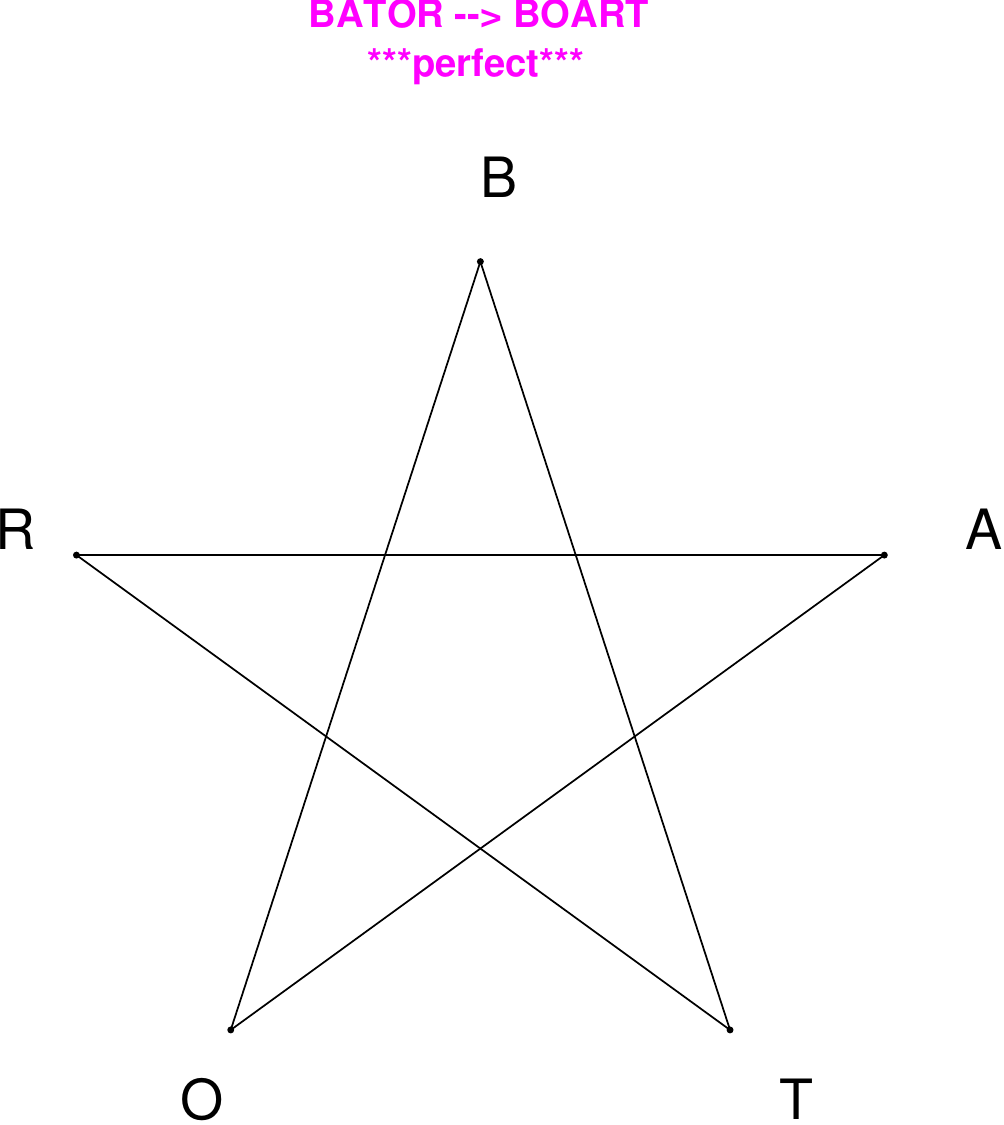}
\end{subfigure}
\hfill
\begin{subfigure}[T]{0.19\textwidth}
\centering
\includegraphics[width=\textwidth]{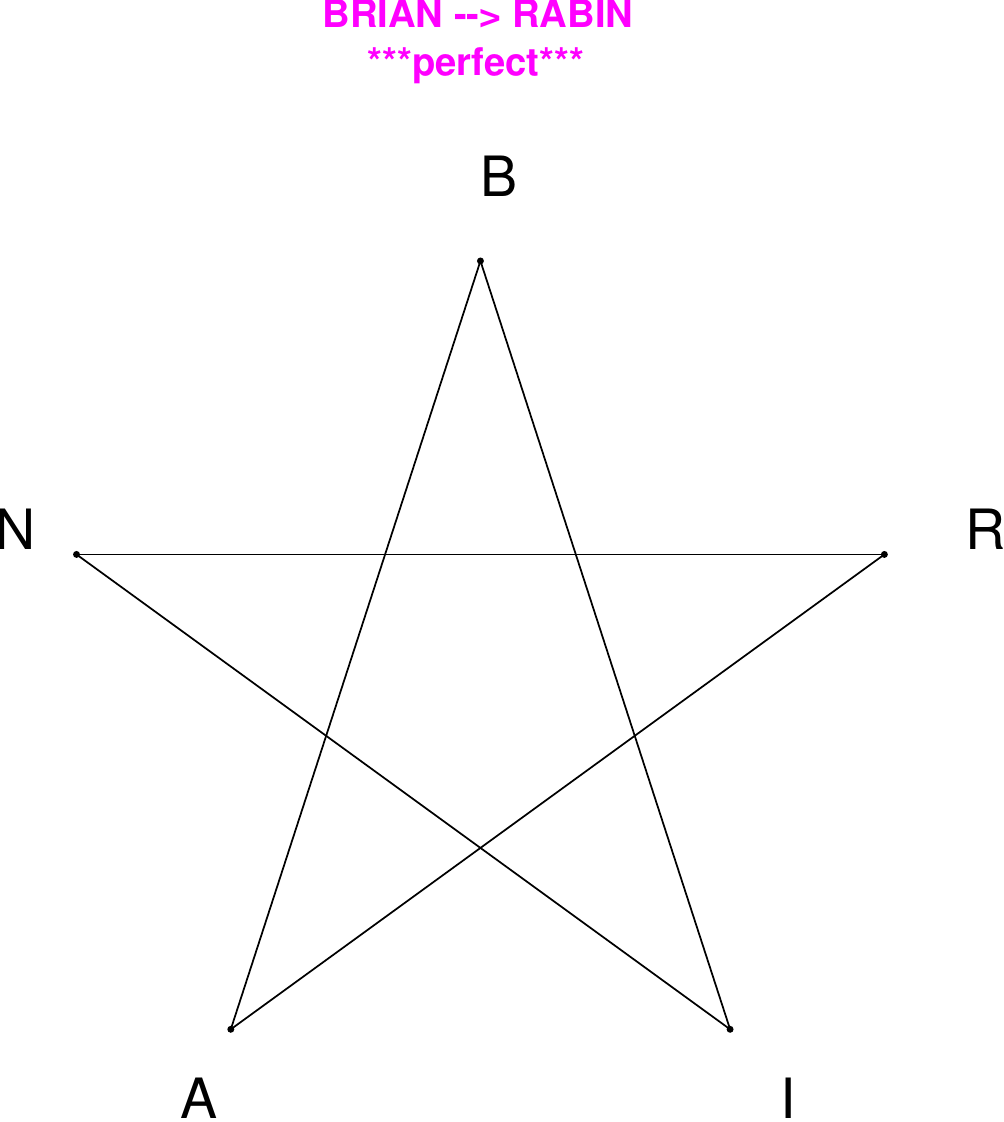}
\end{subfigure}
\hfill
\begin{subfigure}[T]{0.19\textwidth}
\centering
\includegraphics[width=\textwidth]{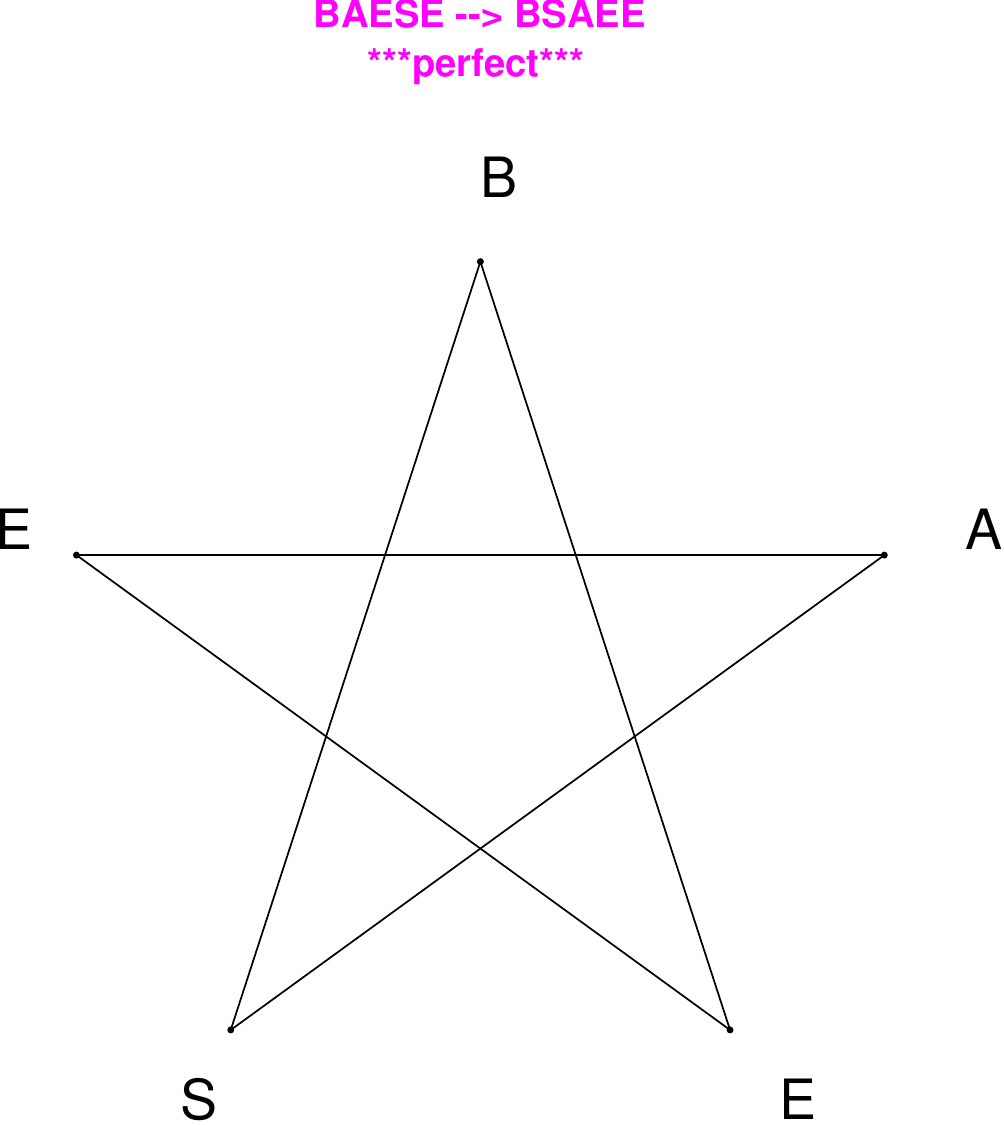}
\end{subfigure}
\end{figure}

\begin{figure}[H]
\centering
\begin{subfigure}[T]{0.19\textwidth}
\centering
\includegraphics[width=\textwidth]{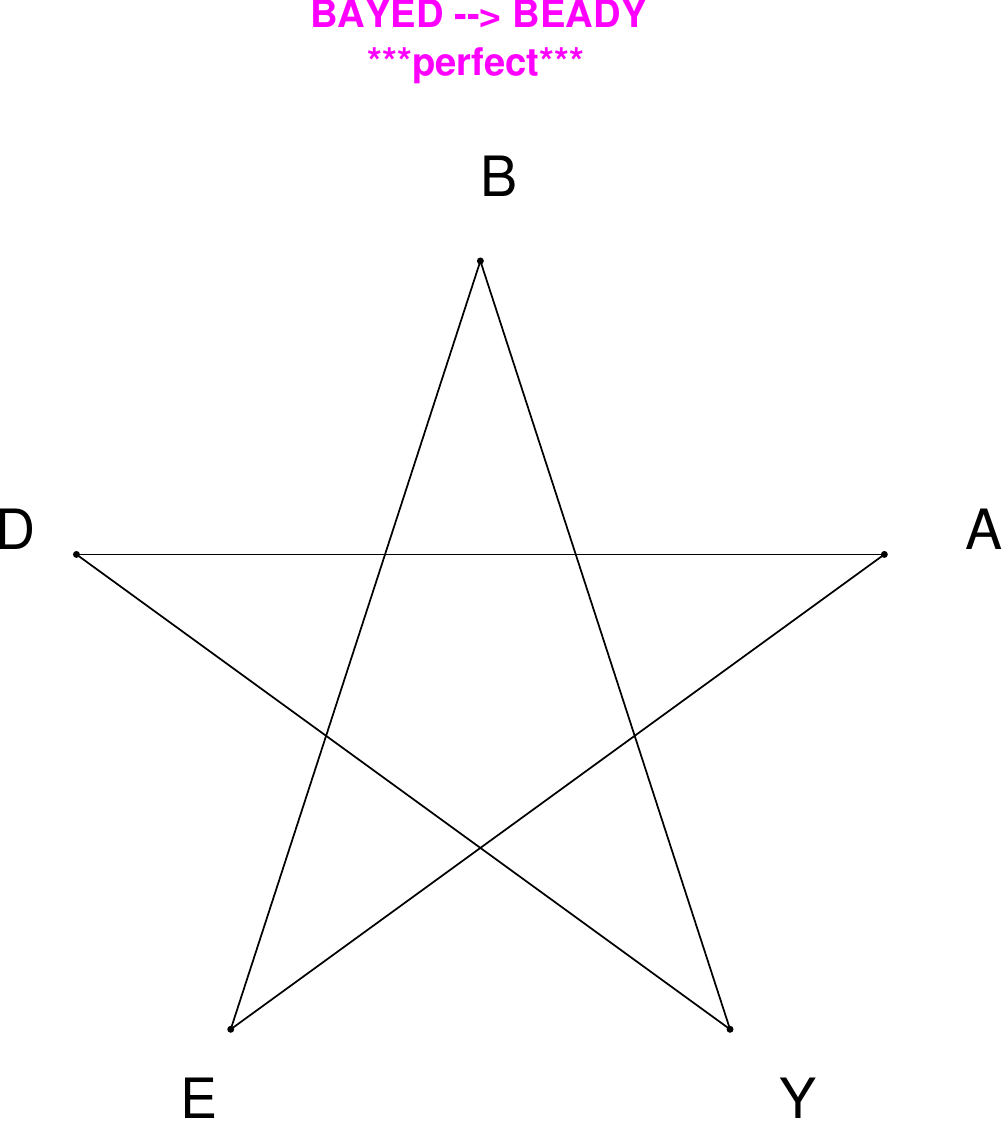}
\end{subfigure}
\hfill
\begin{subfigure}[T]{0.19\textwidth}
\centering
\includegraphics[width=\textwidth]{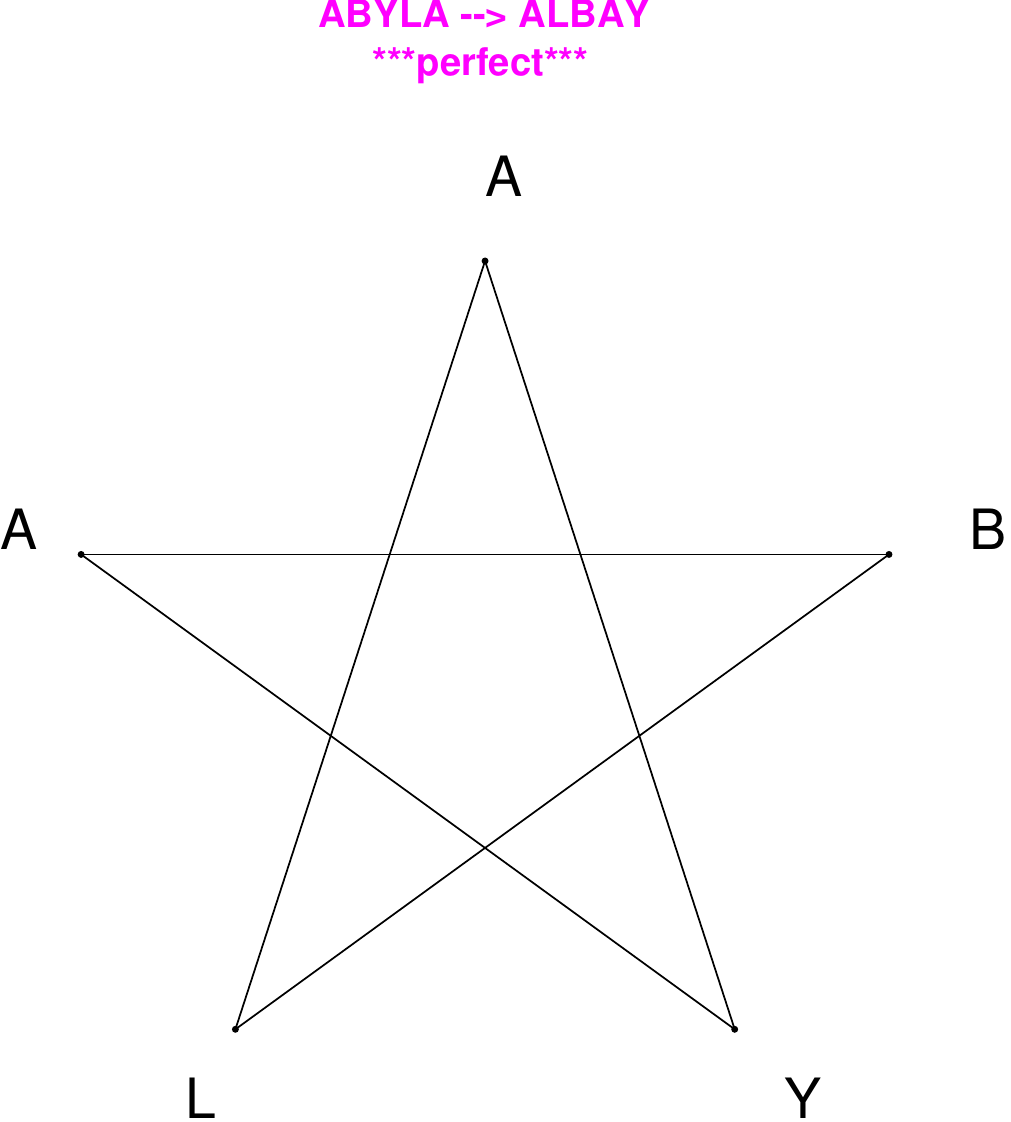}
\end{subfigure}
\hfill
\begin{subfigure}[T]{0.19\textwidth}
\centering
\includegraphics[width=\textwidth]{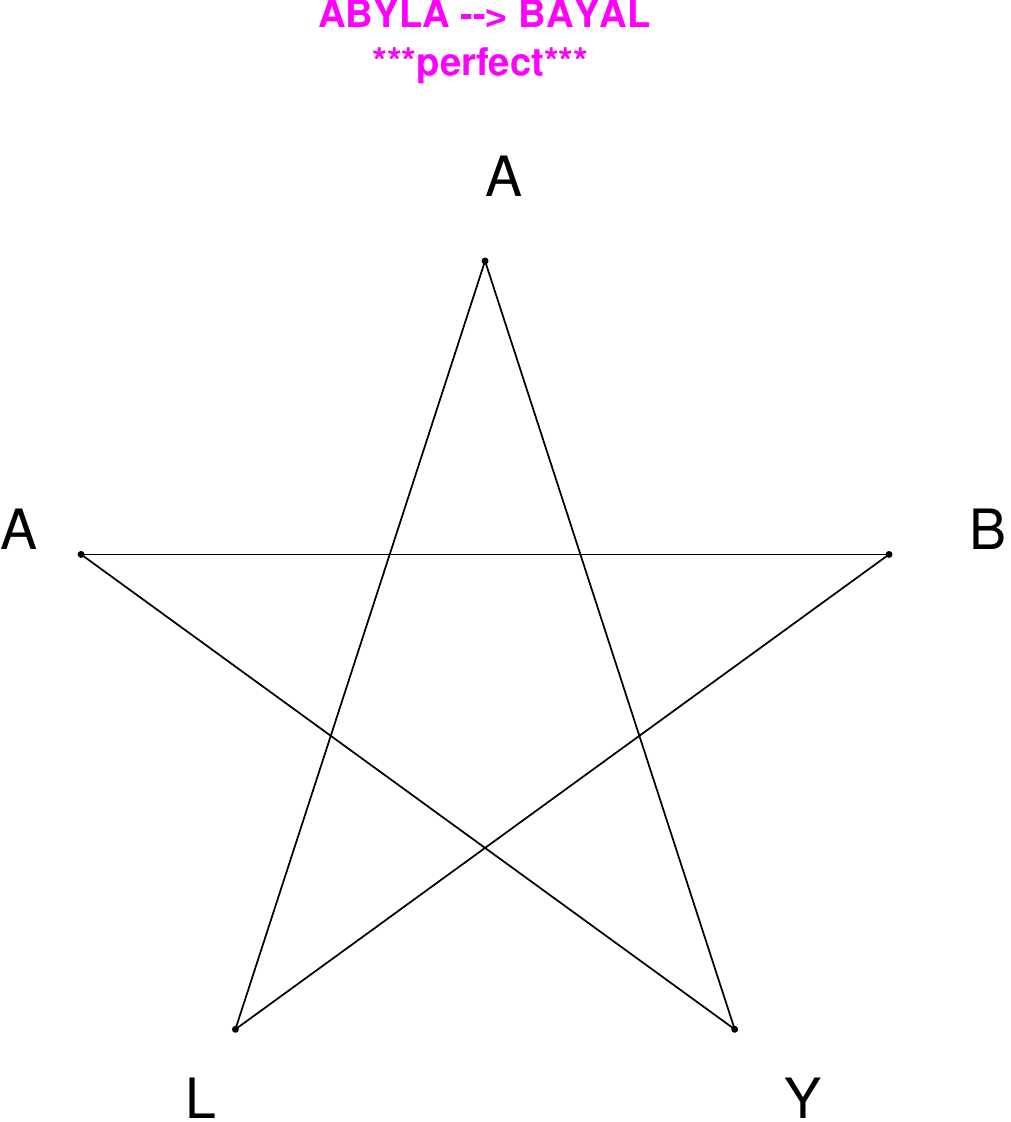}
\end{subfigure}
\hfill
\begin{subfigure}[T]{0.19\textwidth}
\centering
\includegraphics[width=\textwidth]{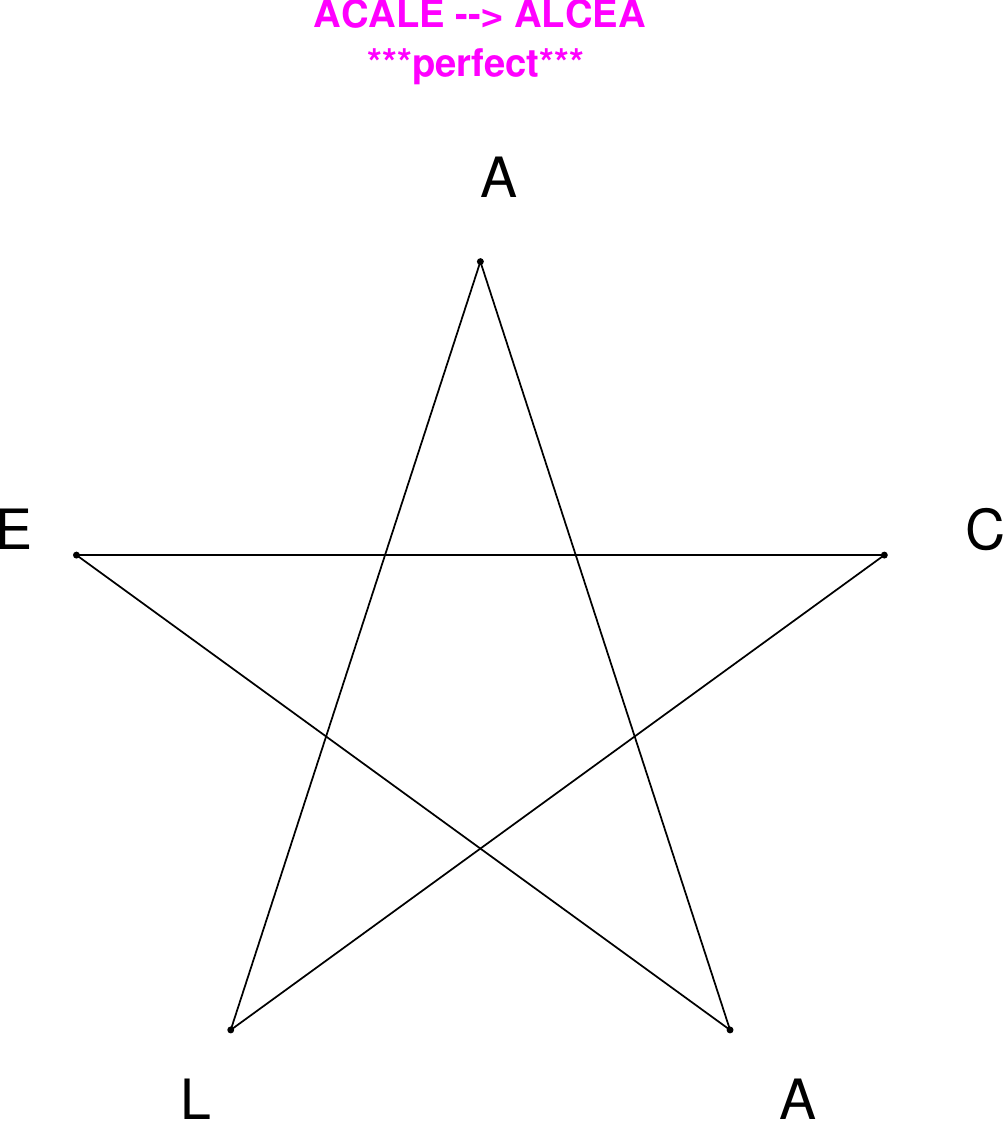}
\end{subfigure}
\hfill
\begin{subfigure}[T]{0.19\textwidth}
\centering
\includegraphics[width=\textwidth]{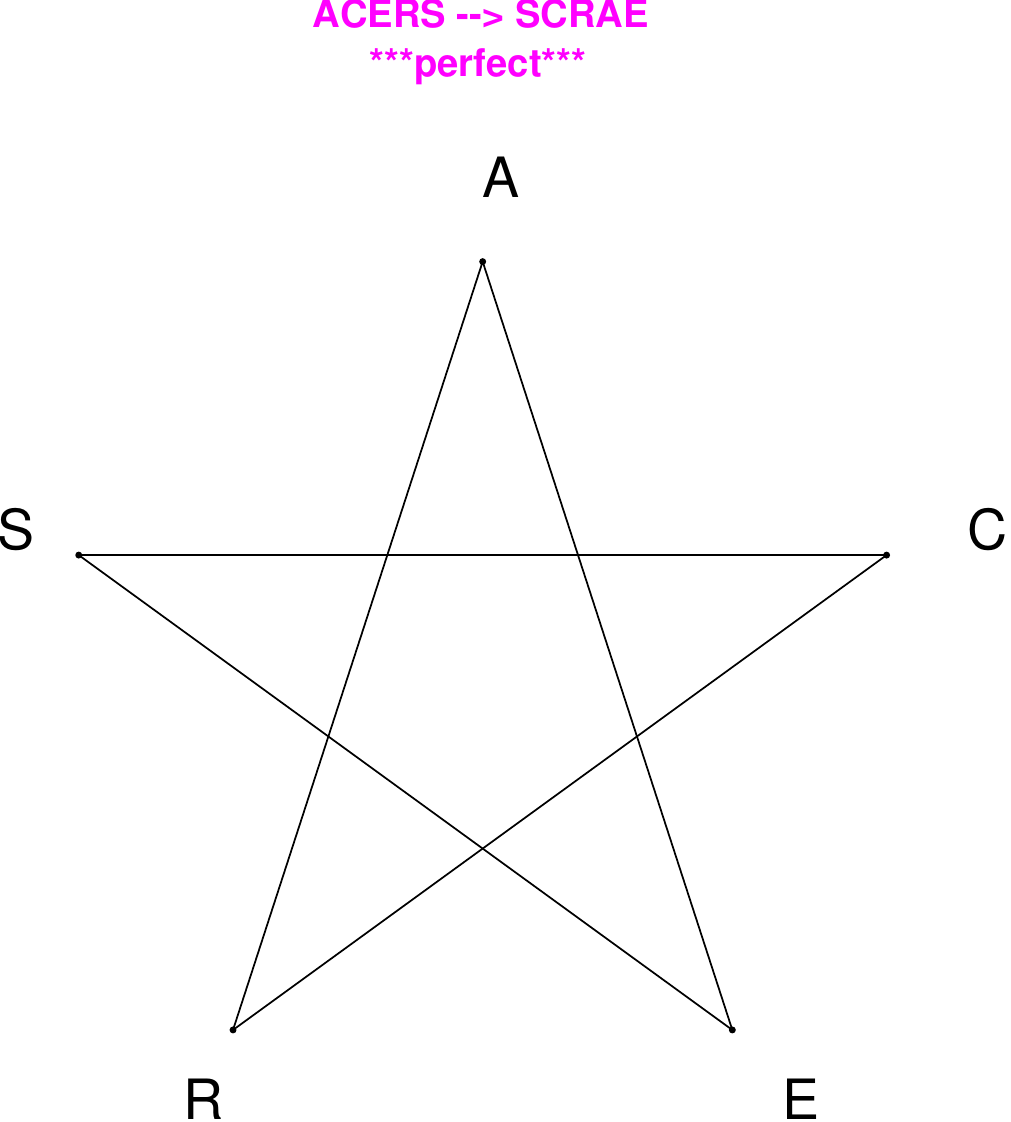}
\end{subfigure}
\end{figure}

\begin{figure}[H]
\centering
\begin{subfigure}[T]{0.19\textwidth}
\centering
\includegraphics[width=\textwidth]{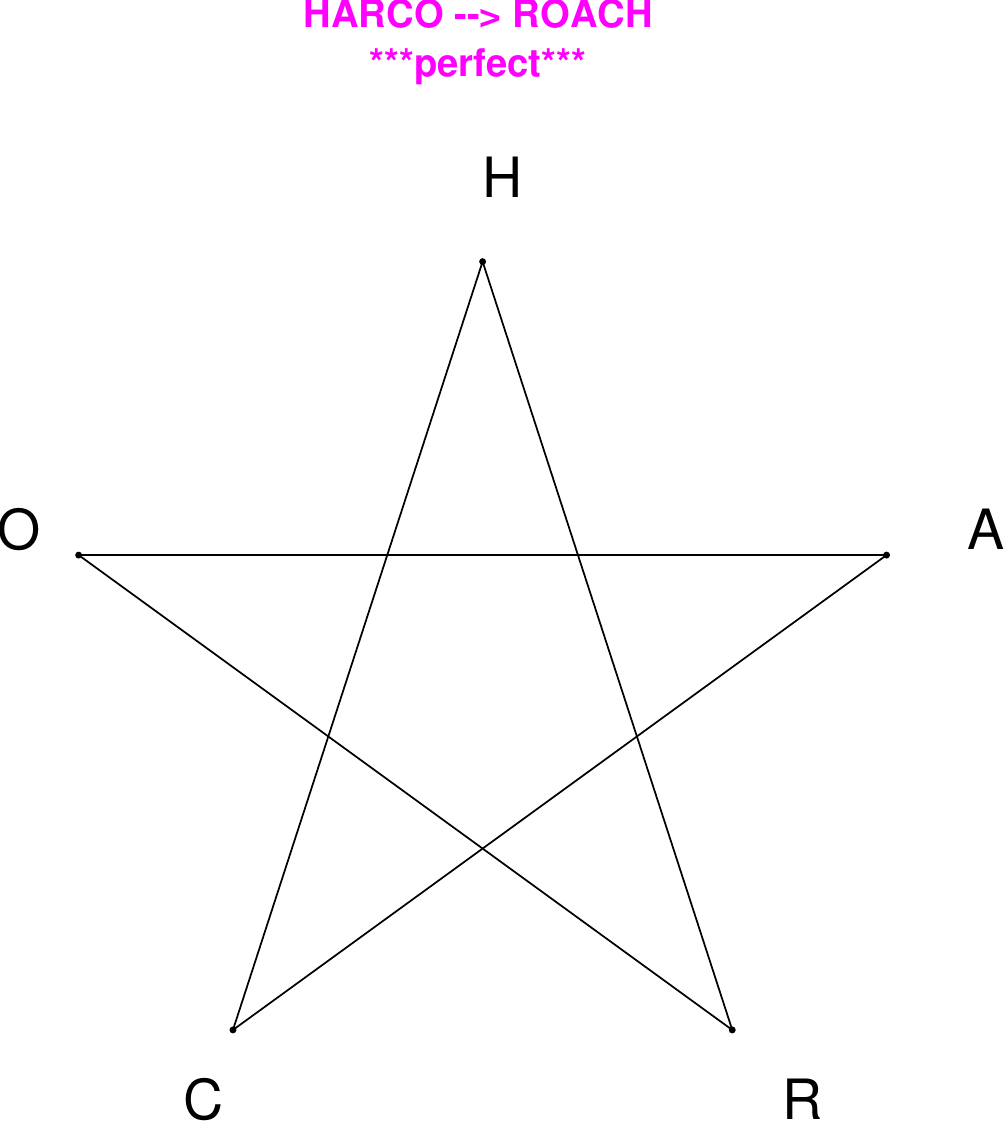}
\end{subfigure}
\hfill
\begin{subfigure}[T]{0.19\textwidth}
\centering
\includegraphics[width=\textwidth]{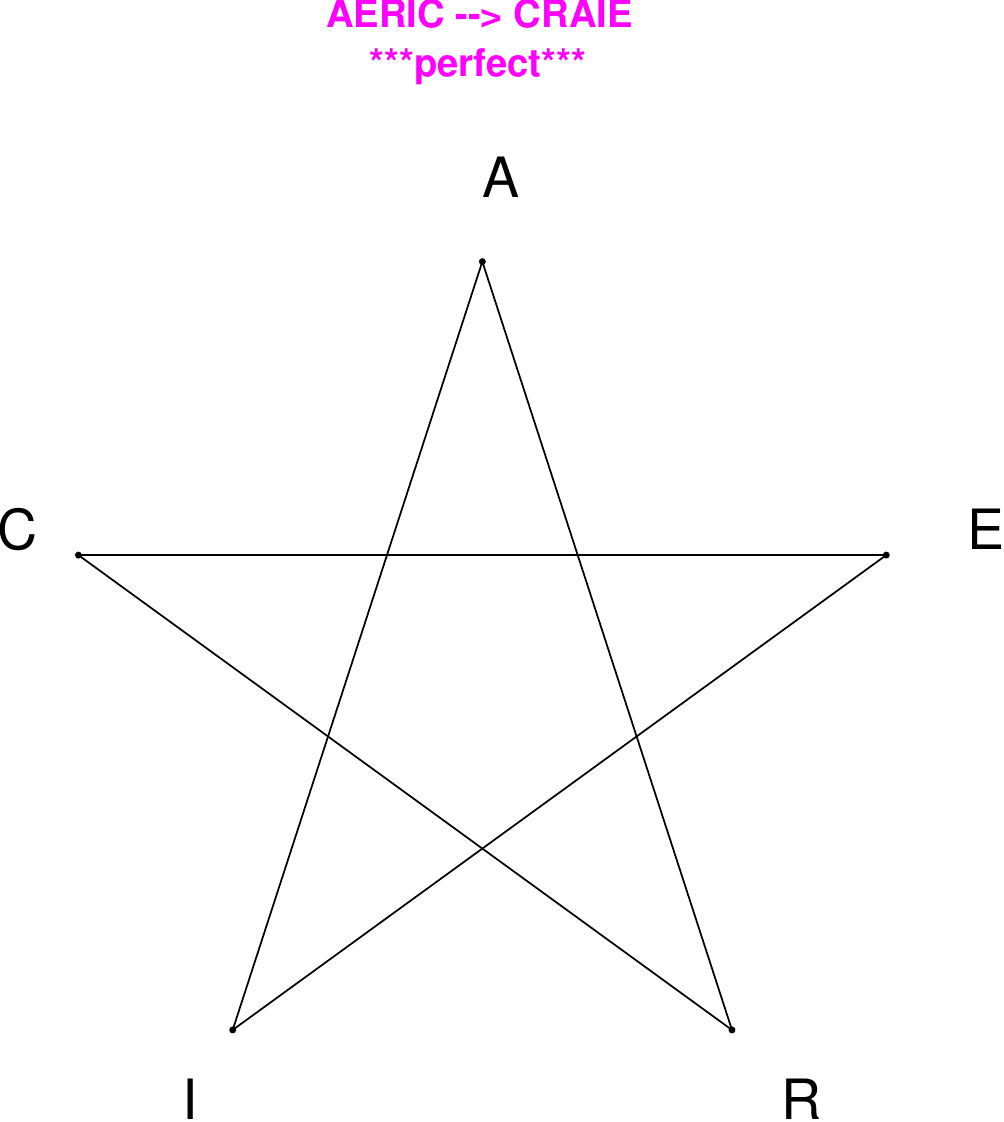}
\end{subfigure}
\hfill
\begin{subfigure}[T]{0.19\textwidth}
\centering
\includegraphics[width=\textwidth]{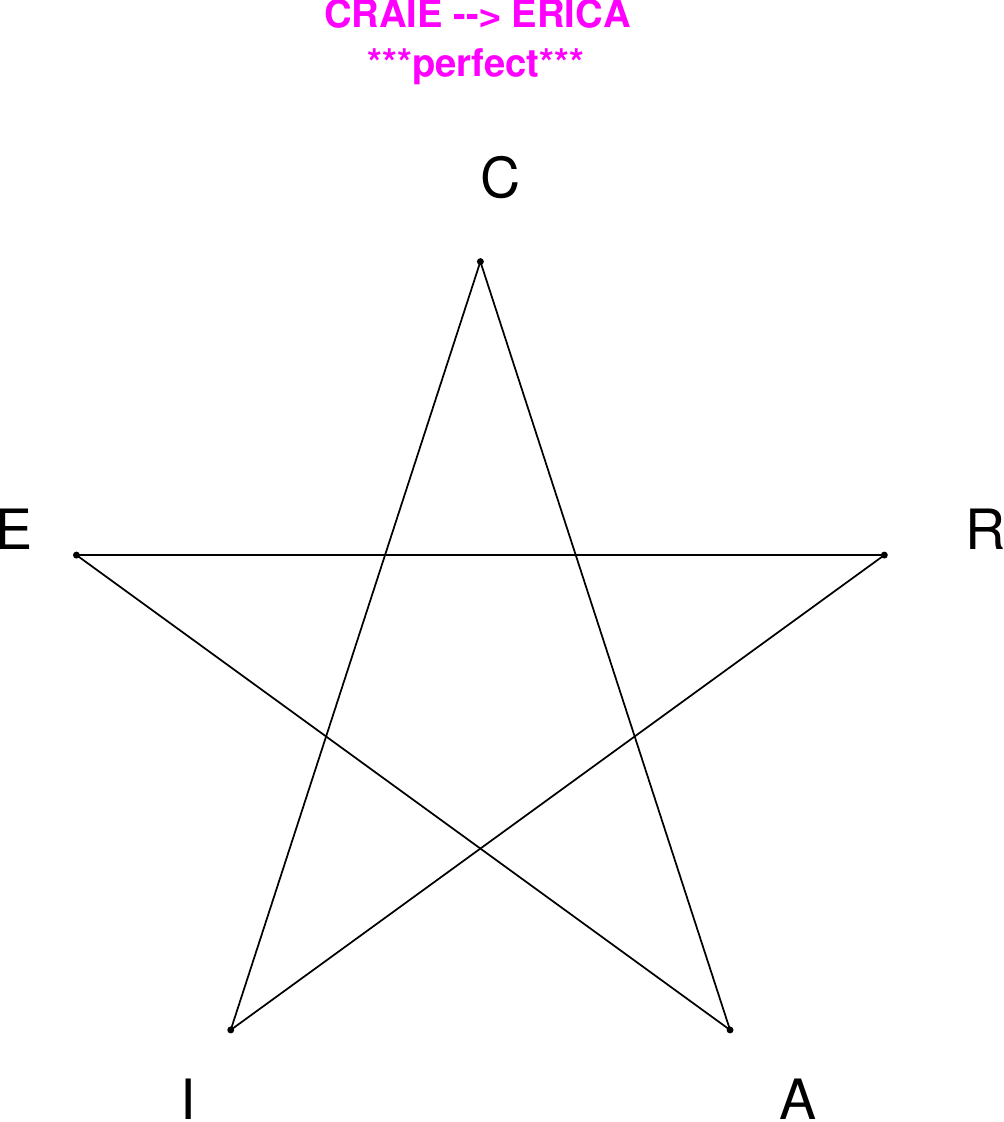}
\end{subfigure}
\hfill
\begin{subfigure}[T]{0.19\textwidth}
\centering
\includegraphics[width=\textwidth]{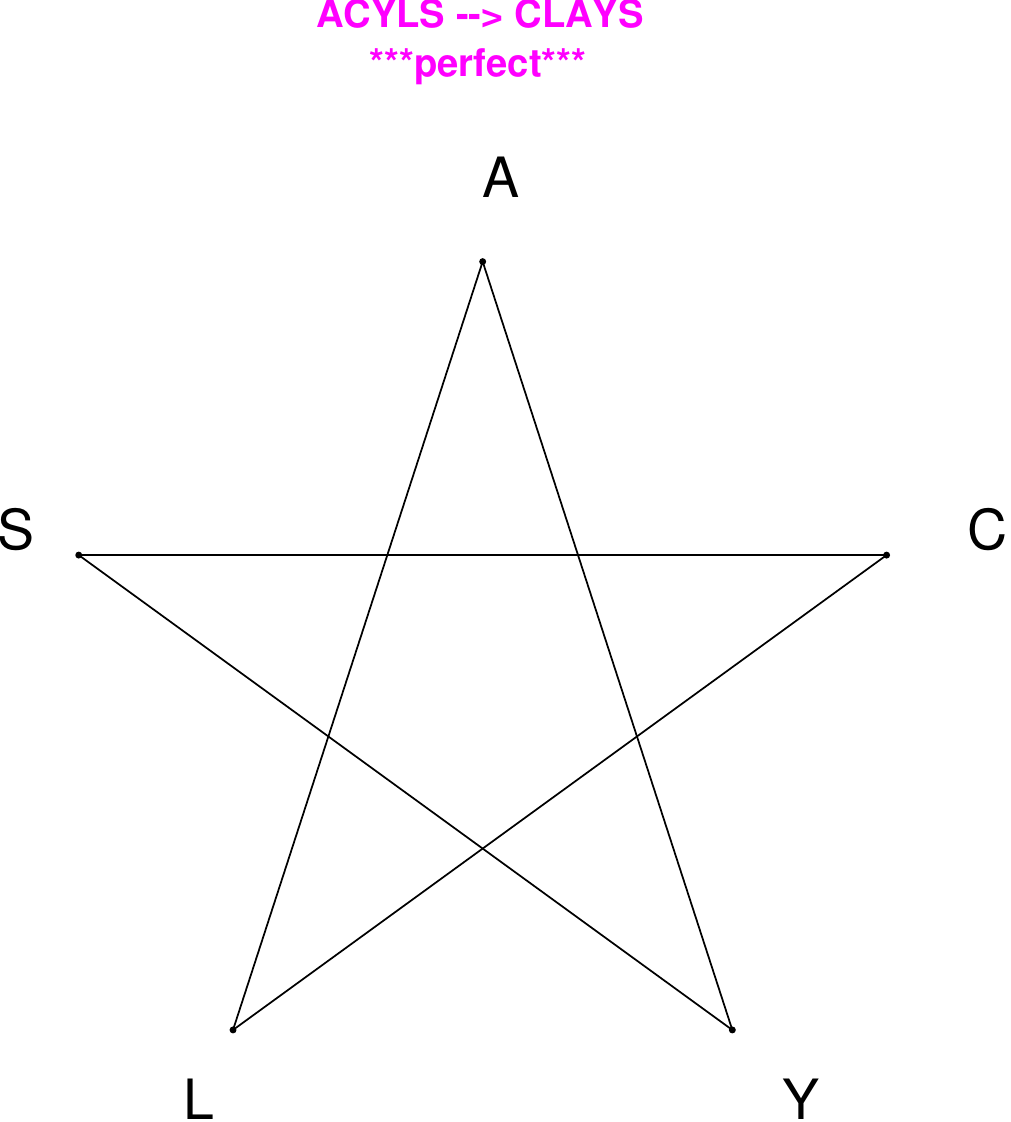}
\end{subfigure}
\hfill
\begin{subfigure}[T]{0.19\textwidth}
\centering
\includegraphics[width=\textwidth]{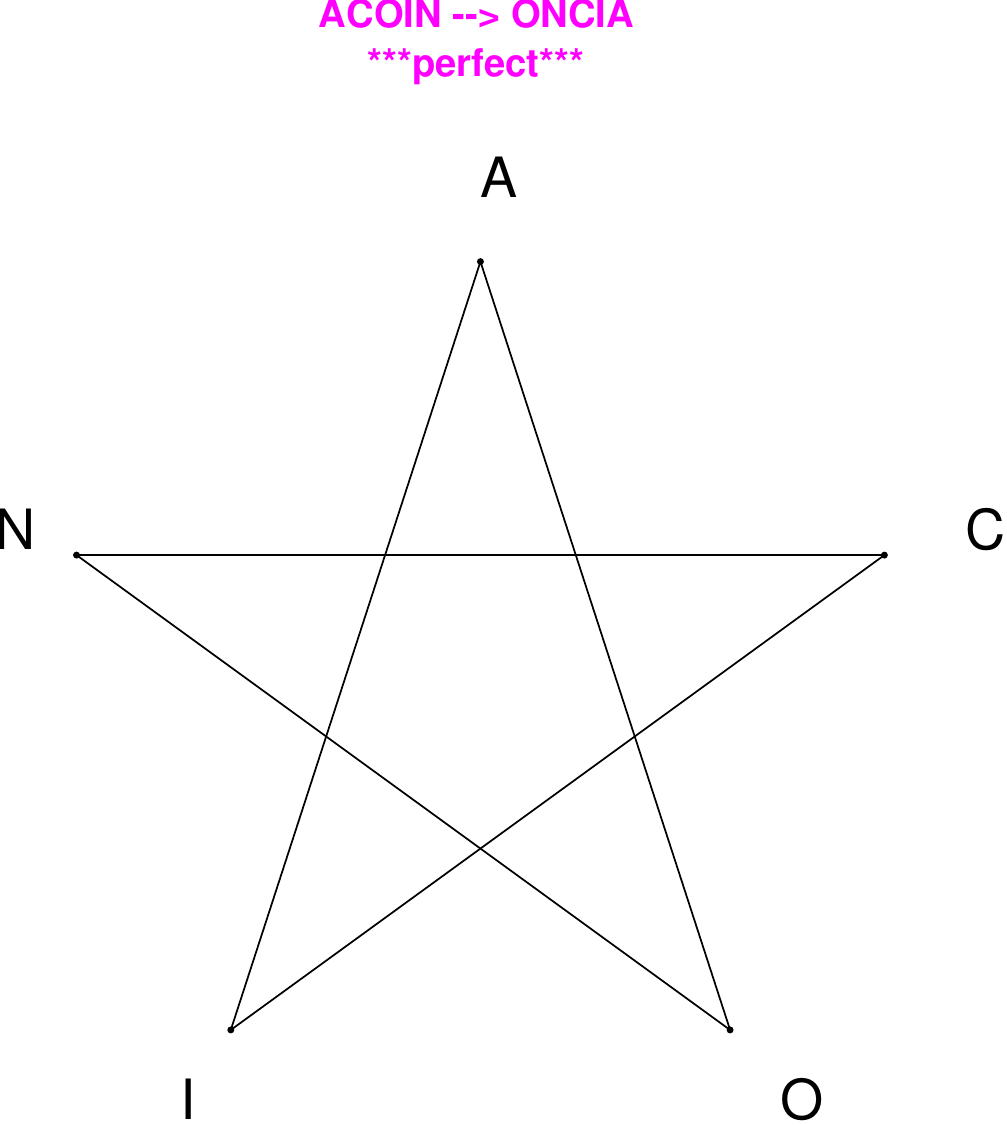}
\end{subfigure}
\end{figure}

\begin{figure}[H]
\centering
\begin{subfigure}[T]{0.19\textwidth}
\centering
\includegraphics[width=\textwidth]{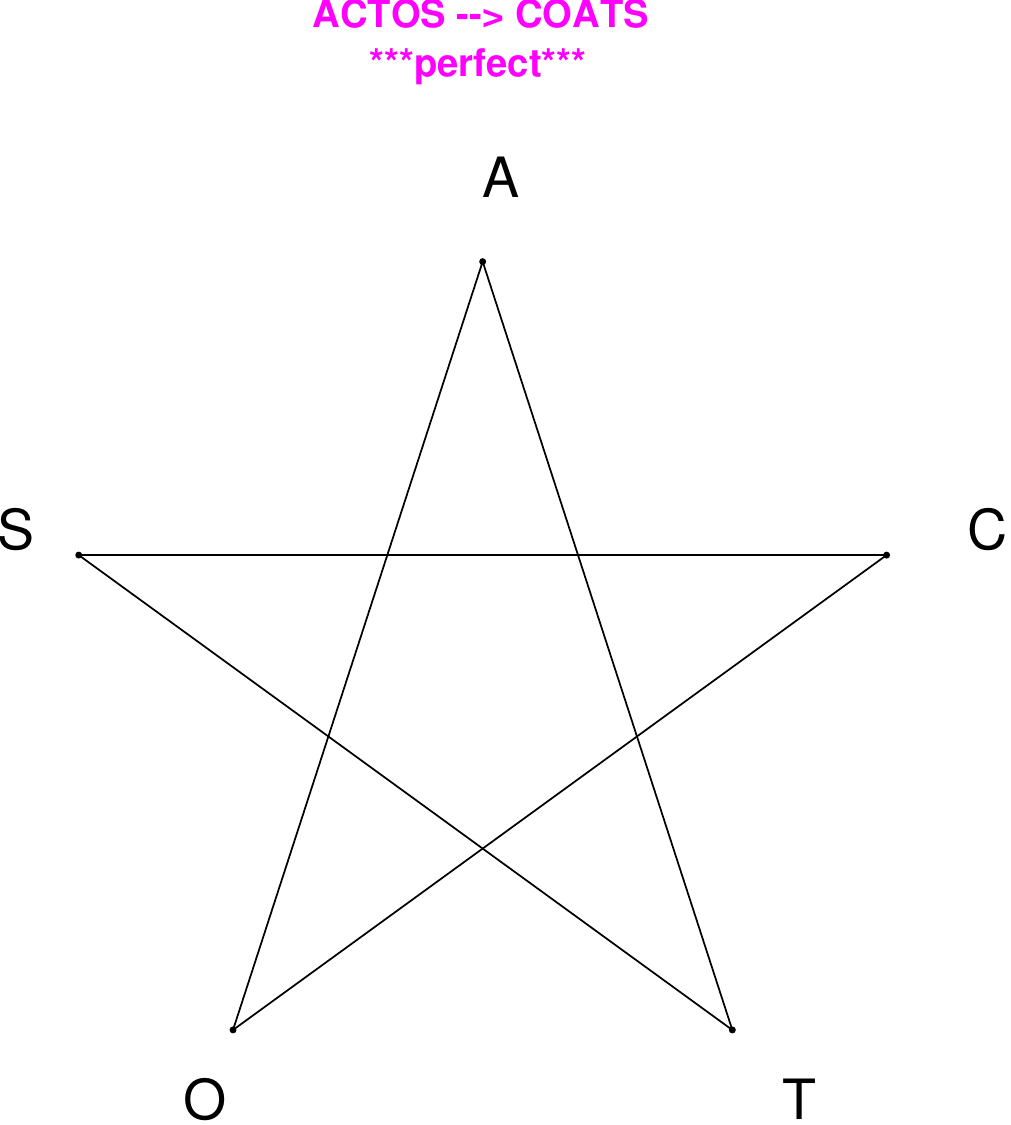}
\end{subfigure}
\hfill
\begin{subfigure}[T]{0.19\textwidth}
\centering
\includegraphics[width=\textwidth]{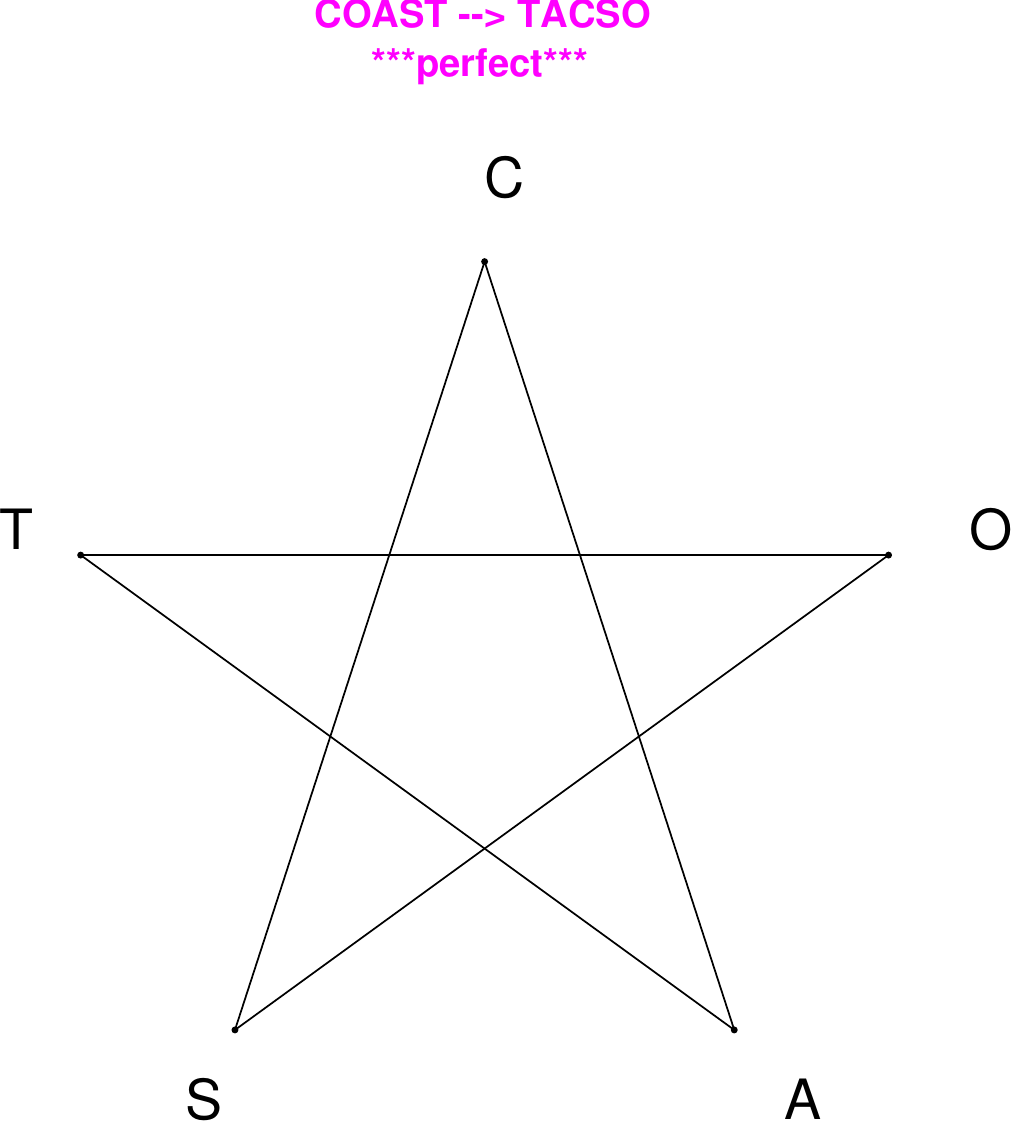}
\end{subfigure}
\hfill
\begin{subfigure}[T]{0.19\textwidth}
\centering
\includegraphics[width=\textwidth]{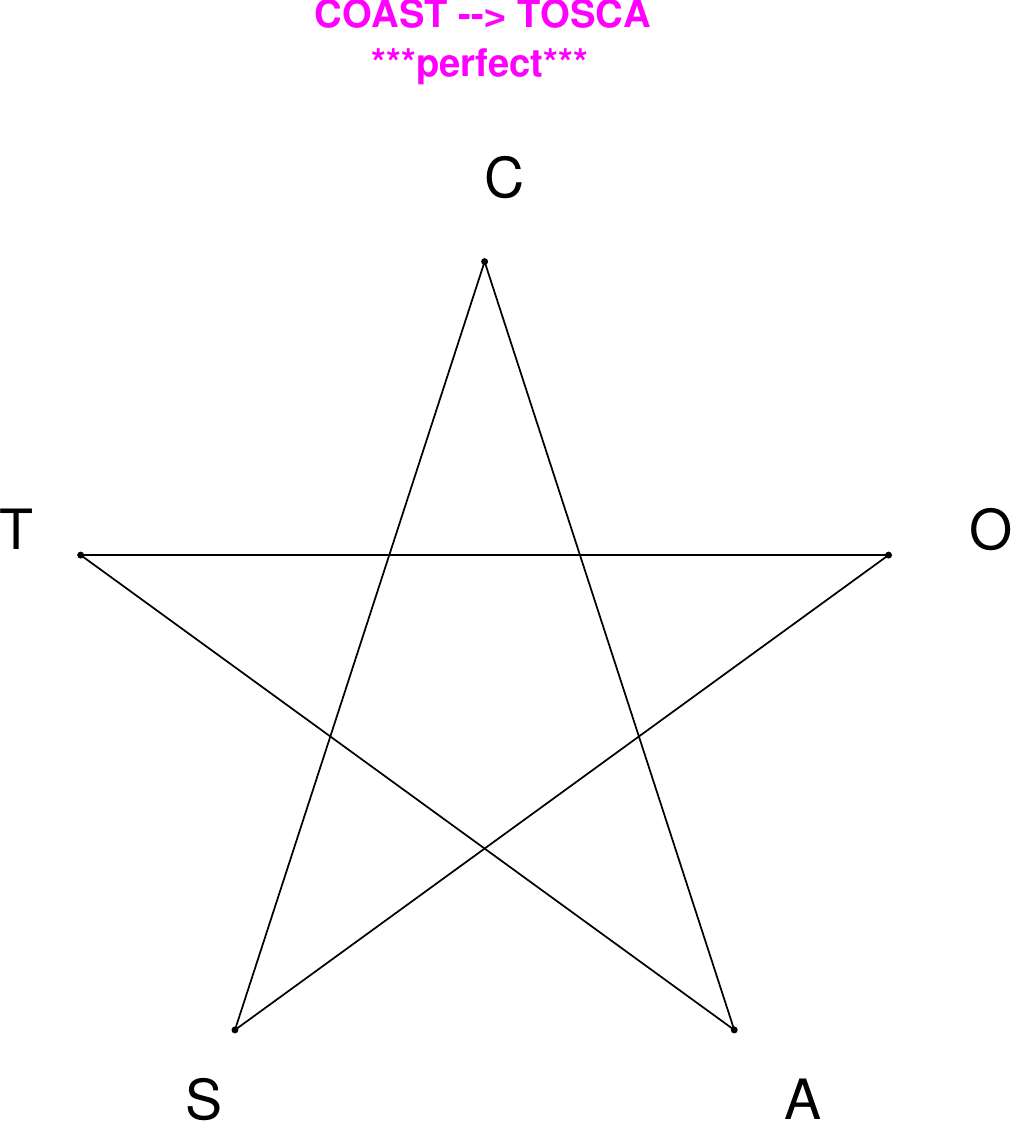}
\end{subfigure}
\hfill
\begin{subfigure}[T]{0.19\textwidth}
\centering
\includegraphics[width=\textwidth]{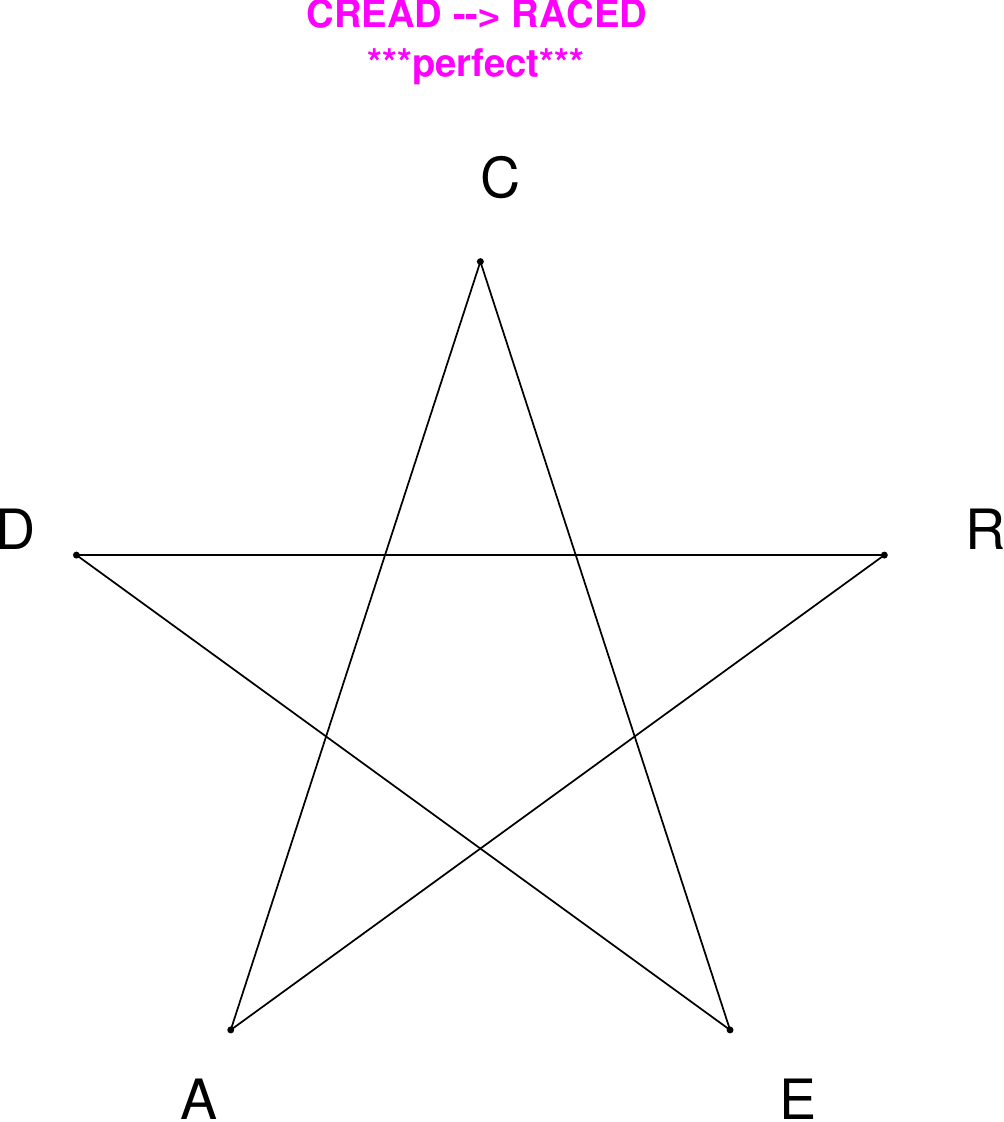}
\end{subfigure}
\hfill
\begin{subfigure}[T]{0.19\textwidth}
\centering
\includegraphics[width=\textwidth]{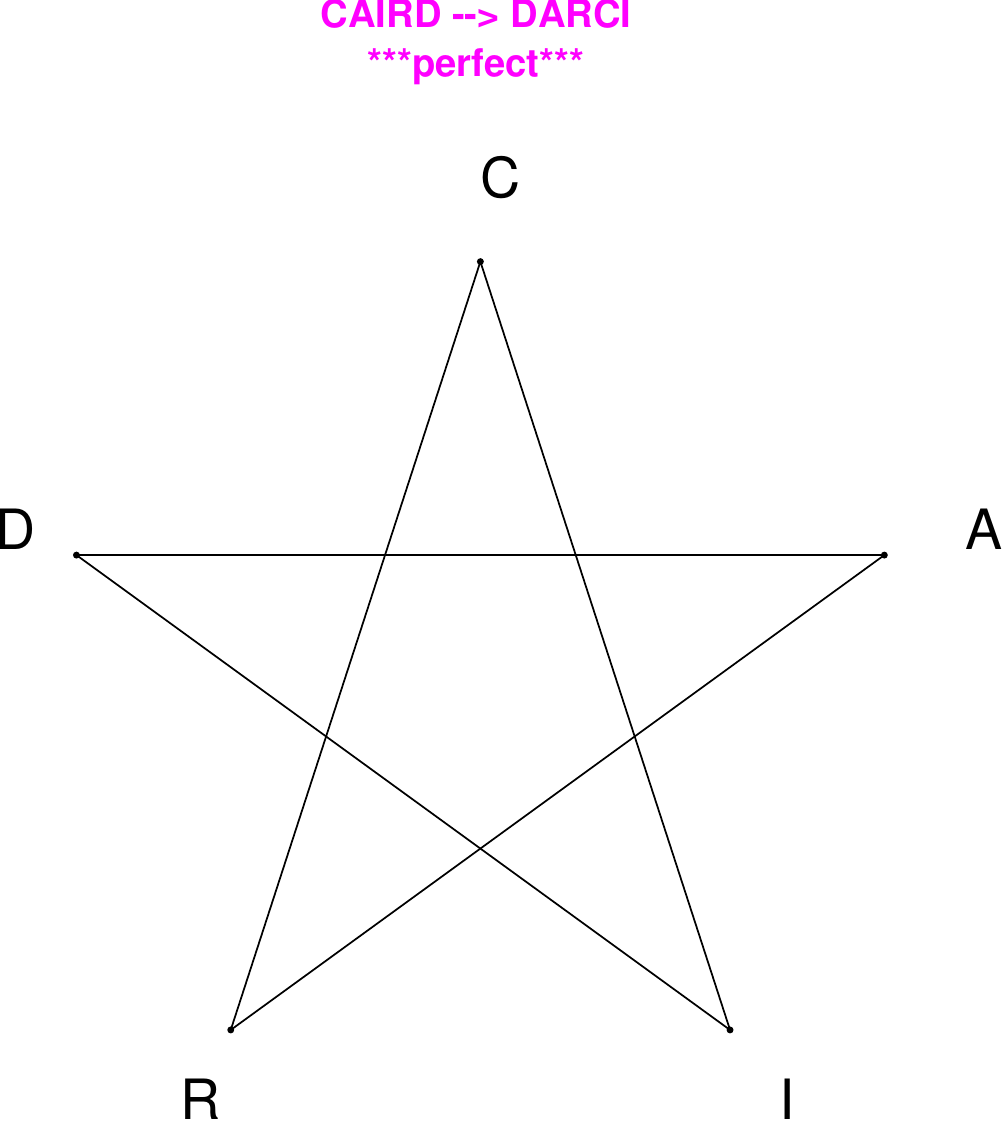}
\end{subfigure}
\end{figure}

\begin{figure}[H]
\centering
\begin{subfigure}[T]{0.19\textwidth}
\centering
\includegraphics[width=\textwidth]{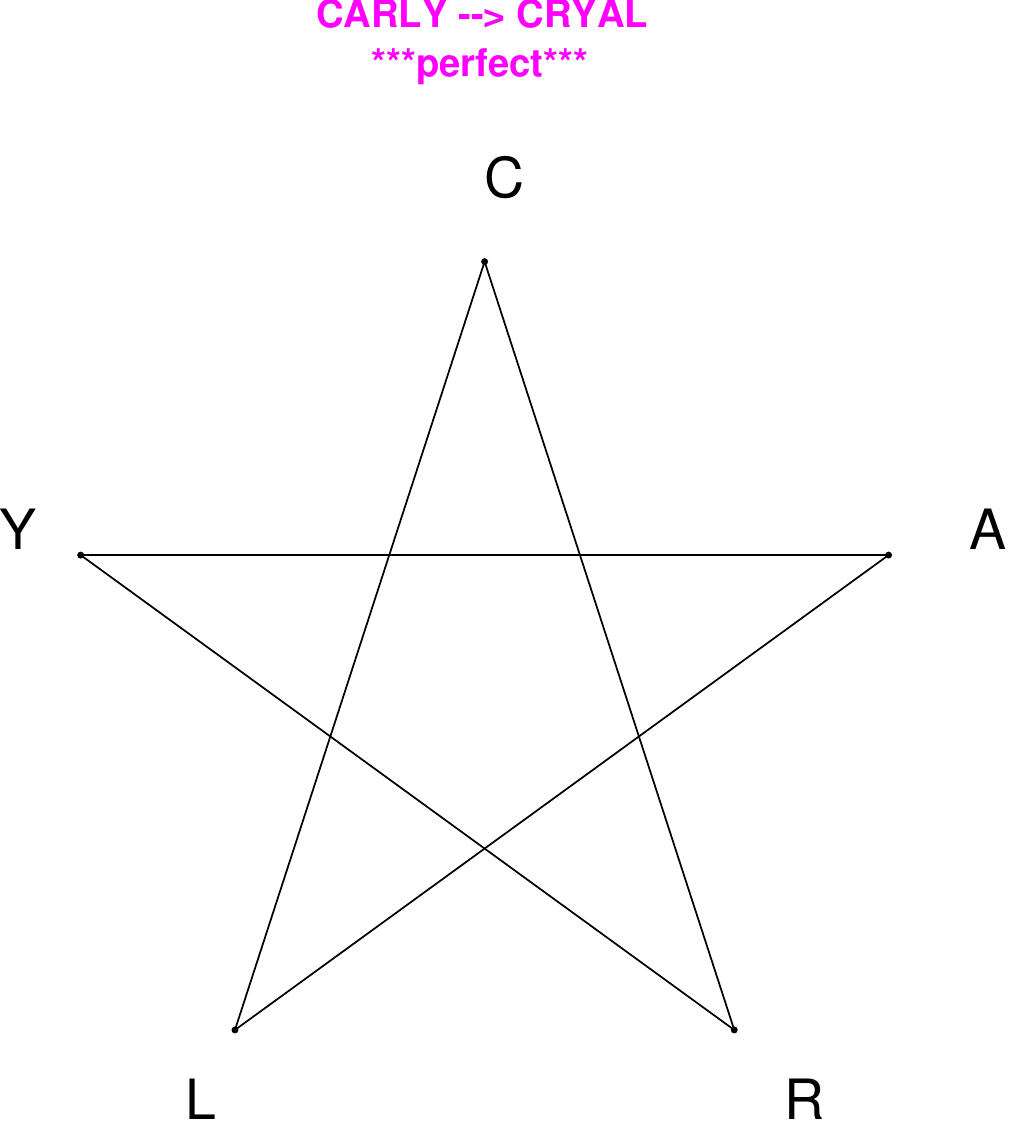}
\end{subfigure}
\hfill
\begin{subfigure}[T]{0.19\textwidth}
\centering
\includegraphics[width=\textwidth]{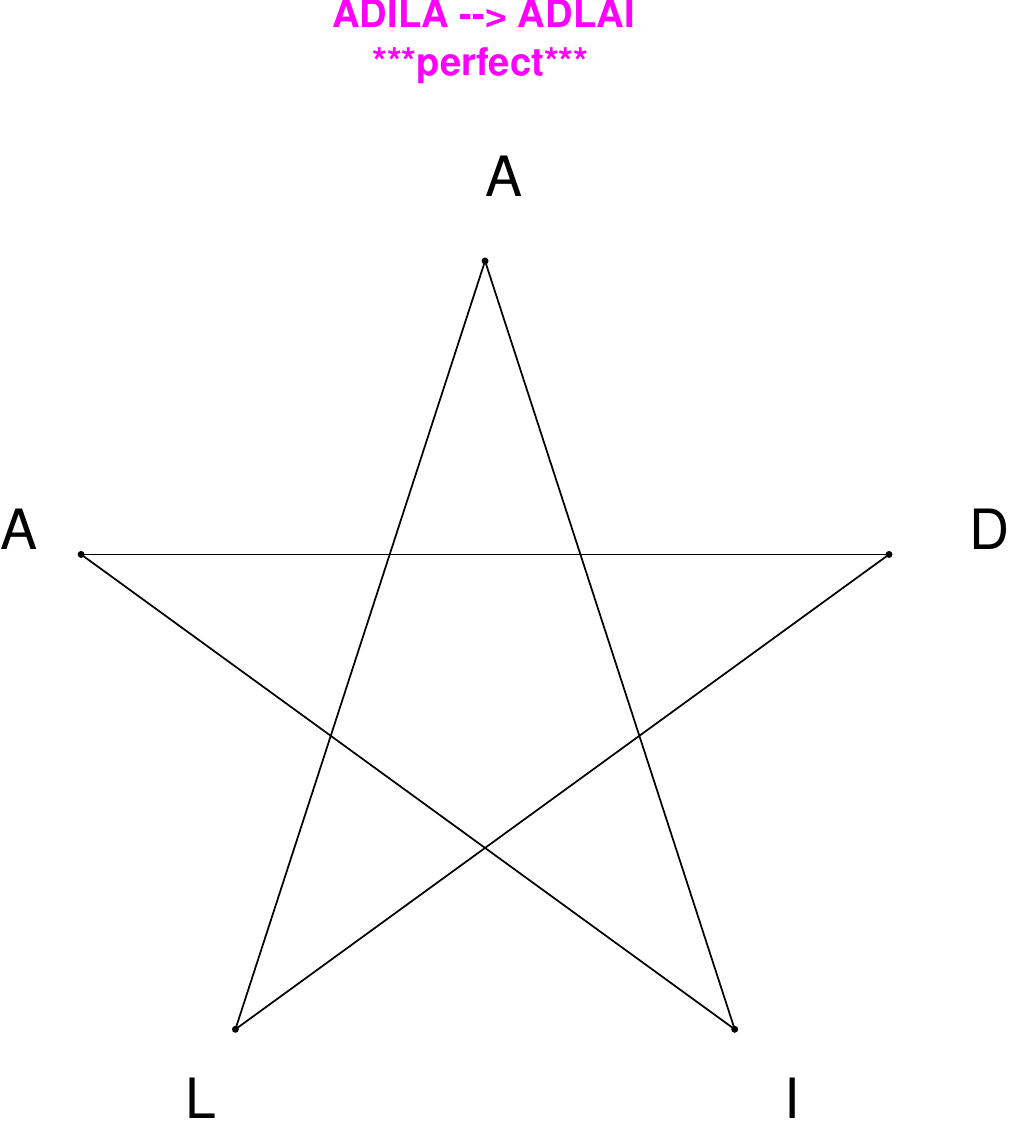}
\end{subfigure}
\hfill
\begin{subfigure}[T]{0.19\textwidth}
\centering
\includegraphics[width=\textwidth]{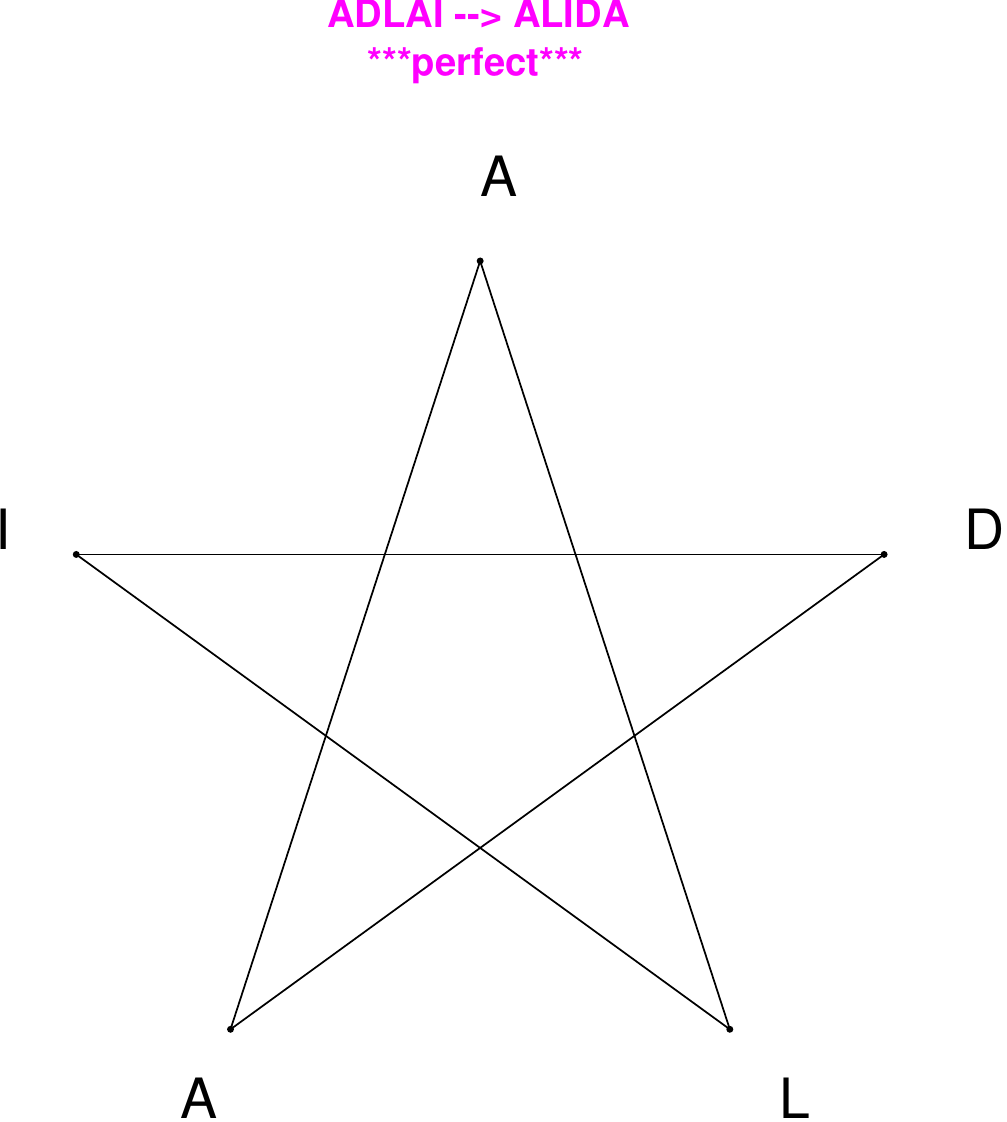}
\end{subfigure}
\hfill
\begin{subfigure}[T]{0.19\textwidth}
\centering
\includegraphics[width=\textwidth]{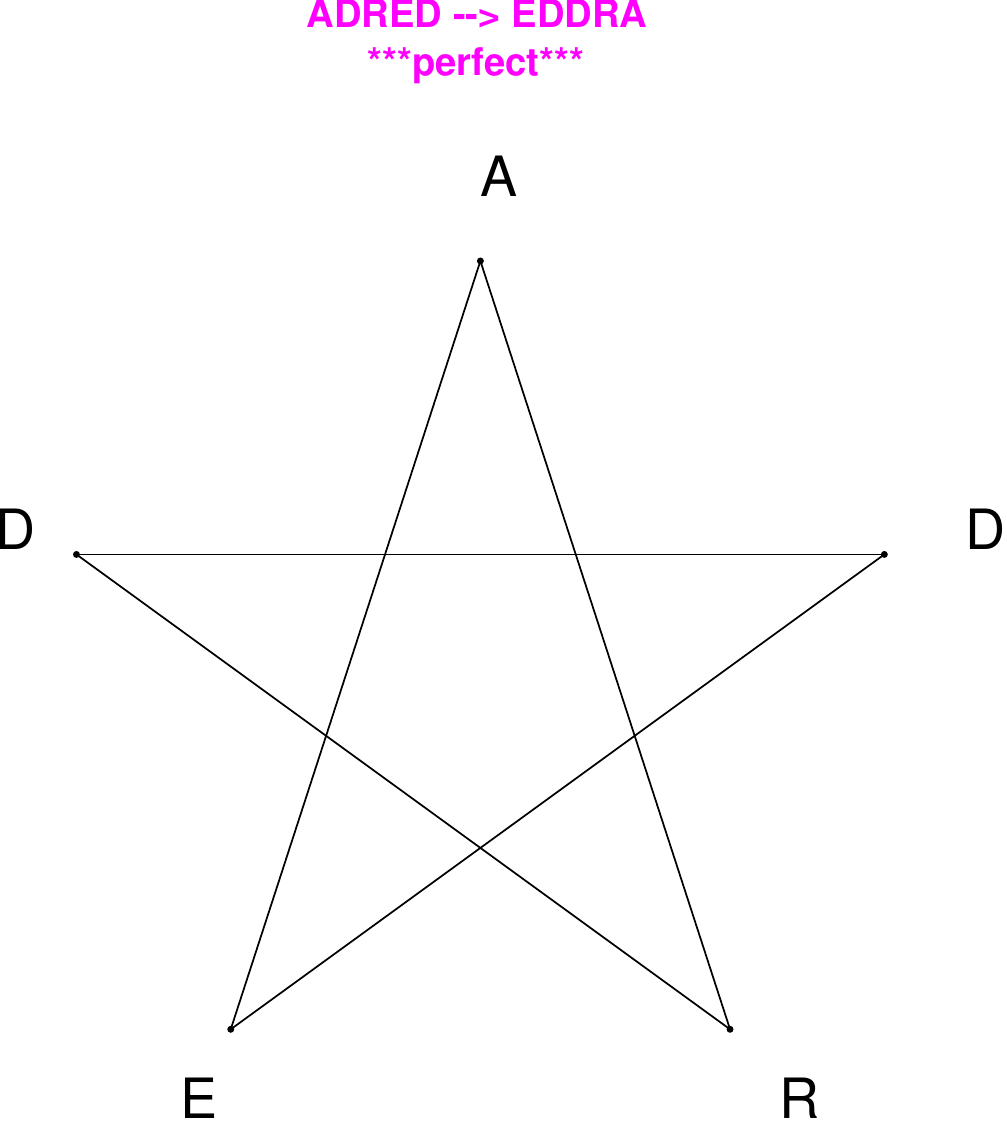}
\end{subfigure}
\hfill
\begin{subfigure}[T]{0.19\textwidth}
\centering
\includegraphics[width=\textwidth]{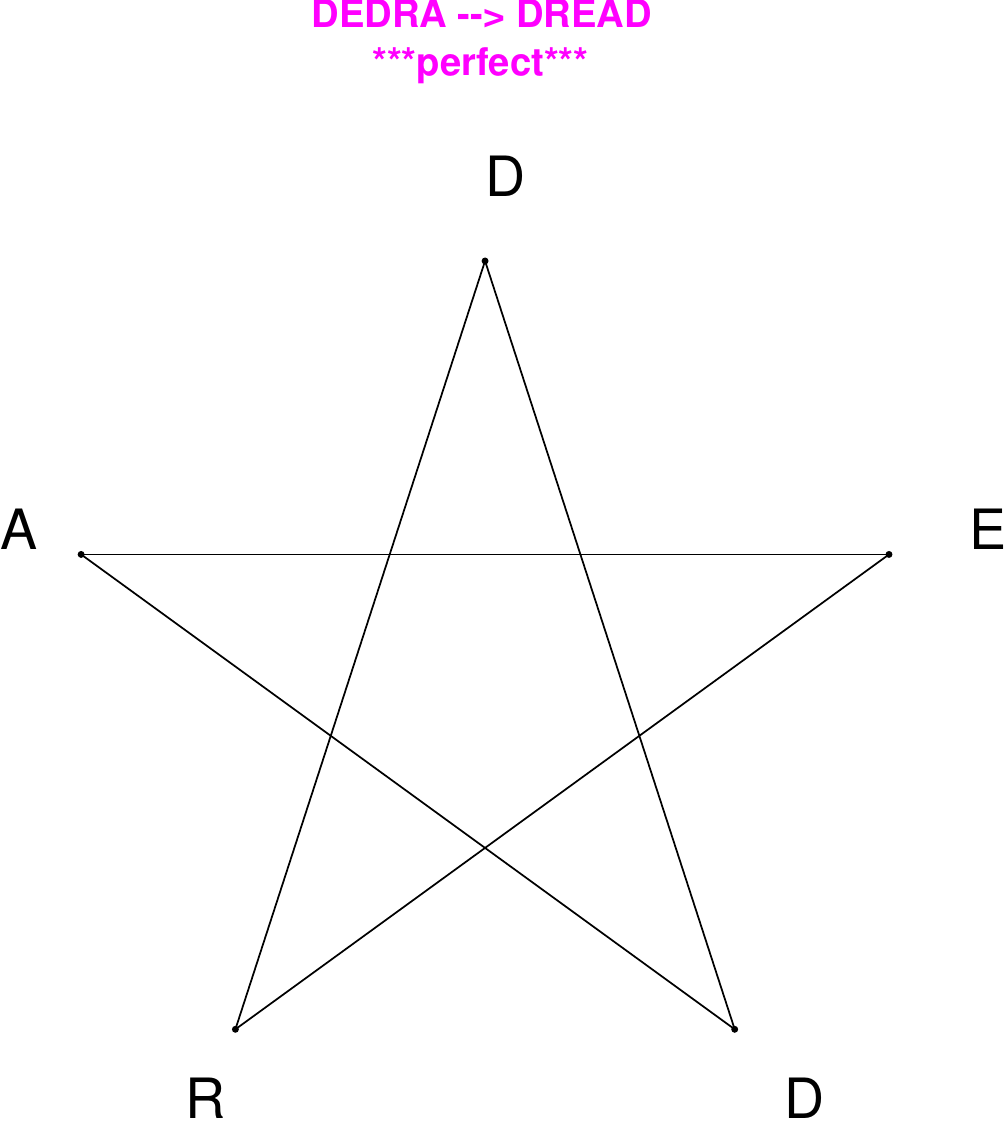}
\end{subfigure}
\end{figure}

\begin{figure}[H]
\centering
\begin{subfigure}[T]{0.19\textwidth}
\centering
\includegraphics[width=\textwidth]{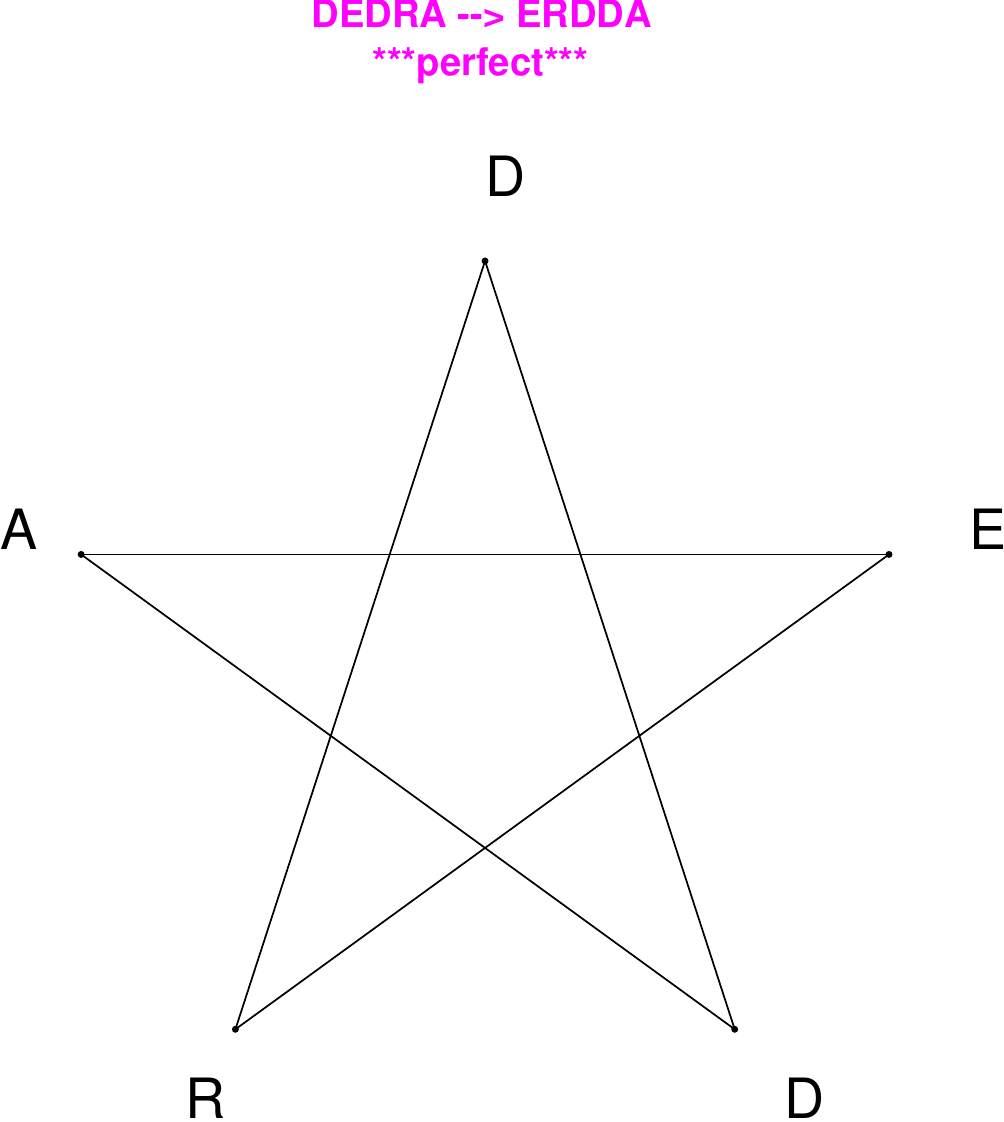}
\end{subfigure}
\hfill
\begin{subfigure}[T]{0.19\textwidth}
\centering
\includegraphics[width=\textwidth]{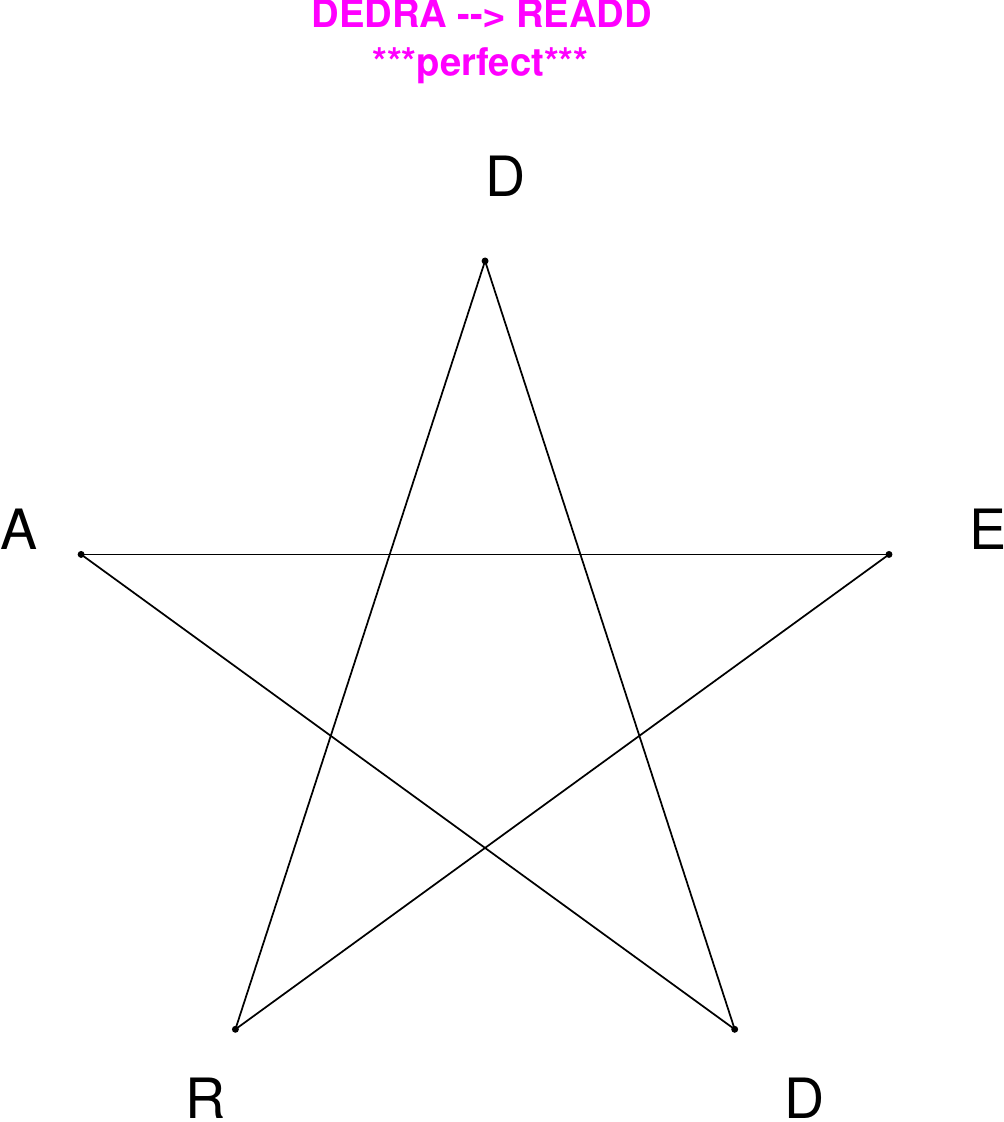}
\end{subfigure}
\hfill
\begin{subfigure}[T]{0.19\textwidth}
\centering
\includegraphics[width=\textwidth]{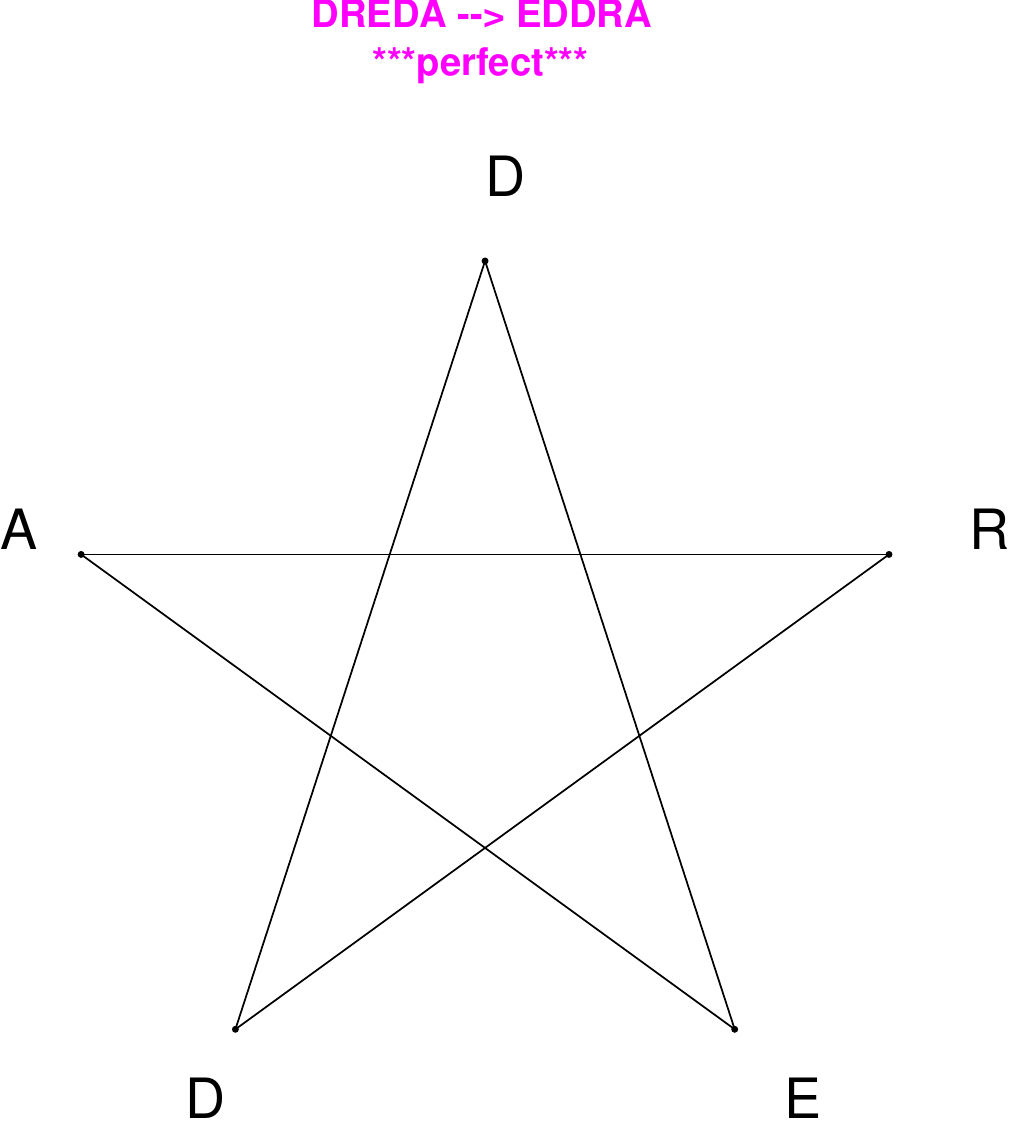}
\end{subfigure}
\hfill
\begin{subfigure}[T]{0.19\textwidth}
\centering
\includegraphics[width=\textwidth]{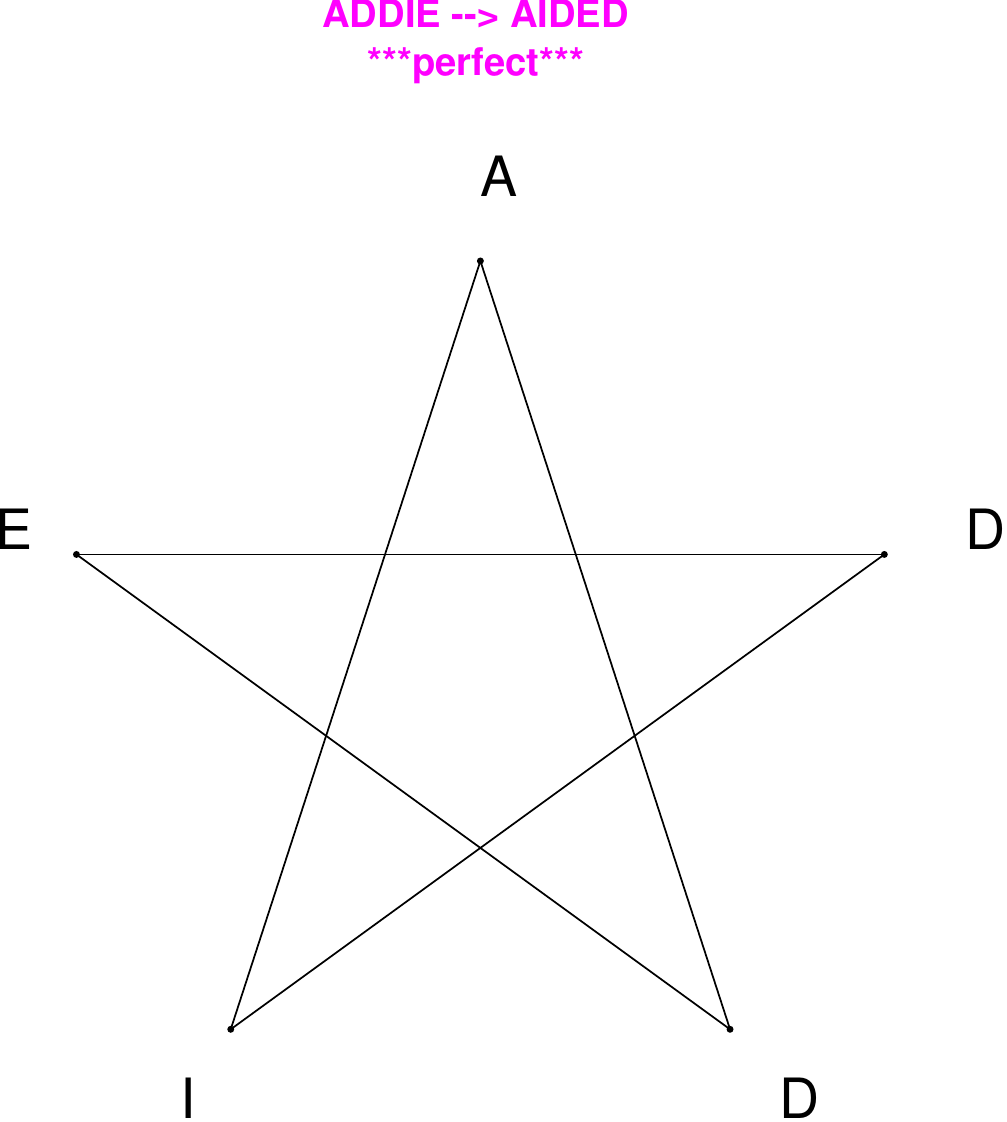}
\end{subfigure}
\hfill
\begin{subfigure}[T]{0.19\textwidth}
\centering
\includegraphics[width=\textwidth]{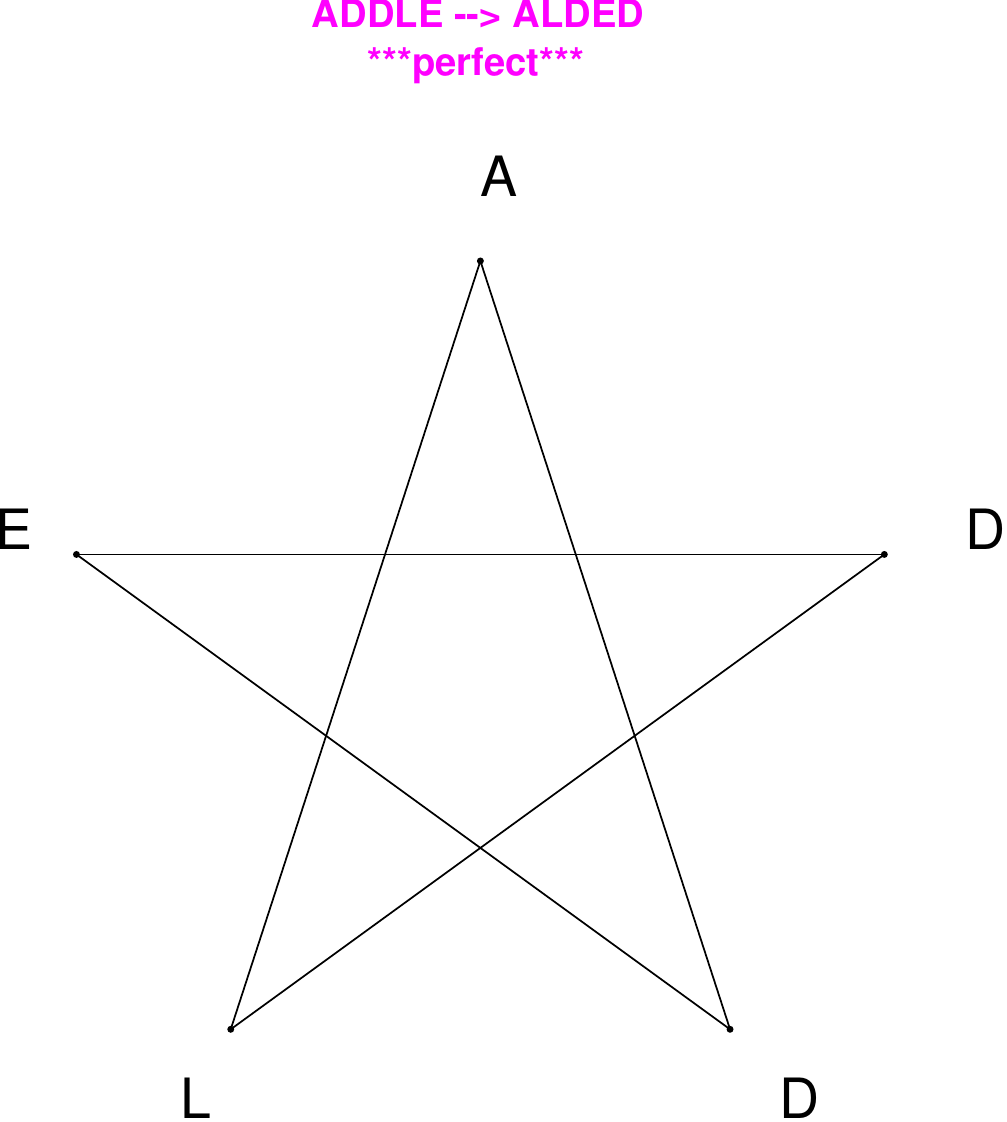}
\end{subfigure}
\end{figure}

\begin{figure}[H]
\centering
\begin{subfigure}[T]{0.19\textwidth}
\centering
\includegraphics[width=\textwidth]{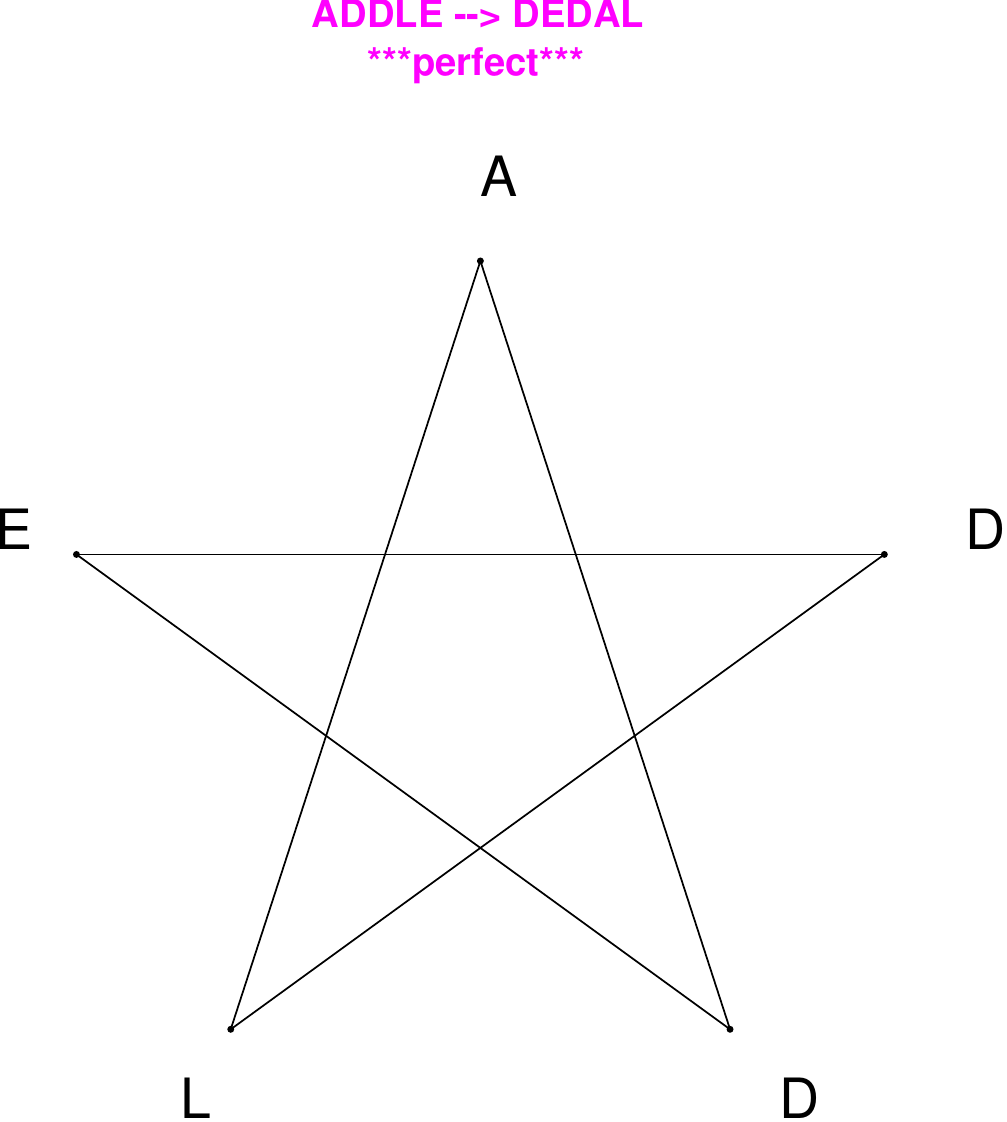}
\end{subfigure}
\hfill
\begin{subfigure}[T]{0.19\textwidth}
\centering
\includegraphics[width=\textwidth]{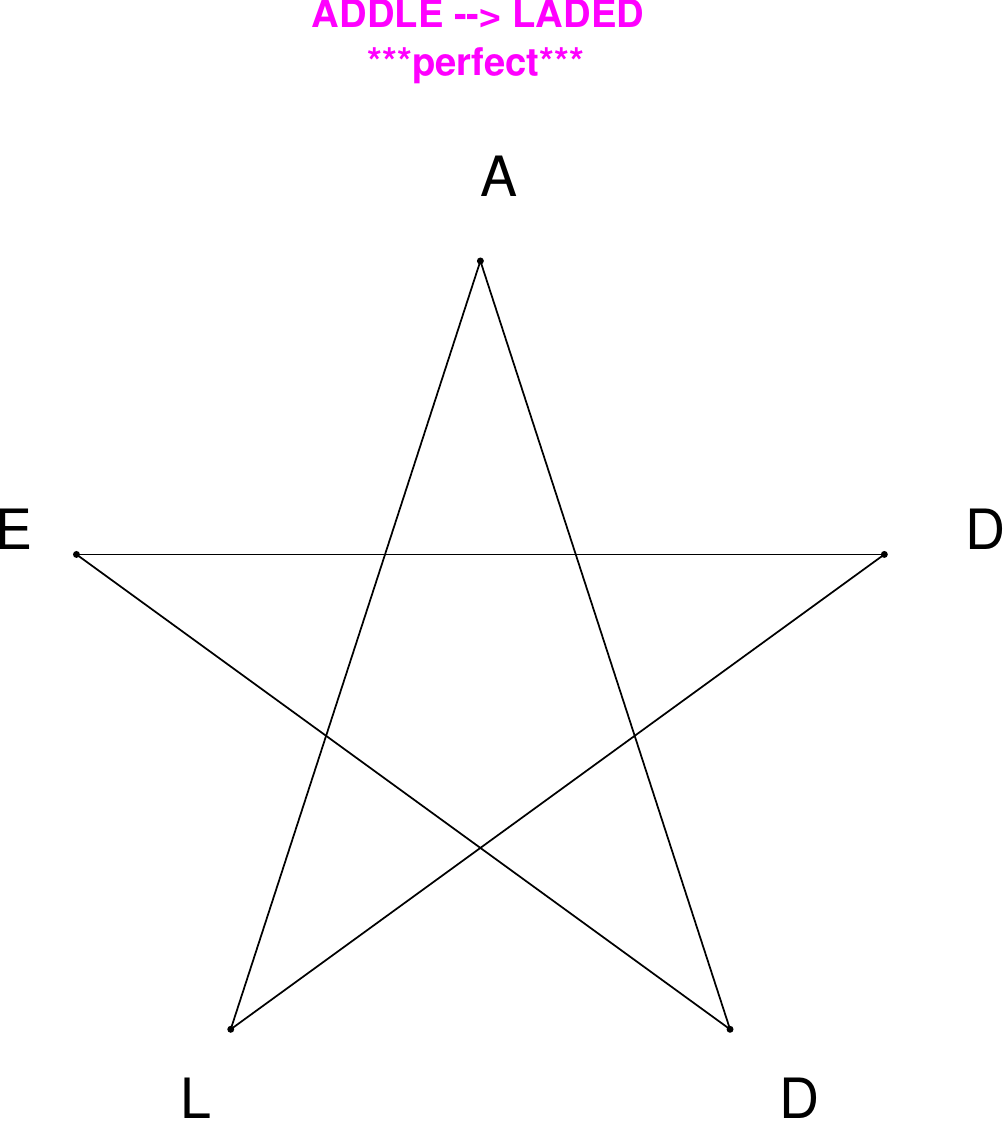}
\end{subfigure}
\hfill
\begin{subfigure}[T]{0.19\textwidth}
\centering
\includegraphics[width=\textwidth]{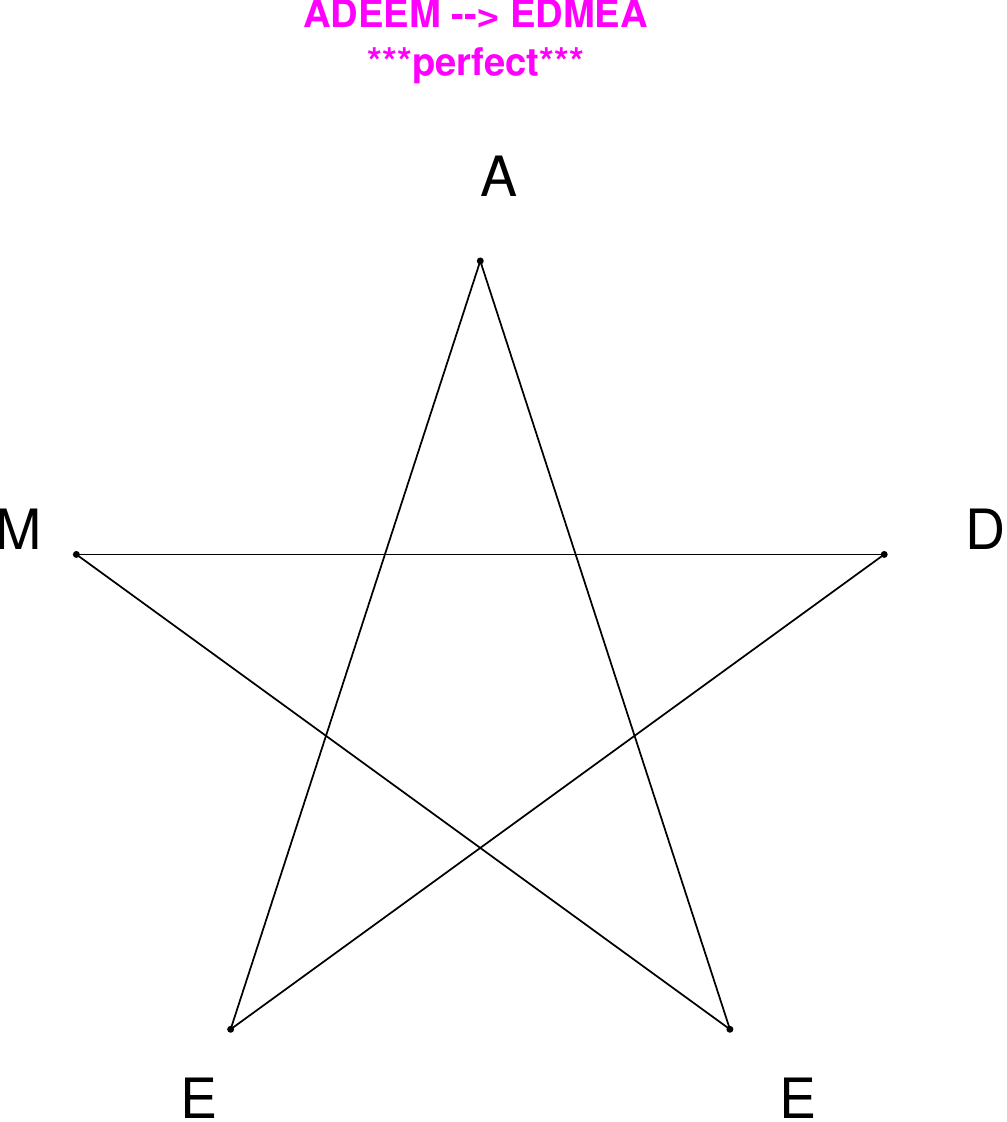}
\end{subfigure}
\hfill
\begin{subfigure}[T]{0.19\textwidth}
\centering
\includegraphics[width=\textwidth]{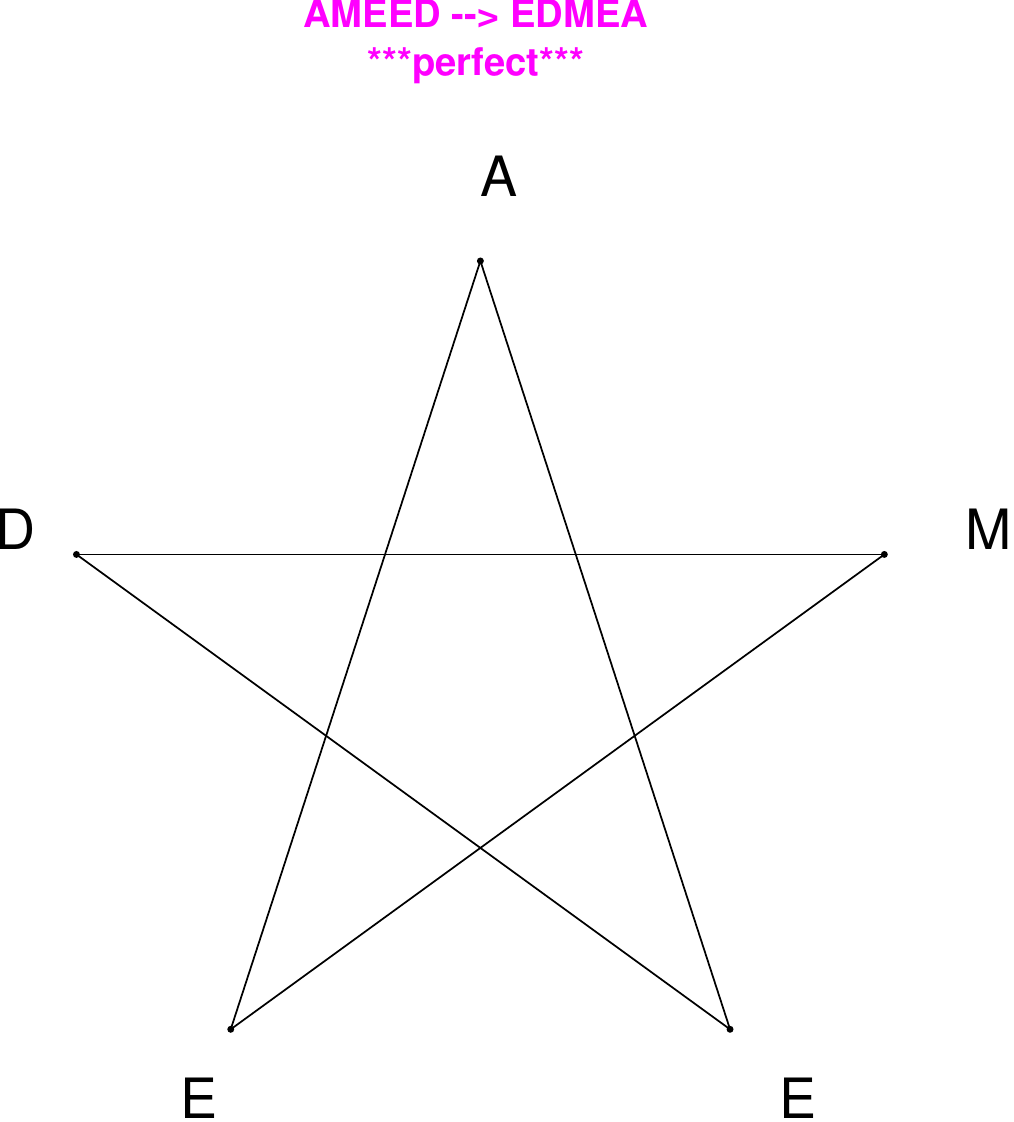}
\end{subfigure}
\hfill
\begin{subfigure}[T]{0.19\textwidth}
\centering
\includegraphics[width=\textwidth]{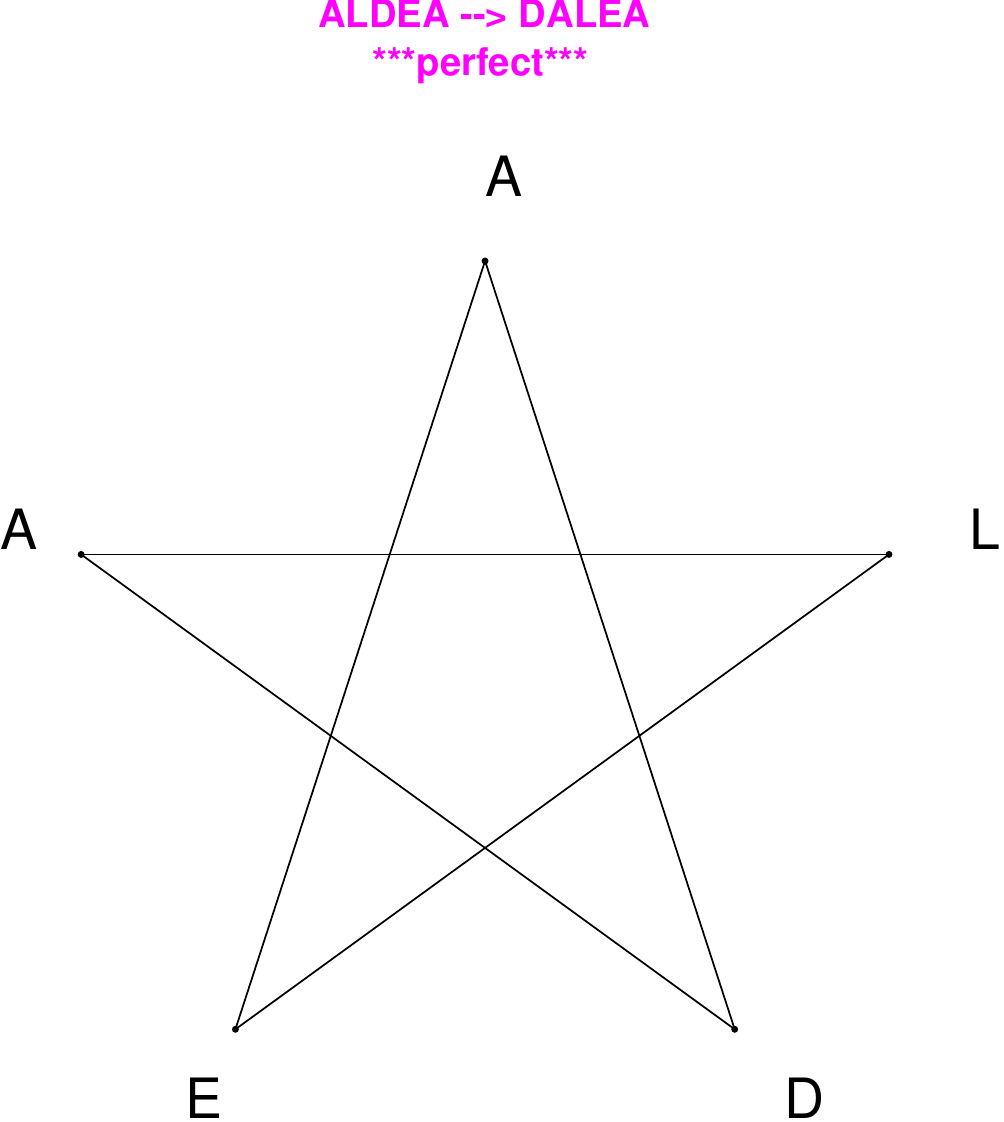}
\end{subfigure}
\end{figure}

\begin{figure}[H]
\centering
\begin{subfigure}[T]{0.19\textwidth}
\centering
\includegraphics[width=\textwidth]{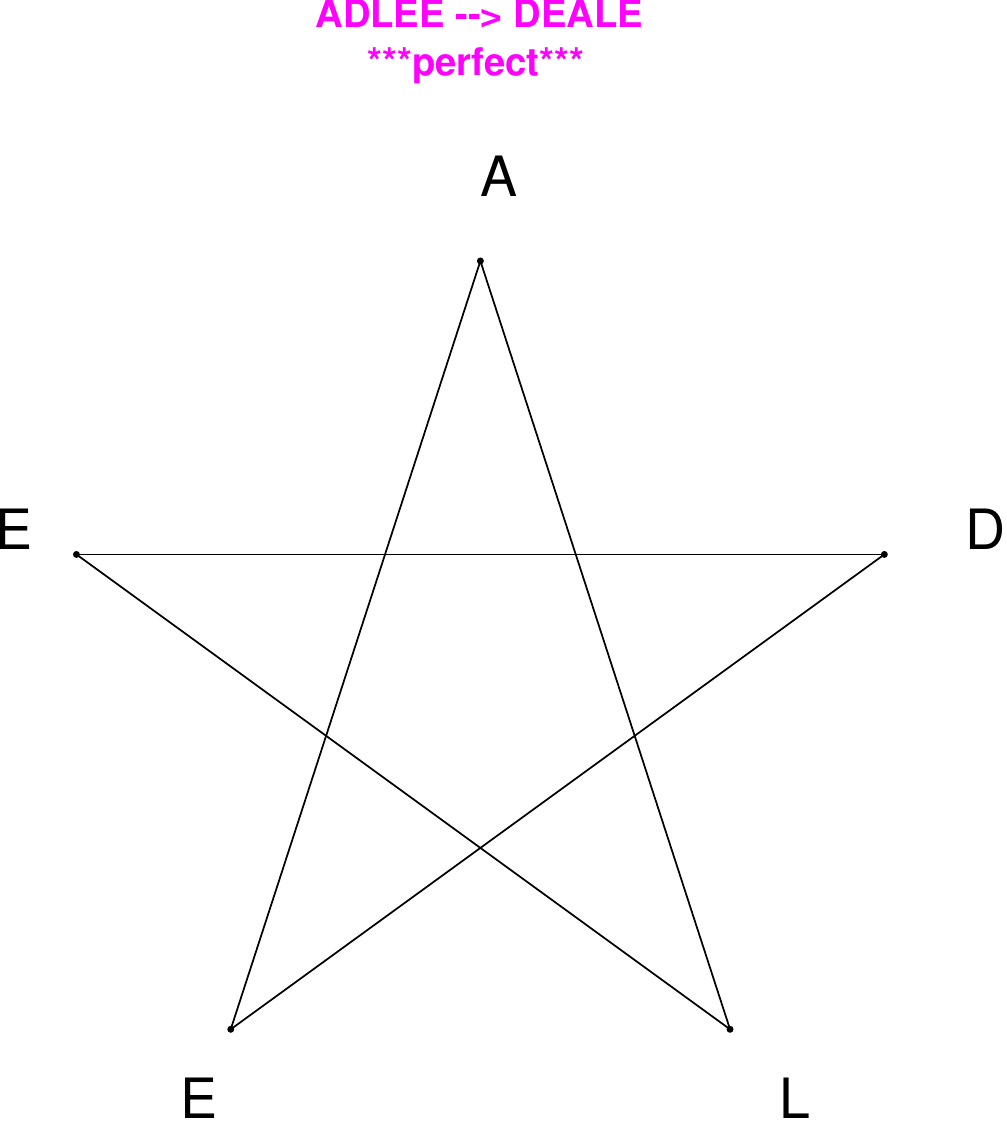}
\end{subfigure}
\hfill
\begin{subfigure}[T]{0.19\textwidth}
\centering
\includegraphics[width=\textwidth]{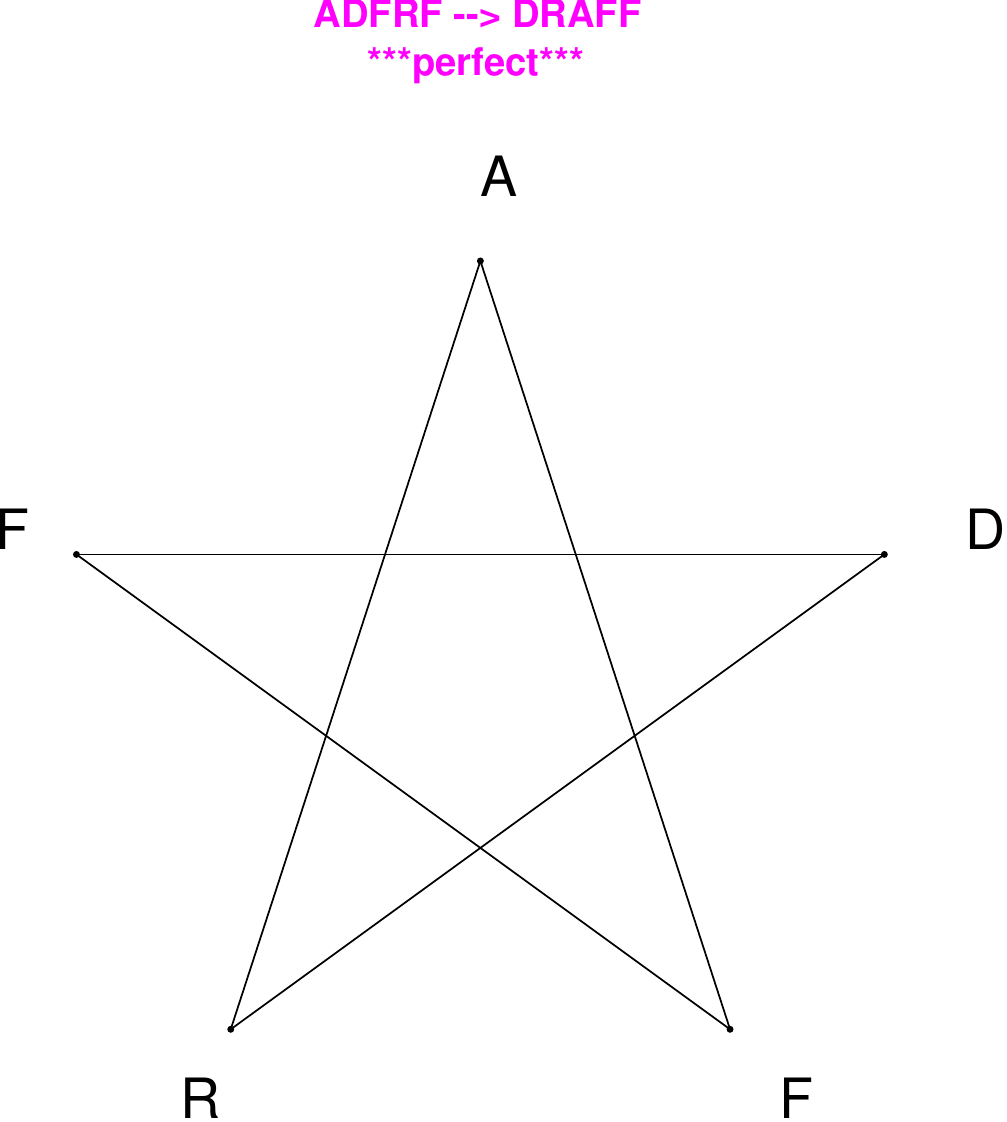}
\end{subfigure}
\hfill
\begin{subfigure}[T]{0.19\textwidth}
\centering
\includegraphics[width=\textwidth]{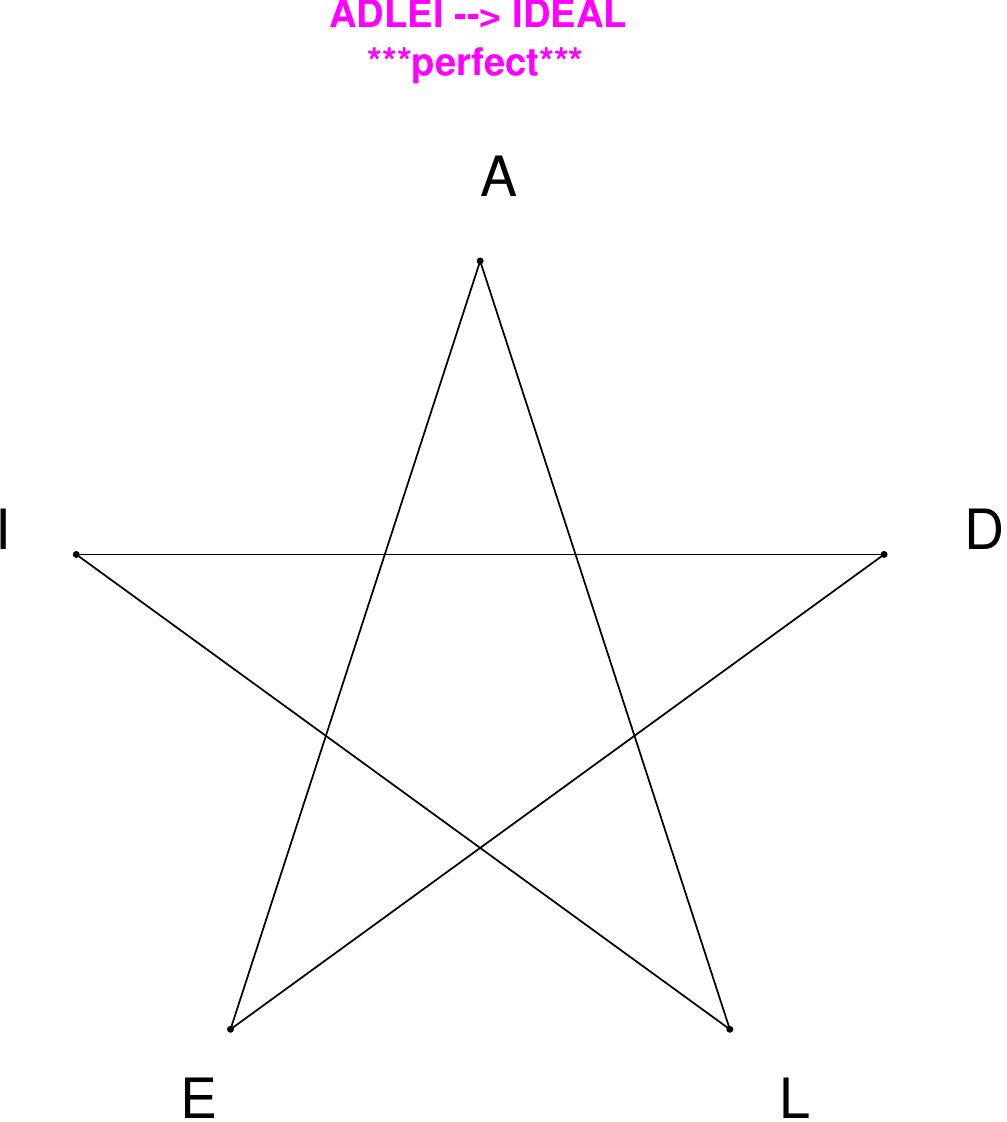}
\end{subfigure}
\hfill
\begin{subfigure}[T]{0.19\textwidth}
\centering
\includegraphics[width=\textwidth]{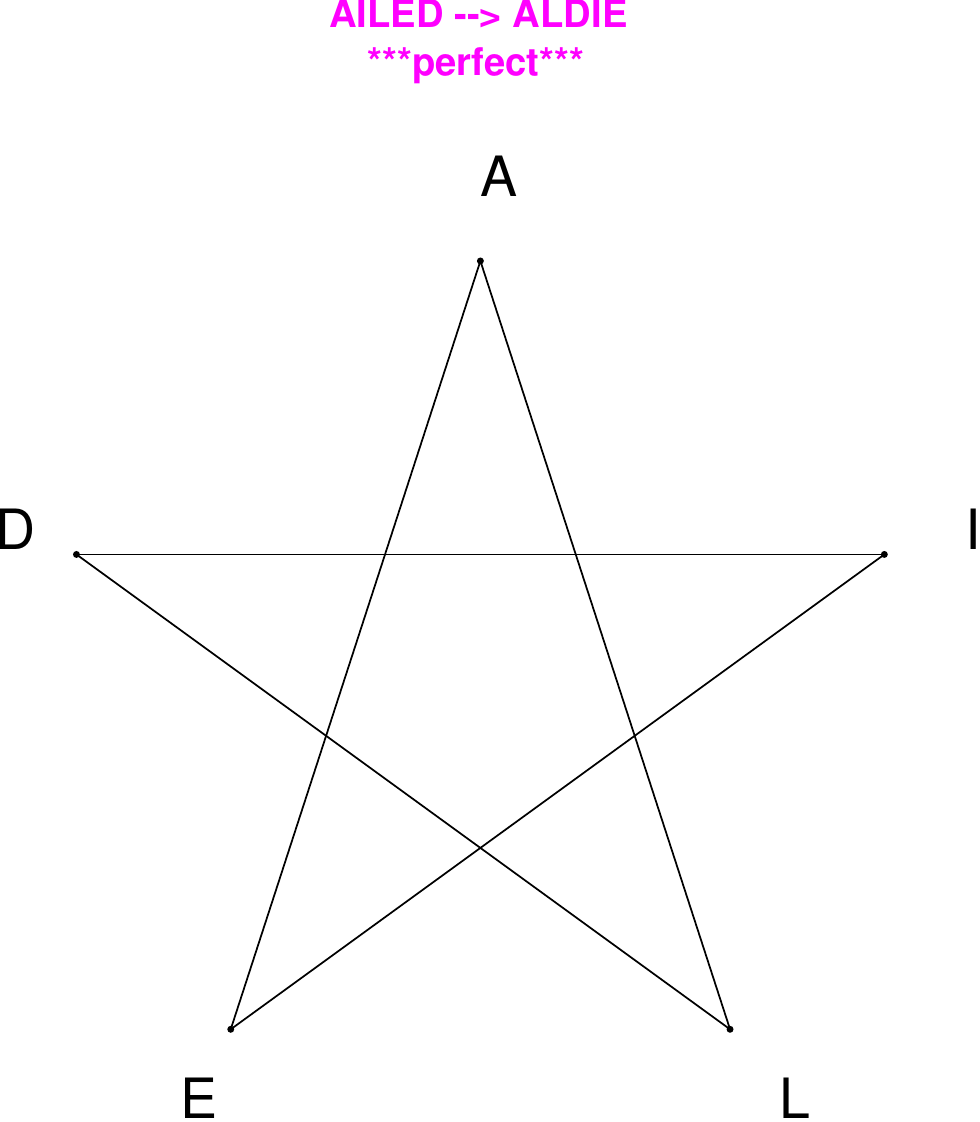}
\end{subfigure}
\hfill
\begin{subfigure}[T]{0.19\textwidth}
\centering
\includegraphics[width=\textwidth]{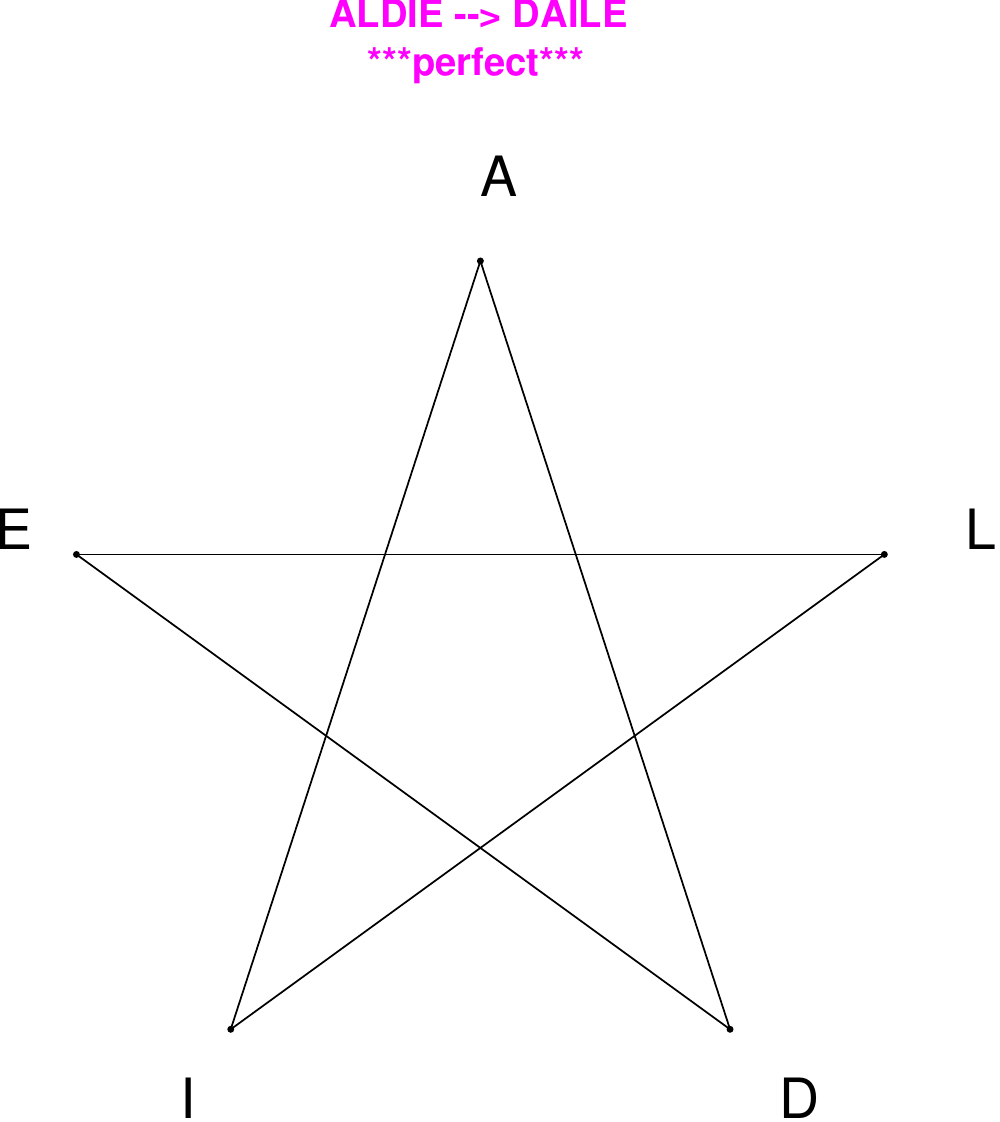}
\end{subfigure}
\end{figure}

\begin{figure}[H]
\centering
\begin{subfigure}[T]{0.19\textwidth}
\centering
\includegraphics[width=\textwidth]{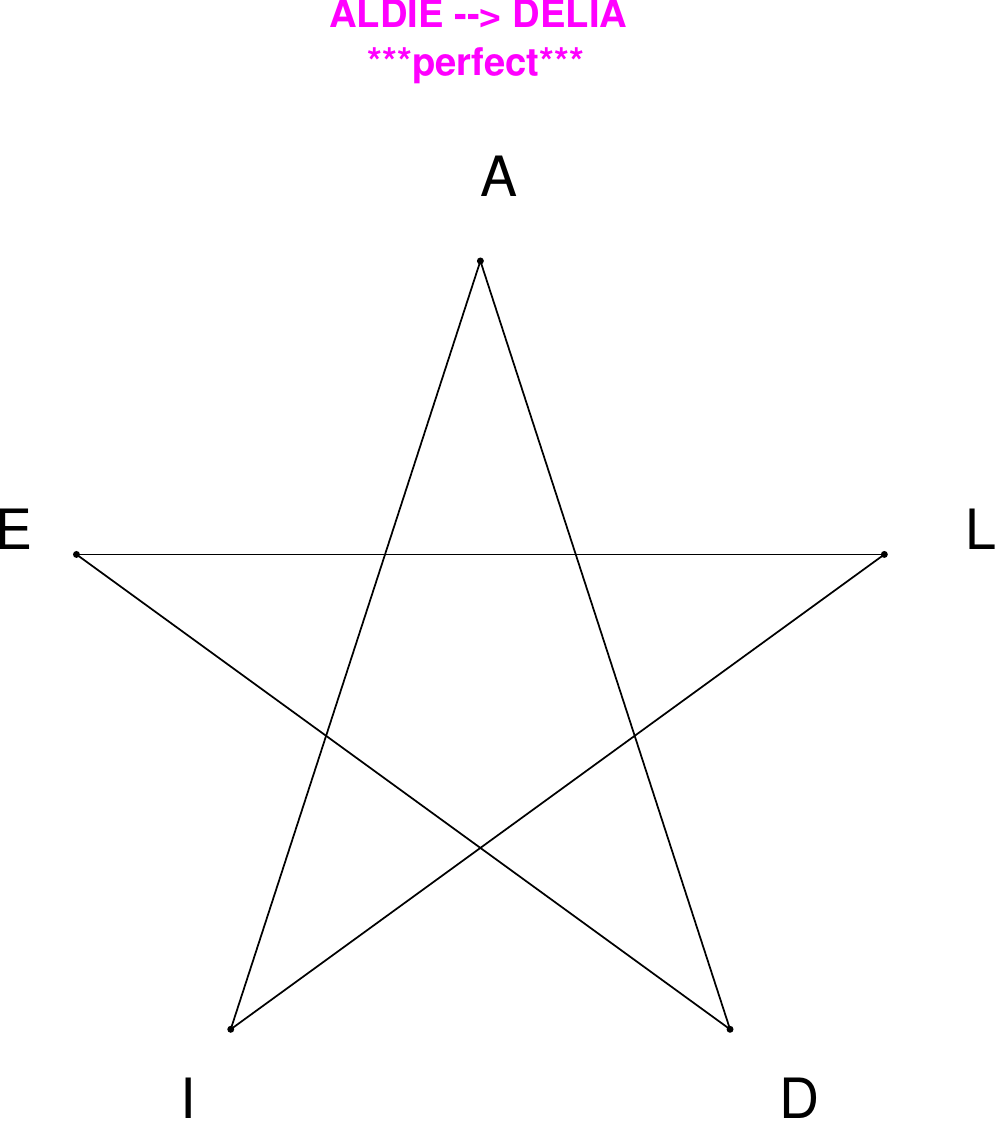}
\end{subfigure}
\hfill
\begin{subfigure}[T]{0.19\textwidth}
\centering
\includegraphics[width=\textwidth]{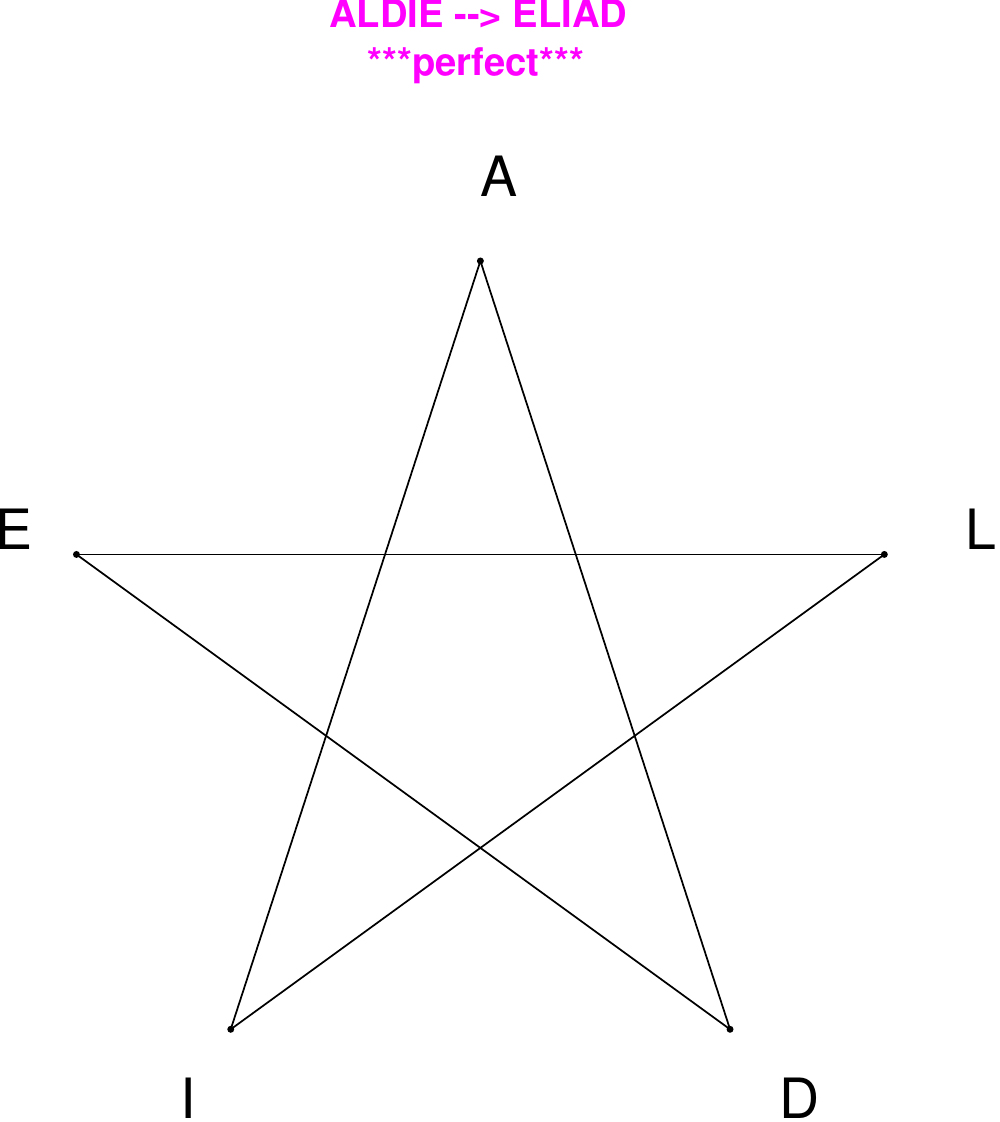}
\end{subfigure}
\hfill
\begin{subfigure}[T]{0.19\textwidth}
\centering
\includegraphics[width=\textwidth]{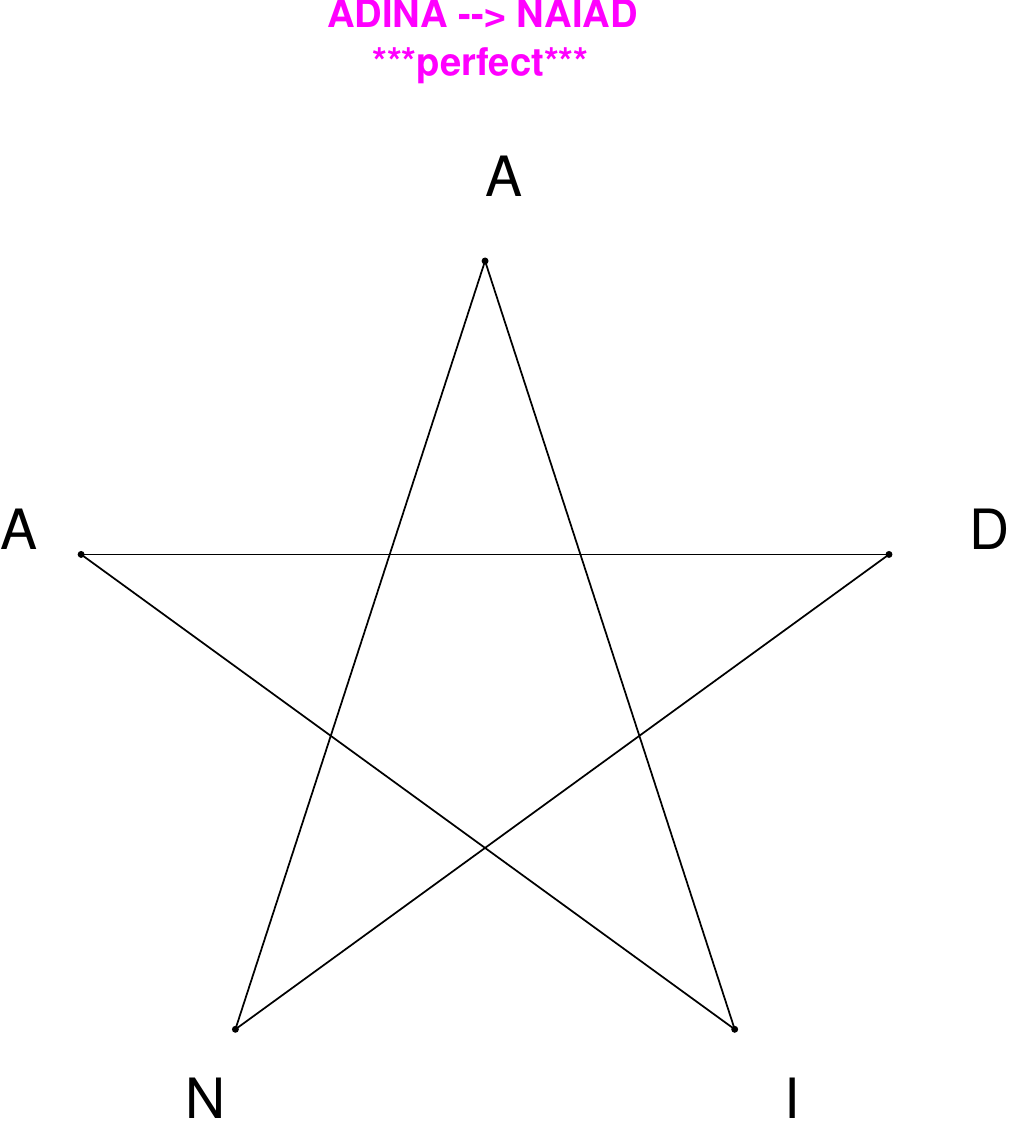}
\end{subfigure}
\hfill
\begin{subfigure}[T]{0.19\textwidth}
\centering
\includegraphics[width=\textwidth]{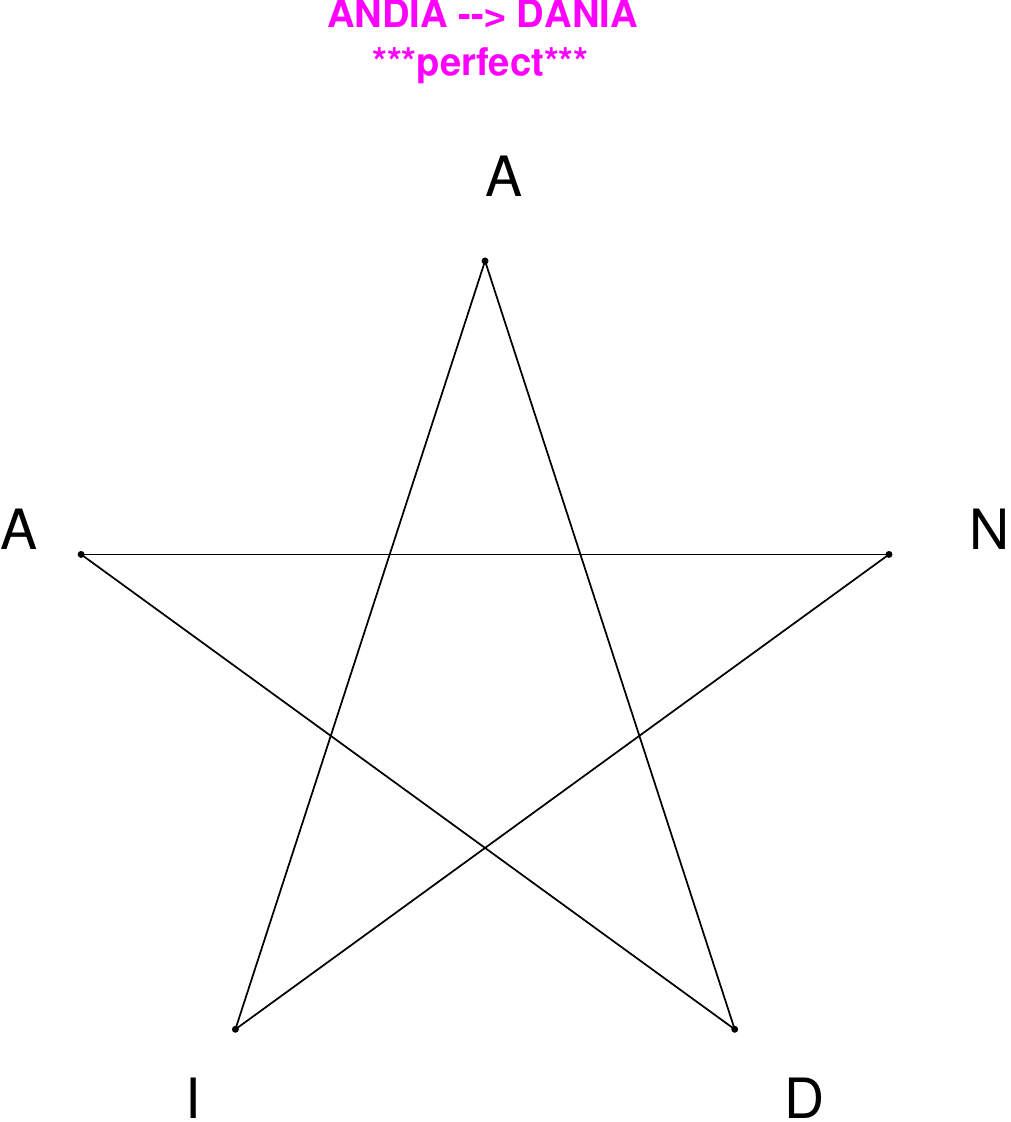}
\end{subfigure}
\hfill
\begin{subfigure}[T]{0.19\textwidth}
\centering
\includegraphics[width=\textwidth]{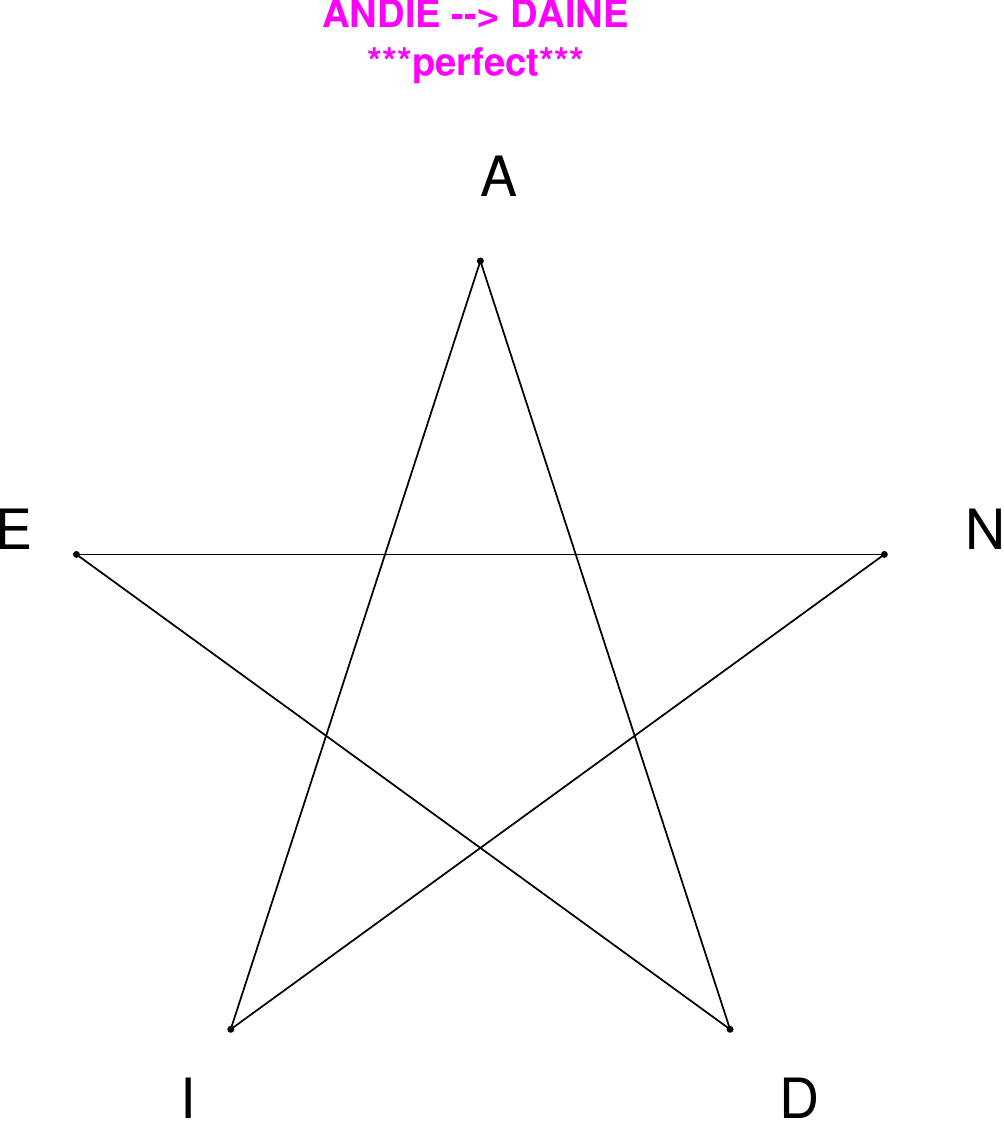}
\end{subfigure}
\end{figure}

\begin{figure}[H]
\centering
\begin{subfigure}[T]{0.19\textwidth}
\centering
\includegraphics[width=\textwidth]{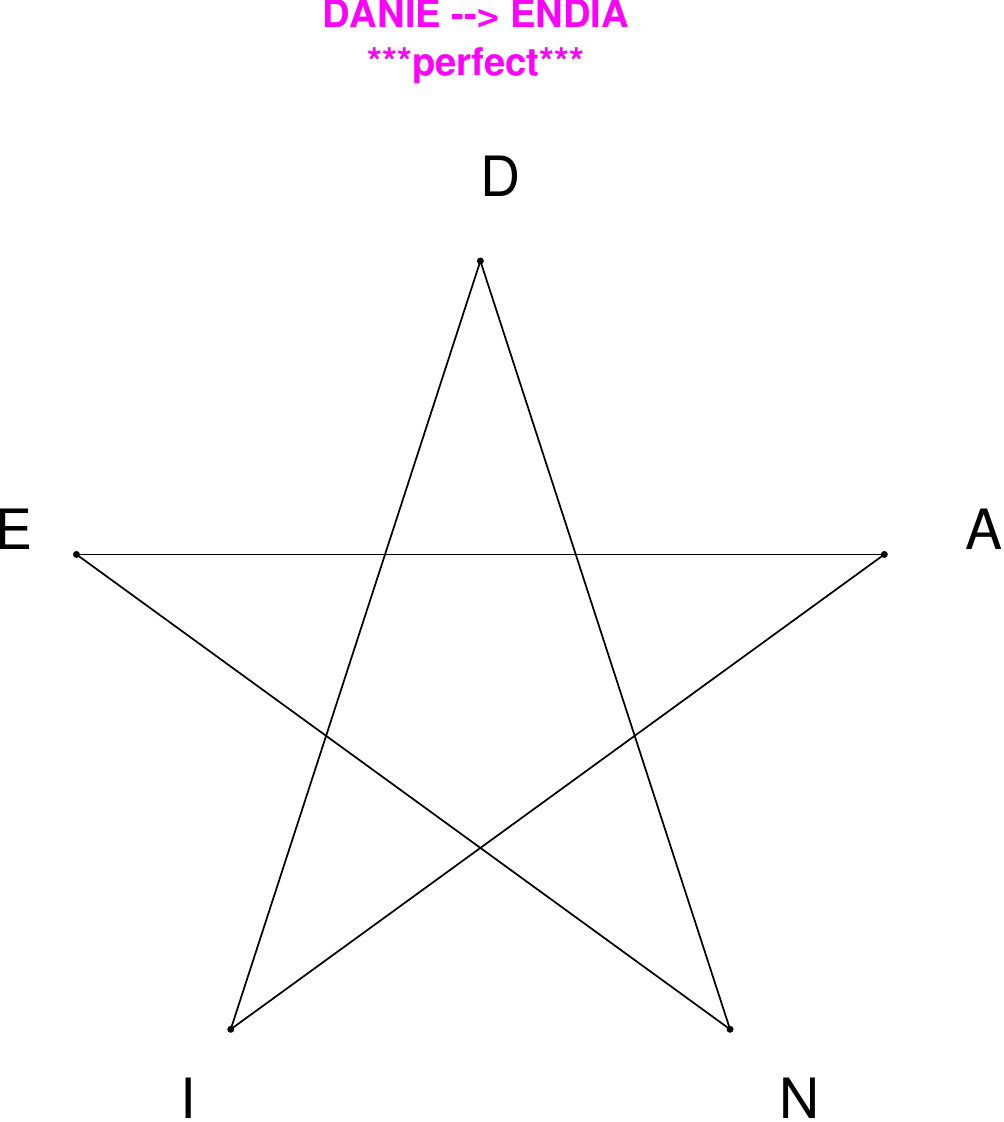}
\end{subfigure}
\hfill
\begin{subfigure}[T]{0.19\textwidth}
\centering
\includegraphics[width=\textwidth]{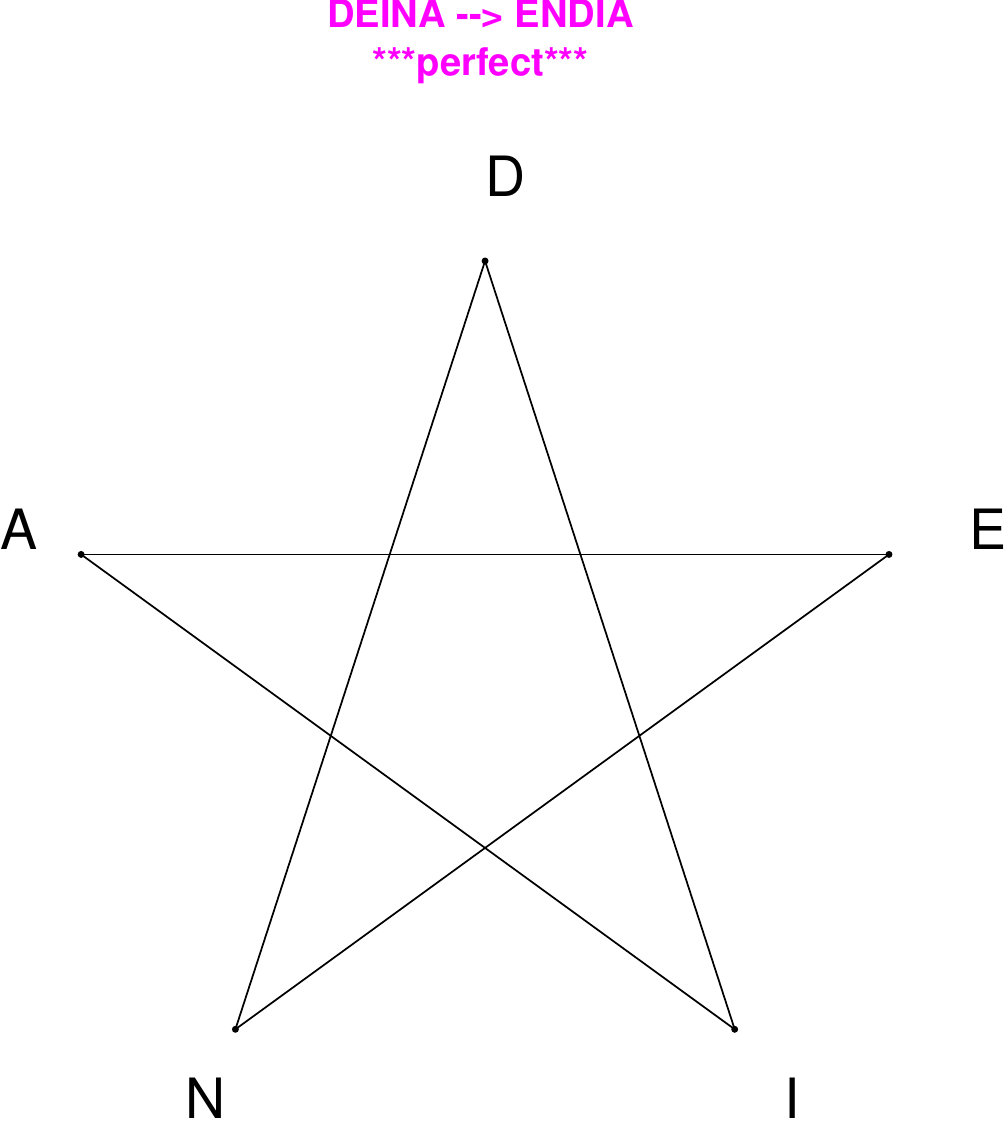}
\end{subfigure}
\hfill
\begin{subfigure}[T]{0.19\textwidth}
\centering
\includegraphics[width=\textwidth]{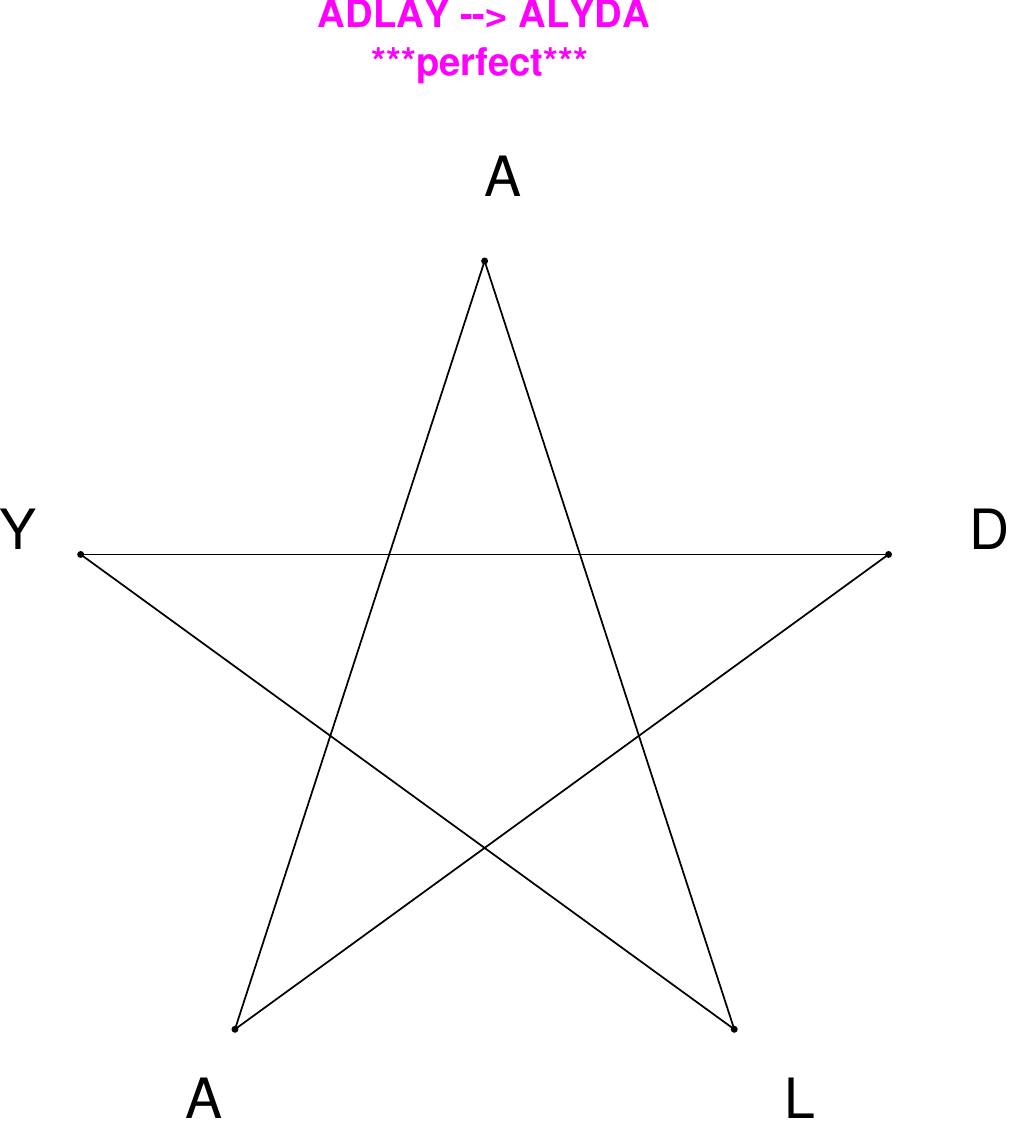}
\end{subfigure}
\hfill
\begin{subfigure}[T]{0.19\textwidth}
\centering
\includegraphics[width=\textwidth]{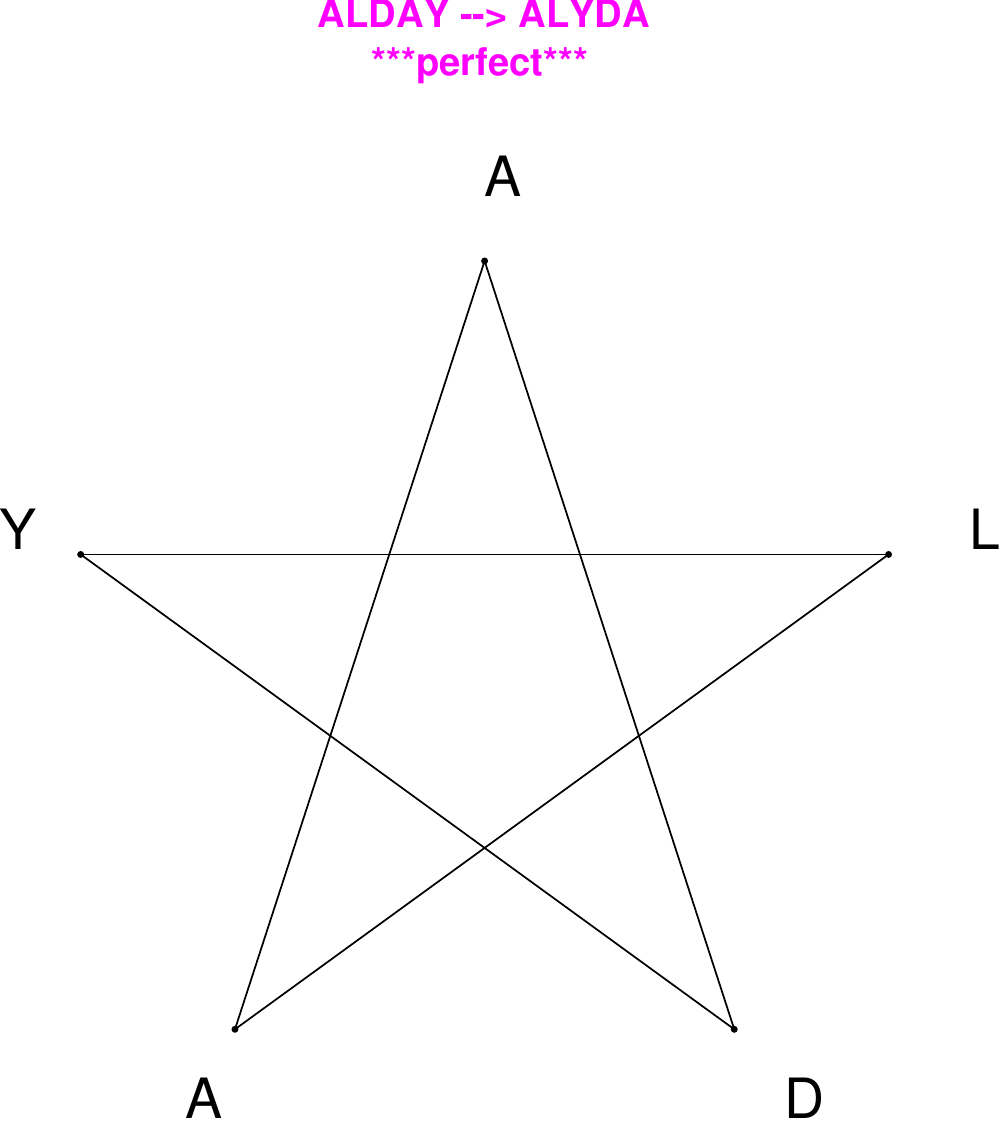}
\end{subfigure}
\hfill
\begin{subfigure}[T]{0.19\textwidth}
\centering
\includegraphics[width=\textwidth]{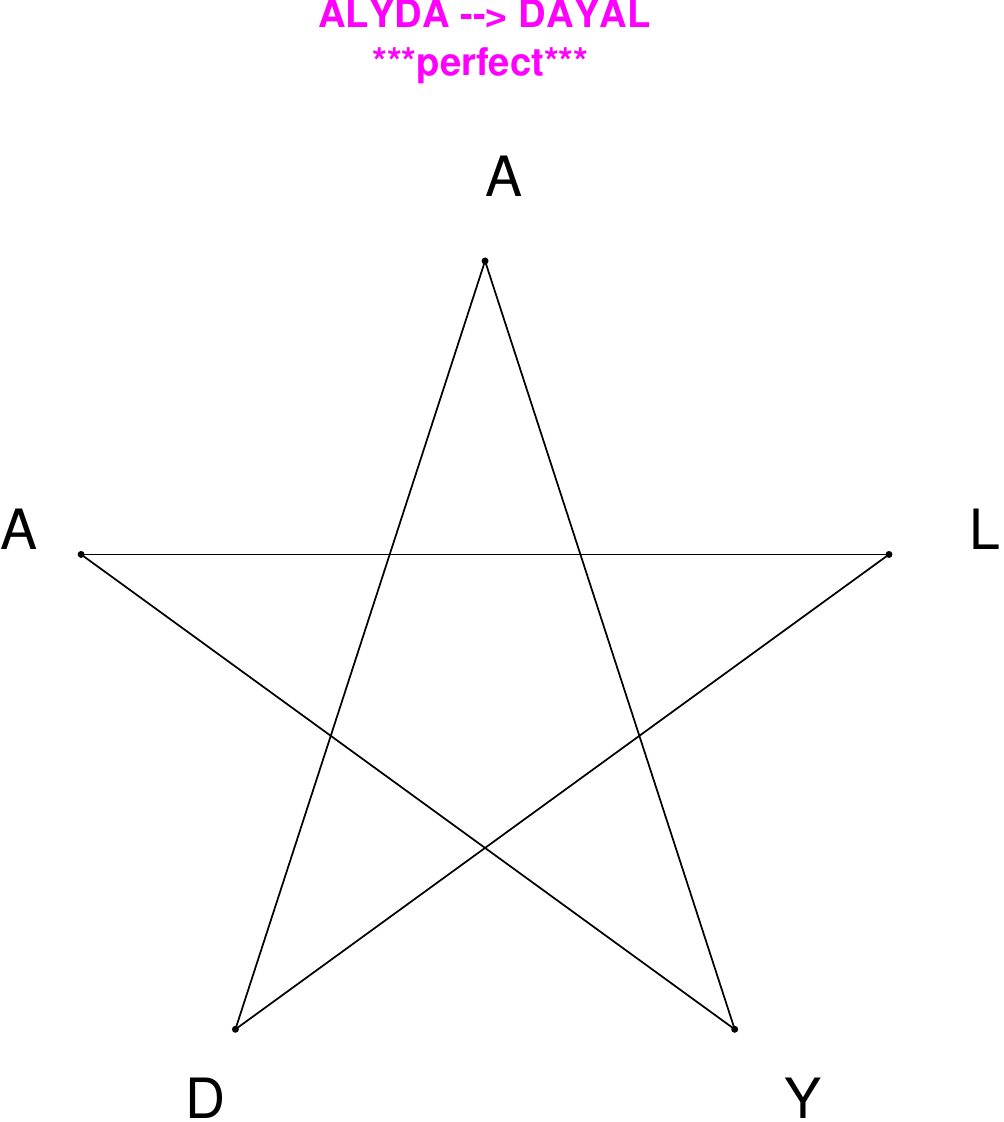}
\end{subfigure}
\end{figure}

\begin{figure}[H]
\centering
\begin{subfigure}[T]{0.19\textwidth}
\centering
\includegraphics[width=\textwidth]{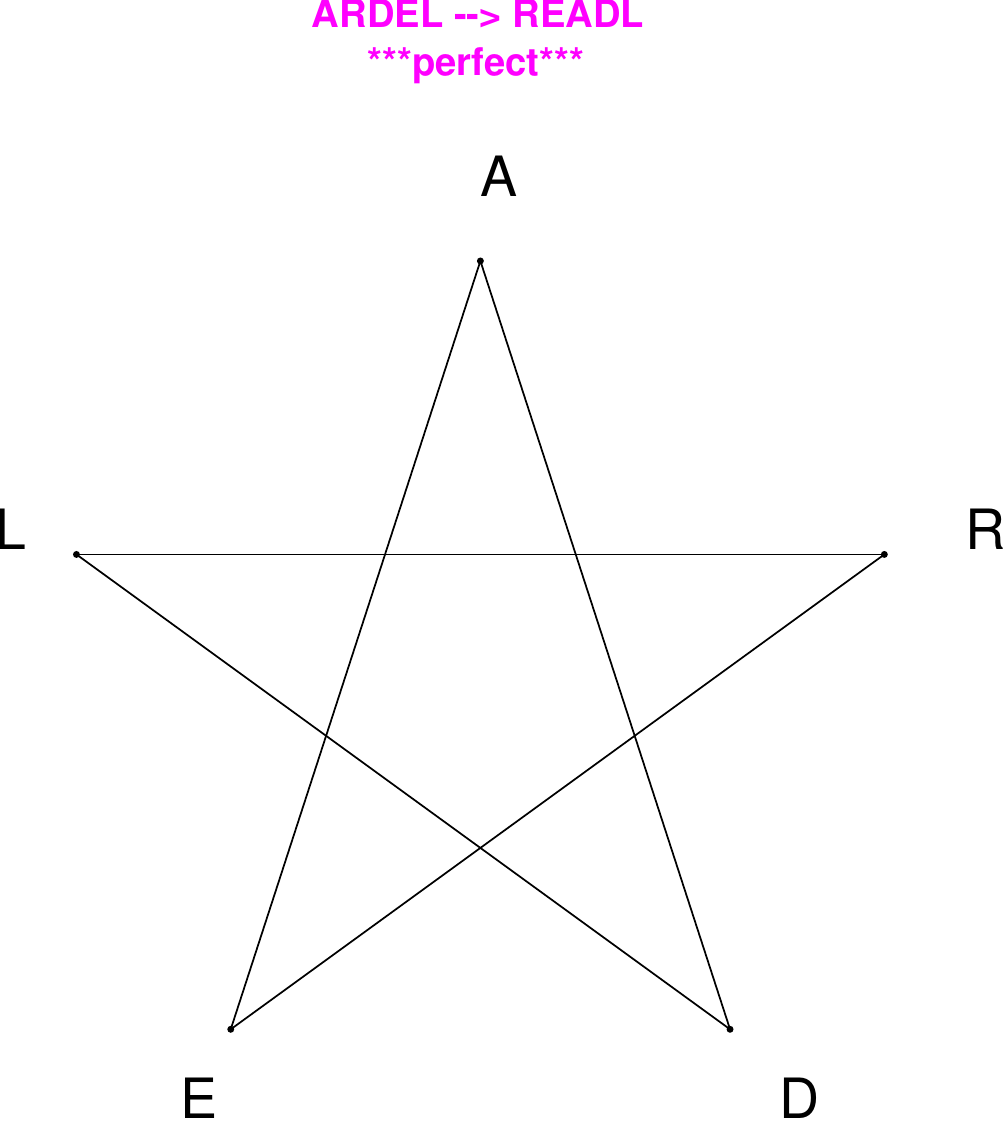}
\end{subfigure}
\hfill
\begin{subfigure}[T]{0.19\textwidth}
\centering
\includegraphics[width=\textwidth]{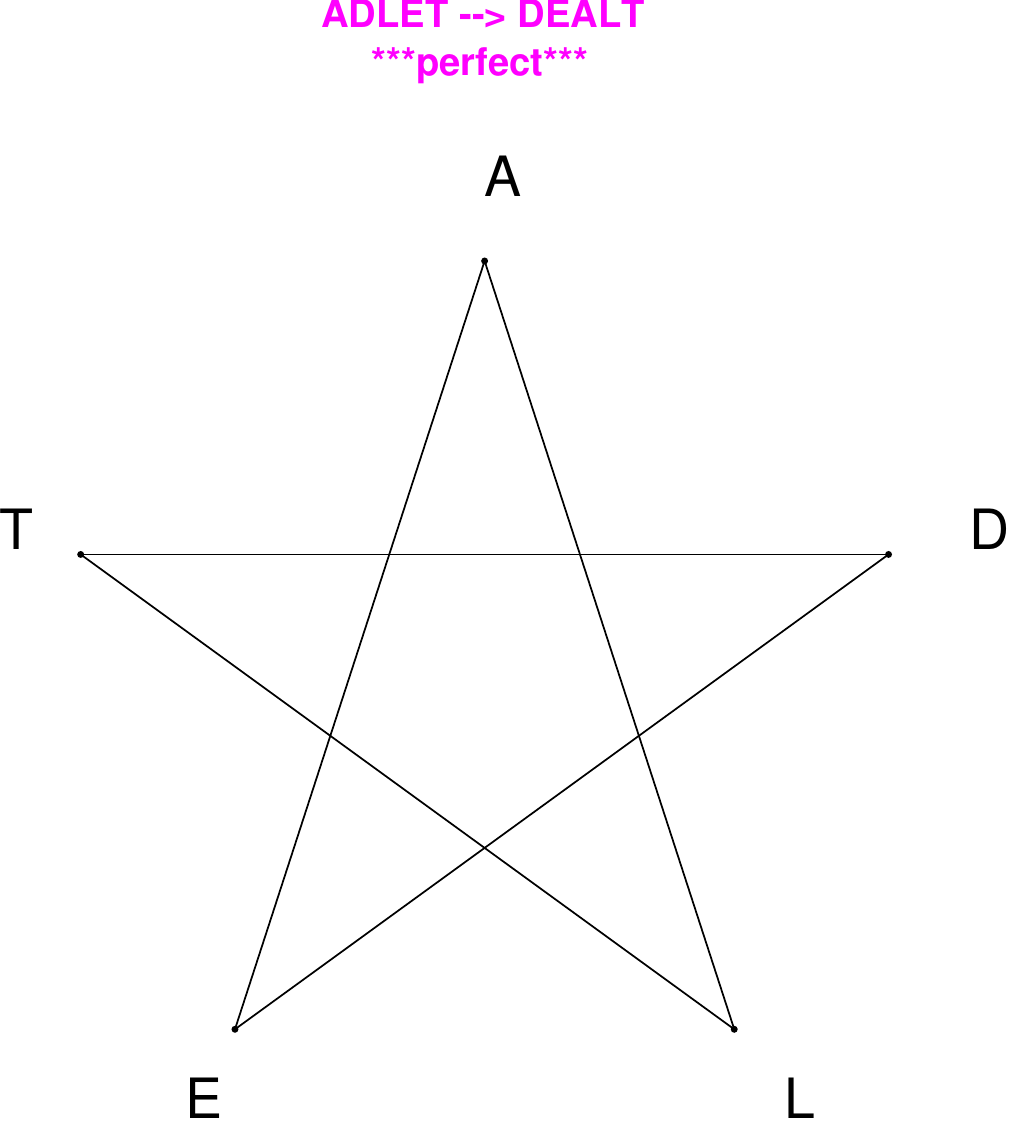}
\end{subfigure}
\hfill
\begin{subfigure}[T]{0.19\textwidth}
\centering
\includegraphics[width=\textwidth]{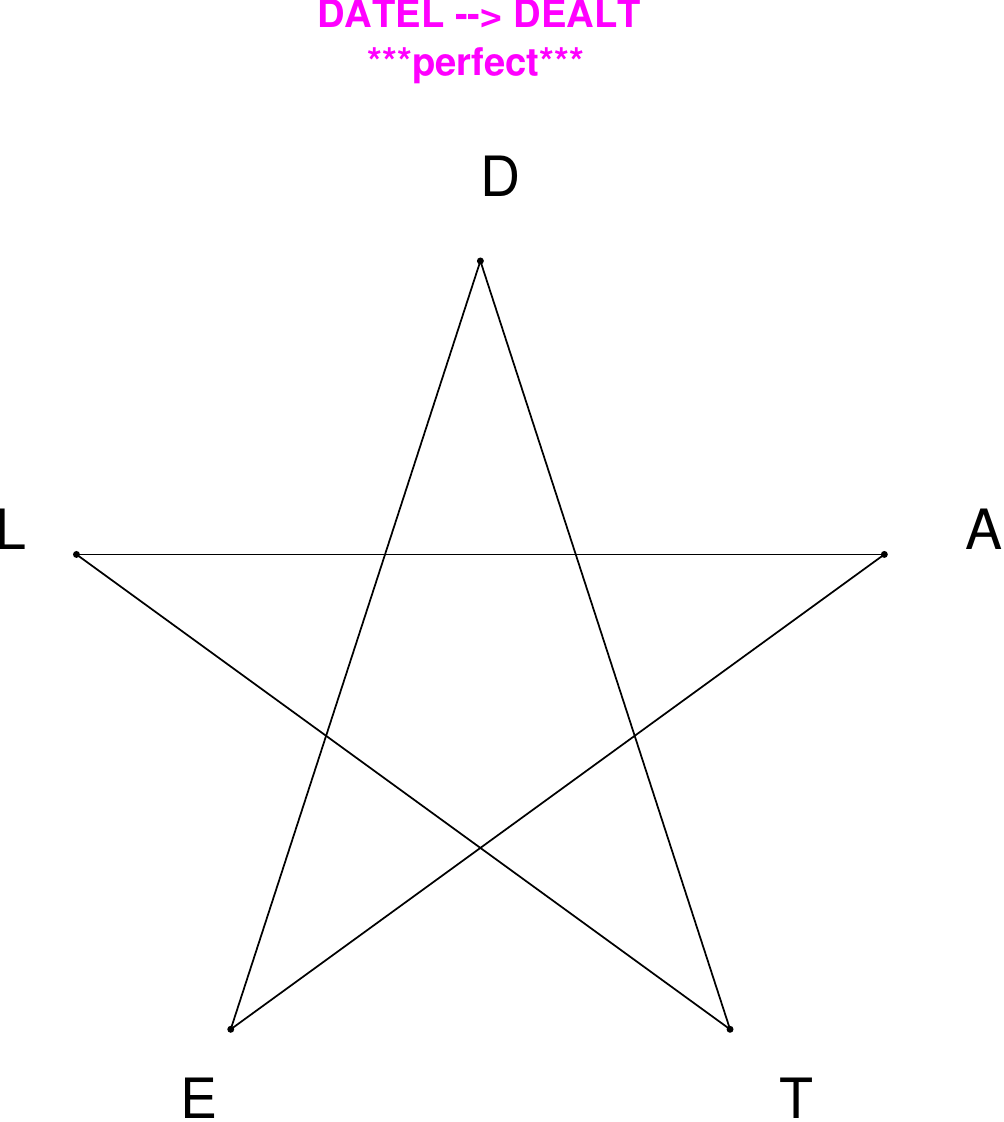}
\end{subfigure}
\hfill
\begin{subfigure}[T]{0.19\textwidth}
\centering
\includegraphics[width=\textwidth]{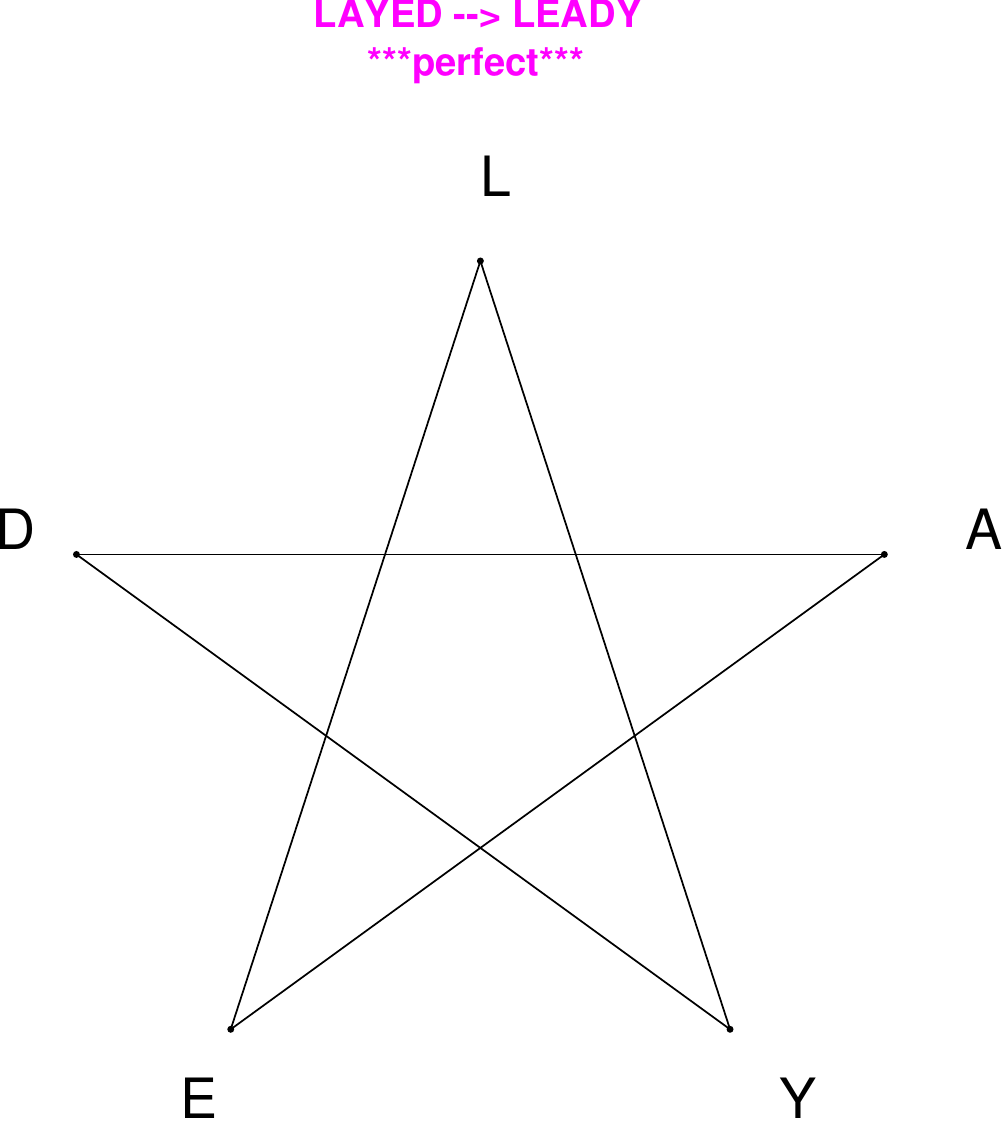}
\end{subfigure}
\hfill
\begin{subfigure}[T]{0.19\textwidth}
\centering
\includegraphics[width=\textwidth]{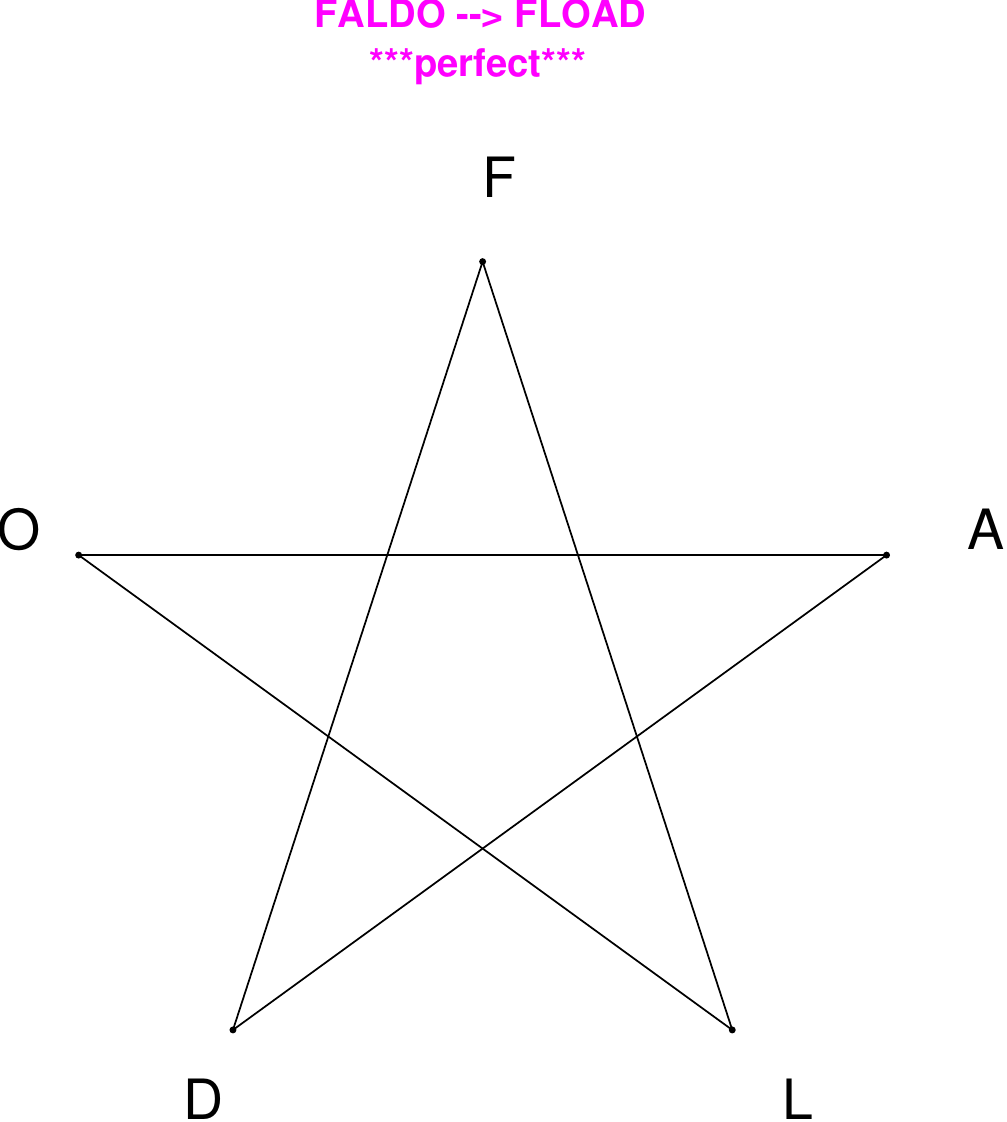}
\end{subfigure}
\end{figure}

\begin{figure}[H]
\centering
\begin{subfigure}[T]{0.19\textwidth}
\centering
\includegraphics[width=\textwidth]{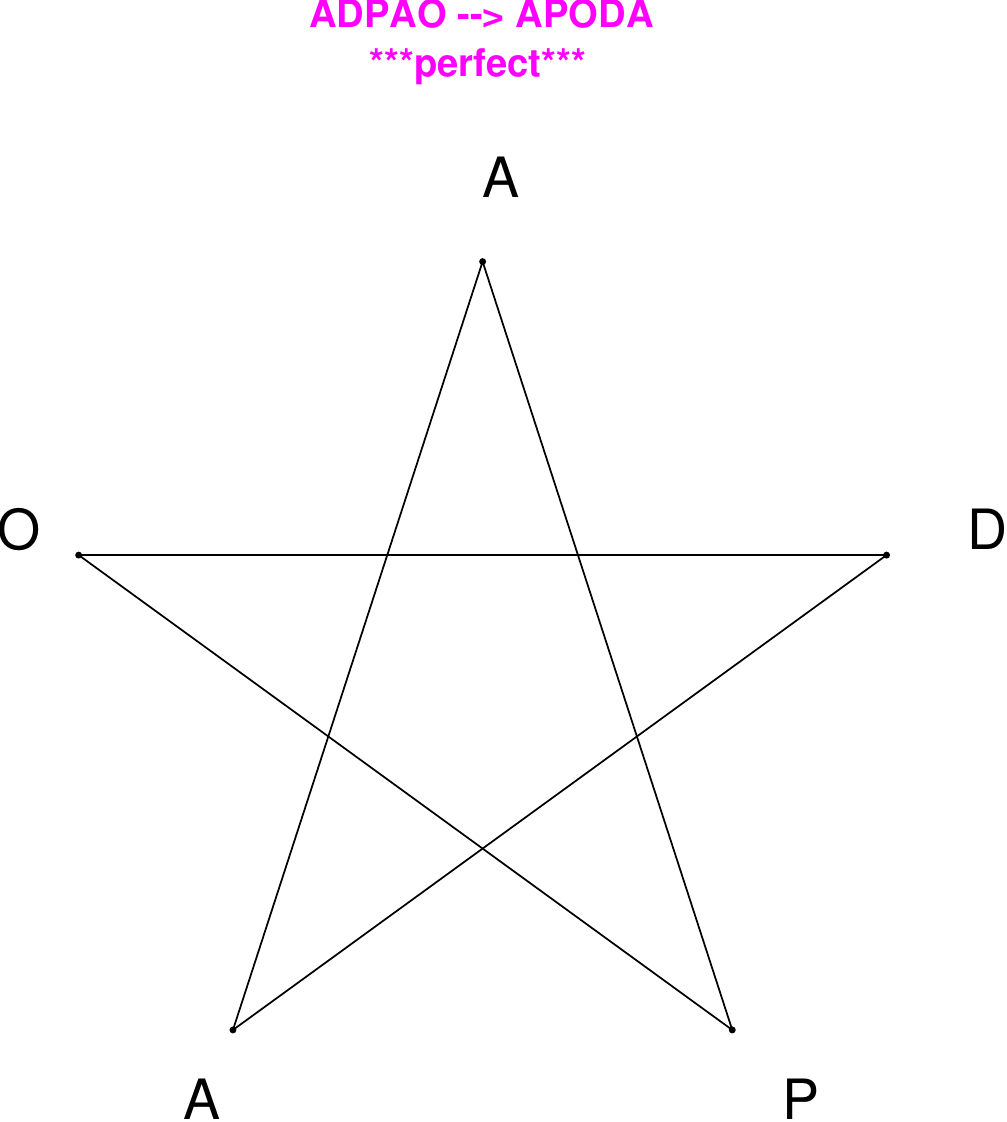}
\end{subfigure}
\hfill
\begin{subfigure}[T]{0.19\textwidth}
\centering
\includegraphics[width=\textwidth]{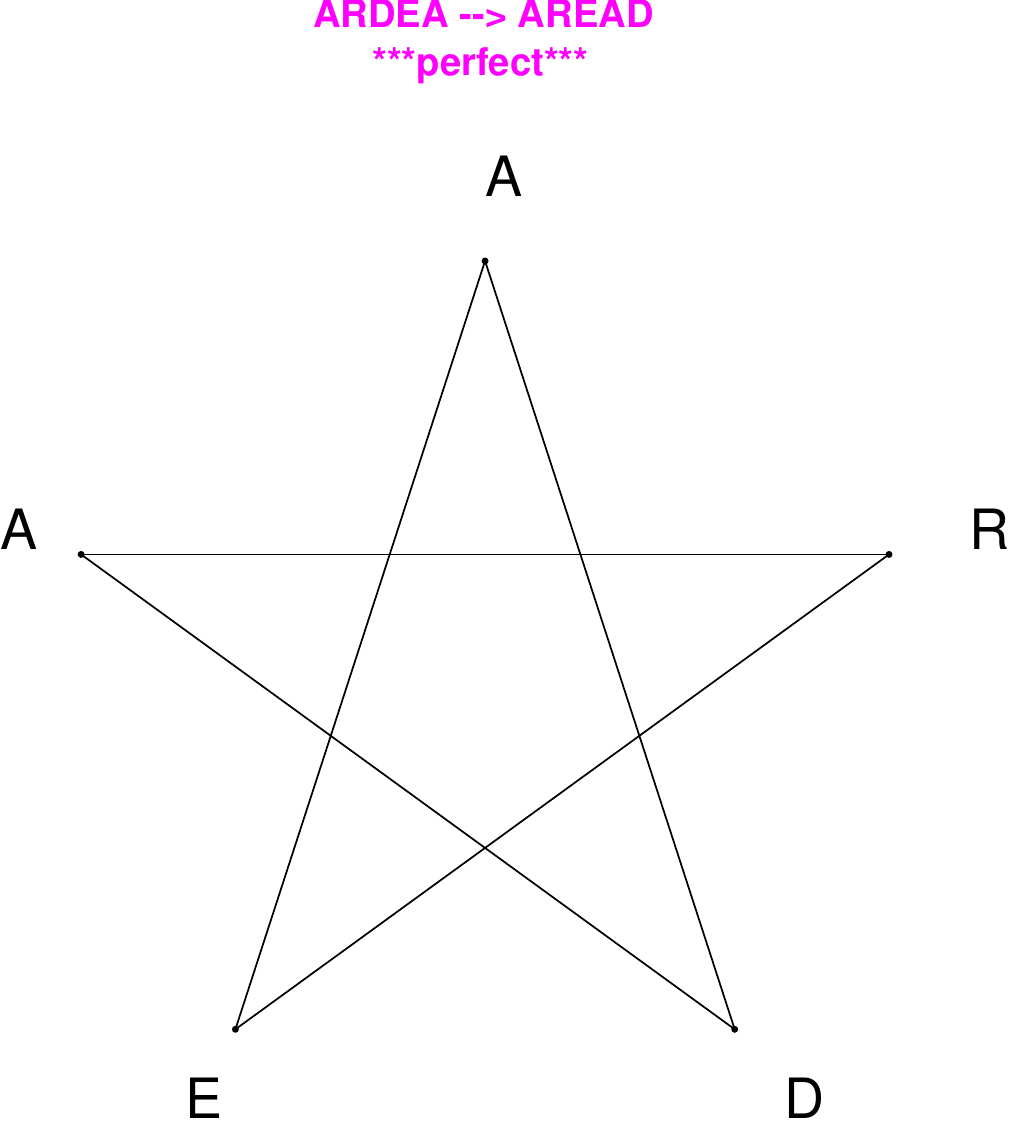}
\end{subfigure}
\hfill
\begin{subfigure}[T]{0.19\textwidth}
\centering
\includegraphics[width=\textwidth]{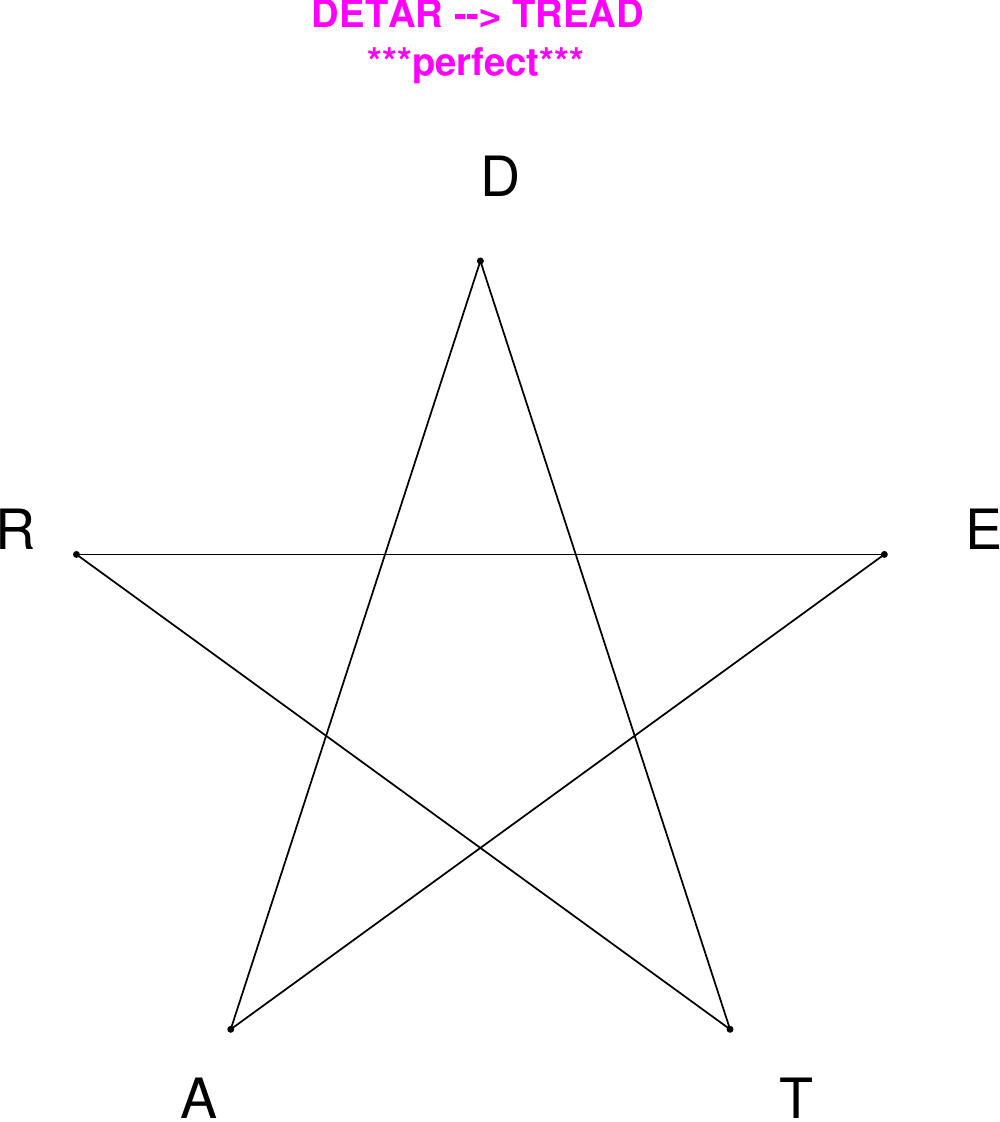}
\end{subfigure}
\hfill
\begin{subfigure}[T]{0.19\textwidth}
\centering
\includegraphics[width=\textwidth]{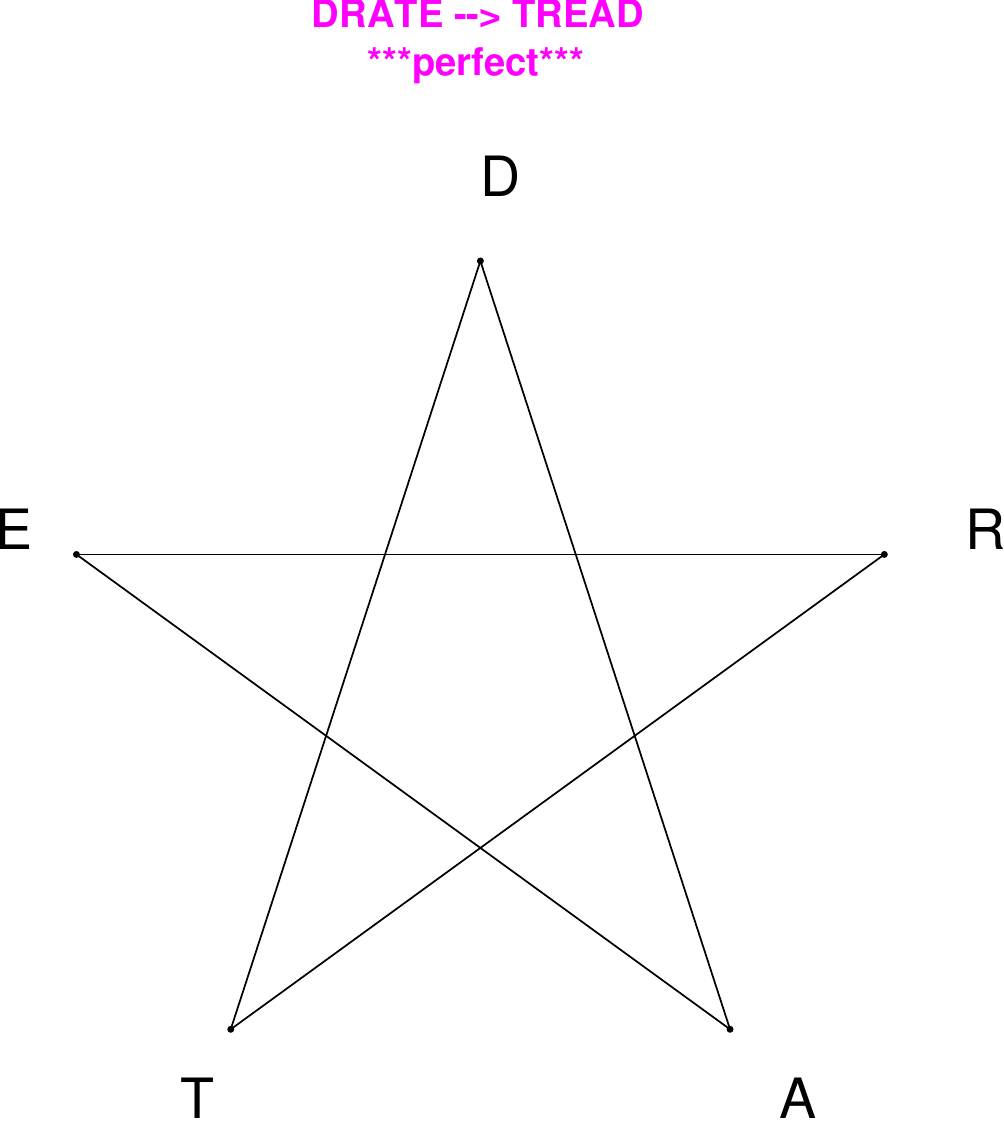}
\end{subfigure}
\hfill
\begin{subfigure}[T]{0.19\textwidth}
\centering
\includegraphics[width=\textwidth]{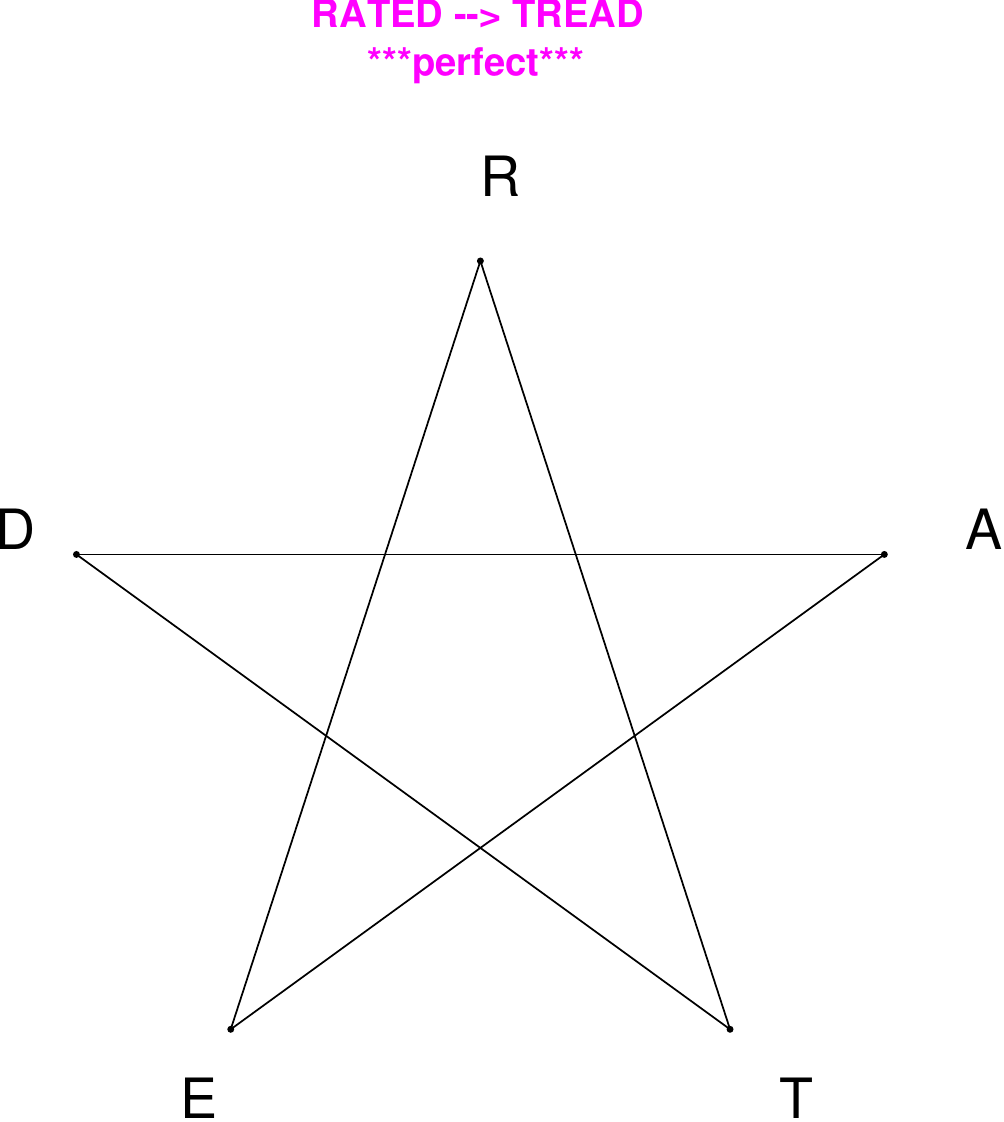}
\end{subfigure}
\end{figure}

\begin{figure}[H]
\centering
\begin{subfigure}[T]{0.19\textwidth}
\centering
\includegraphics[width=\textwidth]{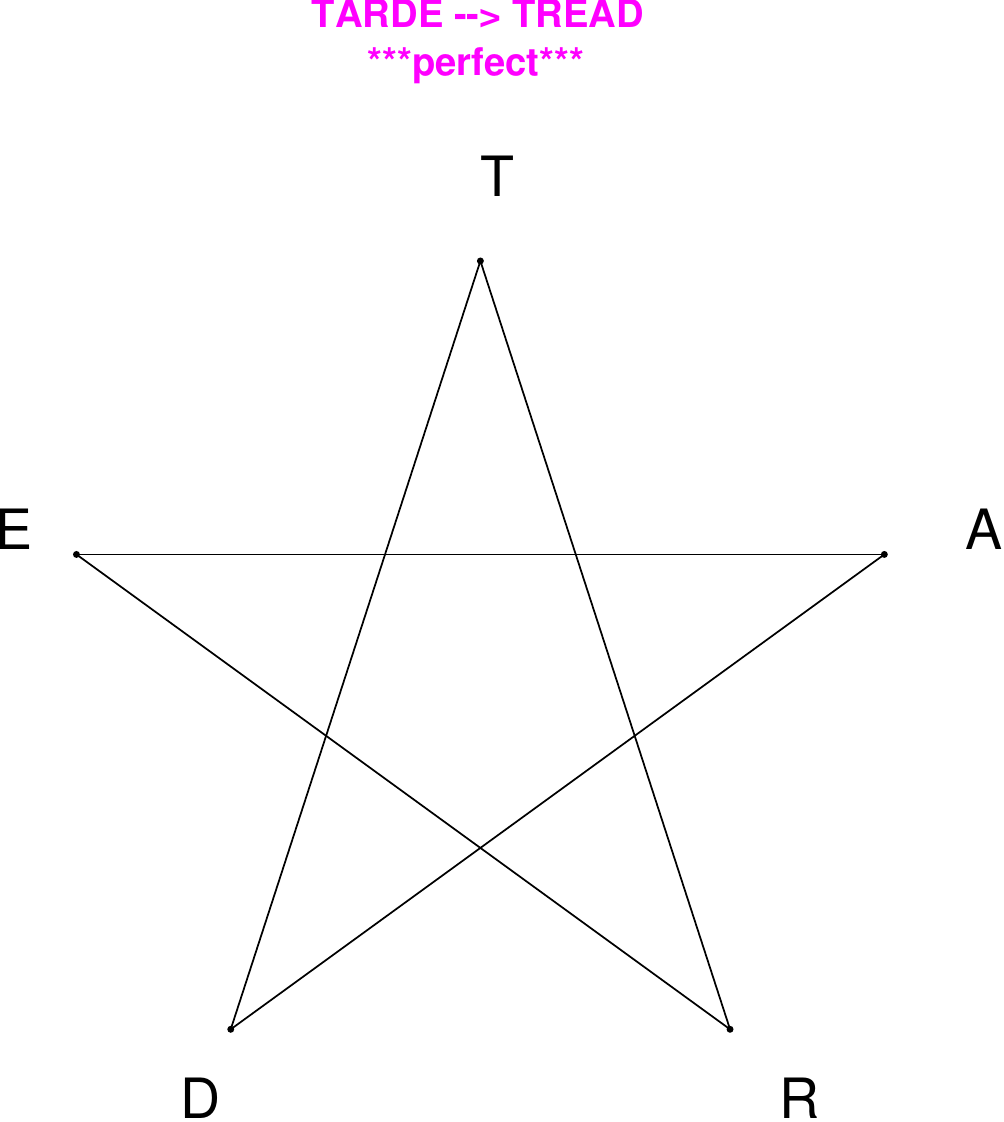}
\end{subfigure}
\hfill
\begin{subfigure}[T]{0.19\textwidth}
\centering
\includegraphics[width=\textwidth]{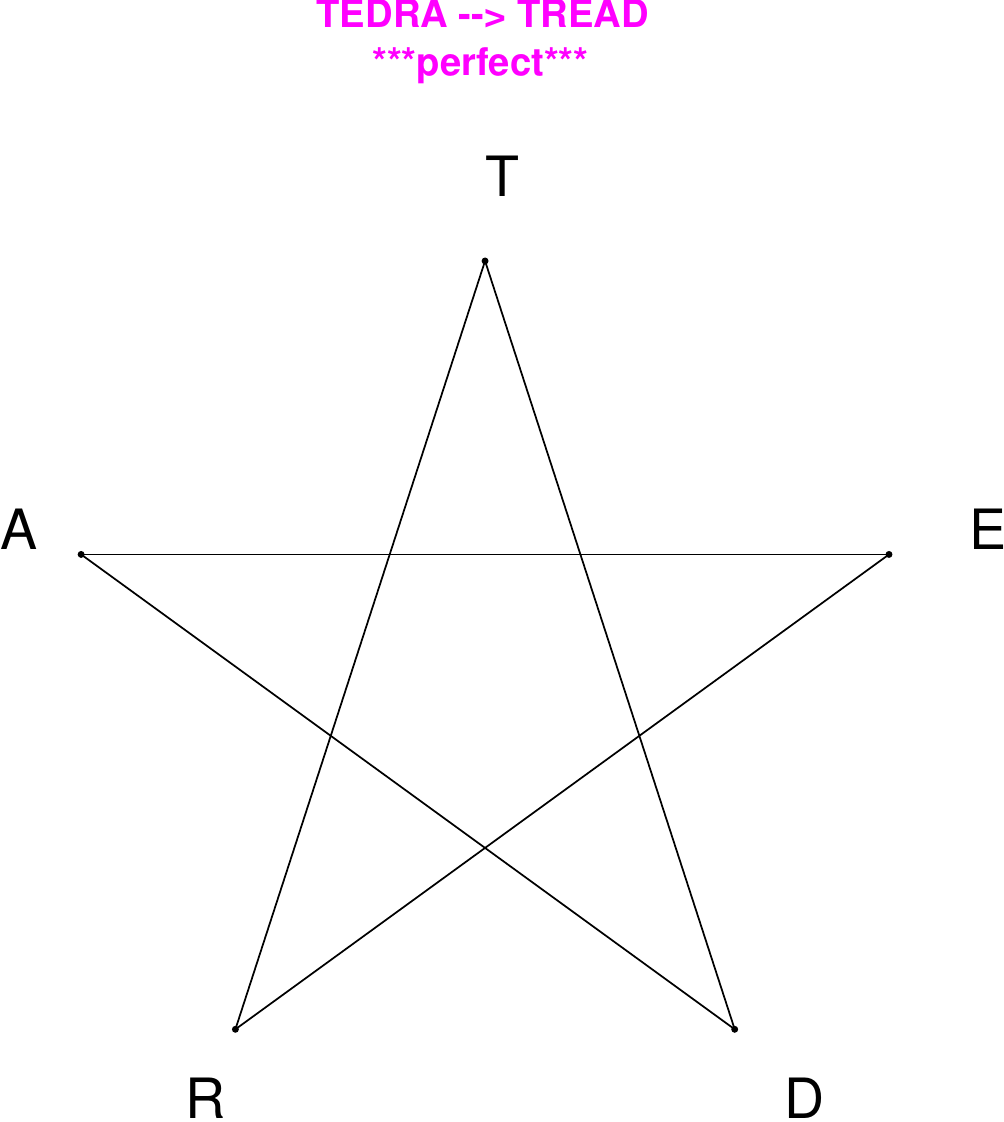}
\end{subfigure}
\hfill
\begin{subfigure}[T]{0.19\textwidth}
\centering
\includegraphics[width=\textwidth]{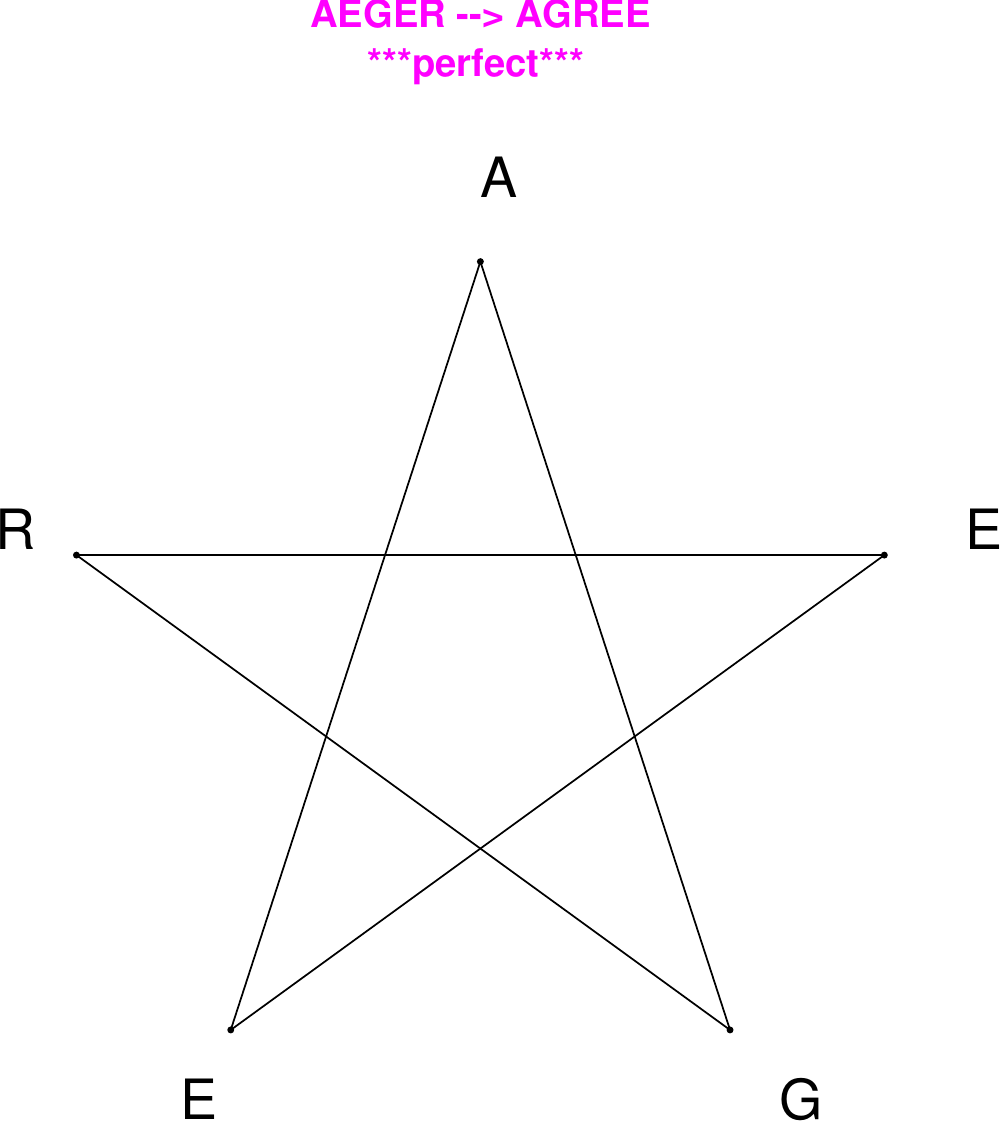}
\end{subfigure}
\hfill
\begin{subfigure}[T]{0.19\textwidth}
\centering
\includegraphics[width=\textwidth]{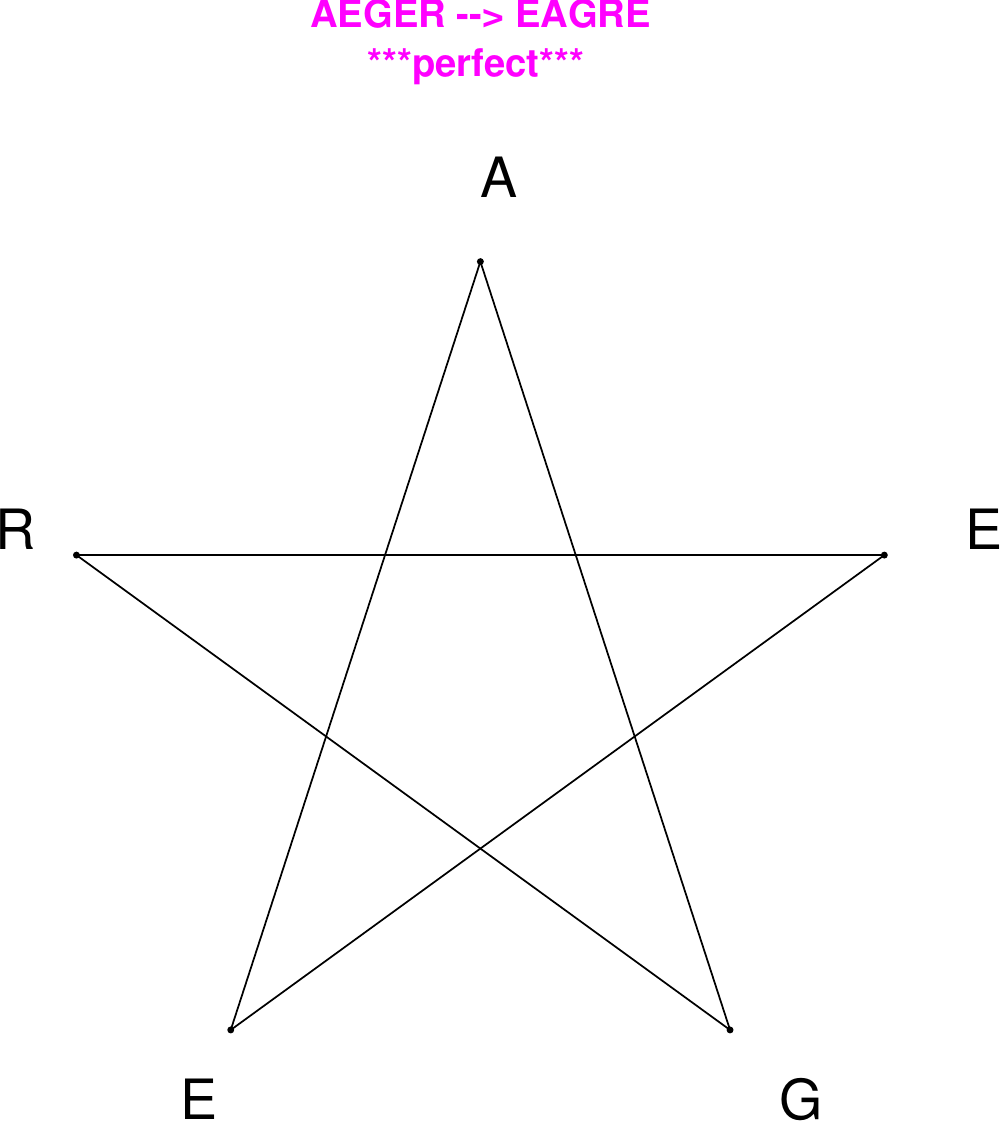}
\end{subfigure}
\hfill
\begin{subfigure}[T]{0.19\textwidth}
\centering
\includegraphics[width=\textwidth]{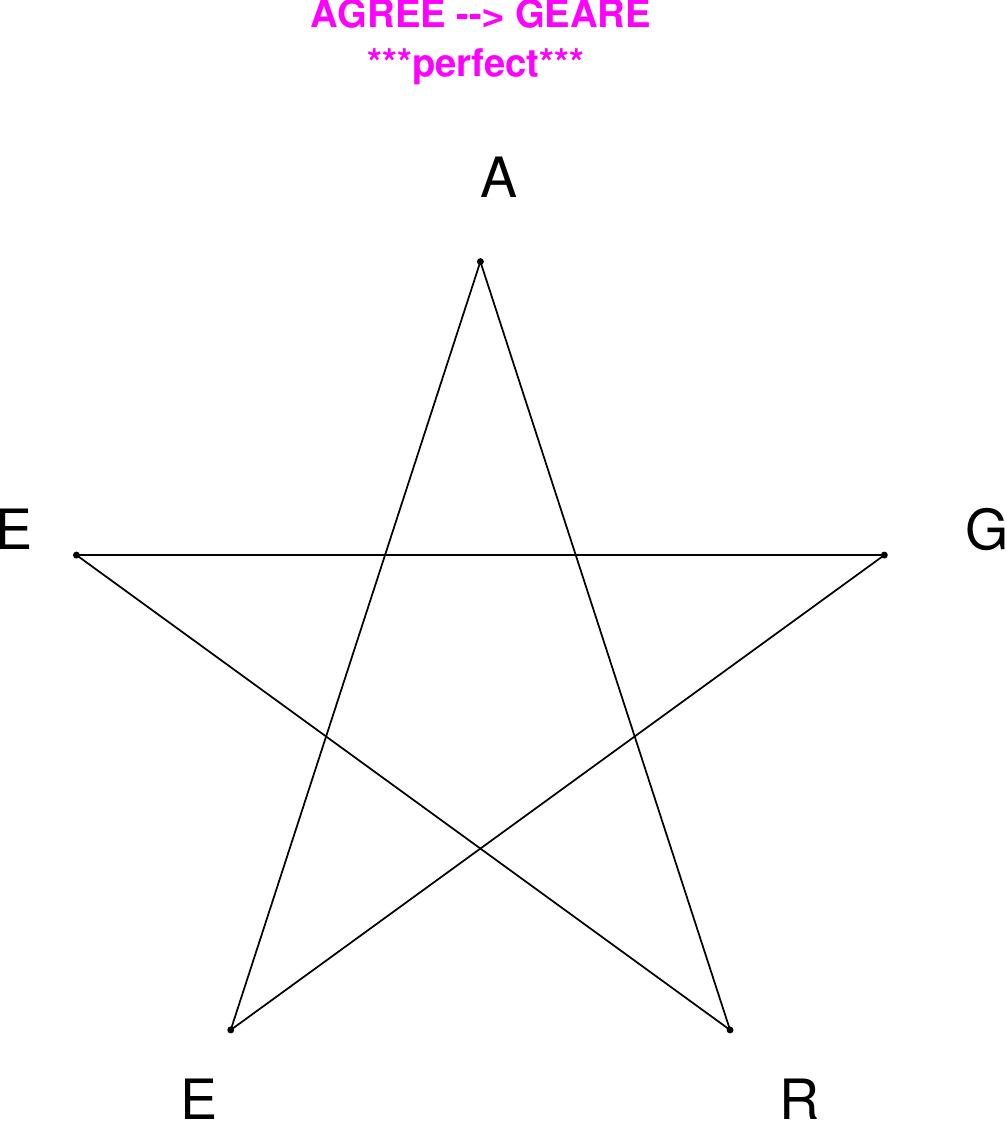}
\end{subfigure}
\end{figure}

\begin{figure}[H]
\centering
\begin{subfigure}[T]{0.19\textwidth}
\centering
\includegraphics[width=\textwidth]{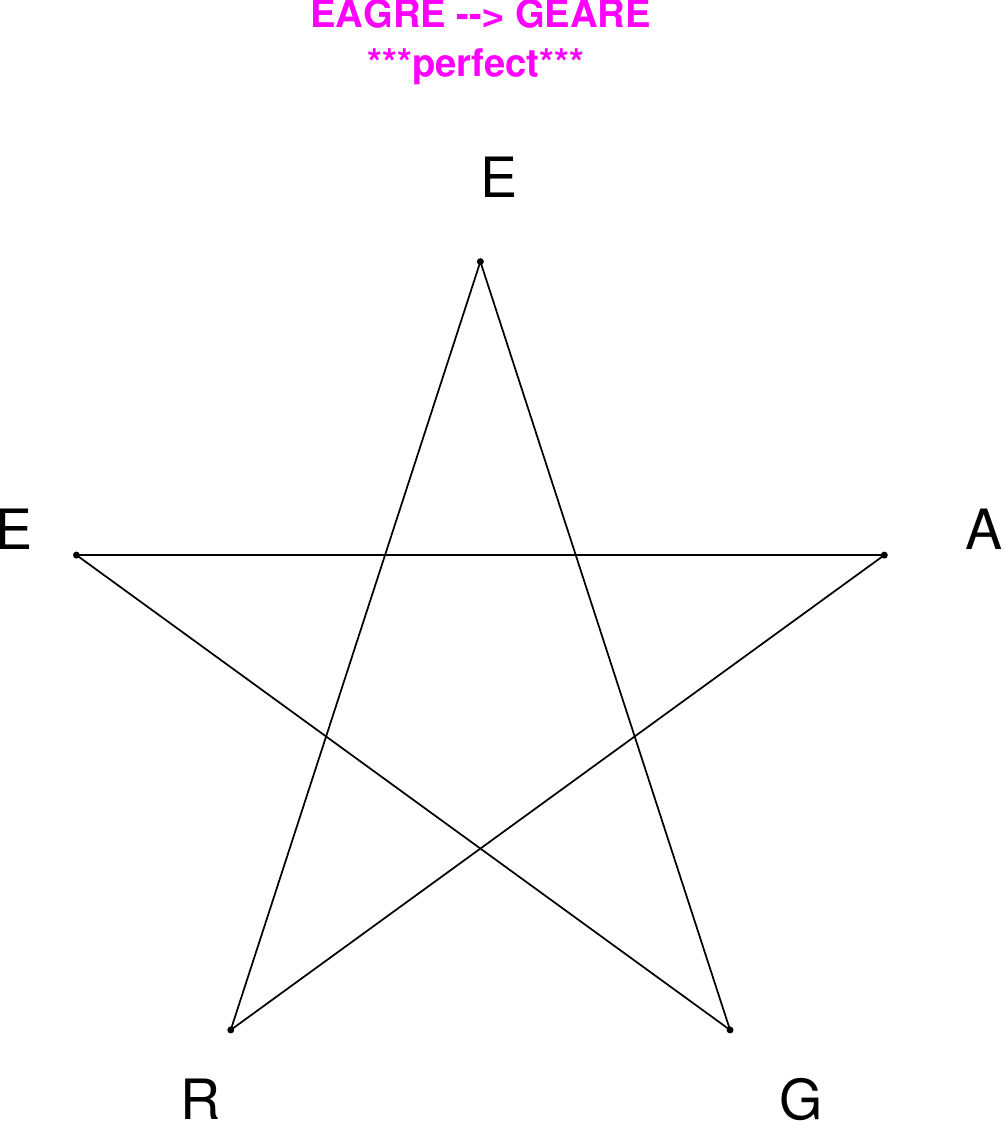}
\end{subfigure}
\hfill
\begin{subfigure}[T]{0.19\textwidth}
\centering
\includegraphics[width=\textwidth]{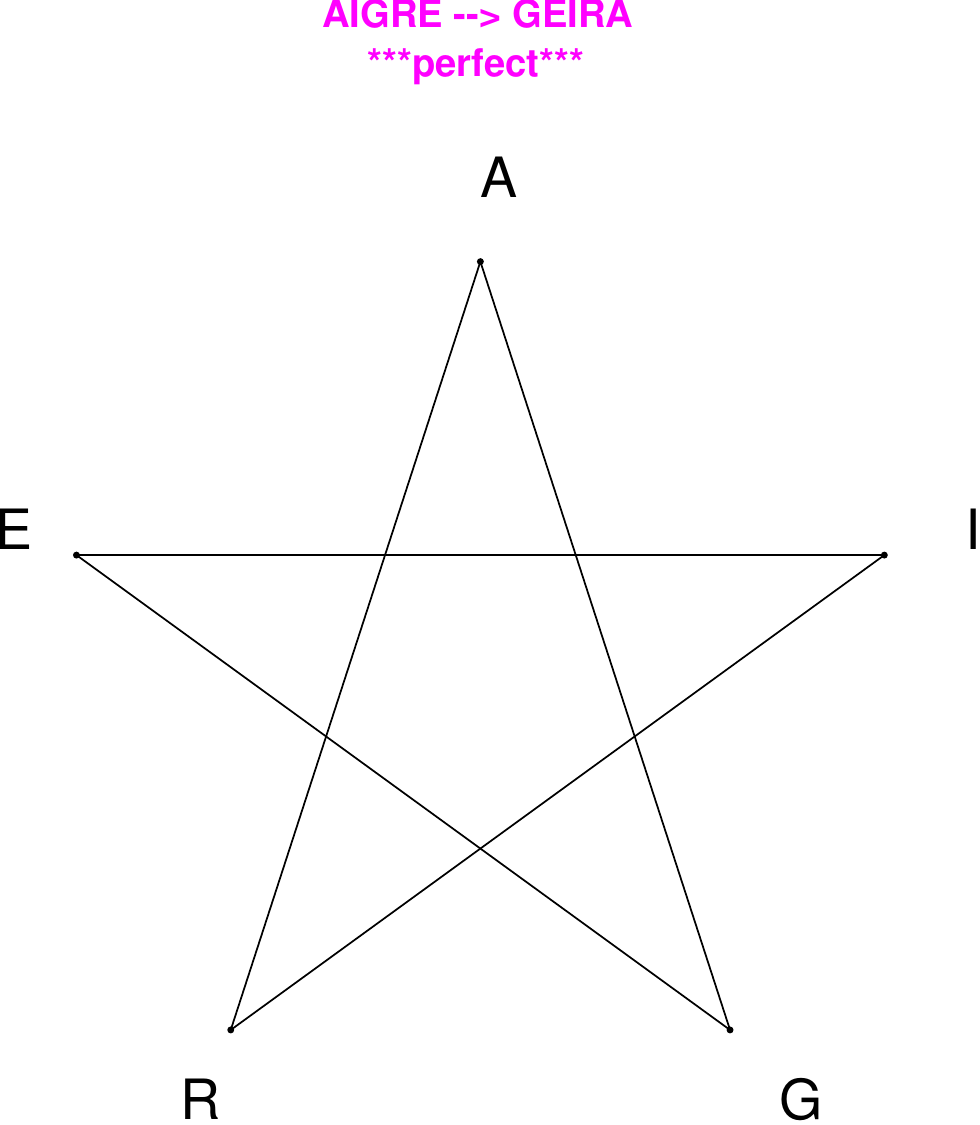}
\end{subfigure}
\hfill
\begin{subfigure}[T]{0.19\textwidth}
\centering
\includegraphics[width=\textwidth]{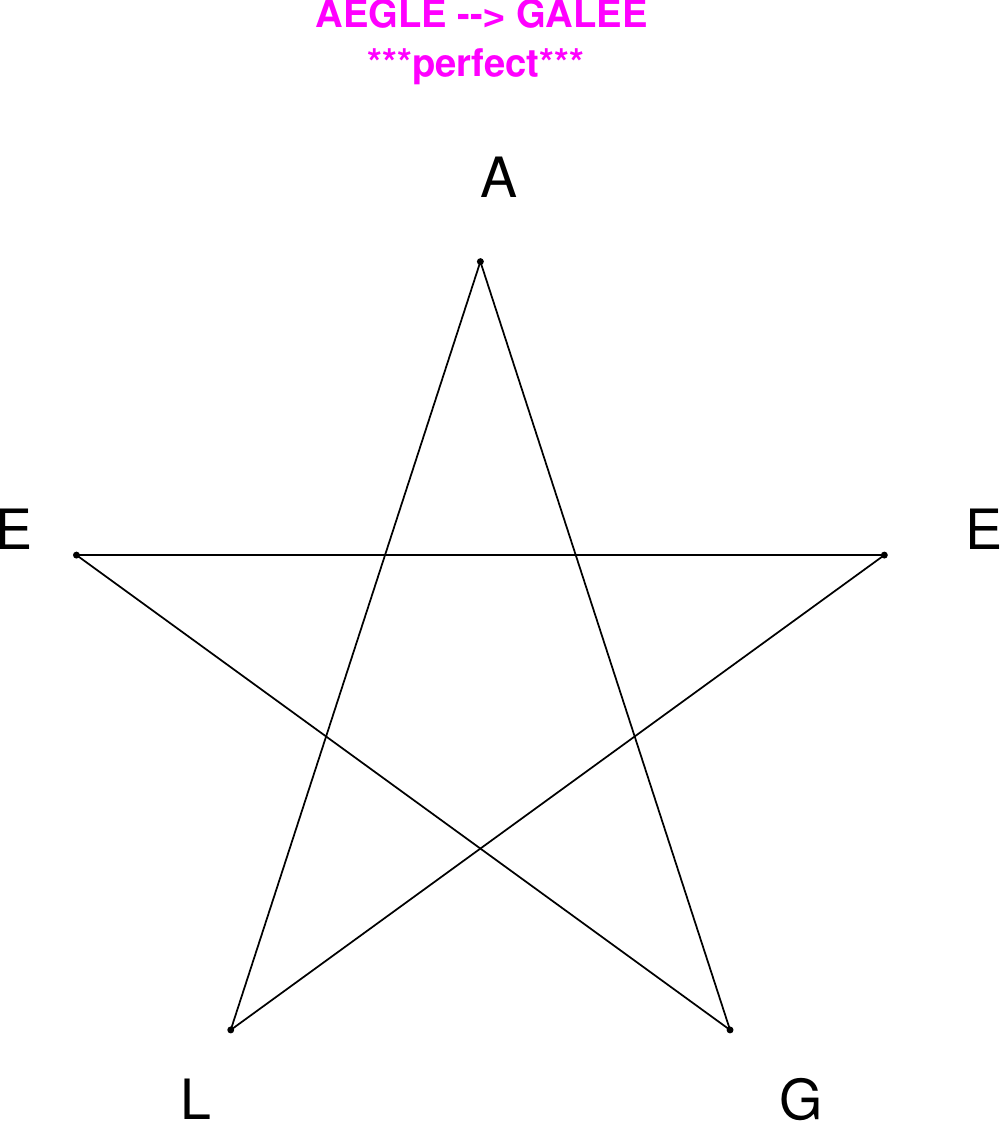}
\end{subfigure}
\hfill
\begin{subfigure}[T]{0.19\textwidth}
\centering
\includegraphics[width=\textwidth]{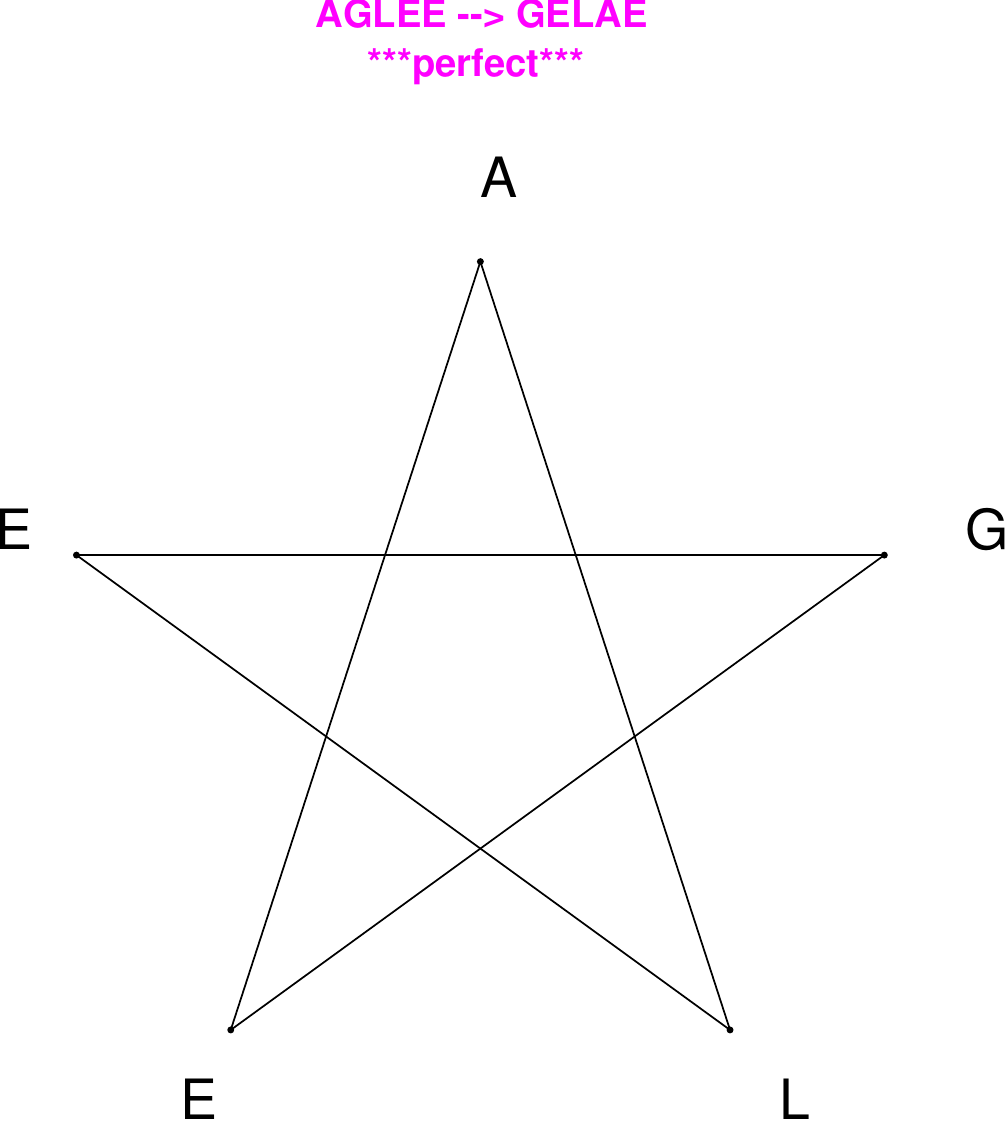}
\end{subfigure}
\hfill
\begin{subfigure}[T]{0.19\textwidth}
\centering
\includegraphics[width=\textwidth]{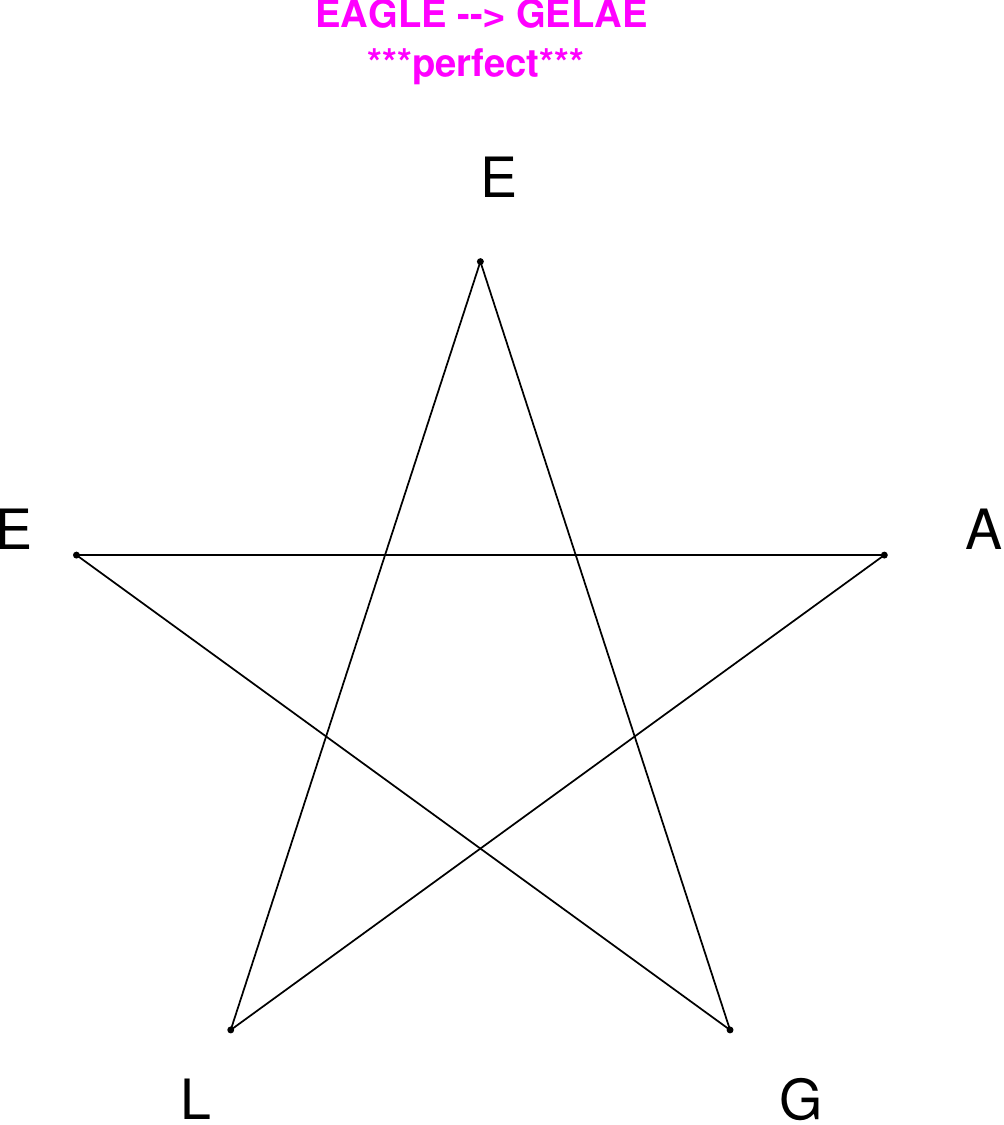}
\end{subfigure}
\end{figure}

\begin{figure}[H]
\centering
\begin{subfigure}[T]{0.19\textwidth}
\centering
\includegraphics[width=\textwidth]{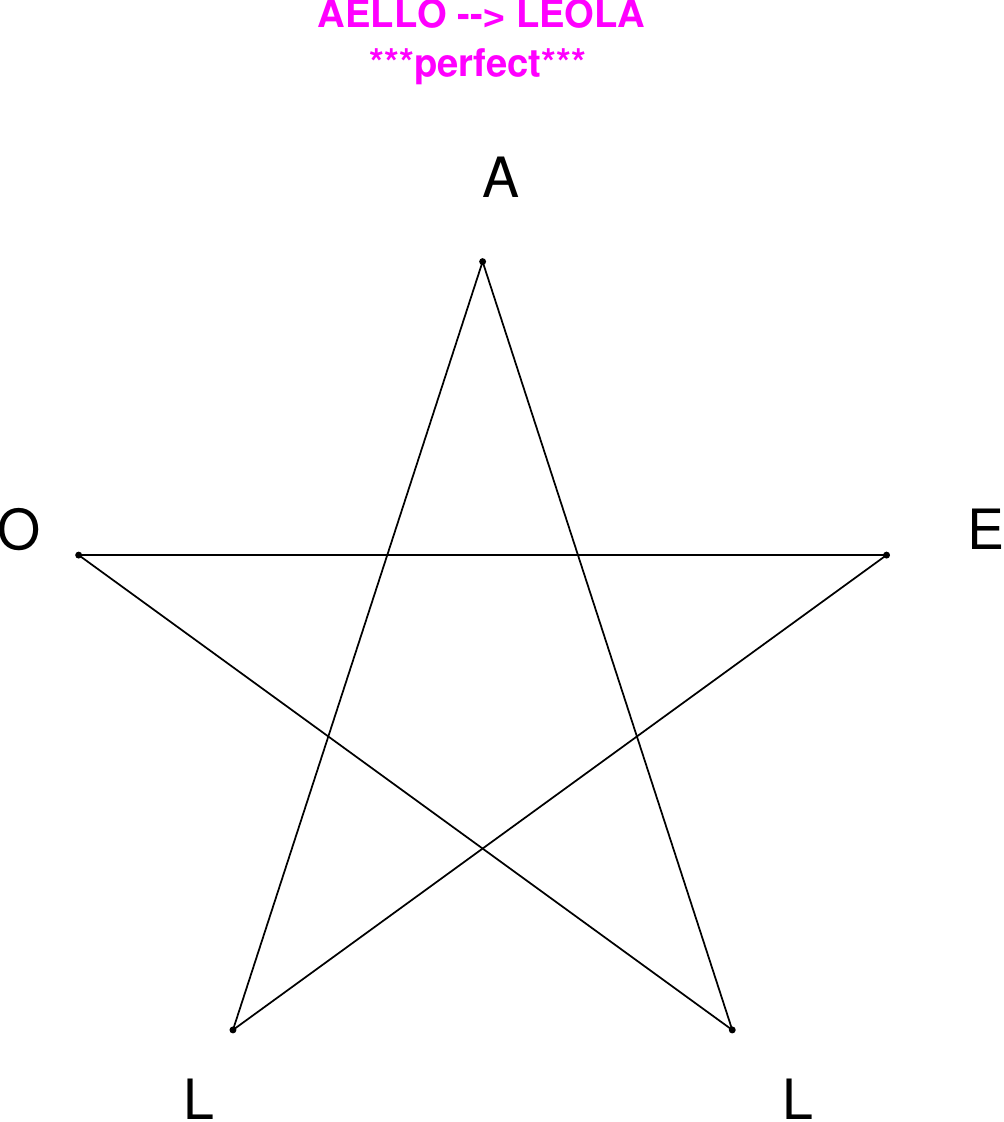}
\end{subfigure}
\hfill
\begin{subfigure}[T]{0.19\textwidth}
\centering
\includegraphics[width=\textwidth]{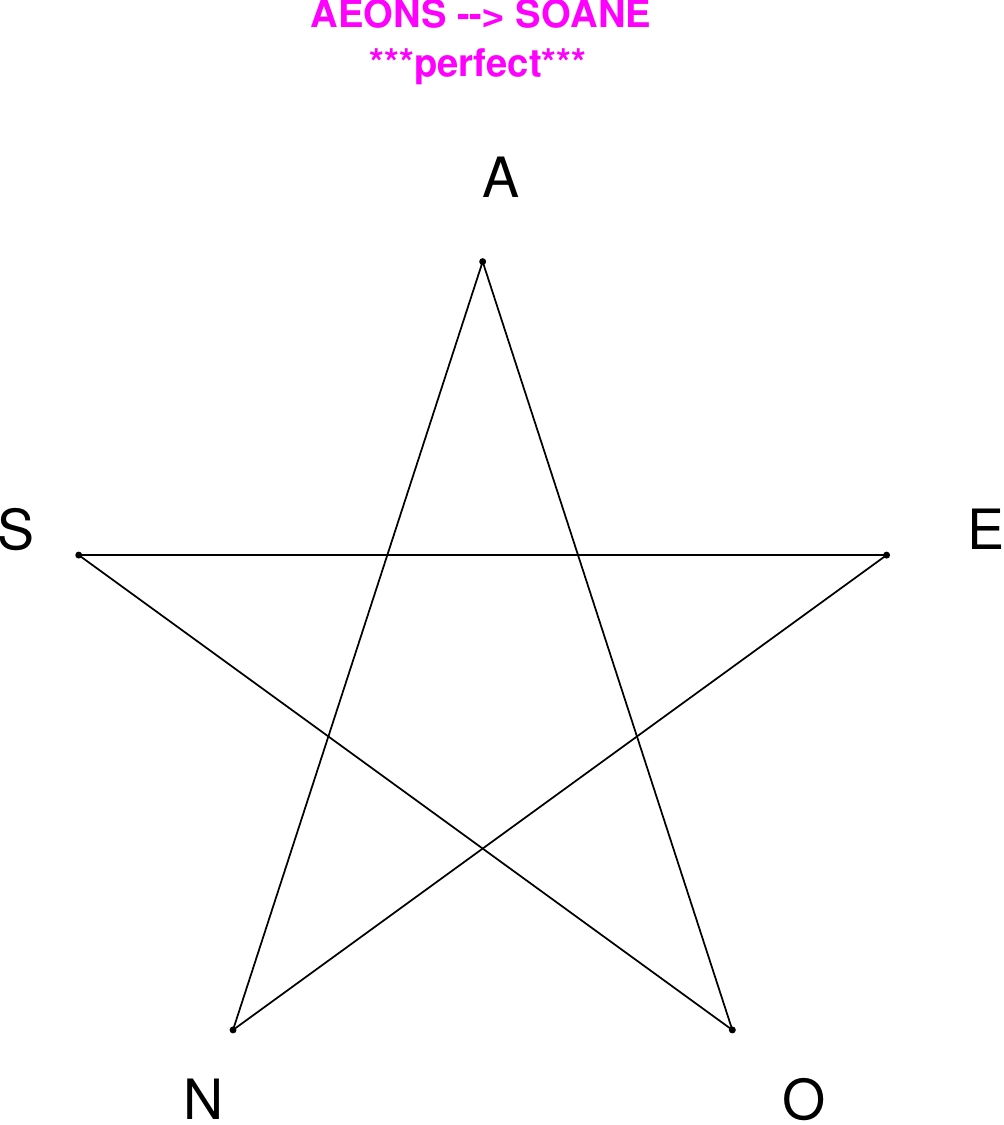}
\end{subfigure}
\hfill
\begin{subfigure}[T]{0.19\textwidth}
\centering
\includegraphics[width=\textwidth]{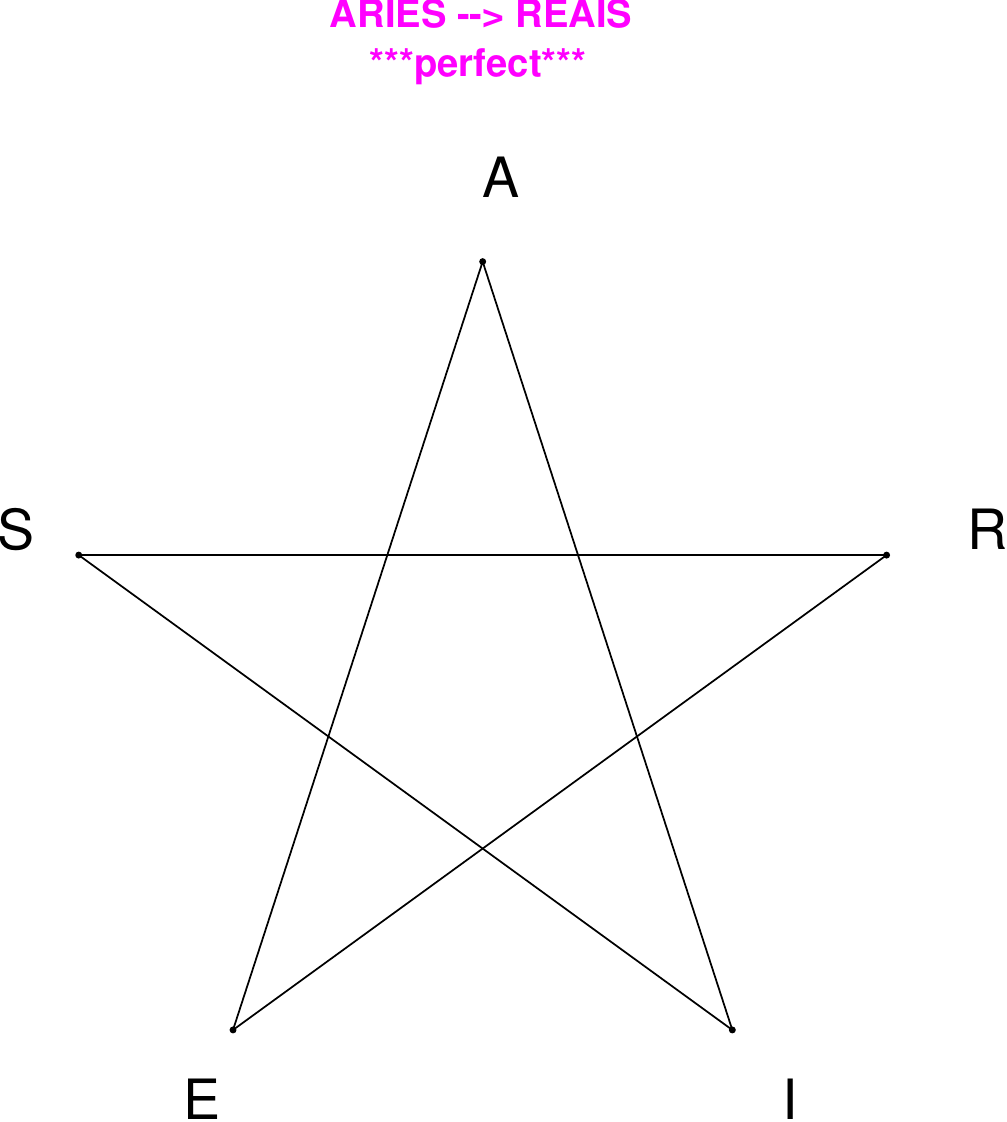}
\end{subfigure}
\hfill
\begin{subfigure}[T]{0.19\textwidth}
\centering
\includegraphics[width=\textwidth]{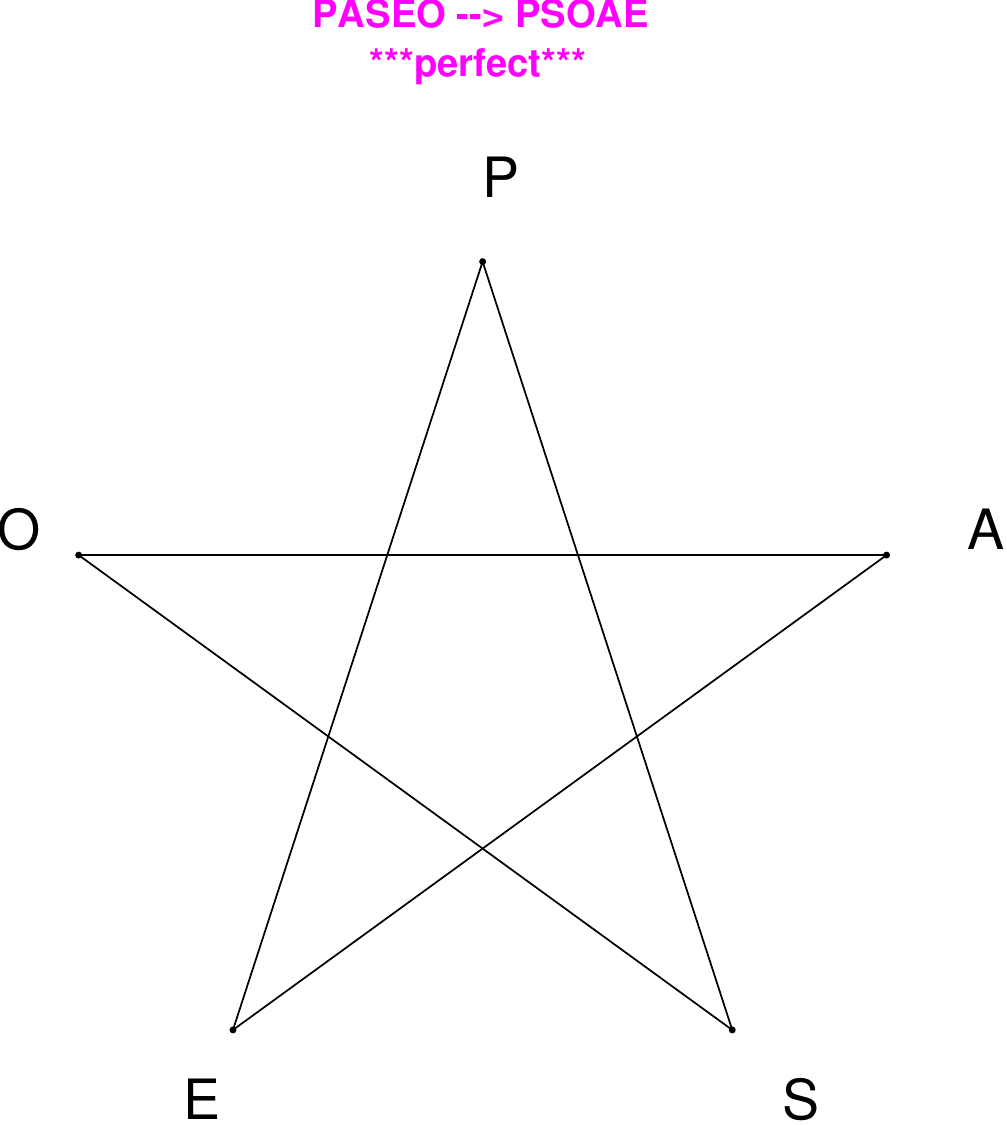}
\end{subfigure}
\hfill
\begin{subfigure}[T]{0.19\textwidth}
\centering
\includegraphics[width=\textwidth]{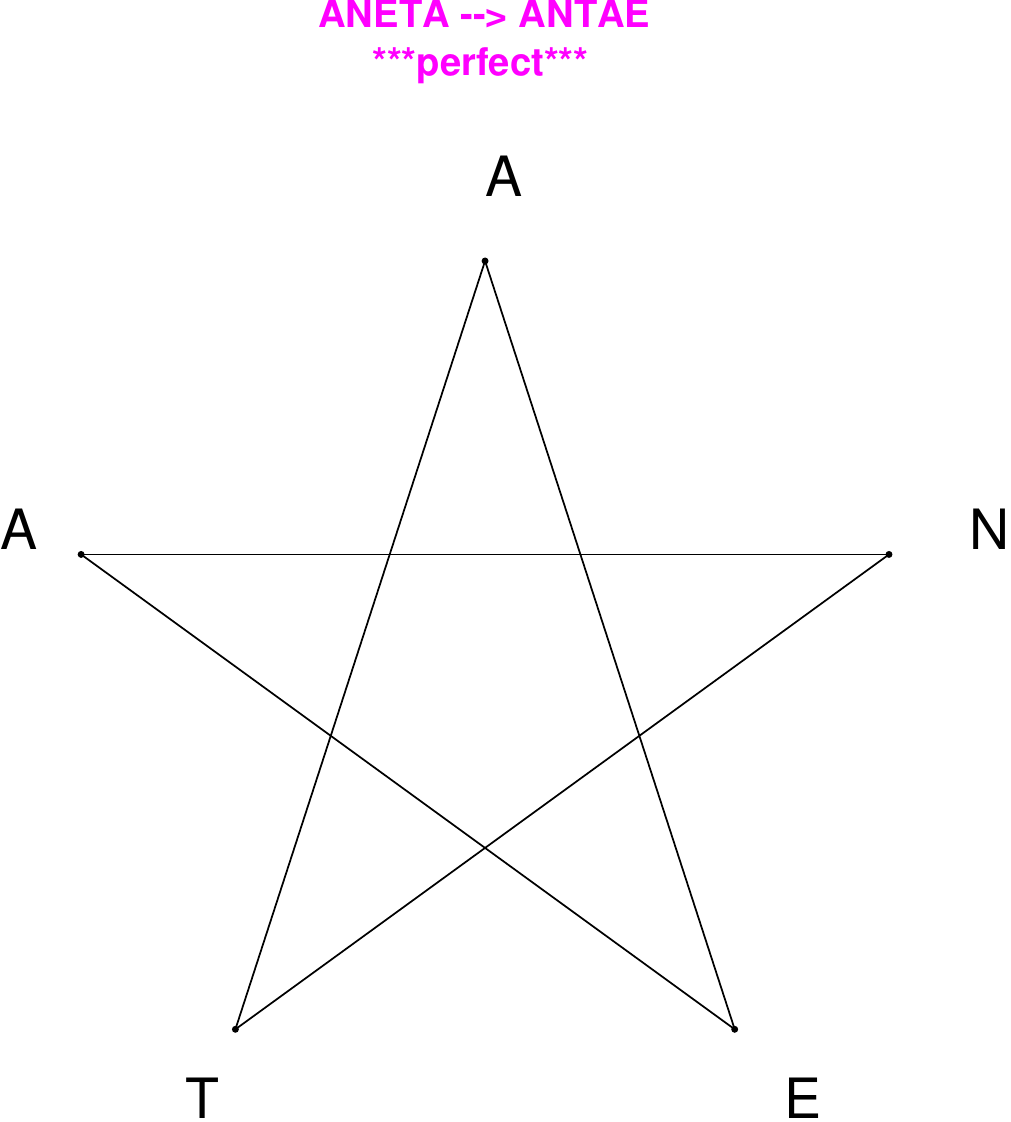}
\end{subfigure}
\end{figure}

\begin{figure}[H]
\centering
\begin{subfigure}[T]{0.19\textwidth}
\centering
\includegraphics[width=\textwidth]{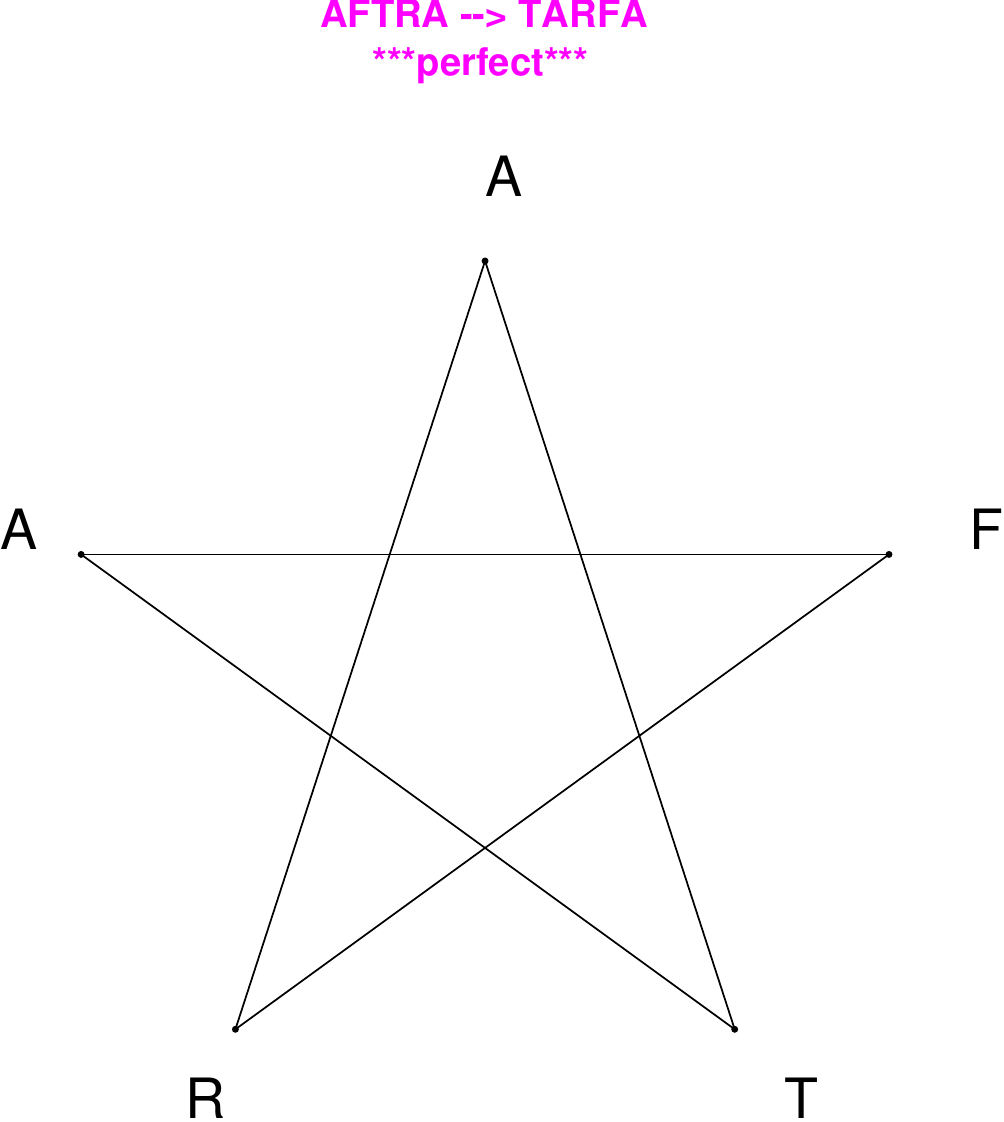}
\end{subfigure}
\hfill
\begin{subfigure}[T]{0.19\textwidth}
\centering
\includegraphics[width=\textwidth]{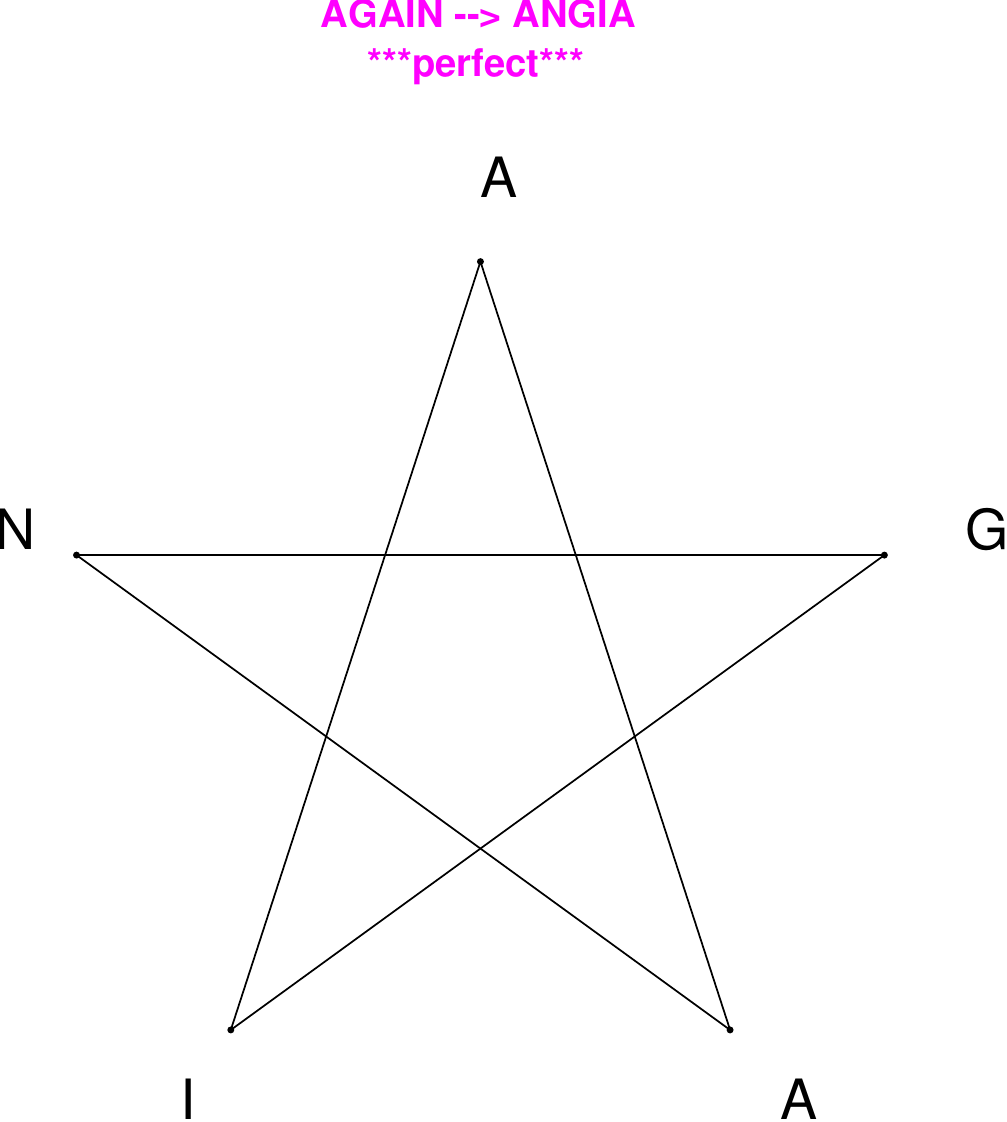}
\end{subfigure}
\hfill
\begin{subfigure}[T]{0.19\textwidth}
\centering
\includegraphics[width=\textwidth]{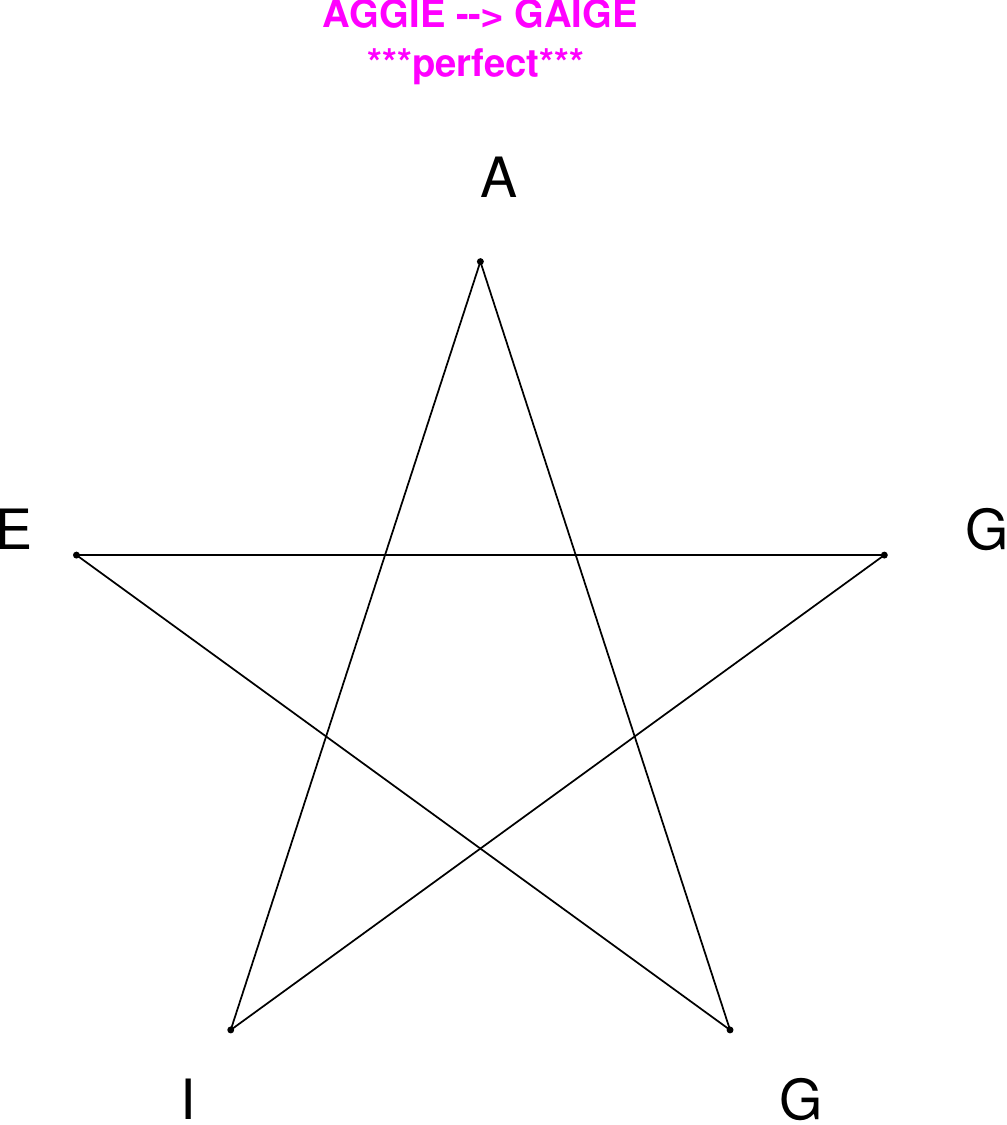}
\end{subfigure}
\hfill
\begin{subfigure}[T]{0.19\textwidth}
\centering
\includegraphics[width=\textwidth]{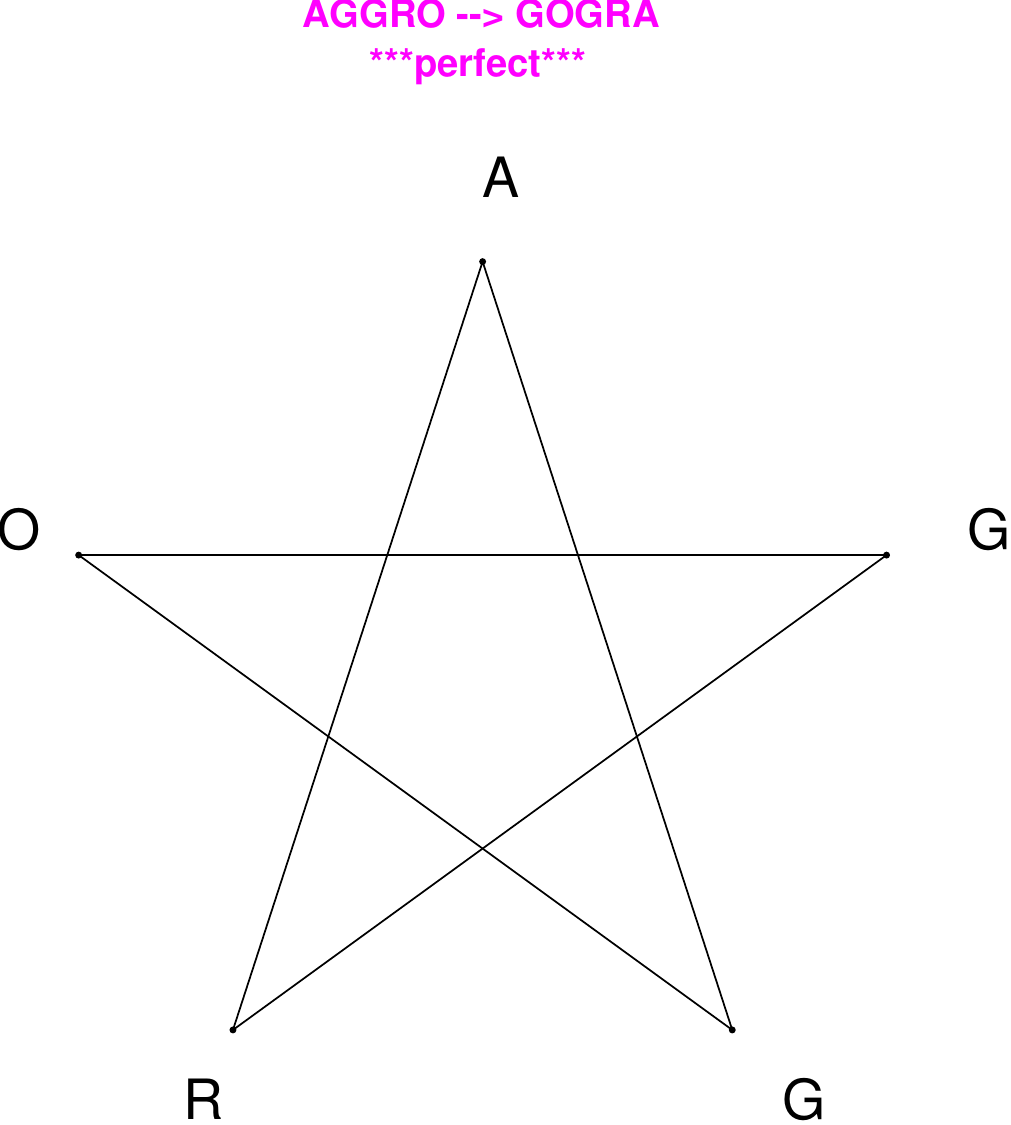}
\end{subfigure}
\hfill
\begin{subfigure}[T]{0.19\textwidth}
\centering
\includegraphics[width=\textwidth]{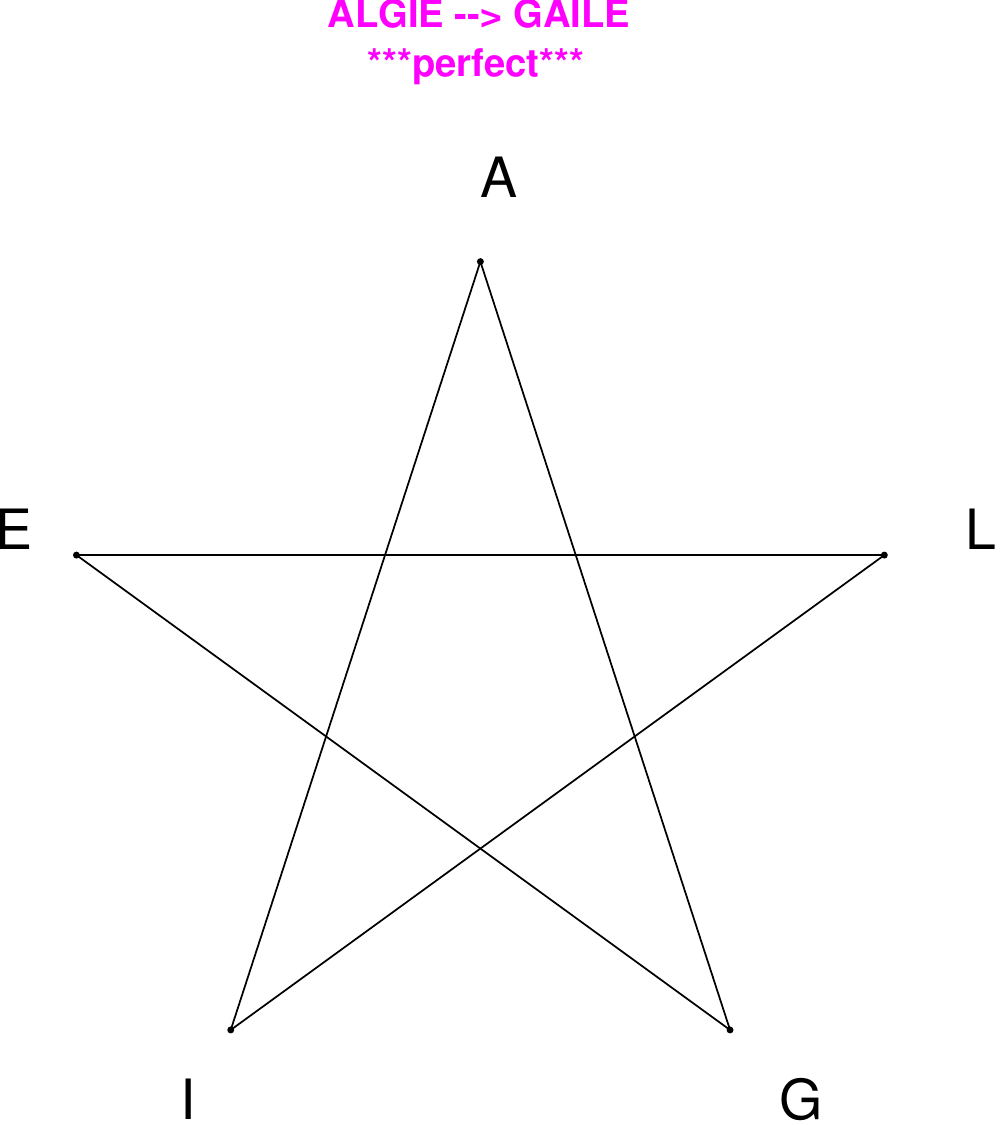}
\end{subfigure}
\end{figure}

\begin{figure}[H]
\centering
\begin{subfigure}[T]{0.19\textwidth}
\centering
\includegraphics[width=\textwidth]{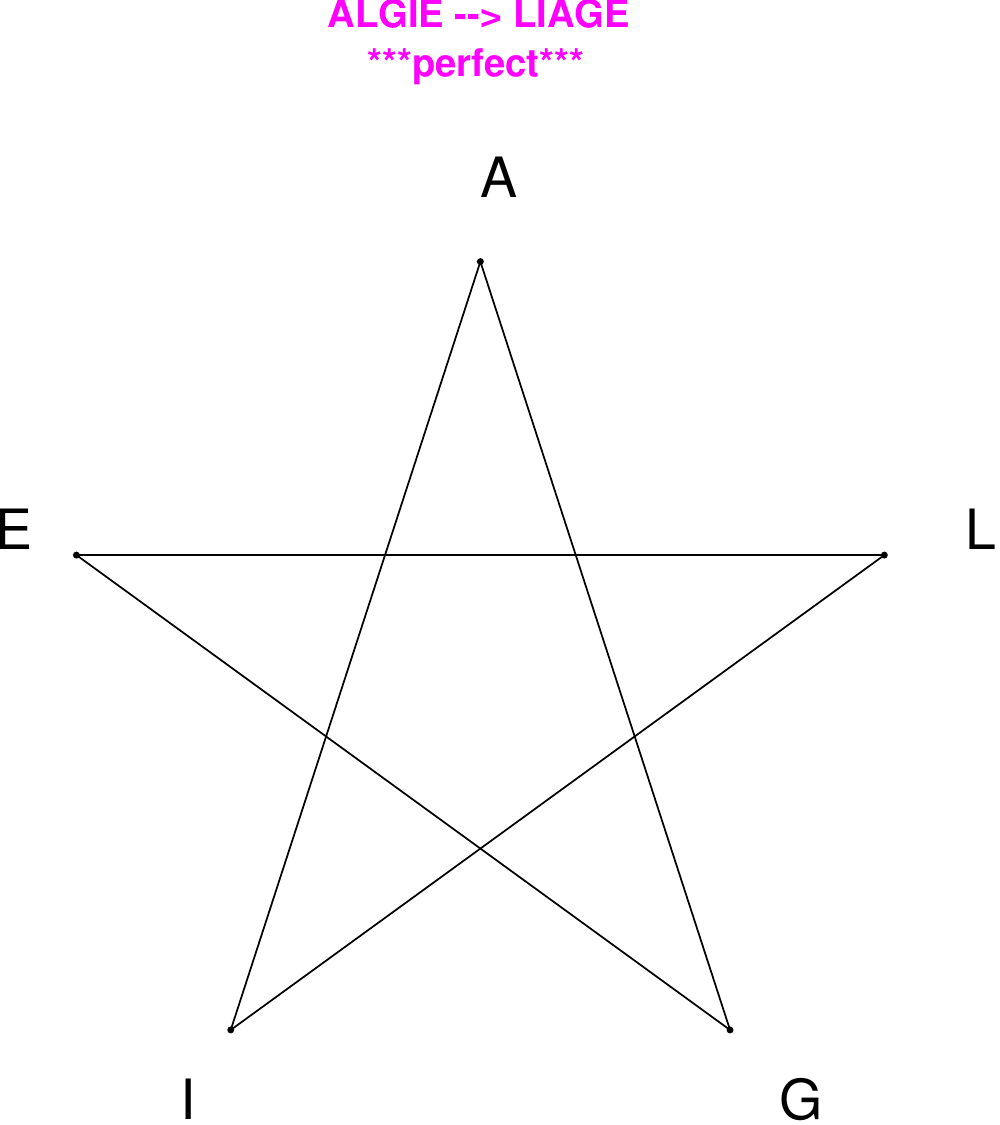}
\end{subfigure}
\hfill
\begin{subfigure}[T]{0.19\textwidth}
\centering
\includegraphics[width=\textwidth]{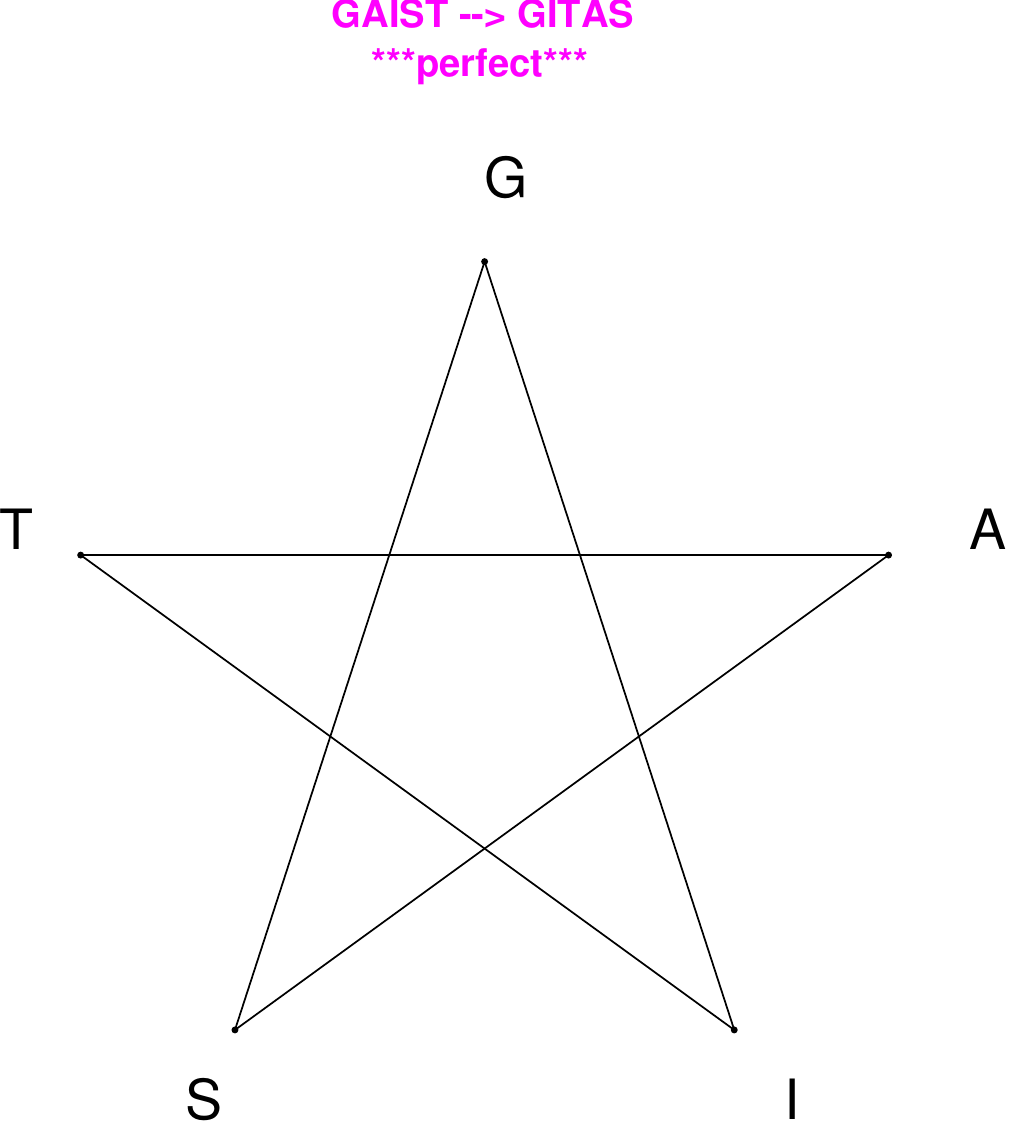}
\end{subfigure}
\hfill
\begin{subfigure}[T]{0.19\textwidth}
\centering
\includegraphics[width=\textwidth]{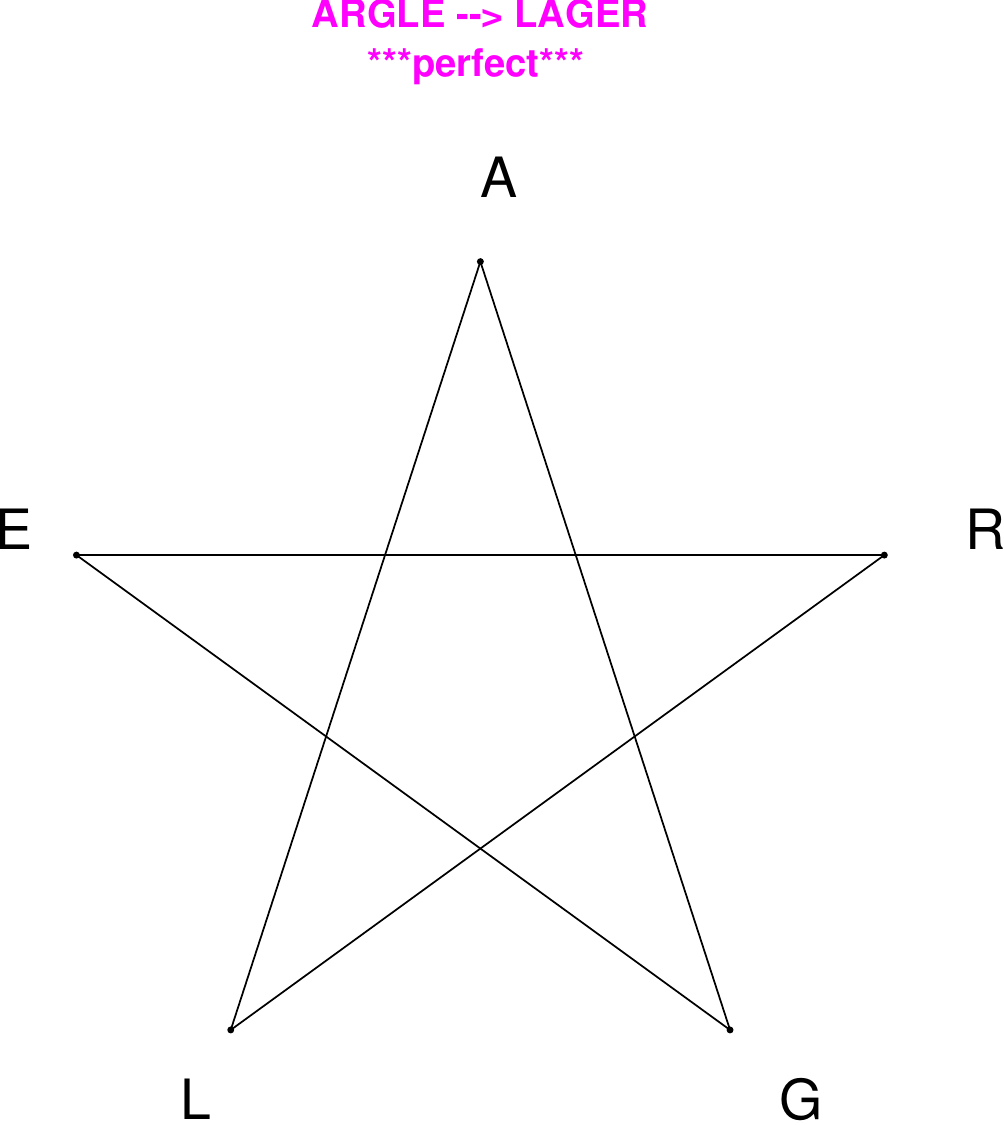}
\end{subfigure}
\hfill
\begin{subfigure}[T]{0.19\textwidth}
\centering
\includegraphics[width=\textwidth]{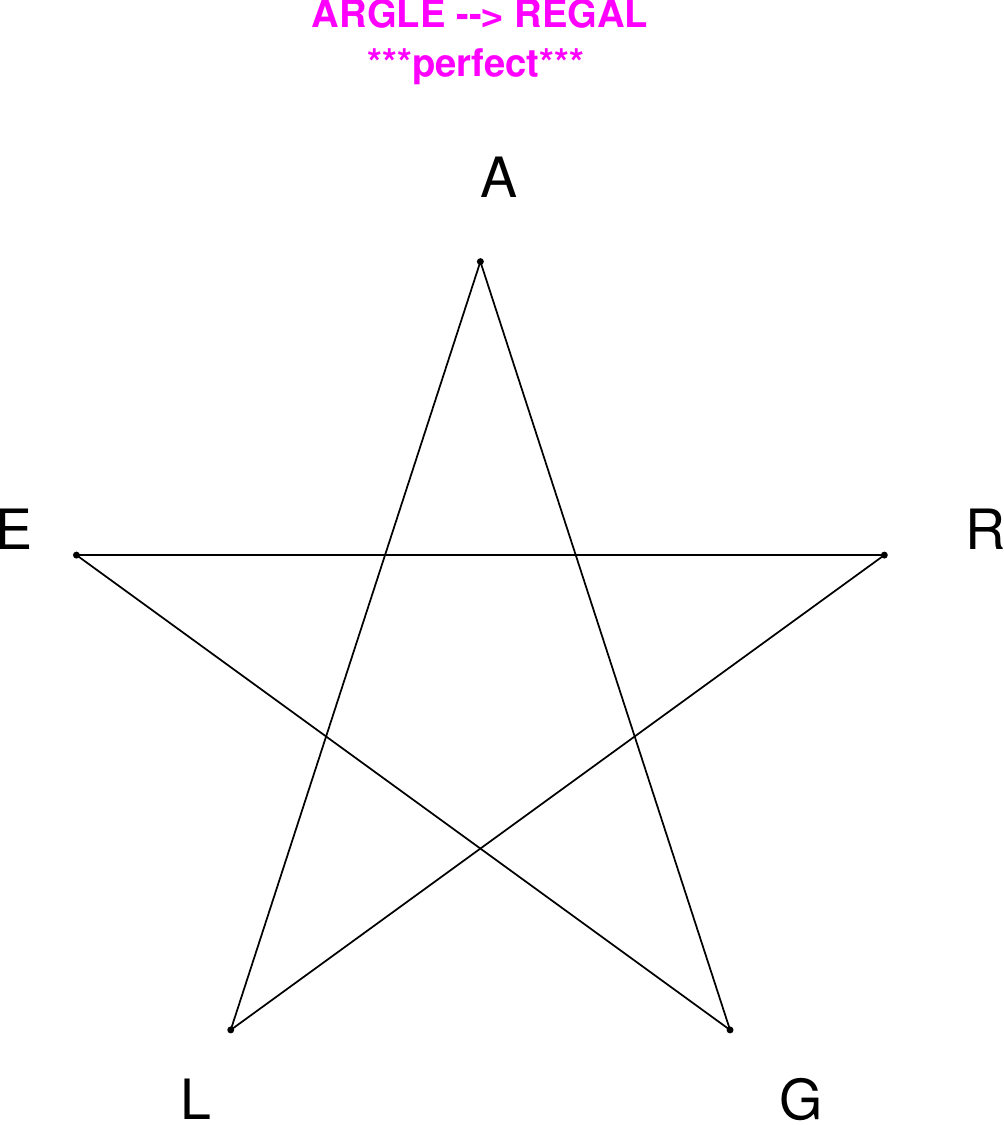}
\end{subfigure}
\hfill
\begin{subfigure}[T]{0.19\textwidth}
\centering
\includegraphics[width=\textwidth]{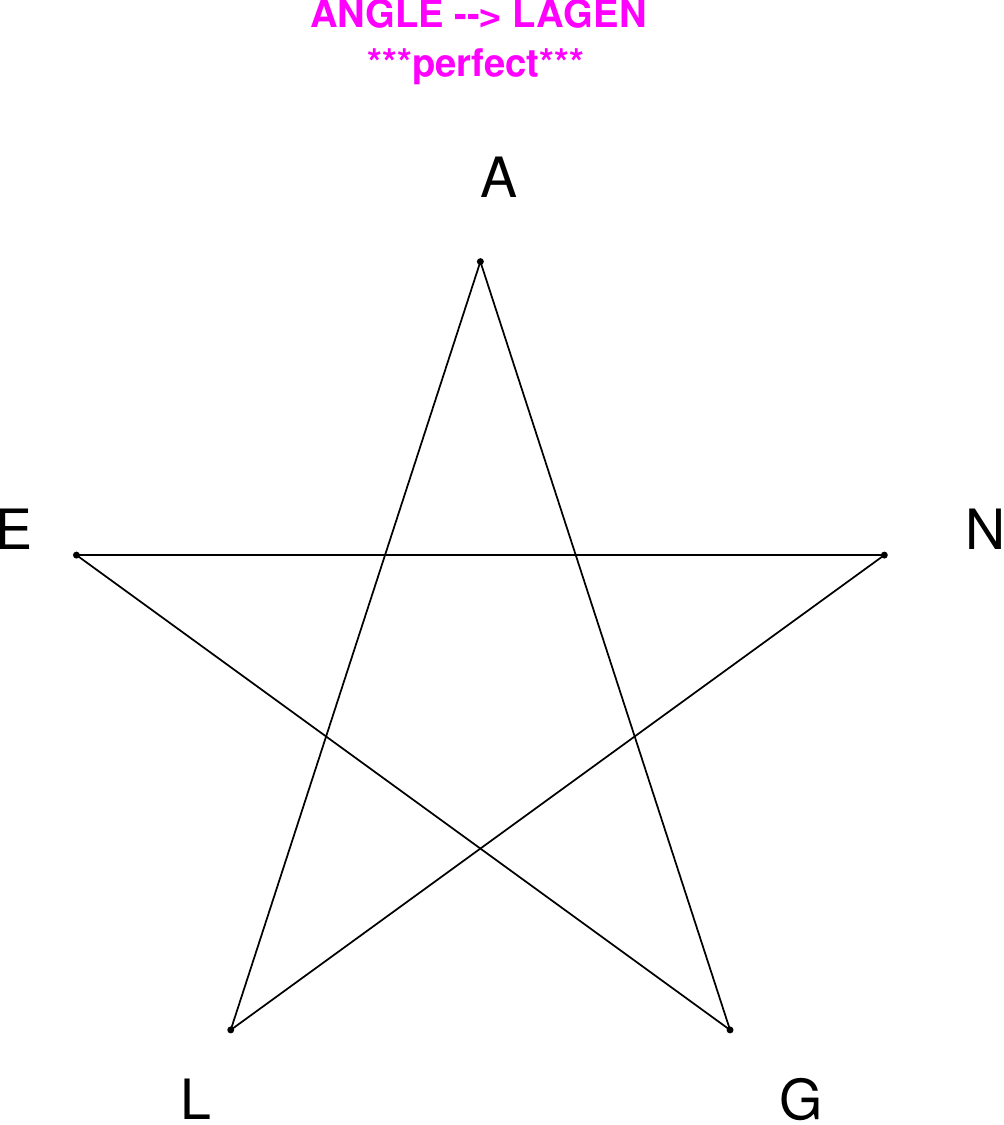}
\end{subfigure}
\end{figure}

\begin{figure}[H]
\centering
\begin{subfigure}[T]{0.19\textwidth}
\centering
\includegraphics[width=\textwidth]{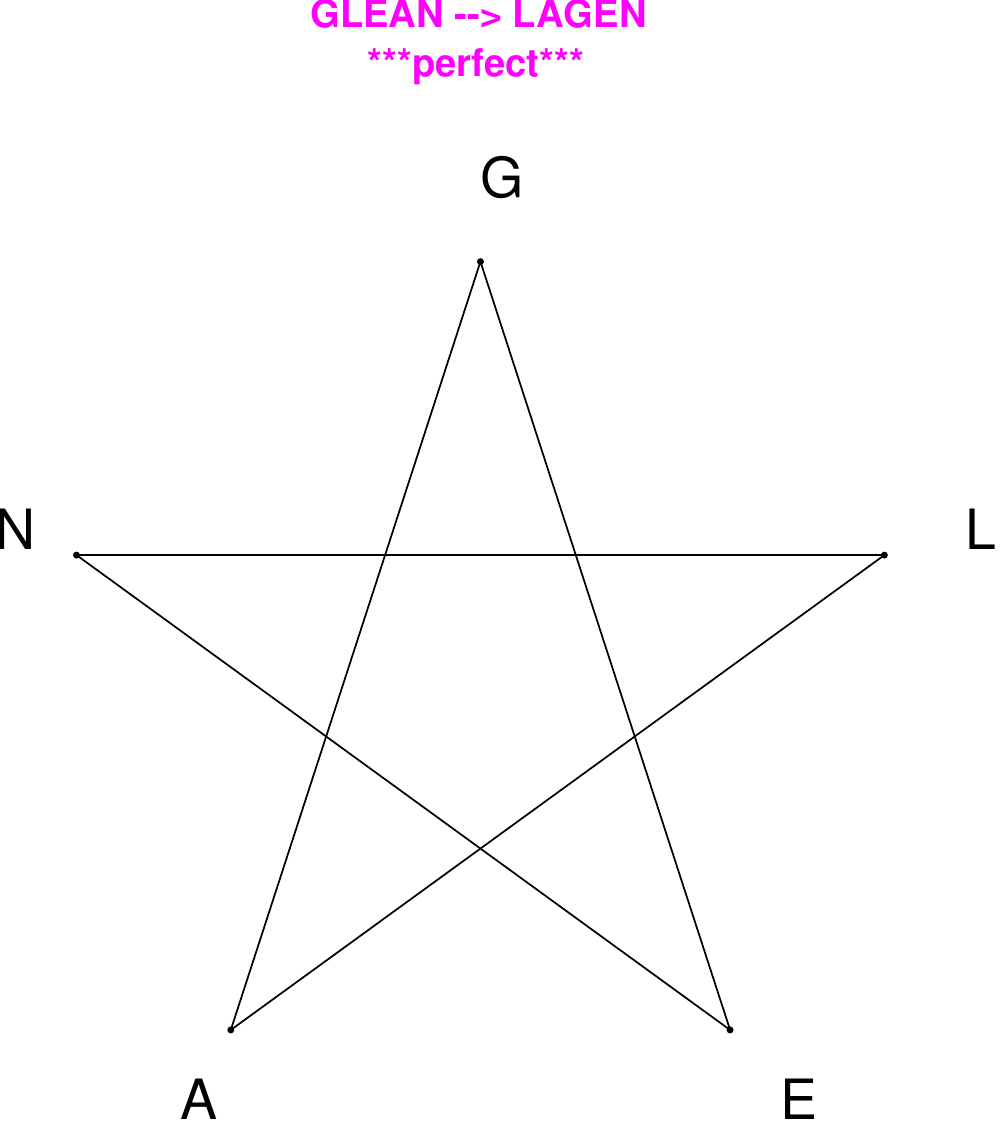}
\end{subfigure}
\hfill
\begin{subfigure}[T]{0.19\textwidth}
\centering
\includegraphics[width=\textwidth]{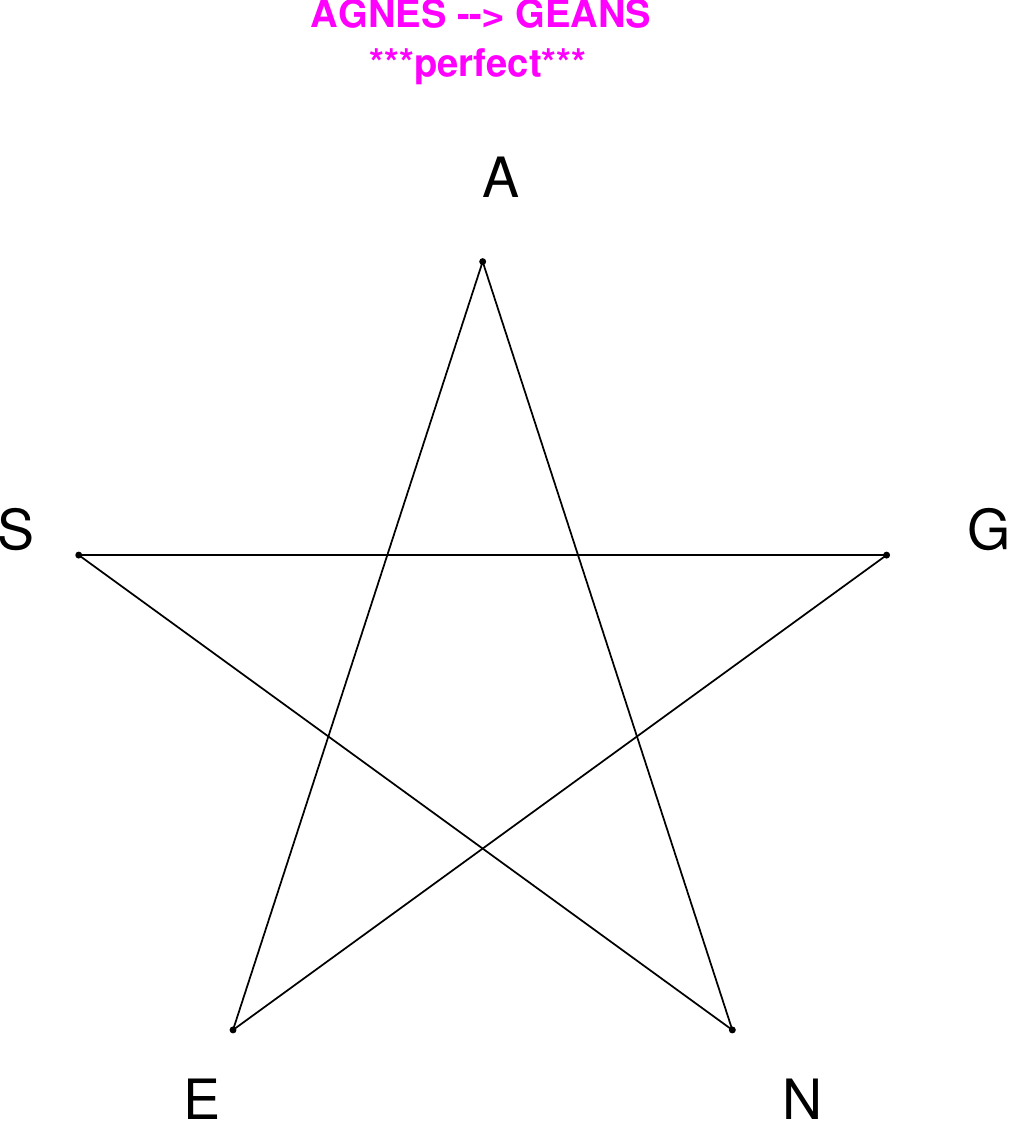}
\end{subfigure}
\hfill
\begin{subfigure}[T]{0.19\textwidth}
\centering
\includegraphics[width=\textwidth]{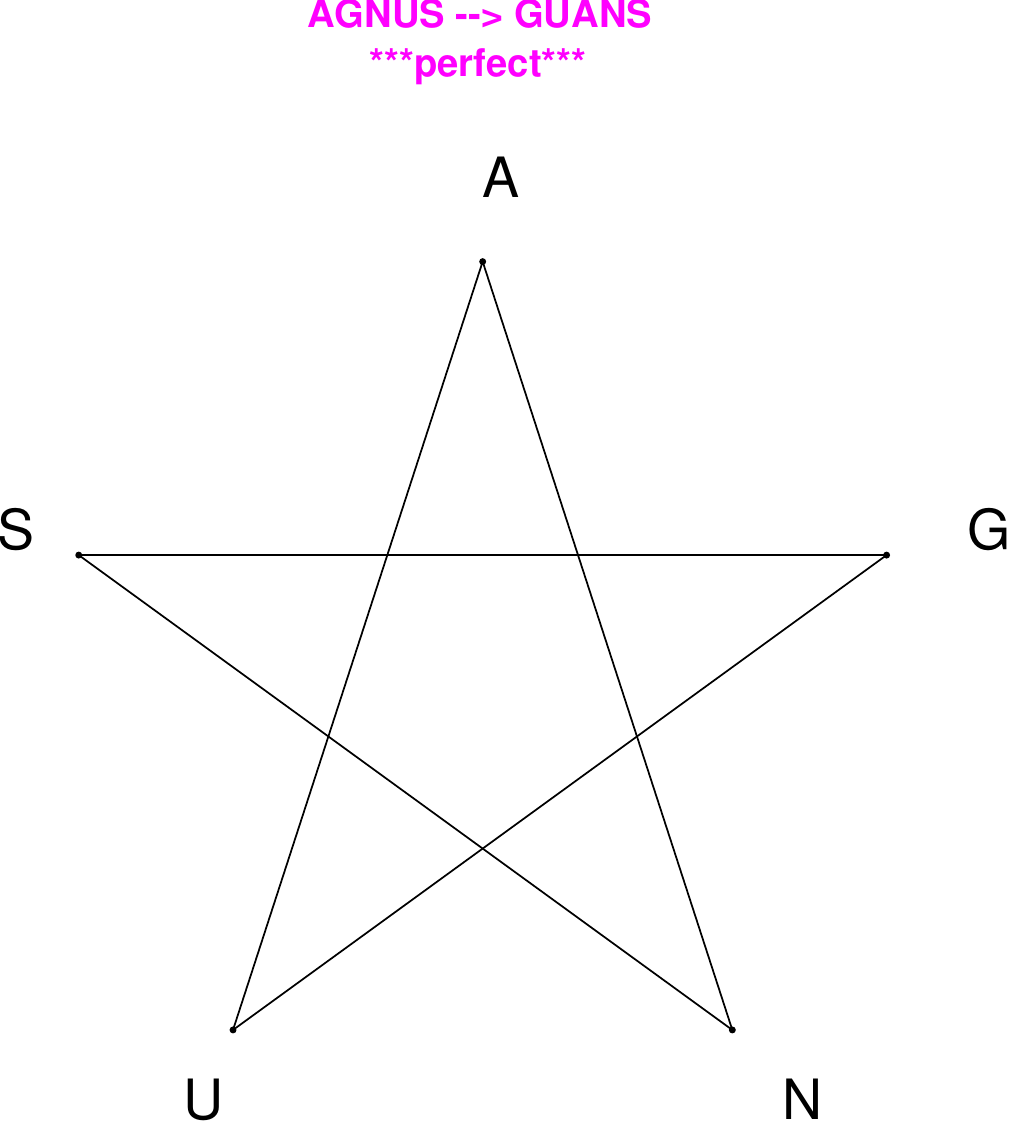}
\end{subfigure}
\hfill
\begin{subfigure}[T]{0.19\textwidth}
\centering
\includegraphics[width=\textwidth]{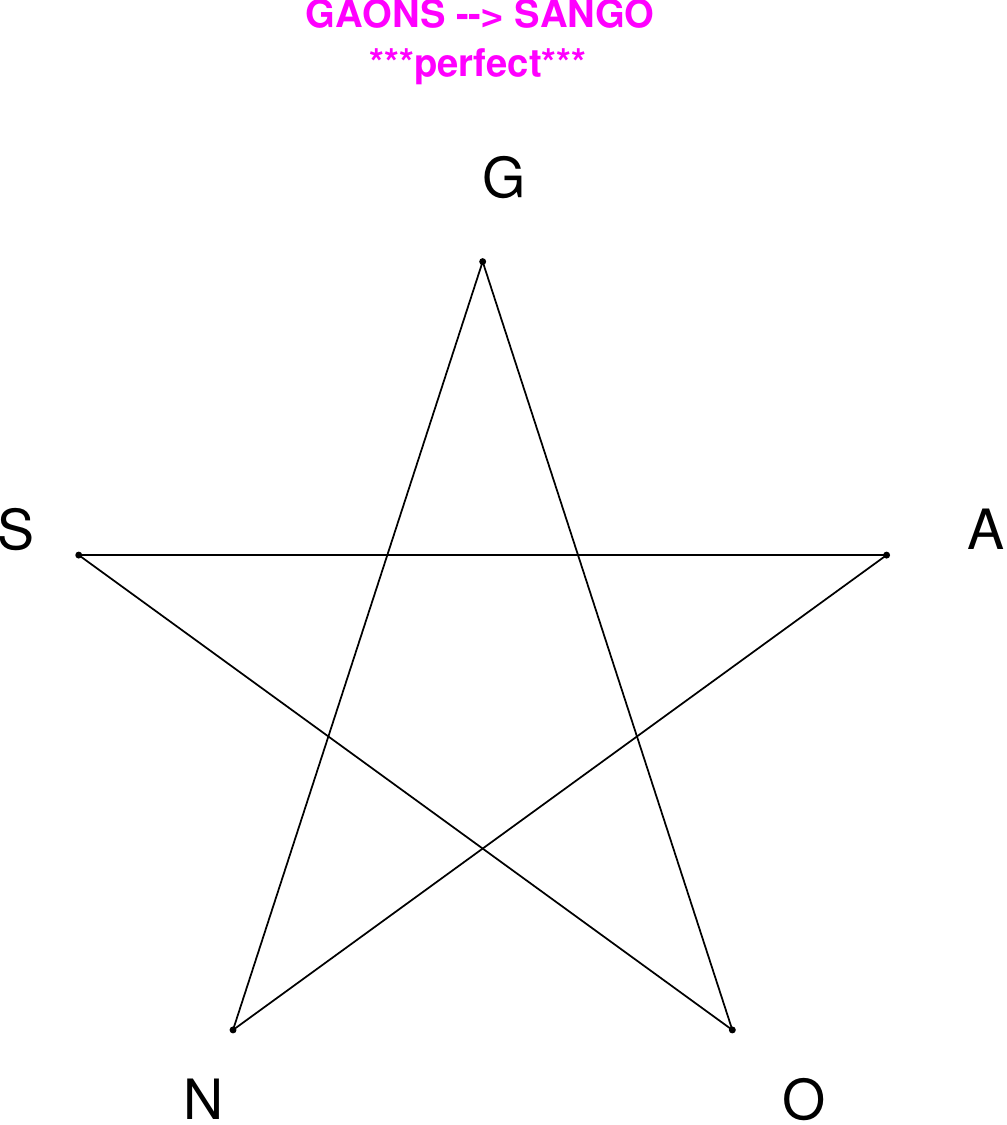}
\end{subfigure}
\hfill
\begin{subfigure}[T]{0.19\textwidth}
\centering
\includegraphics[width=\textwidth]{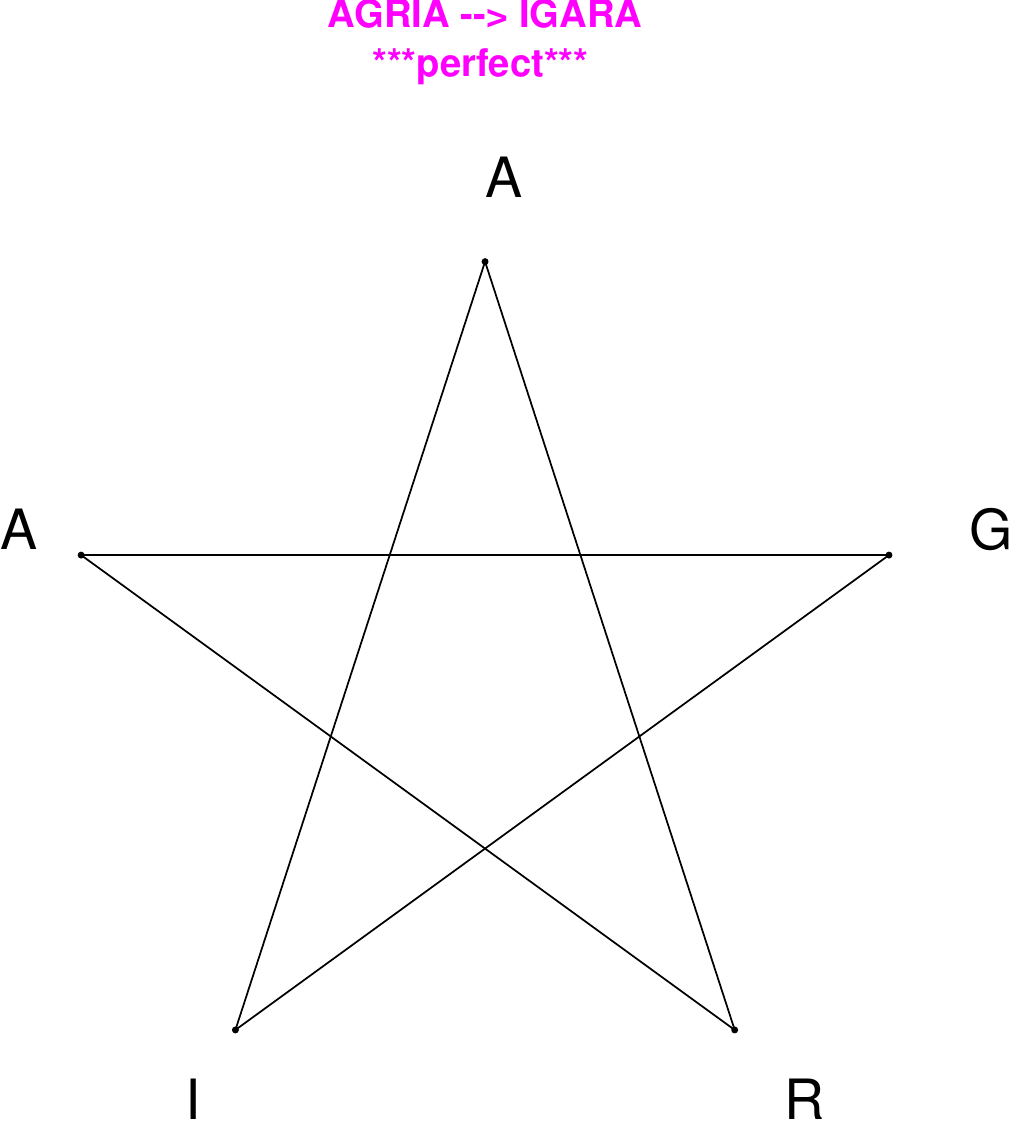}
\end{subfigure}
\end{figure}

\begin{figure}[H]
\centering
\begin{subfigure}[T]{0.19\textwidth}
\centering
\includegraphics[width=\textwidth]{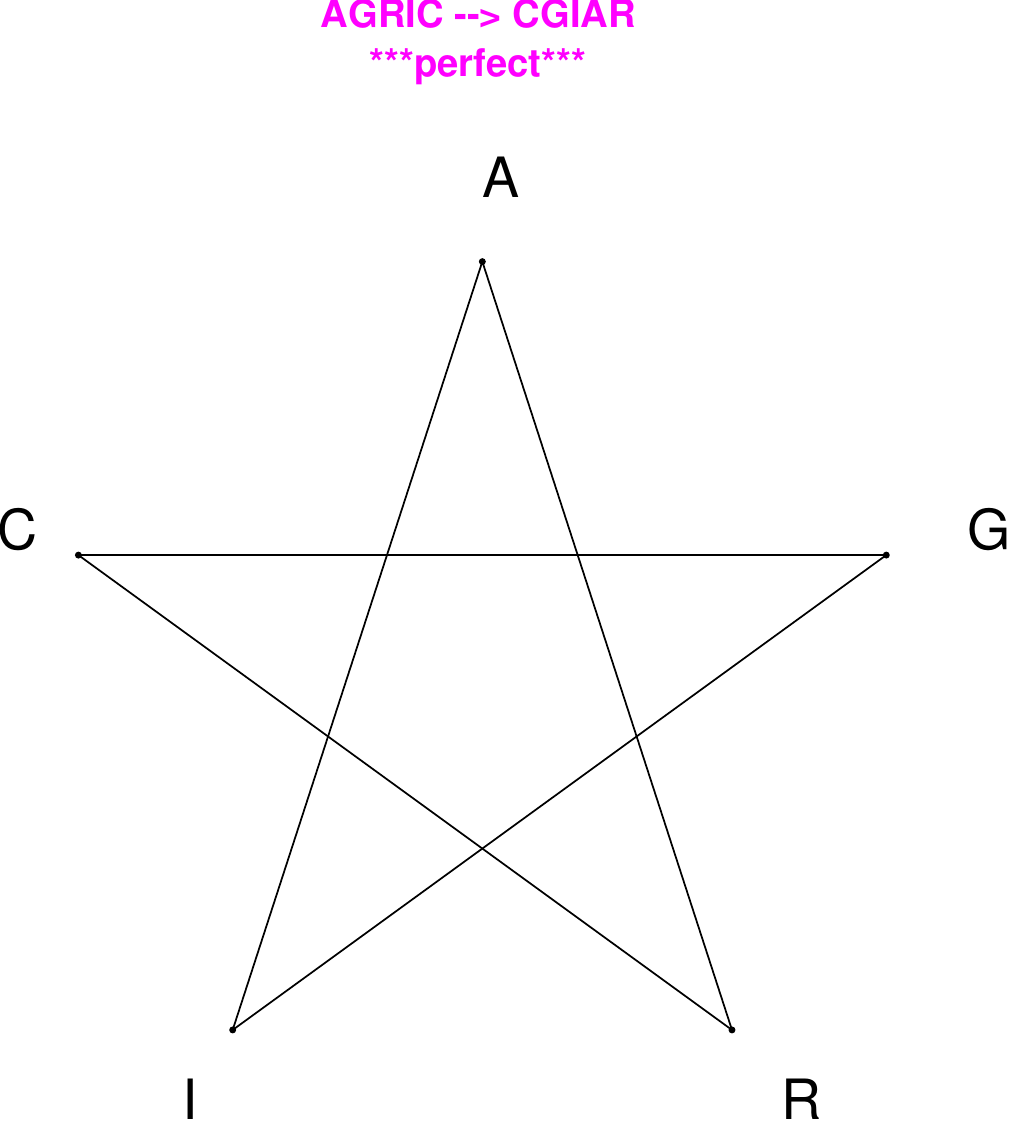}
\end{subfigure}
\hfill
\begin{subfigure}[T]{0.19\textwidth}
\centering
\includegraphics[width=\textwidth]{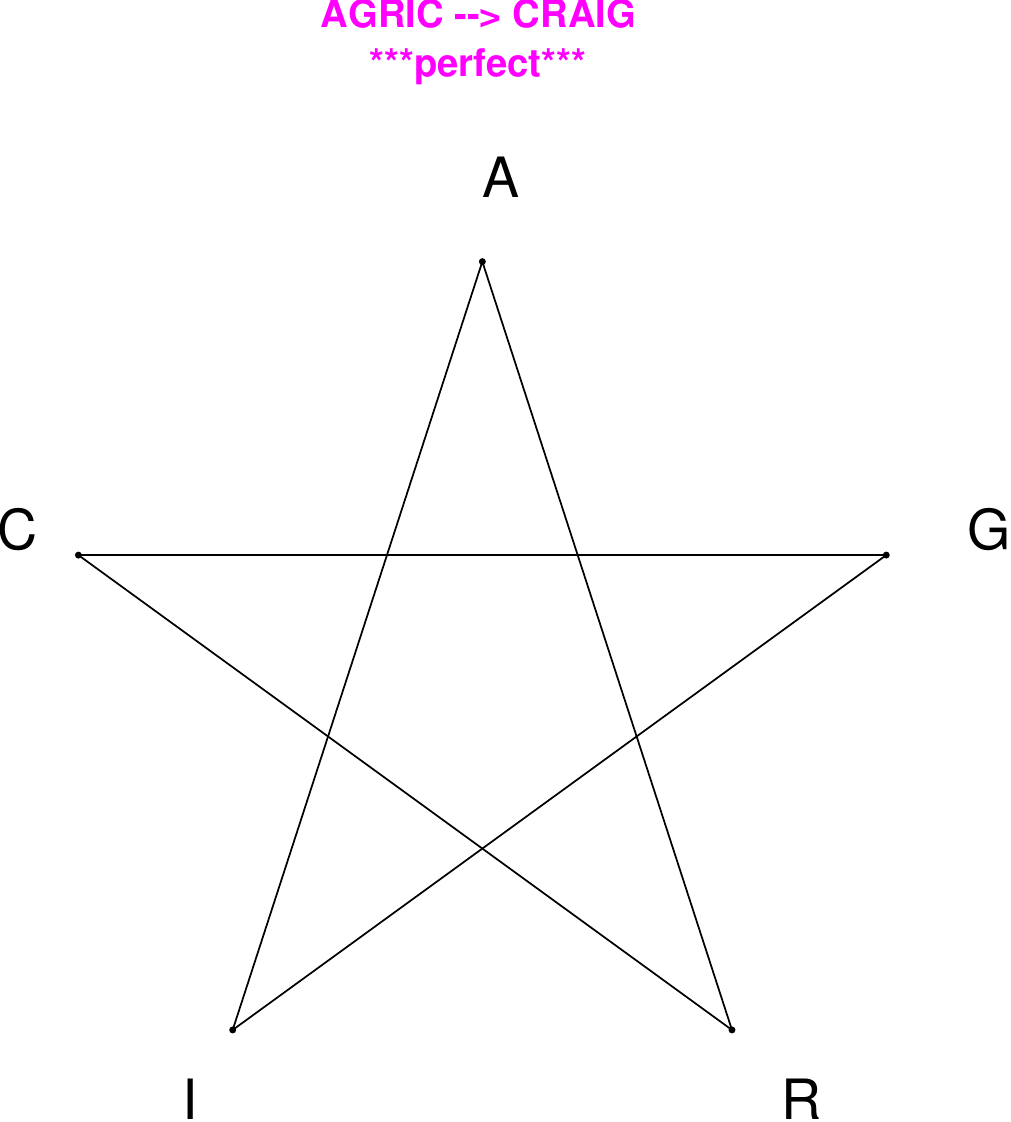}
\end{subfigure}
\hfill
\begin{subfigure}[T]{0.19\textwidth}
\centering
\includegraphics[width=\textwidth]{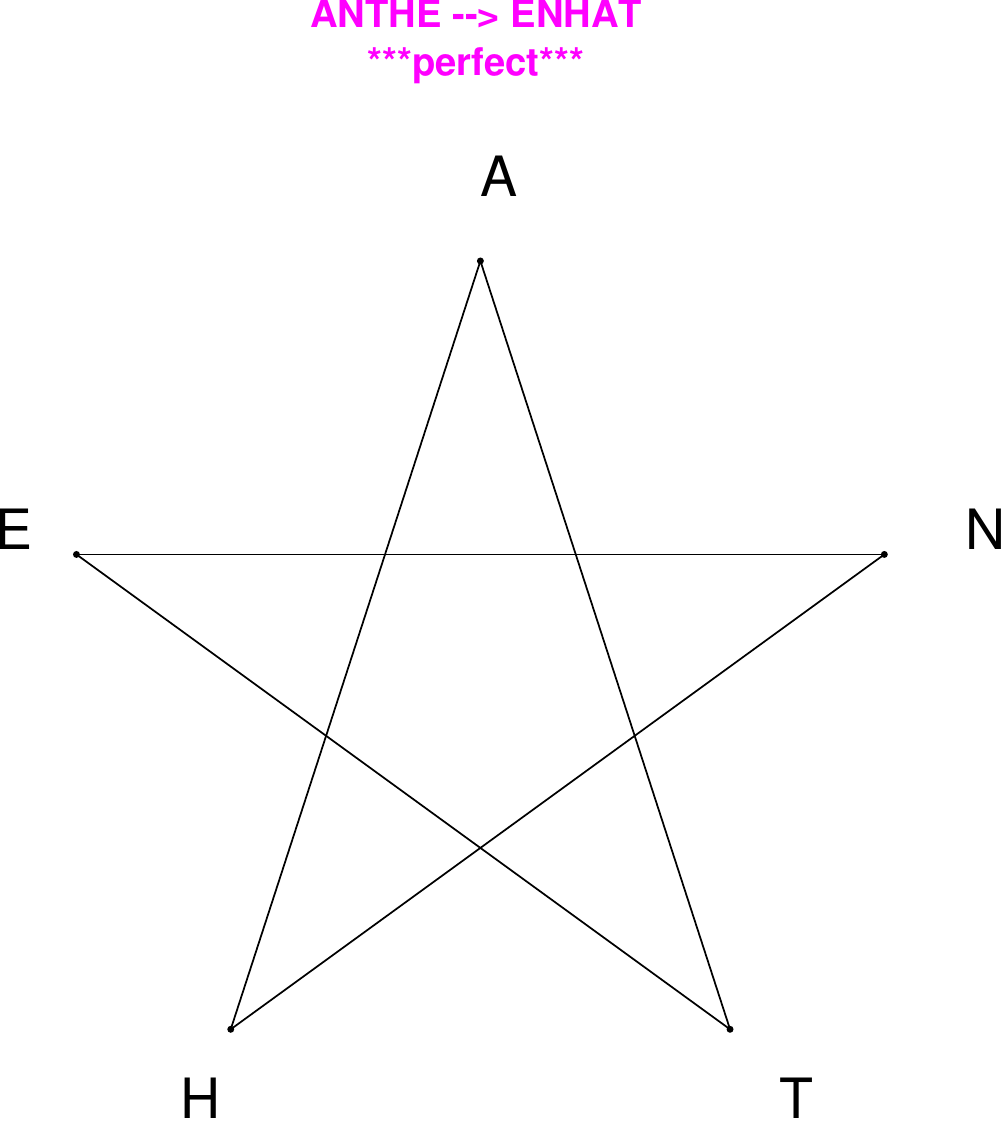}
\end{subfigure}
\hfill
\begin{subfigure}[T]{0.19\textwidth}
\centering
\includegraphics[width=\textwidth]{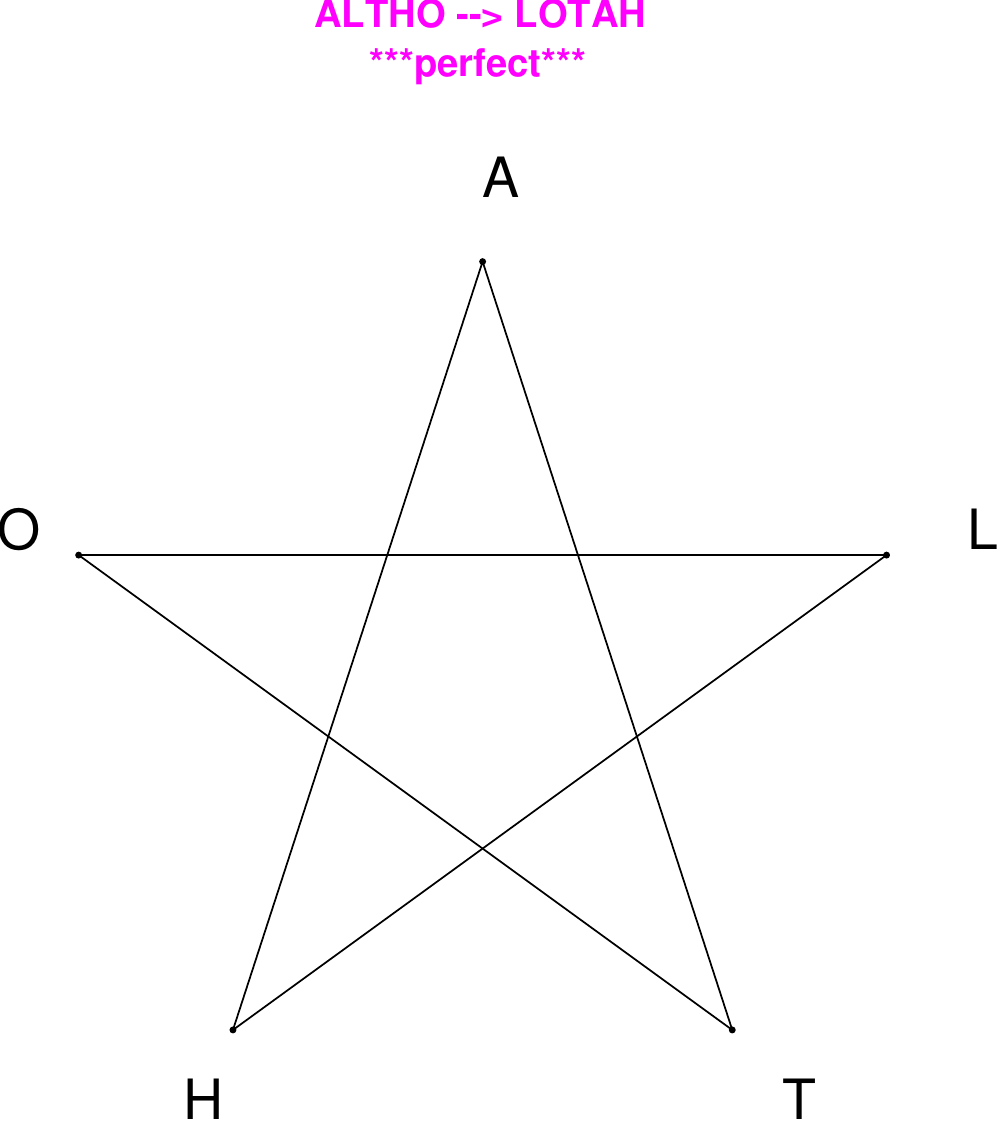}
\end{subfigure}
\hfill
\begin{subfigure}[T]{0.19\textwidth}
\centering
\includegraphics[width=\textwidth]{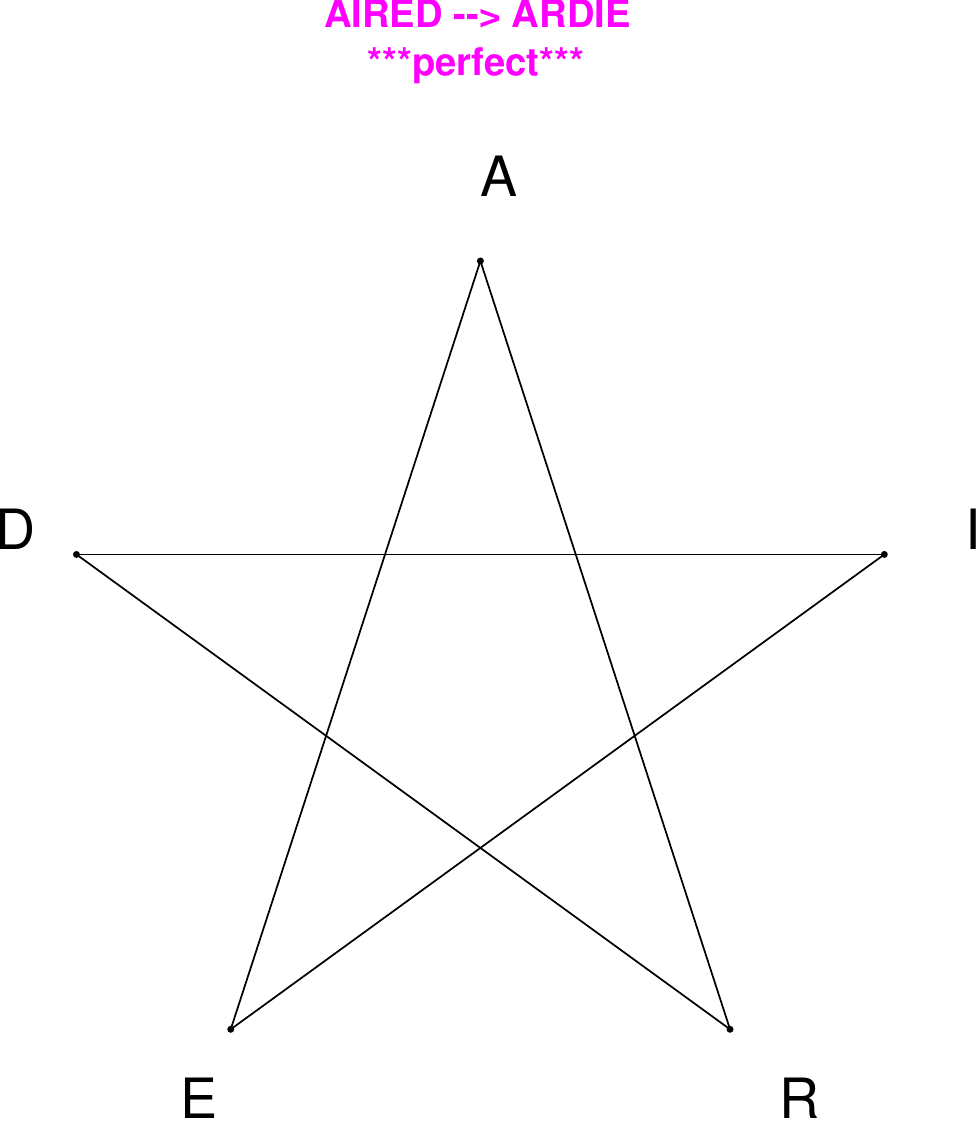}
\end{subfigure}
\end{figure}

\begin{figure}[H]
\centering
\begin{subfigure}[T]{0.19\textwidth}
\centering
\includegraphics[width=\textwidth]{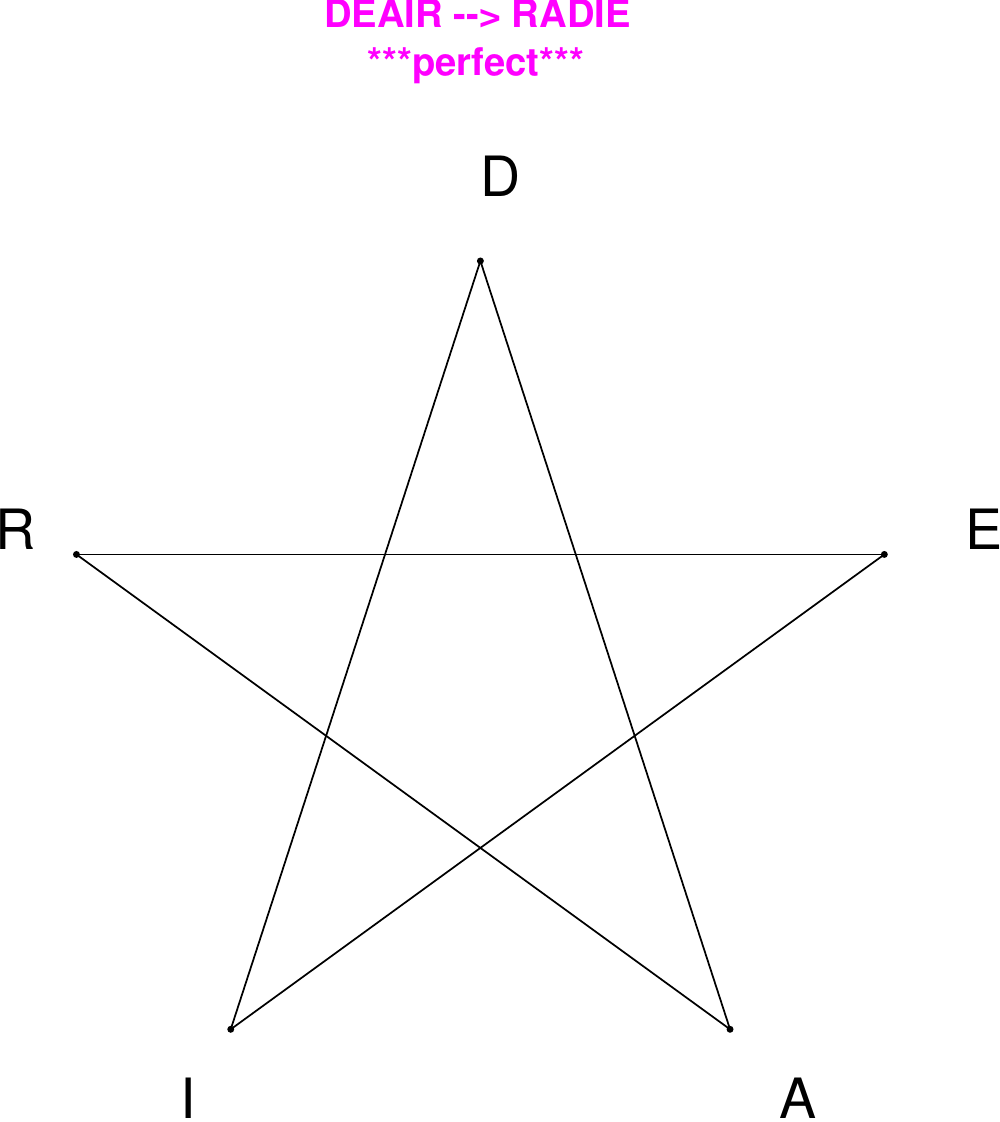}
\end{subfigure}
\hfill
\begin{subfigure}[T]{0.19\textwidth}
\centering
\includegraphics[width=\textwidth]{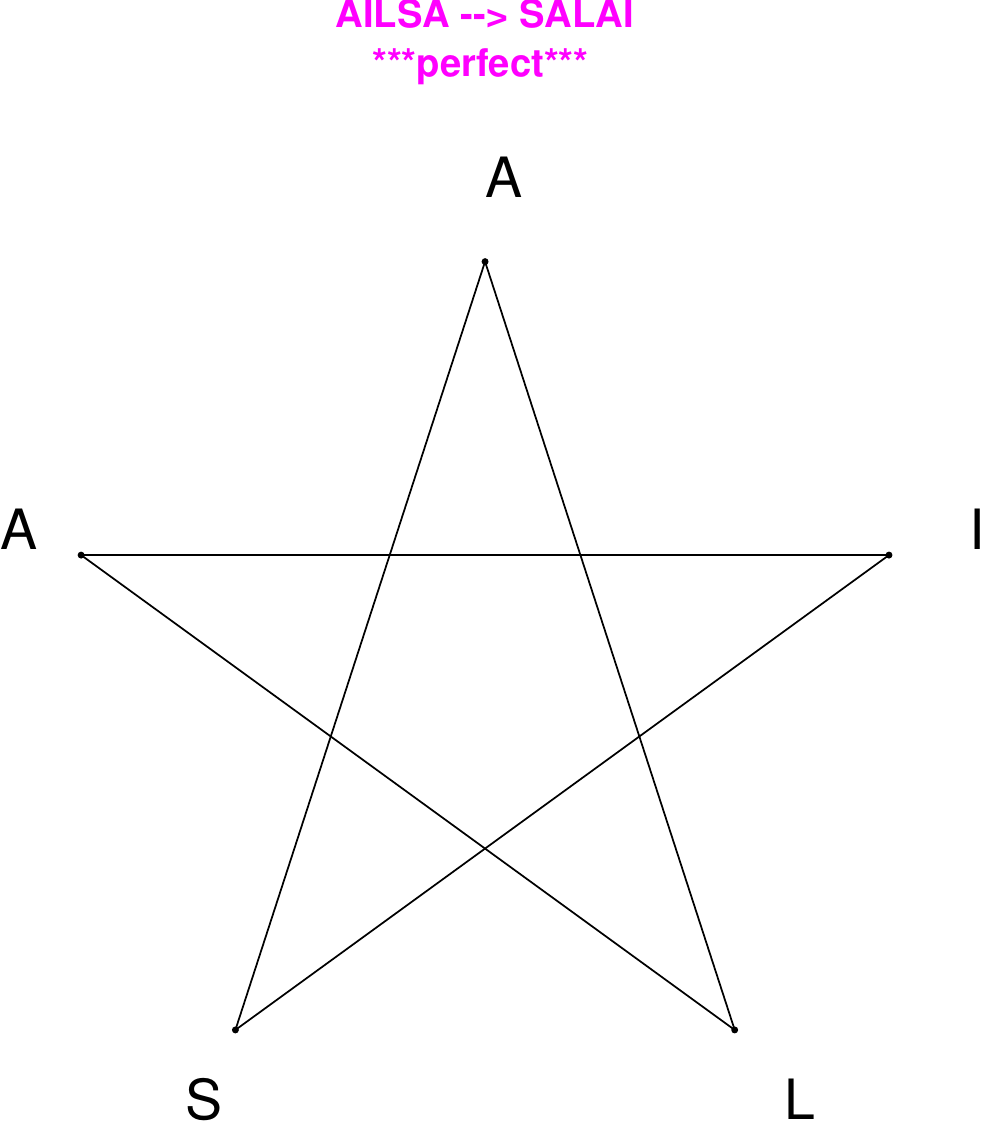}
\end{subfigure}
\hfill
\begin{subfigure}[T]{0.19\textwidth}
\centering
\includegraphics[width=\textwidth]{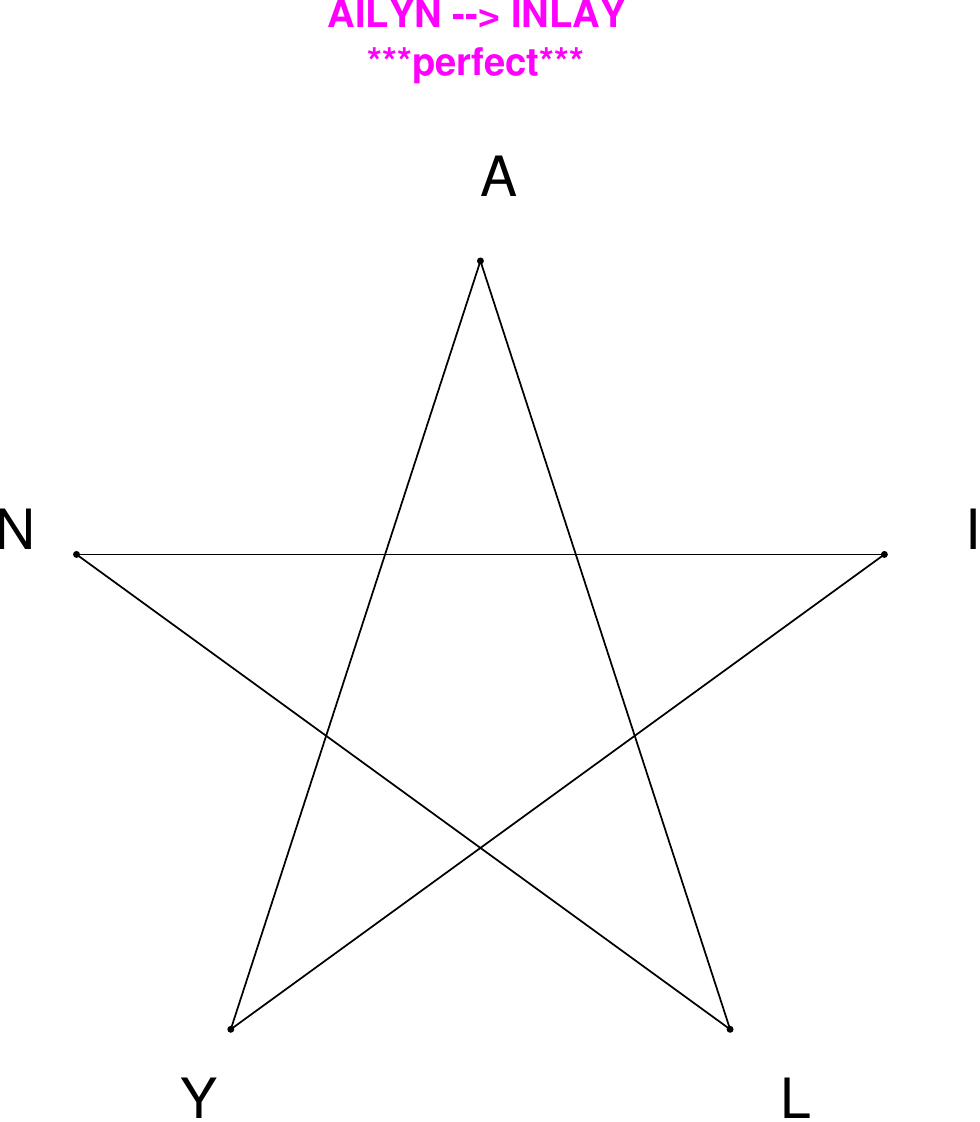}
\end{subfigure}
\hfill
\begin{subfigure}[T]{0.19\textwidth}
\centering
\includegraphics[width=\textwidth]{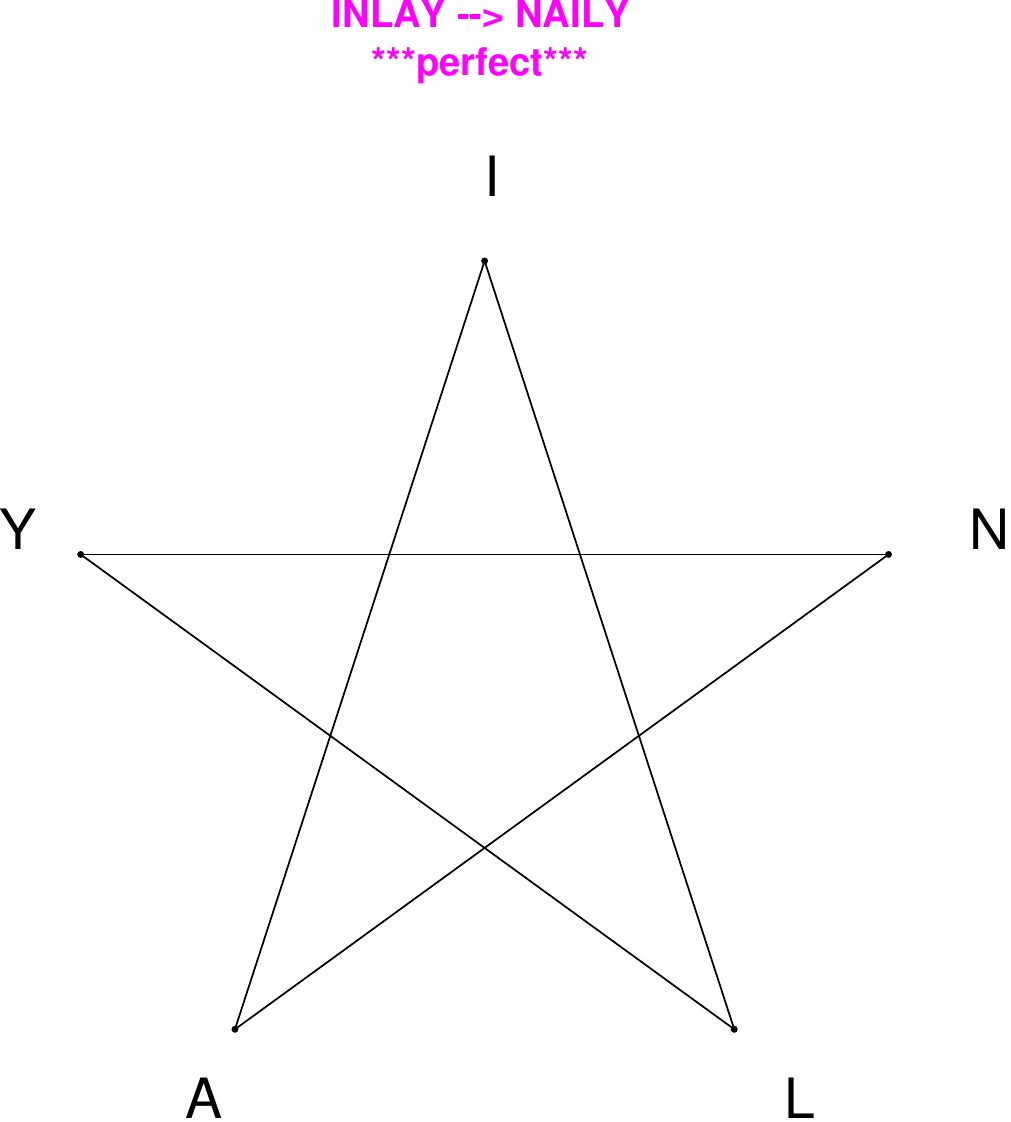}
\end{subfigure}
\hfill
\begin{subfigure}[T]{0.19\textwidth}
\centering
\includegraphics[width=\textwidth]{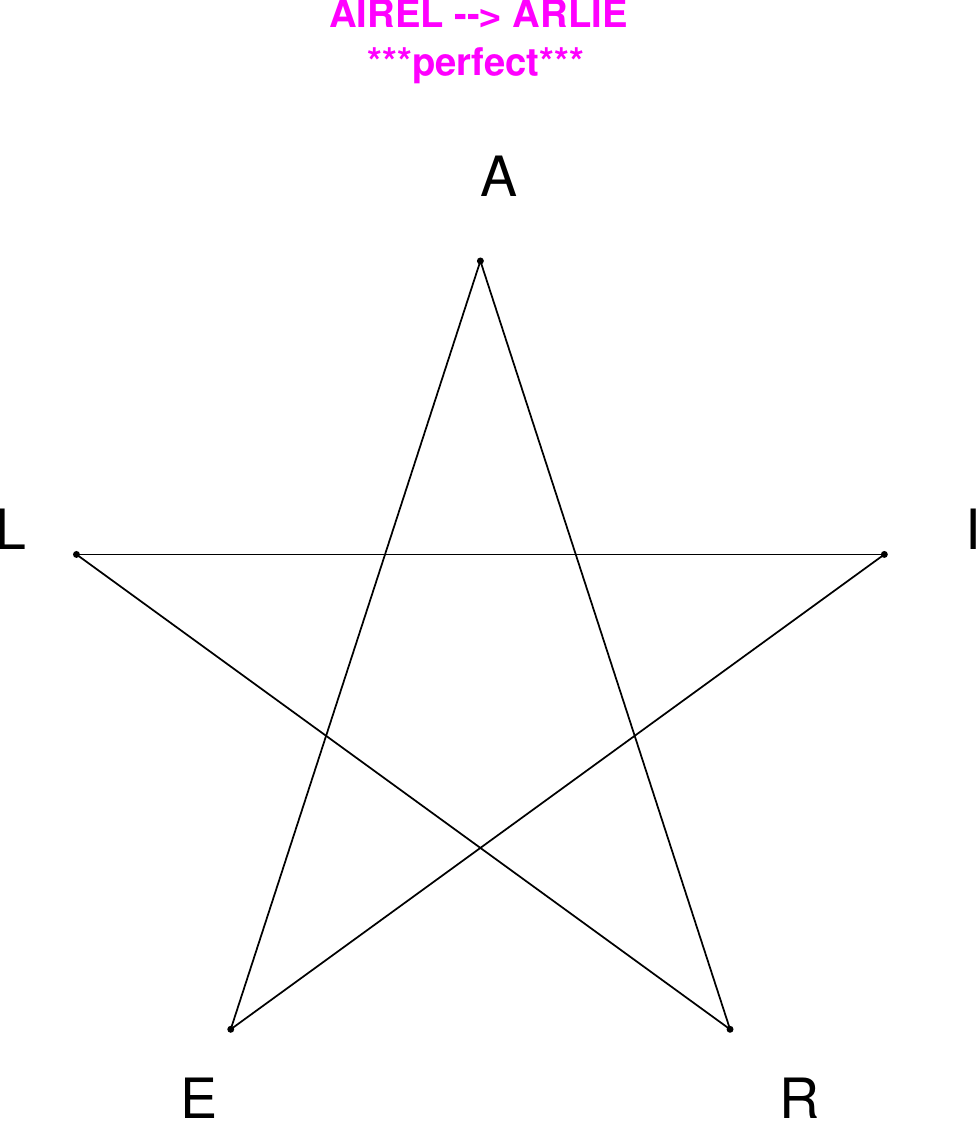}
\end{subfigure}
\end{figure}

\begin{figure}[H]
\centering
\begin{subfigure}[T]{0.19\textwidth}
\centering
\includegraphics[width=\textwidth]{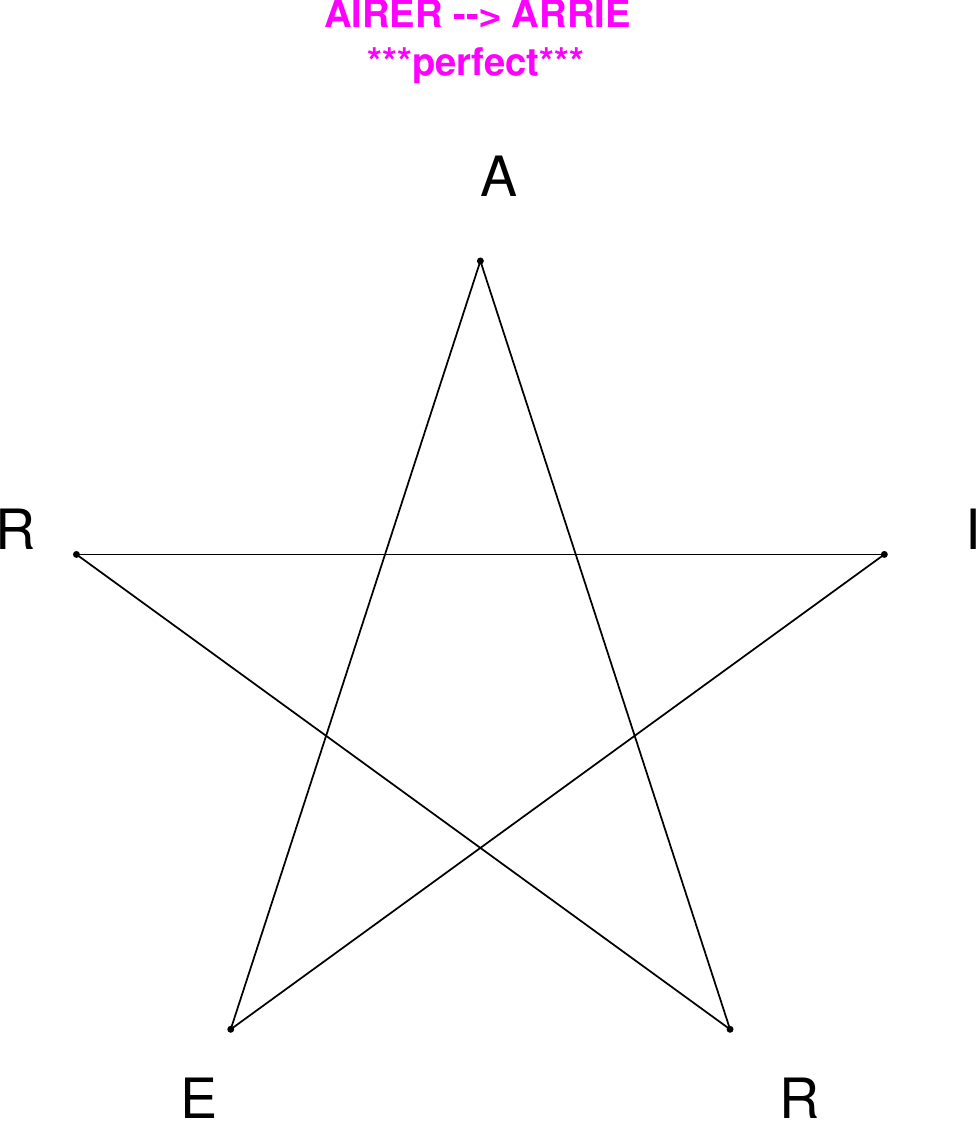}
\end{subfigure}
\hfill
\begin{subfigure}[T]{0.19\textwidth}
\centering
\includegraphics[width=\textwidth]{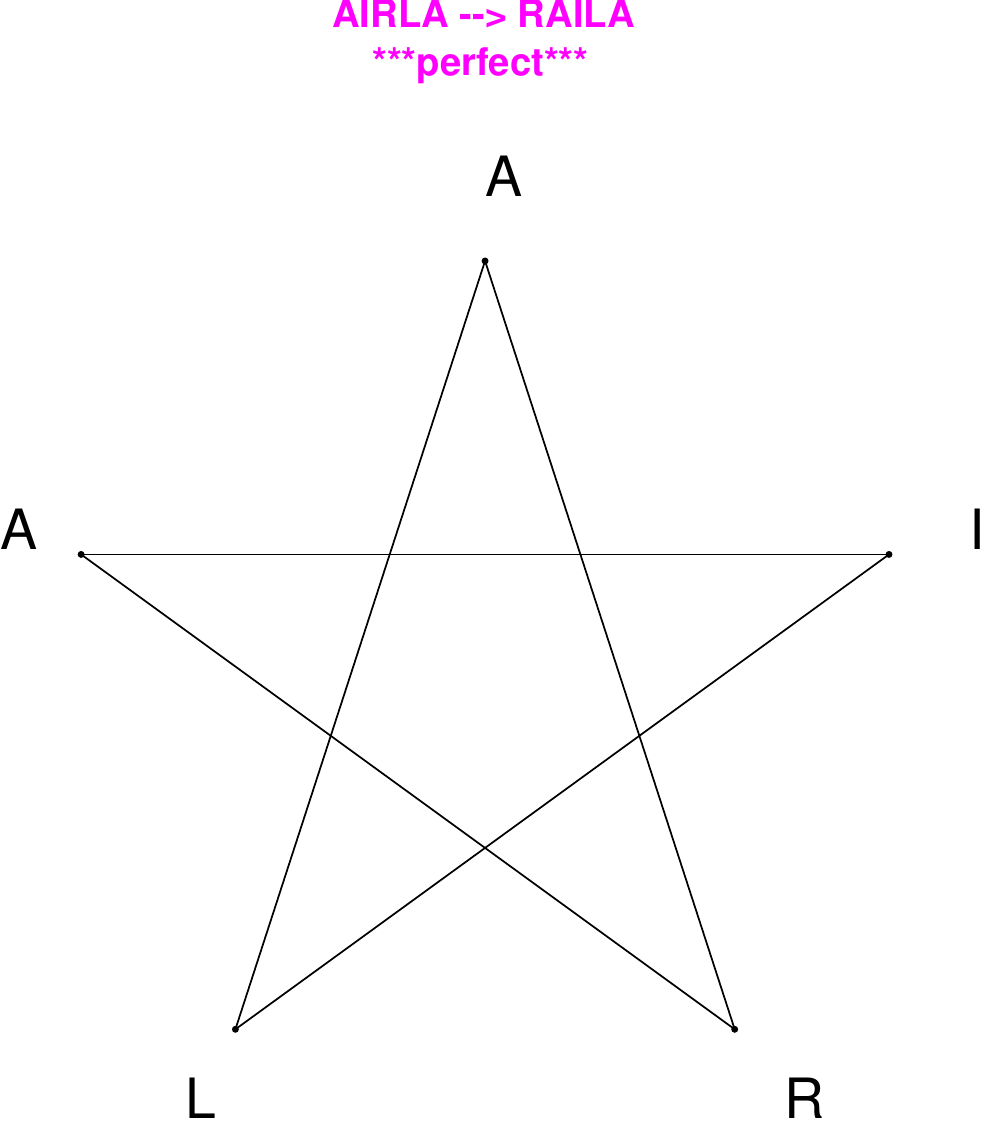}
\end{subfigure}
\hfill
\begin{subfigure}[T]{0.19\textwidth}
\centering
\includegraphics[width=\textwidth]{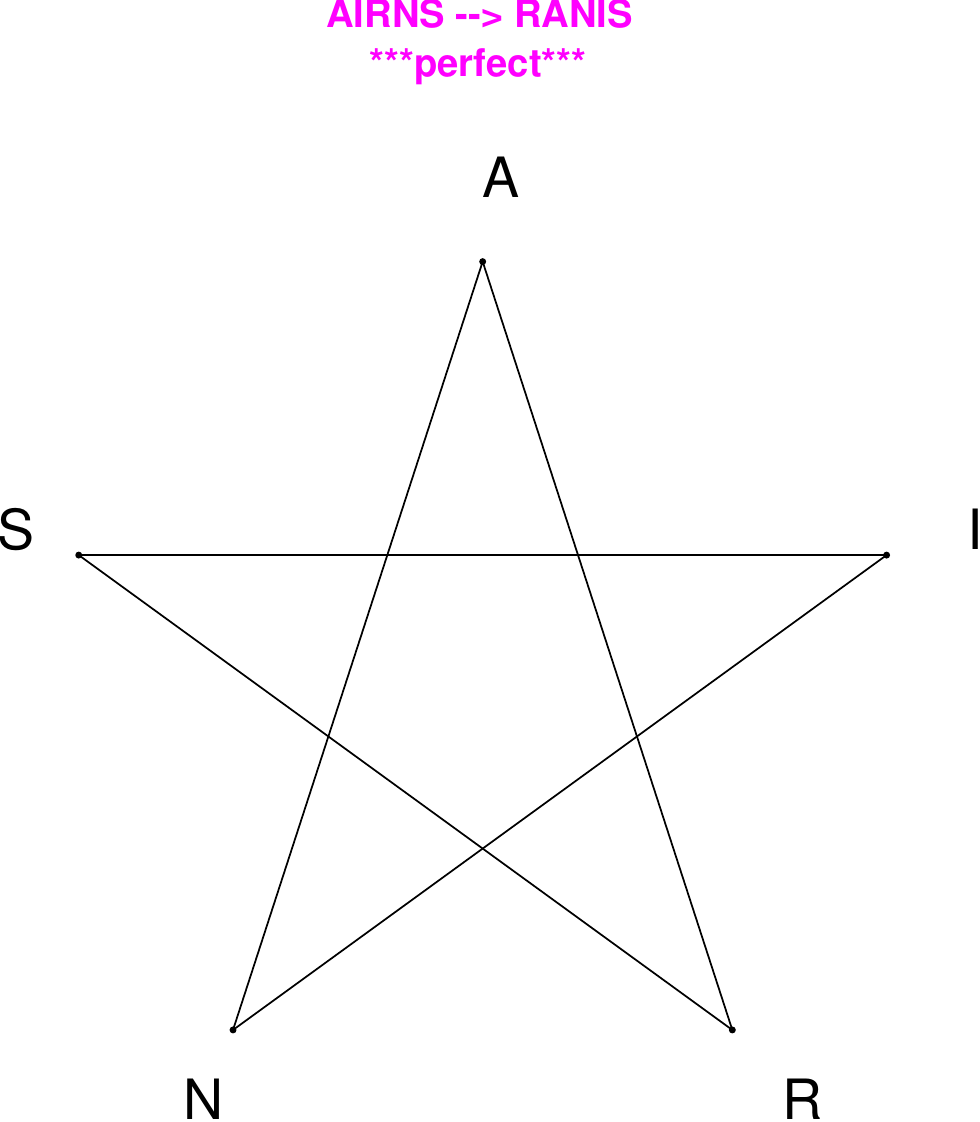}
\end{subfigure}
\hfill
\begin{subfigure}[T]{0.19\textwidth}
\centering
\includegraphics[width=\textwidth]{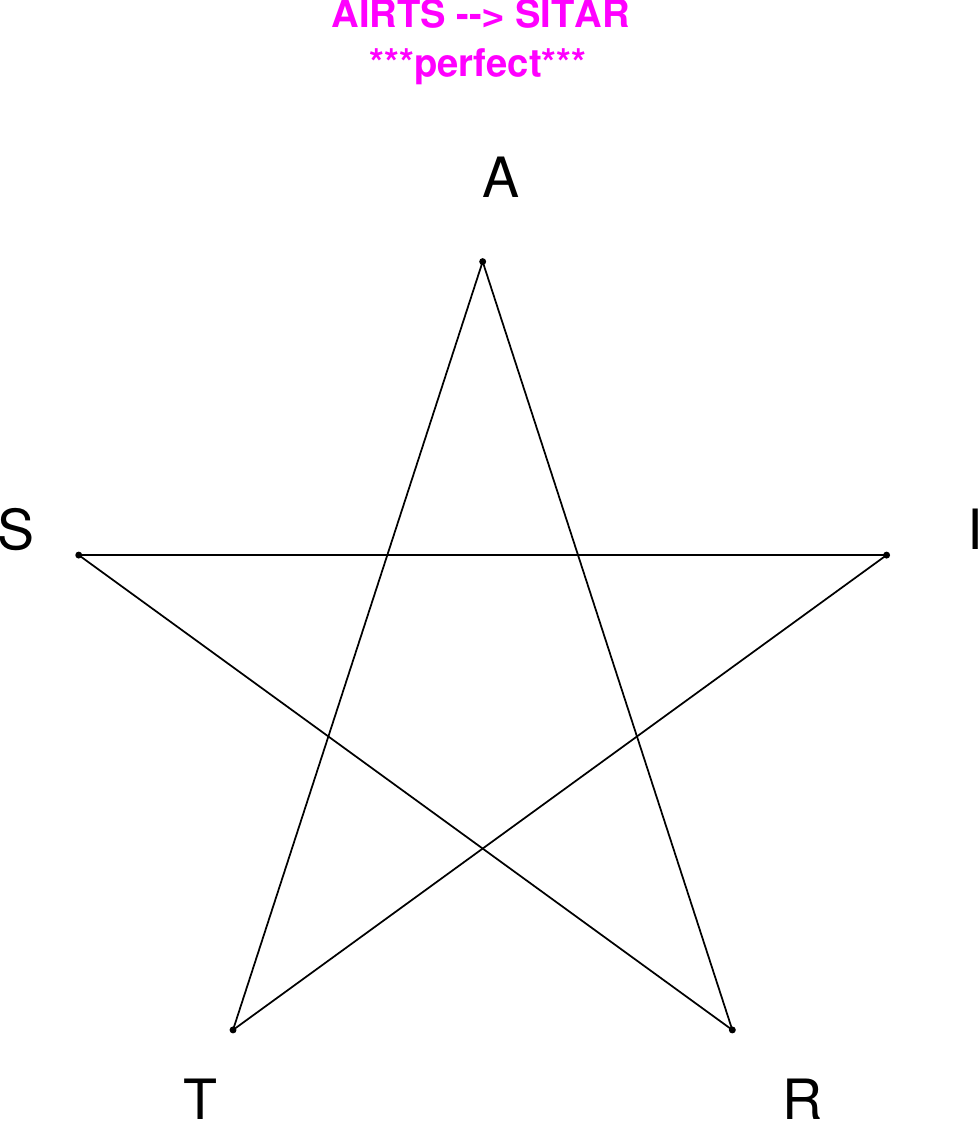}
\end{subfigure}
\hfill
\begin{subfigure}[T]{0.19\textwidth}
\centering
\includegraphics[width=\textwidth]{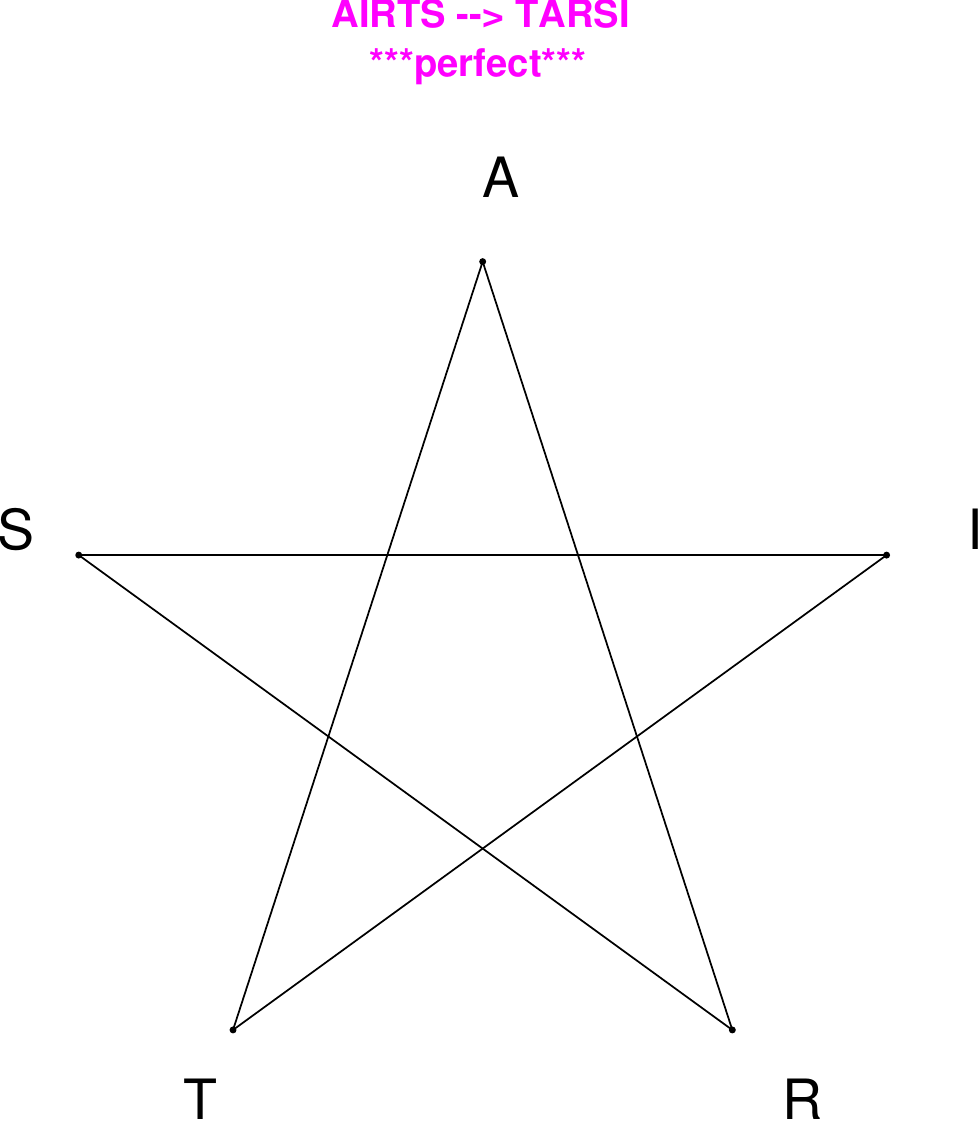}
\end{subfigure}
\end{figure}

\begin{figure}[H]
\centering
\begin{subfigure}[T]{0.19\textwidth}
\centering
\includegraphics[width=\textwidth]{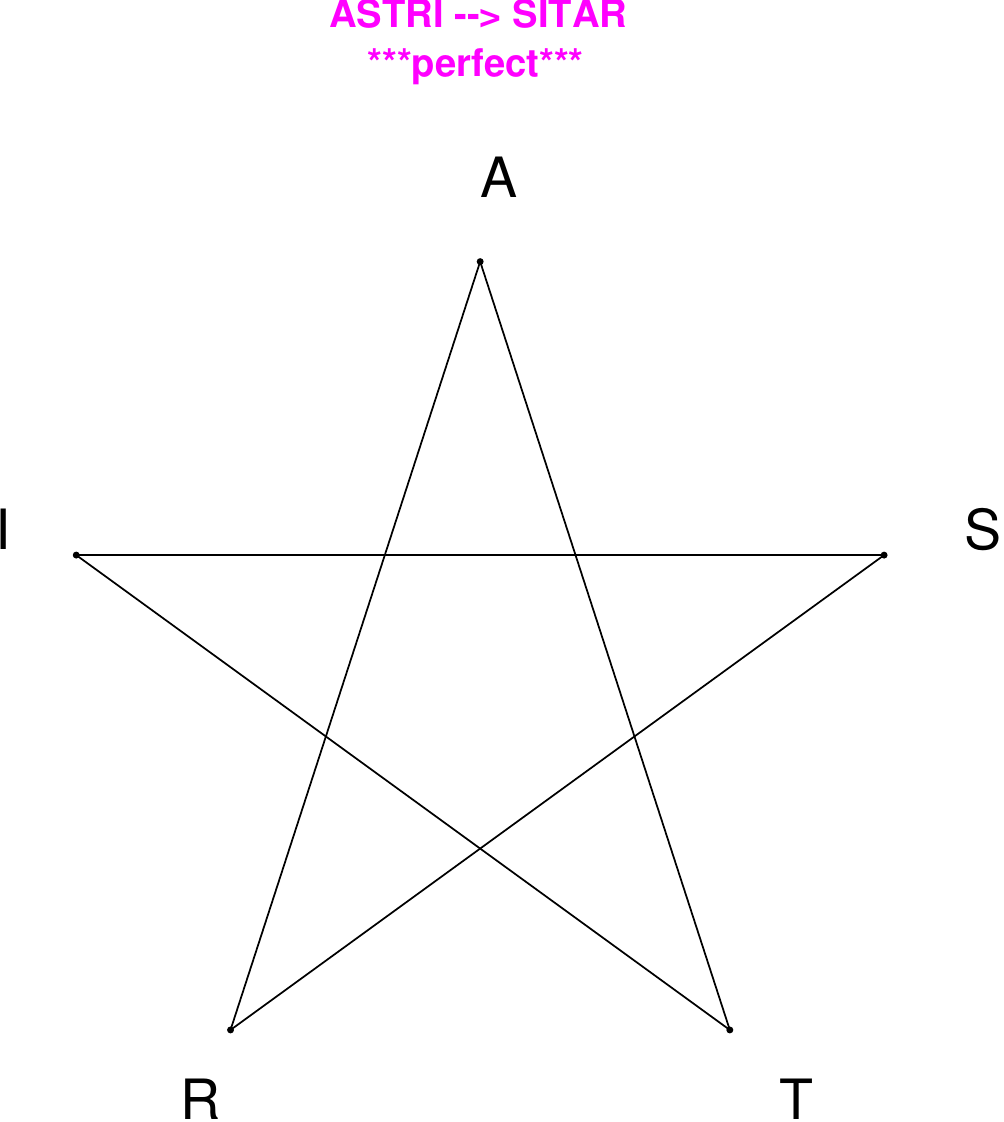}
\end{subfigure}
\hfill
\begin{subfigure}[T]{0.19\textwidth}
\centering
\includegraphics[width=\textwidth]{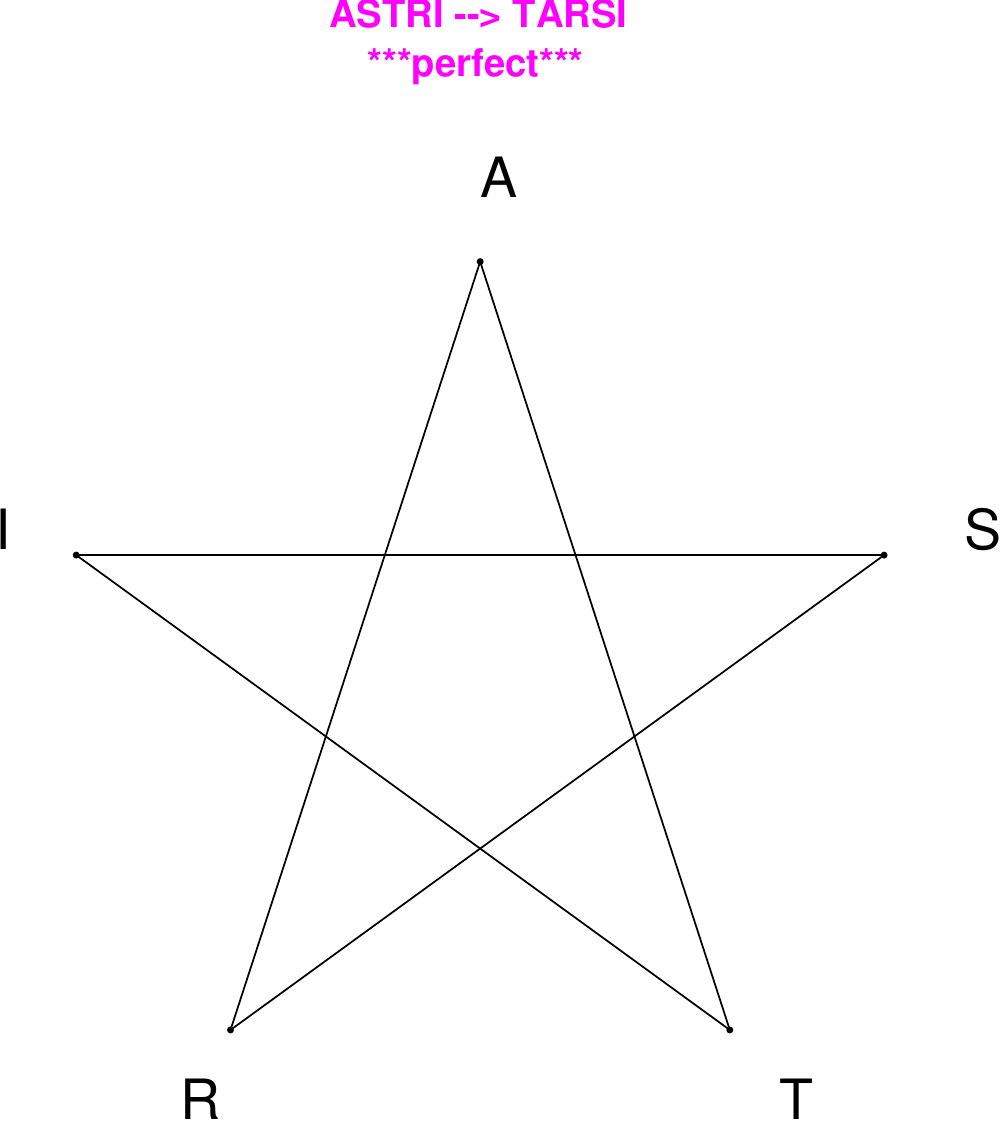}
\end{subfigure}
\hfill
\begin{subfigure}[T]{0.19\textwidth}
\centering
\includegraphics[width=\textwidth]{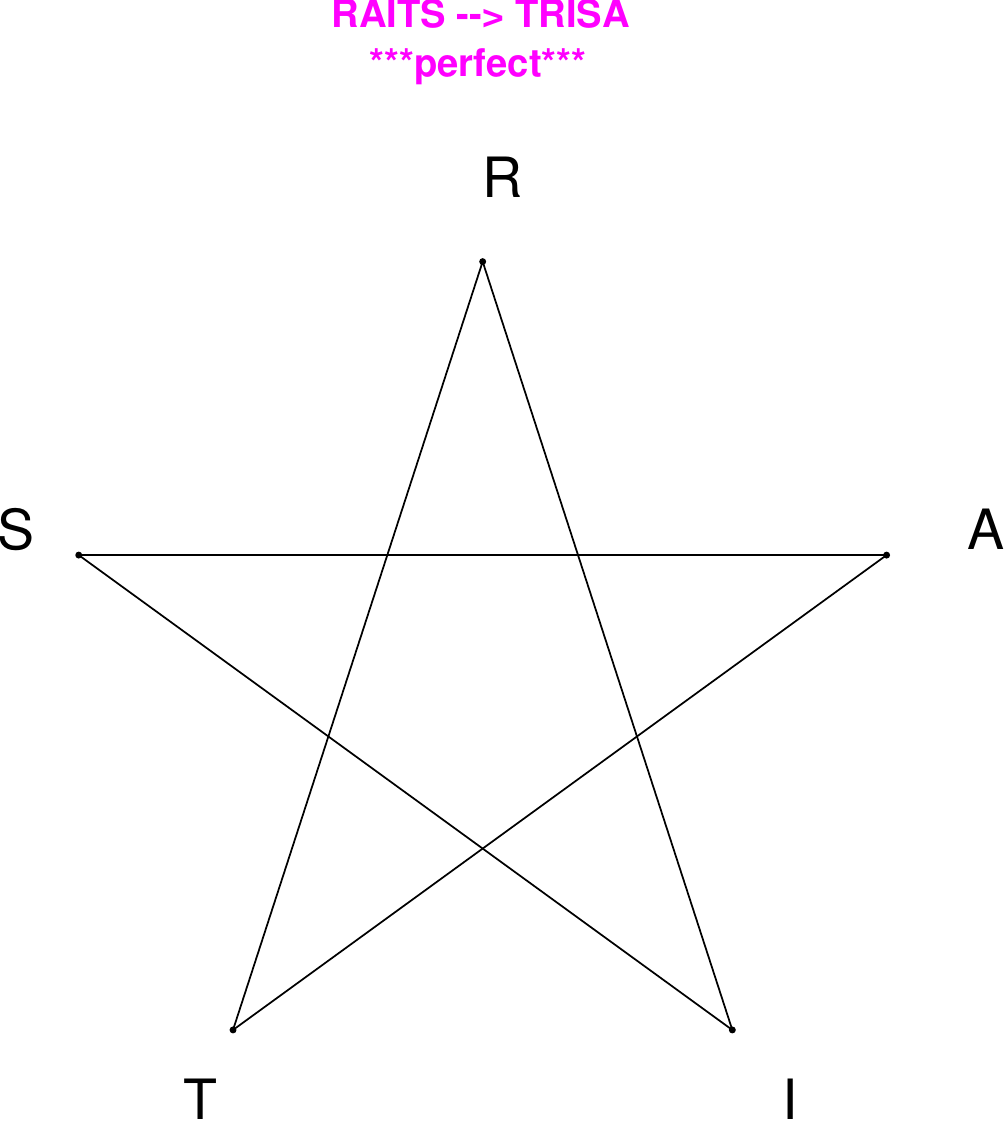}
\end{subfigure}
\hfill
\begin{subfigure}[T]{0.19\textwidth}
\centering
\includegraphics[width=\textwidth]{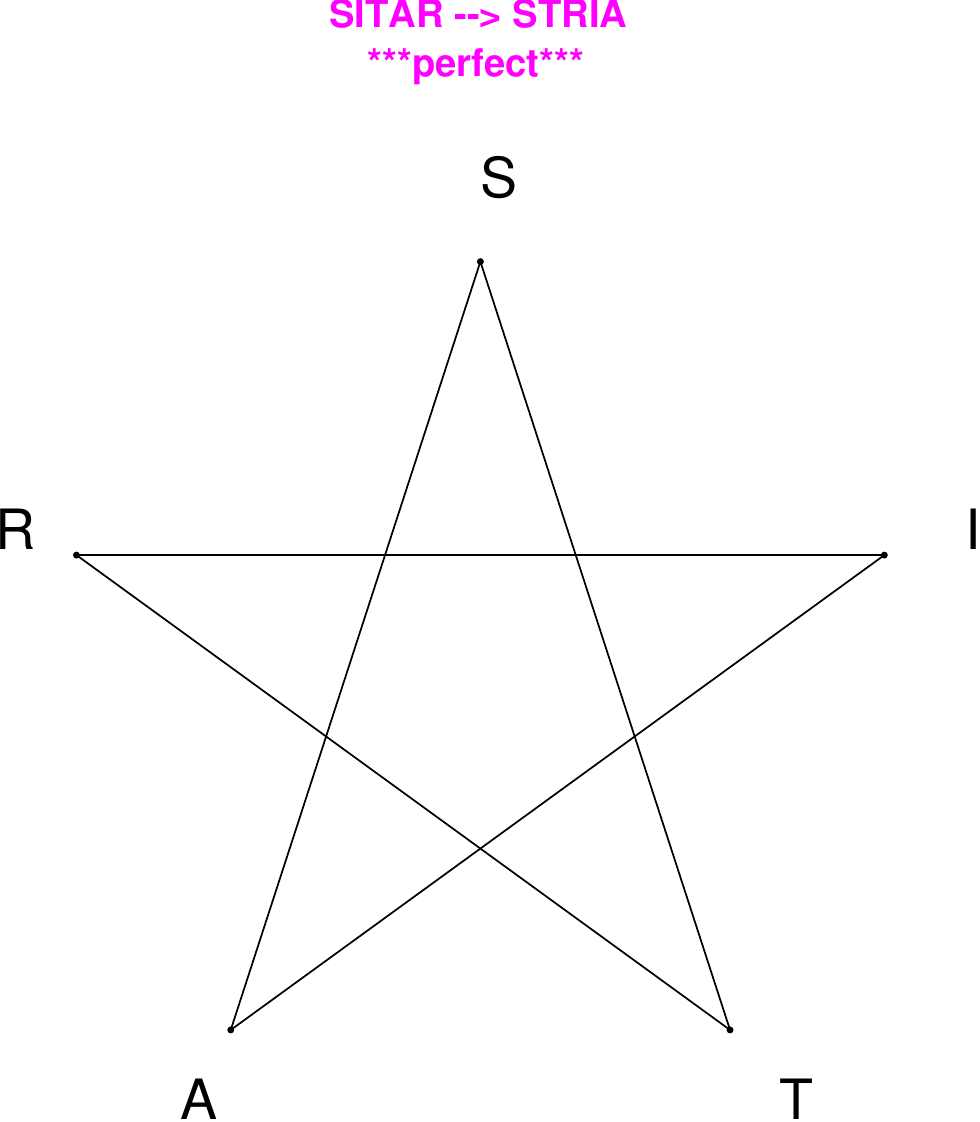}
\end{subfigure}
\hfill
\begin{subfigure}[T]{0.19\textwidth}
\centering
\includegraphics[width=\textwidth]{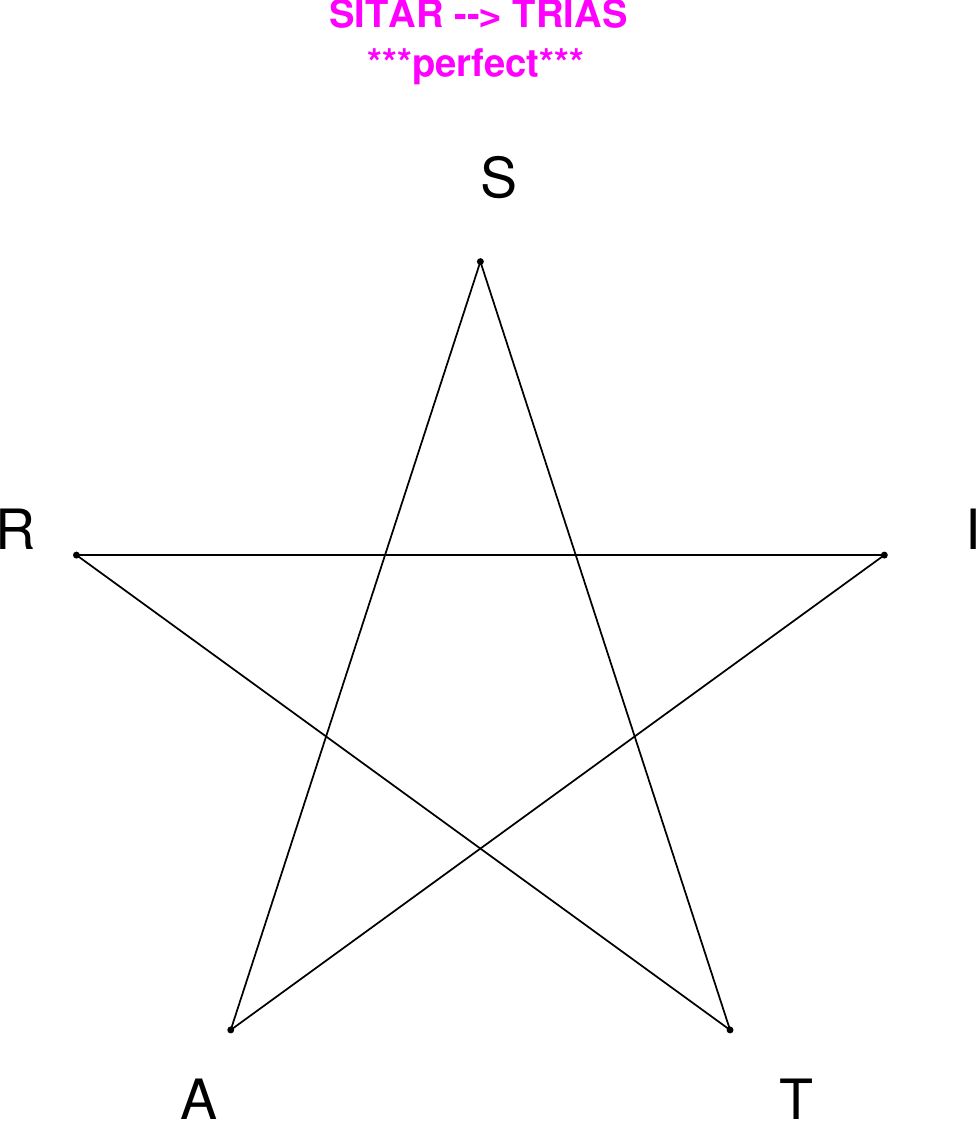}
\end{subfigure}
\end{figure}

\begin{figure}[H]
\centering
\begin{subfigure}[T]{0.19\textwidth}
\centering
\includegraphics[width=\textwidth]{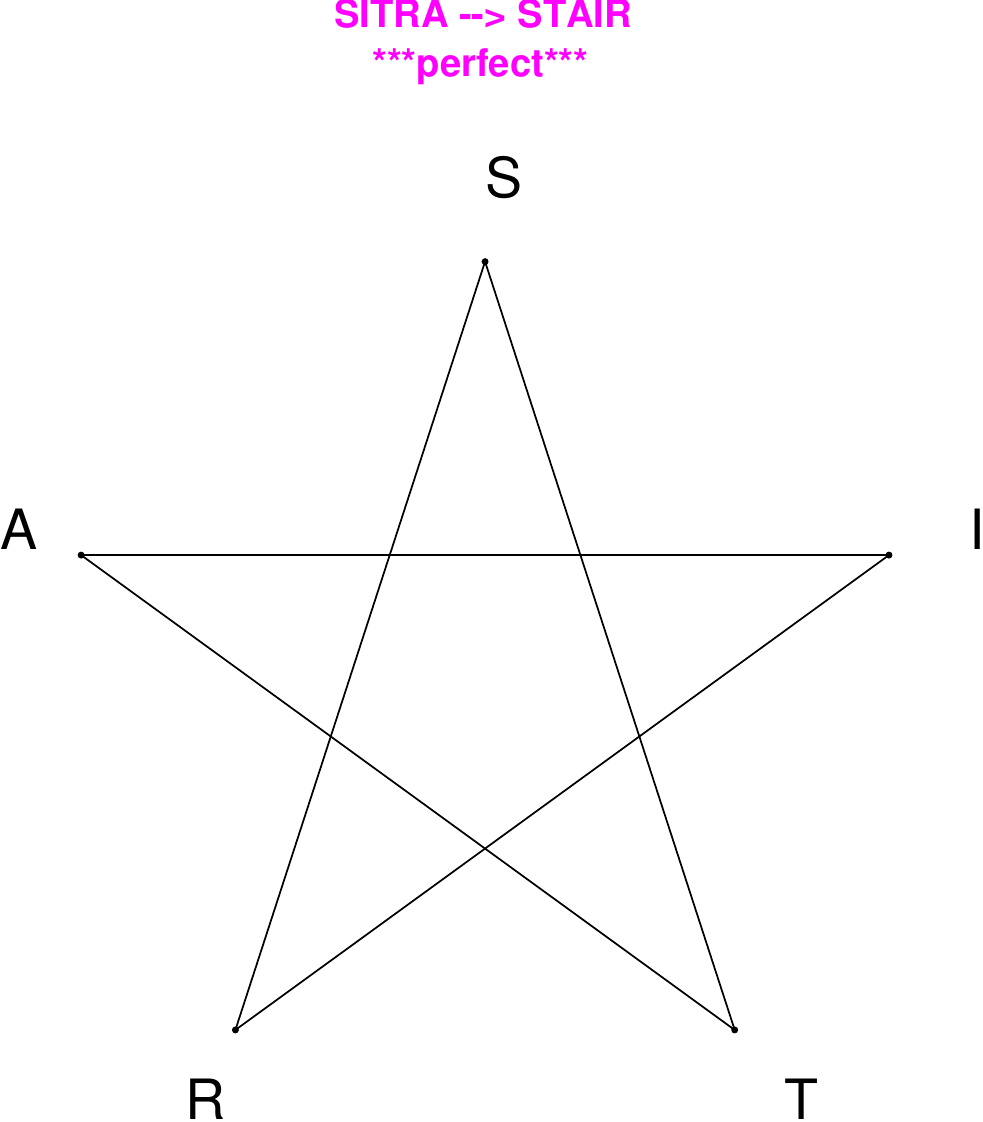}
\end{subfigure}
\hfill
\begin{subfigure}[T]{0.19\textwidth}
\centering
\includegraphics[width=\textwidth]{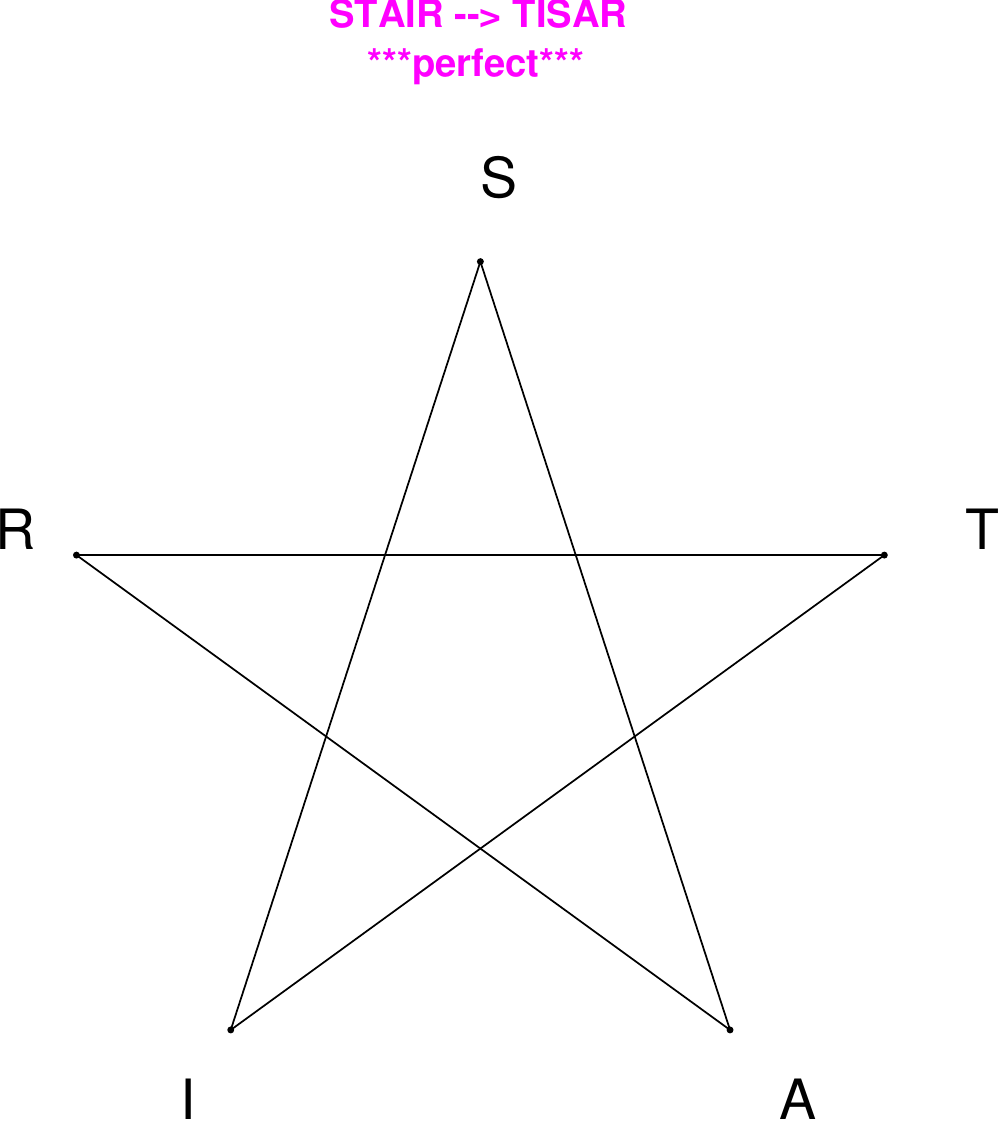}
\end{subfigure}
\hfill
\begin{subfigure}[T]{0.19\textwidth}
\centering
\includegraphics[width=\textwidth]{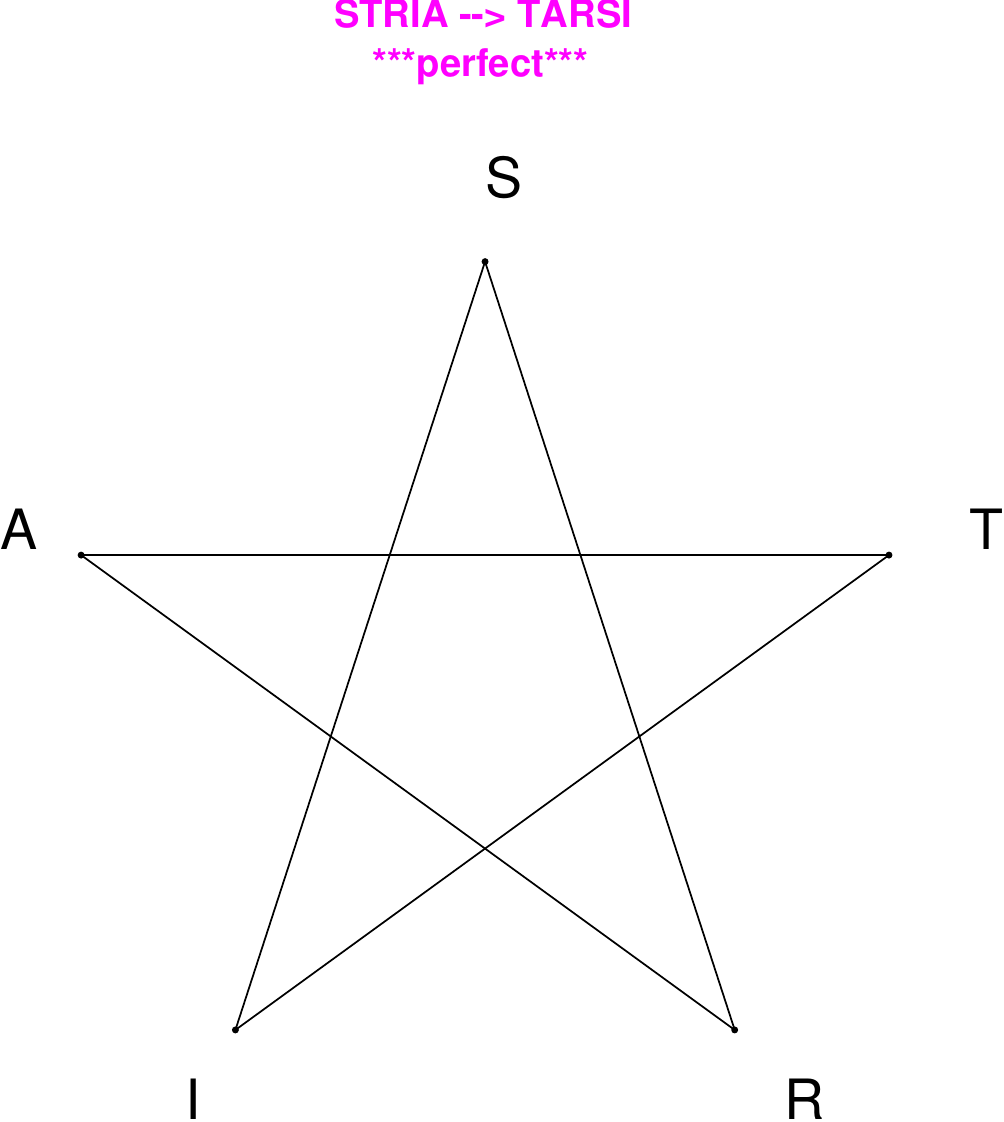}
\end{subfigure}
\hfill
\begin{subfigure}[T]{0.19\textwidth}
\centering
\includegraphics[width=\textwidth]{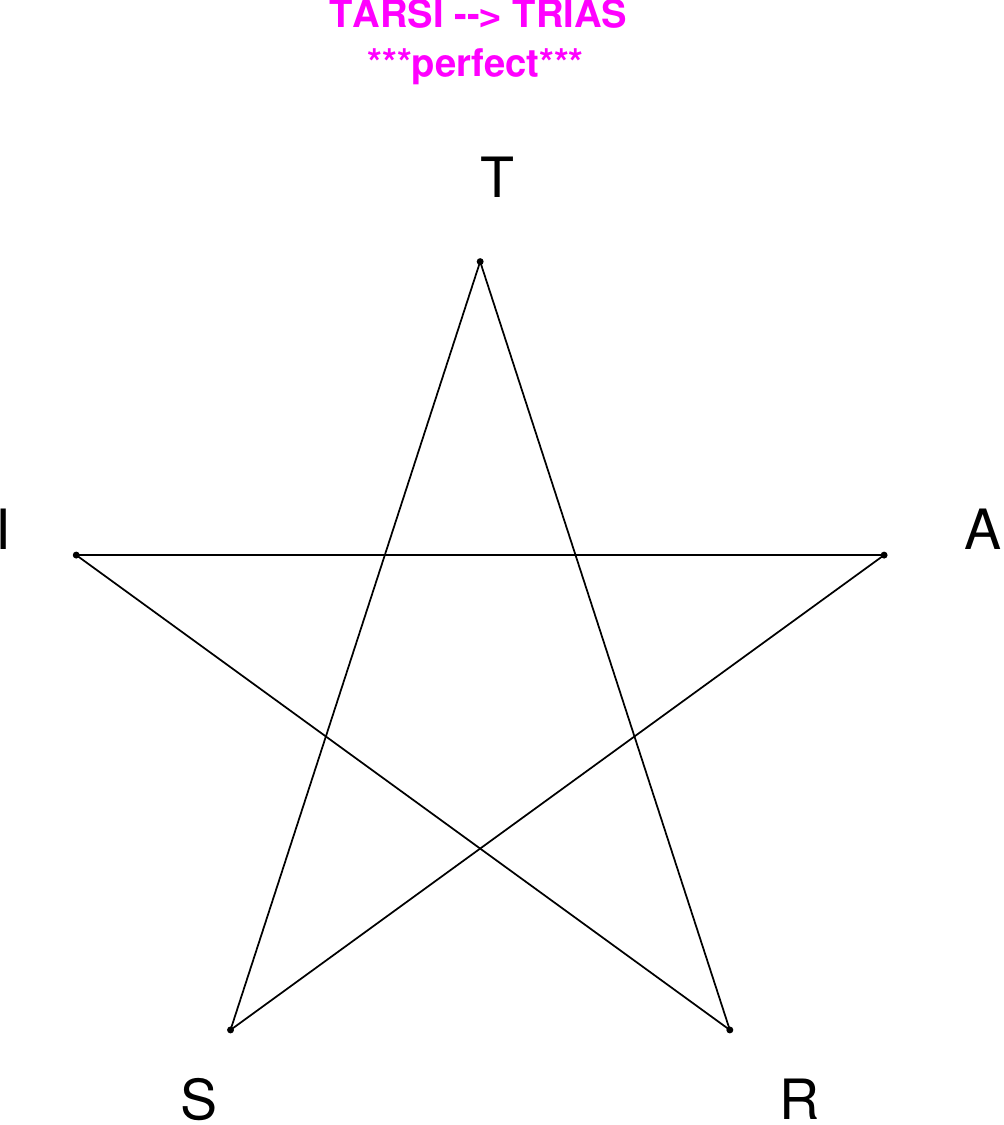}
\end{subfigure}
\hfill
\begin{subfigure}[T]{0.19\textwidth}
\centering
\includegraphics[width=\textwidth]{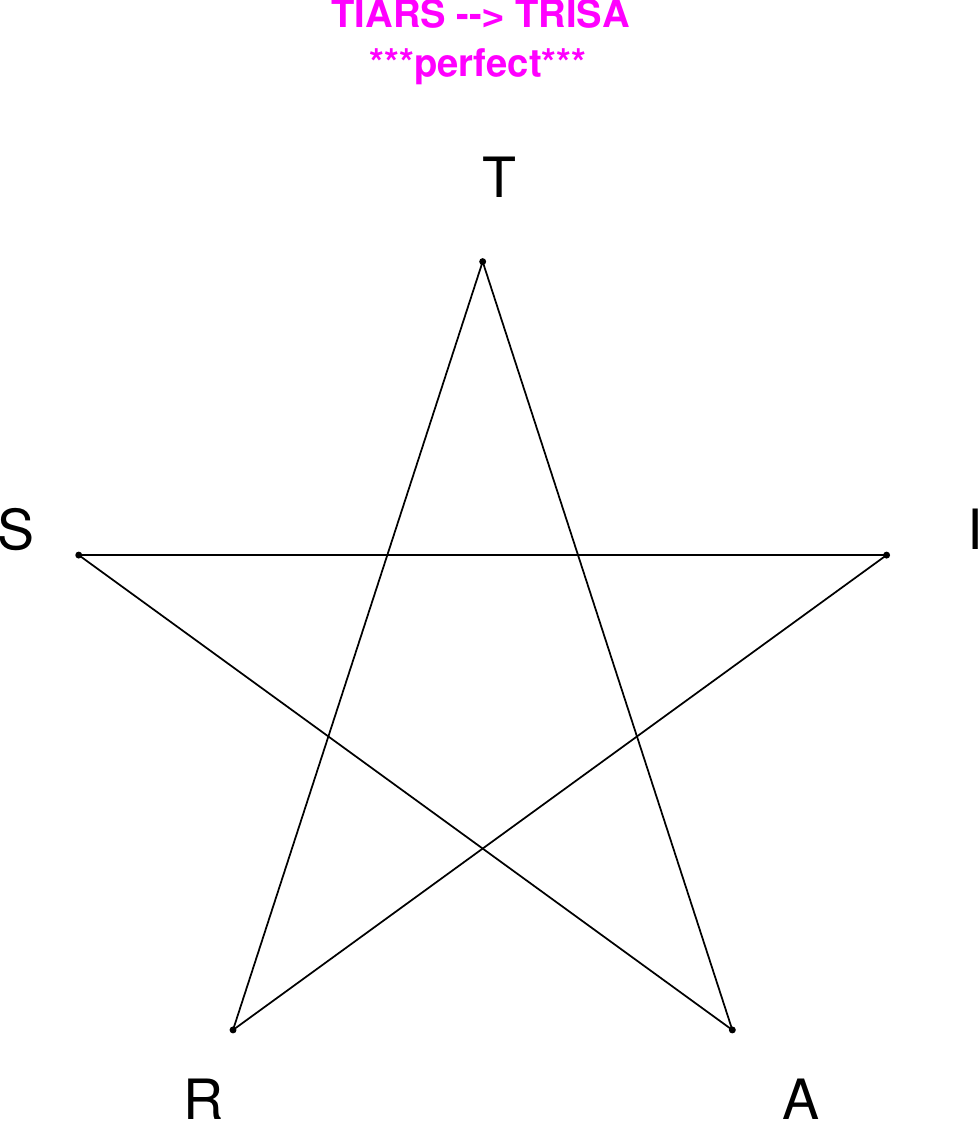}
\end{subfigure}
\end{figure}

\begin{figure}[H]
\centering
\begin{subfigure}[T]{0.19\textwidth}
\centering
\includegraphics[width=\textwidth]{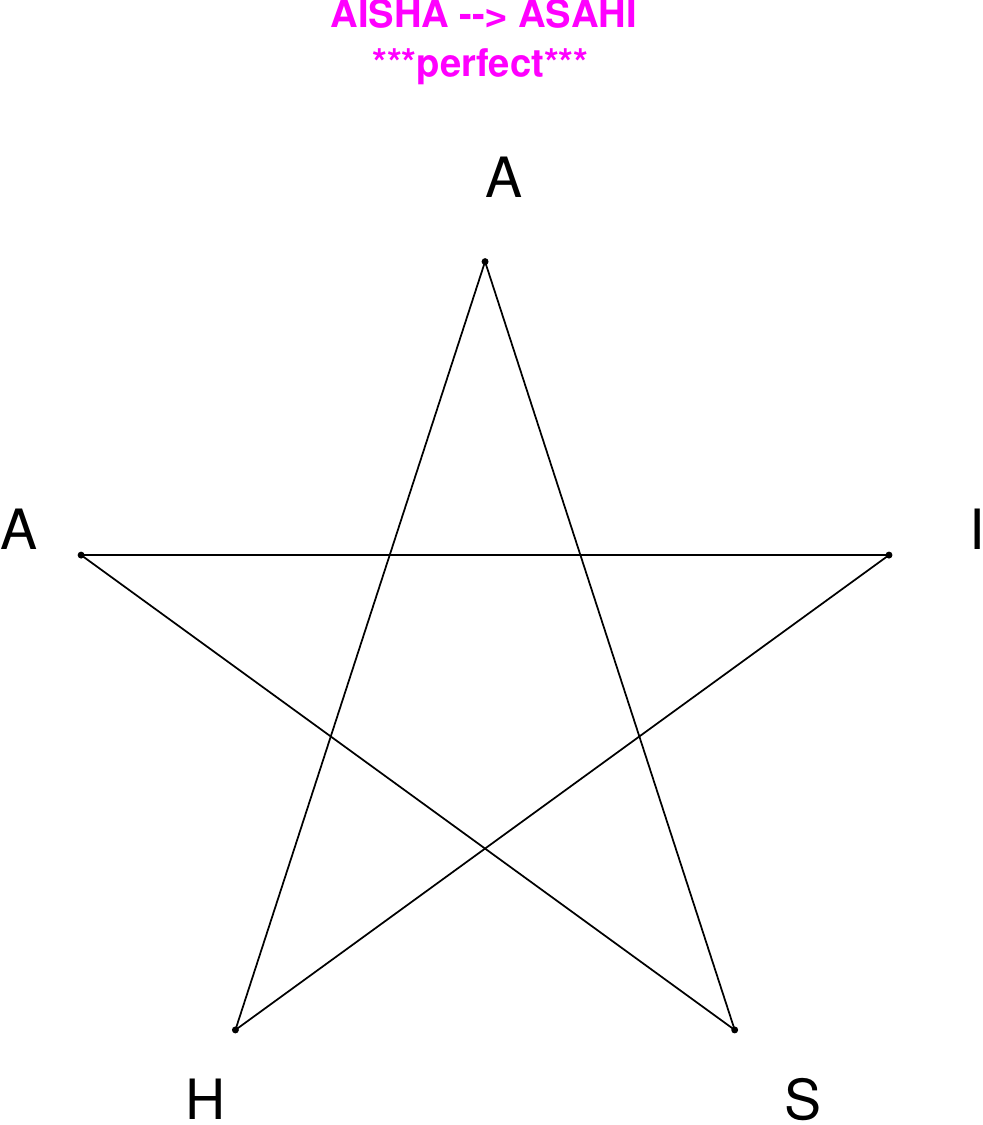}
\end{subfigure}
\hfill
\begin{subfigure}[T]{0.19\textwidth}
\centering
\includegraphics[width=\textwidth]{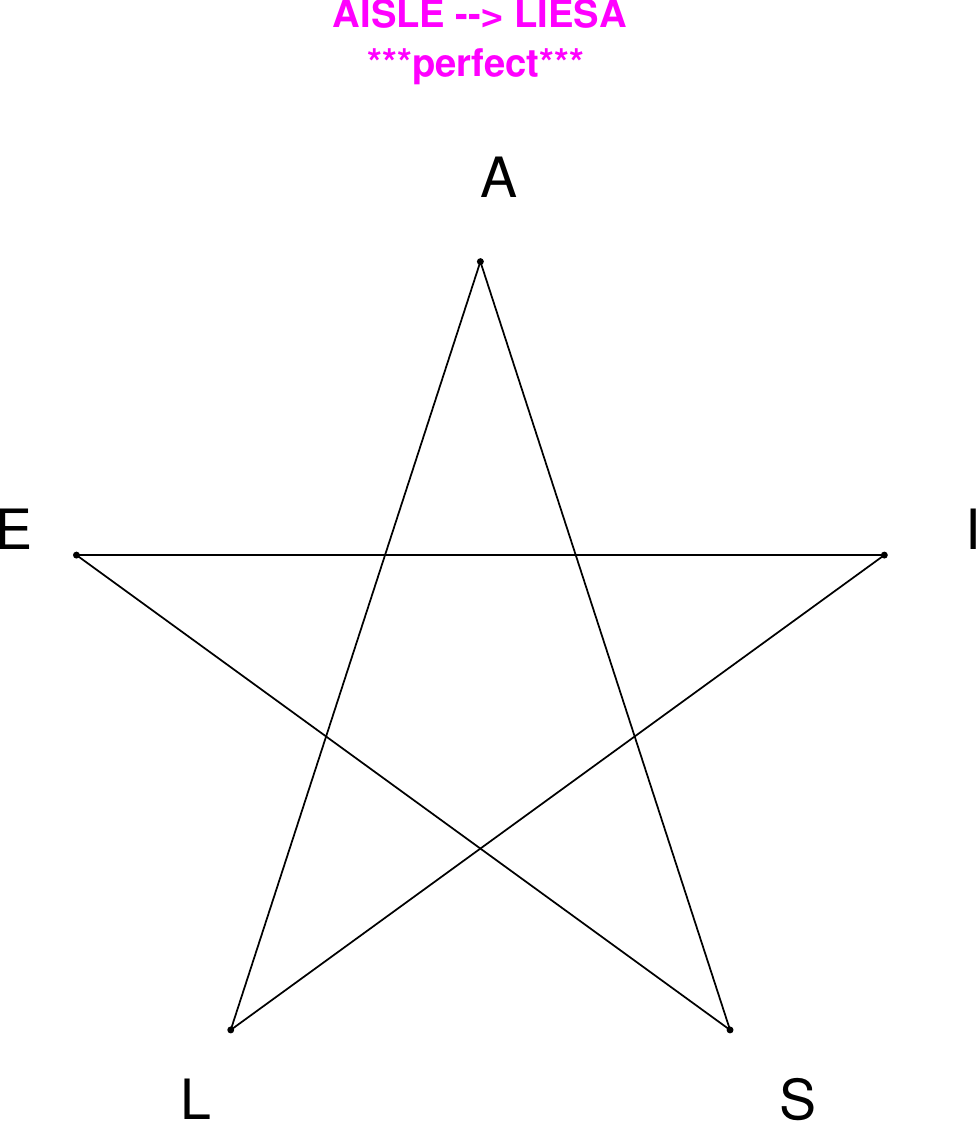}
\end{subfigure}
\hfill
\begin{subfigure}[T]{0.19\textwidth}
\centering
\includegraphics[width=\textwidth]{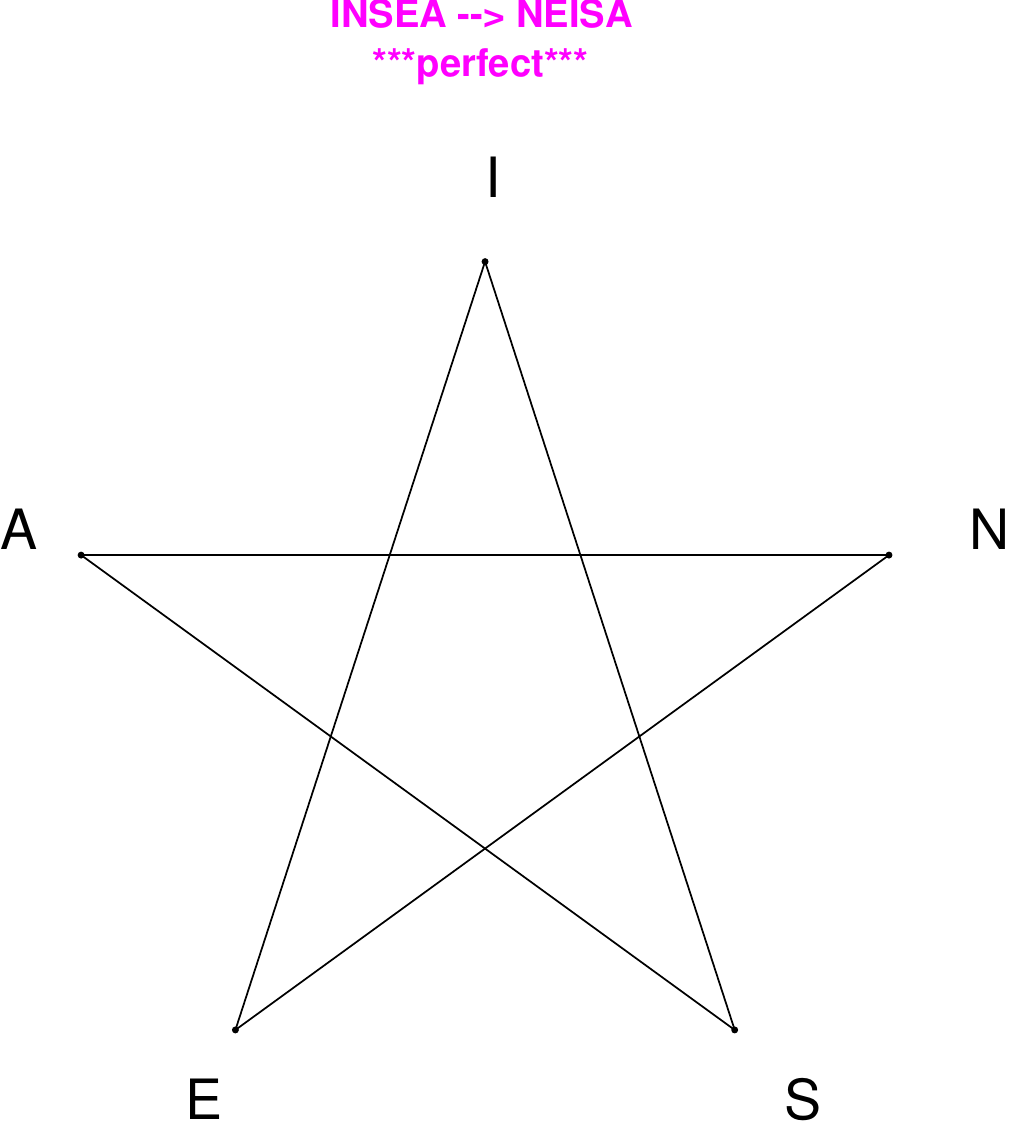}
\end{subfigure}
\hfill
\begin{subfigure}[T]{0.19\textwidth}
\centering
\includegraphics[width=\textwidth]{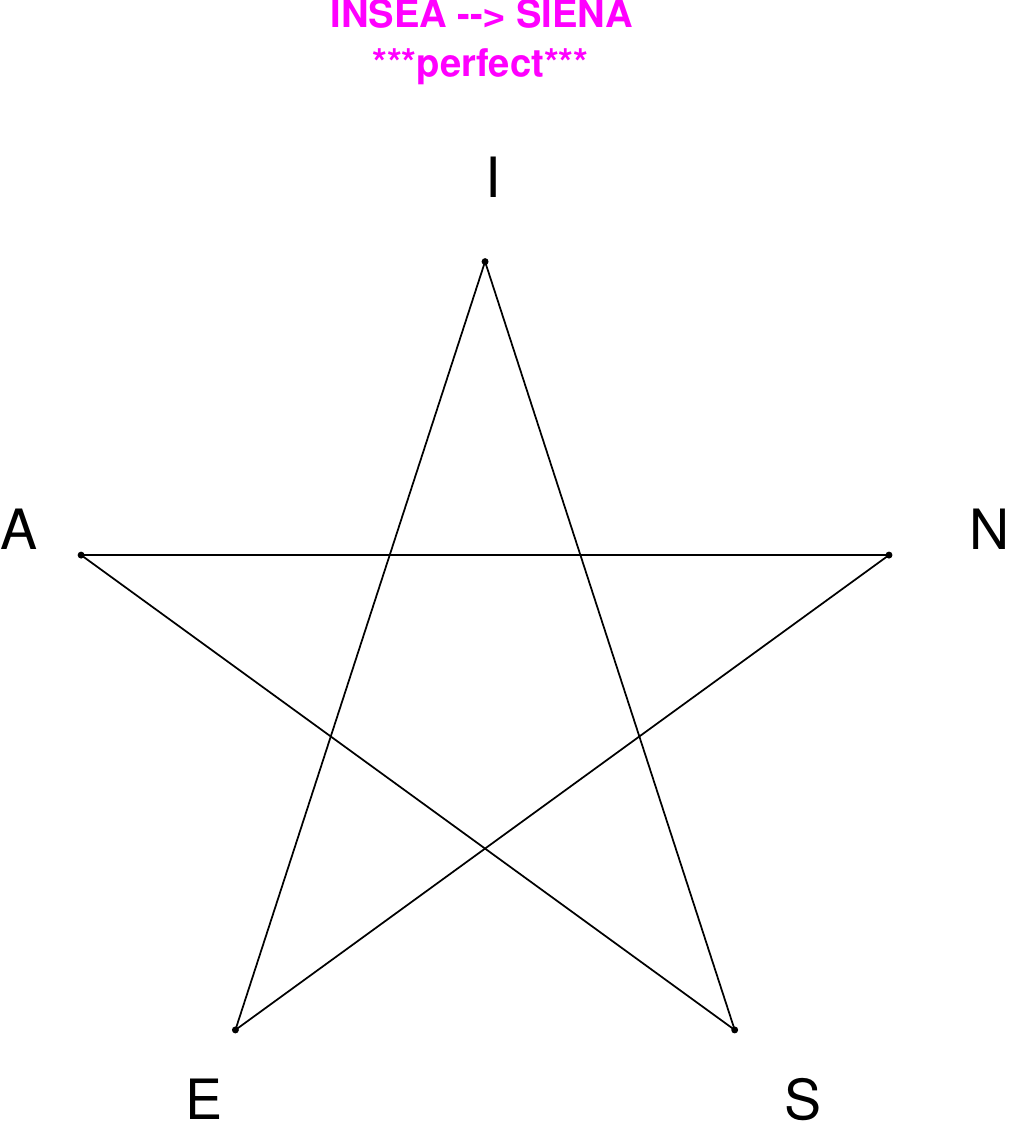}
\end{subfigure}
\hfill
\begin{subfigure}[T]{0.19\textwidth}
\centering
\includegraphics[width=\textwidth]{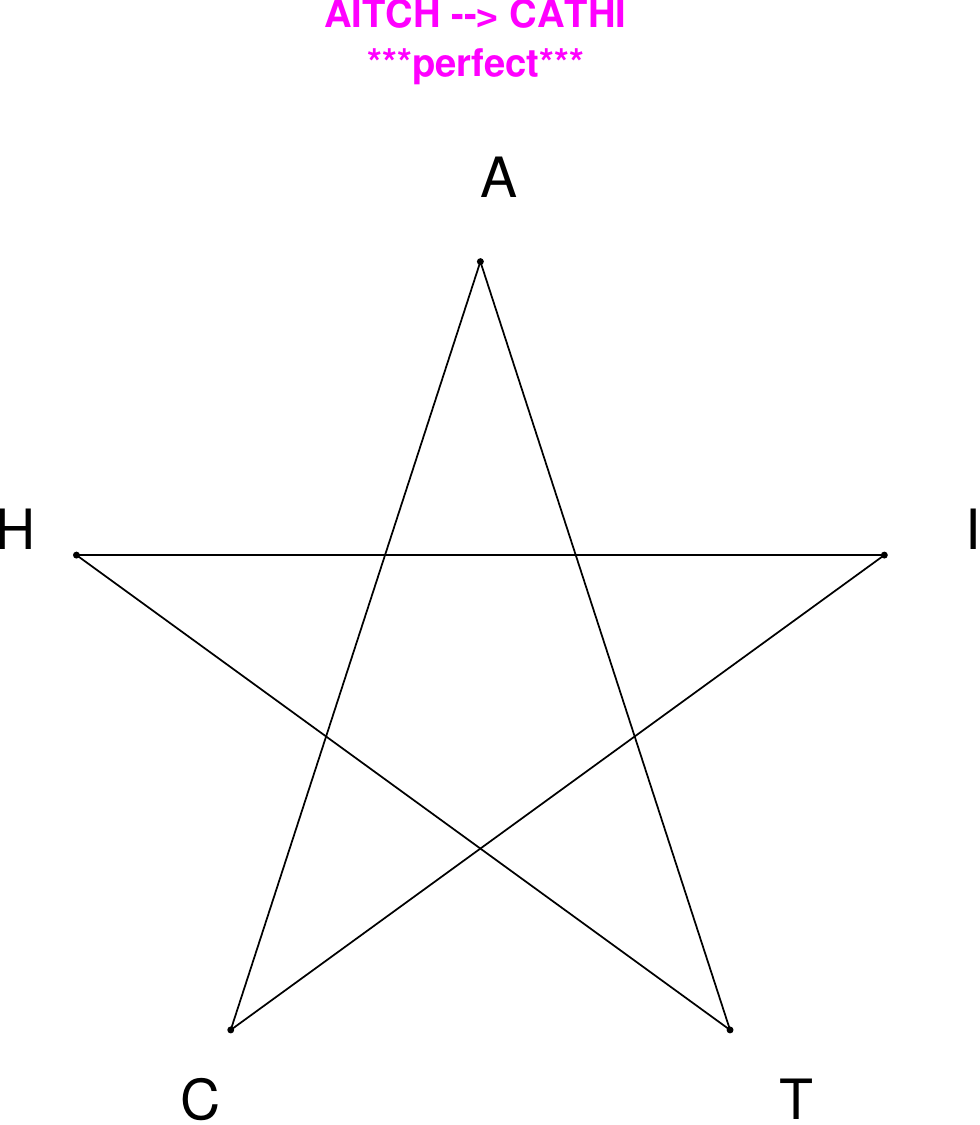}
\end{subfigure}
\end{figure}

\begin{figure}[H]
\centering
\begin{subfigure}[T]{0.19\textwidth}
\centering
\includegraphics[width=\textwidth]{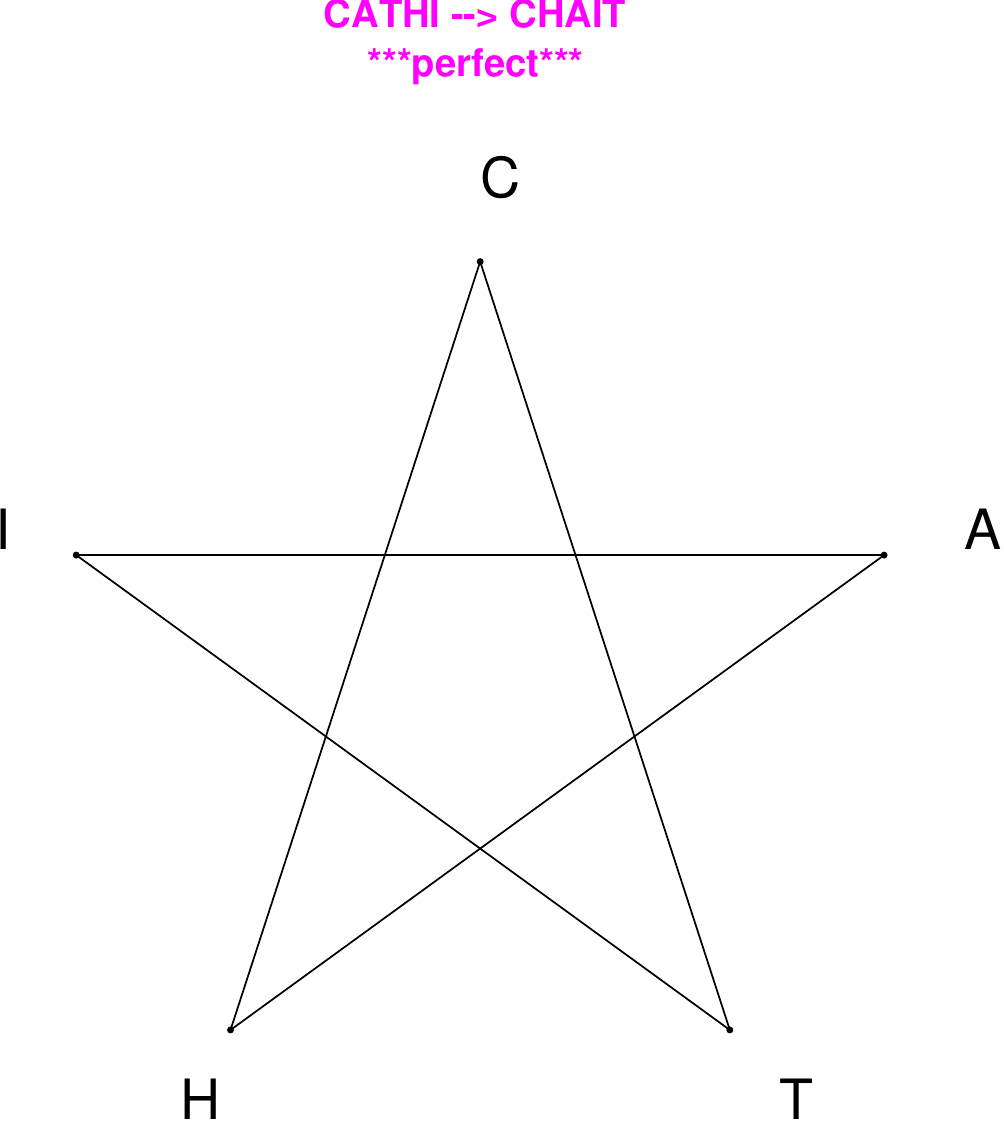}
\end{subfigure}
\hfill
\begin{subfigure}[T]{0.19\textwidth}
\centering
\includegraphics[width=\textwidth]{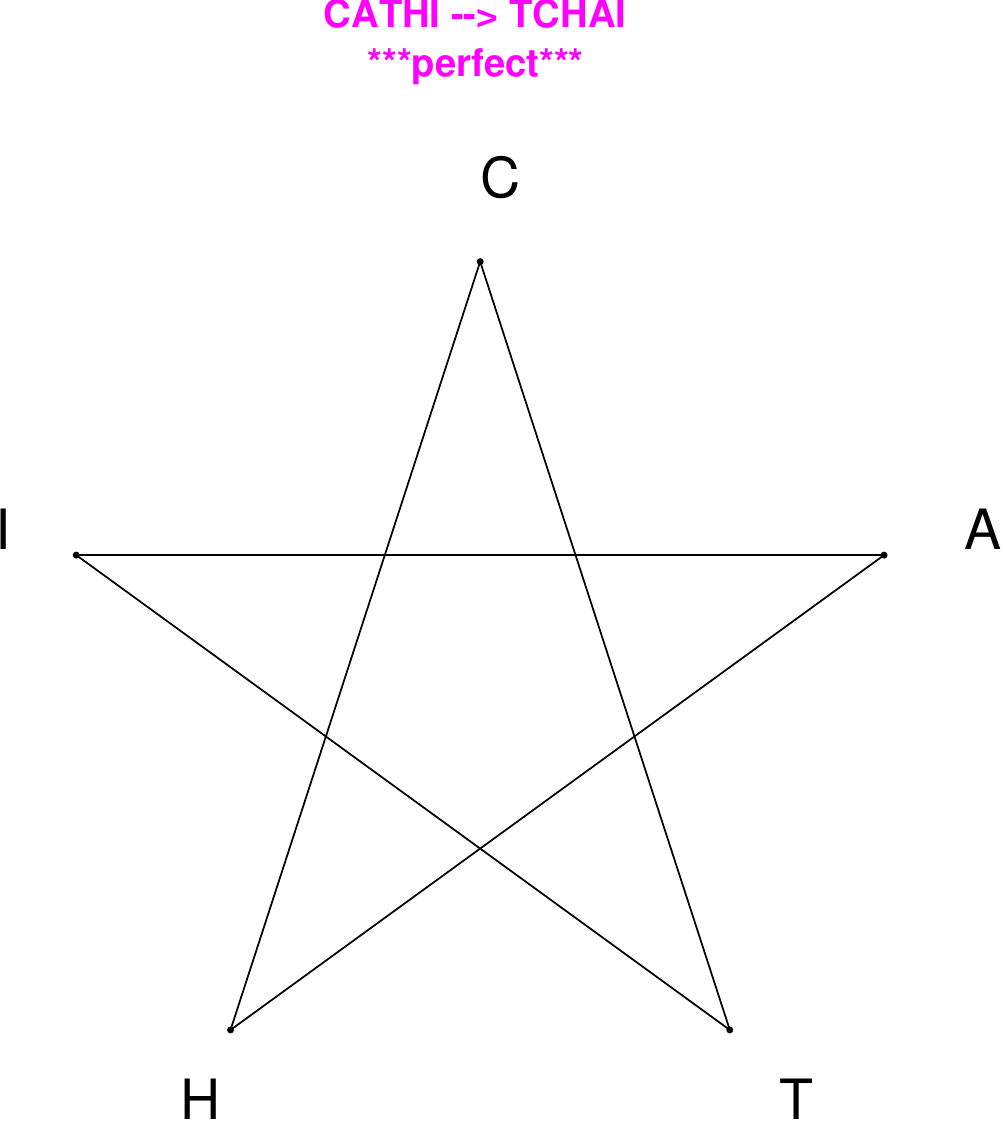}
\end{subfigure}
\hfill
\begin{subfigure}[T]{0.19\textwidth}
\centering
\includegraphics[width=\textwidth]{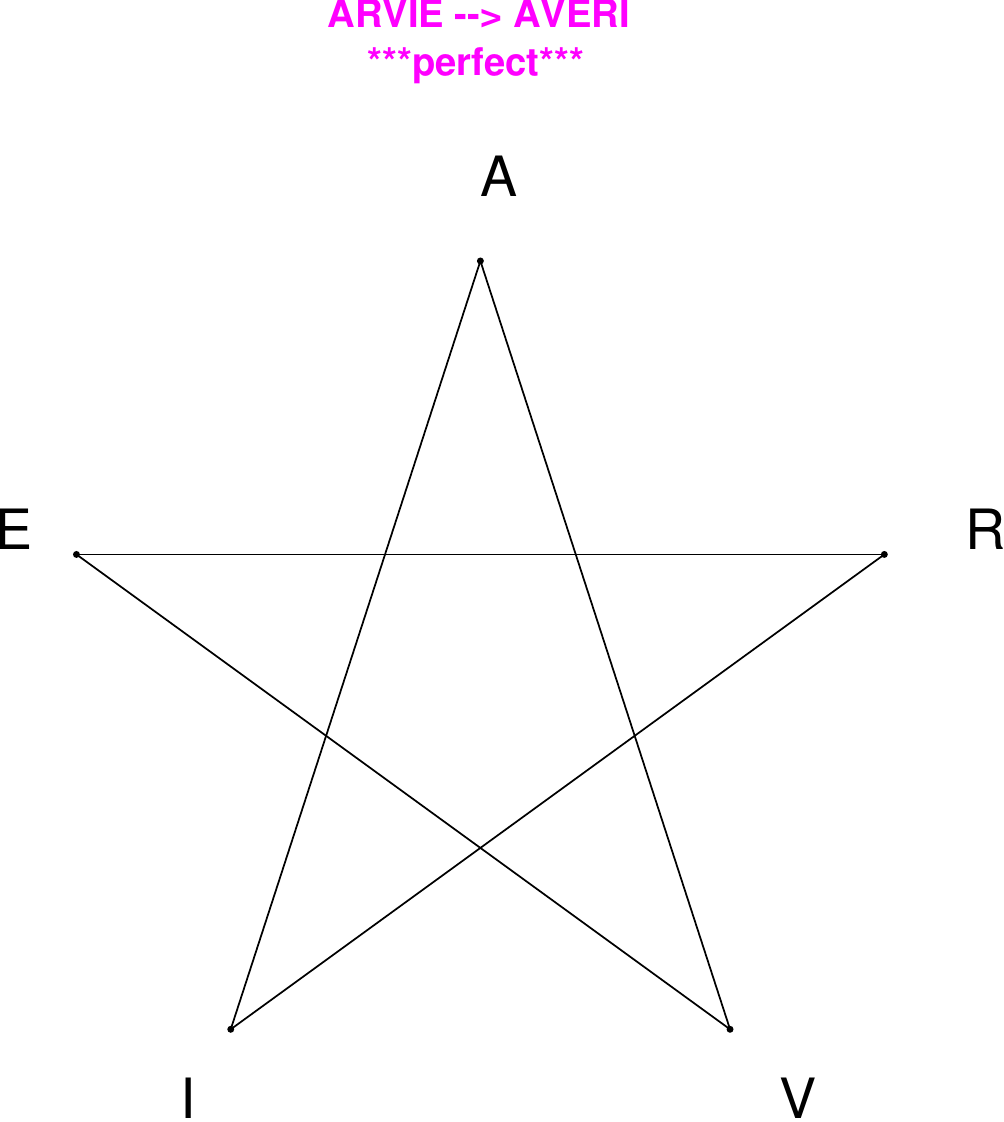}
\end{subfigure}
\hfill
\begin{subfigure}[T]{0.19\textwidth}
\centering
\includegraphics[width=\textwidth]{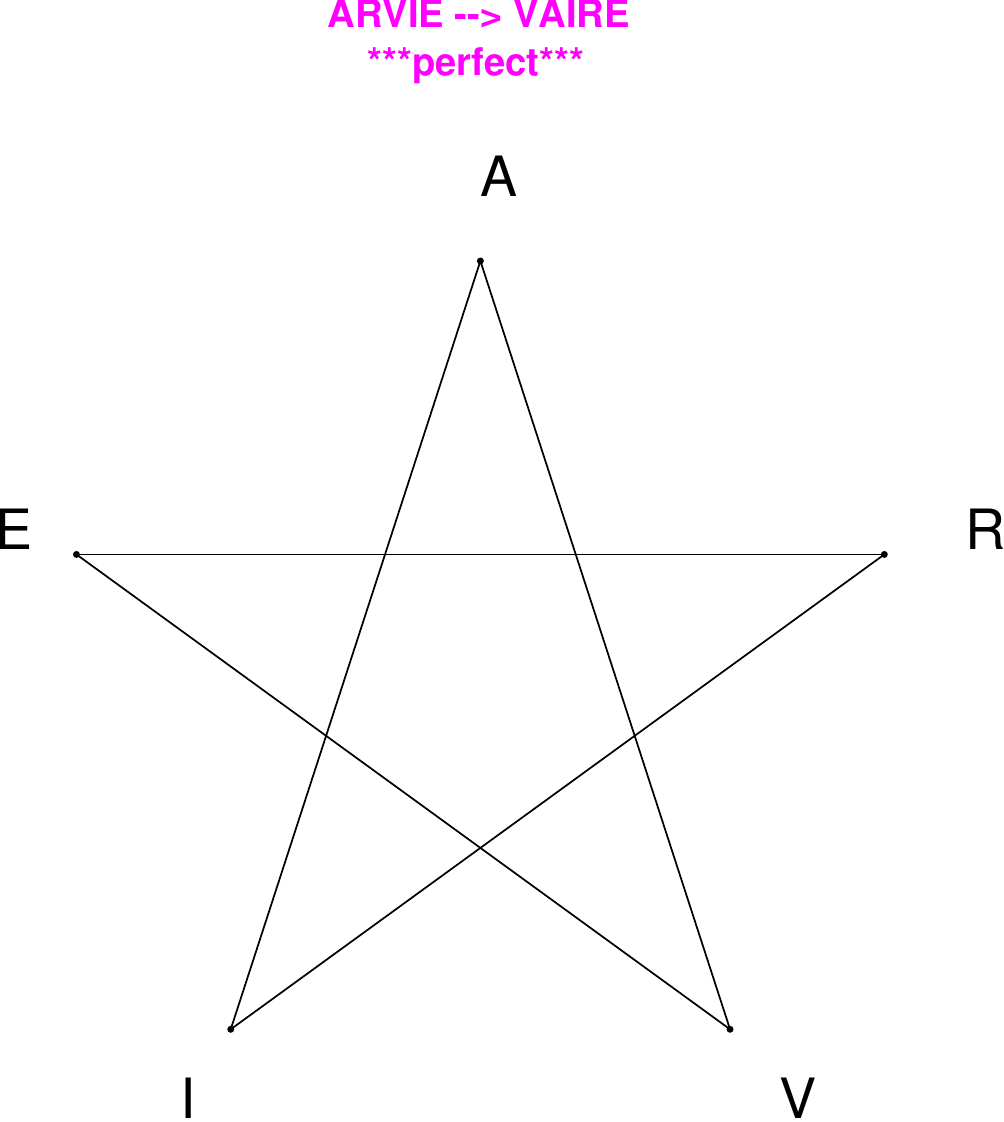}
\end{subfigure}
\hfill
\begin{subfigure}[T]{0.19\textwidth}
\centering
\includegraphics[width=\textwidth]{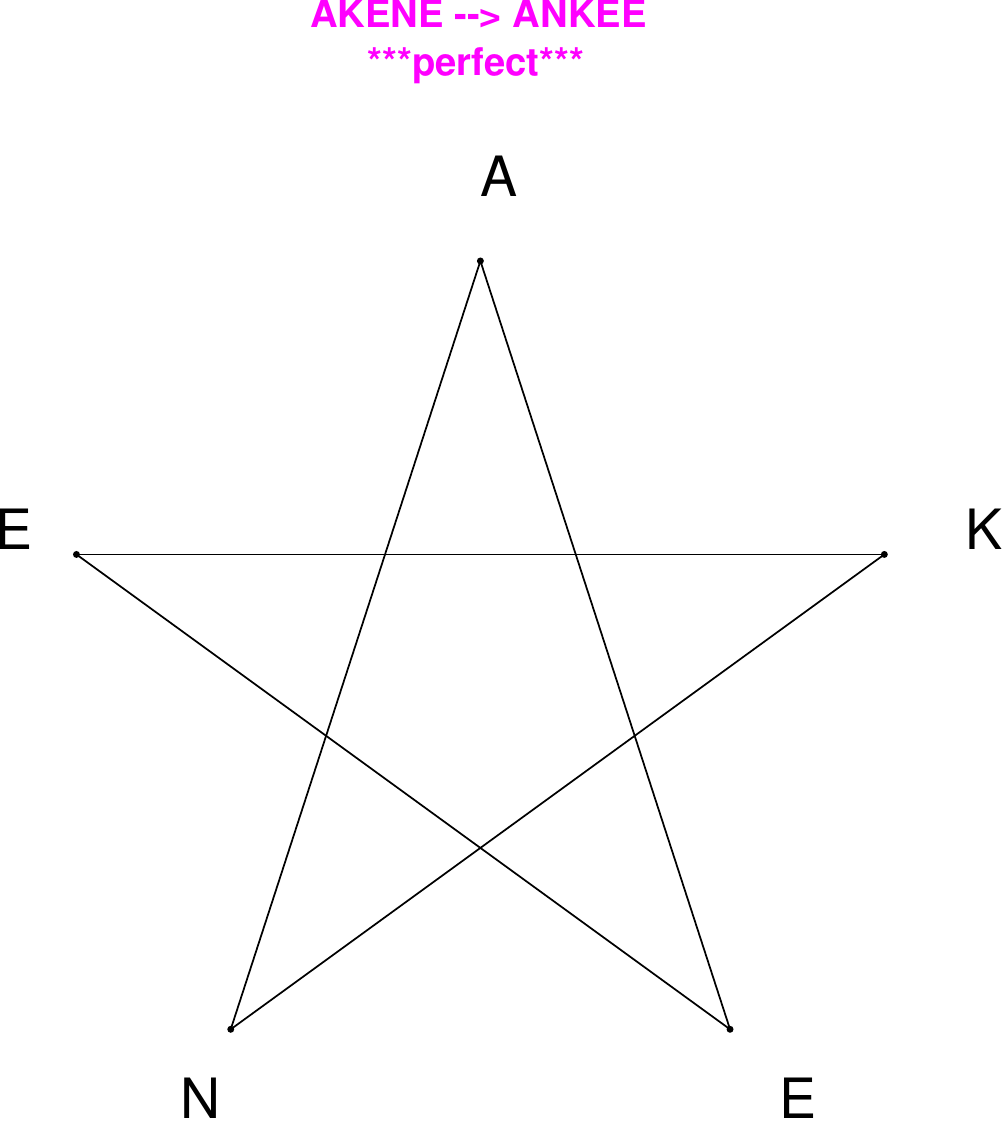}
\end{subfigure}
\end{figure}

\begin{figure}[H]
\centering
\begin{subfigure}[T]{0.19\textwidth}
\centering
\includegraphics[width=\textwidth]{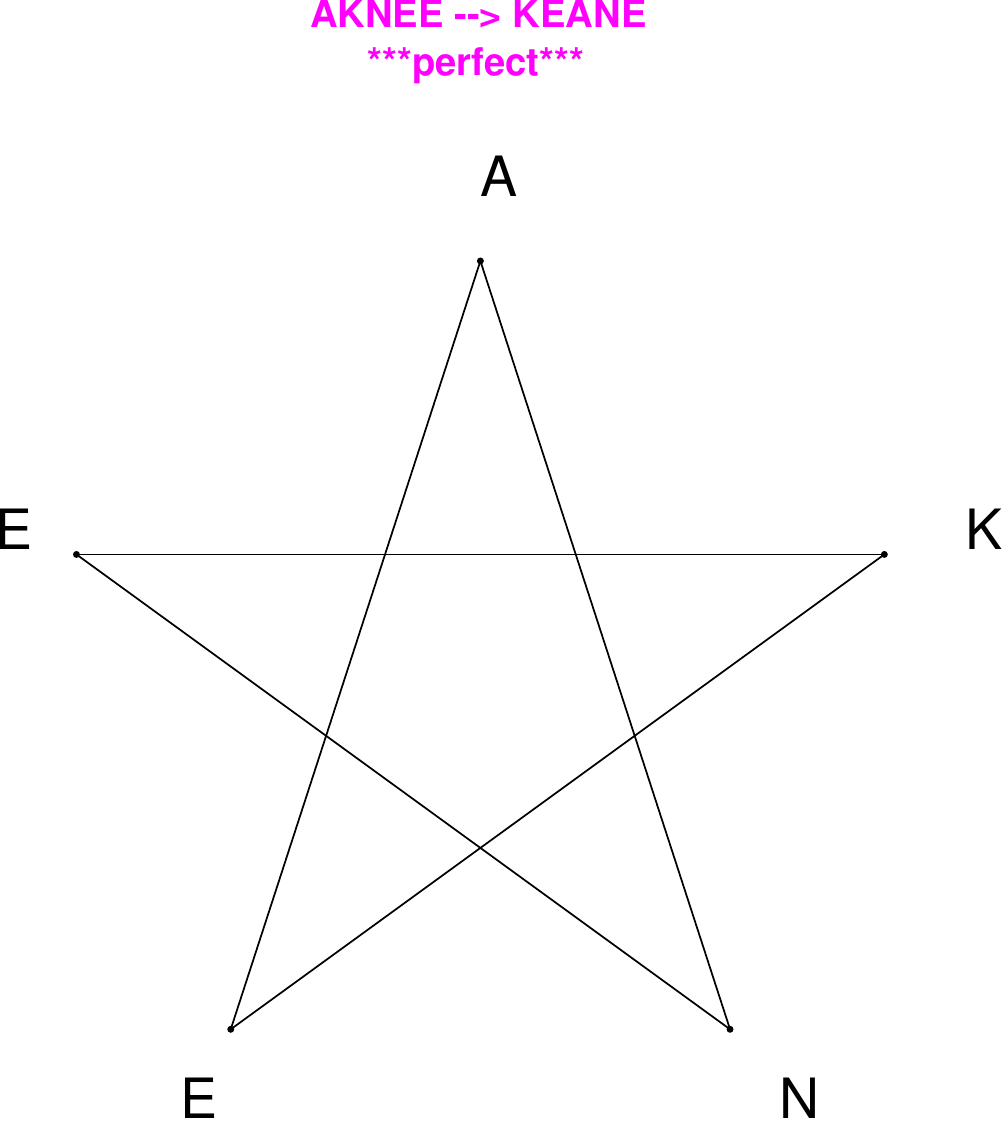}
\end{subfigure}
\hfill
\begin{subfigure}[T]{0.19\textwidth}
\centering
\includegraphics[width=\textwidth]{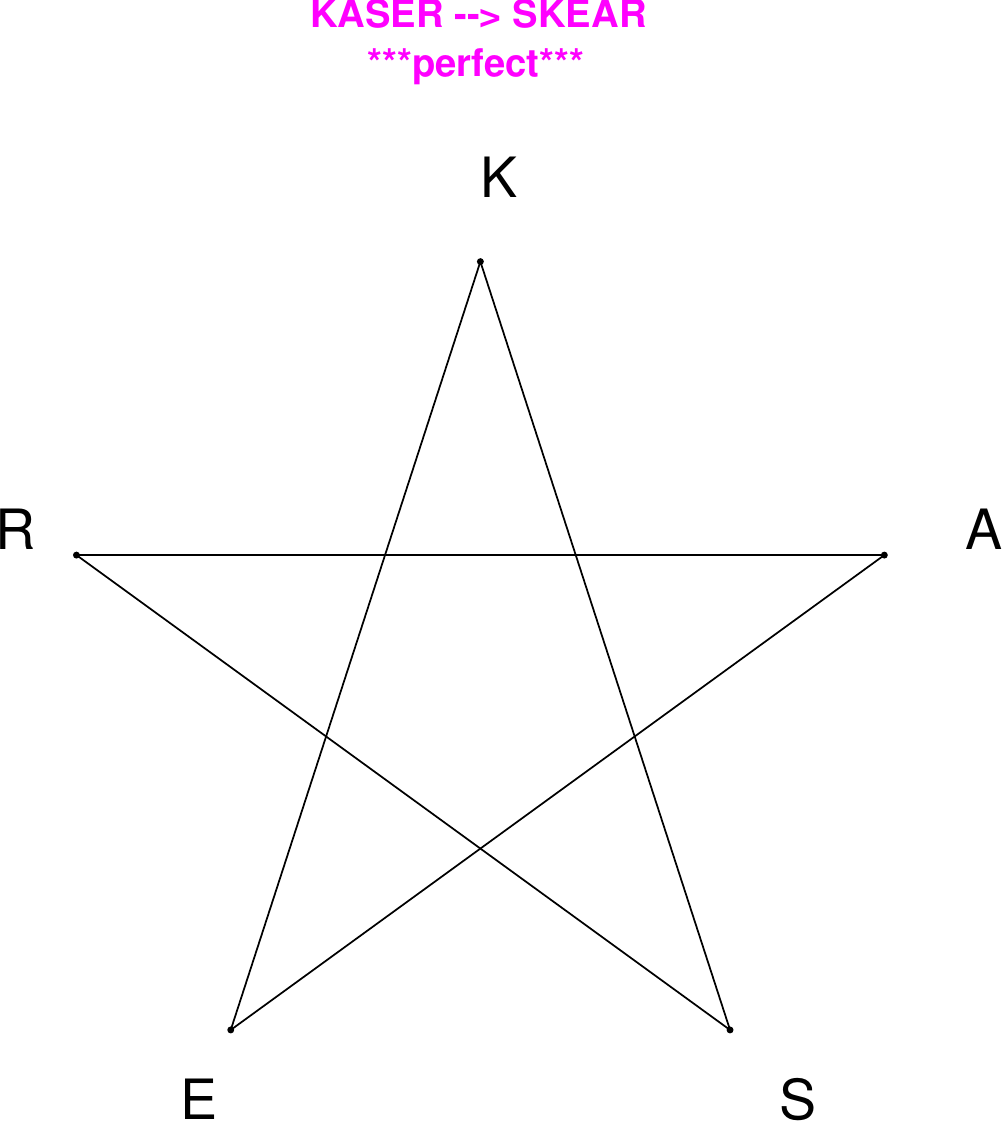}
\end{subfigure}
\hfill
\begin{subfigure}[T]{0.19\textwidth}
\centering
\includegraphics[width=\textwidth]{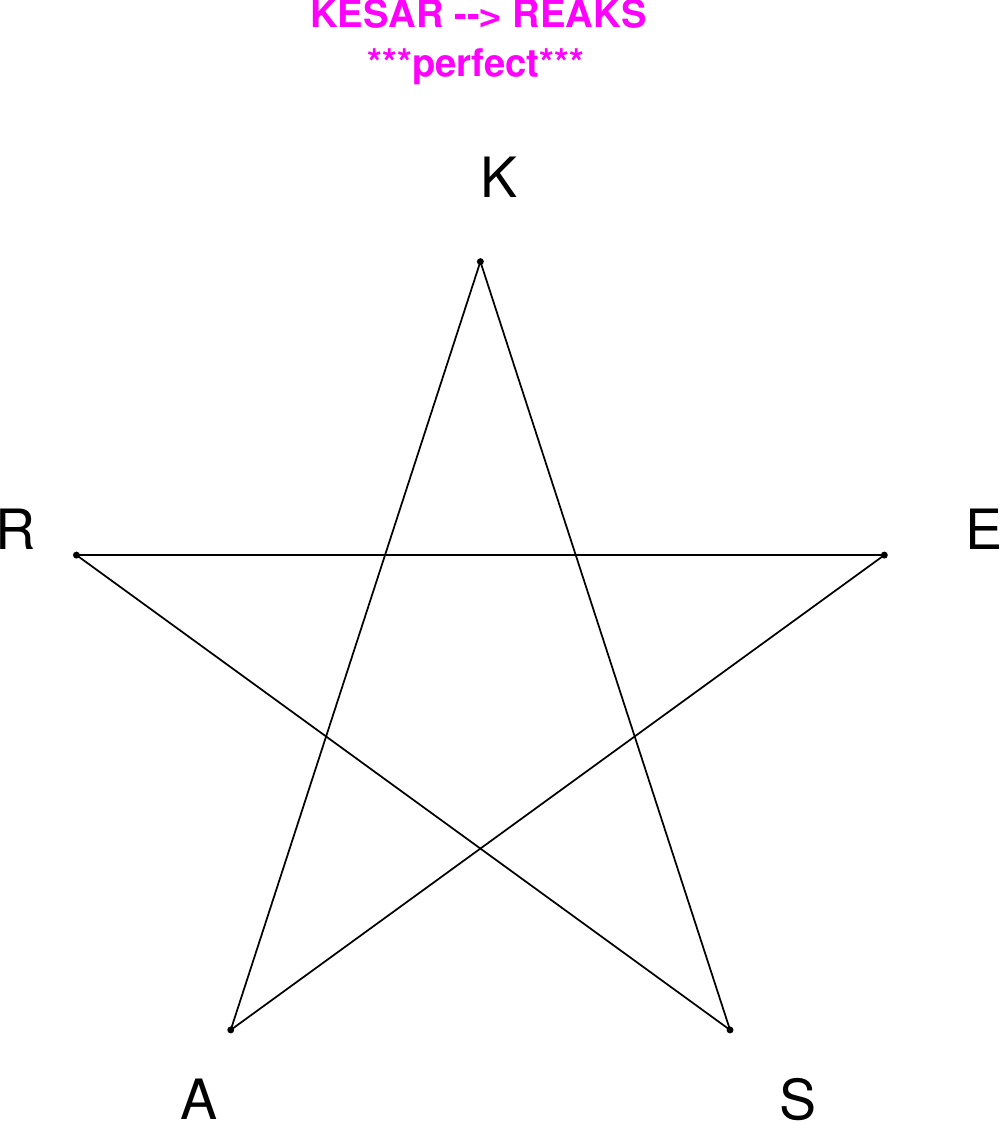}
\end{subfigure}
\hfill
\begin{subfigure}[T]{0.19\textwidth}
\centering
\includegraphics[width=\textwidth]{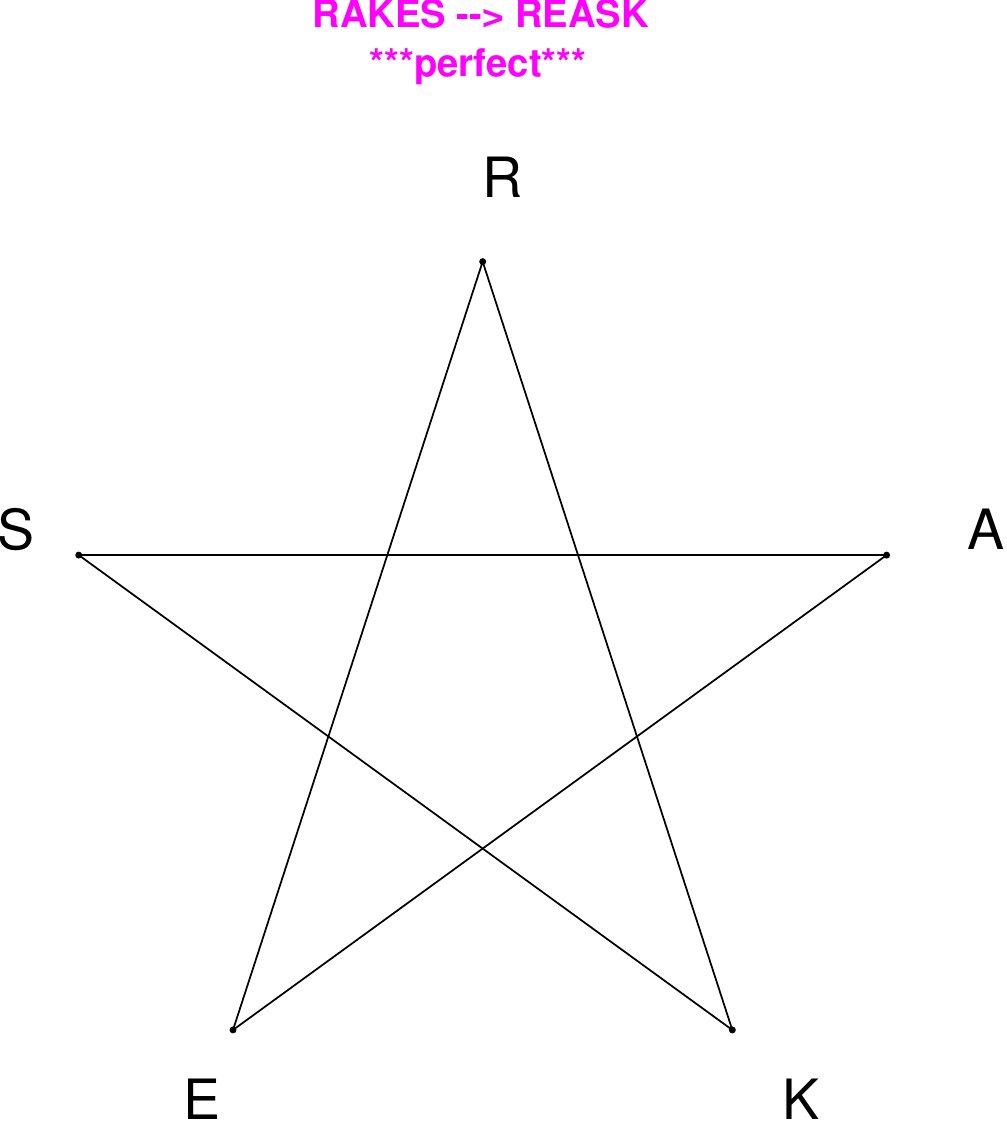}
\end{subfigure}
\hfill
\begin{subfigure}[T]{0.19\textwidth}
\centering
\includegraphics[width=\textwidth]{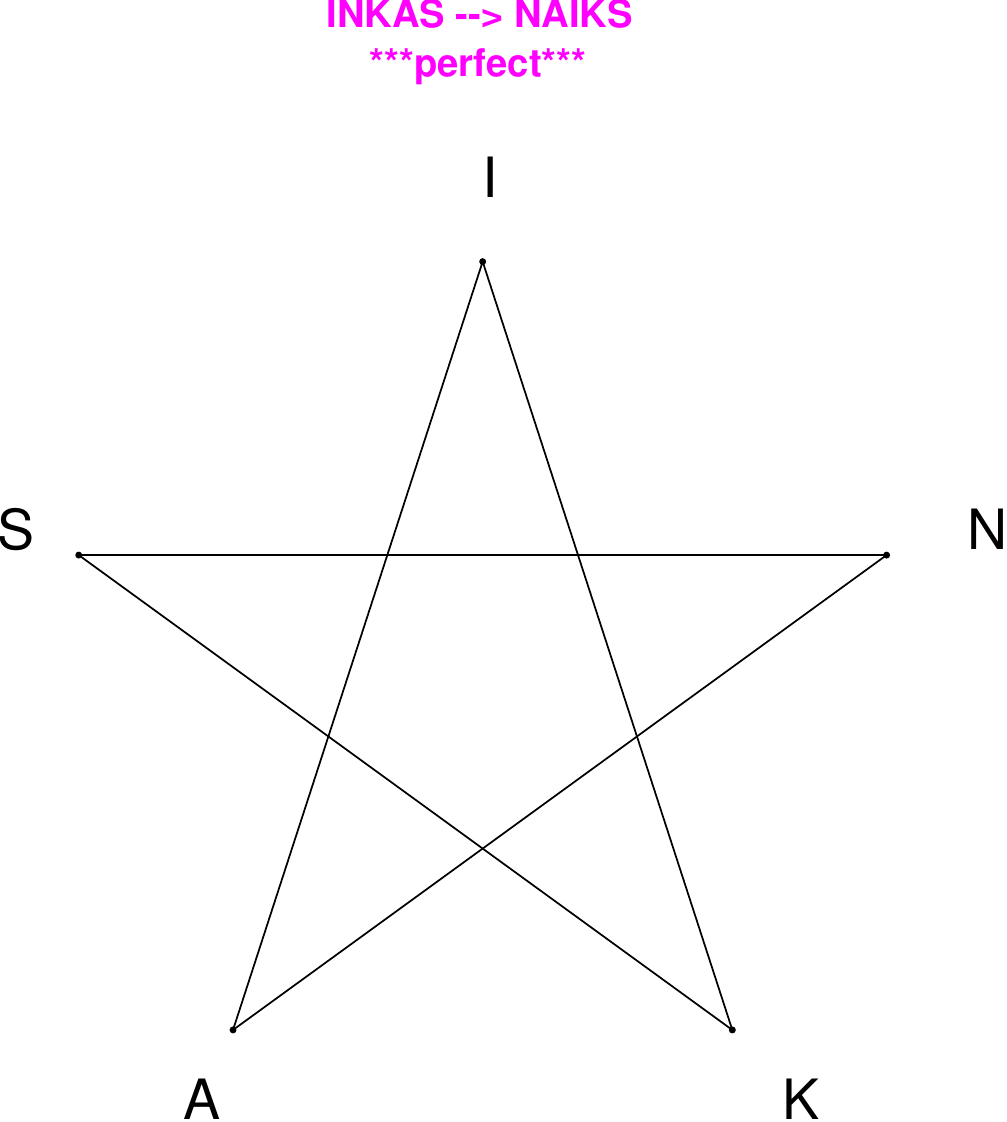}
\end{subfigure}
\end{figure}

\begin{figure}[H]
\centering
\begin{subfigure}[T]{0.19\textwidth}
\centering
\includegraphics[width=\textwidth]{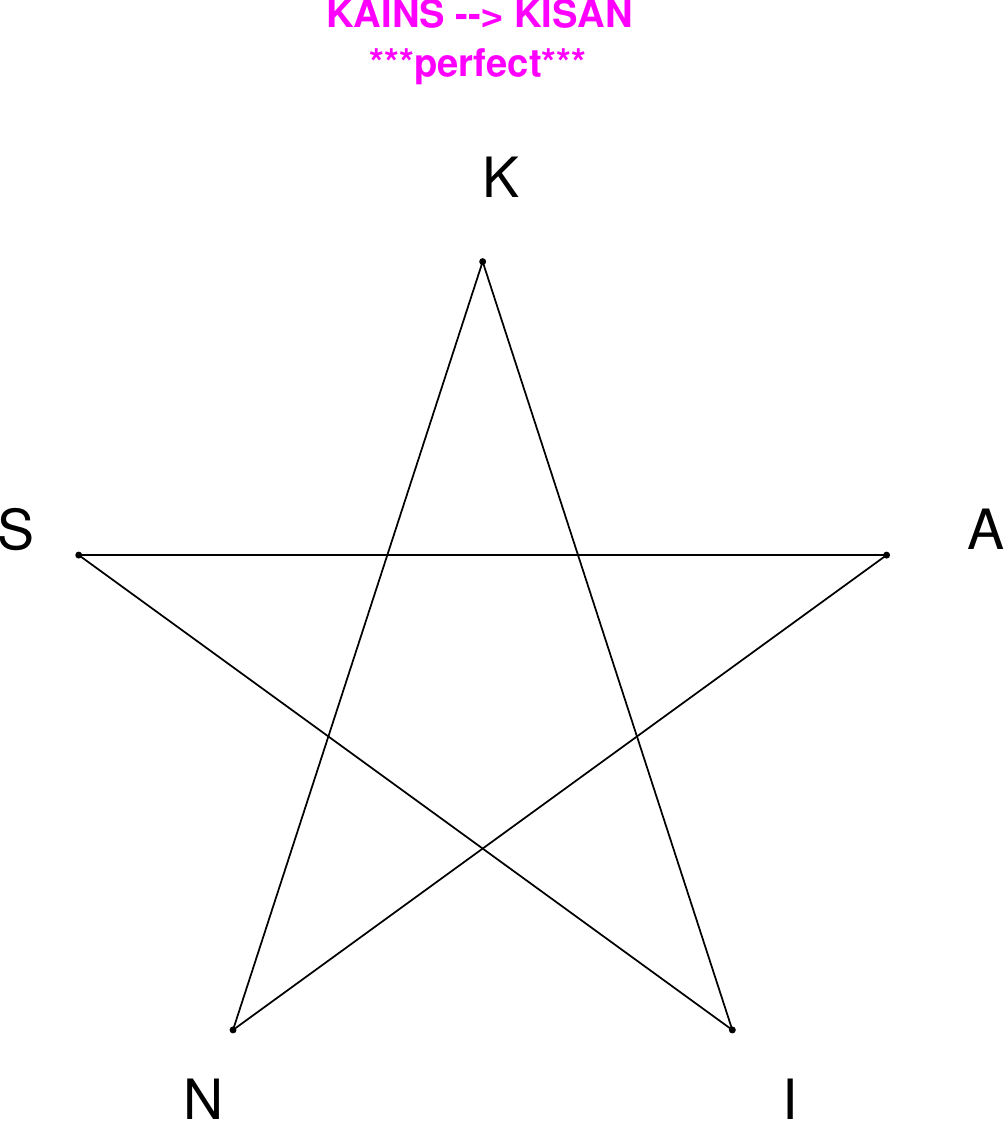}
\end{subfigure}
\hfill
\begin{subfigure}[T]{0.19\textwidth}
\centering
\includegraphics[width=\textwidth]{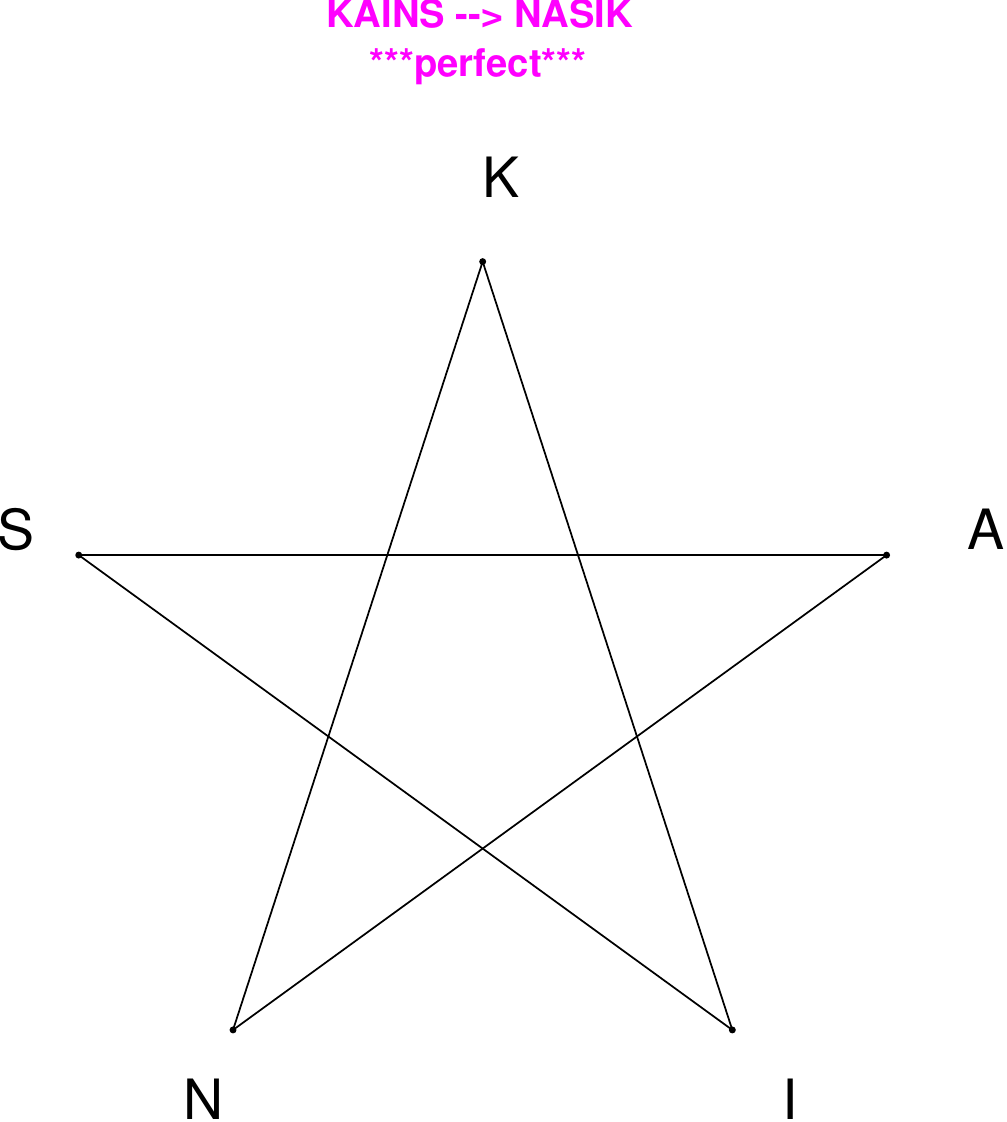}
\end{subfigure}
\hfill
\begin{subfigure}[T]{0.19\textwidth}
\centering
\includegraphics[width=\textwidth]{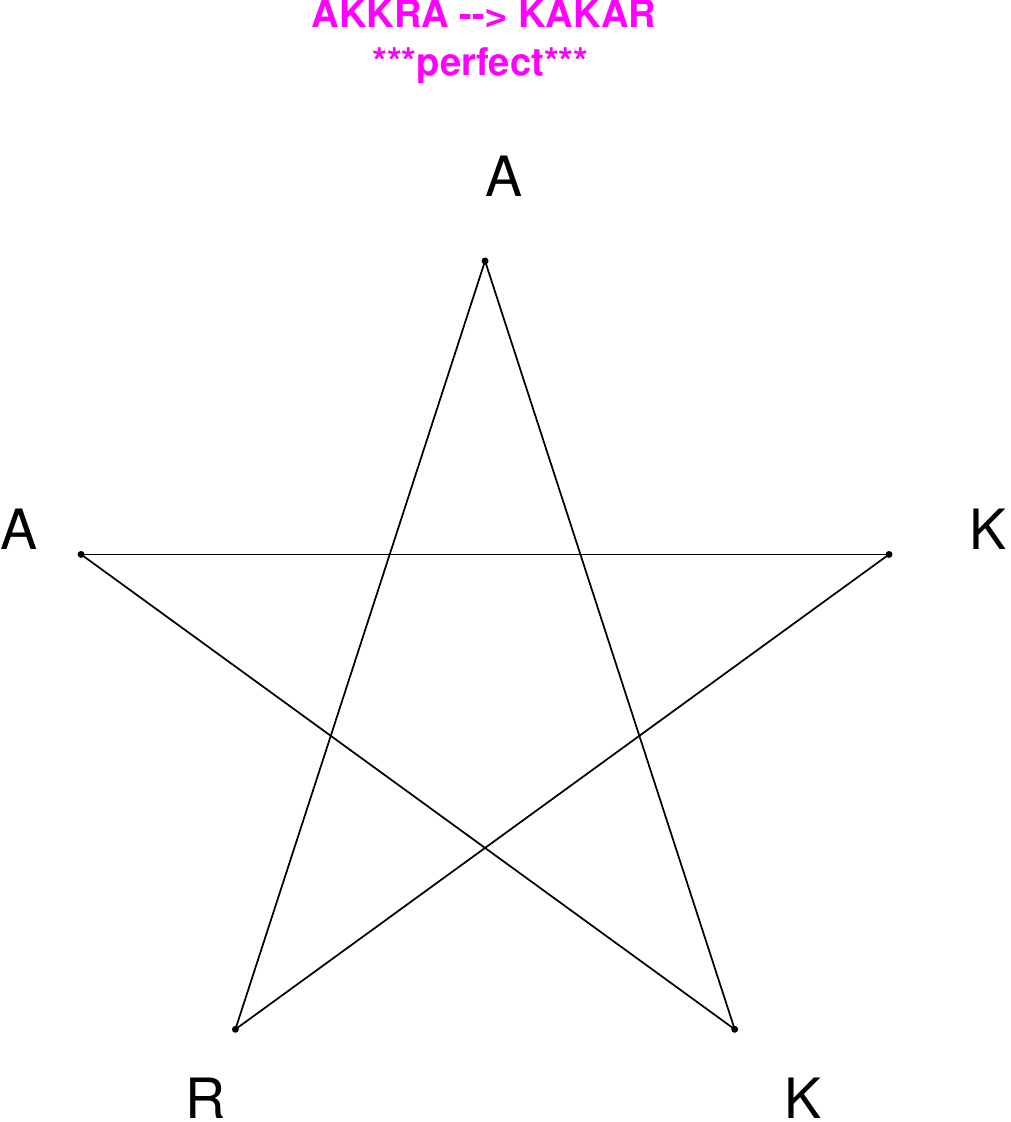}
\end{subfigure}
\hfill
\begin{subfigure}[T]{0.19\textwidth}
\centering
\includegraphics[width=\textwidth]{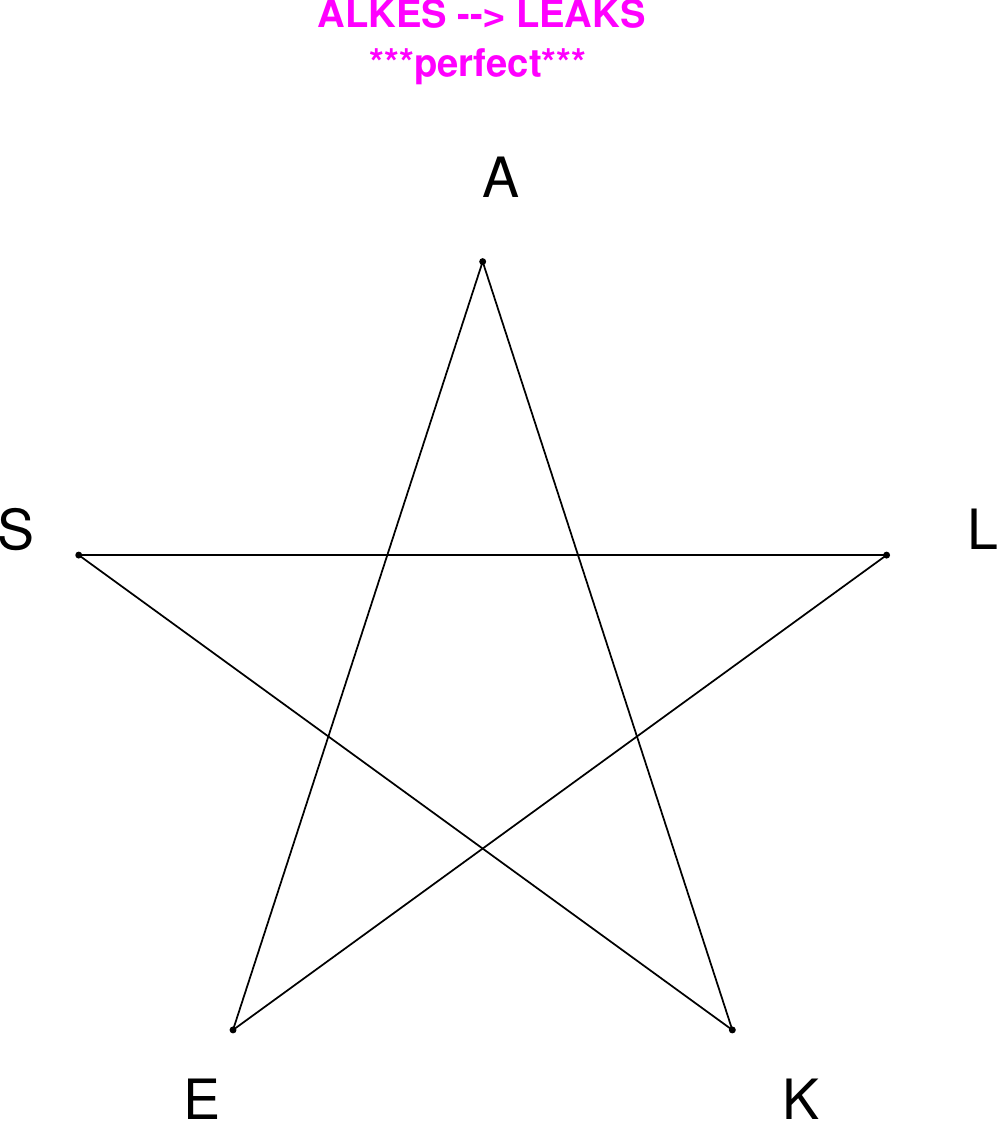}
\end{subfigure}
\hfill
\begin{subfigure}[T]{0.19\textwidth}
\centering
\includegraphics[width=\textwidth]{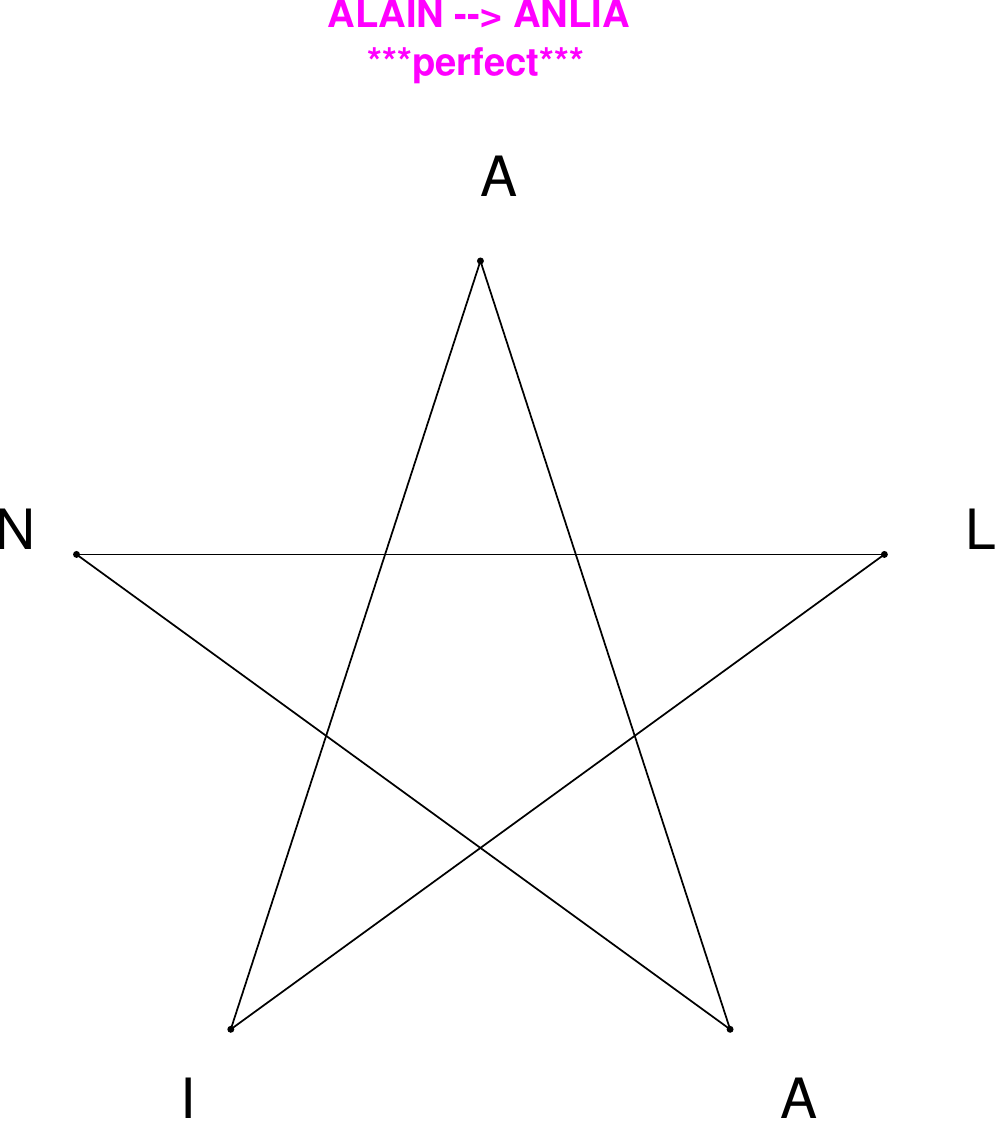}
\end{subfigure}
\end{figure}

\begin{figure}[H]
\centering
\begin{subfigure}[T]{0.19\textwidth}
\centering
\includegraphics[width=\textwidth]{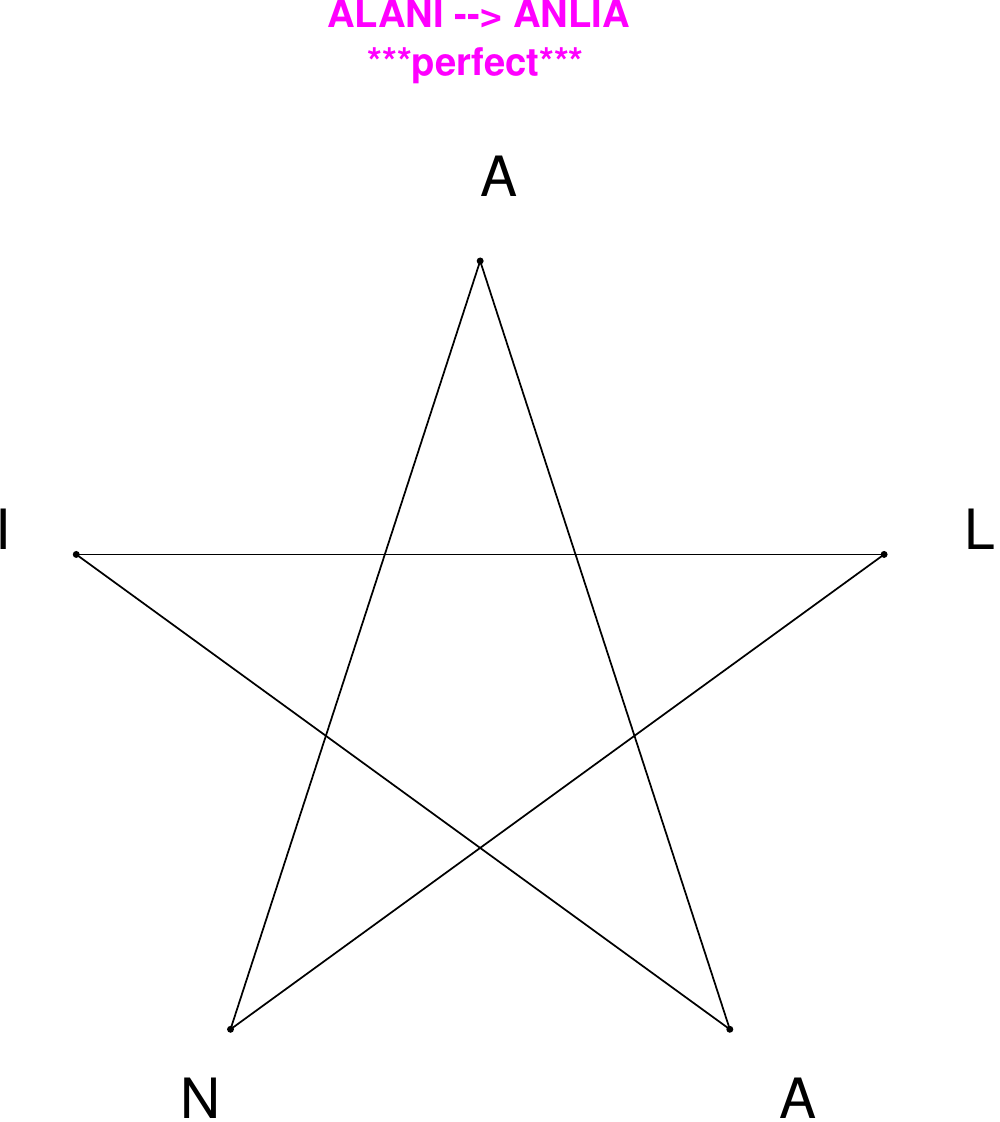}
\end{subfigure}
\hfill
\begin{subfigure}[T]{0.19\textwidth}
\centering
\includegraphics[width=\textwidth]{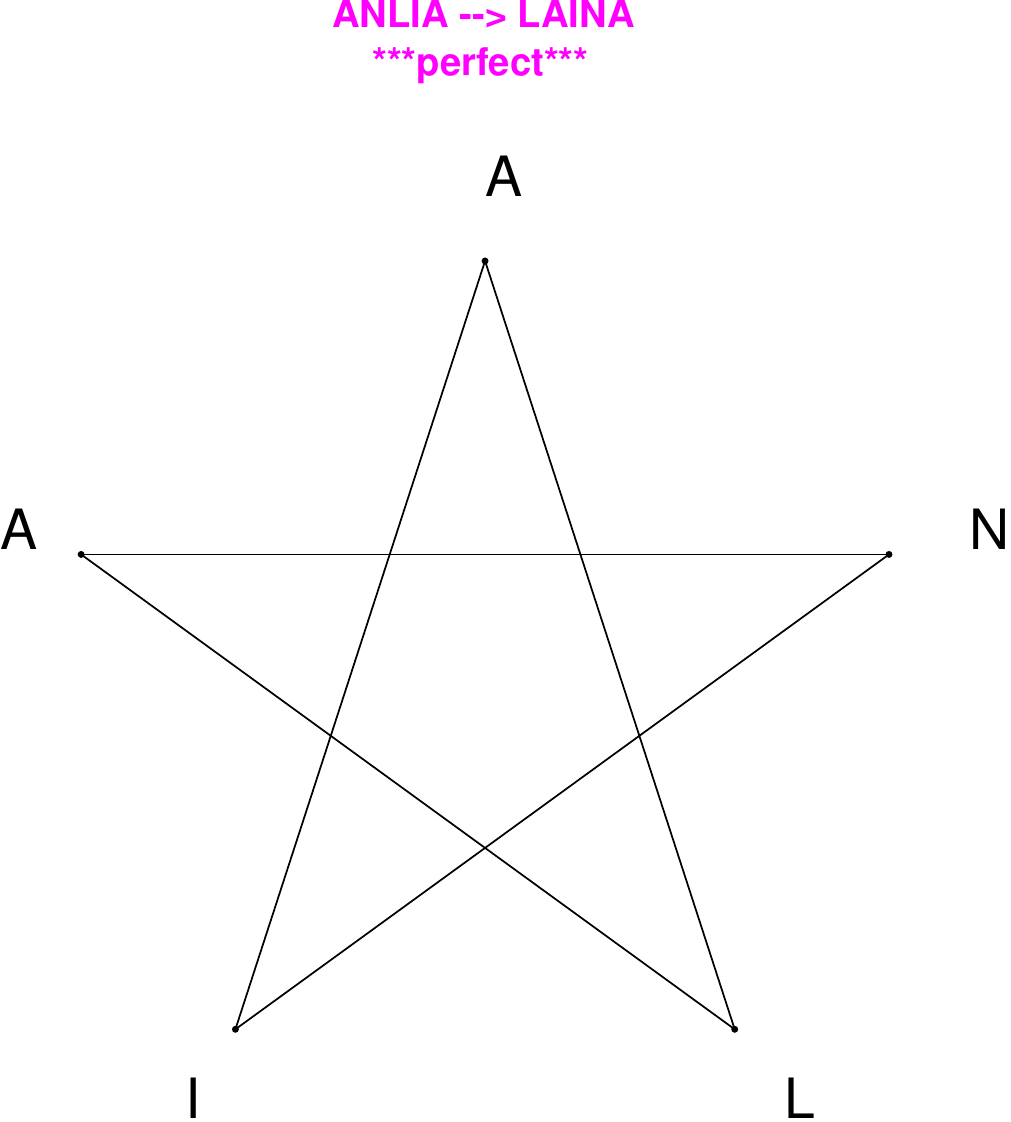}
\end{subfigure}
\hfill
\begin{subfigure}[T]{0.19\textwidth}
\centering
\includegraphics[width=\textwidth]{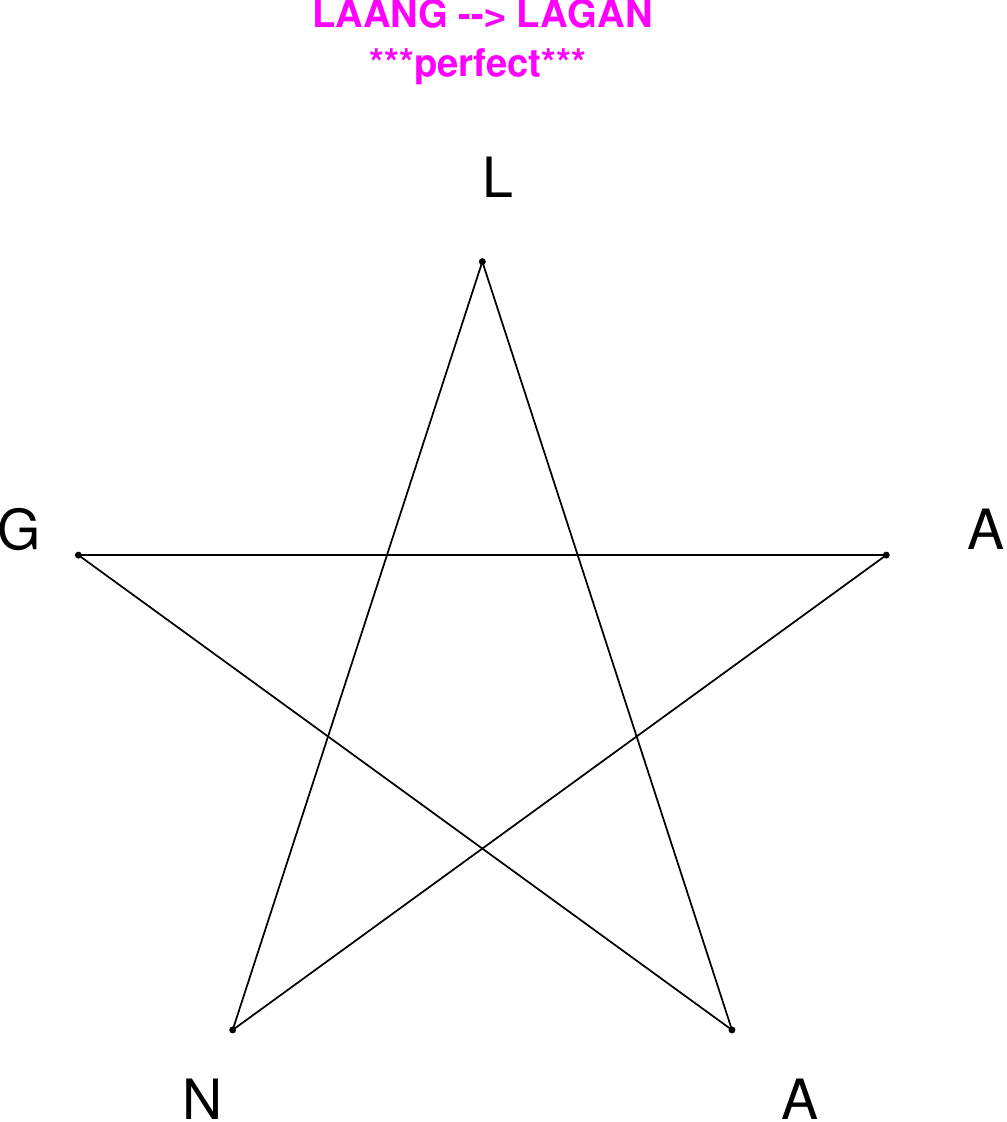}
\end{subfigure}
\hfill
\begin{subfigure}[T]{0.19\textwidth}
\centering
\includegraphics[width=\textwidth]{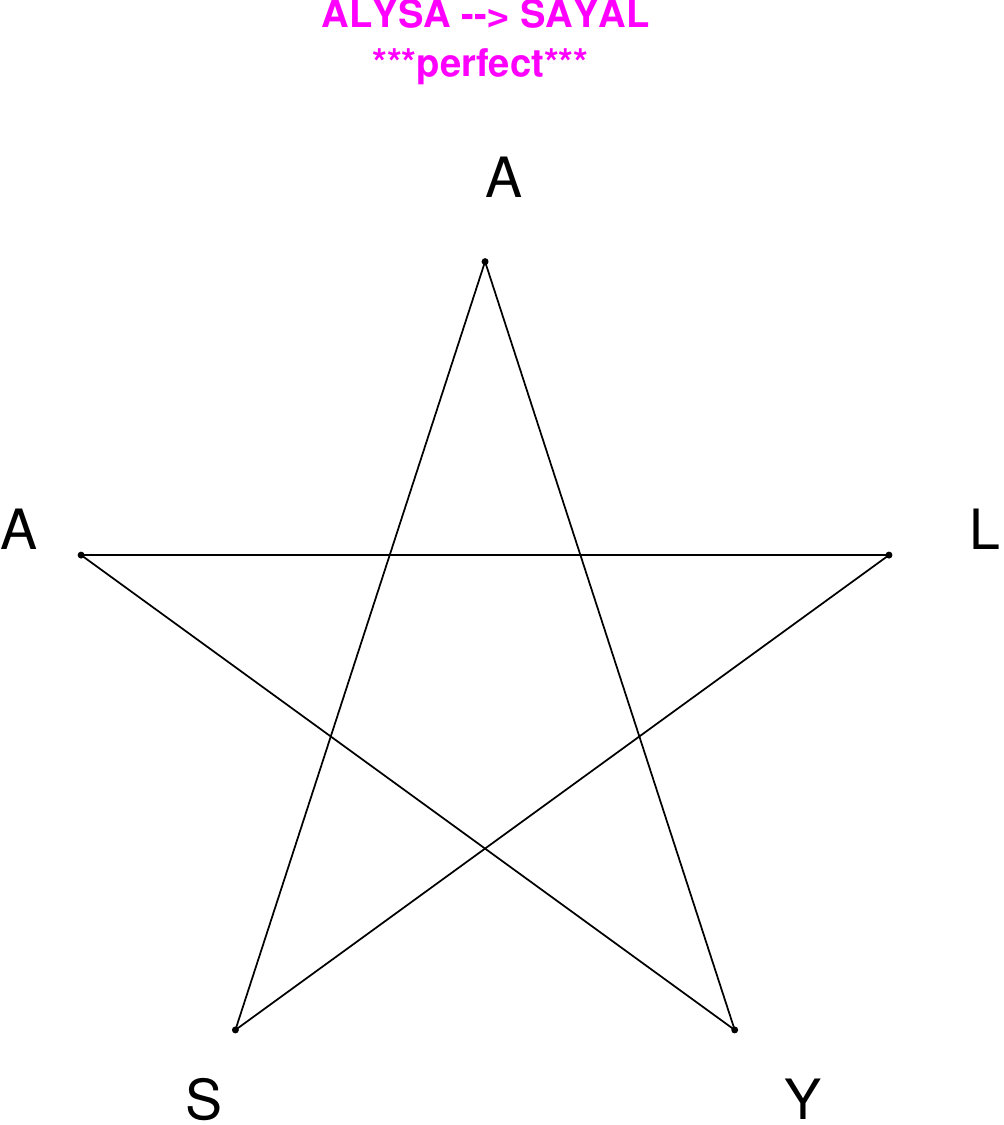}
\end{subfigure}
\hfill
\begin{subfigure}[T]{0.19\textwidth}
\centering
\includegraphics[width=\textwidth]{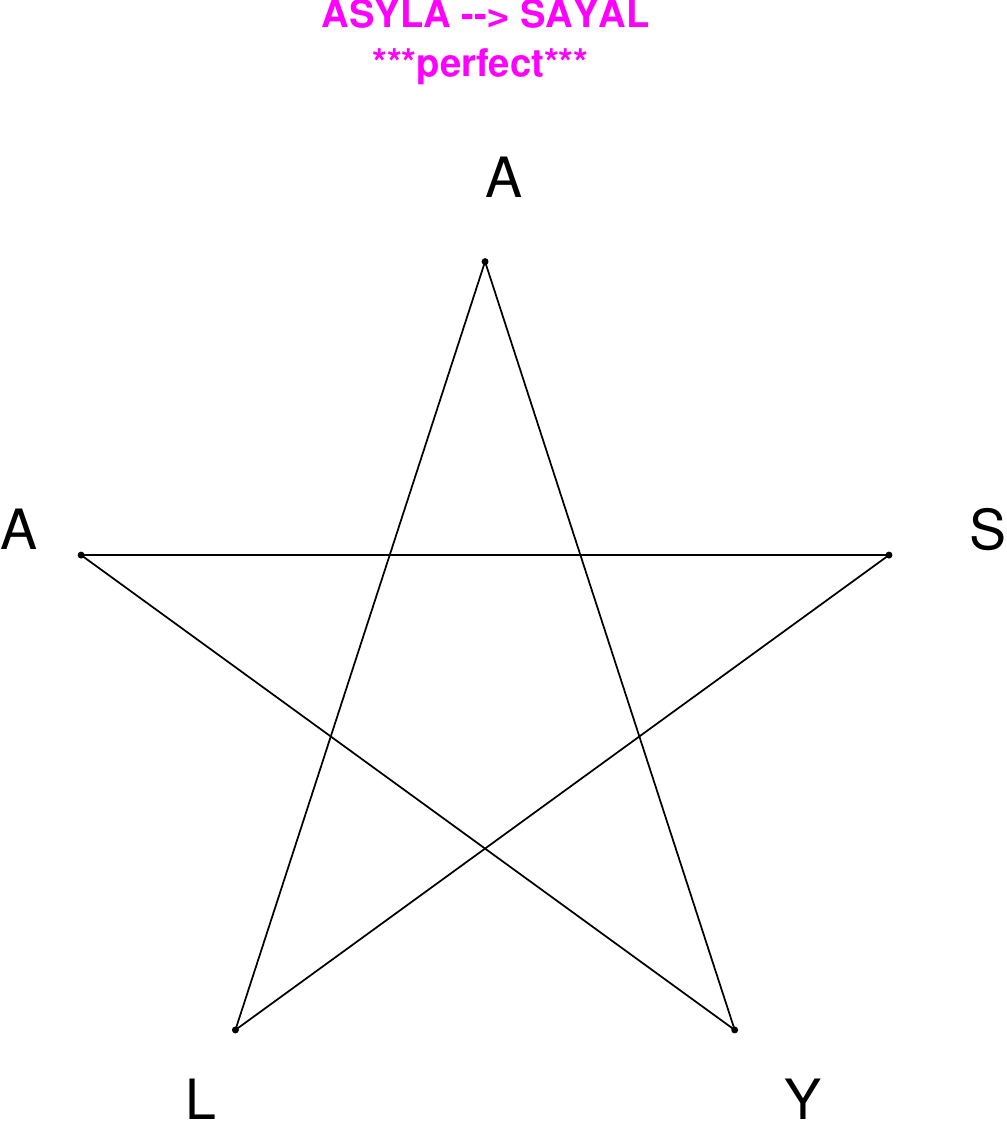}
\end{subfigure}
\end{figure}

\begin{figure}[H]
\centering
\begin{subfigure}[T]{0.19\textwidth}
\centering
\includegraphics[width=\textwidth]{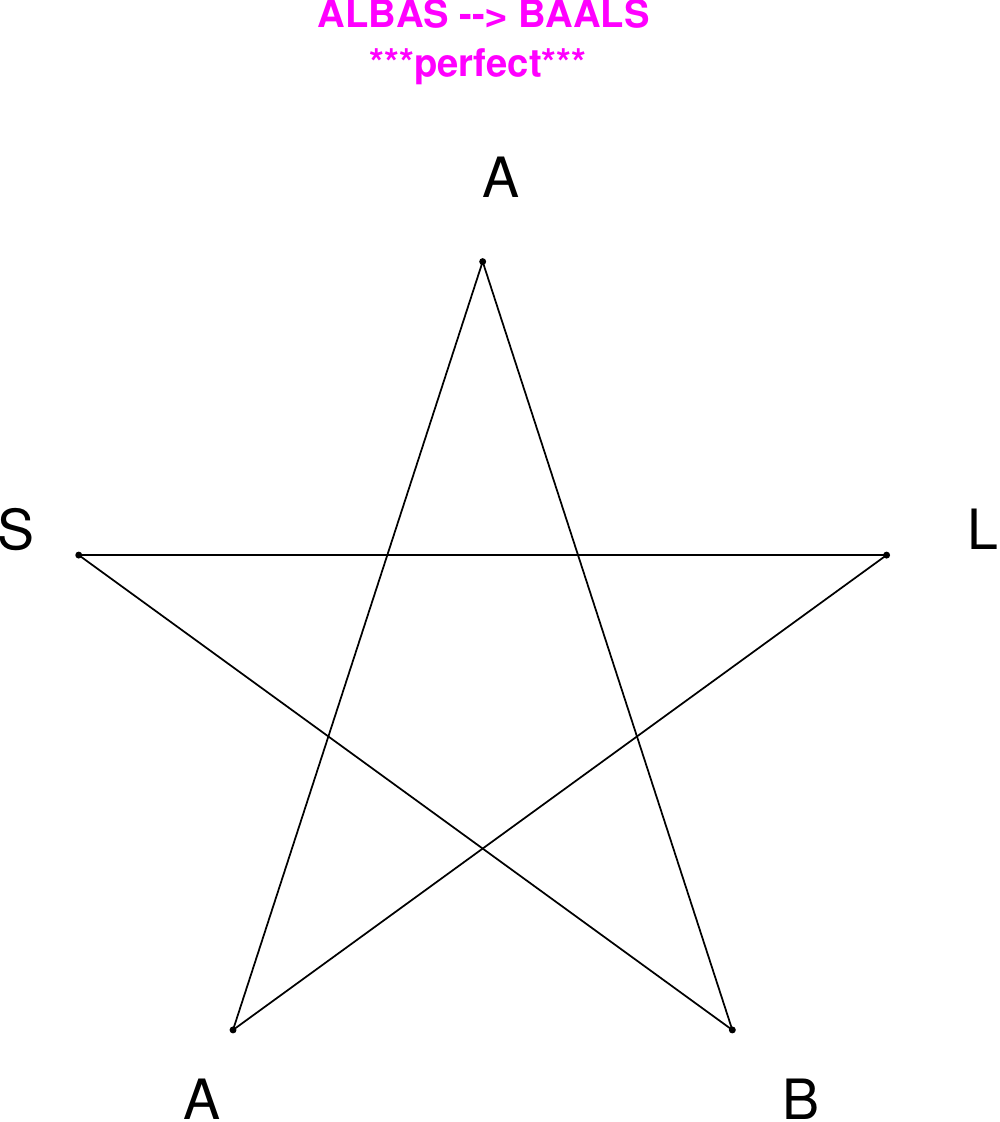}
\end{subfigure}
\hfill
\begin{subfigure}[T]{0.19\textwidth}
\centering
\includegraphics[width=\textwidth]{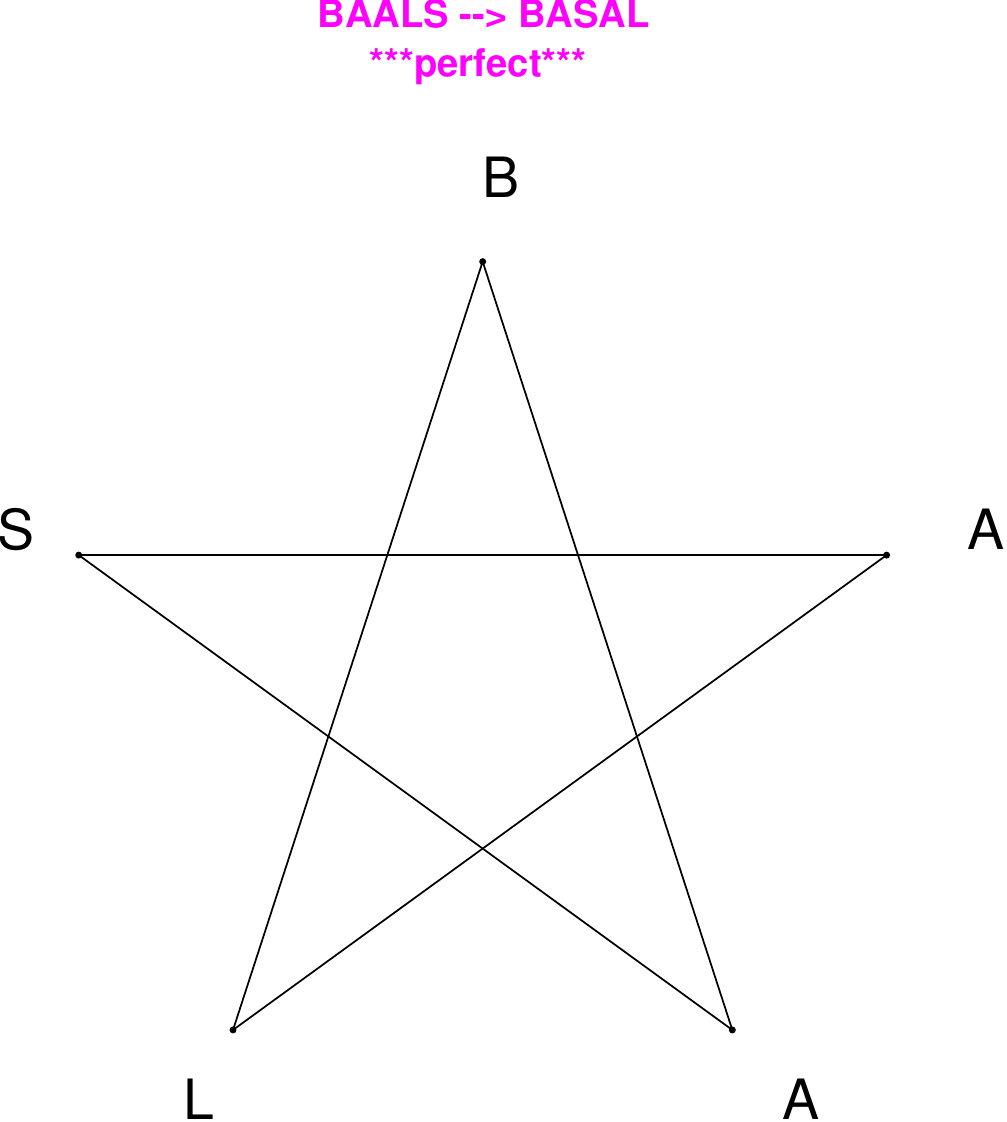}
\end{subfigure}
\hfill
\begin{subfigure}[T]{0.19\textwidth}
\centering
\includegraphics[width=\textwidth]{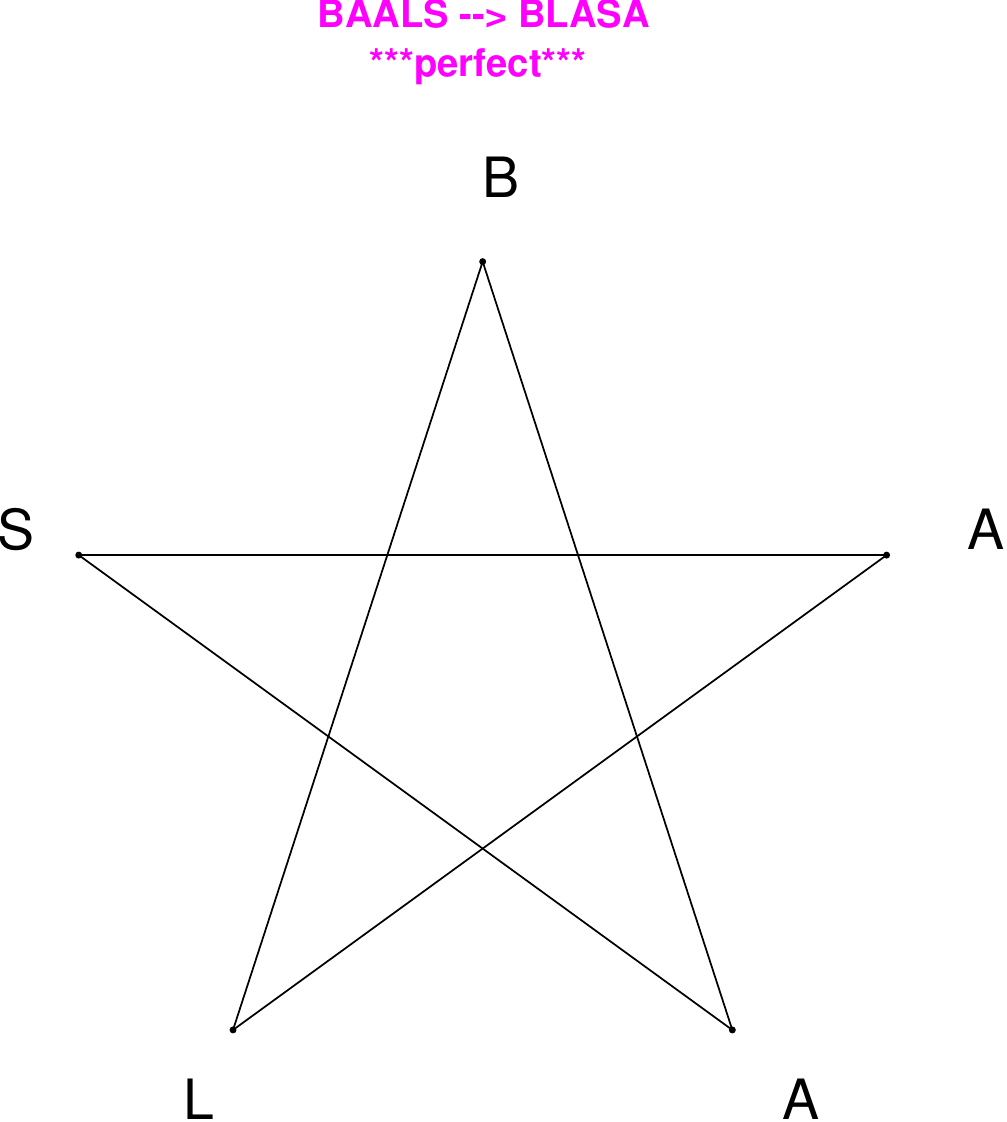}
\end{subfigure}
\hfill
\begin{subfigure}[T]{0.19\textwidth}
\centering
\includegraphics[width=\textwidth]{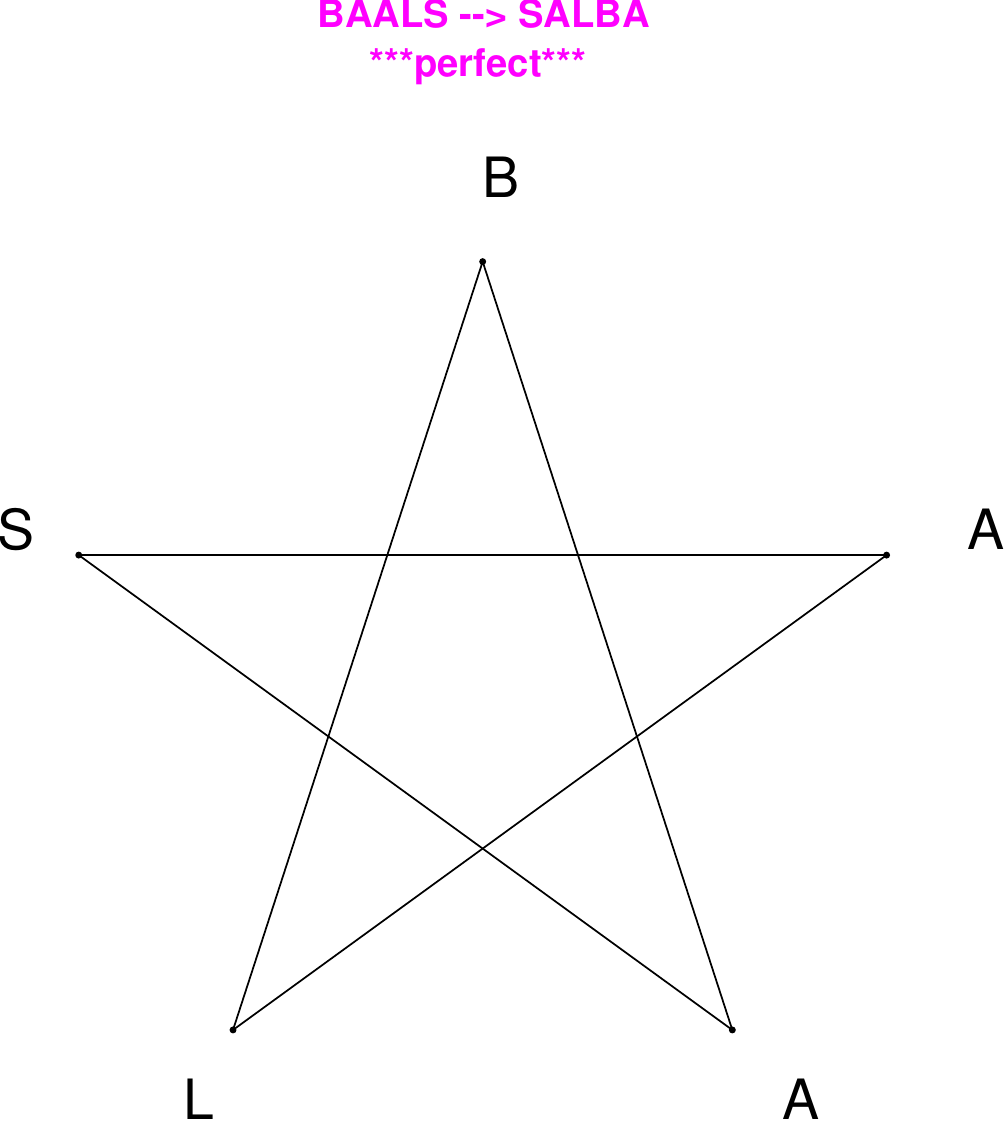}
\end{subfigure}
\hfill
\begin{subfigure}[T]{0.19\textwidth}
\centering
\includegraphics[width=\textwidth]{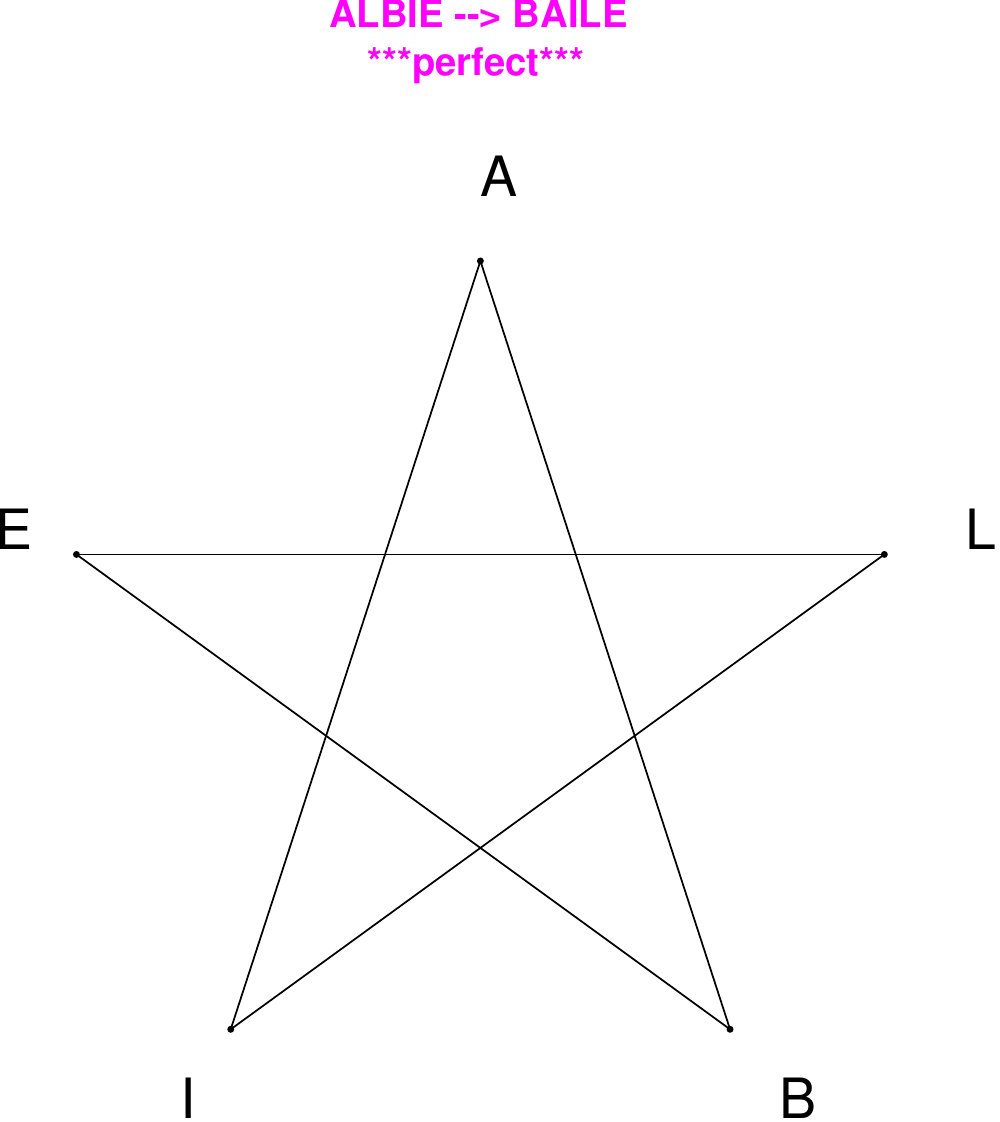}
\end{subfigure}
\end{figure}

\begin{figure}[H]
\centering
\begin{subfigure}[T]{0.19\textwidth}
\centering
\includegraphics[width=\textwidth]{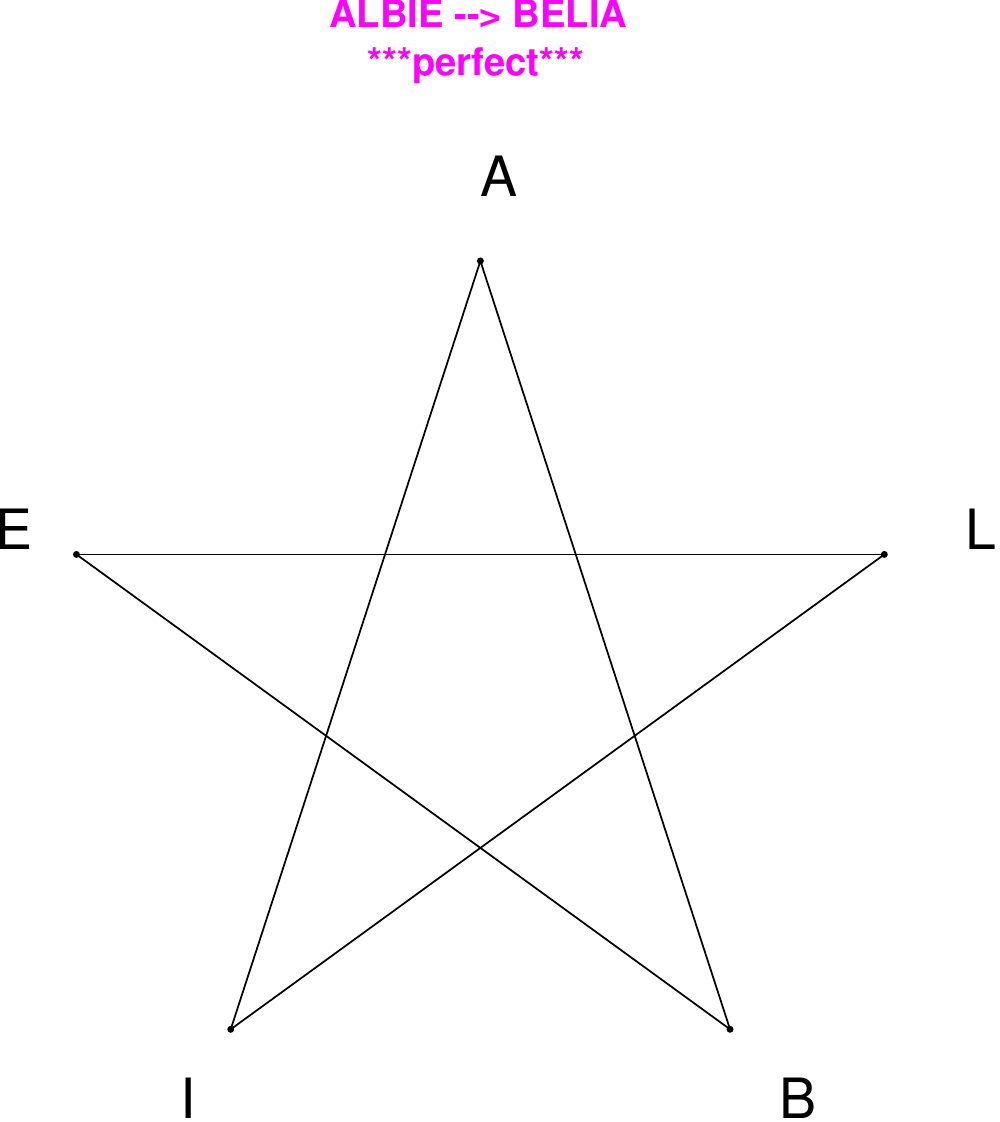}
\end{subfigure}
\hfill
\begin{subfigure}[T]{0.19\textwidth}
\centering
\includegraphics[width=\textwidth]{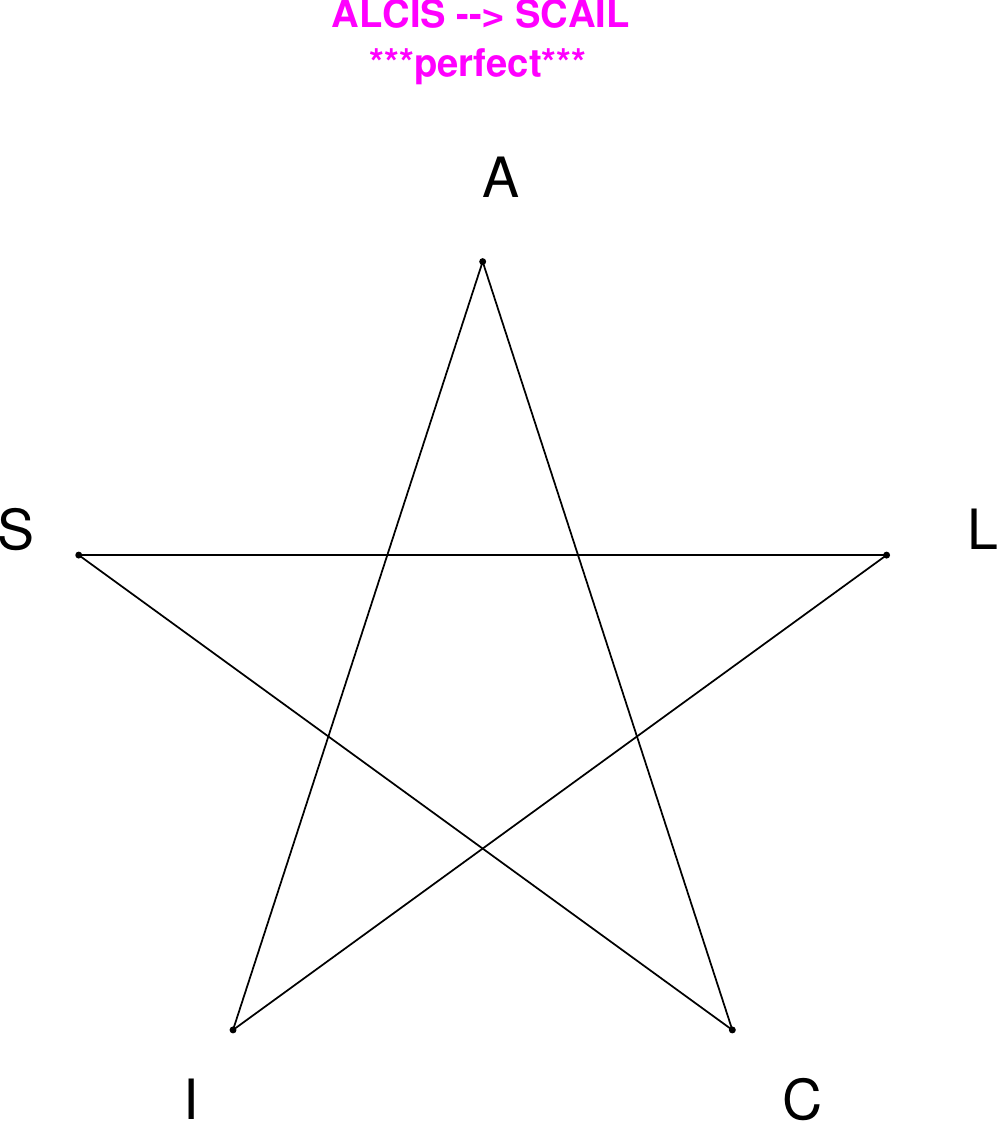}
\end{subfigure}
\hfill
\begin{subfigure}[T]{0.19\textwidth}
\centering
\includegraphics[width=\textwidth]{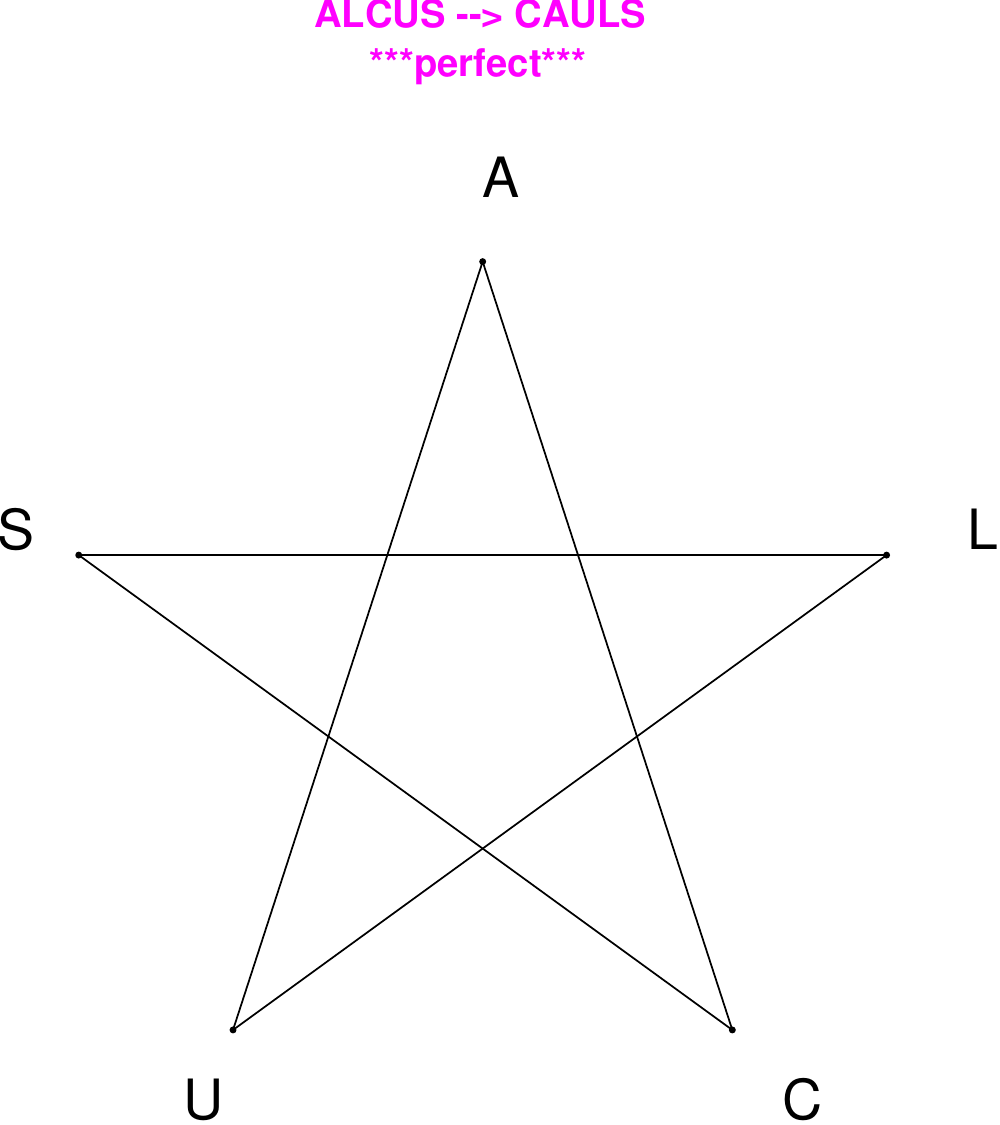}
\end{subfigure}
\hfill
\begin{subfigure}[T]{0.19\textwidth}
\centering
\includegraphics[width=\textwidth]{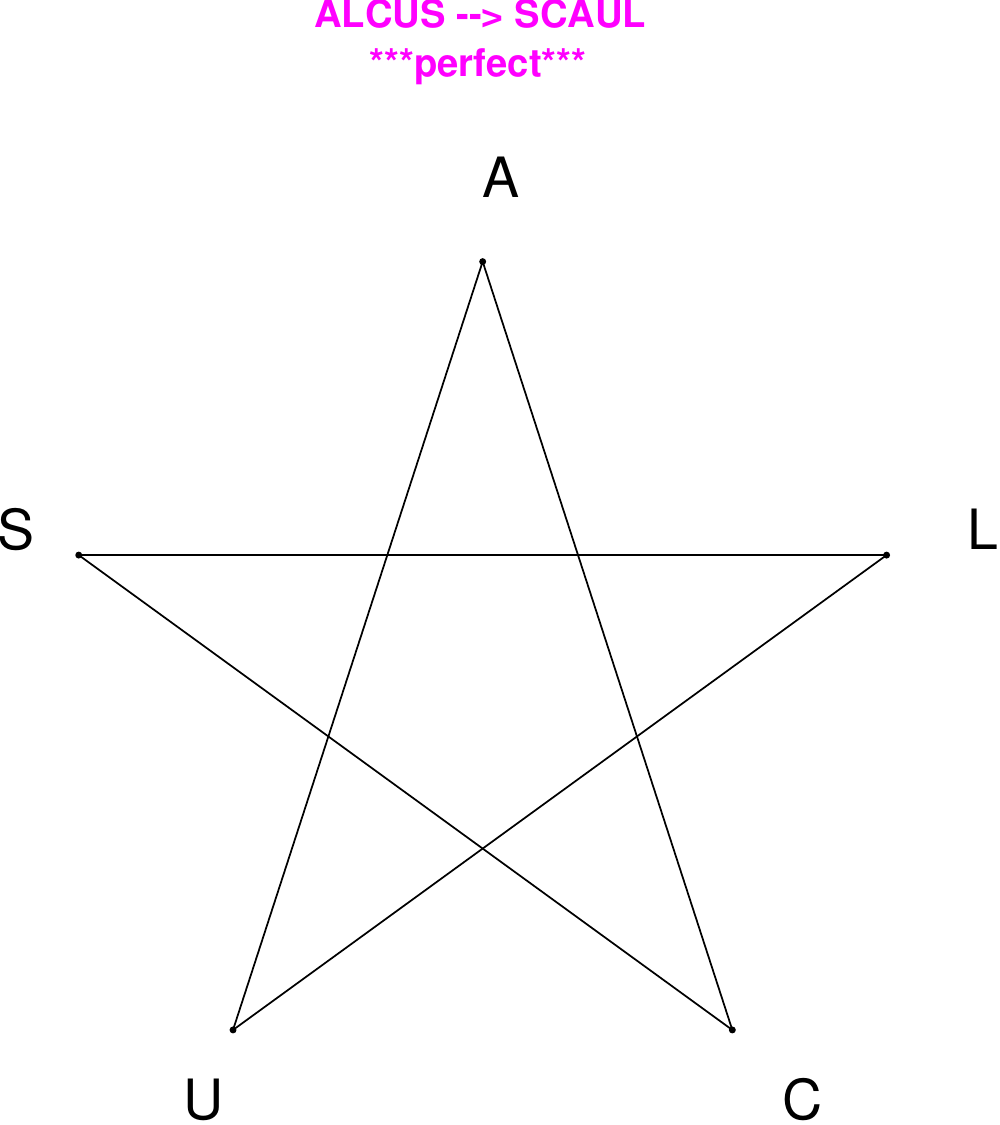}
\end{subfigure}
\hfill
\begin{subfigure}[T]{0.19\textwidth}
\centering
\includegraphics[width=\textwidth]{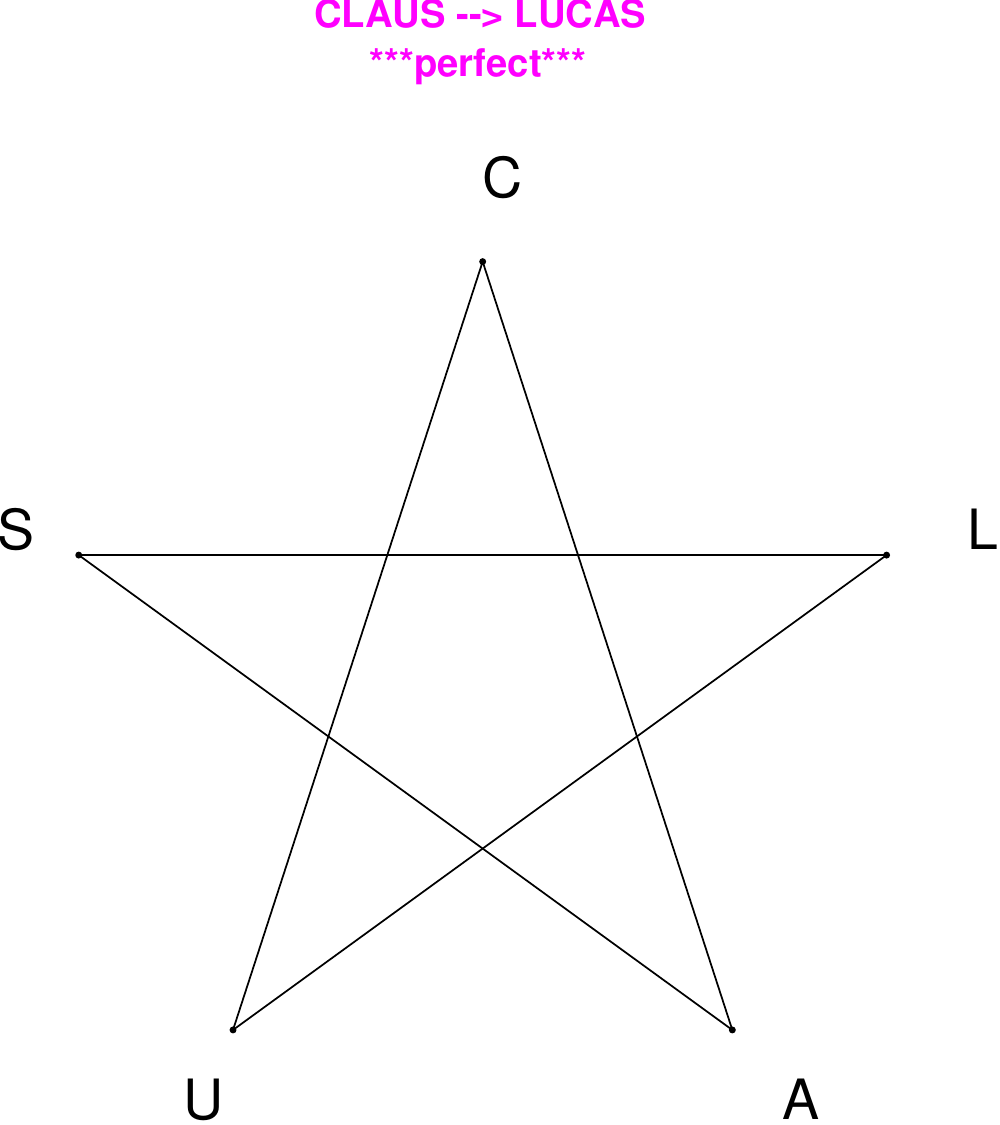}
\end{subfigure}
\end{figure}

\begin{figure}[H]
\centering
\begin{subfigure}[T]{0.19\textwidth}
\centering
\includegraphics[width=\textwidth]{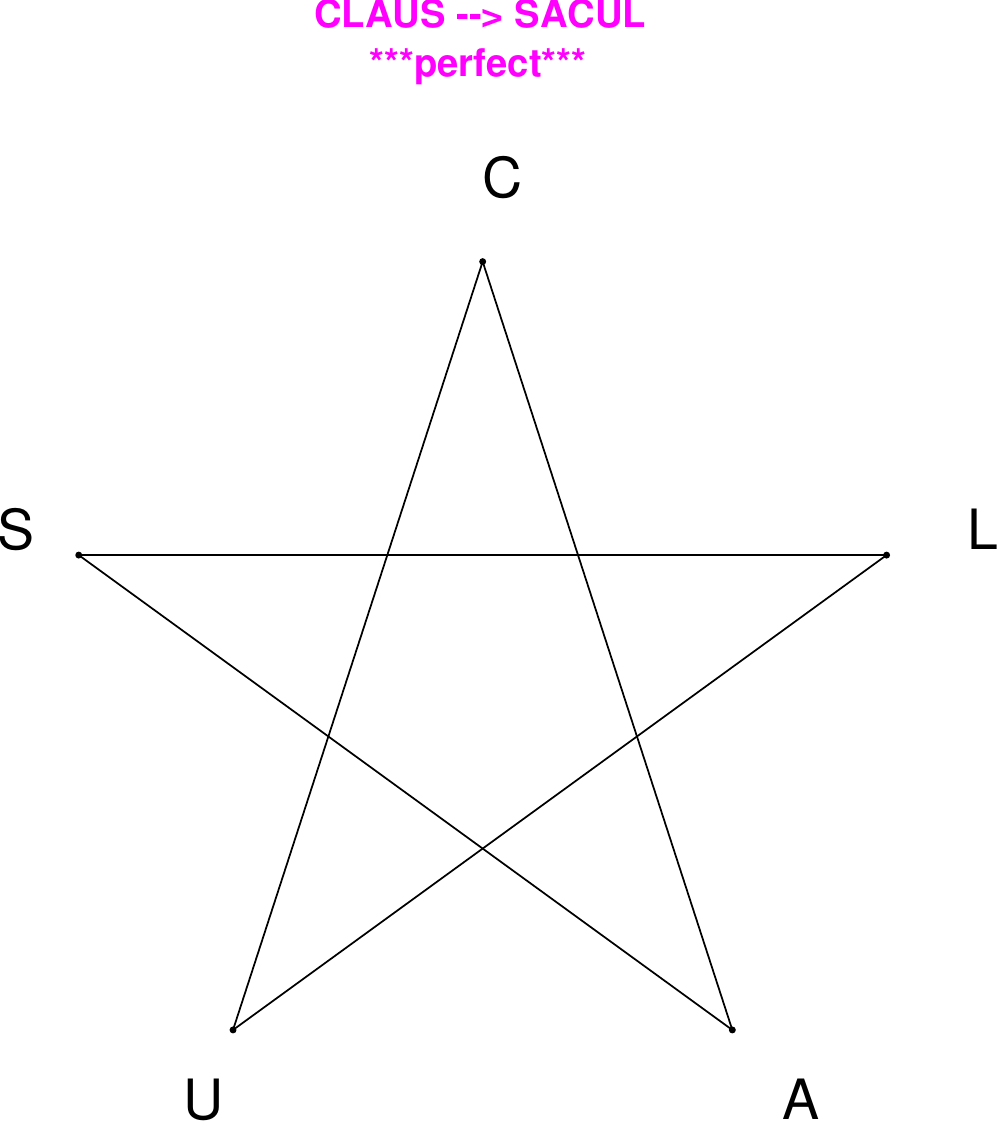}
\end{subfigure}
\hfill
\begin{subfigure}[T]{0.19\textwidth}
\centering
\includegraphics[width=\textwidth]{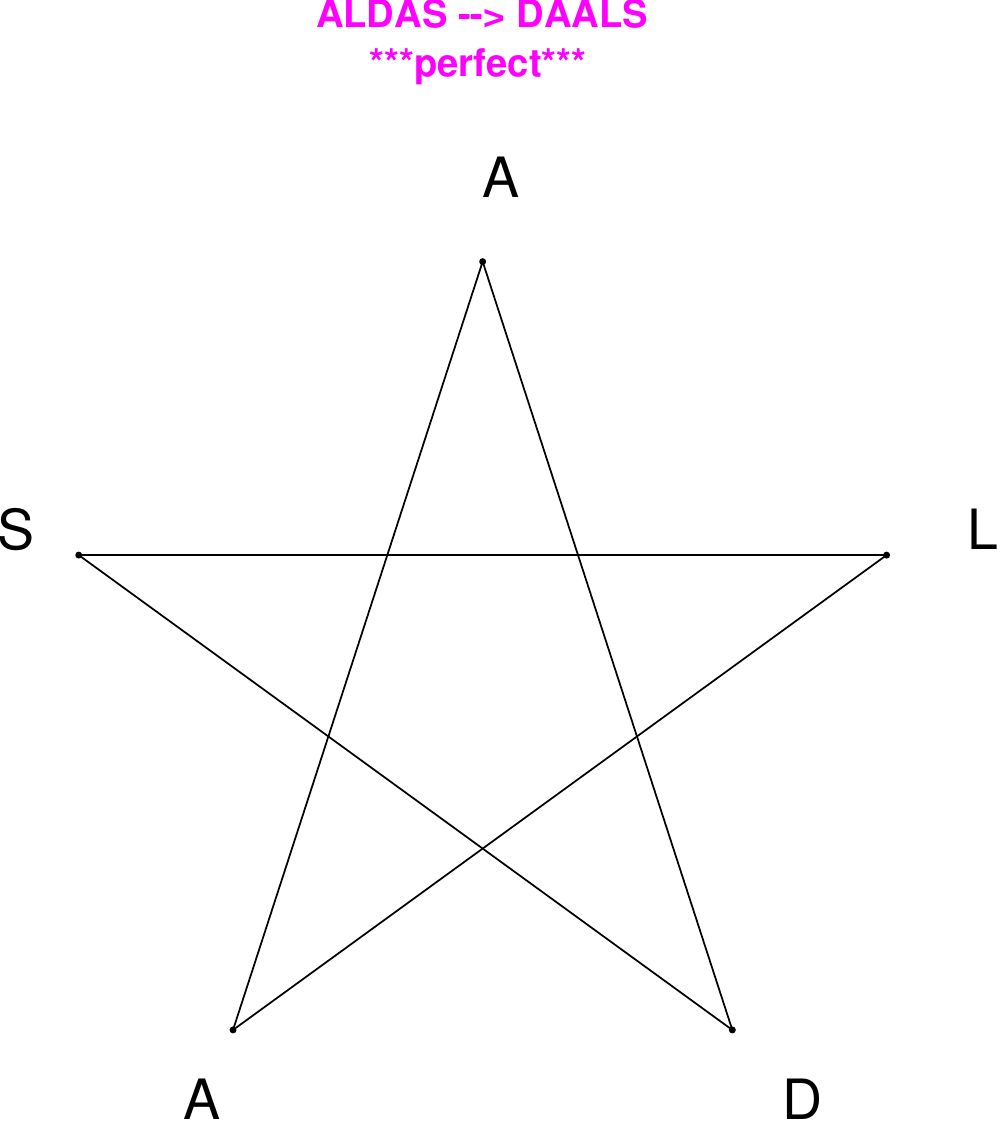}
\end{subfigure}
\hfill
\begin{subfigure}[T]{0.19\textwidth}
\centering
\includegraphics[width=\textwidth]{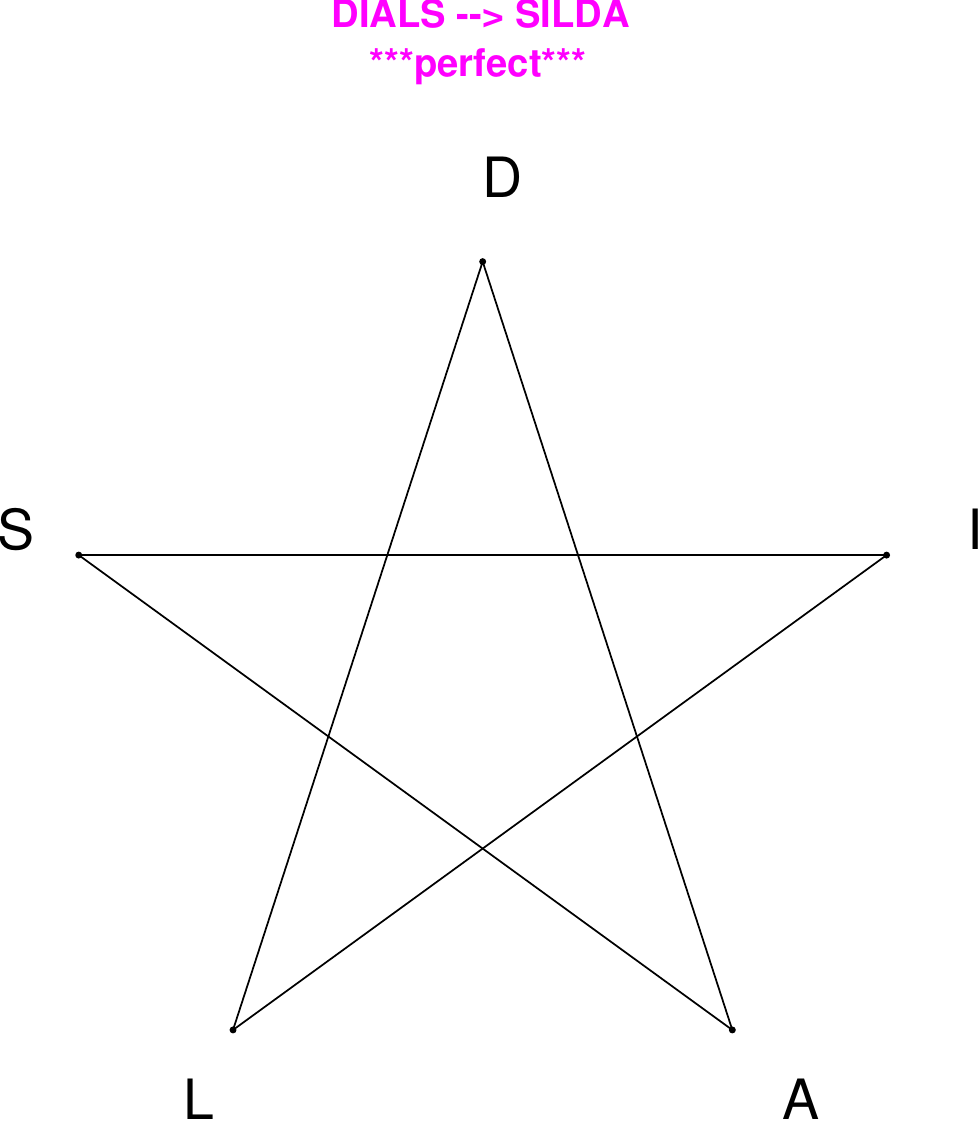}
\end{subfigure}
\hfill
\begin{subfigure}[T]{0.19\textwidth}
\centering
\includegraphics[width=\textwidth]{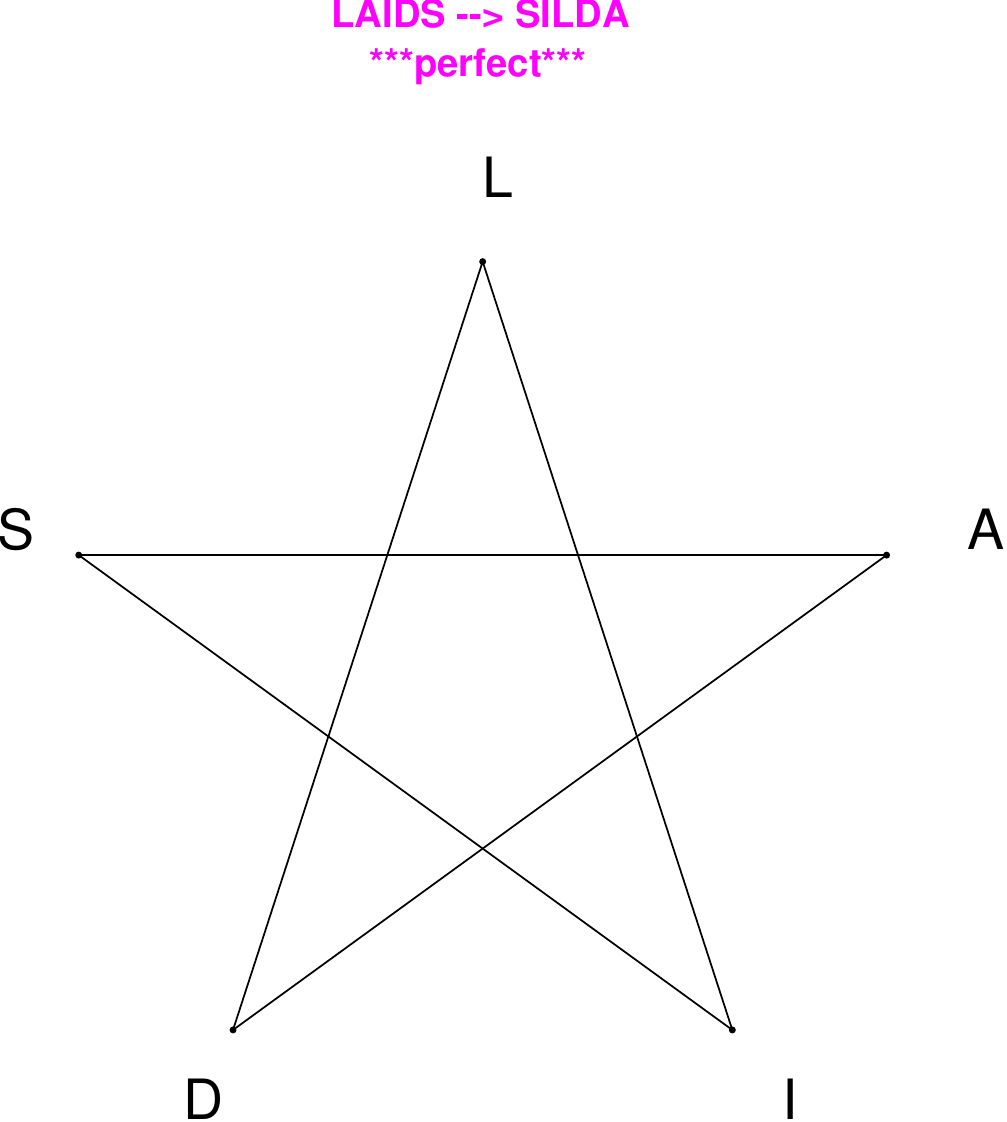}
\end{subfigure}
\hfill
\begin{subfigure}[T]{0.19\textwidth}
\centering
\includegraphics[width=\textwidth]{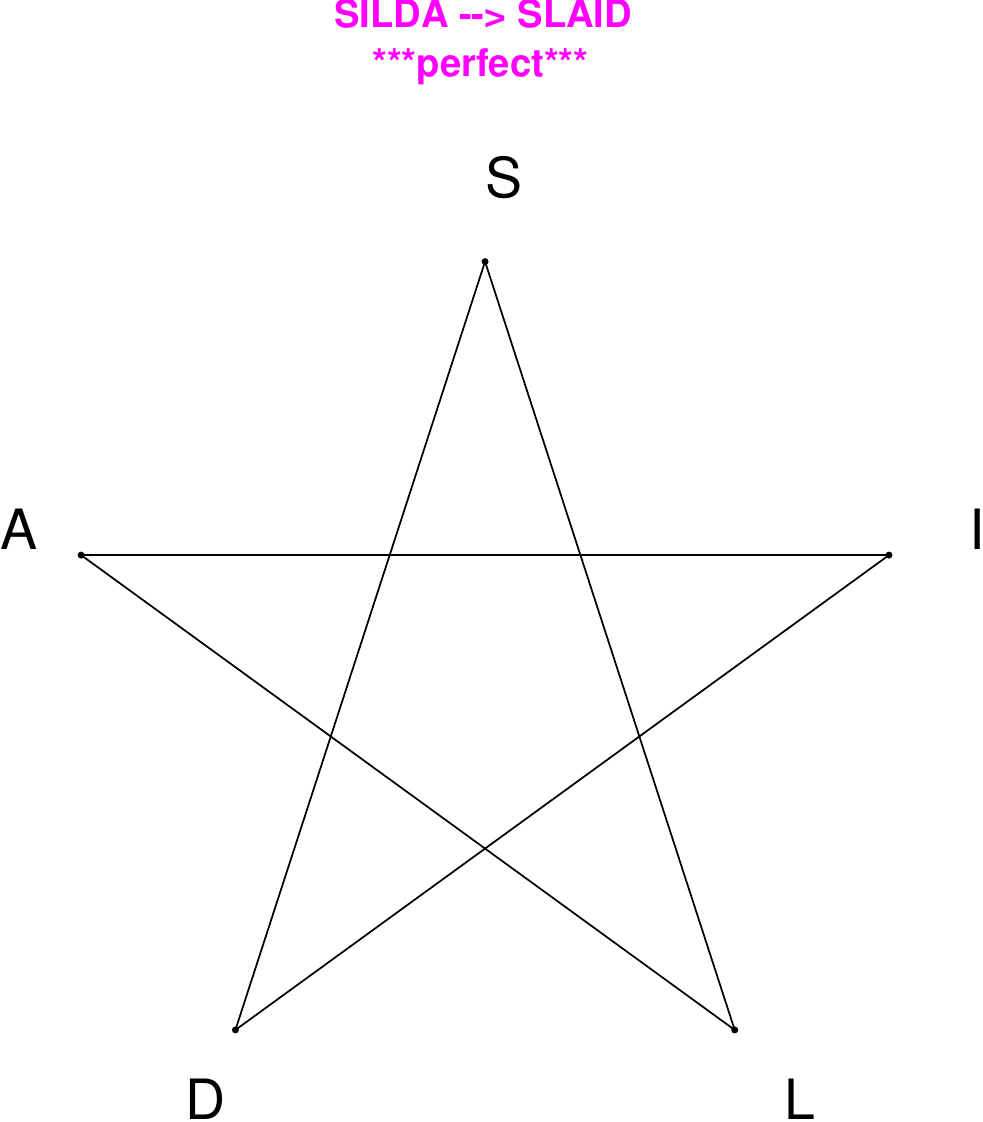}
\end{subfigure}
\end{figure}

\begin{figure}[H]
\centering
\begin{subfigure}[T]{0.19\textwidth}
\centering
\includegraphics[width=\textwidth]{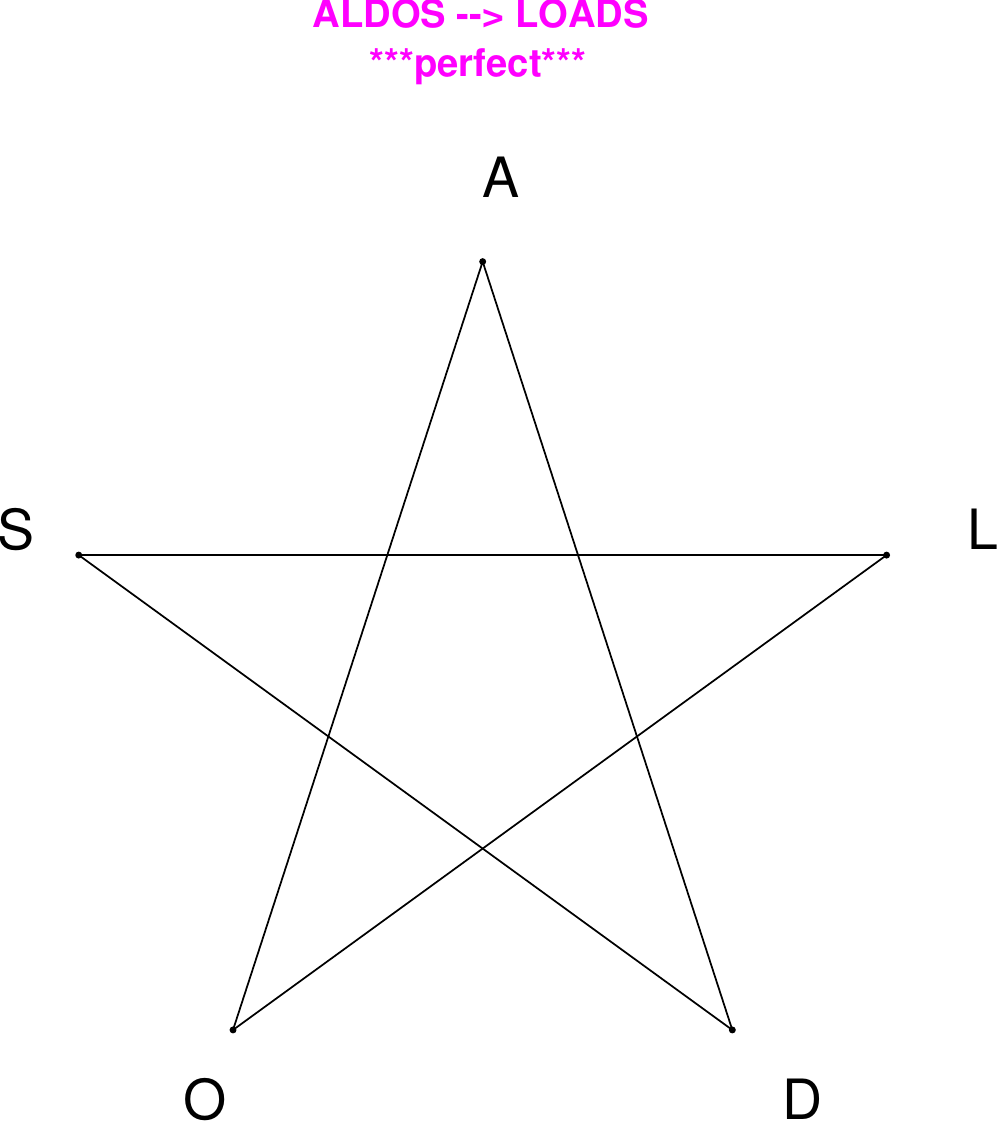}
\end{subfigure}
\hfill
\begin{subfigure}[T]{0.19\textwidth}
\centering
\includegraphics[width=\textwidth]{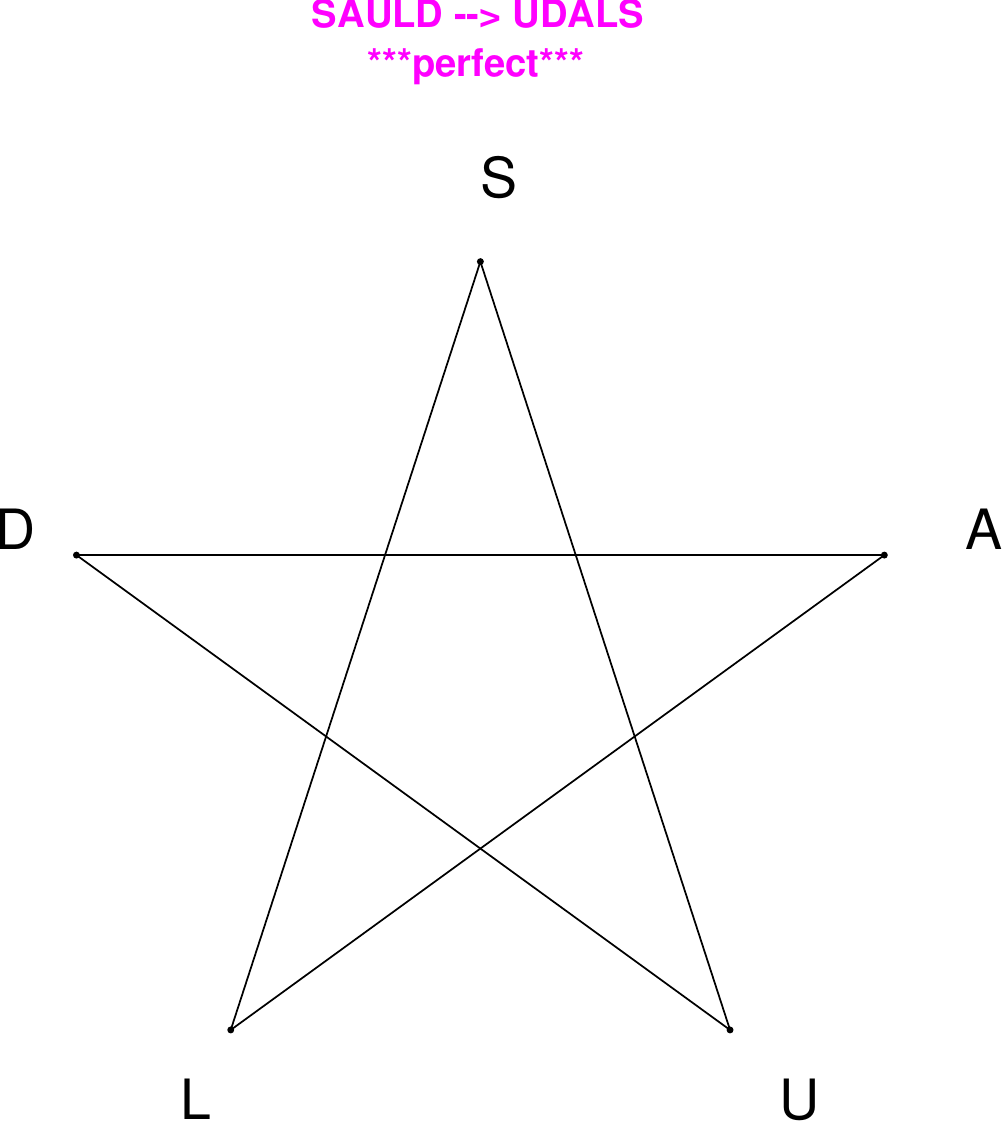}
\end{subfigure}
\hfill
\begin{subfigure}[T]{0.19\textwidth}
\centering
\includegraphics[width=\textwidth]{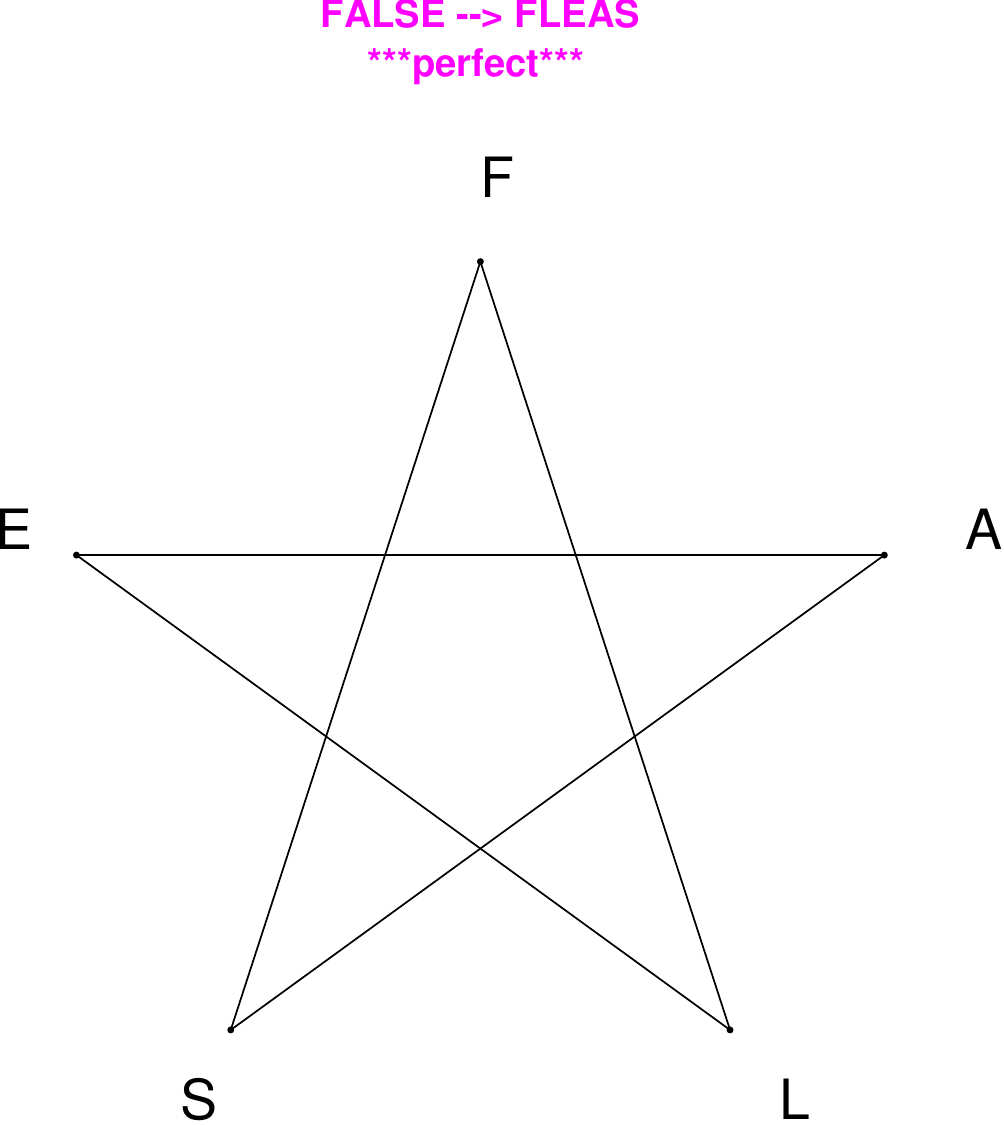}
\end{subfigure}
\hfill
\begin{subfigure}[T]{0.19\textwidth}
\centering
\includegraphics[width=\textwidth]{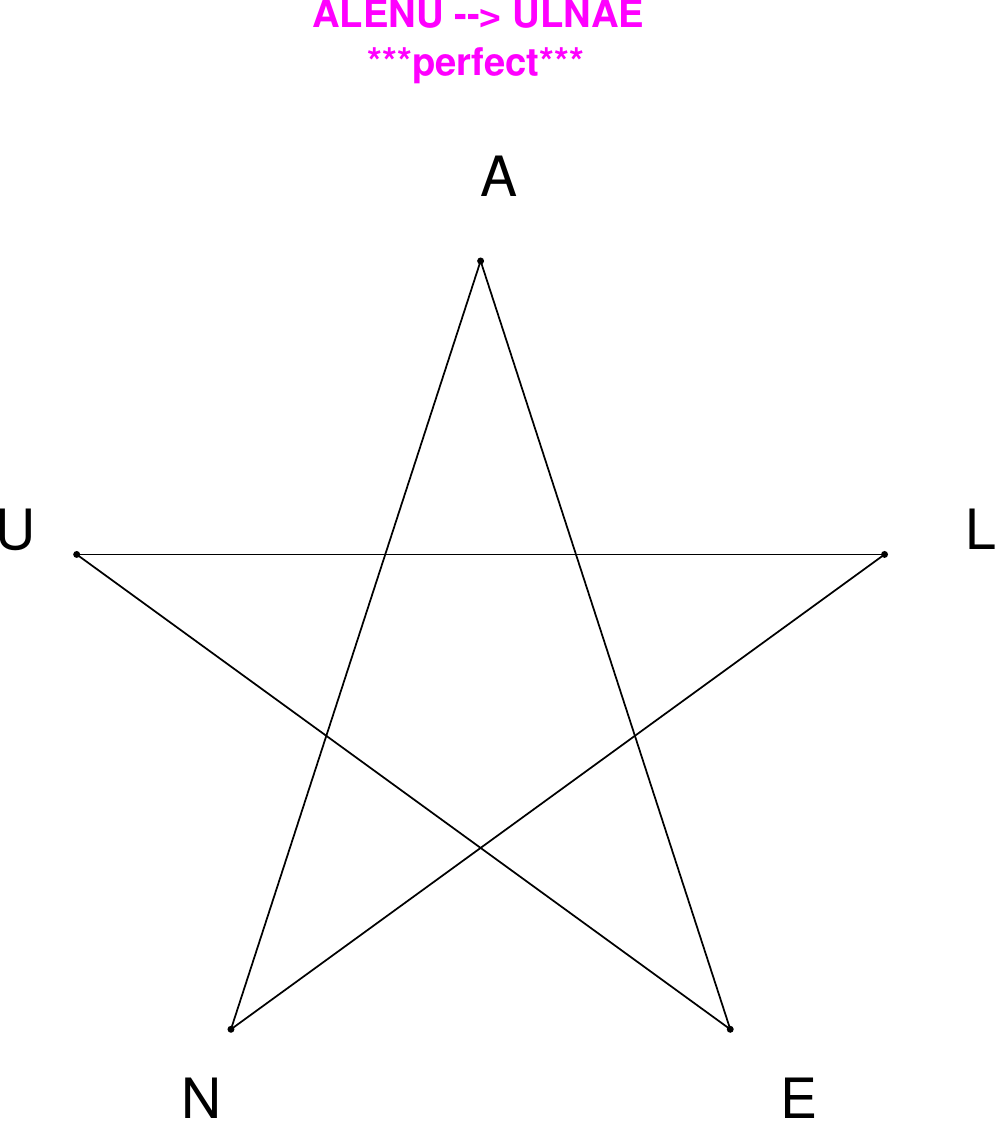}
\end{subfigure}
\hfill
\begin{subfigure}[T]{0.19\textwidth}
\centering
\includegraphics[width=\textwidth]{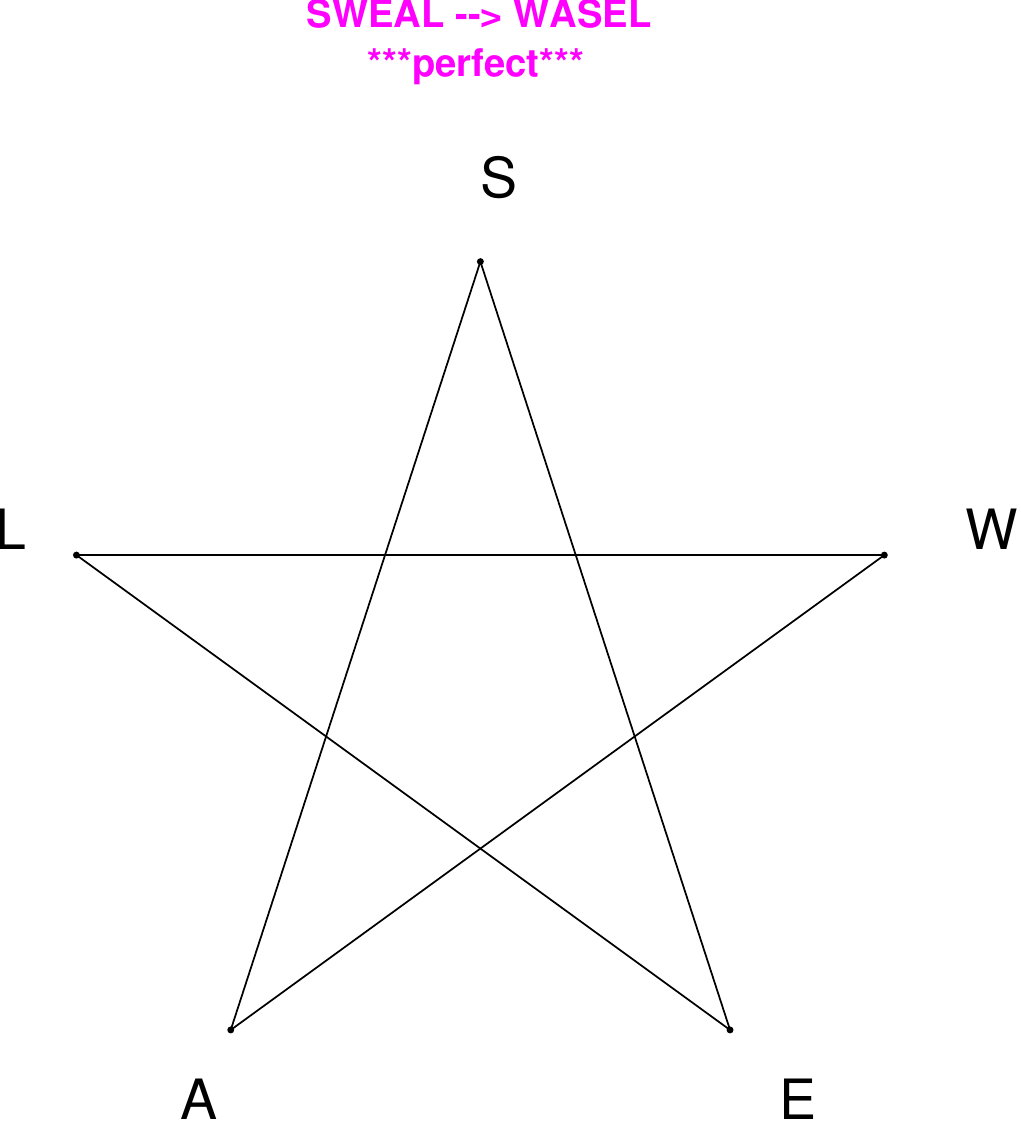}
\end{subfigure}
\end{figure}

\begin{figure}[H]
\centering
\begin{subfigure}[T]{0.19\textwidth}
\centering
\includegraphics[width=\textwidth]{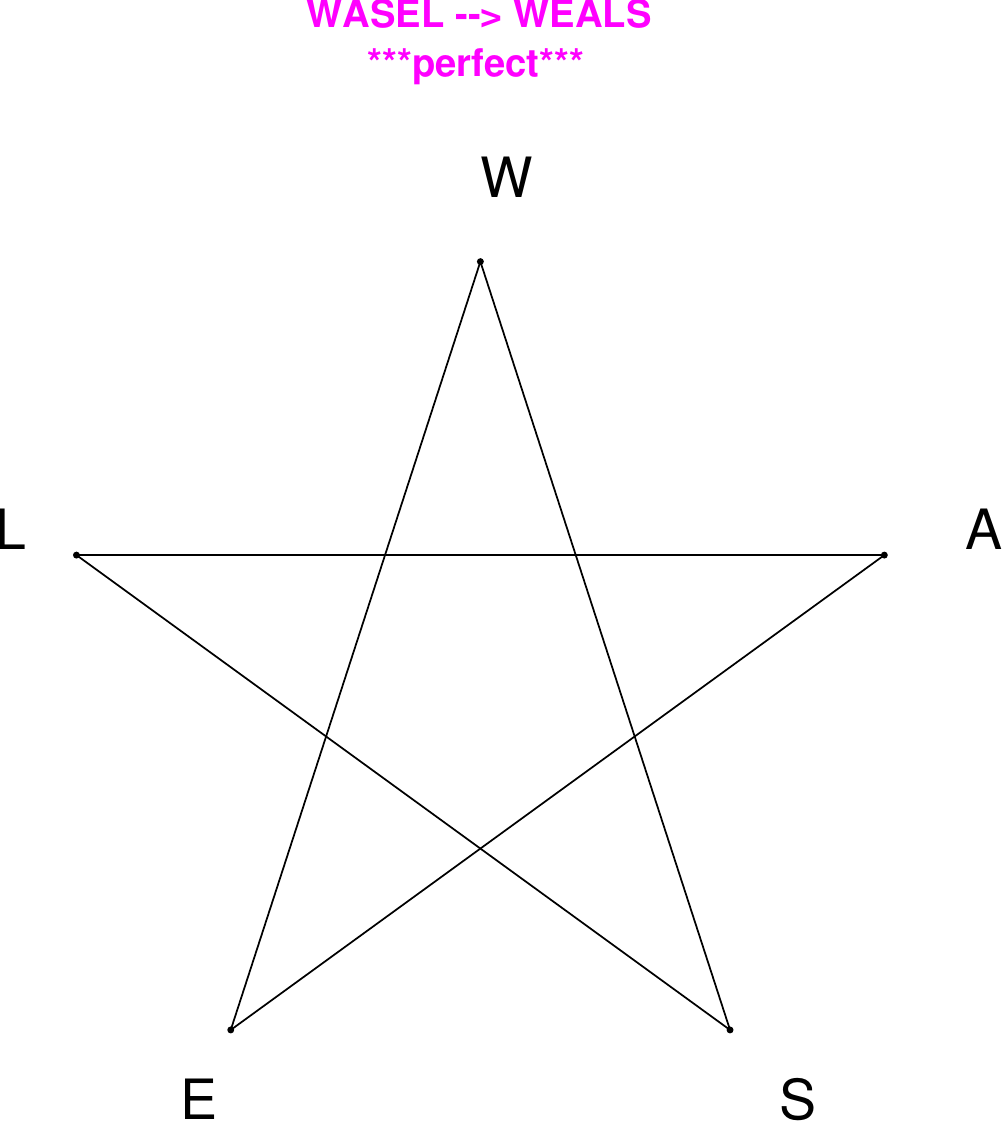}
\end{subfigure}
\hfill
\begin{subfigure}[T]{0.19\textwidth}
\centering
\includegraphics[width=\textwidth]{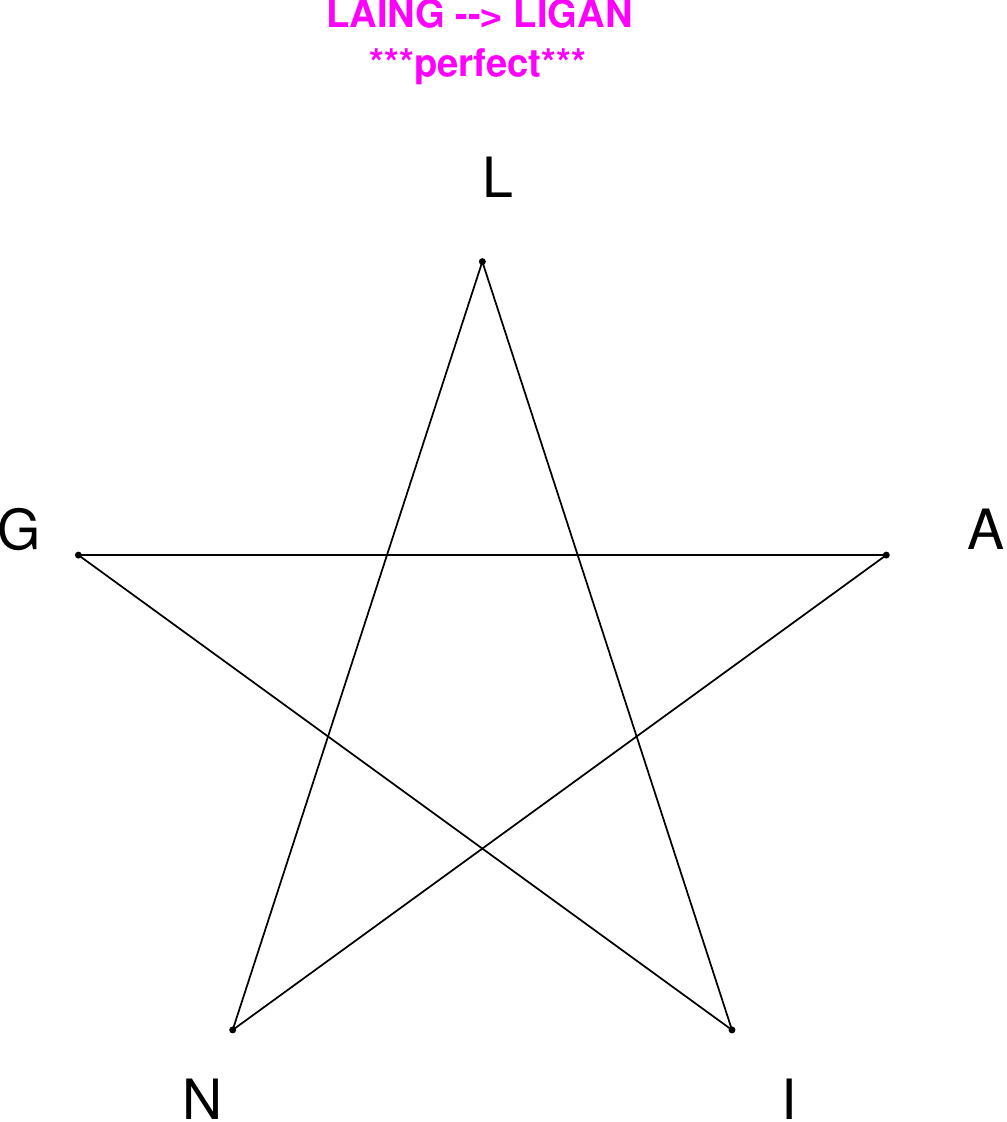}
\end{subfigure}
\hfill
\begin{subfigure}[T]{0.19\textwidth}
\centering
\includegraphics[width=\textwidth]{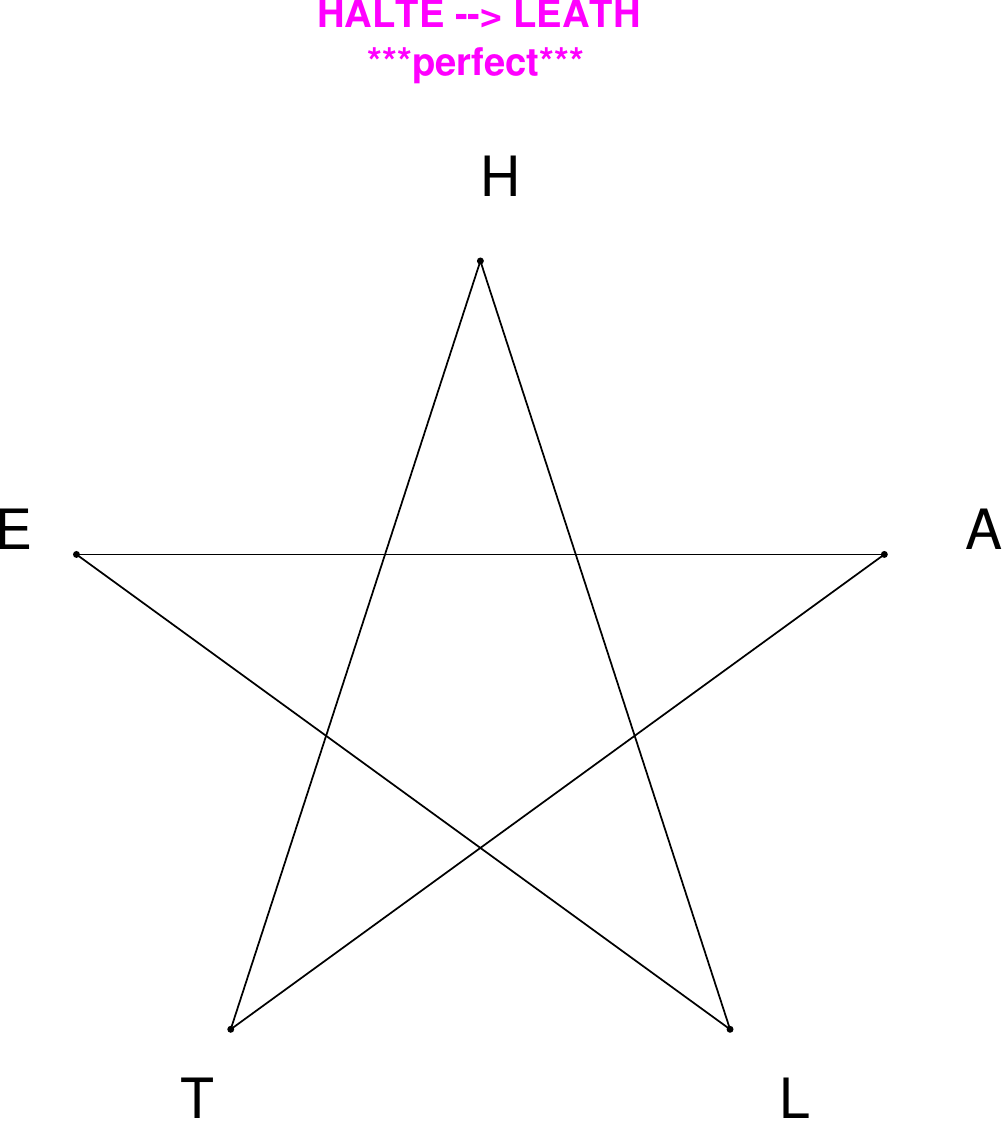}
\end{subfigure}
\hfill
\begin{subfigure}[T]{0.19\textwidth}
\centering
\includegraphics[width=\textwidth]{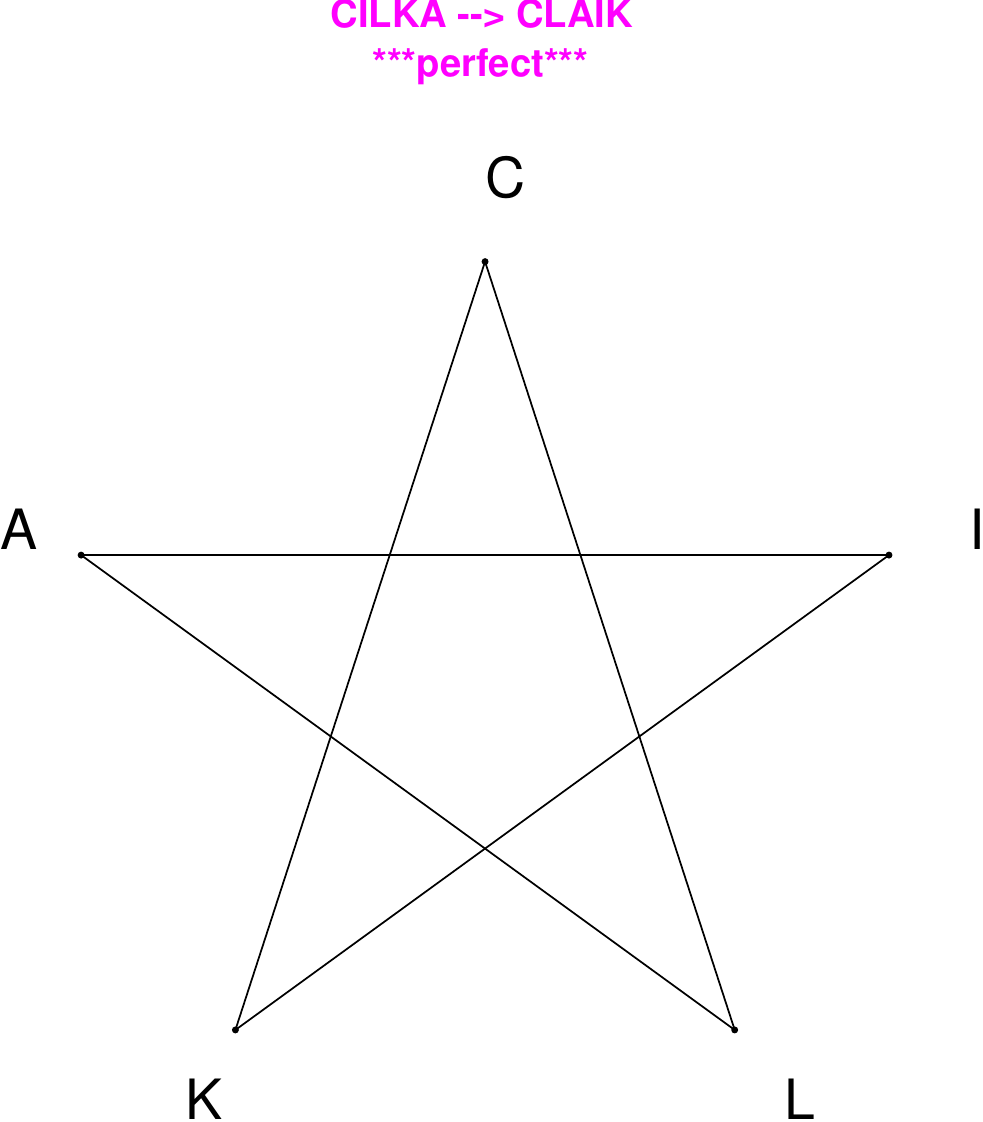}
\end{subfigure}
\hfill
\begin{subfigure}[T]{0.19\textwidth}
\centering
\includegraphics[width=\textwidth]{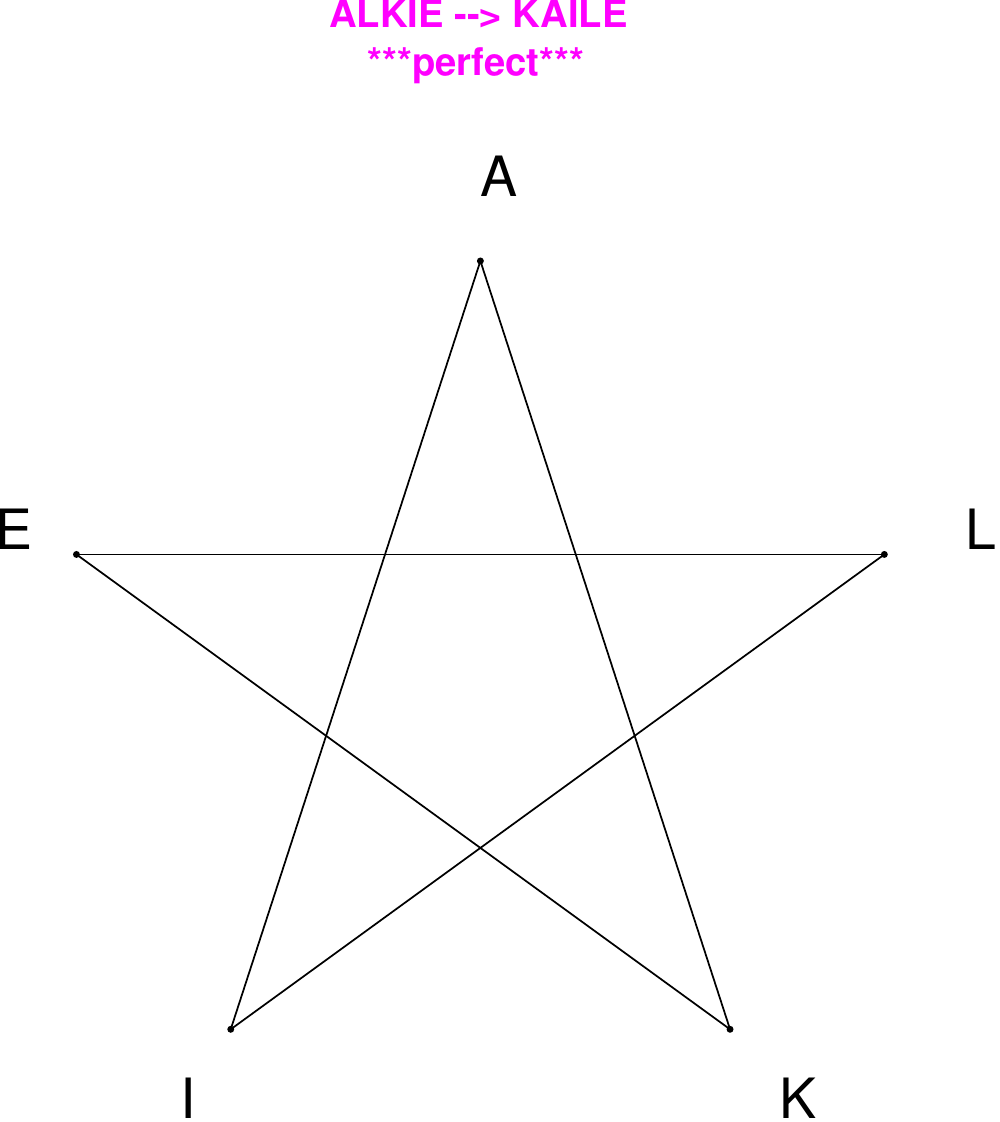}
\end{subfigure}
\end{figure}

\begin{figure}[H]
\centering
\begin{subfigure}[T]{0.19\textwidth}
\centering
\includegraphics[width=\textwidth]{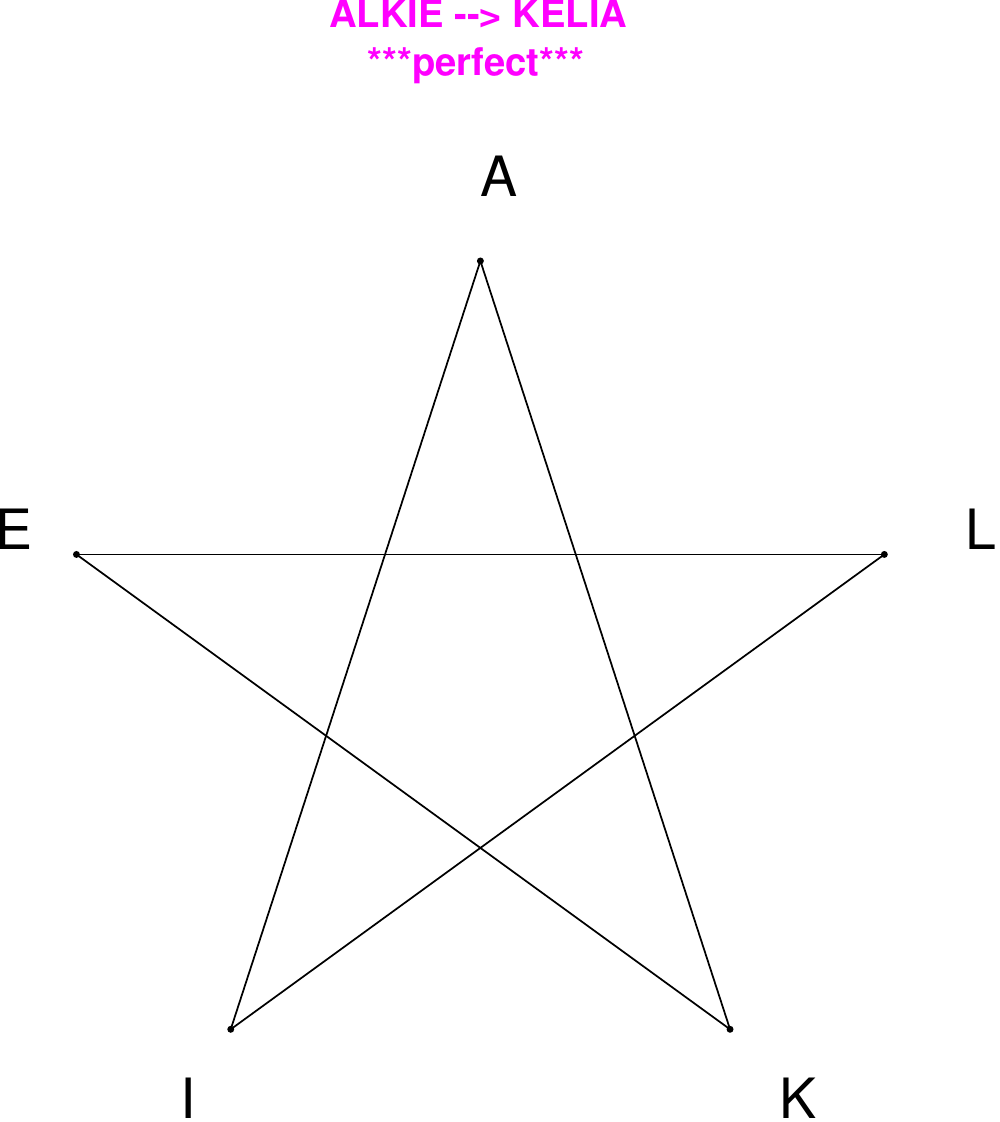}
\end{subfigure}
\hfill
\begin{subfigure}[T]{0.19\textwidth}
\centering
\includegraphics[width=\textwidth]{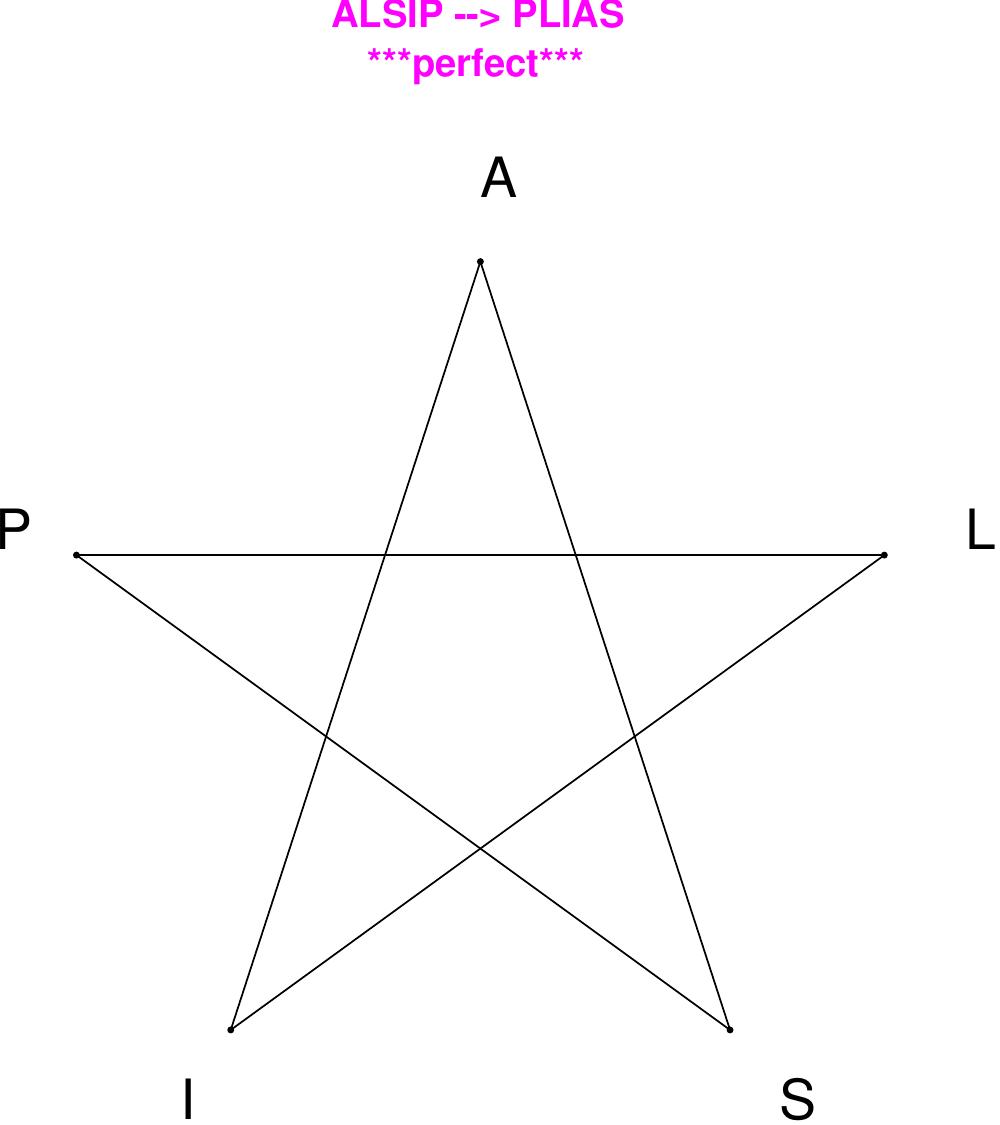}
\end{subfigure}
\hfill
\begin{subfigure}[T]{0.19\textwidth}
\centering
\includegraphics[width=\textwidth]{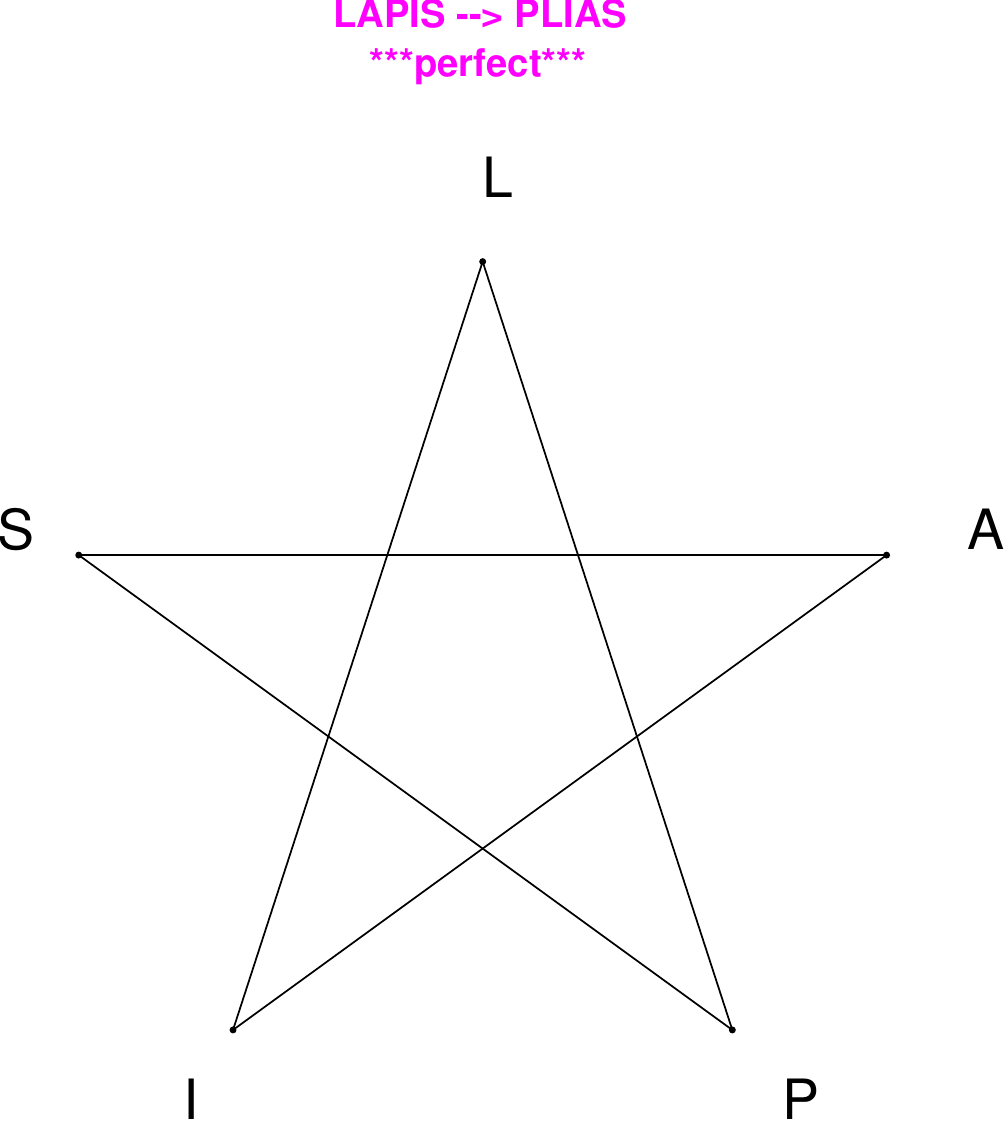}
\end{subfigure}
\hfill
\begin{subfigure}[T]{0.19\textwidth}
\centering
\includegraphics[width=\textwidth]{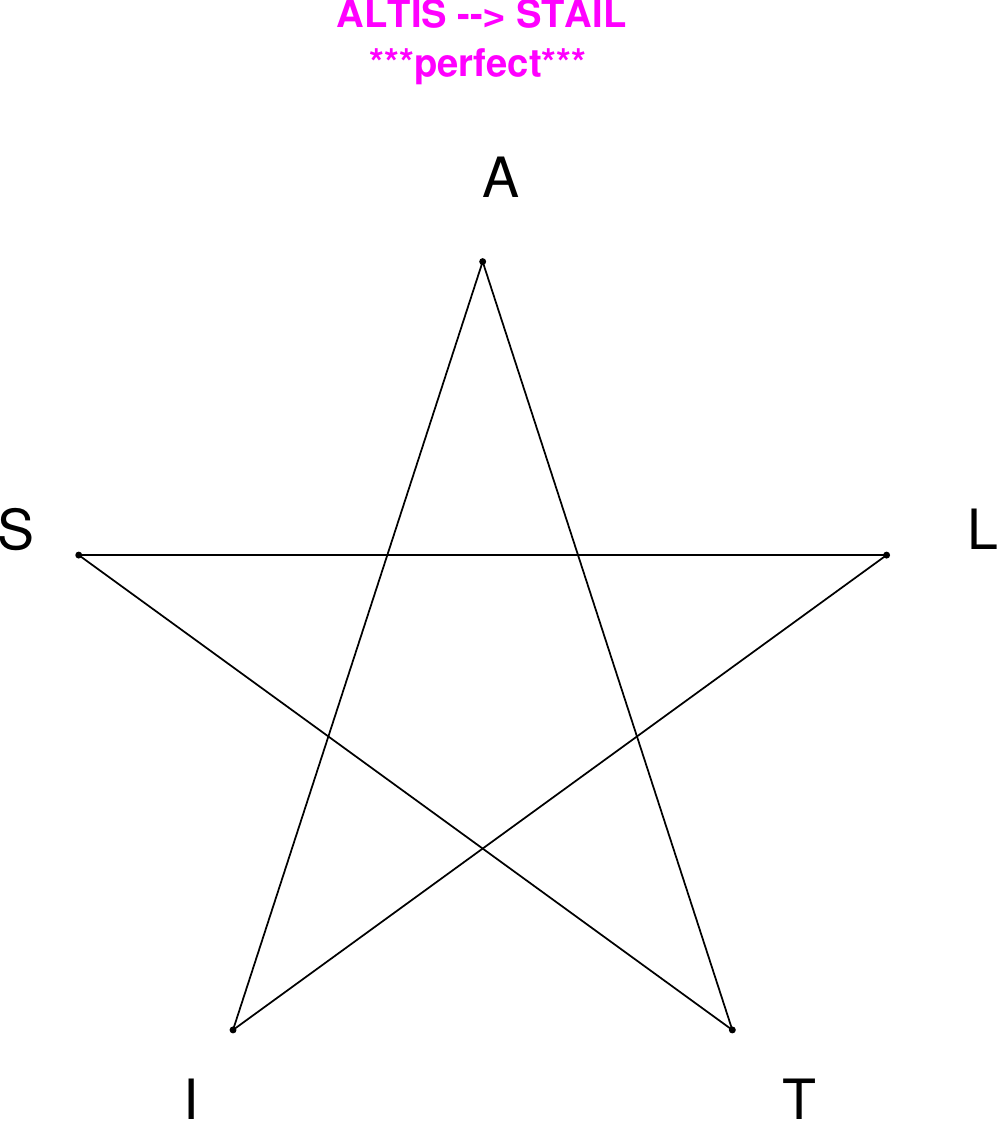}
\end{subfigure}
\hfill
\begin{subfigure}[T]{0.19\textwidth}
\centering
\includegraphics[width=\textwidth]{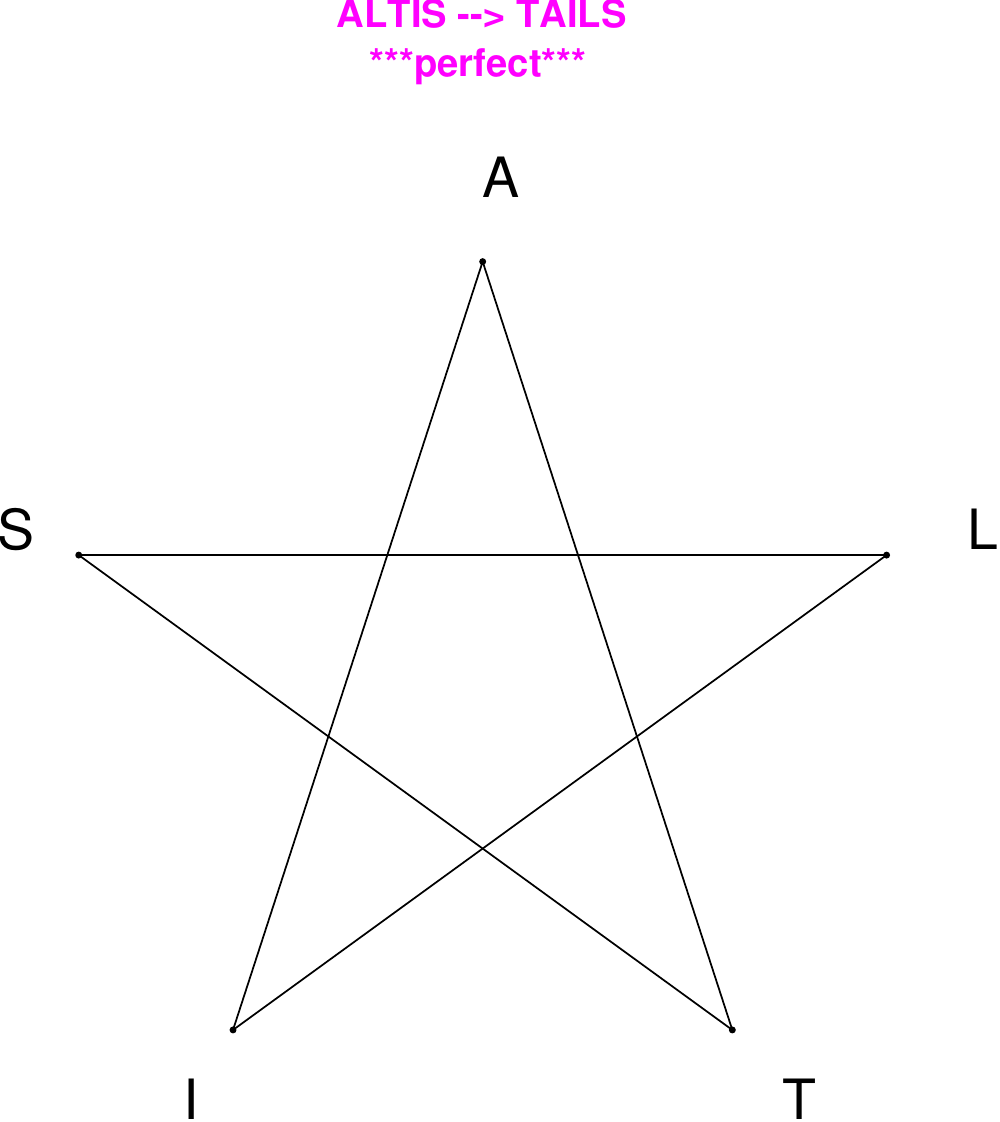}
\end{subfigure}
\end{figure}

\begin{figure}[H]
\centering
\begin{subfigure}[T]{0.19\textwidth}
\centering
\includegraphics[width=\textwidth]{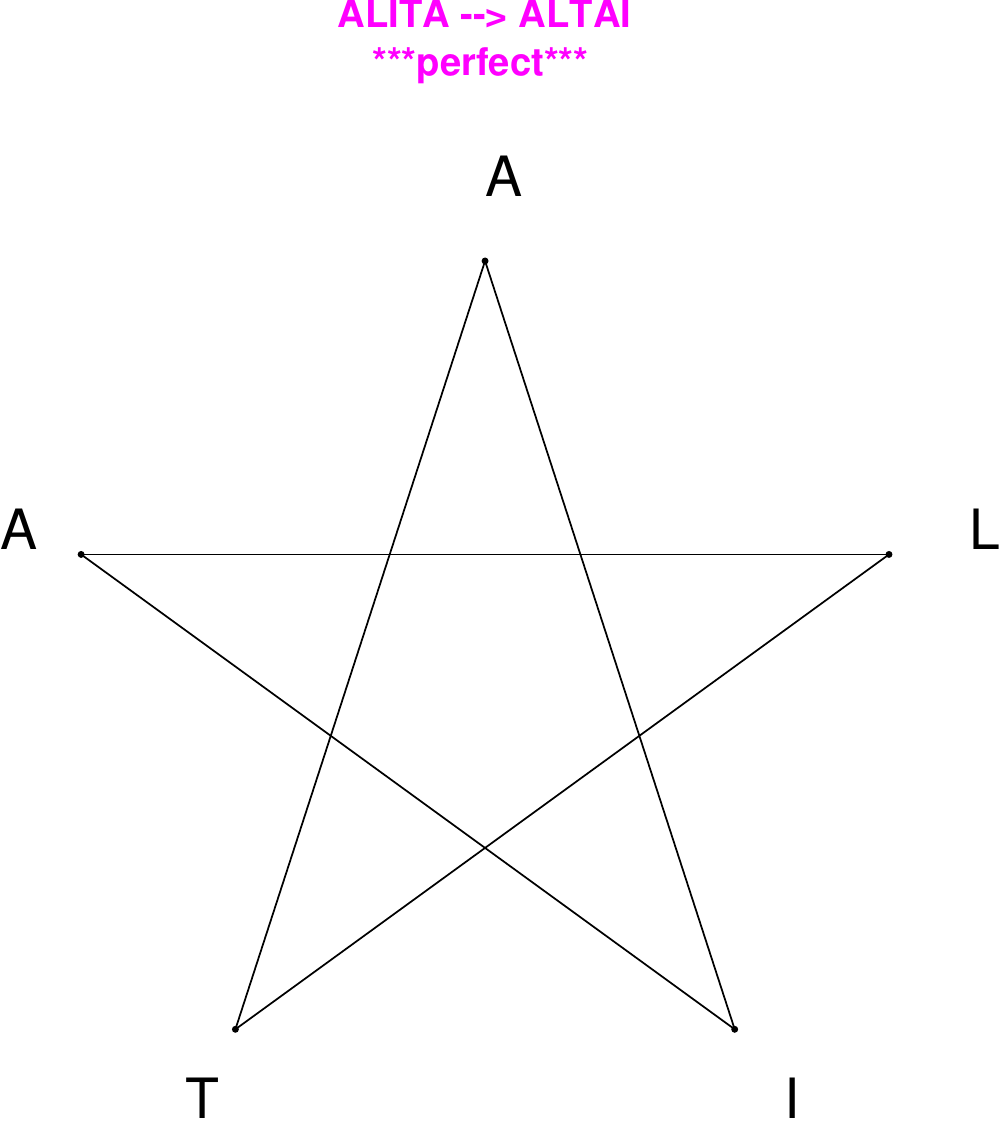}
\end{subfigure}
\hfill
\begin{subfigure}[T]{0.19\textwidth}
\centering
\includegraphics[width=\textwidth]{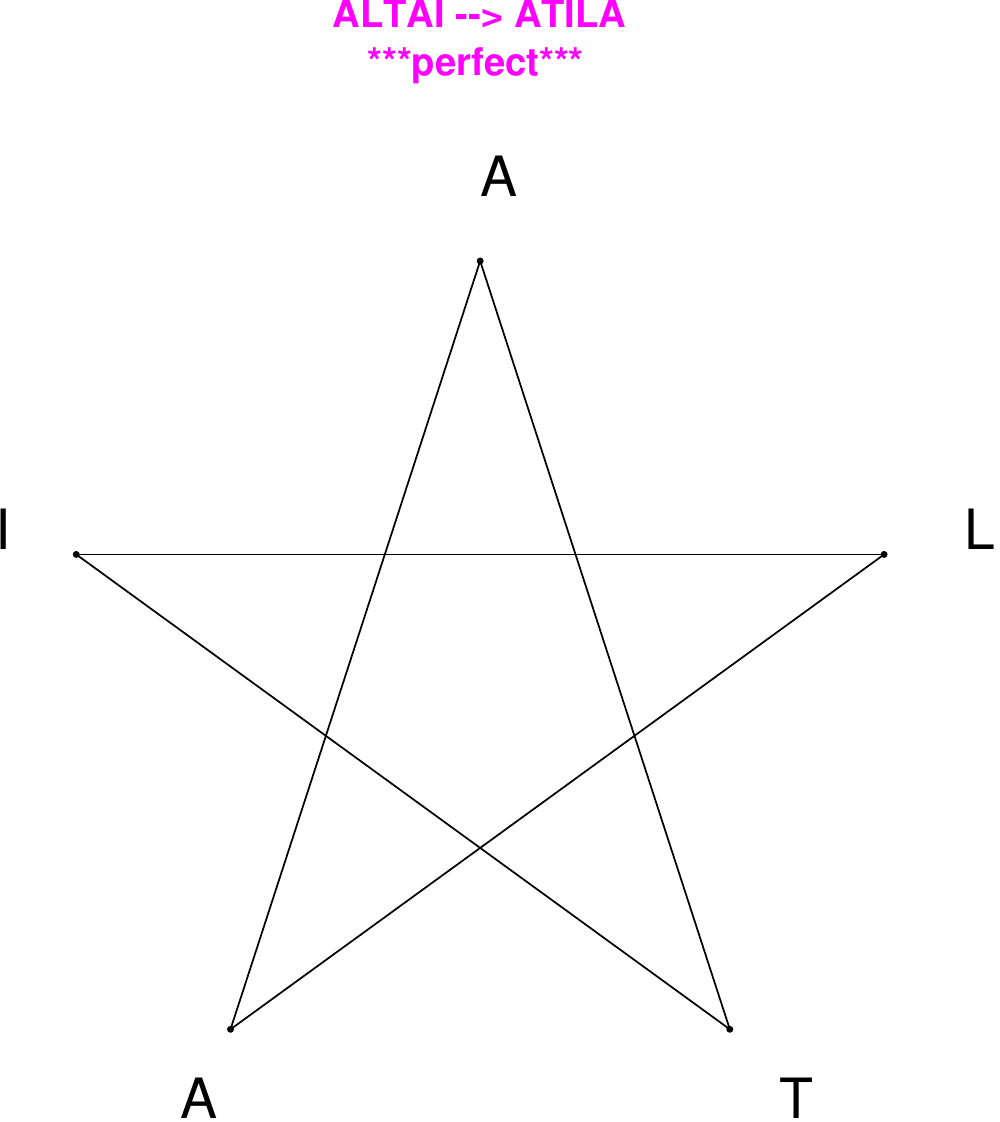}
\end{subfigure}
\hfill
\begin{subfigure}[T]{0.19\textwidth}
\centering
\includegraphics[width=\textwidth]{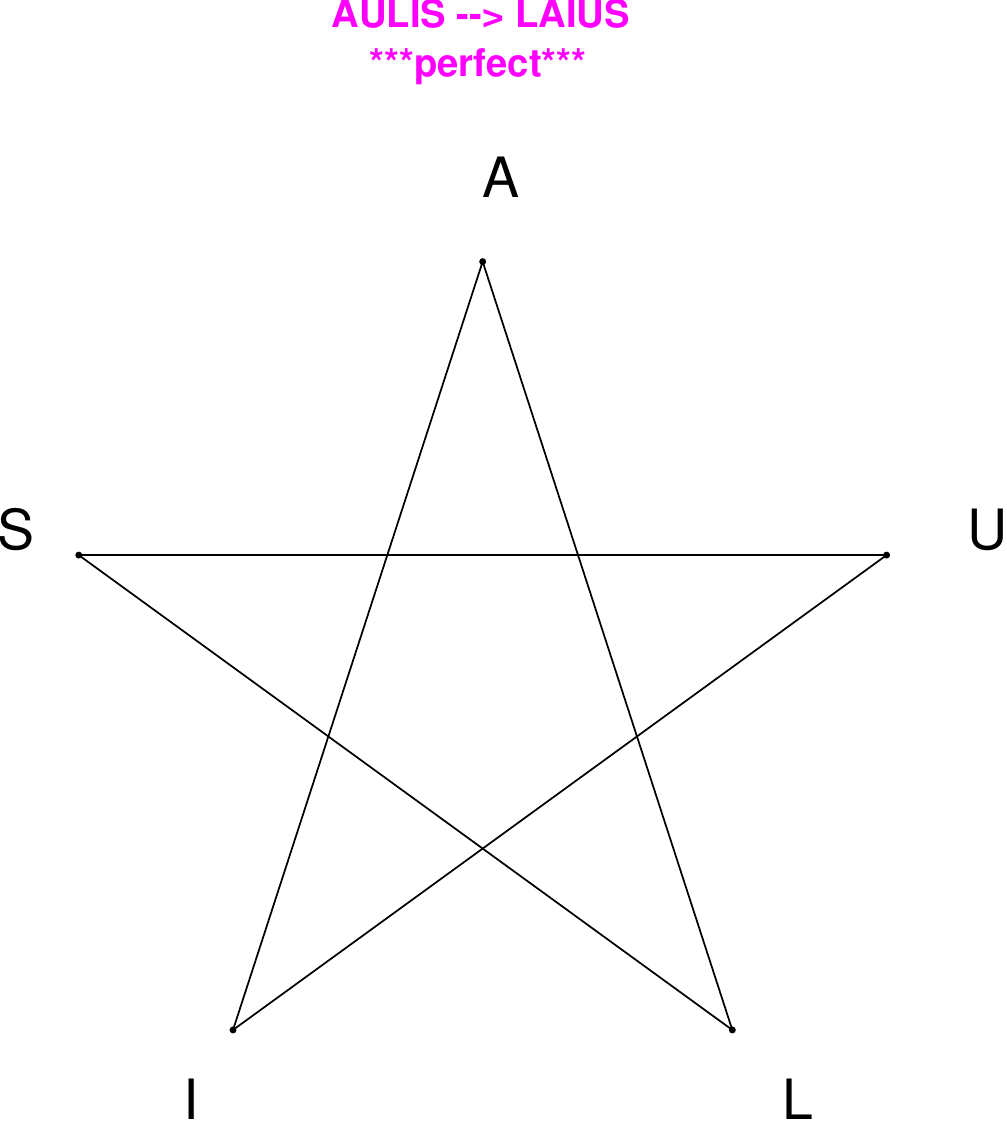}
\end{subfigure}
\hfill
\begin{subfigure}[T]{0.19\textwidth}
\centering
\includegraphics[width=\textwidth]{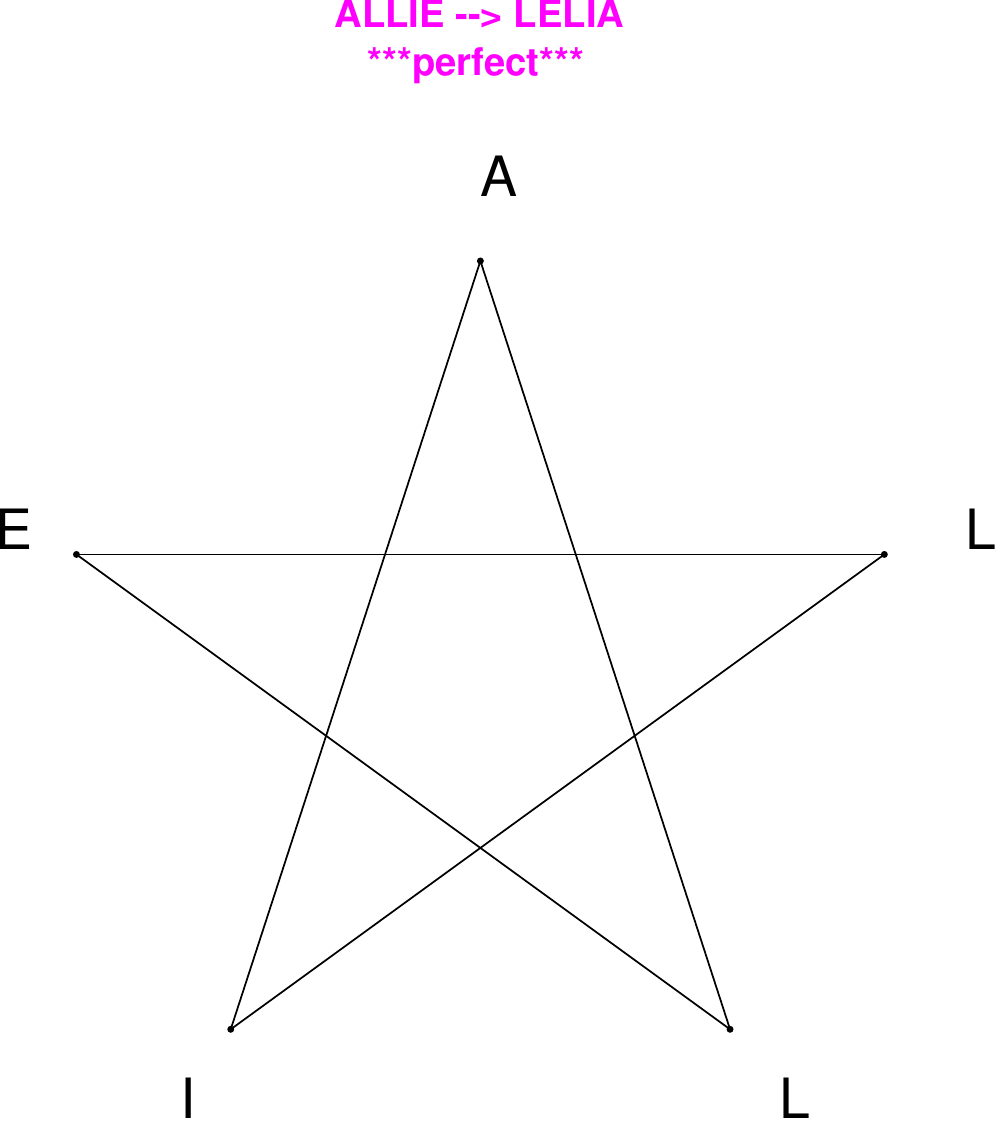}
\end{subfigure}
\hfill
\begin{subfigure}[T]{0.19\textwidth}
\centering
\includegraphics[width=\textwidth]{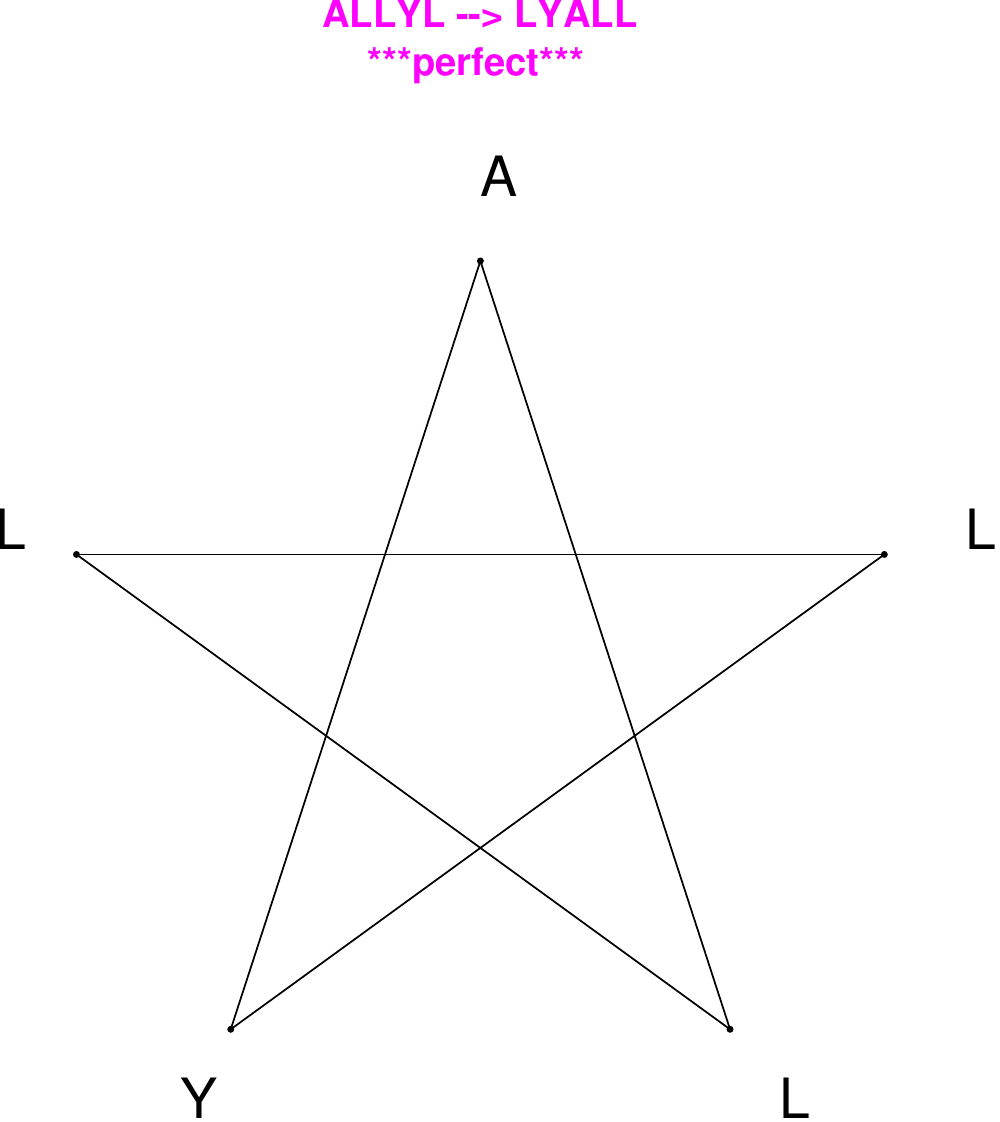}
\end{subfigure}
\end{figure}

\begin{figure}[H]
\centering
\begin{subfigure}[T]{0.19\textwidth}
\centering
\includegraphics[width=\textwidth]{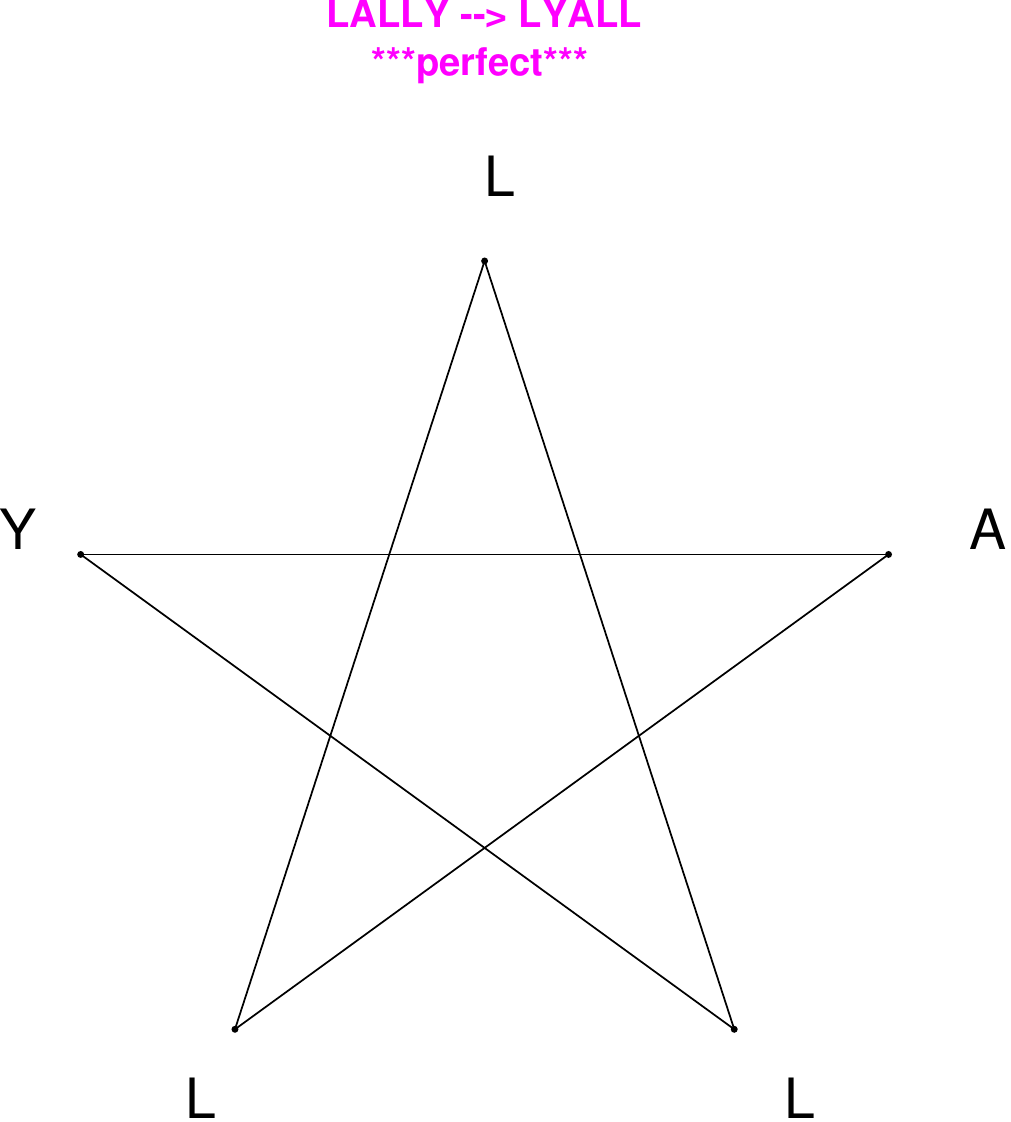}
\end{subfigure}
\hfill
\begin{subfigure}[T]{0.19\textwidth}
\centering
\includegraphics[width=\textwidth]{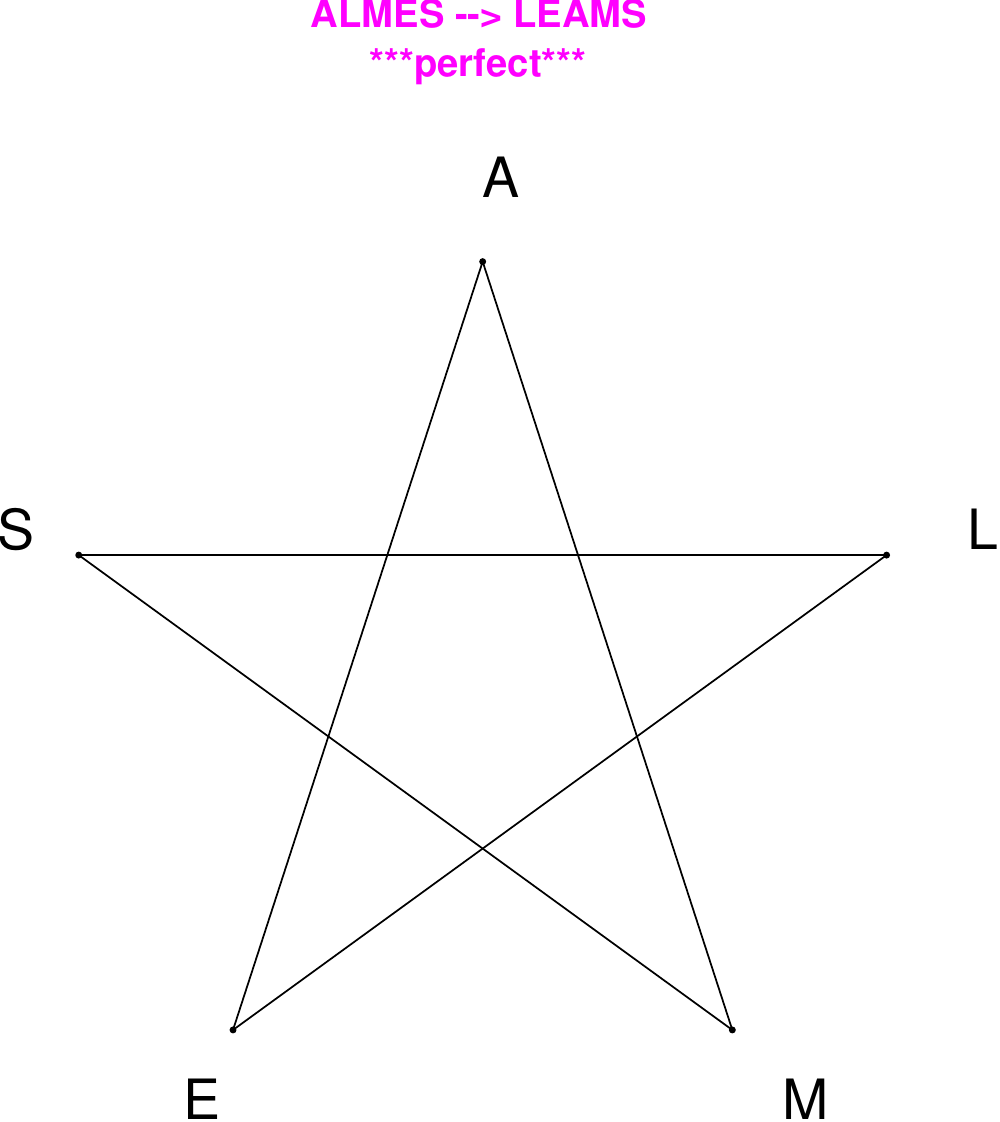}
\end{subfigure}
\hfill
\begin{subfigure}[T]{0.19\textwidth}
\centering
\includegraphics[width=\textwidth]{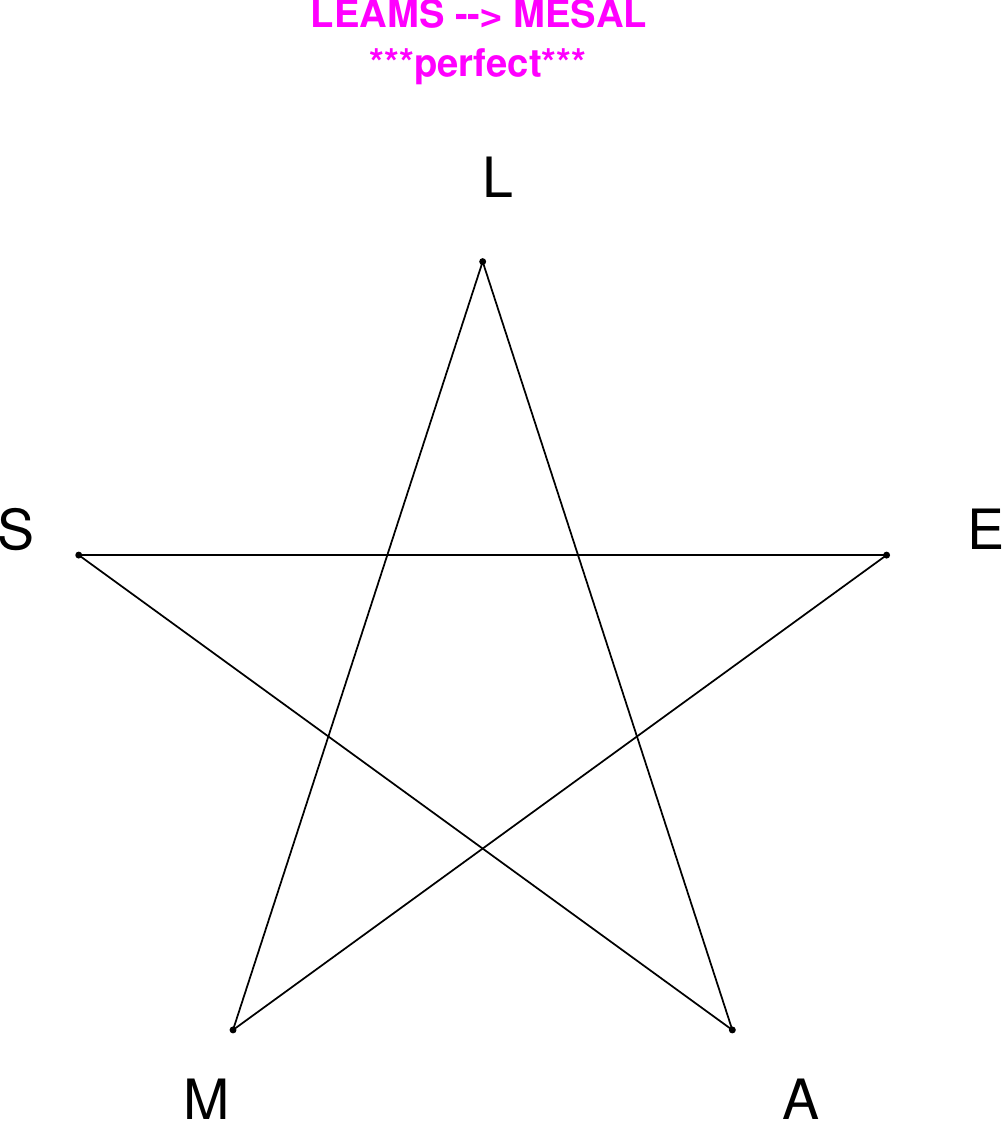}
\end{subfigure}
\hfill
\begin{subfigure}[T]{0.19\textwidth}
\centering
\includegraphics[width=\textwidth]{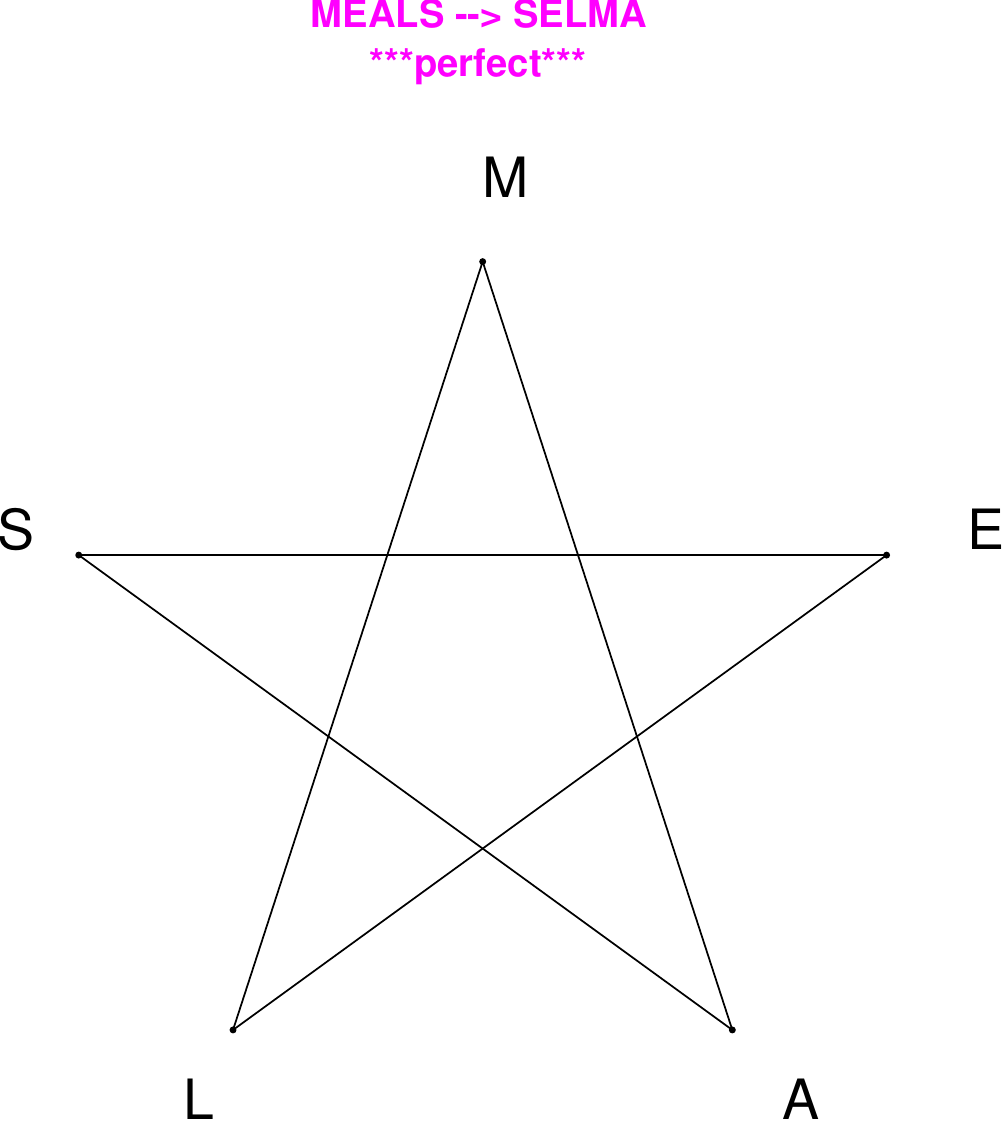}
\end{subfigure}
\hfill
\begin{subfigure}[T]{0.19\textwidth}
\centering
\includegraphics[width=\textwidth]{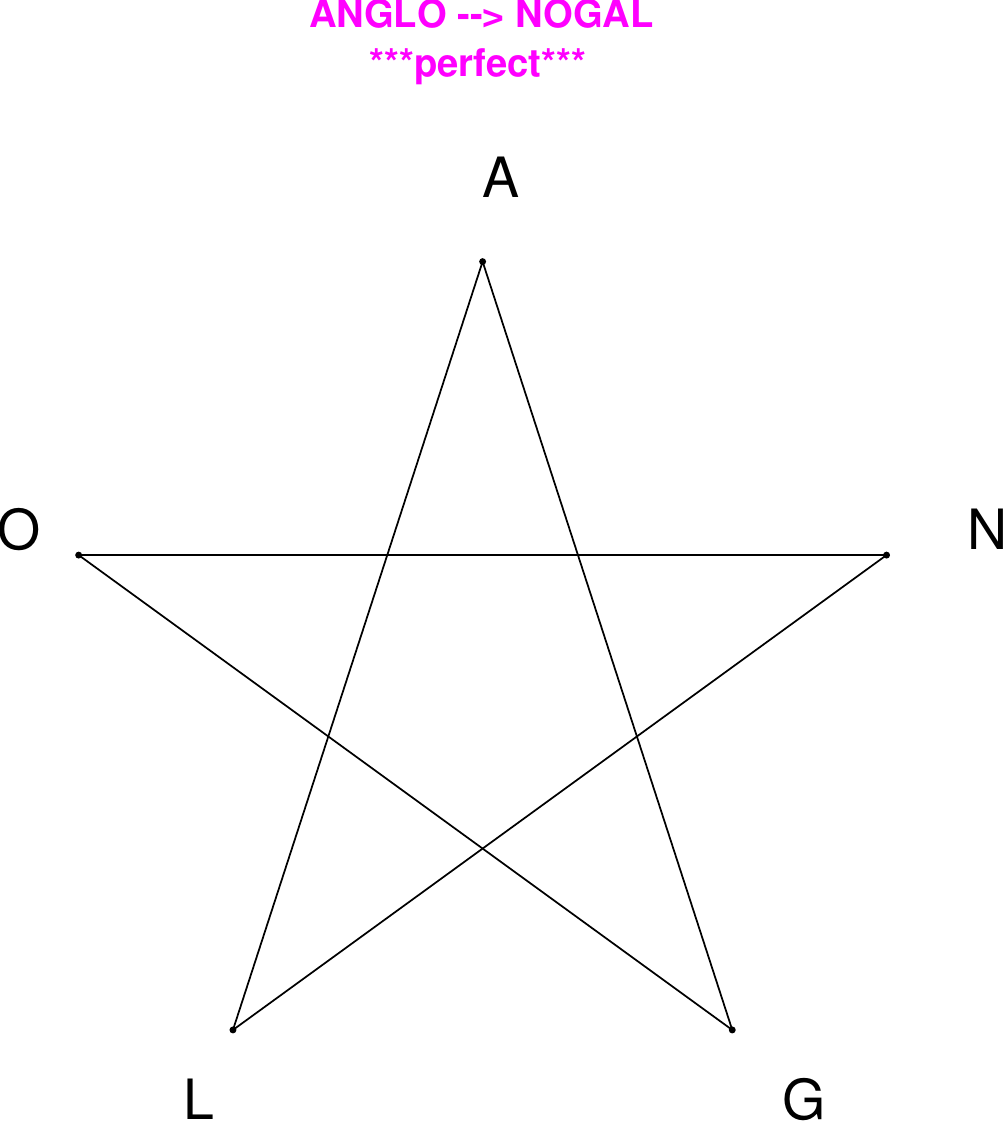}
\end{subfigure}
\end{figure}

\begin{figure}[H]
\centering
\begin{subfigure}[T]{0.19\textwidth}
\centering
\includegraphics[width=\textwidth]{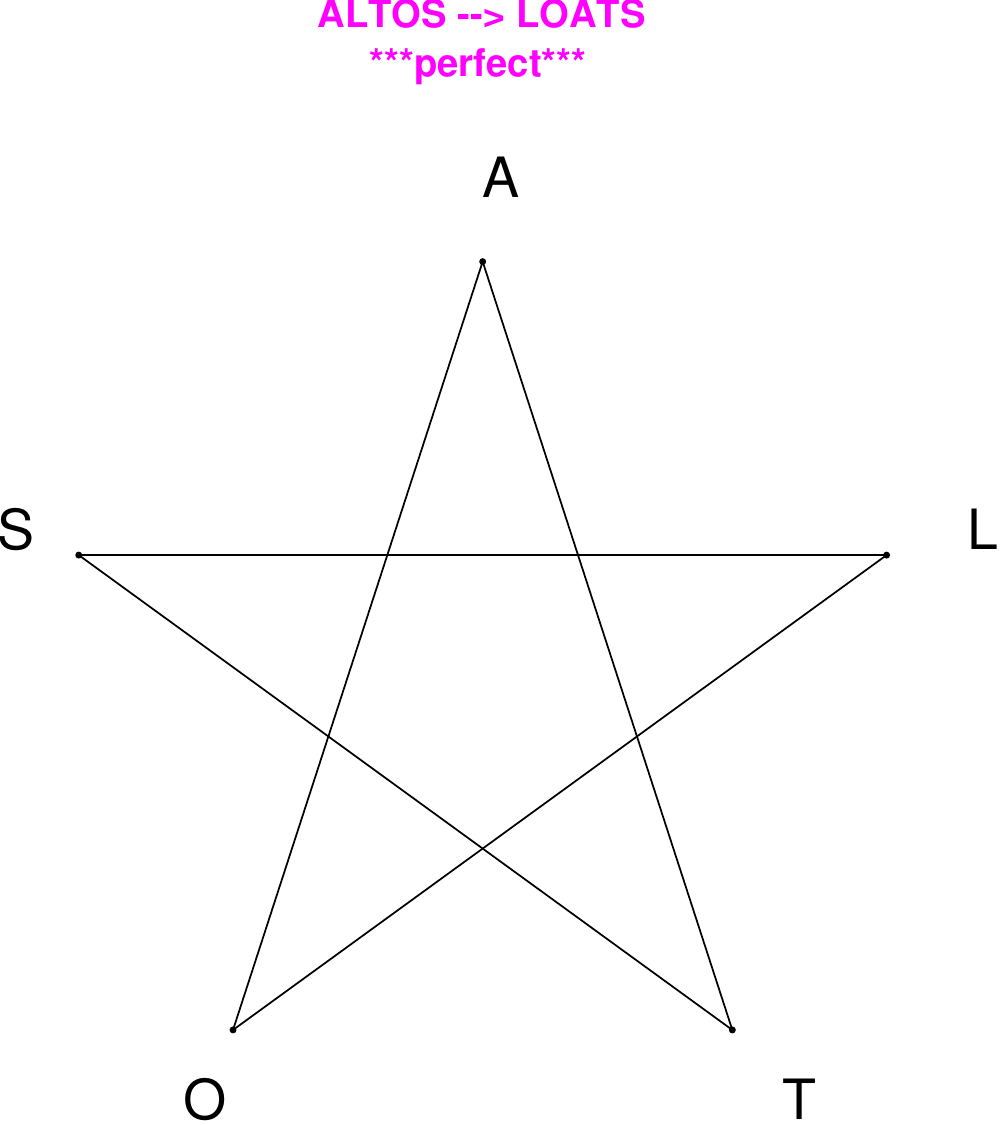}
\end{subfigure}
\hfill
\begin{subfigure}[T]{0.19\textwidth}
\centering
\includegraphics[width=\textwidth]{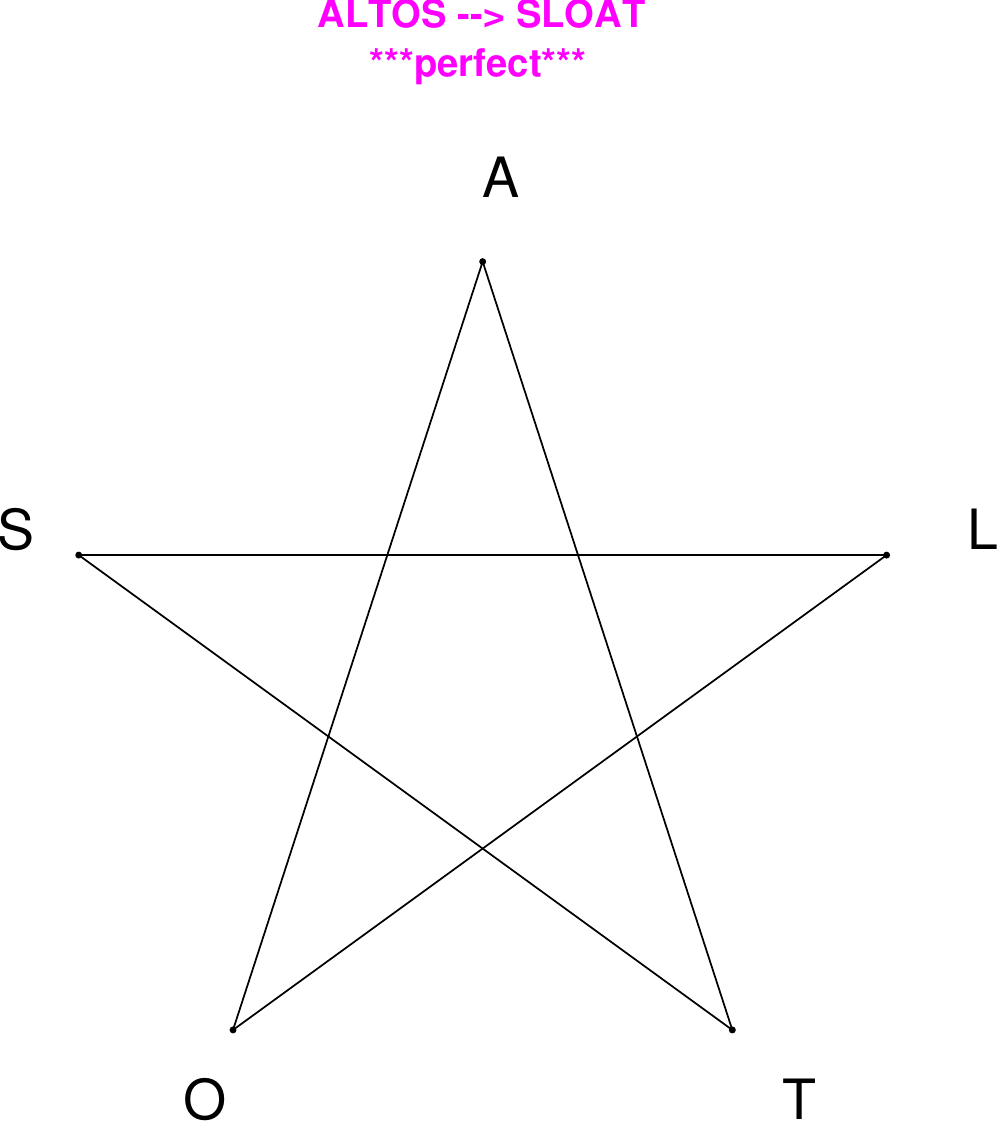}
\end{subfigure}
\hfill
\begin{subfigure}[T]{0.19\textwidth}
\centering
\includegraphics[width=\textwidth]{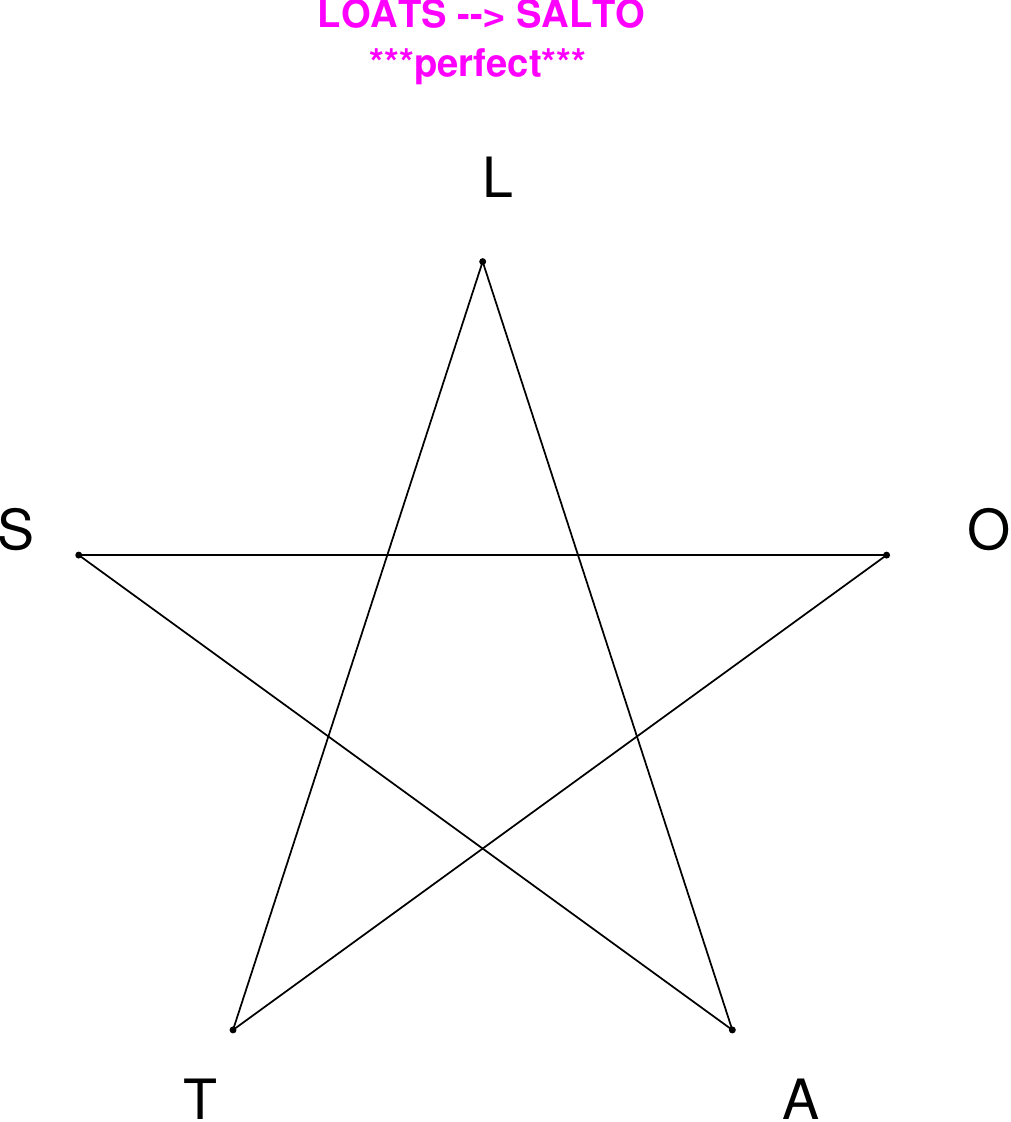}
\end{subfigure}
\hfill
\begin{subfigure}[T]{0.19\textwidth}
\centering
\includegraphics[width=\textwidth]{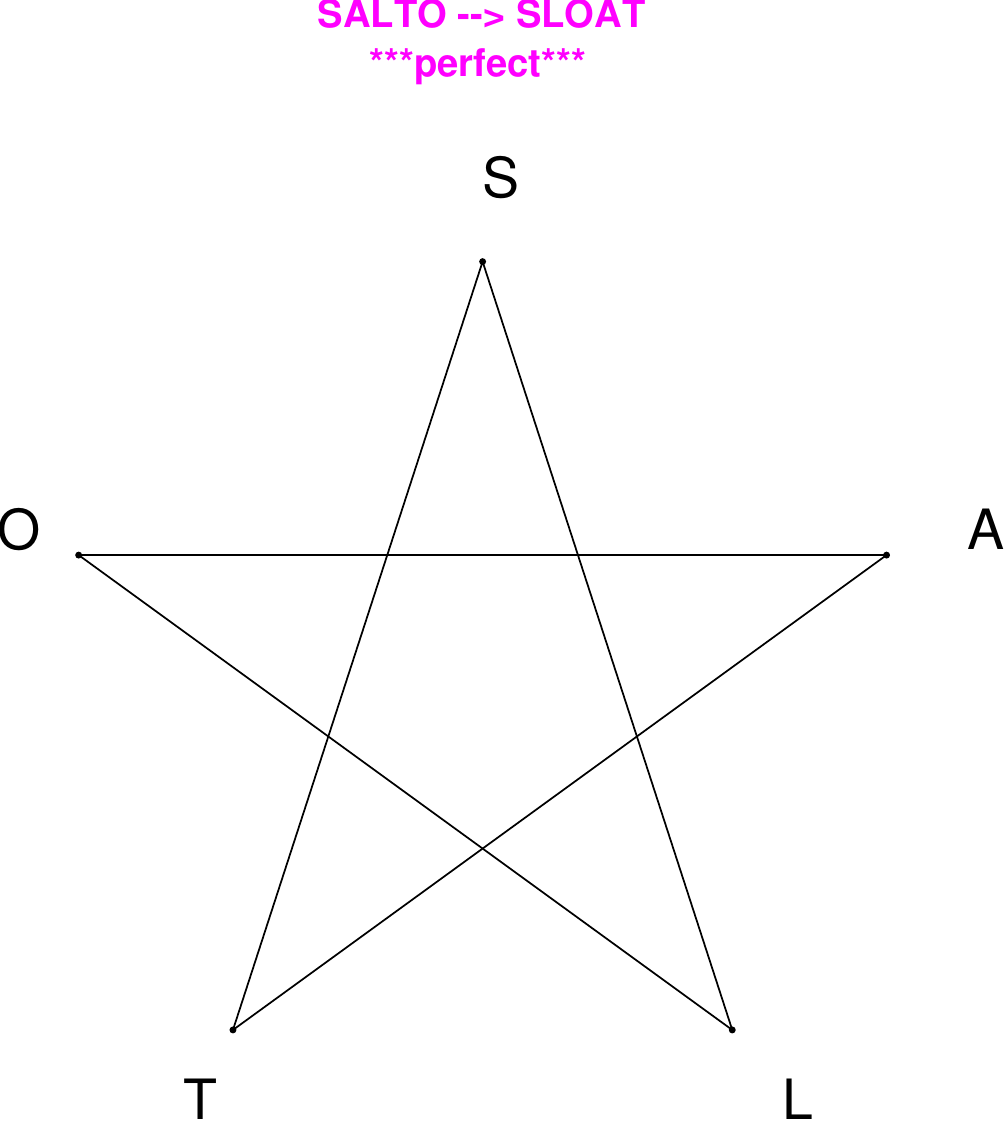}
\end{subfigure}
\hfill
\begin{subfigure}[T]{0.19\textwidth}
\centering
\includegraphics[width=\textwidth]{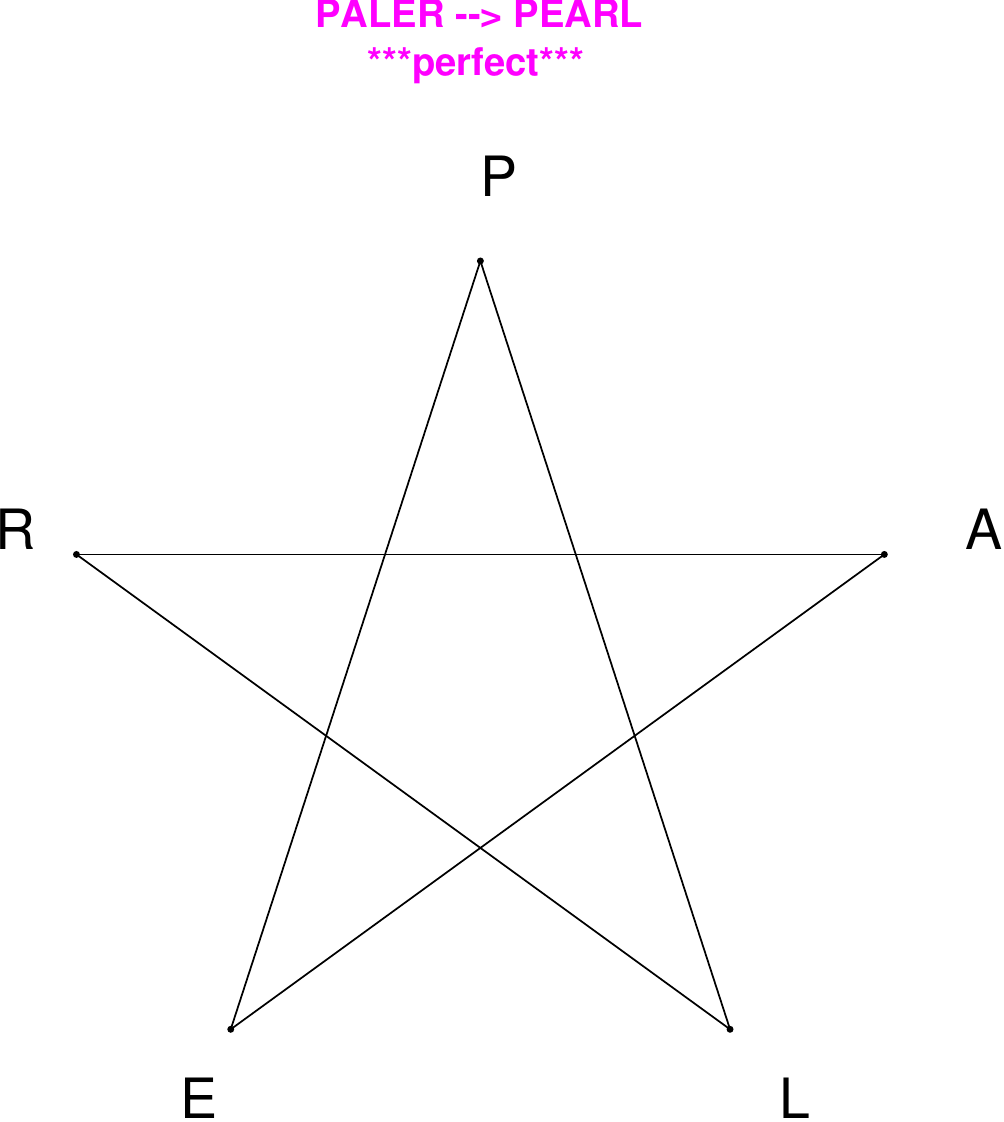}
\end{subfigure}
\end{figure}

\begin{figure}[H]
\centering
\begin{subfigure}[T]{0.19\textwidth}
\centering
\includegraphics[width=\textwidth]{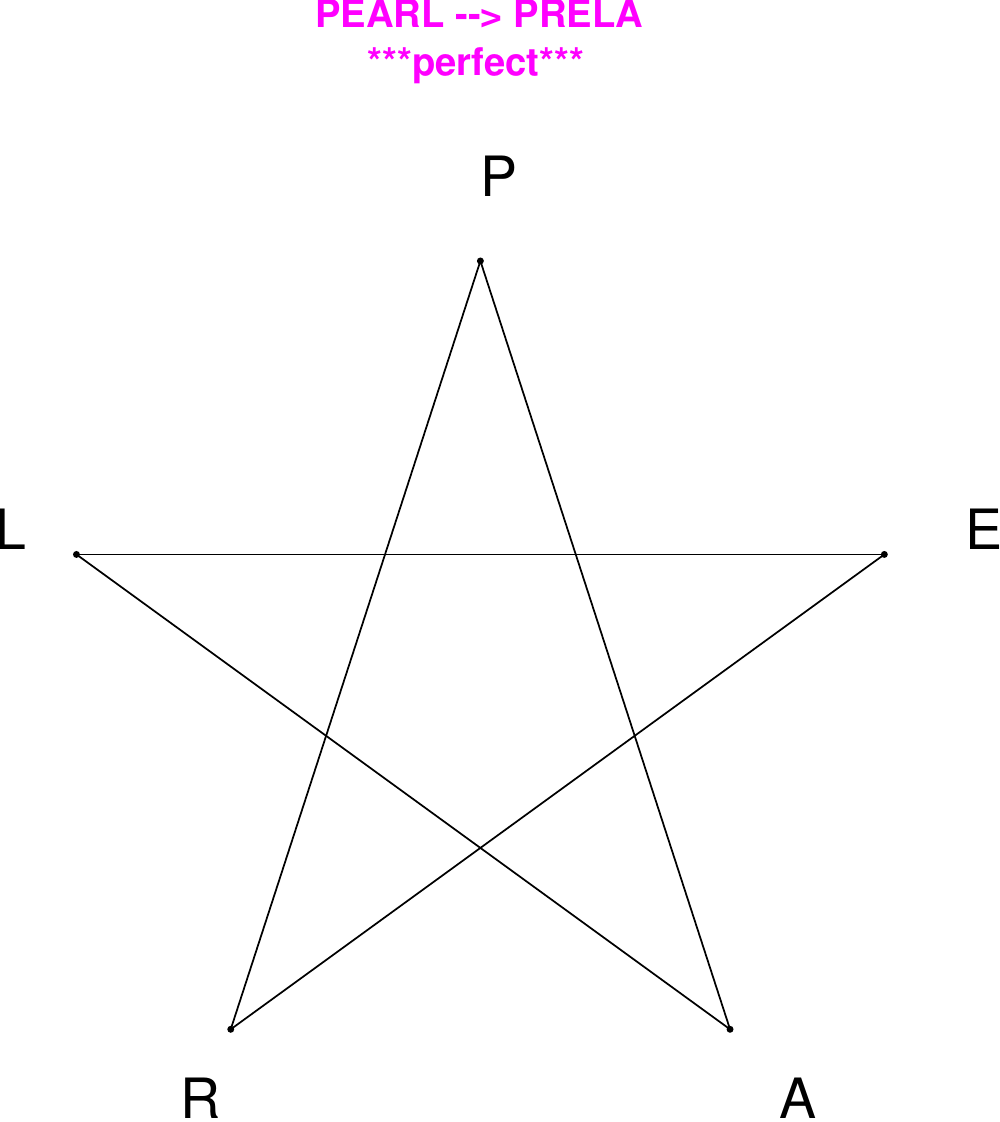}
\end{subfigure}
\hfill
\begin{subfigure}[T]{0.19\textwidth}
\centering
\includegraphics[width=\textwidth]{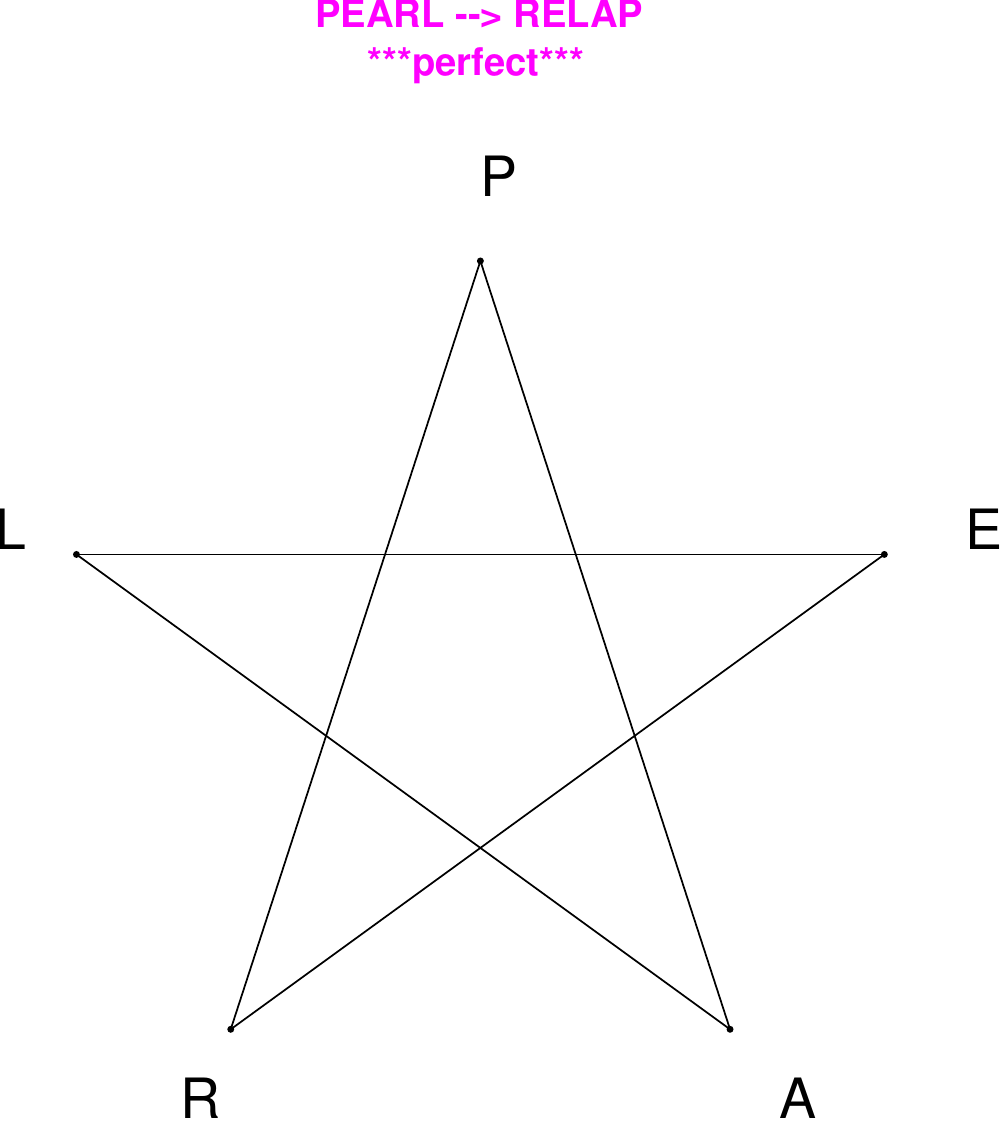}
\end{subfigure}
\hfill
\begin{subfigure}[T]{0.19\textwidth}
\centering
\includegraphics[width=\textwidth]{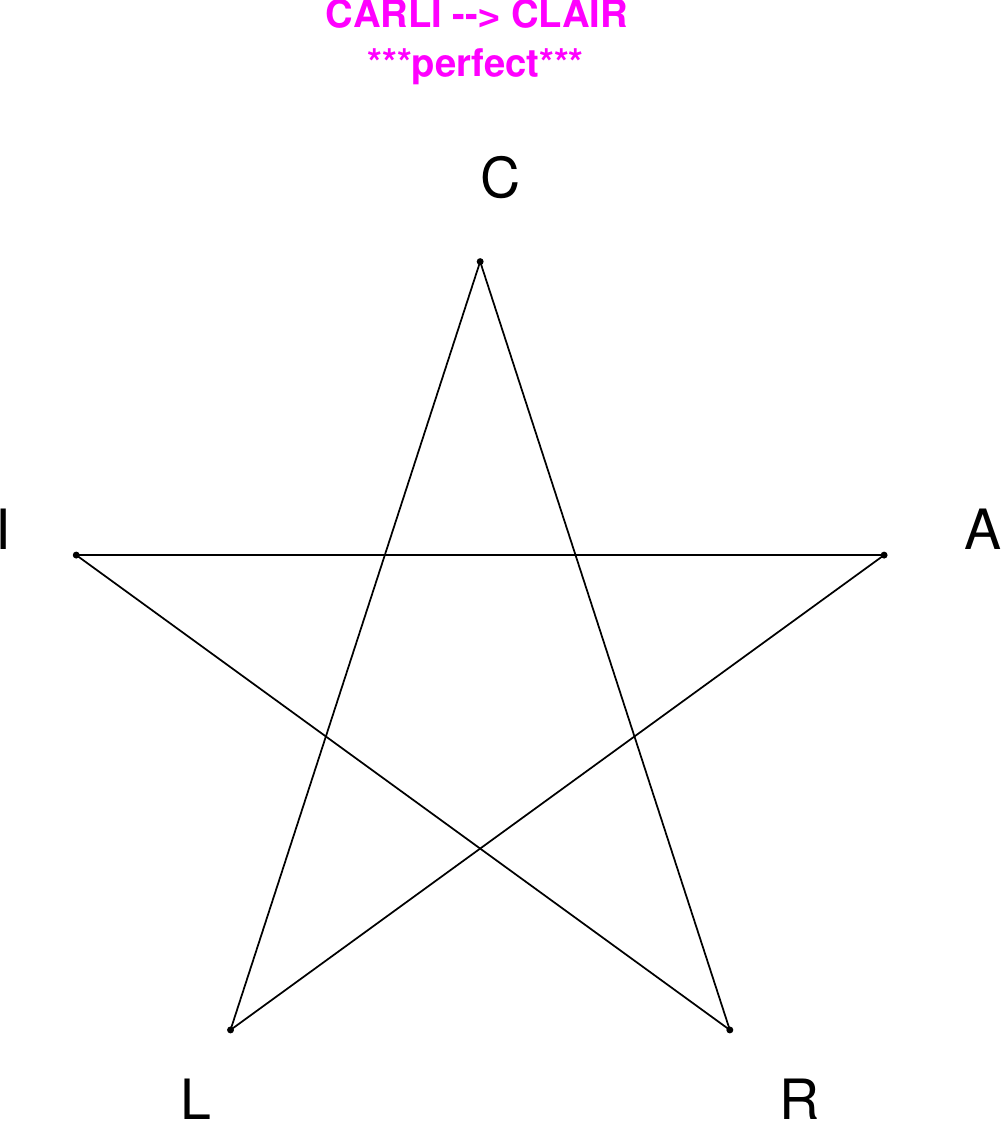}
\end{subfigure}
\hfill
\begin{subfigure}[T]{0.19\textwidth}
\centering
\includegraphics[width=\textwidth]{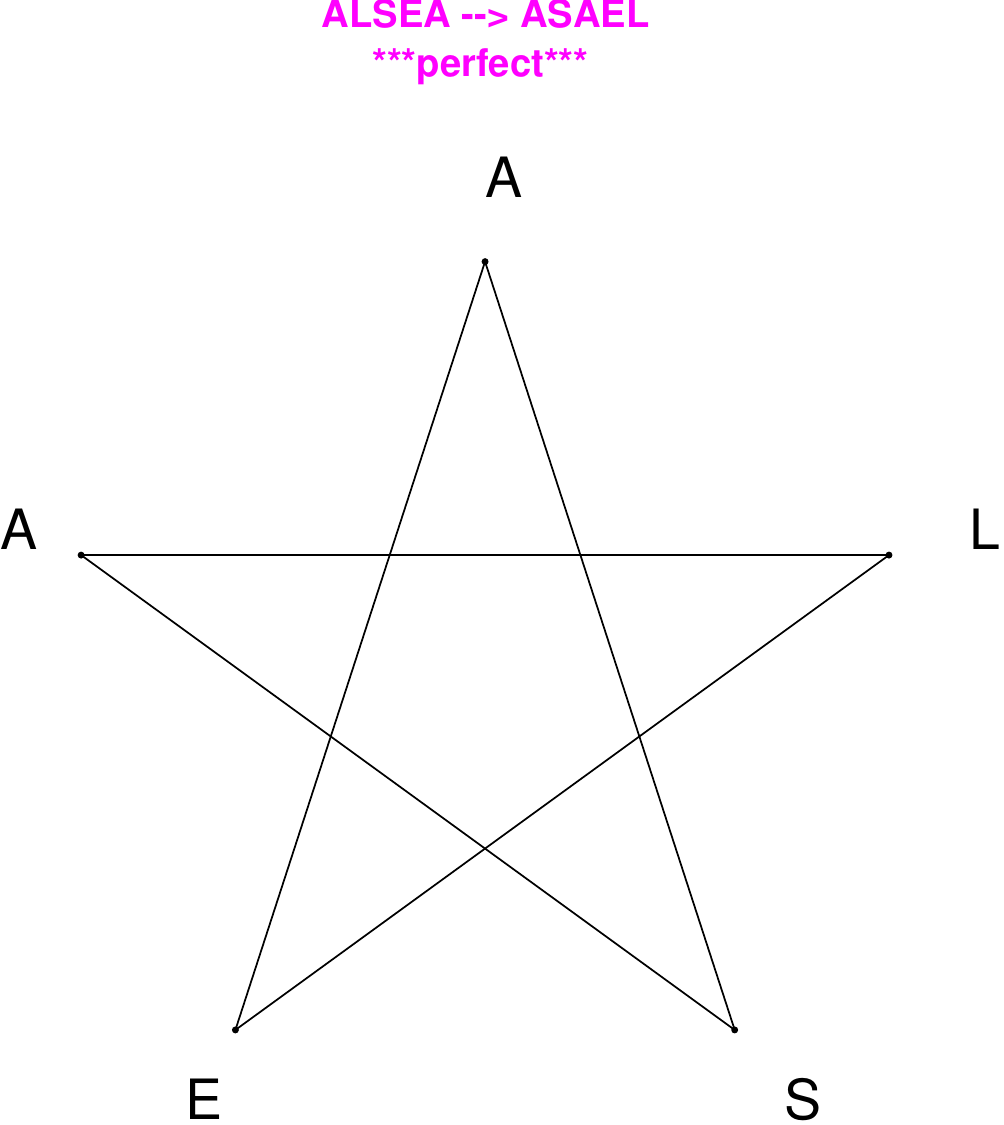}
\end{subfigure}
\hfill
\begin{subfigure}[T]{0.19\textwidth}
\centering
\includegraphics[width=\textwidth]{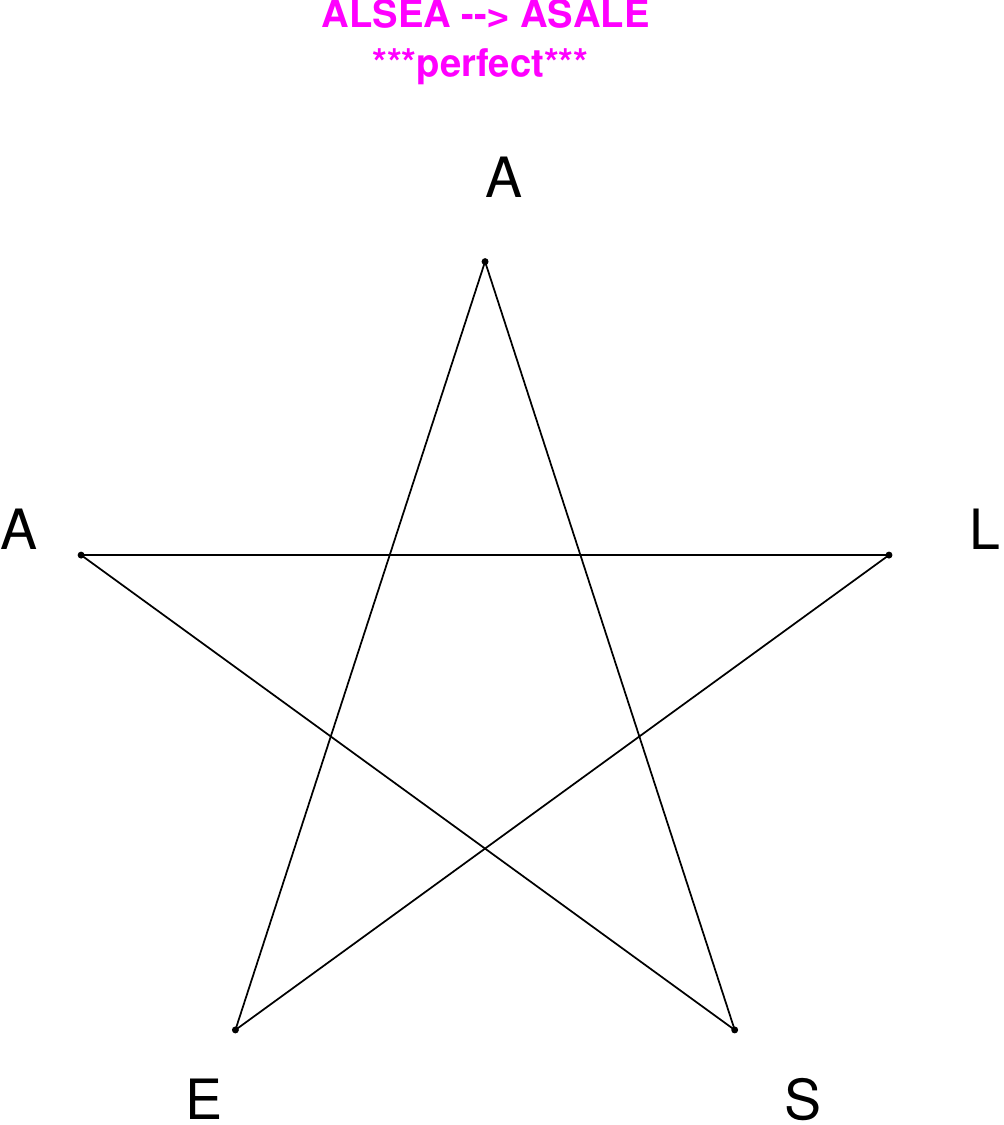}
\end{subfigure}
\end{figure}

\begin{figure}[H]
\centering
\begin{subfigure}[T]{0.19\textwidth}
\centering
\includegraphics[width=\textwidth]{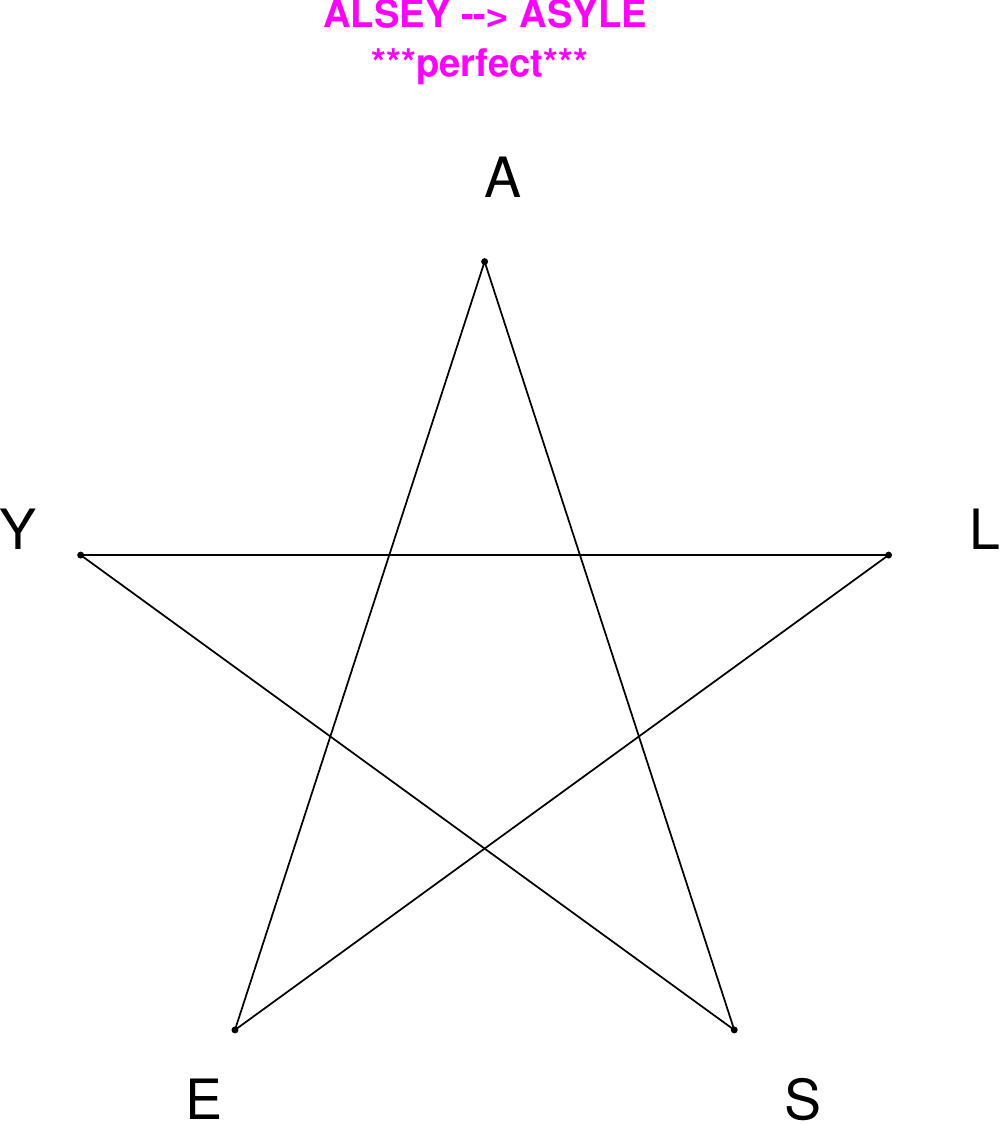}
\end{subfigure}
\hfill
\begin{subfigure}[T]{0.19\textwidth}
\centering
\includegraphics[width=\textwidth]{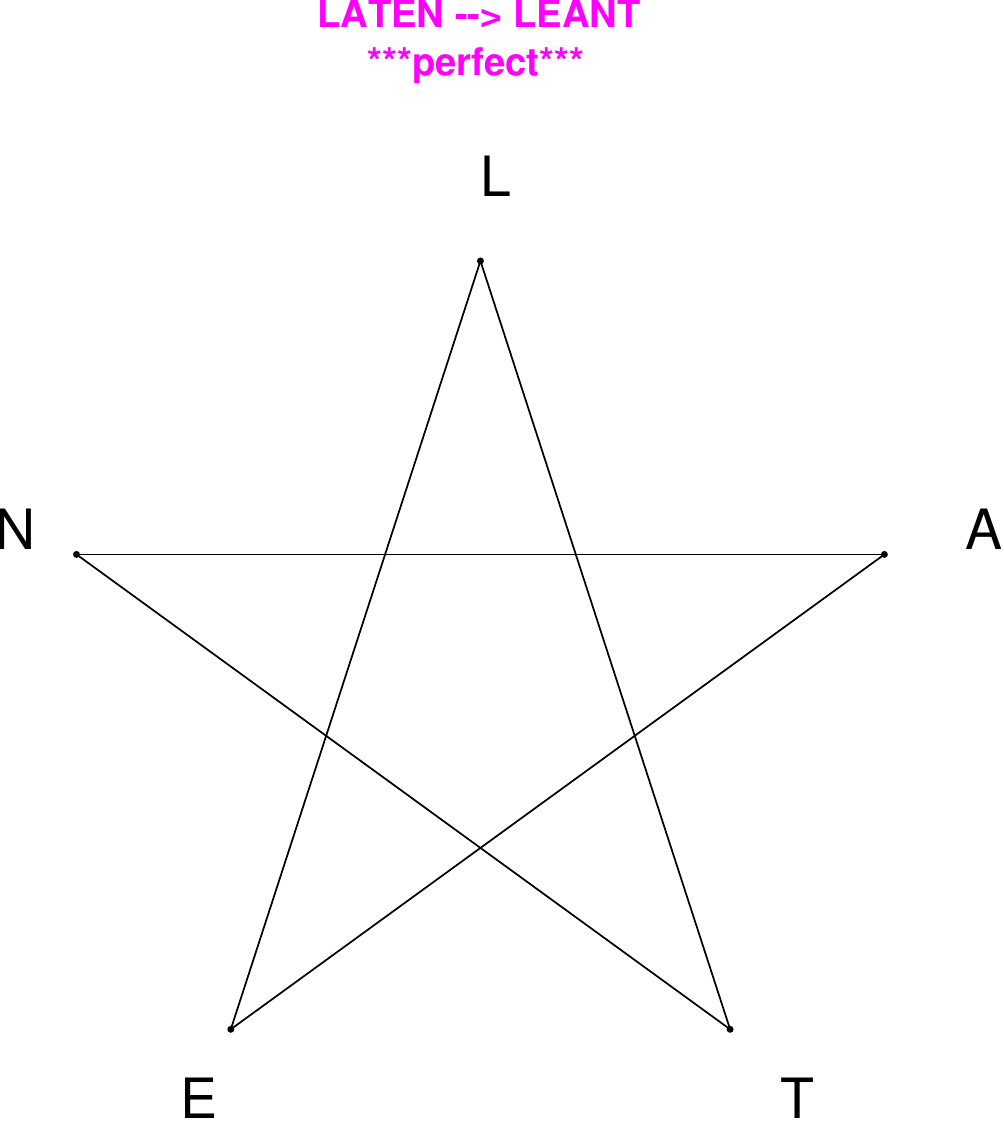}
\end{subfigure}
\hfill
\begin{subfigure}[T]{0.19\textwidth}
\centering
\includegraphics[width=\textwidth]{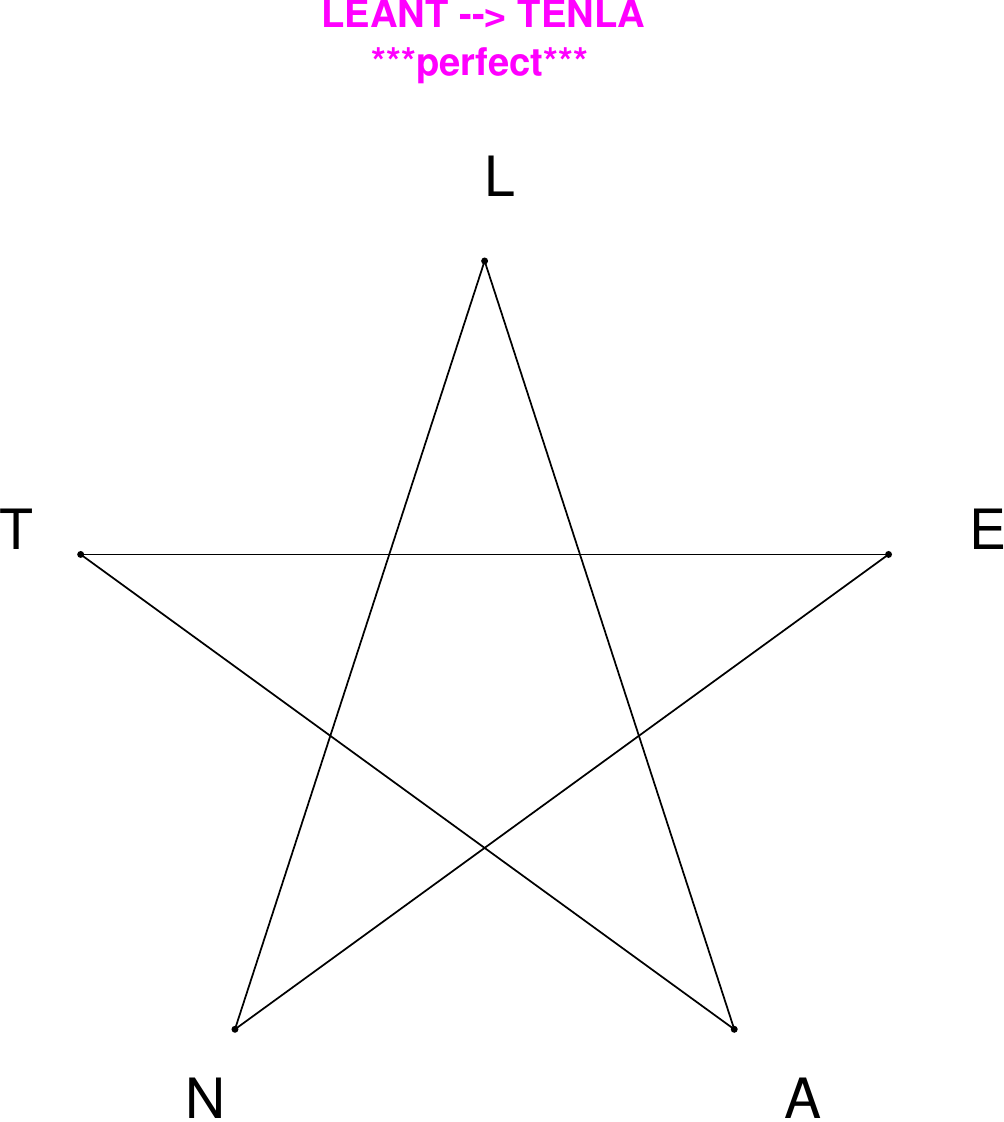}
\end{subfigure}
\hfill
\begin{subfigure}[T]{0.19\textwidth}
\centering
\includegraphics[width=\textwidth]{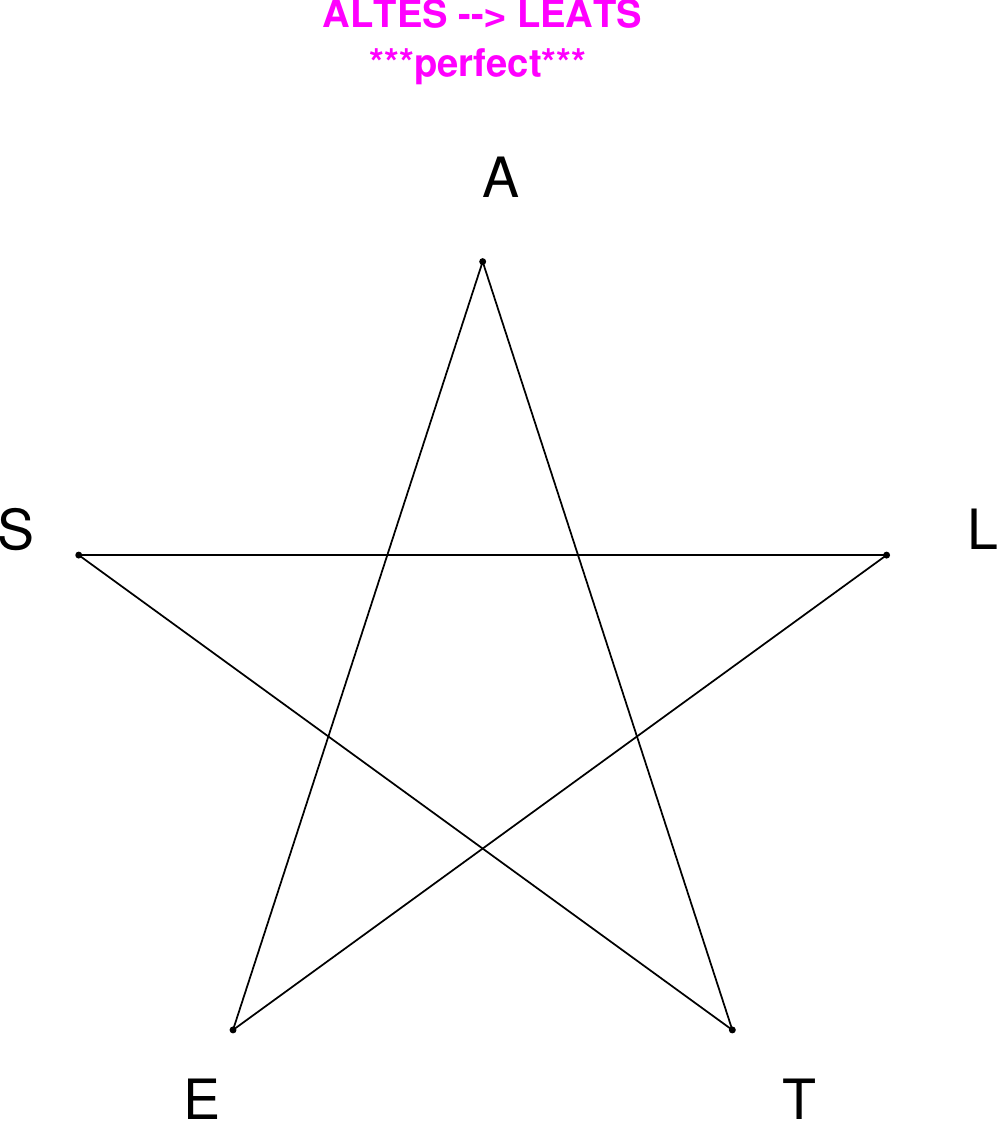}
\end{subfigure}
\hfill
\begin{subfigure}[T]{0.19\textwidth}
\centering
\includegraphics[width=\textwidth]{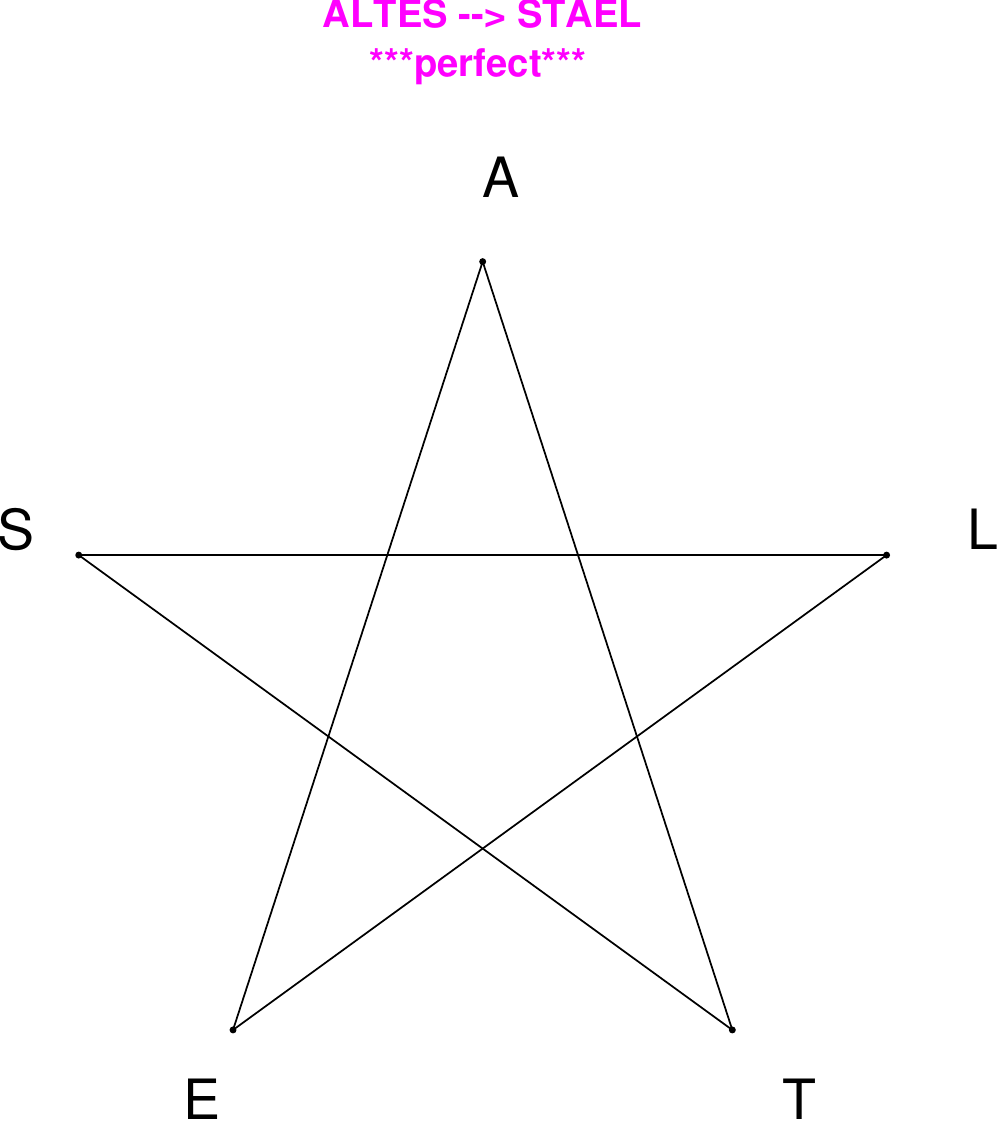}
\end{subfigure}
\end{figure}

\begin{figure}[H]
\centering
\begin{subfigure}[T]{0.19\textwidth}
\centering
\includegraphics[width=\textwidth]{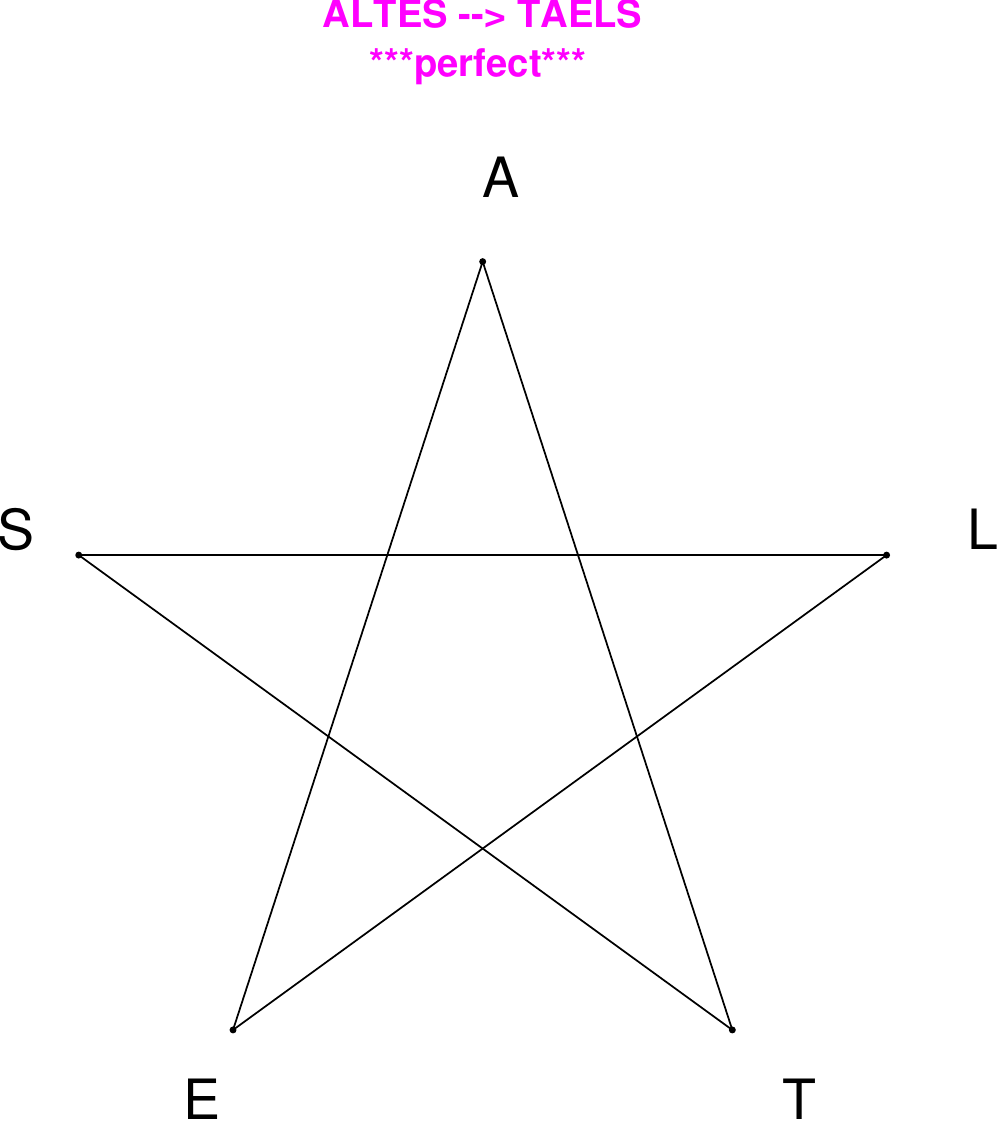}
\end{subfigure}
\hfill
\begin{subfigure}[T]{0.19\textwidth}
\centering
\includegraphics[width=\textwidth]{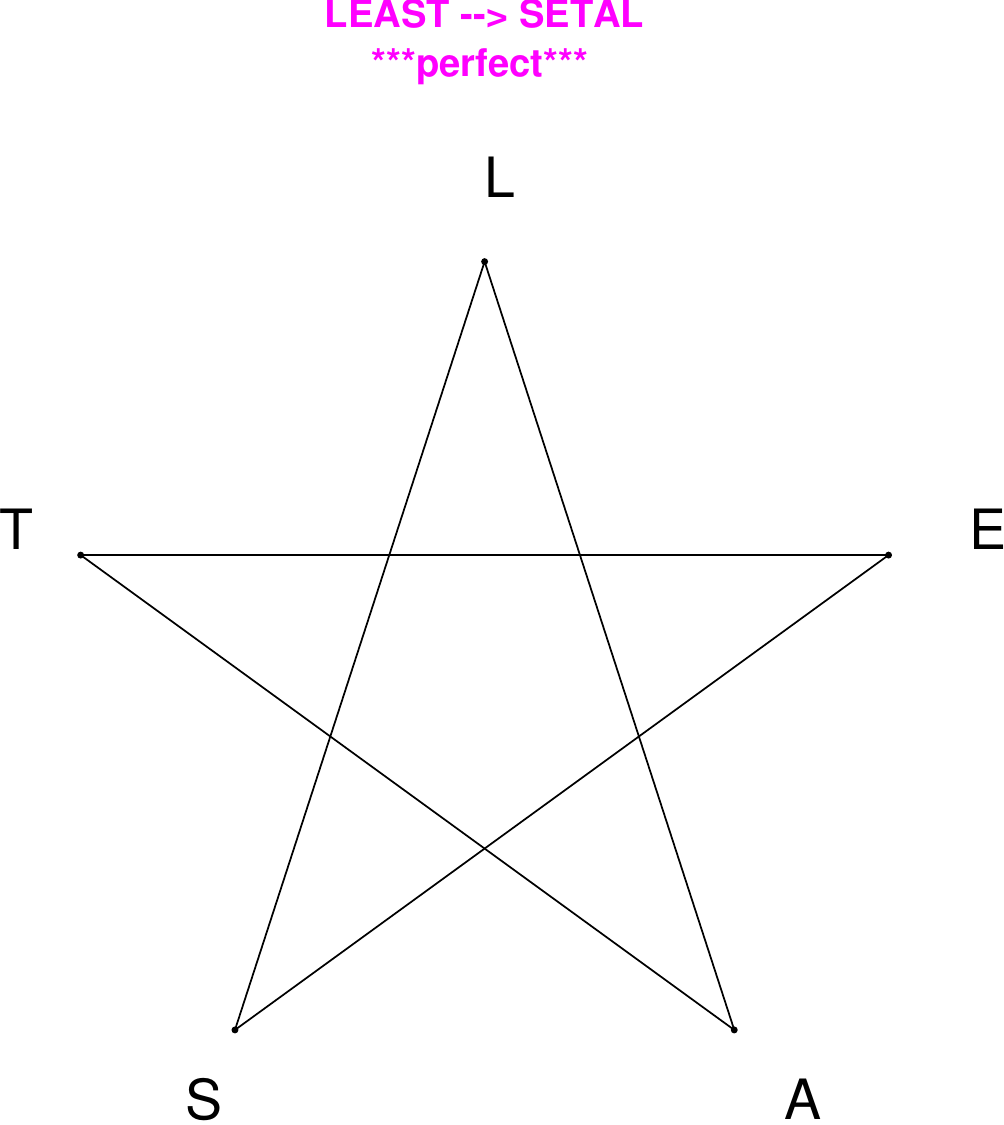}
\end{subfigure}
\hfill
\begin{subfigure}[T]{0.19\textwidth}
\centering
\includegraphics[width=\textwidth]{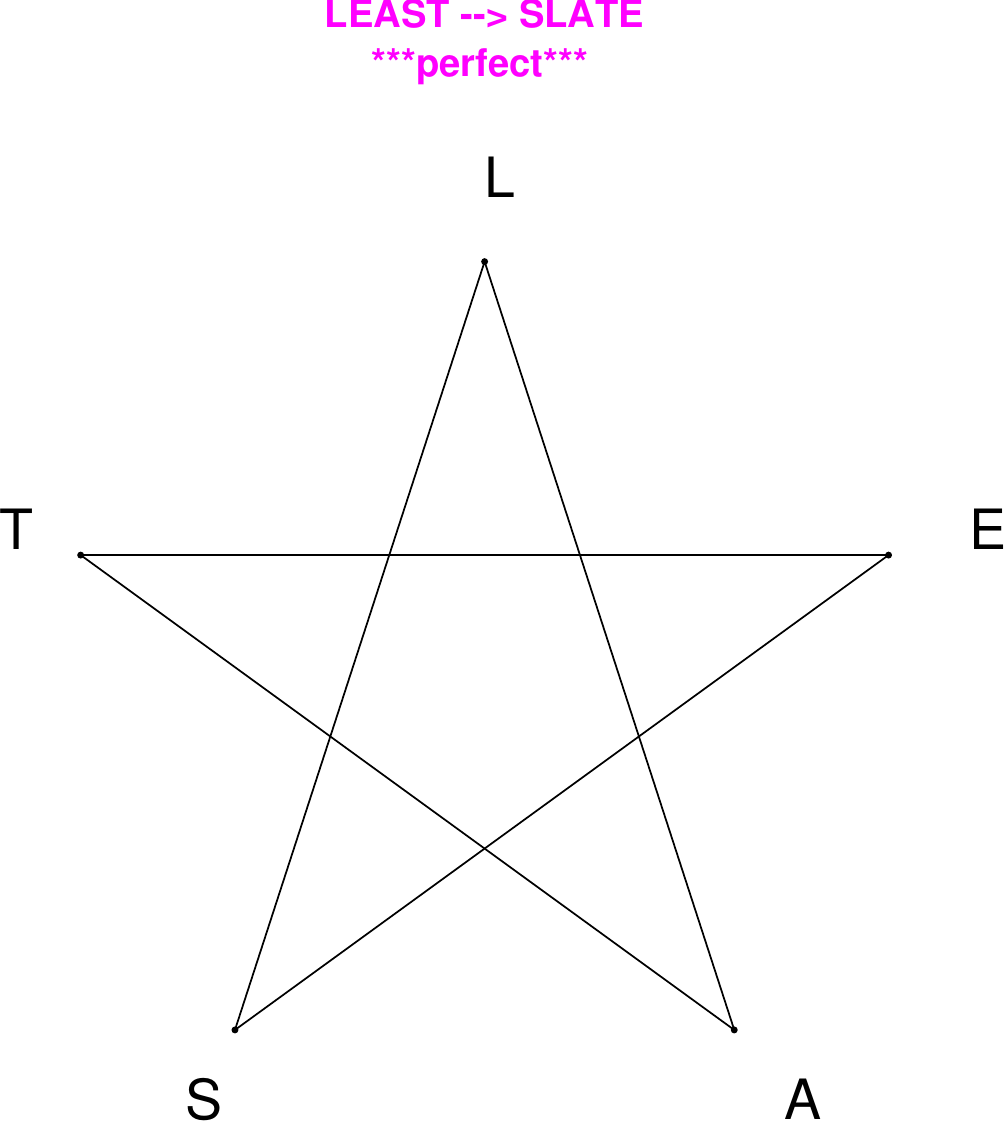}
\end{subfigure}
\hfill
\begin{subfigure}[T]{0.19\textwidth}
\centering
\includegraphics[width=\textwidth]{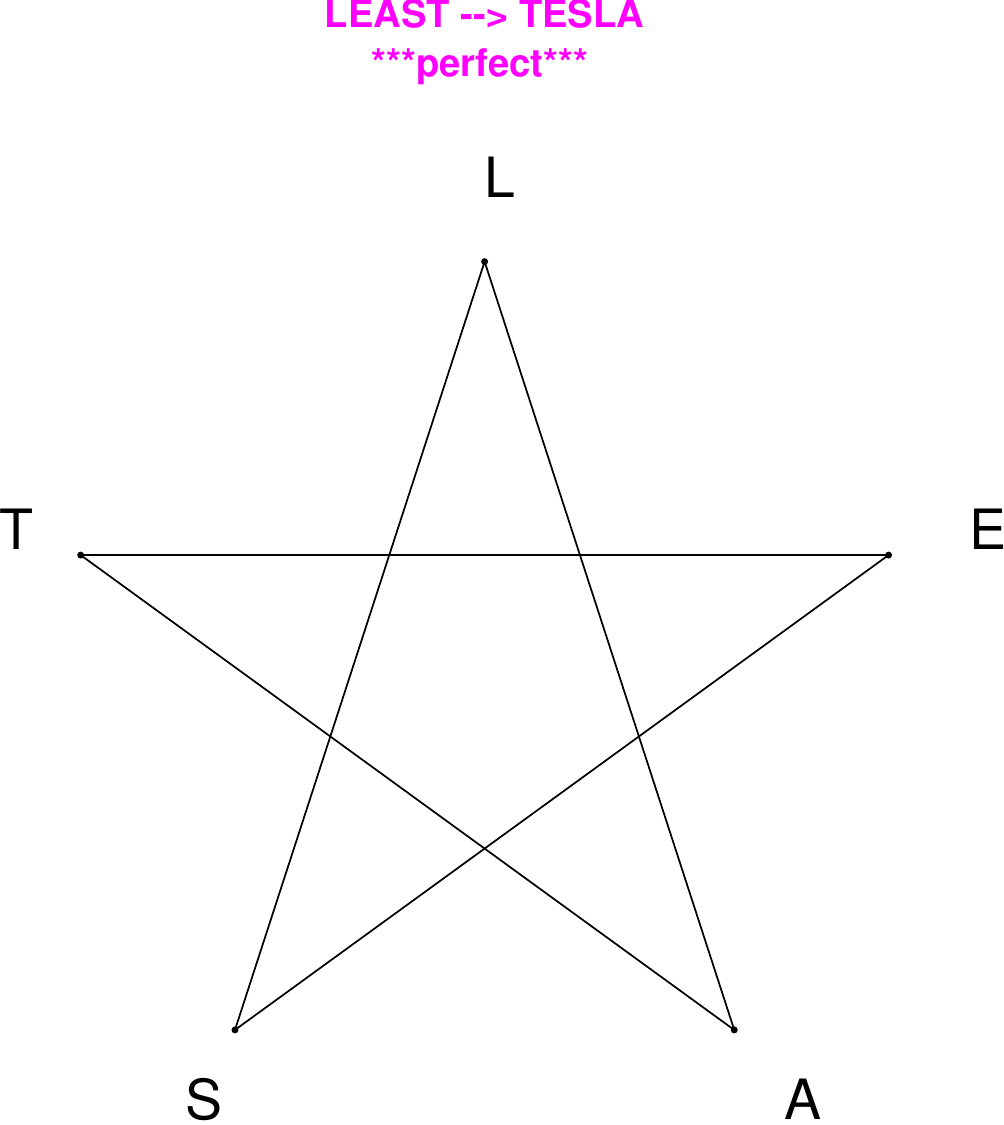}
\end{subfigure}
\hfill
\begin{subfigure}[T]{0.19\textwidth}
\centering
\includegraphics[width=\textwidth]{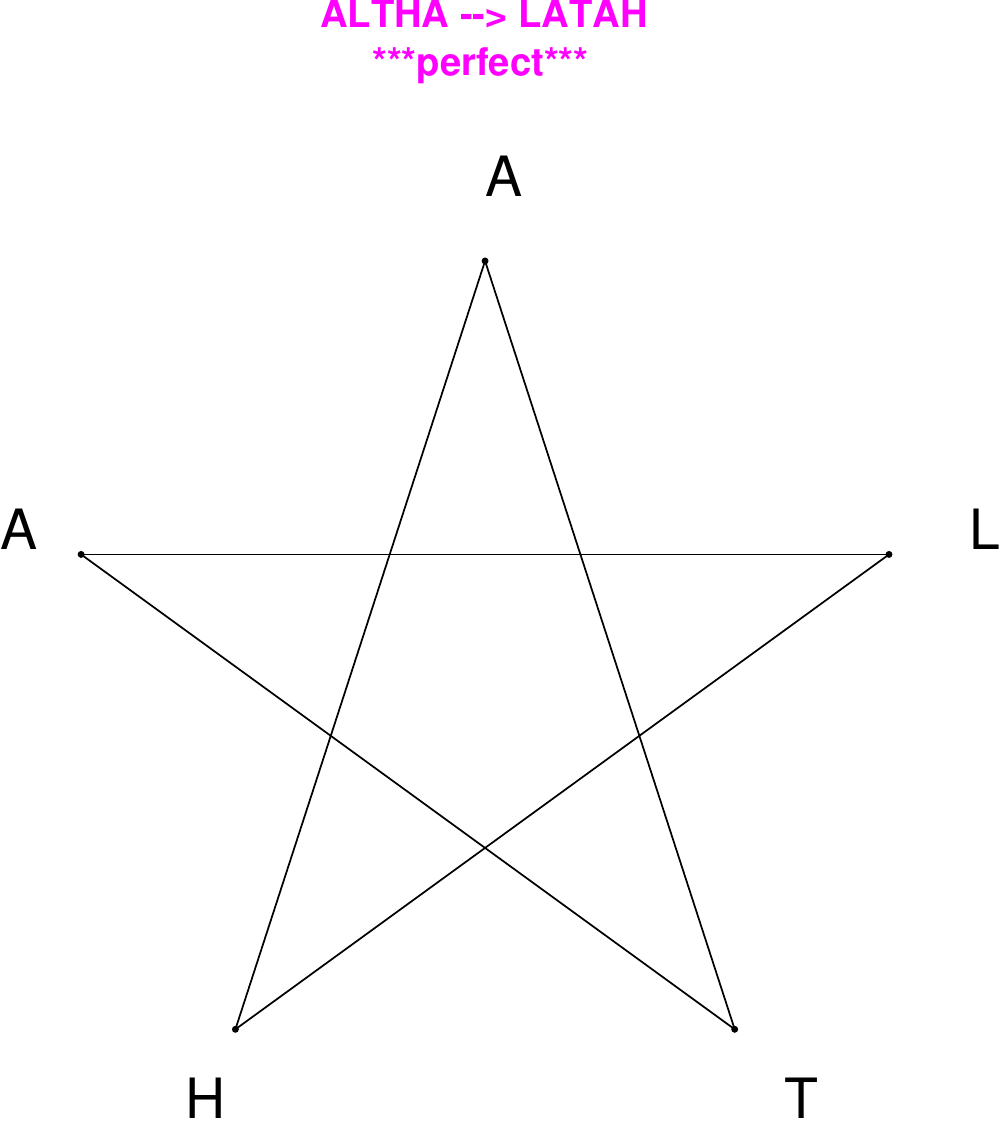}
\end{subfigure}
\end{figure}

\begin{figure}[H]
\centering
\begin{subfigure}[T]{0.19\textwidth}
\centering
\includegraphics[width=\textwidth]{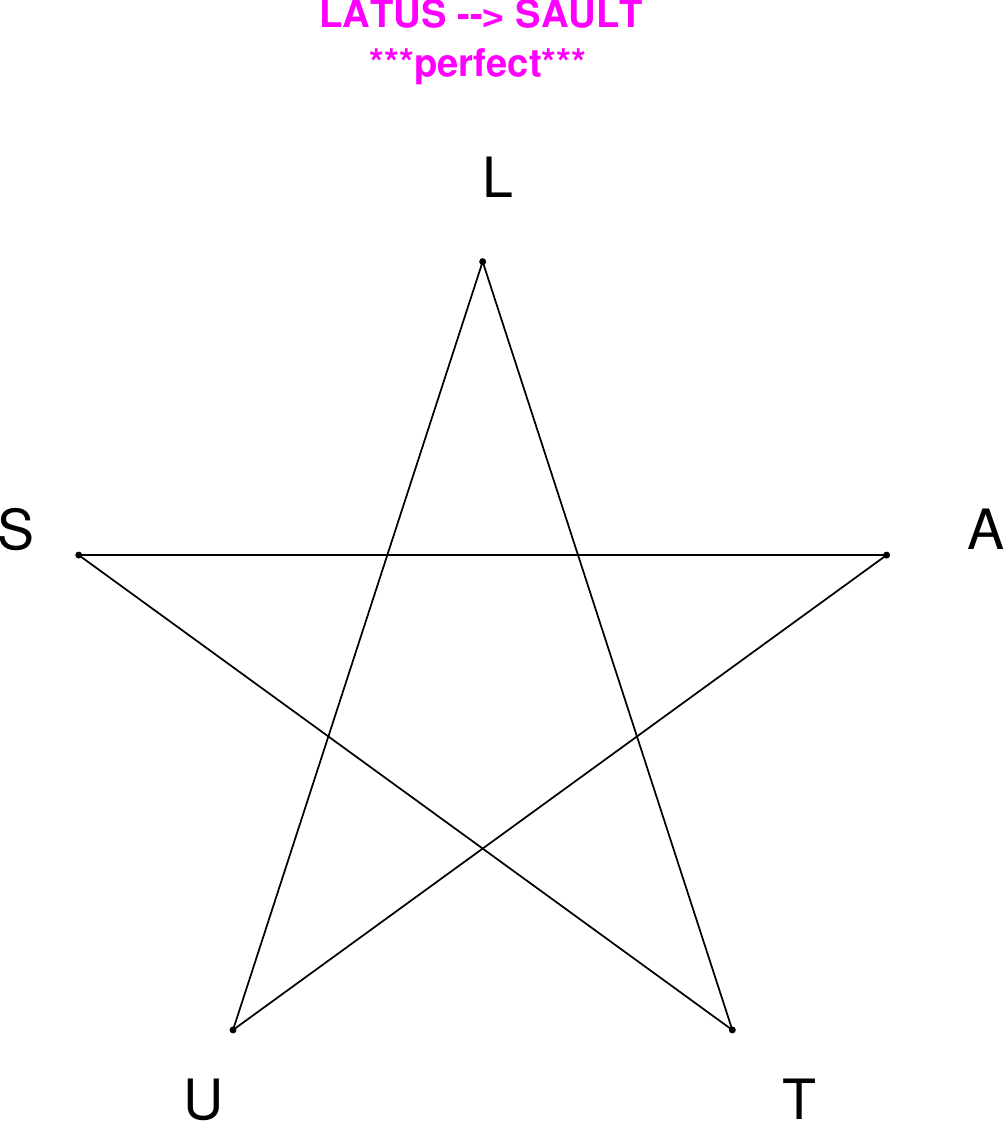}
\end{subfigure}
\hfill
\begin{subfigure}[T]{0.19\textwidth}
\centering
\includegraphics[width=\textwidth]{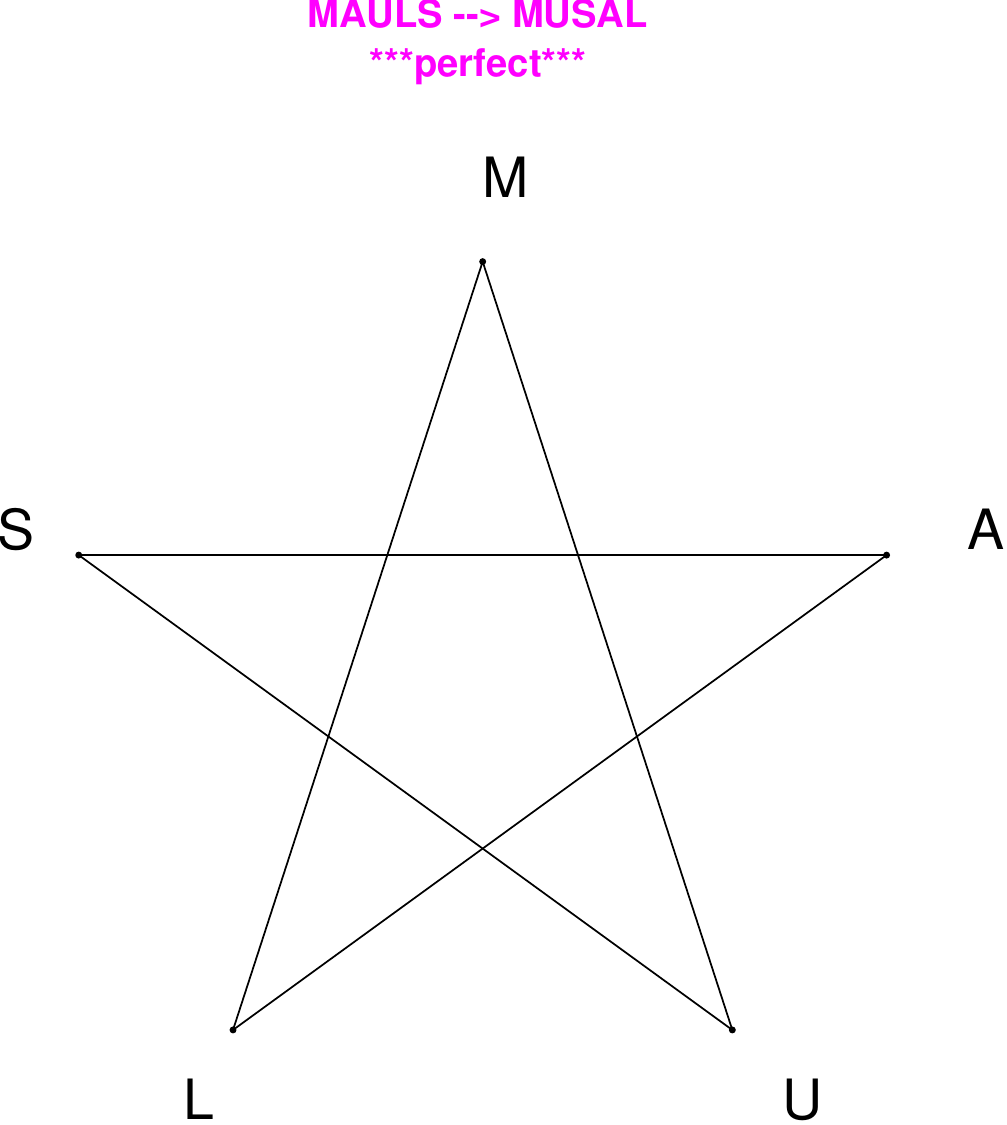}
\end{subfigure}
\hfill
\begin{subfigure}[T]{0.19\textwidth}
\centering
\includegraphics[width=\textwidth]{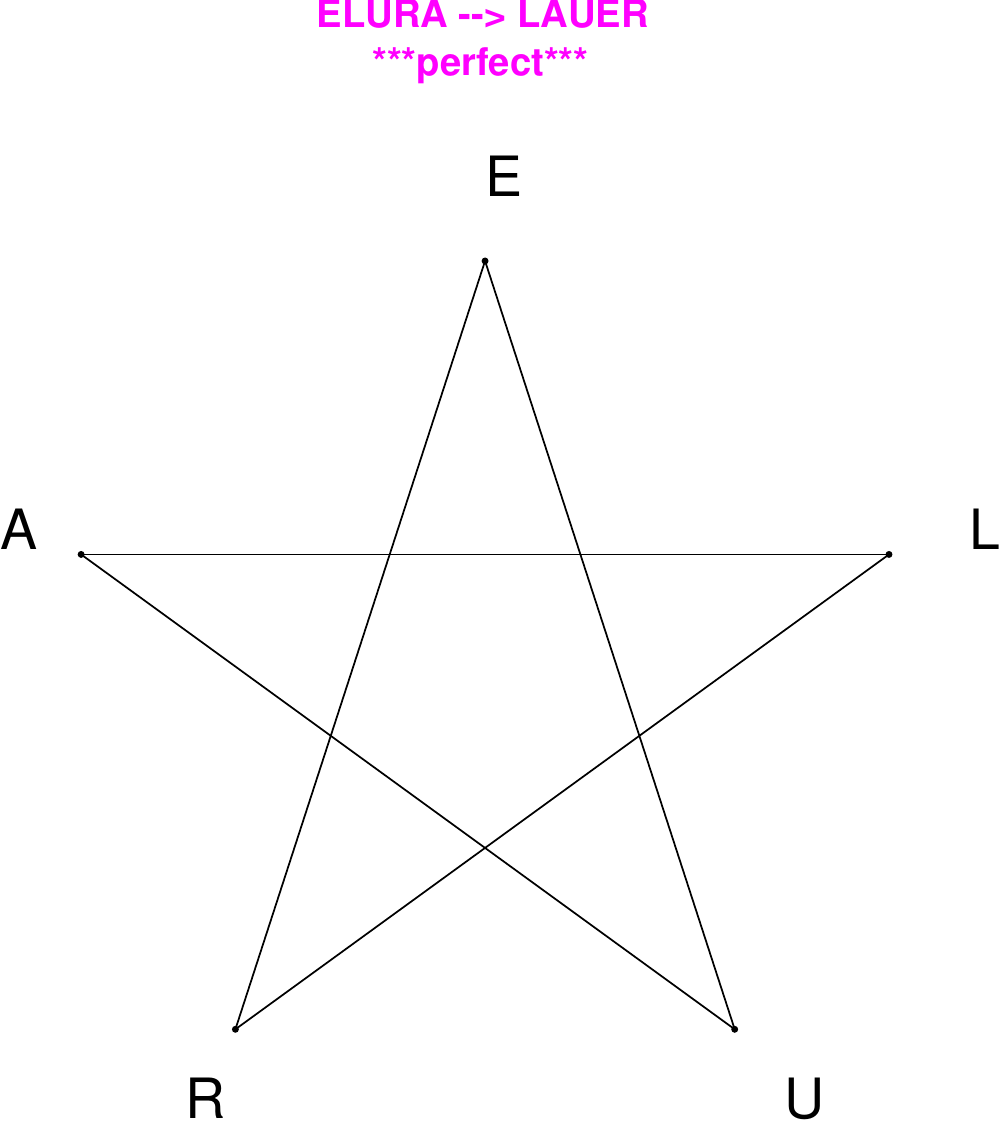}
\end{subfigure}
\hfill
\begin{subfigure}[T]{0.19\textwidth}
\centering
\includegraphics[width=\textwidth]{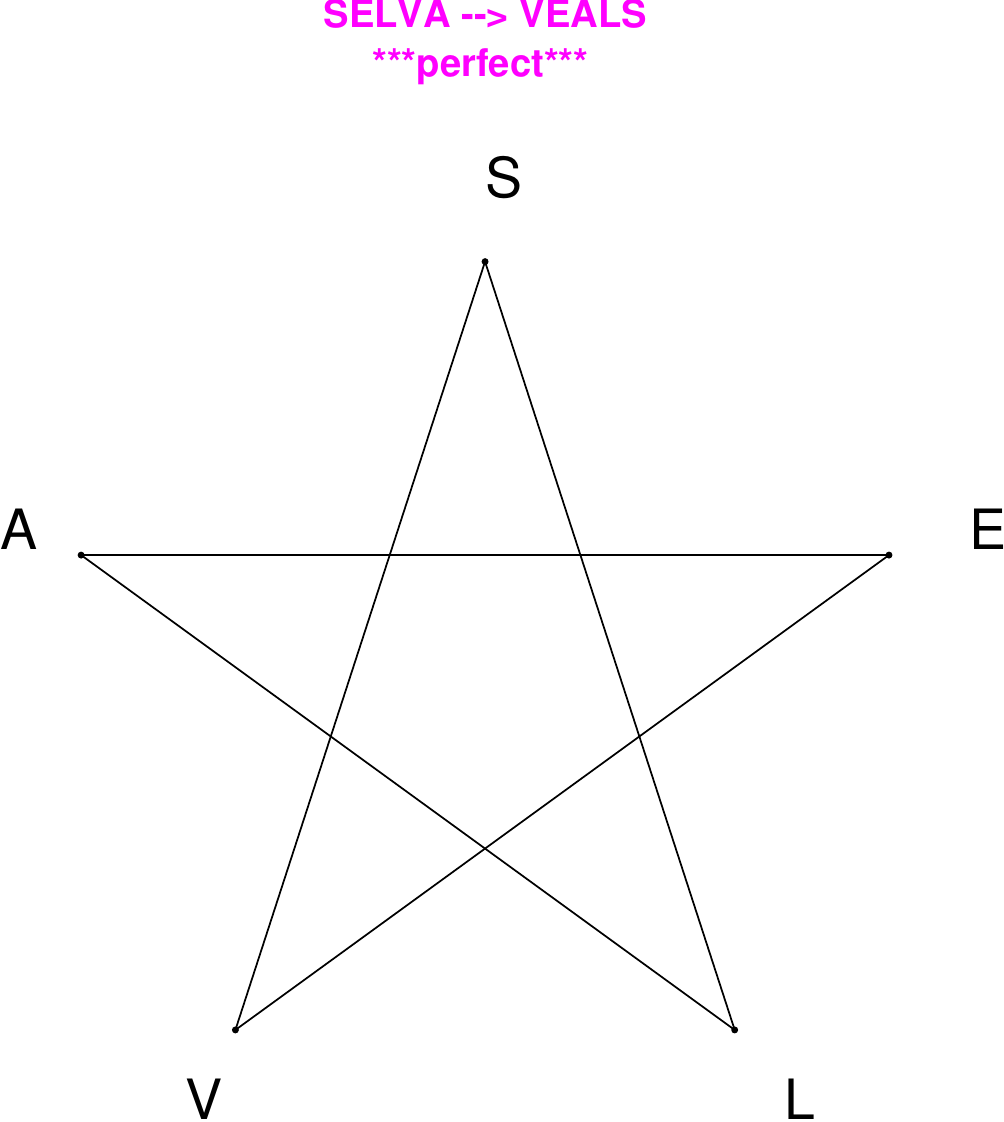}
\end{subfigure}
\hfill
\begin{subfigure}[T]{0.19\textwidth}
\centering
\includegraphics[width=\textwidth]{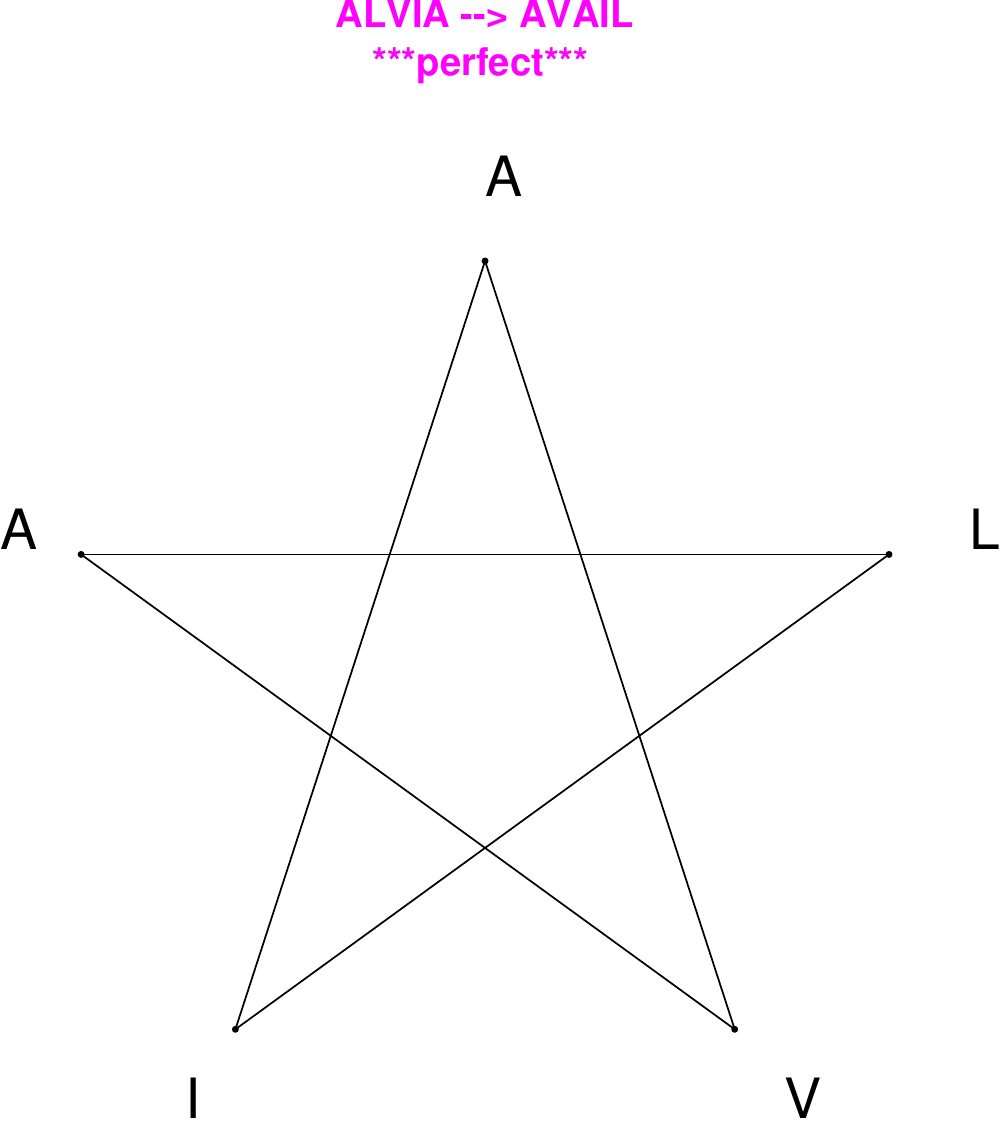}
\end{subfigure}
\end{figure}

\begin{figure}[H]
\centering
\begin{subfigure}[T]{0.19\textwidth}
\centering
\includegraphics[width=\textwidth]{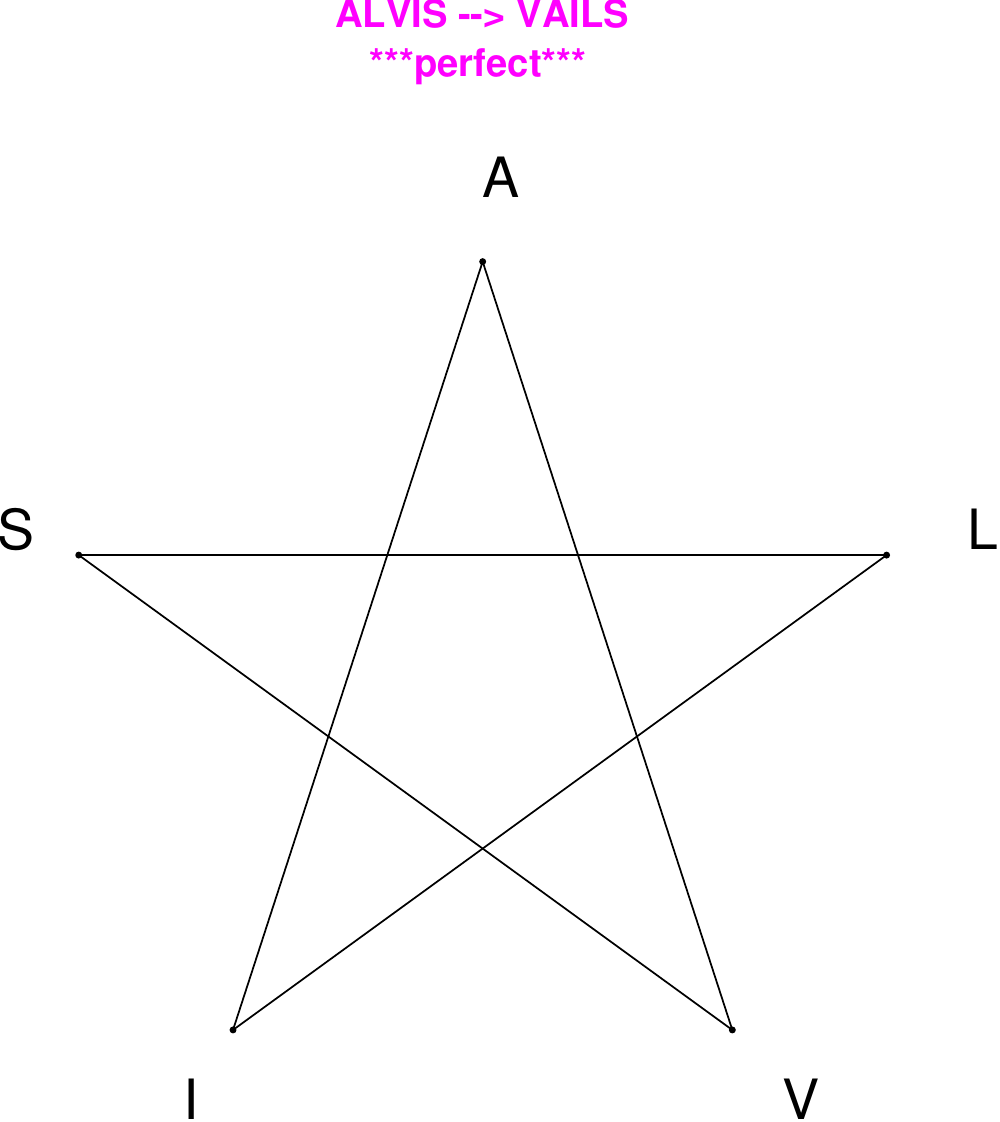}
\end{subfigure}
\hfill
\begin{subfigure}[T]{0.19\textwidth}
\centering
\includegraphics[width=\textwidth]{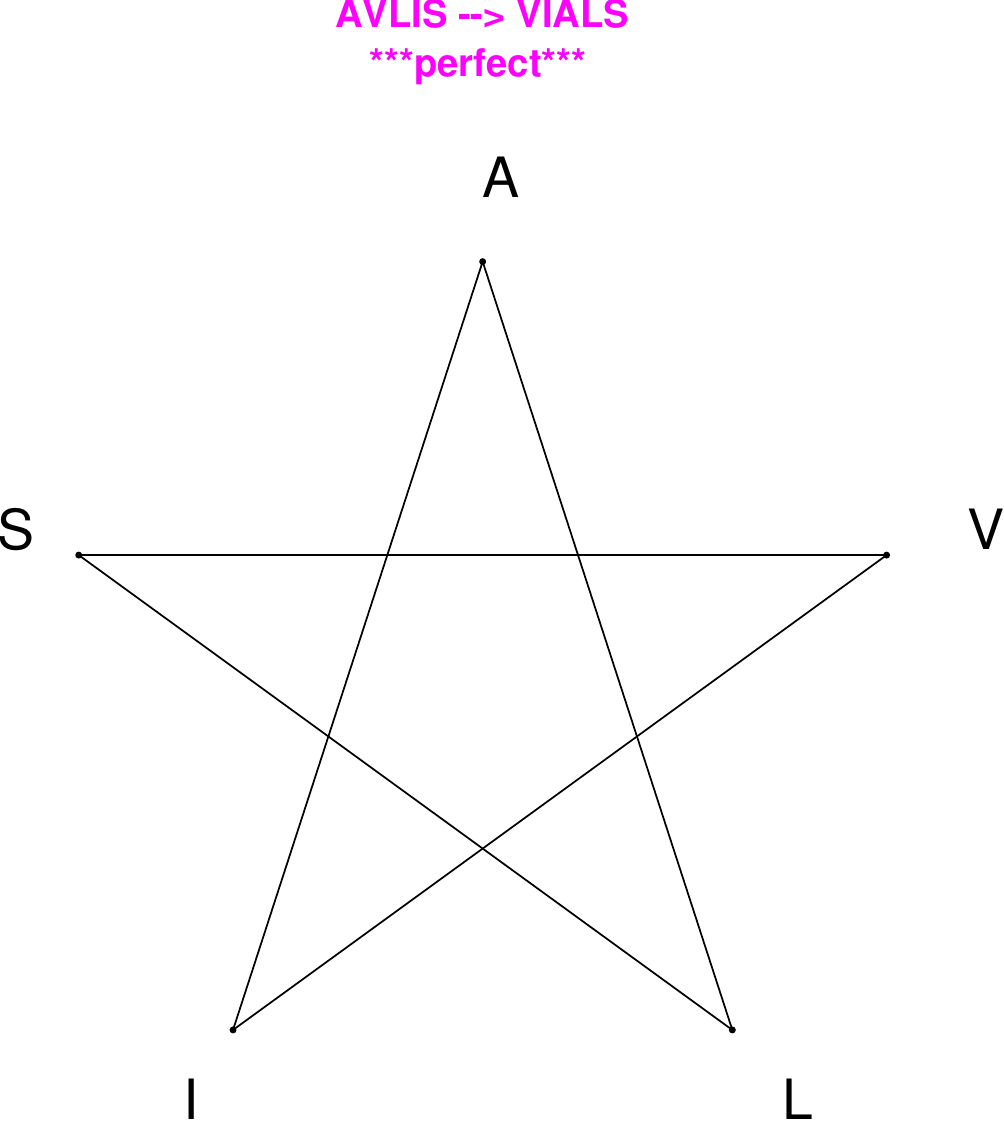}
\end{subfigure}
\hfill
\begin{subfigure}[T]{0.19\textwidth}
\centering
\includegraphics[width=\textwidth]{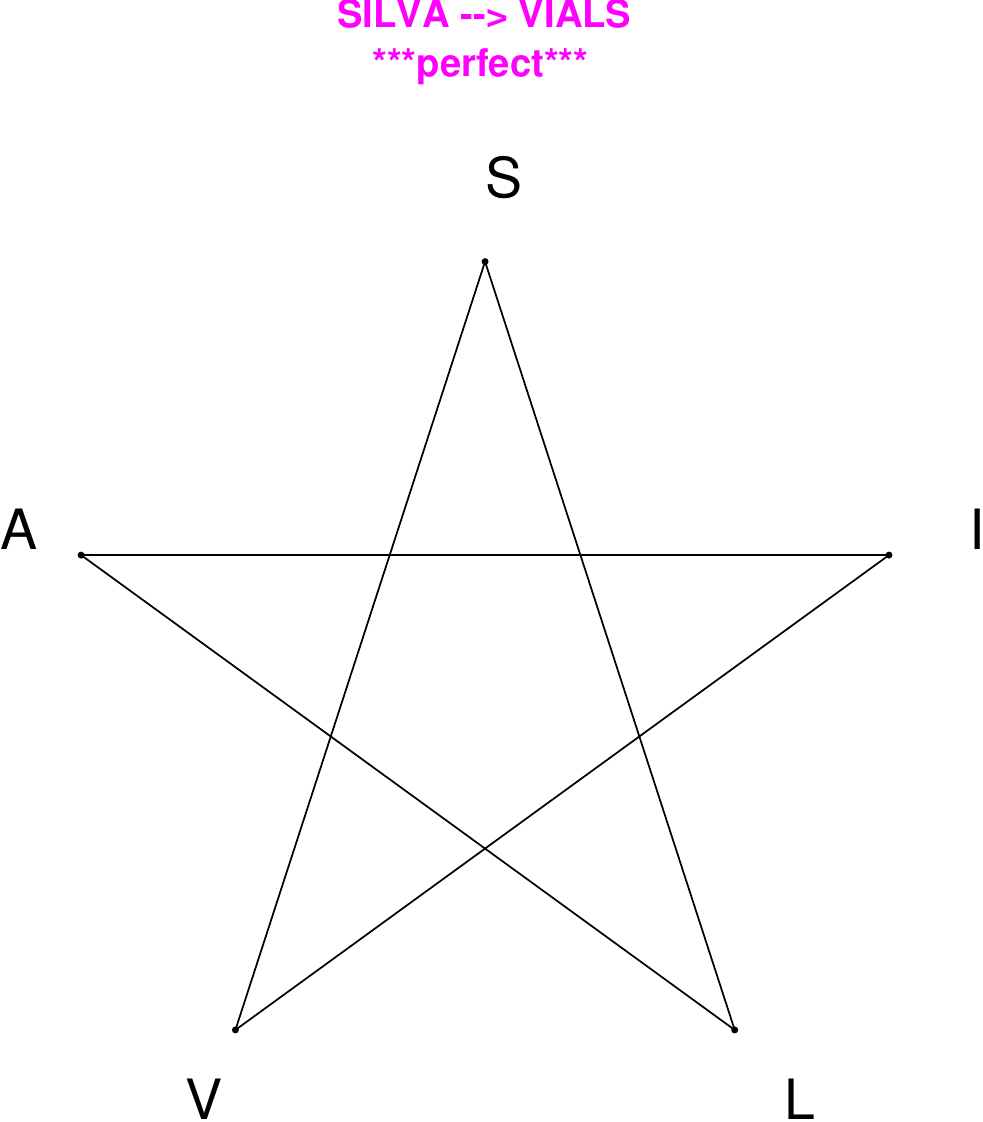}
\end{subfigure}
\hfill
\begin{subfigure}[T]{0.19\textwidth}
\centering
\includegraphics[width=\textwidth]{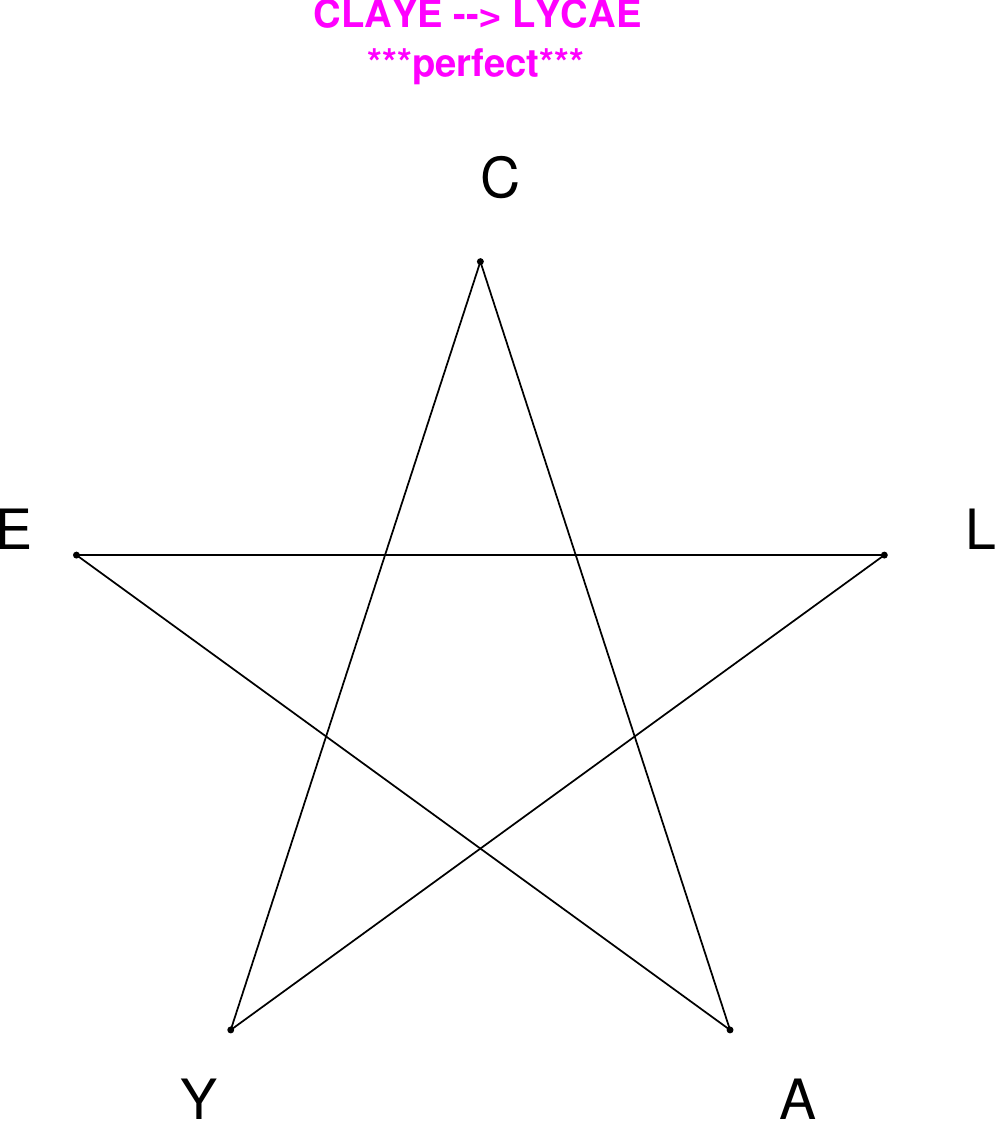}
\end{subfigure}
\hfill
\begin{subfigure}[T]{0.19\textwidth}
\centering
\includegraphics[width=\textwidth]{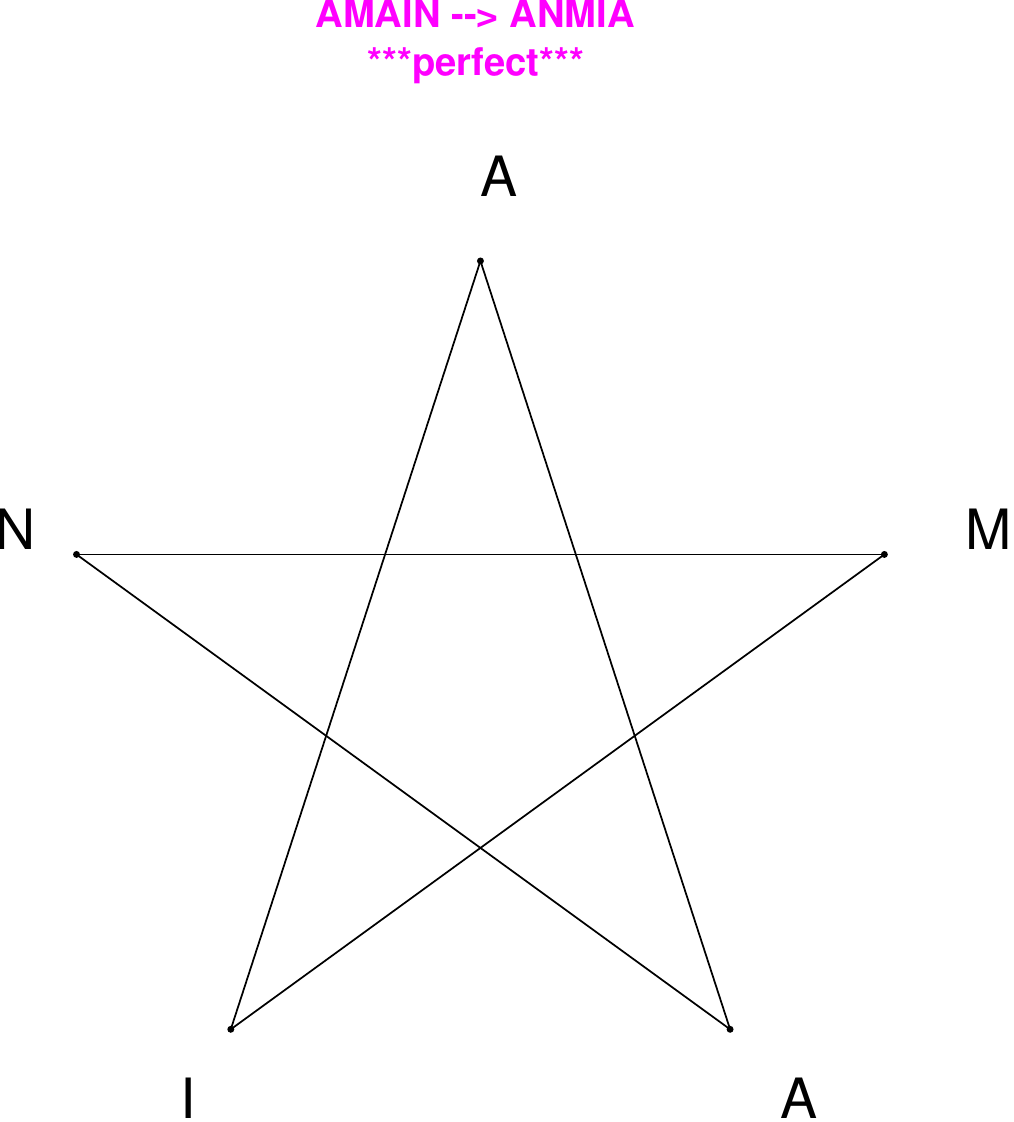}
\end{subfigure}
\end{figure}

\begin{figure}[H]
\centering
\begin{subfigure}[T]{0.19\textwidth}
\centering
\includegraphics[width=\textwidth]{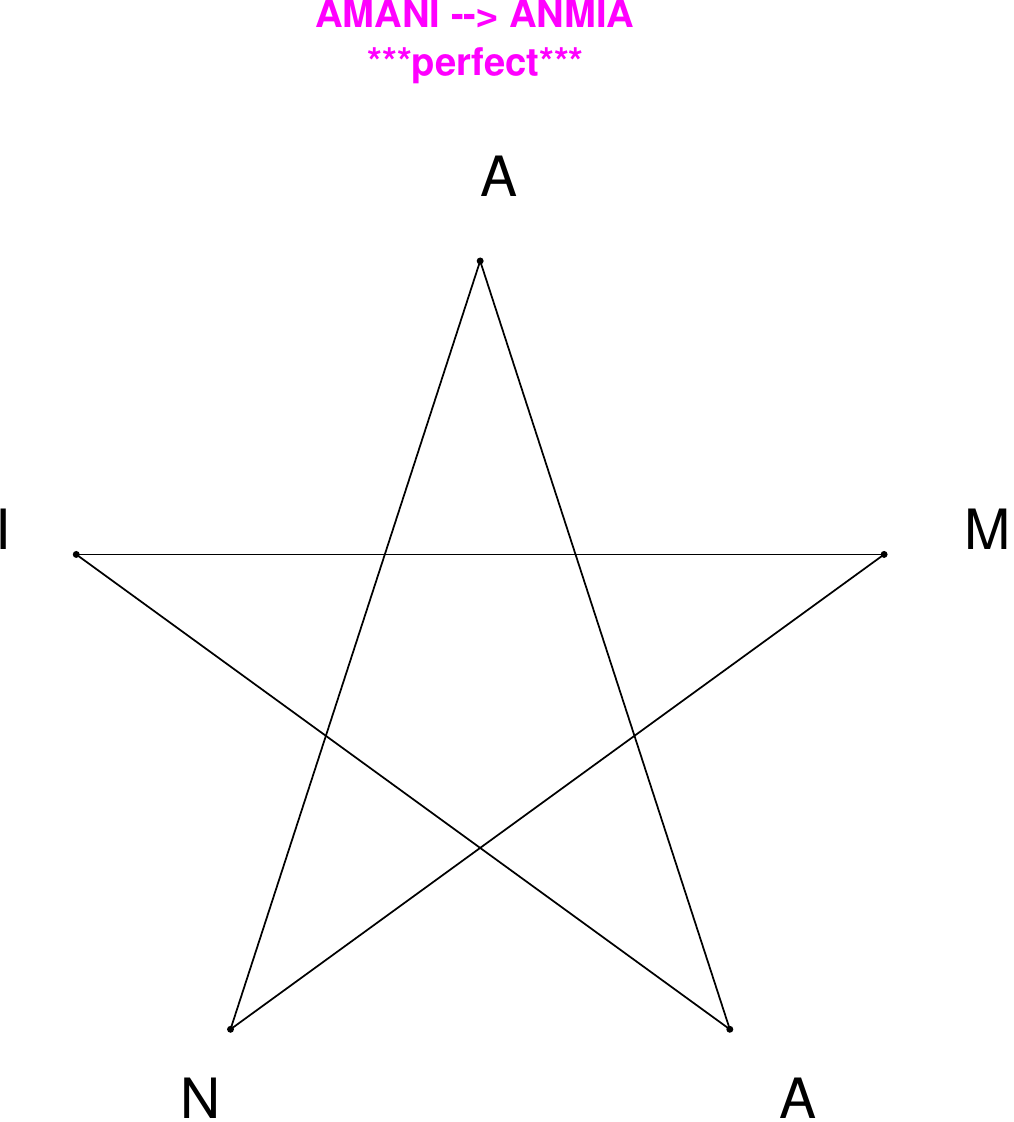}
\end{subfigure}
\hfill
\begin{subfigure}[T]{0.19\textwidth}
\centering
\includegraphics[width=\textwidth]{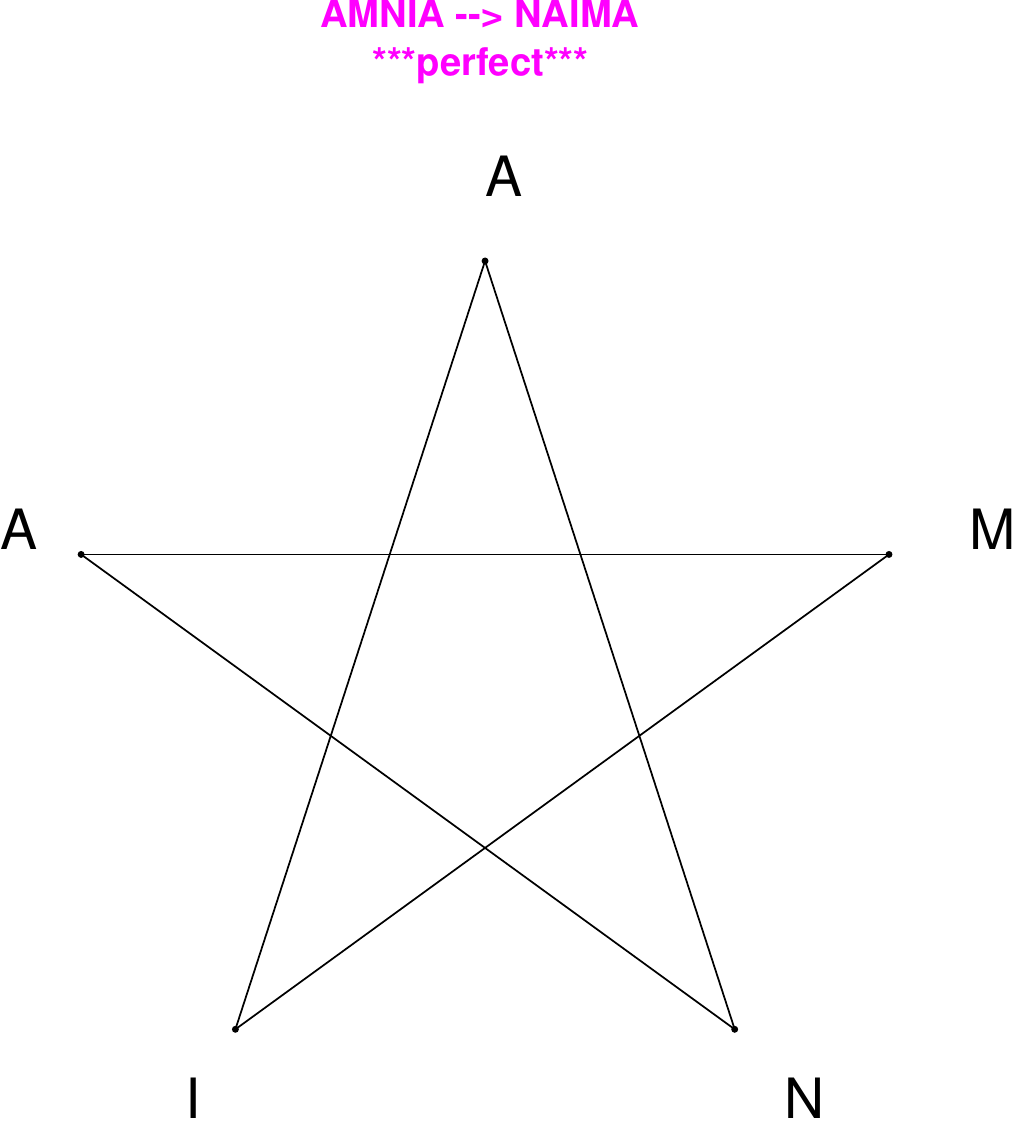}
\end{subfigure}
\hfill
\begin{subfigure}[T]{0.19\textwidth}
\centering
\includegraphics[width=\textwidth]{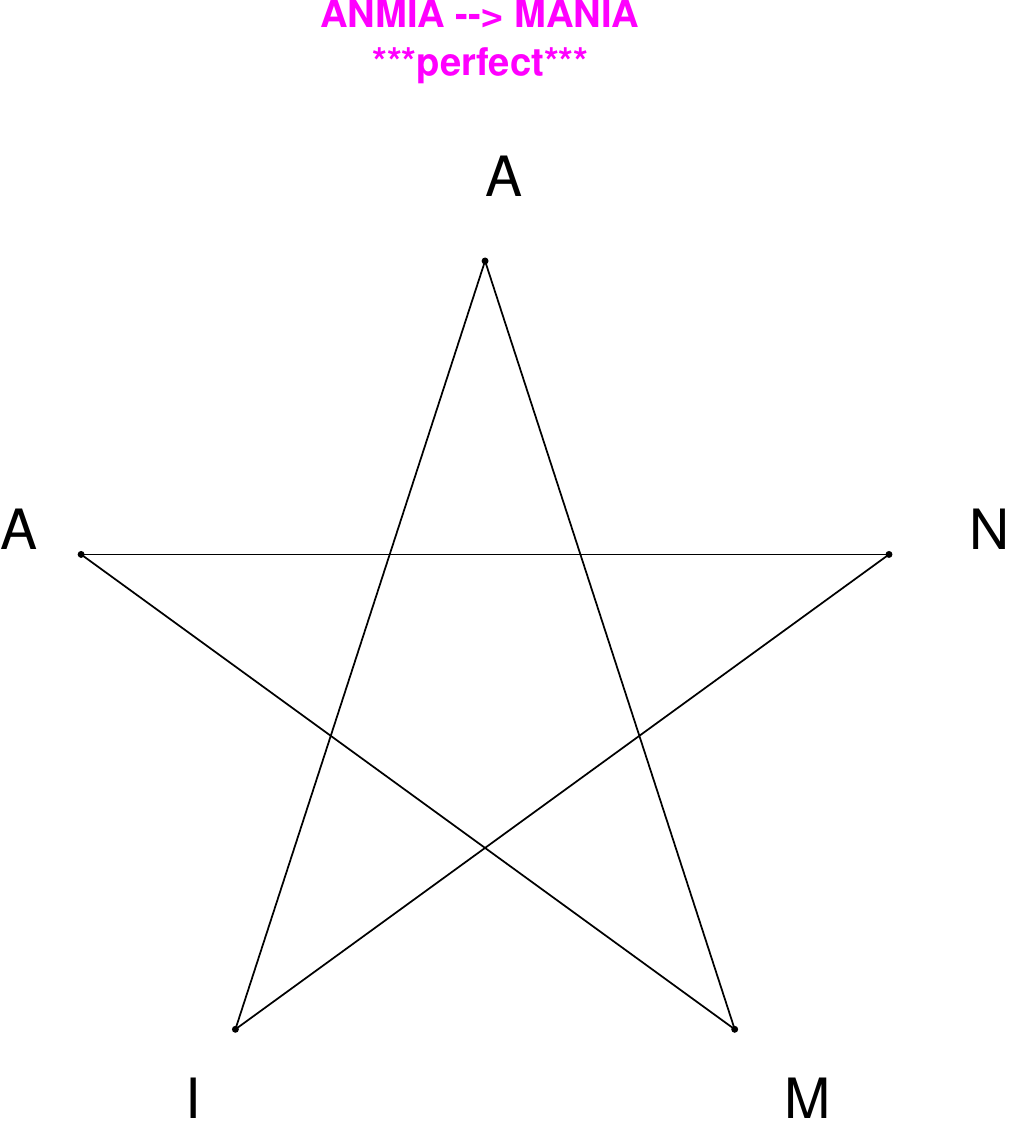}
\end{subfigure}
\hfill
\begin{subfigure}[T]{0.19\textwidth}
\centering
\includegraphics[width=\textwidth]{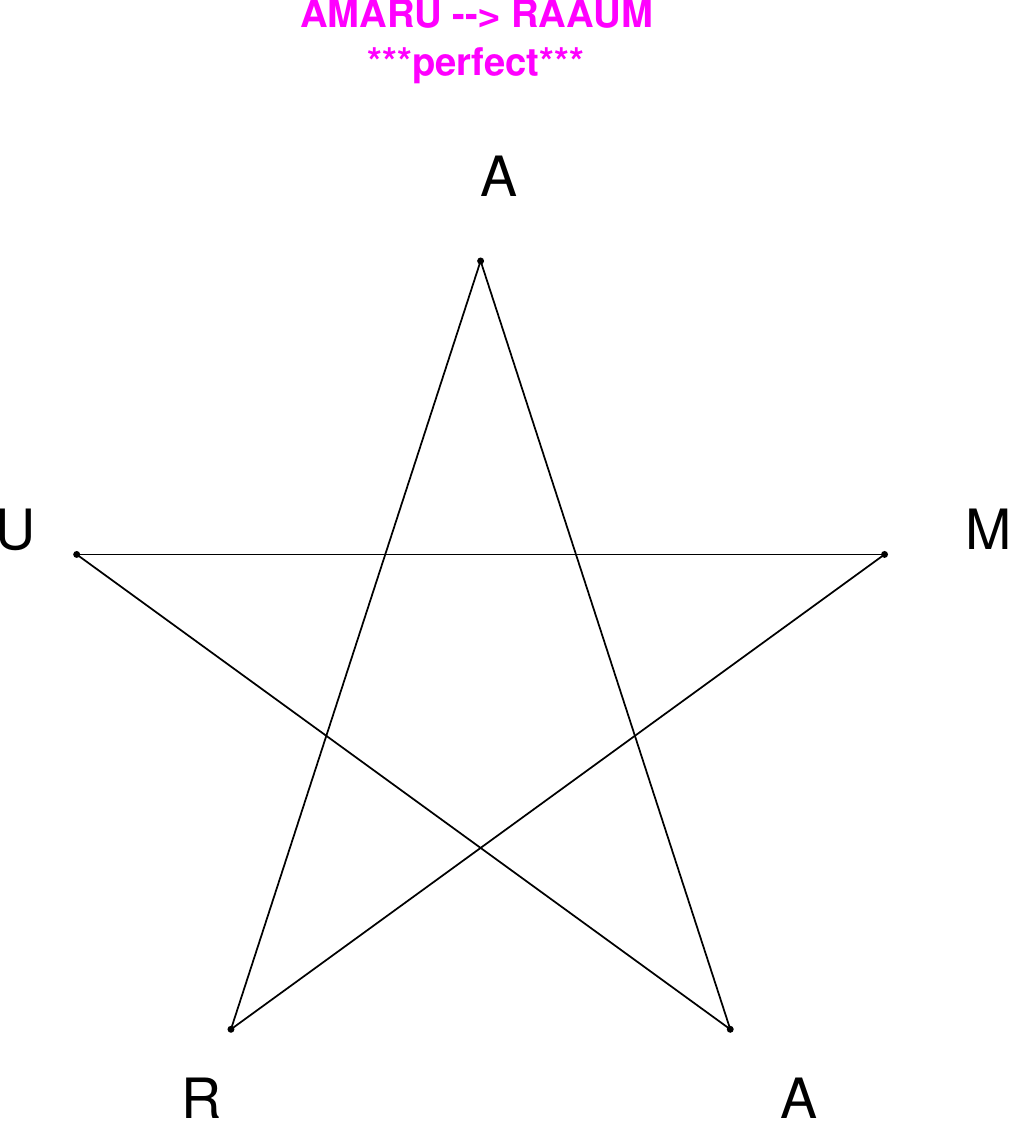}
\end{subfigure}
\hfill
\begin{subfigure}[T]{0.19\textwidth}
\centering
\includegraphics[width=\textwidth]{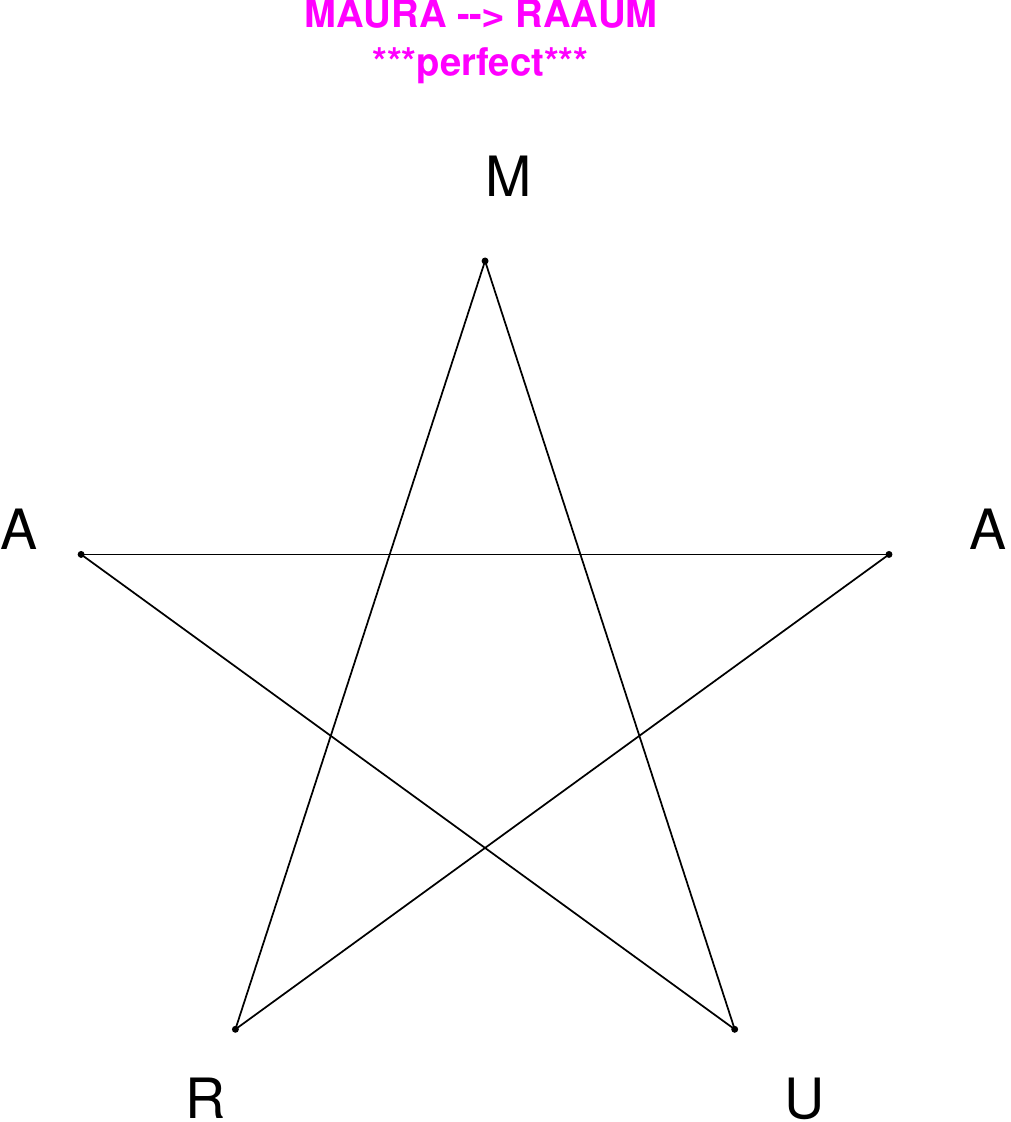}
\end{subfigure}
\end{figure}

\begin{figure}[H]
\centering
\begin{subfigure}[T]{0.19\textwidth}
\centering
\includegraphics[width=\textwidth]{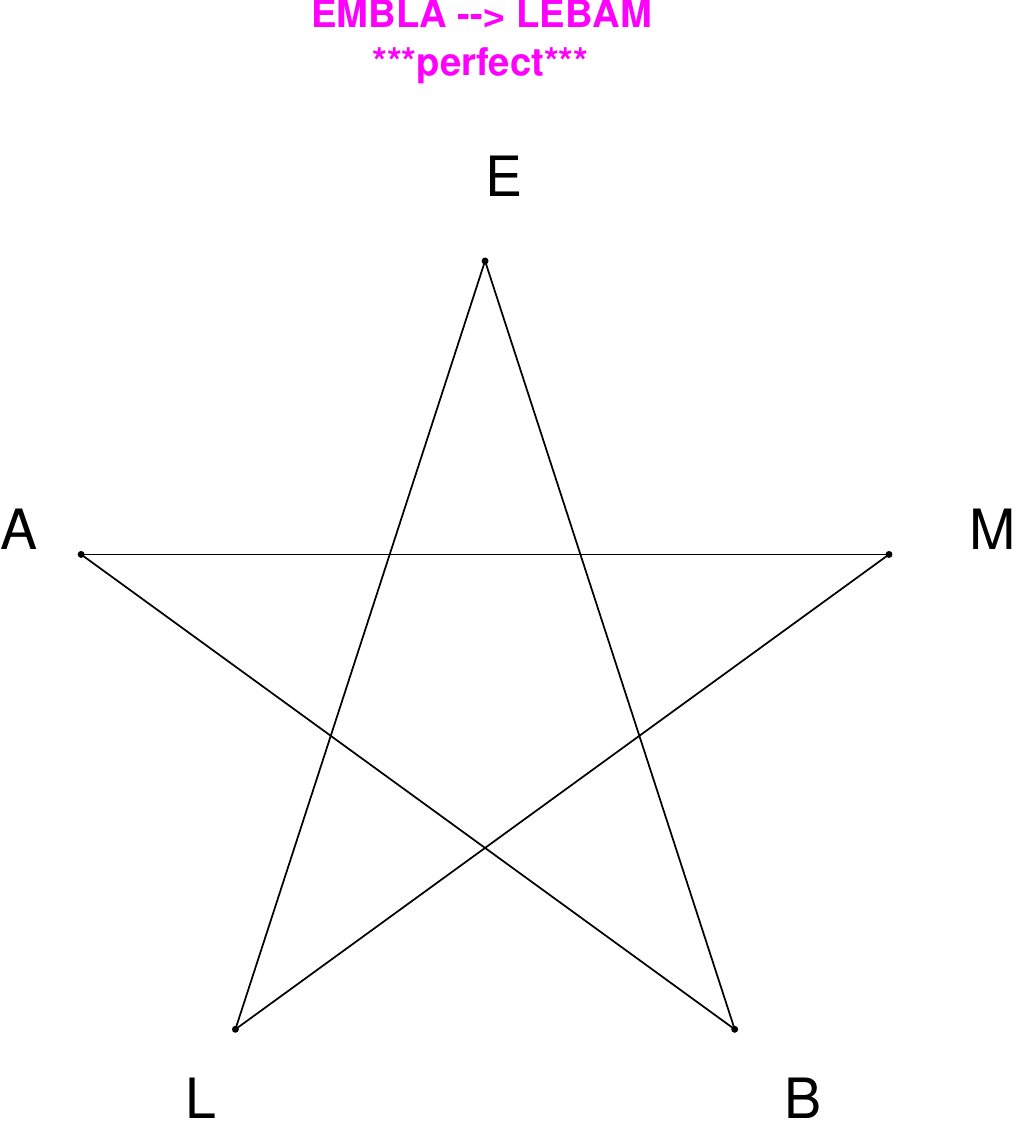}
\end{subfigure}
\hfill
\begin{subfigure}[T]{0.19\textwidth}
\centering
\includegraphics[width=\textwidth]{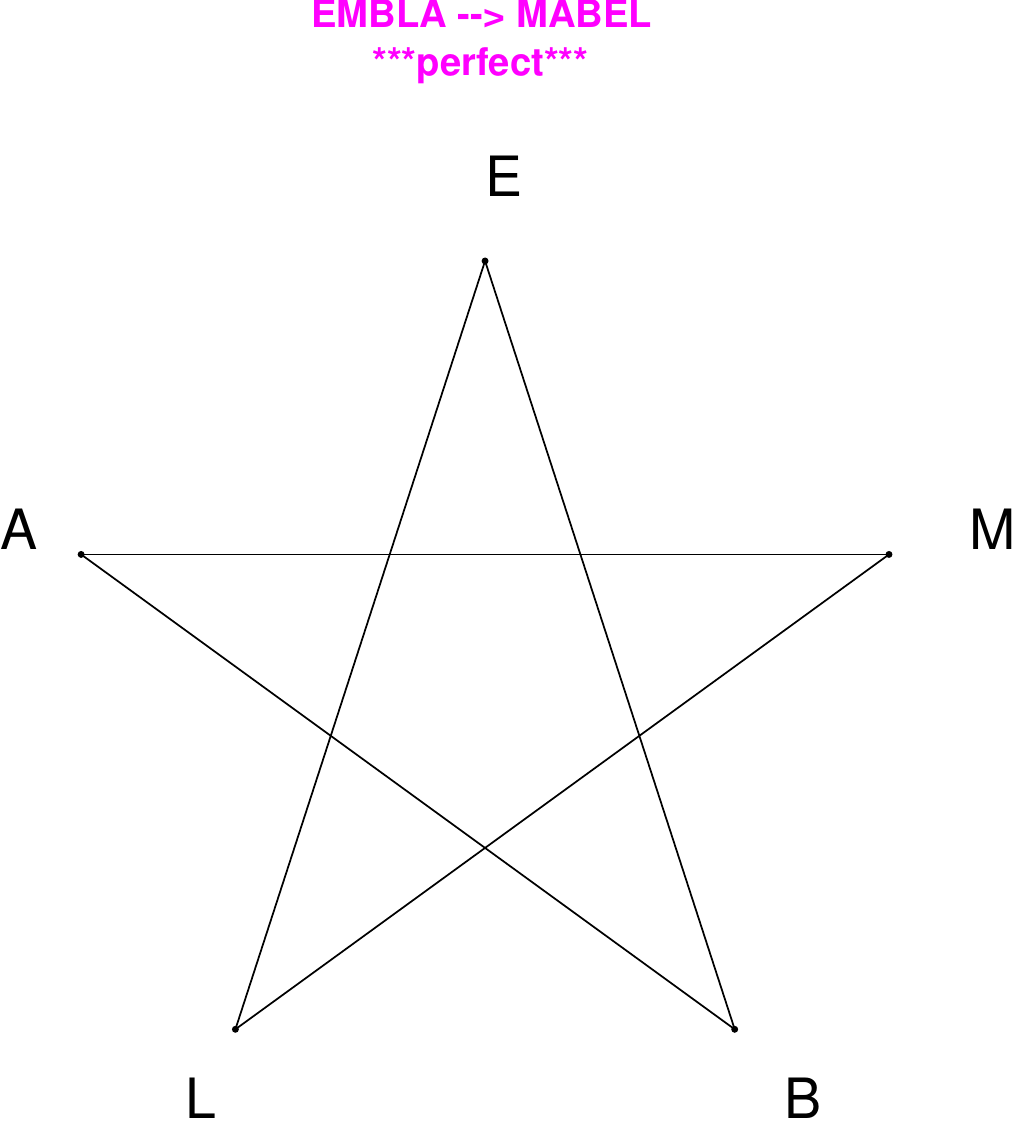}
\end{subfigure}
\hfill
\begin{subfigure}[T]{0.19\textwidth}
\centering
\includegraphics[width=\textwidth]{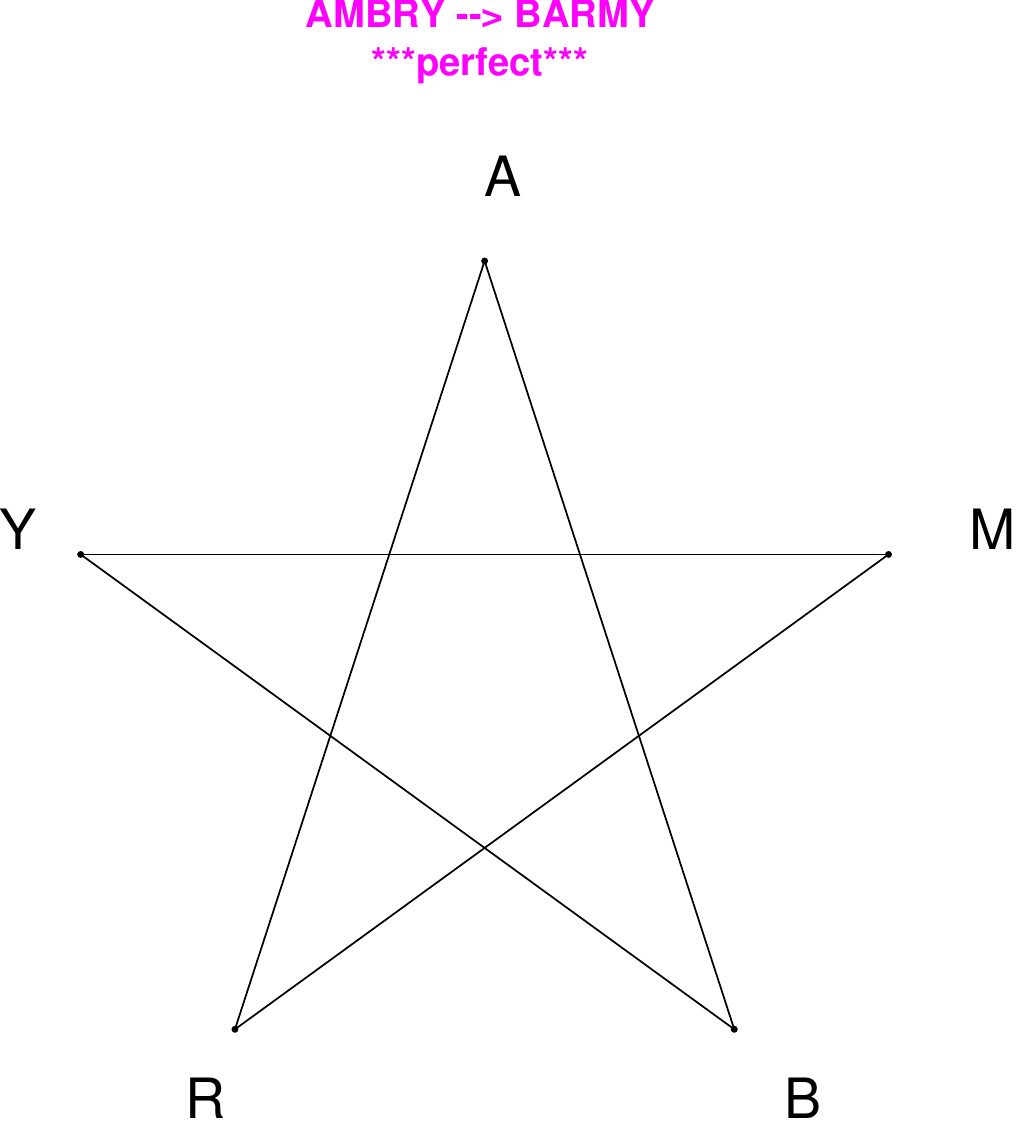}
\end{subfigure}
\hfill
\begin{subfigure}[T]{0.19\textwidth}
\centering
\includegraphics[width=\textwidth]{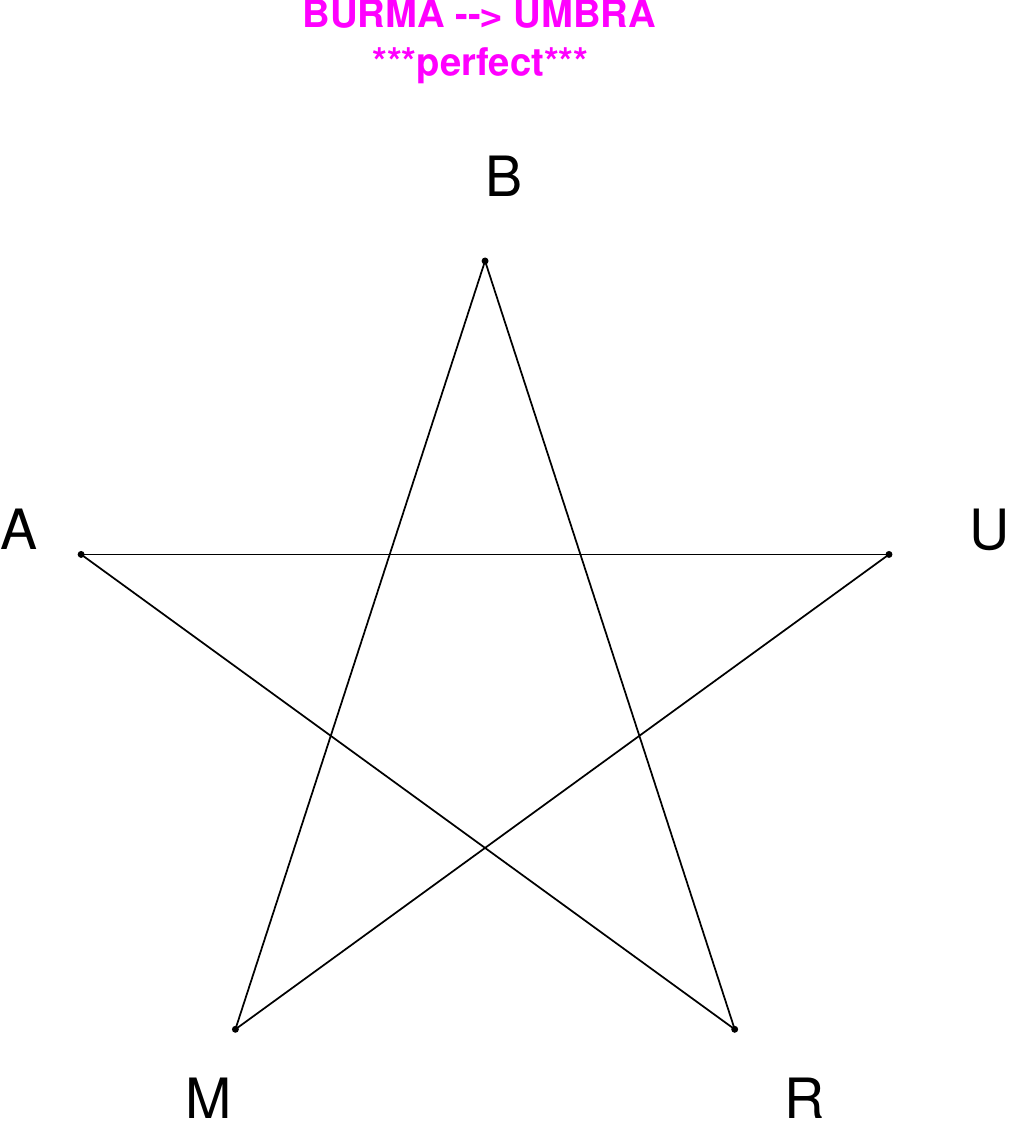}
\end{subfigure}
\hfill
\begin{subfigure}[T]{0.19\textwidth}
\centering
\includegraphics[width=\textwidth]{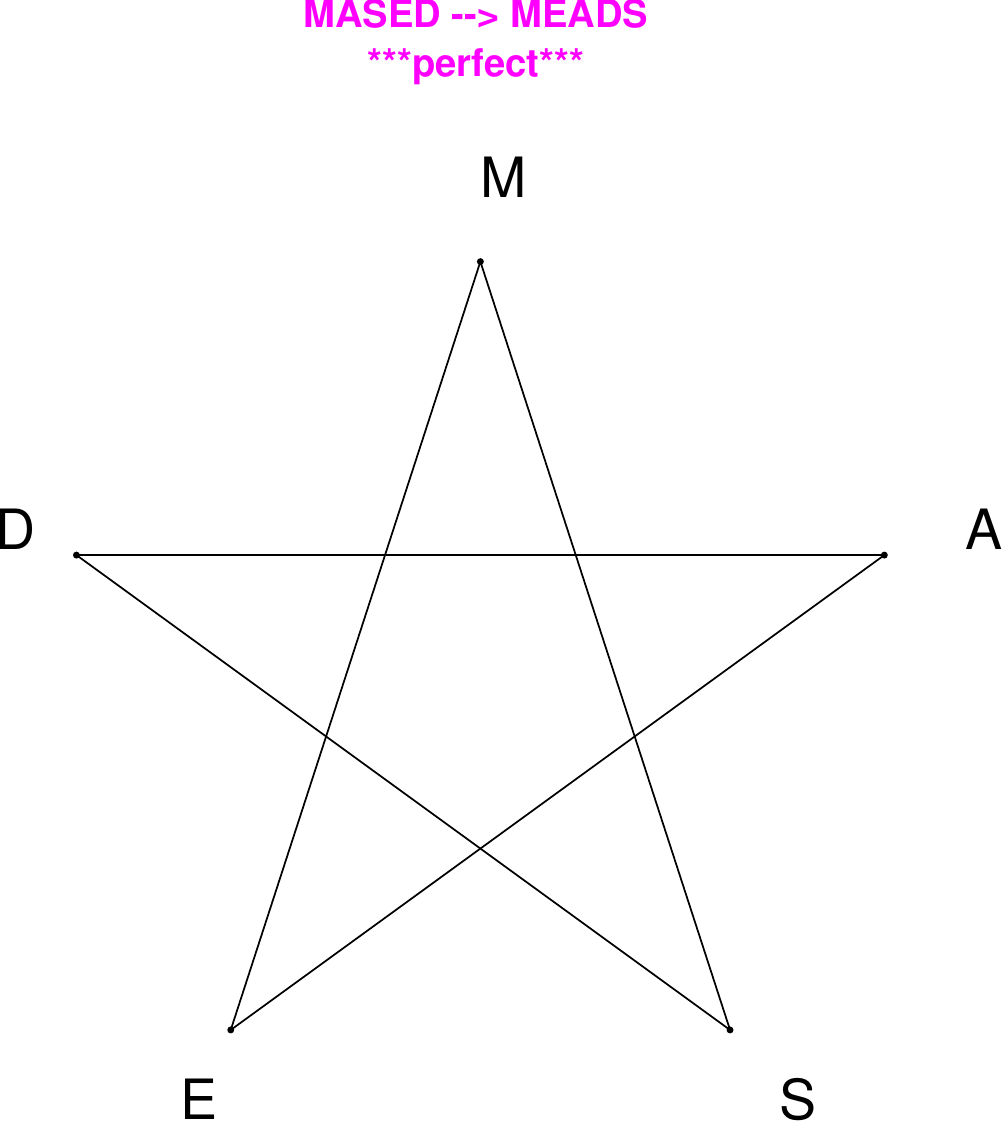}
\end{subfigure}
\end{figure}

\begin{figure}[H]
\centering
\begin{subfigure}[T]{0.19\textwidth}
\centering
\includegraphics[width=\textwidth]{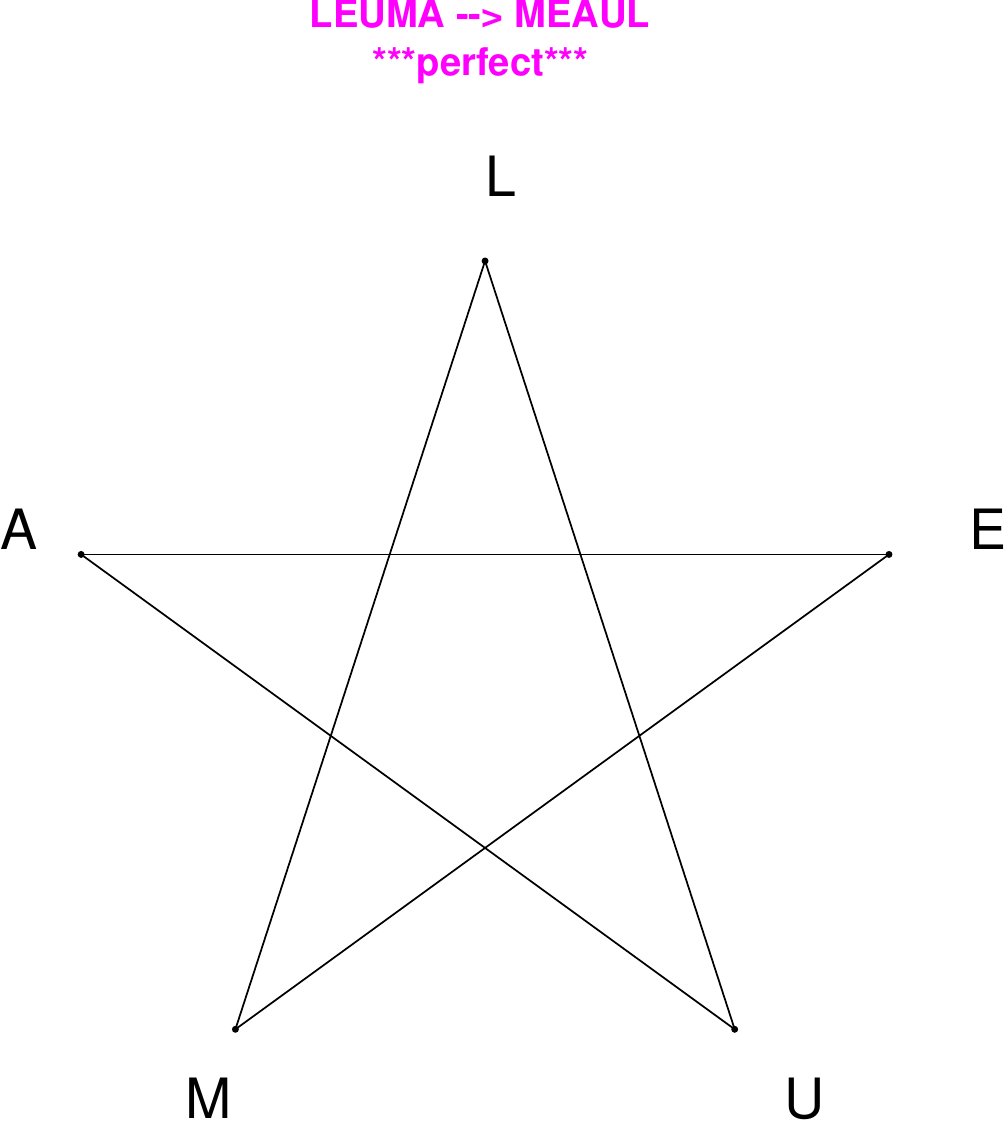}
\end{subfigure}
\hfill
\begin{subfigure}[T]{0.19\textwidth}
\centering
\includegraphics[width=\textwidth]{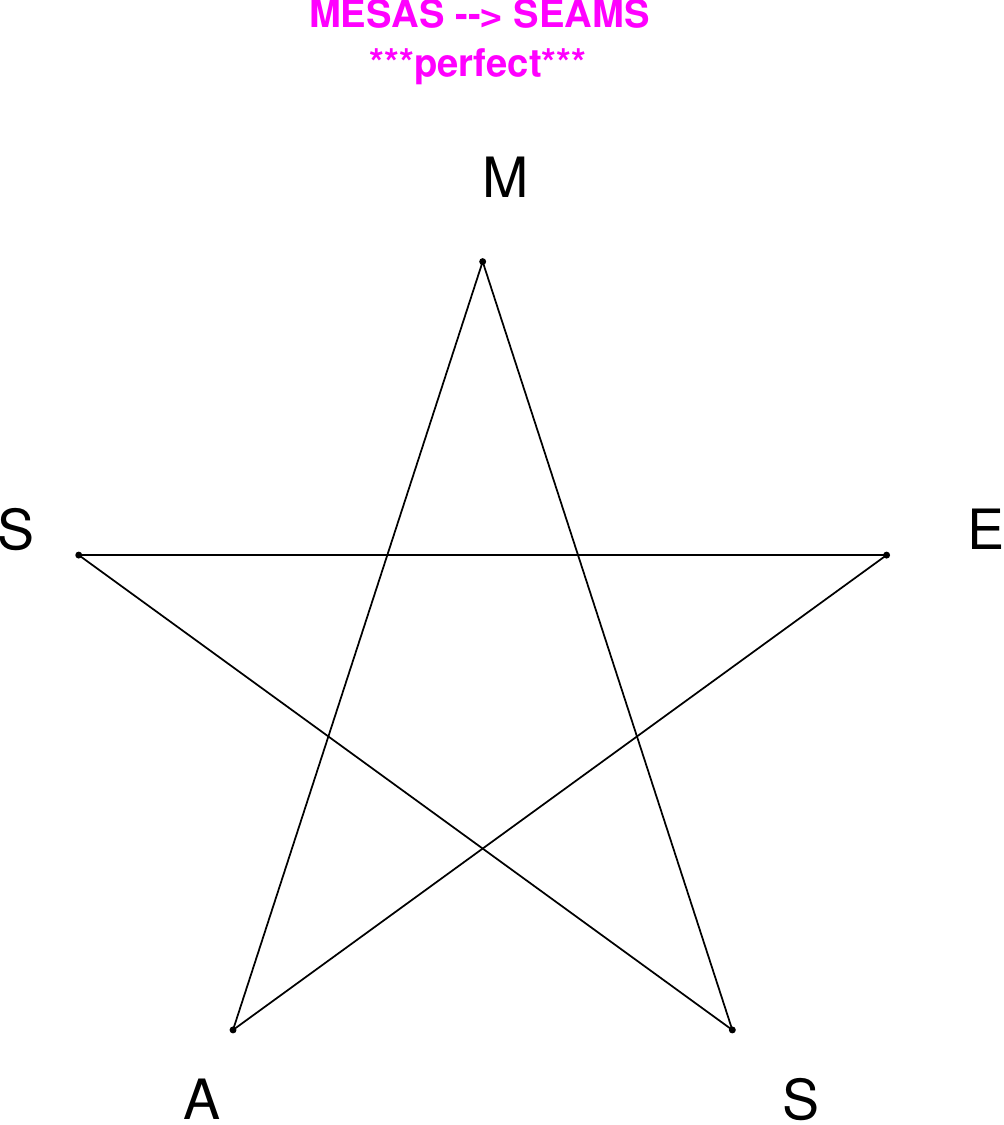}
\end{subfigure}
\hfill
\begin{subfigure}[T]{0.19\textwidth}
\centering
\includegraphics[width=\textwidth]{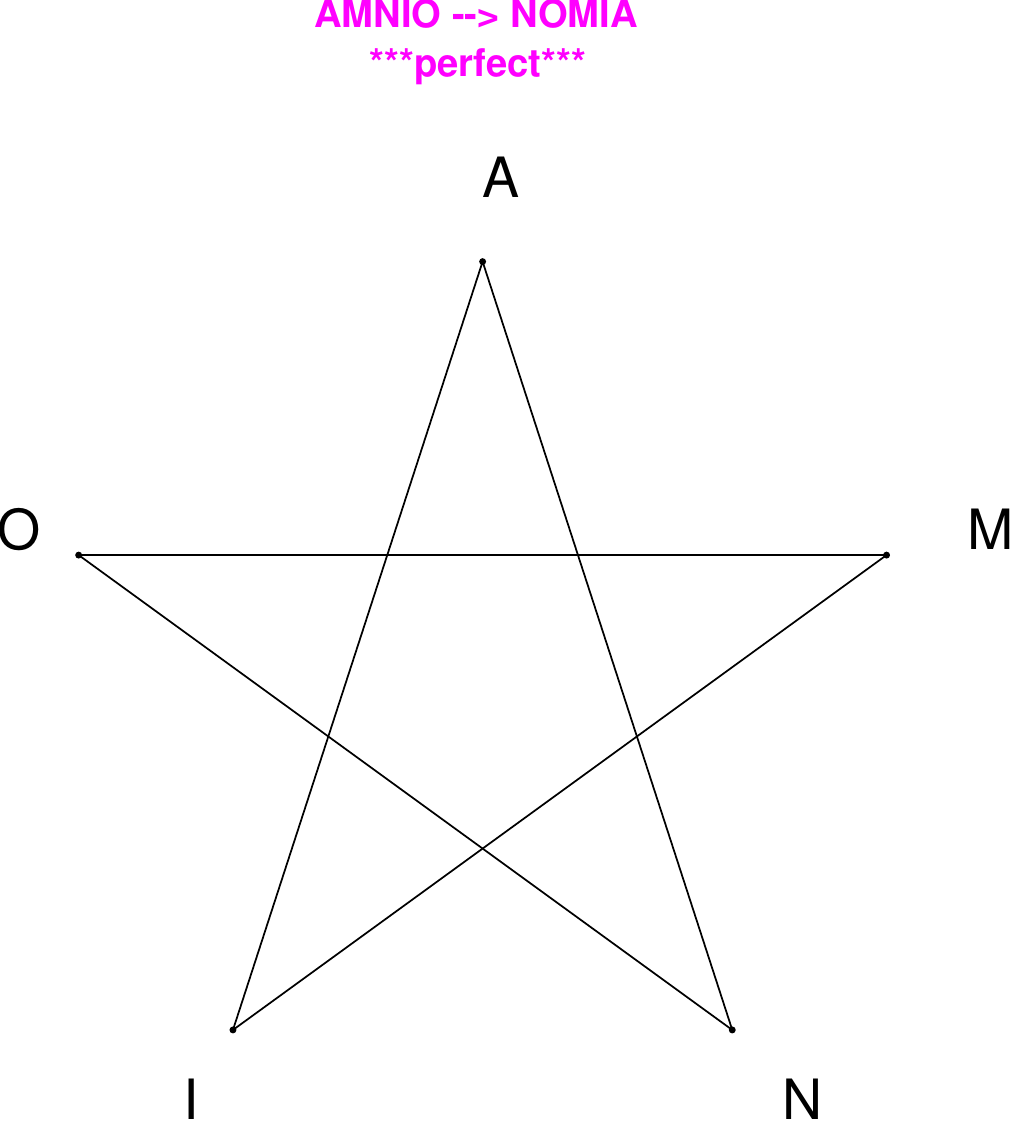}
\end{subfigure}
\hfill
\begin{subfigure}[T]{0.19\textwidth}
\centering
\includegraphics[width=\textwidth]{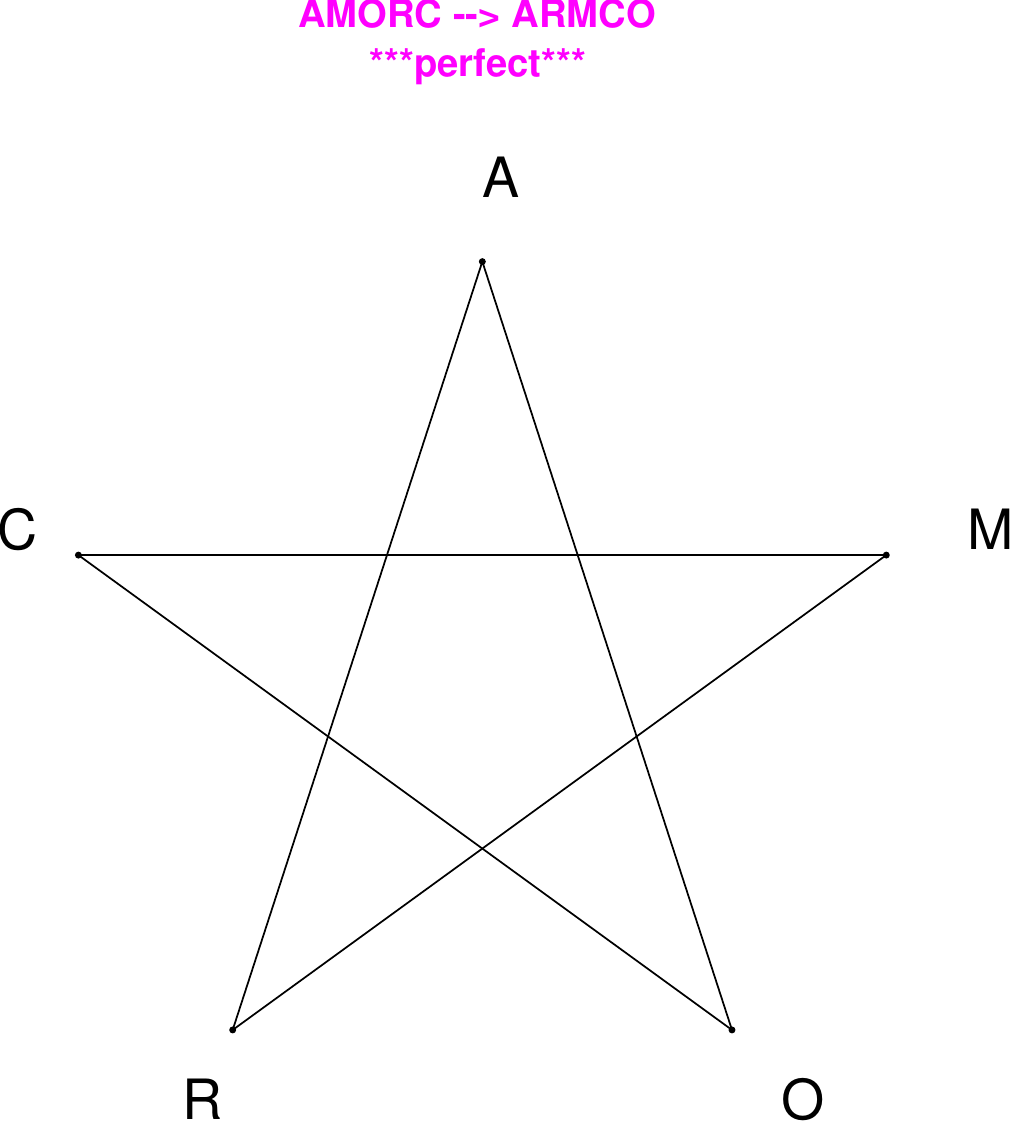}
\end{subfigure}
\hfill
\begin{subfigure}[T]{0.19\textwidth}
\centering
\includegraphics[width=\textwidth]{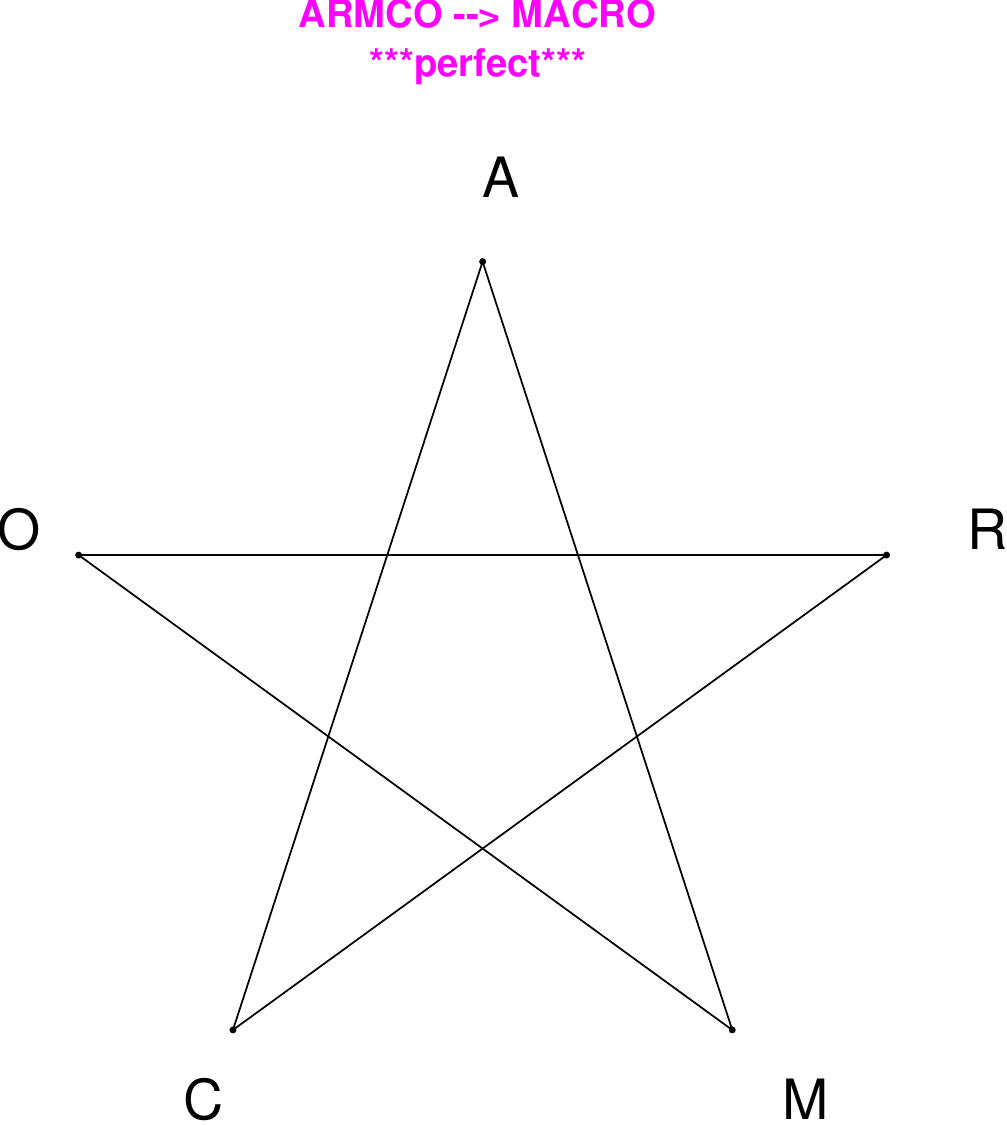}
\end{subfigure}
\end{figure}

\begin{figure}[H]
\centering
\begin{subfigure}[T]{0.19\textwidth}
\centering
\includegraphics[width=\textwidth]{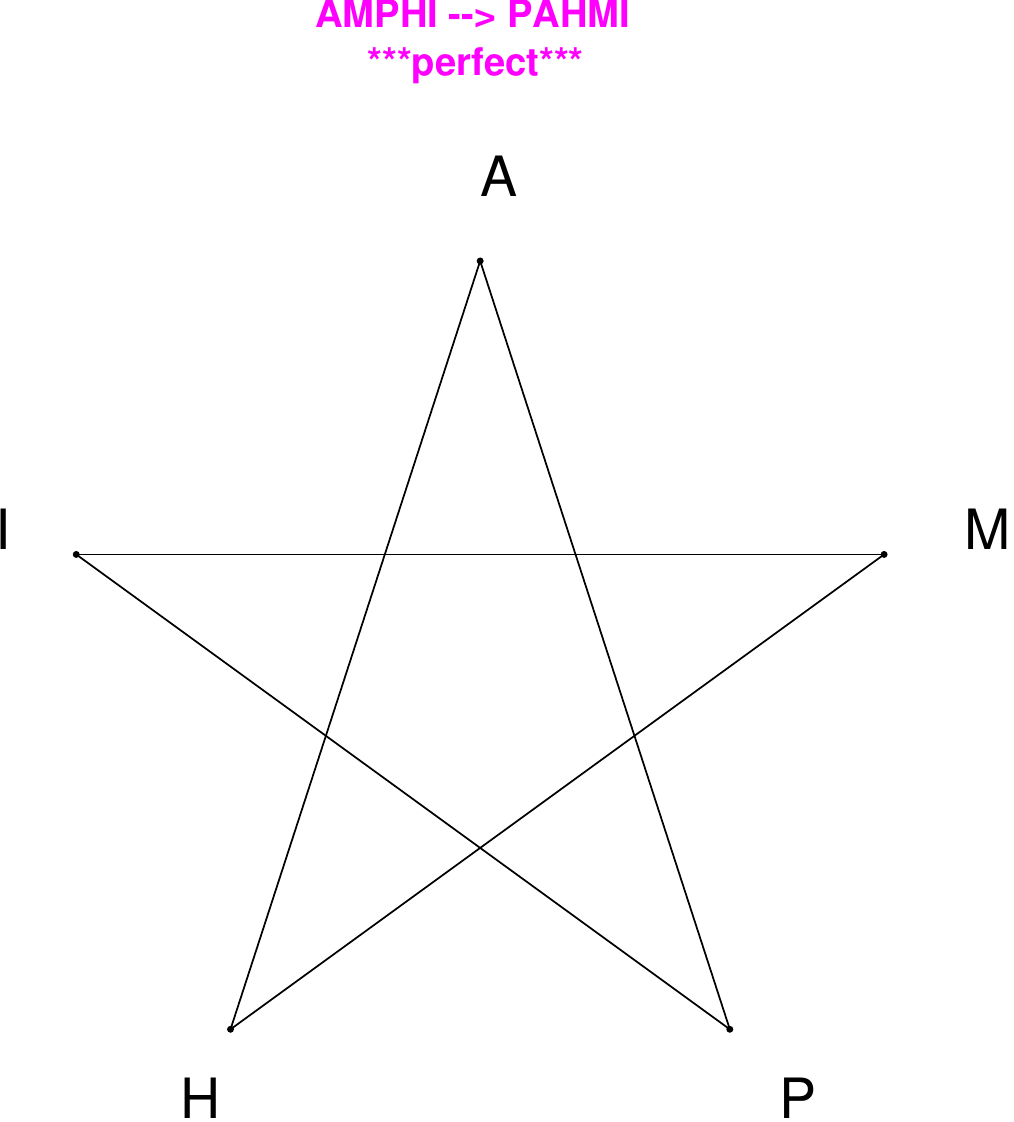}
\end{subfigure}
\hfill
\begin{subfigure}[T]{0.19\textwidth}
\centering
\includegraphics[width=\textwidth]{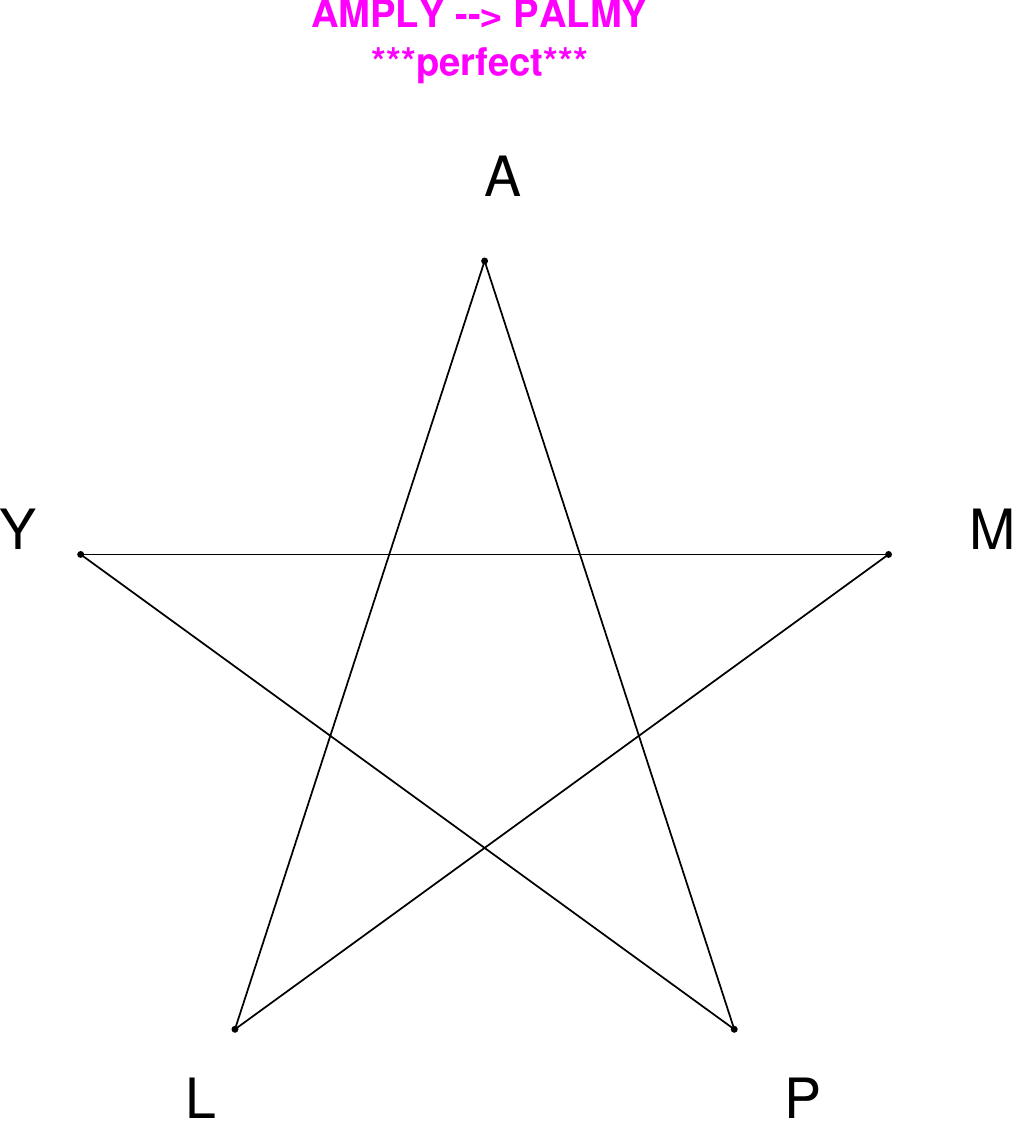}
\end{subfigure}
\hfill
\begin{subfigure}[T]{0.19\textwidth}
\centering
\includegraphics[width=\textwidth]{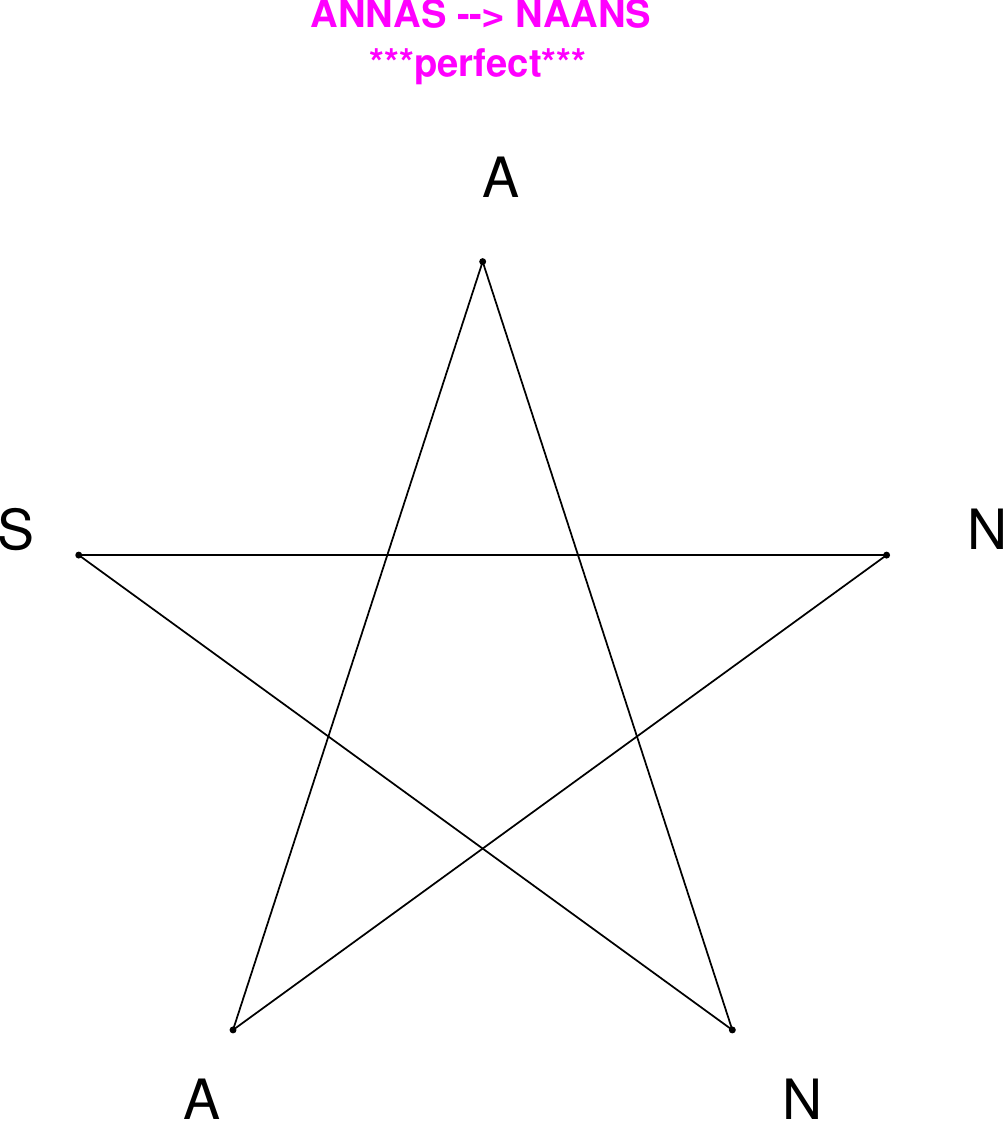}
\end{subfigure}
\hfill
\begin{subfigure}[T]{0.19\textwidth}
\centering
\includegraphics[width=\textwidth]{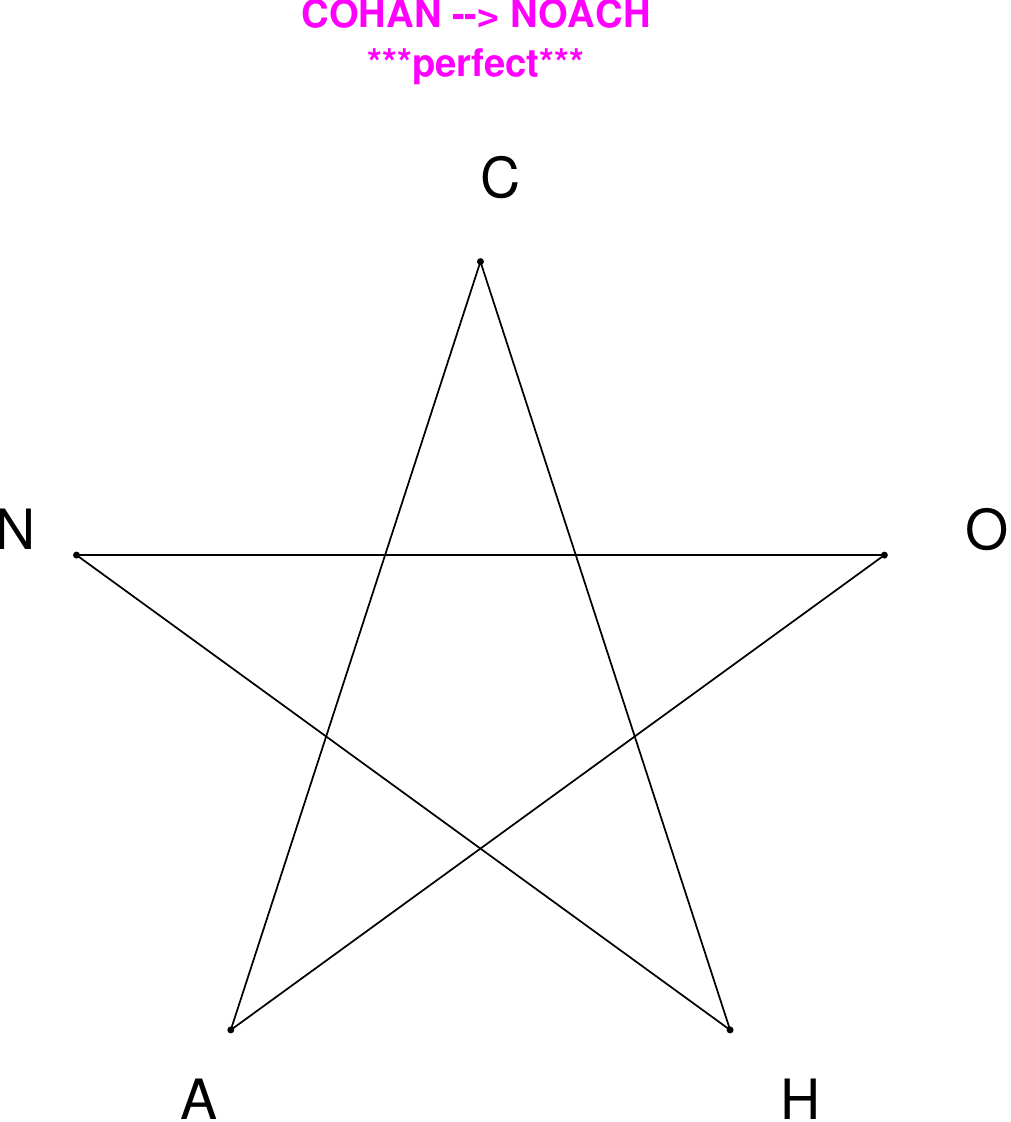}
\end{subfigure}
\hfill
\begin{subfigure}[T]{0.19\textwidth}
\centering
\includegraphics[width=\textwidth]{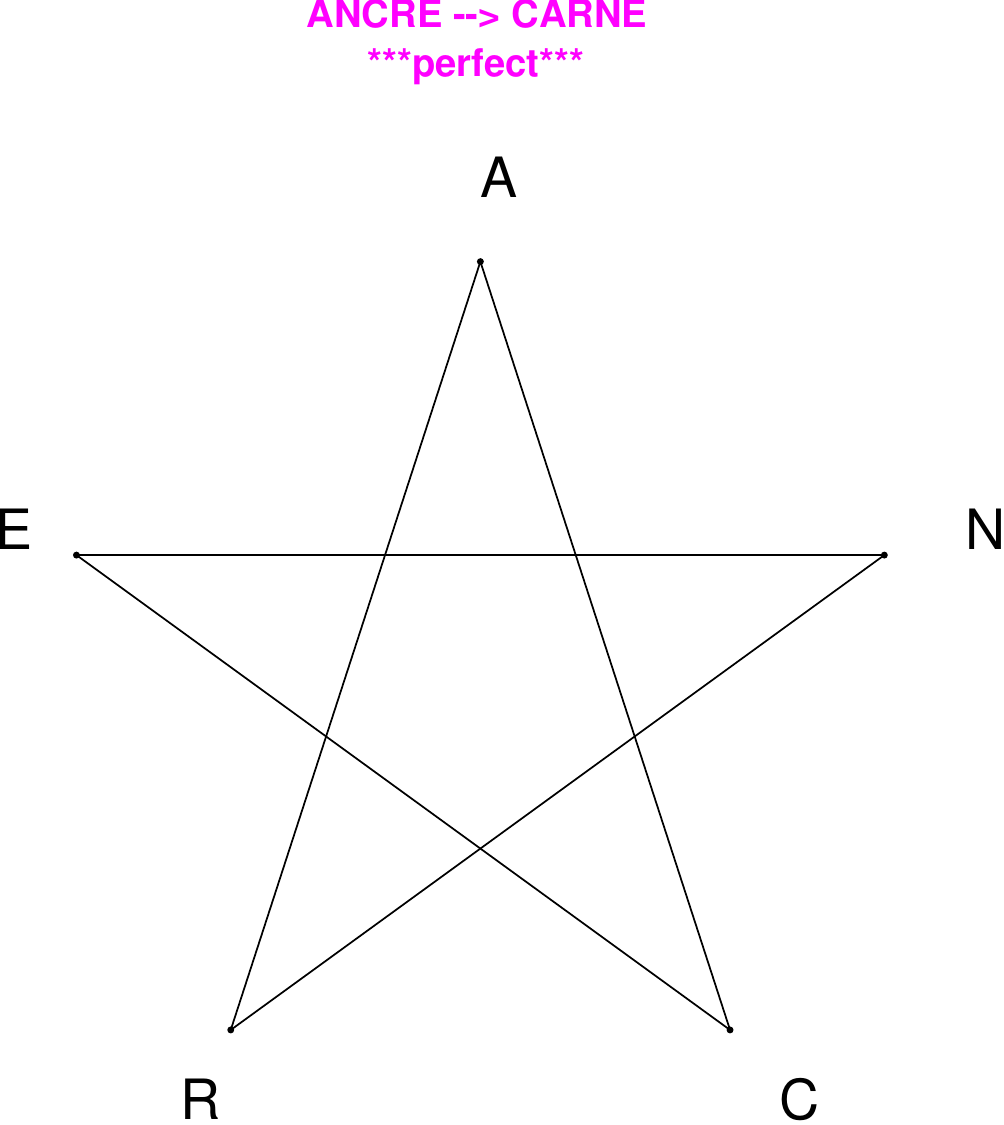}
\end{subfigure}
\end{figure}

\begin{figure}[H]
\centering
\begin{subfigure}[T]{0.19\textwidth}
\centering
\includegraphics[width=\textwidth]{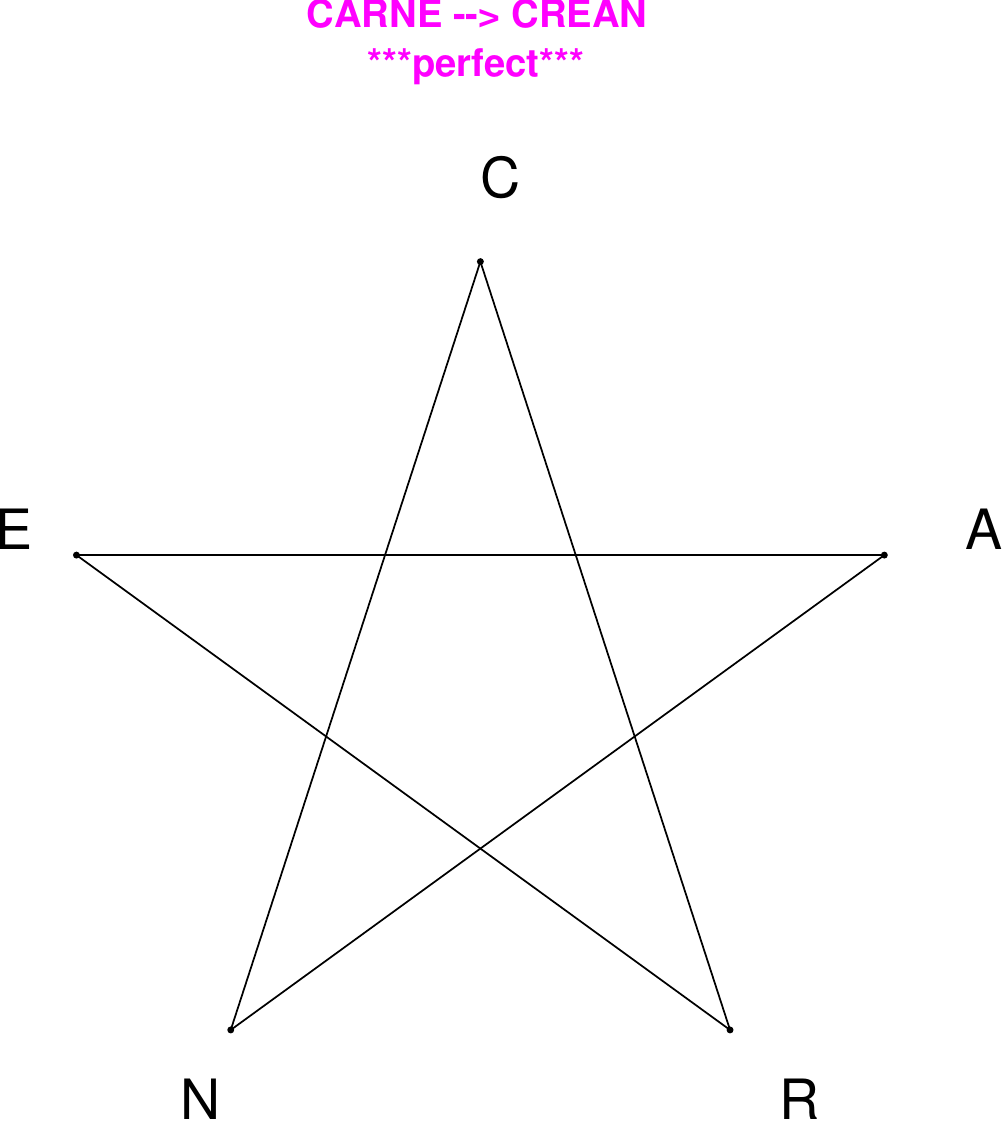}
\end{subfigure}
\hfill
\begin{subfigure}[T]{0.19\textwidth}
\centering
\includegraphics[width=\textwidth]{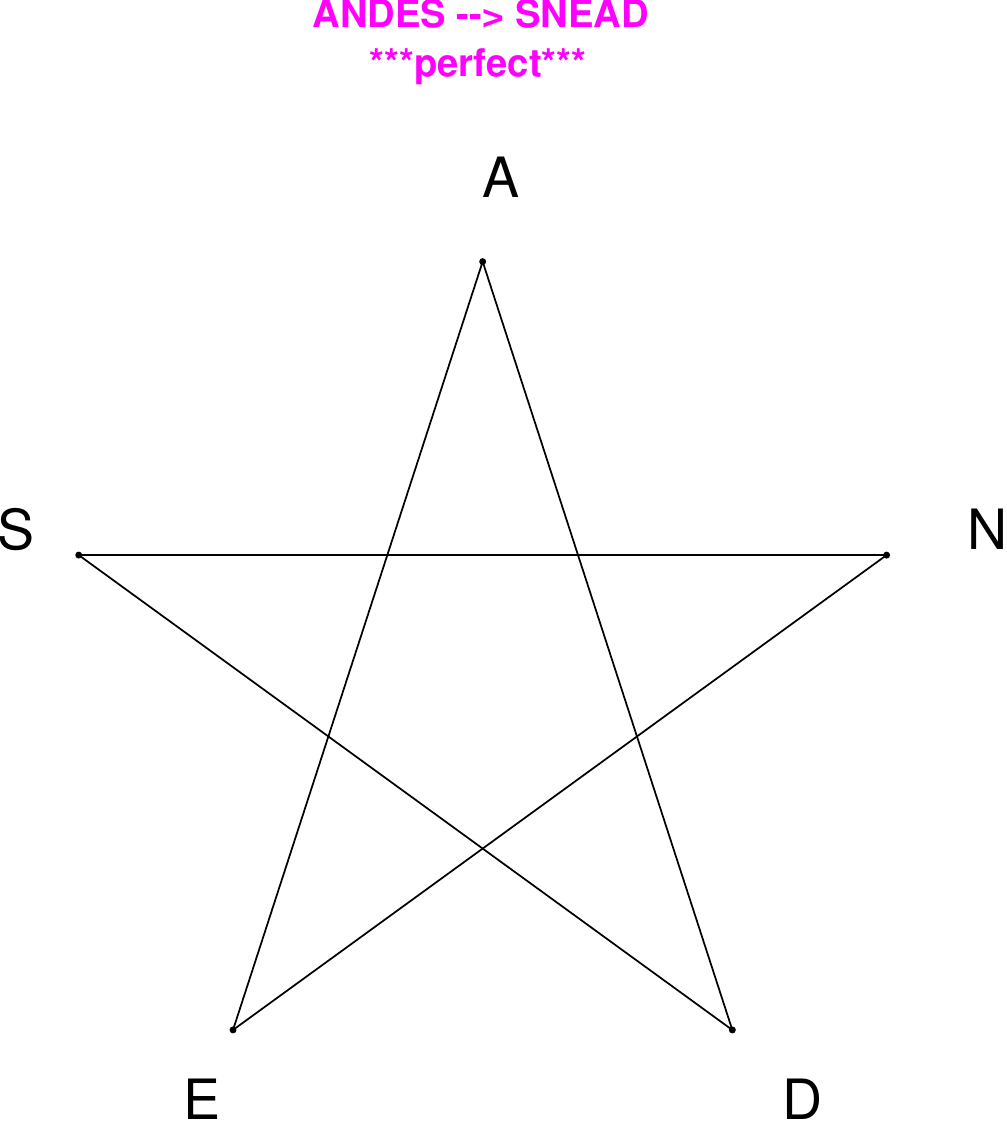}
\end{subfigure}
\hfill
\begin{subfigure}[T]{0.19\textwidth}
\centering
\includegraphics[width=\textwidth]{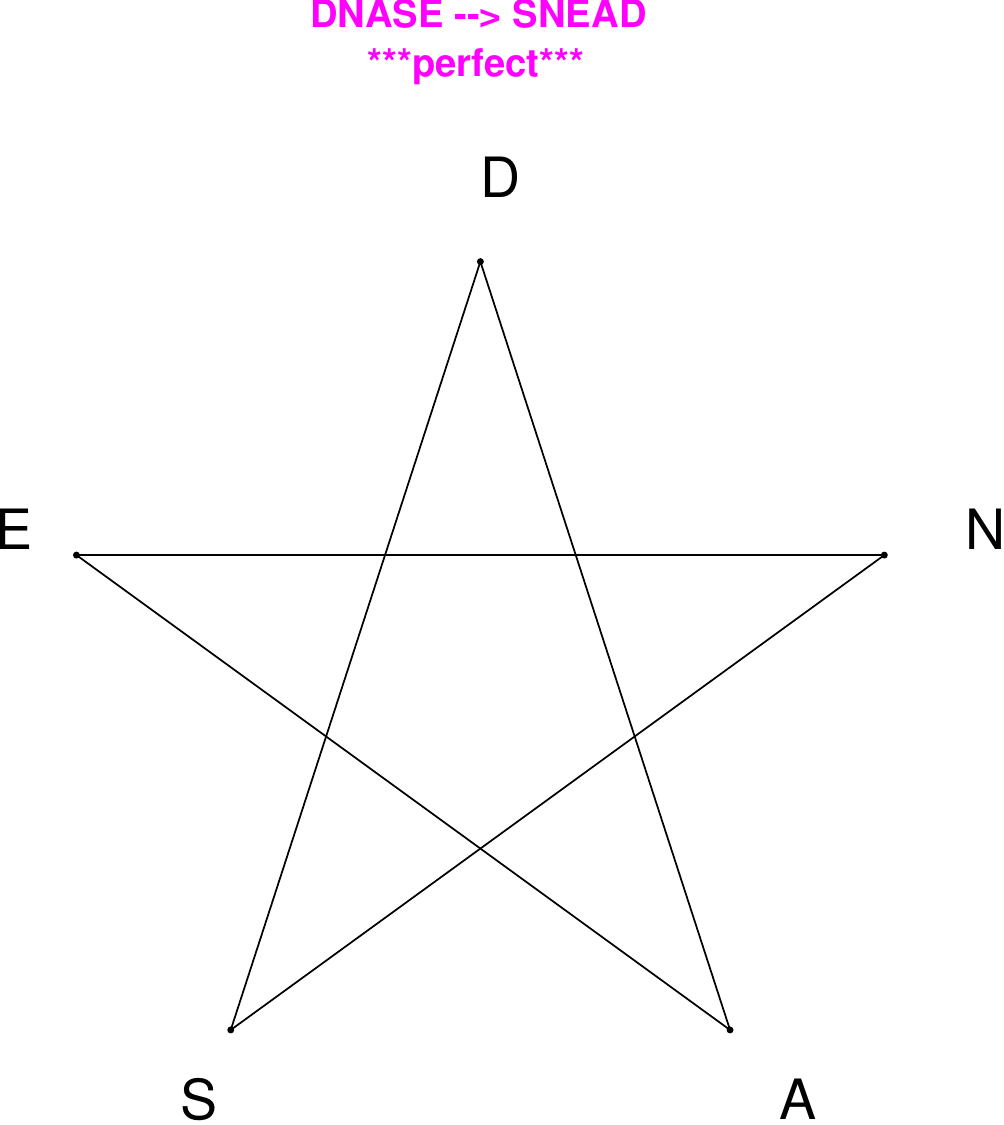}
\end{subfigure}
\hfill
\begin{subfigure}[T]{0.19\textwidth}
\centering
\includegraphics[width=\textwidth]{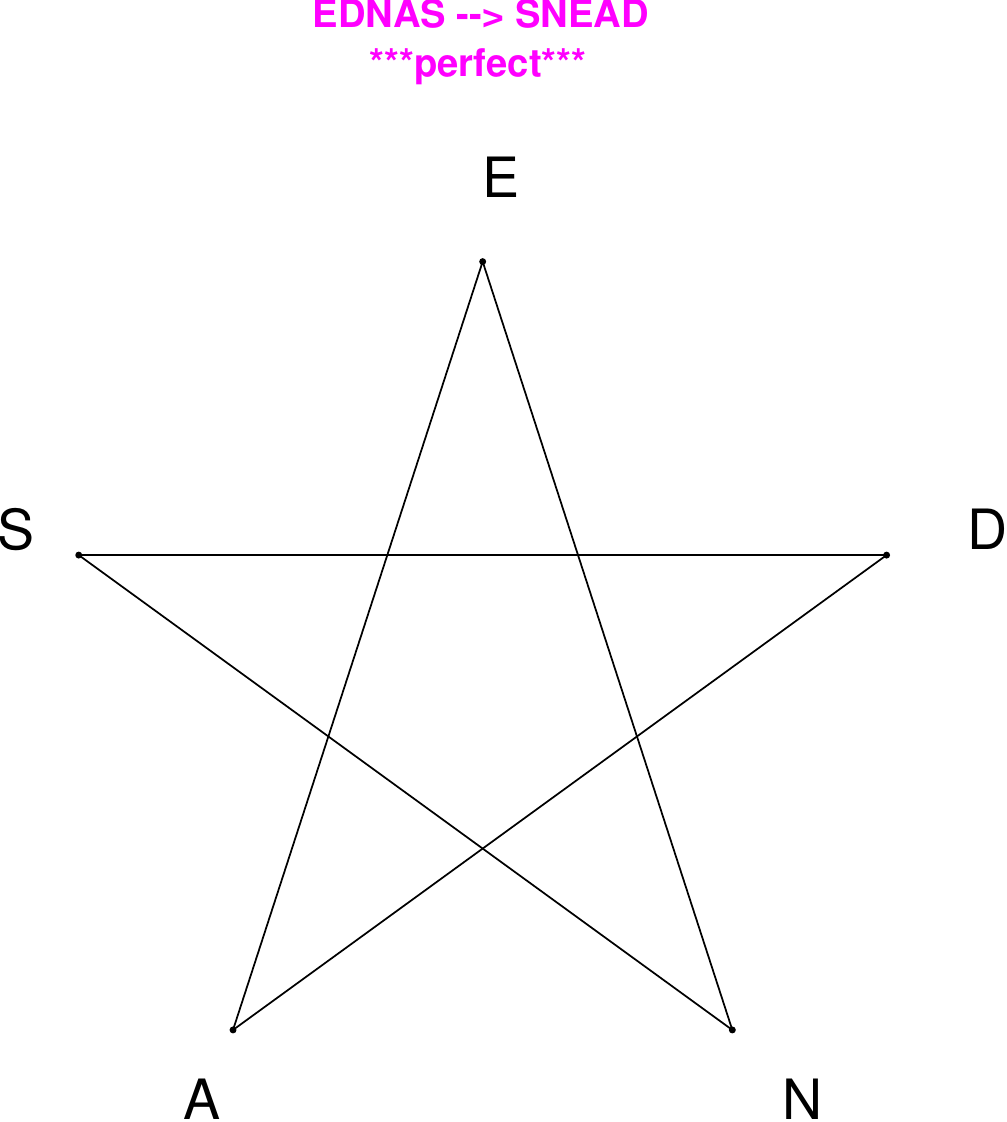}
\end{subfigure}
\hfill
\begin{subfigure}[T]{0.19\textwidth}
\centering
\includegraphics[width=\textwidth]{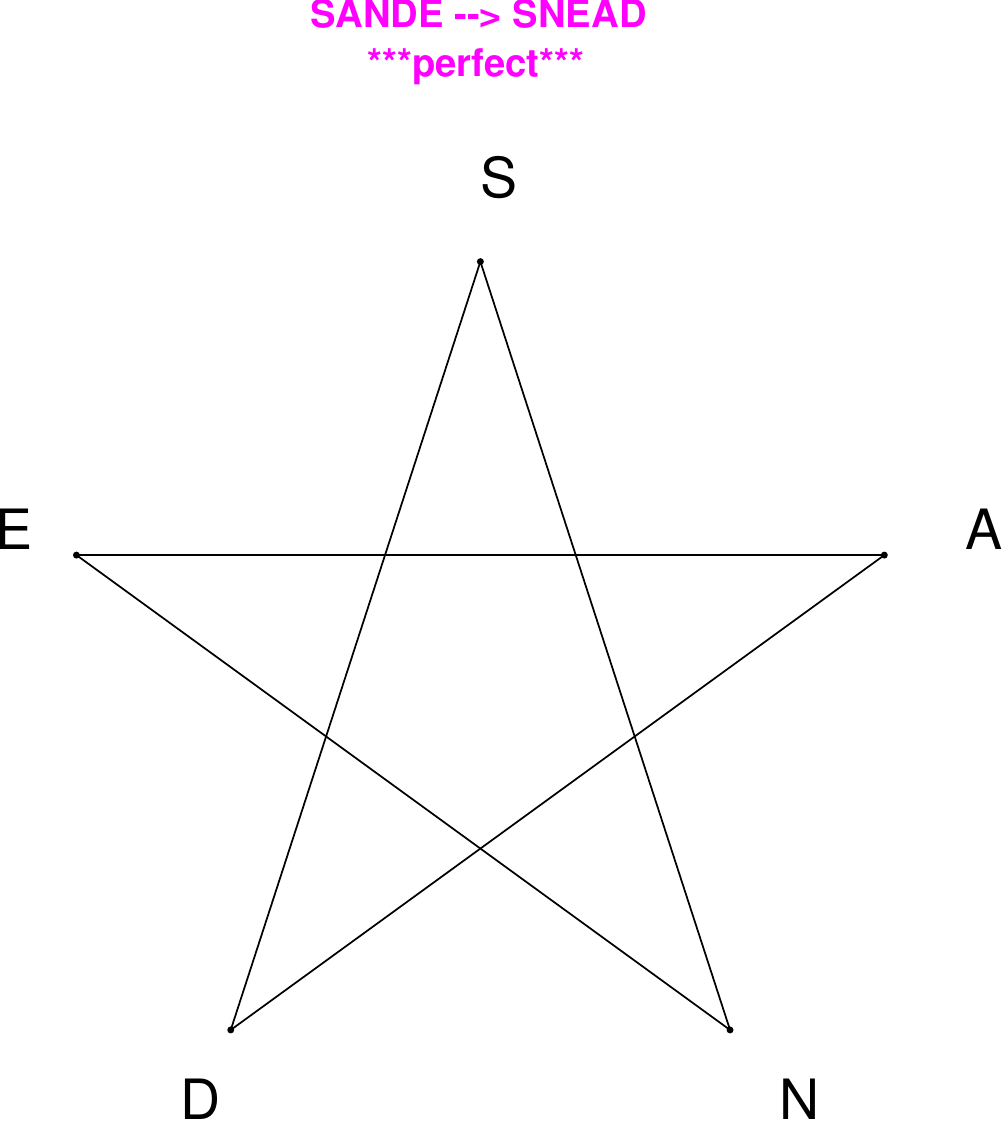}
\end{subfigure}
\end{figure}

\begin{figure}[H]
\centering
\begin{subfigure}[T]{0.19\textwidth}
\centering
\includegraphics[width=\textwidth]{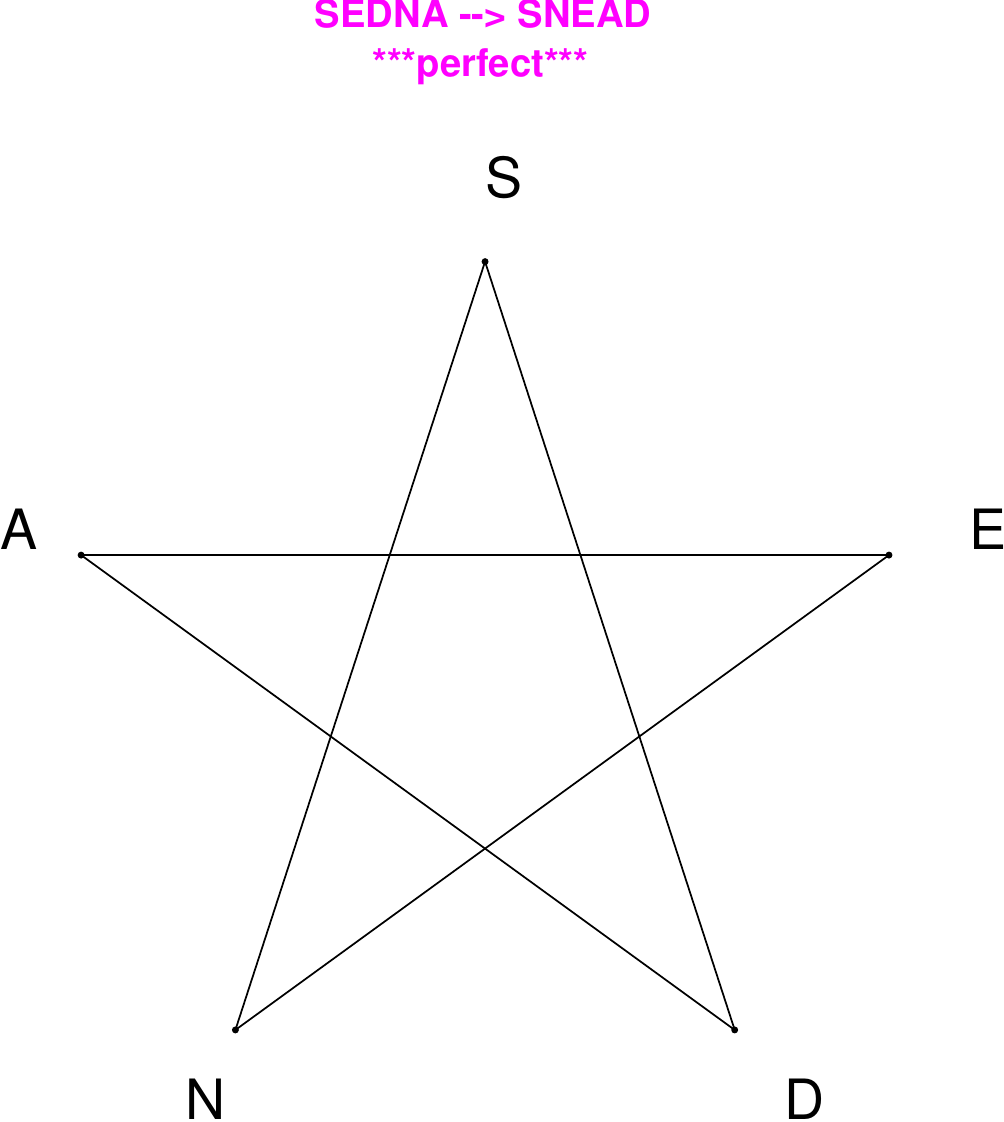}
\end{subfigure}
\hfill
\begin{subfigure}[T]{0.19\textwidth}
\centering
\includegraphics[width=\textwidth]{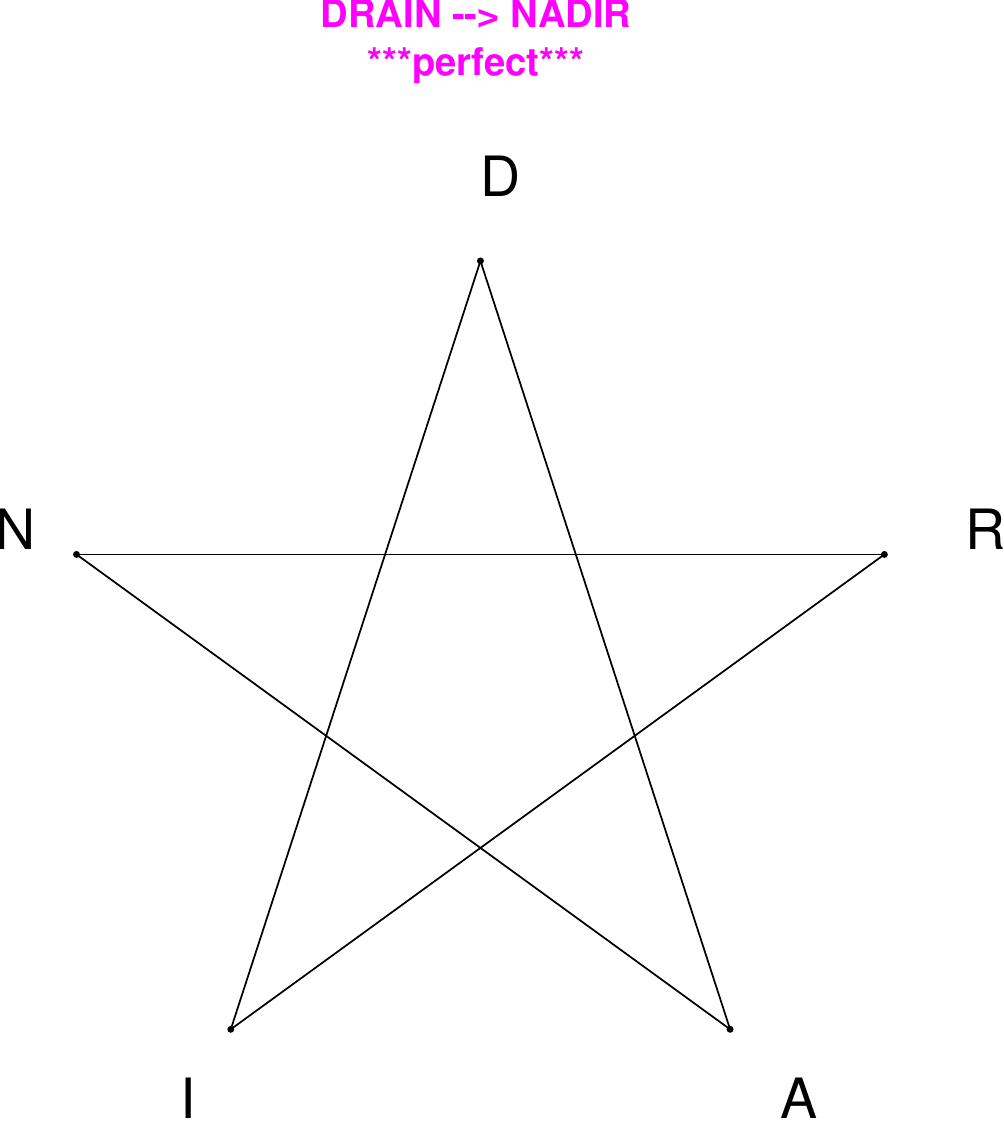}
\end{subfigure}
\hfill
\begin{subfigure}[T]{0.19\textwidth}
\centering
\includegraphics[width=\textwidth]{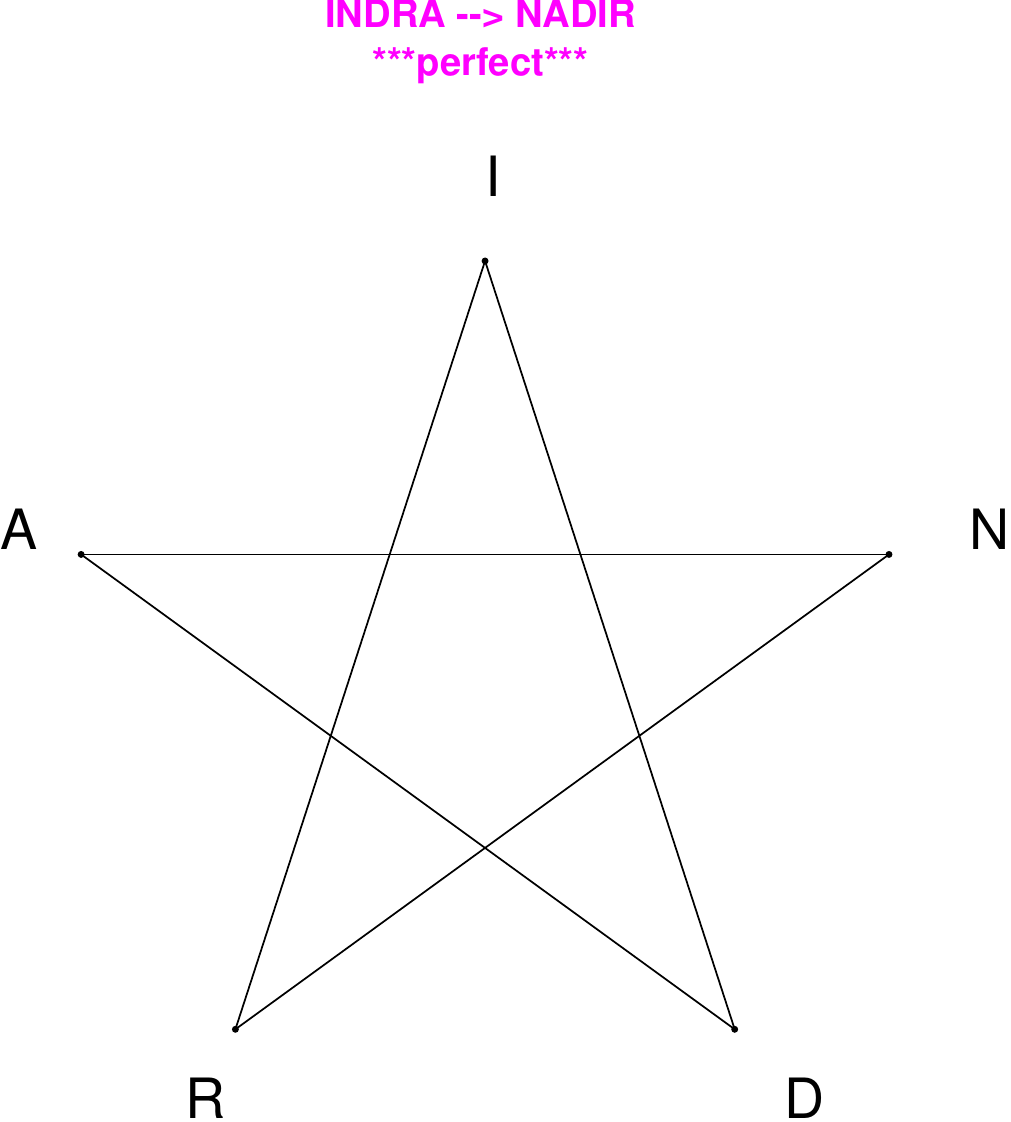}
\end{subfigure}
\hfill
\begin{subfigure}[T]{0.19\textwidth}
\centering
\includegraphics[width=\textwidth]{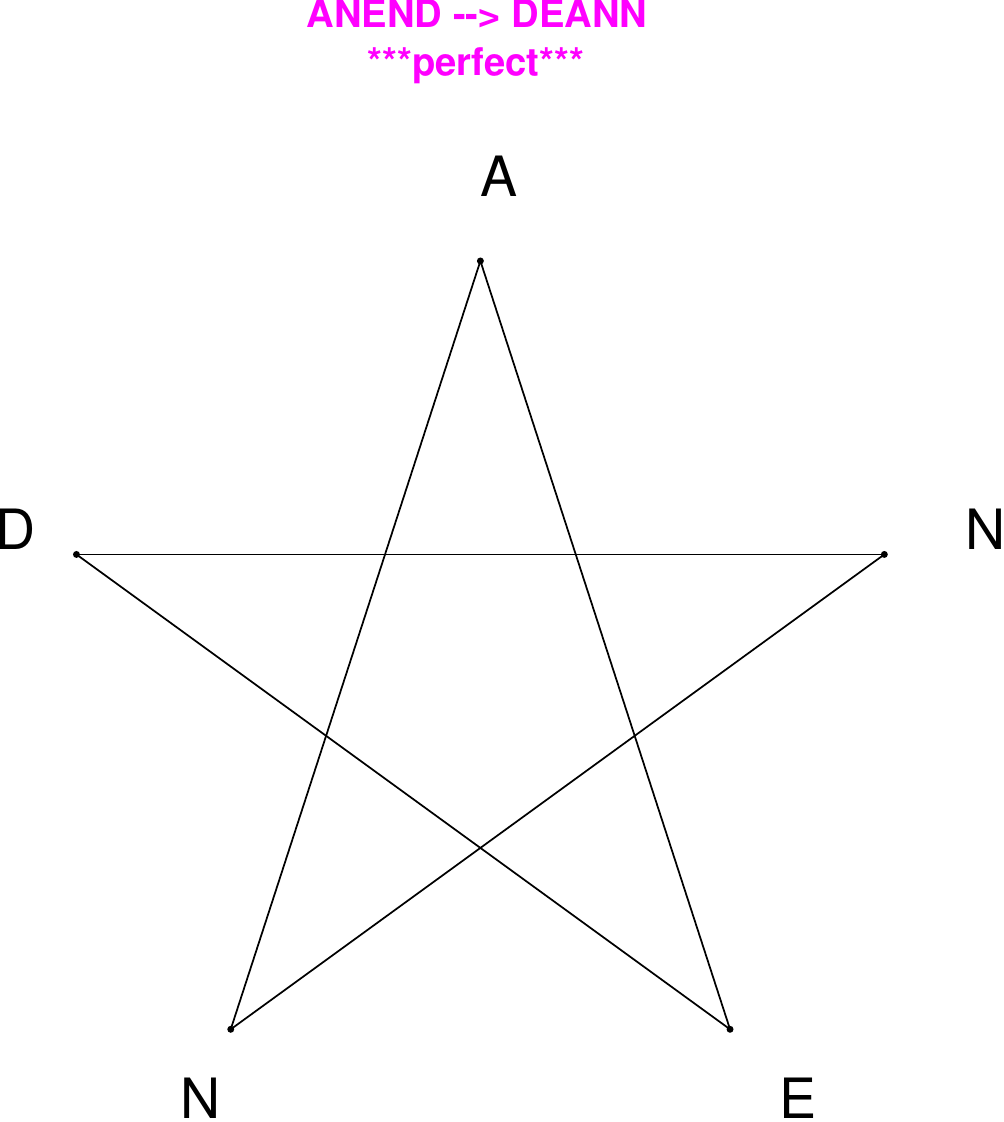}
\end{subfigure}
\hfill
\begin{subfigure}[T]{0.19\textwidth}
\centering
\includegraphics[width=\textwidth]{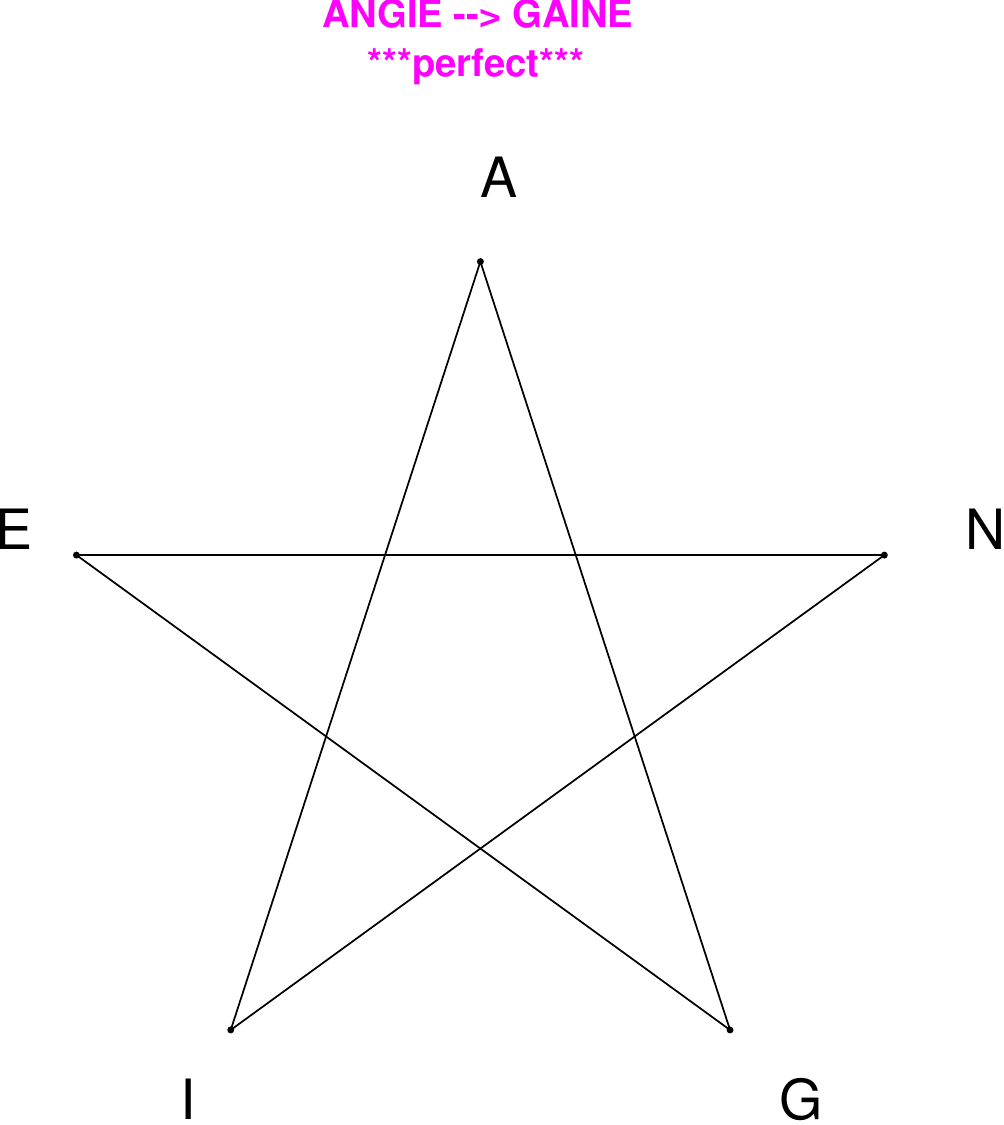}
\end{subfigure}
\end{figure}

\begin{figure}[H]
\centering
\begin{subfigure}[T]{0.19\textwidth}
\centering
\includegraphics[width=\textwidth]{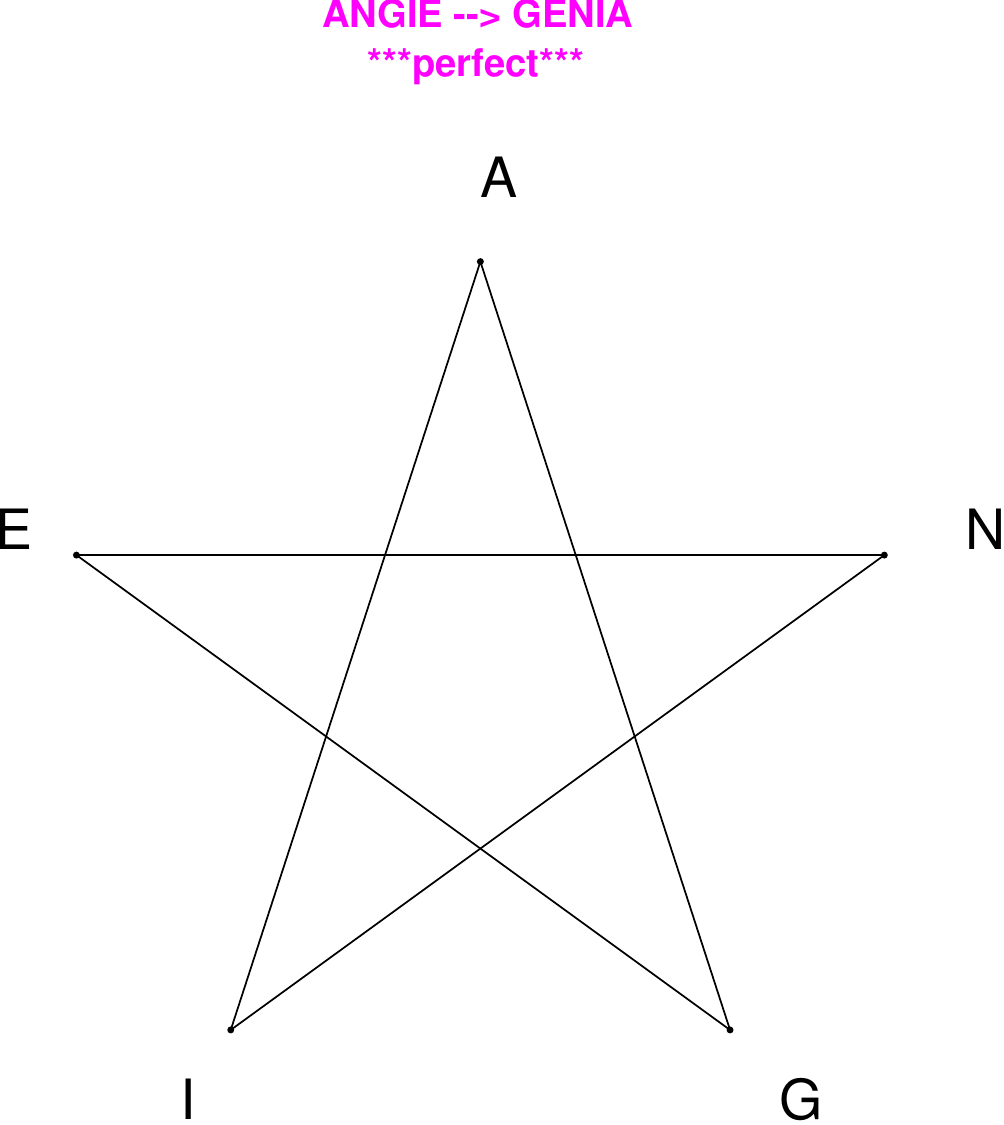}
\end{subfigure}
\hfill
\begin{subfigure}[T]{0.19\textwidth}
\centering
\includegraphics[width=\textwidth]{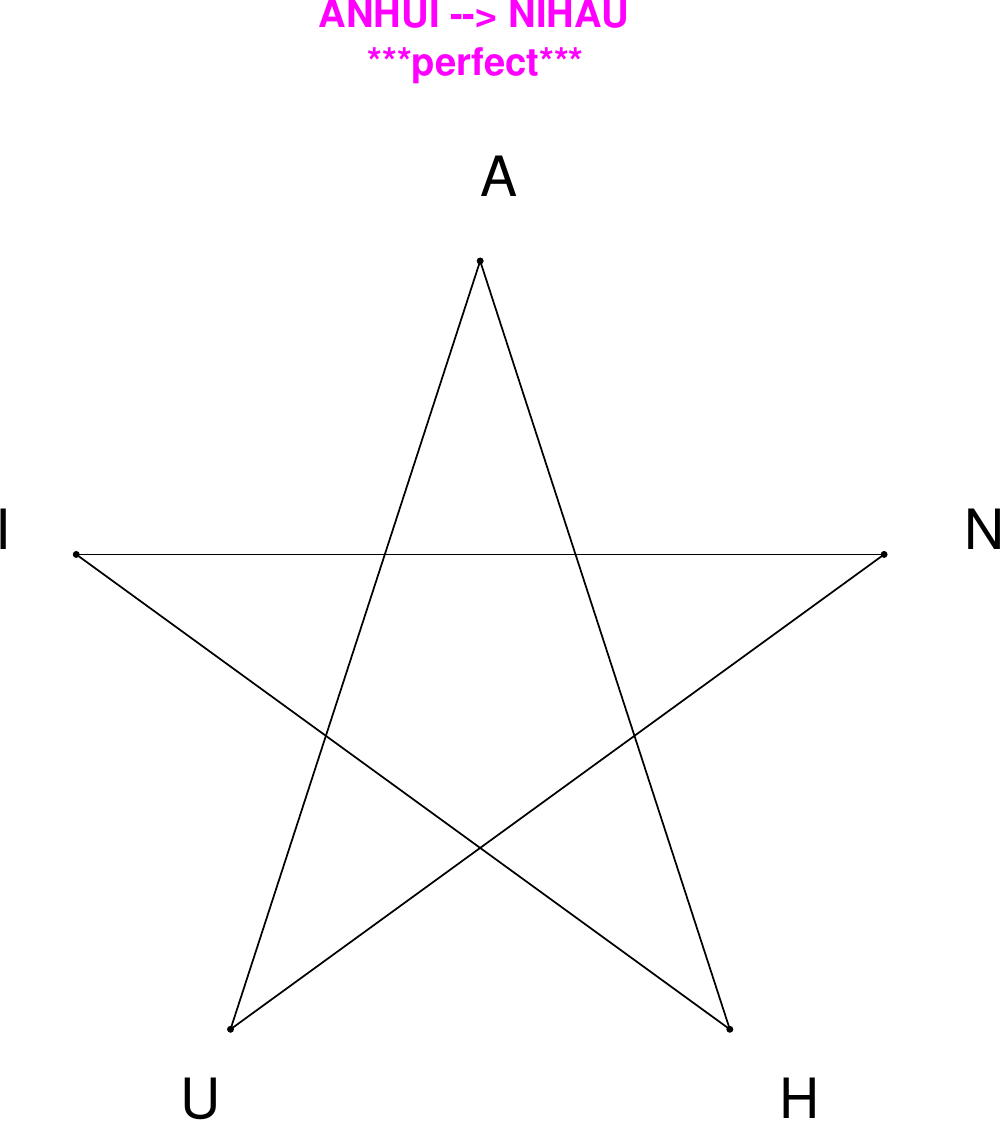}
\end{subfigure}
\hfill
\begin{subfigure}[T]{0.19\textwidth}
\centering
\includegraphics[width=\textwidth]{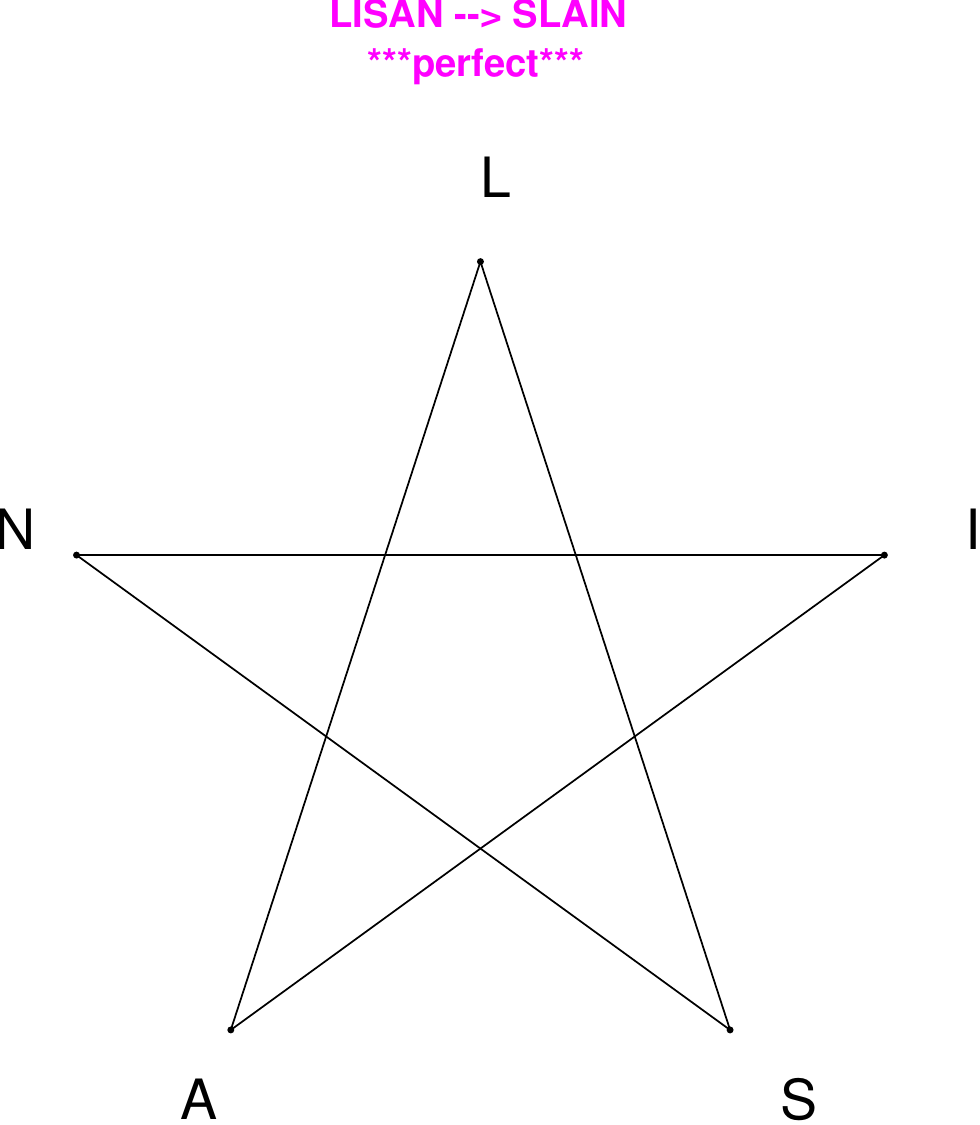}
\end{subfigure}
\hfill
\begin{subfigure}[T]{0.19\textwidth}
\centering
\includegraphics[width=\textwidth]{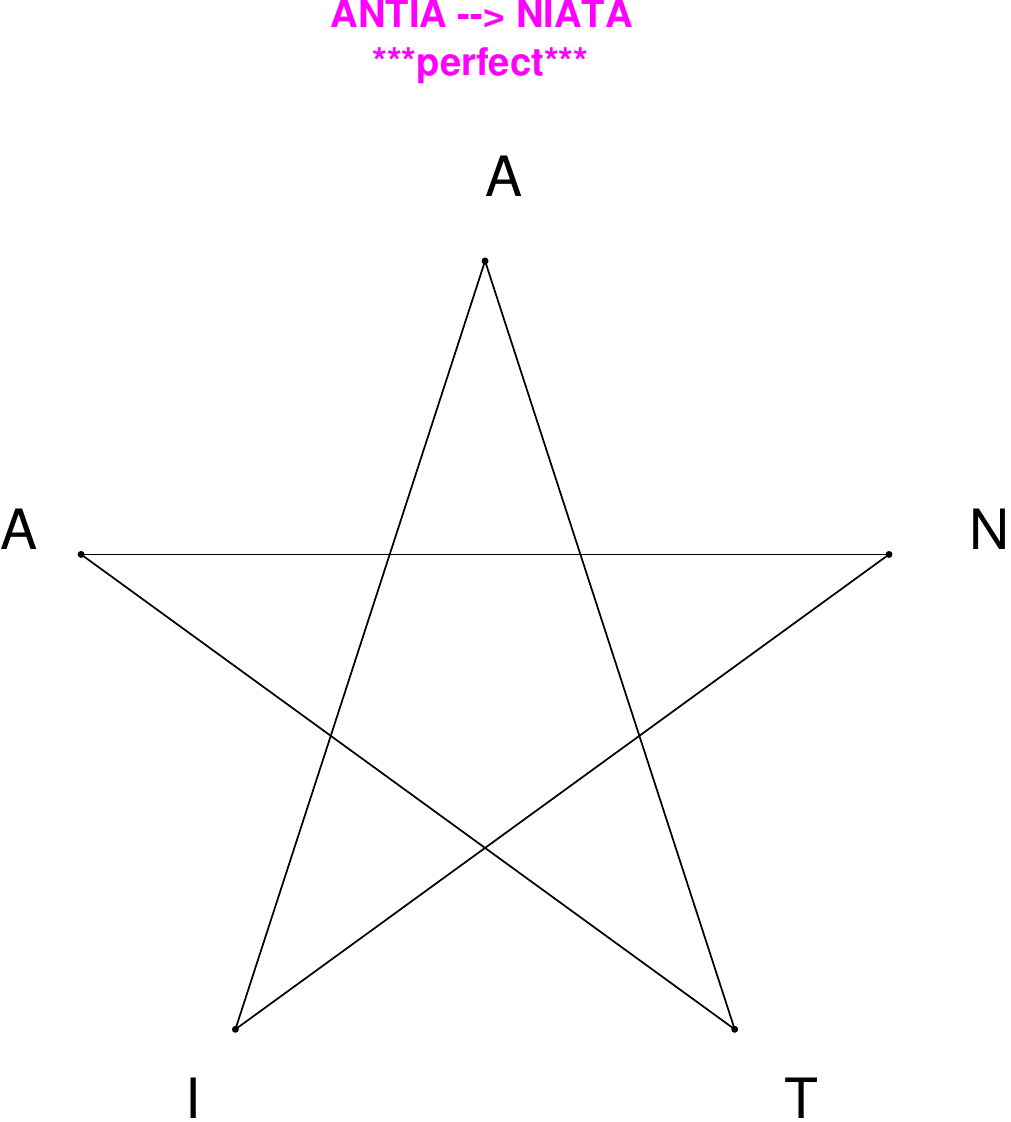}
\end{subfigure}
\hfill
\begin{subfigure}[T]{0.19\textwidth}
\centering
\includegraphics[width=\textwidth]{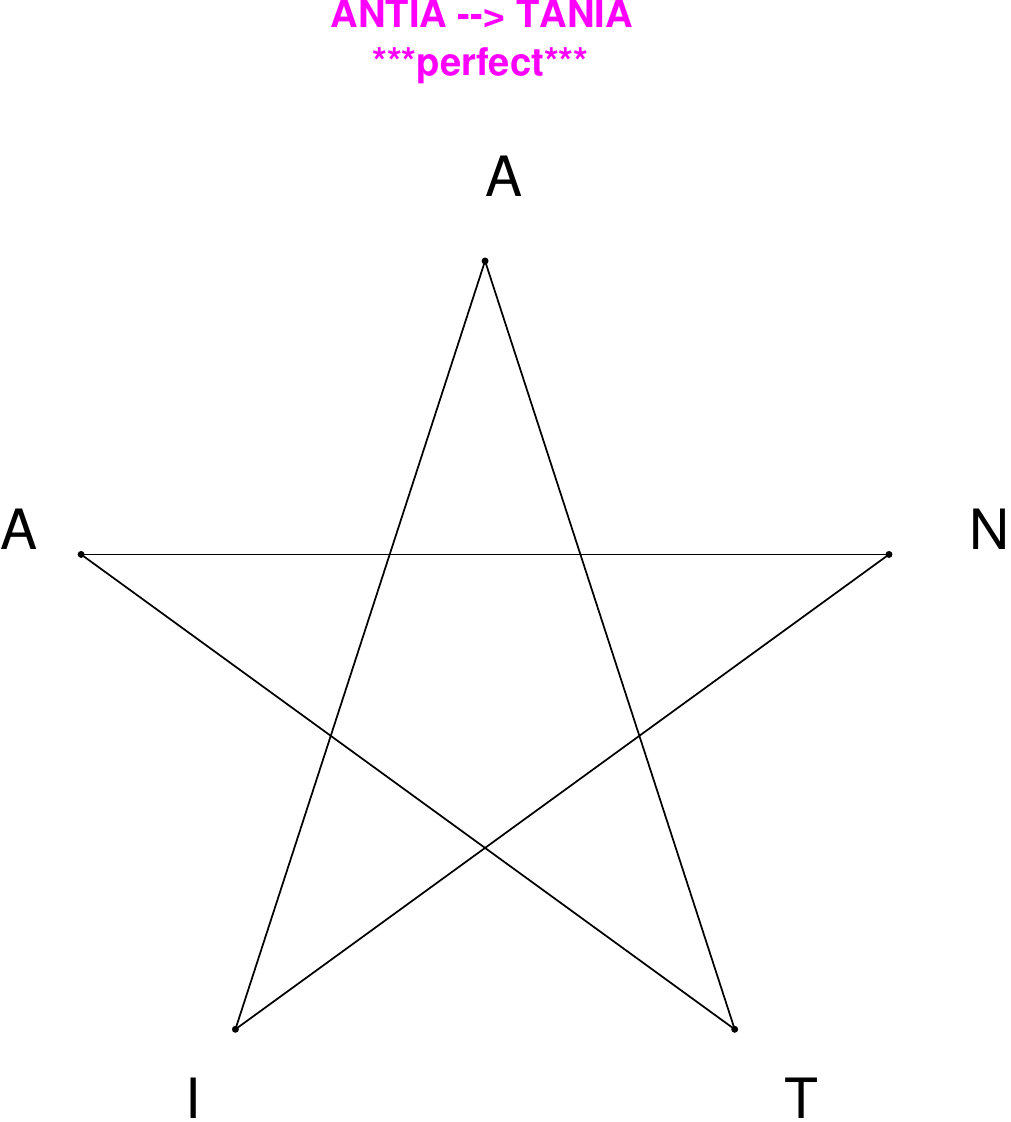}
\end{subfigure}
\end{figure}

\begin{figure}[H]
\centering
\begin{subfigure}[T]{0.19\textwidth}
\centering
\includegraphics[width=\textwidth]{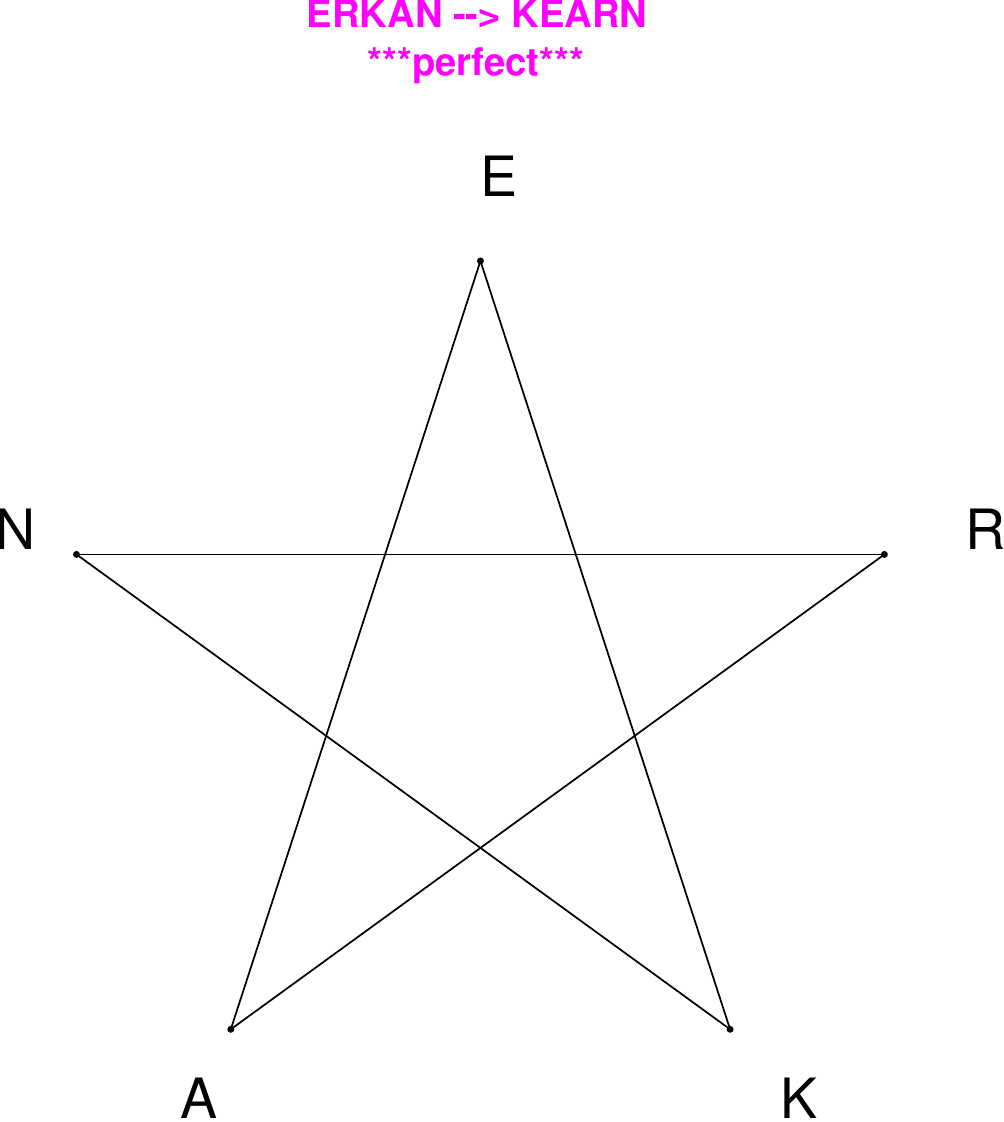}
\end{subfigure}
\hfill
\begin{subfigure}[T]{0.19\textwidth}
\centering
\includegraphics[width=\textwidth]{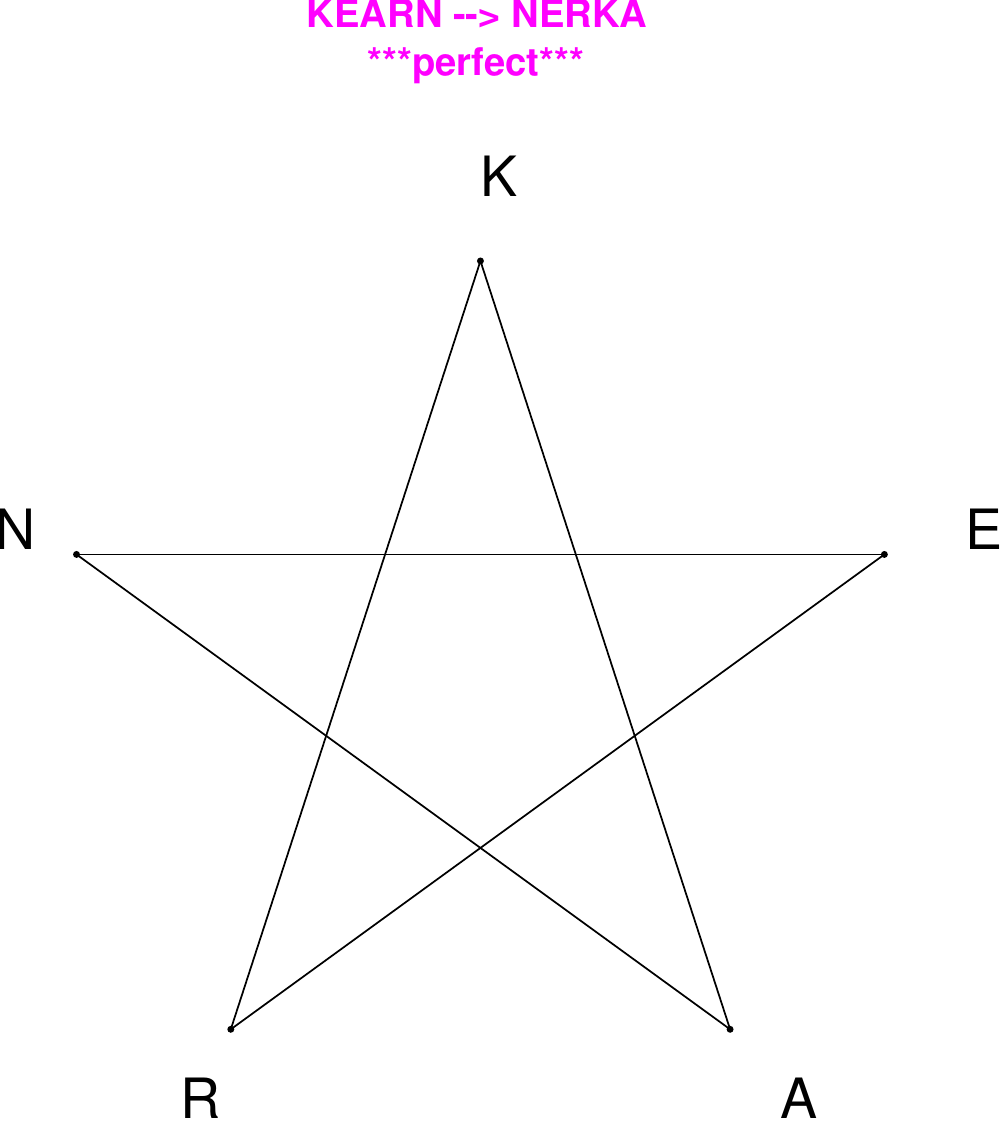}
\end{subfigure}
\hfill
\begin{subfigure}[T]{0.19\textwidth}
\centering
\includegraphics[width=\textwidth]{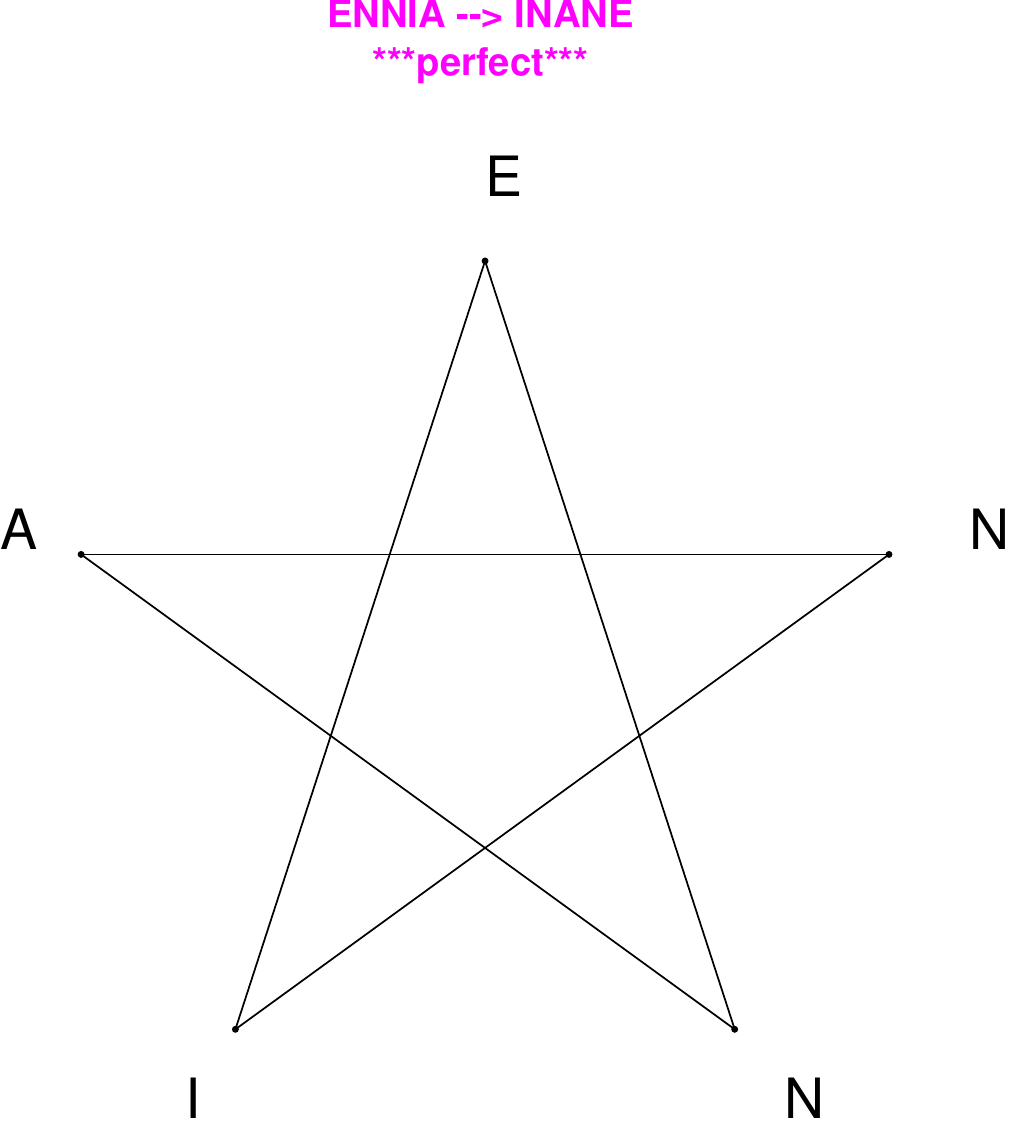}
\end{subfigure}
\hfill
\begin{subfigure}[T]{0.19\textwidth}
\centering
\includegraphics[width=\textwidth]{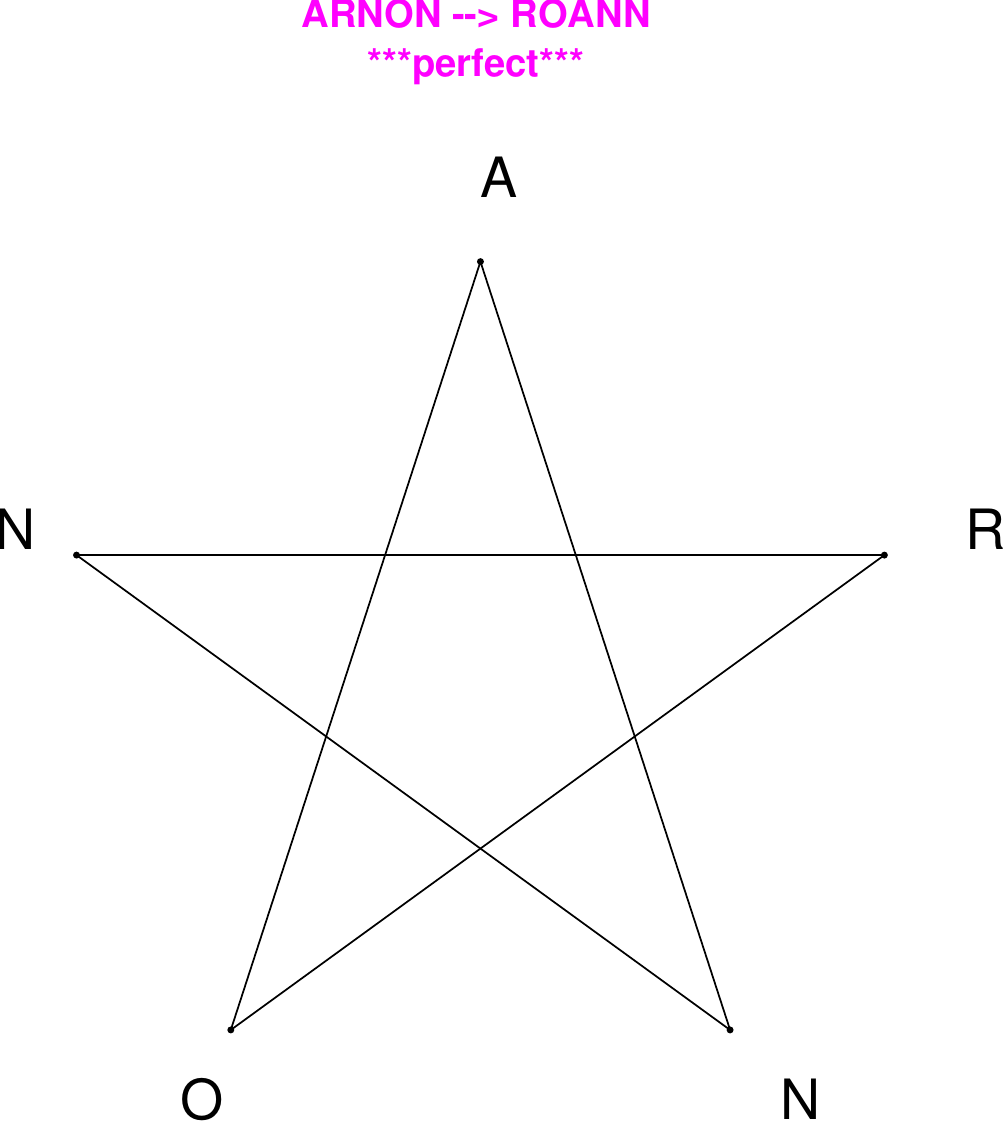}
\end{subfigure}
\hfill
\begin{subfigure}[T]{0.19\textwidth}
\centering
\includegraphics[width=\textwidth]{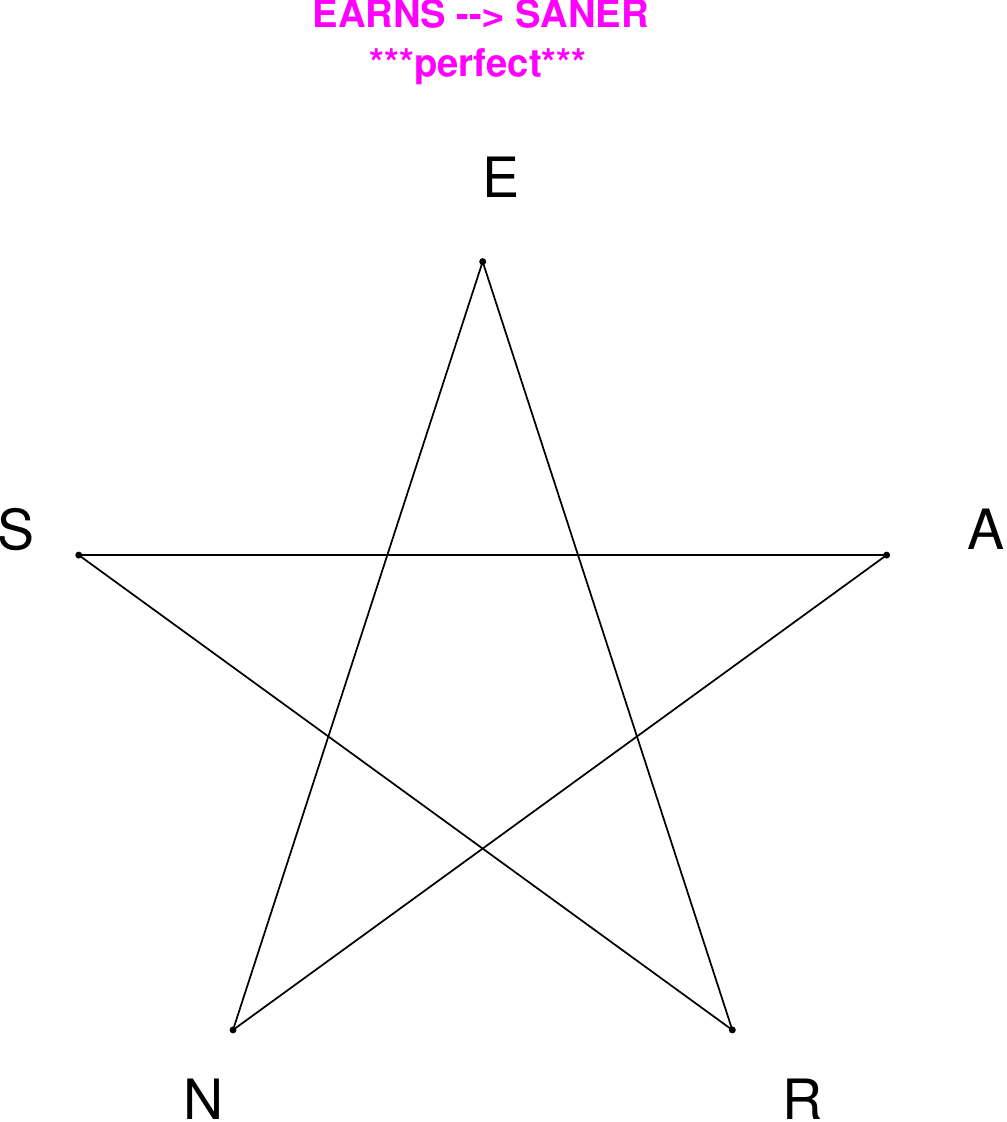}
\end{subfigure}
\end{figure}

\begin{figure}[H]
\centering
\begin{subfigure}[T]{0.19\textwidth}
\centering
\includegraphics[width=\textwidth]{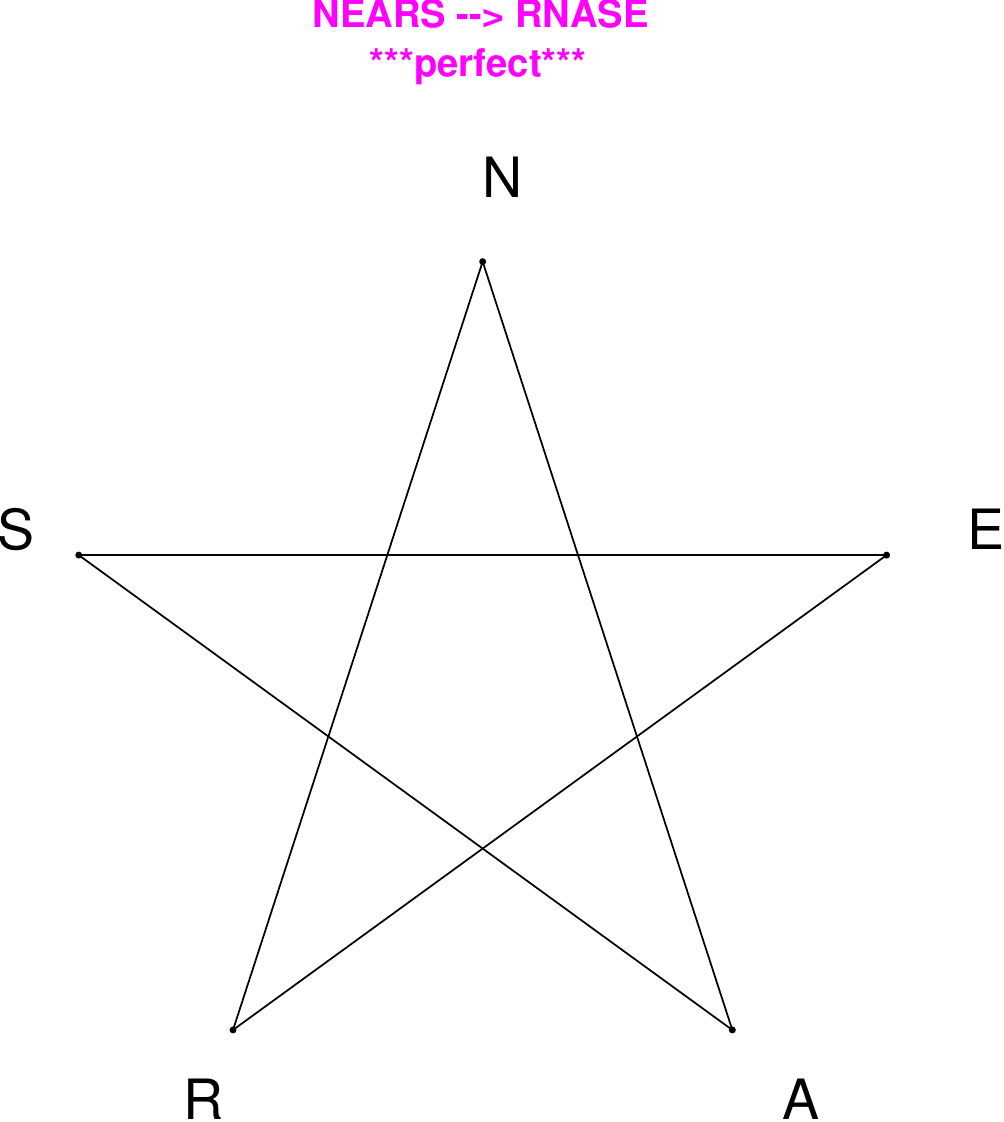}
\end{subfigure}
\hfill
\begin{subfigure}[T]{0.19\textwidth}
\centering
\includegraphics[width=\textwidth]{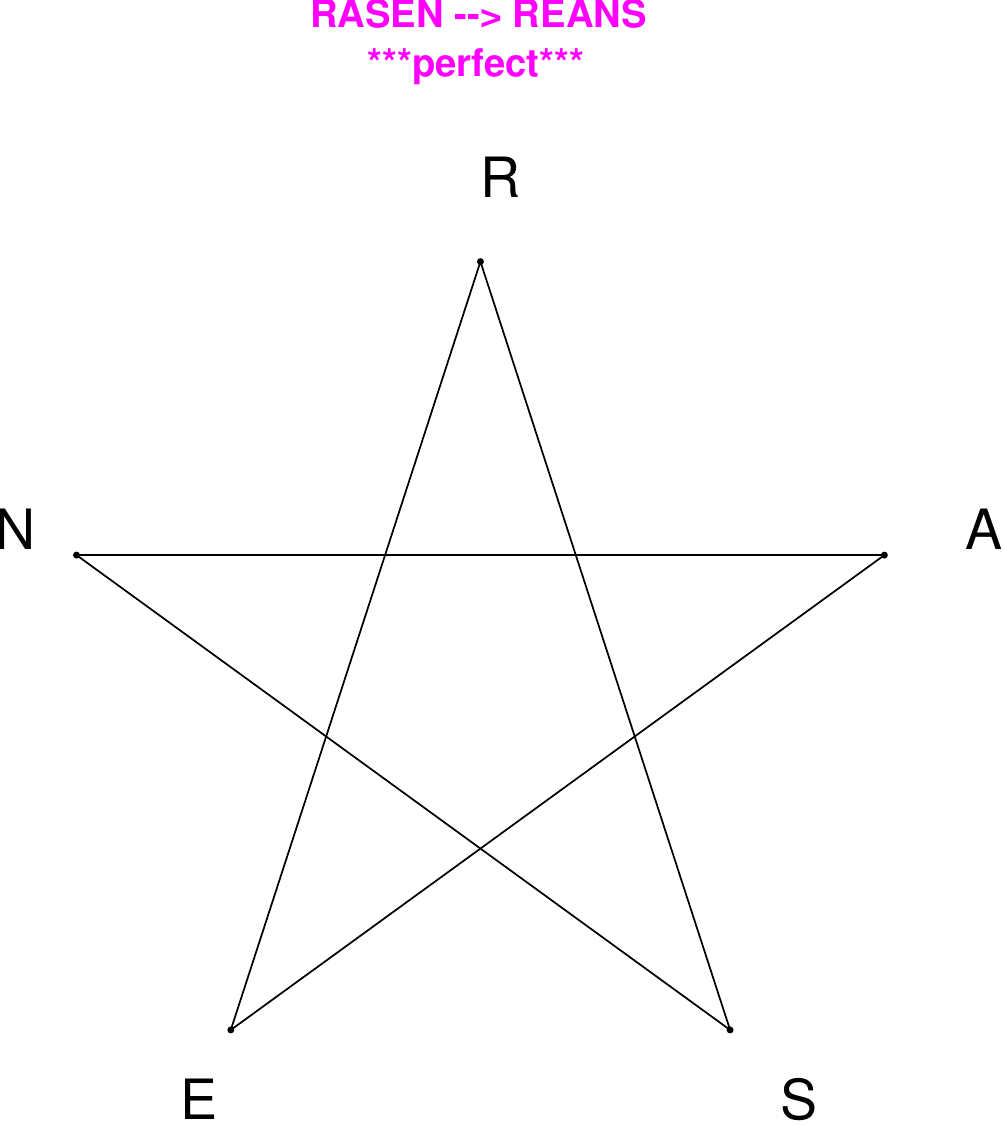}
\end{subfigure}
\hfill
\begin{subfigure}[T]{0.19\textwidth}
\centering
\includegraphics[width=\textwidth]{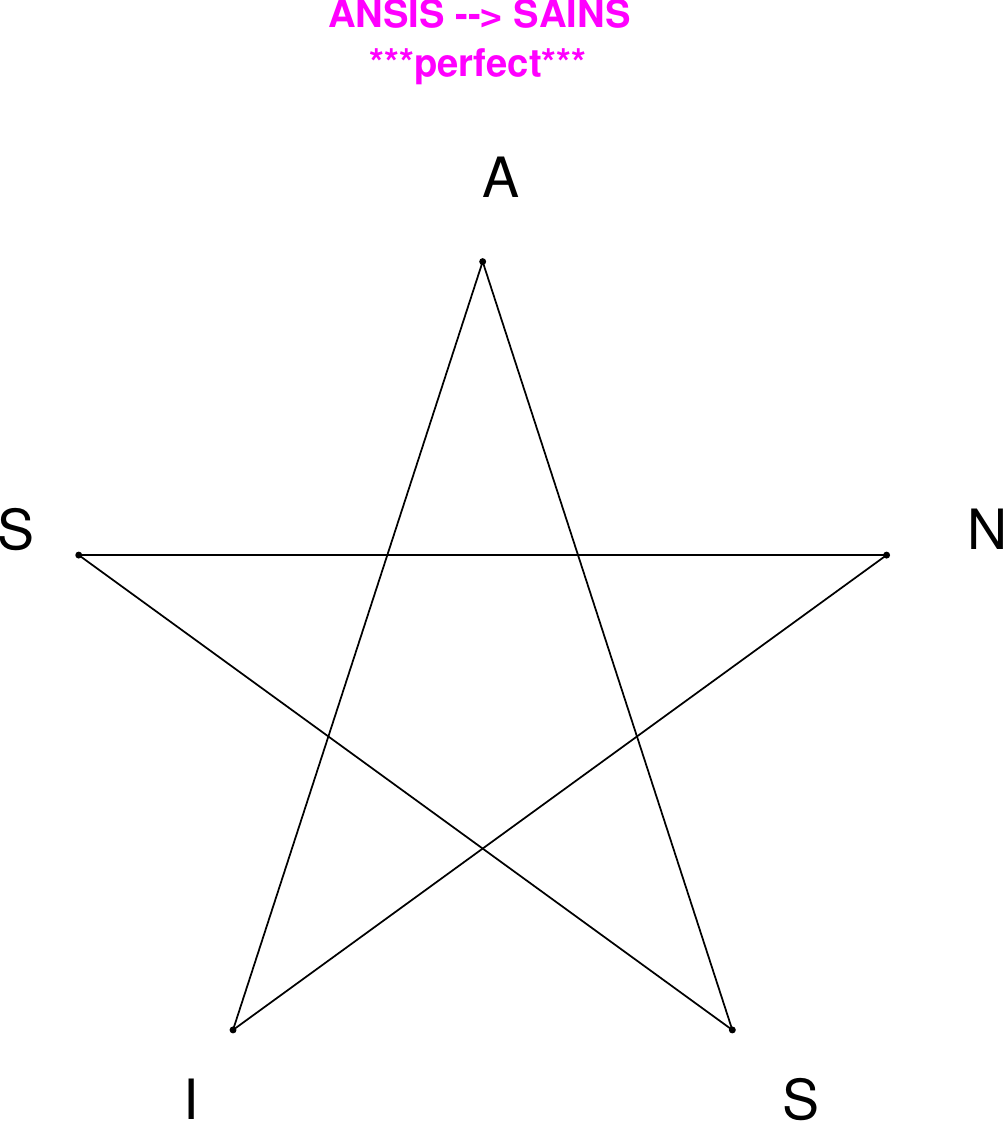}
\end{subfigure}
\hfill
\begin{subfigure}[T]{0.19\textwidth}
\centering
\includegraphics[width=\textwidth]{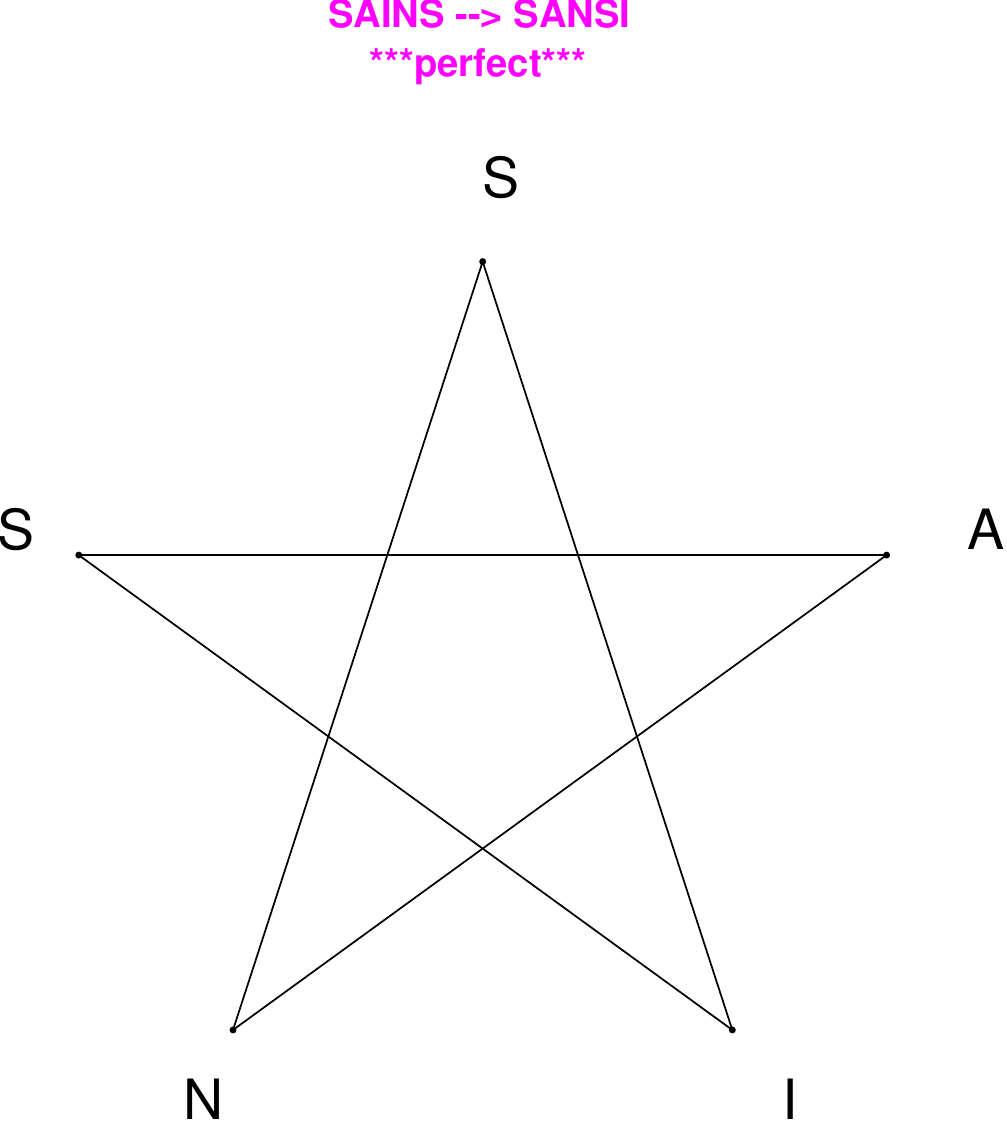}
\end{subfigure}
\hfill
\begin{subfigure}[T]{0.19\textwidth}
\centering
\includegraphics[width=\textwidth]{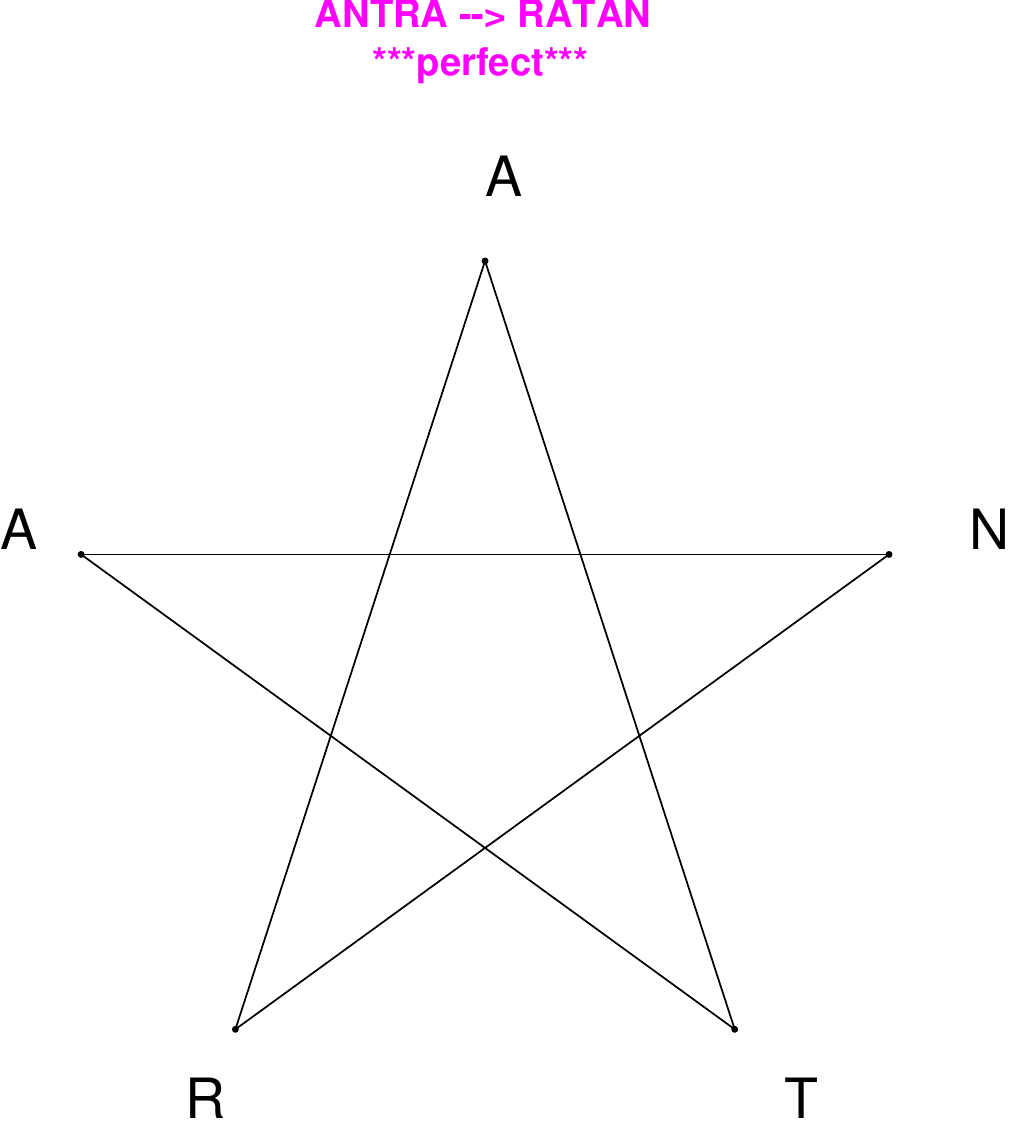}
\end{subfigure}
\end{figure}

\begin{figure}[H]
\centering
\begin{subfigure}[T]{0.19\textwidth}
\centering
\includegraphics[width=\textwidth]{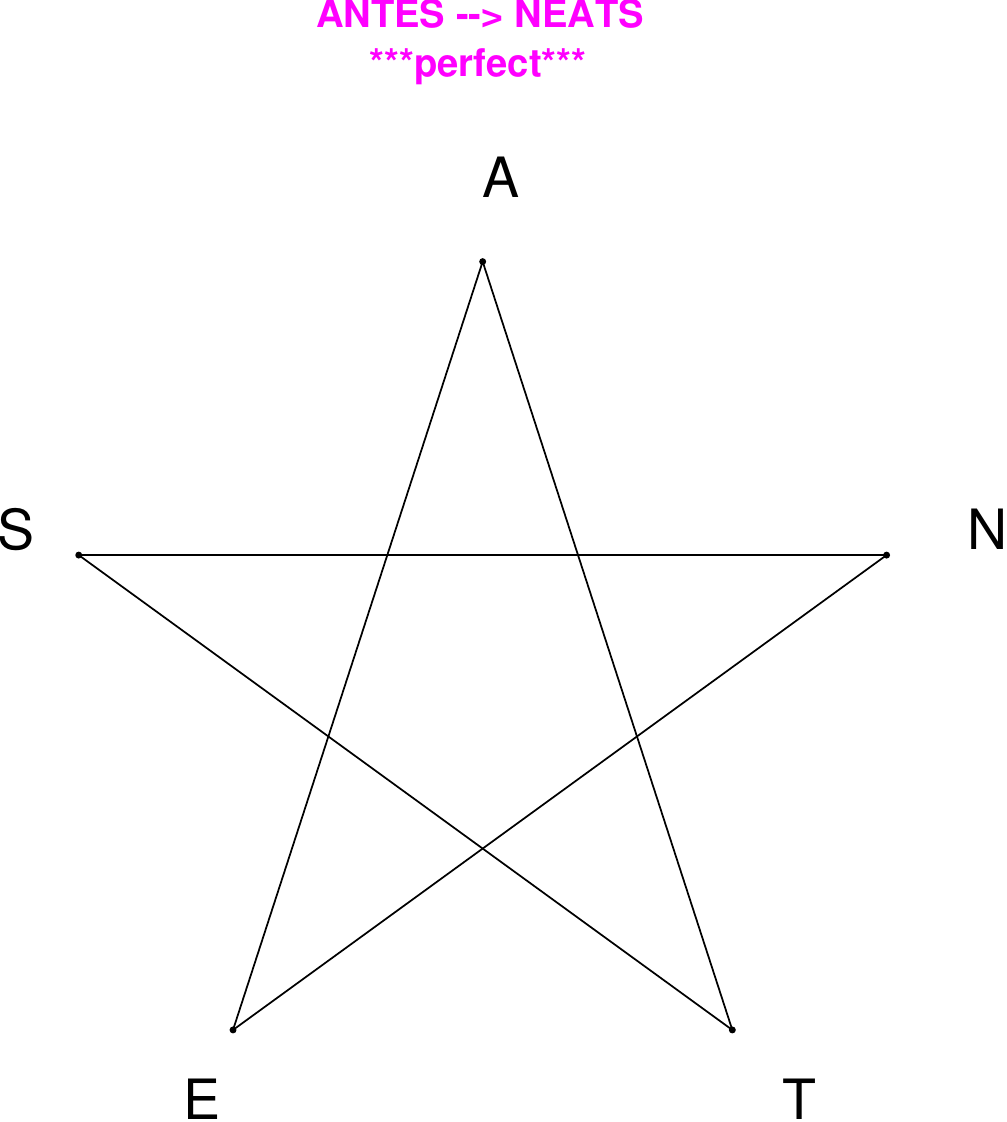}
\end{subfigure}
\hfill
\begin{subfigure}[T]{0.19\textwidth}
\centering
\includegraphics[width=\textwidth]{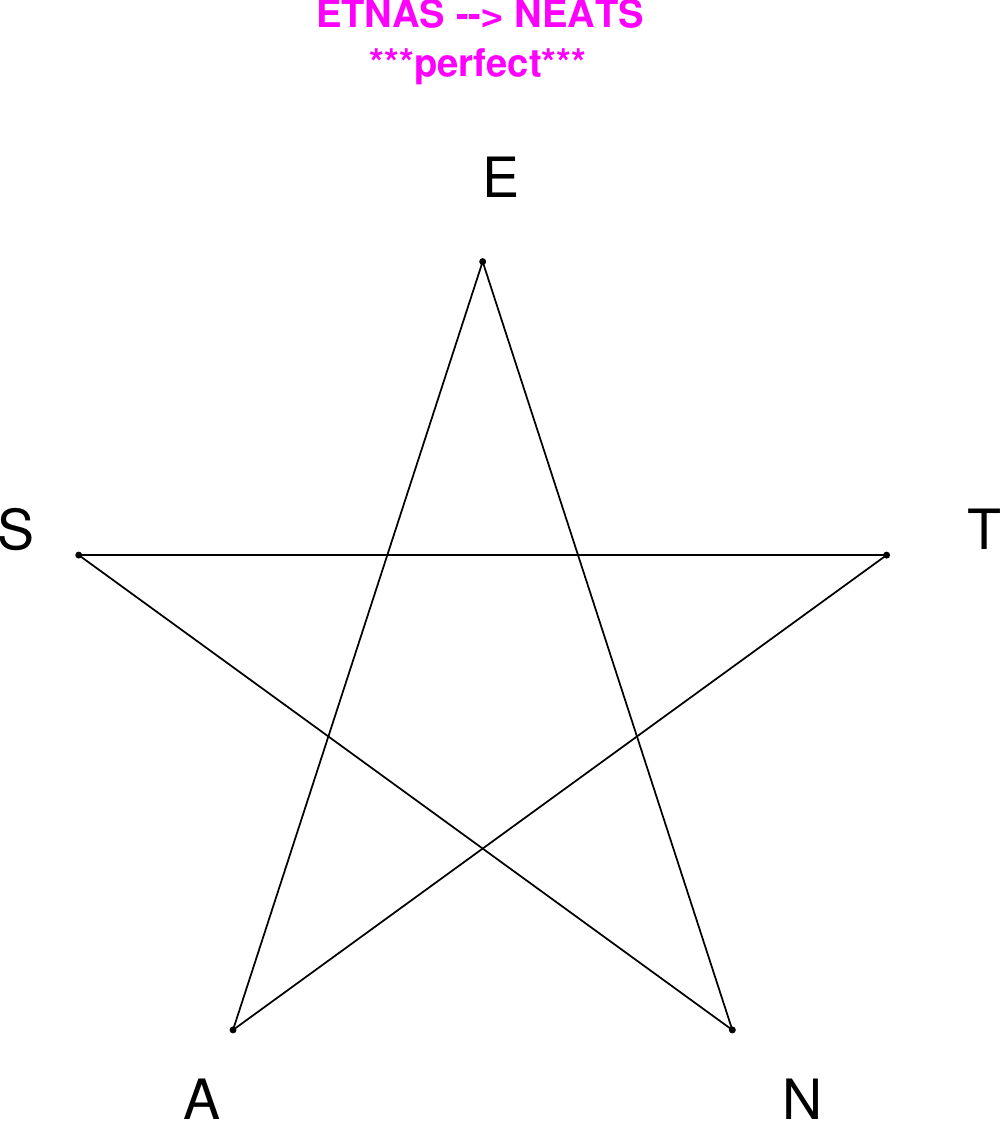}
\end{subfigure}
\hfill
\begin{subfigure}[T]{0.19\textwidth}
\centering
\includegraphics[width=\textwidth]{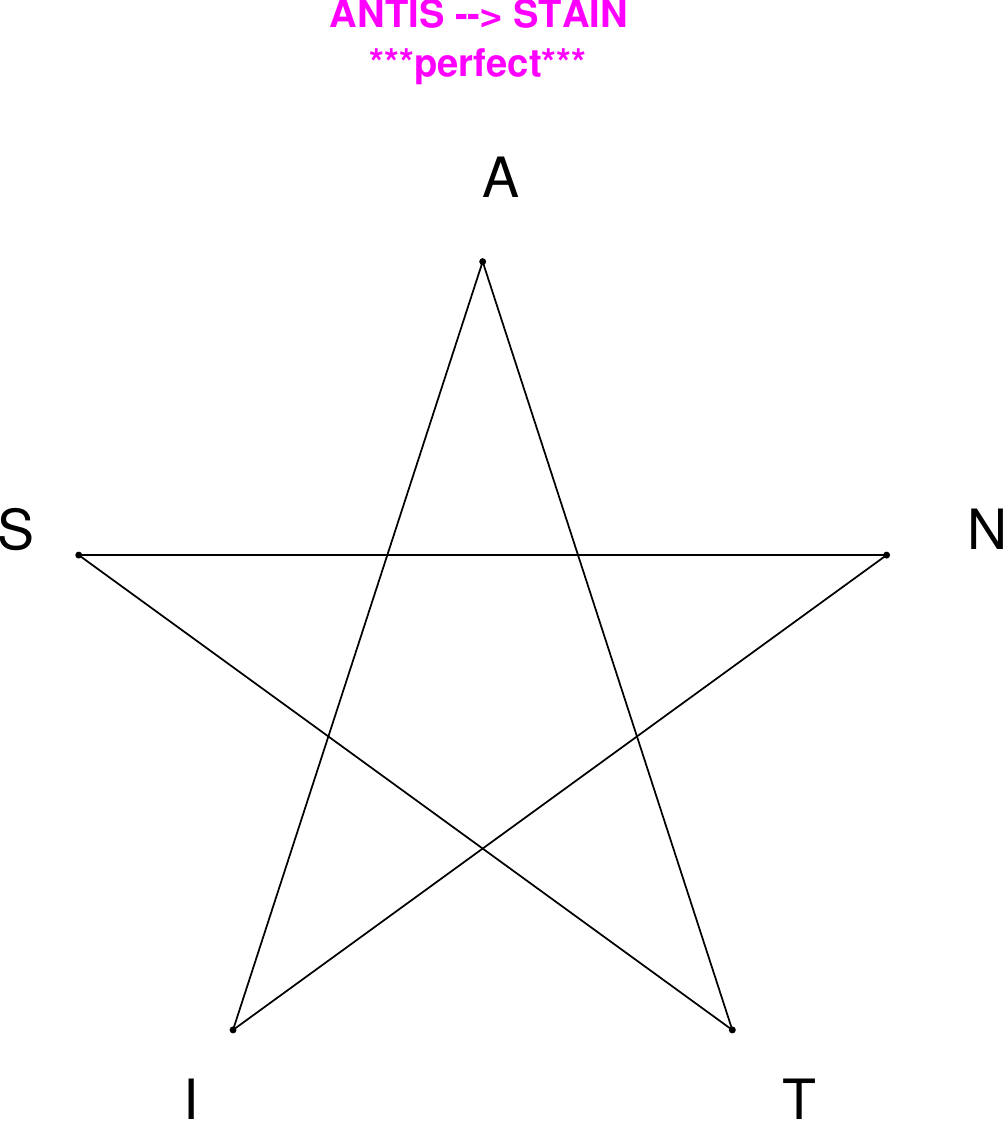}
\end{subfigure}
\hfill
\begin{subfigure}[T]{0.19\textwidth}
\centering
\includegraphics[width=\textwidth]{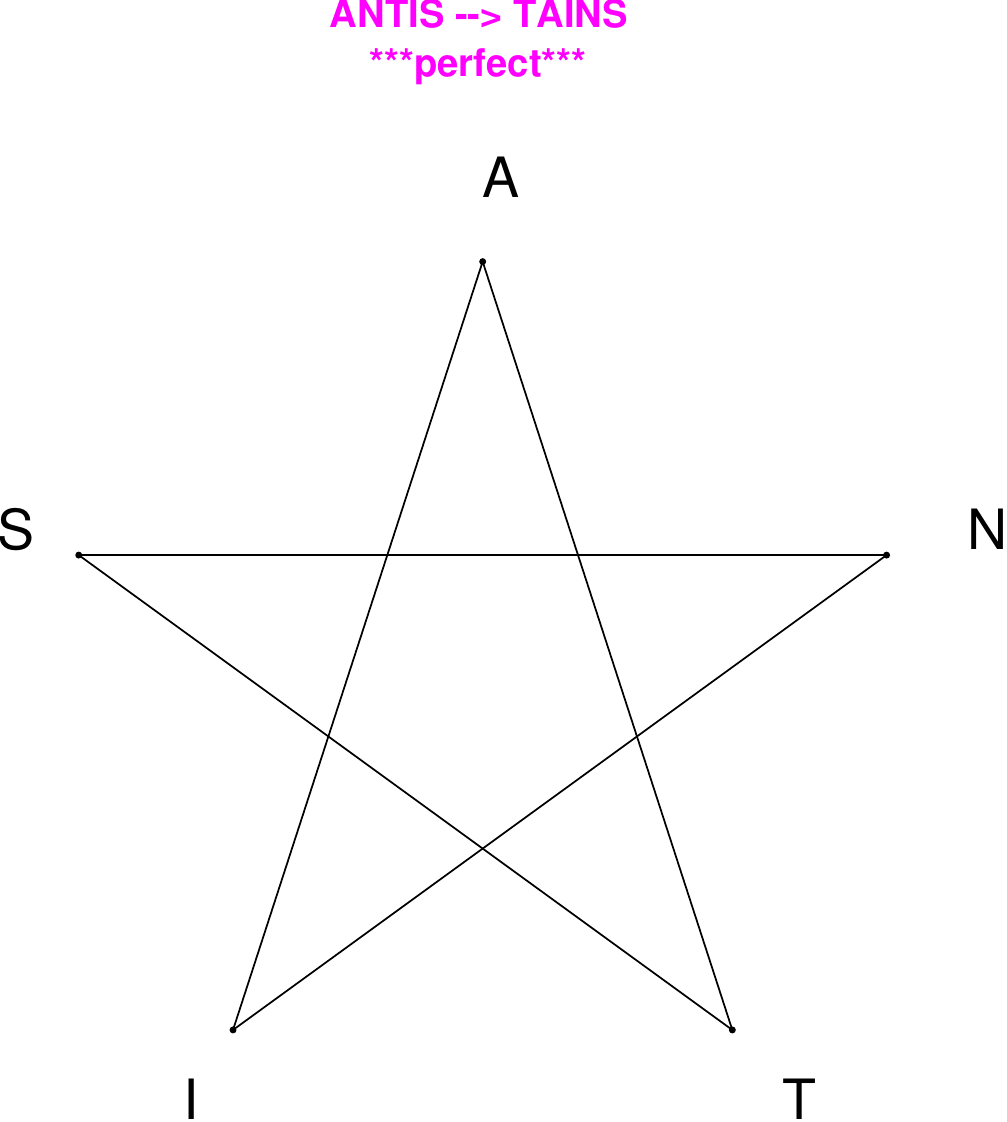}
\end{subfigure}
\hfill
\begin{subfigure}[T]{0.19\textwidth}
\centering
\includegraphics[width=\textwidth]{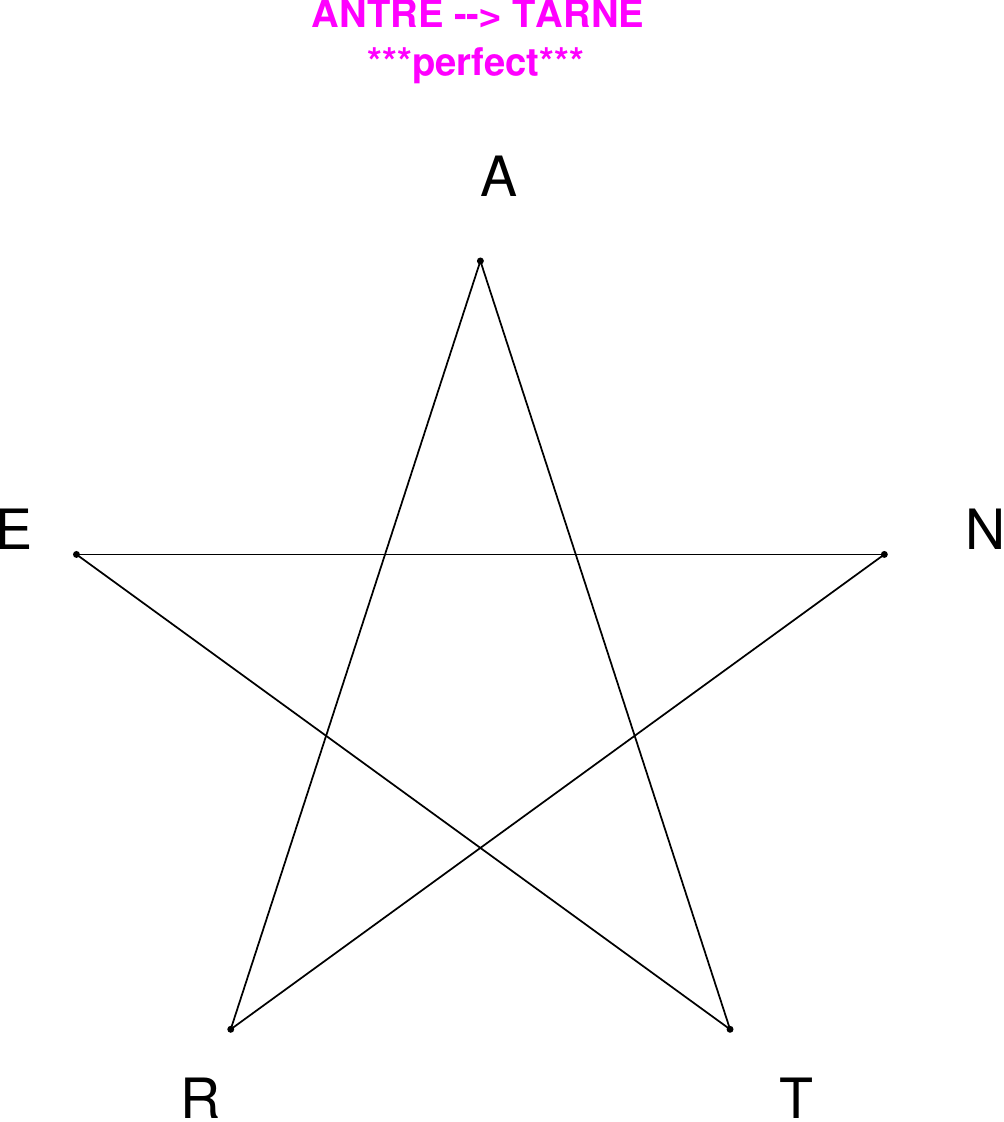}
\end{subfigure}
\end{figure}

\begin{figure}[H]
\centering
\begin{subfigure}[T]{0.19\textwidth}
\centering
\includegraphics[width=\textwidth]{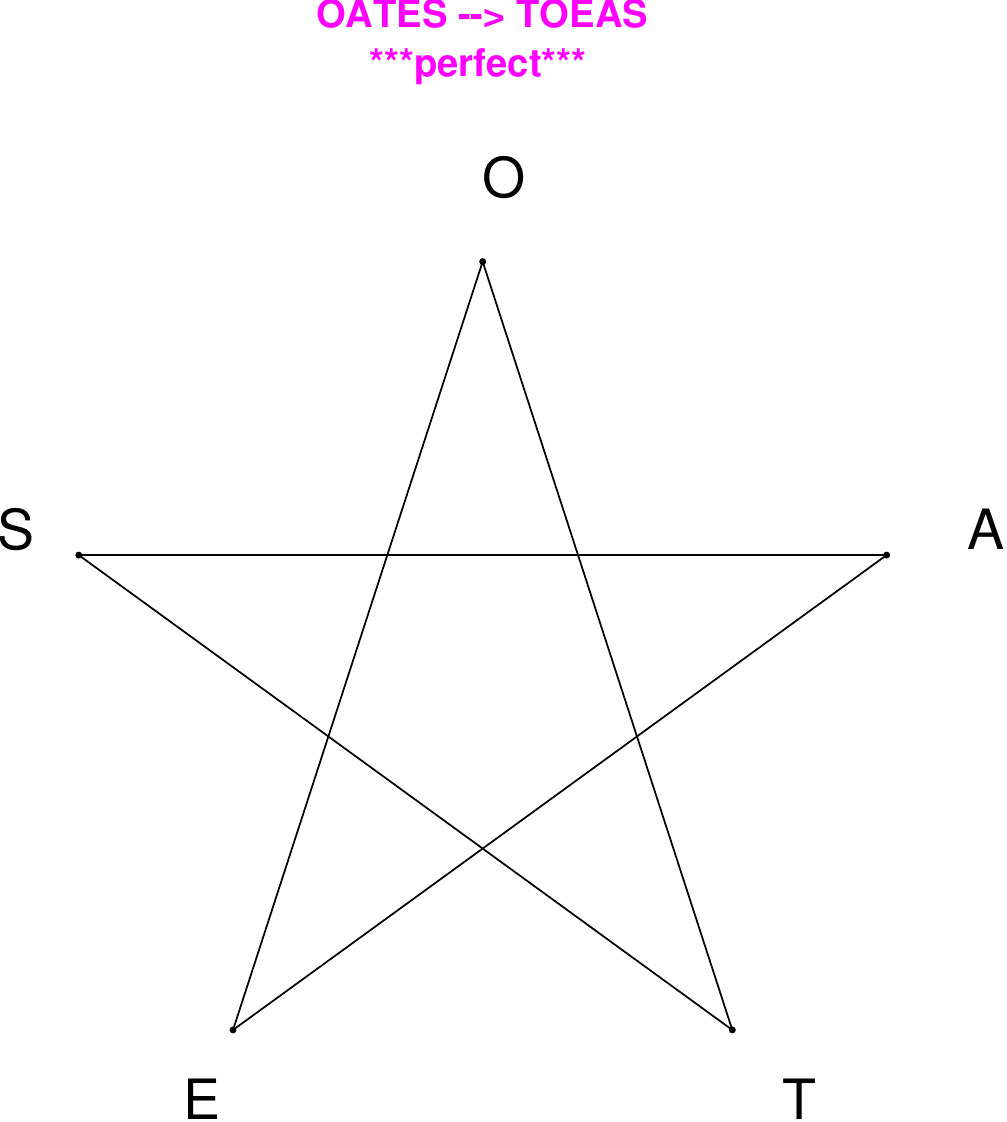}
\end{subfigure}
\hfill
\begin{subfigure}[T]{0.19\textwidth}
\centering
\includegraphics[width=\textwidth]{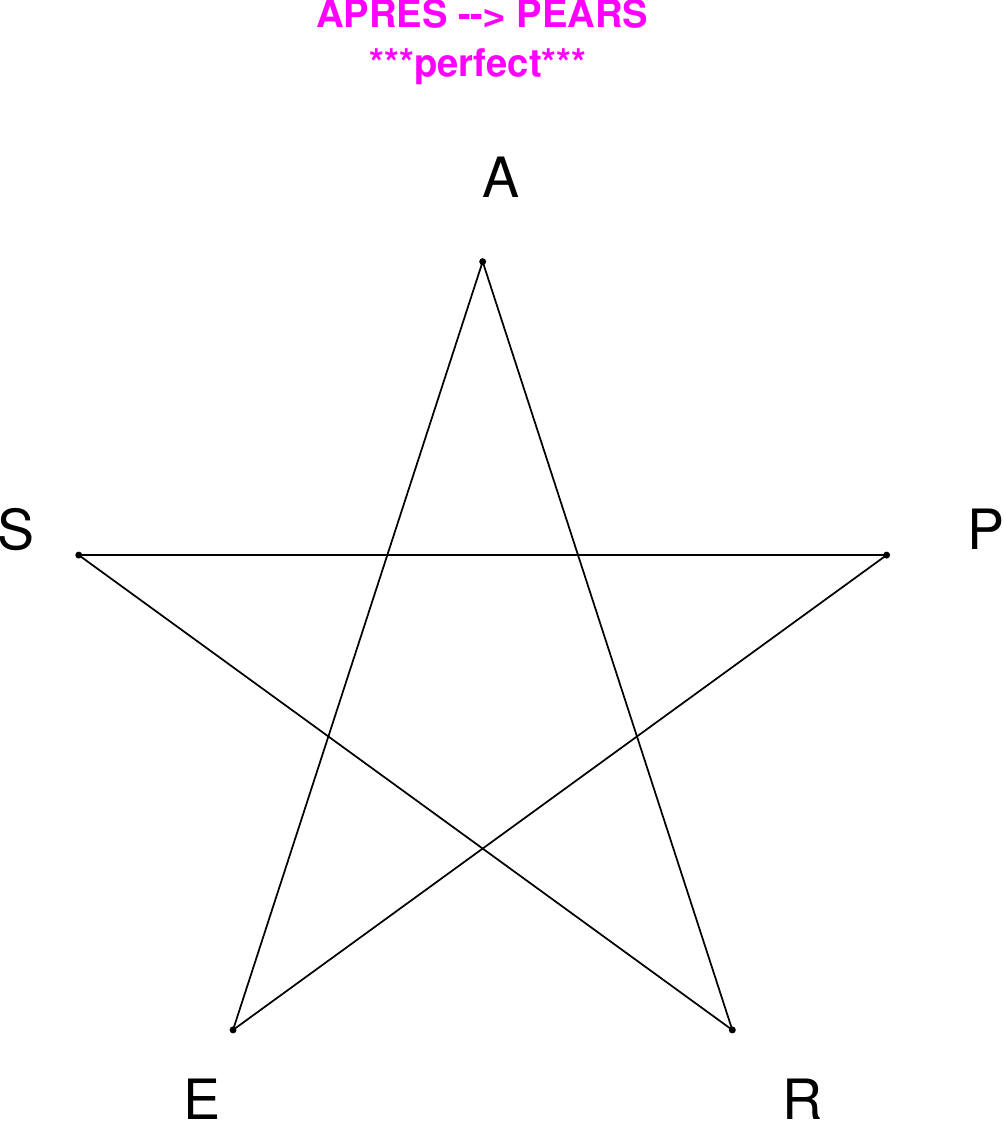}
\end{subfigure}
\hfill
\begin{subfigure}[T]{0.19\textwidth}
\centering
\includegraphics[width=\textwidth]{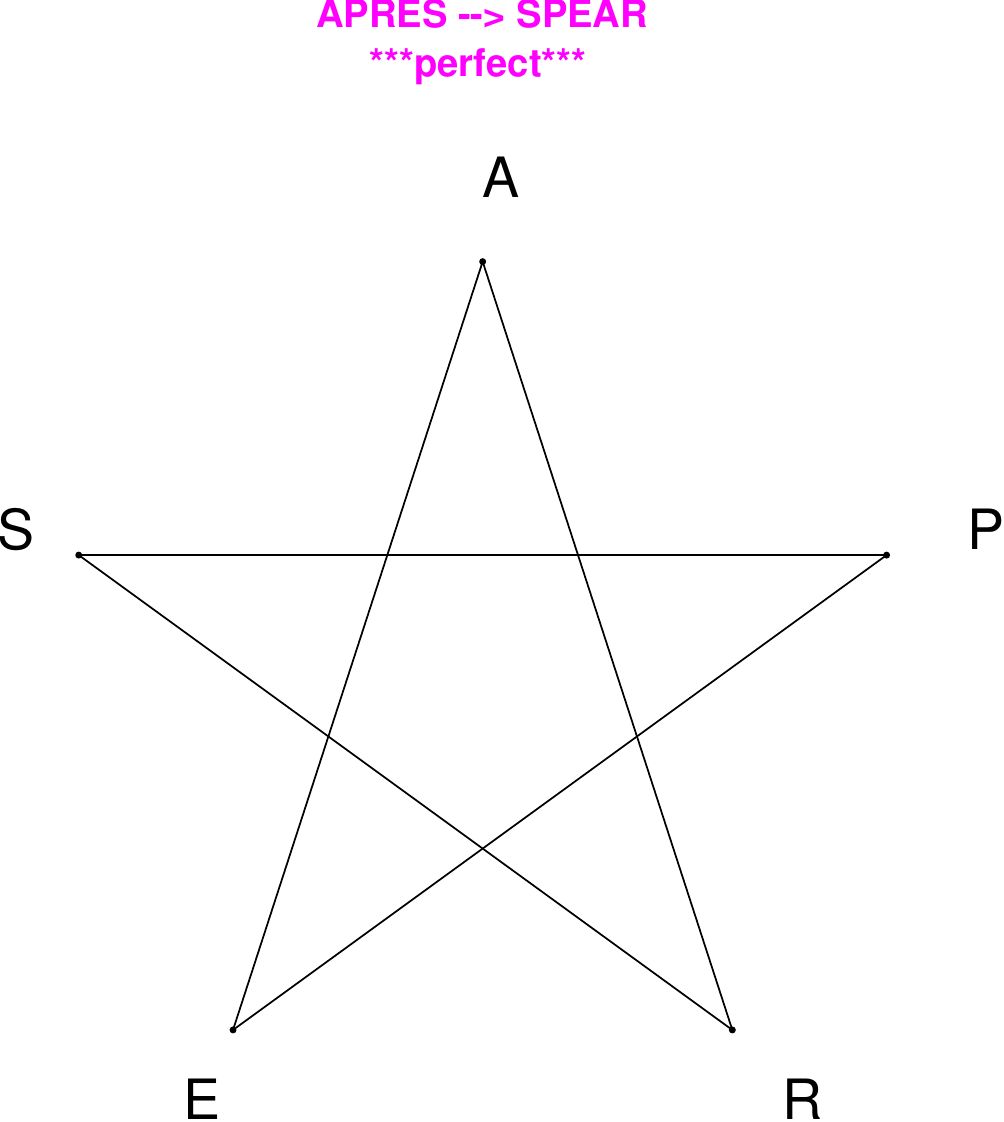}
\end{subfigure}
\hfill
\begin{subfigure}[T]{0.19\textwidth}
\centering
\includegraphics[width=\textwidth]{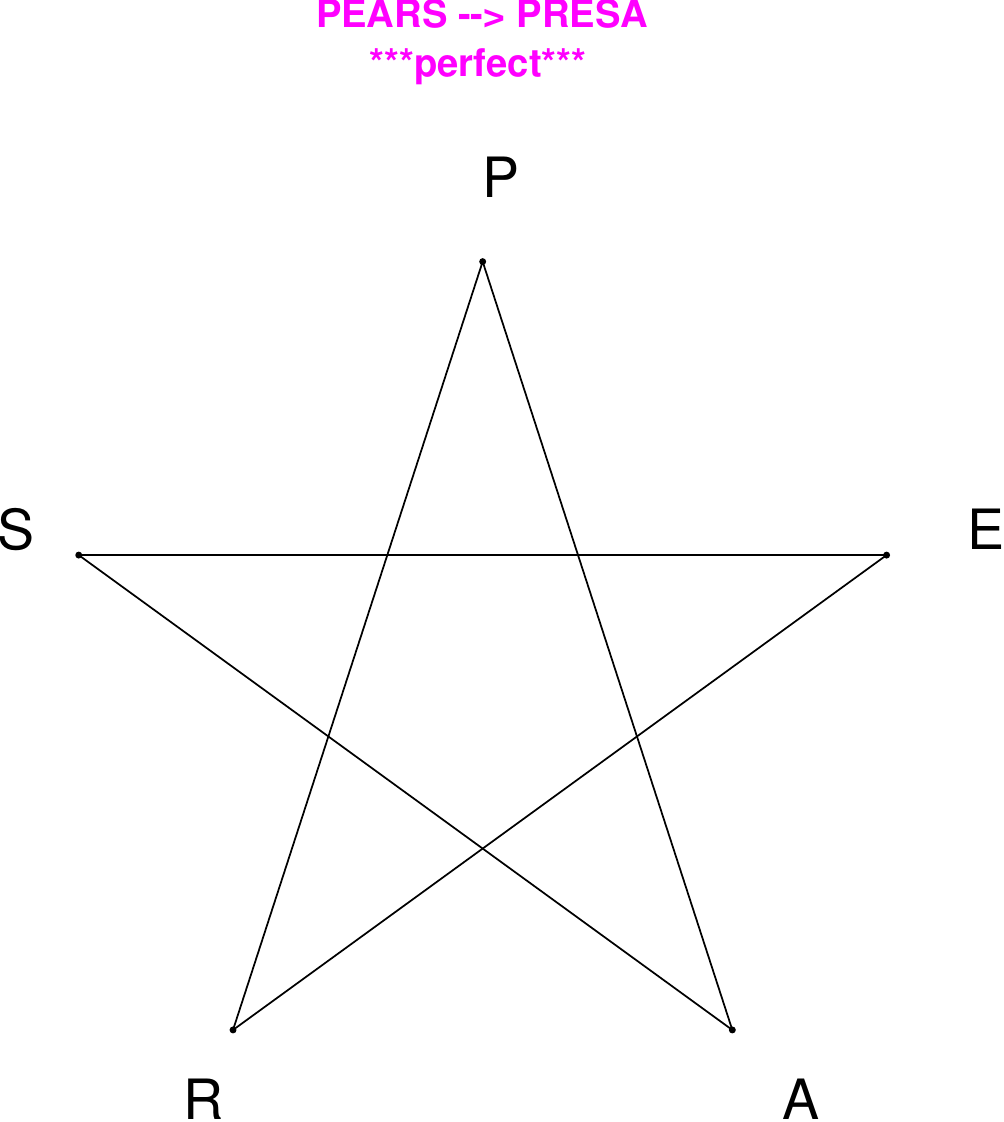}
\end{subfigure}
\hfill
\begin{subfigure}[T]{0.19\textwidth}
\centering
\includegraphics[width=\textwidth]{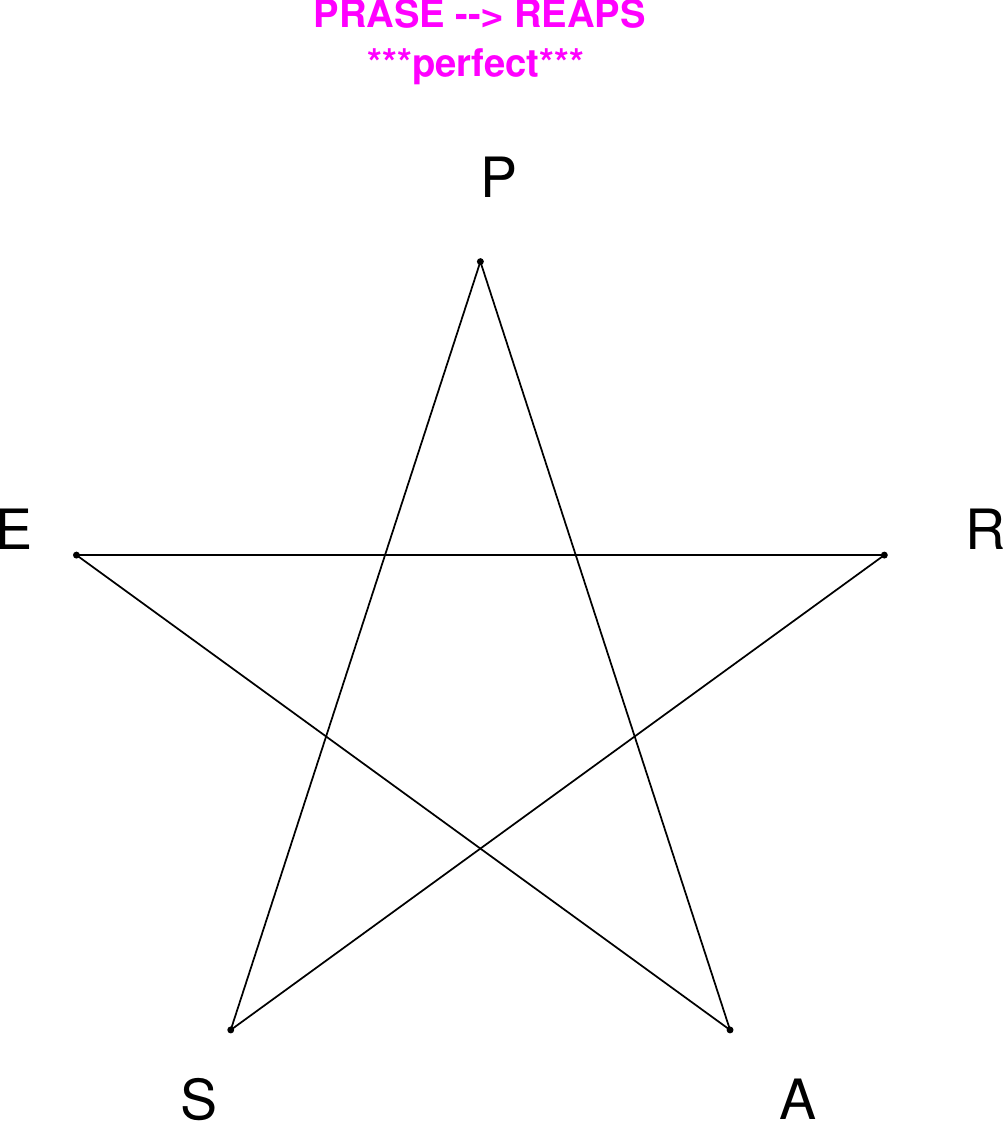}
\end{subfigure}
\end{figure}

\begin{figure}[H]
\centering
\begin{subfigure}[T]{0.19\textwidth}
\centering
\includegraphics[width=\textwidth]{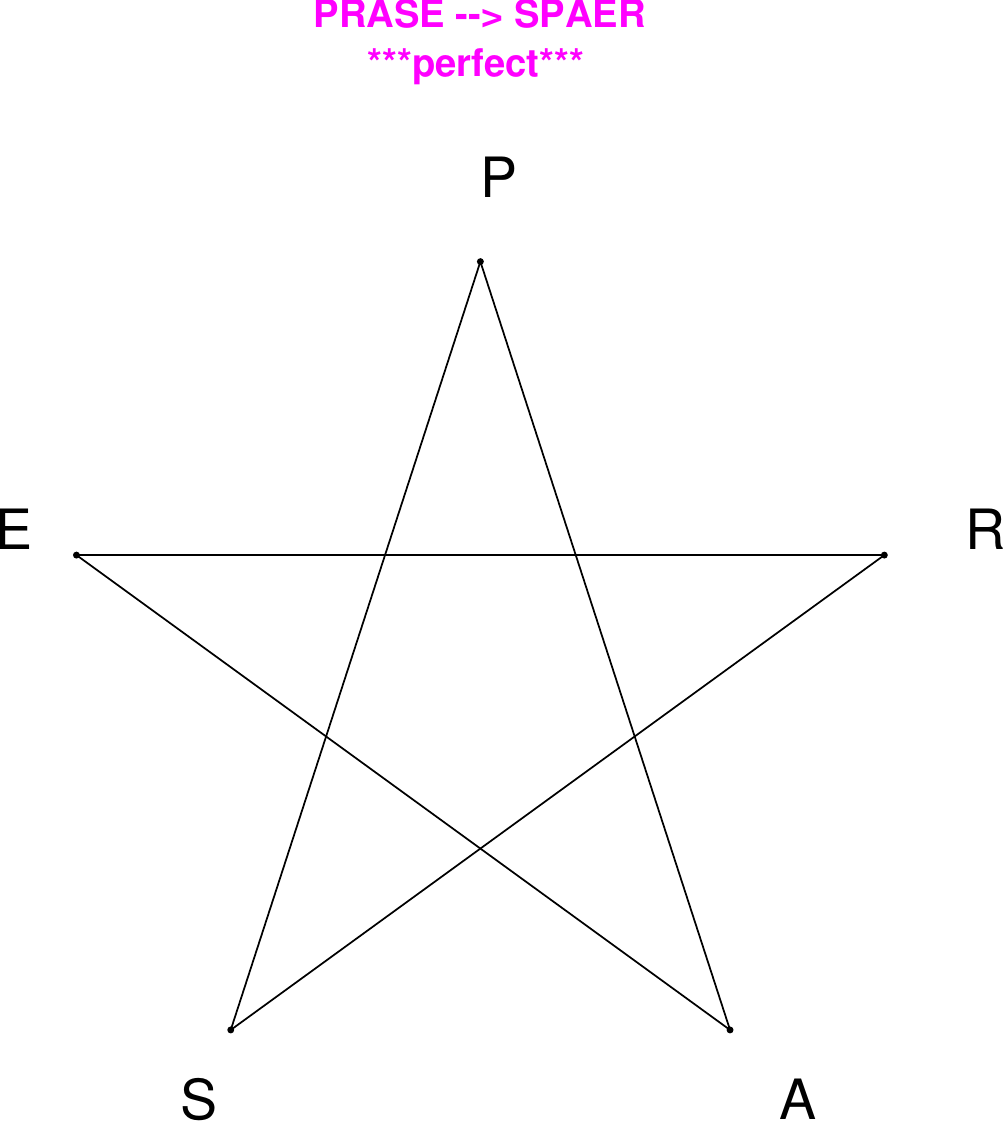}
\end{subfigure}
\hfill
\begin{subfigure}[T]{0.19\textwidth}
\centering
\includegraphics[width=\textwidth]{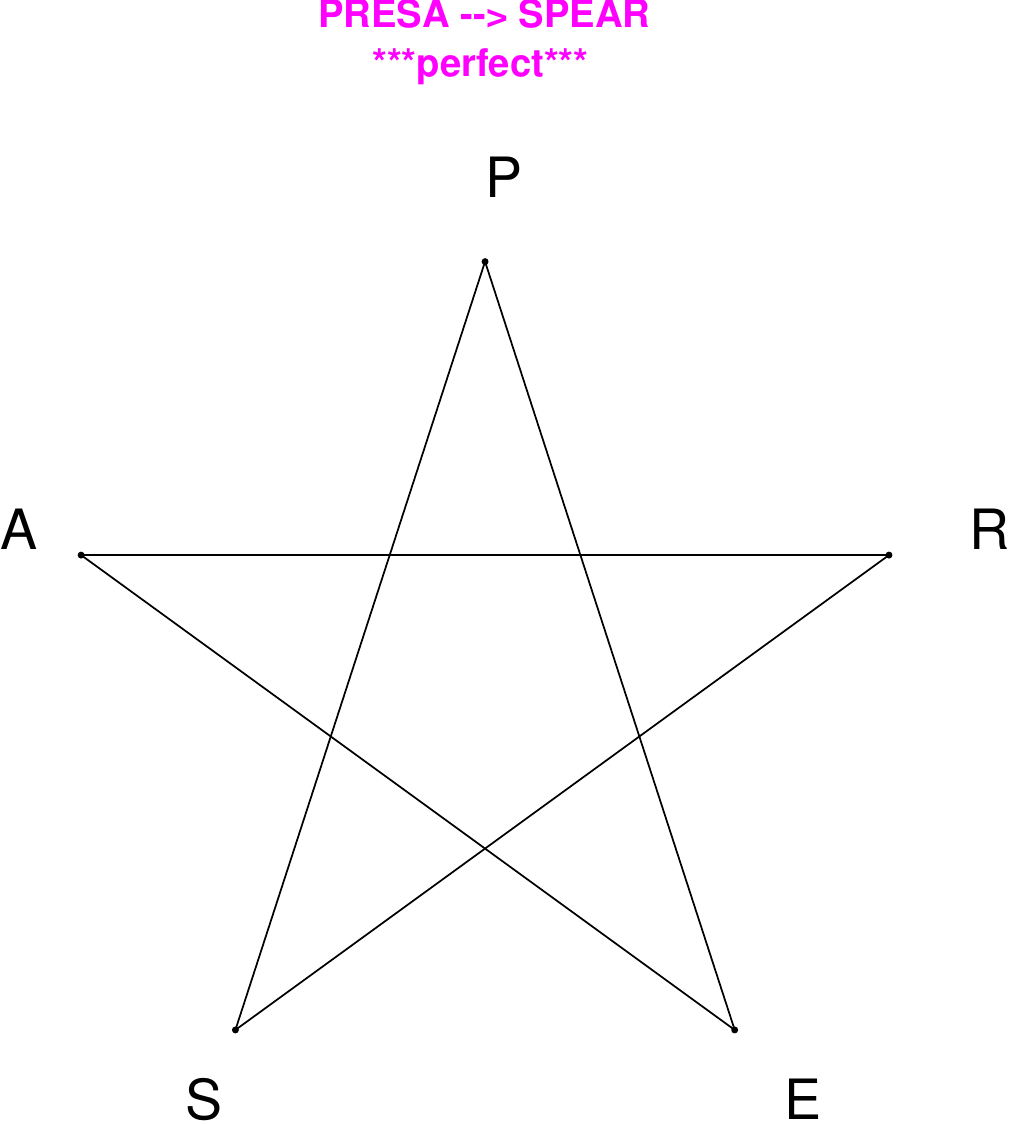}
\end{subfigure}
\hfill
\begin{subfigure}[T]{0.19\textwidth}
\centering
\includegraphics[width=\textwidth]{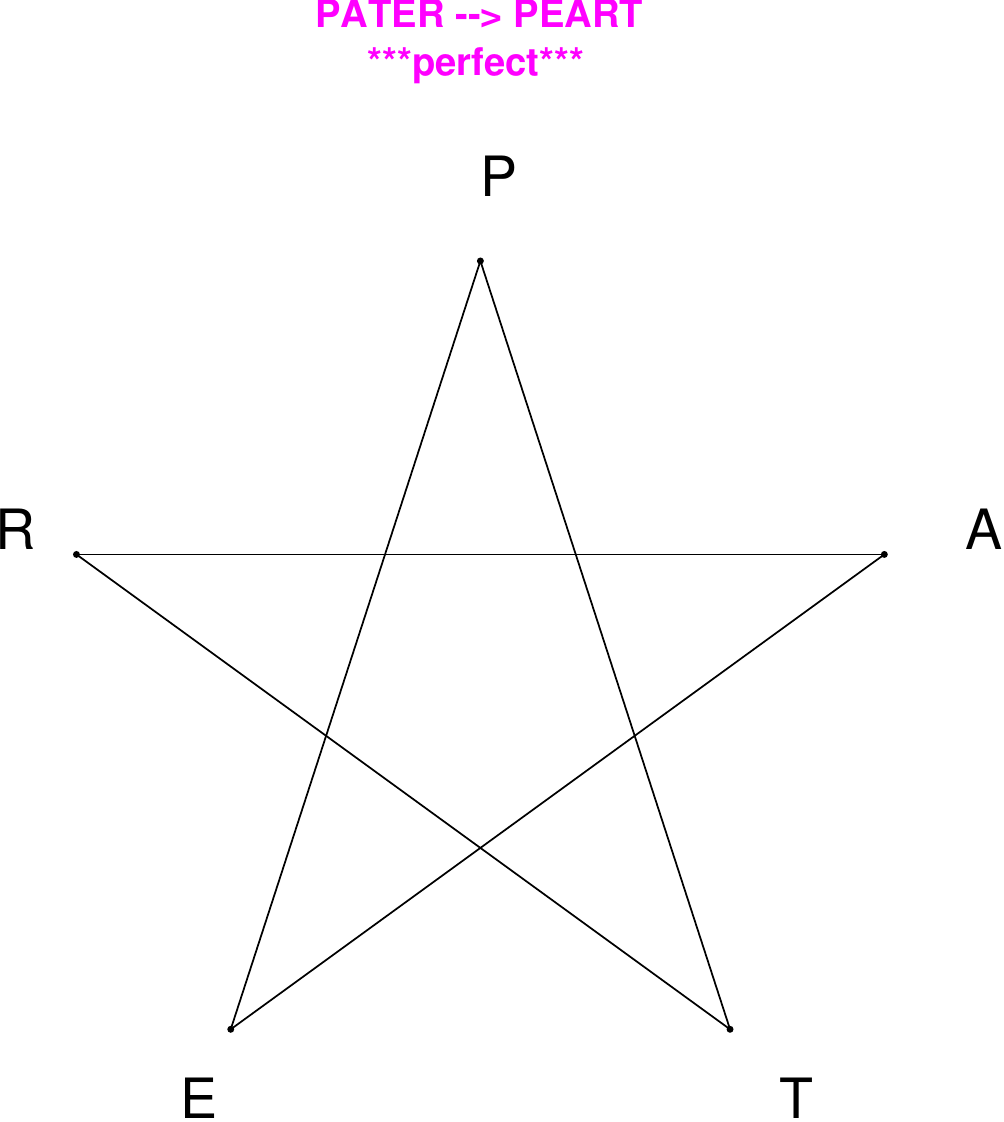}
\end{subfigure}
\hfill
\begin{subfigure}[T]{0.19\textwidth}
\centering
\includegraphics[width=\textwidth]{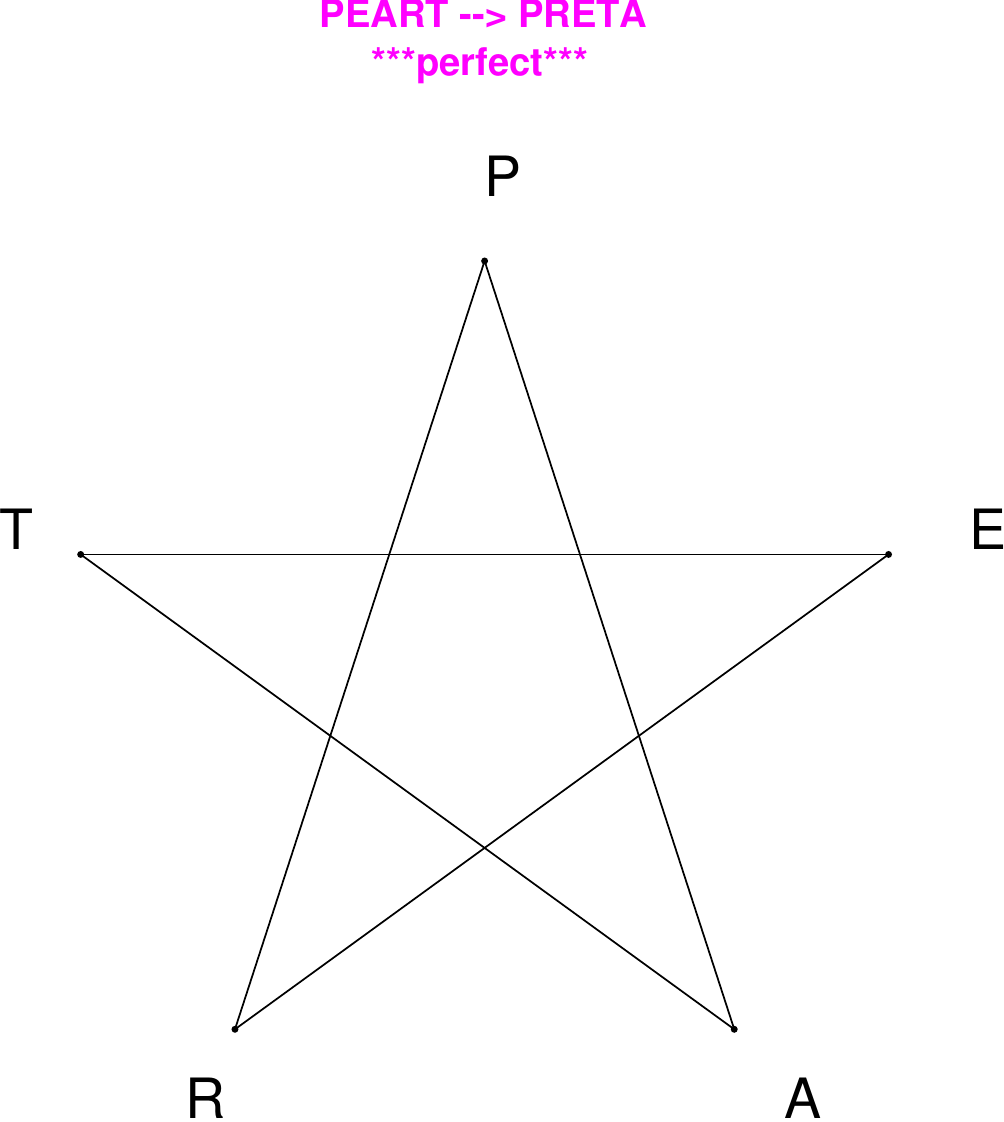}
\end{subfigure}
\hfill
\begin{subfigure}[T]{0.19\textwidth}
\centering
\includegraphics[width=\textwidth]{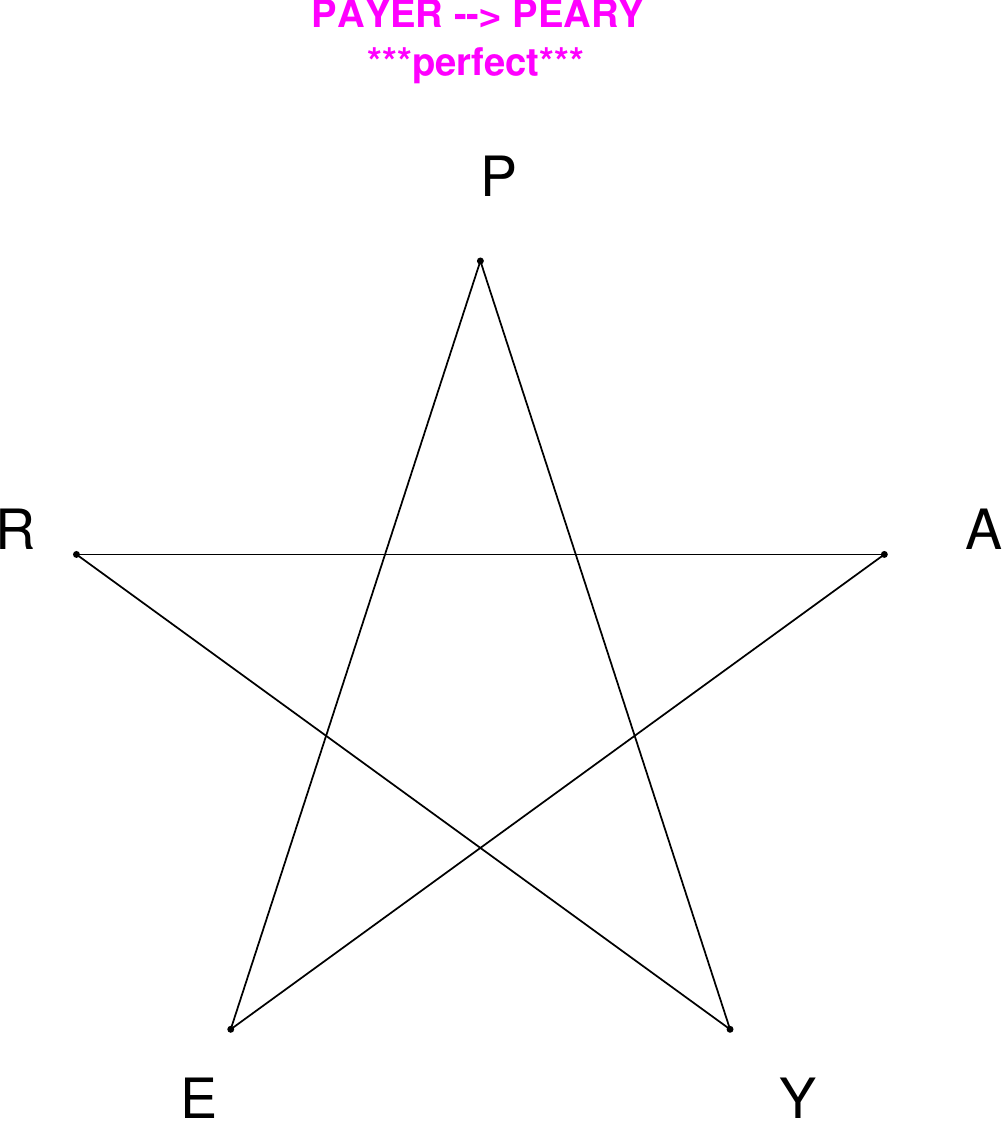}
\end{subfigure}
\end{figure}

\begin{figure}[H]
\centering
\begin{subfigure}[T]{0.19\textwidth}
\centering
\includegraphics[width=\textwidth]{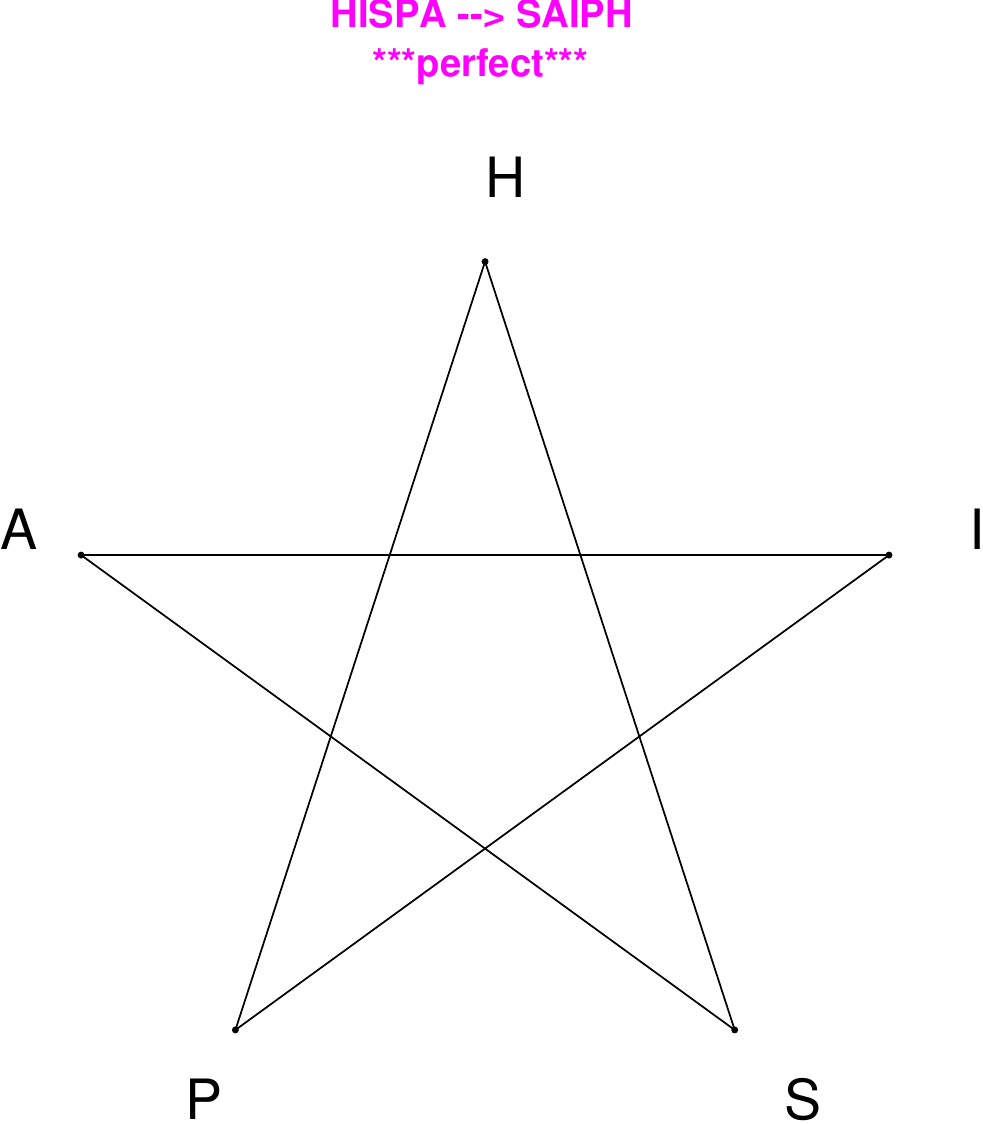}
\end{subfigure}
\hfill
\begin{subfigure}[T]{0.19\textwidth}
\centering
\includegraphics[width=\textwidth]{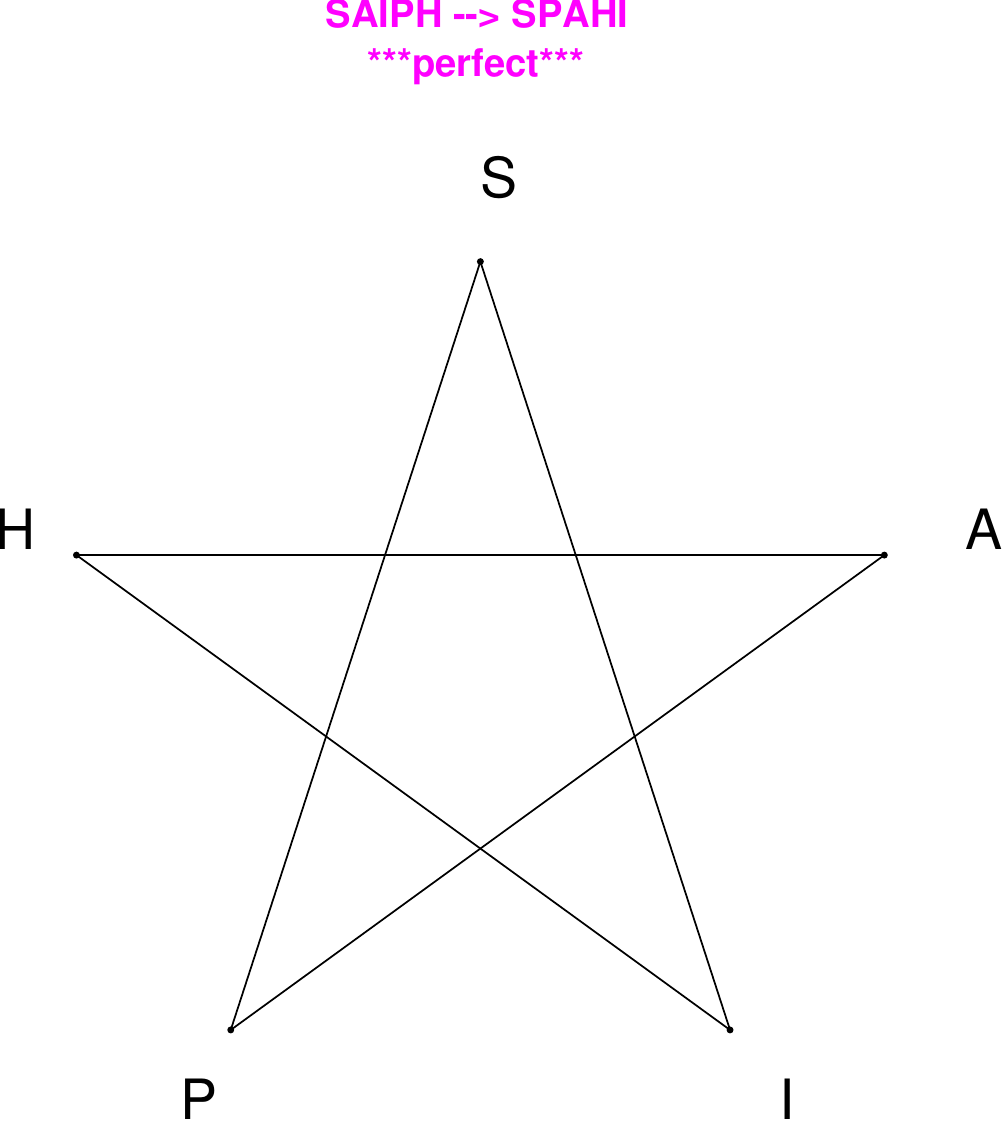}
\end{subfigure}
\hfill
\begin{subfigure}[T]{0.19\textwidth}
\centering
\includegraphics[width=\textwidth]{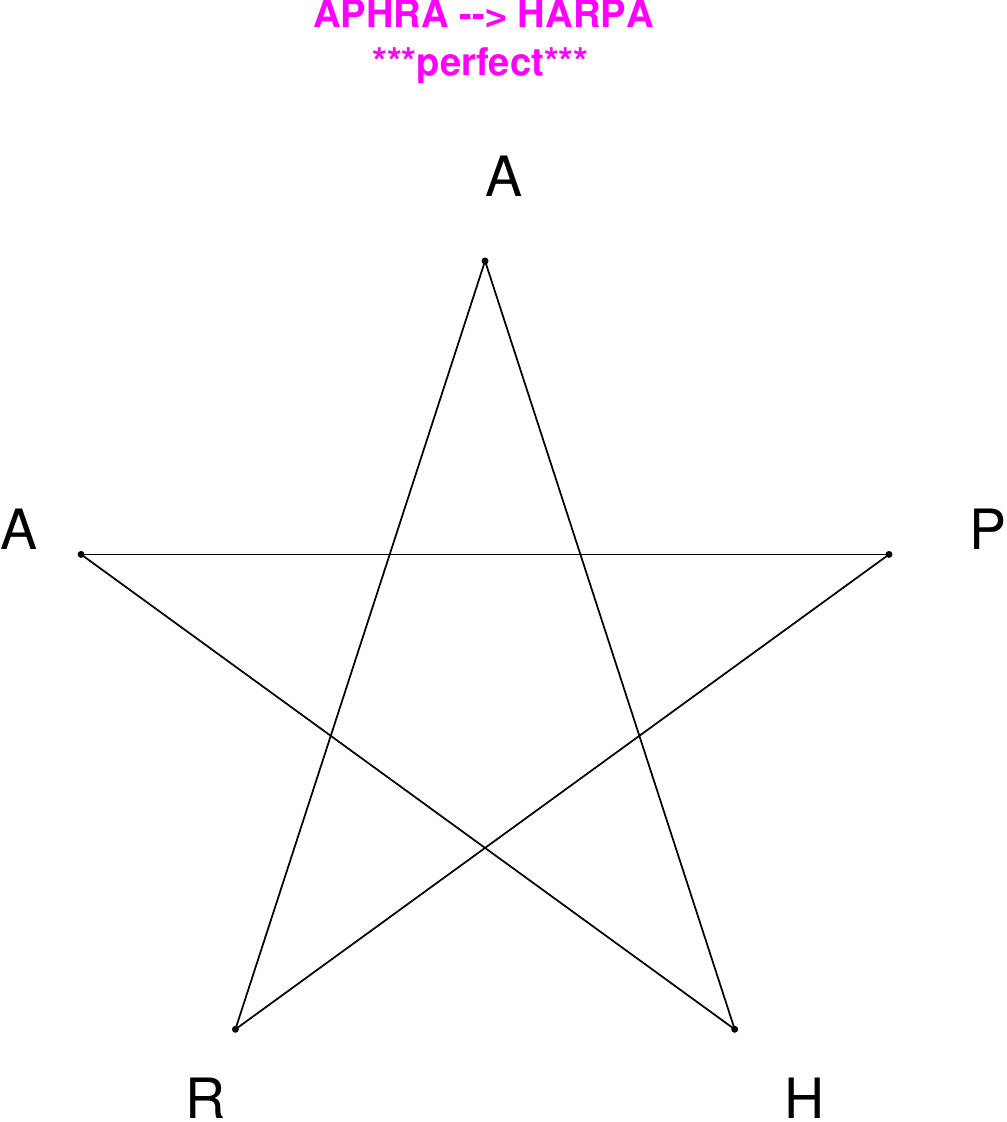}
\end{subfigure}
\hfill
\begin{subfigure}[T]{0.19\textwidth}
\centering
\includegraphics[width=\textwidth]{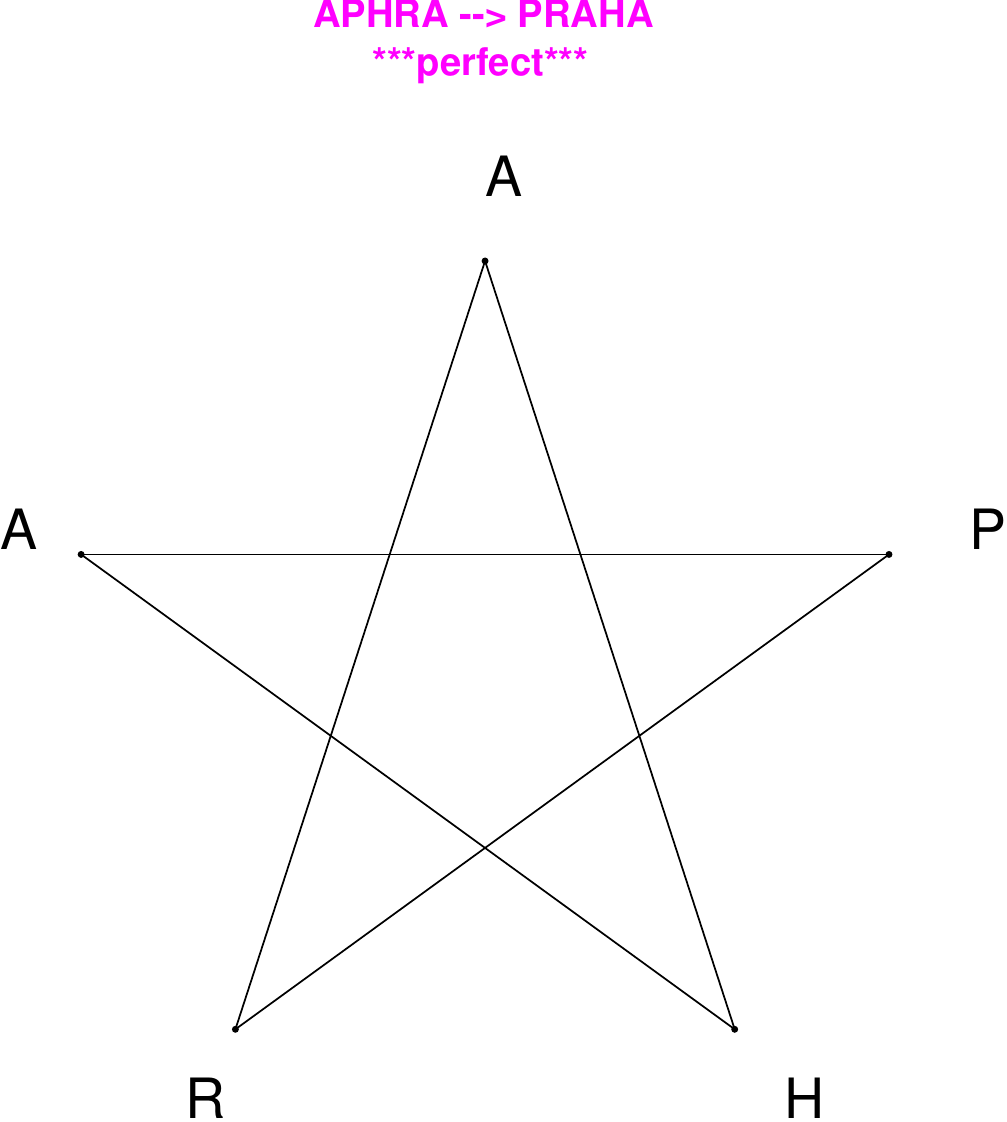}
\end{subfigure}
\hfill
\begin{subfigure}[T]{0.19\textwidth}
\centering
\includegraphics[width=\textwidth]{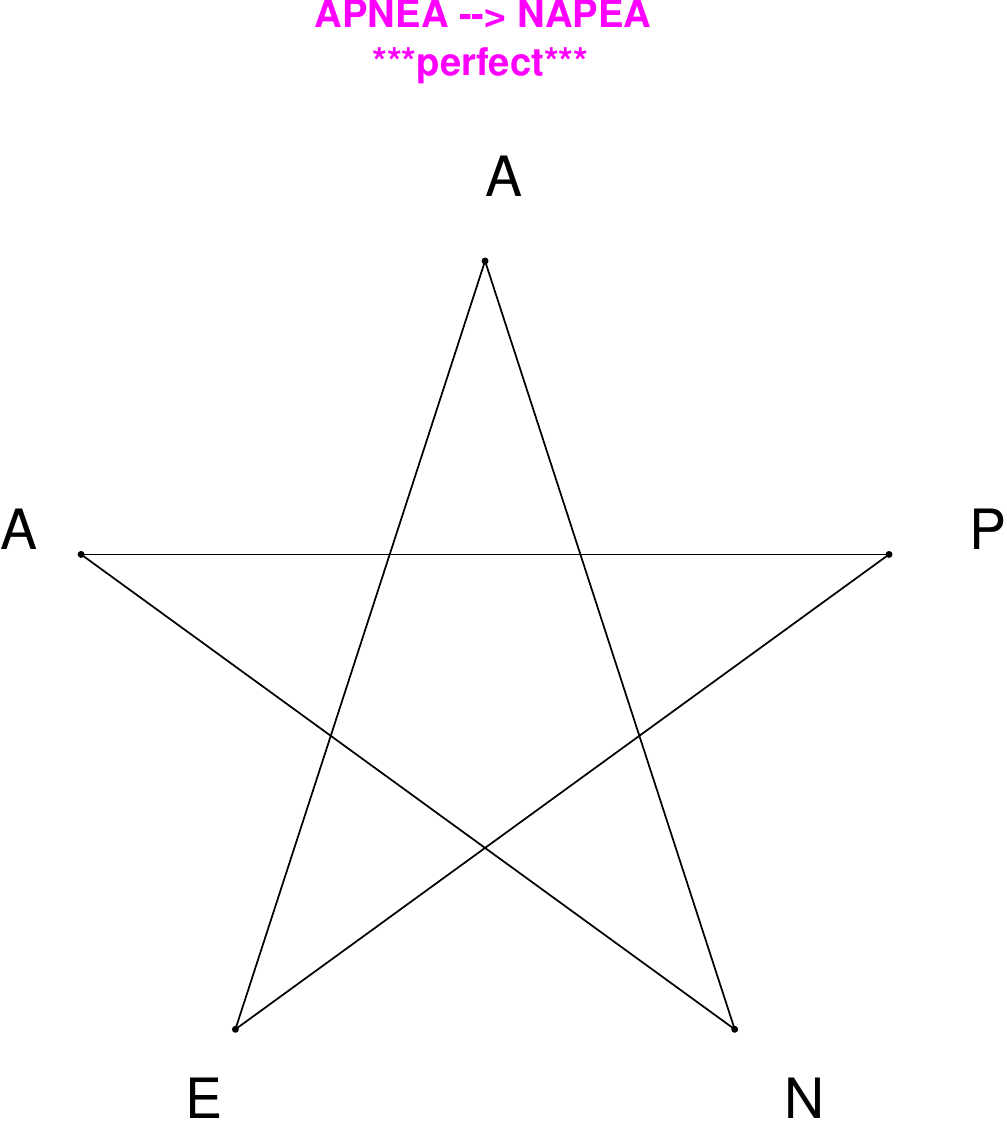}
\end{subfigure}
\end{figure}

\begin{figure}[H]
\centering
\begin{subfigure}[T]{0.19\textwidth}
\centering
\includegraphics[width=\textwidth]{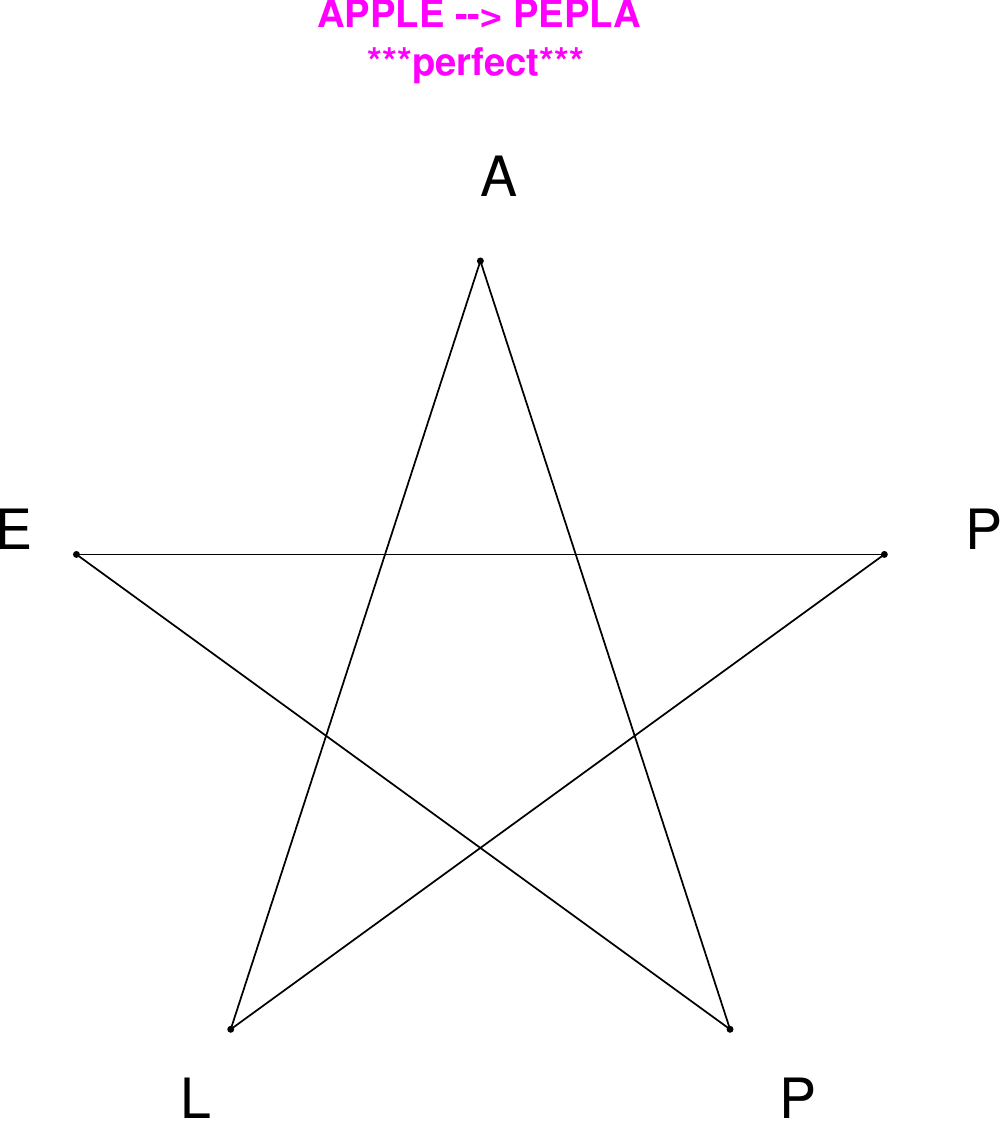}
\end{subfigure}
\hfill
\begin{subfigure}[T]{0.19\textwidth}
\centering
\includegraphics[width=\textwidth]{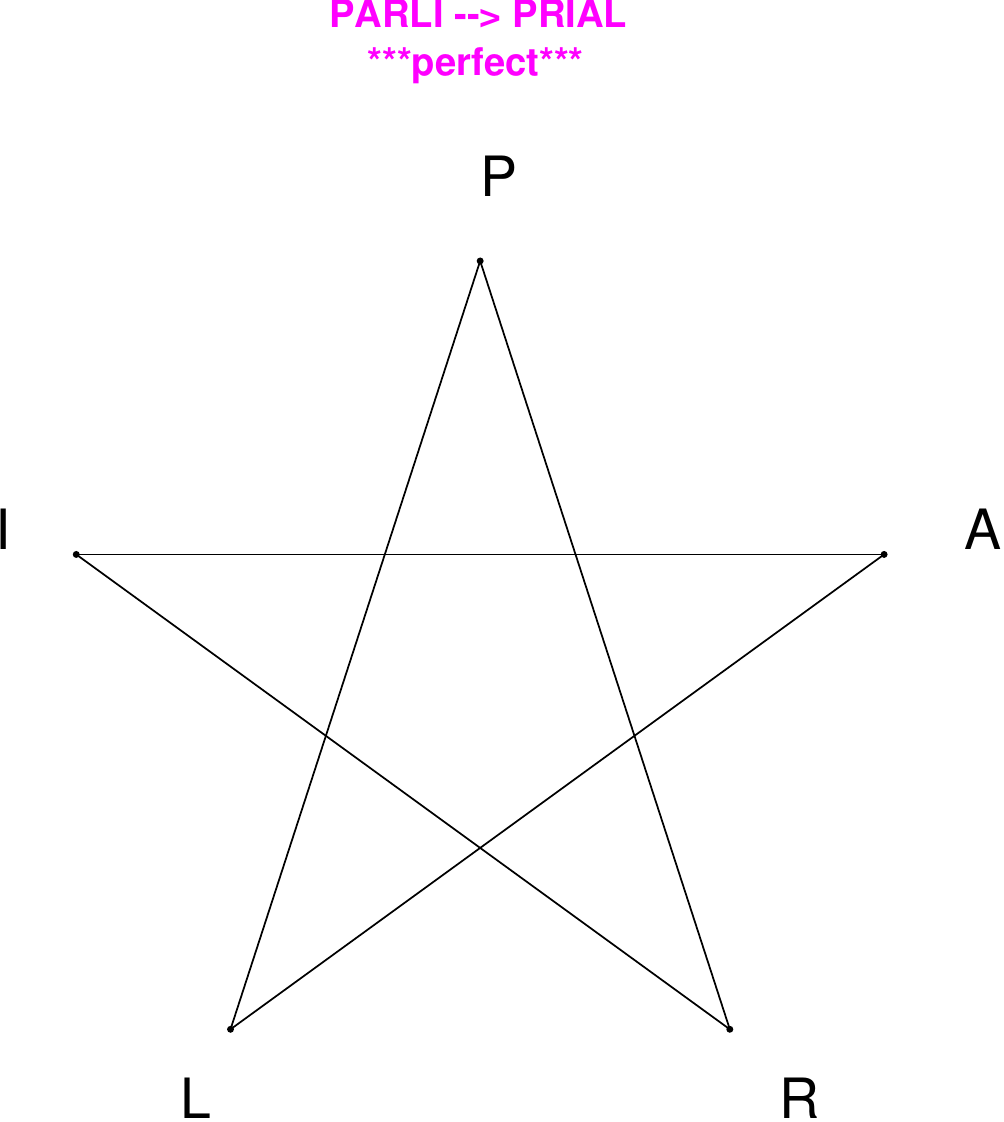}
\end{subfigure}
\hfill
\begin{subfigure}[T]{0.19\textwidth}
\centering
\includegraphics[width=\textwidth]{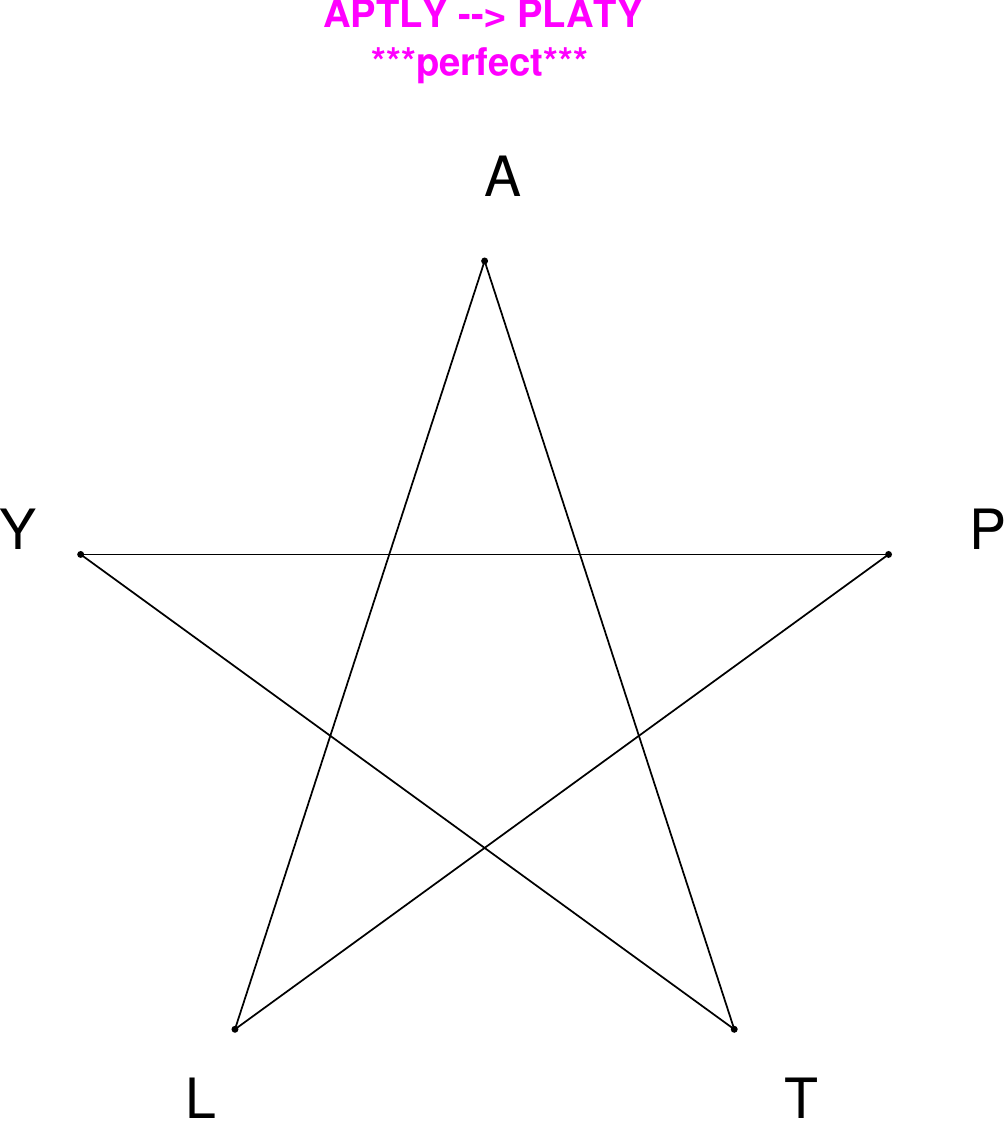}
\end{subfigure}
\hfill
\begin{subfigure}[T]{0.19\textwidth}
\centering
\includegraphics[width=\textwidth]{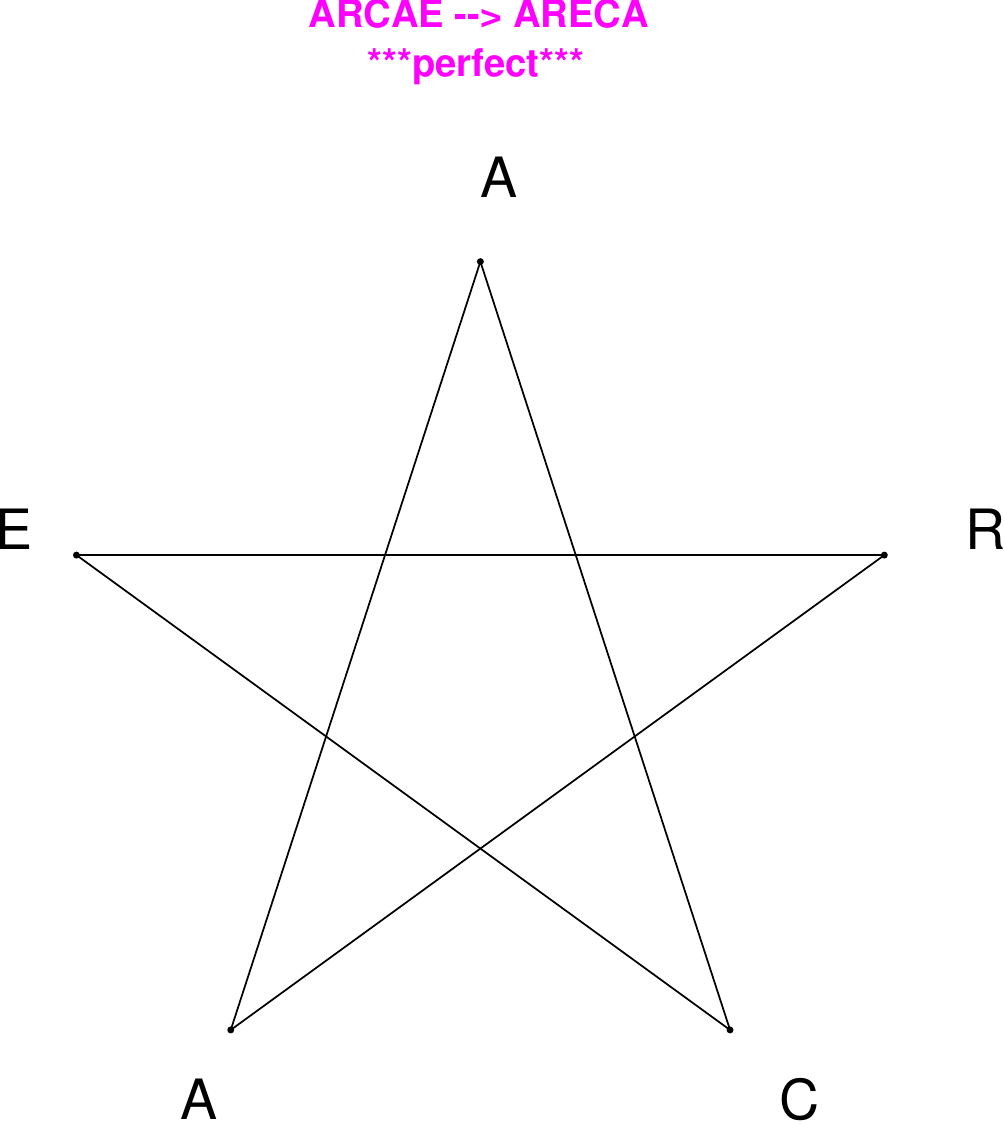}
\end{subfigure}
\hfill
\begin{subfigure}[T]{0.19\textwidth}
\centering
\includegraphics[width=\textwidth]{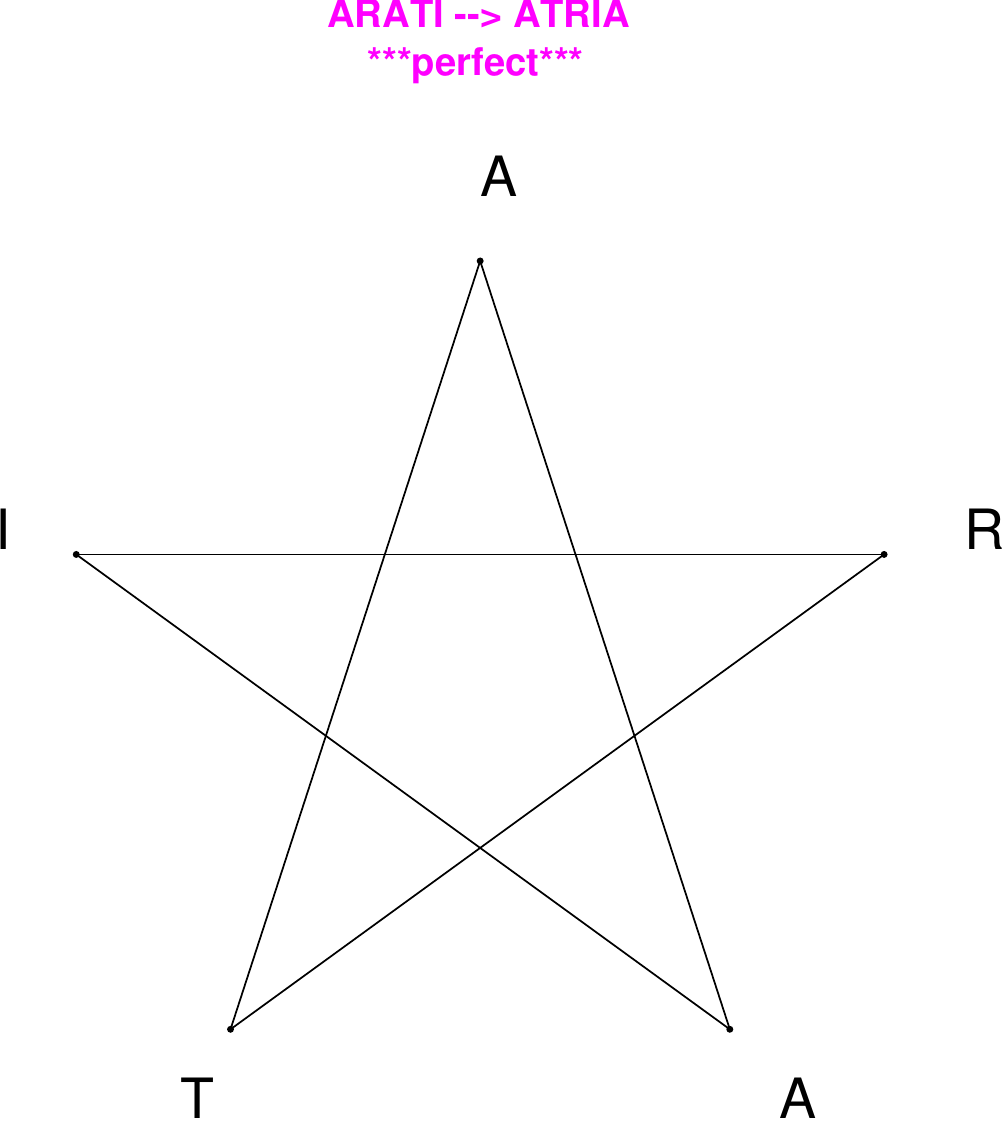}
\end{subfigure}
\end{figure}

\begin{figure}[H]
\centering
\begin{subfigure}[T]{0.19\textwidth}
\centering
\includegraphics[width=\textwidth]{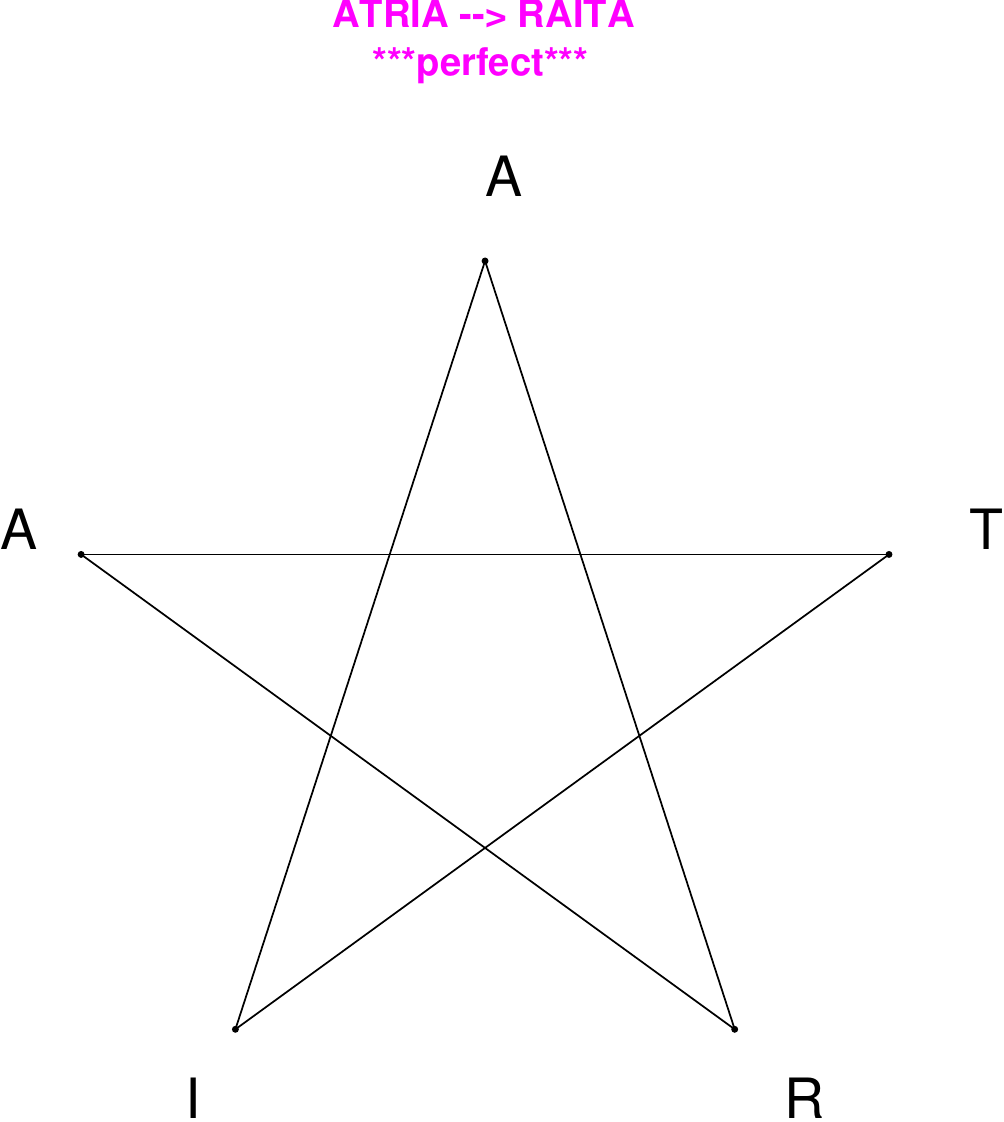}
\end{subfigure}
\hfill
\begin{subfigure}[T]{0.19\textwidth}
\centering
\includegraphics[width=\textwidth]{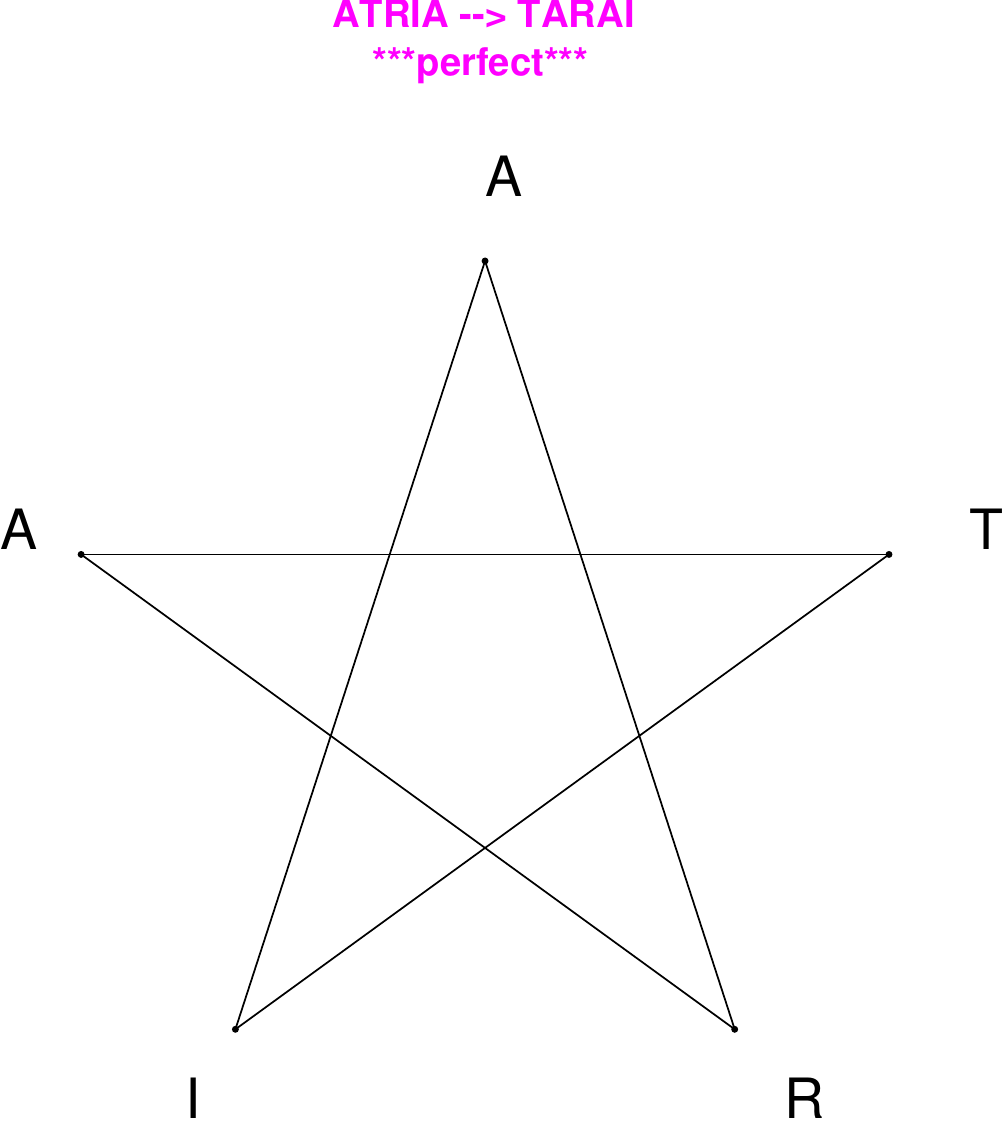}
\end{subfigure}
\hfill
\begin{subfigure}[T]{0.19\textwidth}
\centering
\includegraphics[width=\textwidth]{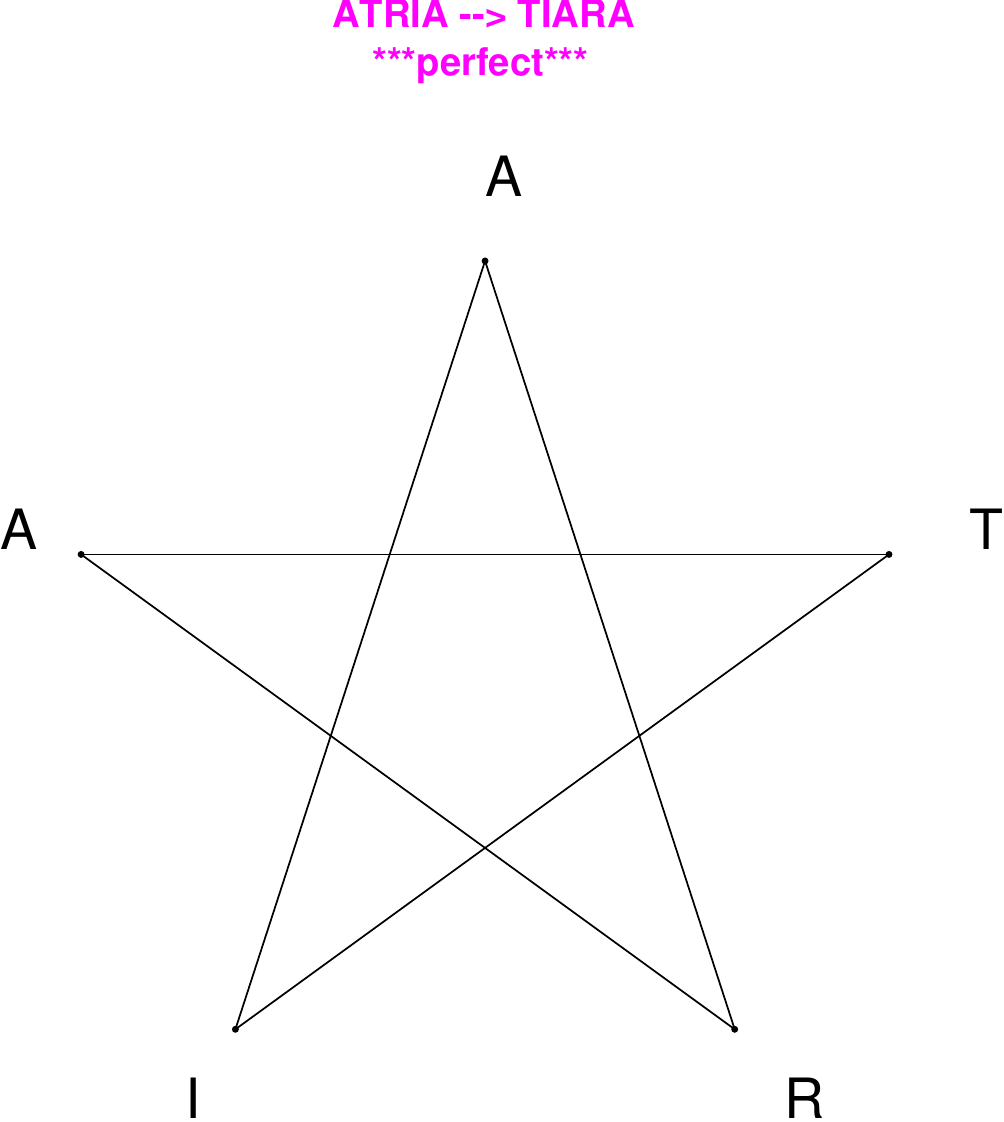}
\end{subfigure}
\hfill
\begin{subfigure}[T]{0.19\textwidth}
\centering
\includegraphics[width=\textwidth]{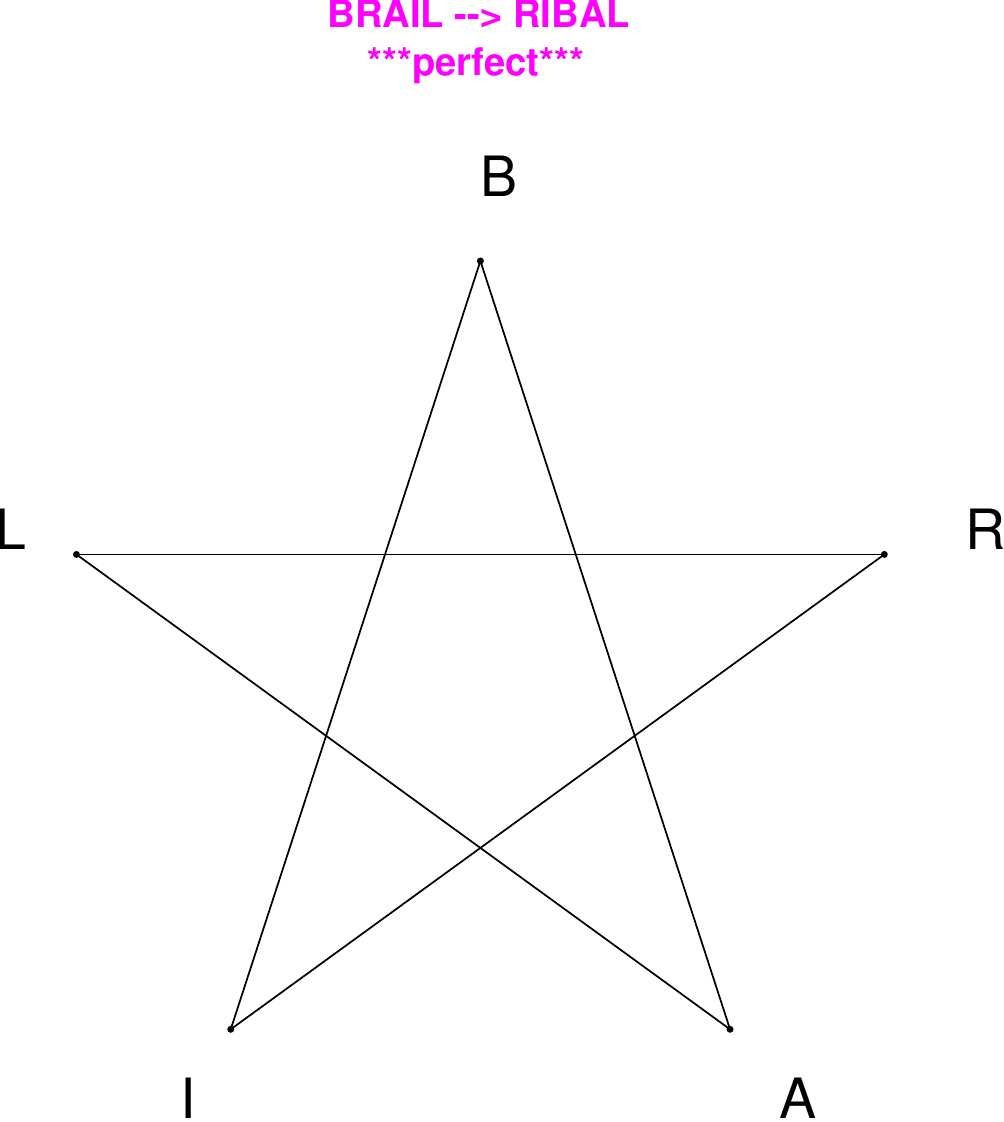}
\end{subfigure}
\hfill
\begin{subfigure}[T]{0.19\textwidth}
\centering
\includegraphics[width=\textwidth]{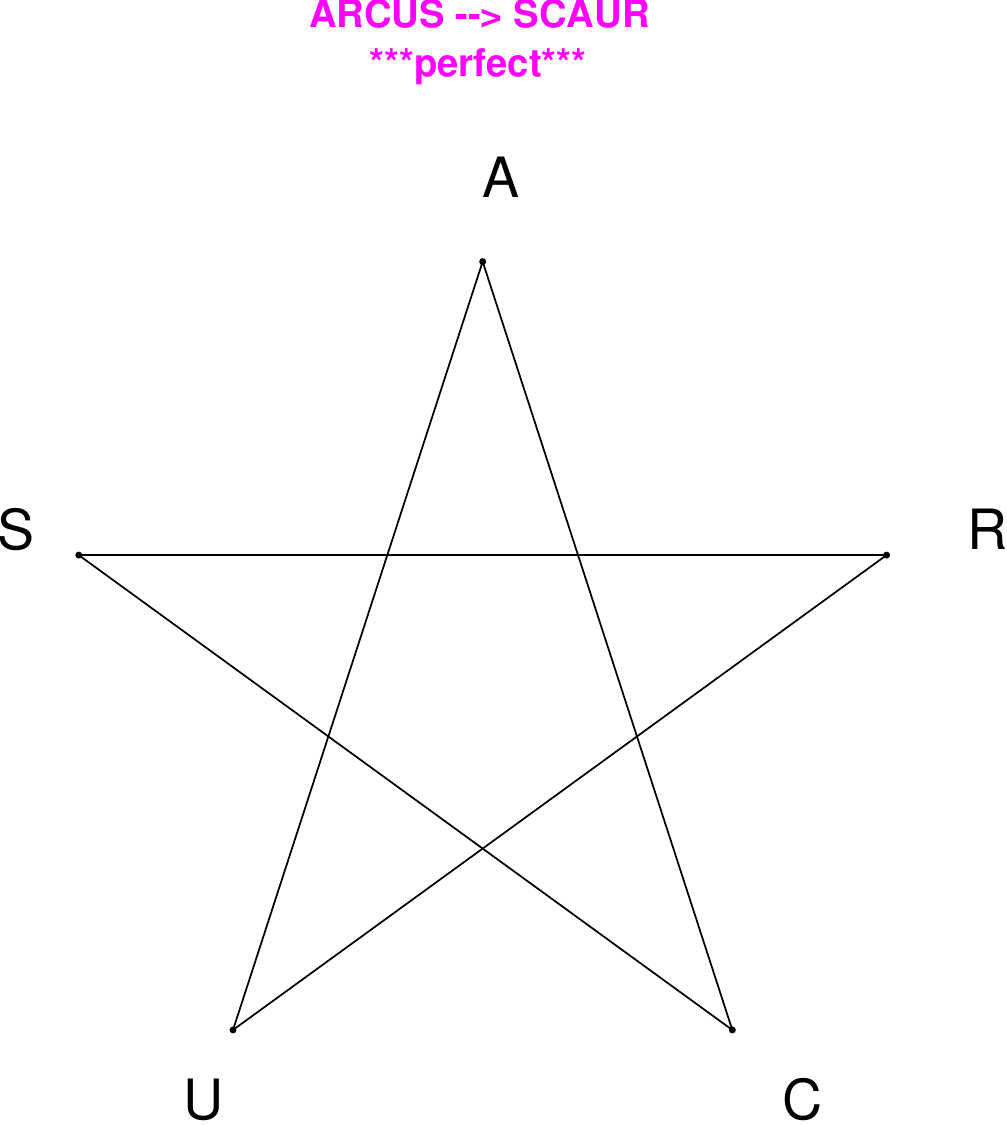}
\end{subfigure}
\end{figure}

\begin{figure}[H]
\centering
\begin{subfigure}[T]{0.19\textwidth}
\centering
\includegraphics[width=\textwidth]{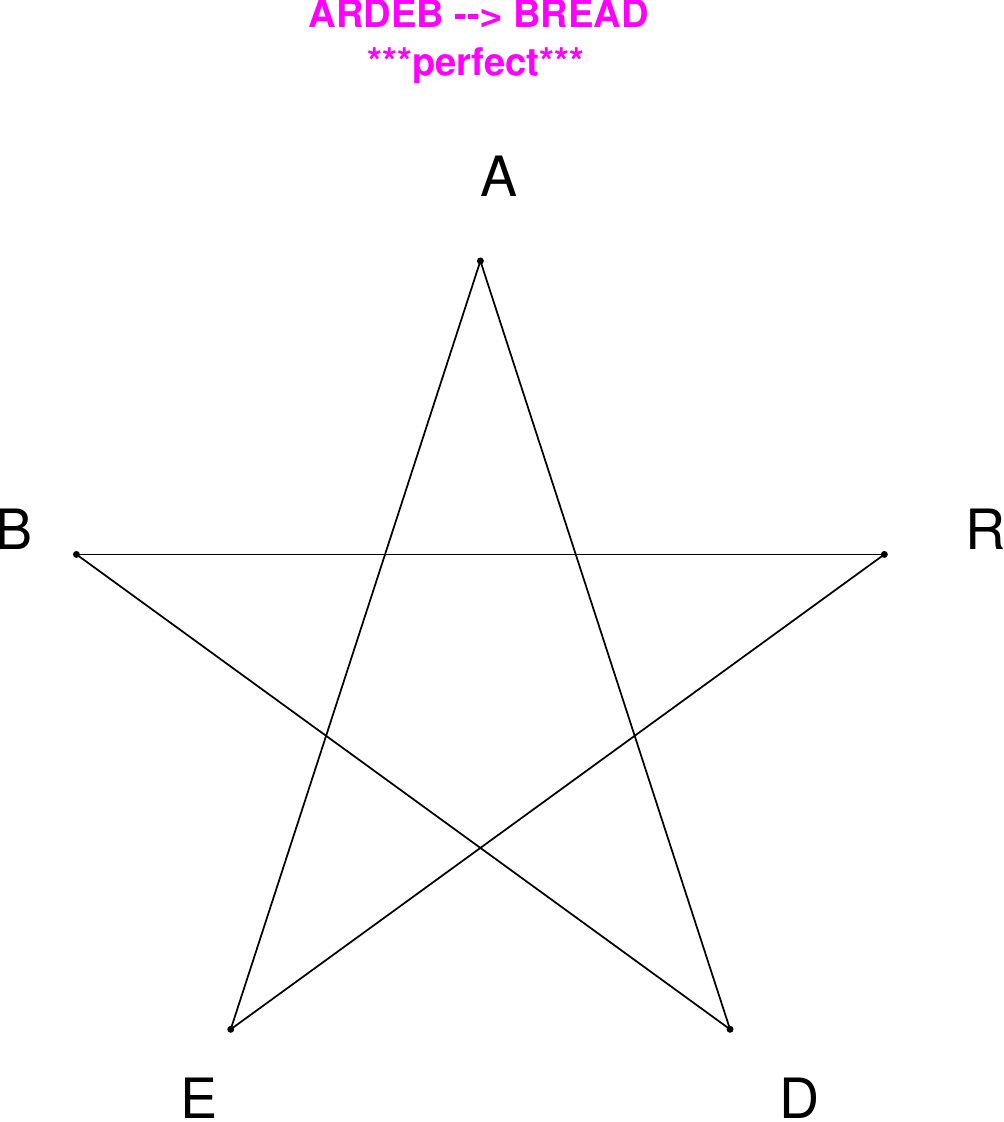}
\end{subfigure}
\hfill
\begin{subfigure}[T]{0.19\textwidth}
\centering
\includegraphics[width=\textwidth]{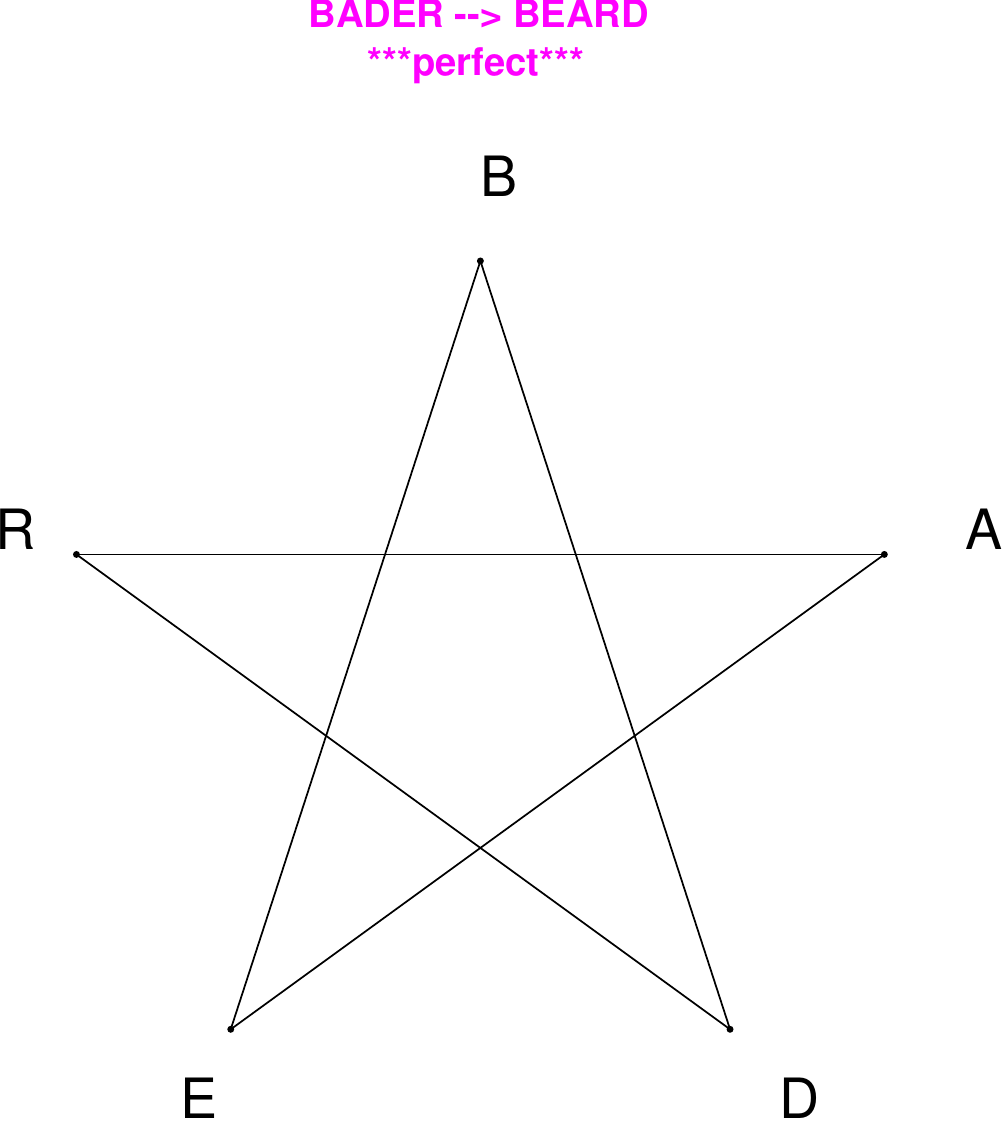}
\end{subfigure}
\hfill
\begin{subfigure}[T]{0.19\textwidth}
\centering
\includegraphics[width=\textwidth]{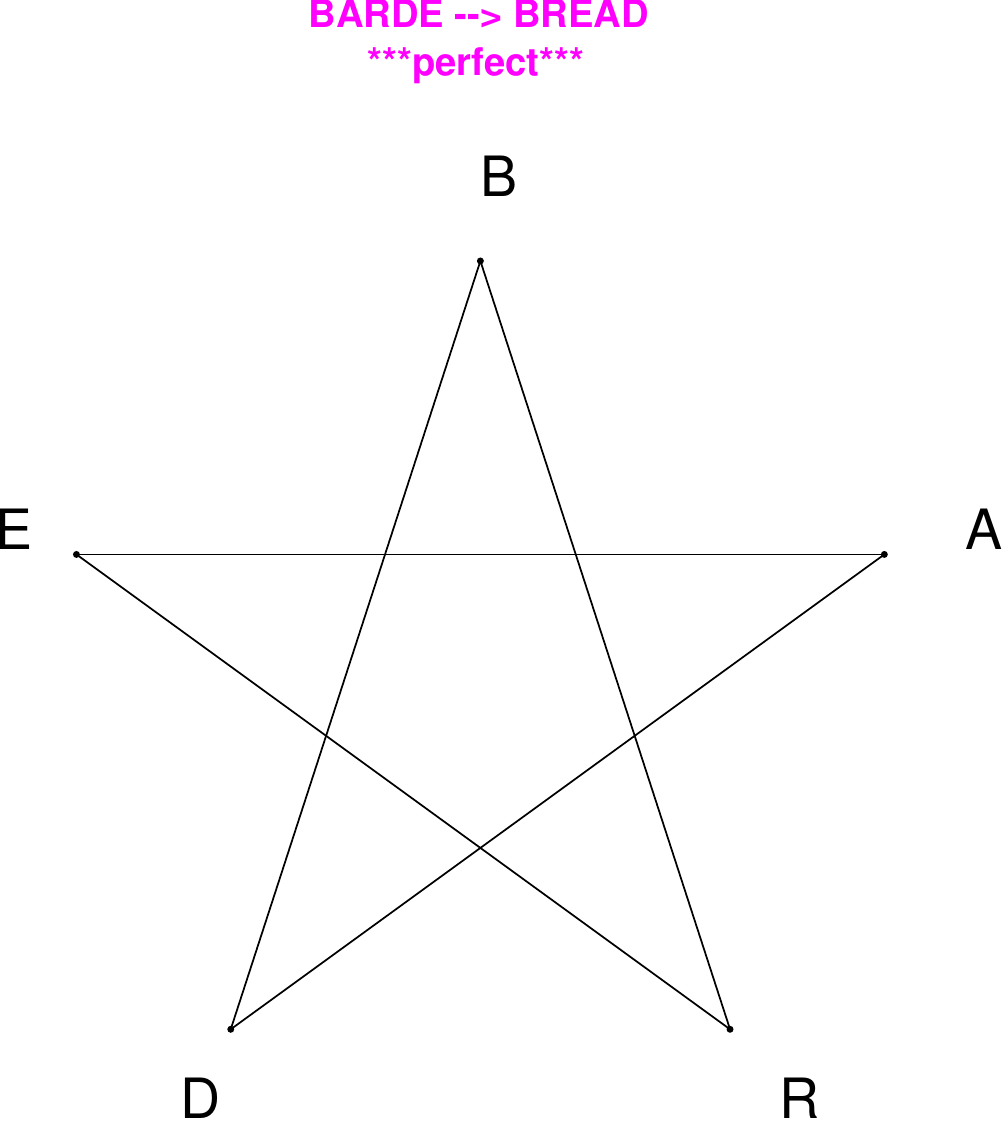}
\end{subfigure}
\hfill
\begin{subfigure}[T]{0.19\textwidth}
\centering
\includegraphics[width=\textwidth]{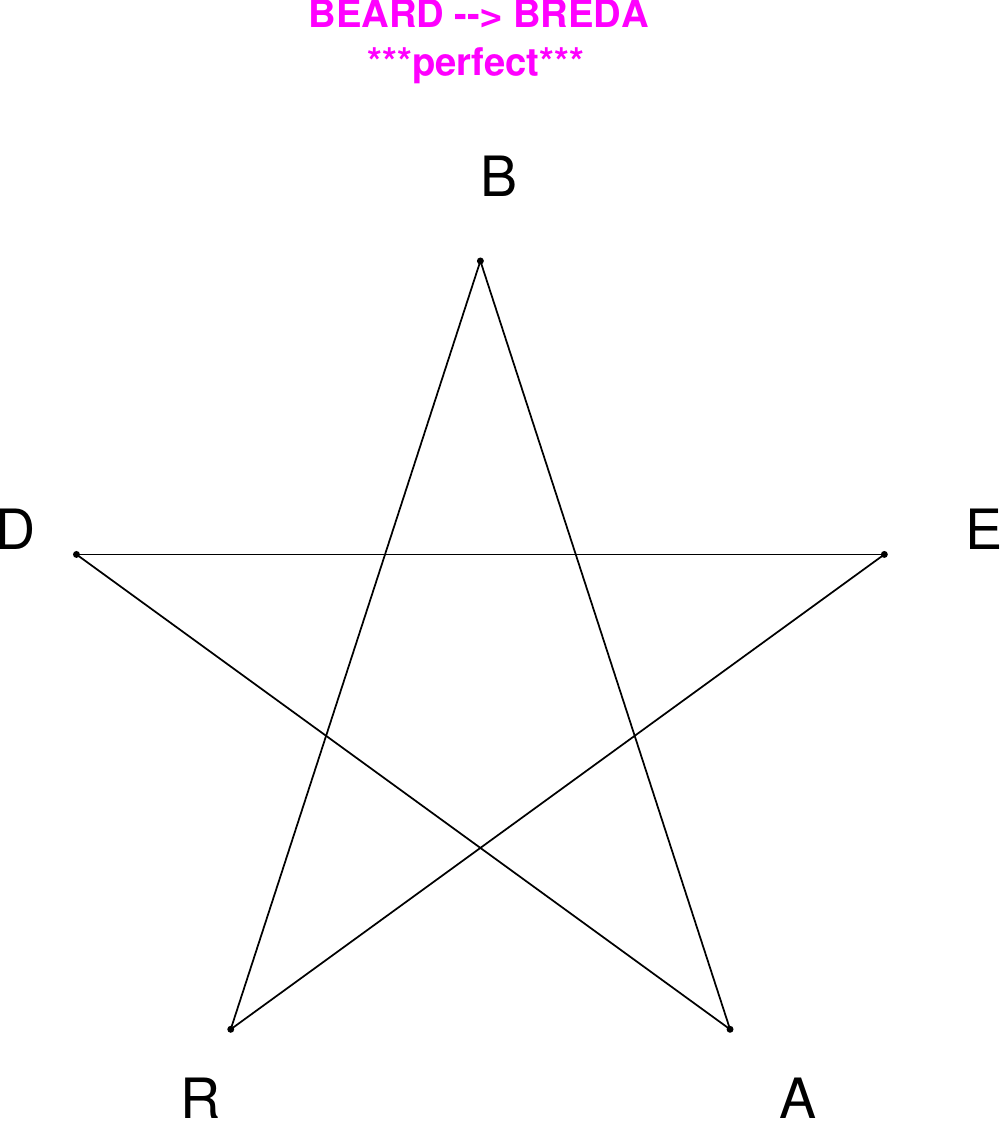}
\end{subfigure}
\hfill
\begin{subfigure}[T]{0.19\textwidth}
\centering
\includegraphics[width=\textwidth]{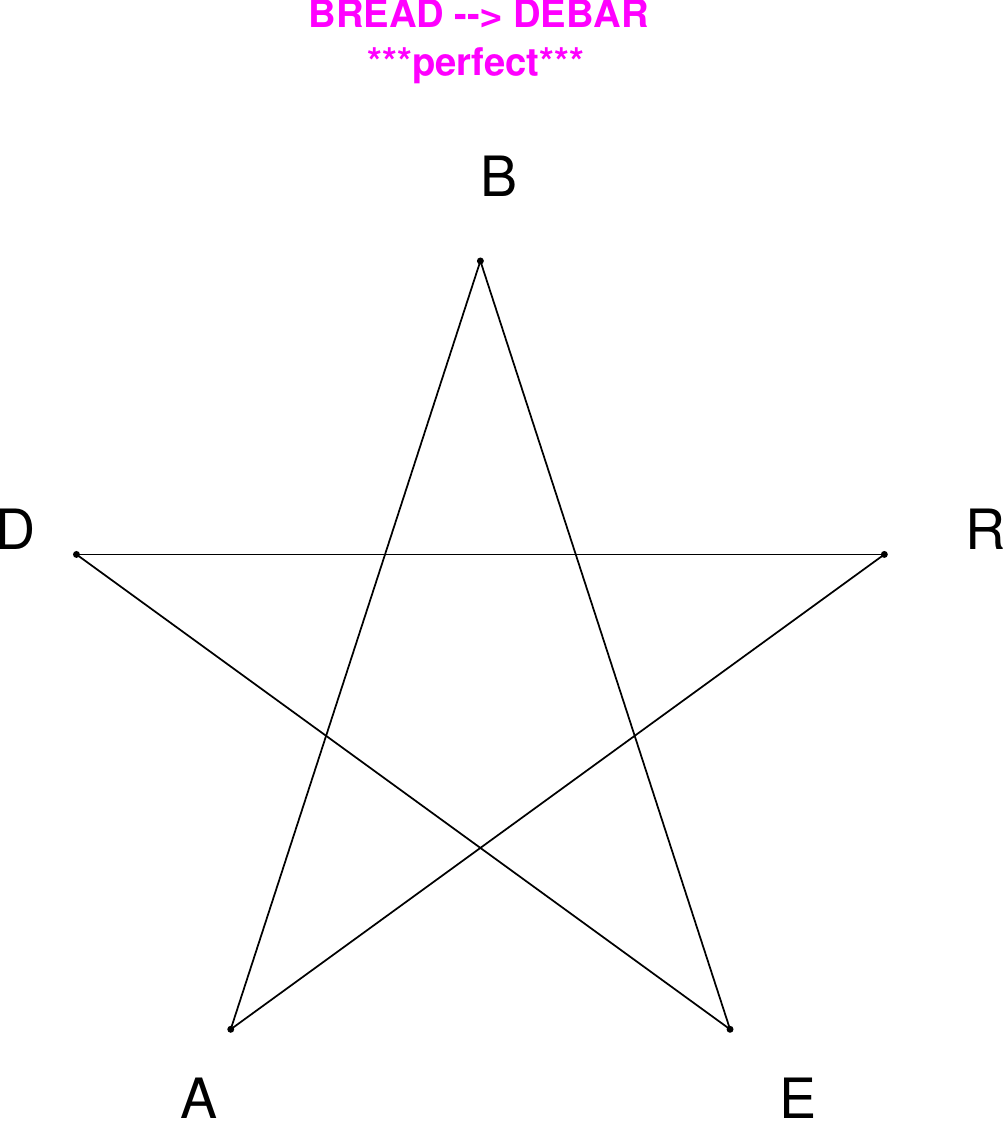}
\end{subfigure}
\end{figure}

\begin{figure}[H]
\centering
\begin{subfigure}[T]{0.19\textwidth}
\centering
\includegraphics[width=\textwidth]{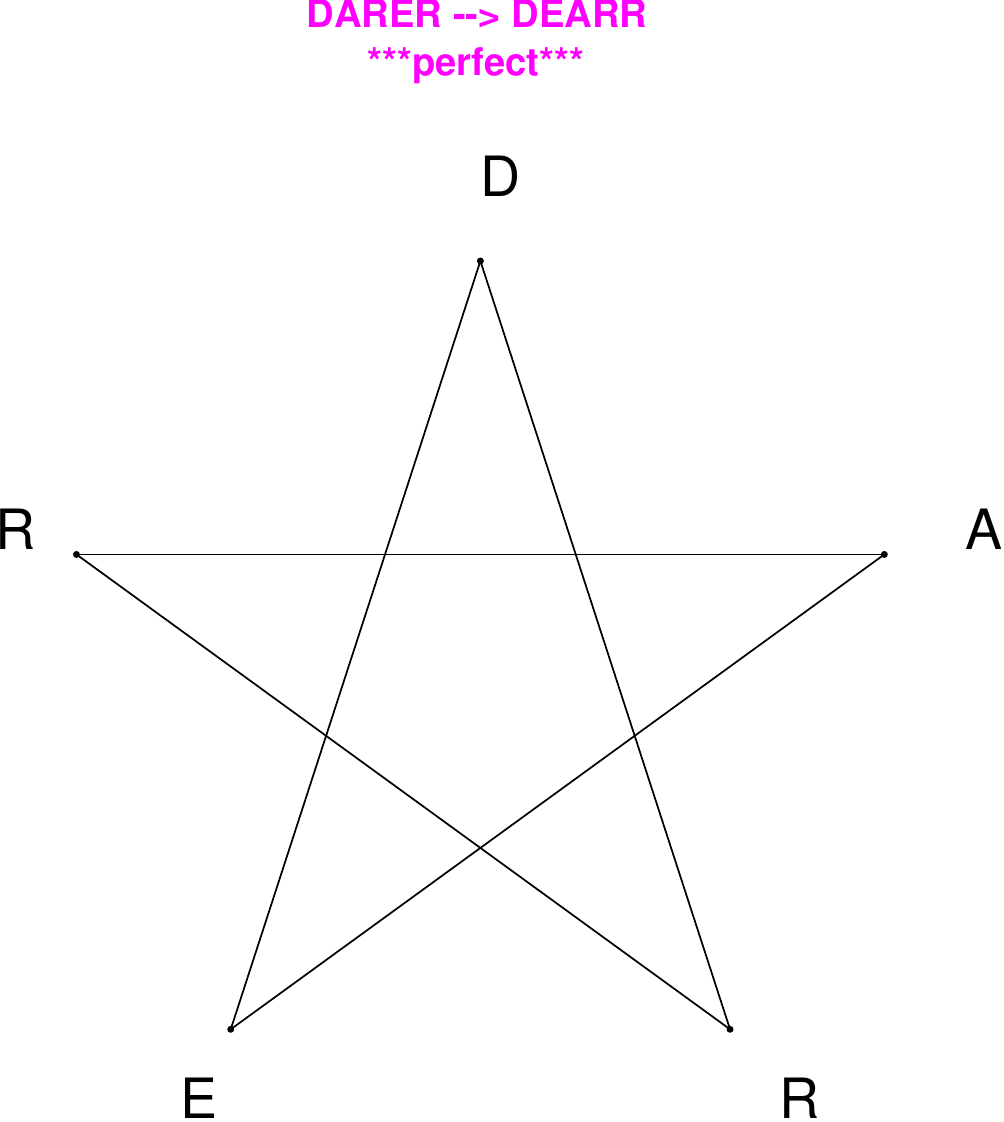}
\end{subfigure}
\hfill
\begin{subfigure}[T]{0.19\textwidth}
\centering
\includegraphics[width=\textwidth]{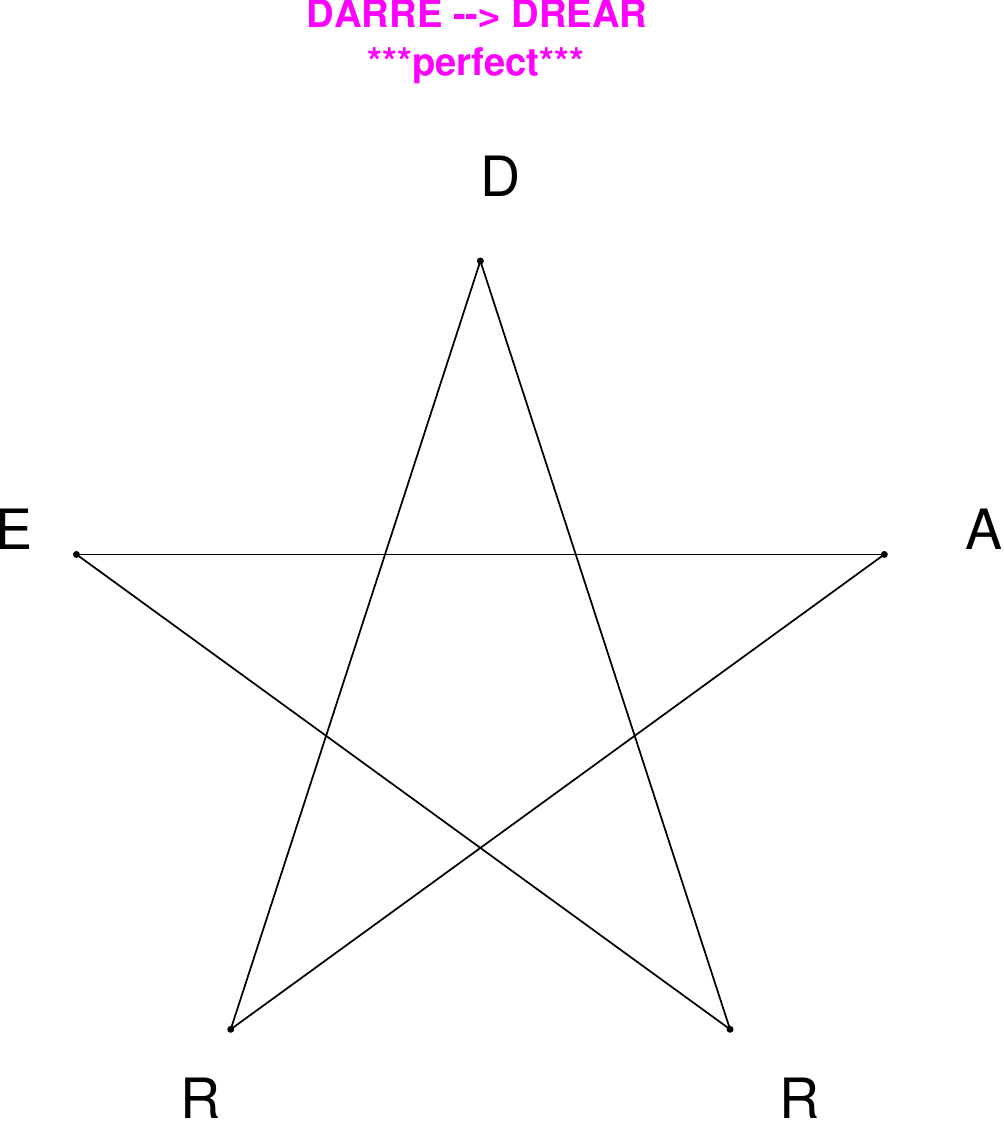}
\end{subfigure}
\hfill
\begin{subfigure}[T]{0.19\textwidth}
\centering
\includegraphics[width=\textwidth]{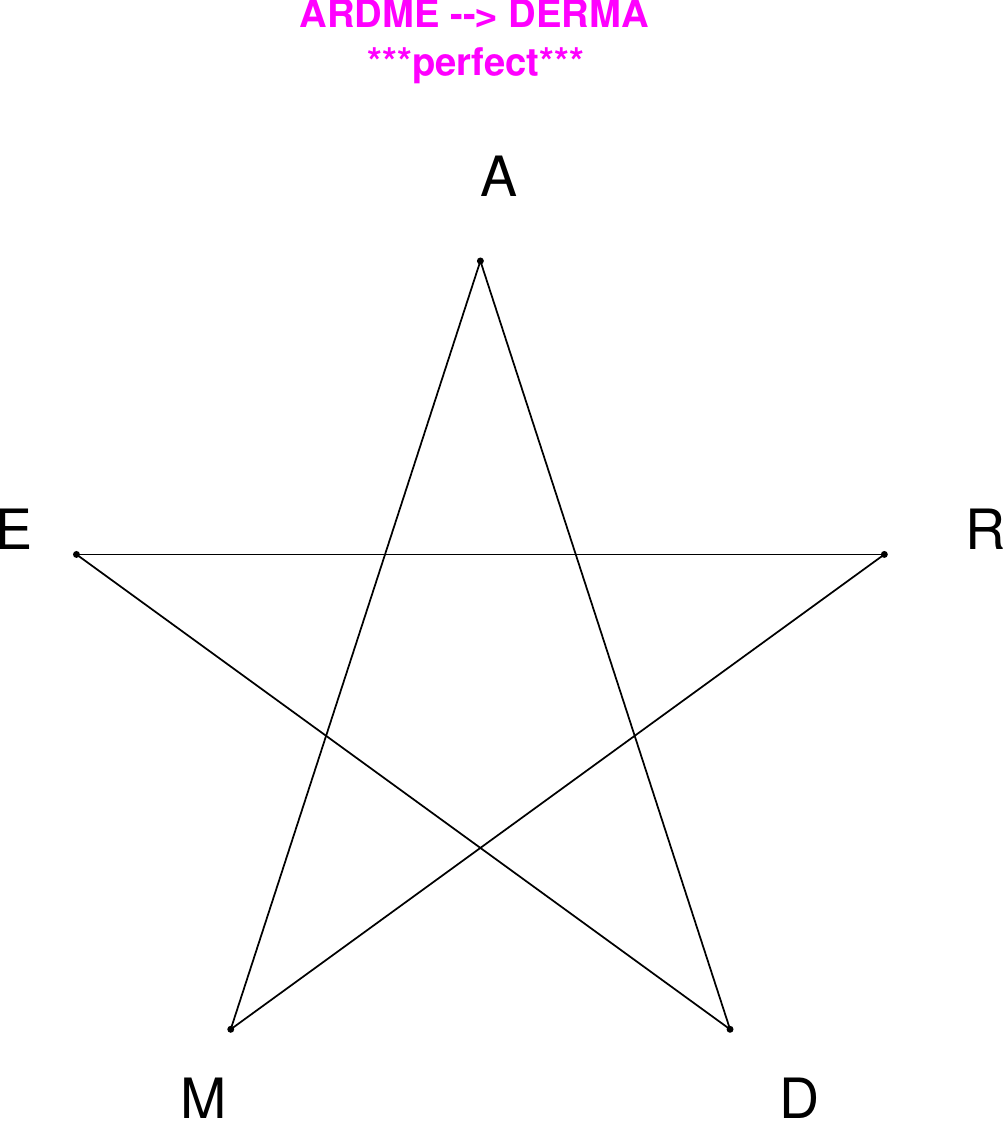}
\end{subfigure}
\hfill
\begin{subfigure}[T]{0.19\textwidth}
\centering
\includegraphics[width=\textwidth]{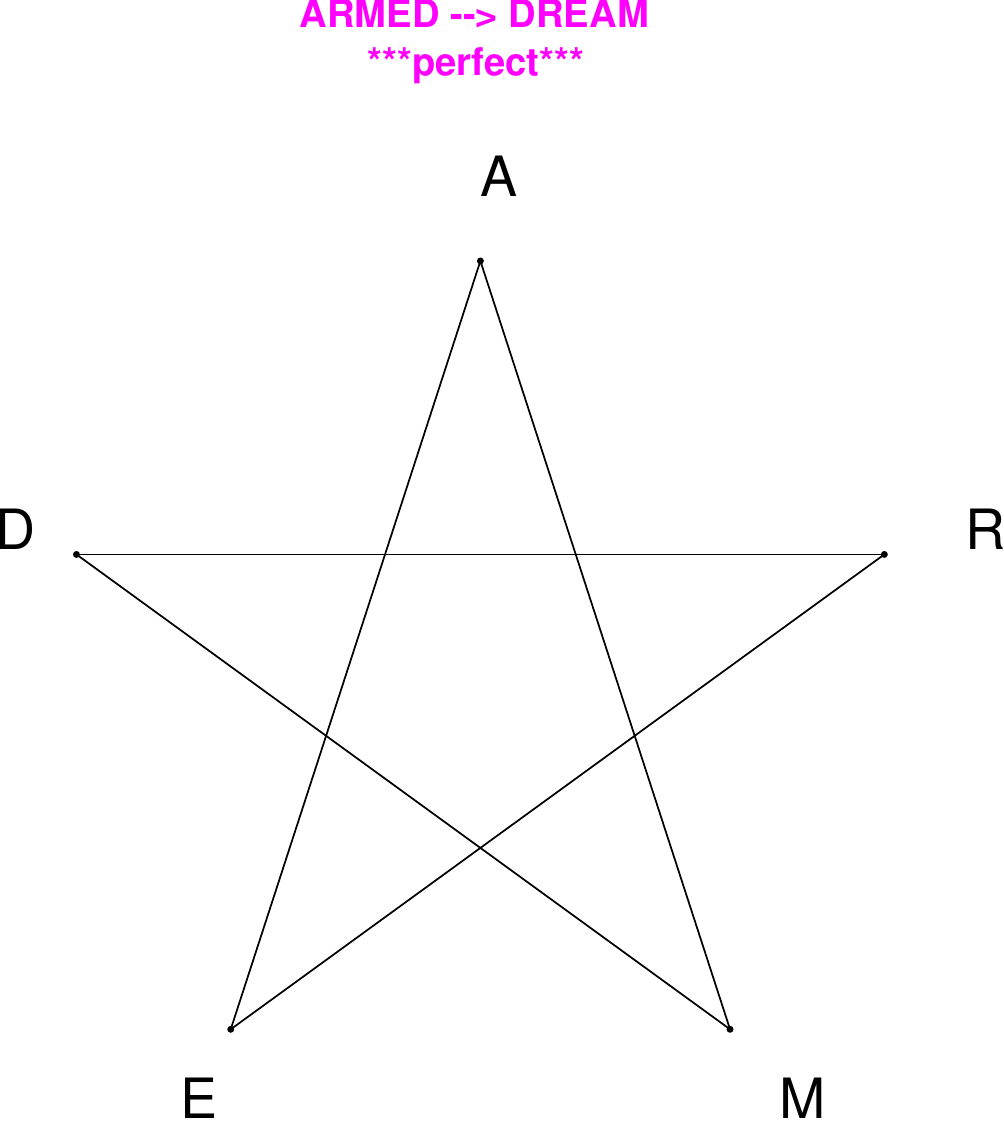}
\end{subfigure}
\hfill
\begin{subfigure}[T]{0.19\textwidth}
\centering
\includegraphics[width=\textwidth]{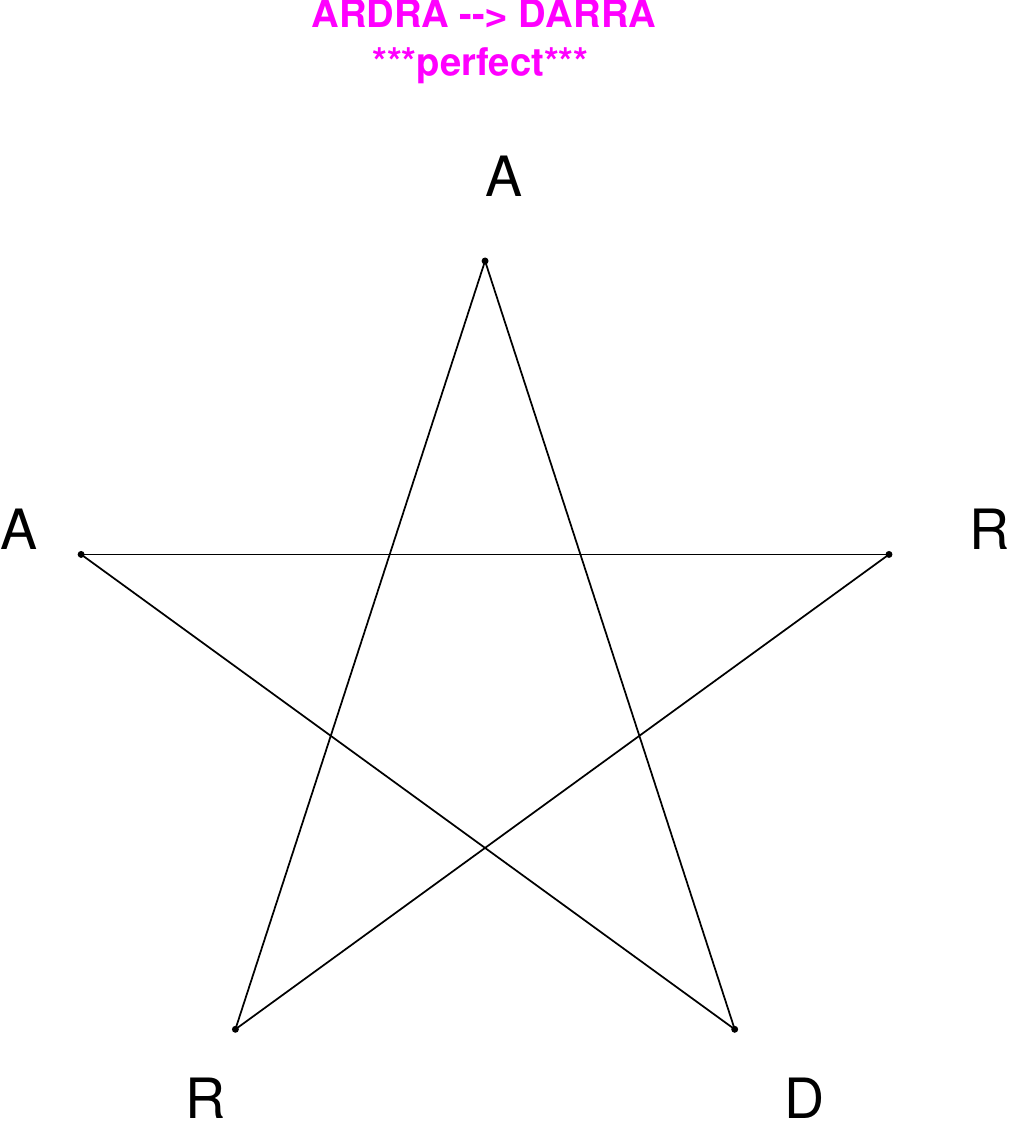}
\end{subfigure}
\end{figure}

\begin{figure}[H]
\centering
\begin{subfigure}[T]{0.19\textwidth}
\centering
\includegraphics[width=\textwidth]{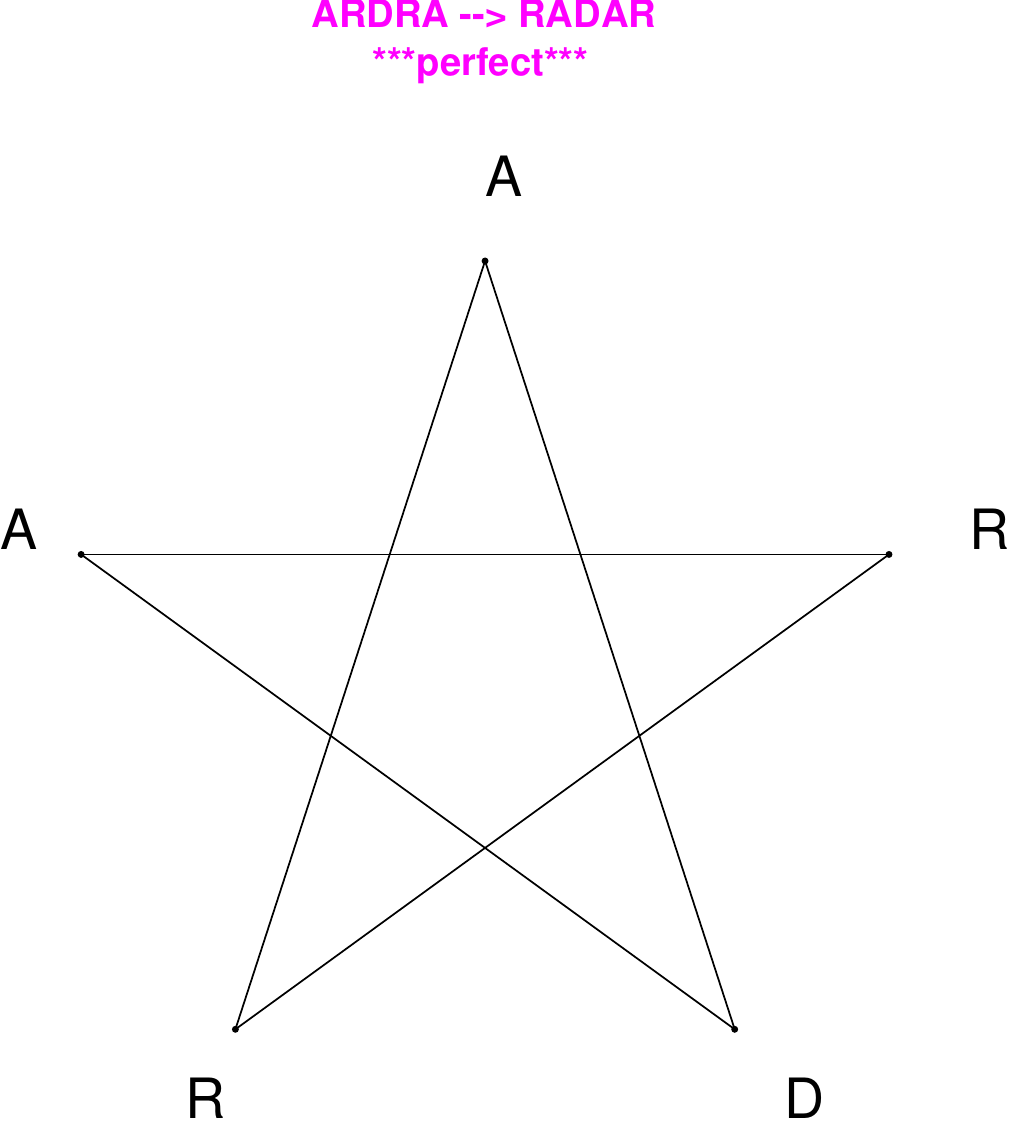}
\end{subfigure}
\hfill
\begin{subfigure}[T]{0.19\textwidth}
\centering
\includegraphics[width=\textwidth]{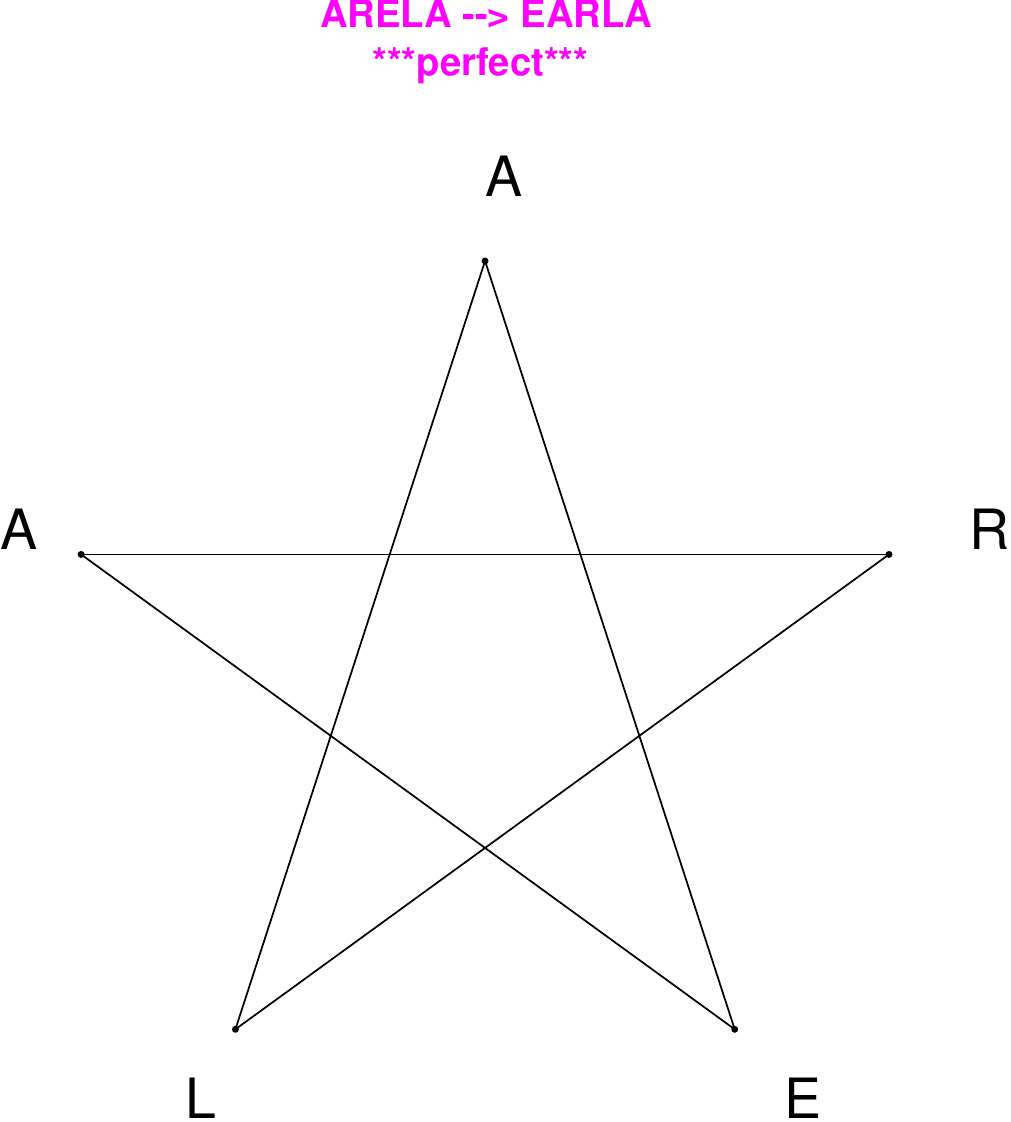}
\end{subfigure}
\hfill
\begin{subfigure}[T]{0.19\textwidth}
\centering
\includegraphics[width=\textwidth]{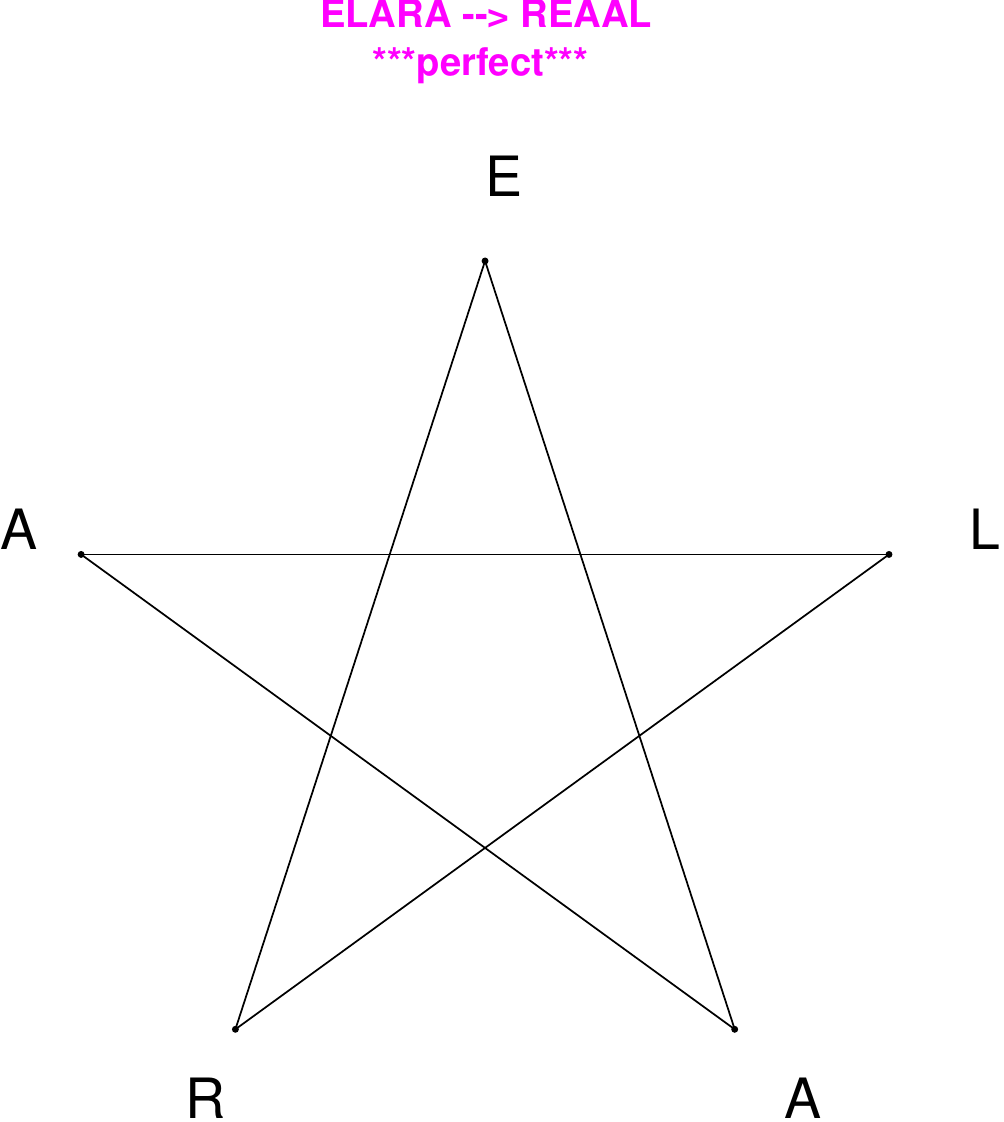}
\end{subfigure}
\hfill
\begin{subfigure}[T]{0.19\textwidth}
\centering
\includegraphics[width=\textwidth]{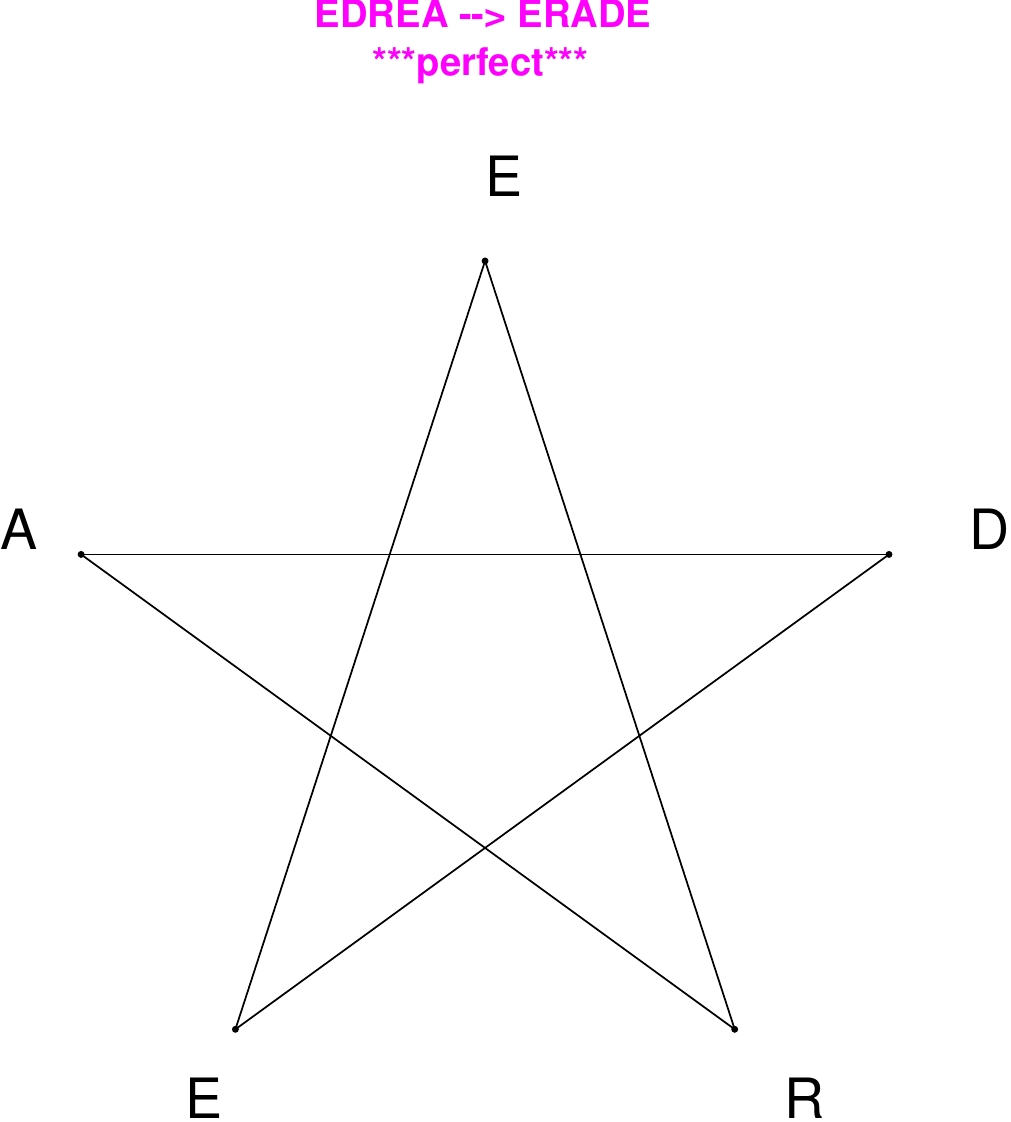}
\end{subfigure}
\hfill
\begin{subfigure}[T]{0.19\textwidth}
\centering
\includegraphics[width=\textwidth]{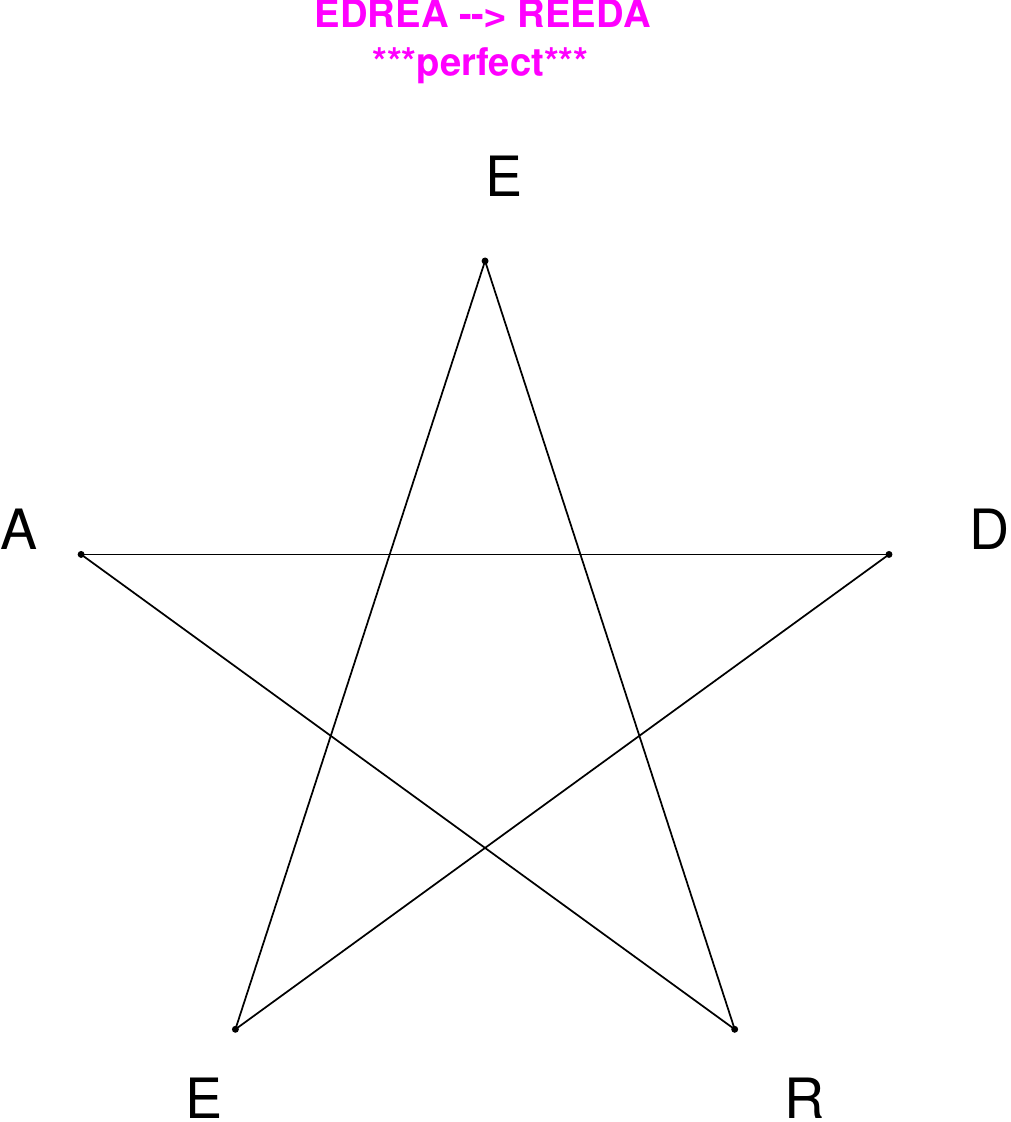}
\end{subfigure}
\end{figure}

\begin{figure}[H]
\centering
\begin{subfigure}[T]{0.19\textwidth}
\centering
\includegraphics[width=\textwidth]{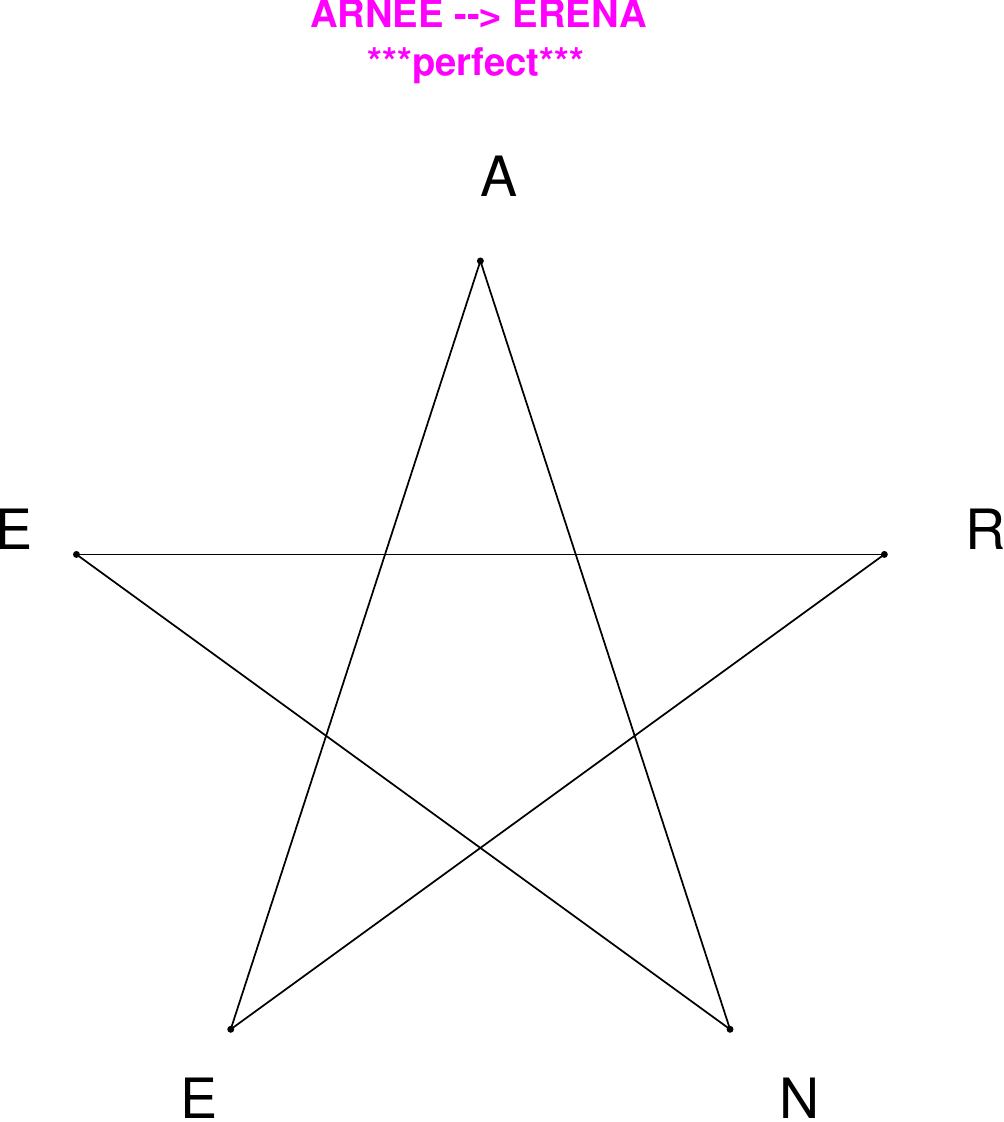}
\end{subfigure}
\hfill
\begin{subfigure}[T]{0.19\textwidth}
\centering
\includegraphics[width=\textwidth]{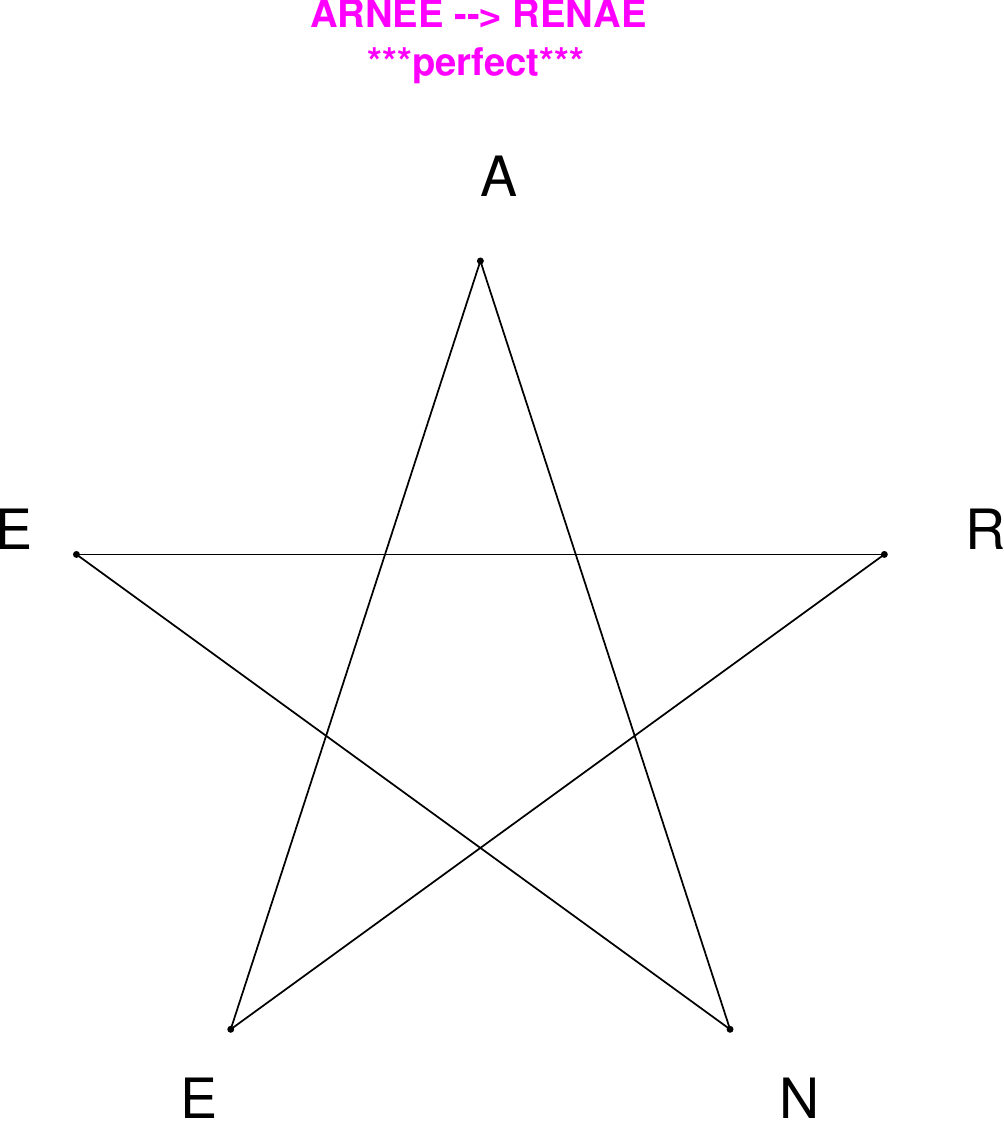}
\end{subfigure}
\hfill
\begin{subfigure}[T]{0.19\textwidth}
\centering
\includegraphics[width=\textwidth]{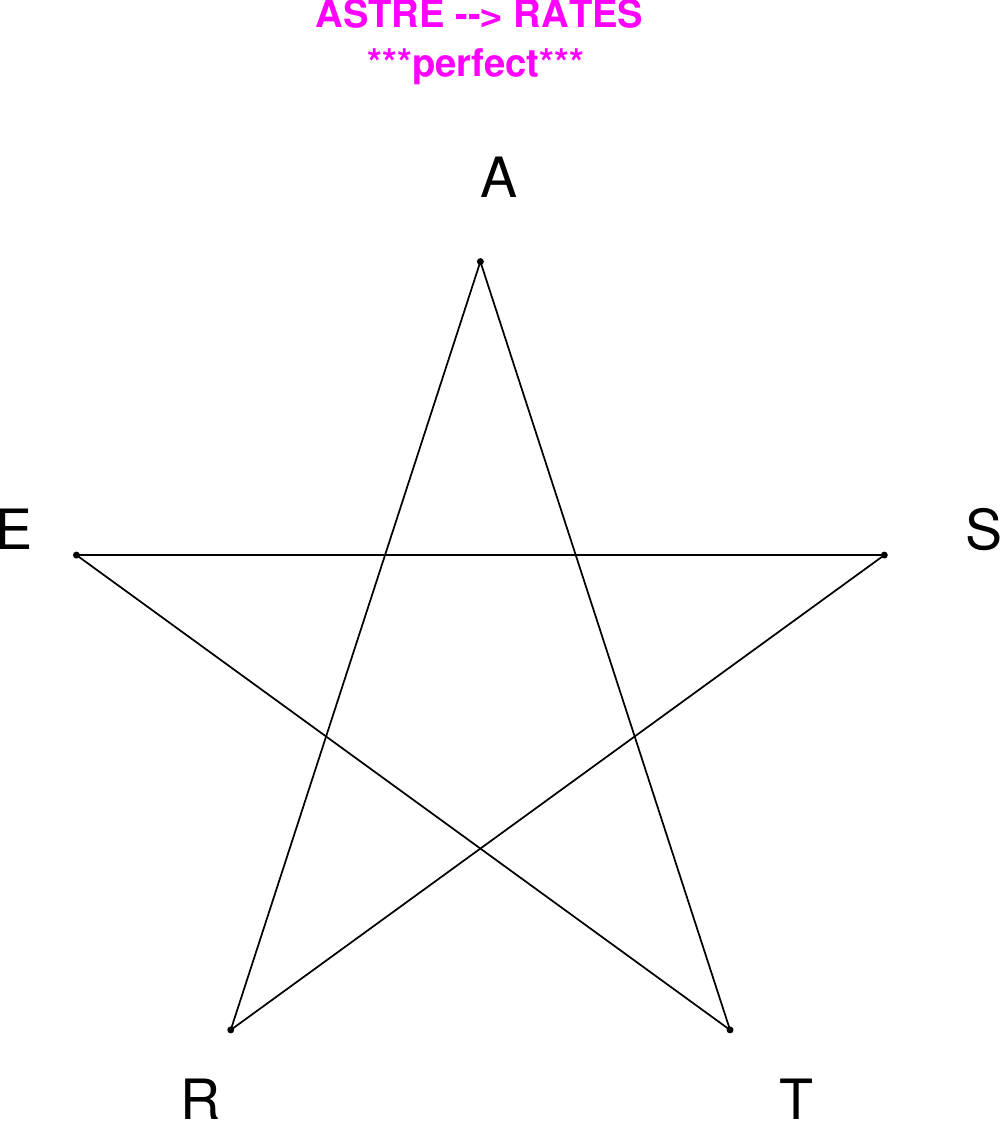}
\end{subfigure}
\hfill
\begin{subfigure}[T]{0.19\textwidth}
\centering
\includegraphics[width=\textwidth]{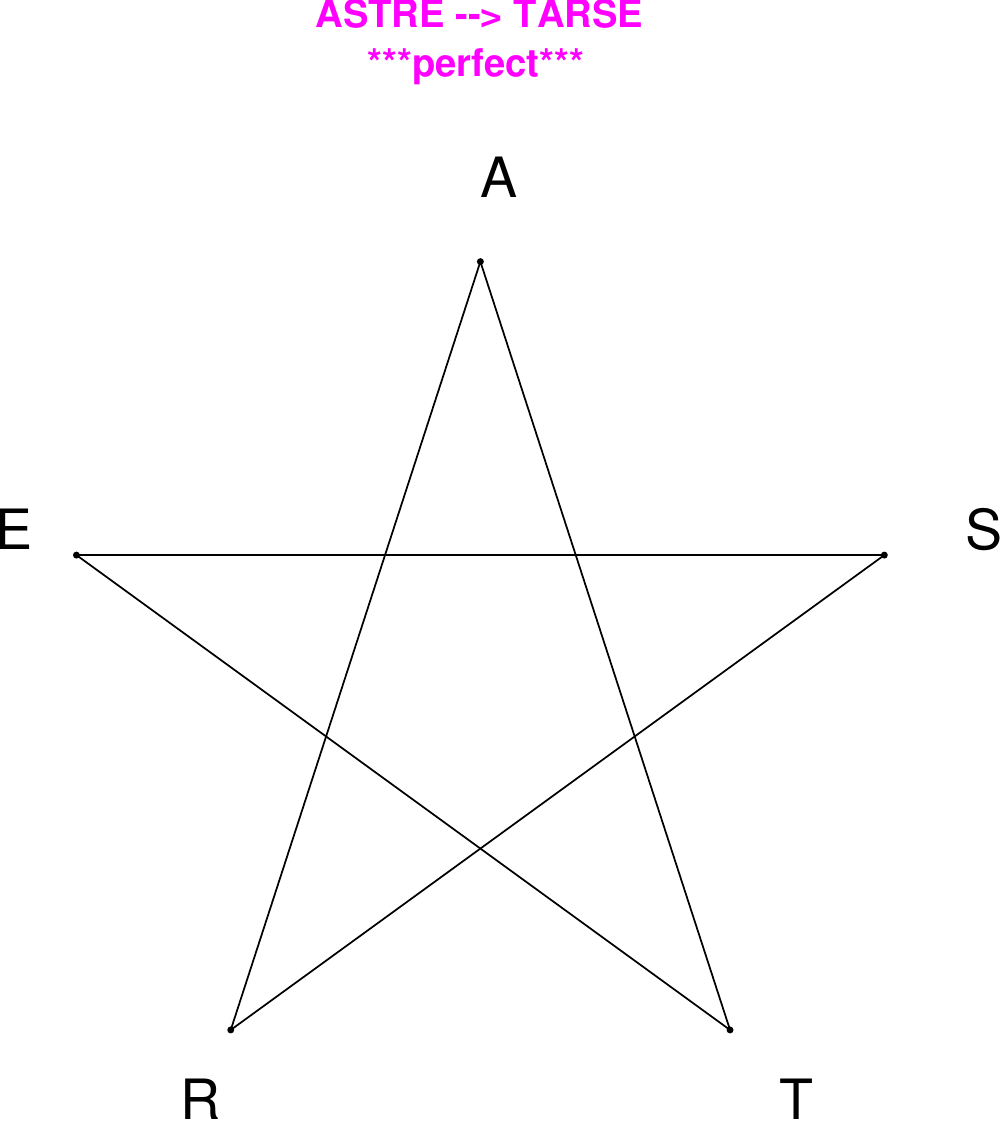}
\end{subfigure}
\hfill
\begin{subfigure}[T]{0.19\textwidth}
\centering
\includegraphics[width=\textwidth]{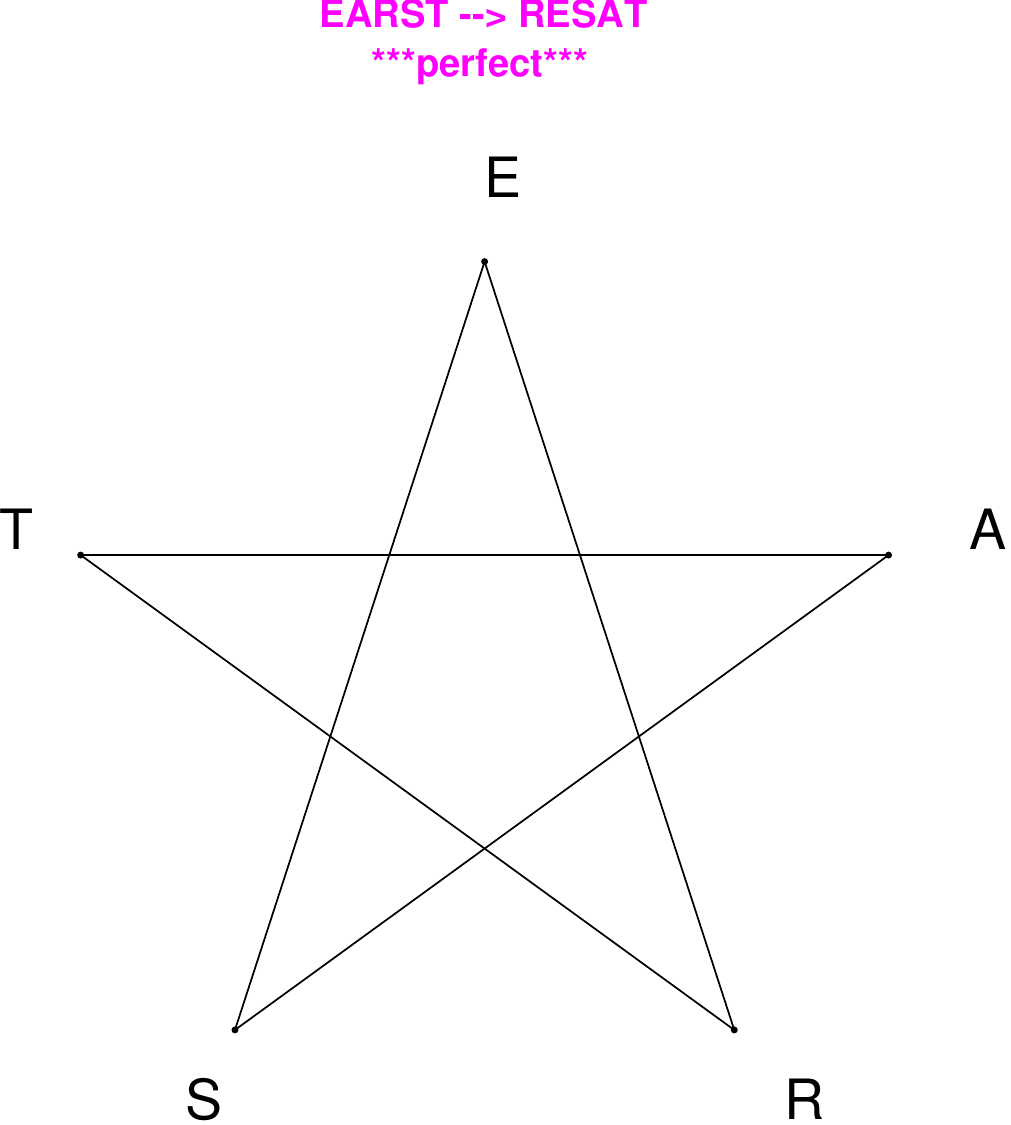}
\end{subfigure}
\end{figure}

\begin{figure}[H]
\centering
\begin{subfigure}[T]{0.19\textwidth}
\centering
\includegraphics[width=\textwidth]{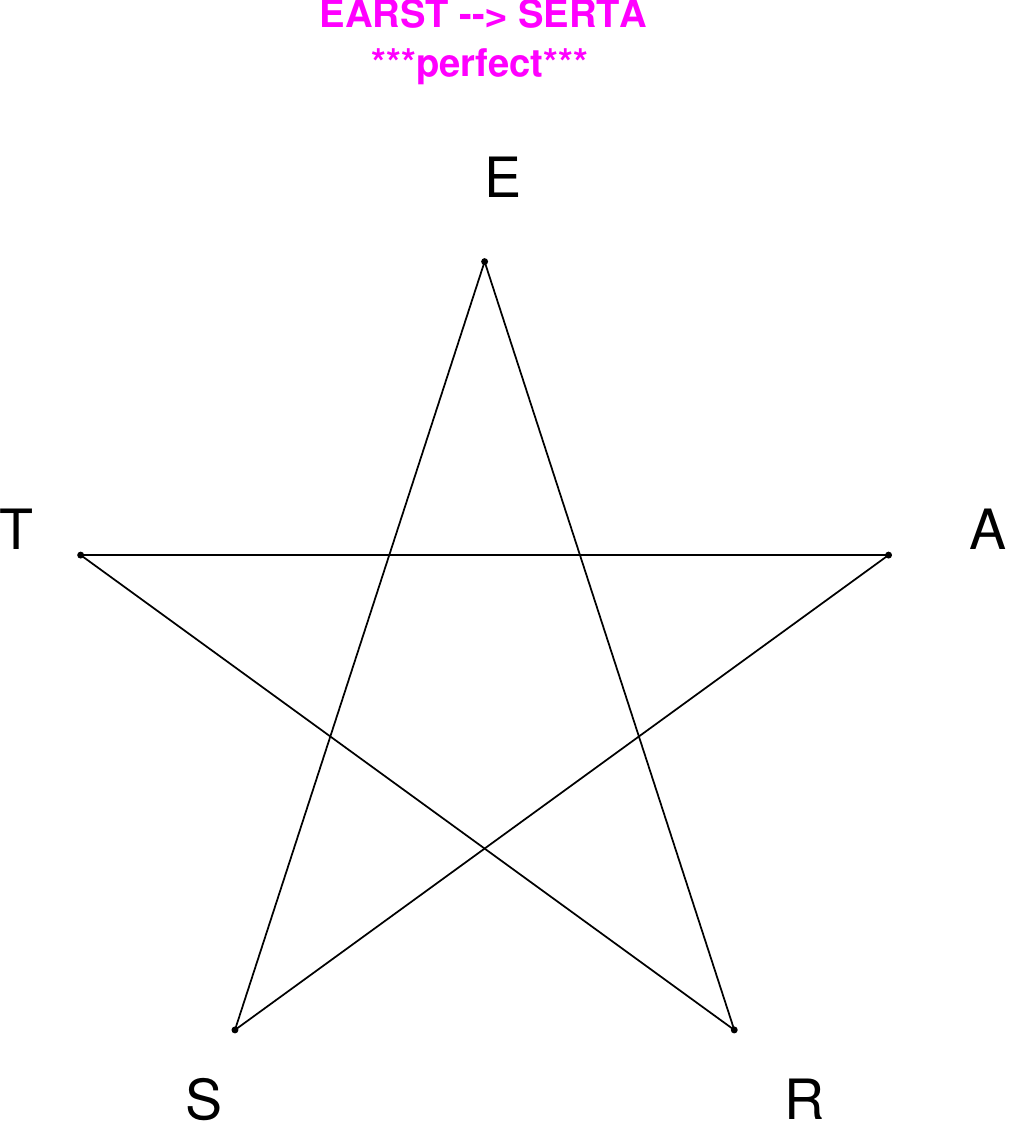}
\end{subfigure}
\hfill
\begin{subfigure}[T]{0.19\textwidth}
\centering
\includegraphics[width=\textwidth]{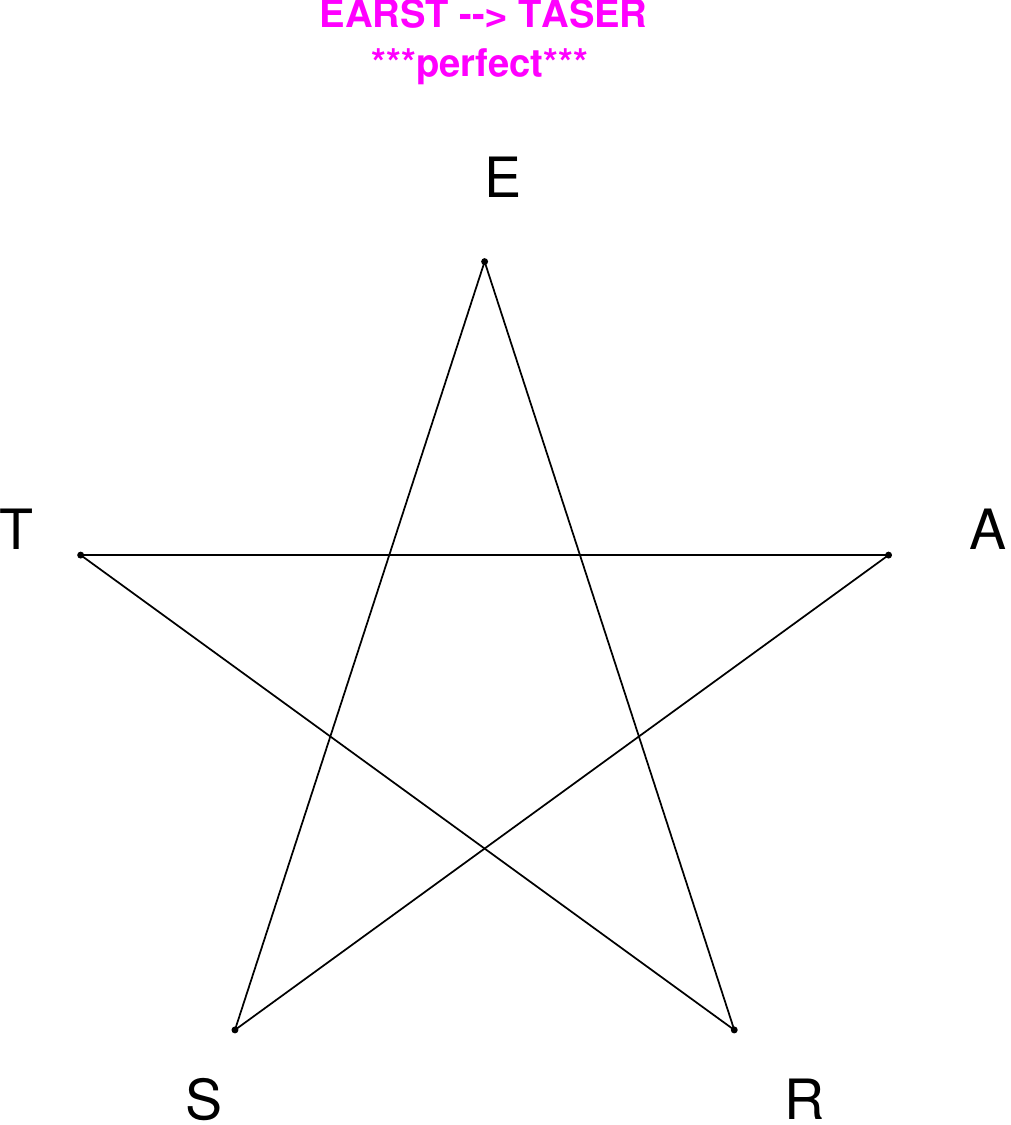}
\end{subfigure}
\hfill
\begin{subfigure}[T]{0.19\textwidth}
\centering
\includegraphics[width=\textwidth]{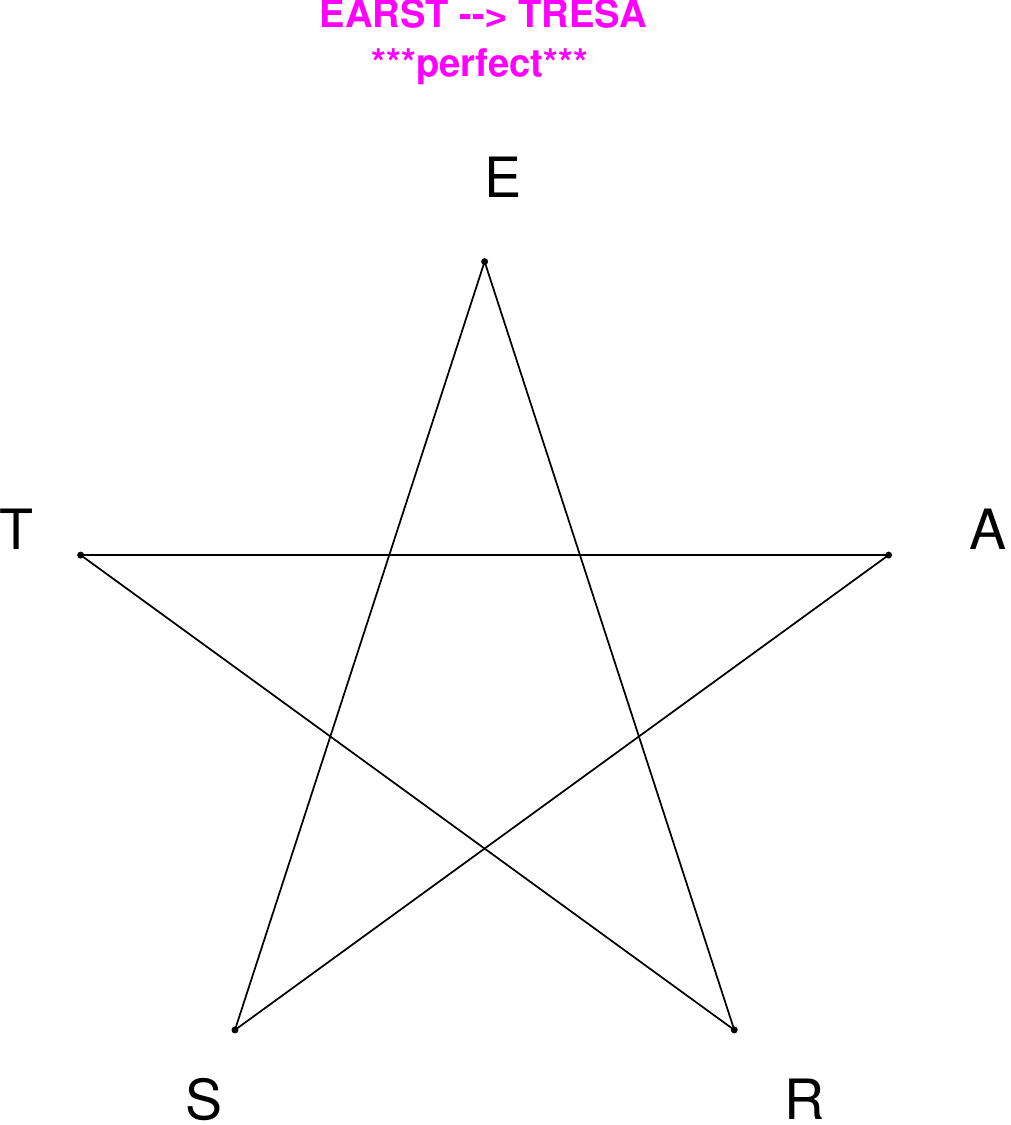}
\end{subfigure}
\hfill
\begin{subfigure}[T]{0.19\textwidth}
\centering
\includegraphics[width=\textwidth]{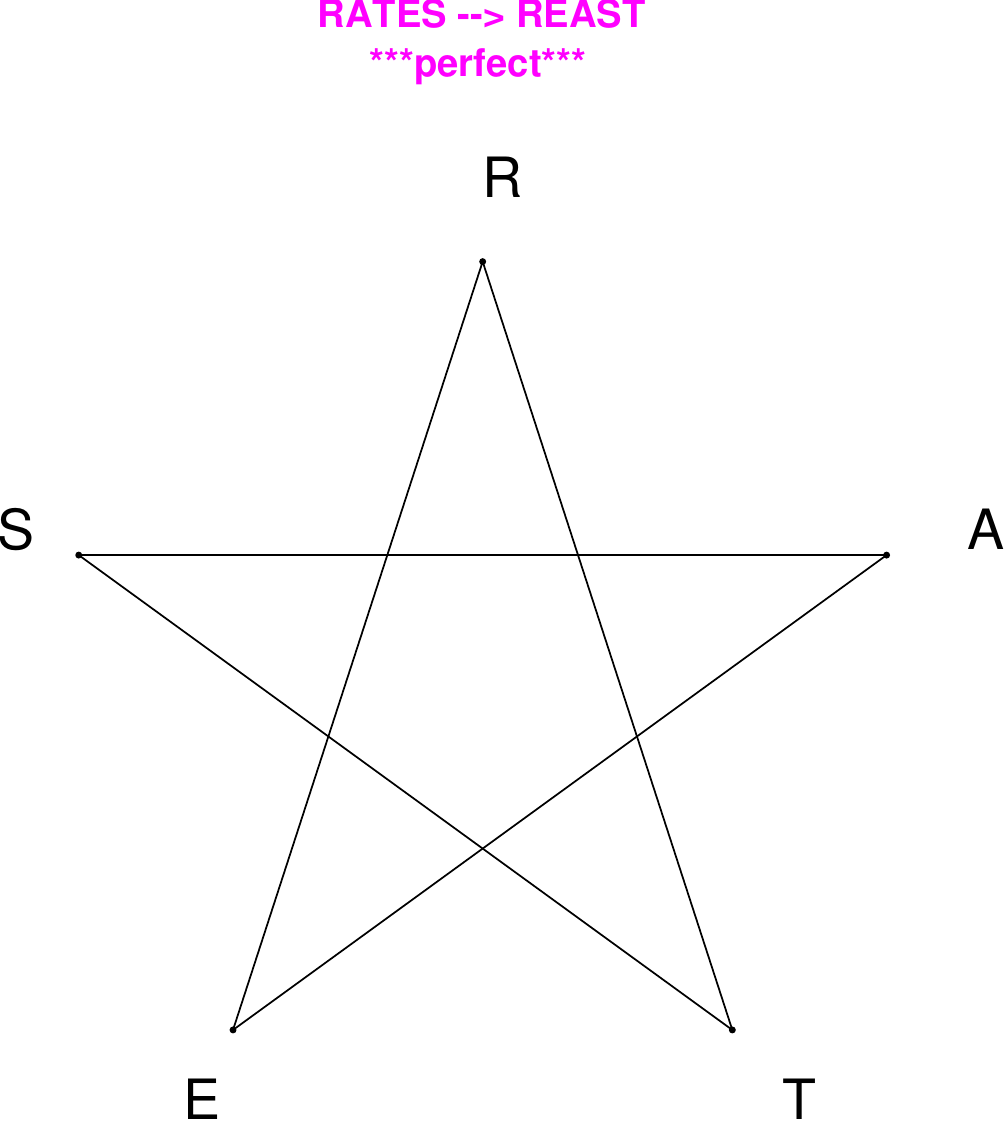}
\end{subfigure}
\hfill
\begin{subfigure}[T]{0.19\textwidth}
\centering
\includegraphics[width=\textwidth]{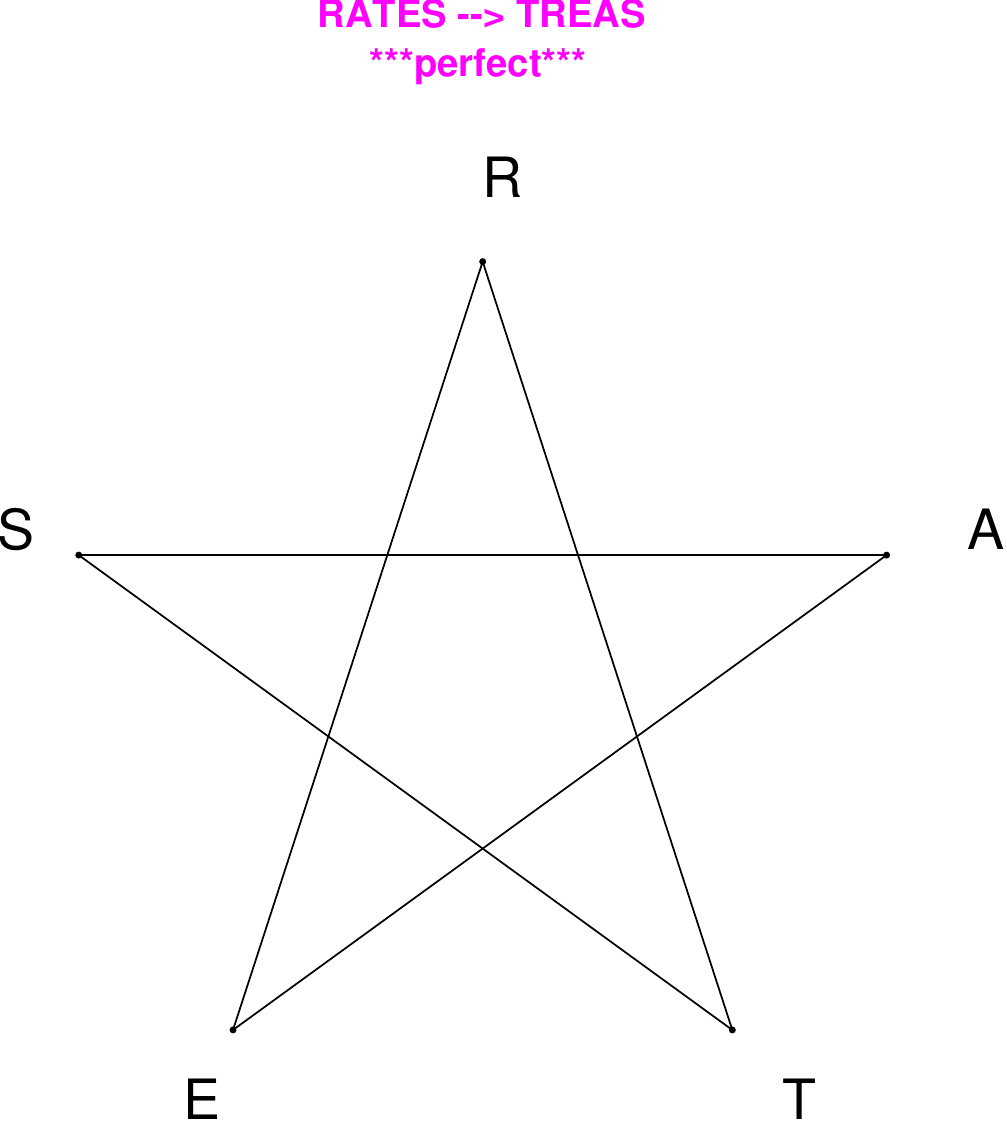}
\end{subfigure}
\end{figure}

\begin{figure}[H]
\centering
\begin{subfigure}[T]{0.19\textwidth}
\centering
\includegraphics[width=\textwidth]{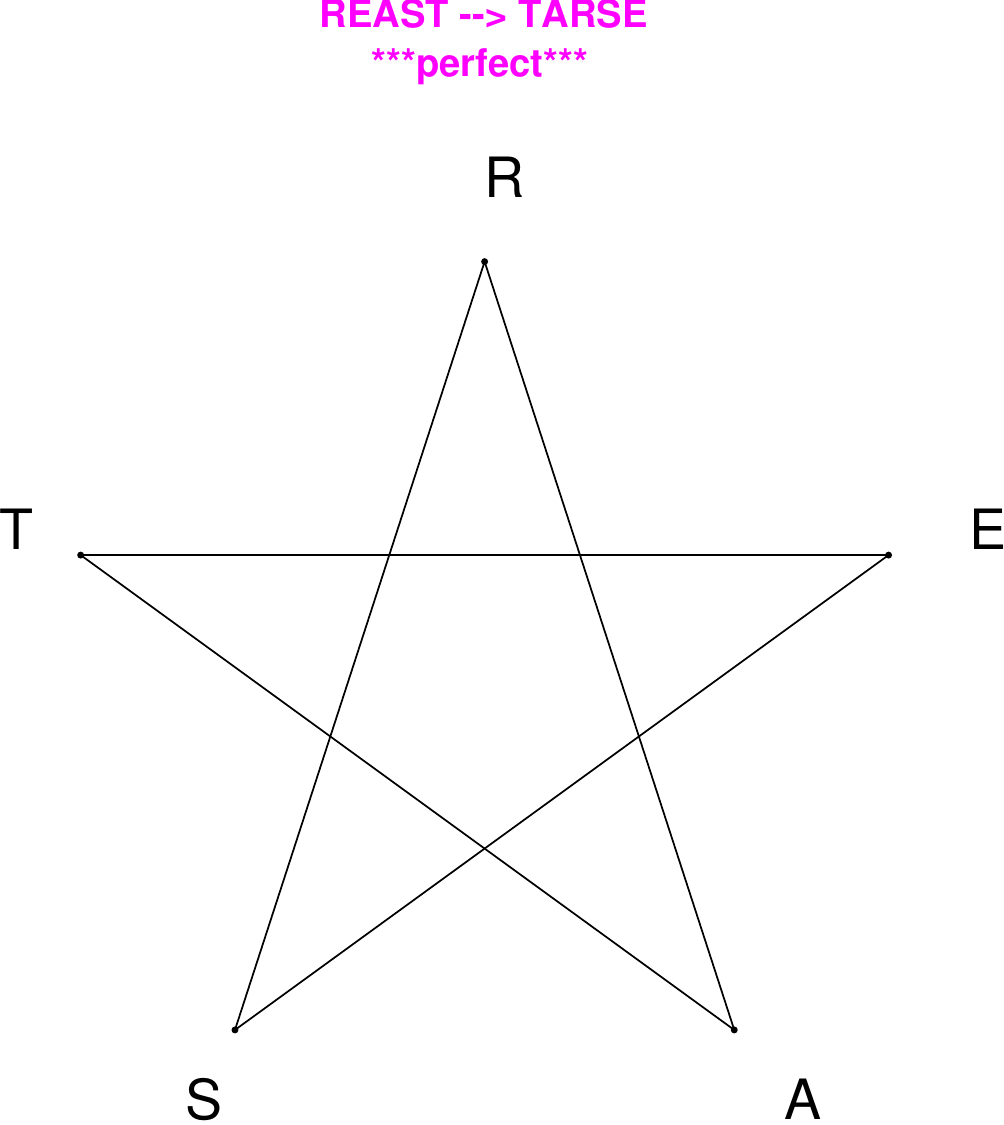}
\end{subfigure}
\hfill
\begin{subfigure}[T]{0.19\textwidth}
\centering
\includegraphics[width=\textwidth]{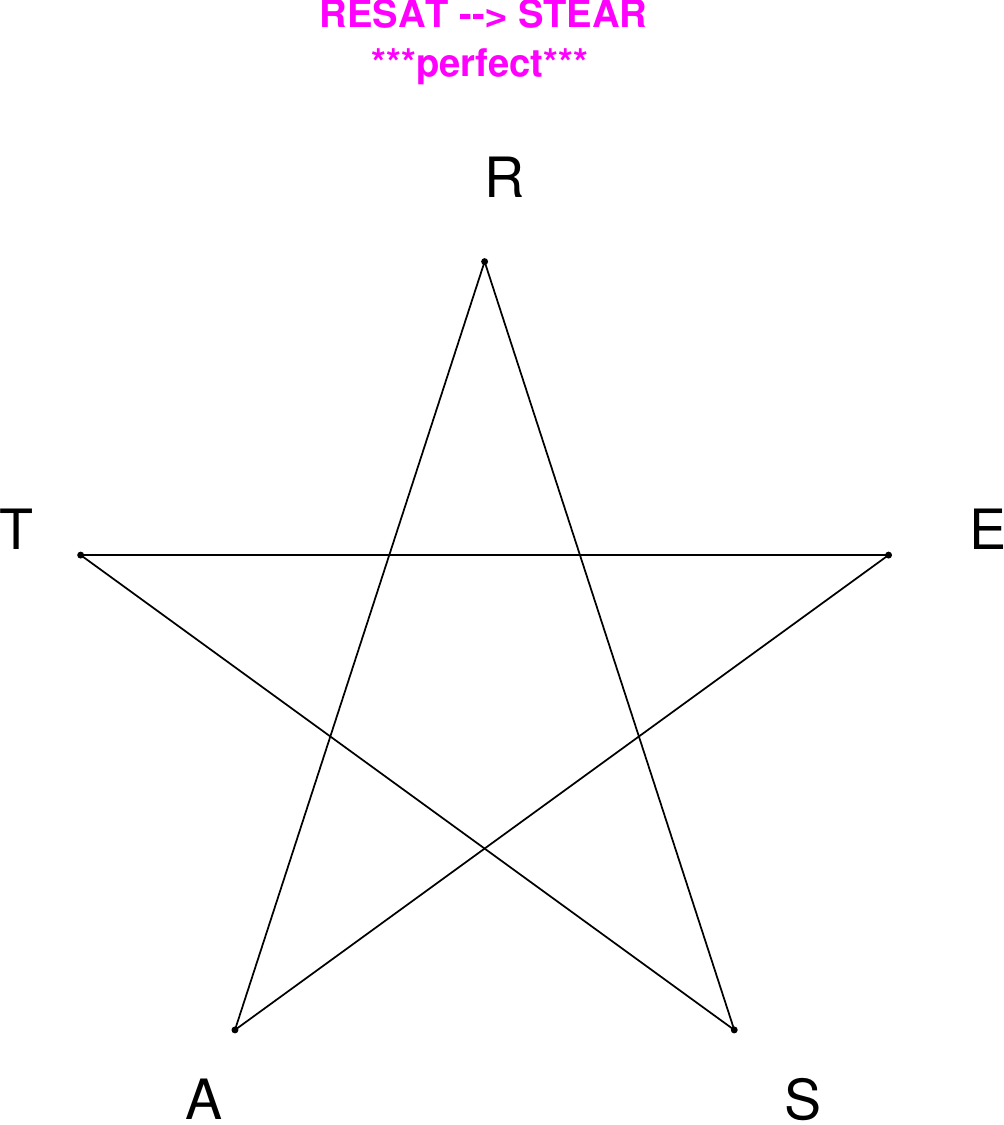}
\end{subfigure}
\hfill
\begin{subfigure}[T]{0.19\textwidth}
\centering
\includegraphics[width=\textwidth]{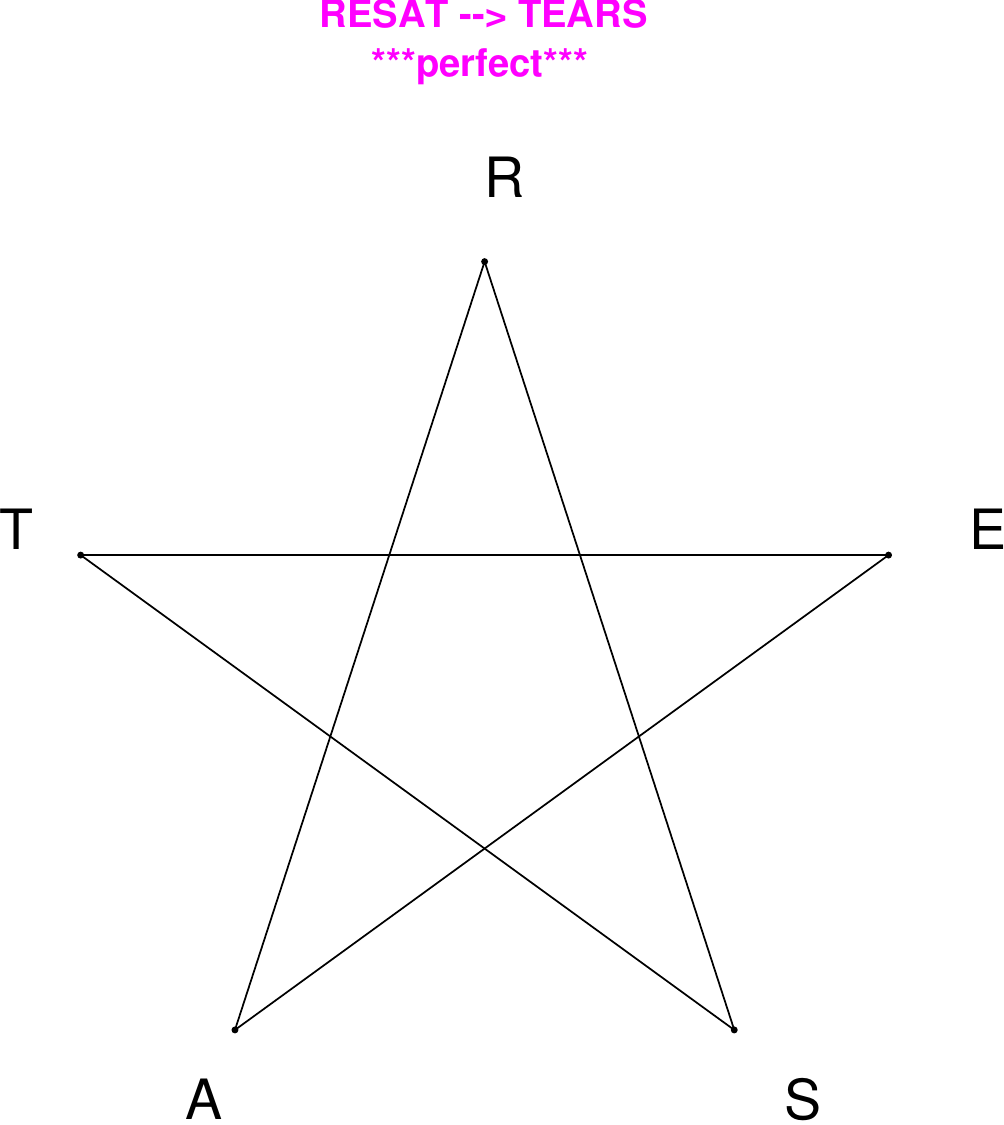}
\end{subfigure}
\hfill
\begin{subfigure}[T]{0.19\textwidth}
\centering
\includegraphics[width=\textwidth]{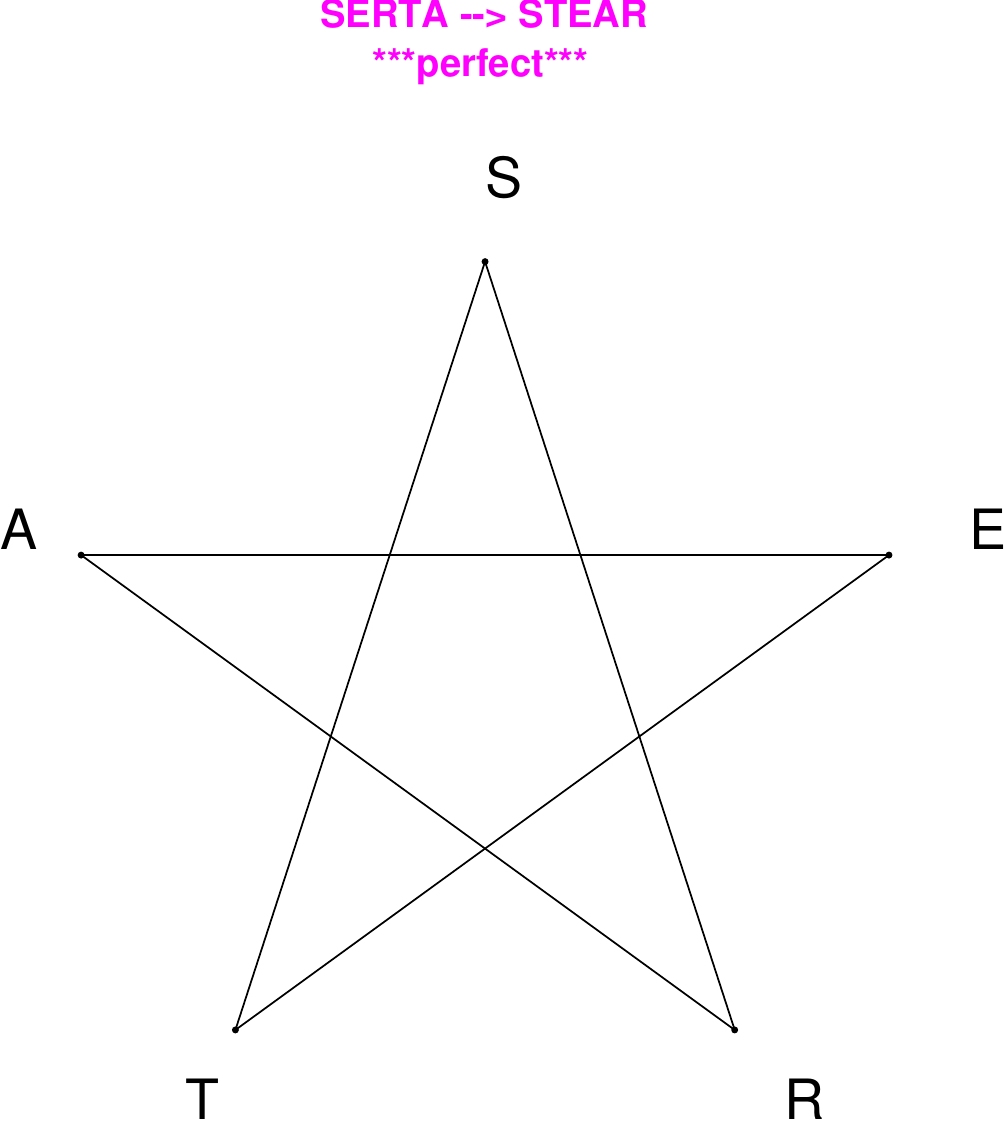}
\end{subfigure}
\hfill
\begin{subfigure}[T]{0.19\textwidth}
\centering
\includegraphics[width=\textwidth]{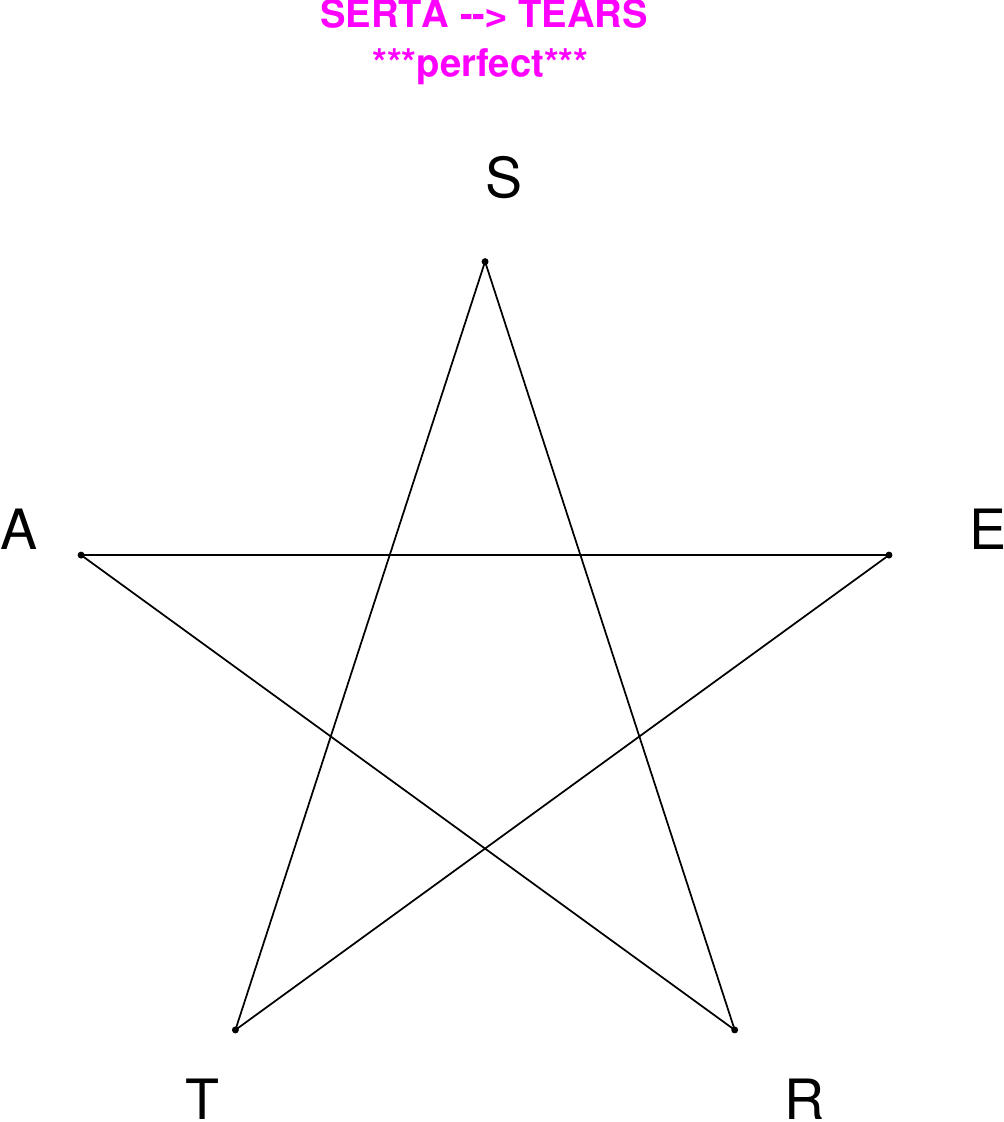}
\end{subfigure}
\end{figure}

\begin{figure}[H]
\centering
\begin{subfigure}[T]{0.19\textwidth}
\centering
\includegraphics[width=\textwidth]{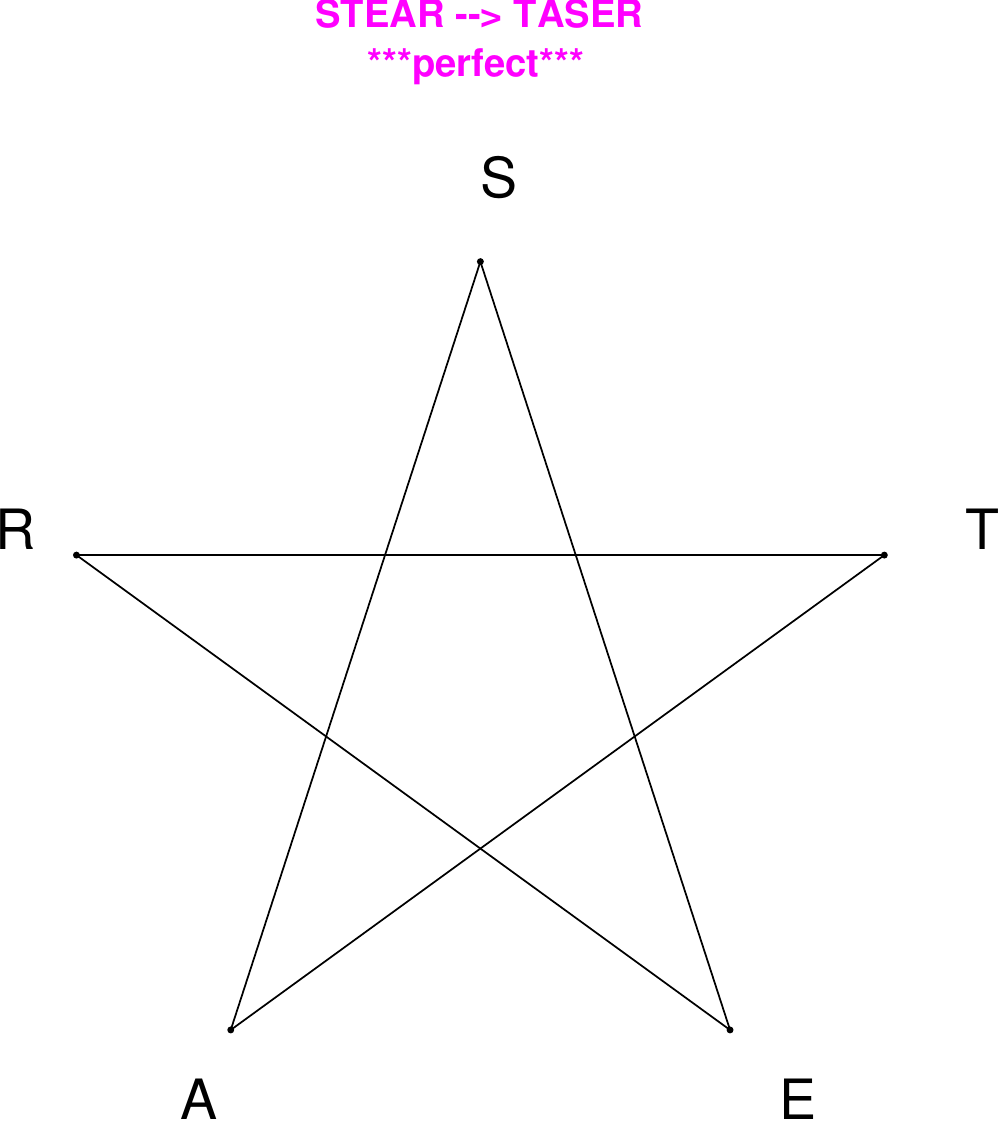}
\end{subfigure}
\hfill
\begin{subfigure}[T]{0.19\textwidth}
\centering
\includegraphics[width=\textwidth]{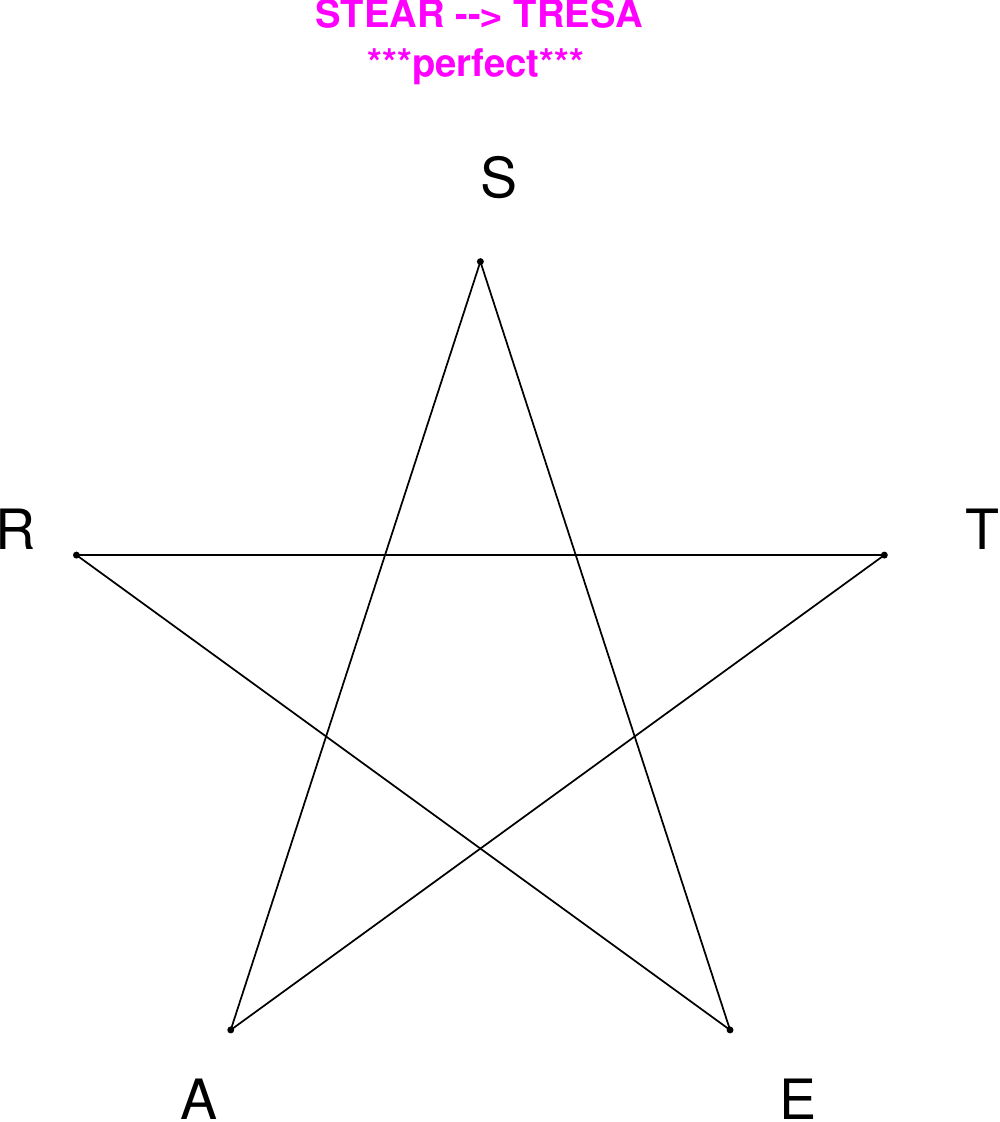}
\end{subfigure}
\hfill
\begin{subfigure}[T]{0.19\textwidth}
\centering
\includegraphics[width=\textwidth]{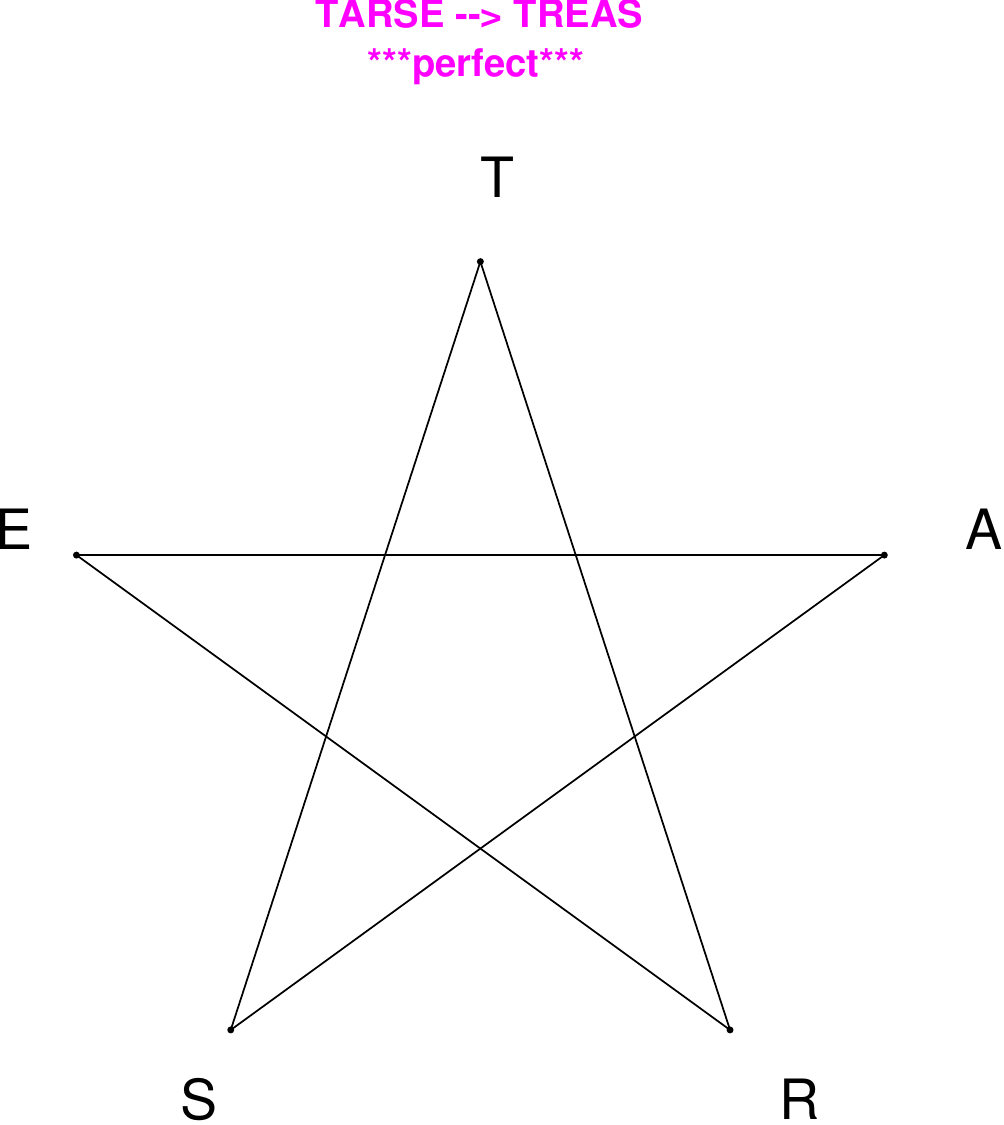}
\end{subfigure}
\hfill
\begin{subfigure}[T]{0.19\textwidth}
\centering
\includegraphics[width=\textwidth]{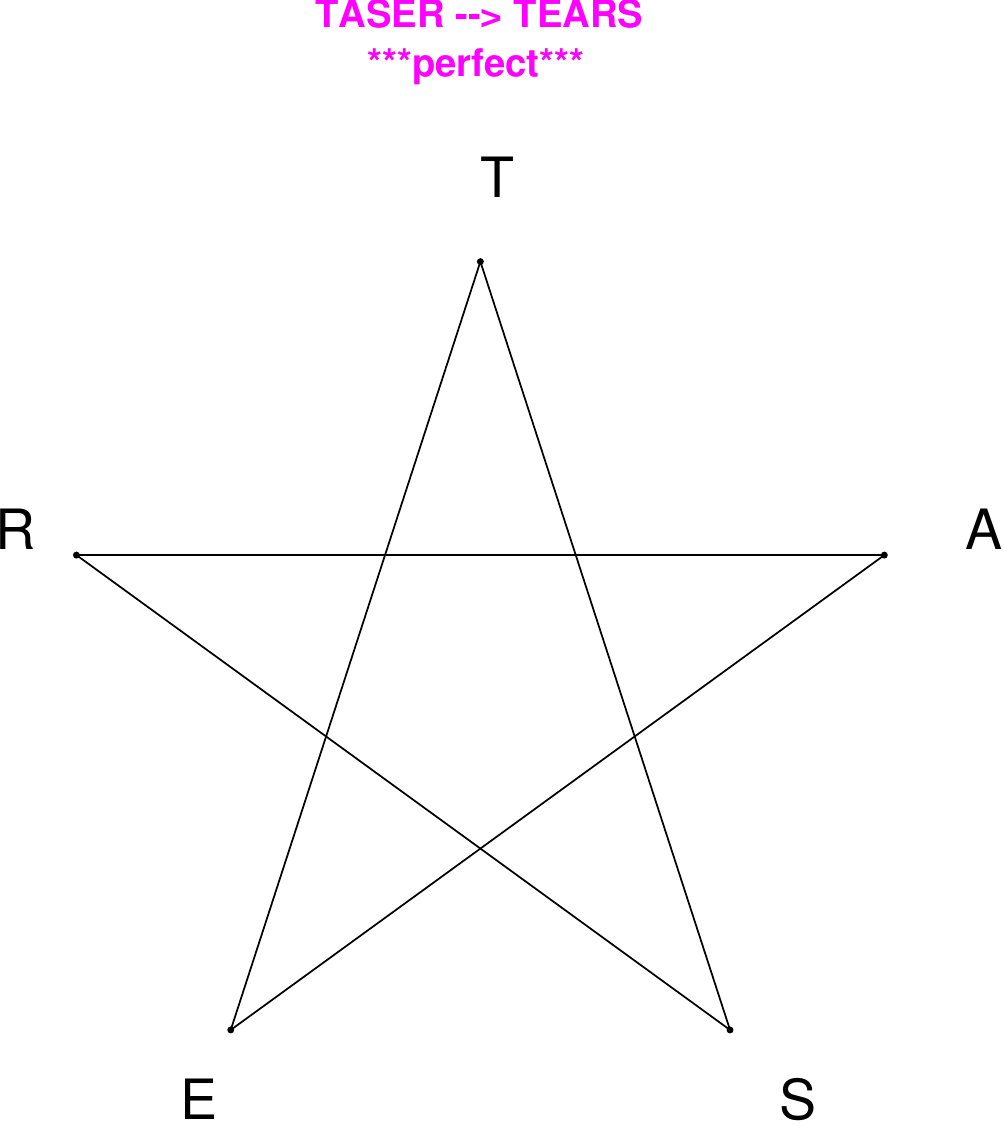}
\end{subfigure}
\hfill
\begin{subfigure}[T]{0.19\textwidth}
\centering
\includegraphics[width=\textwidth]{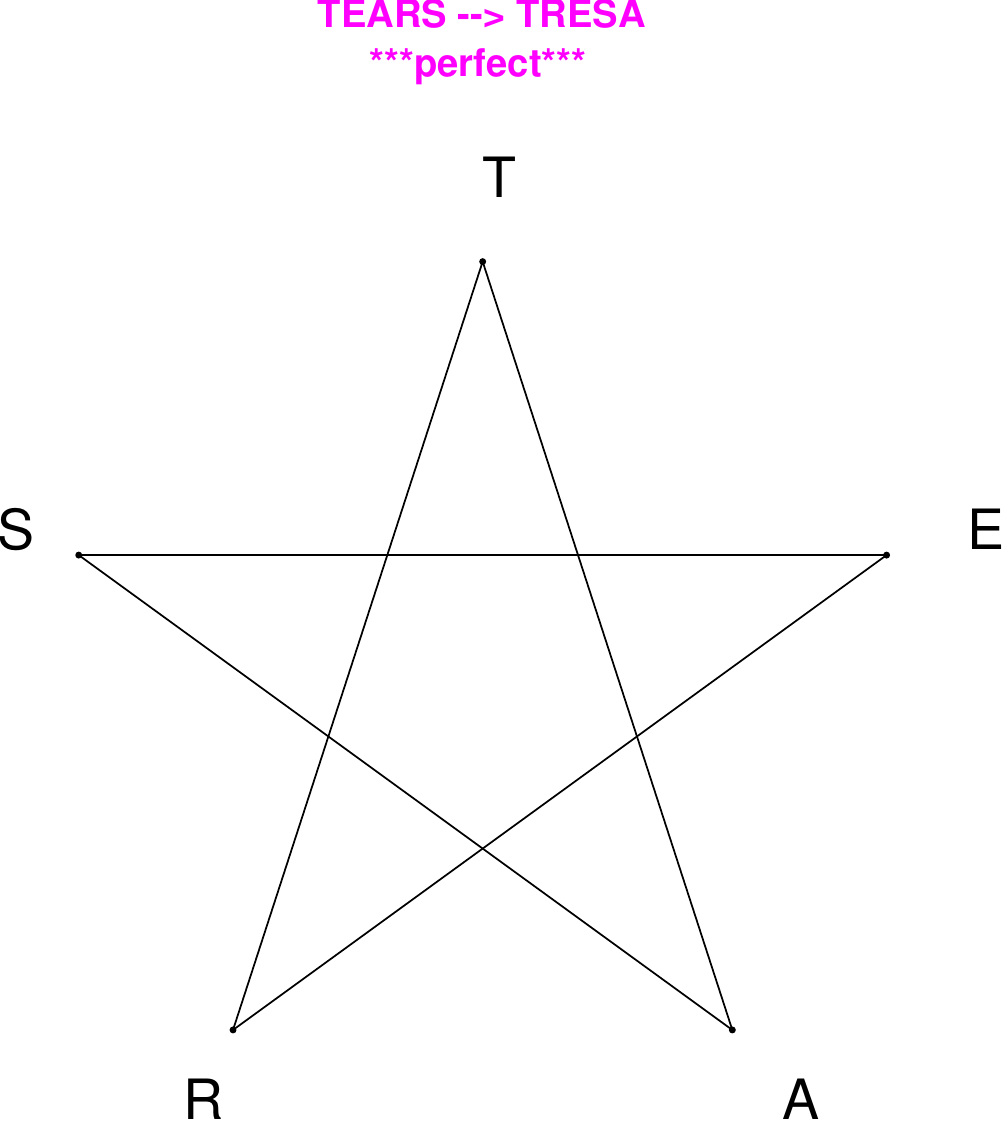}
\end{subfigure}
\end{figure}

\begin{figure}[H]
\centering
\begin{subfigure}[T]{0.19\textwidth}
\centering
\includegraphics[width=\textwidth]{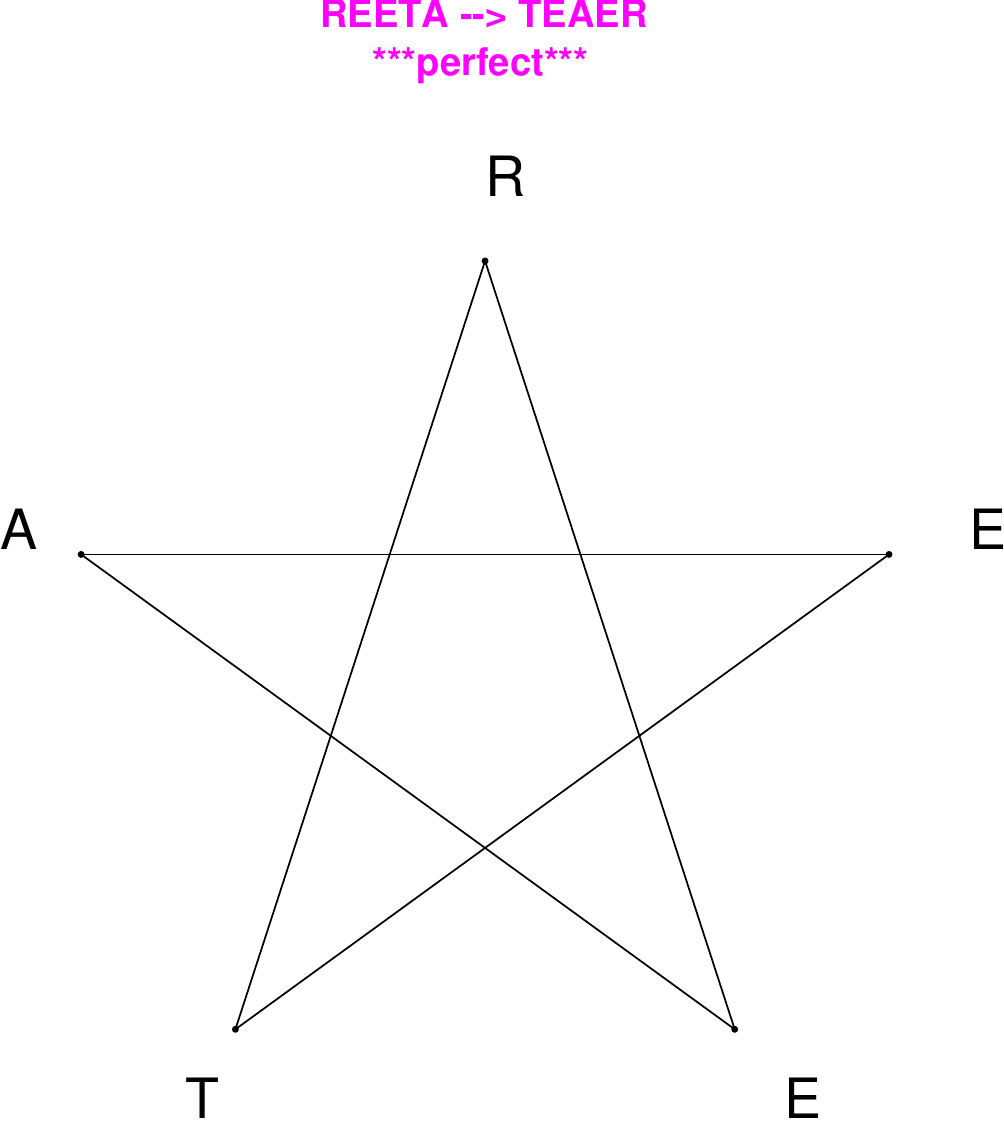}
\end{subfigure}
\hfill
\begin{subfigure}[T]{0.19\textwidth}
\centering
\includegraphics[width=\textwidth]{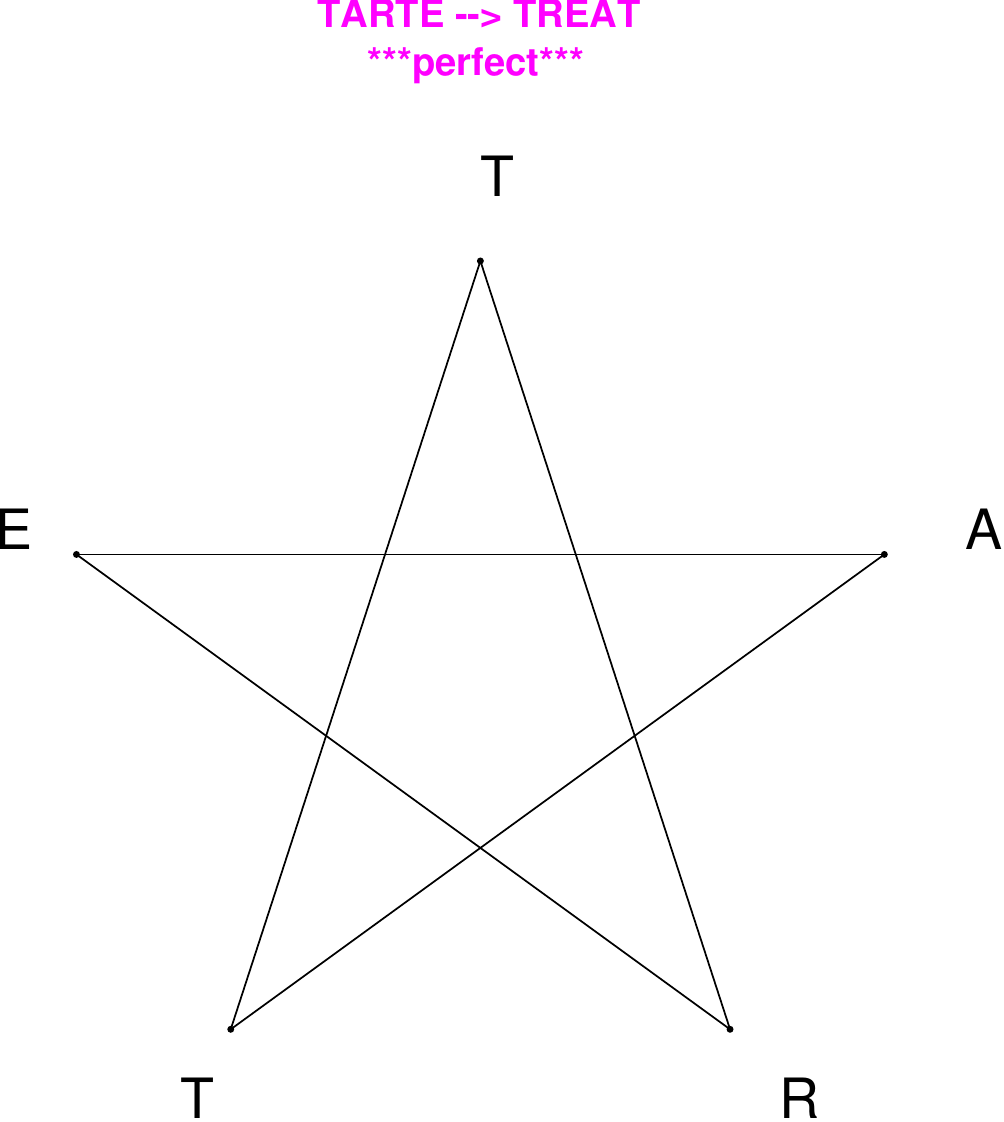}
\end{subfigure}
\hfill
\begin{subfigure}[T]{0.19\textwidth}
\centering
\includegraphics[width=\textwidth]{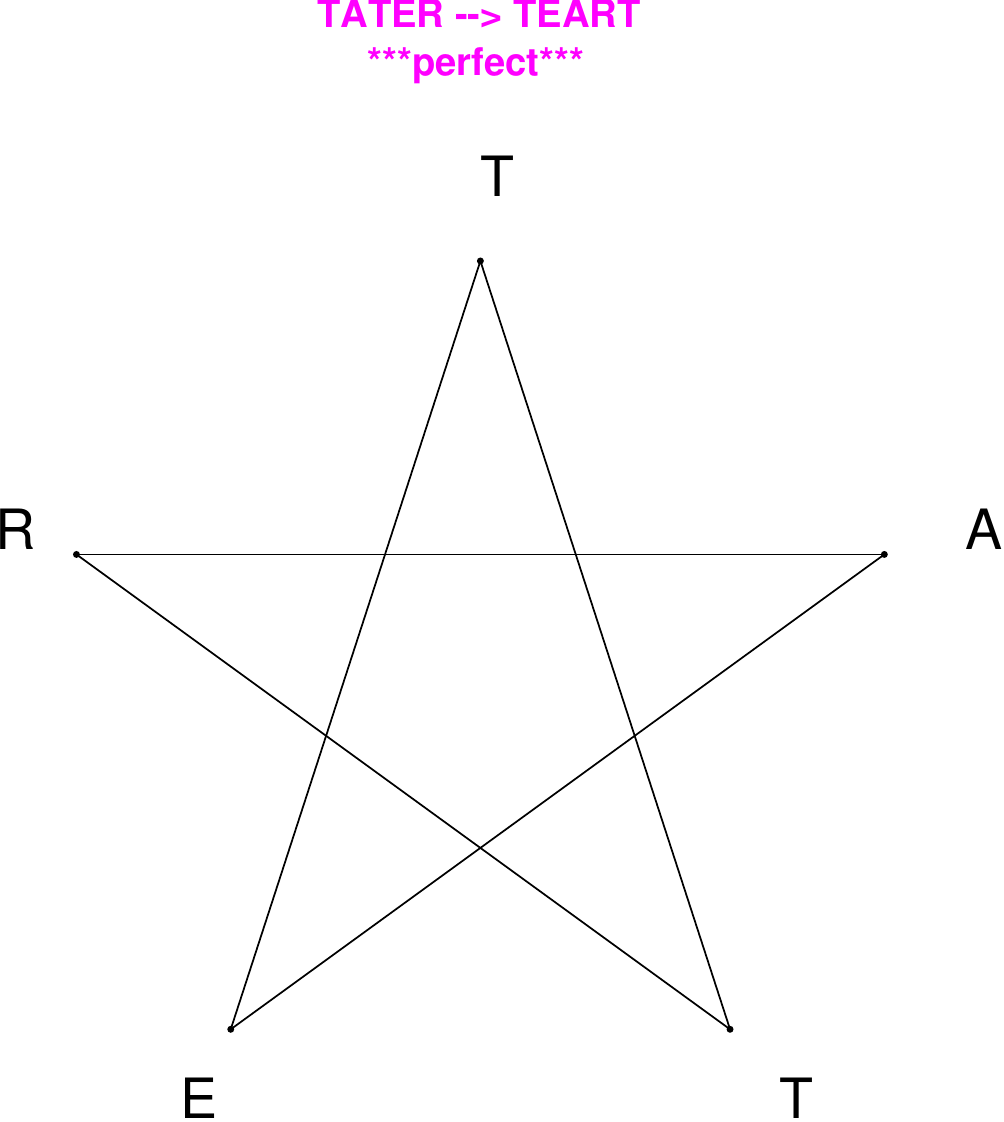}
\end{subfigure}
\hfill
\begin{subfigure}[T]{0.19\textwidth}
\centering
\includegraphics[width=\textwidth]{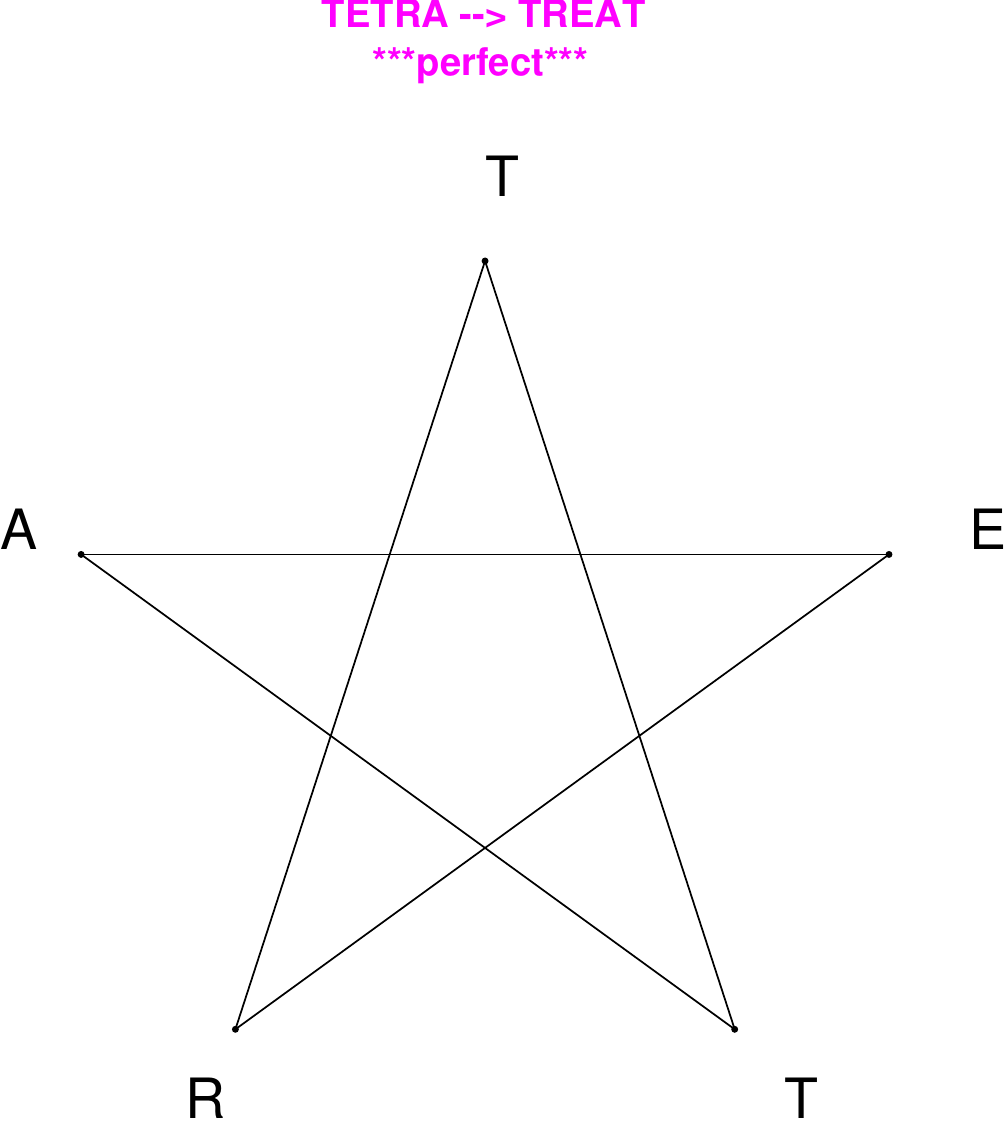}
\end{subfigure}
\hfill
\begin{subfigure}[T]{0.19\textwidth}
\centering
\includegraphics[width=\textwidth]{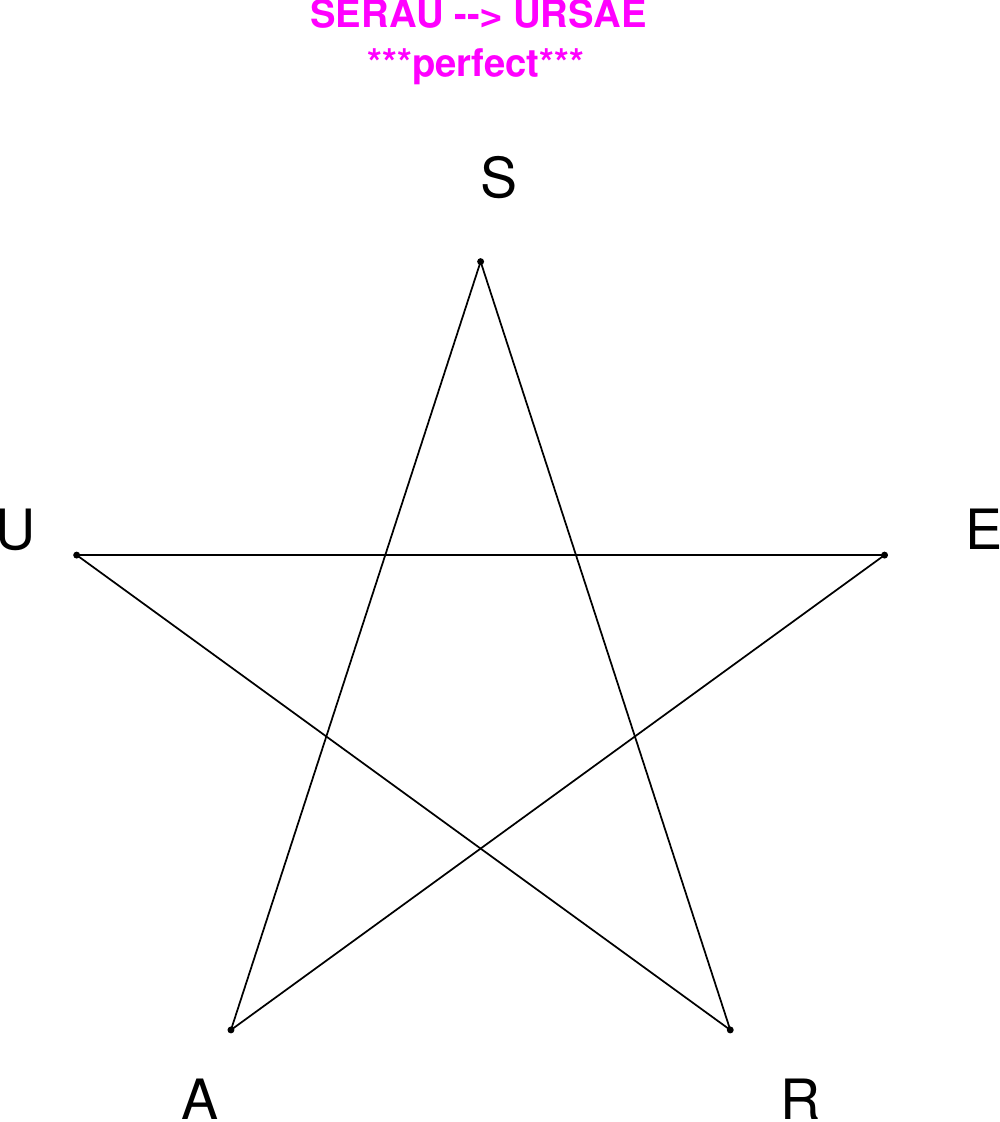}
\end{subfigure}
\end{figure}

\begin{figure}[H]
\centering
\begin{subfigure}[T]{0.19\textwidth}
\centering
\includegraphics[width=\textwidth]{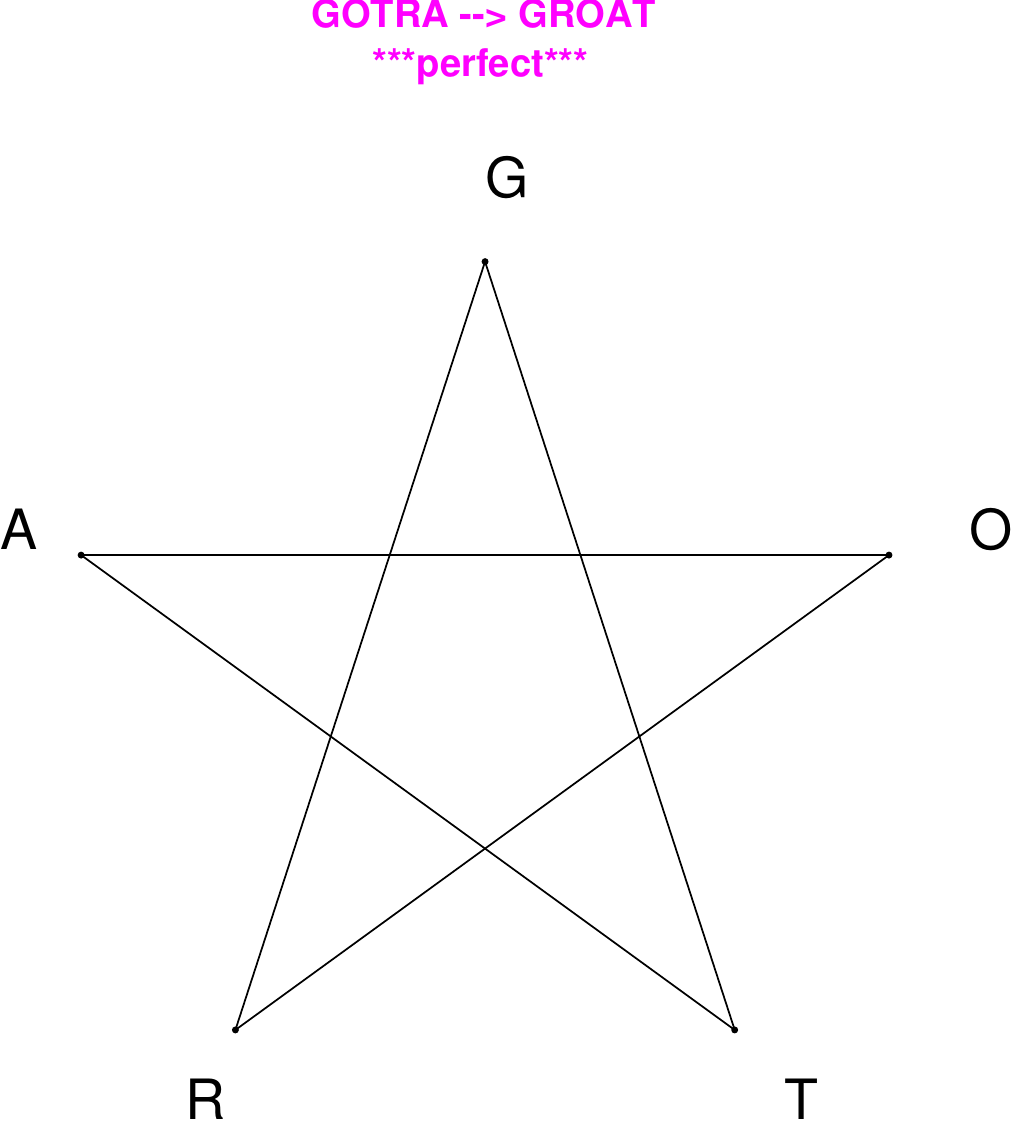}
\end{subfigure}
\hfill
\begin{subfigure}[T]{0.19\textwidth}
\centering
\includegraphics[width=\textwidth]{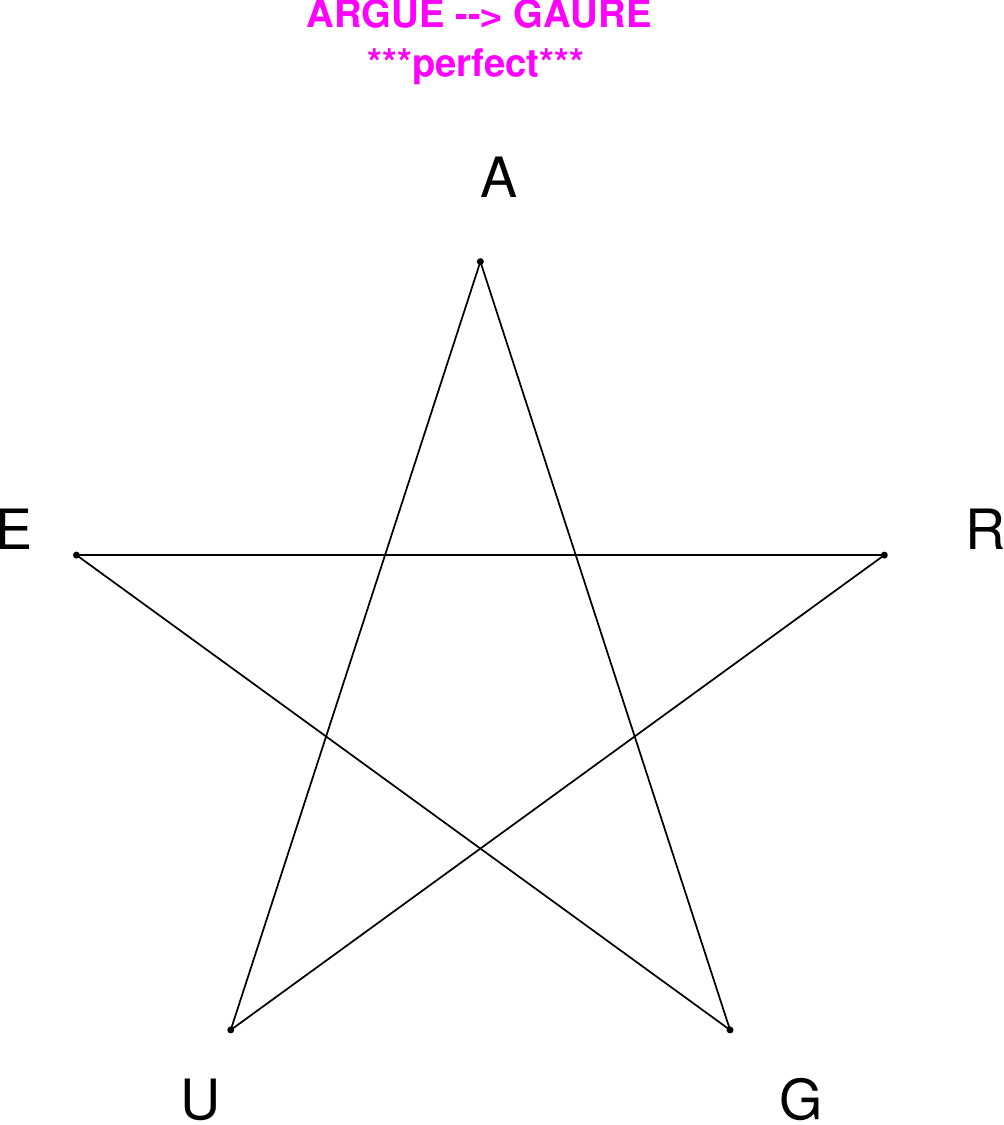}
\end{subfigure}
\hfill
\begin{subfigure}[T]{0.19\textwidth}
\centering
\includegraphics[width=\textwidth]{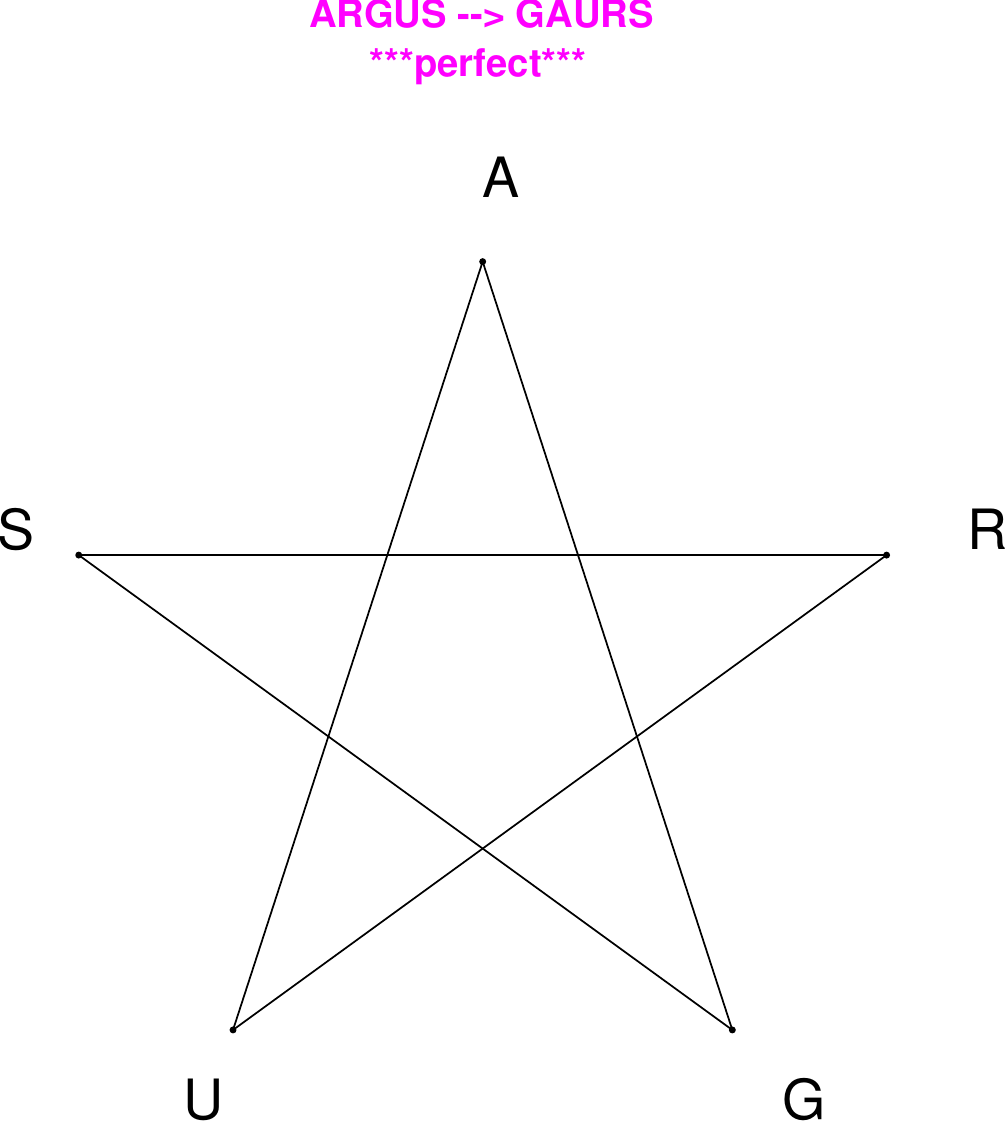}
\end{subfigure}
\hfill
\begin{subfigure}[T]{0.19\textwidth}
\centering
\includegraphics[width=\textwidth]{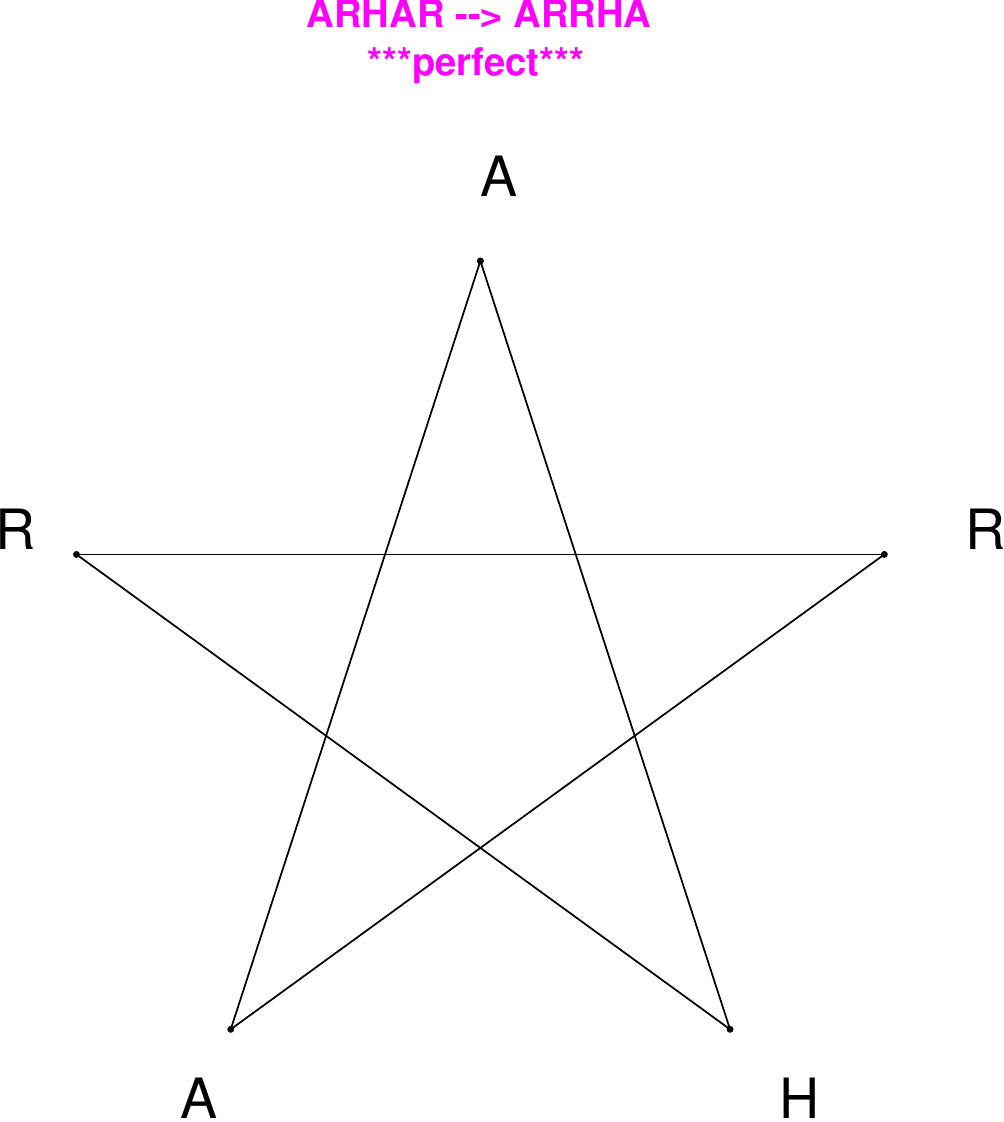}
\end{subfigure}
\hfill
\begin{subfigure}[T]{0.19\textwidth}
\centering
\includegraphics[width=\textwidth]{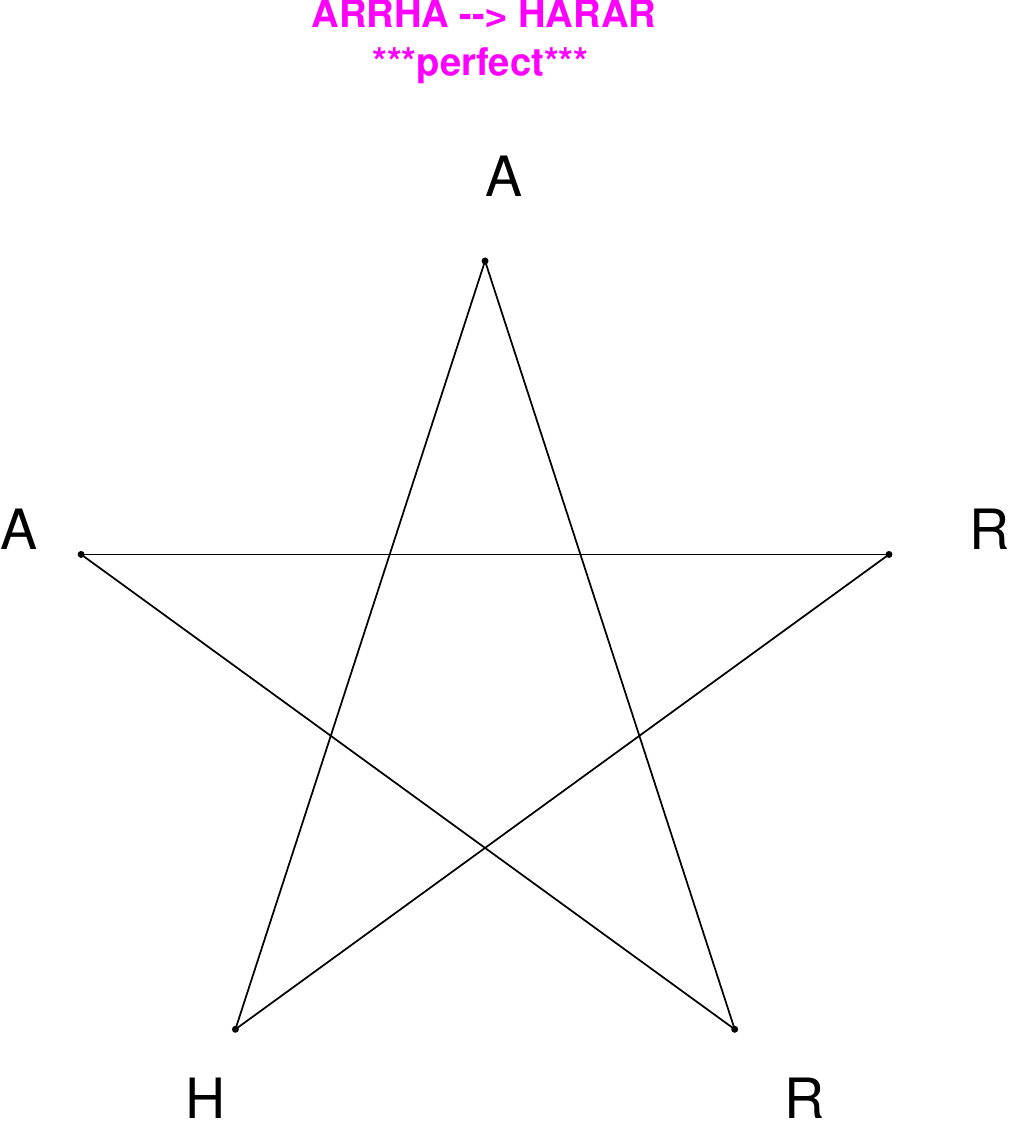}
\end{subfigure}
\end{figure}

\begin{figure}[H]
\centering
\begin{subfigure}[T]{0.19\textwidth}
\centering
\includegraphics[width=\textwidth]{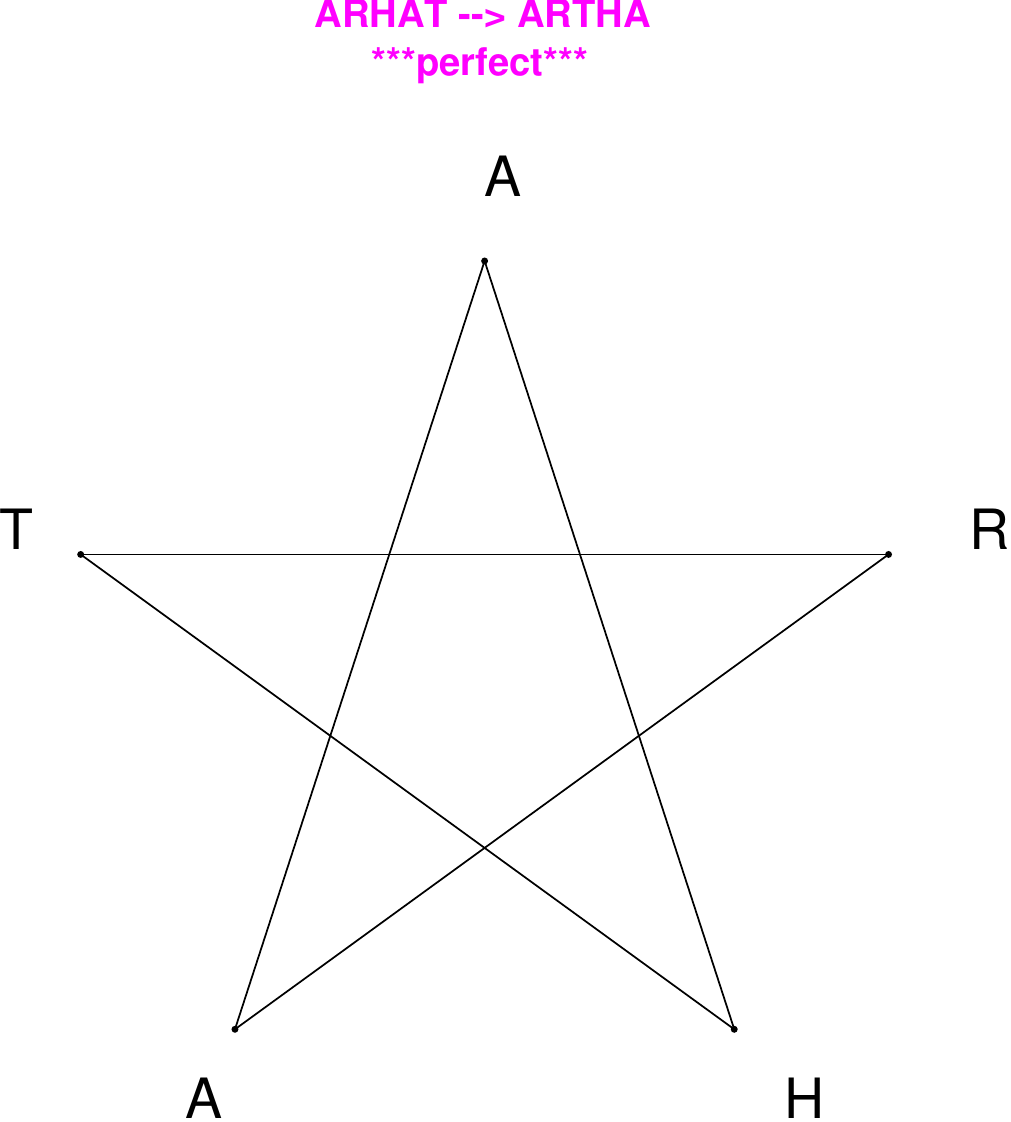}
\end{subfigure}
\hfill
\begin{subfigure}[T]{0.19\textwidth}
\centering
\includegraphics[width=\textwidth]{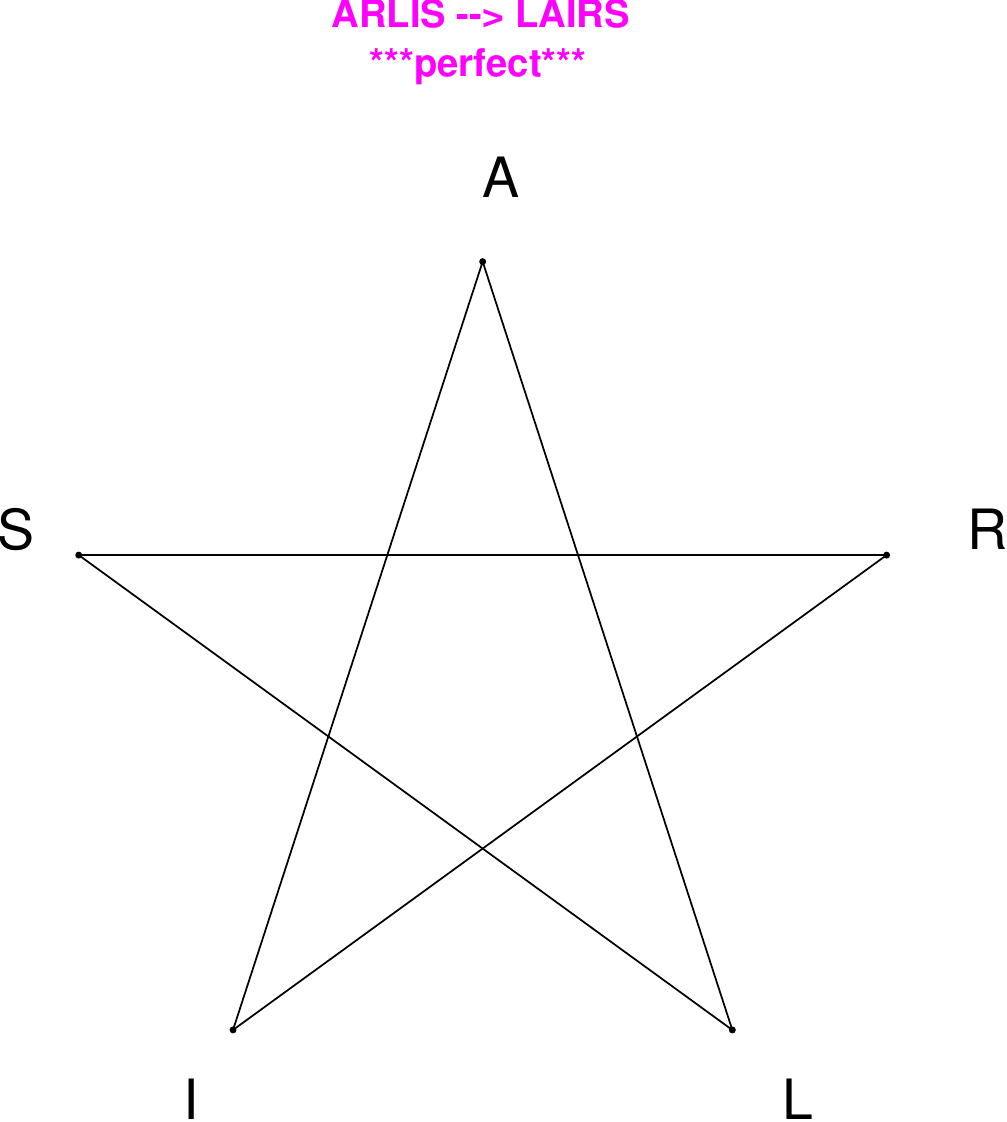}
\end{subfigure}
\hfill
\begin{subfigure}[T]{0.19\textwidth}
\centering
\includegraphics[width=\textwidth]{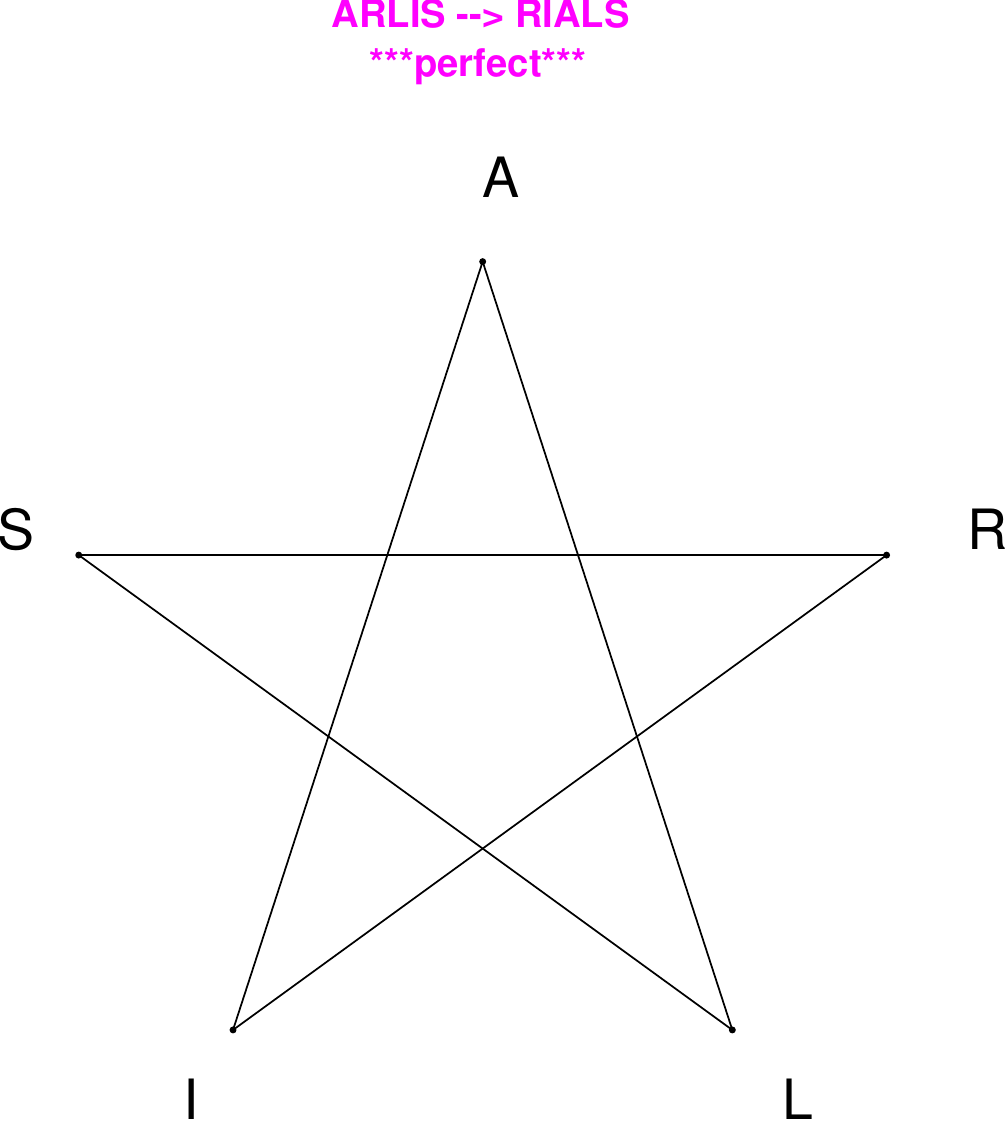}
\end{subfigure}
\hfill
\begin{subfigure}[T]{0.19\textwidth}
\centering
\includegraphics[width=\textwidth]{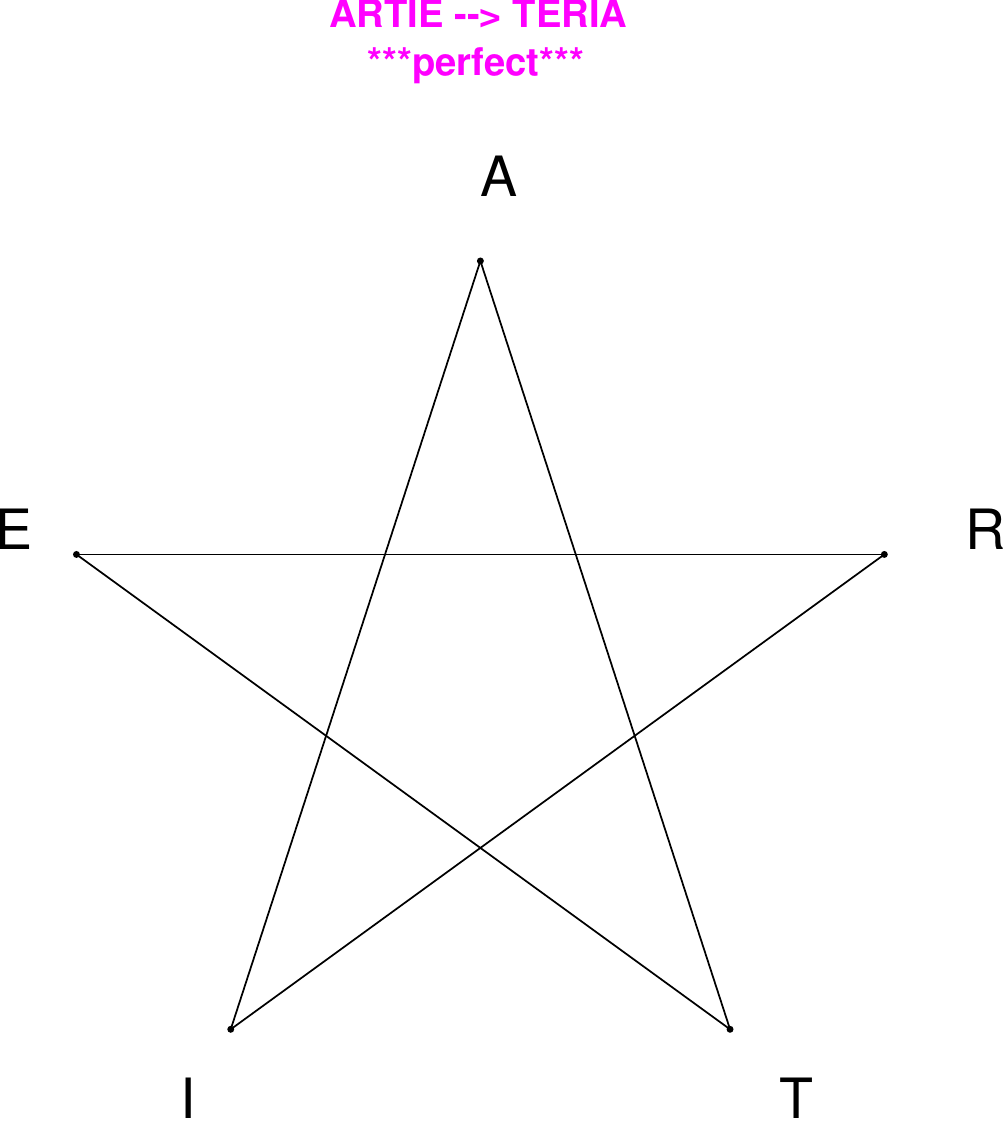}
\end{subfigure}
\hfill
\begin{subfigure}[T]{0.19\textwidth}
\centering
\includegraphics[width=\textwidth]{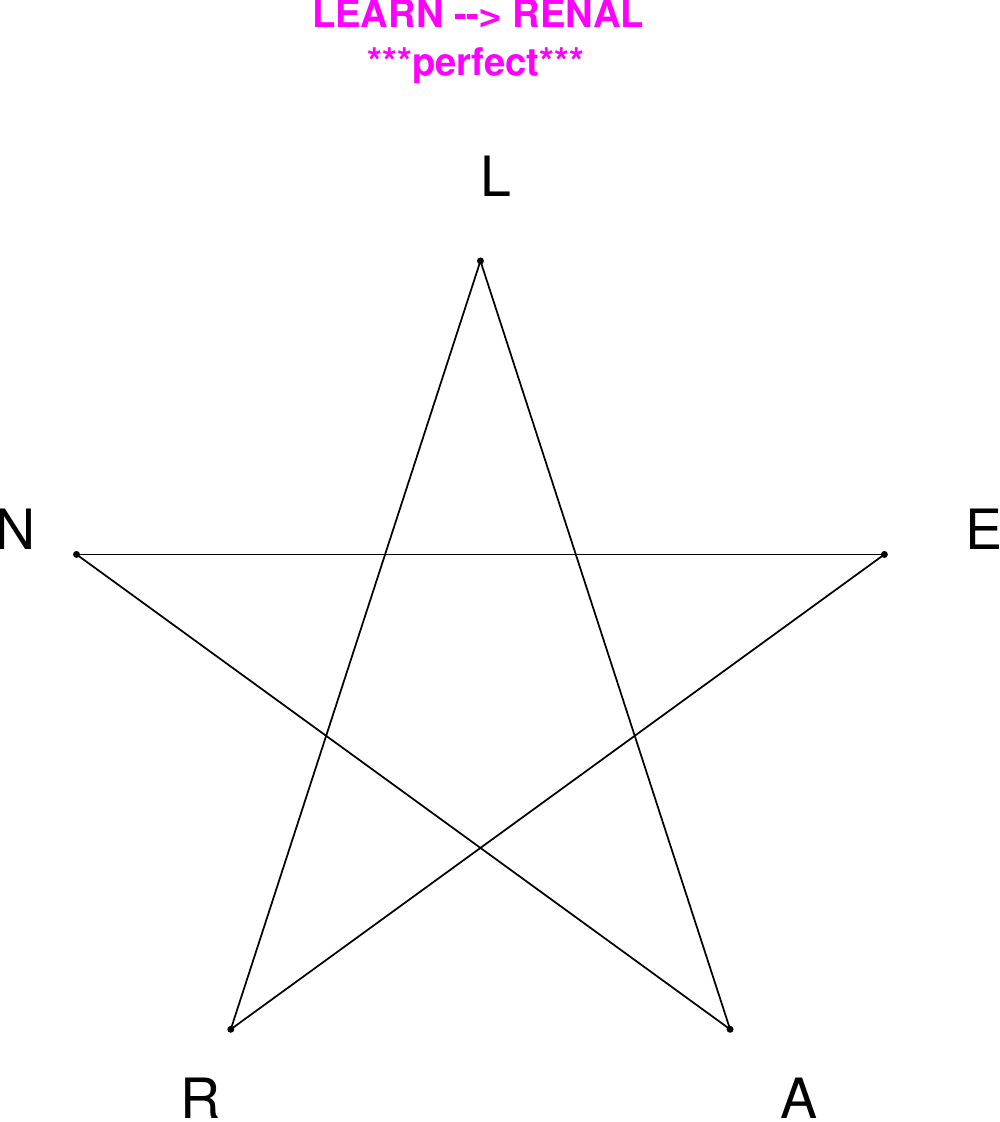}
\end{subfigure}
\end{figure}

\begin{figure}[H]
\centering
\begin{subfigure}[T]{0.19\textwidth}
\centering
\includegraphics[width=\textwidth]{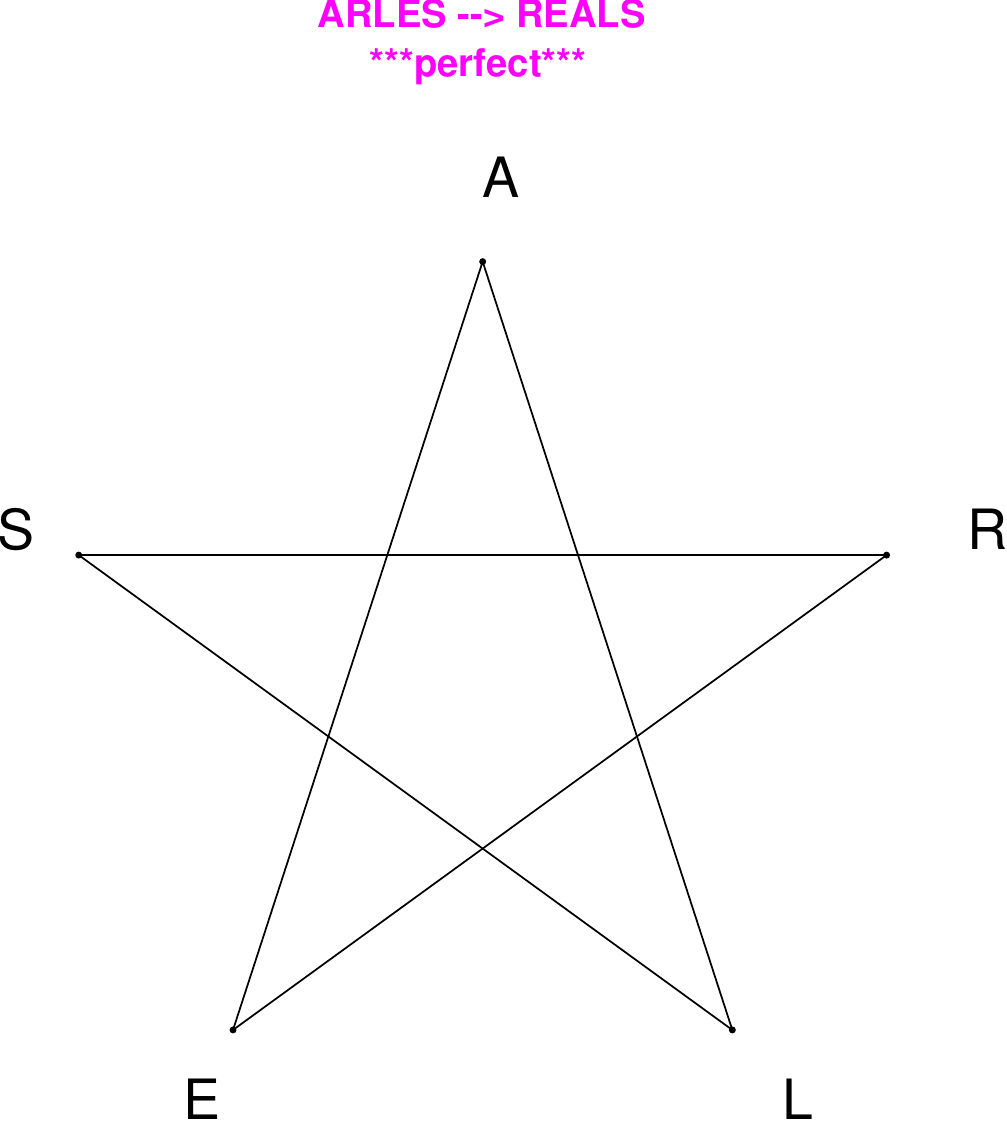}
\end{subfigure}
\hfill
\begin{subfigure}[T]{0.19\textwidth}
\centering
\includegraphics[width=\textwidth]{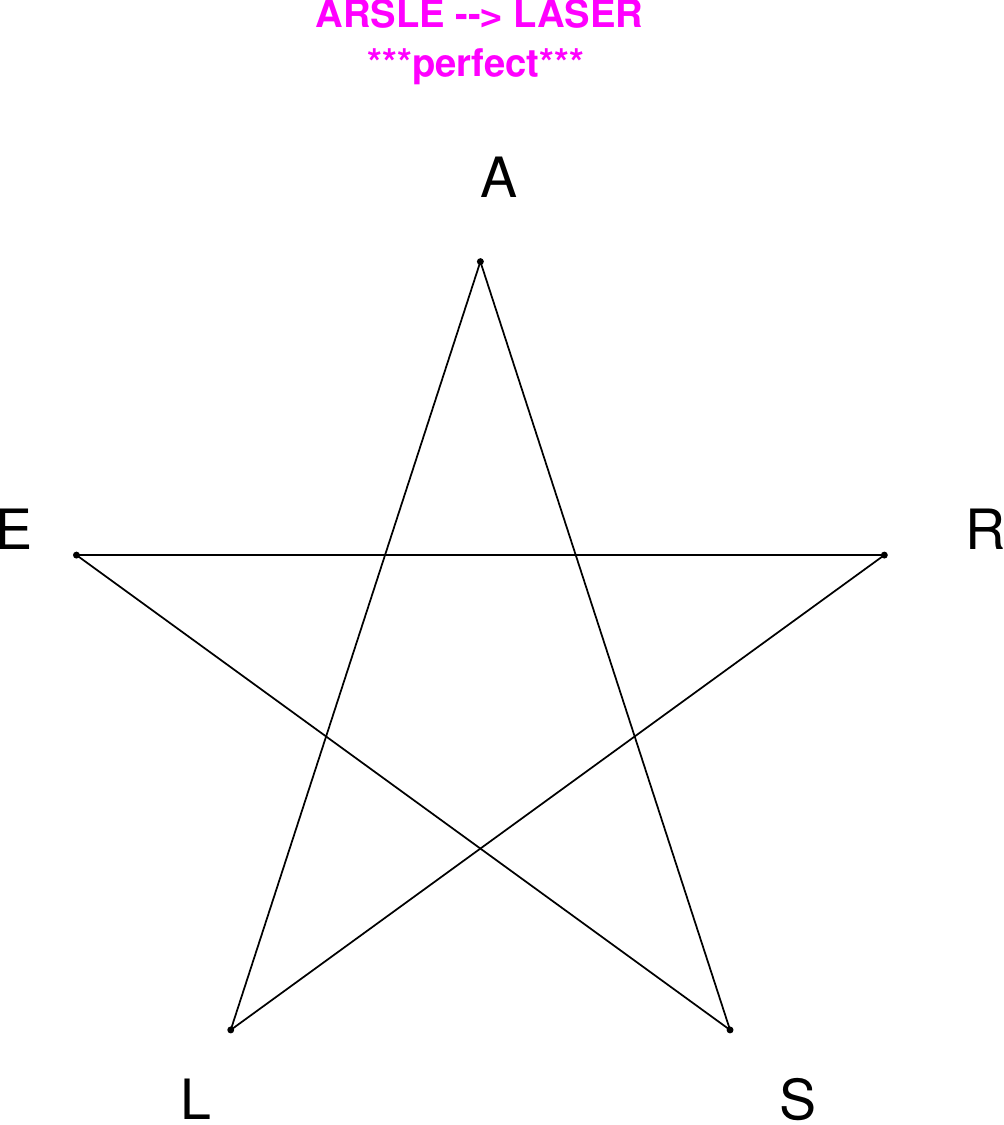}
\end{subfigure}
\hfill
\begin{subfigure}[T]{0.19\textwidth}
\centering
\includegraphics[width=\textwidth]{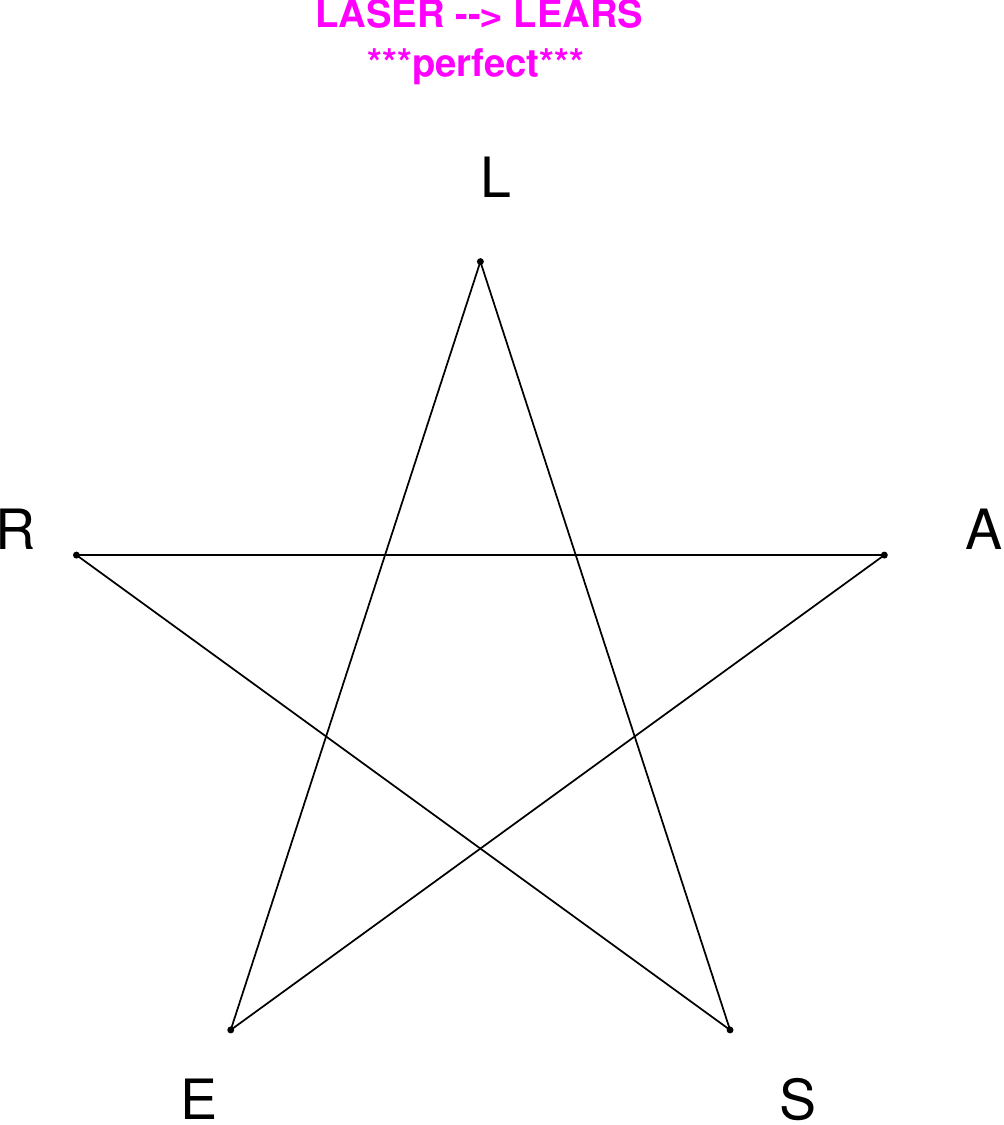}
\end{subfigure}
\hfill
\begin{subfigure}[T]{0.19\textwidth}
\centering
\includegraphics[width=\textwidth]{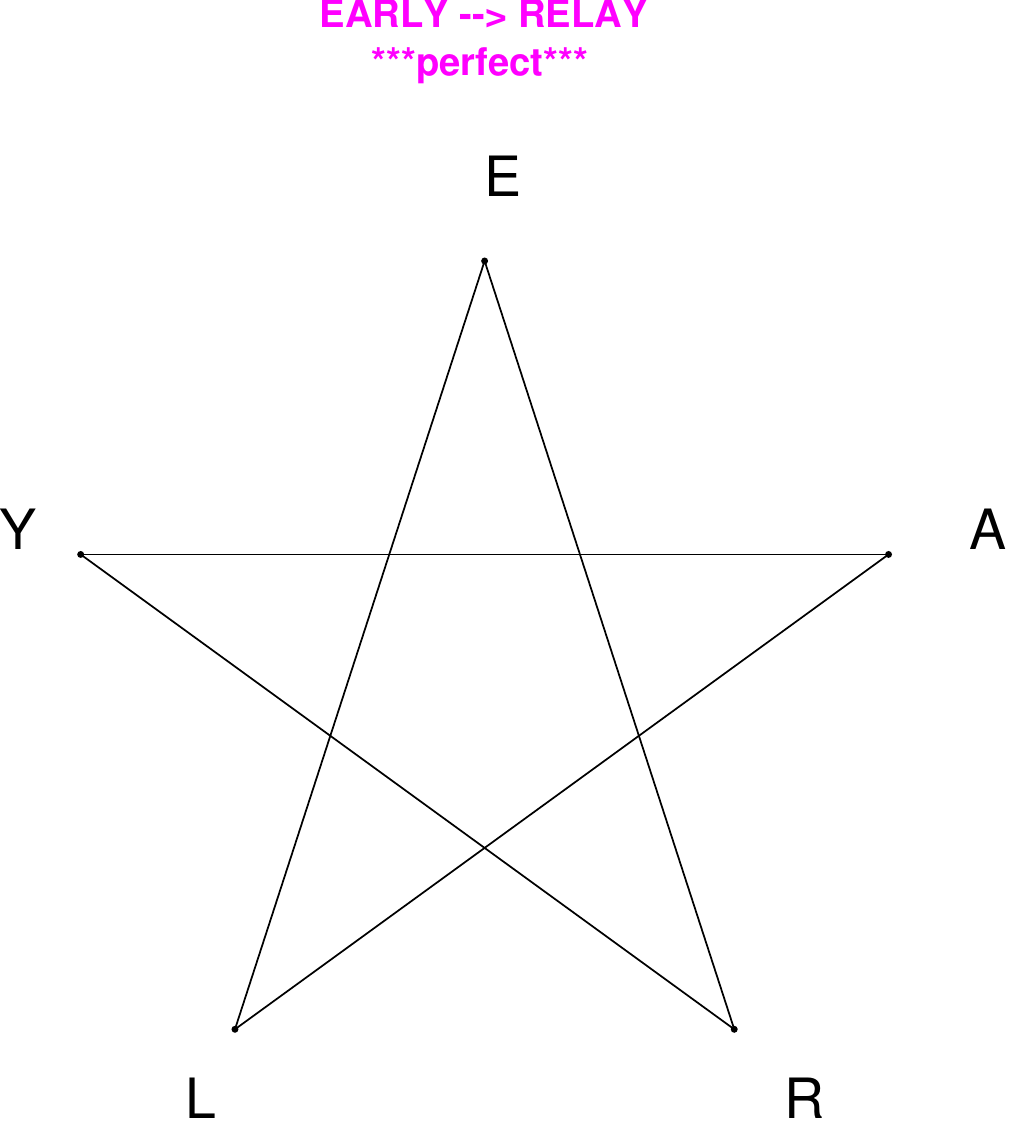}
\end{subfigure}
\hfill
\begin{subfigure}[T]{0.19\textwidth}
\centering
\includegraphics[width=\textwidth]{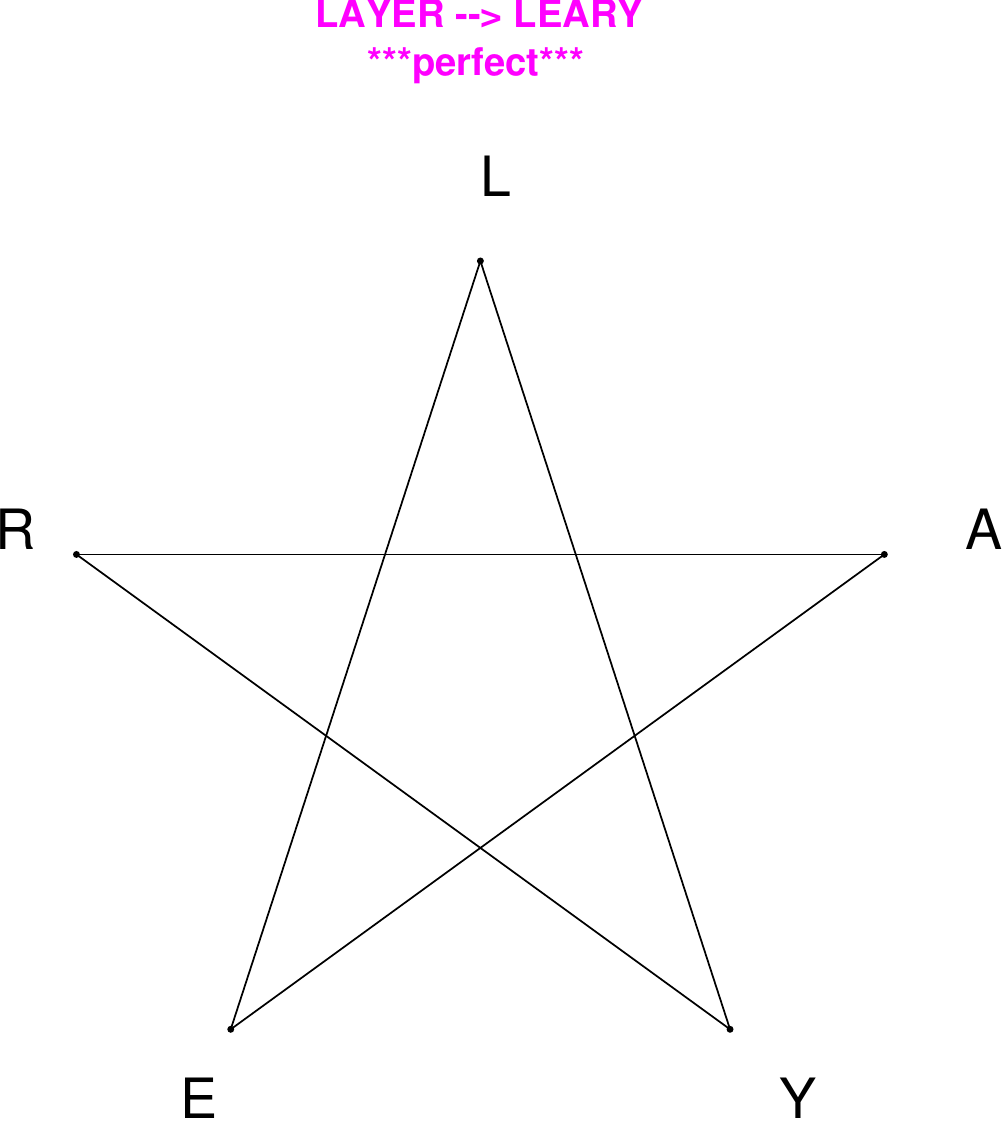}
\end{subfigure}
\end{figure}

\begin{figure}[H]
\centering
\begin{subfigure}[T]{0.19\textwidth}
\centering
\includegraphics[width=\textwidth]{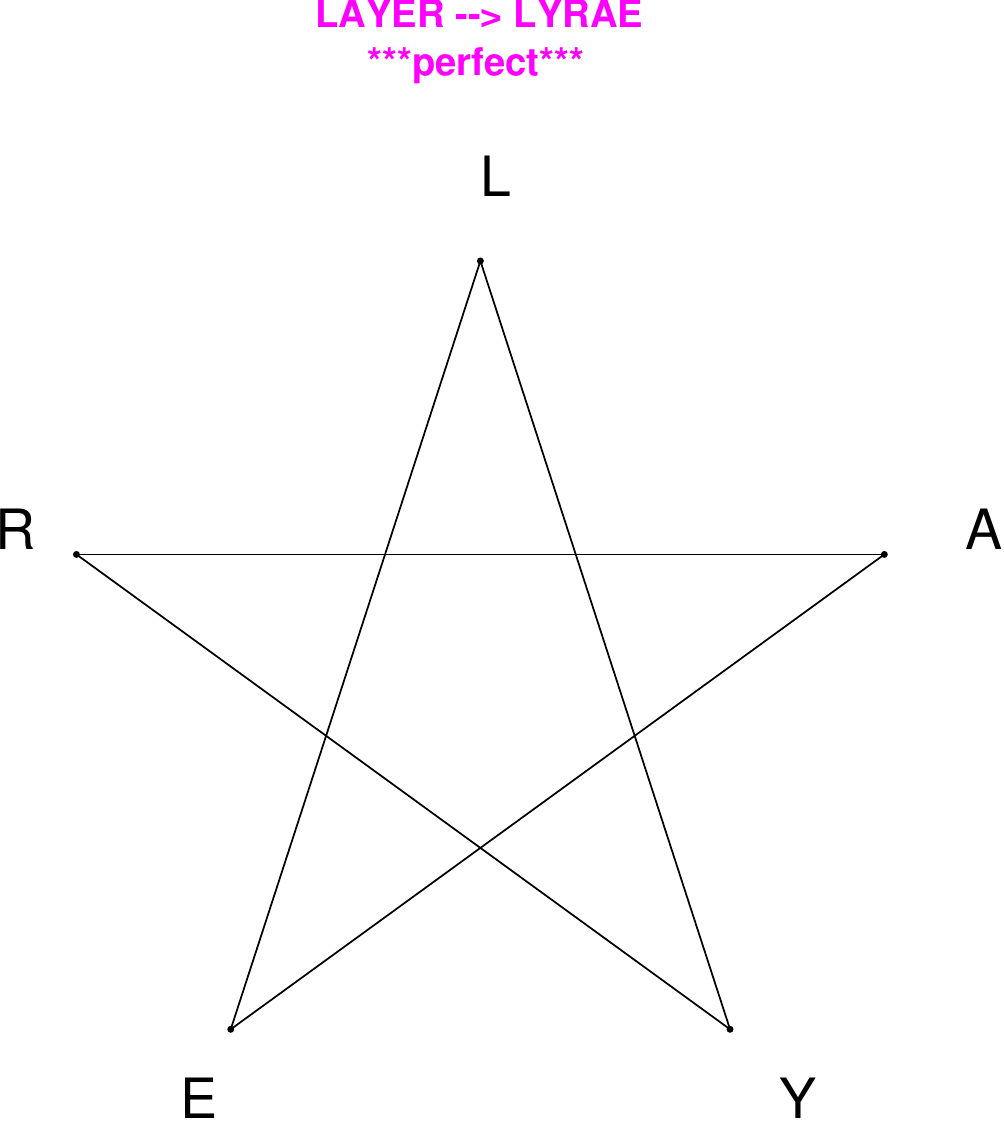}
\end{subfigure}
\hfill
\begin{subfigure}[T]{0.19\textwidth}
\centering
\includegraphics[width=\textwidth]{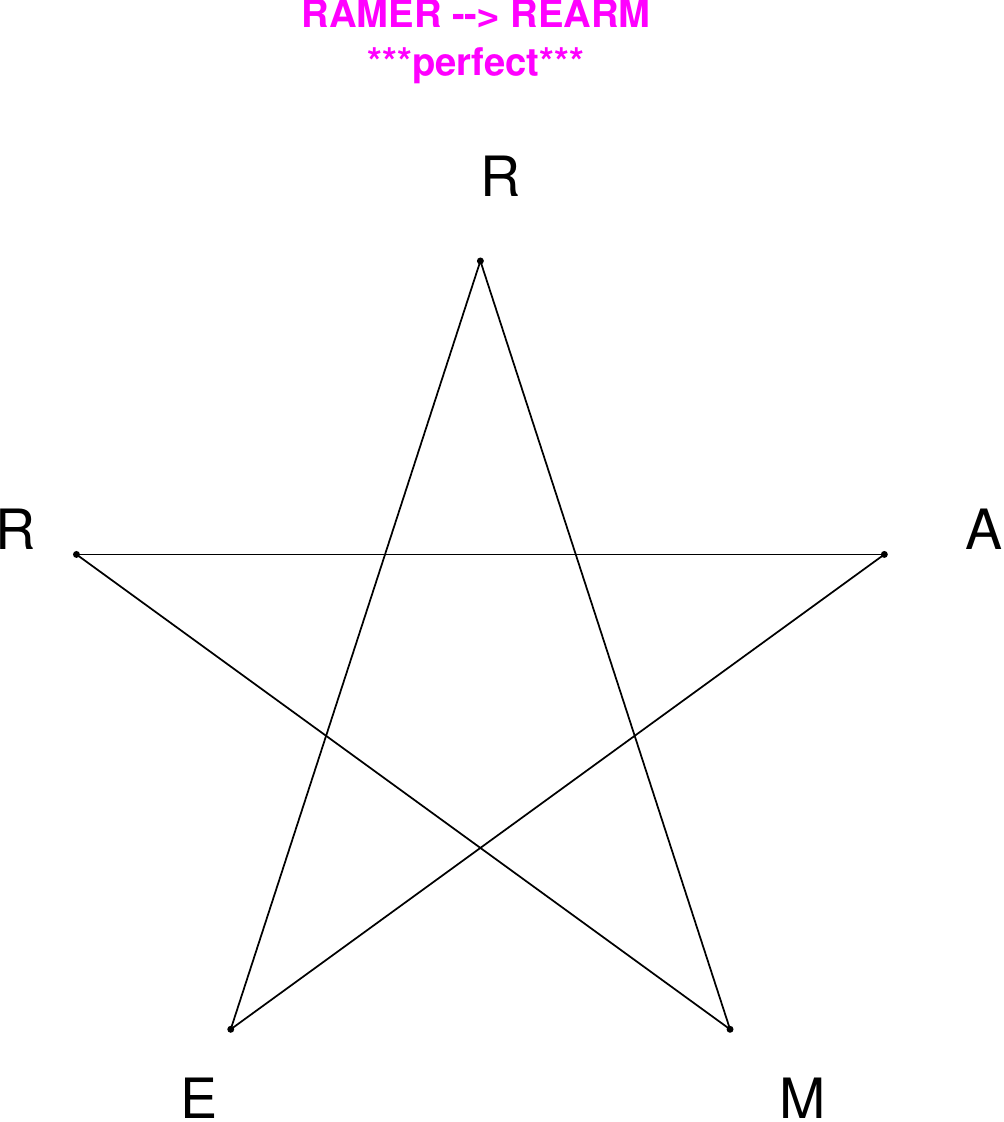}
\end{subfigure}
\hfill
\begin{subfigure}[T]{0.19\textwidth}
\centering
\includegraphics[width=\textwidth]{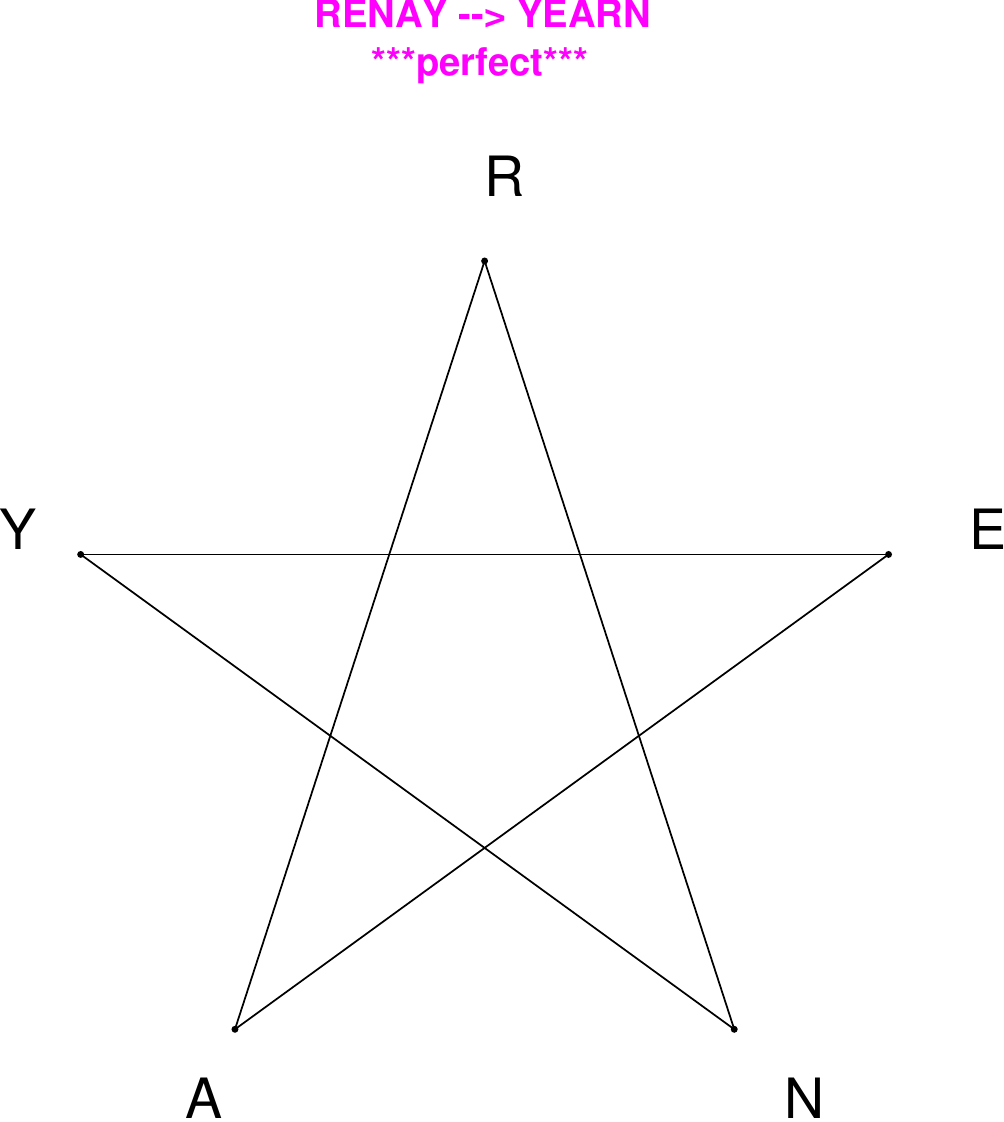}
\end{subfigure}
\hfill
\begin{subfigure}[T]{0.19\textwidth}
\centering
\includegraphics[width=\textwidth]{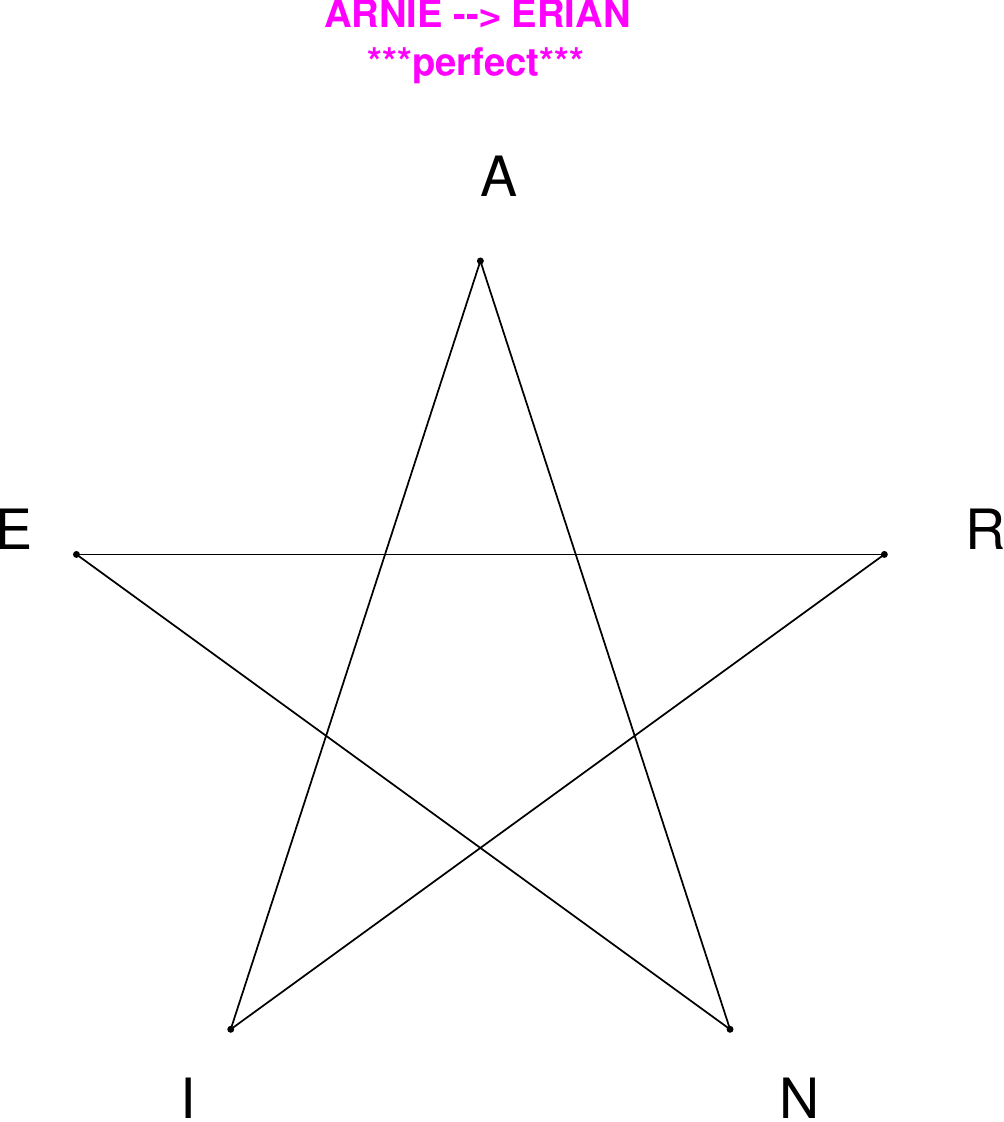}
\end{subfigure}
\hfill
\begin{subfigure}[T]{0.19\textwidth}
\centering
\includegraphics[width=\textwidth]{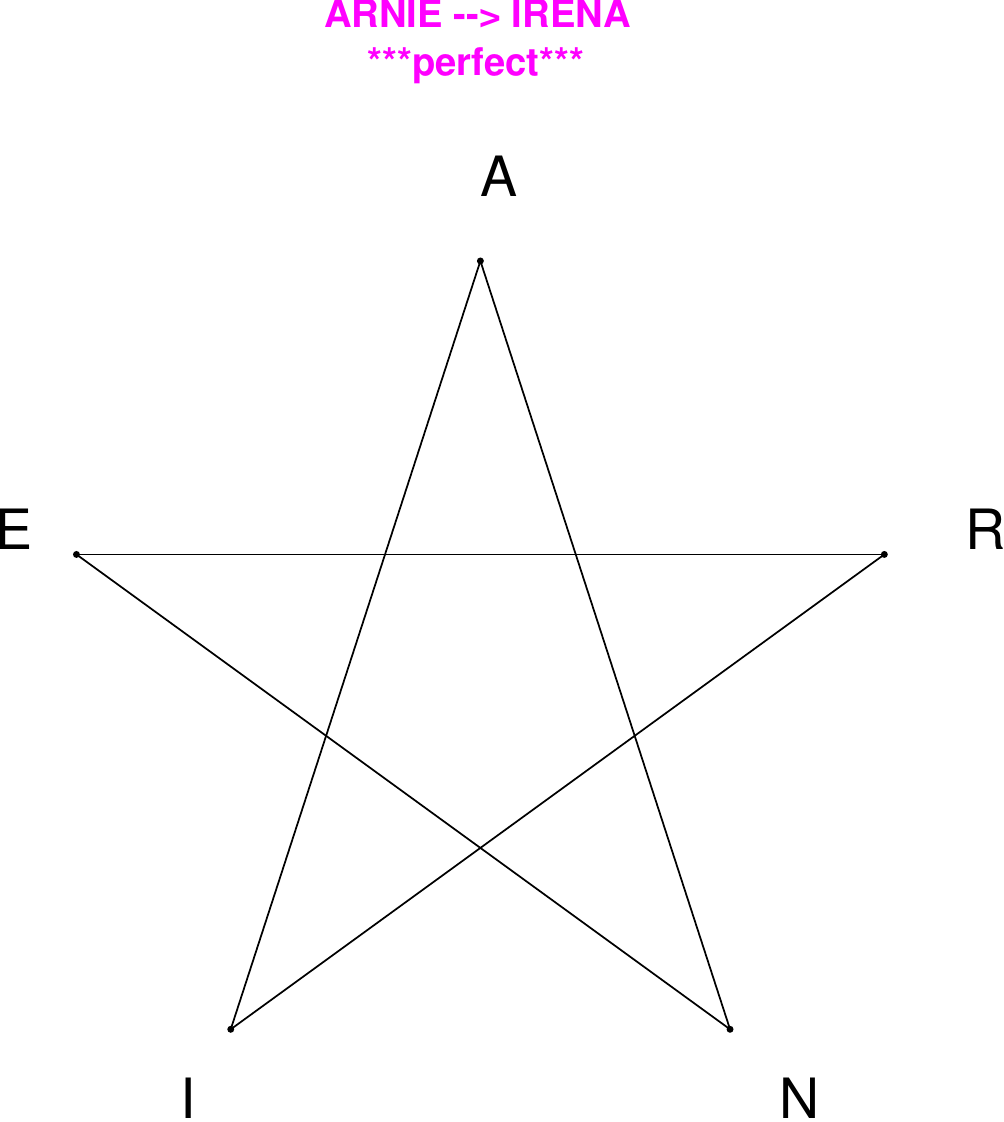}
\end{subfigure}
\end{figure}

\begin{figure}[H]
\centering
\begin{subfigure}[T]{0.19\textwidth}
\centering
\includegraphics[width=\textwidth]{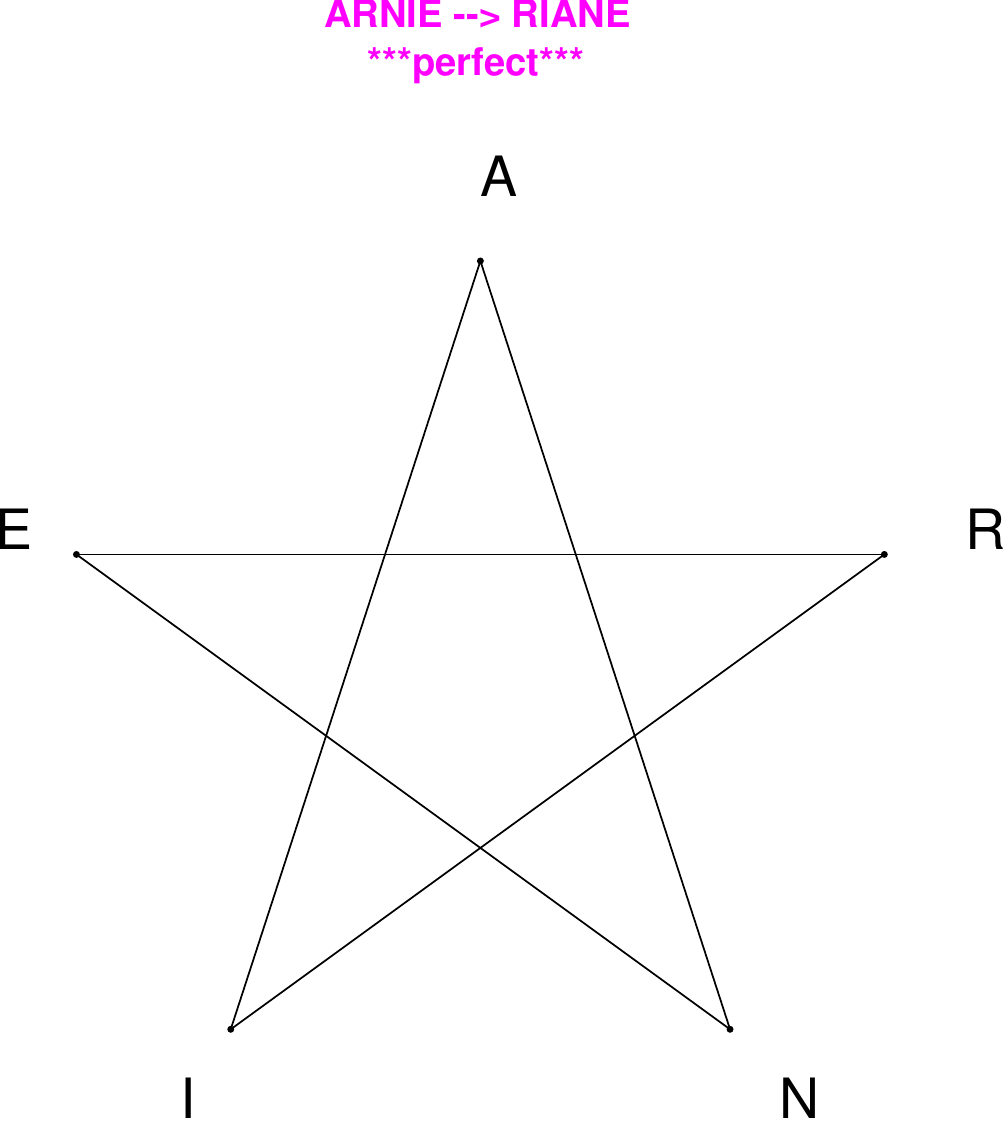}
\end{subfigure}
\hfill
\begin{subfigure}[T]{0.19\textwidth}
\centering
\includegraphics[width=\textwidth]{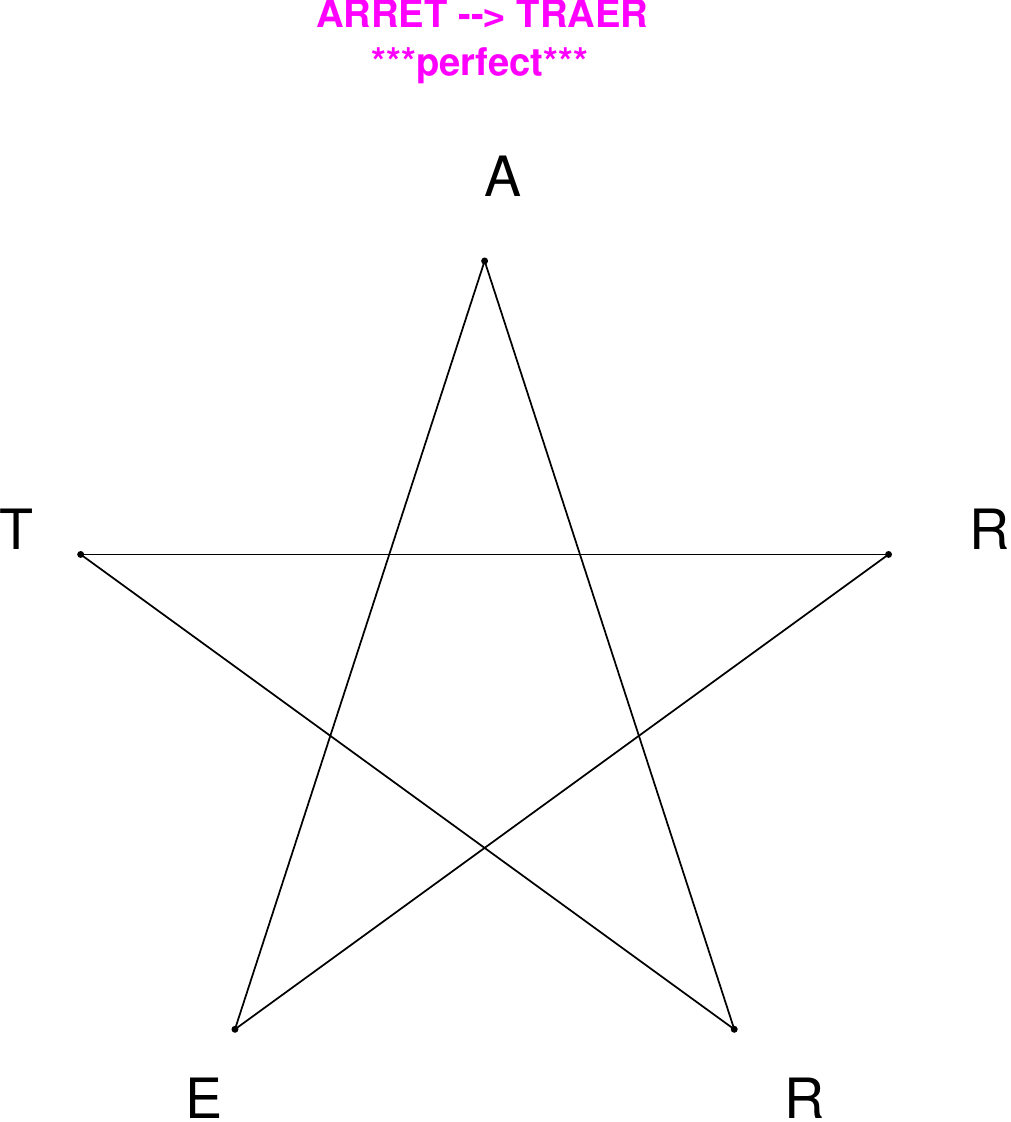}
\end{subfigure}
\hfill
\begin{subfigure}[T]{0.19\textwidth}
\centering
\includegraphics[width=\textwidth]{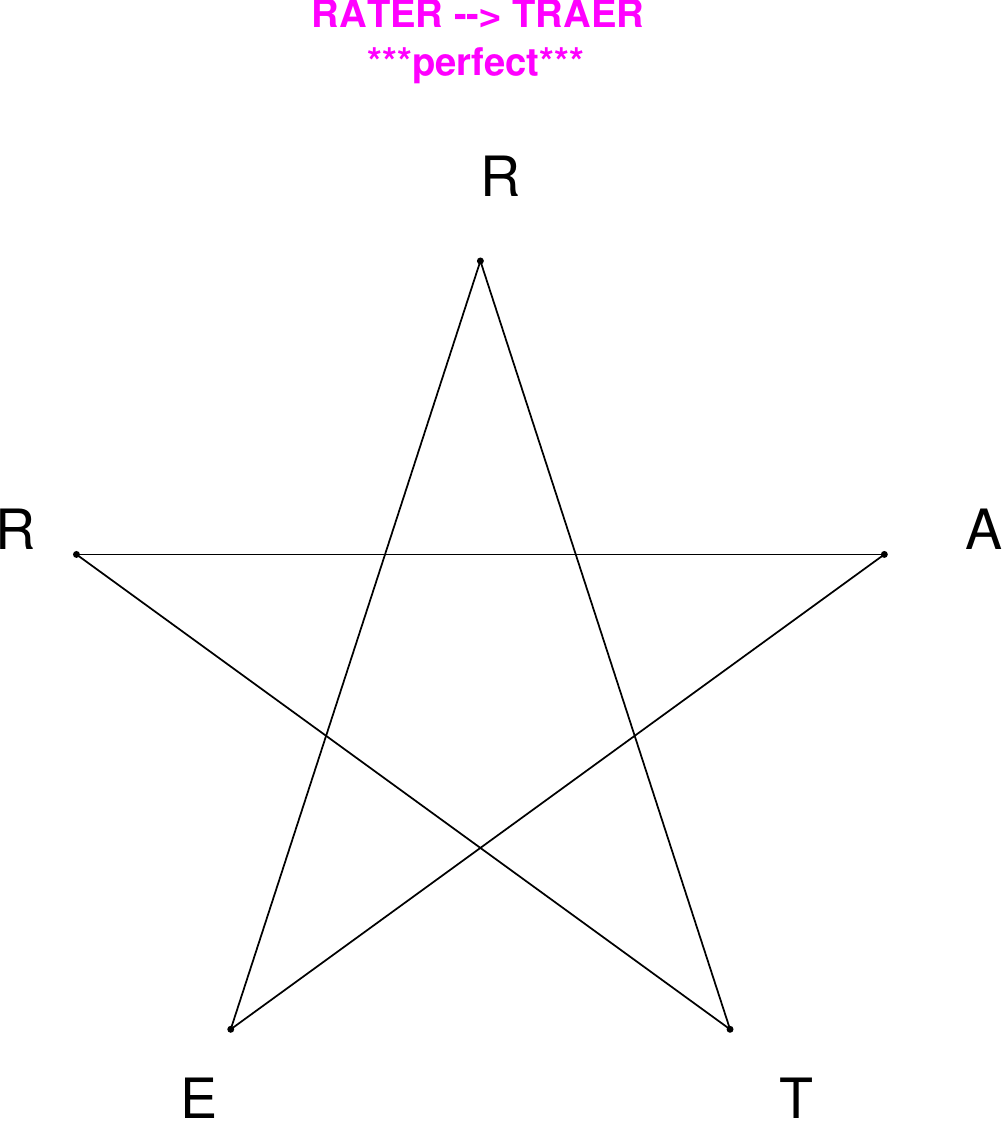}
\end{subfigure}
\hfill
\begin{subfigure}[T]{0.19\textwidth}
\centering
\includegraphics[width=\textwidth]{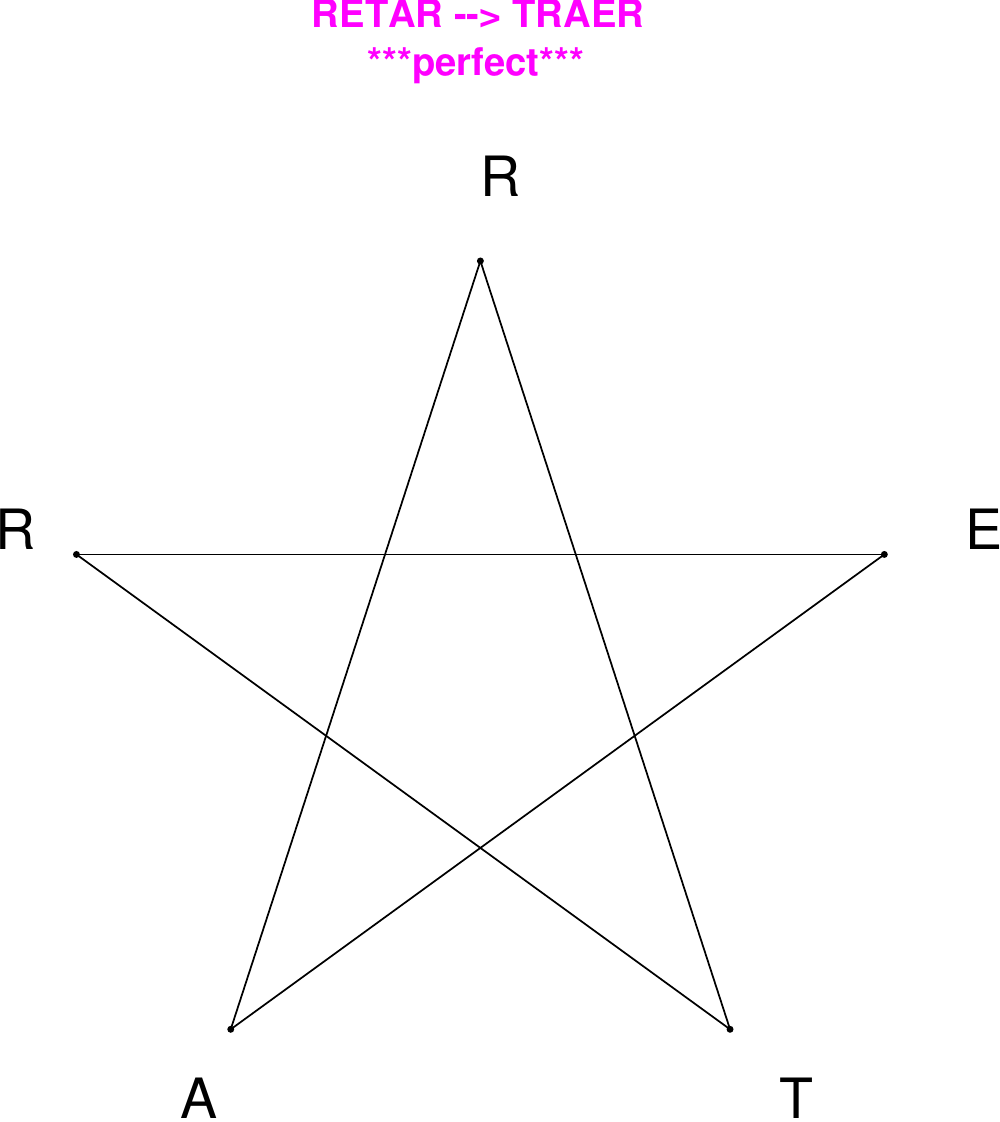}
\end{subfigure}
\hfill
\begin{subfigure}[T]{0.19\textwidth}
\centering
\includegraphics[width=\textwidth]{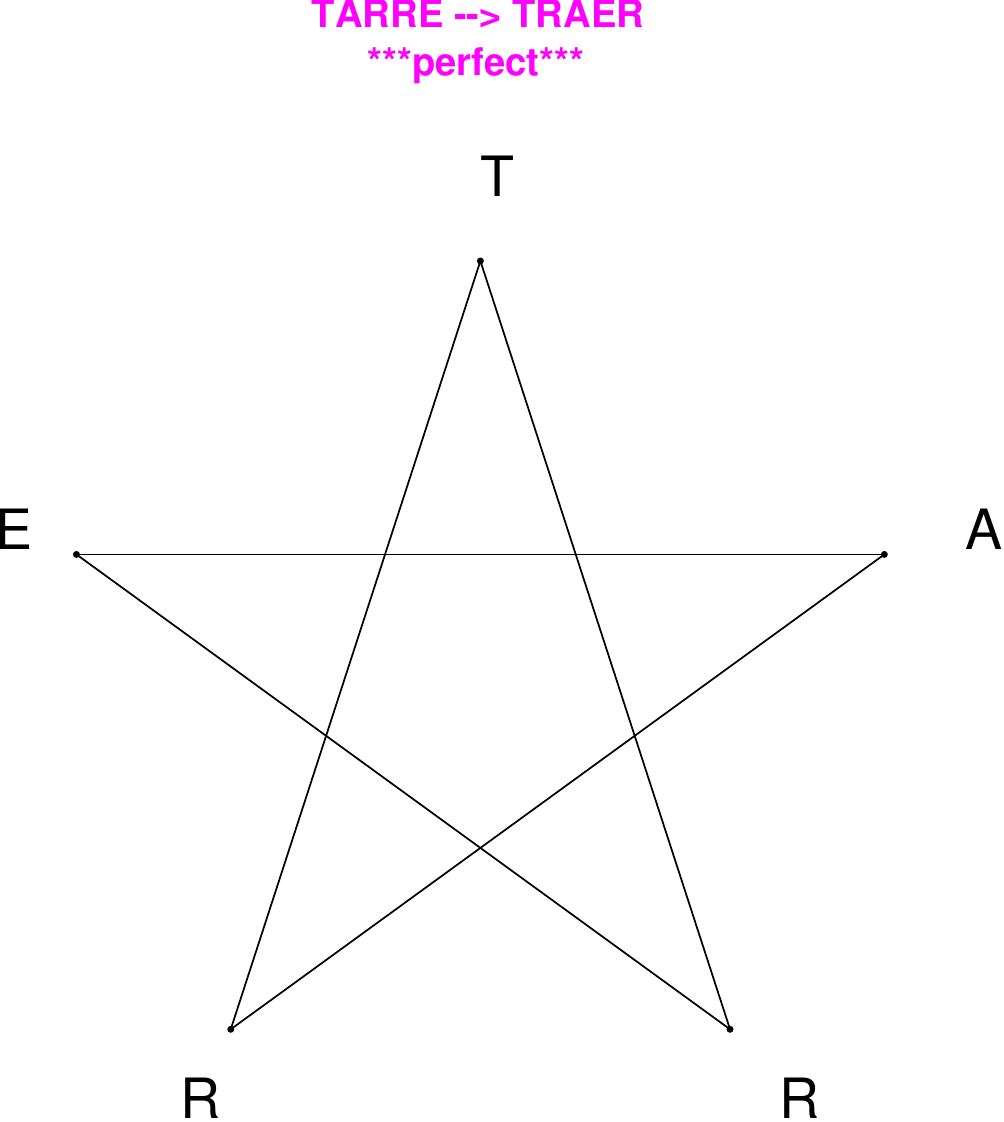}
\end{subfigure}
\end{figure}

\begin{figure}[H]
\centering
\begin{subfigure}[T]{0.19\textwidth}
\centering
\includegraphics[width=\textwidth]{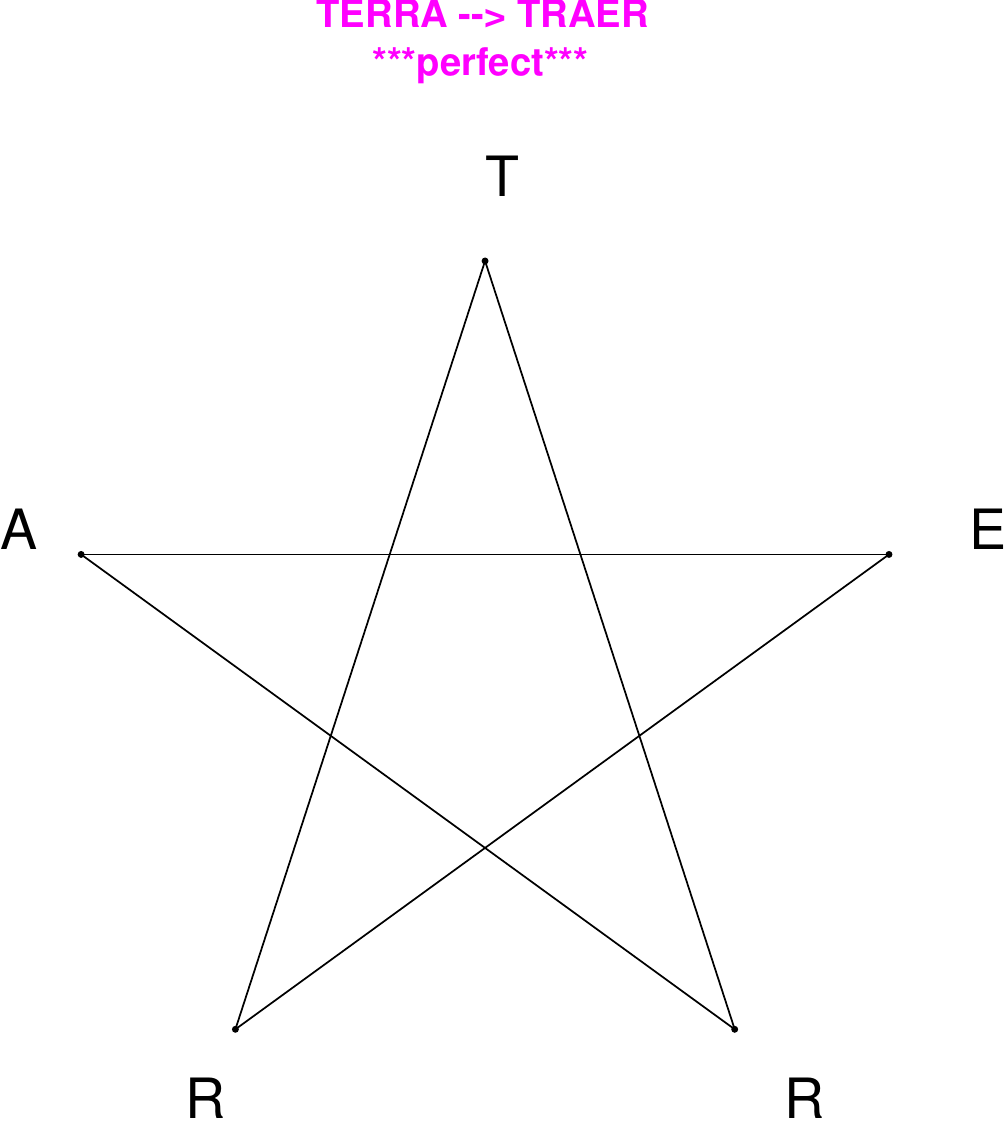}
\end{subfigure}
\hfill
\begin{subfigure}[T]{0.19\textwidth}
\centering
\includegraphics[width=\textwidth]{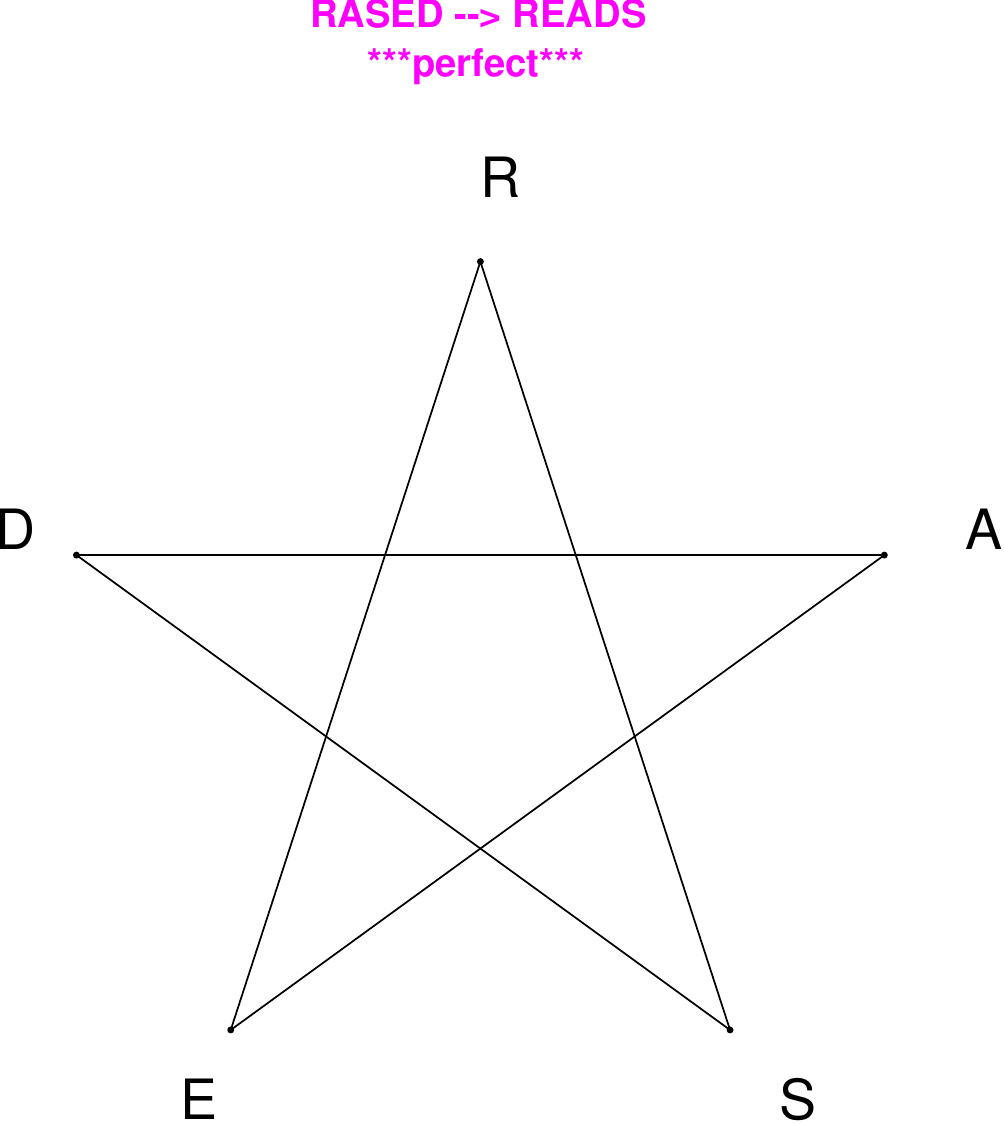}
\end{subfigure}
\hfill
\begin{subfigure}[T]{0.19\textwidth}
\centering
\includegraphics[width=\textwidth]{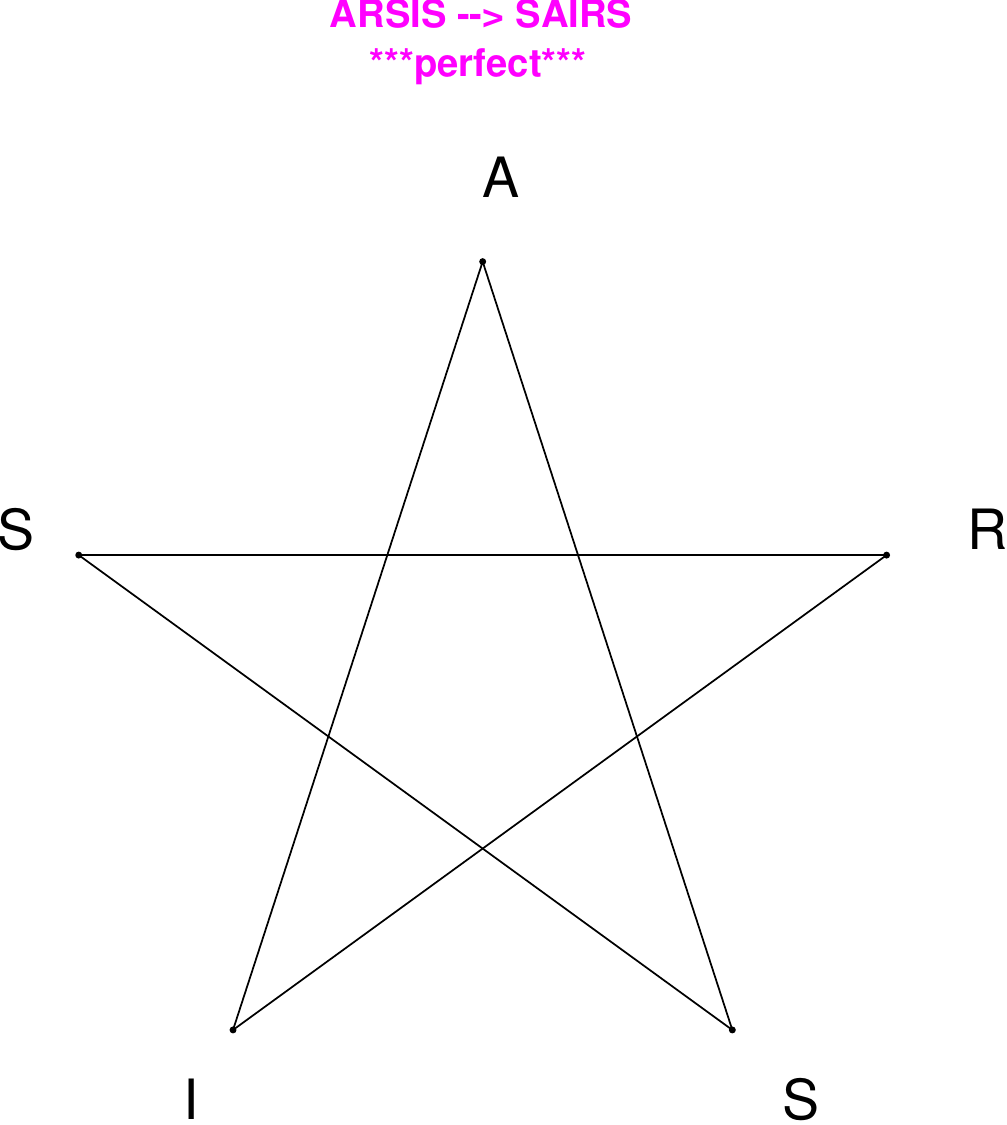}
\end{subfigure}
\hfill
\begin{subfigure}[T]{0.19\textwidth}
\centering
\includegraphics[width=\textwidth]{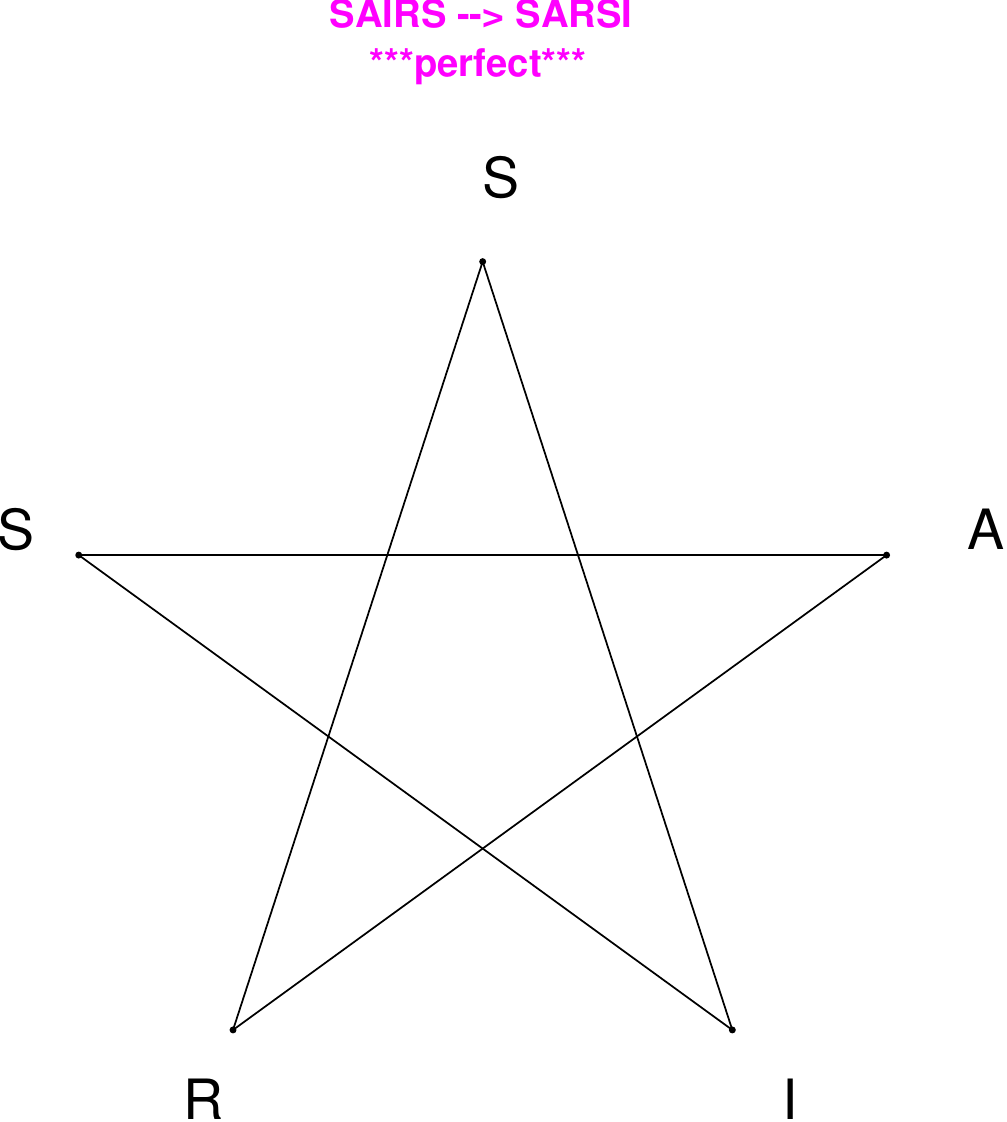}
\end{subfigure}
\hfill
\begin{subfigure}[T]{0.19\textwidth}
\centering
\includegraphics[width=\textwidth]{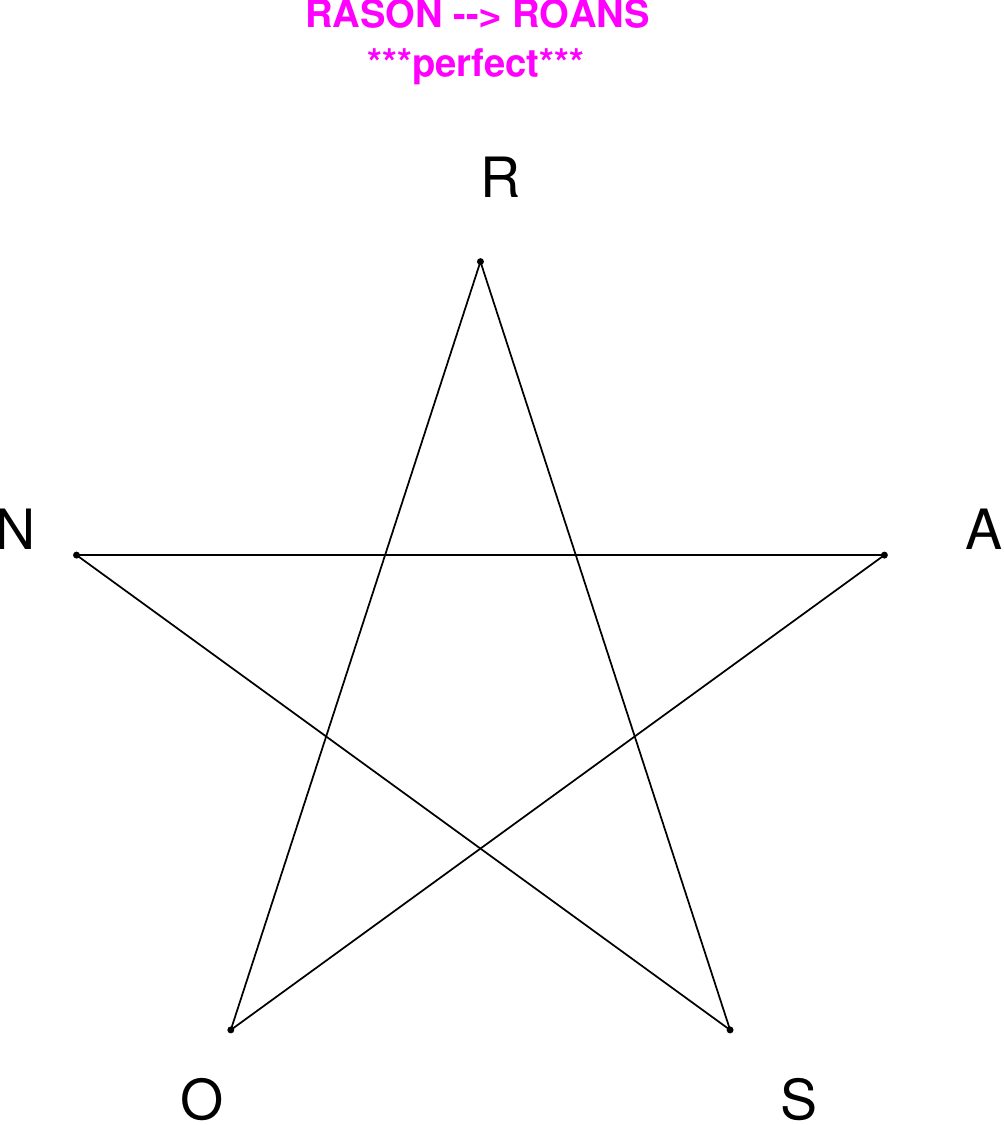}
\end{subfigure}
\end{figure}

\begin{figure}[H]
\centering
\begin{subfigure}[T]{0.19\textwidth}
\centering
\includegraphics[width=\textwidth]{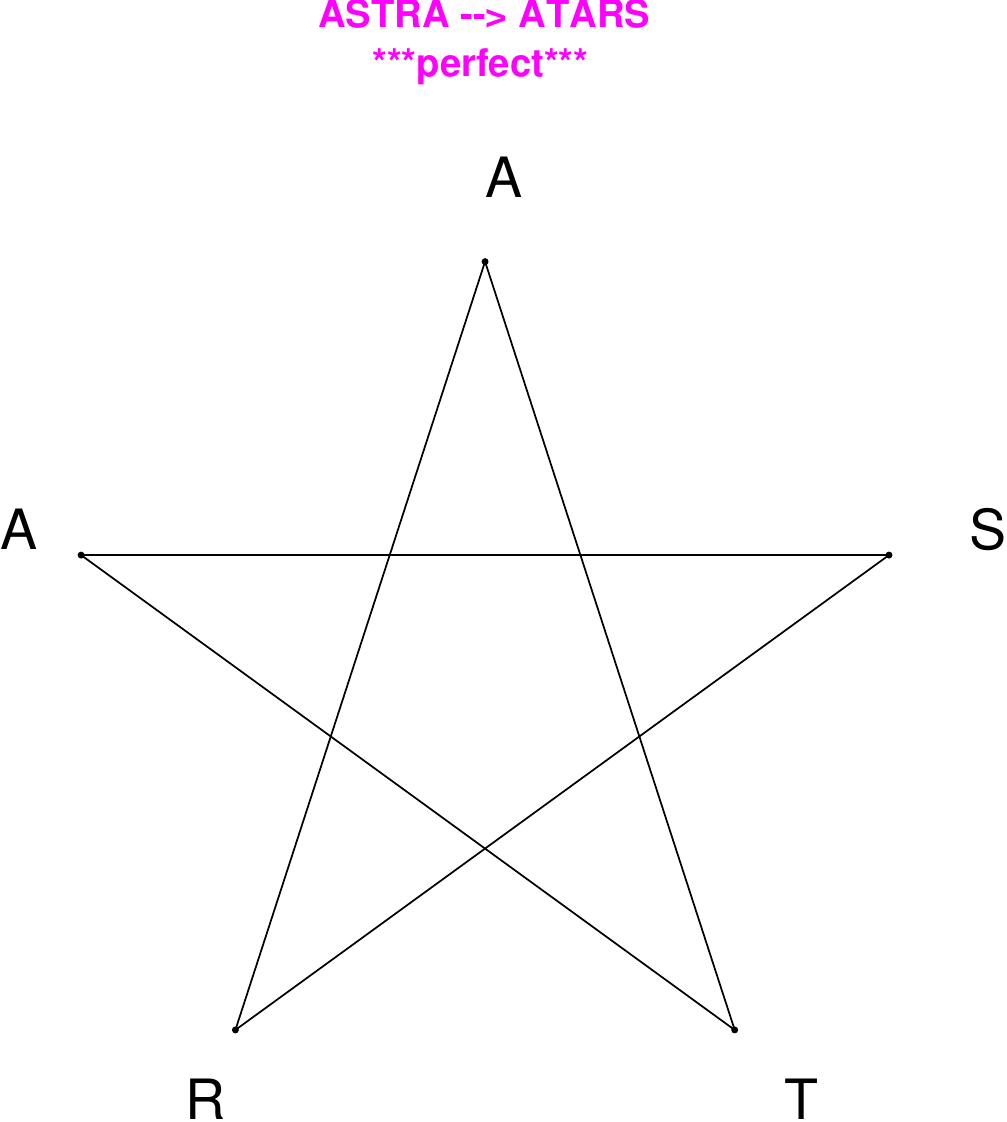}
\end{subfigure}
\hfill
\begin{subfigure}[T]{0.19\textwidth}
\centering
\includegraphics[width=\textwidth]{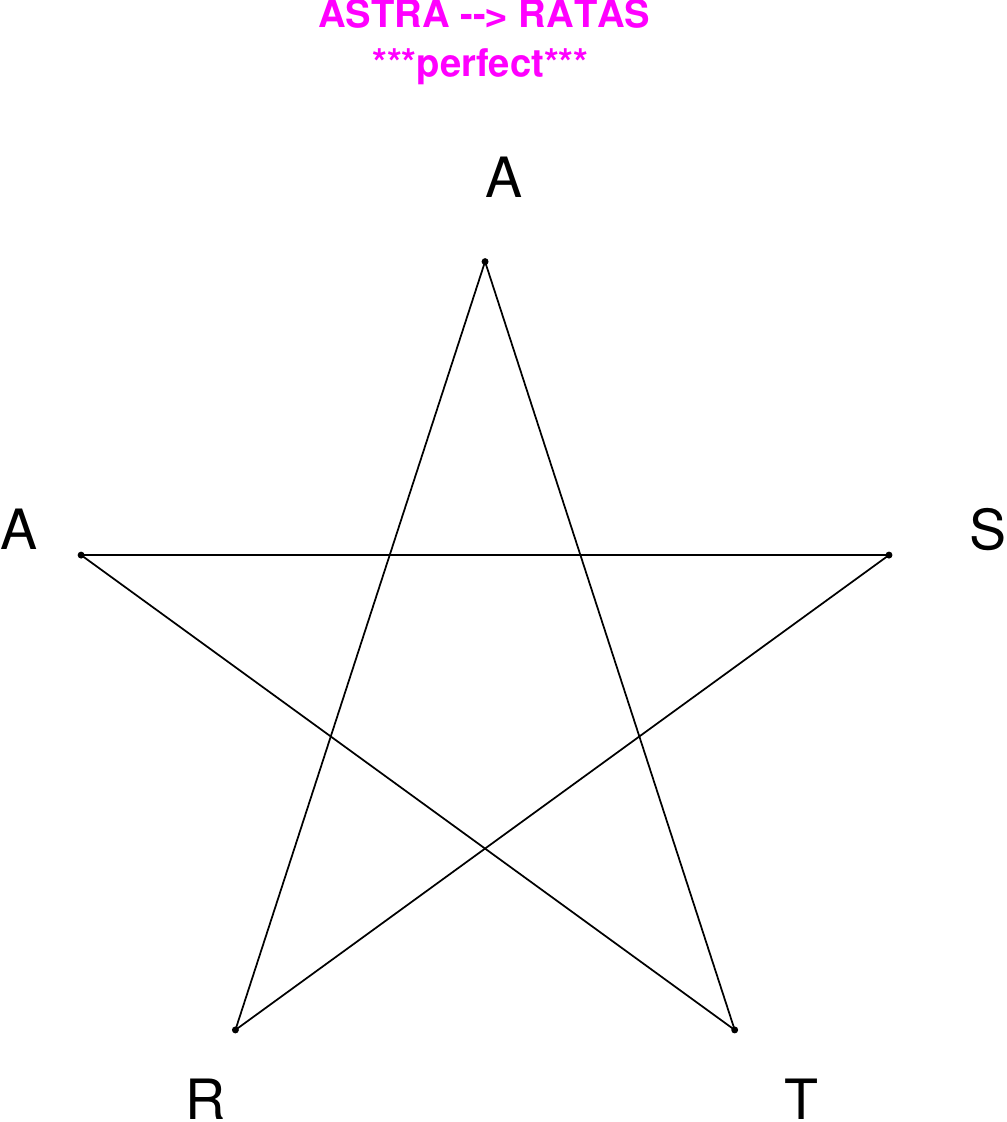}
\end{subfigure}
\hfill
\begin{subfigure}[T]{0.19\textwidth}
\centering
\includegraphics[width=\textwidth]{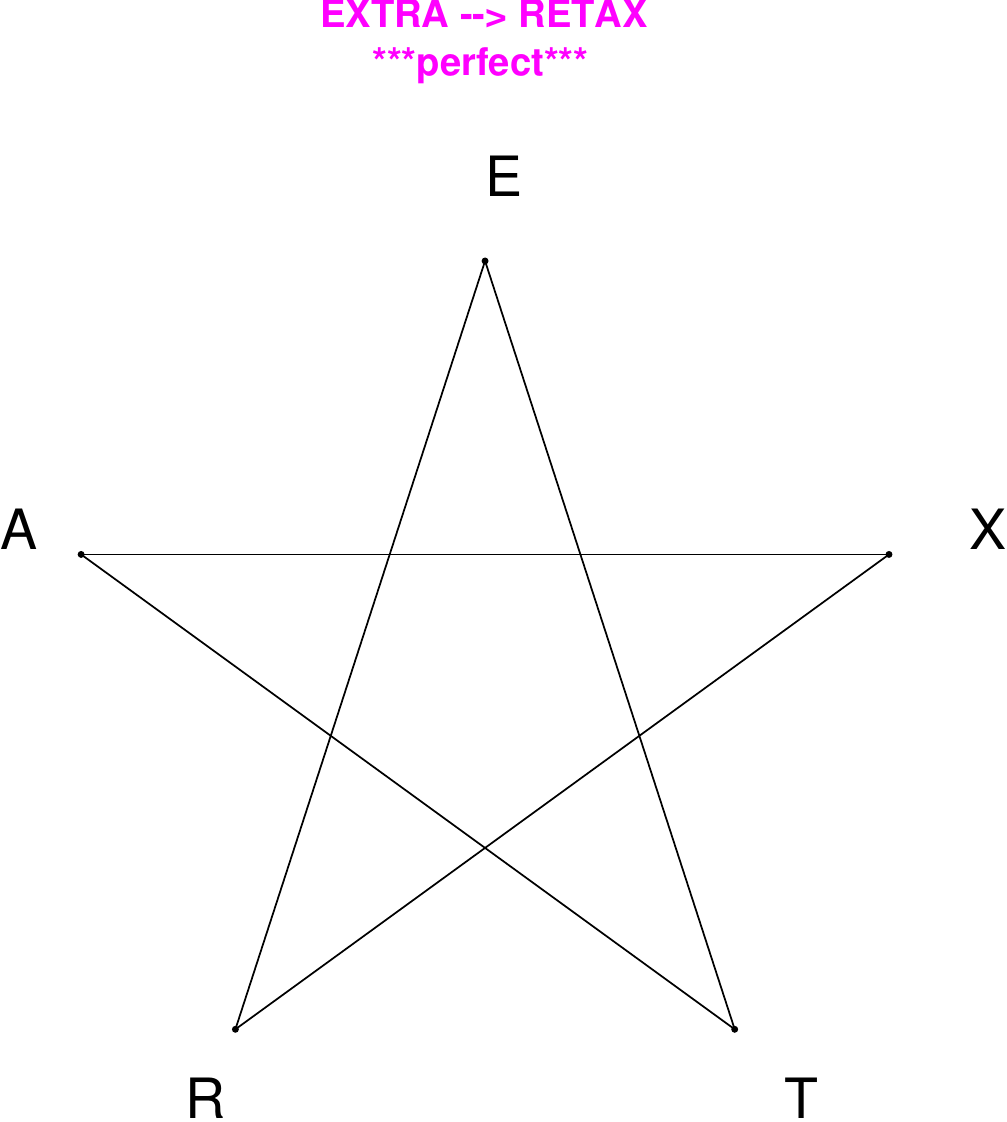}
\end{subfigure}
\hfill
\begin{subfigure}[T]{0.19\textwidth}
\centering
\includegraphics[width=\textwidth]{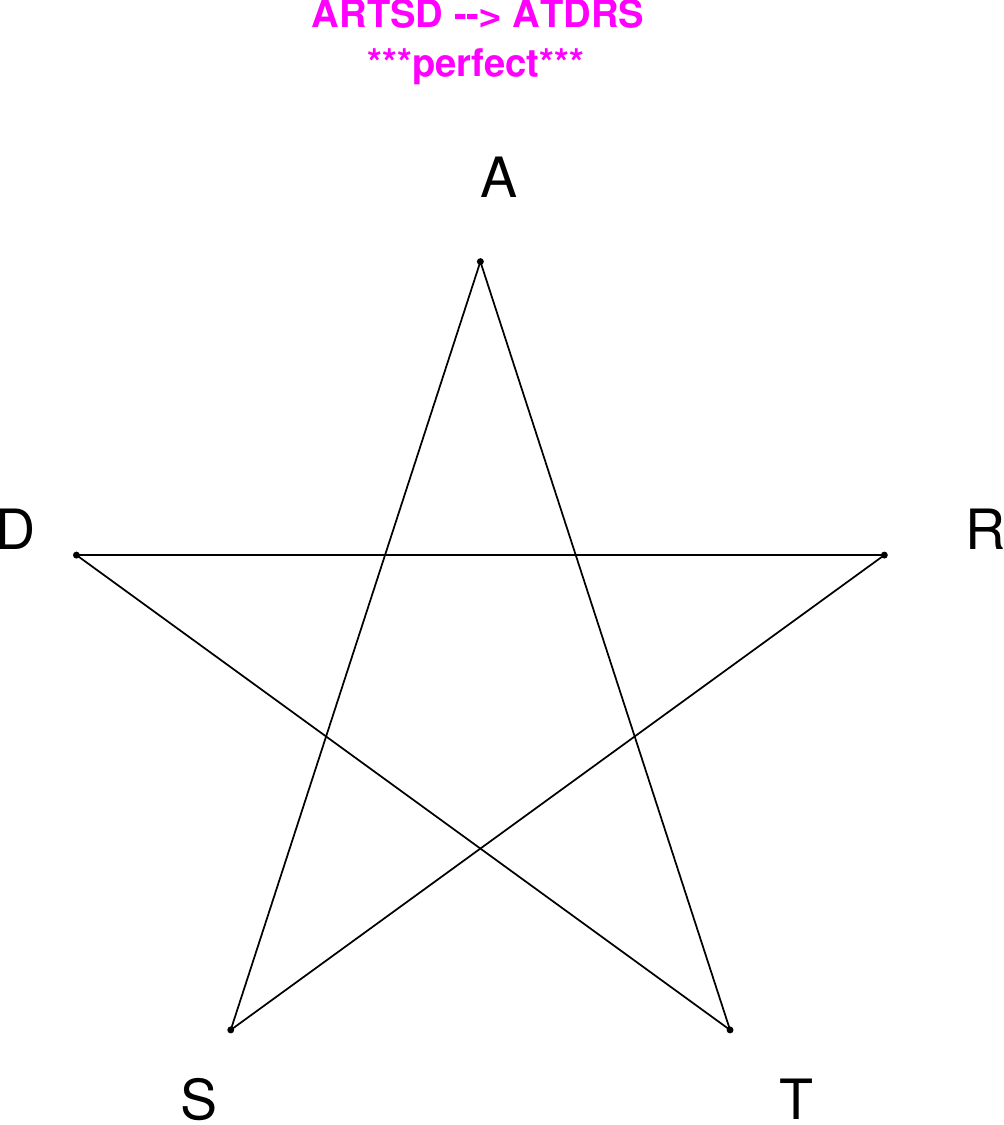}
\end{subfigure}
\hfill
\begin{subfigure}[T]{0.19\textwidth}
\centering
\includegraphics[width=\textwidth]{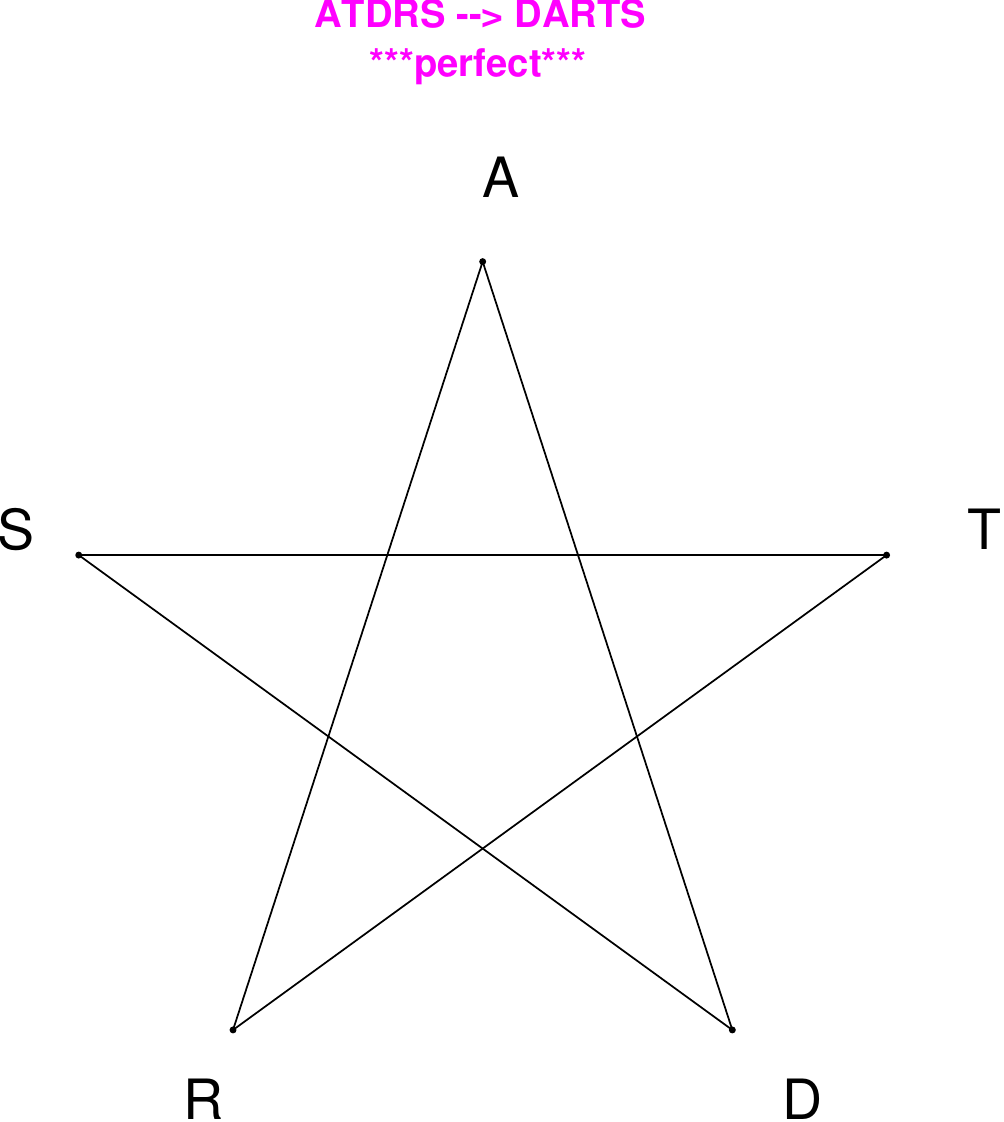}
\end{subfigure}
\end{figure}

\begin{figure}[H]
\centering
\begin{subfigure}[T]{0.19\textwidth}
\centering
\includegraphics[width=\textwidth]{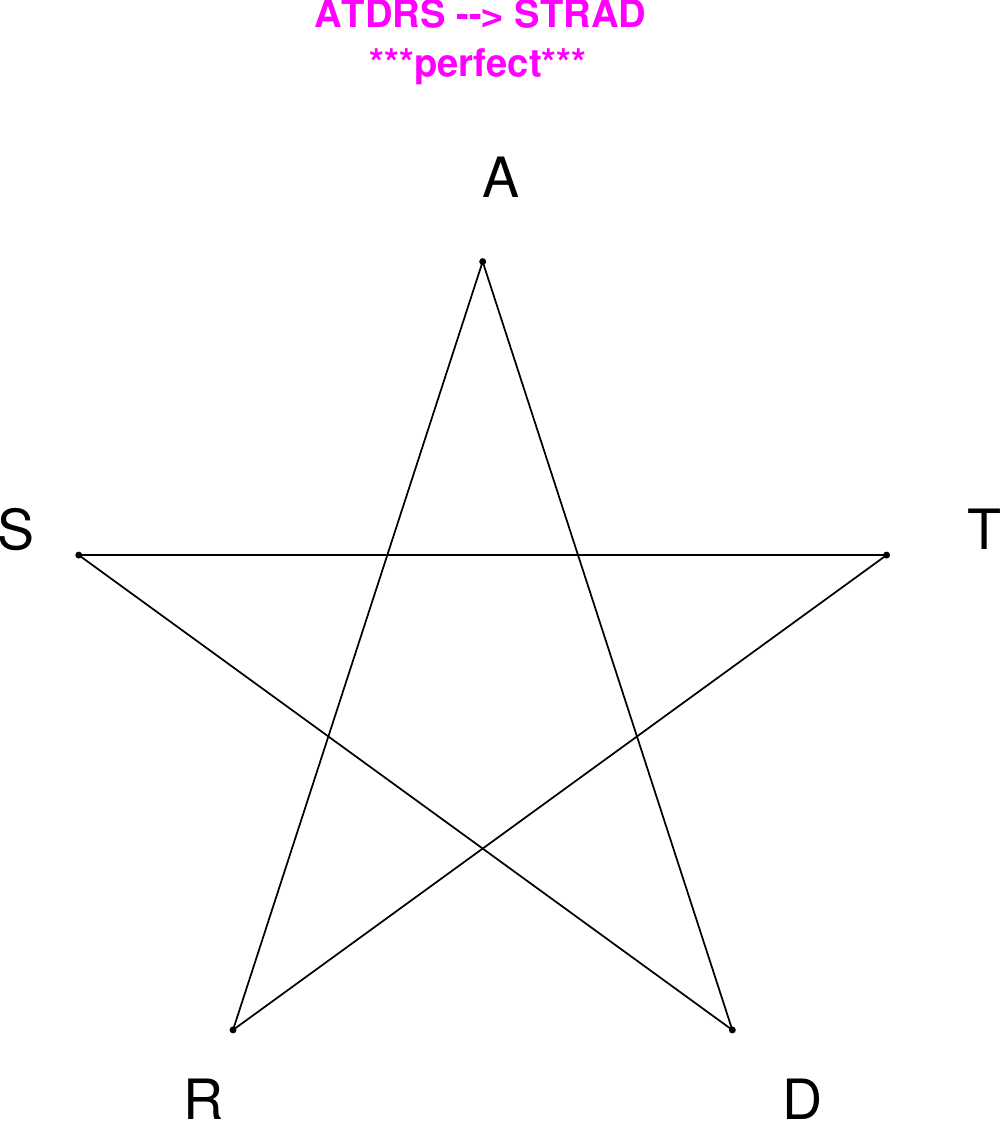}
\end{subfigure}
\hfill
\begin{subfigure}[T]{0.19\textwidth}
\centering
\includegraphics[width=\textwidth]{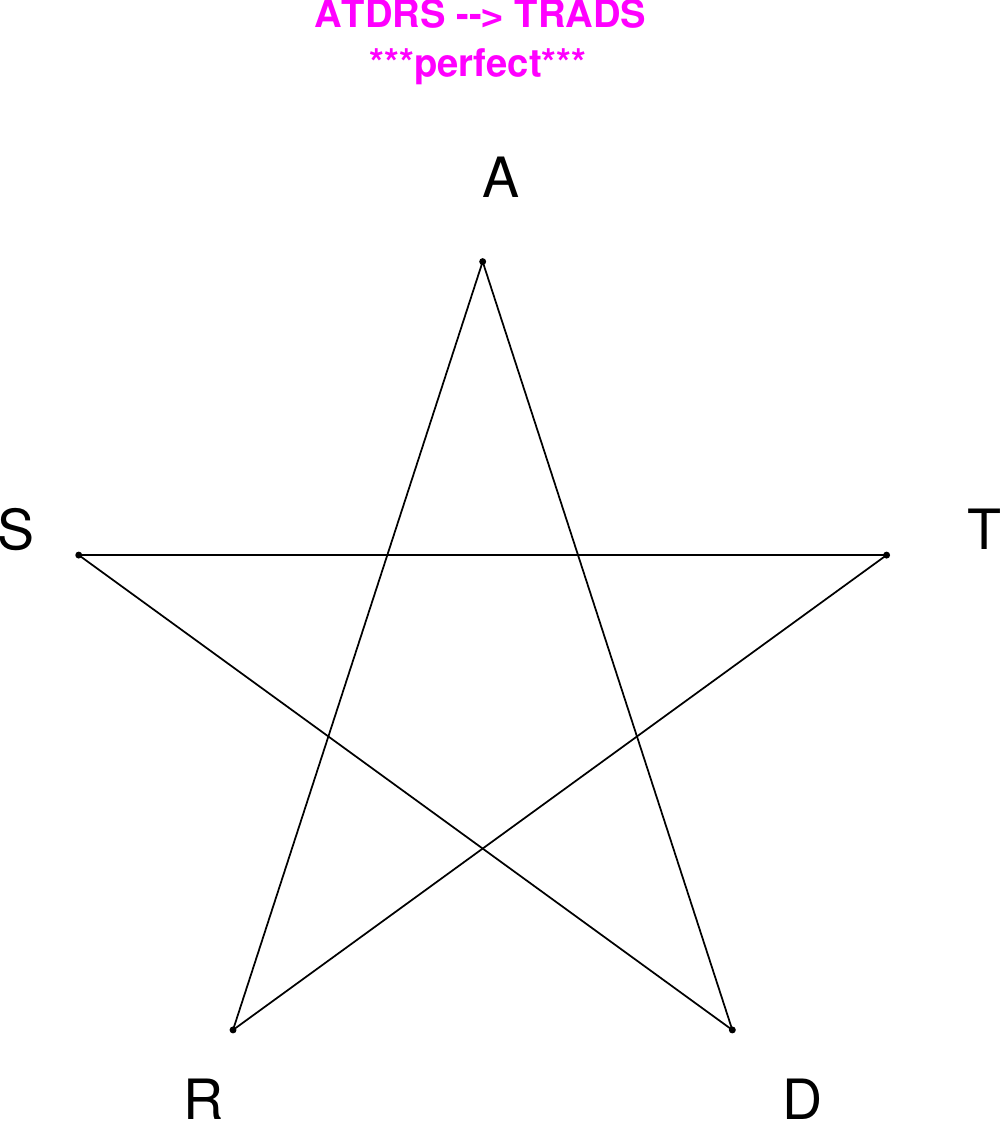}
\end{subfigure}
\hfill
\begin{subfigure}[T]{0.19\textwidth}
\centering
\includegraphics[width=\textwidth]{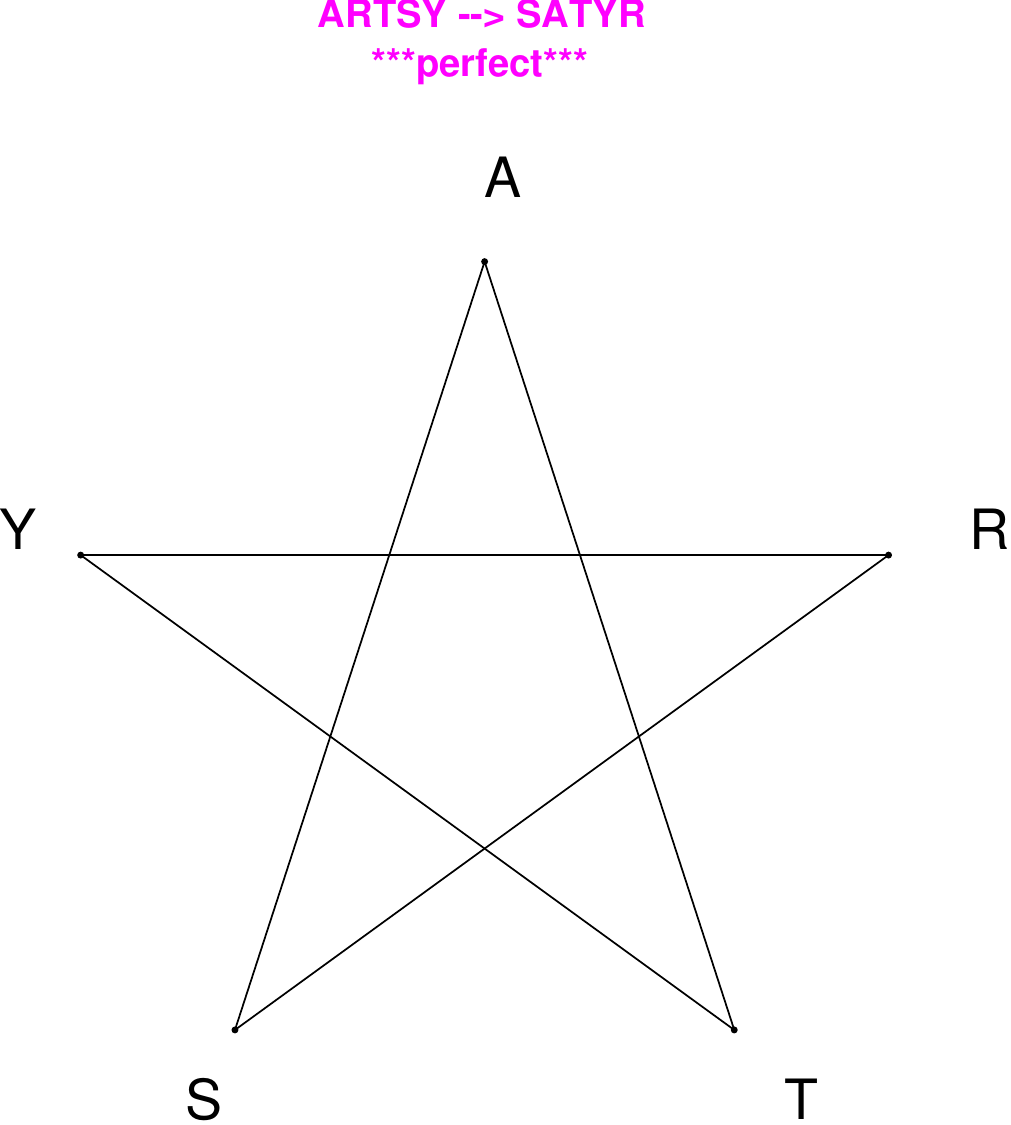}
\end{subfigure}
\hfill
\begin{subfigure}[T]{0.19\textwidth}
\centering
\includegraphics[width=\textwidth]{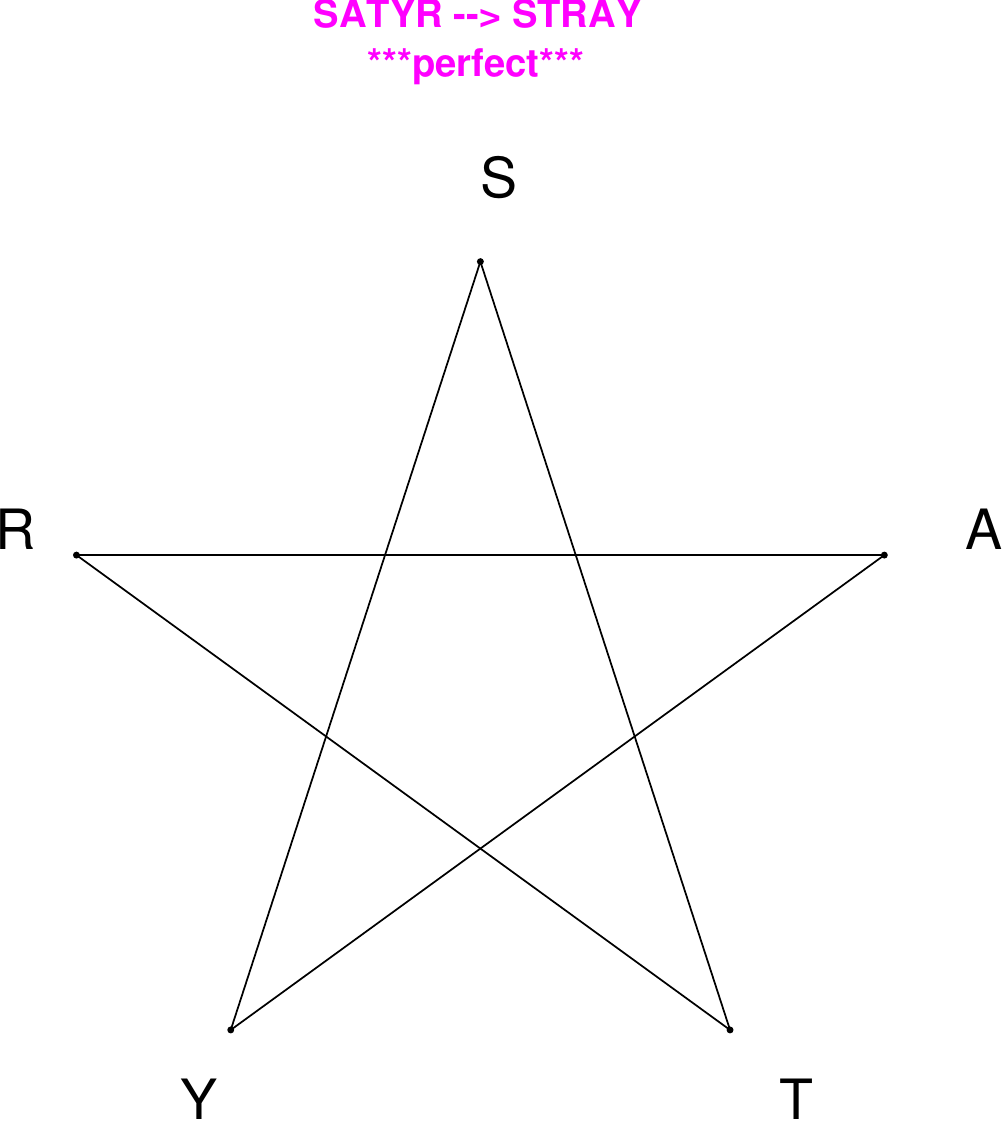}
\end{subfigure}
\hfill
\begin{subfigure}[T]{0.19\textwidth}
\centering
\includegraphics[width=\textwidth]{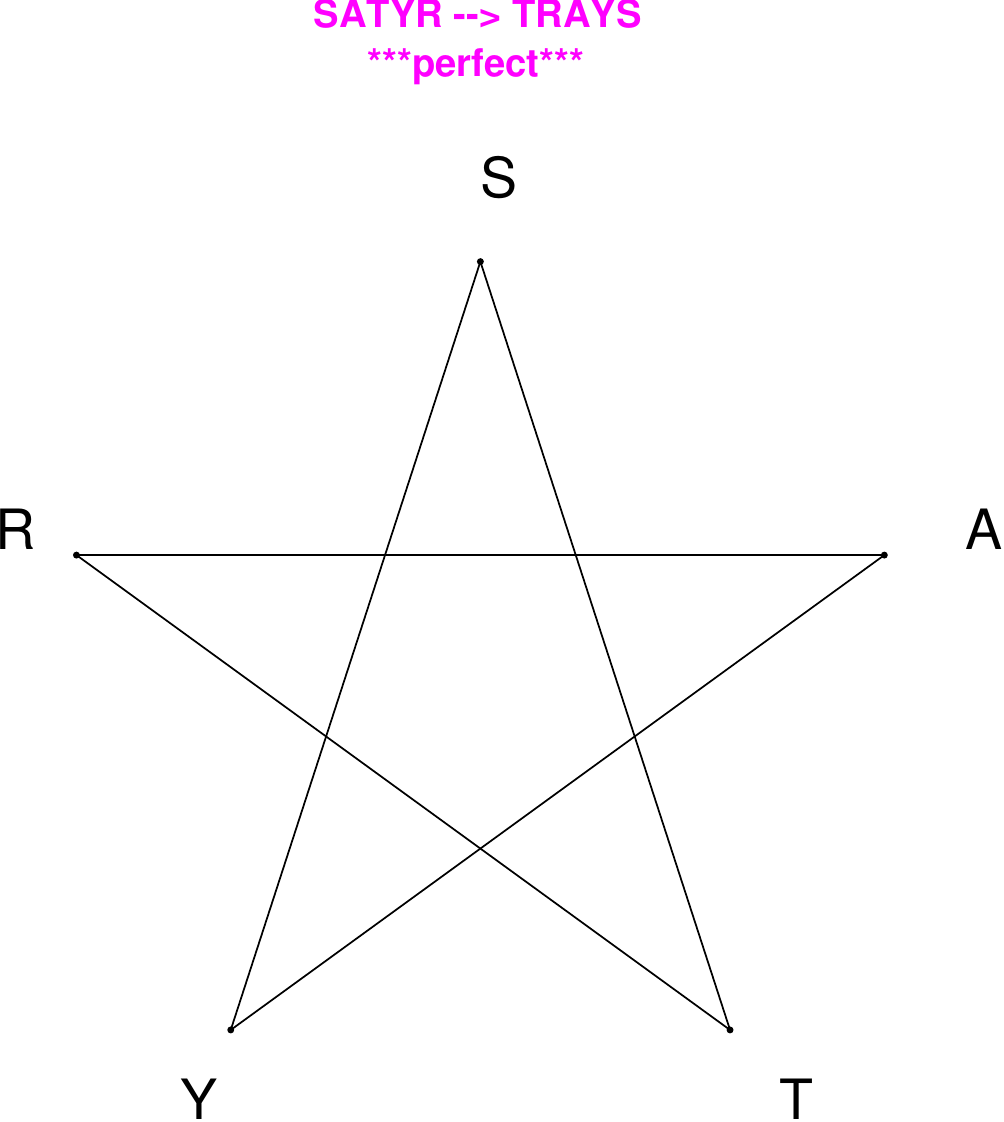}
\end{subfigure}
\end{figure}

\begin{figure}[H]
\centering
\begin{subfigure}[T]{0.19\textwidth}
\centering
\includegraphics[width=\textwidth]{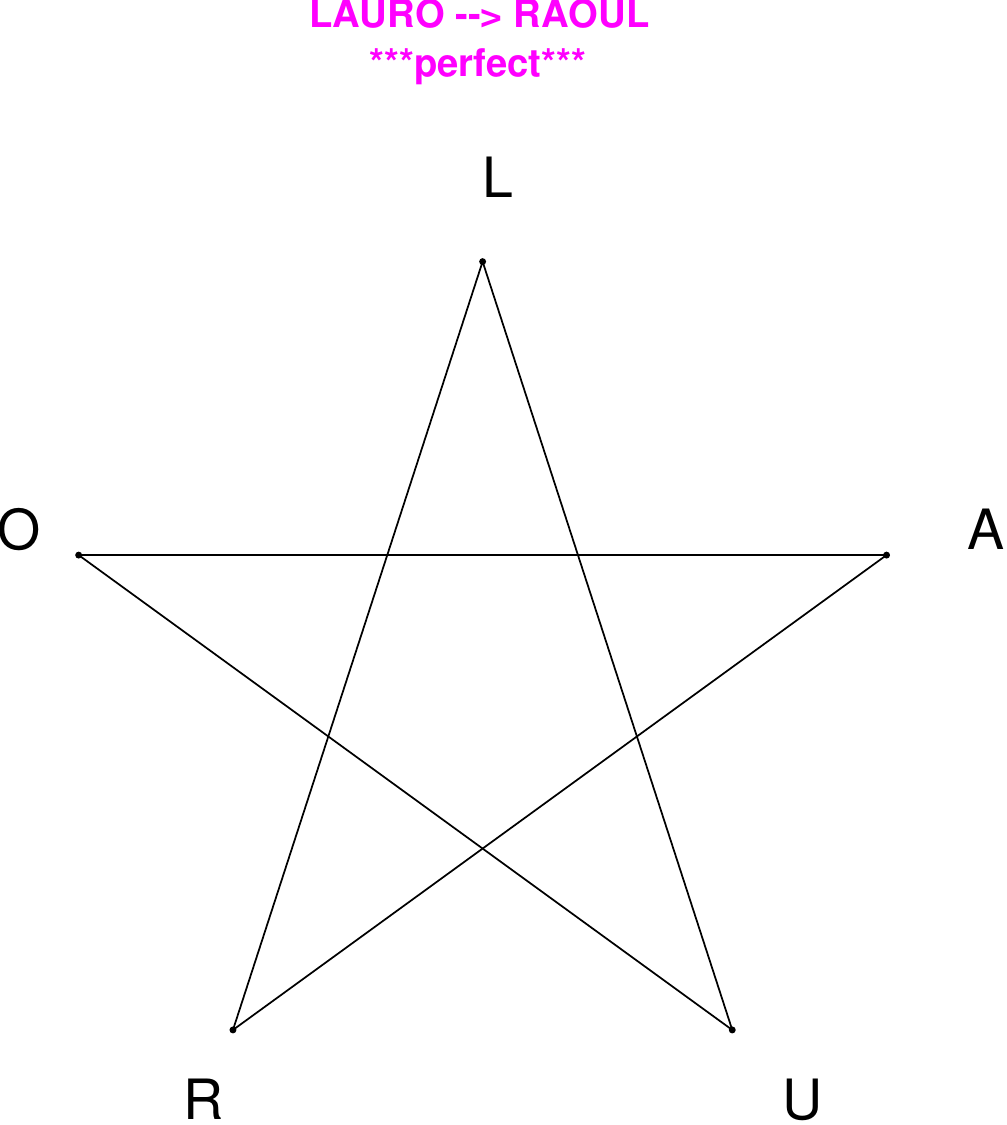}
\end{subfigure}
\hfill
\begin{subfigure}[T]{0.19\textwidth}
\centering
\includegraphics[width=\textwidth]{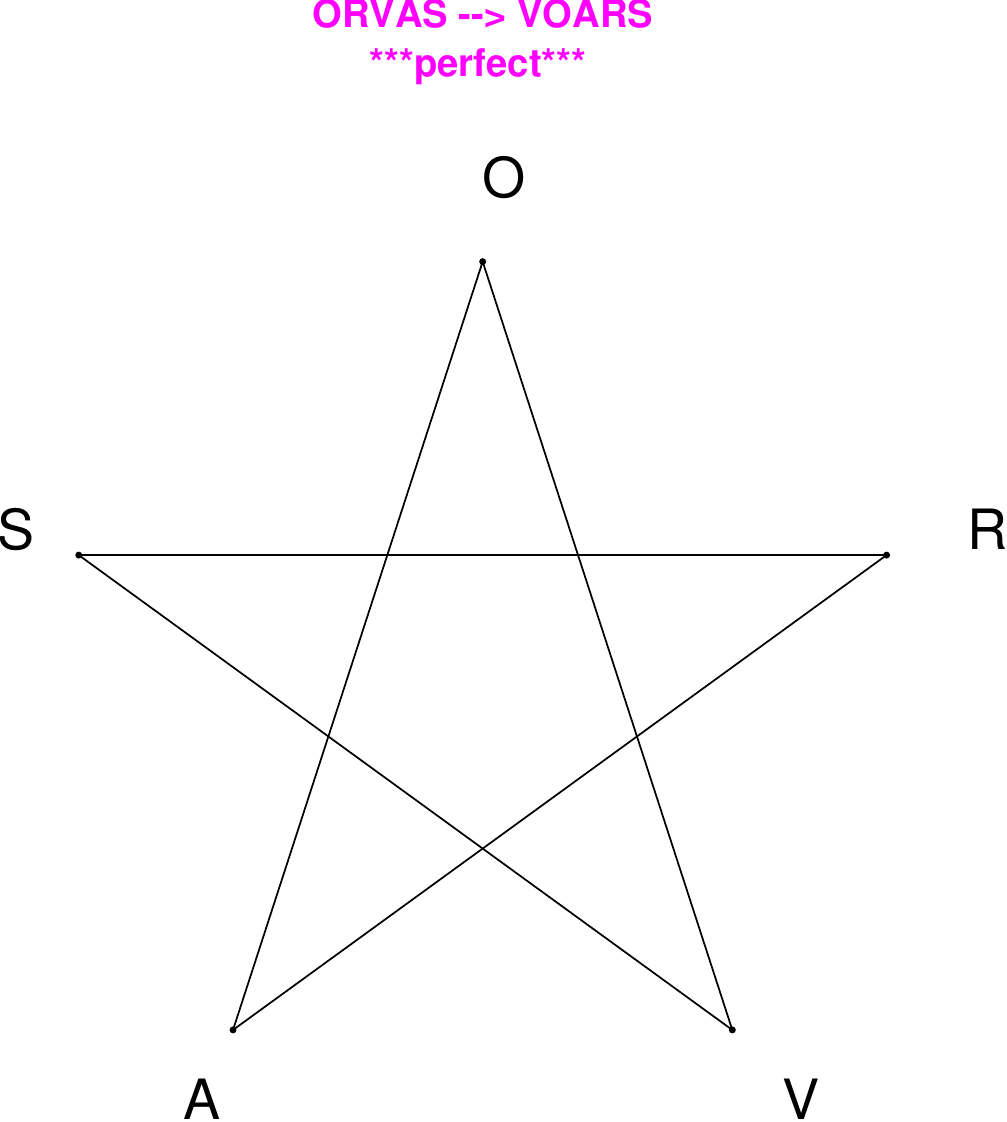}
\end{subfigure}
\hfill
\begin{subfigure}[T]{0.19\textwidth}
\centering
\includegraphics[width=\textwidth]{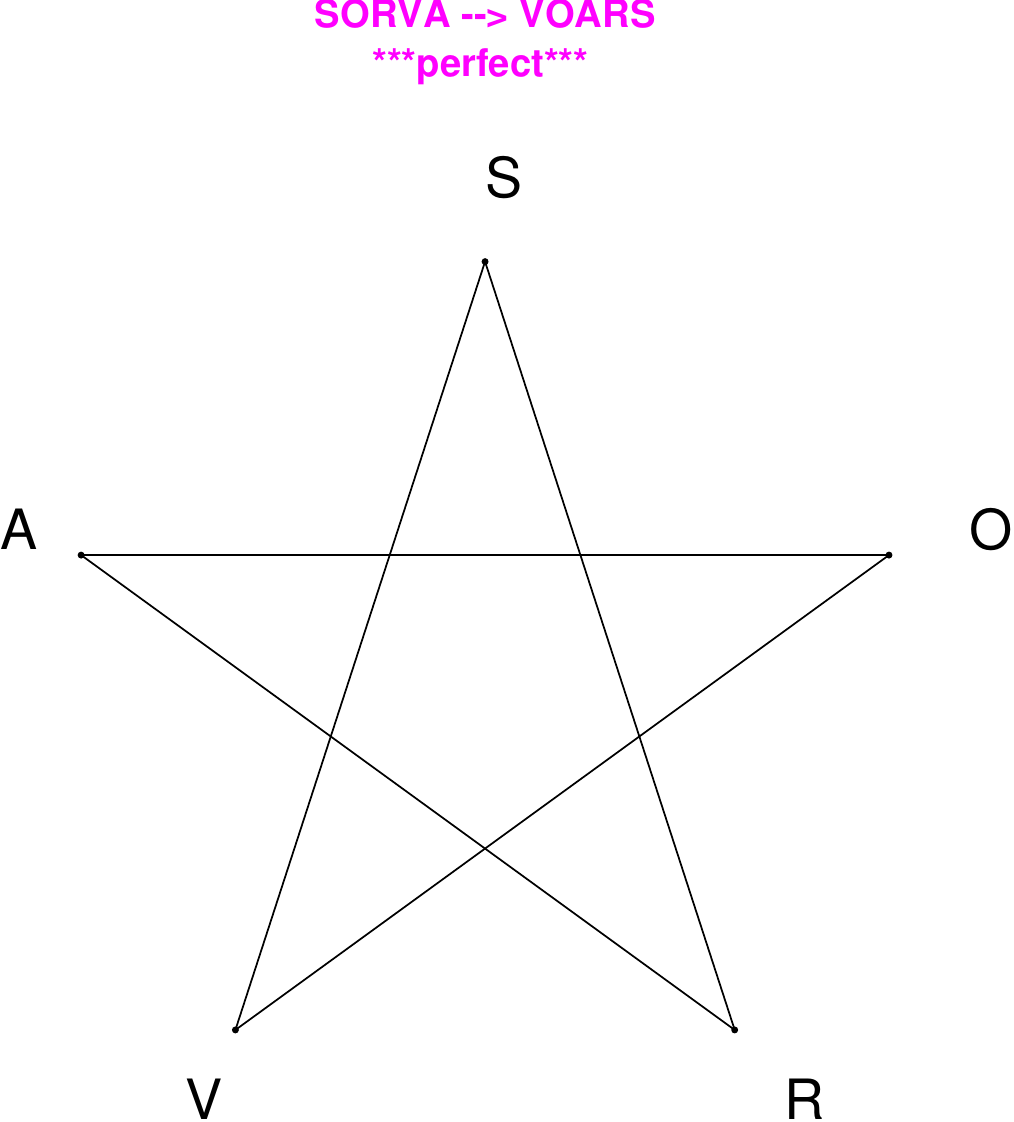}
\end{subfigure}
\hfill
\begin{subfigure}[T]{0.19\textwidth}
\centering
\includegraphics[width=\textwidth]{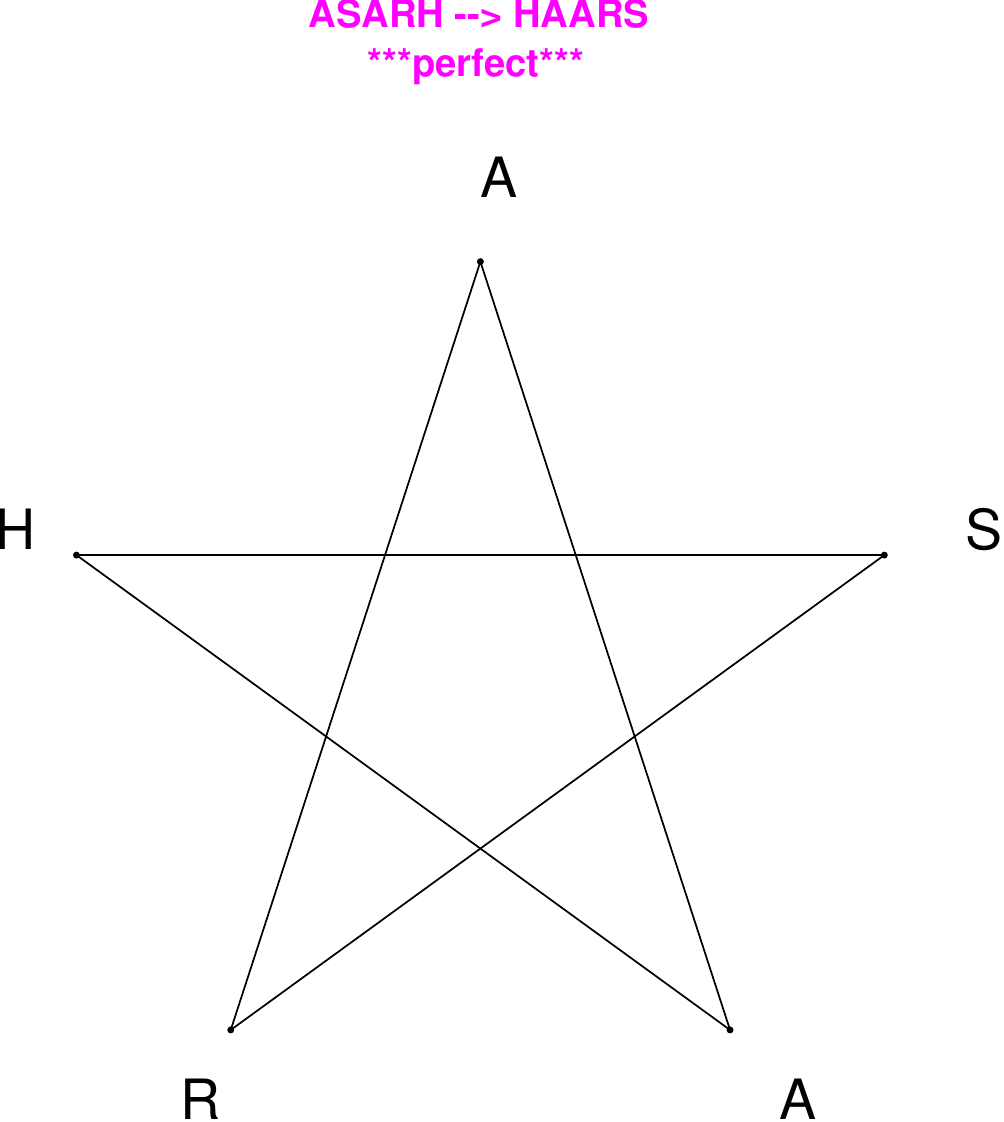}
\end{subfigure}
\hfill
\begin{subfigure}[T]{0.19\textwidth}
\centering
\includegraphics[width=\textwidth]{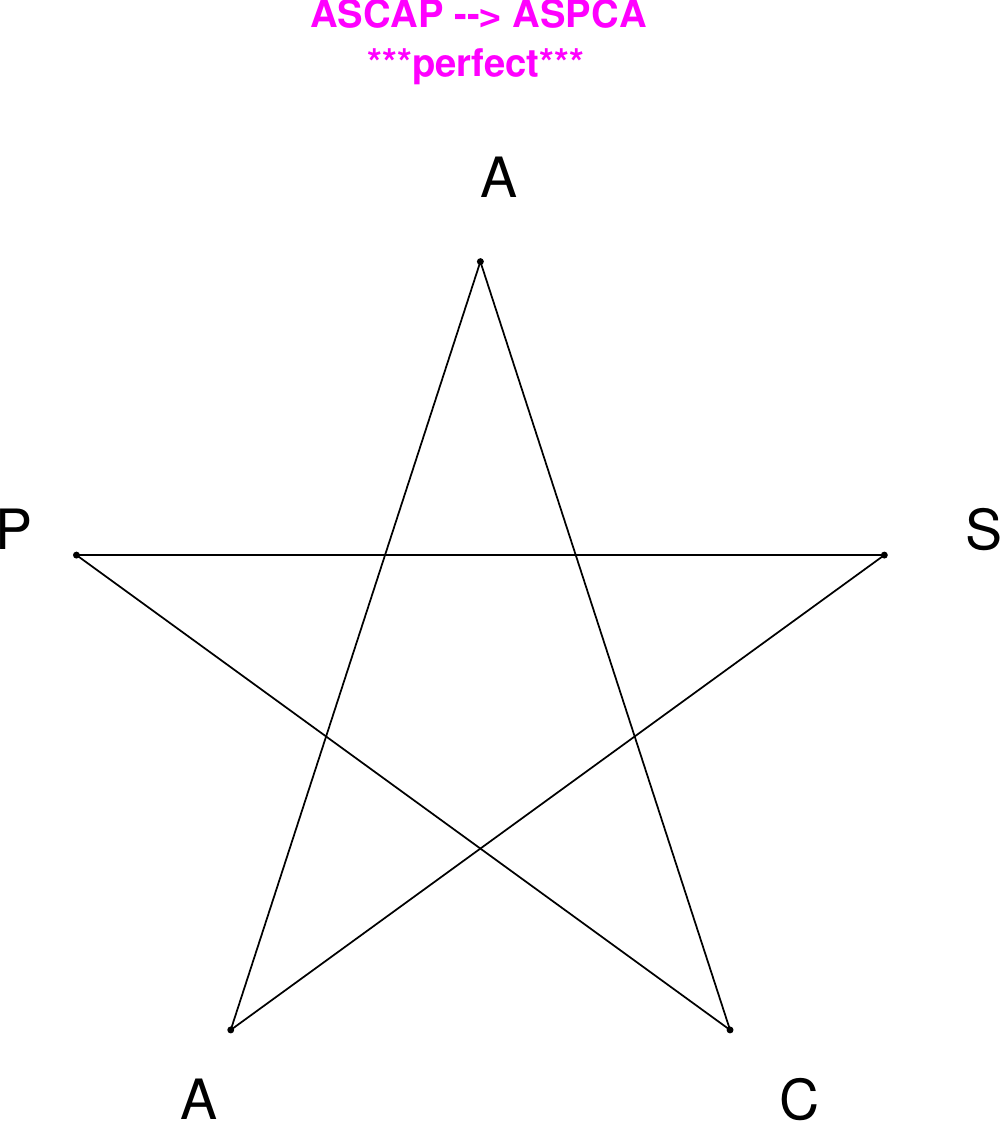}
\end{subfigure}
\end{figure}

\begin{figure}[H]
\centering
\begin{subfigure}[T]{0.19\textwidth}
\centering
\includegraphics[width=\textwidth]{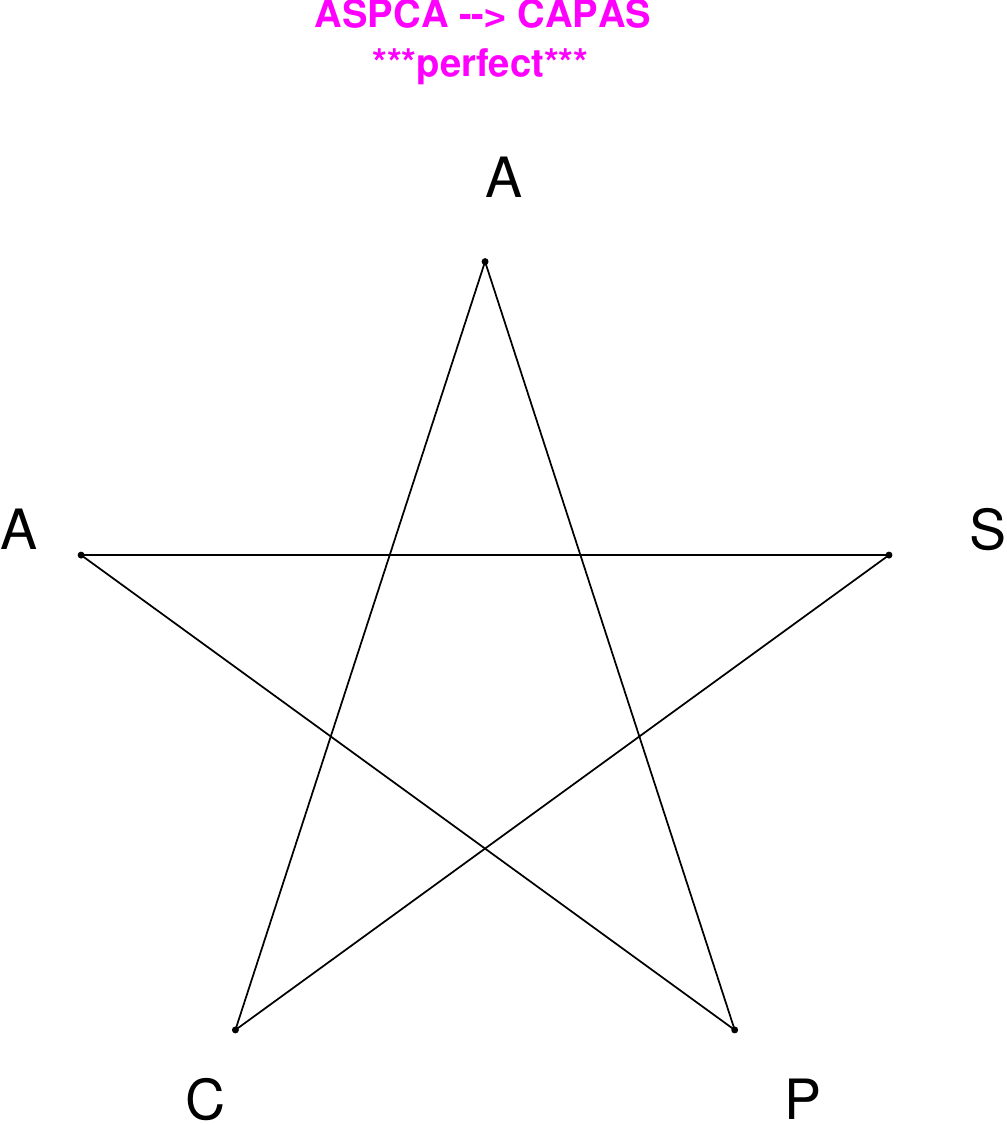}
\end{subfigure}
\hfill
\begin{subfigure}[T]{0.19\textwidth}
\centering
\includegraphics[width=\textwidth]{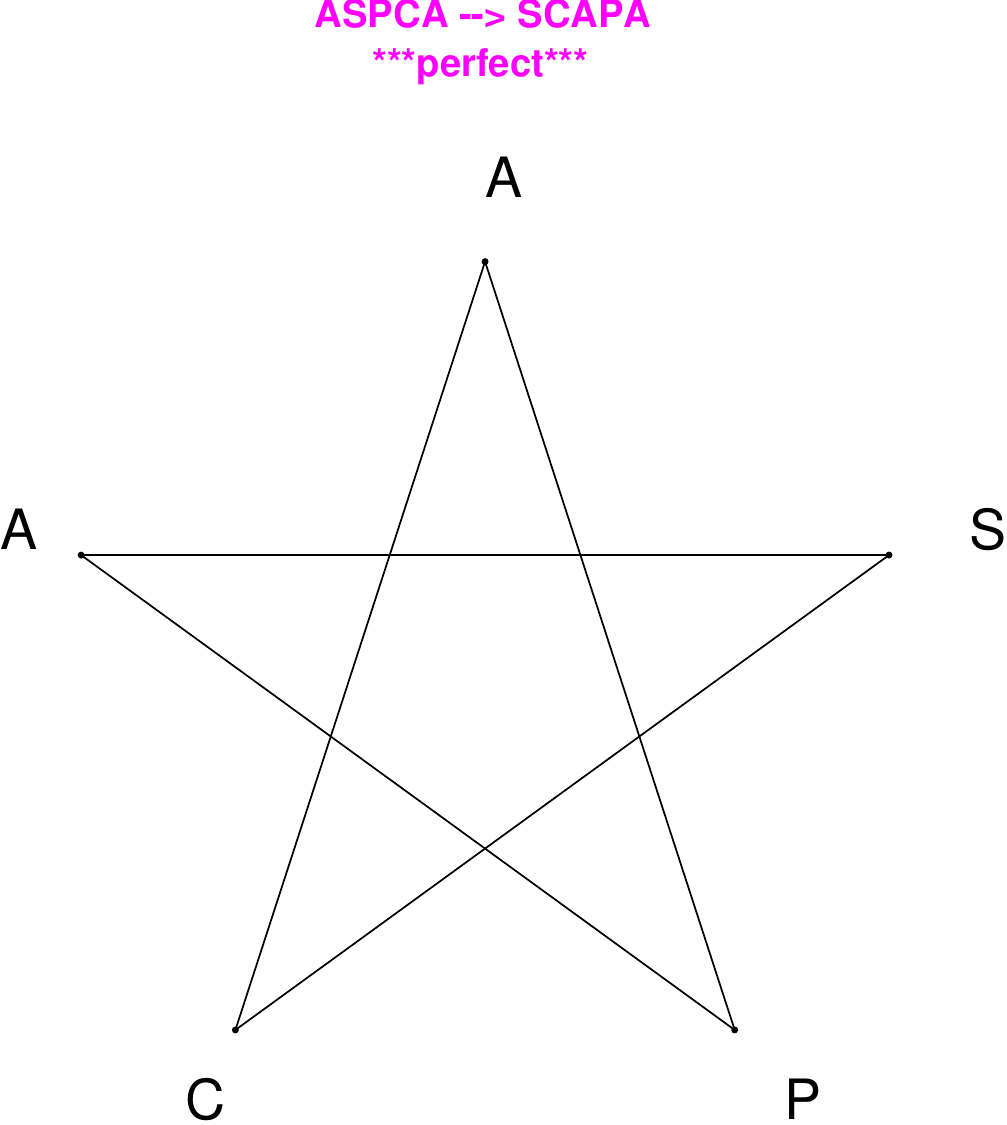}
\end{subfigure}
\hfill
\begin{subfigure}[T]{0.19\textwidth}
\centering
\includegraphics[width=\textwidth]{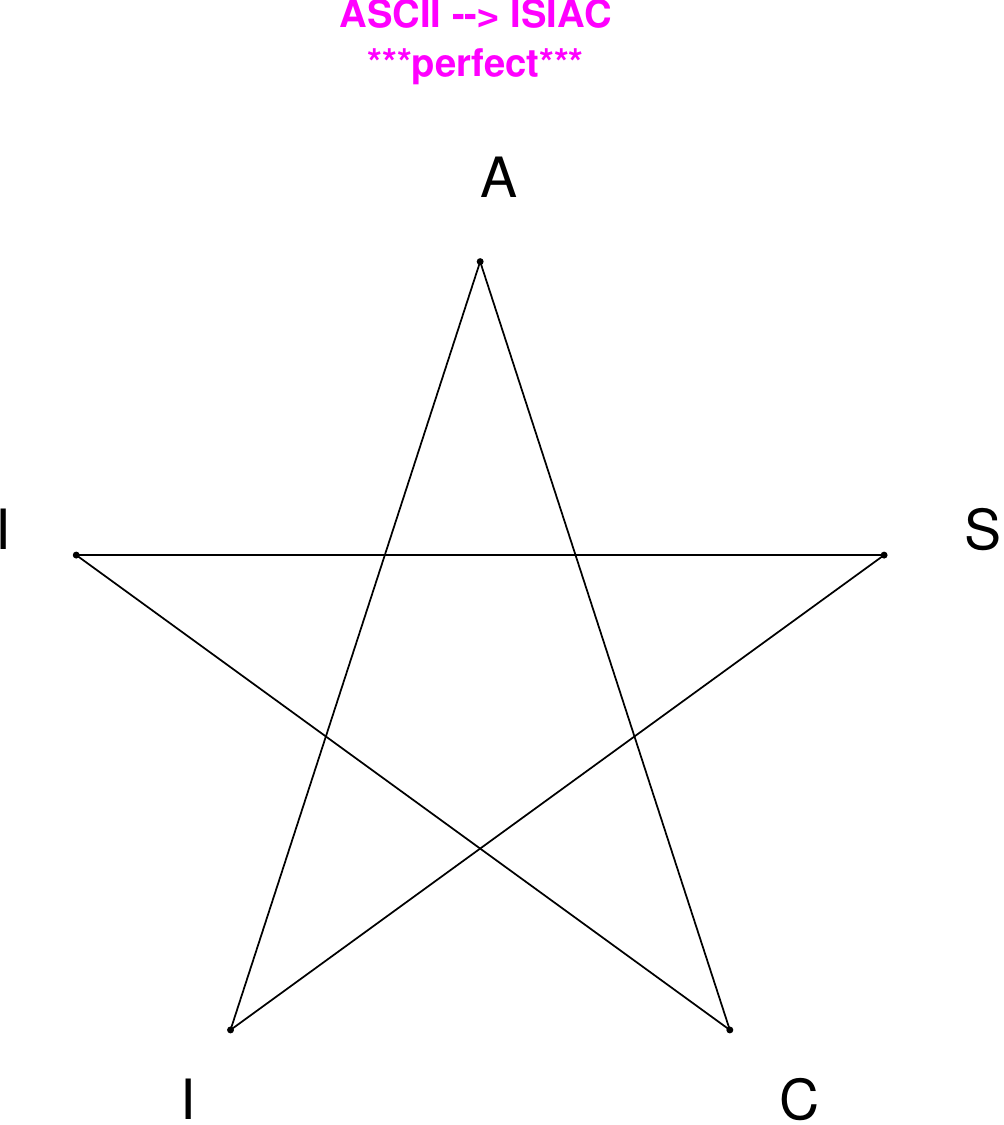}
\end{subfigure}
\hfill
\begin{subfigure}[T]{0.19\textwidth}
\centering
\includegraphics[width=\textwidth]{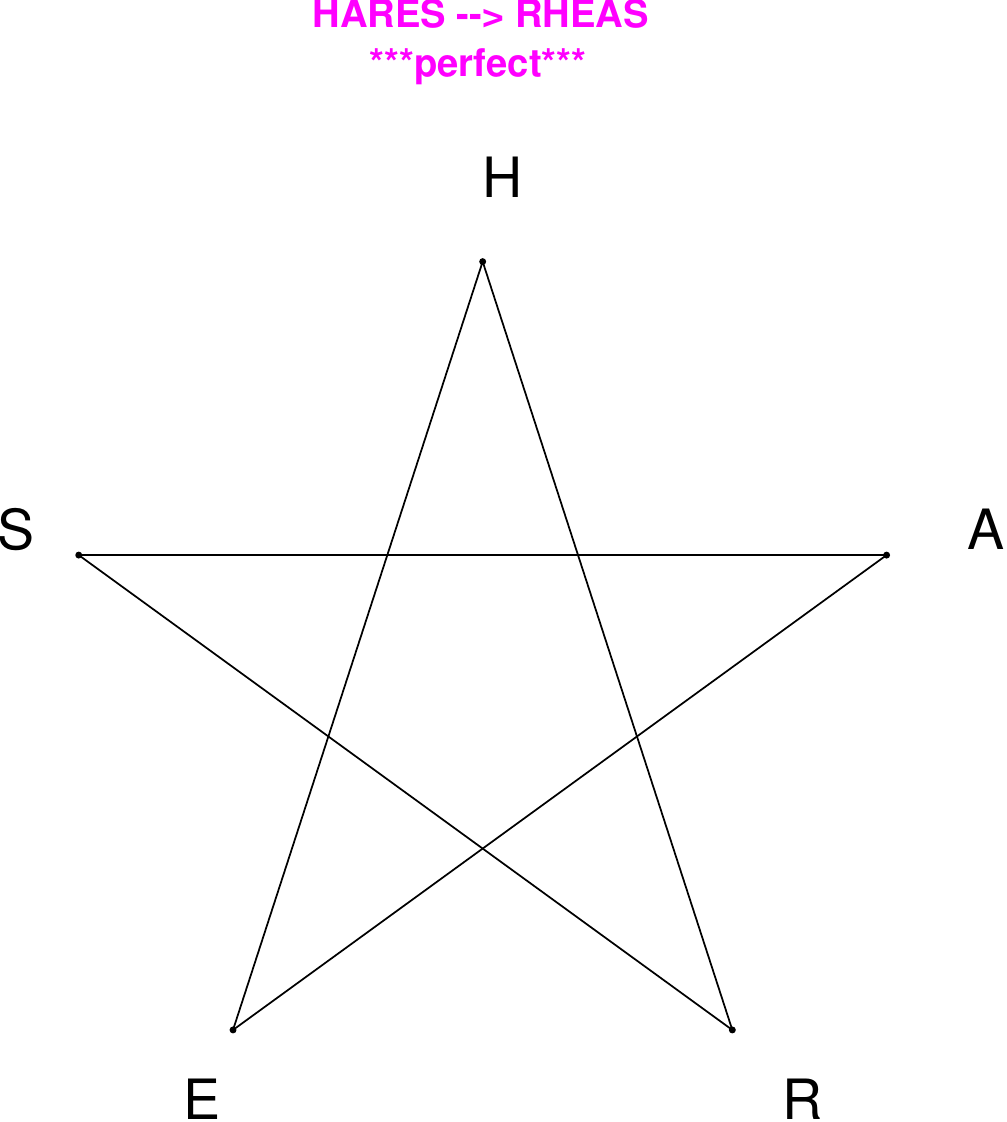}
\end{subfigure}
\hfill
\begin{subfigure}[T]{0.19\textwidth}
\centering
\includegraphics[width=\textwidth]{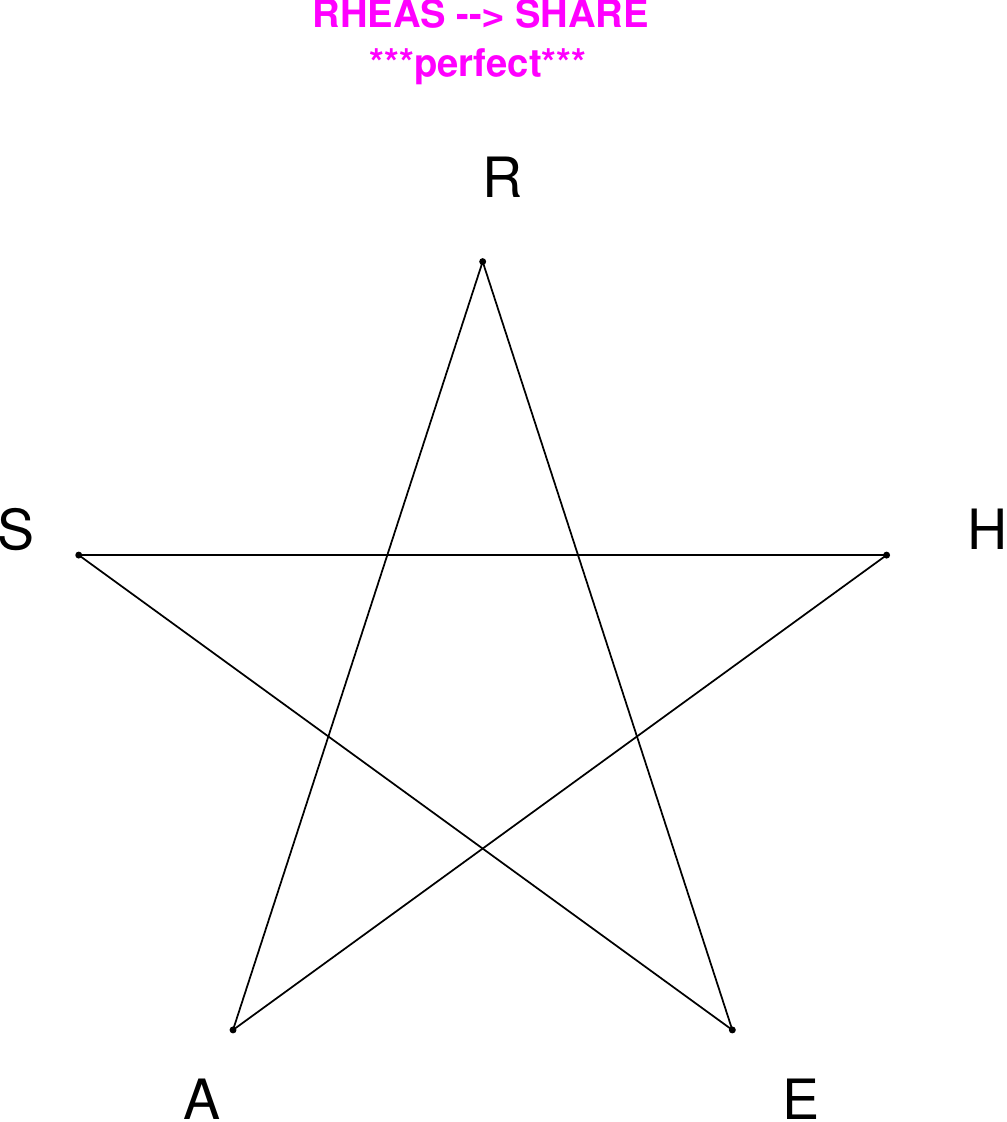}
\end{subfigure}
\end{figure}

\begin{figure}[H]
\centering
\begin{subfigure}[T]{0.19\textwidth}
\centering
\includegraphics[width=\textwidth]{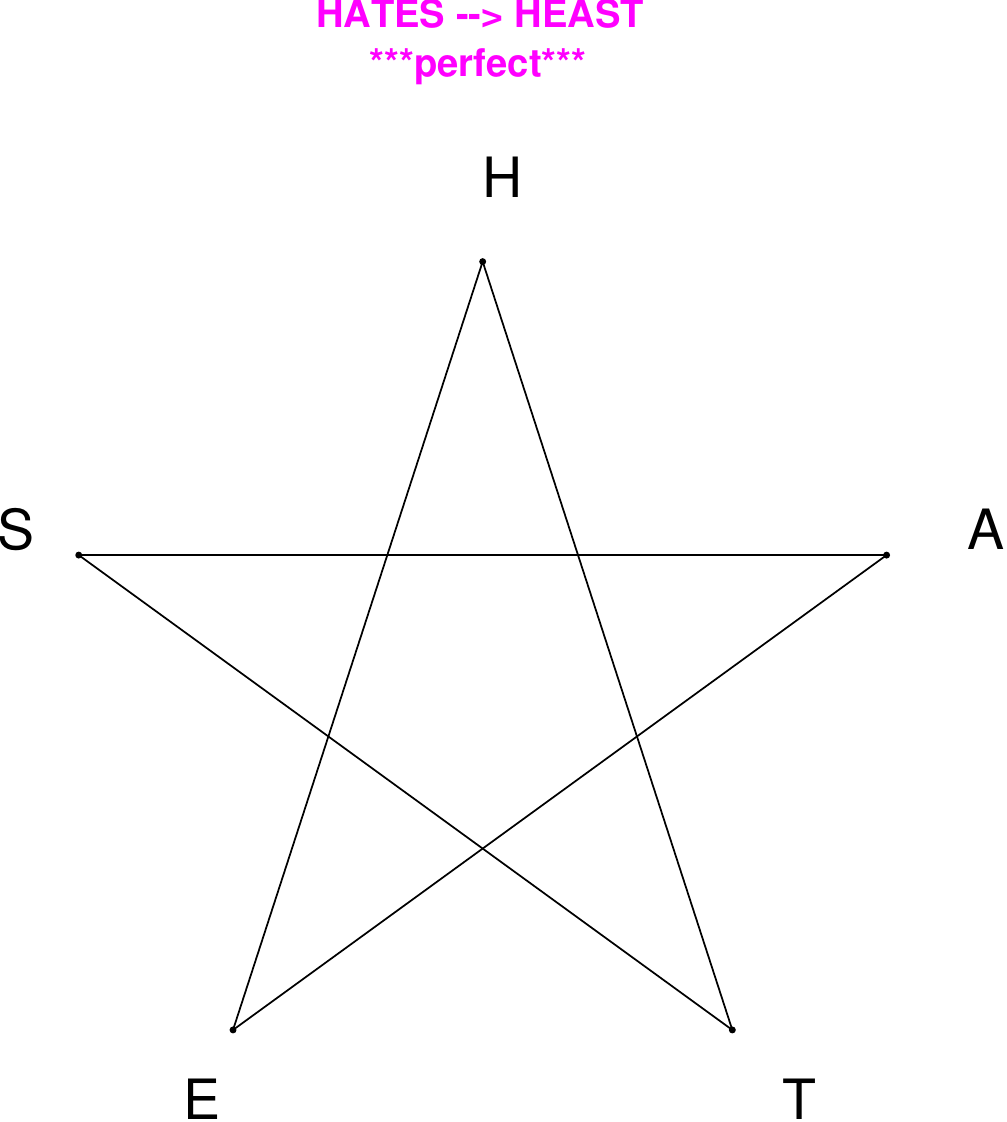}
\end{subfigure}
\hfill
\begin{subfigure}[T]{0.19\textwidth}
\centering
\includegraphics[width=\textwidth]{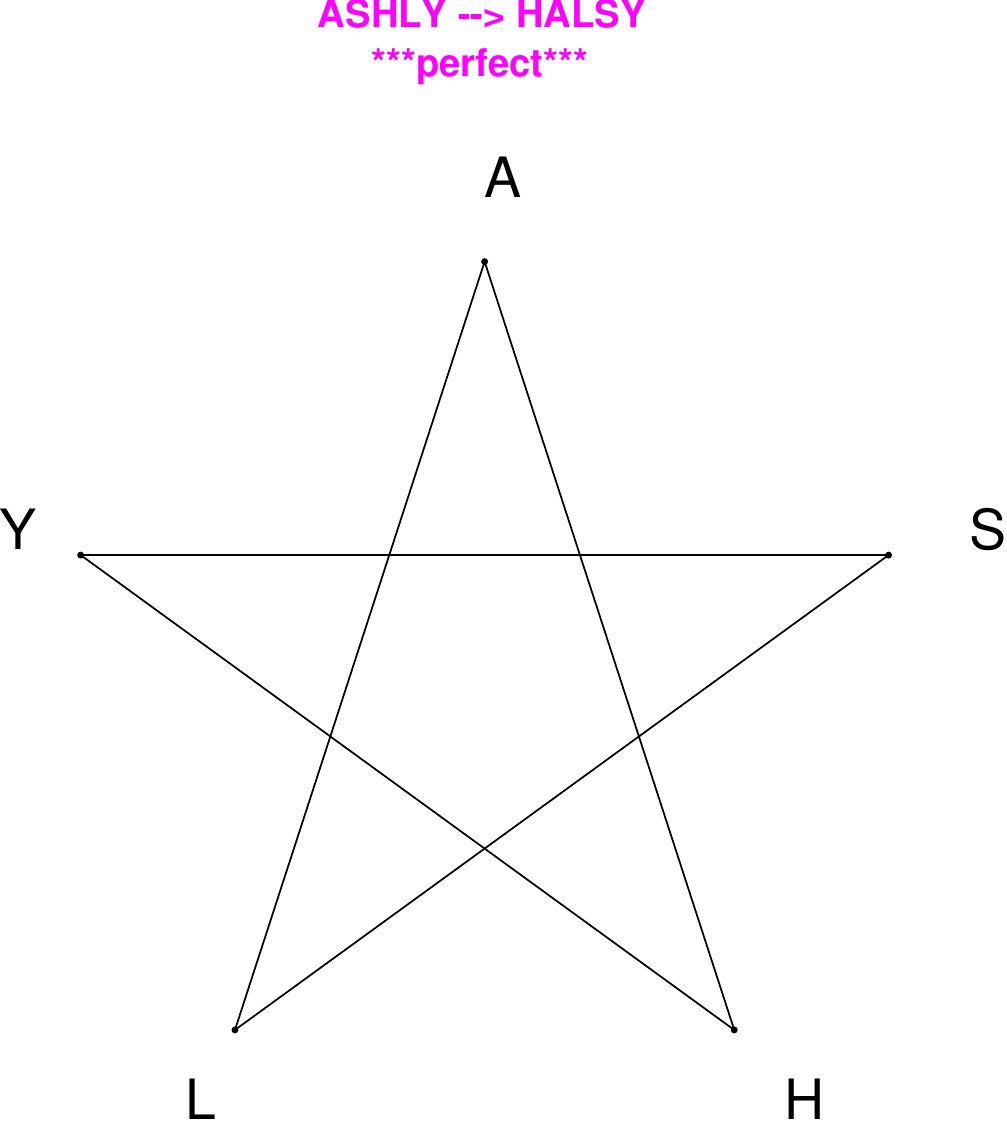}
\end{subfigure}
\hfill
\begin{subfigure}[T]{0.19\textwidth}
\centering
\includegraphics[width=\textwidth]{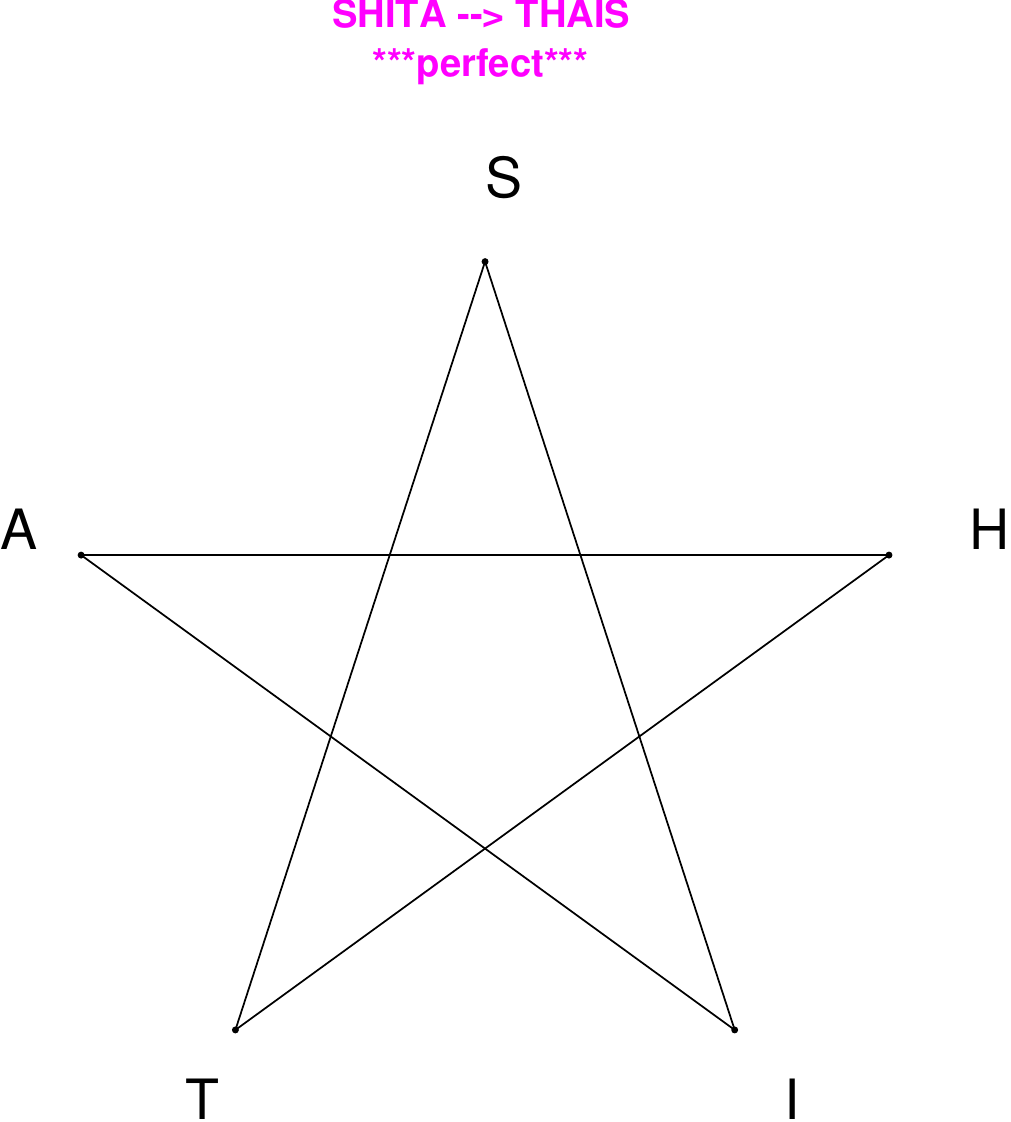}
\end{subfigure}
\hfill
\begin{subfigure}[T]{0.19\textwidth}
\centering
\includegraphics[width=\textwidth]{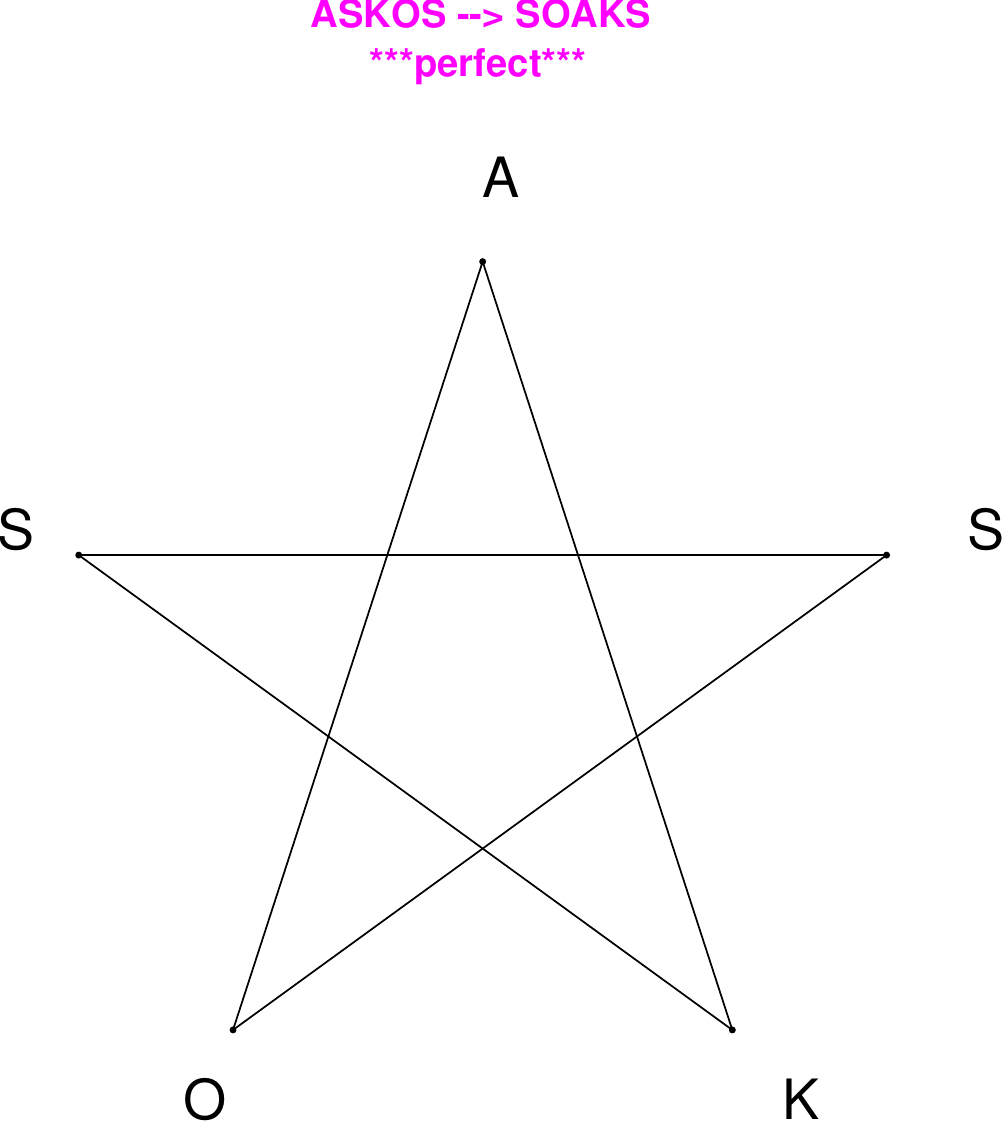}
\end{subfigure}
\hfill
\begin{subfigure}[T]{0.19\textwidth}
\centering
\includegraphics[width=\textwidth]{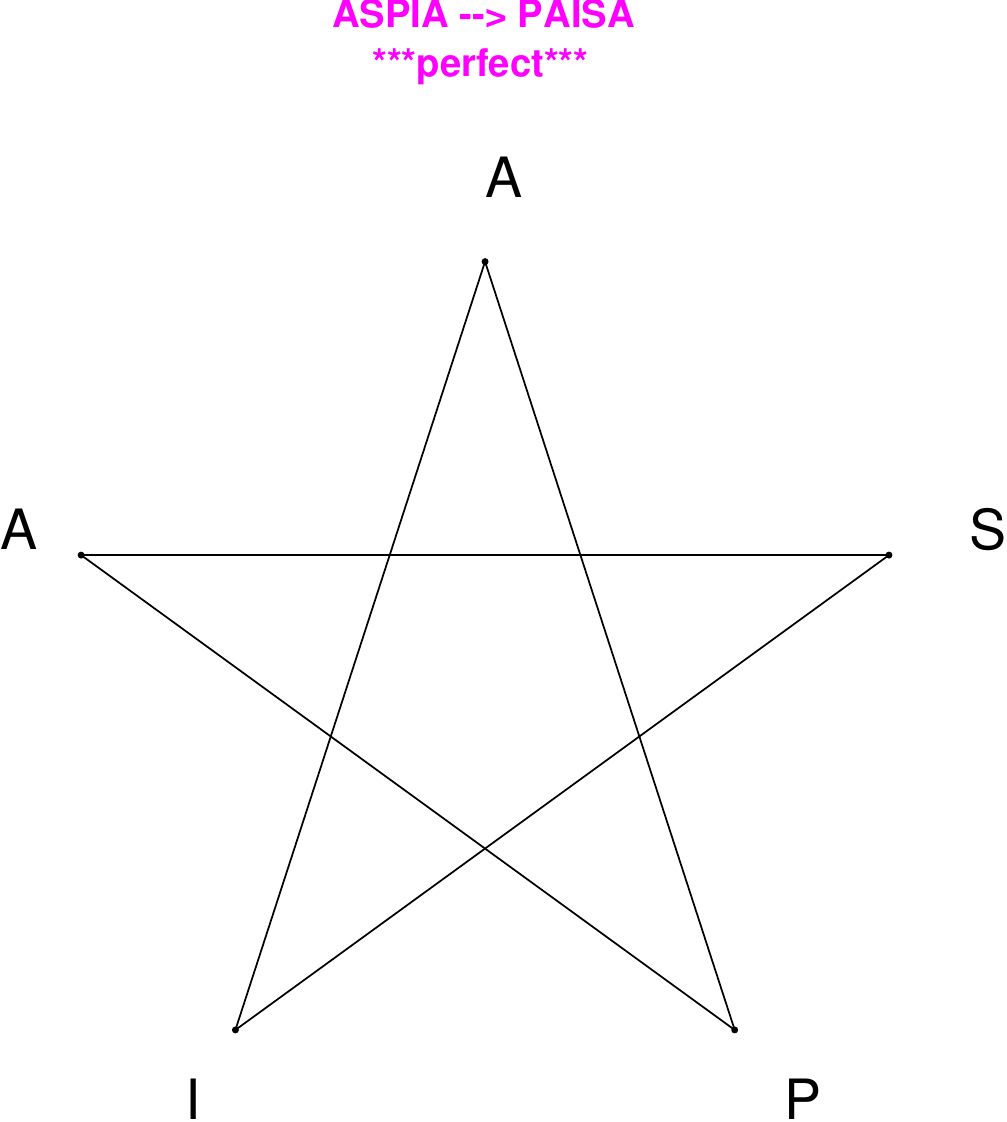}
\end{subfigure}
\end{figure}

\begin{figure}[H]
\centering
\begin{subfigure}[T]{0.19\textwidth}
\centering
\includegraphics[width=\textwidth]{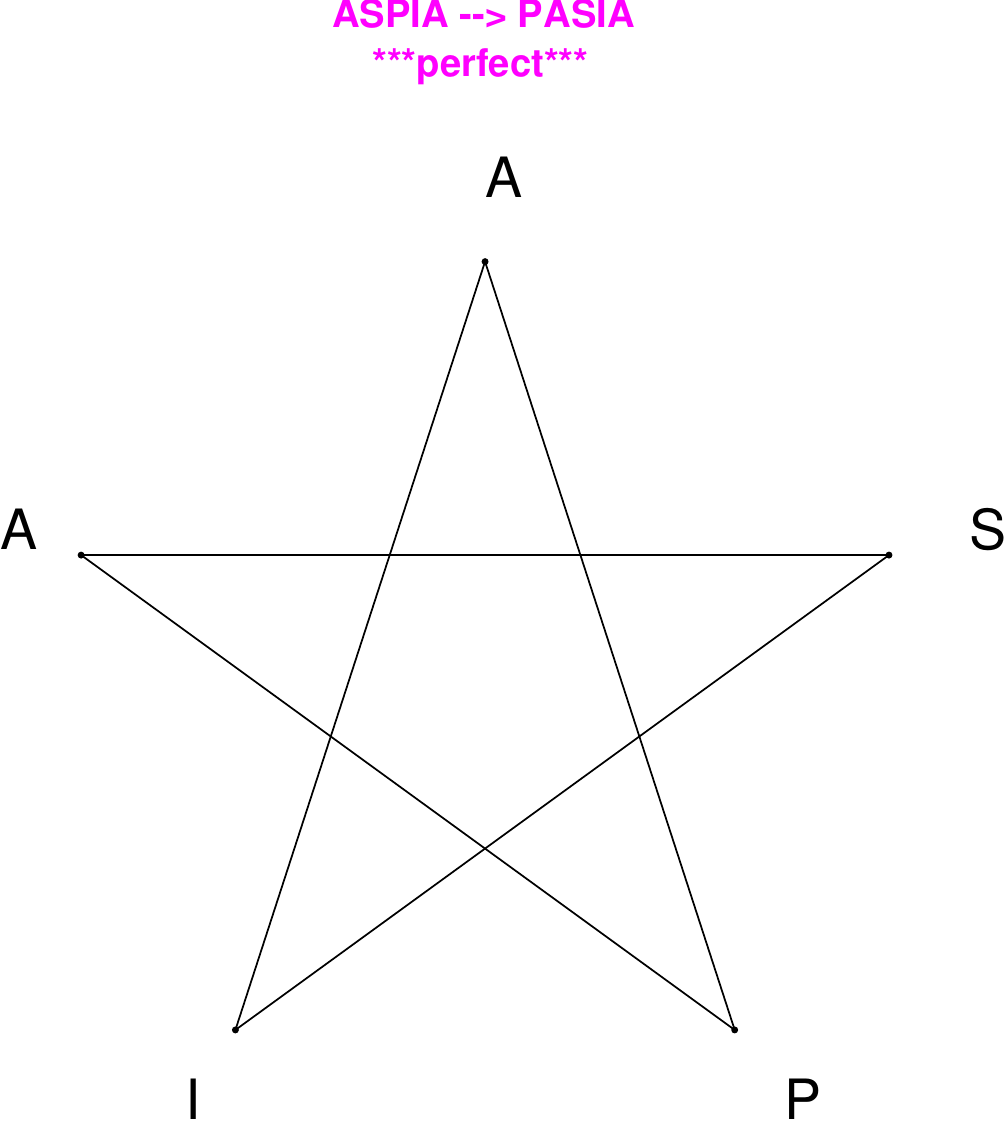}
\end{subfigure}
\hfill
\begin{subfigure}[T]{0.19\textwidth}
\centering
\includegraphics[width=\textwidth]{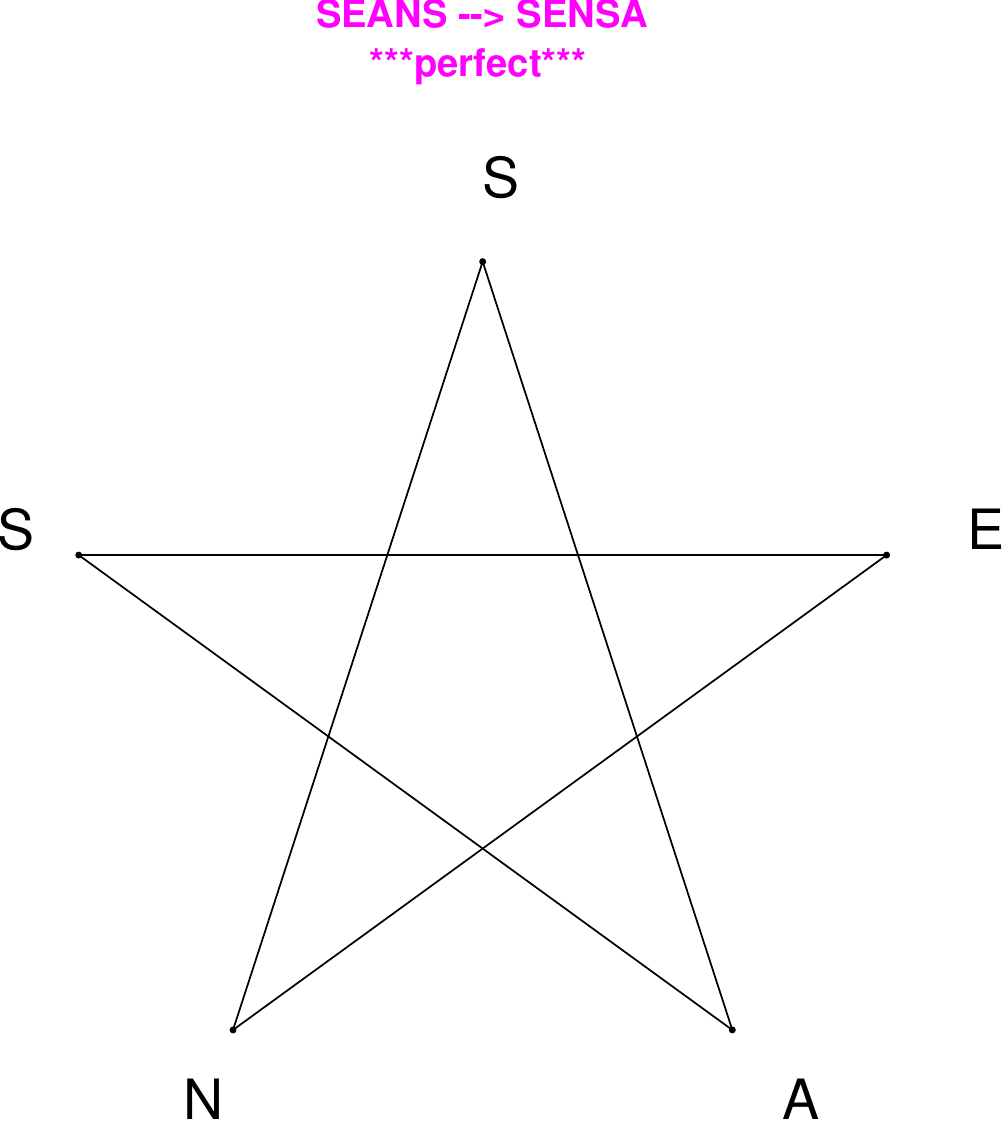}
\end{subfigure}
\hfill
\begin{subfigure}[T]{0.19\textwidth}
\centering
\includegraphics[width=\textwidth]{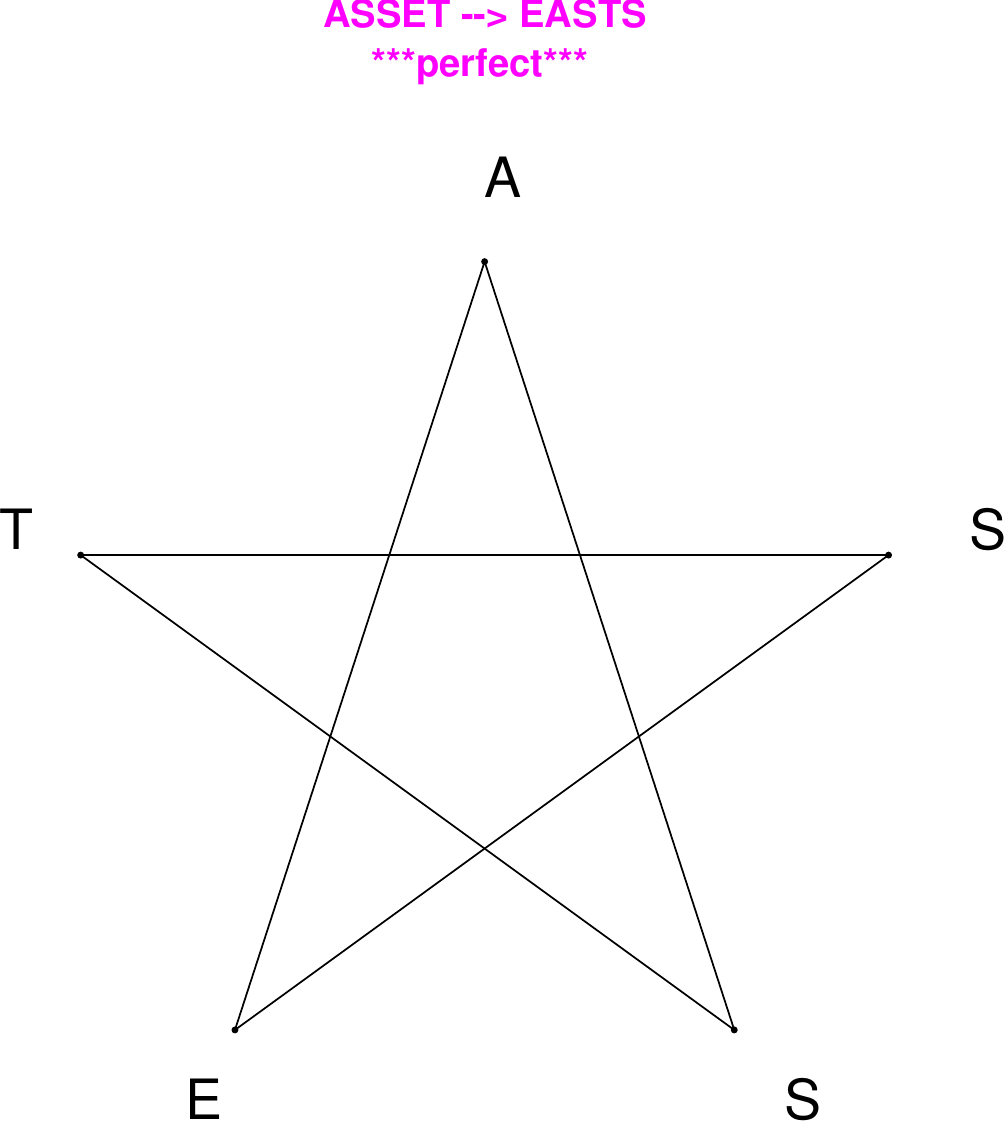}
\end{subfigure}
\hfill
\begin{subfigure}[T]{0.19\textwidth}
\centering
\includegraphics[width=\textwidth]{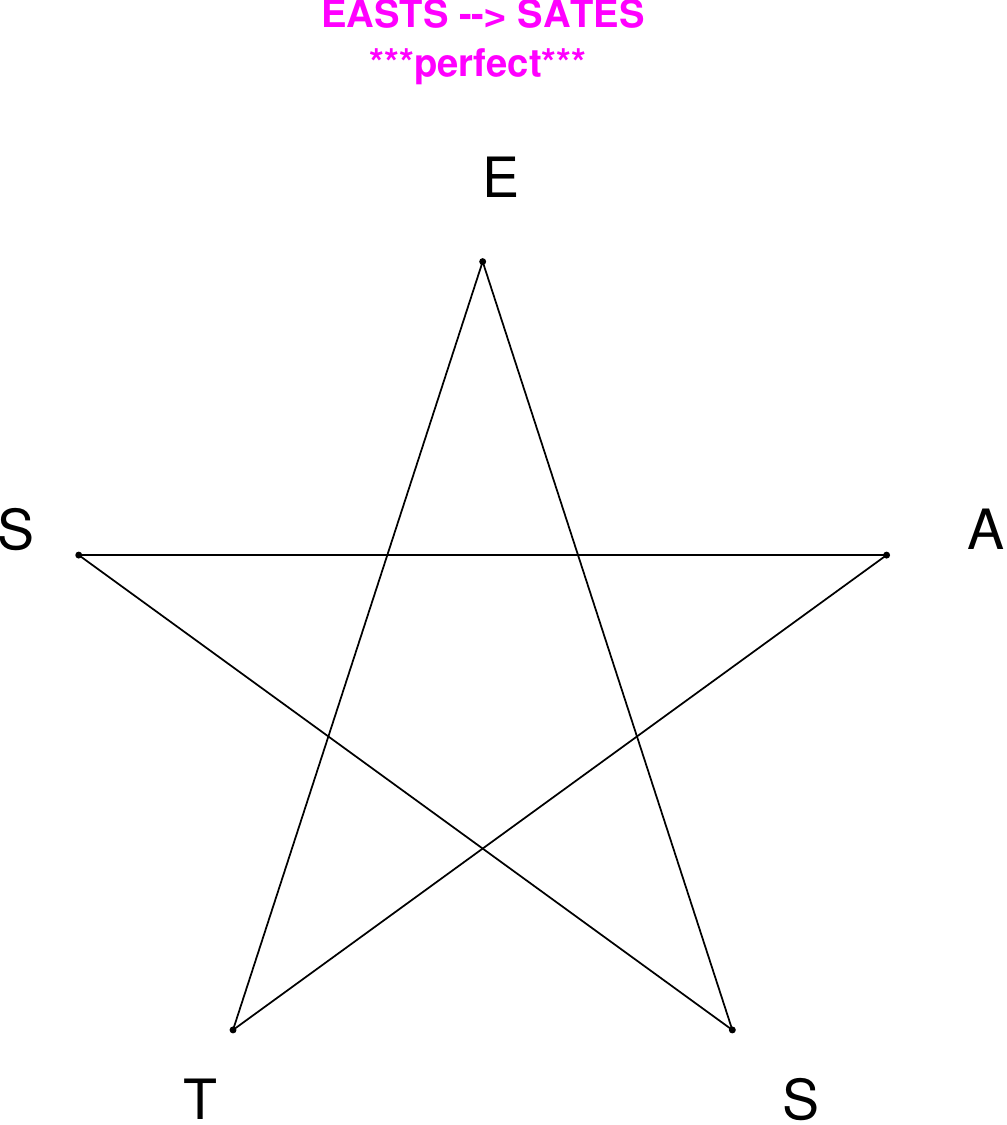}
\end{subfigure}
\hfill
\begin{subfigure}[T]{0.19\textwidth}
\centering
\includegraphics[width=\textwidth]{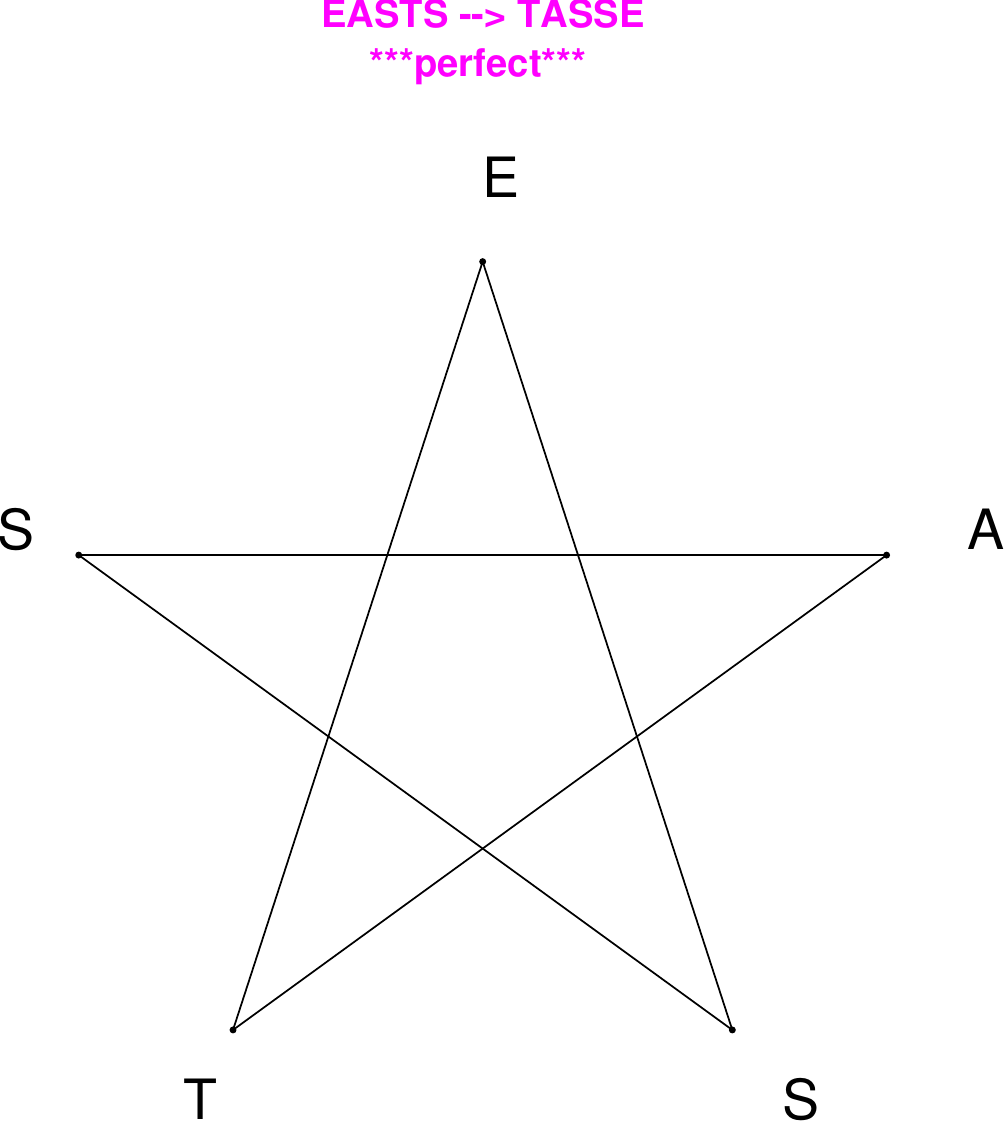}
\end{subfigure}
\end{figure}

\begin{figure}[H]
\centering
\begin{subfigure}[T]{0.19\textwidth}
\centering
\includegraphics[width=\textwidth]{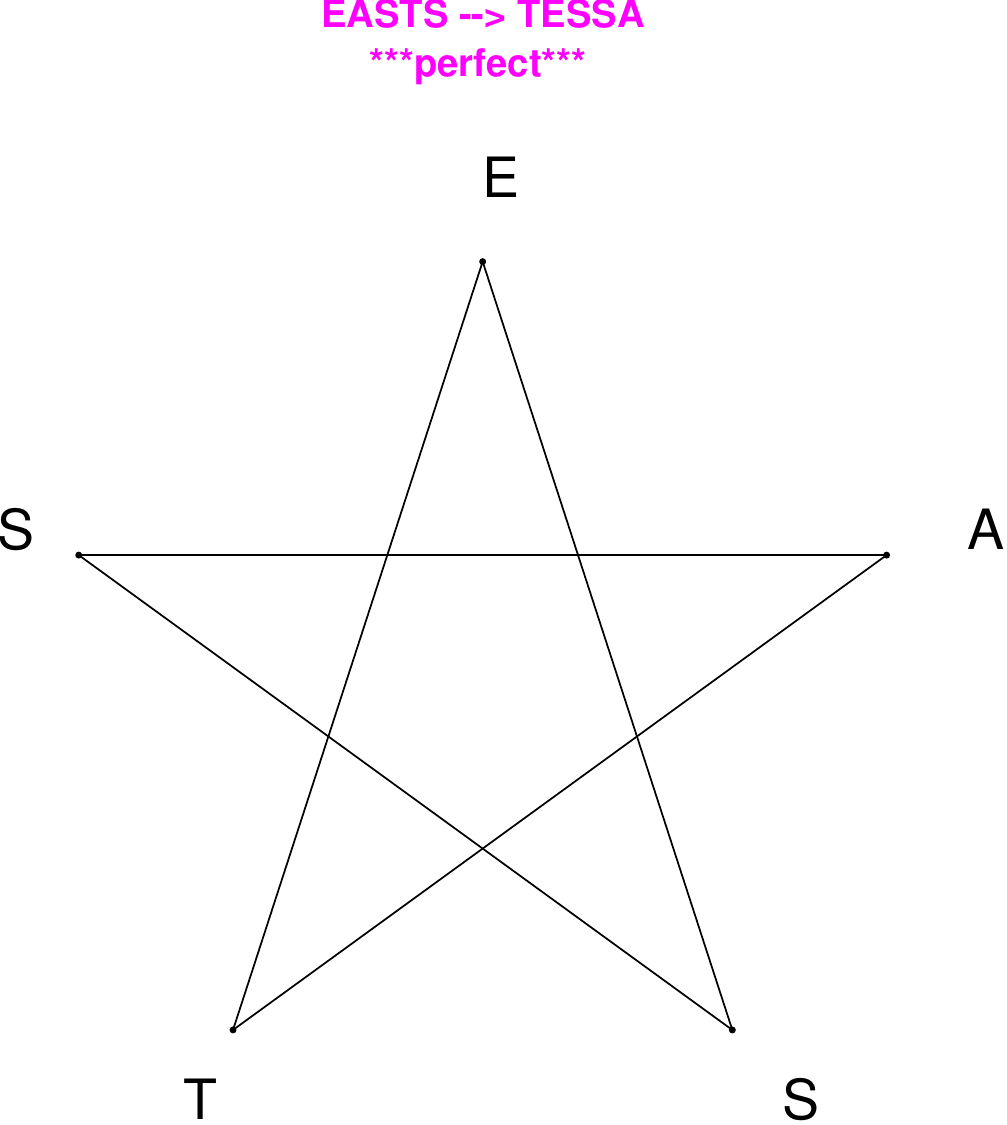}
\end{subfigure}
\hfill
\begin{subfigure}[T]{0.19\textwidth}
\centering
\includegraphics[width=\textwidth]{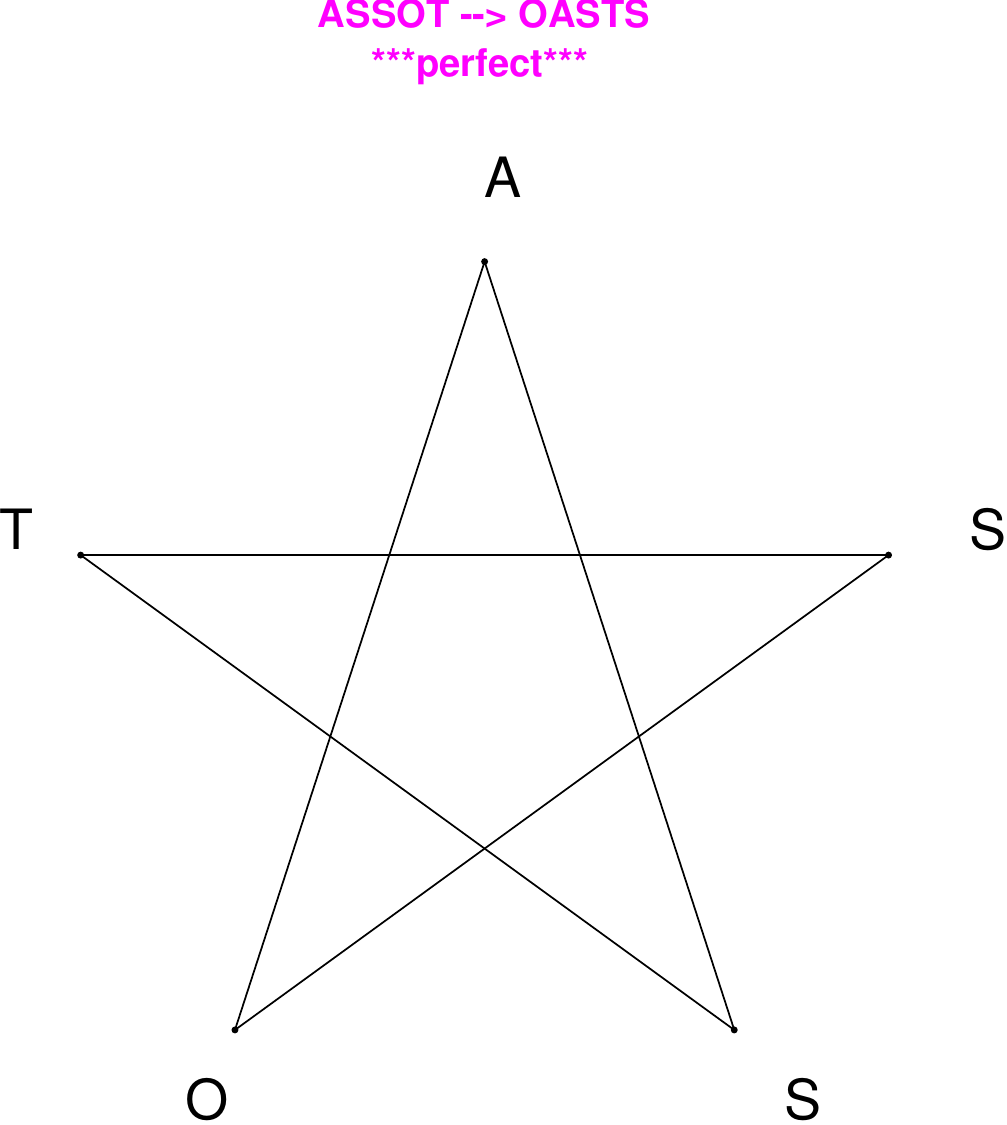}
\end{subfigure}
\hfill
\begin{subfigure}[T]{0.19\textwidth}
\centering
\includegraphics[width=\textwidth]{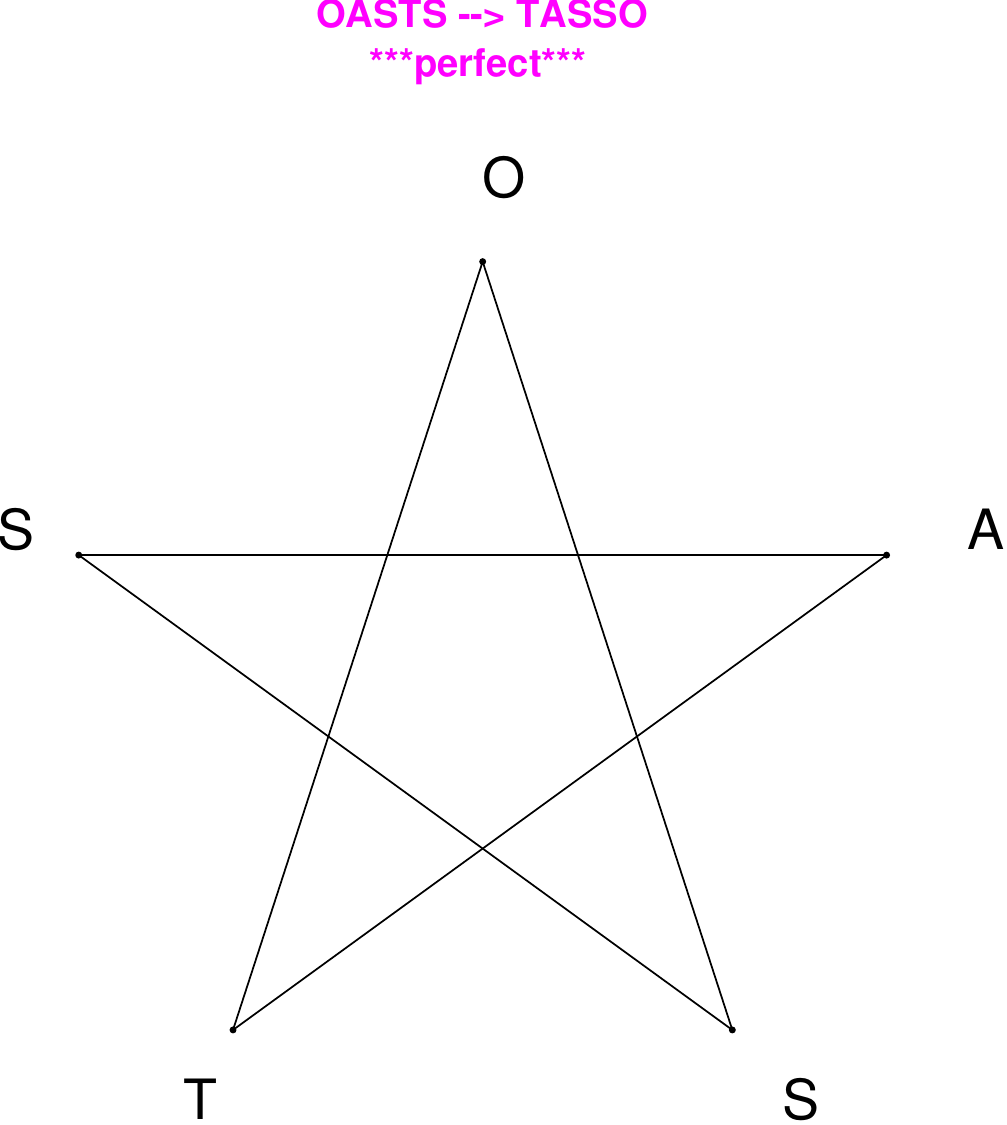}
\end{subfigure}
\hfill
\begin{subfigure}[T]{0.19\textwidth}
\centering
\includegraphics[width=\textwidth]{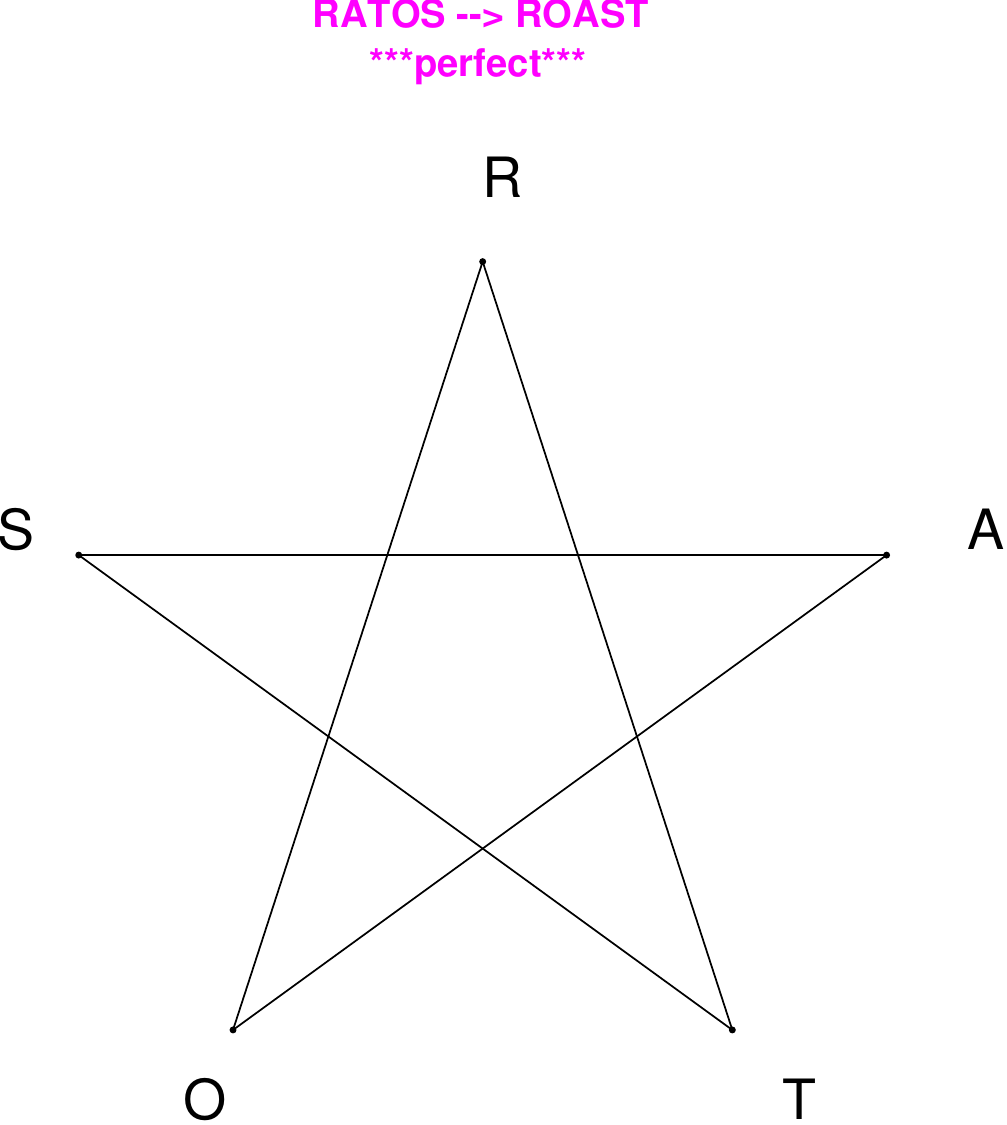}
\end{subfigure}
\hfill
\begin{subfigure}[T]{0.19\textwidth}
\centering
\includegraphics[width=\textwidth]{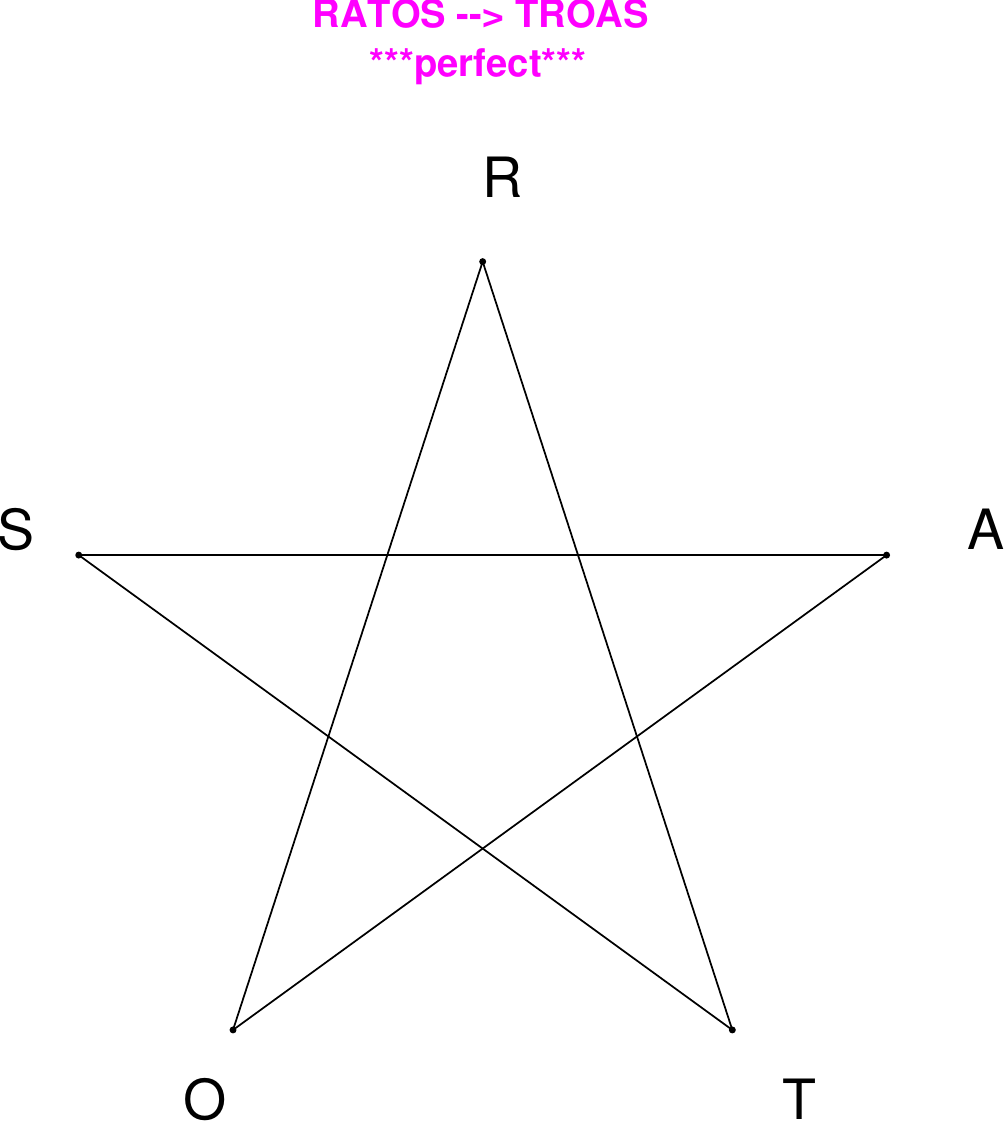}
\end{subfigure}
\end{figure}

\begin{figure}[H]
\centering
\begin{subfigure}[T]{0.19\textwidth}
\centering
\includegraphics[width=\textwidth]{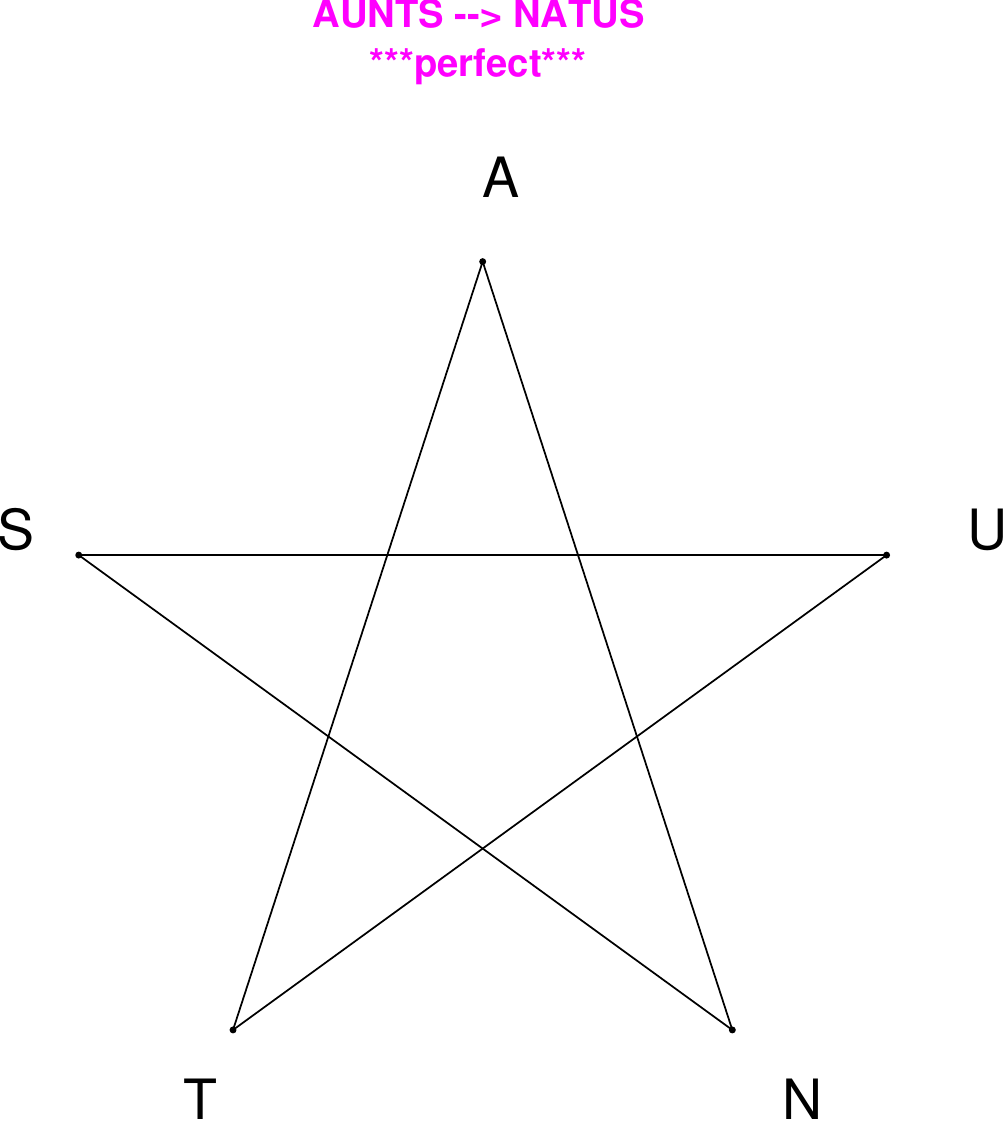}
\end{subfigure}
\hfill
\begin{subfigure}[T]{0.19\textwidth}
\centering
\includegraphics[width=\textwidth]{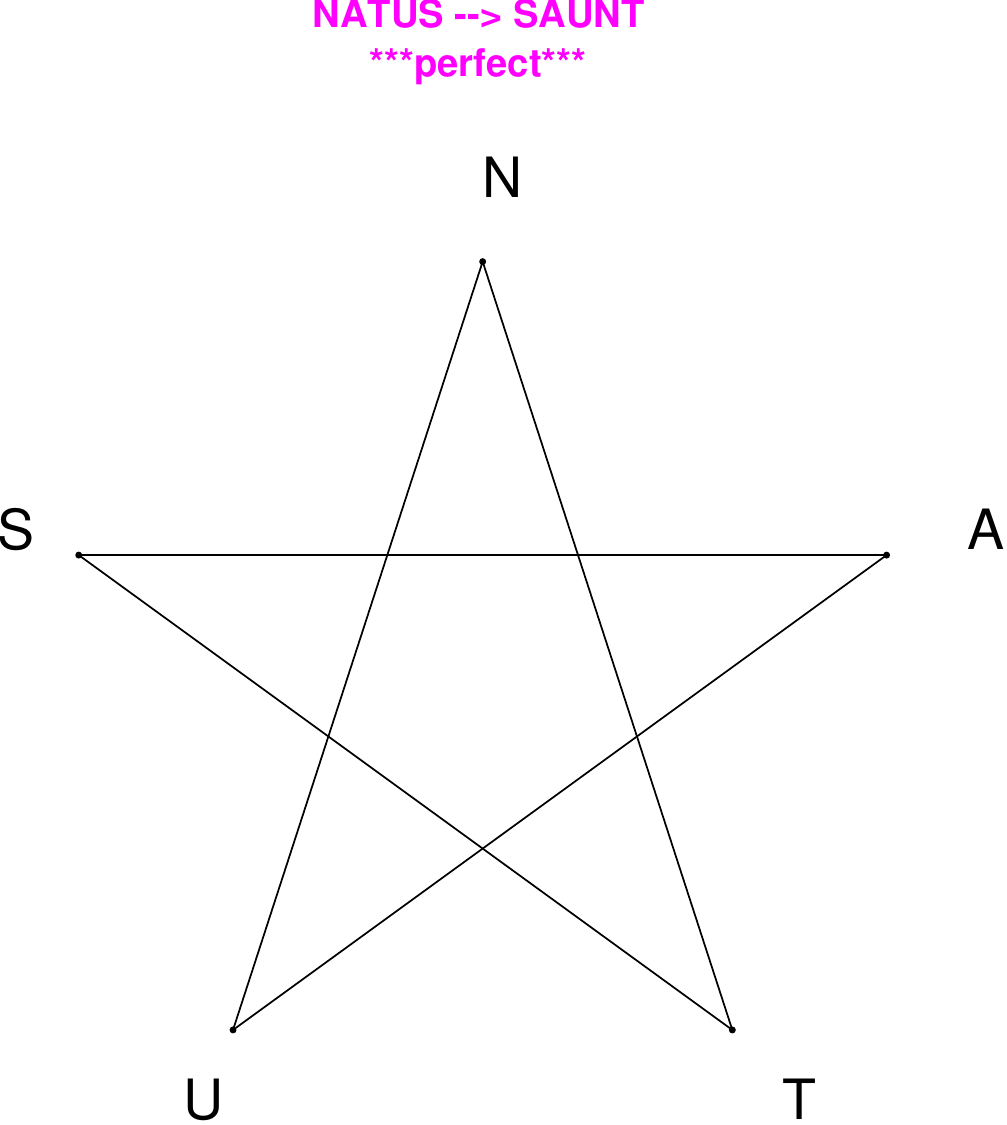}
\end{subfigure}
\hfill
\begin{subfigure}[T]{0.19\textwidth}
\centering
\includegraphics[width=\textwidth]{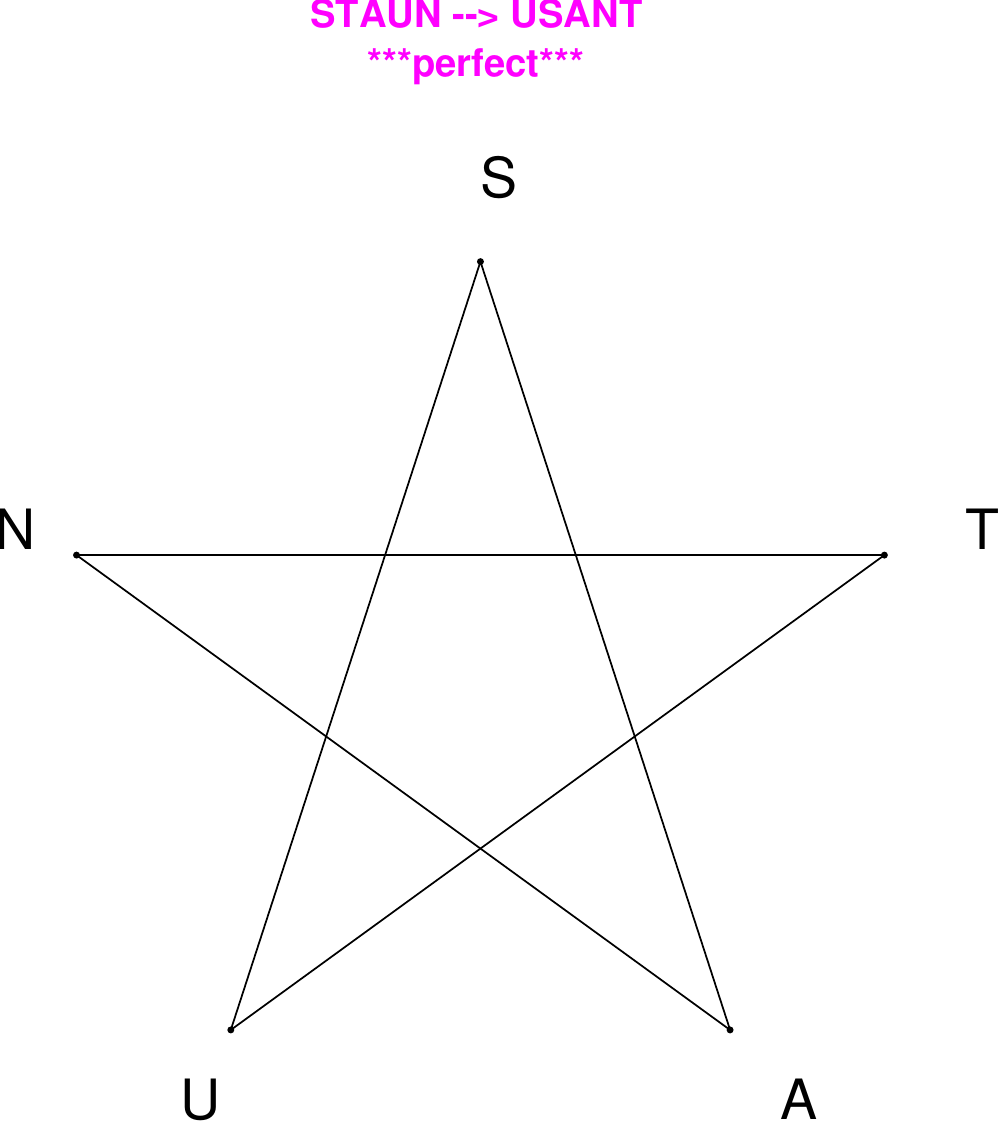}
\end{subfigure}
\hfill
\begin{subfigure}[T]{0.19\textwidth}
\centering
\includegraphics[width=\textwidth]{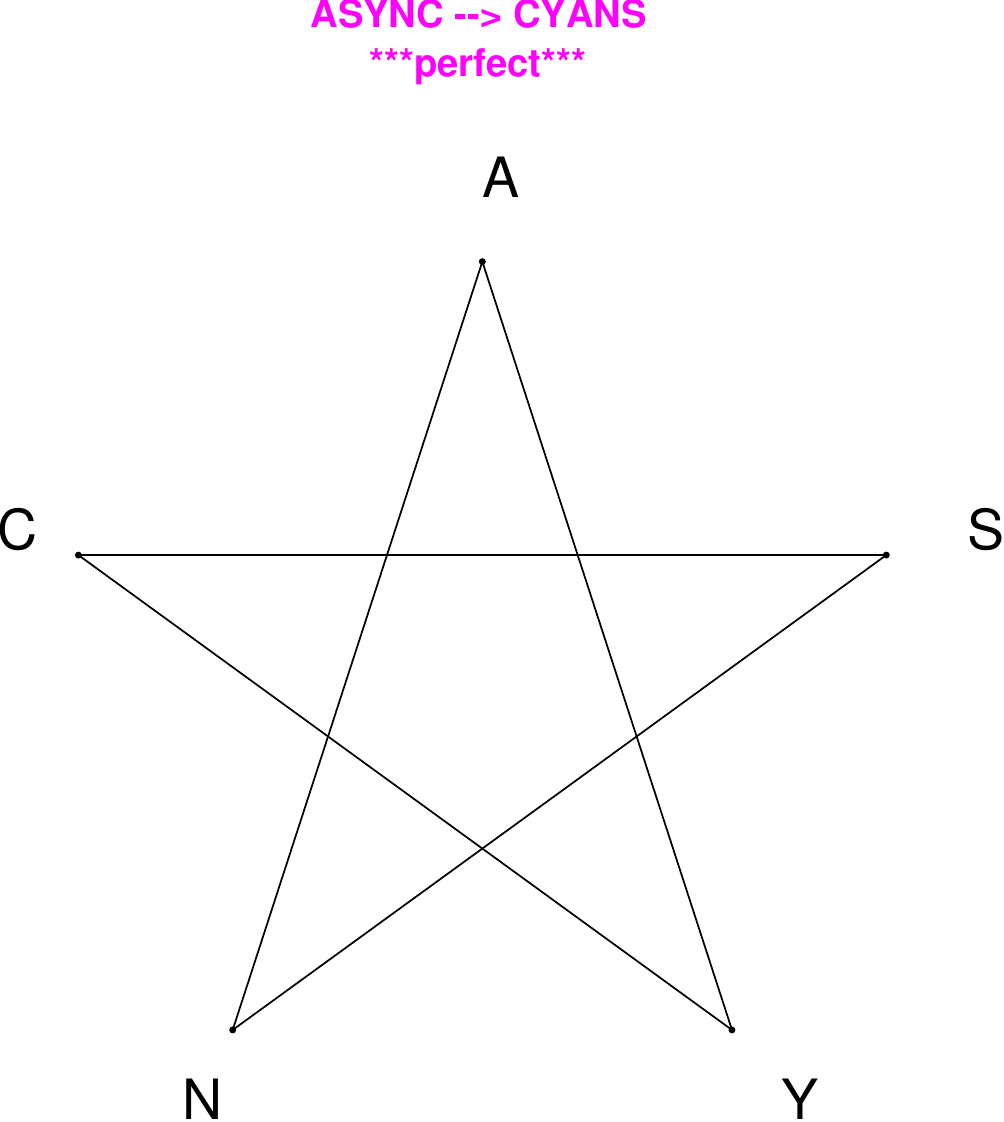}
\end{subfigure}
\hfill
\begin{subfigure}[T]{0.19\textwidth}
\centering
\includegraphics[width=\textwidth]{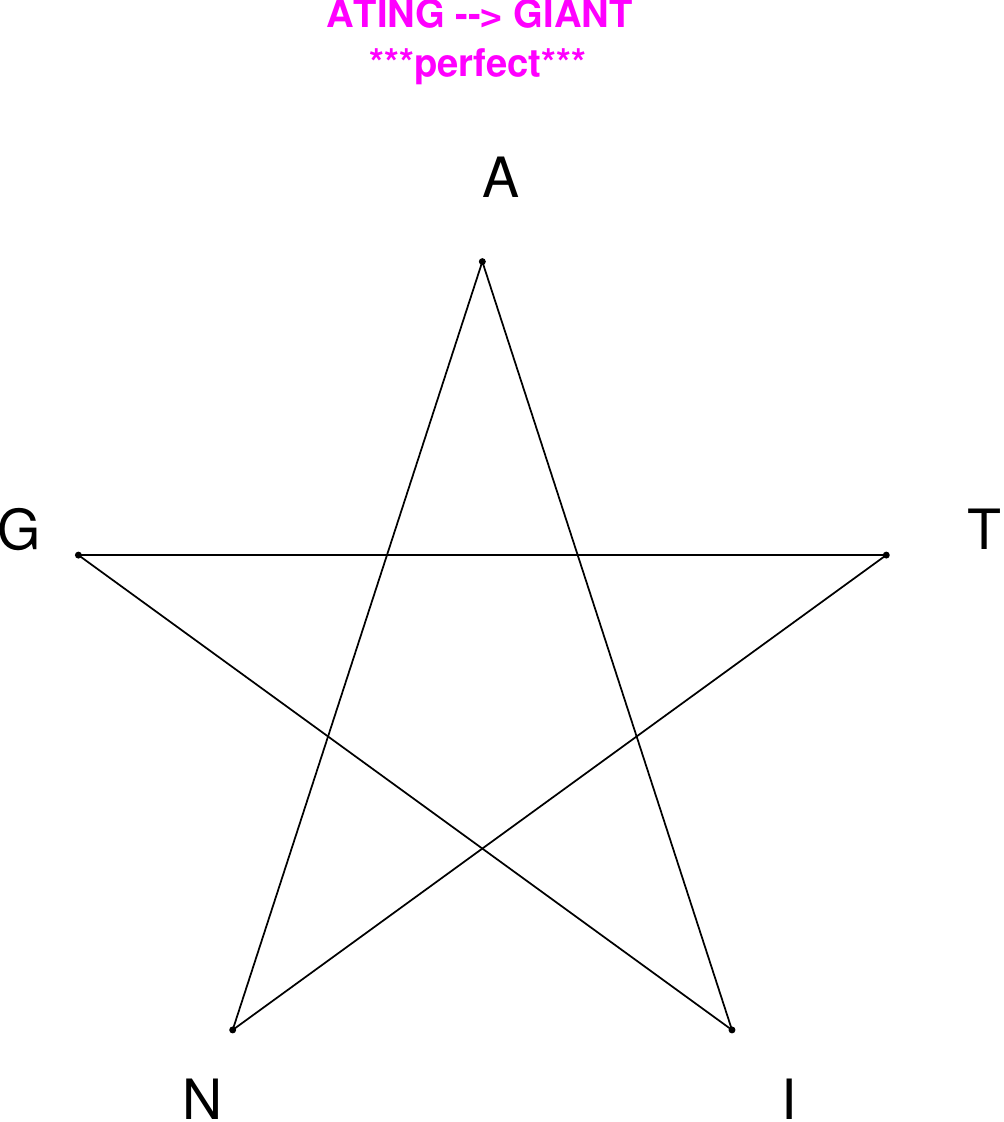}
\end{subfigure}
\end{figure}

\begin{figure}[H]
\centering
\begin{subfigure}[T]{0.19\textwidth}
\centering
\includegraphics[width=\textwidth]{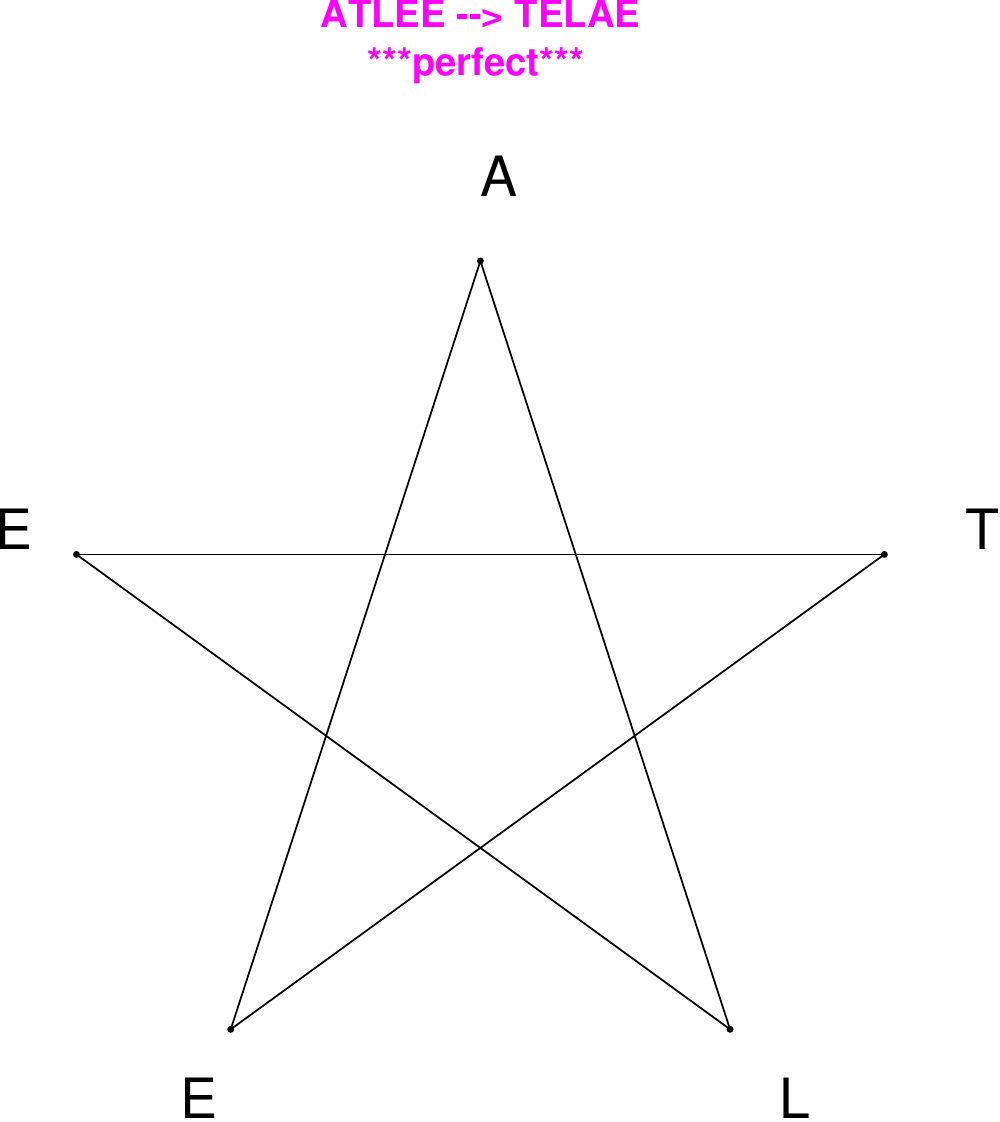}
\end{subfigure}
\hfill
\begin{subfigure}[T]{0.19\textwidth}
\centering
\includegraphics[width=\textwidth]{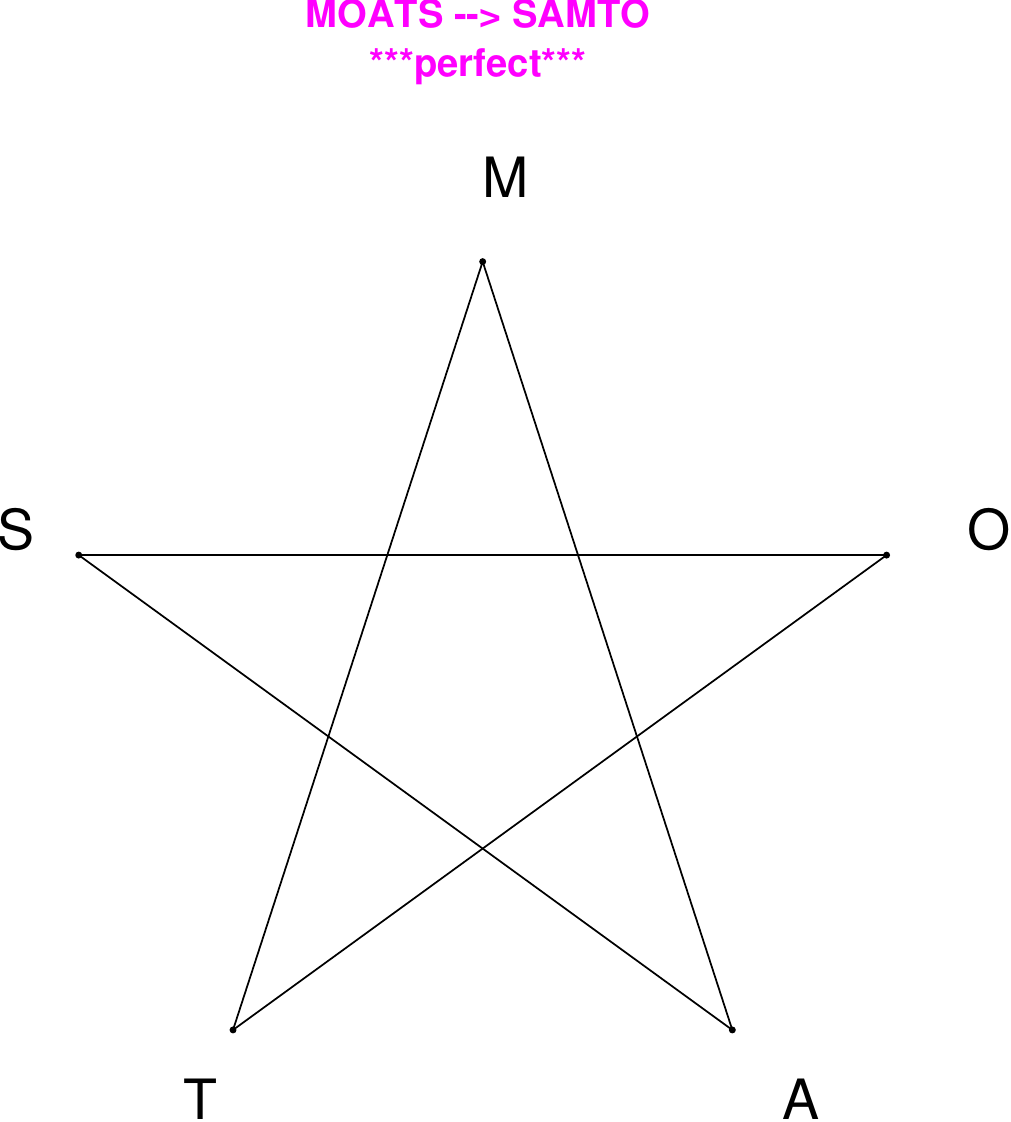}
\end{subfigure}
\hfill
\begin{subfigure}[T]{0.19\textwidth}
\centering
\includegraphics[width=\textwidth]{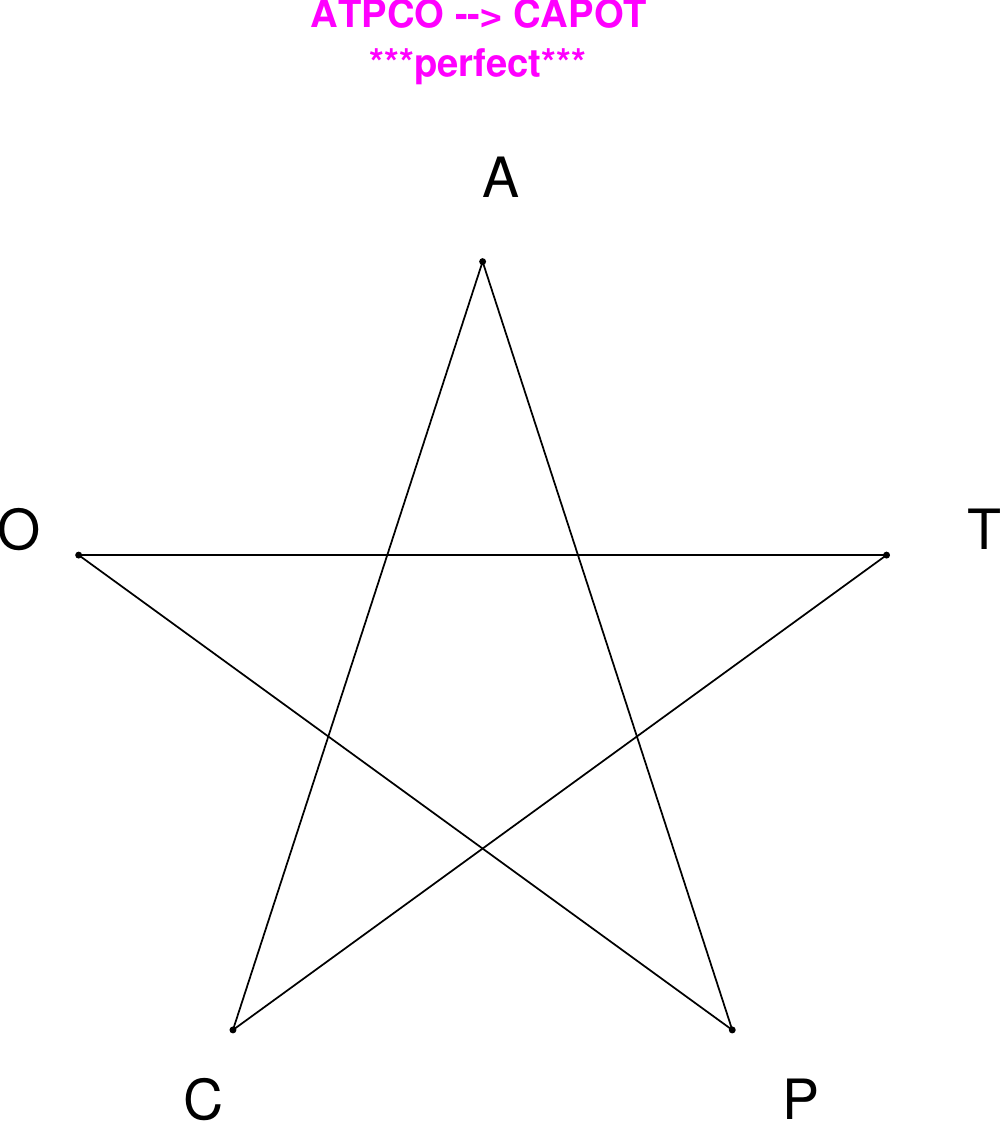}
\end{subfigure}
\hfill
\begin{subfigure}[T]{0.19\textwidth}
\centering
\includegraphics[width=\textwidth]{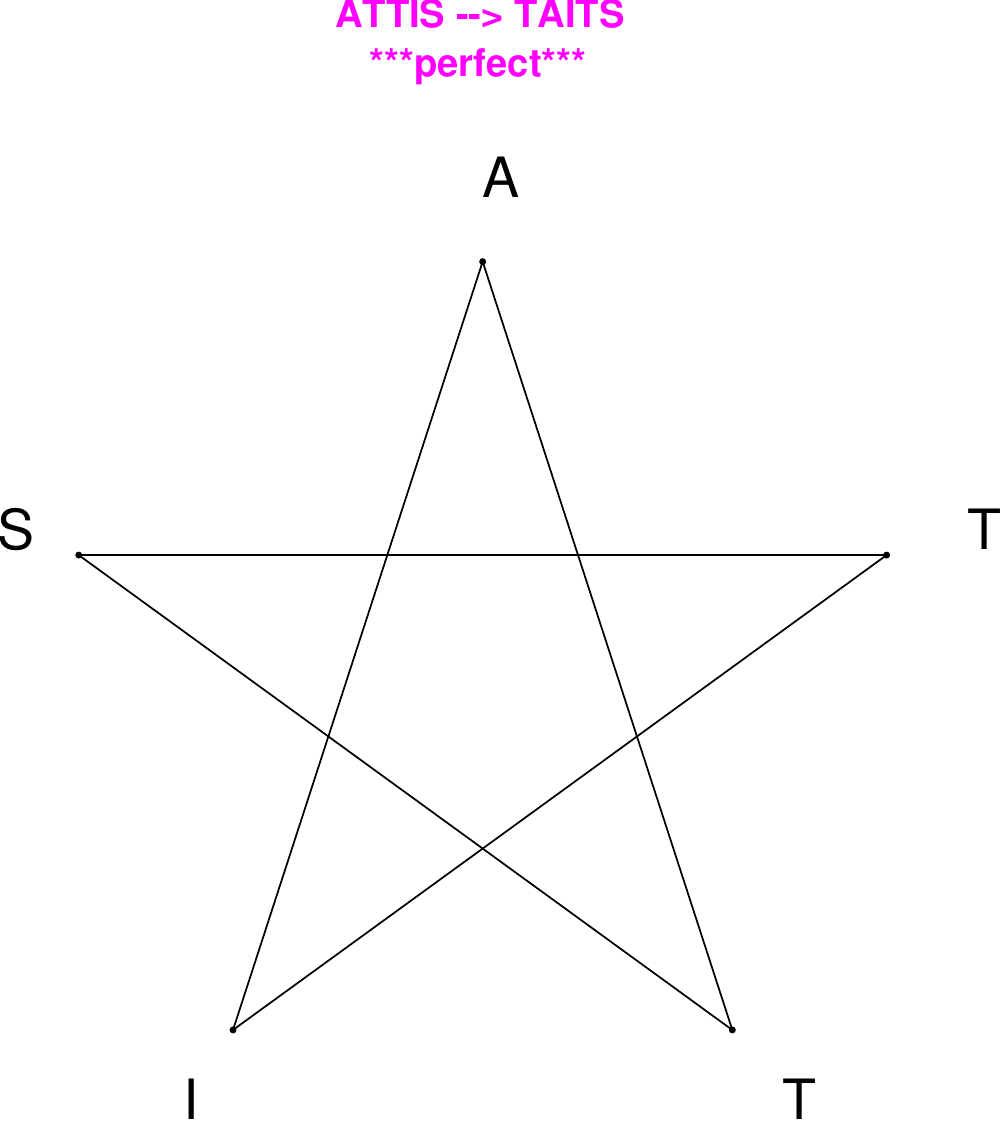}
\end{subfigure}
\hfill
\begin{subfigure}[T]{0.19\textwidth}
\centering
\includegraphics[width=\textwidth]{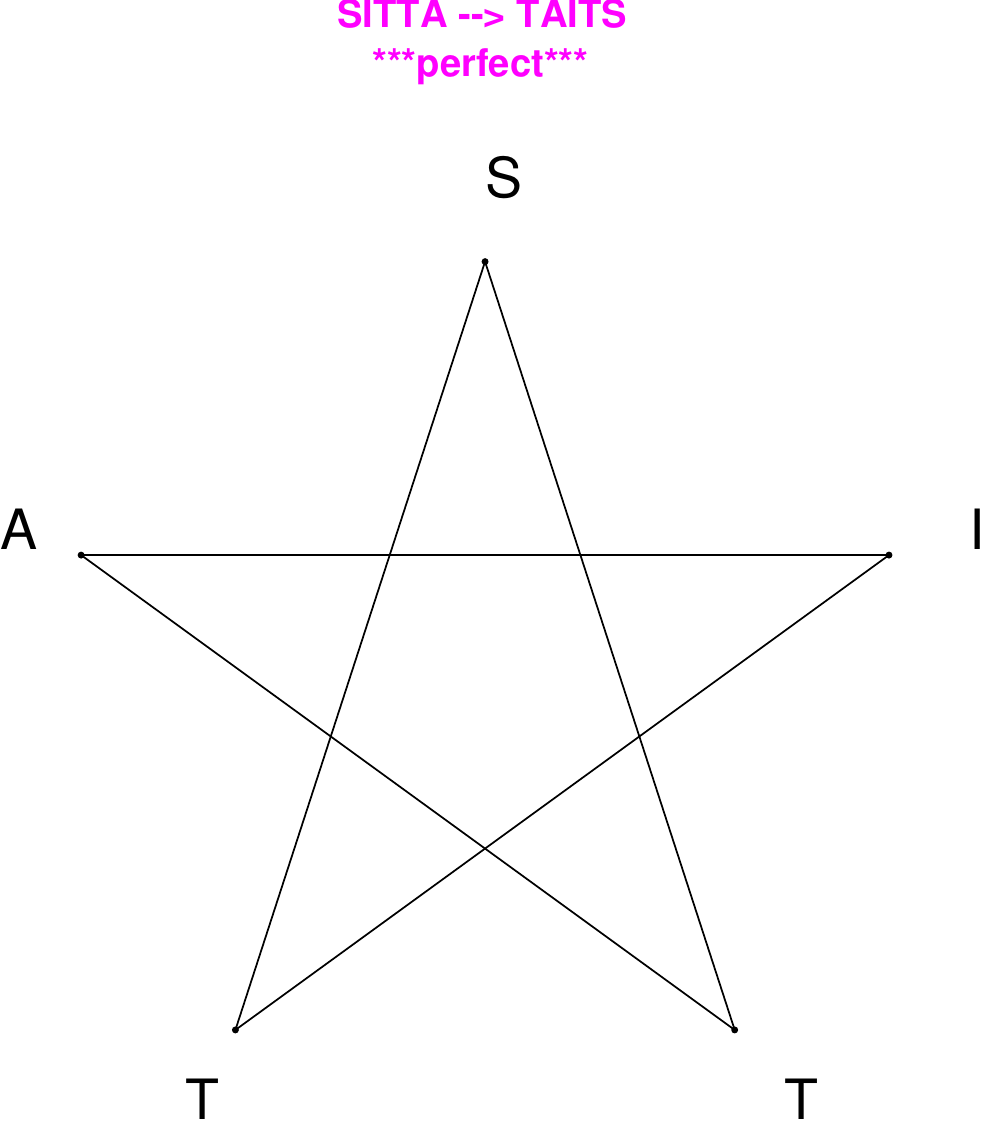}
\end{subfigure}
\end{figure}

\begin{figure}[H]
\centering
\begin{subfigure}[T]{0.19\textwidth}
\centering
\includegraphics[width=\textwidth]{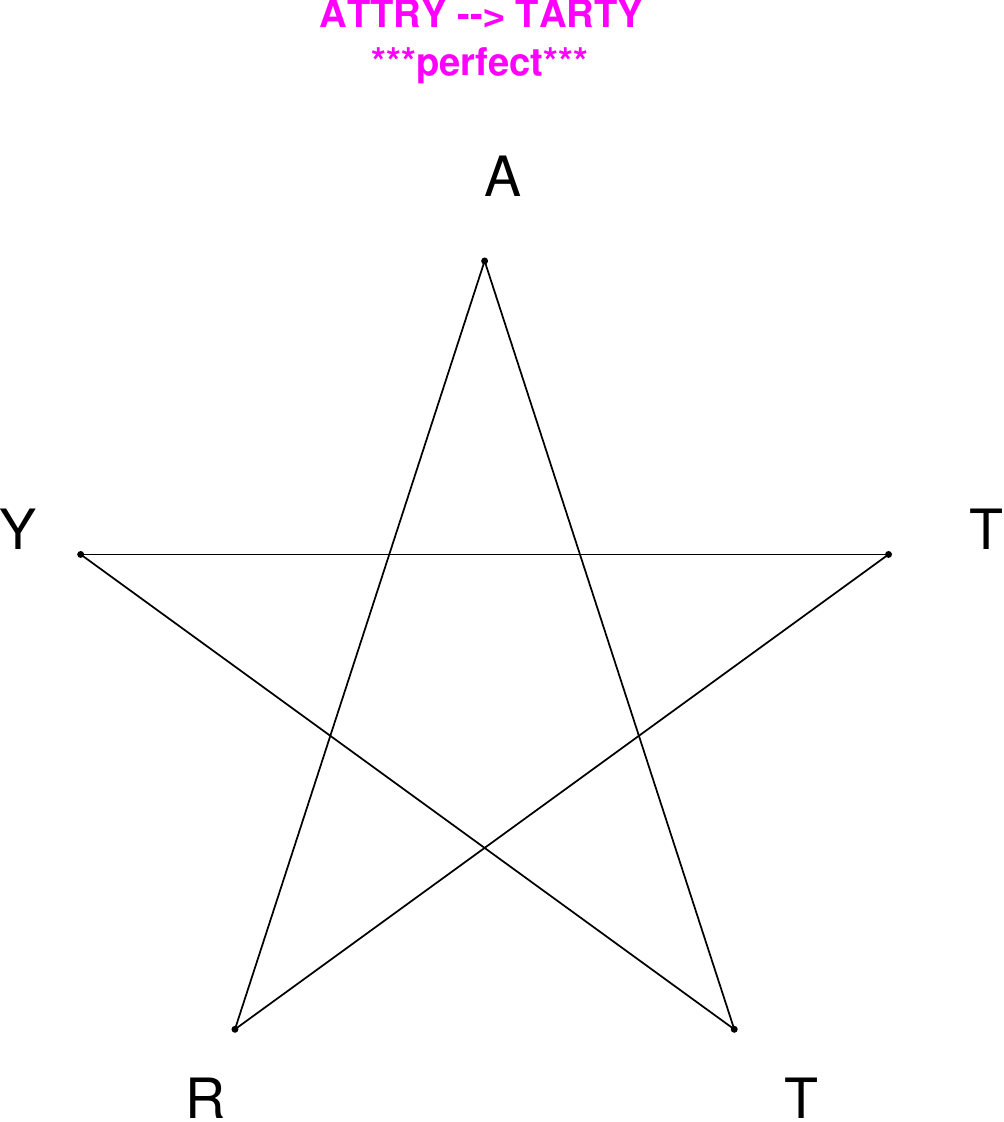}
\end{subfigure}
\hfill
\begin{subfigure}[T]{0.19\textwidth}
\centering
\includegraphics[width=\textwidth]{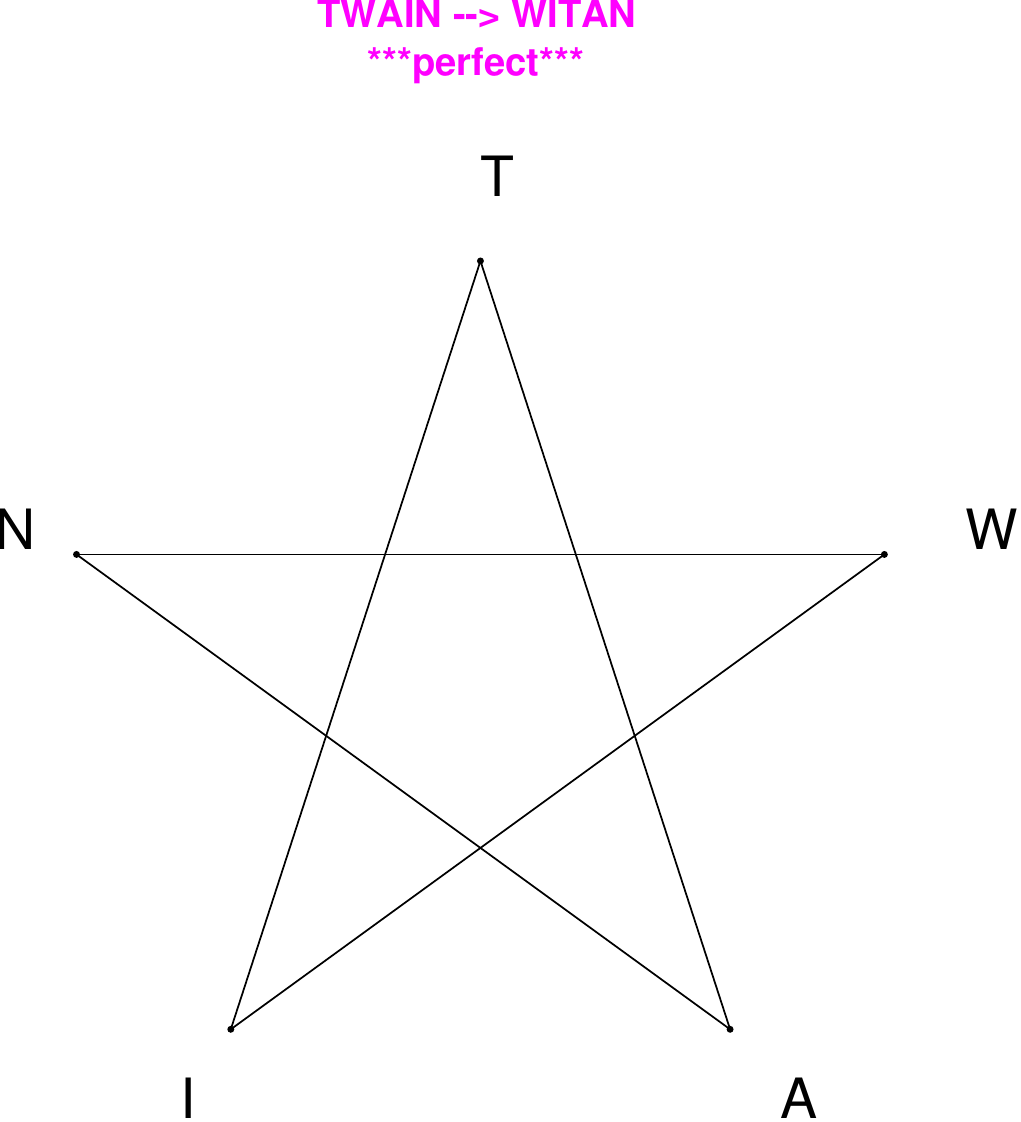}
\end{subfigure}
\hfill
\begin{subfigure}[T]{0.19\textwidth}
\centering
\includegraphics[width=\textwidth]{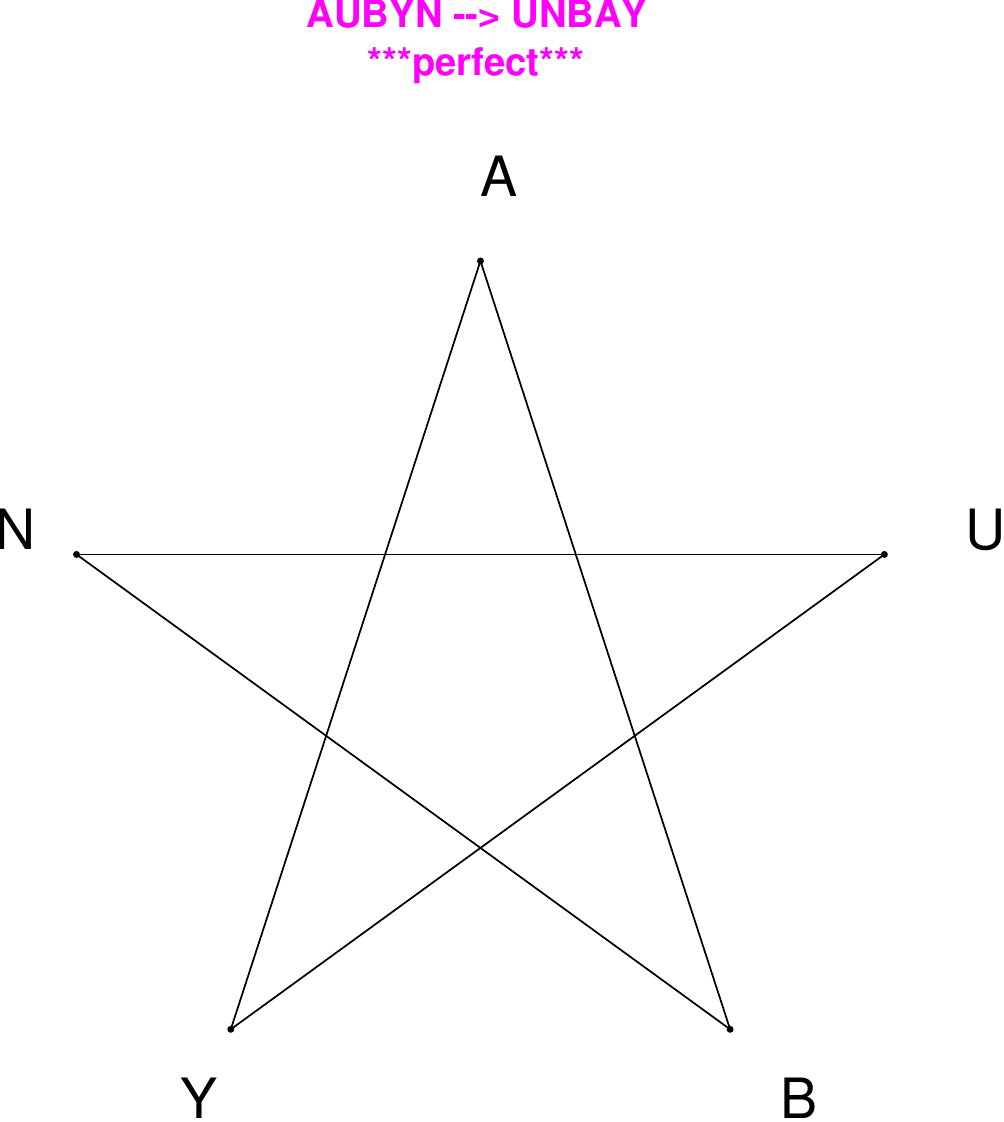}
\end{subfigure}
\hfill
\begin{subfigure}[T]{0.19\textwidth}
\centering
\includegraphics[width=\textwidth]{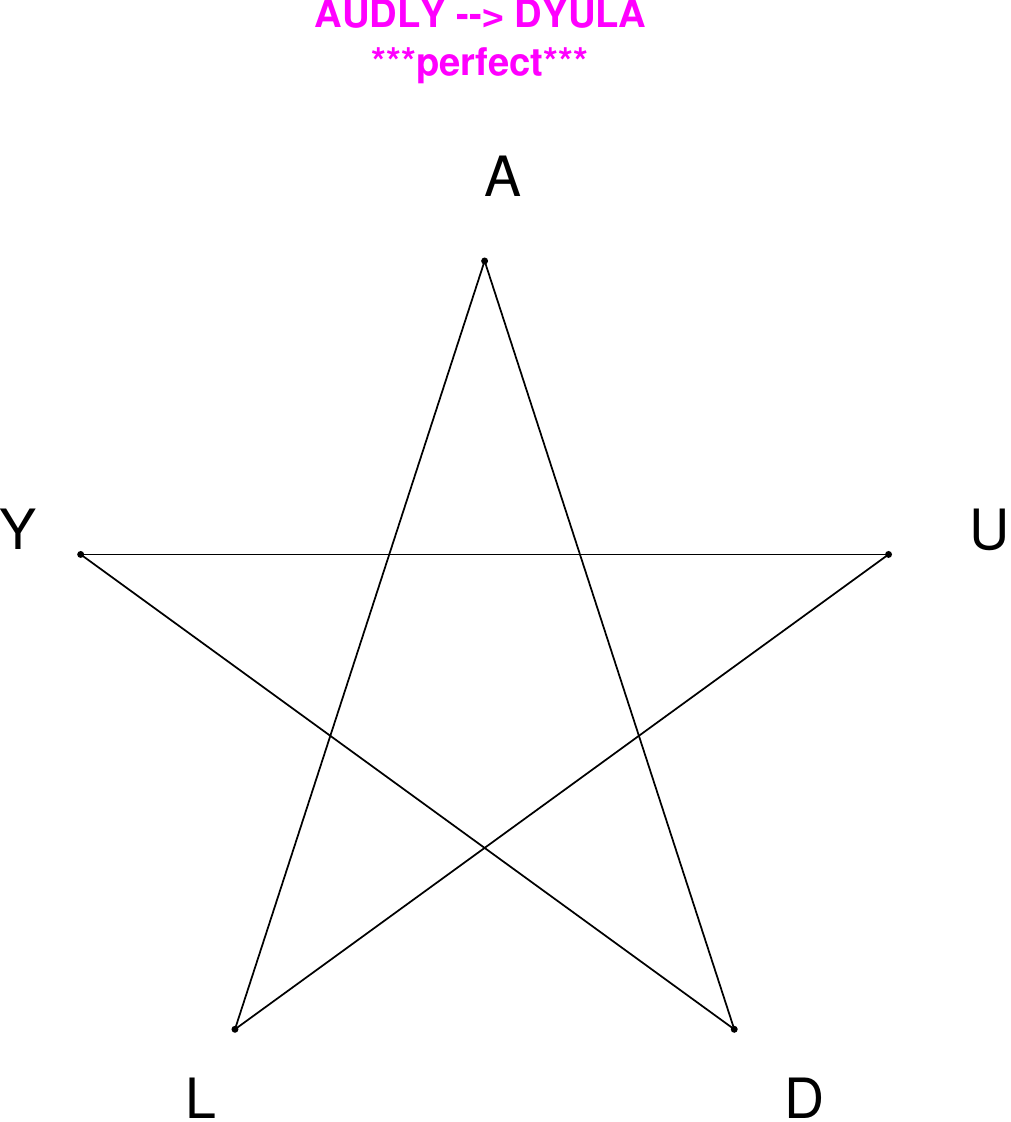}
\end{subfigure}
\hfill
\begin{subfigure}[T]{0.19\textwidth}
\centering
\includegraphics[width=\textwidth]{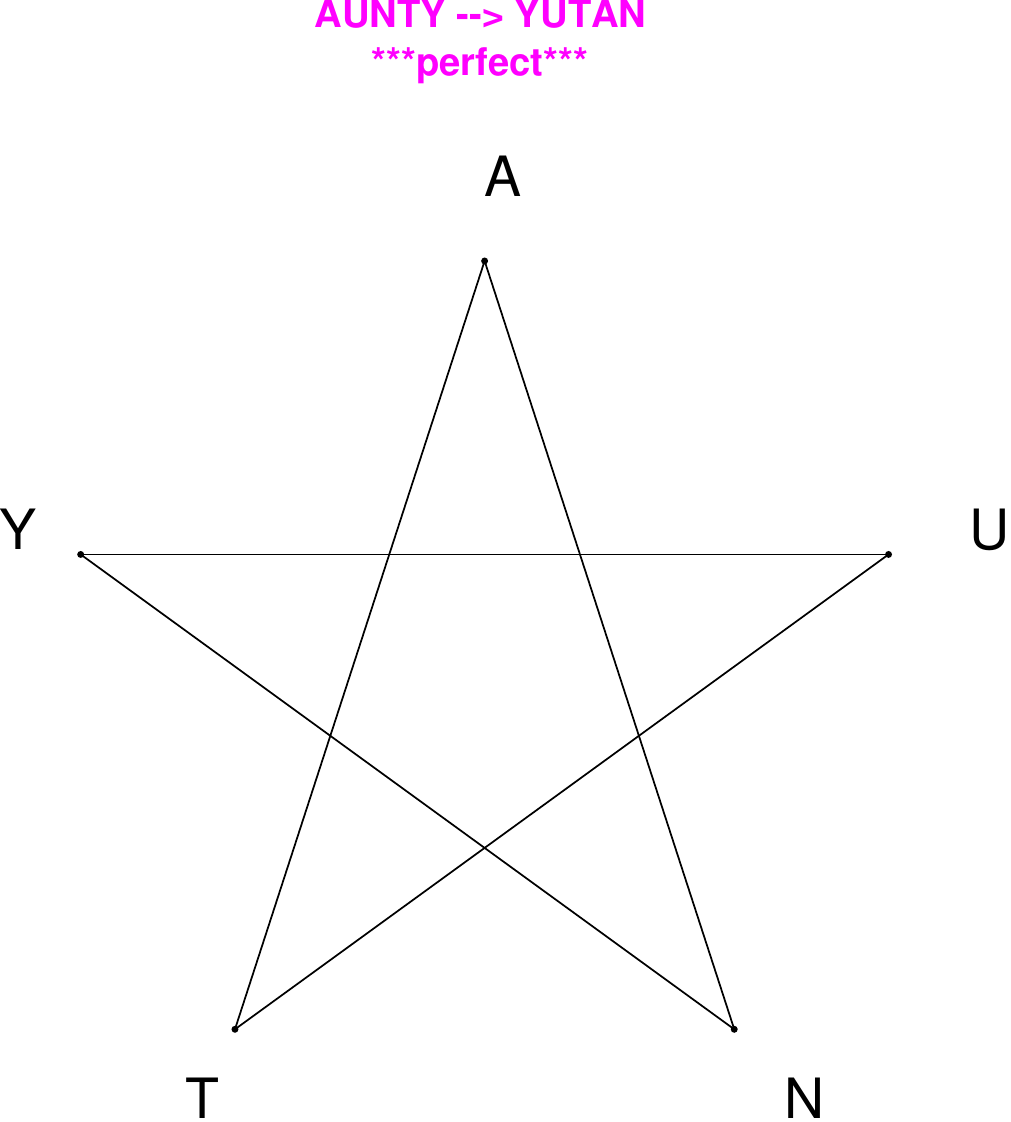}
\end{subfigure}
\end{figure}

\begin{figure}[H]
\centering
\begin{subfigure}[T]{0.19\textwidth}
\centering
\includegraphics[width=\textwidth]{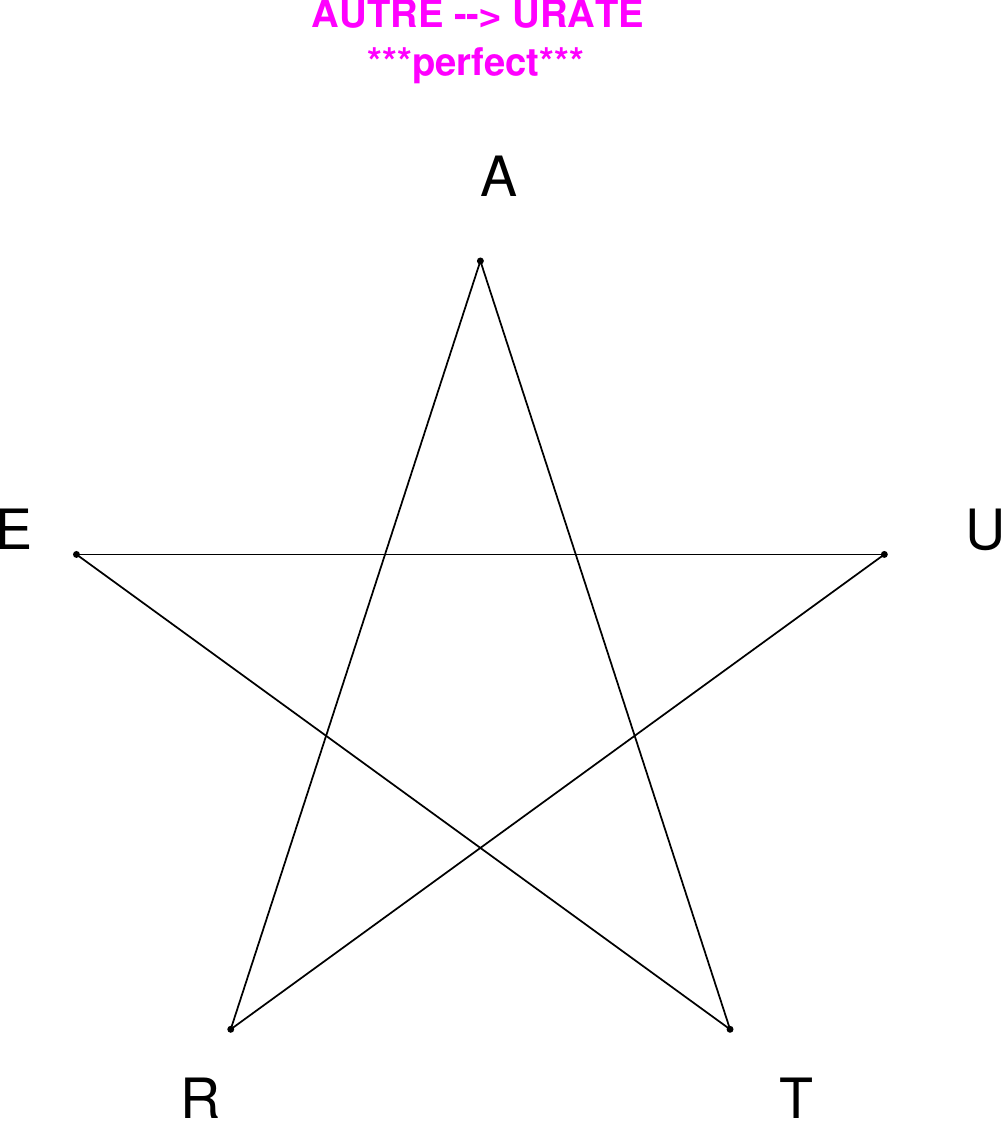}
\end{subfigure}
\hfill
\begin{subfigure}[T]{0.19\textwidth}
\centering
\includegraphics[width=\textwidth]{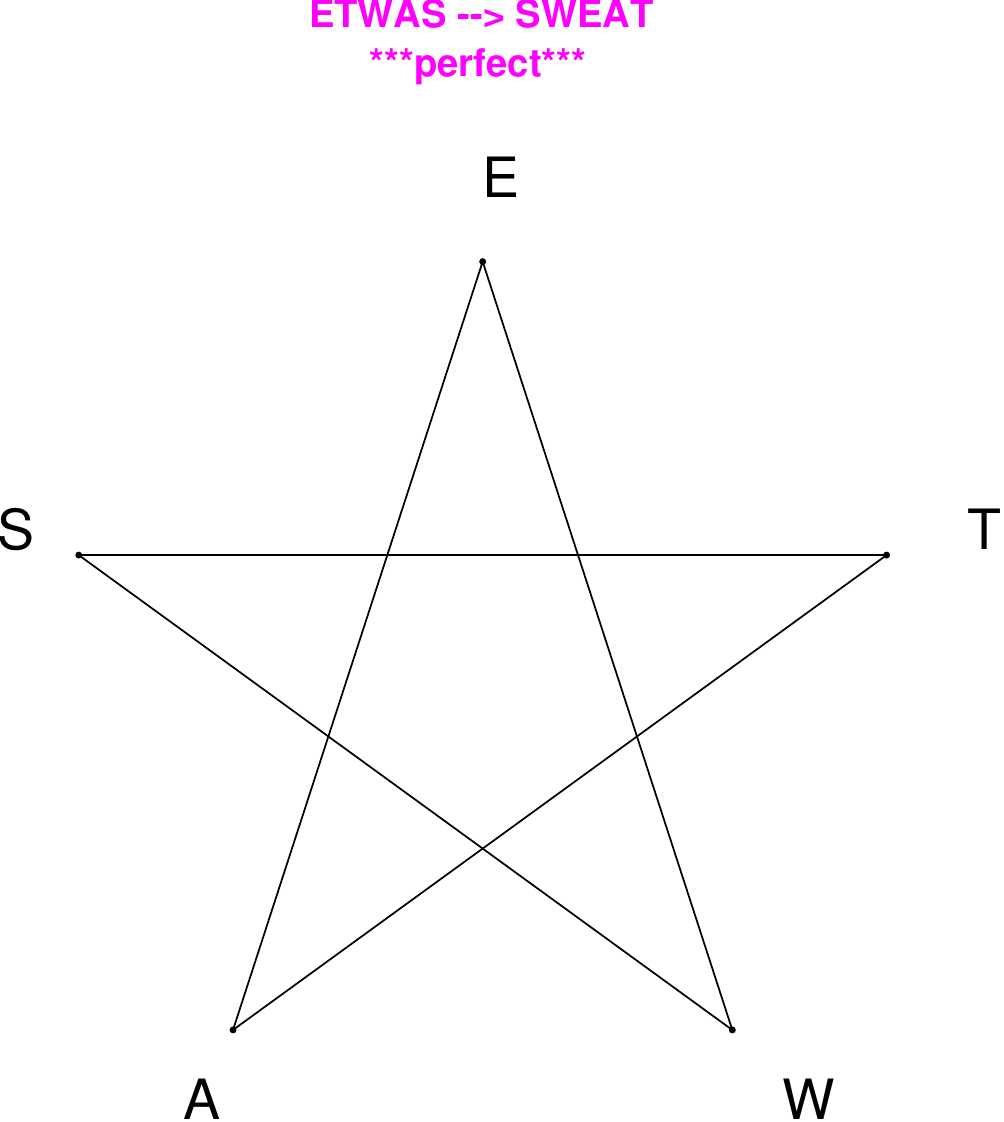}
\end{subfigure}
\hfill
\begin{subfigure}[T]{0.19\textwidth}
\centering
\includegraphics[width=\textwidth]{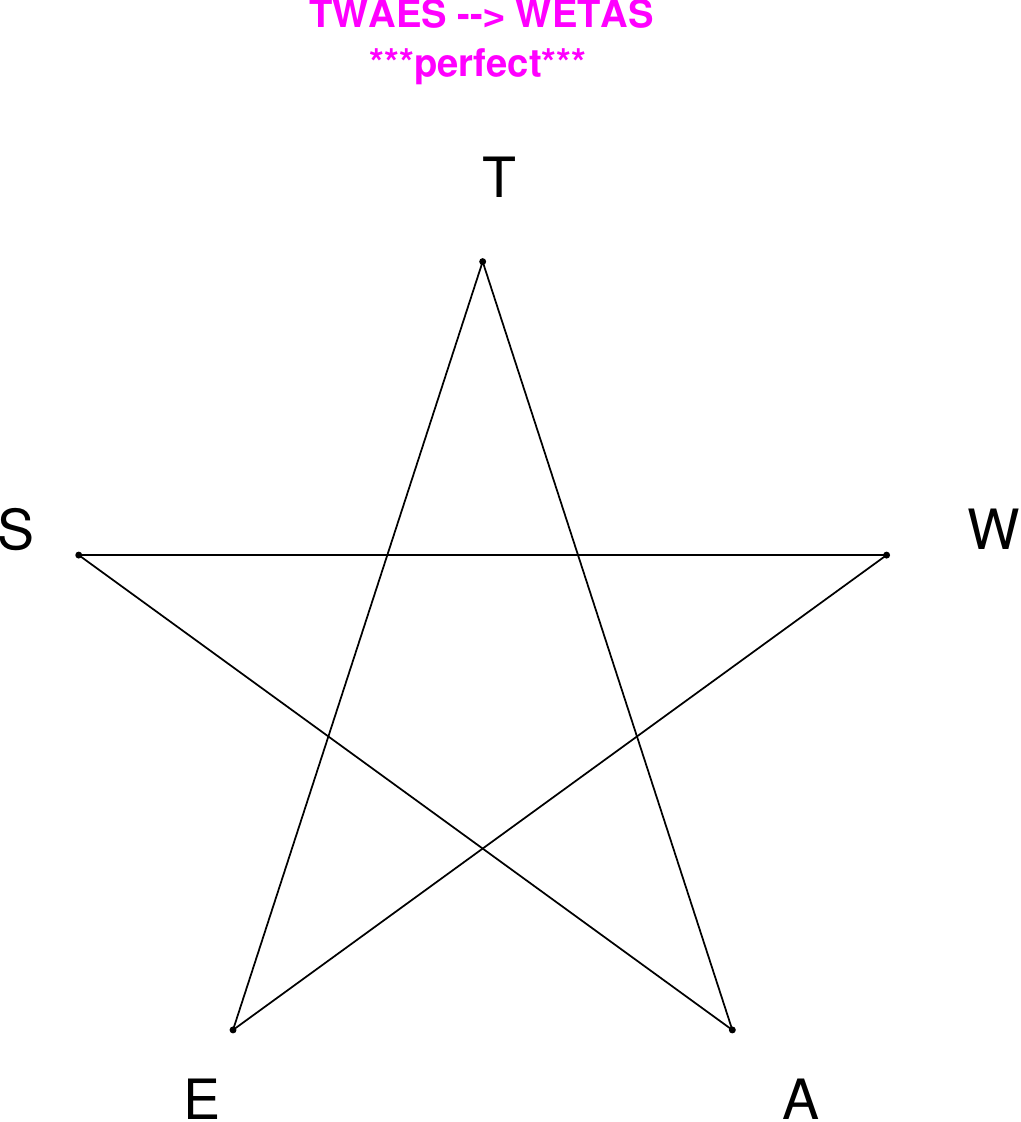}
\end{subfigure}
\hfill
\begin{subfigure}[T]{0.19\textwidth}
\centering
\includegraphics[width=\textwidth]{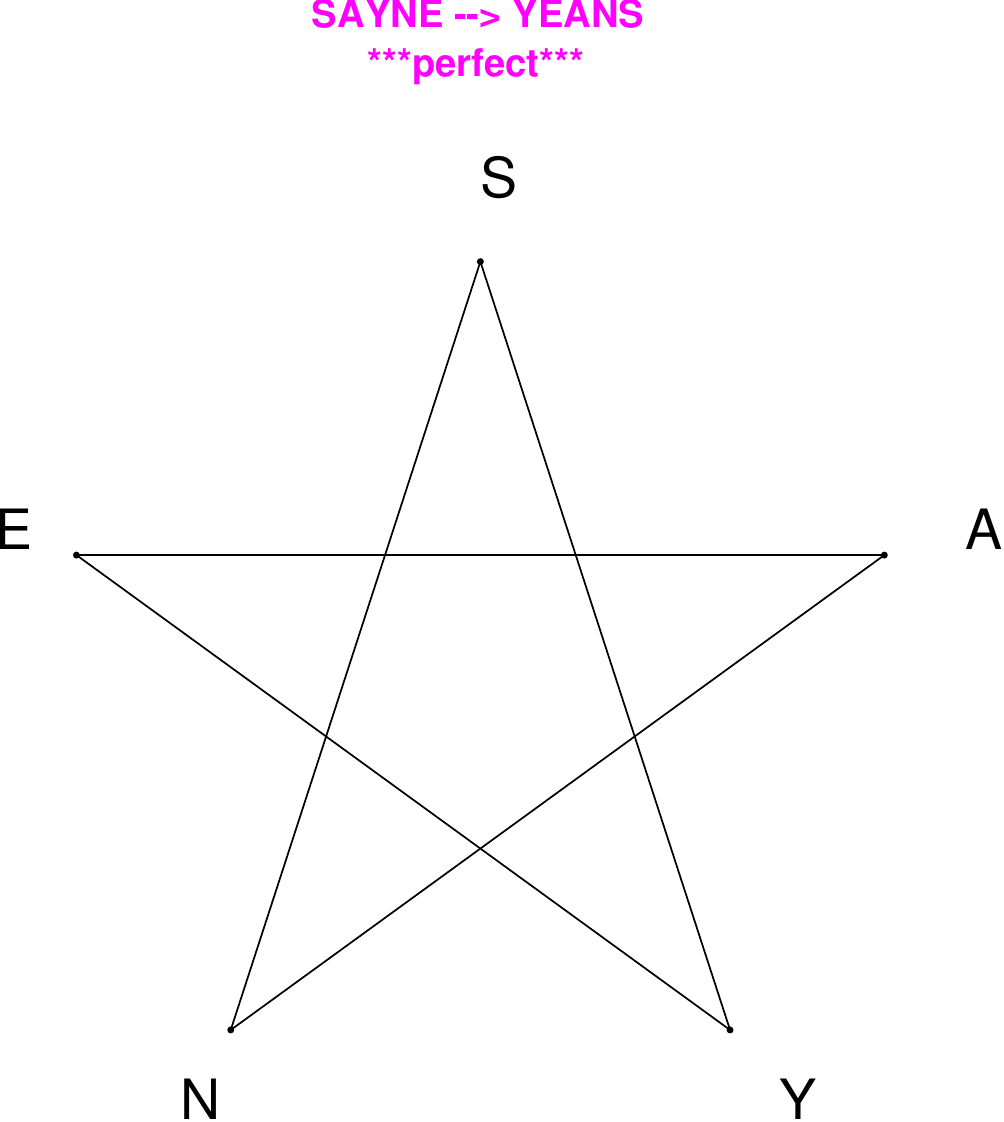}
\end{subfigure}
\hfill
\begin{subfigure}[T]{0.19\textwidth}
\centering
\includegraphics[width=\textwidth]{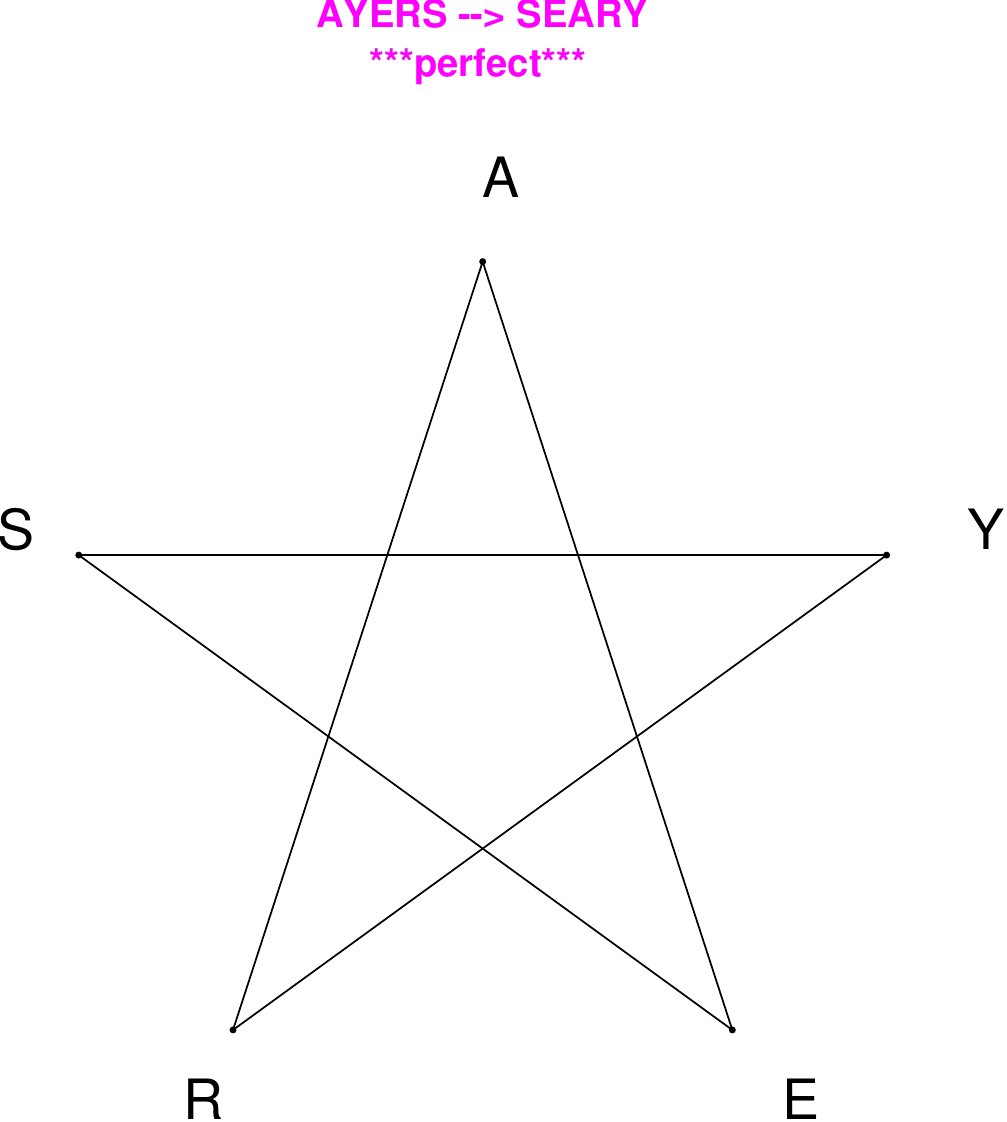}
\end{subfigure}
\end{figure}

\begin{figure}[H]
\centering
\begin{subfigure}[T]{0.19\textwidth}
\centering
\includegraphics[width=\textwidth]{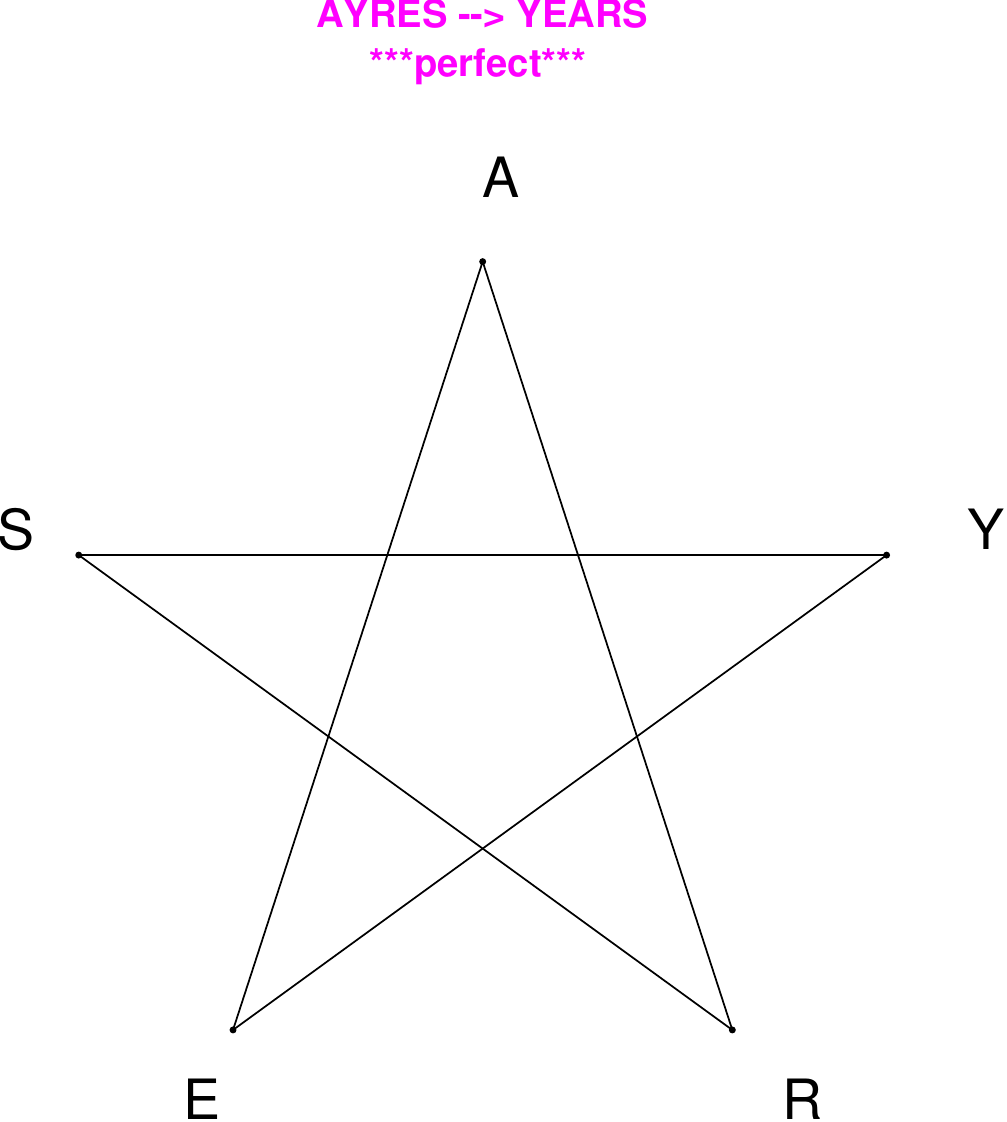}
\end{subfigure}
\hfill
\begin{subfigure}[T]{0.19\textwidth}
\centering
\includegraphics[width=\textwidth]{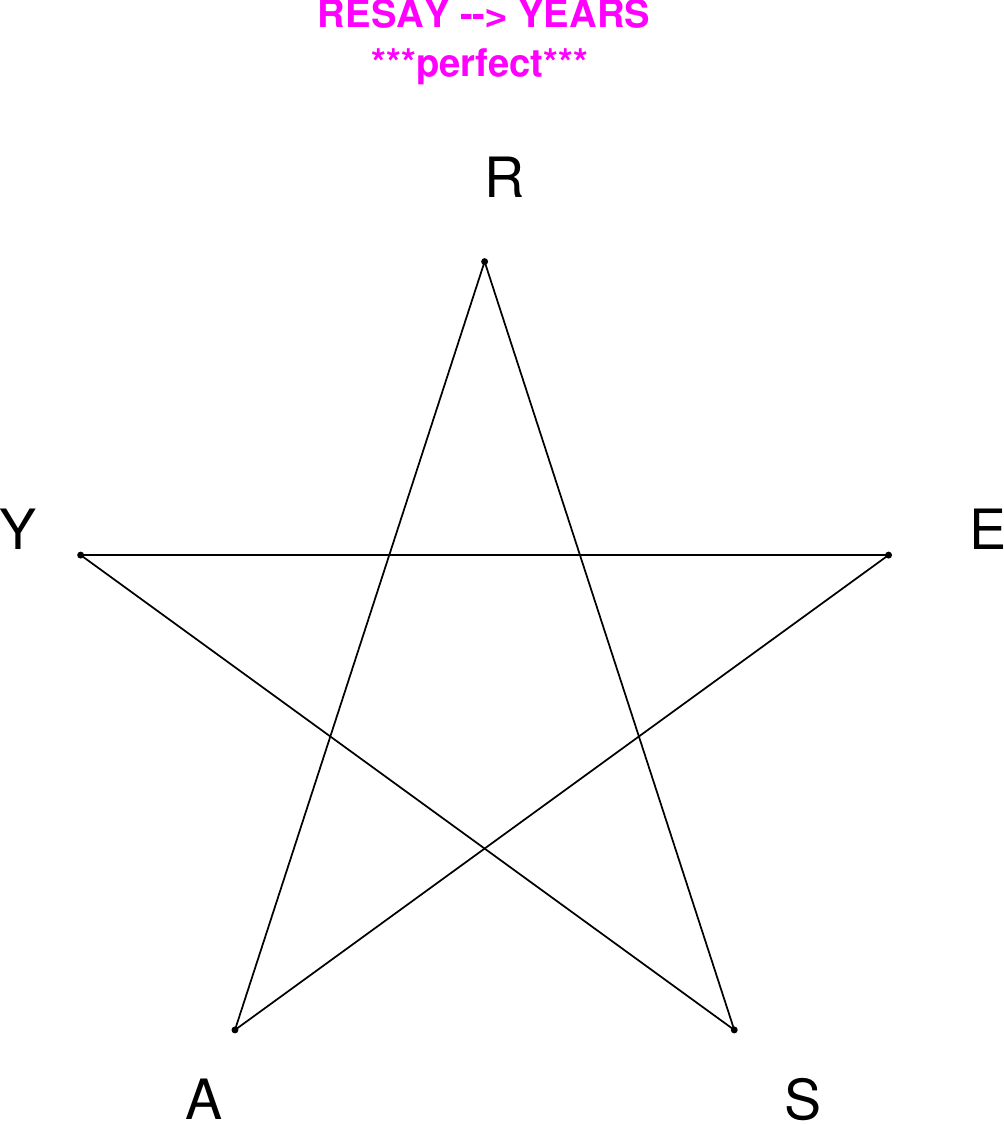}
\end{subfigure}
\hfill
\begin{subfigure}[T]{0.19\textwidth}
\centering
\includegraphics[width=\textwidth]{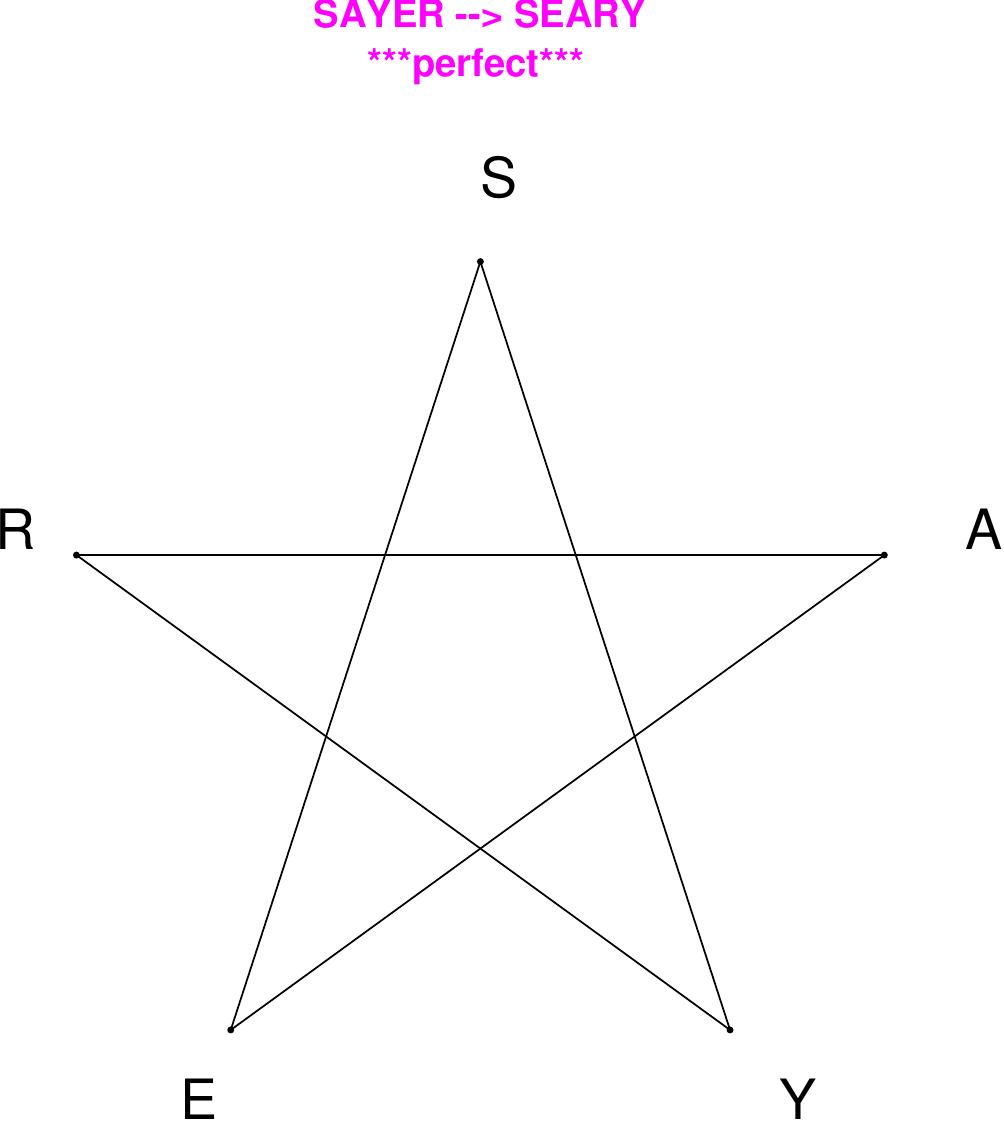}
\end{subfigure}
\hfill
\begin{subfigure}[T]{0.19\textwidth}
\centering
\includegraphics[width=\textwidth]{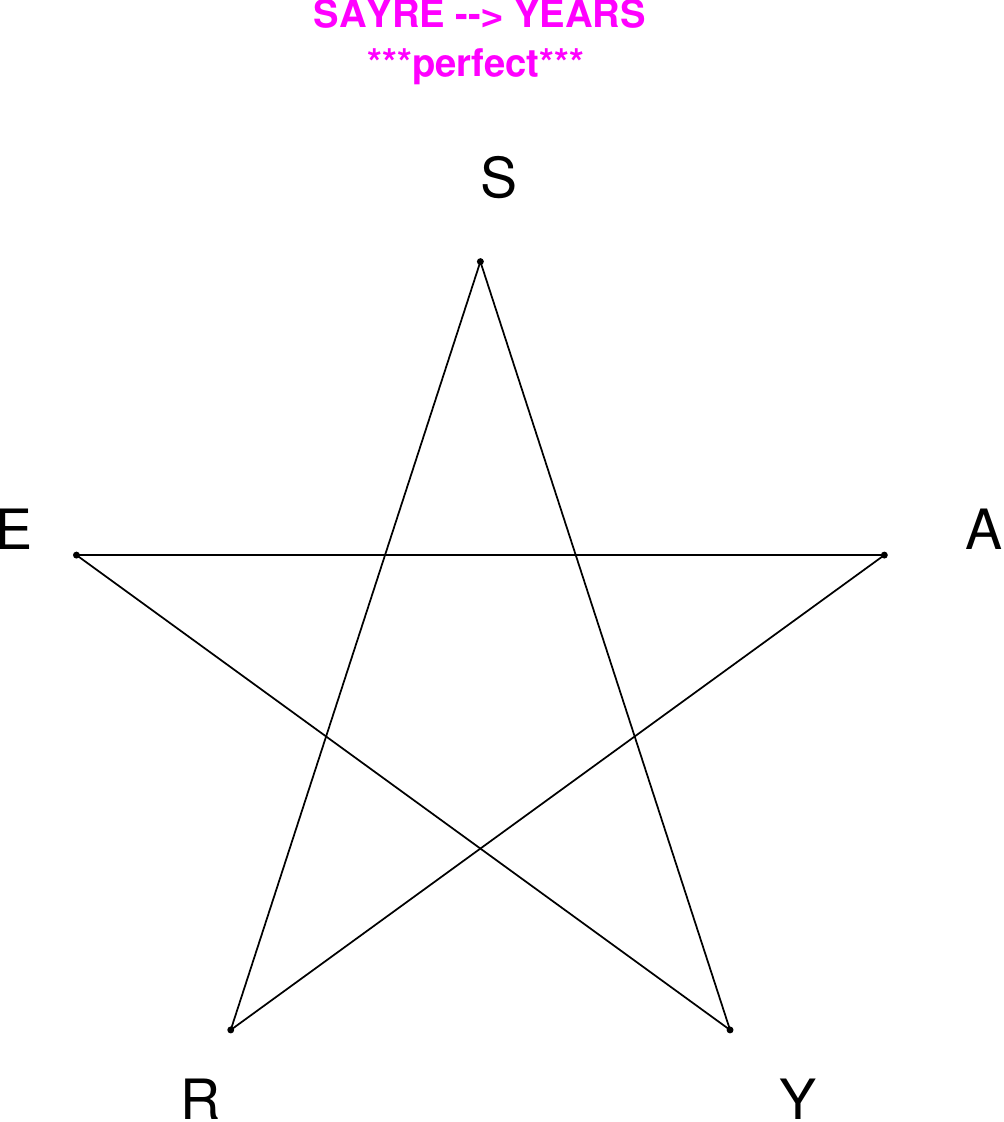}
\end{subfigure}
\hfill
\begin{subfigure}[T]{0.19\textwidth}
\centering
\includegraphics[width=\textwidth]{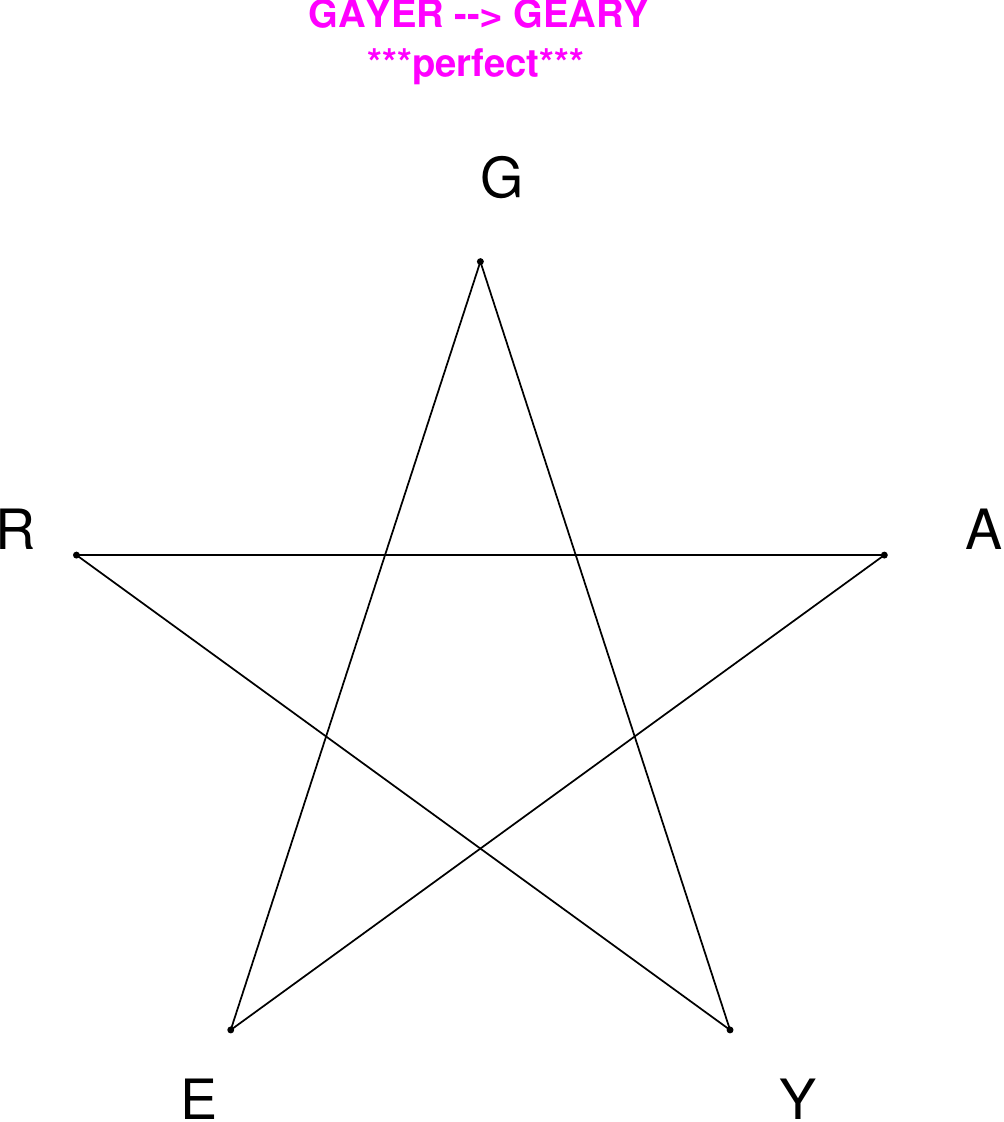}
\end{subfigure}
\end{figure}

\begin{figure}[H]
\centering
\begin{subfigure}[T]{0.19\textwidth}
\centering
\includegraphics[width=\textwidth]{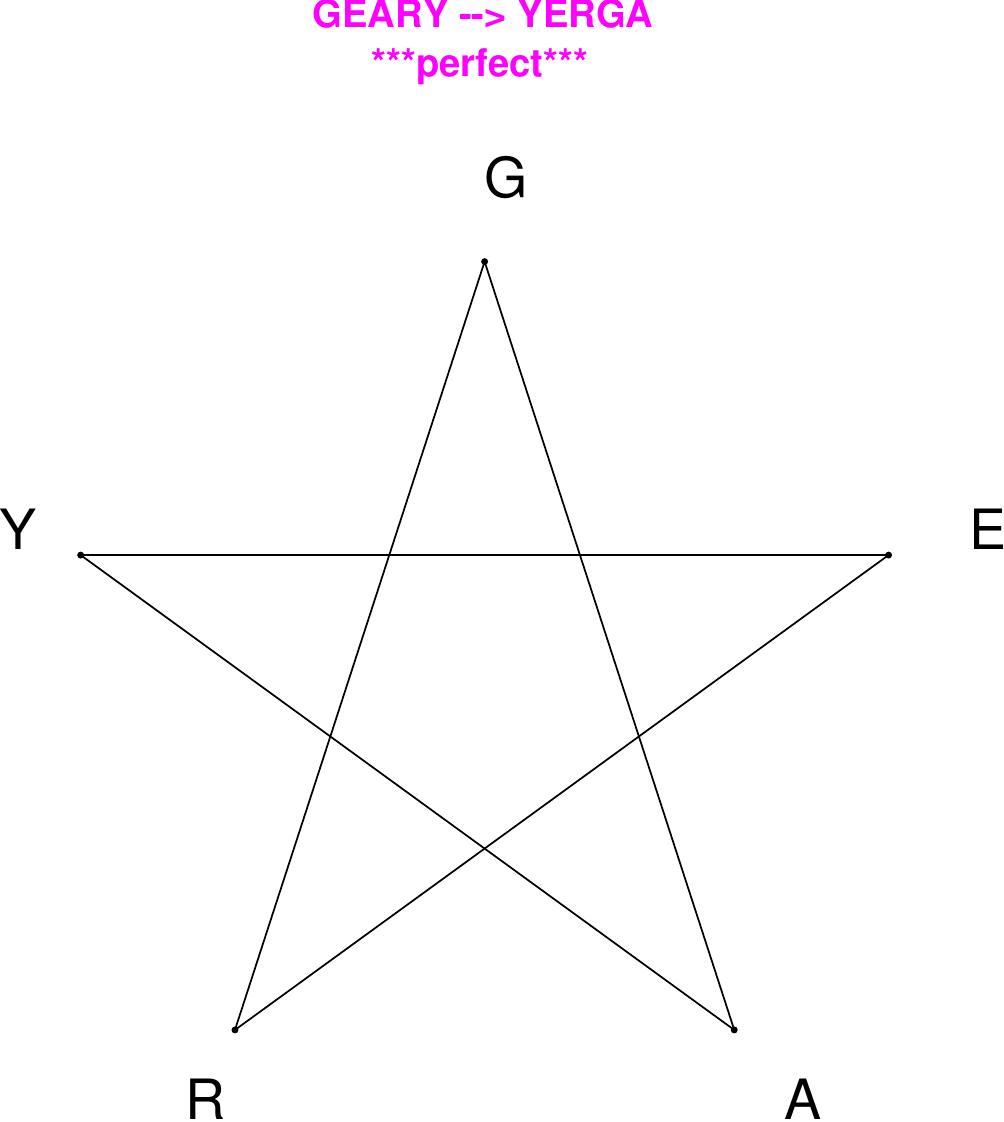}
\end{subfigure}
\hfill
\begin{subfigure}[T]{0.19\textwidth}
\centering
\includegraphics[width=\textwidth]{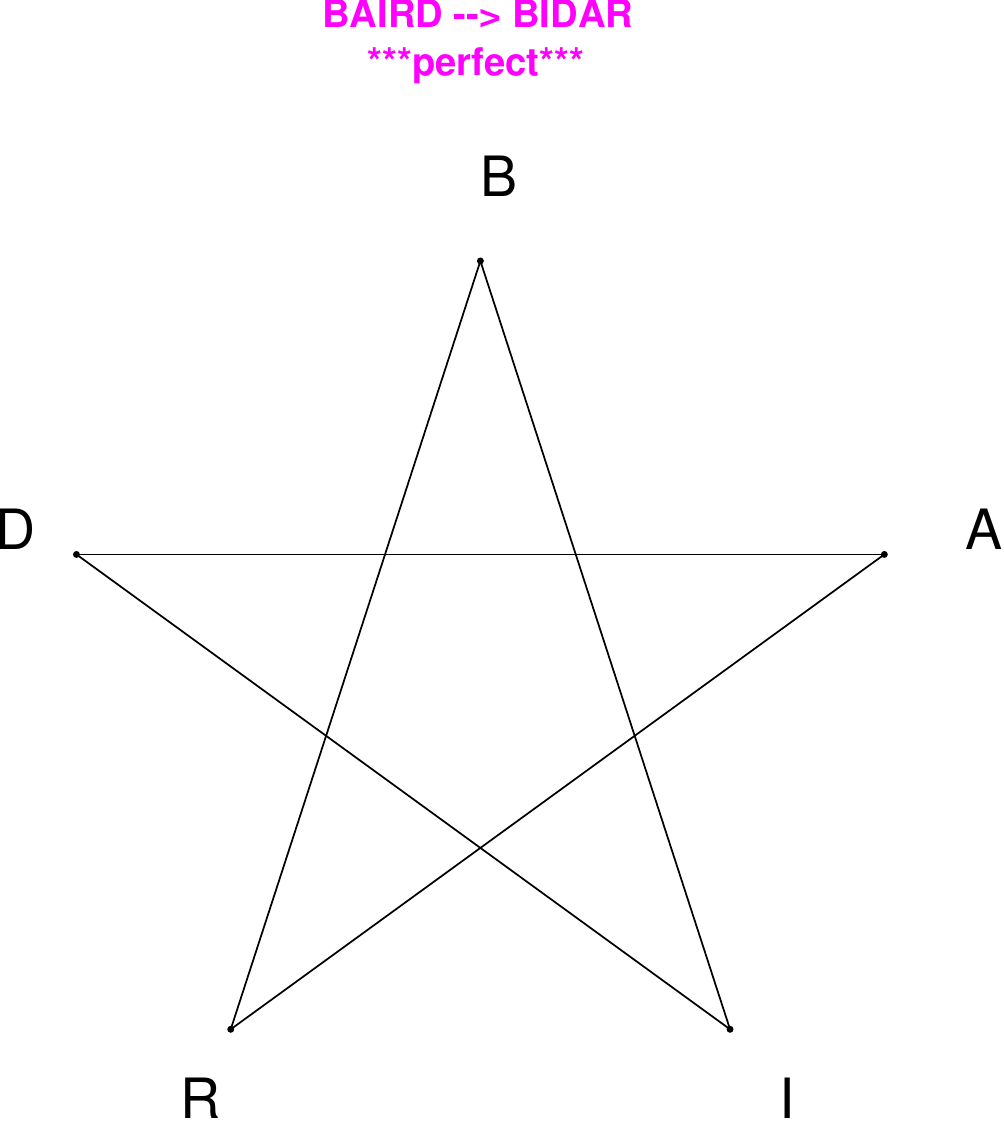}
\end{subfigure}
\hfill
\begin{subfigure}[T]{0.19\textwidth}
\centering
\includegraphics[width=\textwidth]{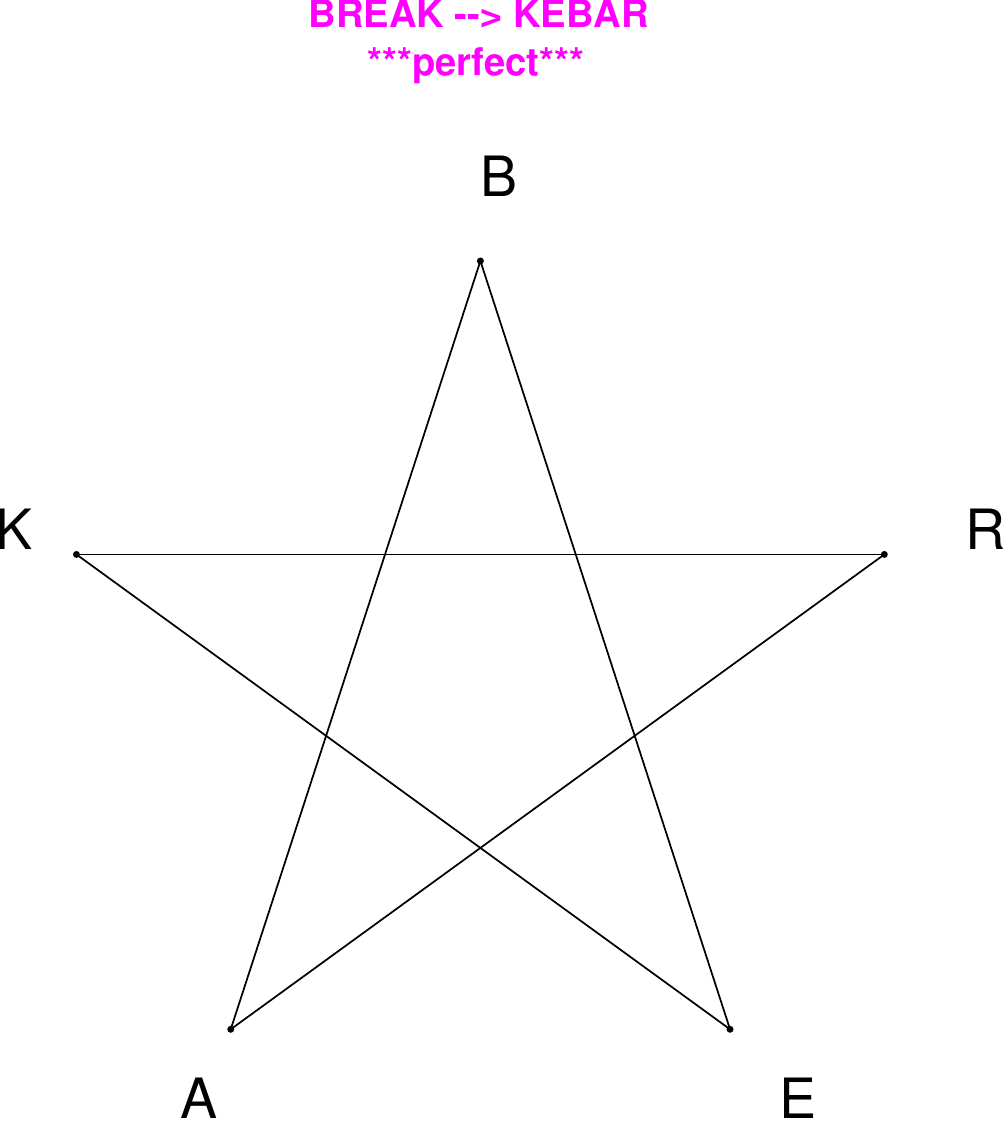}
\end{subfigure}
\hfill
\begin{subfigure}[T]{0.19\textwidth}
\centering
\includegraphics[width=\textwidth]{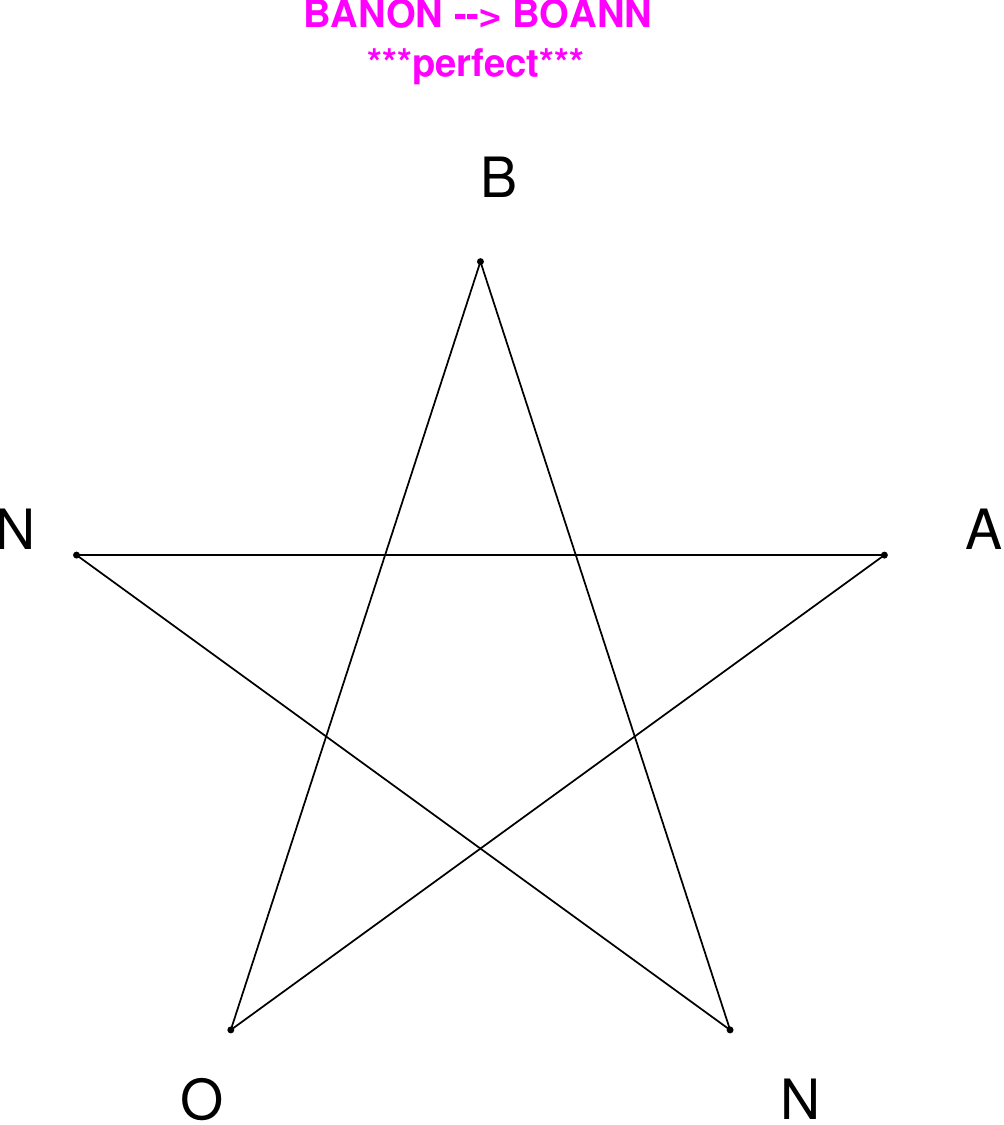}
\end{subfigure}
\hfill
\begin{subfigure}[T]{0.19\textwidth}
\centering
\includegraphics[width=\textwidth]{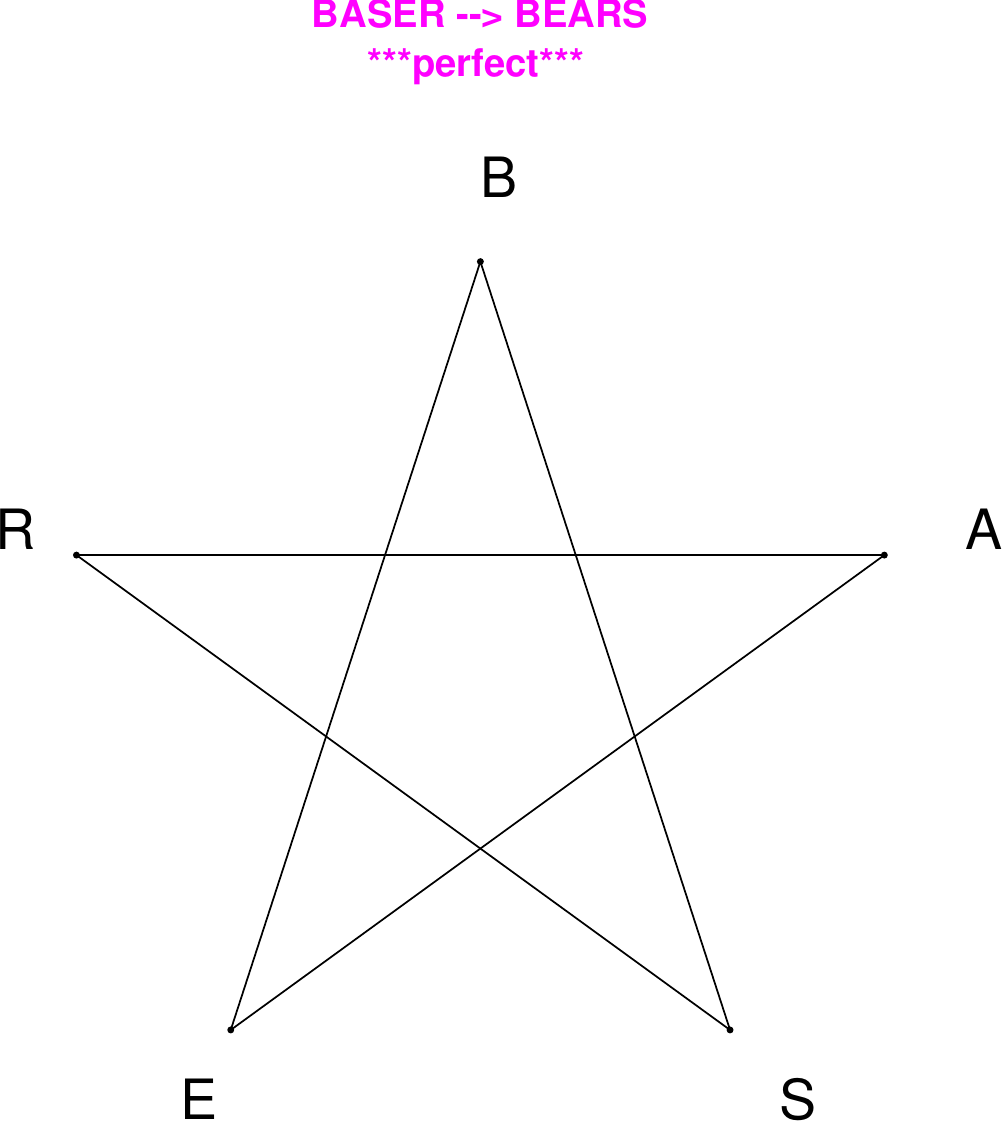}
\end{subfigure}
\end{figure}

\begin{figure}[H]
\centering
\begin{subfigure}[T]{0.19\textwidth}
\centering
\includegraphics[width=\textwidth]{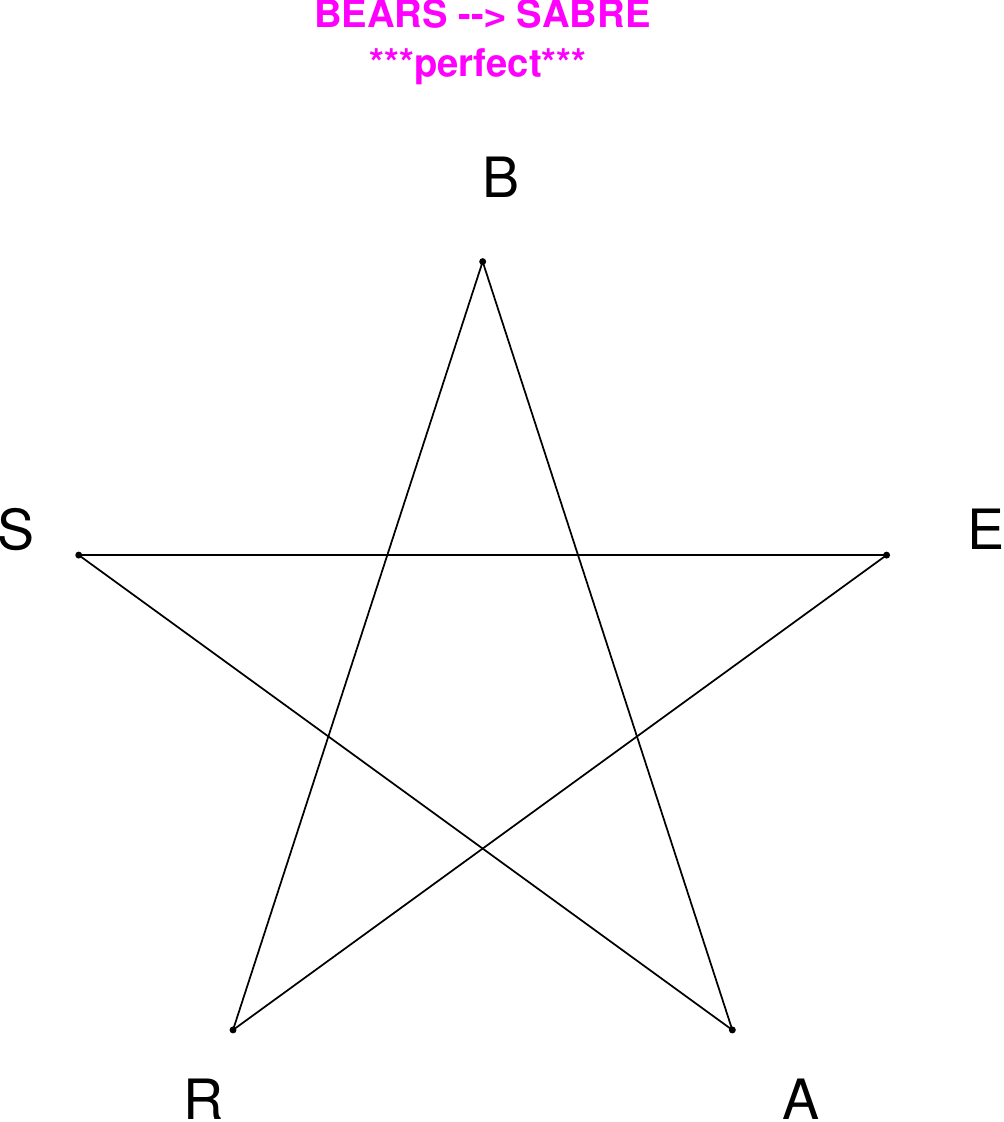}
\end{subfigure}
\hfill
\begin{subfigure}[T]{0.19\textwidth}
\centering
\includegraphics[width=\textwidth]{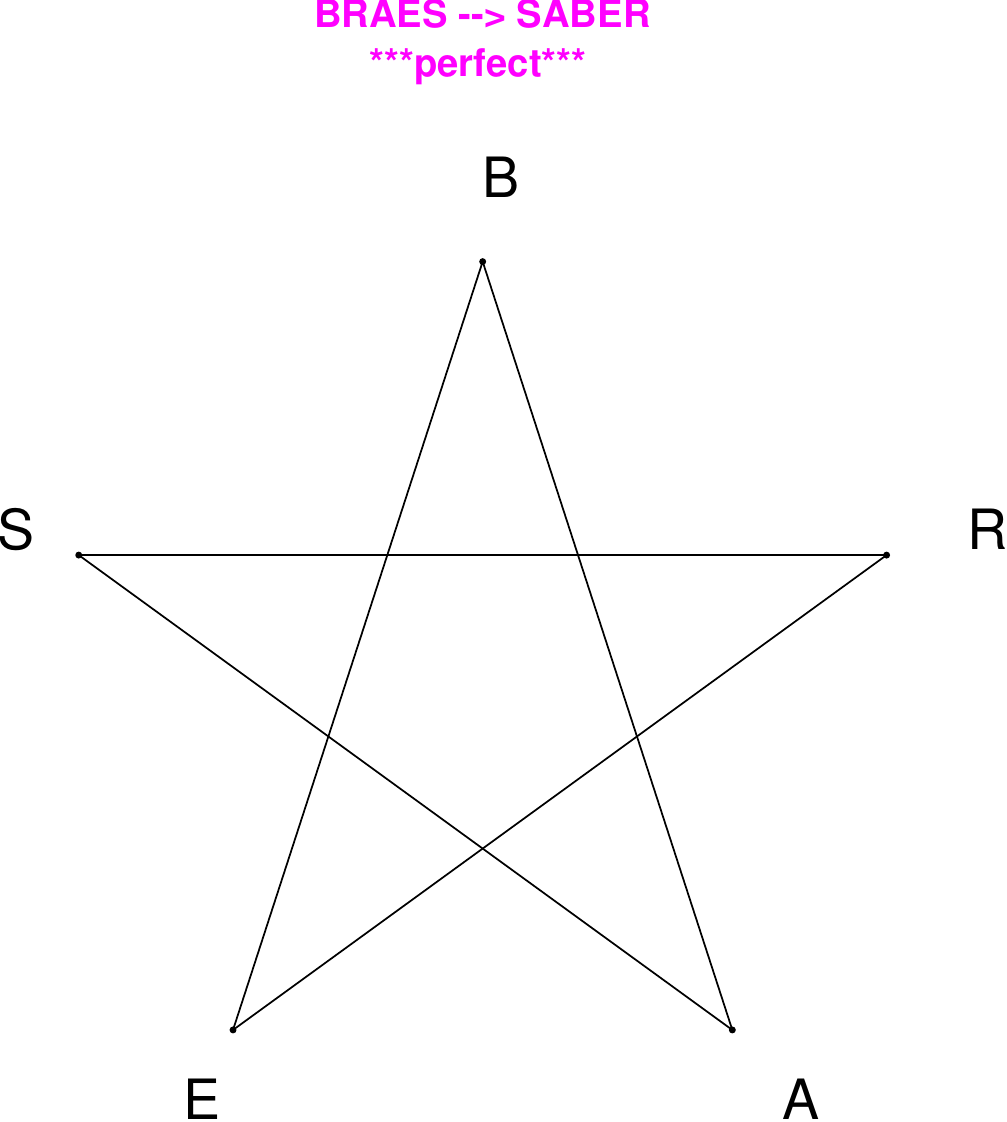}
\end{subfigure}
\hfill
\begin{subfigure}[T]{0.19\textwidth}
\centering
\includegraphics[width=\textwidth]{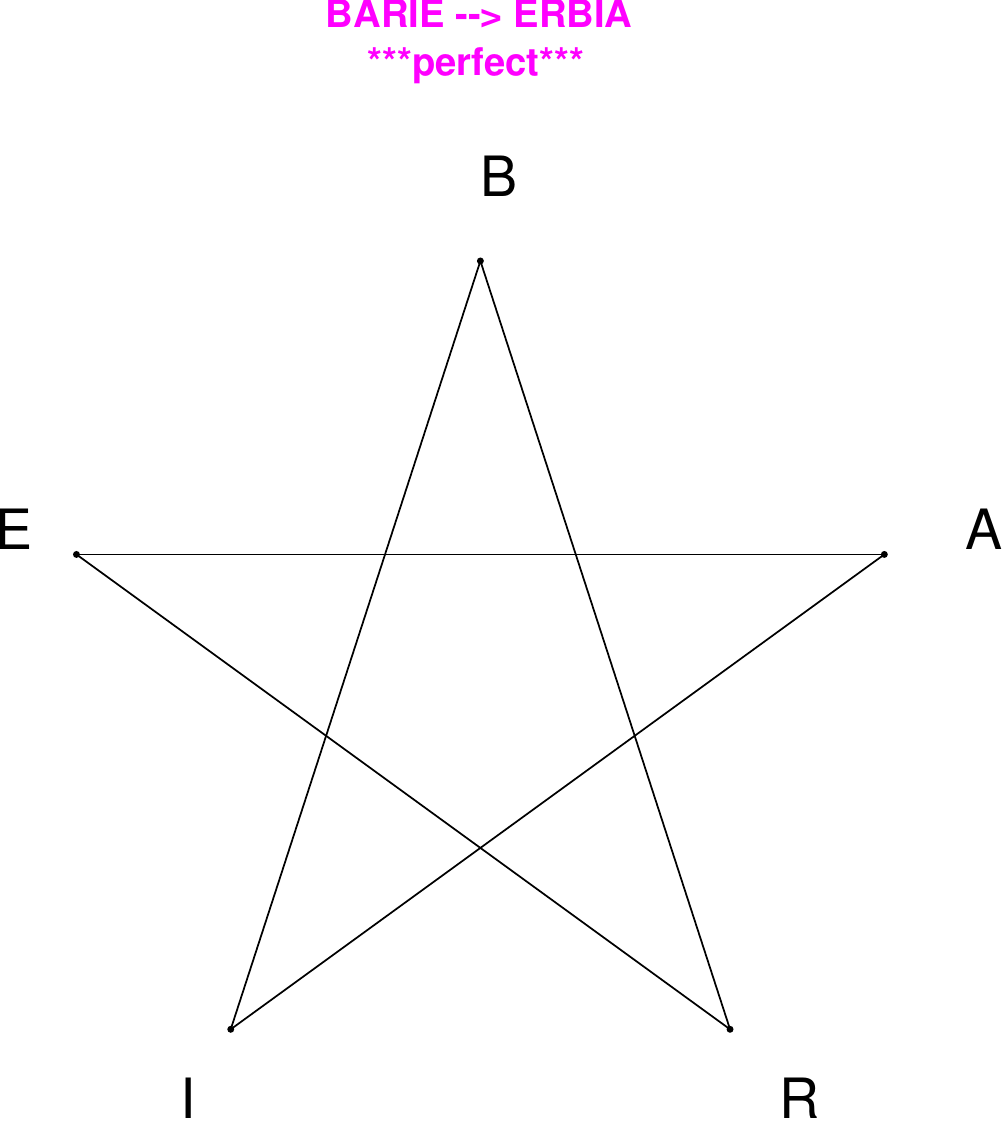}
\end{subfigure}
\hfill
\begin{subfigure}[T]{0.19\textwidth}
\centering
\includegraphics[width=\textwidth]{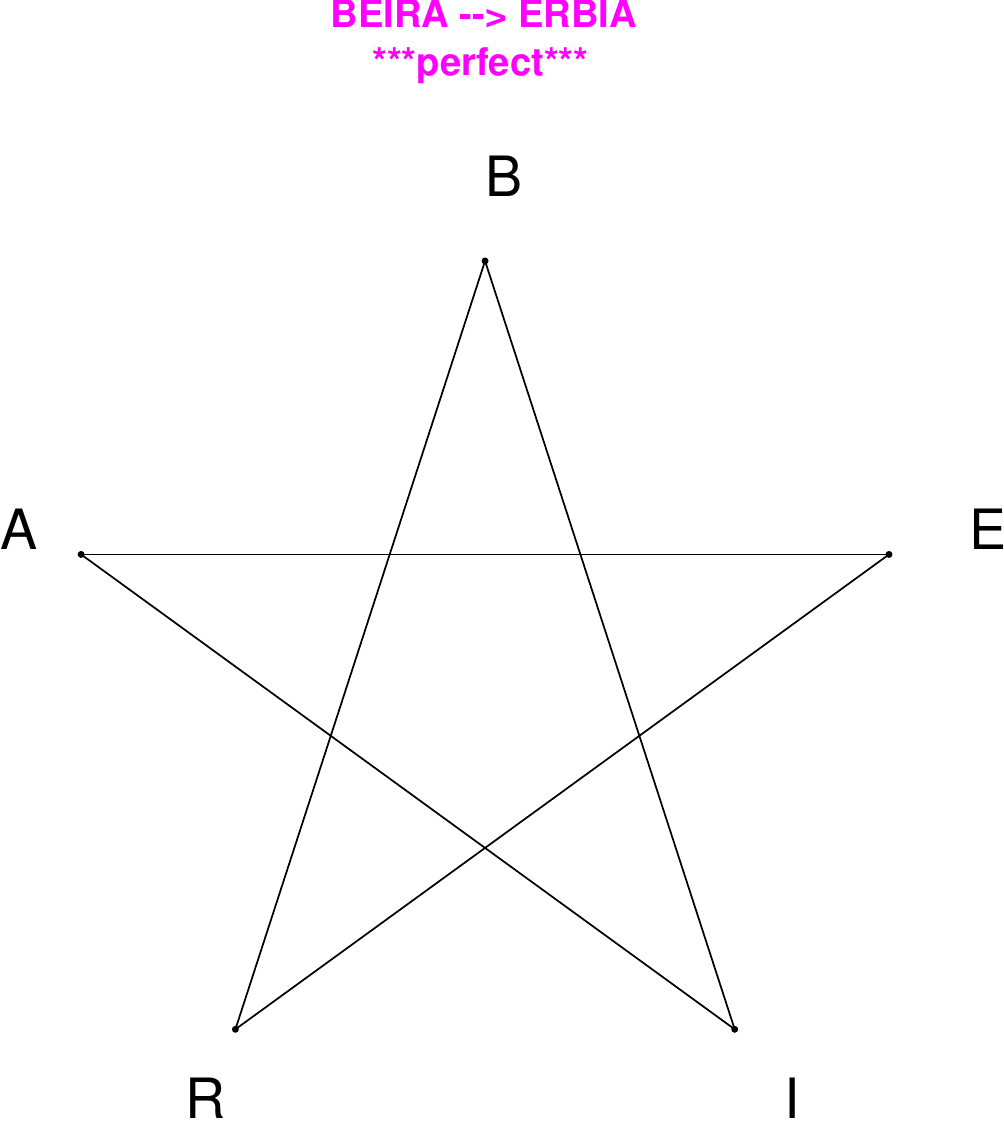}
\end{subfigure}
\hfill
\begin{subfigure}[T]{0.19\textwidth}
\centering
\includegraphics[width=\textwidth]{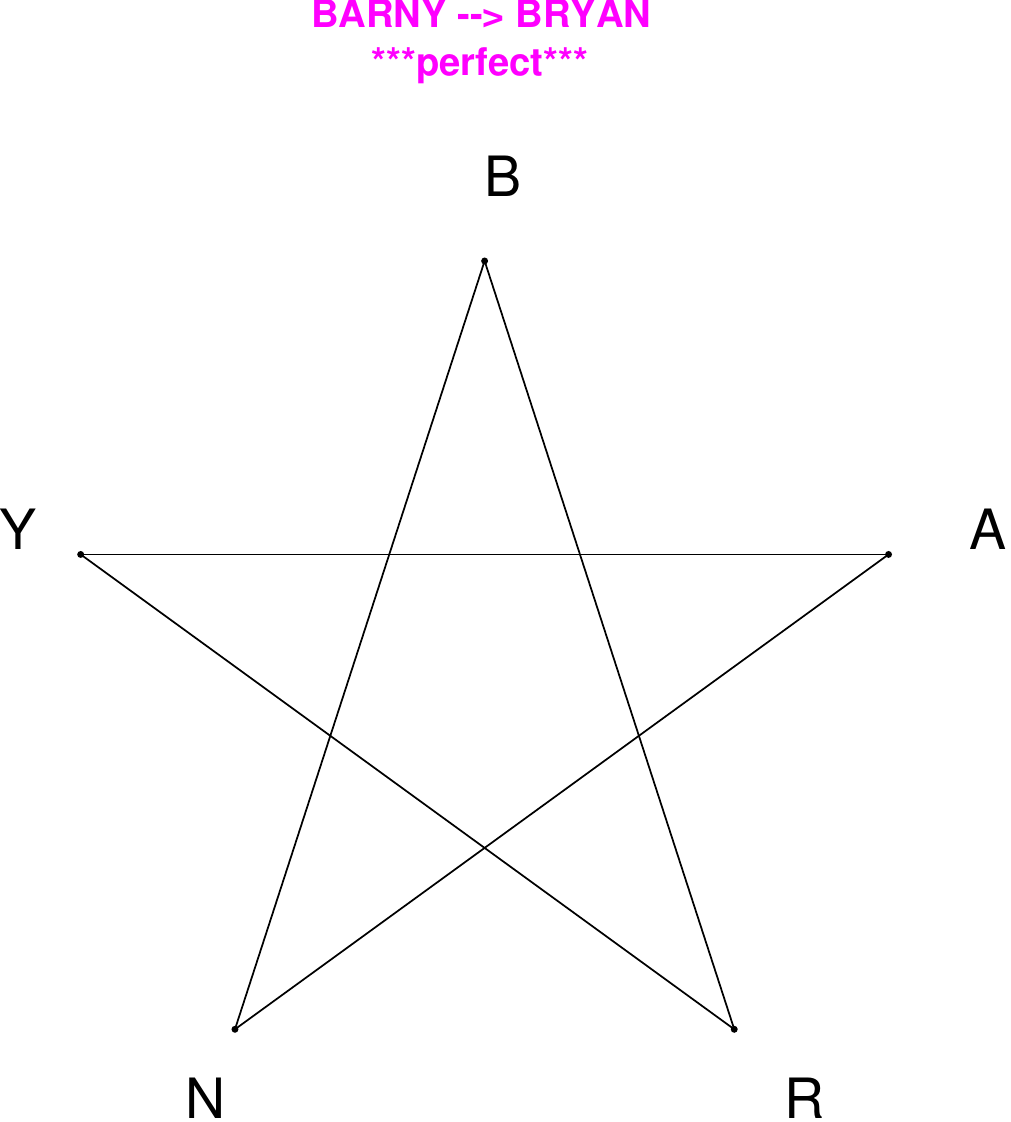}
\end{subfigure}
\end{figure}

\begin{figure}[H]
\centering
\begin{subfigure}[T]{0.19\textwidth}
\centering
\includegraphics[width=\textwidth]{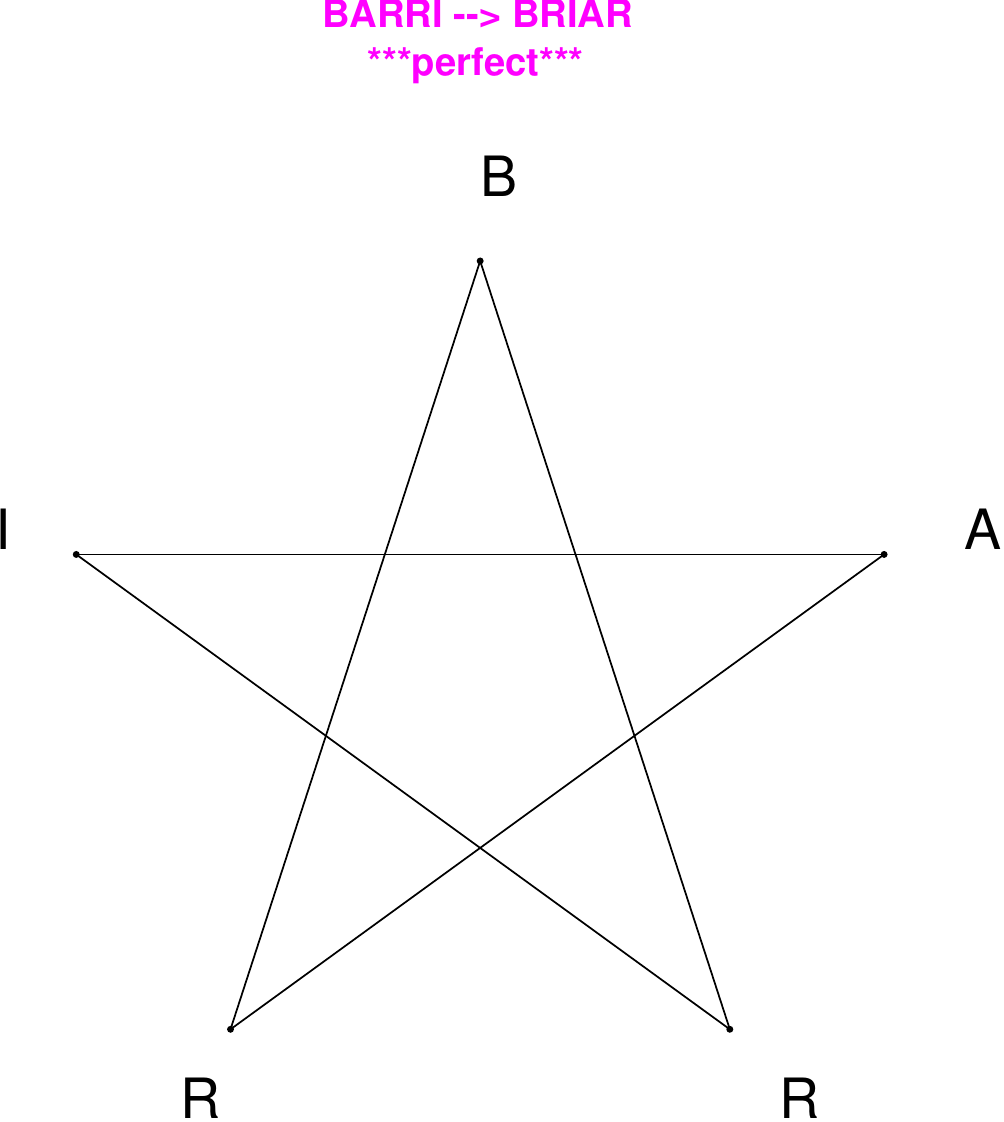}
\end{subfigure}
\hfill
\begin{subfigure}[T]{0.19\textwidth}
\centering
\includegraphics[width=\textwidth]{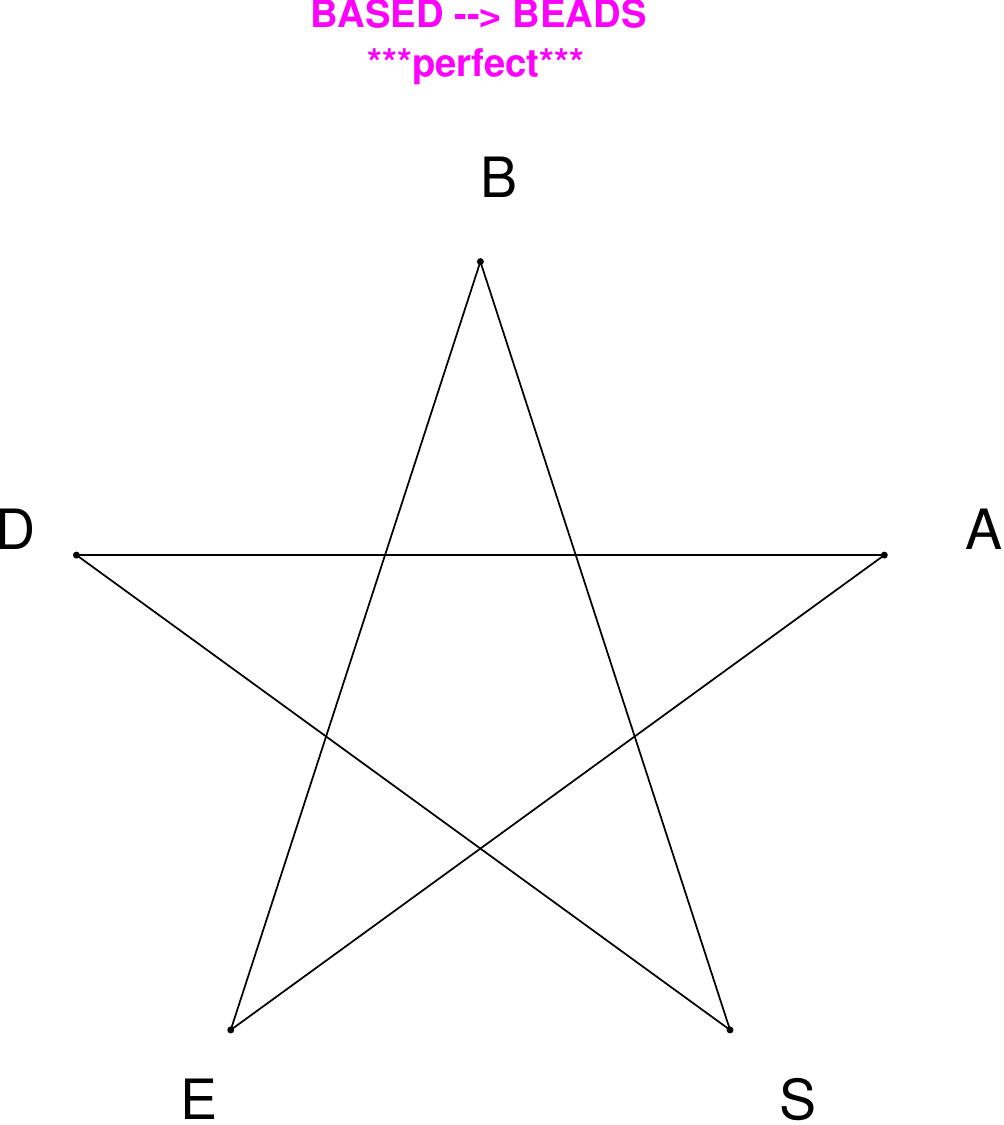}
\end{subfigure}
\hfill
\begin{subfigure}[T]{0.19\textwidth}
\centering
\includegraphics[width=\textwidth]{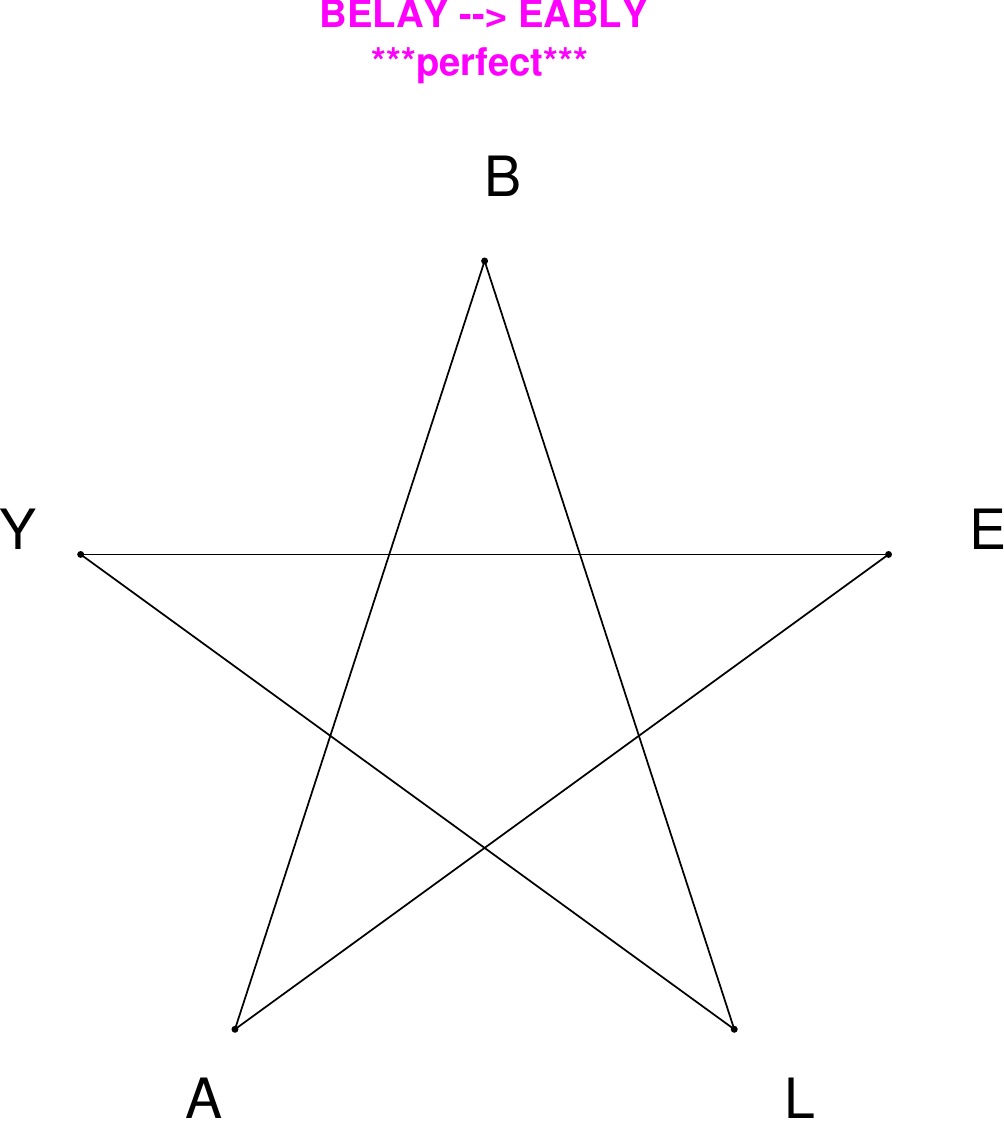}
\end{subfigure}
\hfill
\begin{subfigure}[T]{0.19\textwidth}
\centering
\includegraphics[width=\textwidth]{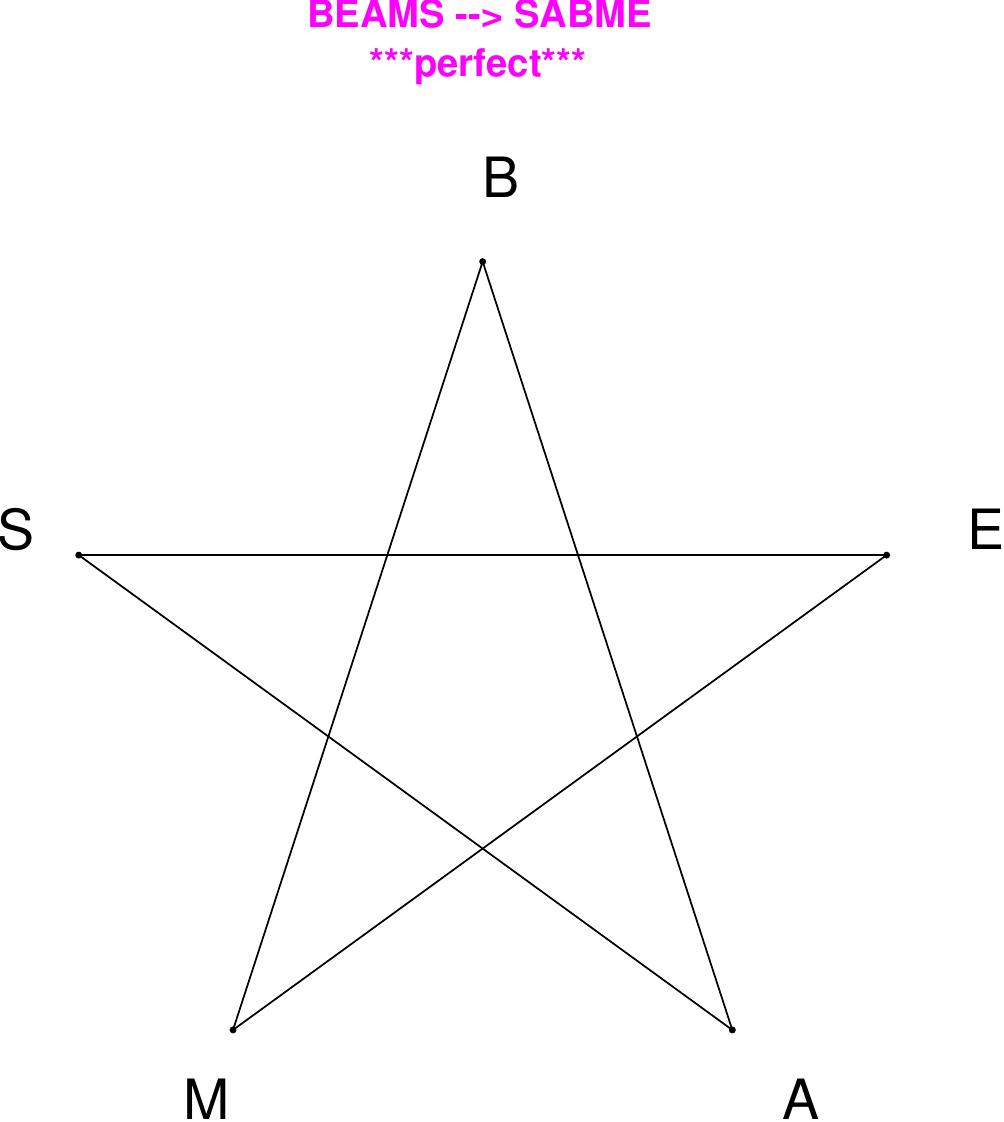}
\end{subfigure}
\hfill
\begin{subfigure}[T]{0.19\textwidth}
\centering
\includegraphics[width=\textwidth]{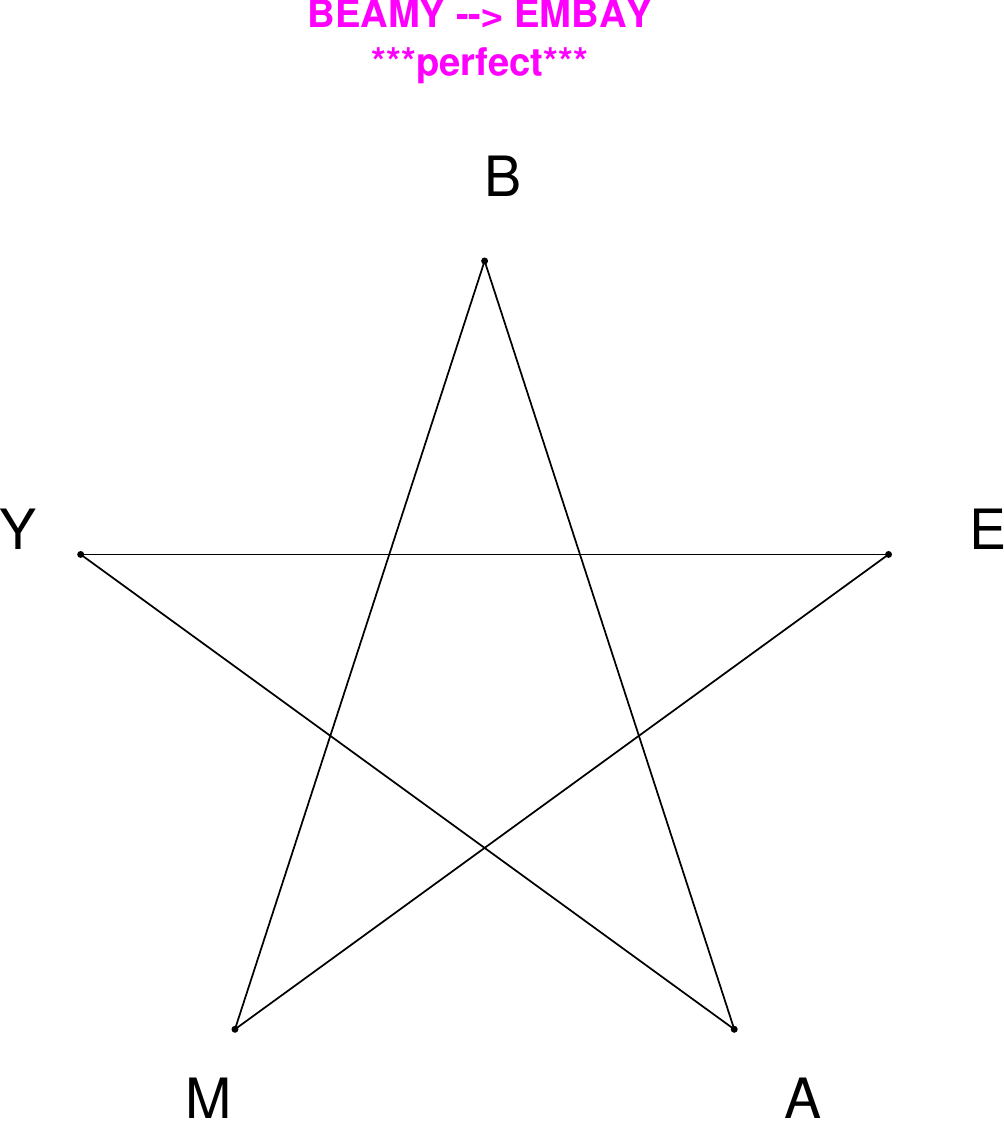}
\end{subfigure}
\end{figure}

\begin{figure}[H]
\centering
\begin{subfigure}[T]{0.19\textwidth}
\centering
\includegraphics[width=\textwidth]{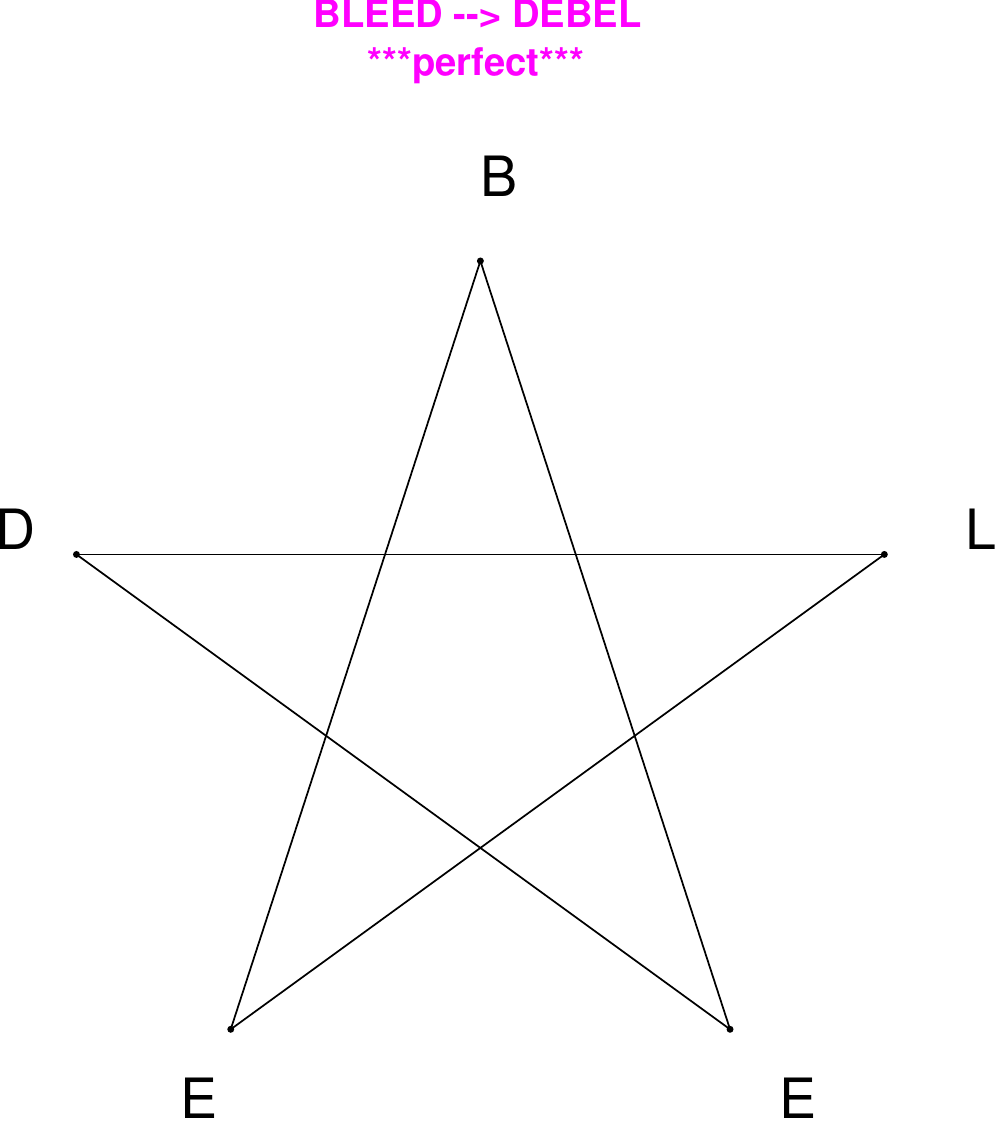}
\end{subfigure}
\hfill
\begin{subfigure}[T]{0.19\textwidth}
\centering
\includegraphics[width=\textwidth]{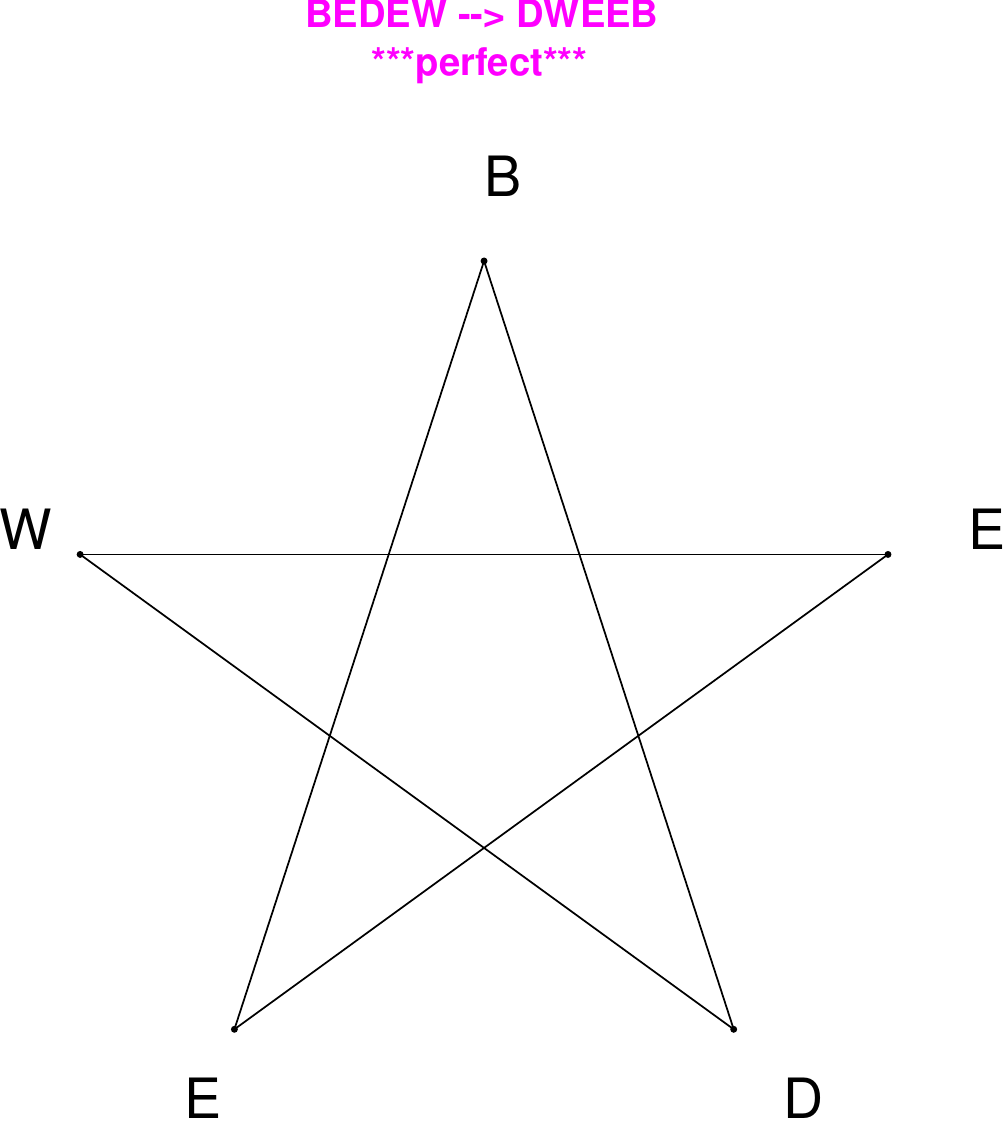}
\end{subfigure}
\hfill
\begin{subfigure}[T]{0.19\textwidth}
\centering
\includegraphics[width=\textwidth]{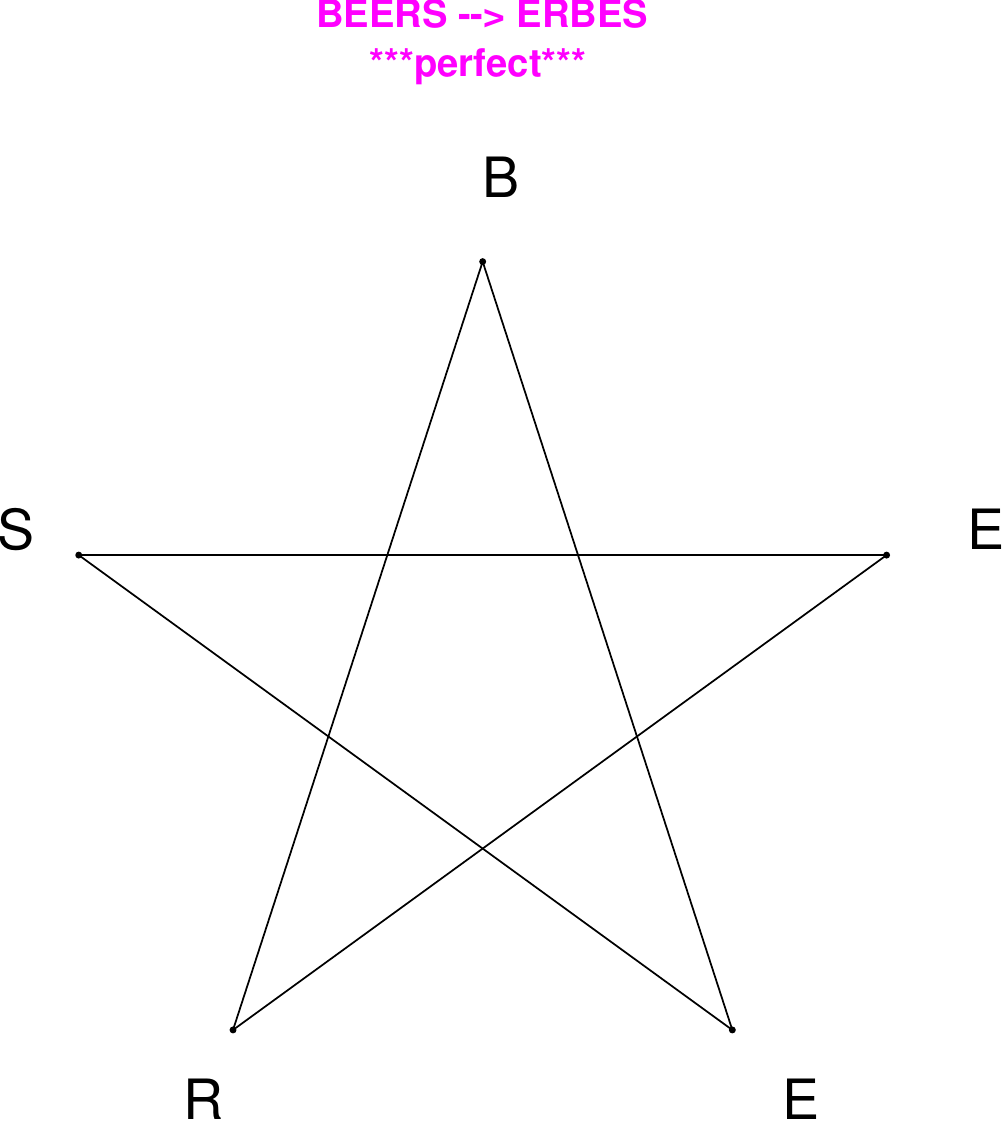}
\end{subfigure}
\hfill
\begin{subfigure}[T]{0.19\textwidth}
\centering
\includegraphics[width=\textwidth]{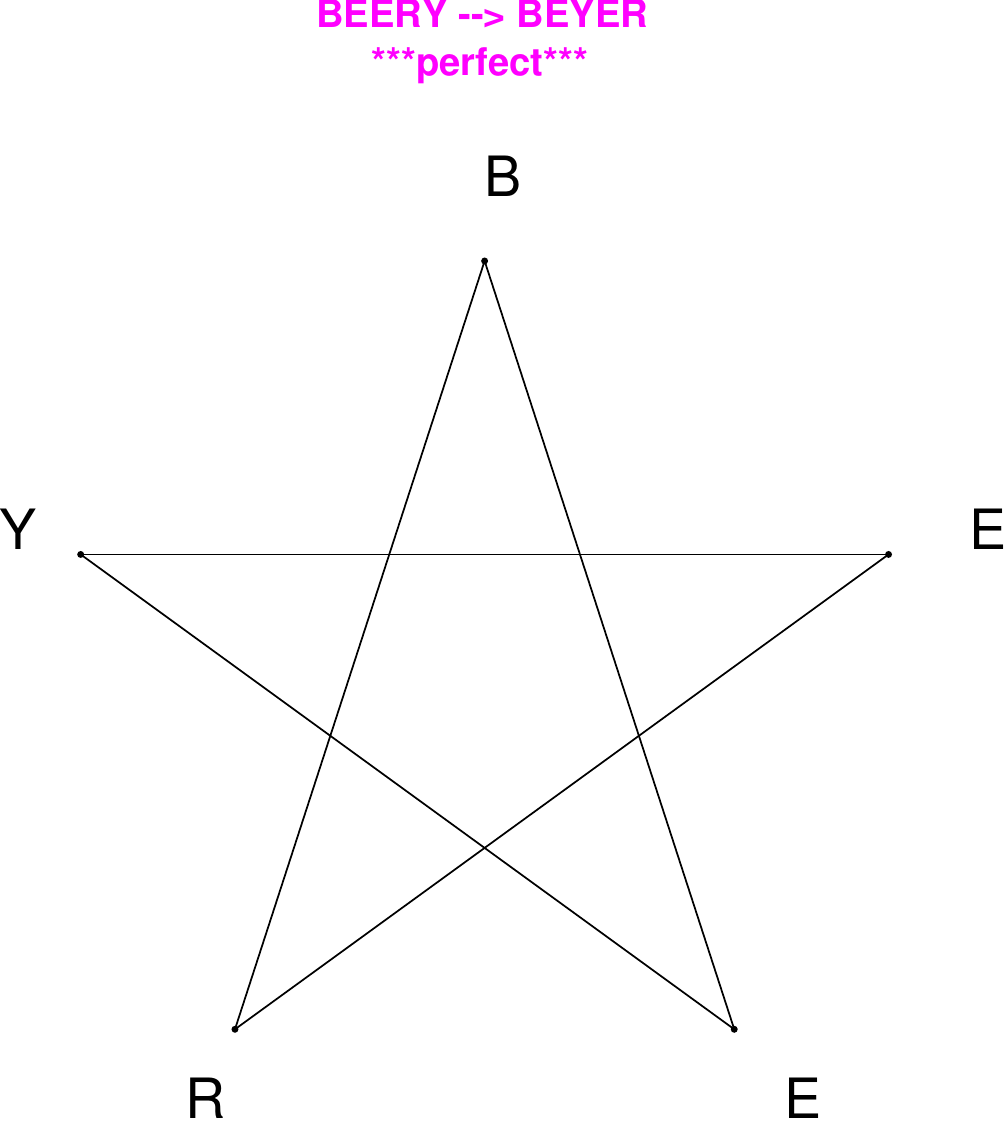}
\end{subfigure}
\hfill
\begin{subfigure}[T]{0.19\textwidth}
\centering
\includegraphics[width=\textwidth]{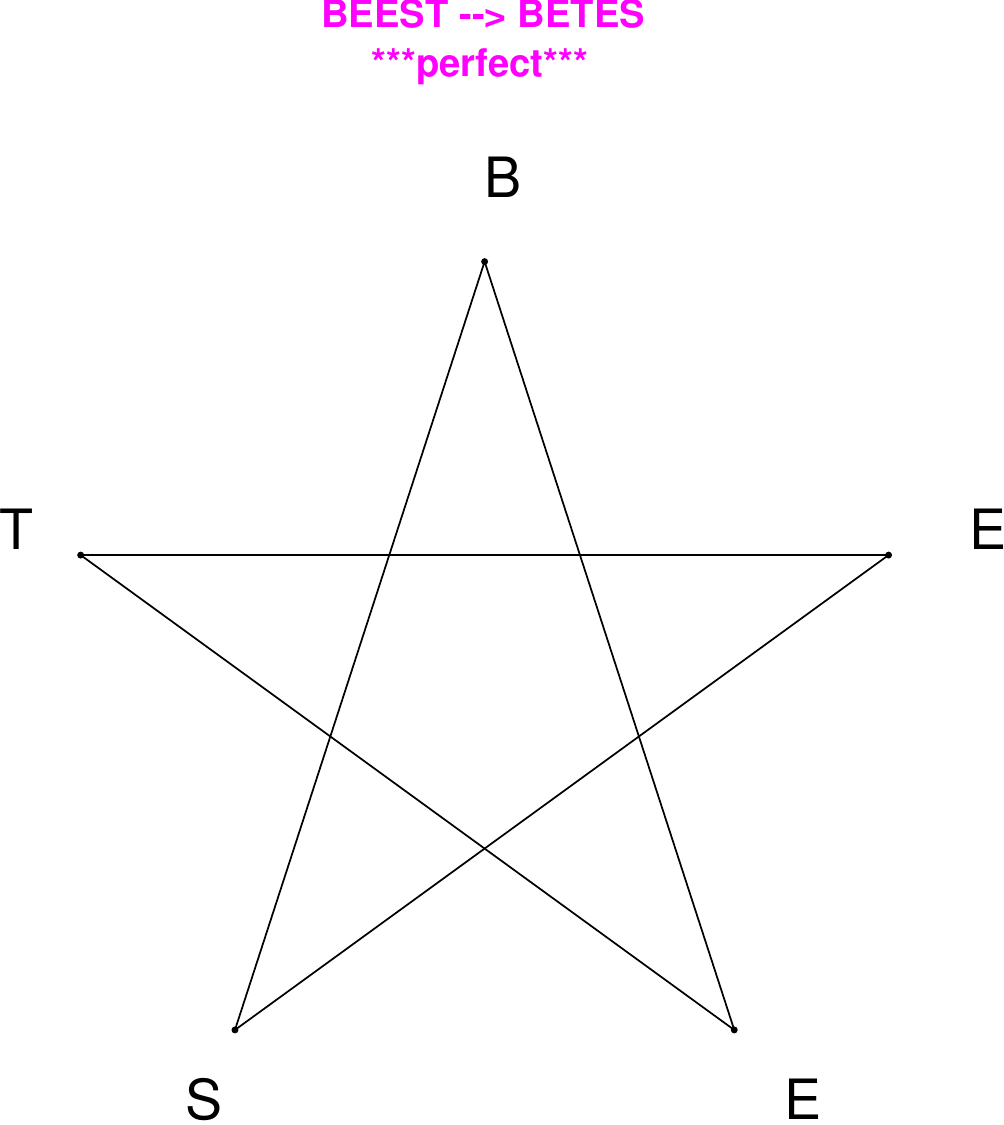}
\end{subfigure}
\end{figure}

\begin{figure}[H]
\centering
\begin{subfigure}[T]{0.19\textwidth}
\centering
\includegraphics[width=\textwidth]{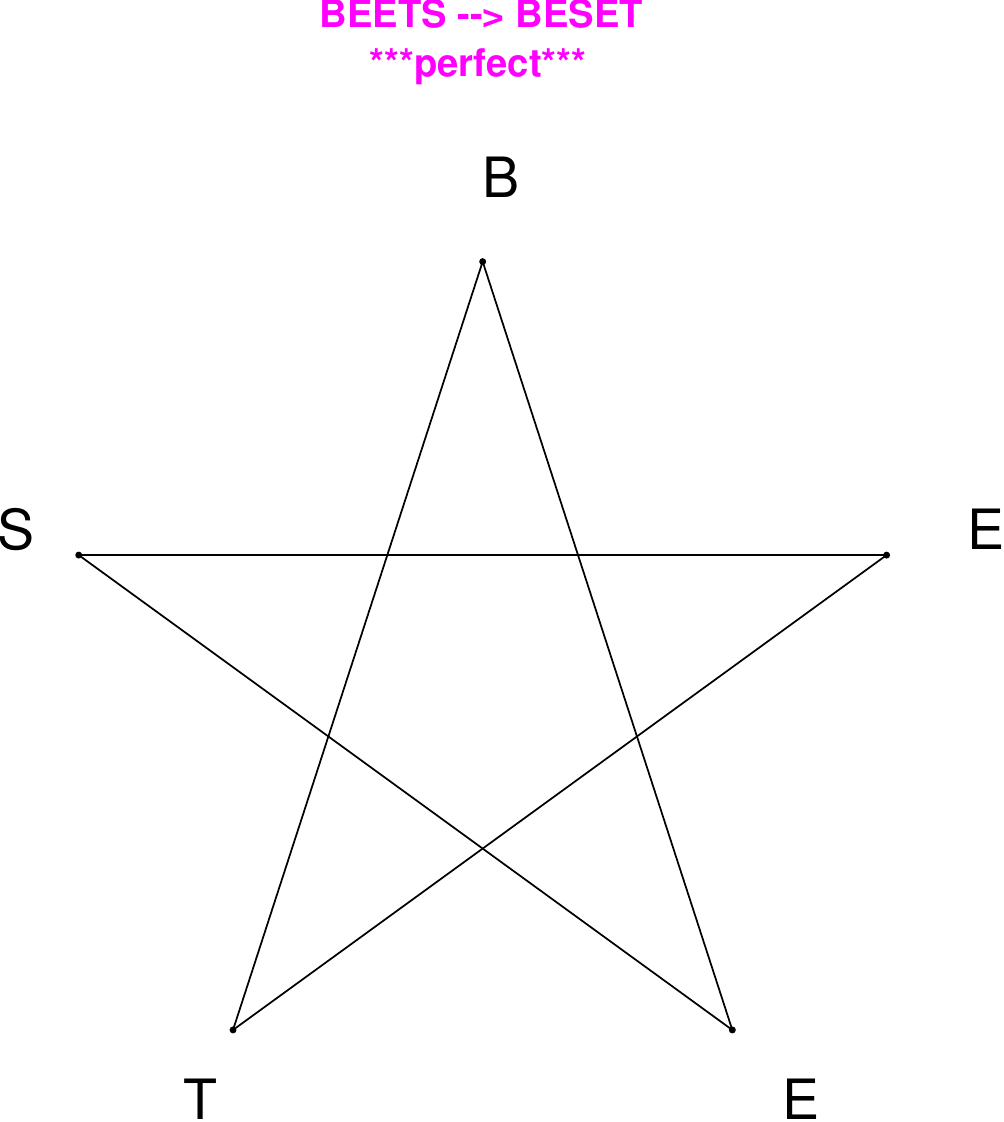}
\end{subfigure}
\hfill
\begin{subfigure}[T]{0.19\textwidth}
\centering
\includegraphics[width=\textwidth]{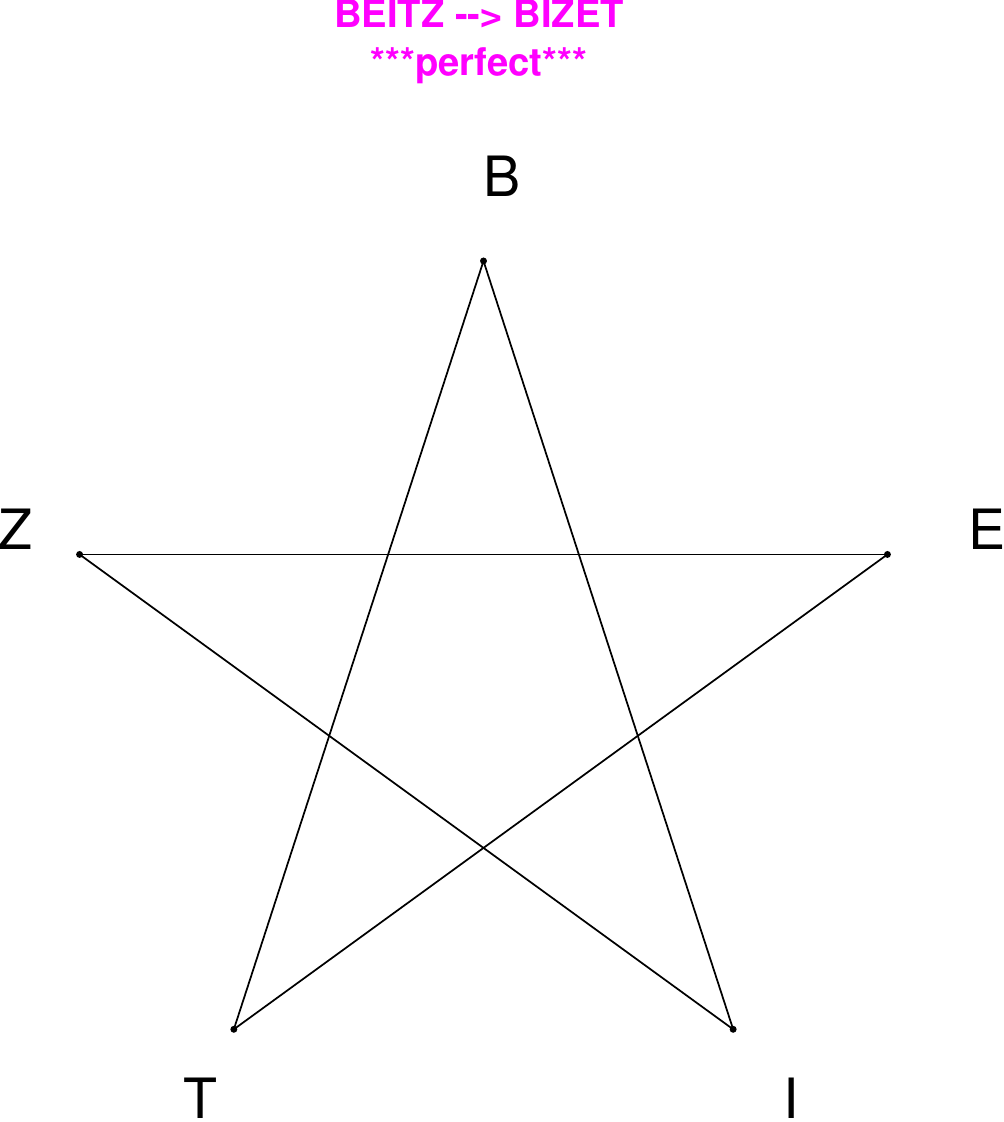}
\end{subfigure}
\hfill
\begin{subfigure}[T]{0.19\textwidth}
\centering
\includegraphics[width=\textwidth]{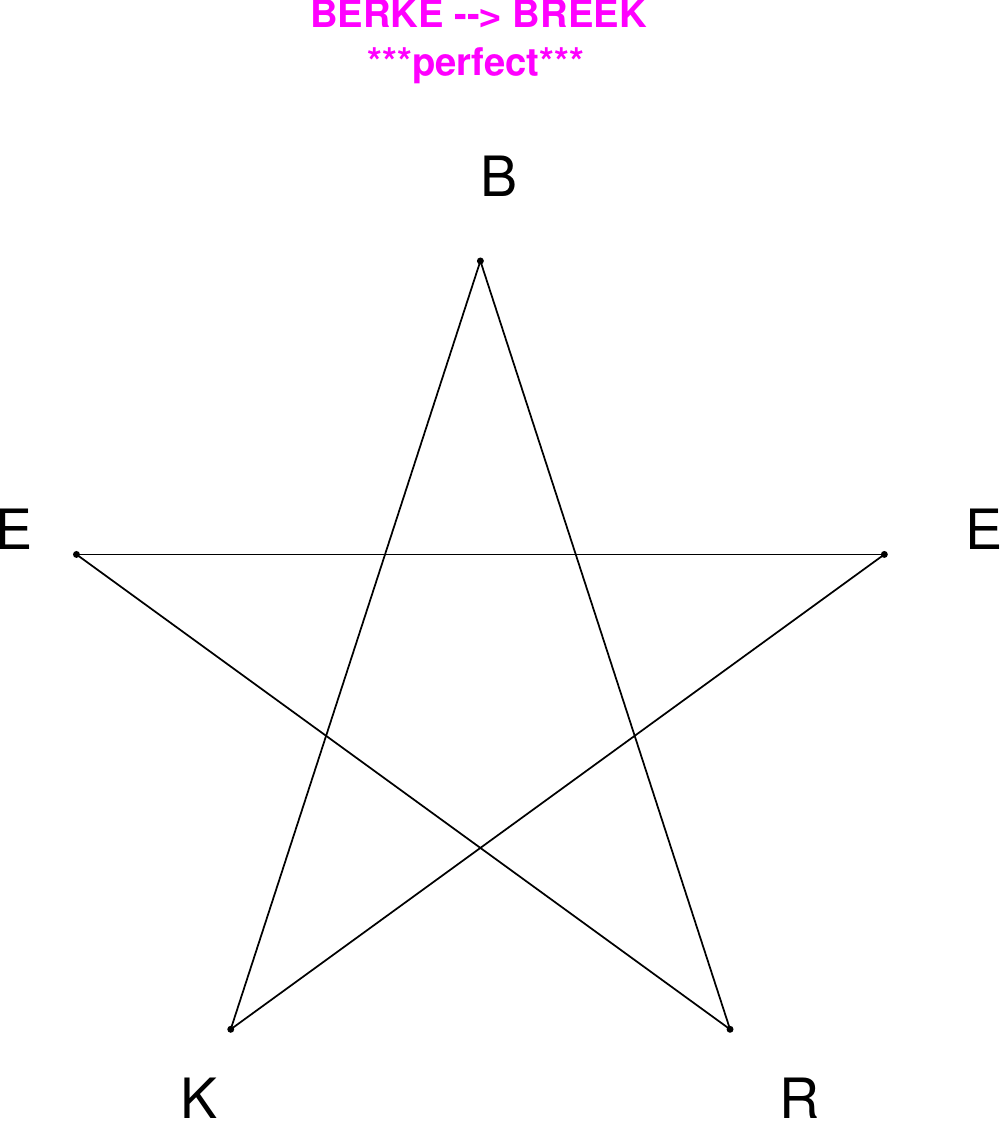}
\end{subfigure}
\hfill
\begin{subfigure}[T]{0.19\textwidth}
\centering
\includegraphics[width=\textwidth]{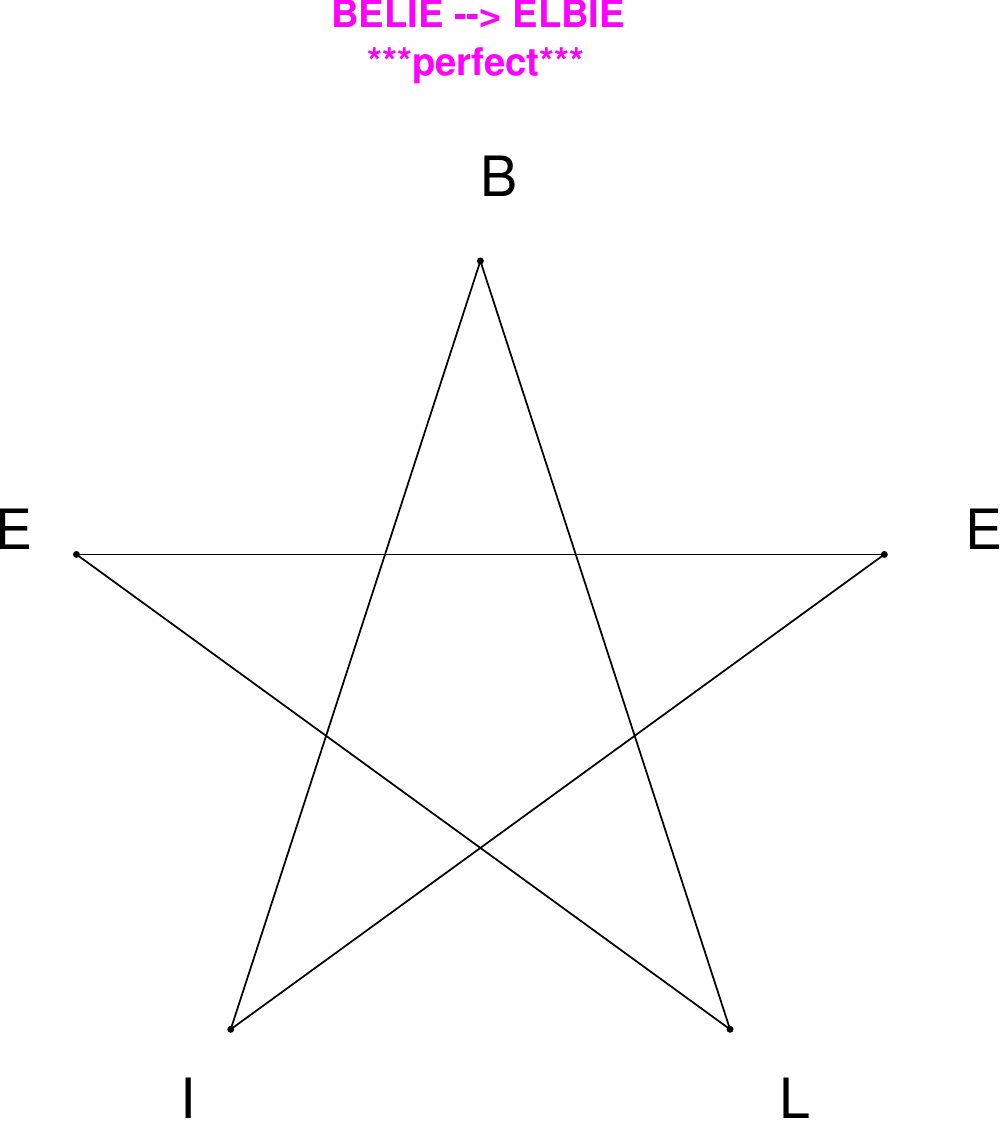}
\end{subfigure}
\hfill
\begin{subfigure}[T]{0.19\textwidth}
\centering
\includegraphics[width=\textwidth]{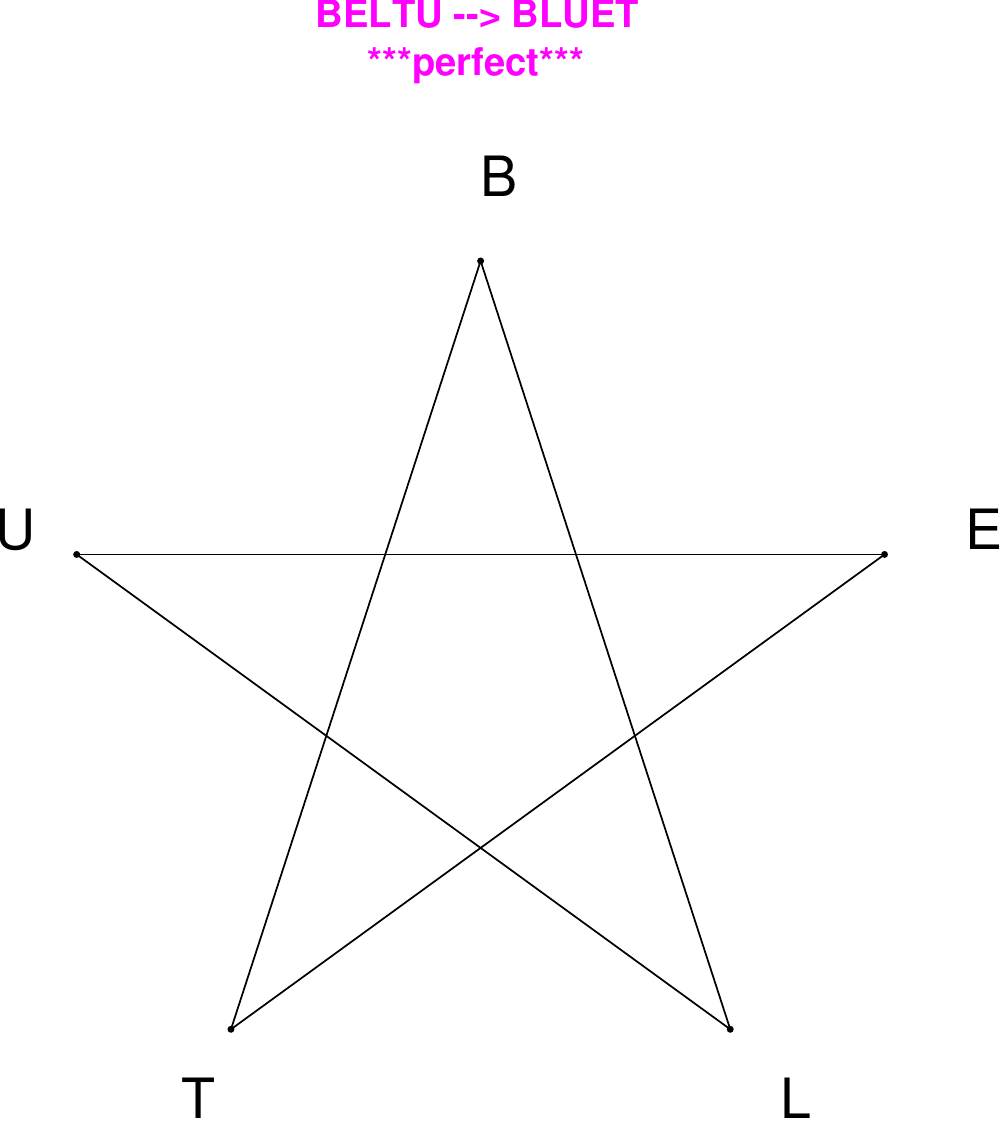}
\end{subfigure}
\end{figure}

\begin{figure}[H]
\centering
\begin{subfigure}[T]{0.19\textwidth}
\centering
\includegraphics[width=\textwidth]{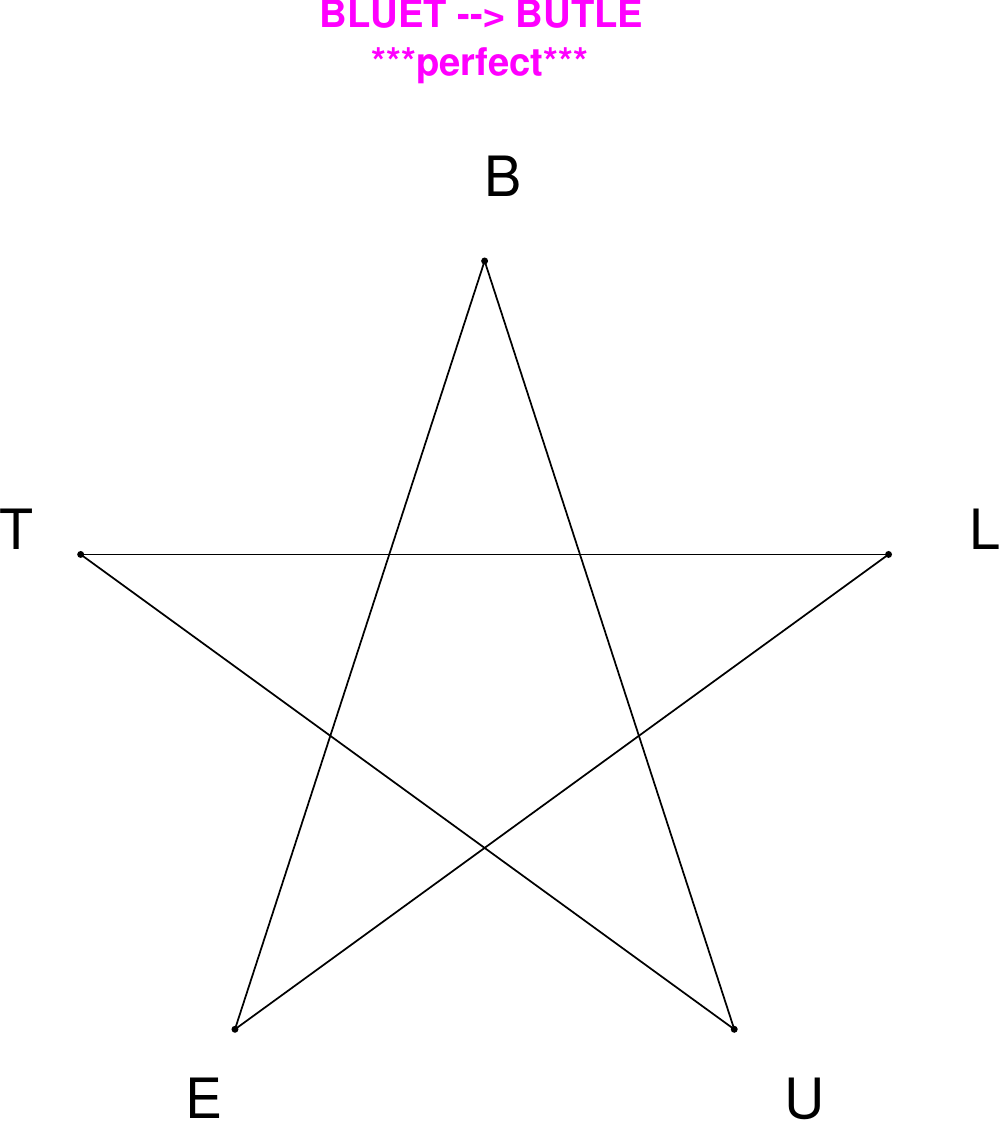}
\end{subfigure}
\hfill
\begin{subfigure}[T]{0.19\textwidth}
\centering
\includegraphics[width=\textwidth]{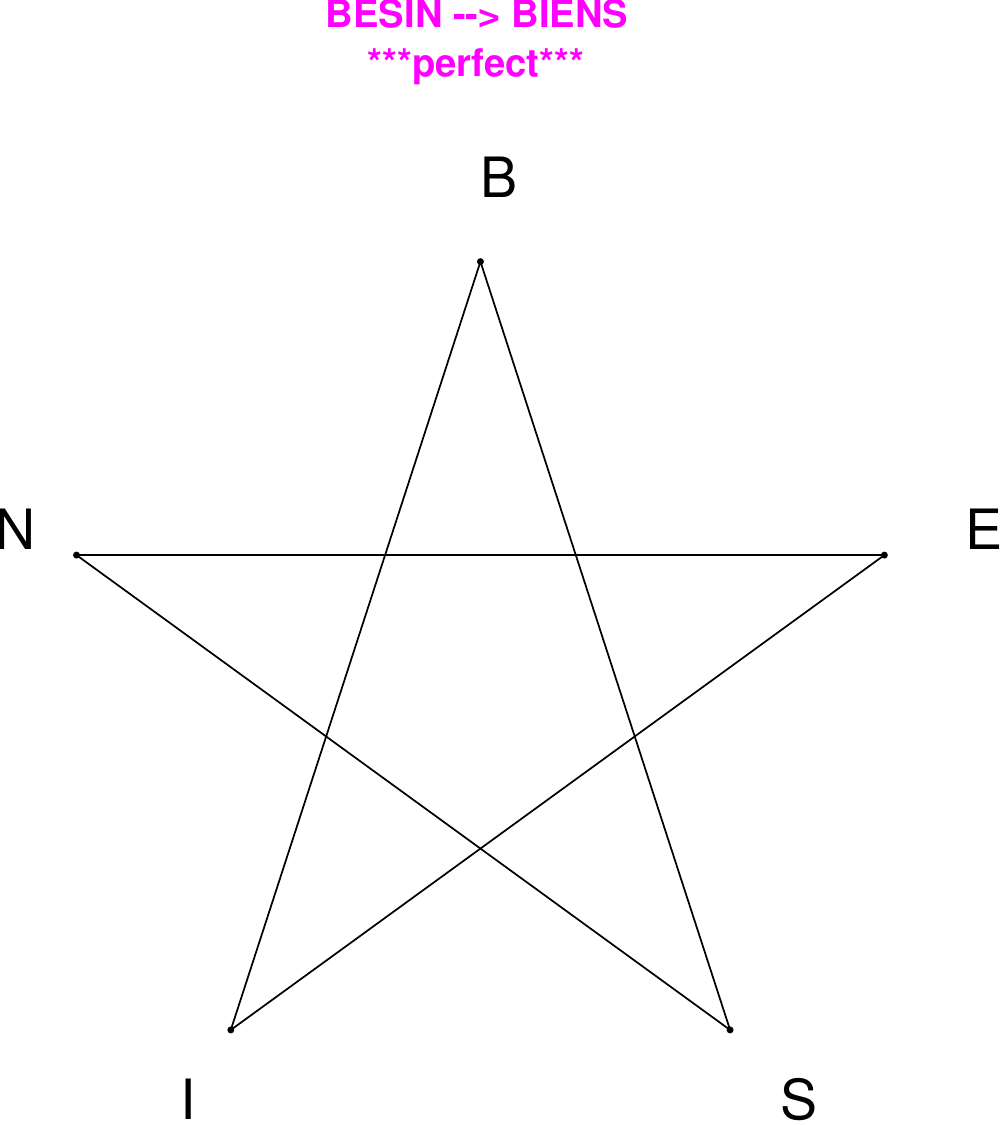}
\end{subfigure}
\hfill
\begin{subfigure}[T]{0.19\textwidth}
\centering
\includegraphics[width=\textwidth]{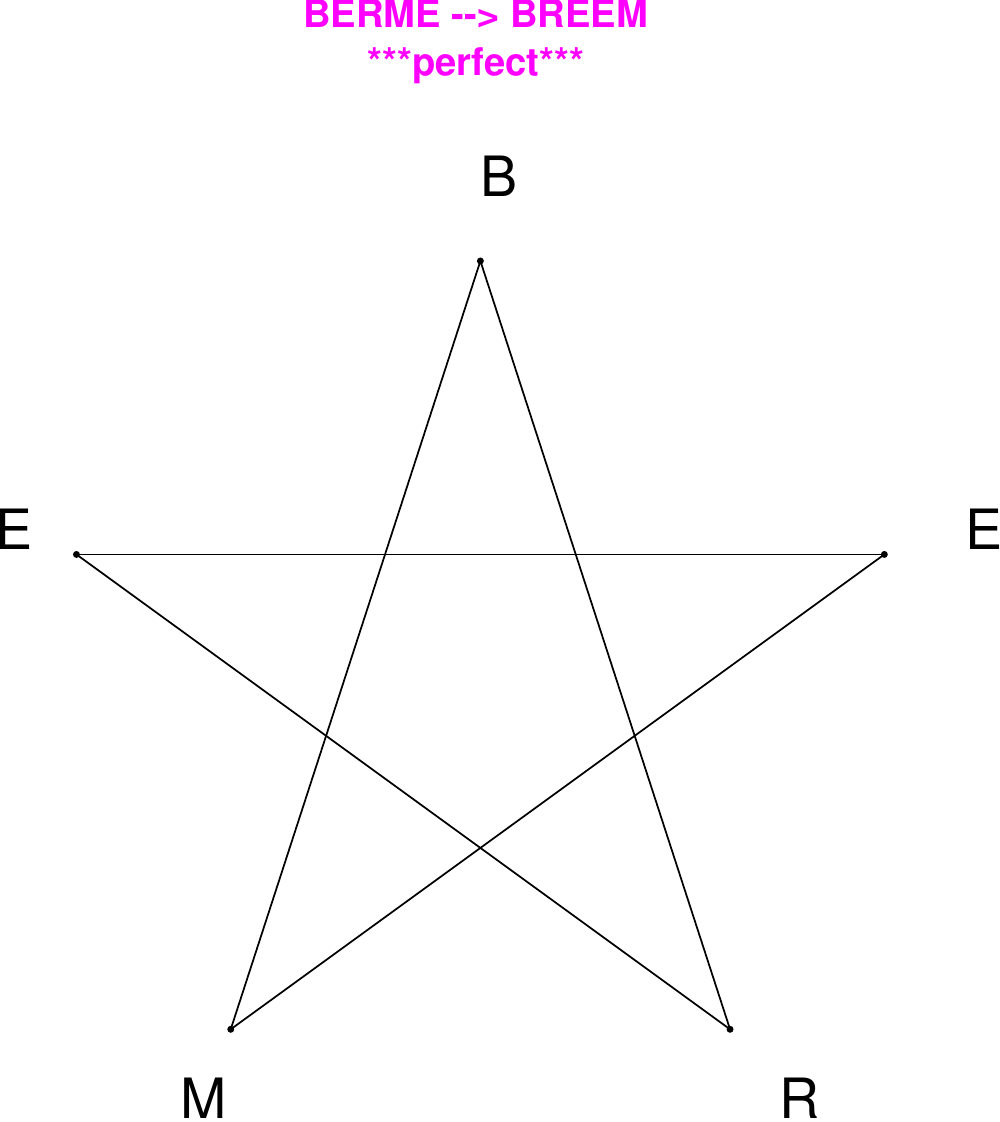}
\end{subfigure}
\hfill
\begin{subfigure}[T]{0.19\textwidth}
\centering
\includegraphics[width=\textwidth]{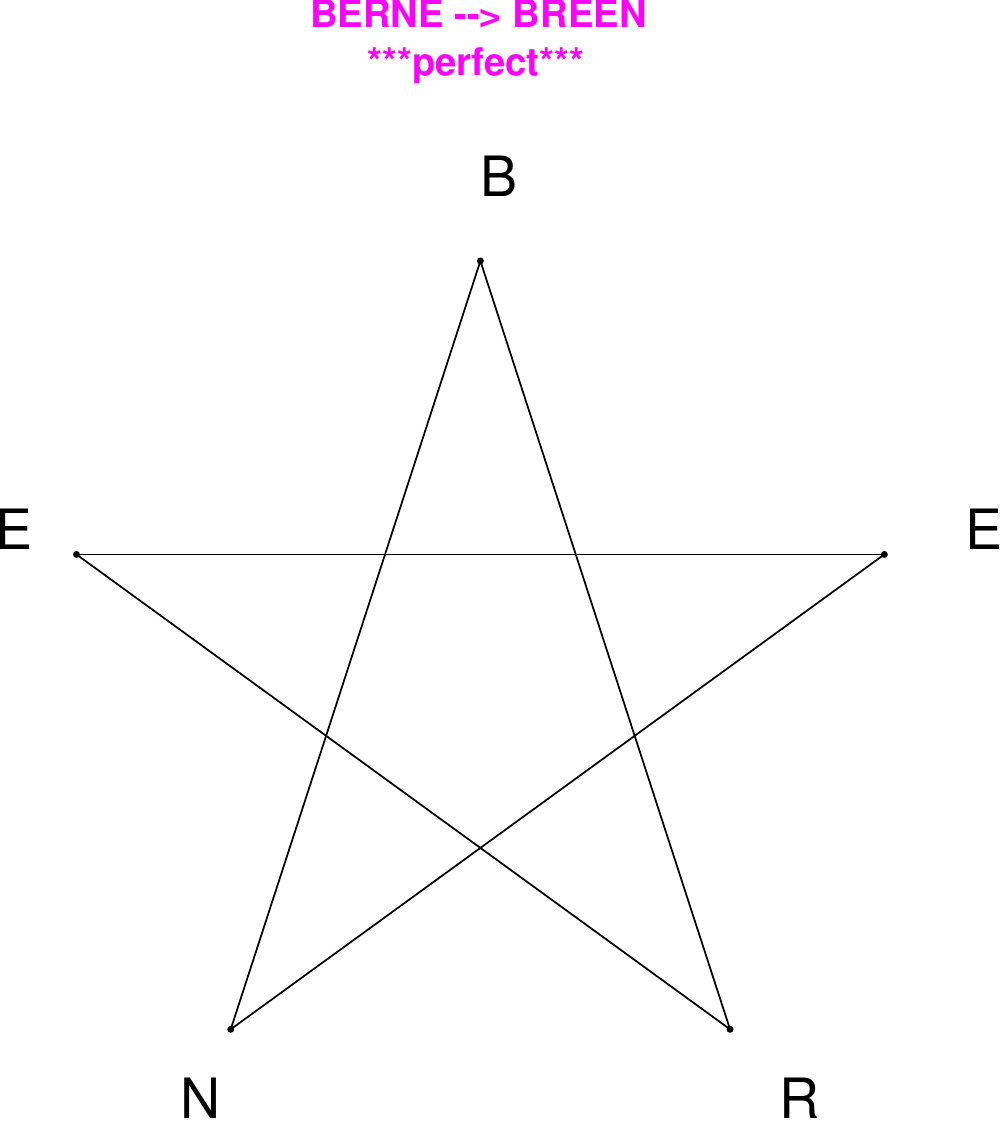}
\end{subfigure}
\hfill
\begin{subfigure}[T]{0.19\textwidth}
\centering
\includegraphics[width=\textwidth]{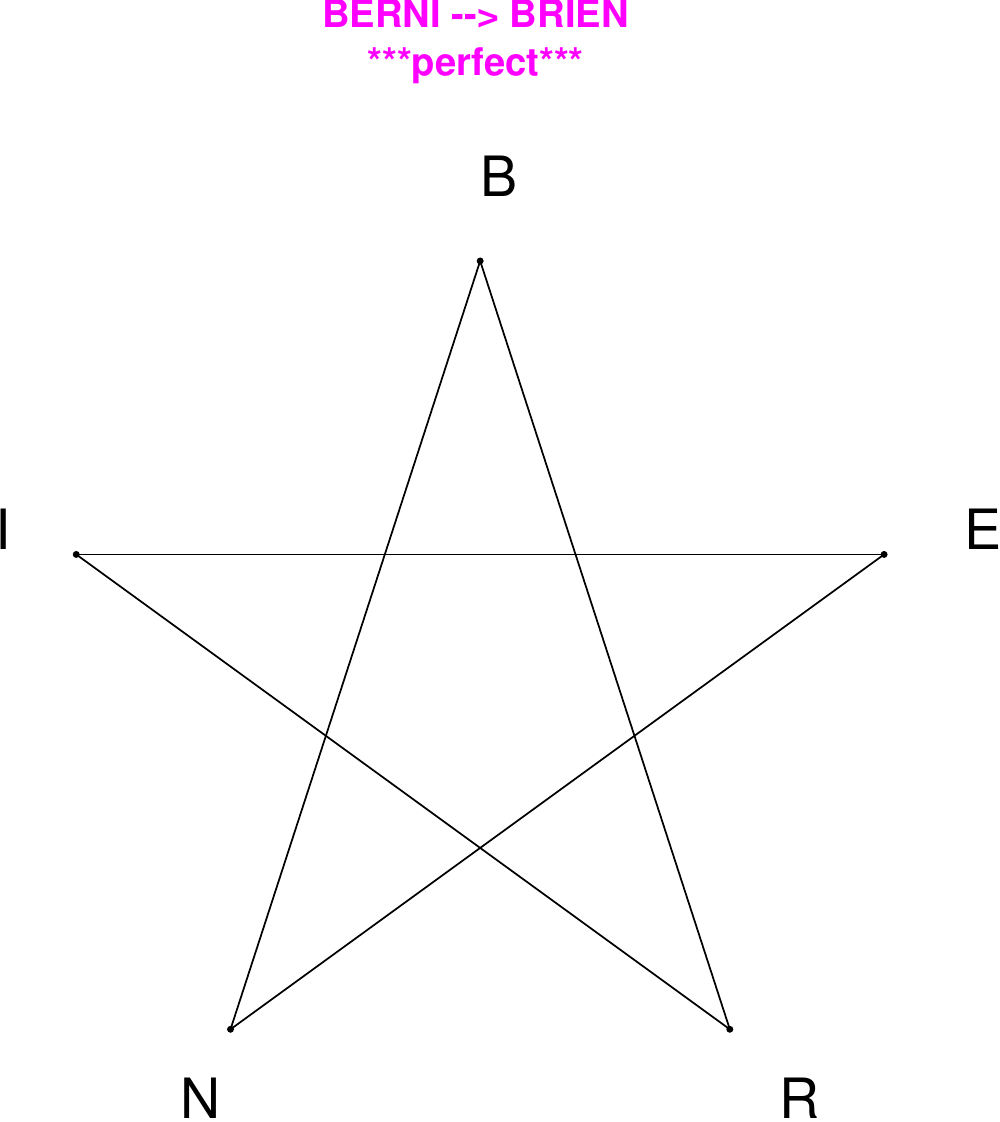}
\end{subfigure}
\end{figure}

\begin{figure}[H]
\centering
\begin{subfigure}[T]{0.19\textwidth}
\centering
\includegraphics[width=\textwidth]{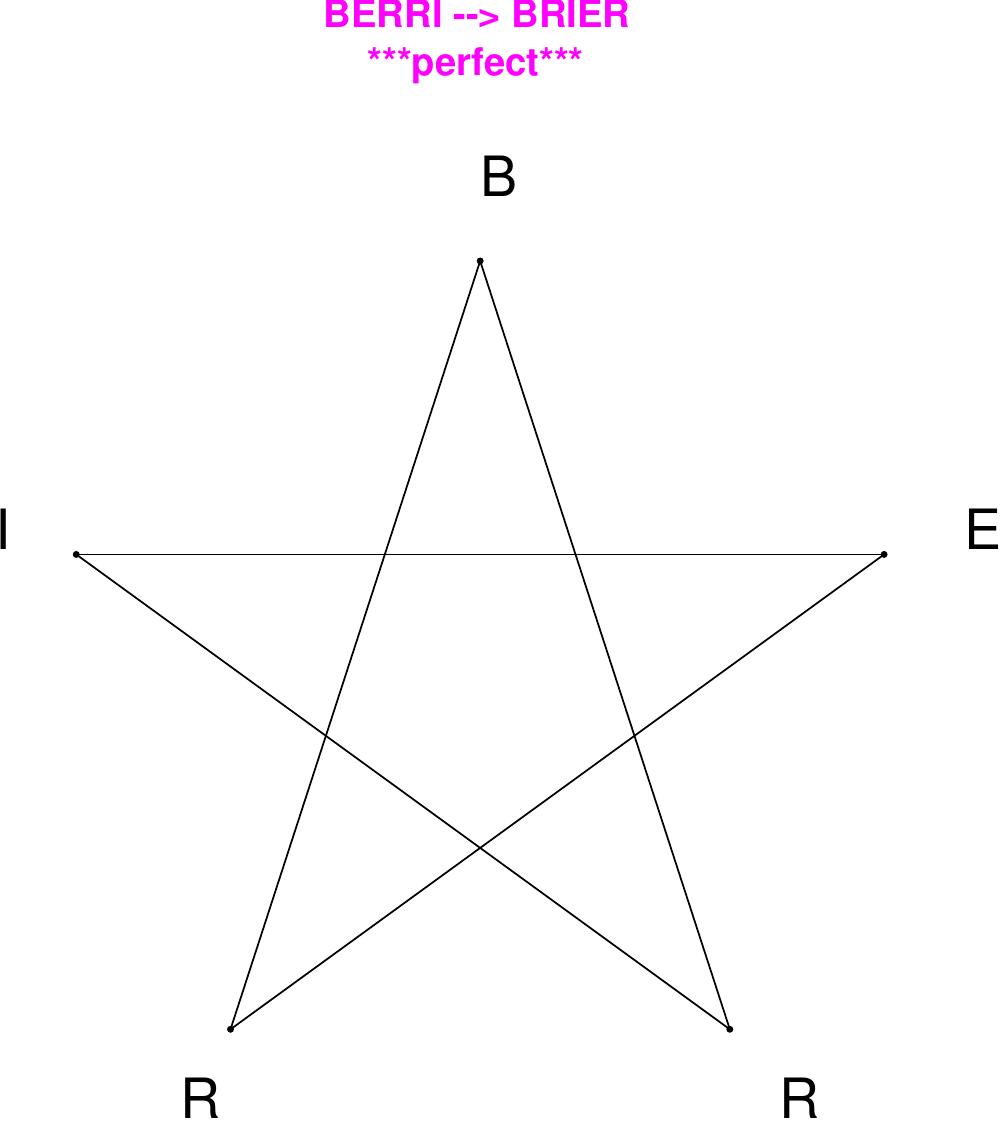}
\end{subfigure}
\hfill
\begin{subfigure}[T]{0.19\textwidth}
\centering
\includegraphics[width=\textwidth]{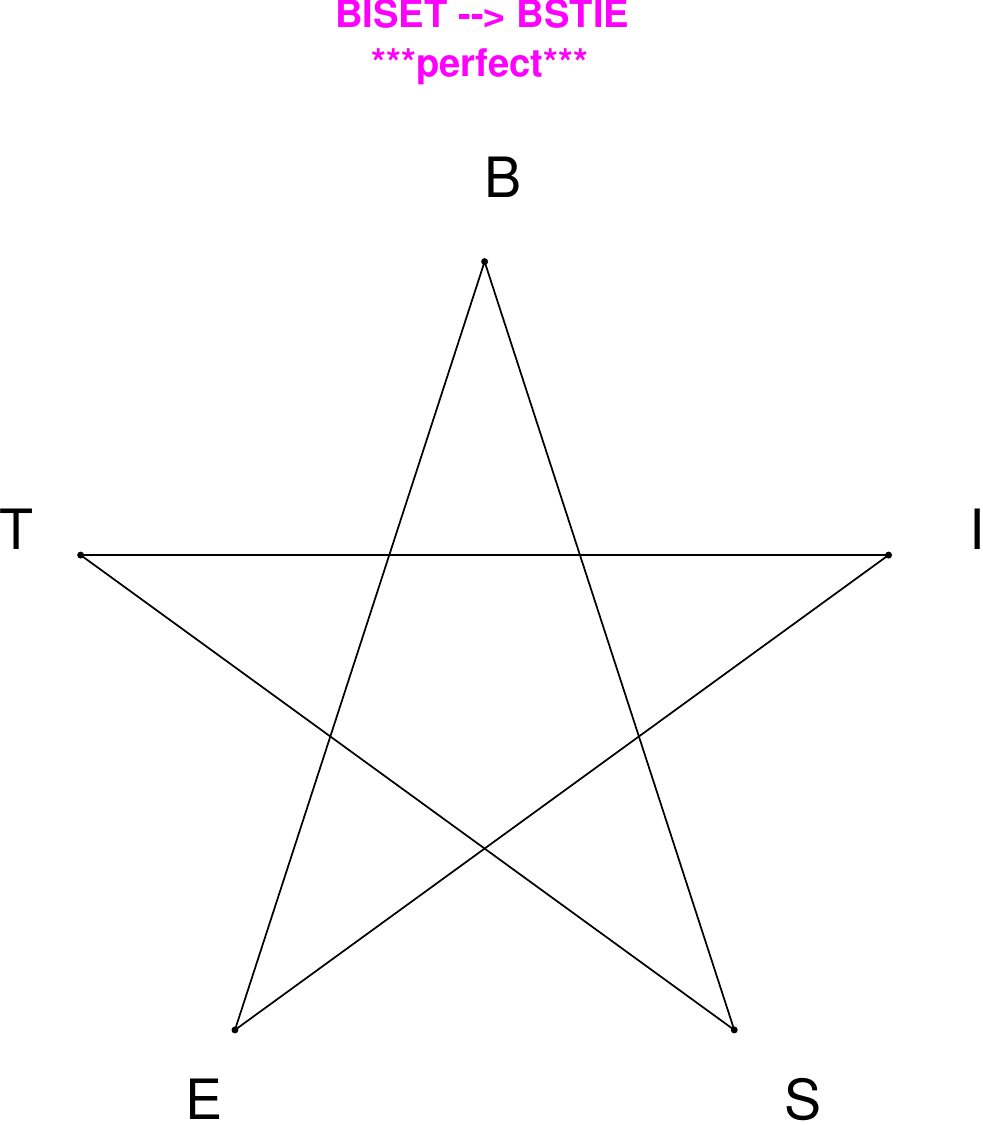}
\end{subfigure}
\hfill
\begin{subfigure}[T]{0.19\textwidth}
\centering
\includegraphics[width=\textwidth]{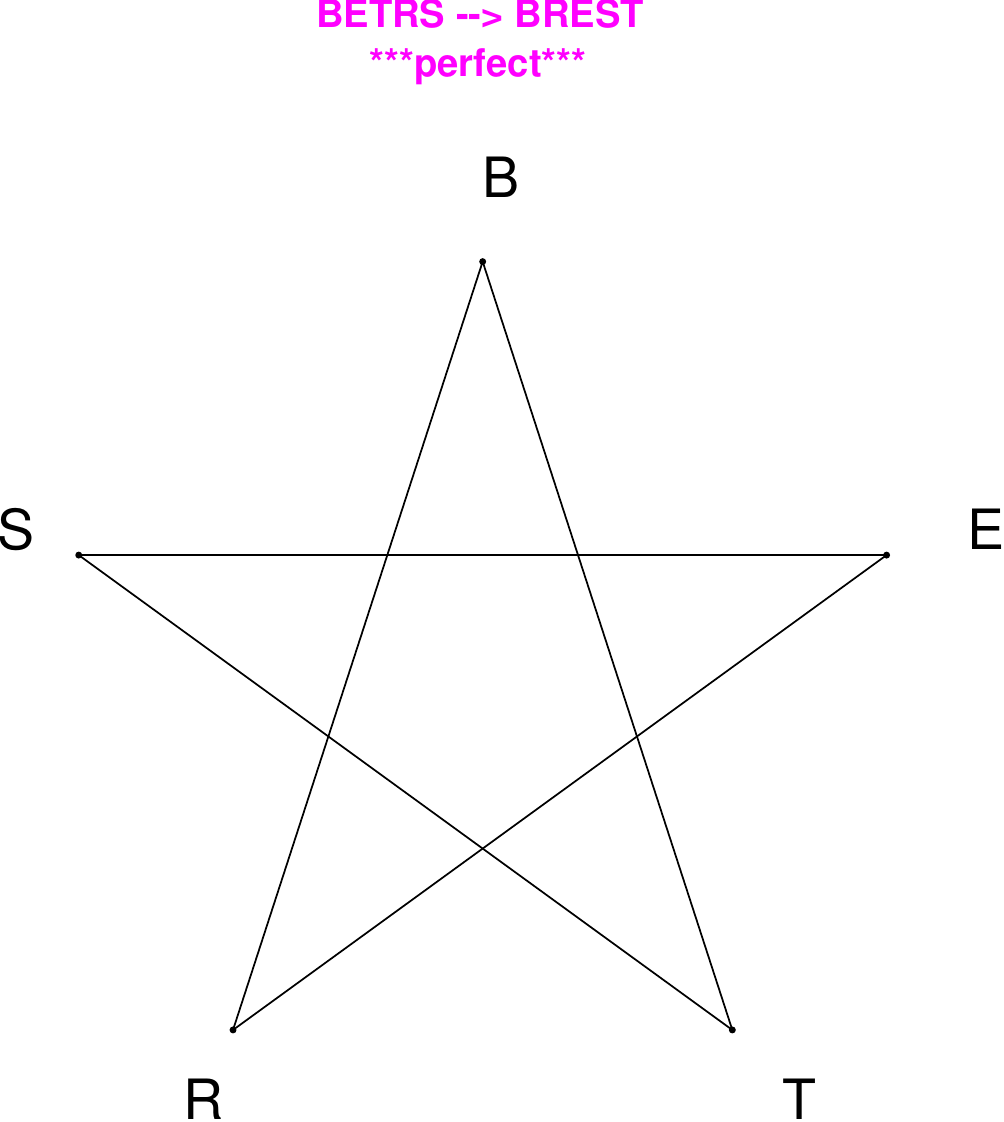}
\end{subfigure}
\hfill
\begin{subfigure}[T]{0.19\textwidth}
\centering
\includegraphics[width=\textwidth]{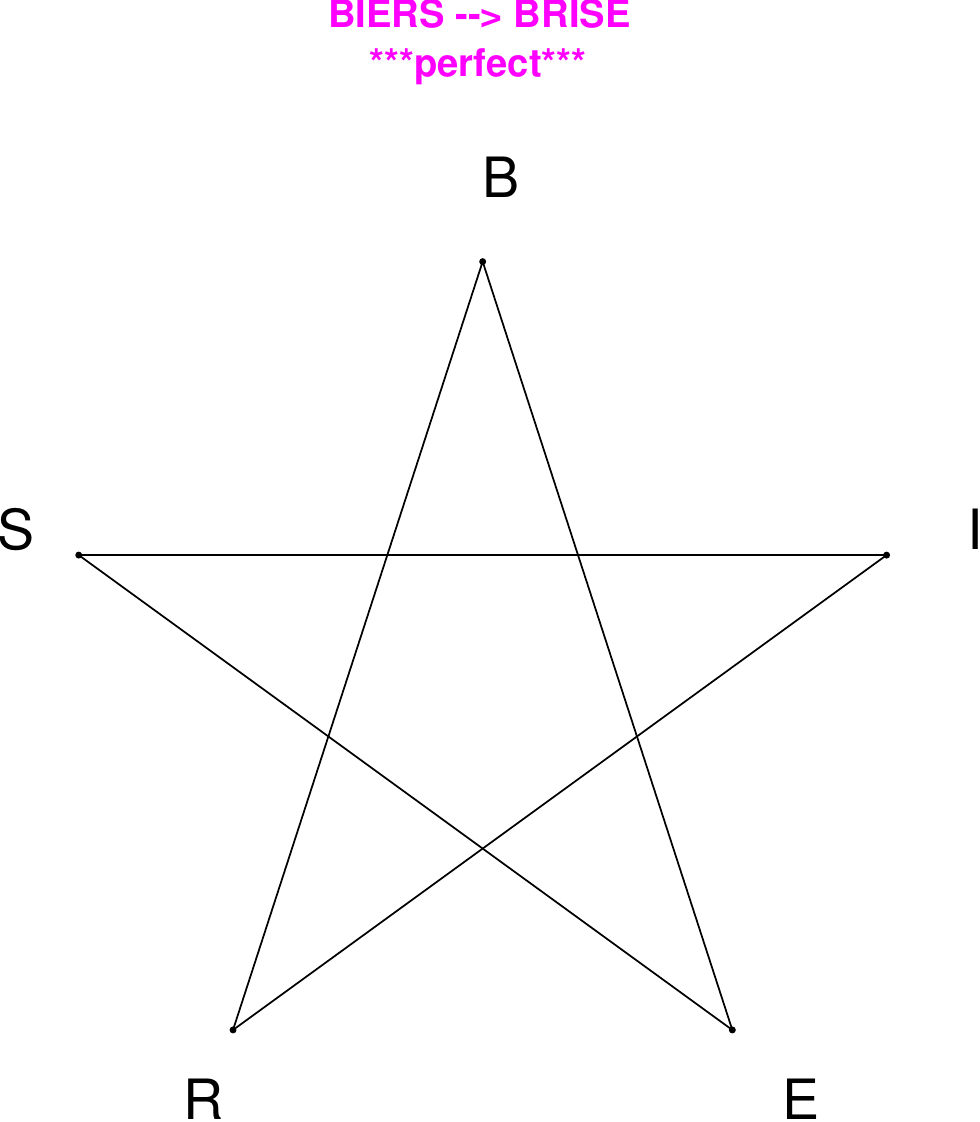}
\end{subfigure}
\hfill
\begin{subfigure}[T]{0.19\textwidth}
\centering
\includegraphics[width=\textwidth]{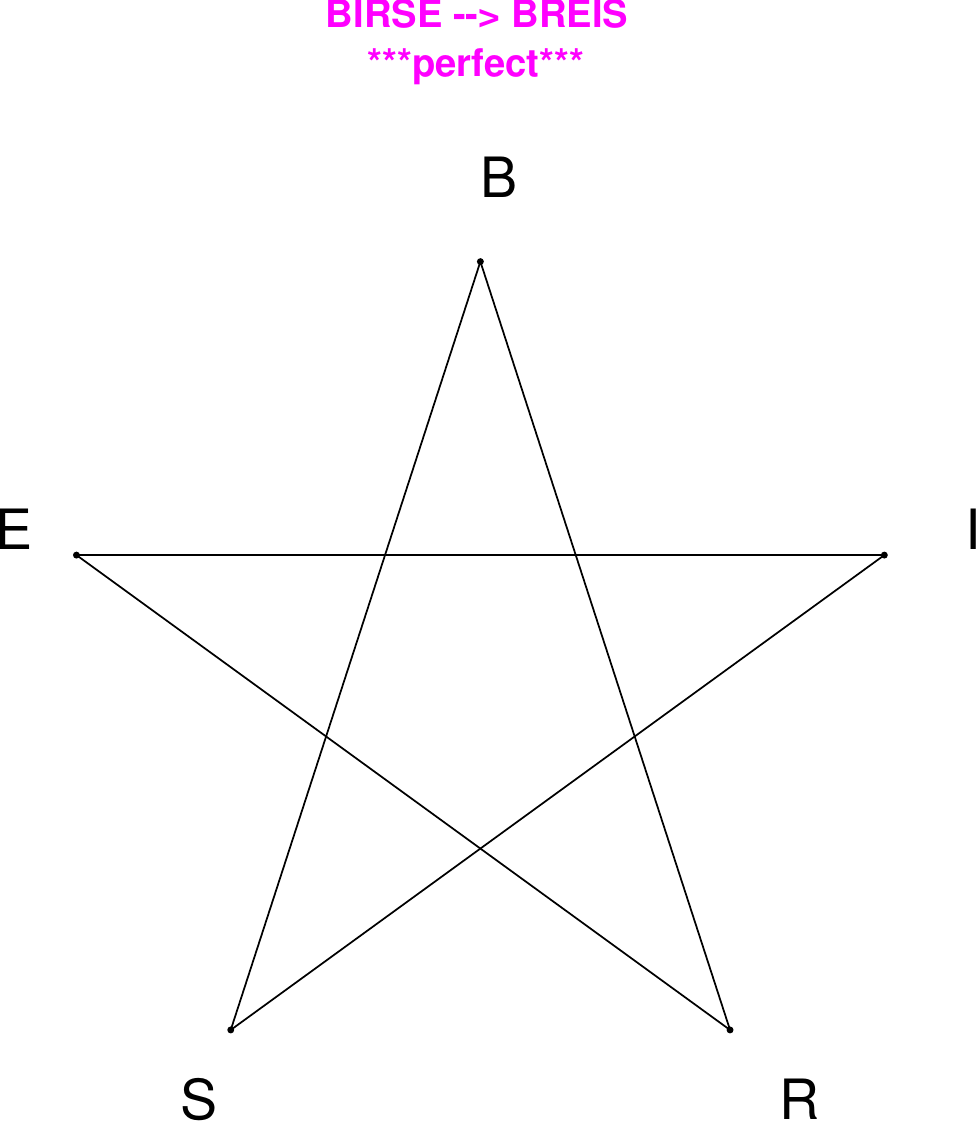}
\end{subfigure}
\end{figure}

\begin{figure}[H]
\centering
\begin{subfigure}[T]{0.19\textwidth}
\centering
\includegraphics[width=\textwidth]{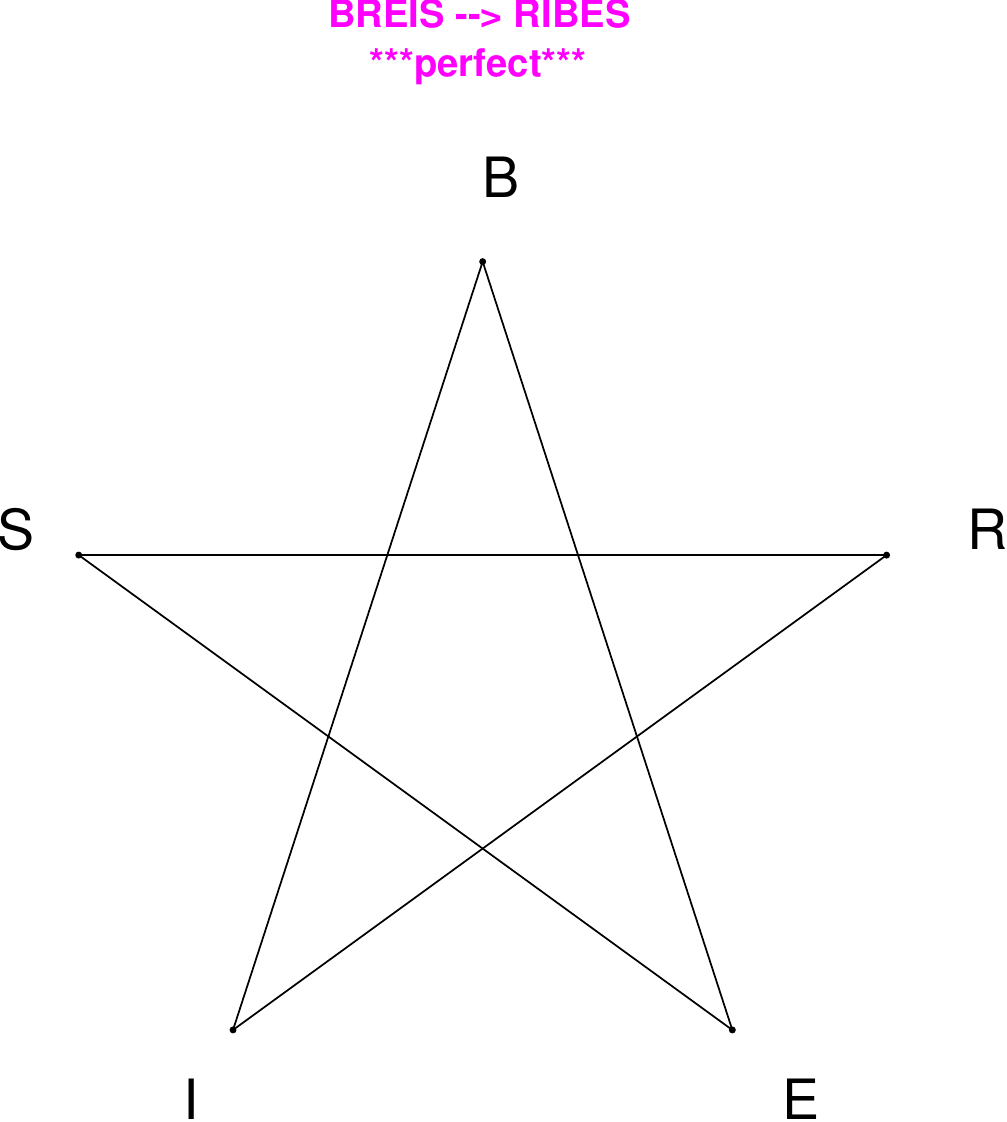}
\end{subfigure}
\hfill
\begin{subfigure}[T]{0.19\textwidth}
\centering
\includegraphics[width=\textwidth]{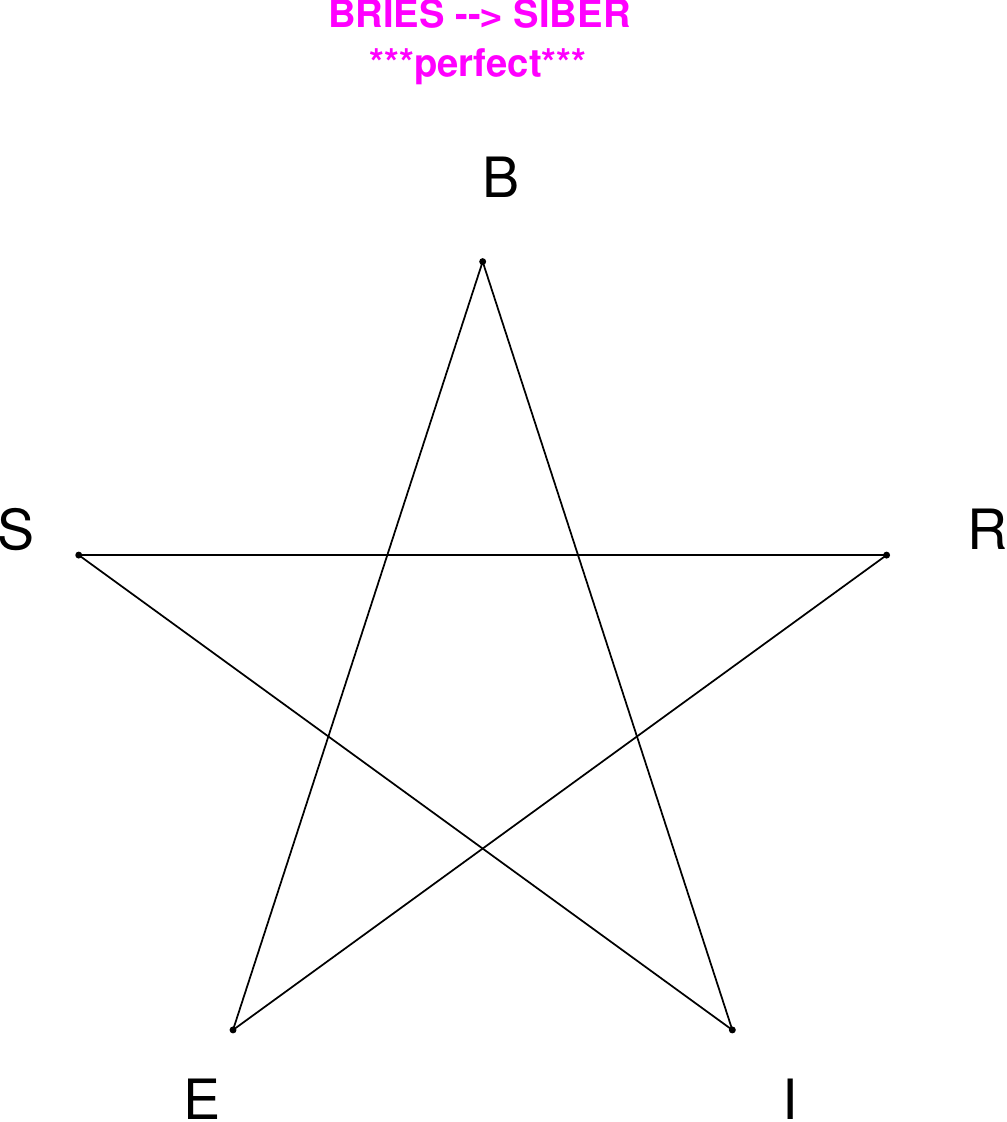}
\end{subfigure}
\hfill
\begin{subfigure}[T]{0.19\textwidth}
\centering
\includegraphics[width=\textwidth]{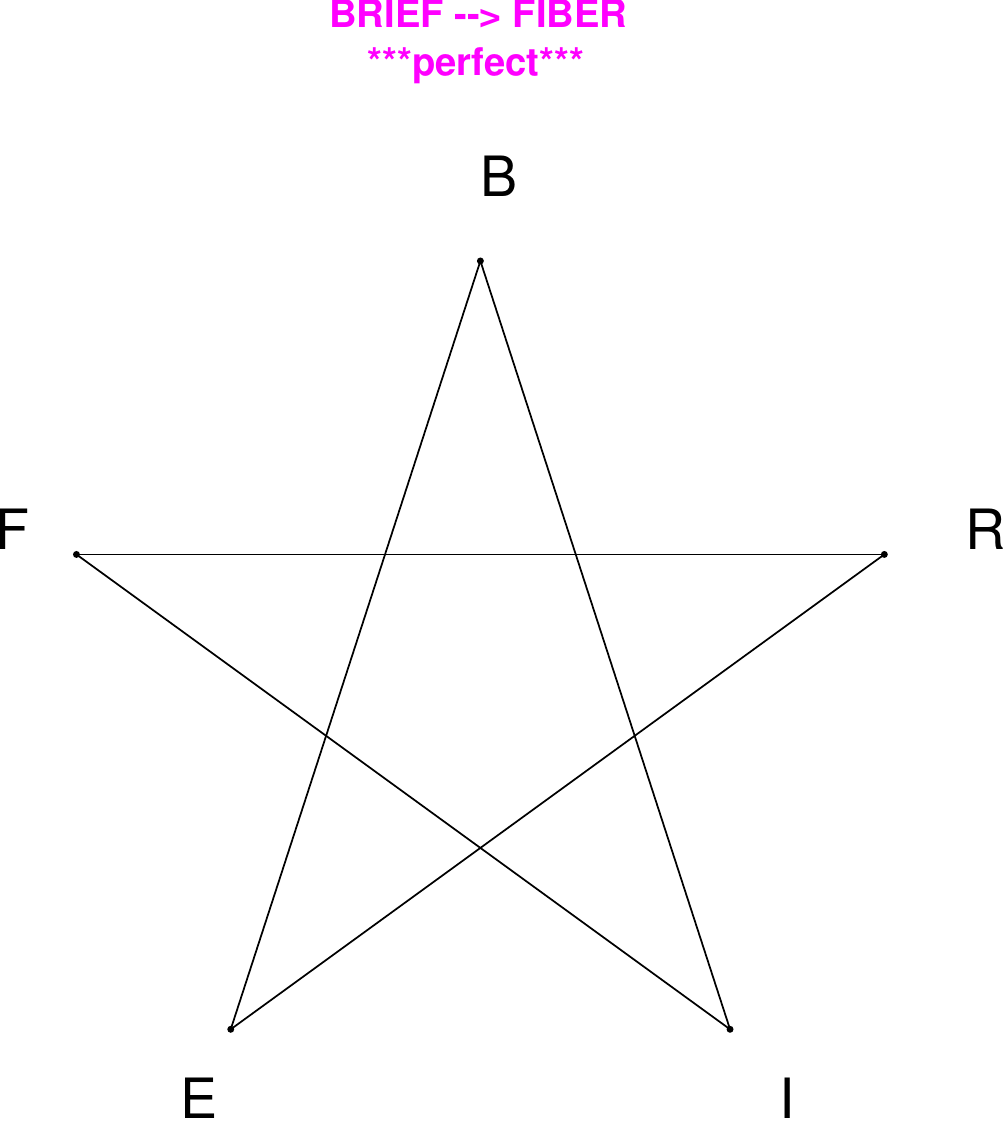}
\end{subfigure}
\hfill
\begin{subfigure}[T]{0.19\textwidth}
\centering
\includegraphics[width=\textwidth]{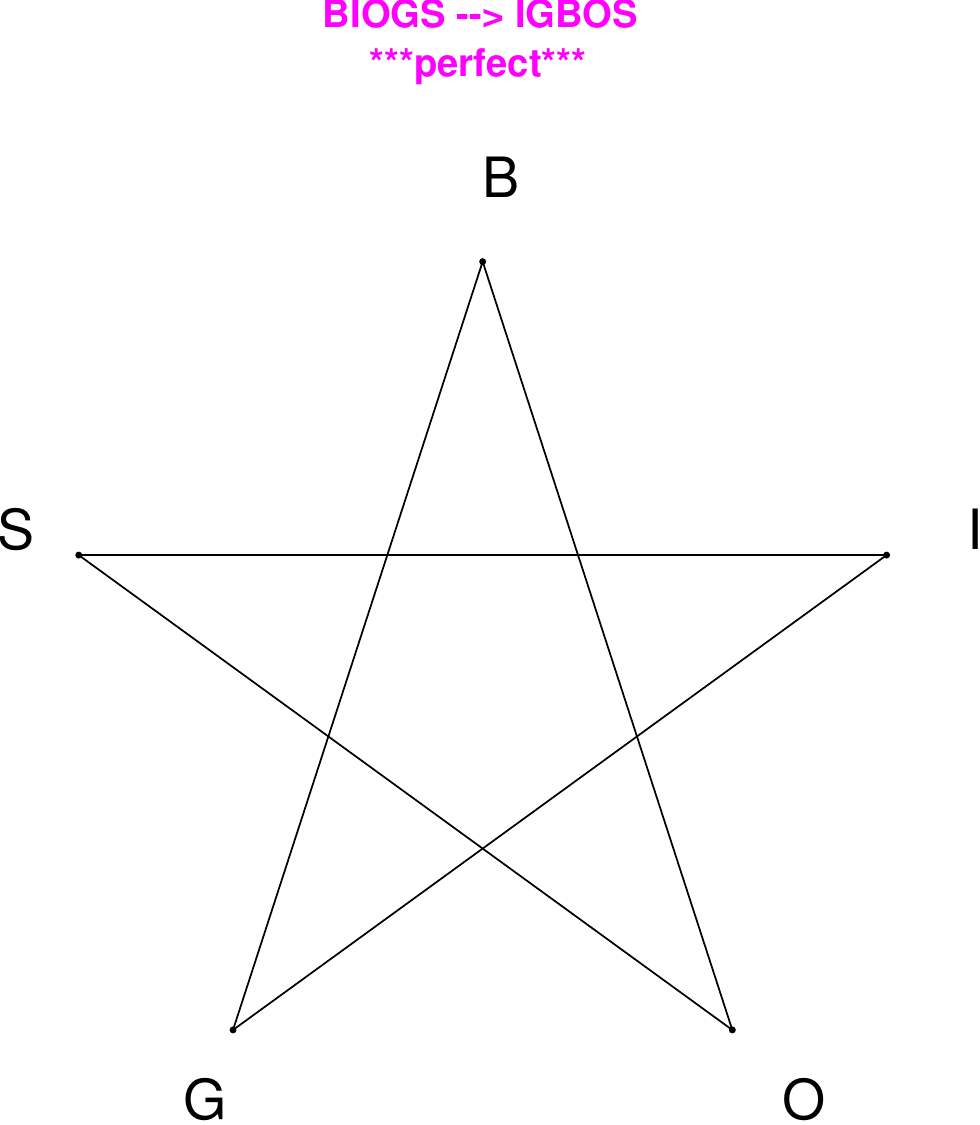}
\end{subfigure}
\hfill
\begin{subfigure}[T]{0.19\textwidth}
\centering
\includegraphics[width=\textwidth]{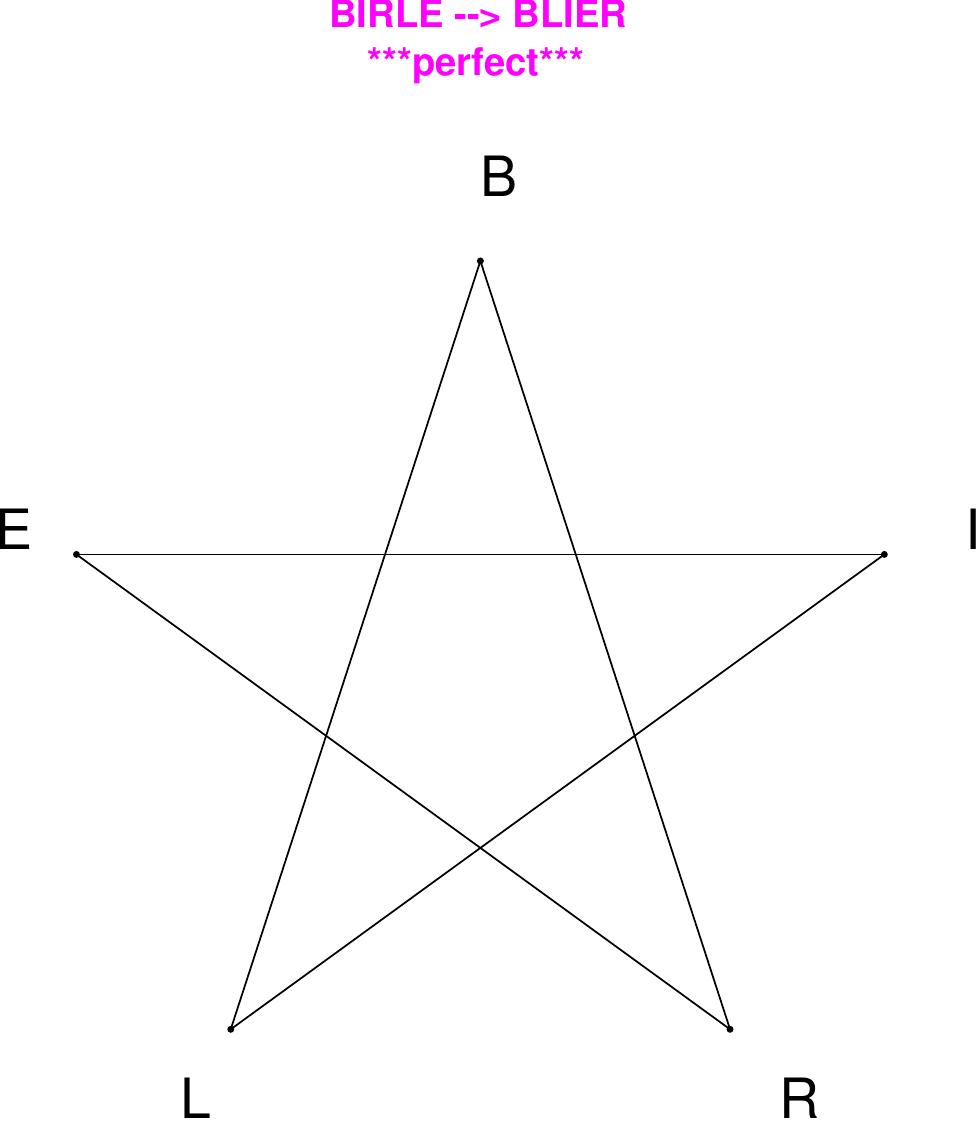}
\end{subfigure}
\end{figure}

\begin{figure}[H]
\centering
\begin{subfigure}[T]{0.19\textwidth}
\centering
\includegraphics[width=\textwidth]{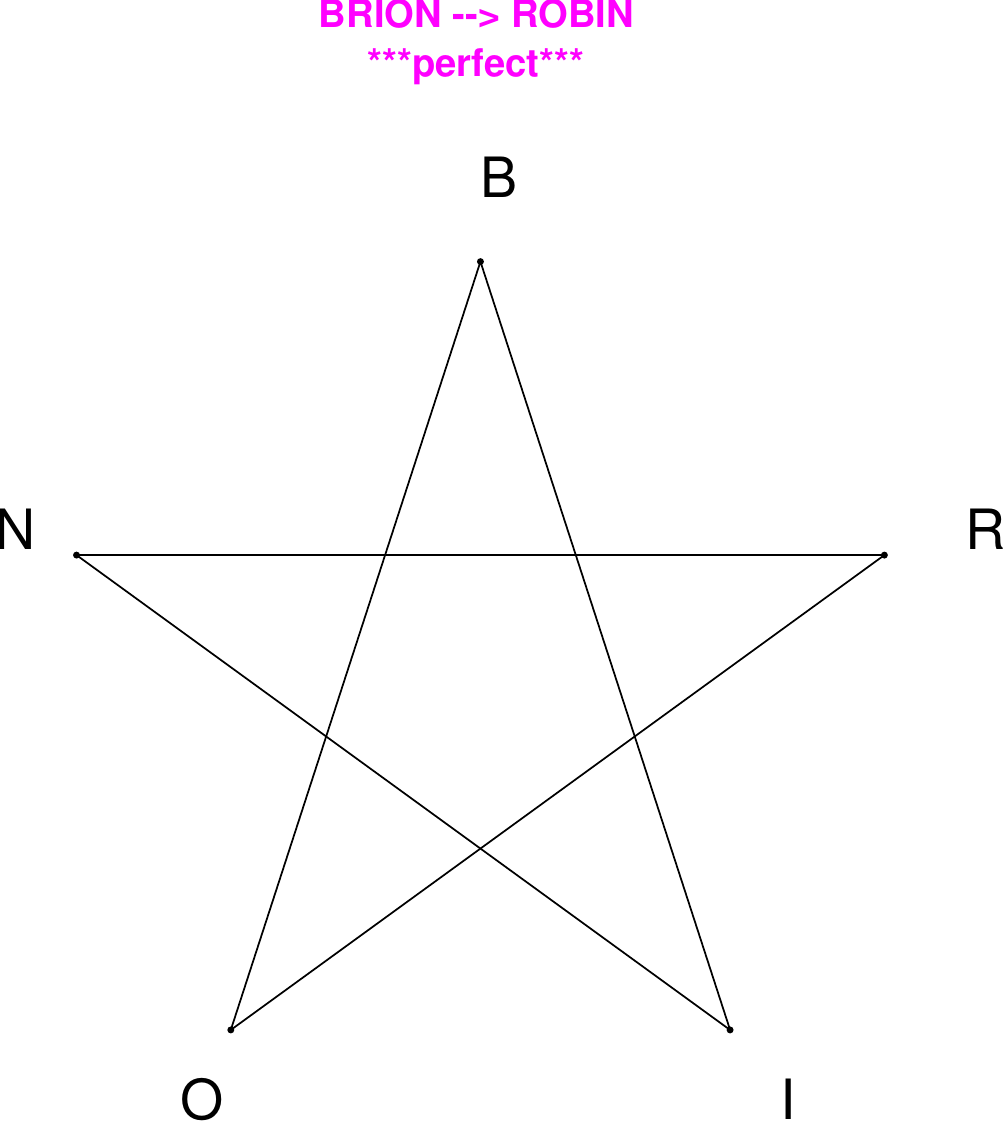}
\end{subfigure}
\hfill
\begin{subfigure}[T]{0.19\textwidth}
\centering
\includegraphics[width=\textwidth]{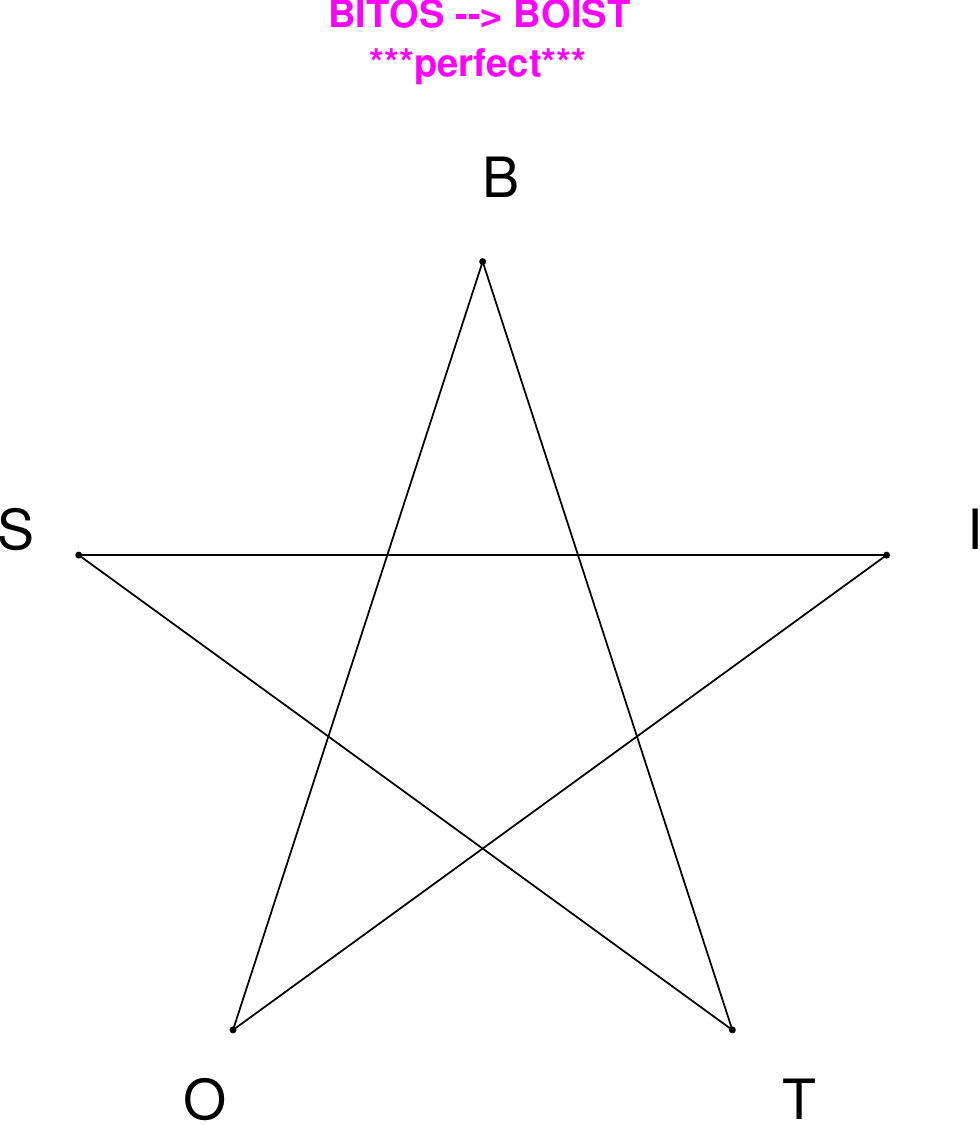}
\end{subfigure}
\hfill
\begin{subfigure}[T]{0.19\textwidth}
\centering
\includegraphics[width=\textwidth]{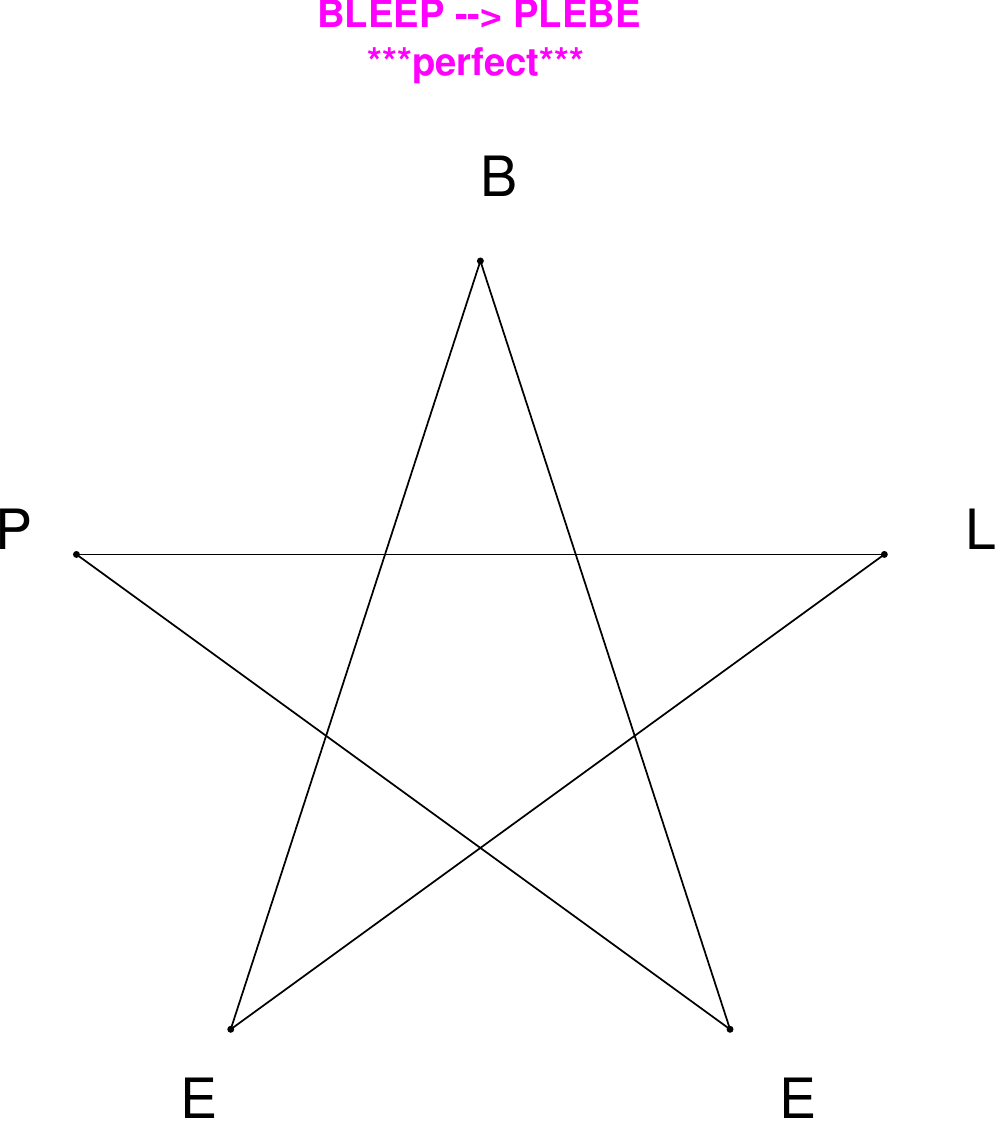}
\end{subfigure}
\hfill
\begin{subfigure}[T]{0.19\textwidth}
\centering
\includegraphics[width=\textwidth]{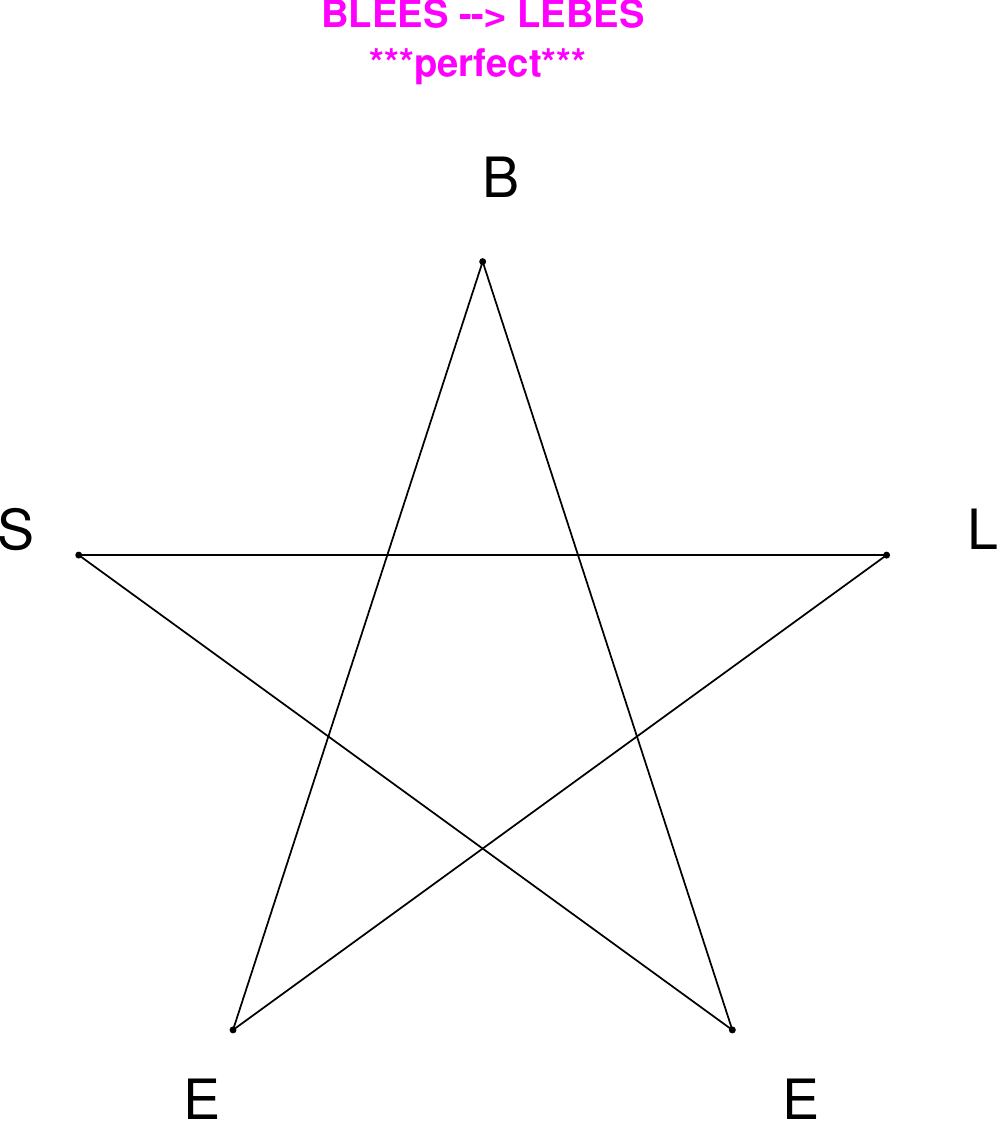}
\end{subfigure}
\hfill
\begin{subfigure}[T]{0.19\textwidth}
\centering
\includegraphics[width=\textwidth]{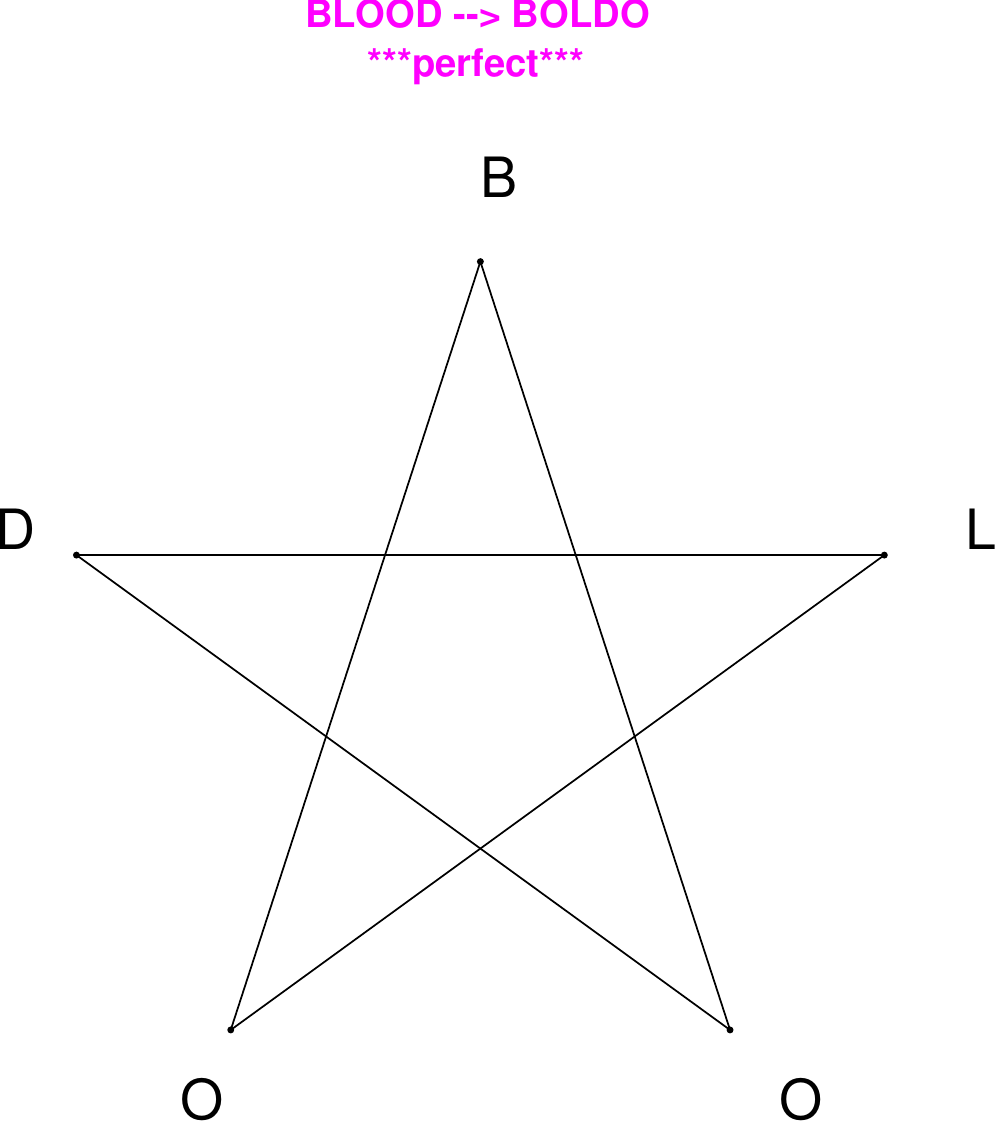}
\end{subfigure}
\end{figure}

\begin{figure}[H]
\centering
\begin{subfigure}[T]{0.19\textwidth}
\centering
\includegraphics[width=\textwidth]{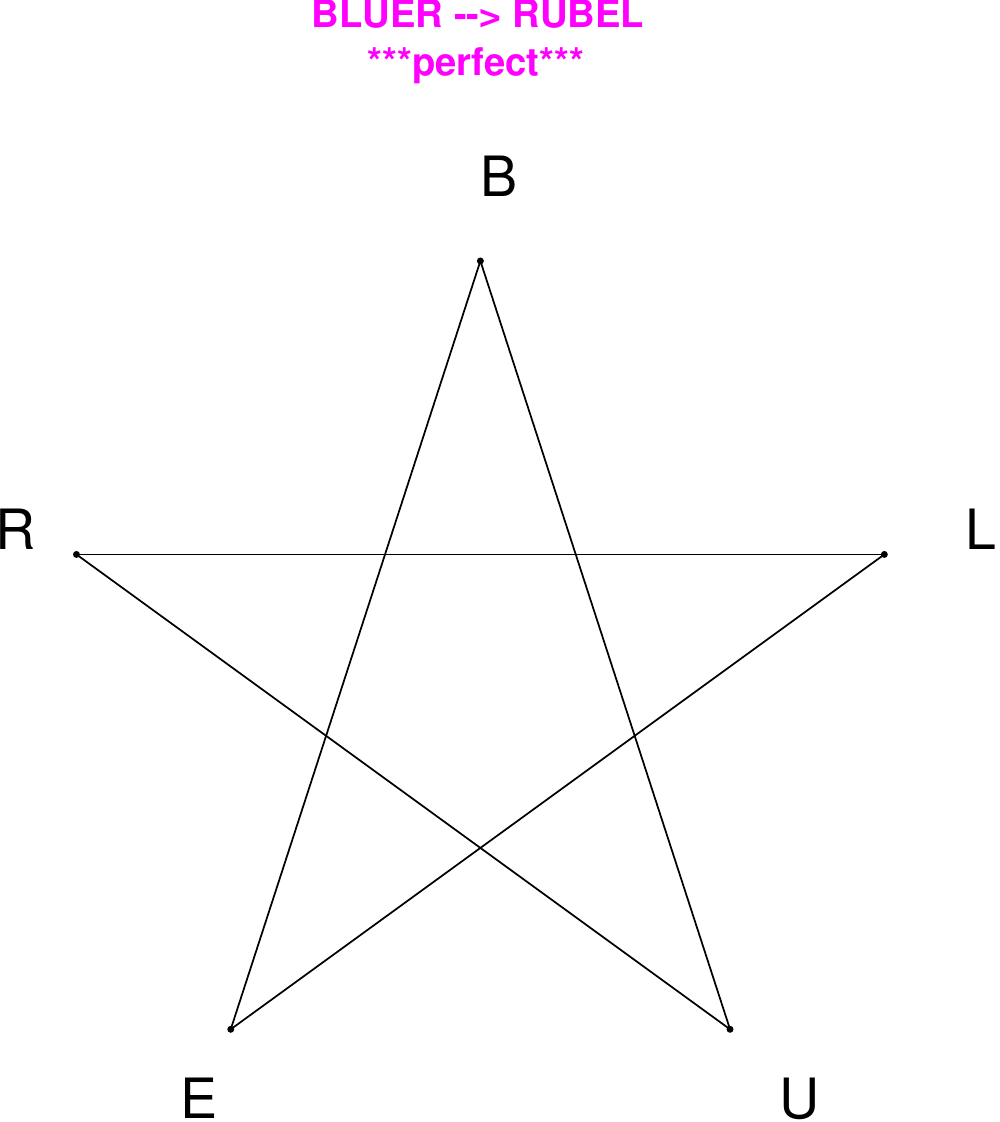}
\end{subfigure}
\hfill
\begin{subfigure}[T]{0.19\textwidth}
\centering
\includegraphics[width=\textwidth]{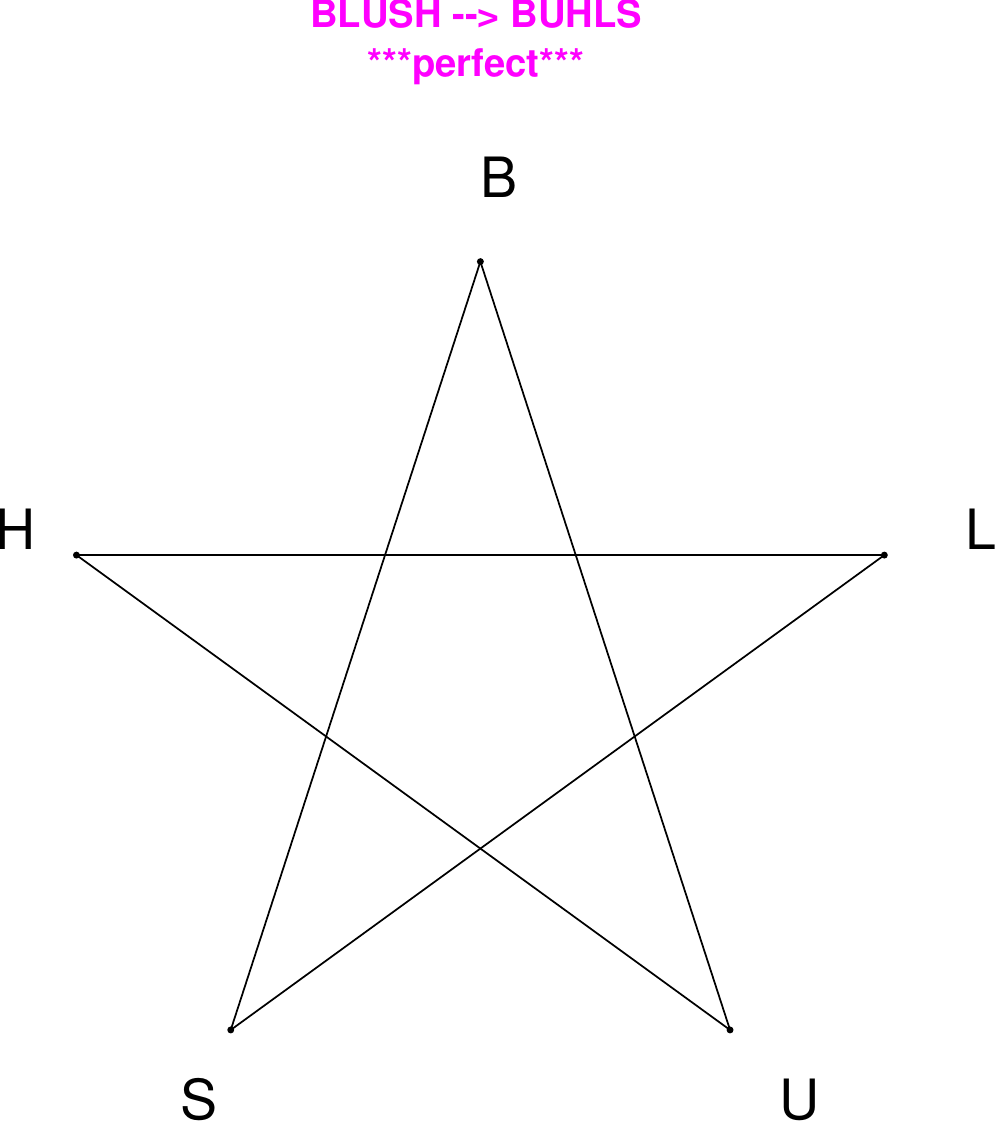}
\end{subfigure}
\hfill
\begin{subfigure}[T]{0.19\textwidth}
\centering
\includegraphics[width=\textwidth]{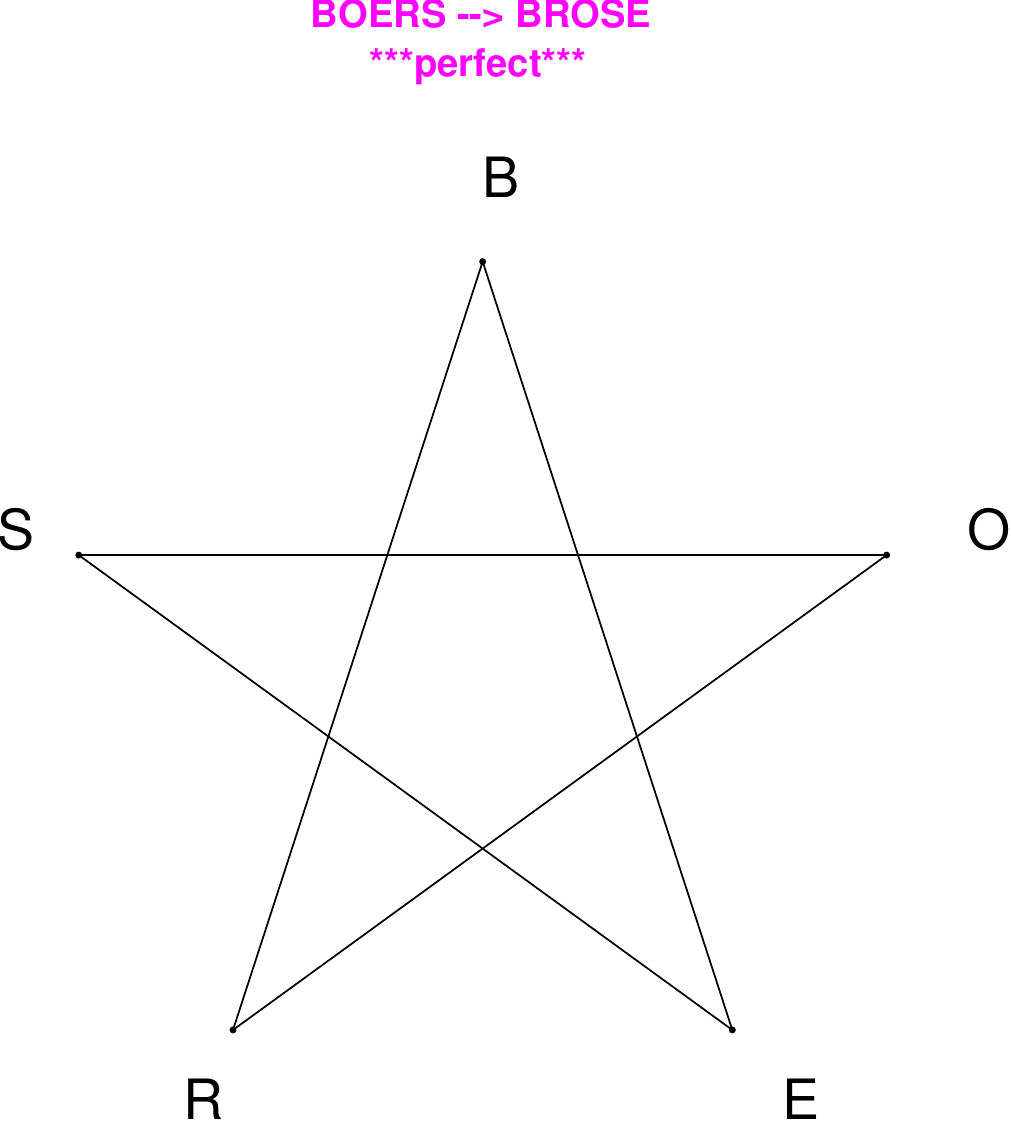}
\end{subfigure}
\hfill
\begin{subfigure}[T]{0.19\textwidth}
\centering
\includegraphics[width=\textwidth]{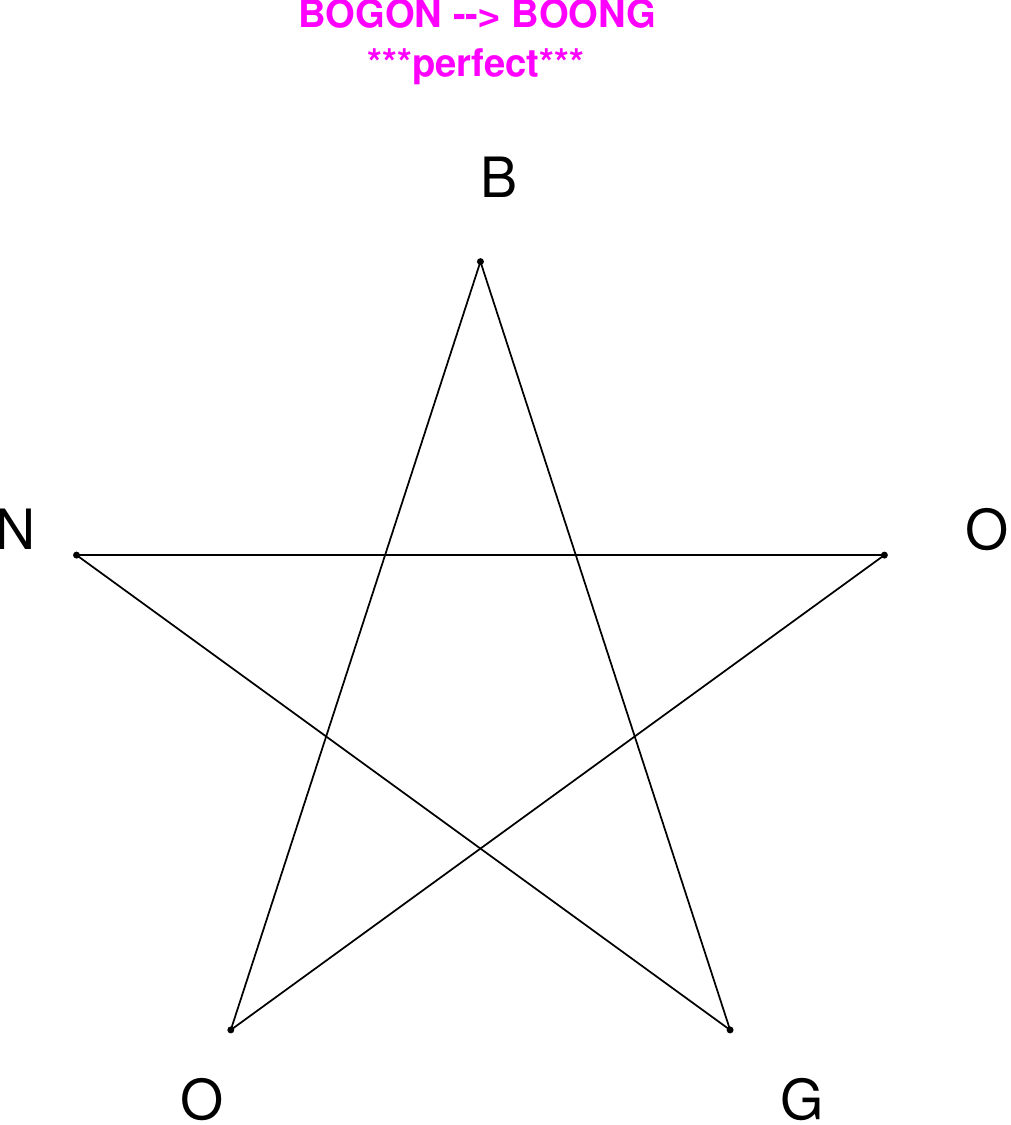}
\end{subfigure}
\hfill
\begin{subfigure}[T]{0.19\textwidth}
\centering
\includegraphics[width=\textwidth]{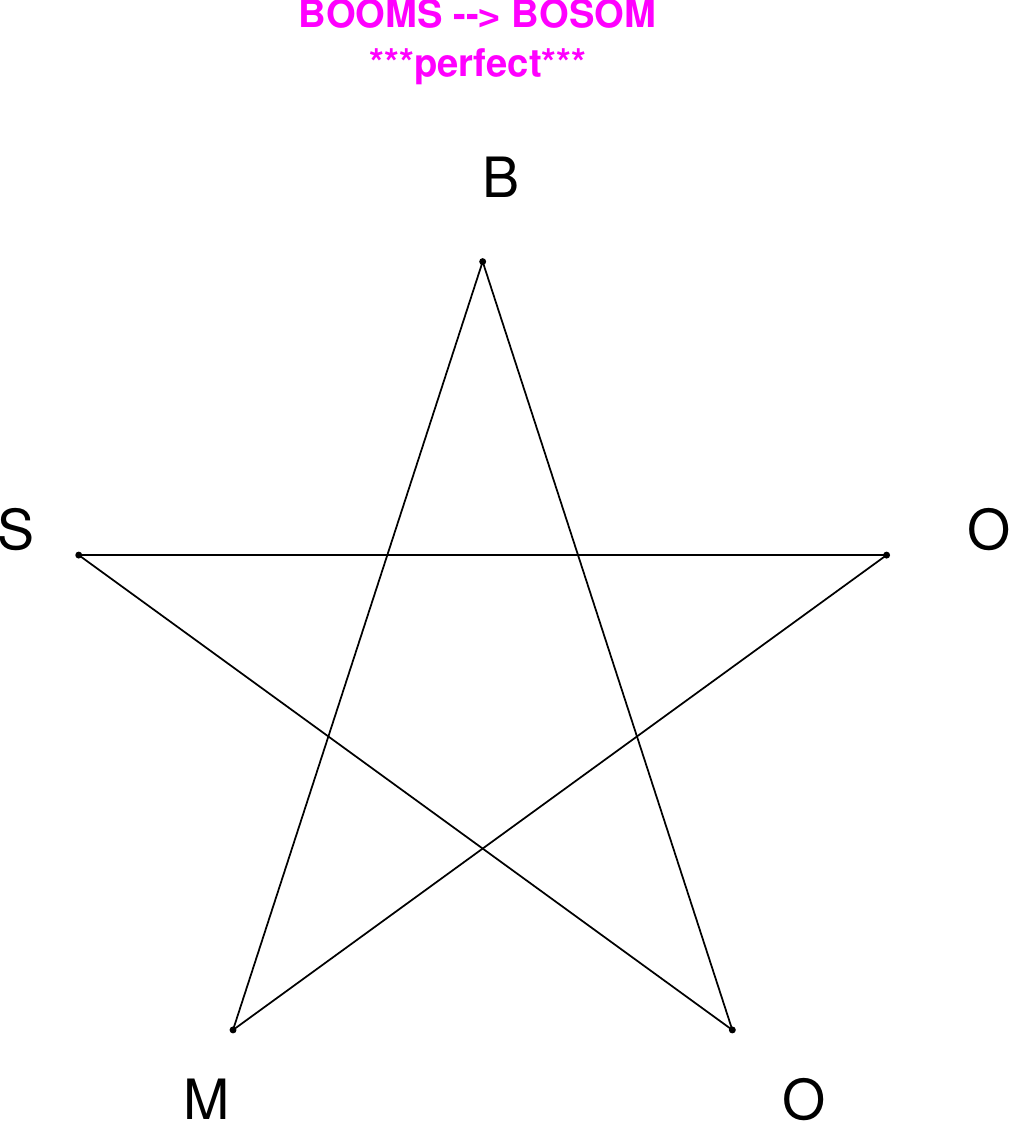}
\end{subfigure}
\end{figure}

\begin{figure}[H]
\centering
\begin{subfigure}[T]{0.19\textwidth}
\centering
\includegraphics[width=\textwidth]{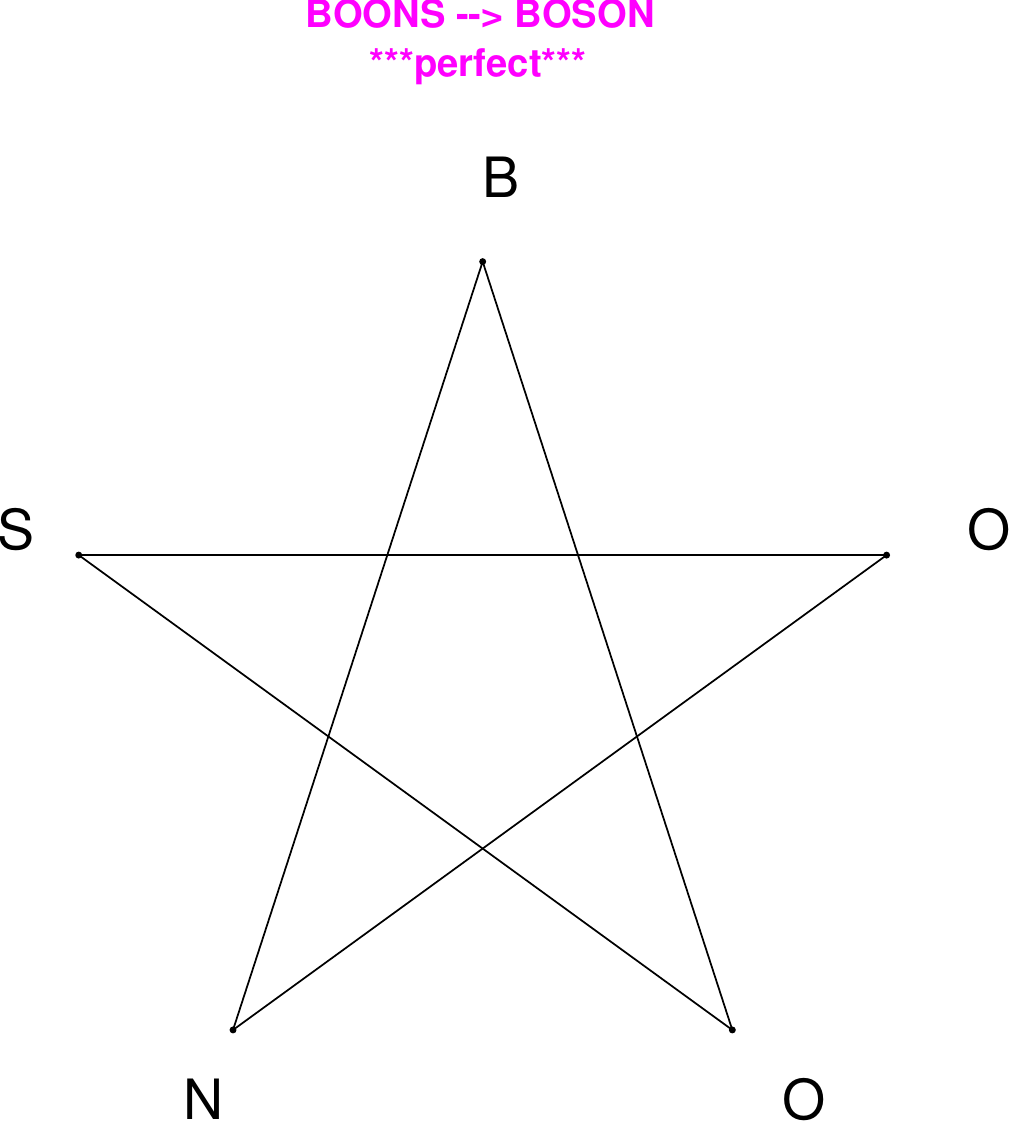}
\end{subfigure}
\hfill
\begin{subfigure}[T]{0.19\textwidth}
\centering
\includegraphics[width=\textwidth]{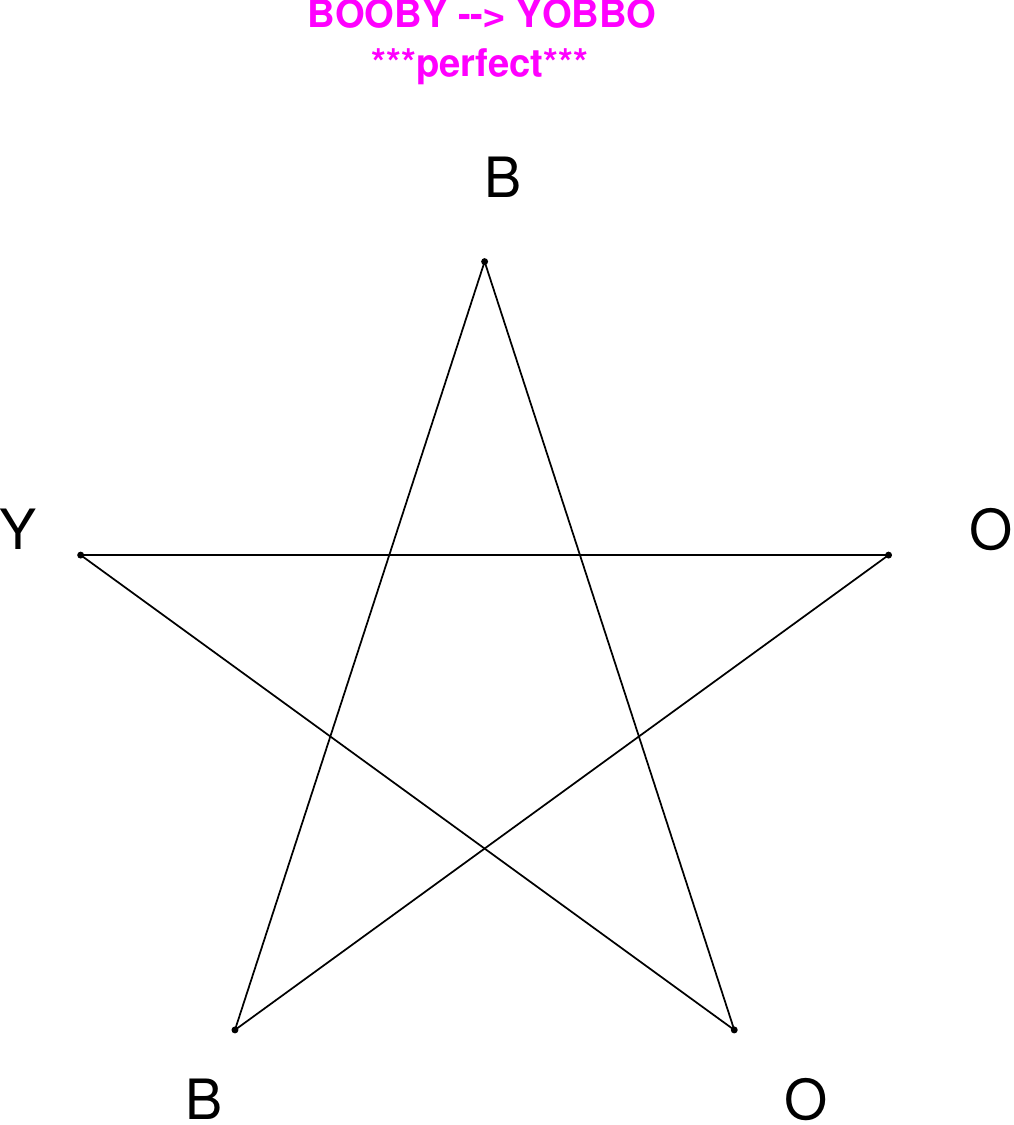}
\end{subfigure}
\hfill
\begin{subfigure}[T]{0.19\textwidth}
\centering
\includegraphics[width=\textwidth]{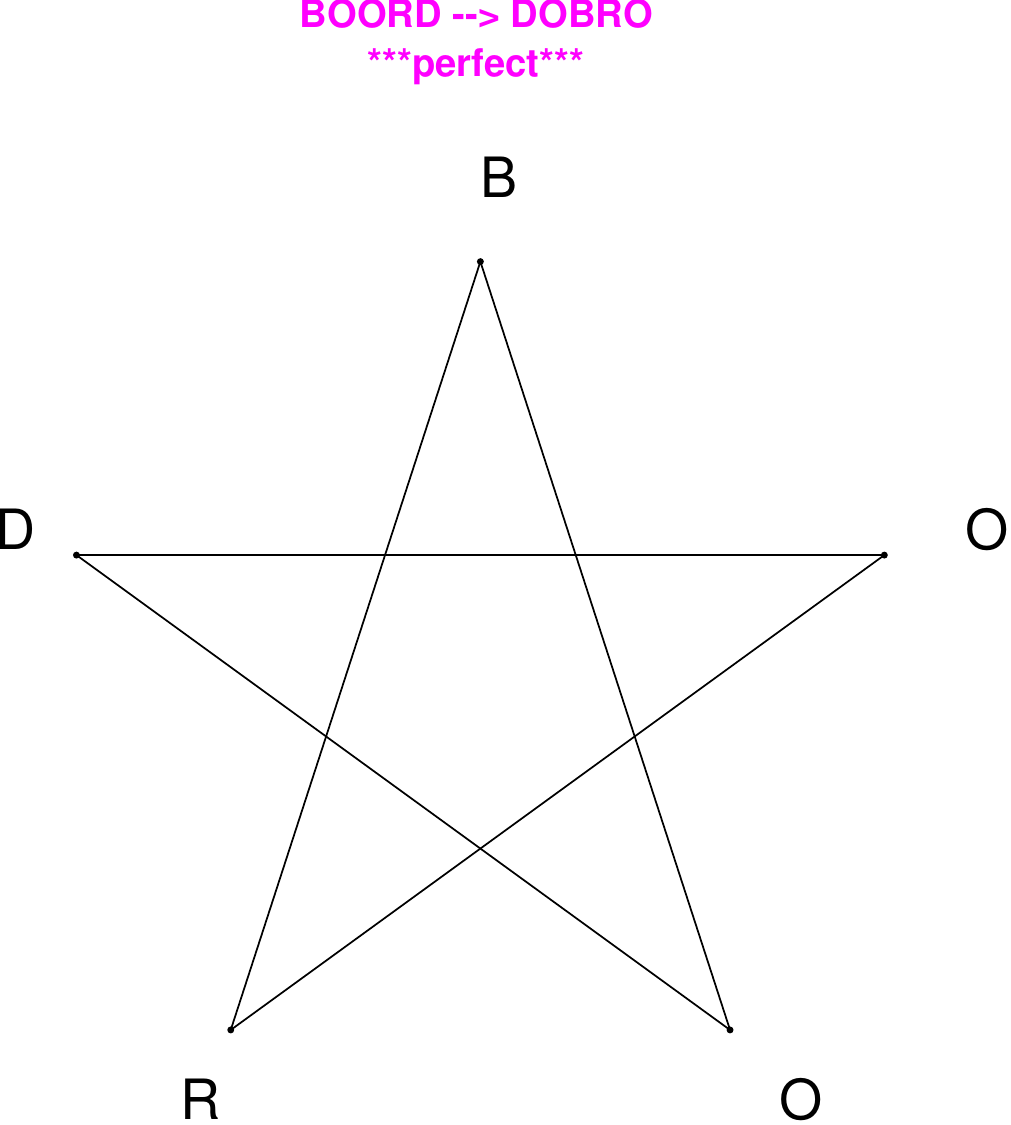}
\end{subfigure}
\hfill
\begin{subfigure}[T]{0.19\textwidth}
\centering
\includegraphics[width=\textwidth]{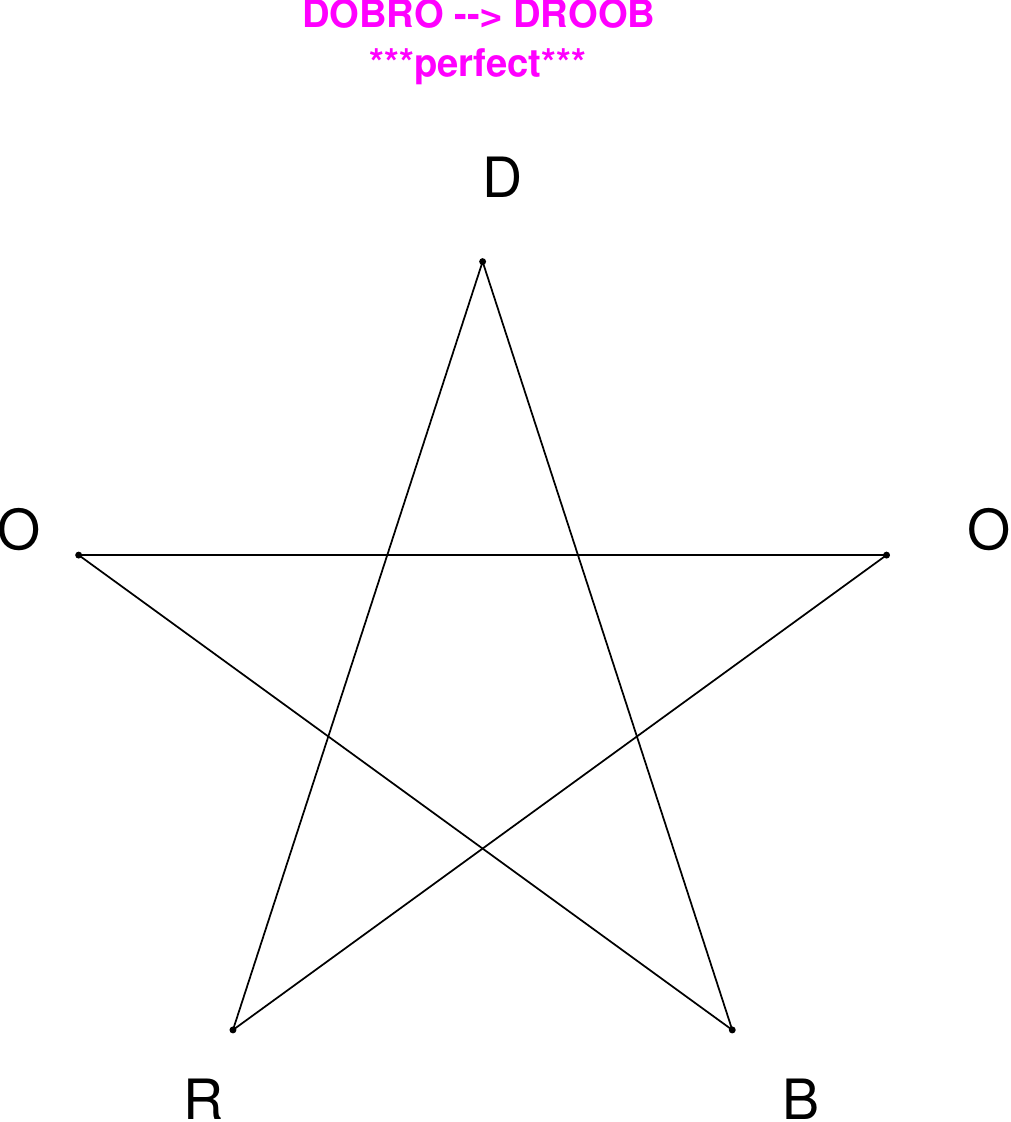}
\end{subfigure}
\hfill
\begin{subfigure}[T]{0.19\textwidth}
\centering
\includegraphics[width=\textwidth]{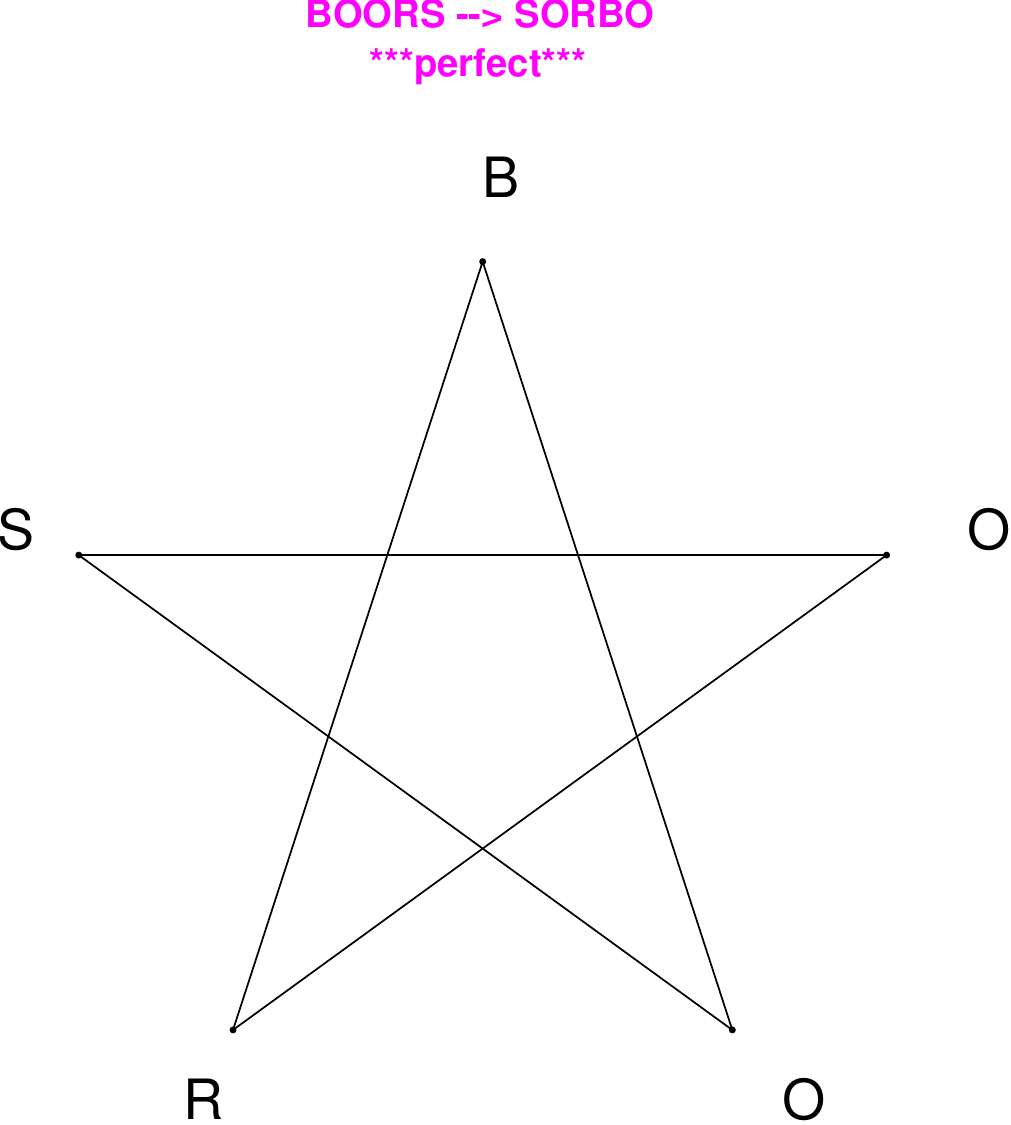}
\end{subfigure}
\end{figure}

\begin{figure}[H]
\centering
\begin{subfigure}[T]{0.19\textwidth}
\centering
\includegraphics[width=\textwidth]{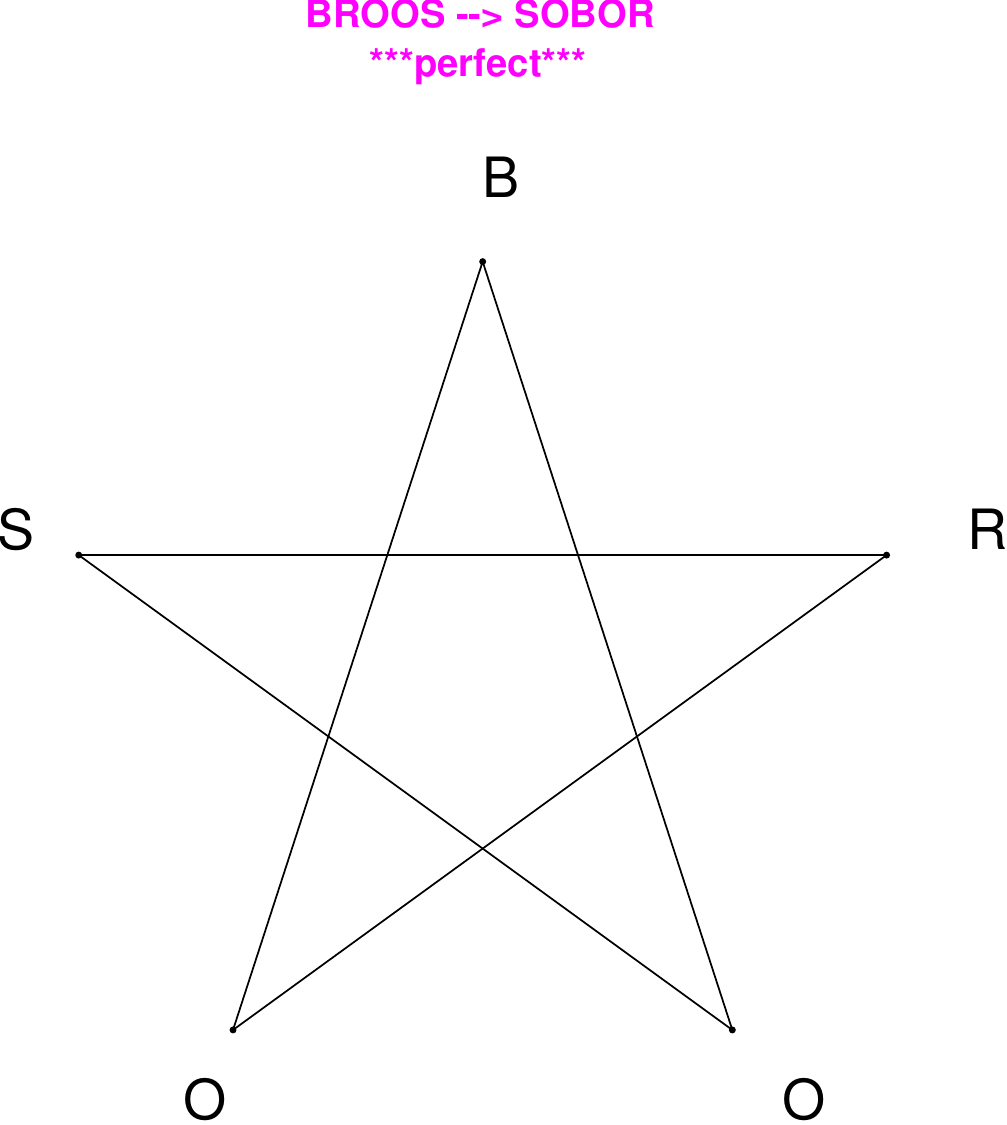}
\end{subfigure}
\hfill
\begin{subfigure}[T]{0.19\textwidth}
\centering
\includegraphics[width=\textwidth]{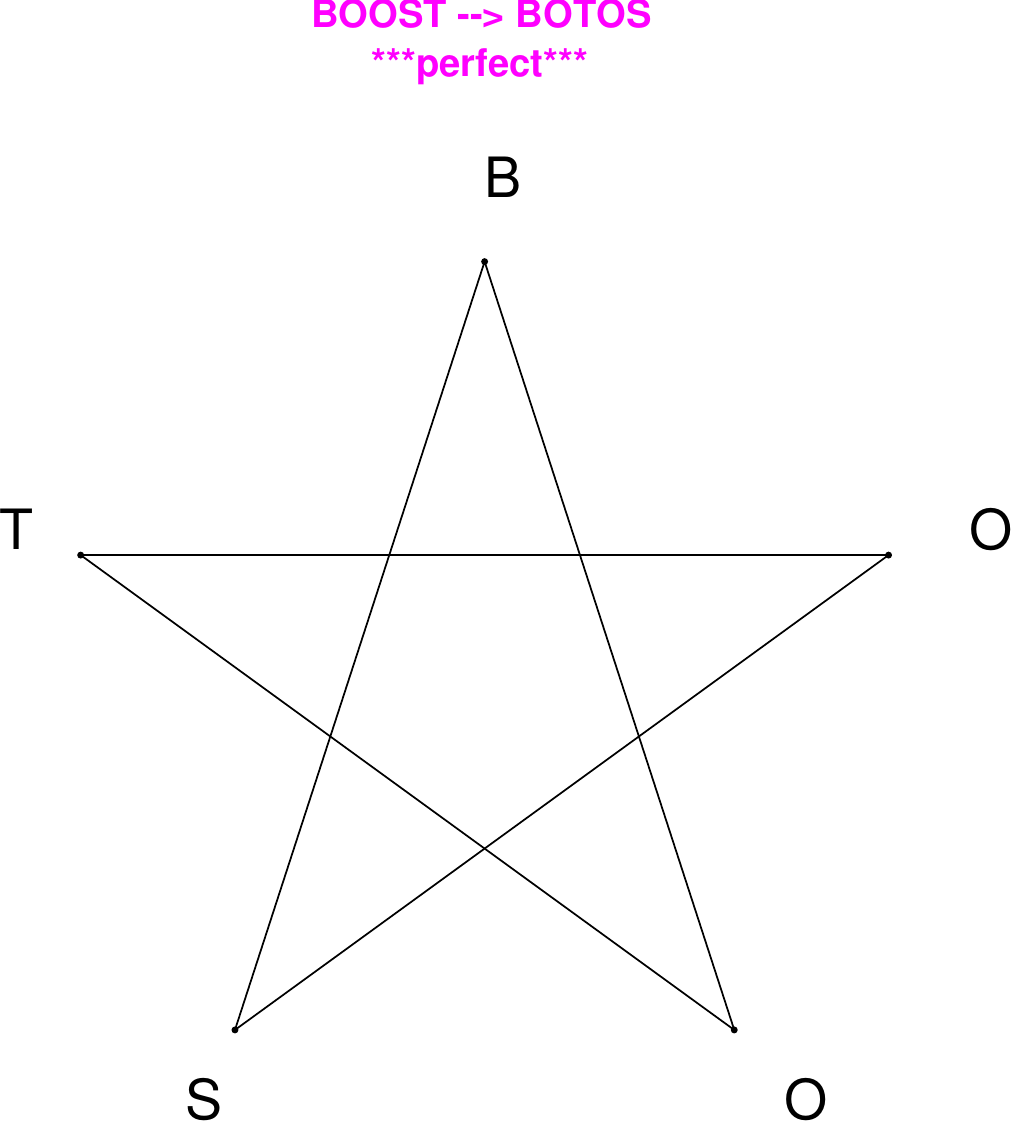}
\end{subfigure}
\hfill
\begin{subfigure}[T]{0.19\textwidth}
\centering
\includegraphics[width=\textwidth]{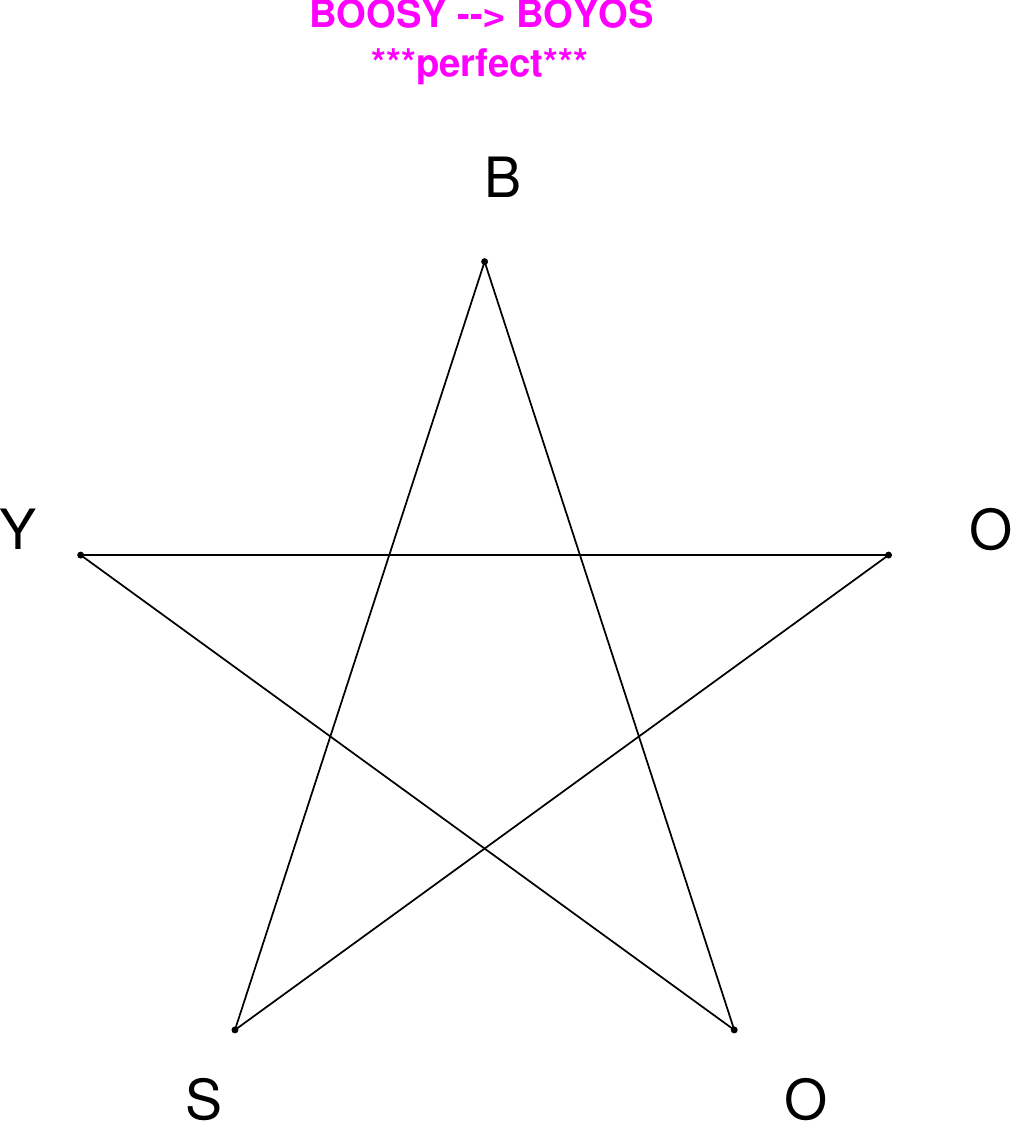}
\end{subfigure}
\hfill
\begin{subfigure}[T]{0.19\textwidth}
\centering
\includegraphics[width=\textwidth]{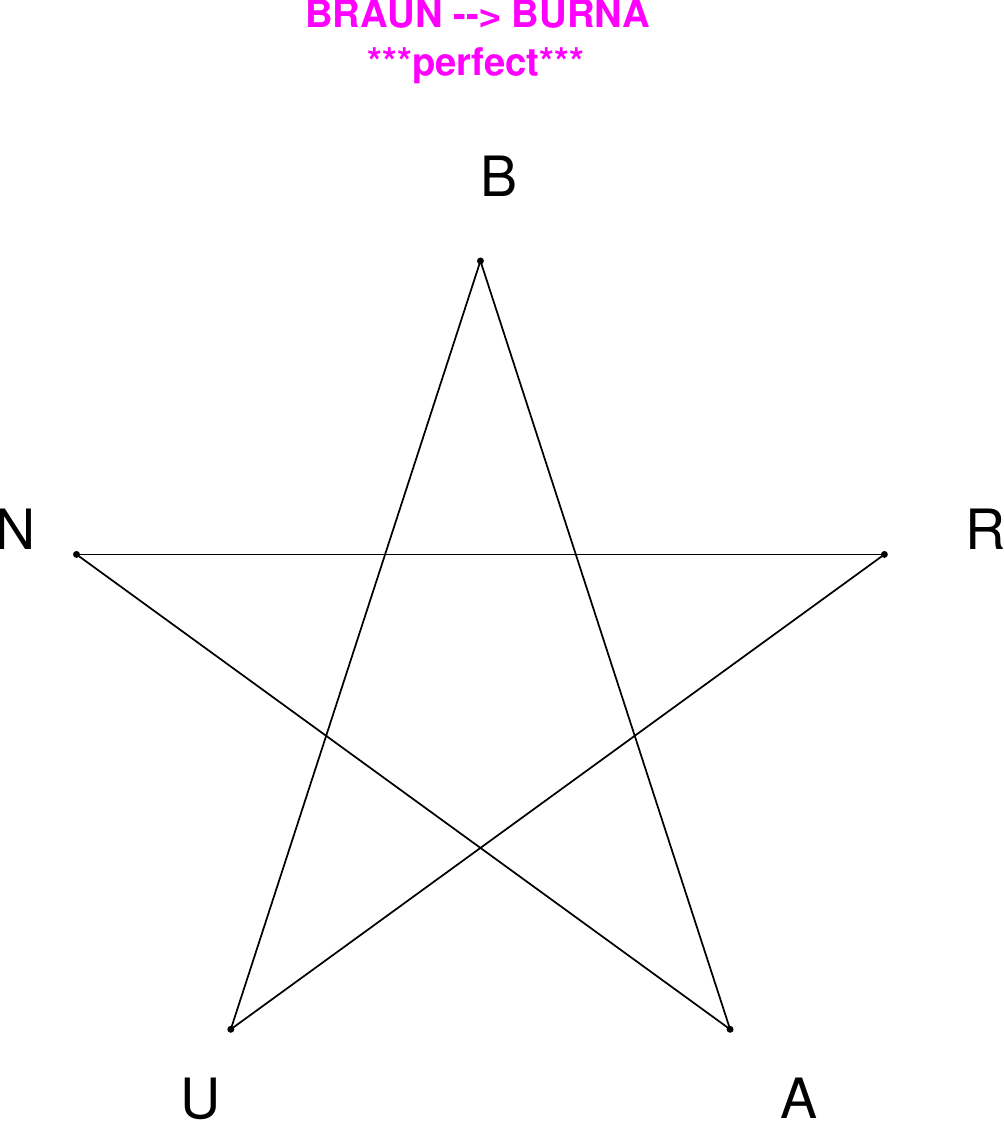}
\end{subfigure}
\hfill
\begin{subfigure}[T]{0.19\textwidth}
\centering
\includegraphics[width=\textwidth]{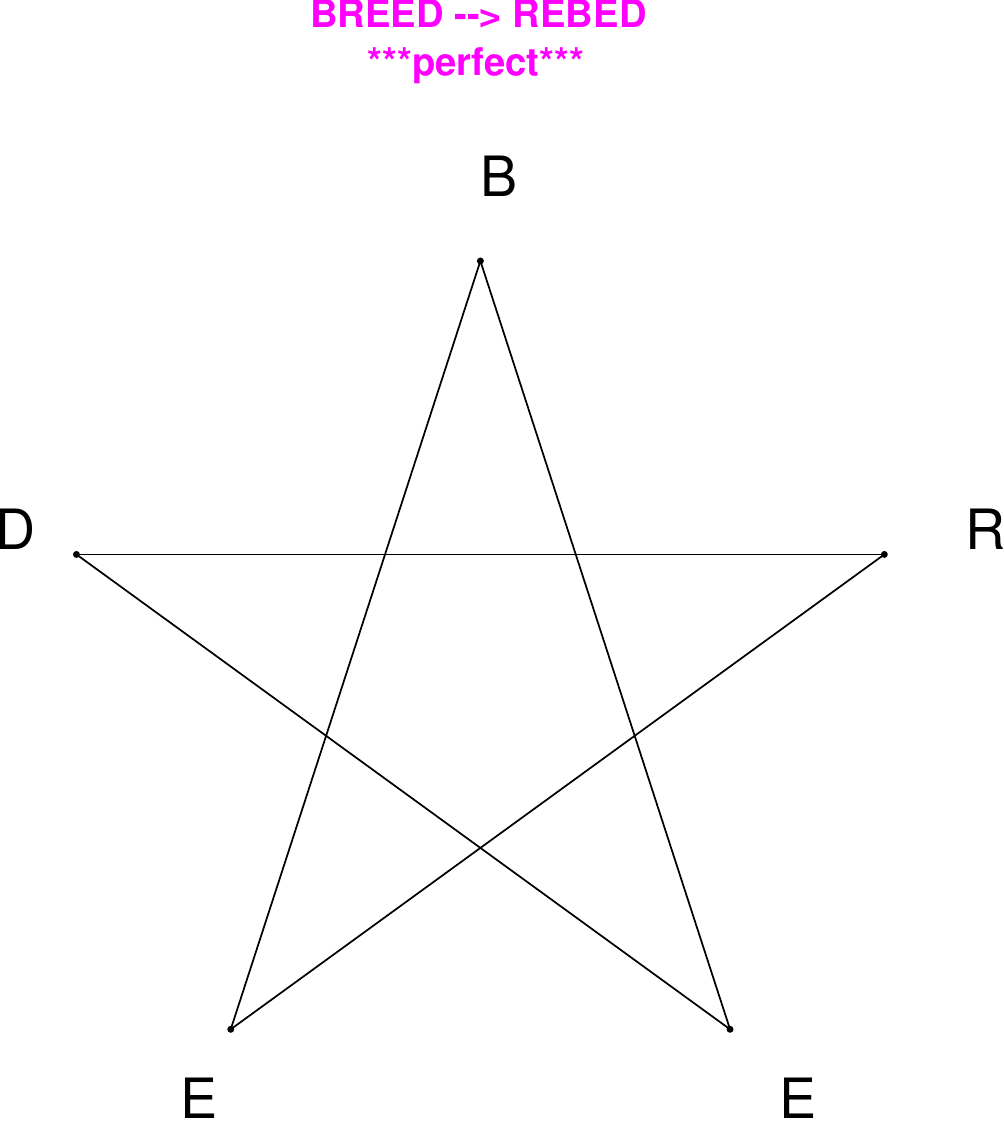}
\end{subfigure}
\end{figure}

\begin{figure}[H]
\centering
\begin{subfigure}[T]{0.19\textwidth}
\centering
\includegraphics[width=\textwidth]{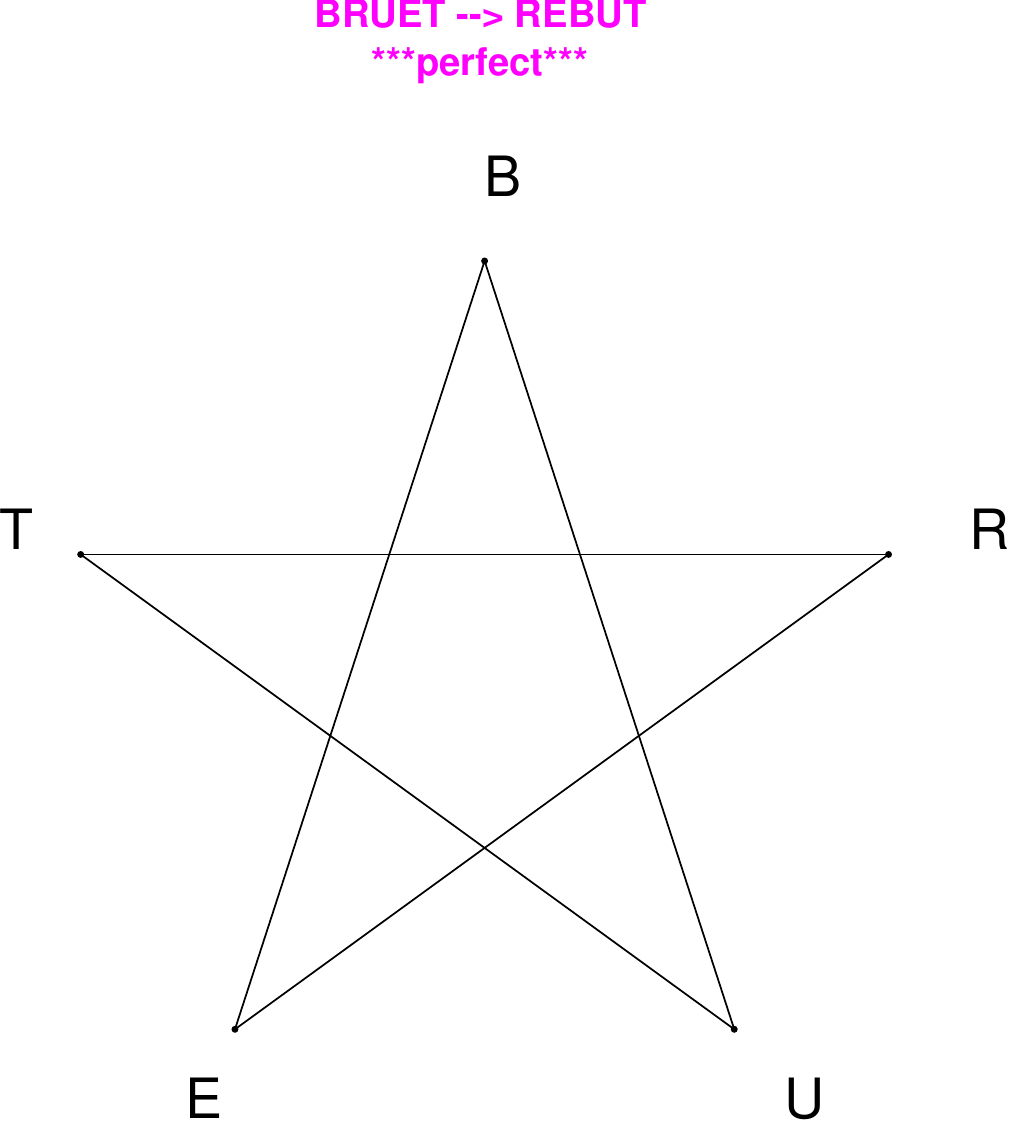}
\end{subfigure}
\hfill
\begin{subfigure}[T]{0.19\textwidth}
\centering
\includegraphics[width=\textwidth]{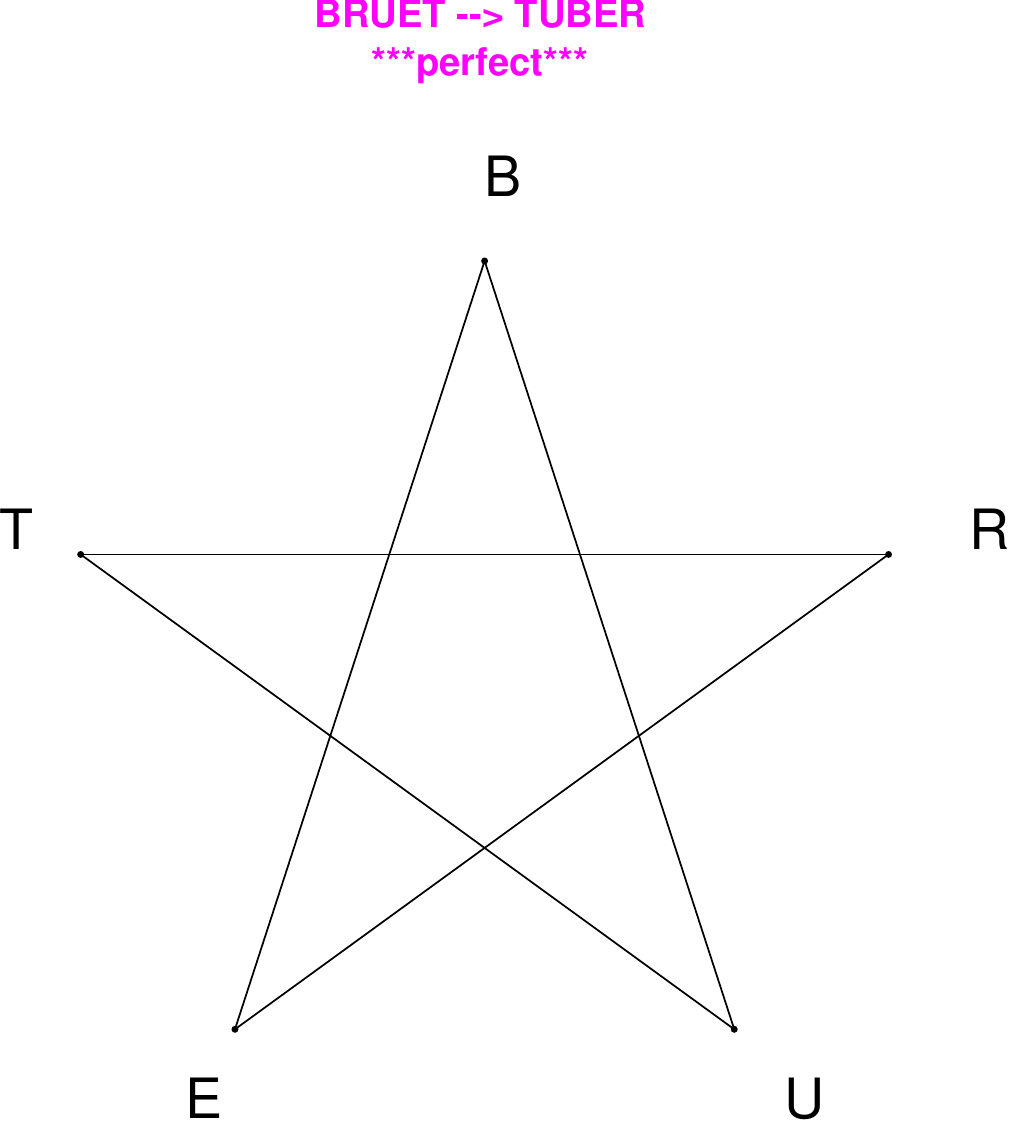}
\end{subfigure}
\hfill
\begin{subfigure}[T]{0.19\textwidth}
\centering
\includegraphics[width=\textwidth]{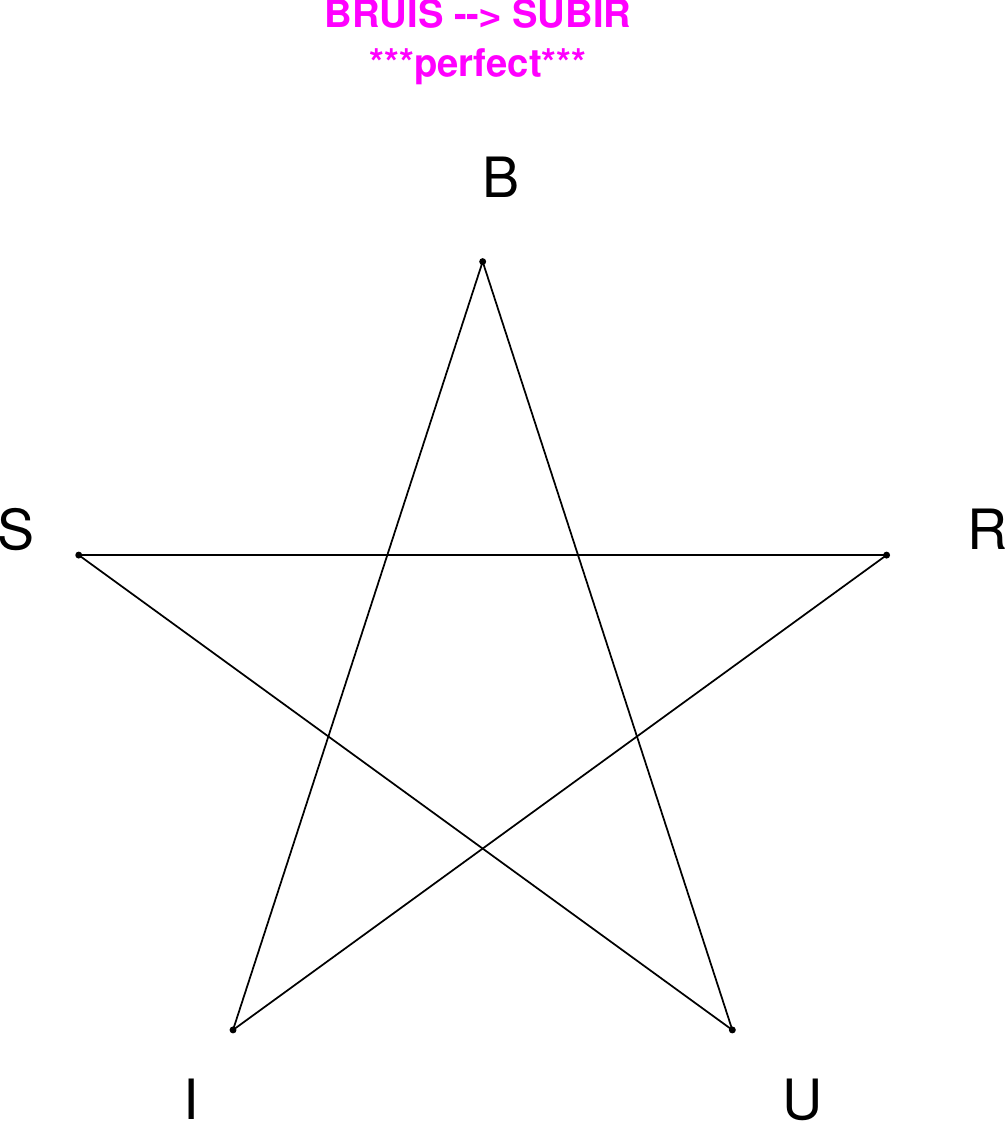}
\end{subfigure}
\hfill
\begin{subfigure}[T]{0.19\textwidth}
\centering
\includegraphics[width=\textwidth]{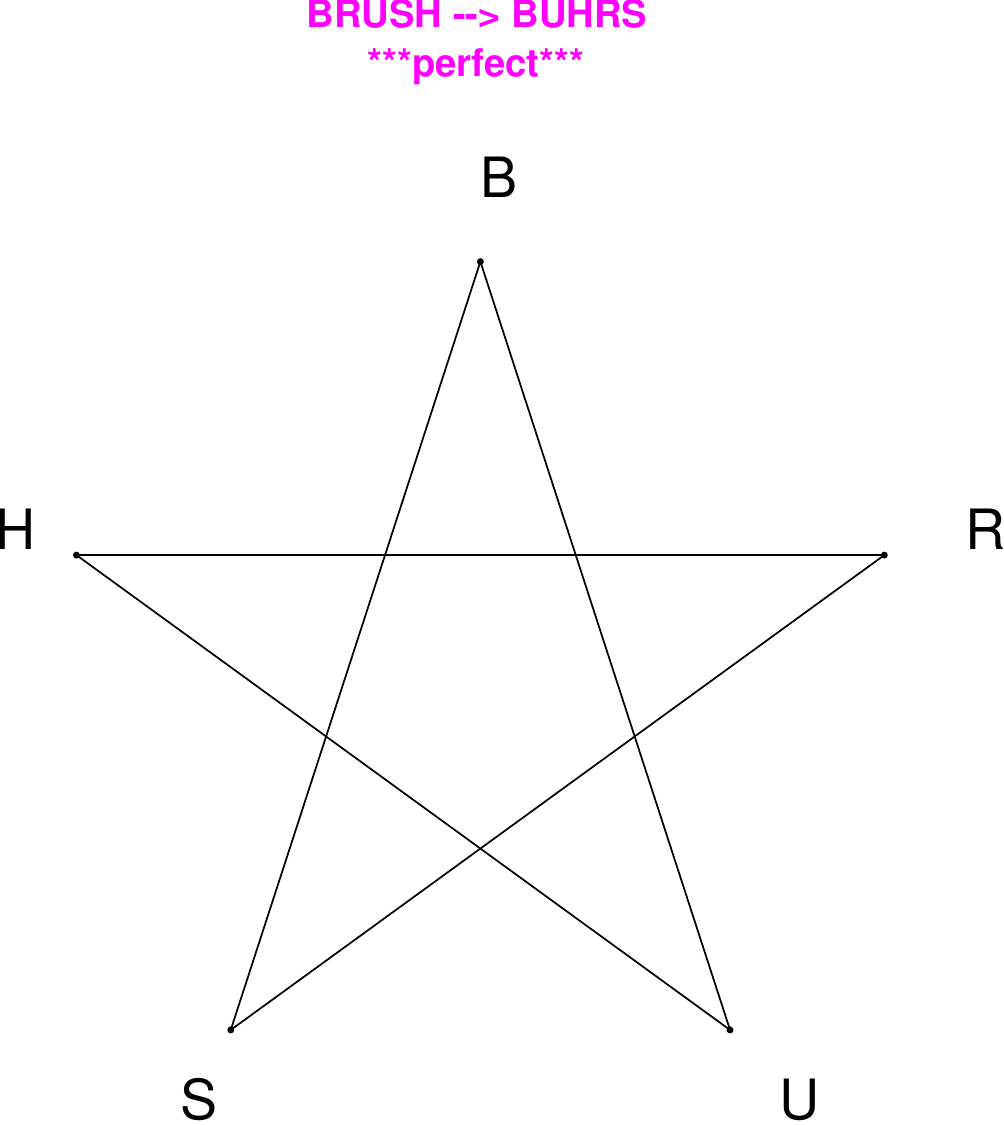}
\end{subfigure}
\hfill
\begin{subfigure}[T]{0.19\textwidth}
\centering
\includegraphics[width=\textwidth]{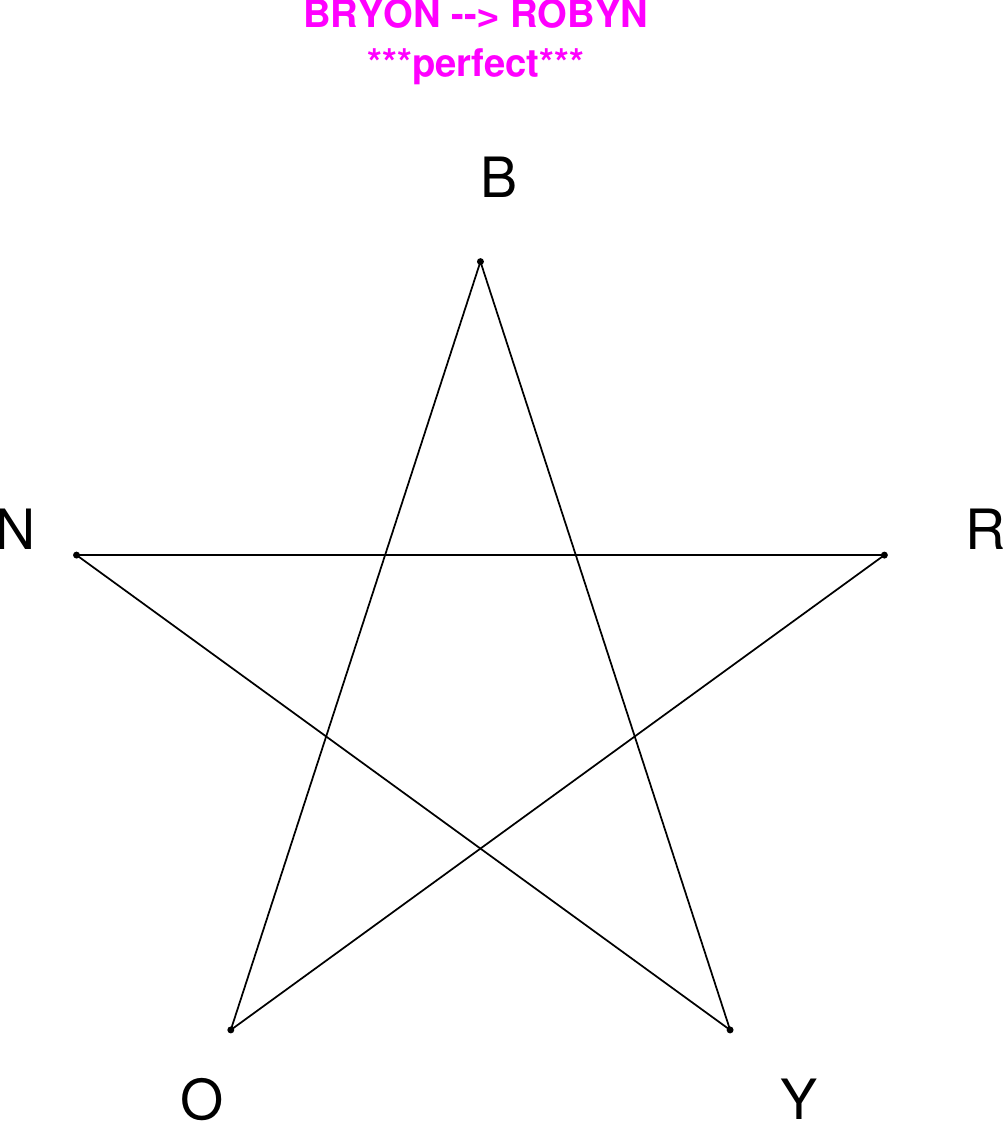}
\end{subfigure}
\end{figure}

\begin{figure}[H]
\centering
\begin{subfigure}[T]{0.19\textwidth}
\centering
\includegraphics[width=\textwidth]{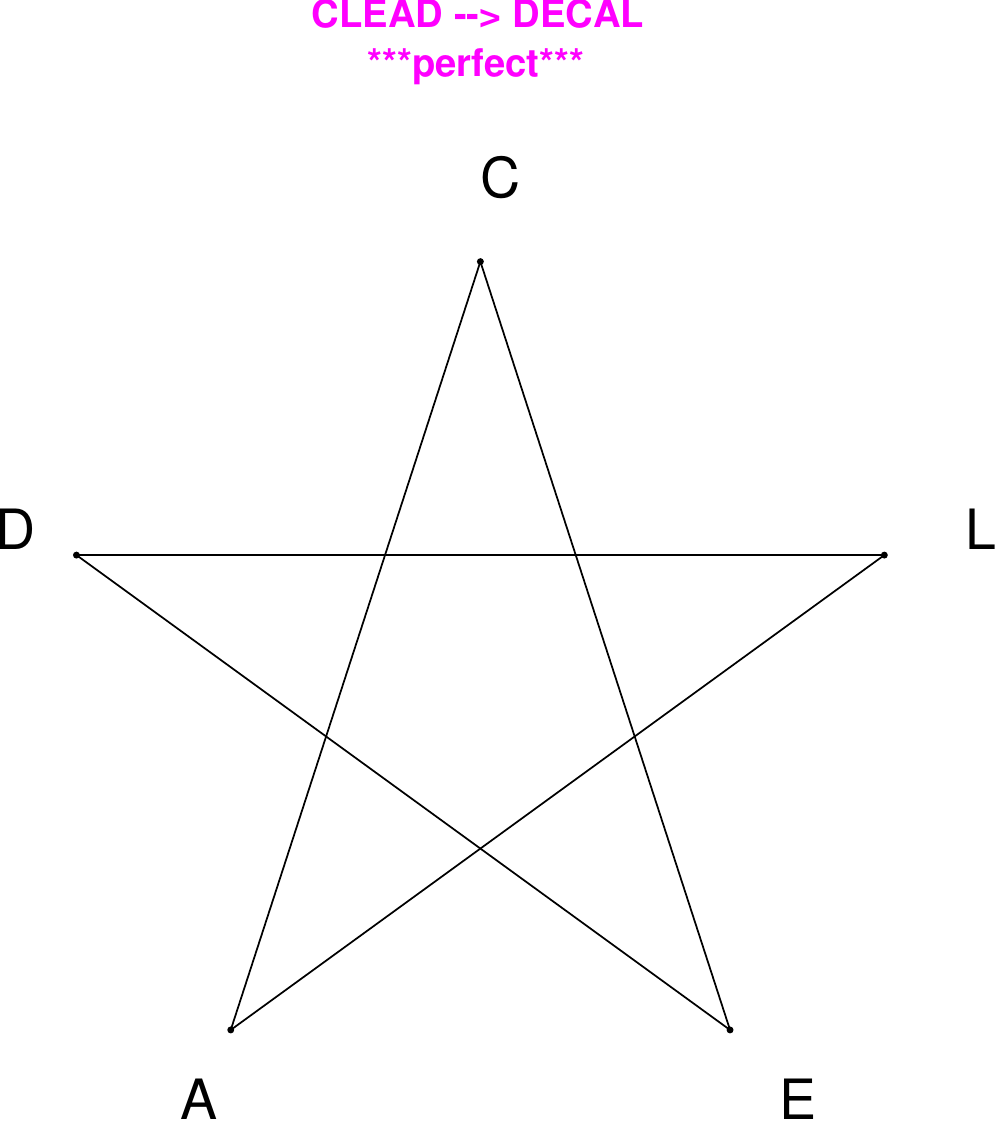}
\end{subfigure}
\hfill
\begin{subfigure}[T]{0.19\textwidth}
\centering
\includegraphics[width=\textwidth]{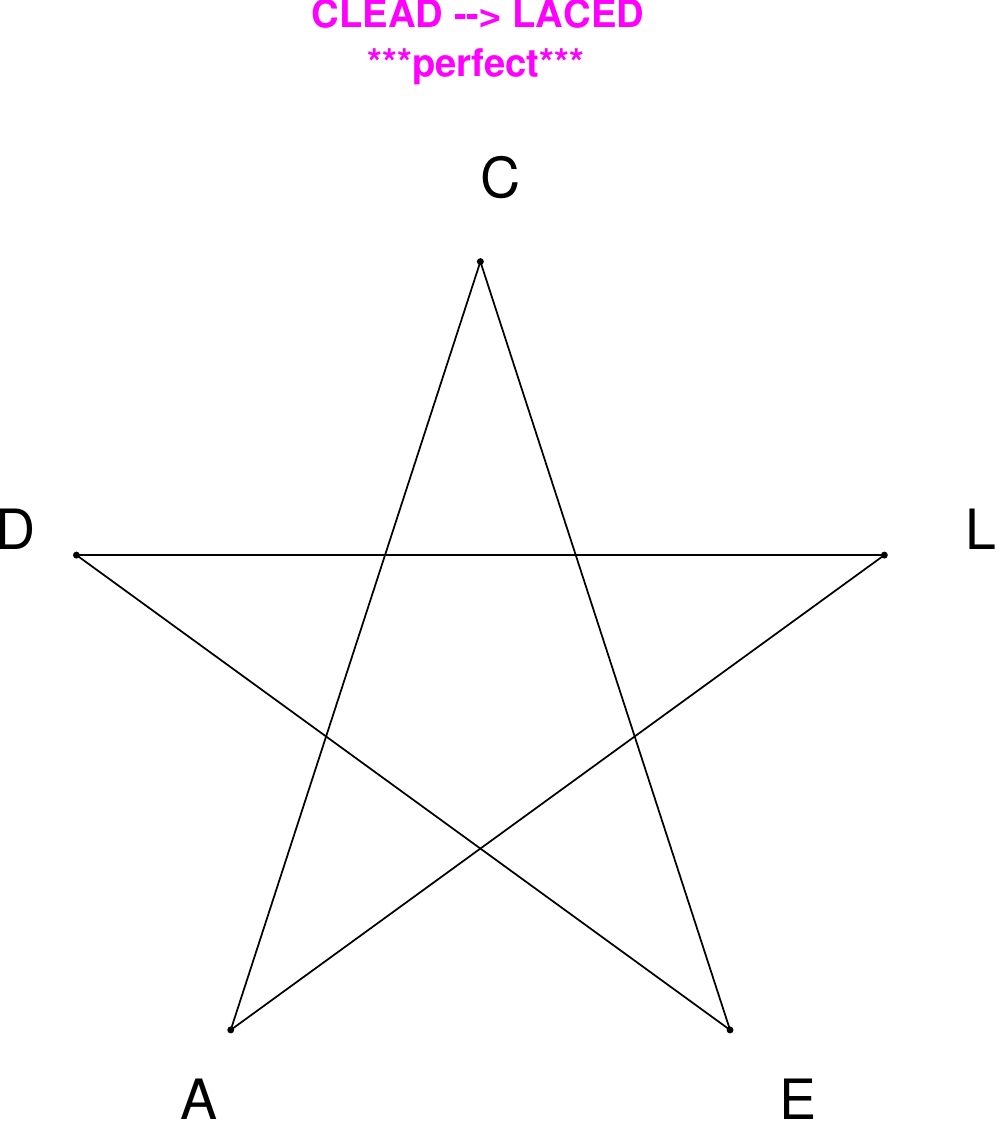}
\end{subfigure}
\hfill
\begin{subfigure}[T]{0.19\textwidth}
\centering
\includegraphics[width=\textwidth]{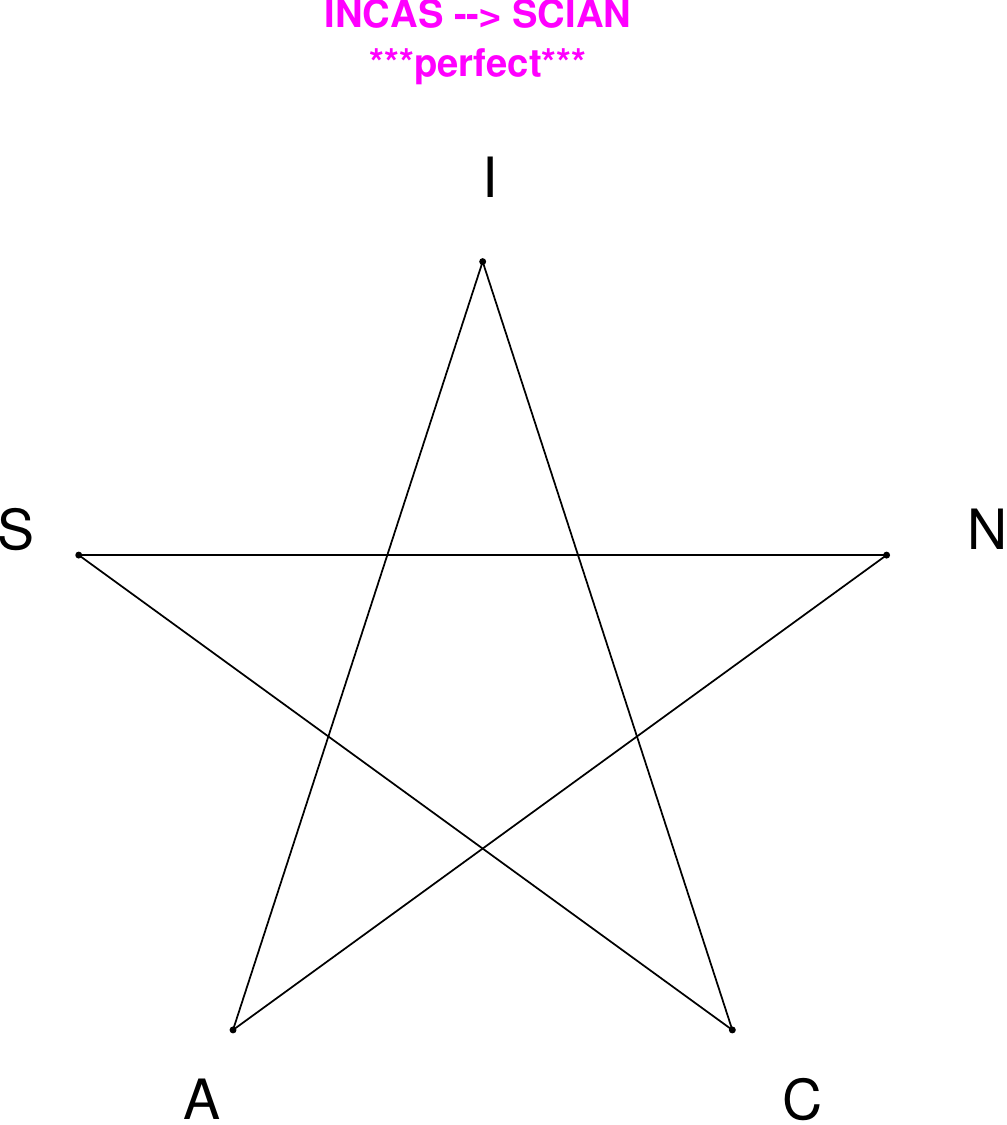}
\end{subfigure}
\hfill
\begin{subfigure}[T]{0.19\textwidth}
\centering
\includegraphics[width=\textwidth]{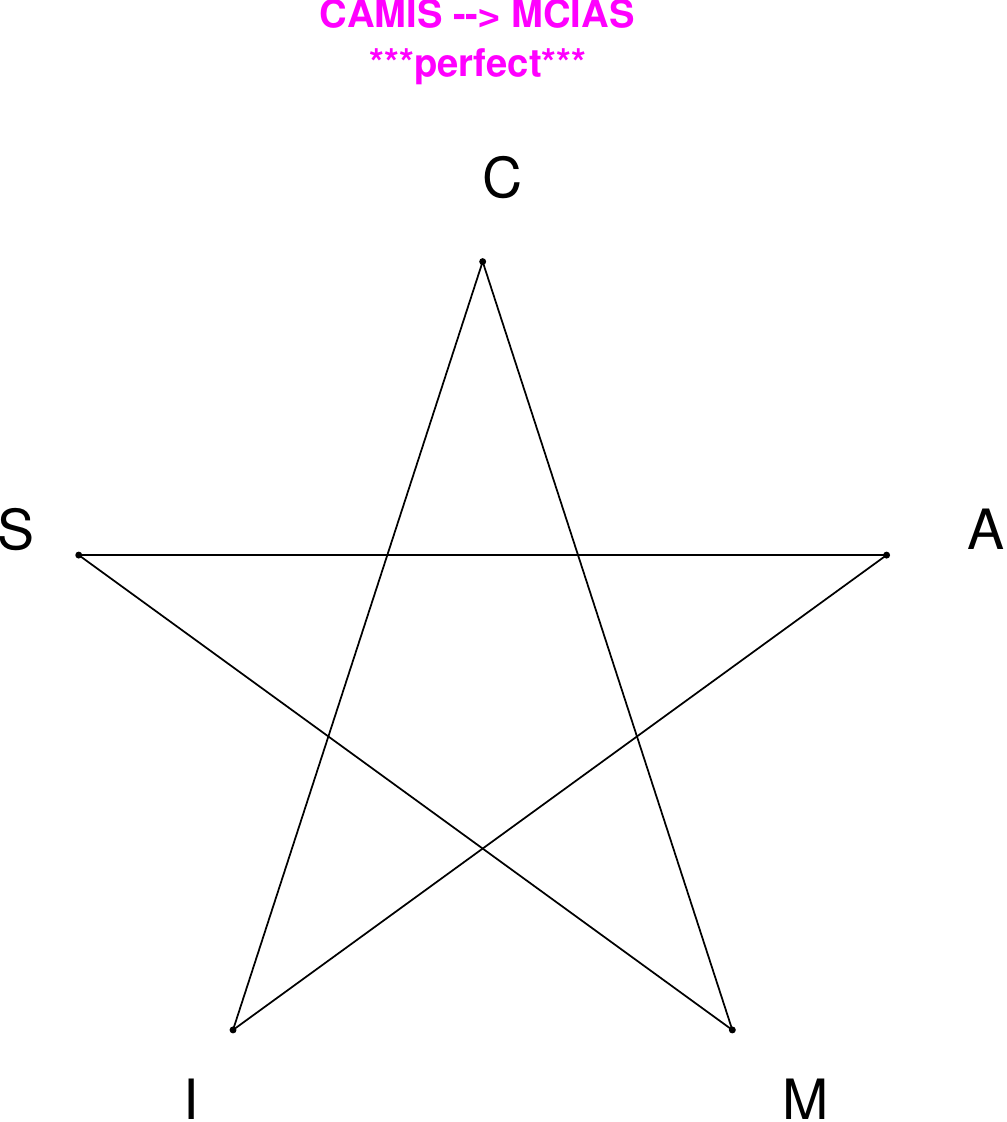}
\end{subfigure}
\hfill
\begin{subfigure}[T]{0.19\textwidth}
\centering
\includegraphics[width=\textwidth]{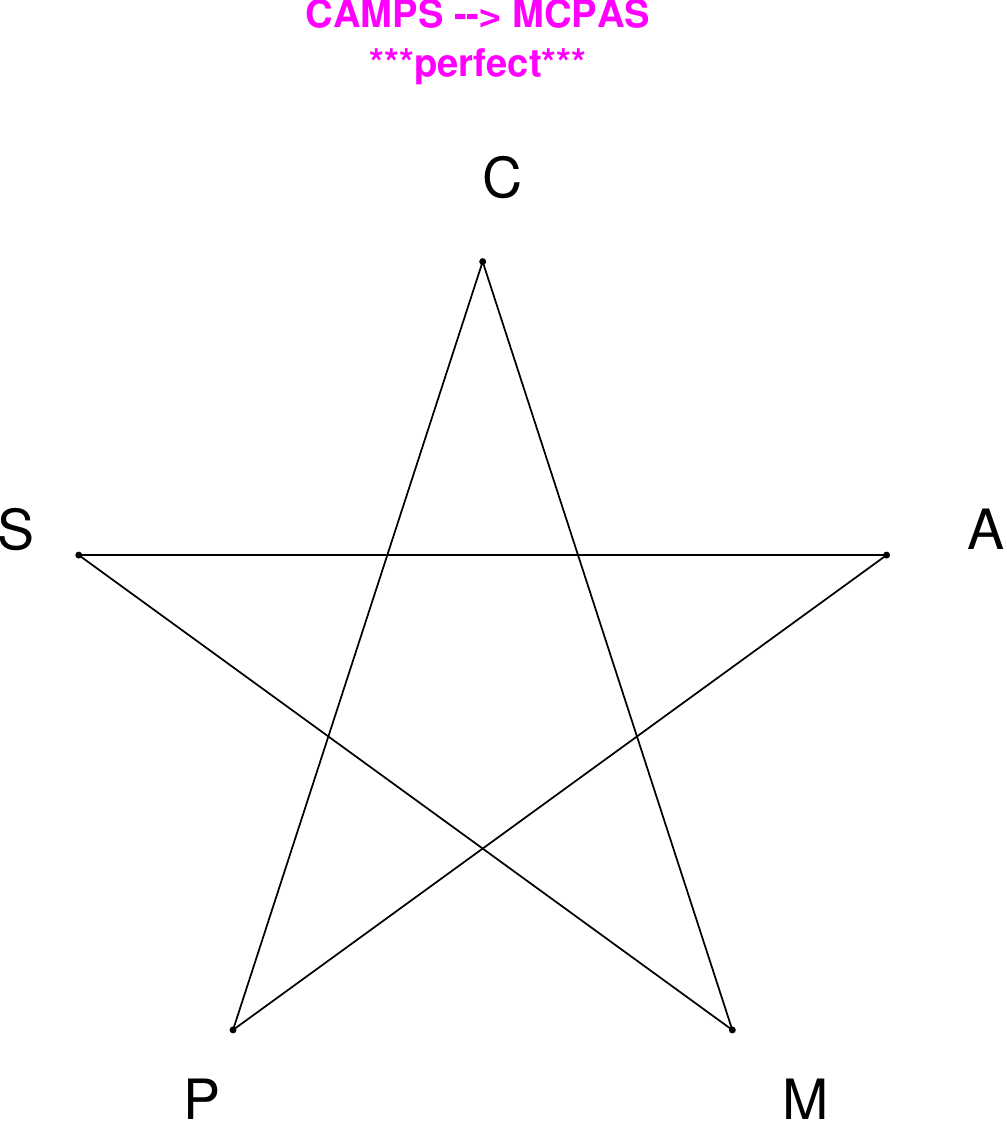}
\end{subfigure}
\end{figure}

\begin{figure}[H]
\centering
\begin{subfigure}[T]{0.19\textwidth}
\centering
\includegraphics[width=\textwidth]{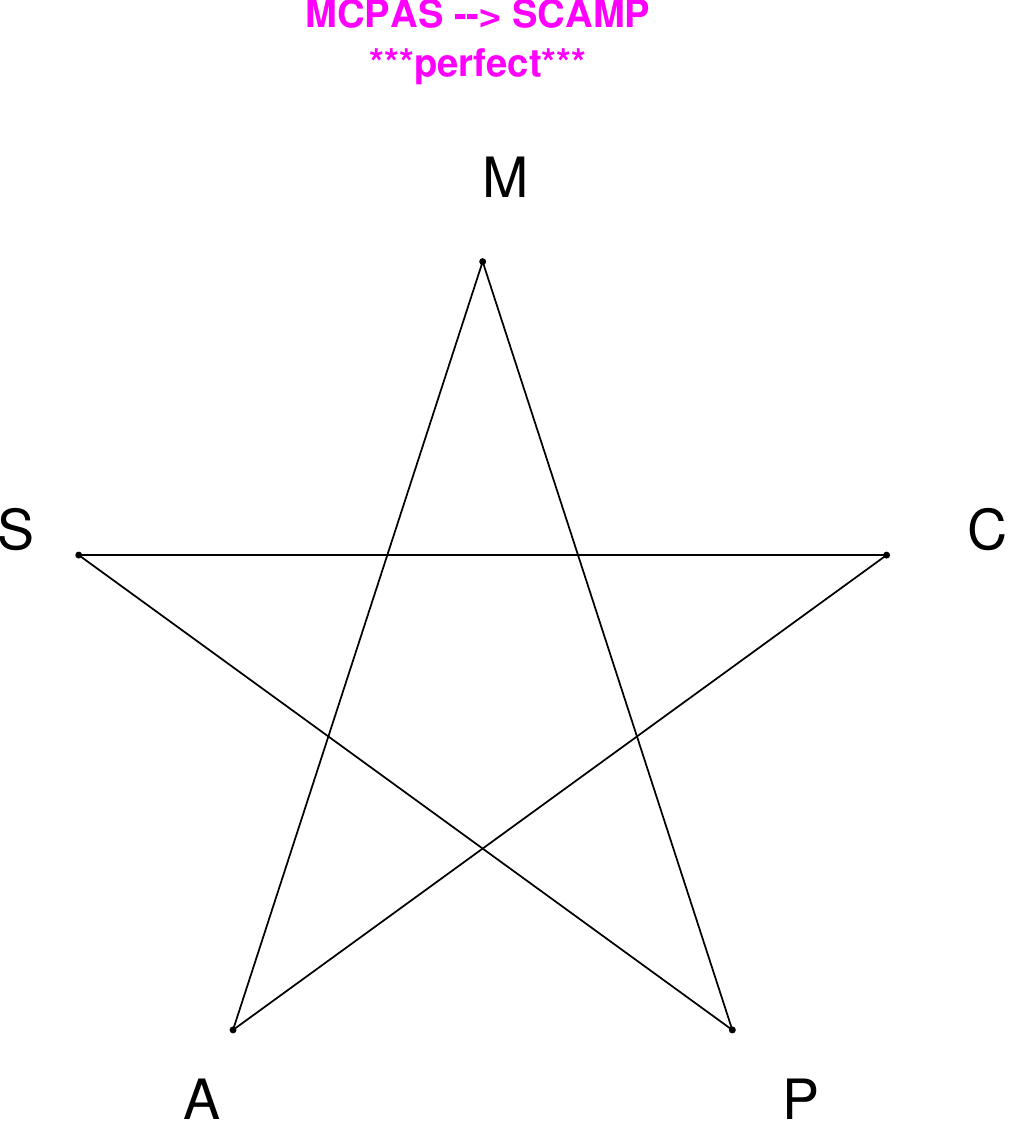}
\end{subfigure}
\hfill
\begin{subfigure}[T]{0.19\textwidth}
\centering
\includegraphics[width=\textwidth]{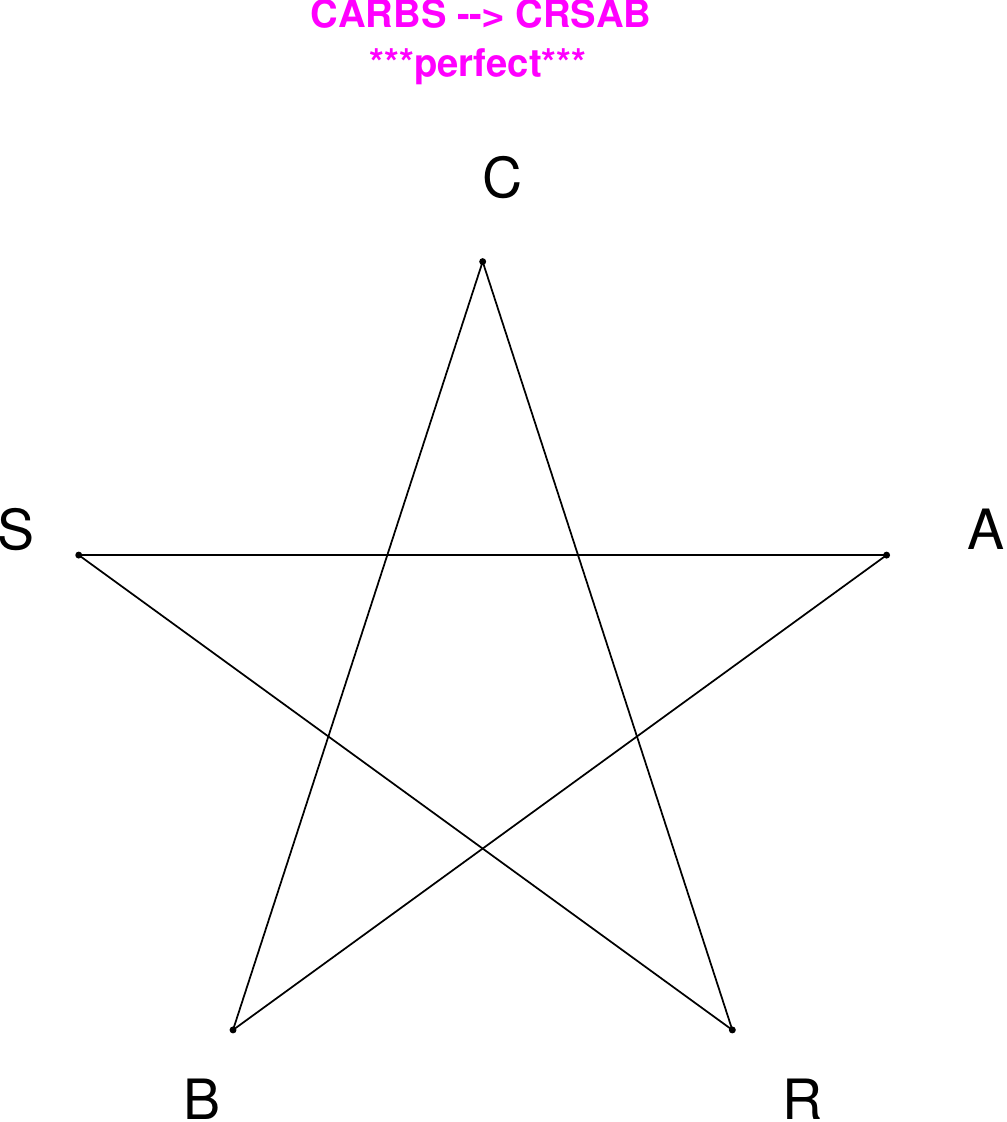}
\end{subfigure}
\hfill
\begin{subfigure}[T]{0.19\textwidth}
\centering
\includegraphics[width=\textwidth]{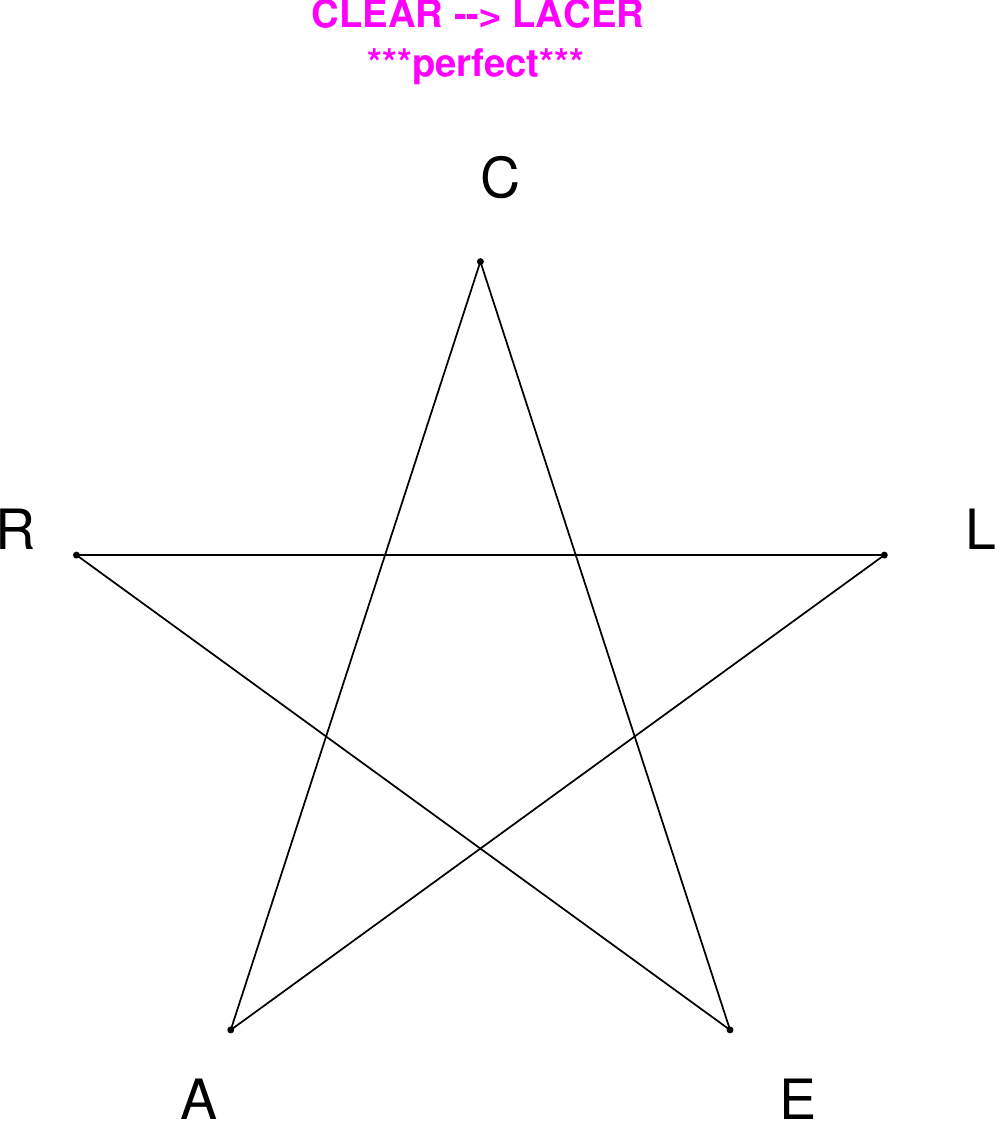}
\end{subfigure}
\hfill
\begin{subfigure}[T]{0.19\textwidth}
\centering
\includegraphics[width=\textwidth]{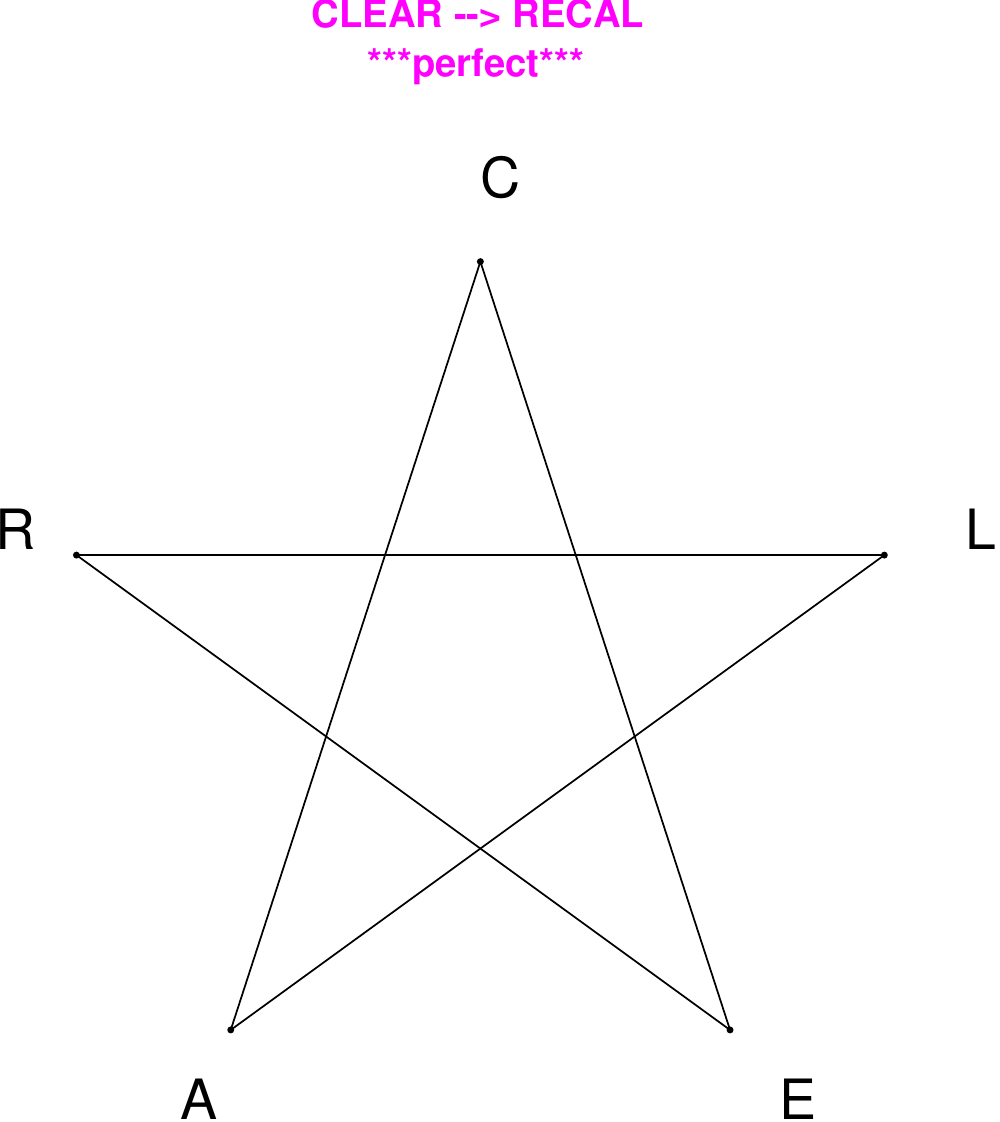}
\end{subfigure}
\hfill
\begin{subfigure}[T]{0.19\textwidth}
\centering
\includegraphics[width=\textwidth]{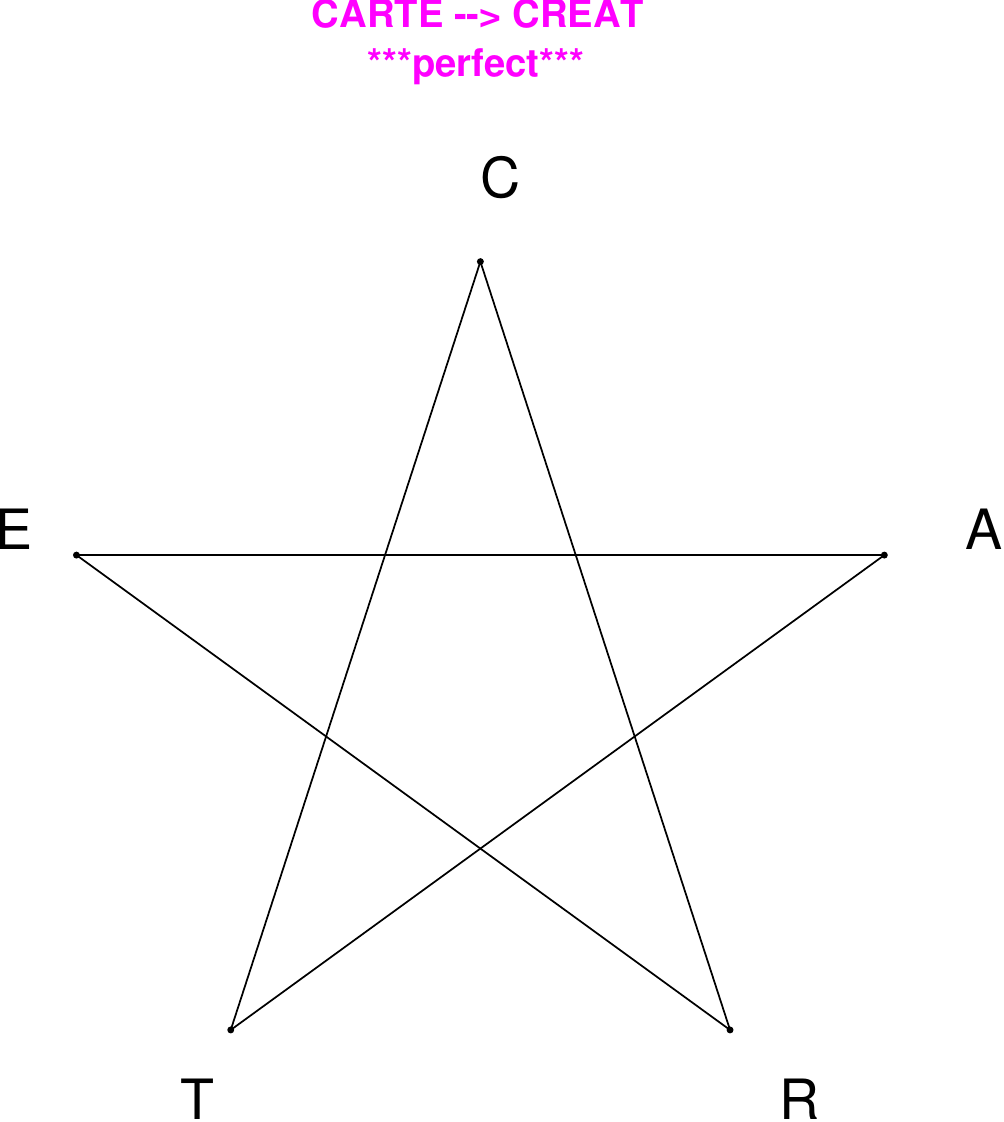}
\end{subfigure}
\end{figure}

\begin{figure}[H]
\centering
\begin{subfigure}[T]{0.19\textwidth}
\centering
\includegraphics[width=\textwidth]{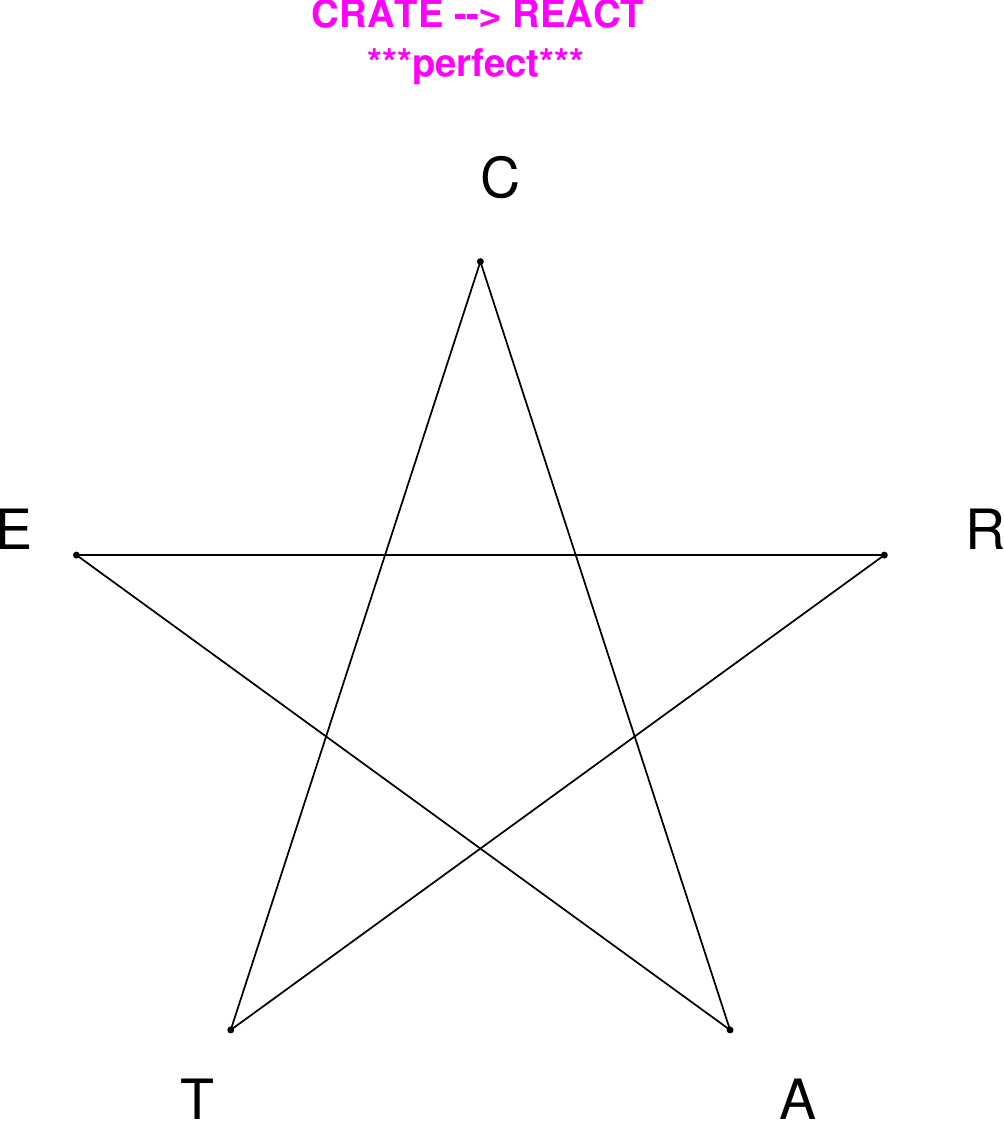}
\end{subfigure}
\hfill
\begin{subfigure}[T]{0.19\textwidth}
\centering
\includegraphics[width=\textwidth]{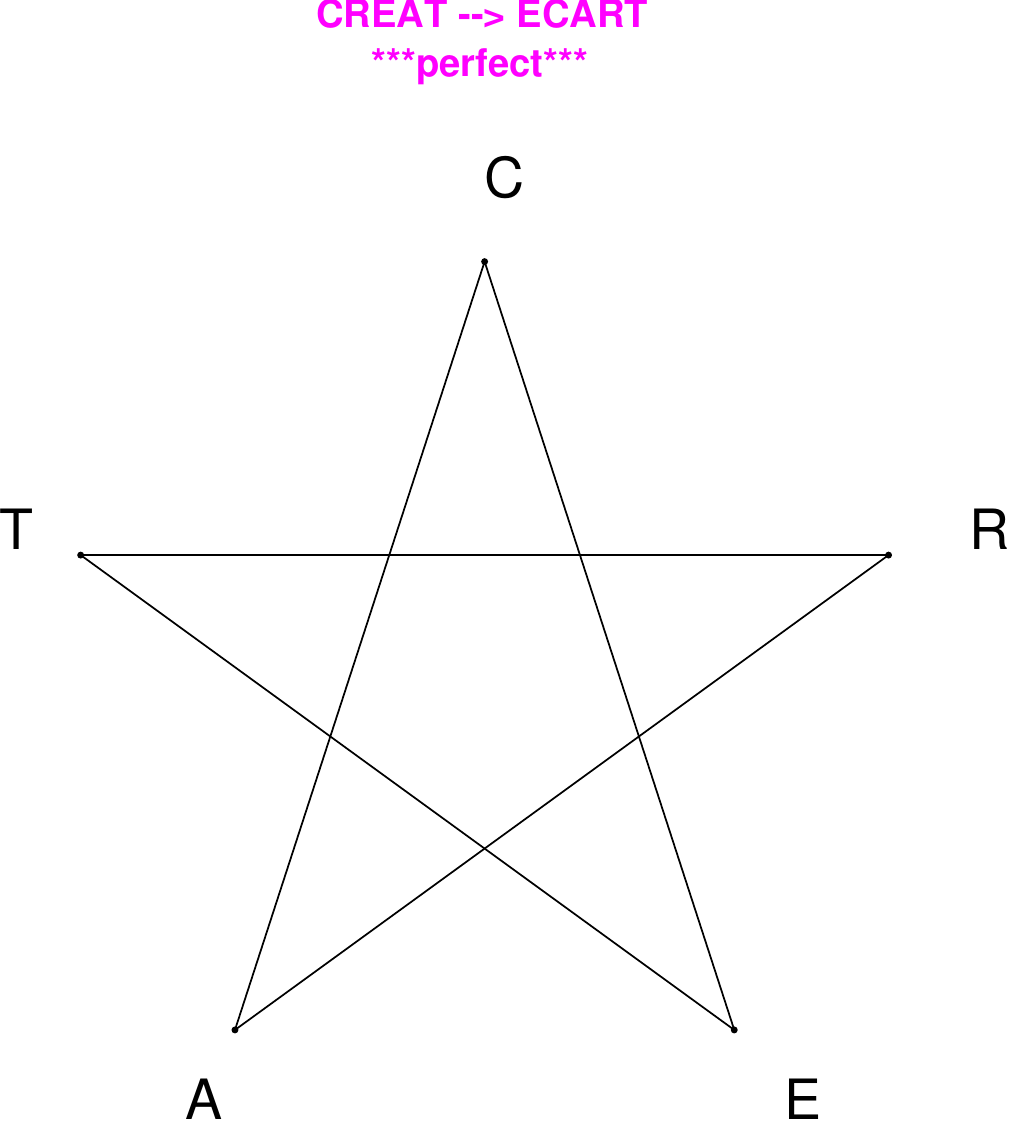}
\end{subfigure}
\hfill
\begin{subfigure}[T]{0.19\textwidth}
\centering
\includegraphics[width=\textwidth]{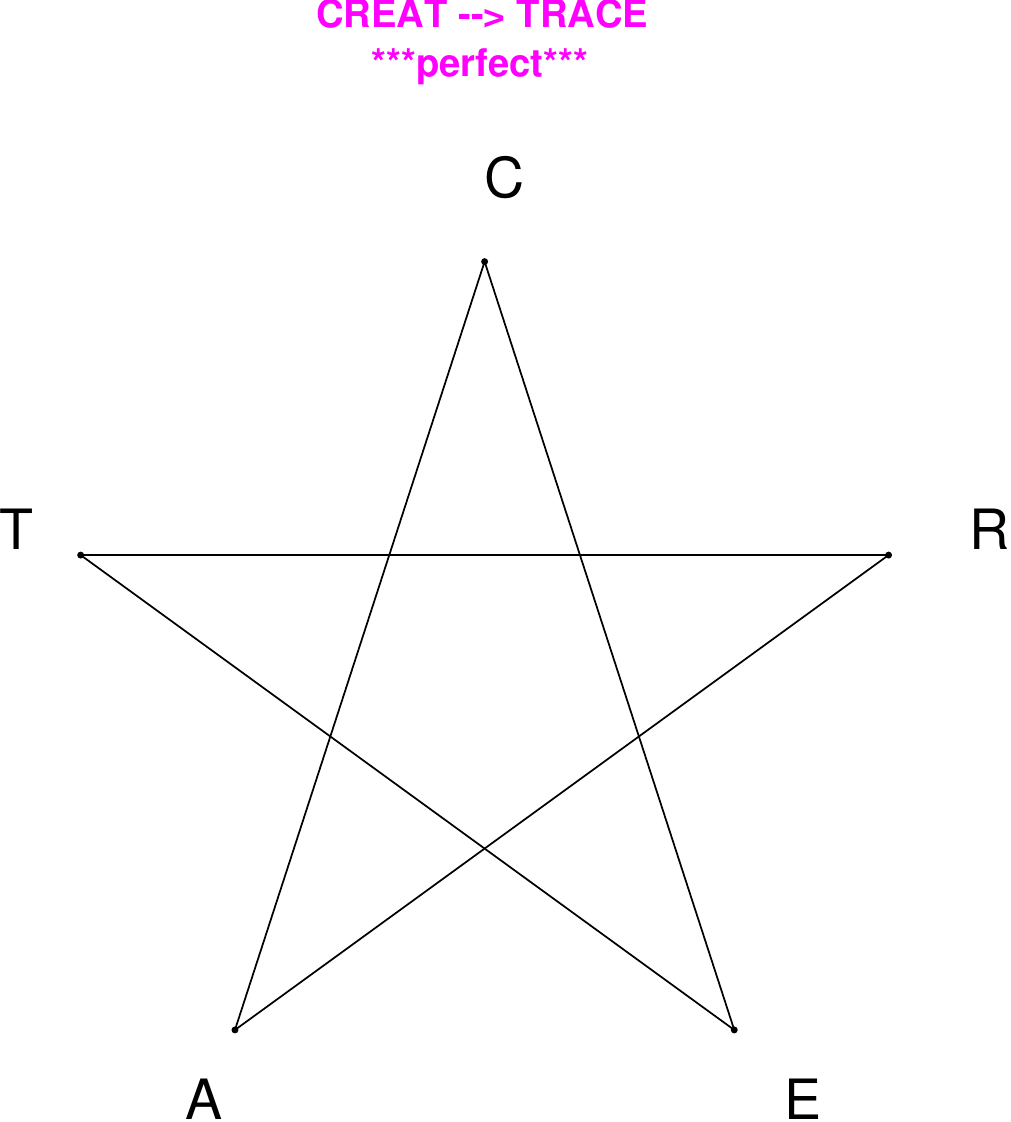}
\end{subfigure}
\hfill
\begin{subfigure}[T]{0.19\textwidth}
\centering
\includegraphics[width=\textwidth]{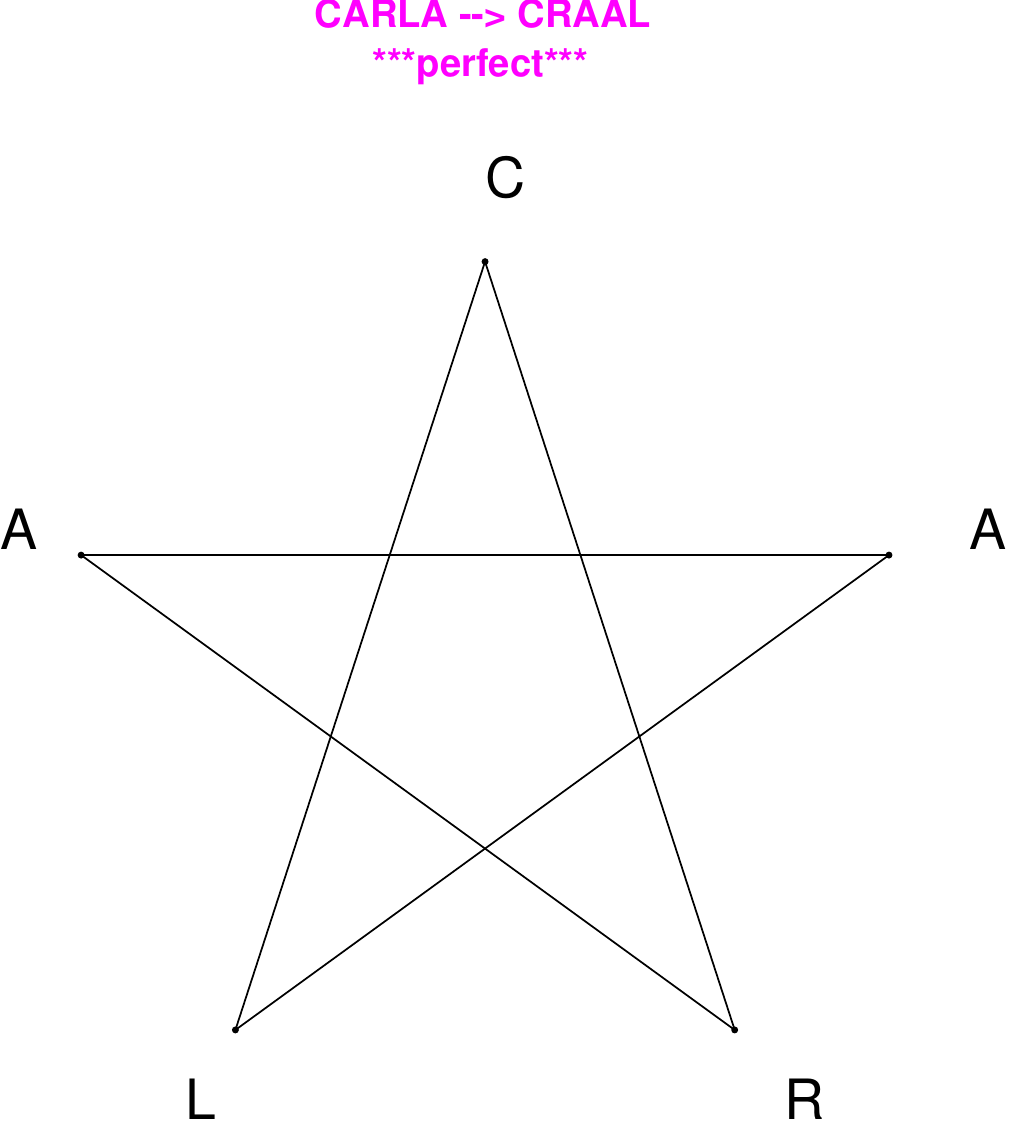}
\end{subfigure}
\hfill
\begin{subfigure}[T]{0.19\textwidth}
\centering
\includegraphics[width=\textwidth]{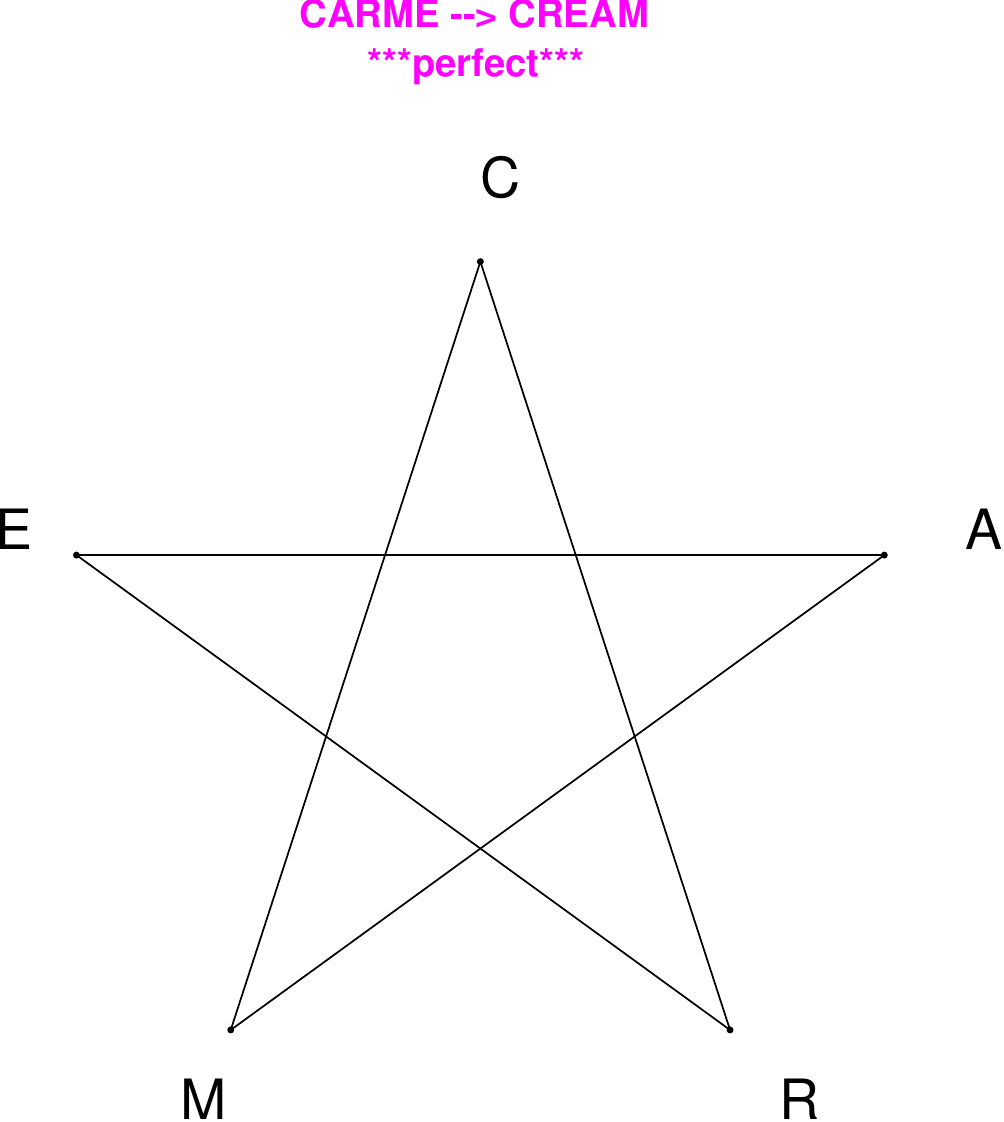}
\end{subfigure}
\end{figure}

\begin{figure}[H]
\centering
\begin{subfigure}[T]{0.19\textwidth}
\centering
\includegraphics[width=\textwidth]{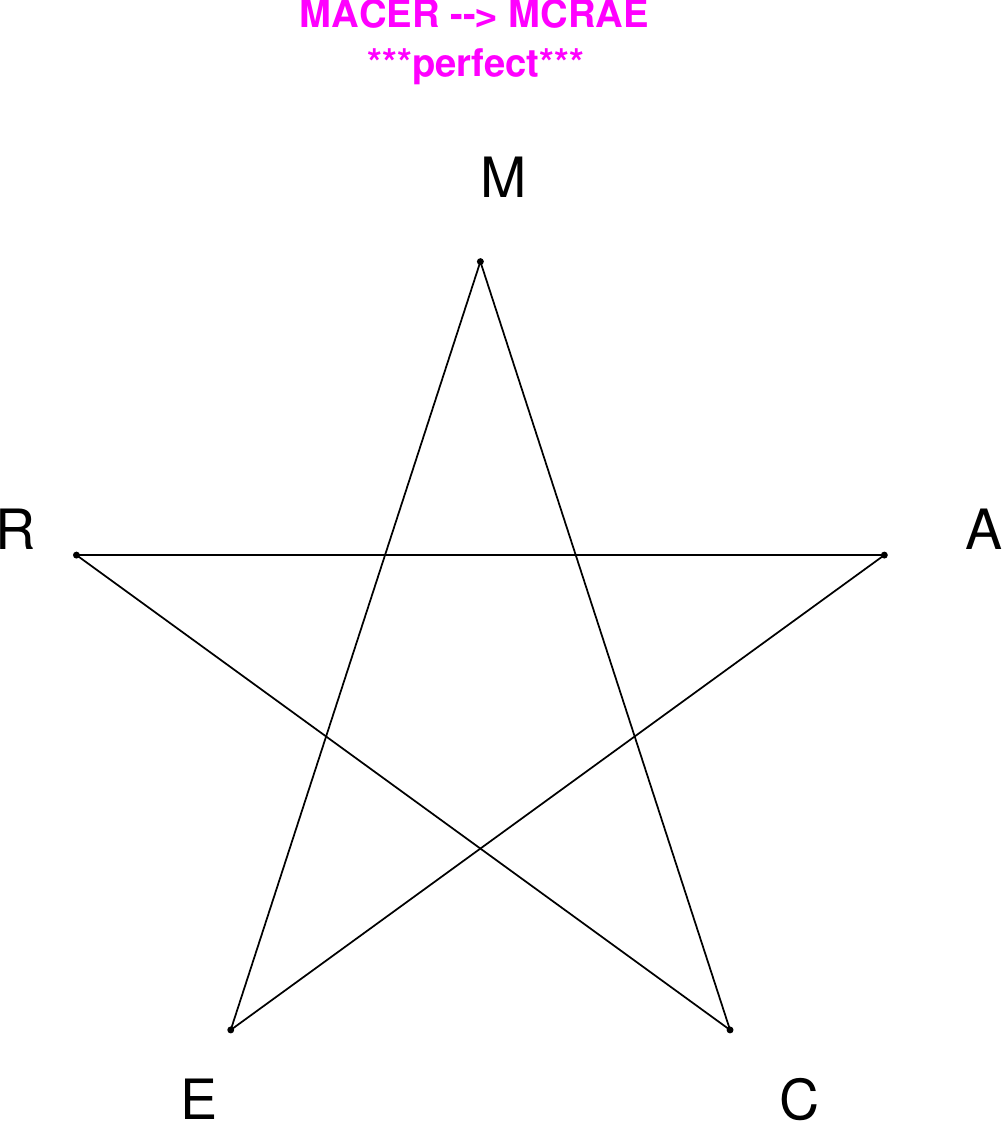}
\end{subfigure}
\hfill
\begin{subfigure}[T]{0.19\textwidth}
\centering
\includegraphics[width=\textwidth]{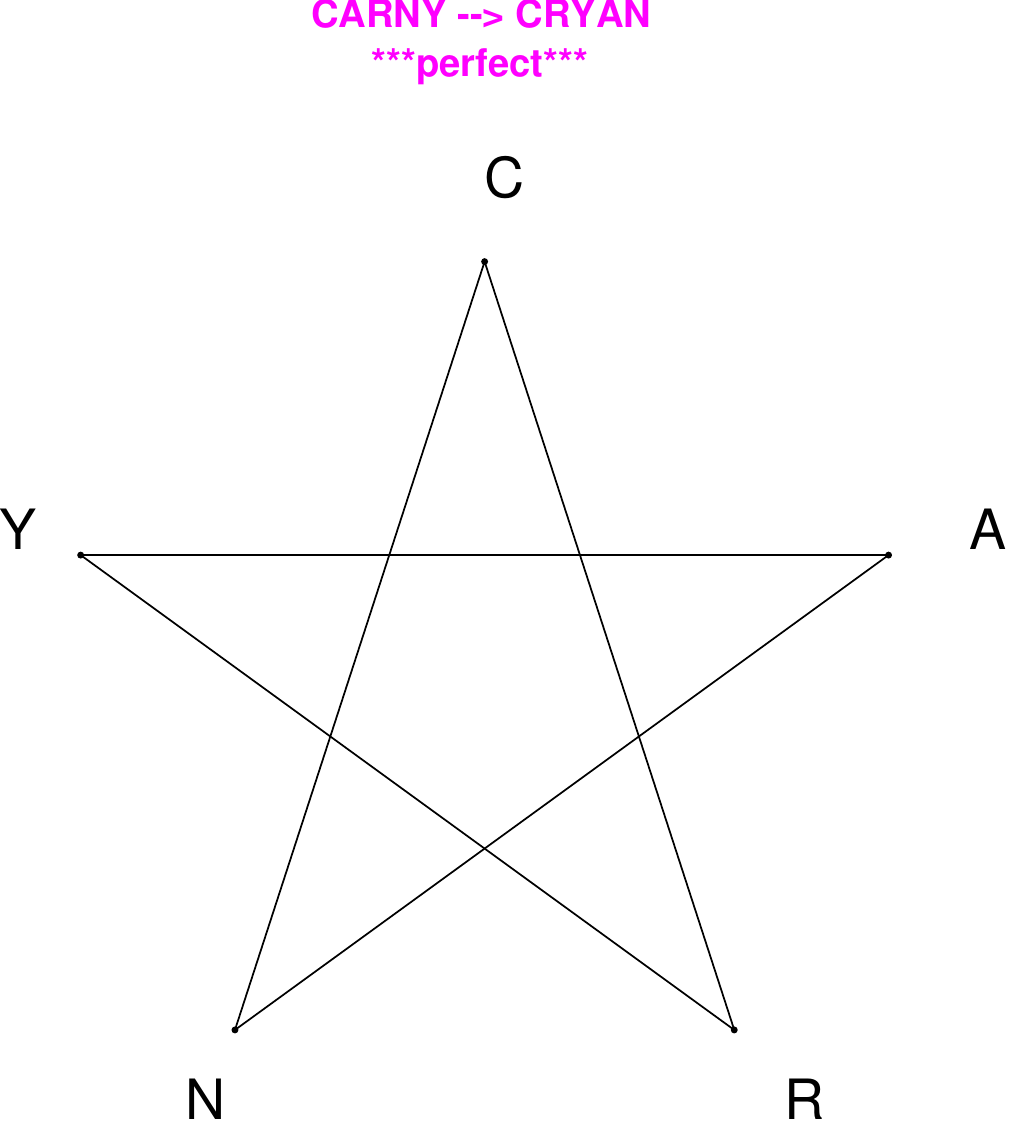}
\end{subfigure}
\hfill
\begin{subfigure}[T]{0.19\textwidth}
\centering
\includegraphics[width=\textwidth]{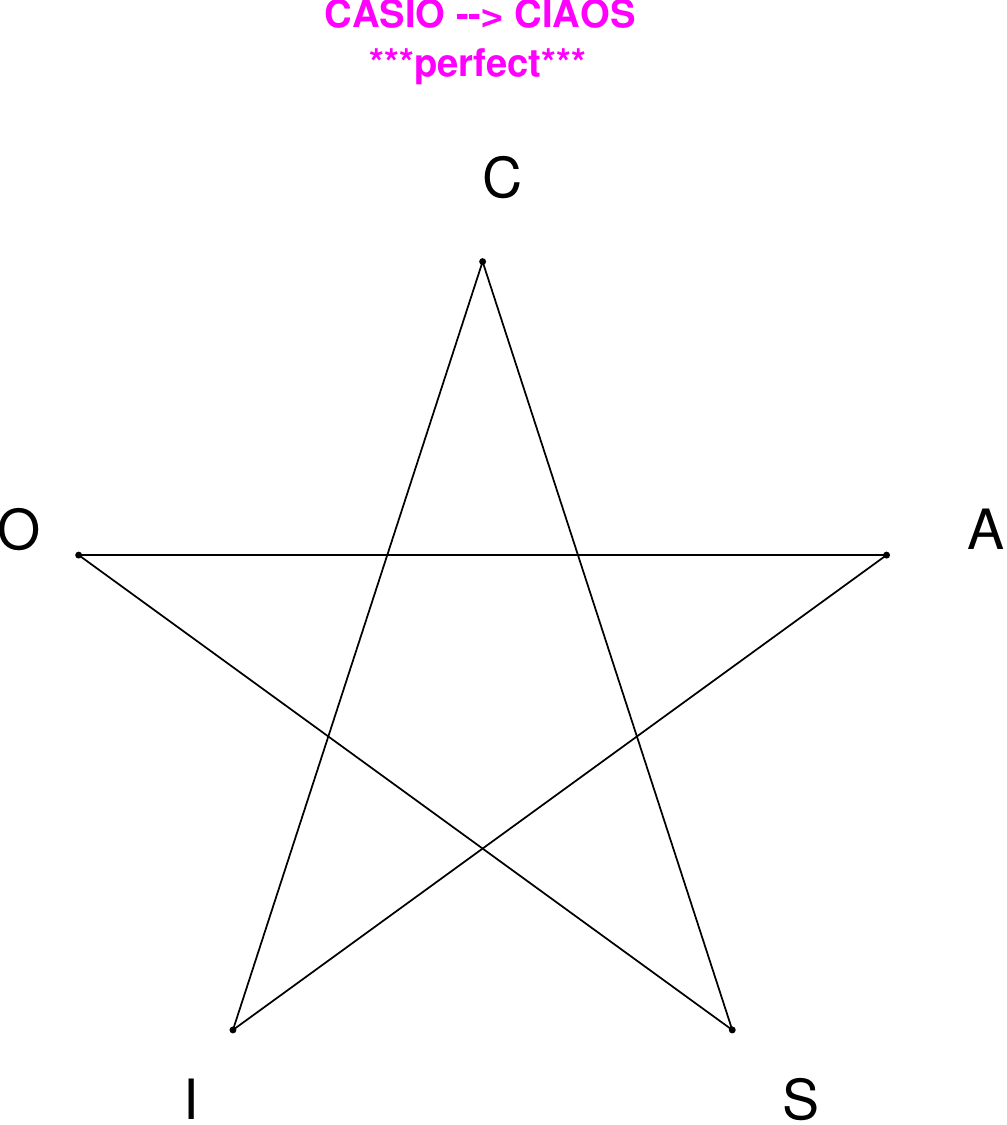}
\end{subfigure}
\hfill
\begin{subfigure}[T]{0.19\textwidth}
\centering
\includegraphics[width=\textwidth]{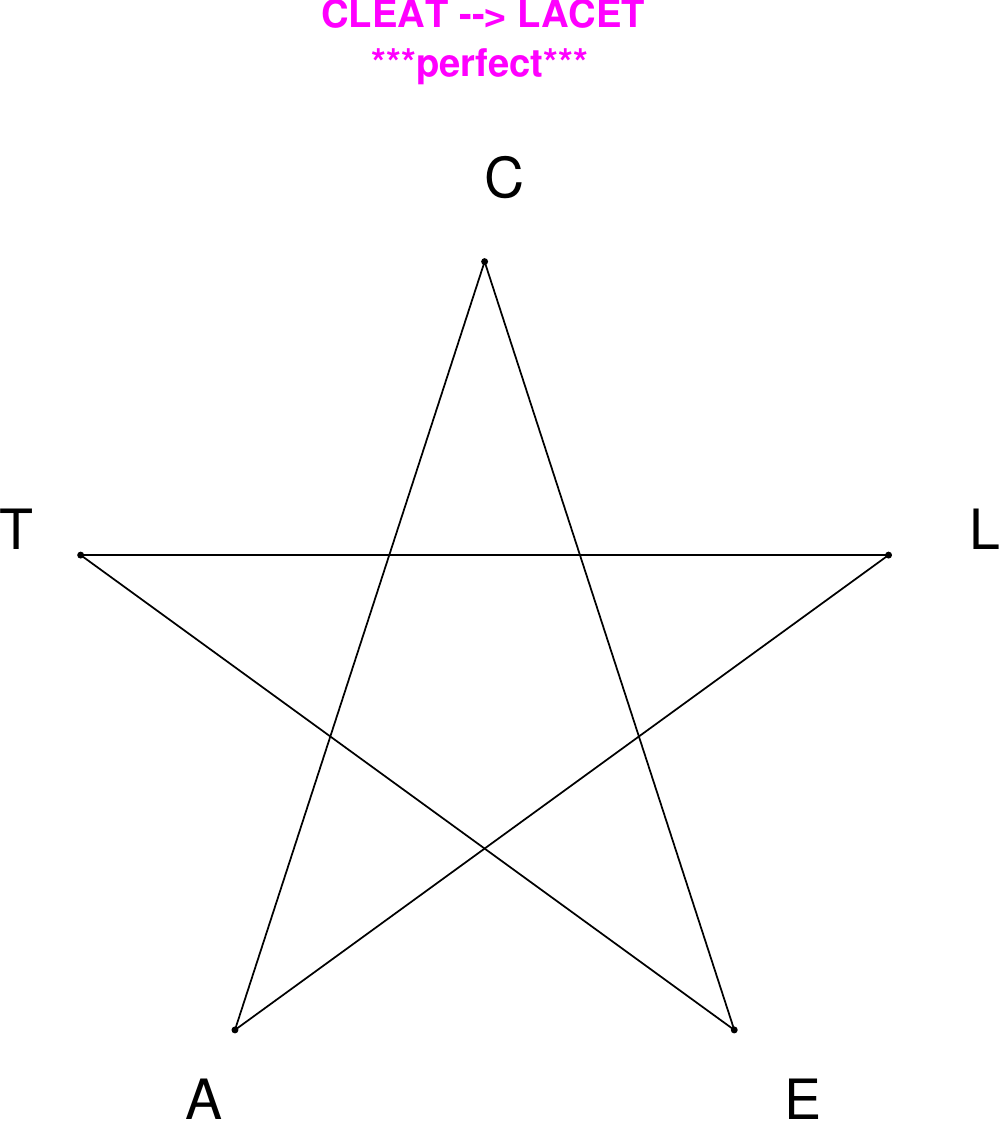}
\end{subfigure}
\hfill
\begin{subfigure}[T]{0.19\textwidth}
\centering
\includegraphics[width=\textwidth]{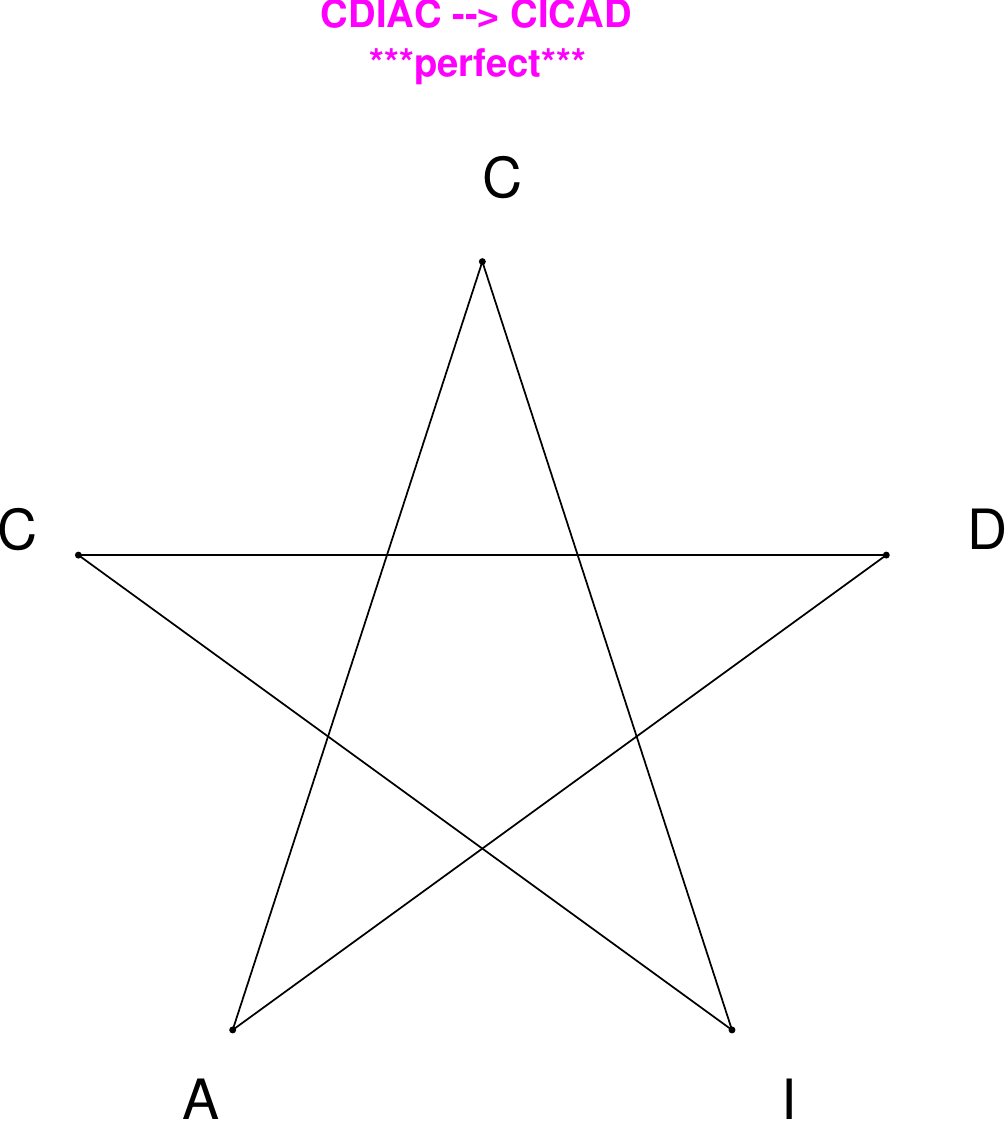}
\end{subfigure}
\end{figure}

\begin{figure}[H]
\centering
\begin{subfigure}[T]{0.19\textwidth}
\centering
\includegraphics[width=\textwidth]{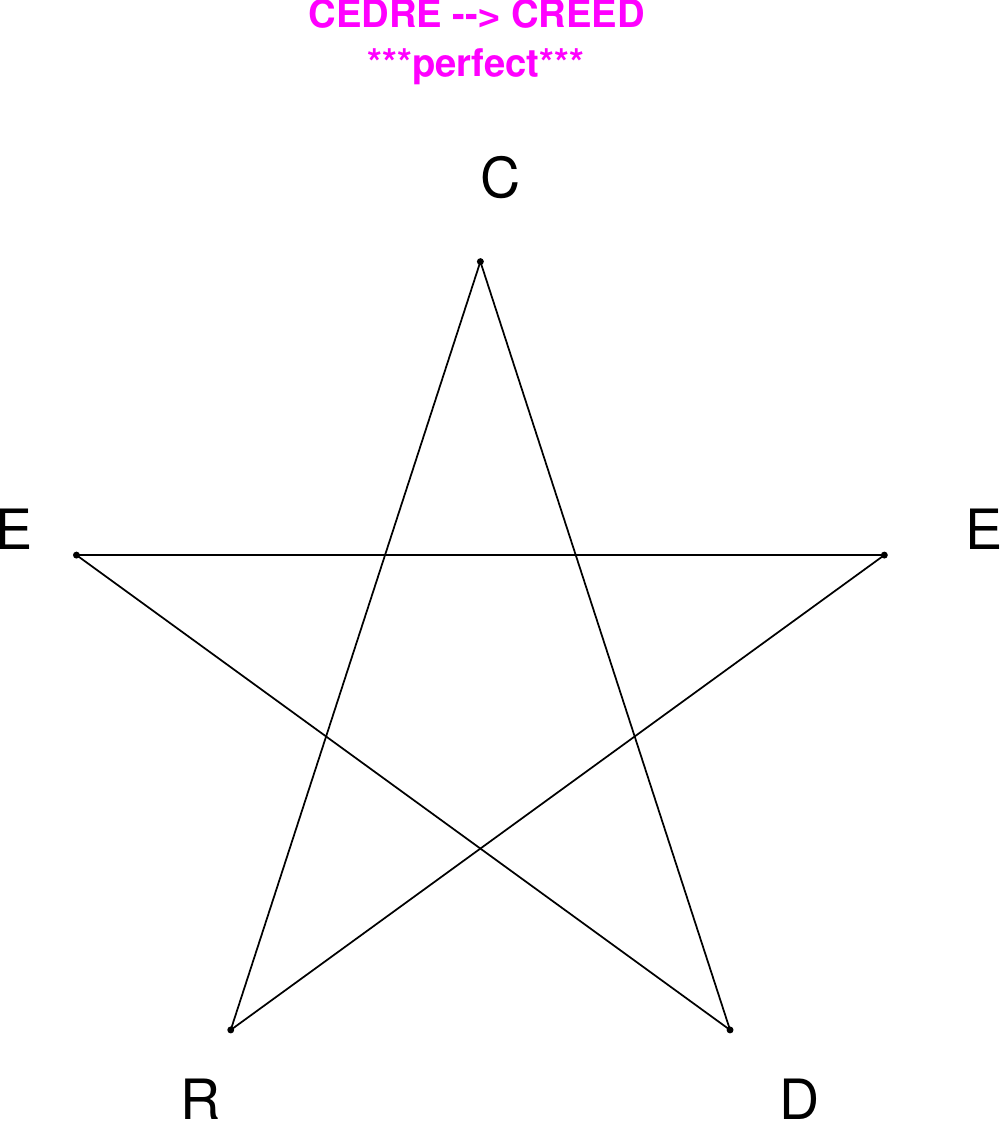}
\end{subfigure}
\hfill
\begin{subfigure}[T]{0.19\textwidth}
\centering
\includegraphics[width=\textwidth]{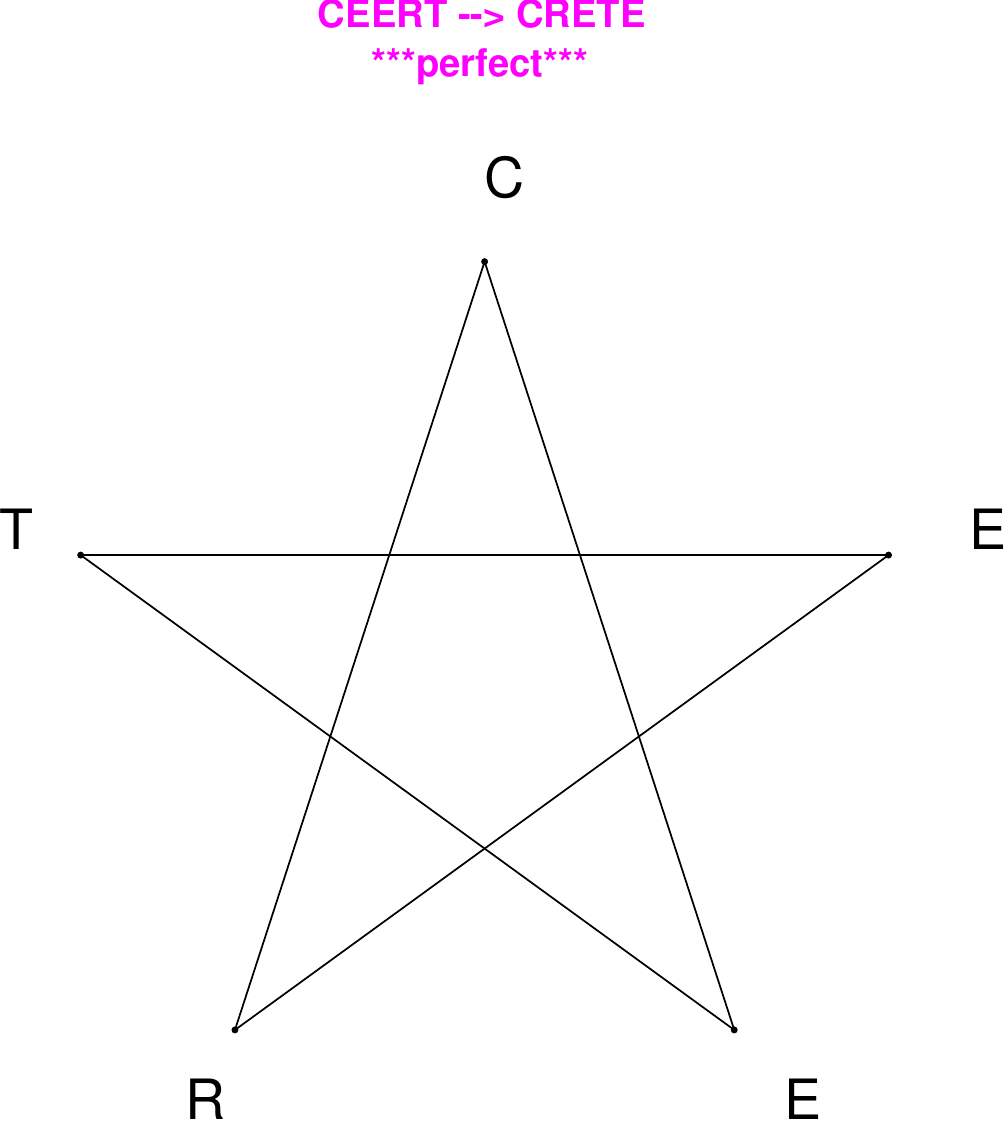}
\end{subfigure}
\hfill
\begin{subfigure}[T]{0.19\textwidth}
\centering
\includegraphics[width=\textwidth]{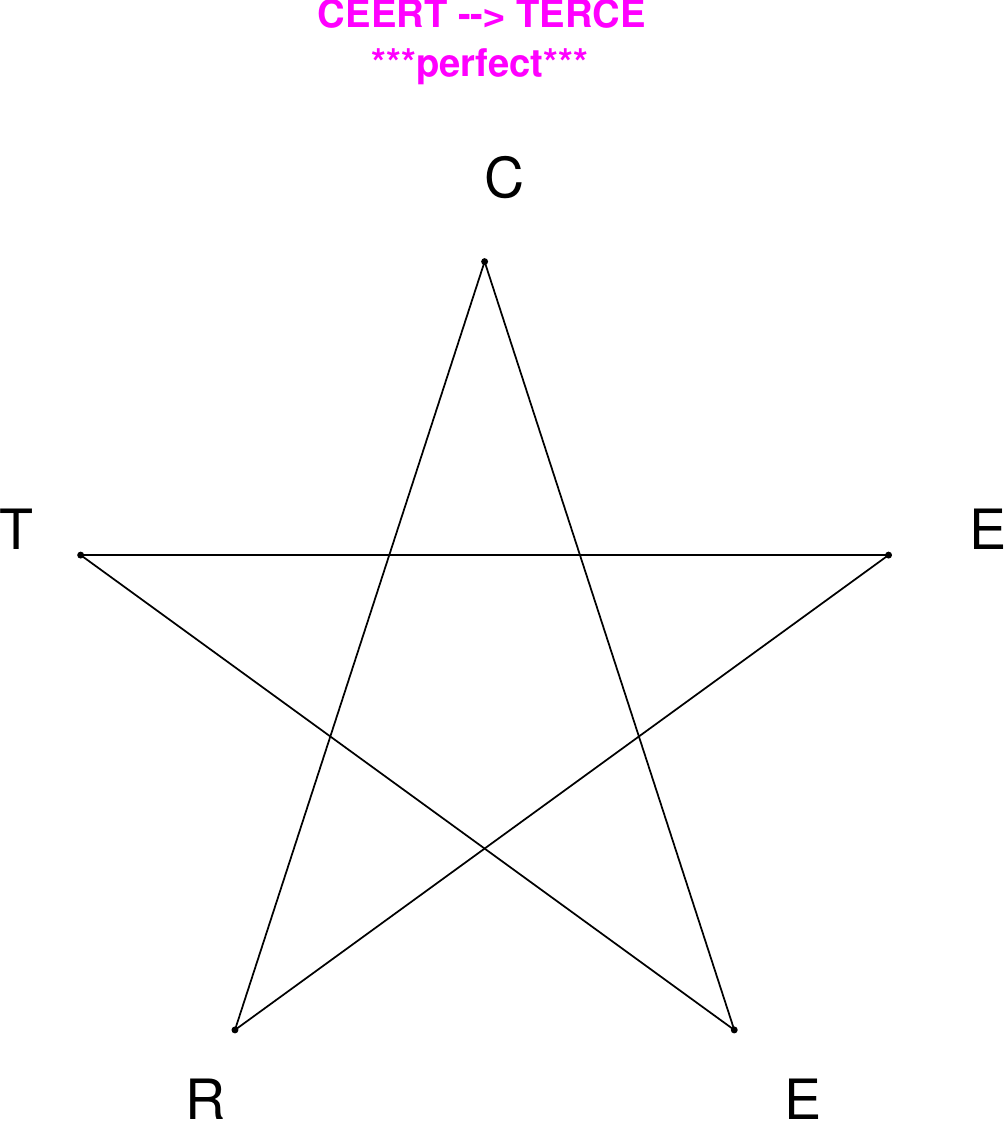}
\end{subfigure}
\hfill
\begin{subfigure}[T]{0.19\textwidth}
\centering
\includegraphics[width=\textwidth]{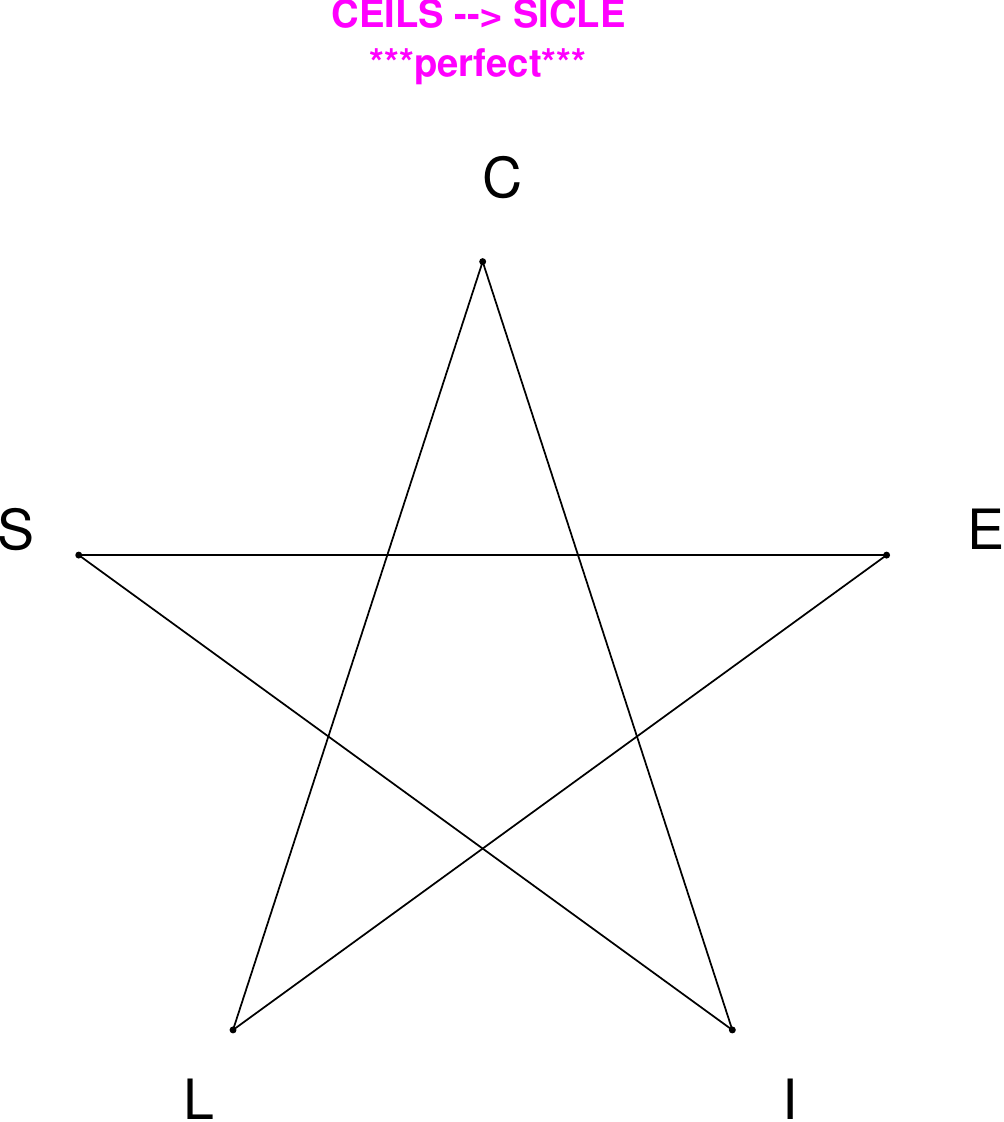}
\end{subfigure}
\hfill
\begin{subfigure}[T]{0.19\textwidth}
\centering
\includegraphics[width=\textwidth]{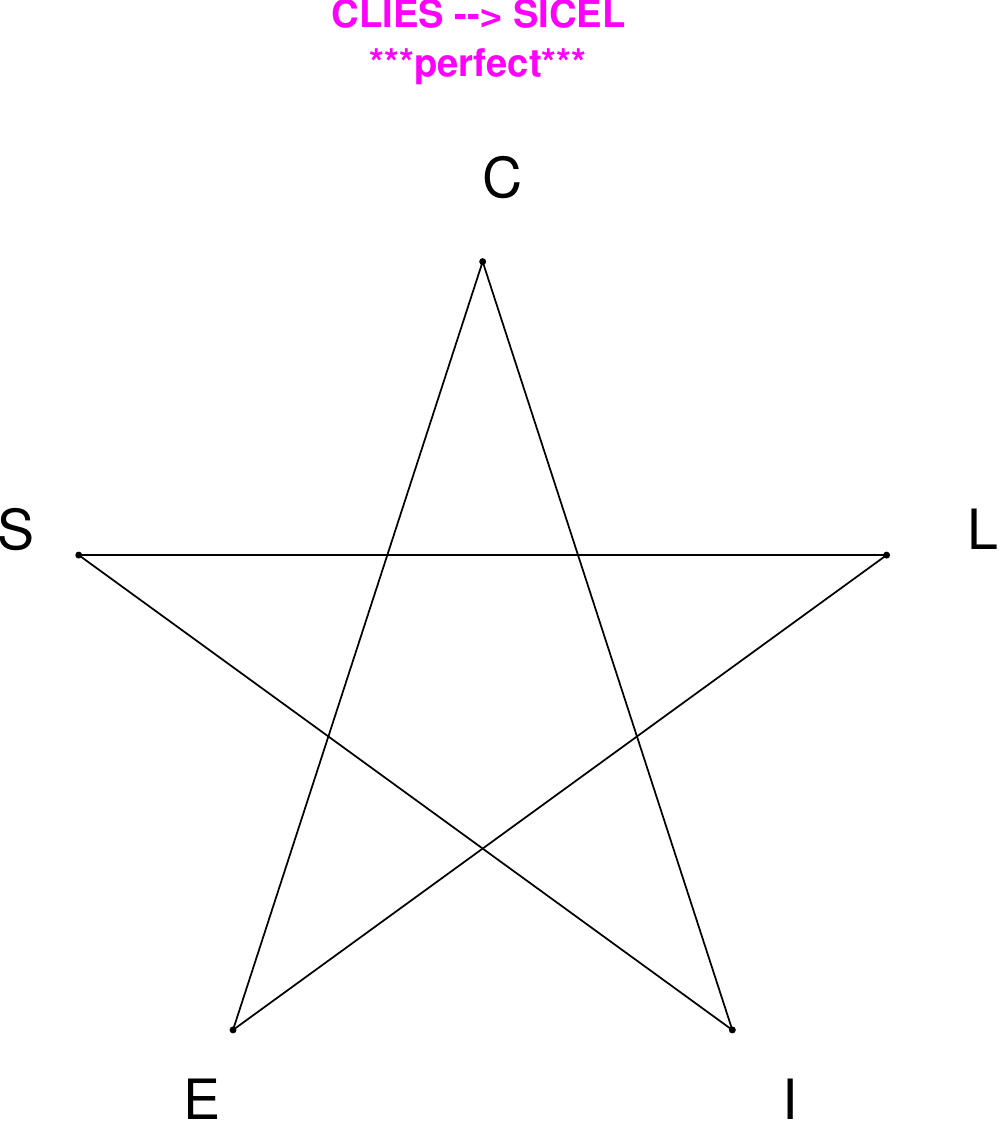}
\end{subfigure}
\end{figure}

\begin{figure}[H]
\centering
\begin{subfigure}[T]{0.19\textwidth}
\centering
\includegraphics[width=\textwidth]{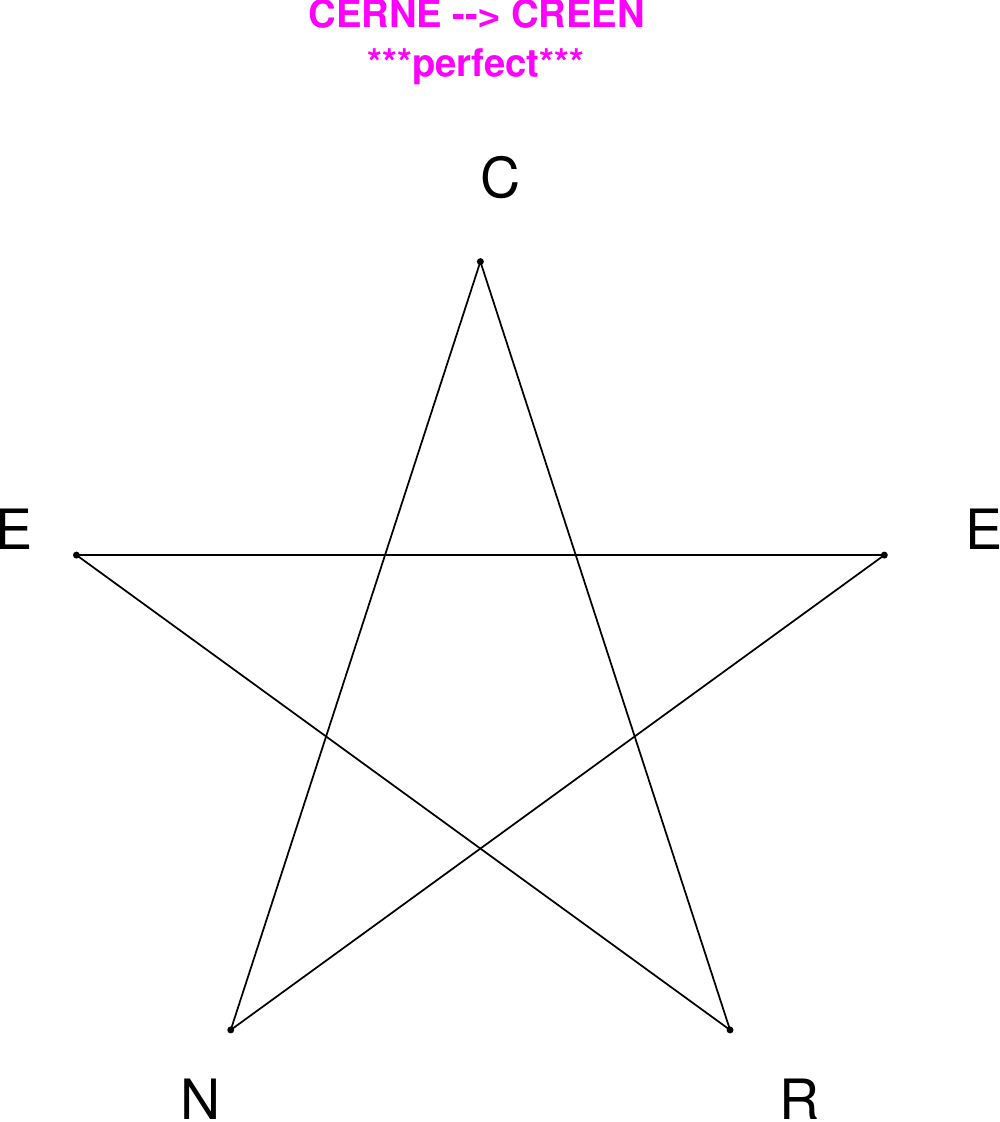}
\end{subfigure}
\hfill
\begin{subfigure}[T]{0.19\textwidth}
\centering
\includegraphics[width=\textwidth]{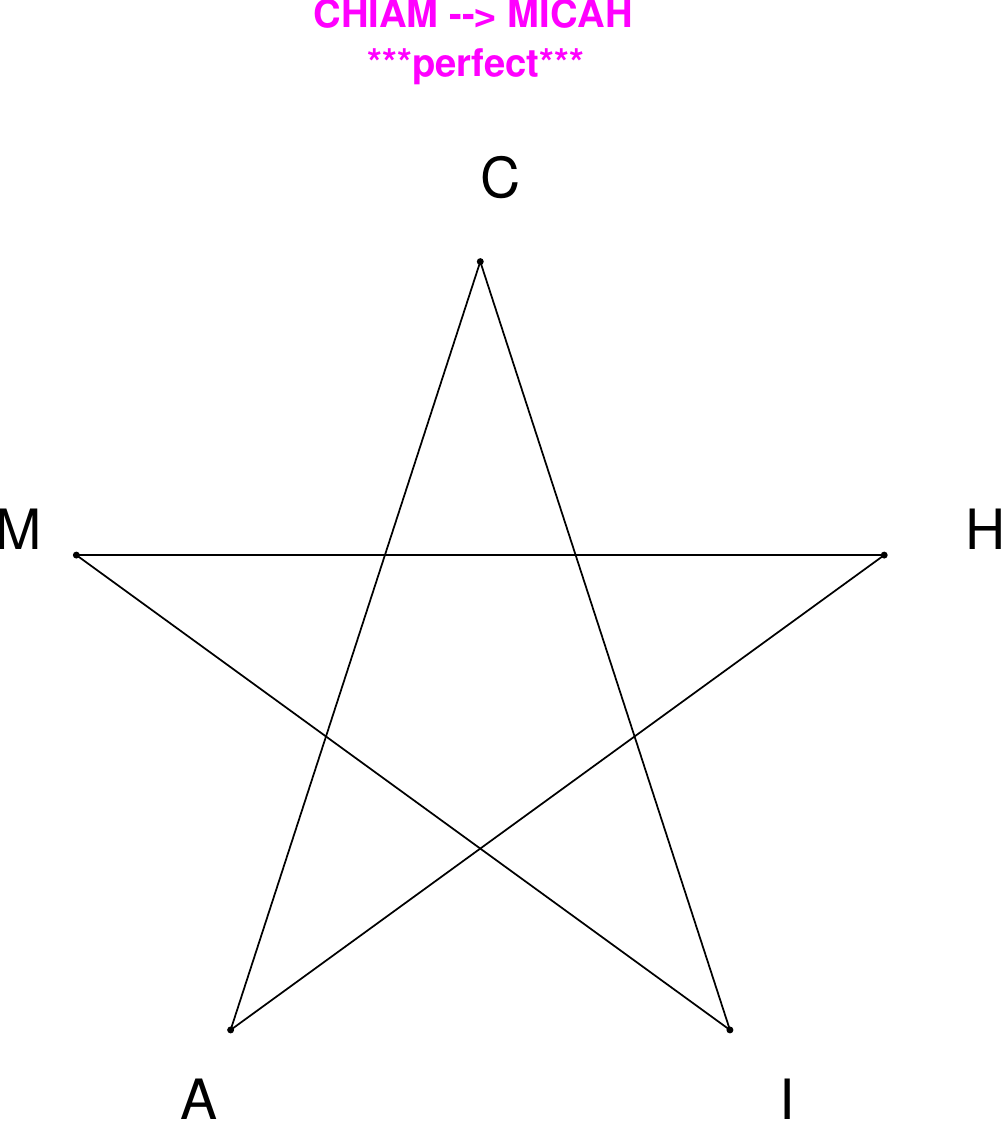}
\end{subfigure}
\hfill
\begin{subfigure}[T]{0.19\textwidth}
\centering
\includegraphics[width=\textwidth]{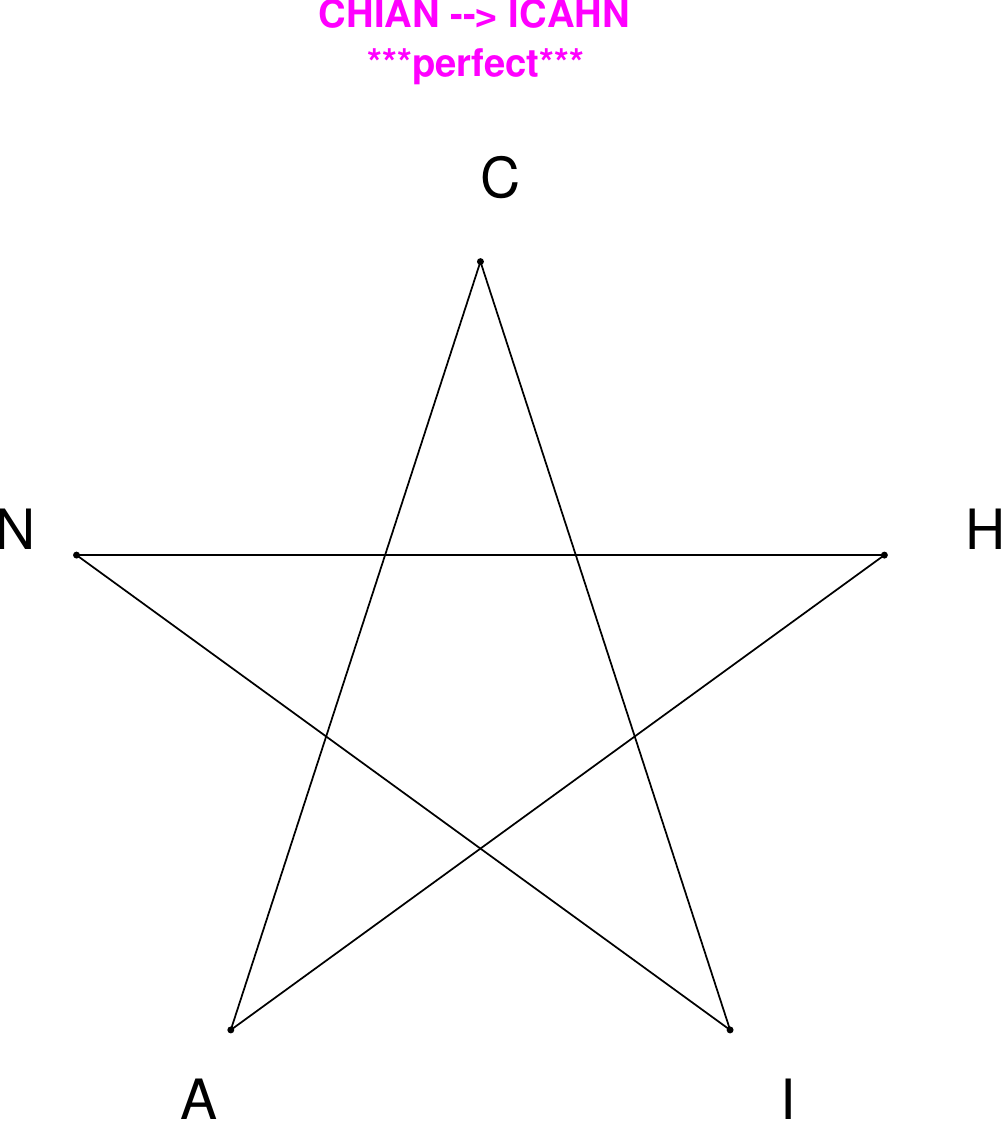}
\end{subfigure}
\hfill
\begin{subfigure}[T]{0.19\textwidth}
\centering
\includegraphics[width=\textwidth]{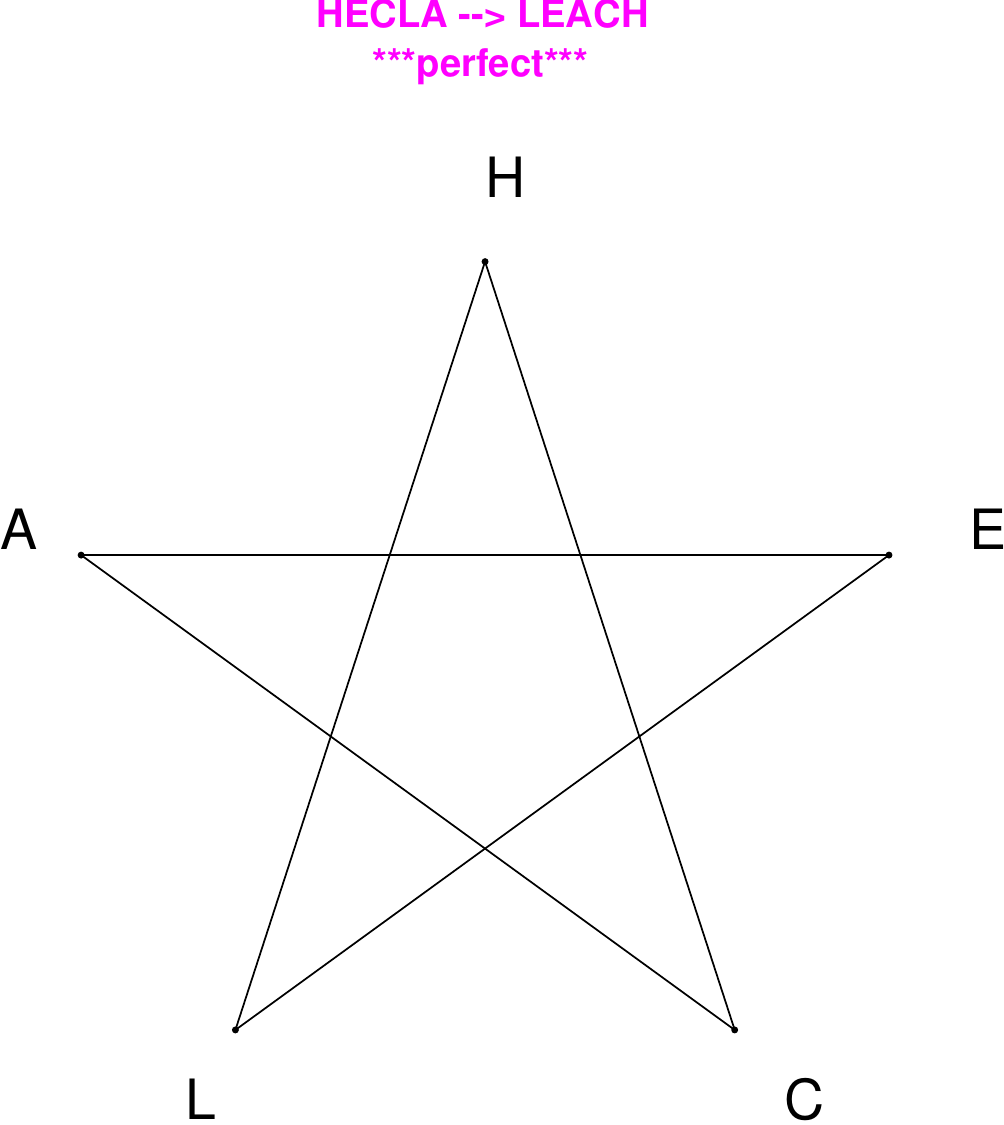}
\end{subfigure}
\hfill
\begin{subfigure}[T]{0.19\textwidth}
\centering
\includegraphics[width=\textwidth]{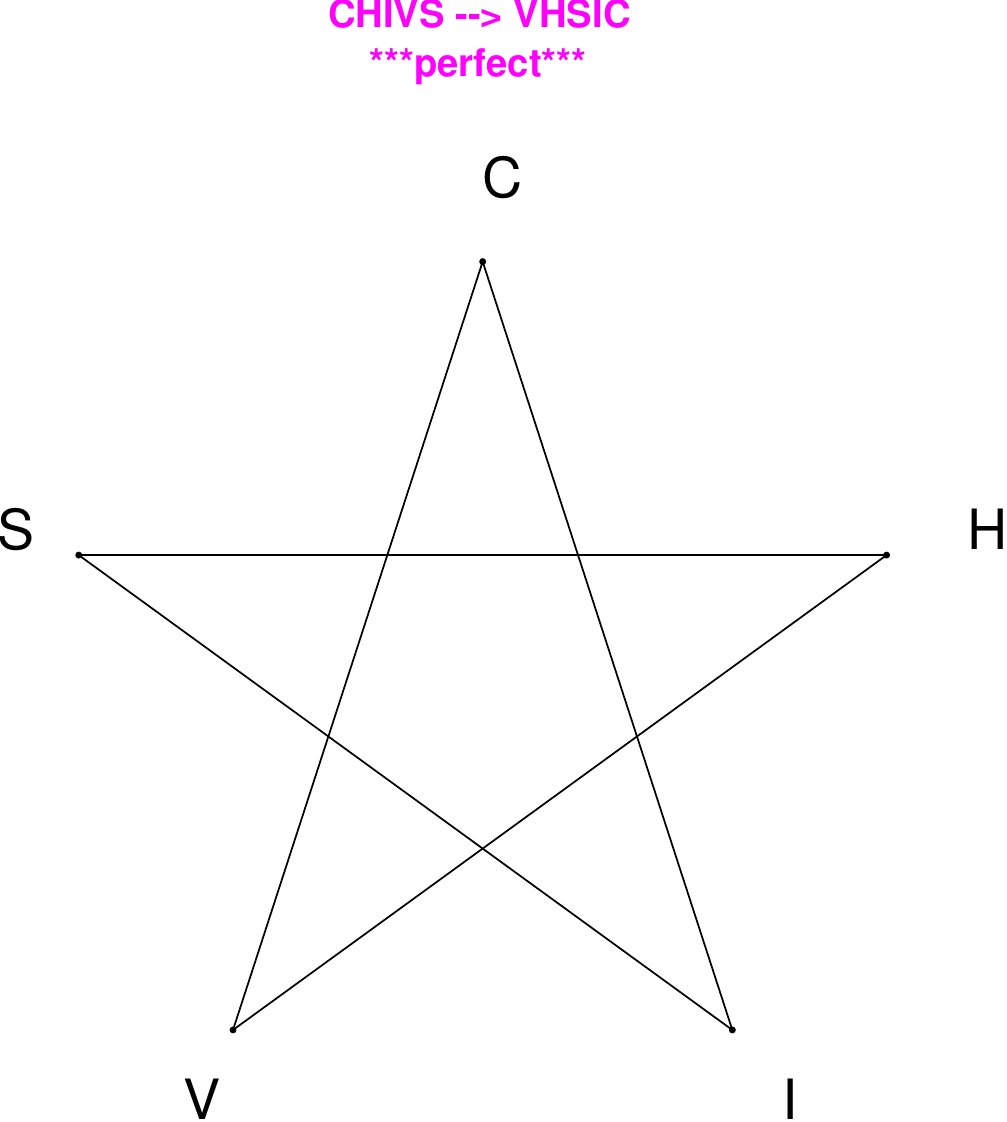}
\end{subfigure}
\end{figure}

\begin{figure}[H]
\centering
\begin{subfigure}[T]{0.19\textwidth}
\centering
\includegraphics[width=\textwidth]{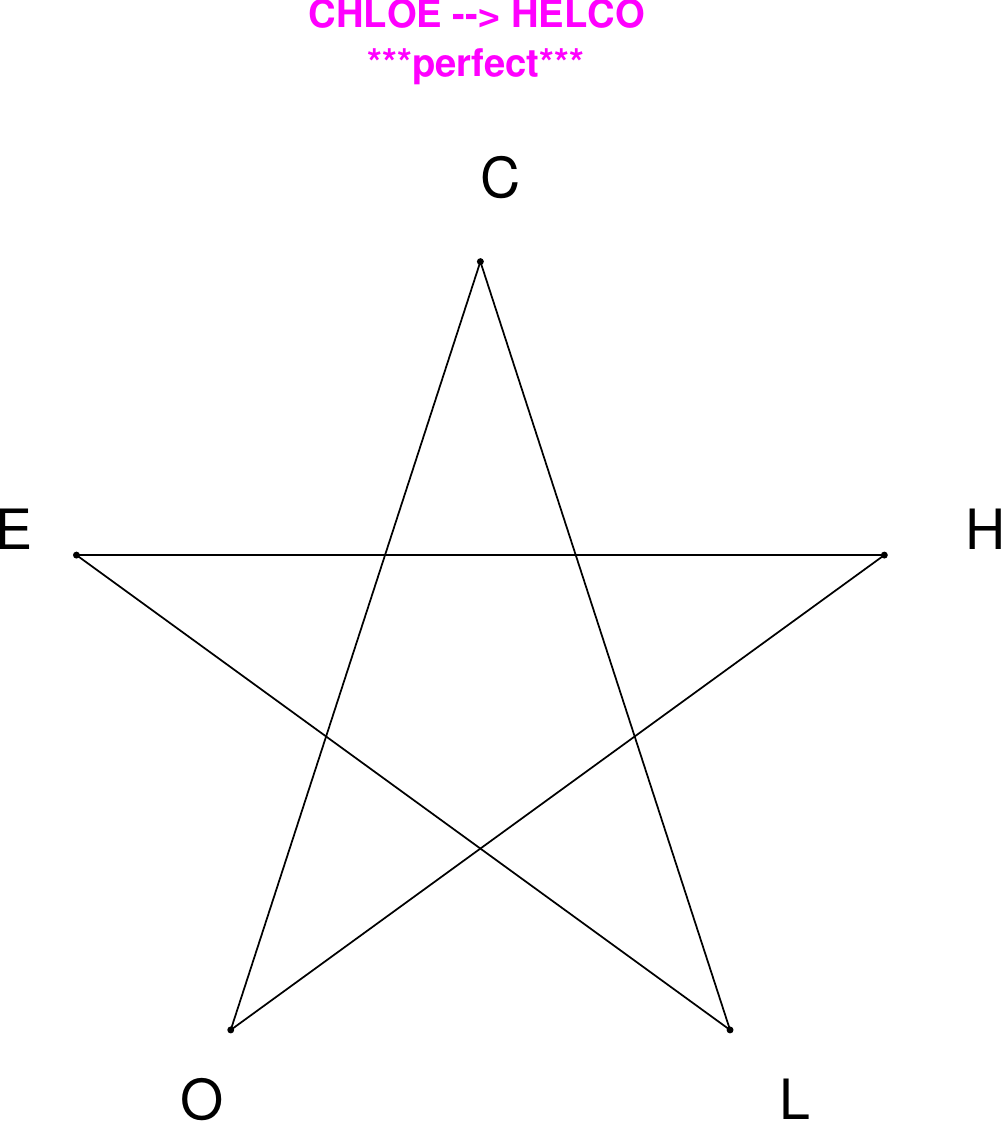}
\end{subfigure}
\hfill
\begin{subfigure}[T]{0.19\textwidth}
\centering
\includegraphics[width=\textwidth]{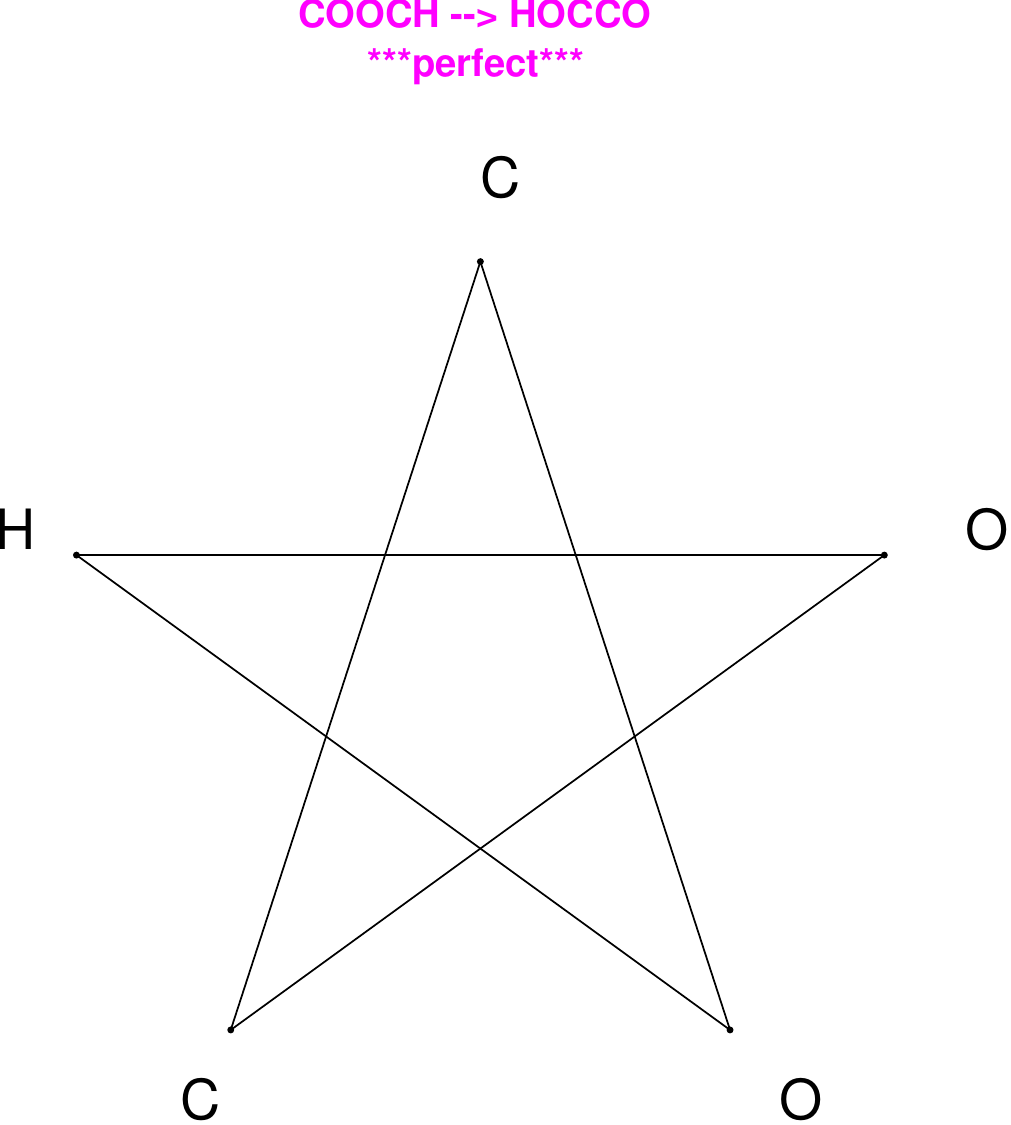}
\end{subfigure}
\hfill
\begin{subfigure}[T]{0.19\textwidth}
\centering
\includegraphics[width=\textwidth]{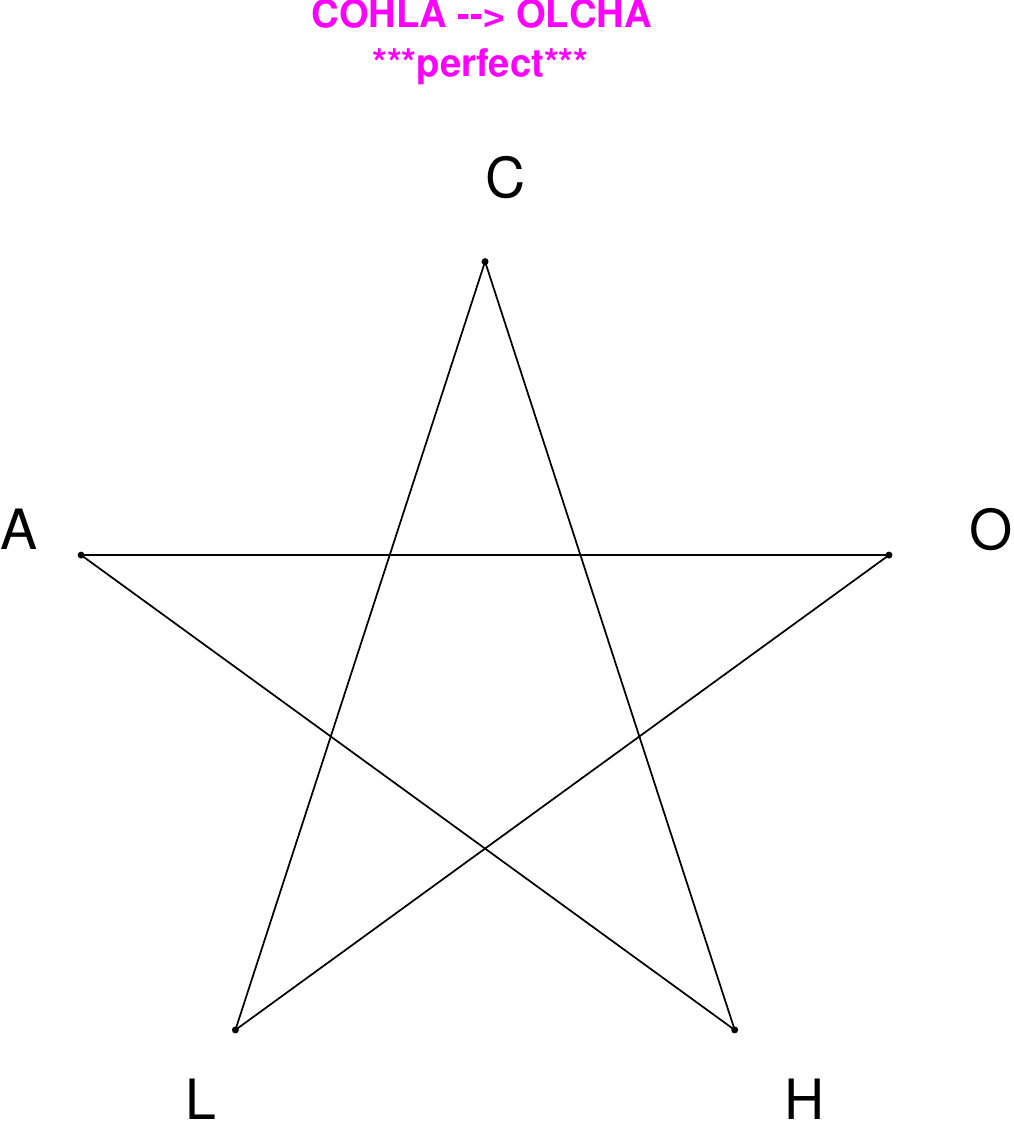}
\end{subfigure}
\hfill
\begin{subfigure}[T]{0.19\textwidth}
\centering
\includegraphics[width=\textwidth]{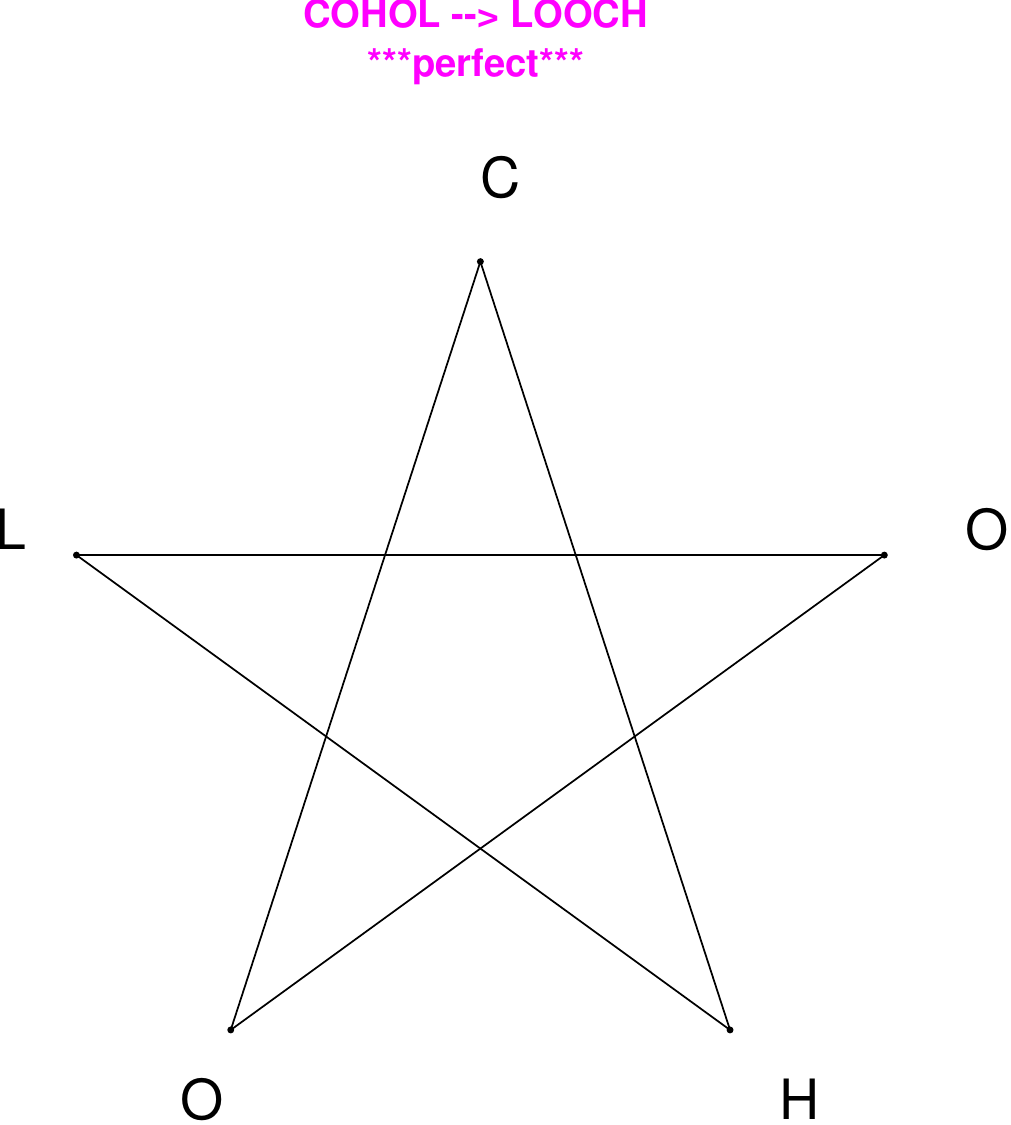}
\end{subfigure}
\hfill
\begin{subfigure}[T]{0.19\textwidth}
\centering
\includegraphics[width=\textwidth]{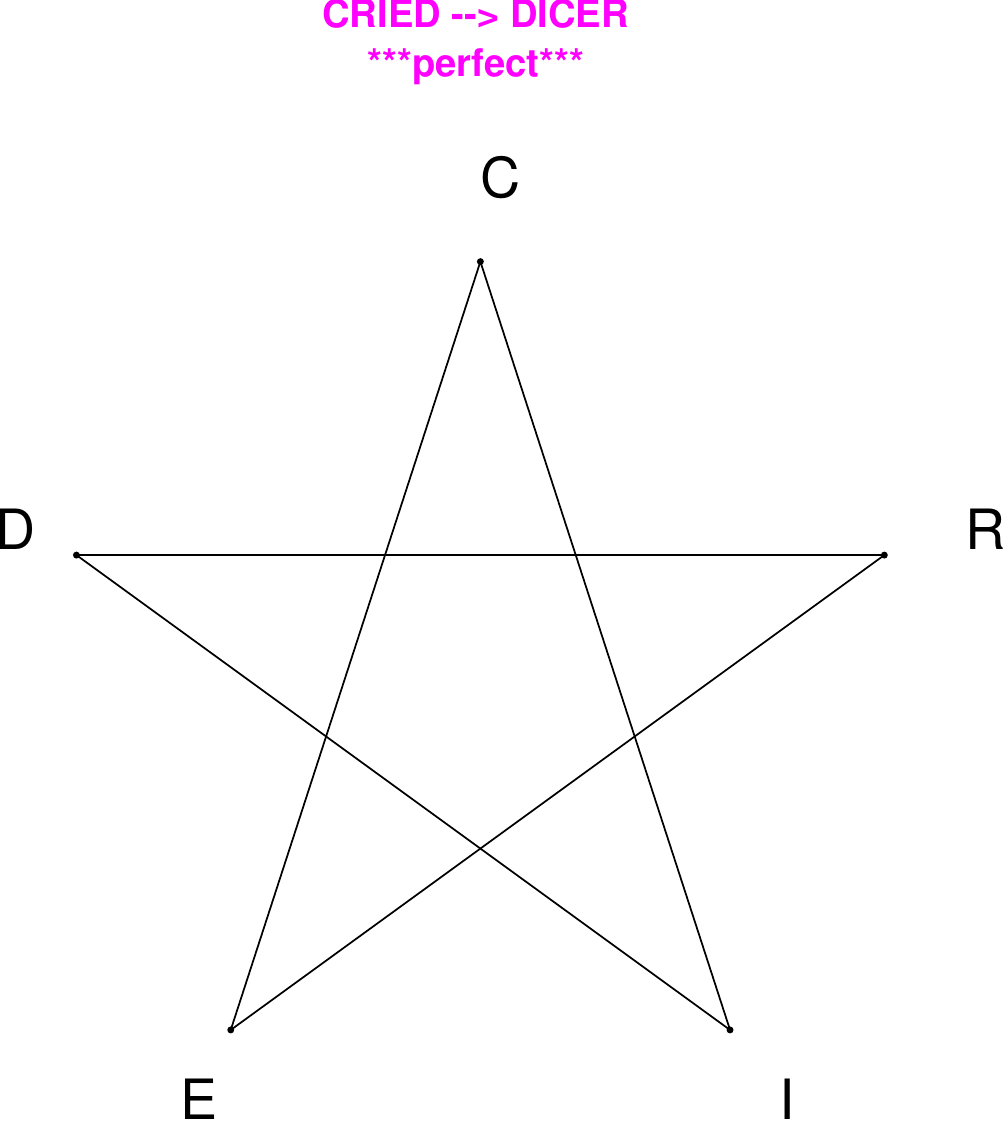}
\end{subfigure}
\end{figure}

\begin{figure}[H]
\centering
\begin{subfigure}[T]{0.19\textwidth}
\centering
\includegraphics[width=\textwidth]{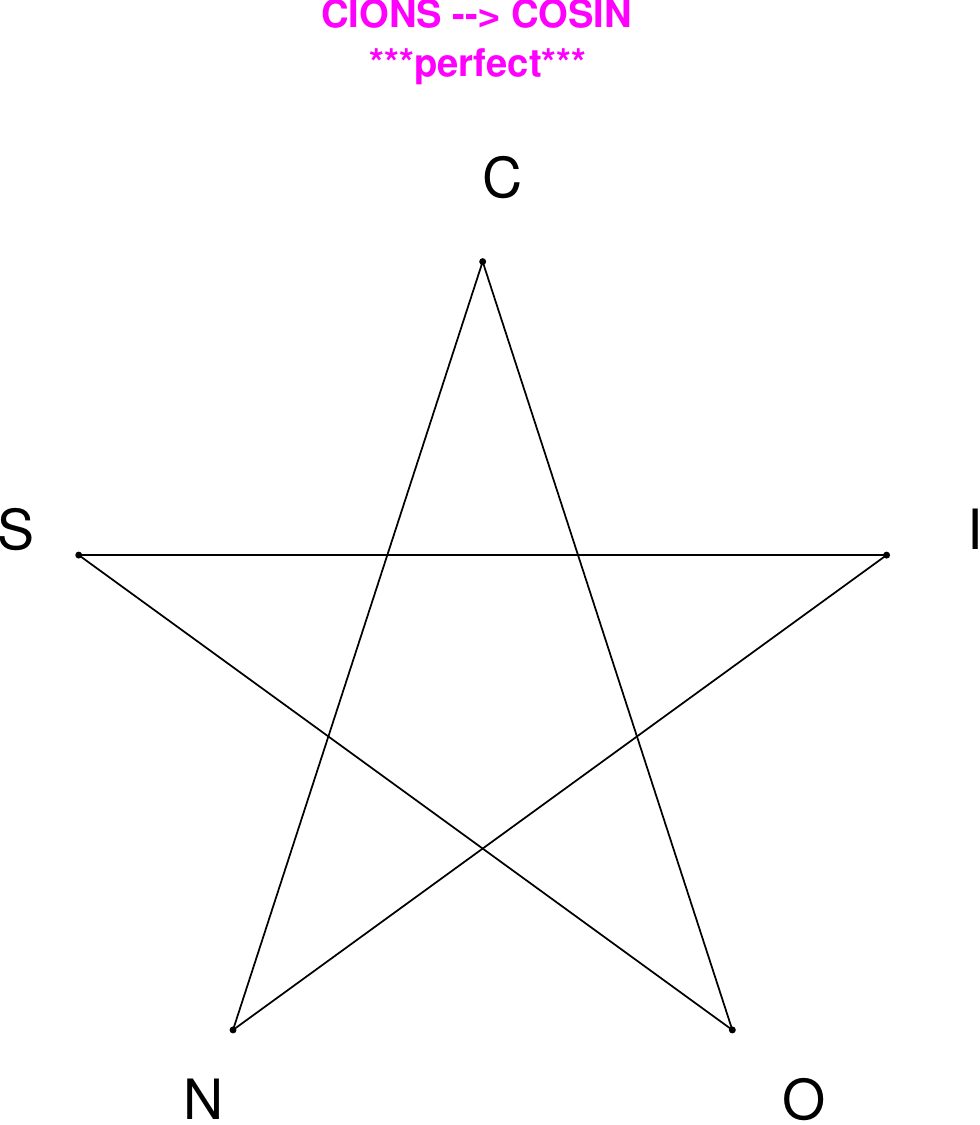}
\end{subfigure}
\hfill
\begin{subfigure}[T]{0.19\textwidth}
\centering
\includegraphics[width=\textwidth]{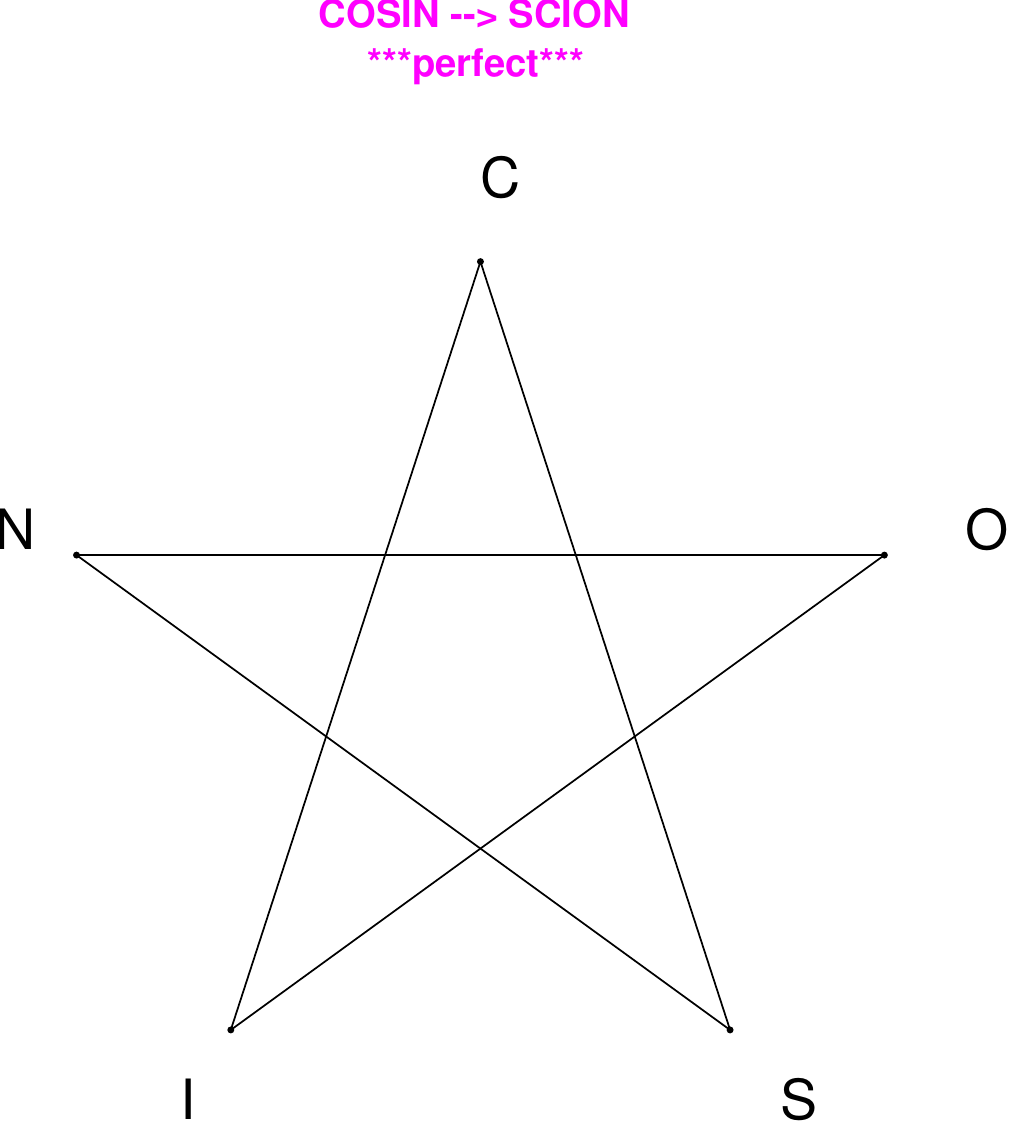}
\end{subfigure}
\hfill
\begin{subfigure}[T]{0.19\textwidth}
\centering
\includegraphics[width=\textwidth]{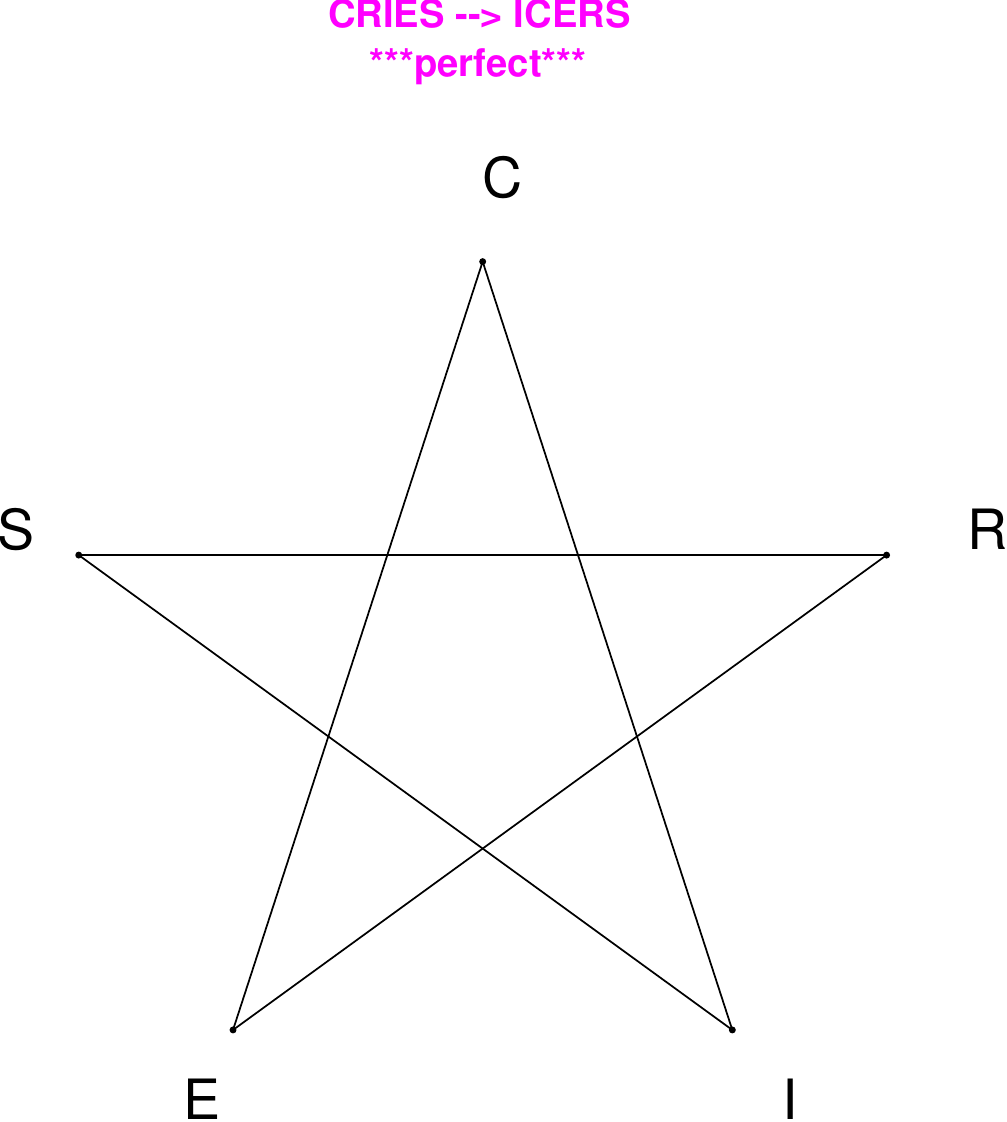}
\end{subfigure}
\hfill
\begin{subfigure}[T]{0.19\textwidth}
\centering
\includegraphics[width=\textwidth]{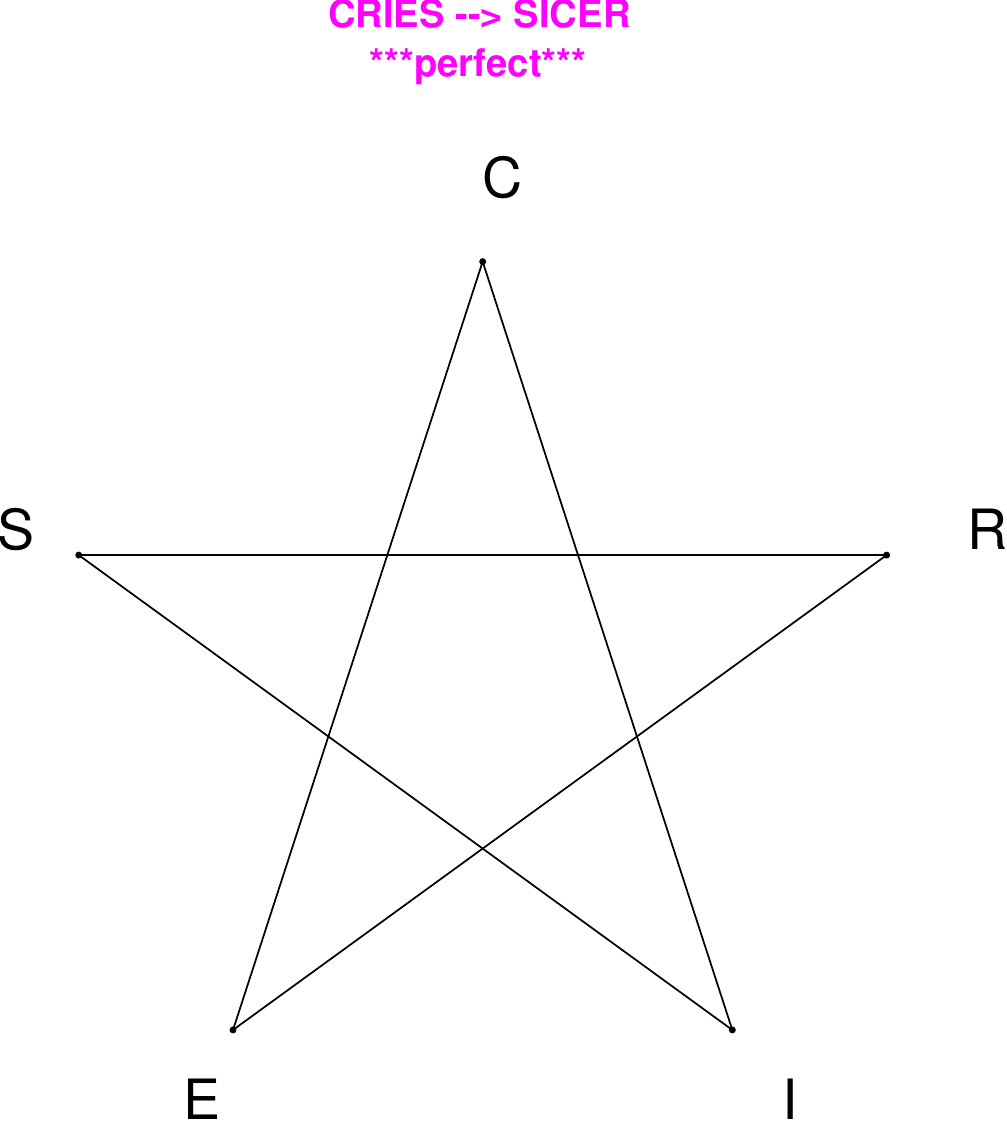}
\end{subfigure}
\hfill
\begin{subfigure}[T]{0.19\textwidth}
\centering
\includegraphics[width=\textwidth]{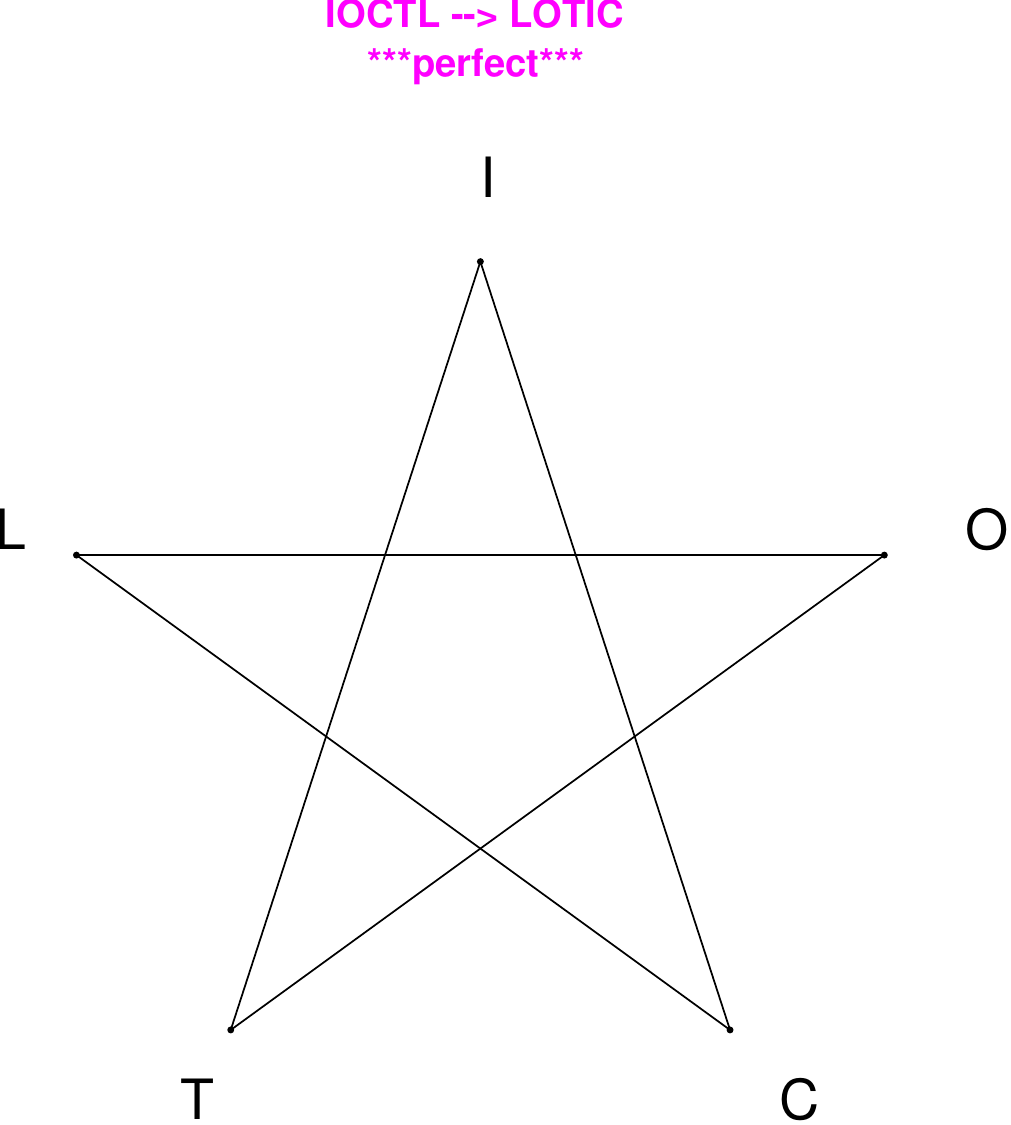}
\end{subfigure}
\end{figure}

\begin{figure}[H]
\centering
\begin{subfigure}[T]{0.19\textwidth}
\centering
\includegraphics[width=\textwidth]{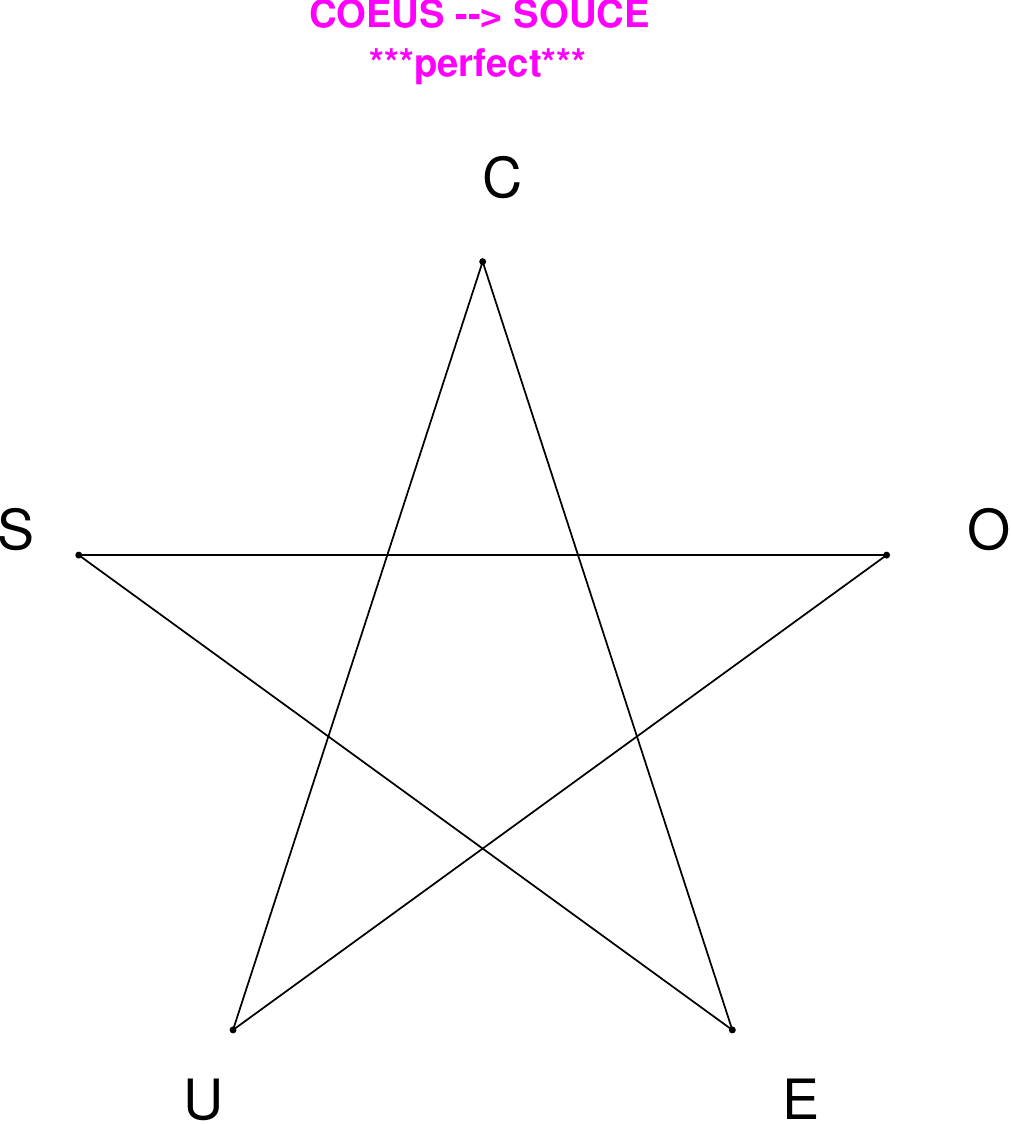}
\end{subfigure}
\hfill
\begin{subfigure}[T]{0.19\textwidth}
\centering
\includegraphics[width=\textwidth]{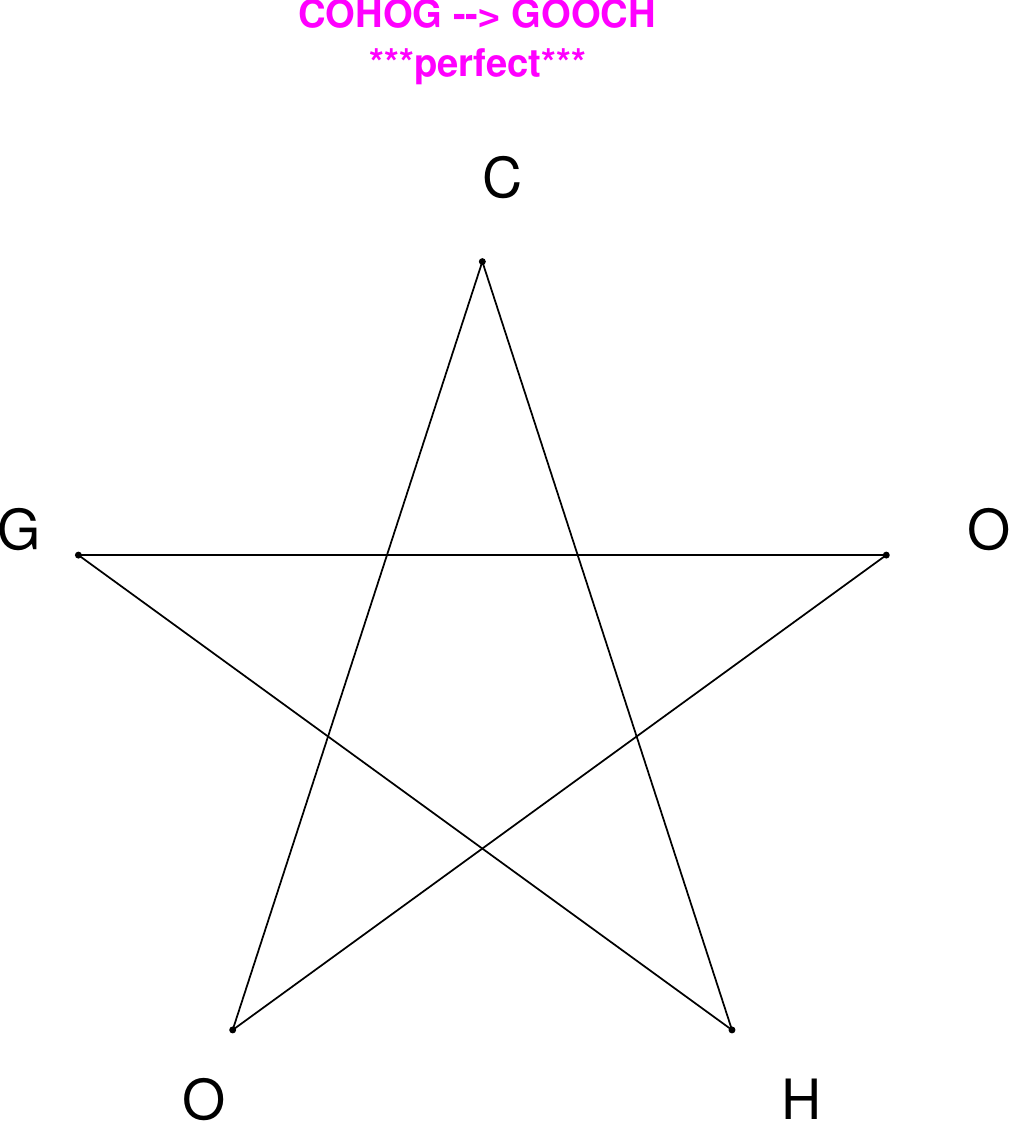}
\end{subfigure}
\hfill
\begin{subfigure}[T]{0.19\textwidth}
\centering
\includegraphics[width=\textwidth]{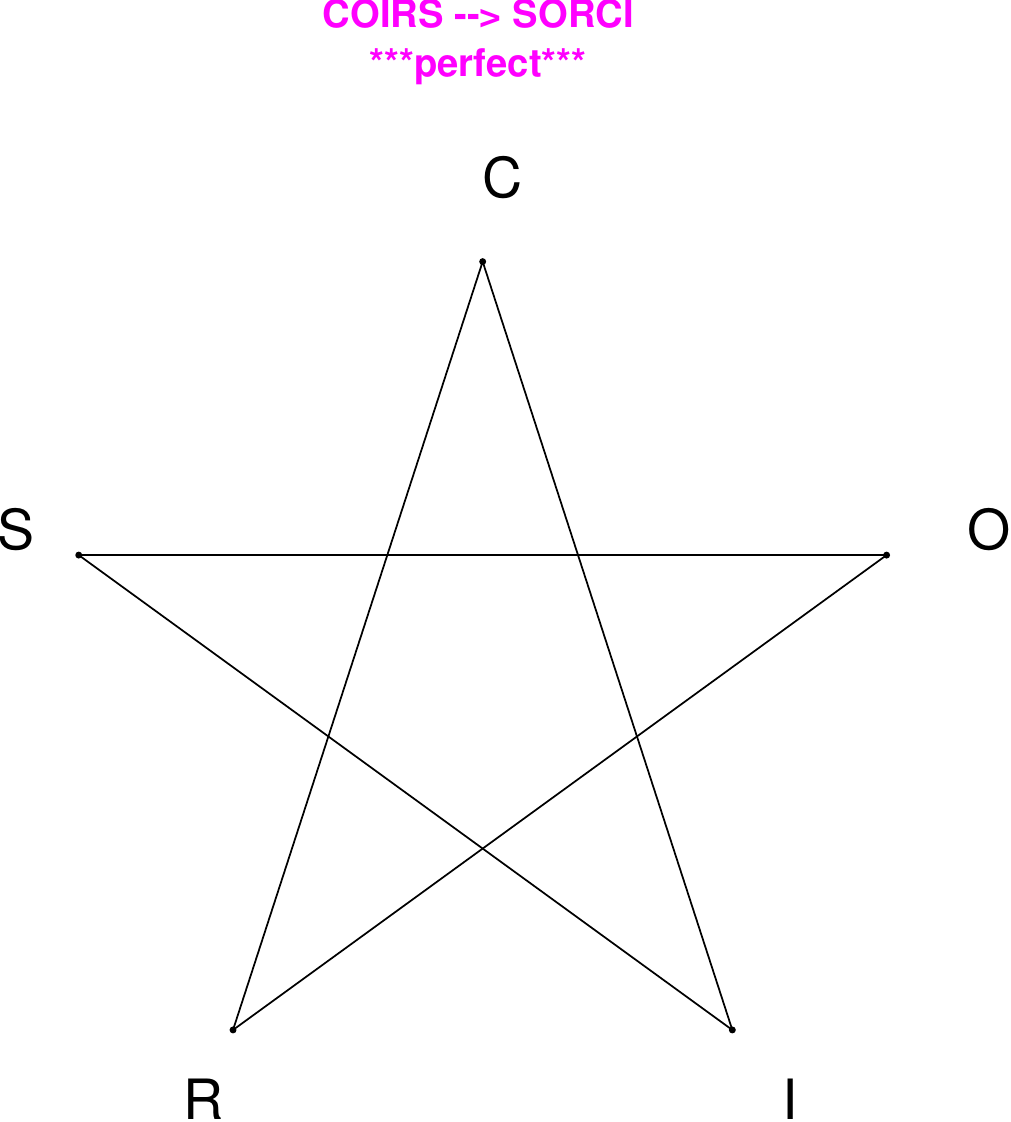}
\end{subfigure}
\hfill
\begin{subfigure}[T]{0.19\textwidth}
\centering
\includegraphics[width=\textwidth]{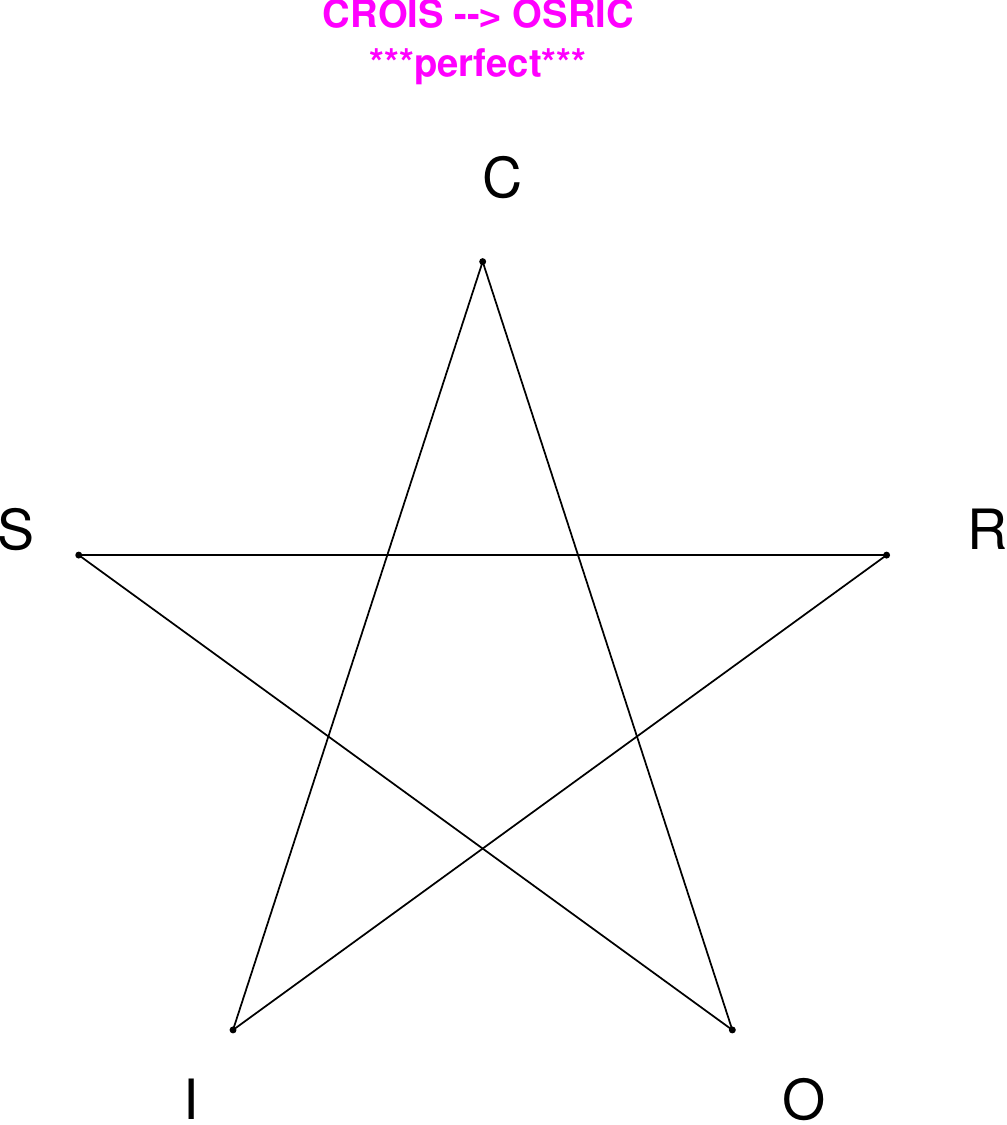}
\end{subfigure}
\hfill
\begin{subfigure}[T]{0.19\textwidth}
\centering
\includegraphics[width=\textwidth]{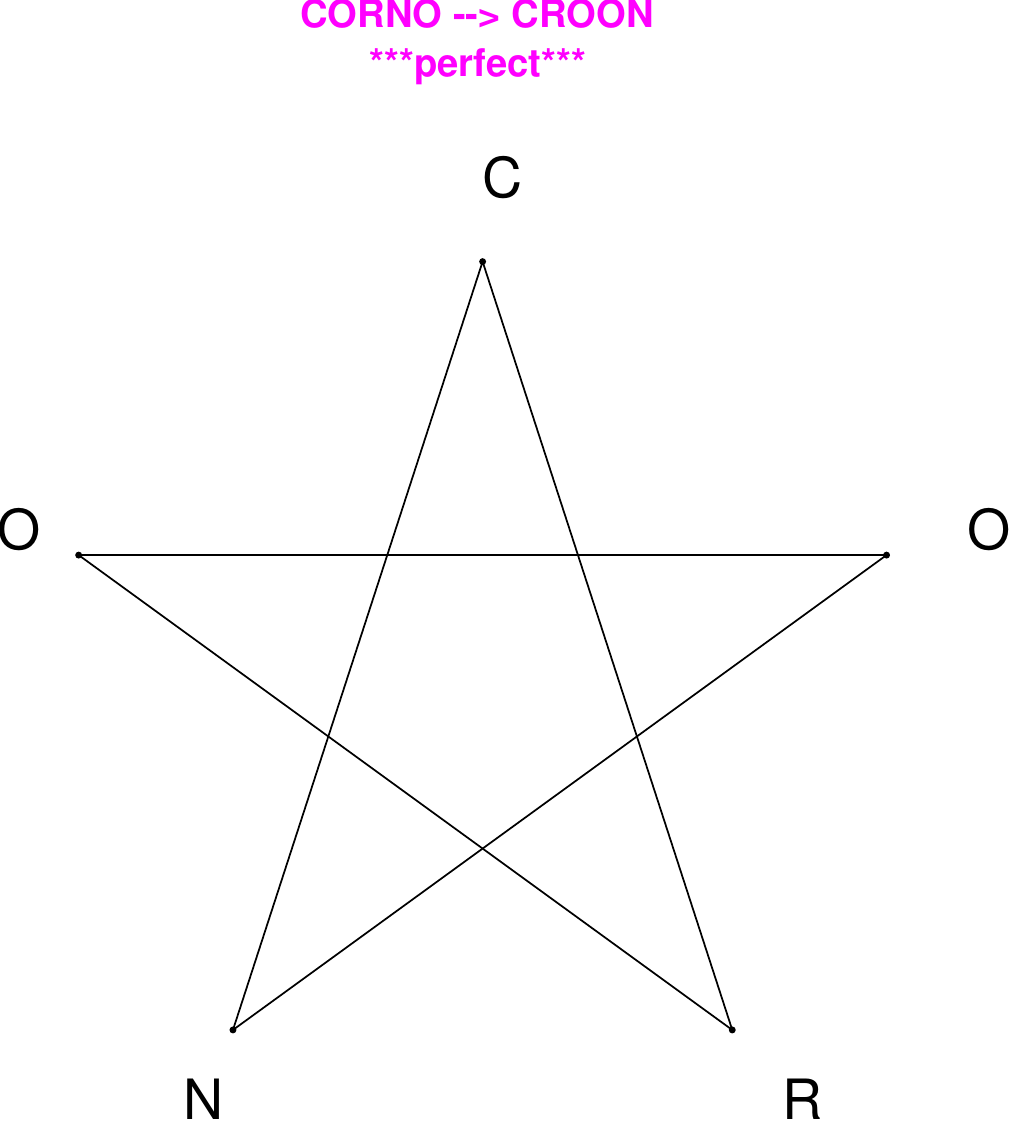}
\end{subfigure}
\end{figure}

\begin{figure}[H]
\centering
\begin{subfigure}[T]{0.19\textwidth}
\centering
\includegraphics[width=\textwidth]{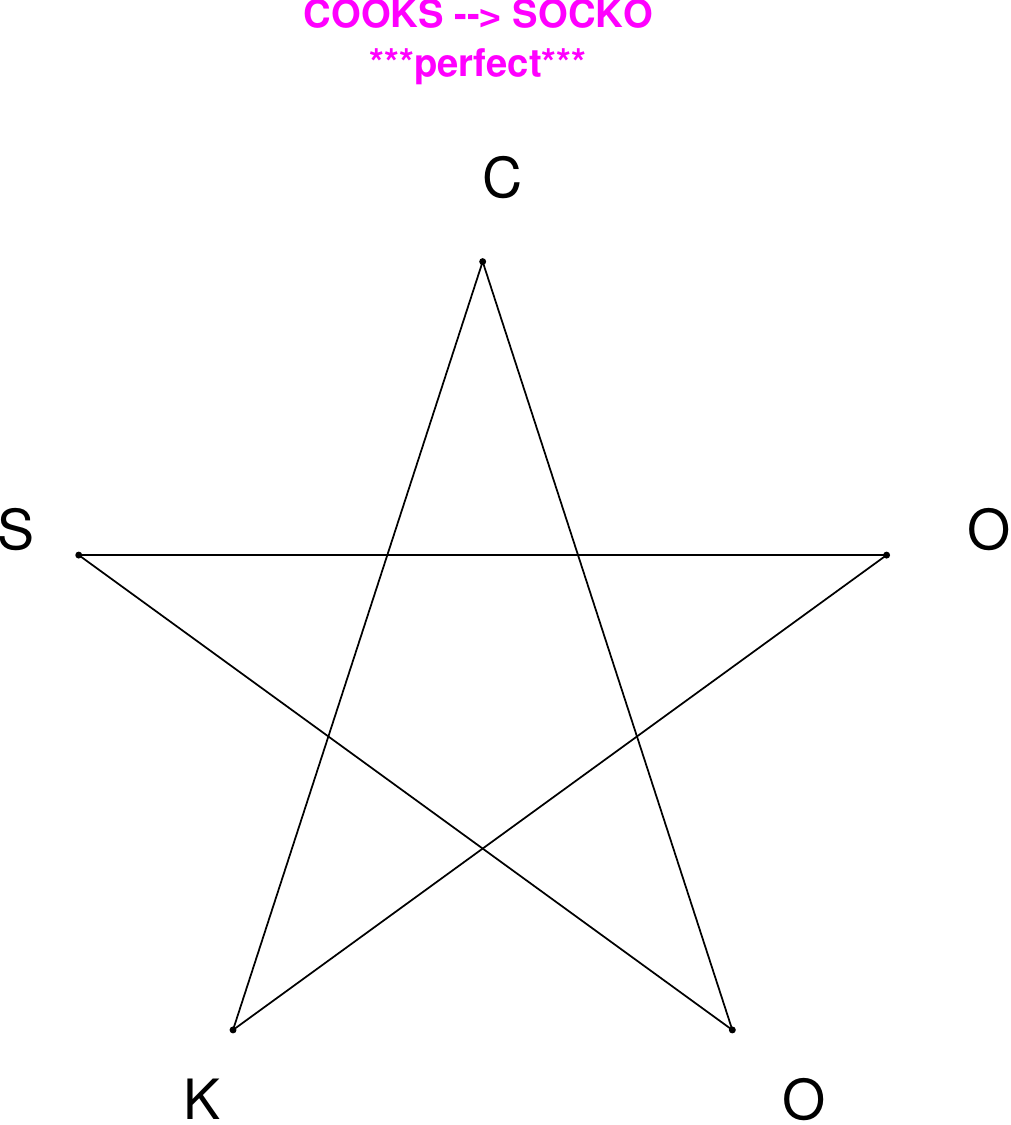}
\end{subfigure}
\hfill
\begin{subfigure}[T]{0.19\textwidth}
\centering
\includegraphics[width=\textwidth]{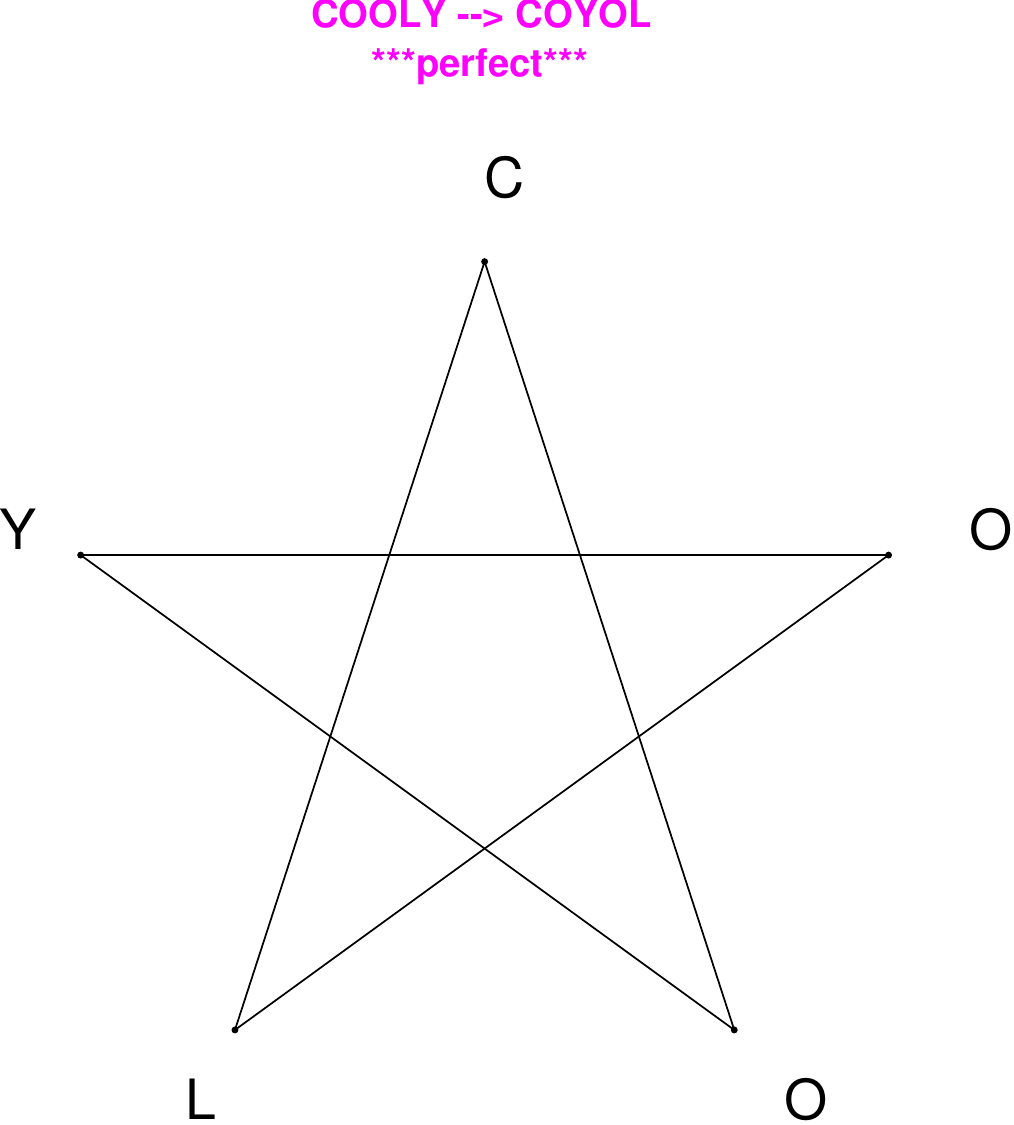}
\end{subfigure}
\hfill
\begin{subfigure}[T]{0.19\textwidth}
\centering
\includegraphics[width=\textwidth]{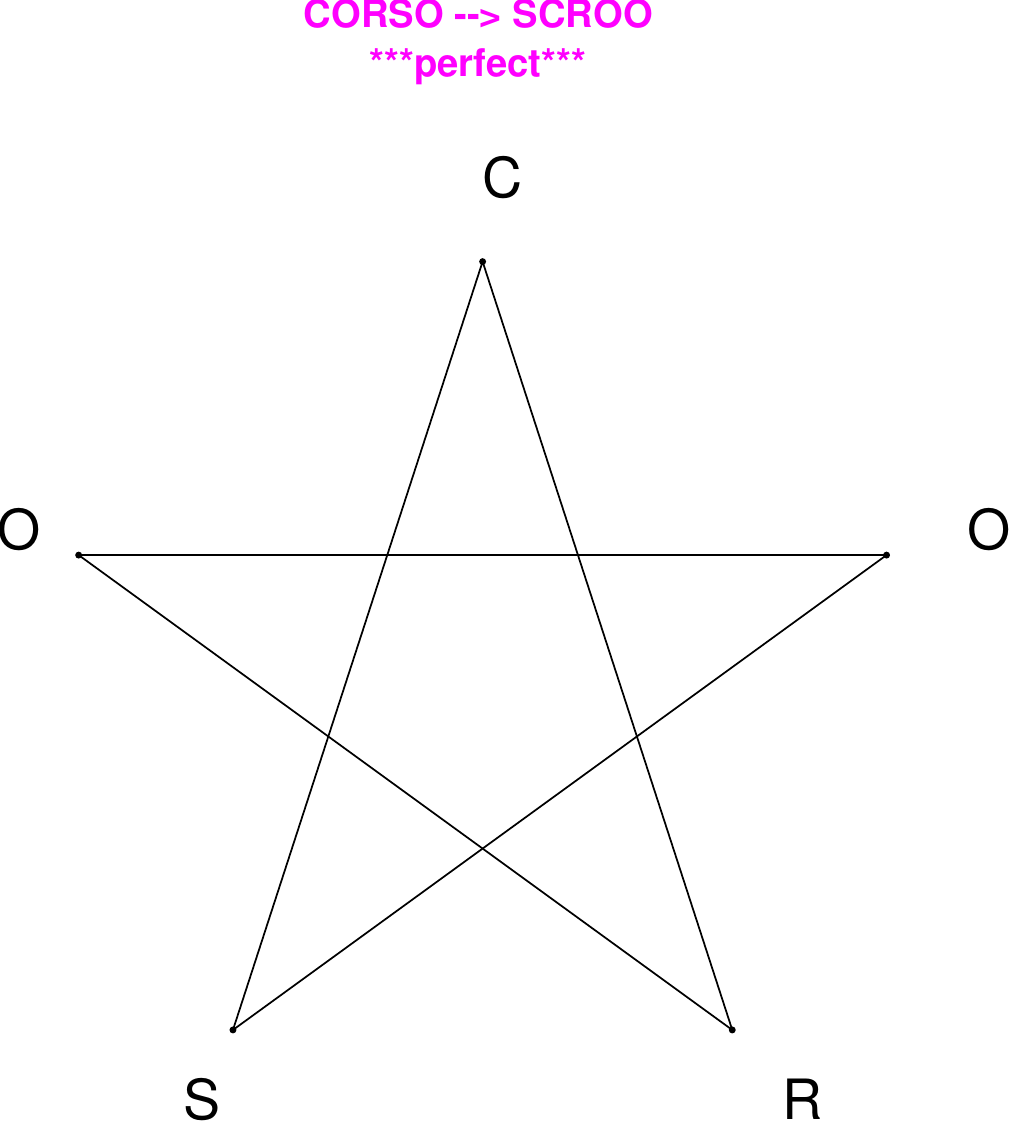}
\end{subfigure}
\hfill
\begin{subfigure}[T]{0.19\textwidth}
\centering
\includegraphics[width=\textwidth]{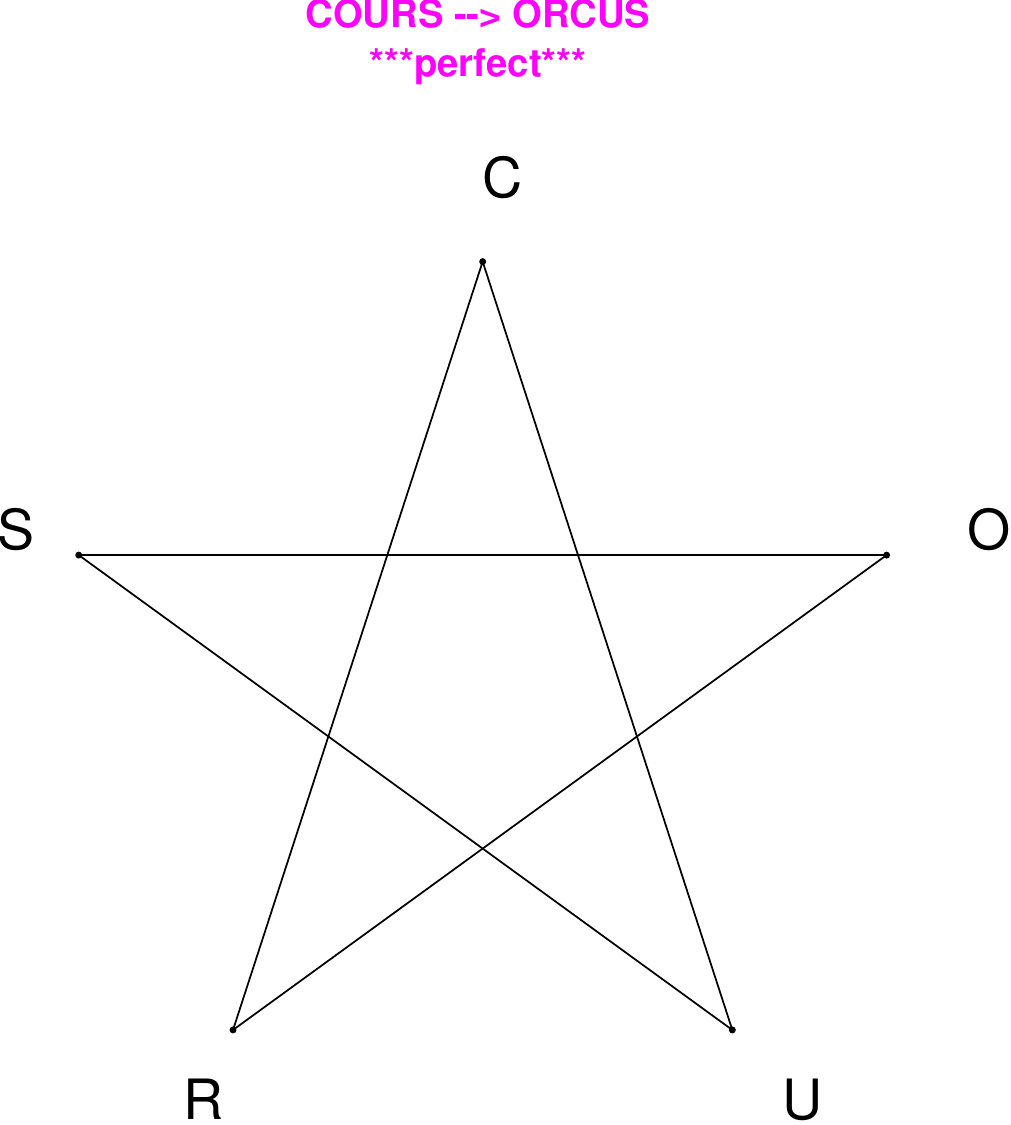}
\end{subfigure}
\hfill
\begin{subfigure}[T]{0.19\textwidth}
\centering
\includegraphics[width=\textwidth]{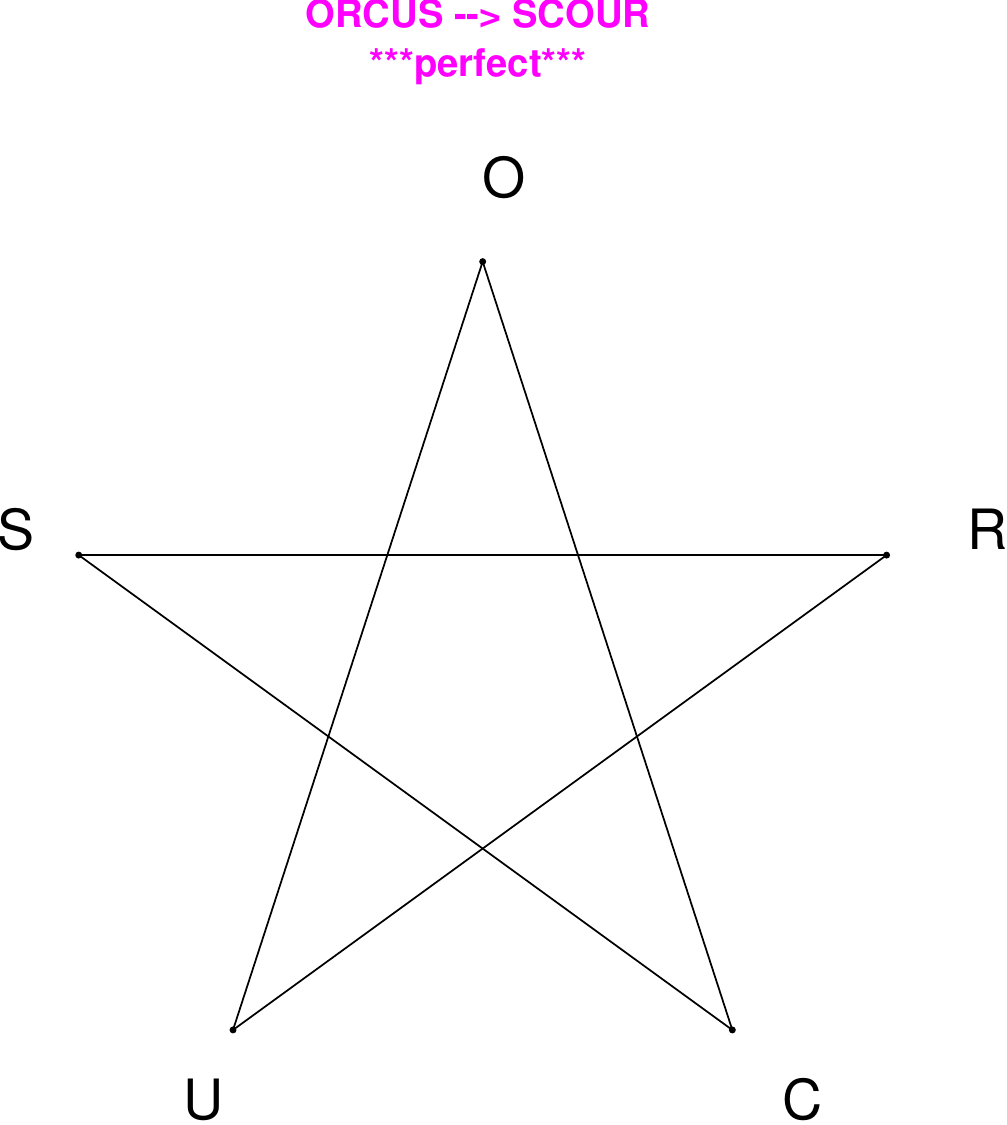}
\end{subfigure}
\end{figure}

\begin{figure}[H]
\centering
\begin{subfigure}[T]{0.19\textwidth}
\centering
\includegraphics[width=\textwidth]{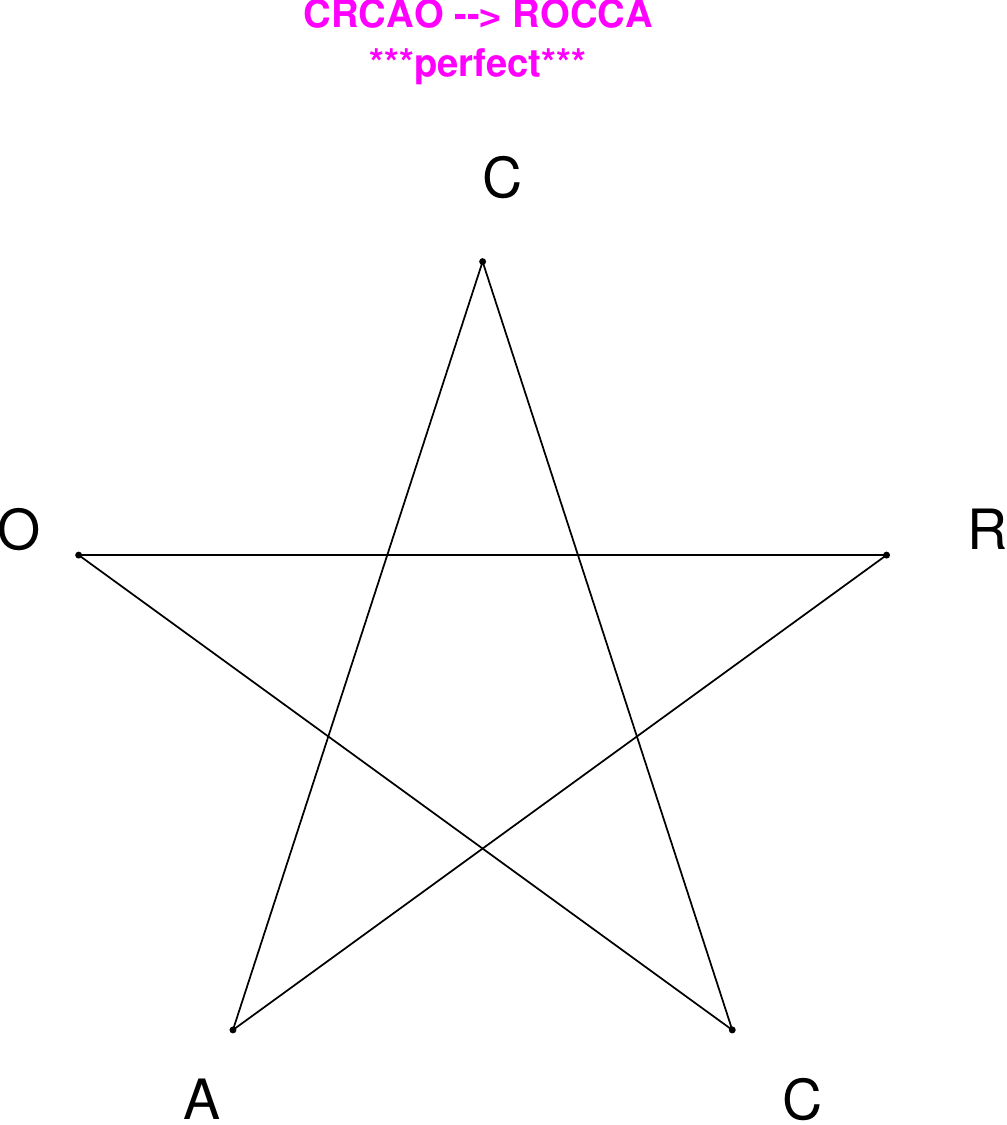}
\end{subfigure}
\hfill
\begin{subfigure}[T]{0.19\textwidth}
\centering
\includegraphics[width=\textwidth]{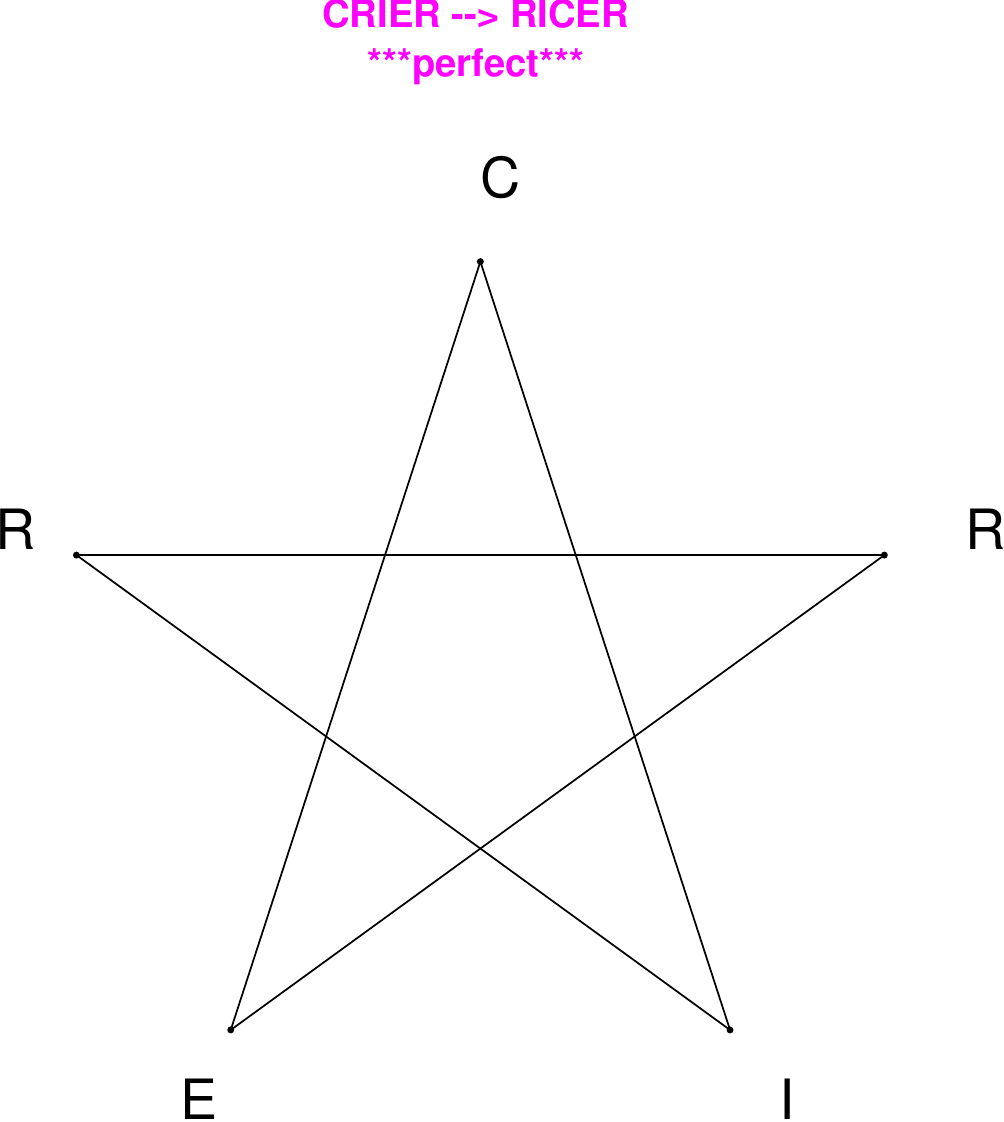}
\end{subfigure}
\hfill
\begin{subfigure}[T]{0.19\textwidth}
\centering
\includegraphics[width=\textwidth]{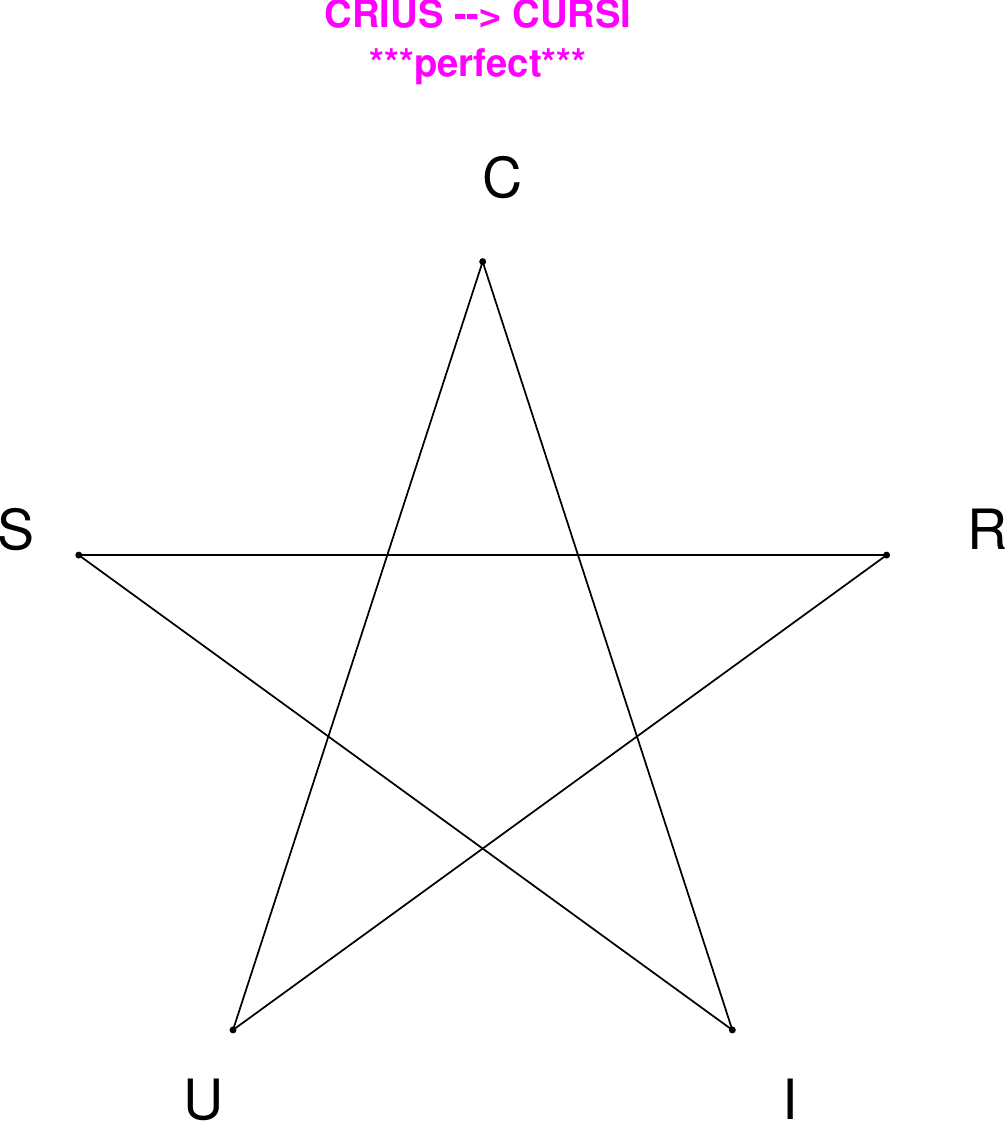}
\end{subfigure}
\hfill
\begin{subfigure}[T]{0.19\textwidth}
\centering
\includegraphics[width=\textwidth]{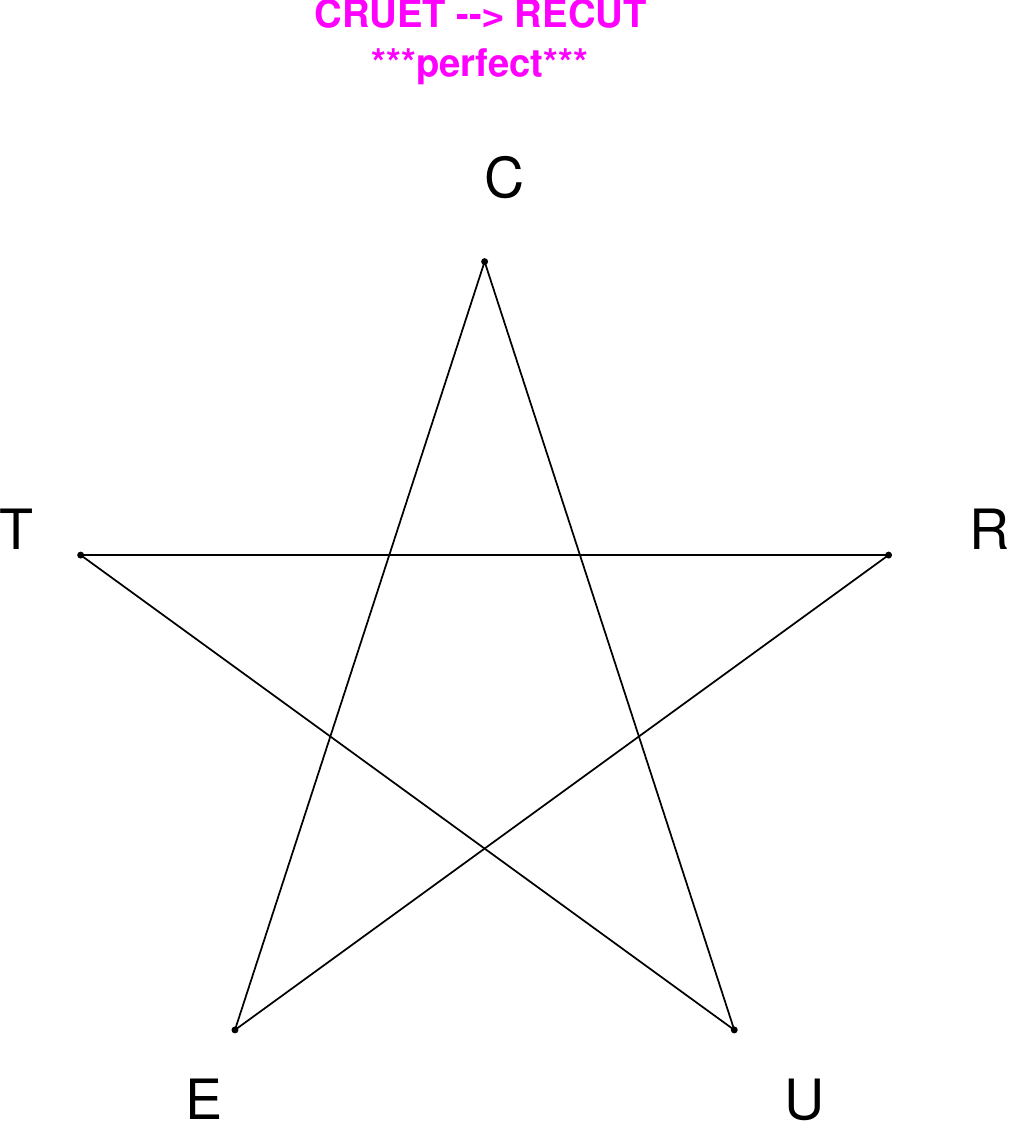}
\end{subfigure}
\hfill
\begin{subfigure}[T]{0.19\textwidth}
\centering
\includegraphics[width=\textwidth]{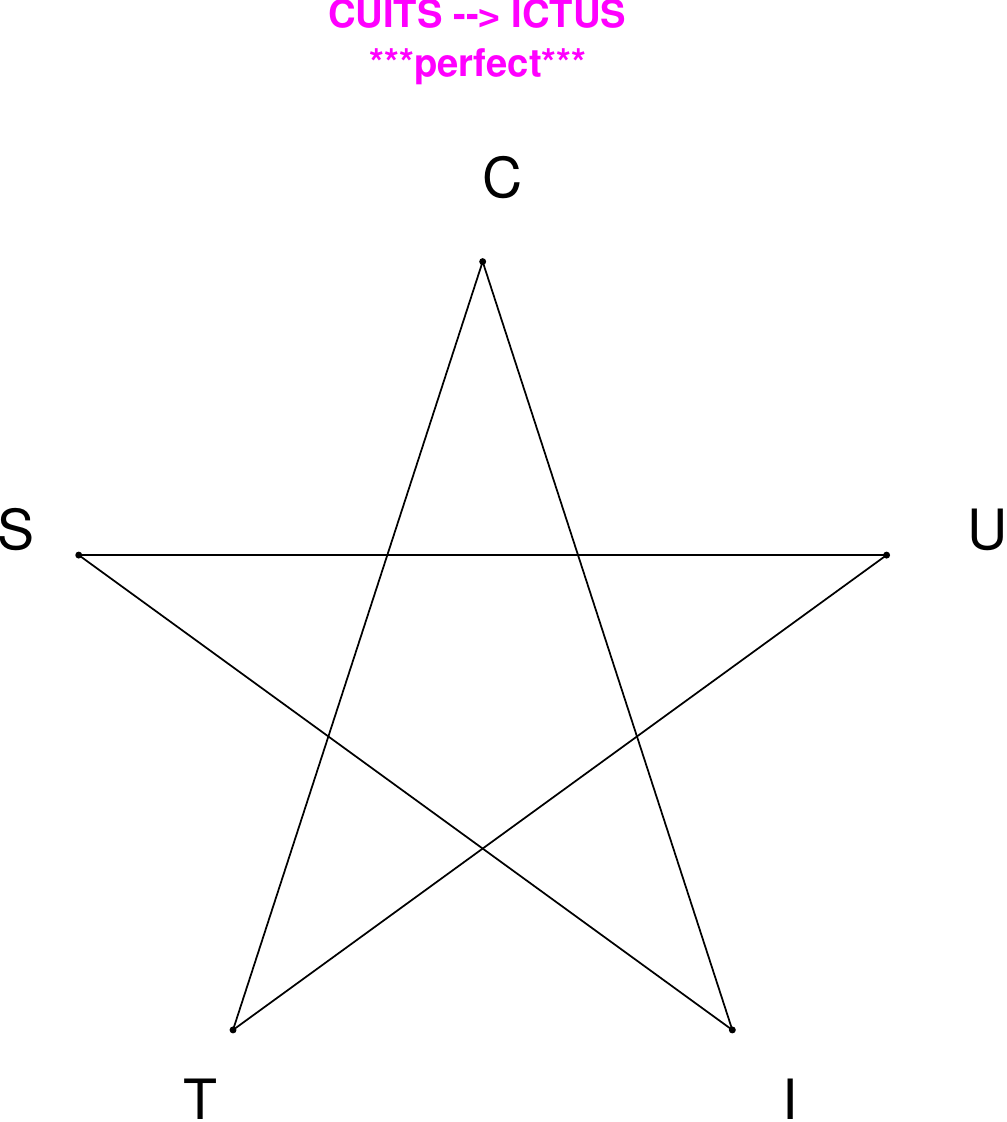}
\end{subfigure}
\end{figure}

\begin{figure}[H]
\centering
\begin{subfigure}[T]{0.19\textwidth}
\centering
\includegraphics[width=\textwidth]{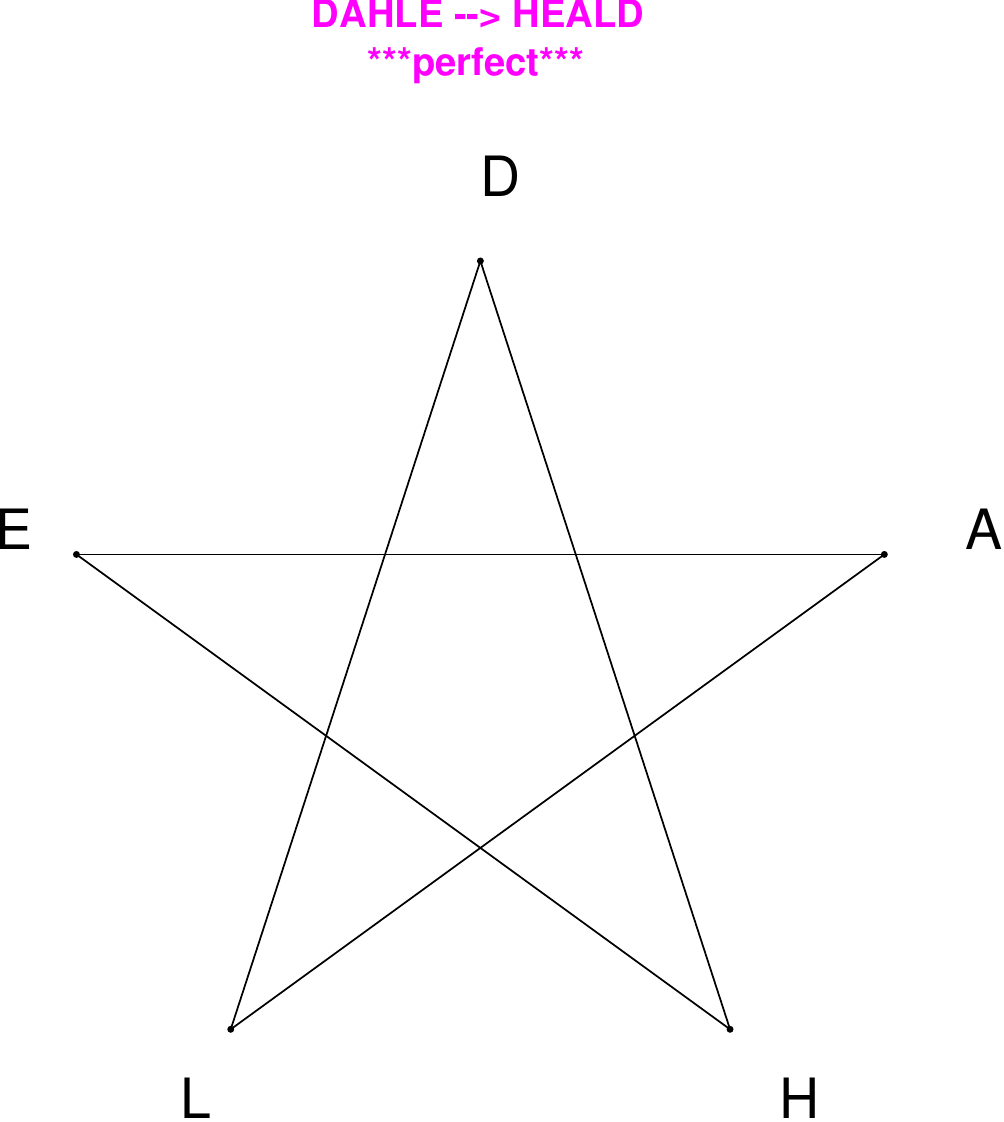}
\end{subfigure}
\hfill
\begin{subfigure}[T]{0.19\textwidth}
\centering
\includegraphics[width=\textwidth]{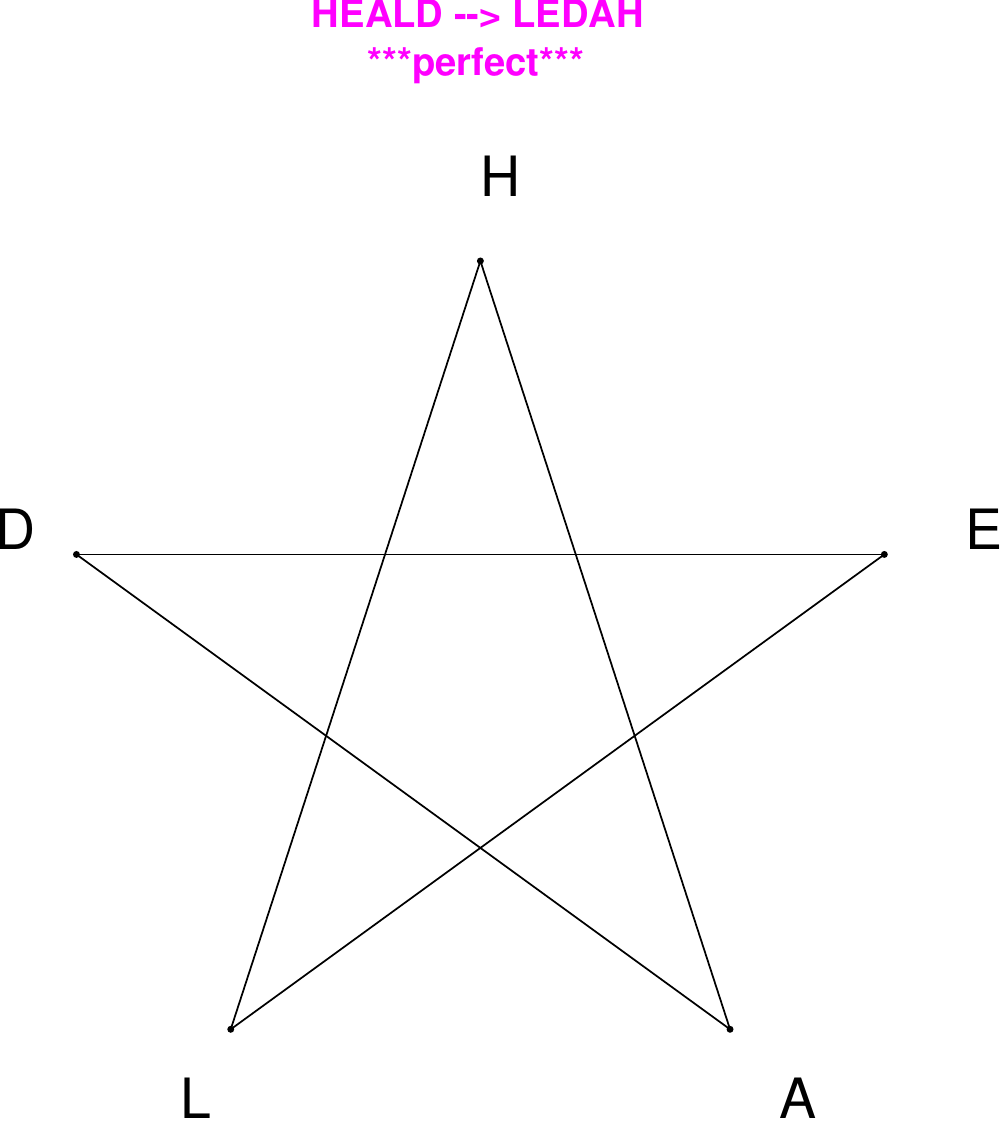}
\end{subfigure}
\hfill
\begin{subfigure}[T]{0.19\textwidth}
\centering
\includegraphics[width=\textwidth]{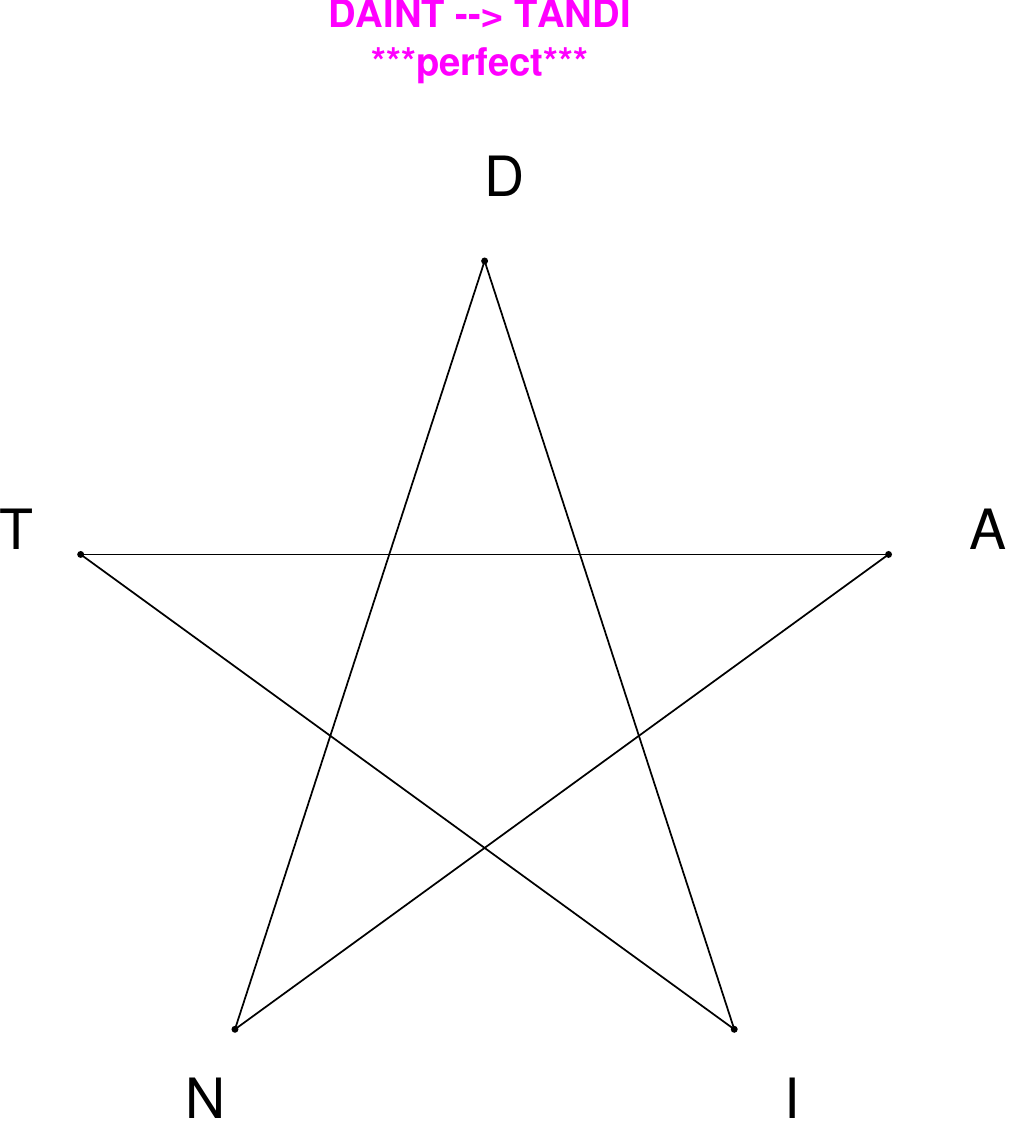}
\end{subfigure}
\hfill
\begin{subfigure}[T]{0.19\textwidth}
\centering
\includegraphics[width=\textwidth]{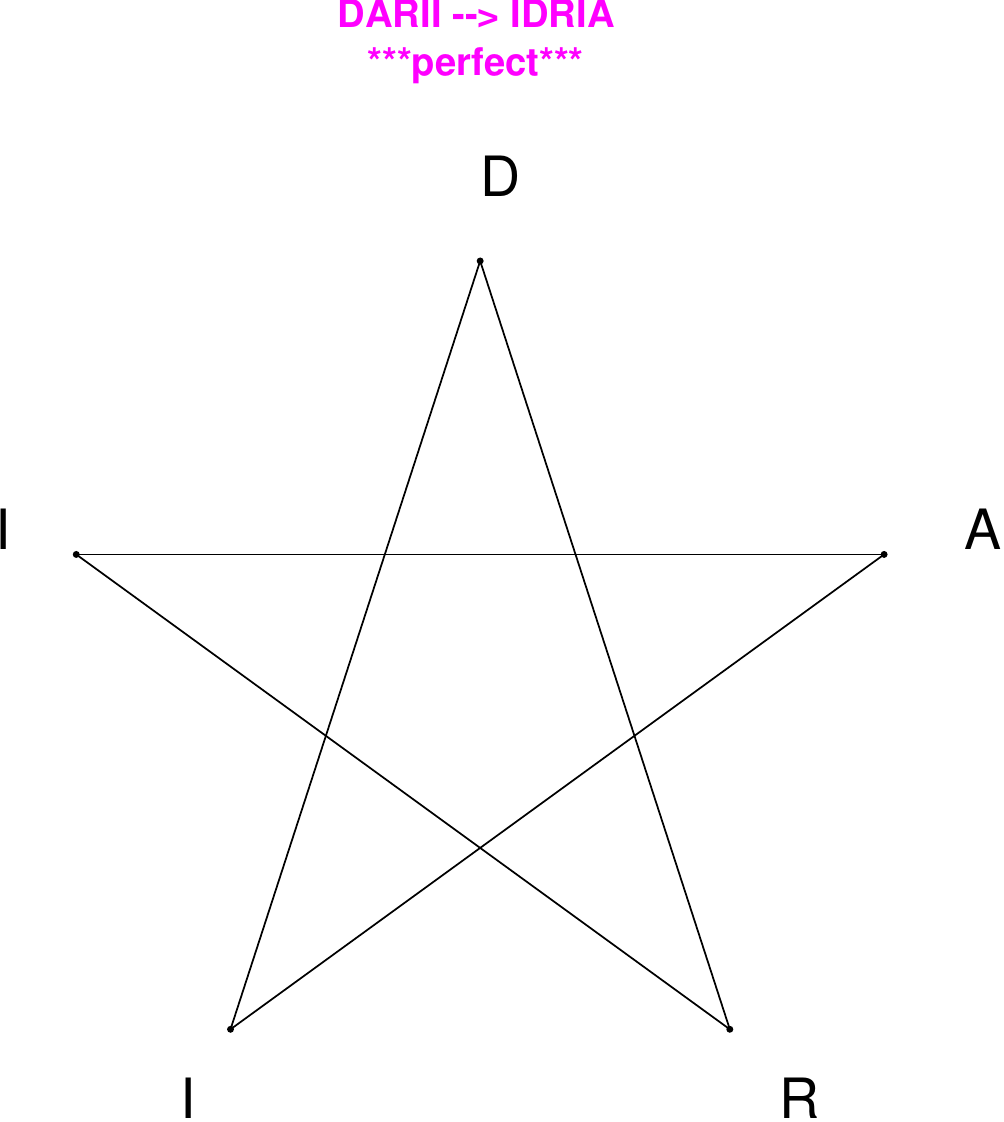}
\end{subfigure}
\hfill
\begin{subfigure}[T]{0.19\textwidth}
\centering
\includegraphics[width=\textwidth]{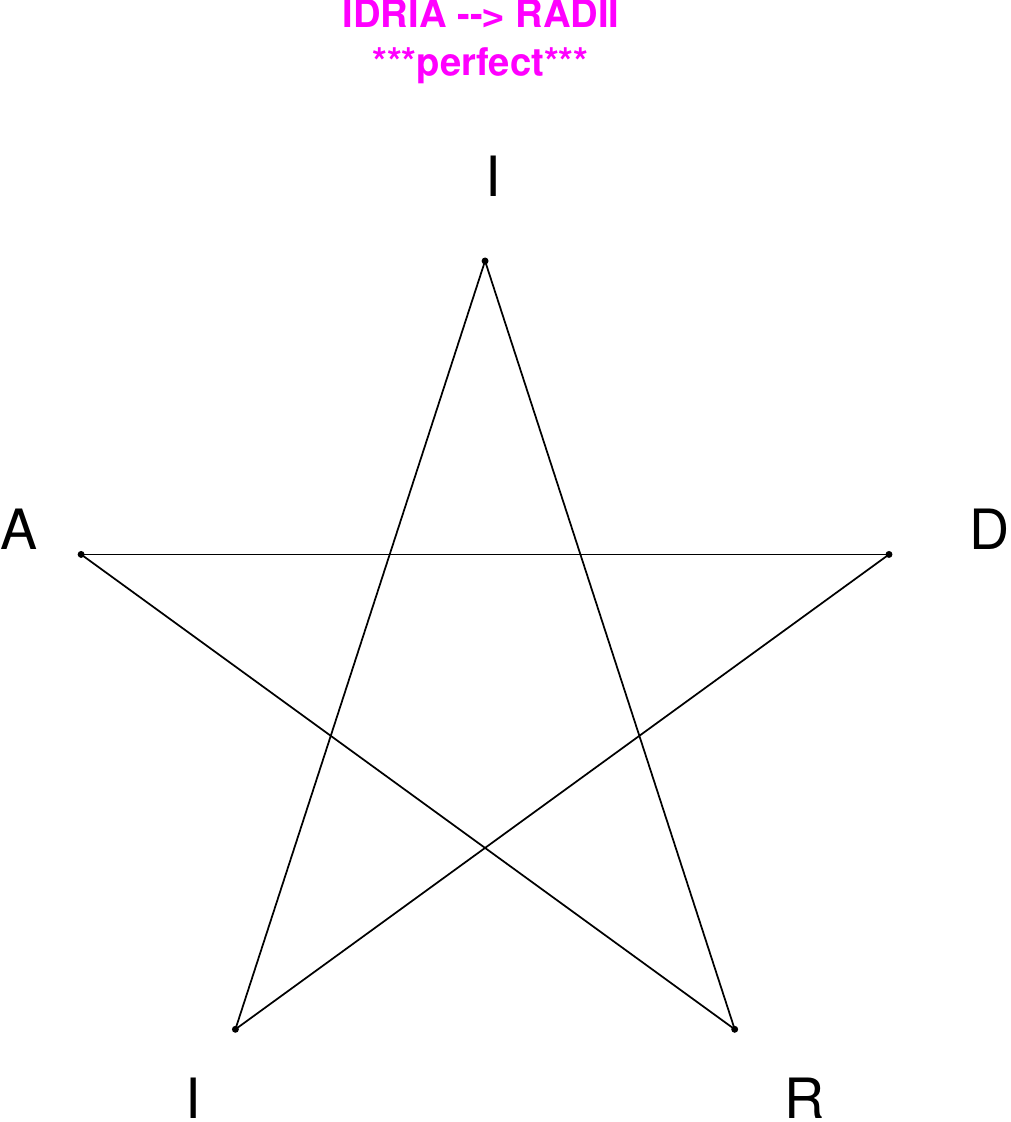}
\end{subfigure}
\end{figure}

\begin{figure}[H]
\centering
\begin{subfigure}[T]{0.19\textwidth}
\centering
\includegraphics[width=\textwidth]{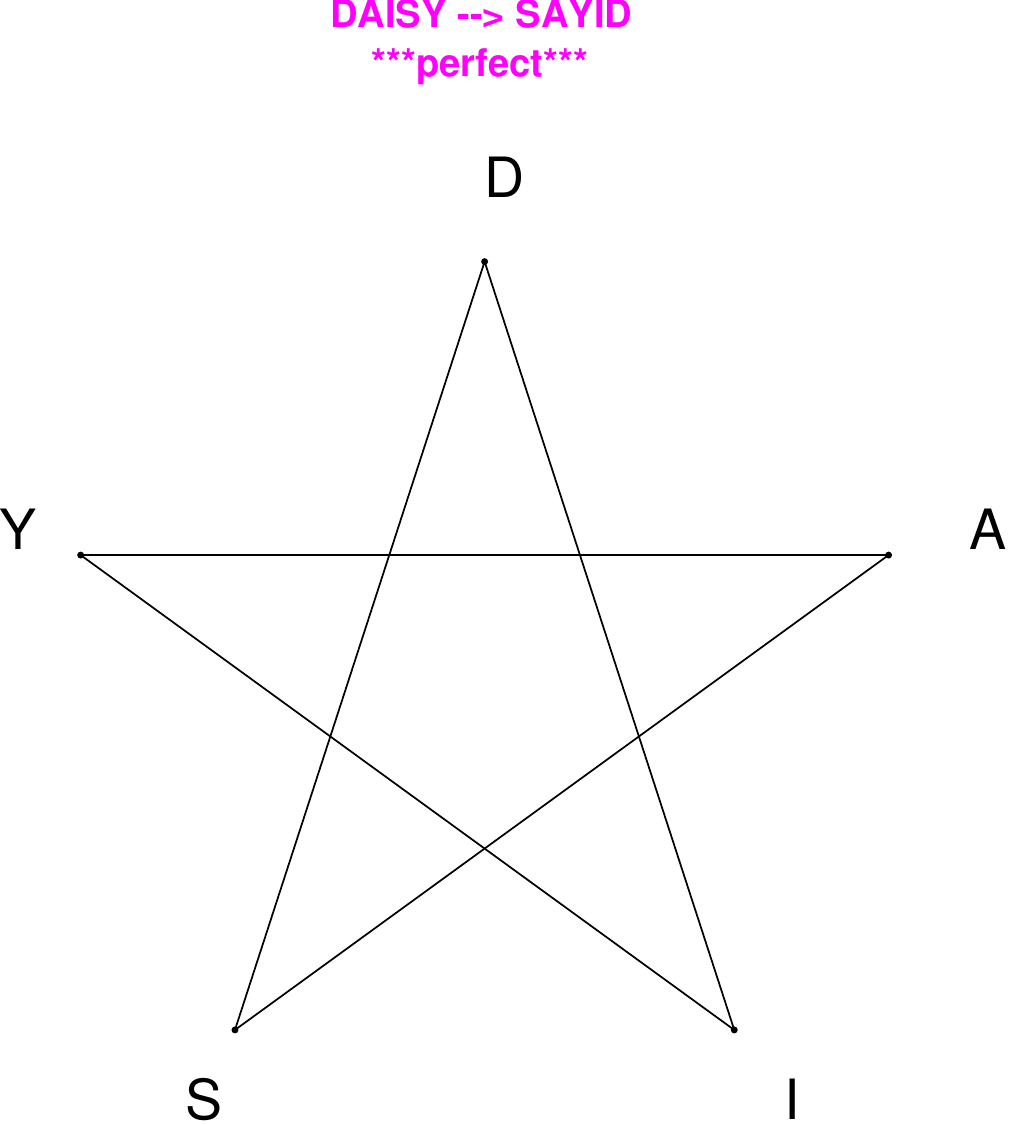}
\end{subfigure}
\hfill
\begin{subfigure}[T]{0.19\textwidth}
\centering
\includegraphics[width=\textwidth]{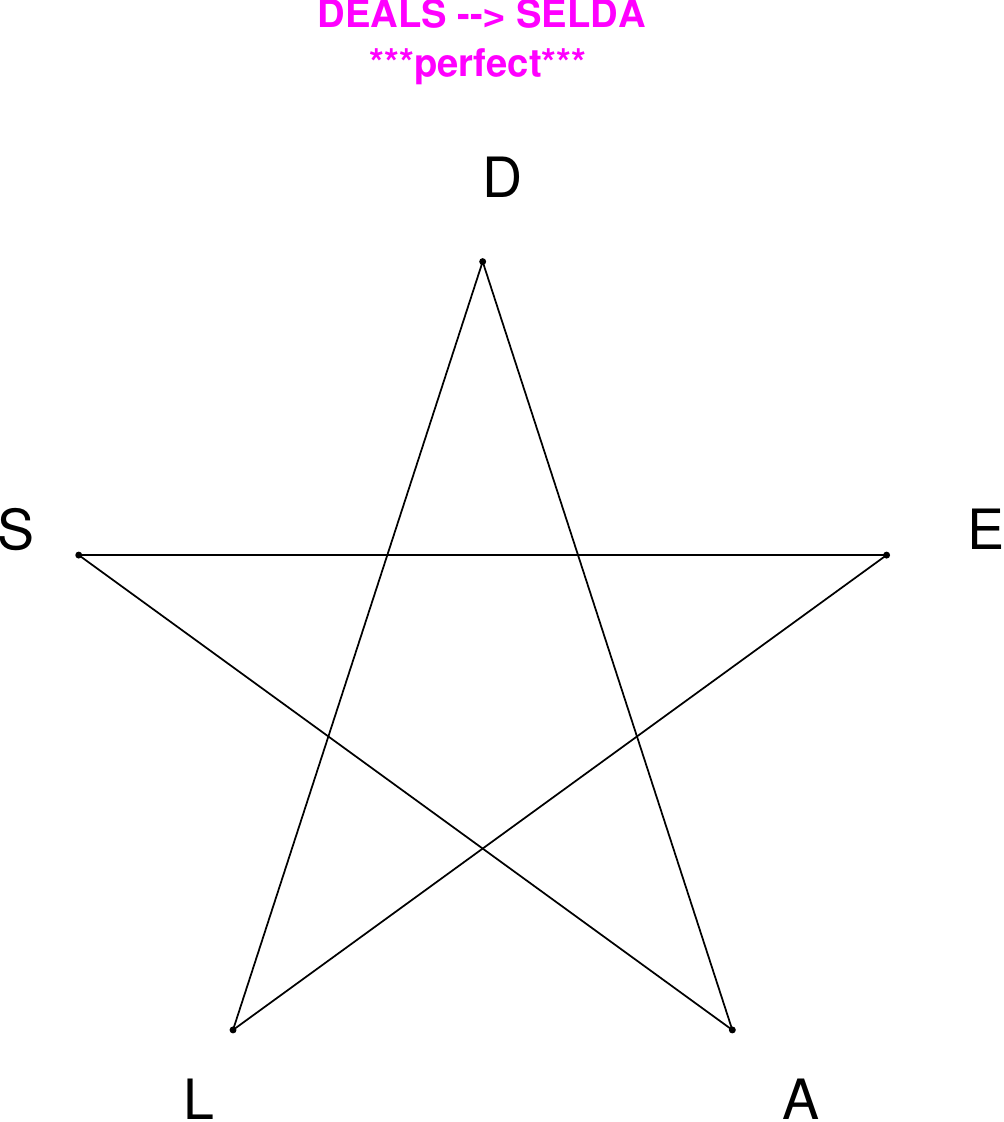}
\end{subfigure}
\hfill
\begin{subfigure}[T]{0.19\textwidth}
\centering
\includegraphics[width=\textwidth]{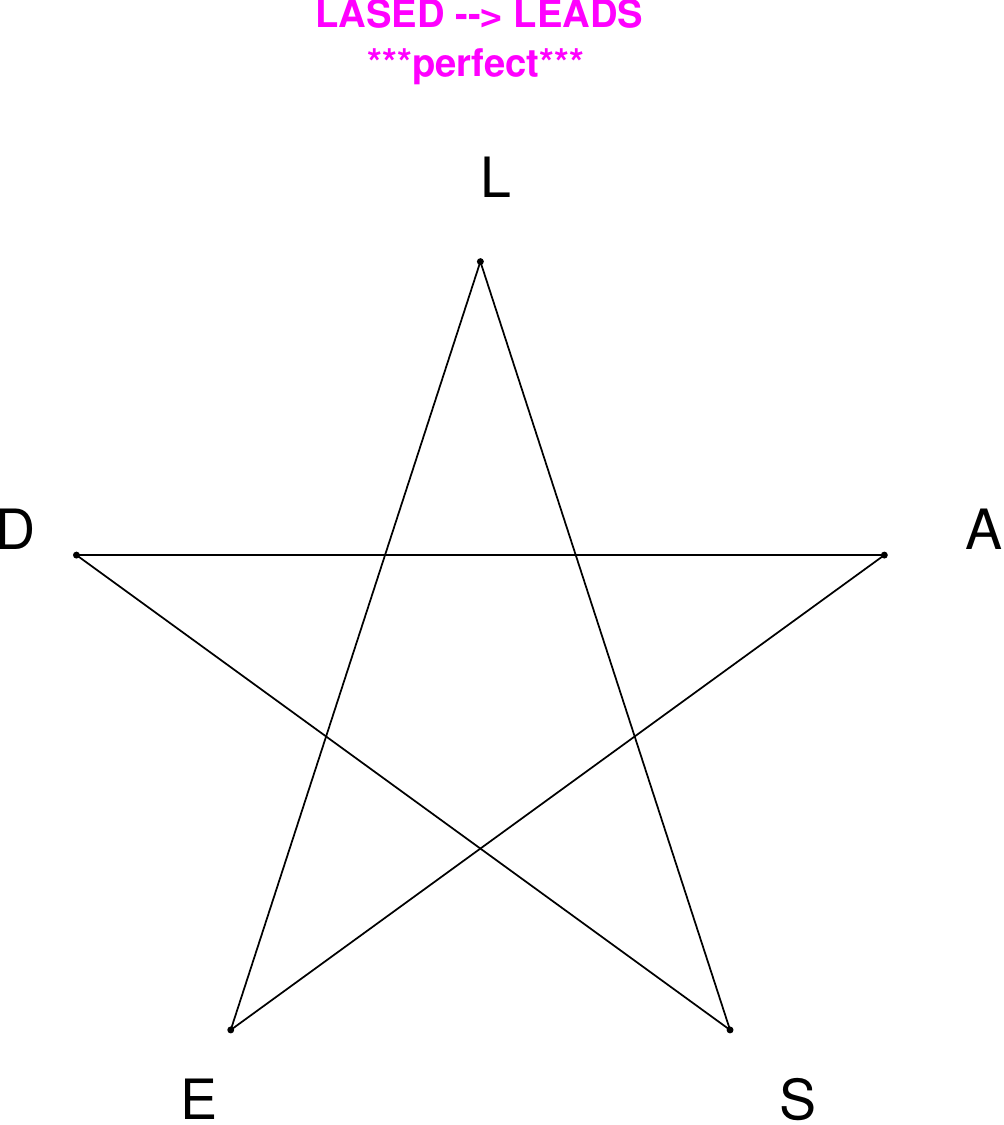}
\end{subfigure}
\hfill
\begin{subfigure}[T]{0.19\textwidth}
\centering
\includegraphics[width=\textwidth]{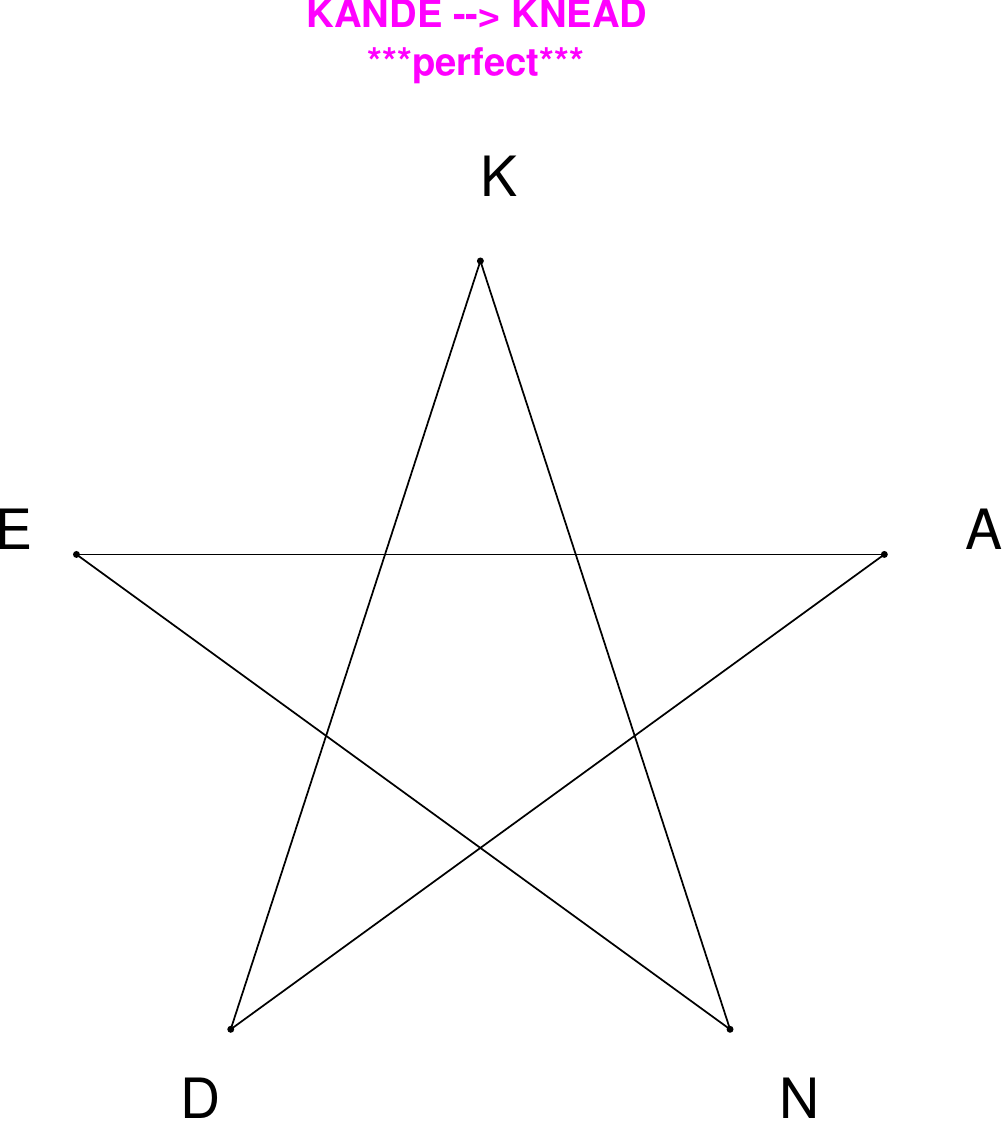}
\end{subfigure}
\hfill
\begin{subfigure}[T]{0.19\textwidth}
\centering
\includegraphics[width=\textwidth]{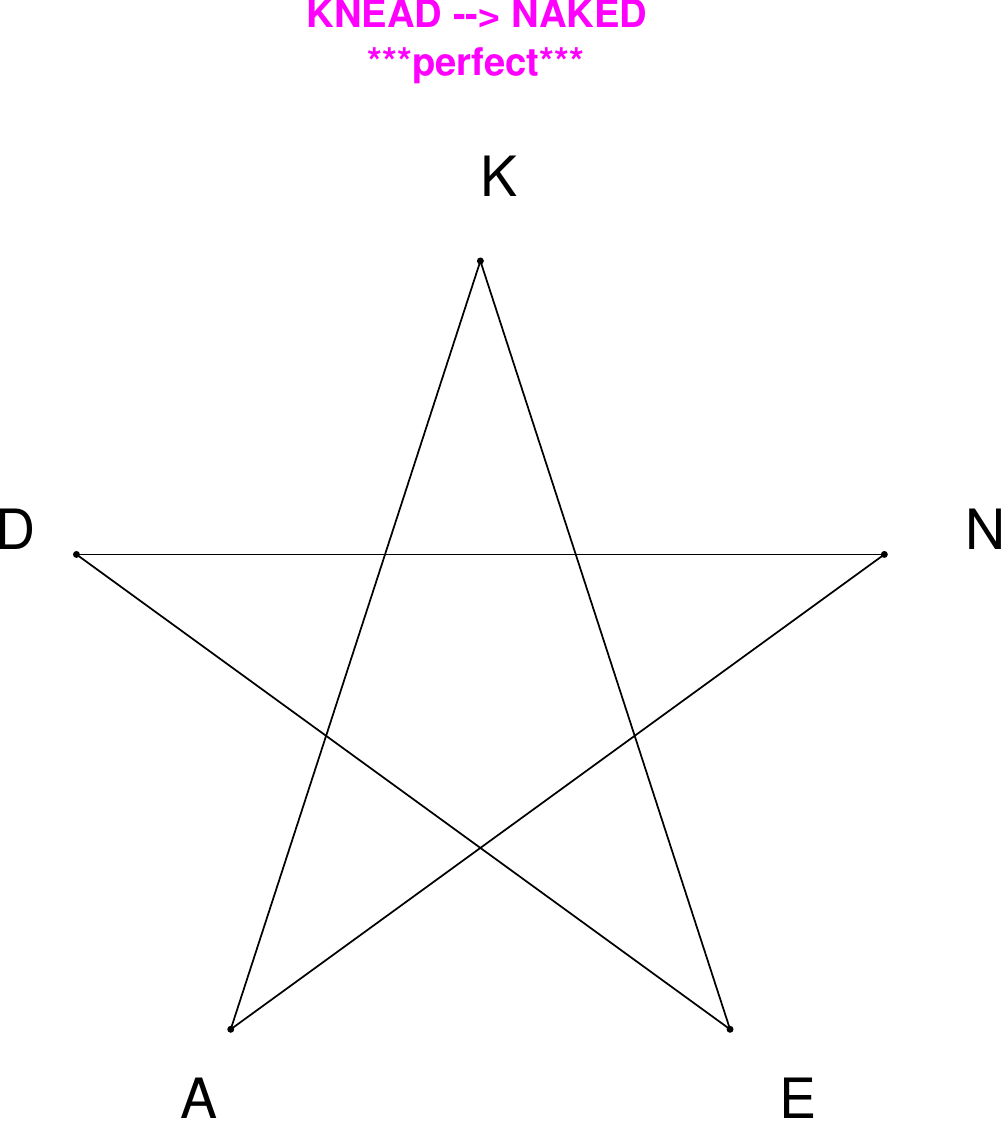}
\end{subfigure}
\end{figure}

\begin{figure}[H]
\centering
\begin{subfigure}[T]{0.19\textwidth}
\centering
\includegraphics[width=\textwidth]{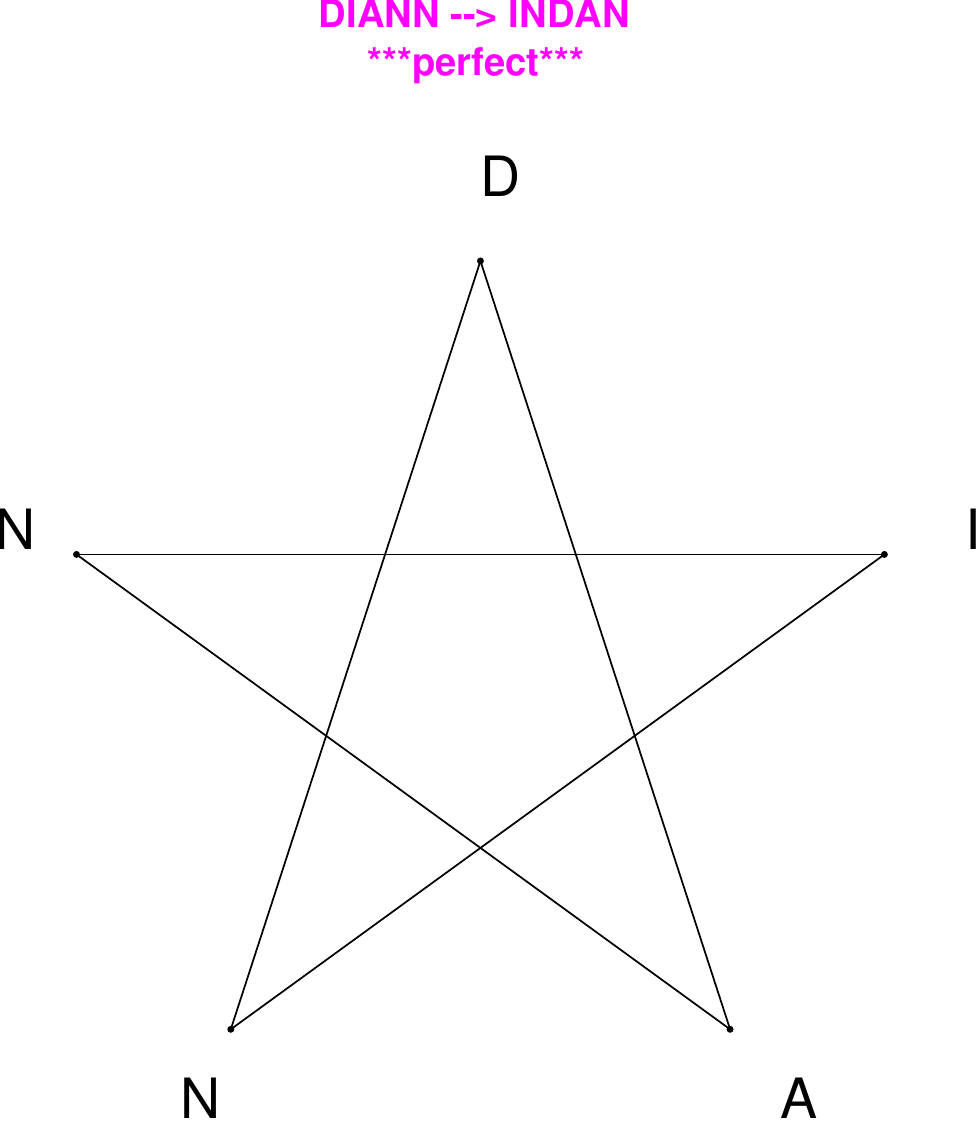}
\end{subfigure}
\hfill
\begin{subfigure}[T]{0.19\textwidth}
\centering
\includegraphics[width=\textwidth]{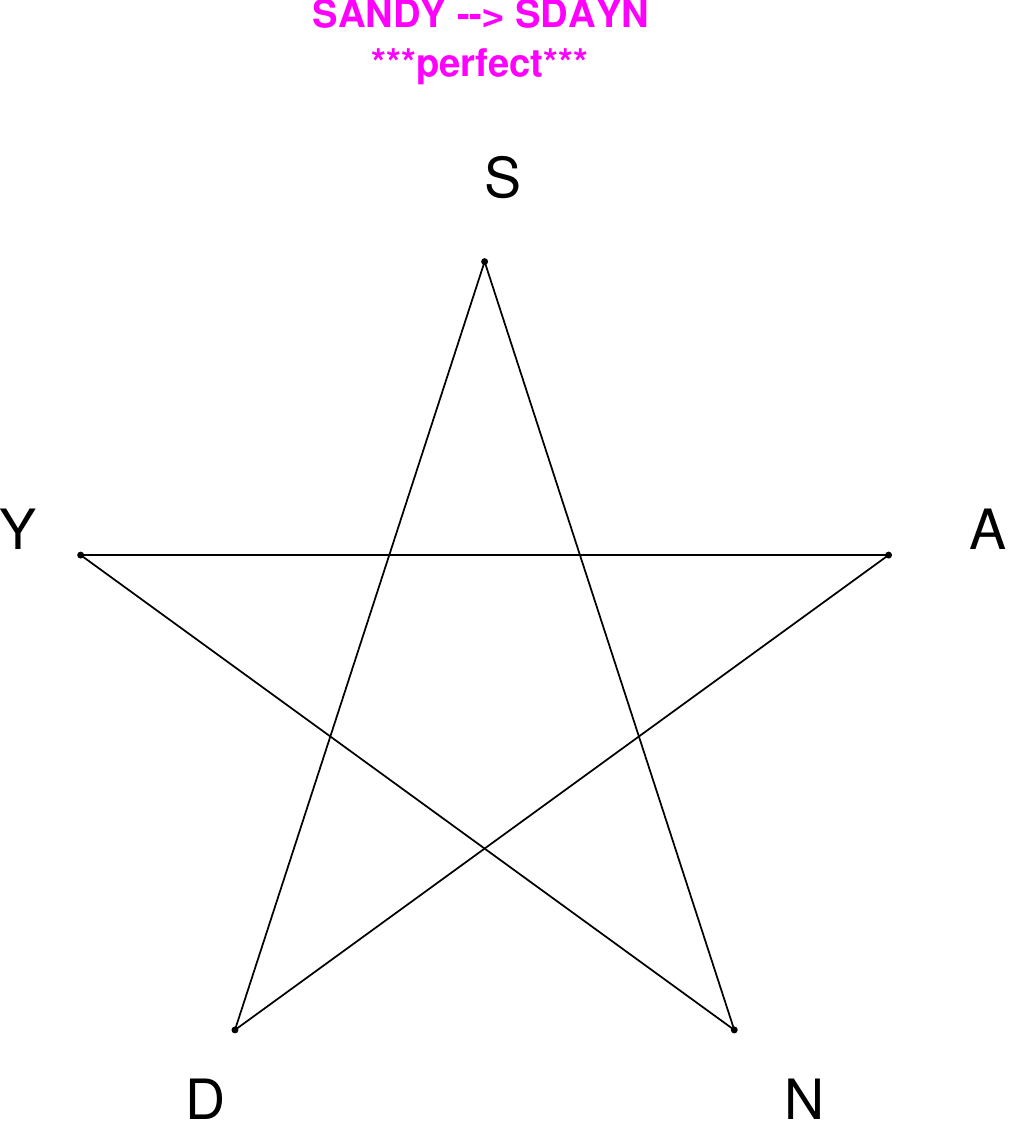}
\end{subfigure}
\hfill
\begin{subfigure}[T]{0.19\textwidth}
\centering
\includegraphics[width=\textwidth]{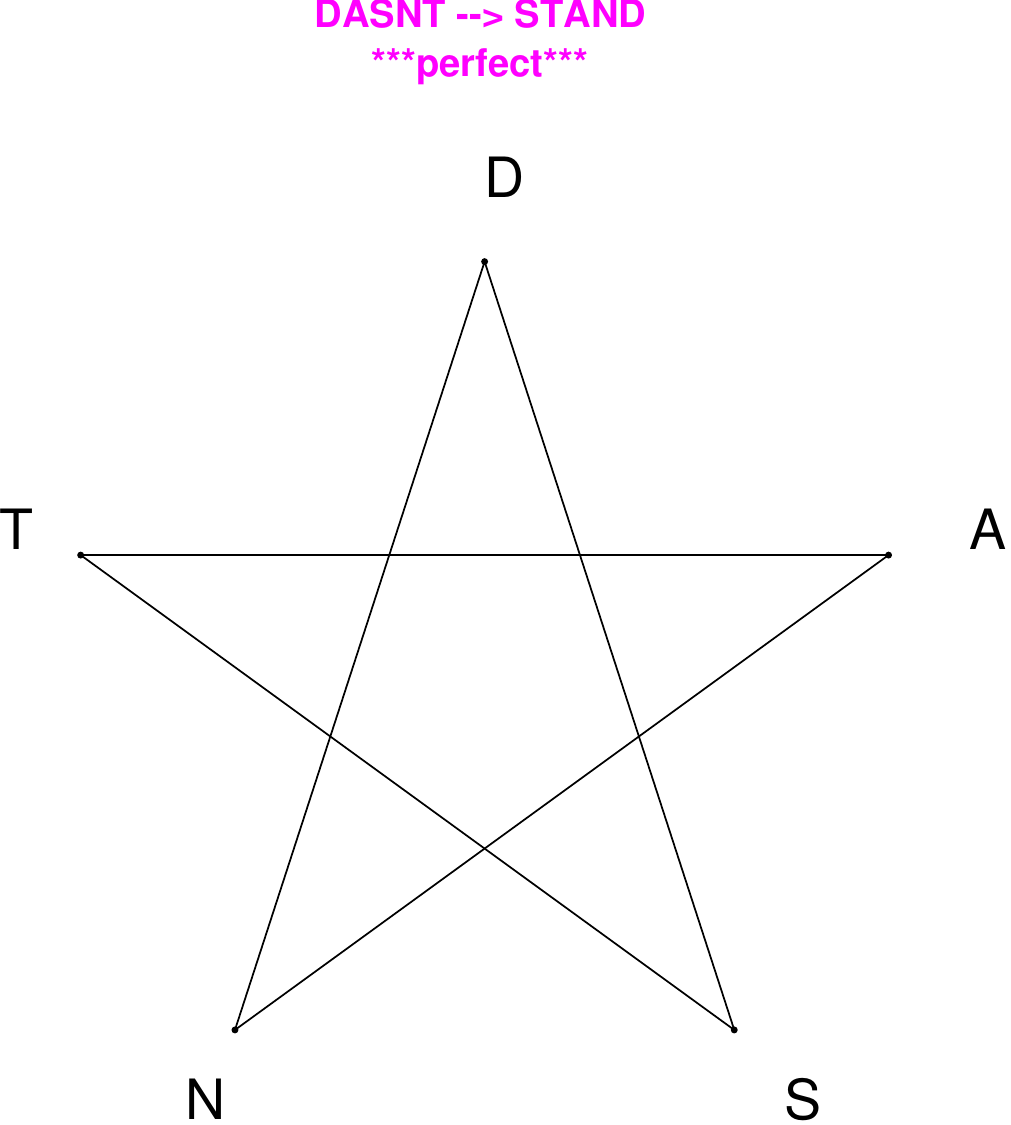}
\end{subfigure}
\hfill
\begin{subfigure}[T]{0.19\textwidth}
\centering
\includegraphics[width=\textwidth]{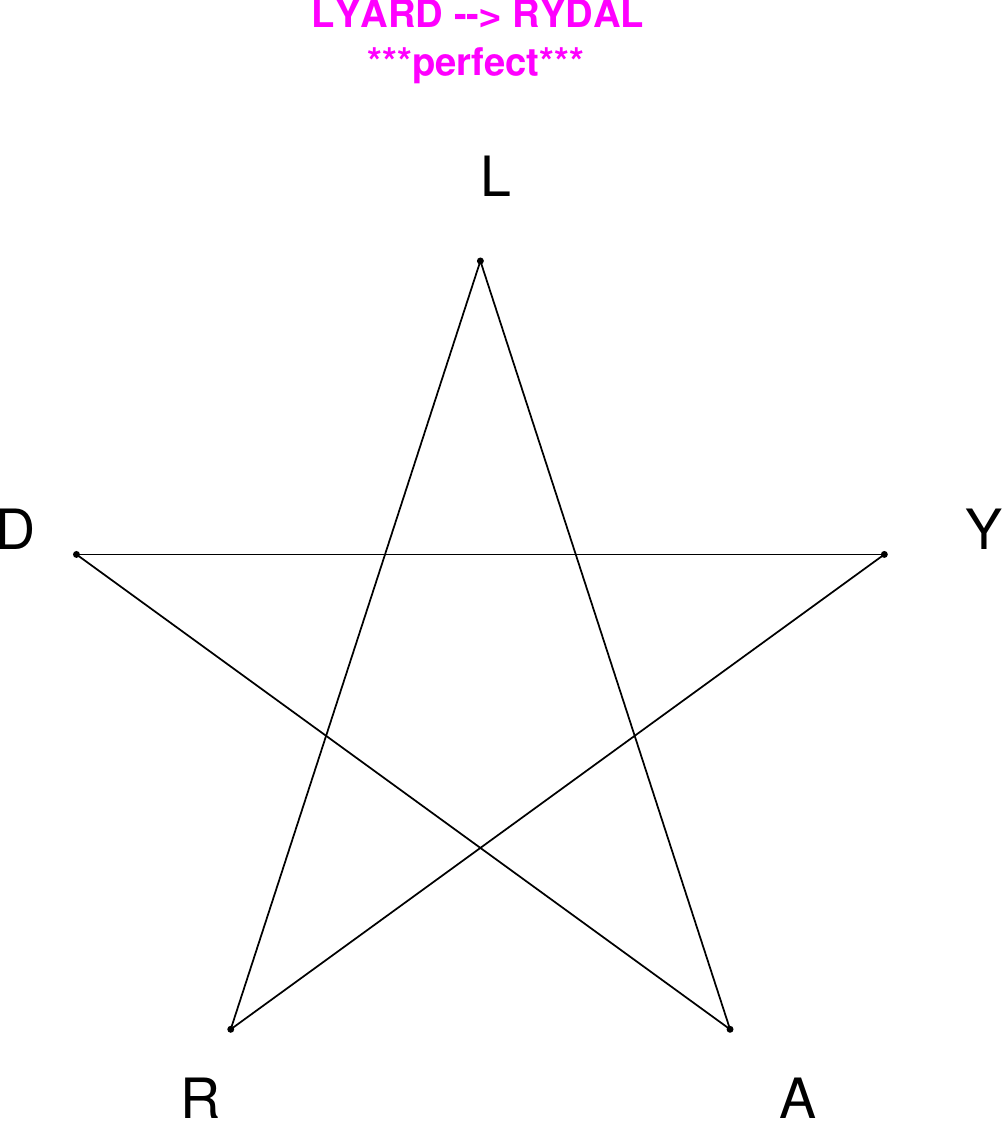}
\end{subfigure}
\hfill
\begin{subfigure}[T]{0.19\textwidth}
\centering
\includegraphics[width=\textwidth]{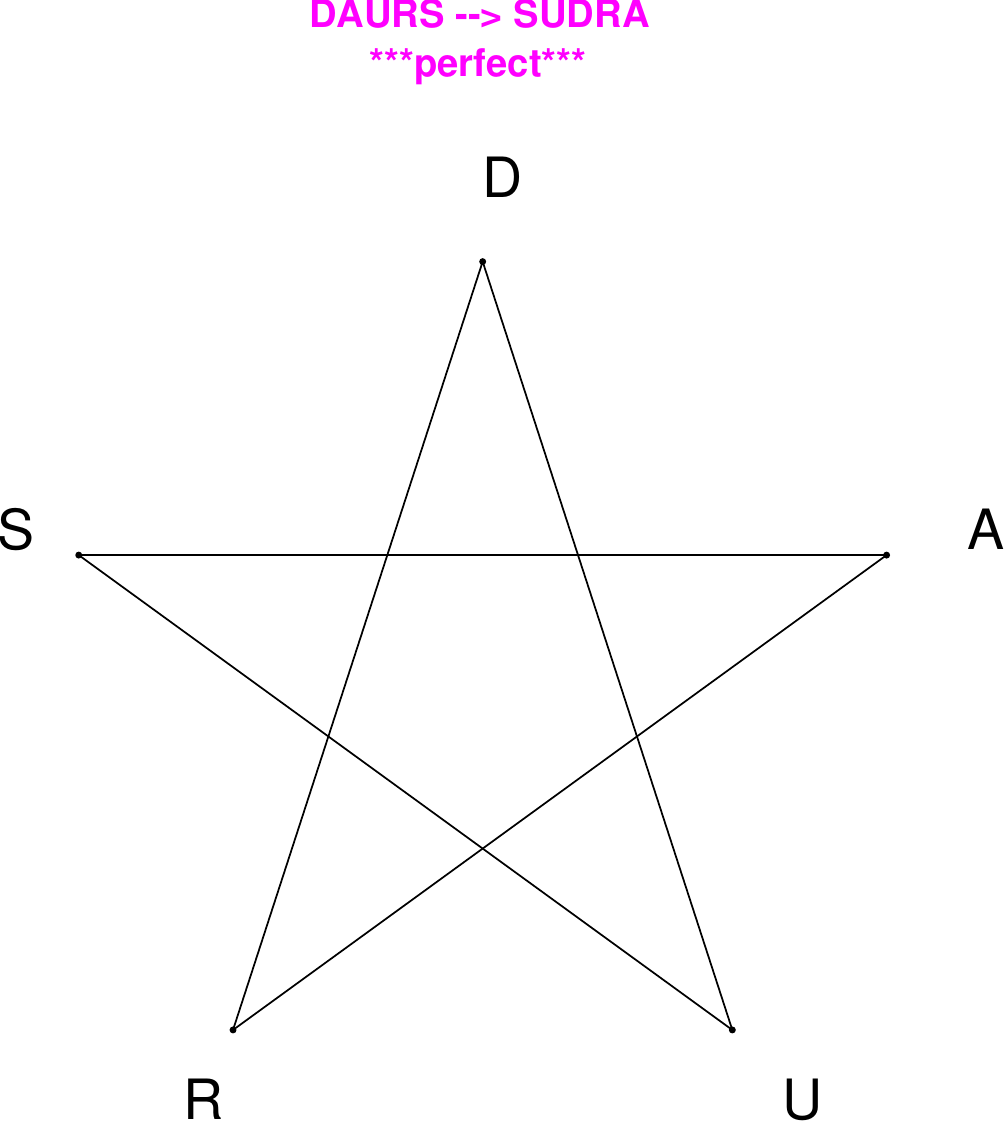}
\end{subfigure}
\end{figure}

\begin{figure}[H]
\centering
\begin{subfigure}[T]{0.19\textwidth}
\centering
\includegraphics[width=\textwidth]{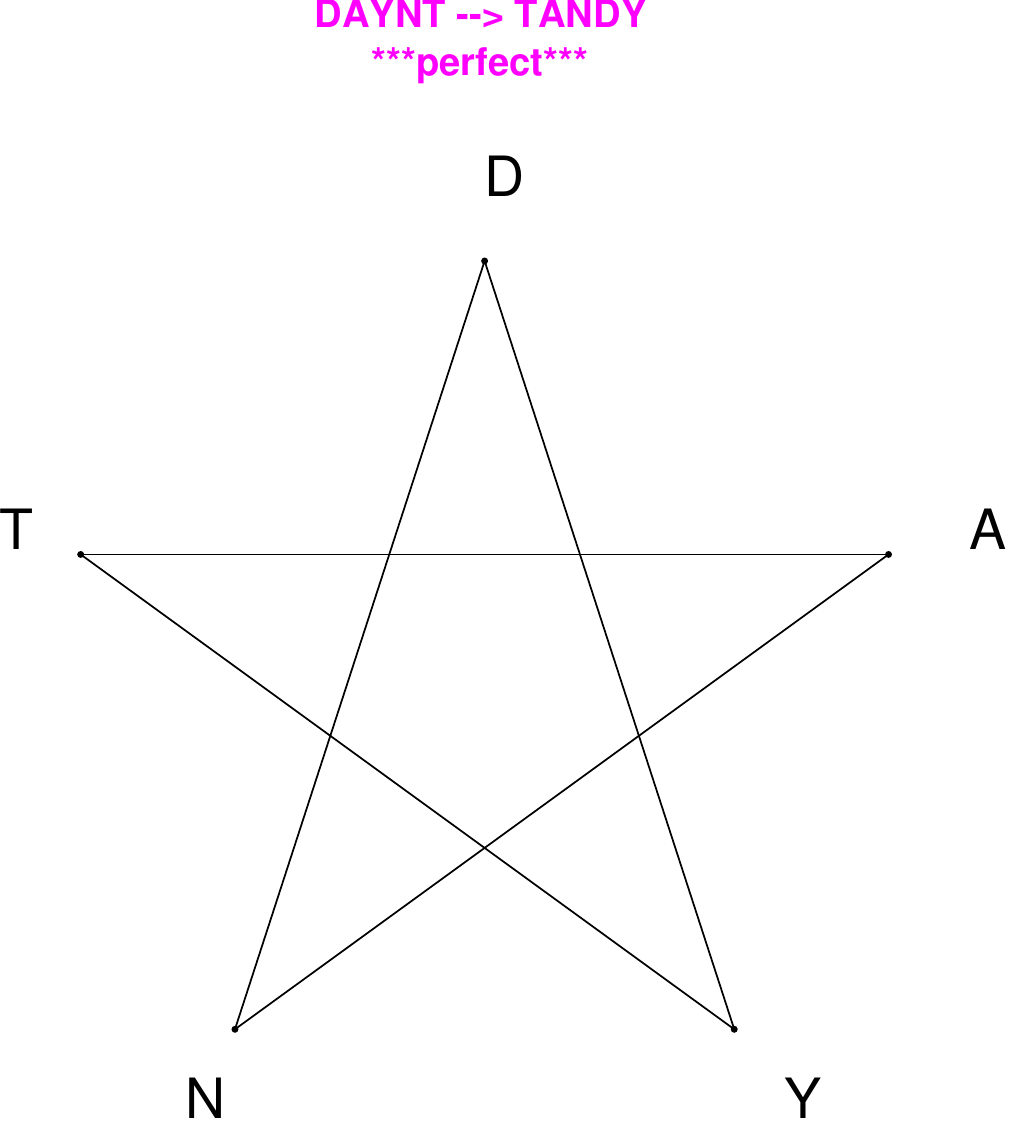}
\end{subfigure}
\hfill
\begin{subfigure}[T]{0.19\textwidth}
\centering
\includegraphics[width=\textwidth]{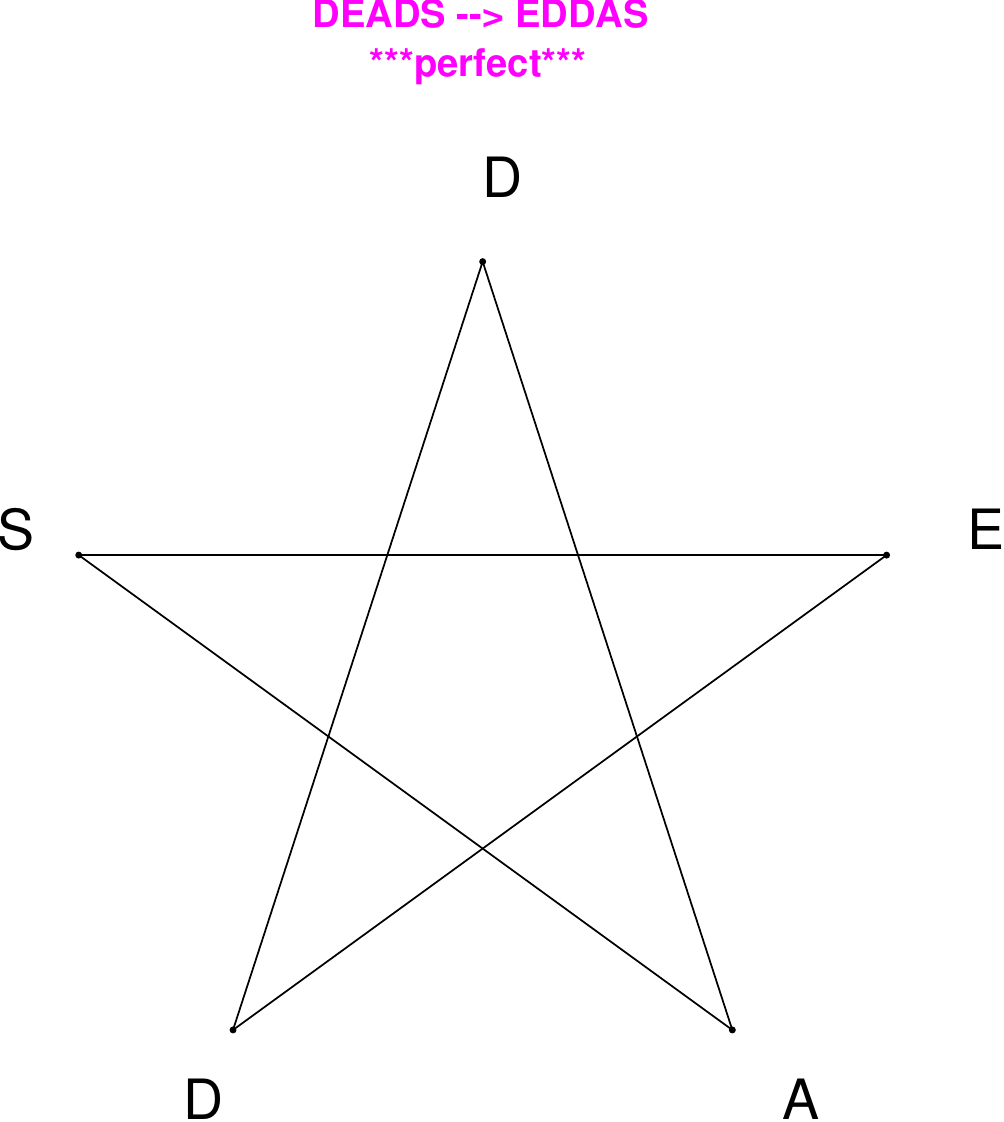}
\end{subfigure}
\hfill
\begin{subfigure}[T]{0.19\textwidth}
\centering
\includegraphics[width=\textwidth]{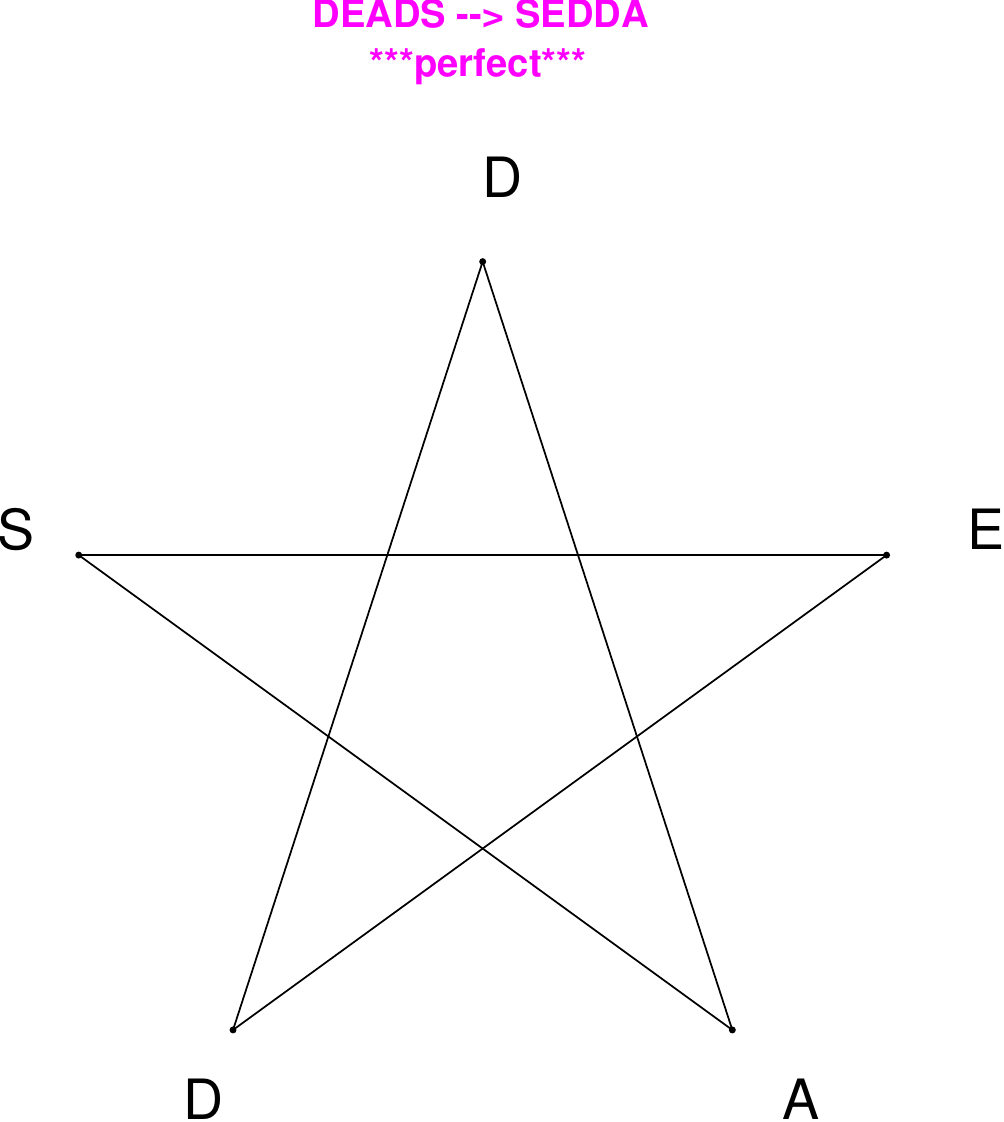}
\end{subfigure}
\hfill
\begin{subfigure}[T]{0.19\textwidth}
\centering
\includegraphics[width=\textwidth]{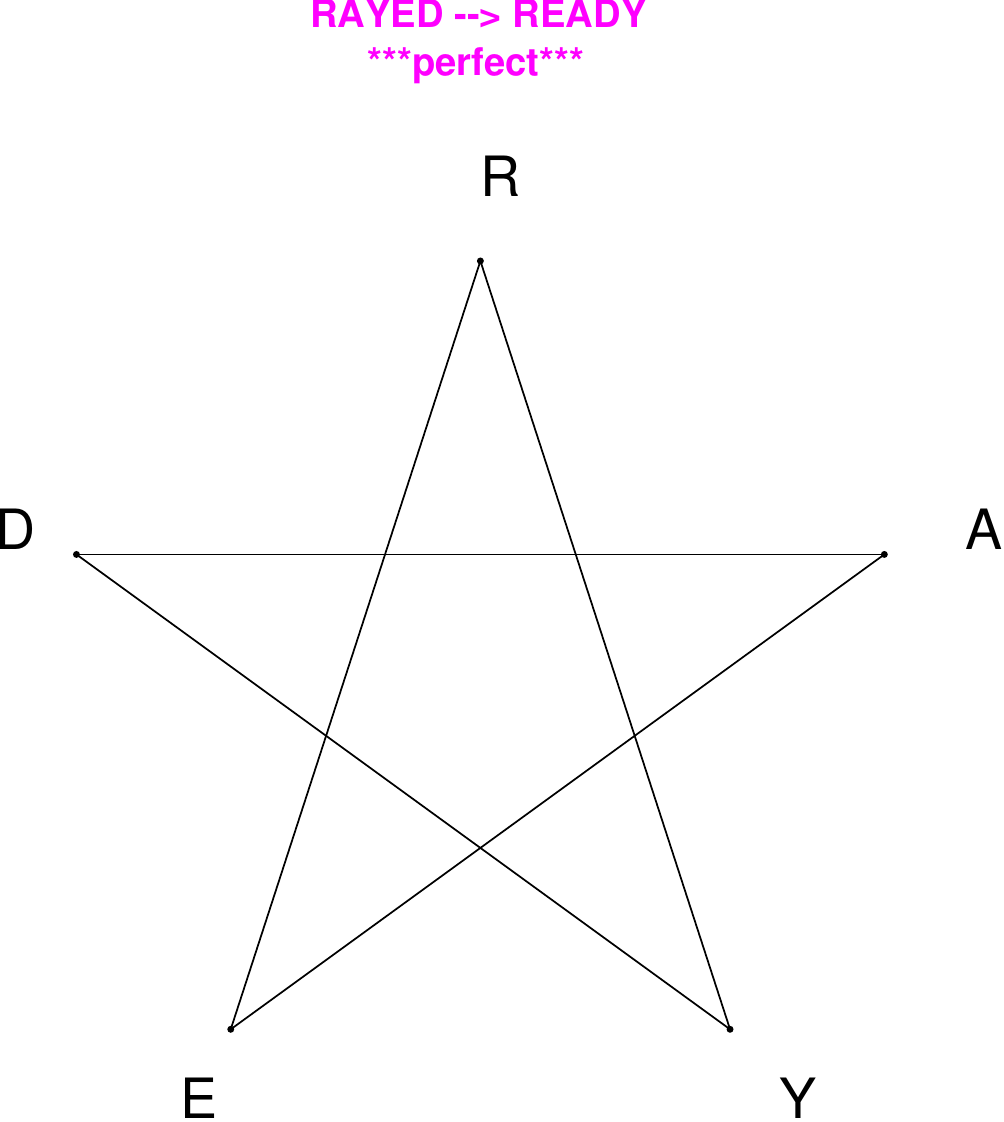}
\end{subfigure}
\hfill
\begin{subfigure}[T]{0.19\textwidth}
\centering
\includegraphics[width=\textwidth]{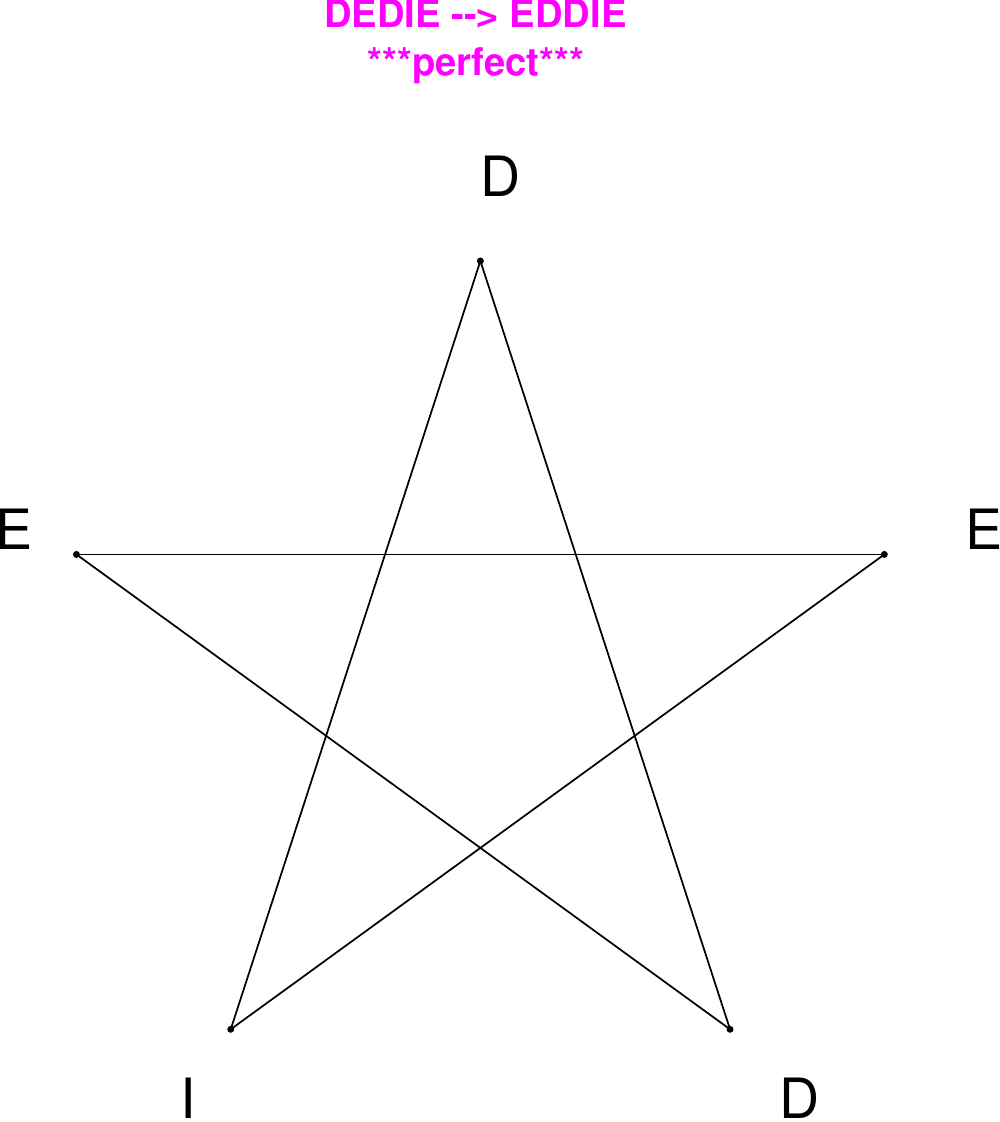}
\end{subfigure}
\end{figure}

\begin{figure}[H]
\centering
\begin{subfigure}[T]{0.19\textwidth}
\centering
\includegraphics[width=\textwidth]{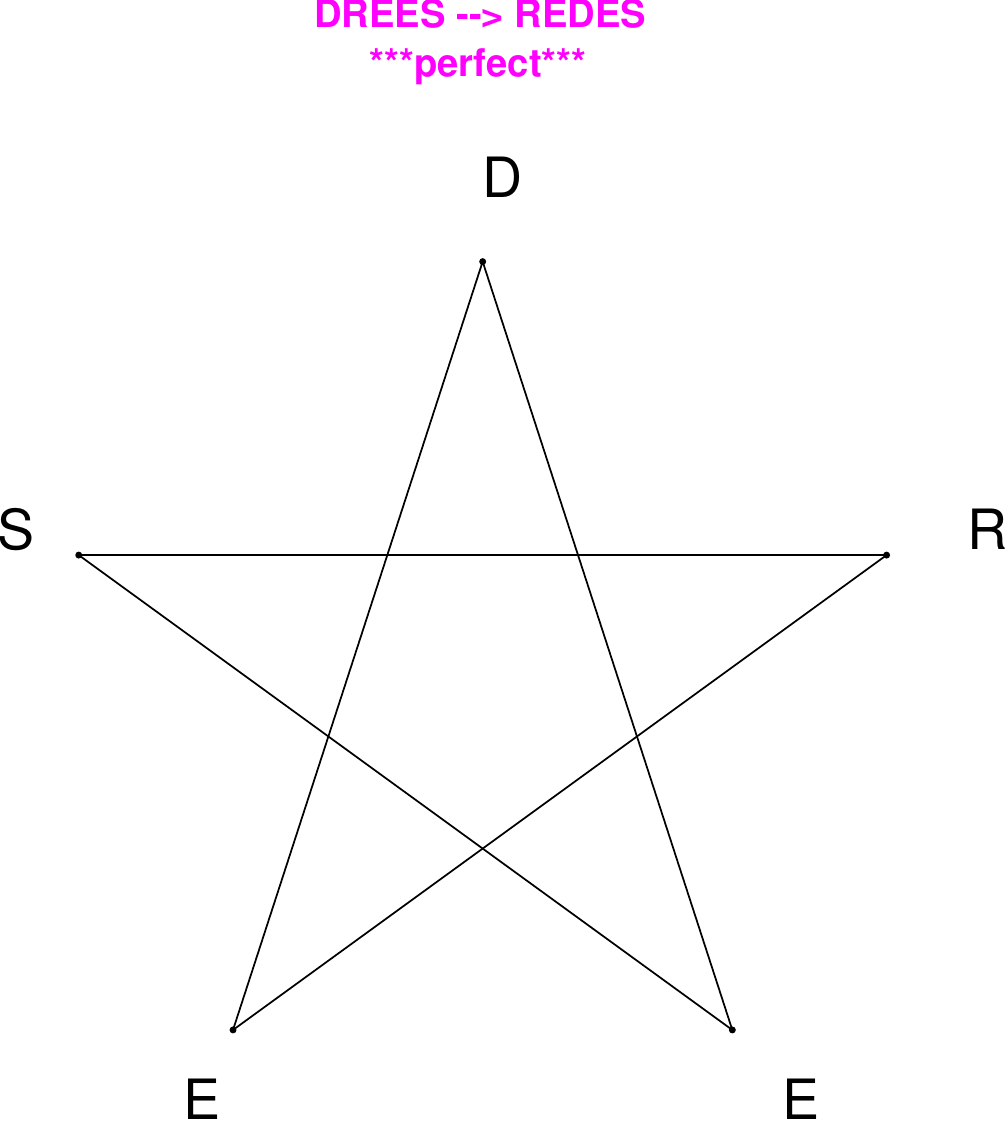}
\end{subfigure}
\hfill
\begin{subfigure}[T]{0.19\textwidth}
\centering
\includegraphics[width=\textwidth]{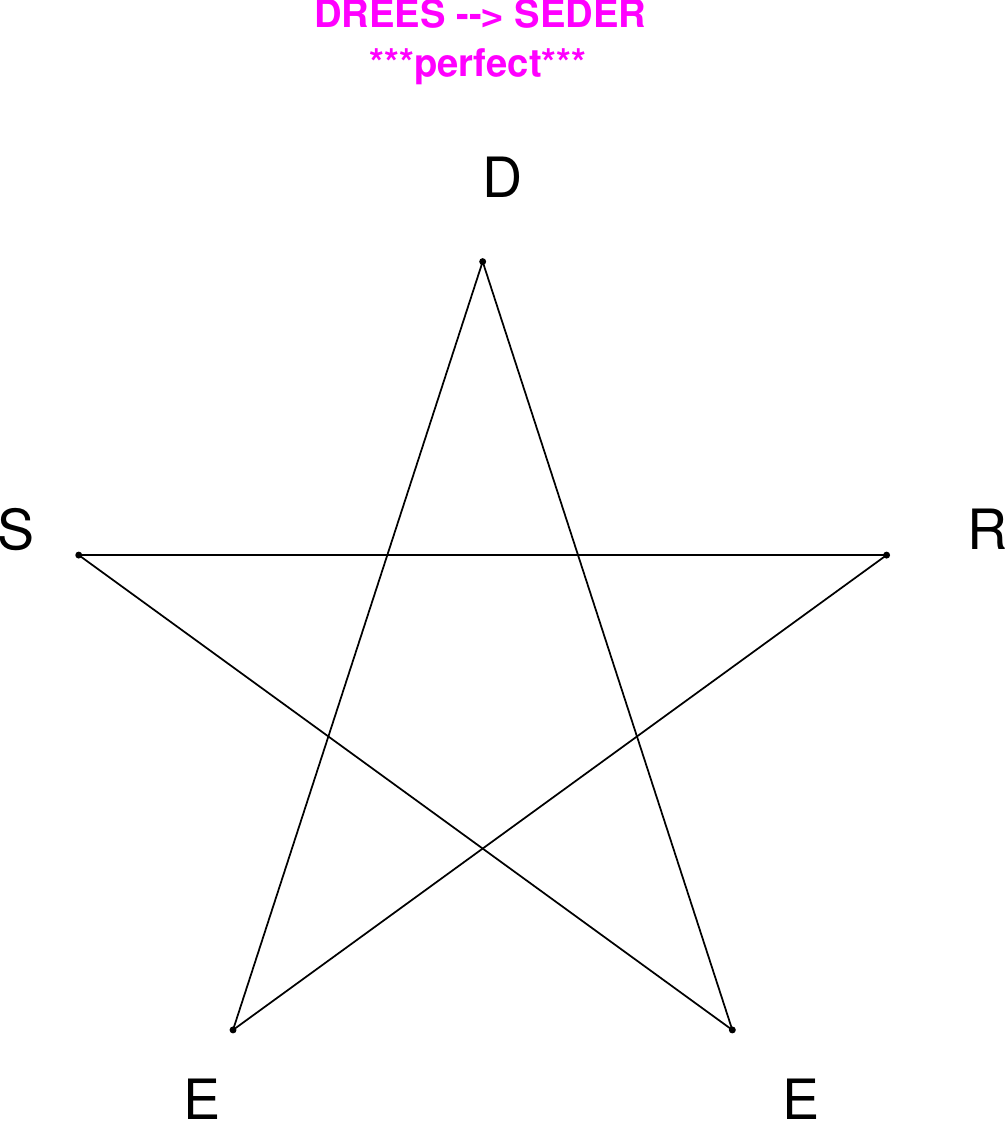}
\end{subfigure}
\hfill
\begin{subfigure}[T]{0.19\textwidth}
\centering
\includegraphics[width=\textwidth]{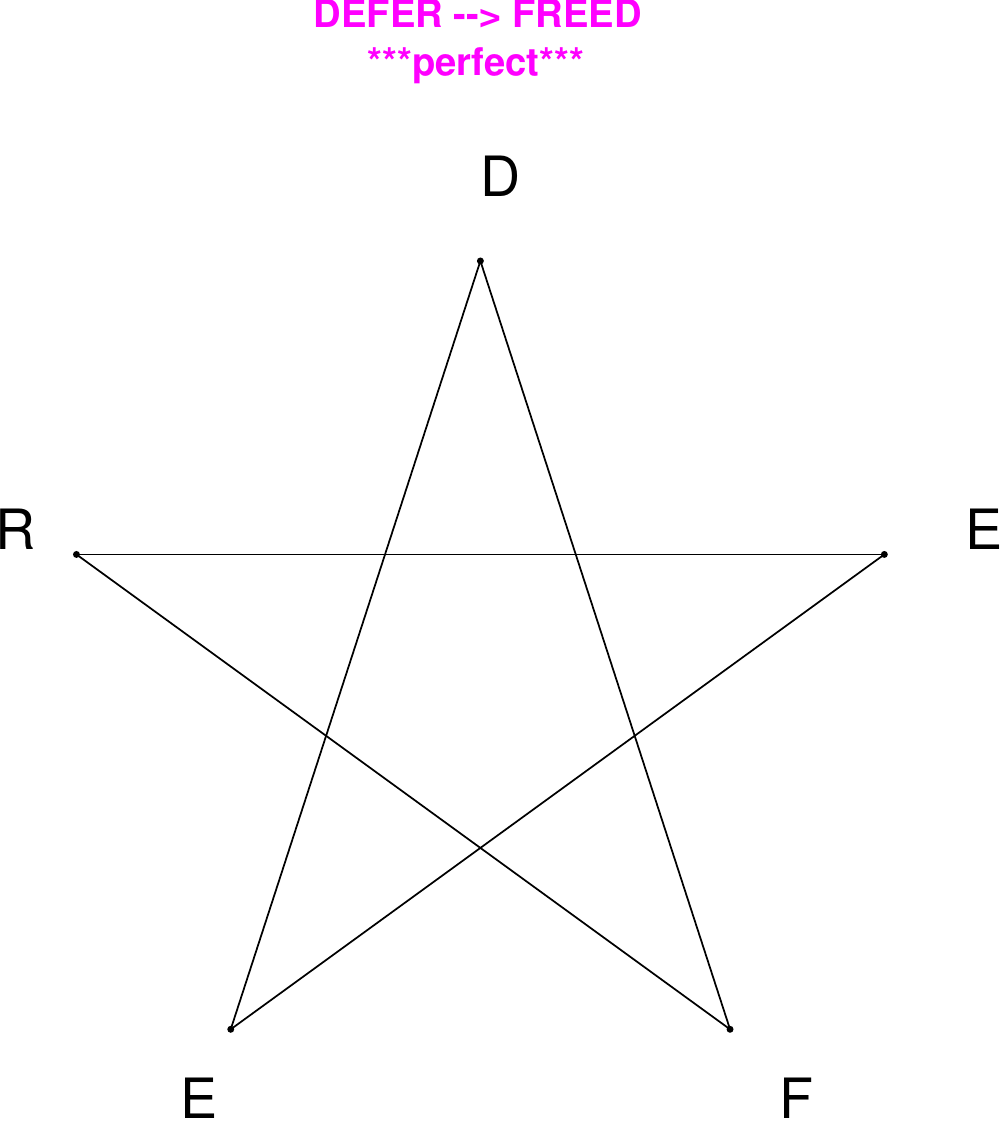}
\end{subfigure}
\hfill
\begin{subfigure}[T]{0.19\textwidth}
\centering
\includegraphics[width=\textwidth]{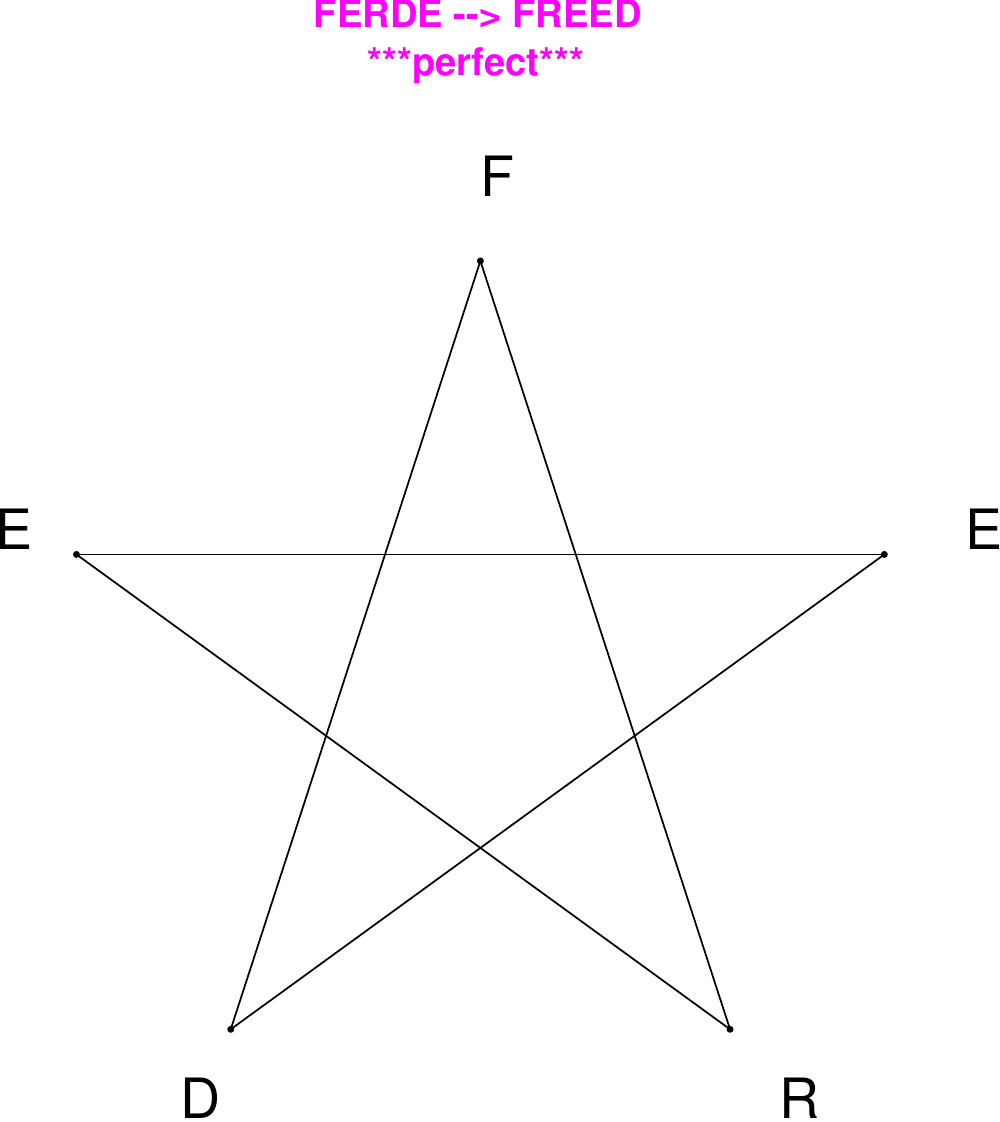}
\end{subfigure}
\hfill
\begin{subfigure}[T]{0.19\textwidth}
\centering
\includegraphics[width=\textwidth]{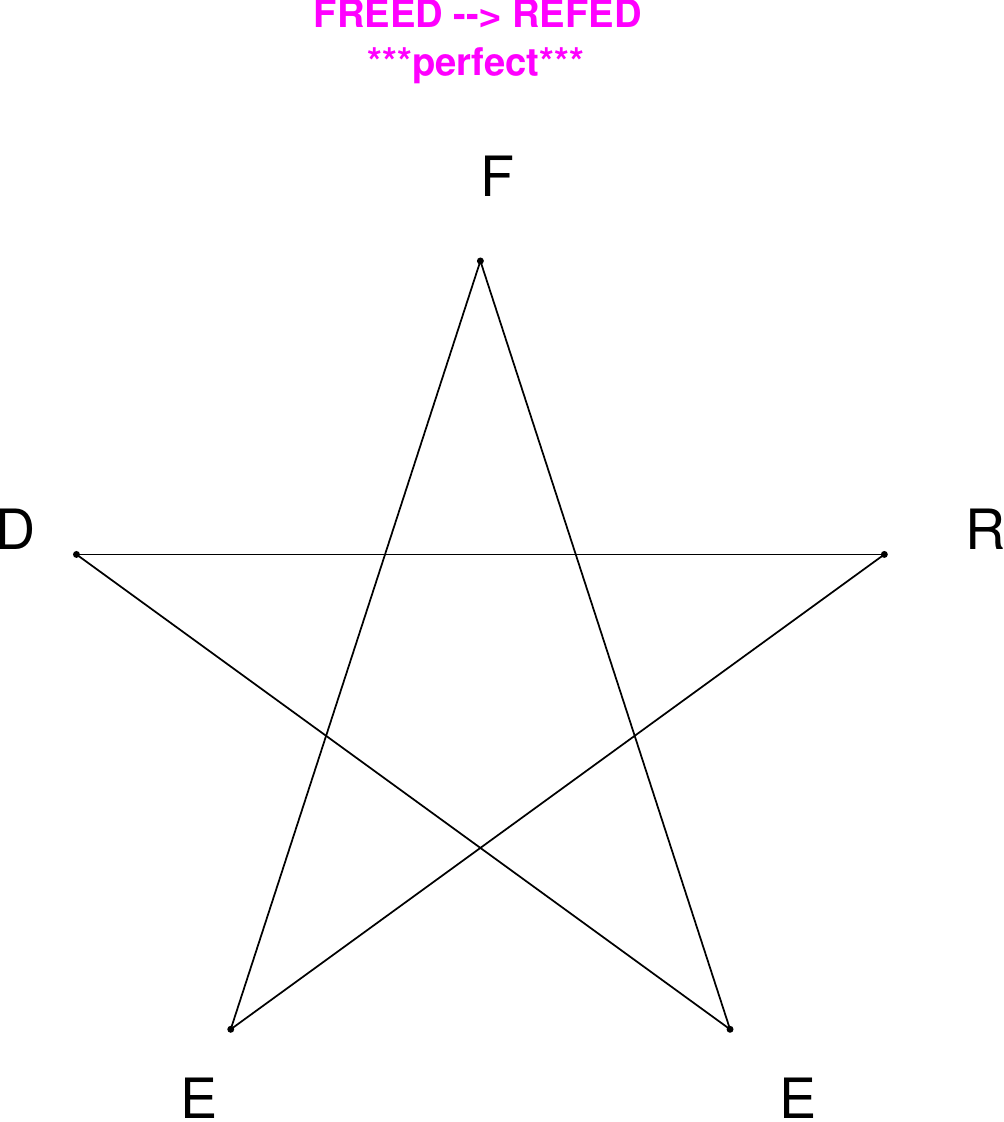}
\end{subfigure}
\end{figure}

\begin{figure}[H]
\centering
\begin{subfigure}[T]{0.19\textwidth}
\centering
\includegraphics[width=\textwidth]{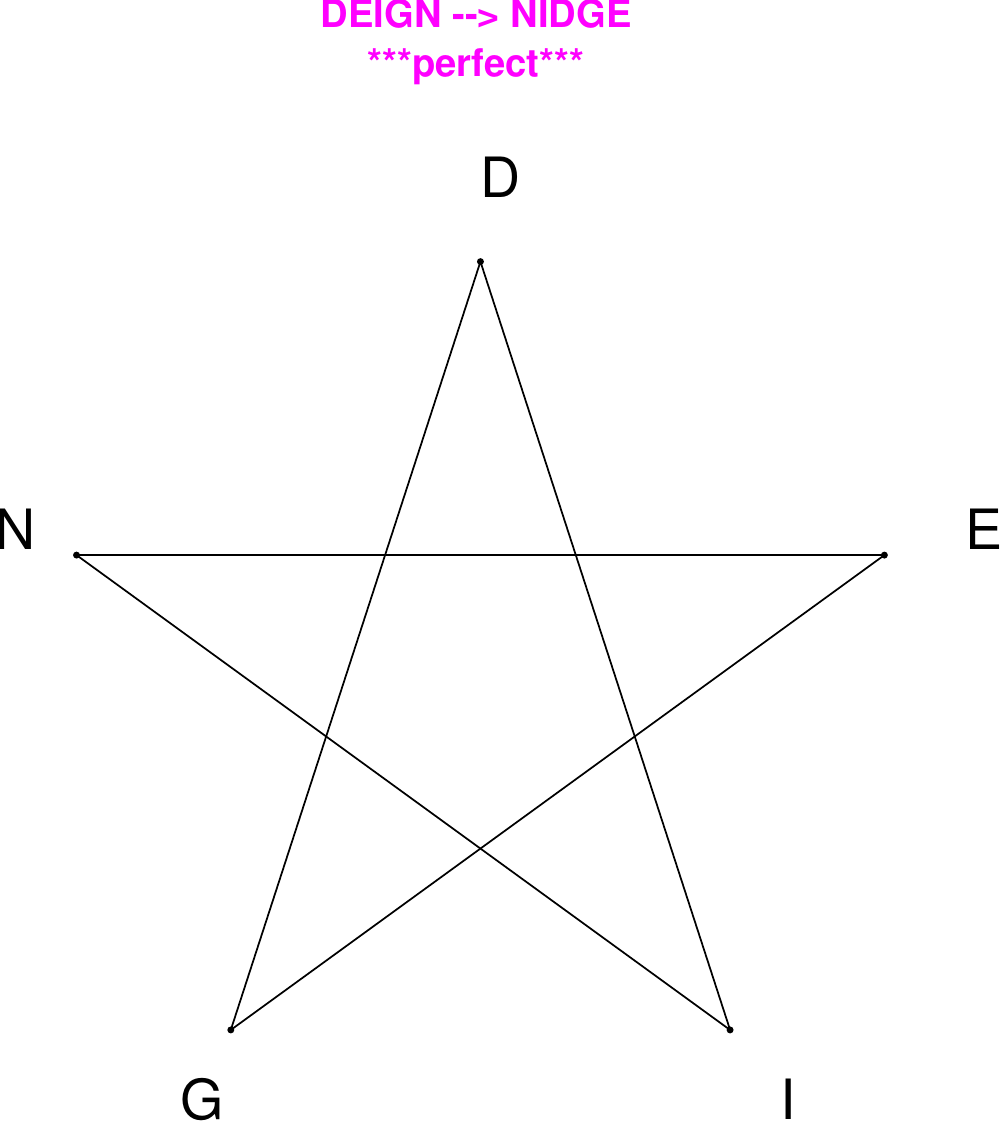}
\end{subfigure}
\hfill
\begin{subfigure}[T]{0.19\textwidth}
\centering
\includegraphics[width=\textwidth]{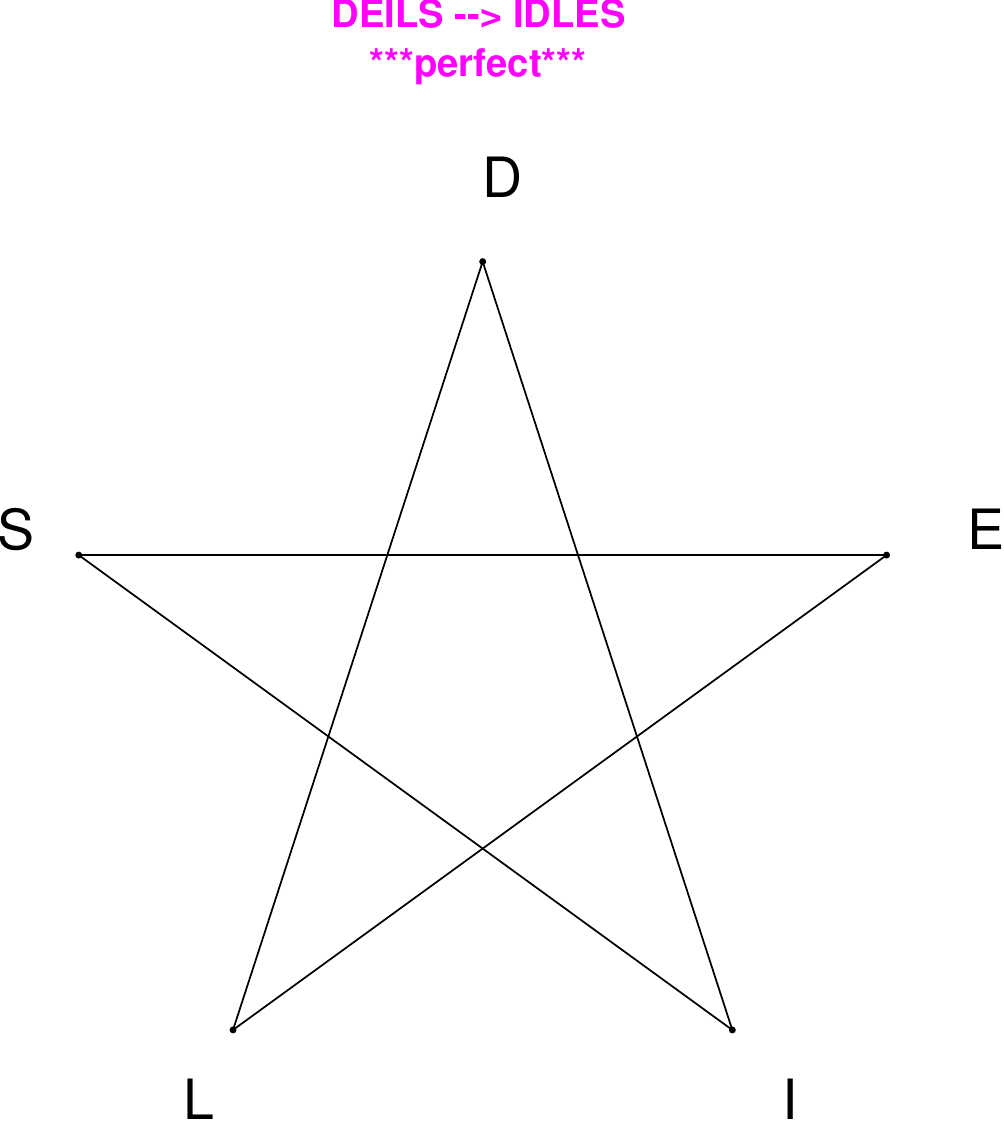}
\end{subfigure}
\hfill
\begin{subfigure}[T]{0.19\textwidth}
\centering
\includegraphics[width=\textwidth]{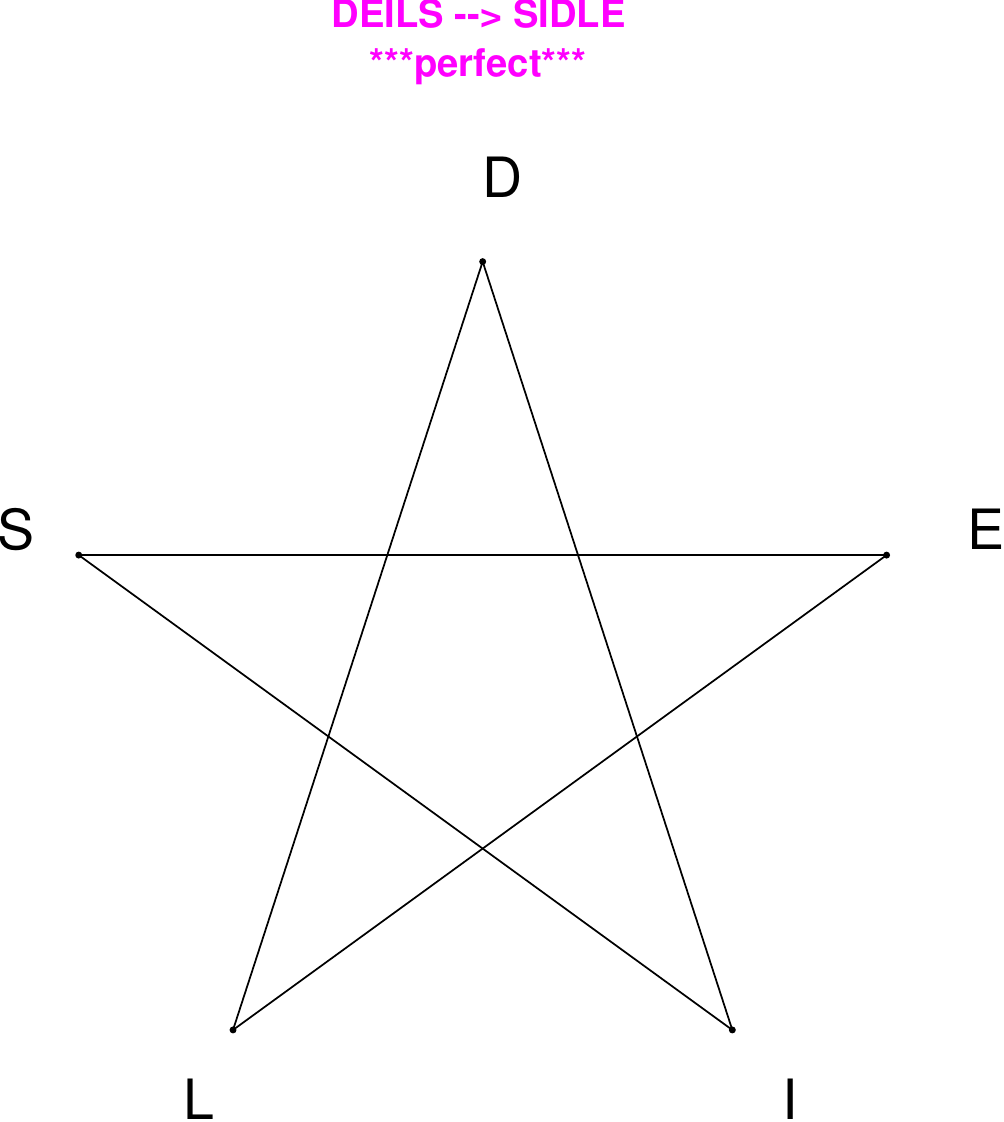}
\end{subfigure}
\hfill
\begin{subfigure}[T]{0.19\textwidth}
\centering
\includegraphics[width=\textwidth]{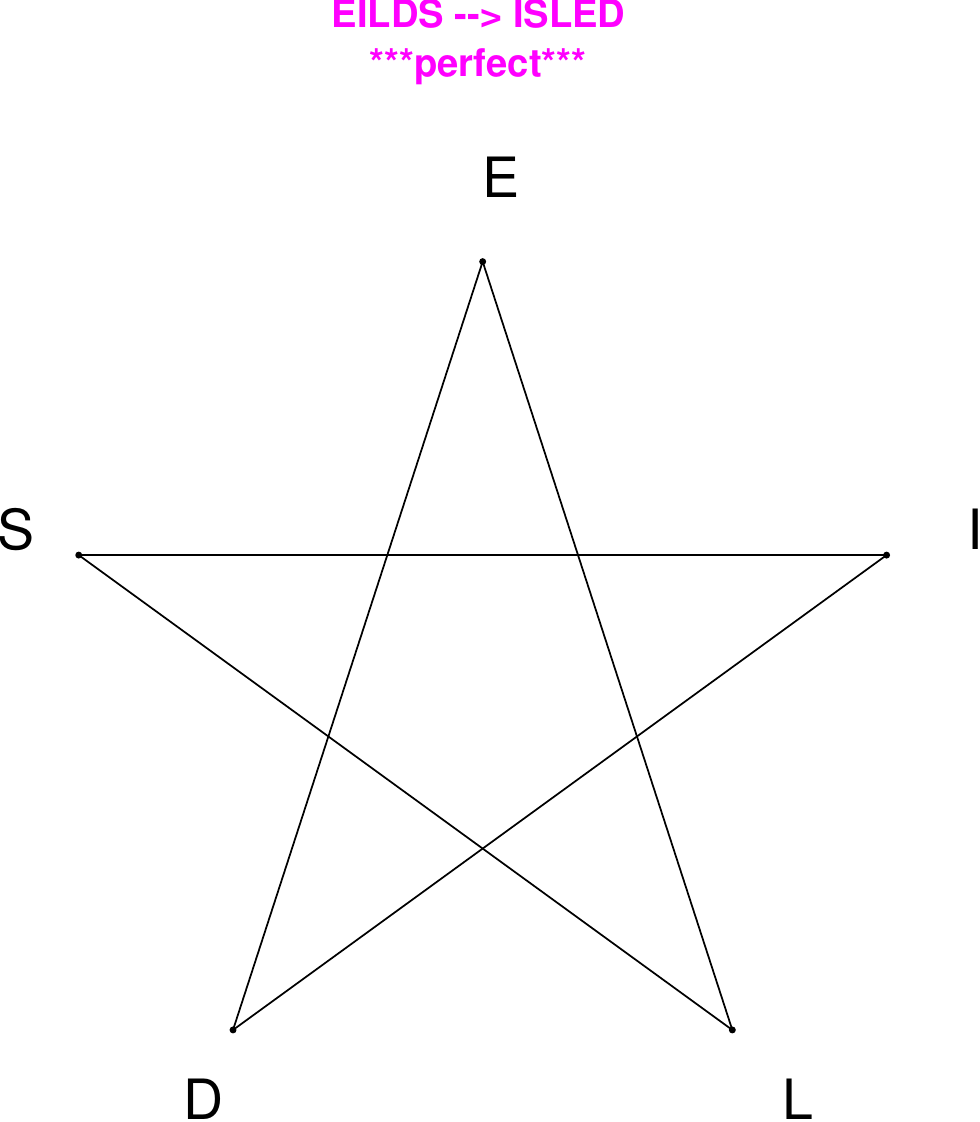}
\end{subfigure}
\hfill
\begin{subfigure}[T]{0.19\textwidth}
\centering
\includegraphics[width=\textwidth]{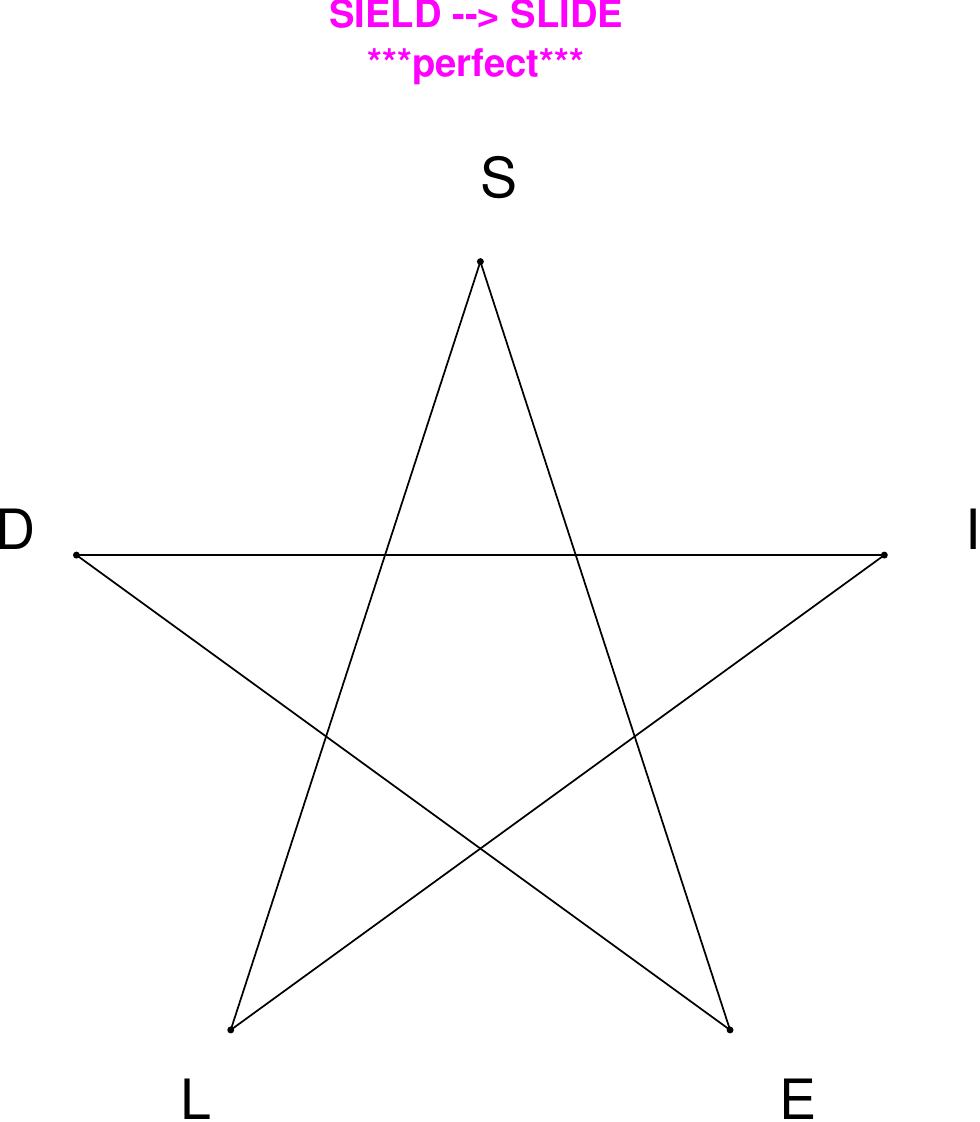}
\end{subfigure}
\end{figure}

\begin{figure}[H]
\centering
\begin{subfigure}[T]{0.19\textwidth}
\centering
\includegraphics[width=\textwidth]{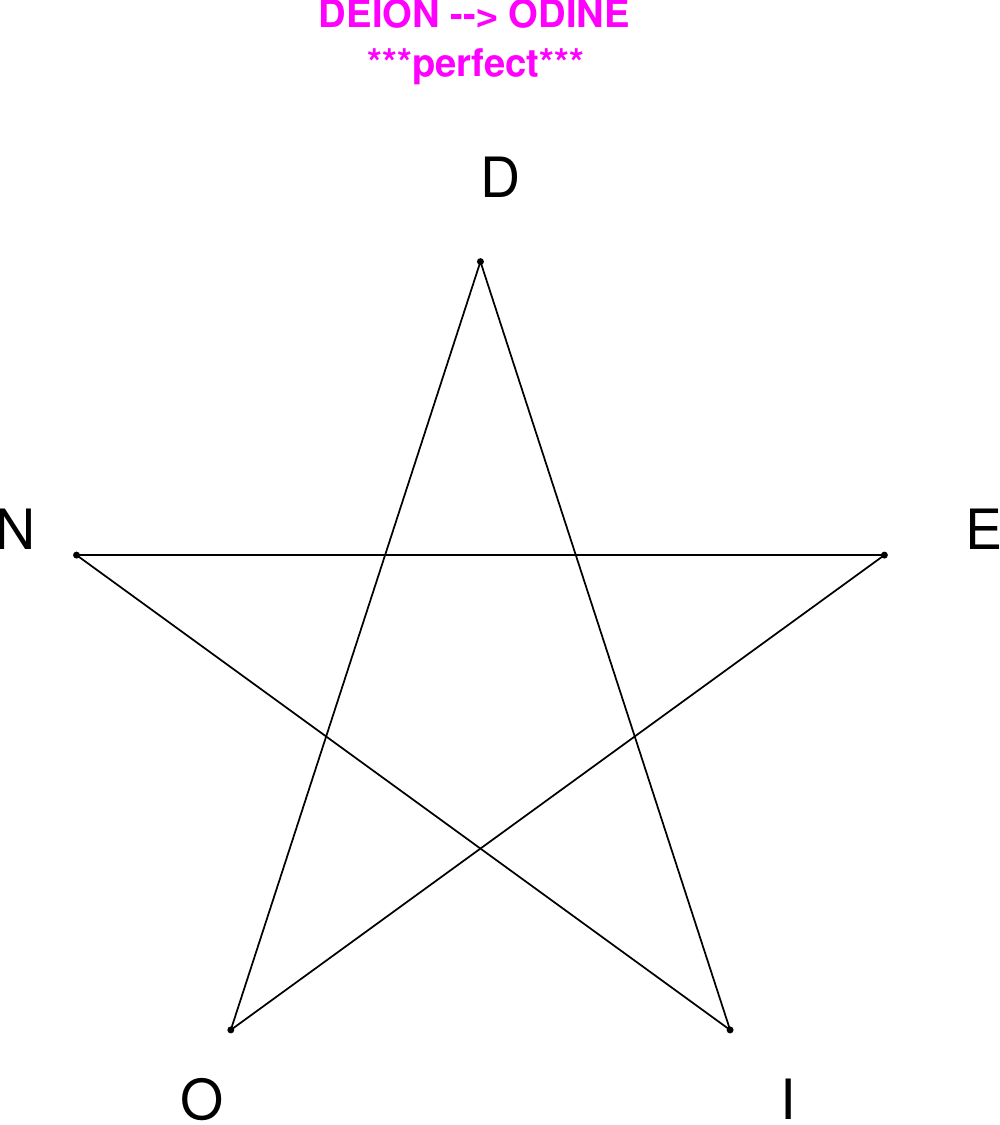}
\end{subfigure}
\hfill
\begin{subfigure}[T]{0.19\textwidth}
\centering
\includegraphics[width=\textwidth]{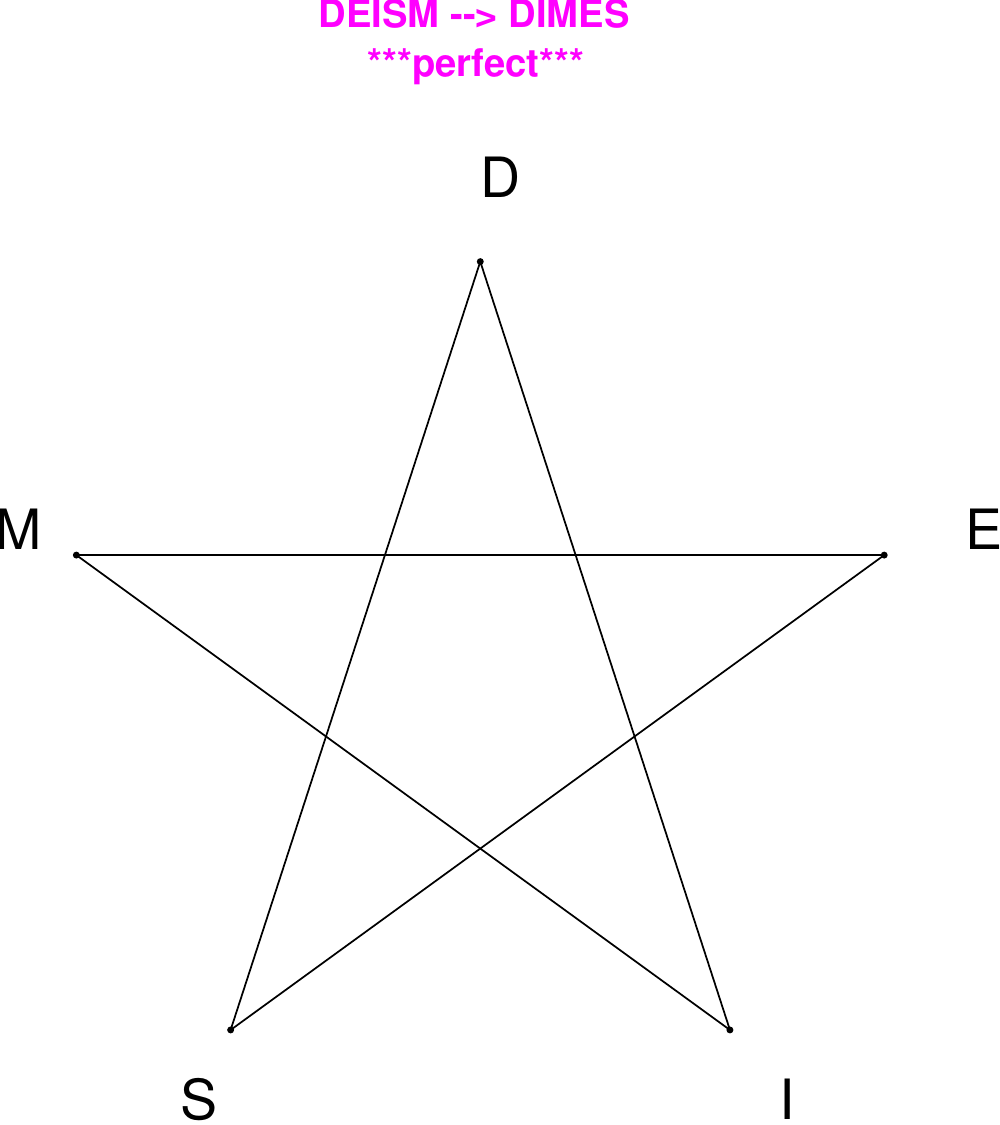}
\end{subfigure}
\hfill
\begin{subfigure}[T]{0.19\textwidth}
\centering
\includegraphics[width=\textwidth]{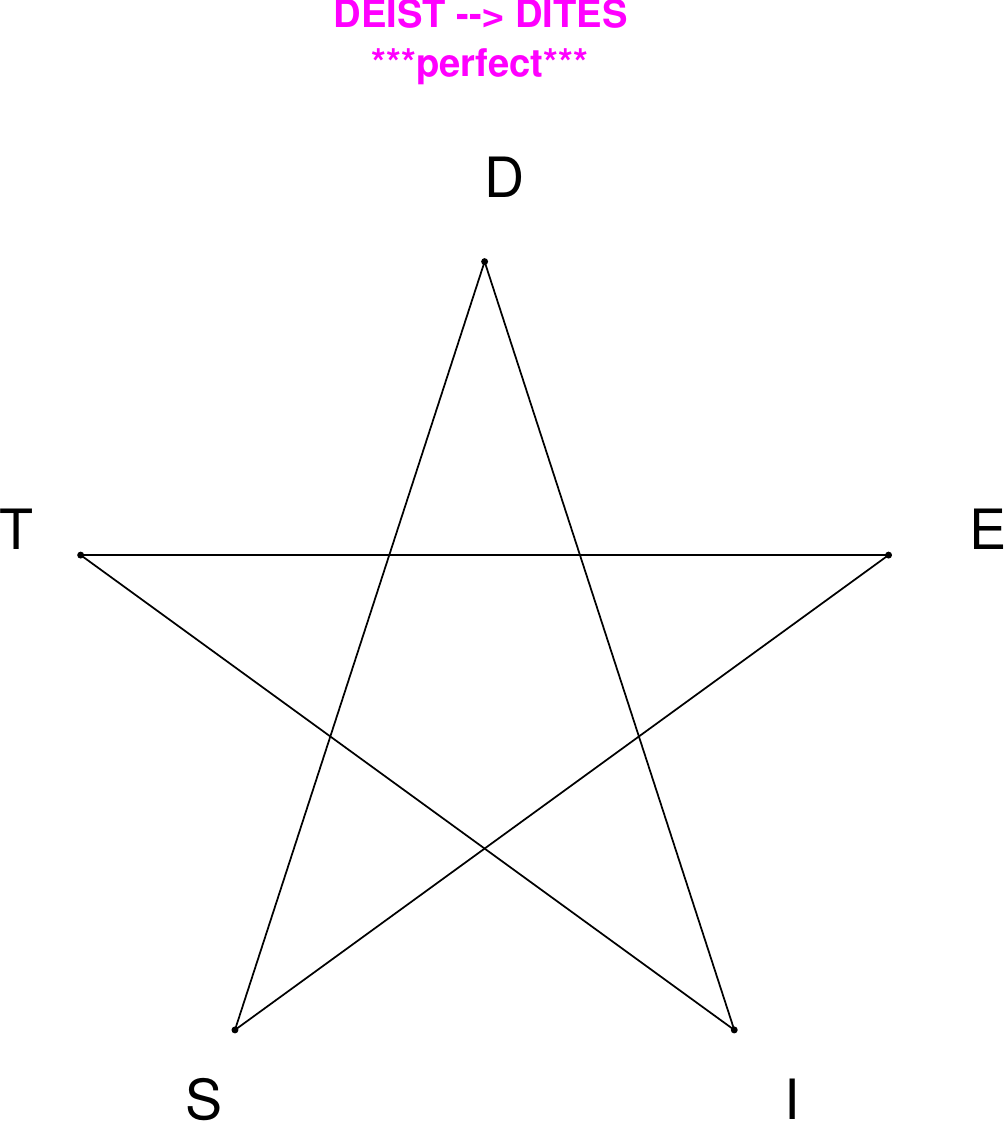}
\end{subfigure}
\hfill
\begin{subfigure}[T]{0.19\textwidth}
\centering
\includegraphics[width=\textwidth]{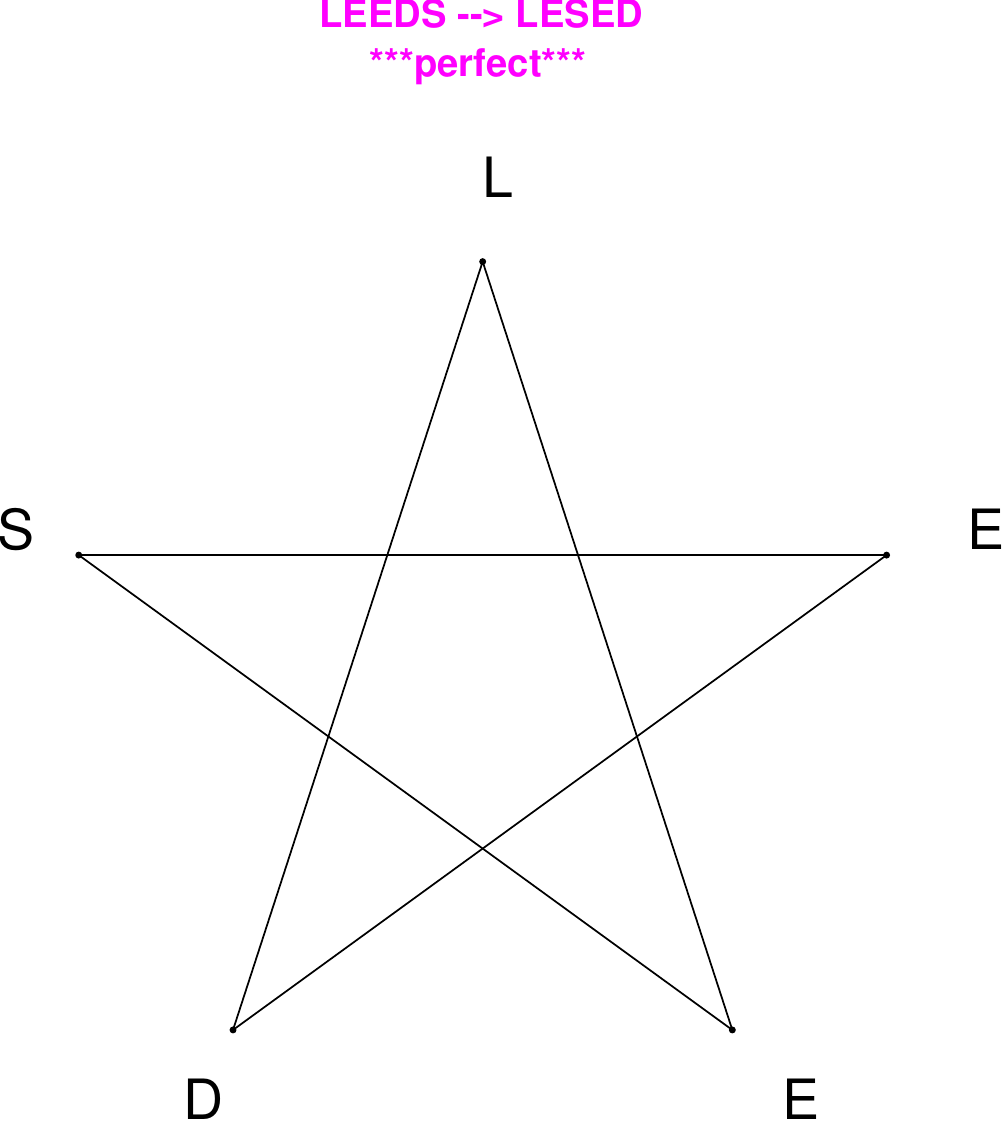}
\end{subfigure}
\hfill
\begin{subfigure}[T]{0.19\textwidth}
\centering
\includegraphics[width=\textwidth]{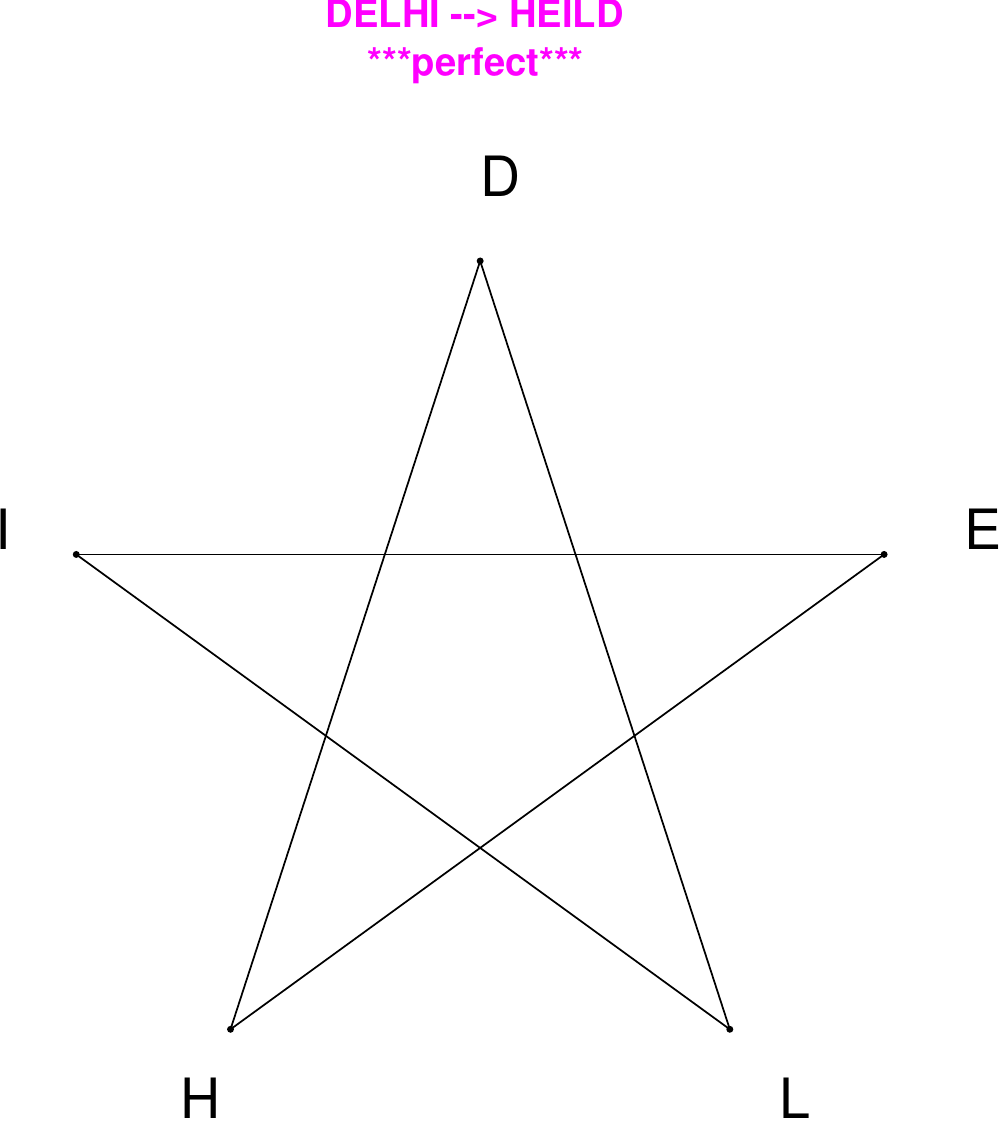}
\end{subfigure}
\end{figure}

\begin{figure}[H]
\centering
\begin{subfigure}[T]{0.19\textwidth}
\centering
\includegraphics[width=\textwidth]{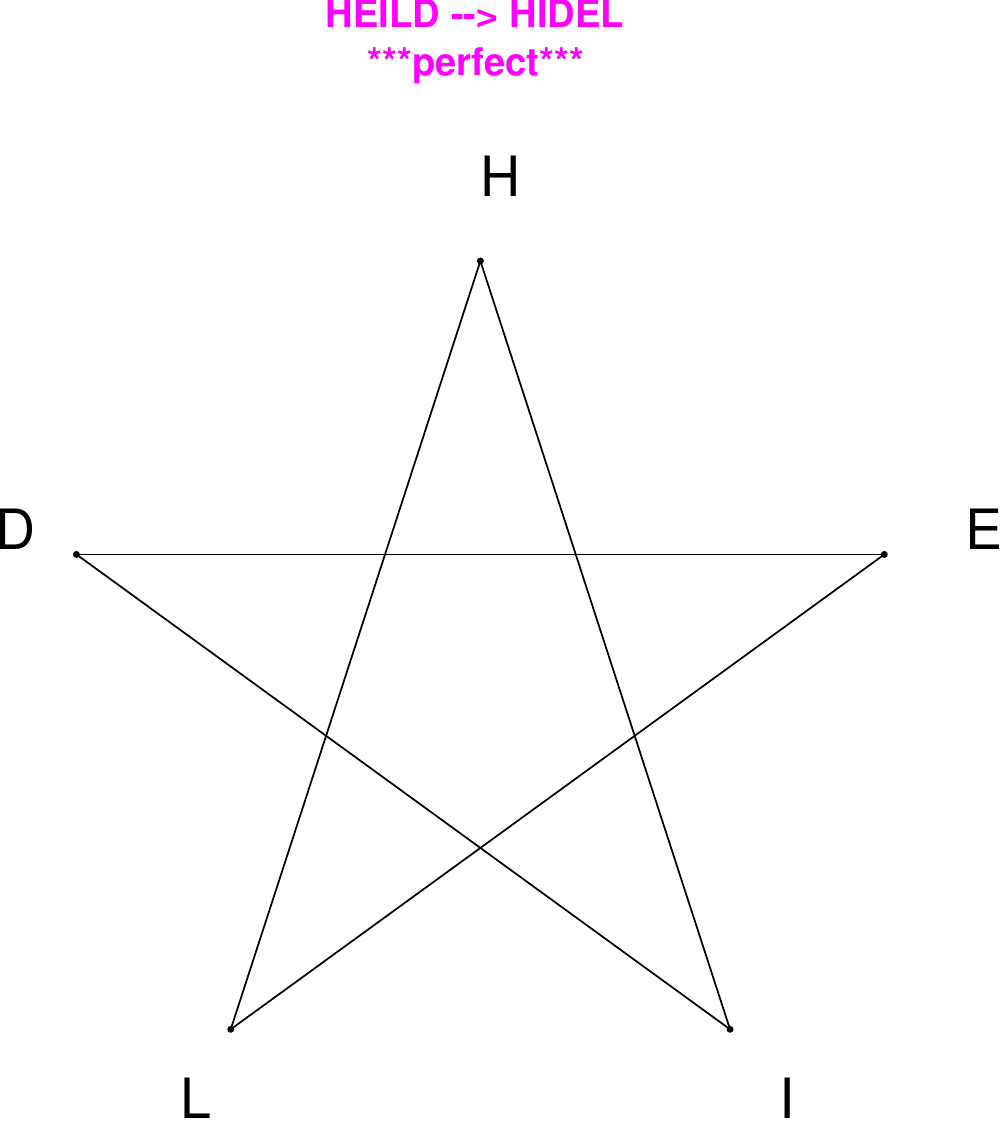}
\end{subfigure}
\hfill
\begin{subfigure}[T]{0.19\textwidth}
\centering
\includegraphics[width=\textwidth]{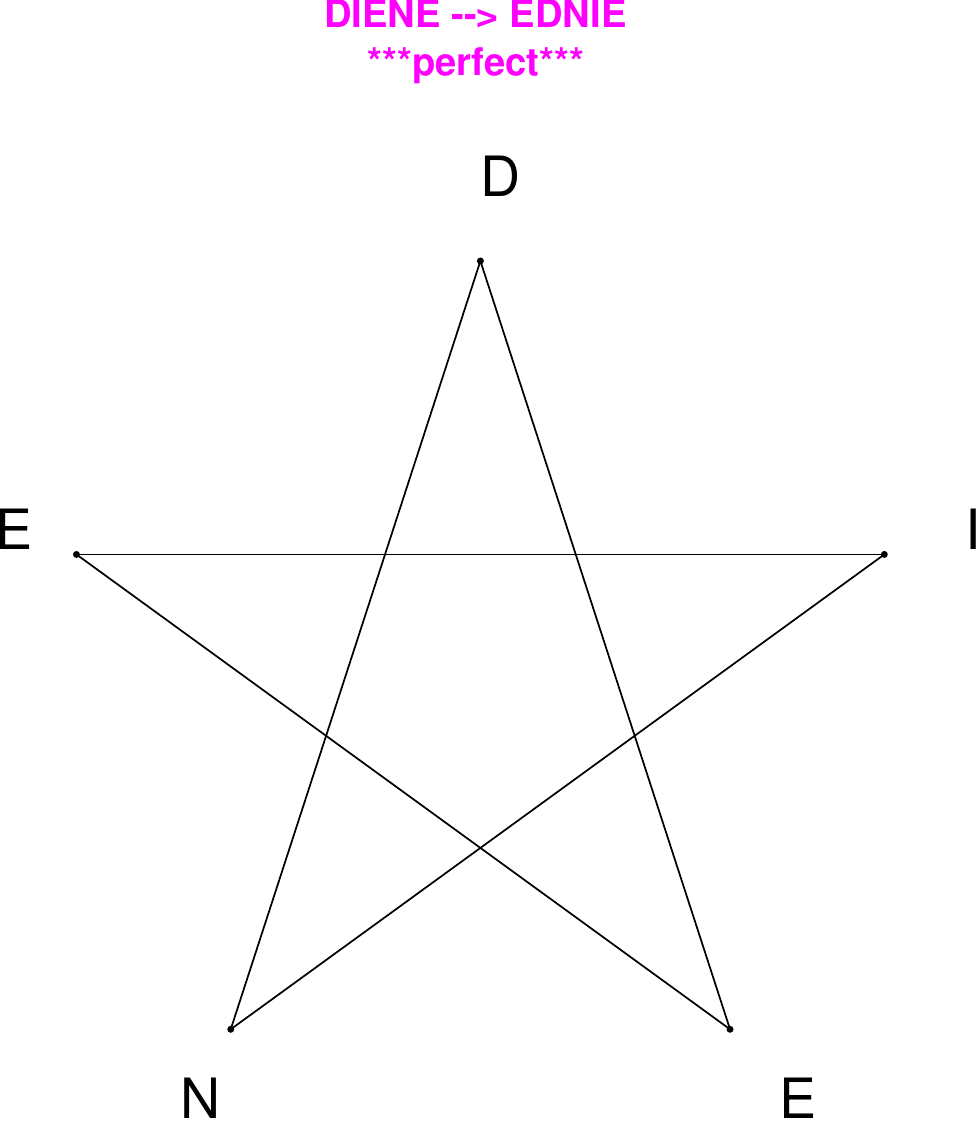}
\end{subfigure}
\hfill
\begin{subfigure}[T]{0.19\textwidth}
\centering
\includegraphics[width=\textwidth]{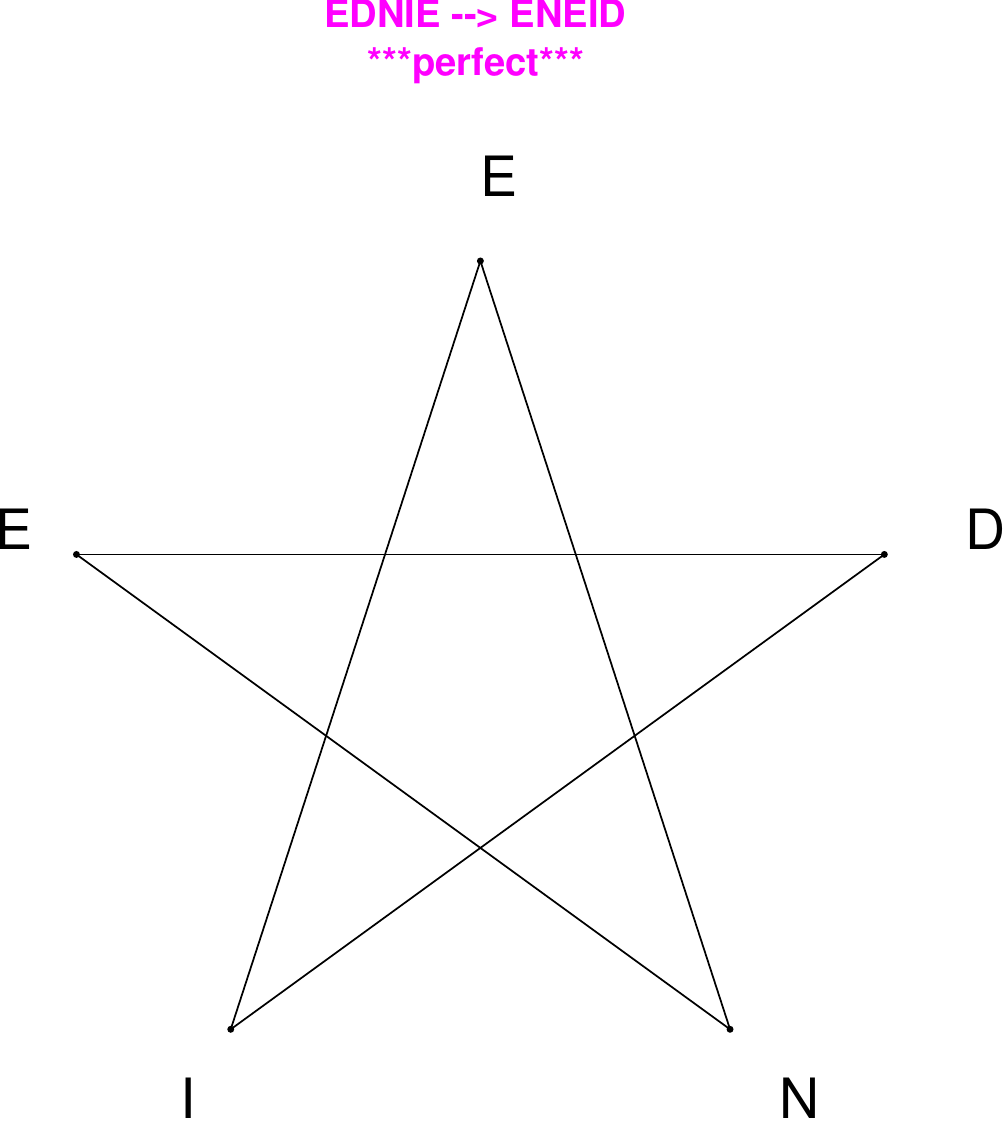}
\end{subfigure}
\hfill
\begin{subfigure}[T]{0.19\textwidth}
\centering
\includegraphics[width=\textwidth]{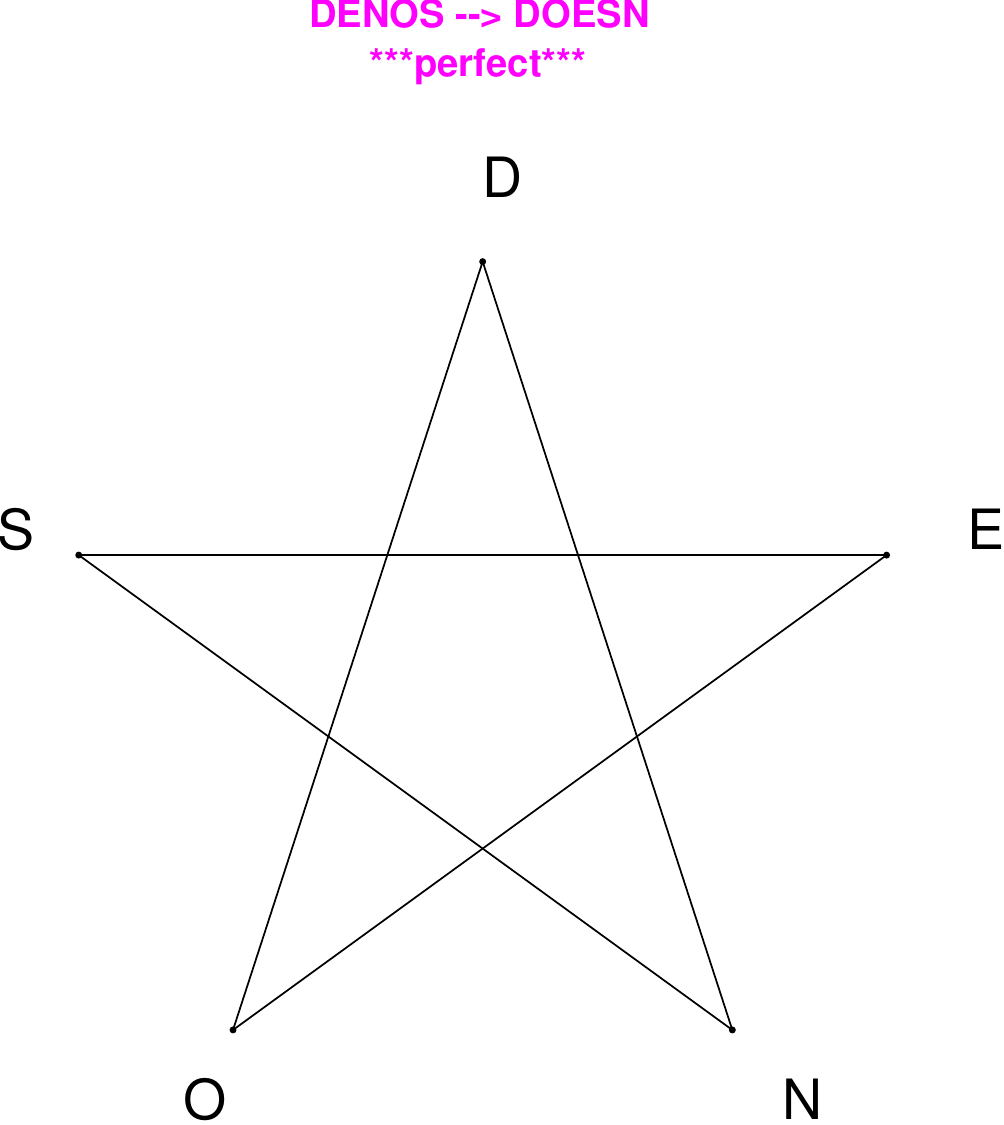}
\end{subfigure}
\hfill
\begin{subfigure}[T]{0.19\textwidth}
\centering
\includegraphics[width=\textwidth]{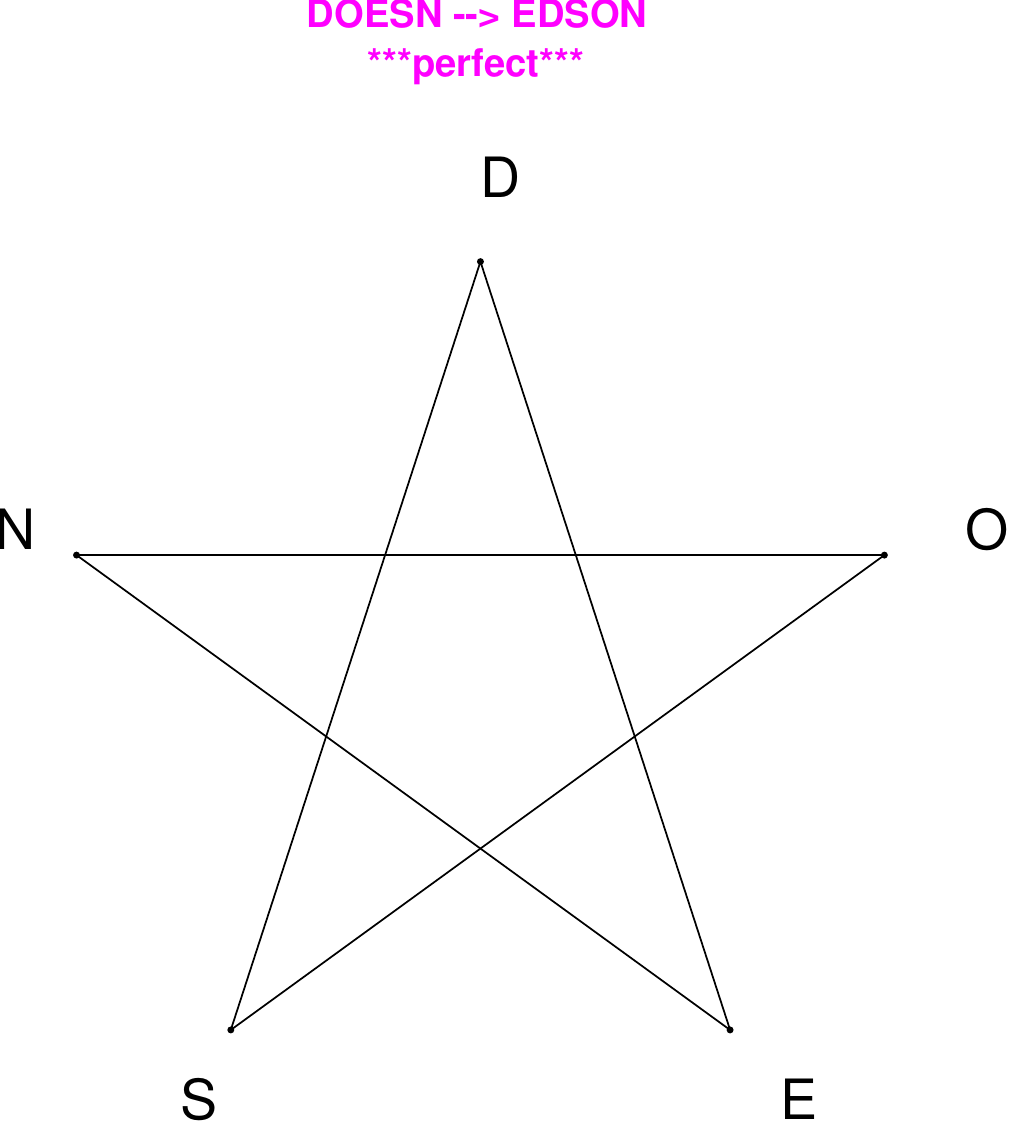}
\end{subfigure}
\end{figure}

\begin{figure}[H]
\centering
\begin{subfigure}[T]{0.19\textwidth}
\centering
\includegraphics[width=\textwidth]{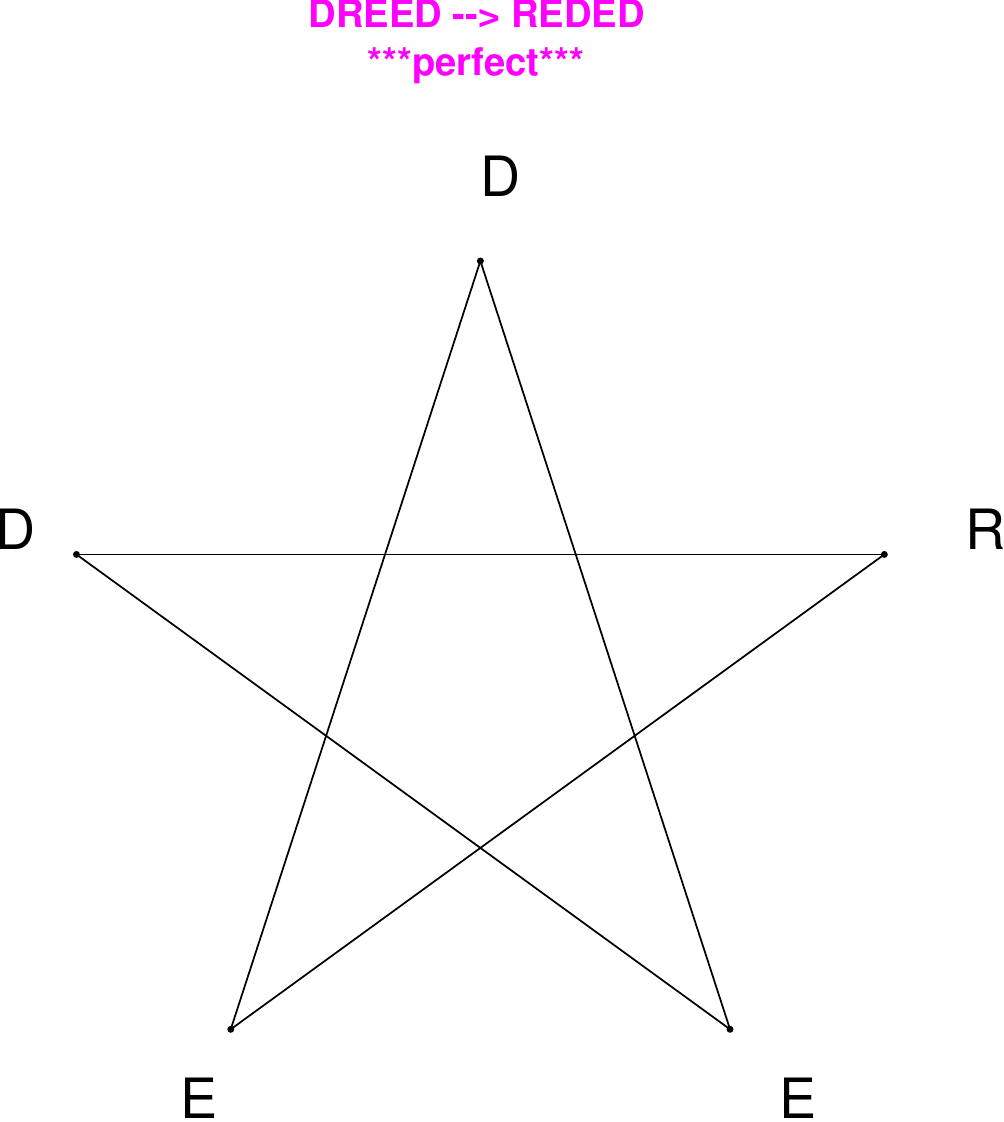}
\end{subfigure}
\hfill
\begin{subfigure}[T]{0.19\textwidth}
\centering
\includegraphics[width=\textwidth]{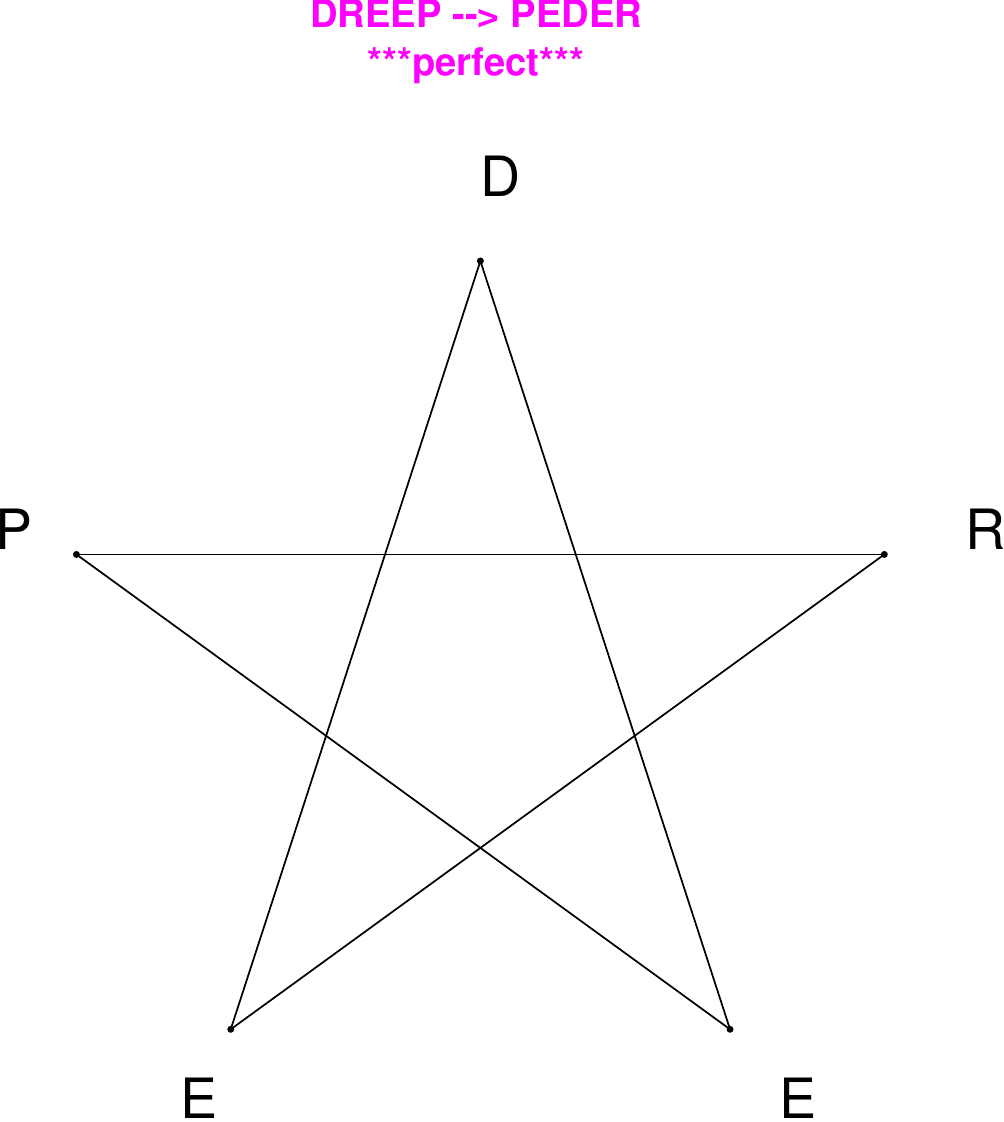}
\end{subfigure}
\hfill
\begin{subfigure}[T]{0.19\textwidth}
\centering
\includegraphics[width=\textwidth]{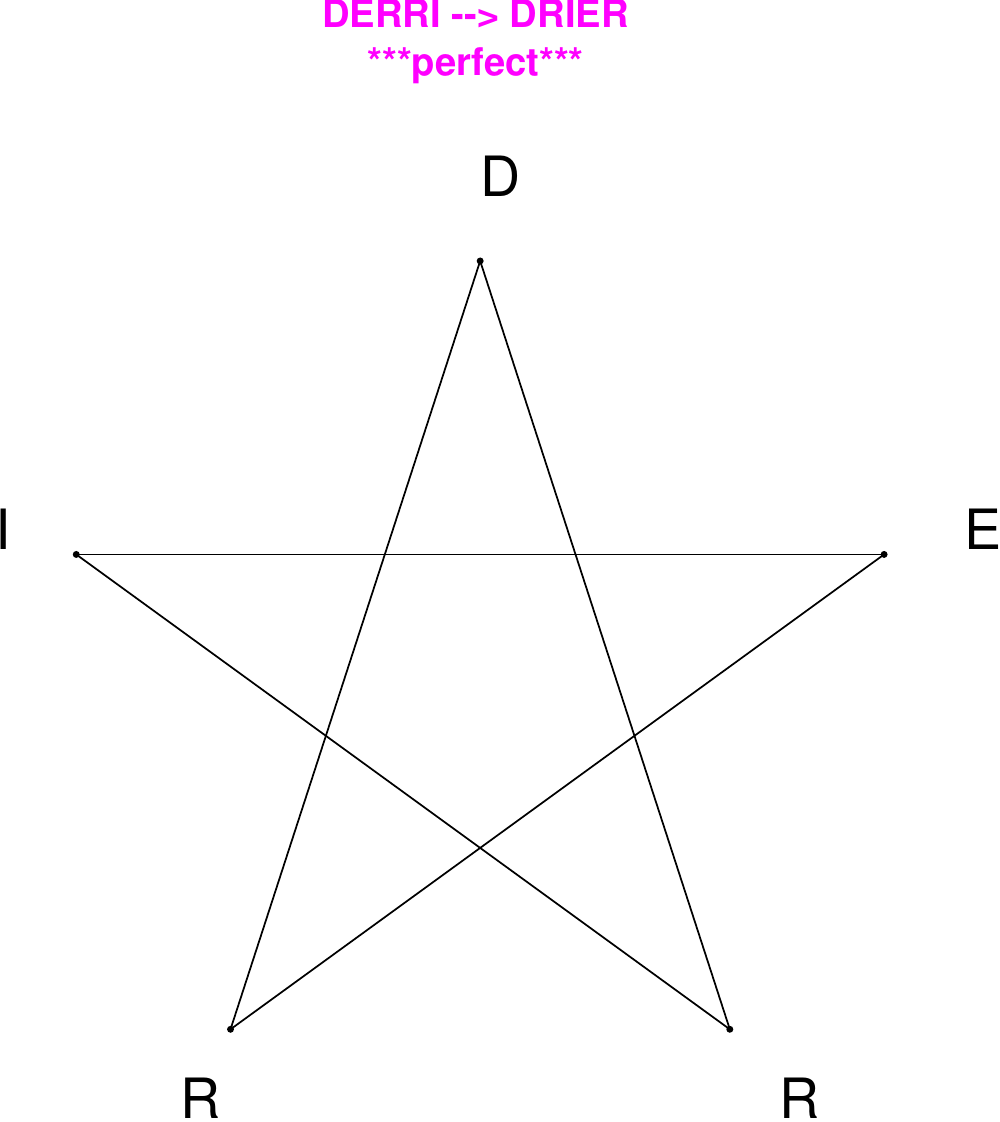}
\end{subfigure}
\hfill
\begin{subfigure}[T]{0.19\textwidth}
\centering
\includegraphics[width=\textwidth]{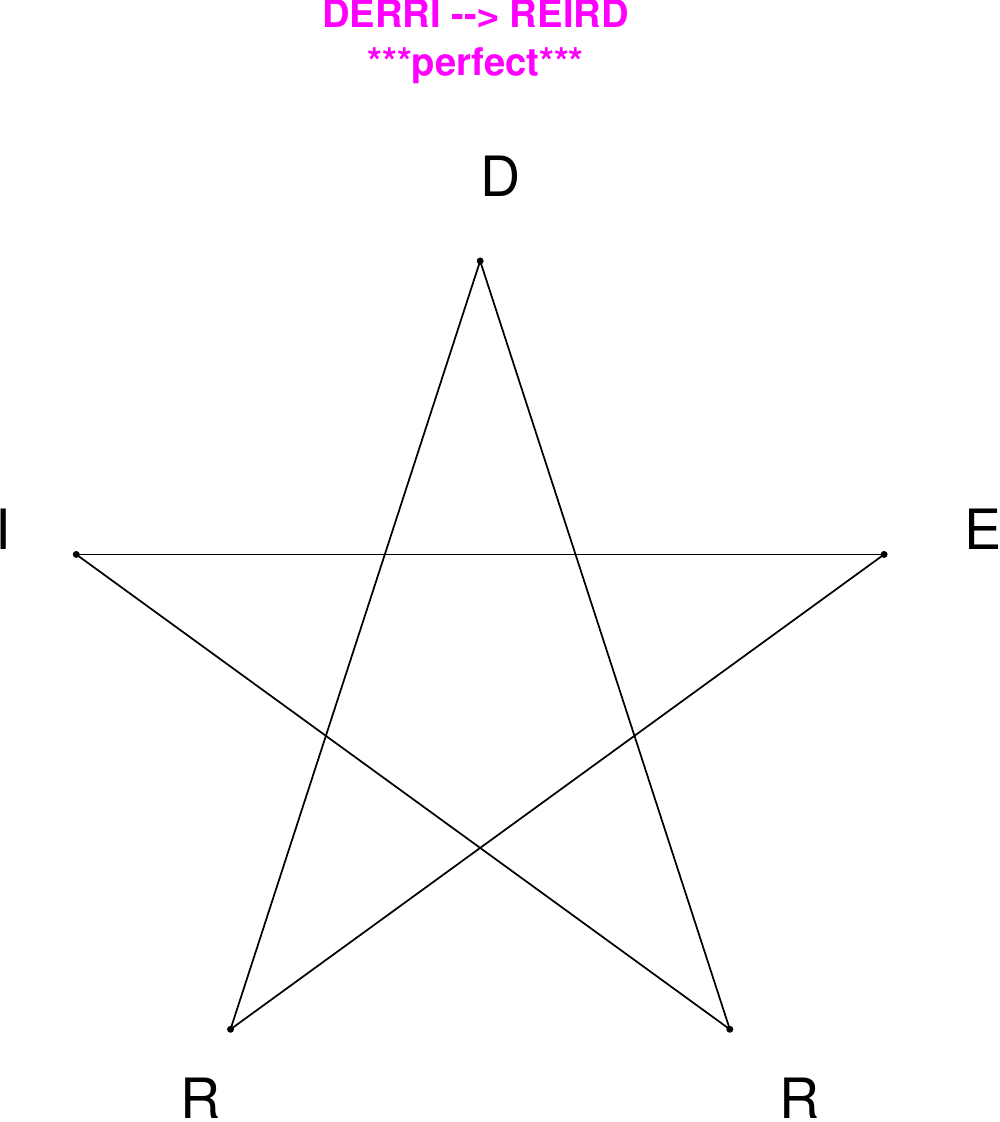}
\end{subfigure}
\hfill
\begin{subfigure}[T]{0.19\textwidth}
\centering
\includegraphics[width=\textwidth]{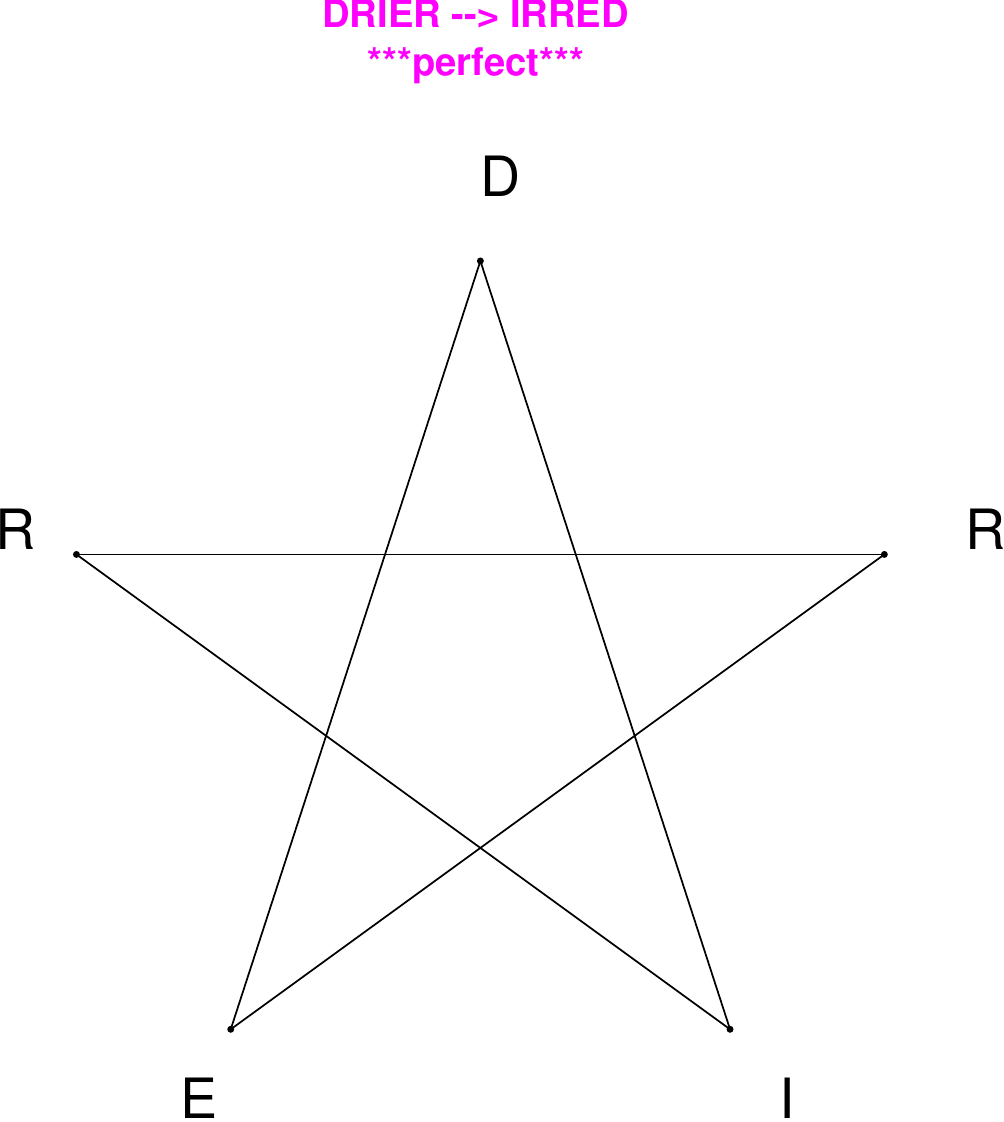}
\end{subfigure}
\end{figure}

\begin{figure}[H]
\centering
\begin{subfigure}[T]{0.19\textwidth}
\centering
\includegraphics[width=\textwidth]{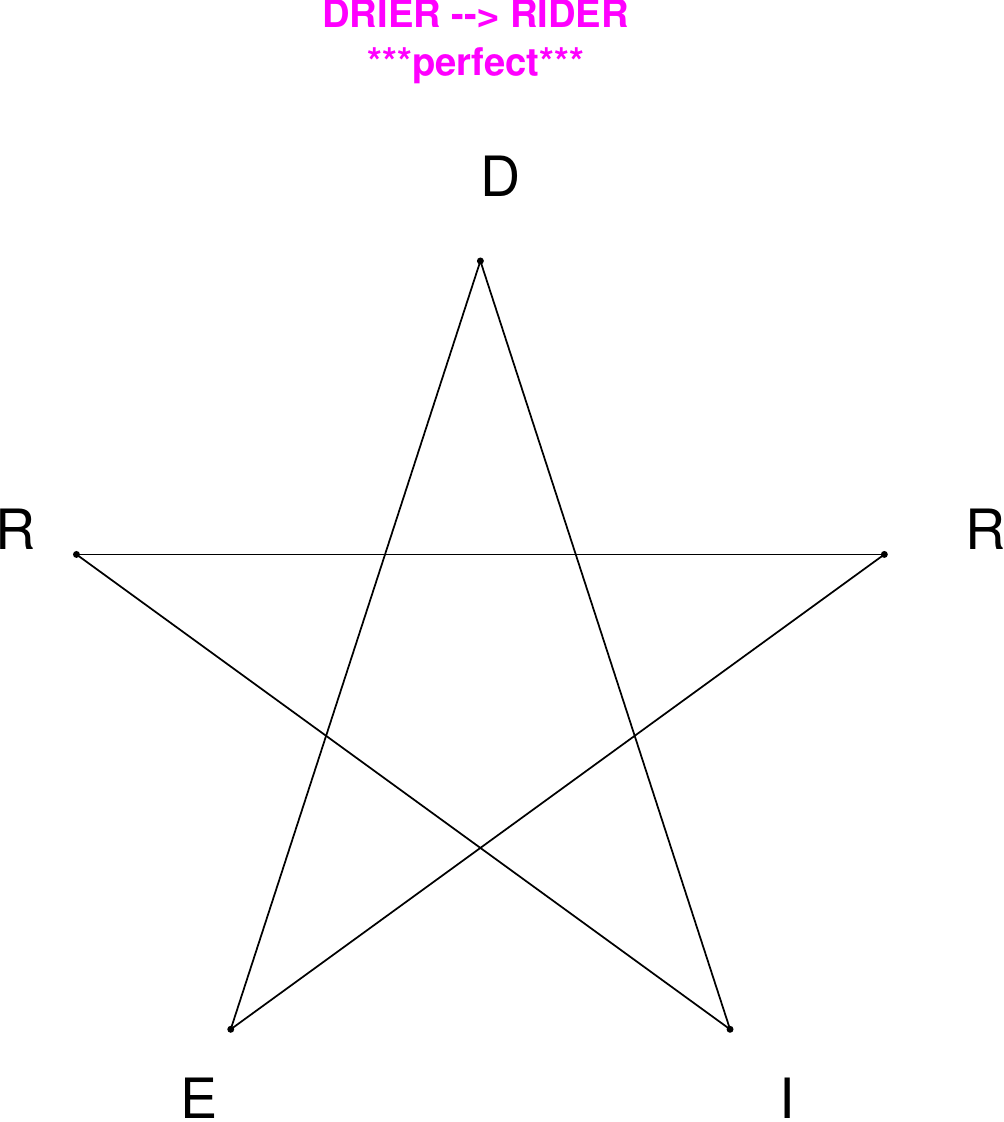}
\end{subfigure}
\hfill
\begin{subfigure}[T]{0.19\textwidth}
\centering
\includegraphics[width=\textwidth]{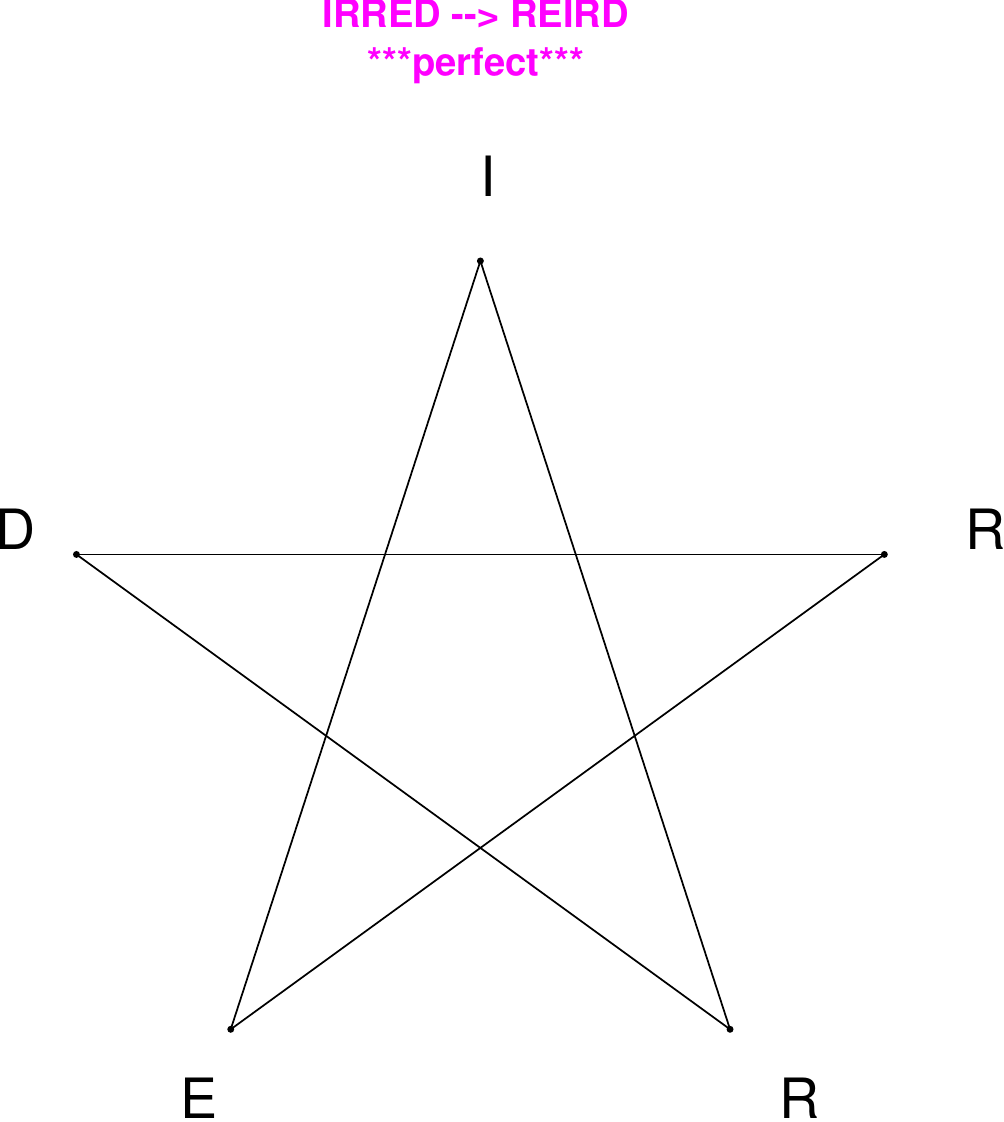}
\end{subfigure}
\hfill
\begin{subfigure}[T]{0.19\textwidth}
\centering
\includegraphics[width=\textwidth]{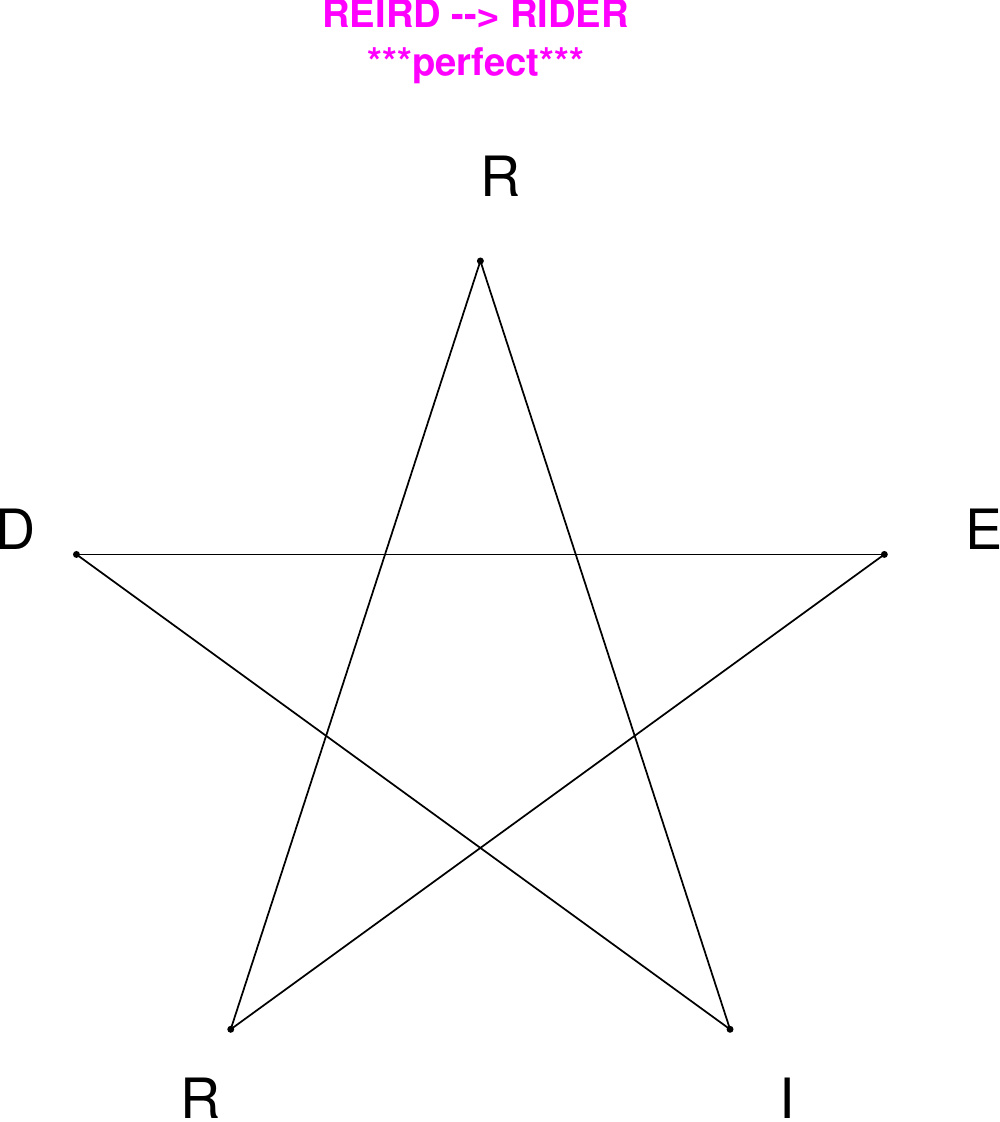}
\end{subfigure}
\hfill
\begin{subfigure}[T]{0.19\textwidth}
\centering
\includegraphics[width=\textwidth]{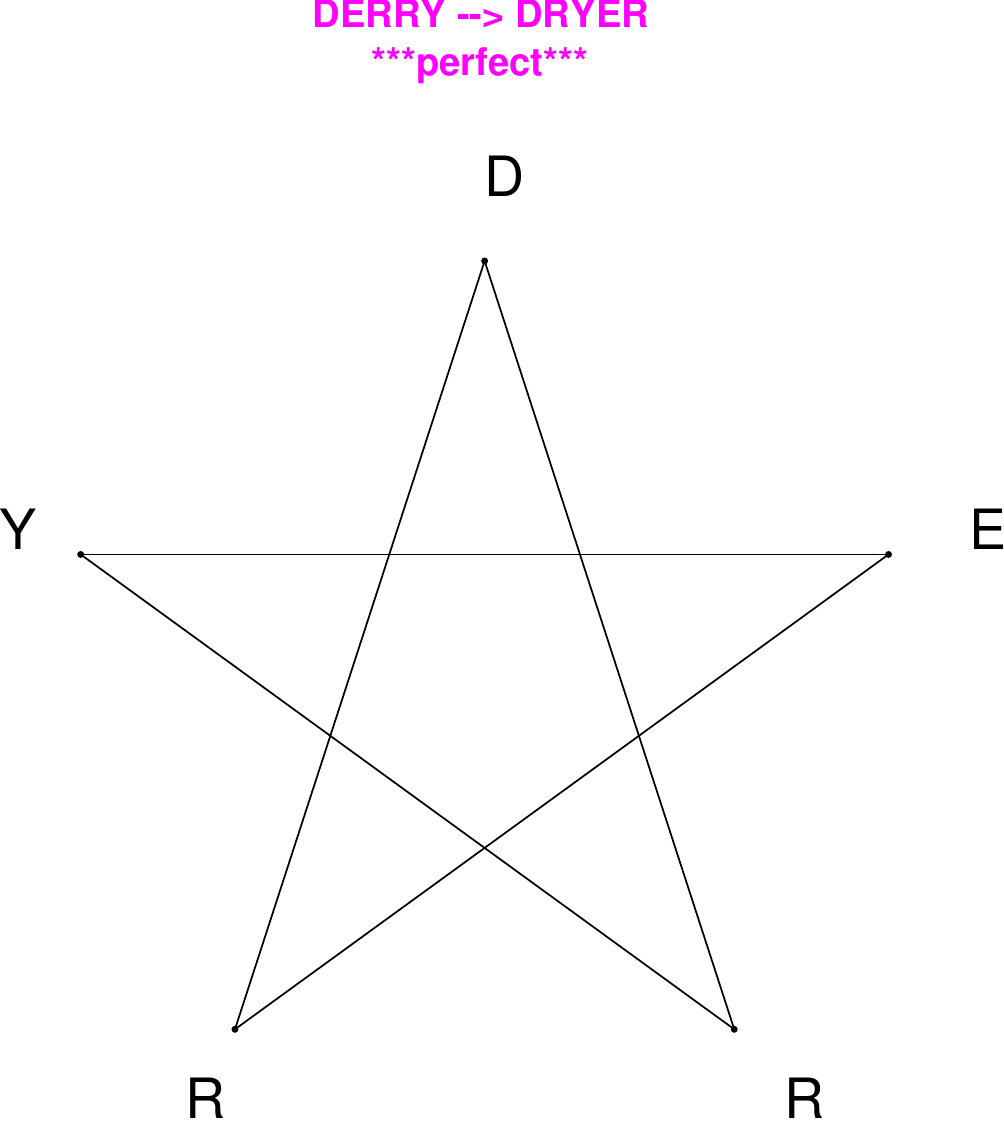}
\end{subfigure}
\hfill
\begin{subfigure}[T]{0.19\textwidth}
\centering
\includegraphics[width=\textwidth]{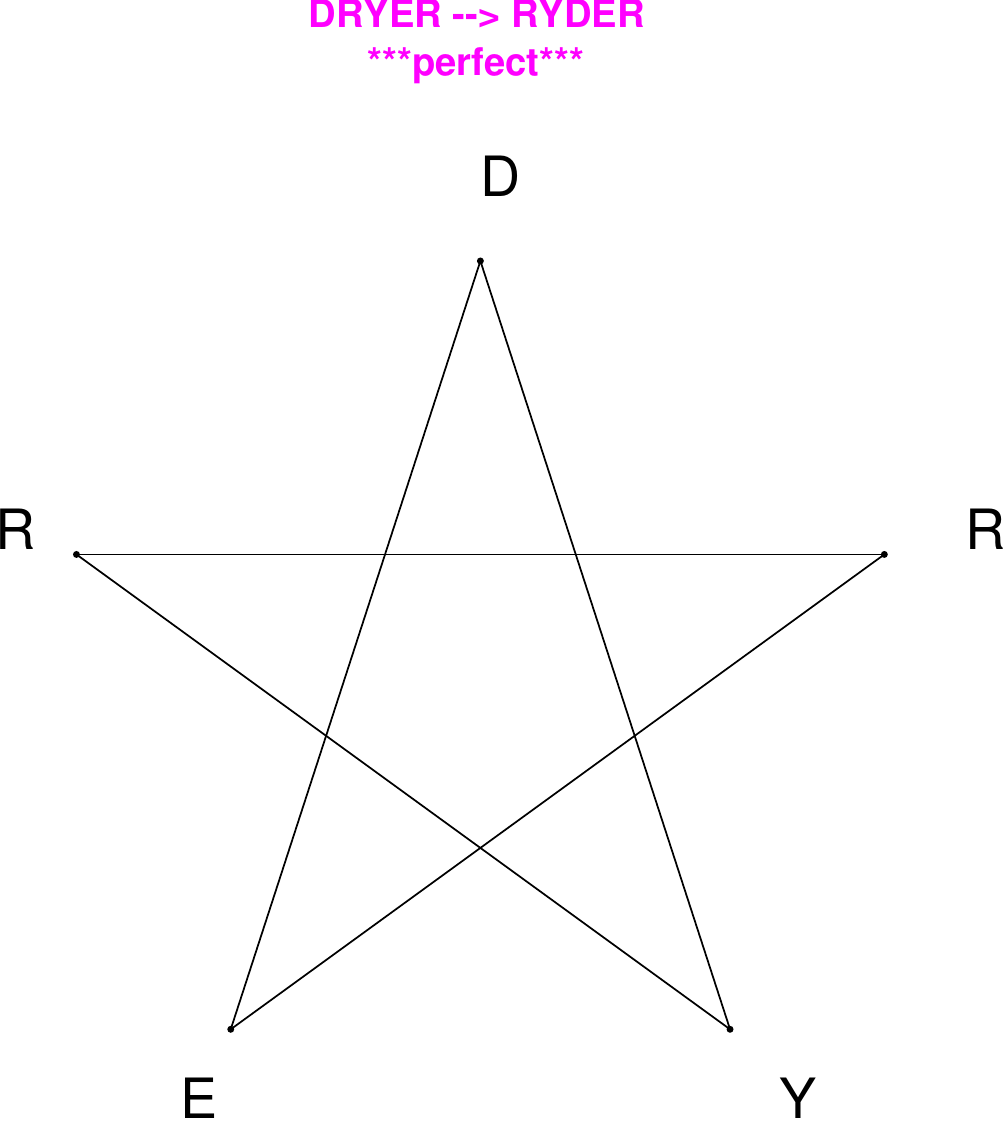}
\end{subfigure}
\end{figure}

\begin{figure}[H]
\centering
\begin{subfigure}[T]{0.19\textwidth}
\centering
\includegraphics[width=\textwidth]{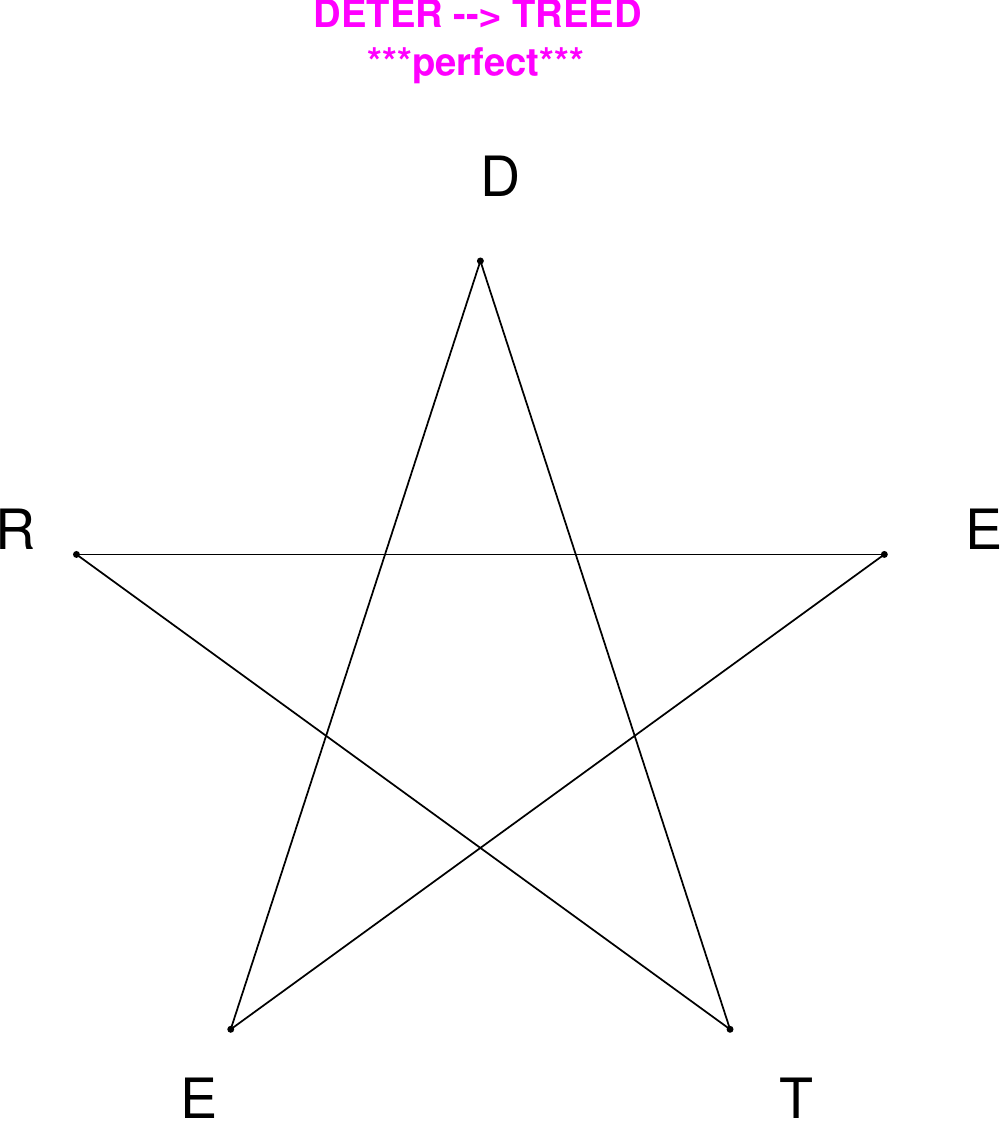}
\end{subfigure}
\hfill
\begin{subfigure}[T]{0.19\textwidth}
\centering
\includegraphics[width=\textwidth]{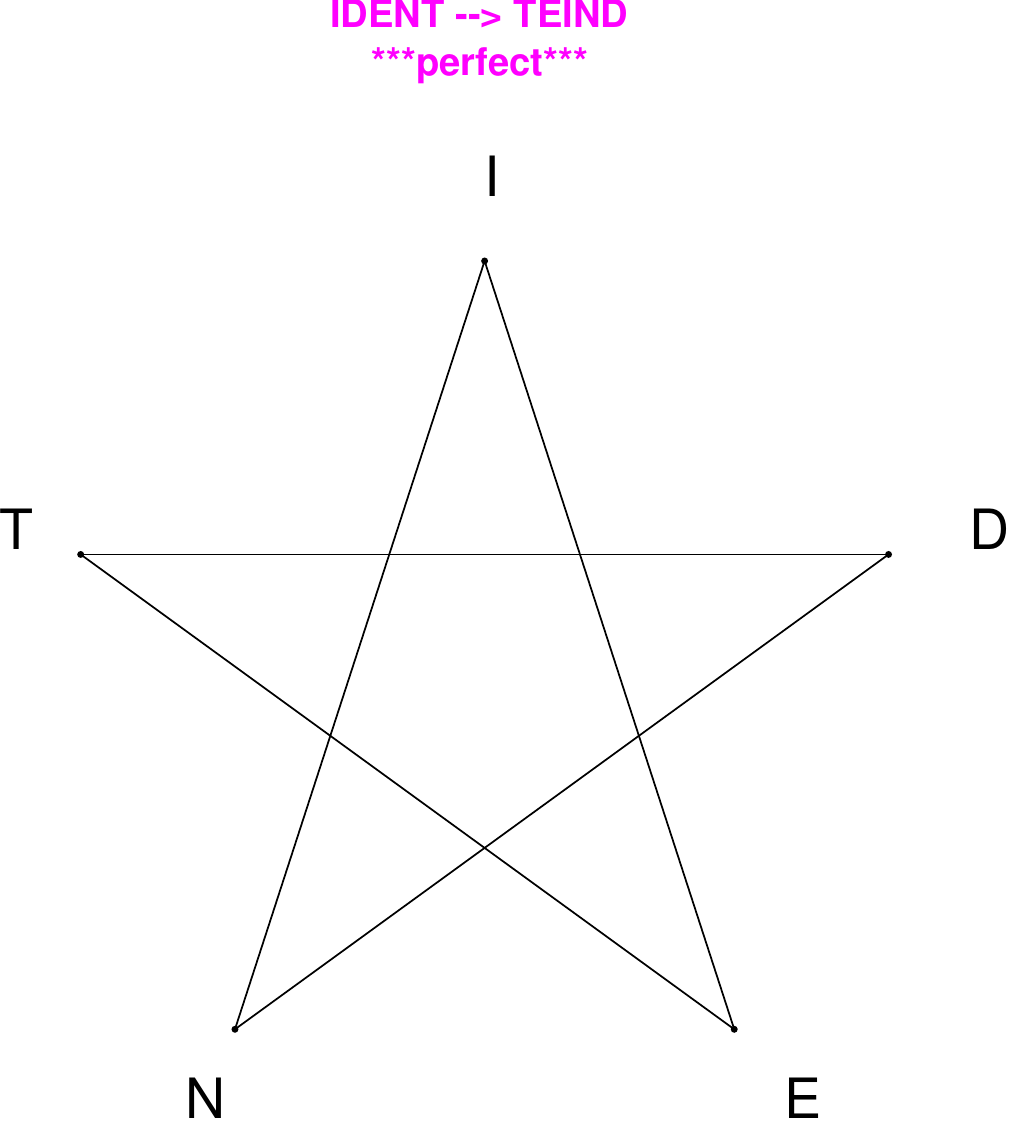}
\end{subfigure}
\hfill
\begin{subfigure}[T]{0.19\textwidth}
\centering
\includegraphics[width=\textwidth]{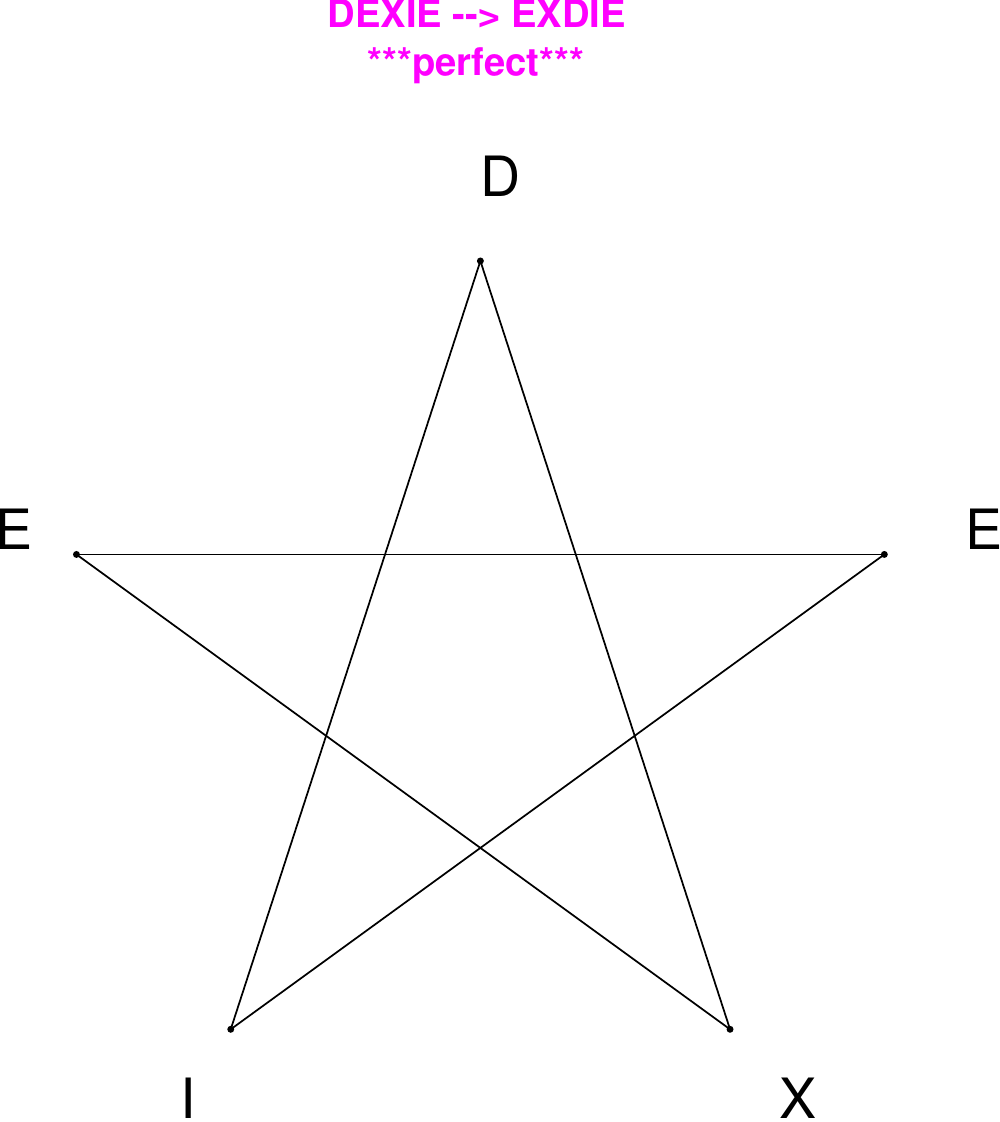}
\end{subfigure}
\hfill
\begin{subfigure}[T]{0.19\textwidth}
\centering
\includegraphics[width=\textwidth]{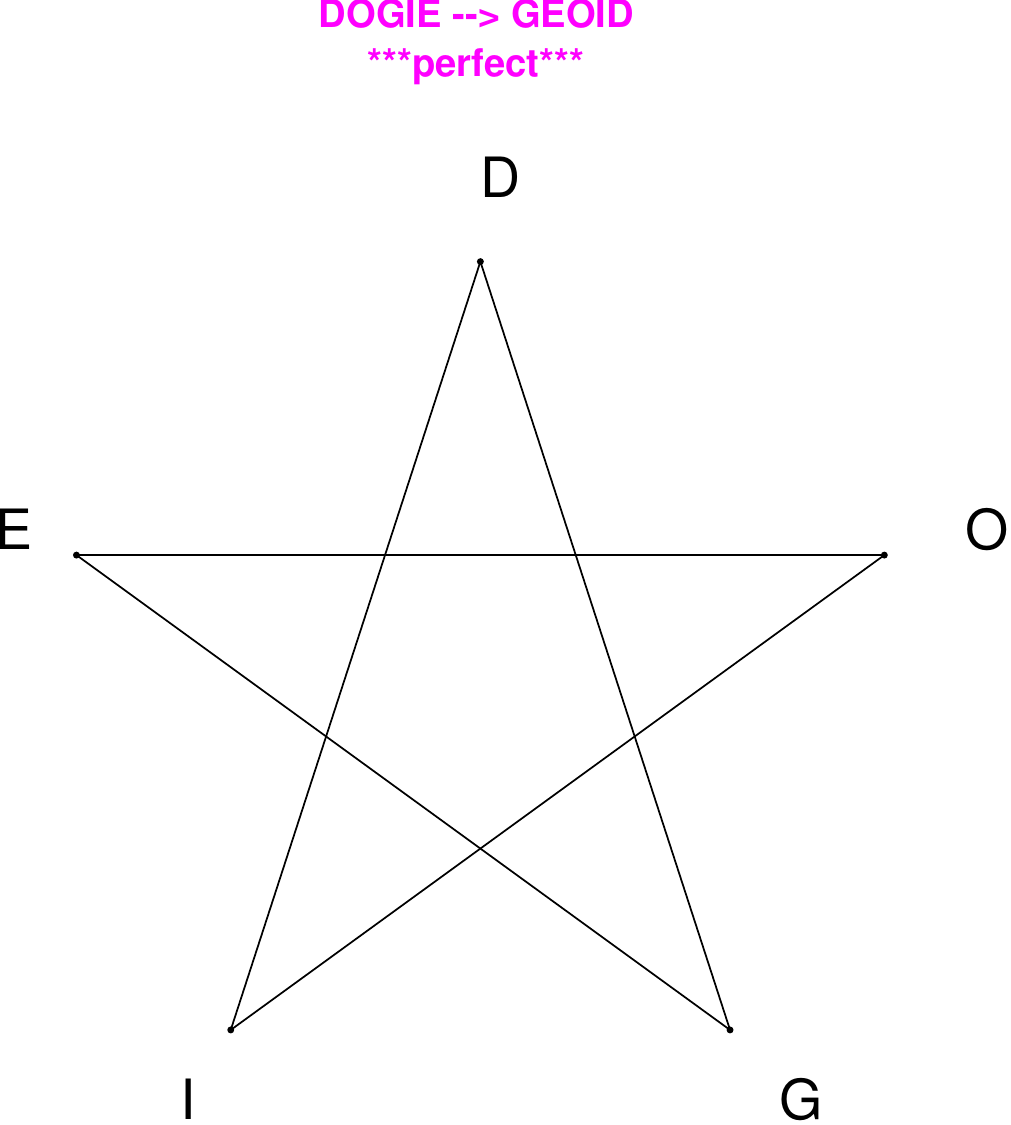}
\end{subfigure}
\hfill
\begin{subfigure}[T]{0.19\textwidth}
\centering
\includegraphics[width=\textwidth]{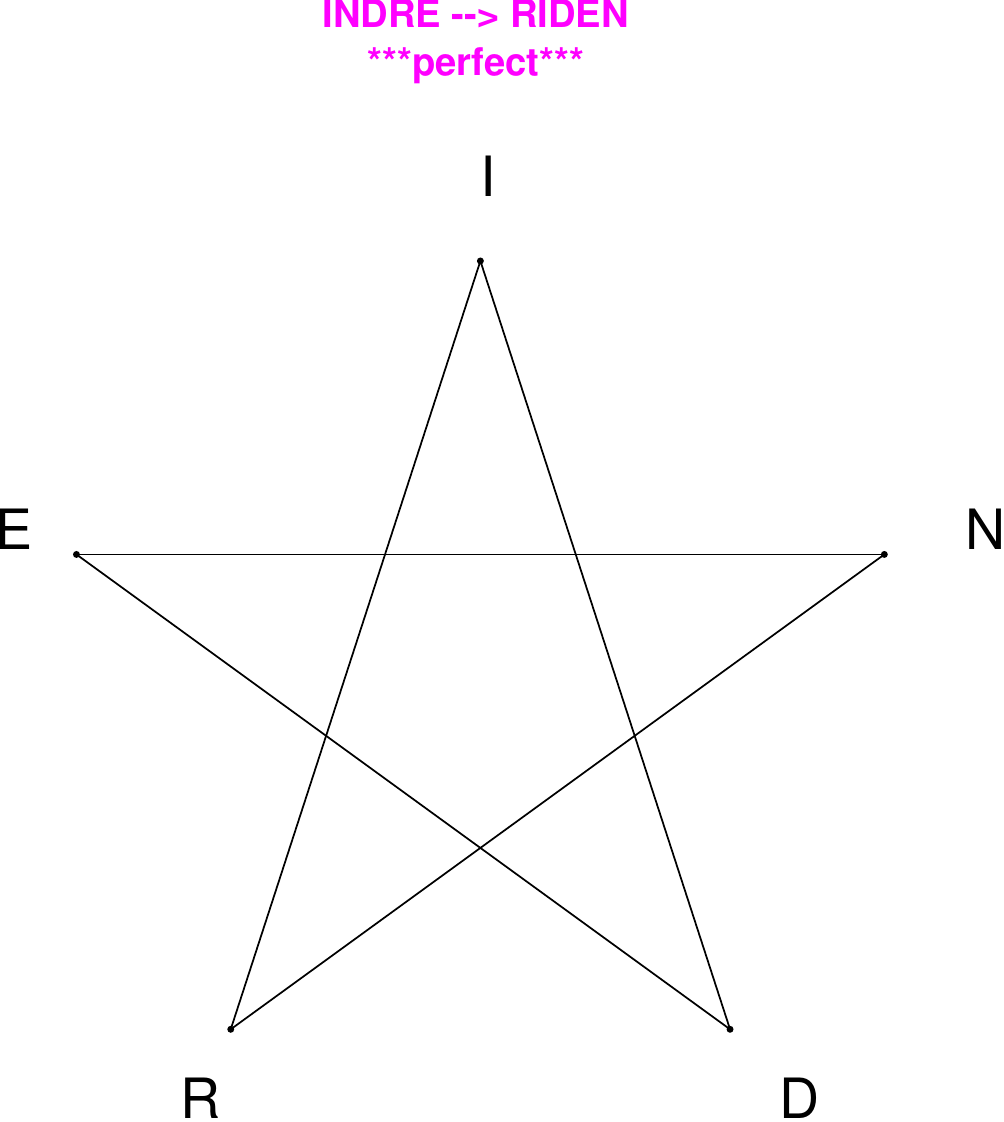}
\end{subfigure}
\end{figure}

\begin{figure}[H]
\centering
\begin{subfigure}[T]{0.19\textwidth}
\centering
\includegraphics[width=\textwidth]{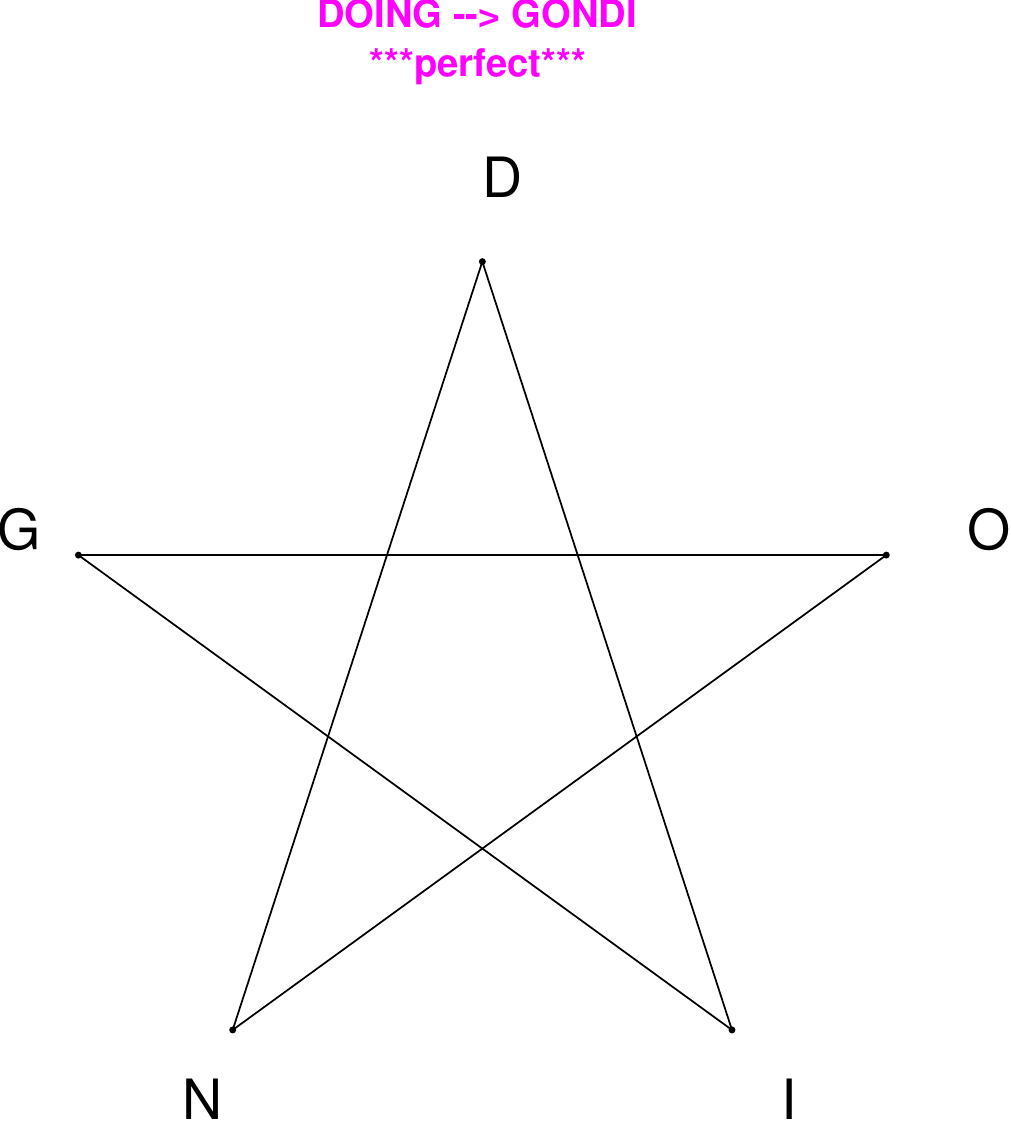}
\end{subfigure}
\hfill
\begin{subfigure}[T]{0.19\textwidth}
\centering
\includegraphics[width=\textwidth]{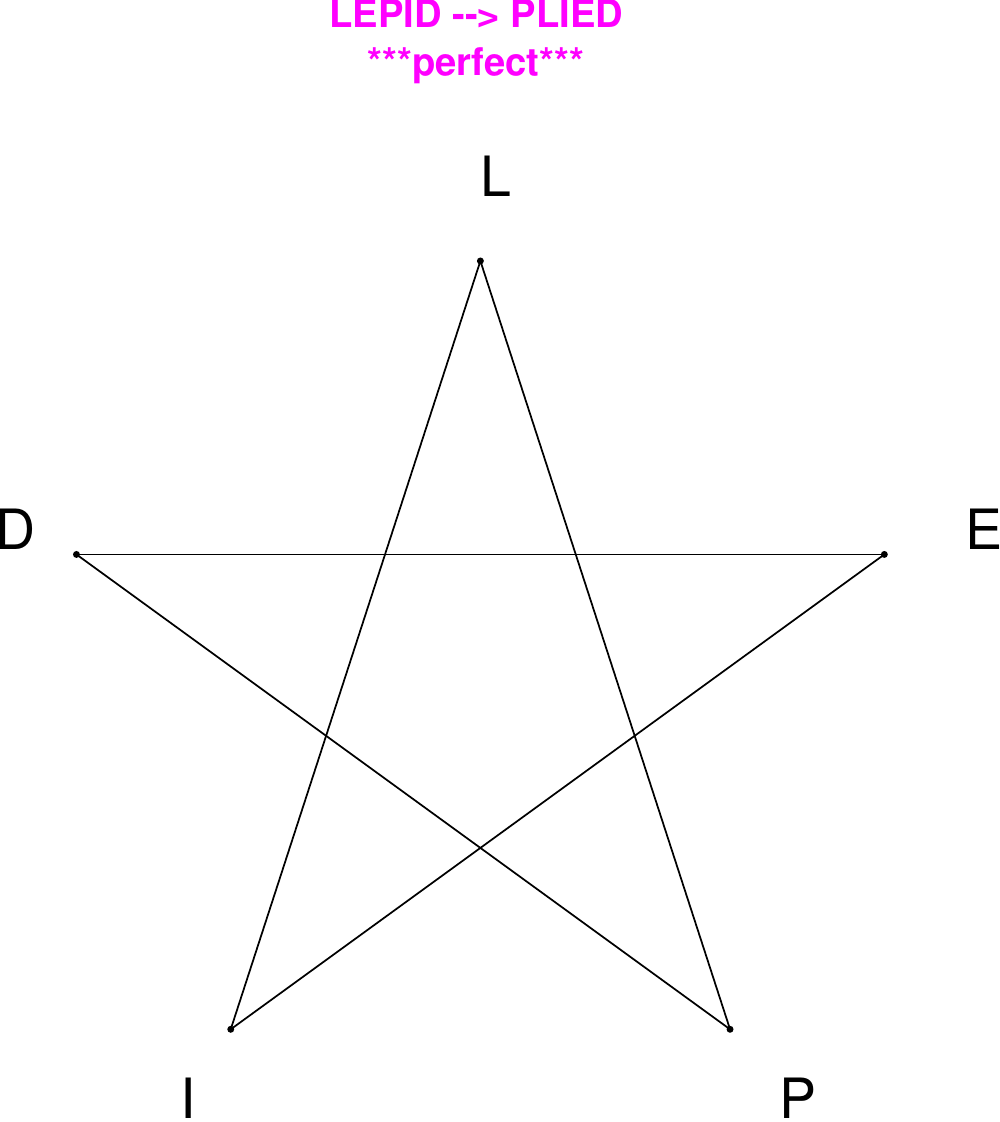}
\end{subfigure}
\hfill
\begin{subfigure}[T]{0.19\textwidth}
\centering
\includegraphics[width=\textwidth]{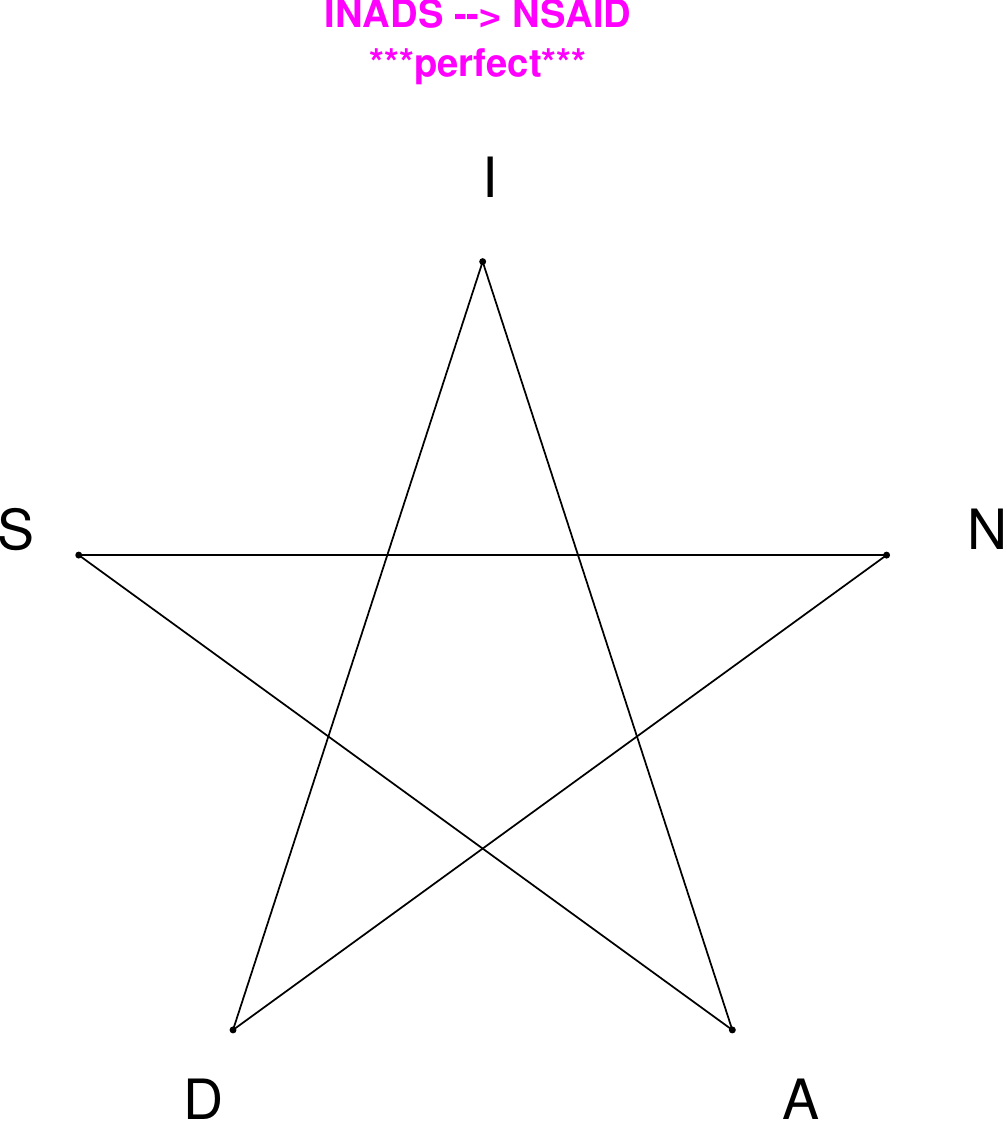}
\end{subfigure}
\hfill
\begin{subfigure}[T]{0.19\textwidth}
\centering
\includegraphics[width=\textwidth]{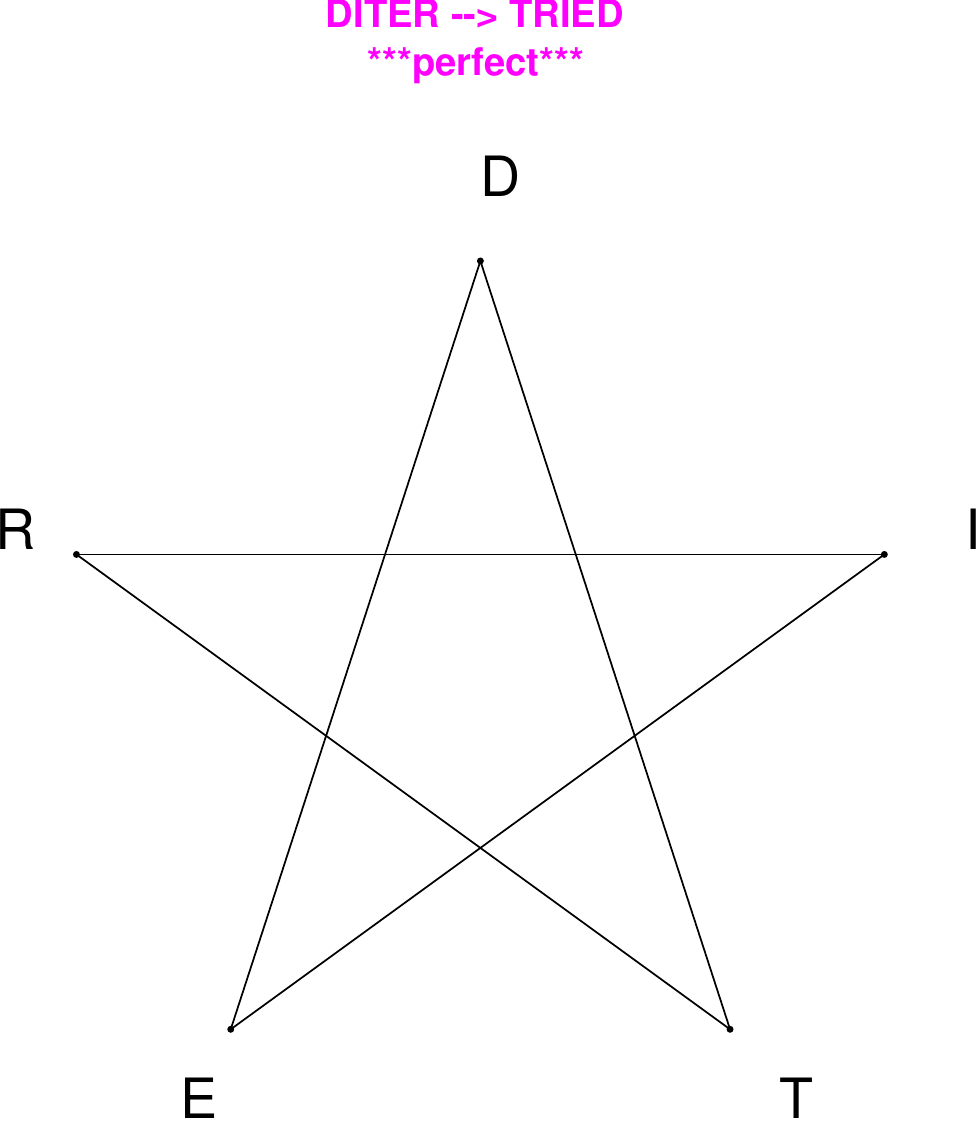}
\end{subfigure}
\hfill
\begin{subfigure}[T]{0.19\textwidth}
\centering
\includegraphics[width=\textwidth]{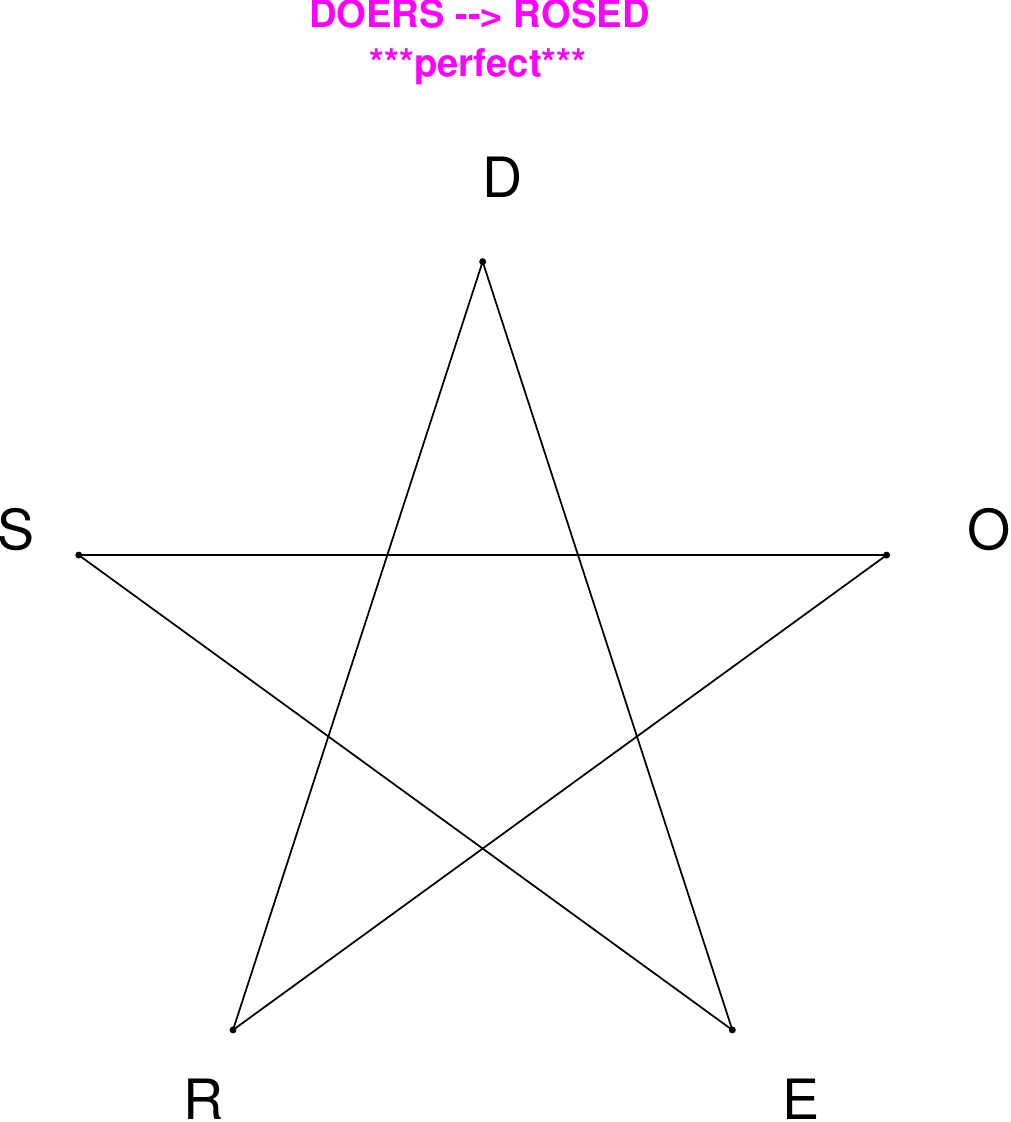}
\end{subfigure}
\end{figure}

\begin{figure}[H]
\centering
\begin{subfigure}[T]{0.19\textwidth}
\centering
\includegraphics[width=\textwidth]{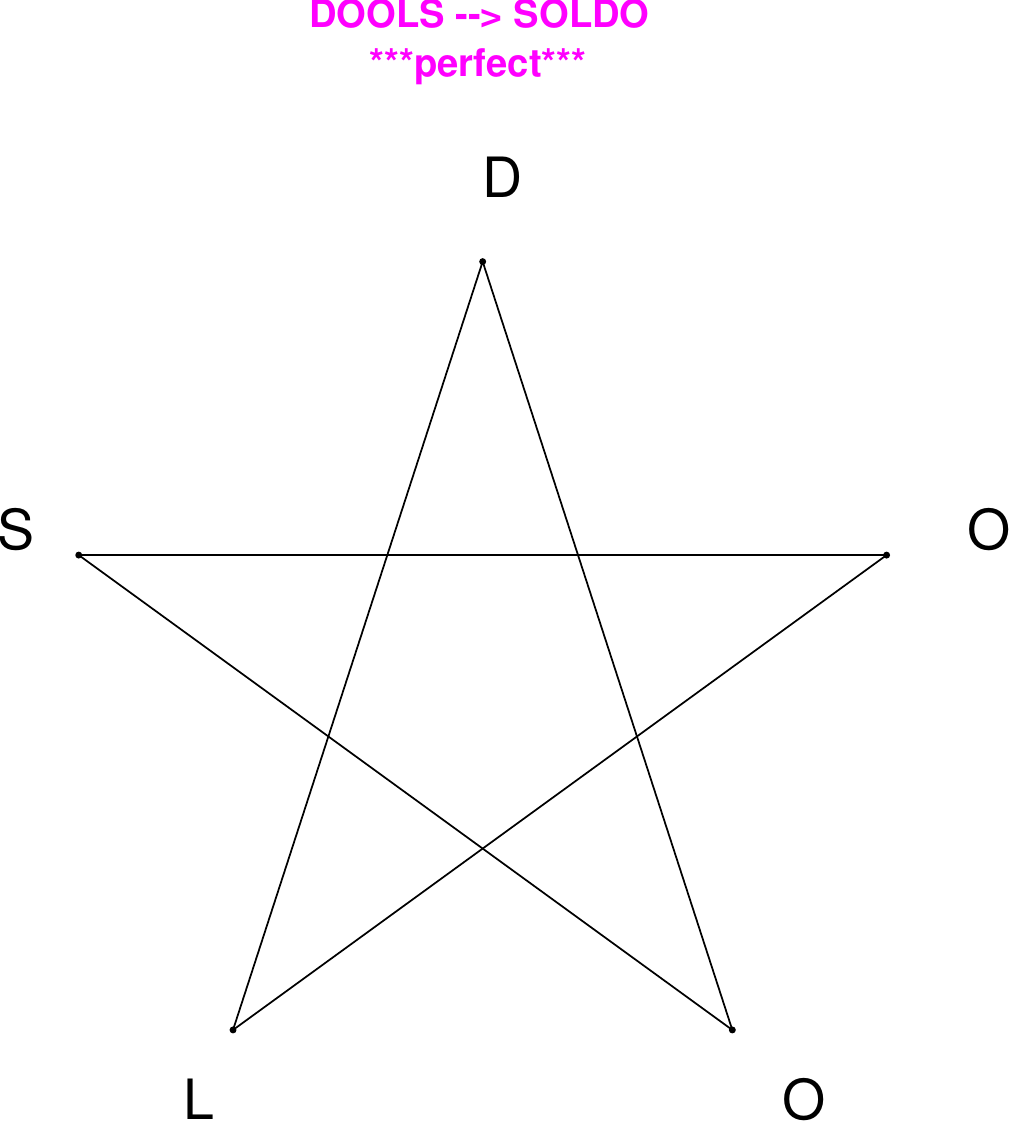}
\end{subfigure}
\hfill
\begin{subfigure}[T]{0.19\textwidth}
\centering
\includegraphics[width=\textwidth]{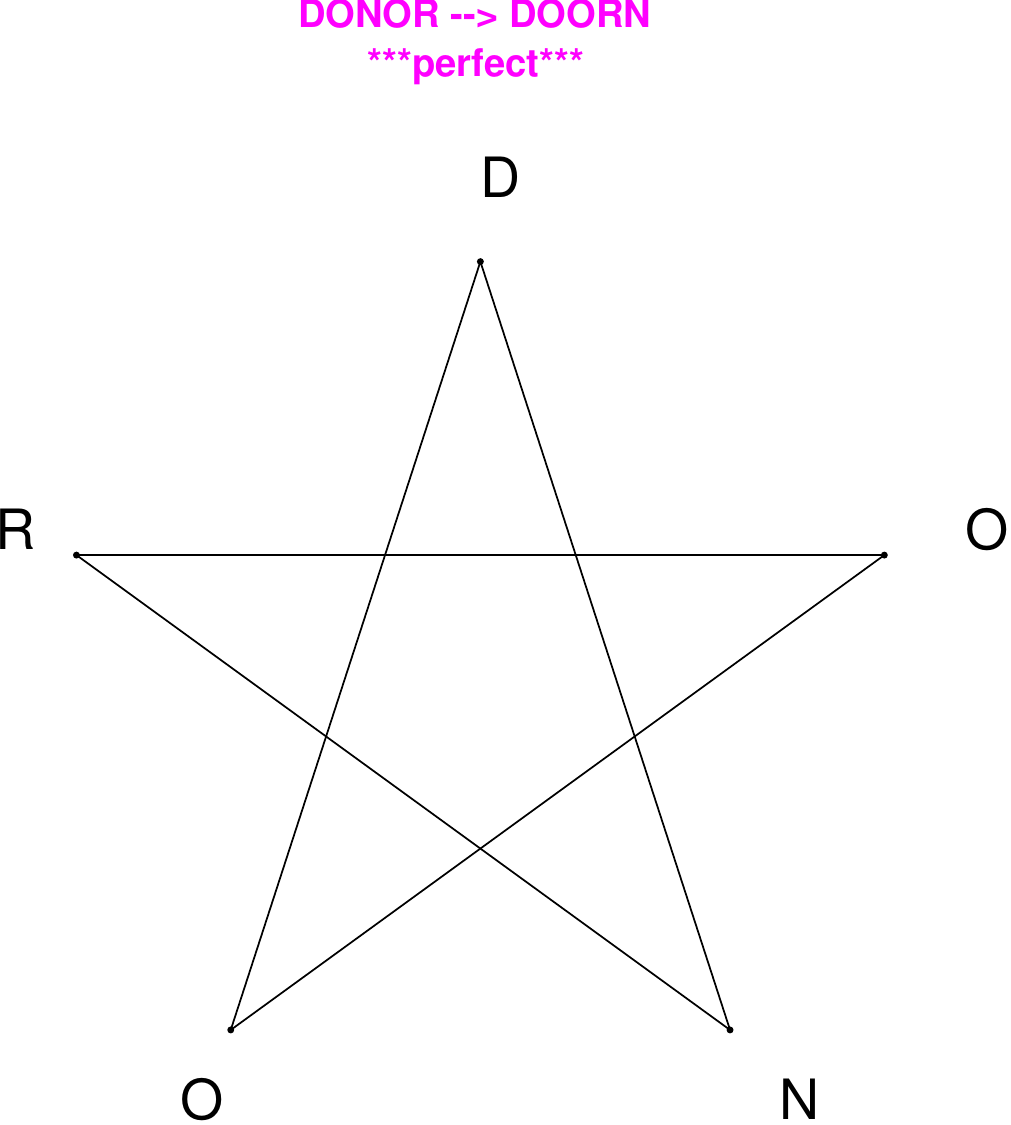}
\end{subfigure}
\hfill
\begin{subfigure}[T]{0.19\textwidth}
\centering
\includegraphics[width=\textwidth]{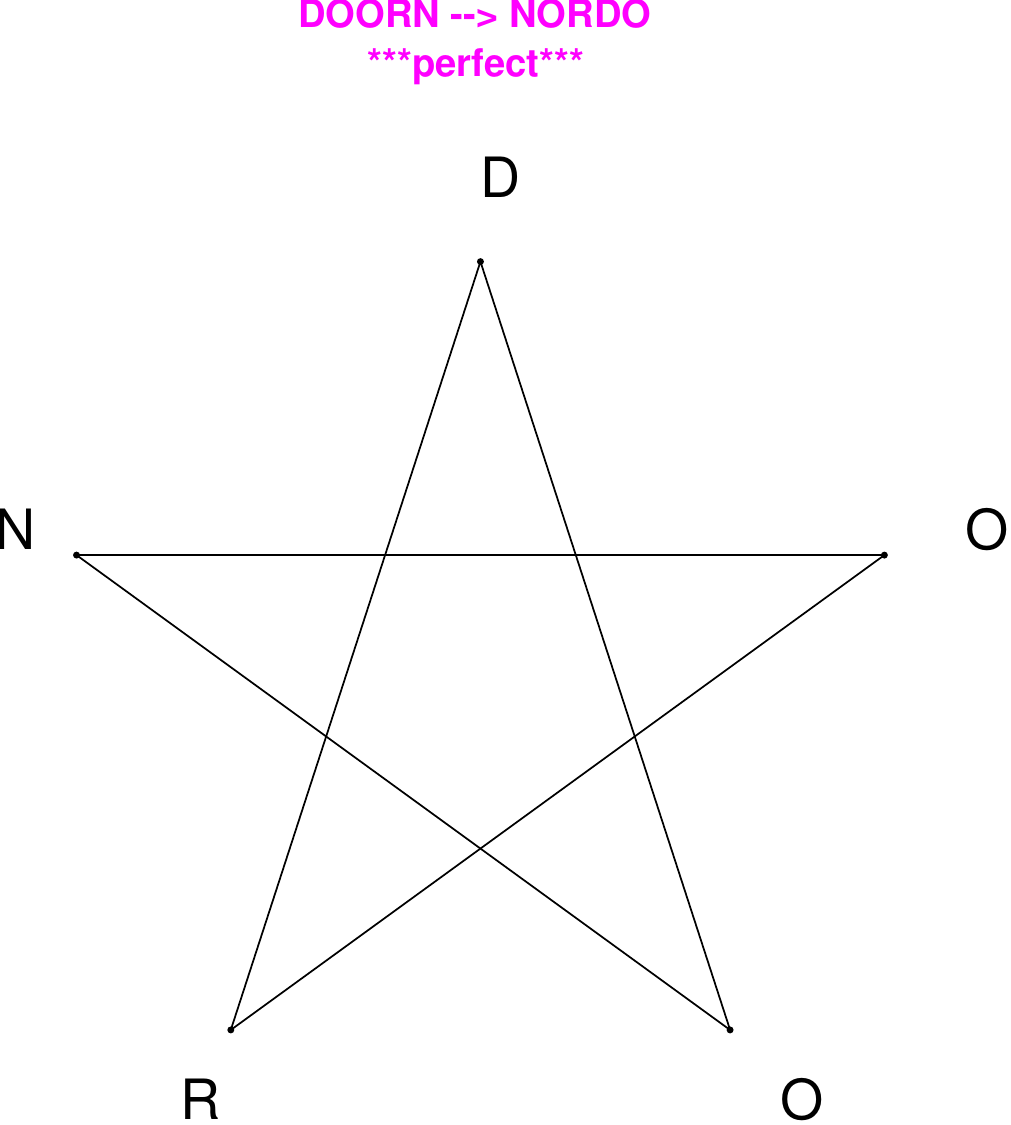}
\end{subfigure}
\hfill
\begin{subfigure}[T]{0.19\textwidth}
\centering
\includegraphics[width=\textwidth]{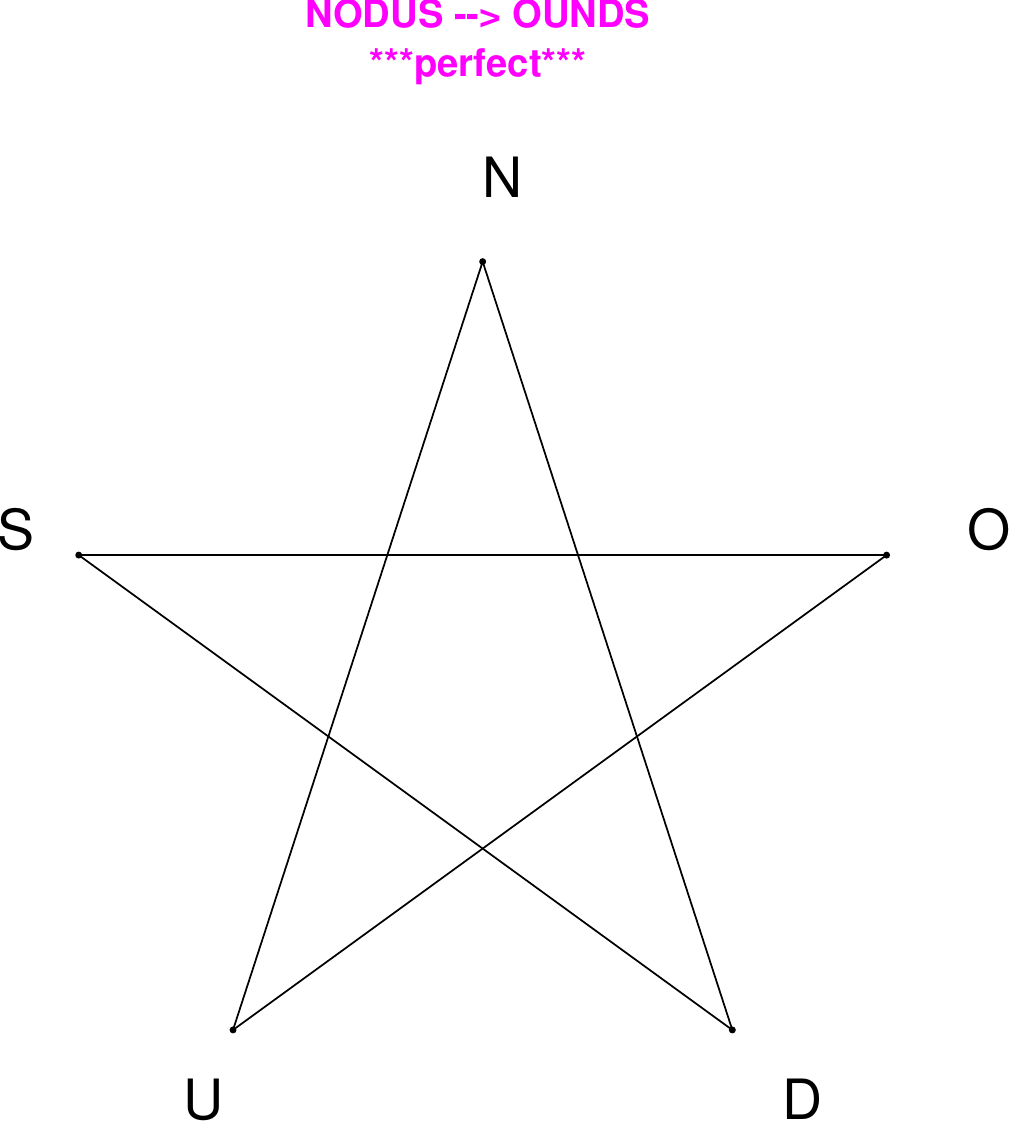}
\end{subfigure}
\hfill
\begin{subfigure}[T]{0.19\textwidth}
\centering
\includegraphics[width=\textwidth]{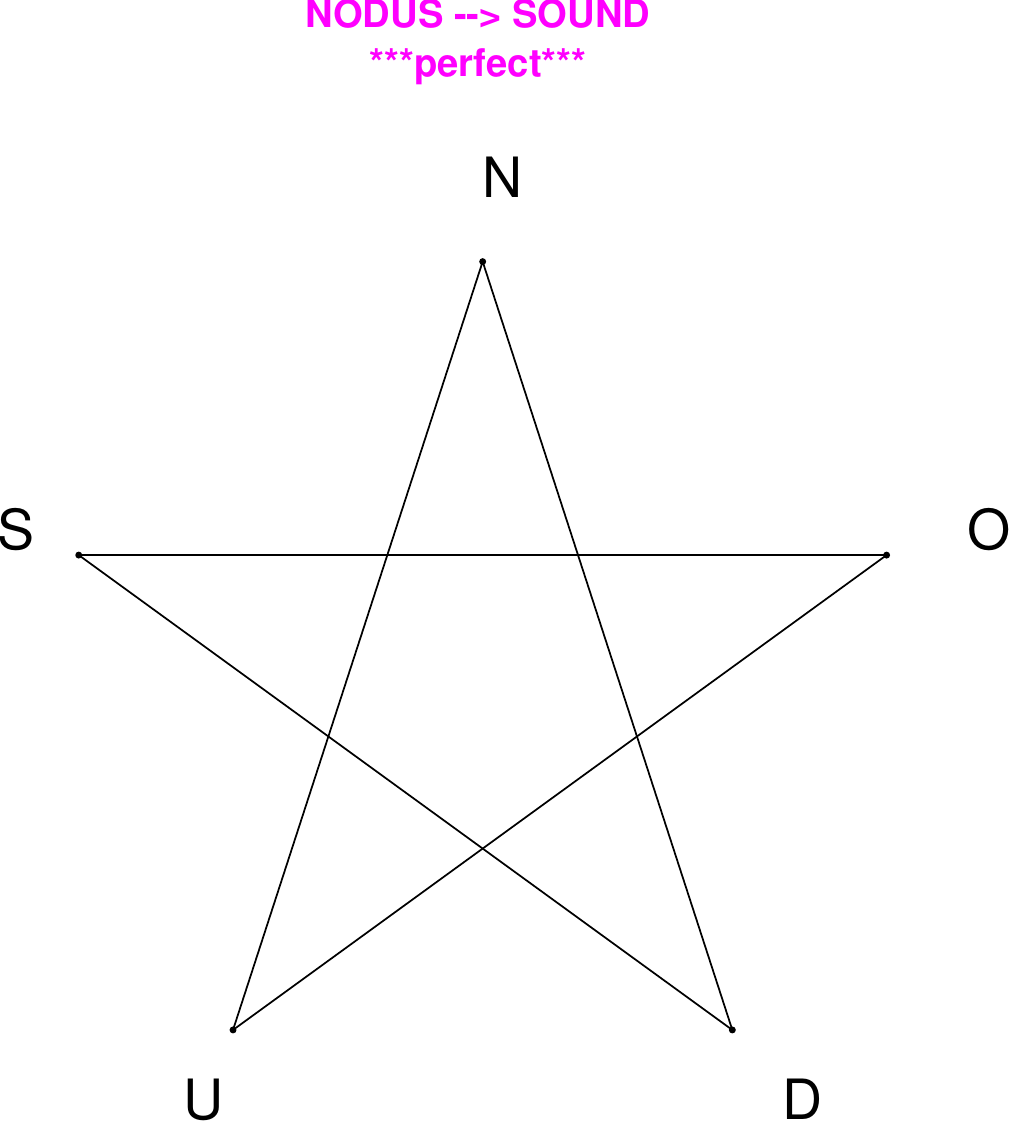}
\end{subfigure}
\end{figure}

\begin{figure}[H]
\centering
\begin{subfigure}[T]{0.19\textwidth}
\centering
\includegraphics[width=\textwidth]{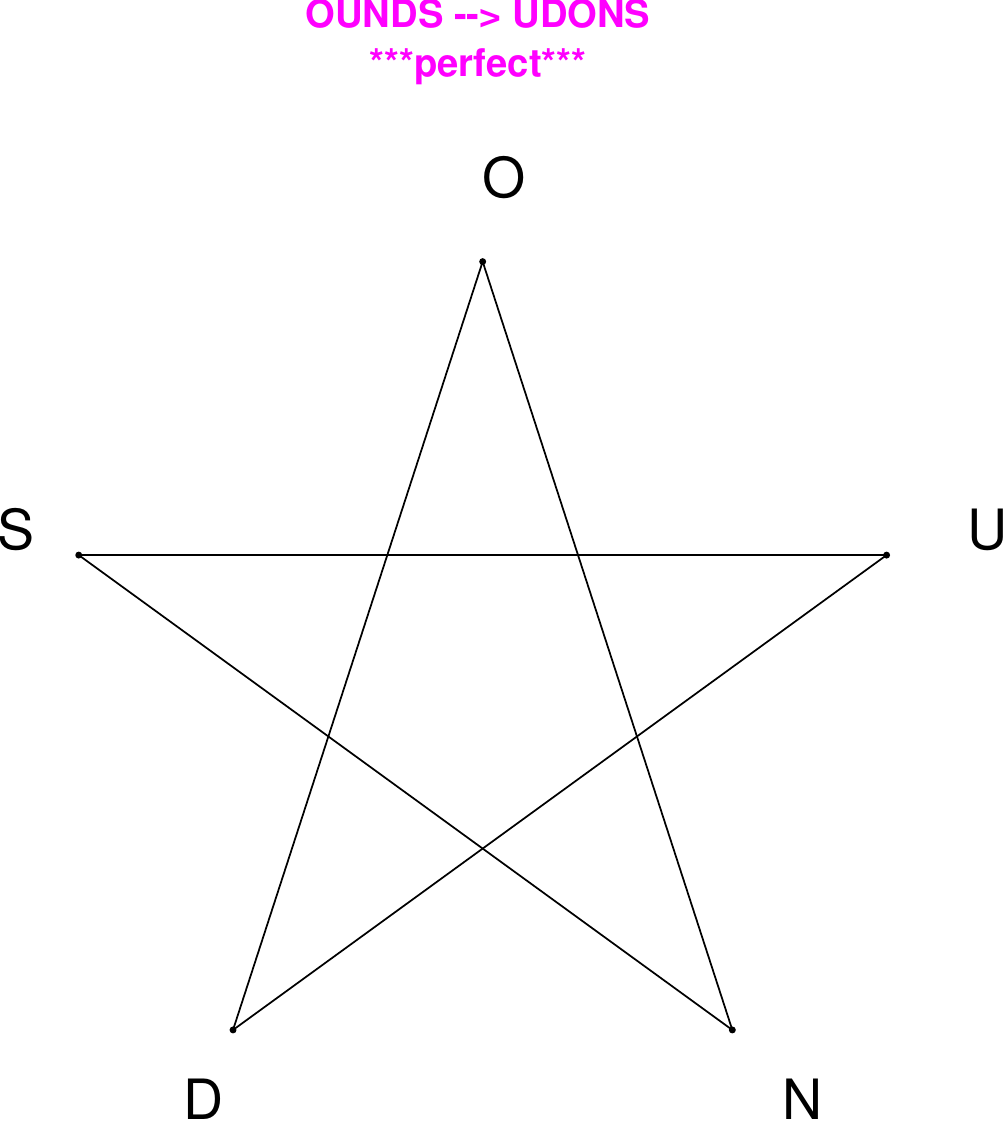}
\end{subfigure}
\hfill
\begin{subfigure}[T]{0.19\textwidth}
\centering
\includegraphics[width=\textwidth]{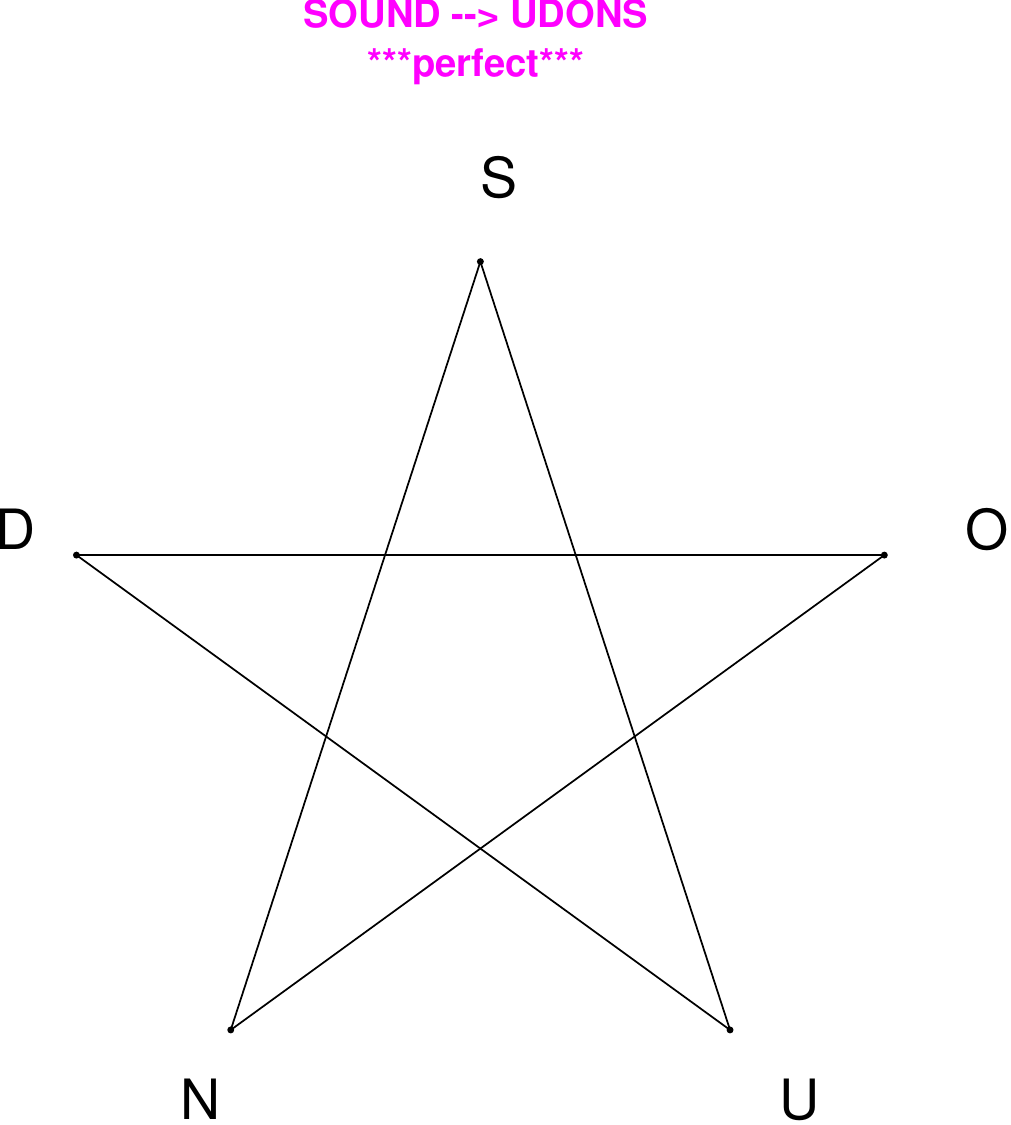}
\end{subfigure}
\hfill
\begin{subfigure}[T]{0.19\textwidth}
\centering
\includegraphics[width=\textwidth]{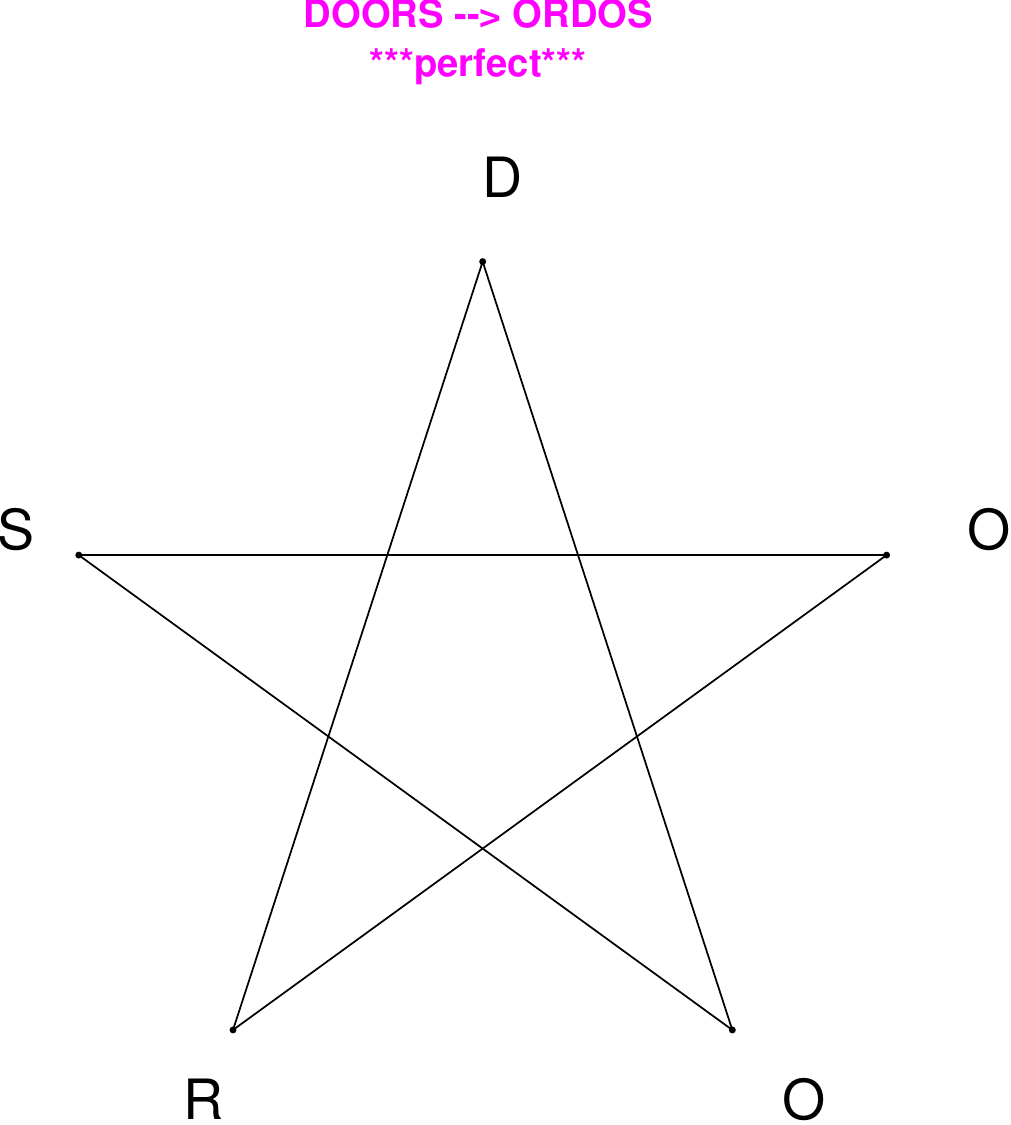}
\end{subfigure}
\hfill
\begin{subfigure}[T]{0.19\textwidth}
\centering
\includegraphics[width=\textwidth]{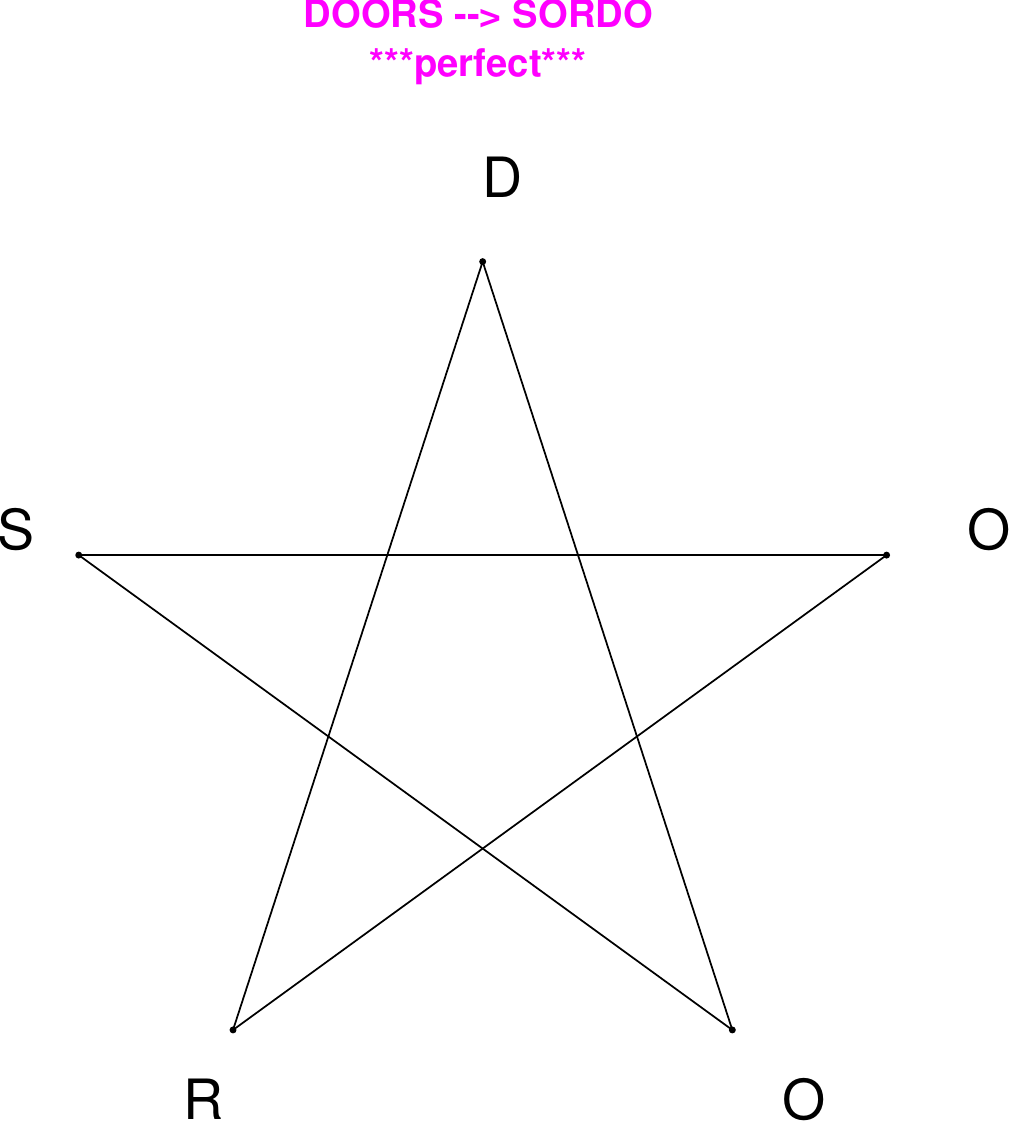}
\end{subfigure}
\hfill
\begin{subfigure}[T]{0.19\textwidth}
\centering
\includegraphics[width=\textwidth]{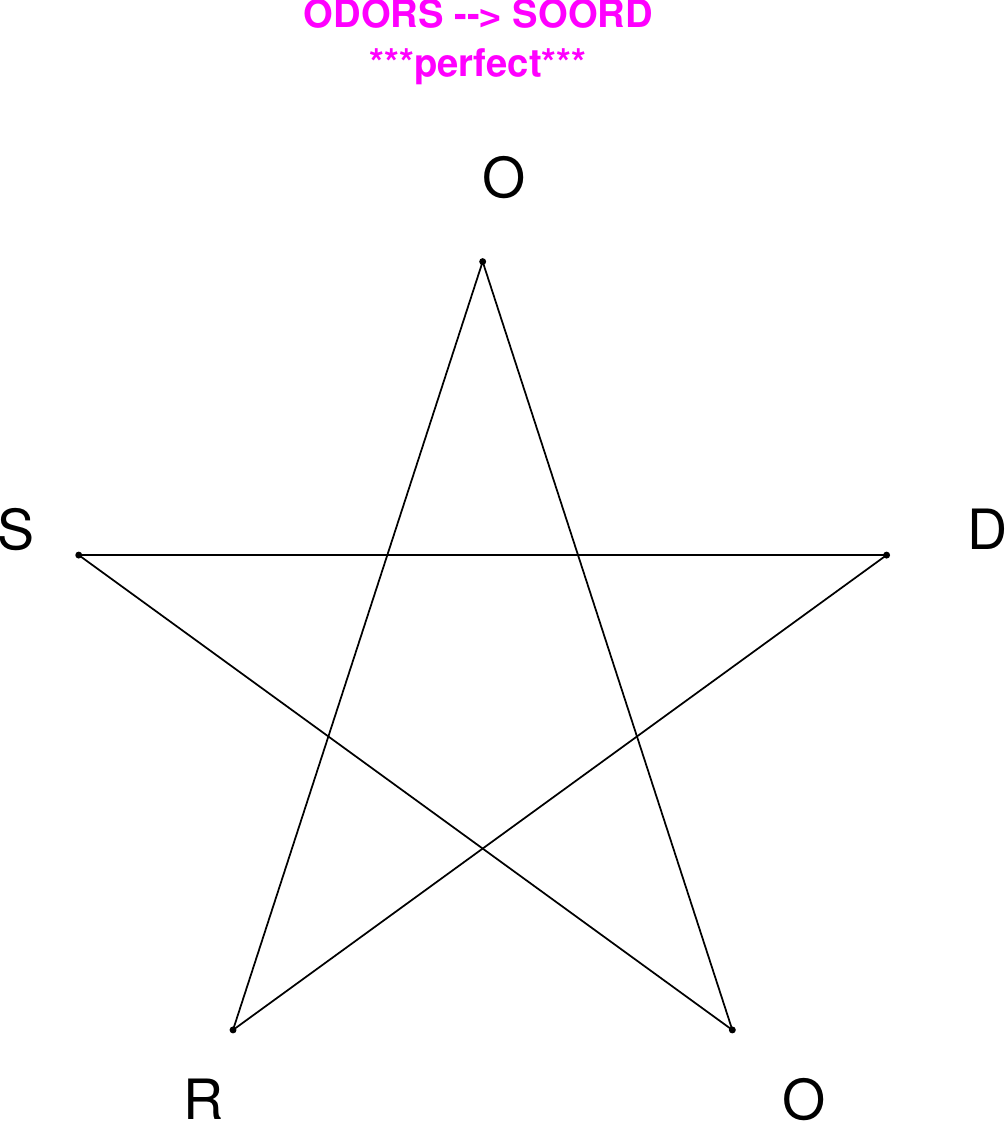}
\end{subfigure}
\end{figure}

\begin{figure}[H]
\centering
\begin{subfigure}[T]{0.19\textwidth}
\centering
\includegraphics[width=\textwidth]{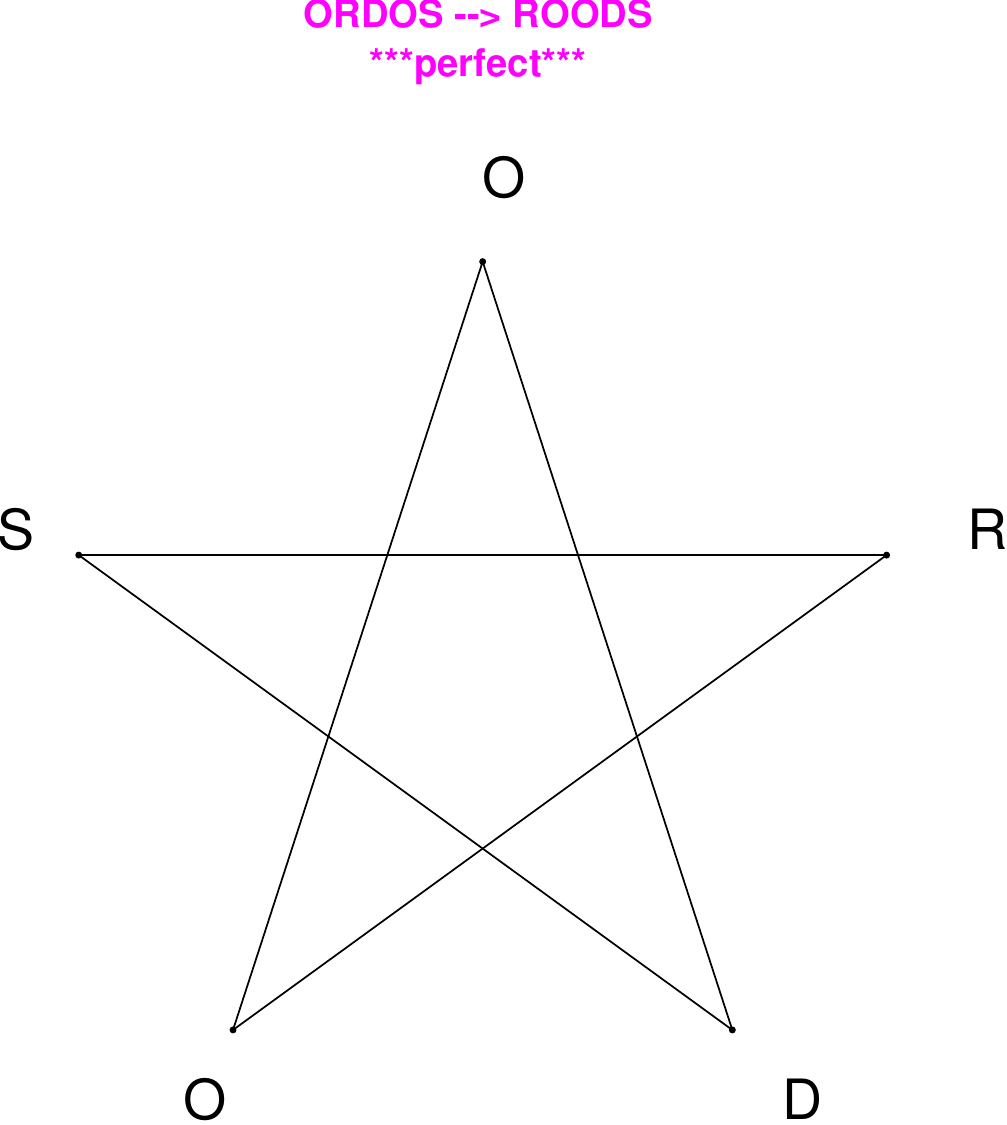}
\end{subfigure}
\hfill
\begin{subfigure}[T]{0.19\textwidth}
\centering
\includegraphics[width=\textwidth]{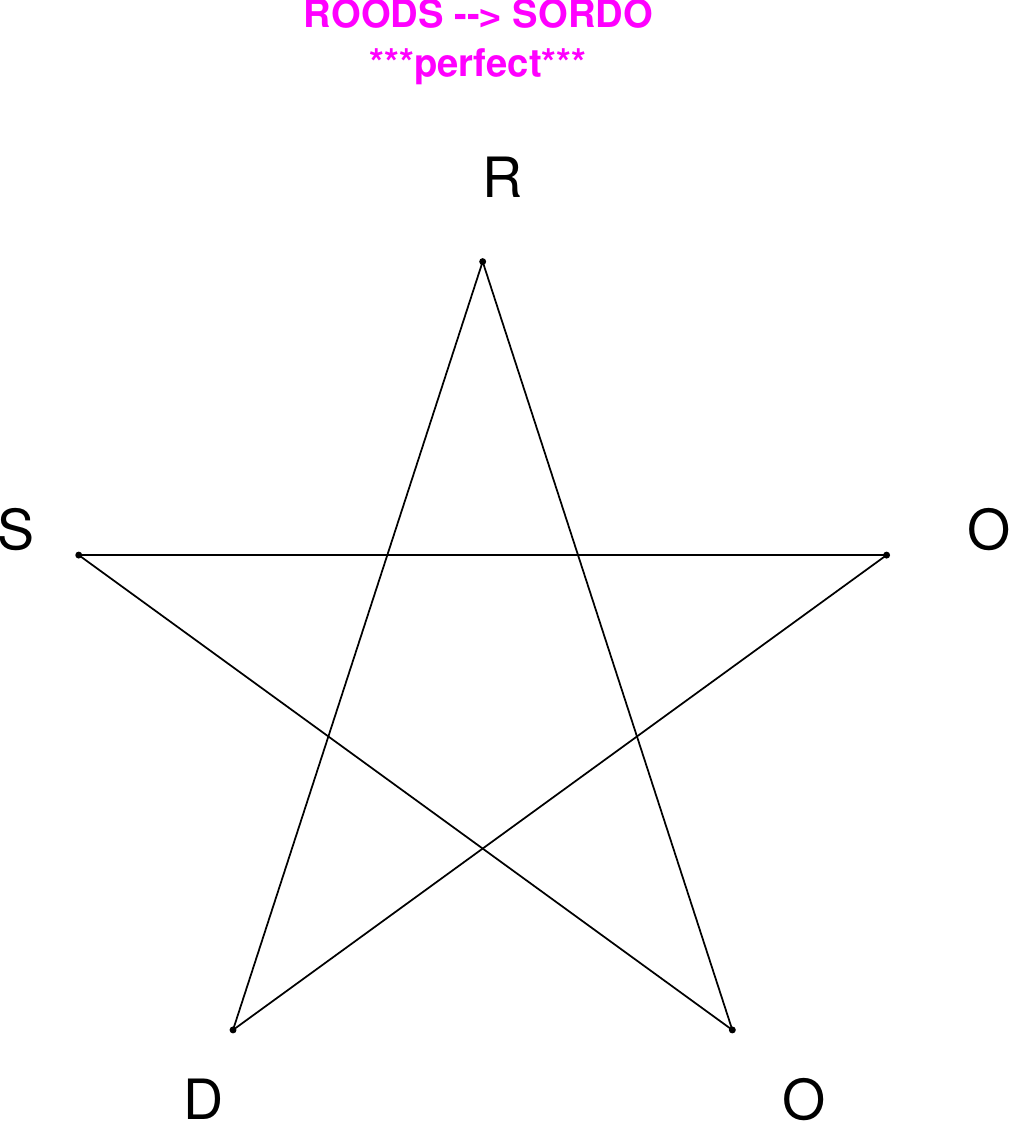}
\end{subfigure}
\hfill
\begin{subfigure}[T]{0.19\textwidth}
\centering
\includegraphics[width=\textwidth]{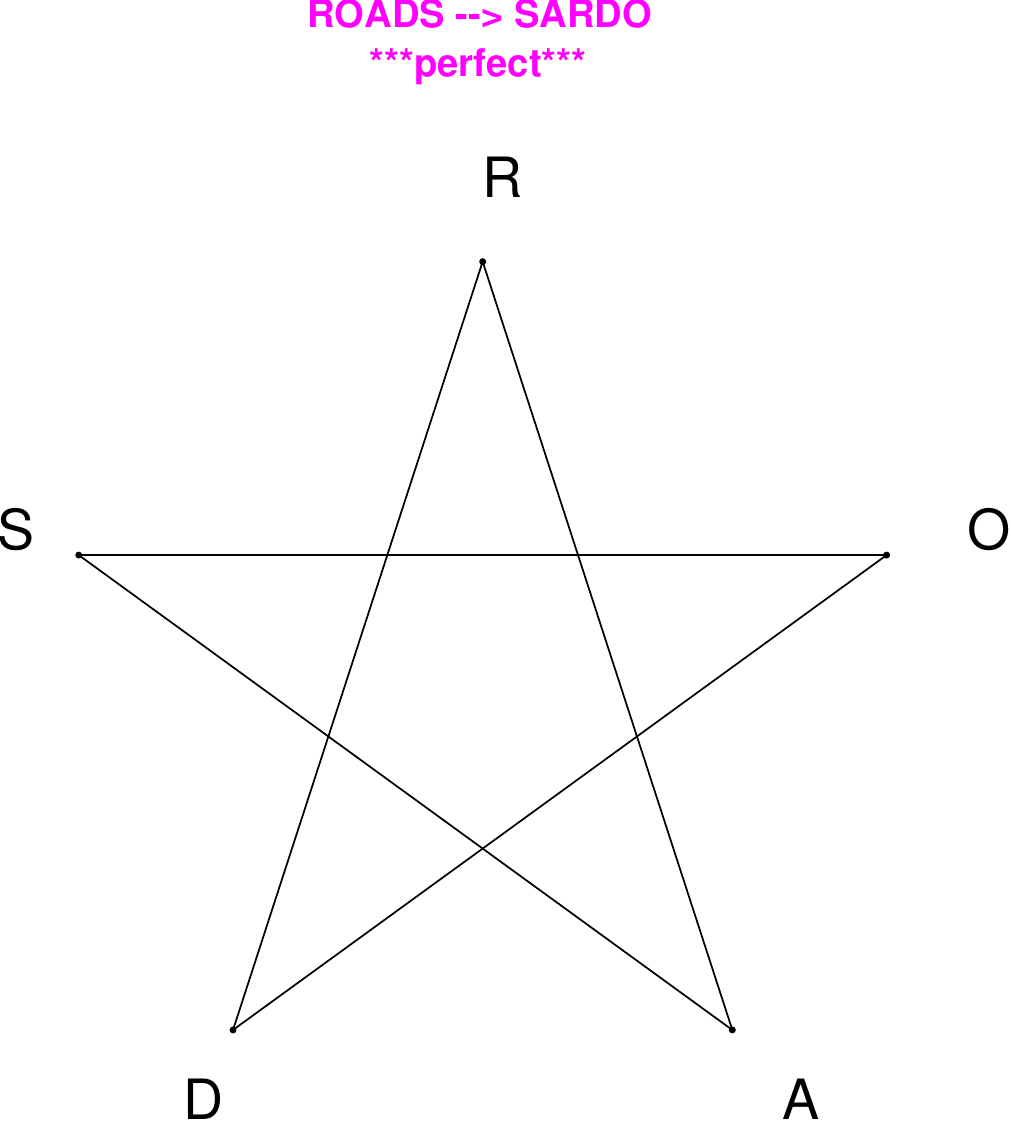}
\end{subfigure}
\hfill
\begin{subfigure}[T]{0.19\textwidth}
\centering
\includegraphics[width=\textwidth]{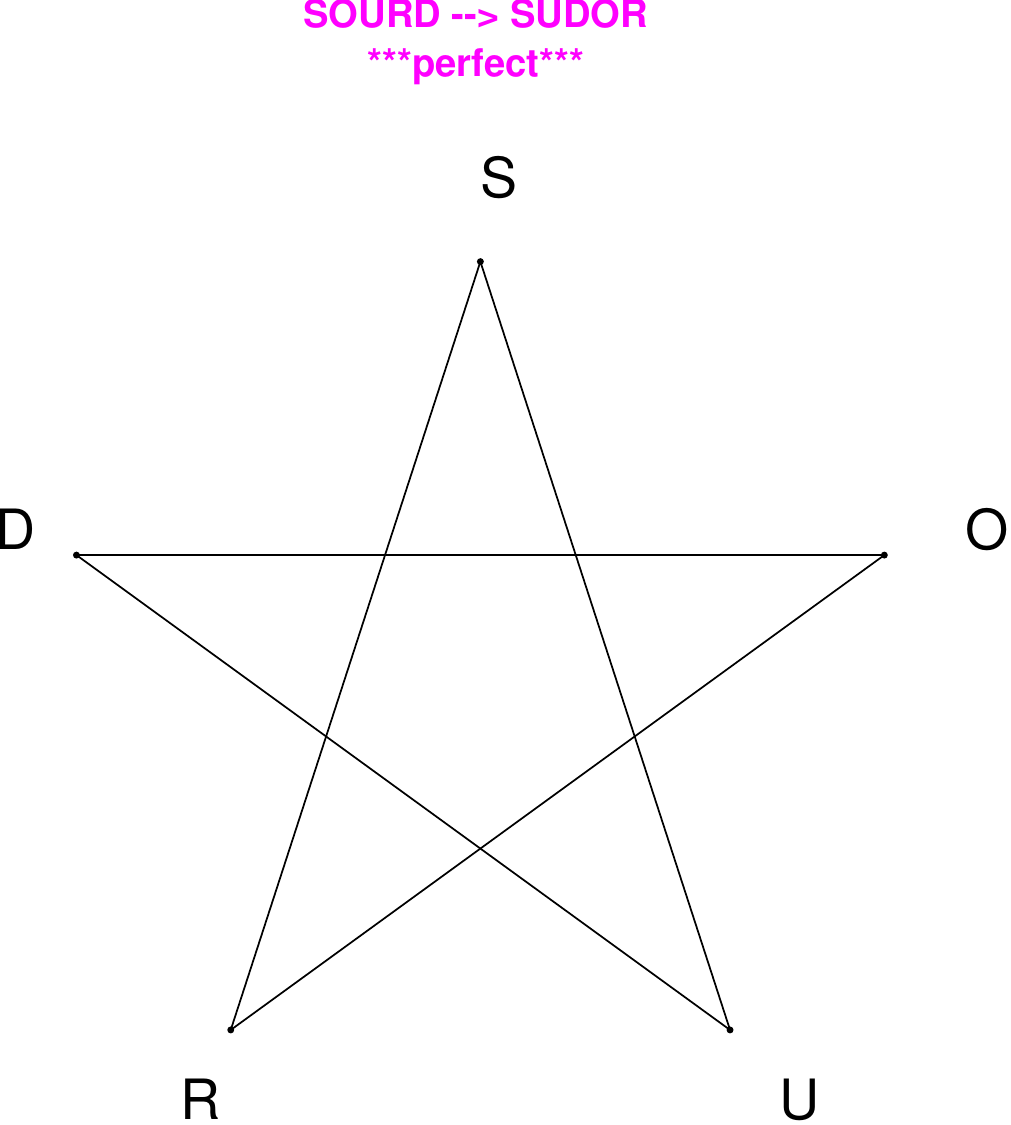}
\end{subfigure}
\hfill
\begin{subfigure}[T]{0.19\textwidth}
\centering
\includegraphics[width=\textwidth]{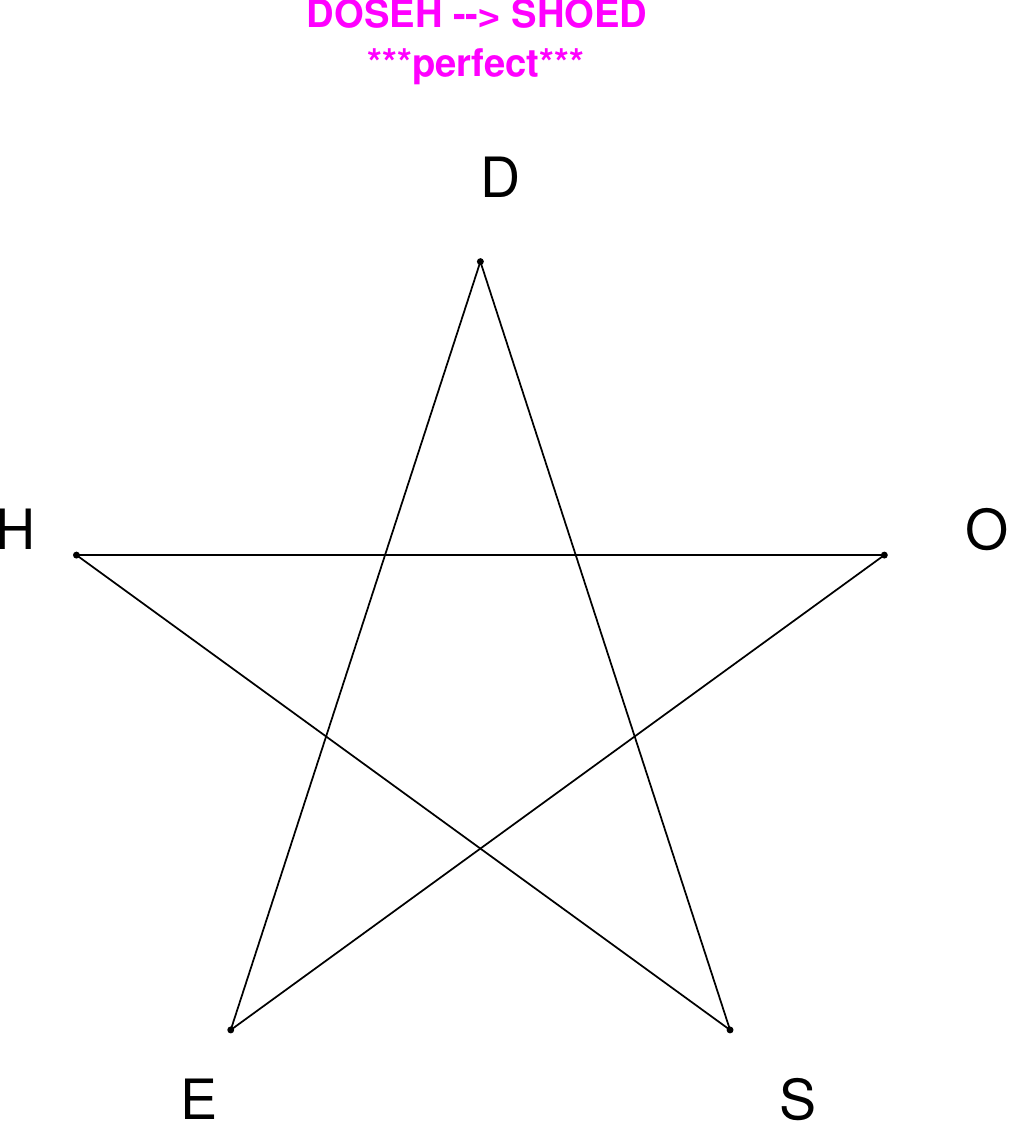}
\end{subfigure}
\end{figure}

\begin{figure}[H]
\centering
\begin{subfigure}[T]{0.19\textwidth}
\centering
\includegraphics[width=\textwidth]{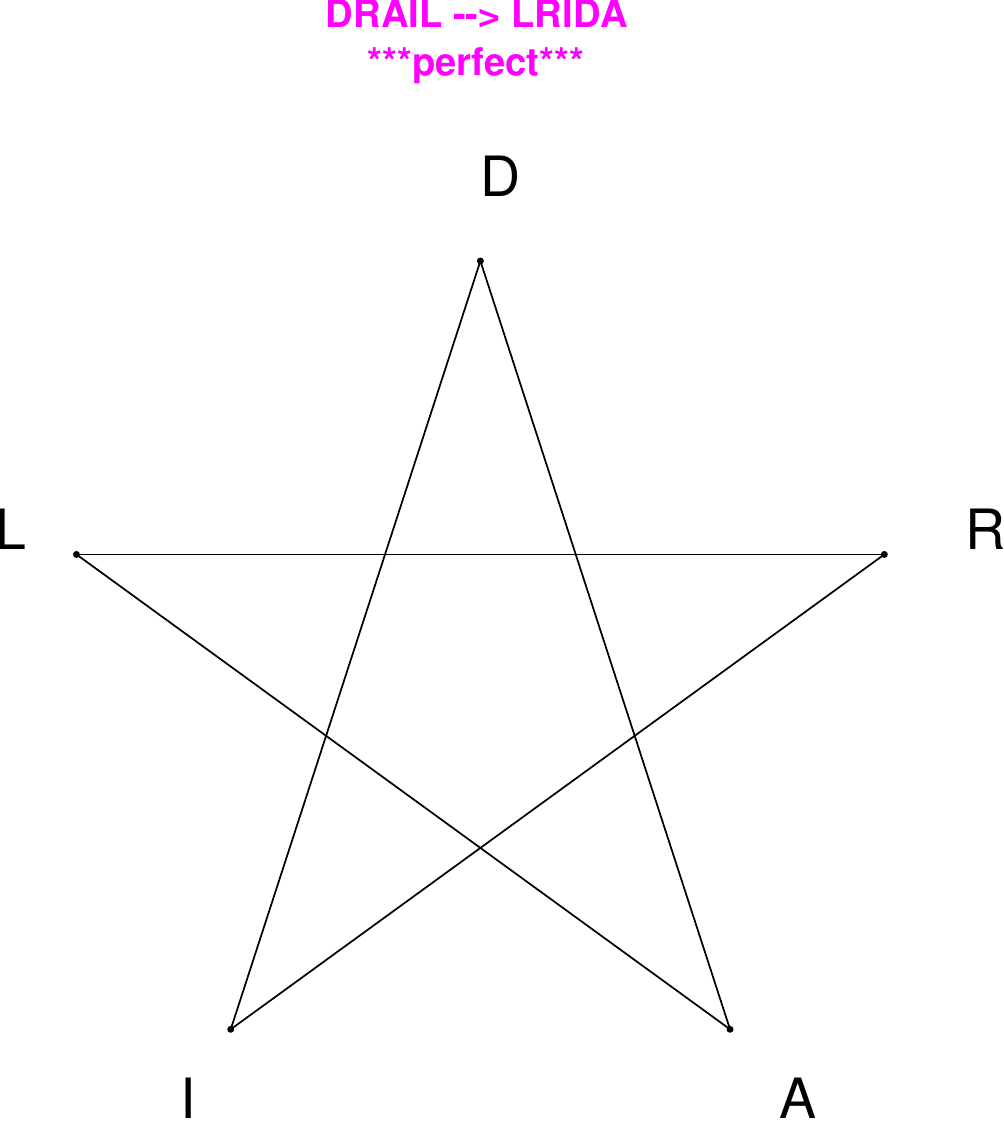}
\end{subfigure}
\hfill
\begin{subfigure}[T]{0.19\textwidth}
\centering
\includegraphics[width=\textwidth]{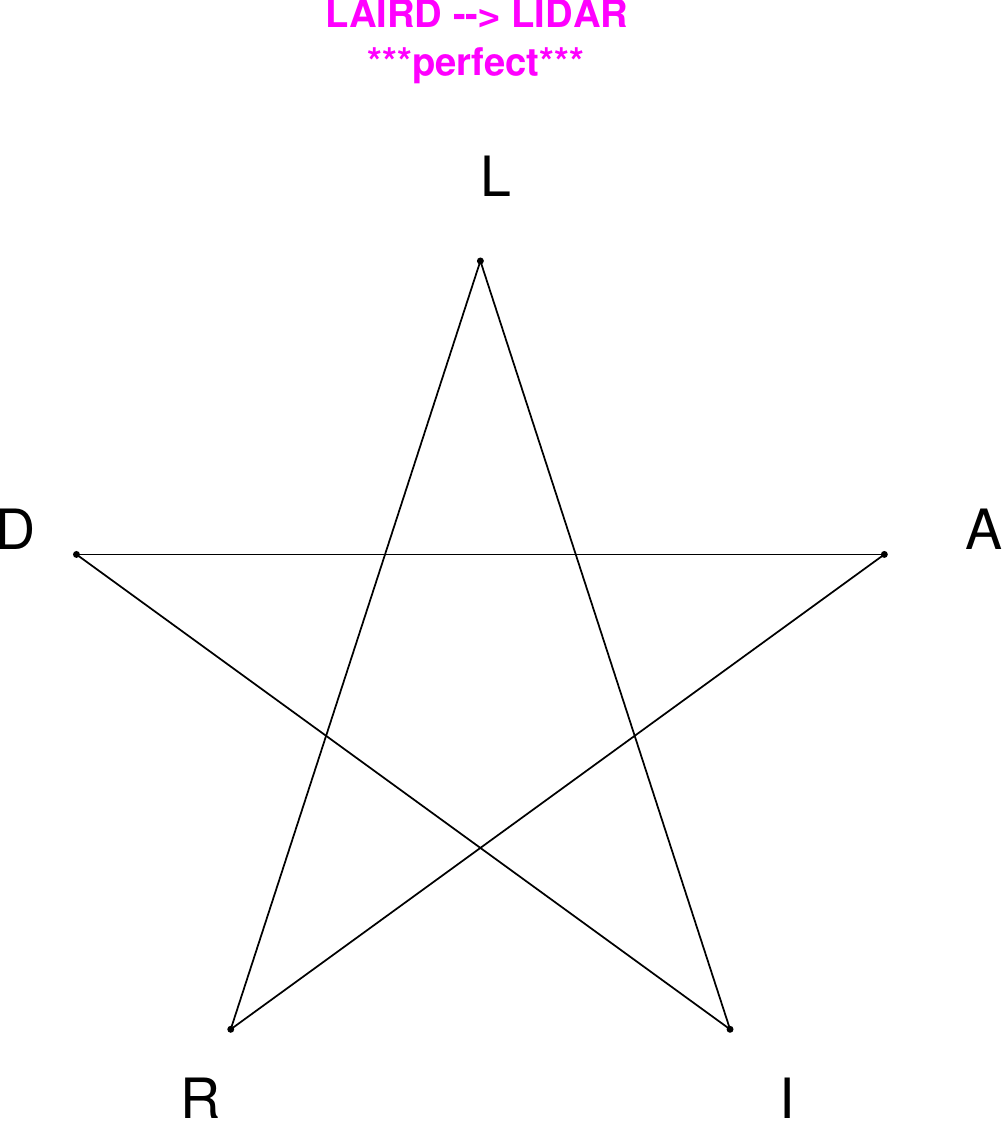}
\end{subfigure}
\hfill
\begin{subfigure}[T]{0.19\textwidth}
\centering
\includegraphics[width=\textwidth]{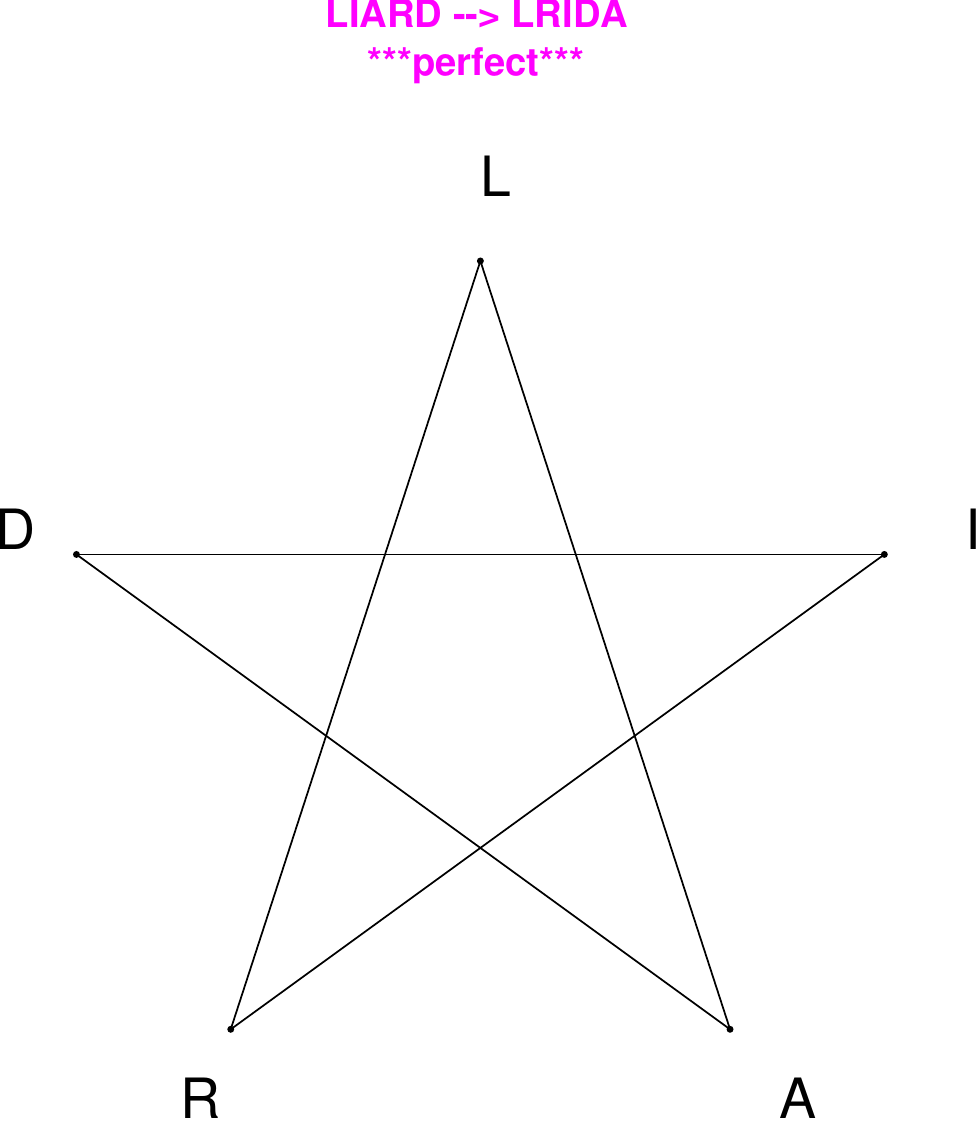}
\end{subfigure}
\hfill
\begin{subfigure}[T]{0.19\textwidth}
\centering
\includegraphics[width=\textwidth]{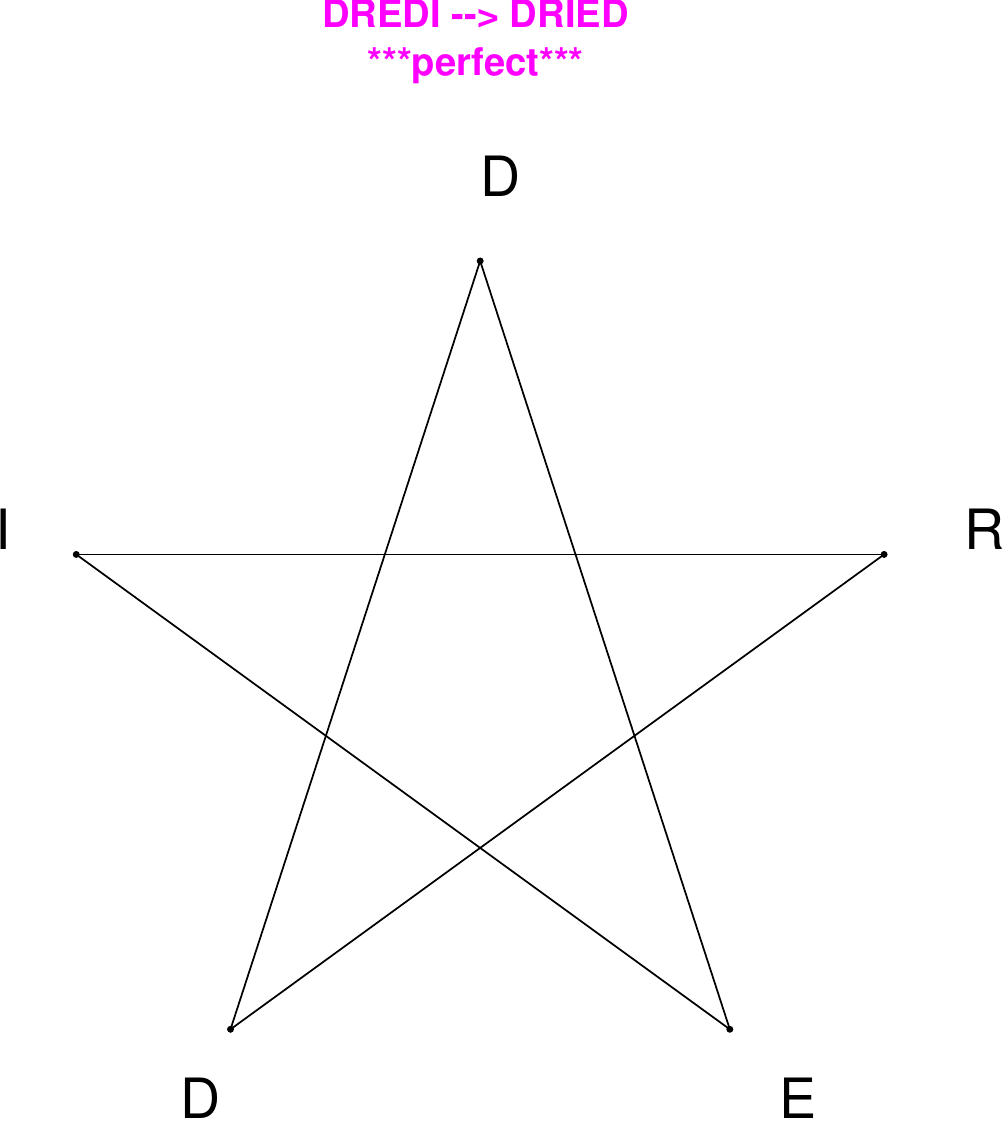}
\end{subfigure}
\hfill
\begin{subfigure}[T]{0.19\textwidth}
\centering
\includegraphics[width=\textwidth]{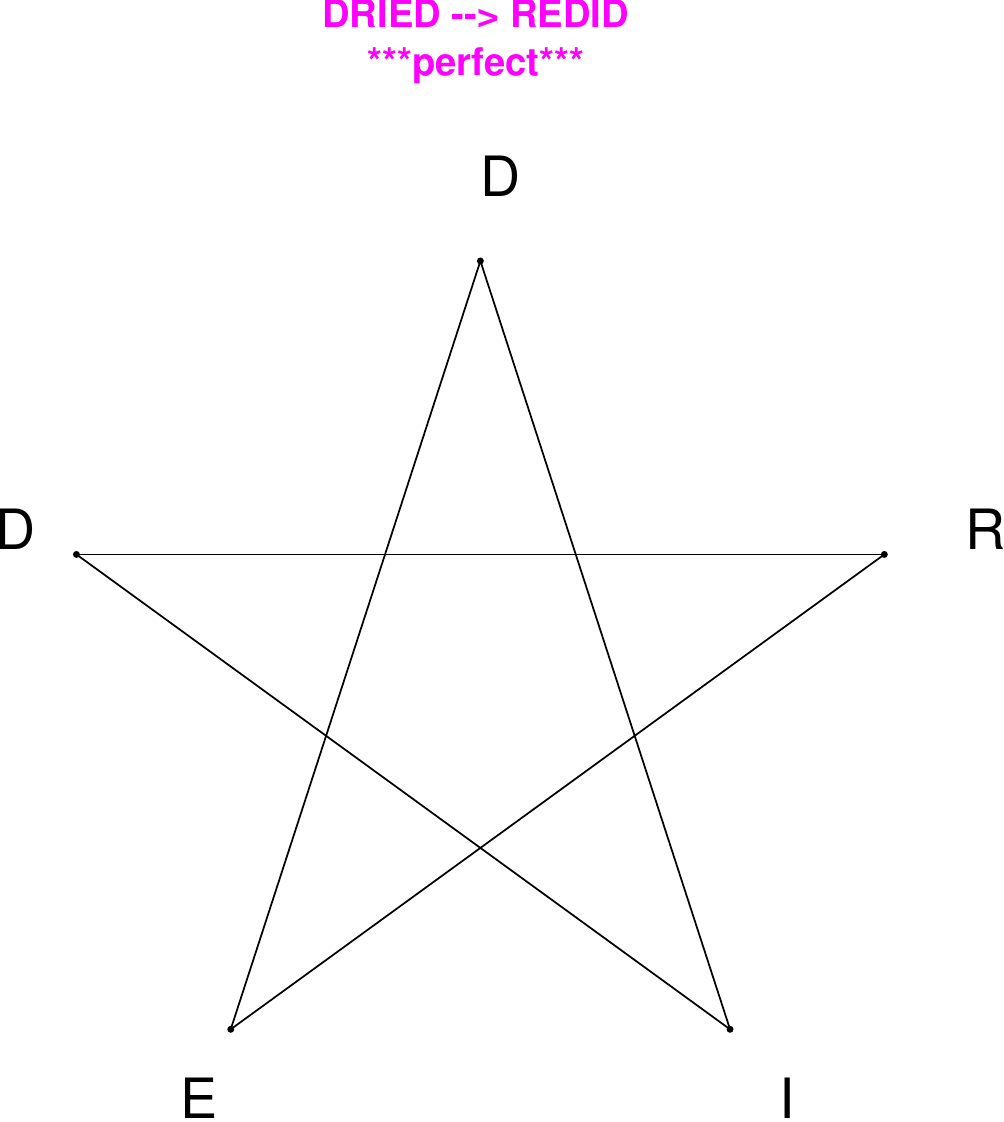}
\end{subfigure}
\end{figure}

\begin{figure}[H]
\centering
\begin{subfigure}[T]{0.19\textwidth}
\centering
\includegraphics[width=\textwidth]{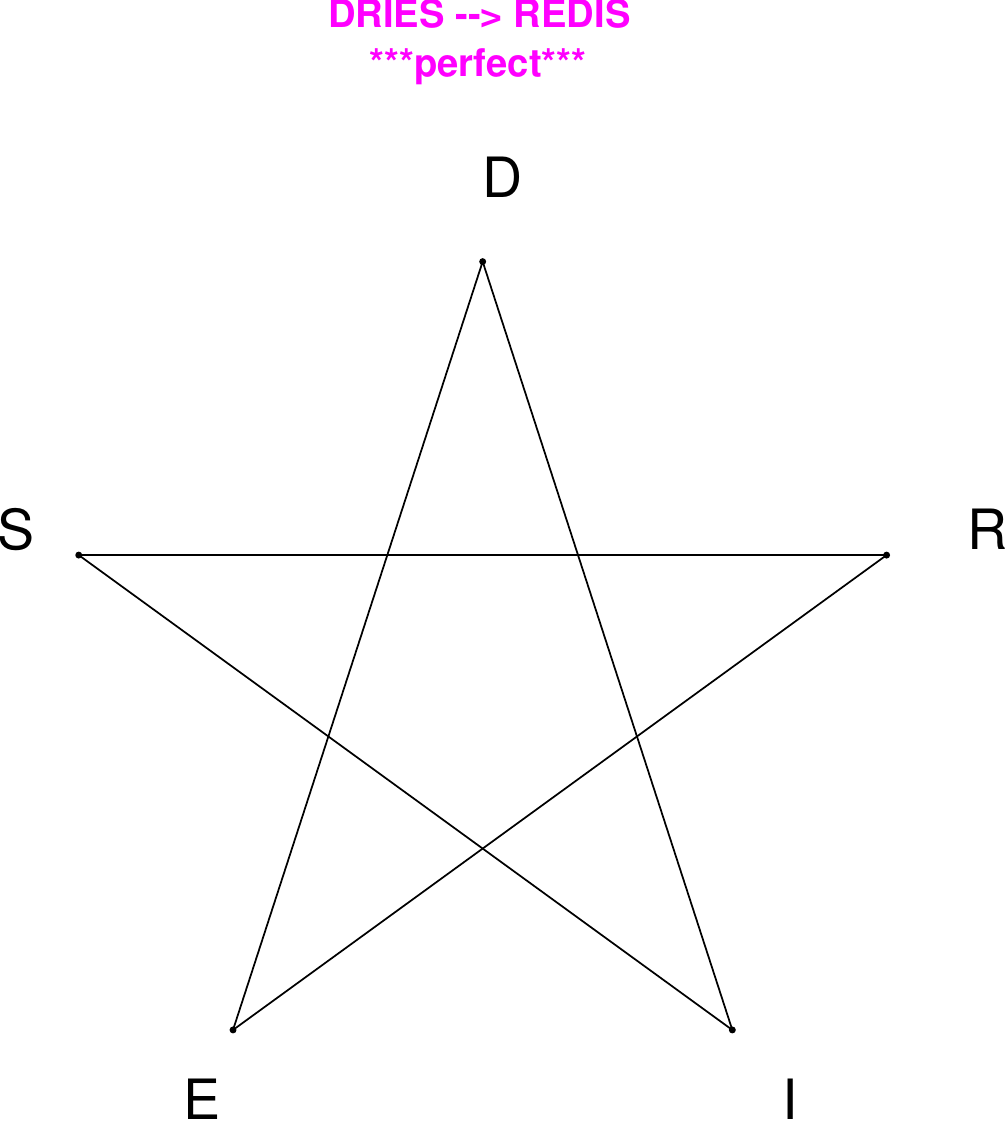}
\end{subfigure}
\hfill
\begin{subfigure}[T]{0.19\textwidth}
\centering
\includegraphics[width=\textwidth]{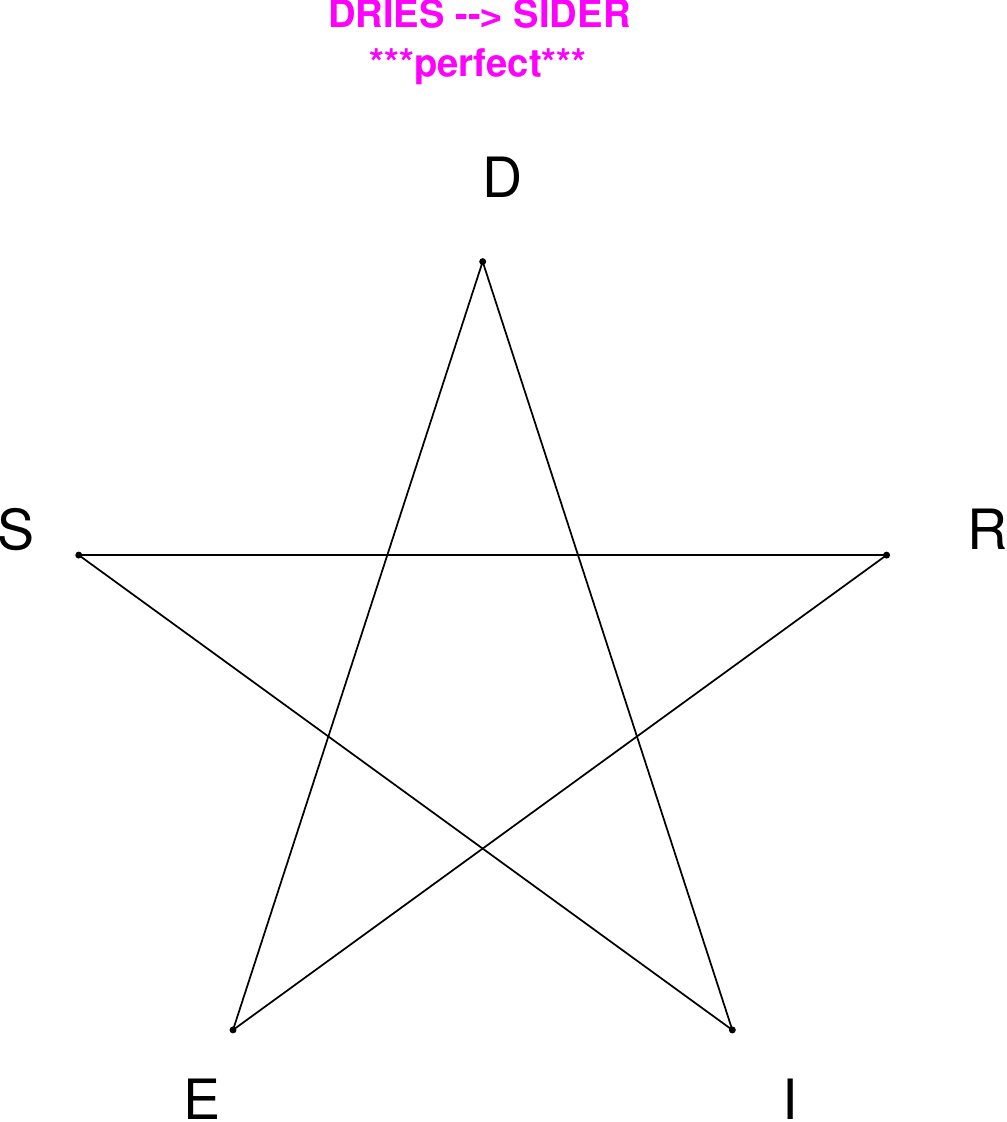}
\end{subfigure}
\hfill
\begin{subfigure}[T]{0.19\textwidth}
\centering
\includegraphics[width=\textwidth]{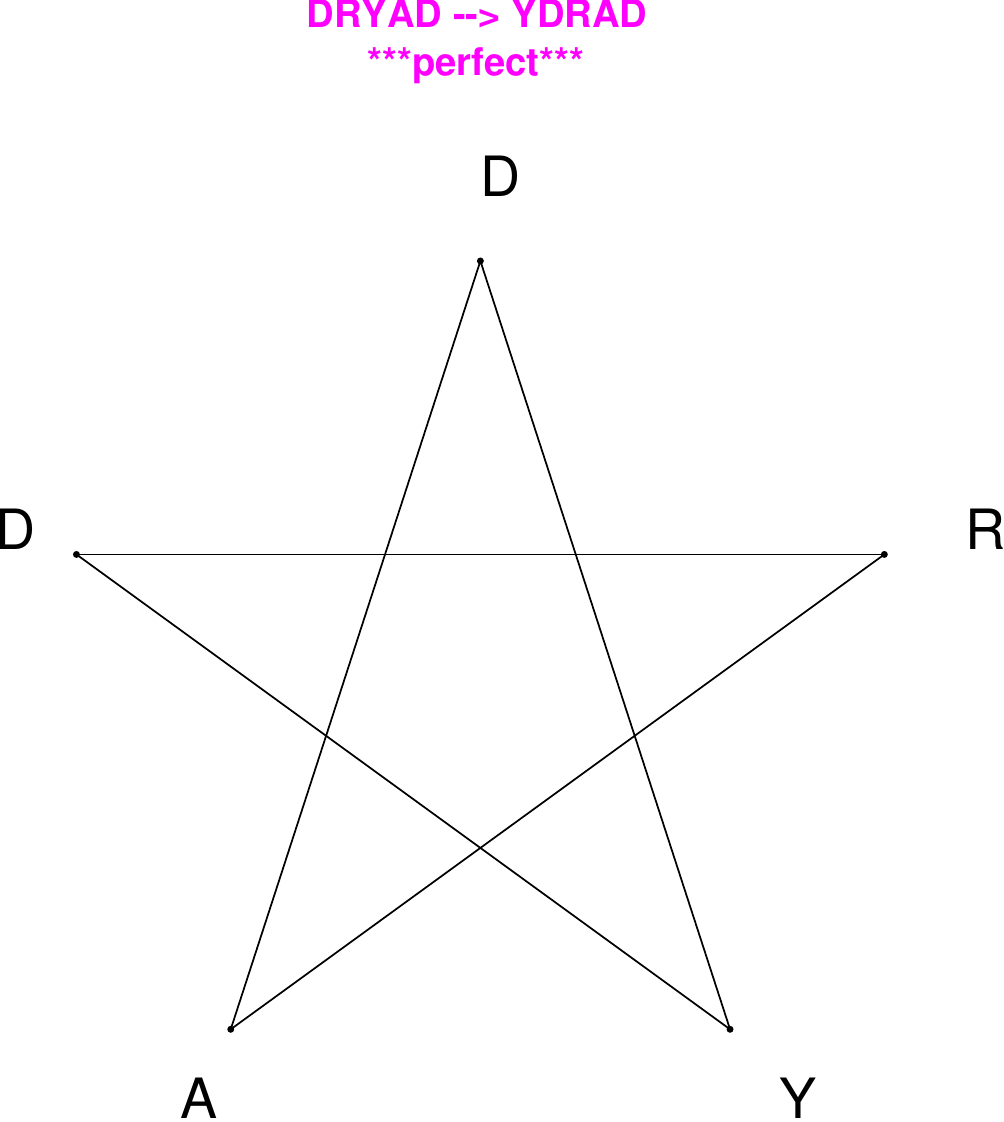}
\end{subfigure}
\hfill
\begin{subfigure}[T]{0.19\textwidth}
\centering
\includegraphics[width=\textwidth]{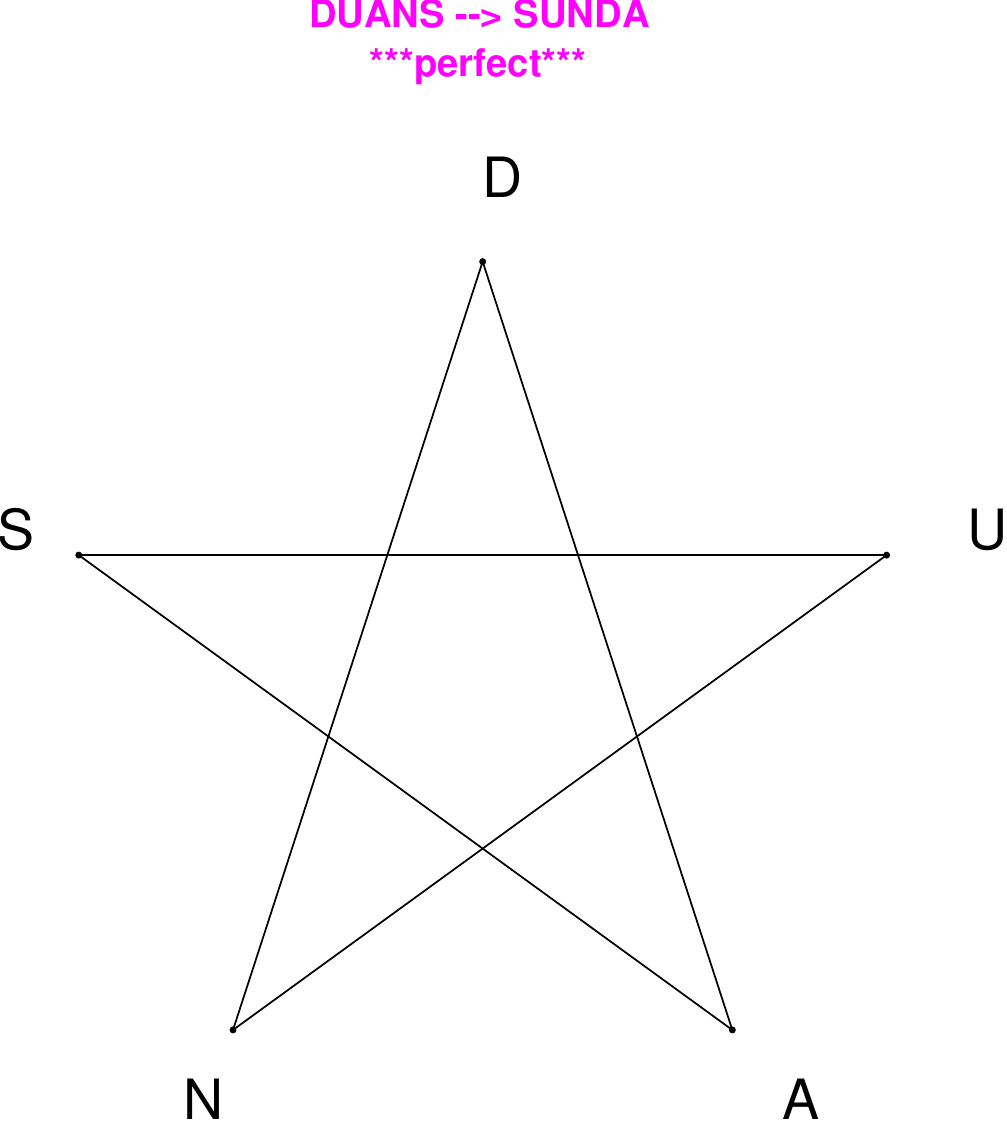}
\end{subfigure}
\hfill
\begin{subfigure}[T]{0.19\textwidth}
\centering
\includegraphics[width=\textwidth]{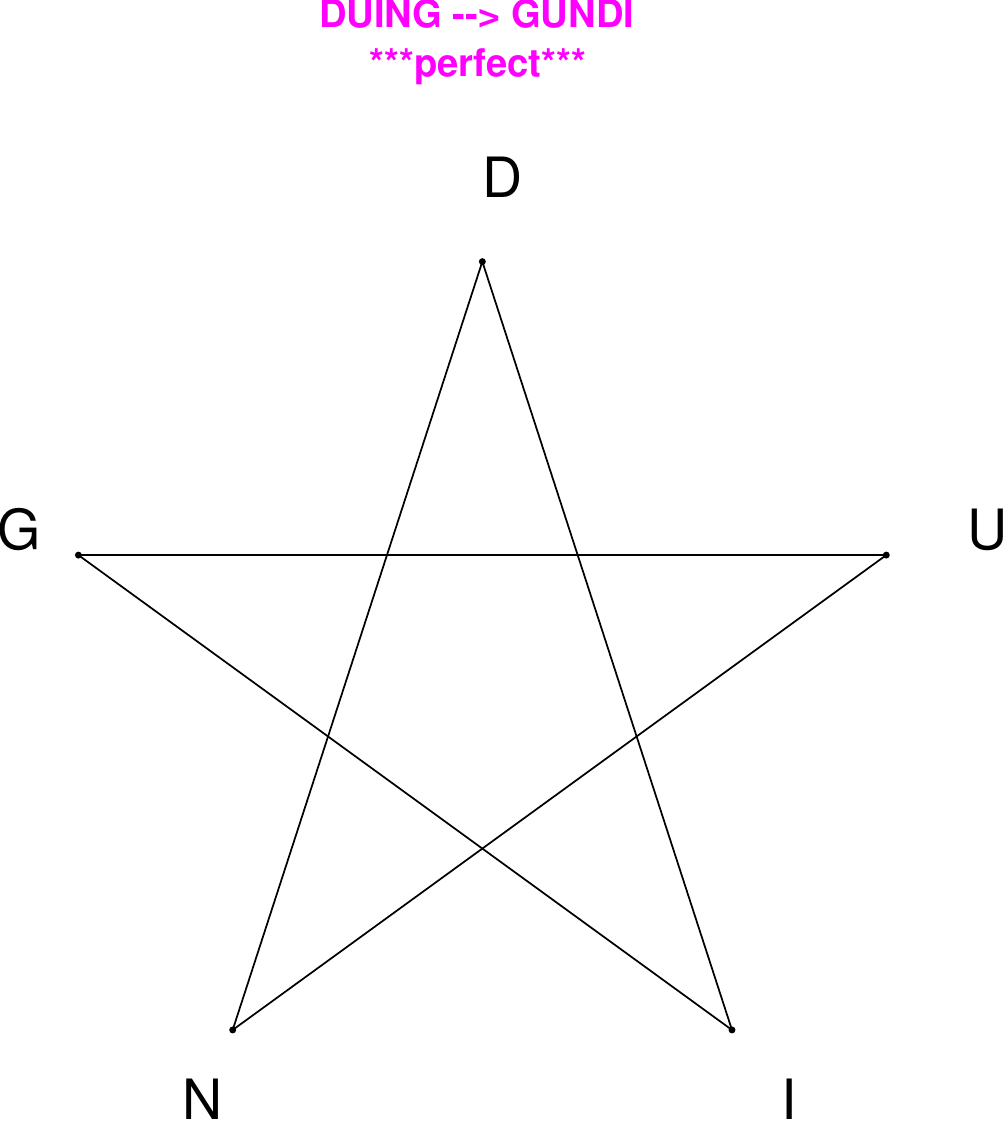}
\end{subfigure}
\end{figure}

\begin{figure}[H]
\centering
\begin{subfigure}[T]{0.19\textwidth}
\centering
\includegraphics[width=\textwidth]{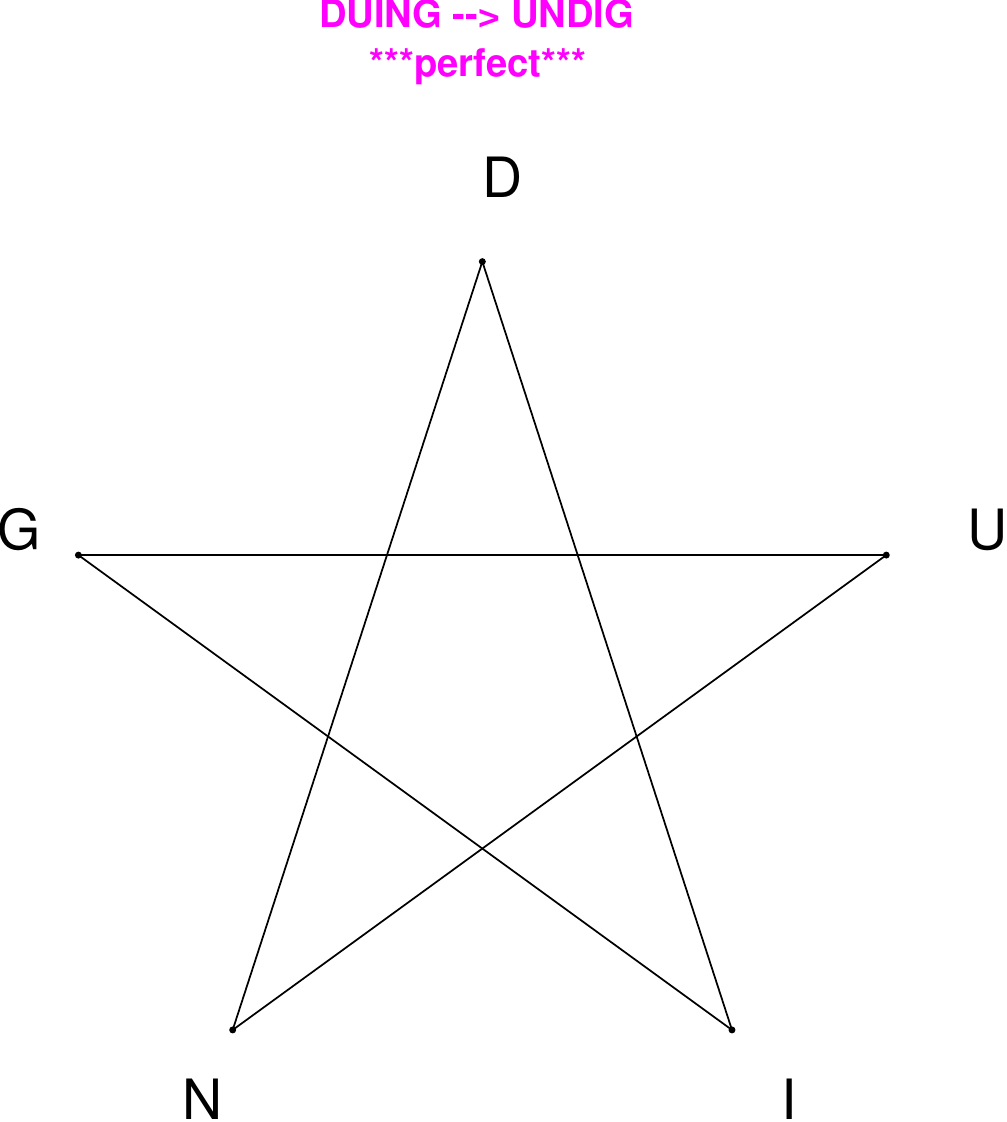}
\end{subfigure}
\hfill
\begin{subfigure}[T]{0.19\textwidth}
\centering
\includegraphics[width=\textwidth]{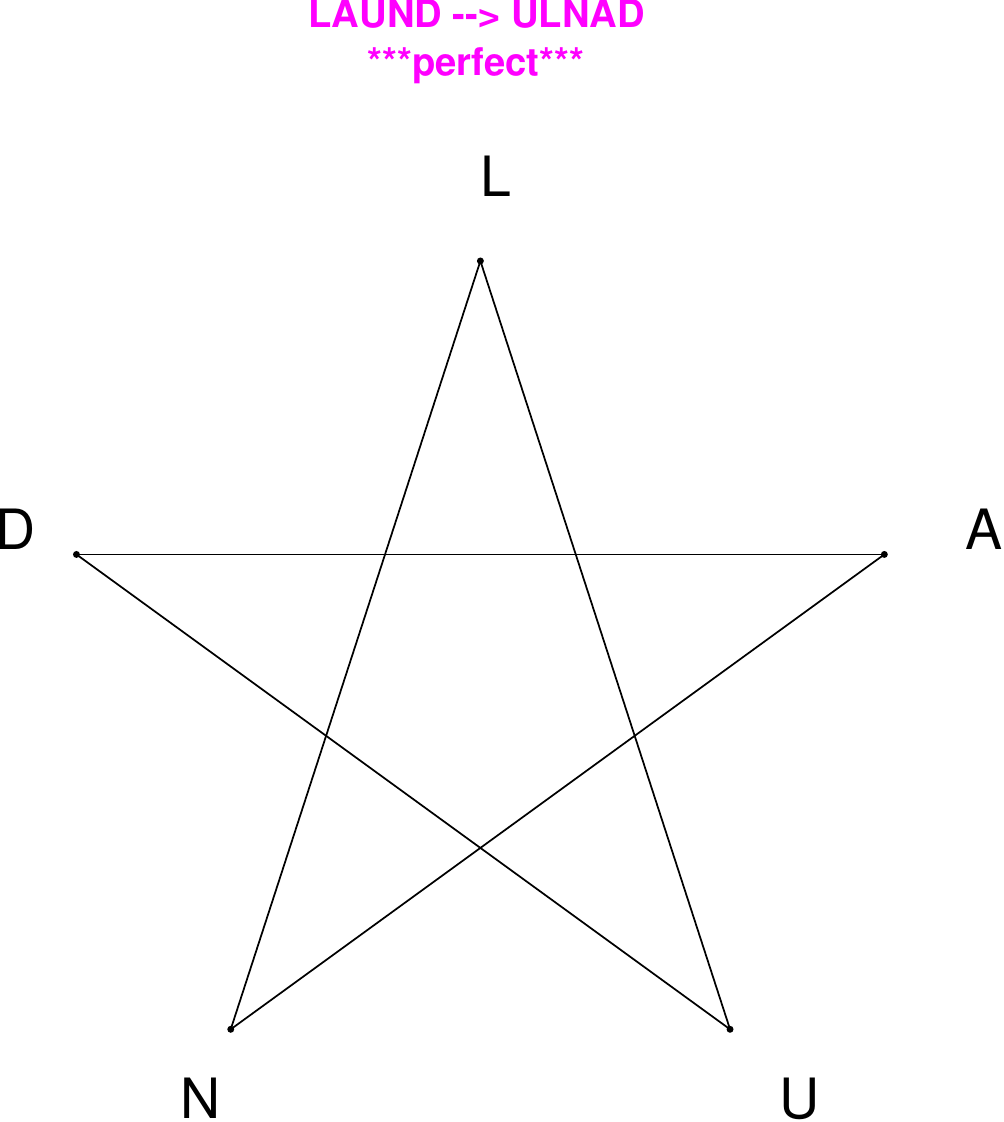}
\end{subfigure}
\hfill
\begin{subfigure}[T]{0.19\textwidth}
\centering
\includegraphics[width=\textwidth]{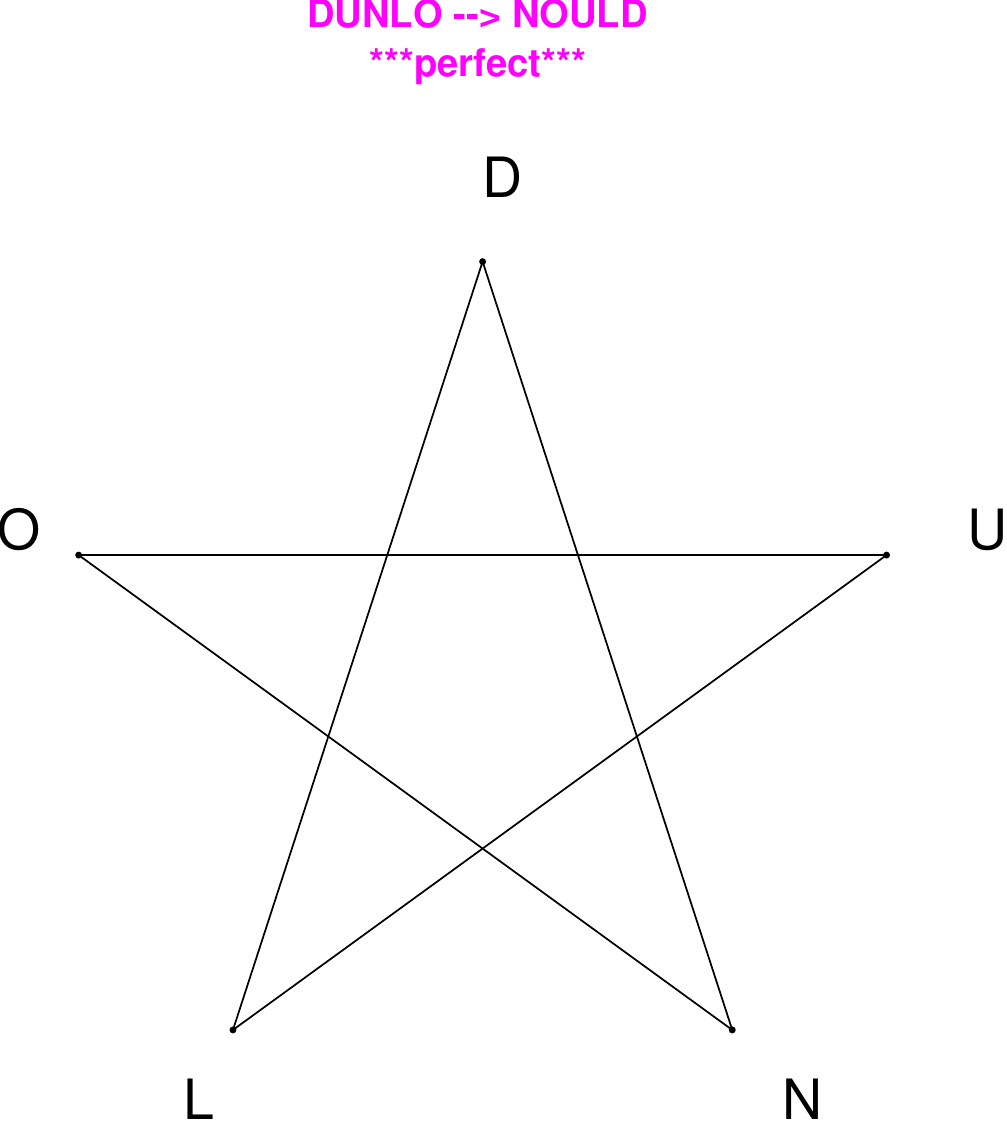}
\end{subfigure}
\hfill
\begin{subfigure}[T]{0.19\textwidth}
\centering
\includegraphics[width=\textwidth]{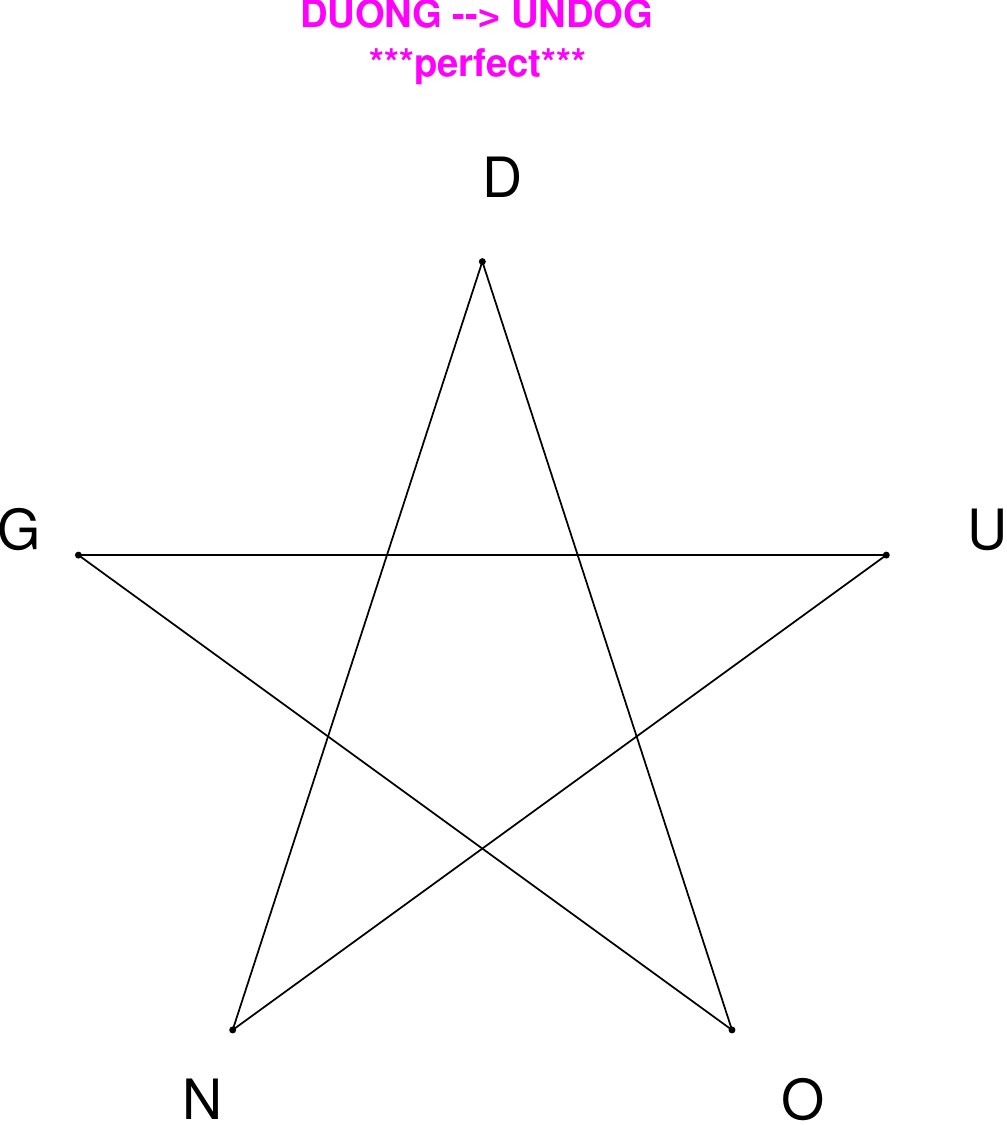}
\end{subfigure}
\hfill
\begin{subfigure}[T]{0.19\textwidth}
\centering
\includegraphics[width=\textwidth]{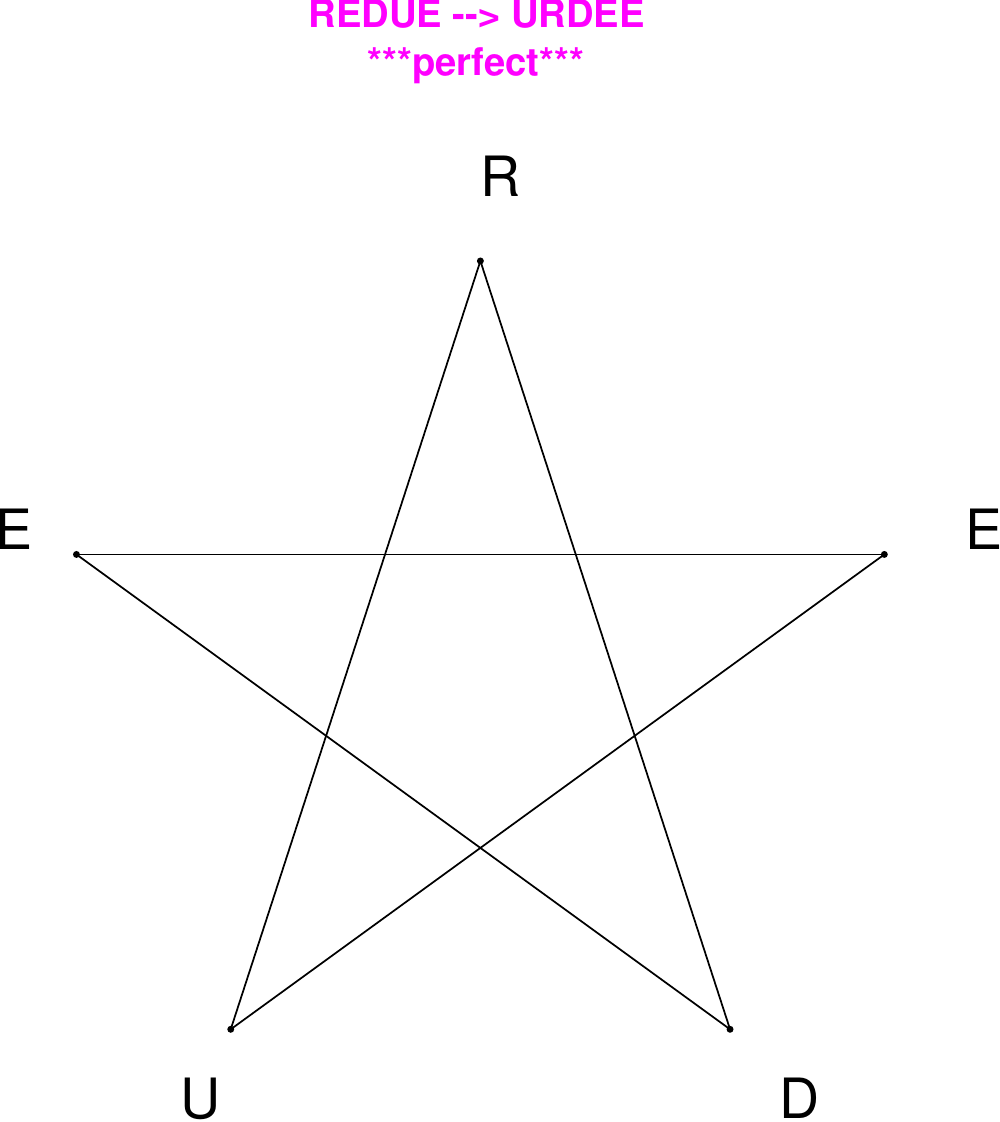}
\end{subfigure}
\end{figure}

\begin{figure}[H]
\centering
\begin{subfigure}[T]{0.19\textwidth}
\centering
\includegraphics[width=\textwidth]{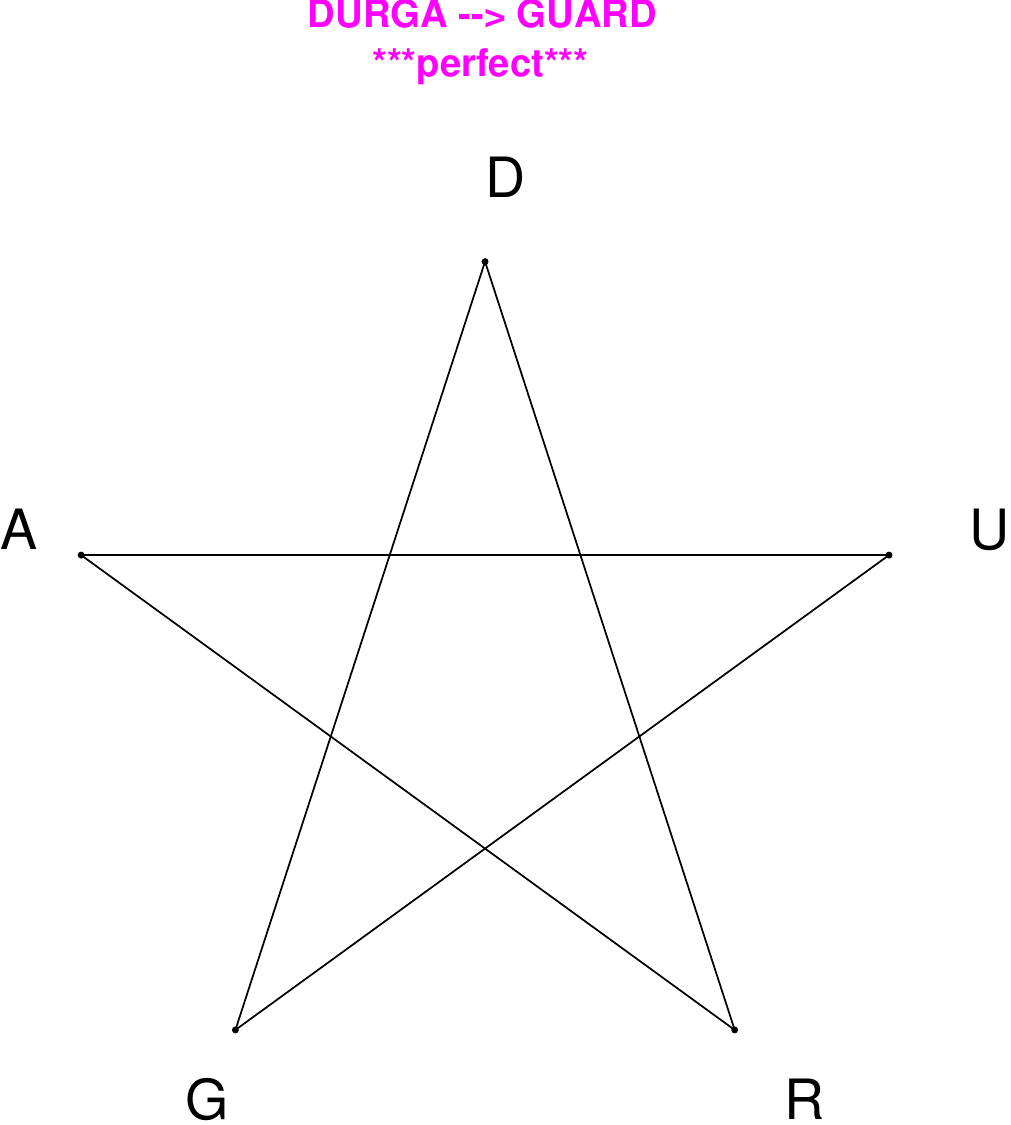}
\end{subfigure}
\hfill
\begin{subfigure}[T]{0.19\textwidth}
\centering
\includegraphics[width=\textwidth]{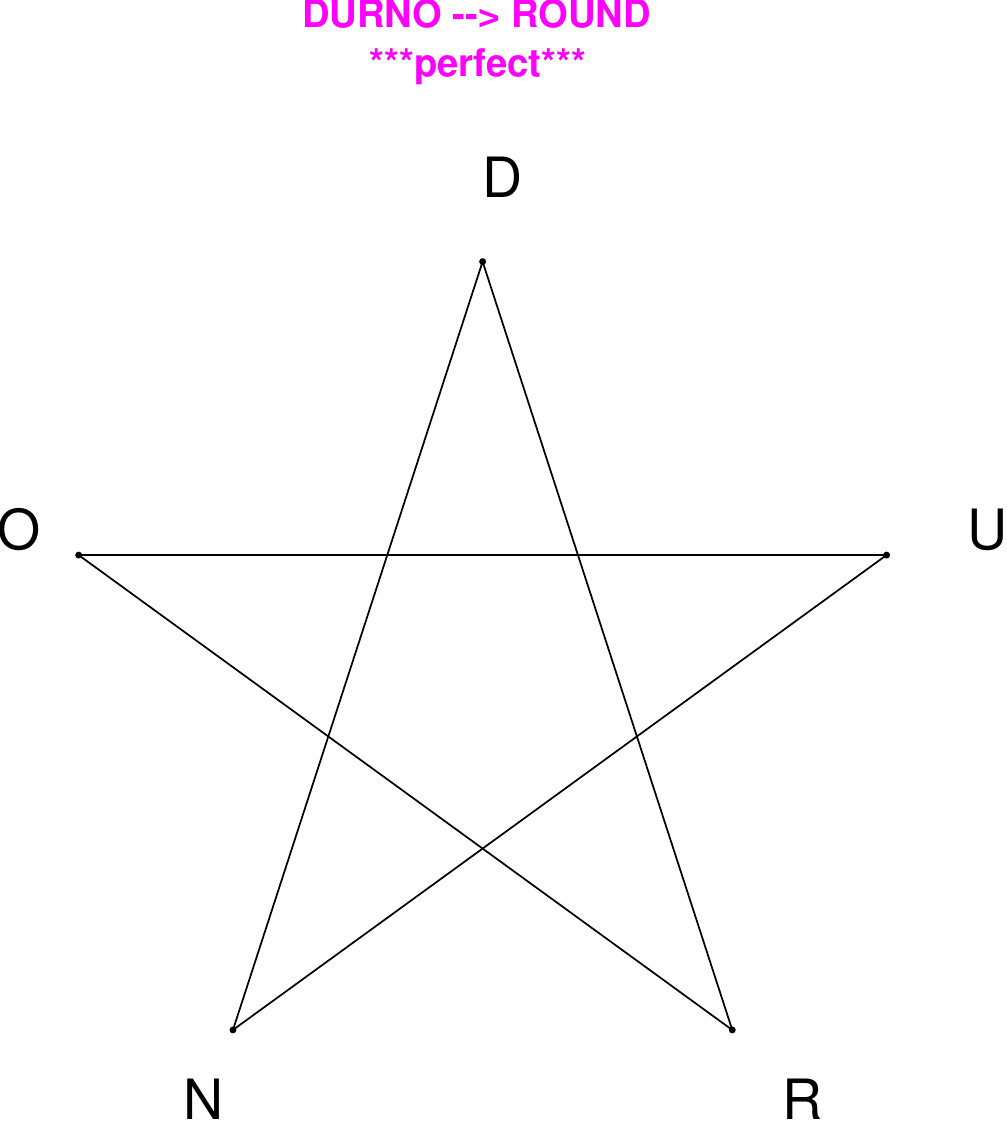}
\end{subfigure}
\hfill
\begin{subfigure}[T]{0.19\textwidth}
\centering
\includegraphics[width=\textwidth]{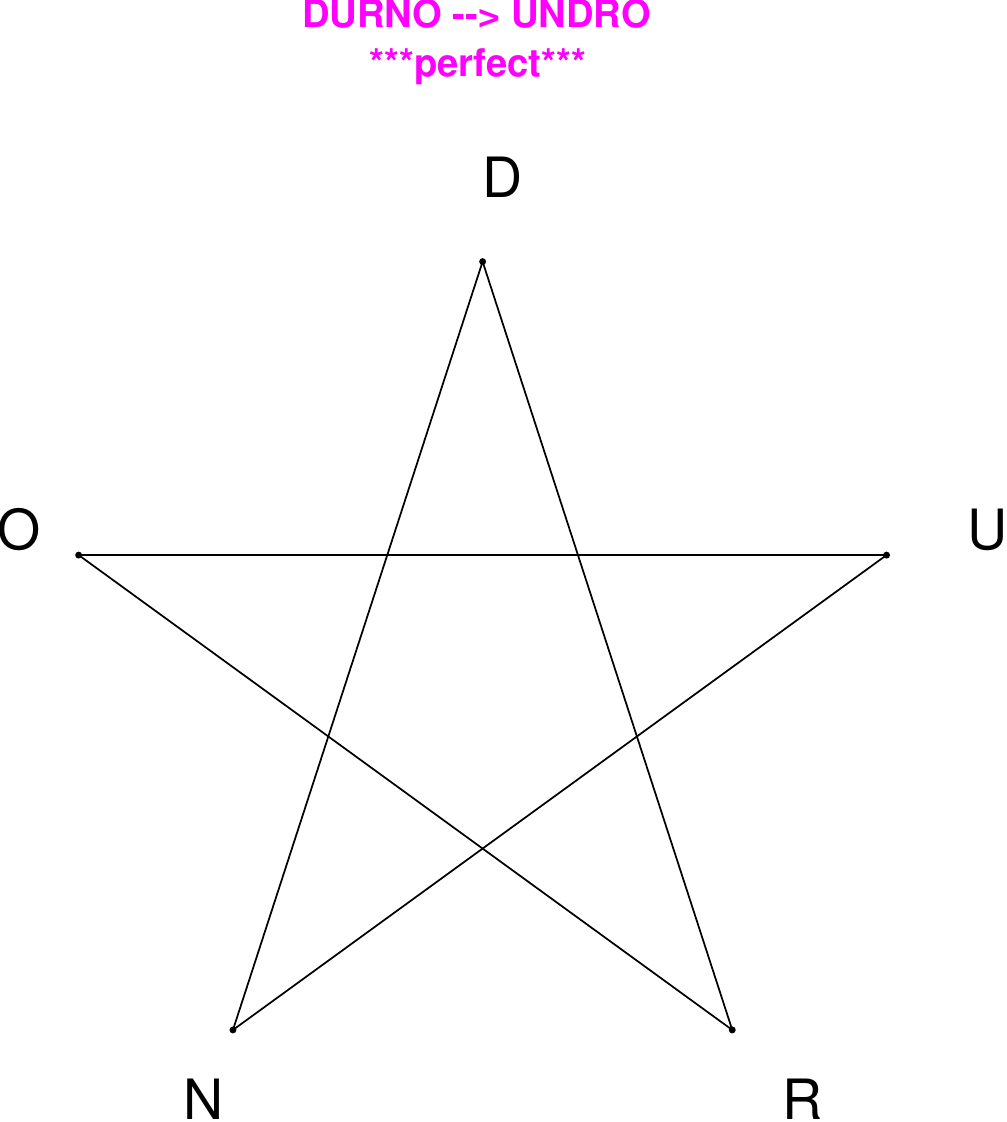}
\end{subfigure}
\hfill
\begin{subfigure}[T]{0.19\textwidth}
\centering
\includegraphics[width=\textwidth]{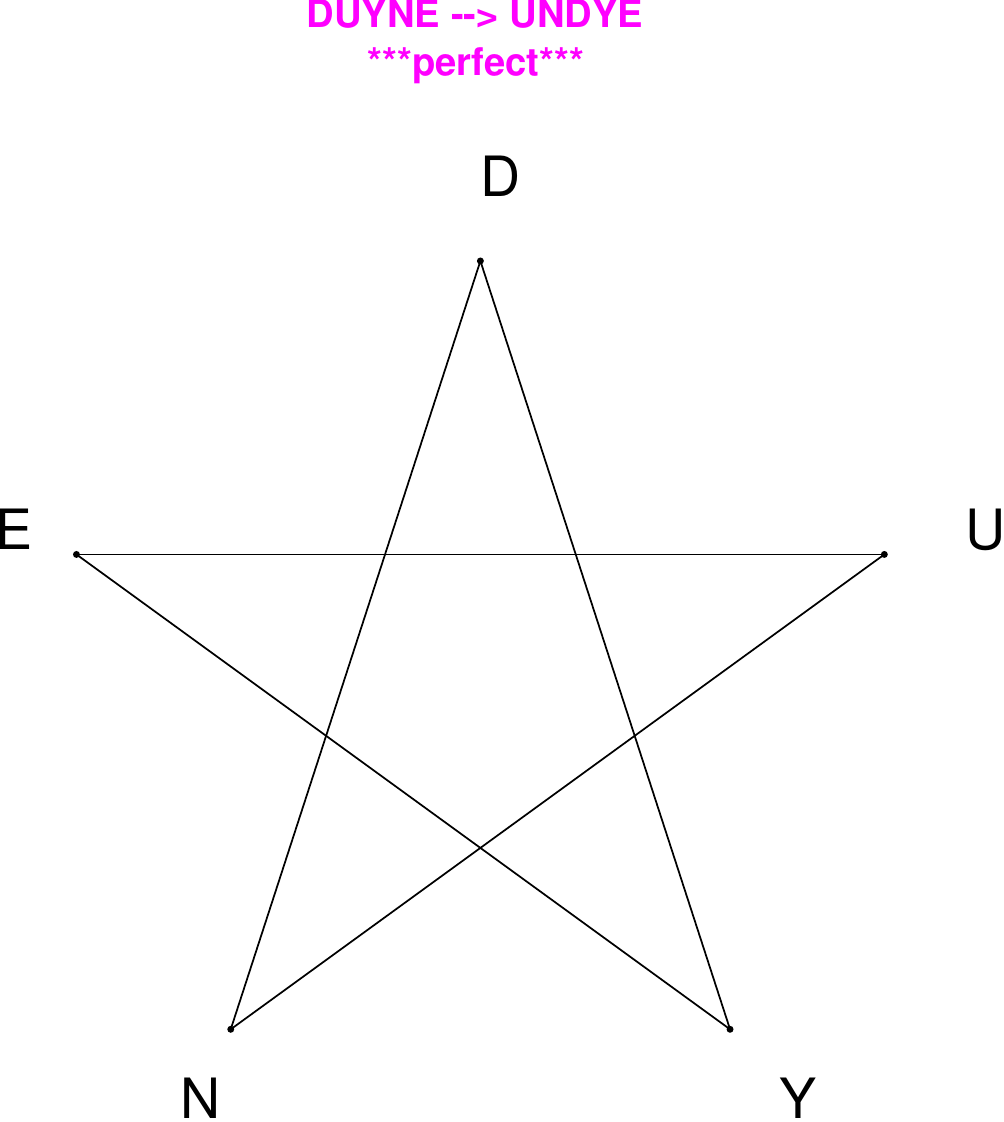}
\end{subfigure}
\hfill
\begin{subfigure}[T]{0.19\textwidth}
\centering
\includegraphics[width=\textwidth]{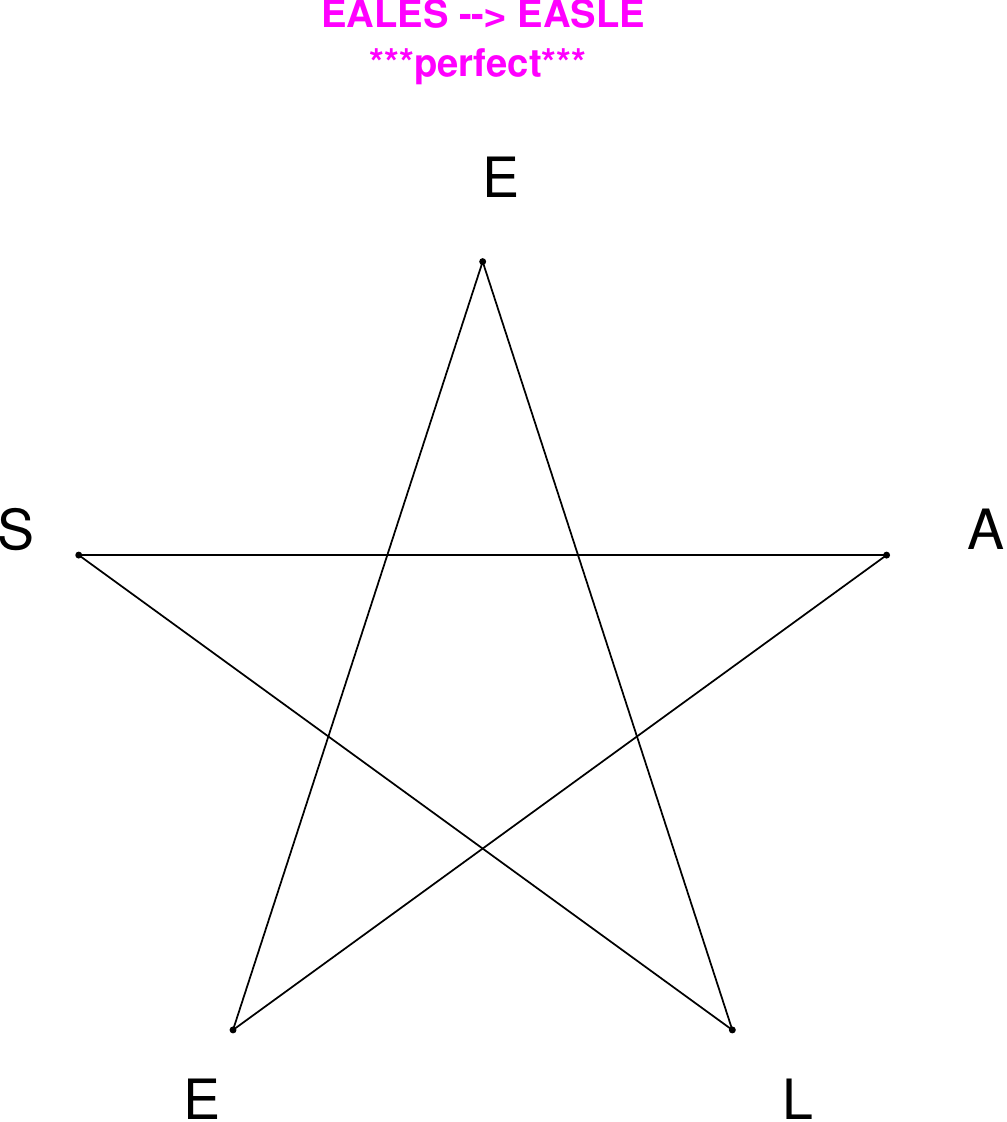}
\end{subfigure}
\end{figure}

\begin{figure}[H]
\centering
\begin{subfigure}[T]{0.19\textwidth}
\centering
\includegraphics[width=\textwidth]{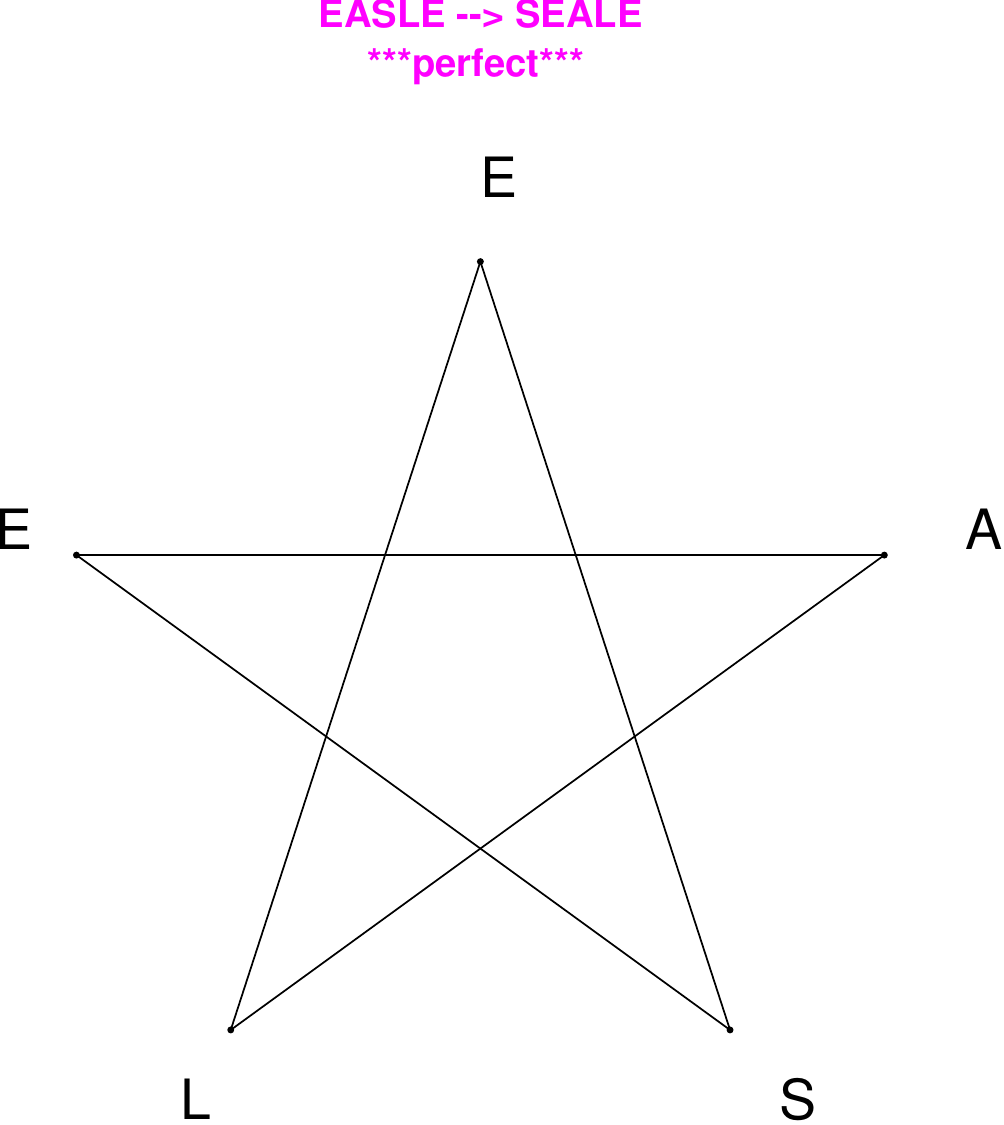}
\end{subfigure}
\hfill
\begin{subfigure}[T]{0.19\textwidth}
\centering
\includegraphics[width=\textwidth]{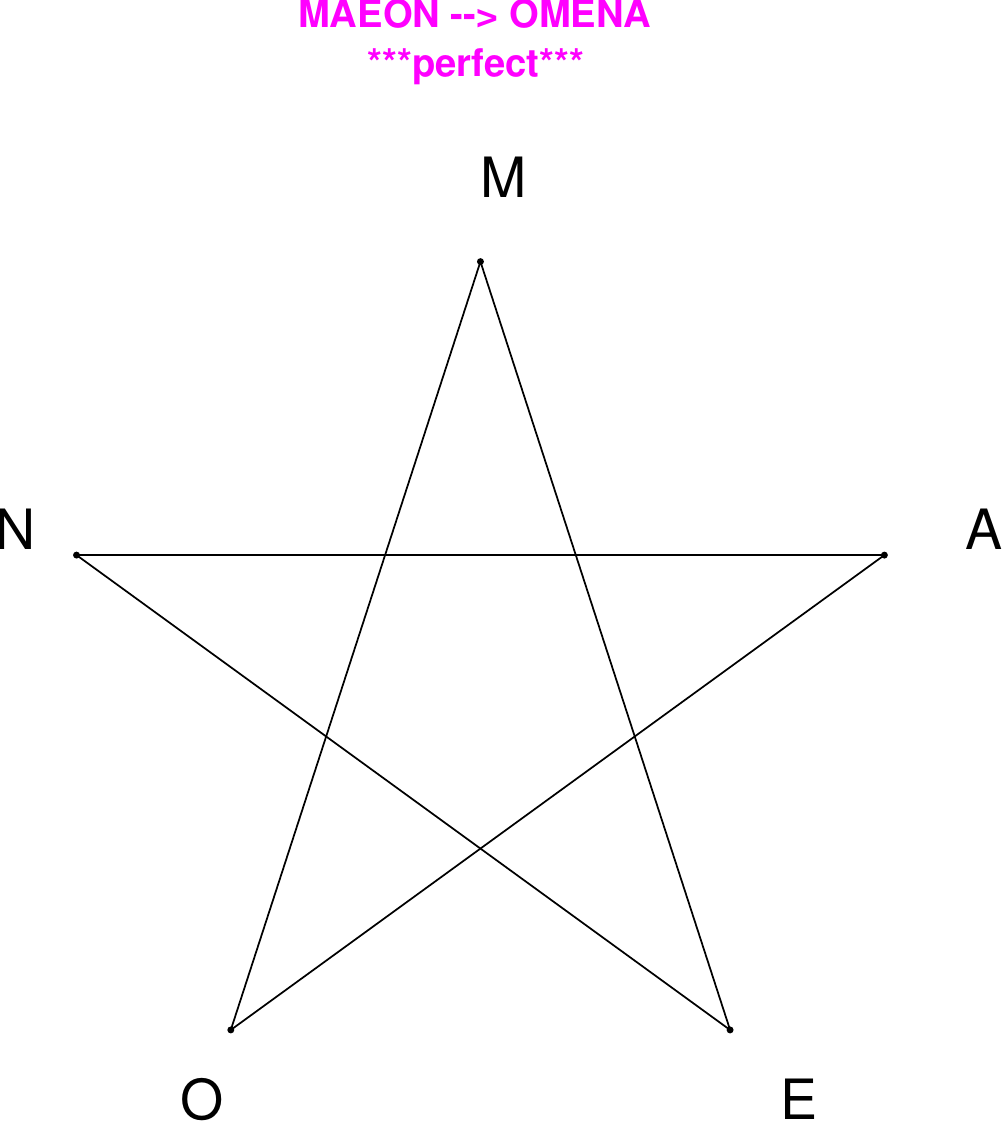}
\end{subfigure}
\hfill
\begin{subfigure}[T]{0.19\textwidth}
\centering
\includegraphics[width=\textwidth]{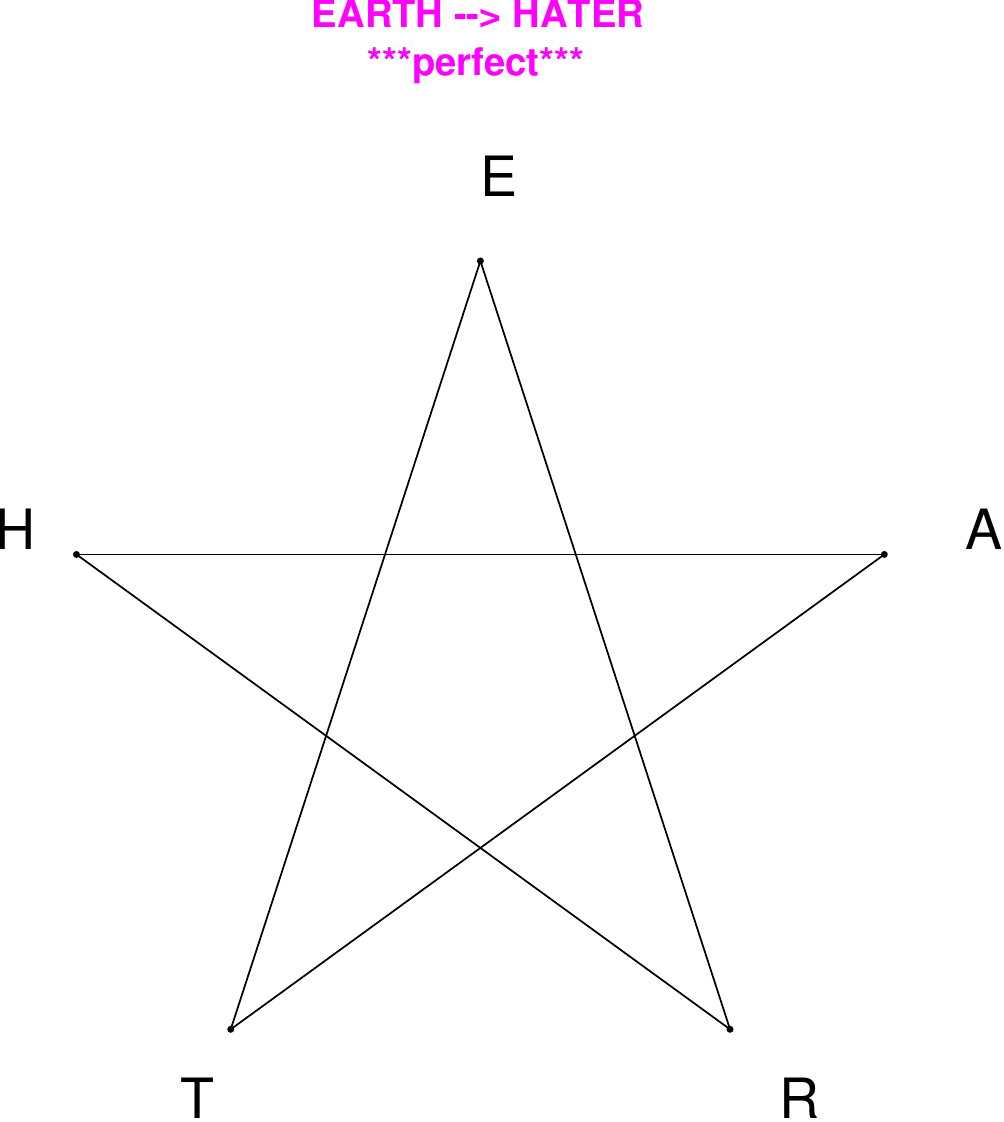}
\end{subfigure}
\hfill
\begin{subfigure}[T]{0.19\textwidth}
\centering
\includegraphics[width=\textwidth]{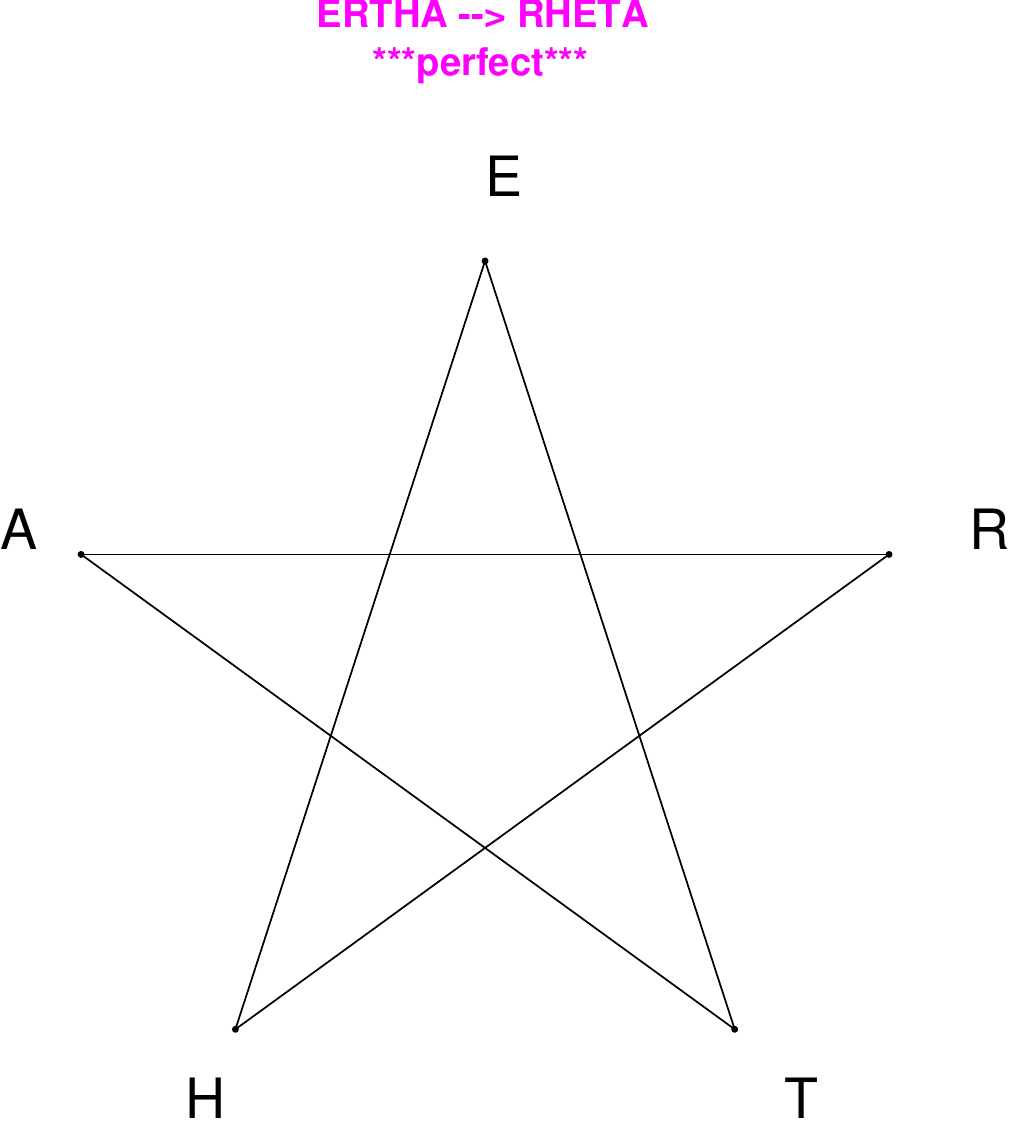}
\end{subfigure}
\hfill
\begin{subfigure}[T]{0.19\textwidth}
\centering
\includegraphics[width=\textwidth]{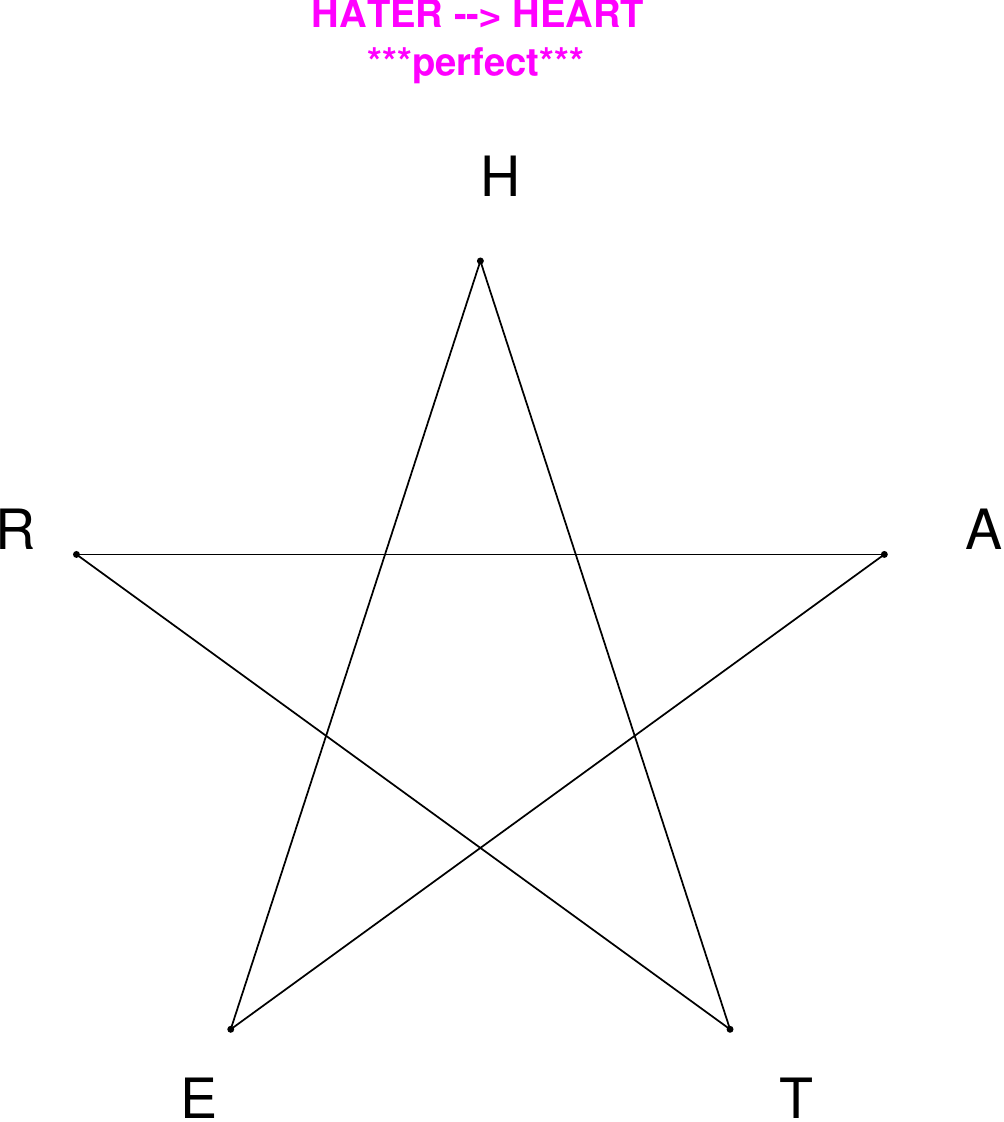}
\end{subfigure}
\end{figure}

\begin{figure}[H]
\centering
\begin{subfigure}[T]{0.19\textwidth}
\centering
\includegraphics[width=\textwidth]{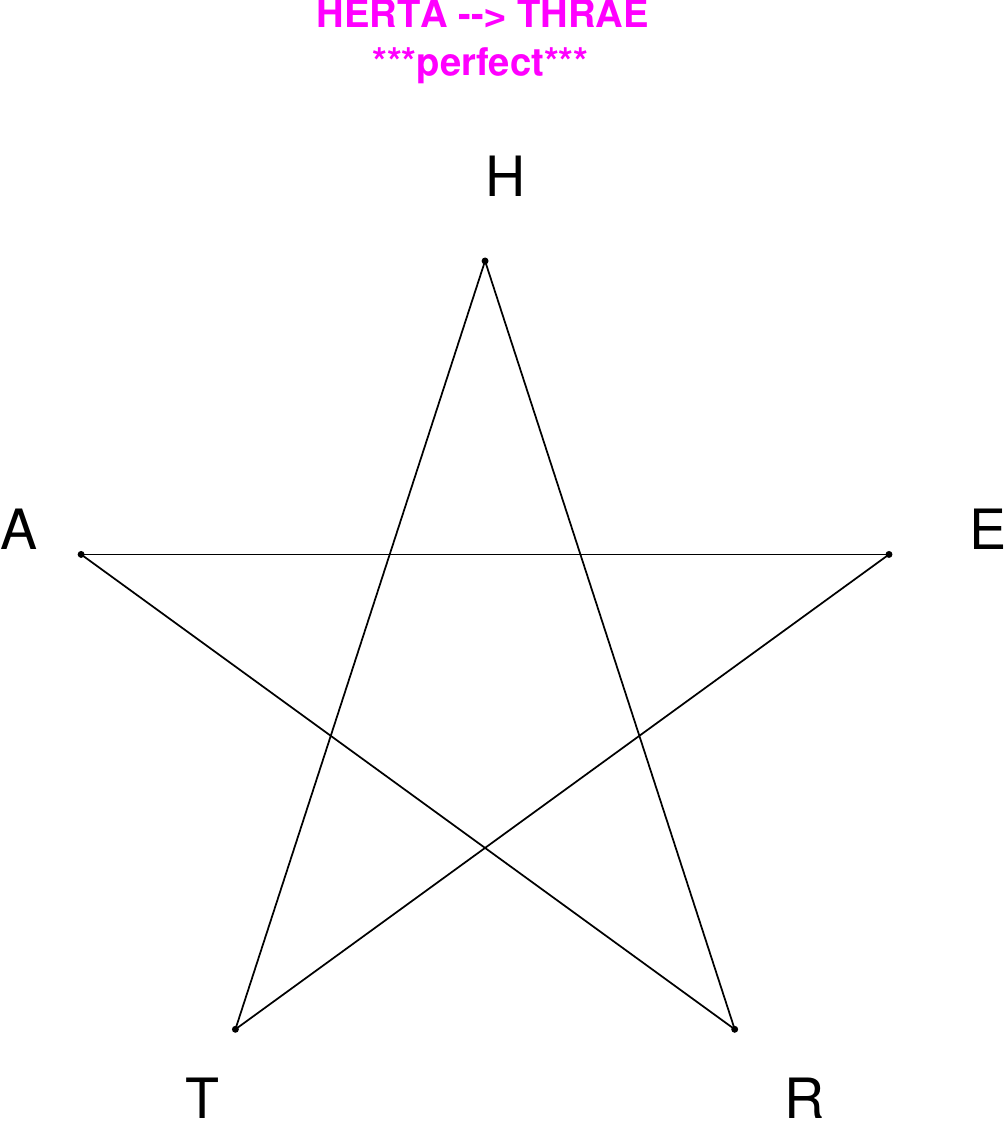}
\end{subfigure}
\hfill
\begin{subfigure}[T]{0.19\textwidth}
\centering
\includegraphics[width=\textwidth]{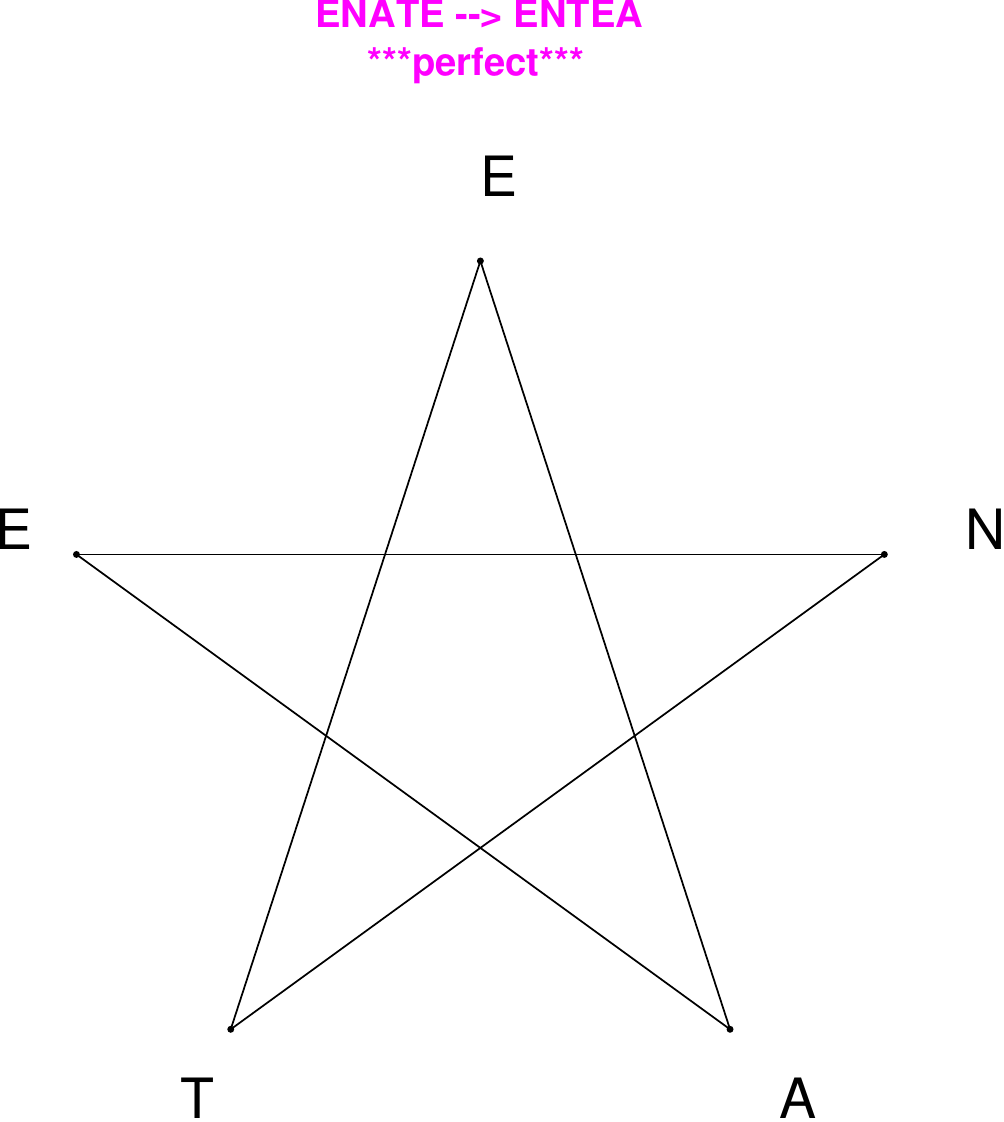}
\end{subfigure}
\hfill
\begin{subfigure}[T]{0.19\textwidth}
\centering
\includegraphics[width=\textwidth]{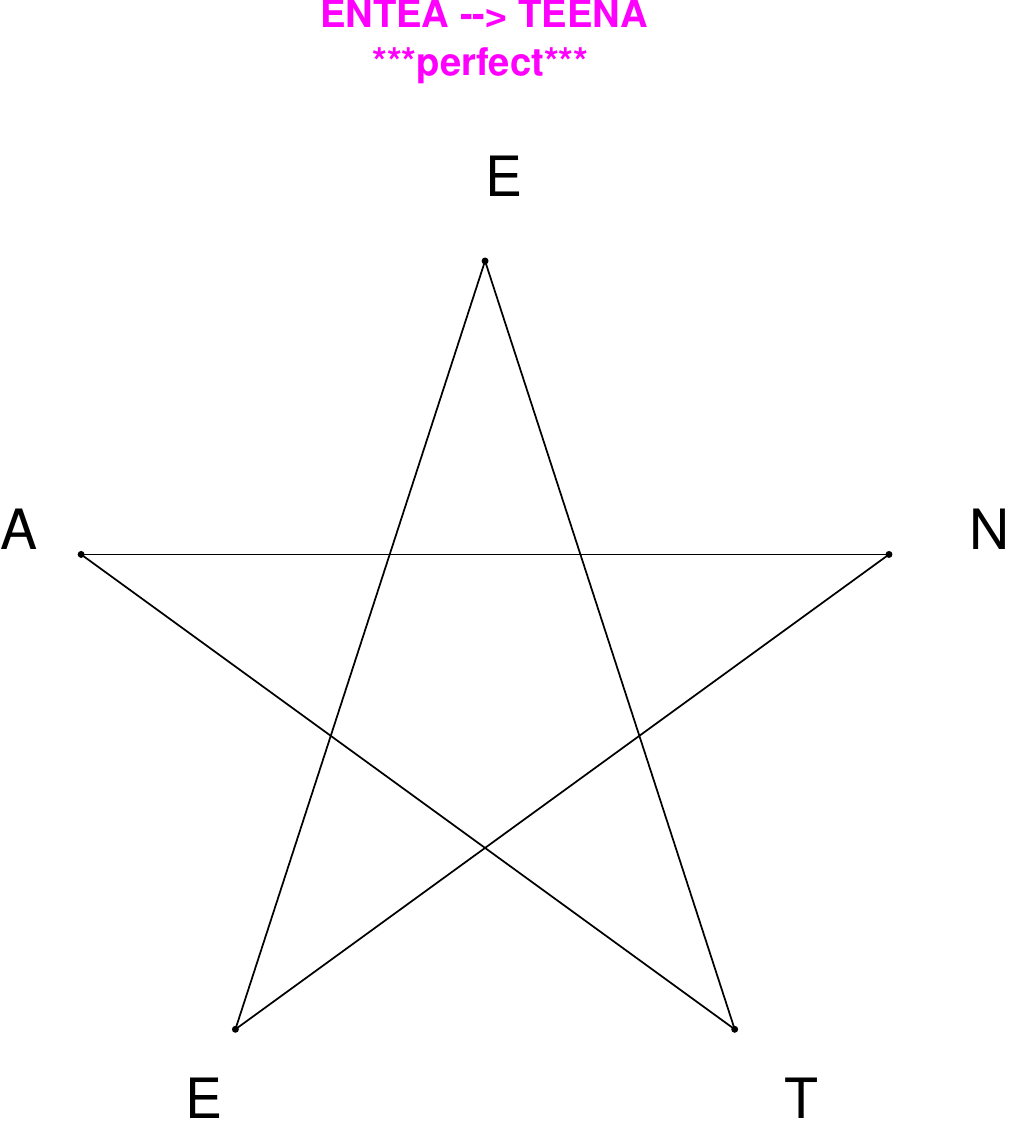}
\end{subfigure}
\hfill
\begin{subfigure}[T]{0.19\textwidth}
\centering
\includegraphics[width=\textwidth]{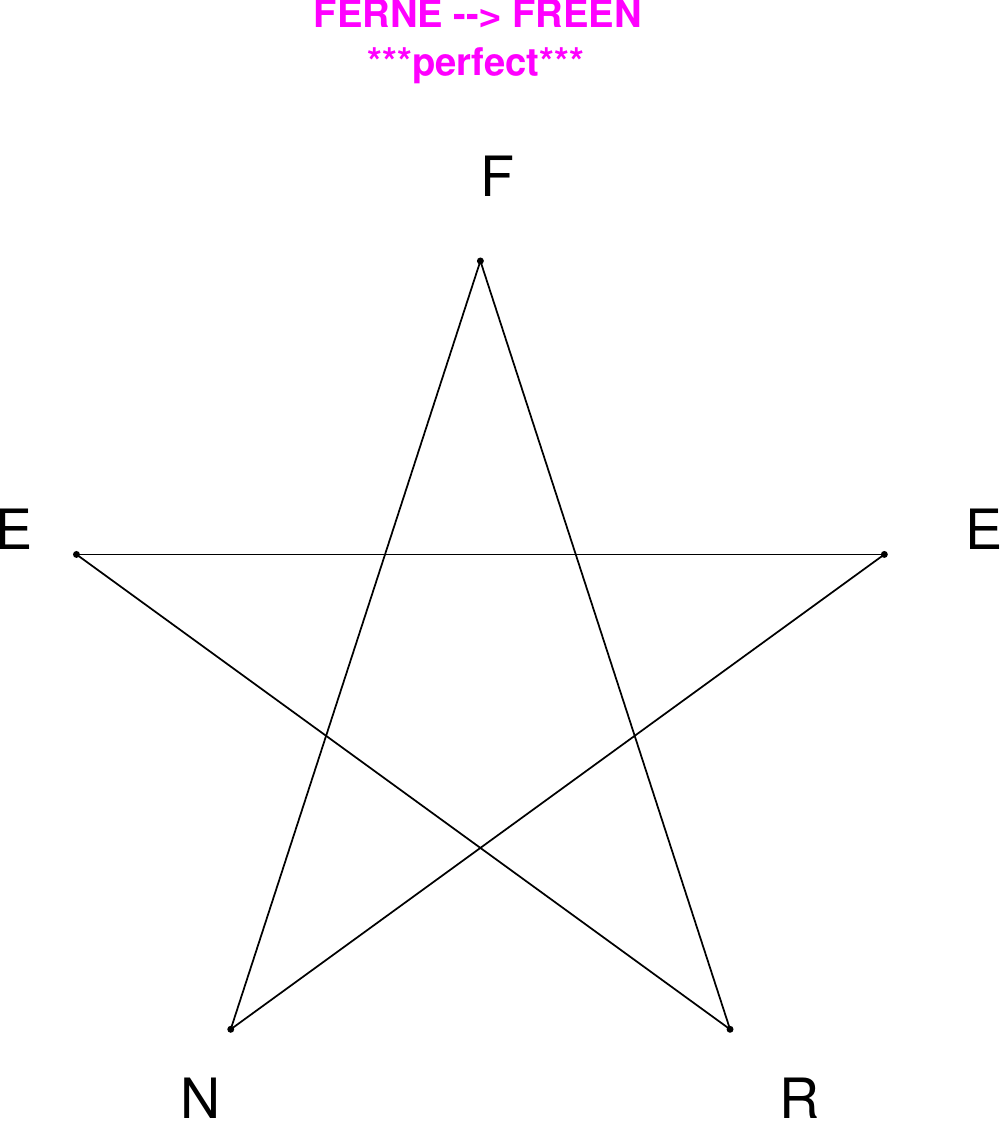}
\end{subfigure}
\hfill
\begin{subfigure}[T]{0.19\textwidth}
\centering
\includegraphics[width=\textwidth]{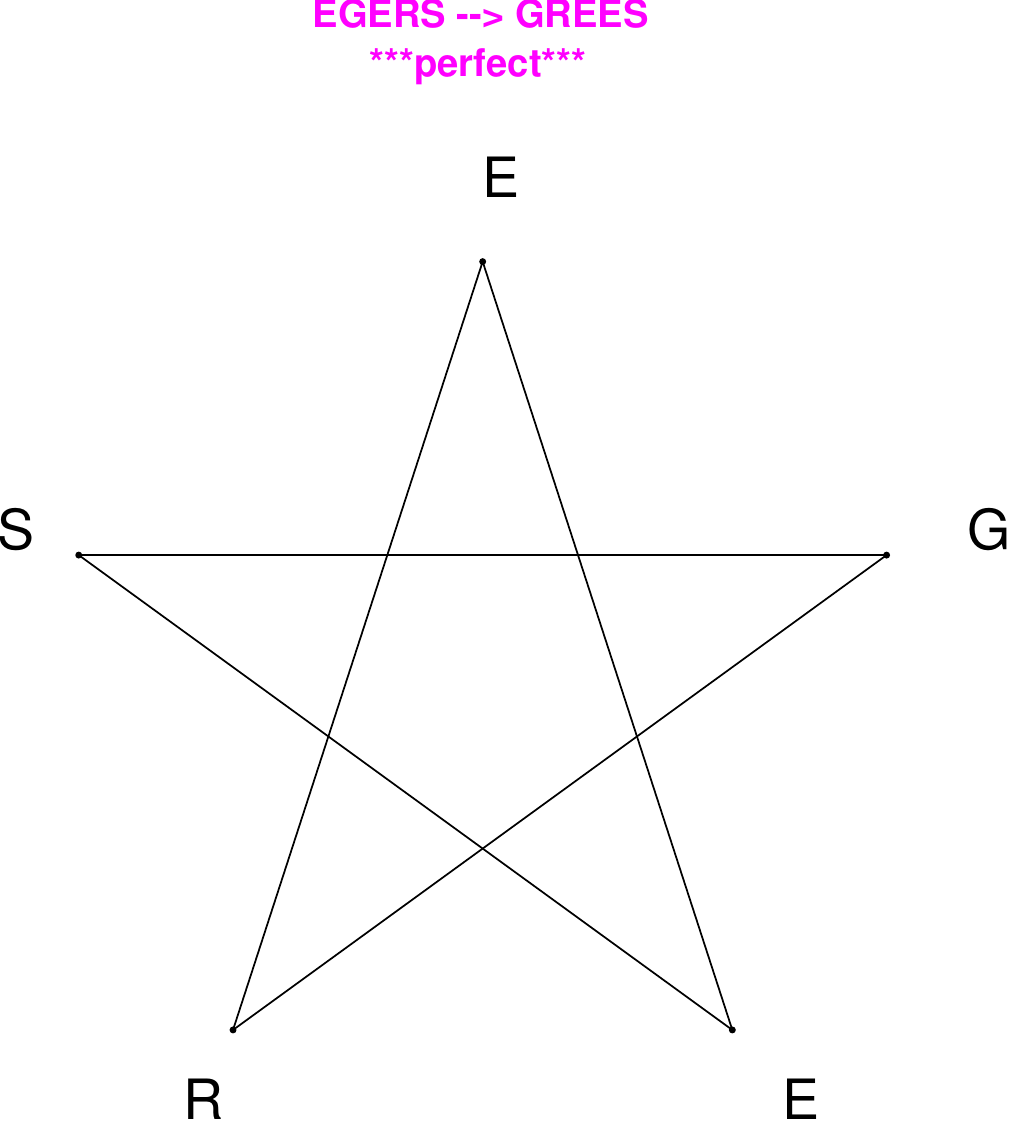}
\end{subfigure}
\end{figure}

\begin{figure}[H]
\centering
\begin{subfigure}[T]{0.19\textwidth}
\centering
\includegraphics[width=\textwidth]{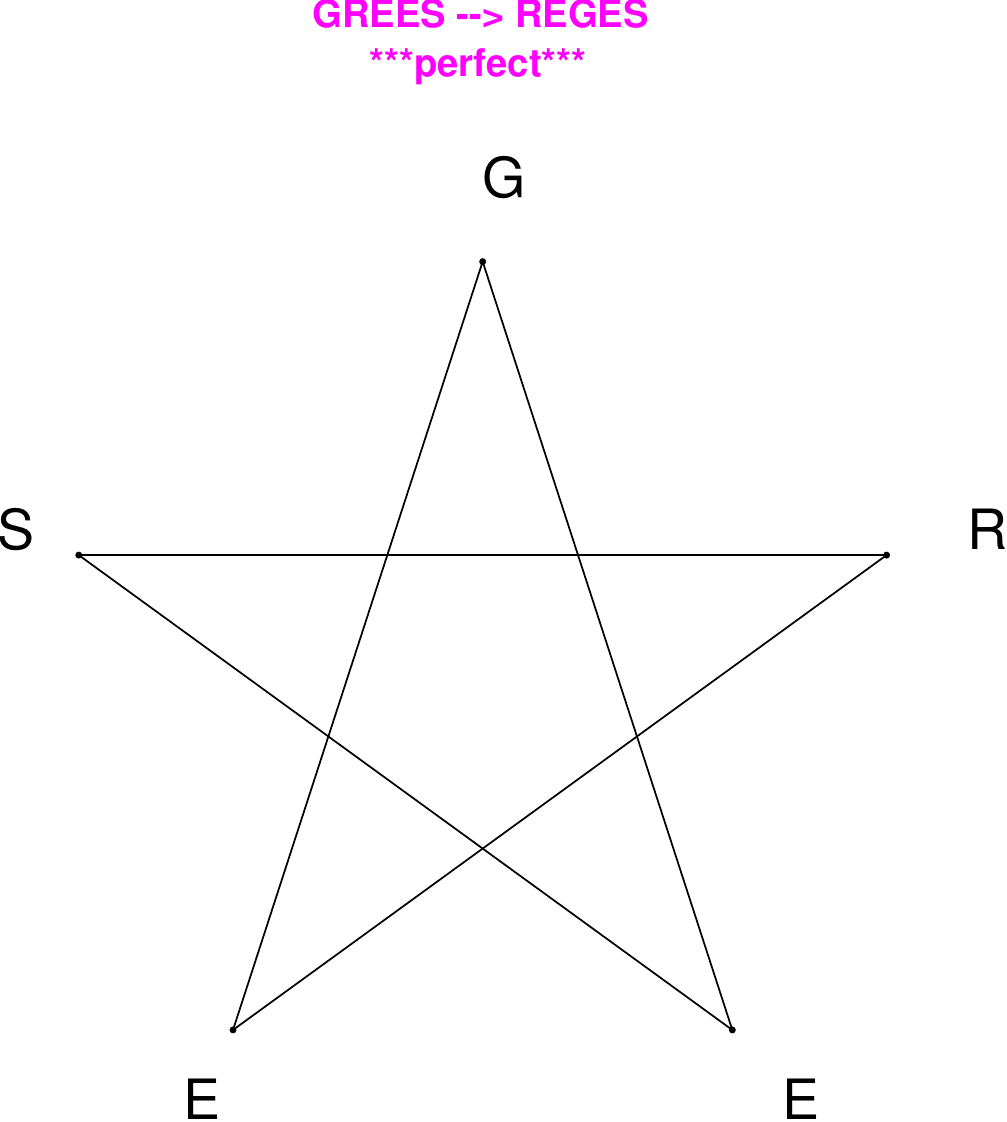}
\end{subfigure}
\hfill
\begin{subfigure}[T]{0.19\textwidth}
\centering
\includegraphics[width=\textwidth]{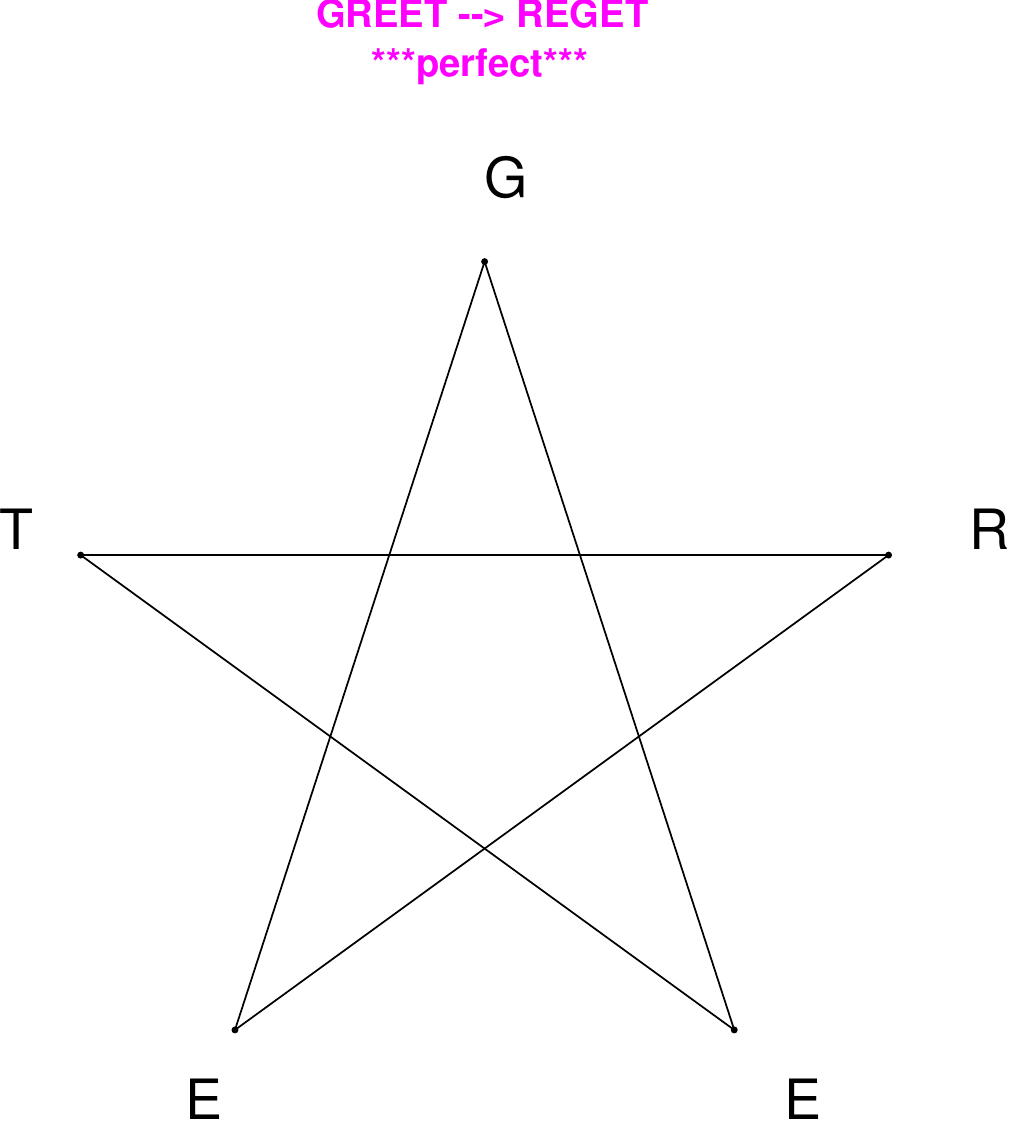}
\end{subfigure}
\hfill
\begin{subfigure}[T]{0.19\textwidth}
\centering
\includegraphics[width=\textwidth]{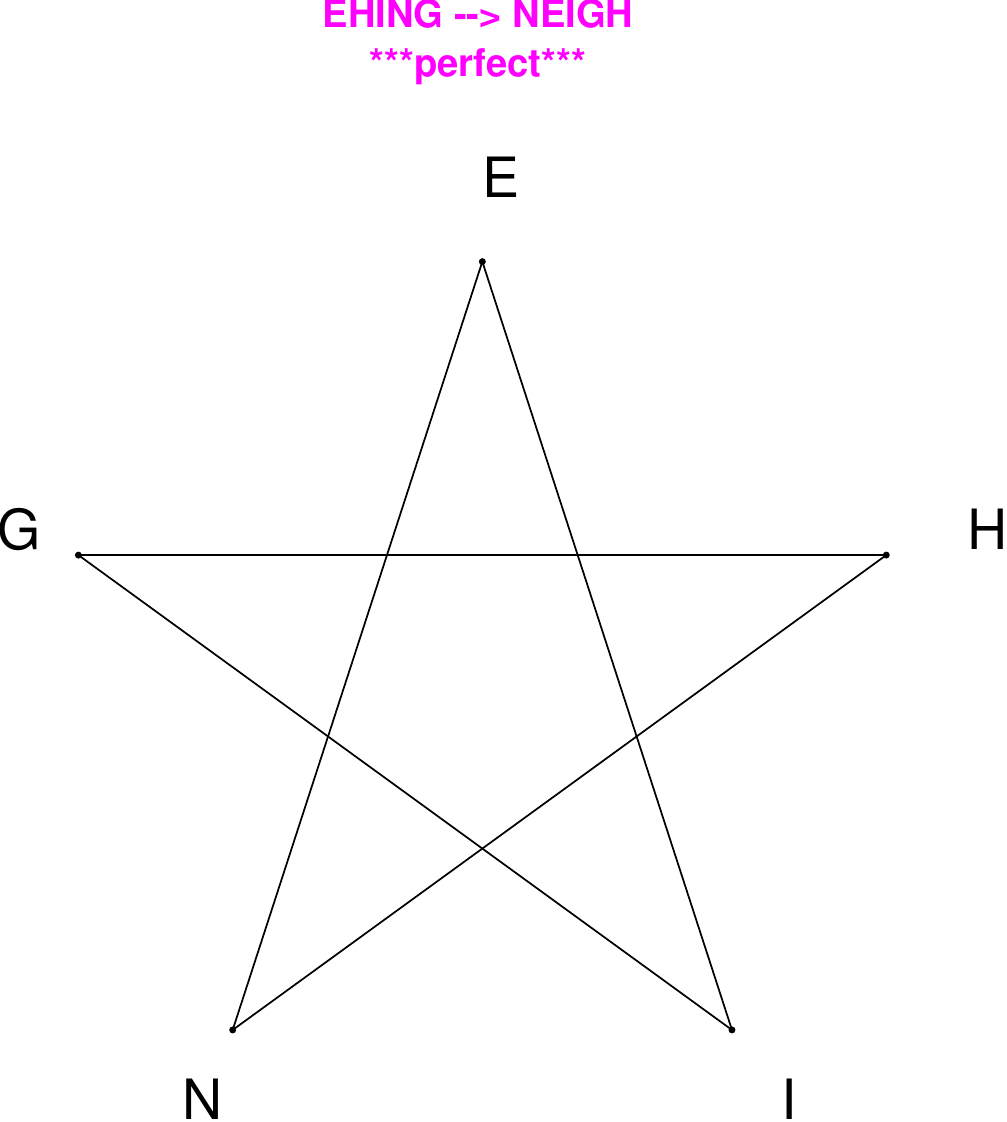}
\end{subfigure}
\hfill
\begin{subfigure}[T]{0.19\textwidth}
\centering
\includegraphics[width=\textwidth]{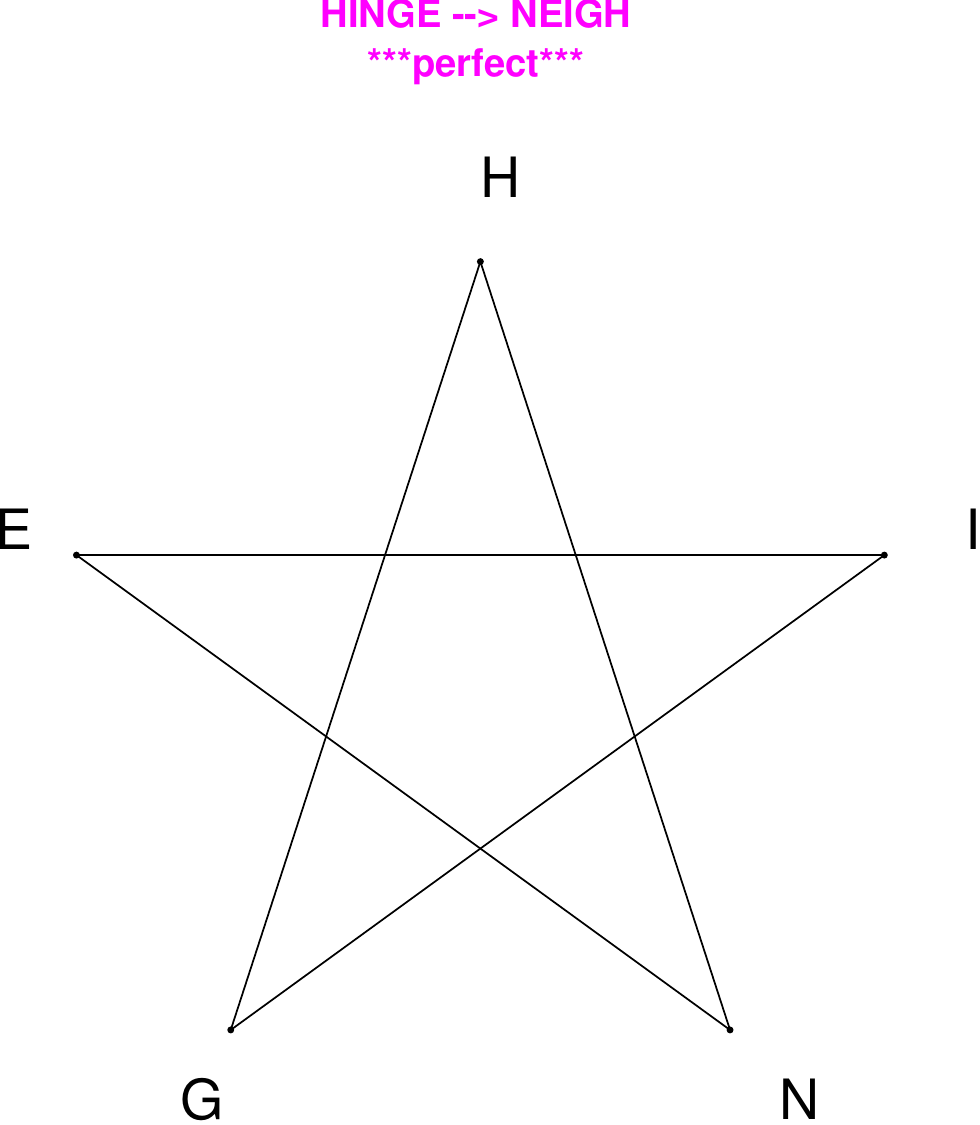}
\end{subfigure}
\hfill
\begin{subfigure}[T]{0.19\textwidth}
\centering
\includegraphics[width=\textwidth]{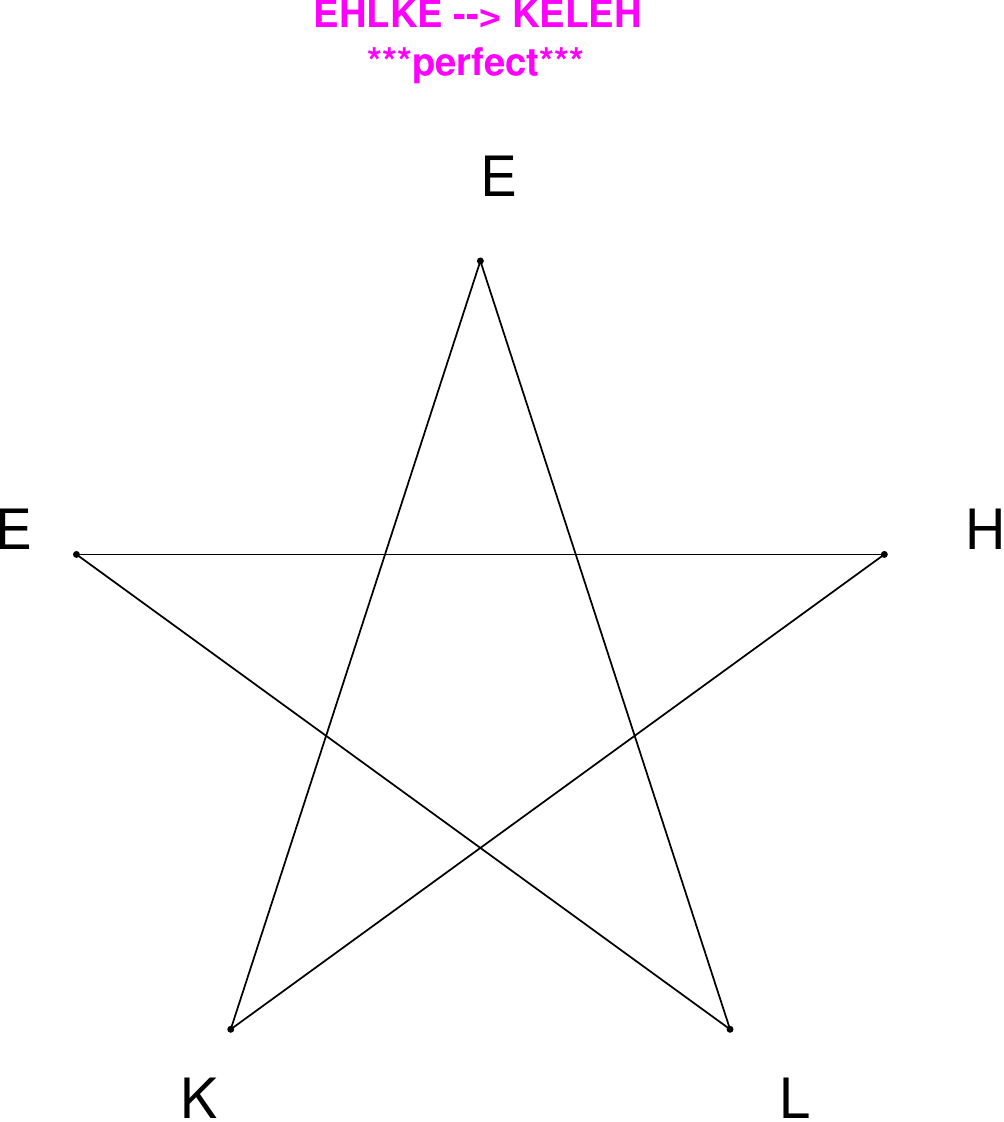}
\end{subfigure}
\end{figure}

\begin{figure}[H]
\centering
\begin{subfigure}[T]{0.19\textwidth}
\centering
\includegraphics[width=\textwidth]{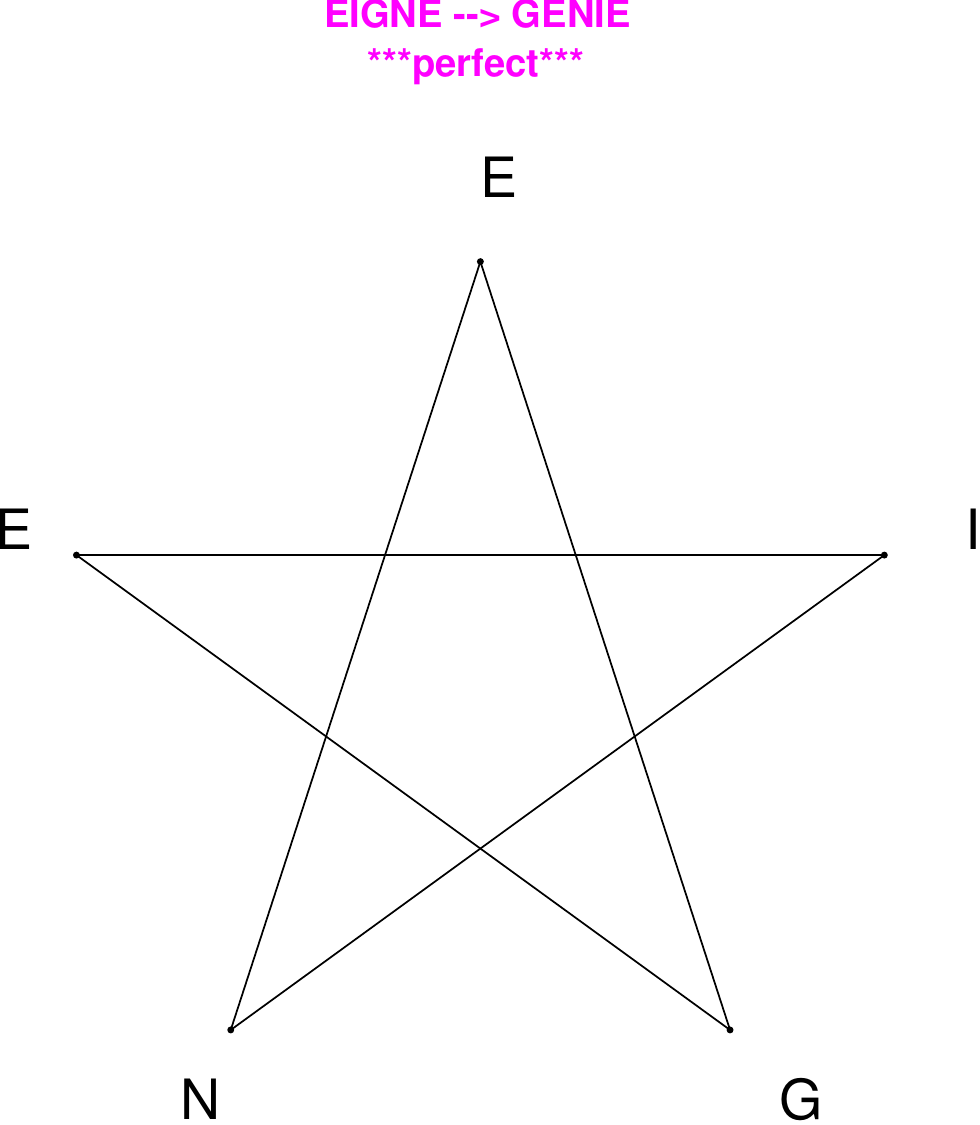}
\end{subfigure}
\hfill
\begin{subfigure}[T]{0.19\textwidth}
\centering
\includegraphics[width=\textwidth]{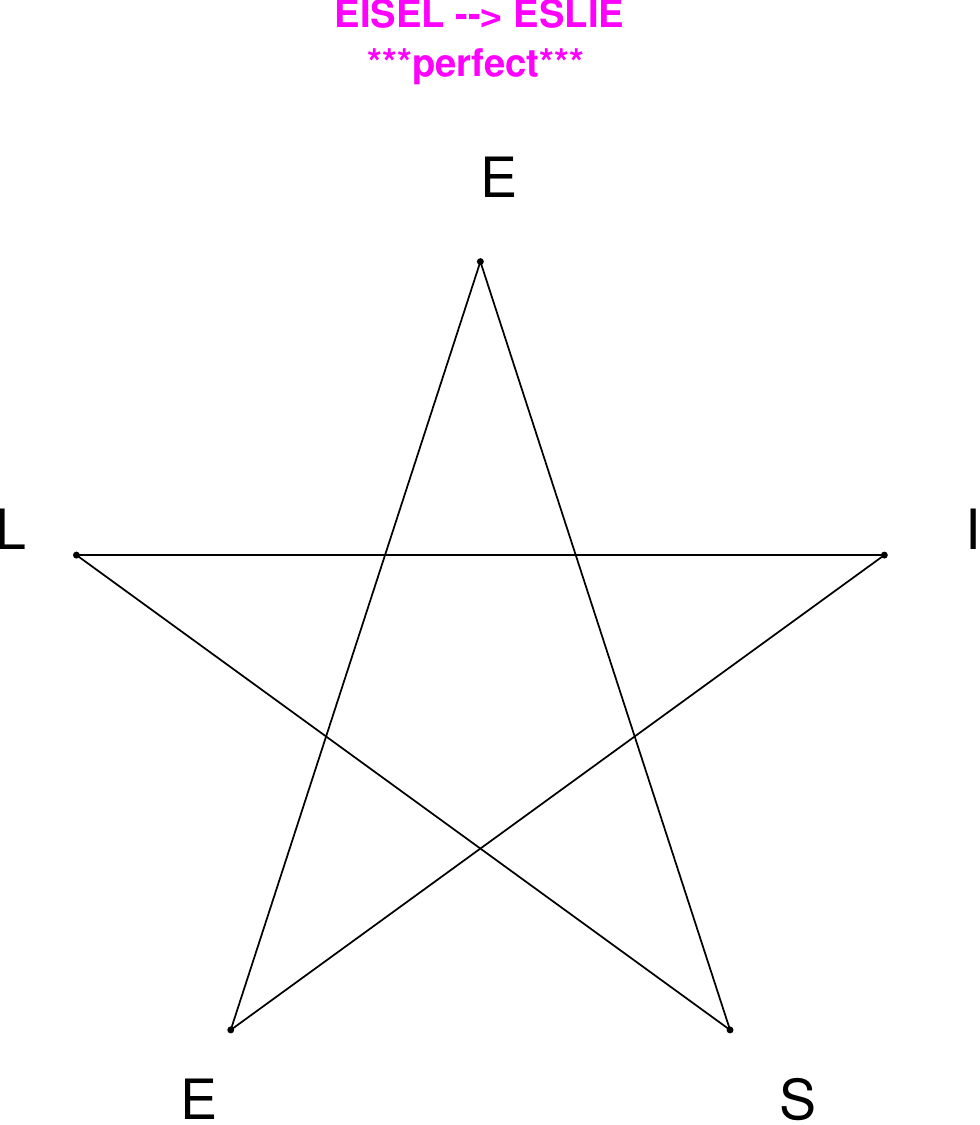}
\end{subfigure}
\hfill
\begin{subfigure}[T]{0.19\textwidth}
\centering
\includegraphics[width=\textwidth]{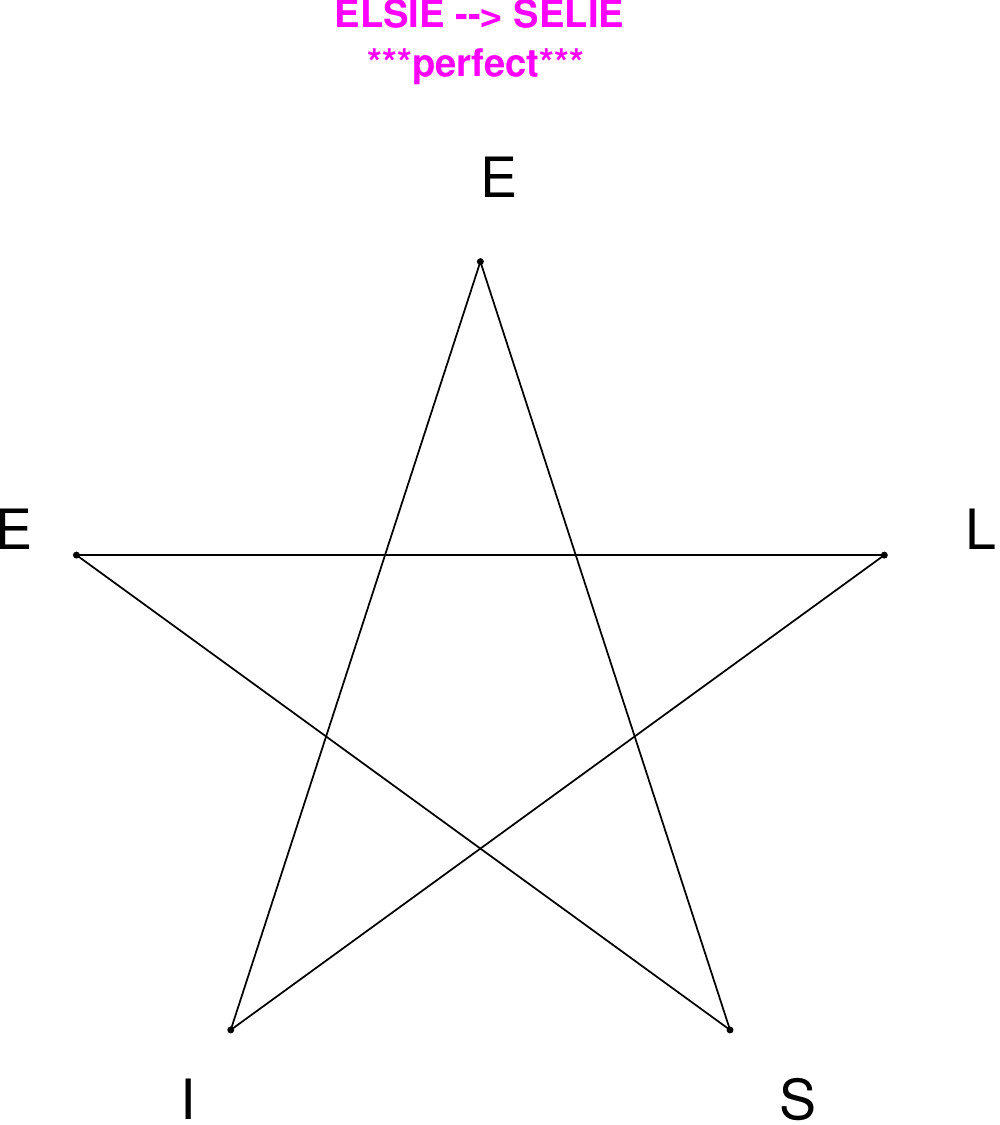}
\end{subfigure}
\hfill
\begin{subfigure}[T]{0.19\textwidth}
\centering
\includegraphics[width=\textwidth]{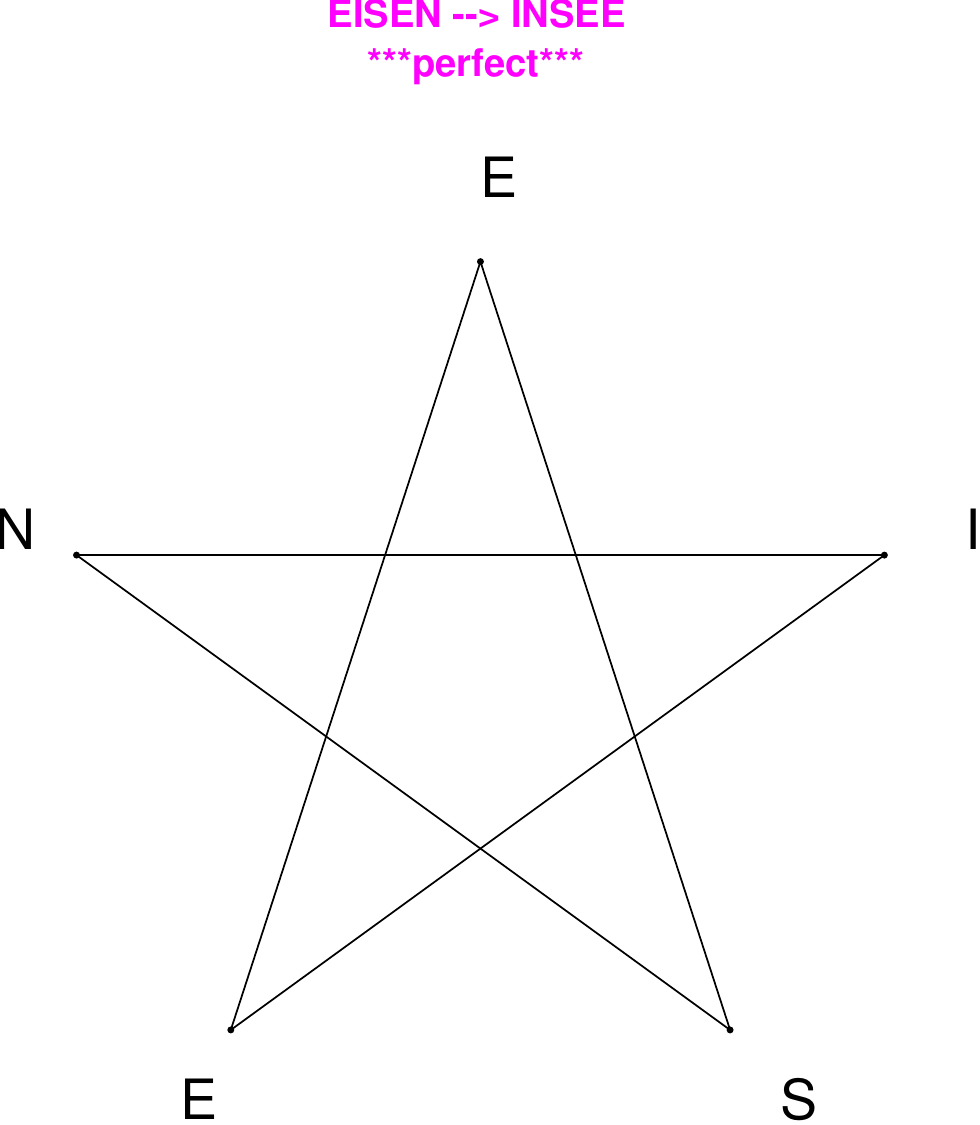}
\end{subfigure}
\hfill
\begin{subfigure}[T]{0.19\textwidth}
\centering
\includegraphics[width=\textwidth]{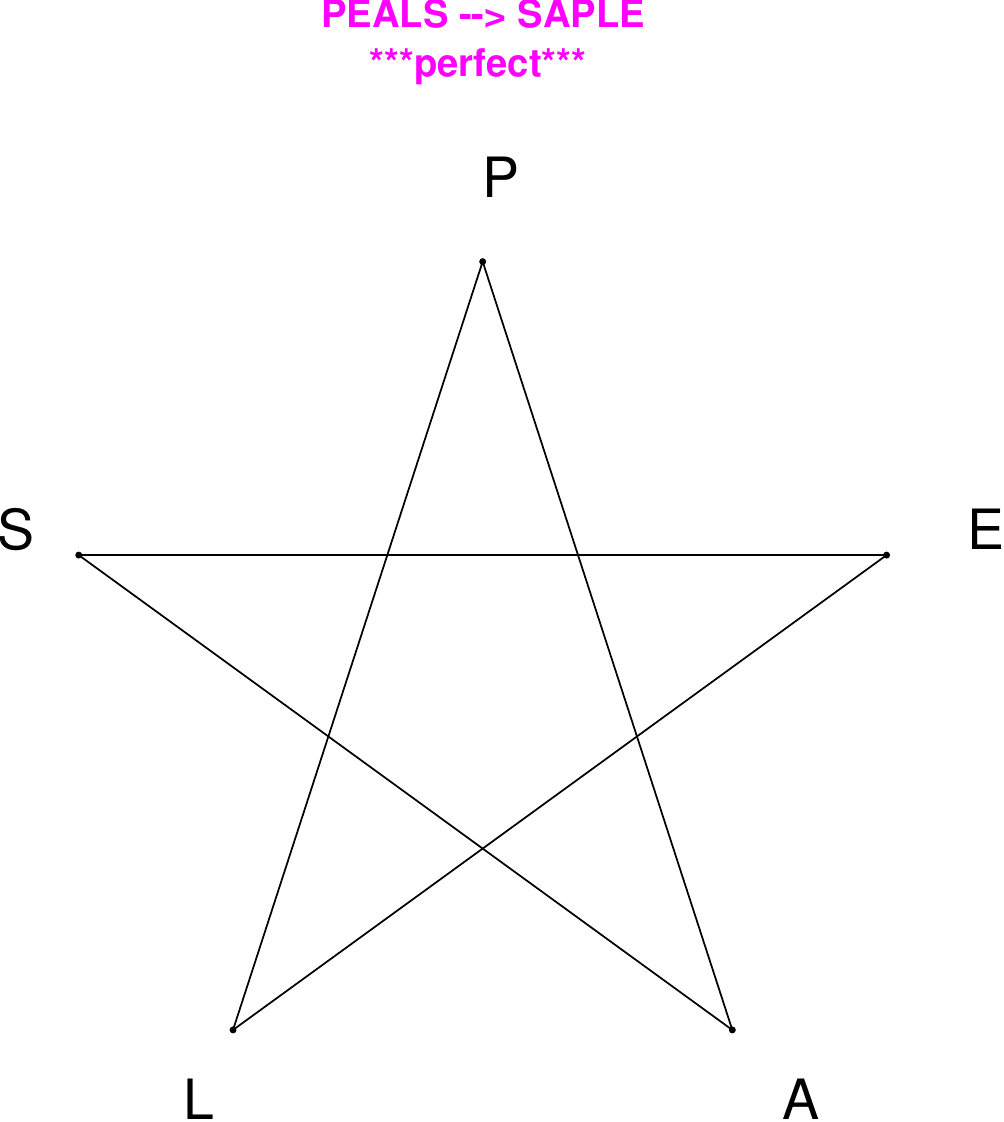}
\end{subfigure}
\end{figure}

\begin{figure}[H]
\centering
\begin{subfigure}[T]{0.19\textwidth}
\centering
\includegraphics[width=\textwidth]{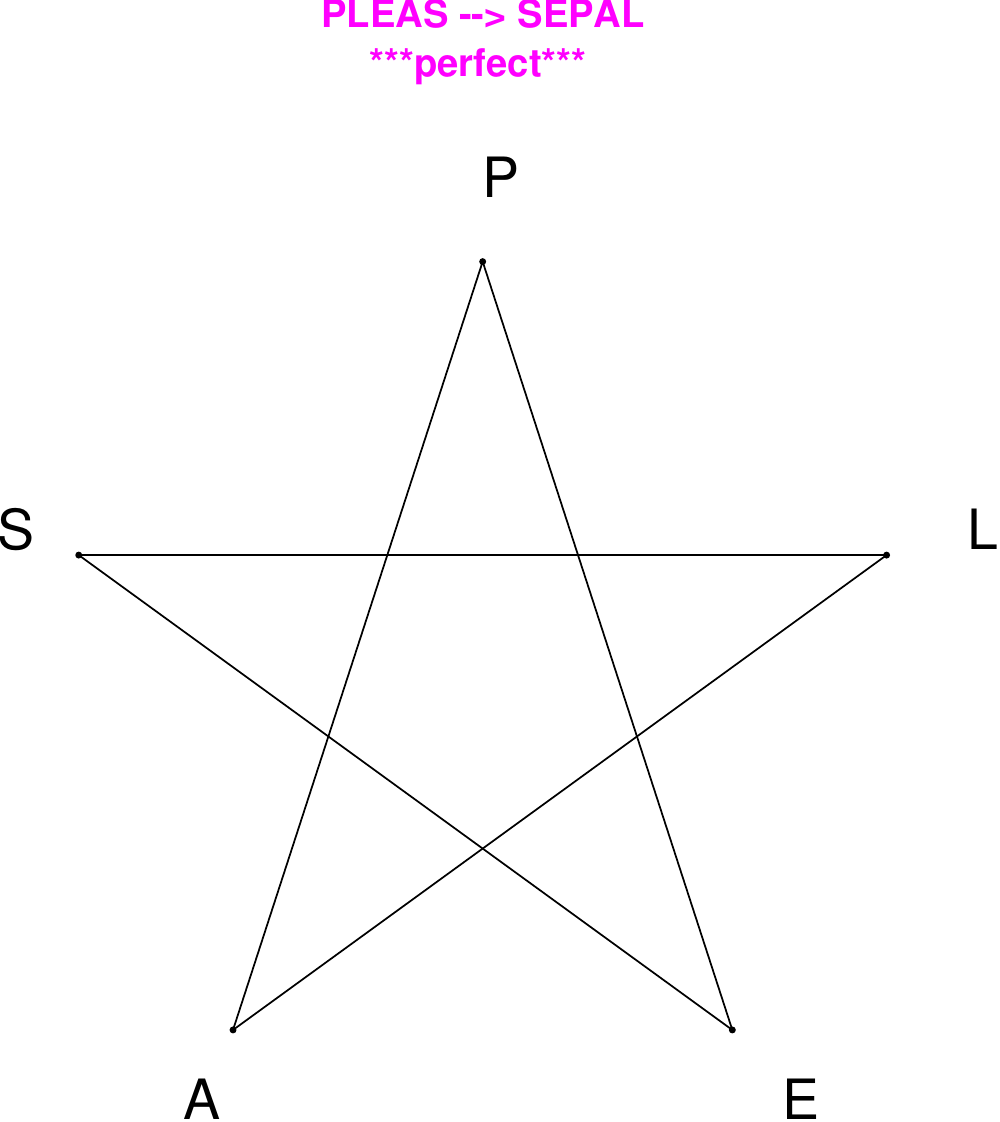}
\end{subfigure}
\hfill
\begin{subfigure}[T]{0.19\textwidth}
\centering
\includegraphics[width=\textwidth]{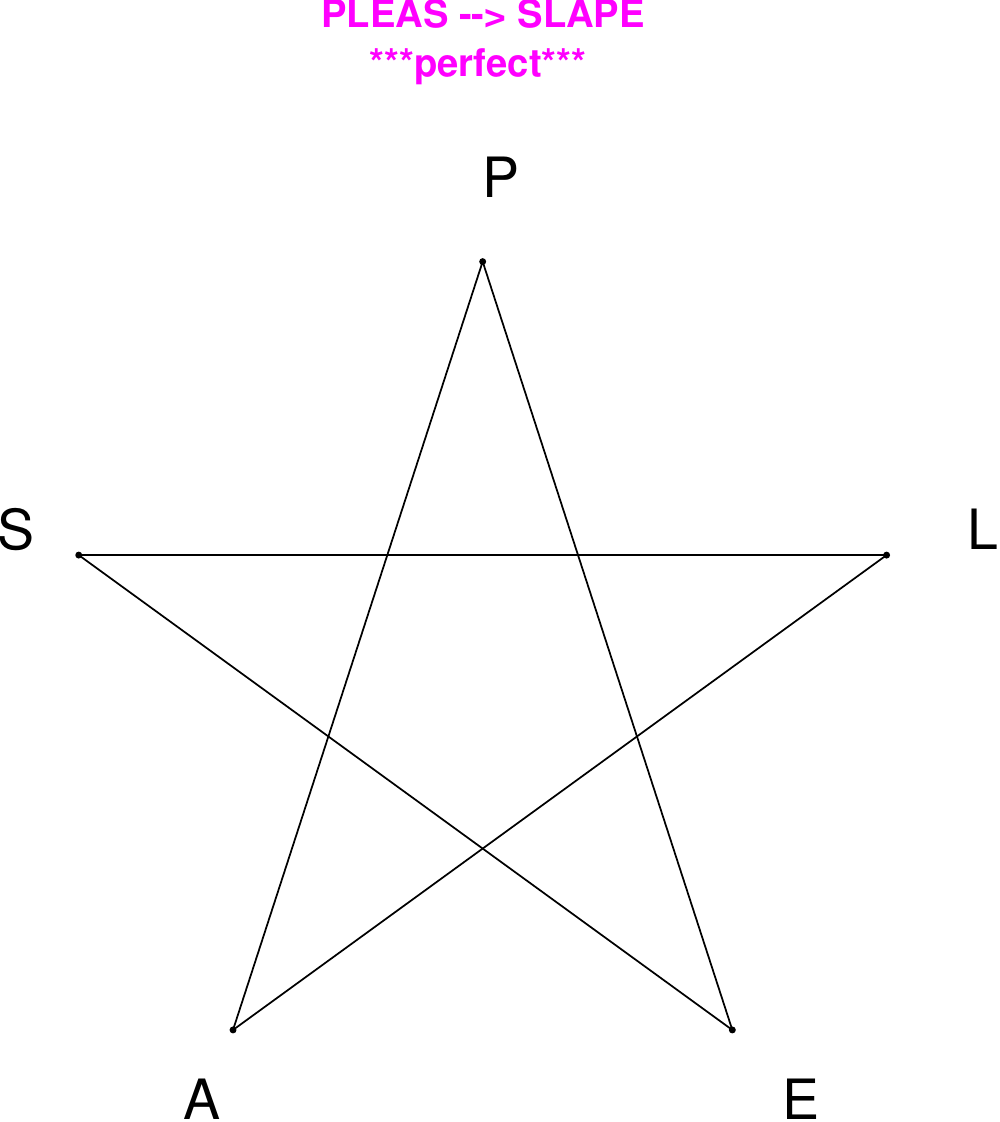}
\end{subfigure}
\hfill
\begin{subfigure}[T]{0.19\textwidth}
\centering
\includegraphics[width=\textwidth]{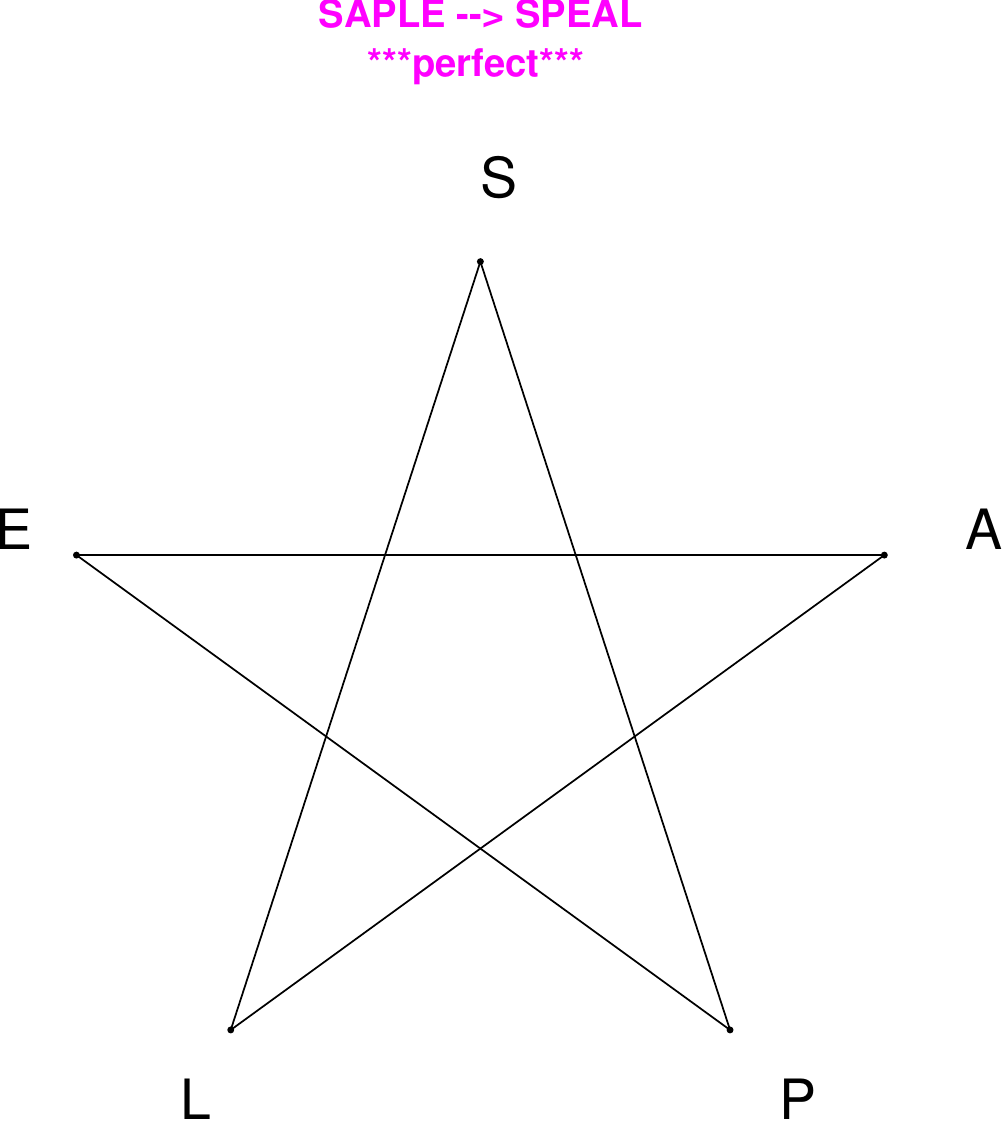}
\end{subfigure}
\hfill
\begin{subfigure}[T]{0.19\textwidth}
\centering
\includegraphics[width=\textwidth]{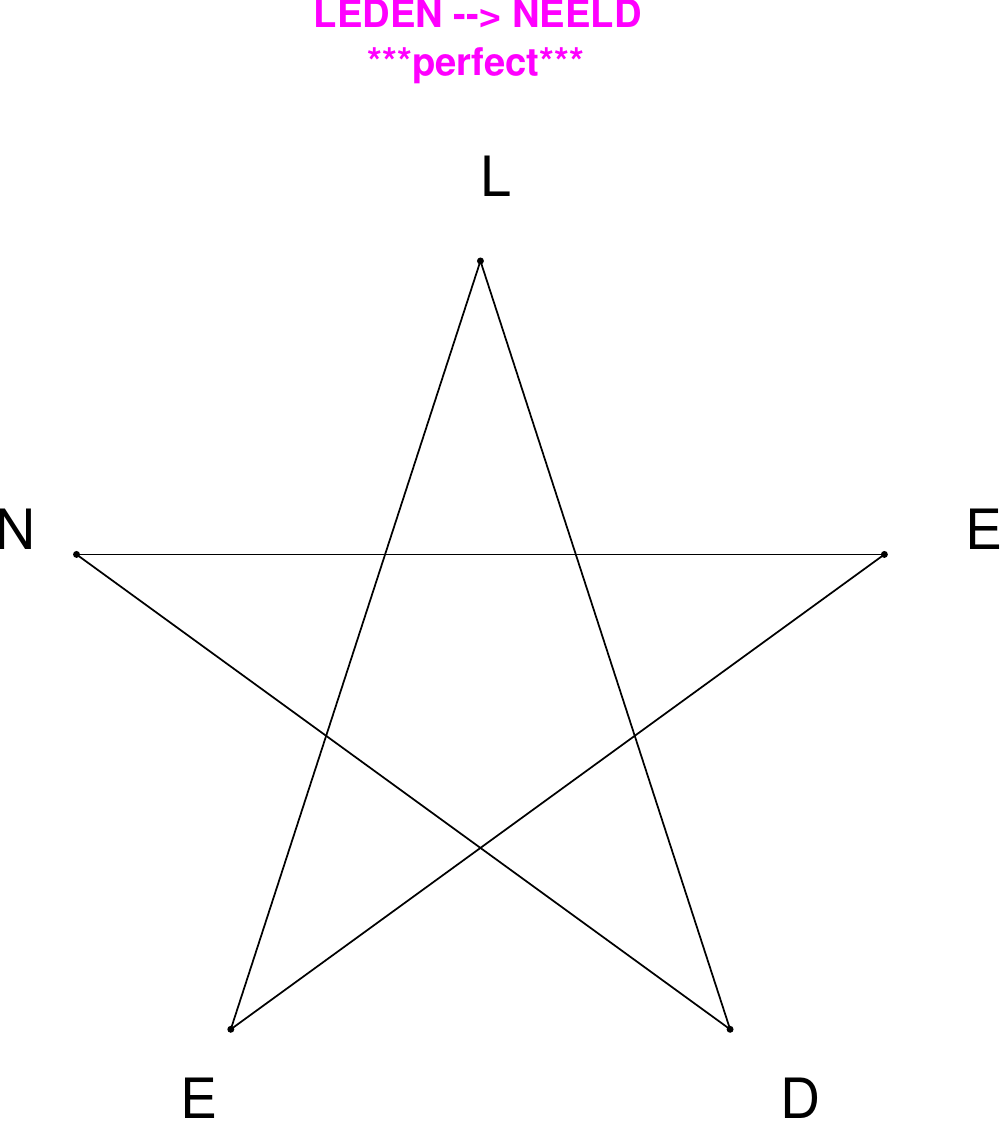}
\end{subfigure}
\hfill
\begin{subfigure}[T]{0.19\textwidth}
\centering
\includegraphics[width=\textwidth]{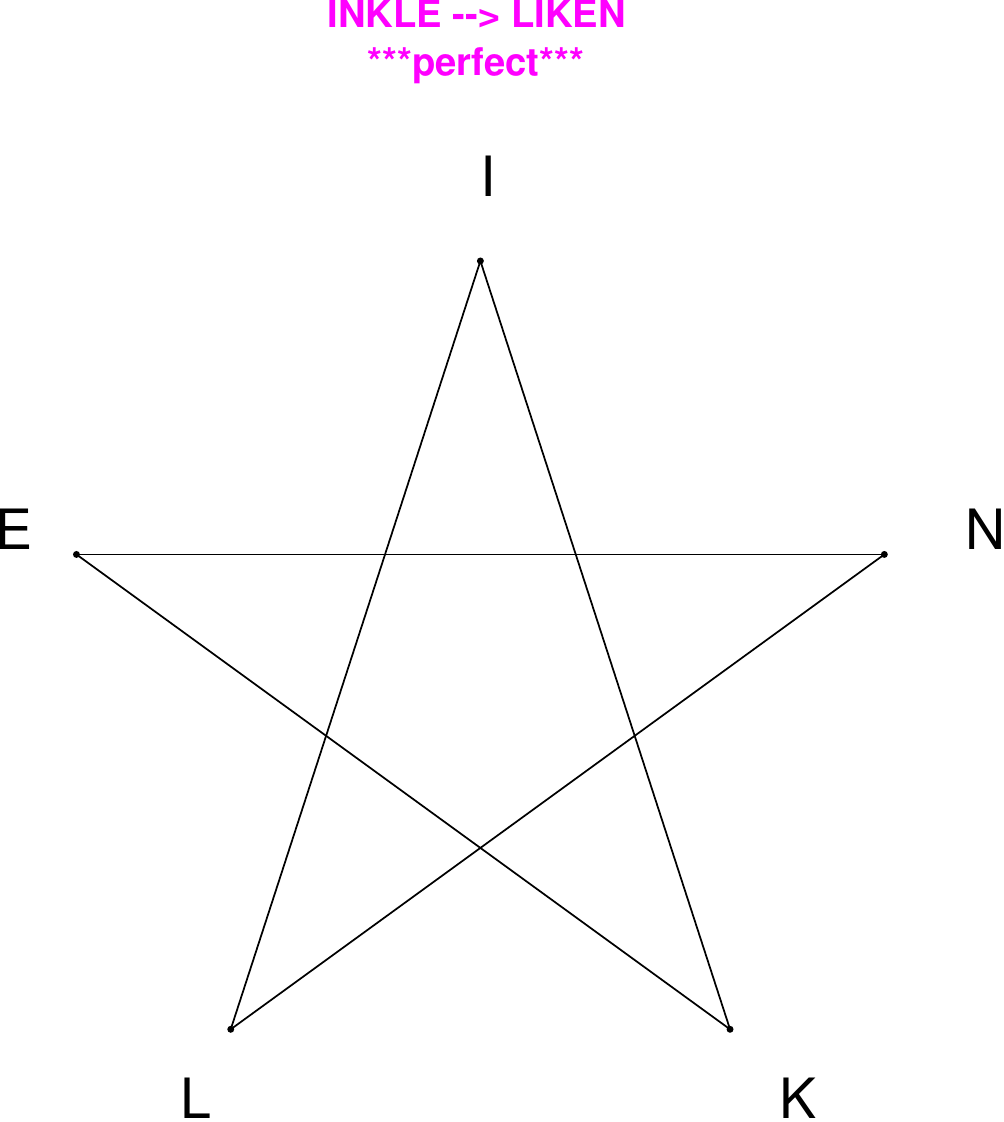}
\end{subfigure}
\end{figure}

\begin{figure}[H]
\centering
\begin{subfigure}[T]{0.19\textwidth}
\centering
\includegraphics[width=\textwidth]{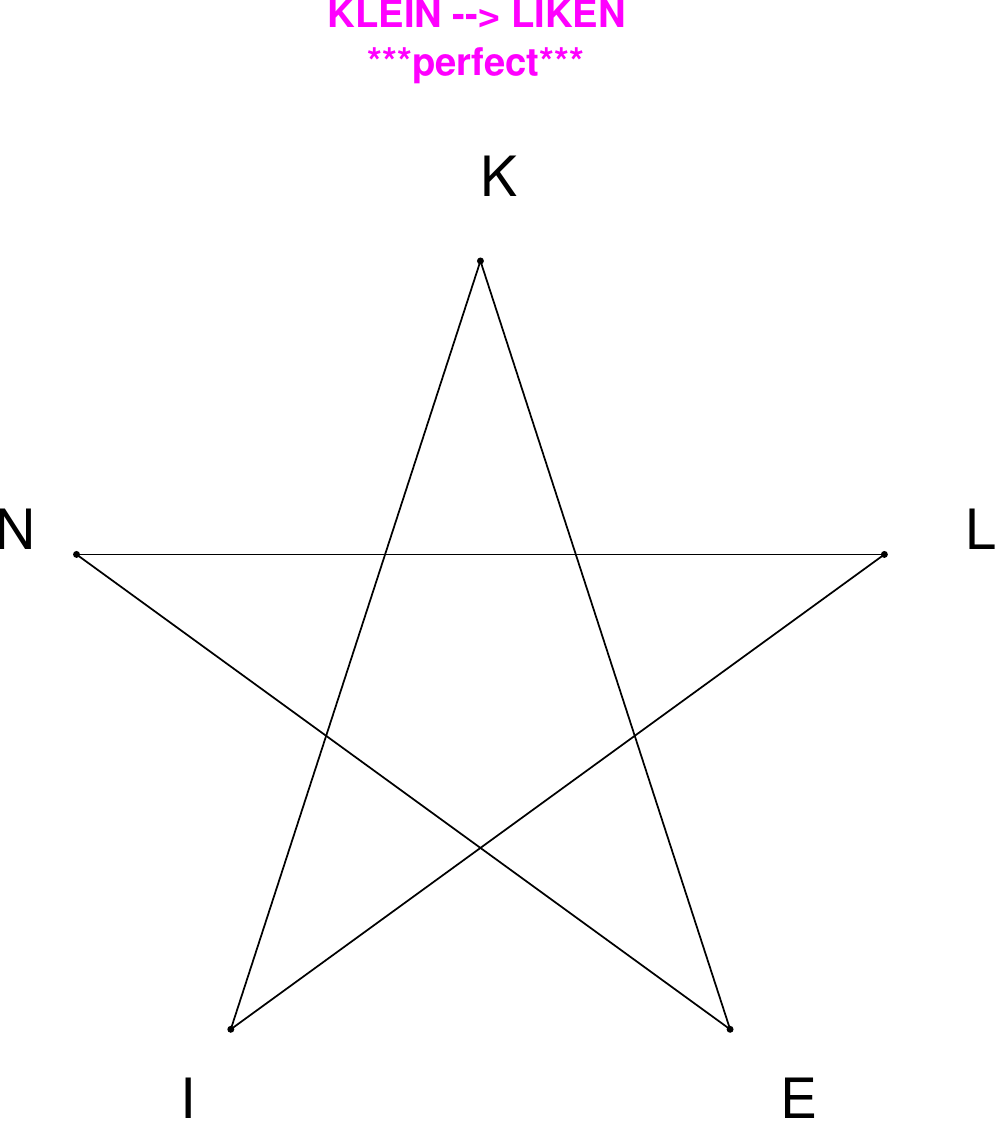}
\end{subfigure}
\hfill
\begin{subfigure}[T]{0.19\textwidth}
\centering
\includegraphics[width=\textwidth]{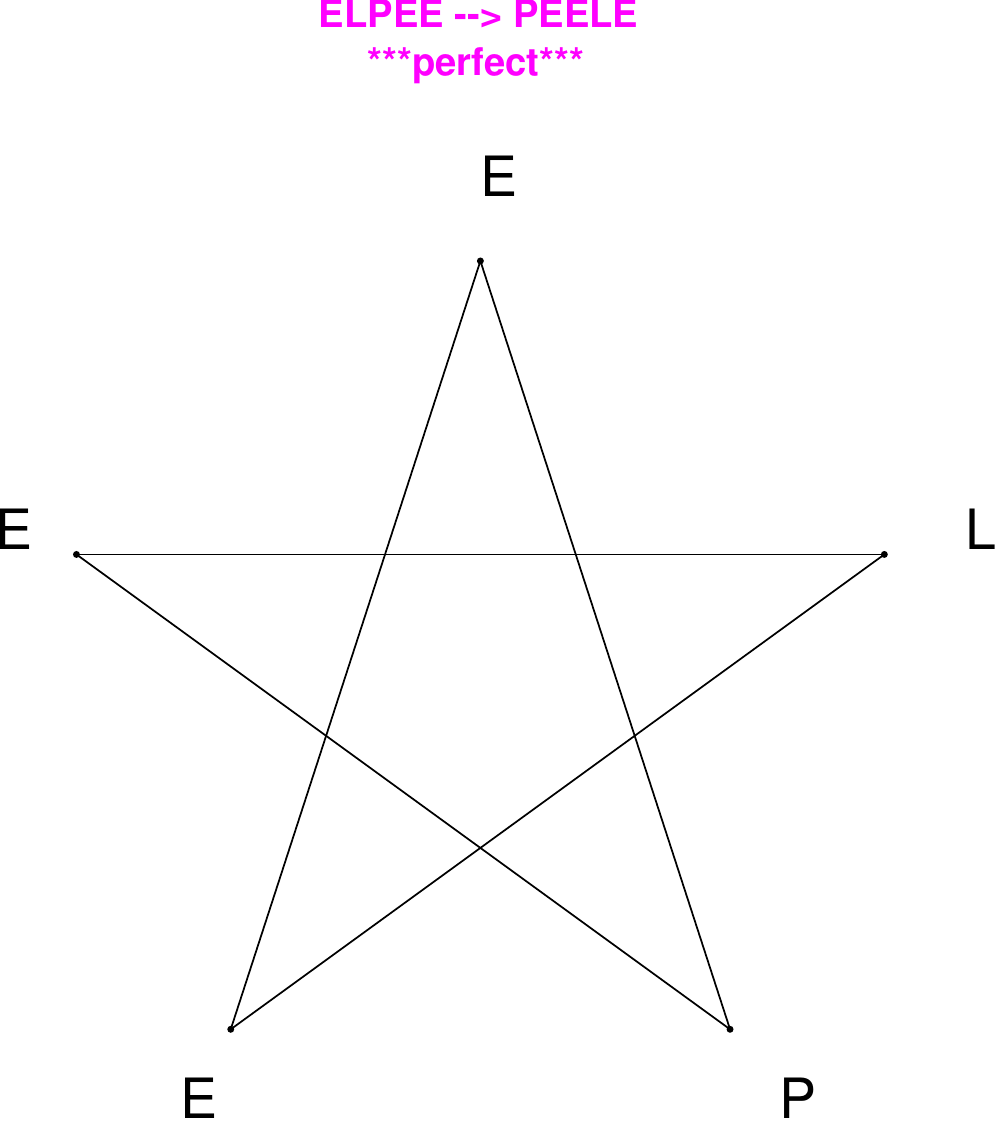}
\end{subfigure}
\hfill
\begin{subfigure}[T]{0.19\textwidth}
\centering
\includegraphics[width=\textwidth]{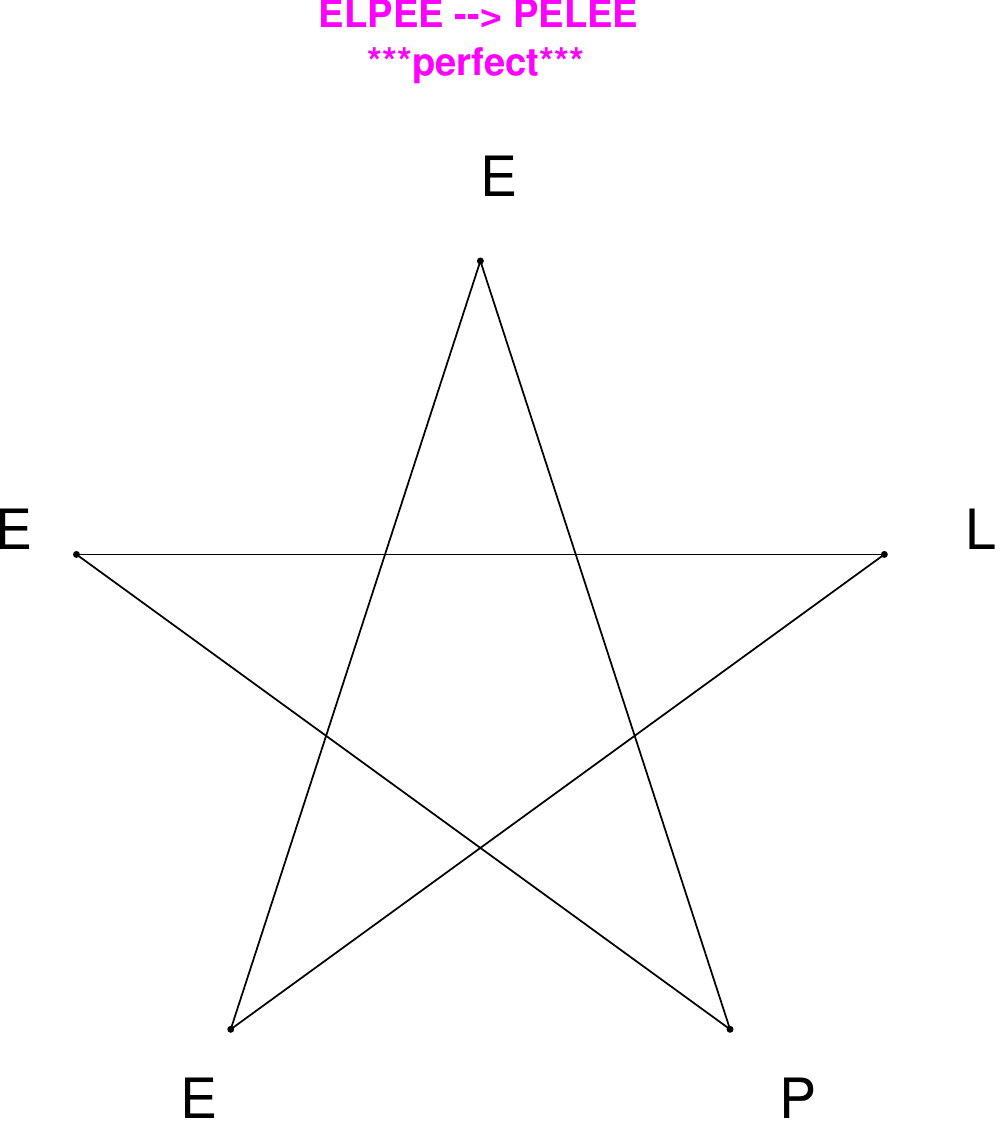}
\end{subfigure}
\hfill
\begin{subfigure}[T]{0.19\textwidth}
\centering
\includegraphics[width=\textwidth]{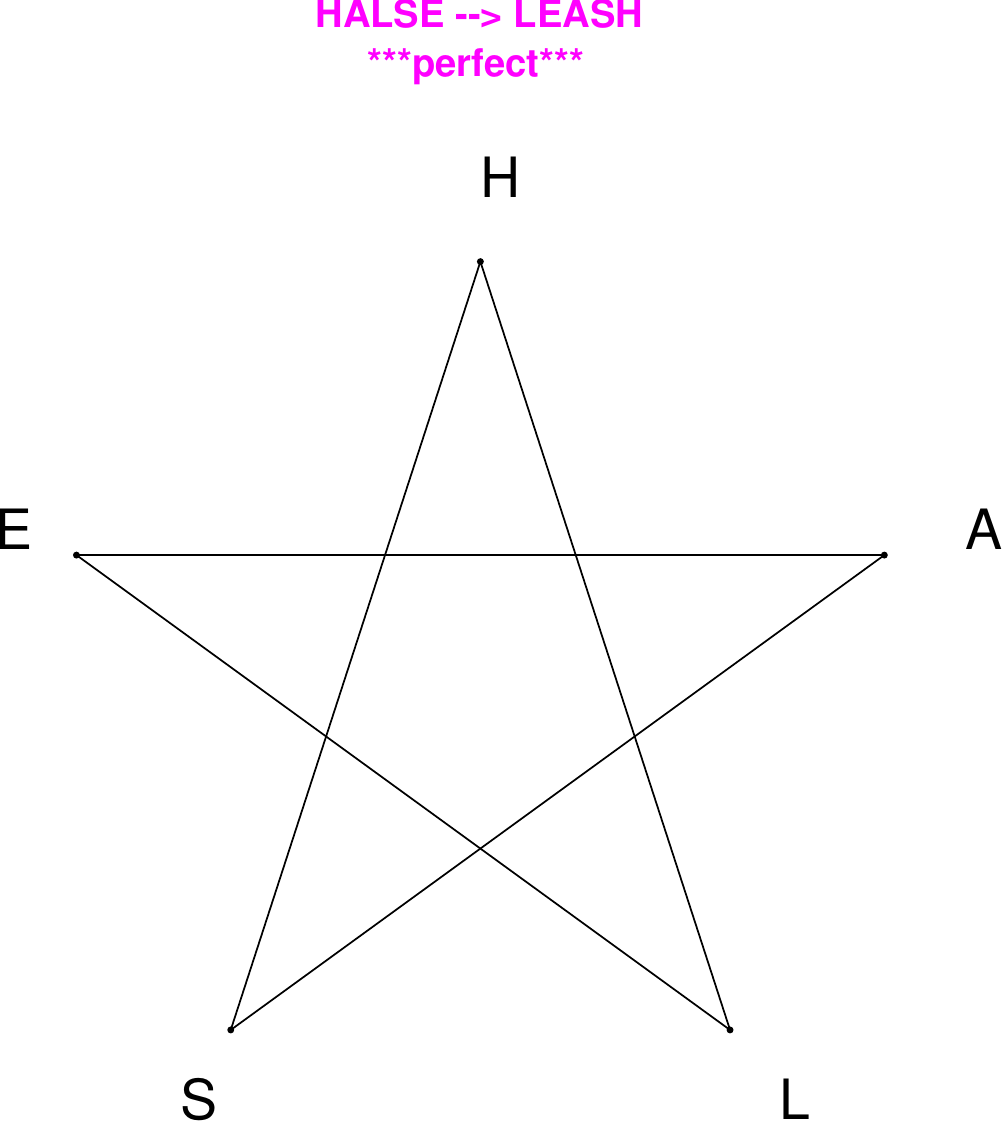}
\end{subfigure}
\hfill
\begin{subfigure}[T]{0.19\textwidth}
\centering
\includegraphics[width=\textwidth]{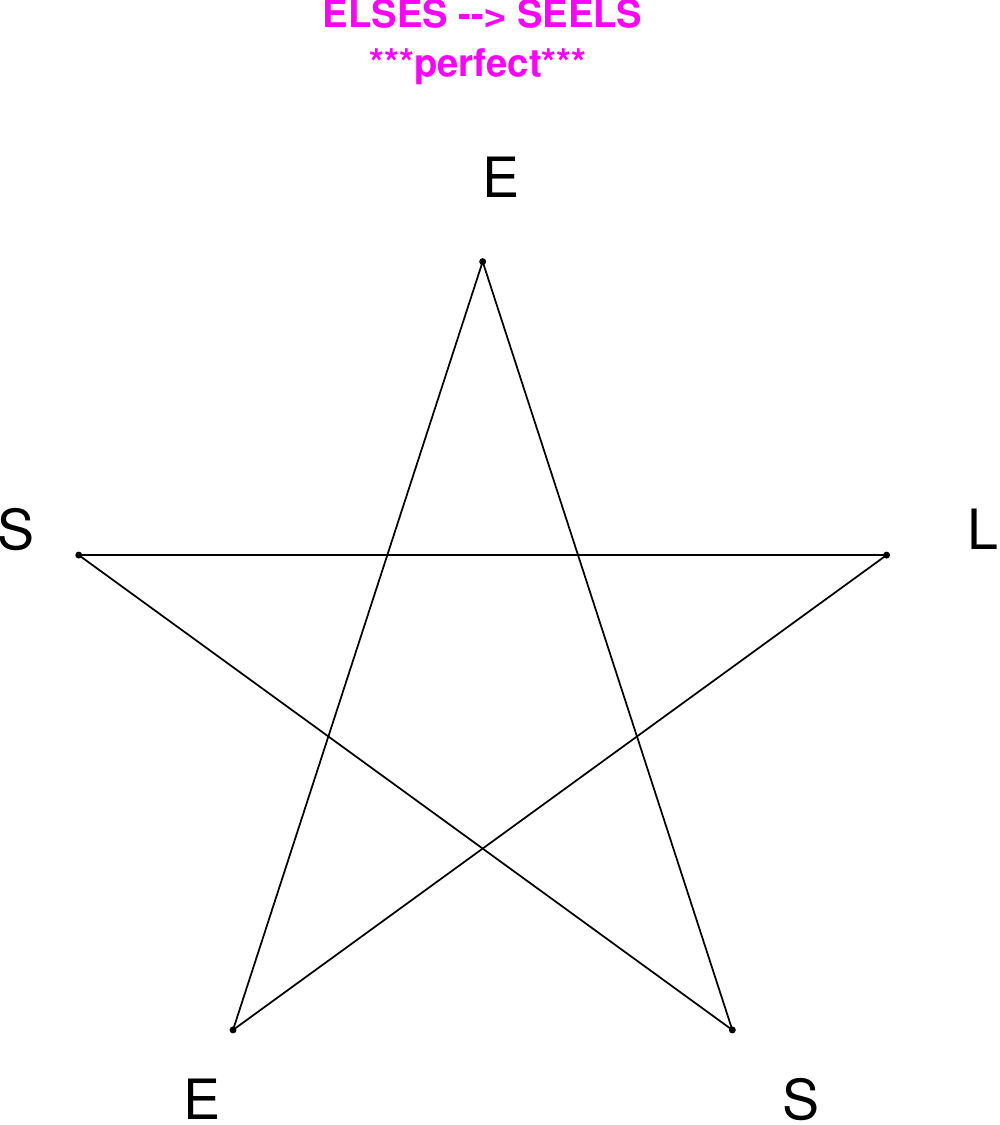}
\end{subfigure}
\end{figure}

\begin{figure}[H]
\centering
\begin{subfigure}[T]{0.19\textwidth}
\centering
\includegraphics[width=\textwidth]{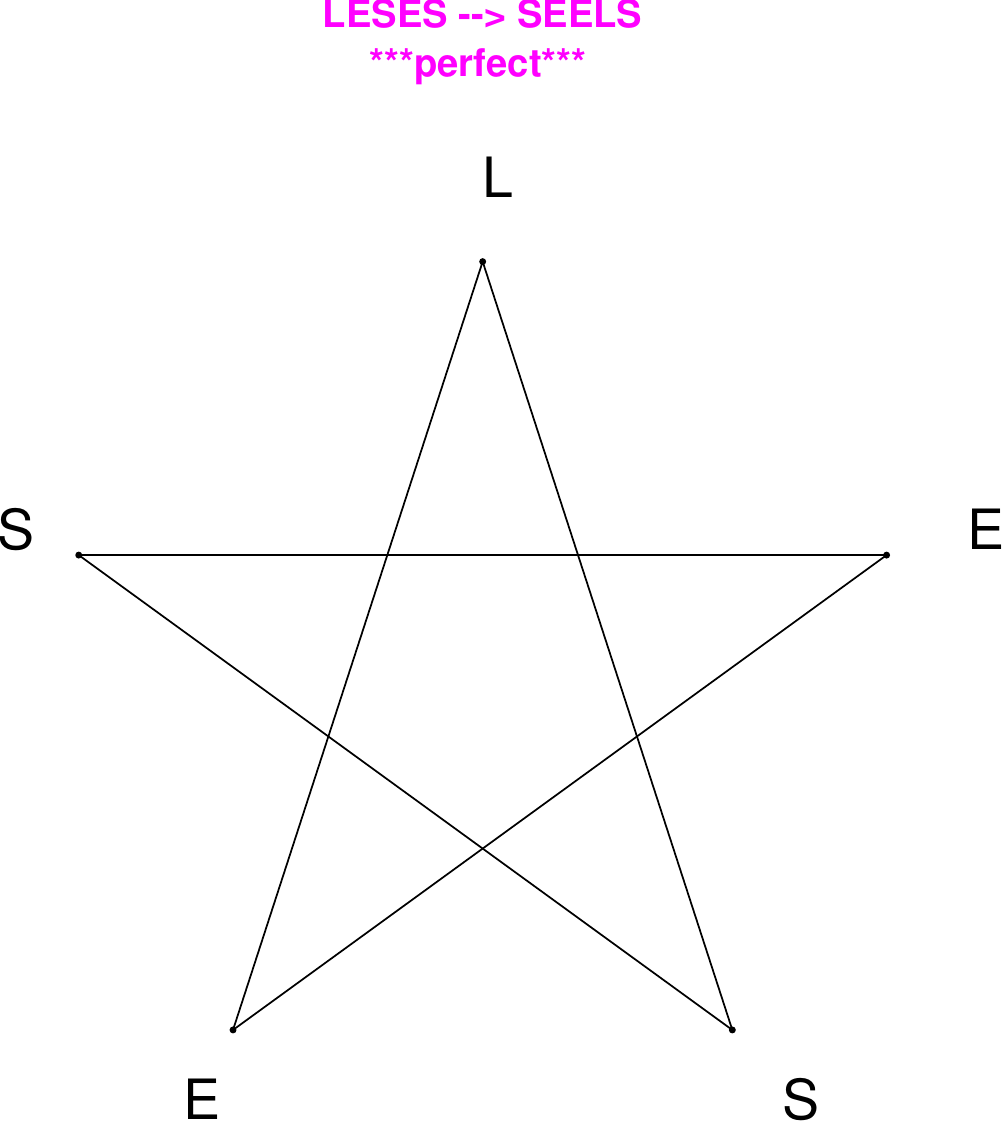}
\end{subfigure}
\hfill
\begin{subfigure}[T]{0.19\textwidth}
\centering
\includegraphics[width=\textwidth]{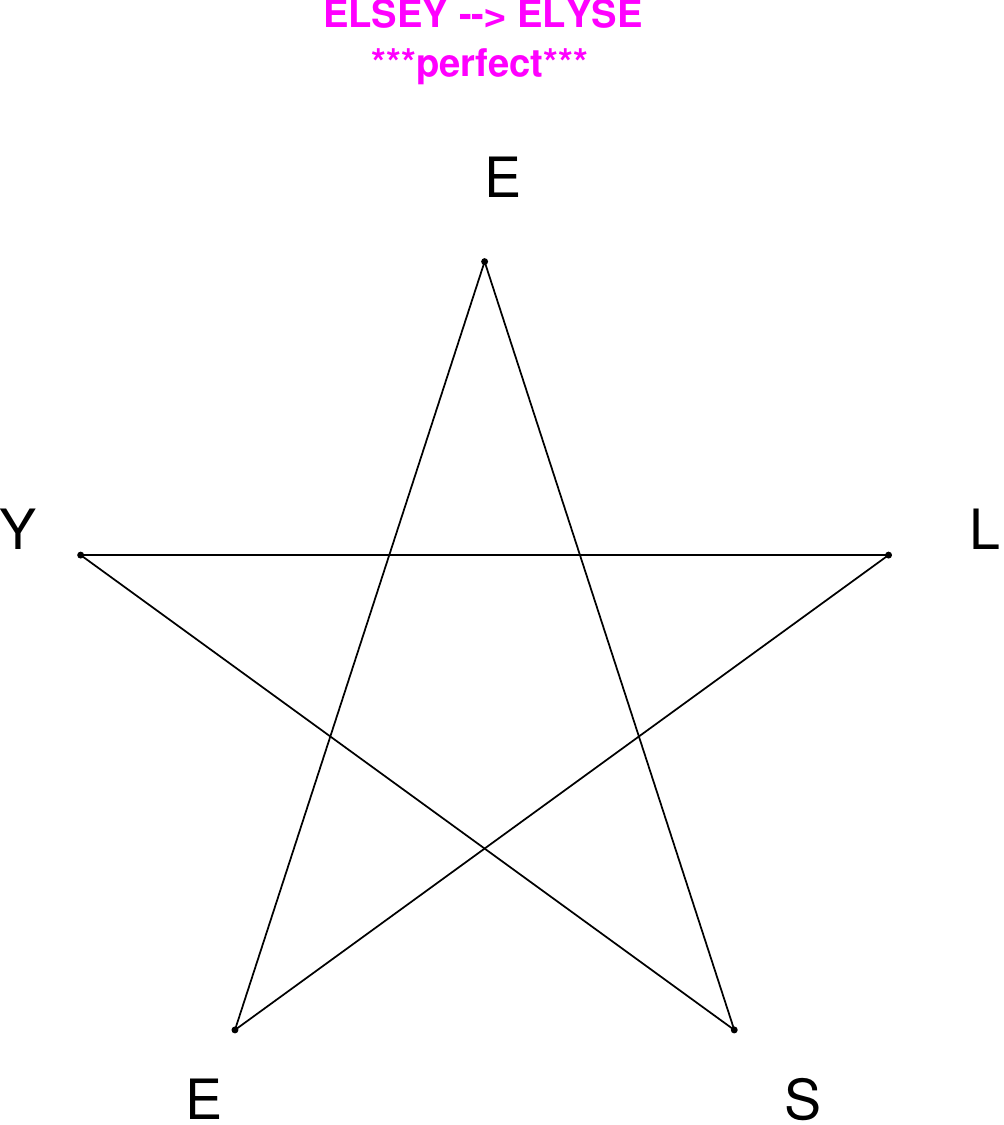}
\end{subfigure}
\hfill
\begin{subfigure}[T]{0.19\textwidth}
\centering
\includegraphics[width=\textwidth]{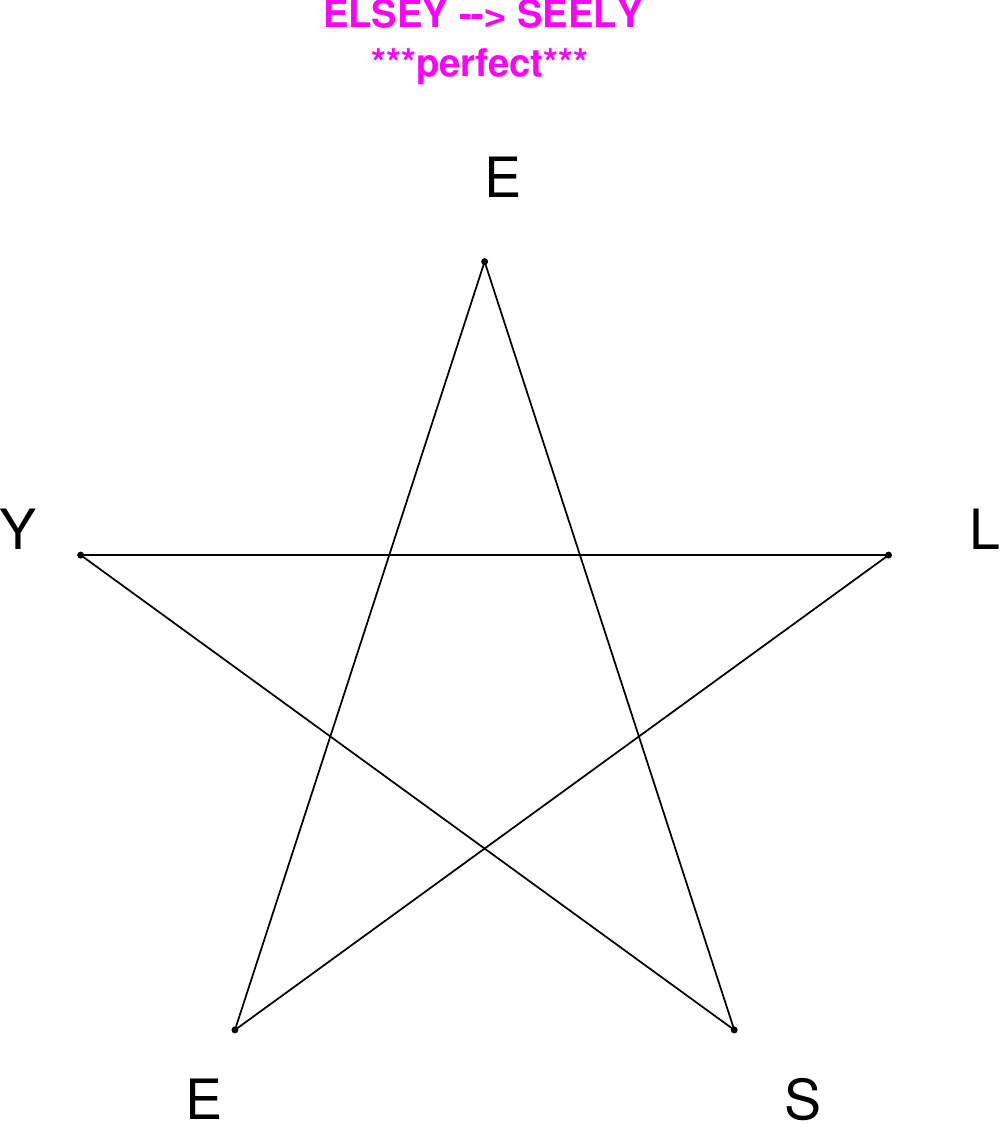}
\end{subfigure}
\hfill
\begin{subfigure}[T]{0.19\textwidth}
\centering
\includegraphics[width=\textwidth]{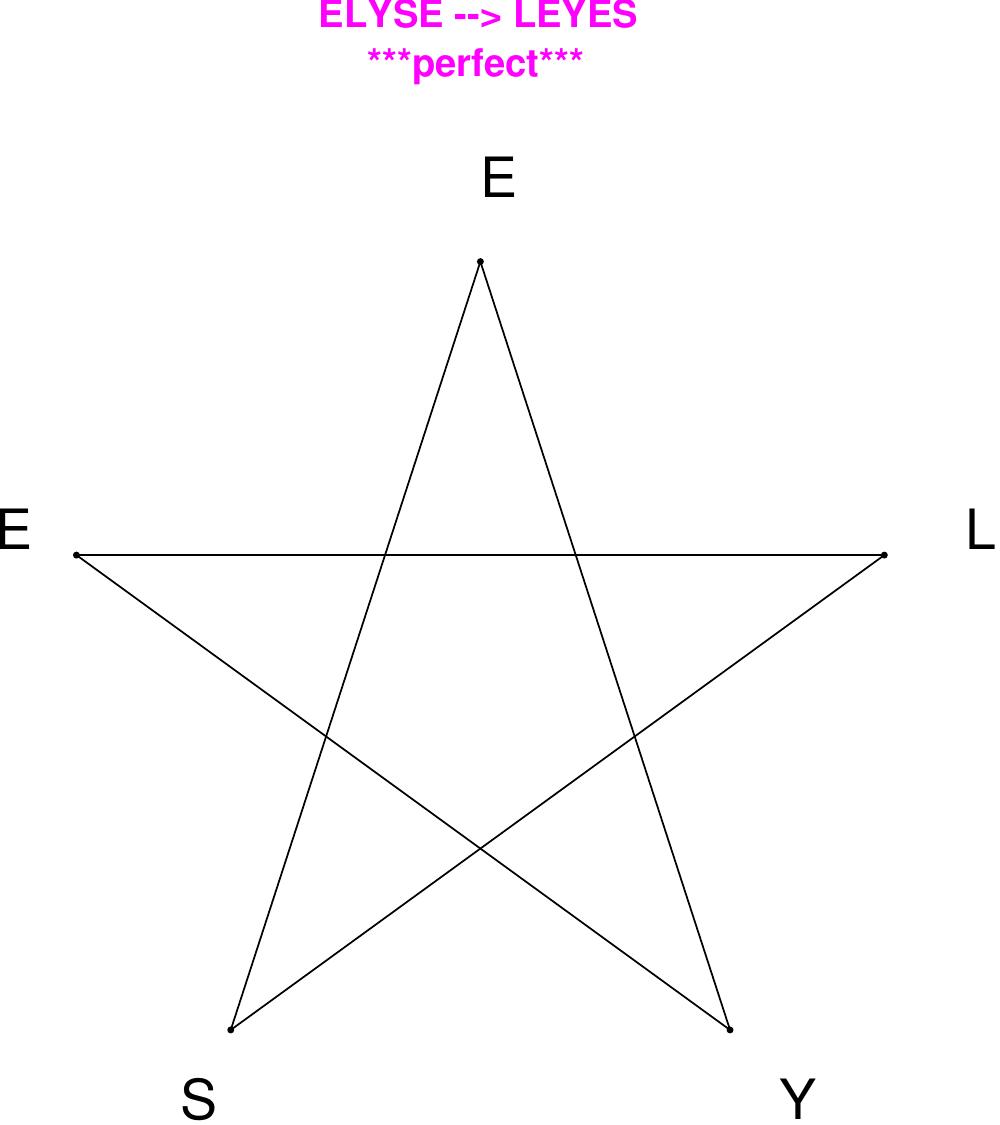}
\end{subfigure}
\hfill
\begin{subfigure}[T]{0.19\textwidth}
\centering
\includegraphics[width=\textwidth]{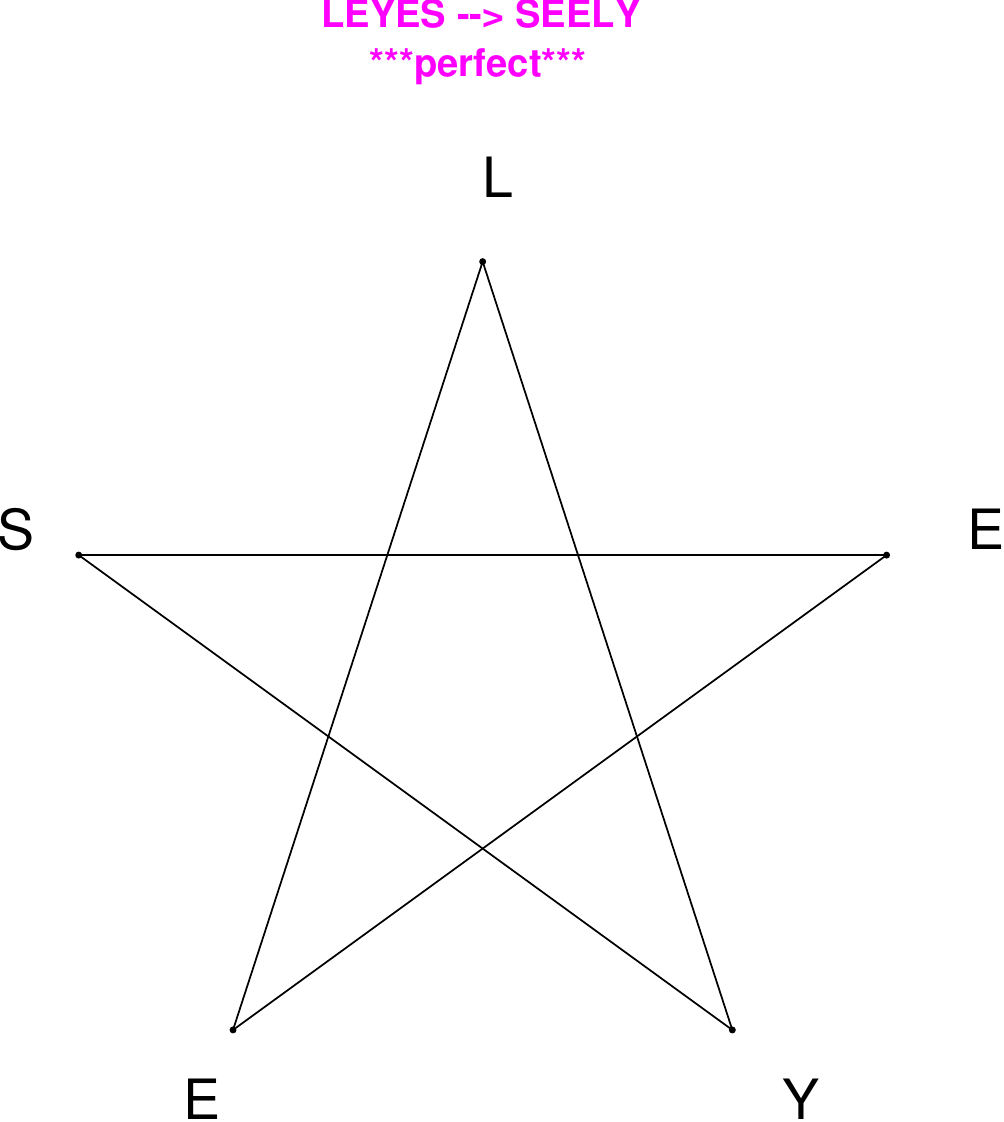}
\end{subfigure}
\end{figure}

\begin{figure}[H]
\centering
\begin{subfigure}[T]{0.19\textwidth}
\centering
\includegraphics[width=\textwidth]{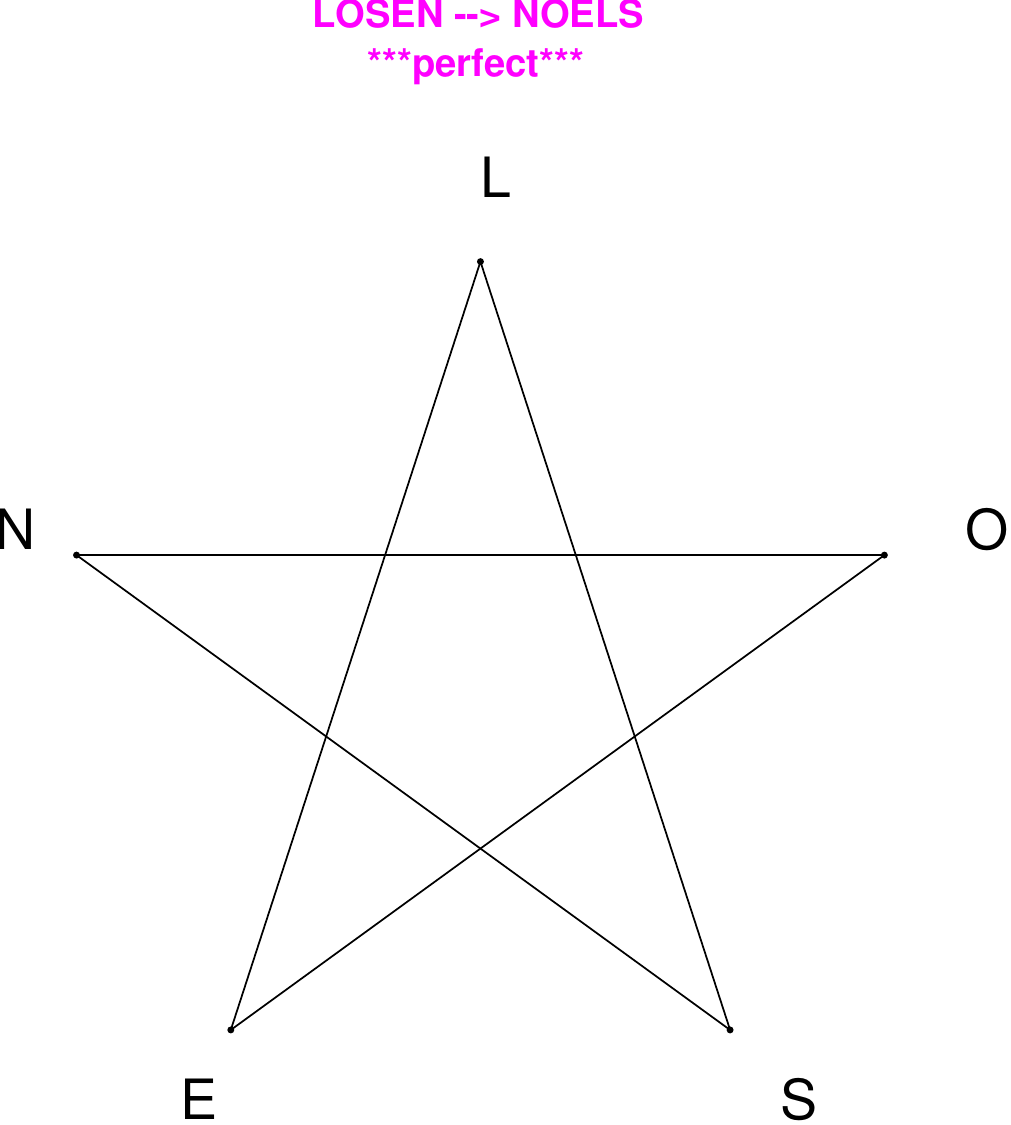}
\end{subfigure}
\hfill
\begin{subfigure}[T]{0.19\textwidth}
\centering
\includegraphics[width=\textwidth]{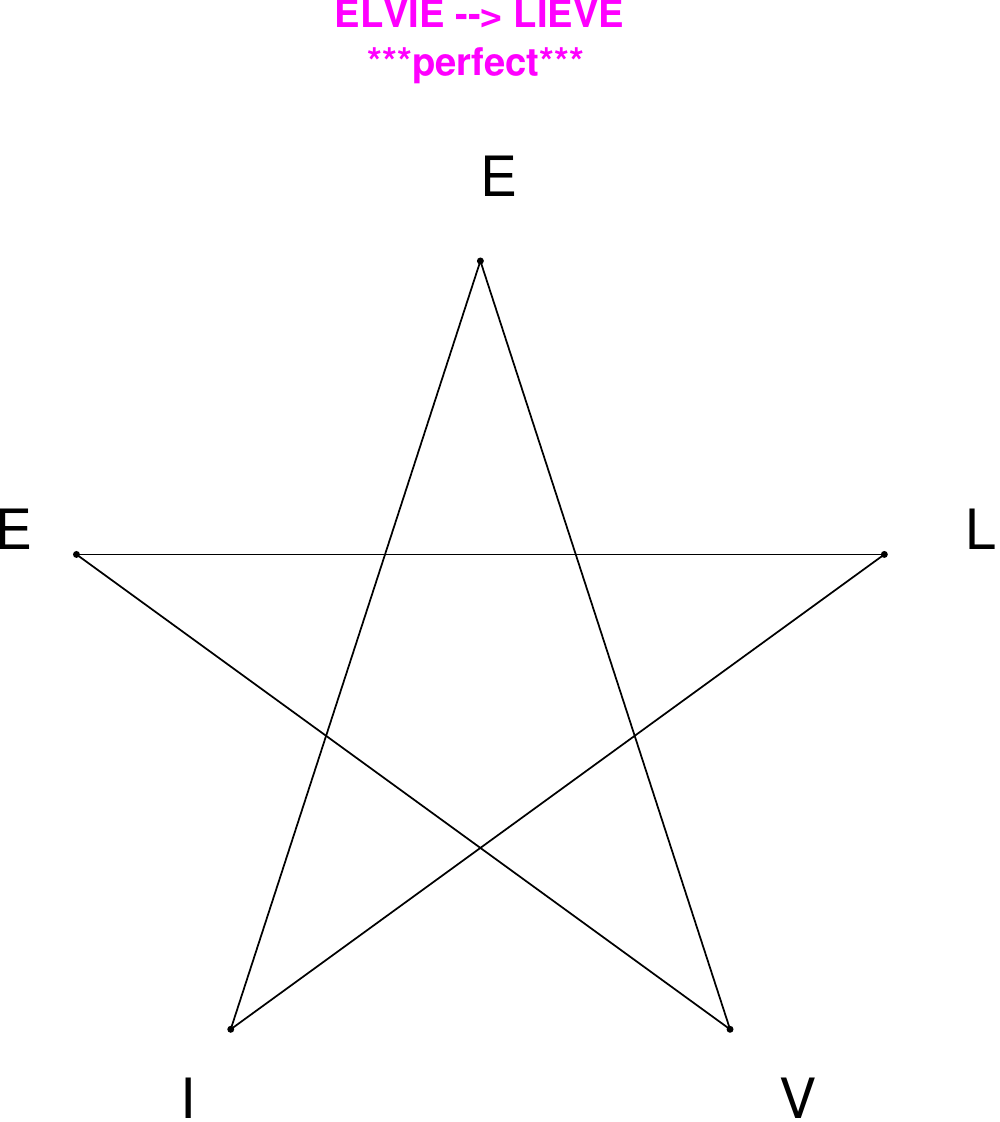}
\end{subfigure}
\hfill
\begin{subfigure}[T]{0.19\textwidth}
\centering
\includegraphics[width=\textwidth]{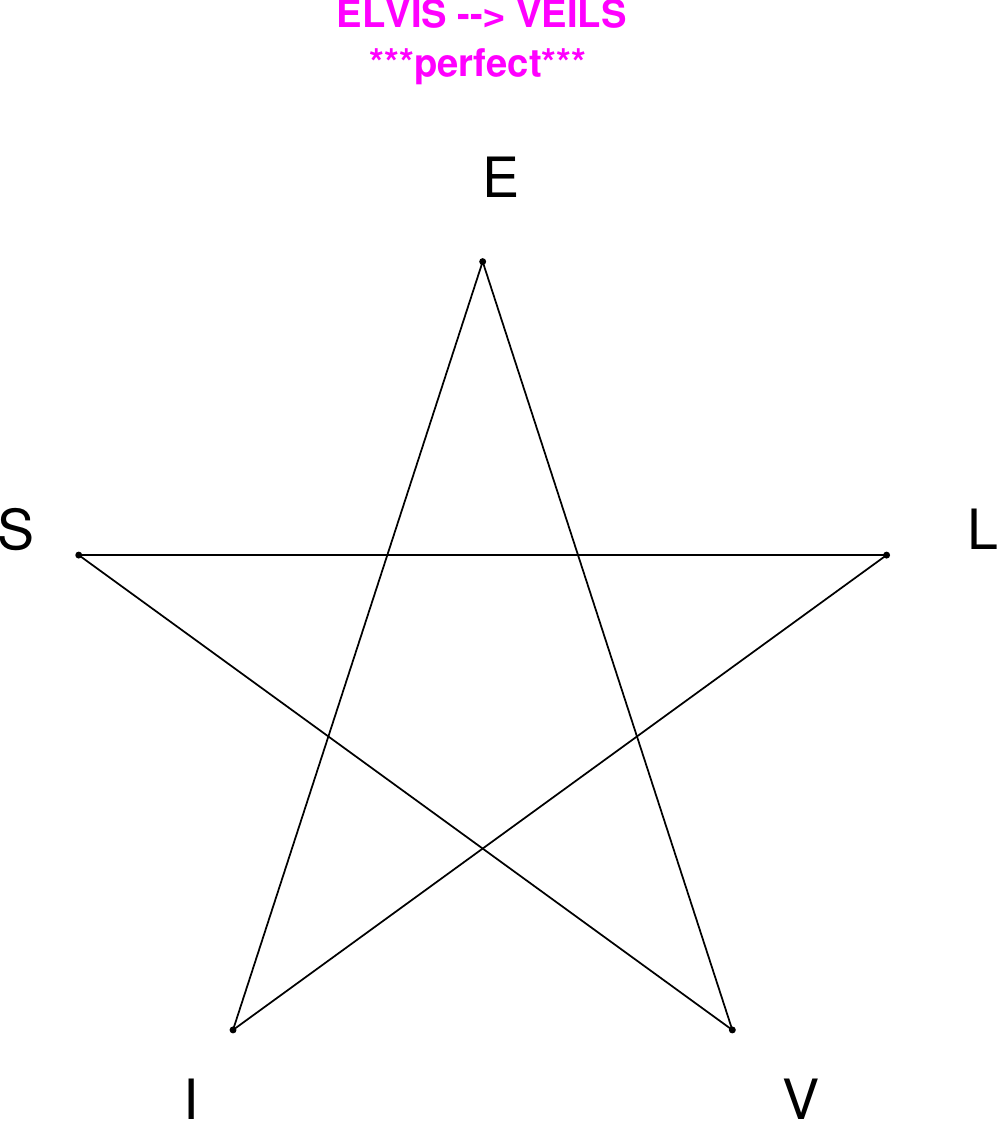}
\end{subfigure}
\hfill
\begin{subfigure}[T]{0.19\textwidth}
\centering
\includegraphics[width=\textwidth]{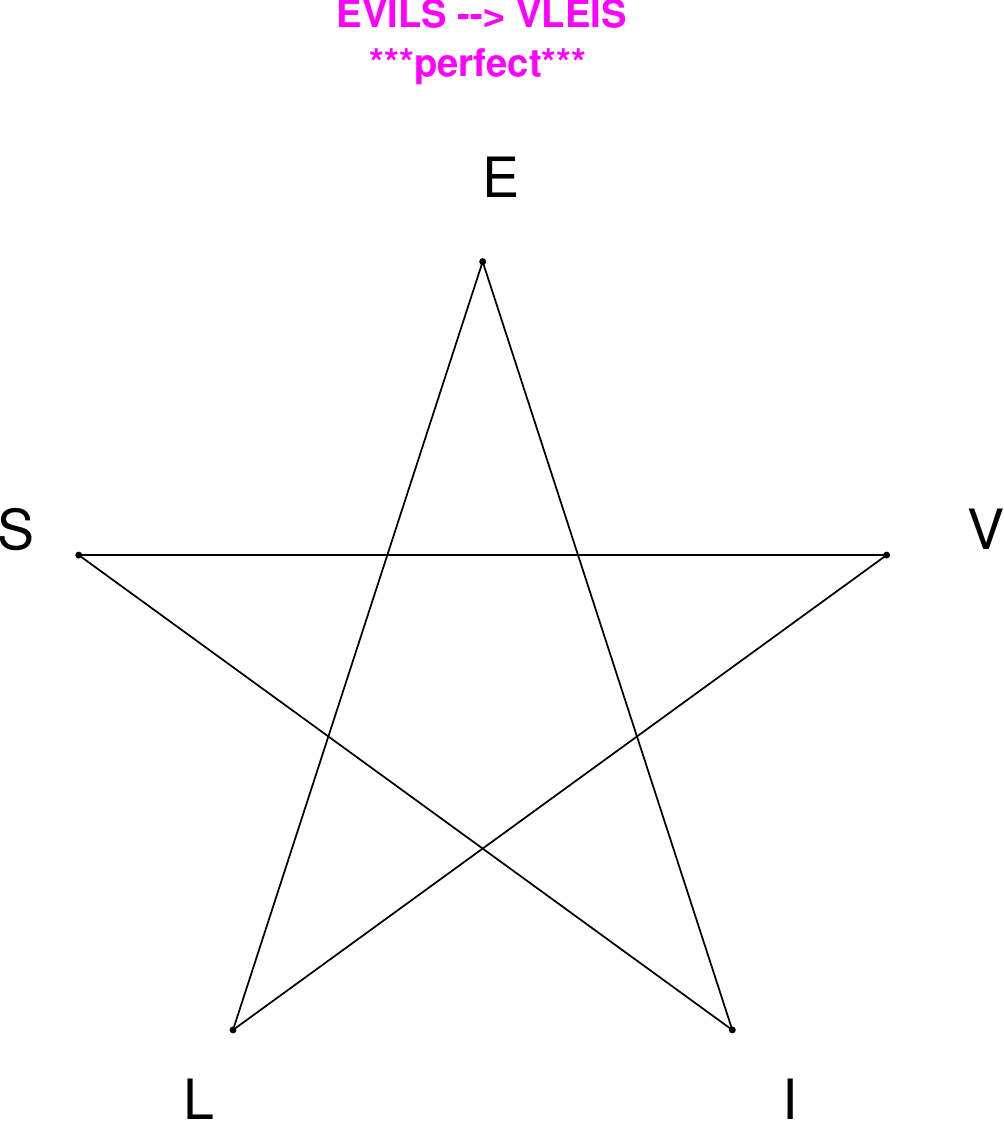}
\end{subfigure}
\hfill
\begin{subfigure}[T]{0.19\textwidth}
\centering
\includegraphics[width=\textwidth]{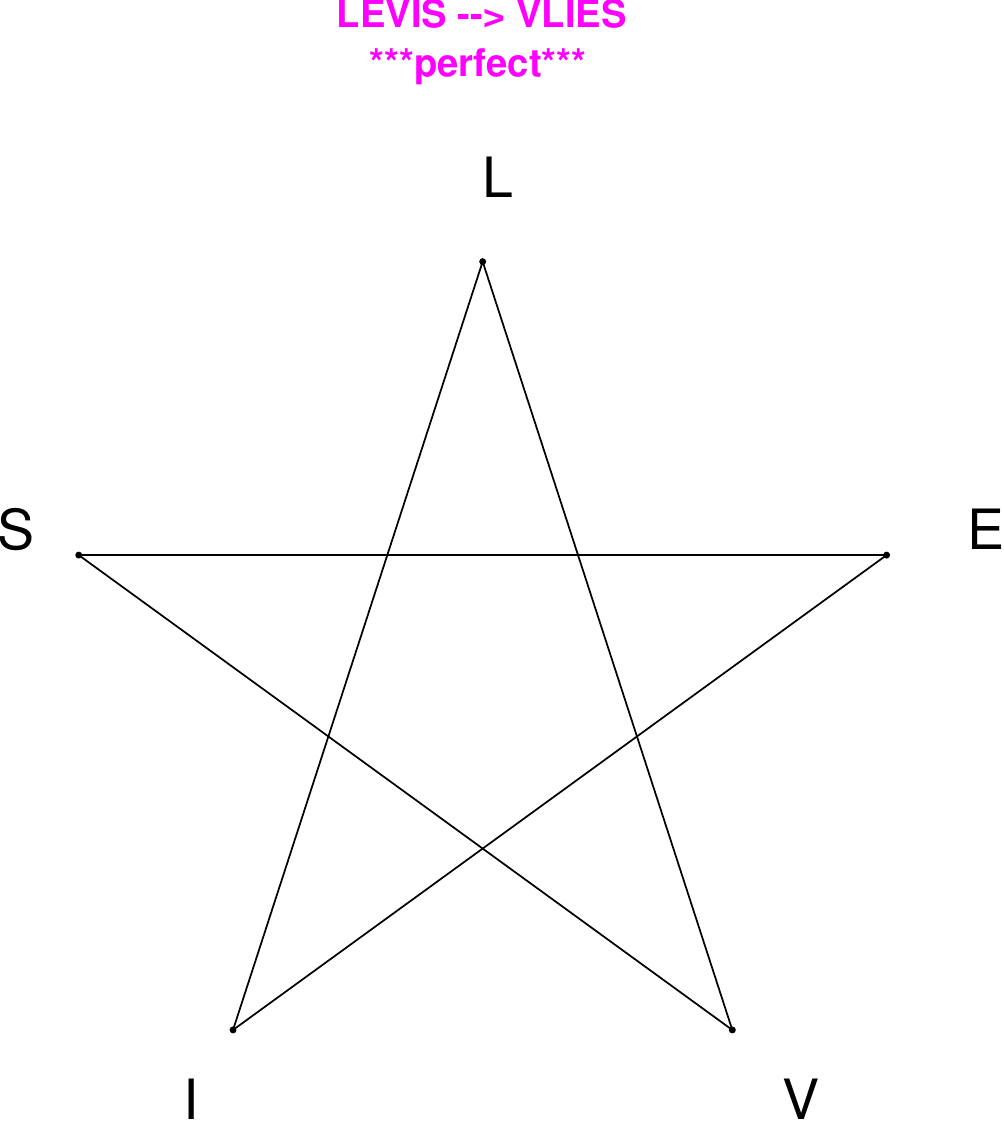}
\end{subfigure}
\end{figure}

\begin{figure}[H]
\centering
\begin{subfigure}[T]{0.19\textwidth}
\centering
\includegraphics[width=\textwidth]{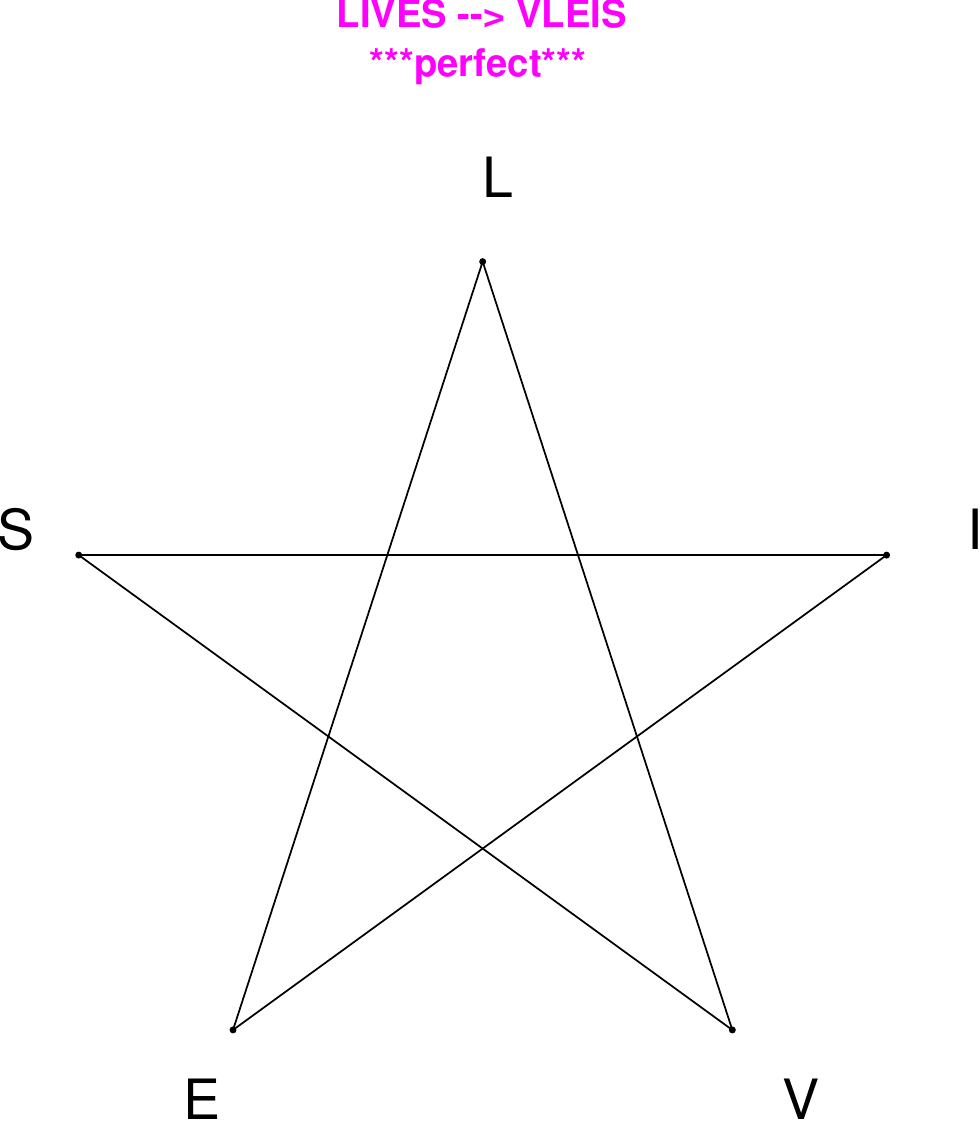}
\end{subfigure}
\hfill
\begin{subfigure}[T]{0.19\textwidth}
\centering
\includegraphics[width=\textwidth]{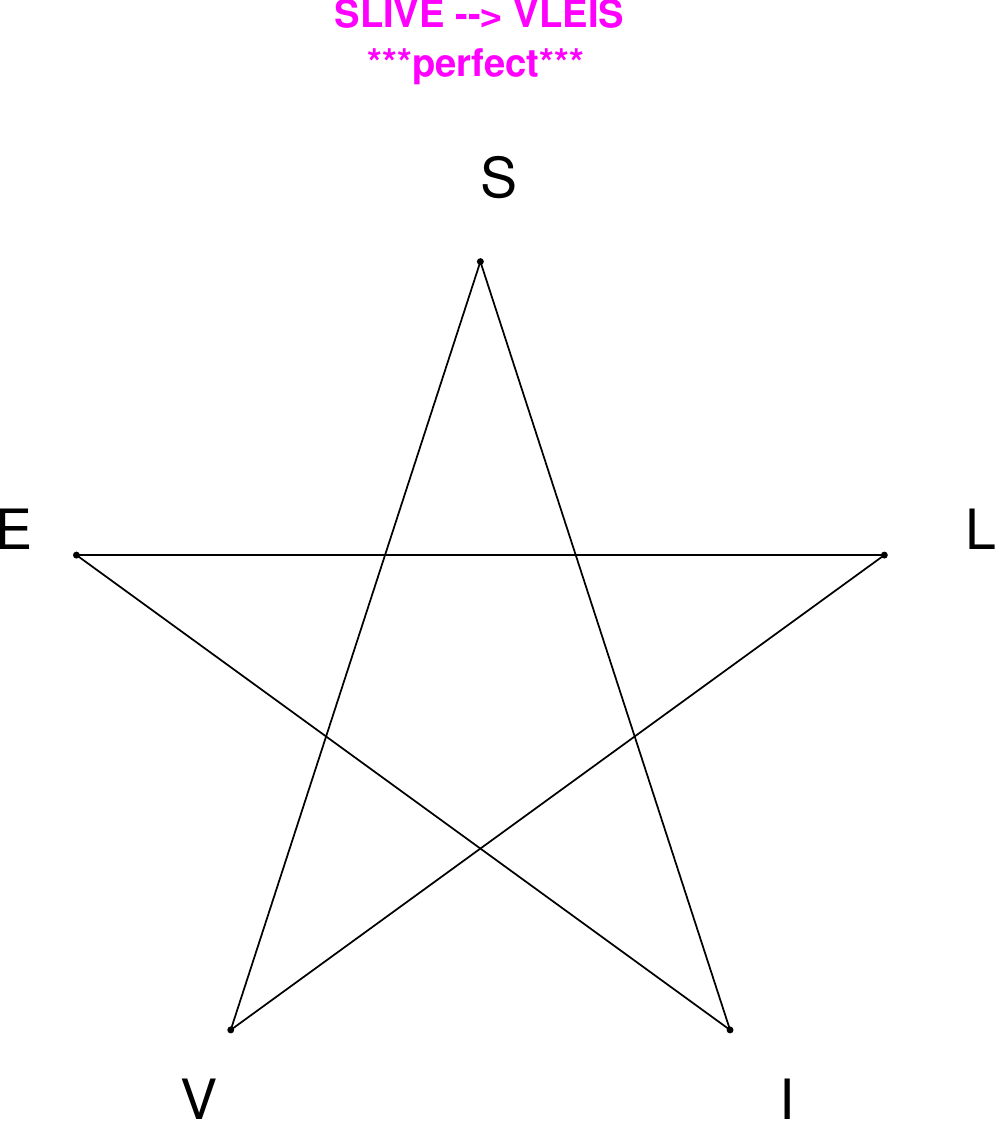}
\end{subfigure}
\hfill
\begin{subfigure}[T]{0.19\textwidth}
\centering
\includegraphics[width=\textwidth]{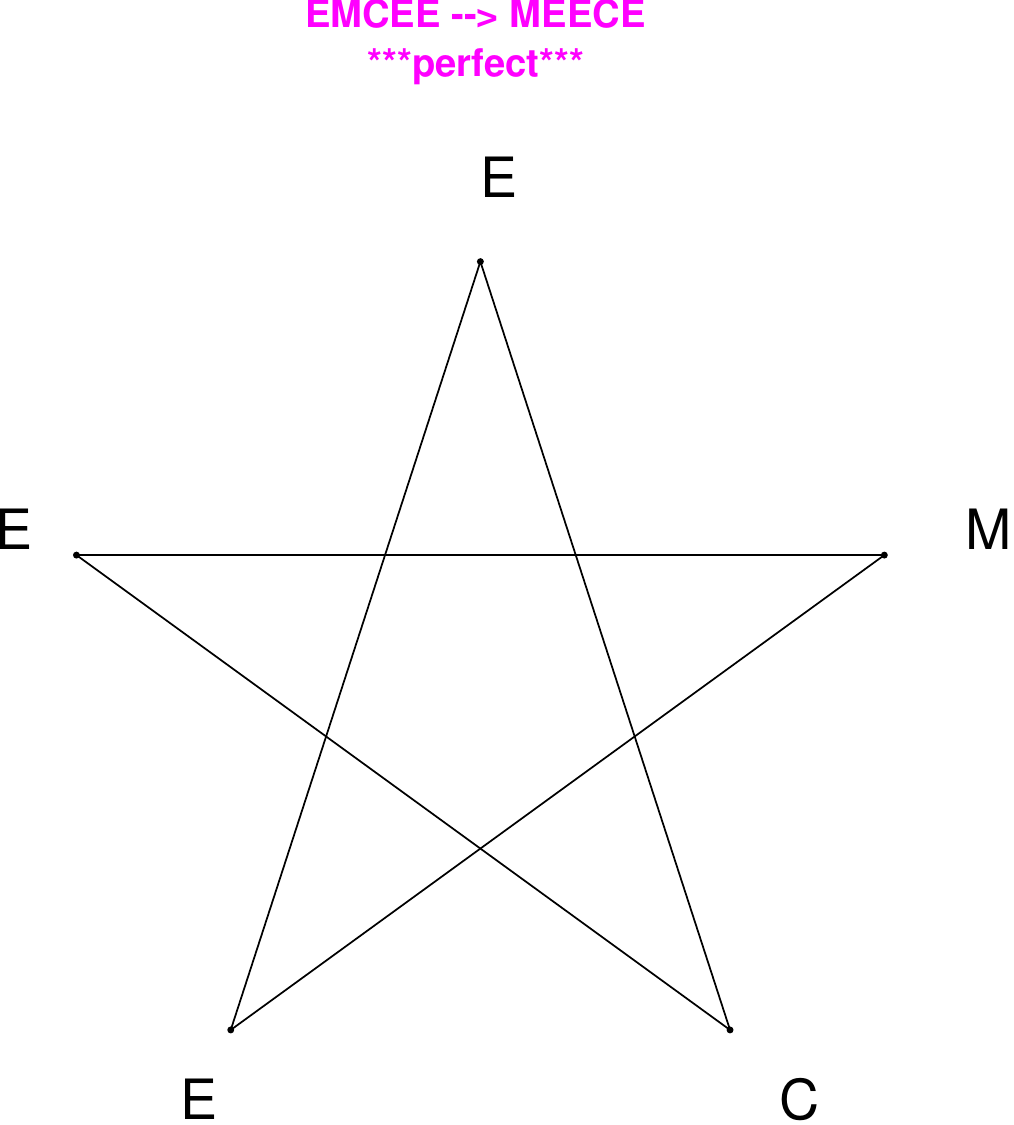}
\end{subfigure}
\hfill
\begin{subfigure}[T]{0.19\textwidth}
\centering
\includegraphics[width=\textwidth]{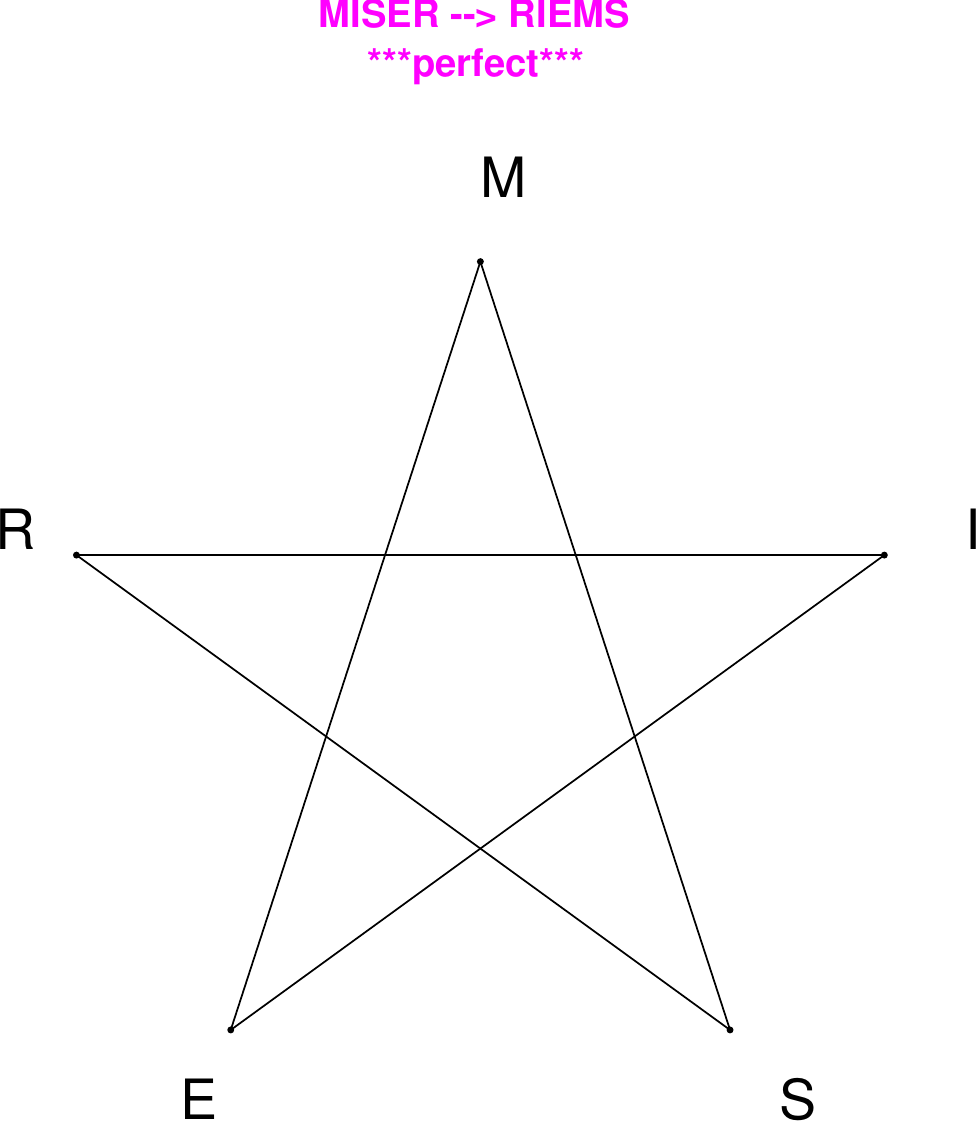}
\end{subfigure}
\hfill
\begin{subfigure}[T]{0.19\textwidth}
\centering
\includegraphics[width=\textwidth]{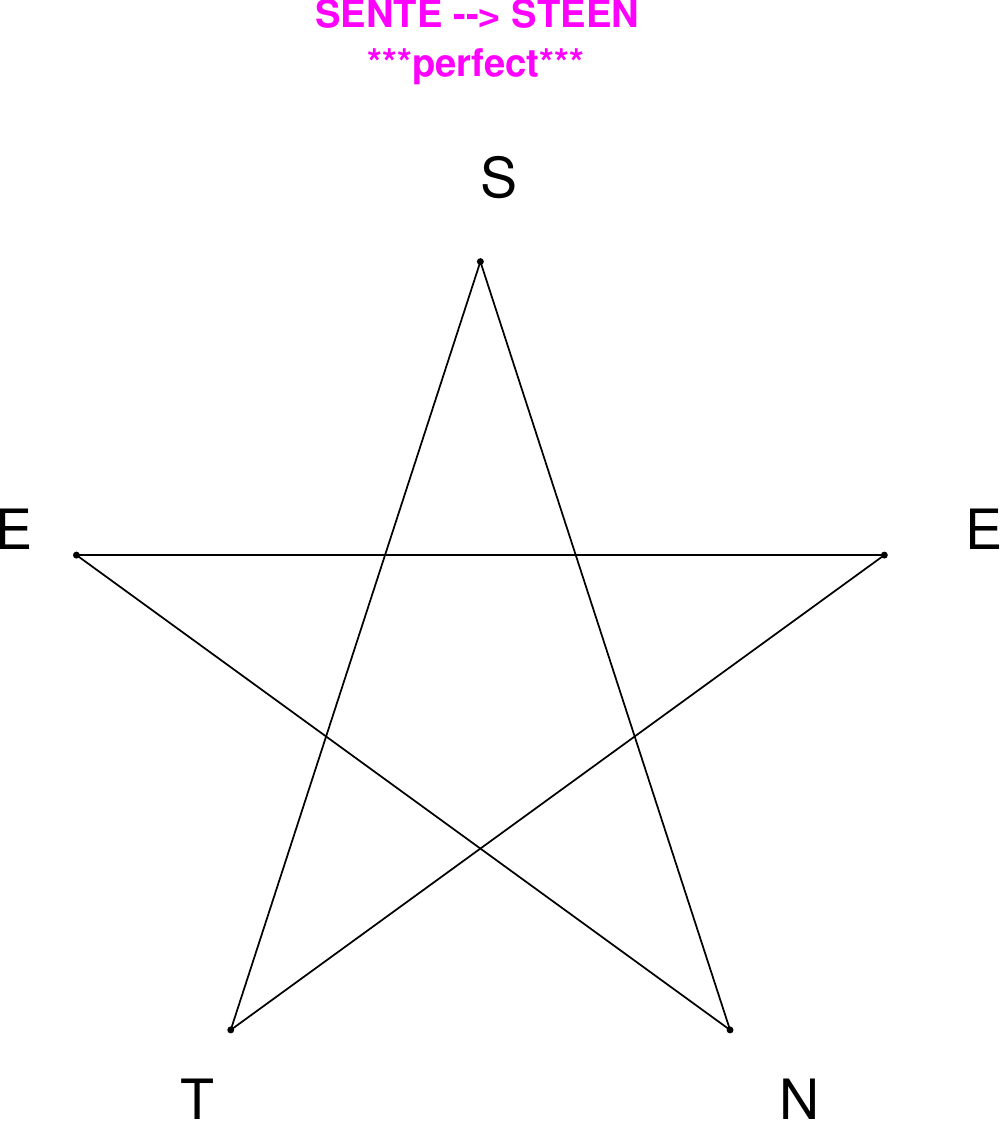}
\end{subfigure}
\end{figure}

\begin{figure}[H]
\centering
\begin{subfigure}[T]{0.19\textwidth}
\centering
\includegraphics[width=\textwidth]{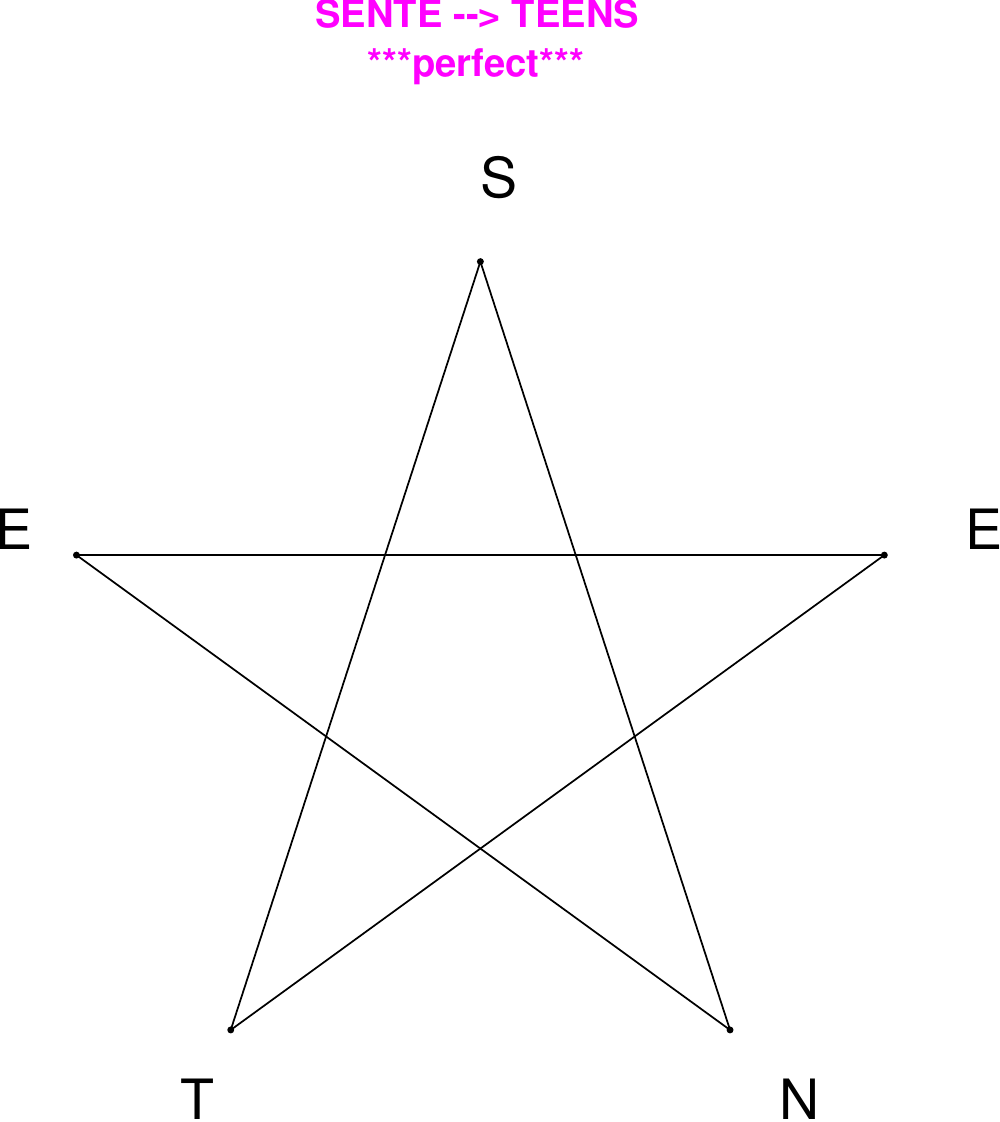}
\end{subfigure}
\hfill
\begin{subfigure}[T]{0.19\textwidth}
\centering
\includegraphics[width=\textwidth]{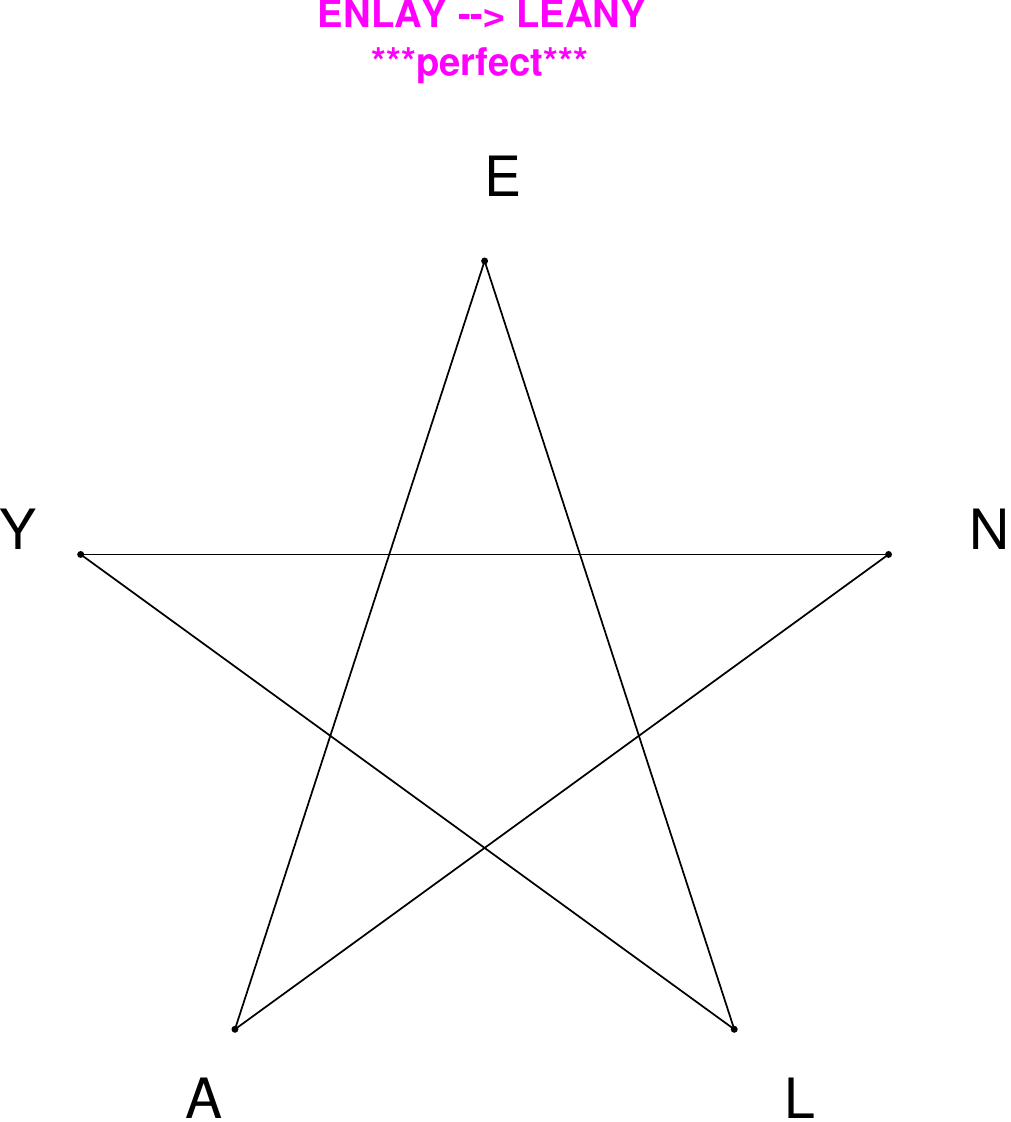}
\end{subfigure}
\hfill
\begin{subfigure}[T]{0.19\textwidth}
\centering
\includegraphics[width=\textwidth]{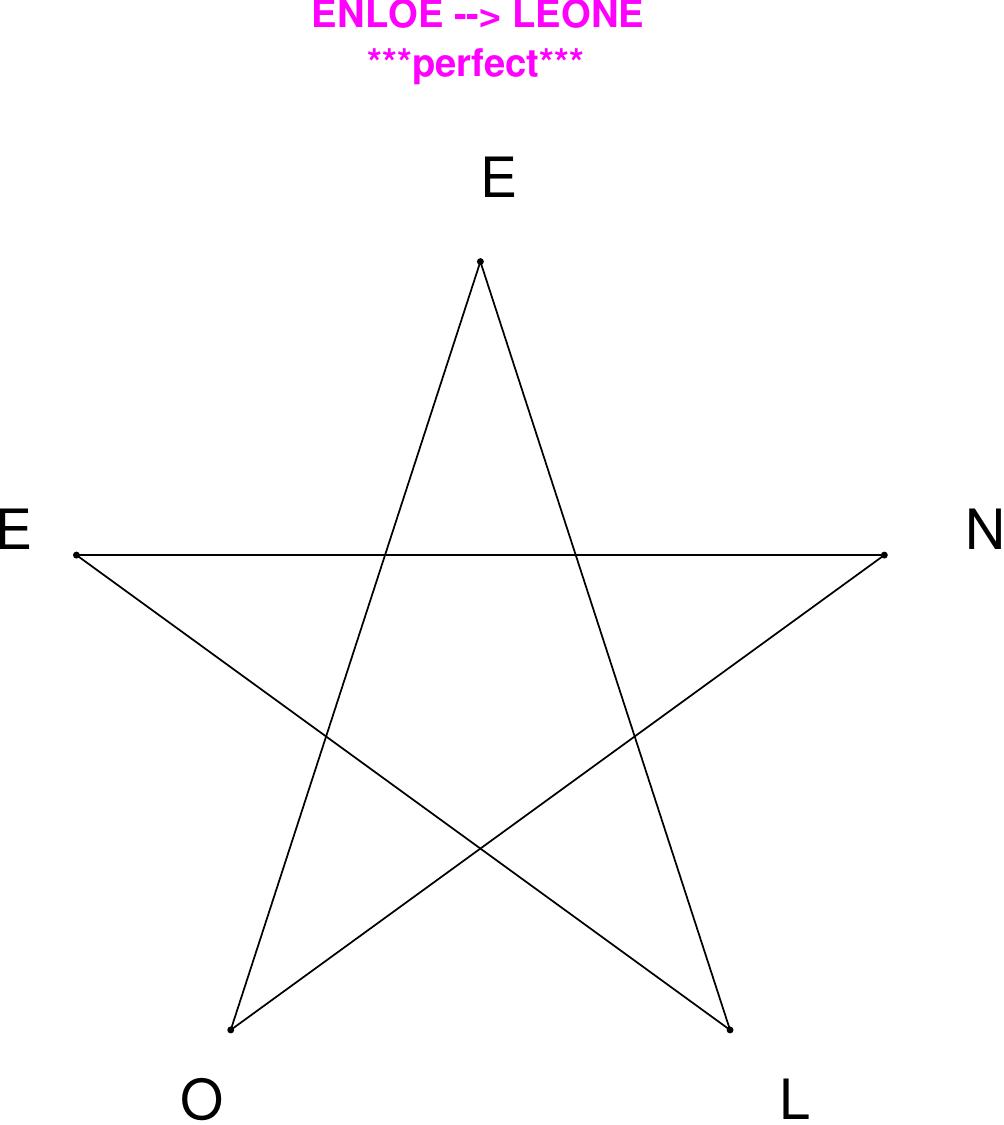}
\end{subfigure}
\hfill
\begin{subfigure}[T]{0.19\textwidth}
\centering
\includegraphics[width=\textwidth]{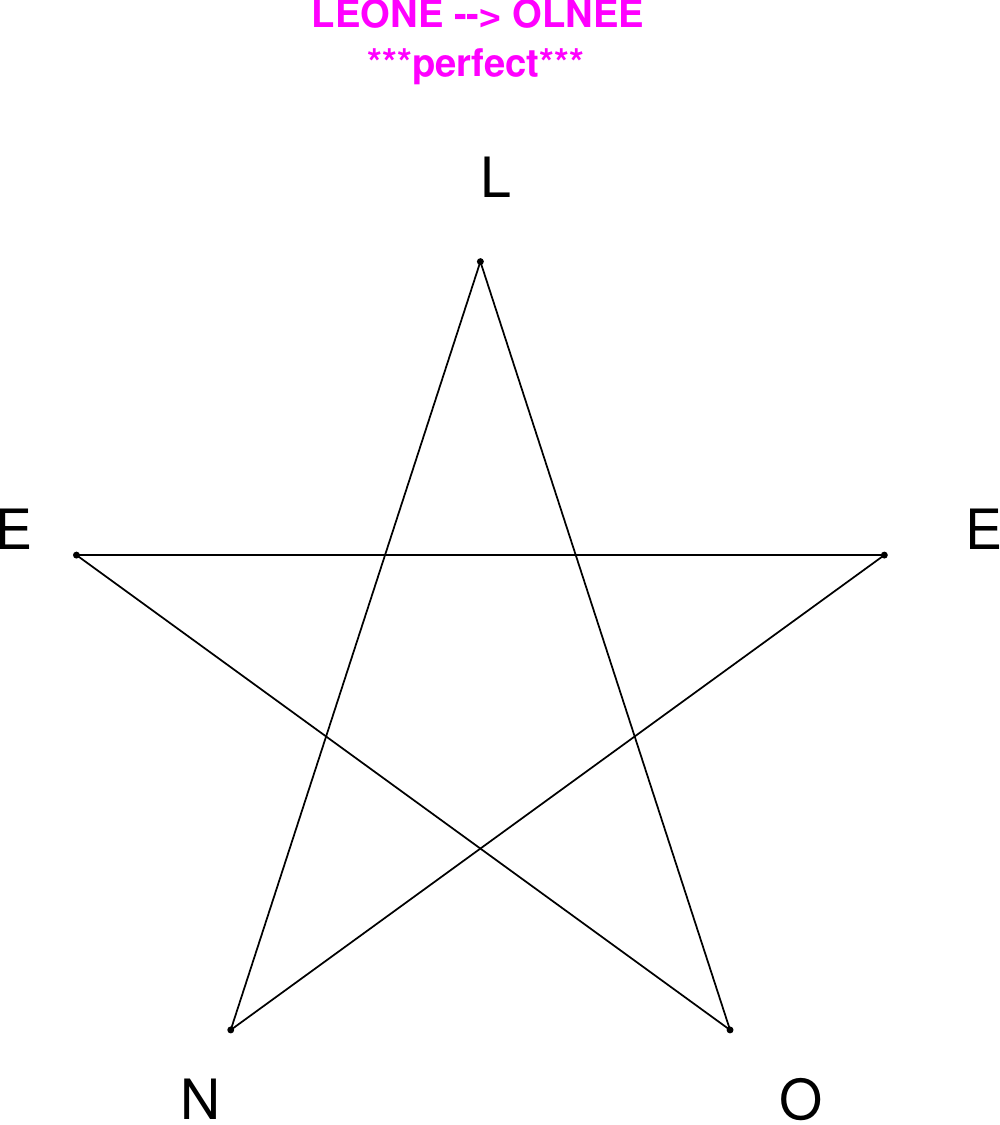}
\end{subfigure}
\hfill
\begin{subfigure}[T]{0.19\textwidth}
\centering
\includegraphics[width=\textwidth]{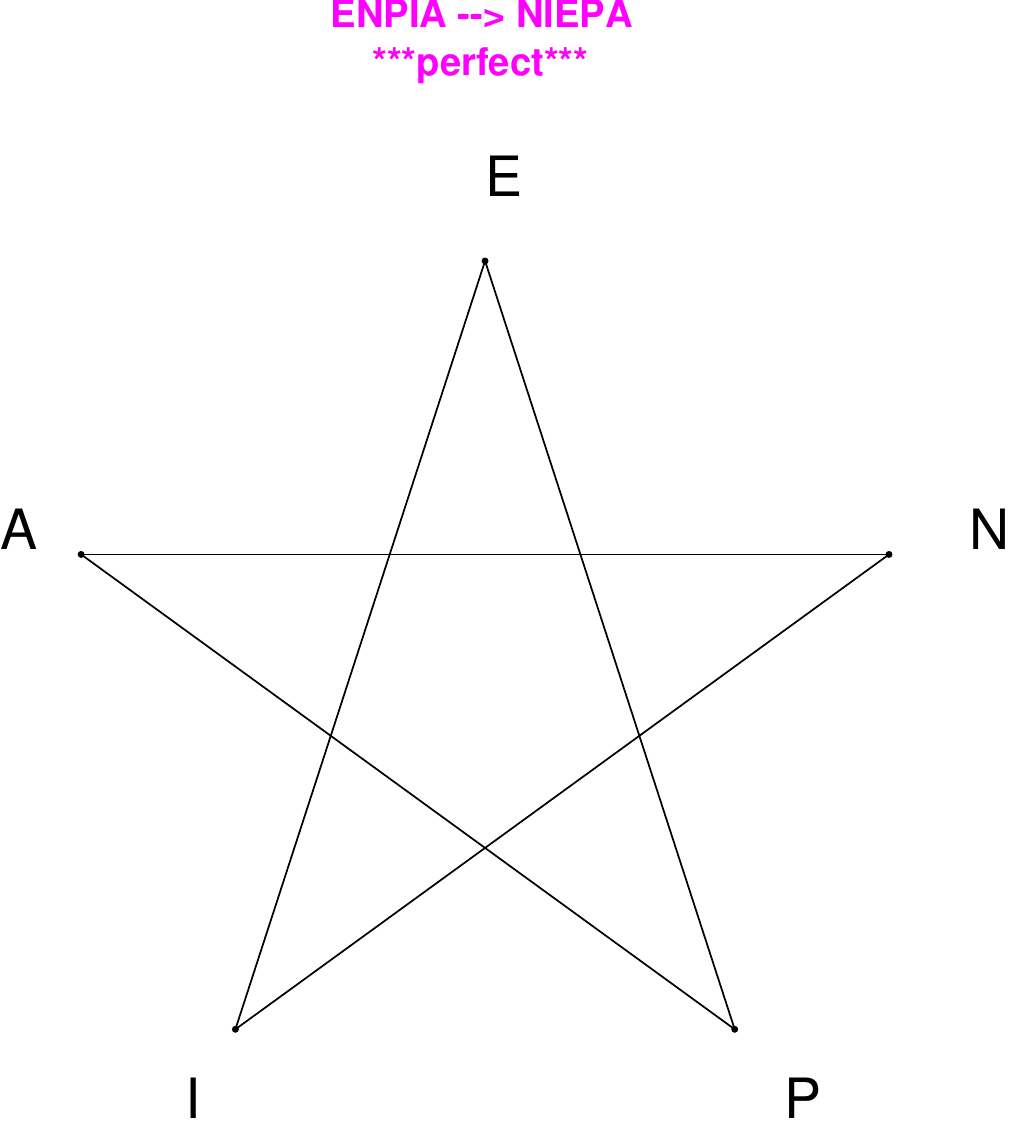}
\end{subfigure}
\end{figure}

\begin{figure}[H]
\centering
\begin{subfigure}[T]{0.19\textwidth}
\centering
\includegraphics[width=\textwidth]{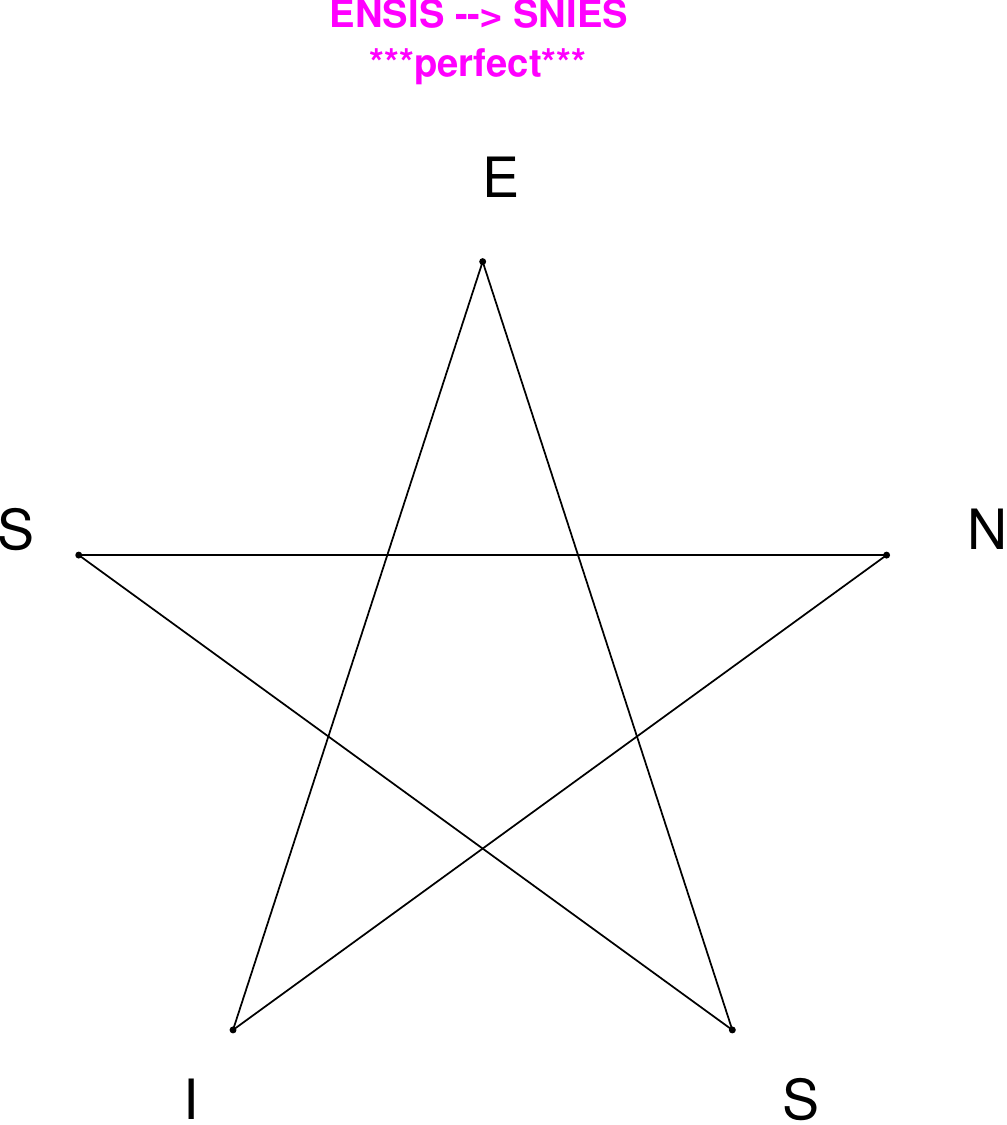}
\end{subfigure}
\hfill
\begin{subfigure}[T]{0.19\textwidth}
\centering
\includegraphics[width=\textwidth]{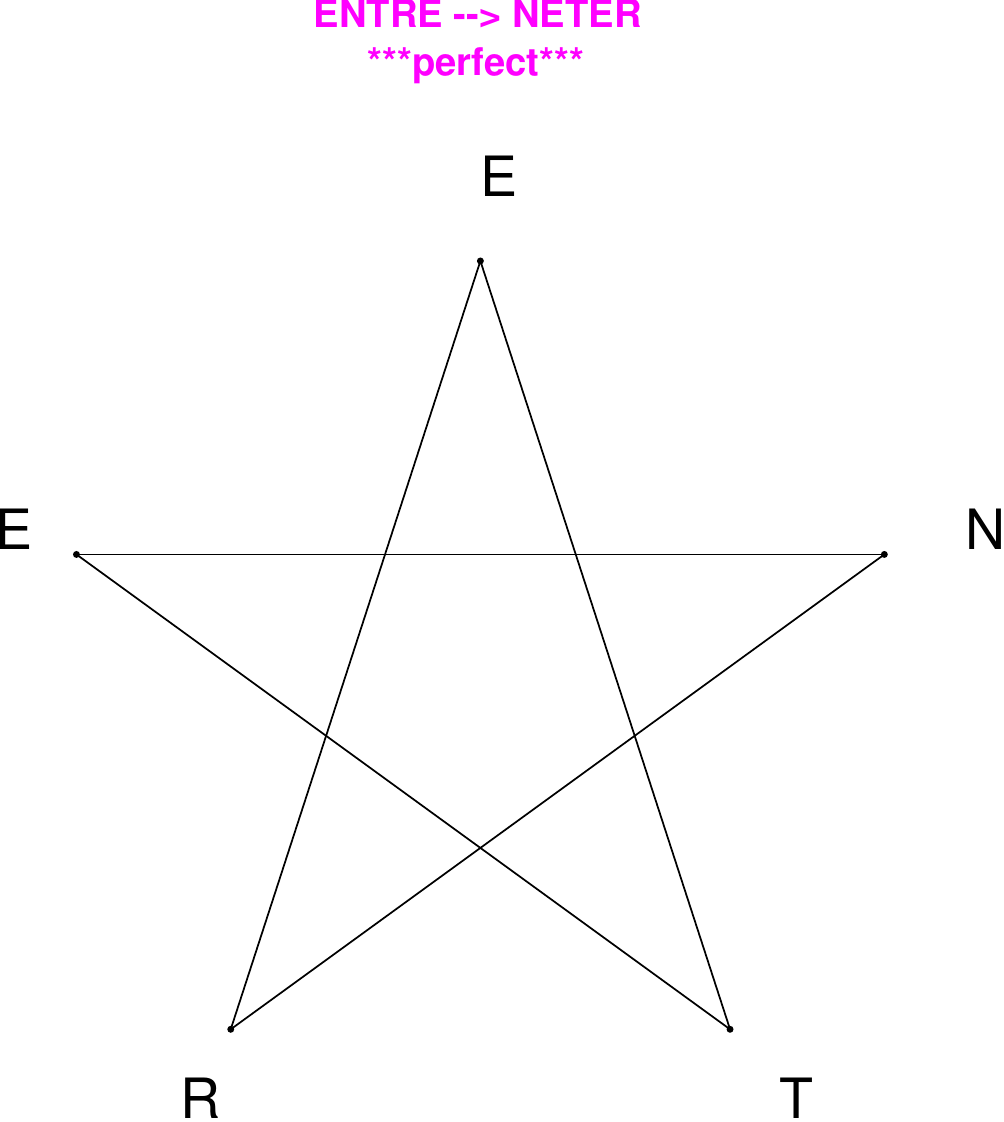}
\end{subfigure}
\hfill
\begin{subfigure}[T]{0.19\textwidth}
\centering
\includegraphics[width=\textwidth]{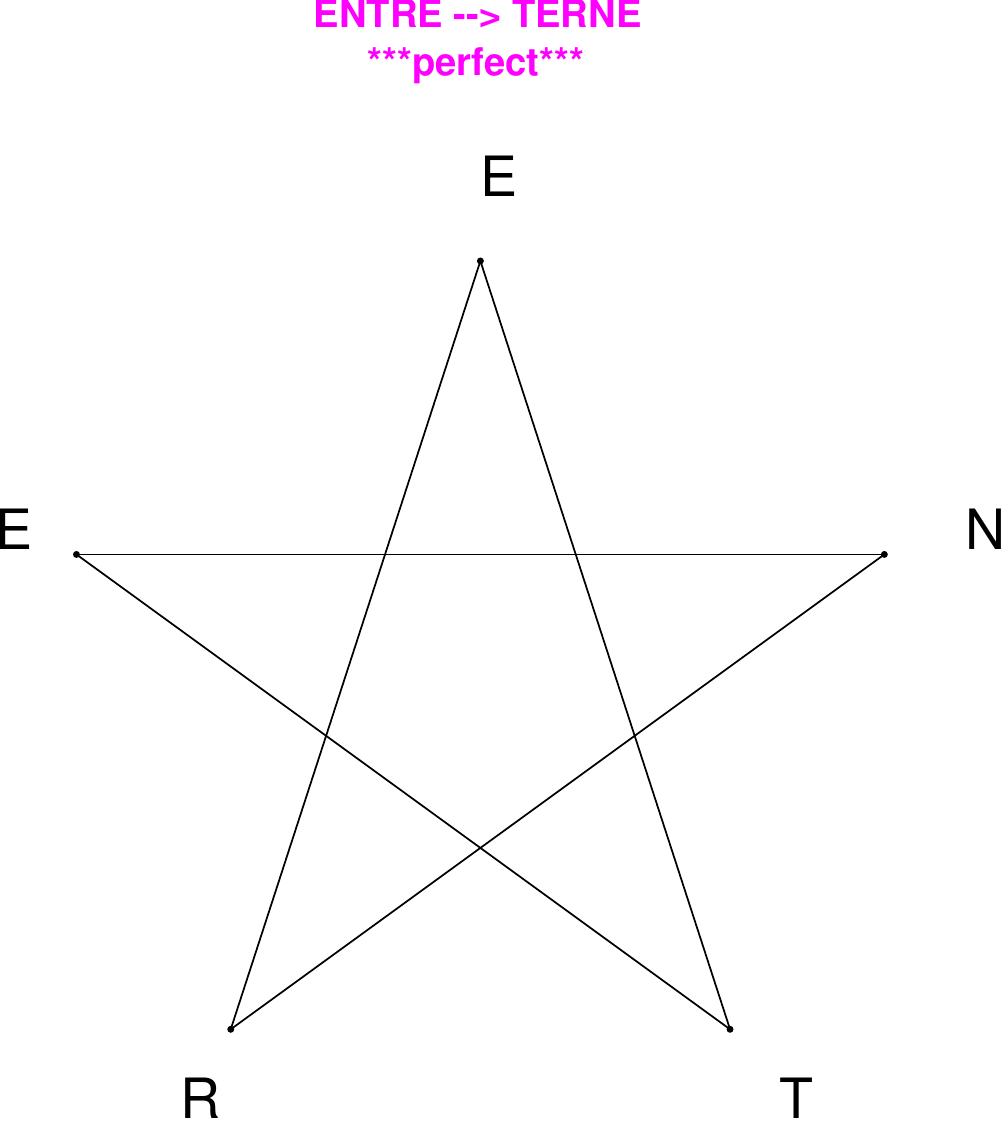}
\end{subfigure}
\hfill
\begin{subfigure}[T]{0.19\textwidth}
\centering
\includegraphics[width=\textwidth]{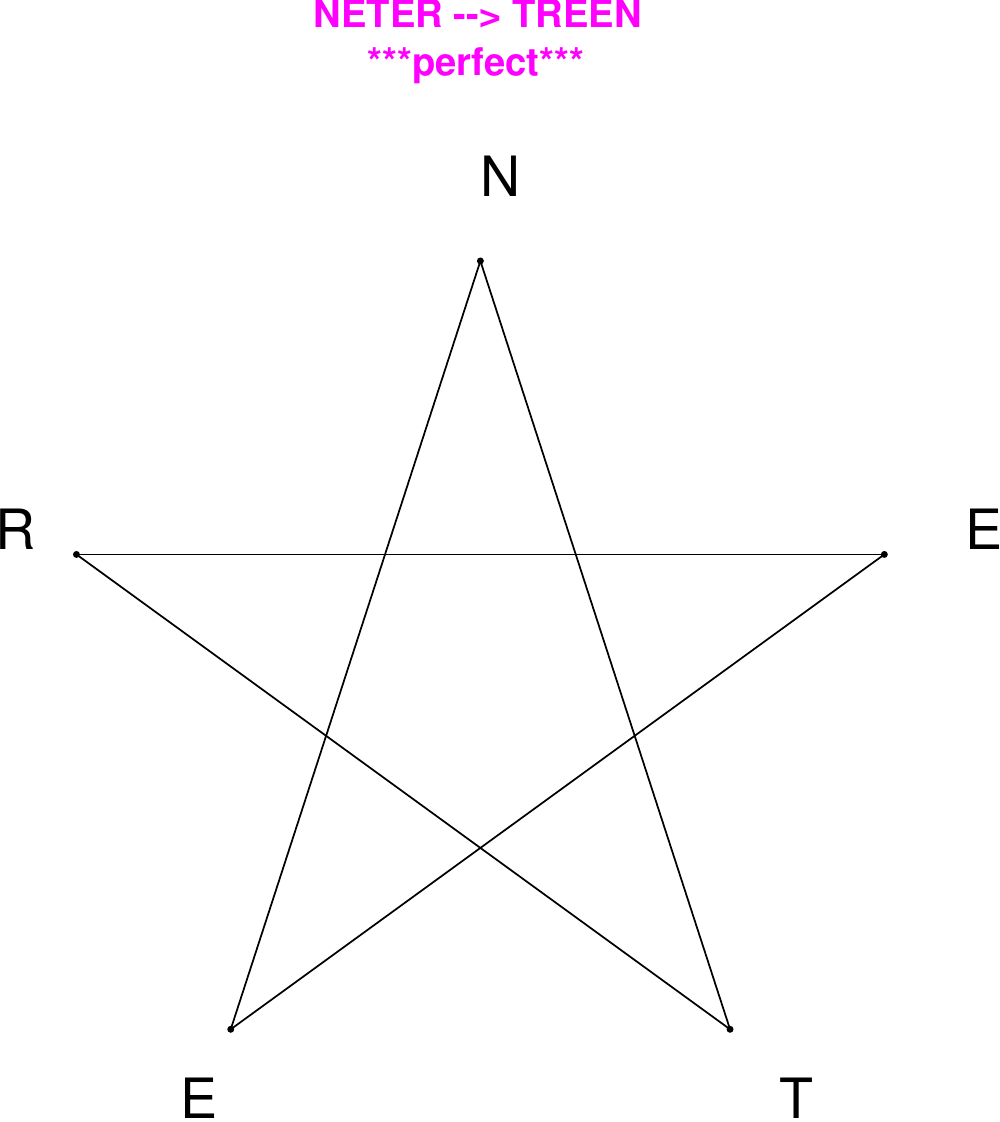}
\end{subfigure}
\hfill
\begin{subfigure}[T]{0.19\textwidth}
\centering
\includegraphics[width=\textwidth]{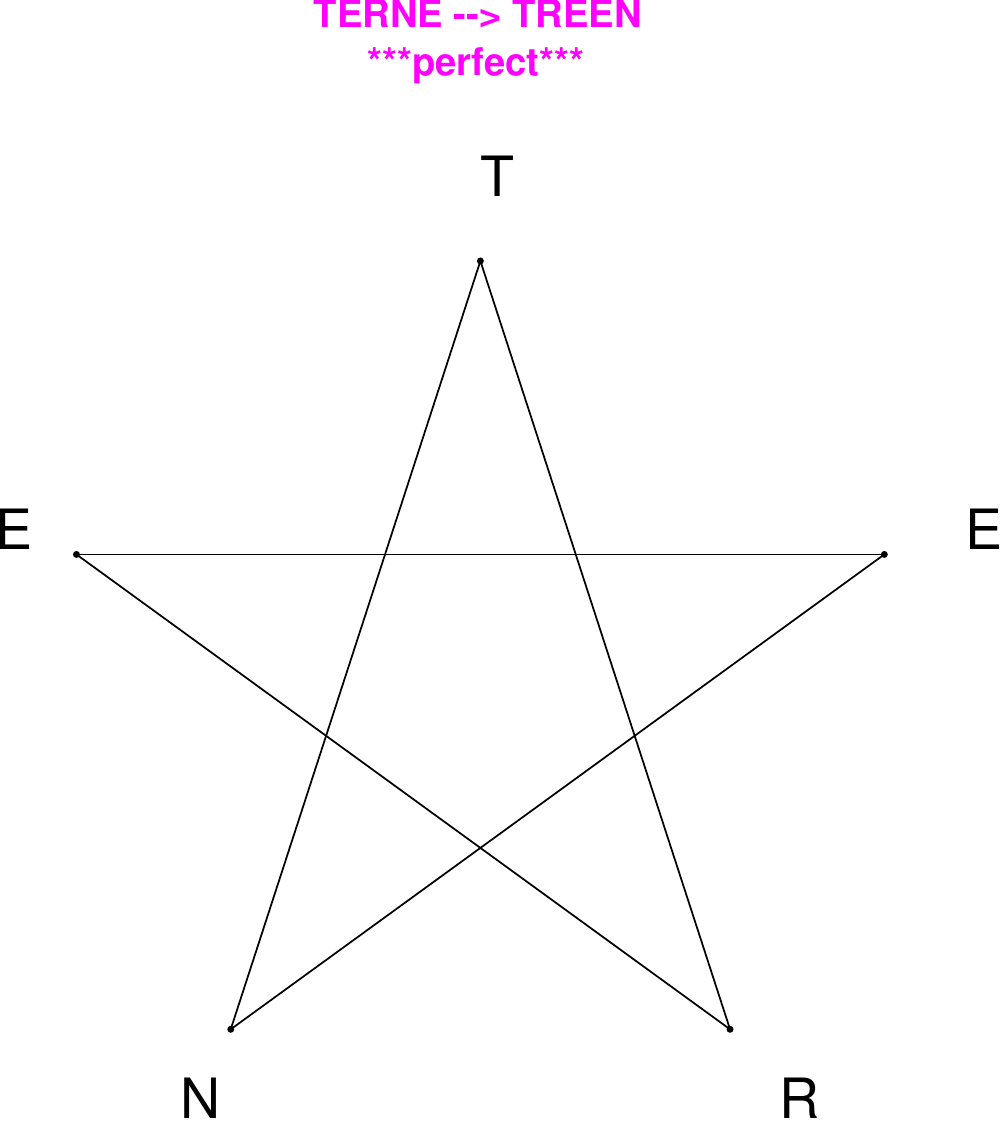}
\end{subfigure}
\end{figure}

\begin{figure}[H]
\centering
\begin{subfigure}[T]{0.19\textwidth}
\centering
\includegraphics[width=\textwidth]{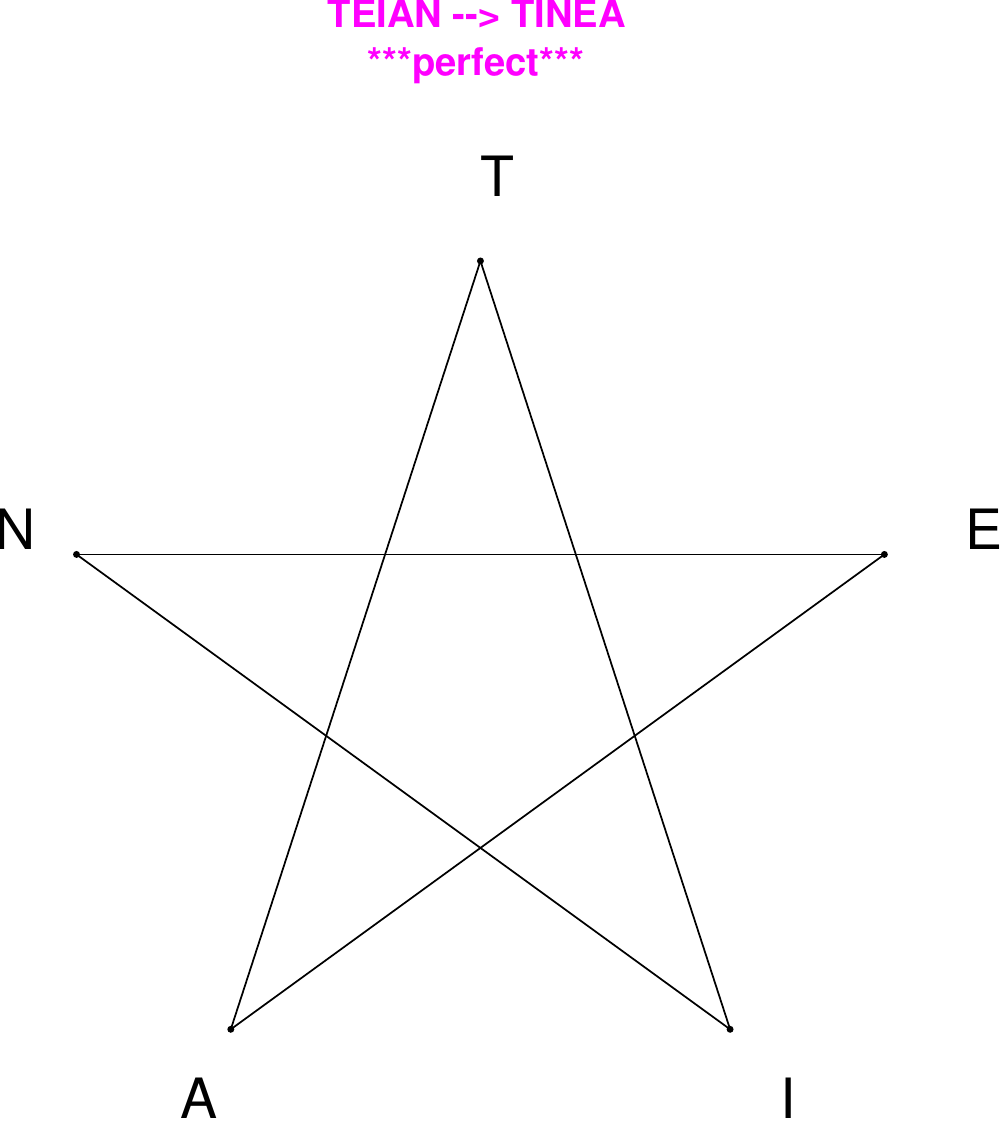}
\end{subfigure}
\hfill
\begin{subfigure}[T]{0.19\textwidth}
\centering
\includegraphics[width=\textwidth]{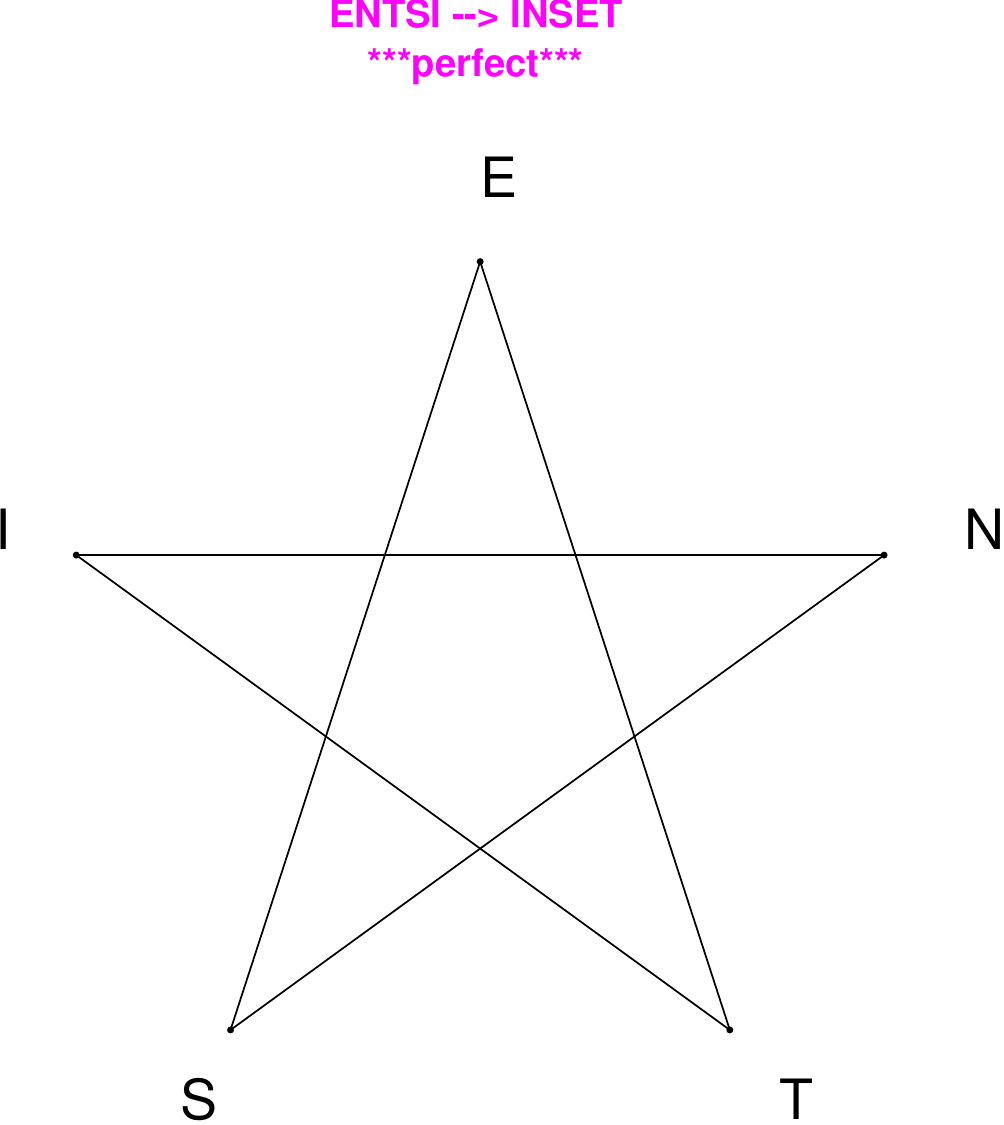}
\end{subfigure}
\hfill
\begin{subfigure}[T]{0.19\textwidth}
\centering
\includegraphics[width=\textwidth]{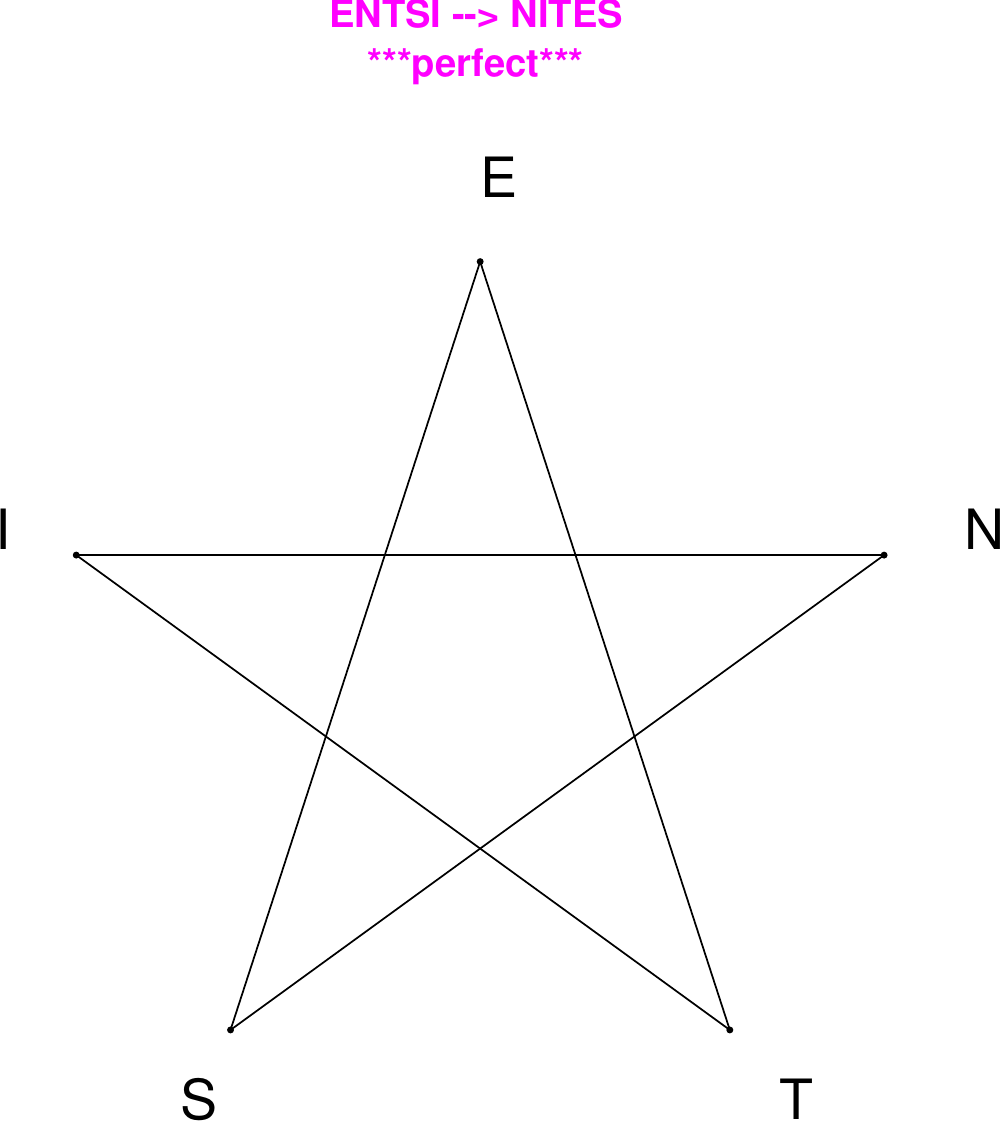}
\end{subfigure}
\hfill
\begin{subfigure}[T]{0.19\textwidth}
\centering
\includegraphics[width=\textwidth]{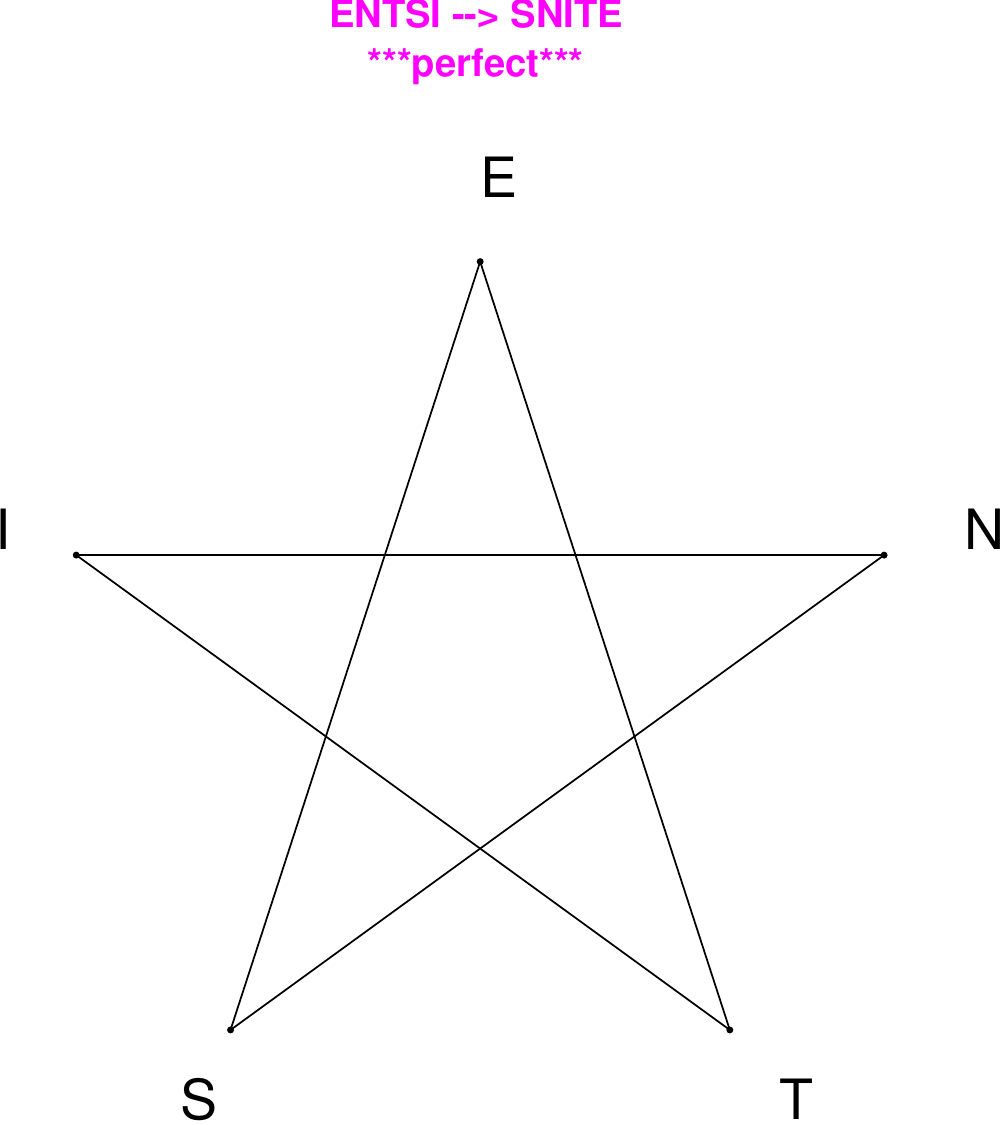}
\end{subfigure}
\hfill
\begin{subfigure}[T]{0.19\textwidth}
\centering
\includegraphics[width=\textwidth]{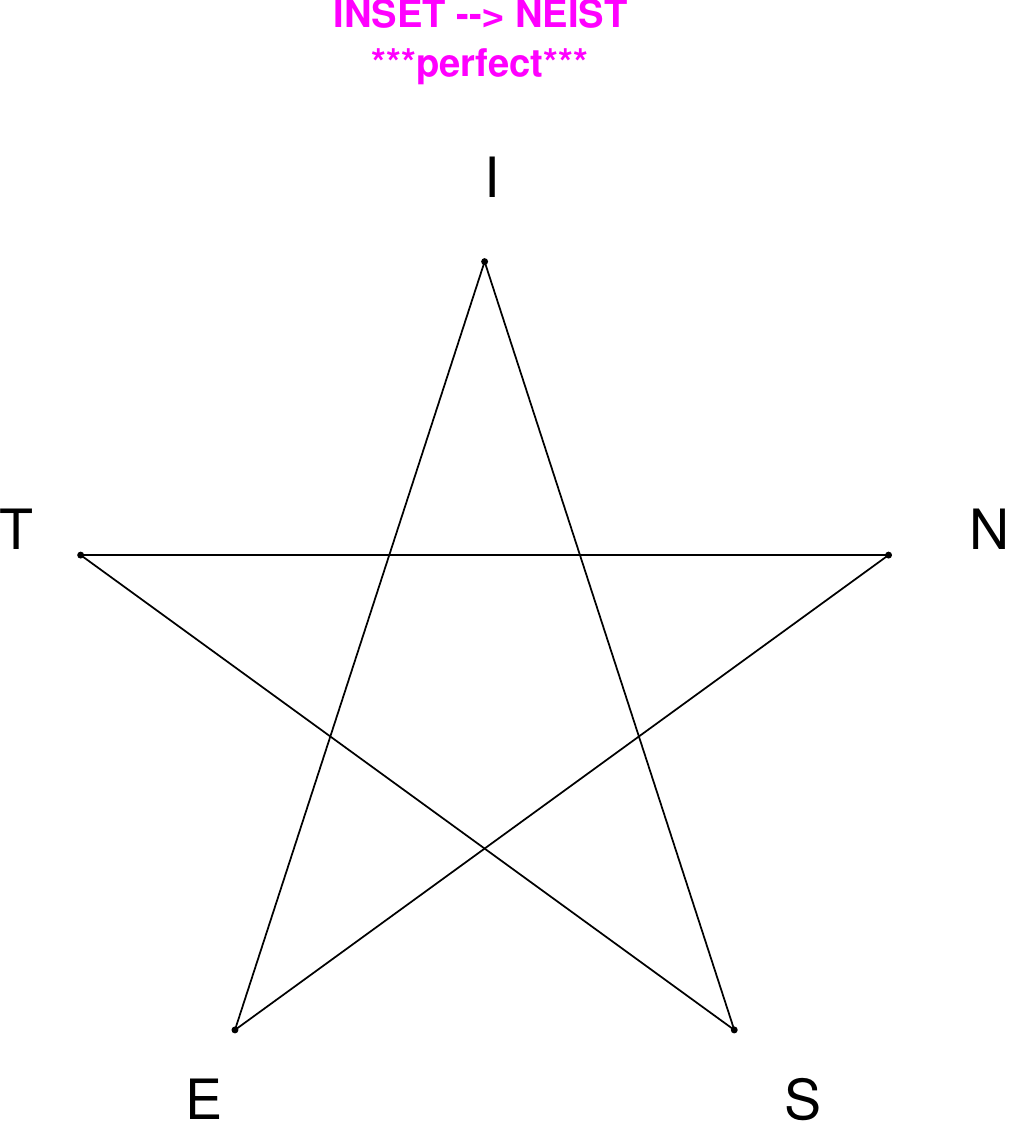}
\end{subfigure}
\end{figure}

\begin{figure}[H]
\centering
\begin{subfigure}[T]{0.19\textwidth}
\centering
\includegraphics[width=\textwidth]{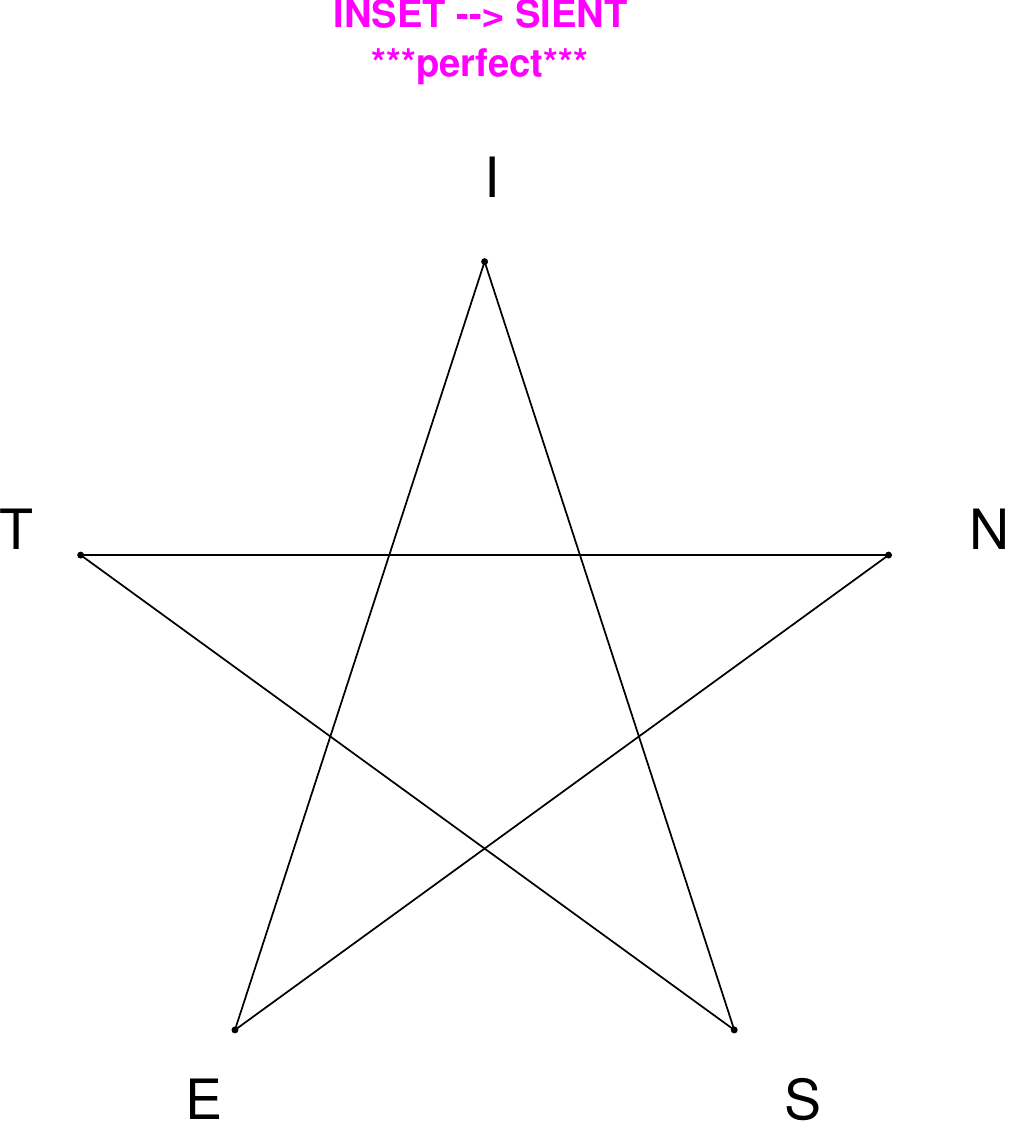}
\end{subfigure}
\hfill
\begin{subfigure}[T]{0.19\textwidth}
\centering
\includegraphics[width=\textwidth]{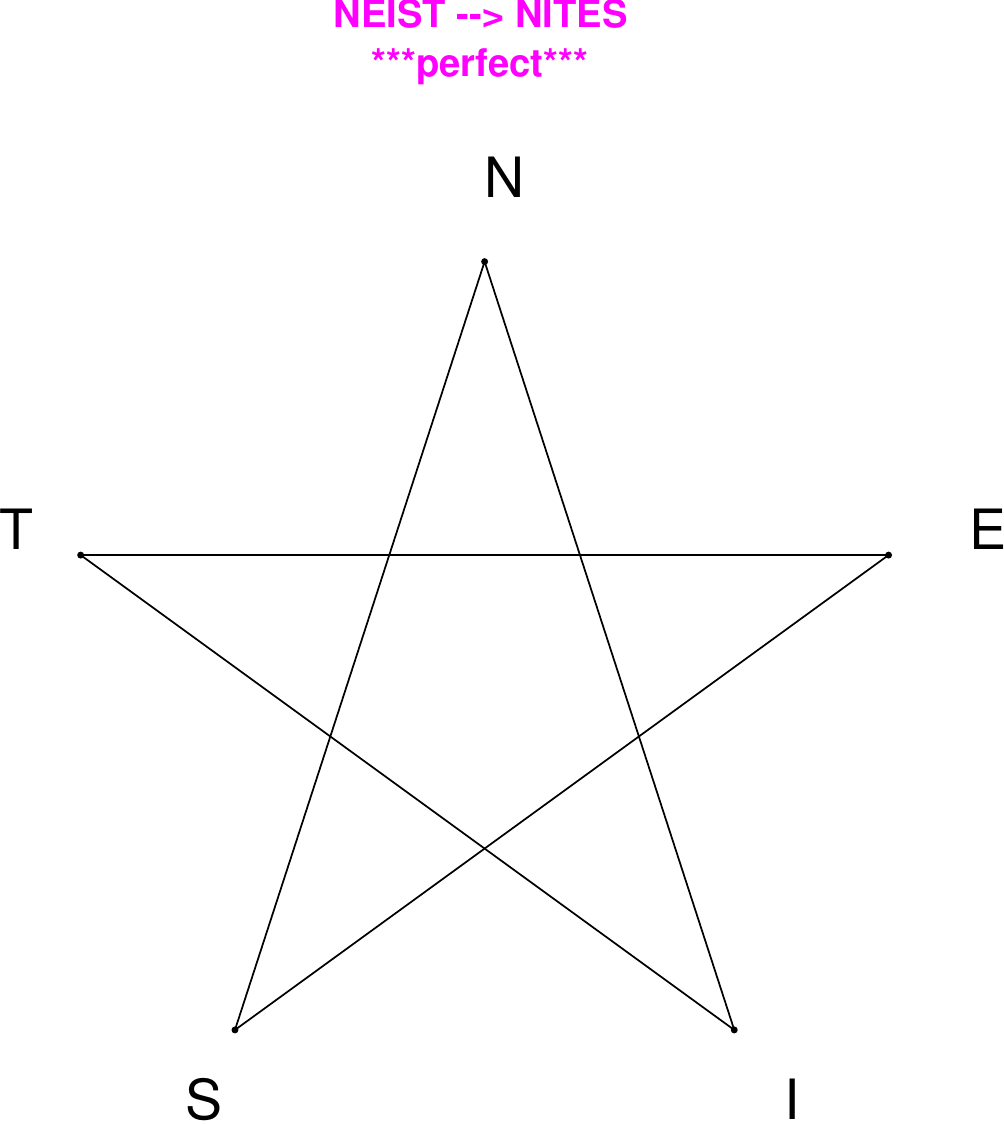}
\end{subfigure}
\hfill
\begin{subfigure}[T]{0.19\textwidth}
\centering
\includegraphics[width=\textwidth]{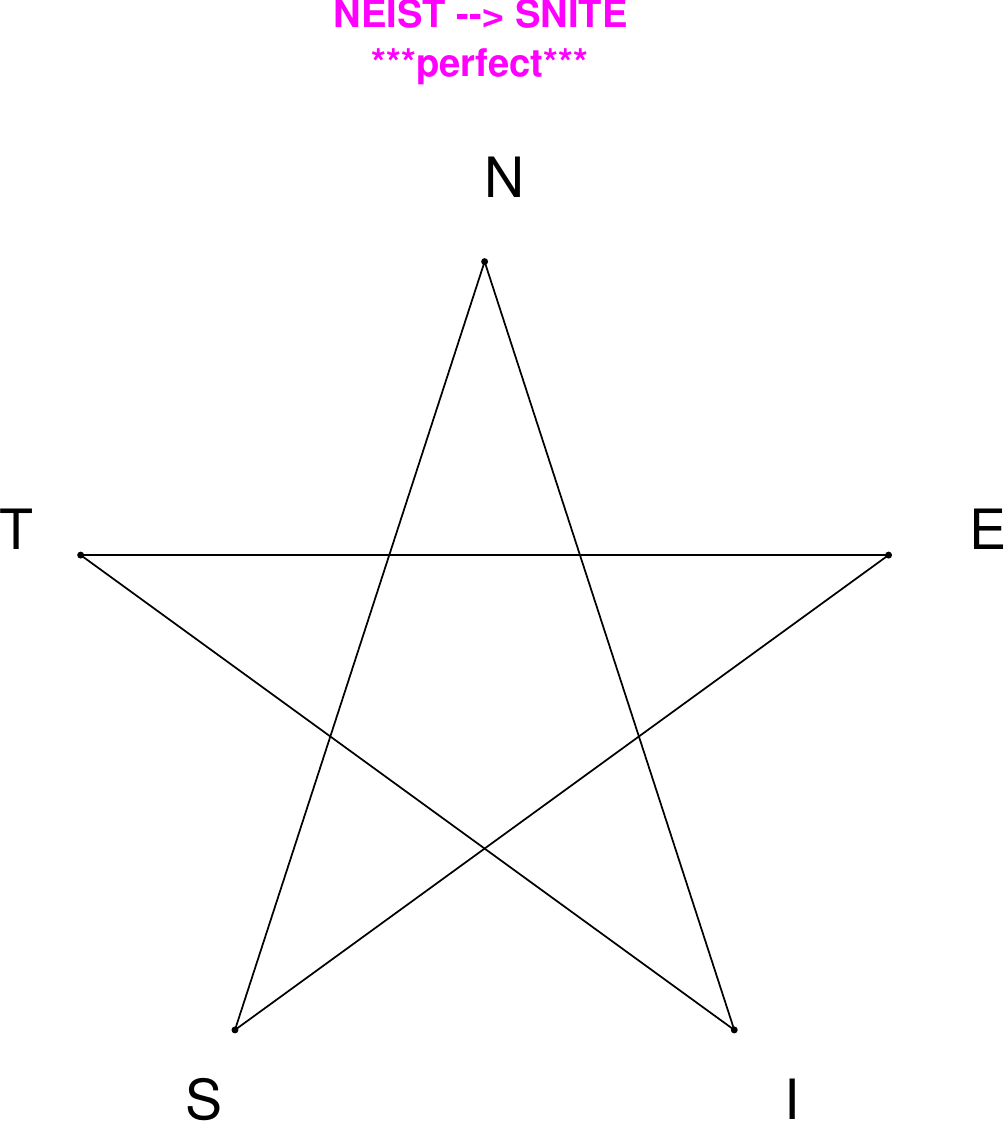}
\end{subfigure}
\hfill
\begin{subfigure}[T]{0.19\textwidth}
\centering
\includegraphics[width=\textwidth]{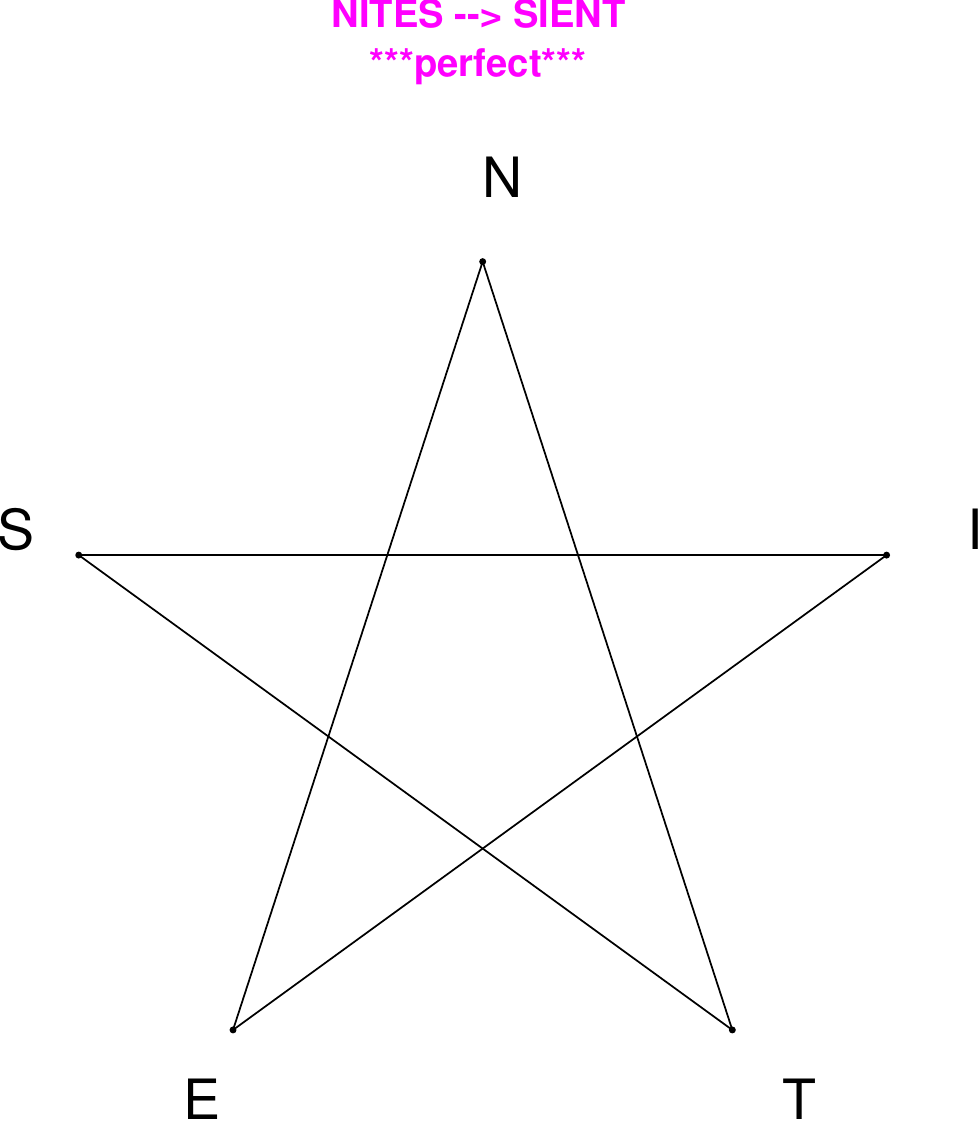}
\end{subfigure}
\hfill
\begin{subfigure}[T]{0.19\textwidth}
\centering
\includegraphics[width=\textwidth]{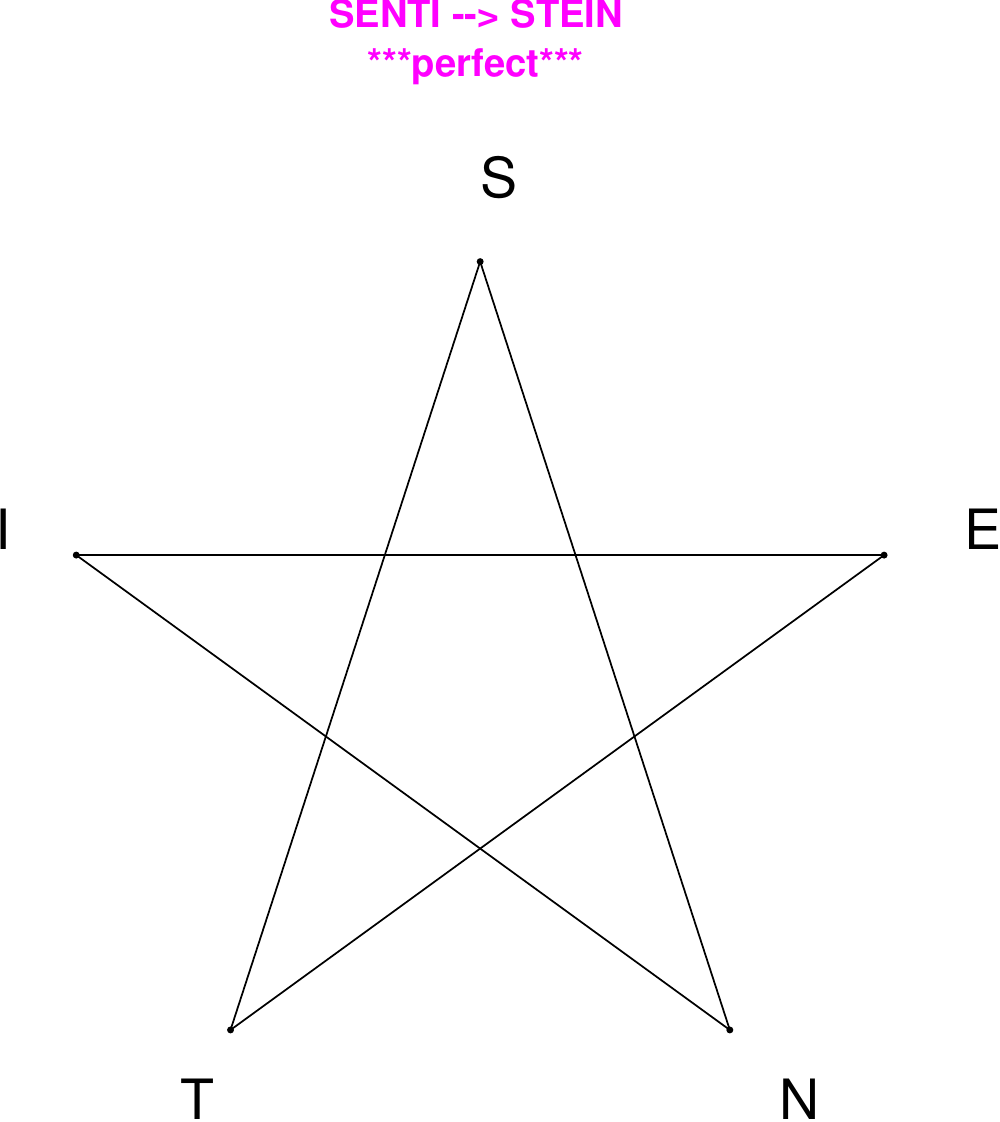}
\end{subfigure}
\end{figure}

\begin{figure}[H]
\centering
\begin{subfigure}[T]{0.19\textwidth}
\centering
\includegraphics[width=\textwidth]{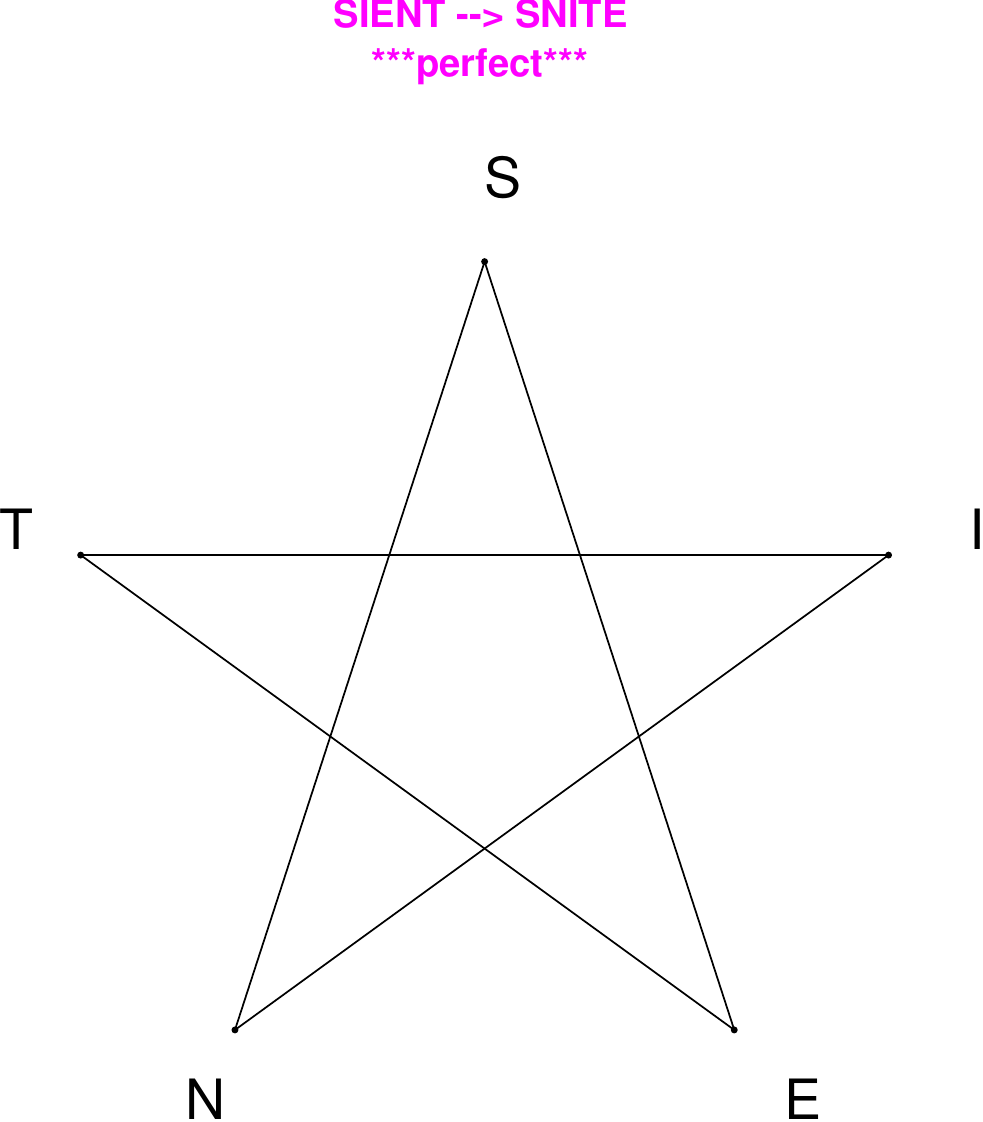}
\end{subfigure}
\hfill
\begin{subfigure}[T]{0.19\textwidth}
\centering
\includegraphics[width=\textwidth]{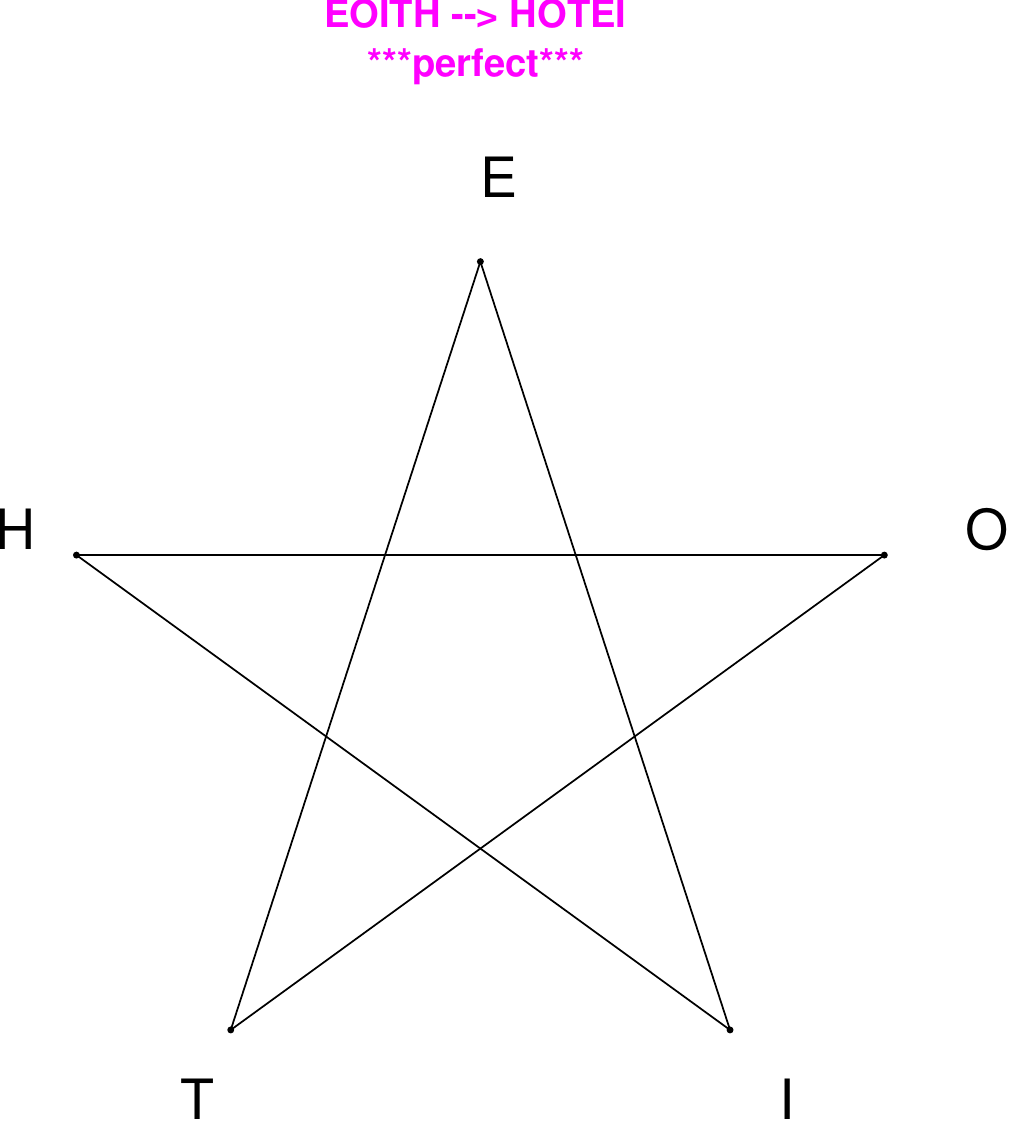}
\end{subfigure}
\hfill
\begin{subfigure}[T]{0.19\textwidth}
\centering
\includegraphics[width=\textwidth]{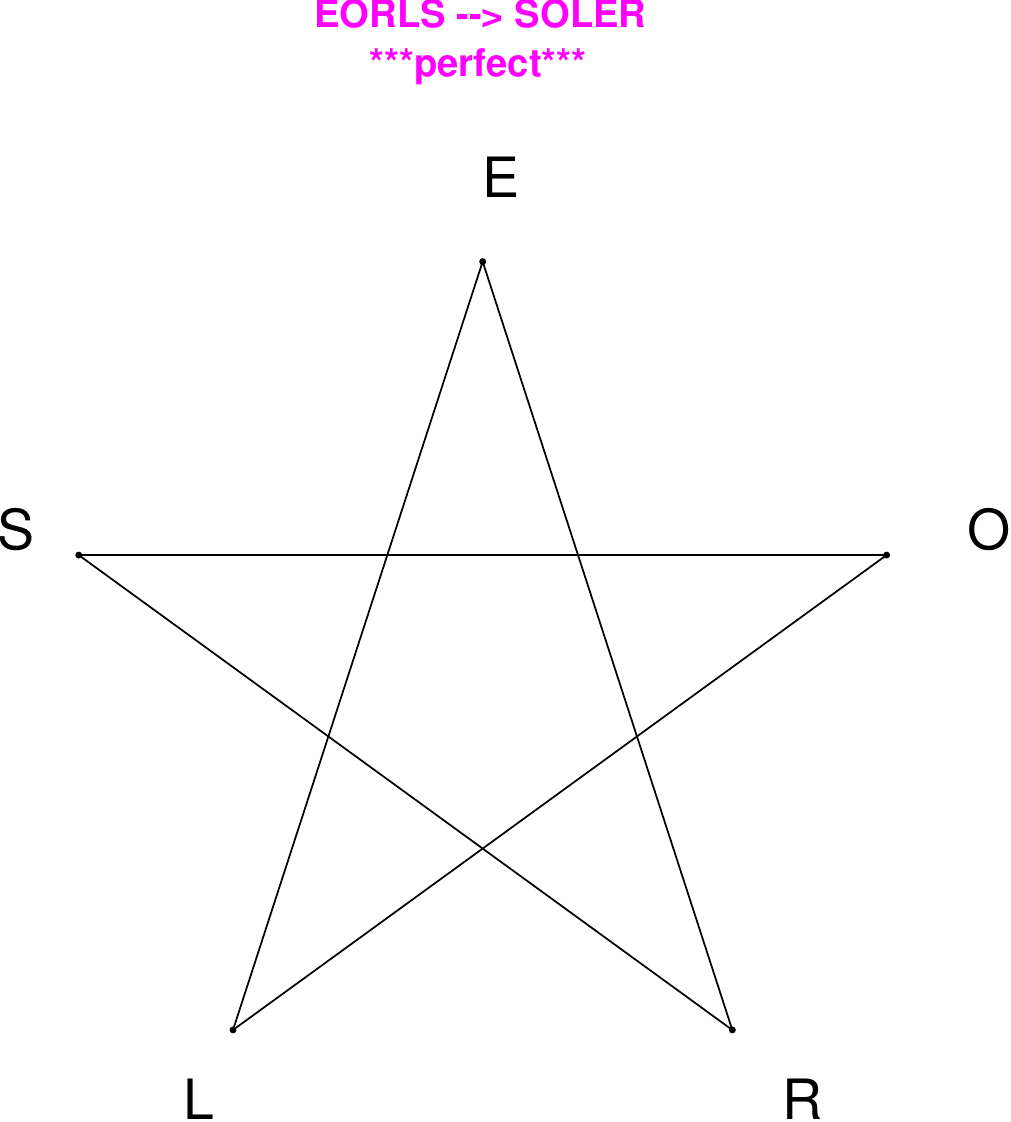}
\end{subfigure}
\hfill
\begin{subfigure}[T]{0.19\textwidth}
\centering
\includegraphics[width=\textwidth]{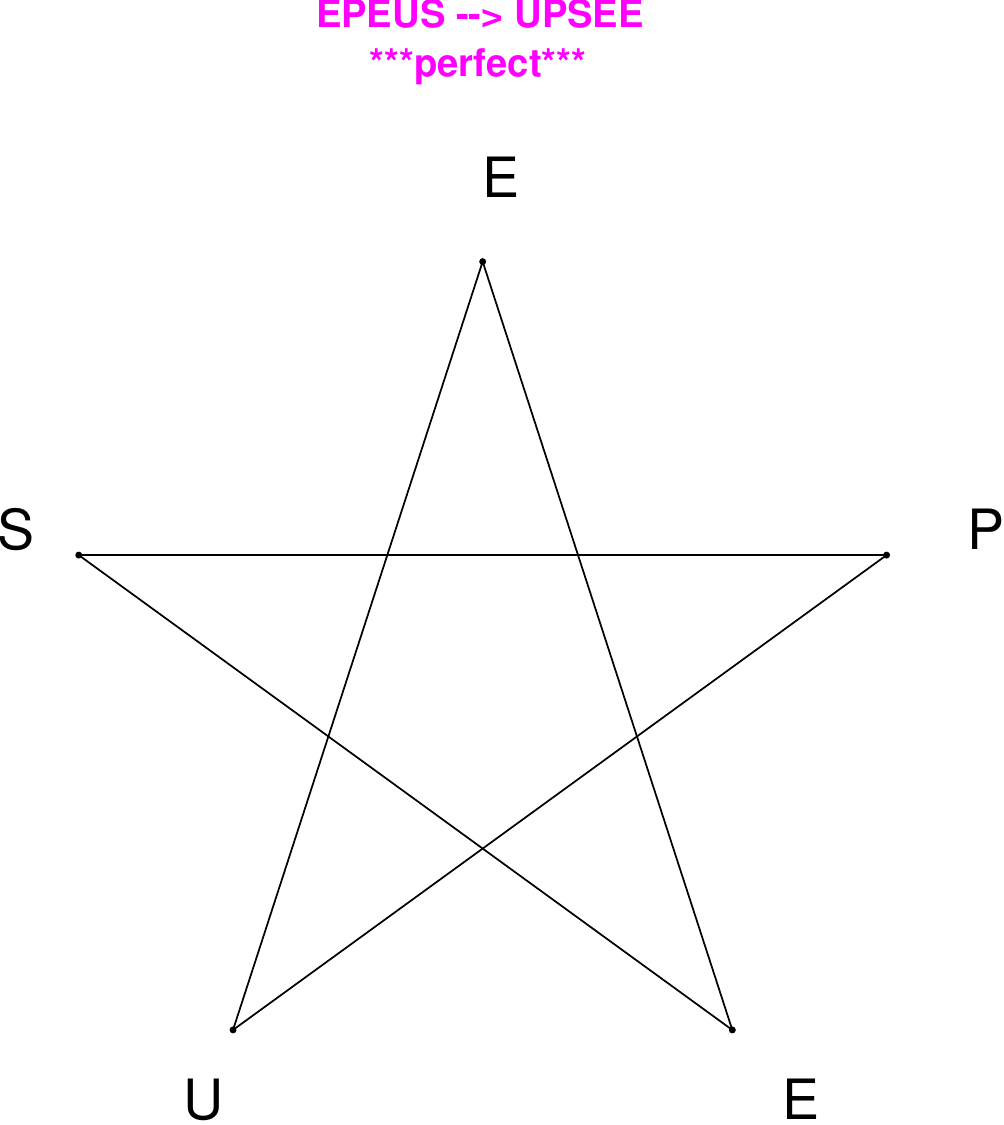}
\end{subfigure}
\hfill
\begin{subfigure}[T]{0.19\textwidth}
\centering
\includegraphics[width=\textwidth]{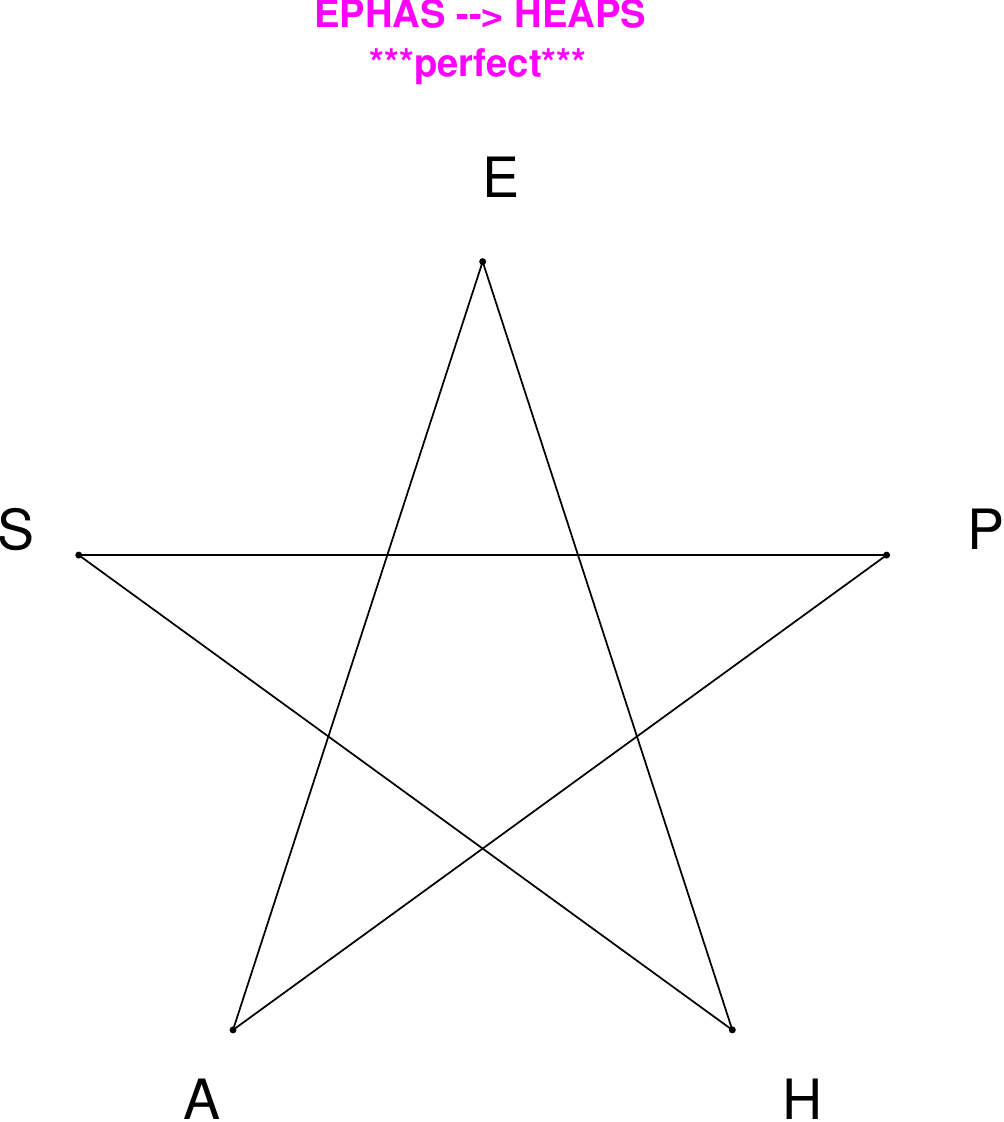}
\end{subfigure}
\end{figure}

\begin{figure}[H]
\centering
\begin{subfigure}[T]{0.19\textwidth}
\centering
\includegraphics[width=\textwidth]{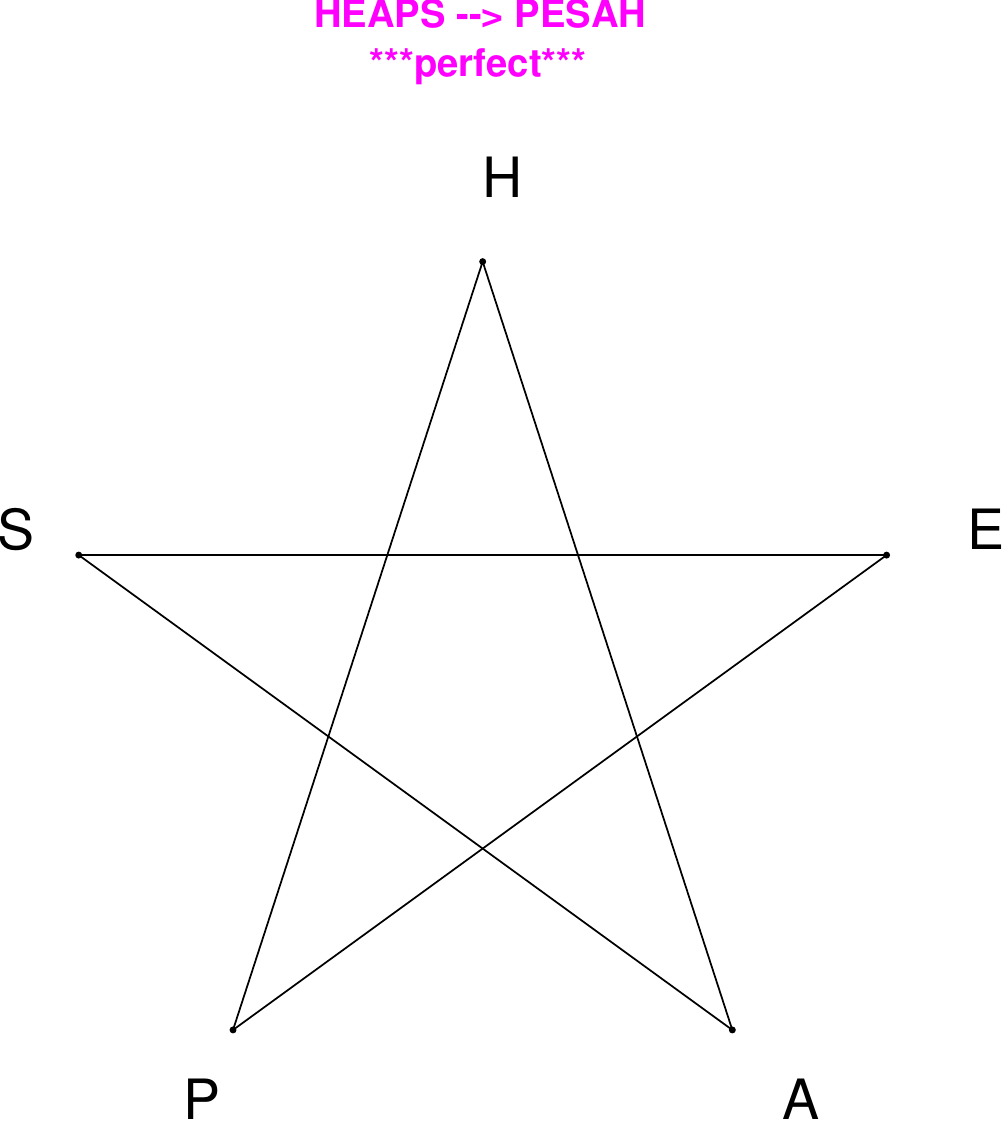}
\end{subfigure}
\hfill
\begin{subfigure}[T]{0.19\textwidth}
\centering
\includegraphics[width=\textwidth]{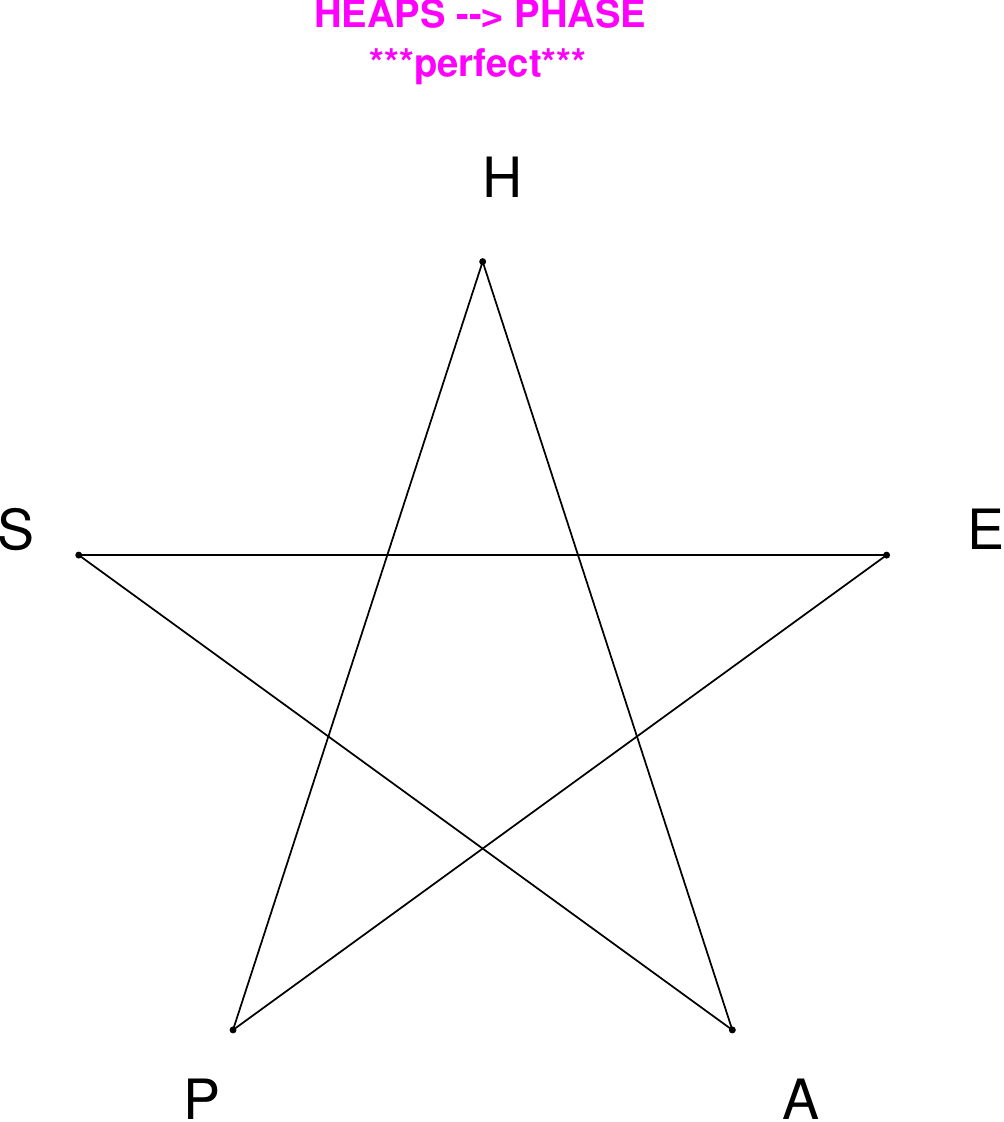}
\end{subfigure}
\hfill
\begin{subfigure}[T]{0.19\textwidth}
\centering
\includegraphics[width=\textwidth]{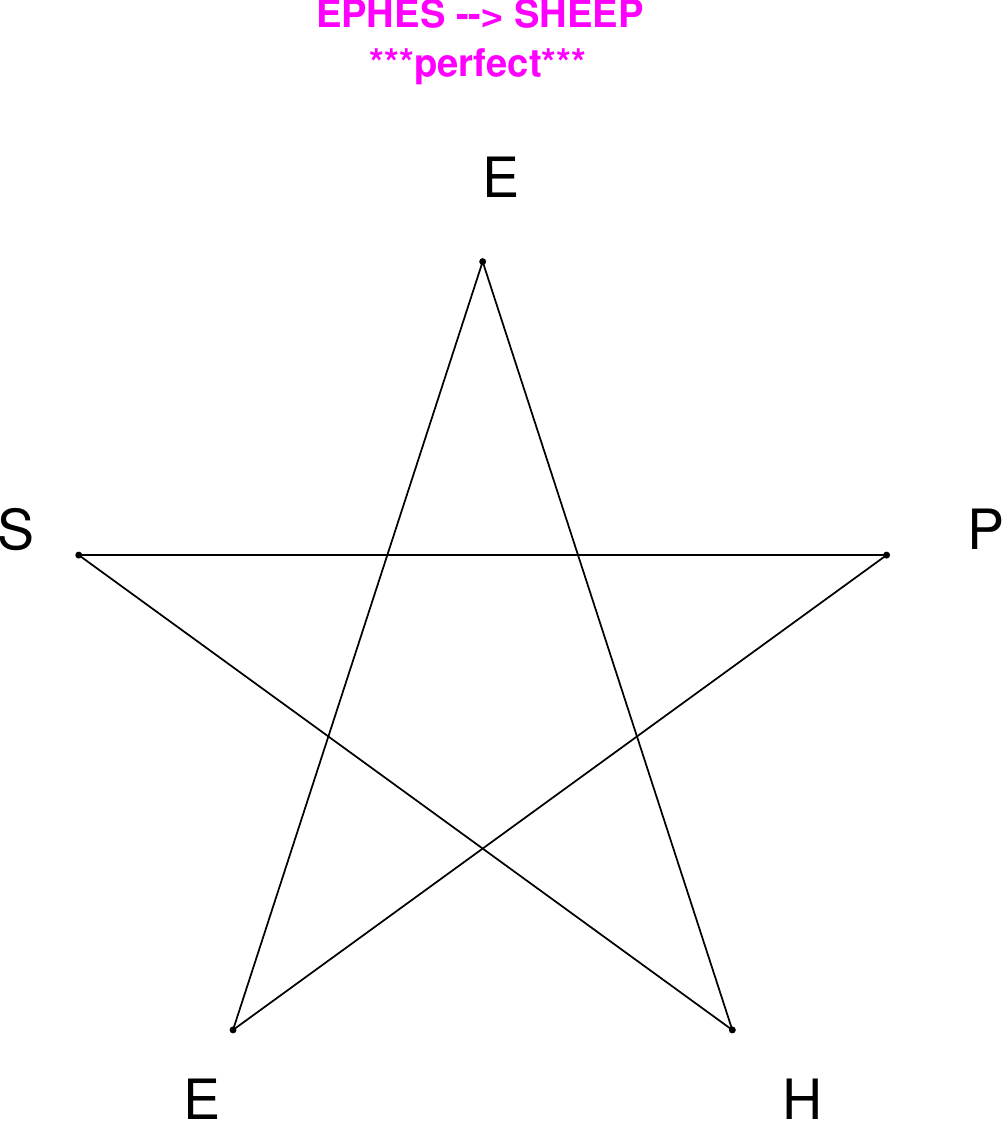}
\end{subfigure}
\hfill
\begin{subfigure}[T]{0.19\textwidth}
\centering
\includegraphics[width=\textwidth]{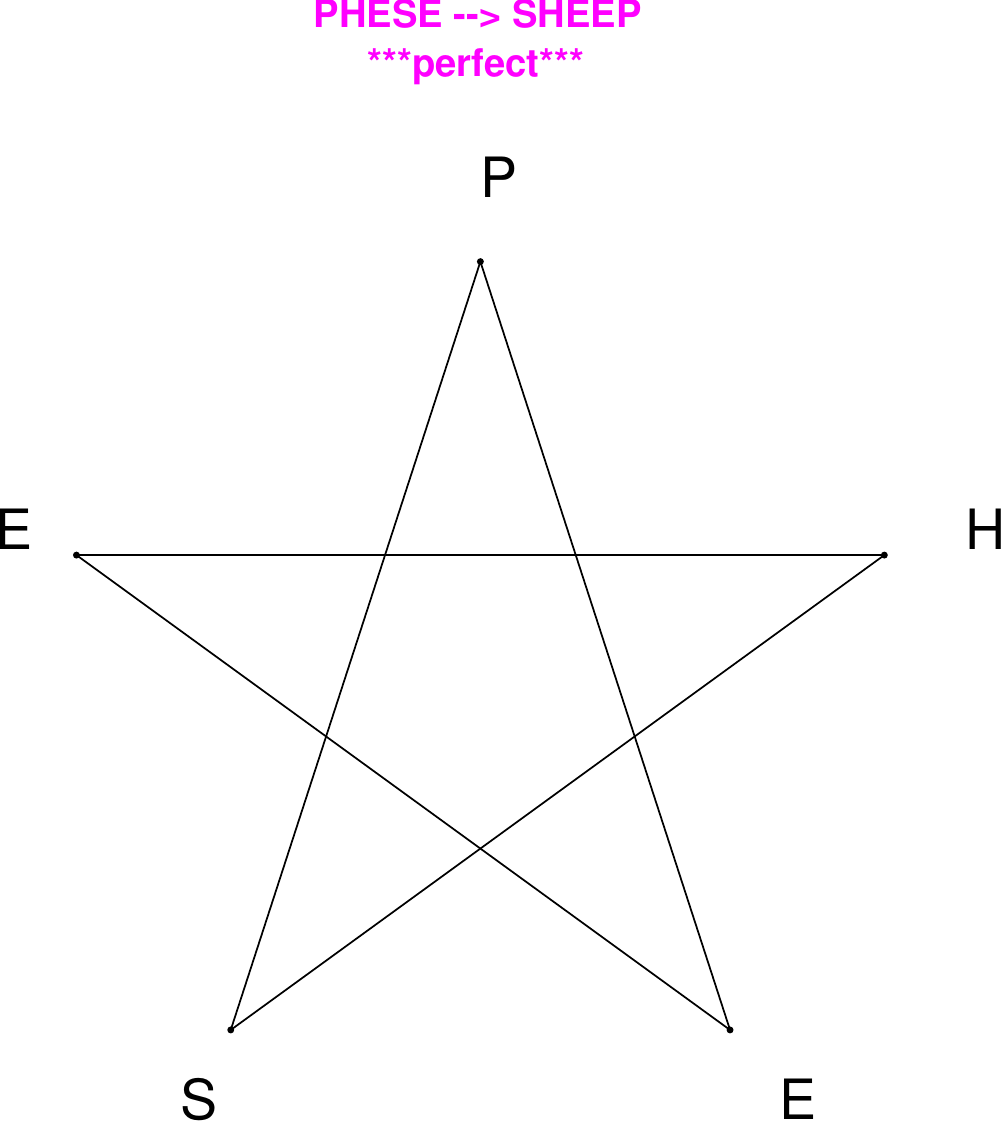}
\end{subfigure}
\hfill
\begin{subfigure}[T]{0.19\textwidth}
\centering
\includegraphics[width=\textwidth]{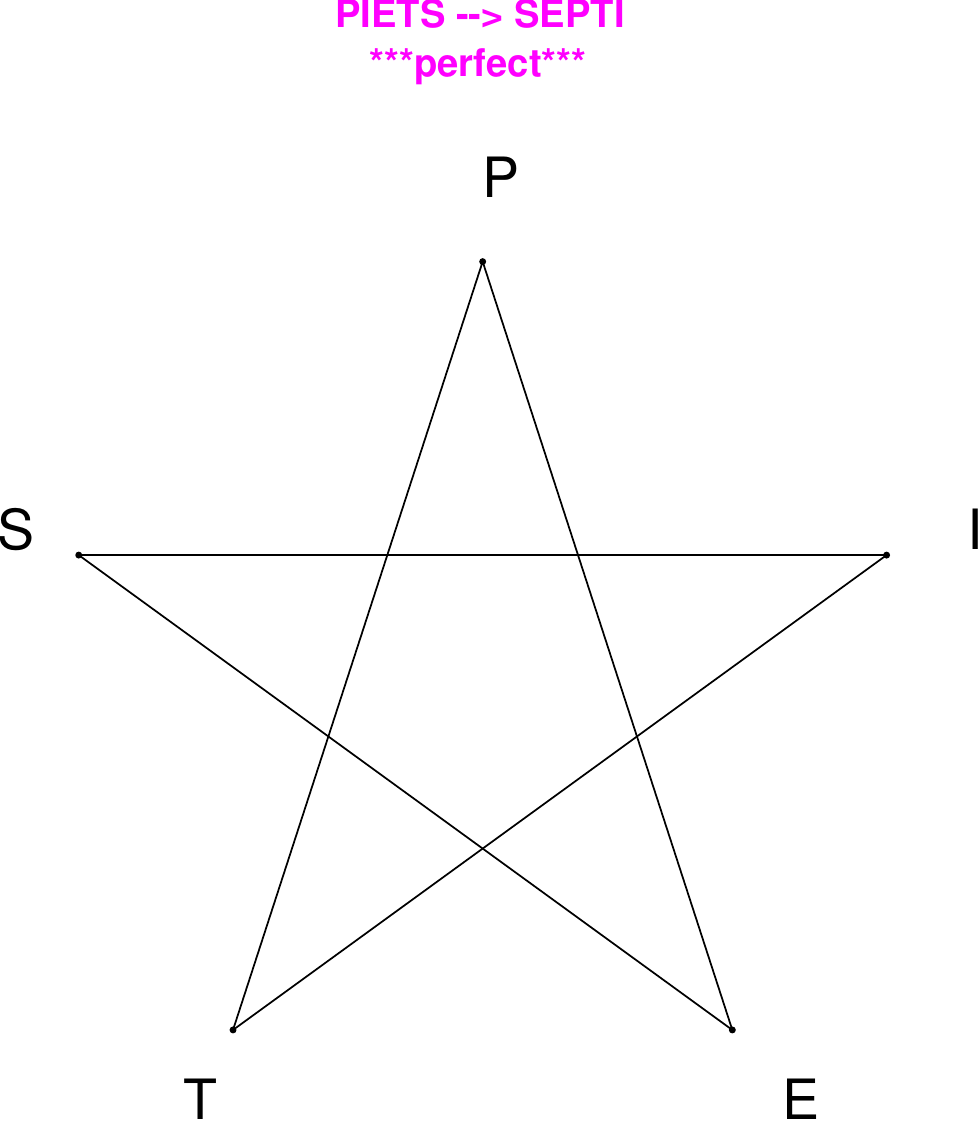}
\end{subfigure}
\end{figure}

\begin{figure}[H]
\centering
\begin{subfigure}[T]{0.19\textwidth}
\centering
\includegraphics[width=\textwidth]{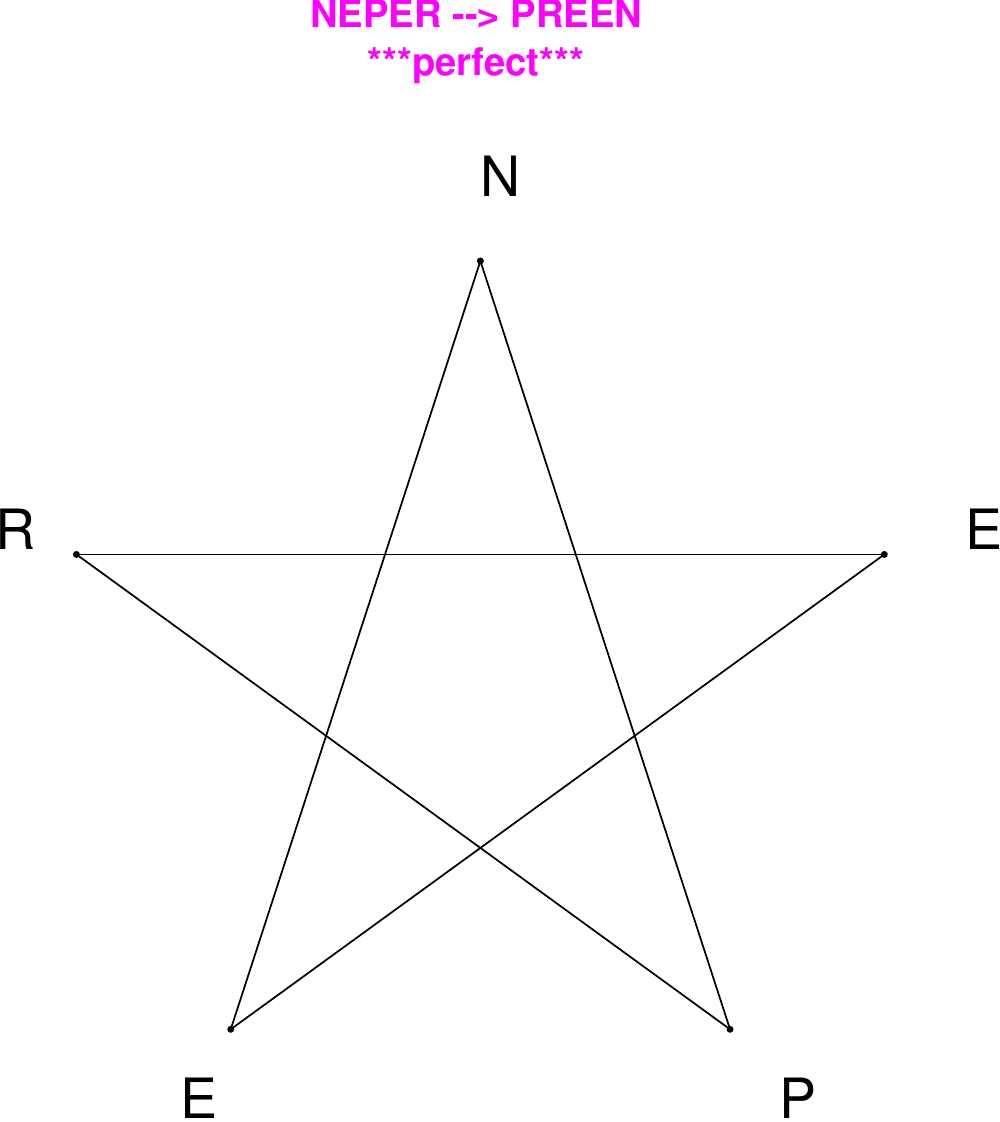}
\end{subfigure}
\hfill
\begin{subfigure}[T]{0.19\textwidth}
\centering
\includegraphics[width=\textwidth]{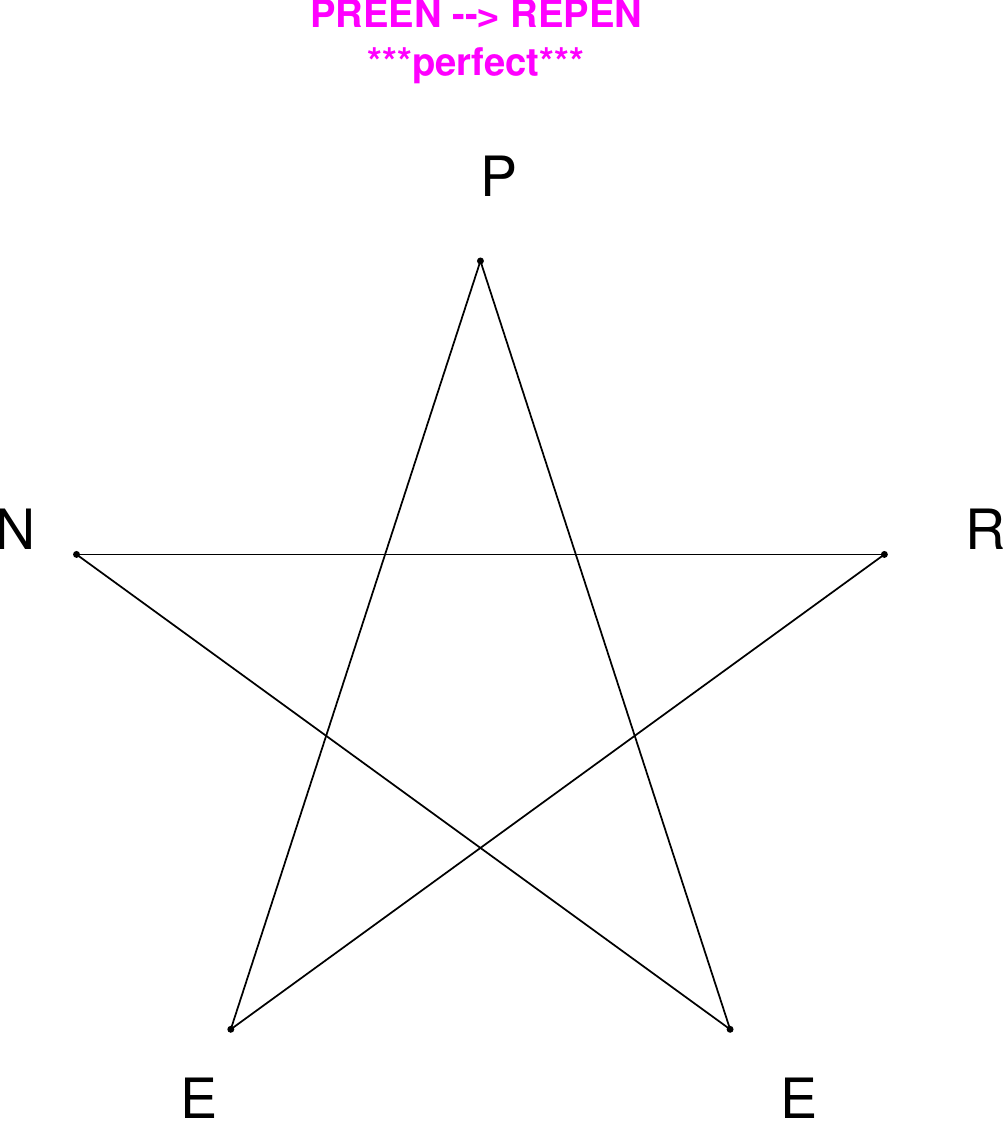}
\end{subfigure}
\hfill
\begin{subfigure}[T]{0.19\textwidth}
\centering
\includegraphics[width=\textwidth]{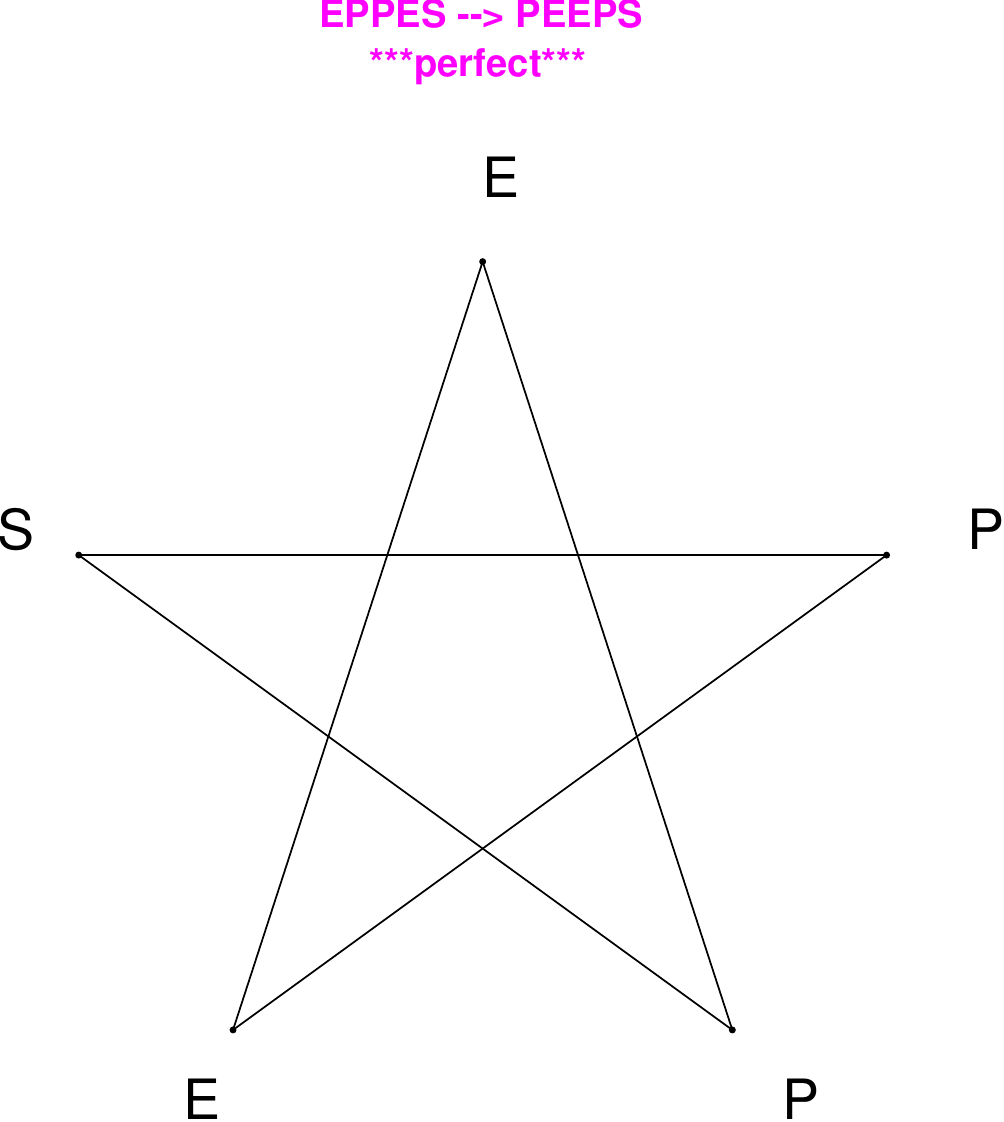}
\end{subfigure}
\hfill
\begin{subfigure}[T]{0.19\textwidth}
\centering
\includegraphics[width=\textwidth]{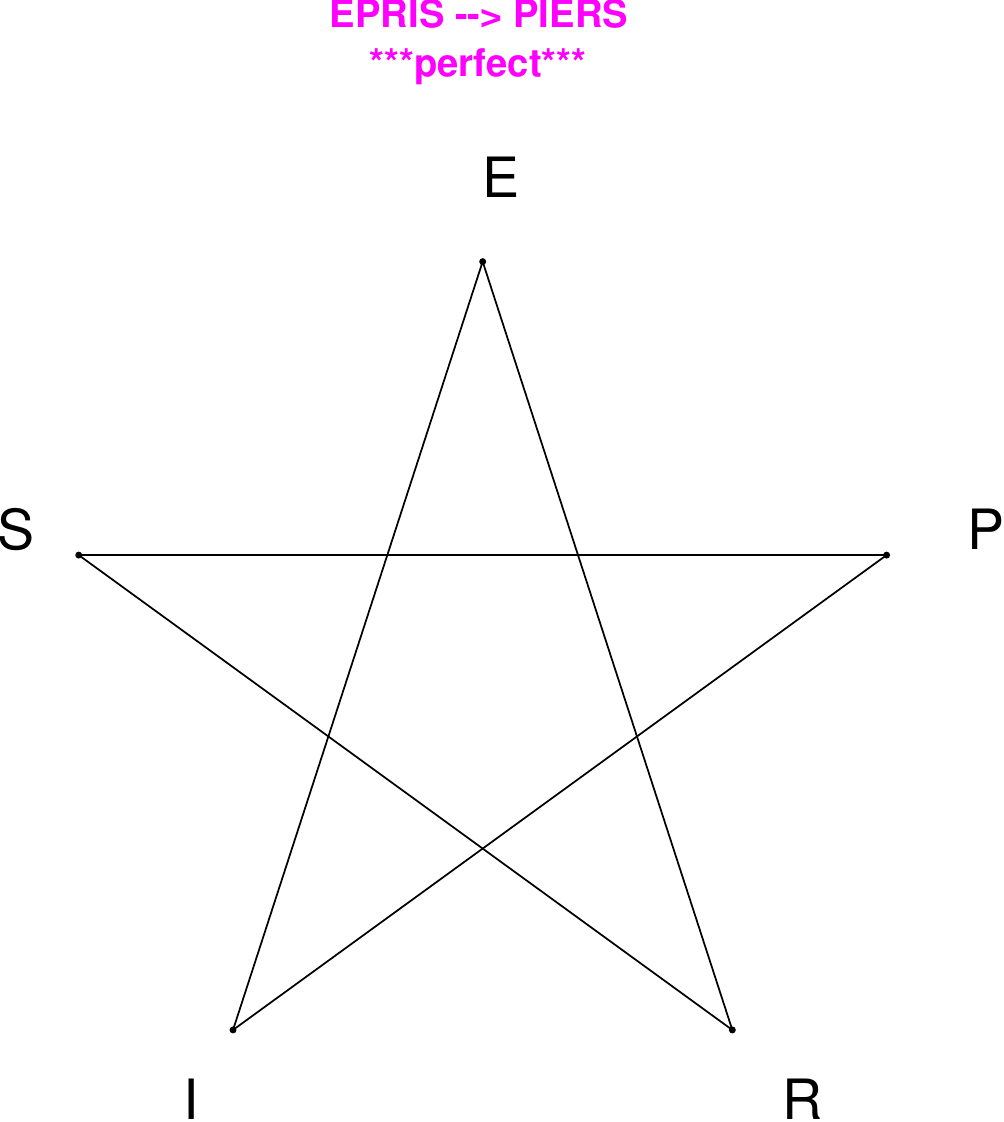}
\end{subfigure}
\hfill
\begin{subfigure}[T]{0.19\textwidth}
\centering
\includegraphics[width=\textwidth]{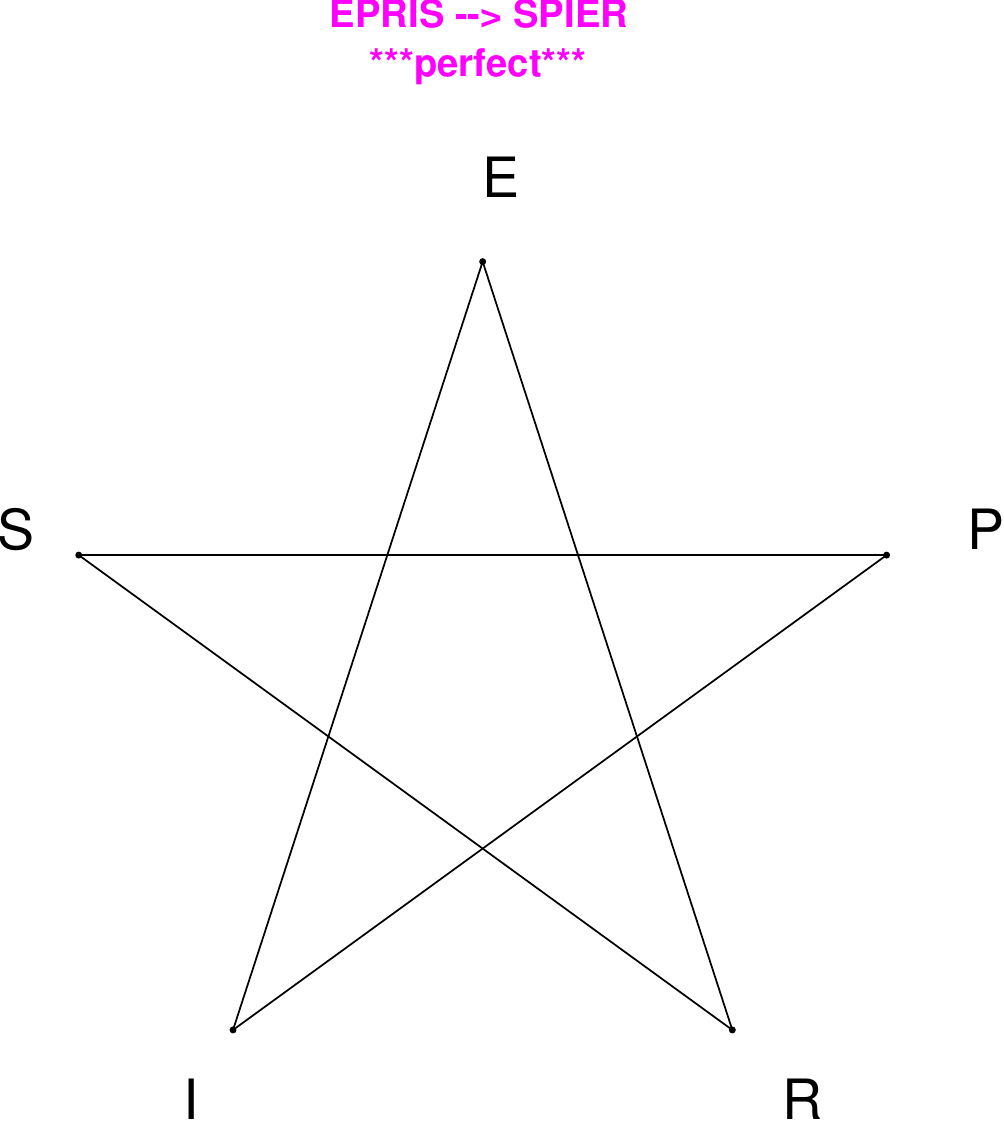}
\end{subfigure}
\end{figure}

\begin{figure}[H]
\centering
\begin{subfigure}[T]{0.19\textwidth}
\centering
\includegraphics[width=\textwidth]{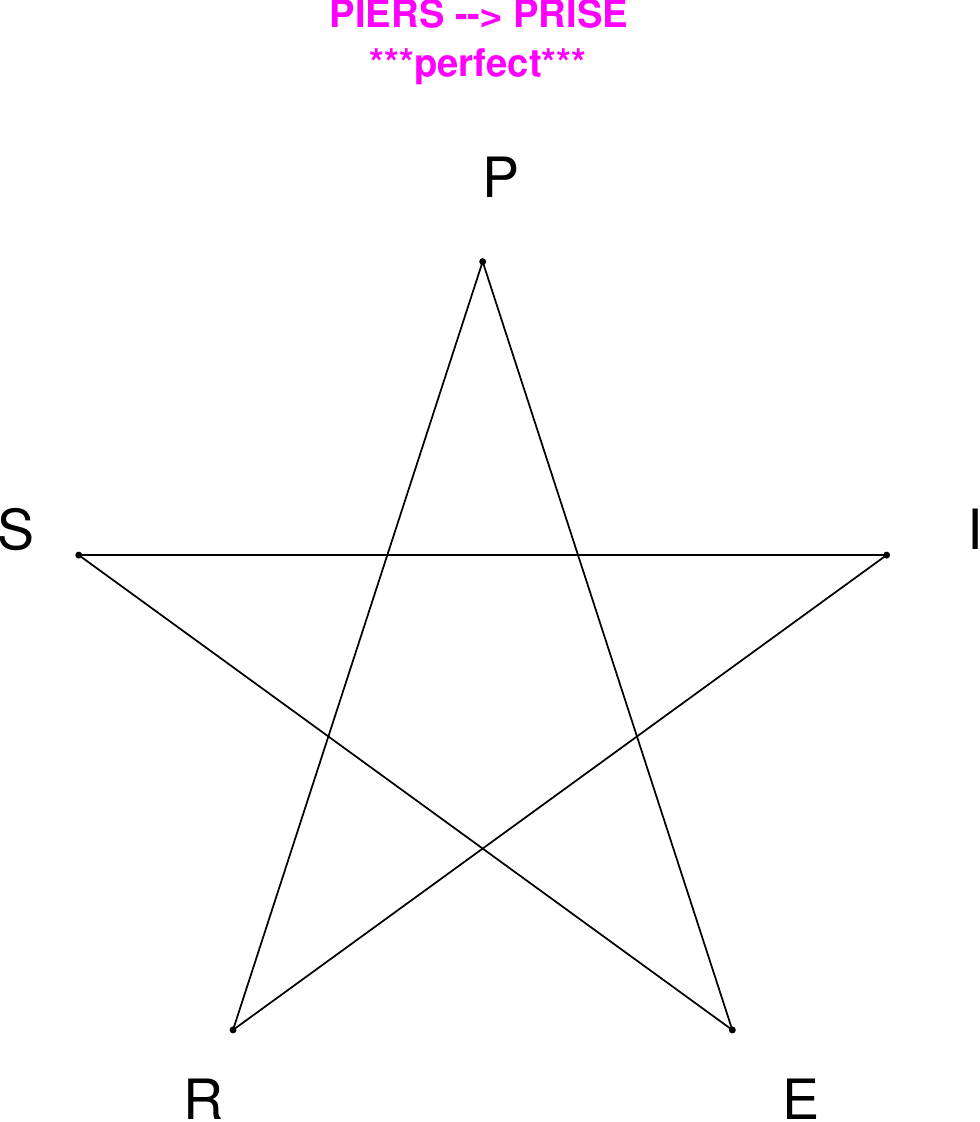}
\end{subfigure}
\hfill
\begin{subfigure}[T]{0.19\textwidth}
\centering
\includegraphics[width=\textwidth]{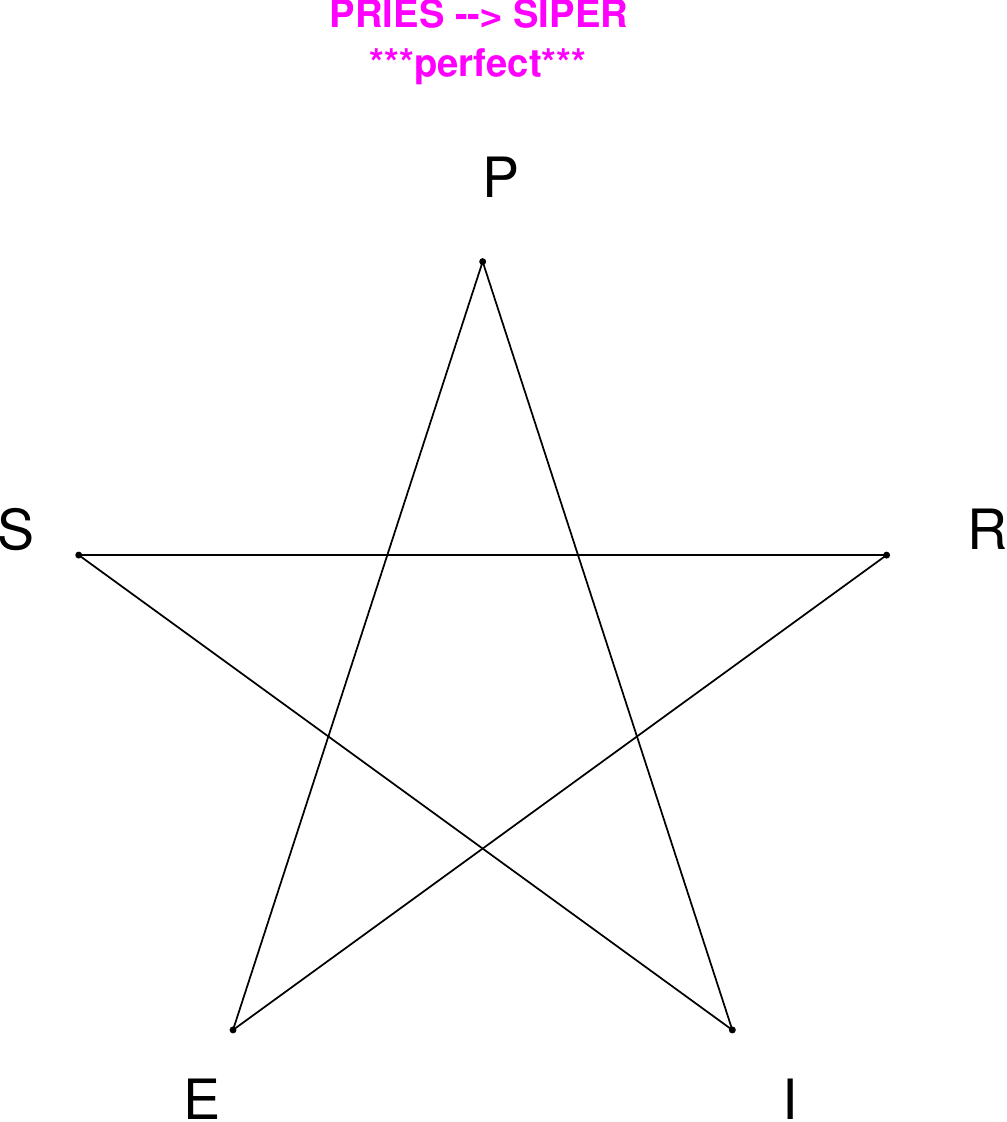}
\end{subfigure}
\hfill
\begin{subfigure}[T]{0.19\textwidth}
\centering
\includegraphics[width=\textwidth]{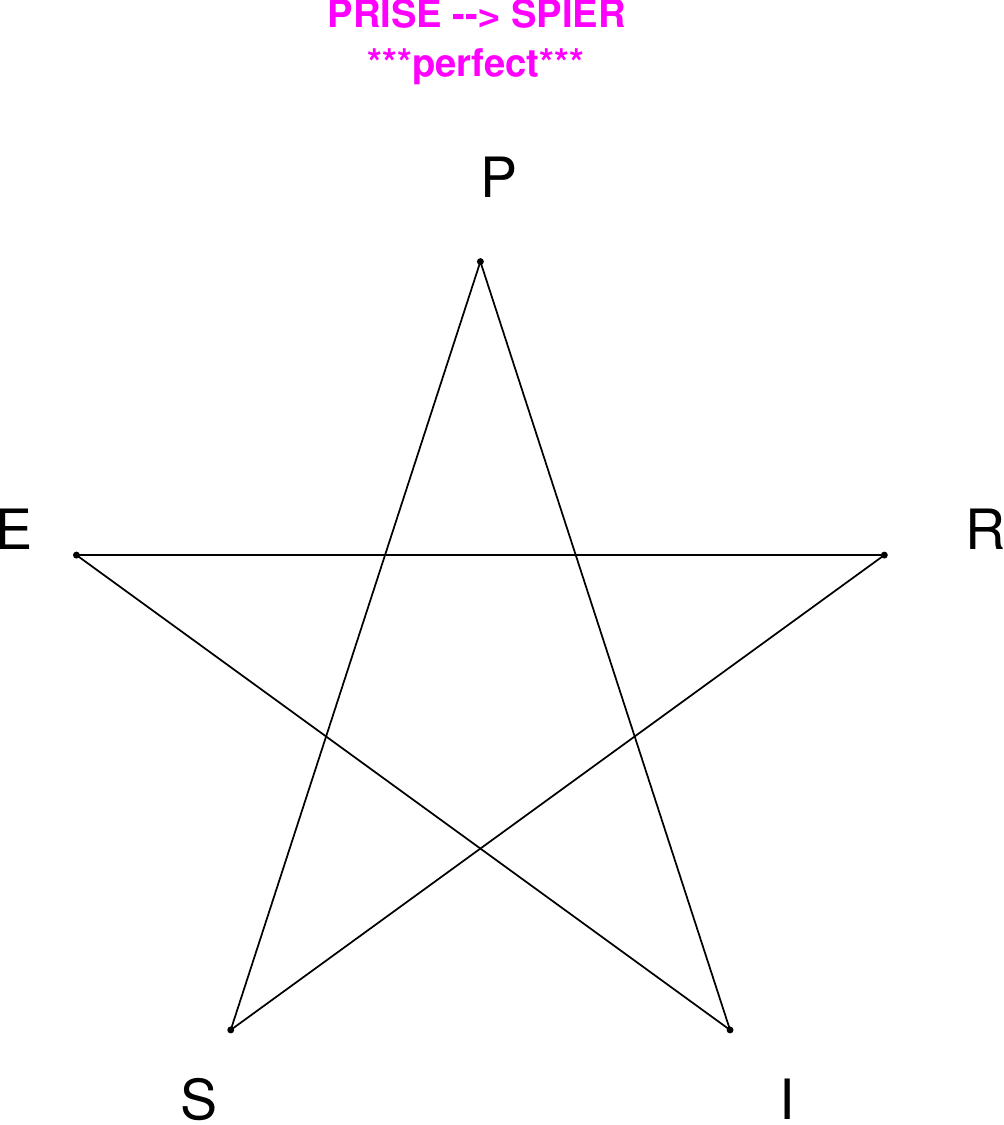}
\end{subfigure}
\hfill
\begin{subfigure}[T]{0.19\textwidth}
\centering
\includegraphics[width=\textwidth]{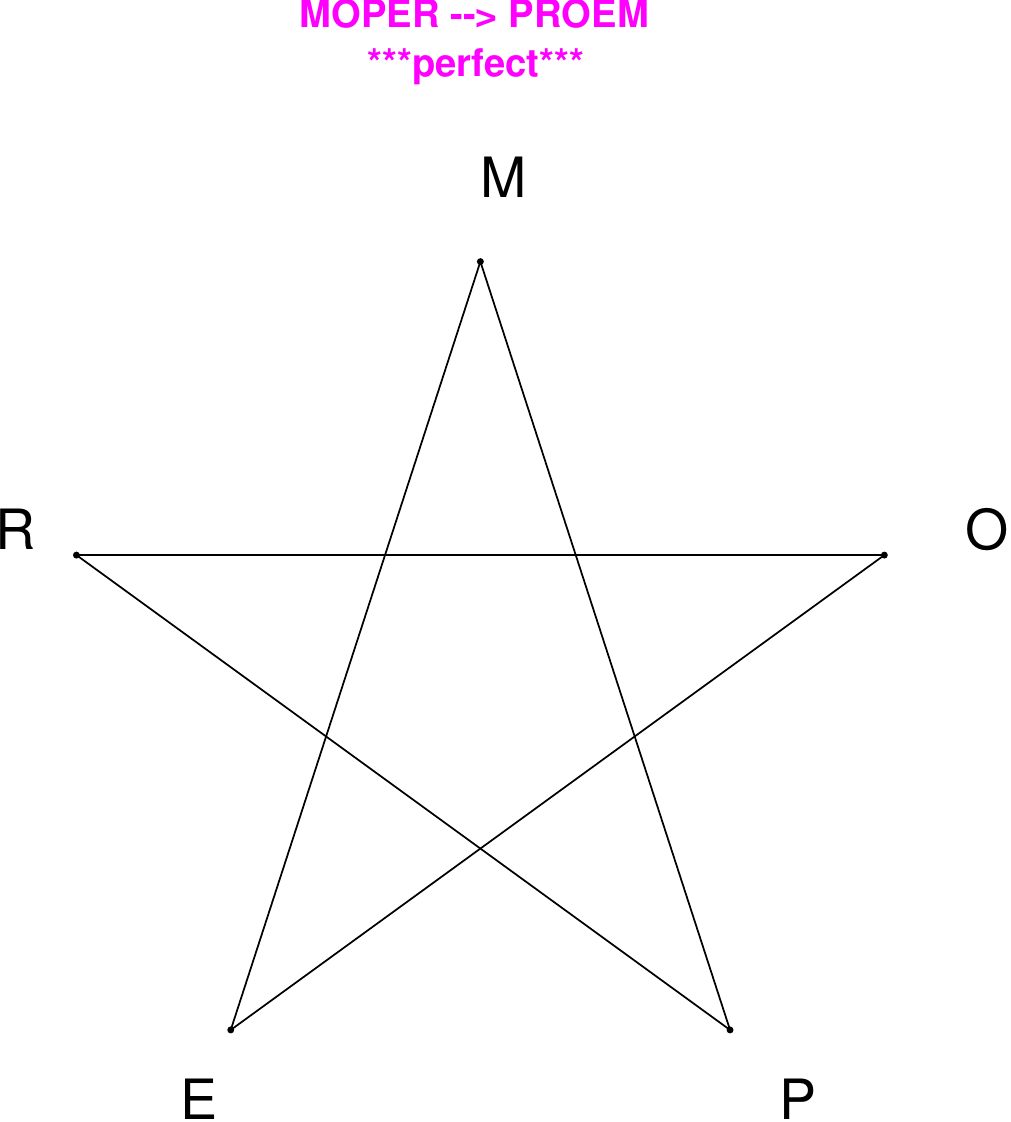}
\end{subfigure}
\hfill
\begin{subfigure}[T]{0.19\textwidth}
\centering
\includegraphics[width=\textwidth]{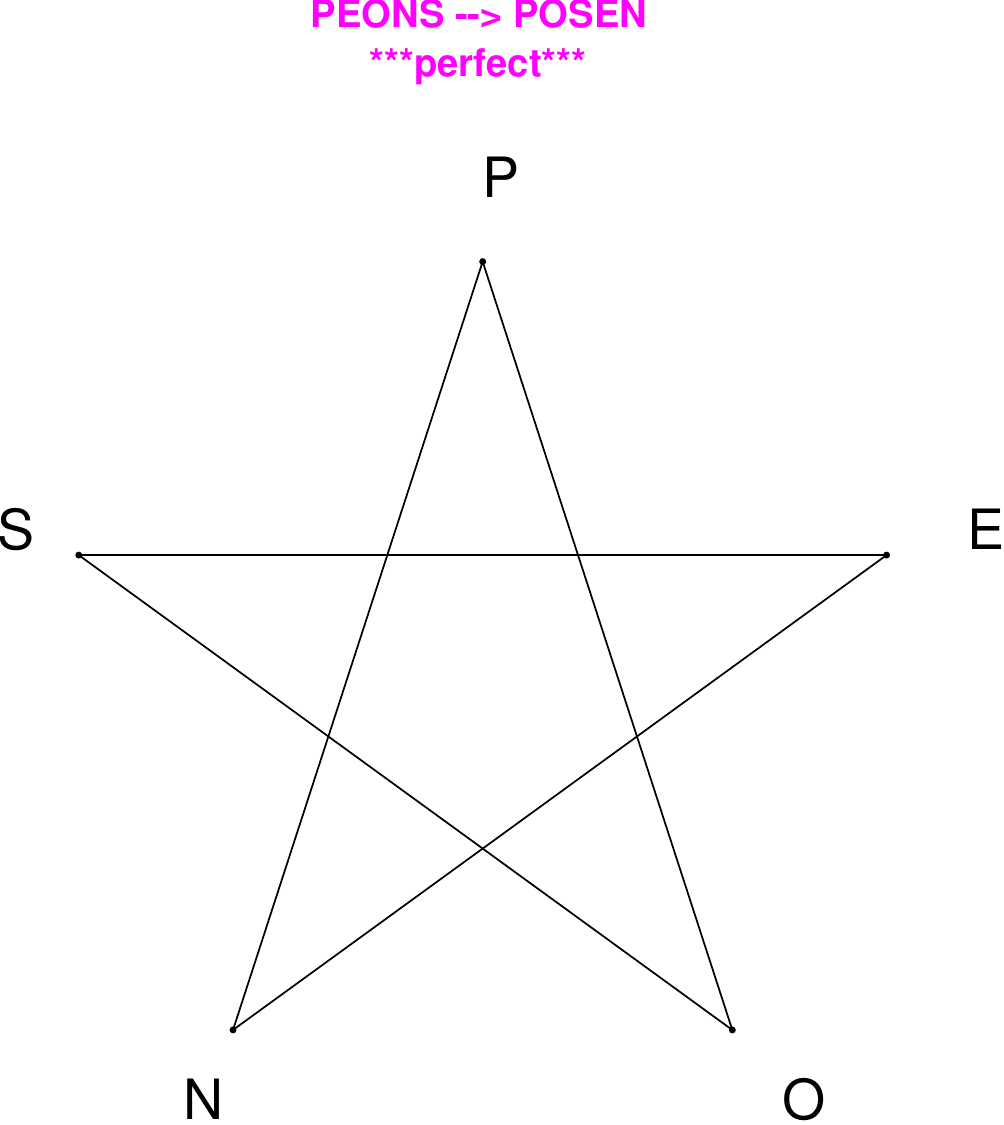}
\end{subfigure}
\end{figure}

\begin{figure}[H]
\centering
\begin{subfigure}[T]{0.19\textwidth}
\centering
\includegraphics[width=\textwidth]{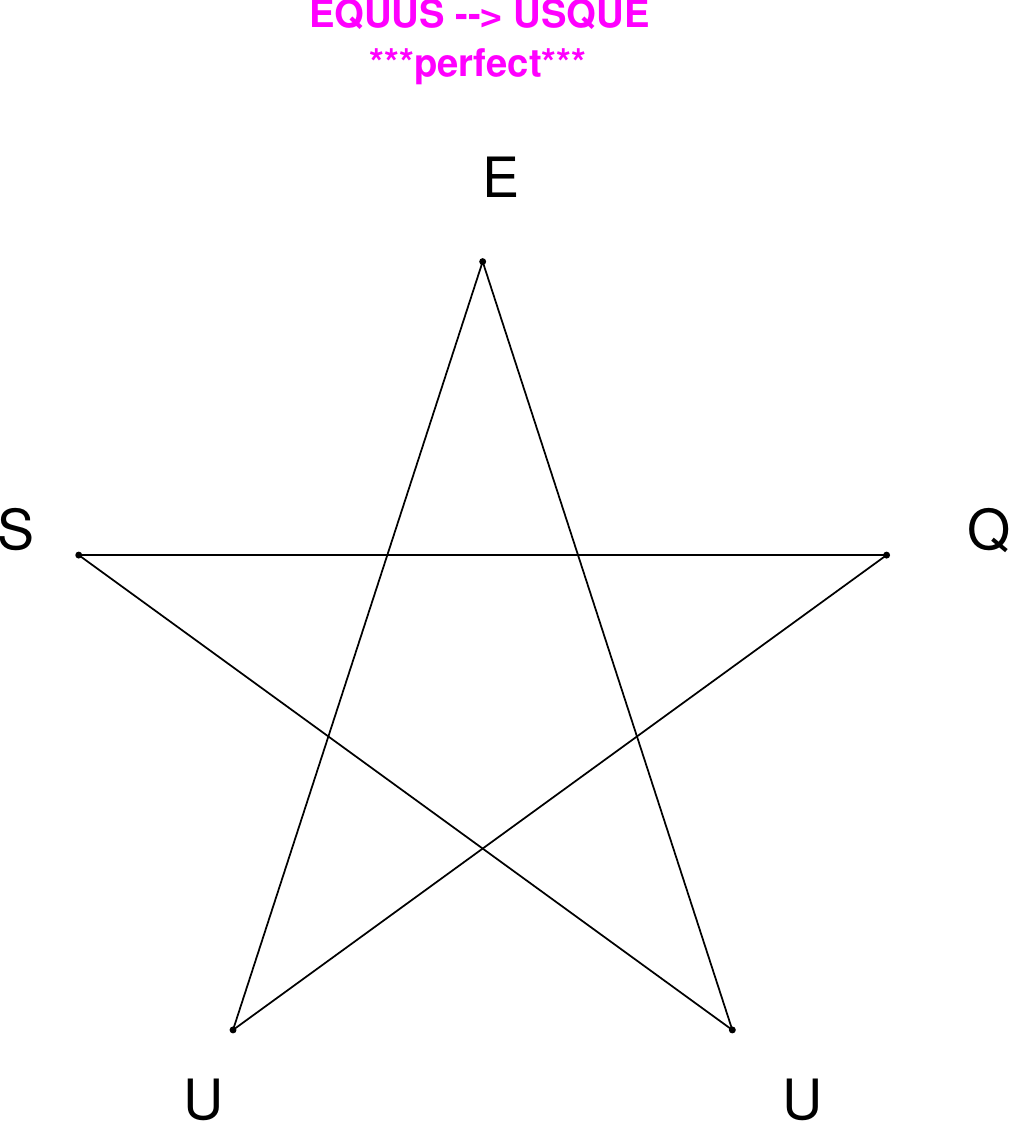}
\end{subfigure}
\hfill
\begin{subfigure}[T]{0.19\textwidth}
\centering
\includegraphics[width=\textwidth]{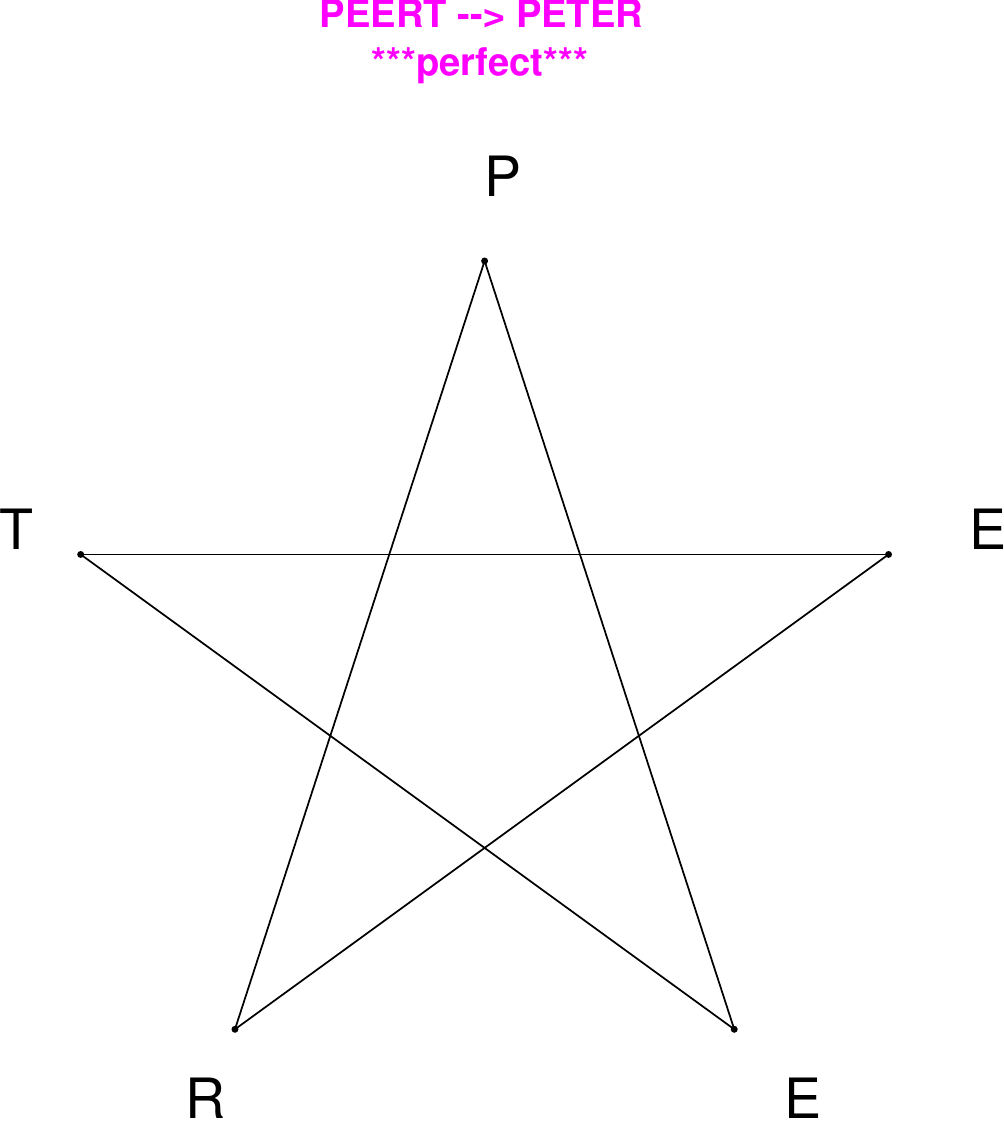}
\end{subfigure}
\hfill
\begin{subfigure}[T]{0.19\textwidth}
\centering
\includegraphics[width=\textwidth]{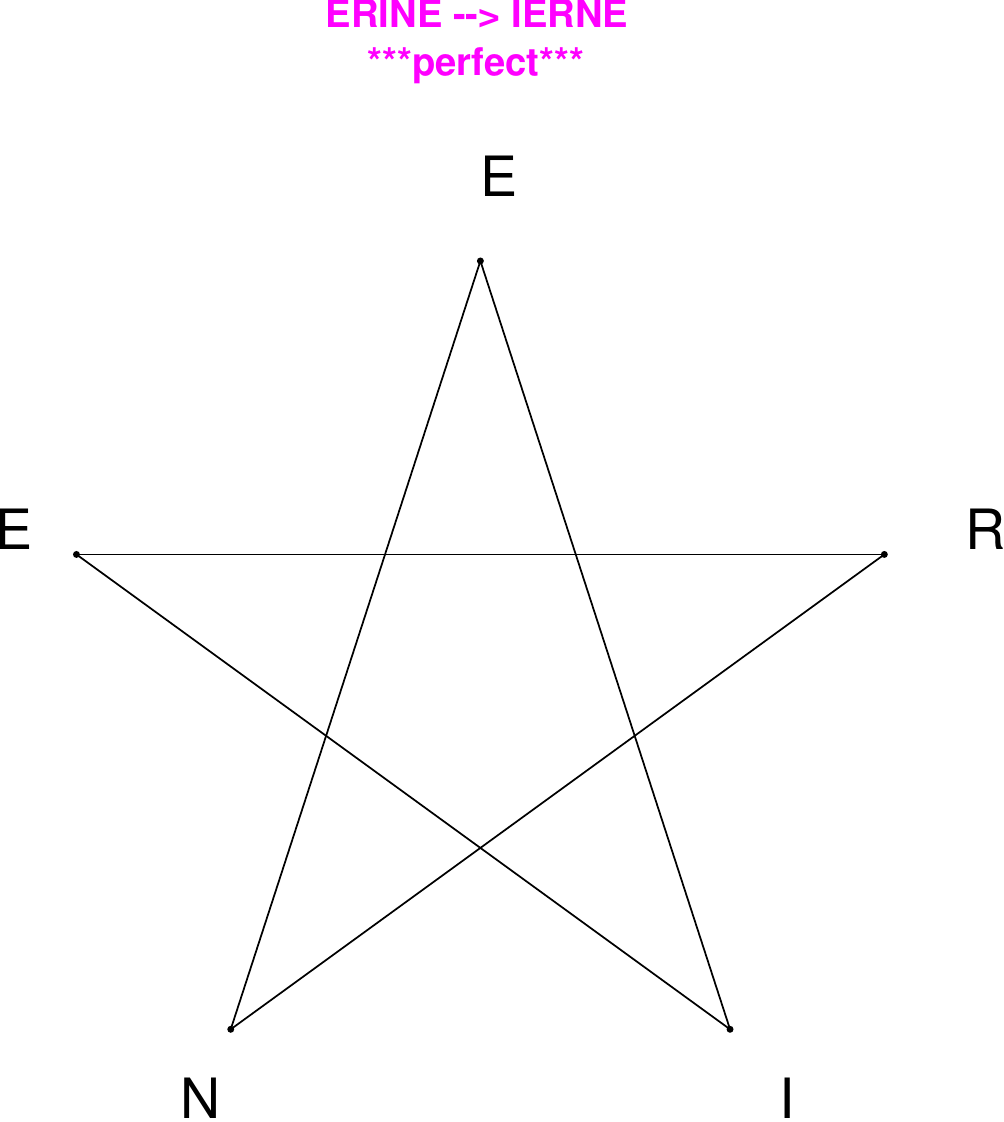}
\end{subfigure}
\hfill
\begin{subfigure}[T]{0.19\textwidth}
\centering
\includegraphics[width=\textwidth]{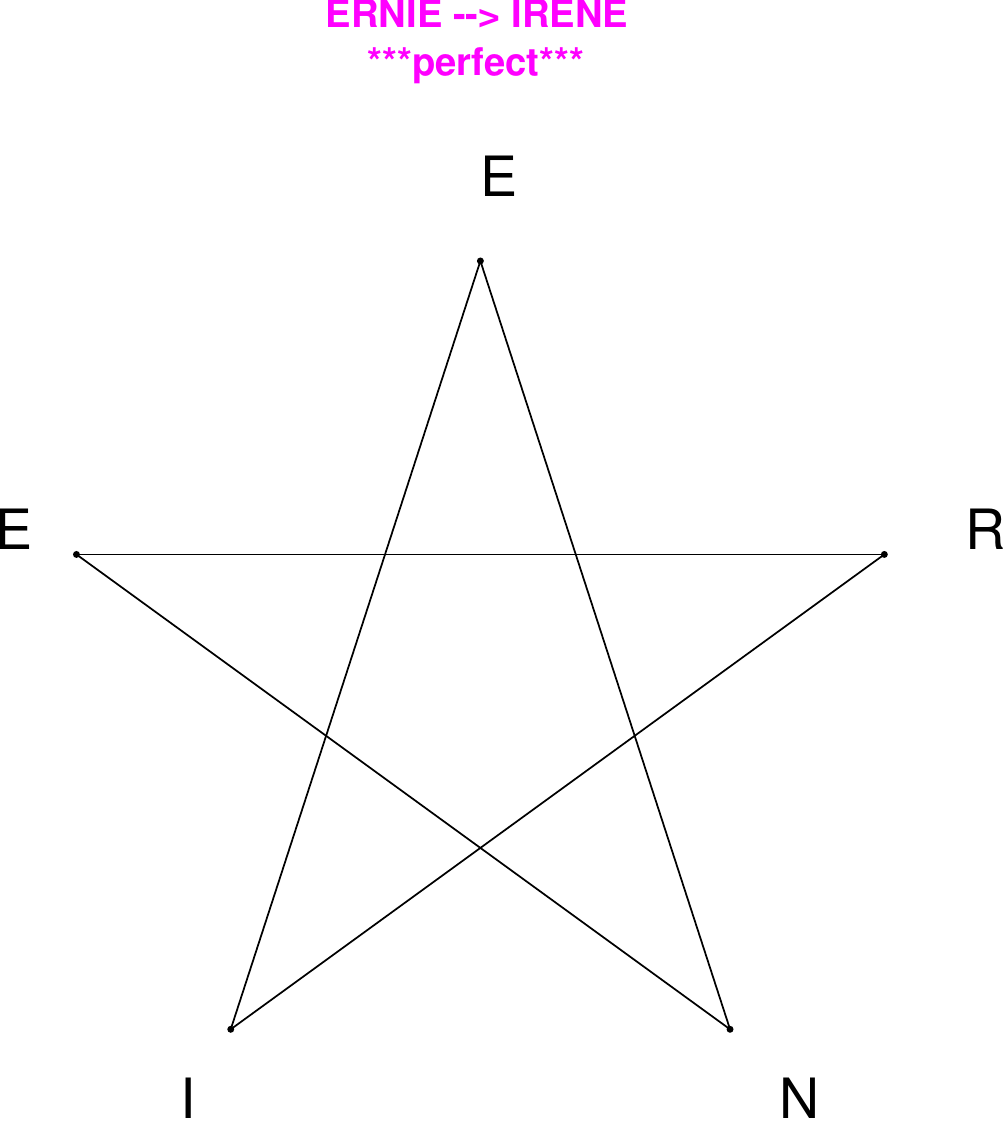}
\end{subfigure}
\hfill
\begin{subfigure}[T]{0.19\textwidth}
\centering
\includegraphics[width=\textwidth]{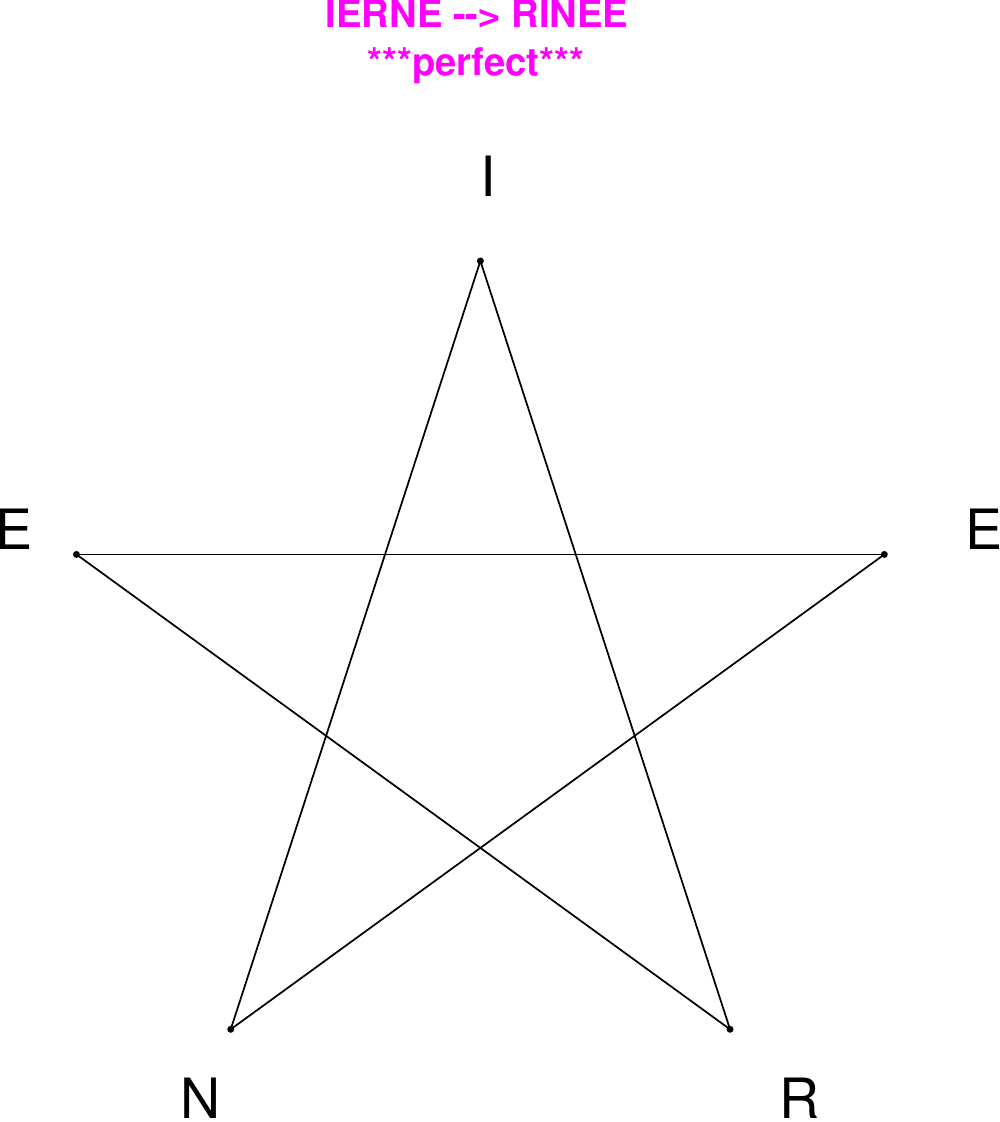}
\end{subfigure}
\end{figure}

\begin{figure}[H]
\centering
\begin{subfigure}[T]{0.19\textwidth}
\centering
\includegraphics[width=\textwidth]{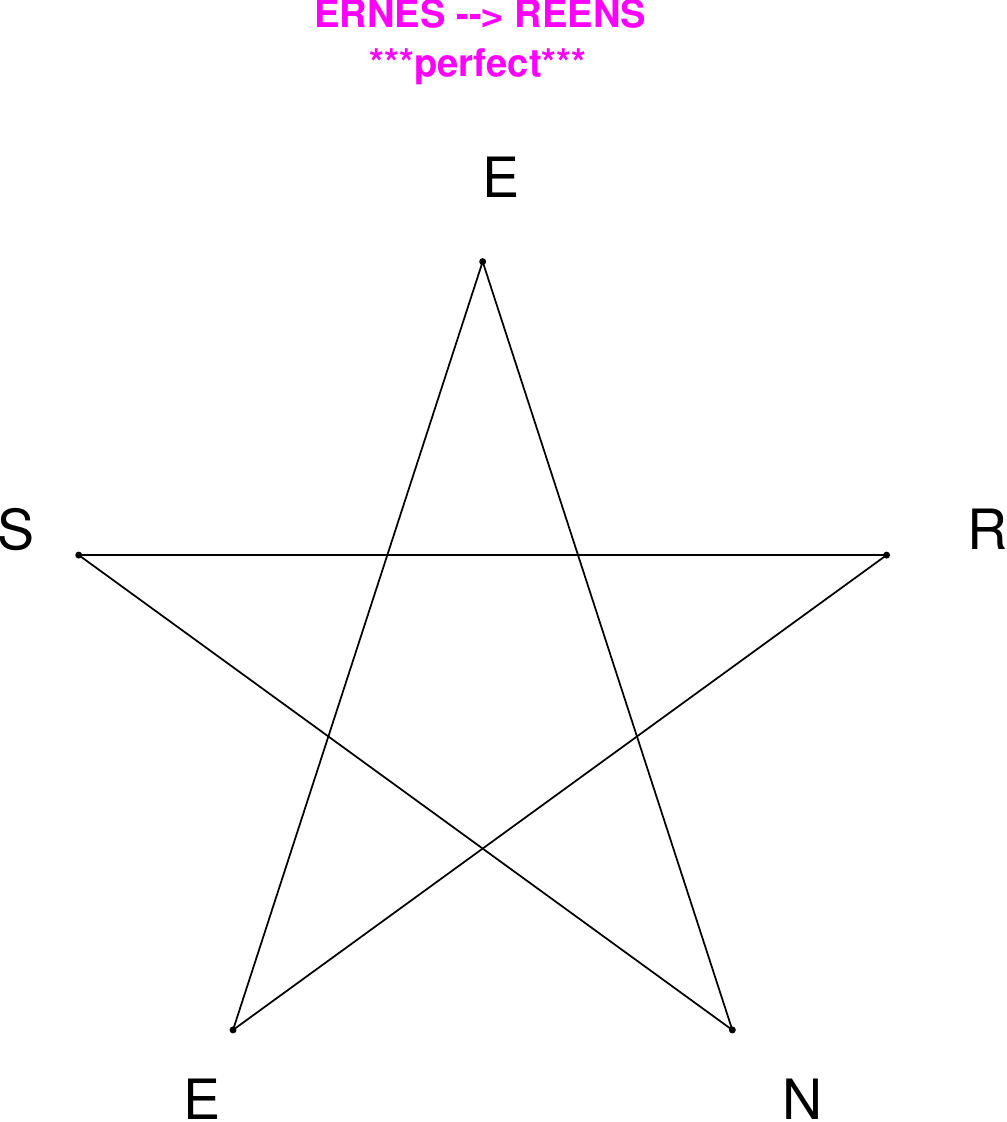}
\end{subfigure}
\hfill
\begin{subfigure}[T]{0.19\textwidth}
\centering
\includegraphics[width=\textwidth]{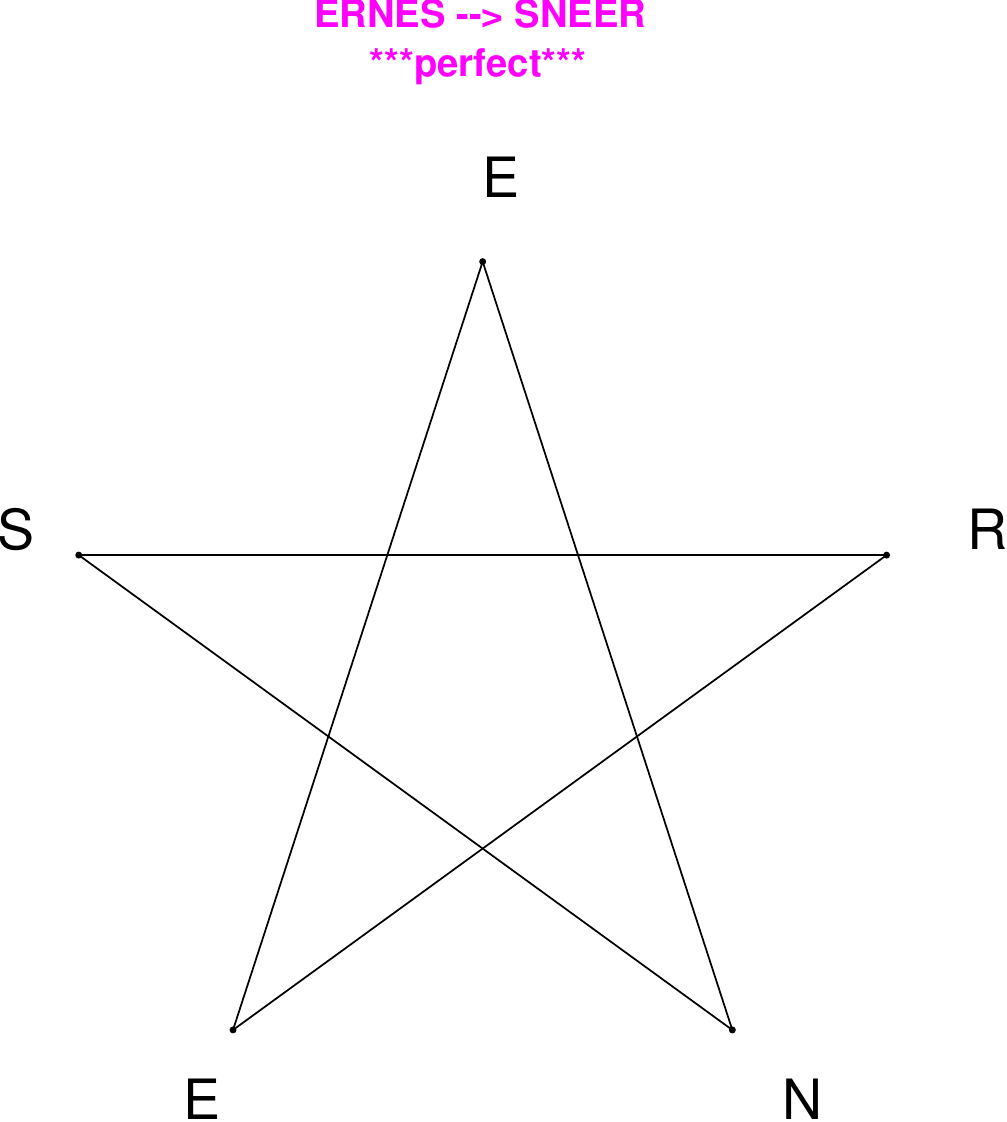}
\end{subfigure}
\hfill
\begin{subfigure}[T]{0.19\textwidth}
\centering
\includegraphics[width=\textwidth]{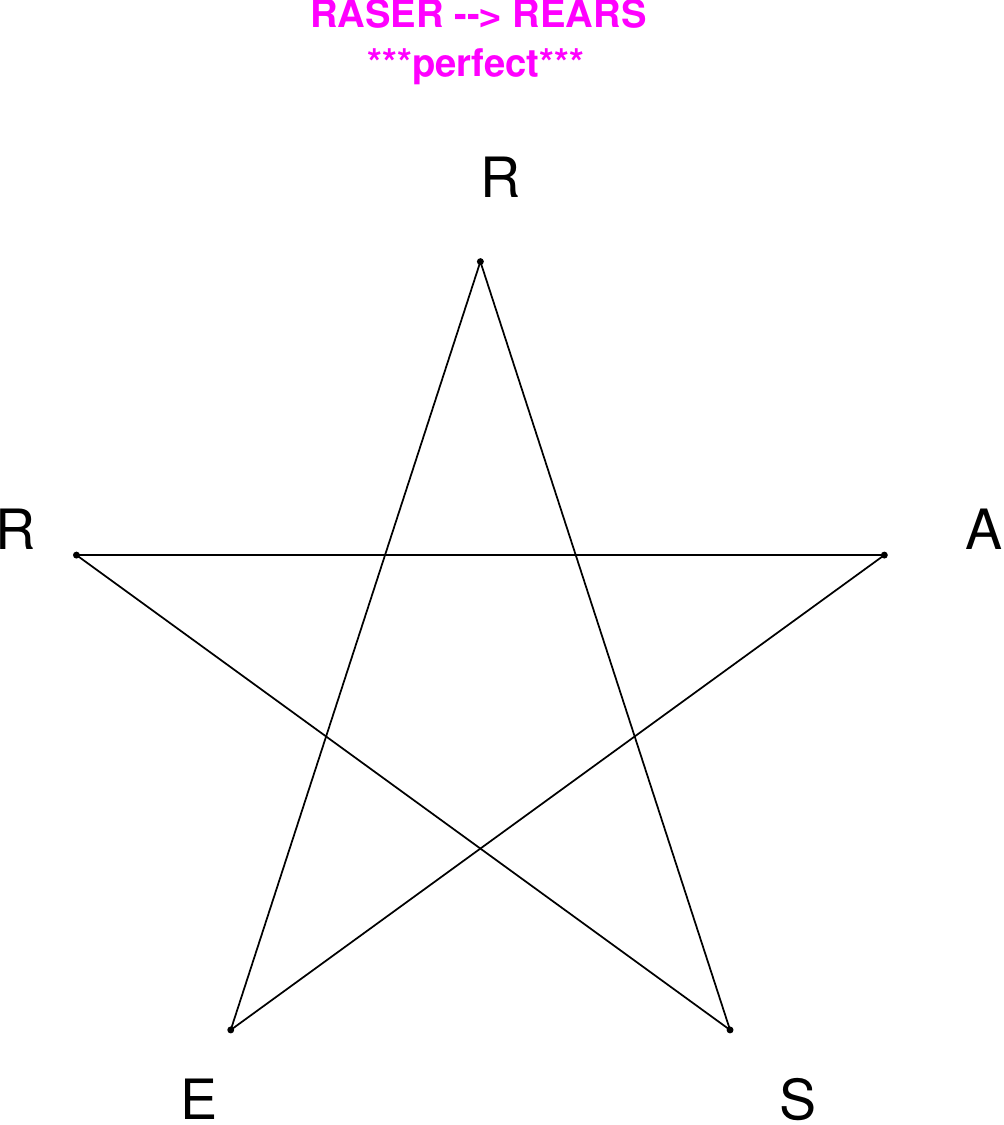}
\end{subfigure}
\hfill
\begin{subfigure}[T]{0.19\textwidth}
\centering
\includegraphics[width=\textwidth]{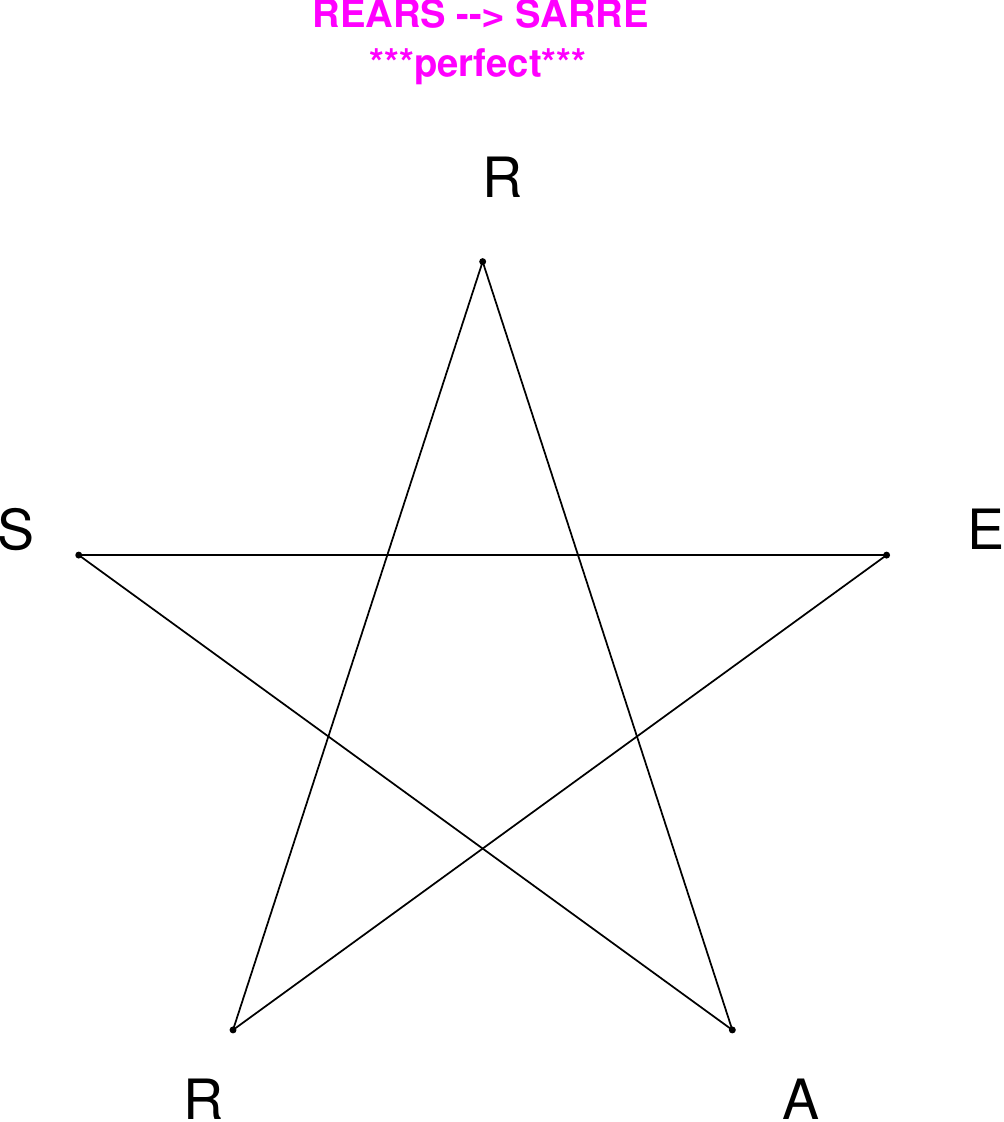}
\end{subfigure}
\hfill
\begin{subfigure}[T]{0.19\textwidth}
\centering
\includegraphics[width=\textwidth]{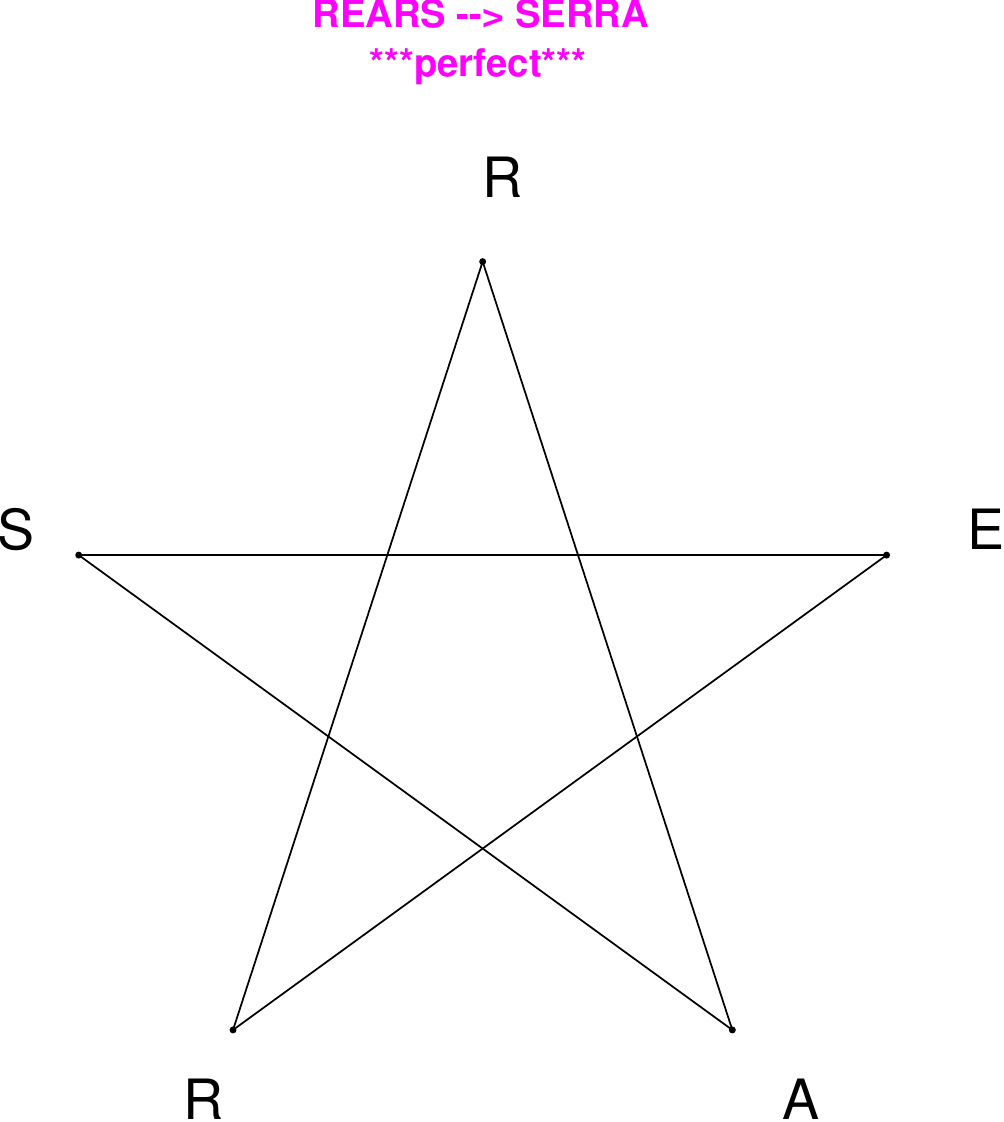}
\end{subfigure}
\end{figure}

\begin{figure}[H]
\centering
\begin{subfigure}[T]{0.19\textwidth}
\centering
\includegraphics[width=\textwidth]{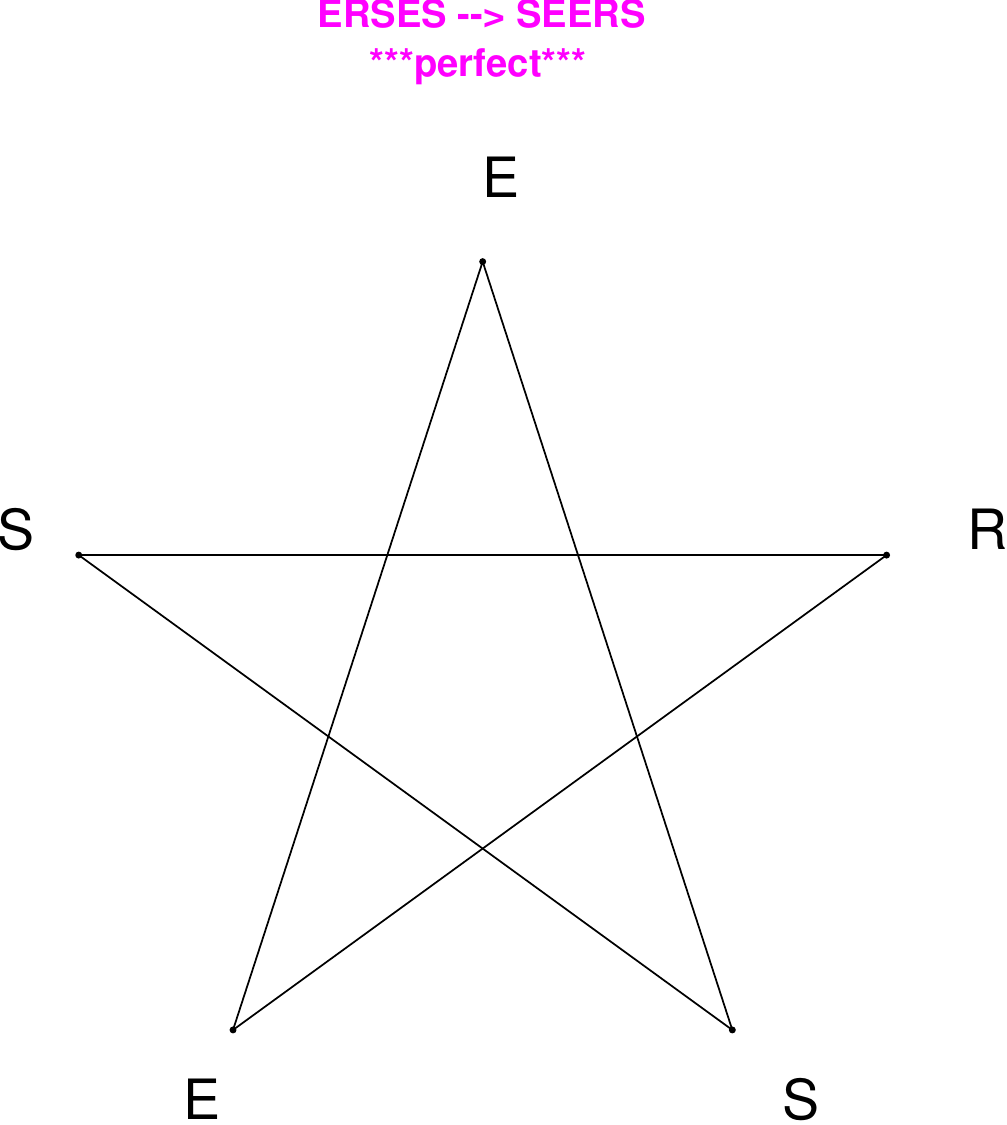}
\end{subfigure}
\hfill
\begin{subfigure}[T]{0.19\textwidth}
\centering
\includegraphics[width=\textwidth]{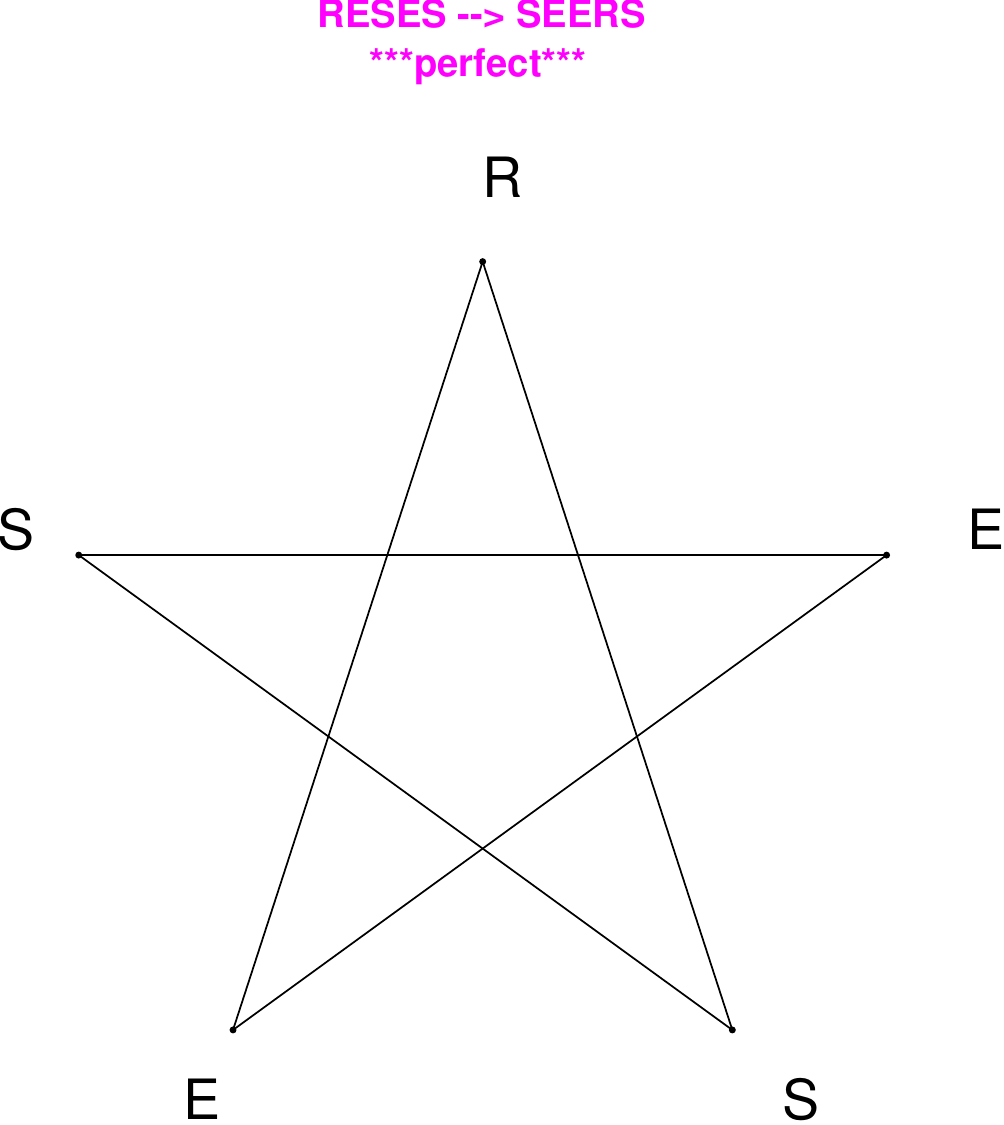}
\end{subfigure}
\hfill
\begin{subfigure}[T]{0.19\textwidth}
\centering
\includegraphics[width=\textwidth]{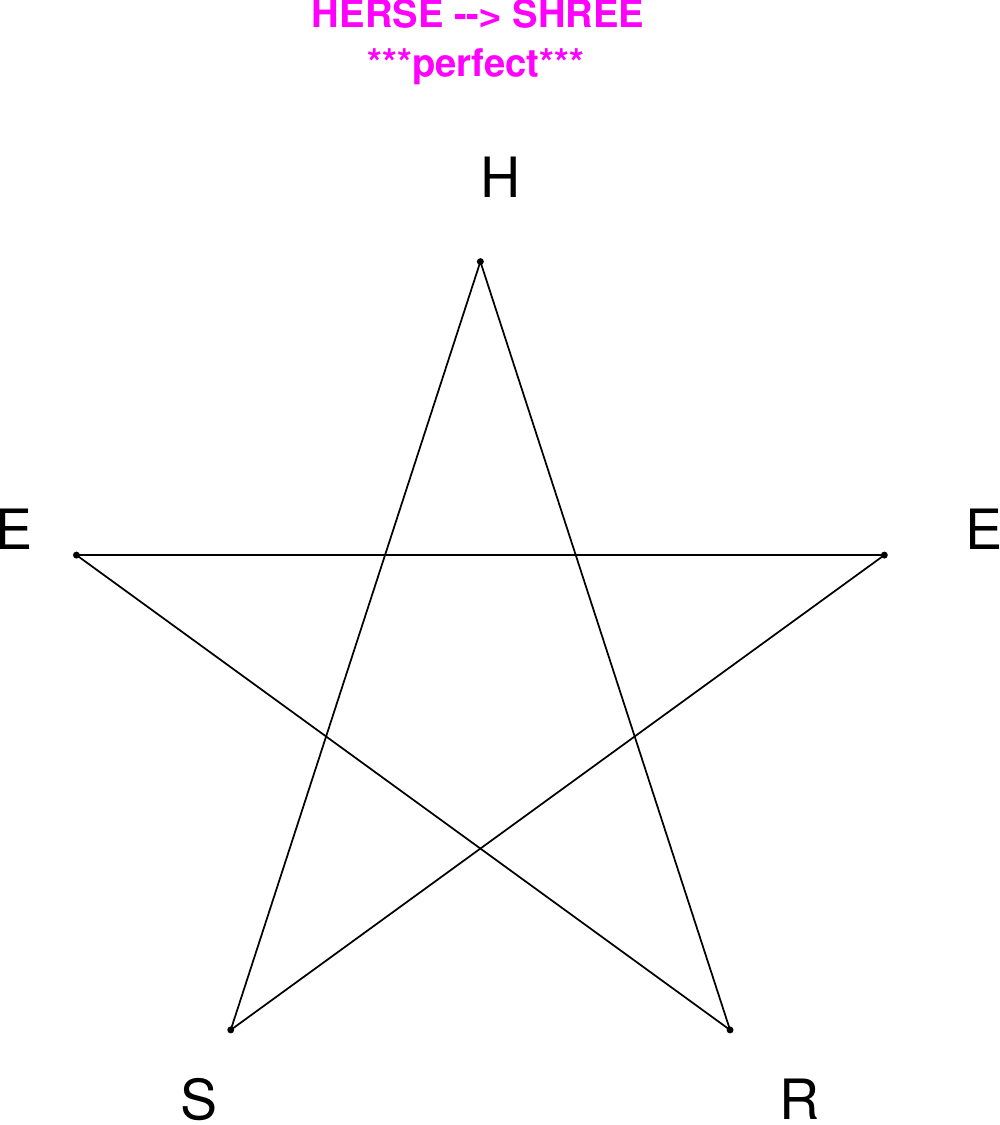}
\end{subfigure}
\hfill
\begin{subfigure}[T]{0.19\textwidth}
\centering
\includegraphics[width=\textwidth]{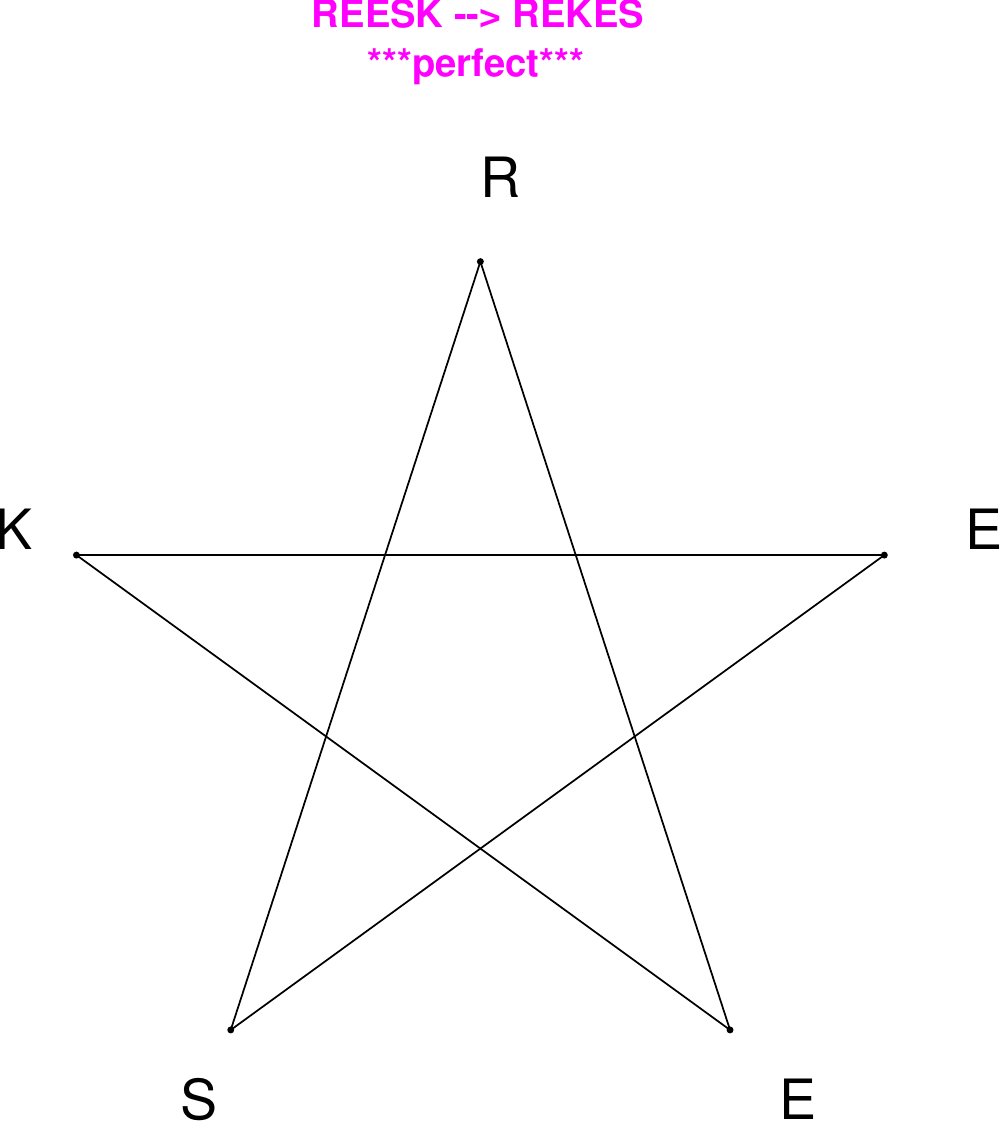}
\end{subfigure}
\hfill
\begin{subfigure}[T]{0.19\textwidth}
\centering
\includegraphics[width=\textwidth]{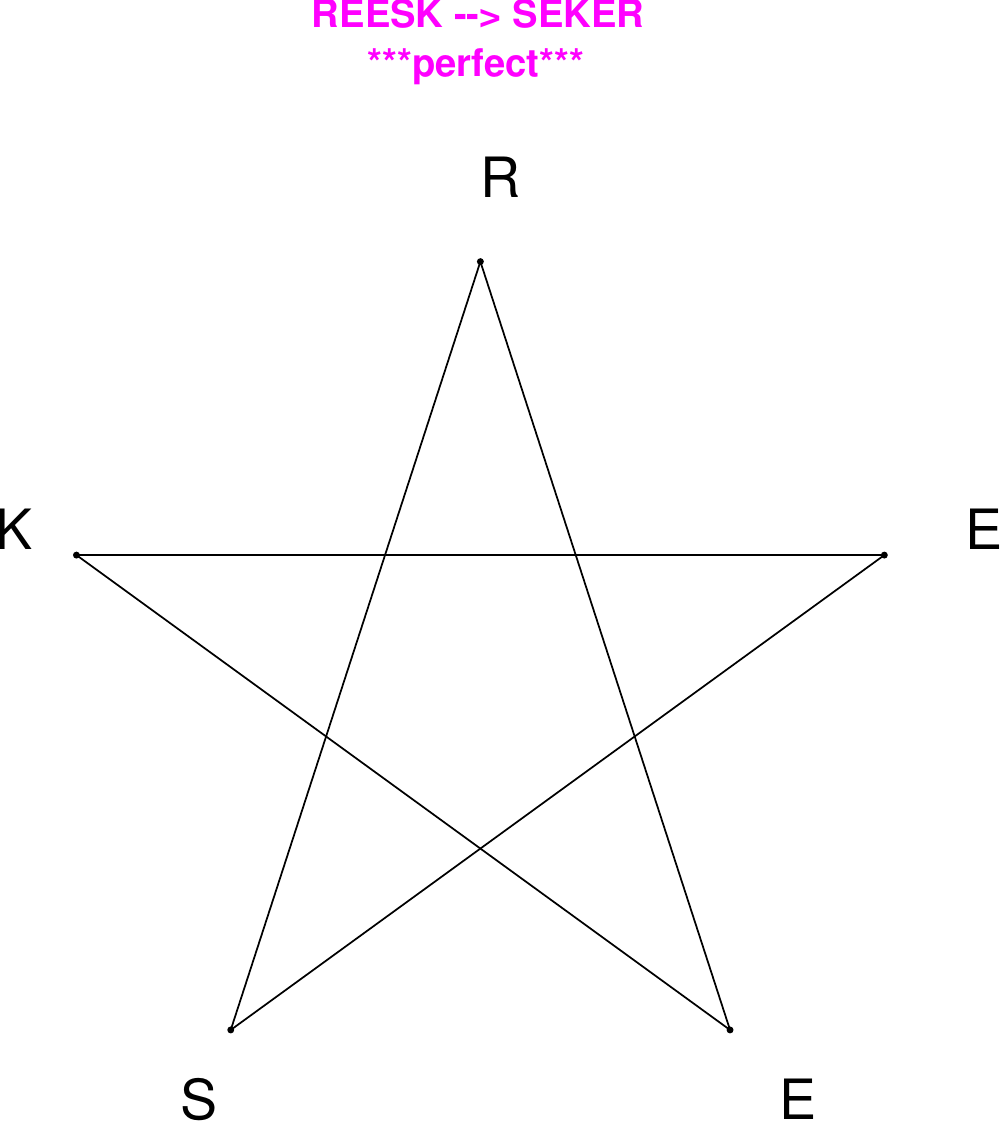}
\end{subfigure}
\end{figure}

\begin{figure}[H]
\centering
\begin{subfigure}[T]{0.19\textwidth}
\centering
\includegraphics[width=\textwidth]{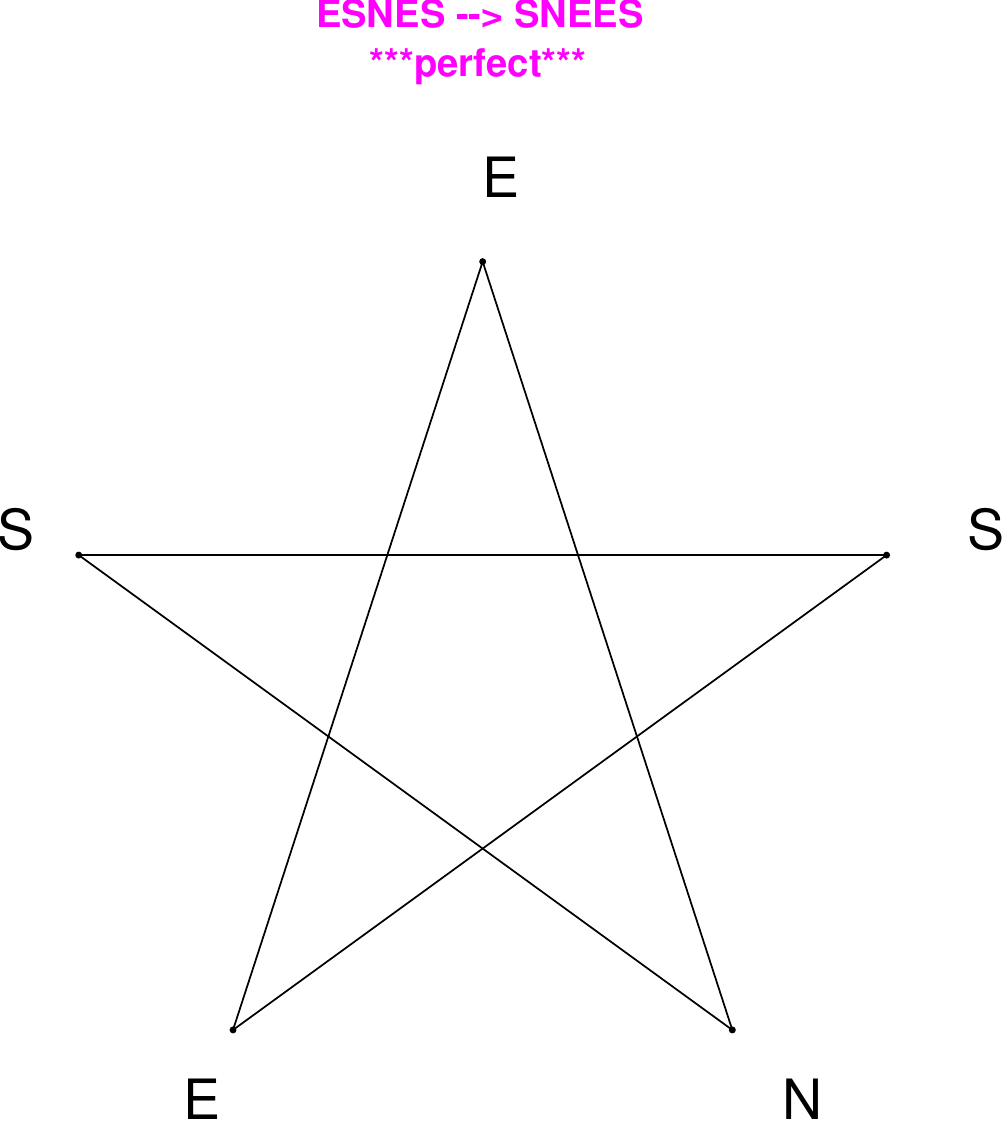}
\end{subfigure}
\hfill
\begin{subfigure}[T]{0.19\textwidth}
\centering
\includegraphics[width=\textwidth]{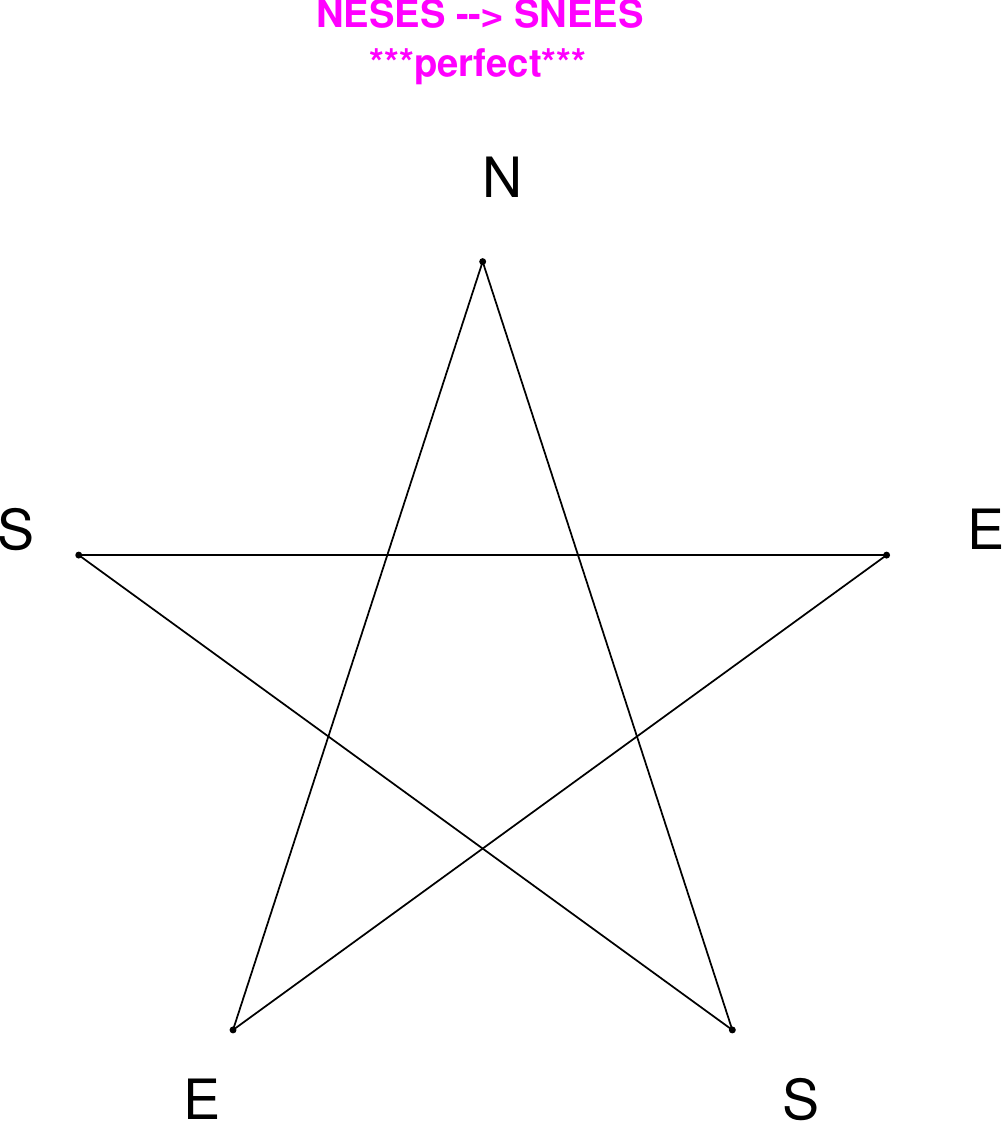}
\end{subfigure}
\hfill
\begin{subfigure}[T]{0.19\textwidth}
\centering
\includegraphics[width=\textwidth]{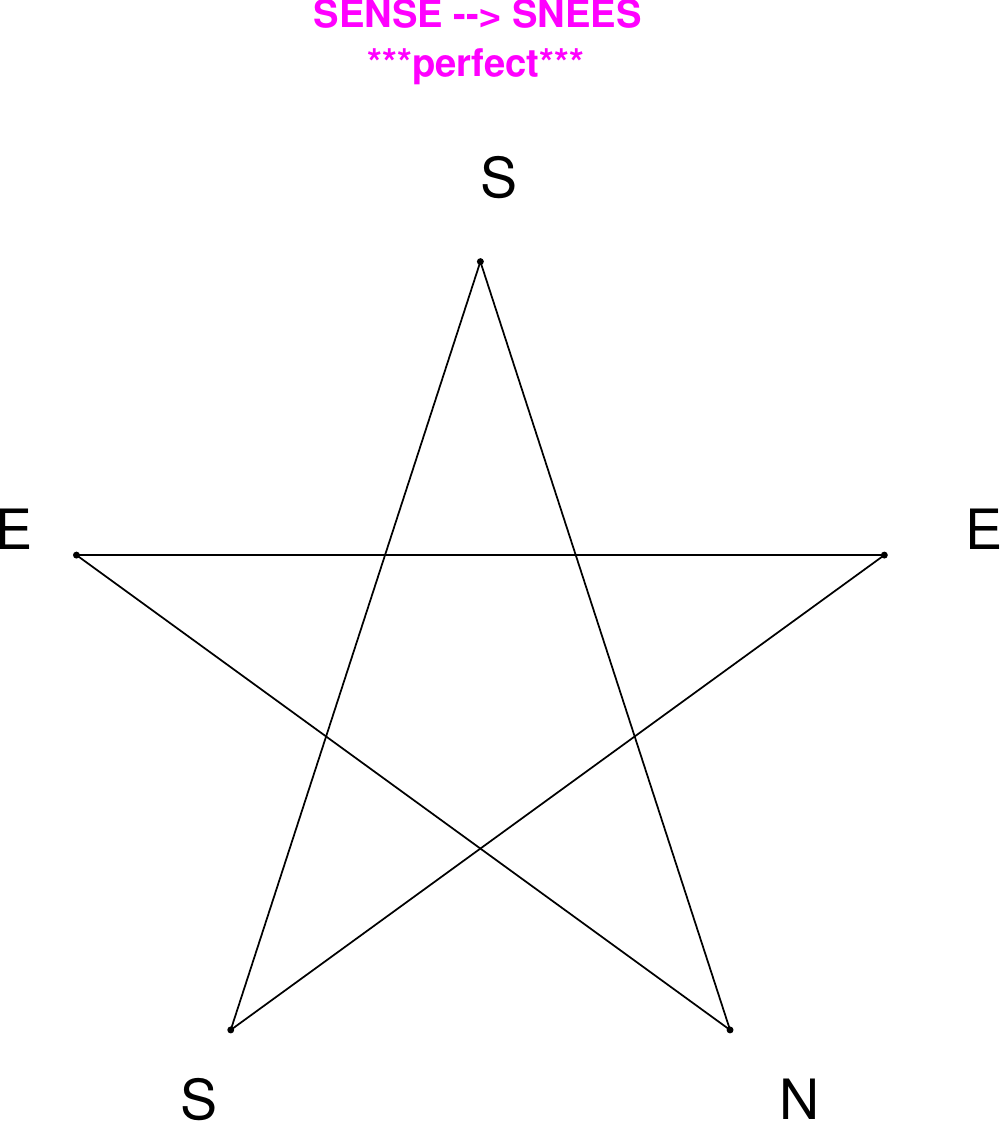}
\end{subfigure}
\hfill
\begin{subfigure}[T]{0.19\textwidth}
\centering
\includegraphics[width=\textwidth]{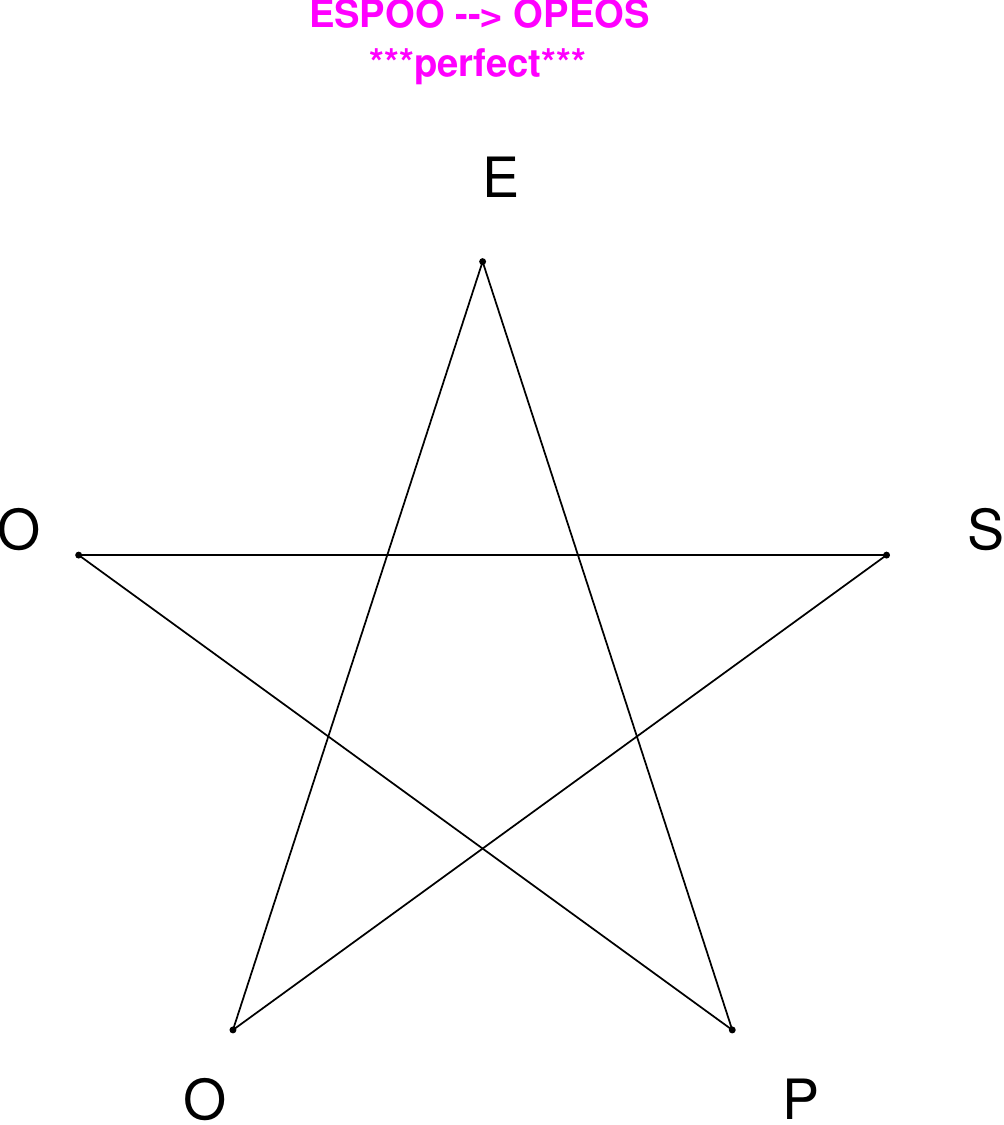}
\end{subfigure}
\hfill
\begin{subfigure}[T]{0.19\textwidth}
\centering
\includegraphics[width=\textwidth]{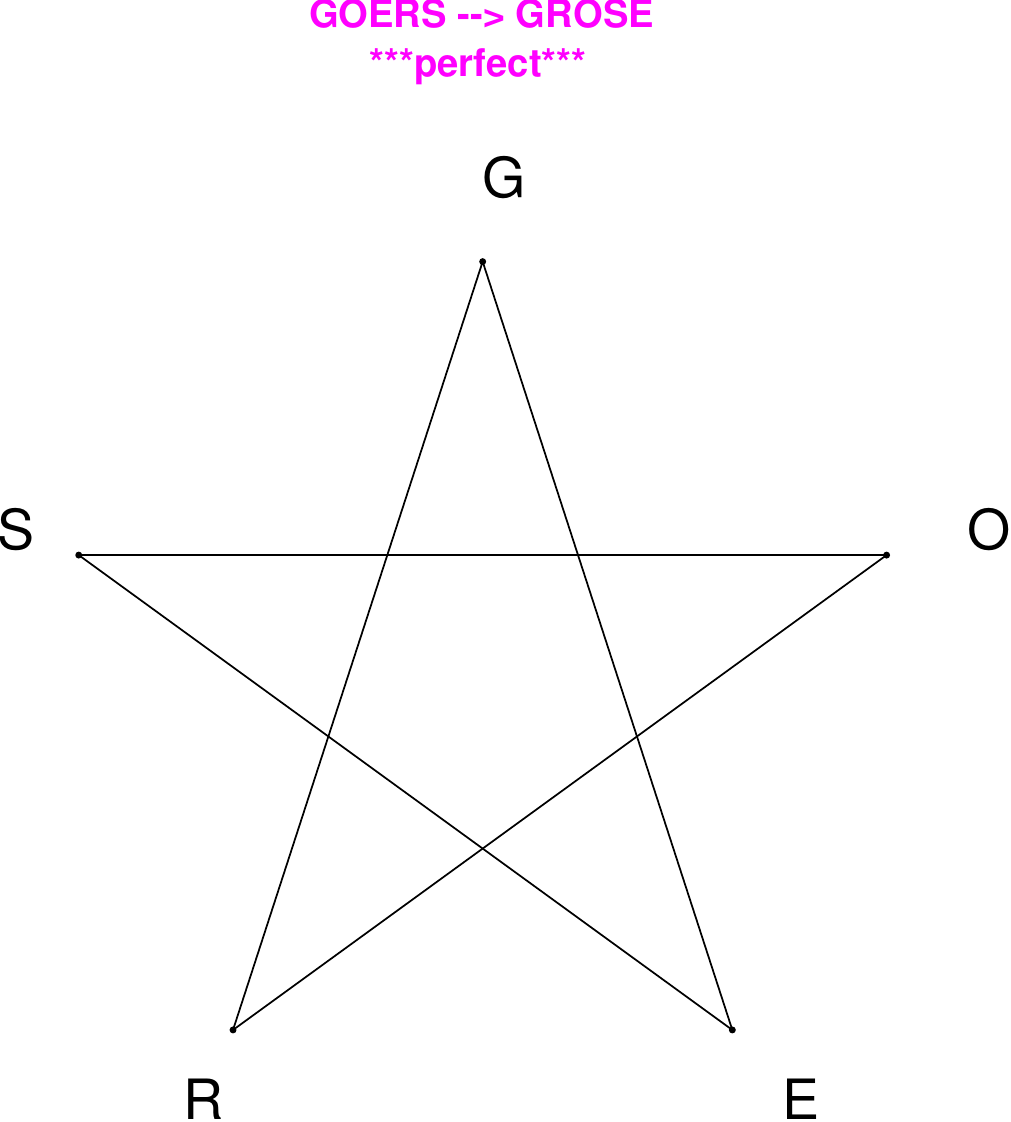}
\end{subfigure}
\end{figure}

\begin{figure}[H]
\centering
\begin{subfigure}[T]{0.19\textwidth}
\centering
\includegraphics[width=\textwidth]{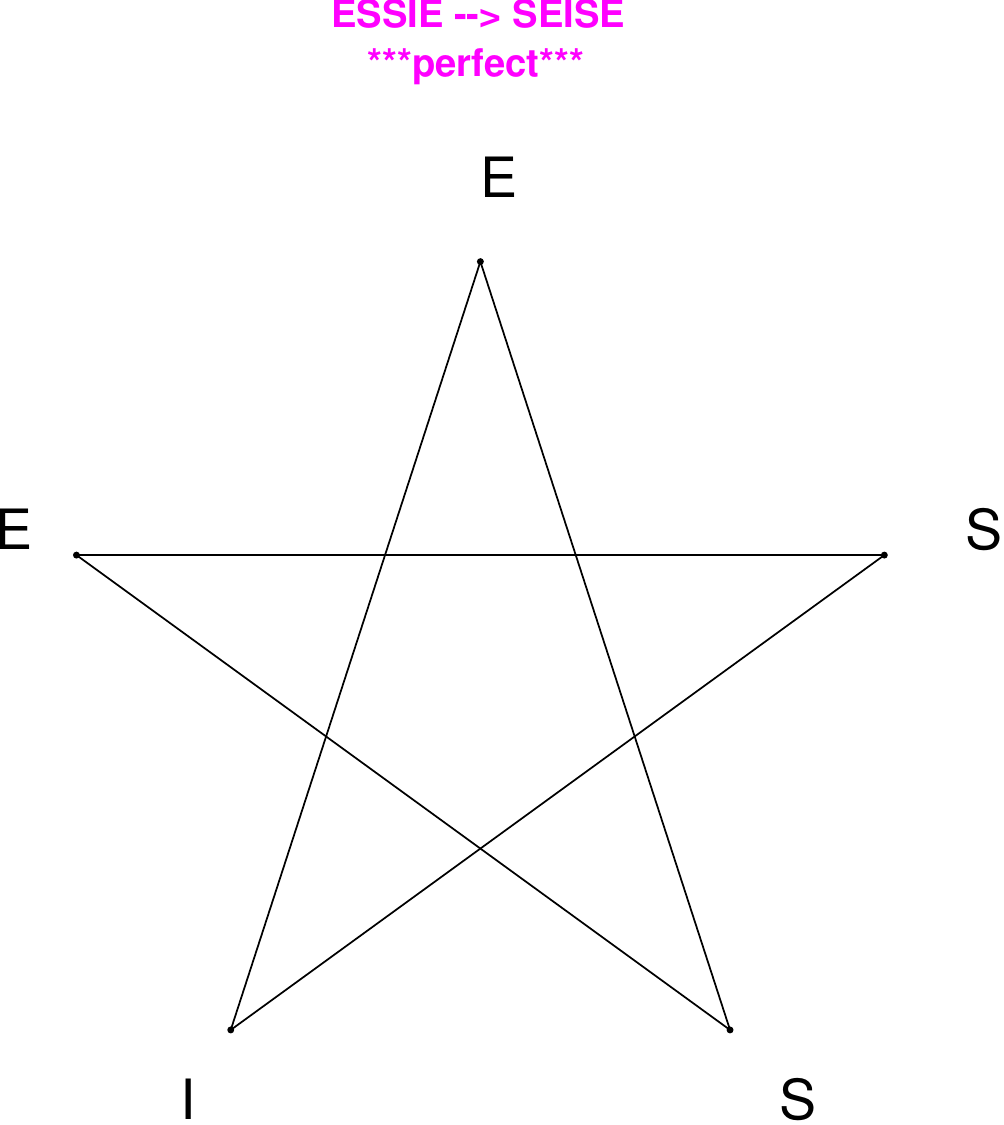}
\end{subfigure}
\hfill
\begin{subfigure}[T]{0.19\textwidth}
\centering
\includegraphics[width=\textwidth]{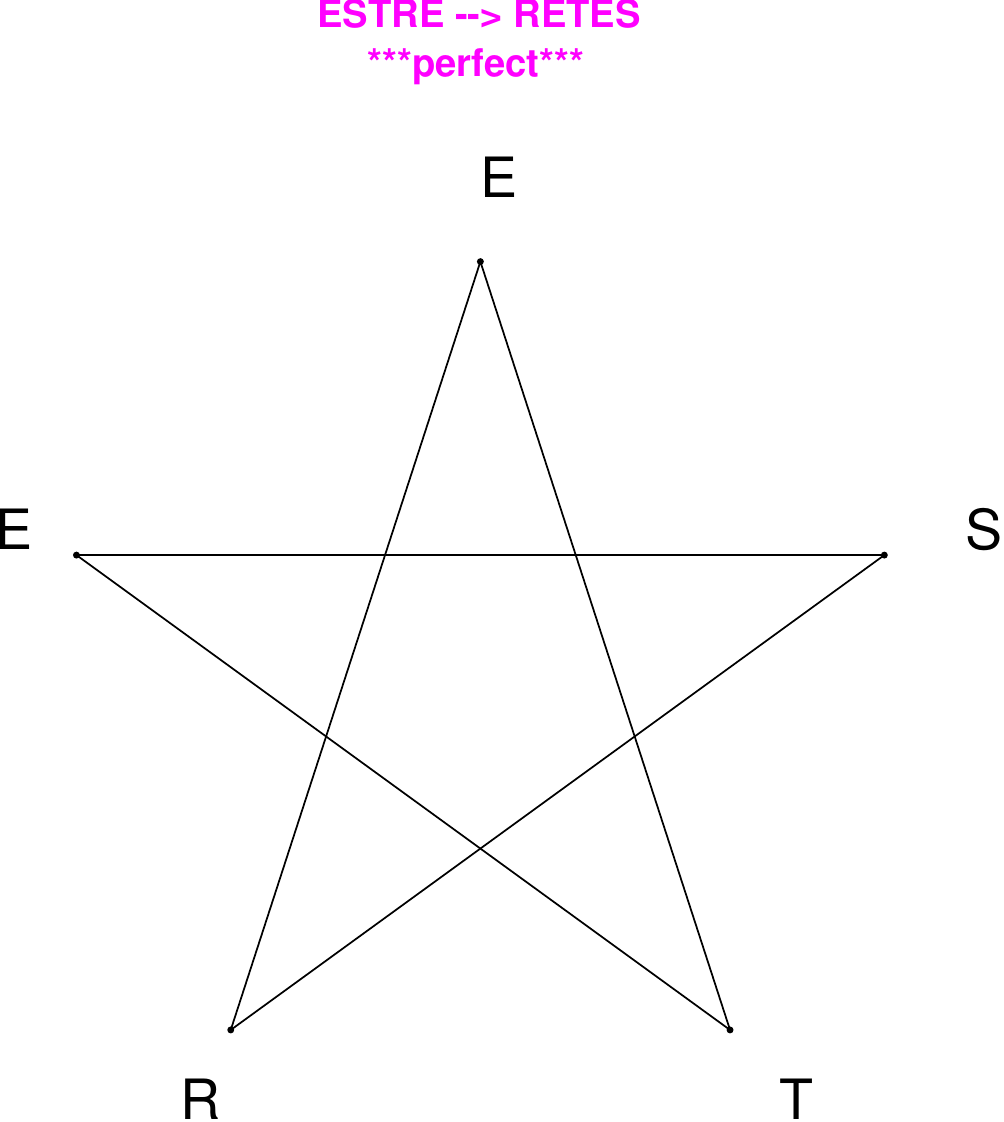}
\end{subfigure}
\hfill
\begin{subfigure}[T]{0.19\textwidth}
\centering
\includegraphics[width=\textwidth]{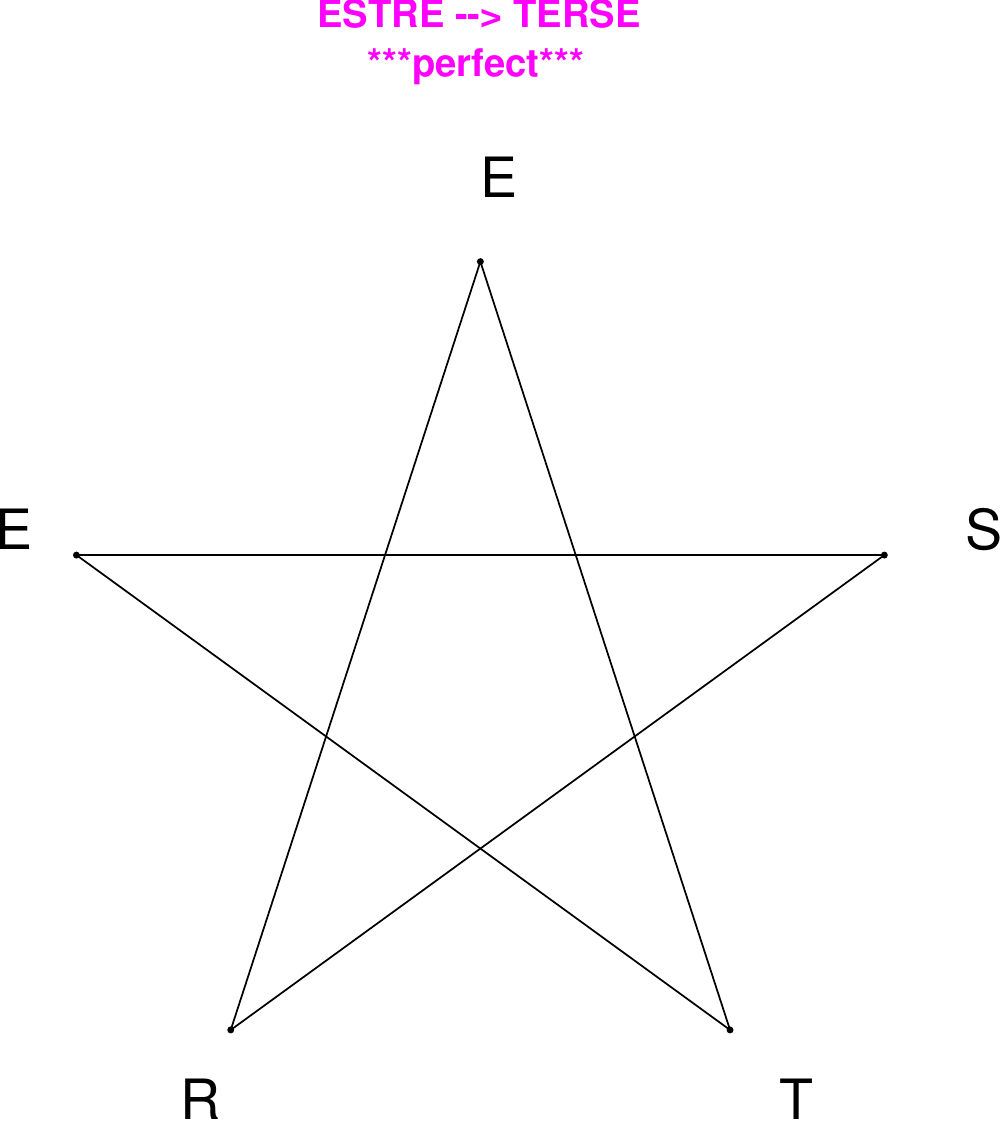}
\end{subfigure}
\hfill
\begin{subfigure}[T]{0.19\textwidth}
\centering
\includegraphics[width=\textwidth]{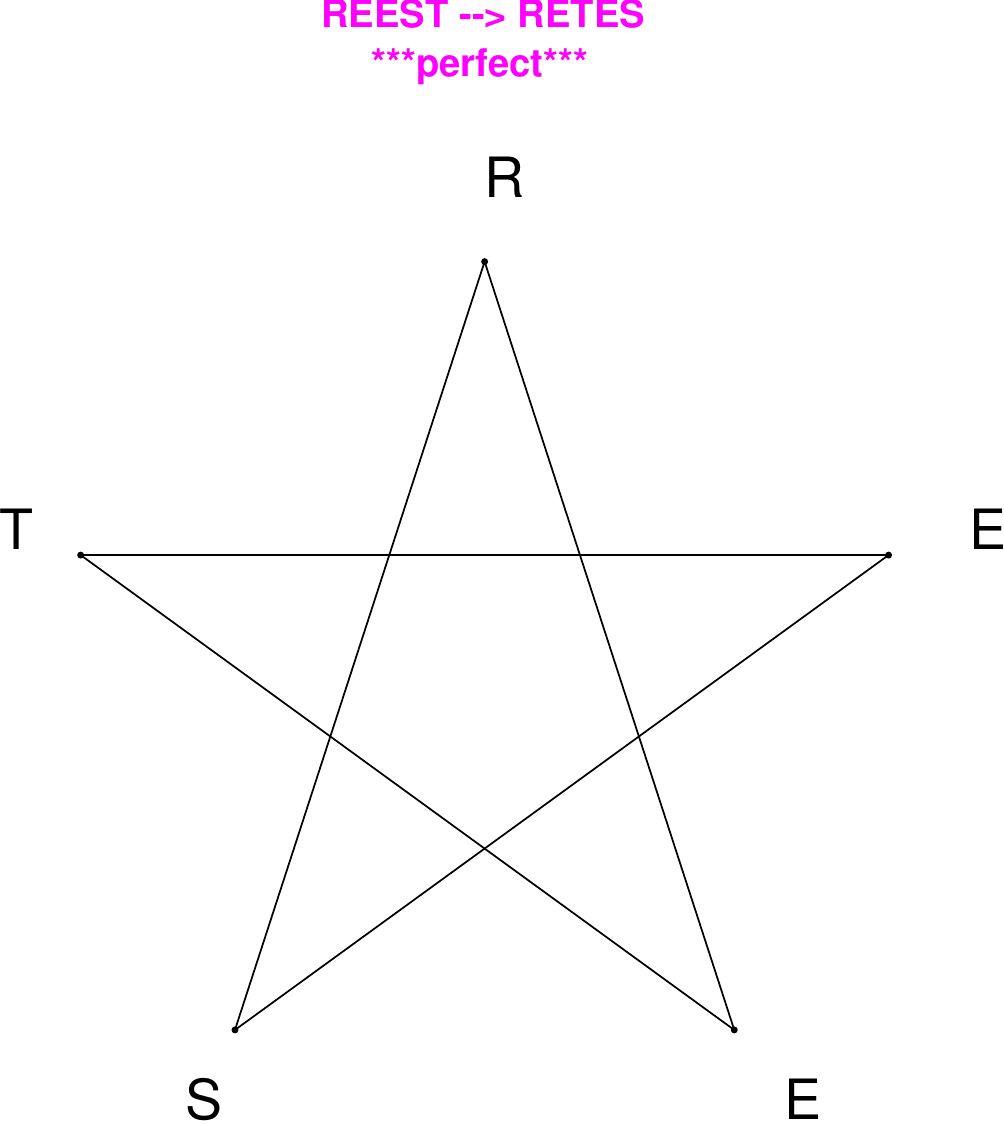}
\end{subfigure}
\hfill
\begin{subfigure}[T]{0.19\textwidth}
\centering
\includegraphics[width=\textwidth]{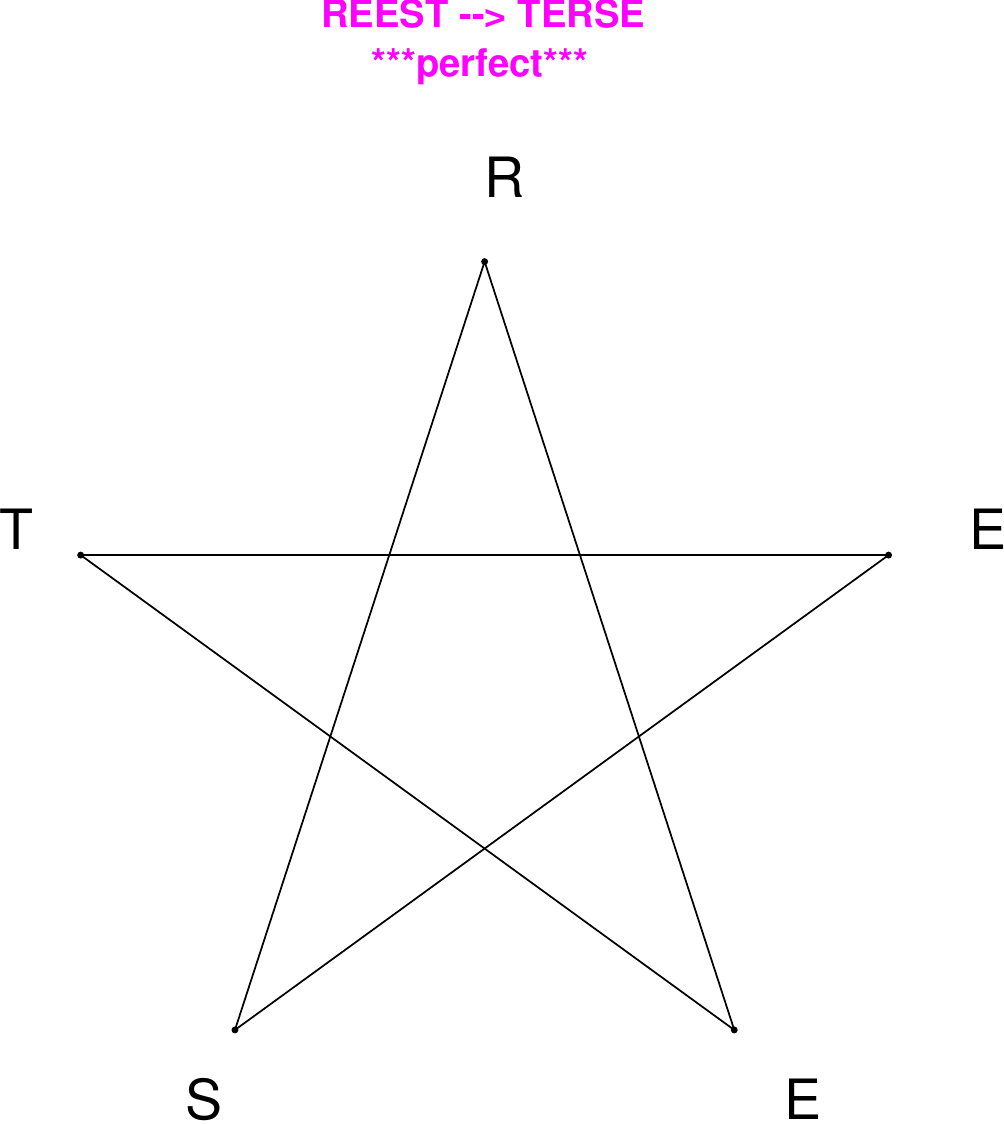}
\end{subfigure}
\end{figure}

\begin{figure}[H]
\centering
\begin{subfigure}[T]{0.19\textwidth}
\centering
\includegraphics[width=\textwidth]{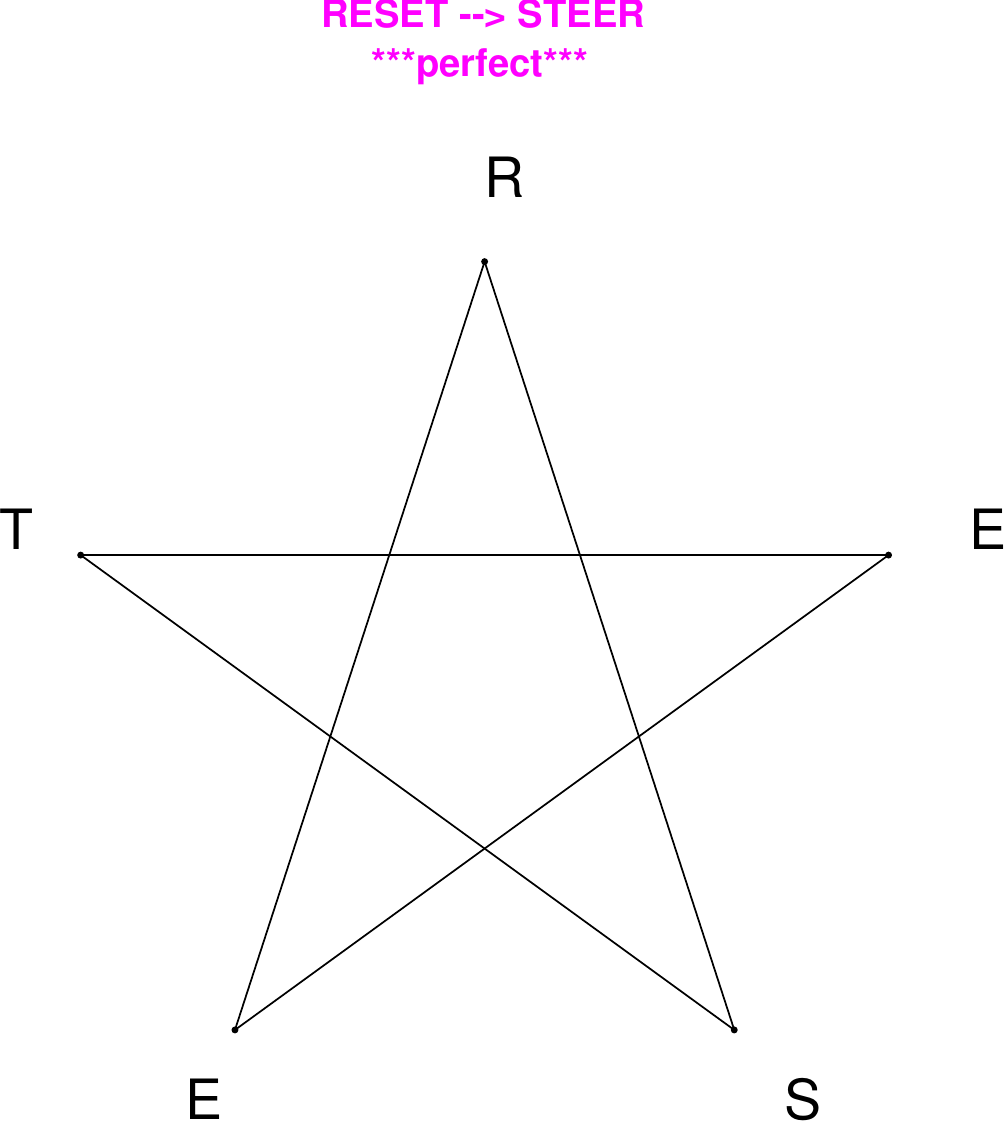}
\end{subfigure}
\hfill
\begin{subfigure}[T]{0.19\textwidth}
\centering
\includegraphics[width=\textwidth]{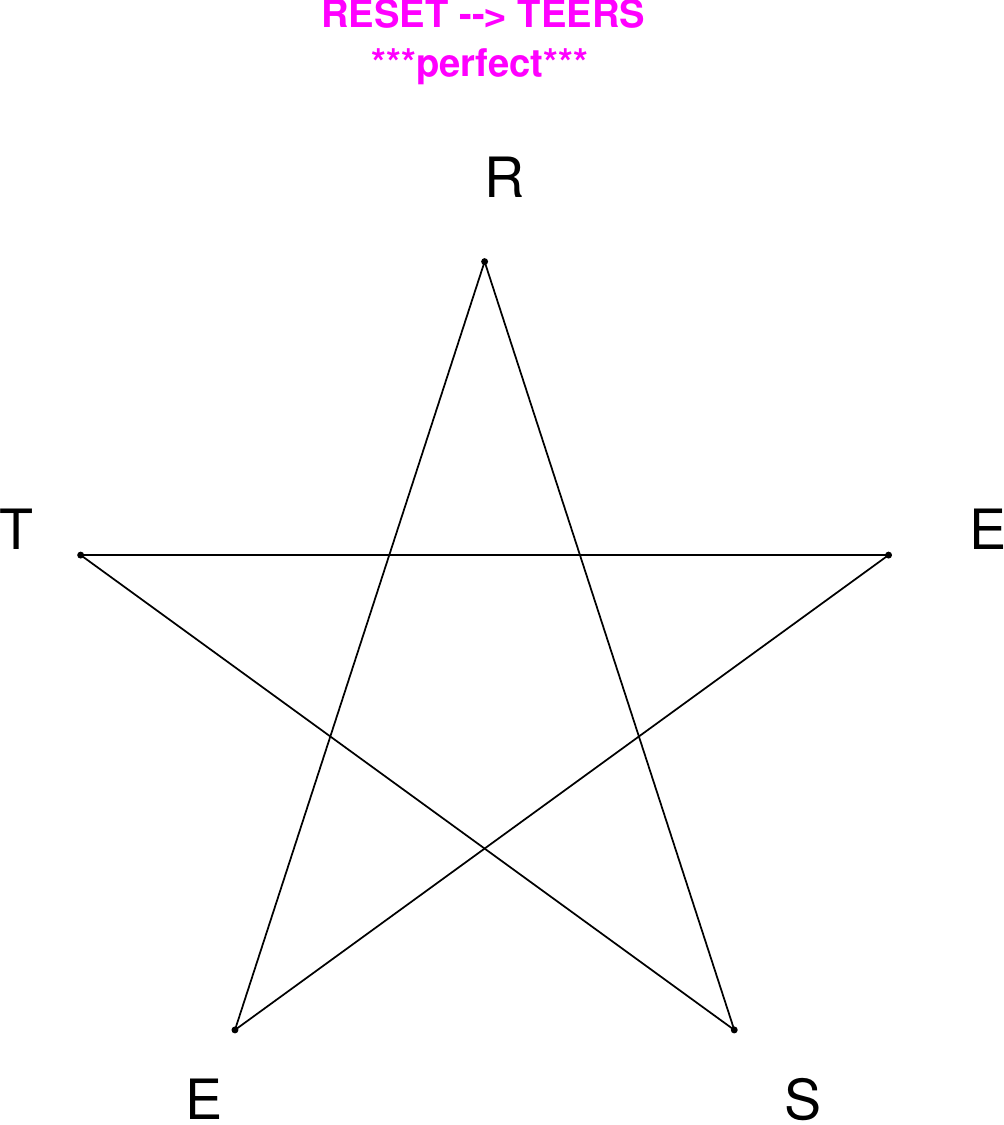}
\end{subfigure}
\hfill
\begin{subfigure}[T]{0.19\textwidth}
\centering
\includegraphics[width=\textwidth]{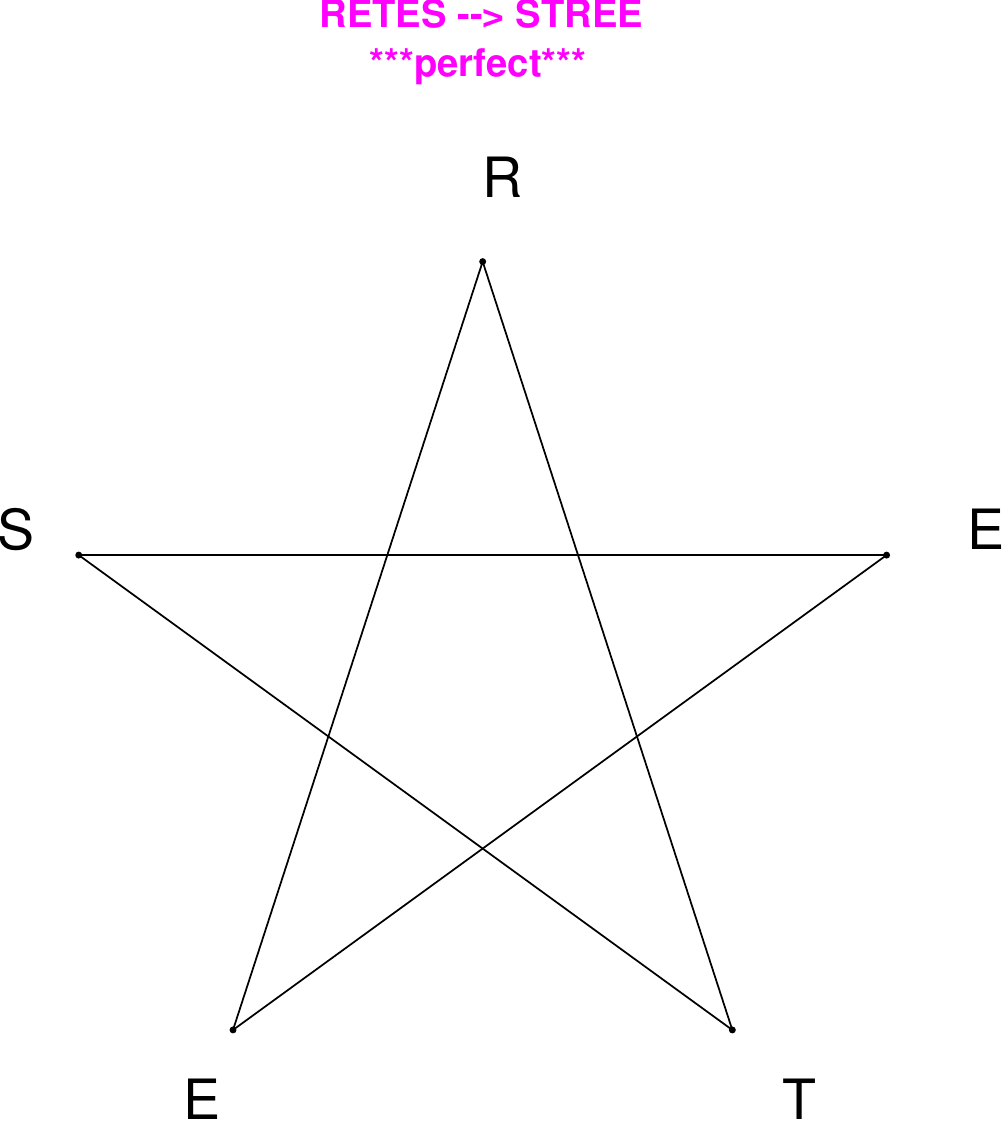}
\end{subfigure}
\hfill
\begin{subfigure}[T]{0.19\textwidth}
\centering
\includegraphics[width=\textwidth]{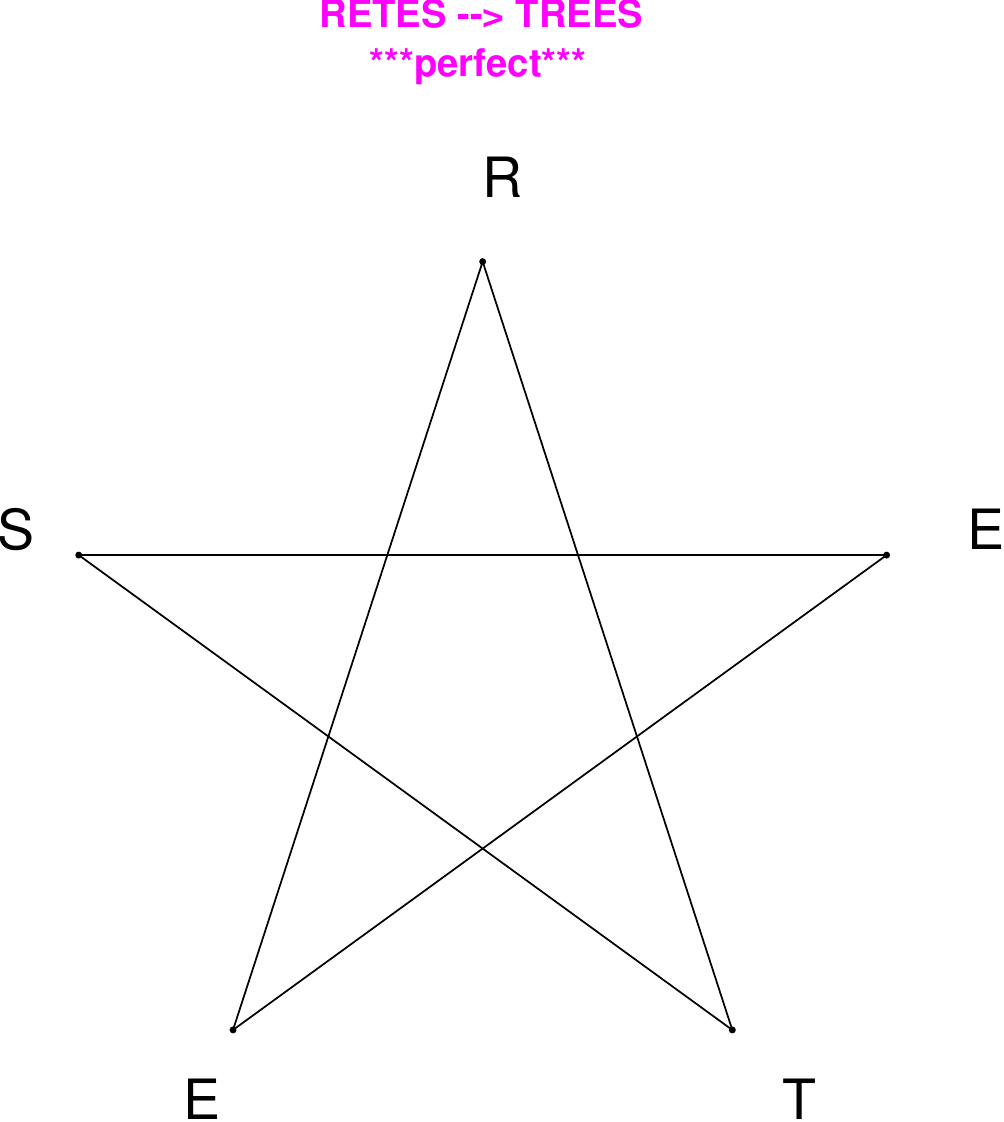}
\end{subfigure}
\hfill
\begin{subfigure}[T]{0.19\textwidth}
\centering
\includegraphics[width=\textwidth]{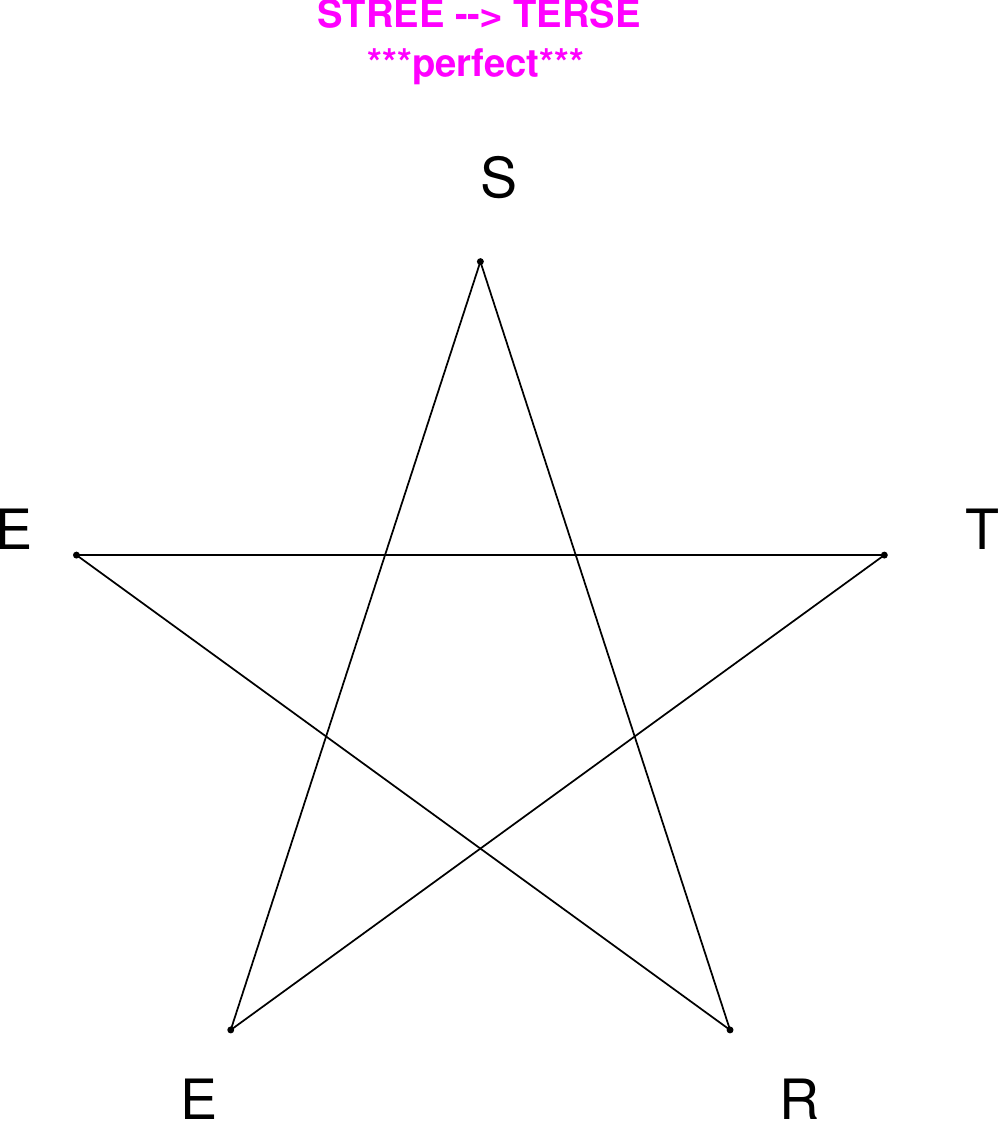}
\end{subfigure}
\end{figure}

\begin{figure}[H]
\centering
\begin{subfigure}[T]{0.19\textwidth}
\centering
\includegraphics[width=\textwidth]{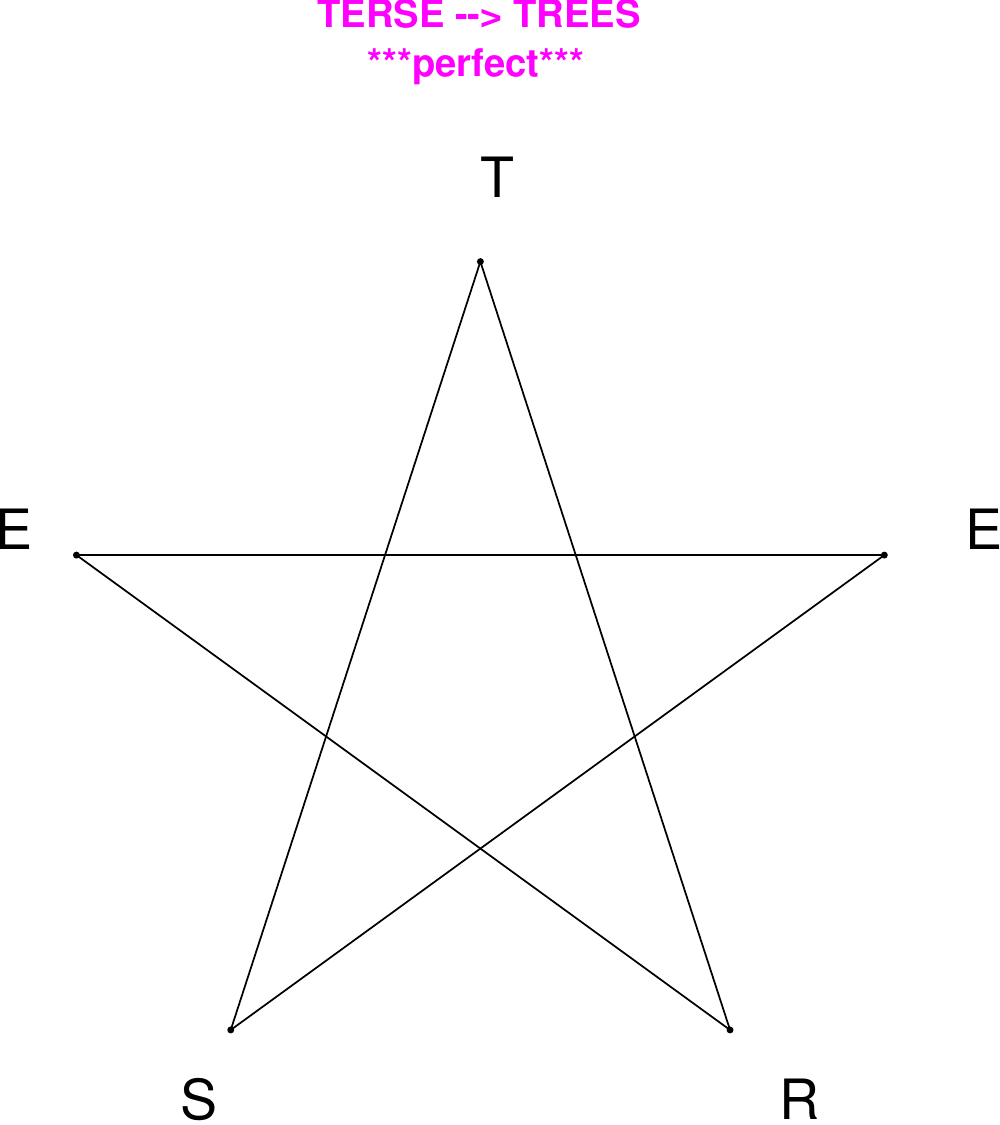}
\end{subfigure}
\hfill
\begin{subfigure}[T]{0.19\textwidth}
\centering
\includegraphics[width=\textwidth]{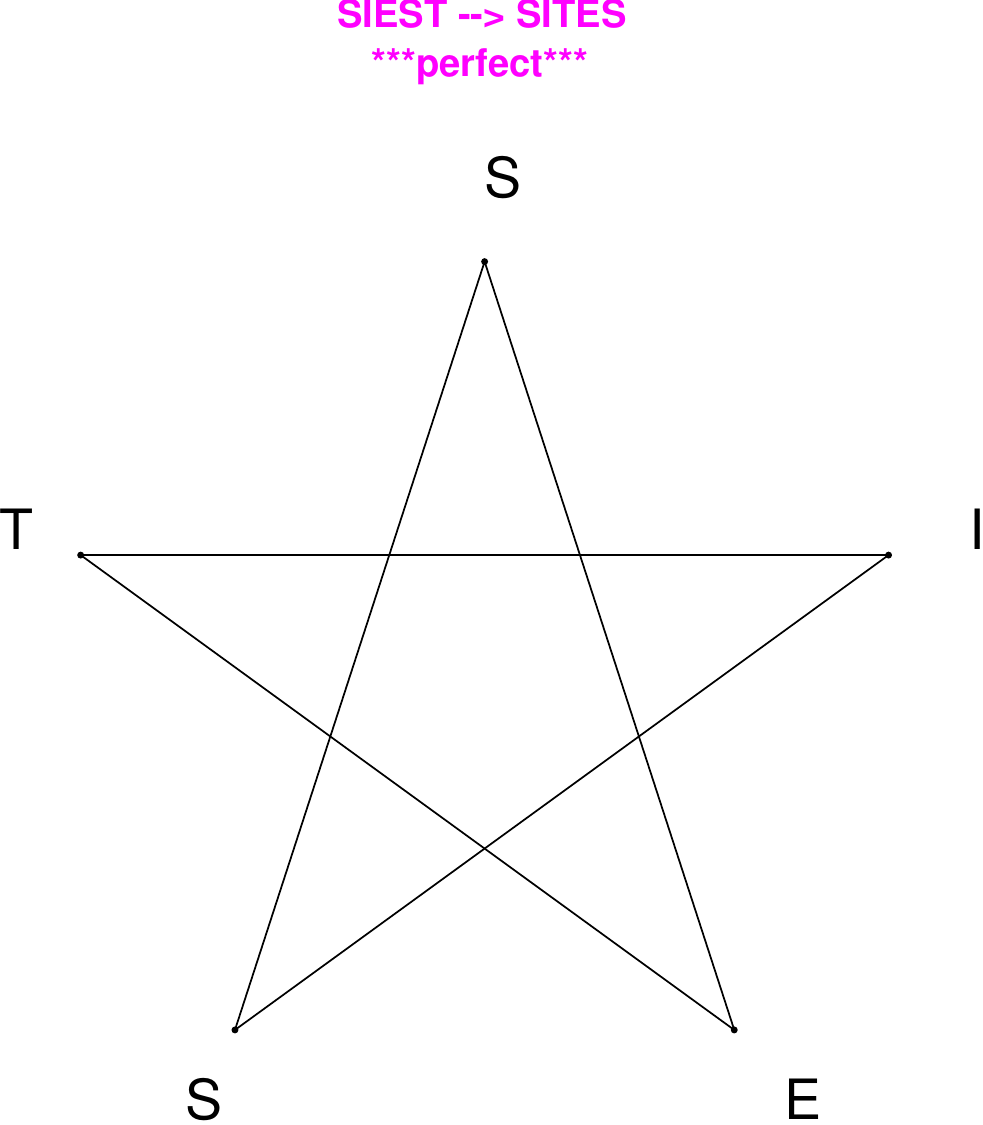}
\end{subfigure}
\hfill
\begin{subfigure}[T]{0.19\textwidth}
\centering
\includegraphics[width=\textwidth]{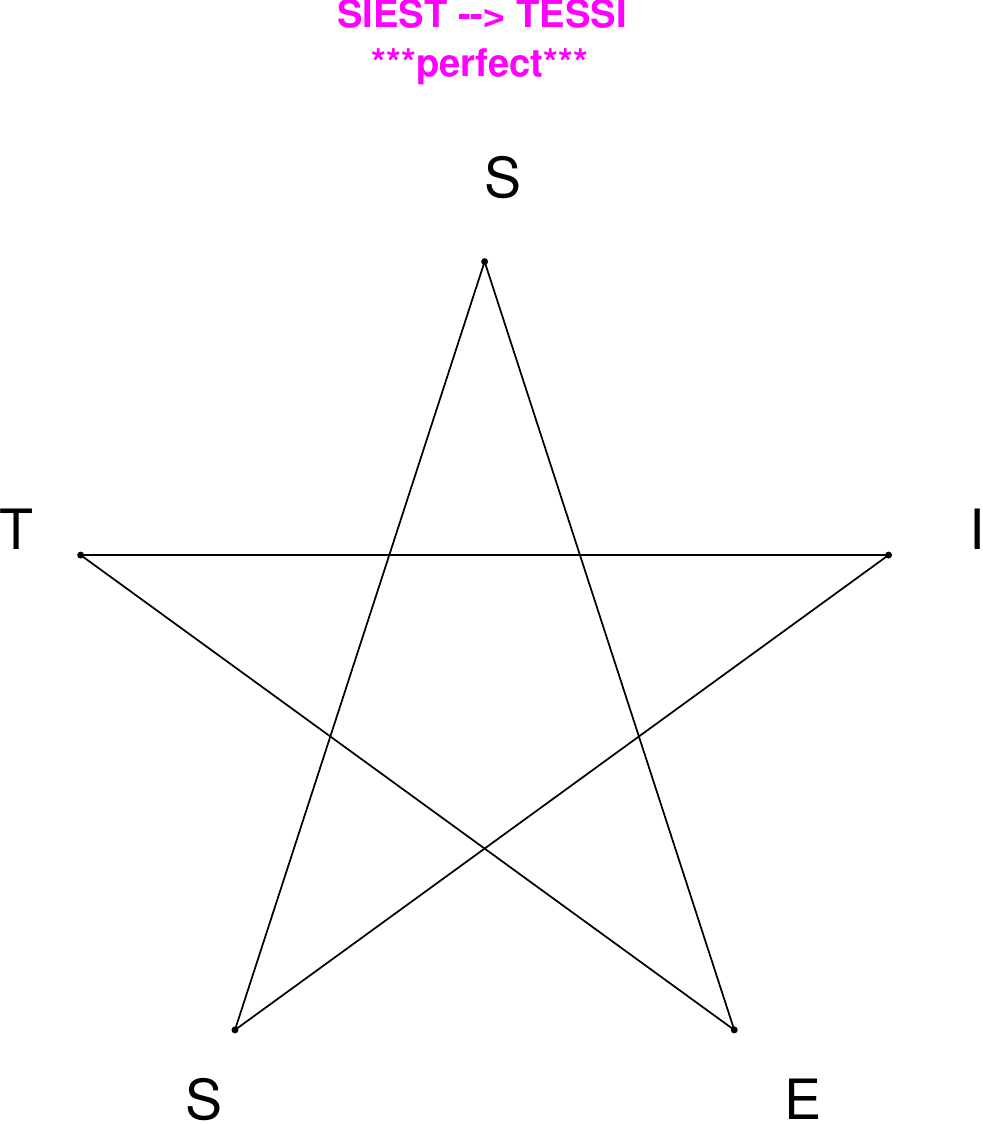}
\end{subfigure}
\hfill
\begin{subfigure}[T]{0.19\textwidth}
\centering
\includegraphics[width=\textwidth]{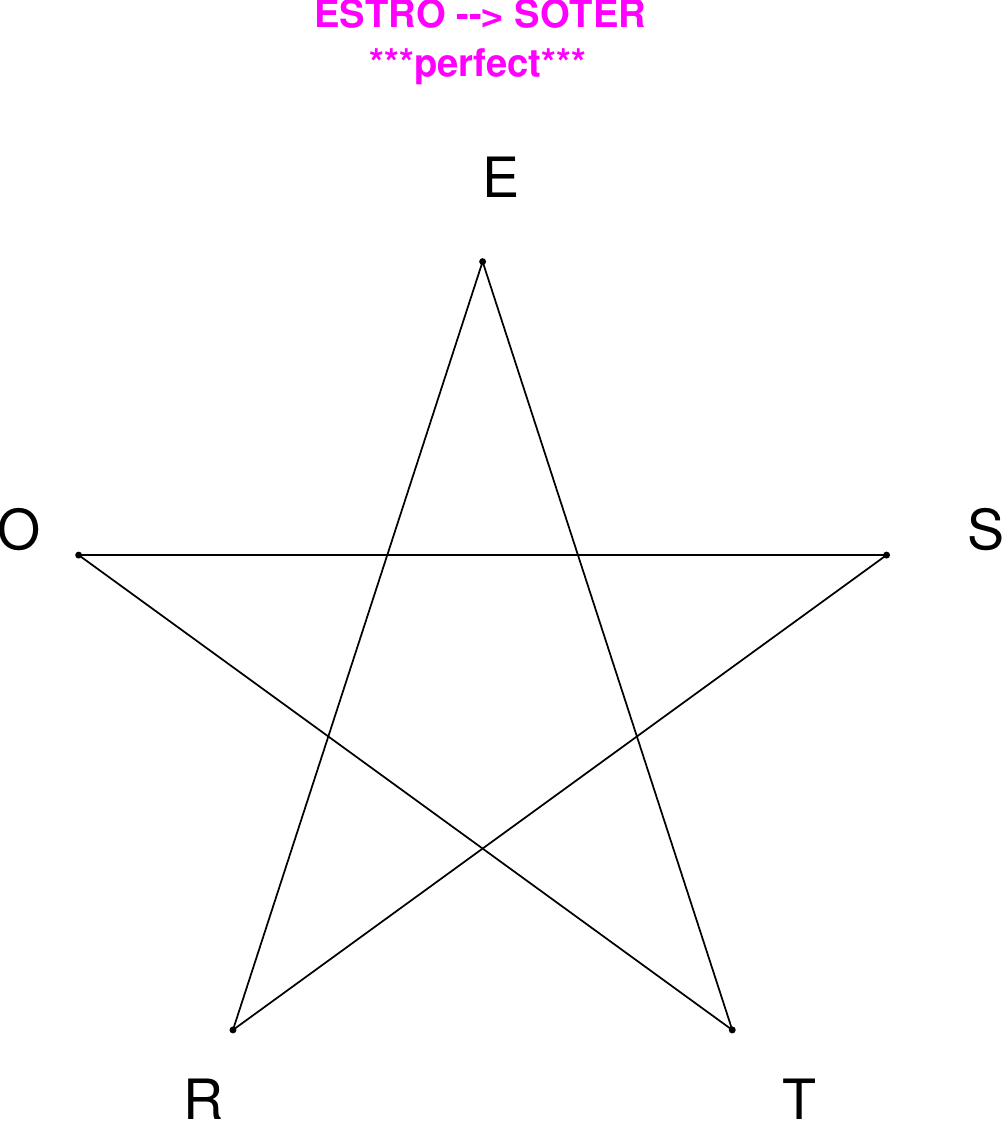}
\end{subfigure}
\hfill
\begin{subfigure}[T]{0.19\textwidth}
\centering
\includegraphics[width=\textwidth]{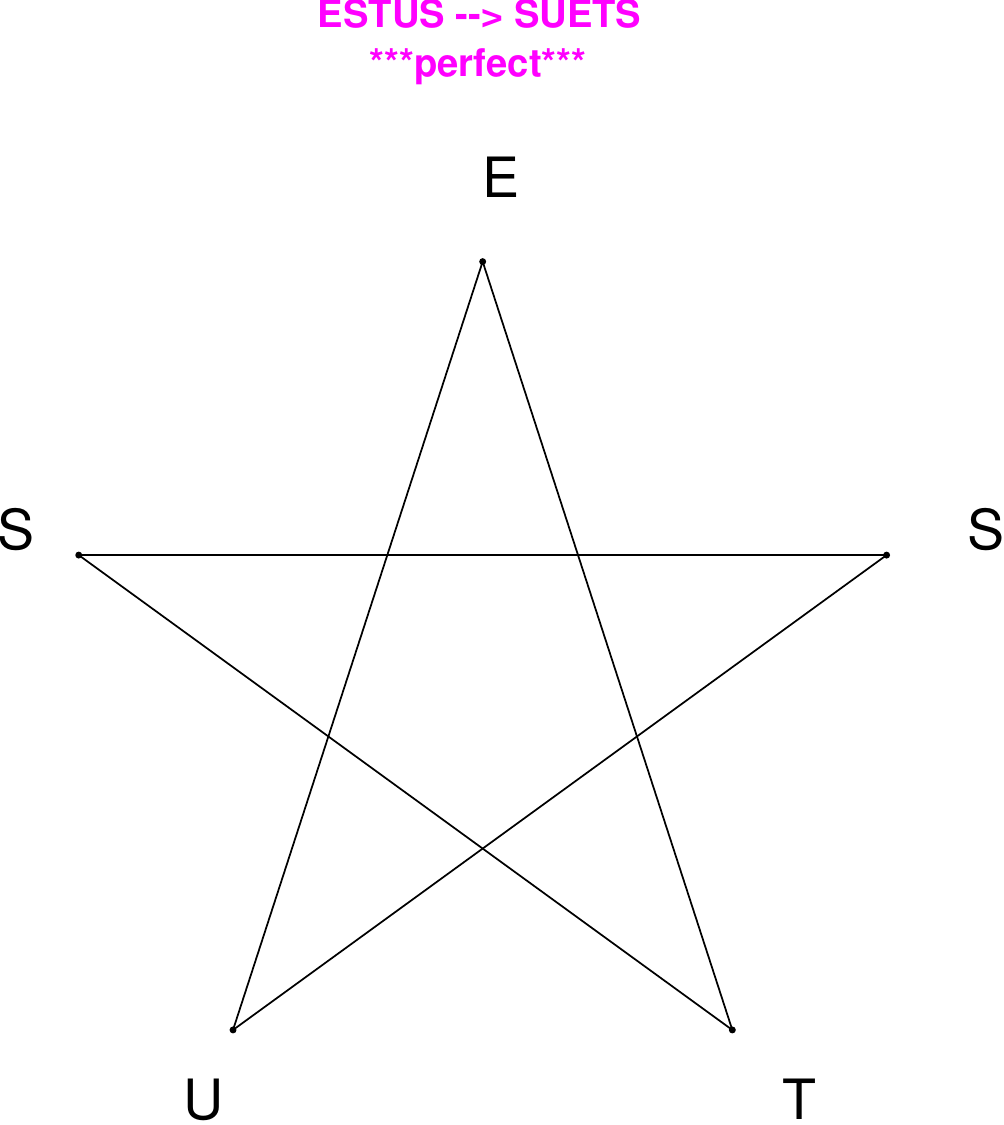}
\end{subfigure}
\end{figure}

\begin{figure}[H]
\centering
\begin{subfigure}[T]{0.19\textwidth}
\centering
\includegraphics[width=\textwidth]{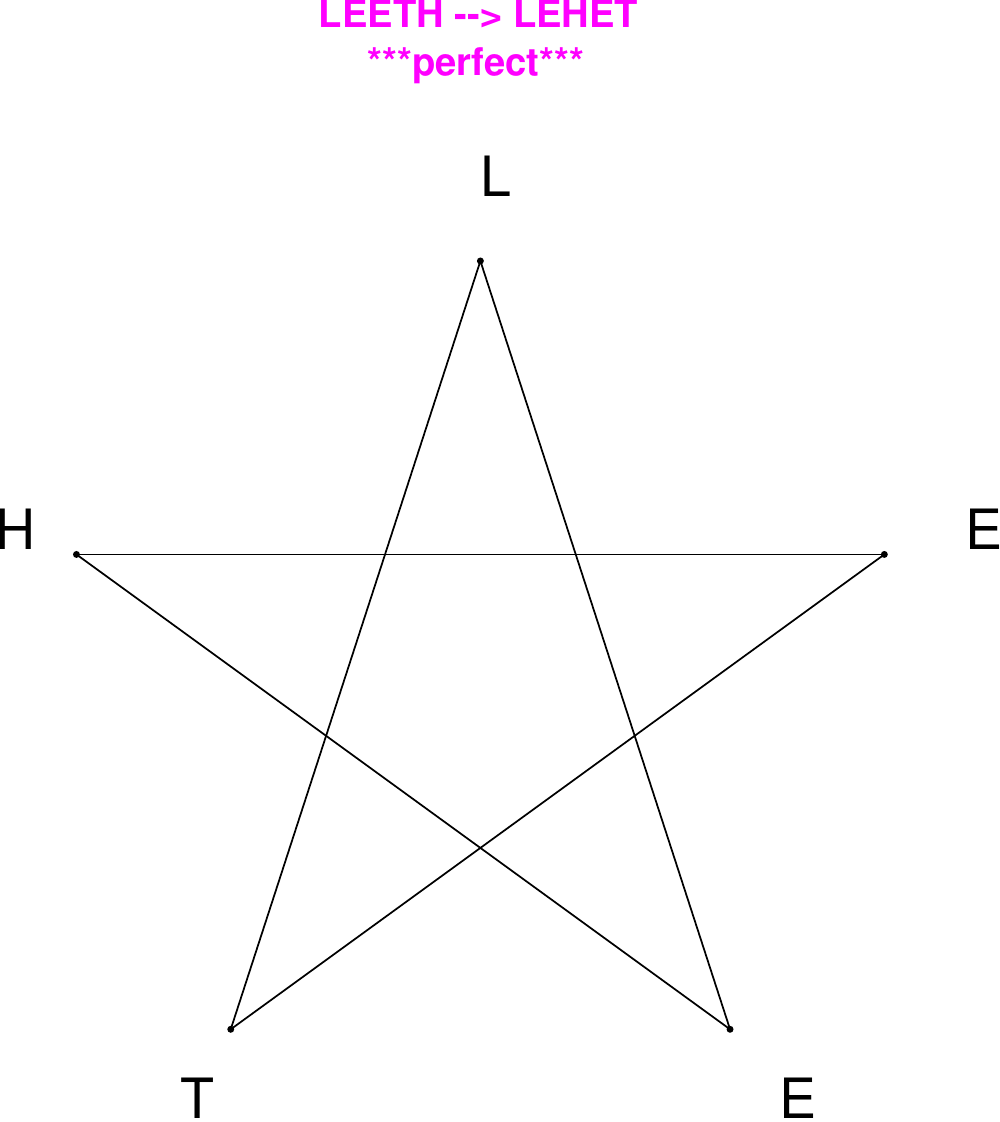}
\end{subfigure}
\hfill
\begin{subfigure}[T]{0.19\textwidth}
\centering
\includegraphics[width=\textwidth]{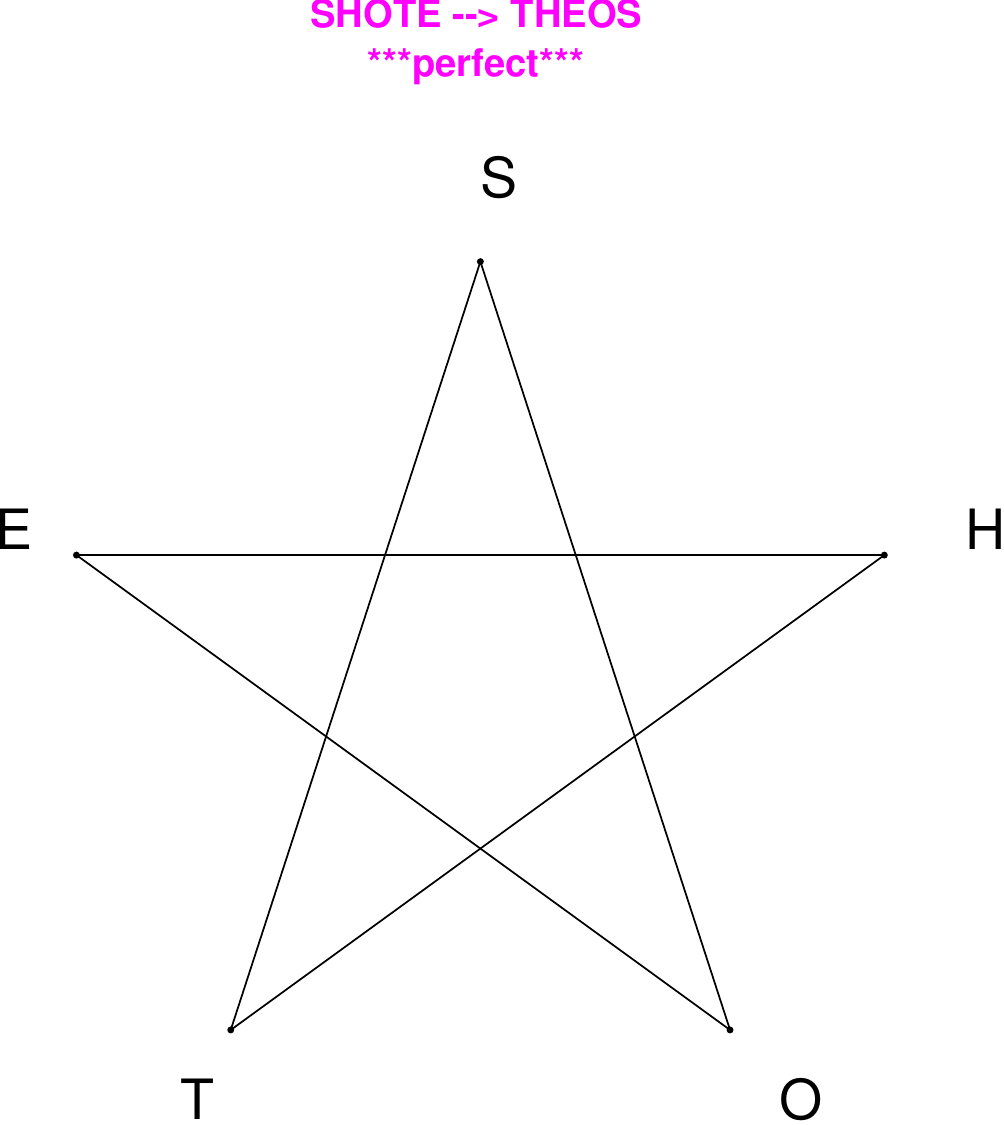}
\end{subfigure}
\hfill
\begin{subfigure}[T]{0.19\textwidth}
\centering
\includegraphics[width=\textwidth]{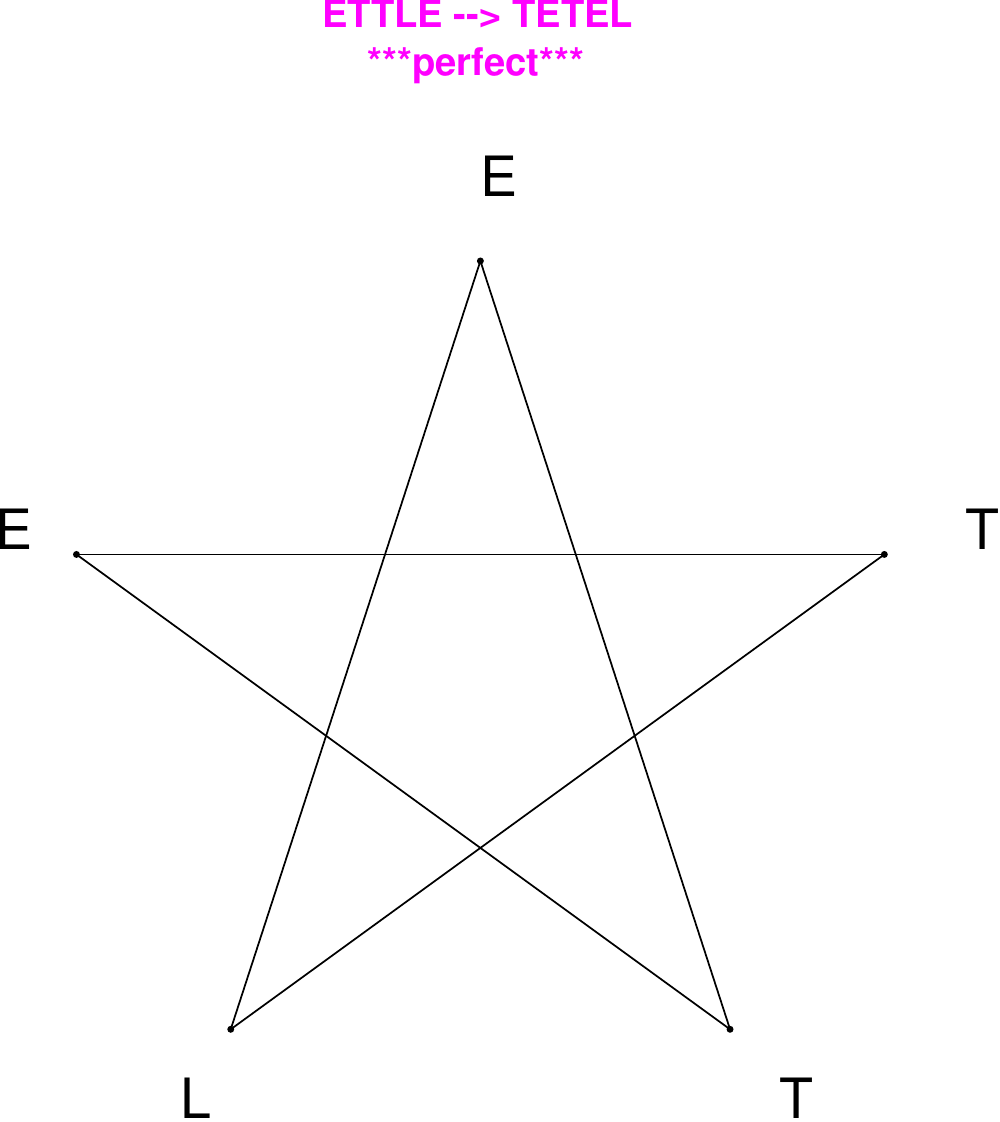}
\end{subfigure}
\hfill
\begin{subfigure}[T]{0.19\textwidth}
\centering
\includegraphics[width=\textwidth]{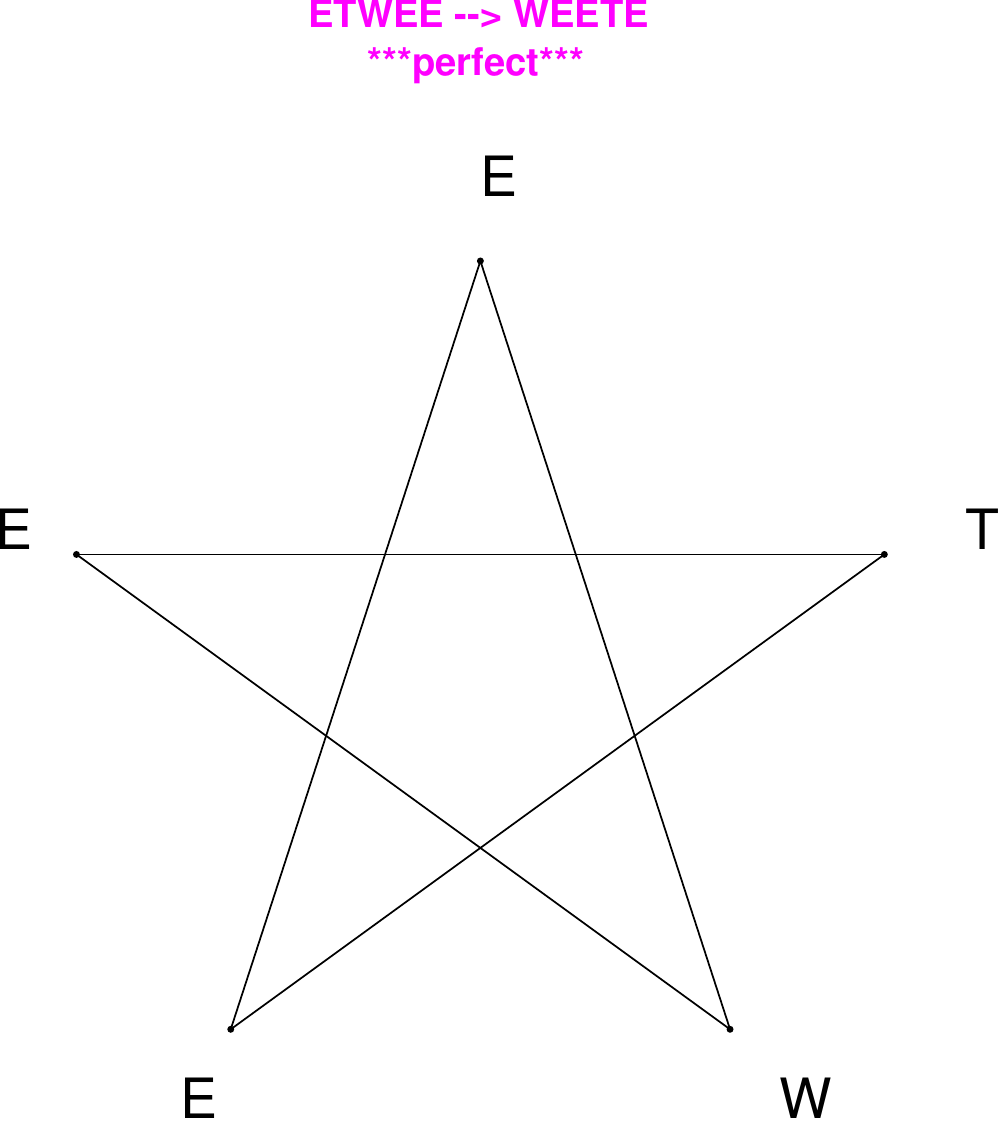}
\end{subfigure}
\hfill
\begin{subfigure}[T]{0.19\textwidth}
\centering
\includegraphics[width=\textwidth]{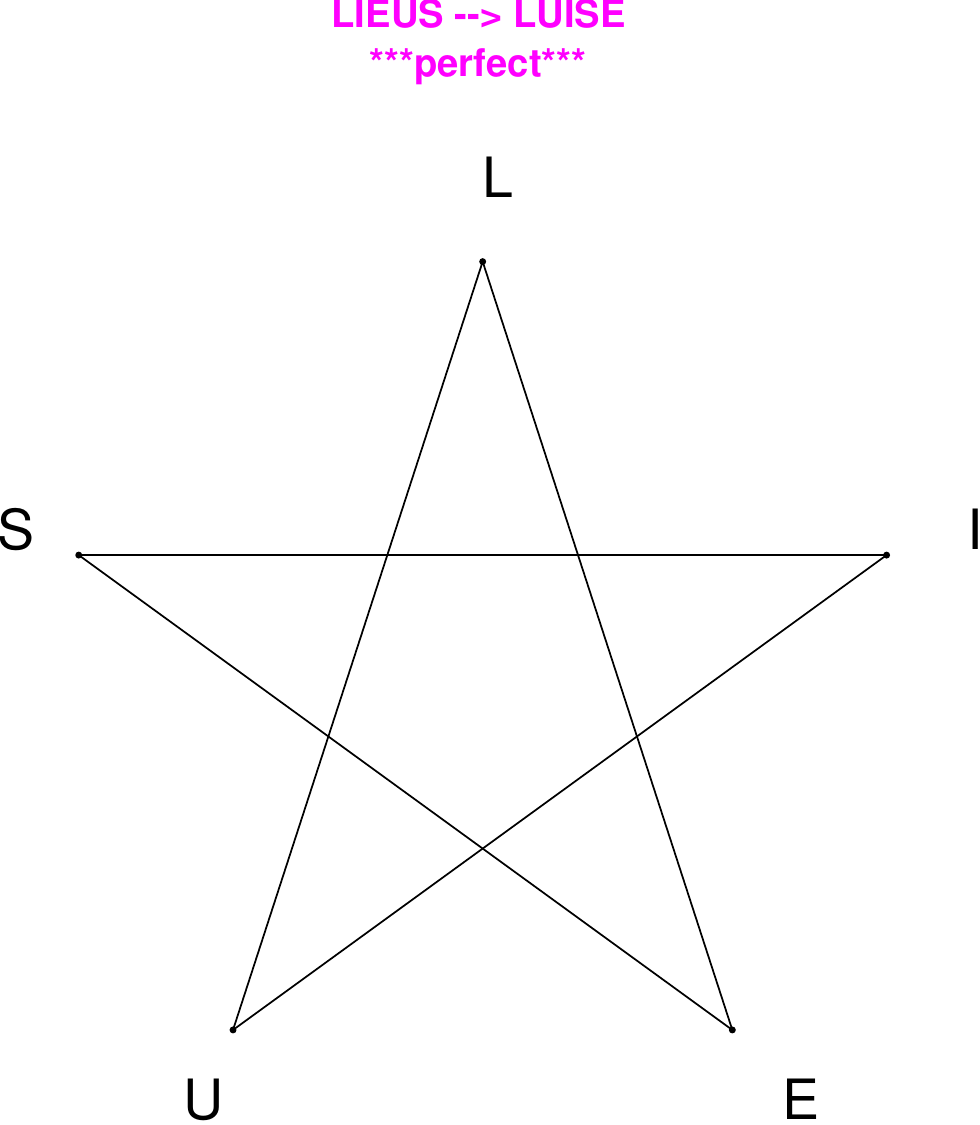}
\end{subfigure}
\end{figure}

\begin{figure}[H]
\centering
\begin{subfigure}[T]{0.19\textwidth}
\centering
\includegraphics[width=\textwidth]{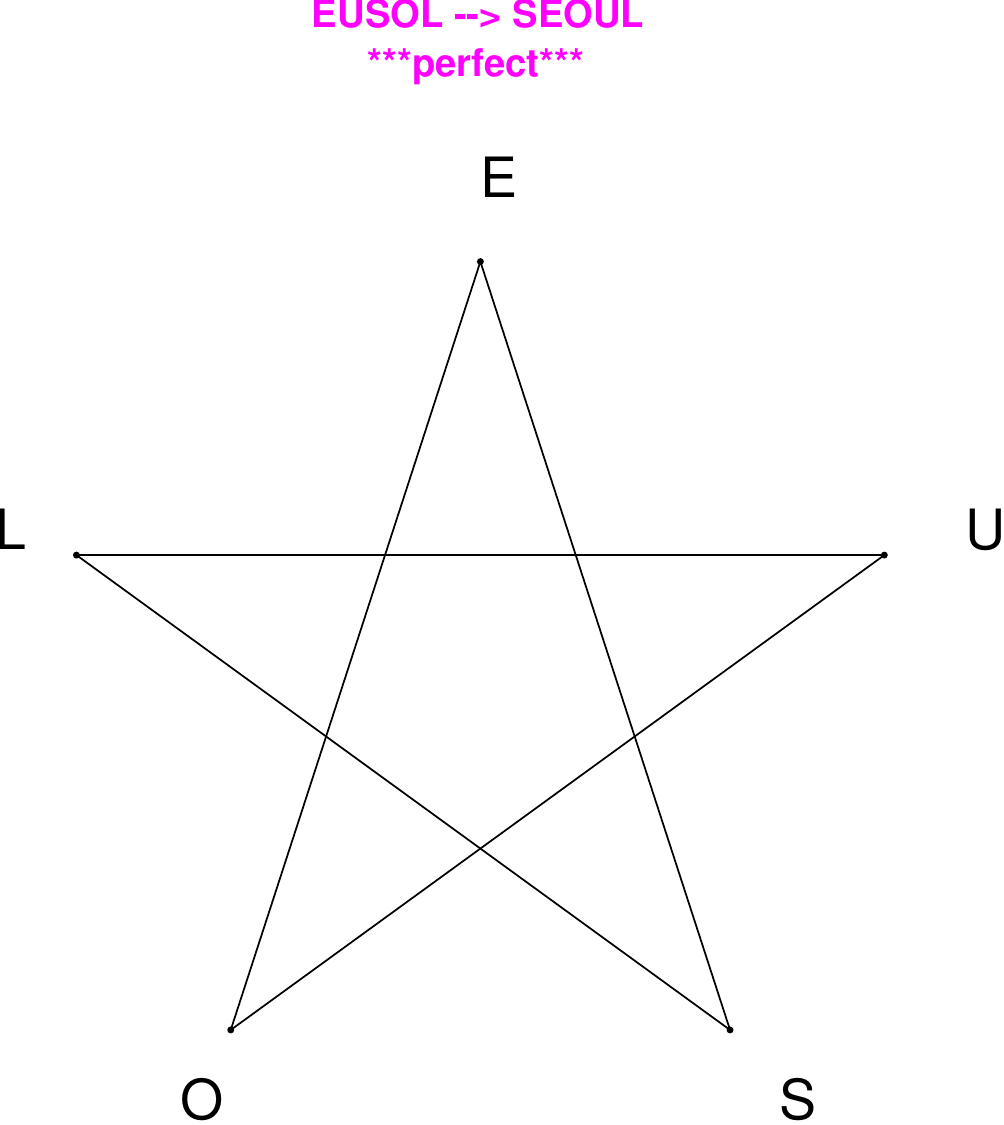}
\end{subfigure}
\hfill
\begin{subfigure}[T]{0.19\textwidth}
\centering
\includegraphics[width=\textwidth]{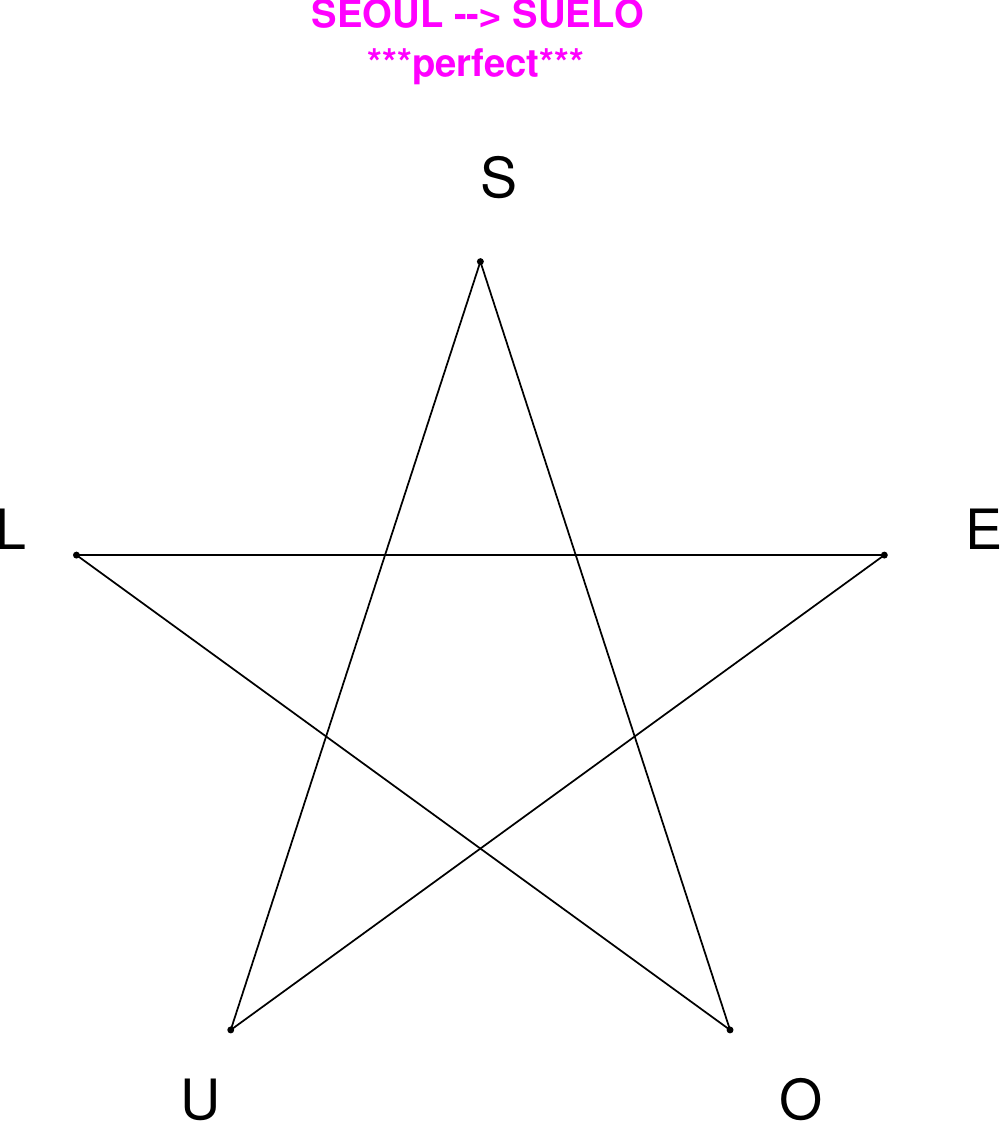}
\end{subfigure}
\hfill
\begin{subfigure}[T]{0.19\textwidth}
\centering
\includegraphics[width=\textwidth]{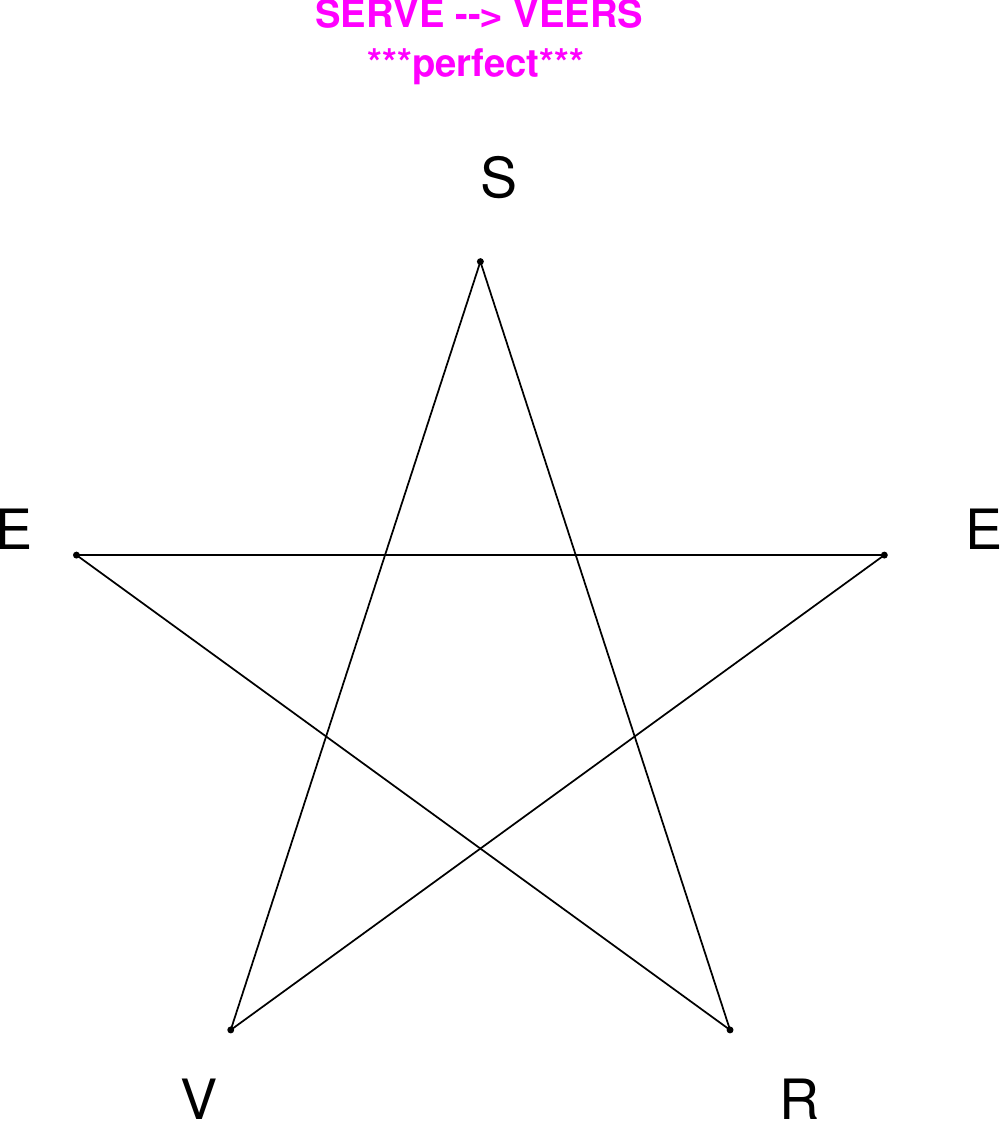}
\end{subfigure}
\hfill
\begin{subfigure}[T]{0.19\textwidth}
\centering
\includegraphics[width=\textwidth]{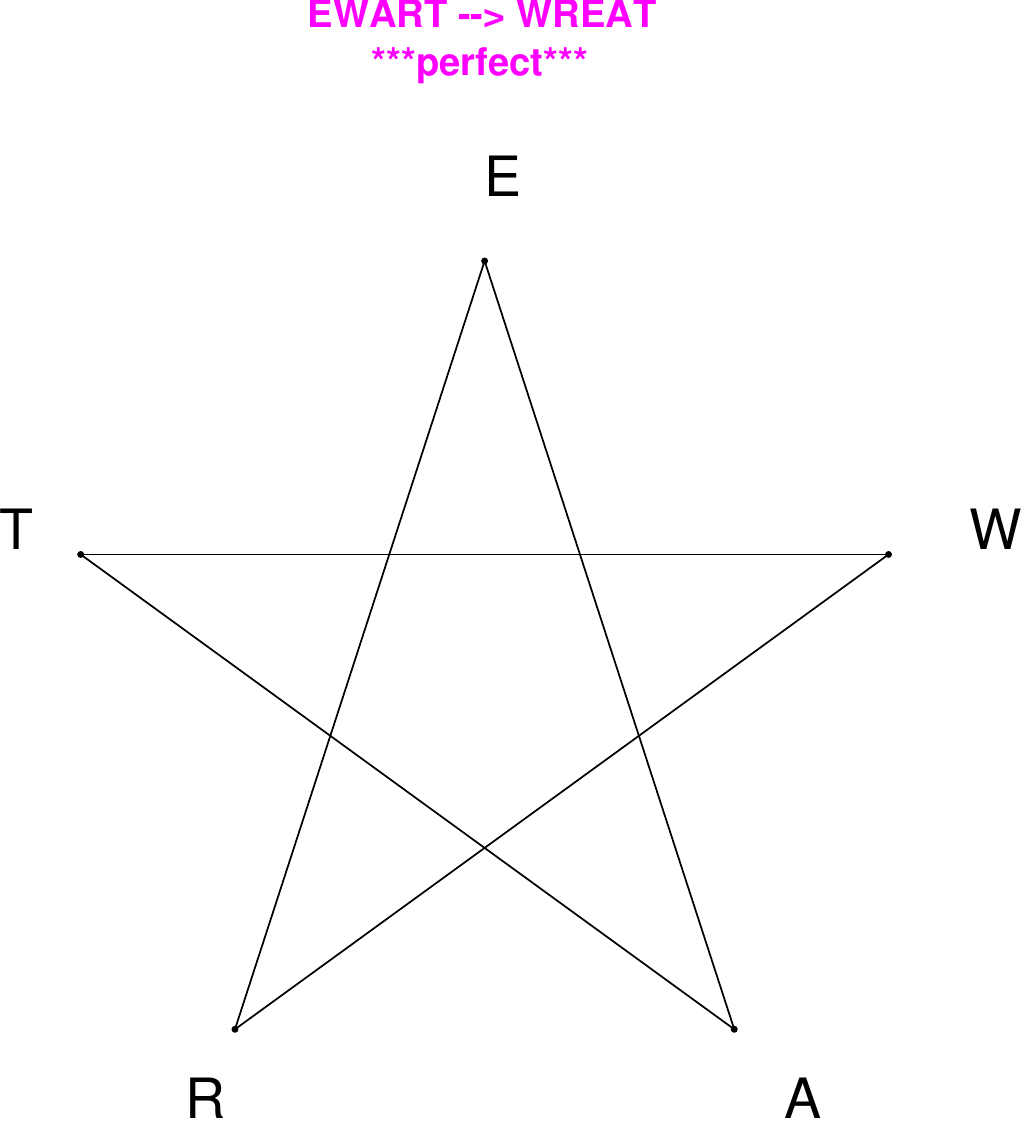}
\end{subfigure}
\hfill
\begin{subfigure}[T]{0.19\textwidth}
\centering
\includegraphics[width=\textwidth]{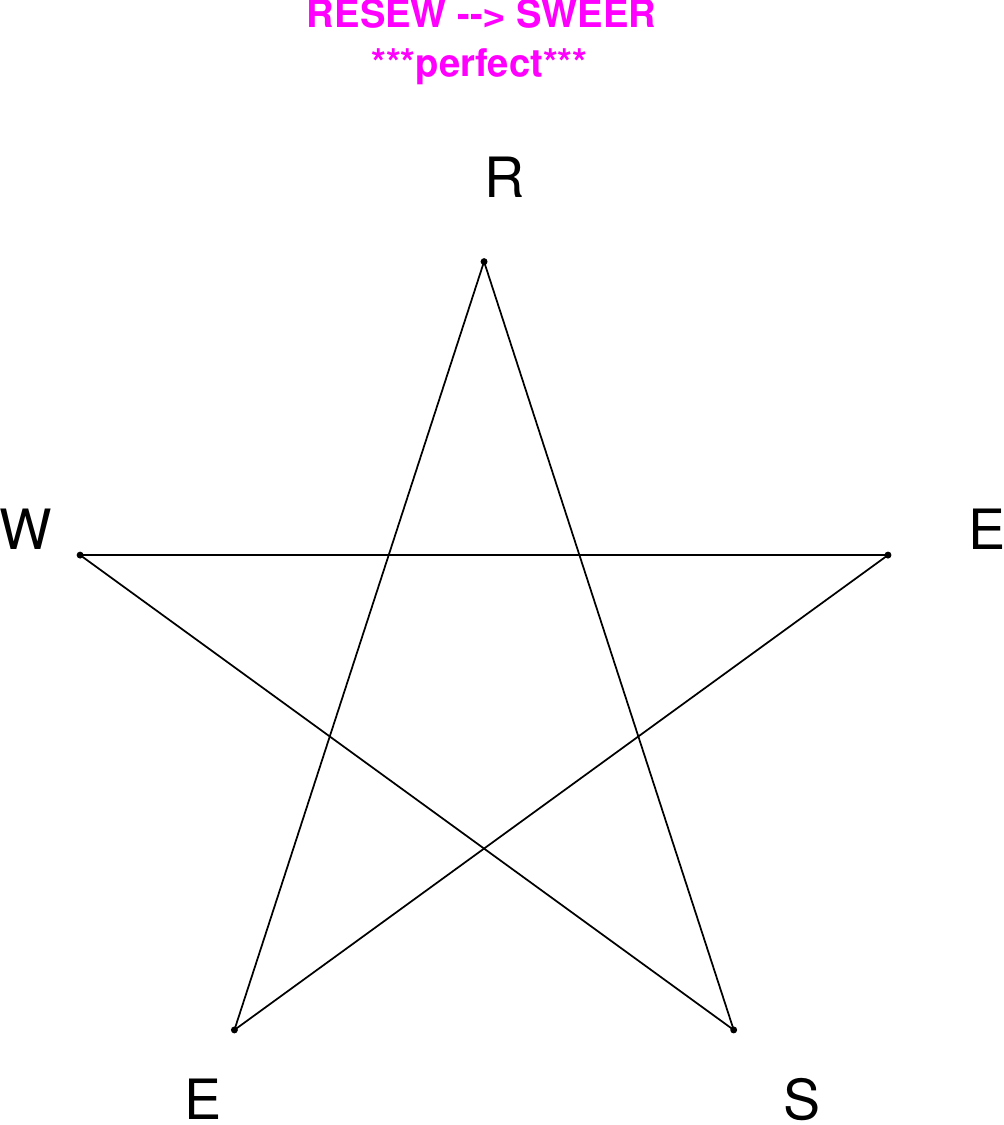}
\end{subfigure}
\end{figure}

\begin{figure}[H]
\centering
\begin{subfigure}[T]{0.19\textwidth}
\centering
\includegraphics[width=\textwidth]{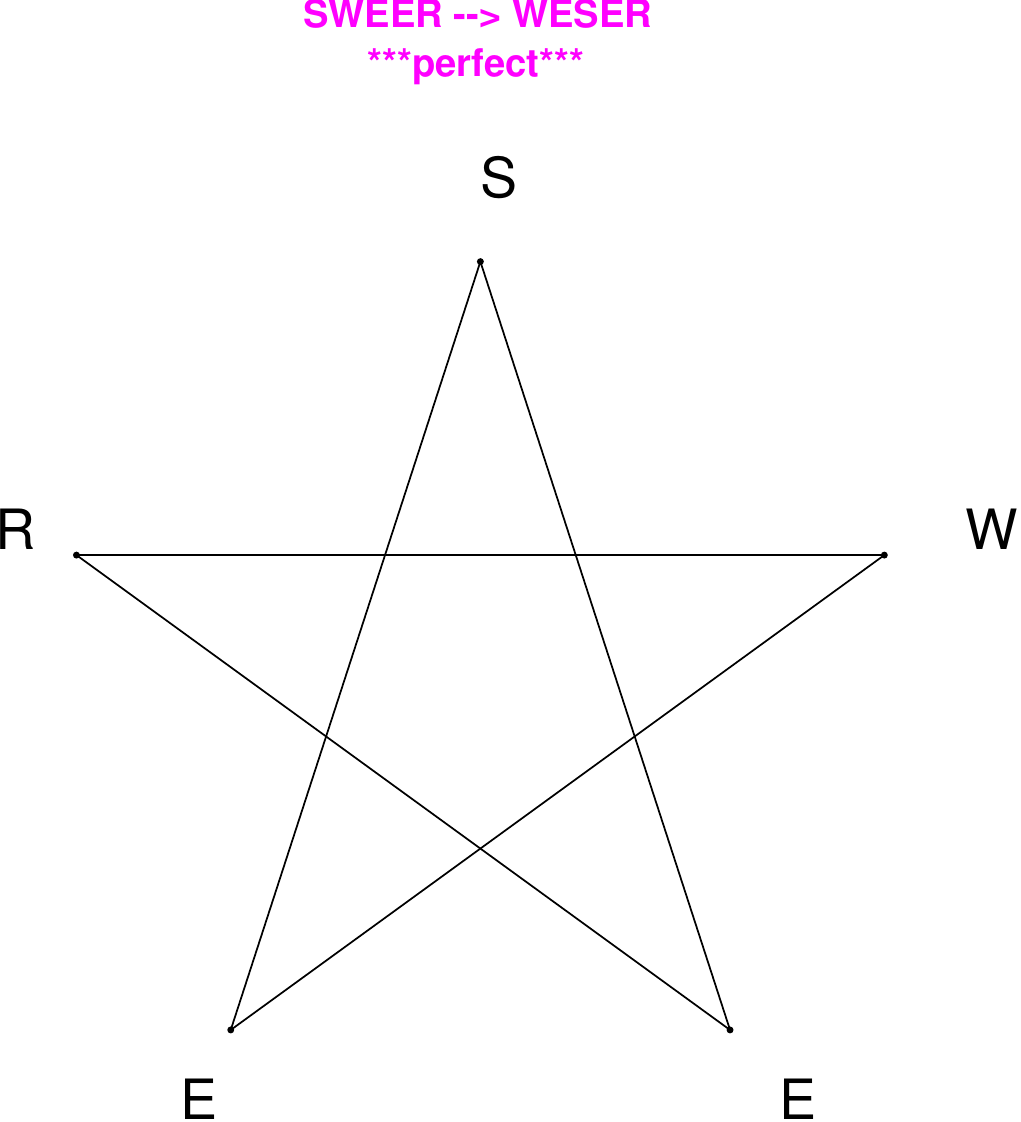}
\end{subfigure}
\hfill
\begin{subfigure}[T]{0.19\textwidth}
\centering
\includegraphics[width=\textwidth]{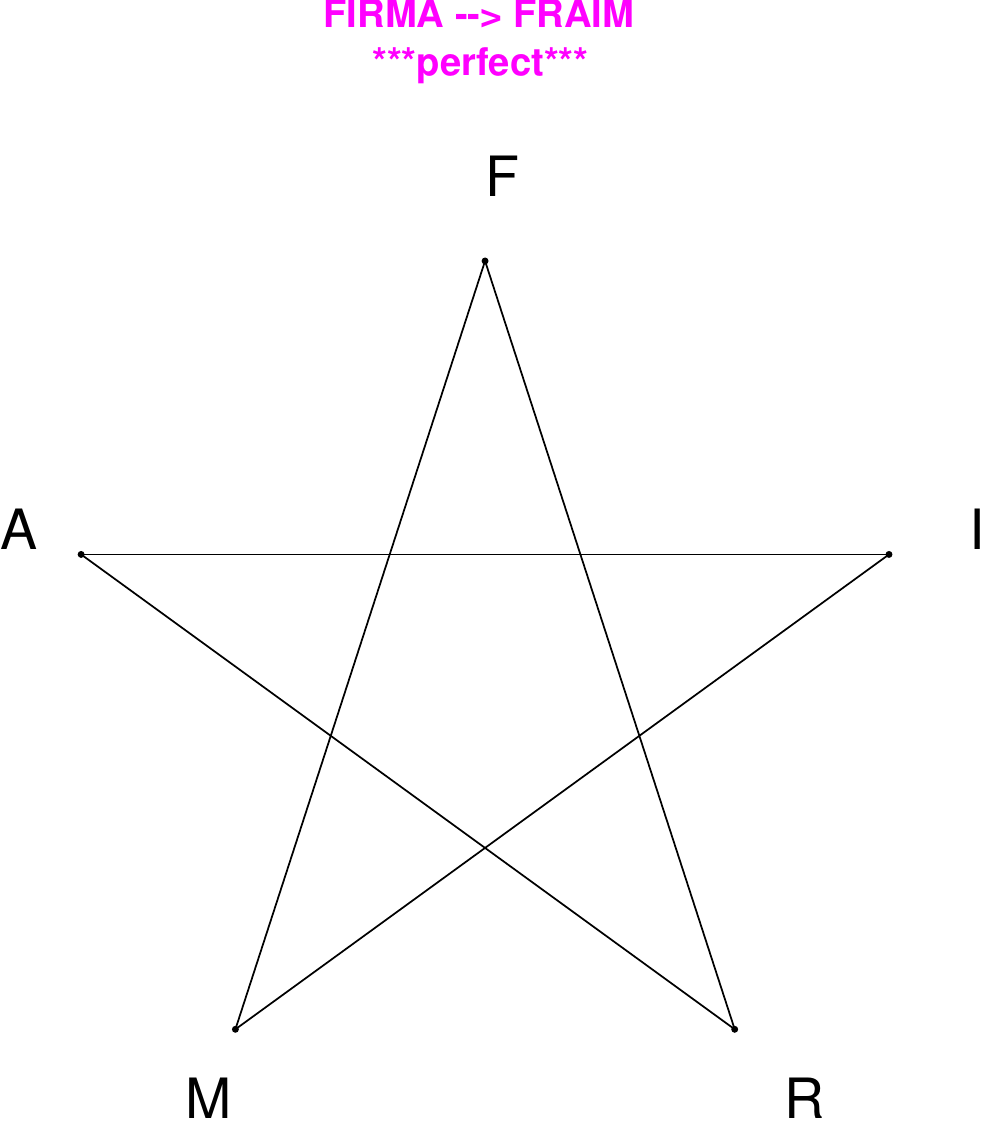}
\end{subfigure}
\hfill
\begin{subfigure}[T]{0.19\textwidth}
\centering
\includegraphics[width=\textwidth]{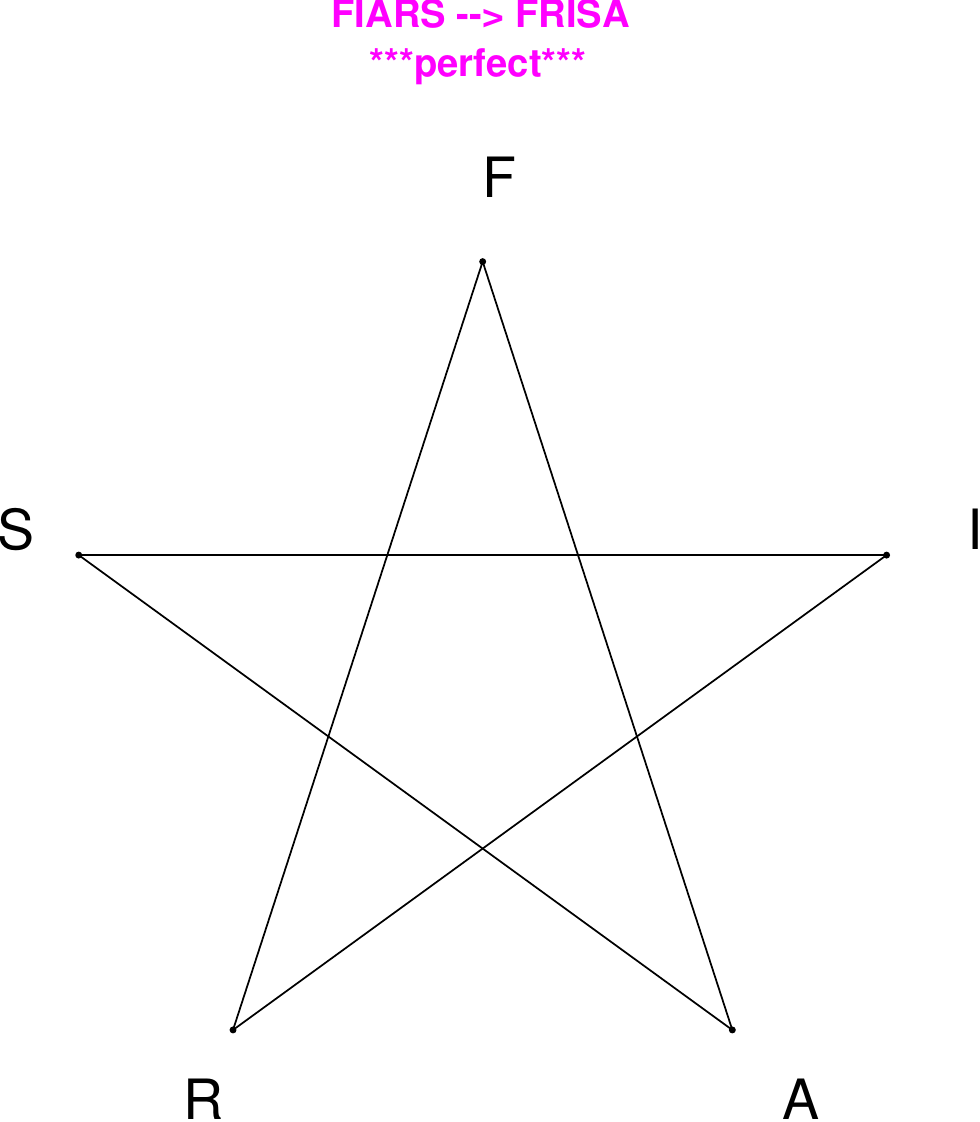}
\end{subfigure}
\hfill
\begin{subfigure}[T]{0.19\textwidth}
\centering
\includegraphics[width=\textwidth]{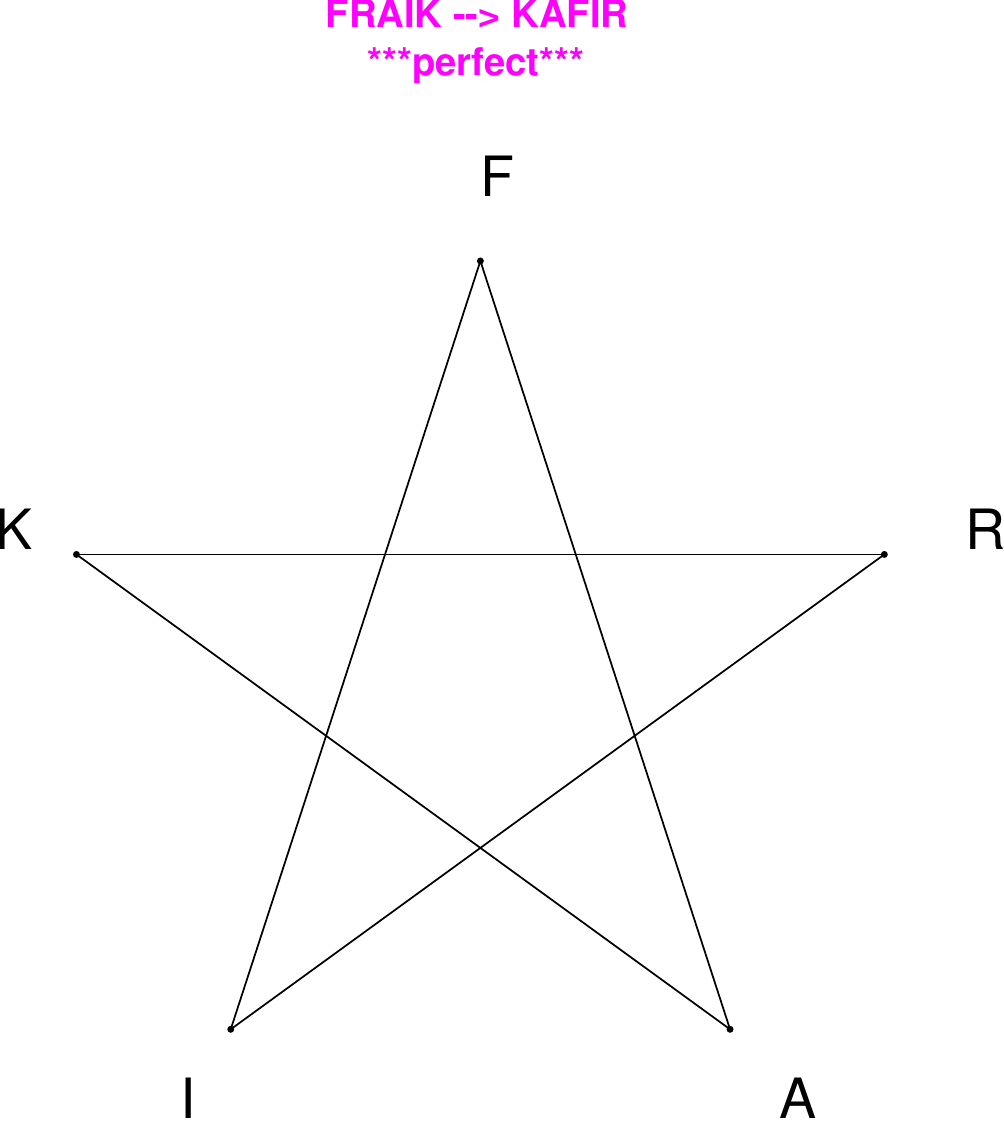}
\end{subfigure}
\hfill
\begin{subfigure}[T]{0.19\textwidth}
\centering
\includegraphics[width=\textwidth]{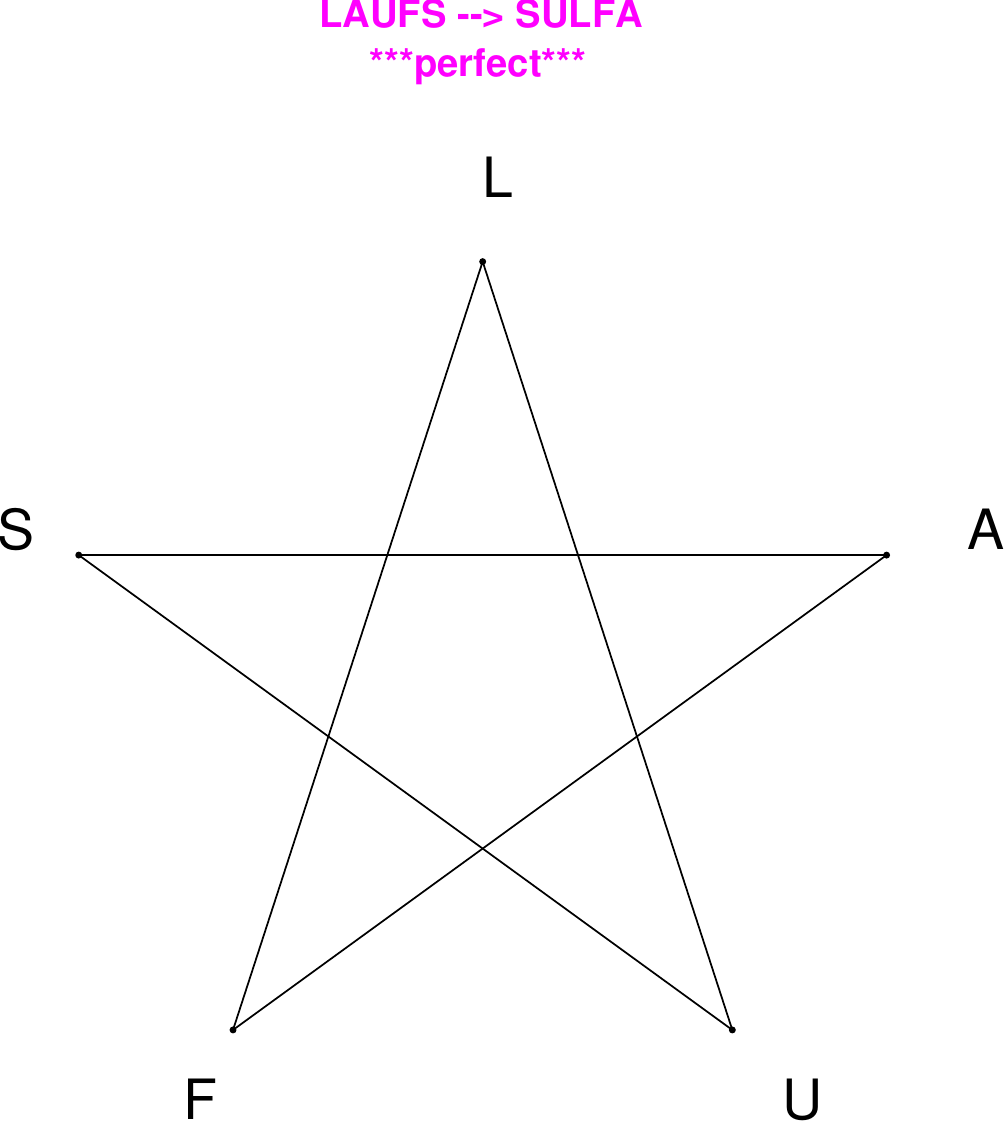}
\end{subfigure}
\end{figure}

\begin{figure}[H]
\centering
\begin{subfigure}[T]{0.19\textwidth}
\centering
\includegraphics[width=\textwidth]{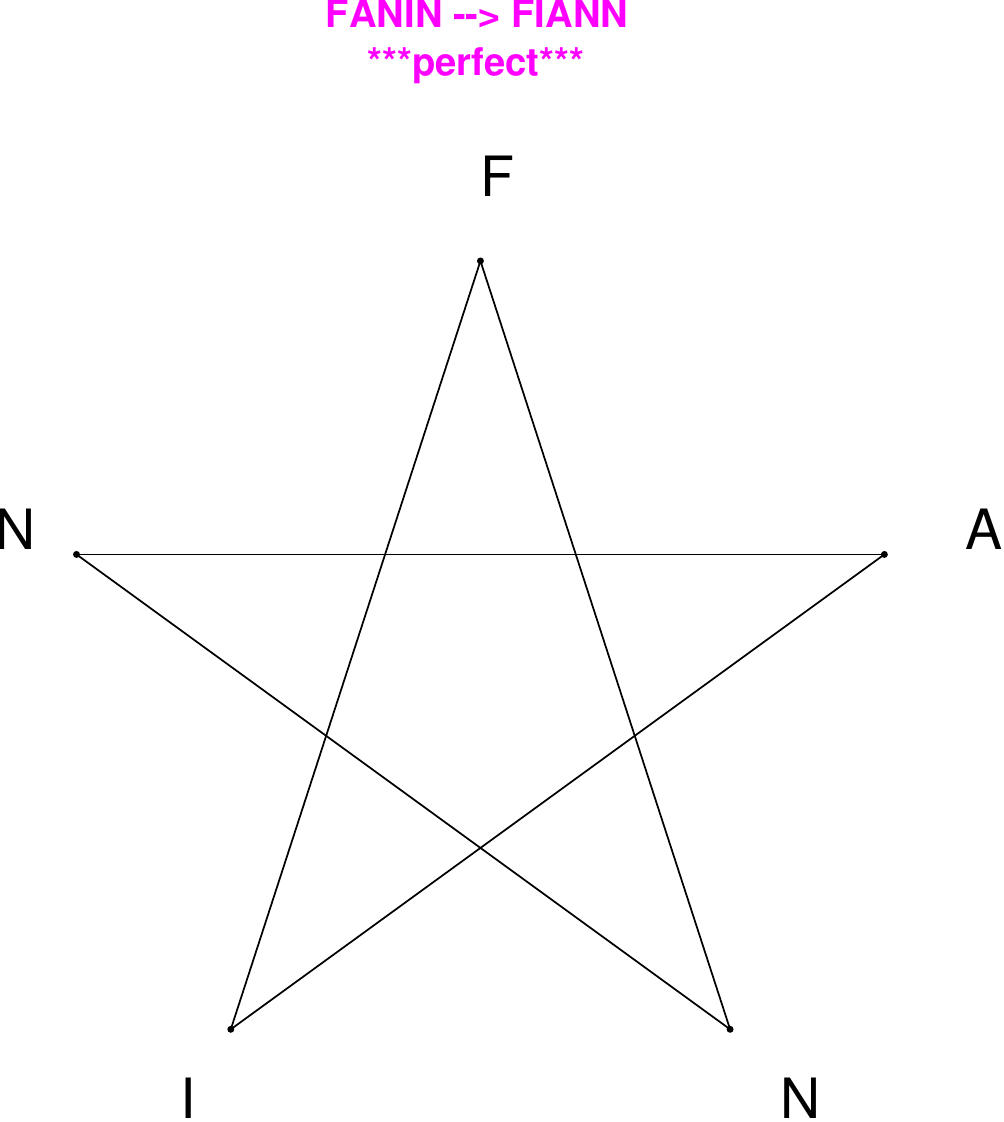}
\end{subfigure}
\hfill
\begin{subfigure}[T]{0.19\textwidth}
\centering
\includegraphics[width=\textwidth]{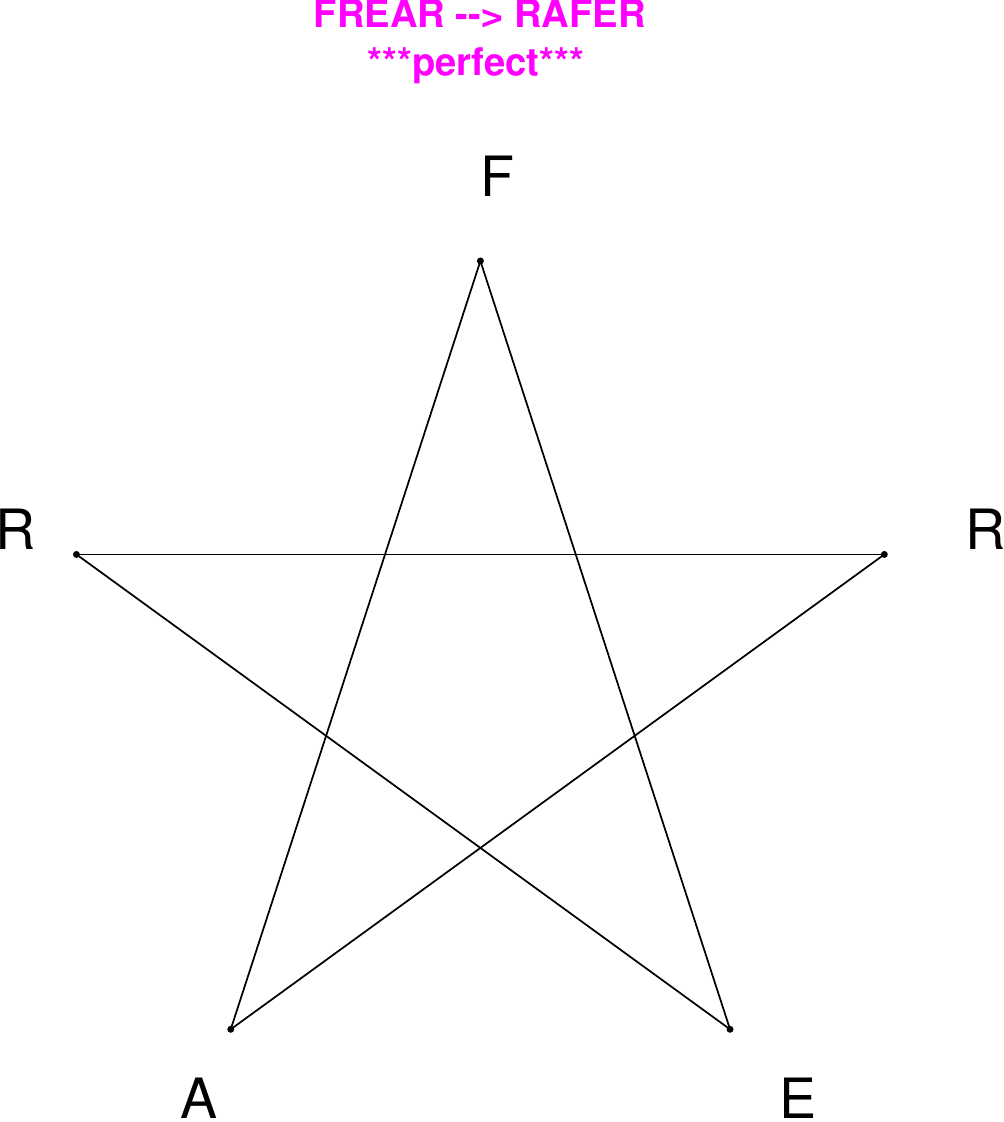}
\end{subfigure}
\hfill
\begin{subfigure}[T]{0.19\textwidth}
\centering
\includegraphics[width=\textwidth]{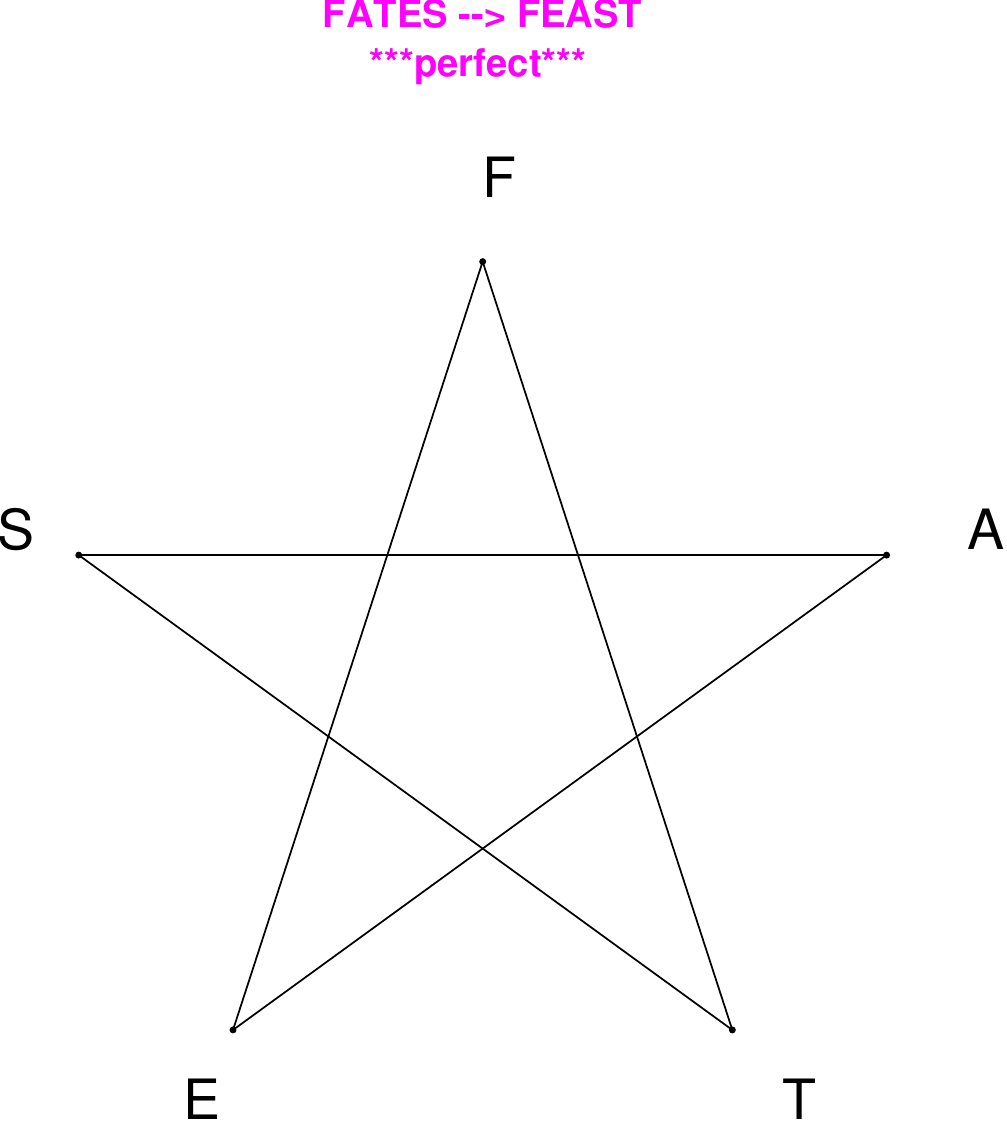}
\end{subfigure}
\hfill
\begin{subfigure}[T]{0.19\textwidth}
\centering
\includegraphics[width=\textwidth]{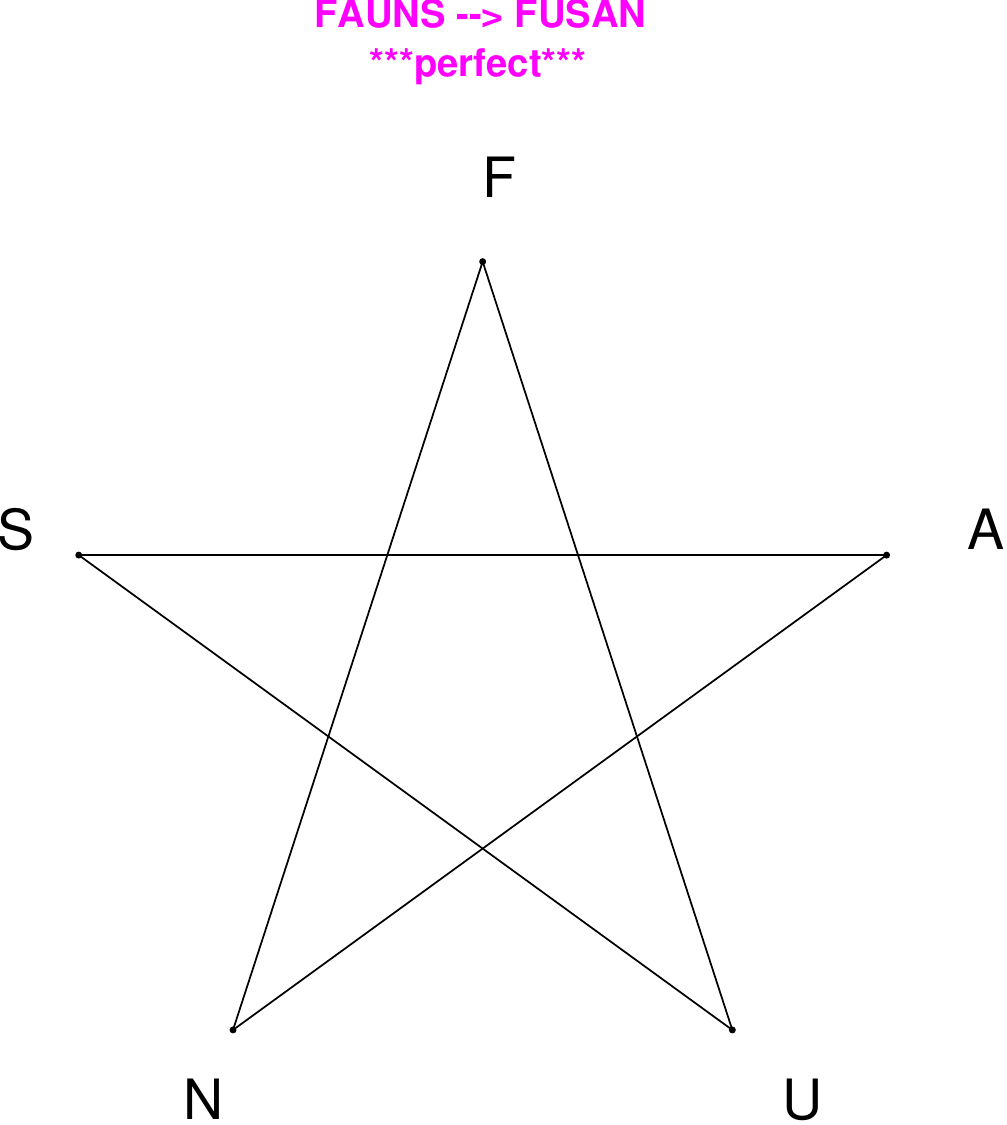}
\end{subfigure}
\hfill
\begin{subfigure}[T]{0.19\textwidth}
\centering
\includegraphics[width=\textwidth]{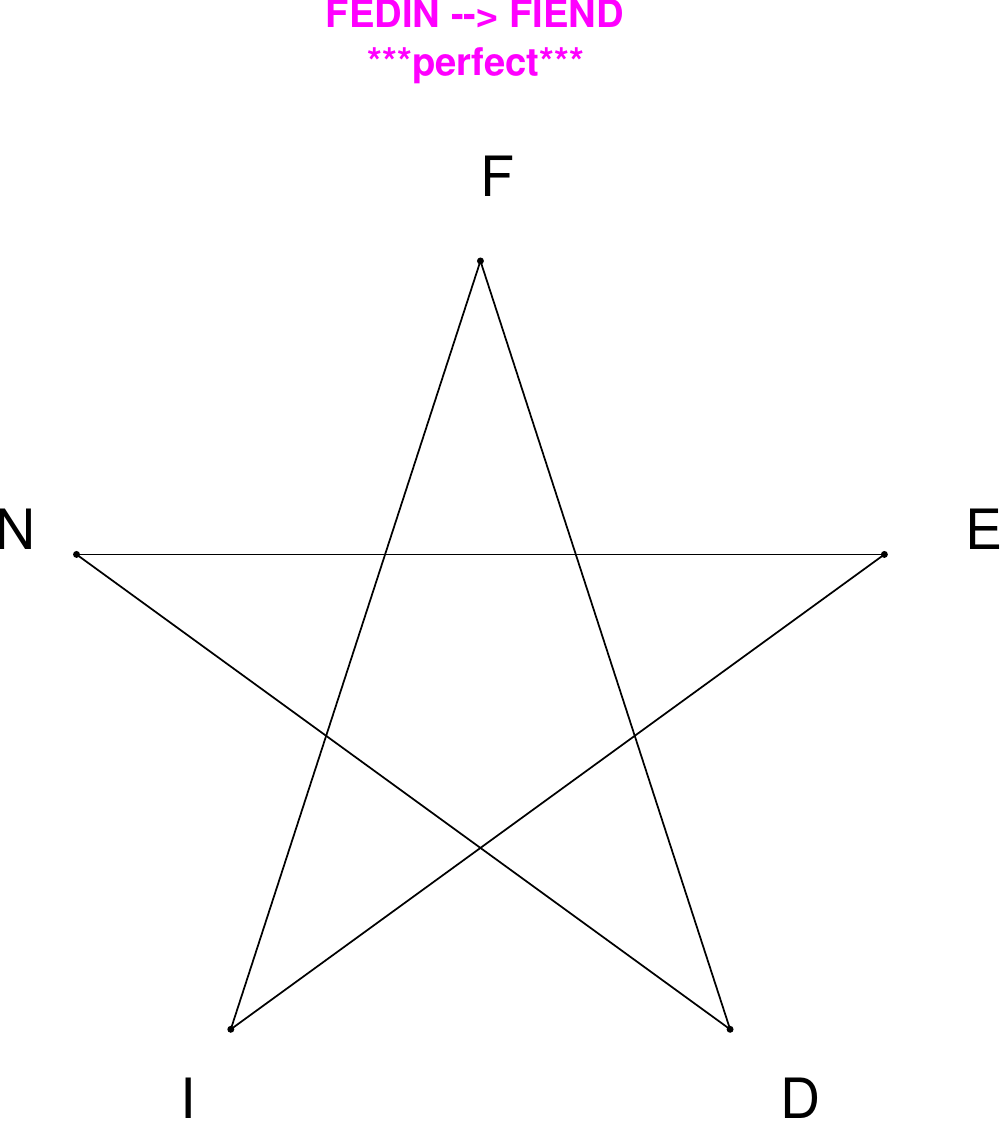}
\end{subfigure}
\end{figure}

\begin{figure}[H]
\centering
\begin{subfigure}[T]{0.19\textwidth}
\centering
\includegraphics[width=\textwidth]{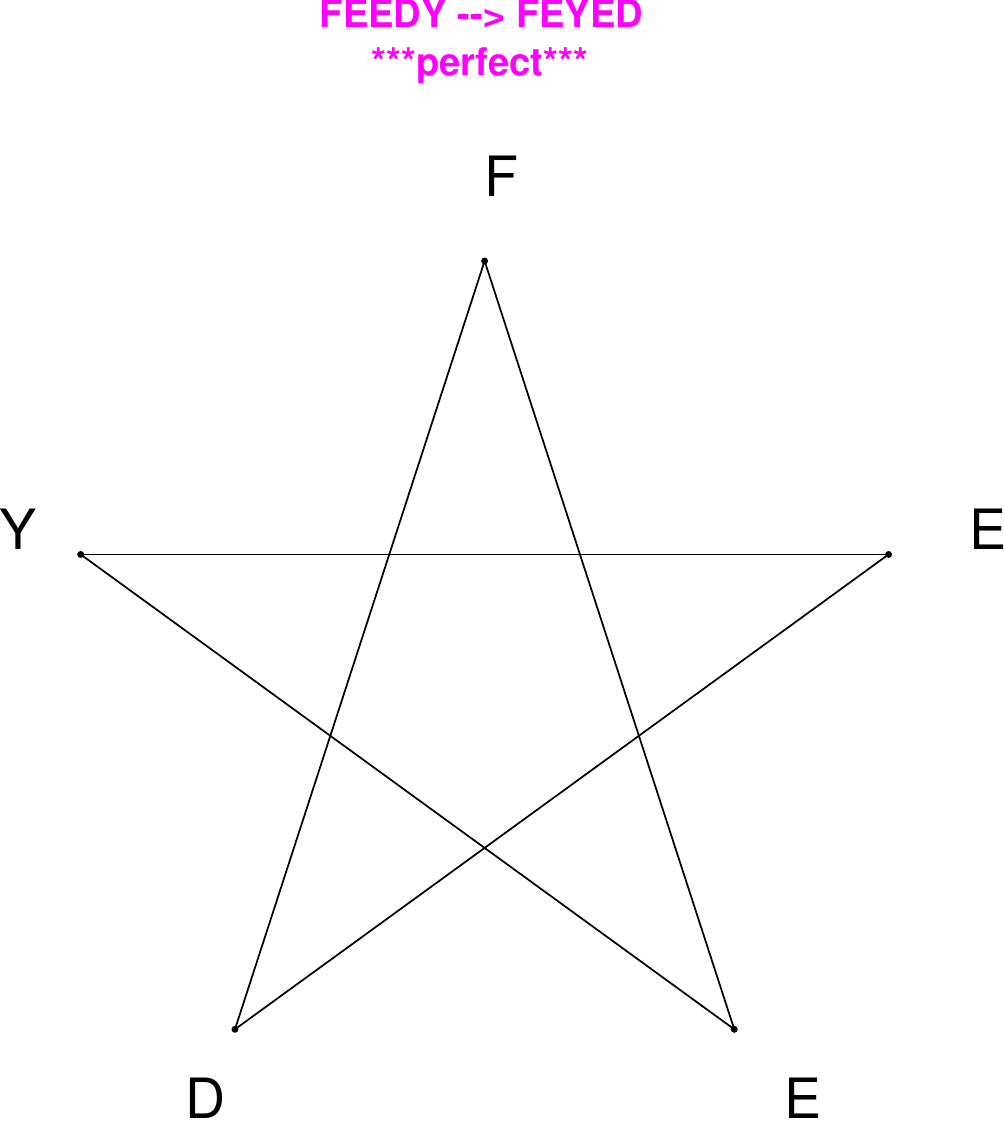}
\end{subfigure}
\hfill
\begin{subfigure}[T]{0.19\textwidth}
\centering
\includegraphics[width=\textwidth]{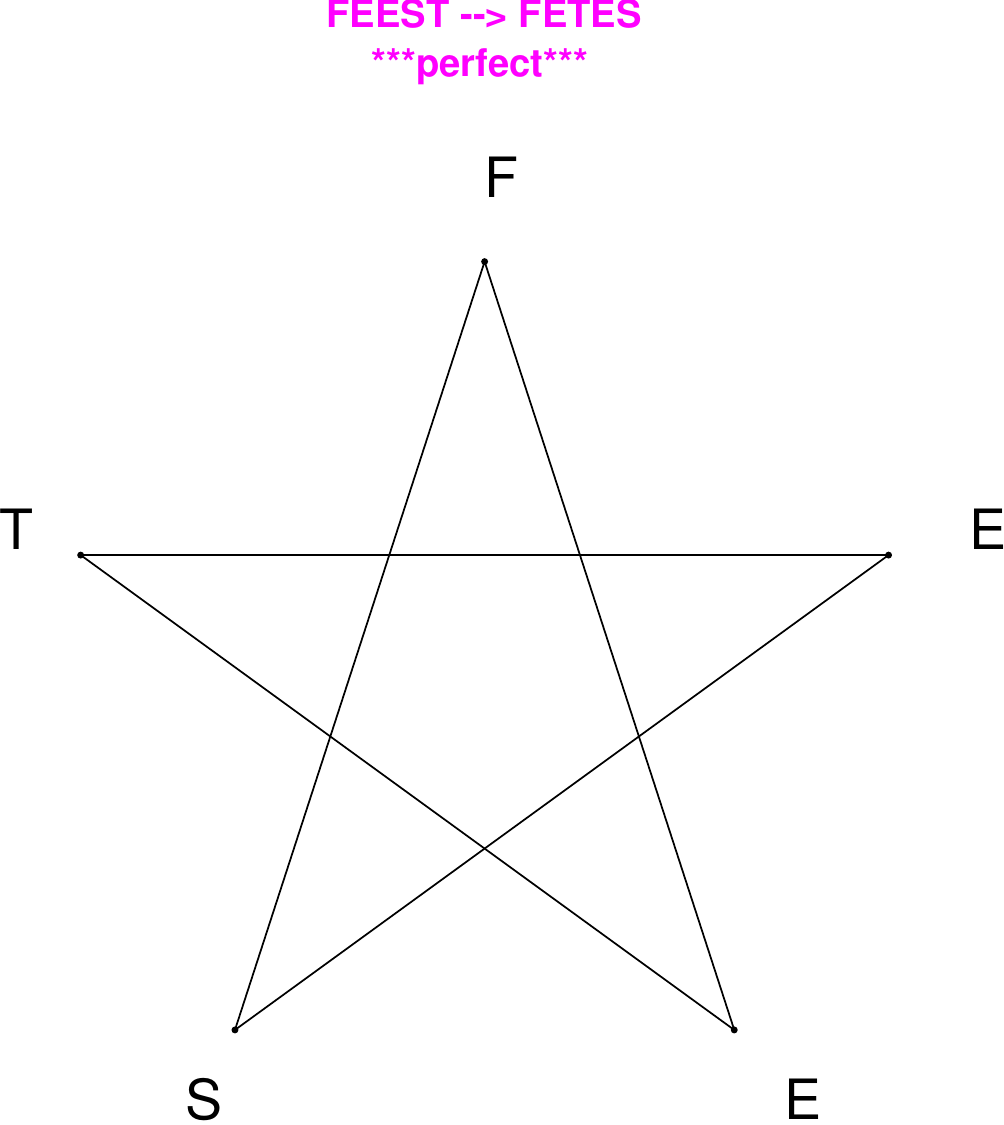}
\end{subfigure}
\hfill
\begin{subfigure}[T]{0.19\textwidth}
\centering
\includegraphics[width=\textwidth]{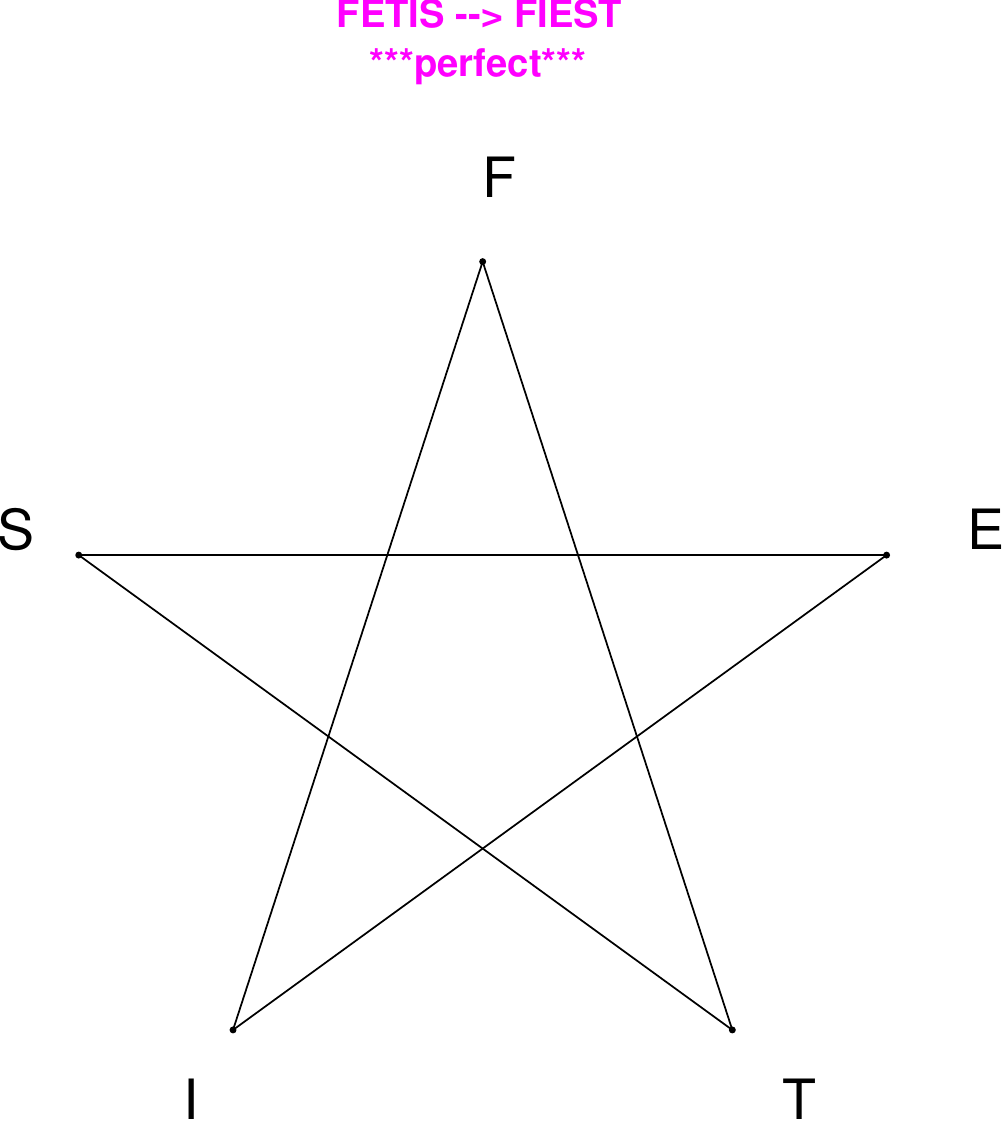}
\end{subfigure}
\hfill
\begin{subfigure}[T]{0.19\textwidth}
\centering
\includegraphics[width=\textwidth]{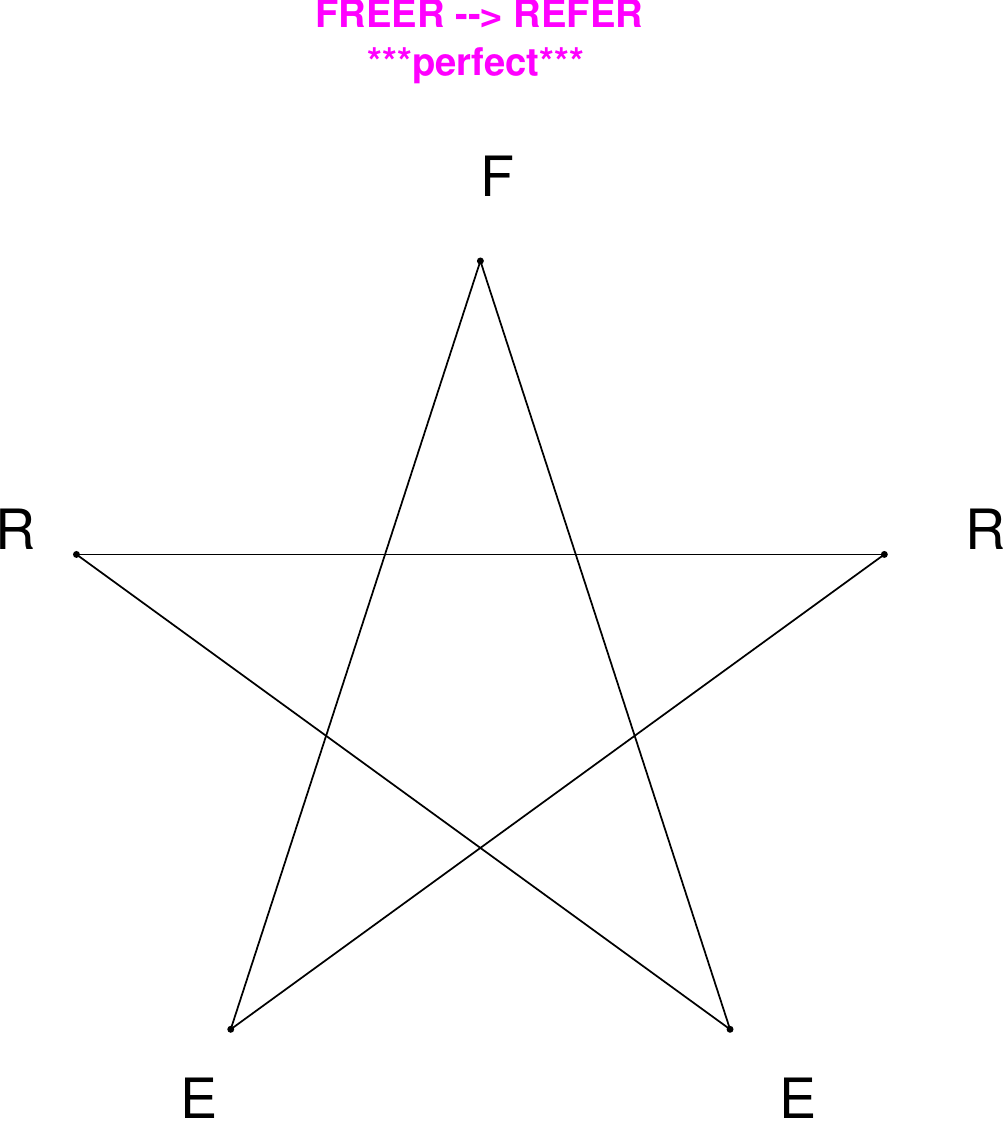}
\end{subfigure}
\hfill
\begin{subfigure}[T]{0.19\textwidth}
\centering
\includegraphics[width=\textwidth]{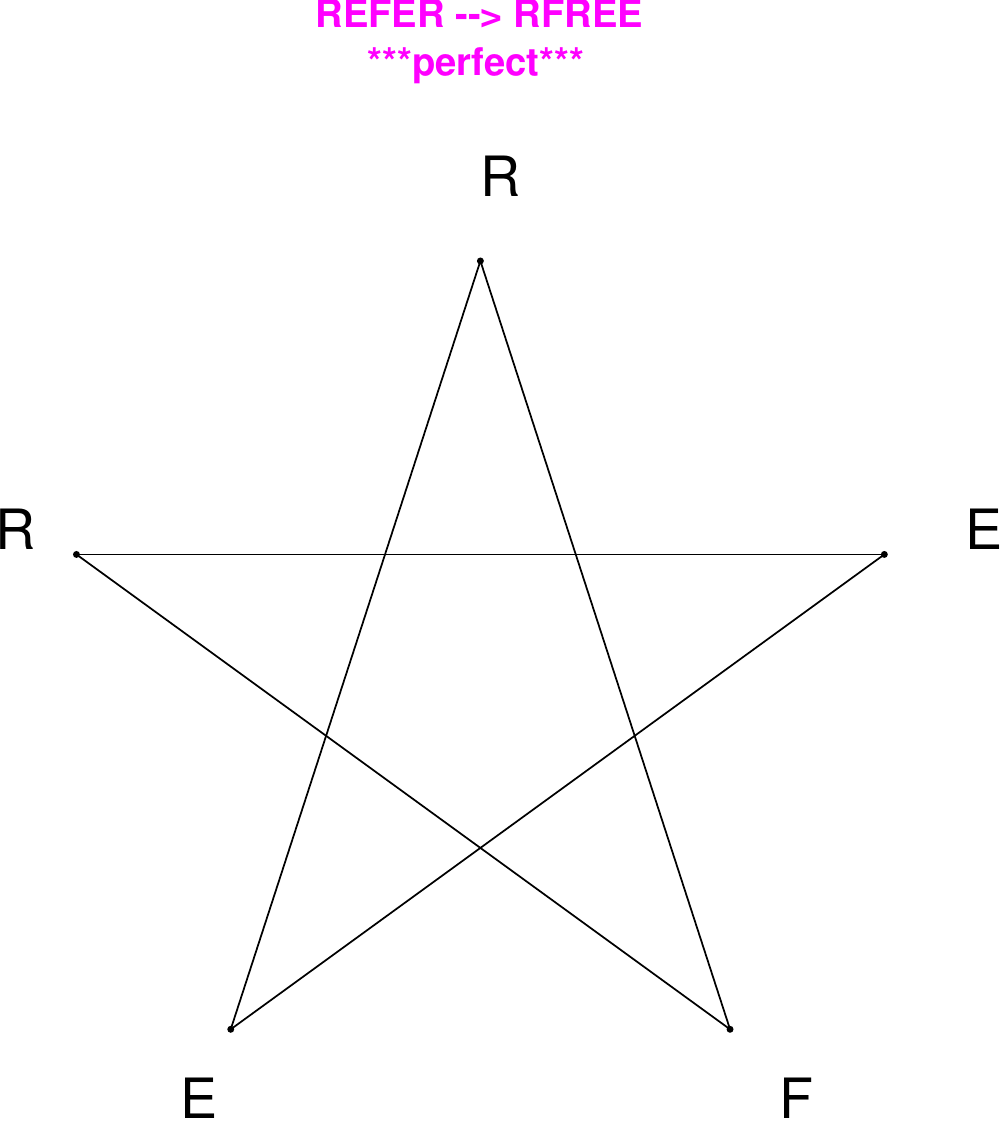}
\end{subfigure}
\end{figure}

\begin{figure}[H]
\centering
\begin{subfigure}[T]{0.19\textwidth}
\centering
\includegraphics[width=\textwidth]{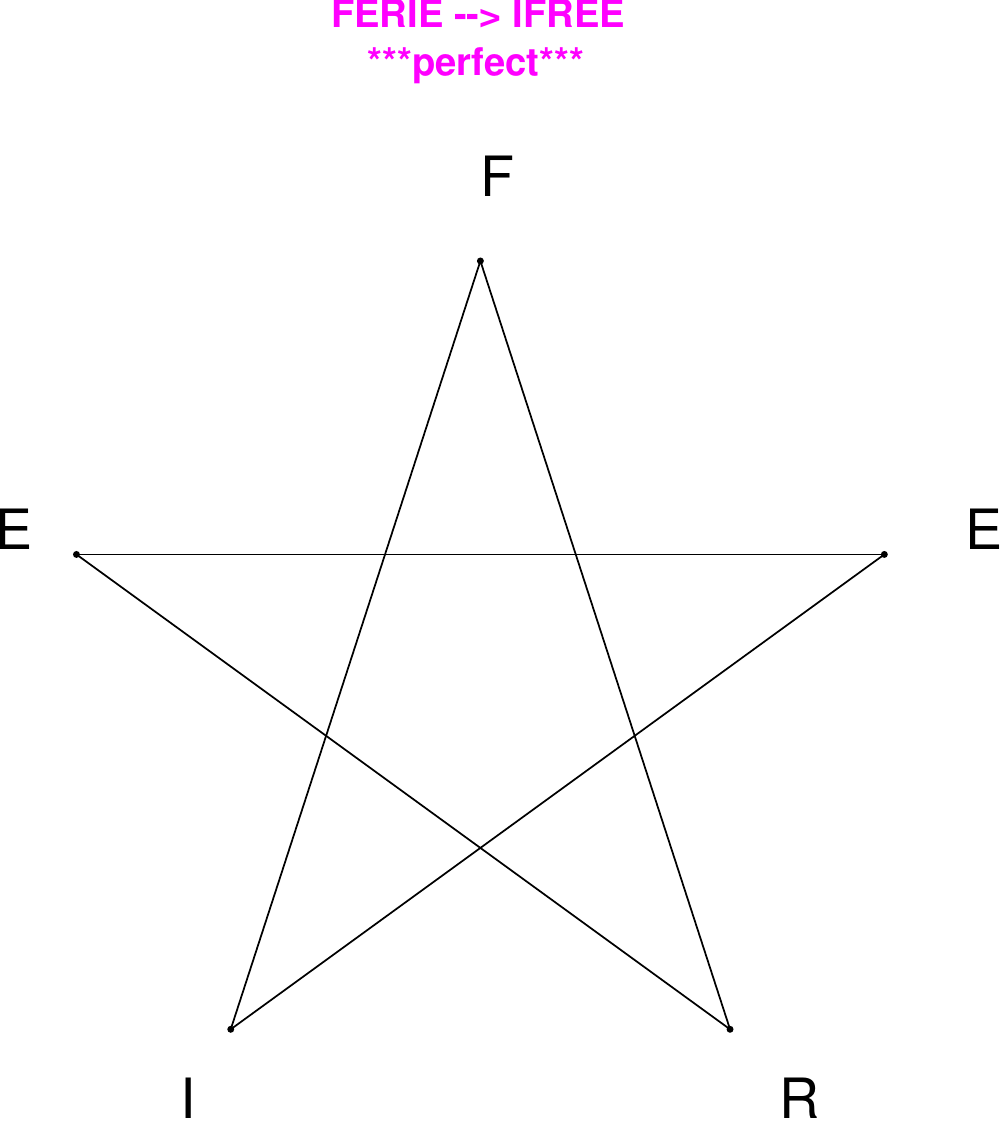}
\end{subfigure}
\hfill
\begin{subfigure}[T]{0.19\textwidth}
\centering
\includegraphics[width=\textwidth]{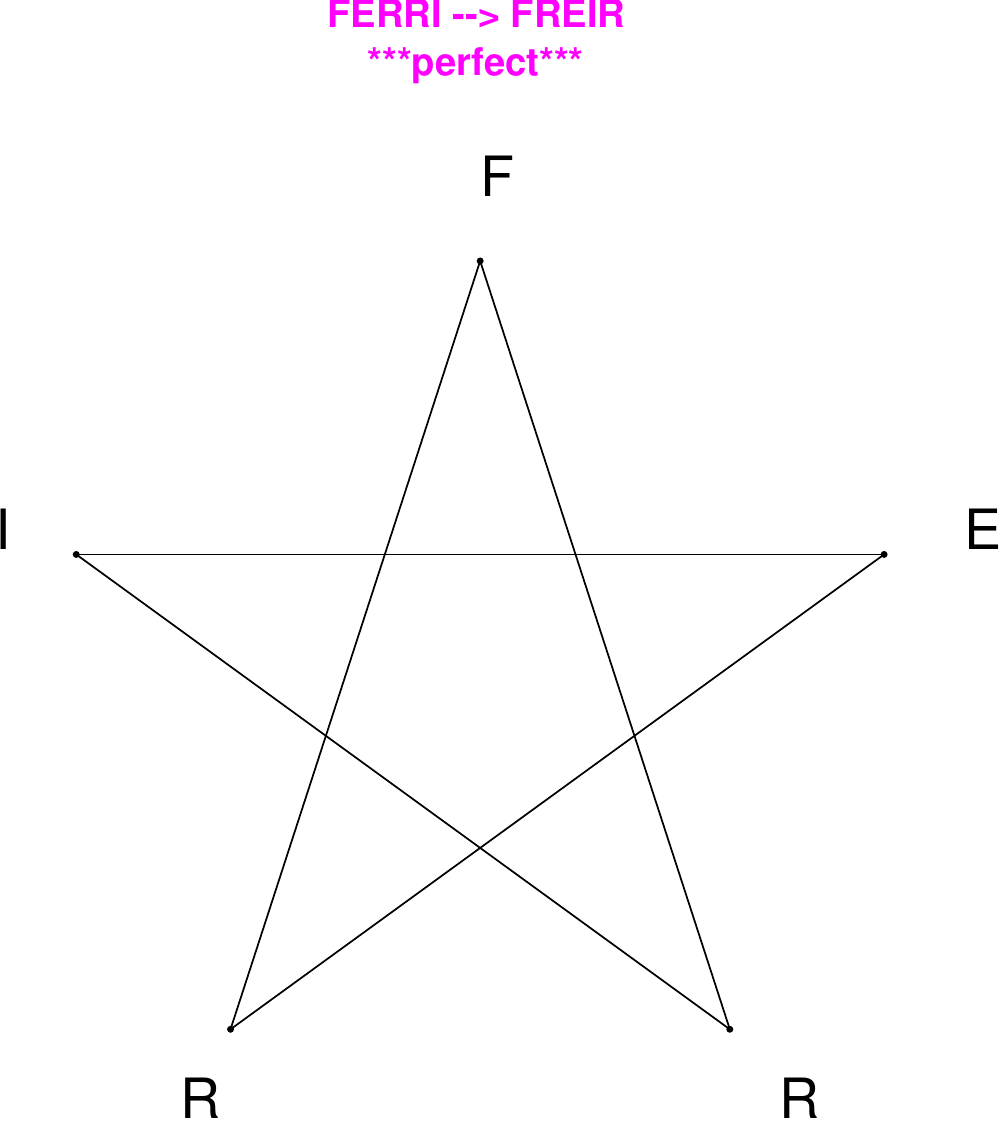}
\end{subfigure}
\hfill
\begin{subfigure}[T]{0.19\textwidth}
\centering
\includegraphics[width=\textwidth]{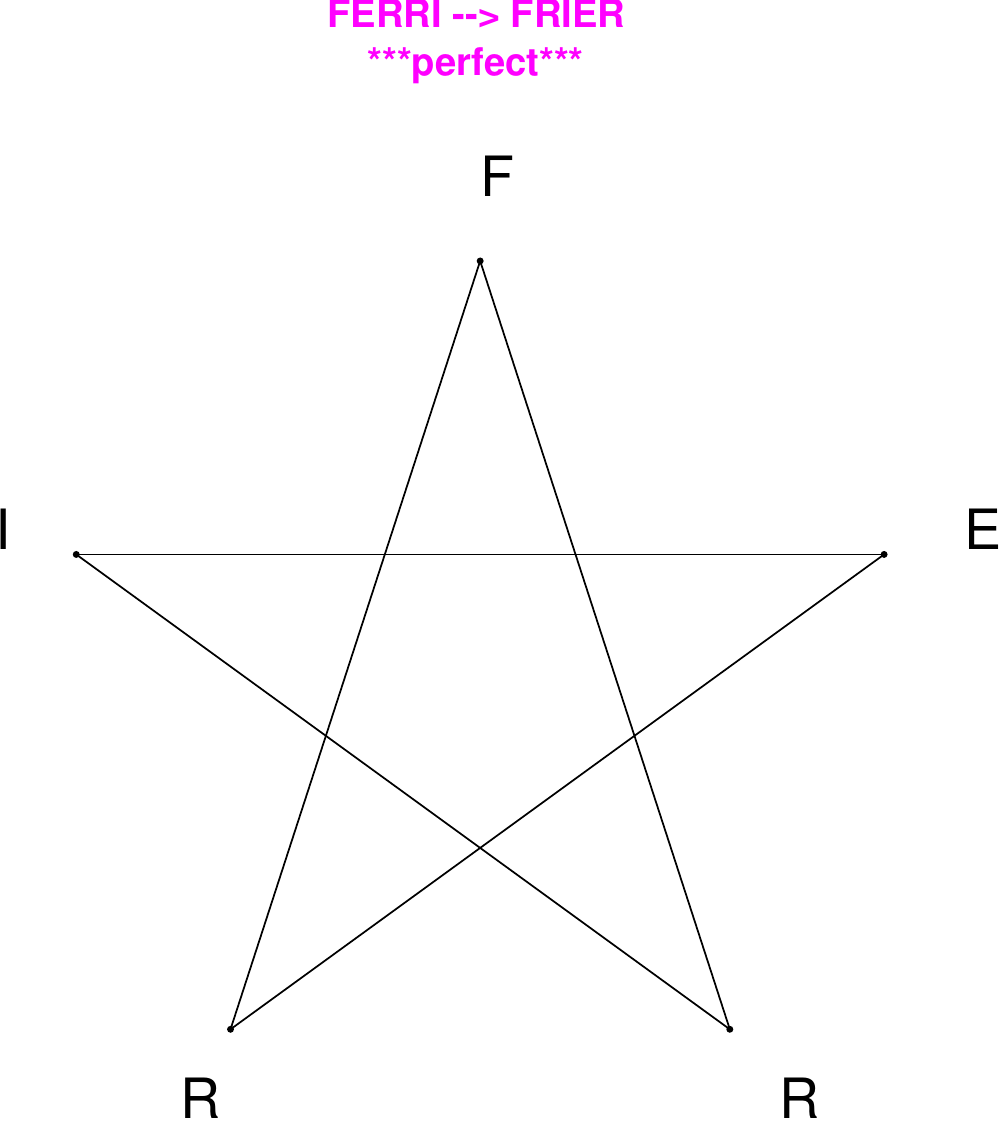}
\end{subfigure}
\hfill
\begin{subfigure}[T]{0.19\textwidth}
\centering
\includegraphics[width=\textwidth]{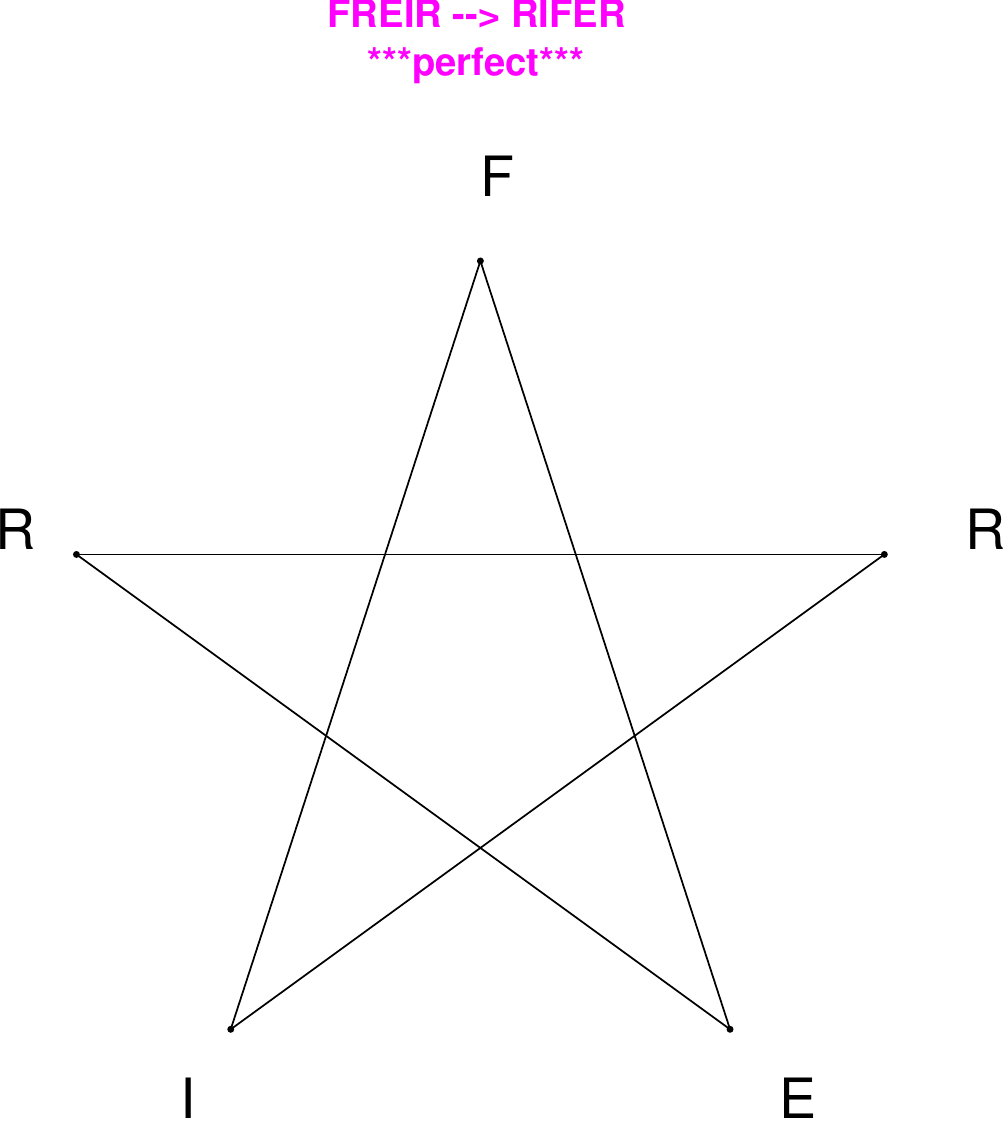}
\end{subfigure}
\hfill
\begin{subfigure}[T]{0.19\textwidth}
\centering
\includegraphics[width=\textwidth]{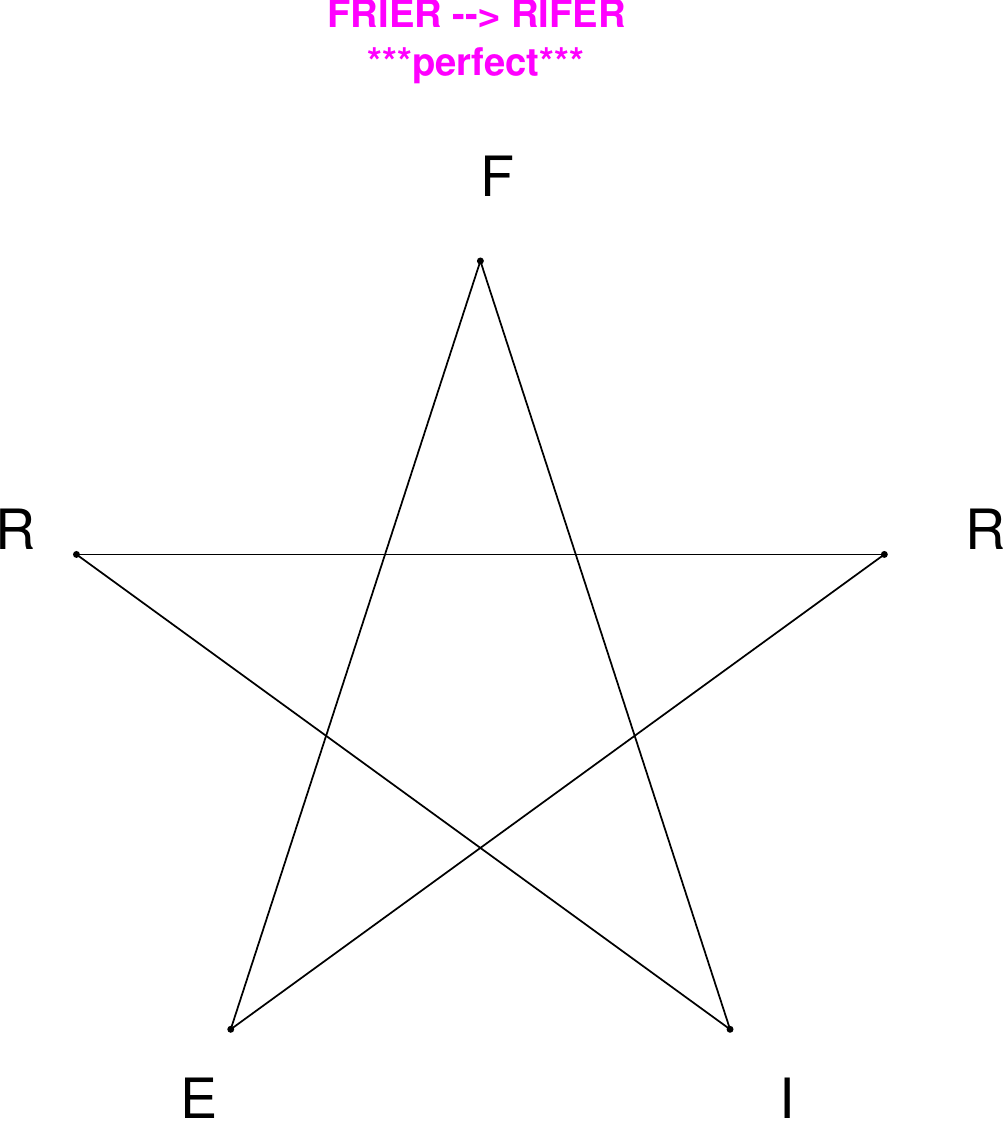}
\end{subfigure}
\end{figure}

\begin{figure}[H]
\centering
\begin{subfigure}[T]{0.19\textwidth}
\centering
\includegraphics[width=\textwidth]{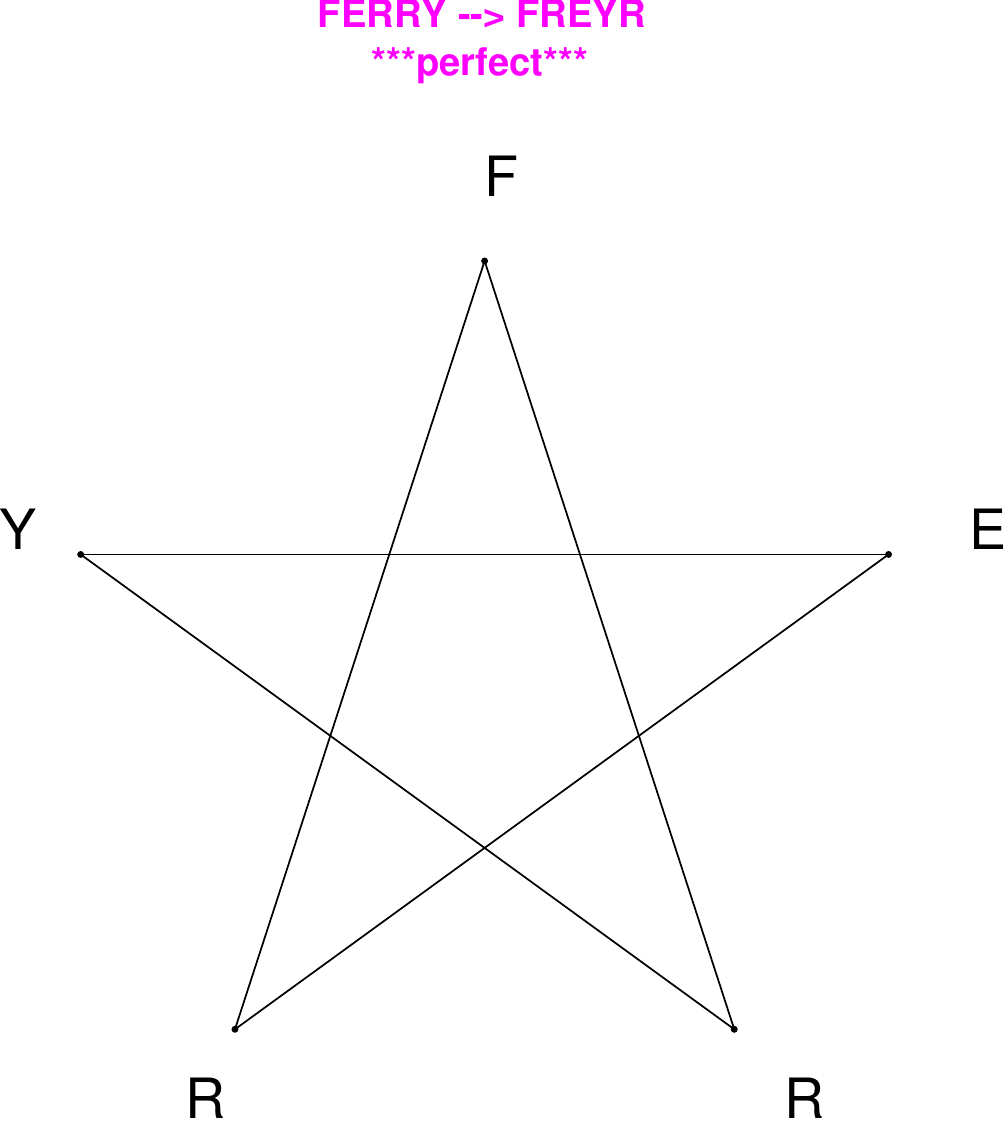}
\end{subfigure}
\hfill
\begin{subfigure}[T]{0.19\textwidth}
\centering
\includegraphics[width=\textwidth]{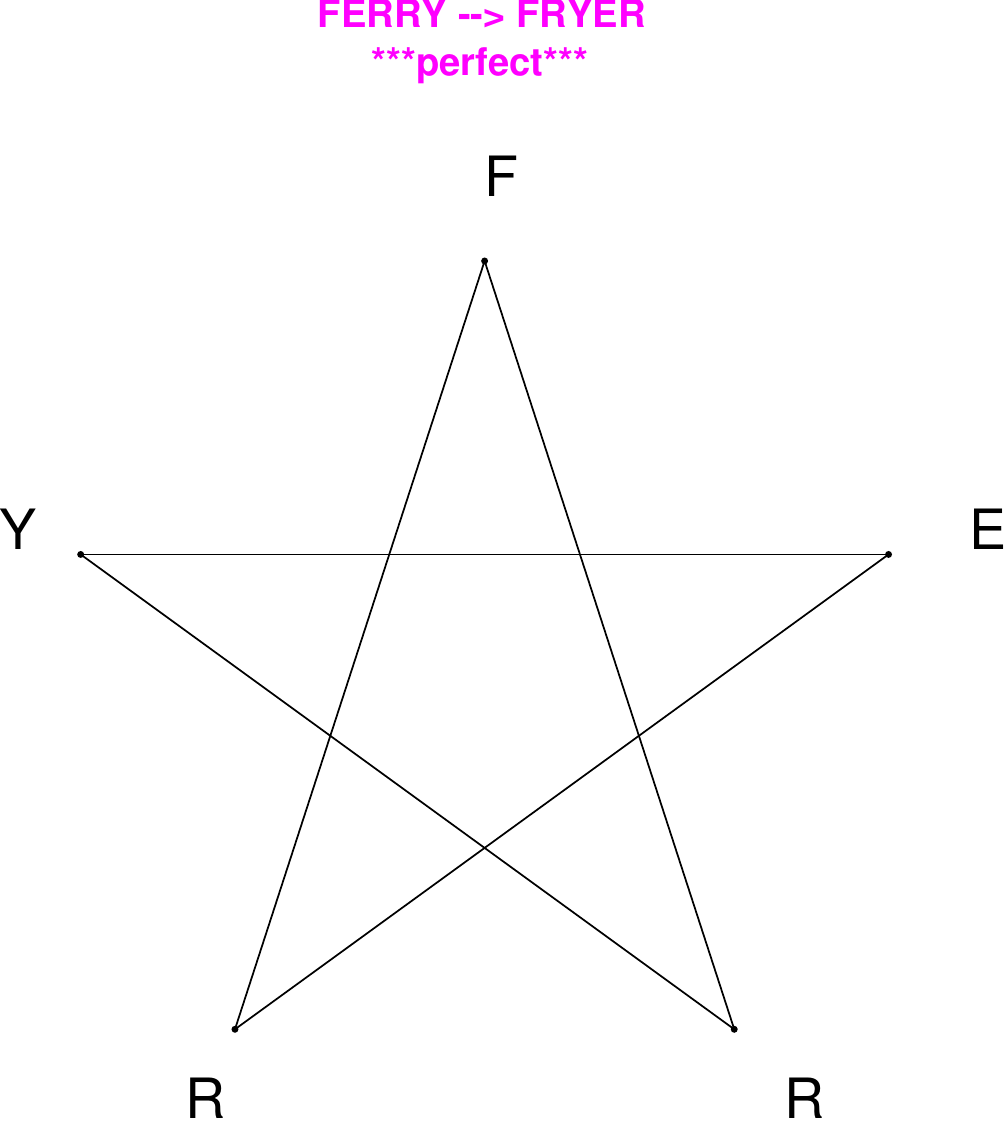}
\end{subfigure}
\hfill
\begin{subfigure}[T]{0.19\textwidth}
\centering
\includegraphics[width=\textwidth]{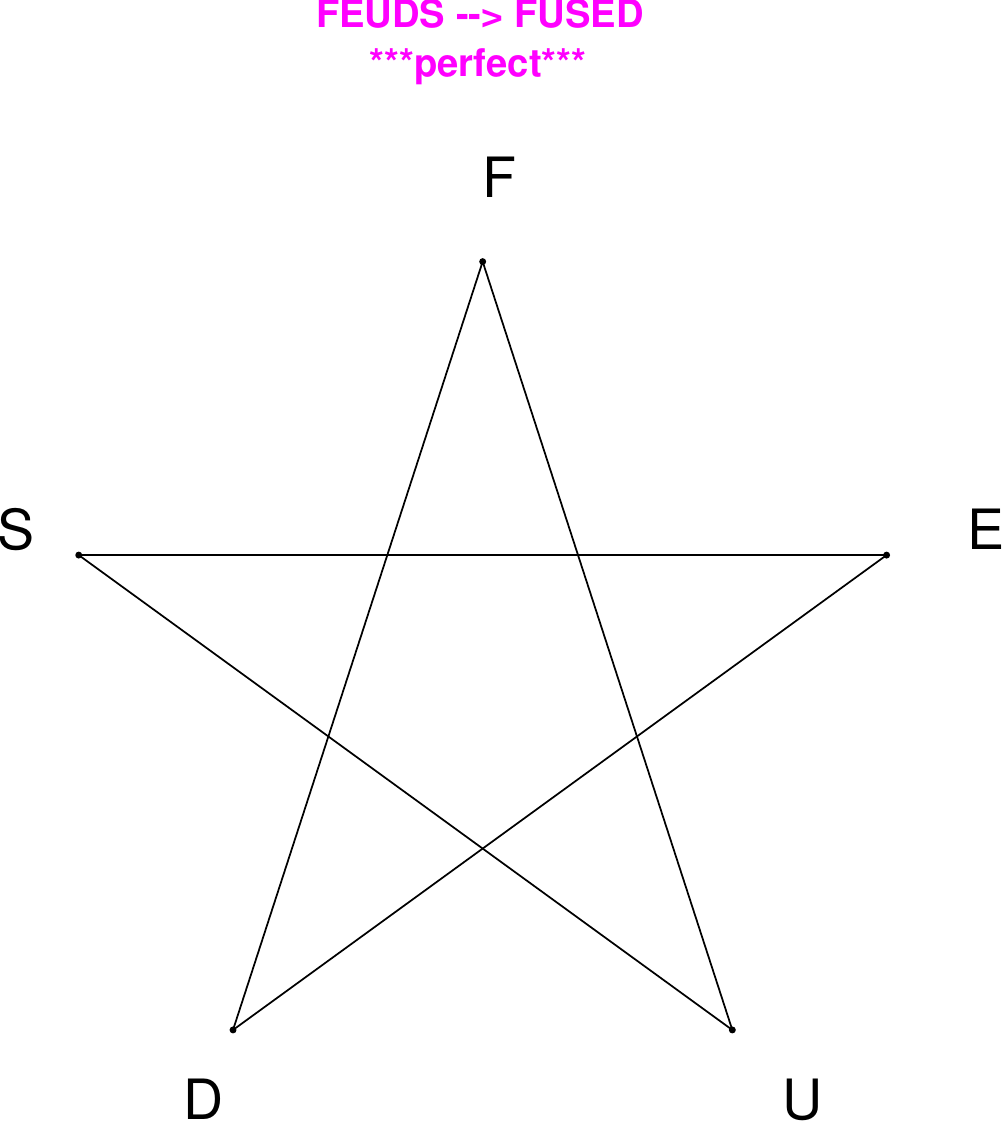}
\end{subfigure}
\hfill
\begin{subfigure}[T]{0.19\textwidth}
\centering
\includegraphics[width=\textwidth]{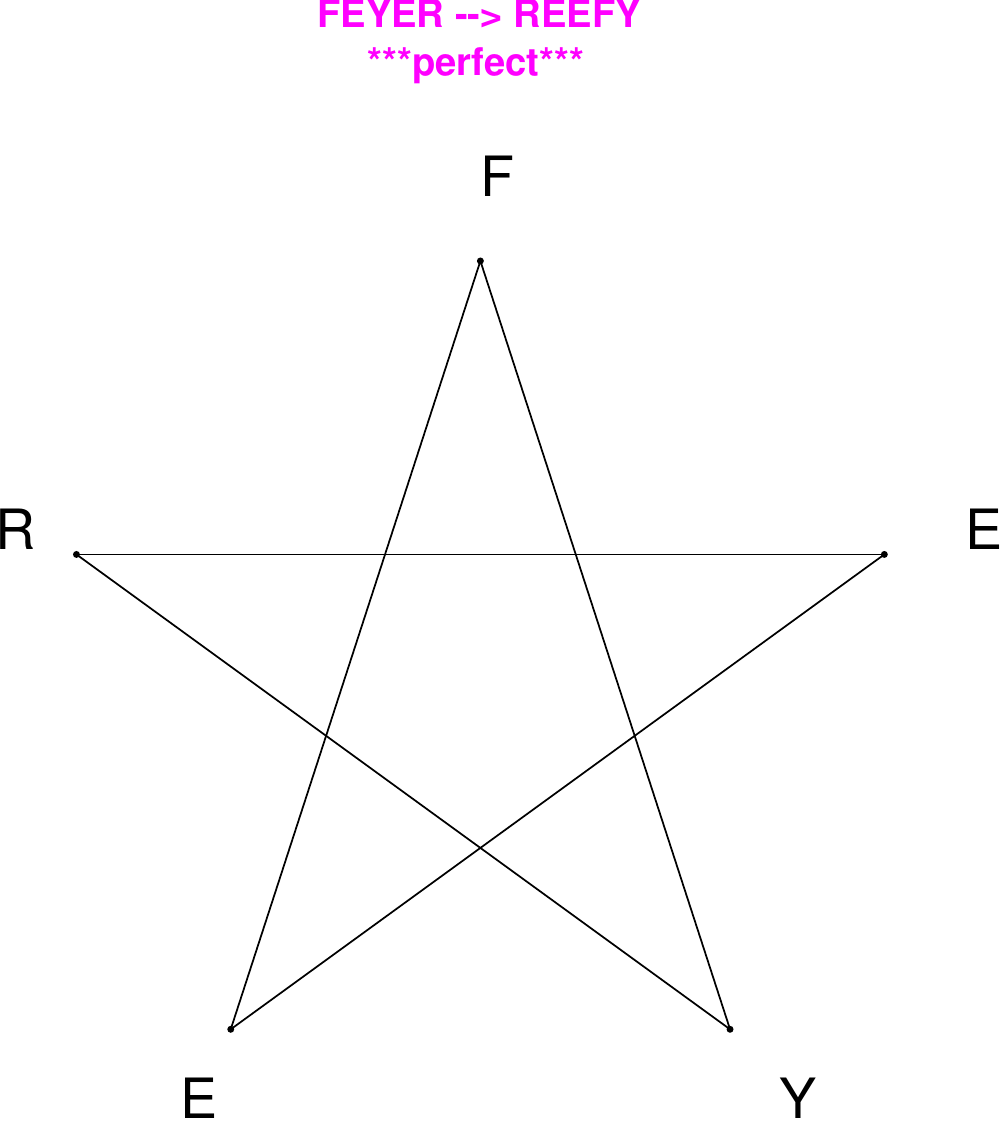}
\end{subfigure}
\hfill
\begin{subfigure}[T]{0.19\textwidth}
\centering
\includegraphics[width=\textwidth]{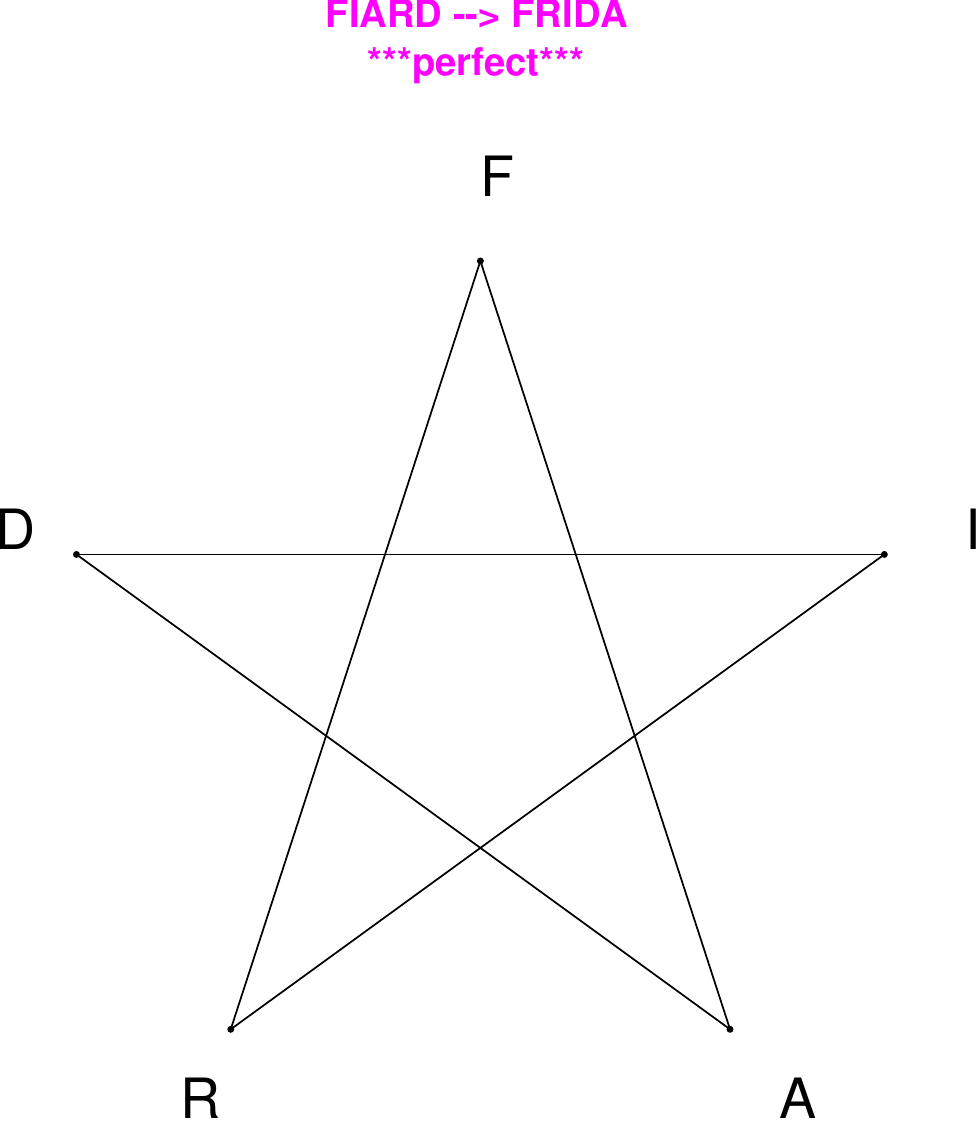}
\end{subfigure}
\end{figure}

\begin{figure}[H]
\centering
\begin{subfigure}[T]{0.19\textwidth}
\centering
\includegraphics[width=\textwidth]{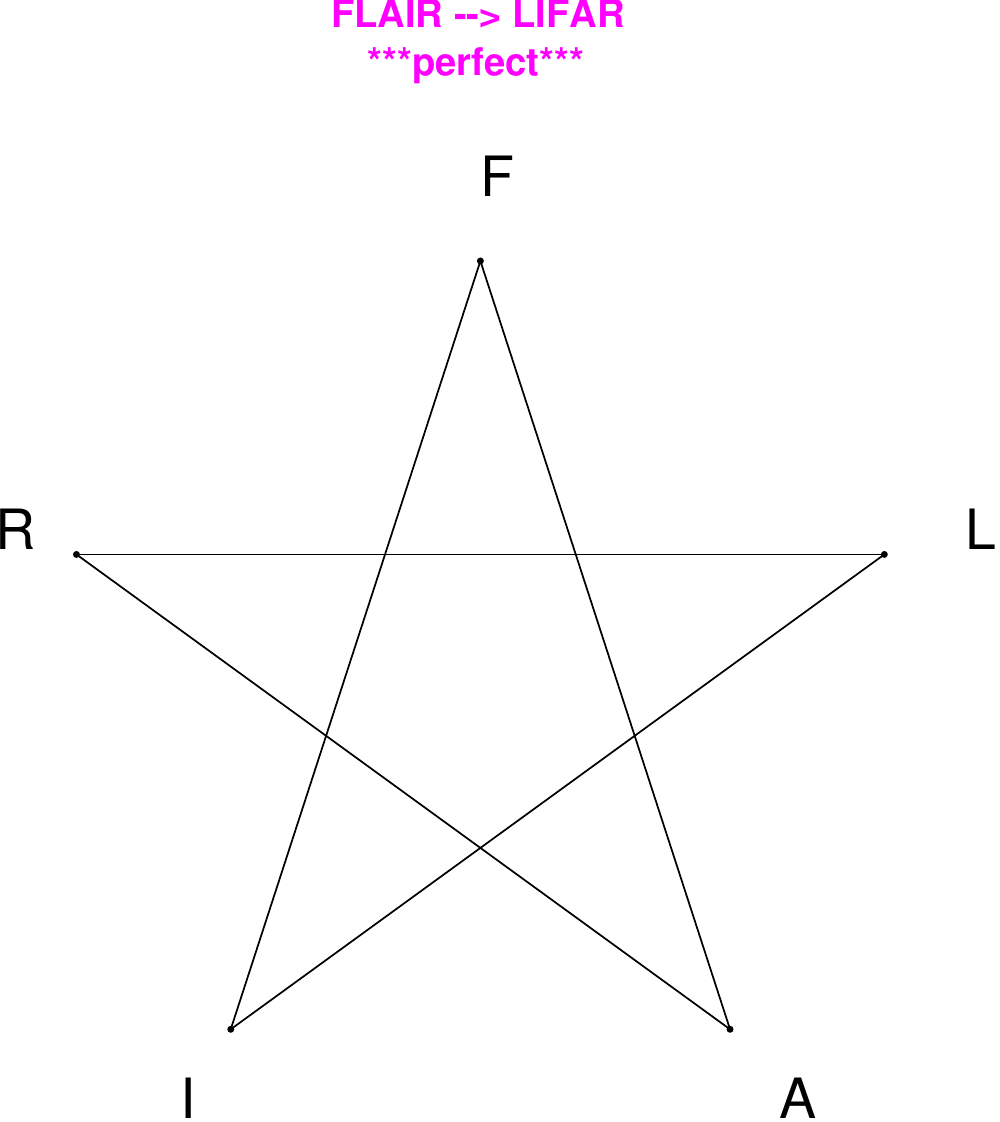}
\end{subfigure}
\hfill
\begin{subfigure}[T]{0.19\textwidth}
\centering
\includegraphics[width=\textwidth]{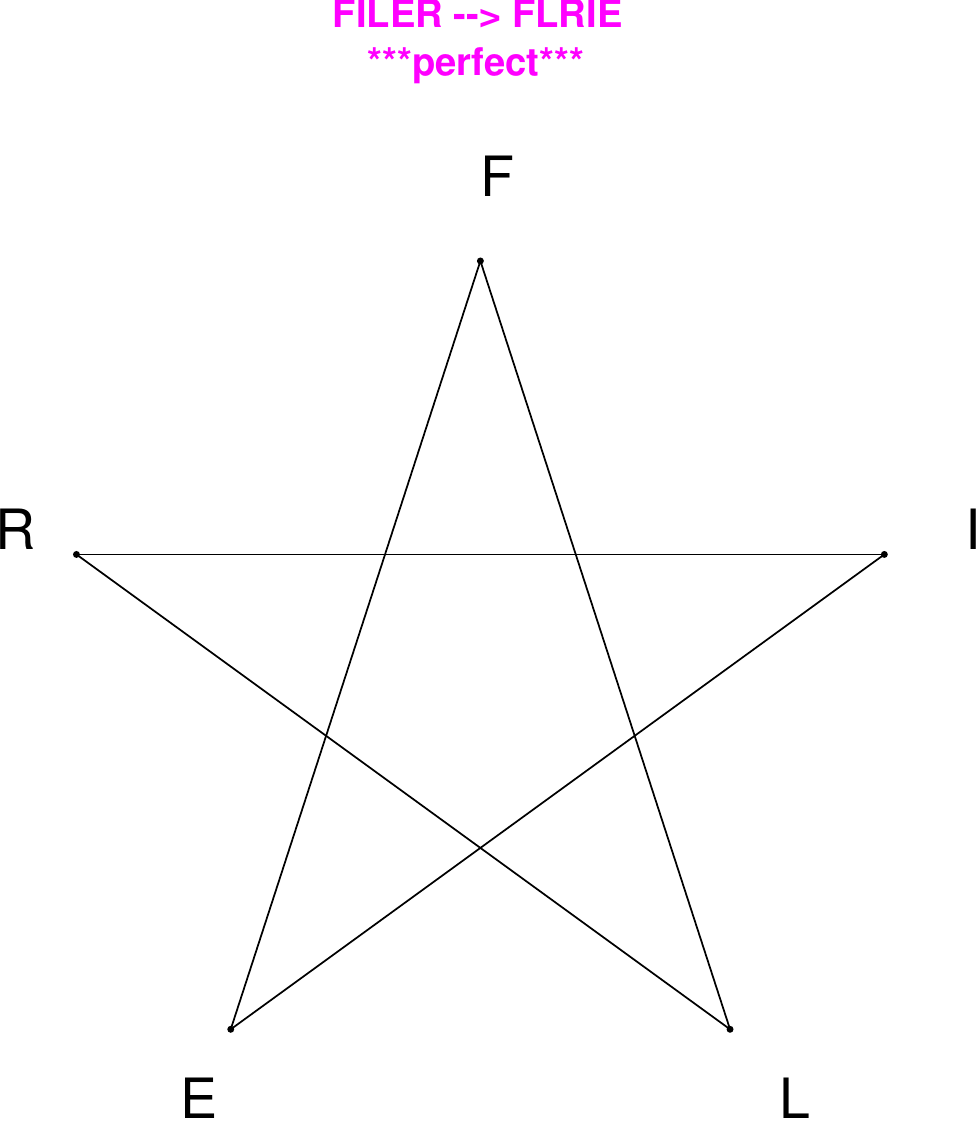}
\end{subfigure}
\hfill
\begin{subfigure}[T]{0.19\textwidth}
\centering
\includegraphics[width=\textwidth]{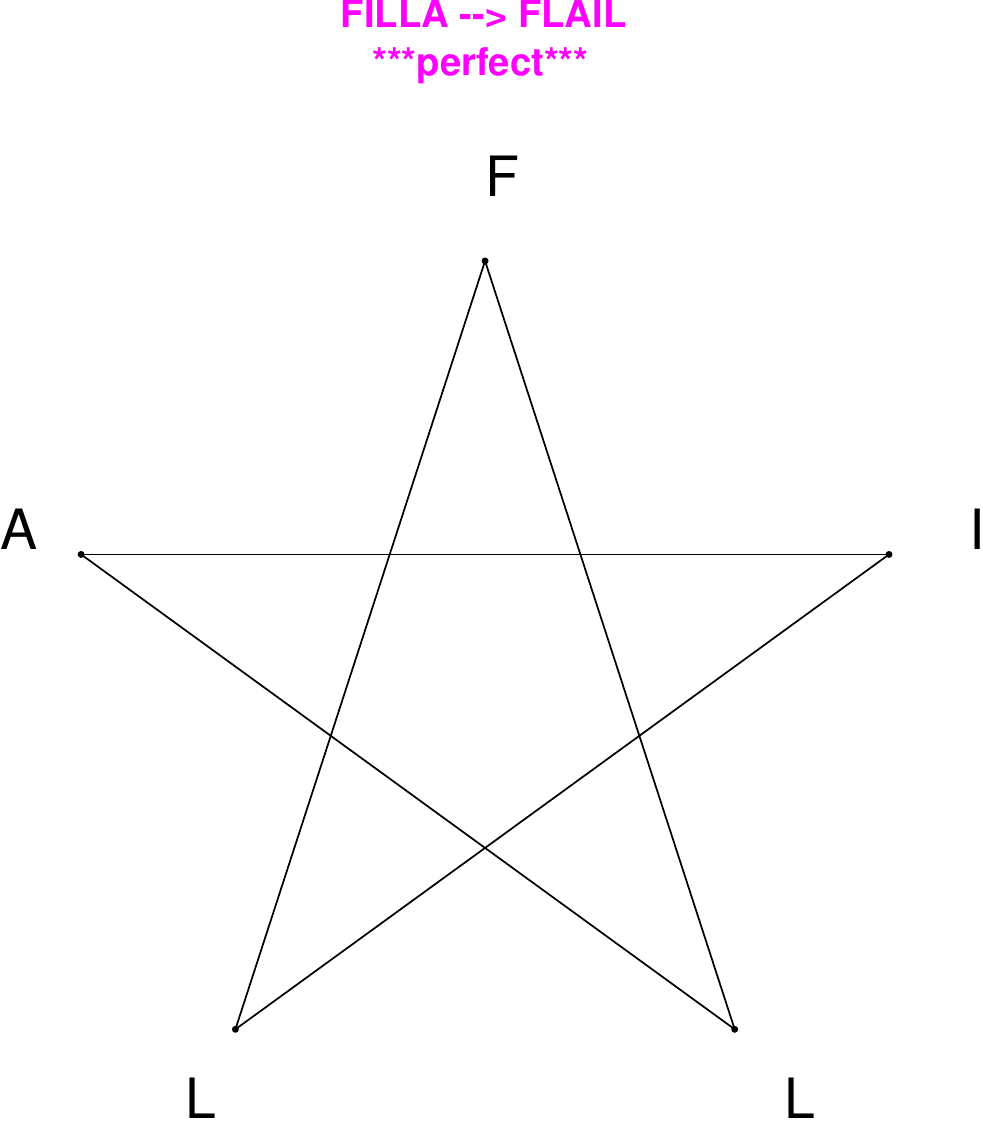}
\end{subfigure}
\hfill
\begin{subfigure}[T]{0.19\textwidth}
\centering
\includegraphics[width=\textwidth]{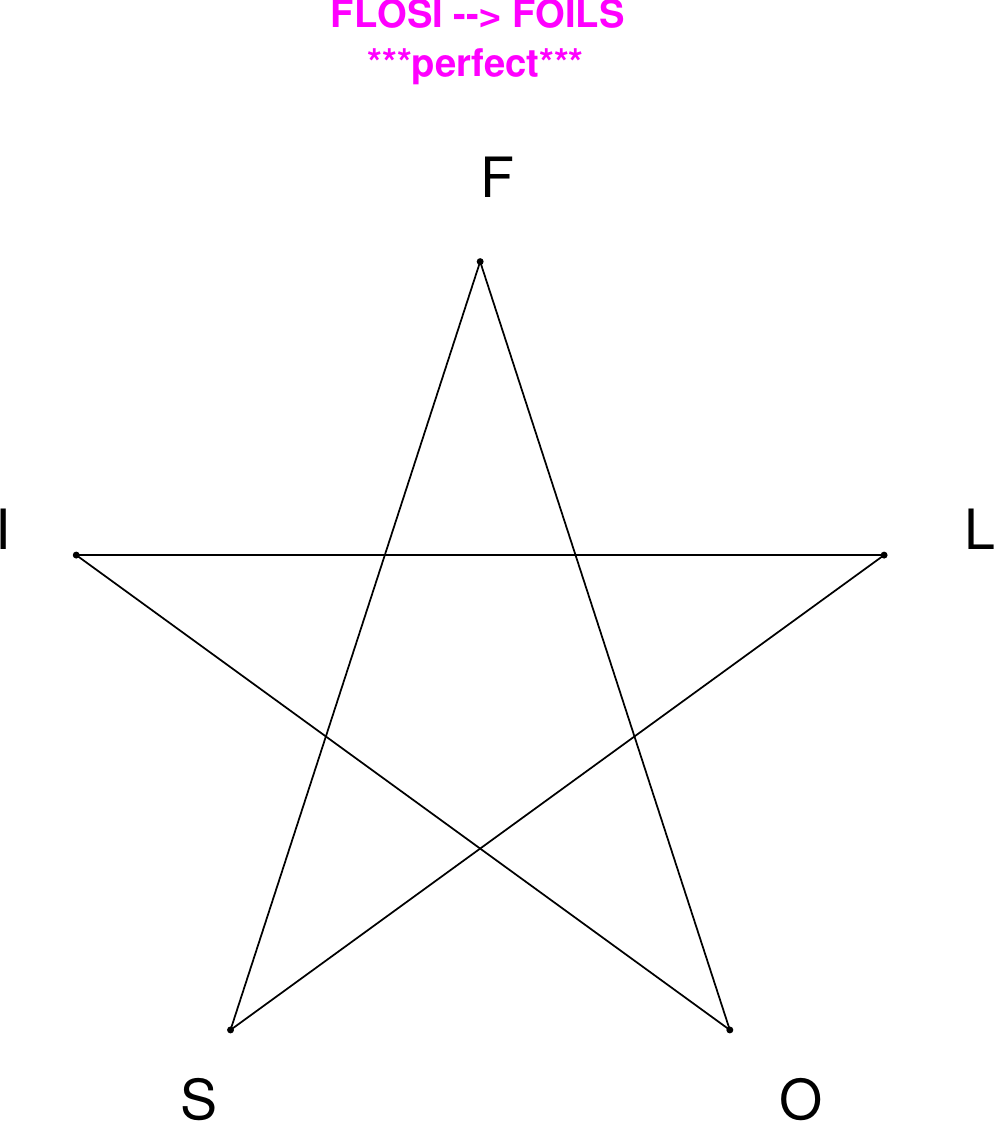}
\end{subfigure}
\hfill
\begin{subfigure}[T]{0.19\textwidth}
\centering
\includegraphics[width=\textwidth]{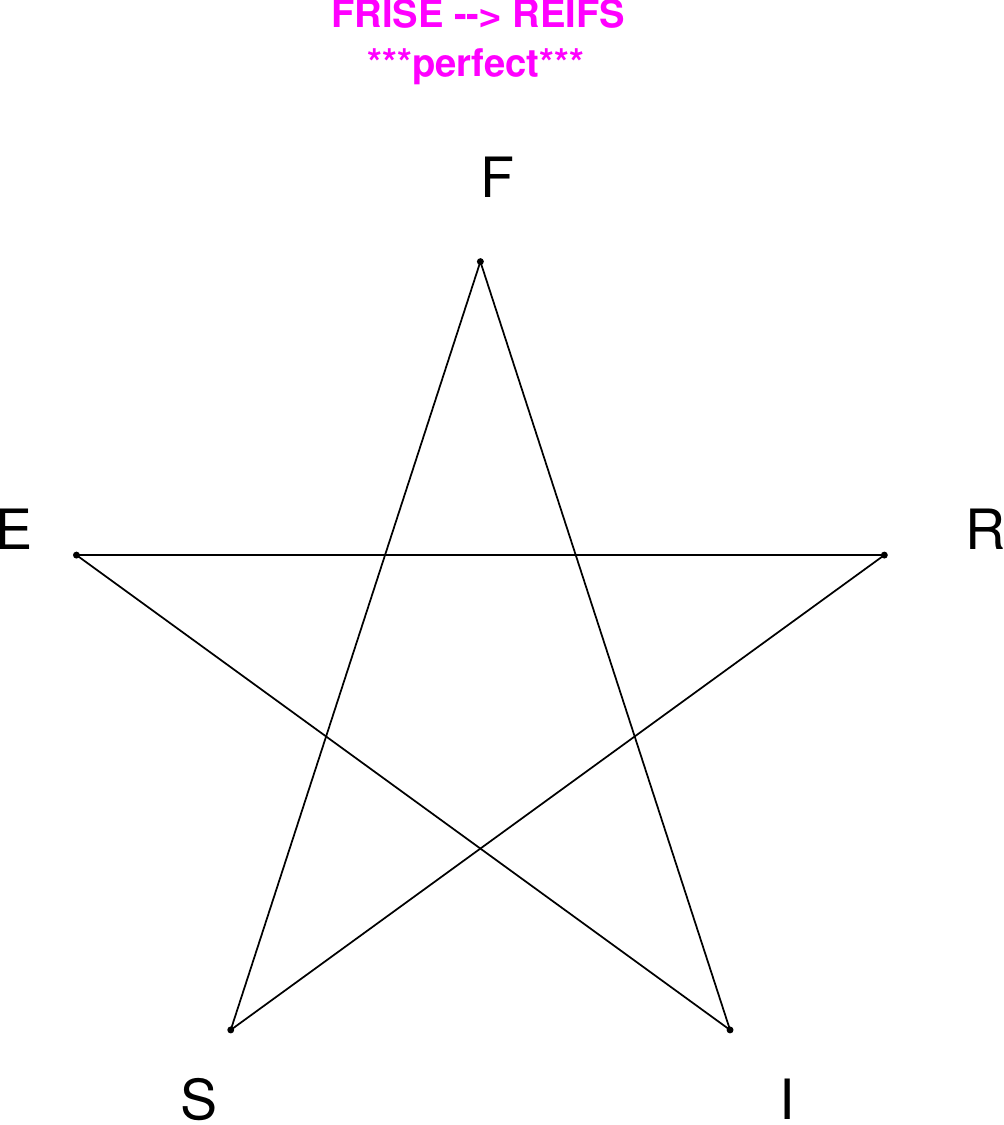}
\end{subfigure}
\end{figure}

\begin{figure}[H]
\centering
\begin{subfigure}[T]{0.19\textwidth}
\centering
\includegraphics[width=\textwidth]{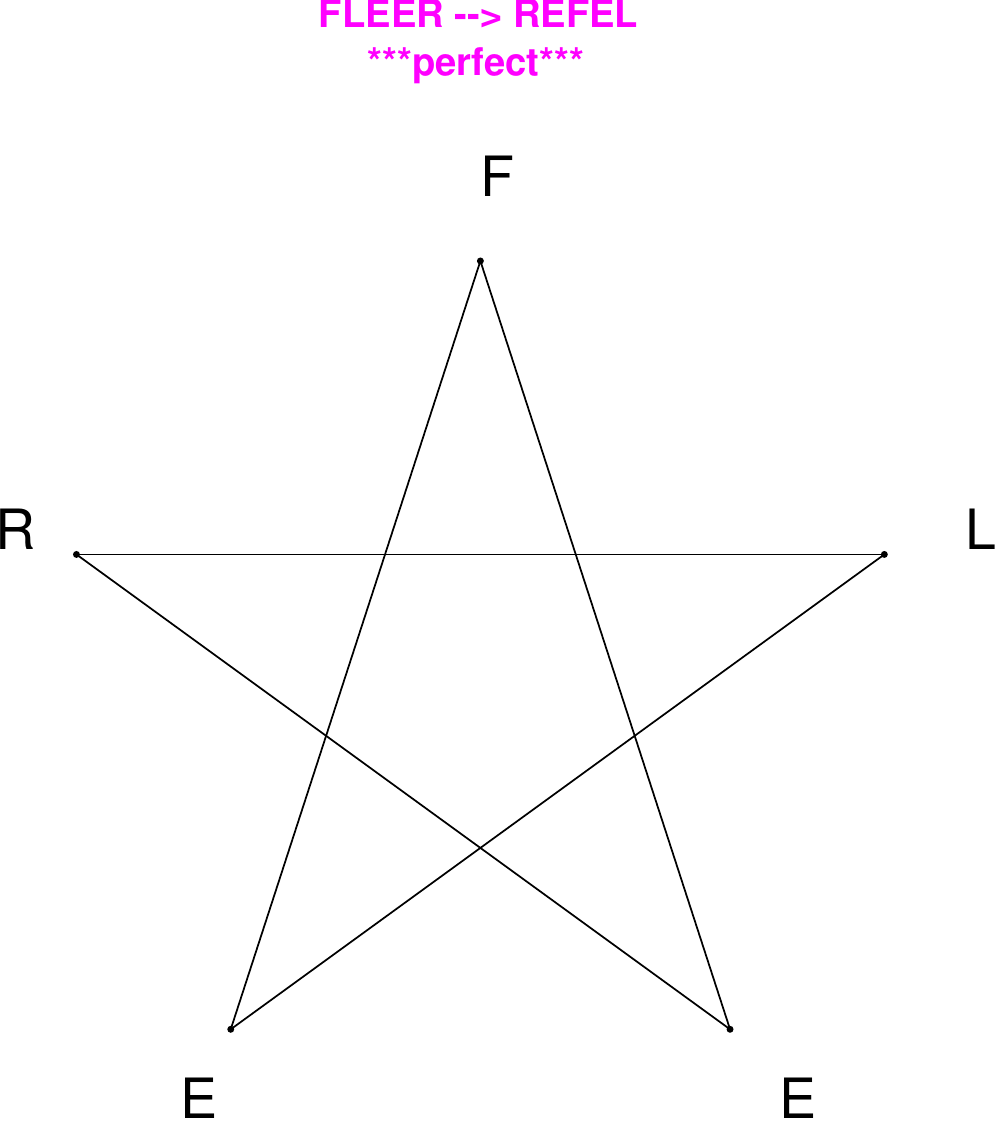}
\end{subfigure}
\hfill
\begin{subfigure}[T]{0.19\textwidth}
\centering
\includegraphics[width=\textwidth]{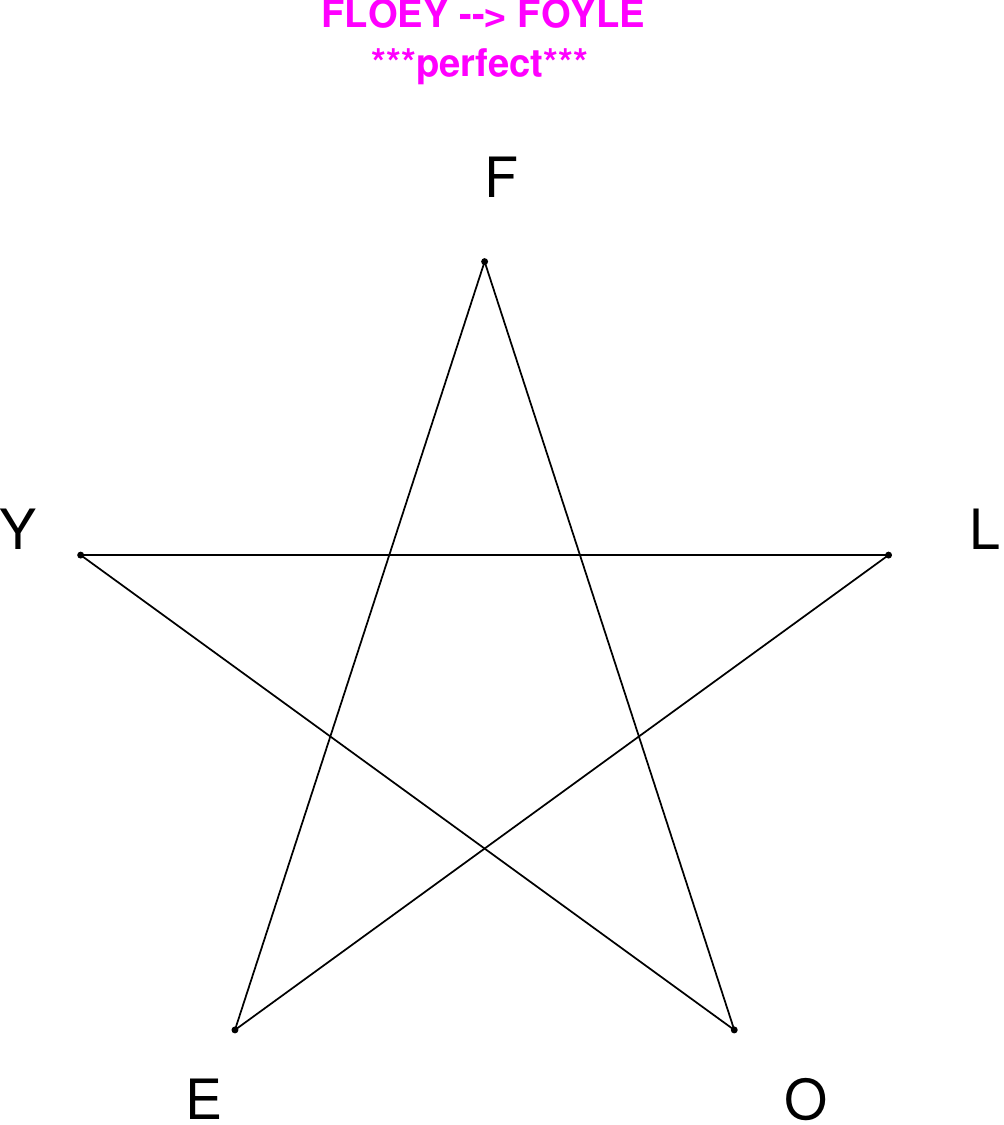}
\end{subfigure}
\hfill
\begin{subfigure}[T]{0.19\textwidth}
\centering
\includegraphics[width=\textwidth]{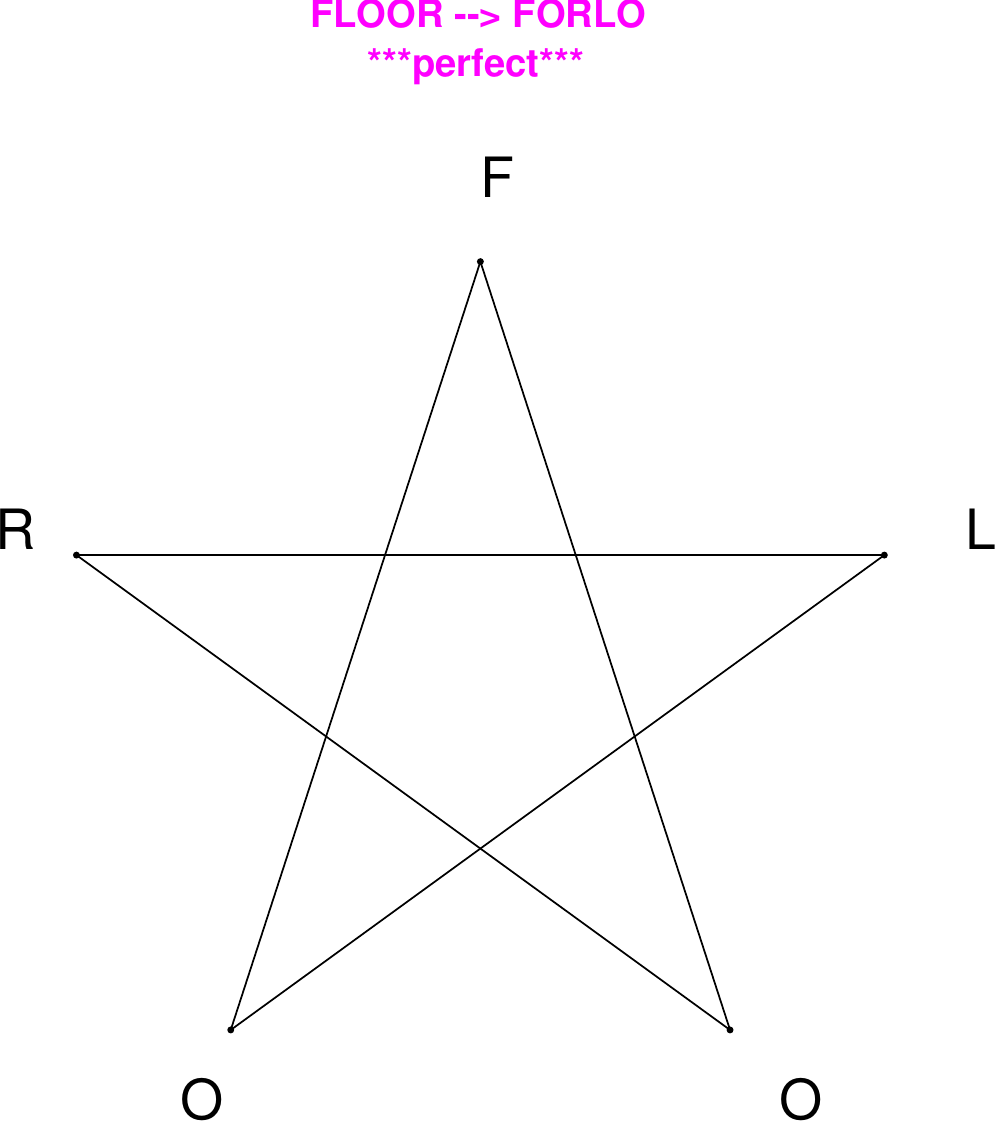}
\end{subfigure}
\hfill
\begin{subfigure}[T]{0.19\textwidth}
\centering
\includegraphics[width=\textwidth]{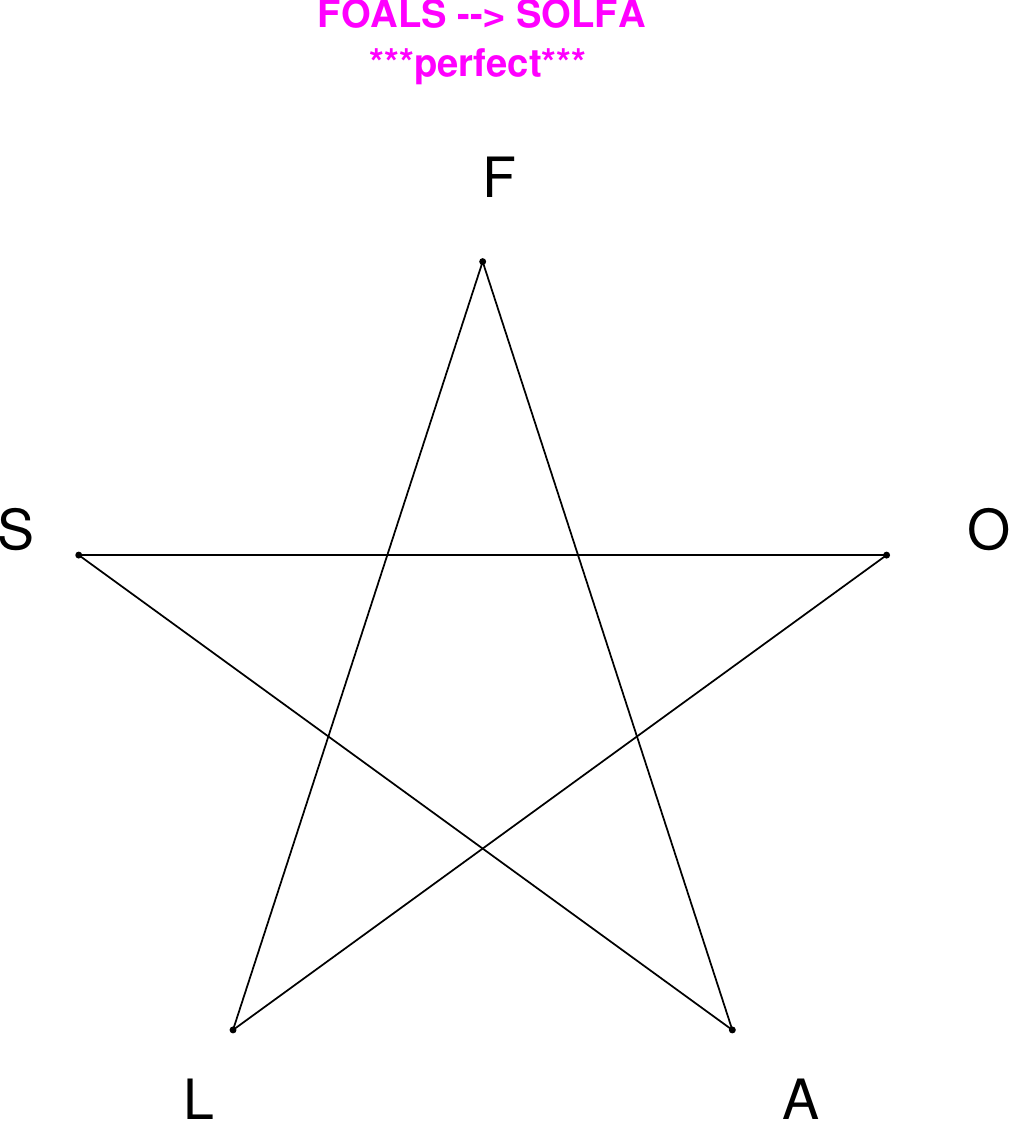}
\end{subfigure}
\hfill
\begin{subfigure}[T]{0.19\textwidth}
\centering
\includegraphics[width=\textwidth]{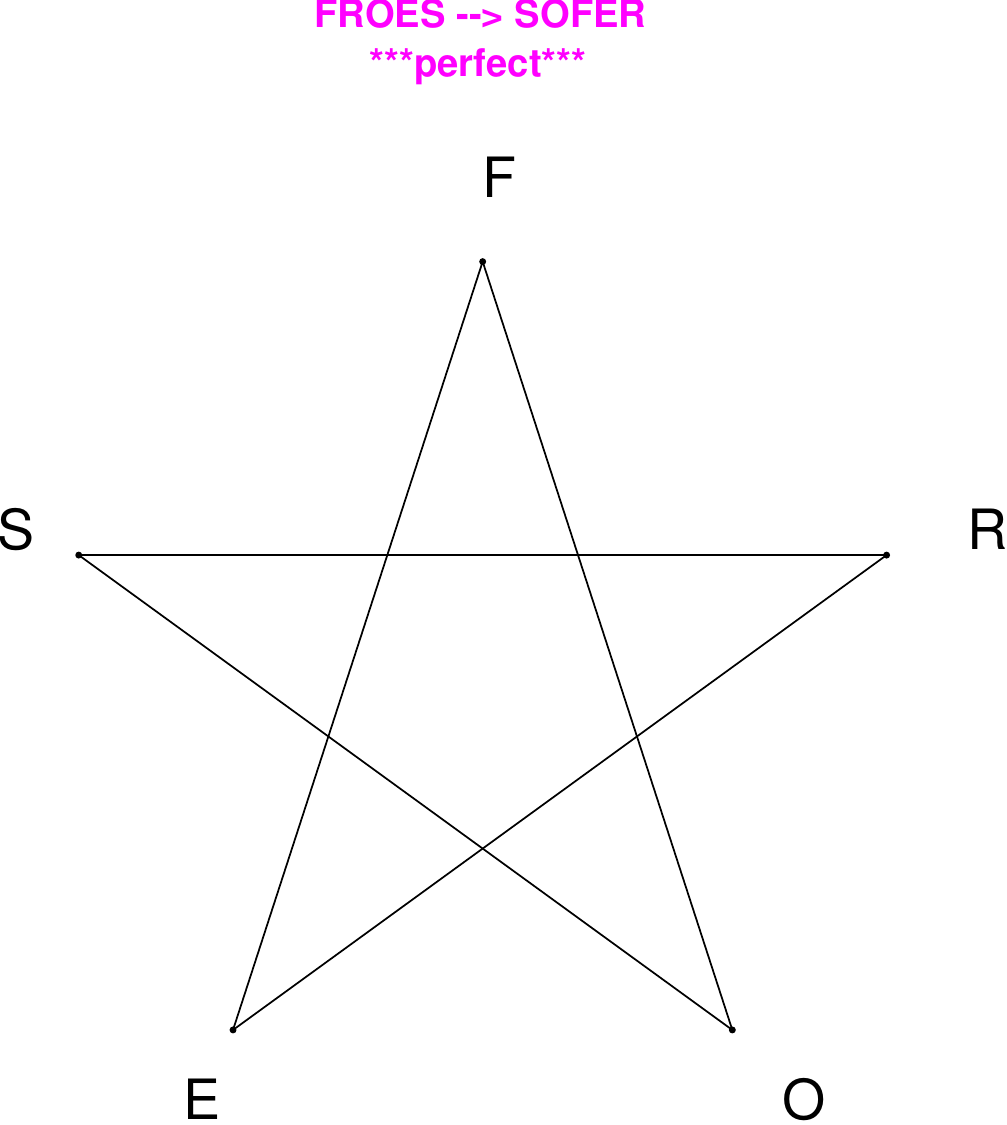}
\end{subfigure}
\end{figure}

\begin{figure}[H]
\centering
\begin{subfigure}[T]{0.19\textwidth}
\centering
\includegraphics[width=\textwidth]{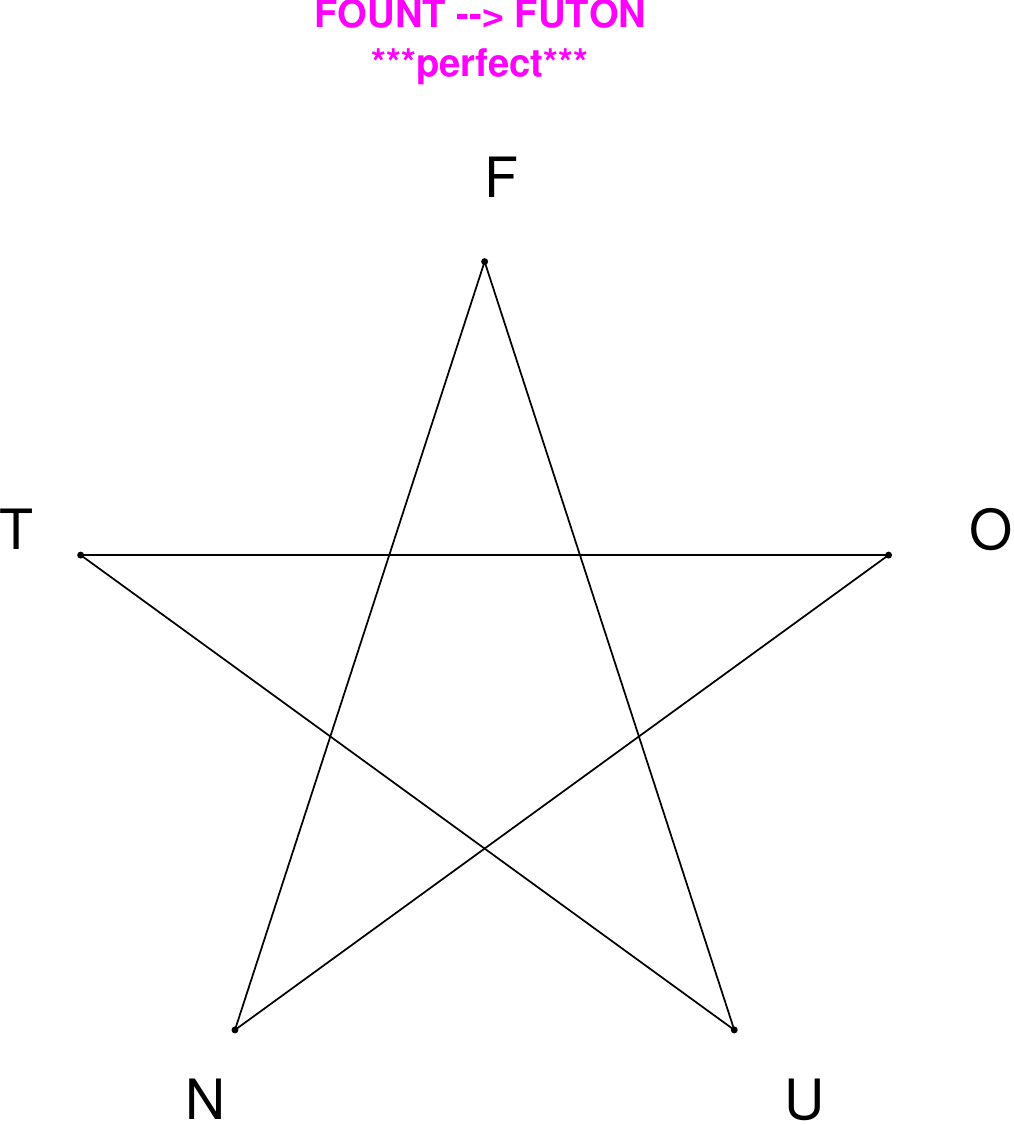}
\end{subfigure}
\hfill
\begin{subfigure}[T]{0.19\textwidth}
\centering
\includegraphics[width=\textwidth]{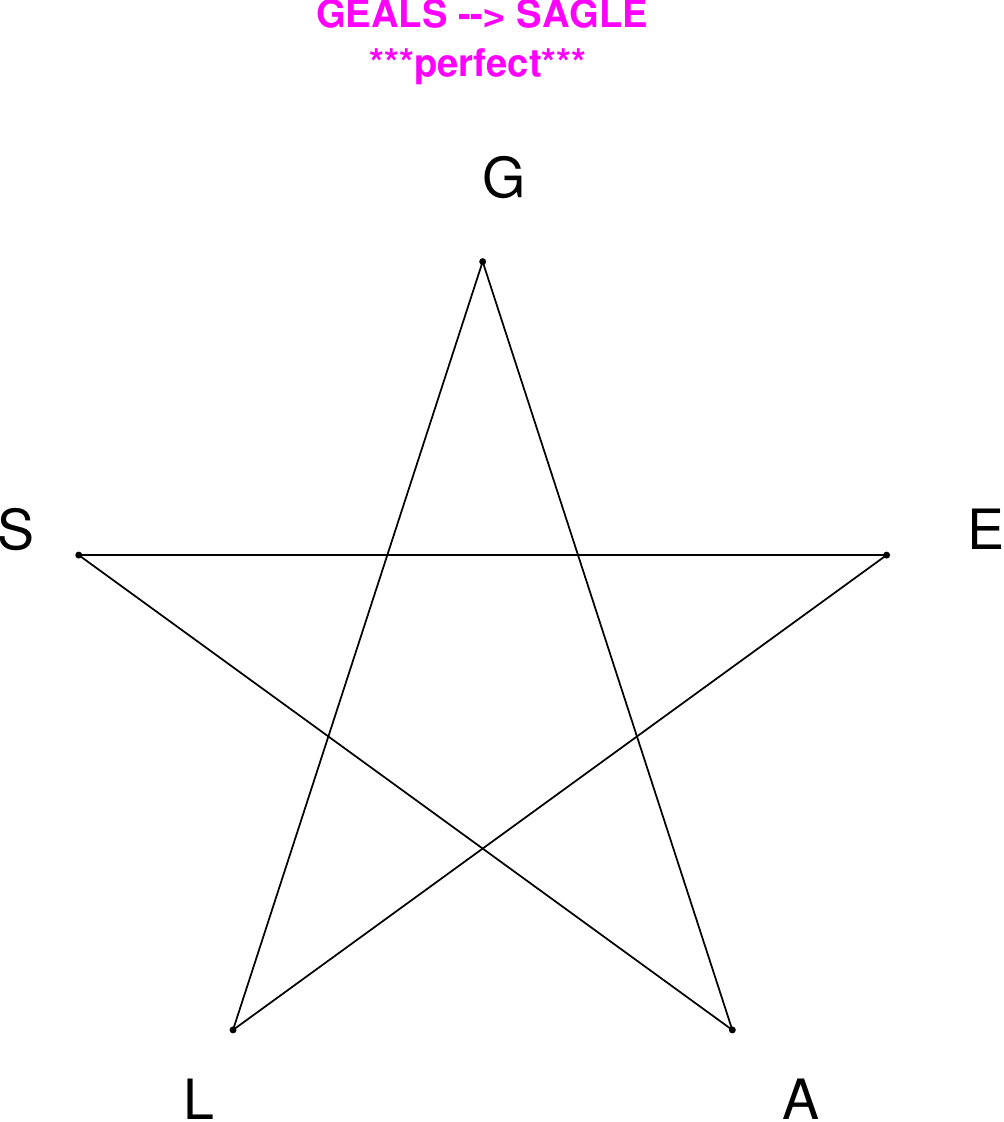}
\end{subfigure}
\hfill
\begin{subfigure}[T]{0.19\textwidth}
\centering
\includegraphics[width=\textwidth]{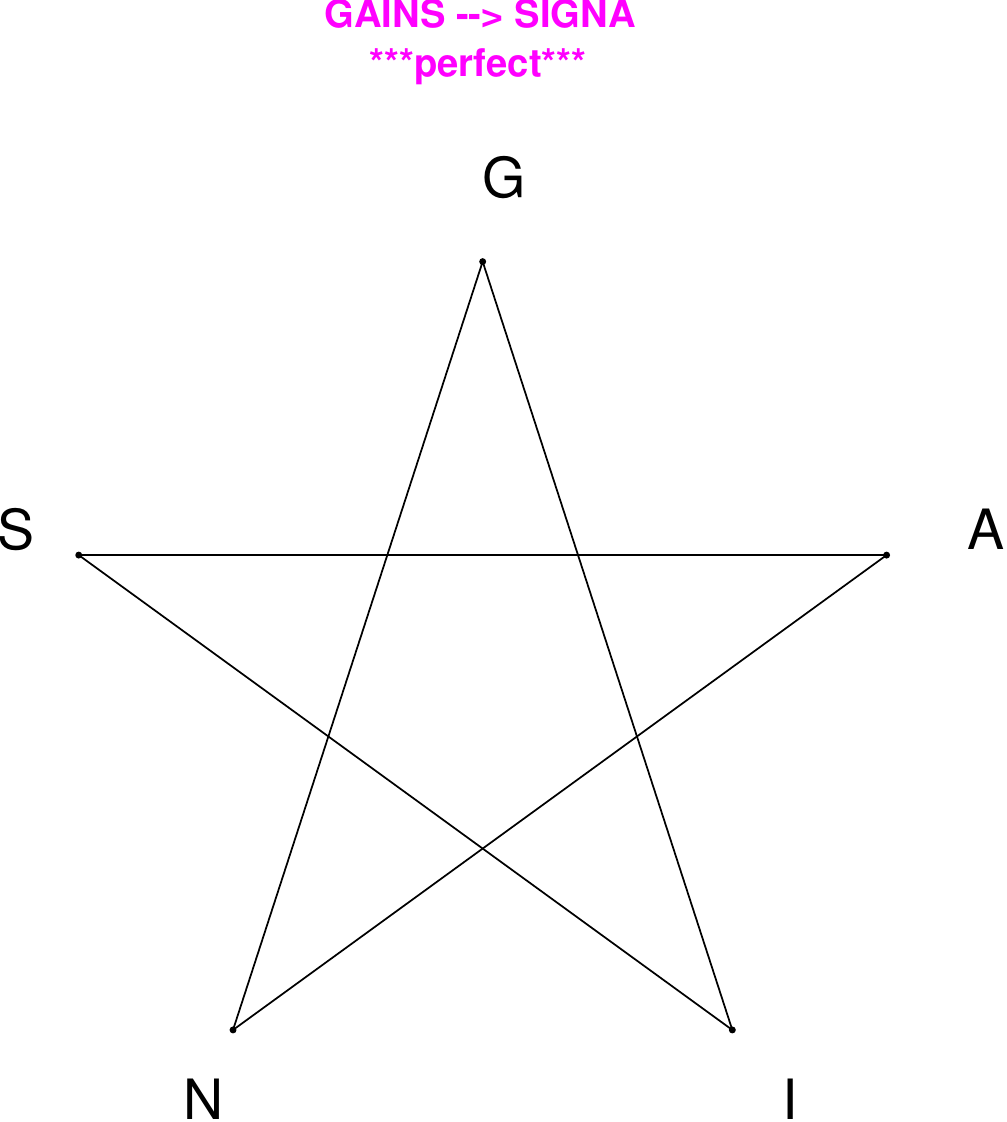}
\end{subfigure}
\hfill
\begin{subfigure}[T]{0.19\textwidth}
\centering
\includegraphics[width=\textwidth]{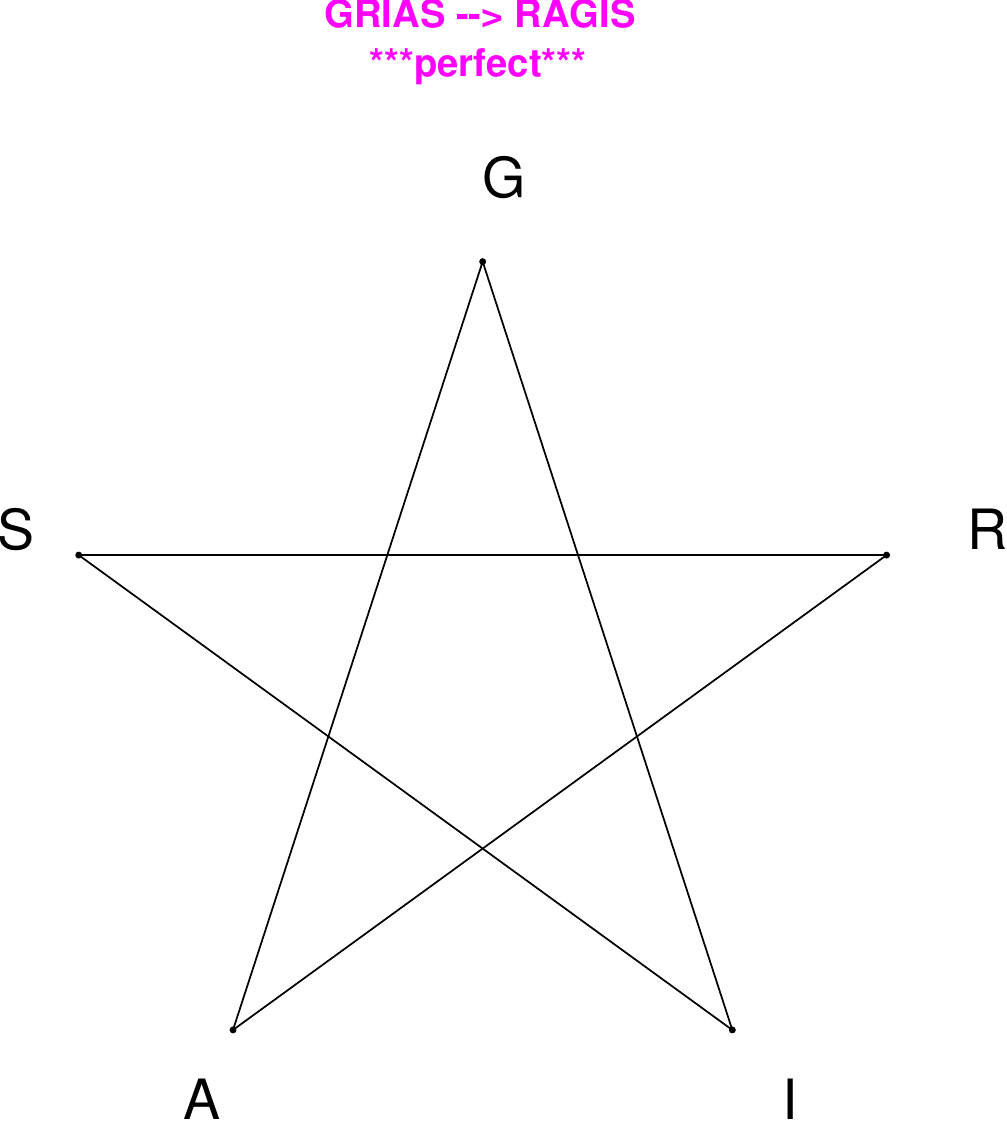}
\end{subfigure}
\hfill
\begin{subfigure}[T]{0.19\textwidth}
\centering
\includegraphics[width=\textwidth]{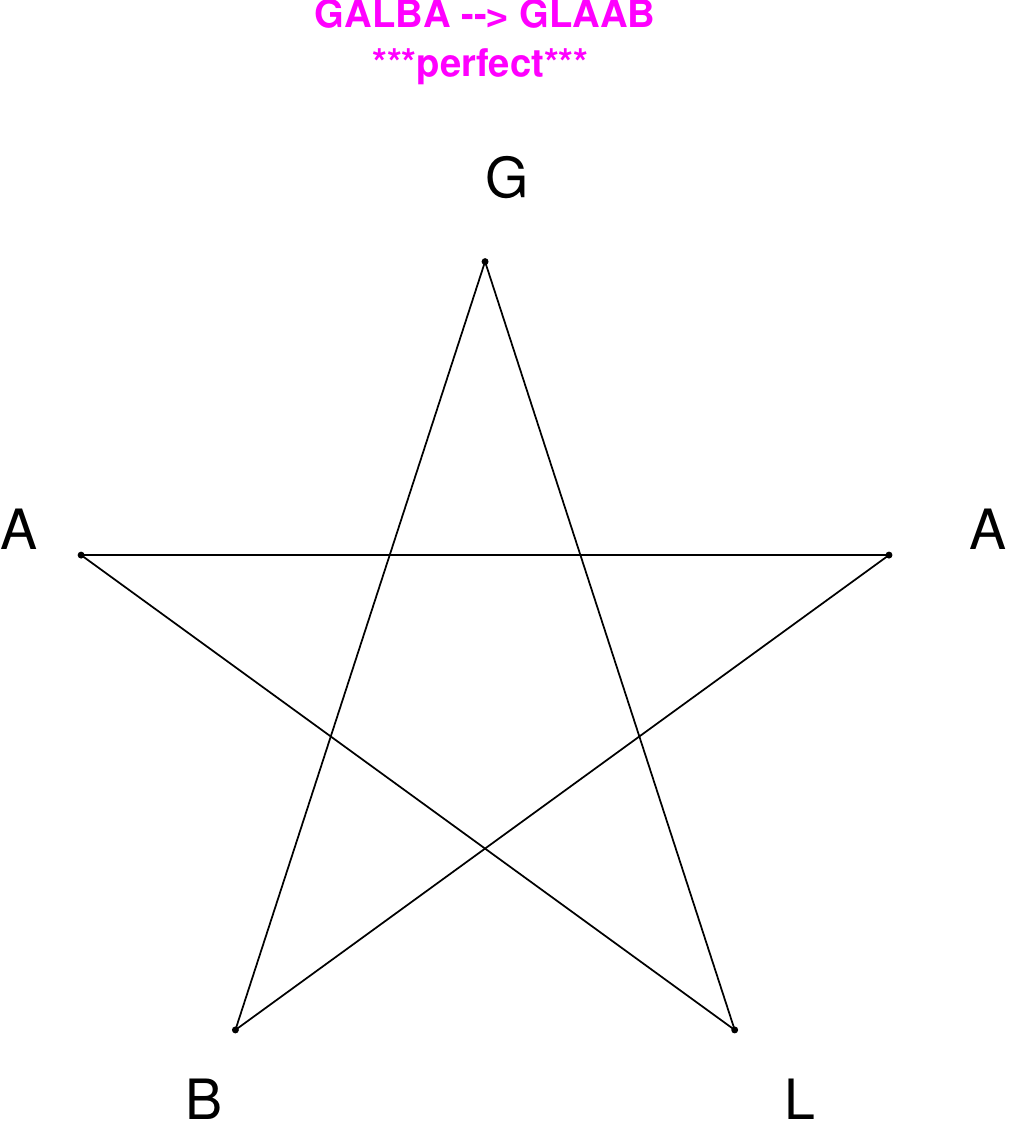}
\end{subfigure}
\end{figure}

\begin{figure}[H]
\centering
\begin{subfigure}[T]{0.19\textwidth}
\centering
\includegraphics[width=\textwidth]{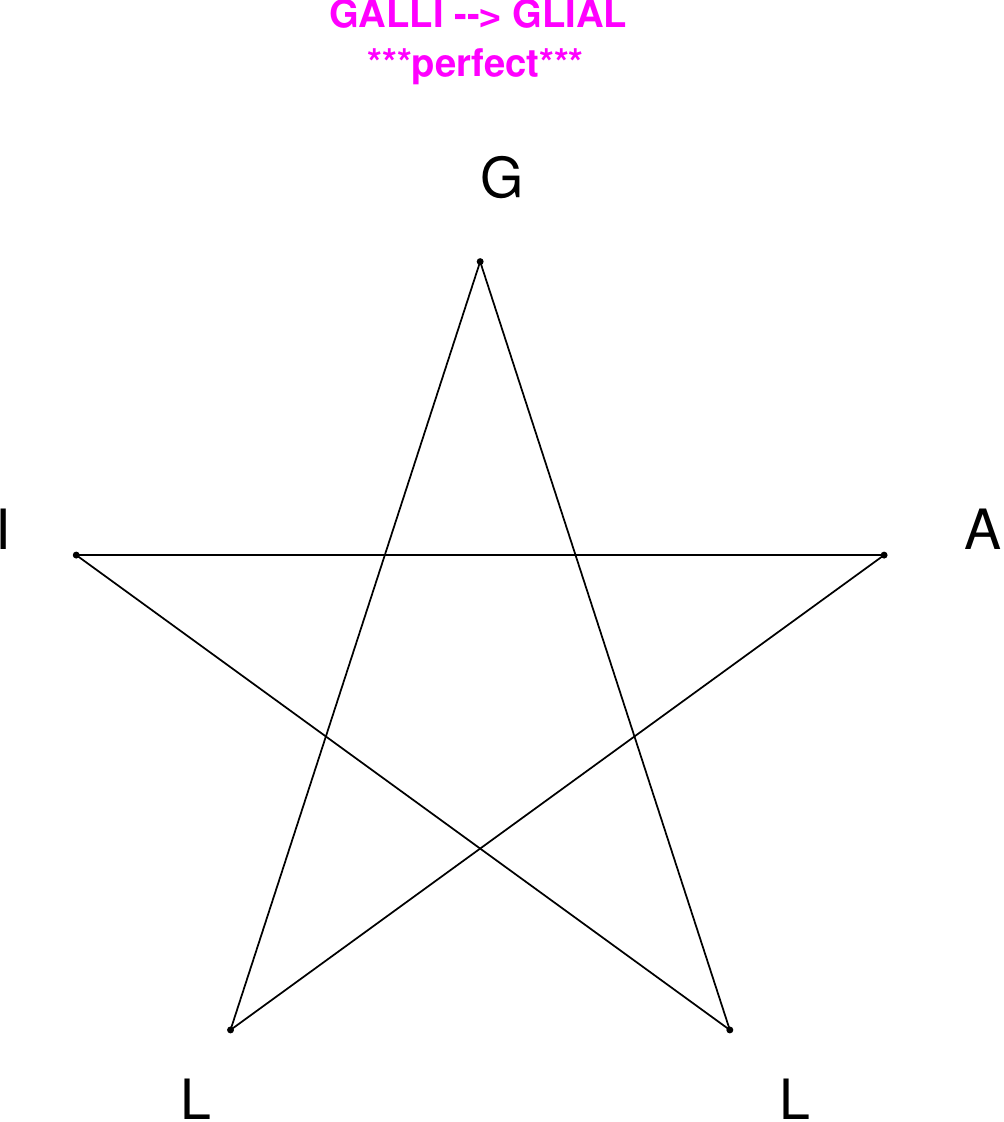}
\end{subfigure}
\hfill
\begin{subfigure}[T]{0.19\textwidth}
\centering
\includegraphics[width=\textwidth]{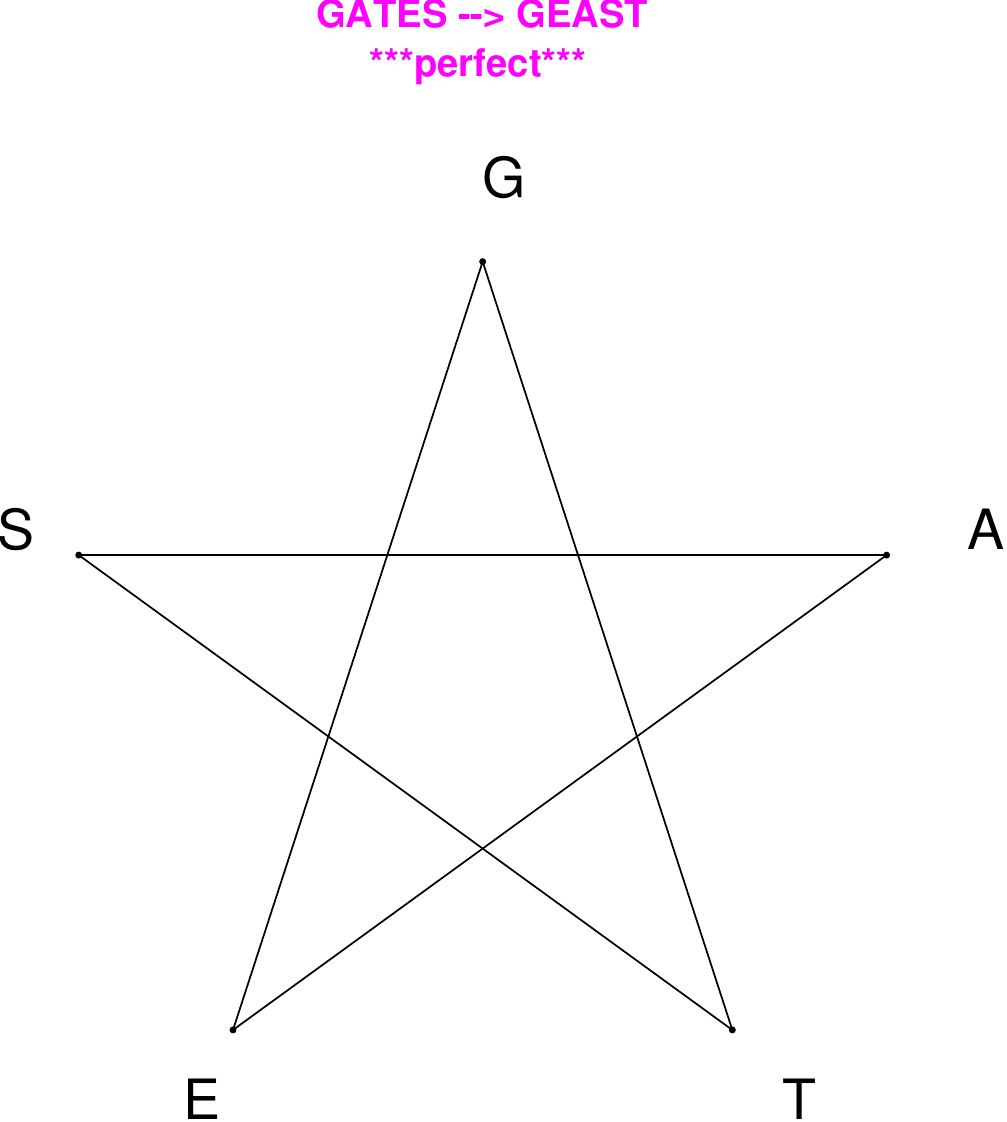}
\end{subfigure}
\hfill
\begin{subfigure}[T]{0.19\textwidth}
\centering
\includegraphics[width=\textwidth]{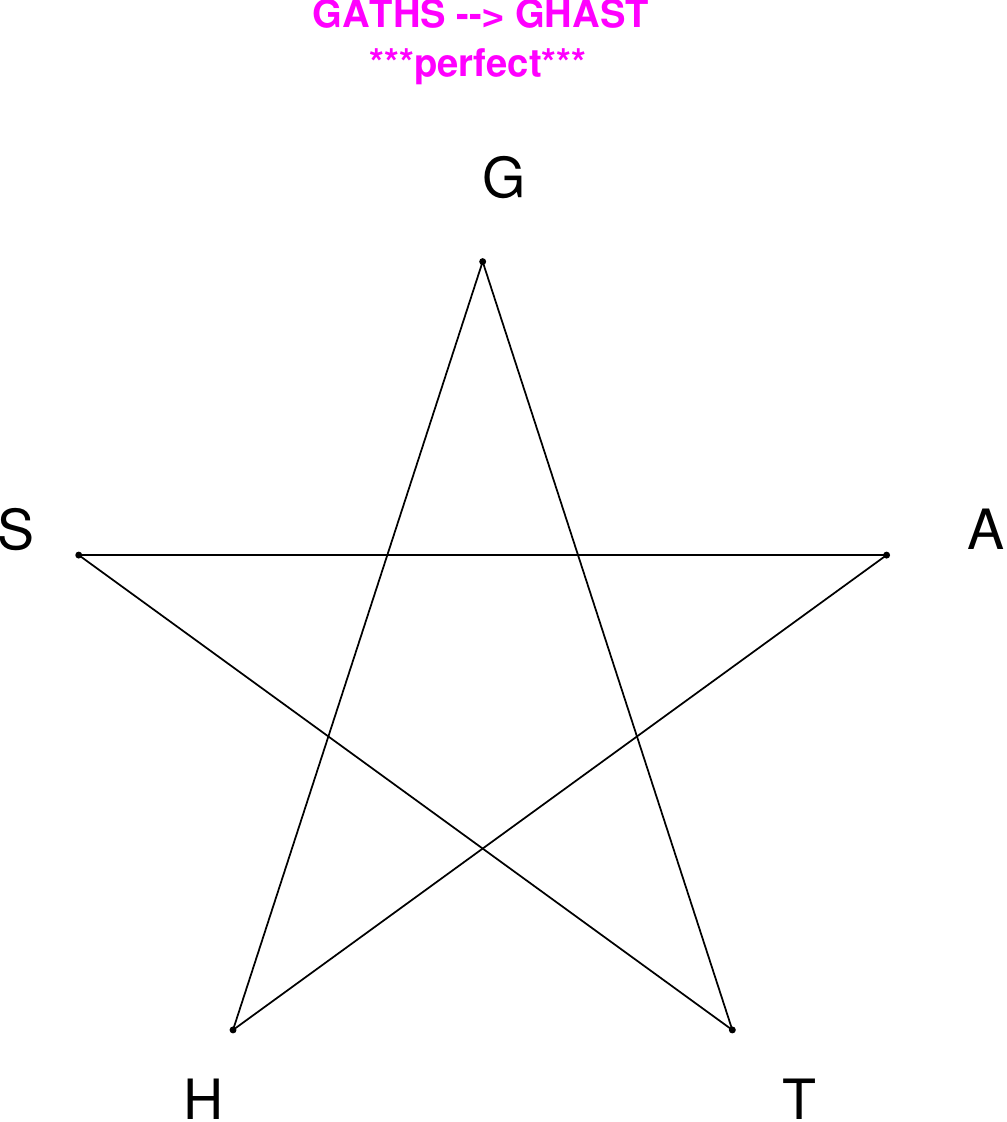}
\end{subfigure}
\hfill
\begin{subfigure}[T]{0.19\textwidth}
\centering
\includegraphics[width=\textwidth]{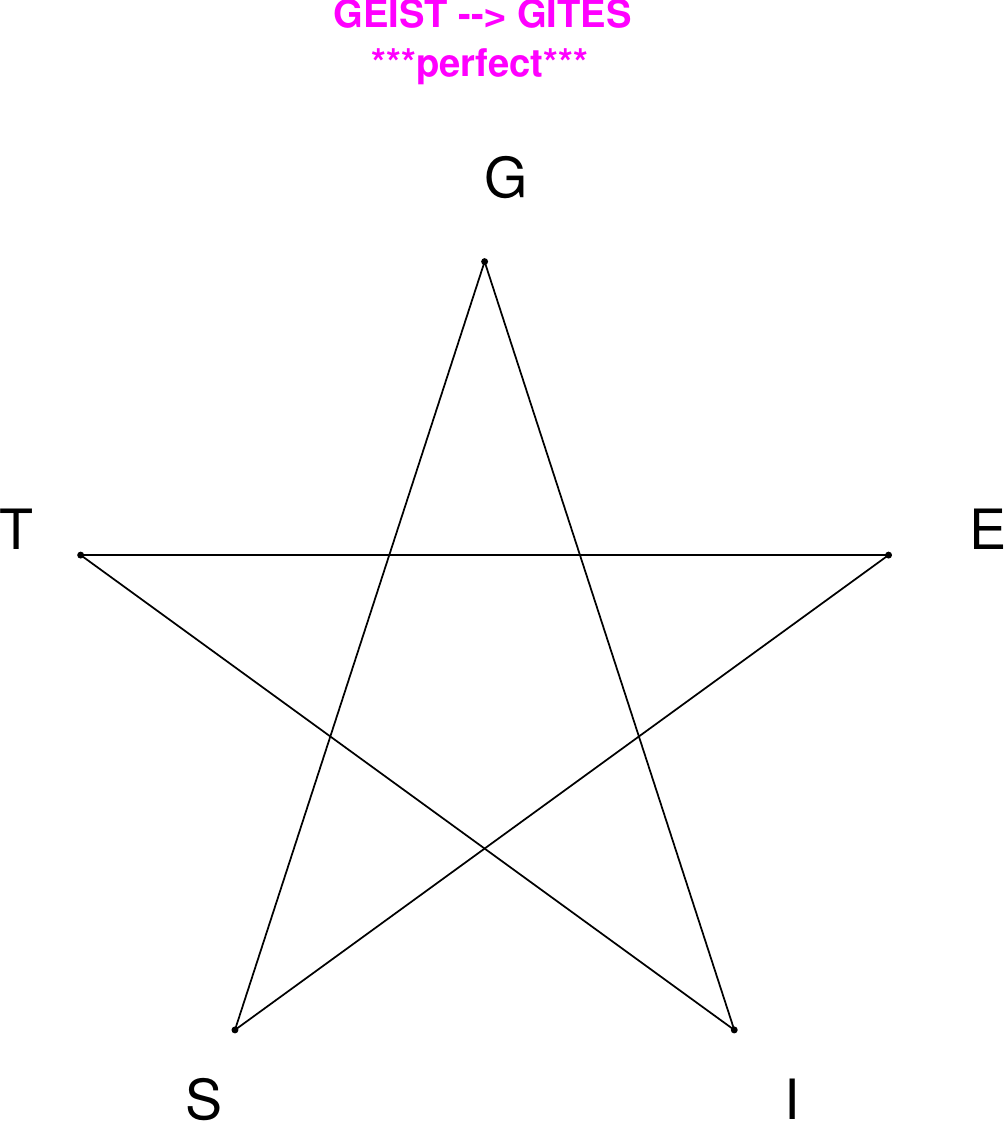}
\end{subfigure}
\hfill
\begin{subfigure}[T]{0.19\textwidth}
\centering
\includegraphics[width=\textwidth]{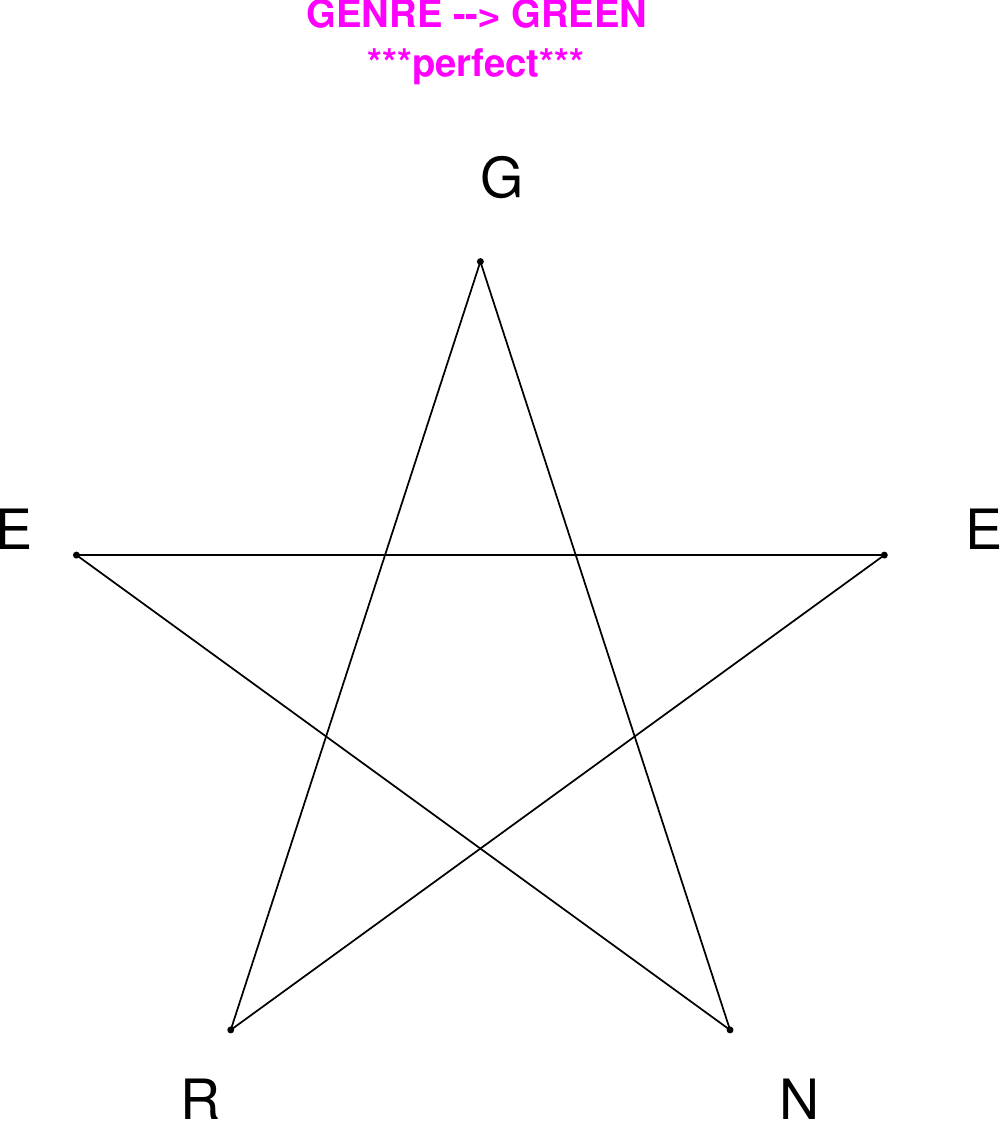}
\end{subfigure}
\end{figure}

\begin{figure}[H]
\centering
\begin{subfigure}[T]{0.19\textwidth}
\centering
\includegraphics[width=\textwidth]{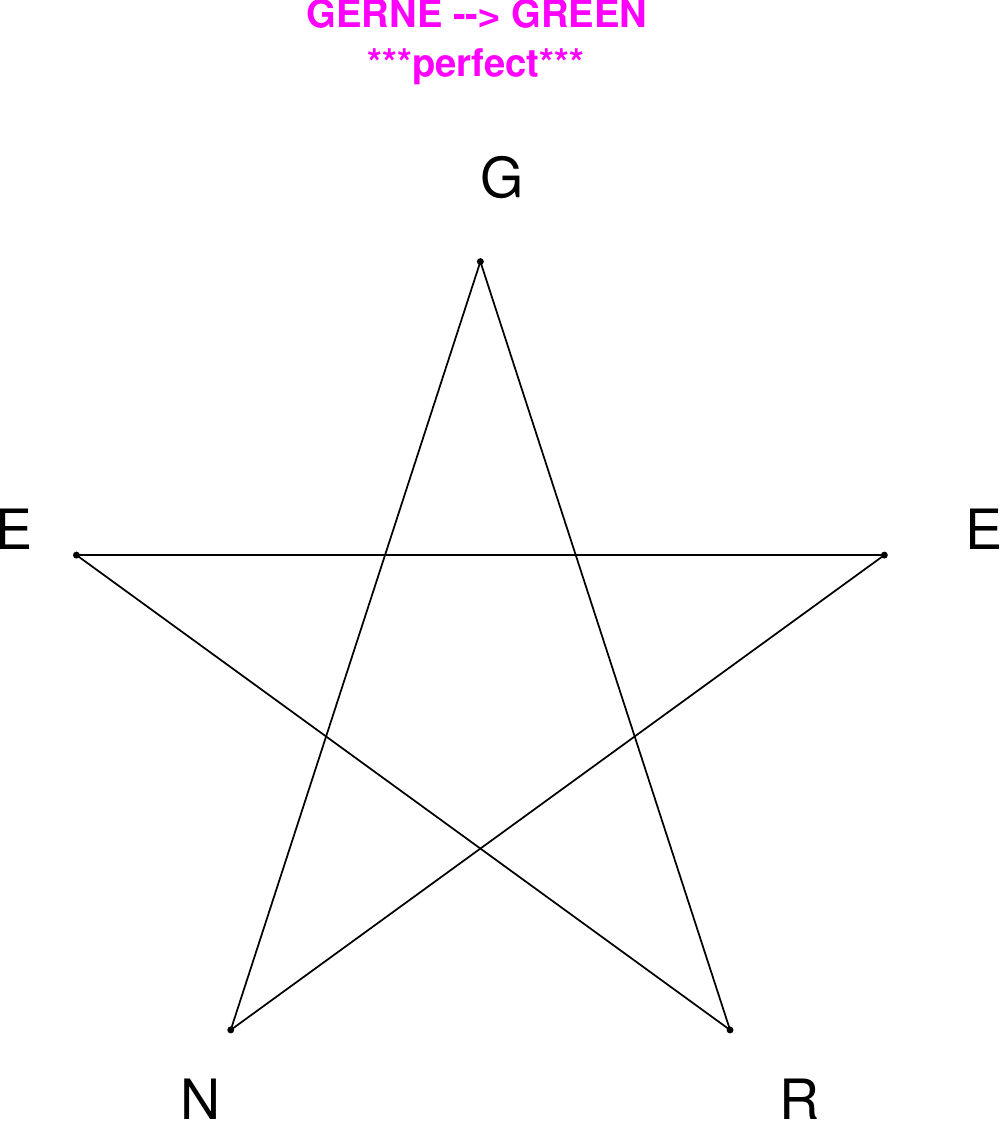}
\end{subfigure}
\hfill
\begin{subfigure}[T]{0.19\textwidth}
\centering
\includegraphics[width=\textwidth]{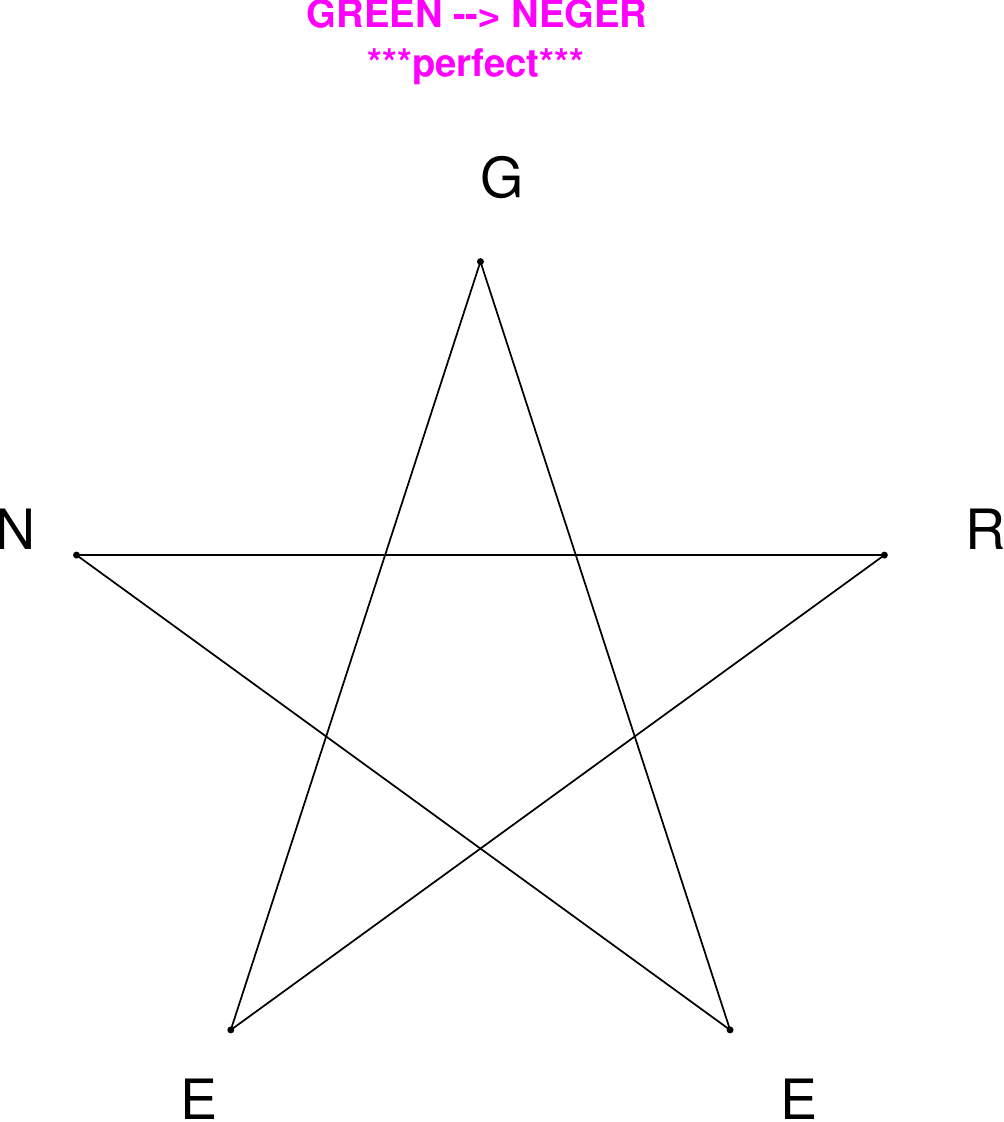}
\end{subfigure}
\hfill
\begin{subfigure}[T]{0.19\textwidth}
\centering
\includegraphics[width=\textwidth]{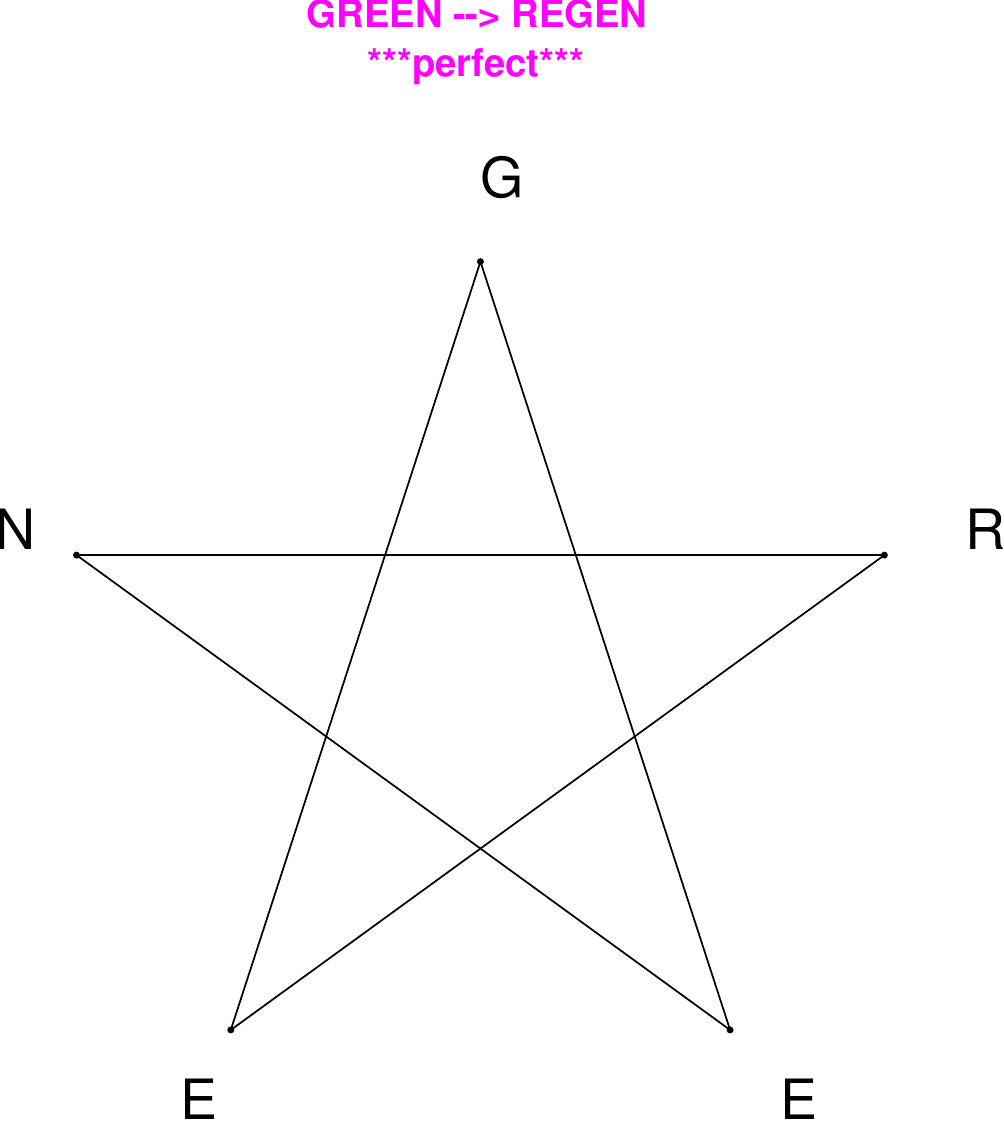}
\end{subfigure}
\hfill
\begin{subfigure}[T]{0.19\textwidth}
\centering
\includegraphics[width=\textwidth]{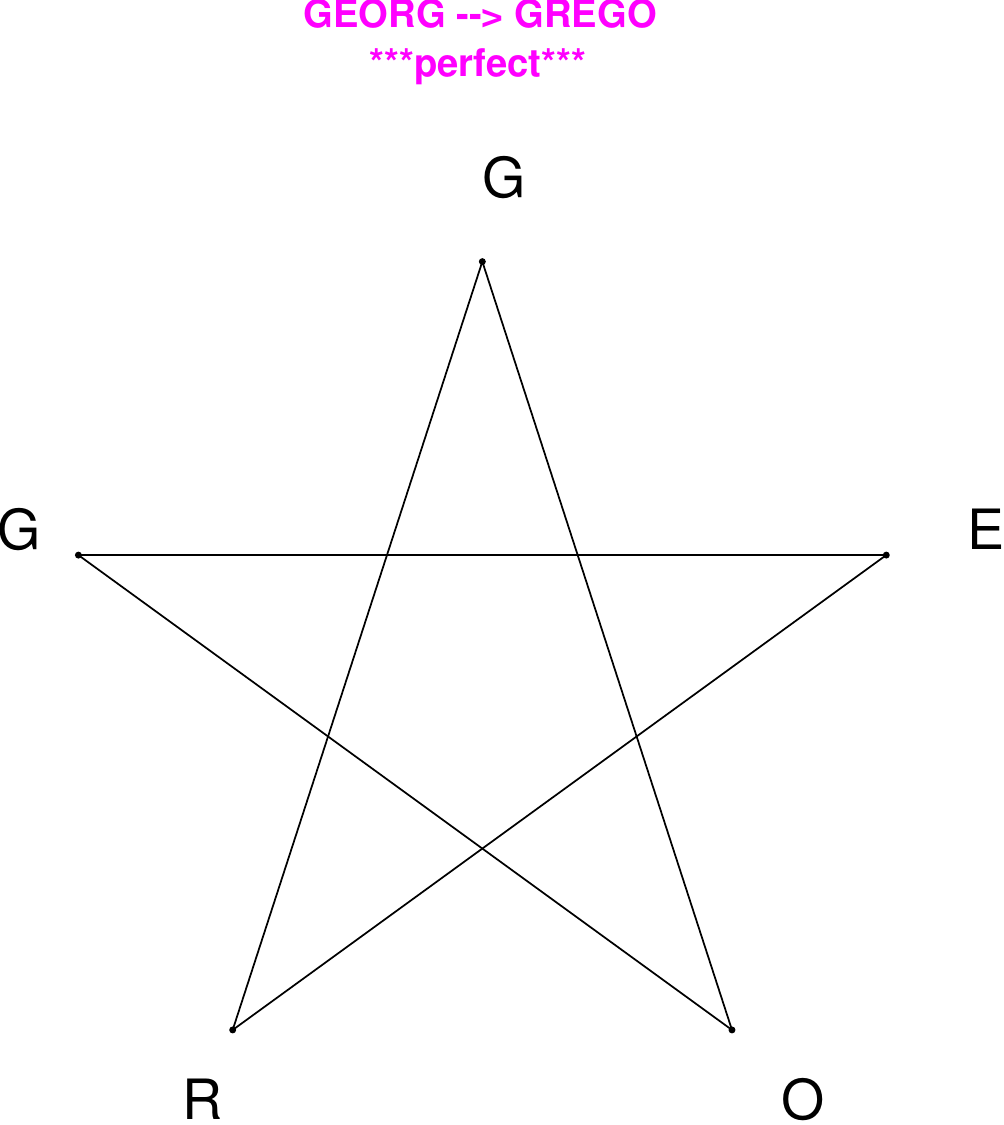}
\end{subfigure}
\hfill
\begin{subfigure}[T]{0.19\textwidth}
\centering
\includegraphics[width=\textwidth]{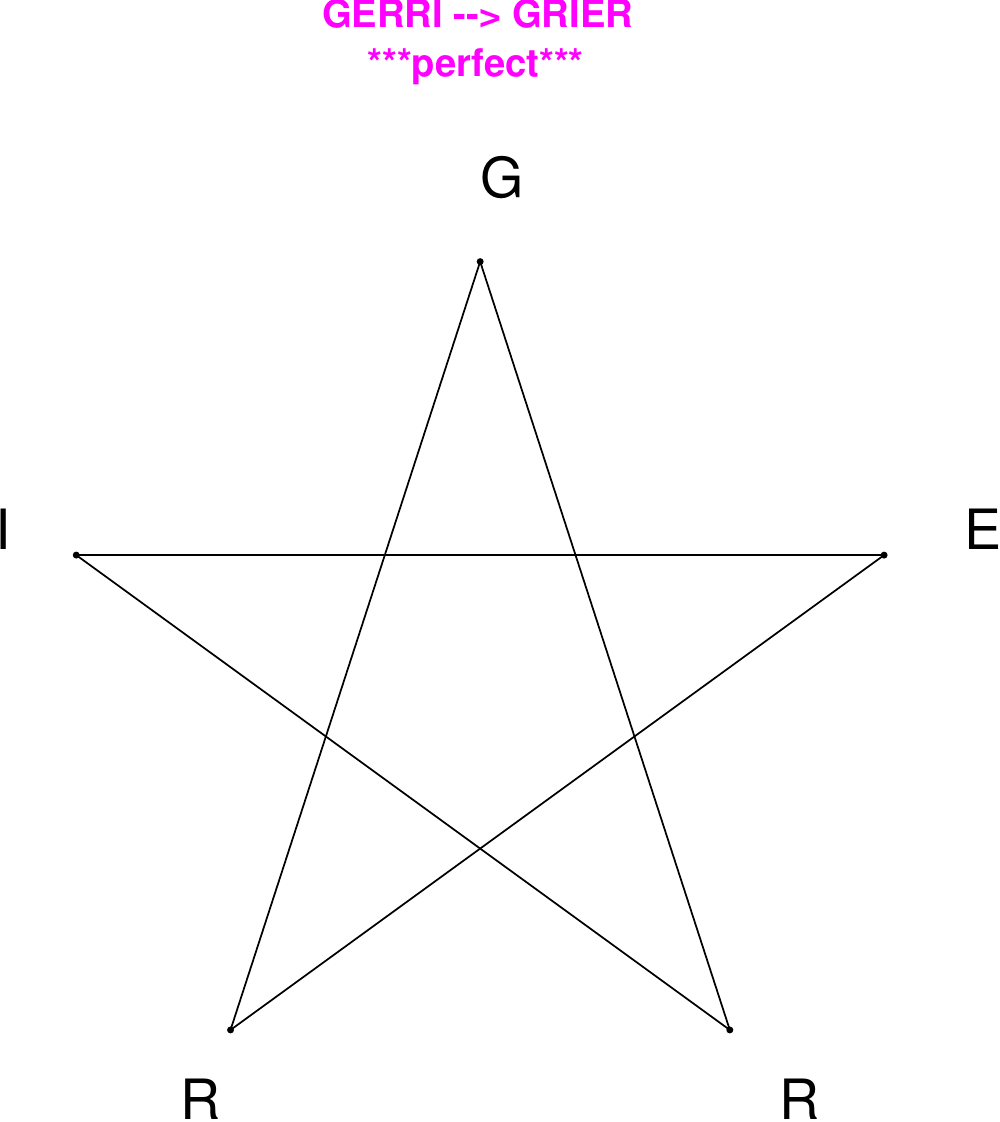}
\end{subfigure}
\end{figure}

\begin{figure}[H]
\centering
\begin{subfigure}[T]{0.19\textwidth}
\centering
\includegraphics[width=\textwidth]{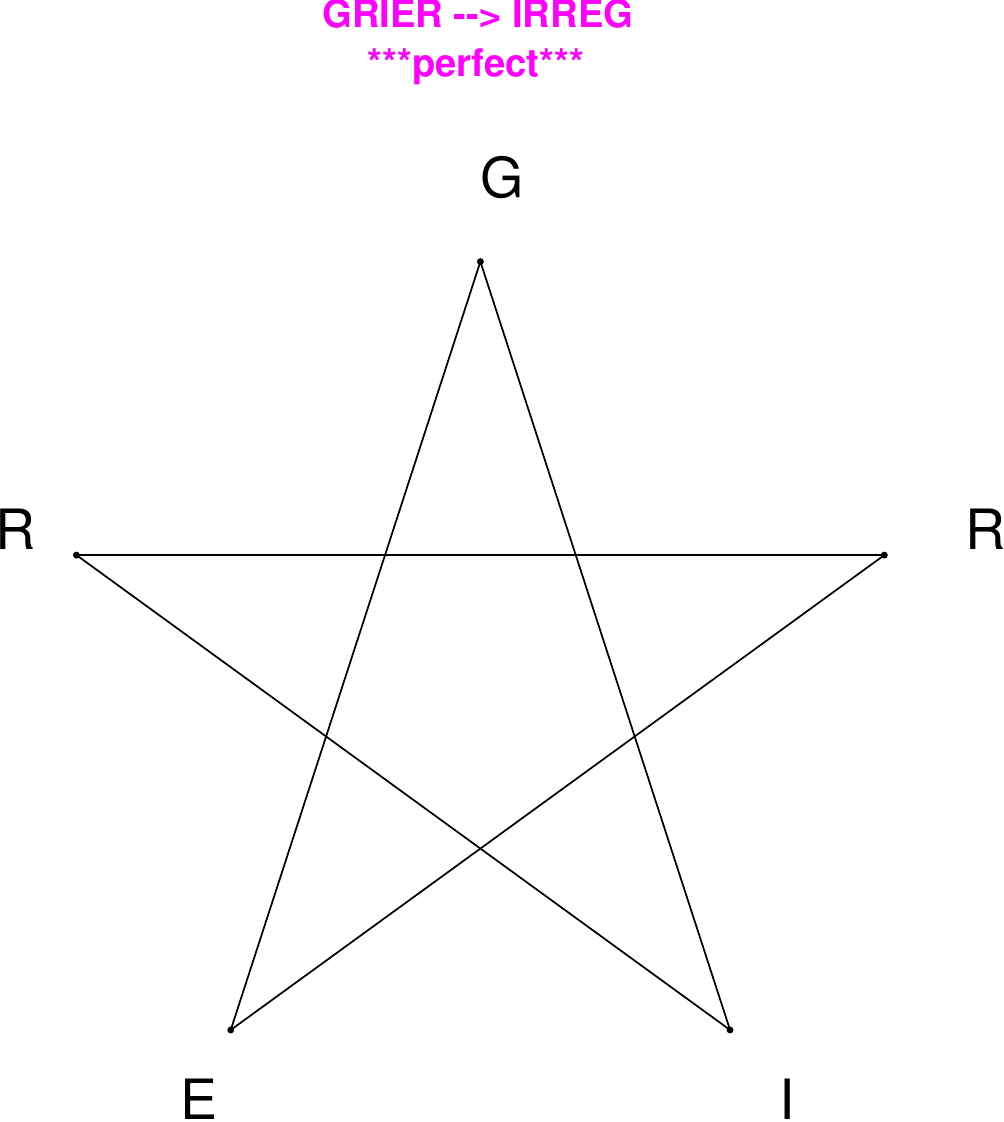}
\end{subfigure}
\hfill
\begin{subfigure}[T]{0.19\textwidth}
\centering
\includegraphics[width=\textwidth]{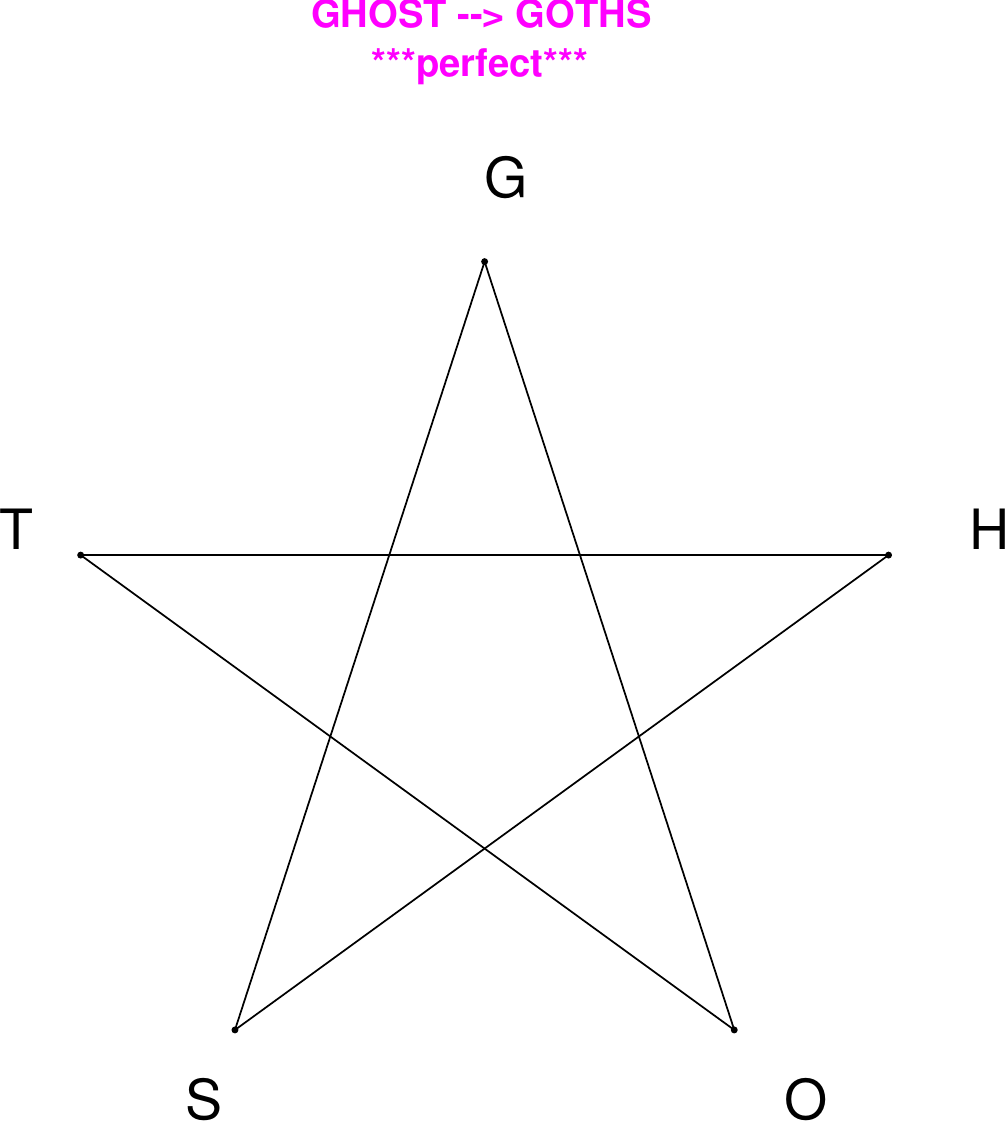}
\end{subfigure}
\hfill
\begin{subfigure}[T]{0.19\textwidth}
\centering
\includegraphics[width=\textwidth]{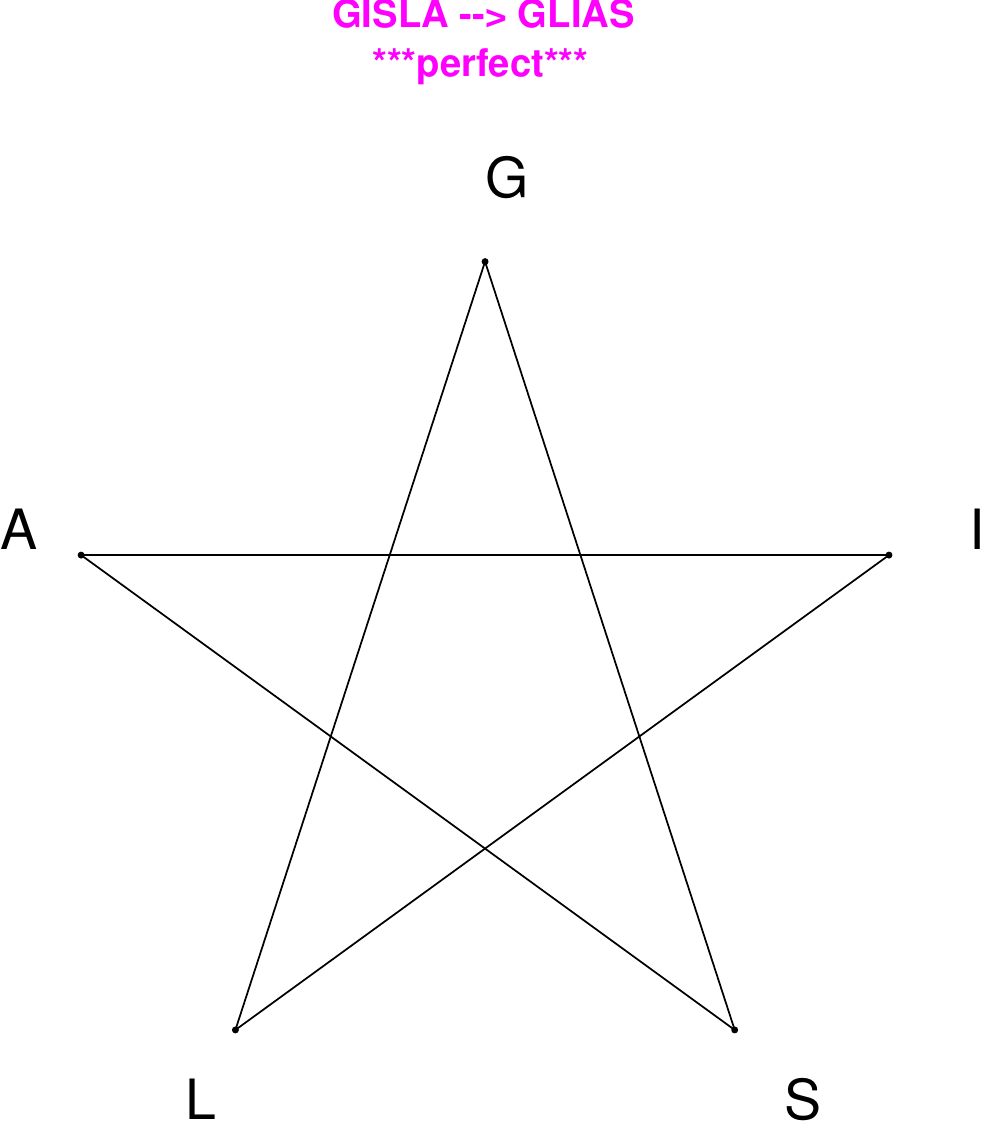}
\end{subfigure}
\hfill
\begin{subfigure}[T]{0.19\textwidth}
\centering
\includegraphics[width=\textwidth]{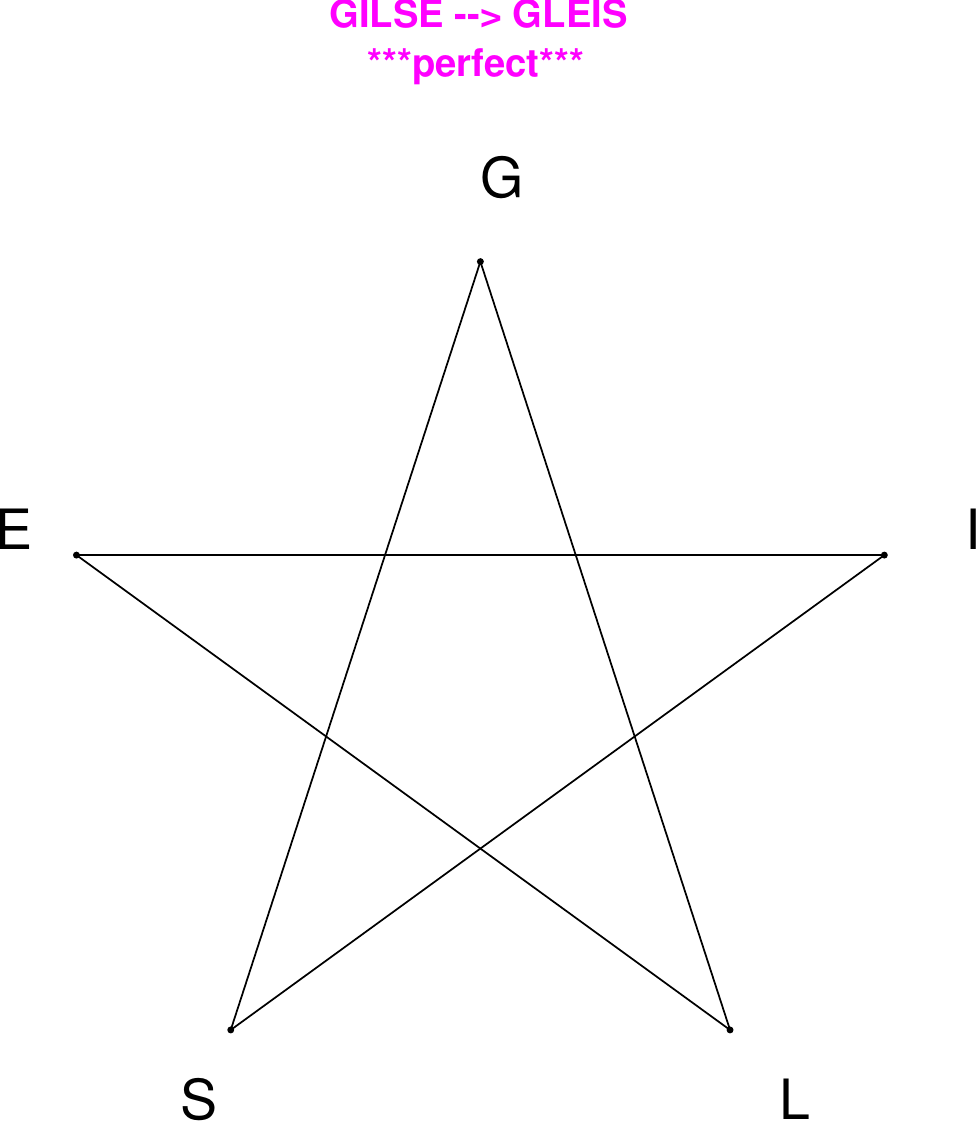}
\end{subfigure}
\hfill
\begin{subfigure}[T]{0.19\textwidth}
\centering
\includegraphics[width=\textwidth]{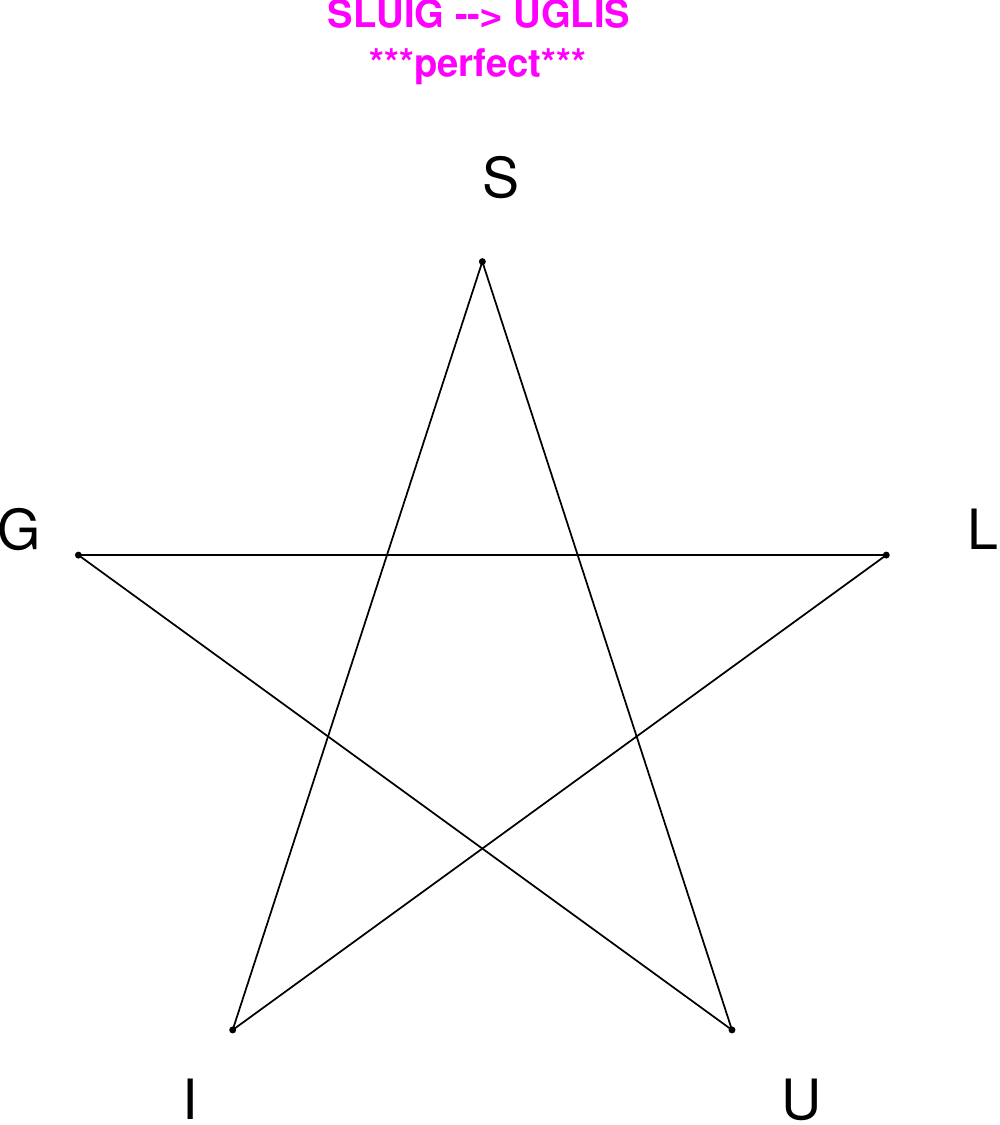}
\end{subfigure}
\end{figure}

\begin{figure}[H]
\centering
\begin{subfigure}[T]{0.19\textwidth}
\centering
\includegraphics[width=\textwidth]{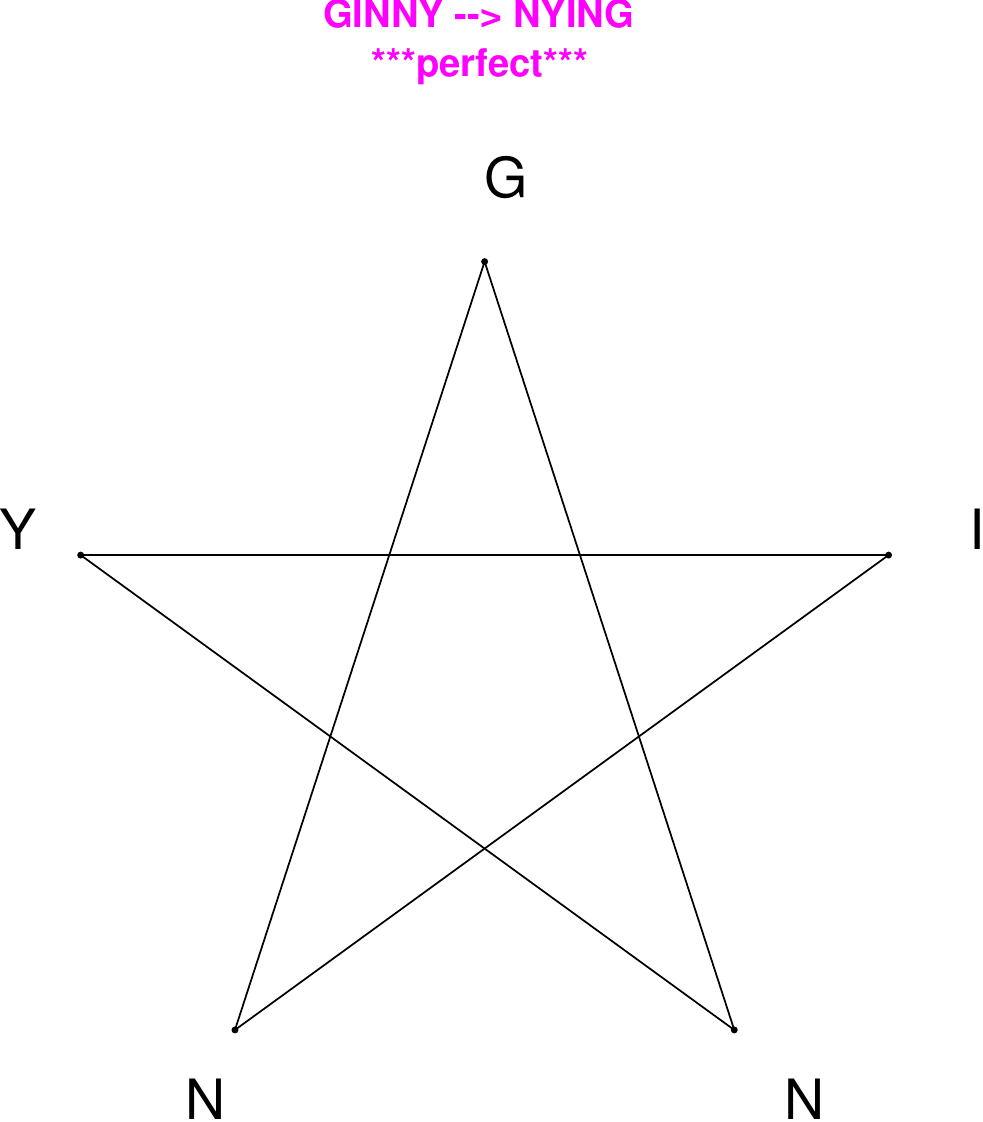}
\end{subfigure}
\hfill
\begin{subfigure}[T]{0.19\textwidth}
\centering
\includegraphics[width=\textwidth]{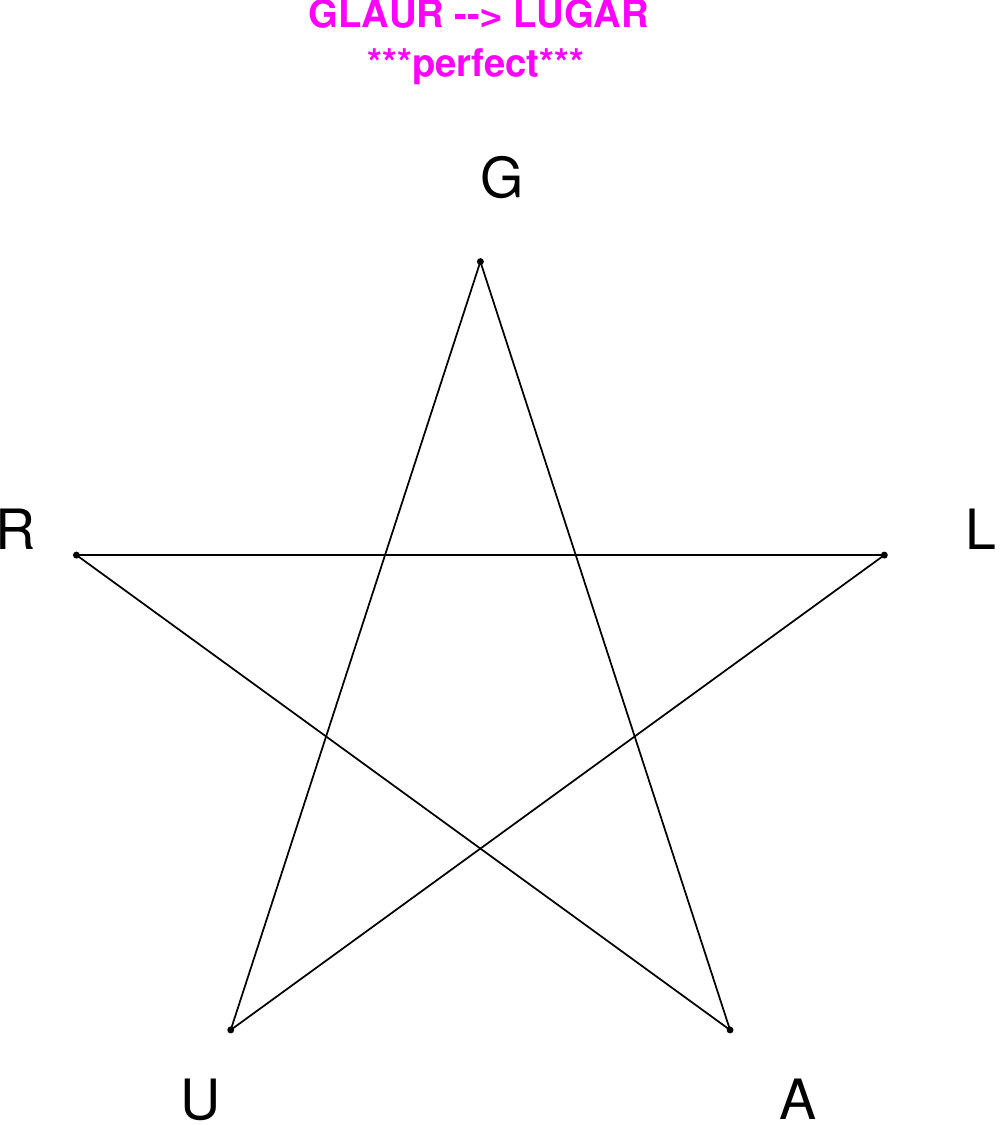}
\end{subfigure}
\hfill
\begin{subfigure}[T]{0.19\textwidth}
\centering
\includegraphics[width=\textwidth]{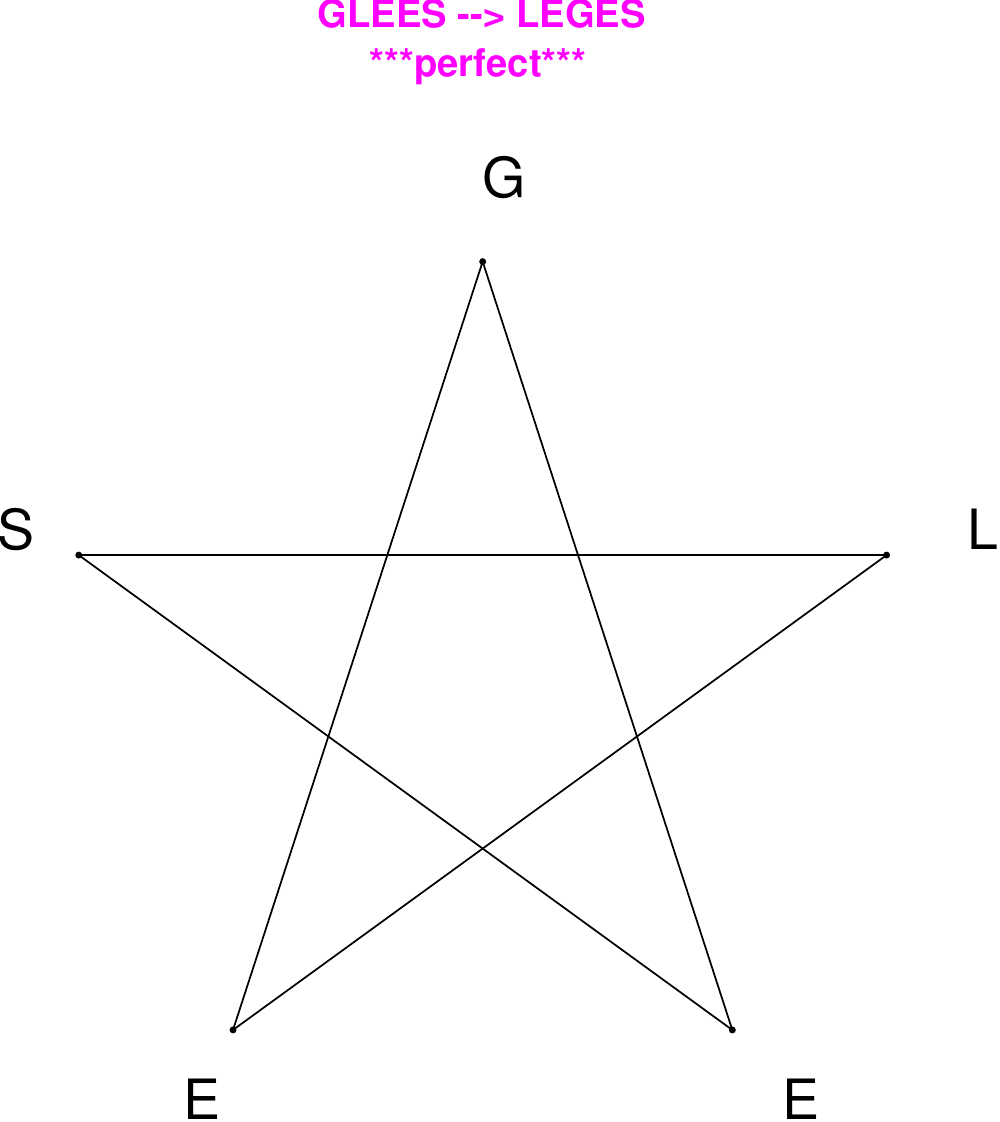}
\end{subfigure}
\hfill
\begin{subfigure}[T]{0.19\textwidth}
\centering
\includegraphics[width=\textwidth]{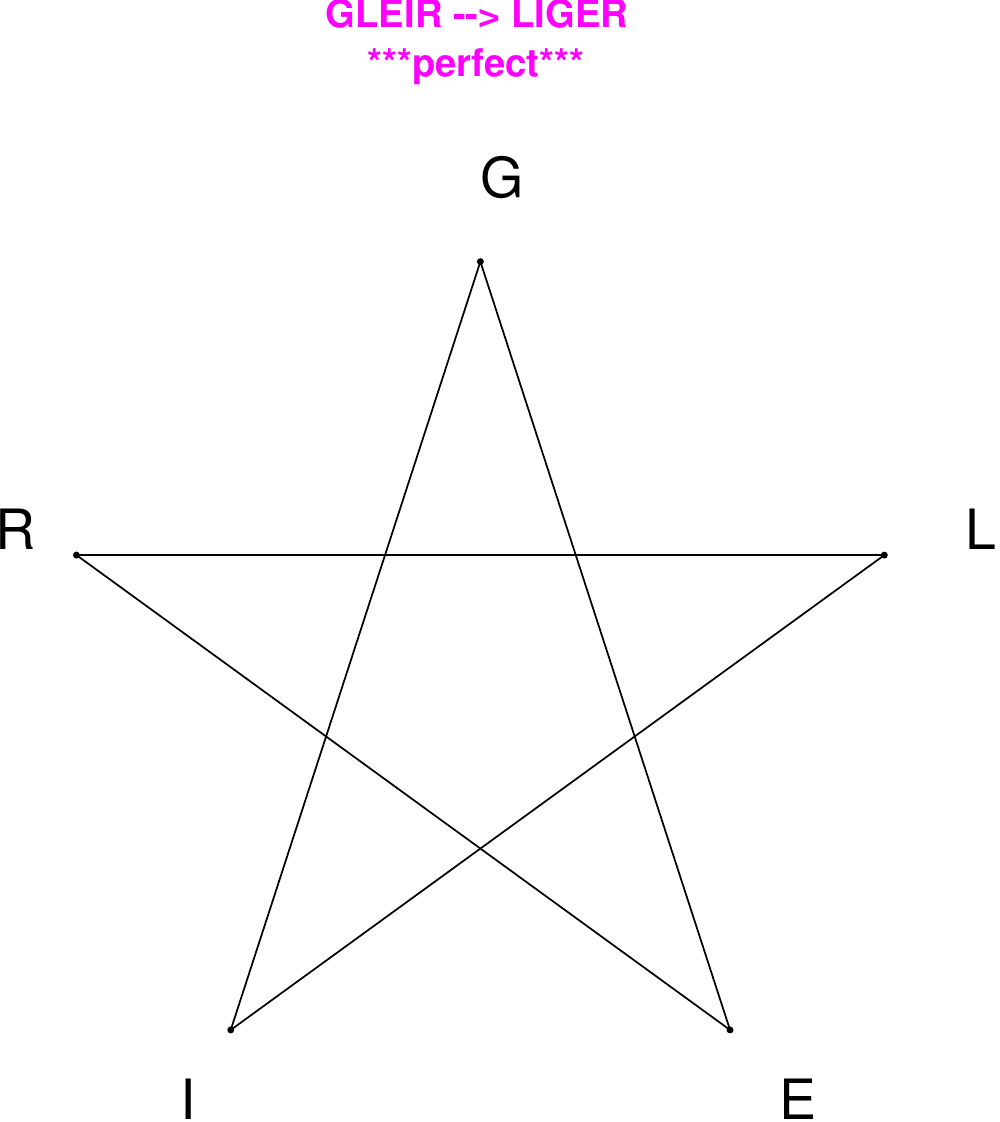}
\end{subfigure}
\hfill
\begin{subfigure}[T]{0.19\textwidth}
\centering
\includegraphics[width=\textwidth]{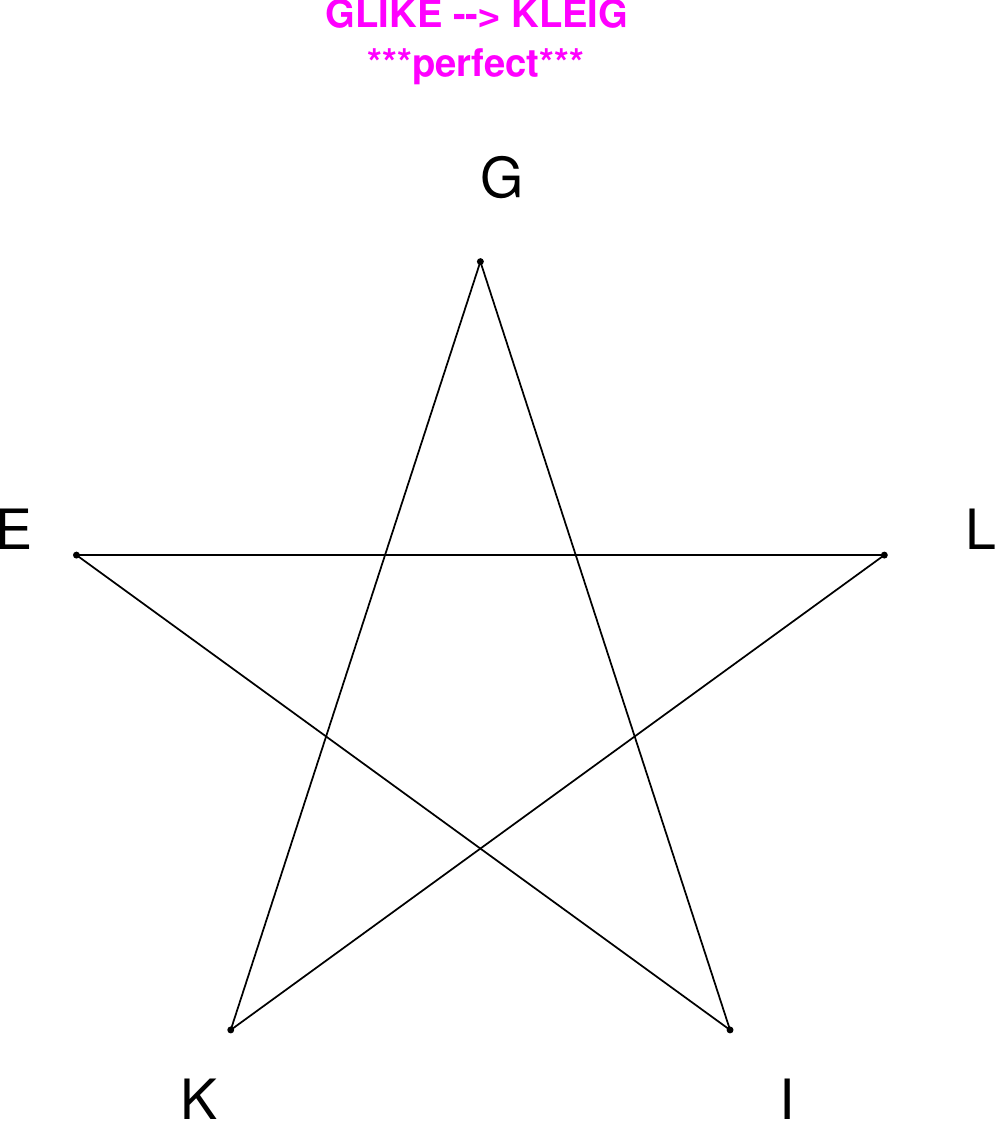}
\end{subfigure}
\end{figure}

\begin{figure}[H]
\centering
\begin{subfigure}[T]{0.19\textwidth}
\centering
\includegraphics[width=\textwidth]{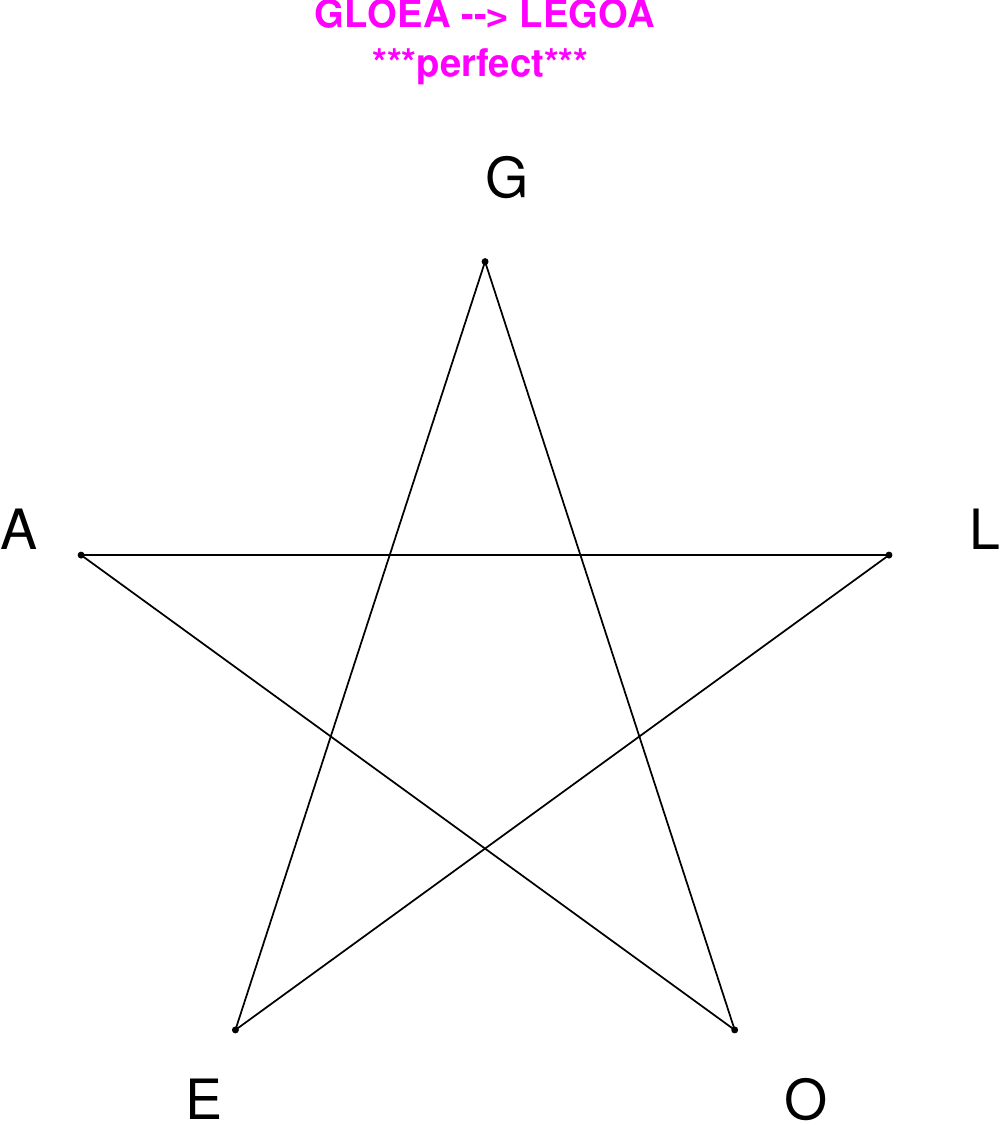}
\end{subfigure}
\hfill
\begin{subfigure}[T]{0.19\textwidth}
\centering
\includegraphics[width=\textwidth]{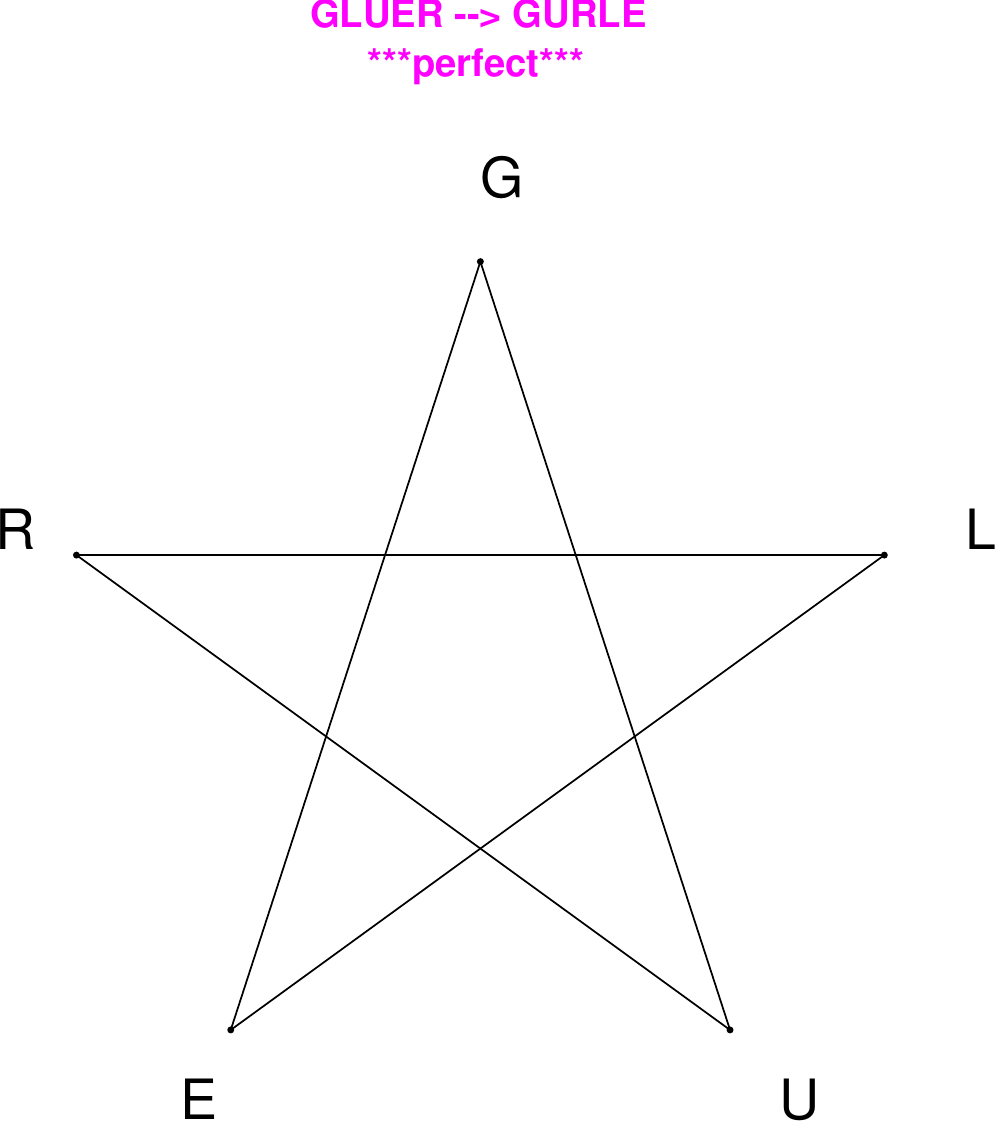}
\end{subfigure}
\hfill
\begin{subfigure}[T]{0.19\textwidth}
\centering
\includegraphics[width=\textwidth]{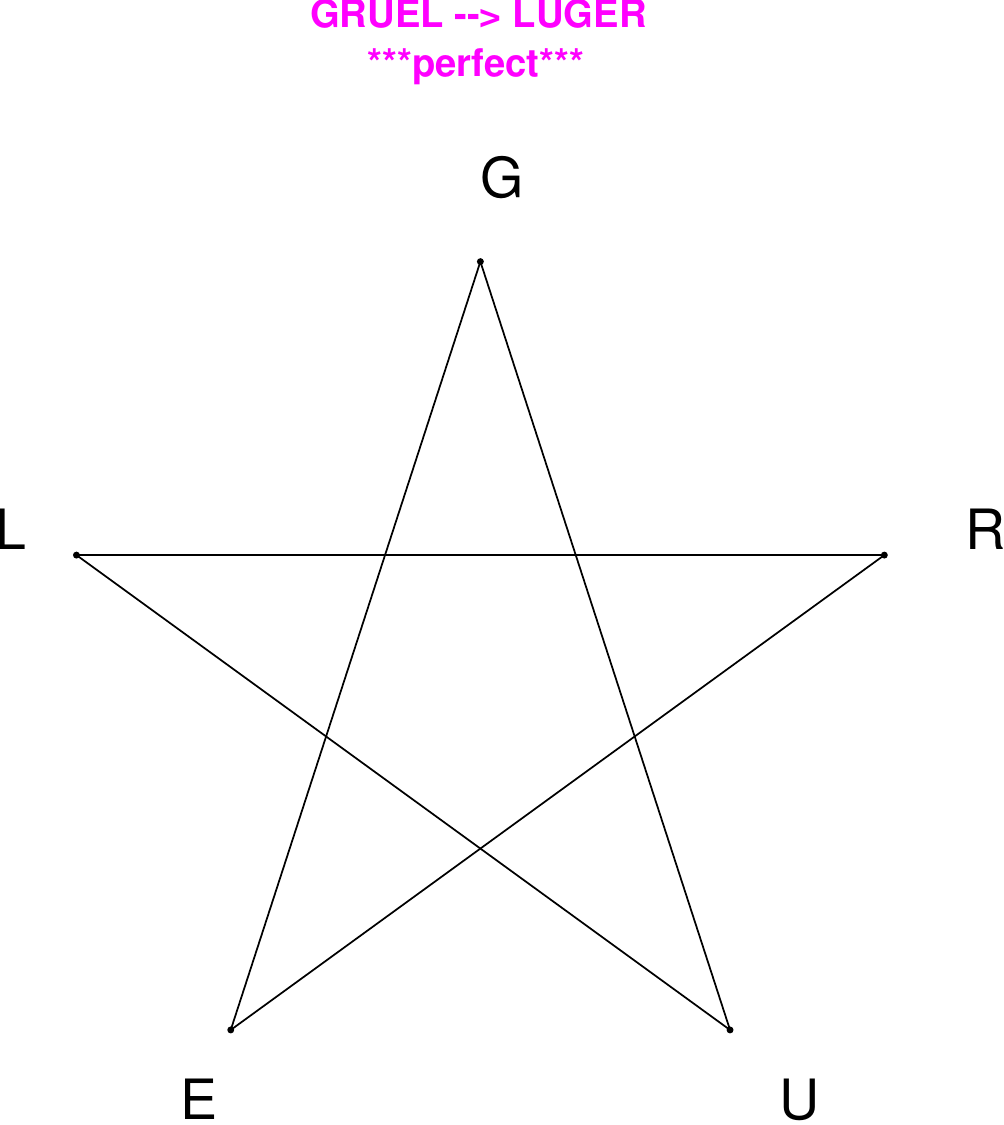}
\end{subfigure}
\hfill
\begin{subfigure}[T]{0.19\textwidth}
\centering
\includegraphics[width=\textwidth]{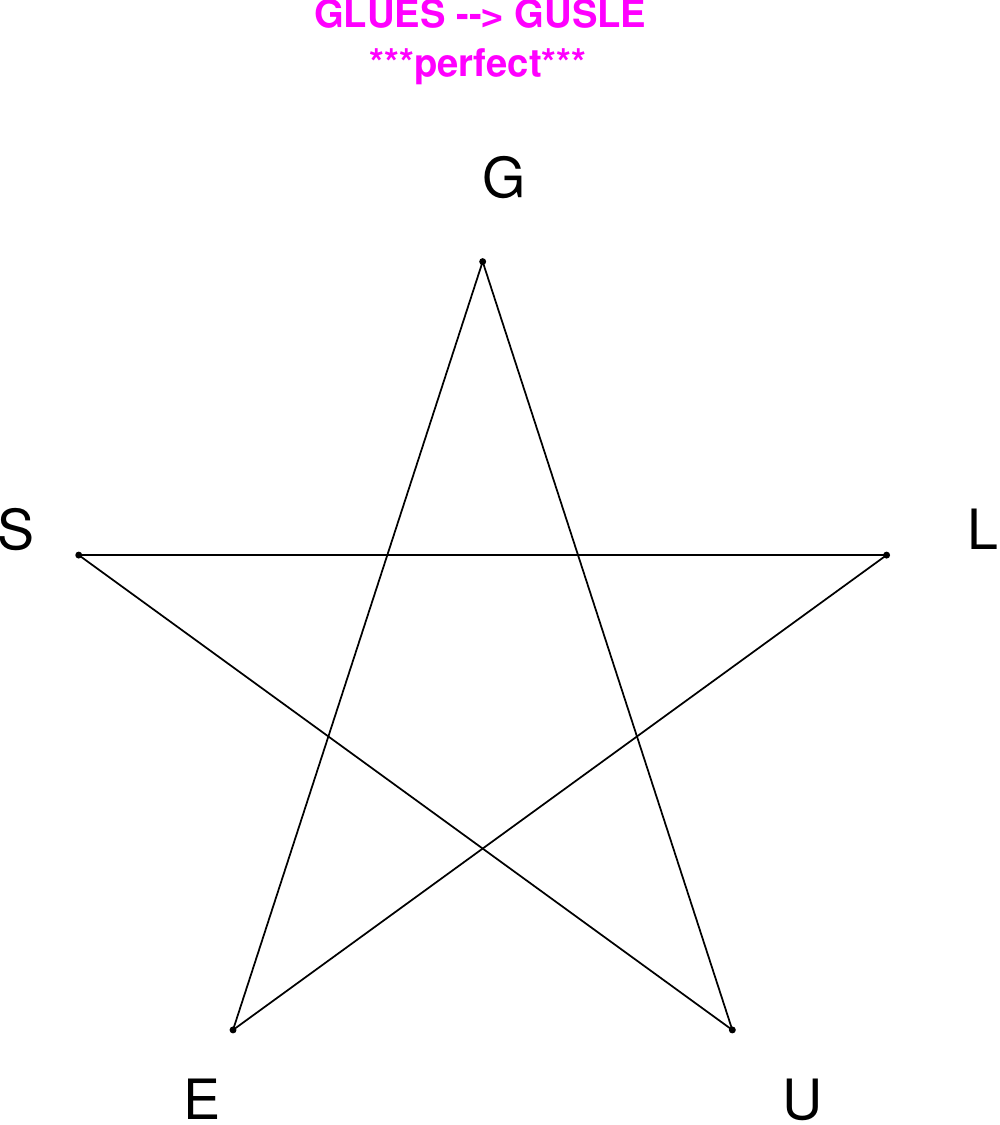}
\end{subfigure}
\hfill
\begin{subfigure}[T]{0.19\textwidth}
\centering
\includegraphics[width=\textwidth]{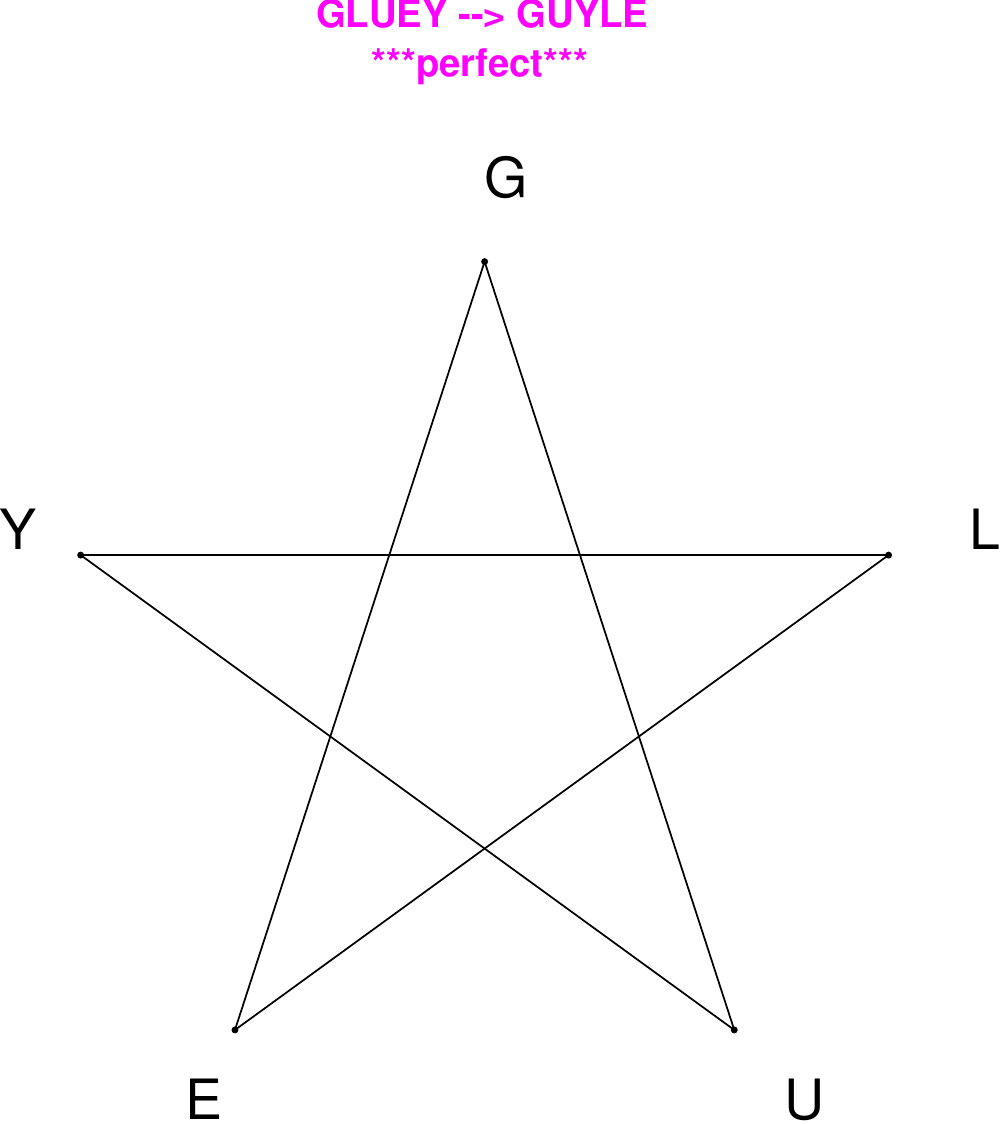}
\end{subfigure}
\end{figure}

\begin{figure}[H]
\centering
\begin{subfigure}[T]{0.19\textwidth}
\centering
\includegraphics[width=\textwidth]{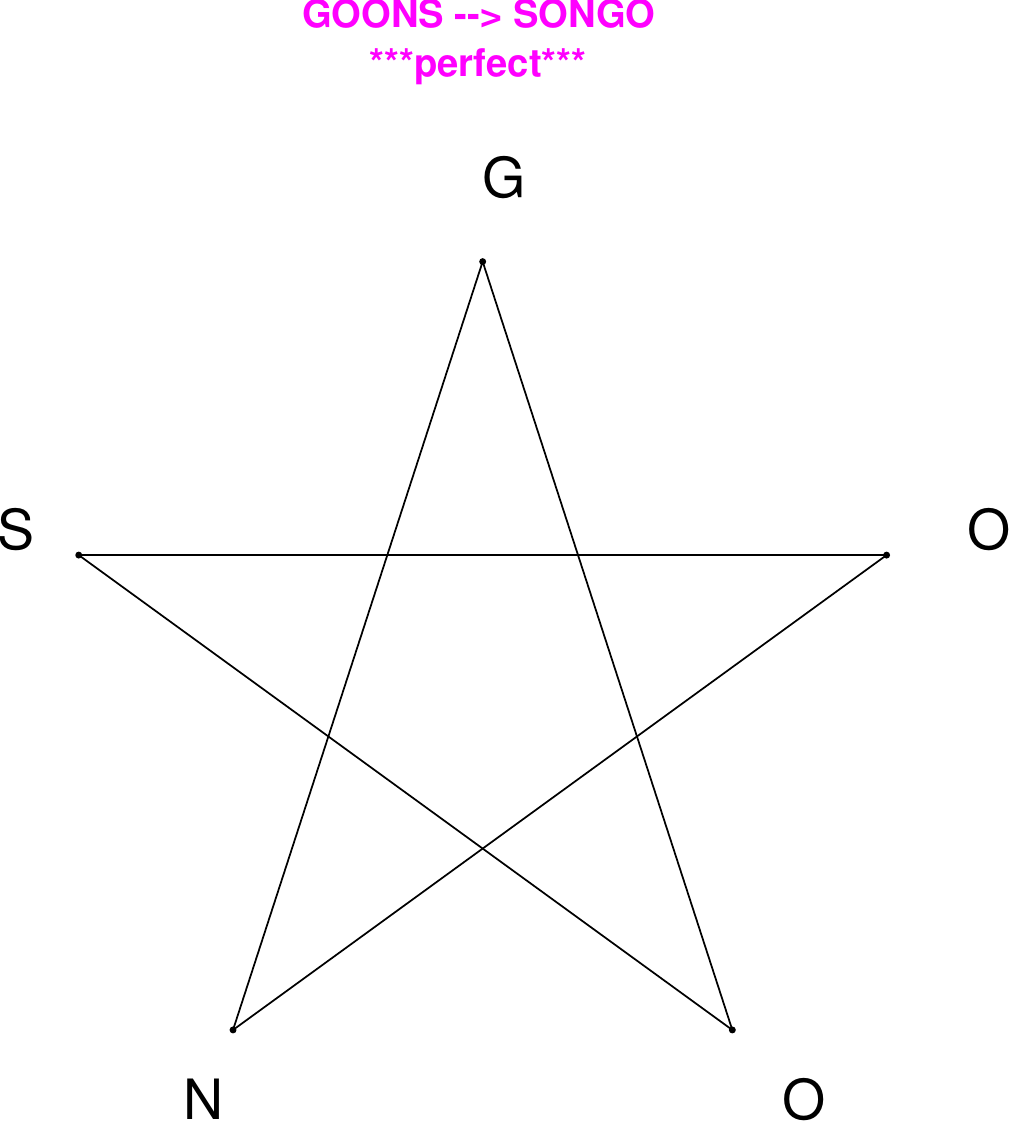}
\end{subfigure}
\hfill
\begin{subfigure}[T]{0.19\textwidth}
\centering
\includegraphics[width=\textwidth]{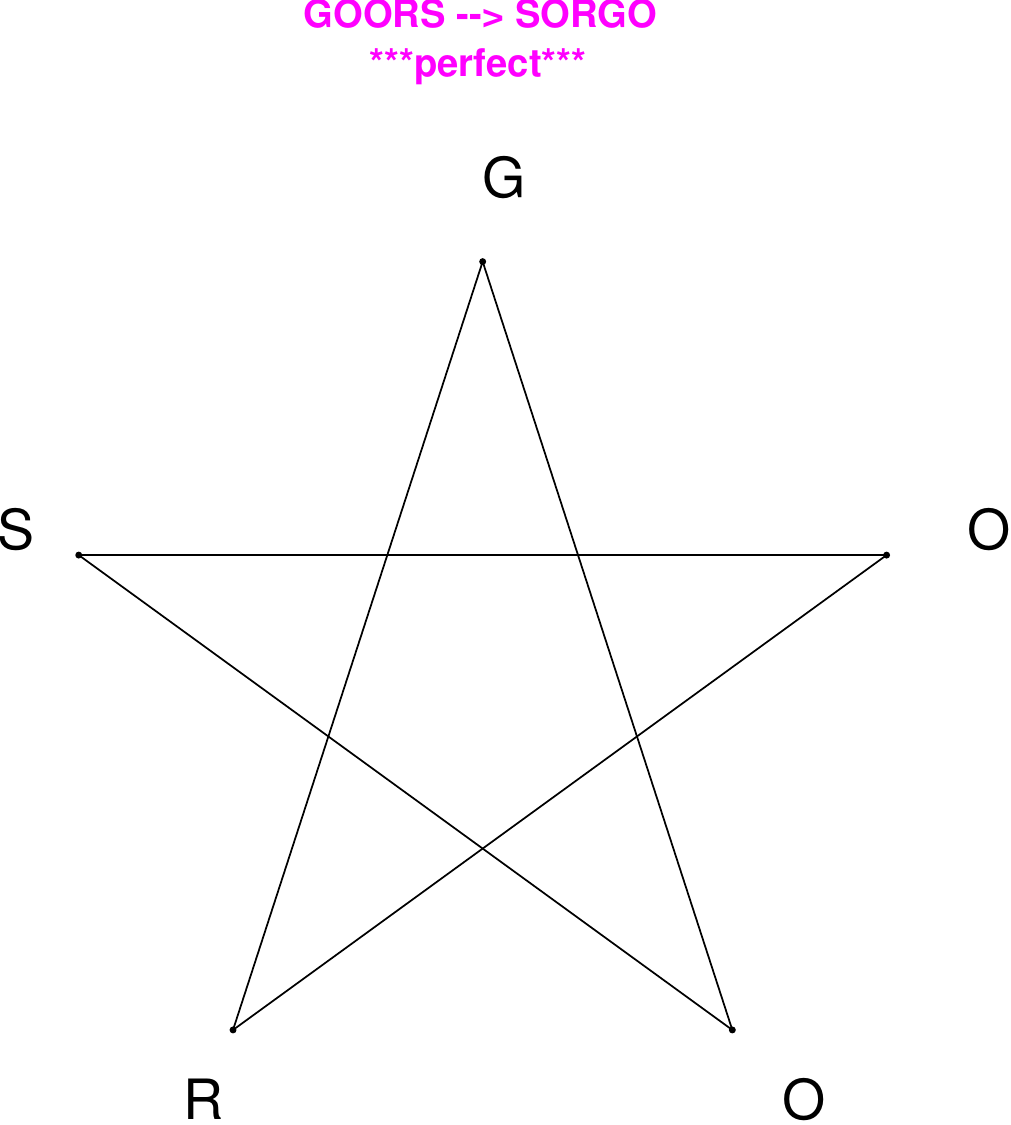}
\end{subfigure}
\hfill
\begin{subfigure}[T]{0.19\textwidth}
\centering
\includegraphics[width=\textwidth]{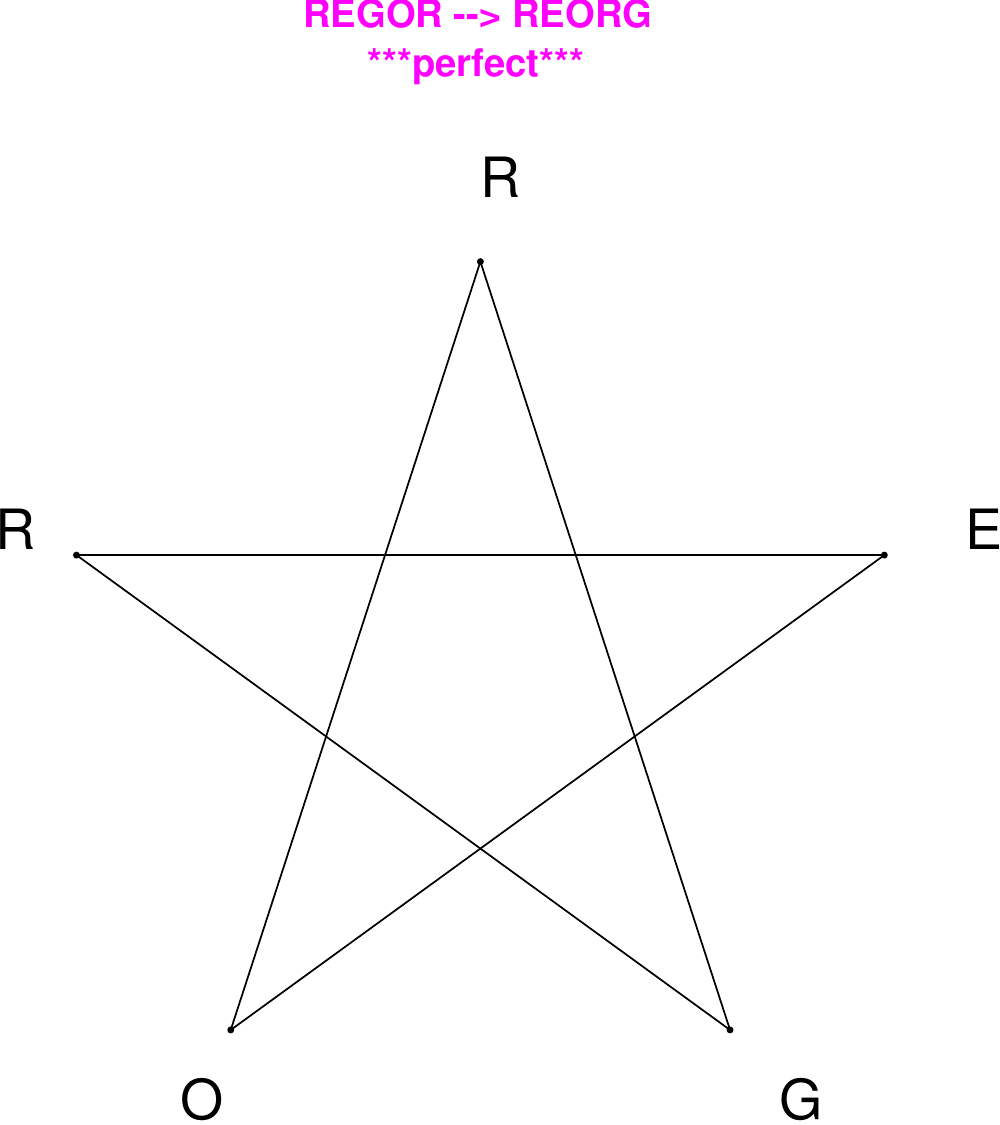}
\end{subfigure}
\hfill
\begin{subfigure}[T]{0.19\textwidth}
\centering
\includegraphics[width=\textwidth]{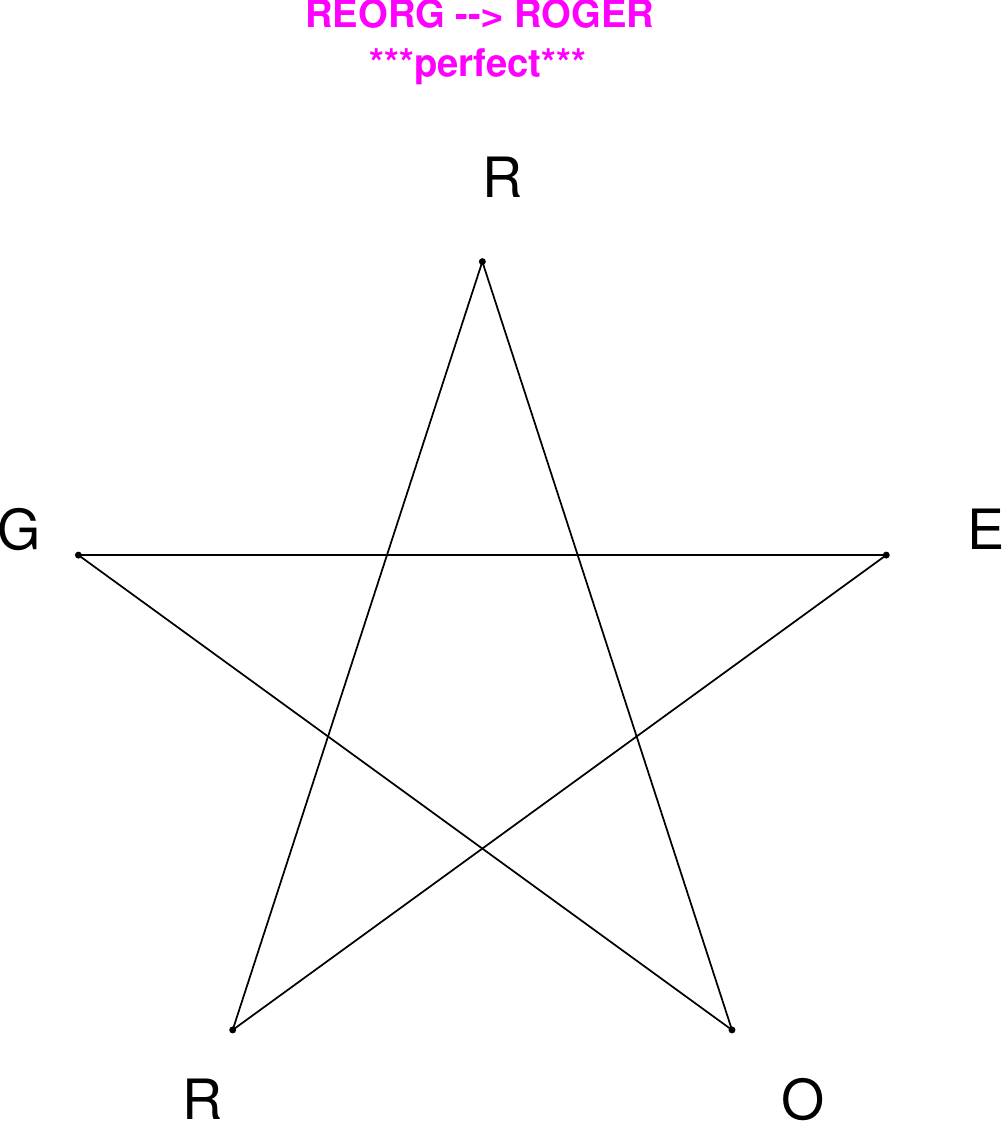}
\end{subfigure}
\hfill
\begin{subfigure}[T]{0.19\textwidth}
\centering
\includegraphics[width=\textwidth]{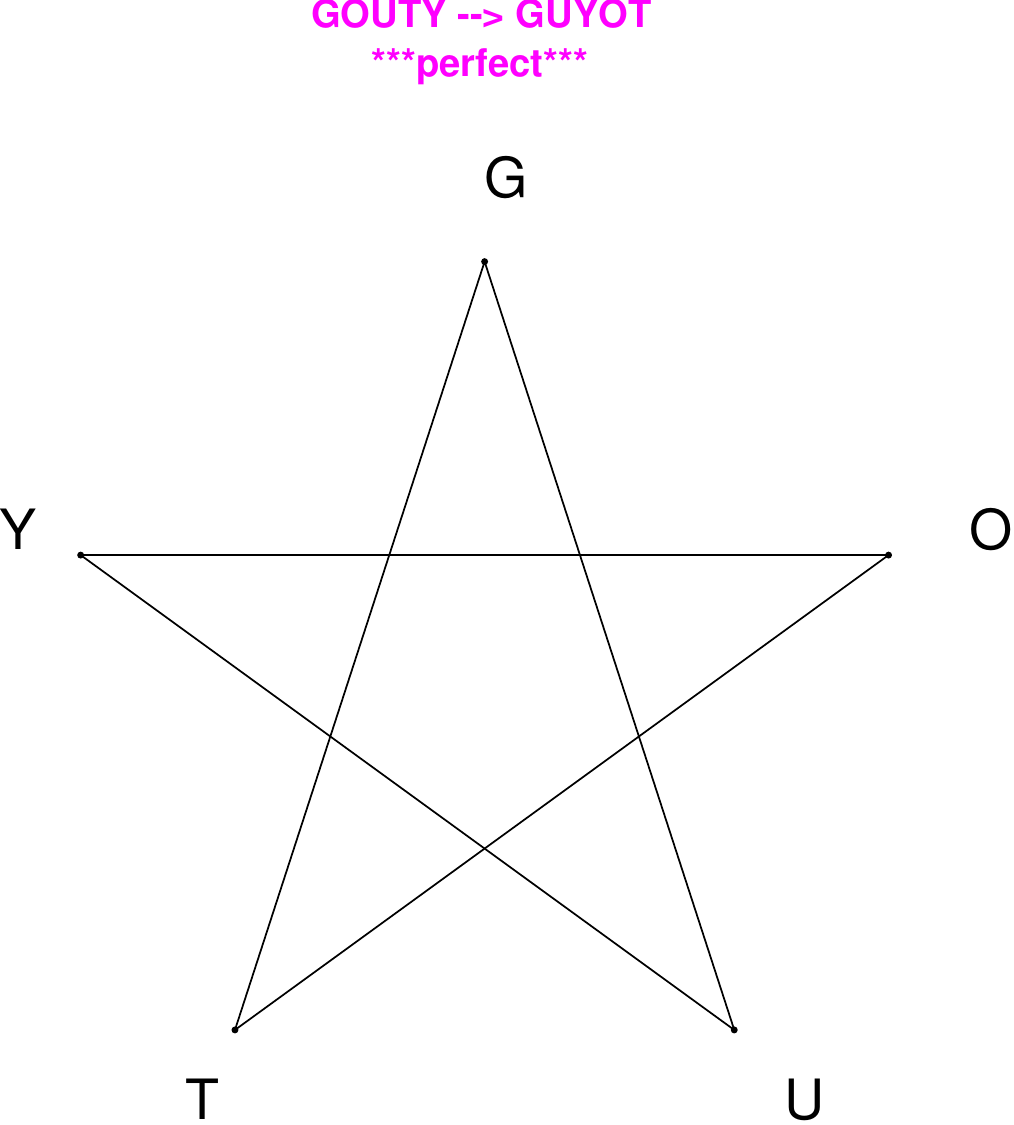}
\end{subfigure}
\end{figure}

\begin{figure}[H]
\centering
\begin{subfigure}[T]{0.19\textwidth}
\centering
\includegraphics[width=\textwidth]{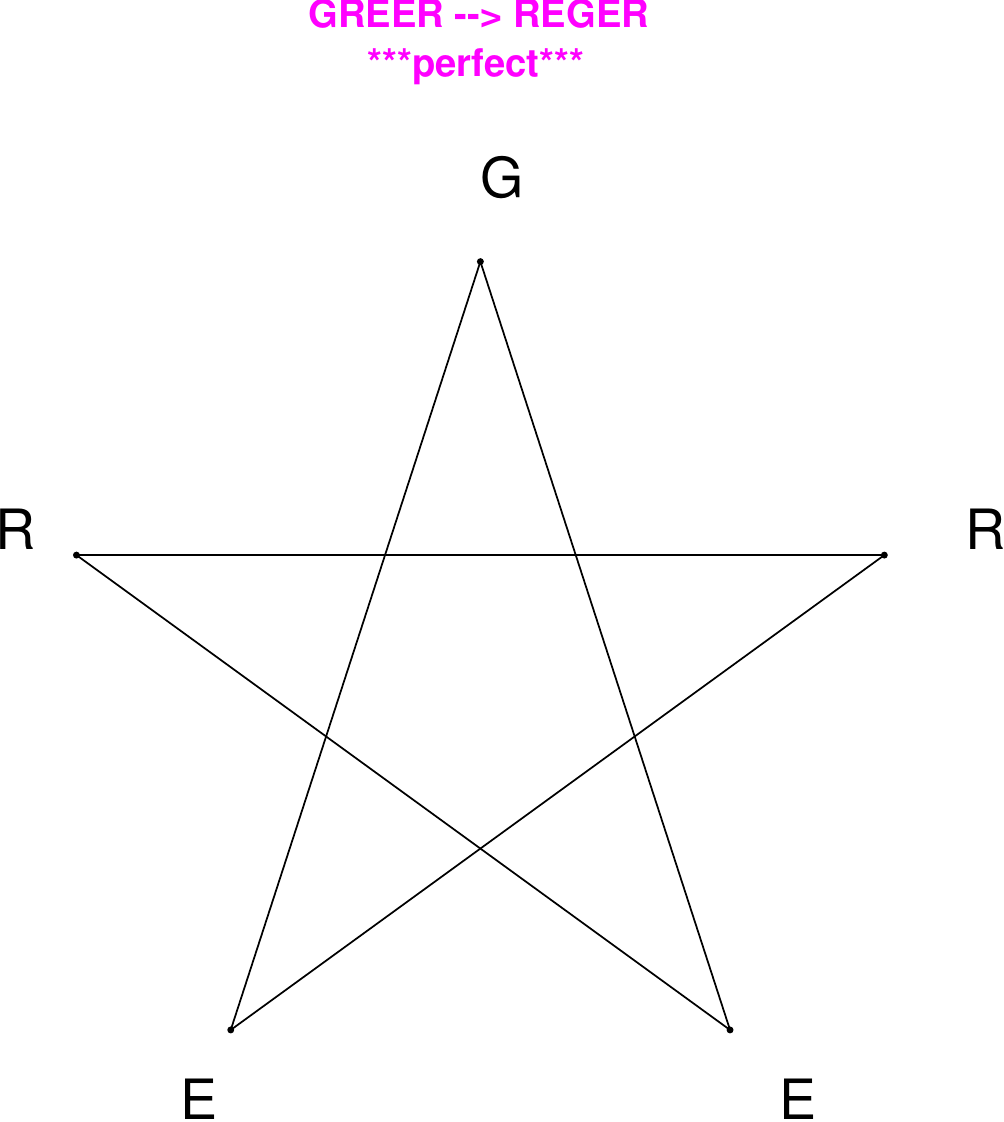}
\end{subfigure}
\hfill
\begin{subfigure}[T]{0.19\textwidth}
\centering
\includegraphics[width=\textwidth]{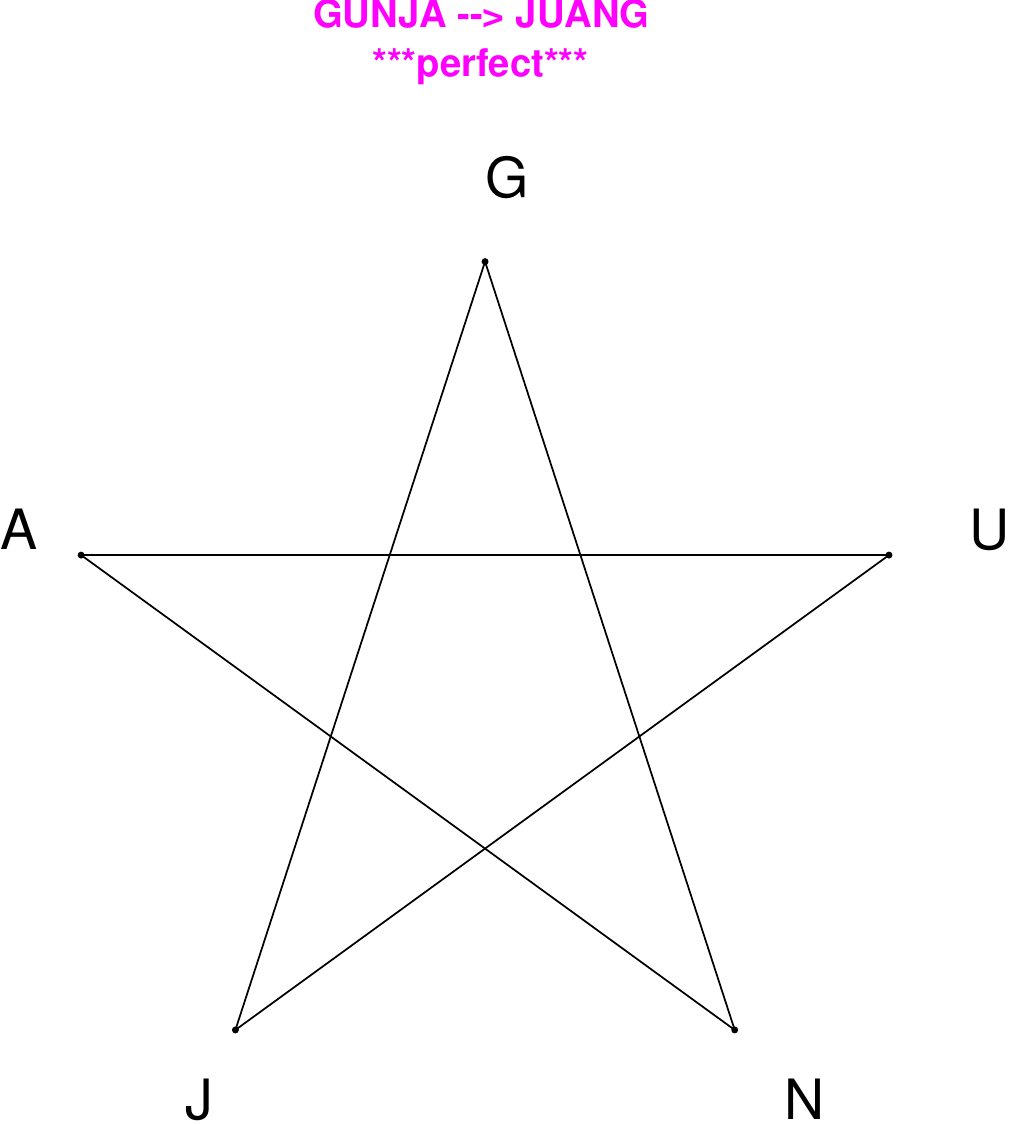}
\end{subfigure}
\hfill
\begin{subfigure}[T]{0.19\textwidth}
\centering
\includegraphics[width=\textwidth]{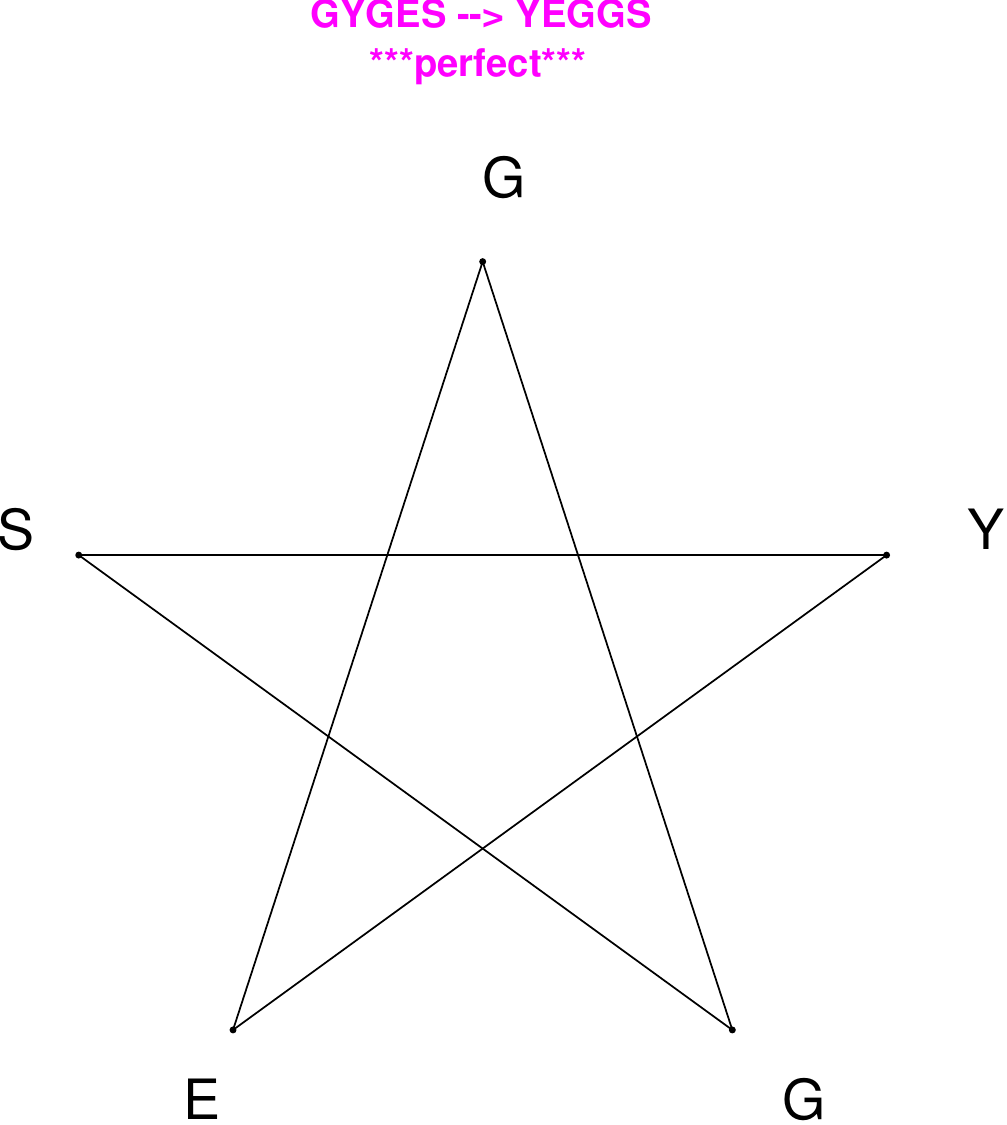}
\end{subfigure}
\hfill
\begin{subfigure}[T]{0.19\textwidth}
\centering
\includegraphics[width=\textwidth]{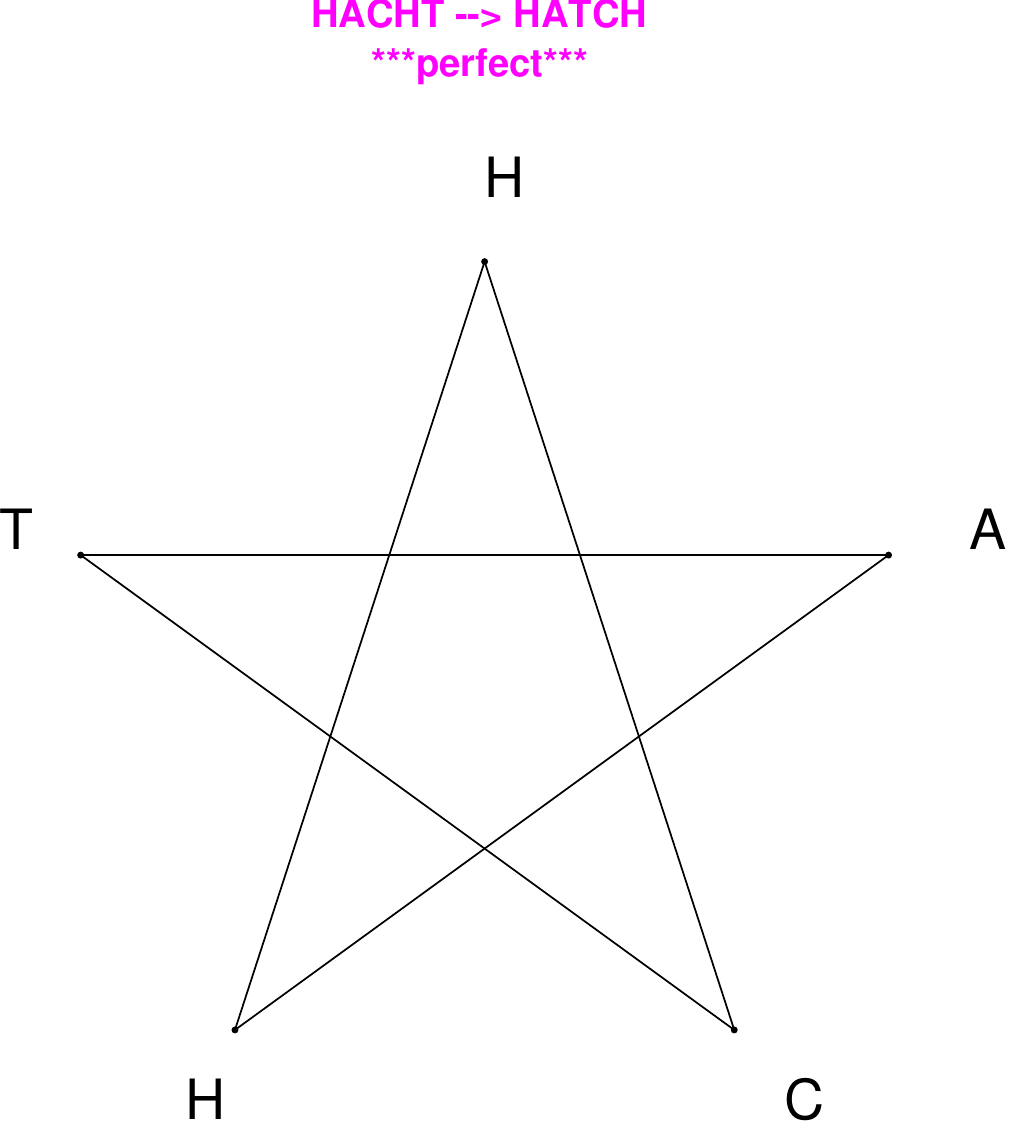}
\end{subfigure}
\hfill
\begin{subfigure}[T]{0.19\textwidth}
\centering
\includegraphics[width=\textwidth]{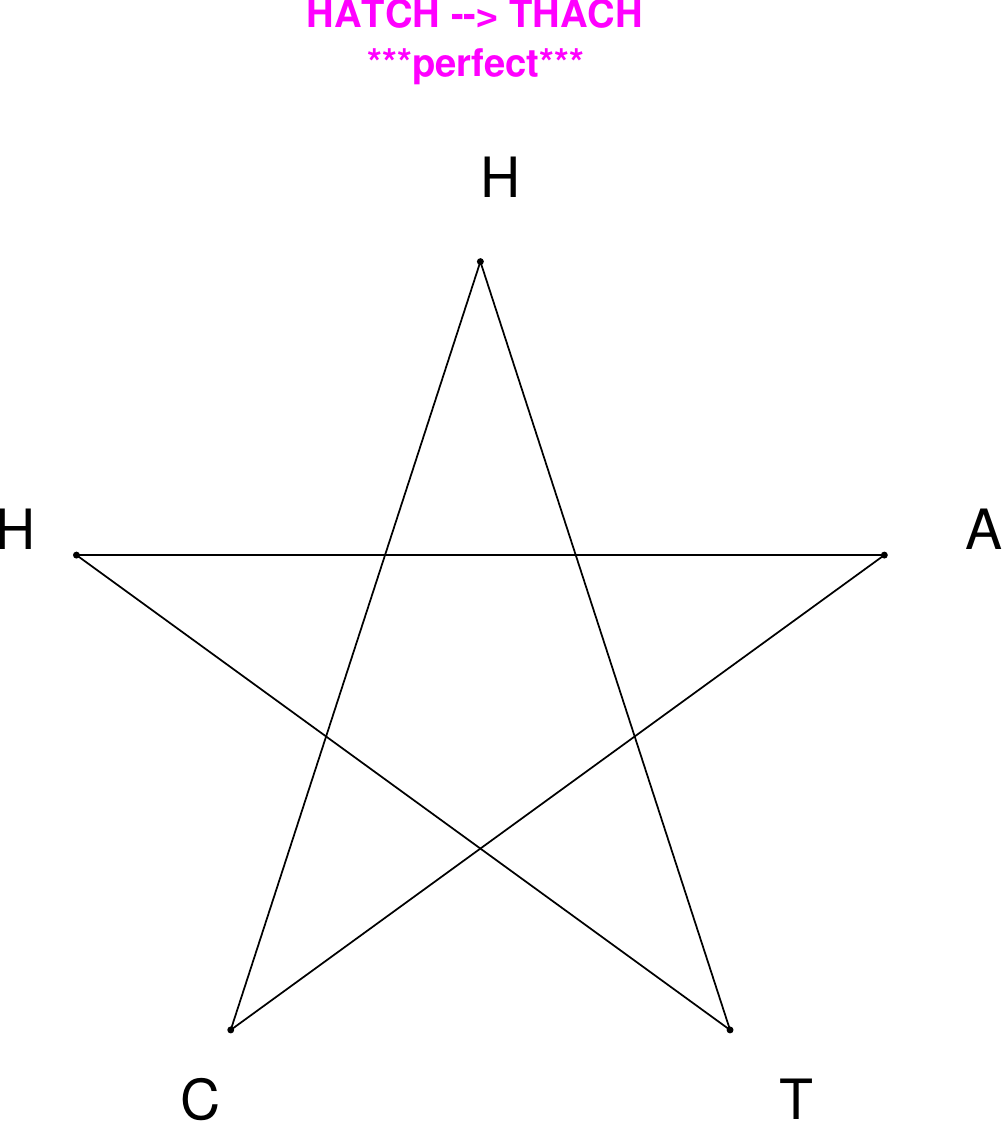}
\end{subfigure}
\end{figure}

\begin{figure}[H]
\centering
\begin{subfigure}[T]{0.19\textwidth}
\centering
\includegraphics[width=\textwidth]{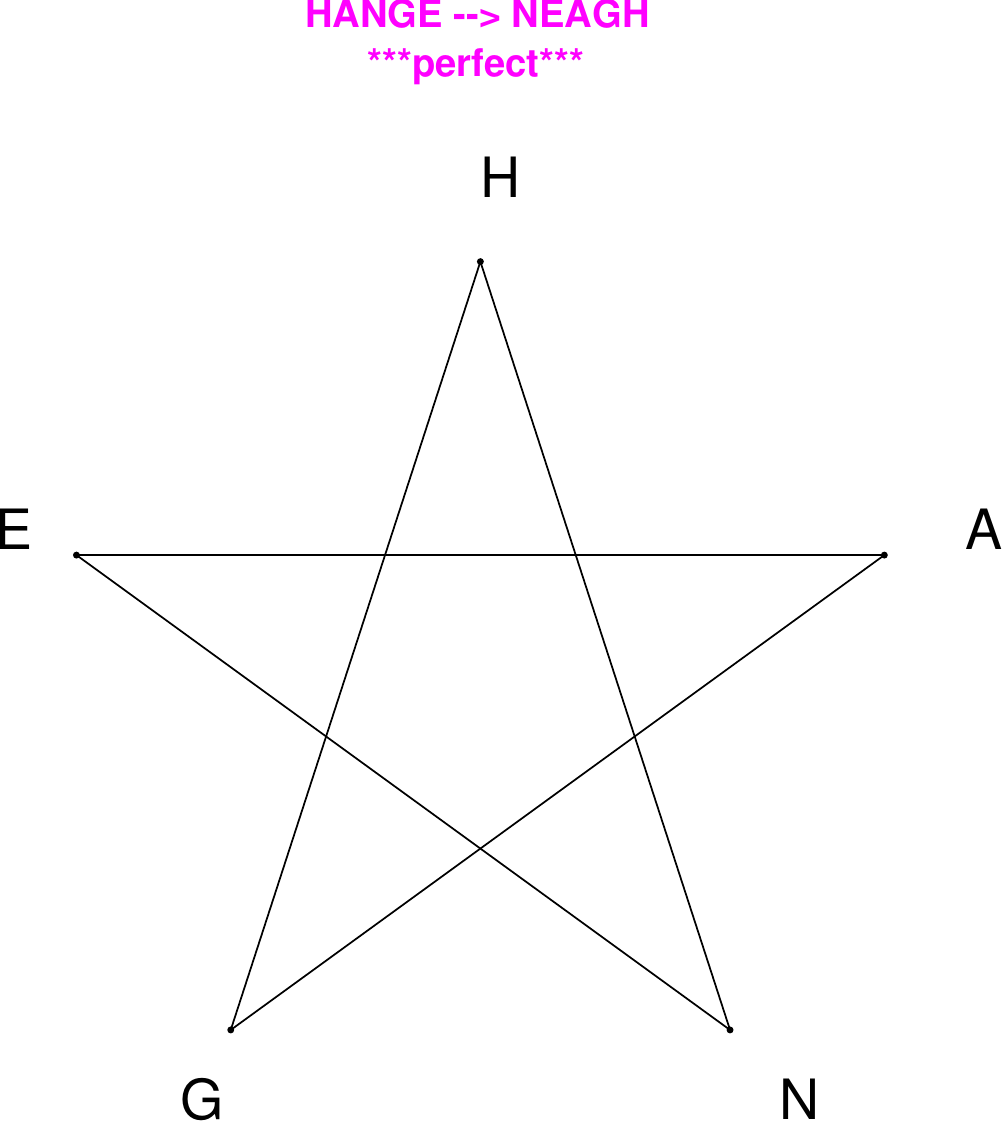}
\end{subfigure}
\hfill
\begin{subfigure}[T]{0.19\textwidth}
\centering
\includegraphics[width=\textwidth]{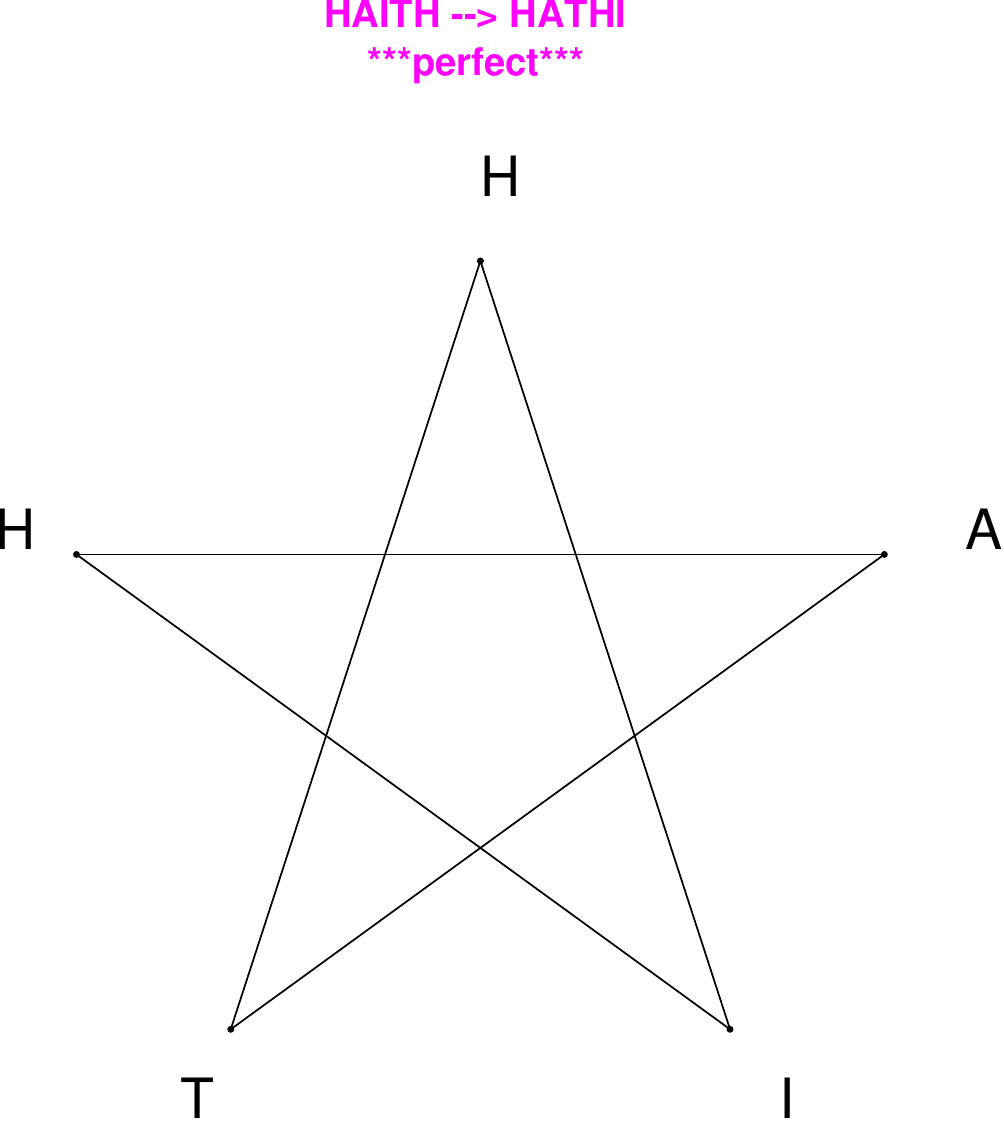}
\end{subfigure}
\hfill
\begin{subfigure}[T]{0.19\textwidth}
\centering
\includegraphics[width=\textwidth]{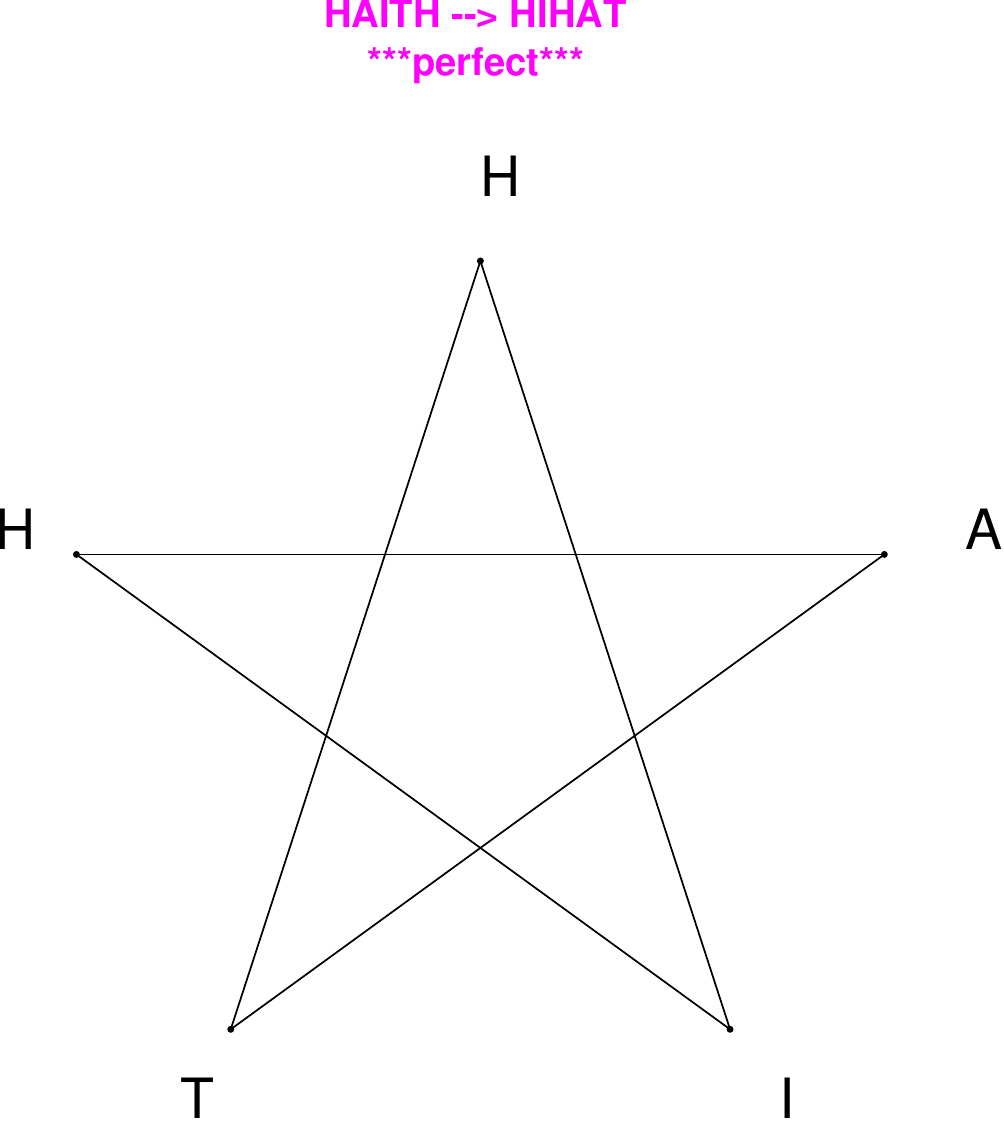}
\end{subfigure}
\hfill
\begin{subfigure}[T]{0.19\textwidth}
\centering
\includegraphics[width=\textwidth]{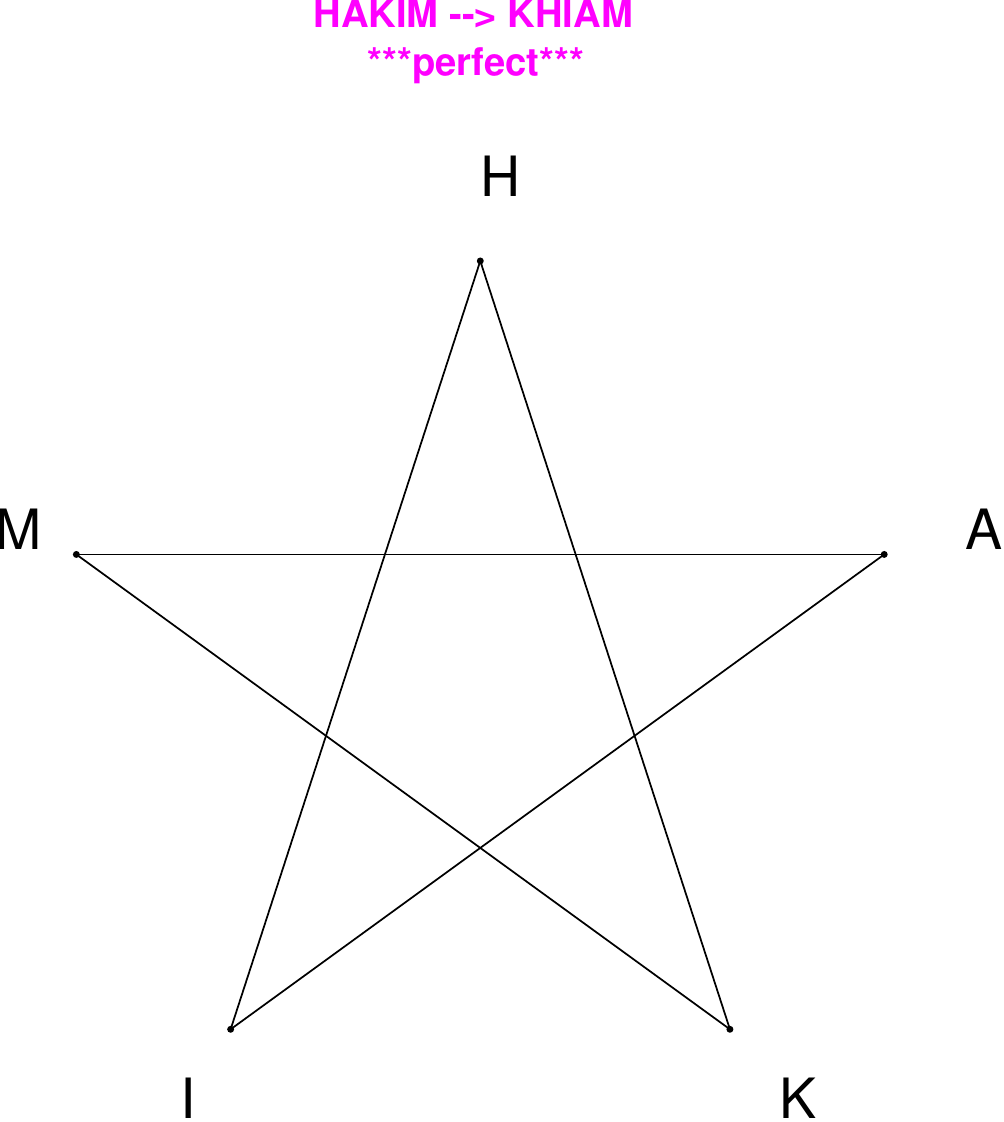}
\end{subfigure}
\hfill
\begin{subfigure}[T]{0.19\textwidth}
\centering
\includegraphics[width=\textwidth]{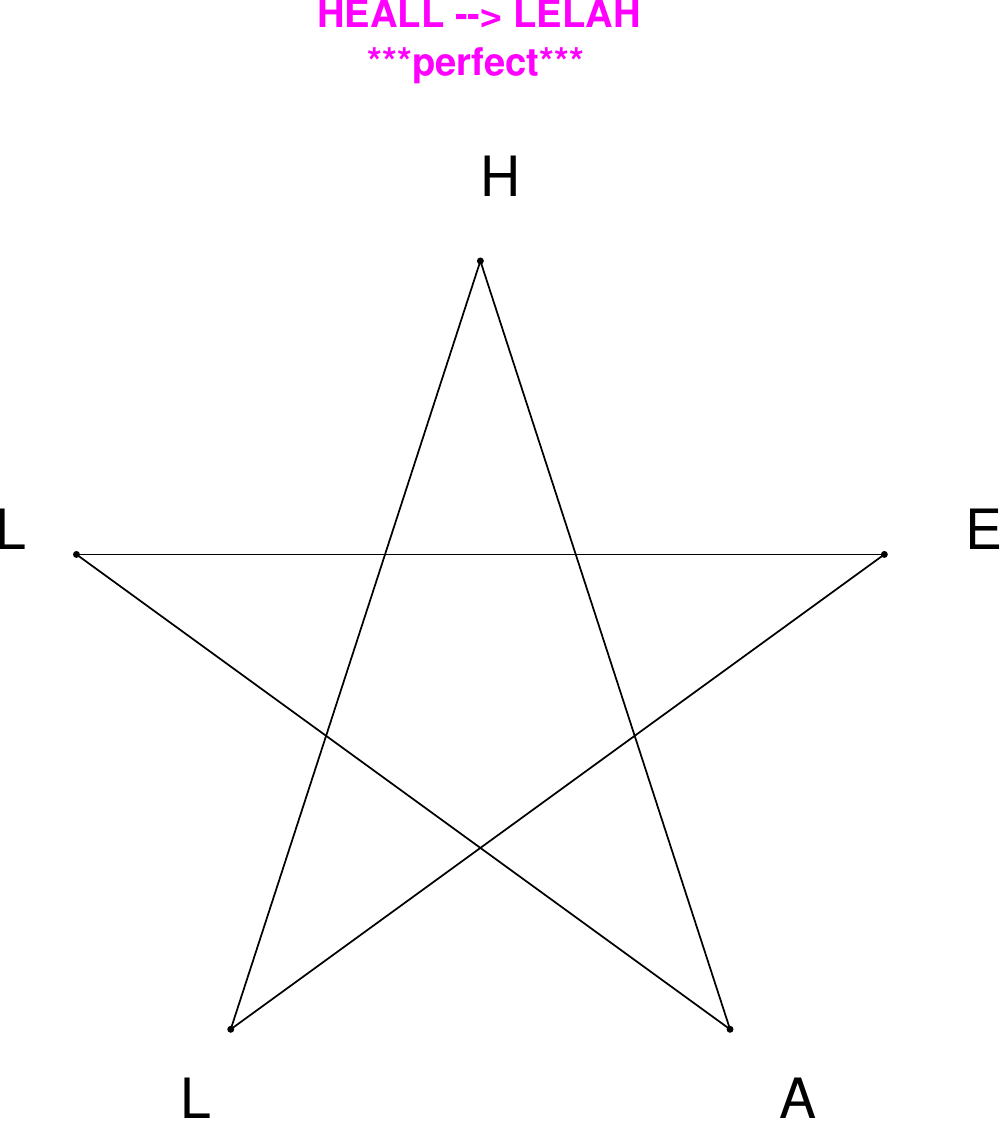}
\end{subfigure}
\end{figure}

\begin{figure}[H]
\centering
\begin{subfigure}[T]{0.19\textwidth}
\centering
\includegraphics[width=\textwidth]{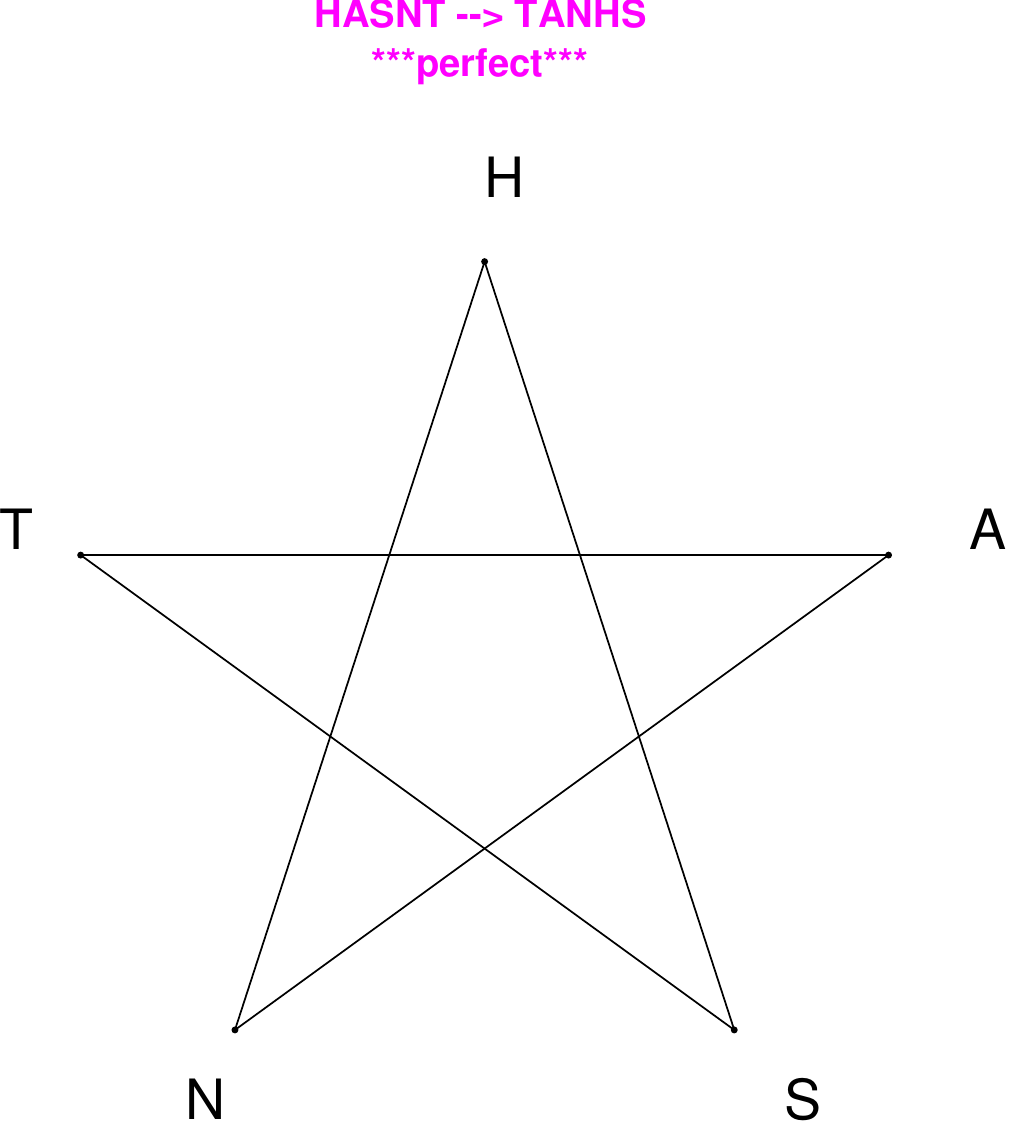}
\end{subfigure}
\hfill
\begin{subfigure}[T]{0.19\textwidth}
\centering
\includegraphics[width=\textwidth]{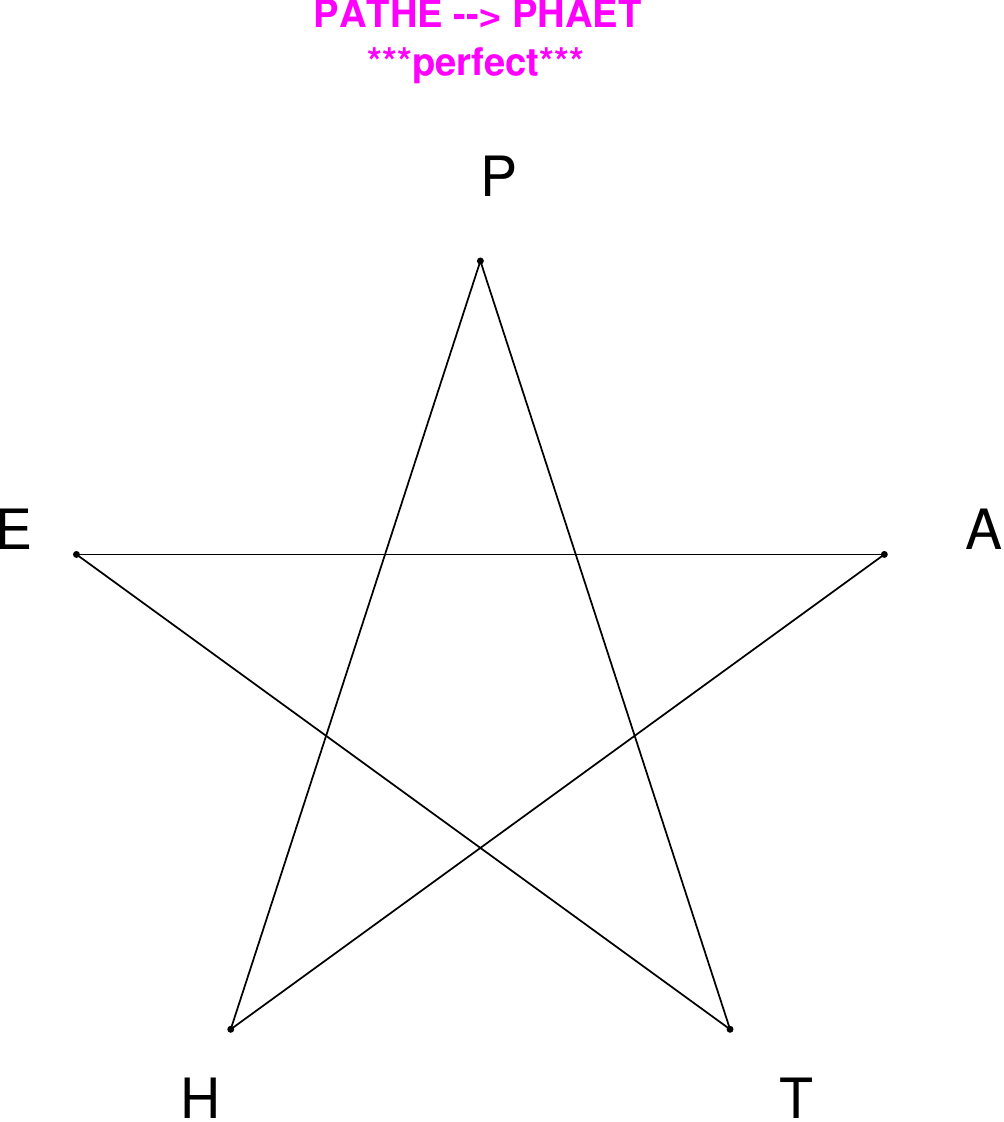}
\end{subfigure}
\hfill
\begin{subfigure}[T]{0.19\textwidth}
\centering
\includegraphics[width=\textwidth]{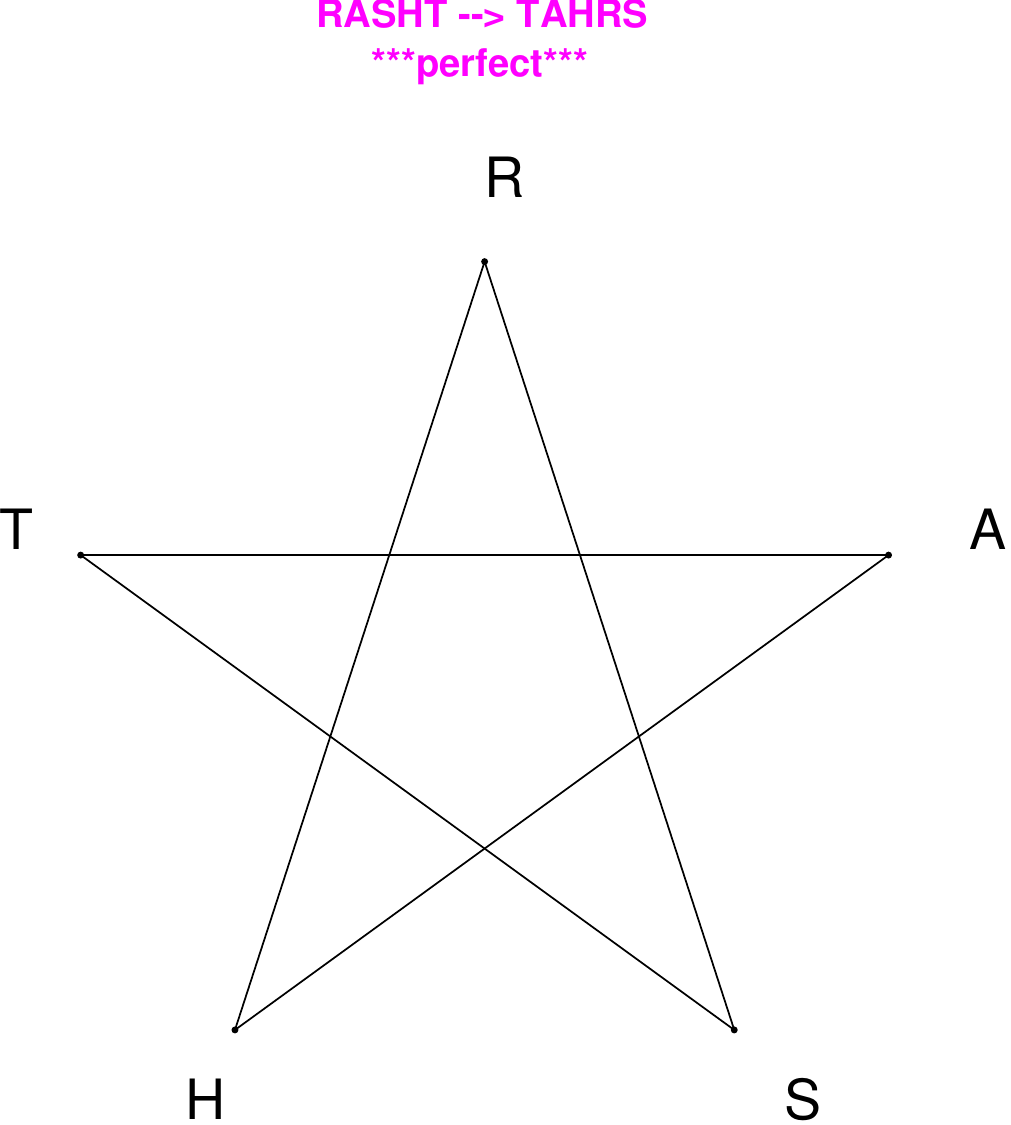}
\end{subfigure}
\hfill
\begin{subfigure}[T]{0.19\textwidth}
\centering
\includegraphics[width=\textwidth]{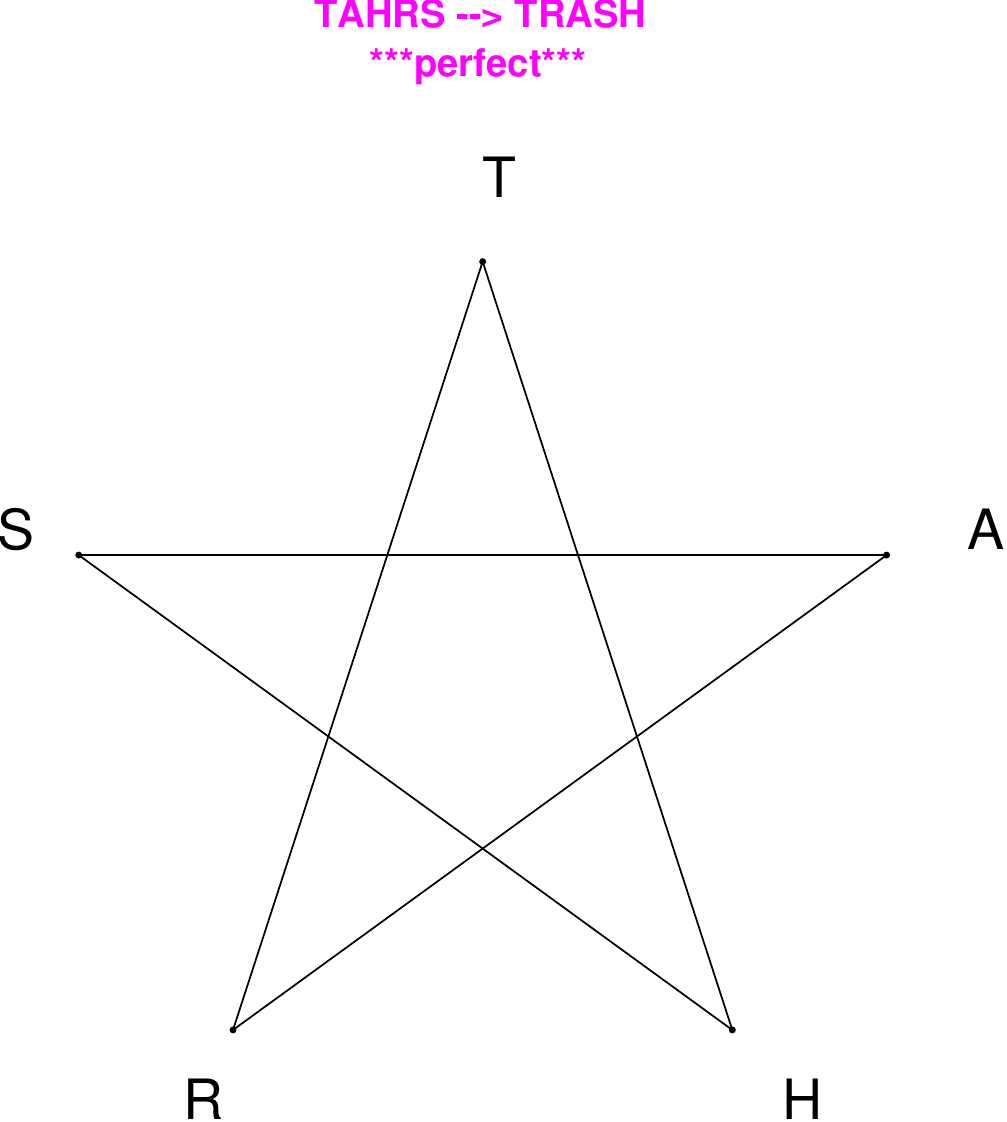}
\end{subfigure}
\hfill
\begin{subfigure}[T]{0.19\textwidth}
\centering
\includegraphics[width=\textwidth]{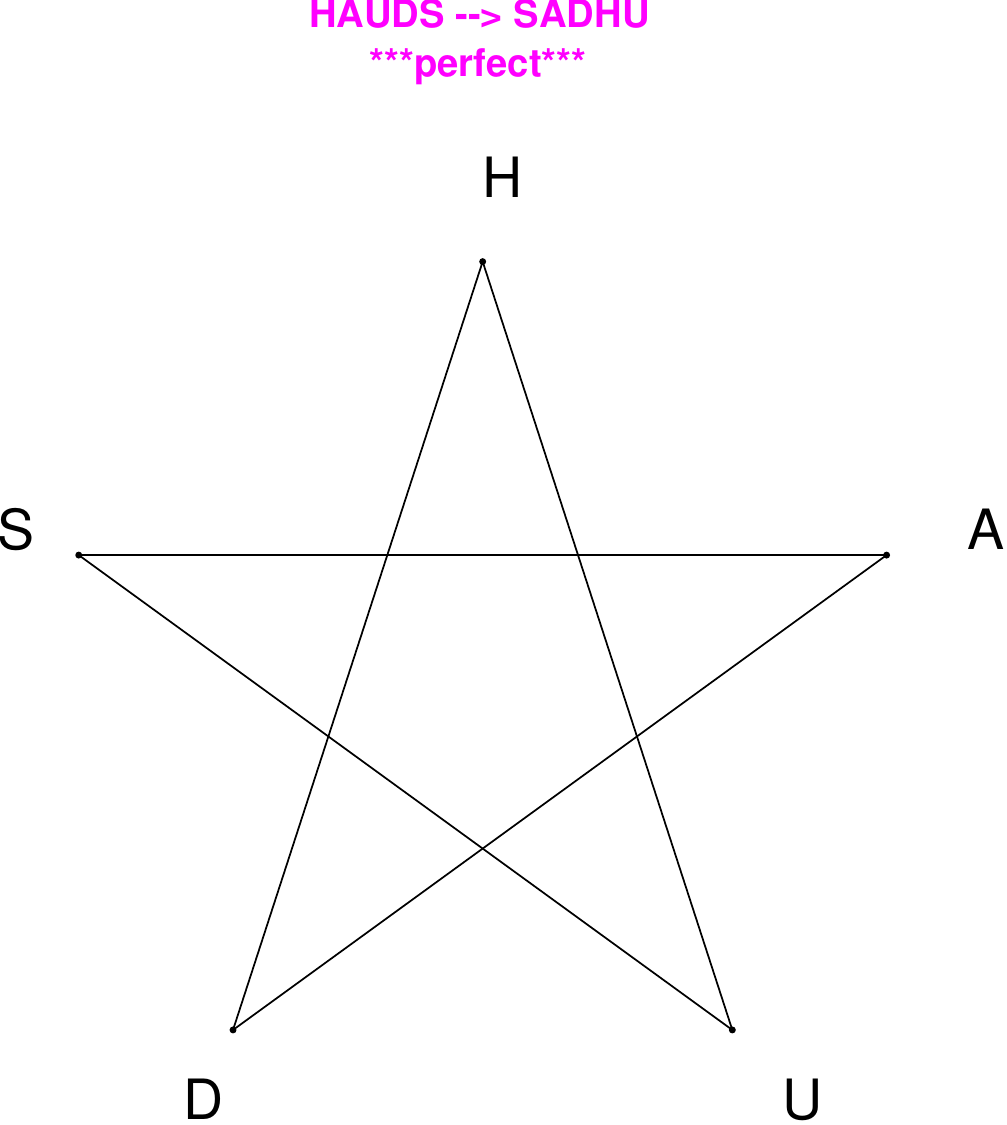}
\end{subfigure}
\end{figure}

\begin{figure}[H]
\centering
\begin{subfigure}[T]{0.19\textwidth}
\centering
\includegraphics[width=\textwidth]{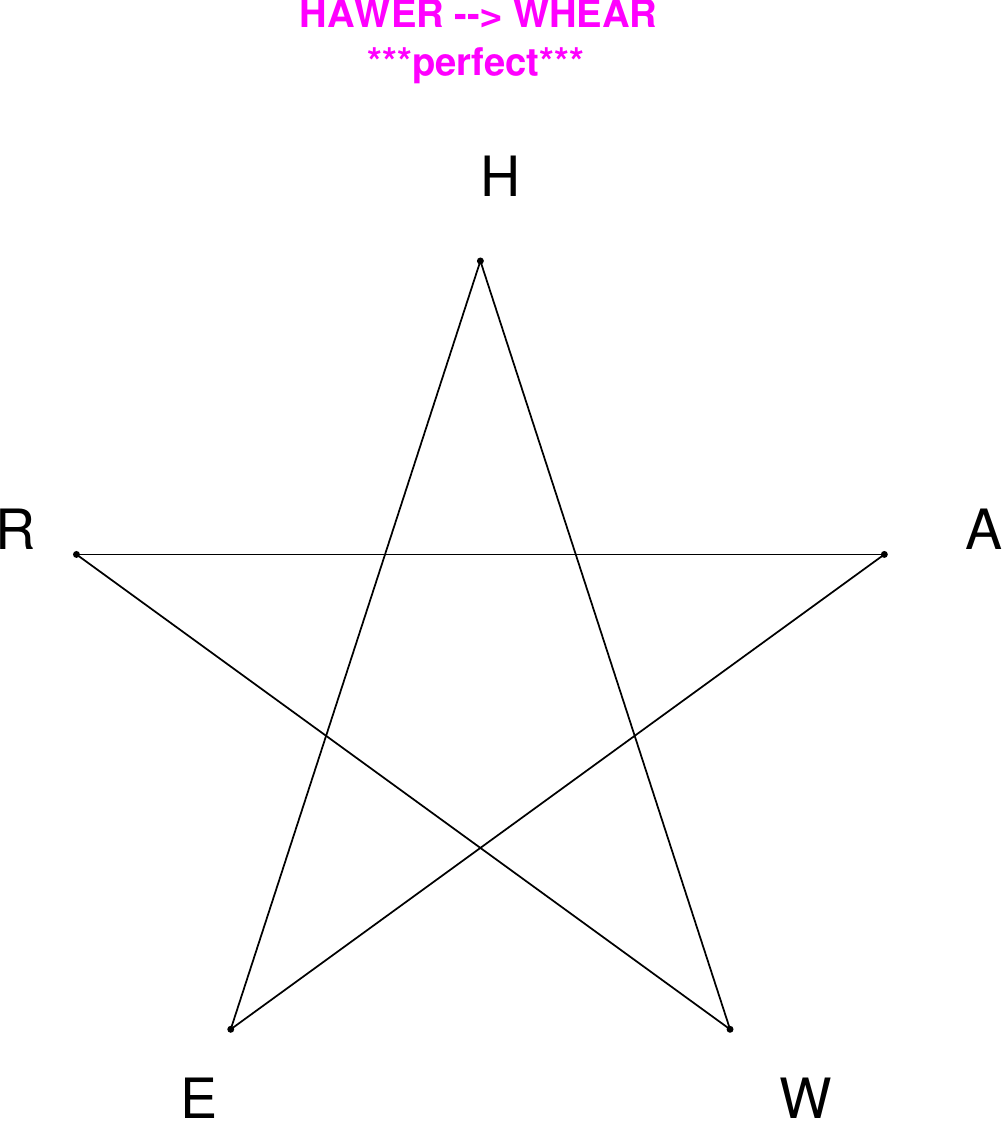}
\end{subfigure}
\hfill
\begin{subfigure}[T]{0.19\textwidth}
\centering
\includegraphics[width=\textwidth]{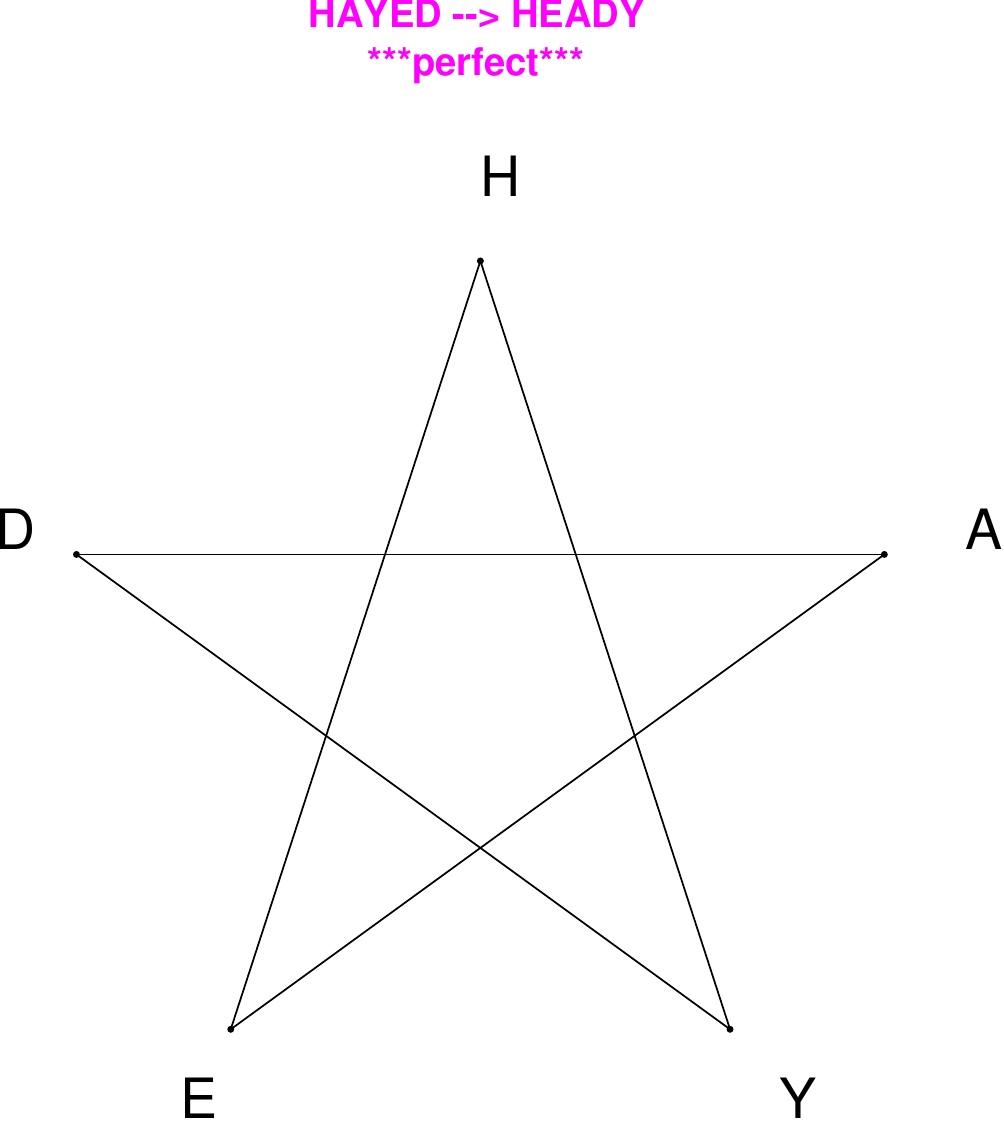}
\end{subfigure}
\hfill
\begin{subfigure}[T]{0.19\textwidth}
\centering
\includegraphics[width=\textwidth]{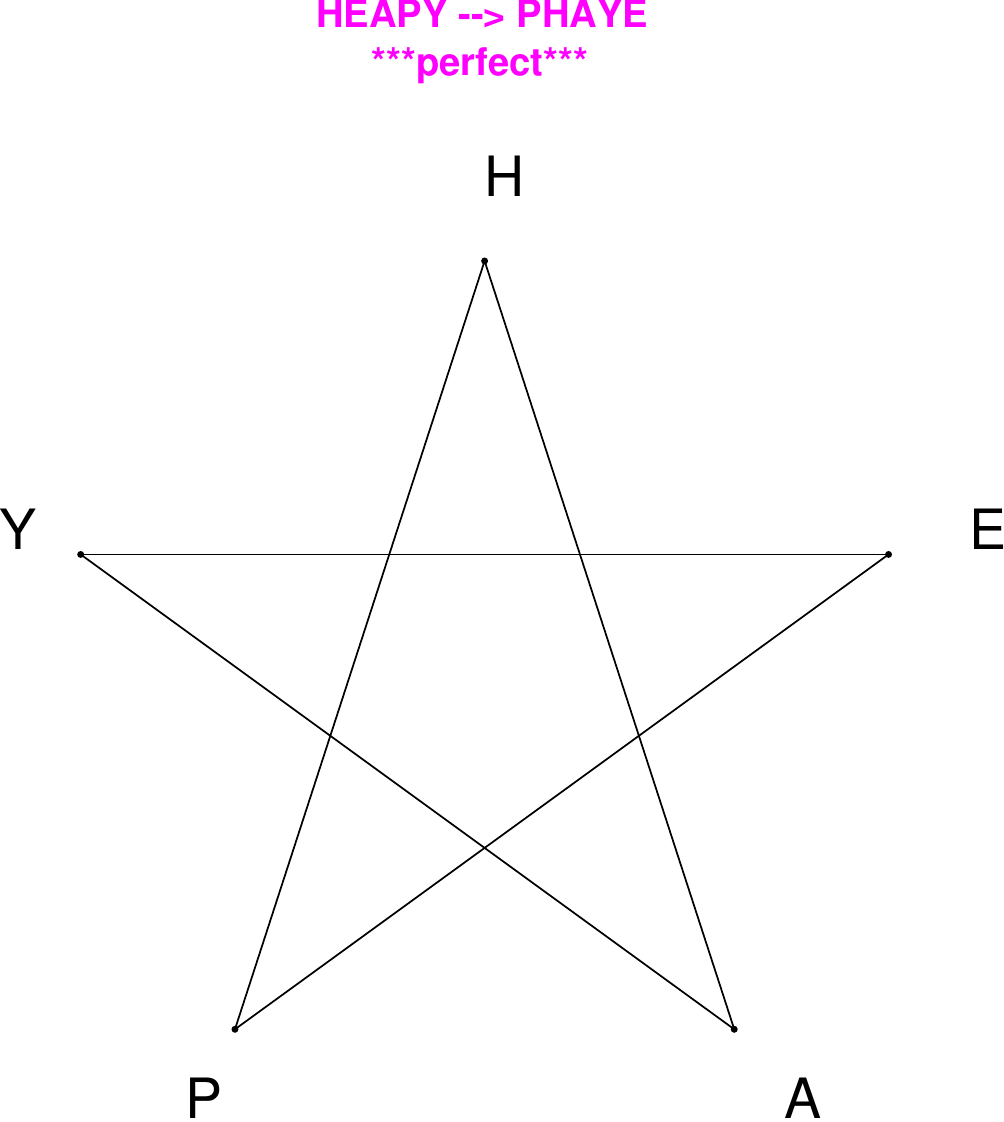}
\end{subfigure}
\hfill
\begin{subfigure}[T]{0.19\textwidth}
\centering
\includegraphics[width=\textwidth]{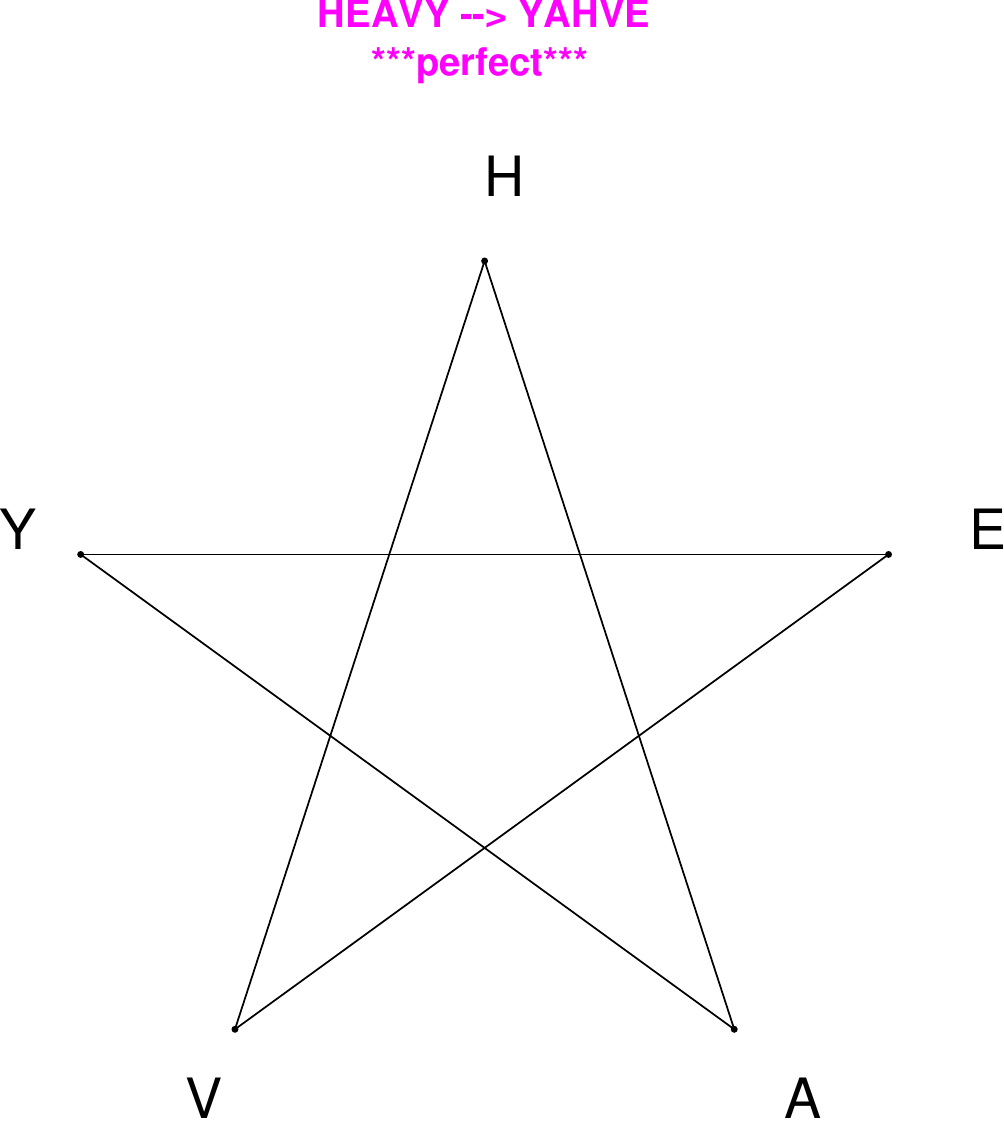}
\end{subfigure}
\hfill
\begin{subfigure}[T]{0.19\textwidth}
\centering
\includegraphics[width=\textwidth]{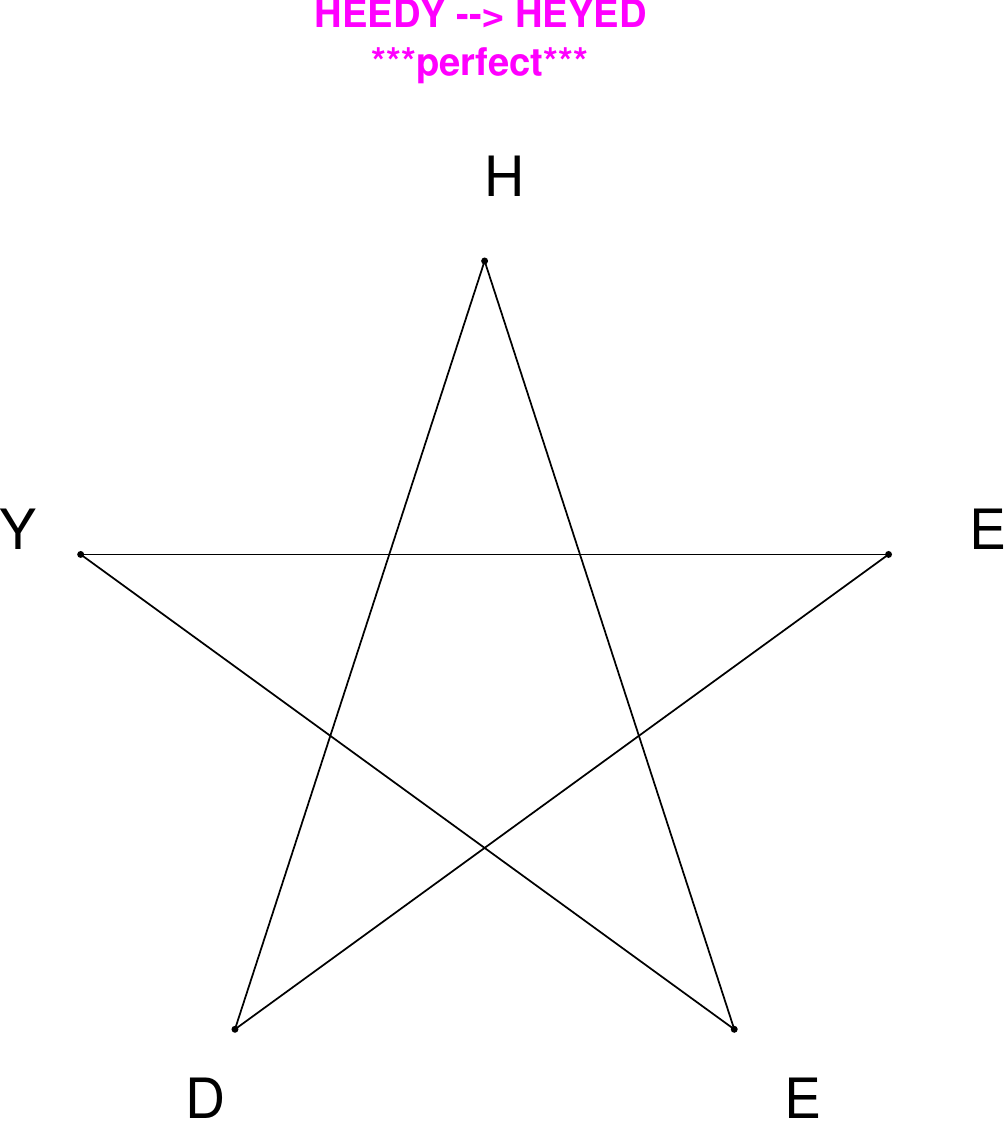}
\end{subfigure}
\end{figure}

\begin{figure}[H]
\centering
\begin{subfigure}[T]{0.19\textwidth}
\centering
\includegraphics[width=\textwidth]{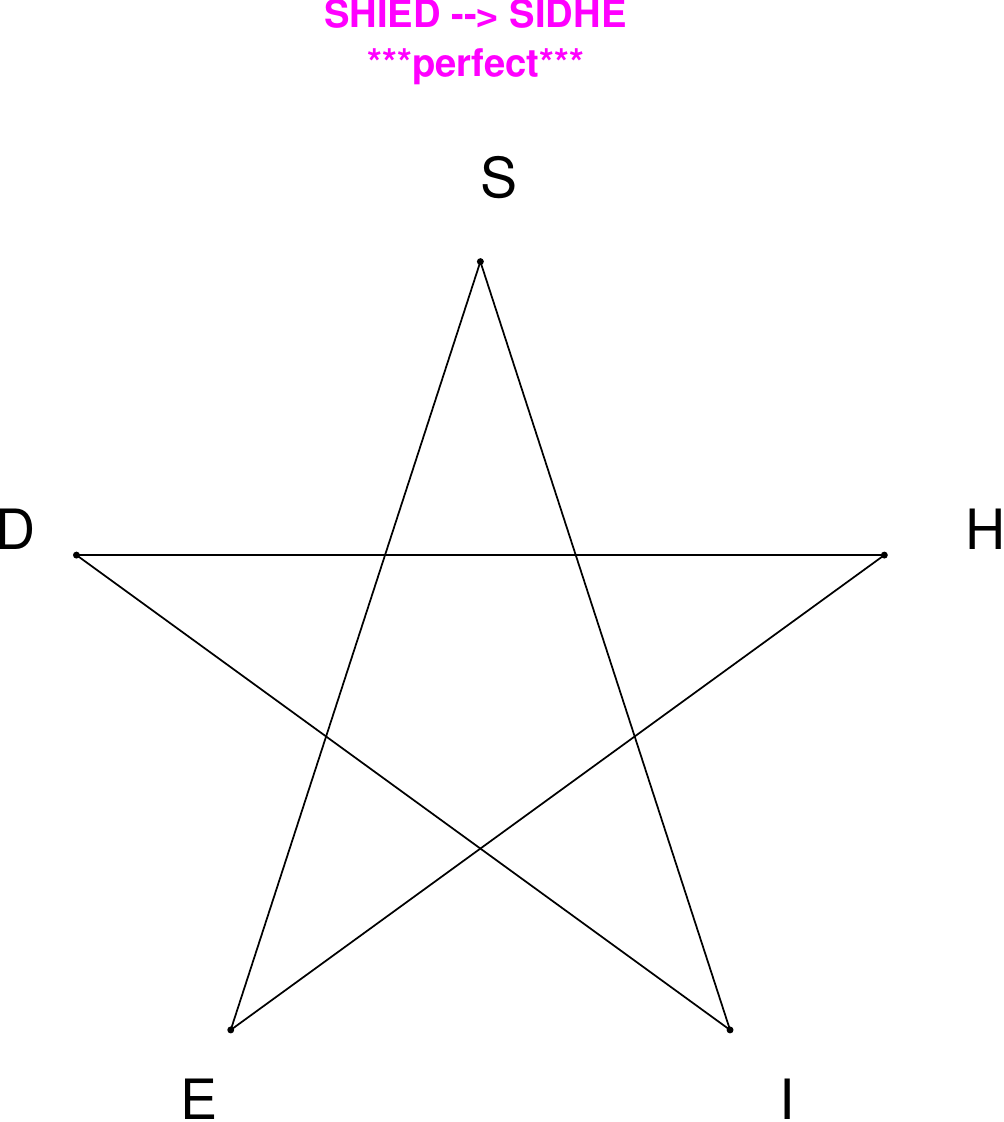}
\end{subfigure}
\hfill
\begin{subfigure}[T]{0.19\textwidth}
\centering
\includegraphics[width=\textwidth]{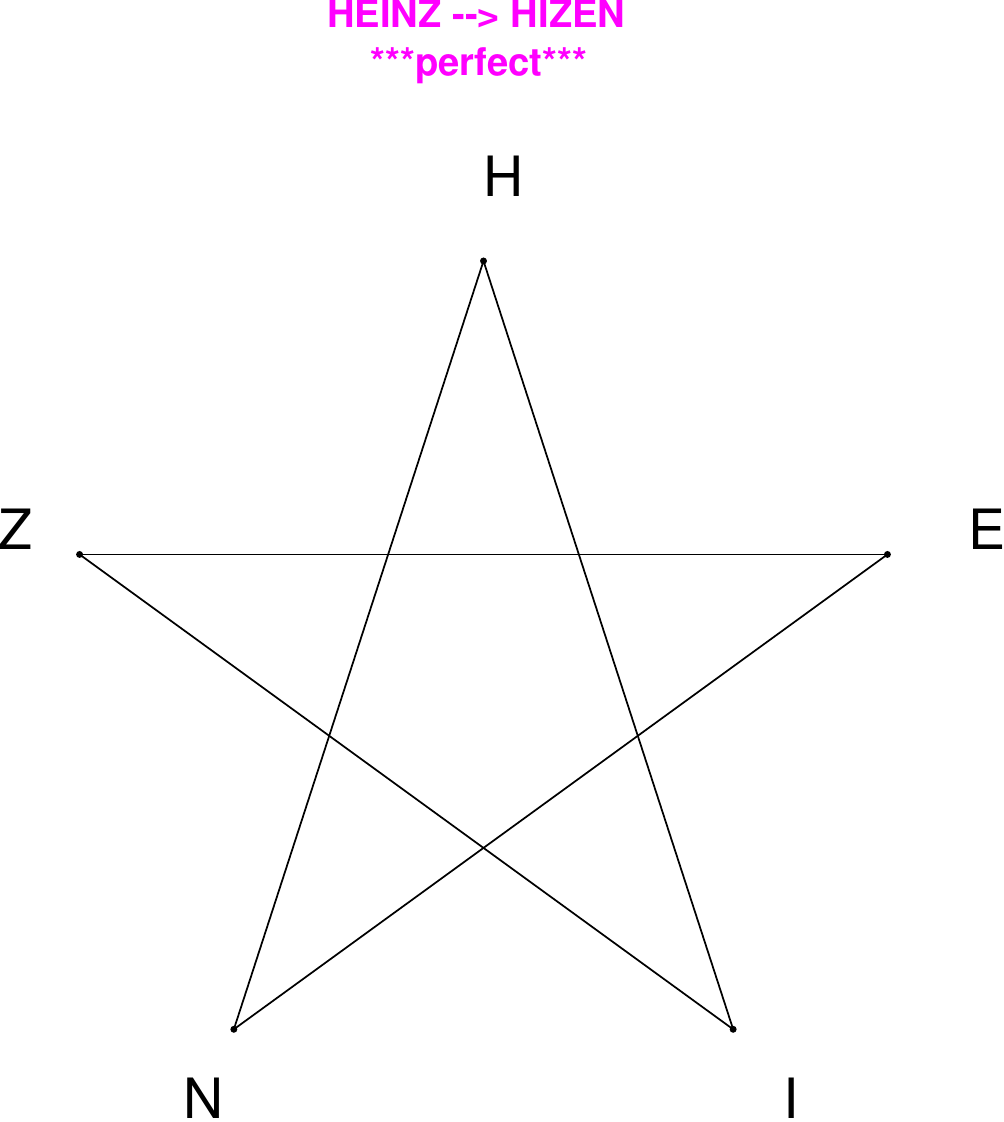}
\end{subfigure}
\hfill
\begin{subfigure}[T]{0.19\textwidth}
\centering
\includegraphics[width=\textwidth]{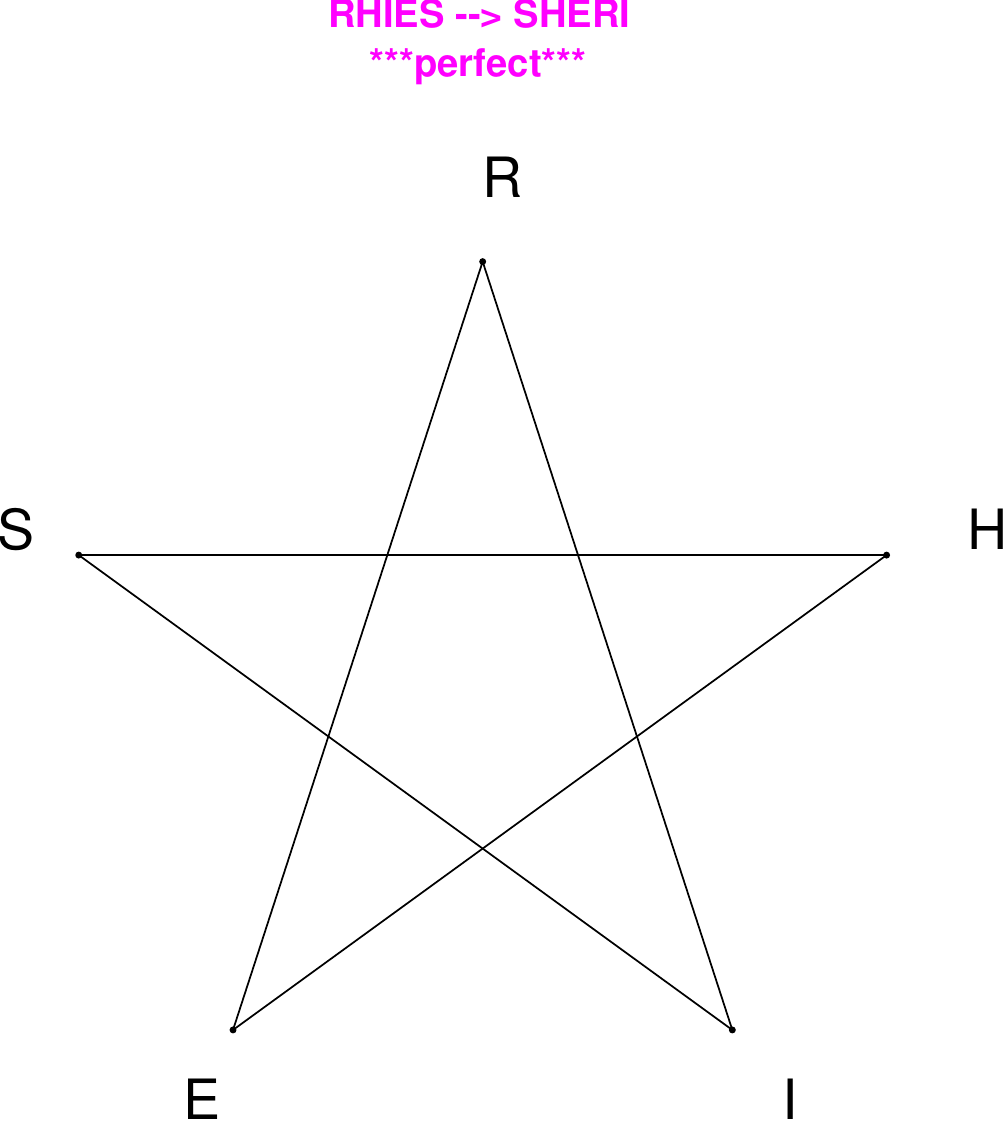}
\end{subfigure}
\hfill
\begin{subfigure}[T]{0.19\textwidth}
\centering
\includegraphics[width=\textwidth]{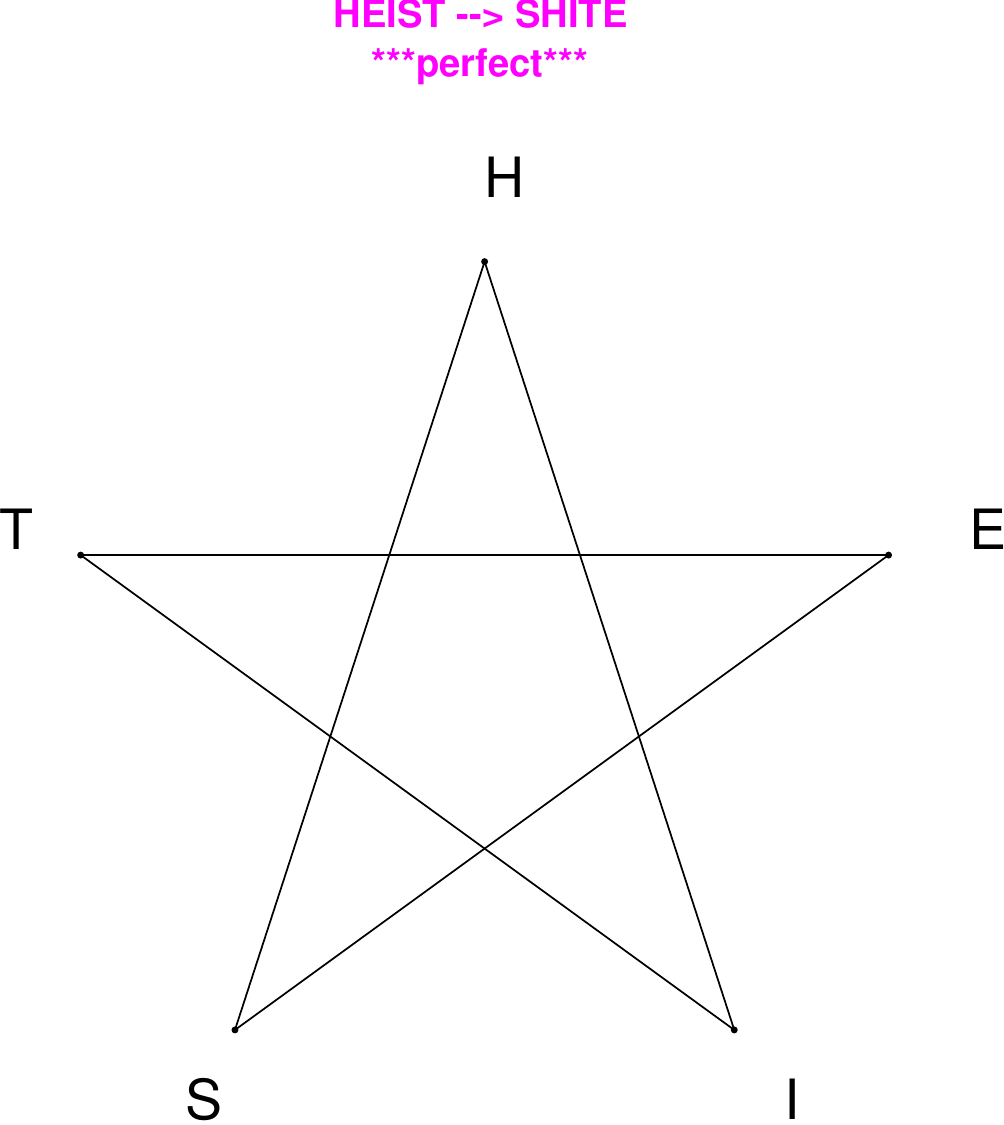}
\end{subfigure}
\hfill
\begin{subfigure}[T]{0.19\textwidth}
\centering
\includegraphics[width=\textwidth]{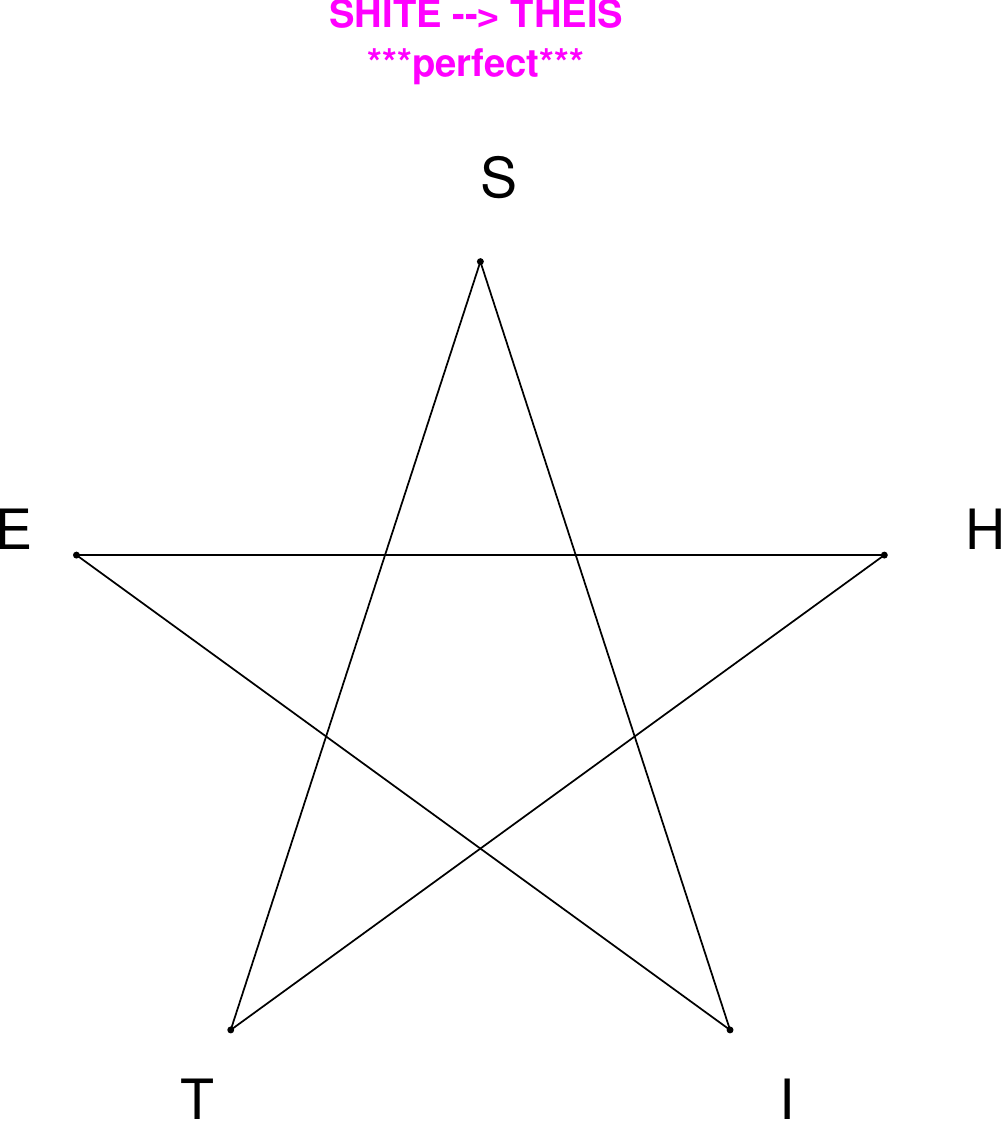}
\end{subfigure}
\end{figure}

\begin{figure}[H]
\centering
\begin{subfigure}[T]{0.19\textwidth}
\centering
\includegraphics[width=\textwidth]{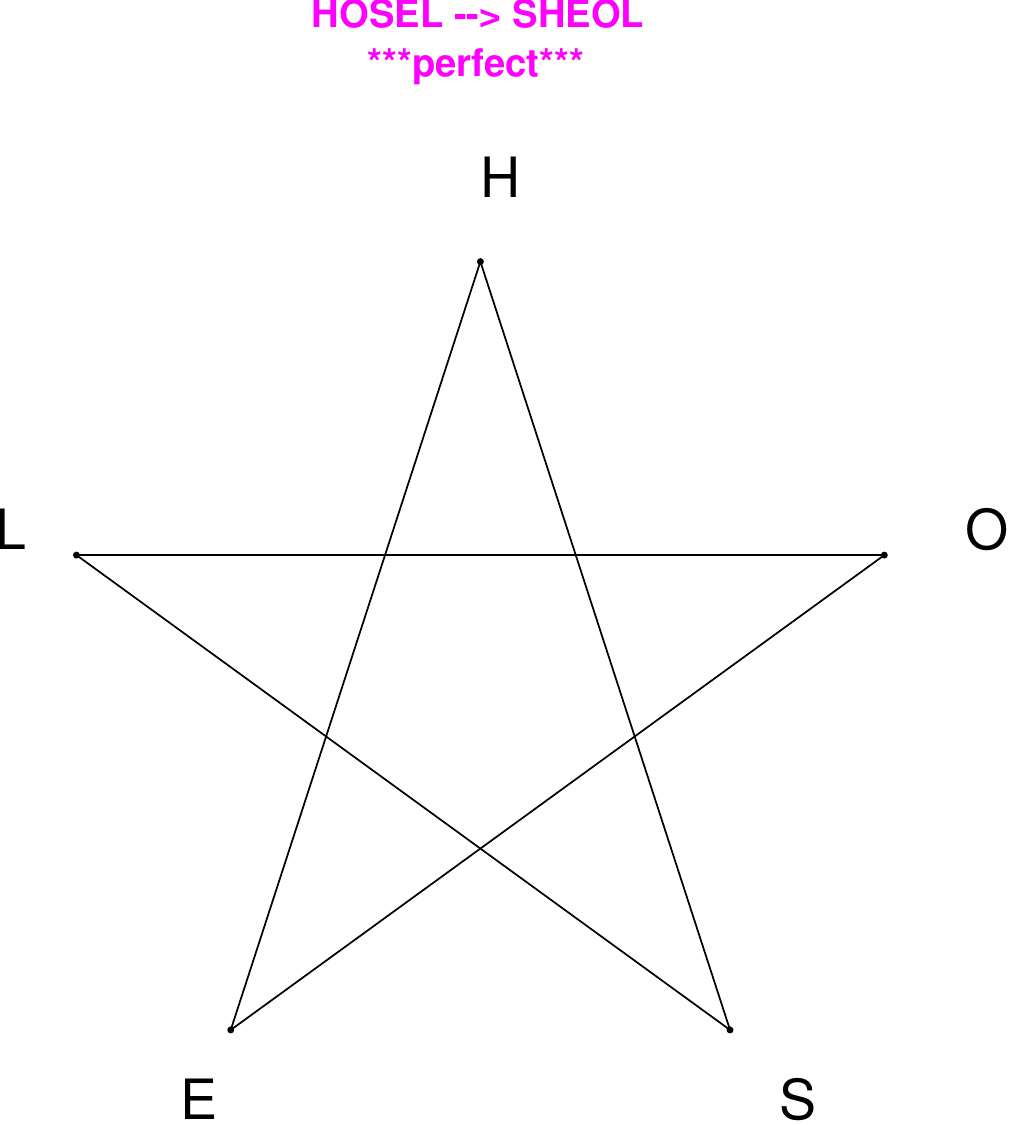}
\end{subfigure}
\hfill
\begin{subfigure}[T]{0.19\textwidth}
\centering
\includegraphics[width=\textwidth]{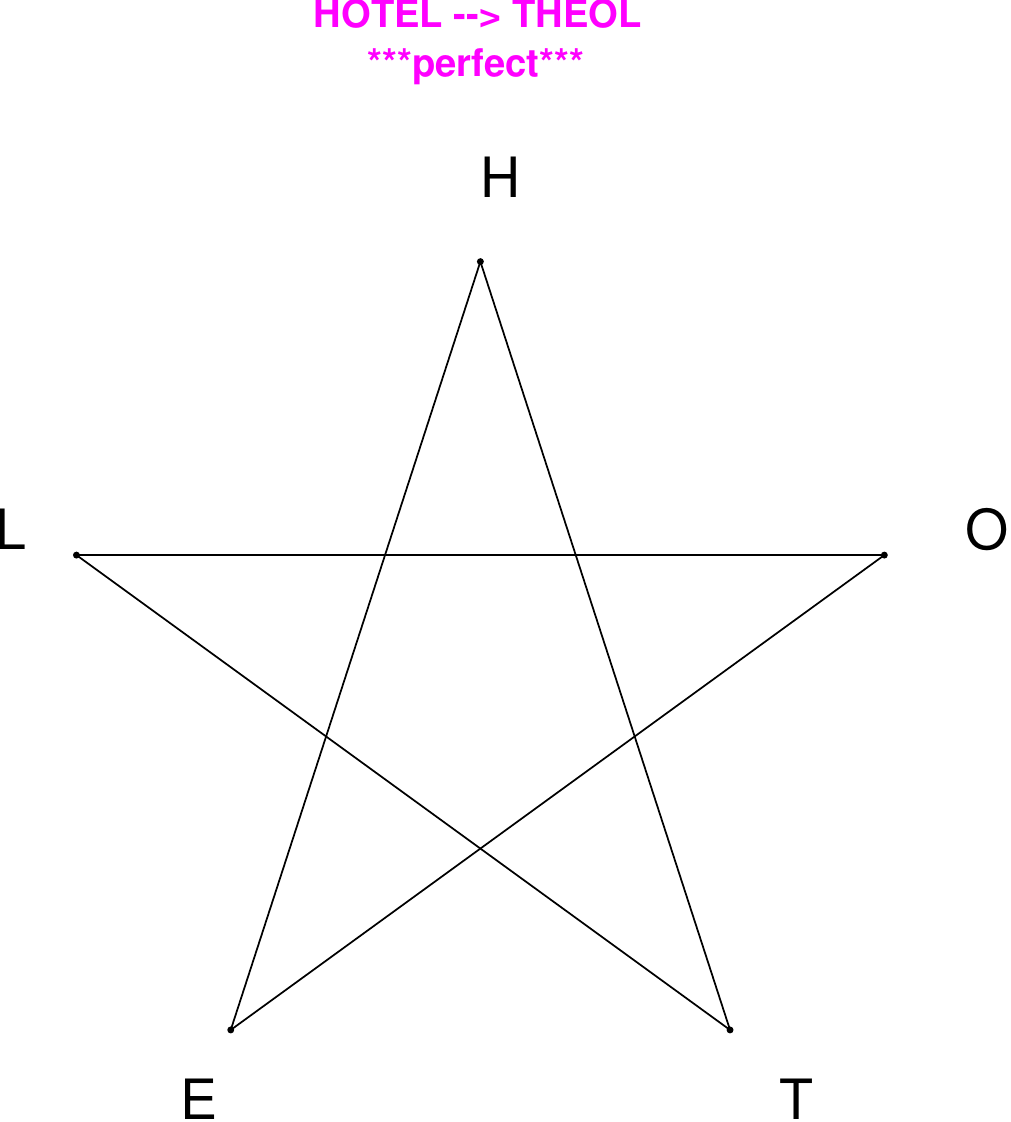}
\end{subfigure}
\hfill
\begin{subfigure}[T]{0.19\textwidth}
\centering
\includegraphics[width=\textwidth]{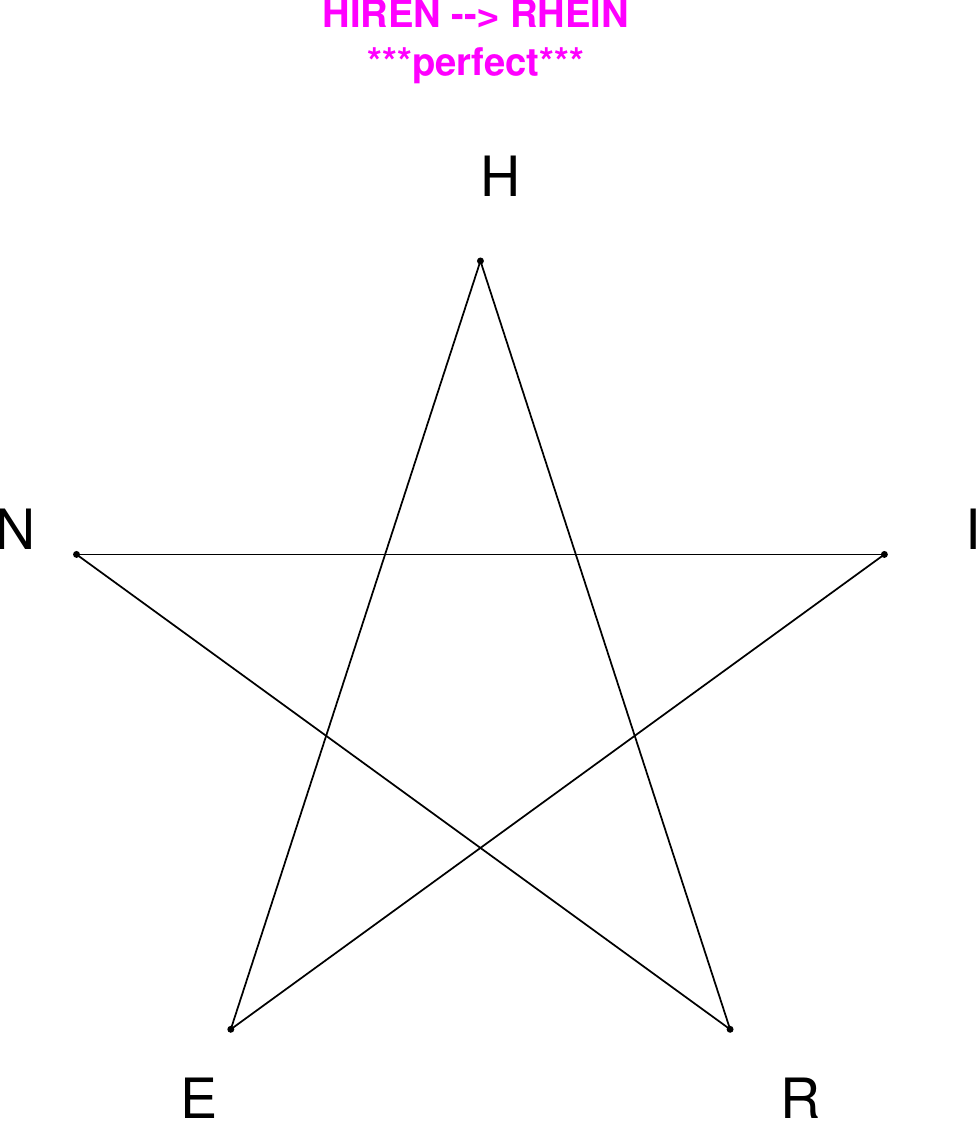}
\end{subfigure}
\hfill
\begin{subfigure}[T]{0.19\textwidth}
\centering
\includegraphics[width=\textwidth]{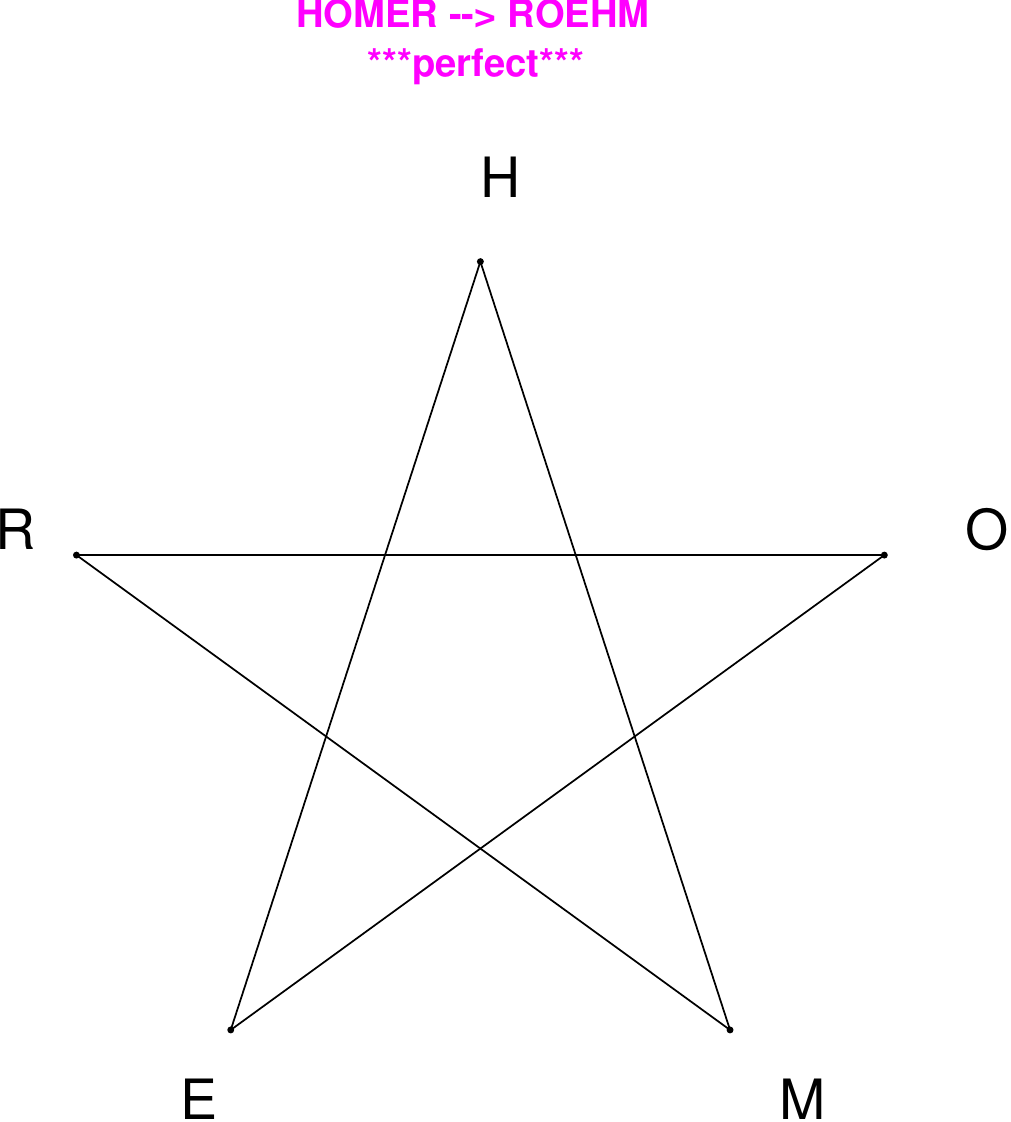}
\end{subfigure}
\hfill
\begin{subfigure}[T]{0.19\textwidth}
\centering
\includegraphics[width=\textwidth]{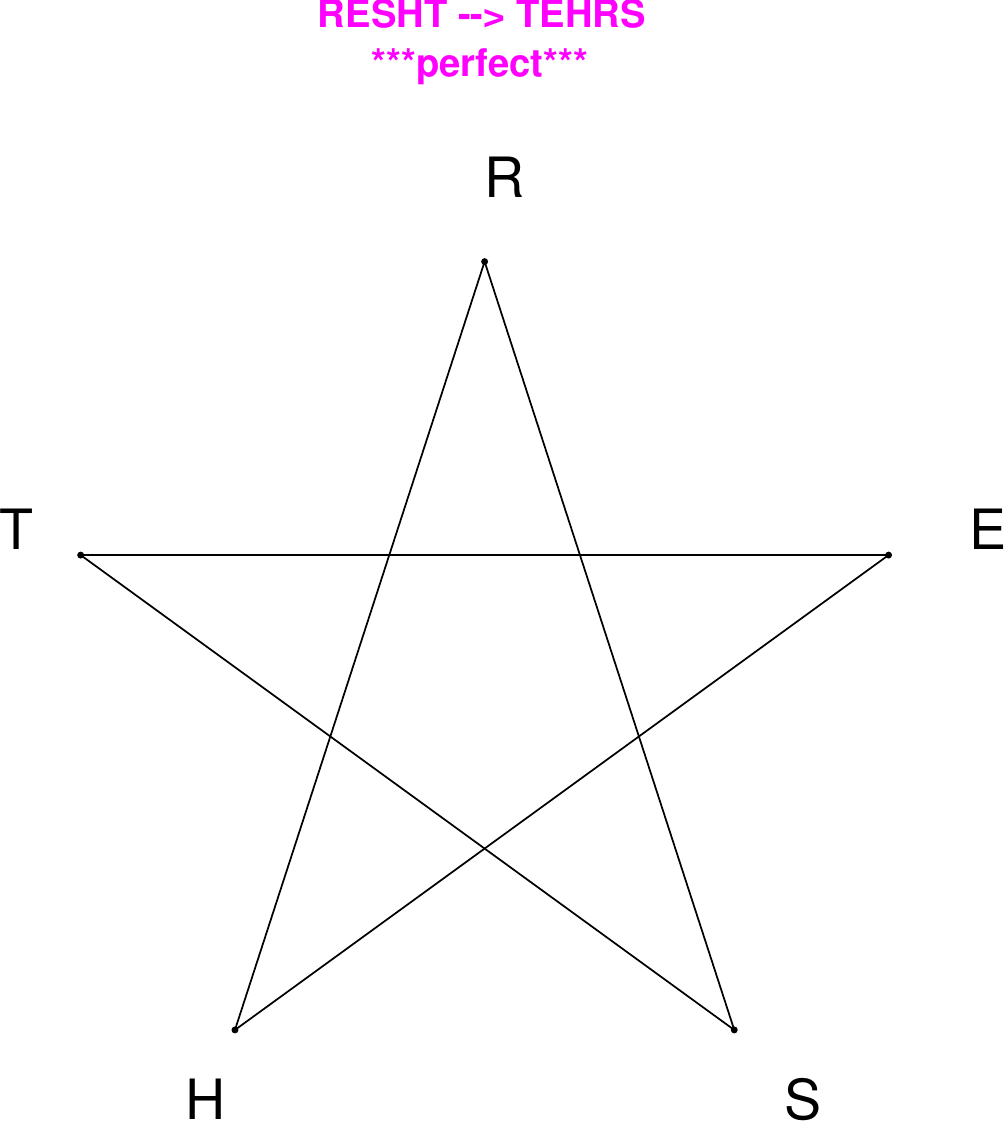}
\end{subfigure}
\end{figure}

\begin{figure}[H]
\centering
\begin{subfigure}[T]{0.19\textwidth}
\centering
\includegraphics[width=\textwidth]{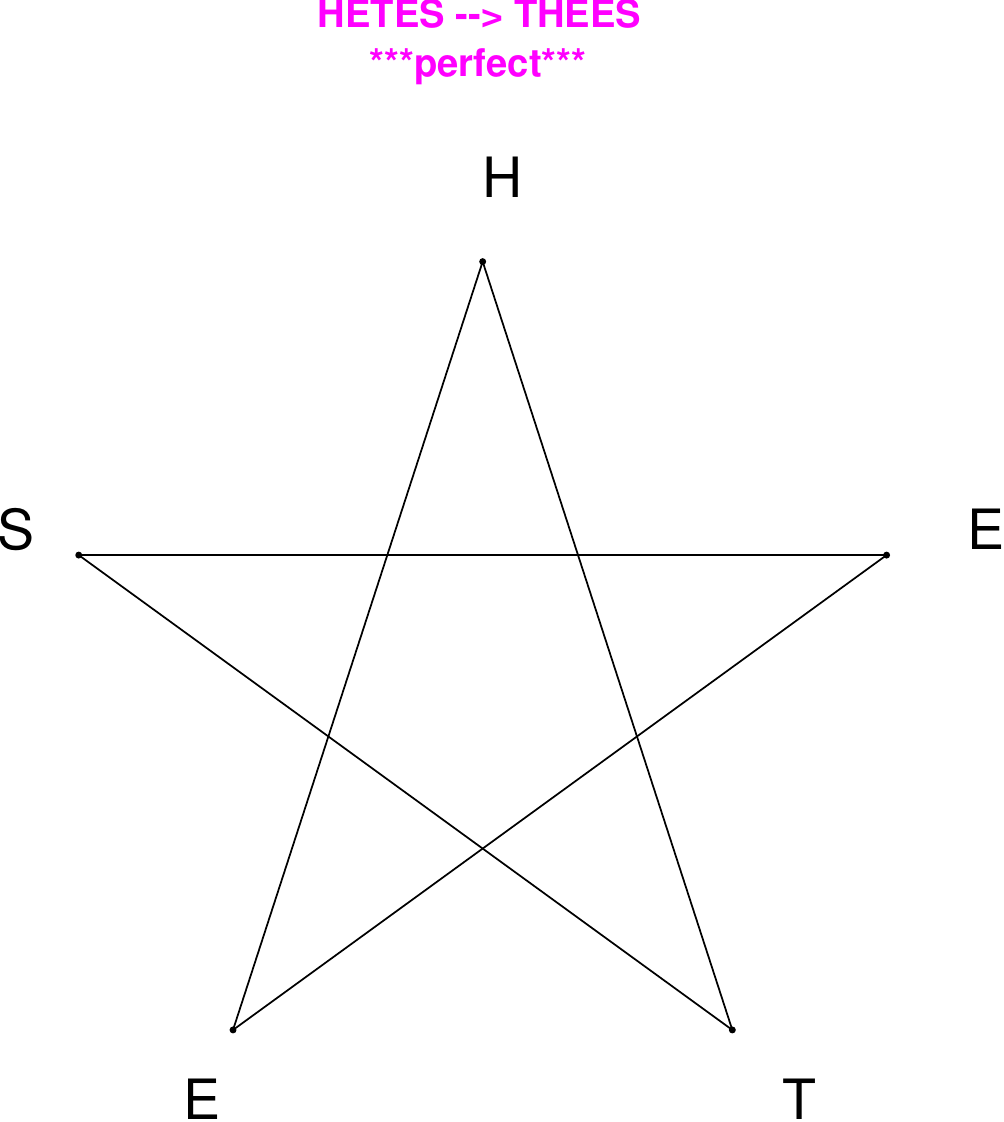}
\end{subfigure}
\hfill
\begin{subfigure}[T]{0.19\textwidth}
\centering
\includegraphics[width=\textwidth]{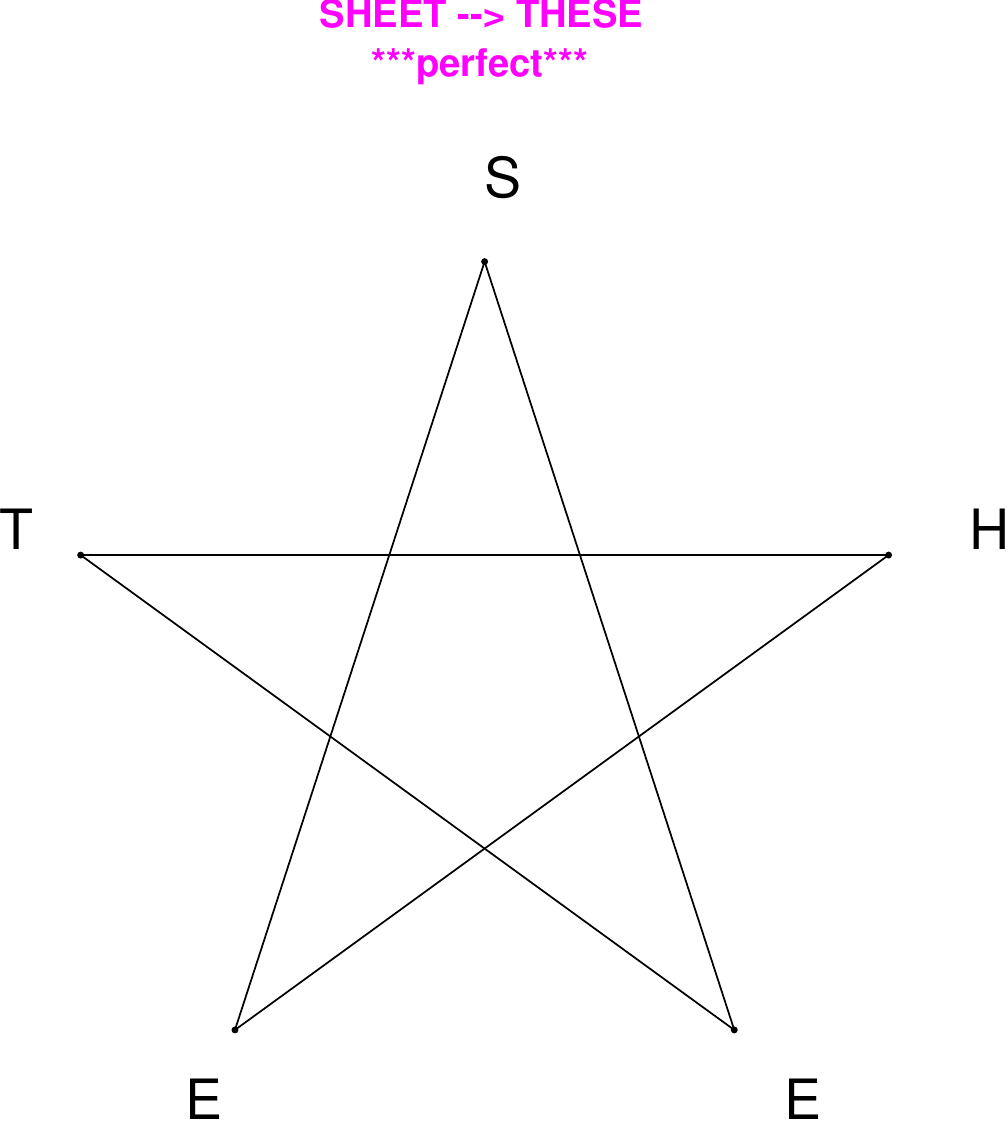}
\end{subfigure}
\hfill
\begin{subfigure}[T]{0.19\textwidth}
\centering
\includegraphics[width=\textwidth]{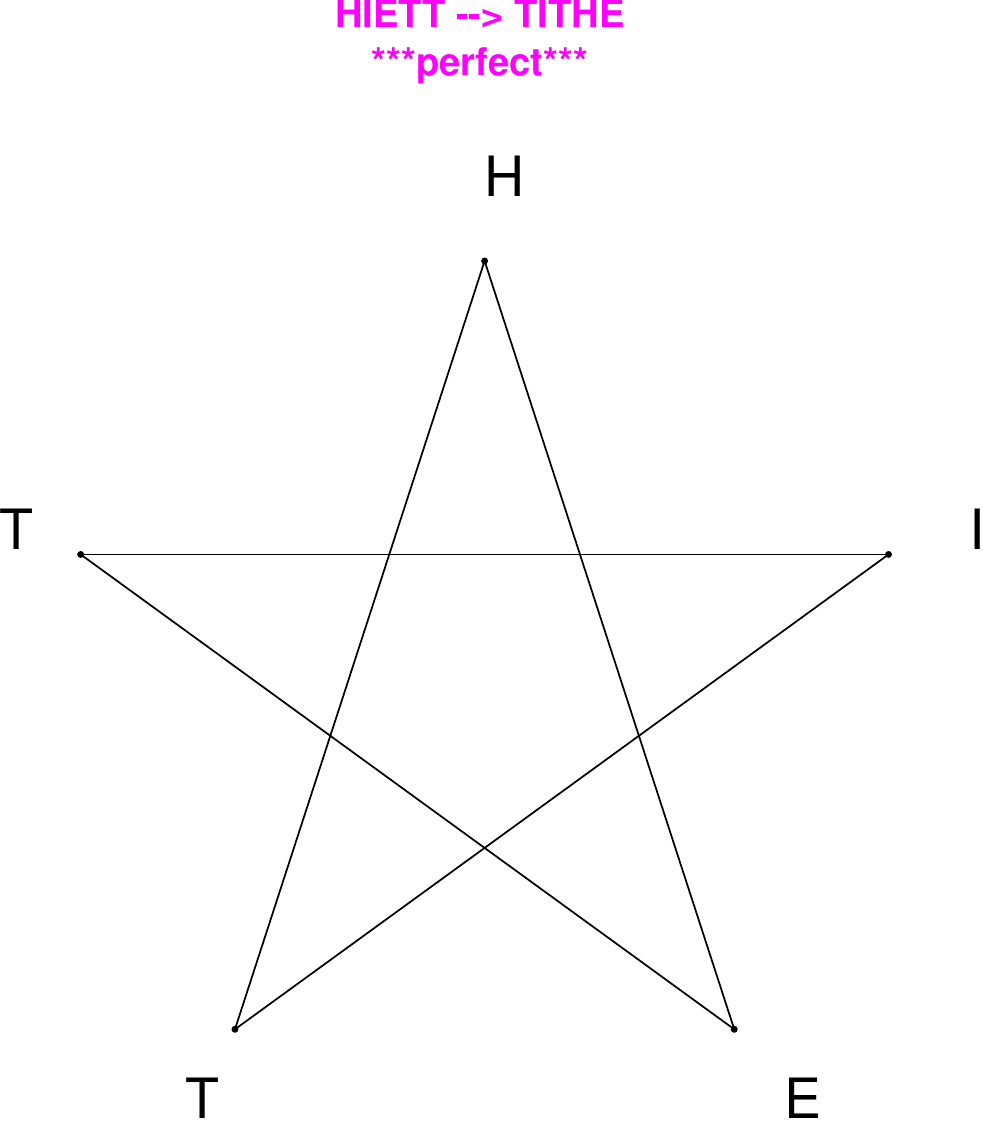}
\end{subfigure}
\hfill
\begin{subfigure}[T]{0.19\textwidth}
\centering
\includegraphics[width=\textwidth]{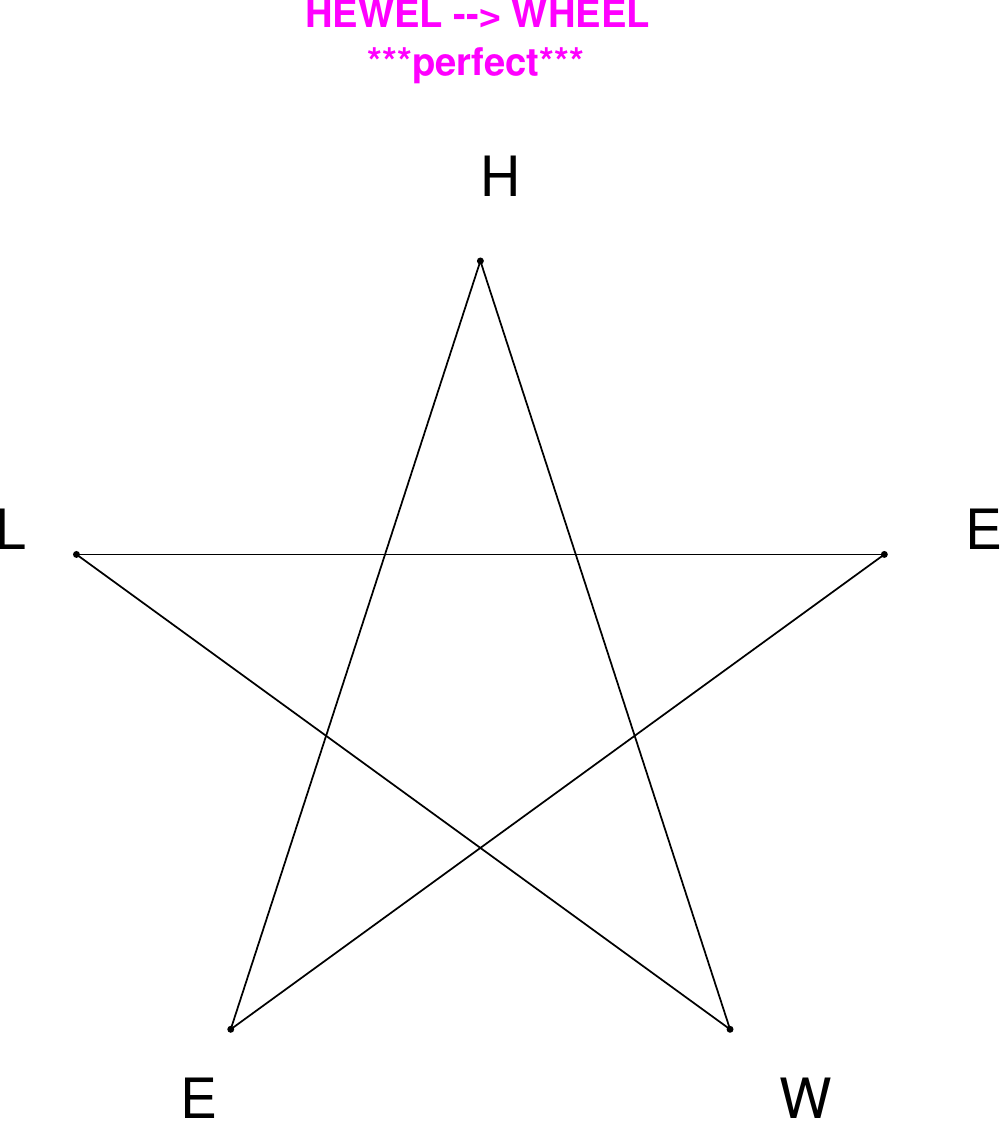}
\end{subfigure}
\hfill
\begin{subfigure}[T]{0.19\textwidth}
\centering
\includegraphics[width=\textwidth]{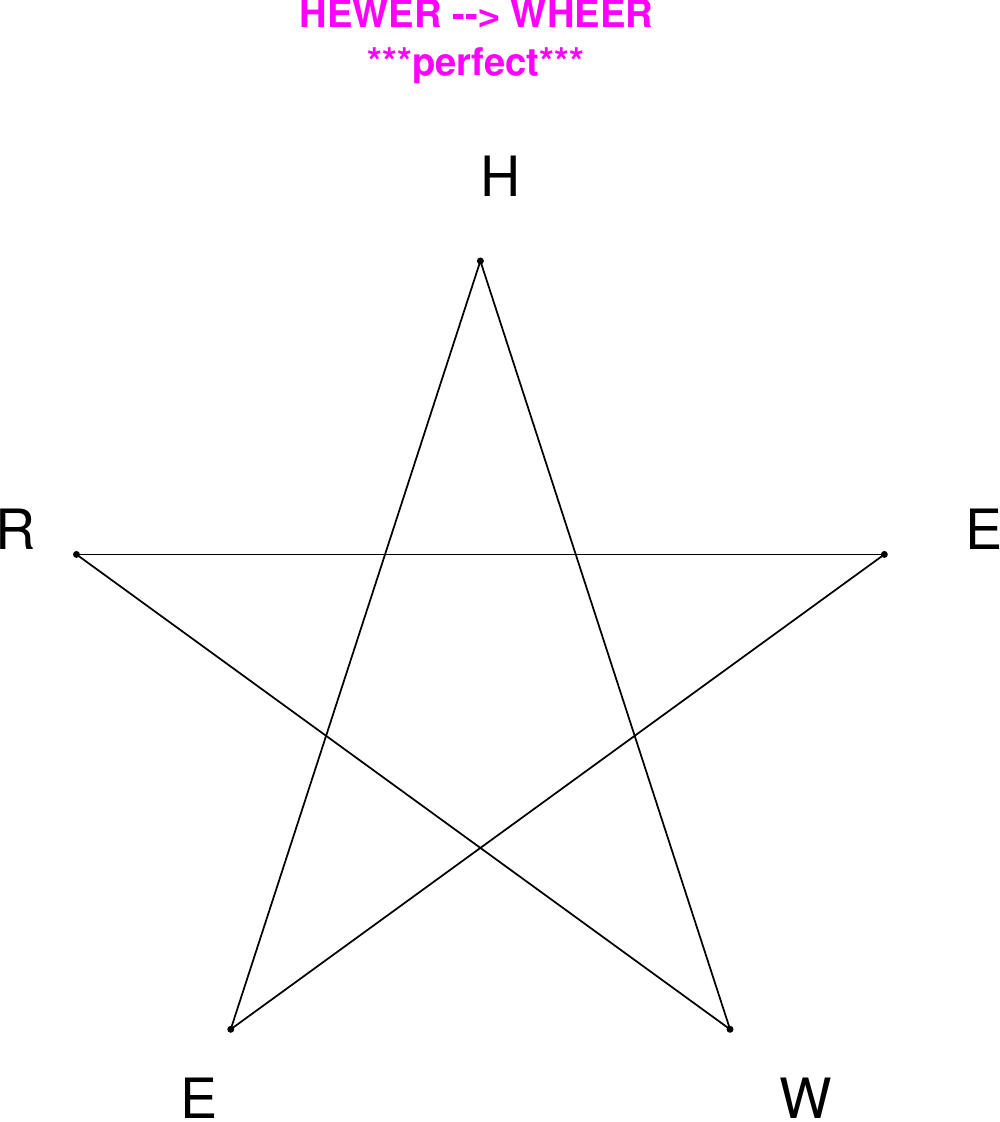}
\end{subfigure}
\end{figure}

\begin{figure}[H]
\centering
\begin{subfigure}[T]{0.19\textwidth}
\centering
\includegraphics[width=\textwidth]{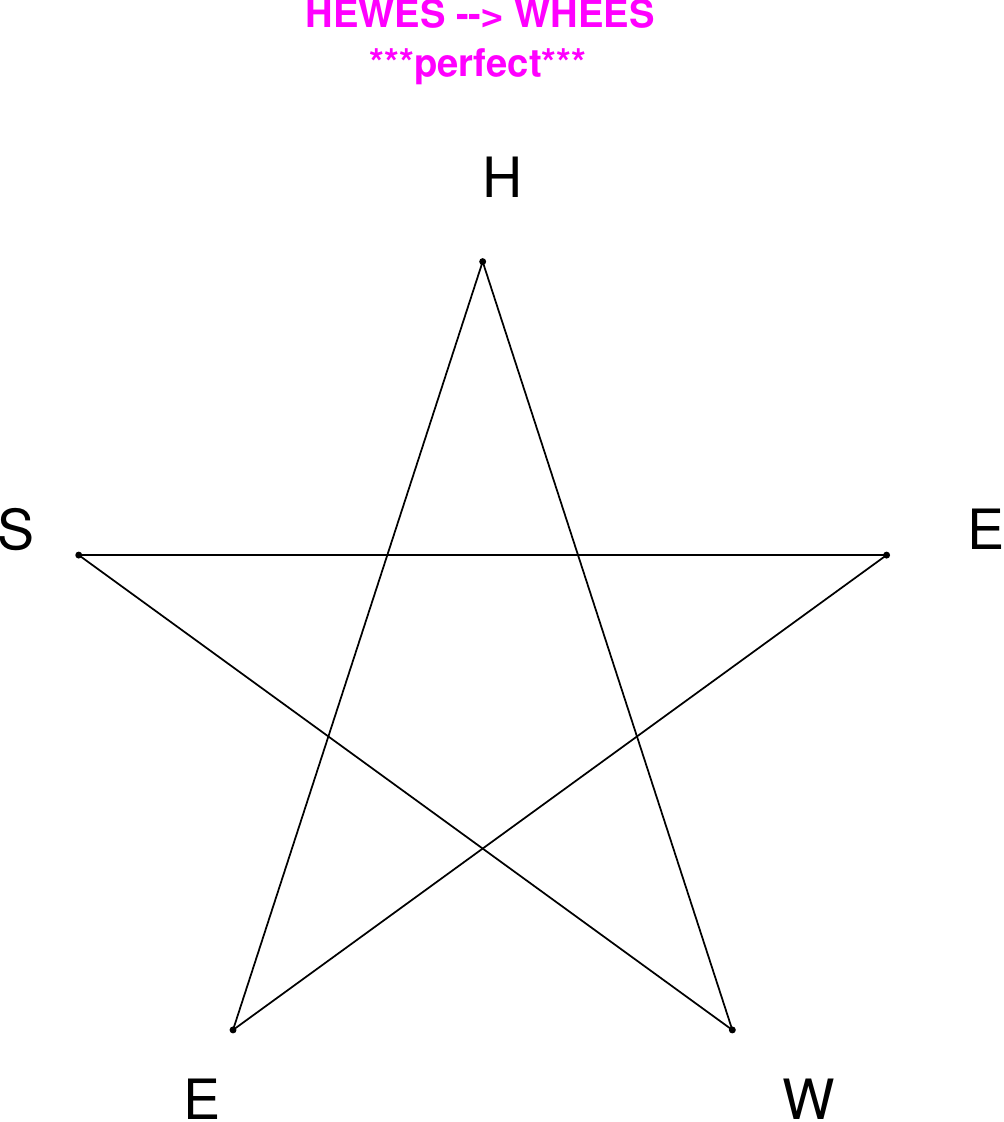}
\end{subfigure}
\hfill
\begin{subfigure}[T]{0.19\textwidth}
\centering
\includegraphics[width=\textwidth]{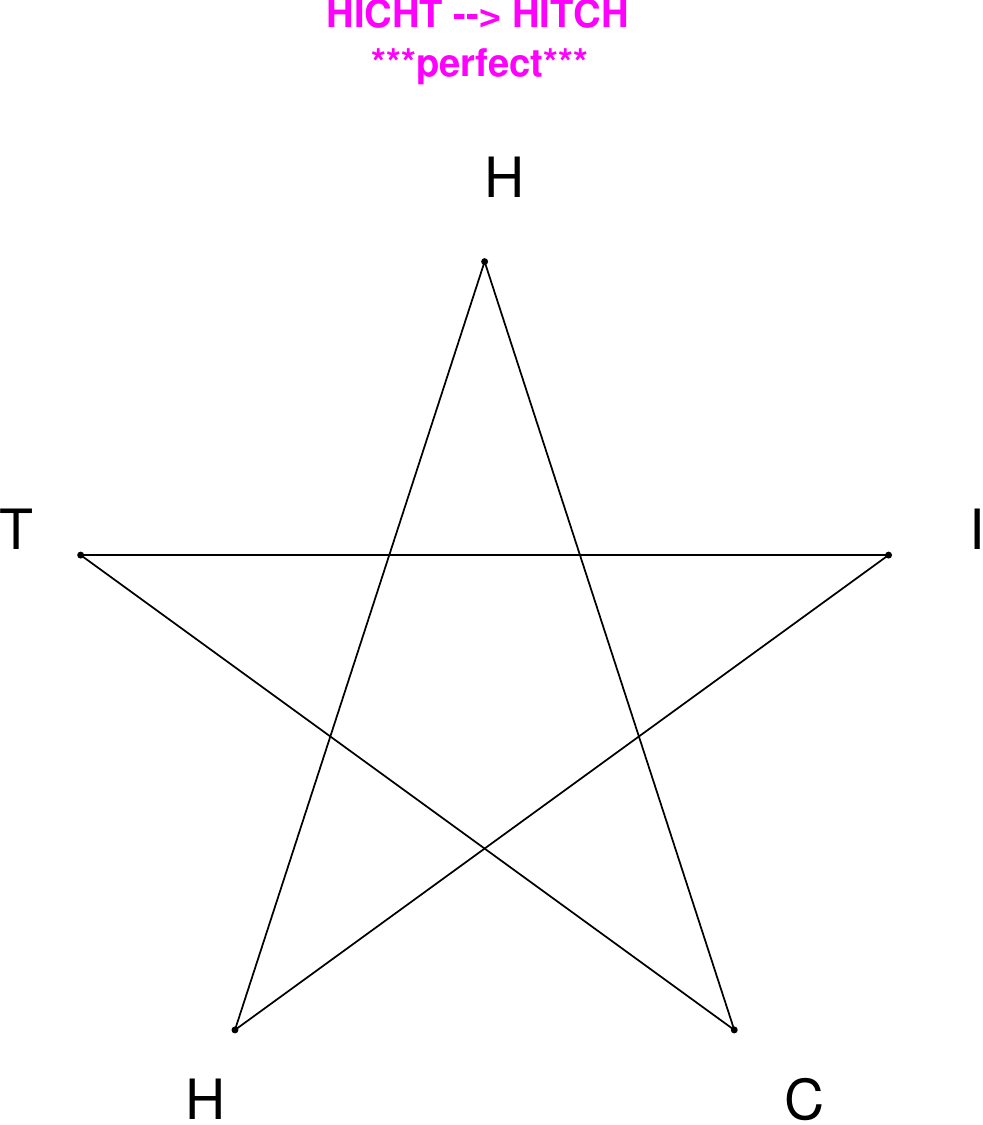}
\end{subfigure}
\hfill
\begin{subfigure}[T]{0.19\textwidth}
\centering
\includegraphics[width=\textwidth]{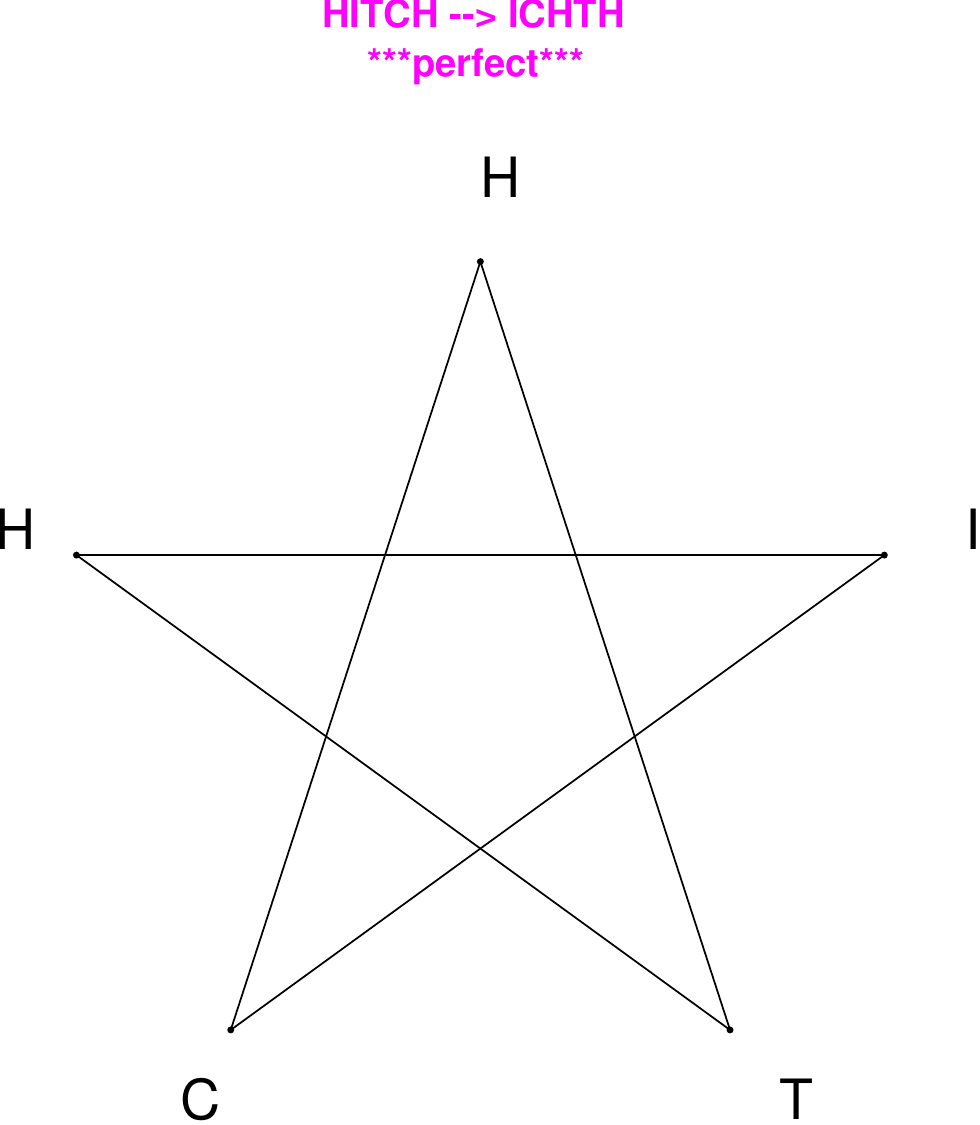}
\end{subfigure}
\hfill
\begin{subfigure}[T]{0.19\textwidth}
\centering
\includegraphics[width=\textwidth]{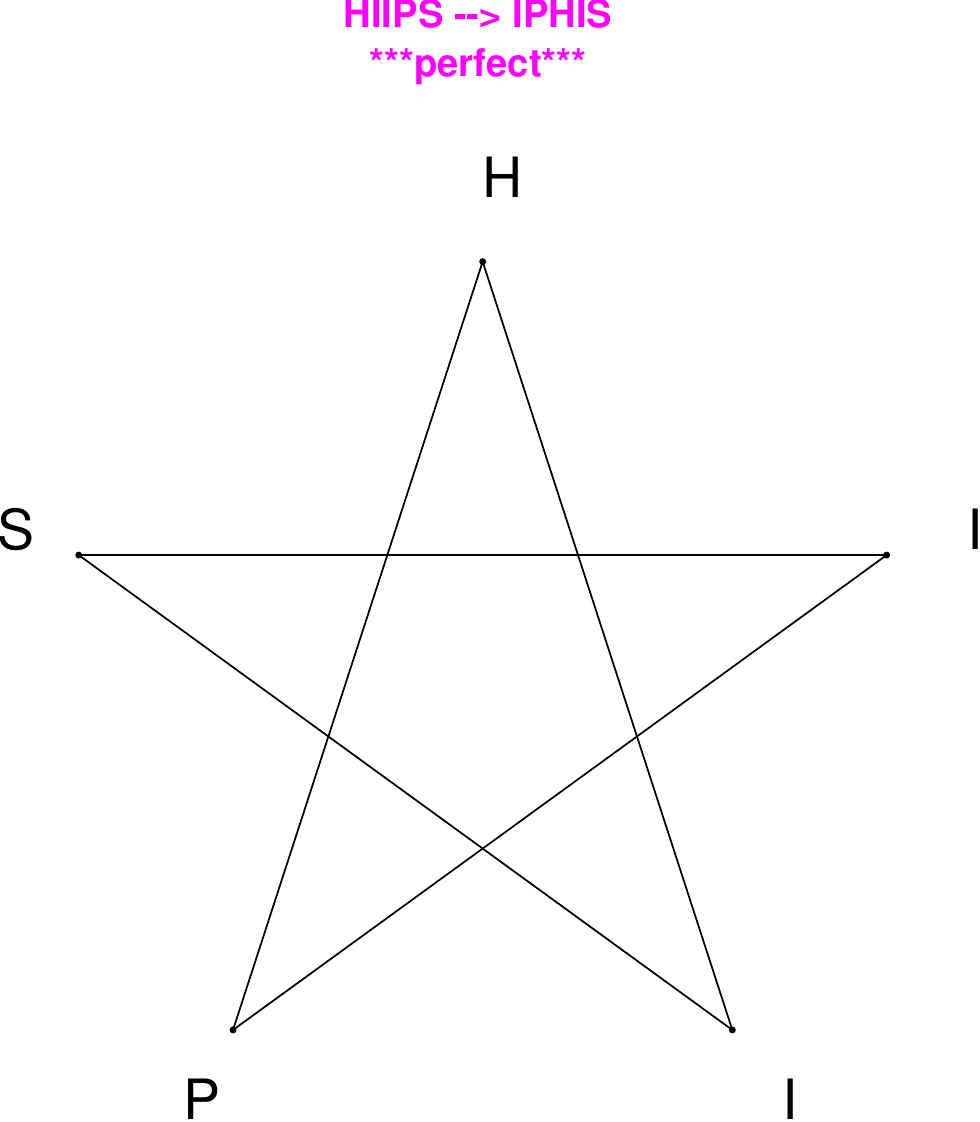}
\end{subfigure}
\hfill
\begin{subfigure}[T]{0.19\textwidth}
\centering
\includegraphics[width=\textwidth]{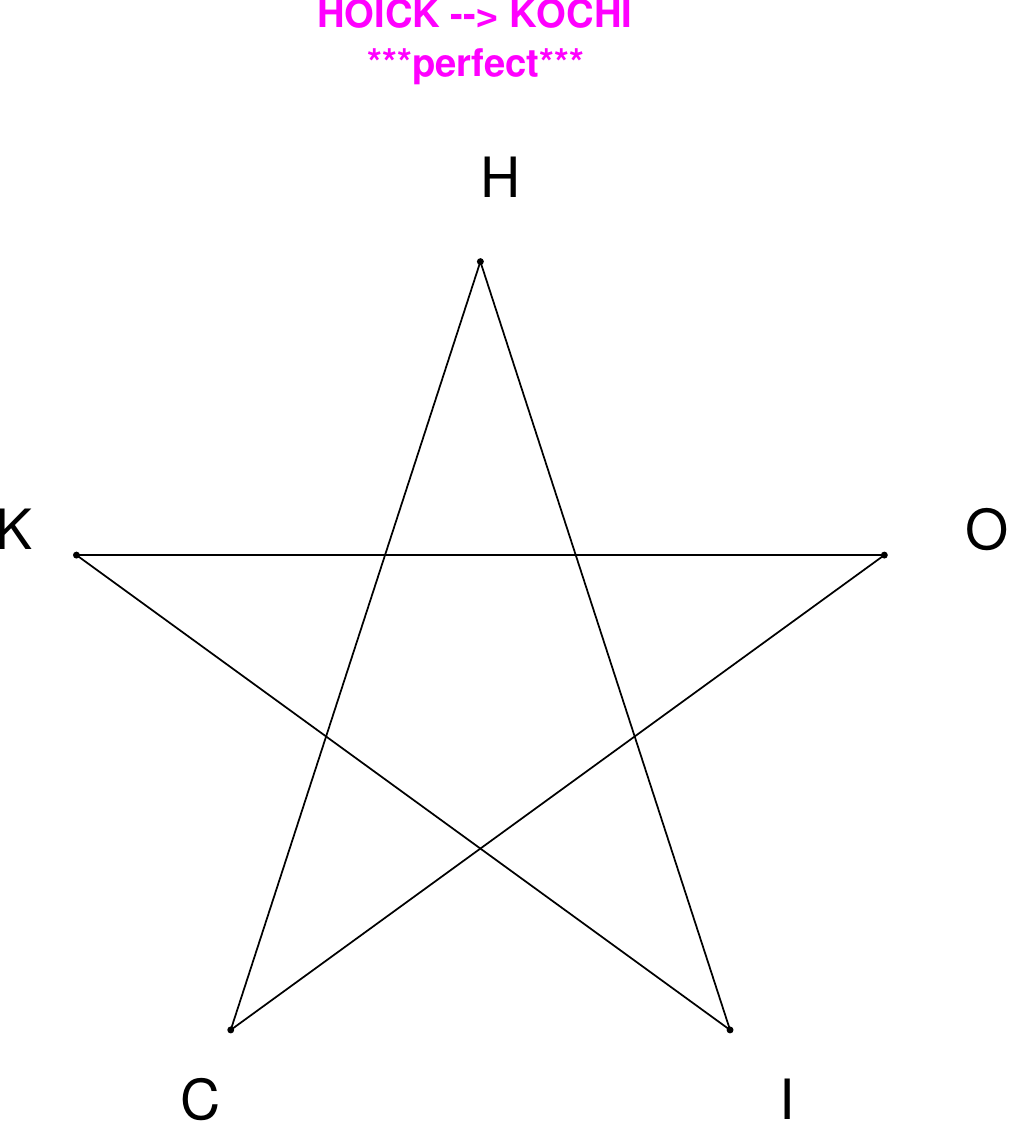}
\end{subfigure}
\end{figure}

\begin{figure}[H]
\centering
\begin{subfigure}[T]{0.19\textwidth}
\centering
\includegraphics[width=\textwidth]{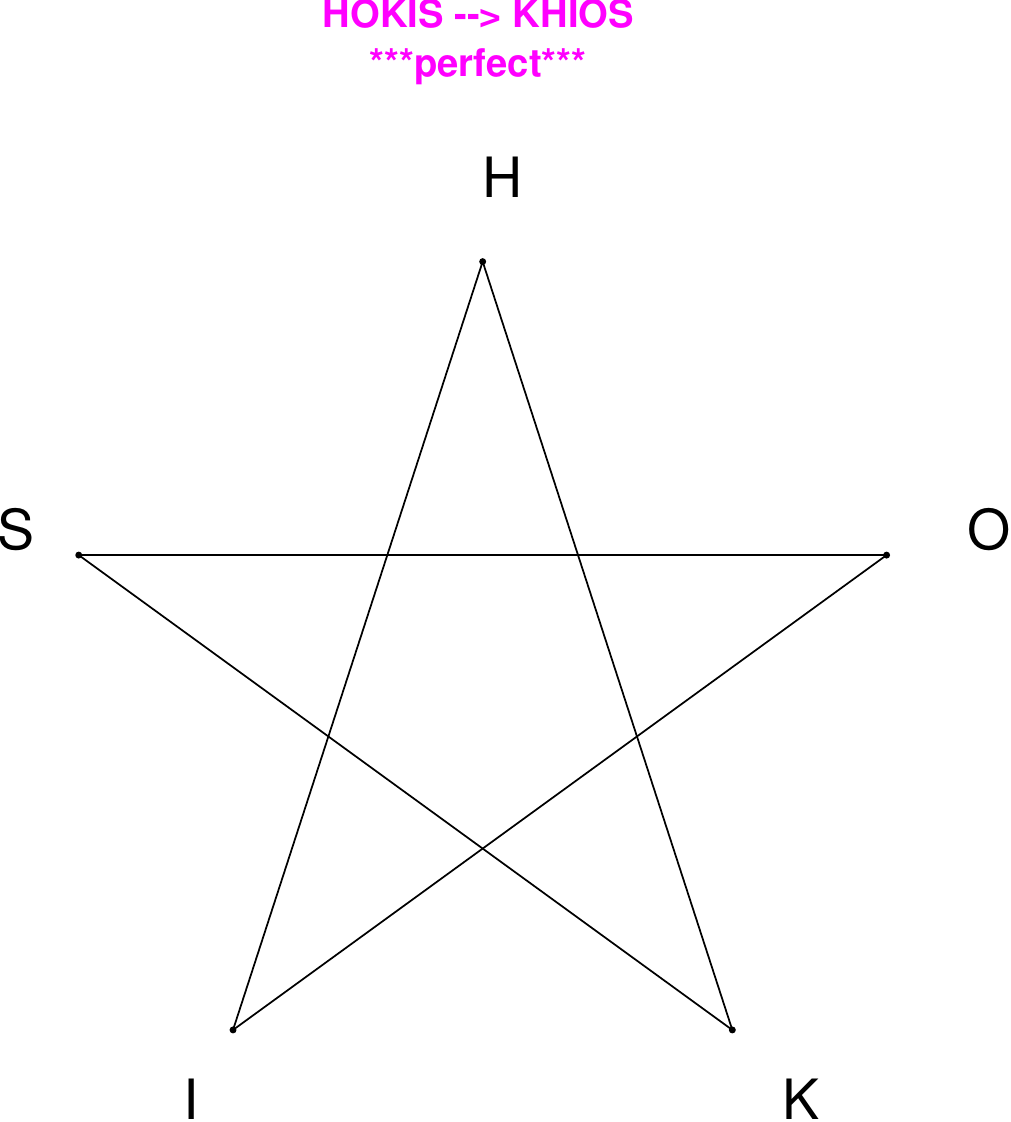}
\end{subfigure}
\hfill
\begin{subfigure}[T]{0.19\textwidth}
\centering
\includegraphics[width=\textwidth]{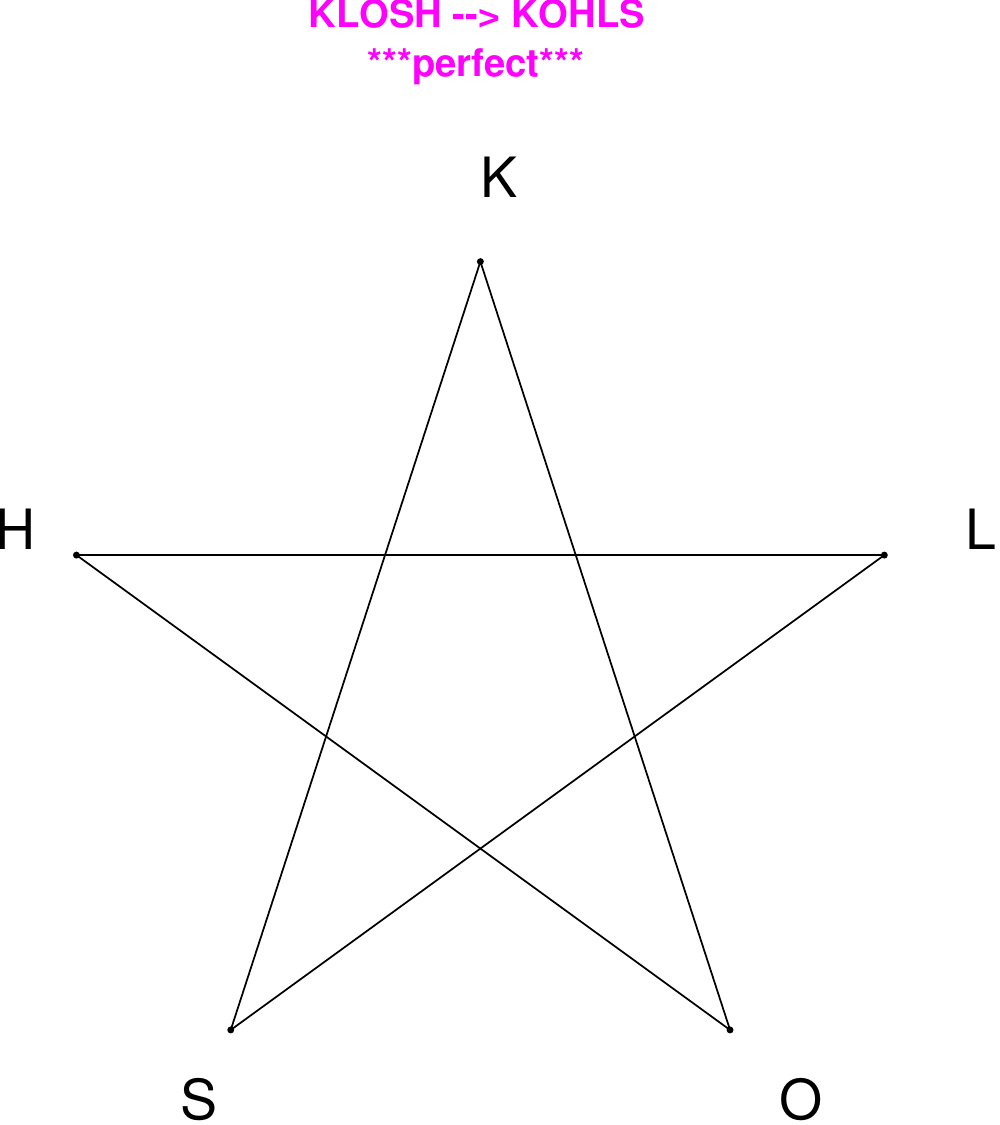}
\end{subfigure}
\hfill
\begin{subfigure}[T]{0.19\textwidth}
\centering
\includegraphics[width=\textwidth]{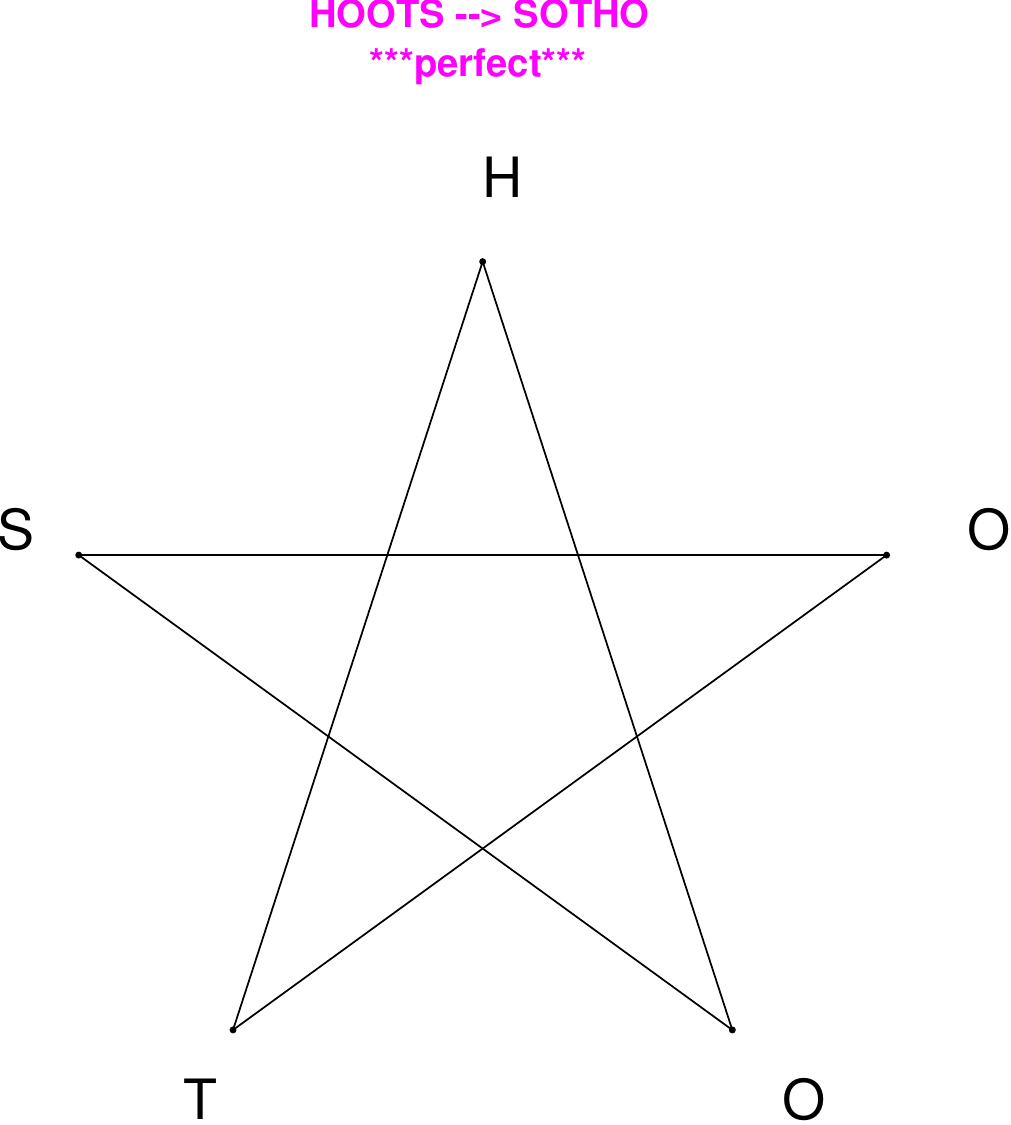}
\end{subfigure}
\hfill
\begin{subfigure}[T]{0.19\textwidth}
\centering
\includegraphics[width=\textwidth]{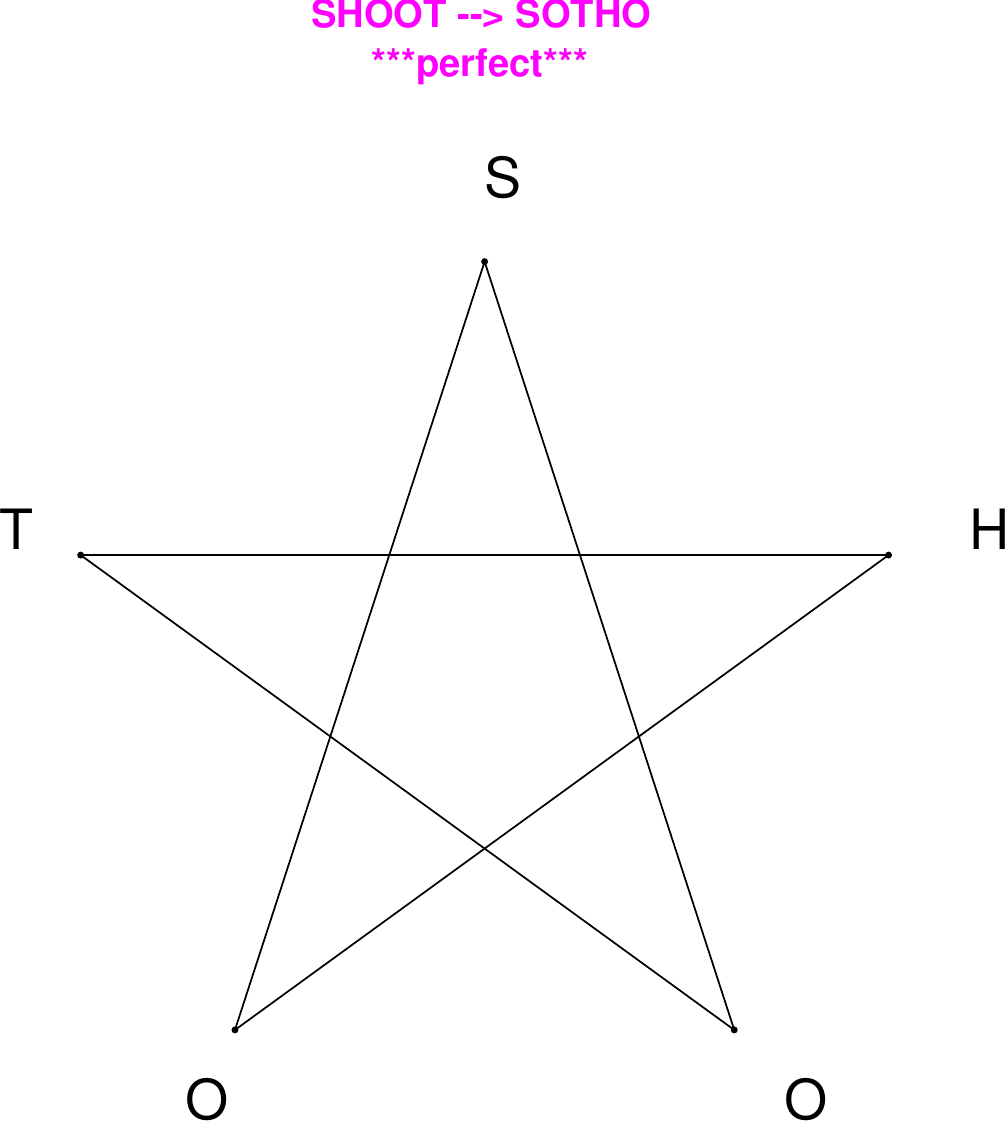}
\end{subfigure}
\hfill
\begin{subfigure}[T]{0.19\textwidth}
\centering
\includegraphics[width=\textwidth]{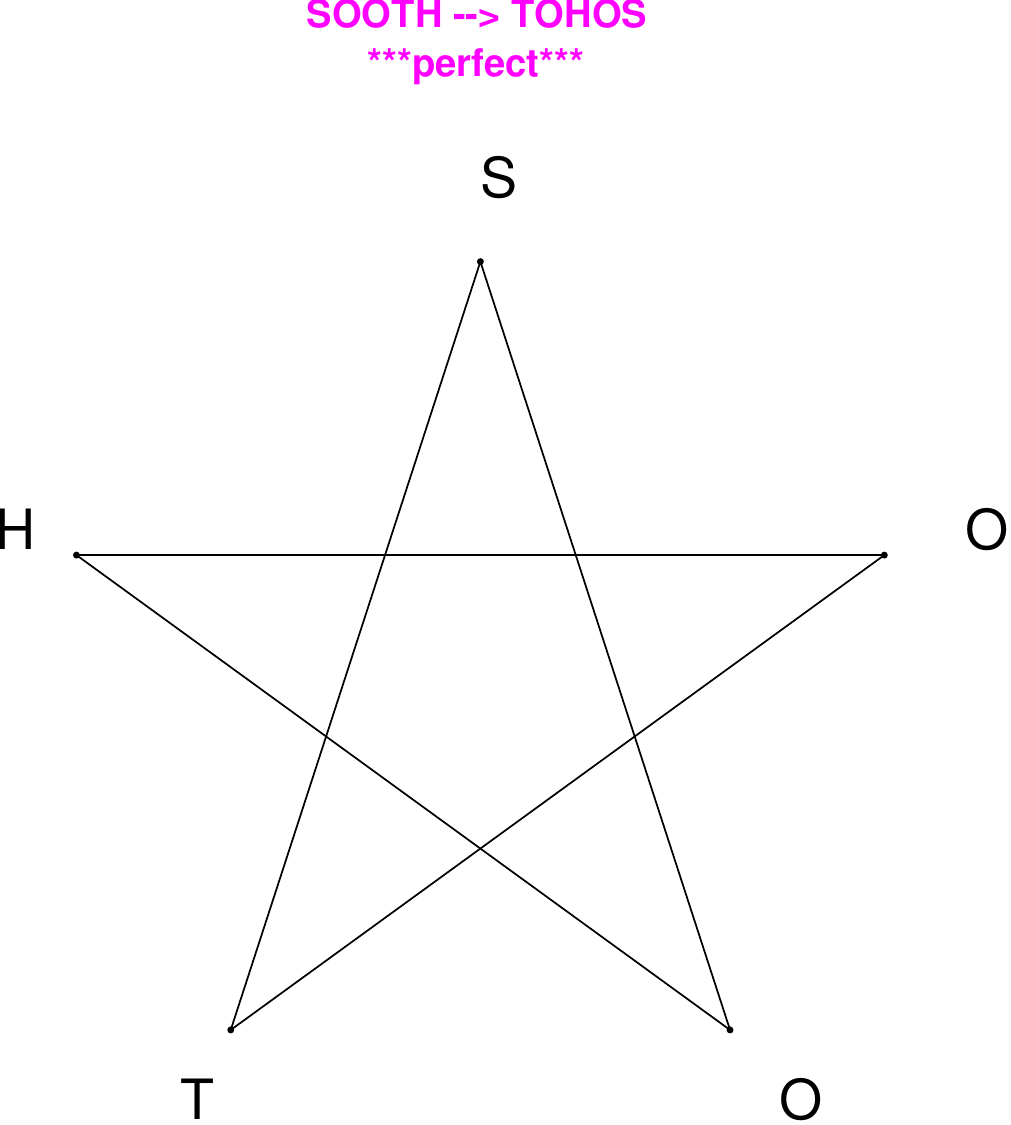}
\end{subfigure}
\end{figure}

\begin{figure}[H]
\centering
\begin{subfigure}[T]{0.19\textwidth}
\centering
\includegraphics[width=\textwidth]{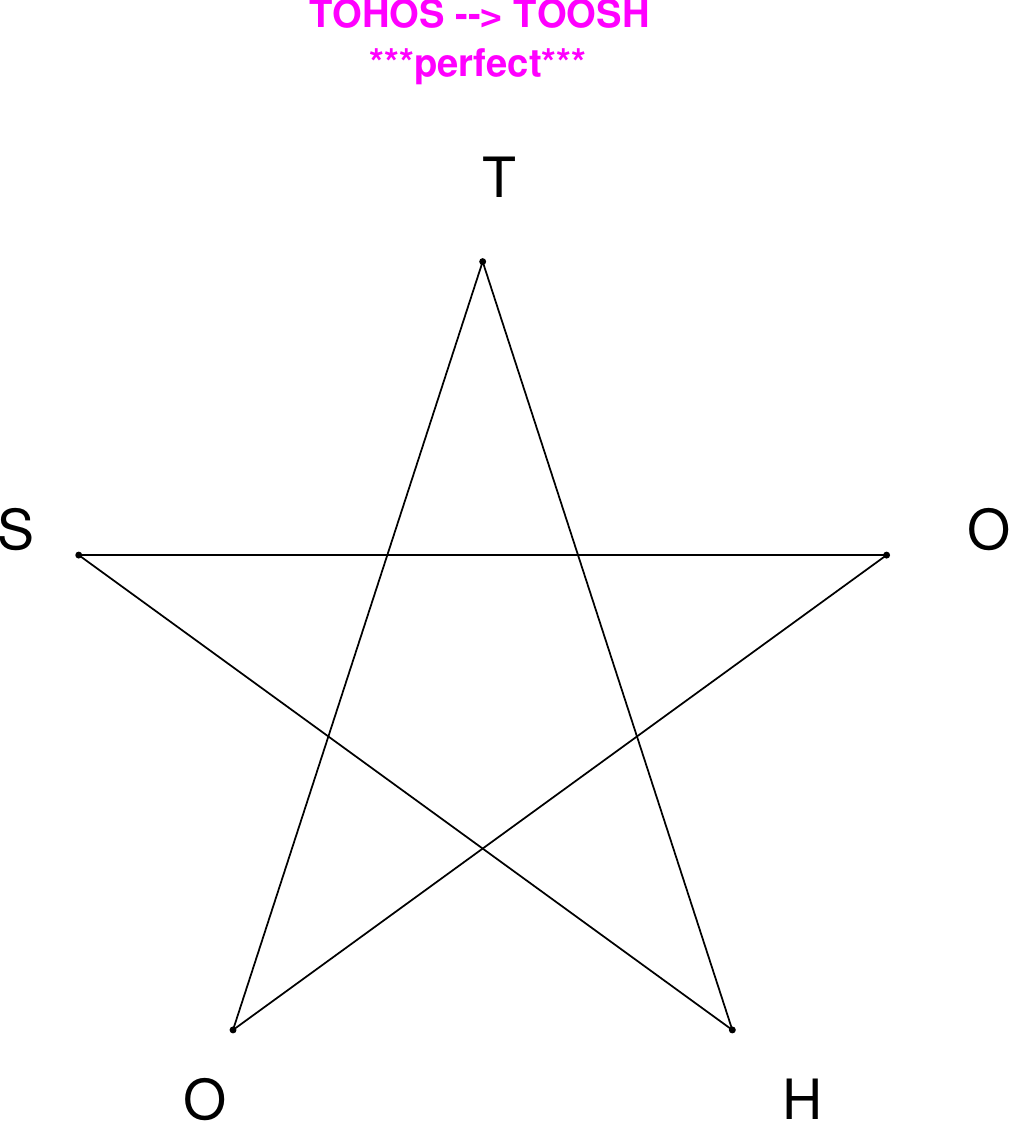}
\end{subfigure}
\hfill
\begin{subfigure}[T]{0.19\textwidth}
\centering
\includegraphics[width=\textwidth]{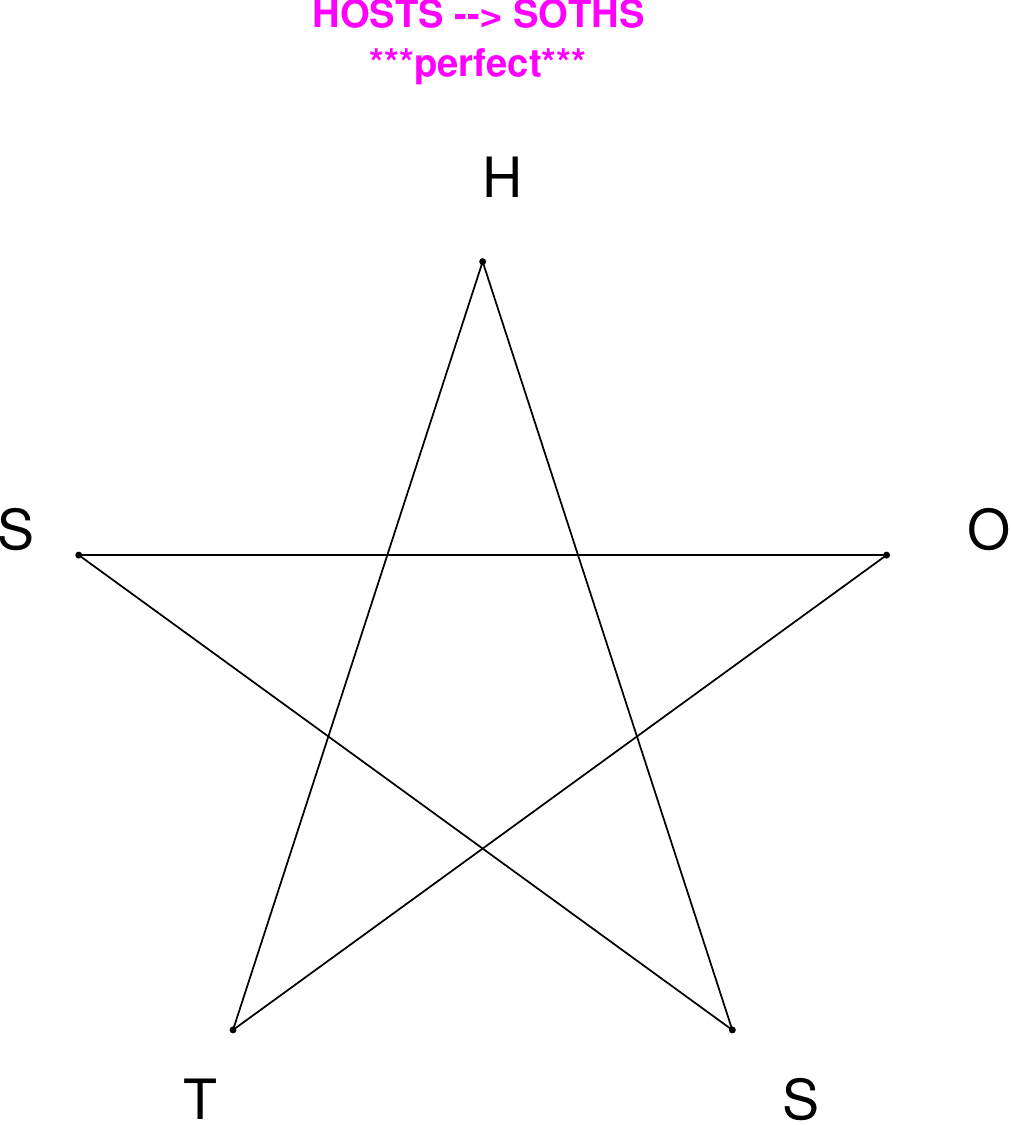}
\end{subfigure}
\hfill
\begin{subfigure}[T]{0.19\textwidth}
\centering
\includegraphics[width=\textwidth]{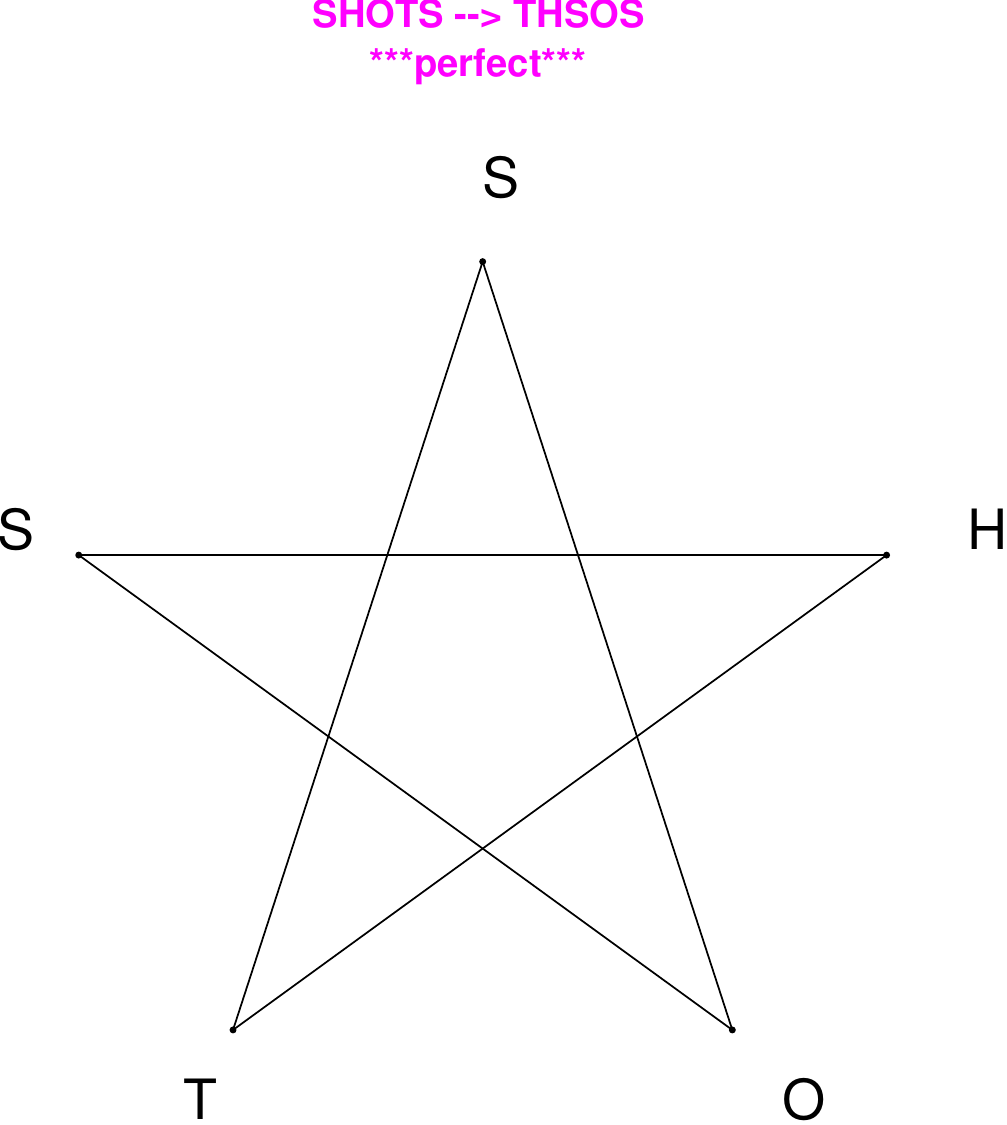}
\end{subfigure}
\hfill
\begin{subfigure}[T]{0.19\textwidth}
\centering
\includegraphics[width=\textwidth]{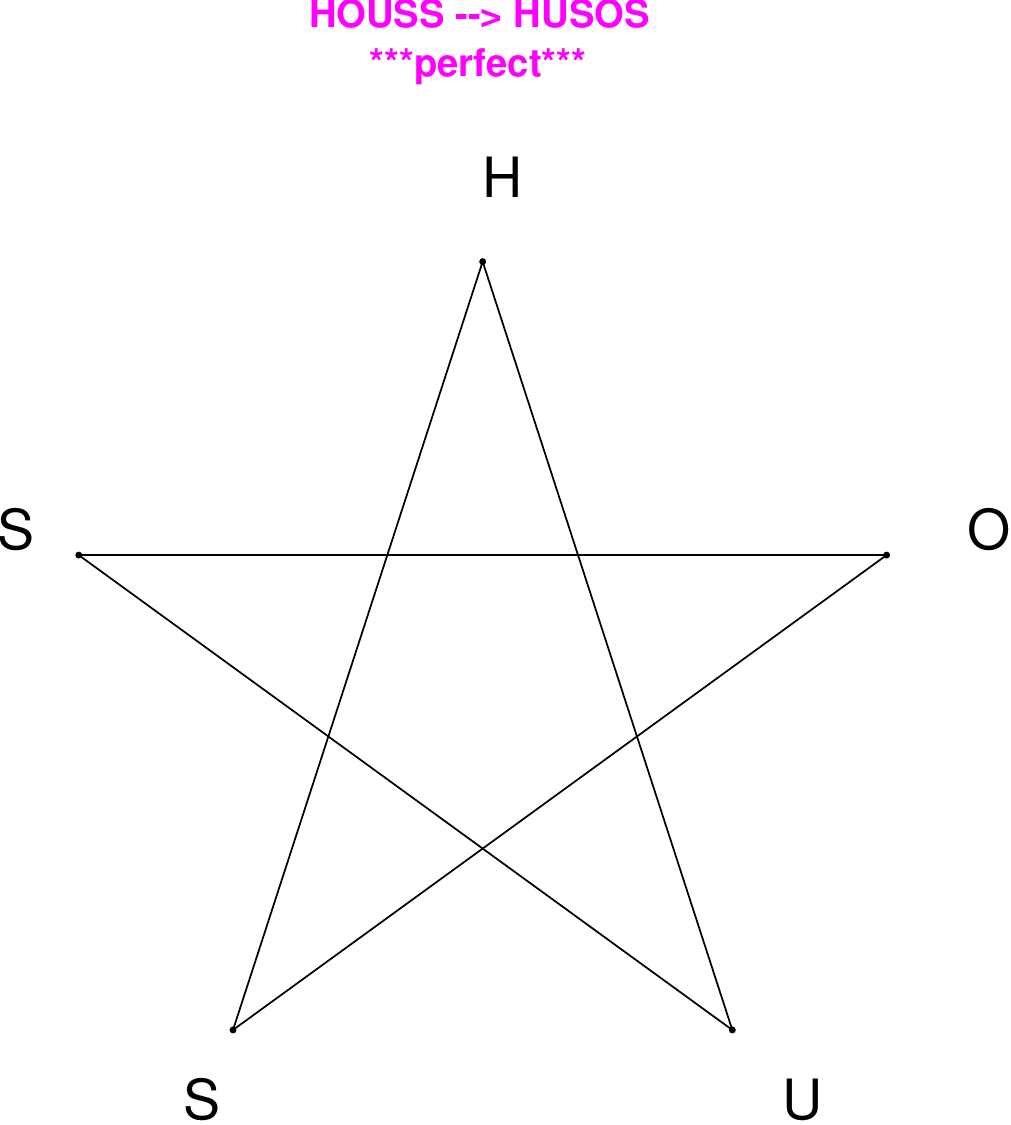}
\end{subfigure}
\hfill
\begin{subfigure}[T]{0.19\textwidth}
\centering
\includegraphics[width=\textwidth]{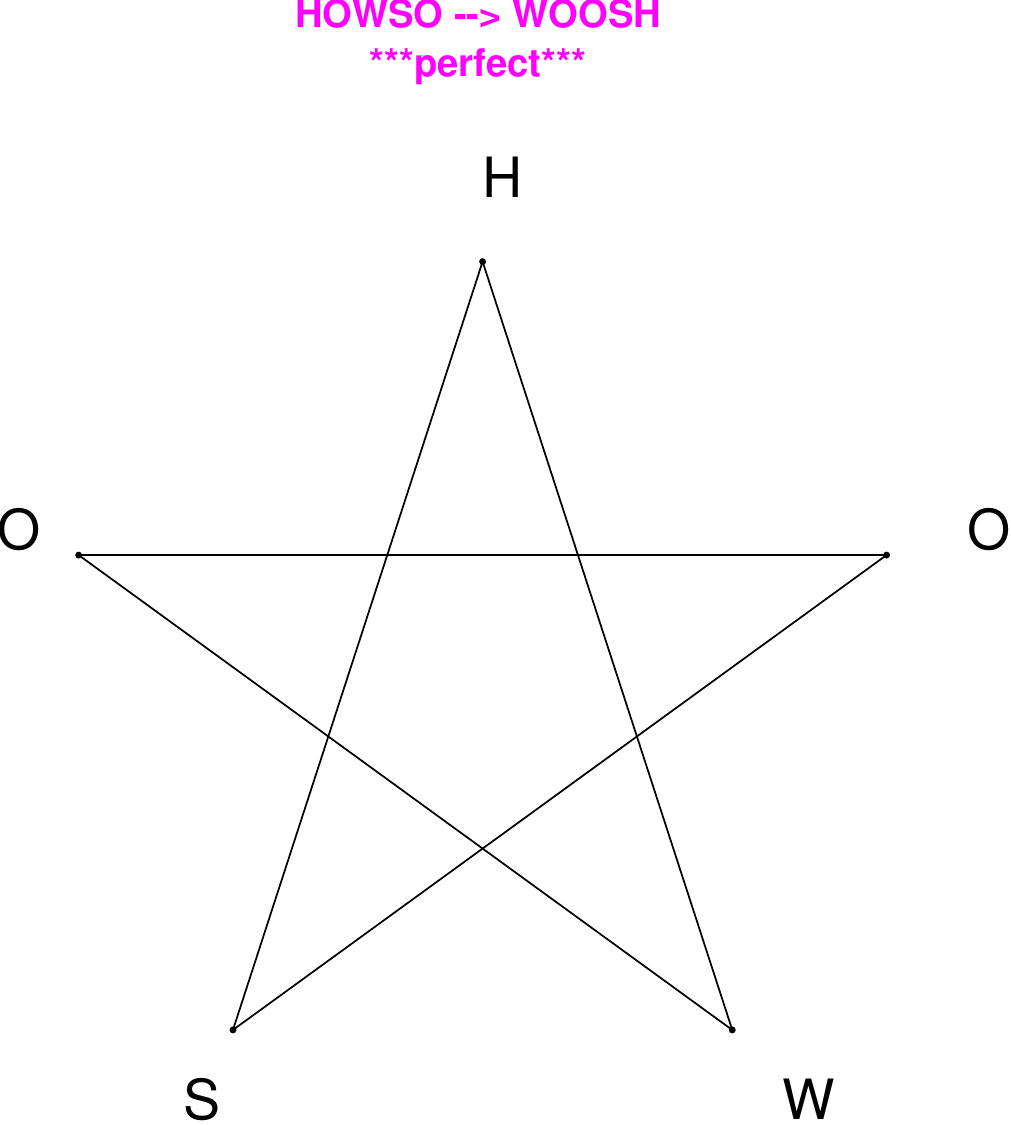}
\end{subfigure}
\end{figure}

\begin{figure}[H]
\centering
\begin{subfigure}[T]{0.19\textwidth}
\centering
\includegraphics[width=\textwidth]{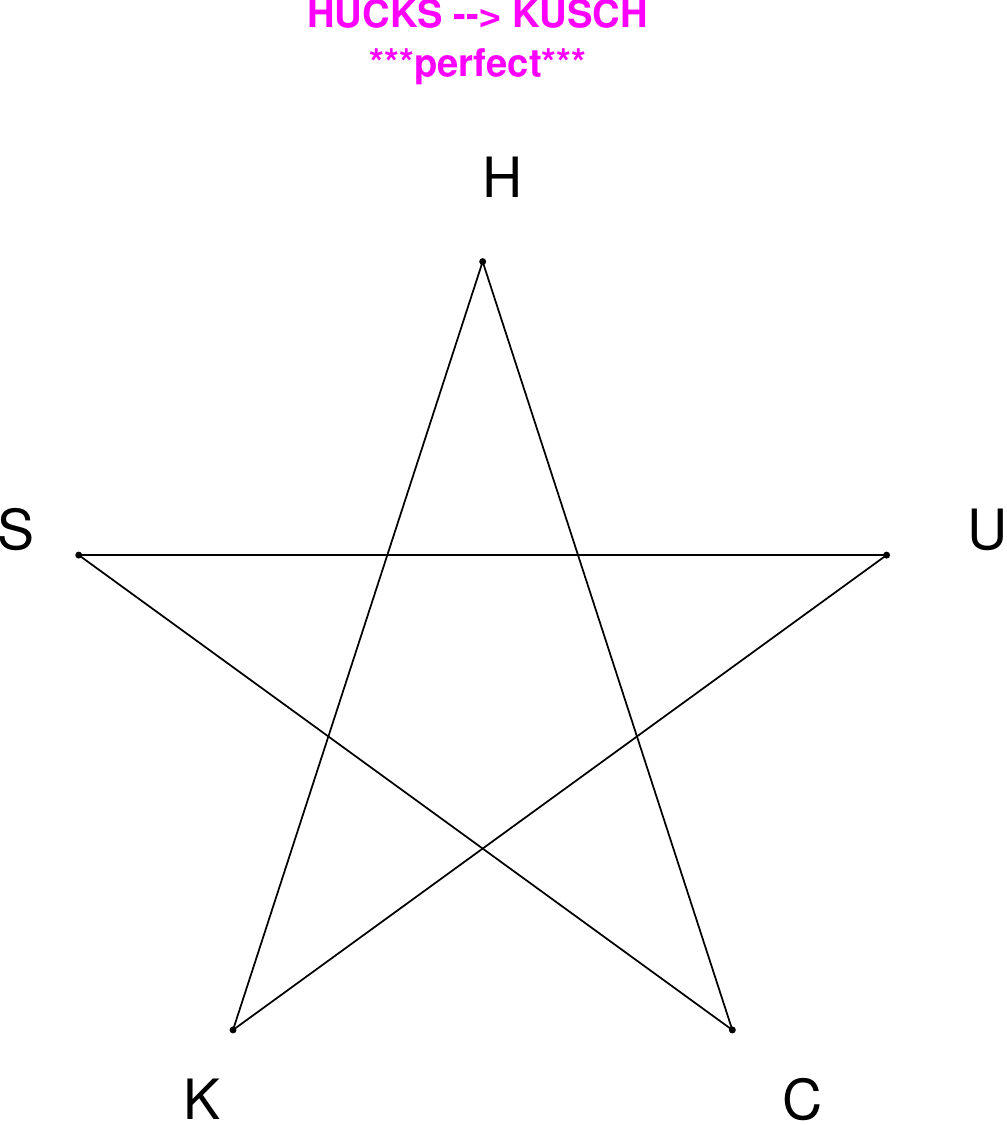}
\end{subfigure}
\hfill
\begin{subfigure}[T]{0.19\textwidth}
\centering
\includegraphics[width=\textwidth]{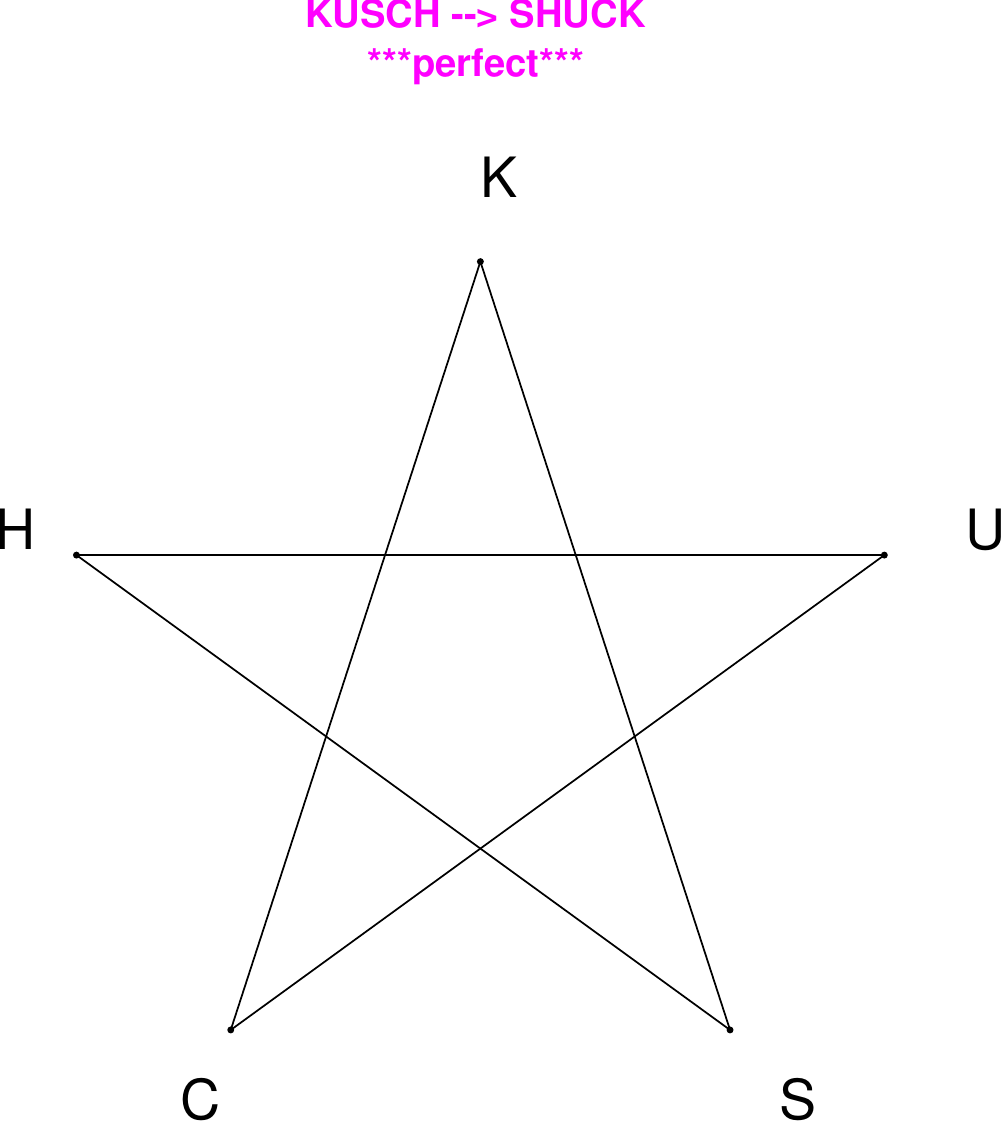}
\end{subfigure}
\hfill
\begin{subfigure}[T]{0.19\textwidth}
\centering
\includegraphics[width=\textwidth]{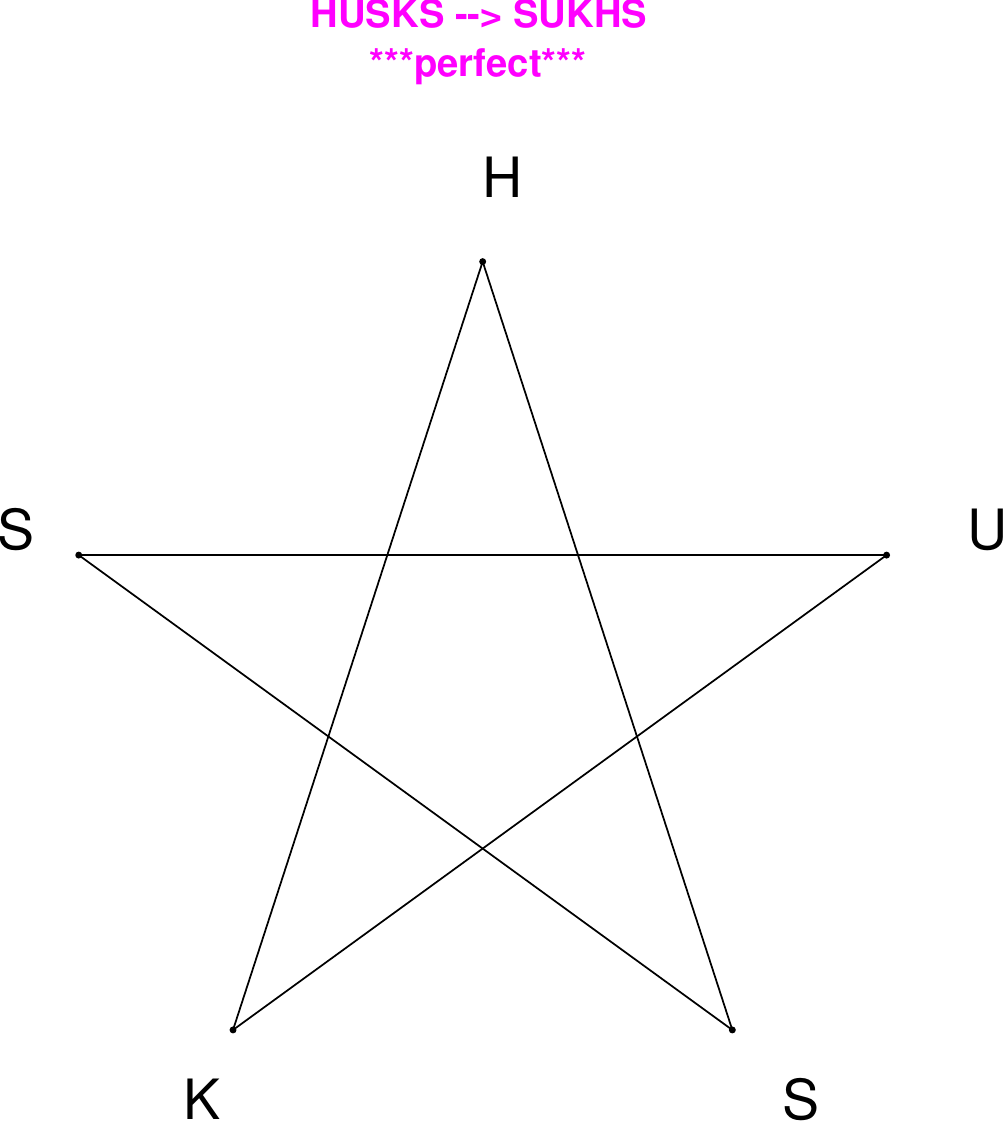}
\end{subfigure}
\hfill
\begin{subfigure}[T]{0.19\textwidth}
\centering
\includegraphics[width=\textwidth]{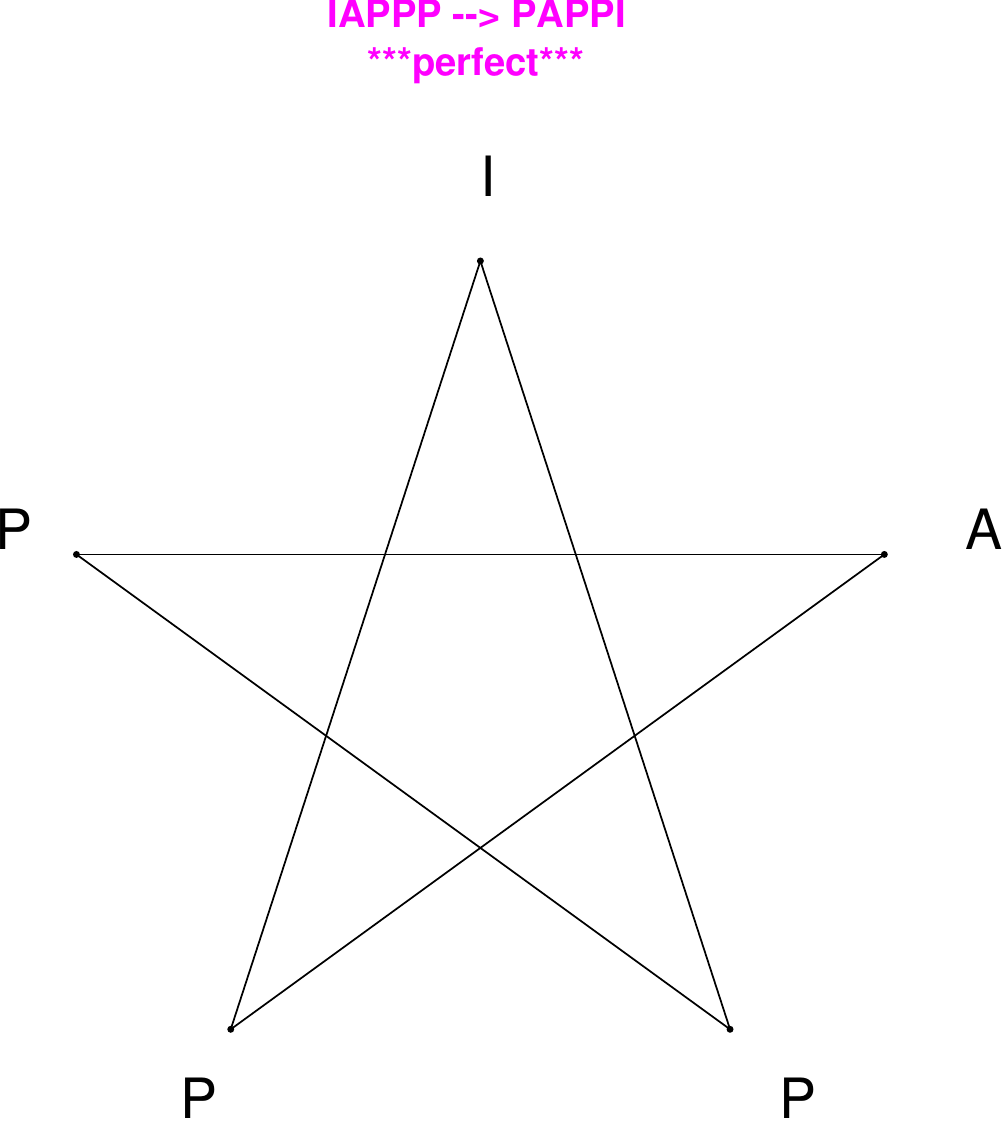}
\end{subfigure}
\hfill
\begin{subfigure}[T]{0.19\textwidth}
\centering
\includegraphics[width=\textwidth]{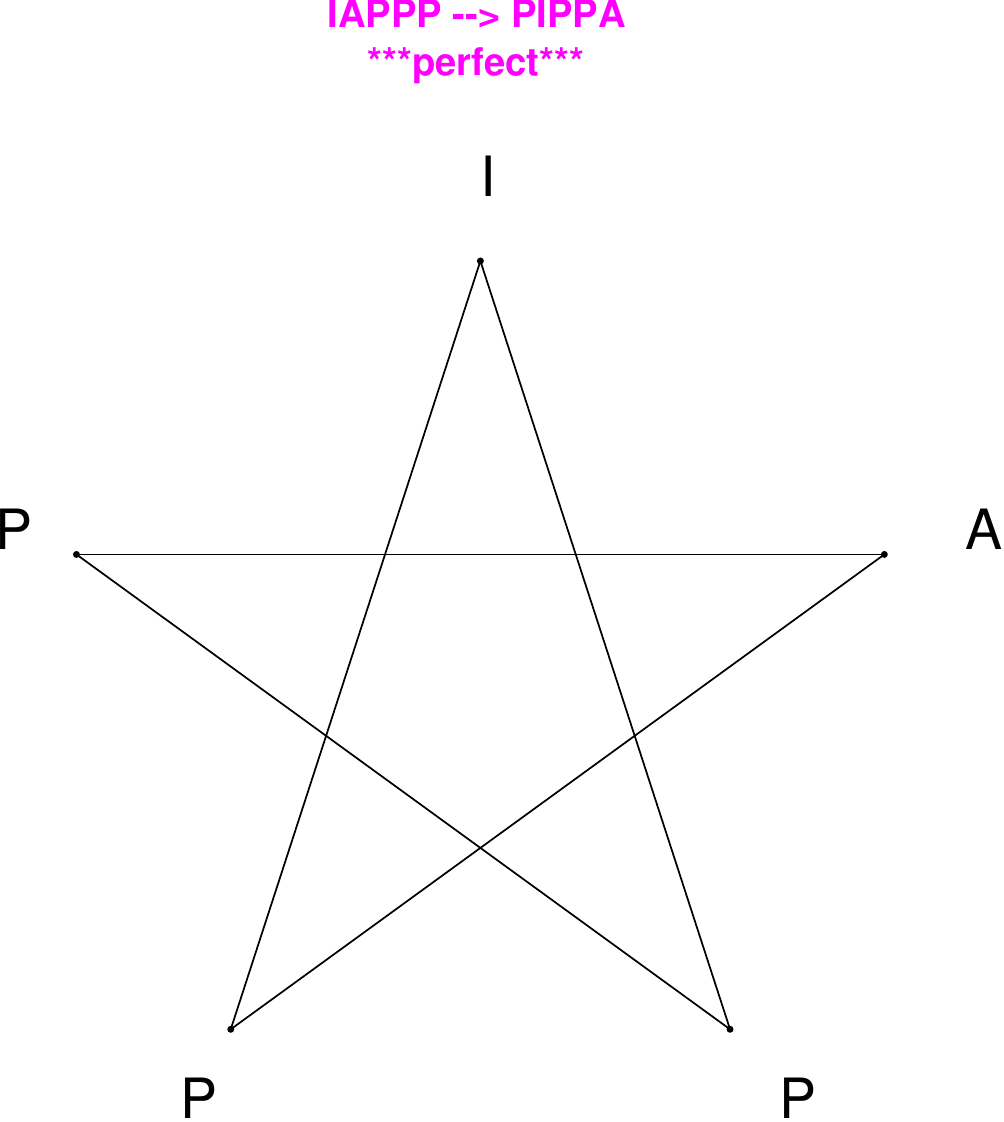}
\end{subfigure}
\end{figure}

\begin{figure}[H]
\centering
\begin{subfigure}[T]{0.19\textwidth}
\centering
\includegraphics[width=\textwidth]{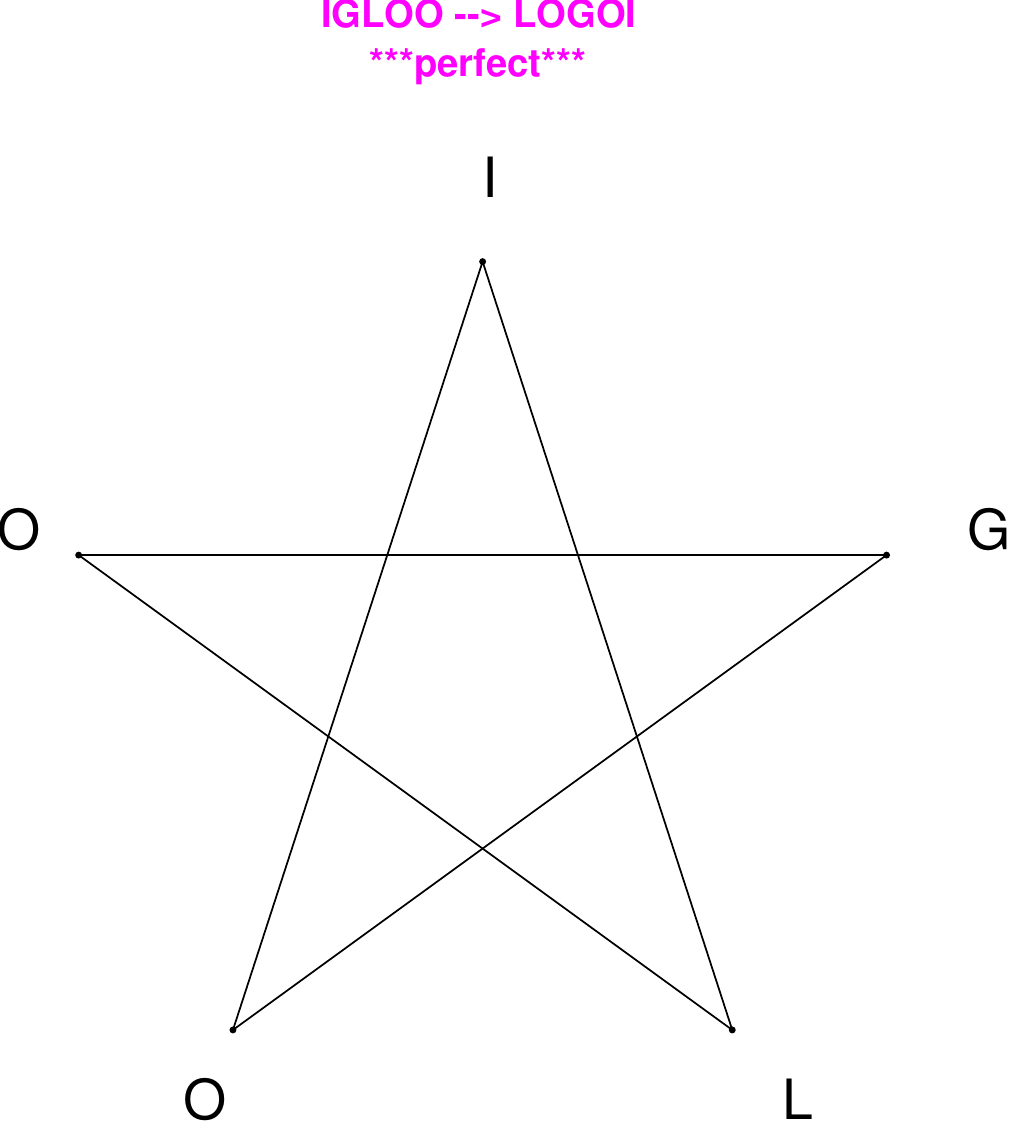}
\end{subfigure}
\hfill
\begin{subfigure}[T]{0.19\textwidth}
\centering
\includegraphics[width=\textwidth]{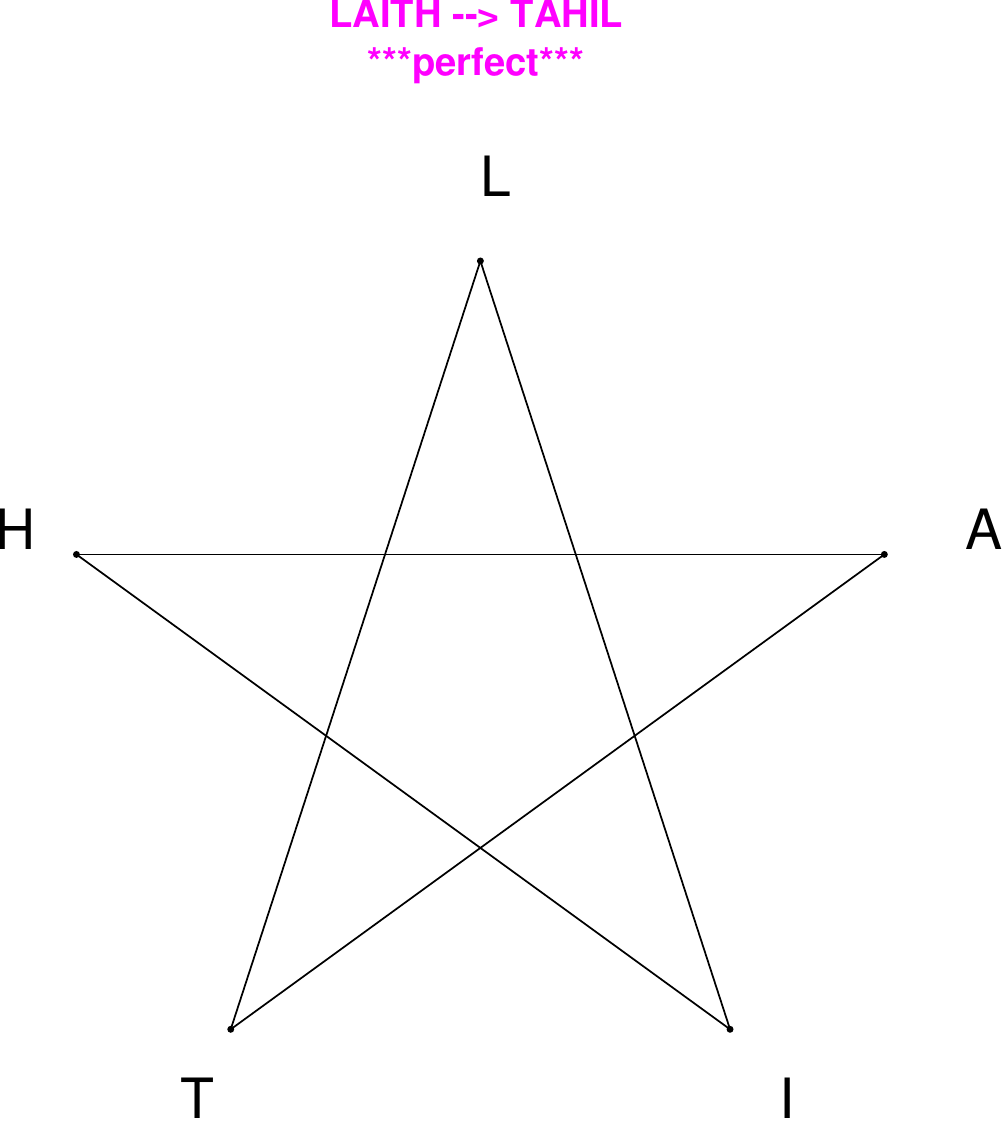}
\end{subfigure}
\hfill
\begin{subfigure}[T]{0.19\textwidth}
\centering
\includegraphics[width=\textwidth]{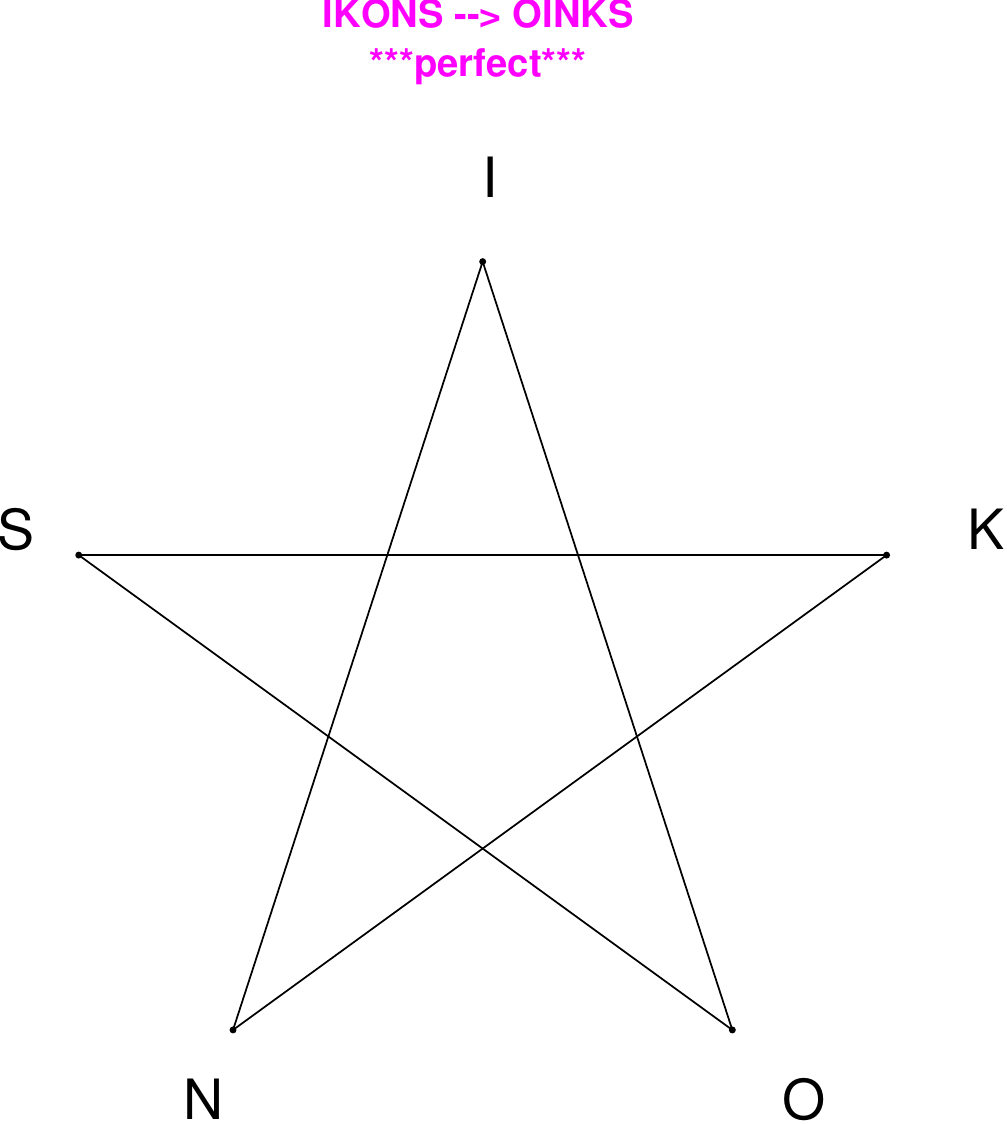}
\end{subfigure}
\hfill
\begin{subfigure}[T]{0.19\textwidth}
\centering
\includegraphics[width=\textwidth]{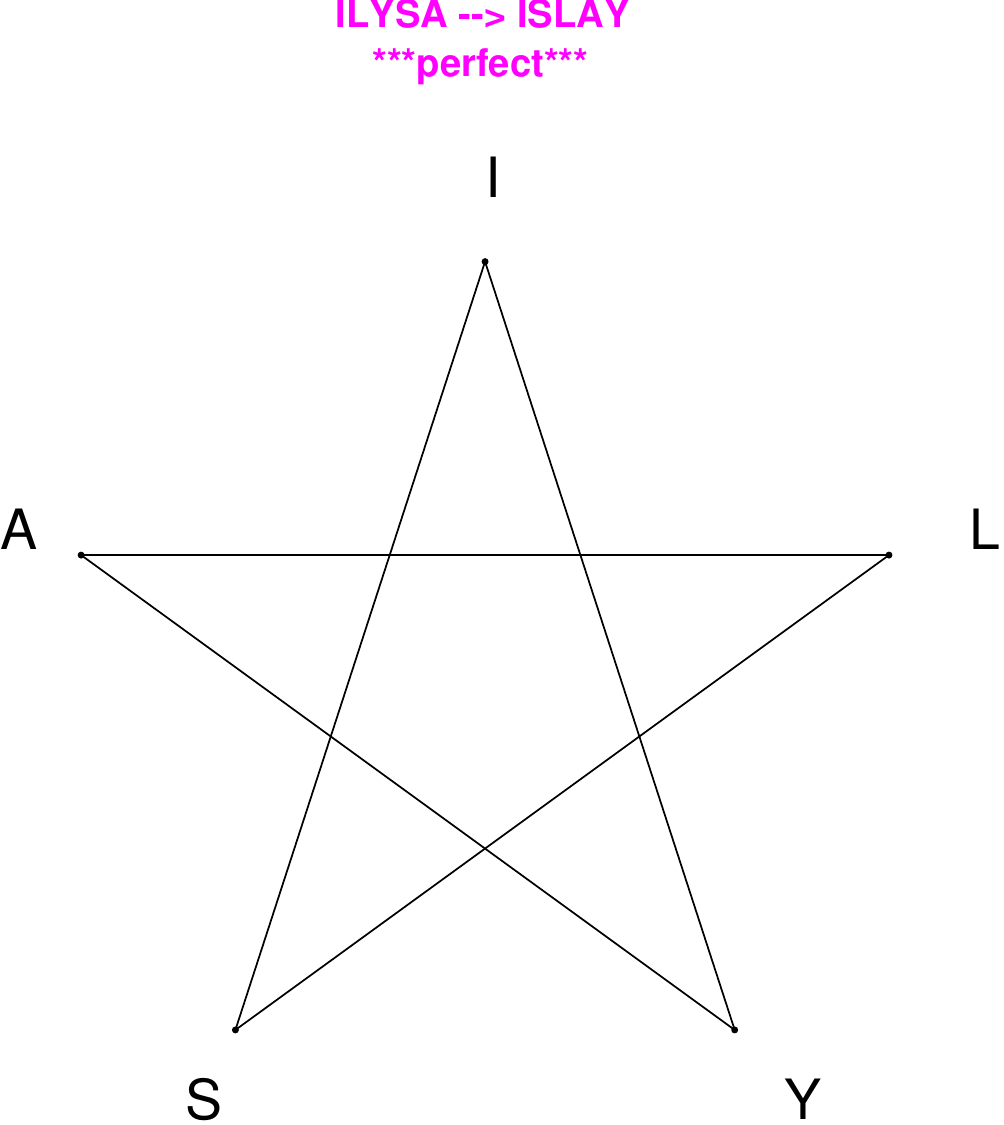}
\end{subfigure}
\hfill
\begin{subfigure}[T]{0.19\textwidth}
\centering
\includegraphics[width=\textwidth]{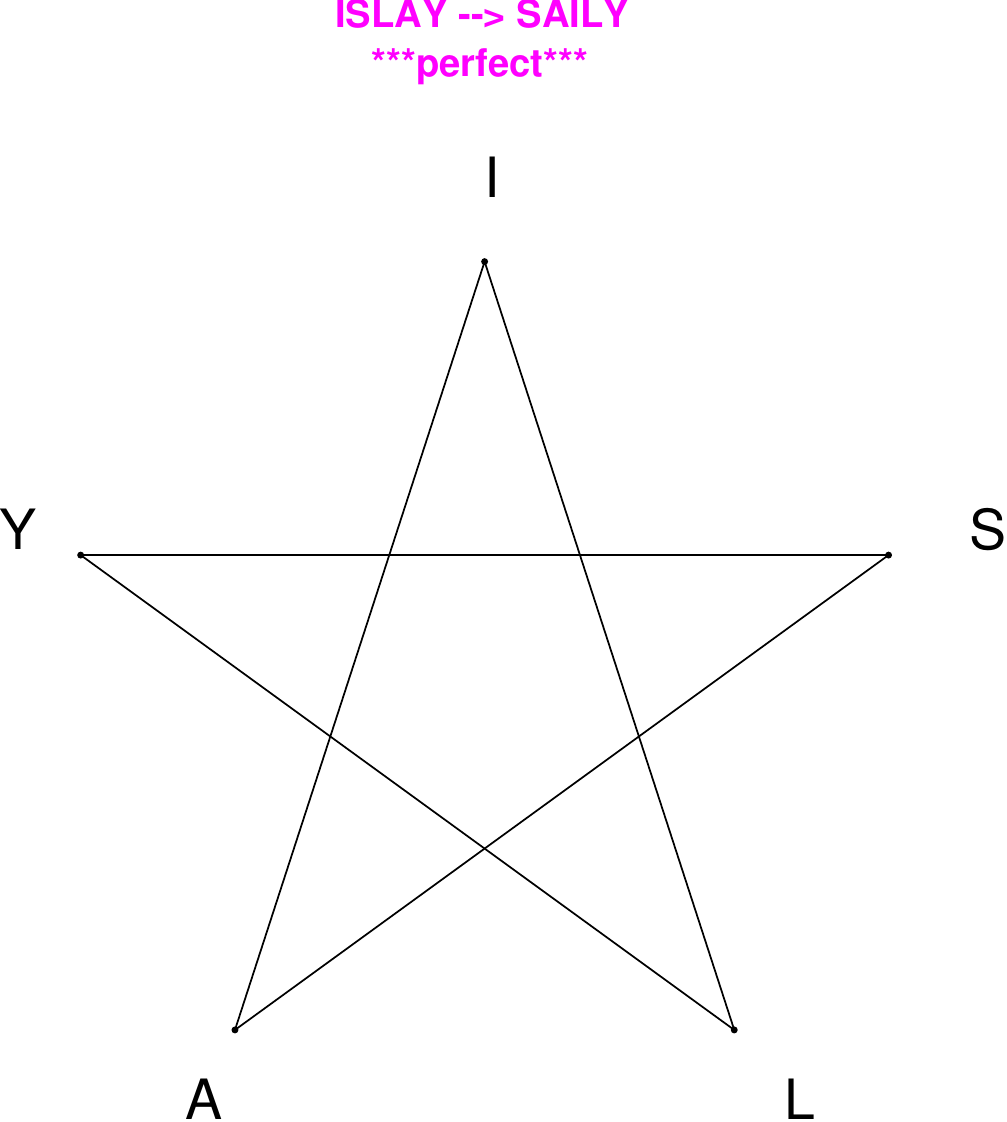}
\end{subfigure}
\end{figure}

\begin{figure}[H]
\centering
\begin{subfigure}[T]{0.19\textwidth}
\centering
\includegraphics[width=\textwidth]{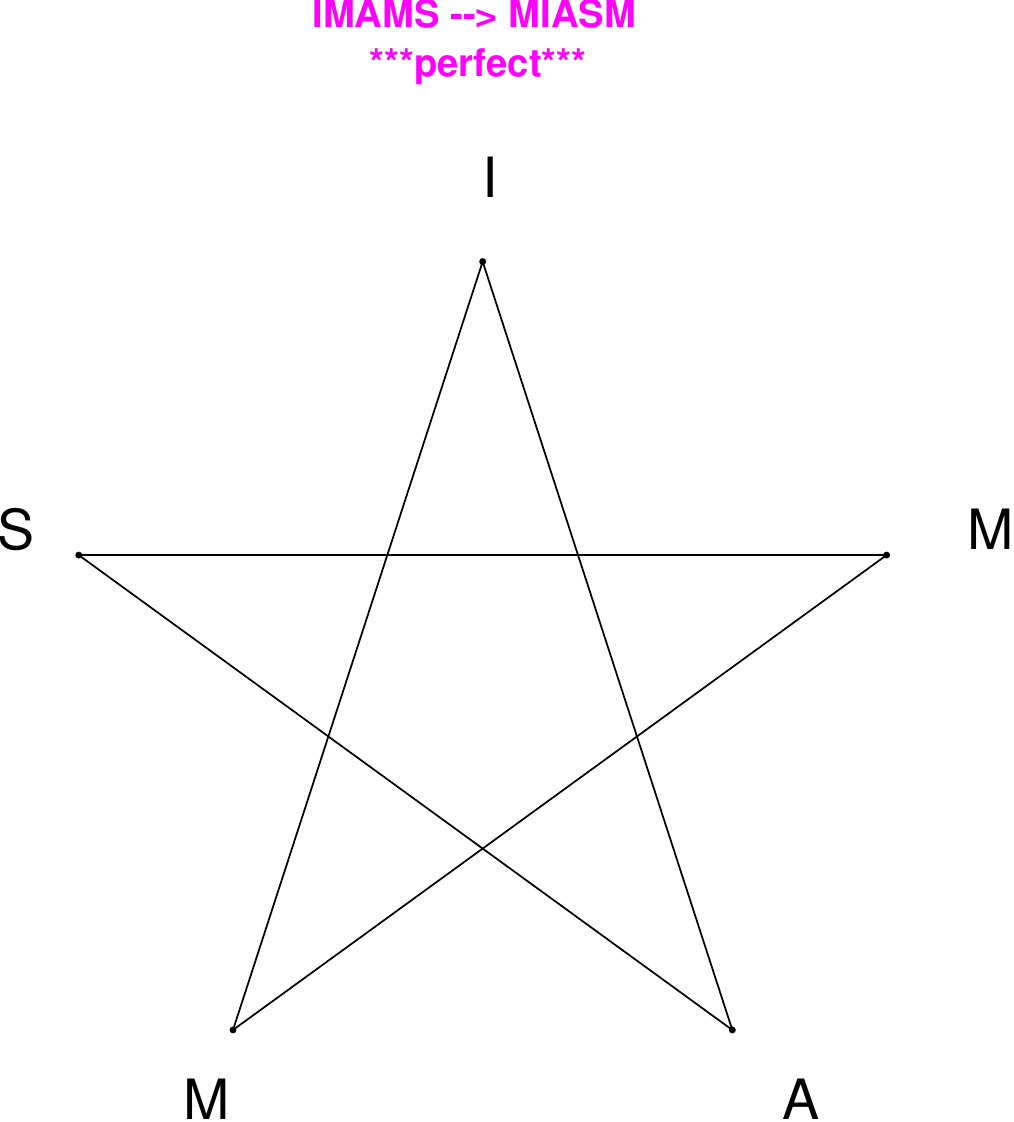}
\end{subfigure}
\hfill
\begin{subfigure}[T]{0.19\textwidth}
\centering
\includegraphics[width=\textwidth]{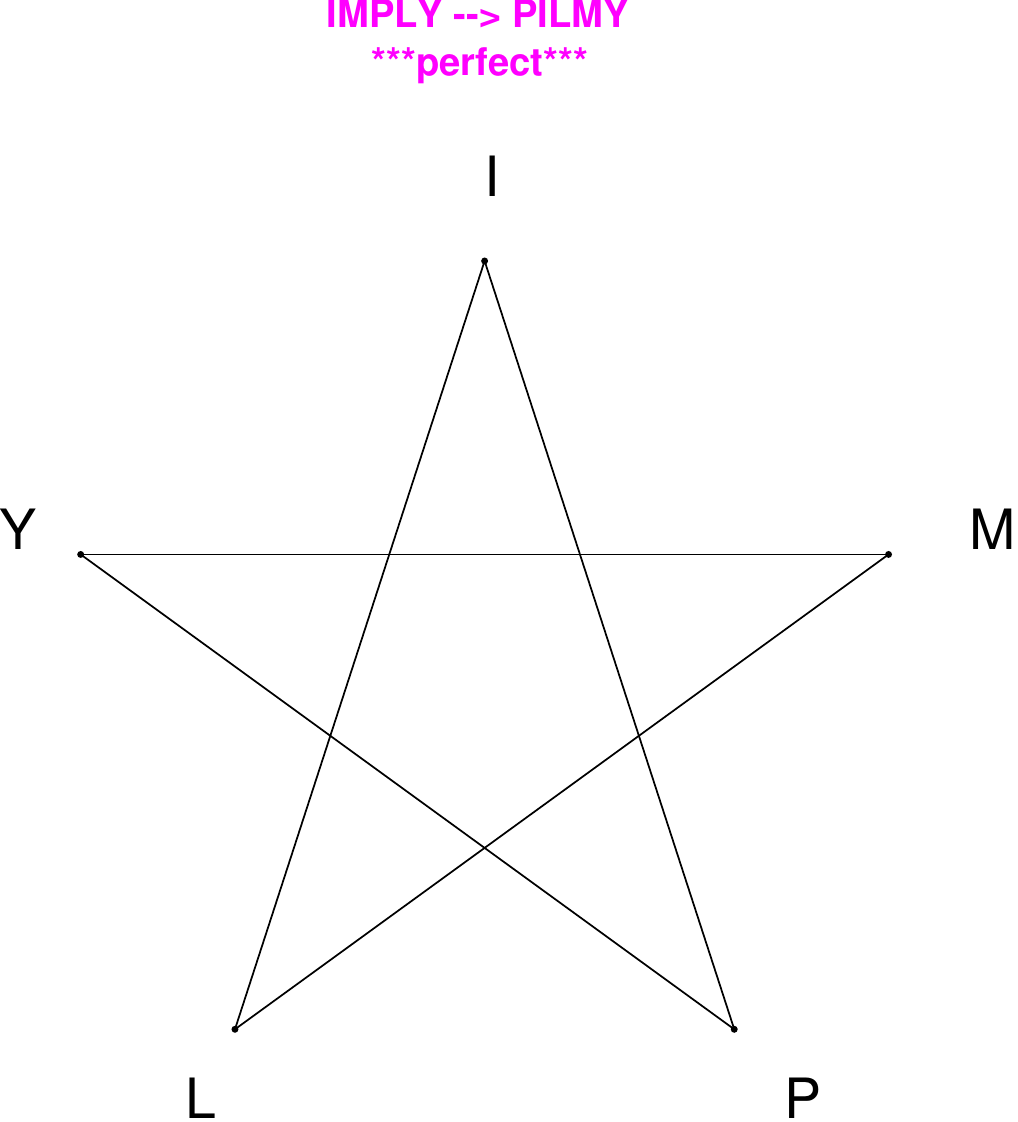}
\end{subfigure}
\hfill
\begin{subfigure}[T]{0.19\textwidth}
\centering
\includegraphics[width=\textwidth]{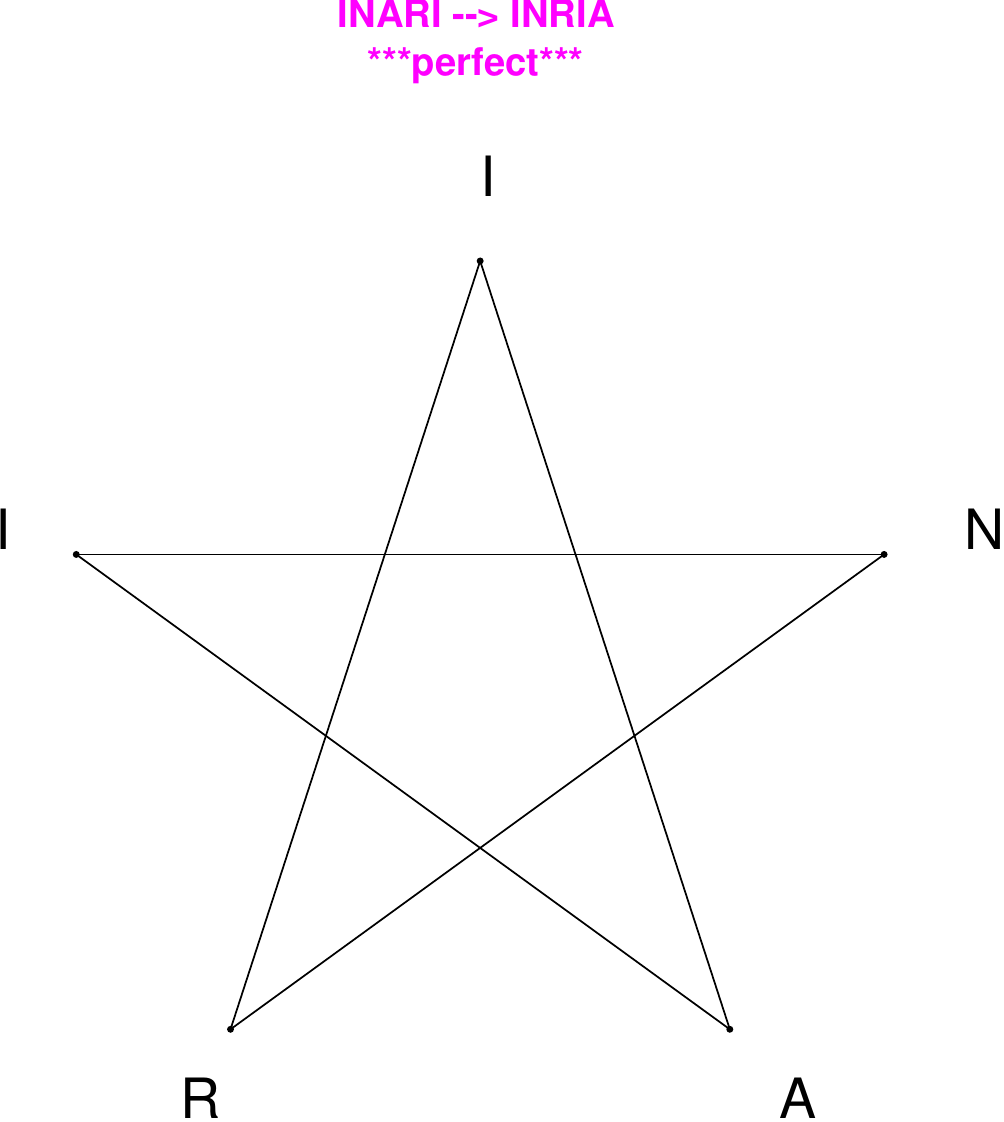}
\end{subfigure}
\hfill
\begin{subfigure}[T]{0.19\textwidth}
\centering
\includegraphics[width=\textwidth]{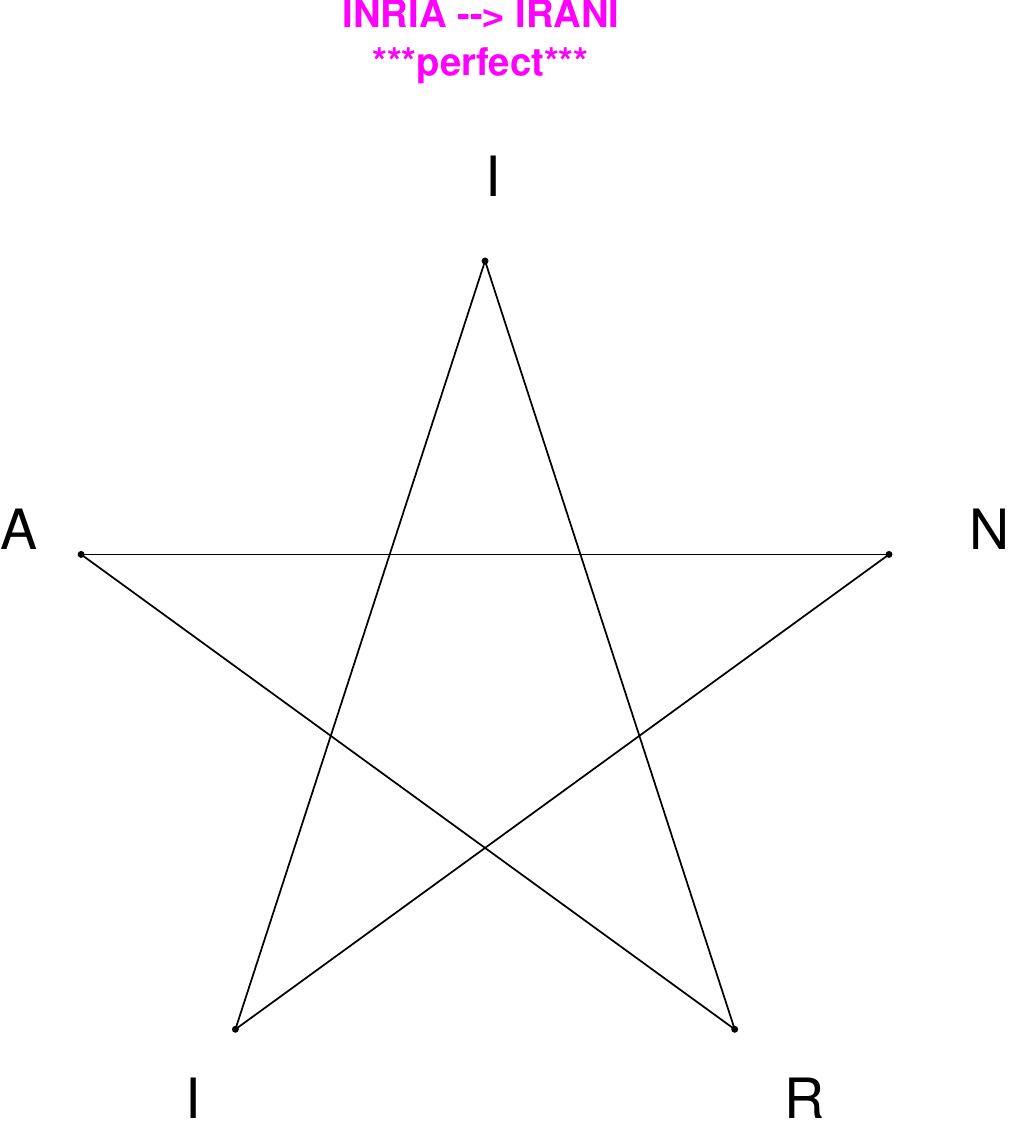}
\end{subfigure}
\hfill
\begin{subfigure}[T]{0.19\textwidth}
\centering
\includegraphics[width=\textwidth]{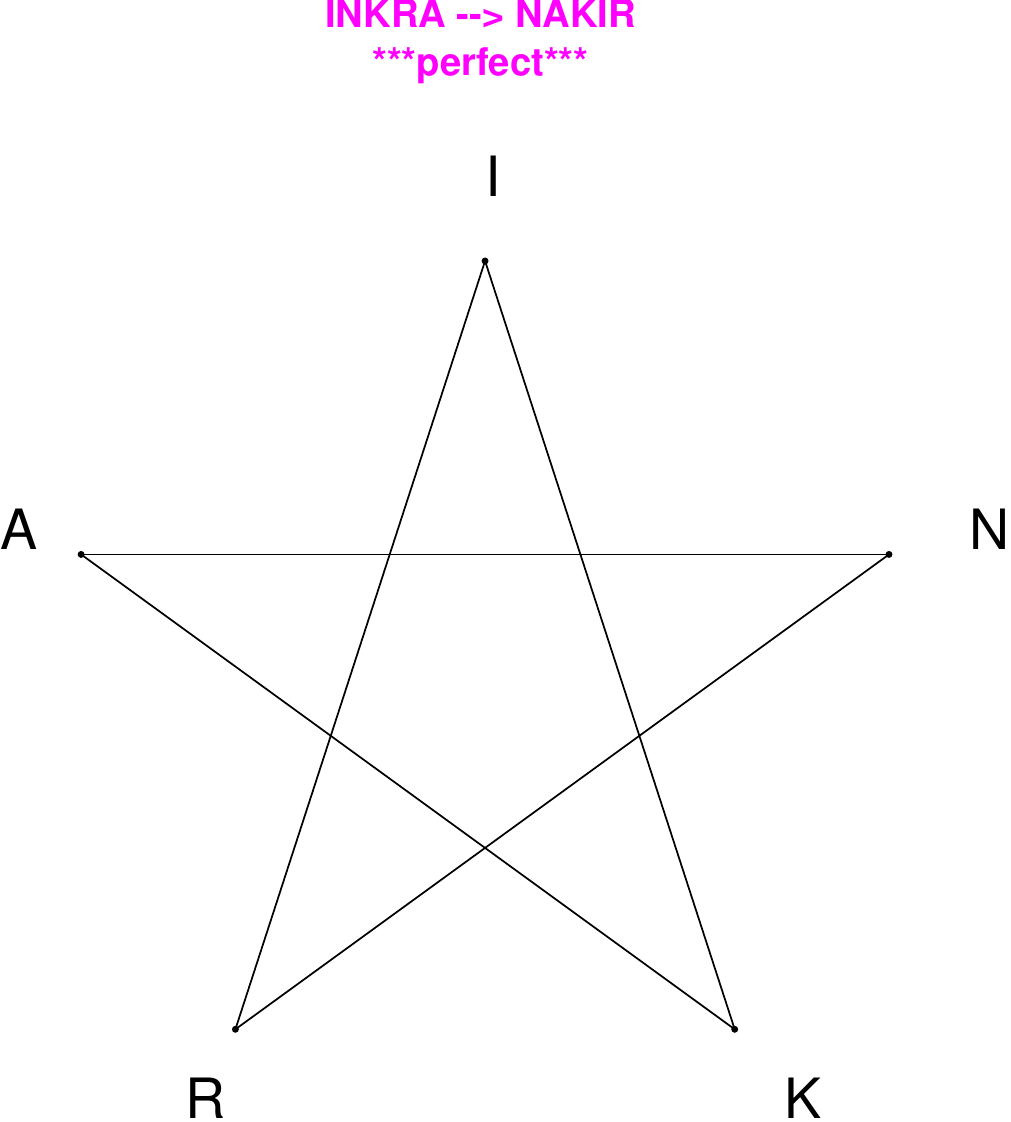}
\end{subfigure}
\end{figure}

\begin{figure}[H]
\centering
\begin{subfigure}[T]{0.19\textwidth}
\centering
\includegraphics[width=\textwidth]{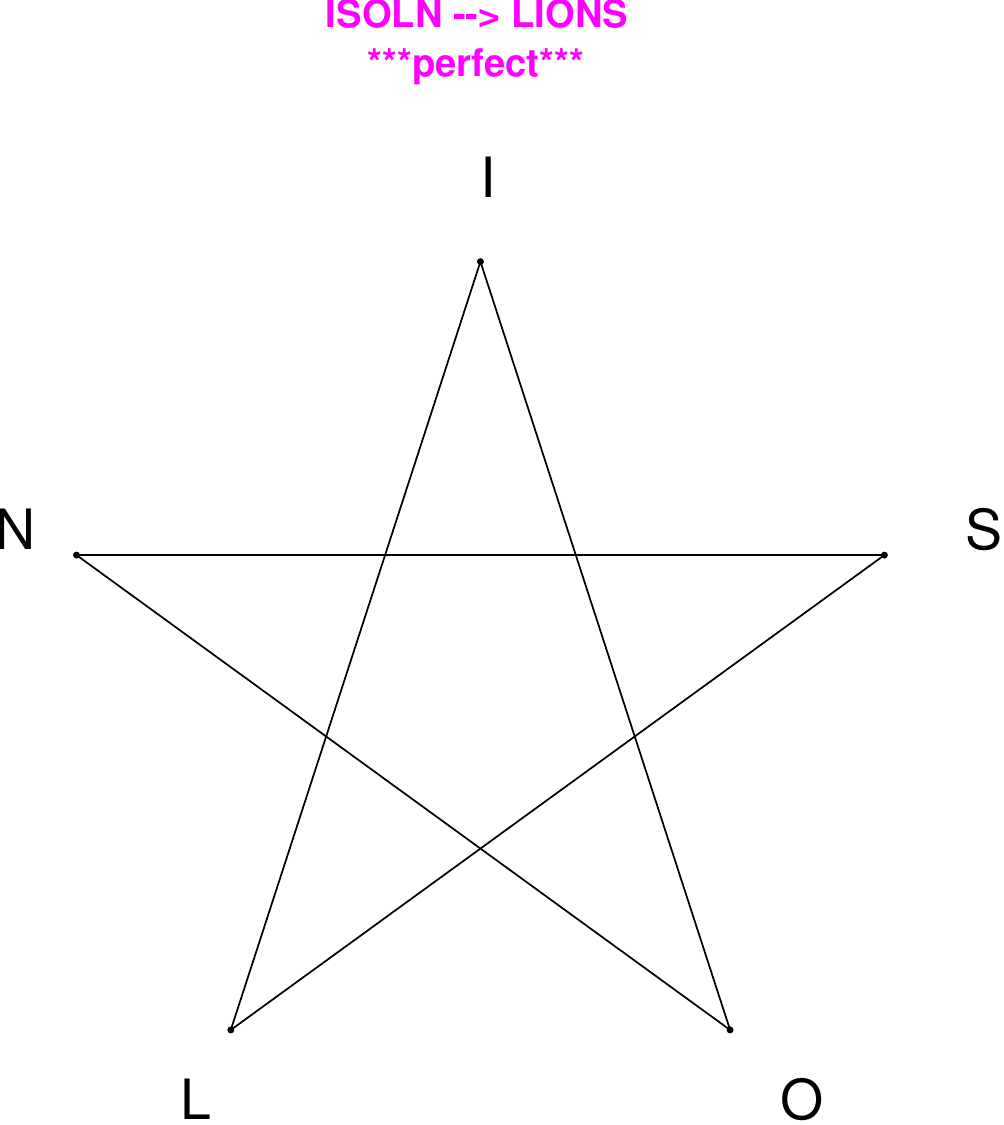}
\end{subfigure}
\hfill
\begin{subfigure}[T]{0.19\textwidth}
\centering
\includegraphics[width=\textwidth]{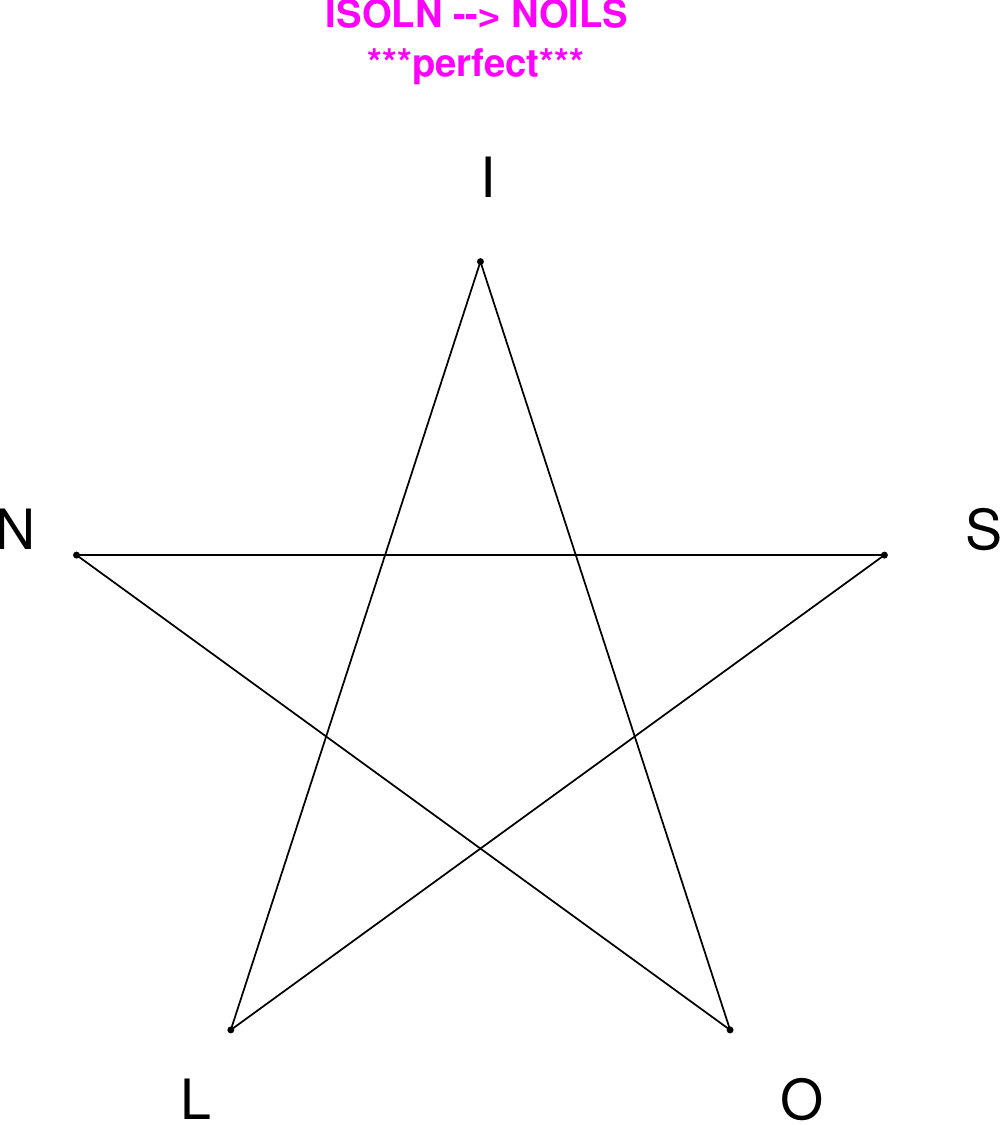}
\end{subfigure}
\hfill
\begin{subfigure}[T]{0.19\textwidth}
\centering
\includegraphics[width=\textwidth]{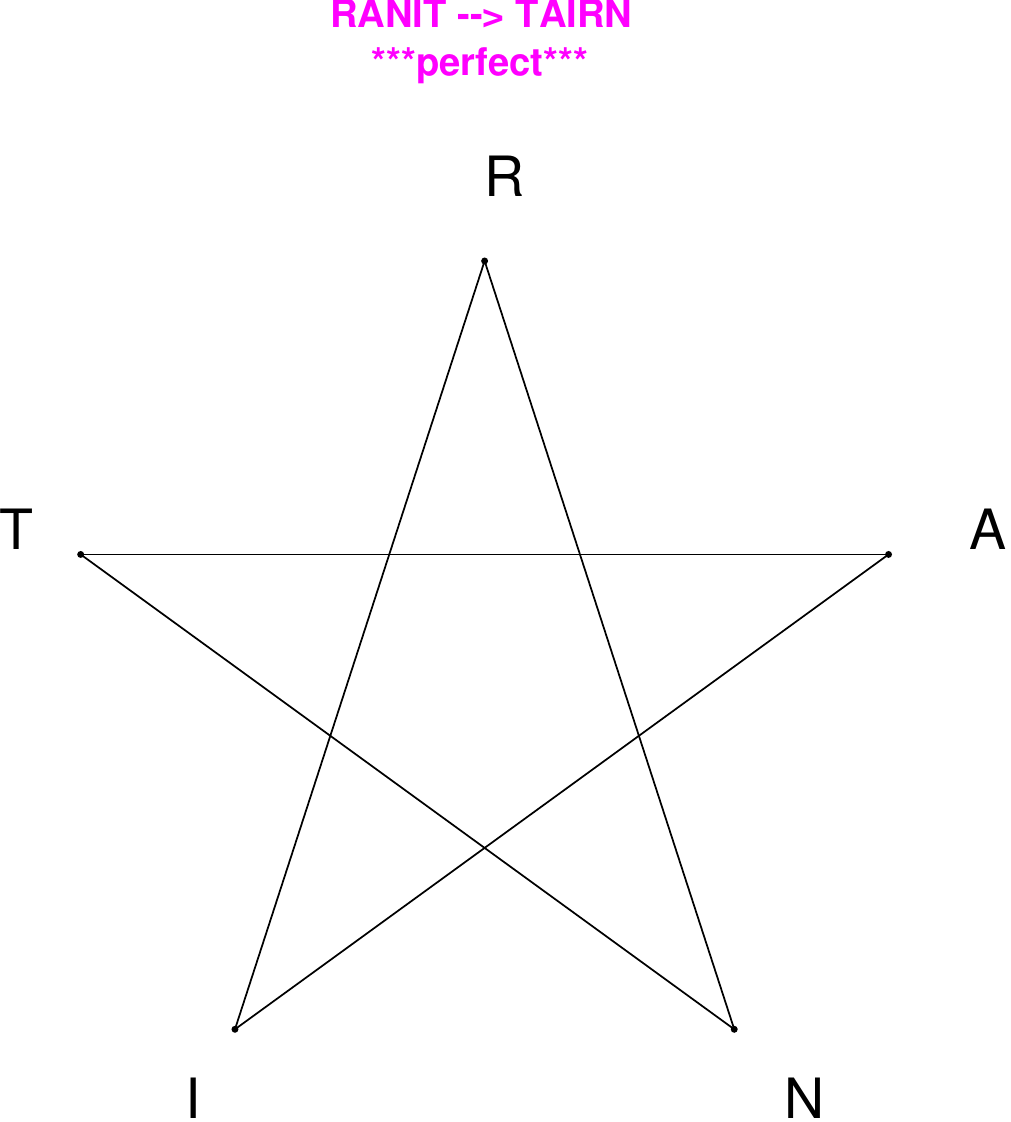}
\end{subfigure}
\hfill
\begin{subfigure}[T]{0.19\textwidth}
\centering
\includegraphics[width=\textwidth]{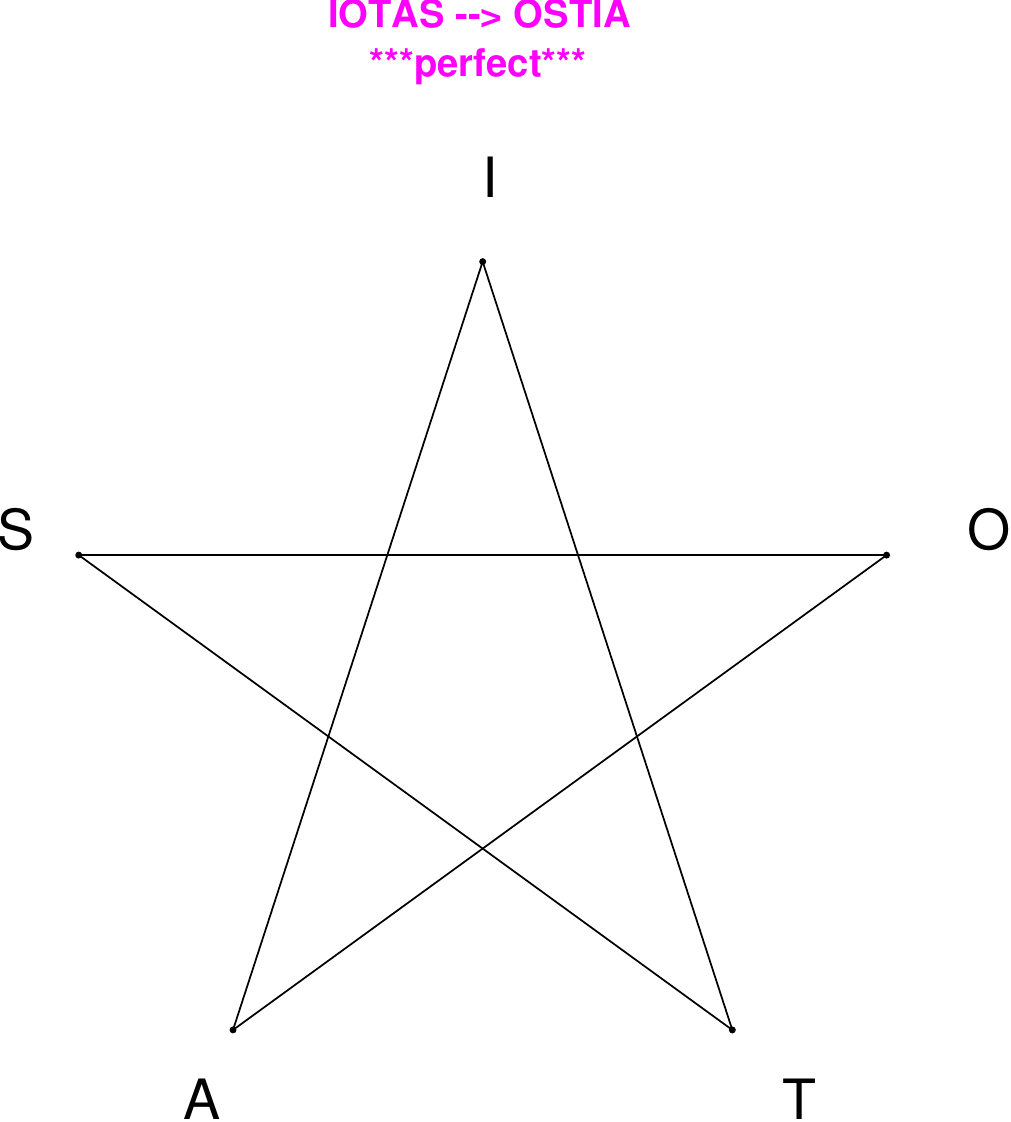}
\end{subfigure}
\hfill
\begin{subfigure}[T]{0.19\textwidth}
\centering
\includegraphics[width=\textwidth]{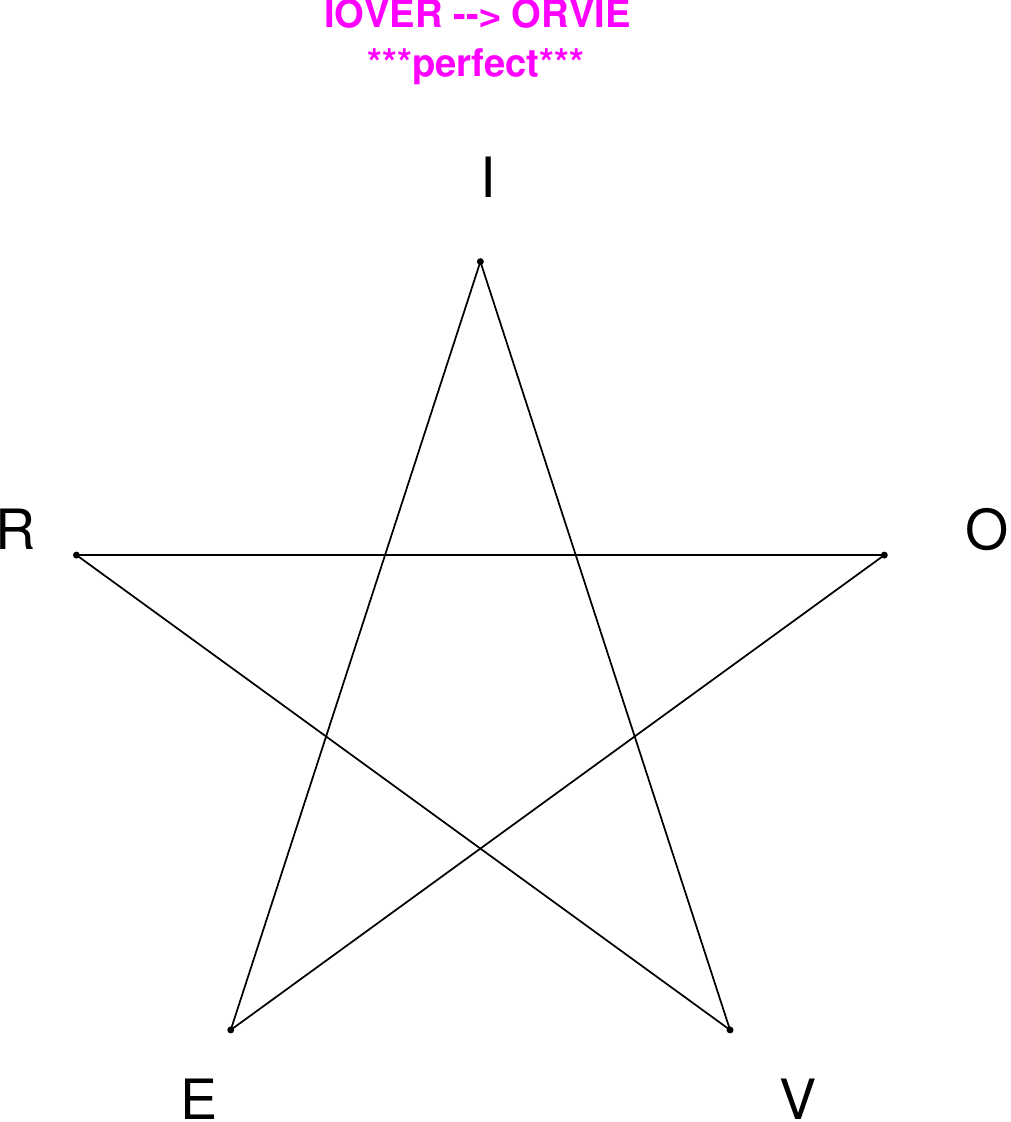}
\end{subfigure}
\end{figure}

\begin{figure}[H]
\centering
\begin{subfigure}[T]{0.19\textwidth}
\centering
\includegraphics[width=\textwidth]{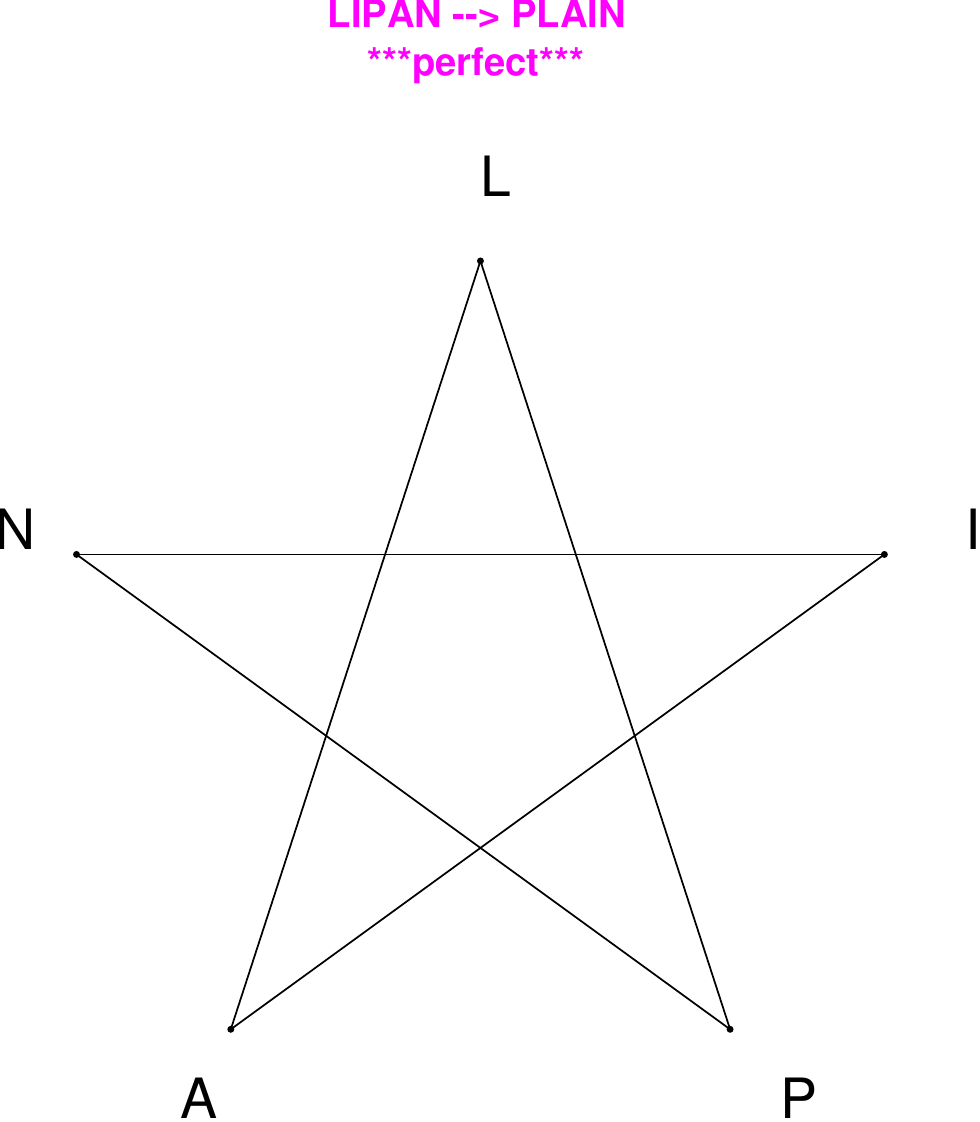}
\end{subfigure}
\hfill
\begin{subfigure}[T]{0.19\textwidth}
\centering
\includegraphics[width=\textwidth]{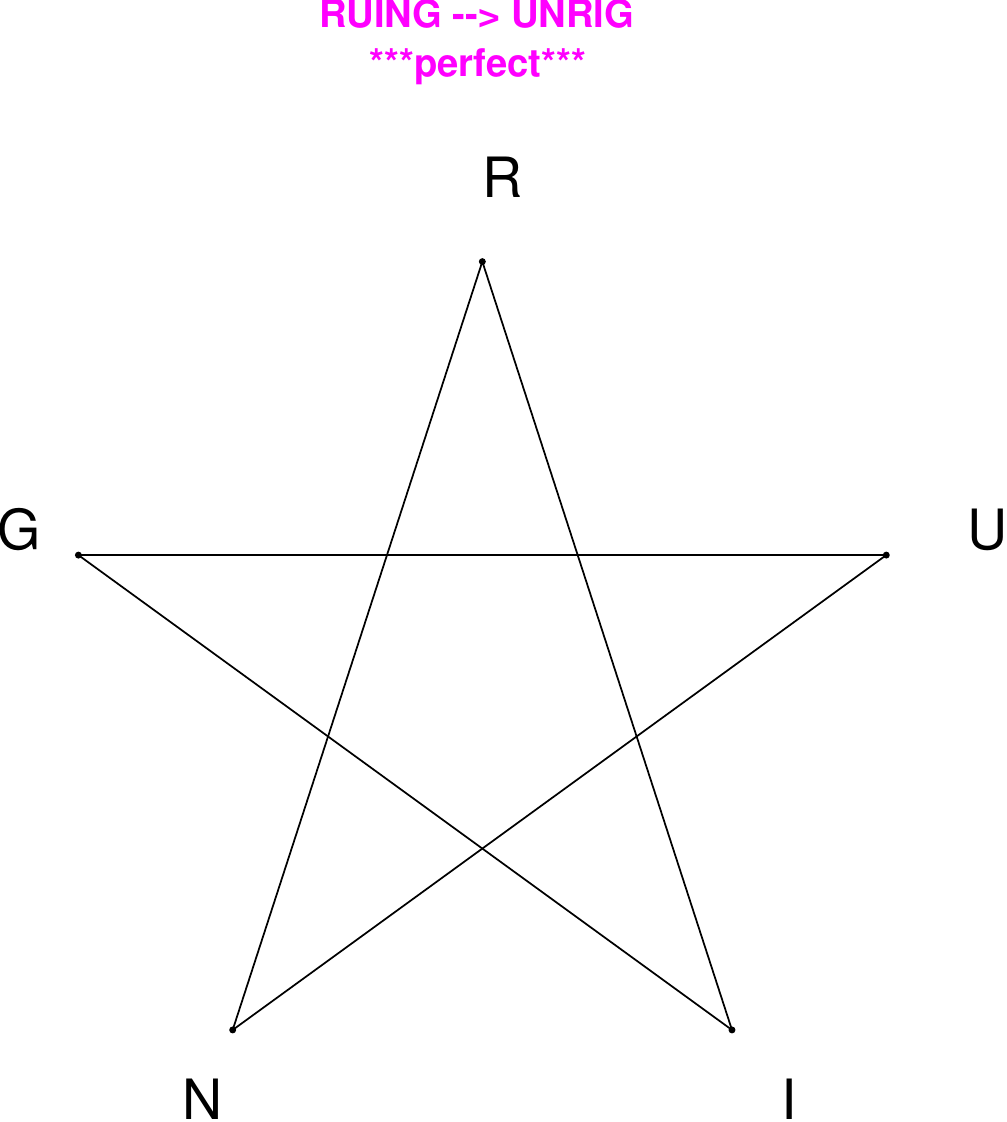}
\end{subfigure}
\hfill
\begin{subfigure}[T]{0.19\textwidth}
\centering
\includegraphics[width=\textwidth]{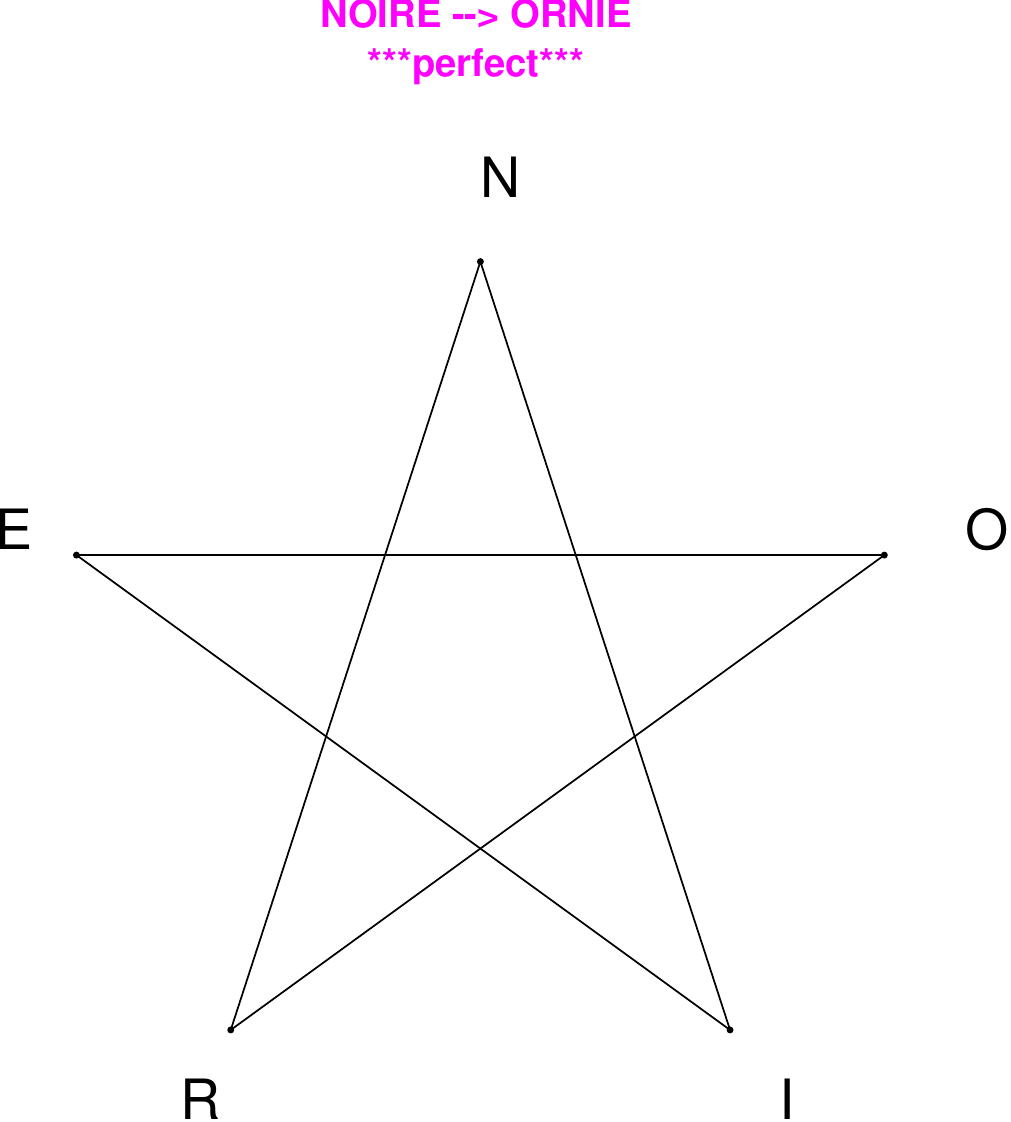}
\end{subfigure}
\hfill
\begin{subfigure}[T]{0.19\textwidth}
\centering
\includegraphics[width=\textwidth]{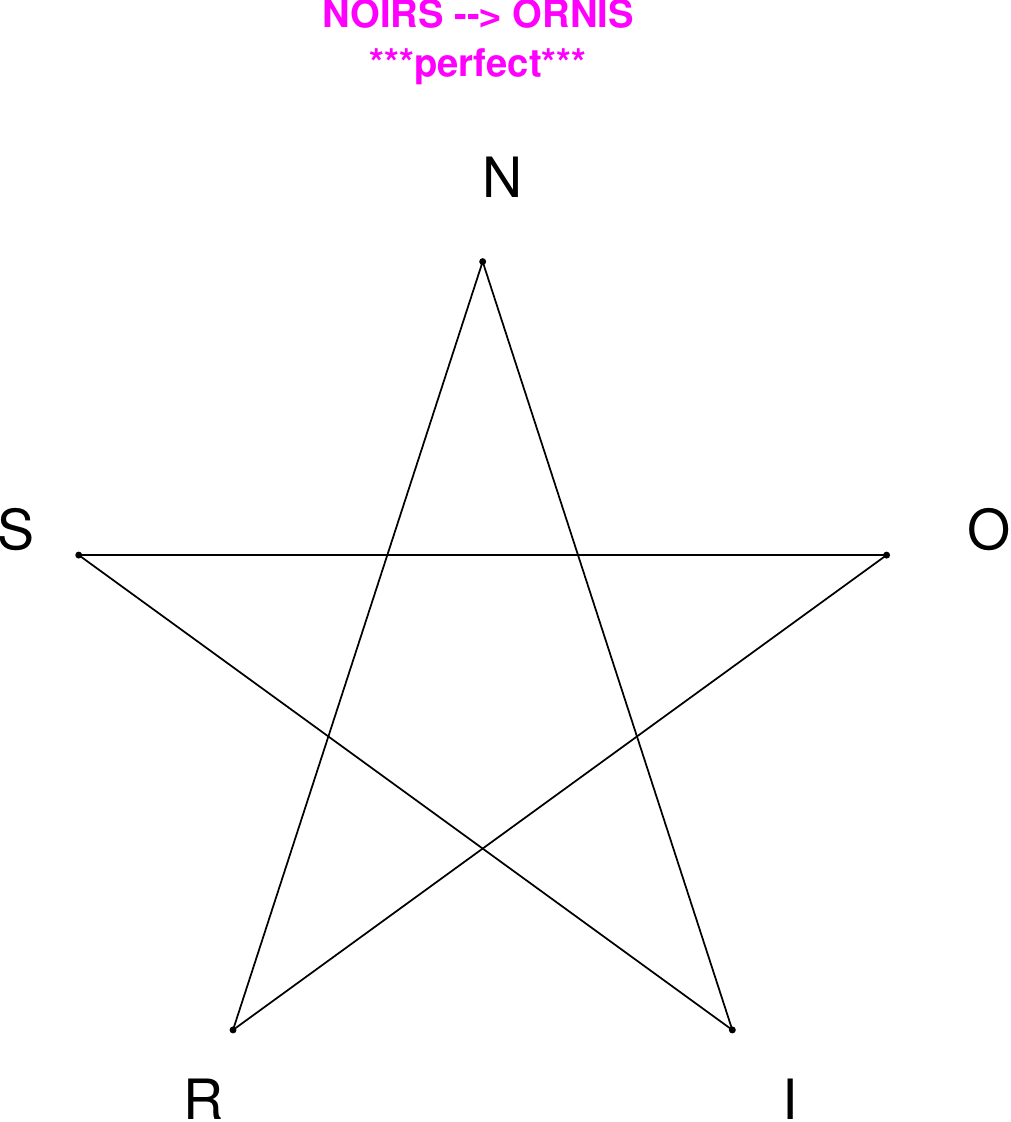}
\end{subfigure}
\hfill
\begin{subfigure}[T]{0.19\textwidth}
\centering
\includegraphics[width=\textwidth]{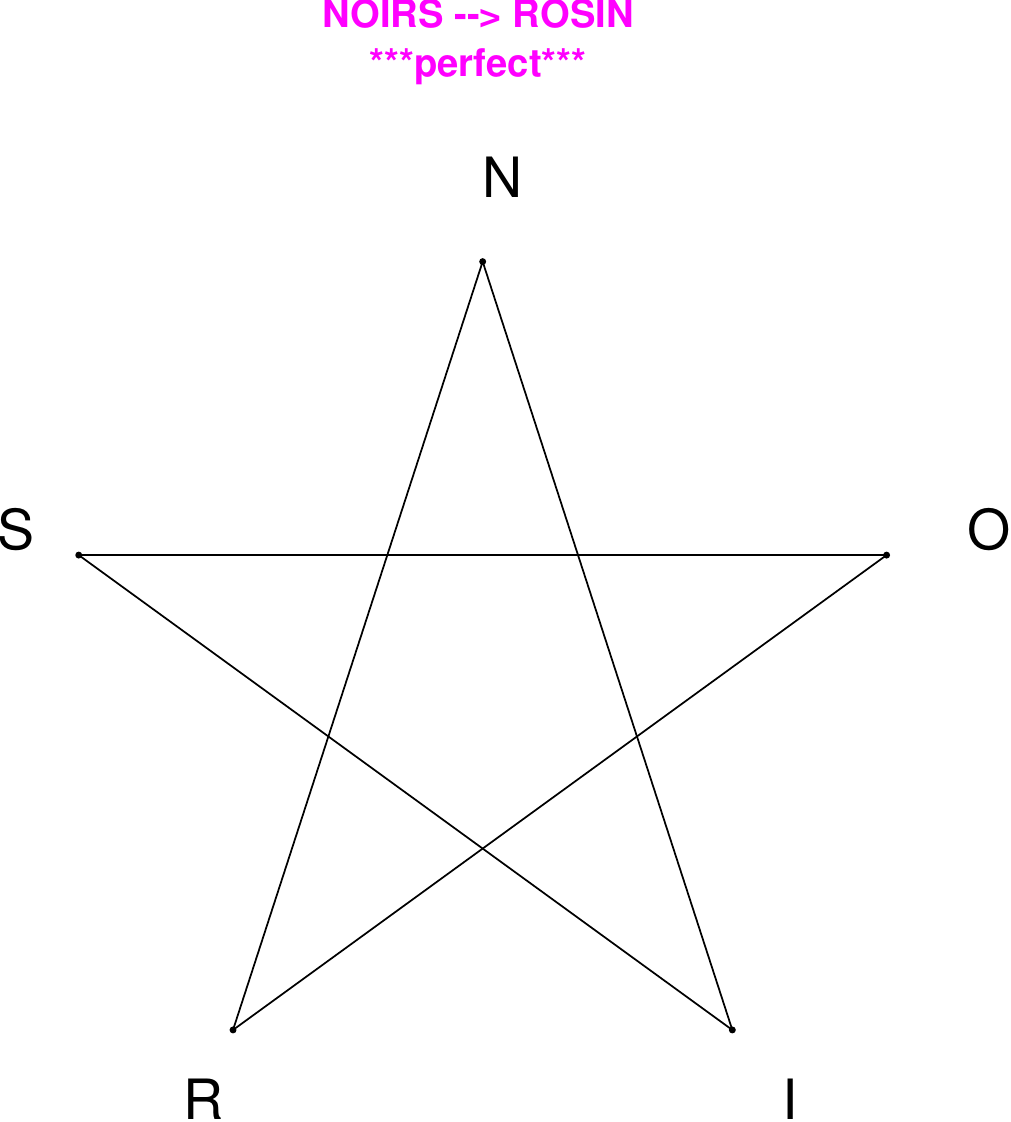}
\end{subfigure}
\end{figure}

\begin{figure}[H]
\centering
\begin{subfigure}[T]{0.19\textwidth}
\centering
\includegraphics[width=\textwidth]{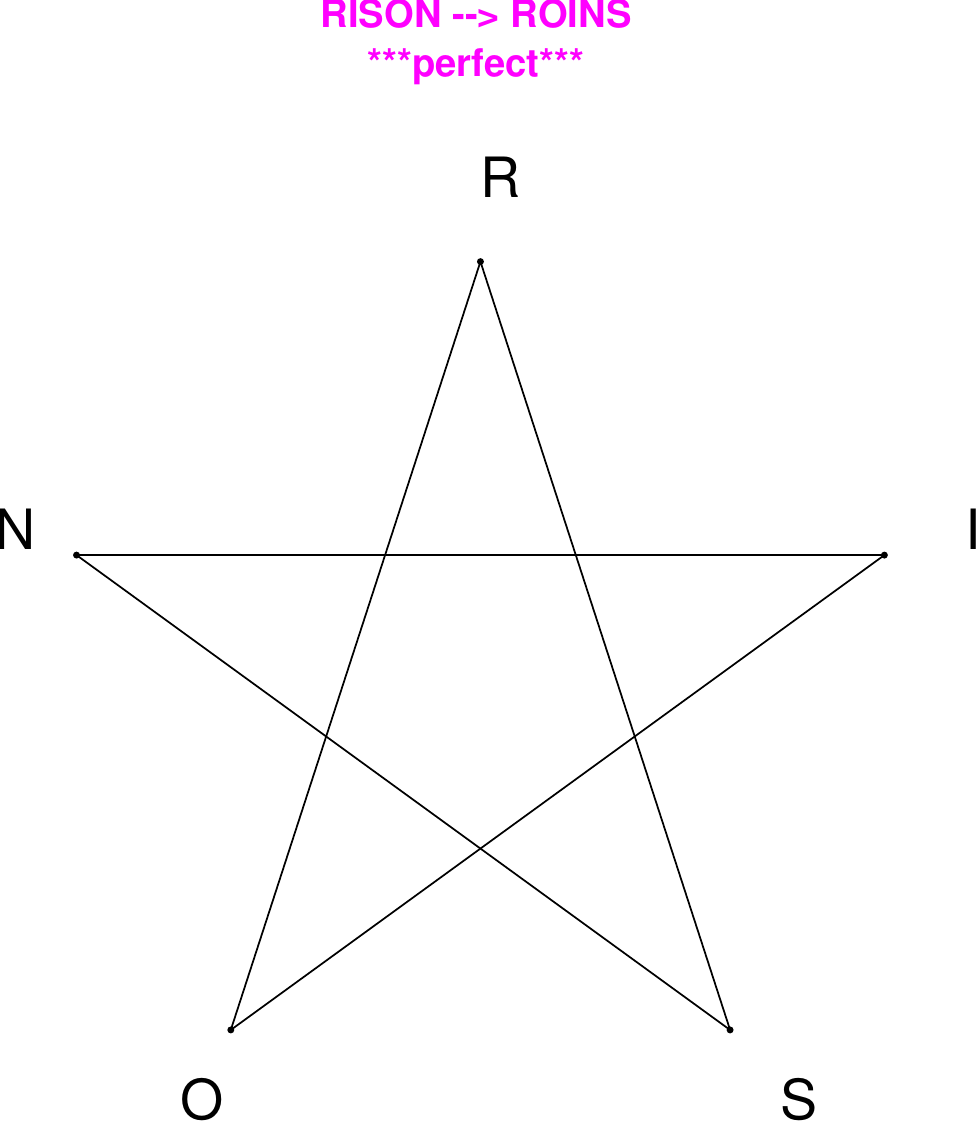}
\end{subfigure}
\hfill
\begin{subfigure}[T]{0.19\textwidth}
\centering
\includegraphics[width=\textwidth]{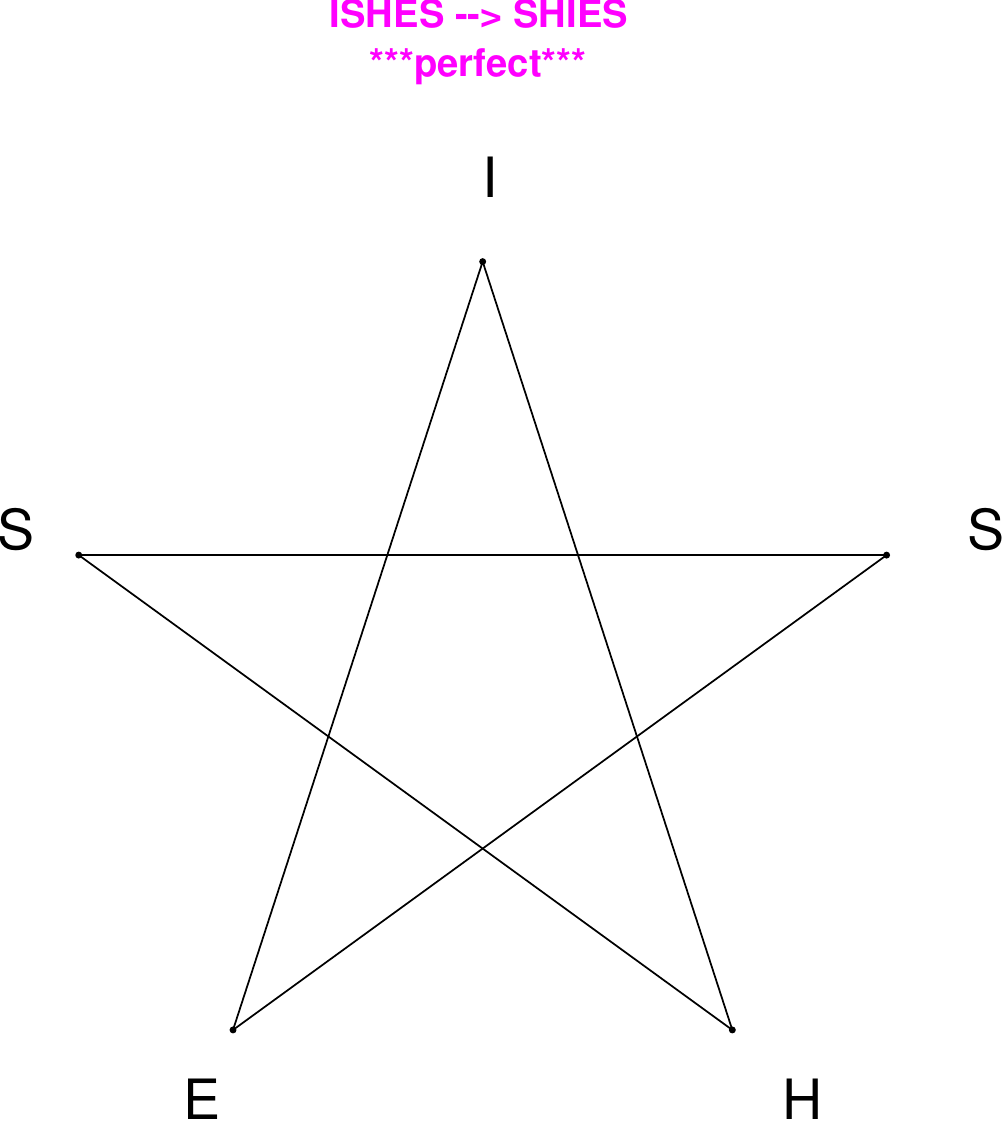}
\end{subfigure}
\hfill
\begin{subfigure}[T]{0.19\textwidth}
\centering
\includegraphics[width=\textwidth]{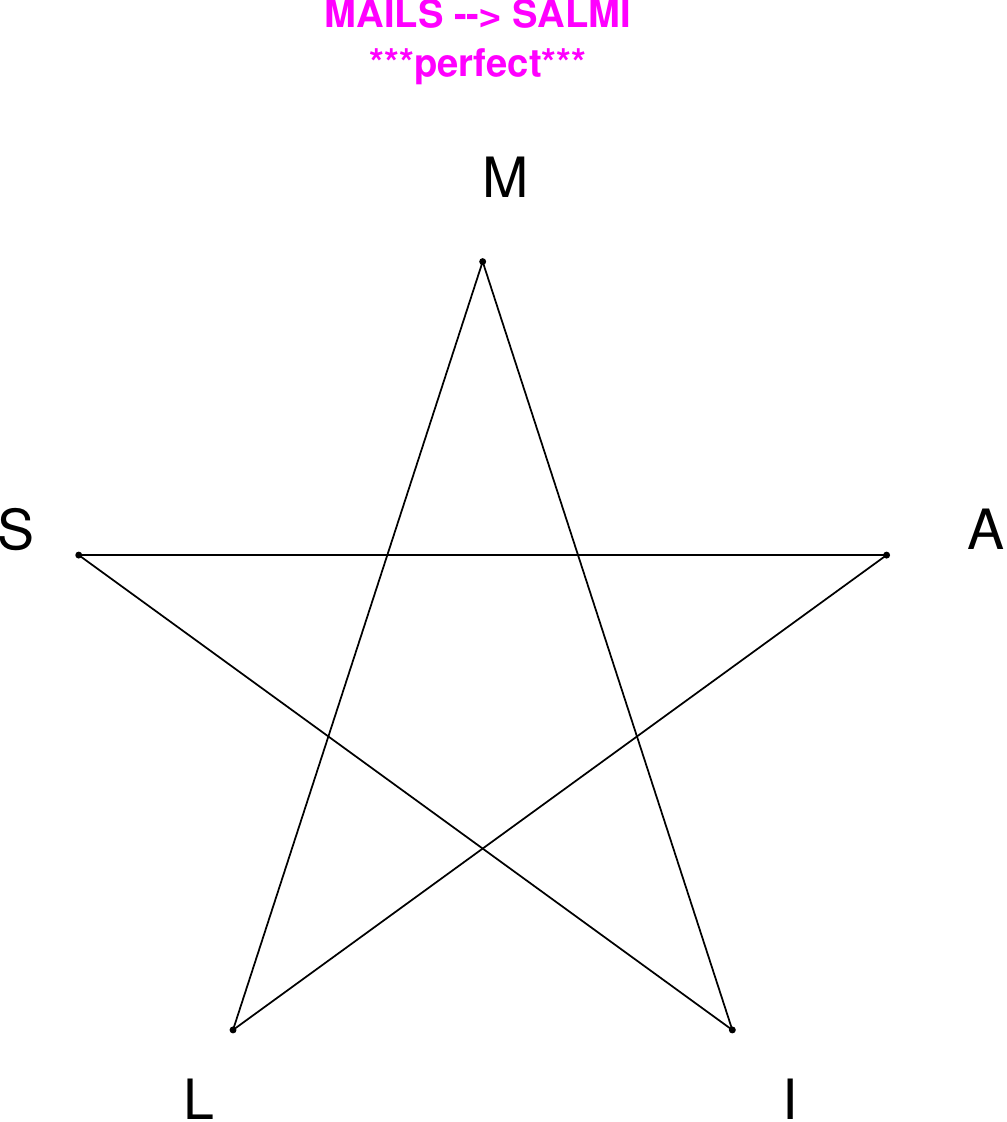}
\end{subfigure}
\hfill
\begin{subfigure}[T]{0.19\textwidth}
\centering
\includegraphics[width=\textwidth]{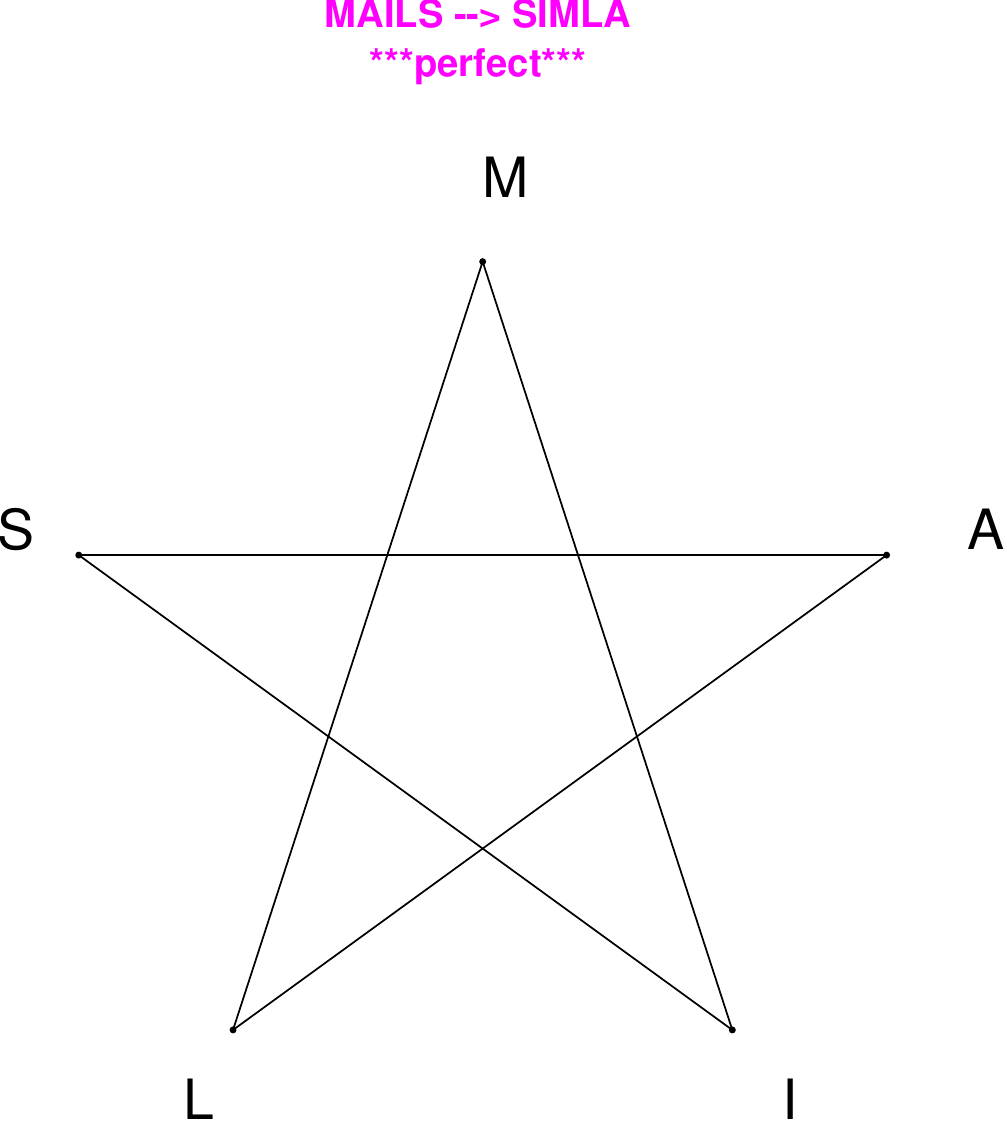}
\end{subfigure}
\hfill
\begin{subfigure}[T]{0.19\textwidth}
\centering
\includegraphics[width=\textwidth]{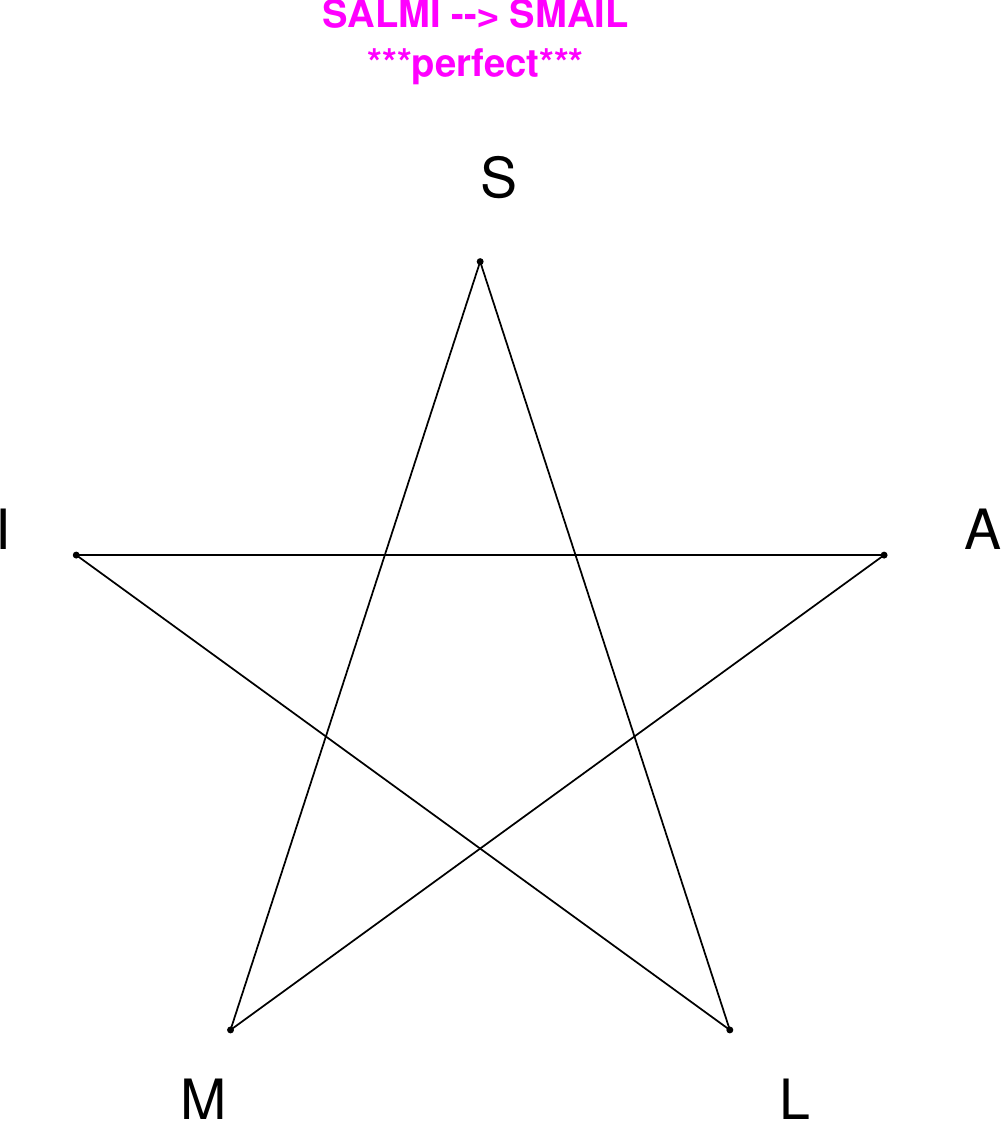}
\end{subfigure}
\end{figure}

\begin{figure}[H]
\centering
\begin{subfigure}[T]{0.19\textwidth}
\centering
\includegraphics[width=\textwidth]{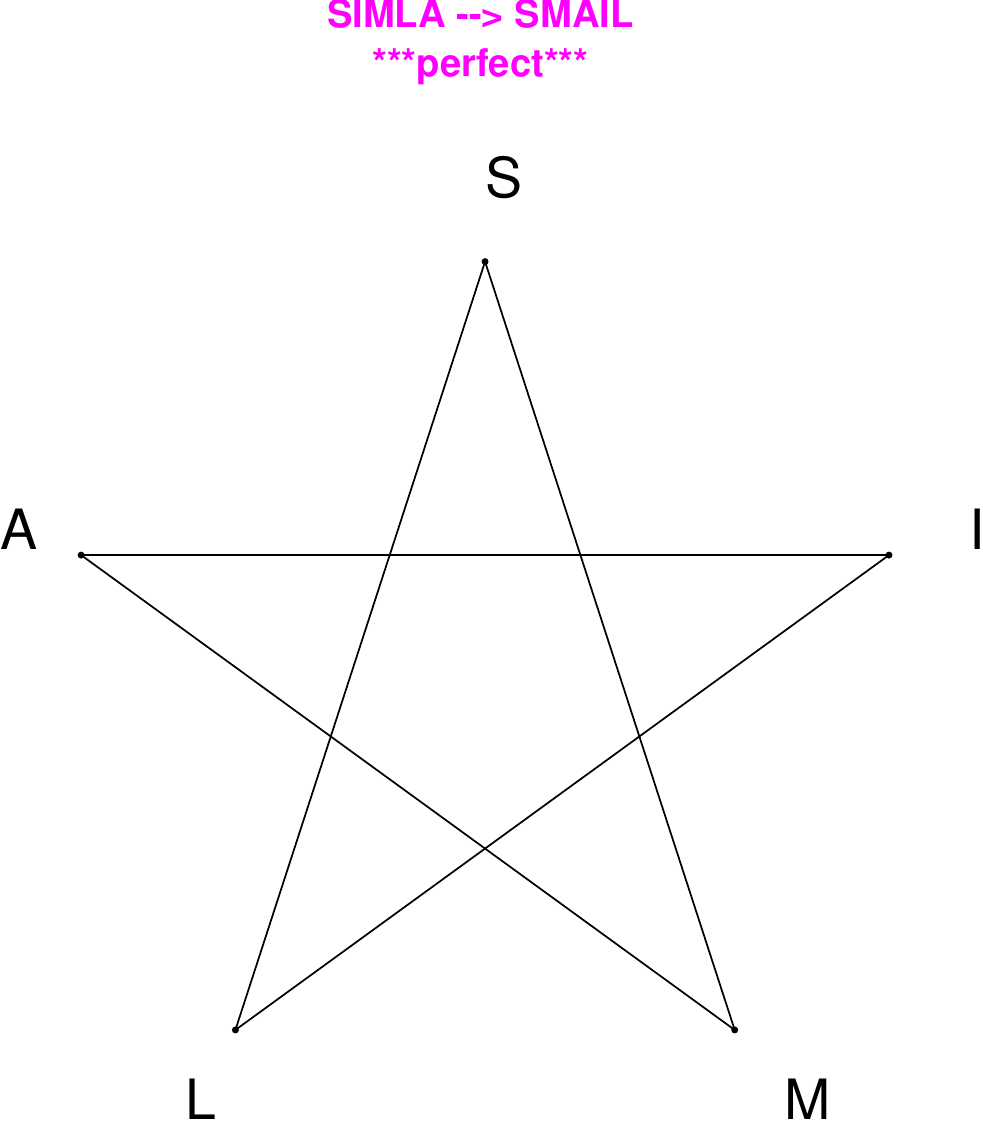}
\end{subfigure}
\hfill
\begin{subfigure}[T]{0.19\textwidth}
\centering
\includegraphics[width=\textwidth]{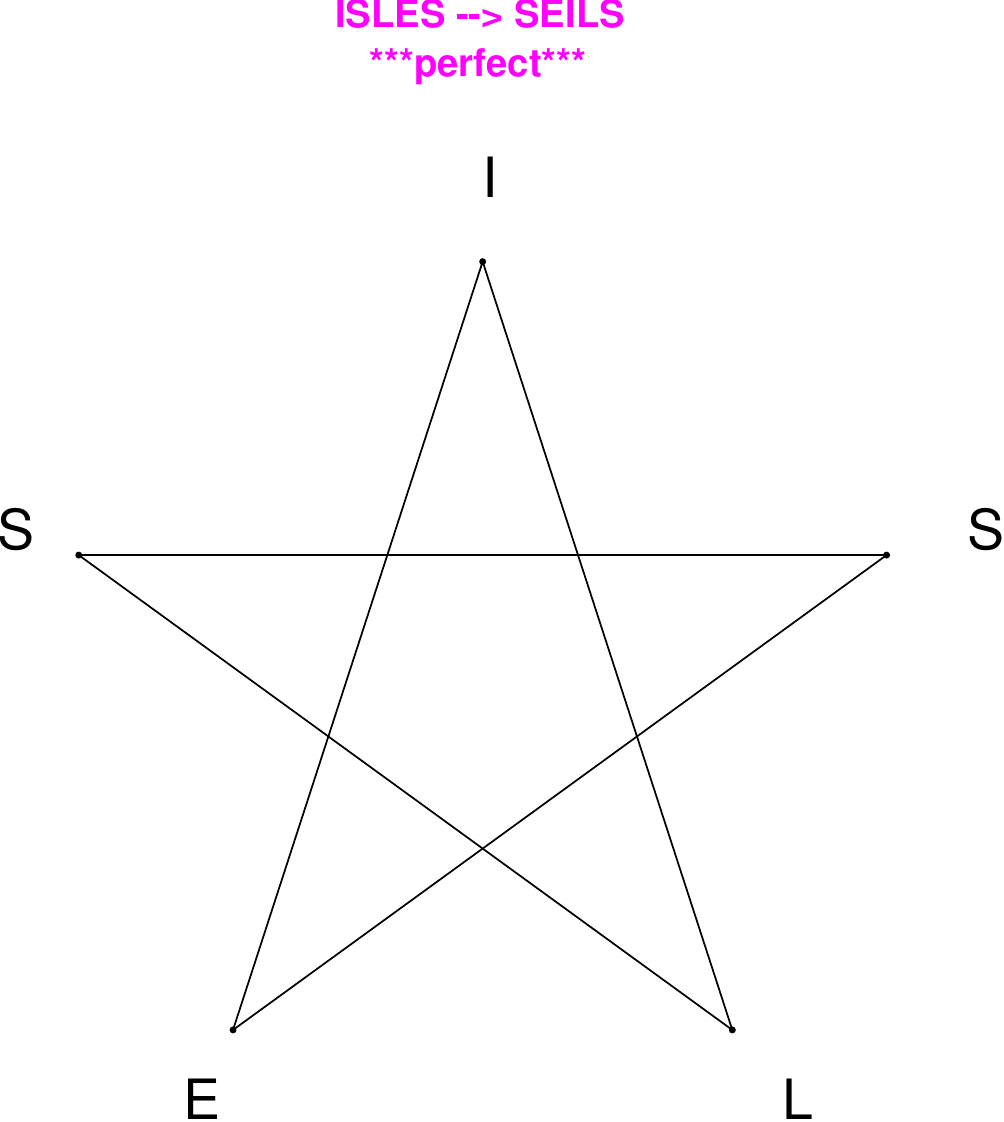}
\end{subfigure}
\hfill
\begin{subfigure}[T]{0.19\textwidth}
\centering
\includegraphics[width=\textwidth]{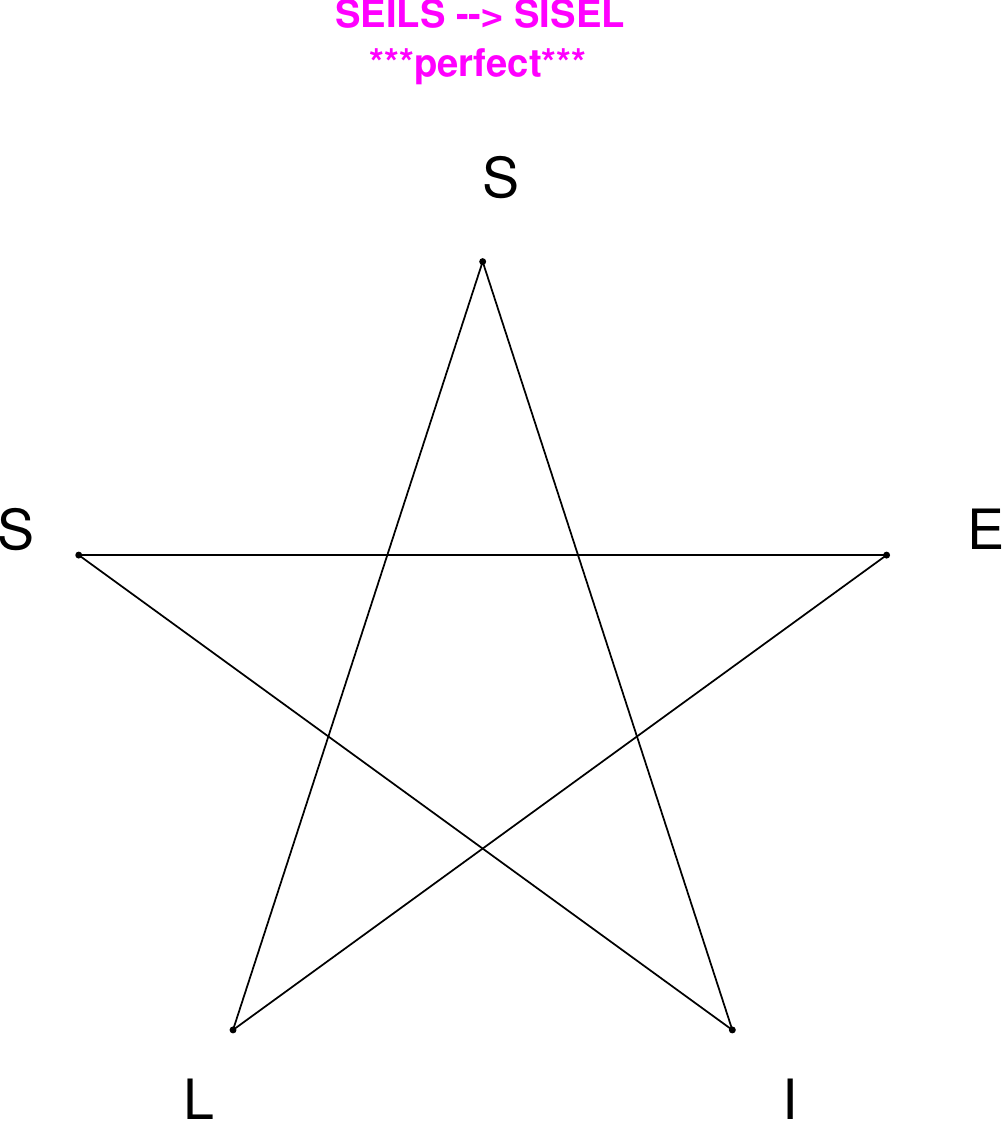}
\end{subfigure}
\hfill
\begin{subfigure}[T]{0.19\textwidth}
\centering
\includegraphics[width=\textwidth]{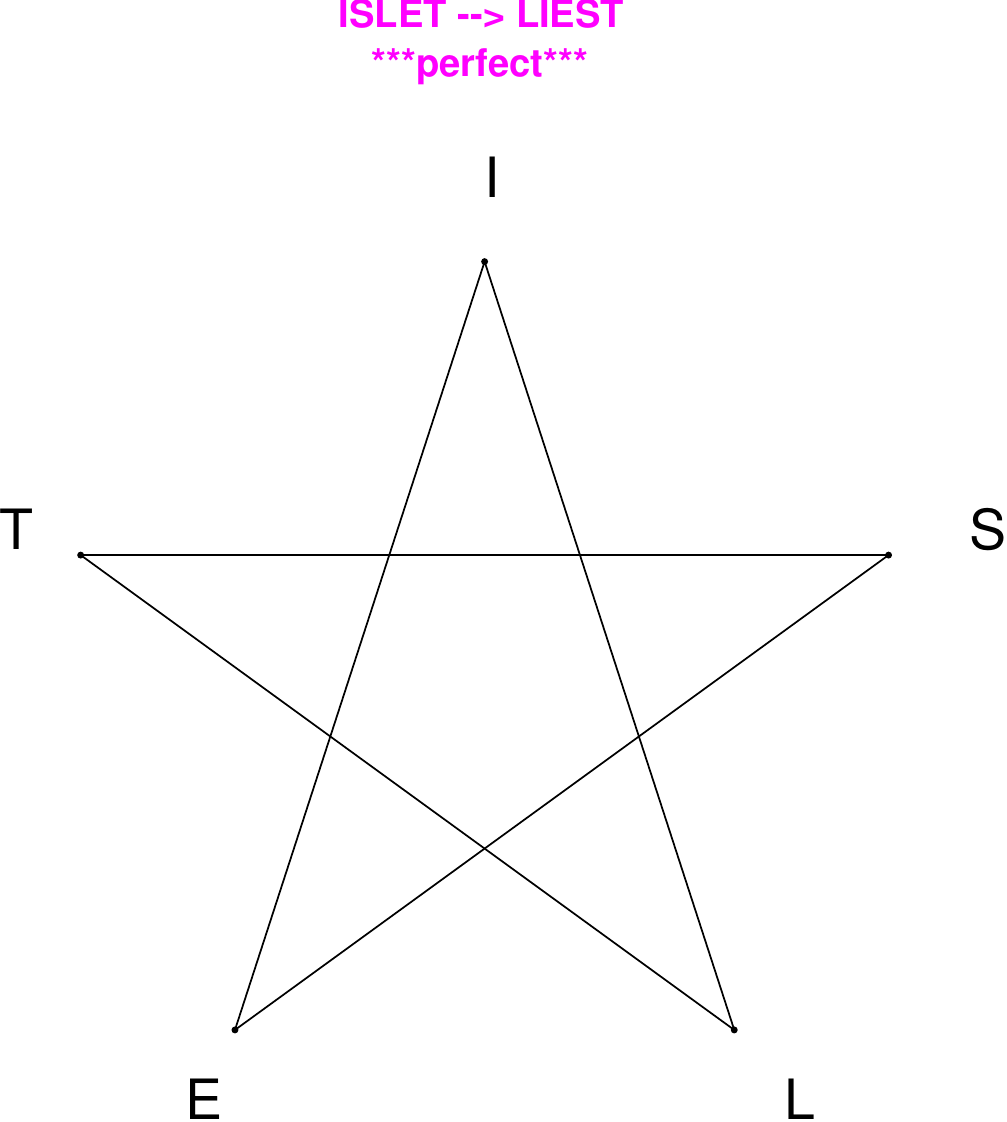}
\end{subfigure}
\hfill
\begin{subfigure}[T]{0.19\textwidth}
\centering
\includegraphics[width=\textwidth]{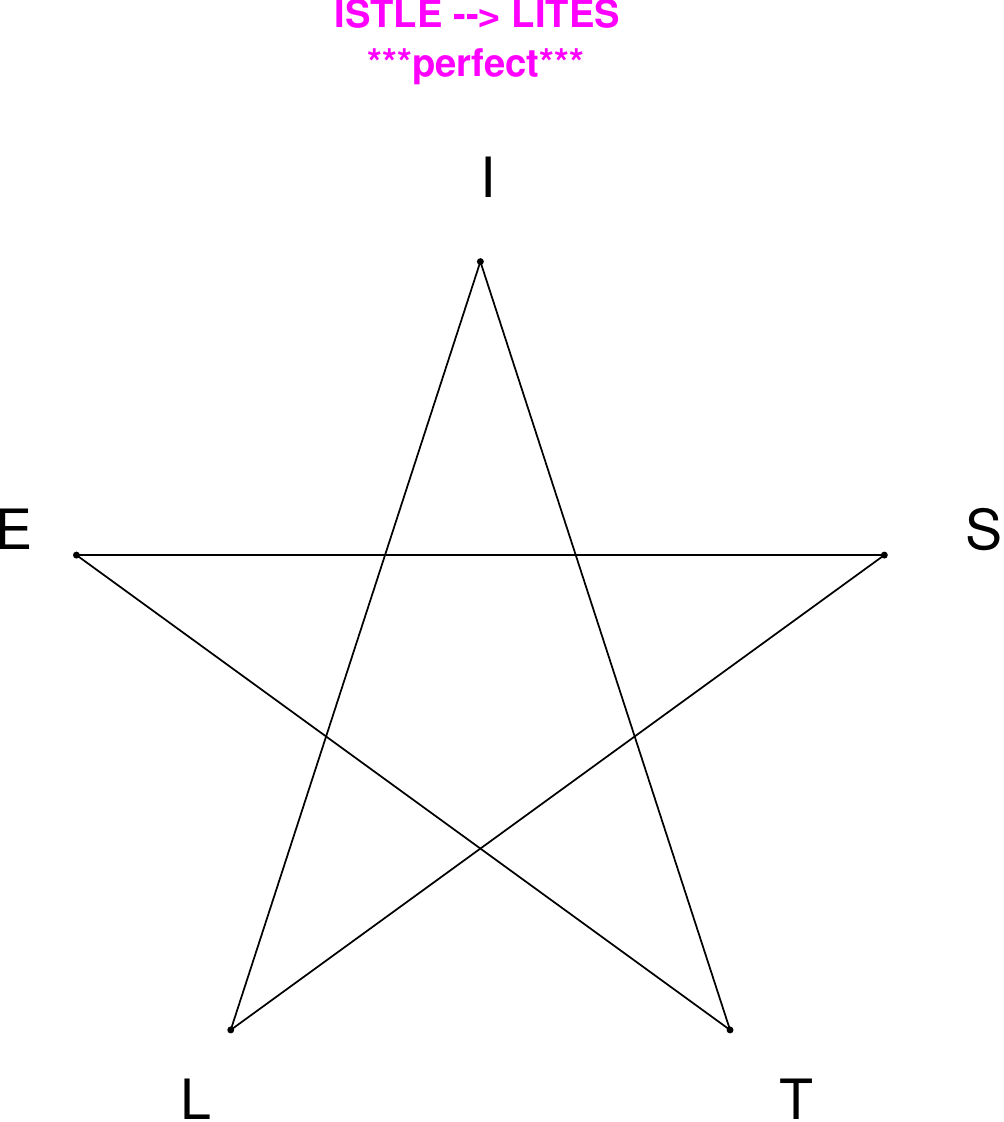}
\end{subfigure}
\end{figure}

\begin{figure}[H]
\centering
\begin{subfigure}[T]{0.19\textwidth}
\centering
\includegraphics[width=\textwidth]{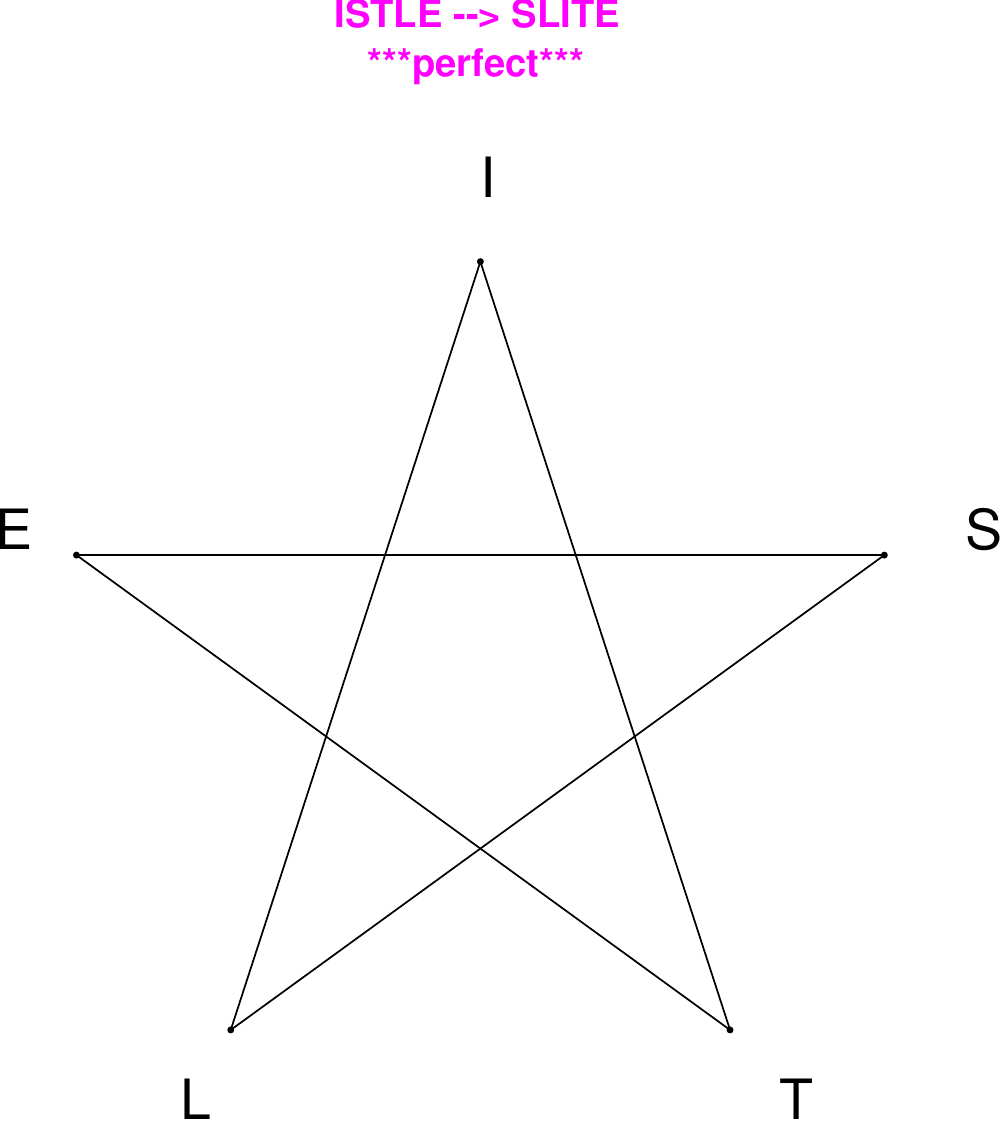}
\end{subfigure}
\hfill
\begin{subfigure}[T]{0.19\textwidth}
\centering
\includegraphics[width=\textwidth]{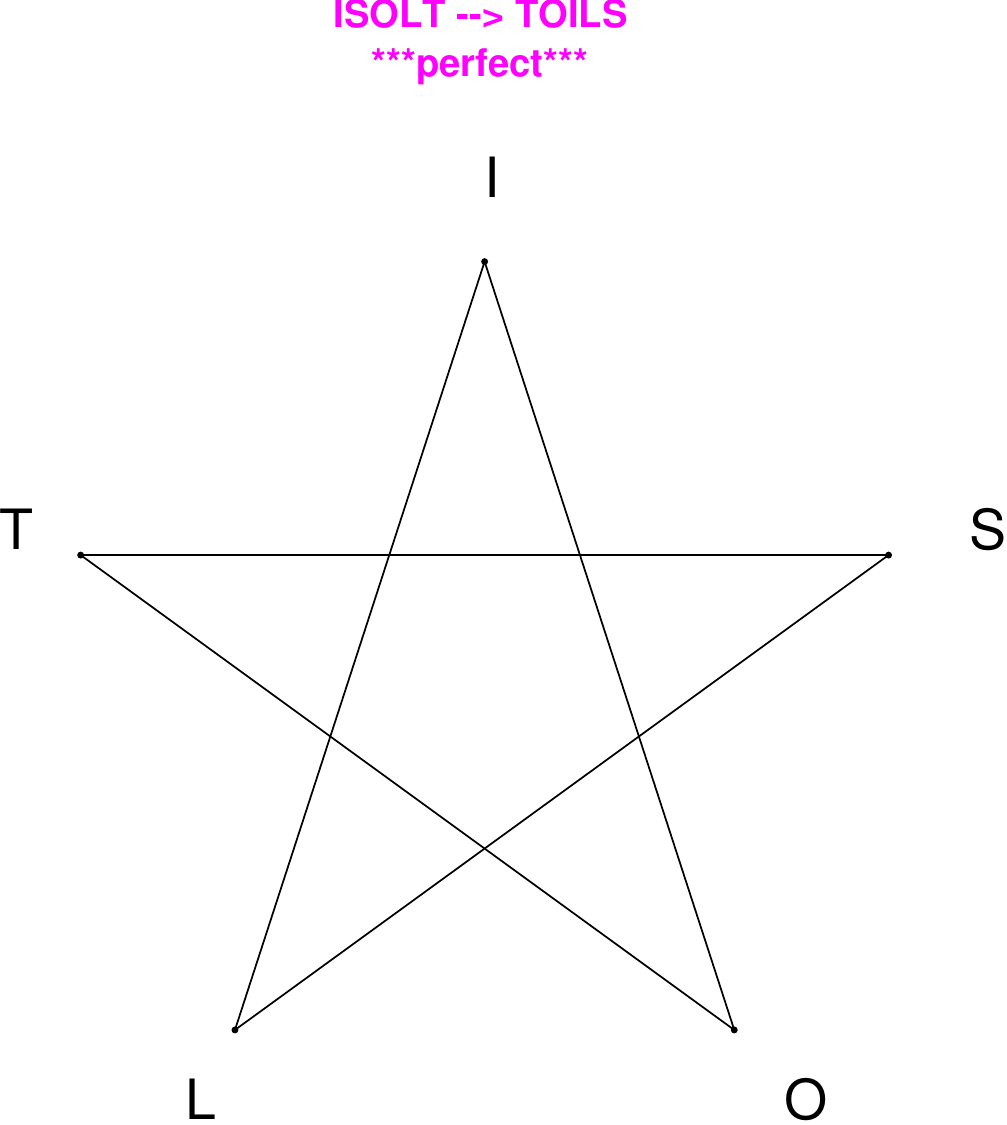}
\end{subfigure}
\hfill
\begin{subfigure}[T]{0.19\textwidth}
\centering
\includegraphics[width=\textwidth]{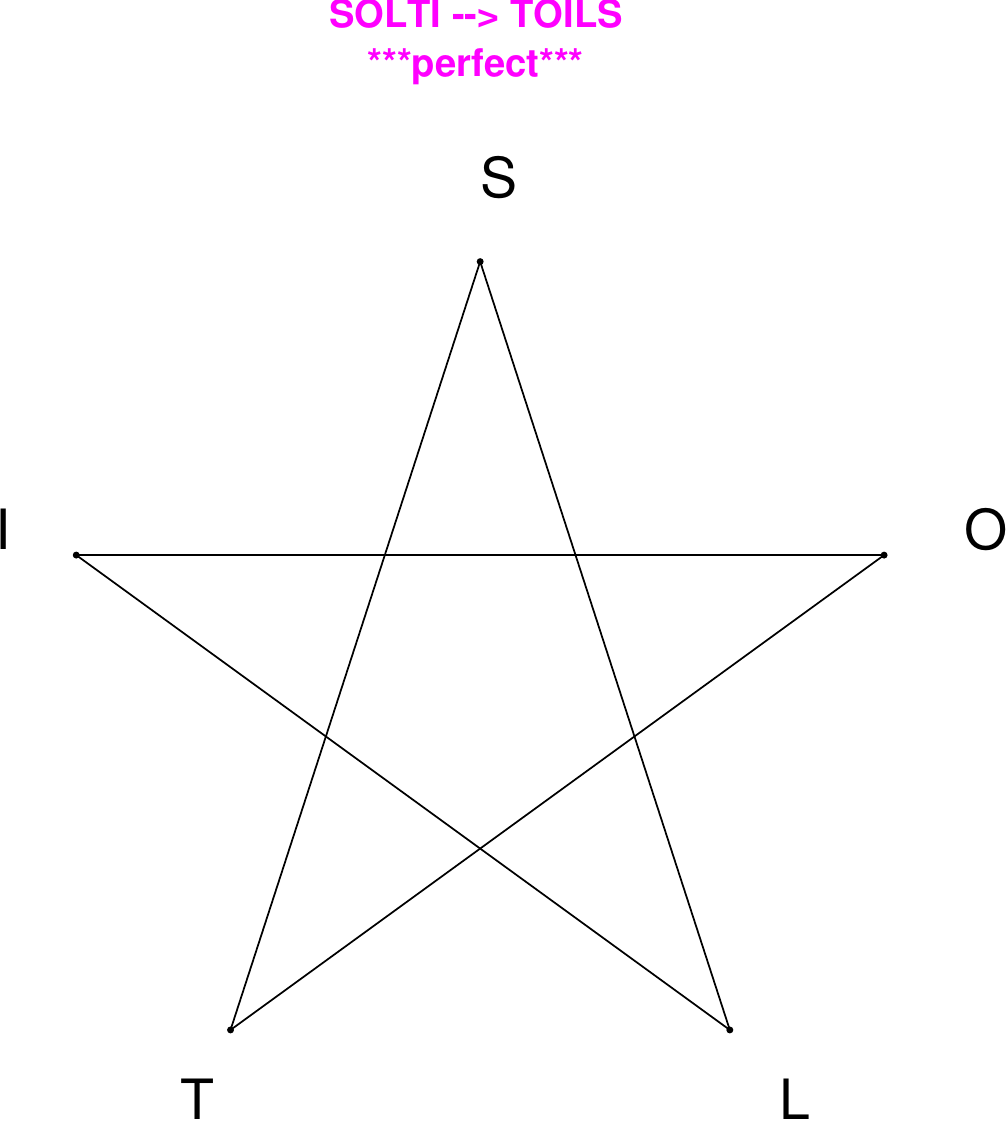}
\end{subfigure}
\hfill
\begin{subfigure}[T]{0.19\textwidth}
\centering
\includegraphics[width=\textwidth]{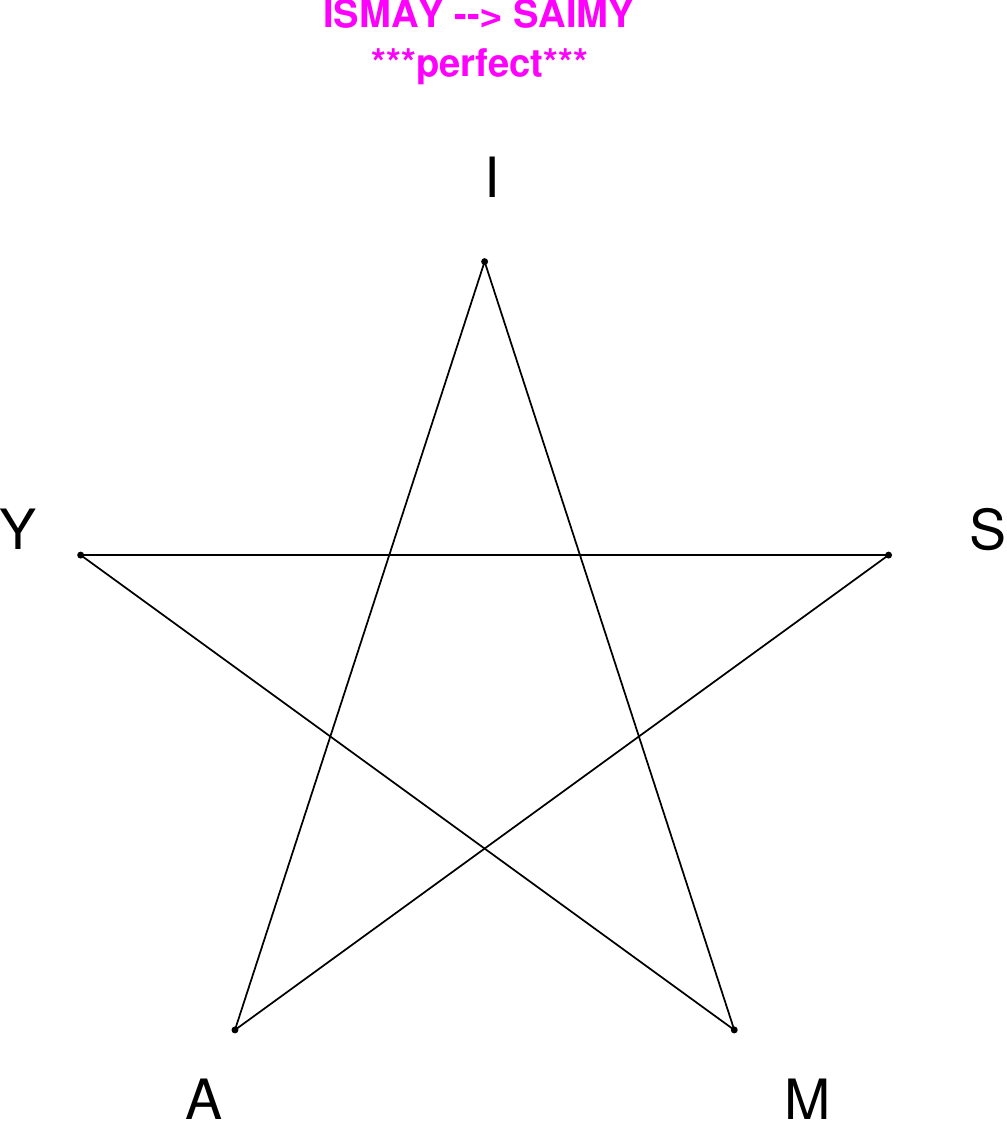}
\end{subfigure}
\hfill
\begin{subfigure}[T]{0.19\textwidth}
\centering
\includegraphics[width=\textwidth]{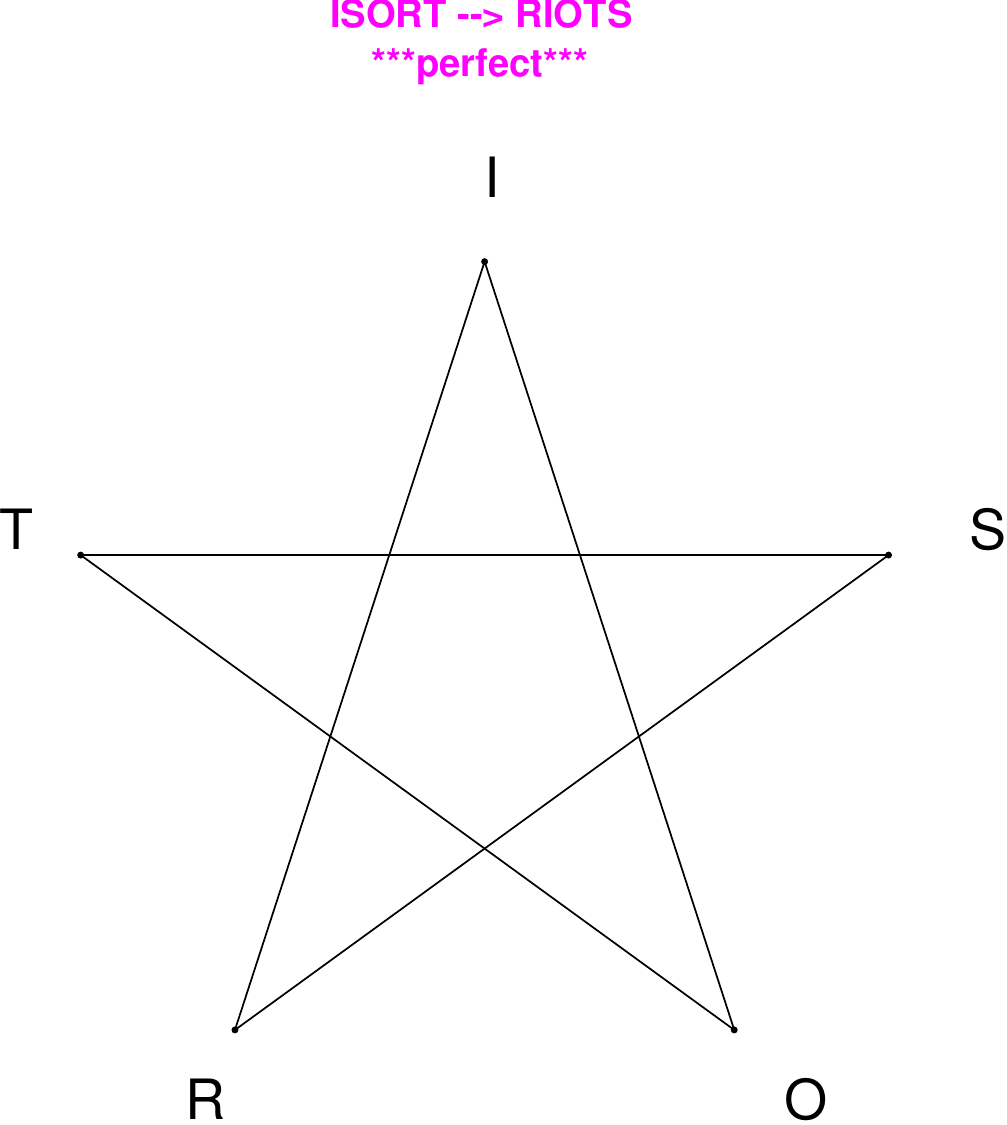}
\end{subfigure}
\end{figure}

\begin{figure}[H]
\centering
\begin{subfigure}[T]{0.19\textwidth}
\centering
\includegraphics[width=\textwidth]{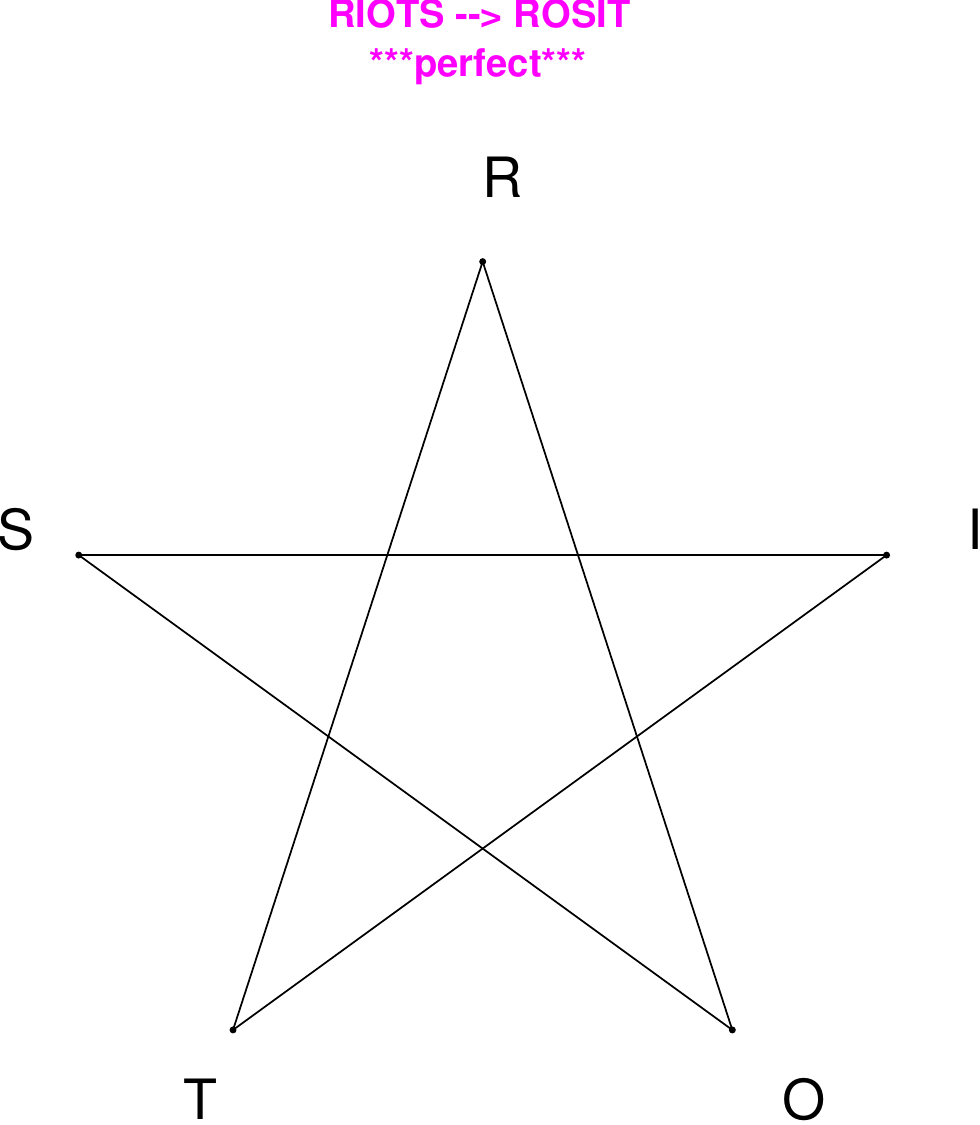}
\end{subfigure}
\hfill
\begin{subfigure}[T]{0.19\textwidth}
\centering
\includegraphics[width=\textwidth]{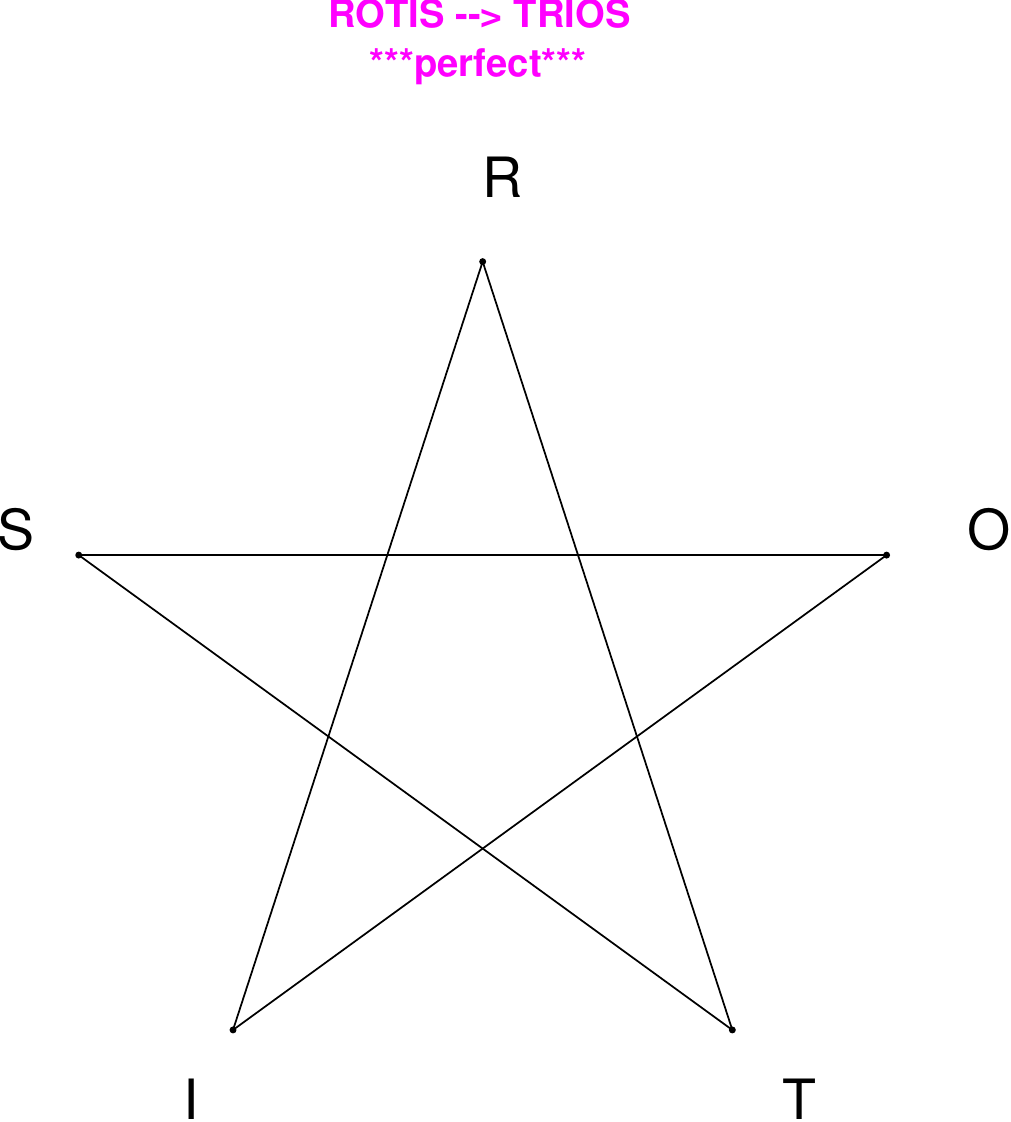}
\end{subfigure}
\hfill
\begin{subfigure}[T]{0.19\textwidth}
\centering
\includegraphics[width=\textwidth]{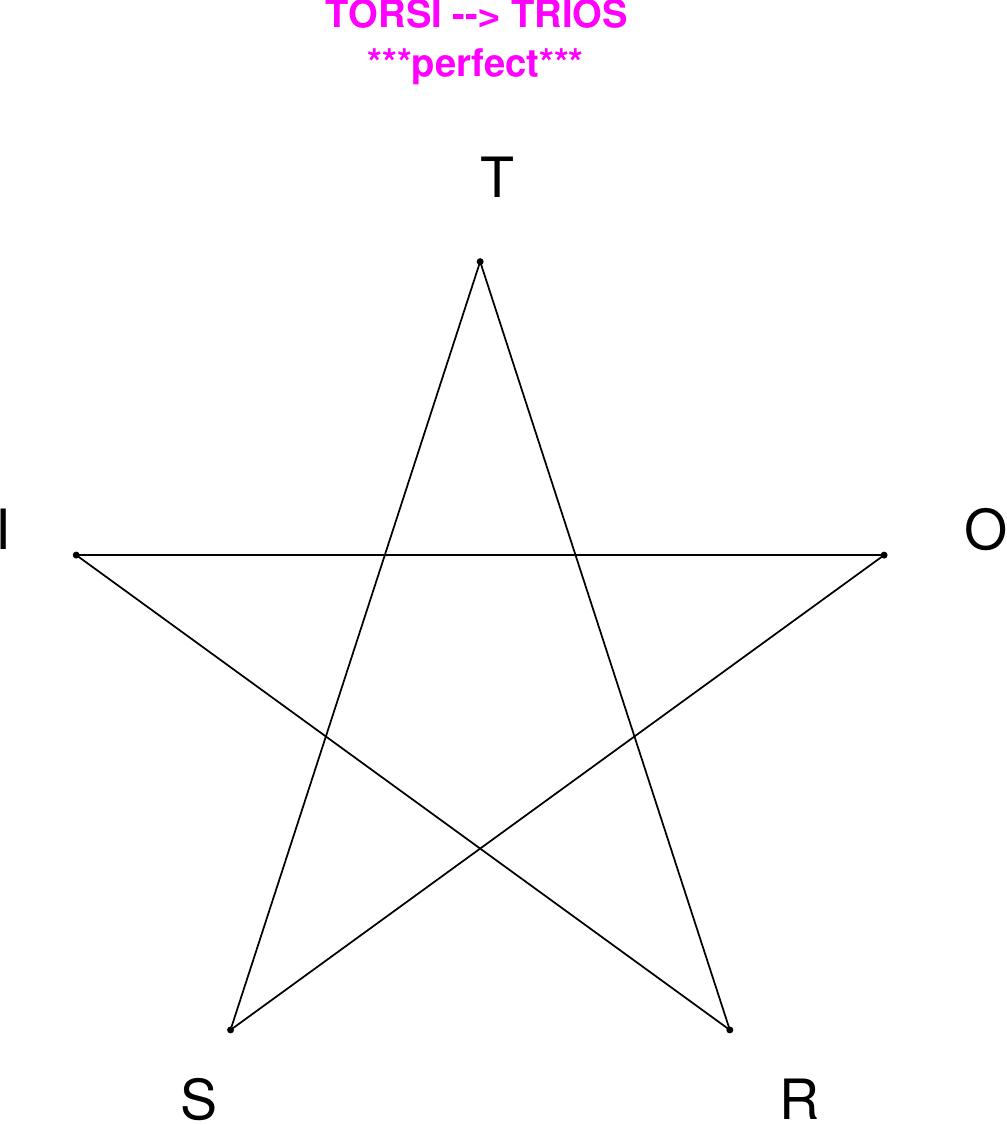}
\end{subfigure}
\hfill
\begin{subfigure}[T]{0.19\textwidth}
\centering
\includegraphics[width=\textwidth]{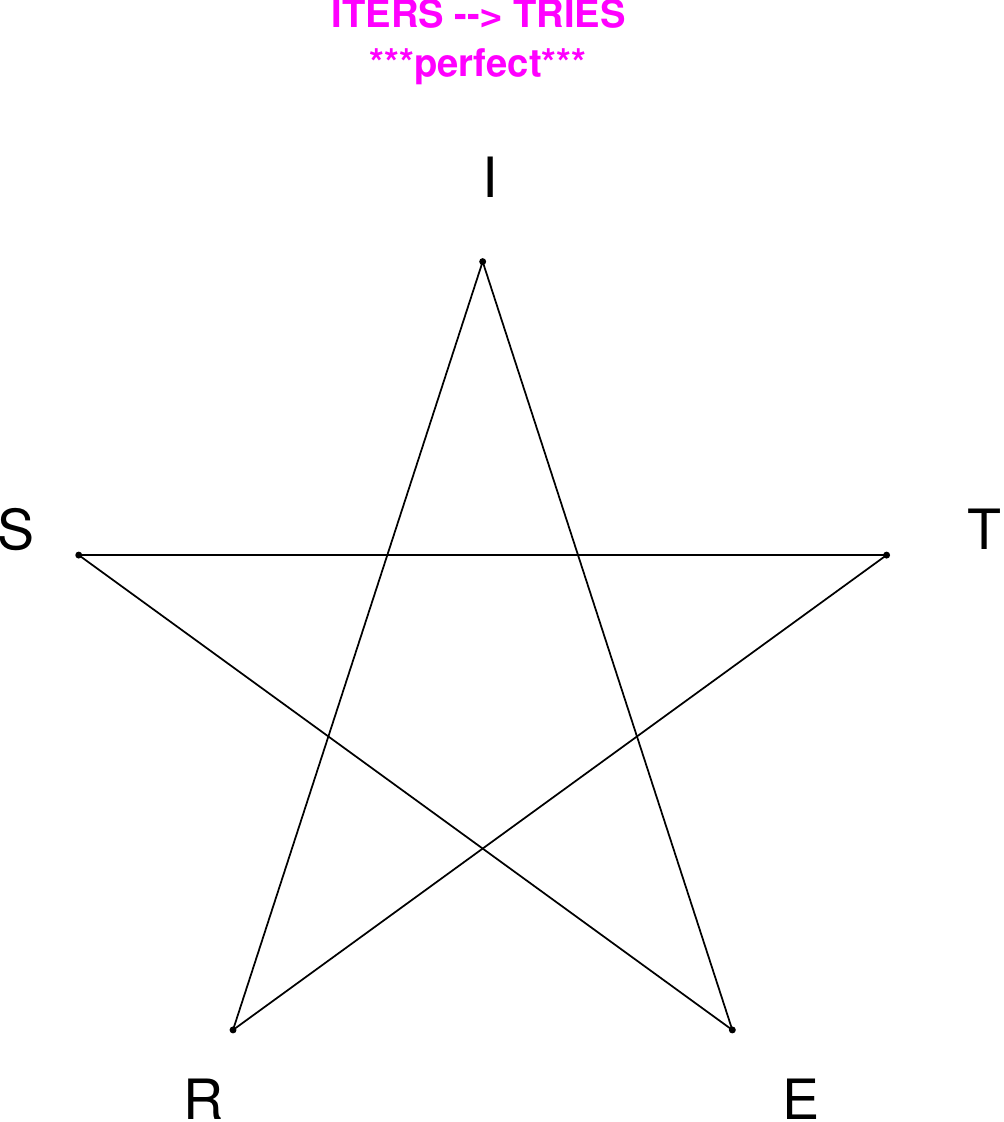}
\end{subfigure}
\hfill
\begin{subfigure}[T]{0.19\textwidth}
\centering
\includegraphics[width=\textwidth]{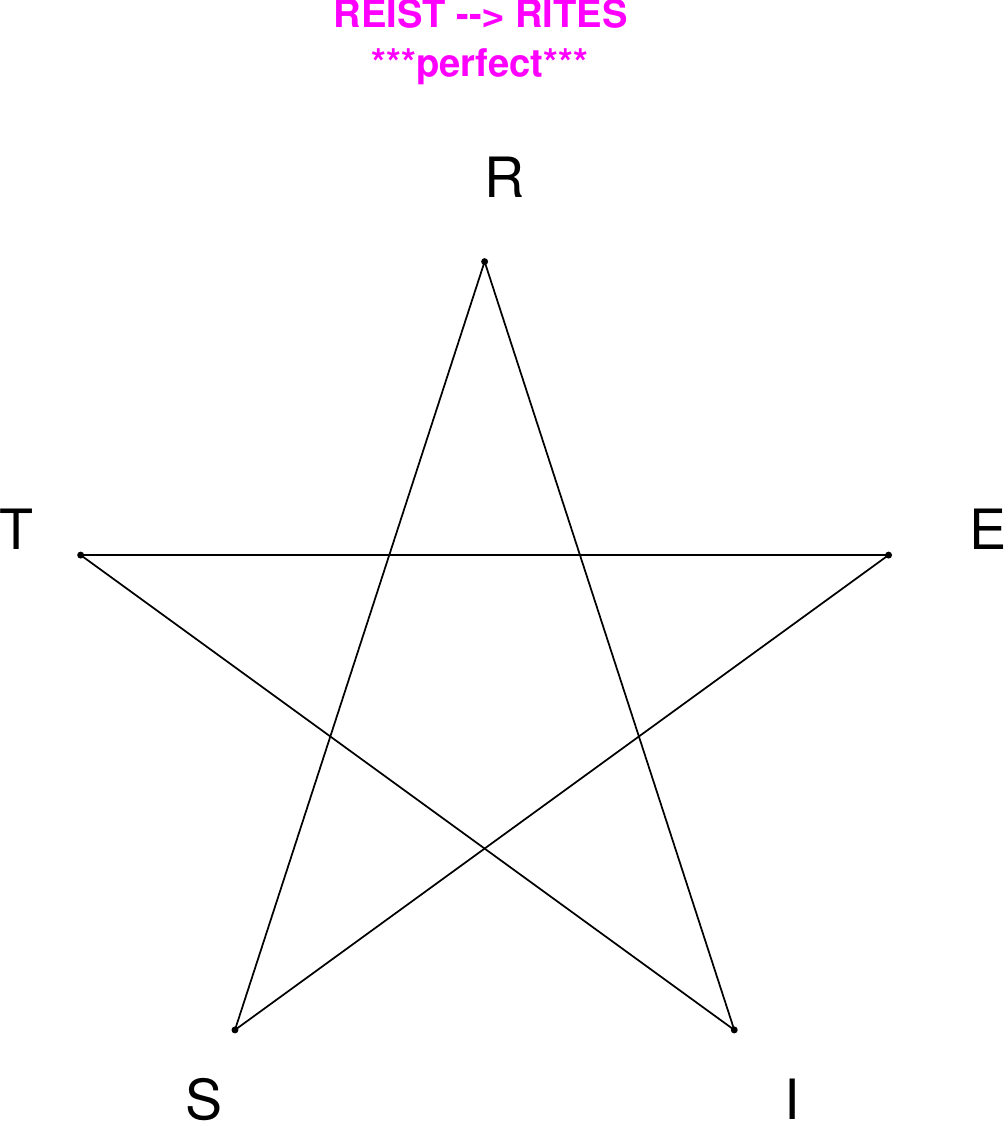}
\end{subfigure}
\end{figure}

\begin{figure}[H]
\centering
\begin{subfigure}[T]{0.19\textwidth}
\centering
\includegraphics[width=\textwidth]{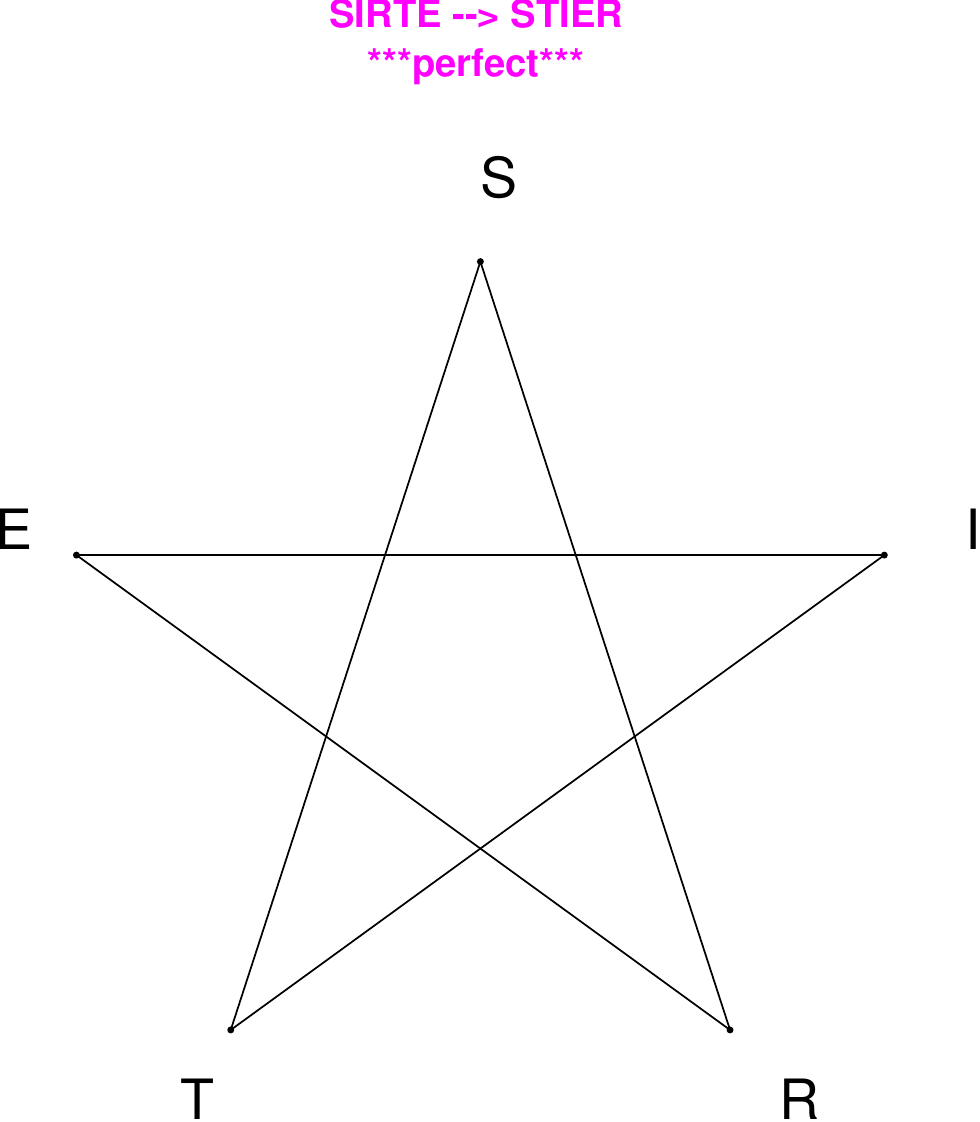}
\end{subfigure}
\hfill
\begin{subfigure}[T]{0.19\textwidth}
\centering
\includegraphics[width=\textwidth]{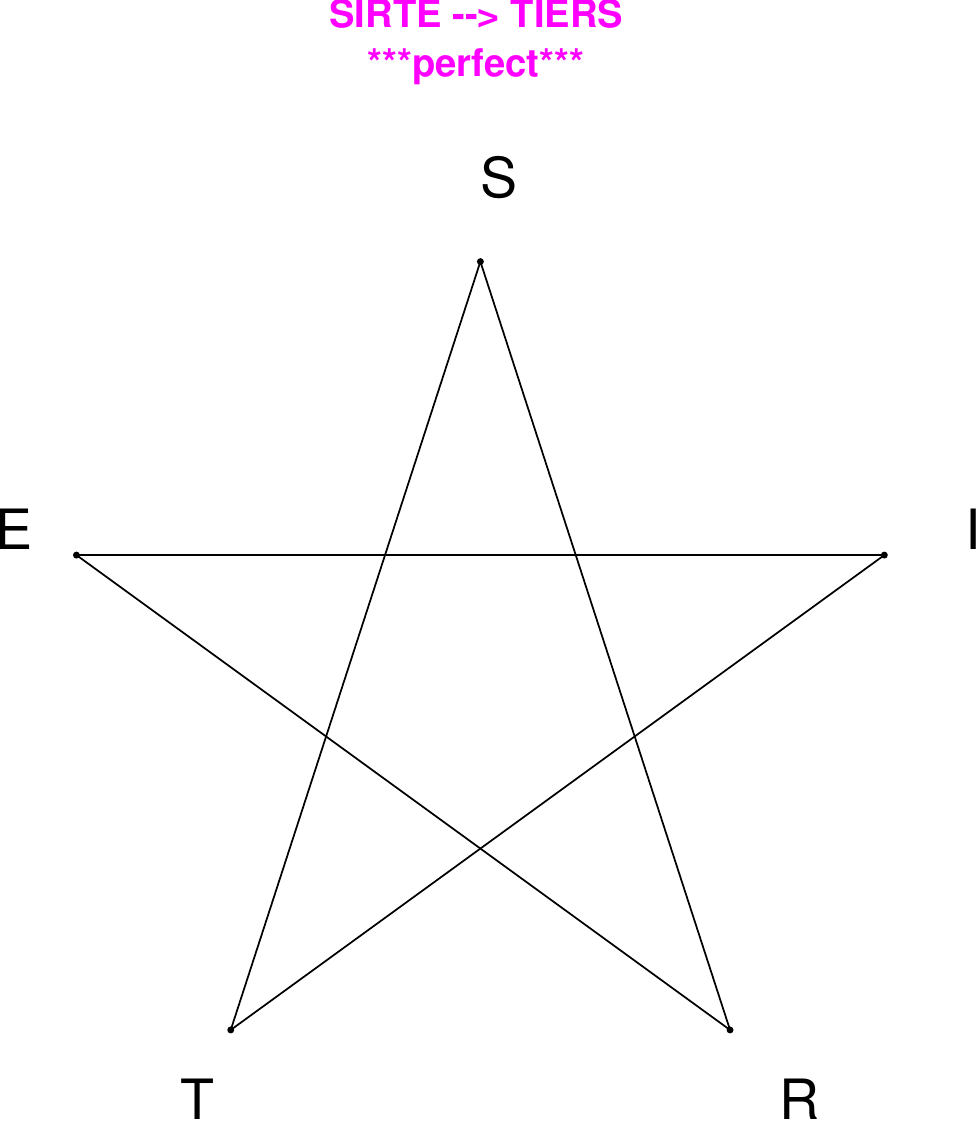}
\end{subfigure}
\hfill
\begin{subfigure}[T]{0.19\textwidth}
\centering
\includegraphics[width=\textwidth]{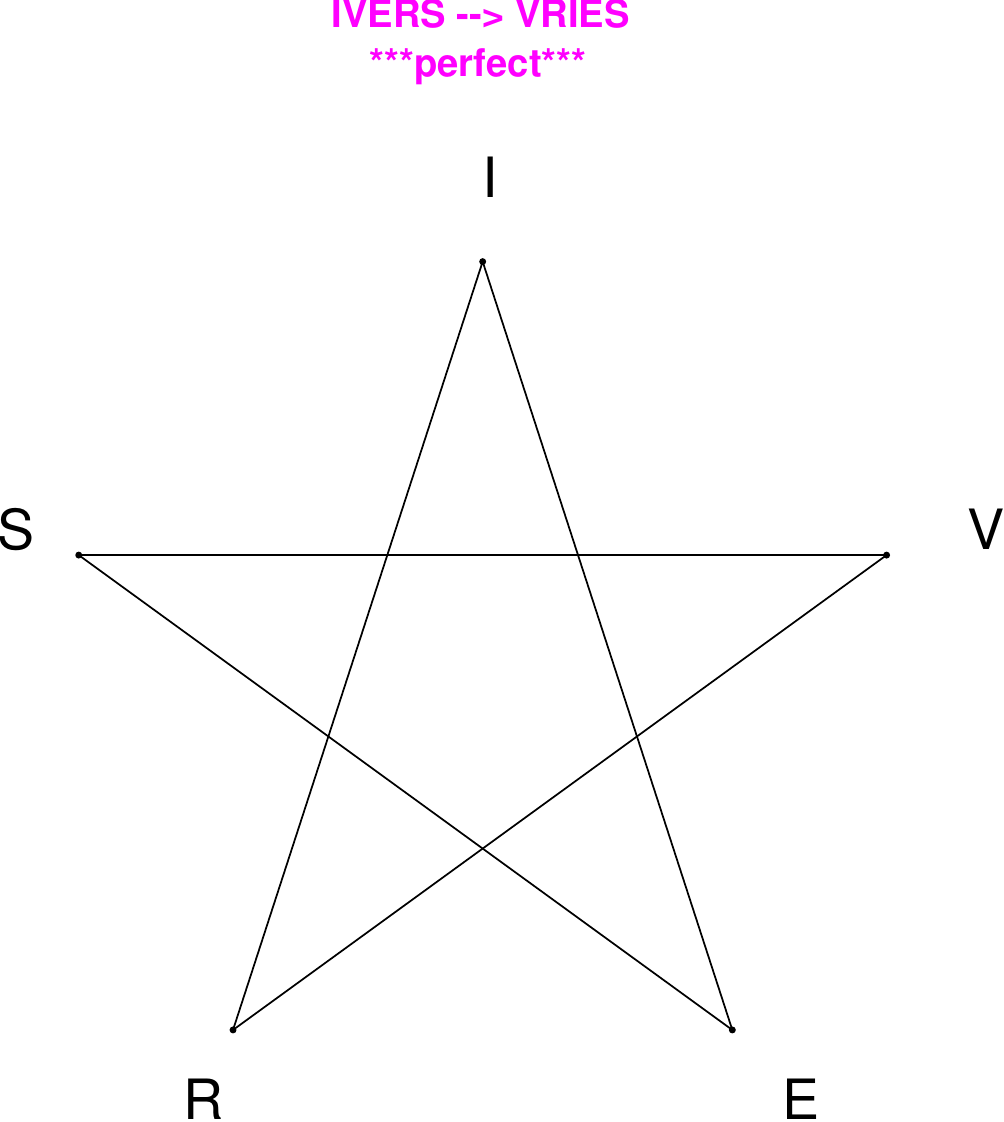}
\end{subfigure}
\hfill
\begin{subfigure}[T]{0.19\textwidth}
\centering
\includegraphics[width=\textwidth]{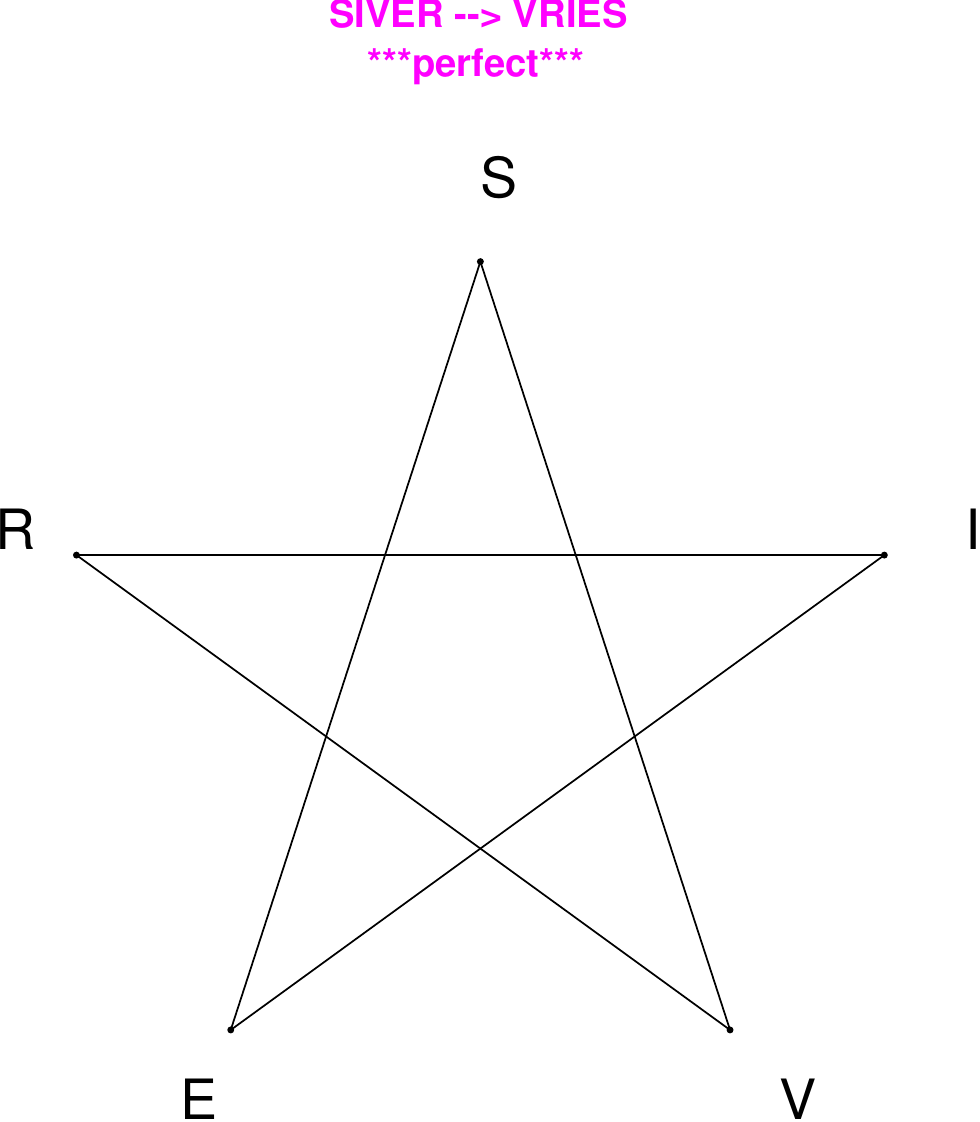}
\end{subfigure}
\hfill
\begin{subfigure}[T]{0.19\textwidth}
\centering
\includegraphics[width=\textwidth]{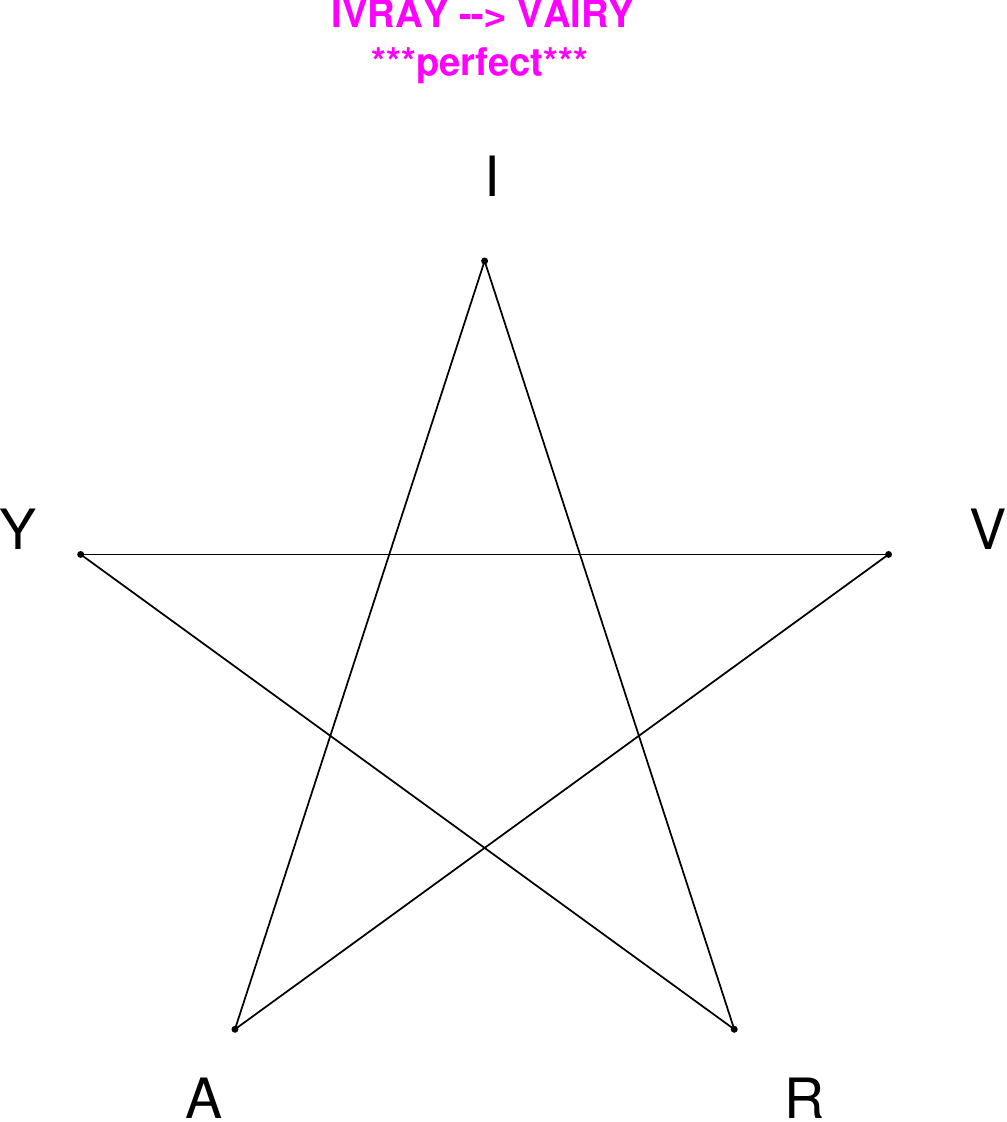}
\end{subfigure}
\end{figure}

\begin{figure}[H]
\centering
\begin{subfigure}[T]{0.19\textwidth}
\centering
\includegraphics[width=\textwidth]{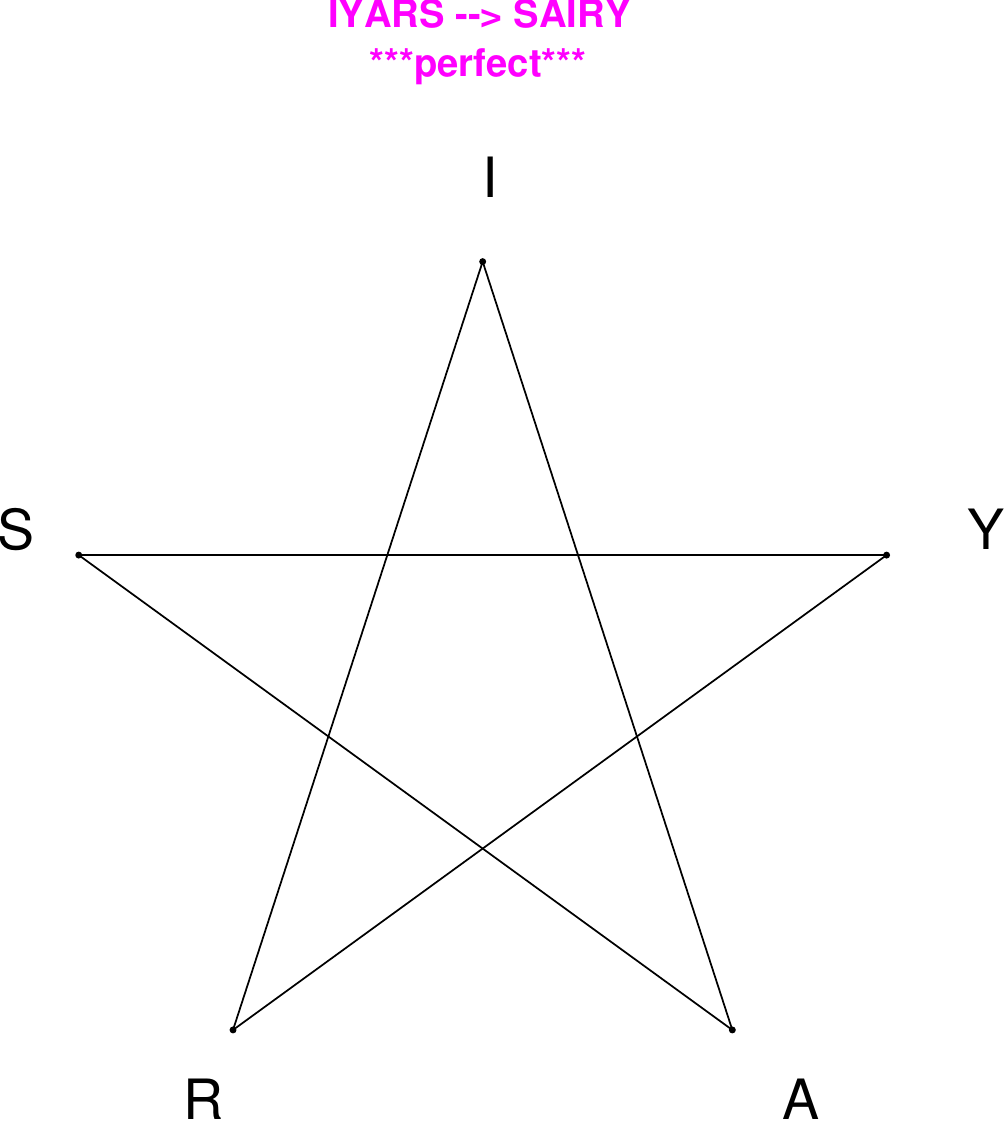}
\end{subfigure}
\hfill
\begin{subfigure}[T]{0.19\textwidth}
\centering
\includegraphics[width=\textwidth]{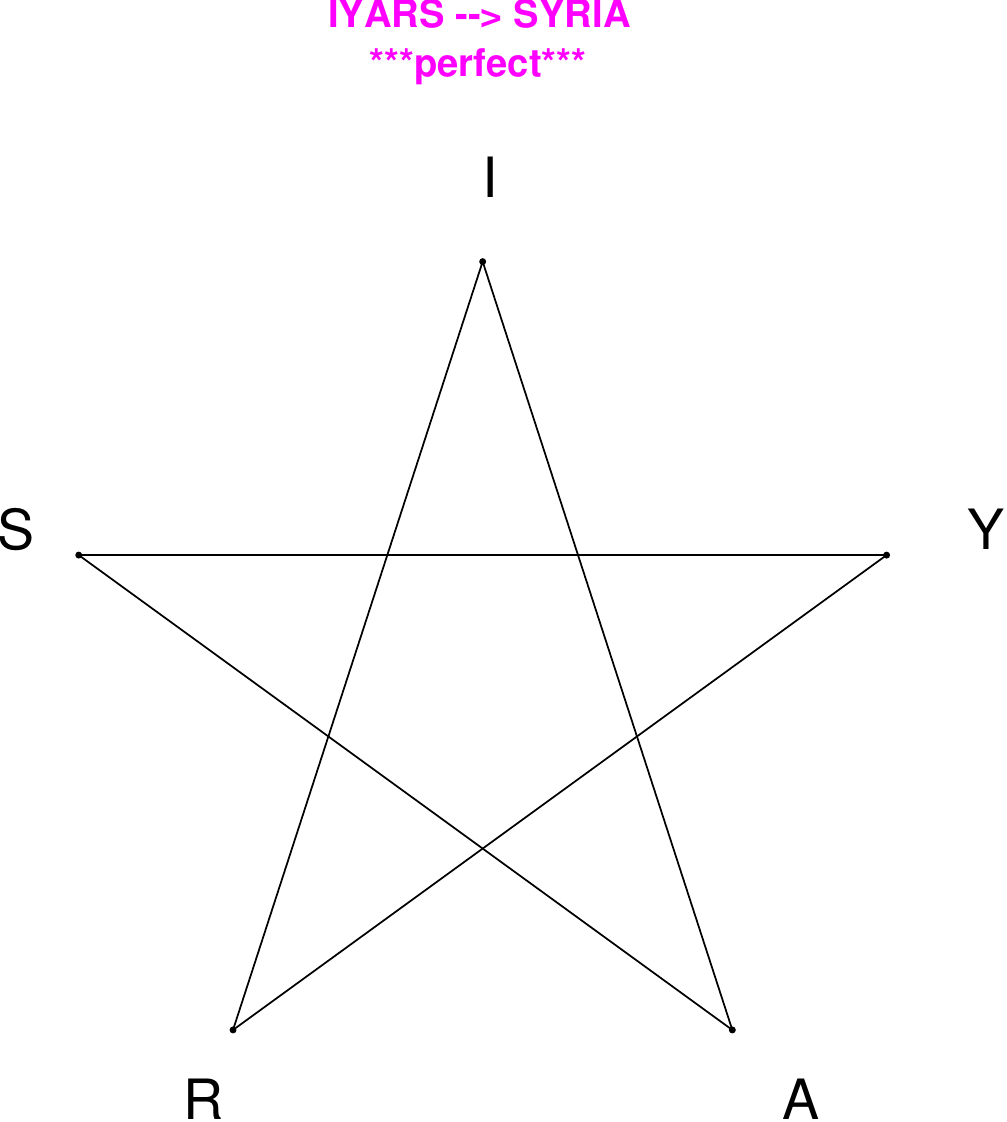}
\end{subfigure}
\hfill
\begin{subfigure}[T]{0.19\textwidth}
\centering
\includegraphics[width=\textwidth]{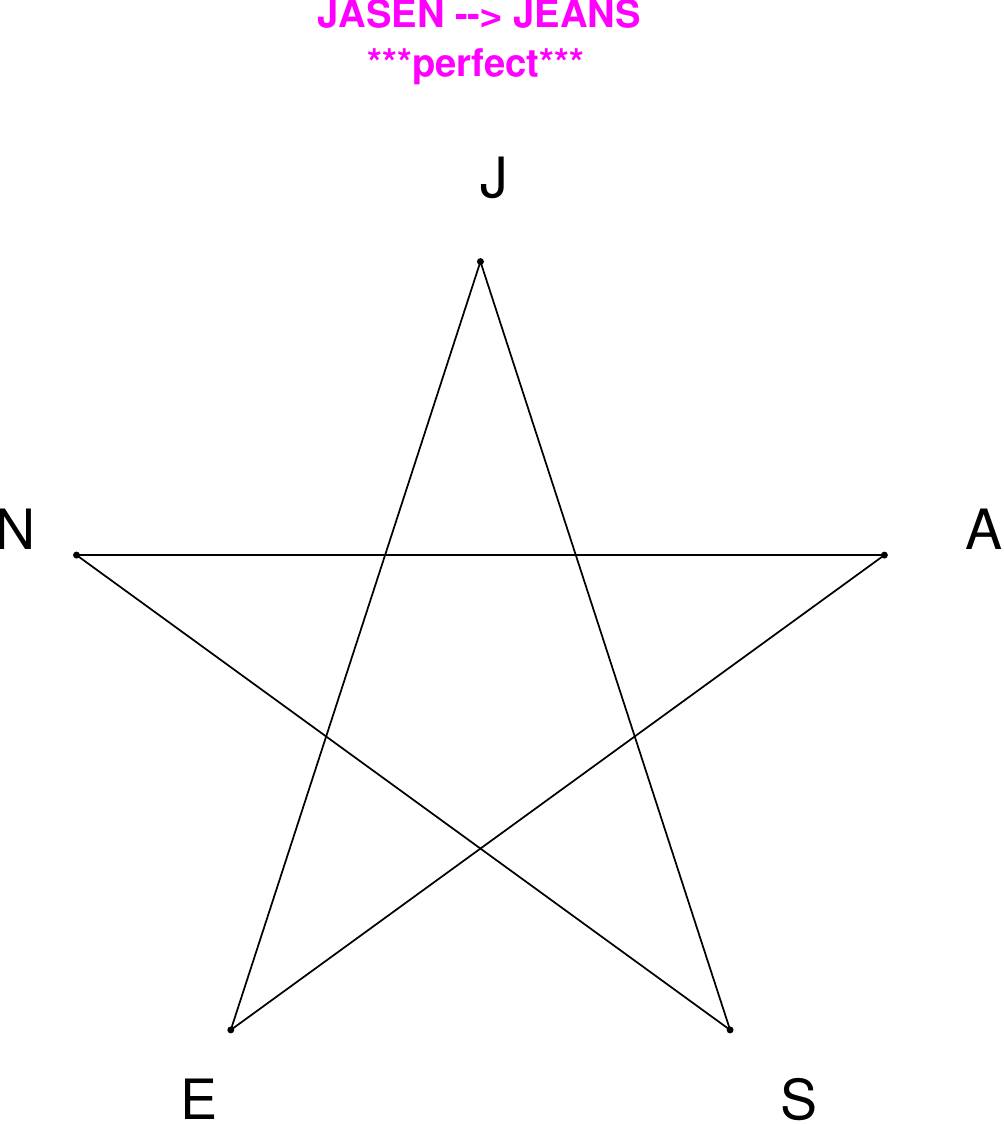}
\end{subfigure}
\hfill
\begin{subfigure}[T]{0.19\textwidth}
\centering
\includegraphics[width=\textwidth]{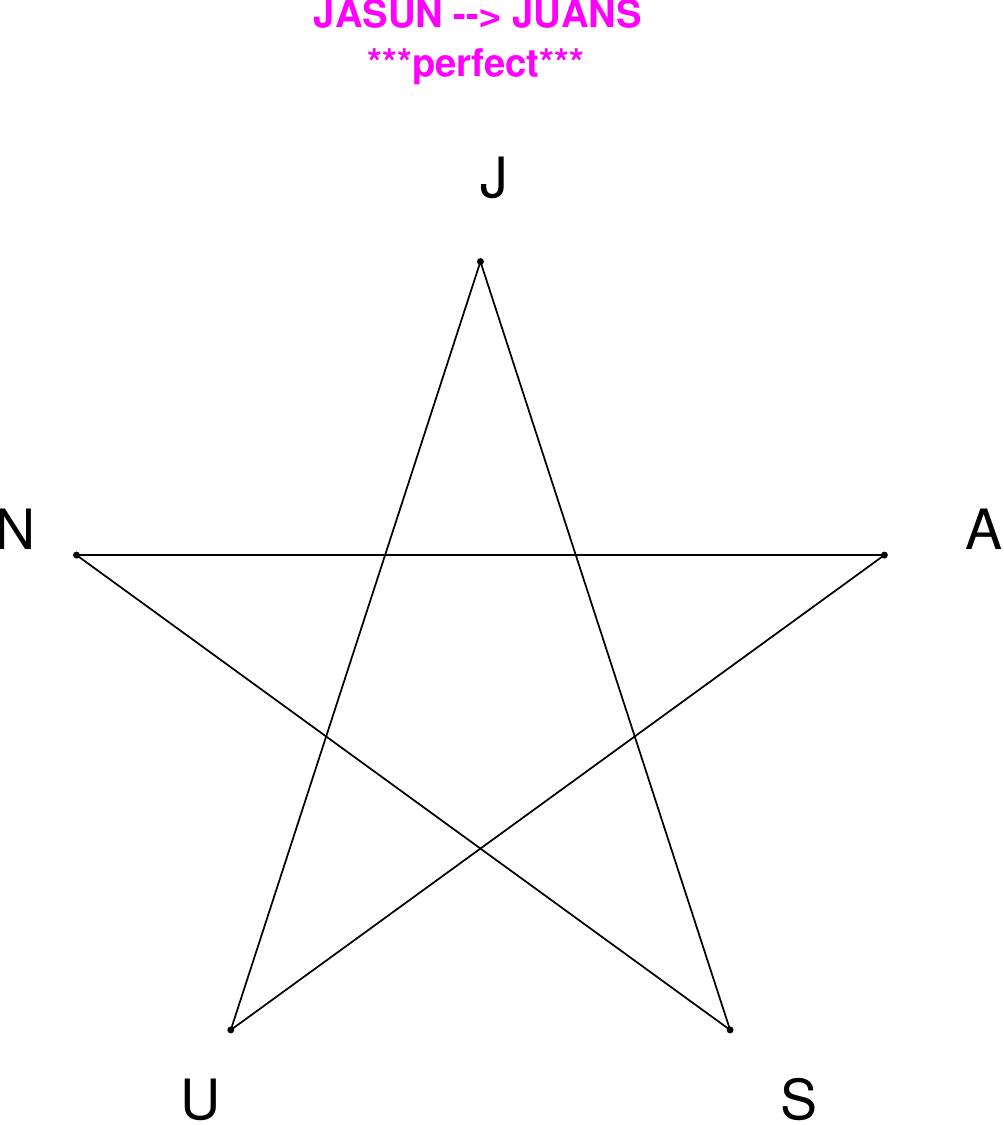}
\end{subfigure}
\hfill
\begin{subfigure}[T]{0.19\textwidth}
\centering
\includegraphics[width=\textwidth]{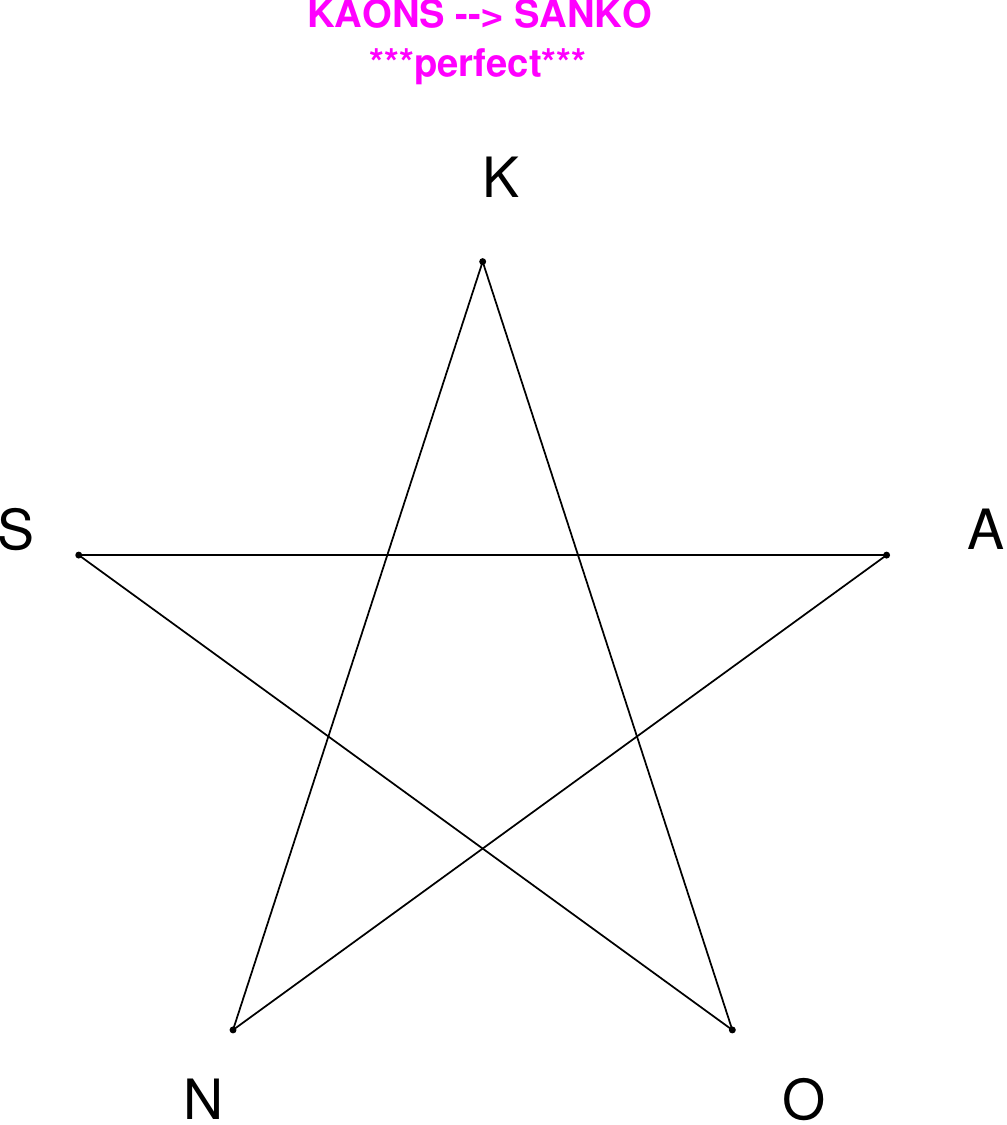}
\end{subfigure}
\end{figure}

\begin{figure}[H]
\centering
\begin{subfigure}[T]{0.19\textwidth}
\centering
\includegraphics[width=\textwidth]{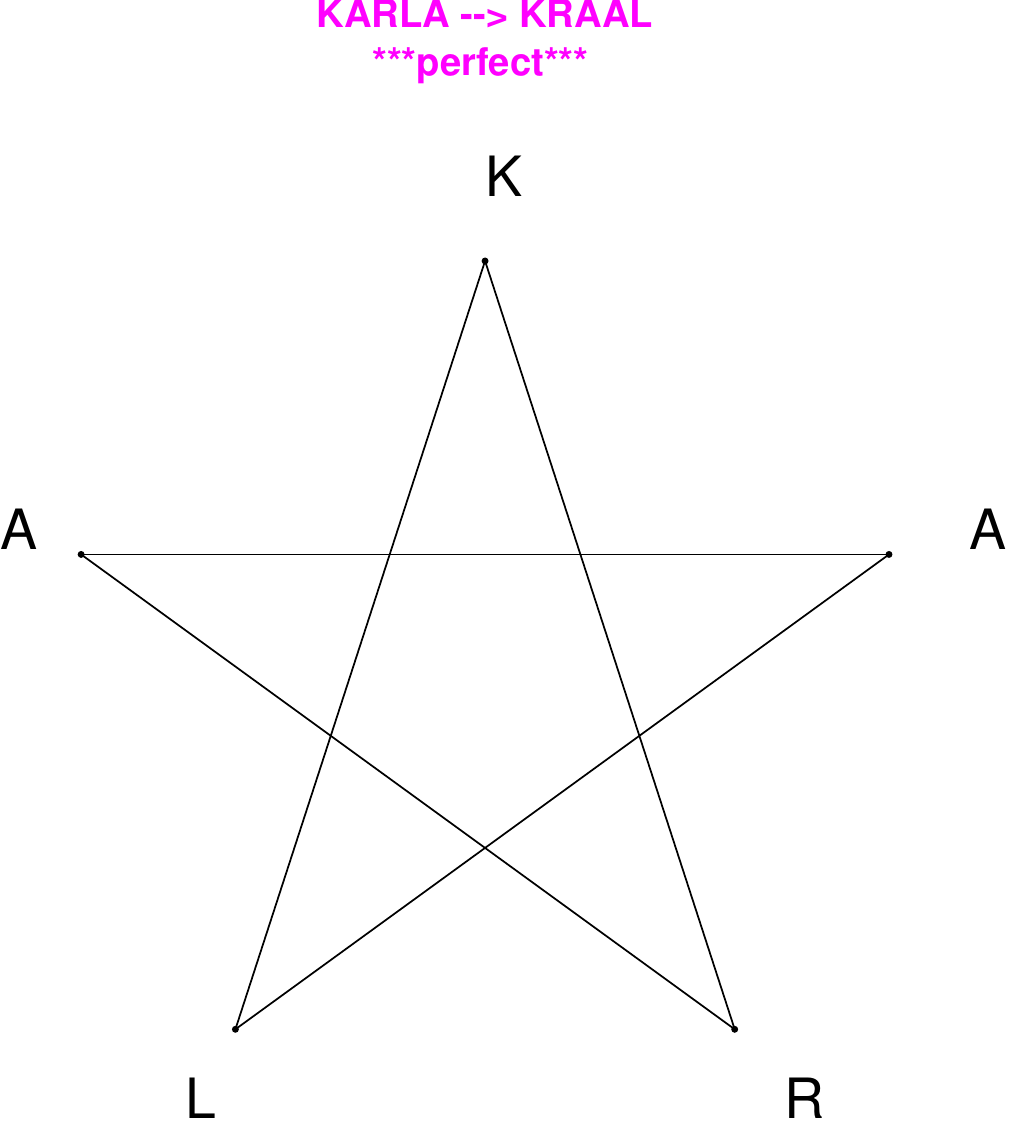}
\end{subfigure}
\hfill
\begin{subfigure}[T]{0.19\textwidth}
\centering
\includegraphics[width=\textwidth]{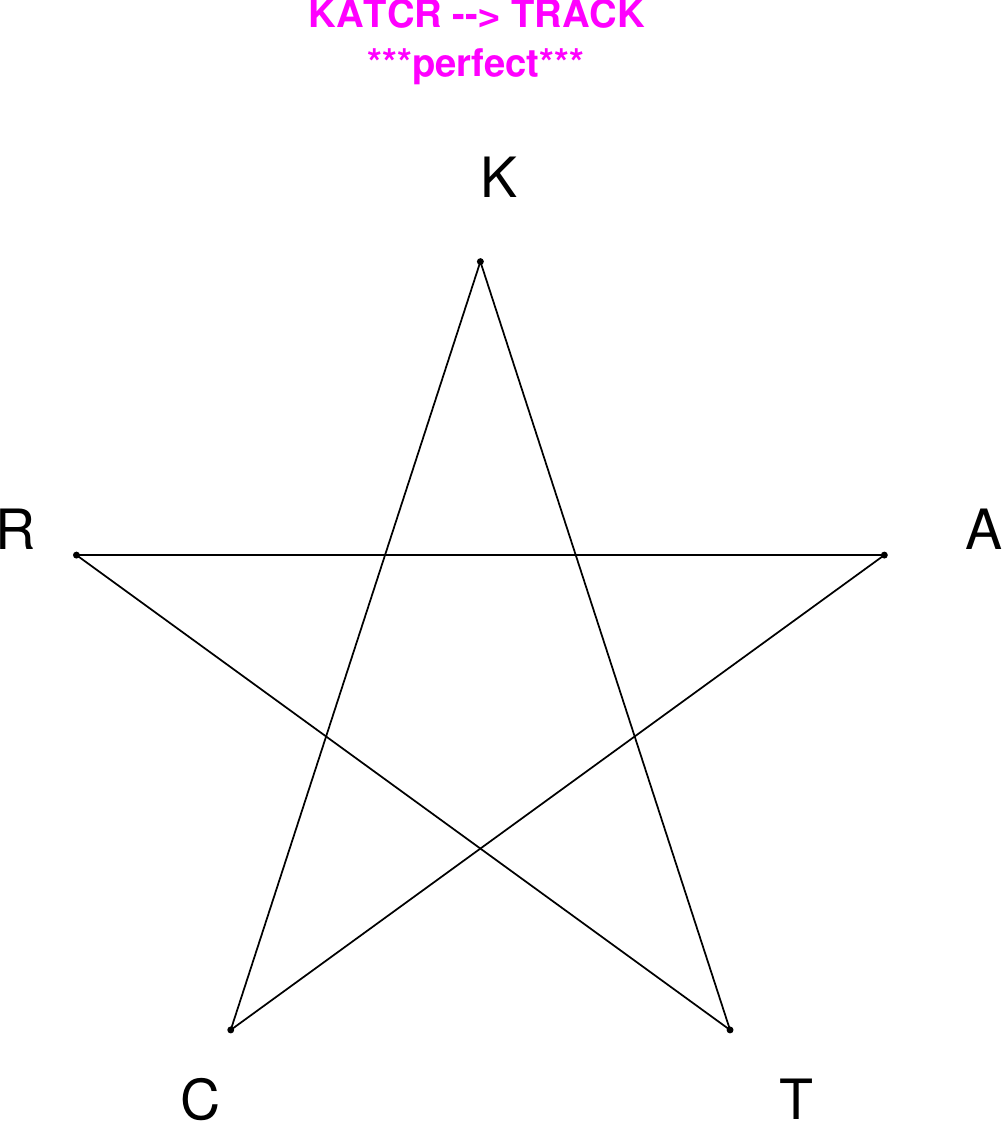}
\end{subfigure}
\hfill
\begin{subfigure}[T]{0.19\textwidth}
\centering
\includegraphics[width=\textwidth]{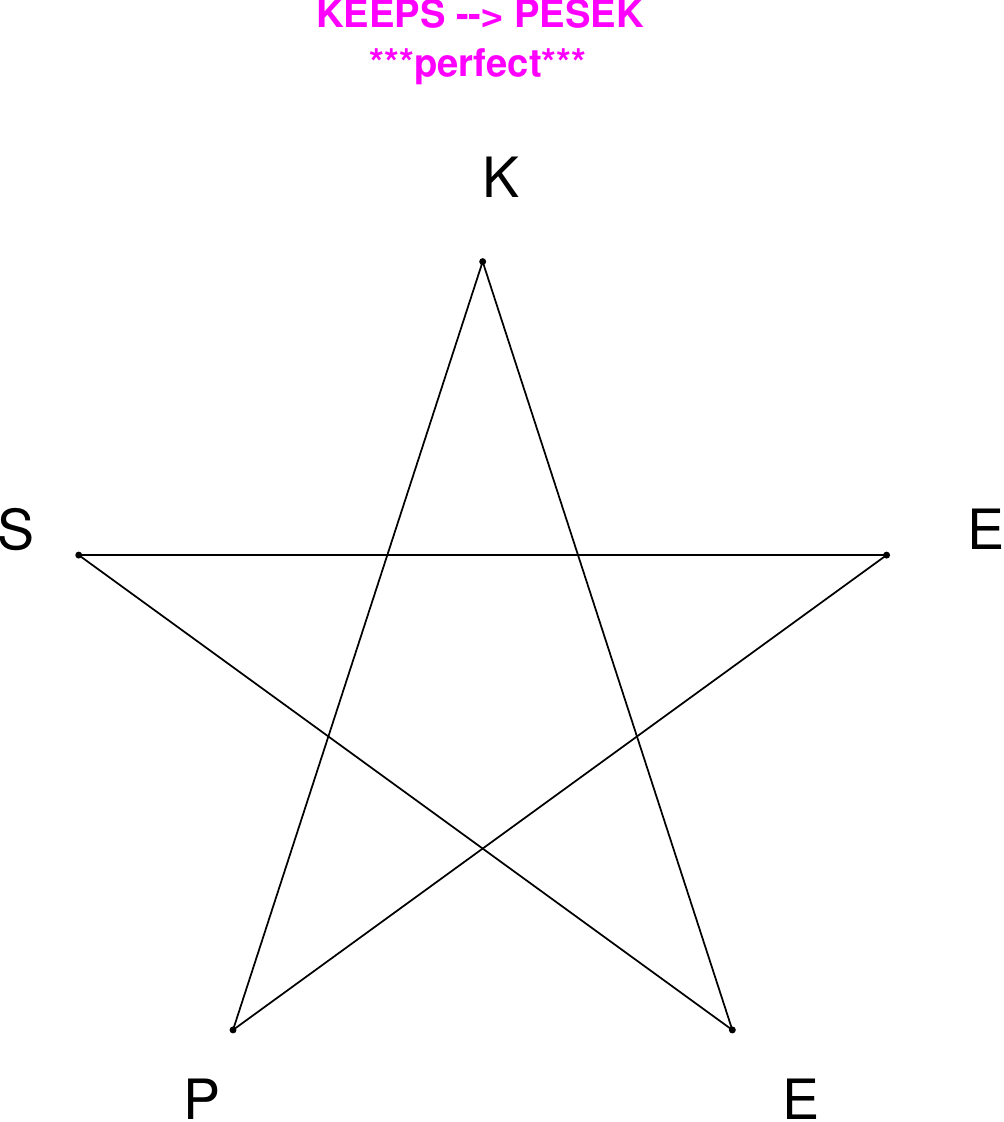}
\end{subfigure}
\hfill
\begin{subfigure}[T]{0.19\textwidth}
\centering
\includegraphics[width=\textwidth]{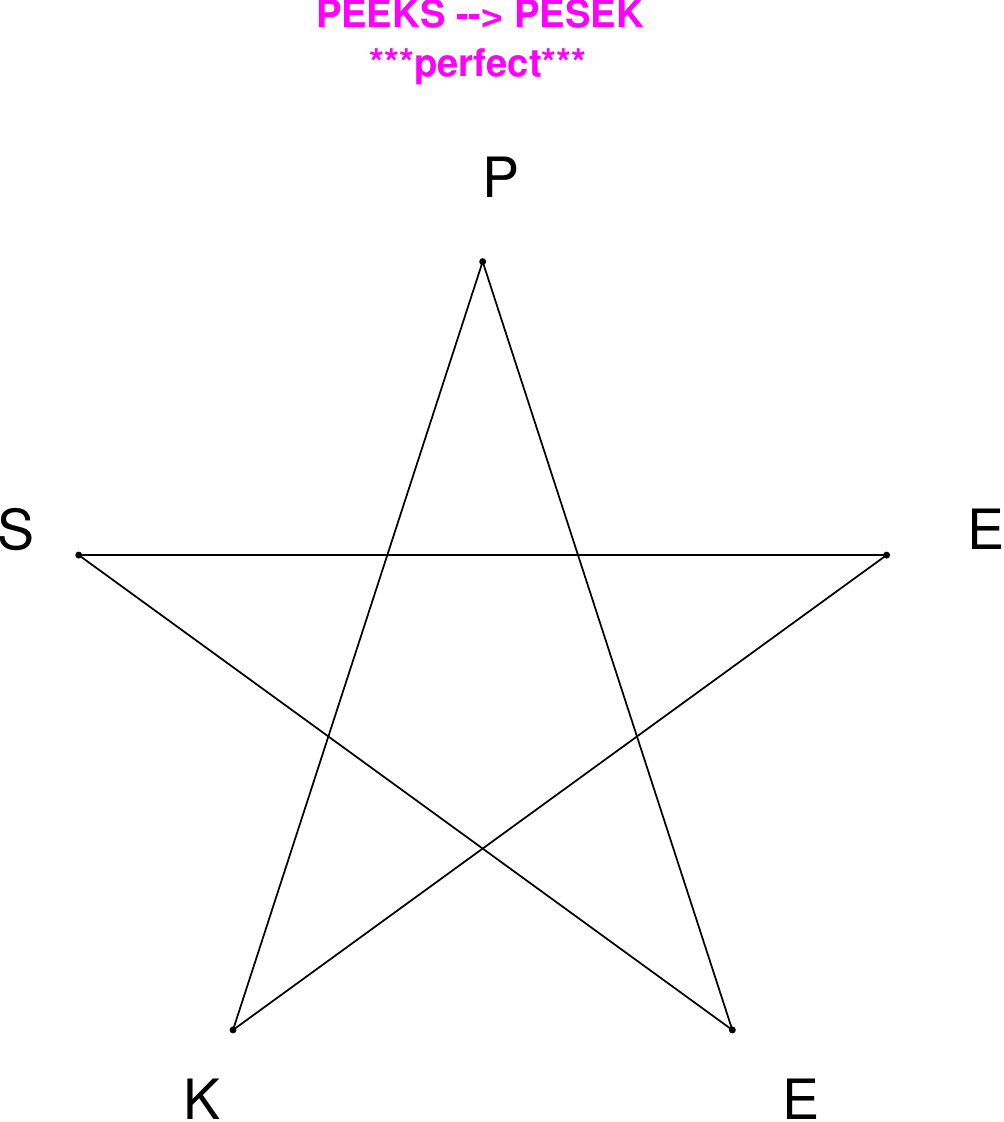}
\end{subfigure}
\hfill
\begin{subfigure}[T]{0.19\textwidth}
\centering
\includegraphics[width=\textwidth]{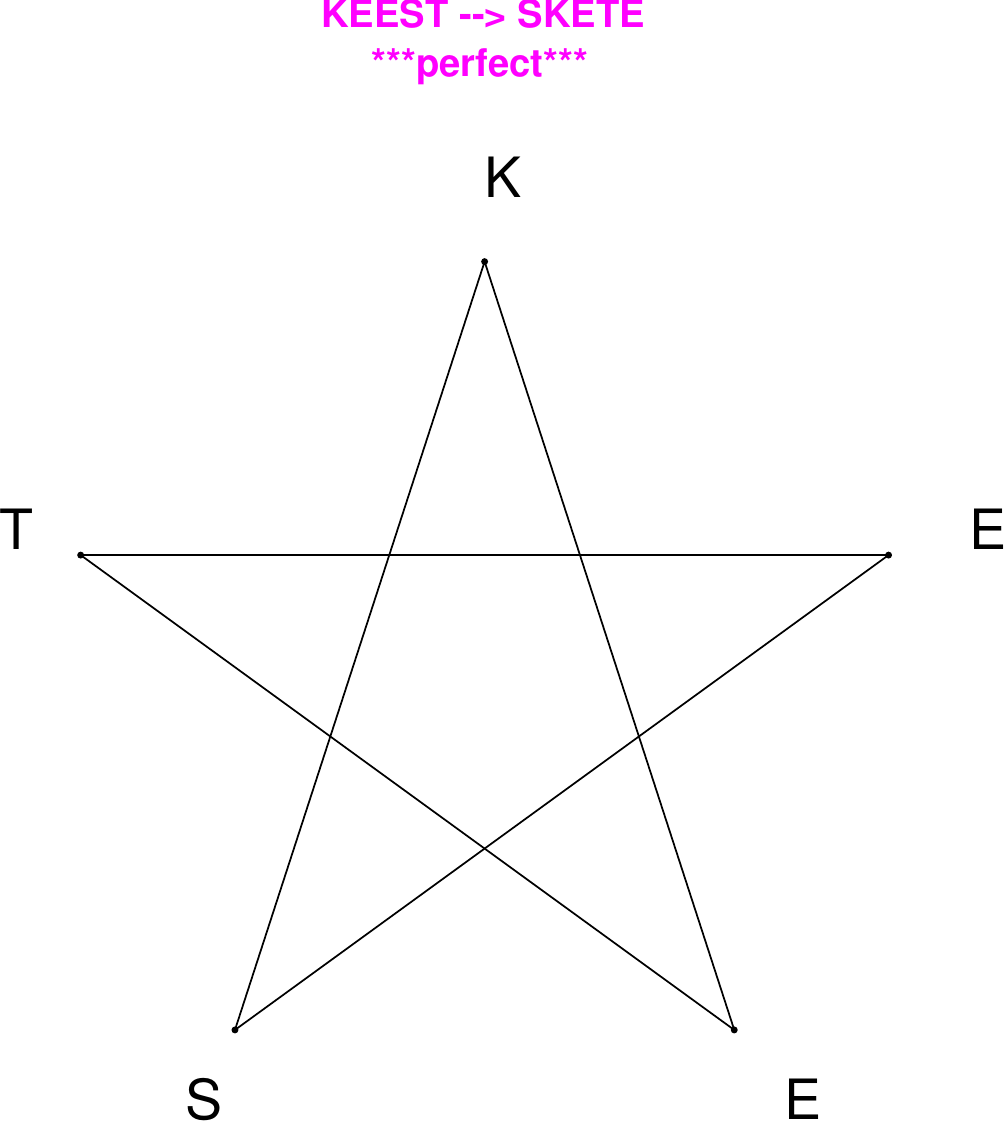}
\end{subfigure}
\end{figure}

\begin{figure}[H]
\centering
\begin{subfigure}[T]{0.19\textwidth}
\centering
\includegraphics[width=\textwidth]{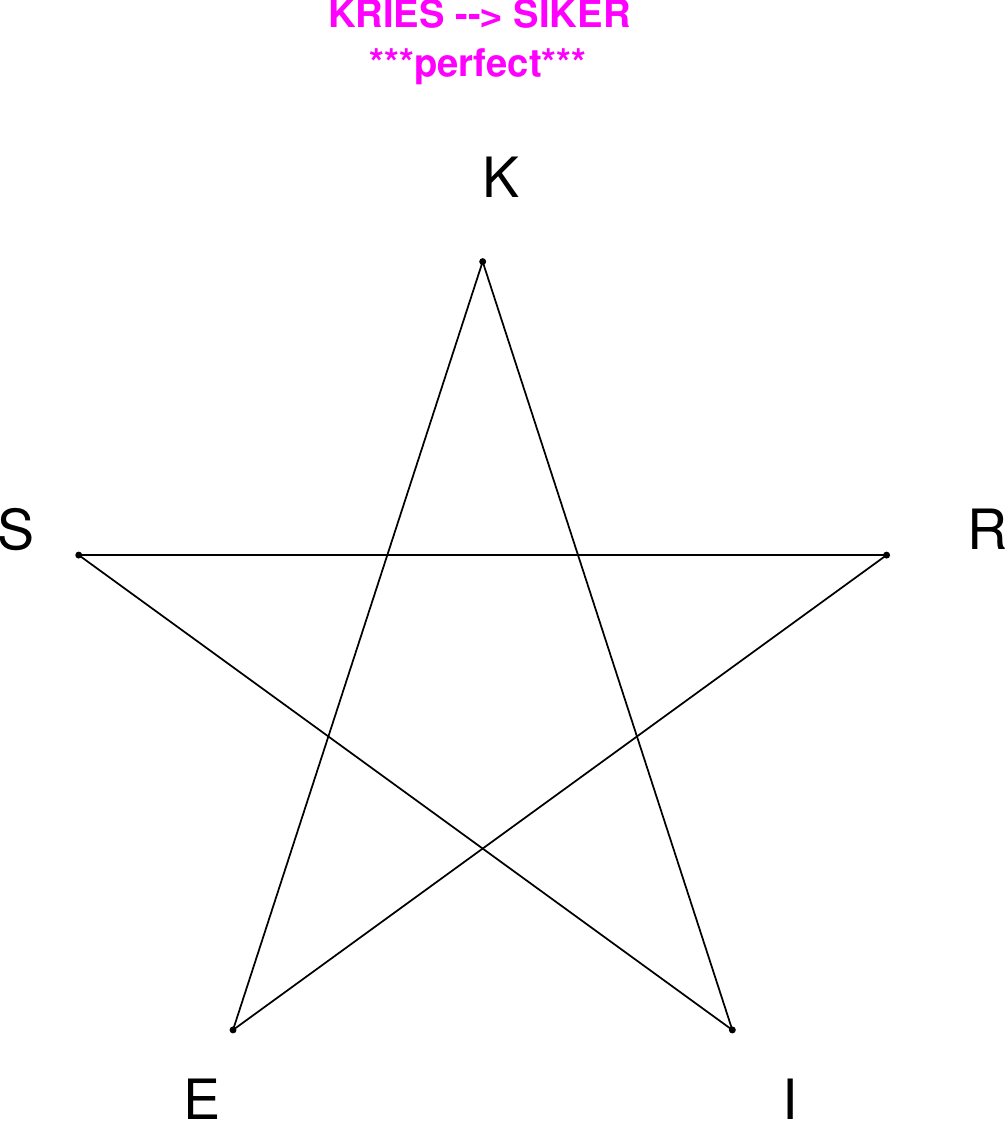}
\end{subfigure}
\hfill
\begin{subfigure}[T]{0.19\textwidth}
\centering
\includegraphics[width=\textwidth]{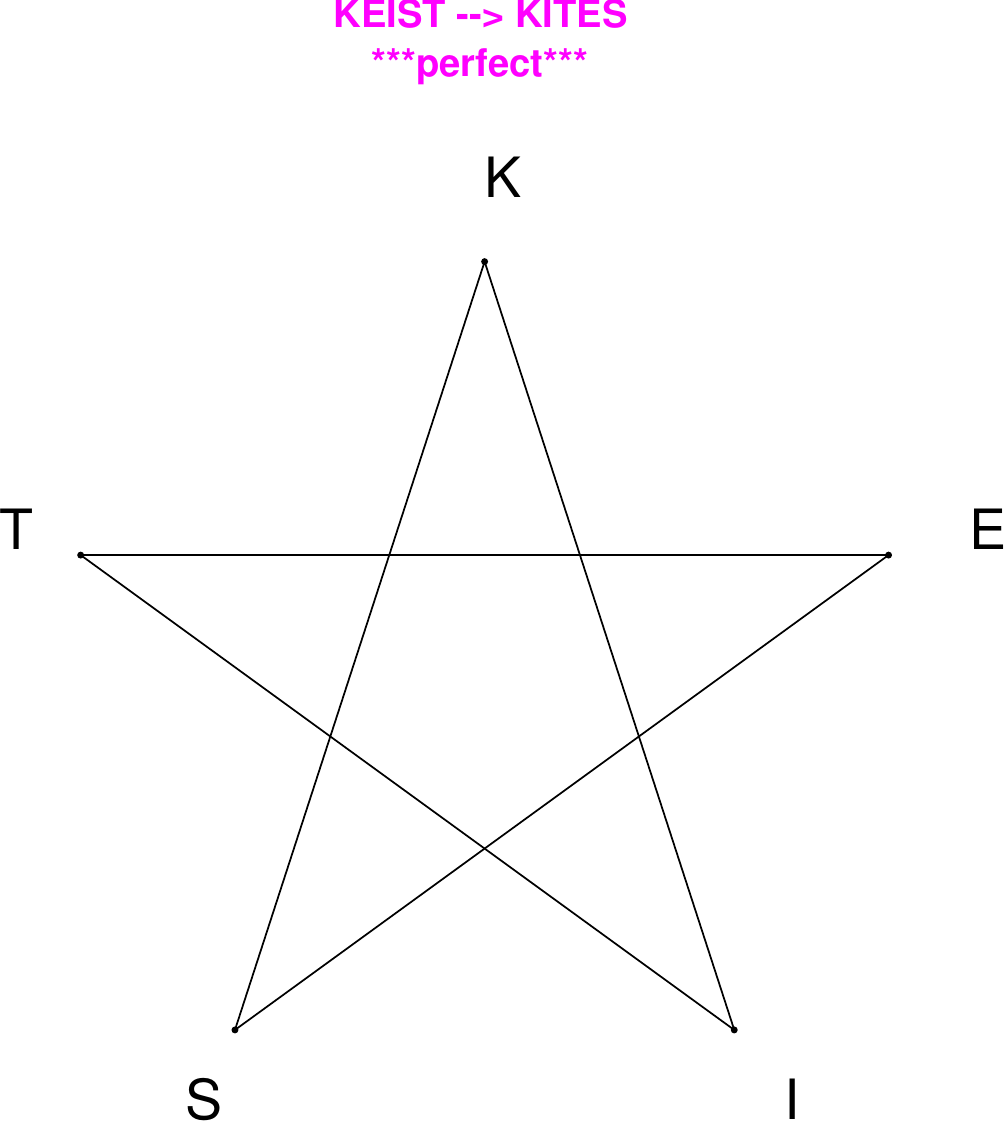}
\end{subfigure}
\hfill
\begin{subfigure}[T]{0.19\textwidth}
\centering
\includegraphics[width=\textwidth]{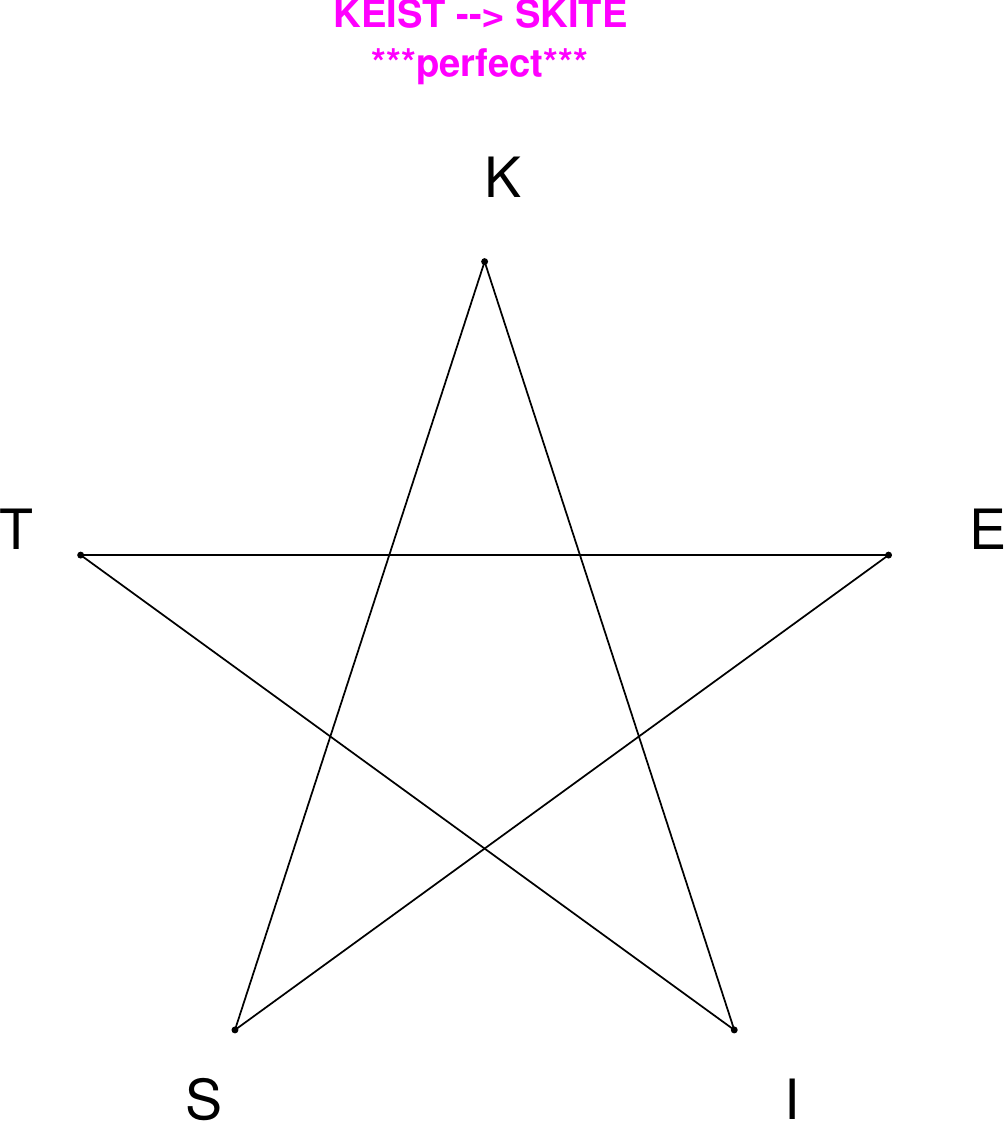}
\end{subfigure}
\hfill
\begin{subfigure}[T]{0.19\textwidth}
\centering
\includegraphics[width=\textwidth]{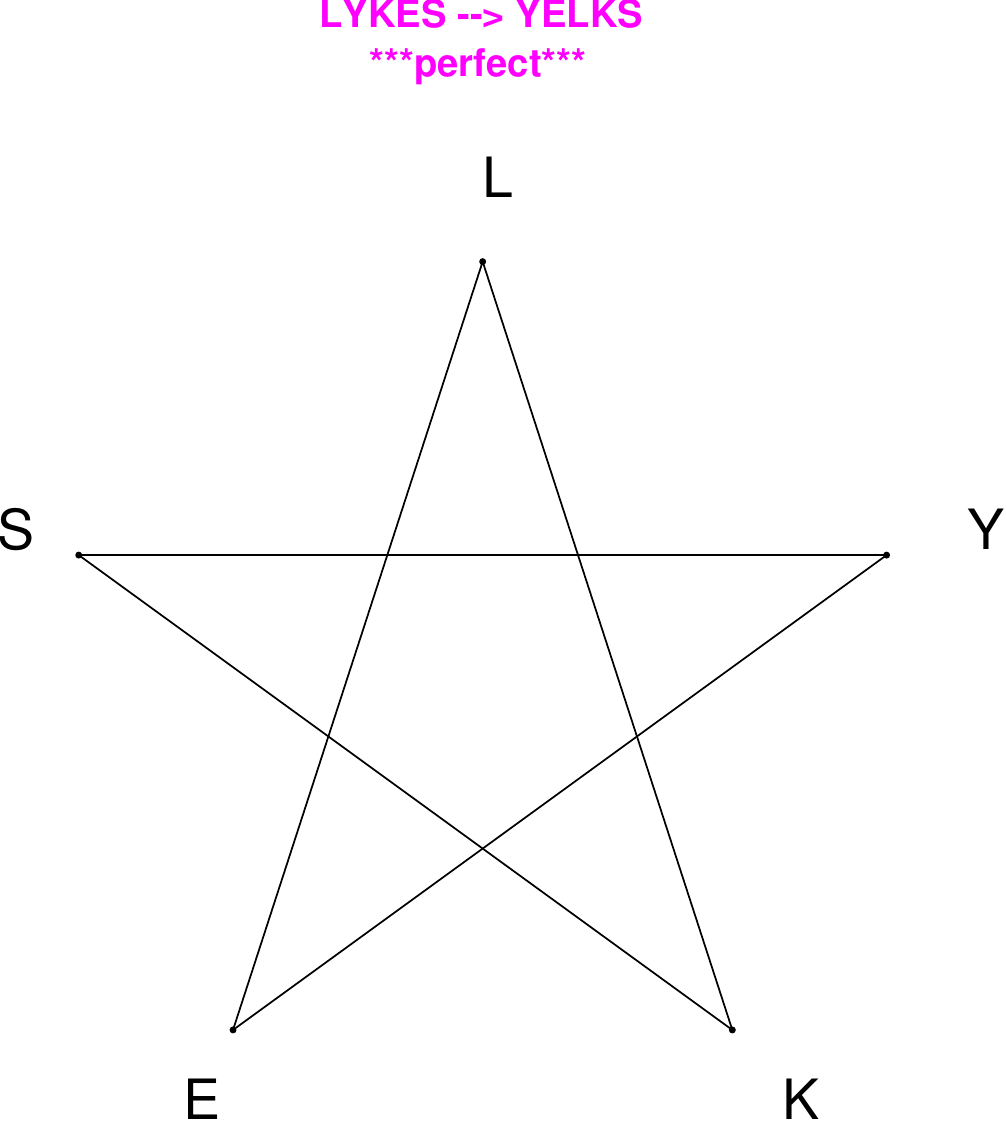}
\end{subfigure}
\hfill
\begin{subfigure}[T]{0.19\textwidth}
\centering
\includegraphics[width=\textwidth]{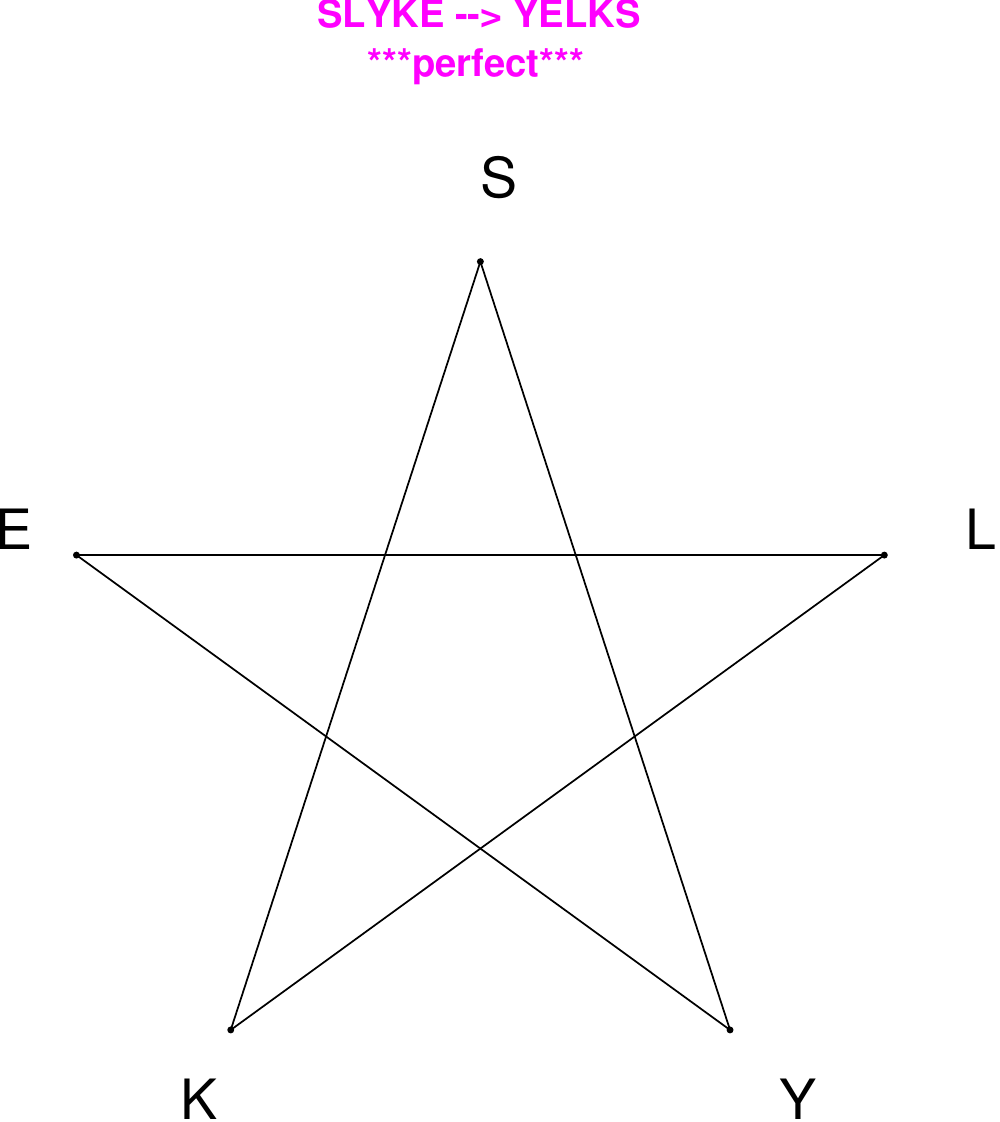}
\end{subfigure}
\end{figure}

\begin{figure}[H]
\centering
\begin{subfigure}[T]{0.19\textwidth}
\centering
\includegraphics[width=\textwidth]{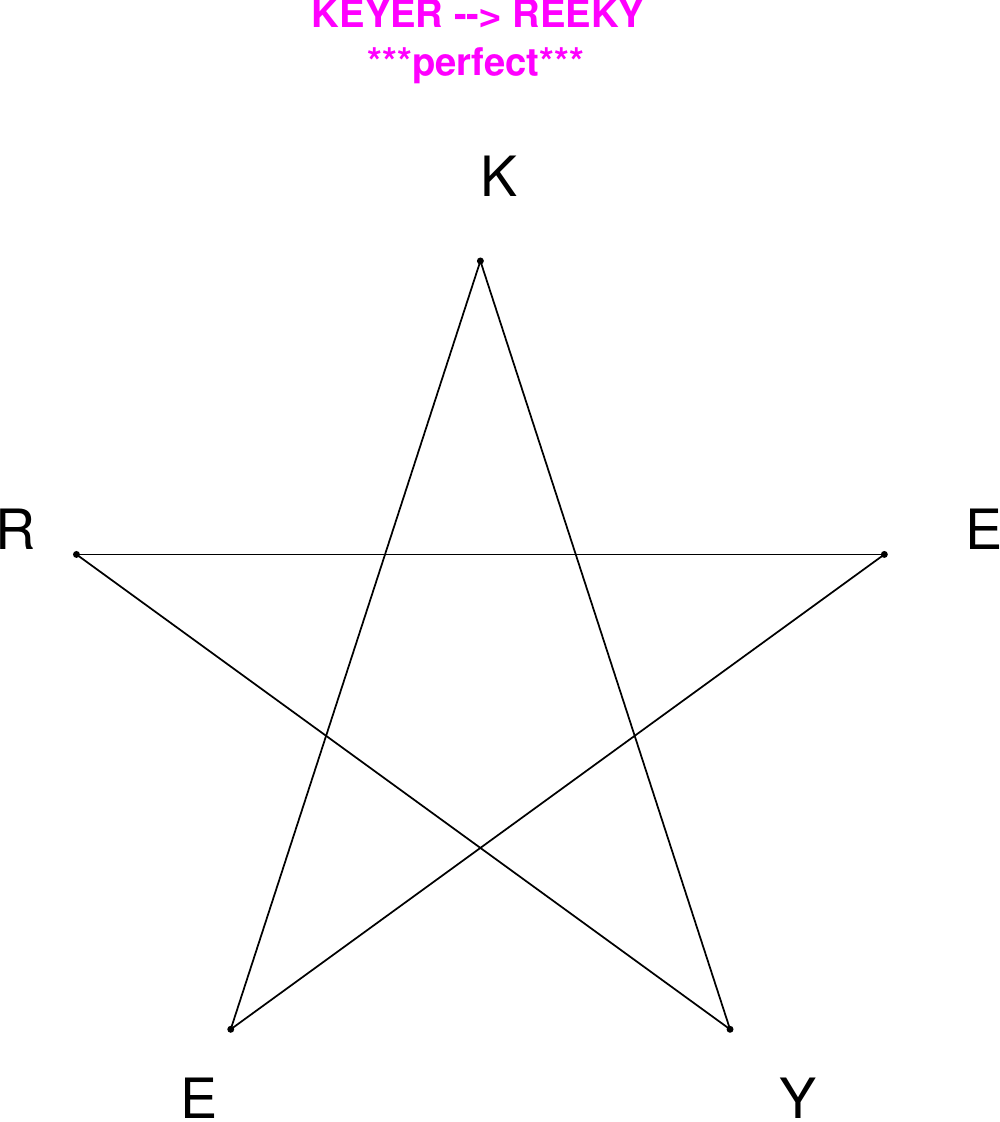}
\end{subfigure}
\hfill
\begin{subfigure}[T]{0.19\textwidth}
\centering
\includegraphics[width=\textwidth]{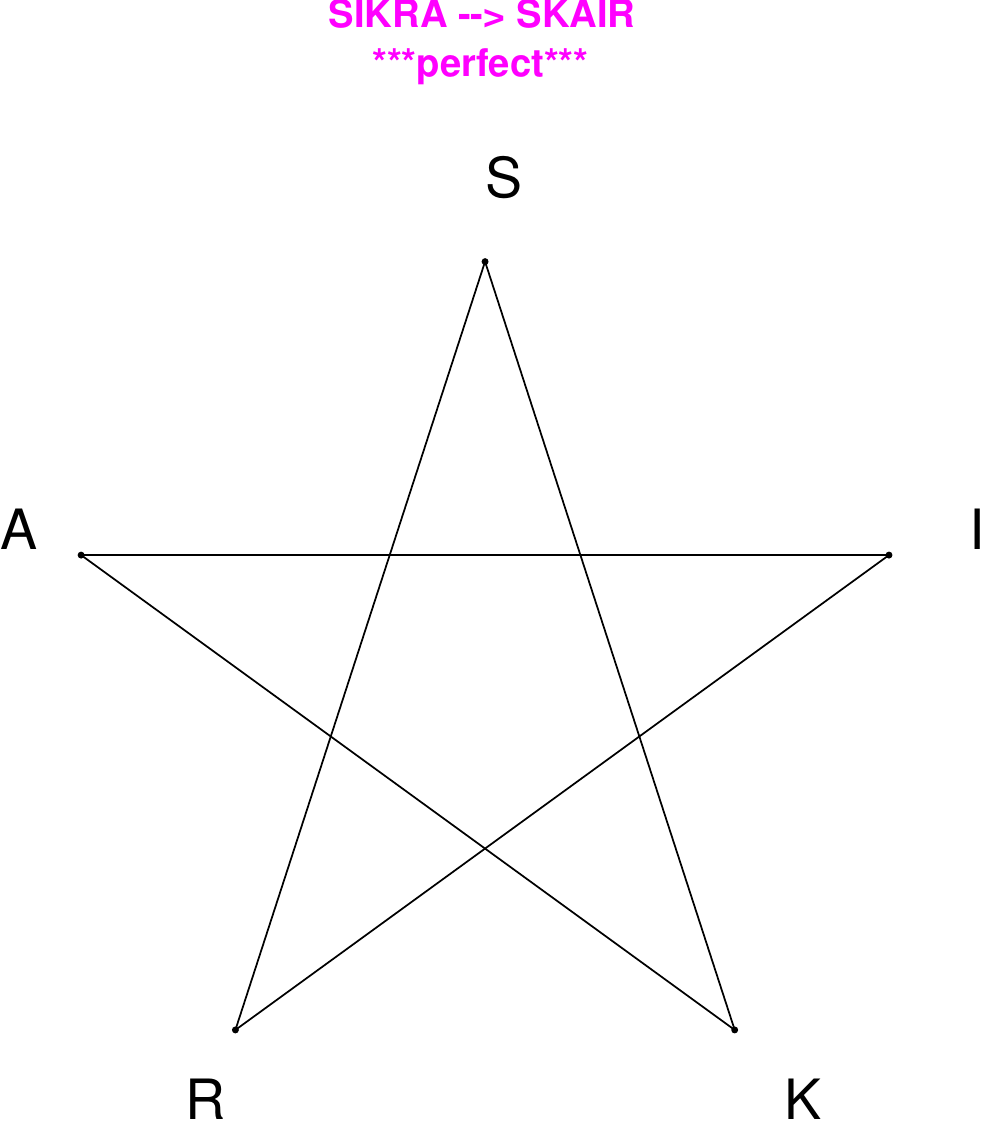}
\end{subfigure}
\hfill
\begin{subfigure}[T]{0.19\textwidth}
\centering
\includegraphics[width=\textwidth]{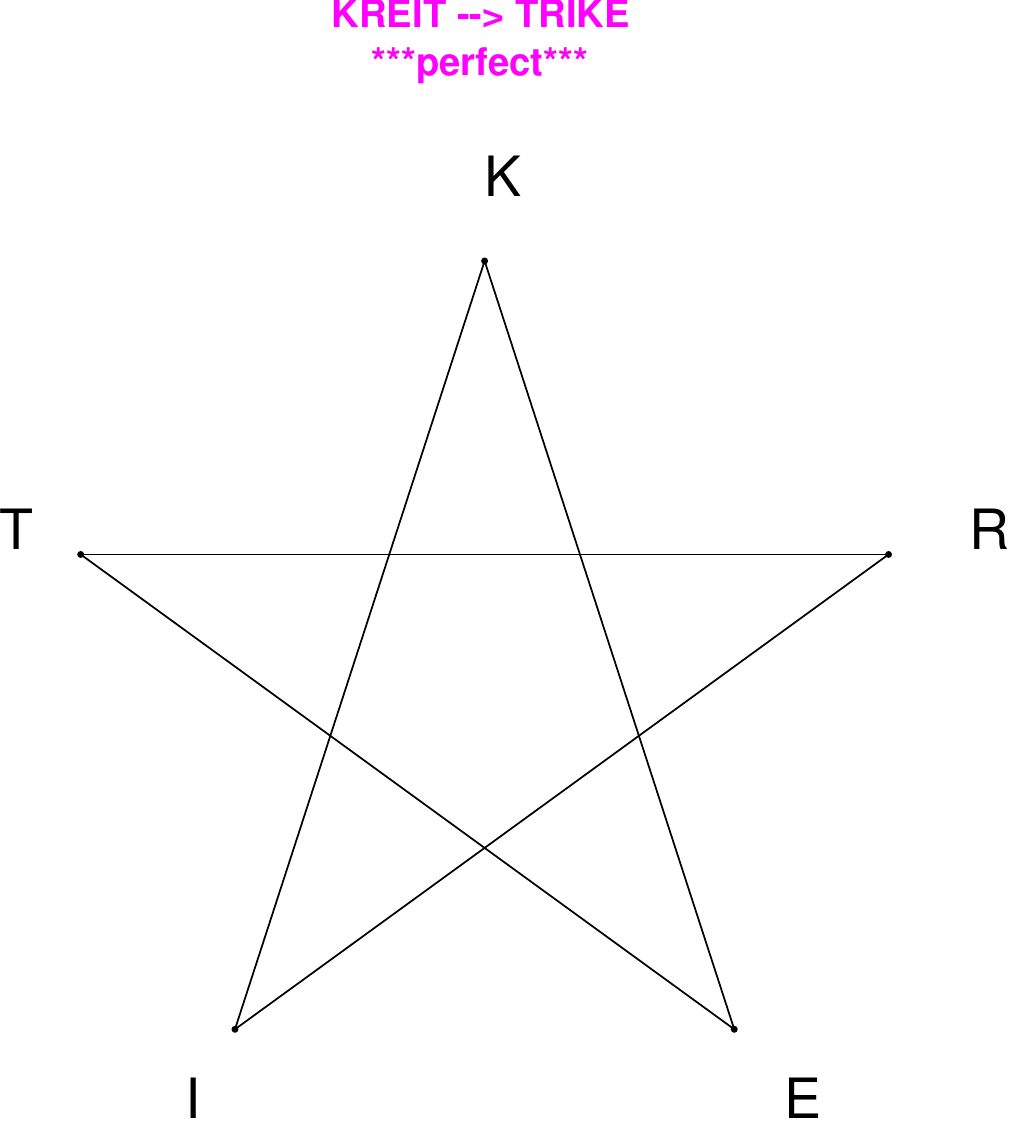}
\end{subfigure}
\hfill
\begin{subfigure}[T]{0.19\textwidth}
\centering
\includegraphics[width=\textwidth]{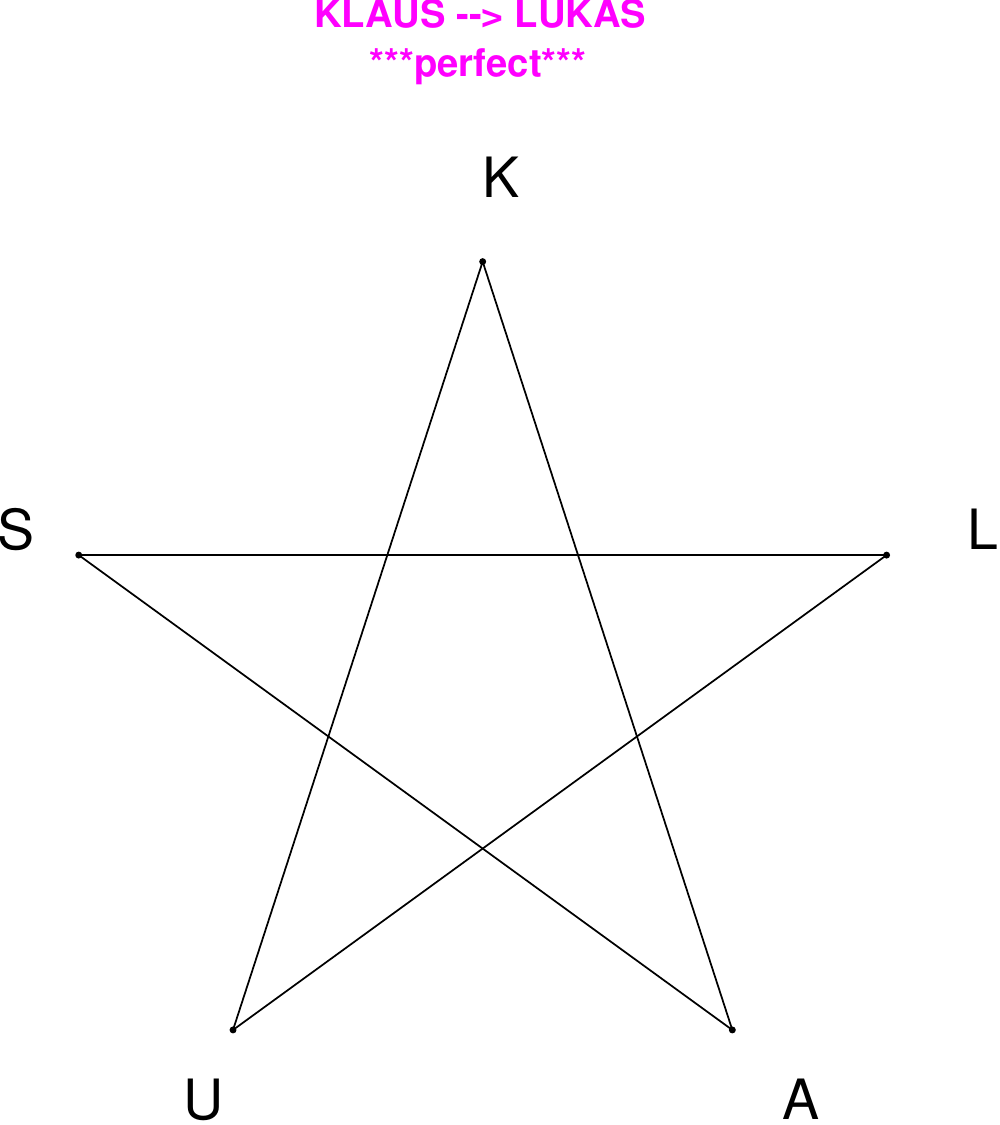}
\end{subfigure}
\hfill
\begin{subfigure}[T]{0.19\textwidth}
\centering
\includegraphics[width=\textwidth]{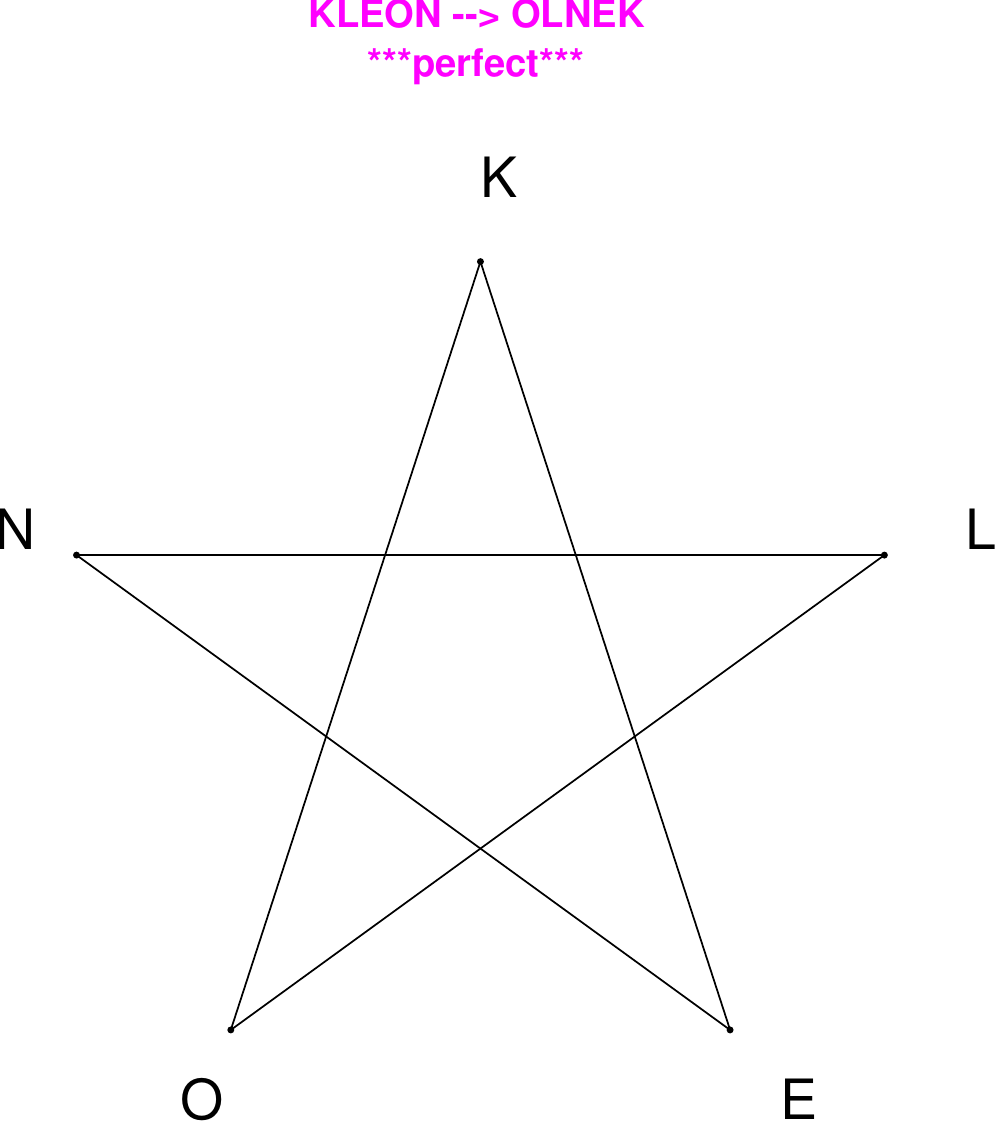}
\end{subfigure}
\end{figure}

\begin{figure}[H]
\centering
\begin{subfigure}[T]{0.19\textwidth}
\centering
\includegraphics[width=\textwidth]{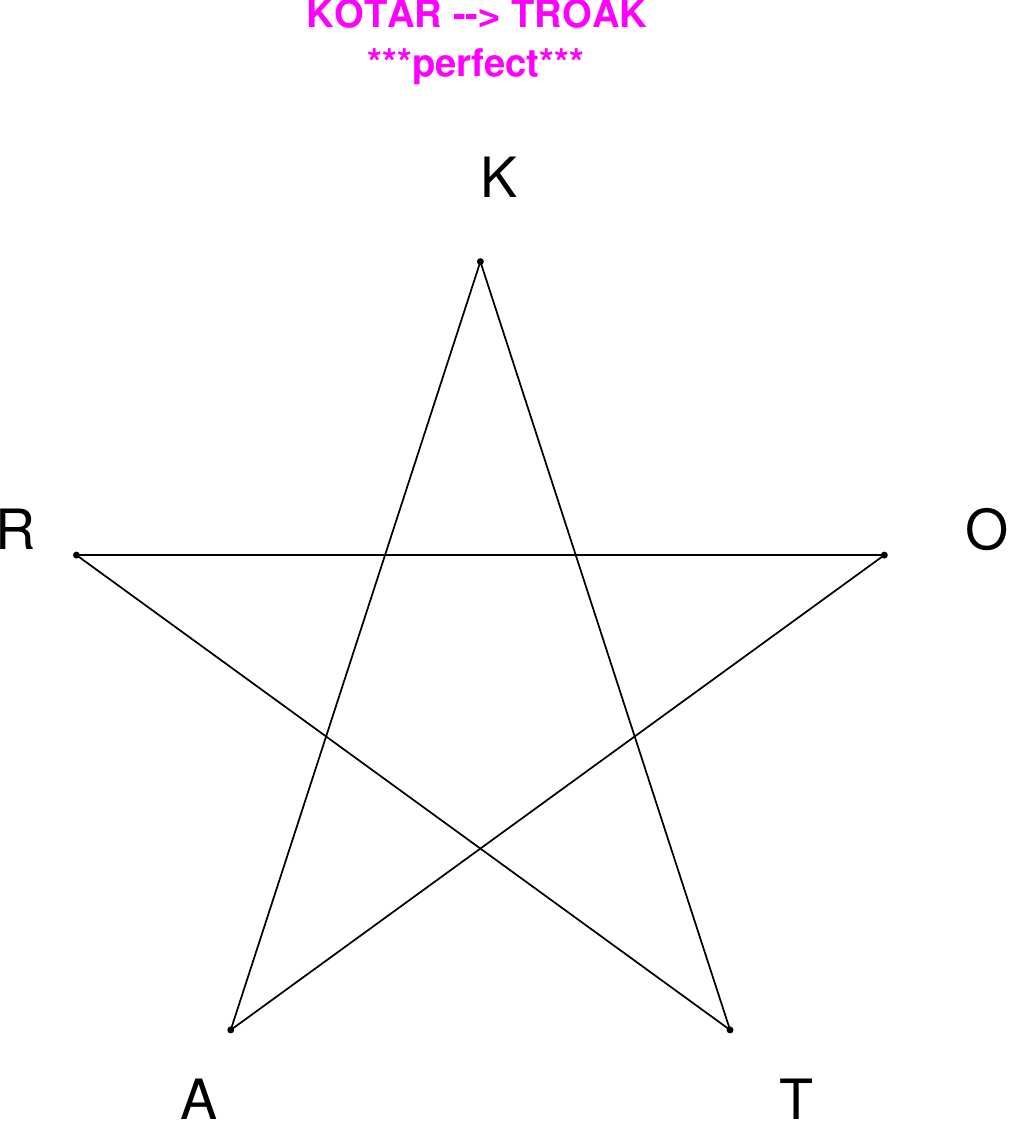}
\end{subfigure}
\hfill
\begin{subfigure}[T]{0.19\textwidth}
\centering
\includegraphics[width=\textwidth]{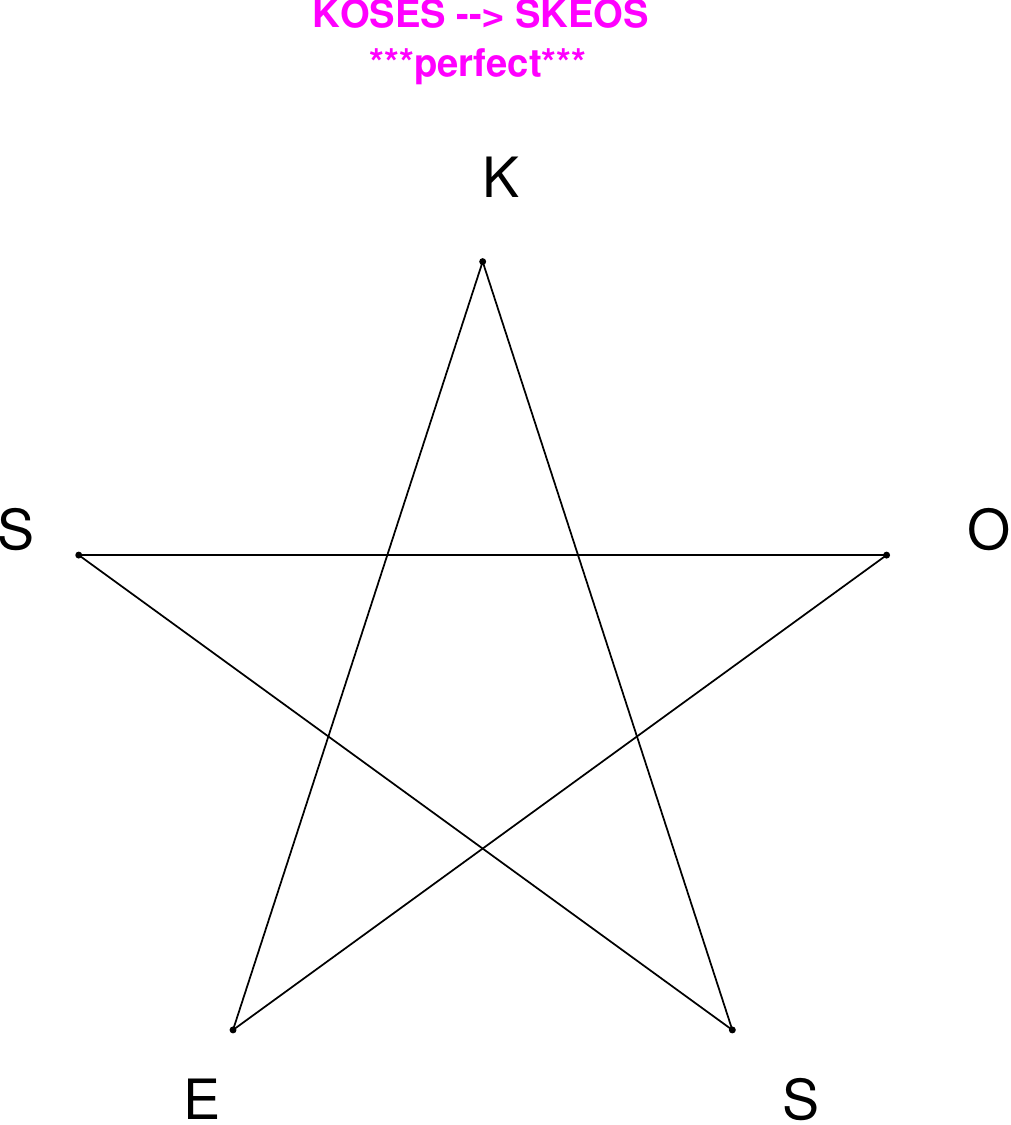}
\end{subfigure}
\hfill
\begin{subfigure}[T]{0.19\textwidth}
\centering
\includegraphics[width=\textwidth]{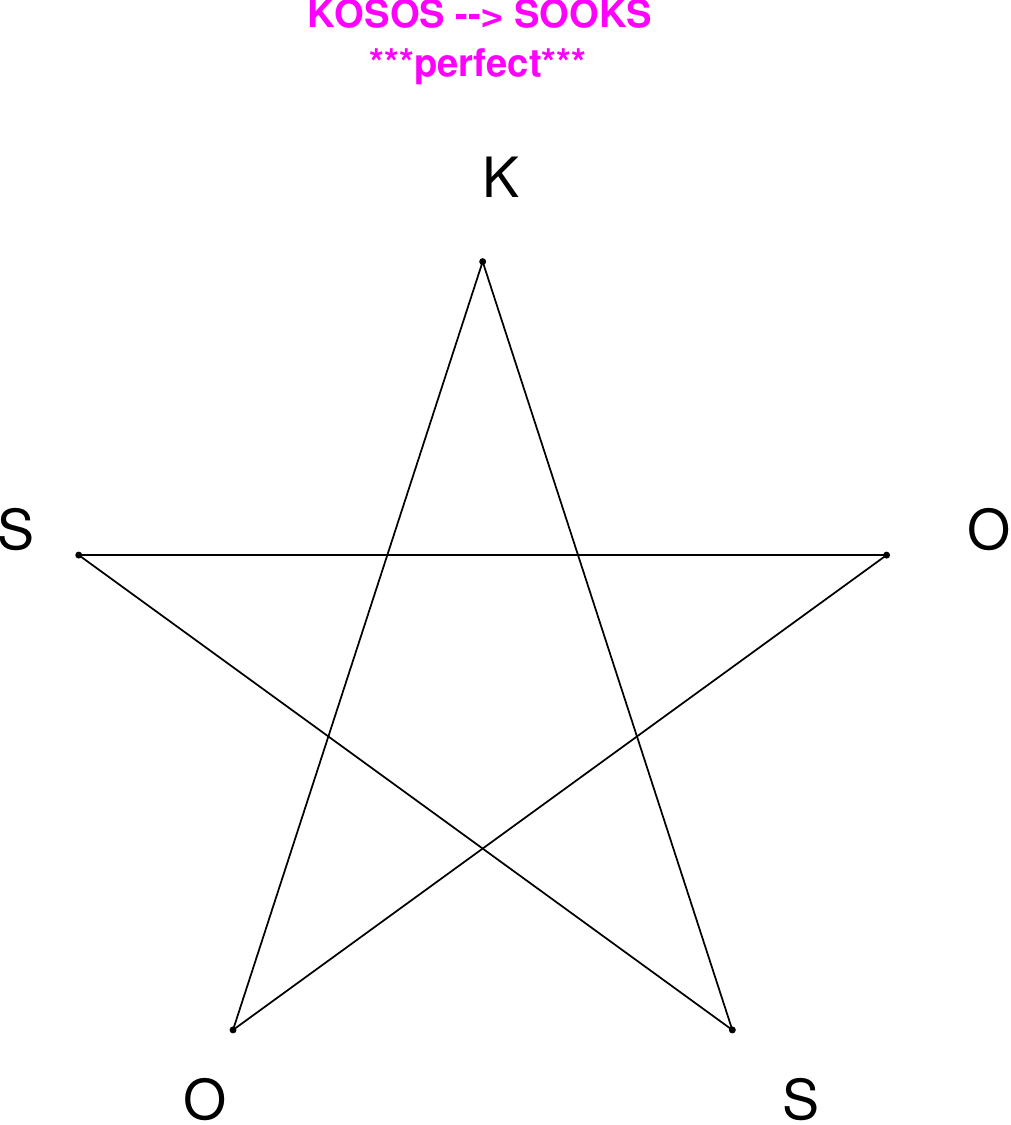}
\end{subfigure}
\hfill
\begin{subfigure}[T]{0.19\textwidth}
\centering
\includegraphics[width=\textwidth]{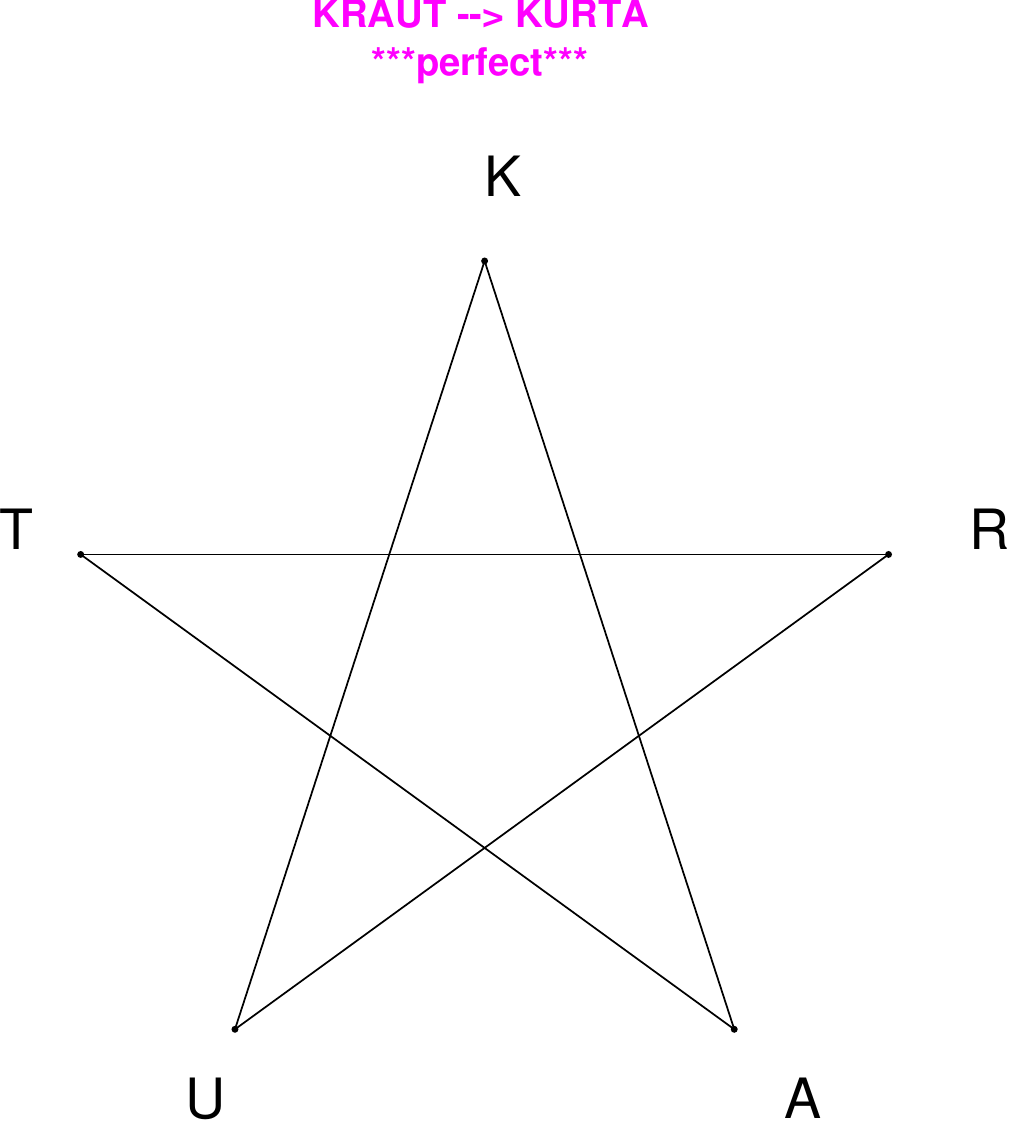}
\end{subfigure}
\hfill
\begin{subfigure}[T]{0.19\textwidth}
\centering
\includegraphics[width=\textwidth]{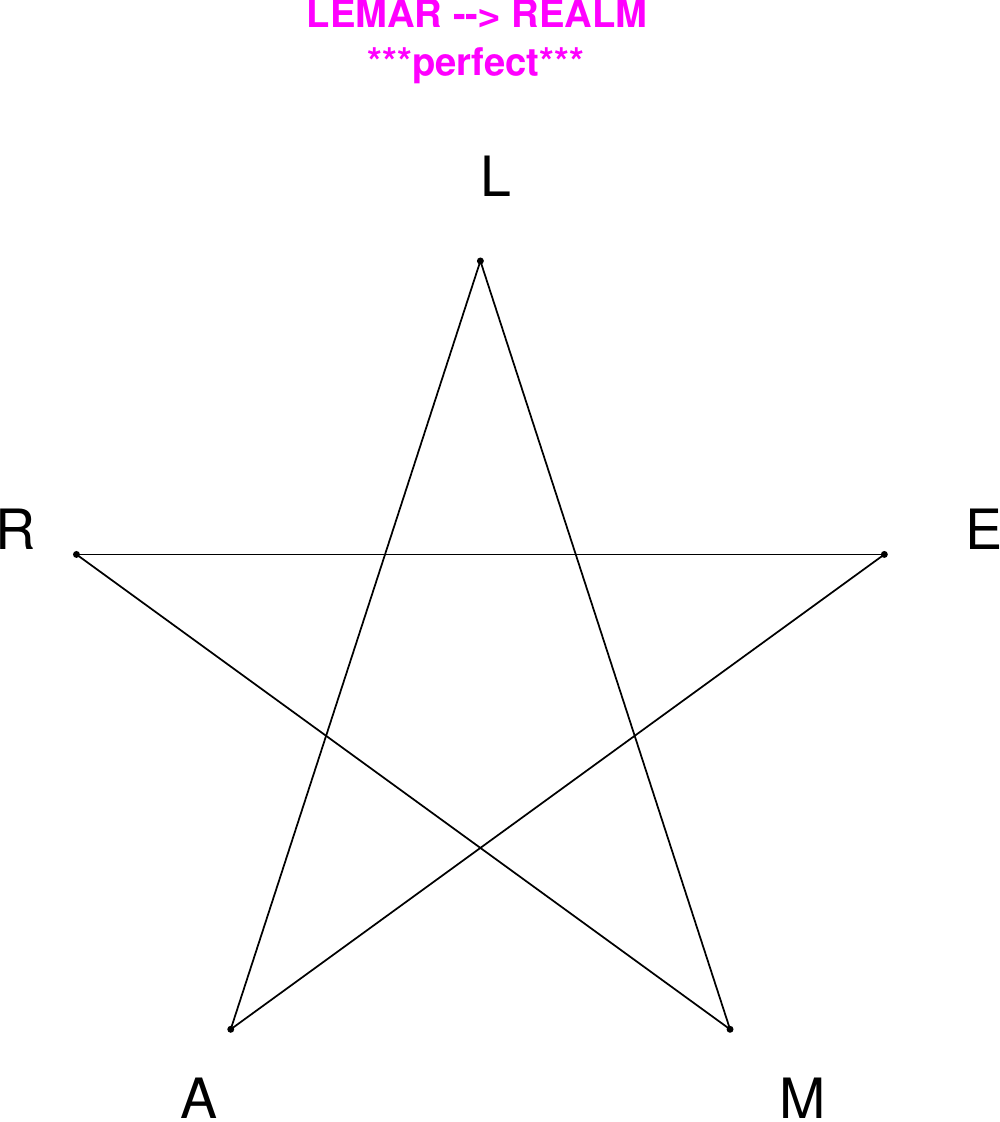}
\end{subfigure}
\end{figure}

\begin{figure}[H]
\centering
\begin{subfigure}[T]{0.19\textwidth}
\centering
\includegraphics[width=\textwidth]{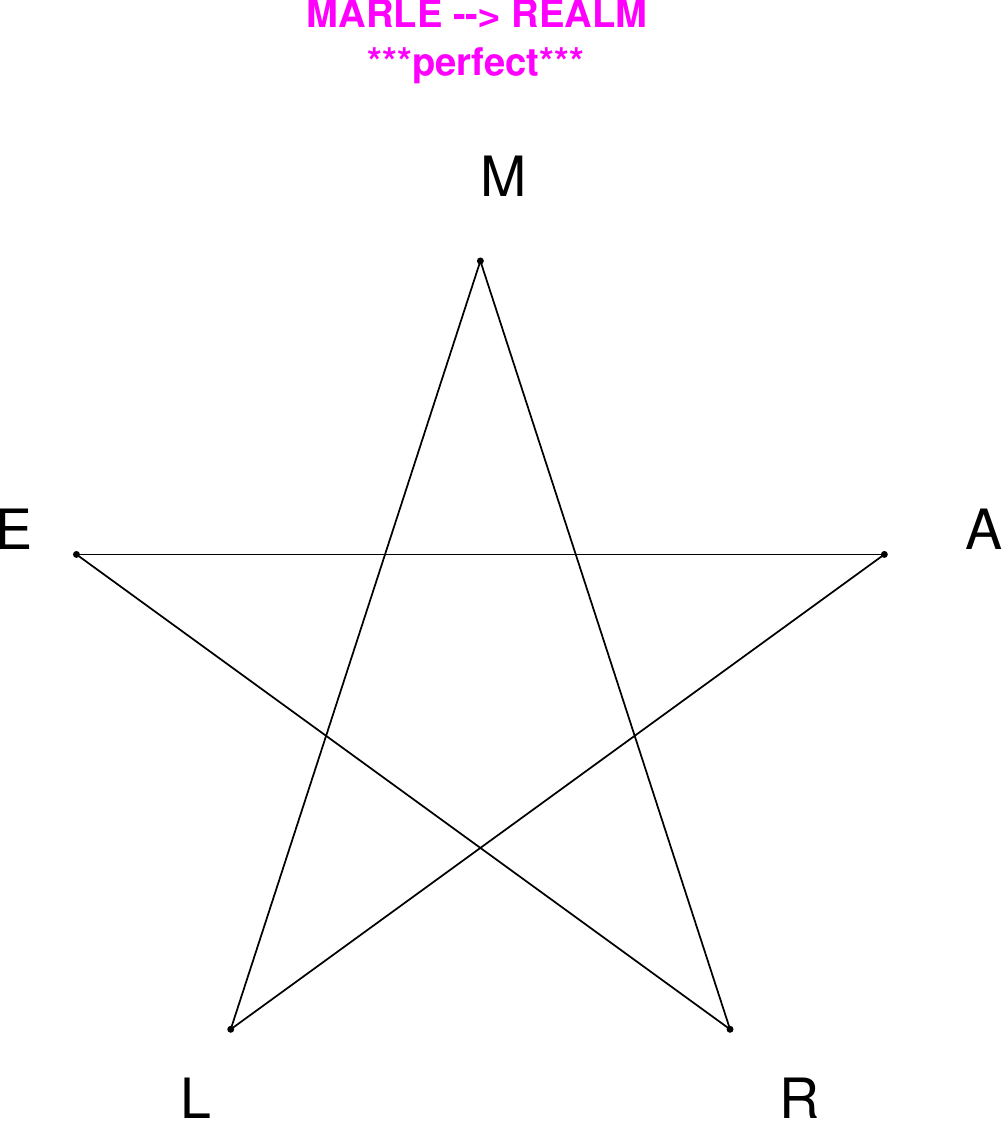}
\end{subfigure}
\hfill
\begin{subfigure}[T]{0.19\textwidth}
\centering
\includegraphics[width=\textwidth]{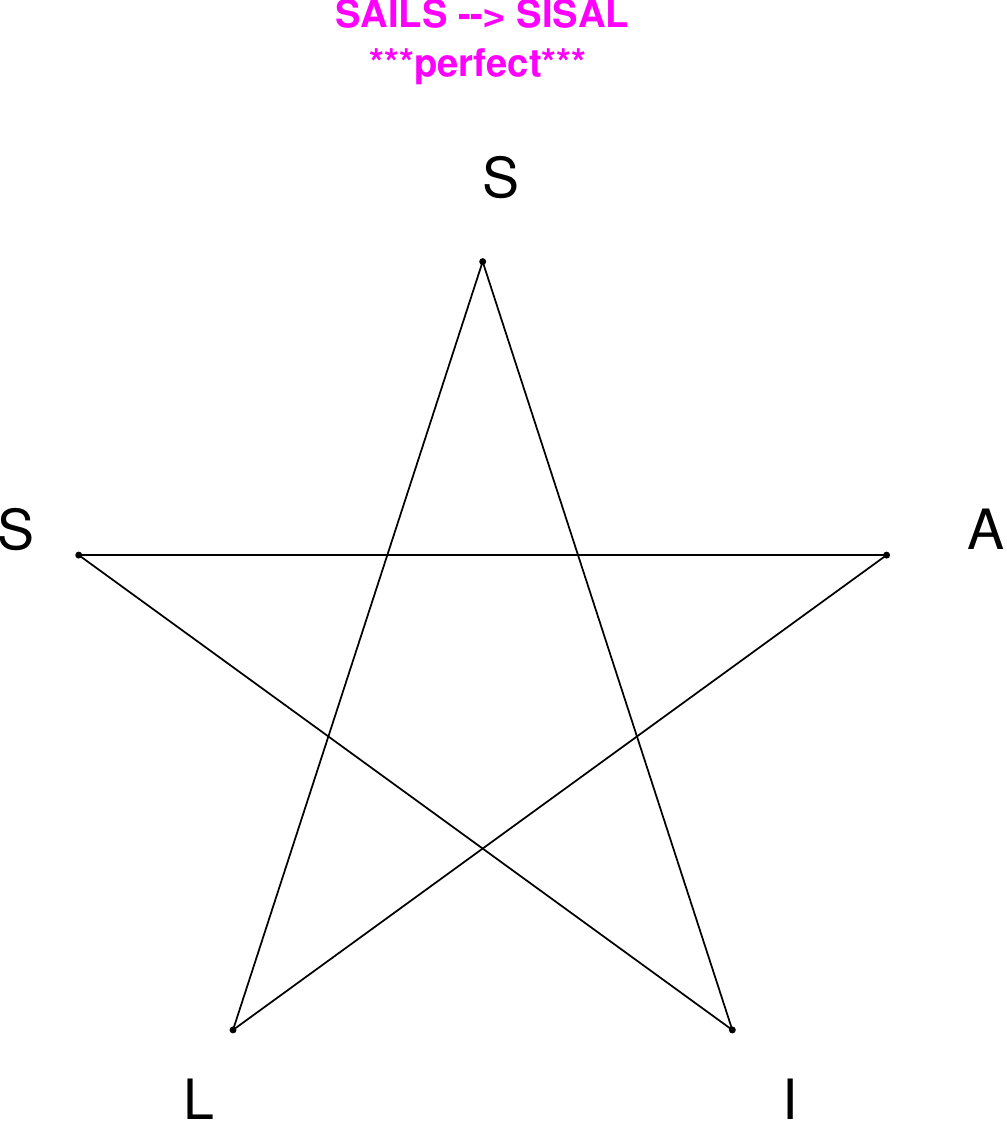}
\end{subfigure}
\hfill
\begin{subfigure}[T]{0.19\textwidth}
\centering
\includegraphics[width=\textwidth]{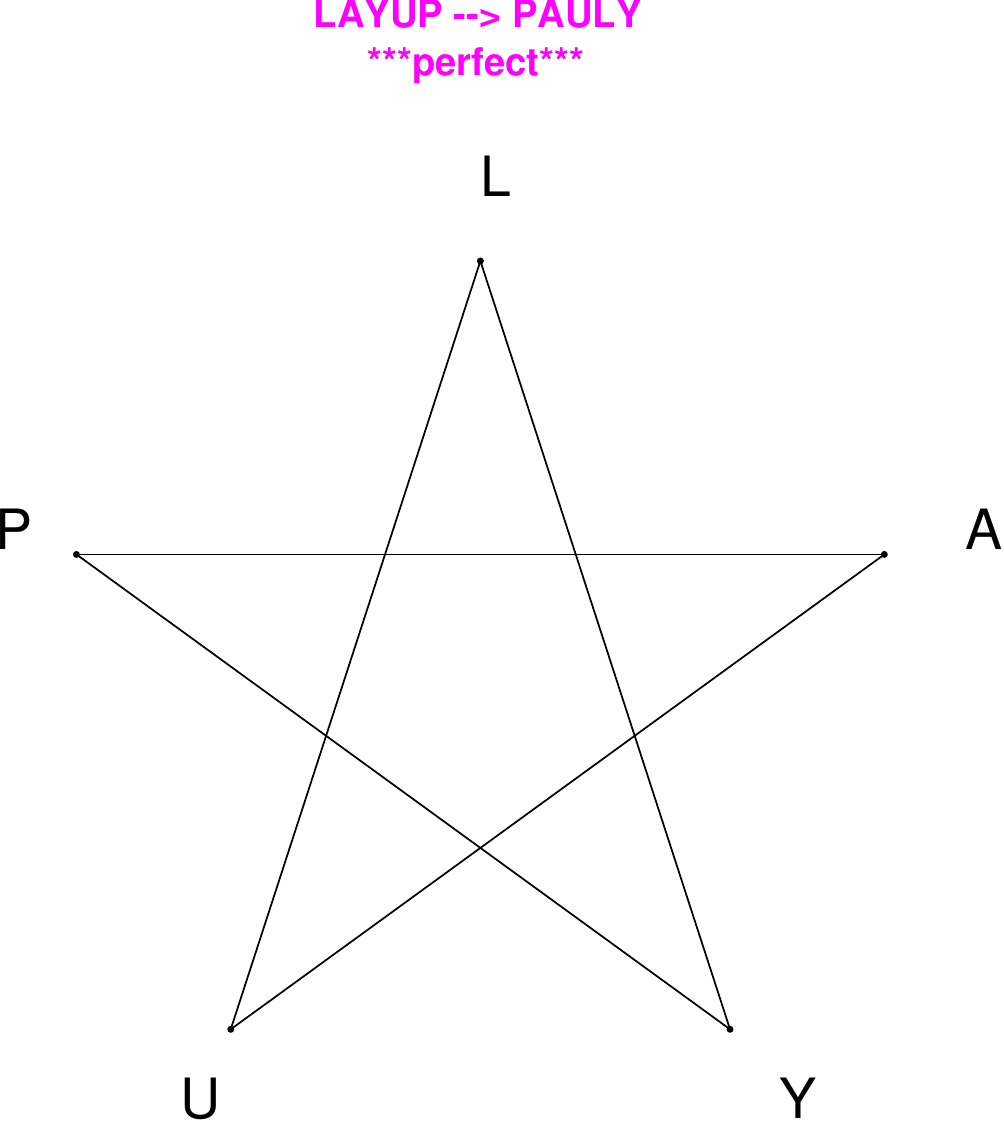}
\end{subfigure}
\hfill
\begin{subfigure}[T]{0.19\textwidth}
\centering
\includegraphics[width=\textwidth]{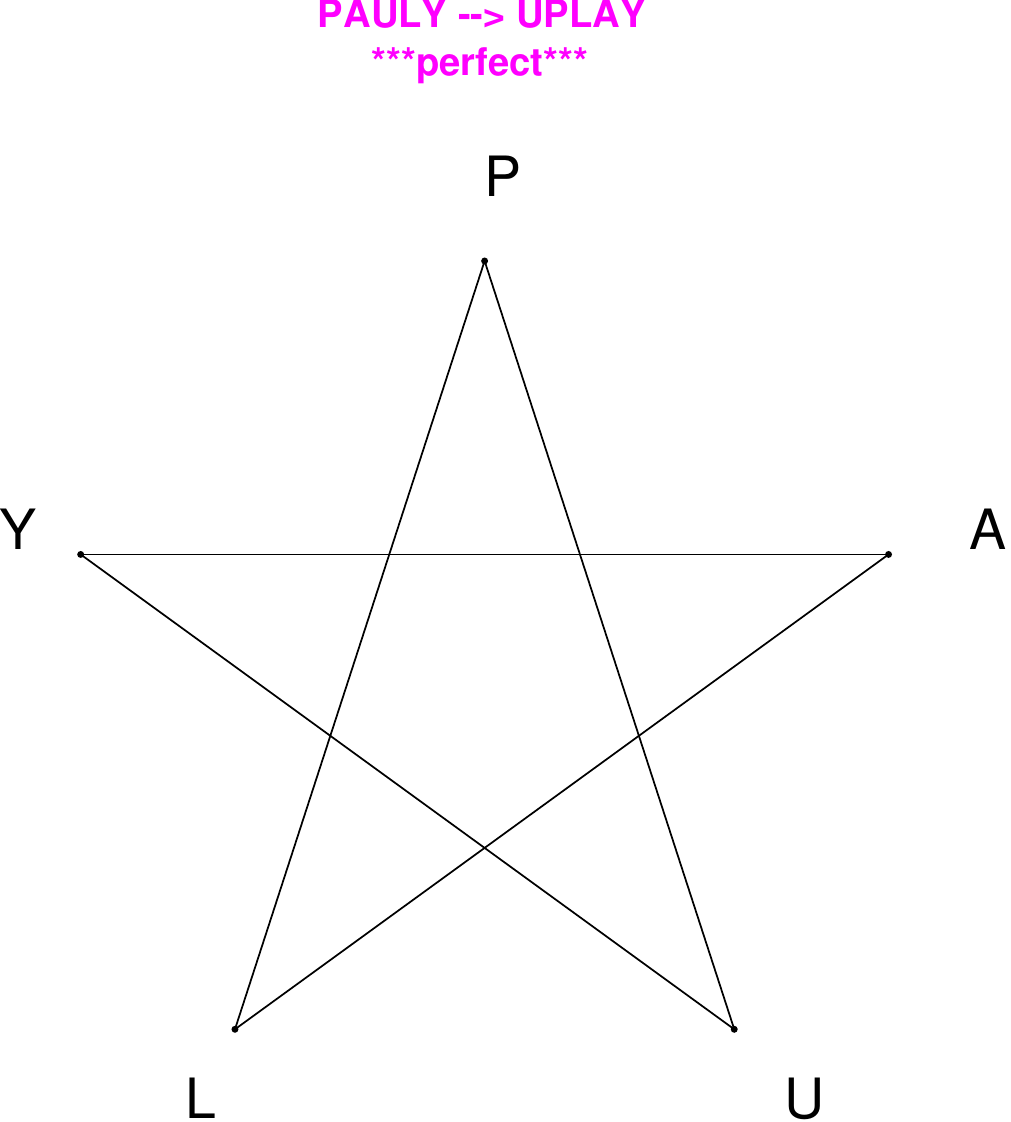}
\end{subfigure}
\hfill
\begin{subfigure}[T]{0.19\textwidth}
\centering
\includegraphics[width=\textwidth]{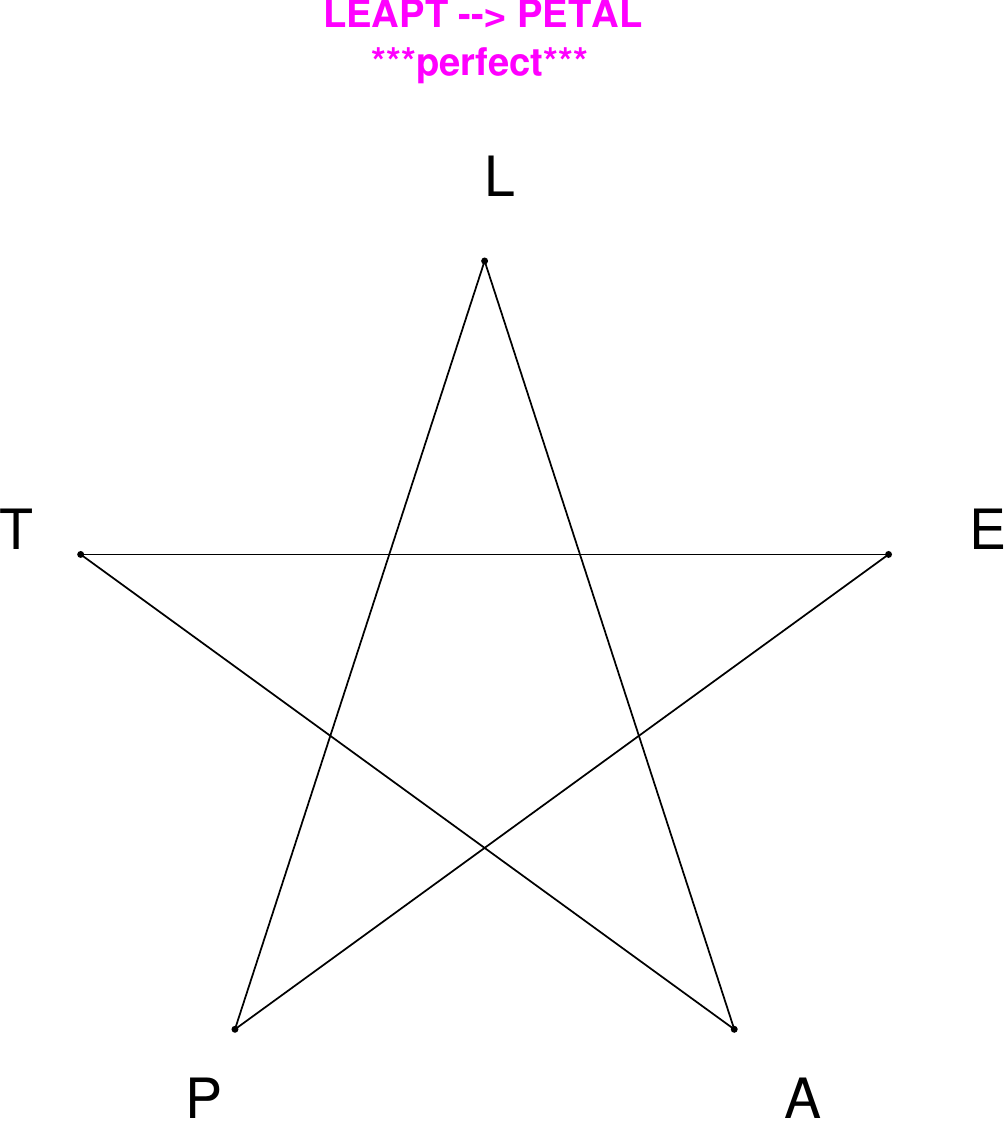}
\end{subfigure}
\end{figure}

\begin{figure}[H]
\centering
\begin{subfigure}[T]{0.19\textwidth}
\centering
\includegraphics[width=\textwidth]{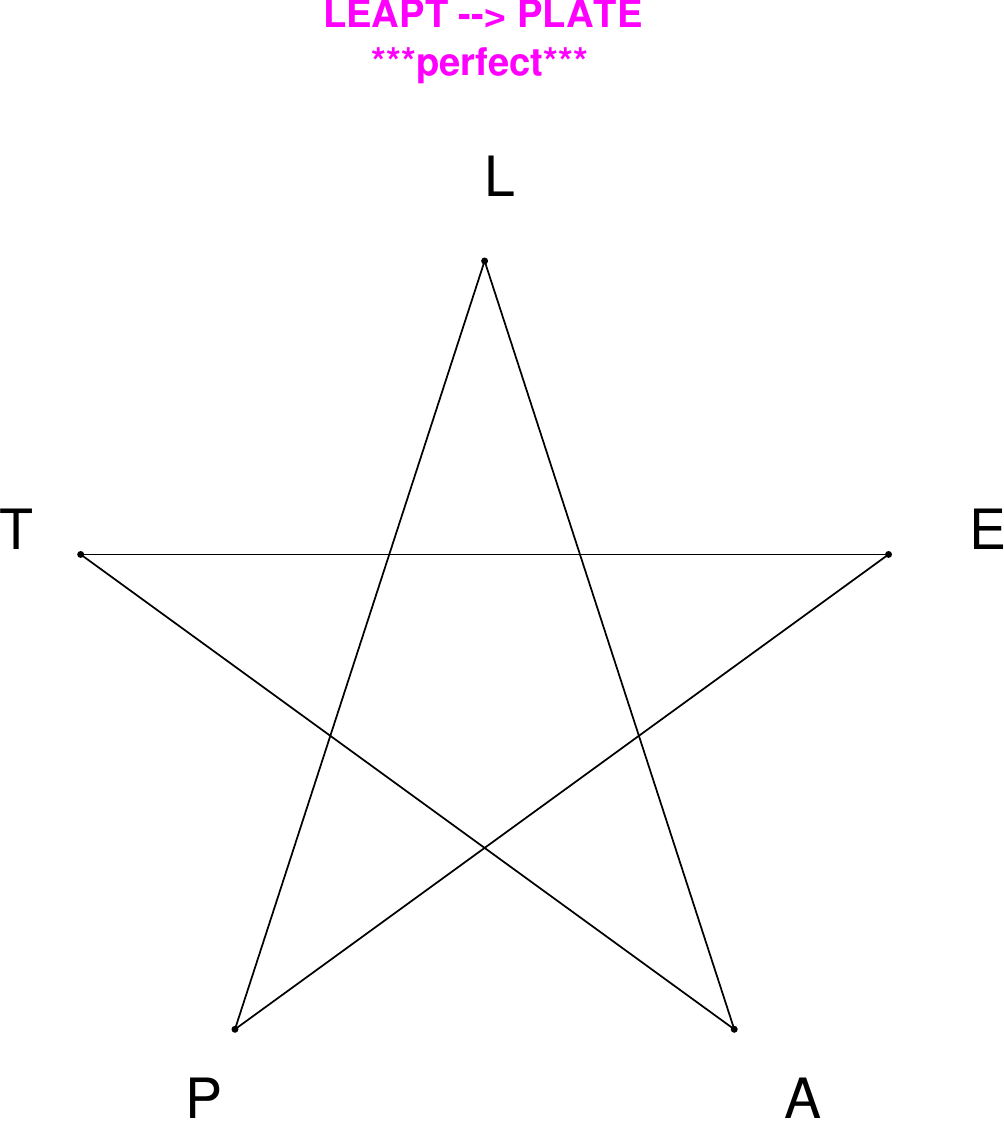}
\end{subfigure}
\hfill
\begin{subfigure}[T]{0.19\textwidth}
\centering
\includegraphics[width=\textwidth]{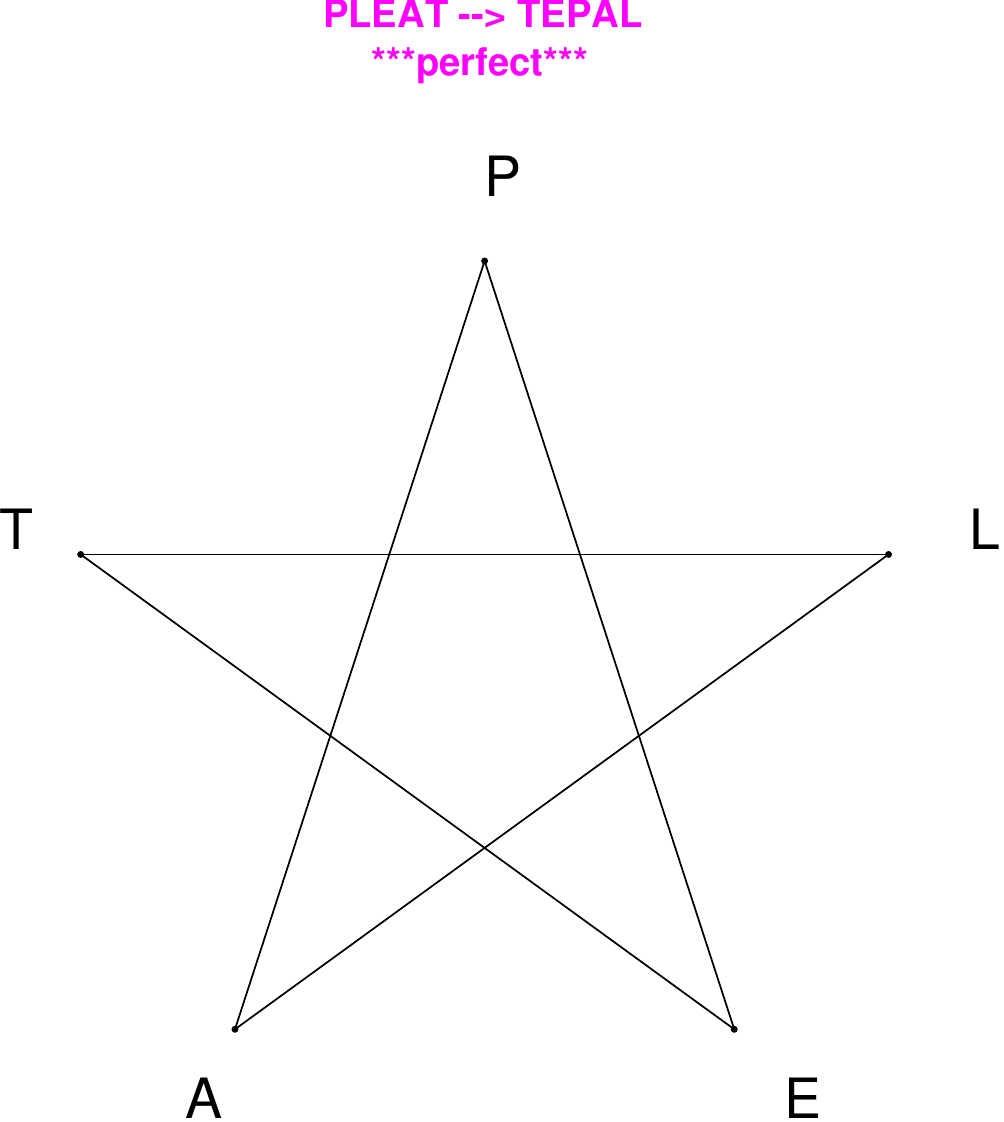}
\end{subfigure}
\hfill
\begin{subfigure}[T]{0.19\textwidth}
\centering
\includegraphics[width=\textwidth]{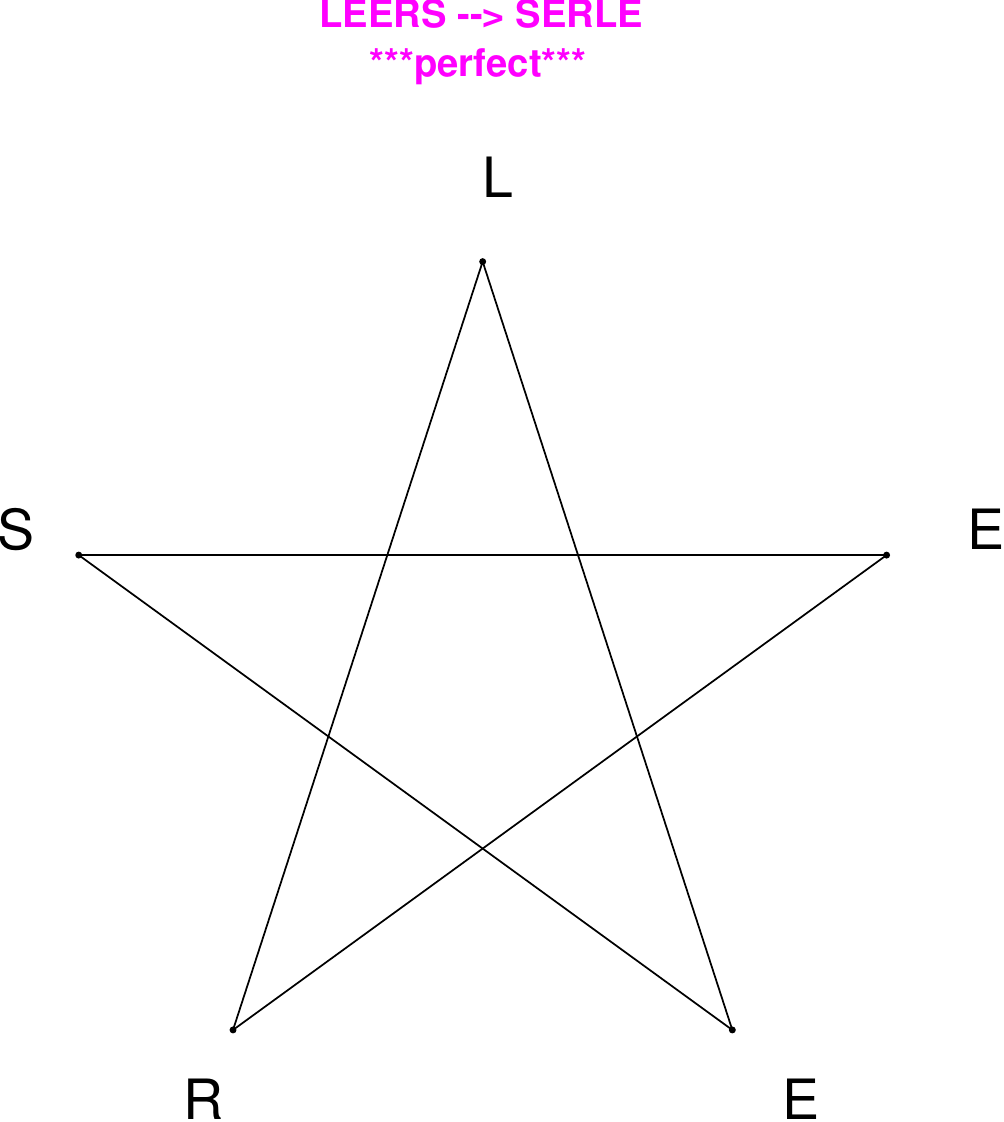}
\end{subfigure}
\hfill
\begin{subfigure}[T]{0.19\textwidth}
\centering
\includegraphics[width=\textwidth]{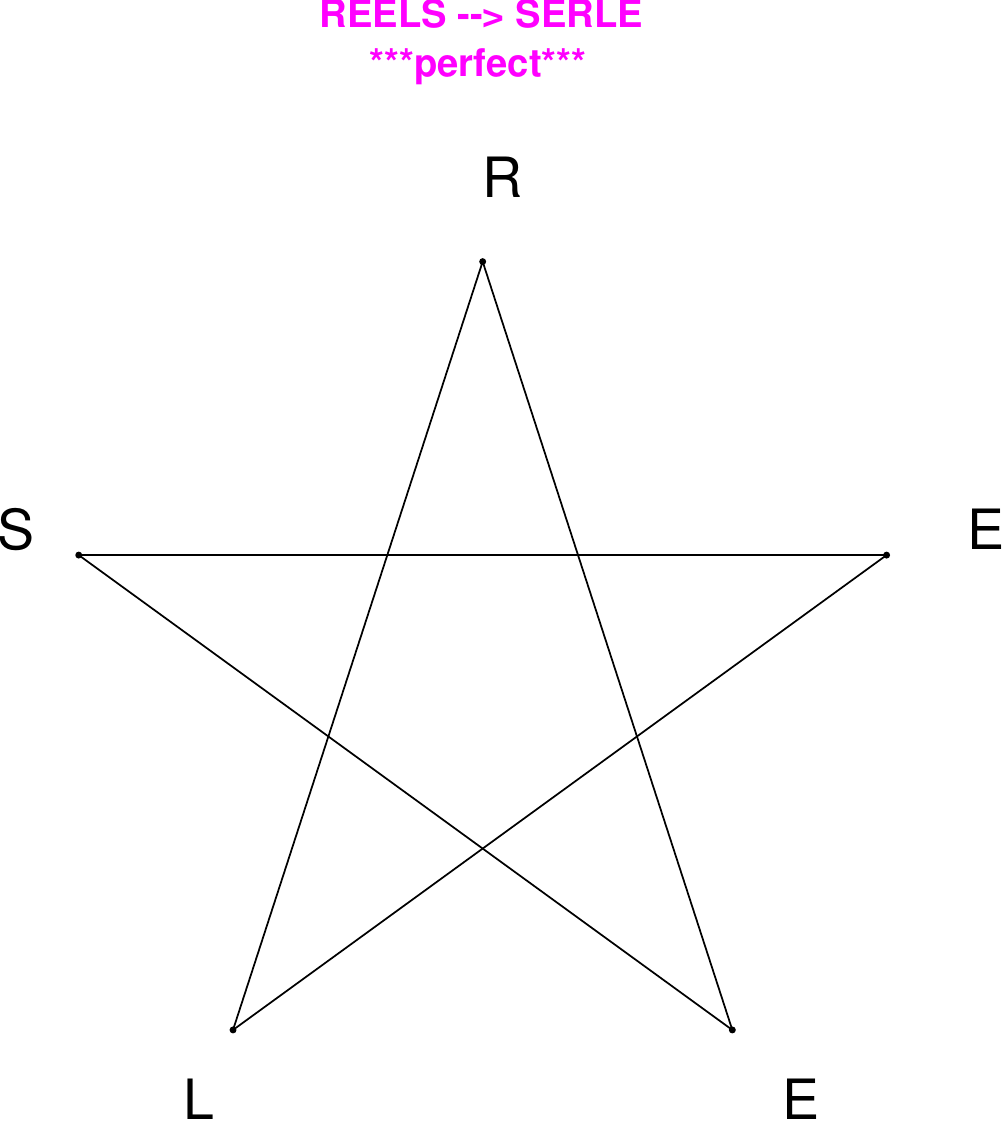}
\end{subfigure}
\hfill
\begin{subfigure}[T]{0.19\textwidth}
\centering
\includegraphics[width=\textwidth]{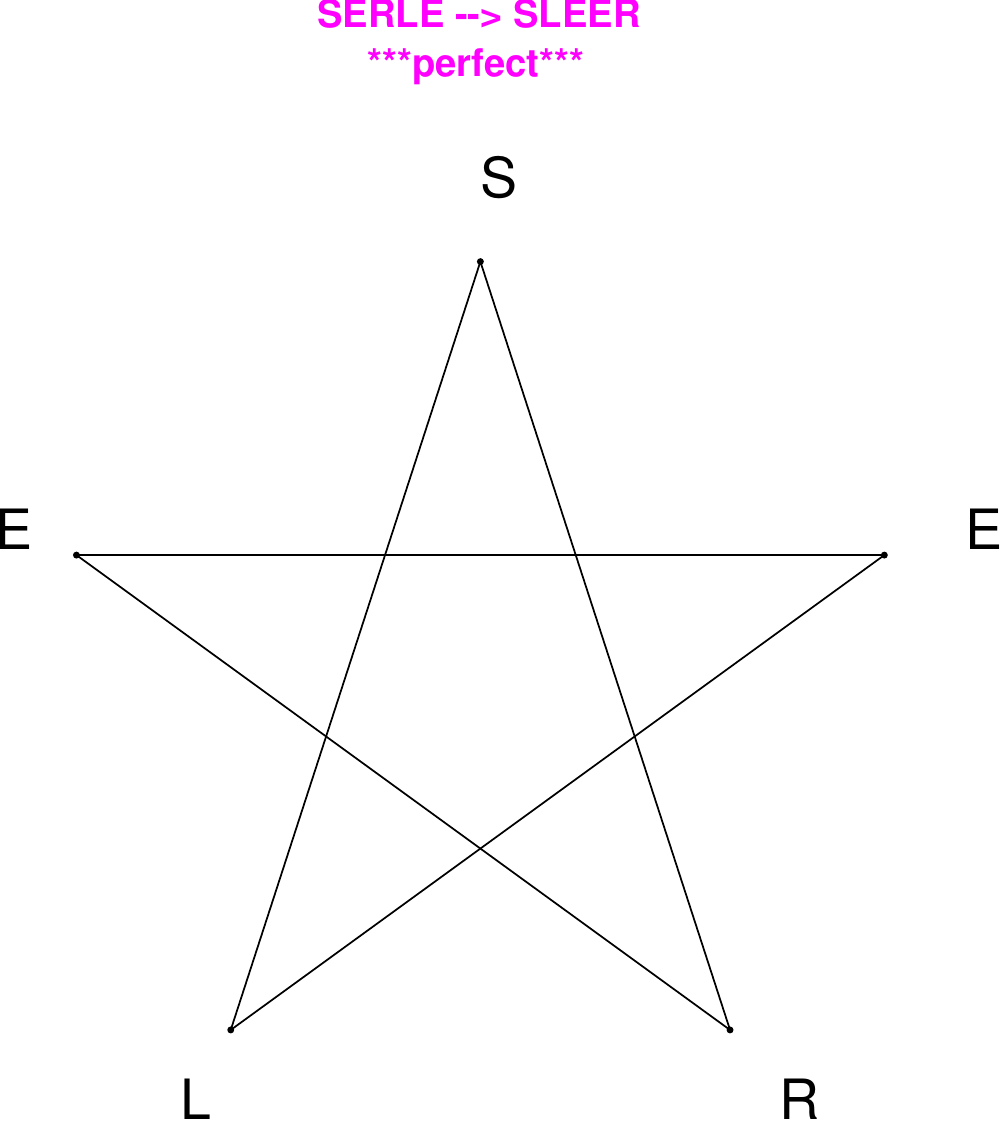}
\end{subfigure}
\end{figure}

\begin{figure}[H]
\centering
\begin{subfigure}[T]{0.19\textwidth}
\centering
\includegraphics[width=\textwidth]{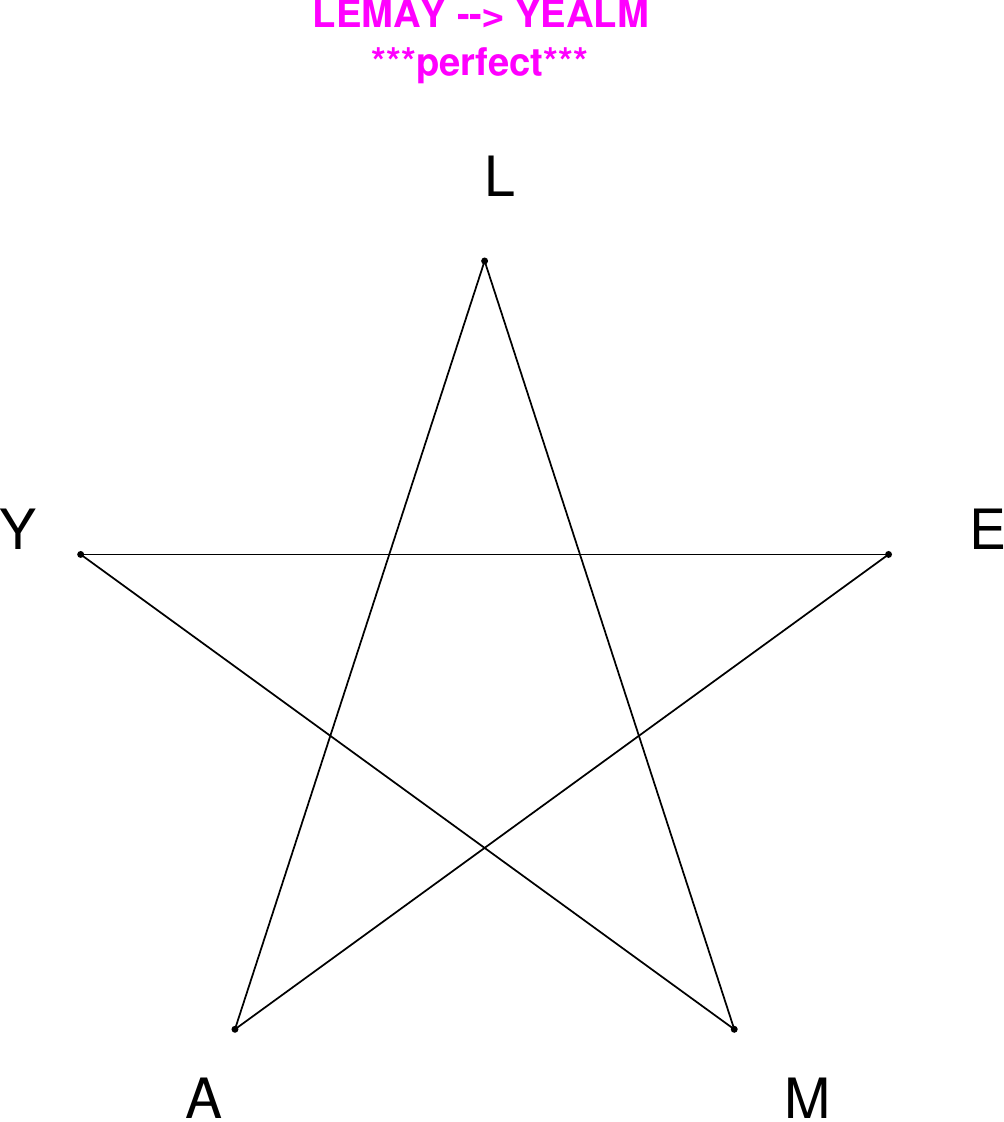}
\end{subfigure}
\hfill
\begin{subfigure}[T]{0.19\textwidth}
\centering
\includegraphics[width=\textwidth]{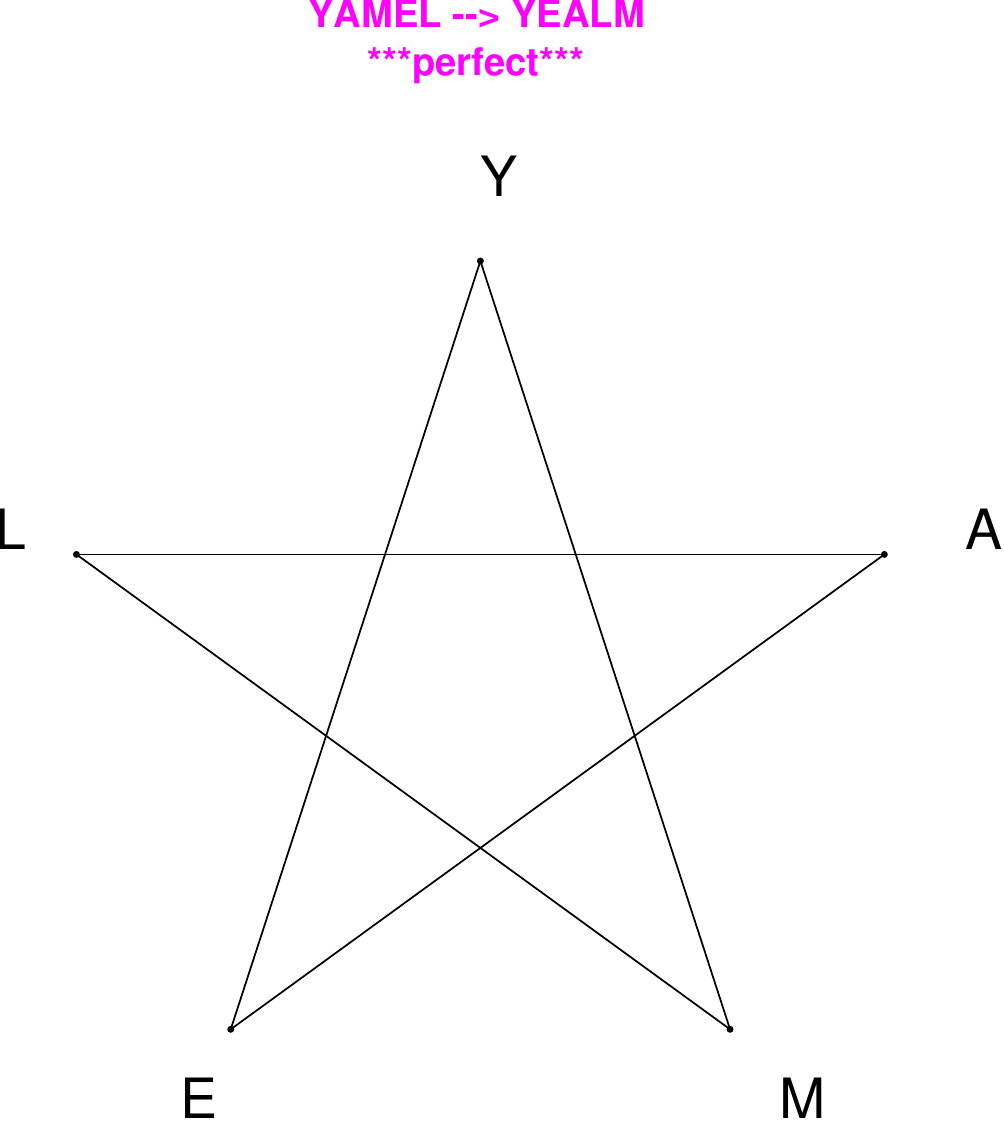}
\end{subfigure}
\hfill
\begin{subfigure}[T]{0.19\textwidth}
\centering
\includegraphics[width=\textwidth]{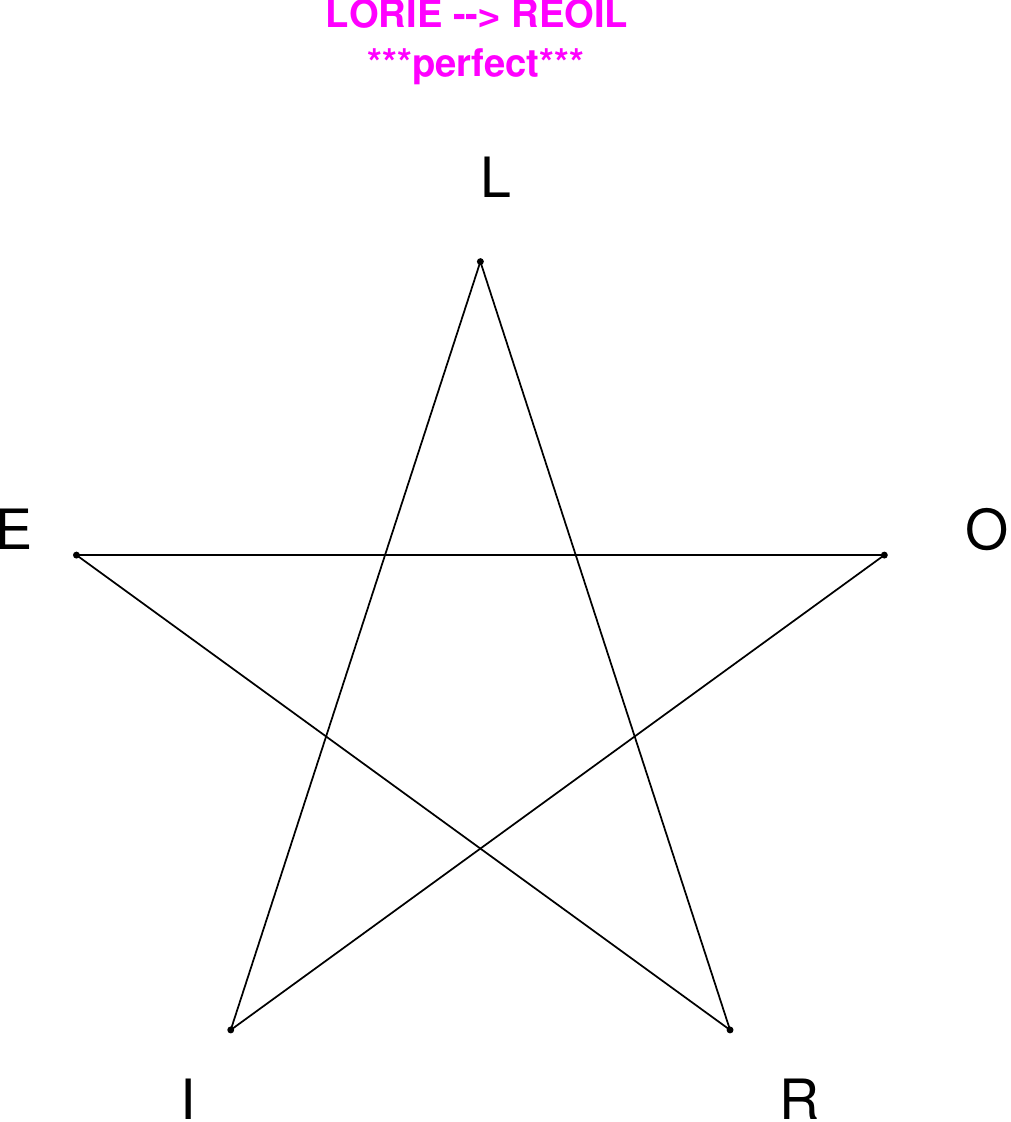}
\end{subfigure}
\hfill
\begin{subfigure}[T]{0.19\textwidth}
\centering
\includegraphics[width=\textwidth]{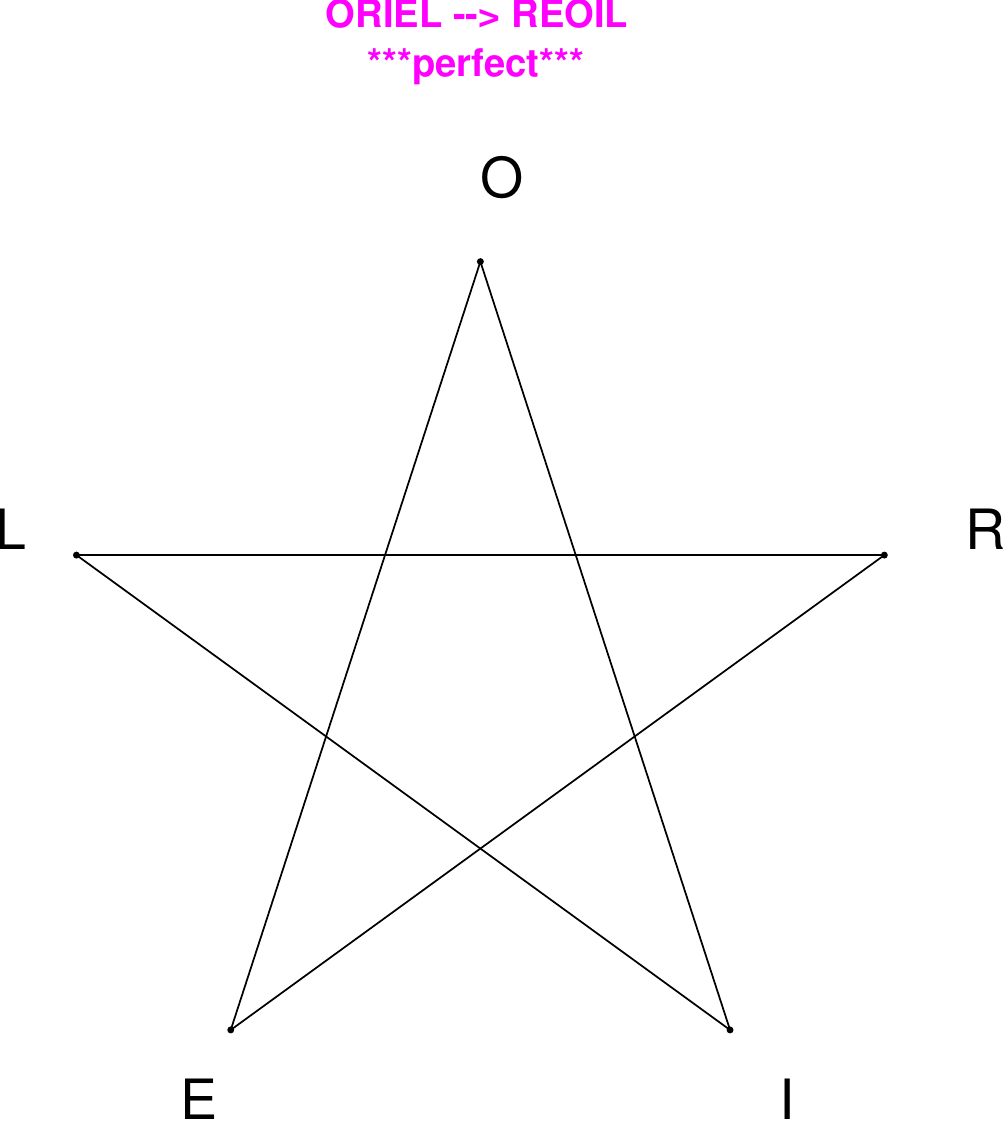}
\end{subfigure}
\hfill
\begin{subfigure}[T]{0.19\textwidth}
\centering
\includegraphics[width=\textwidth]{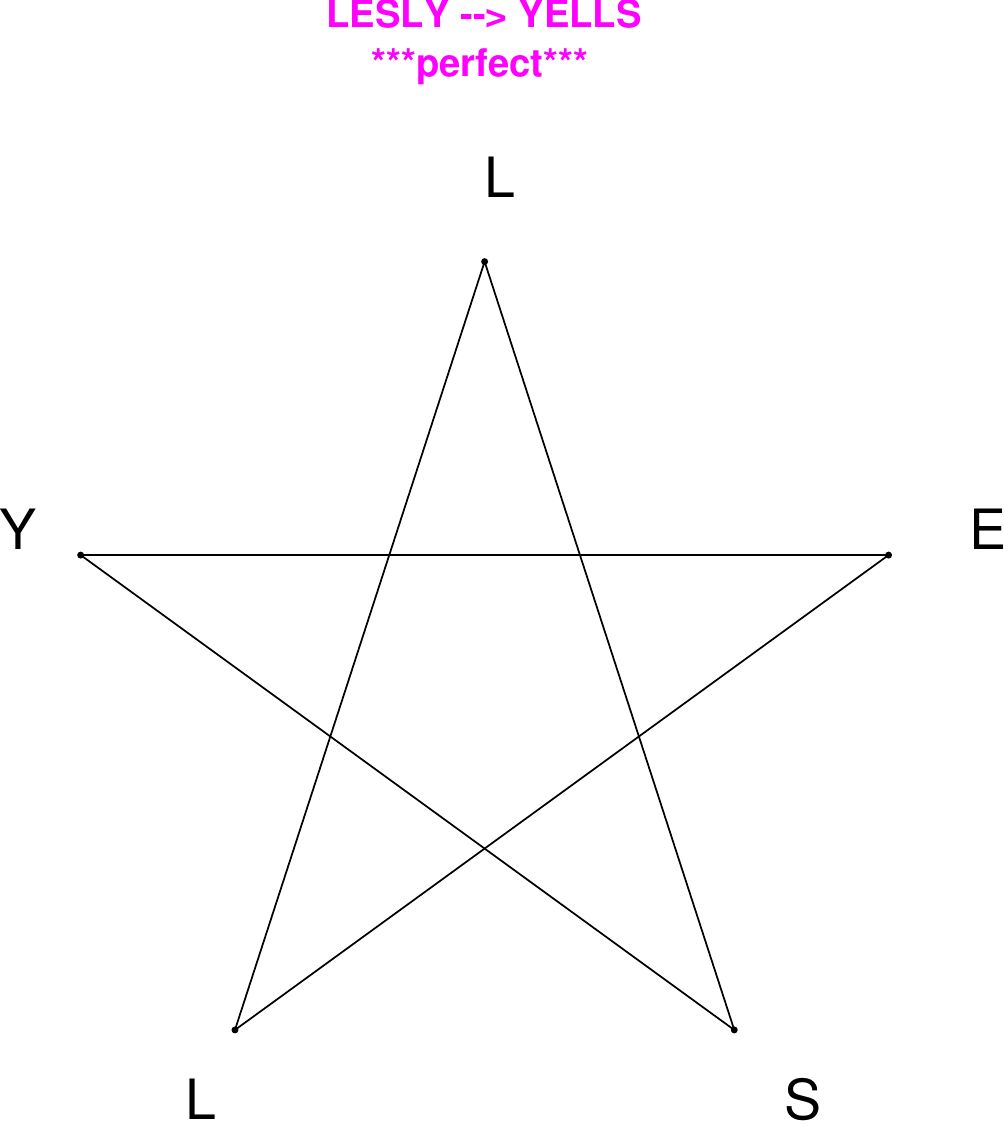}
\end{subfigure}
\end{figure}

\begin{figure}[H]
\centering
\begin{subfigure}[T]{0.19\textwidth}
\centering
\includegraphics[width=\textwidth]{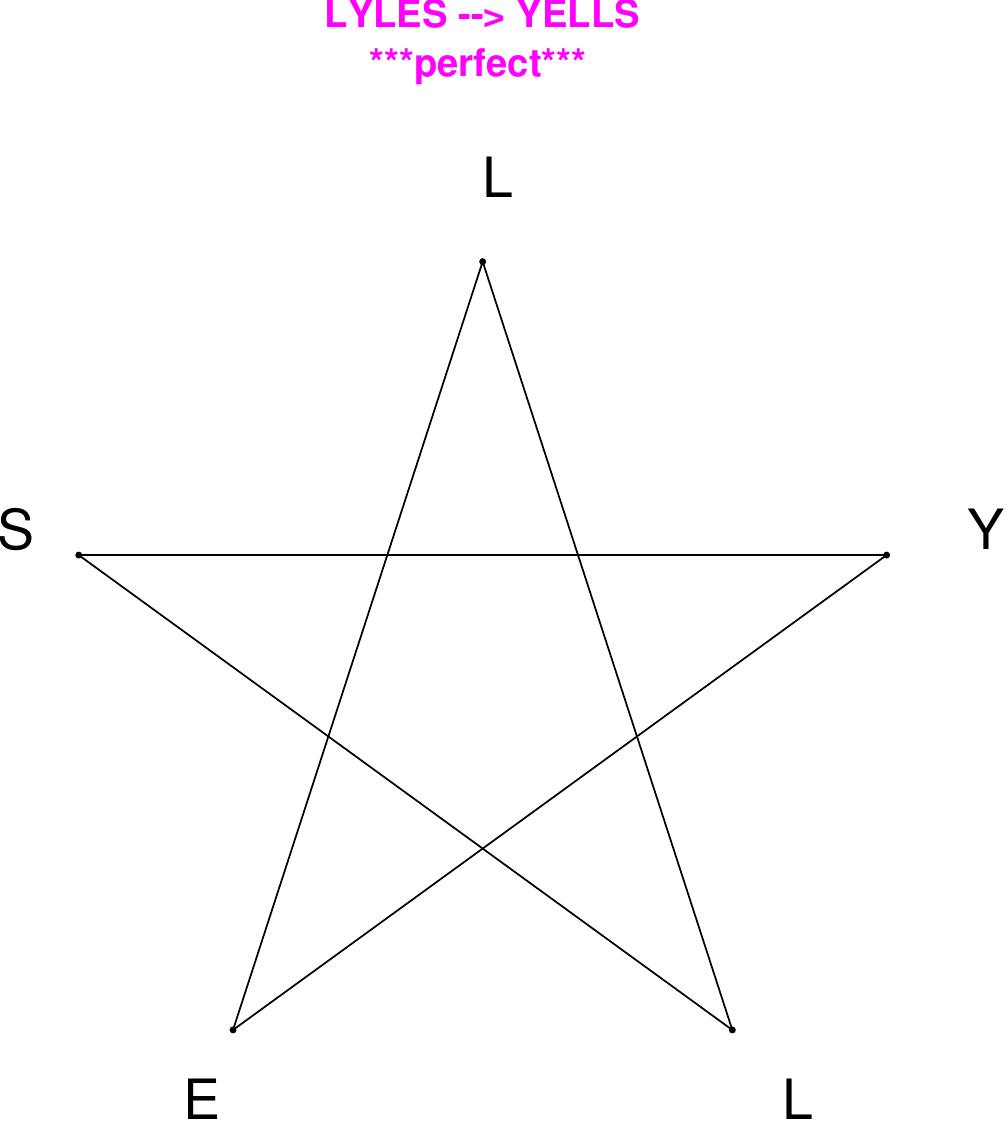}
\end{subfigure}
\hfill
\begin{subfigure}[T]{0.19\textwidth}
\centering
\includegraphics[width=\textwidth]{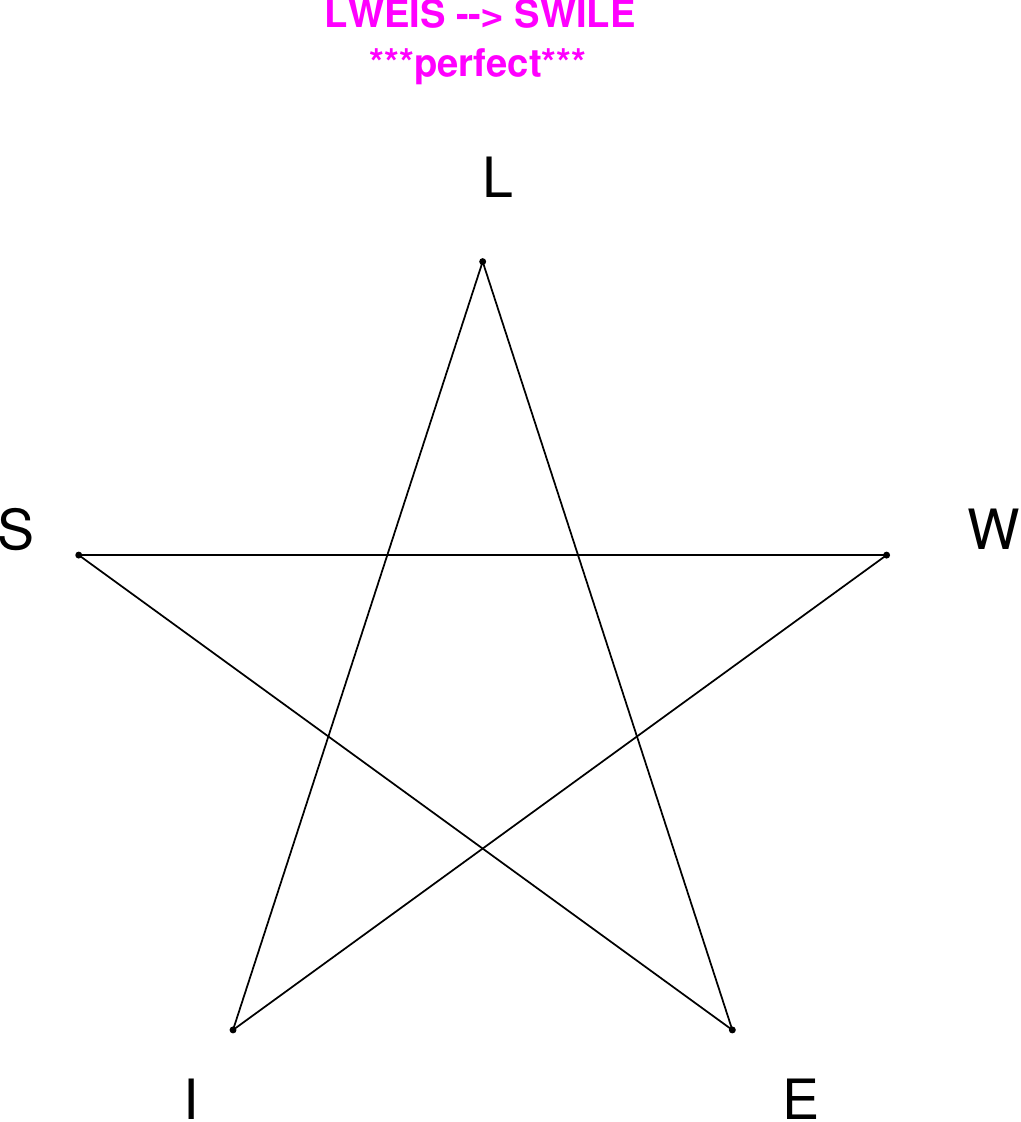}
\end{subfigure}
\hfill
\begin{subfigure}[T]{0.19\textwidth}
\centering
\includegraphics[width=\textwidth]{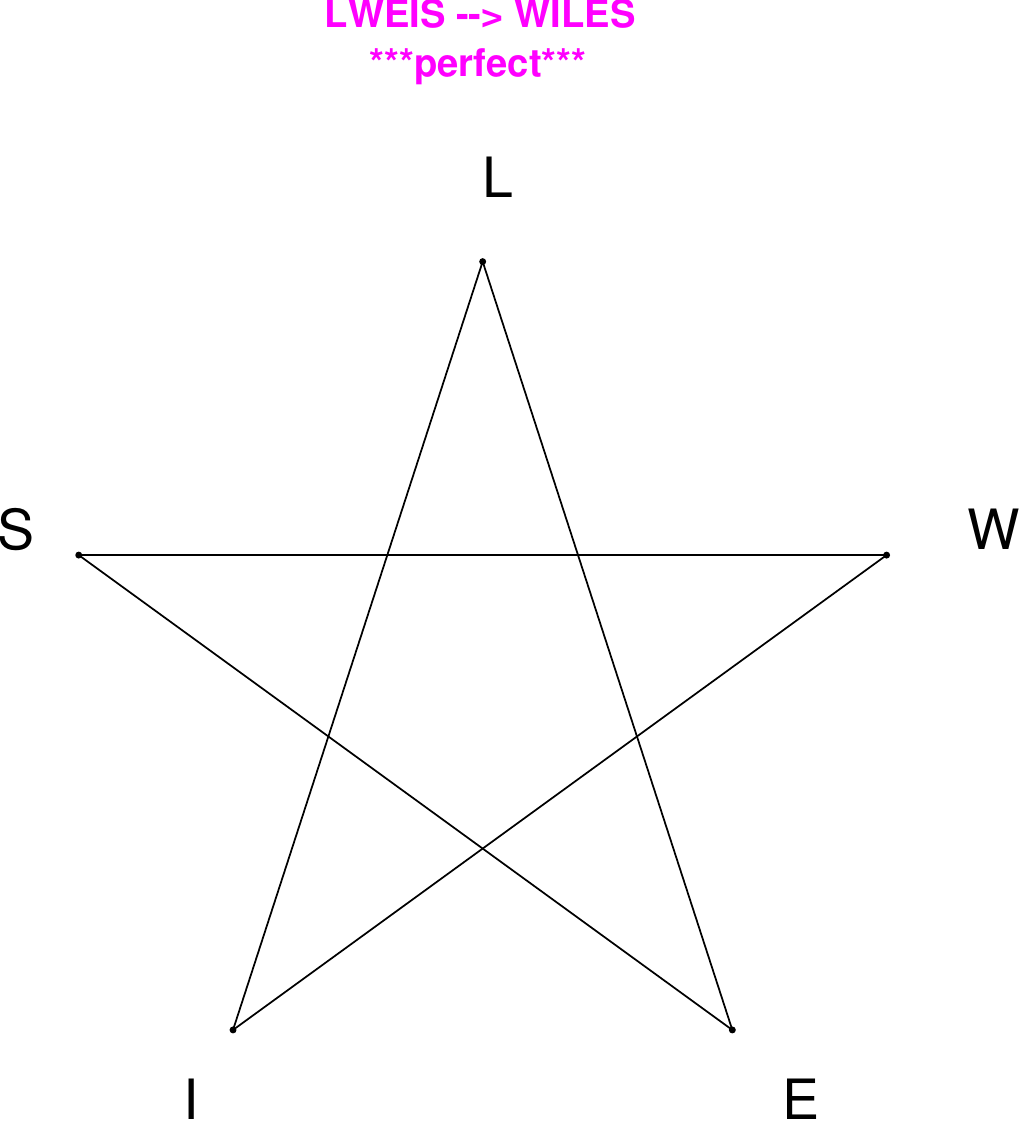}
\end{subfigure}
\hfill
\begin{subfigure}[T]{0.19\textwidth}
\centering
\includegraphics[width=\textwidth]{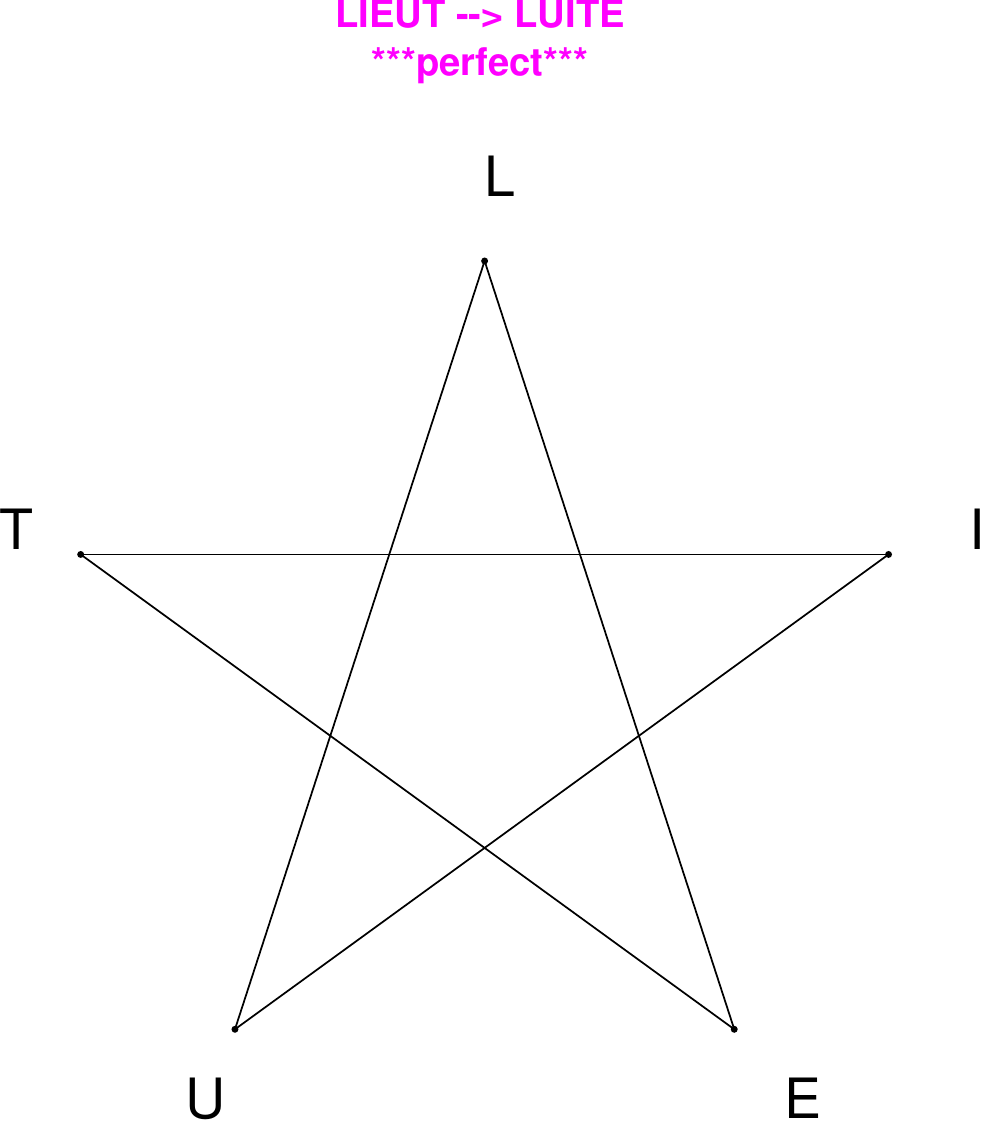}
\end{subfigure}
\hfill
\begin{subfigure}[T]{0.19\textwidth}
\centering
\includegraphics[width=\textwidth]{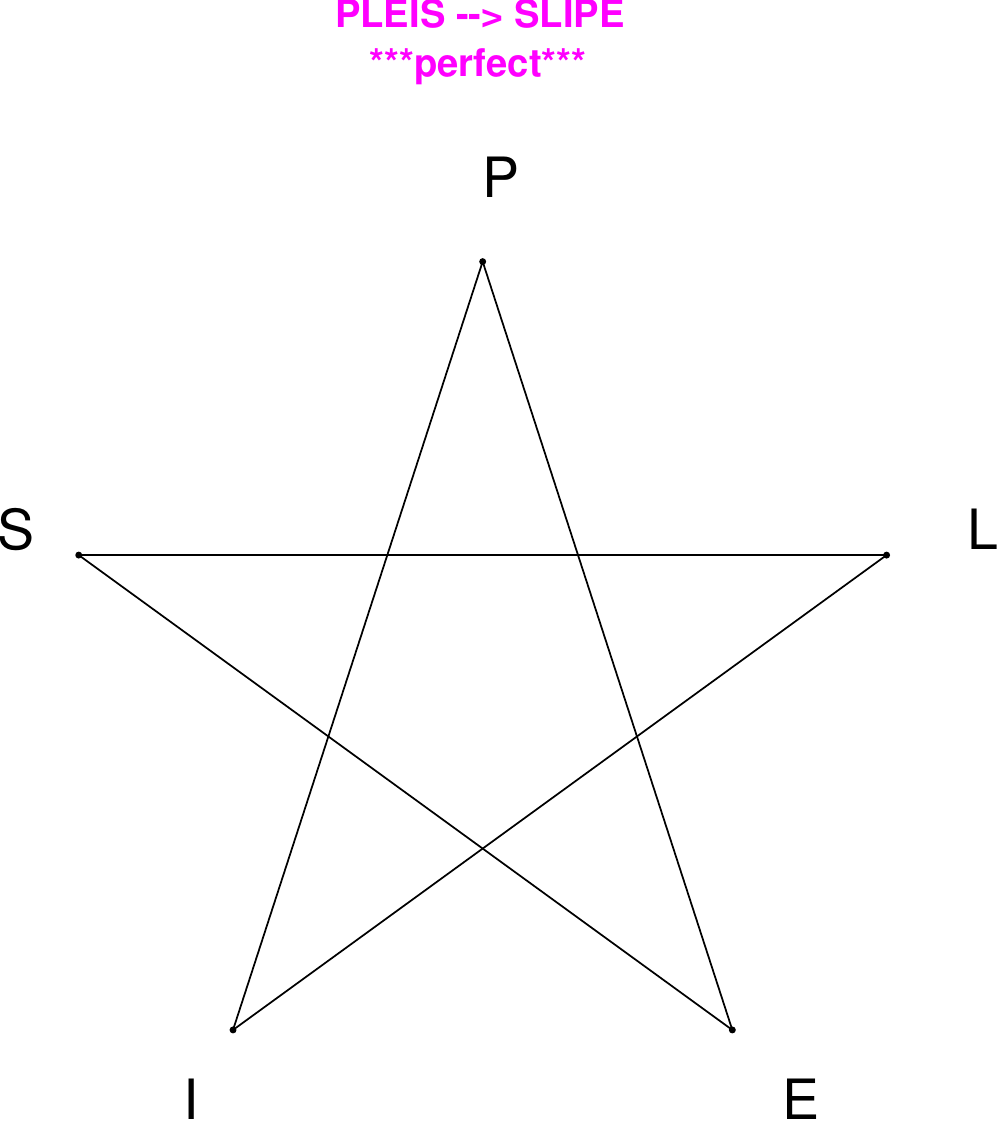}
\end{subfigure}
\end{figure}

\begin{figure}[H]
\centering
\begin{subfigure}[T]{0.19\textwidth}
\centering
\includegraphics[width=\textwidth]{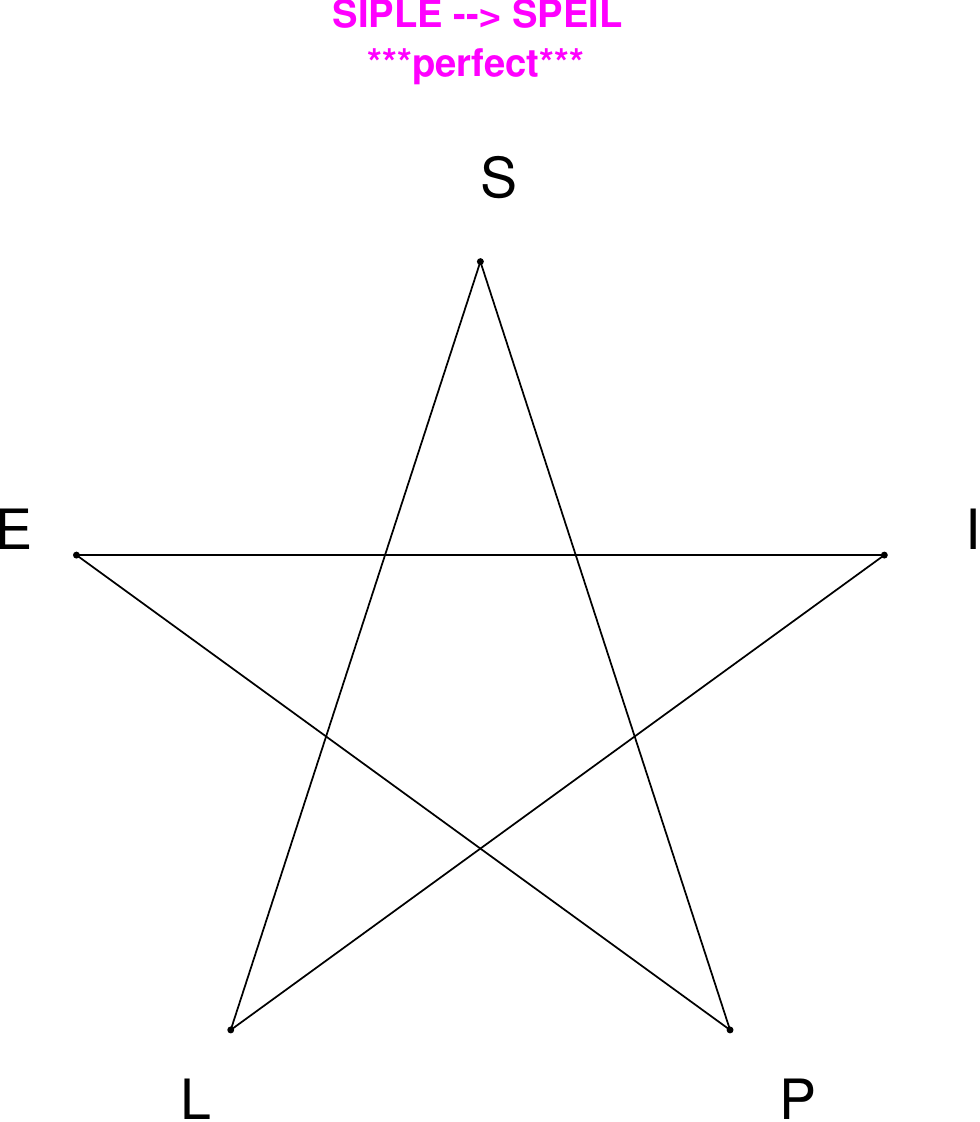}
\end{subfigure}
\hfill
\begin{subfigure}[T]{0.19\textwidth}
\centering
\includegraphics[width=\textwidth]{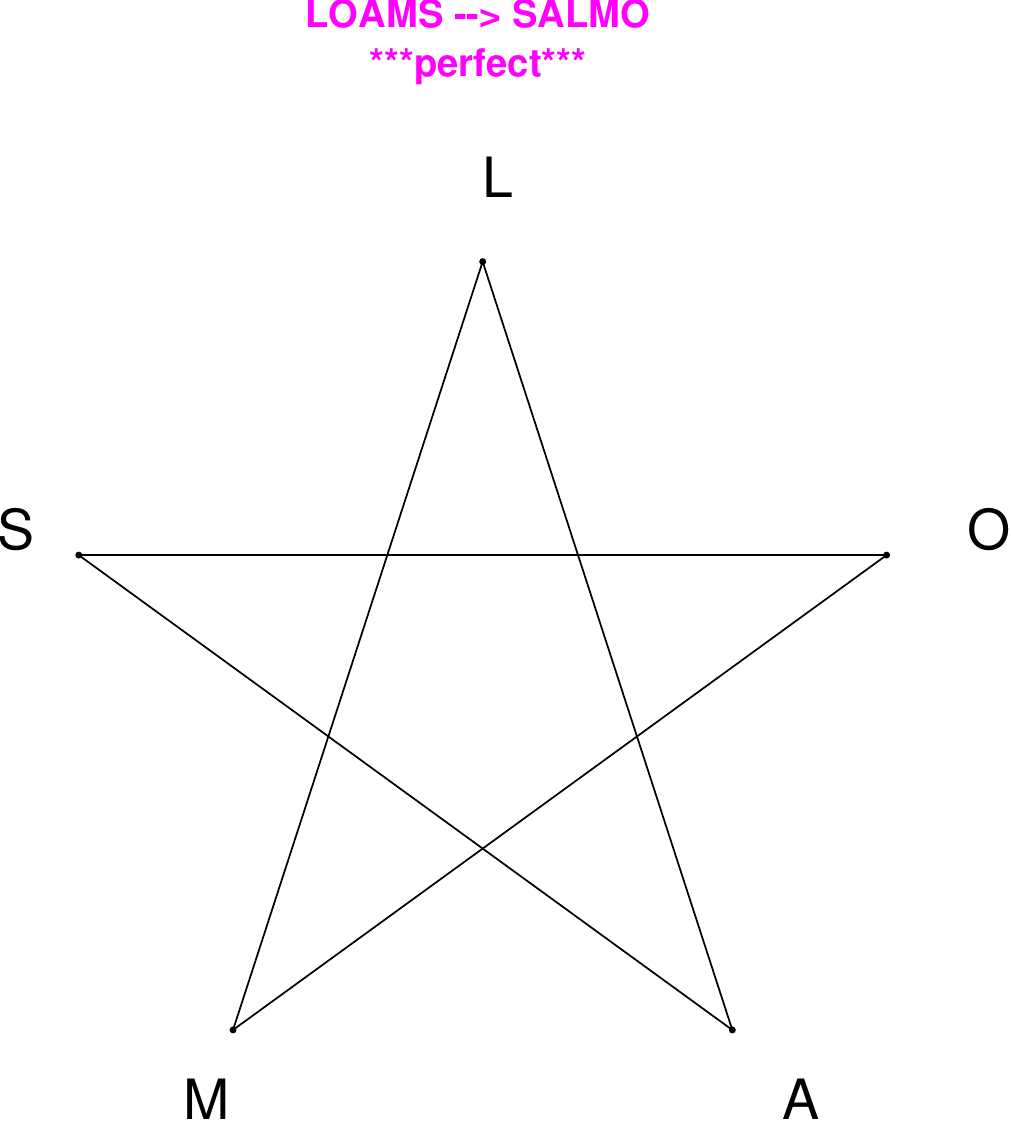}
\end{subfigure}
\hfill
\begin{subfigure}[T]{0.19\textwidth}
\centering
\includegraphics[width=\textwidth]{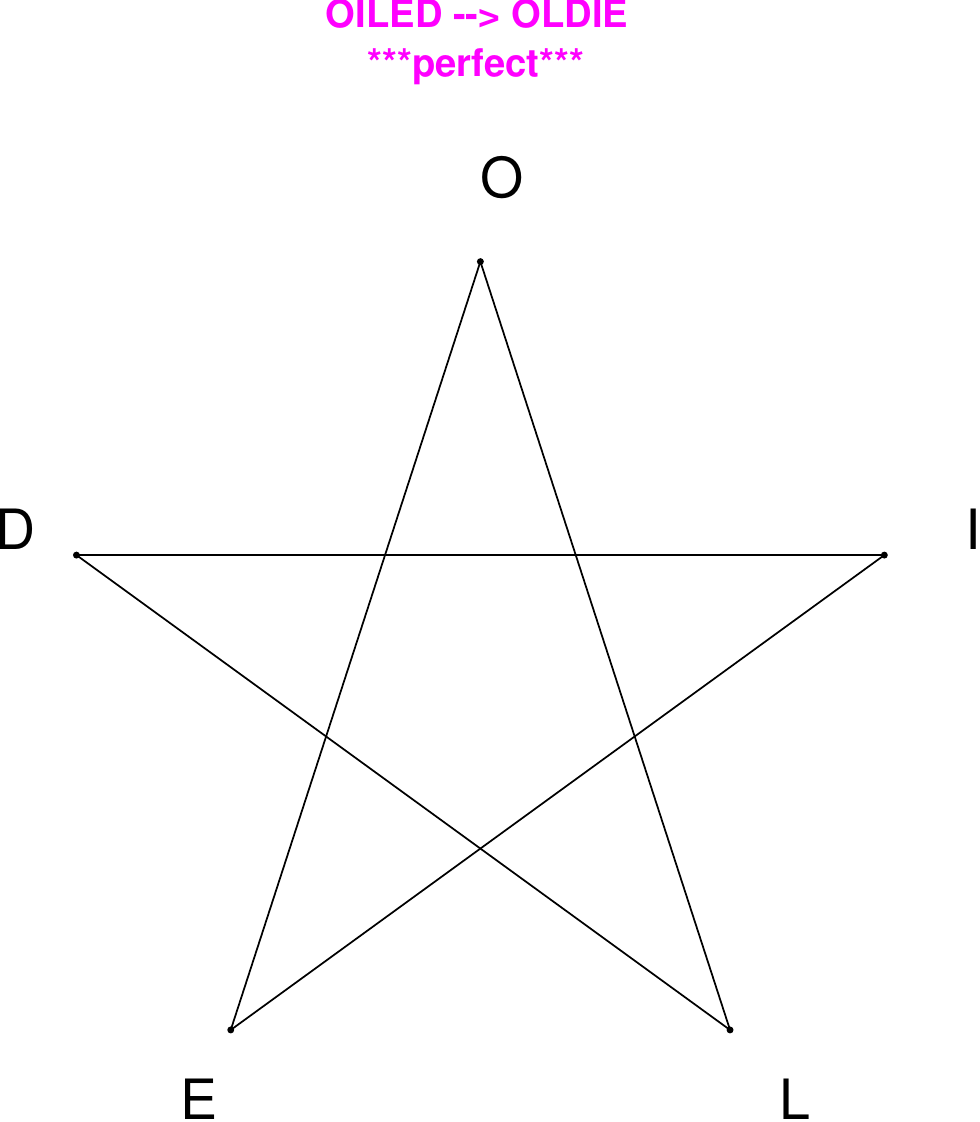}
\end{subfigure}
\hfill
\begin{subfigure}[T]{0.19\textwidth}
\centering
\includegraphics[width=\textwidth]{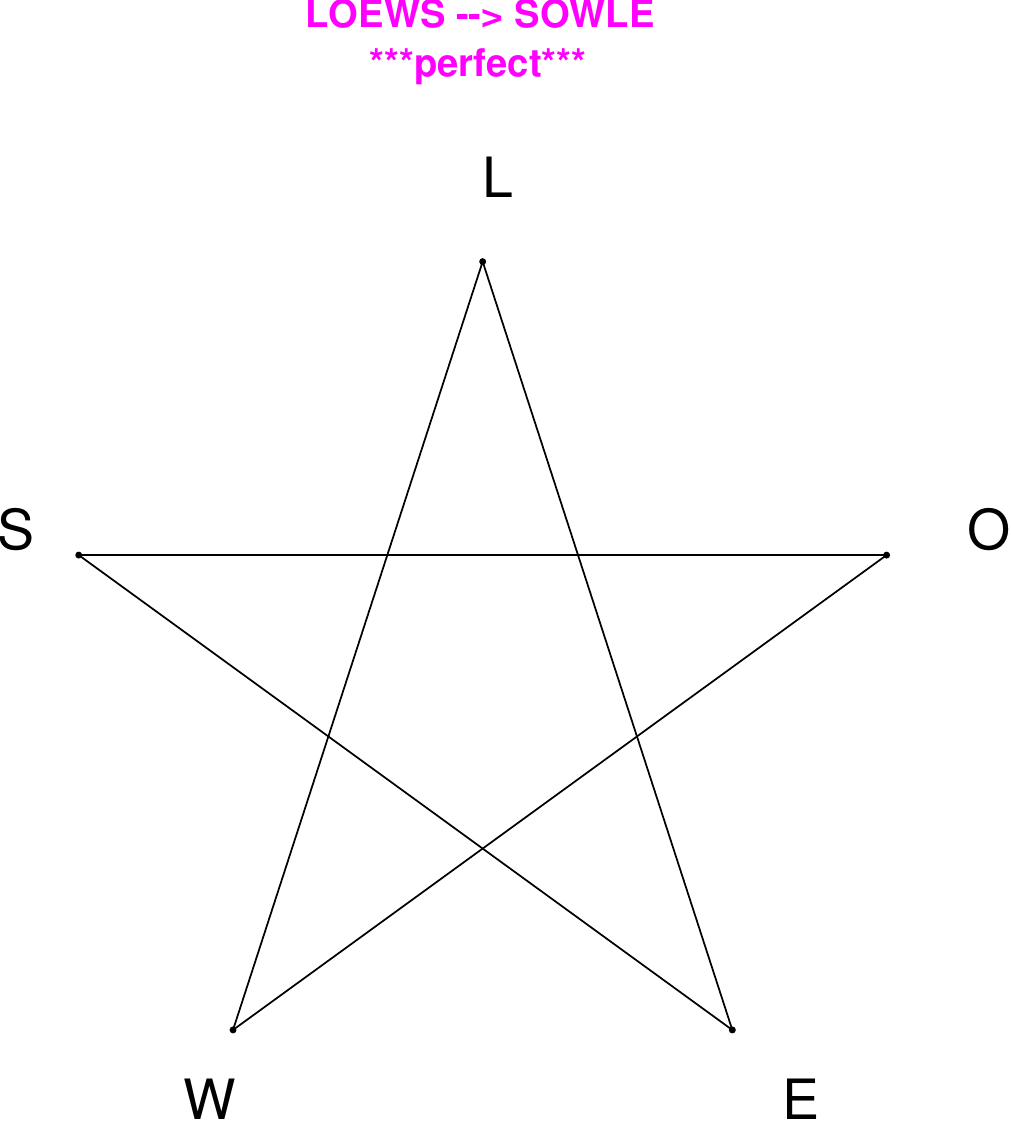}
\end{subfigure}
\hfill
\begin{subfigure}[T]{0.19\textwidth}
\centering
\includegraphics[width=\textwidth]{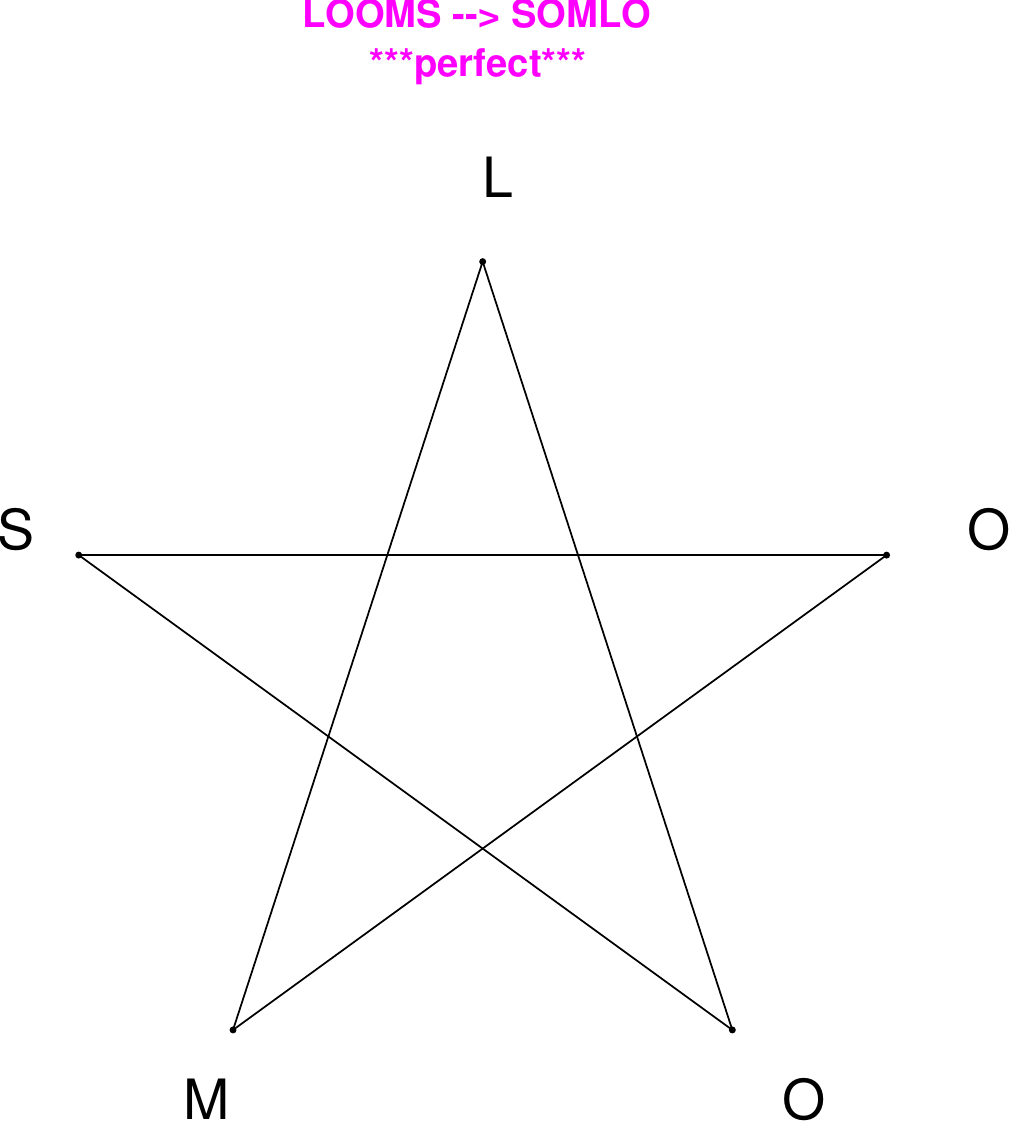}
\end{subfigure}
\end{figure}

\begin{figure}[H]
\centering
\begin{subfigure}[T]{0.19\textwidth}
\centering
\includegraphics[width=\textwidth]{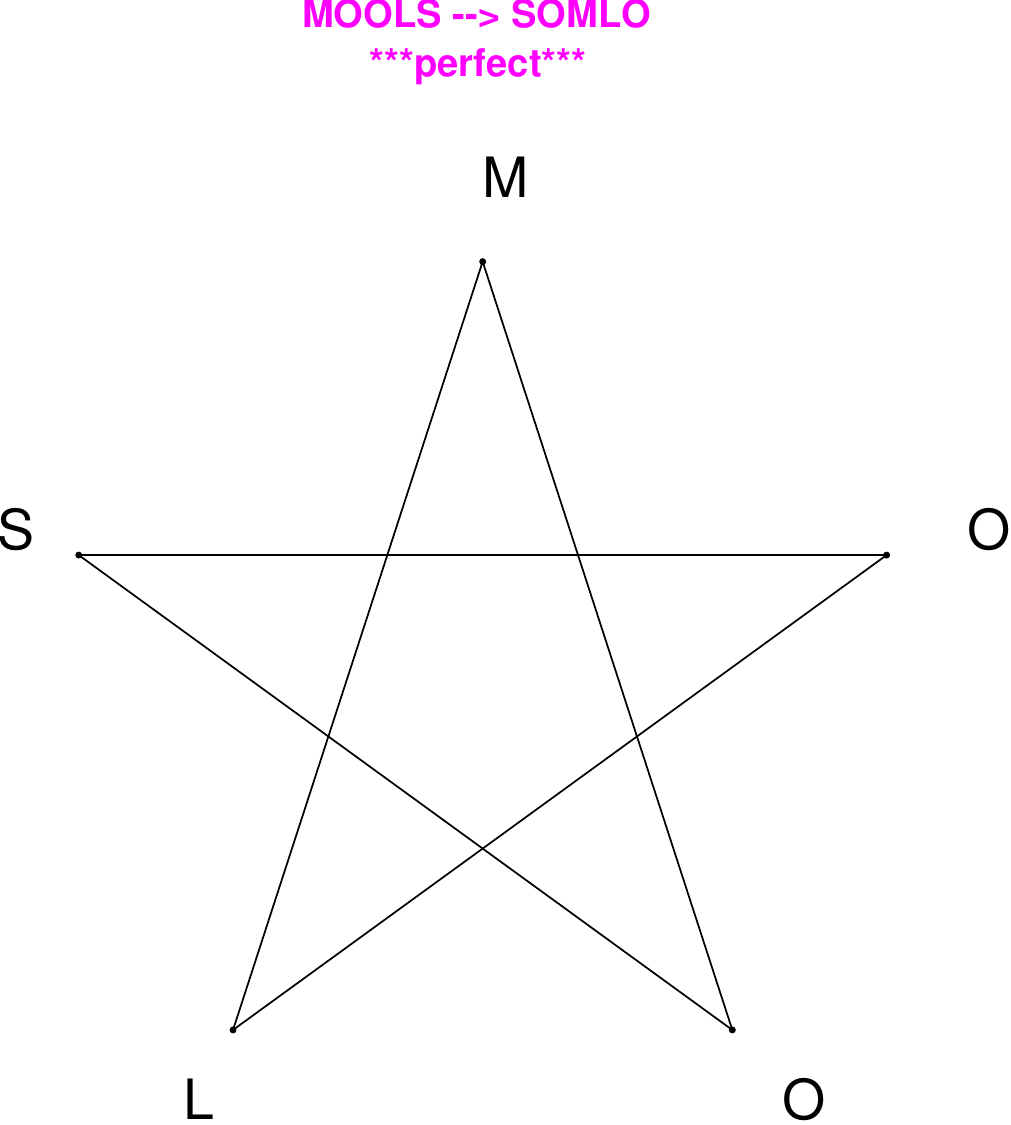}
\end{subfigure}
\hfill
\begin{subfigure}[T]{0.19\textwidth}
\centering
\includegraphics[width=\textwidth]{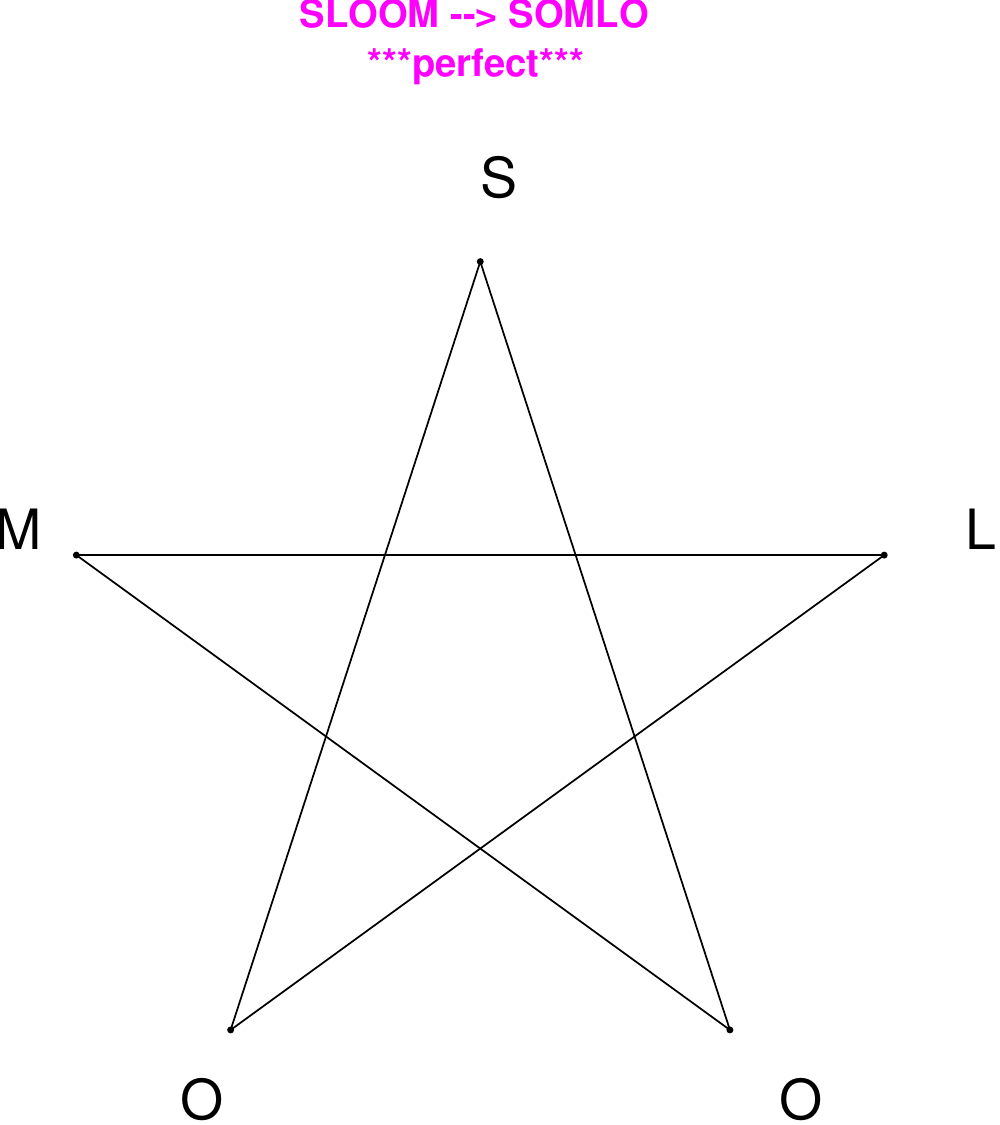}
\end{subfigure}
\hfill
\begin{subfigure}[T]{0.19\textwidth}
\centering
\includegraphics[width=\textwidth]{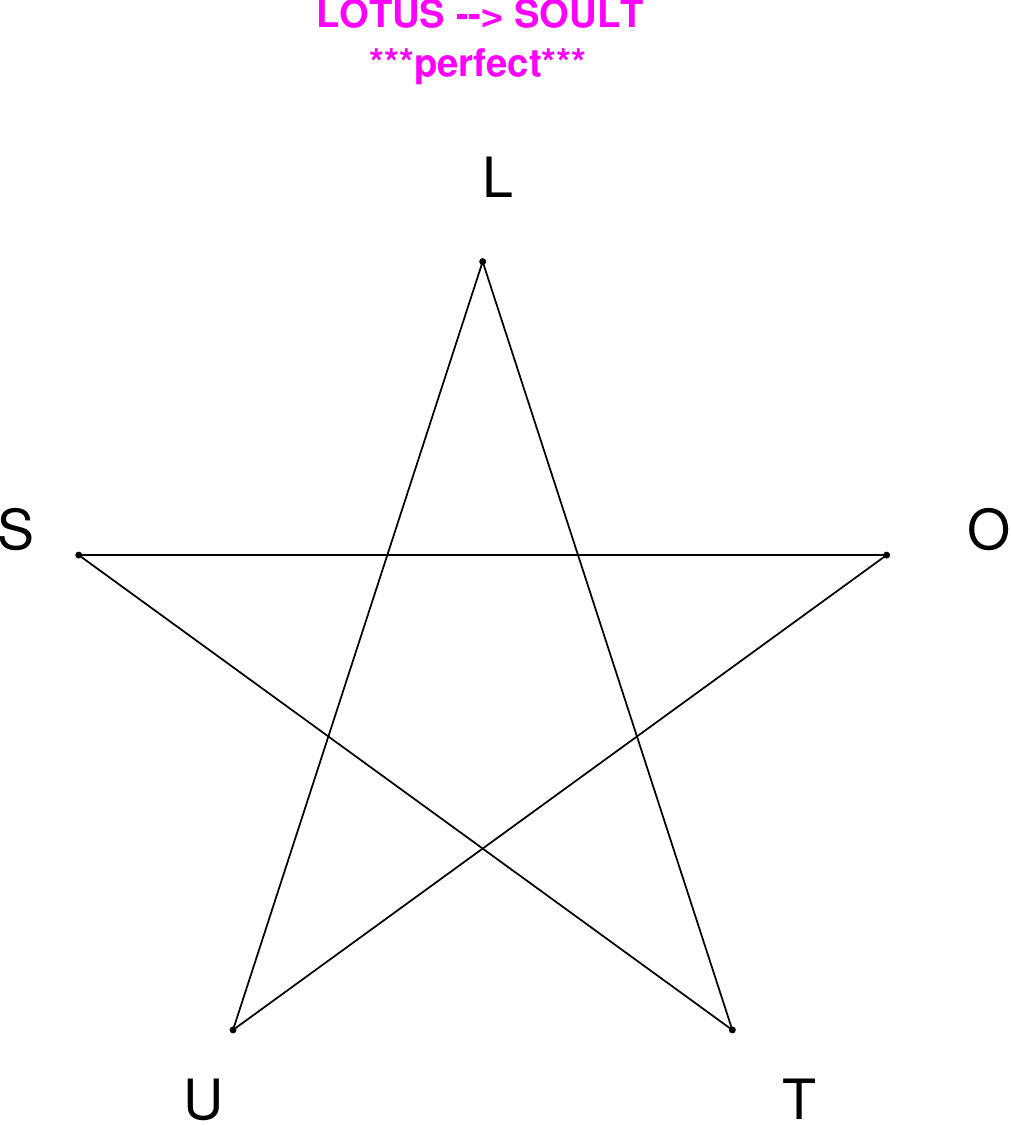}
\end{subfigure}
\hfill
\begin{subfigure}[T]{0.19\textwidth}
\centering
\includegraphics[width=\textwidth]{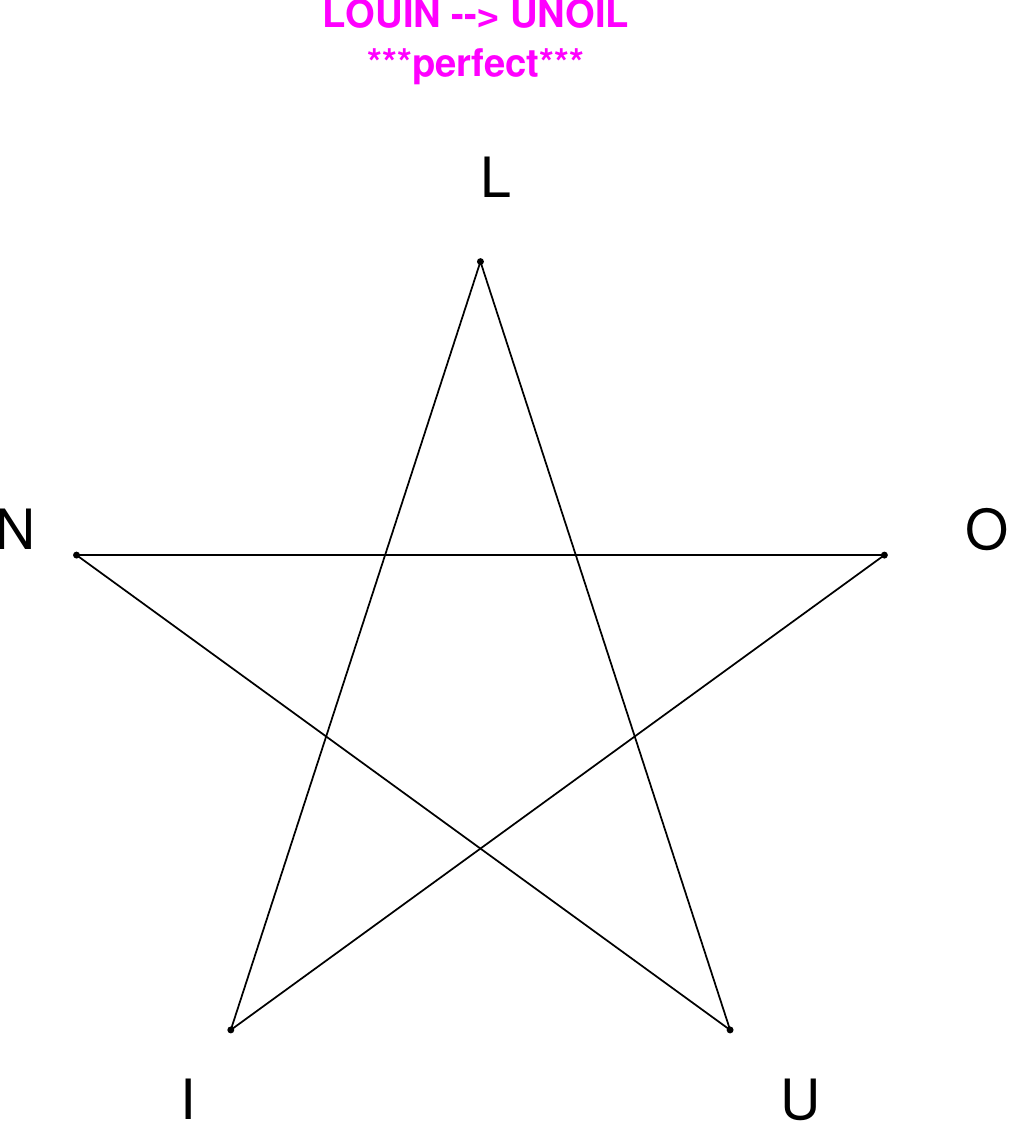}
\end{subfigure}
\hfill
\begin{subfigure}[T]{0.19\textwidth}
\centering
\includegraphics[width=\textwidth]{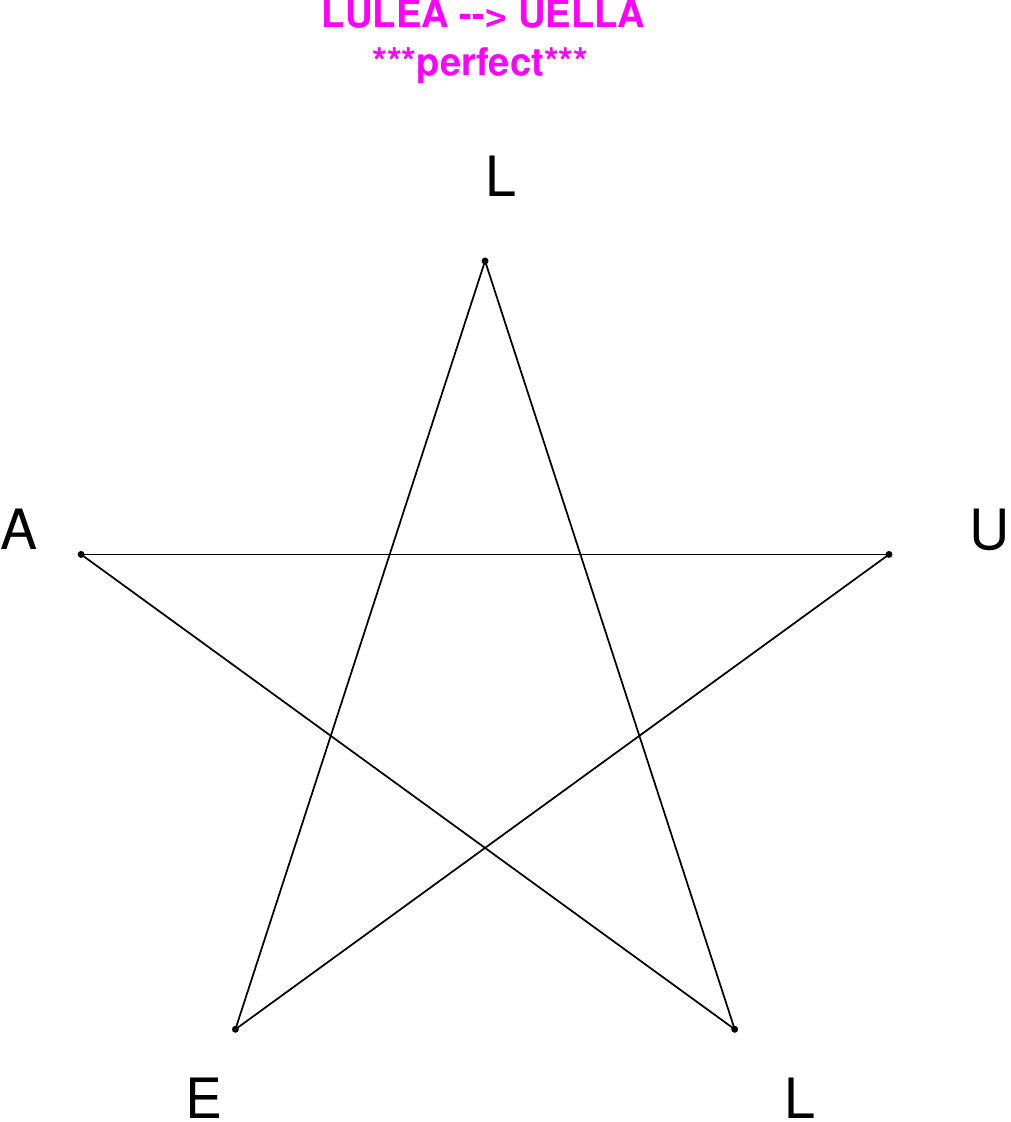}
\end{subfigure}
\end{figure}

\begin{figure}[H]
\centering
\begin{subfigure}[T]{0.19\textwidth}
\centering
\includegraphics[width=\textwidth]{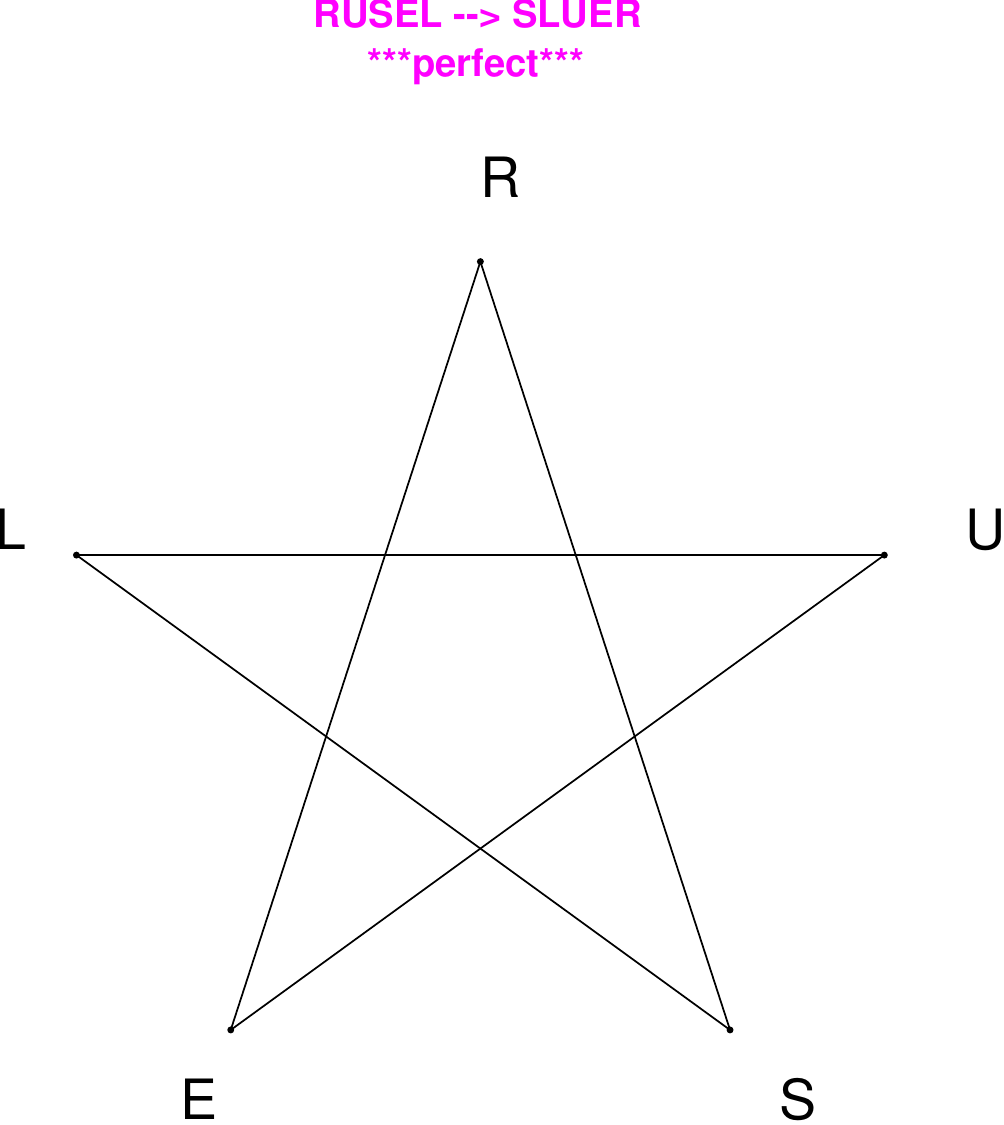}
\end{subfigure}
\hfill
\begin{subfigure}[T]{0.19\textwidth}
\centering
\includegraphics[width=\textwidth]{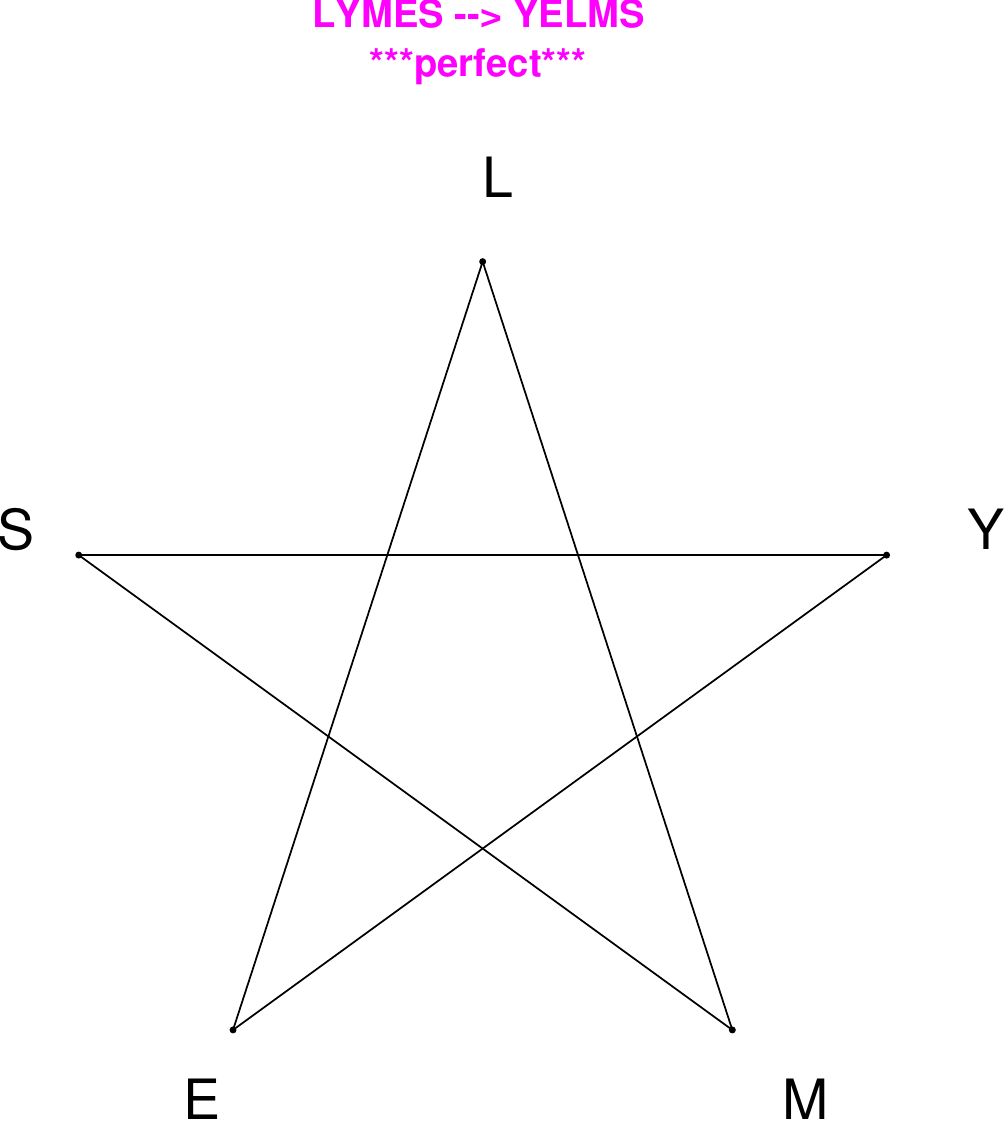}
\end{subfigure}
\hfill
\begin{subfigure}[T]{0.19\textwidth}
\centering
\includegraphics[width=\textwidth]{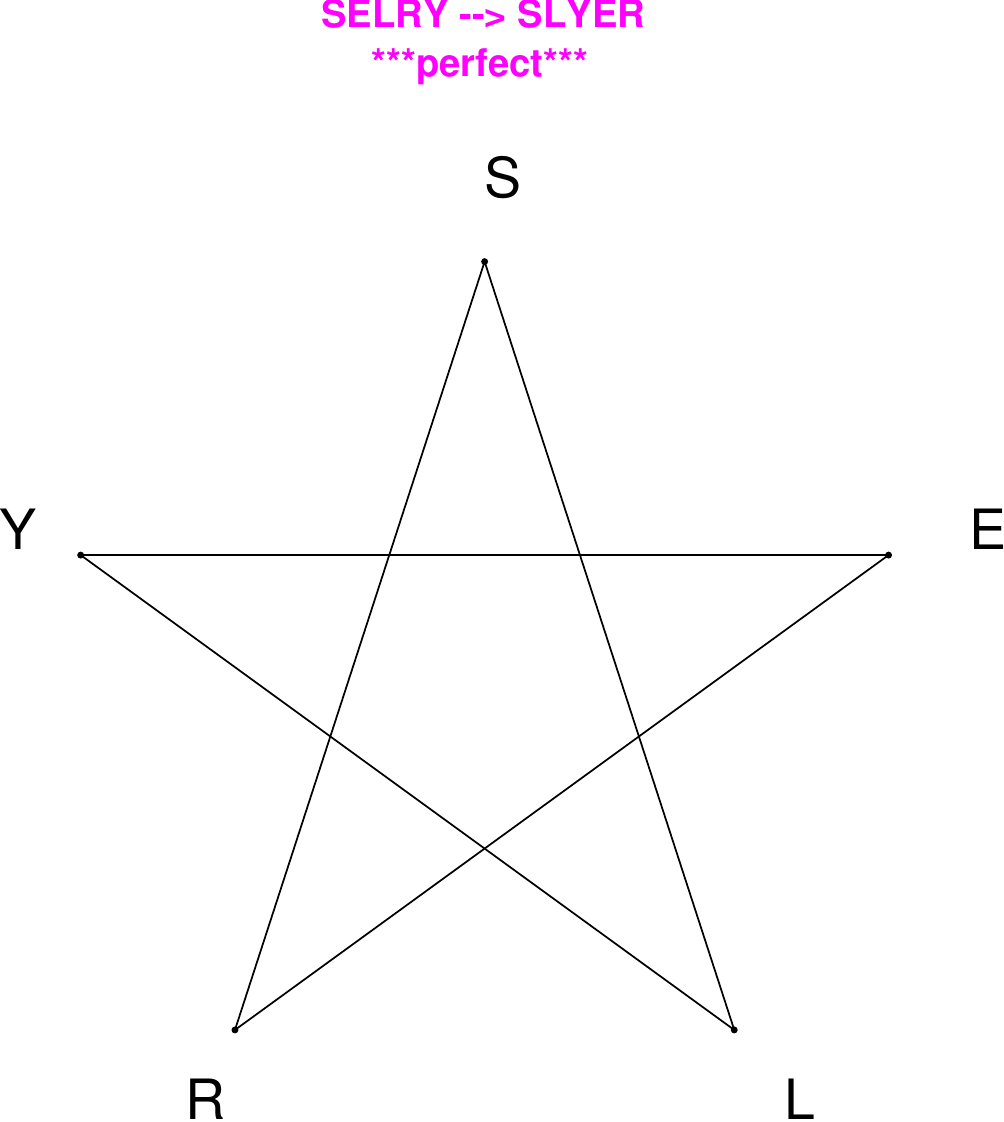}
\end{subfigure}
\hfill
\begin{subfigure}[T]{0.19\textwidth}
\centering
\includegraphics[width=\textwidth]{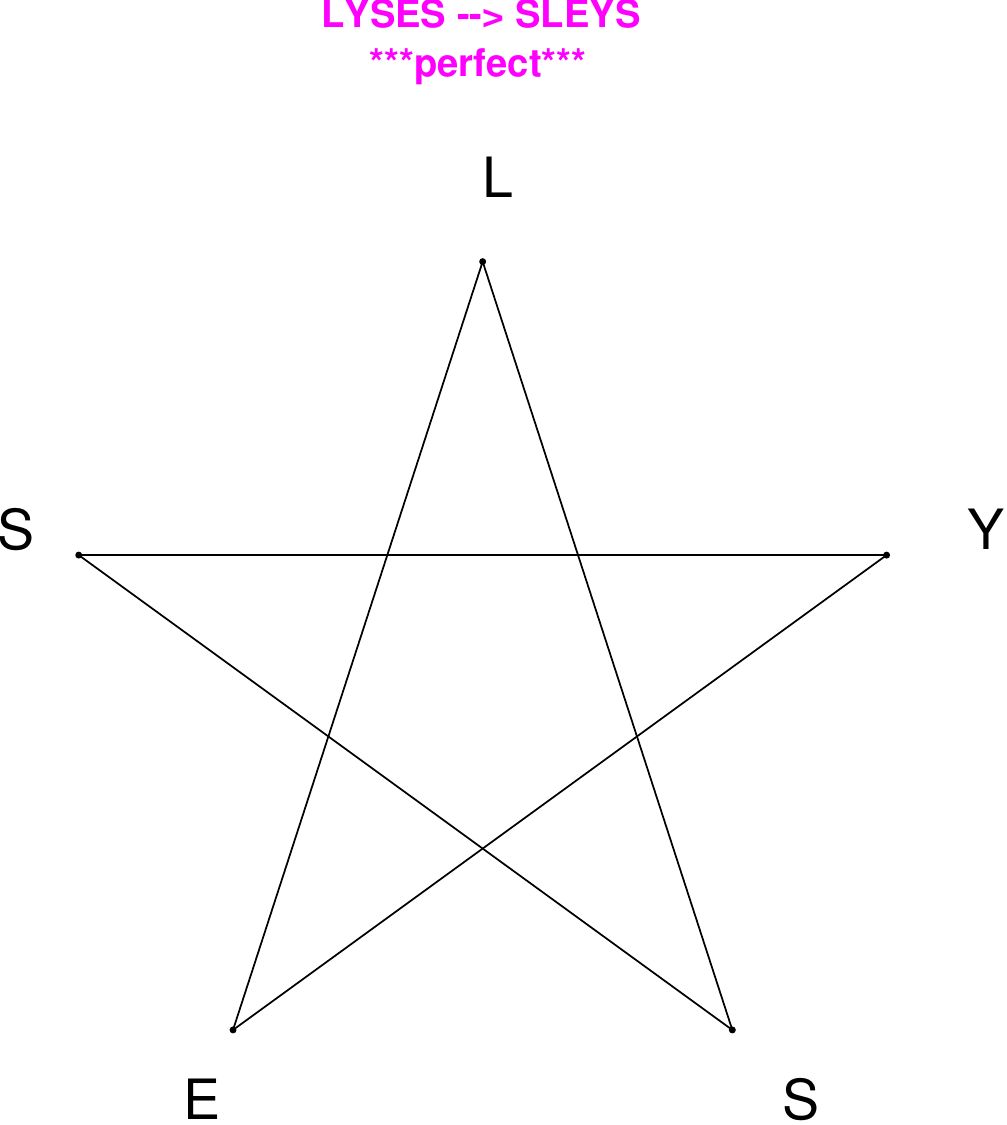}
\end{subfigure}
\hfill
\begin{subfigure}[T]{0.19\textwidth}
\centering
\includegraphics[width=\textwidth]{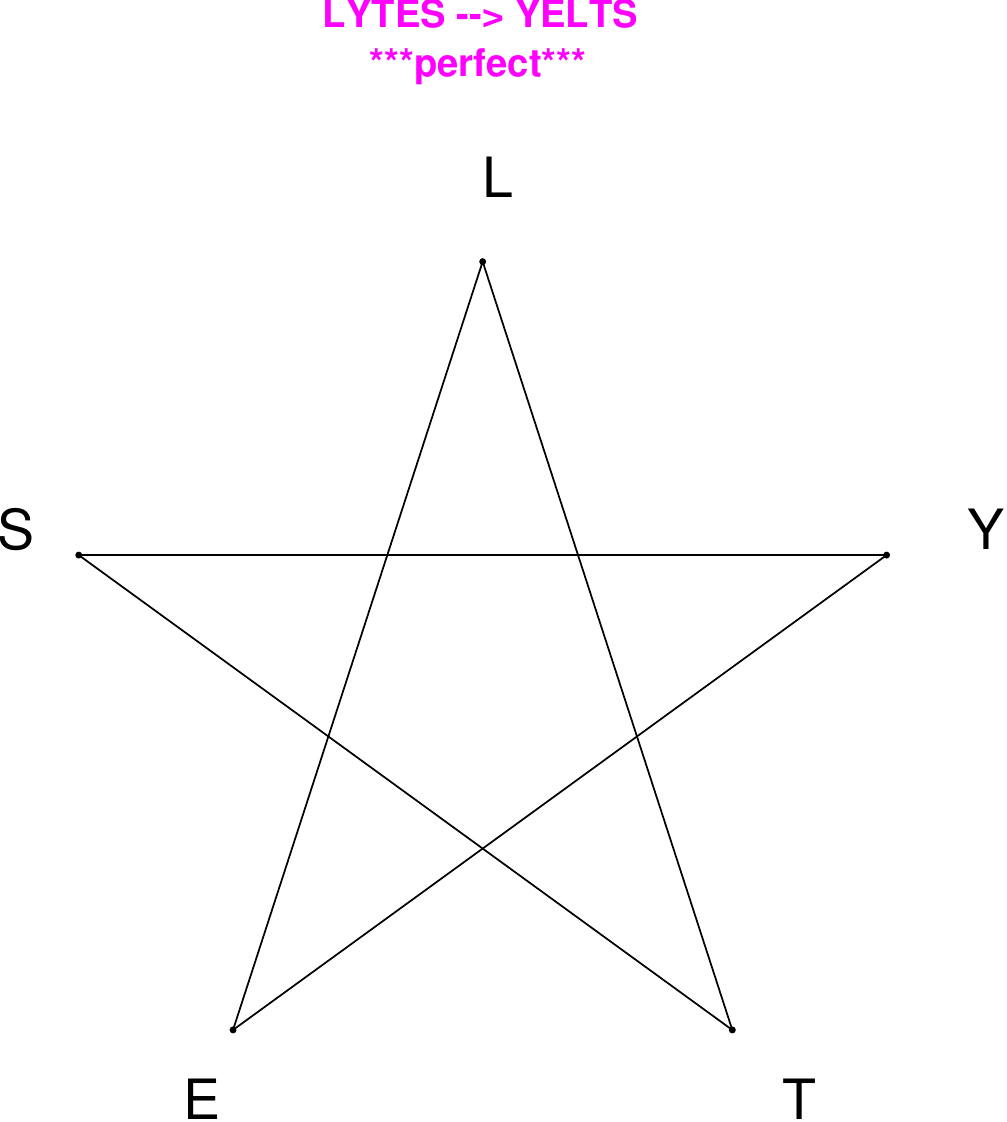}
\end{subfigure}
\end{figure}

\begin{figure}[H]
\centering
\begin{subfigure}[T]{0.19\textwidth}
\centering
\includegraphics[width=\textwidth]{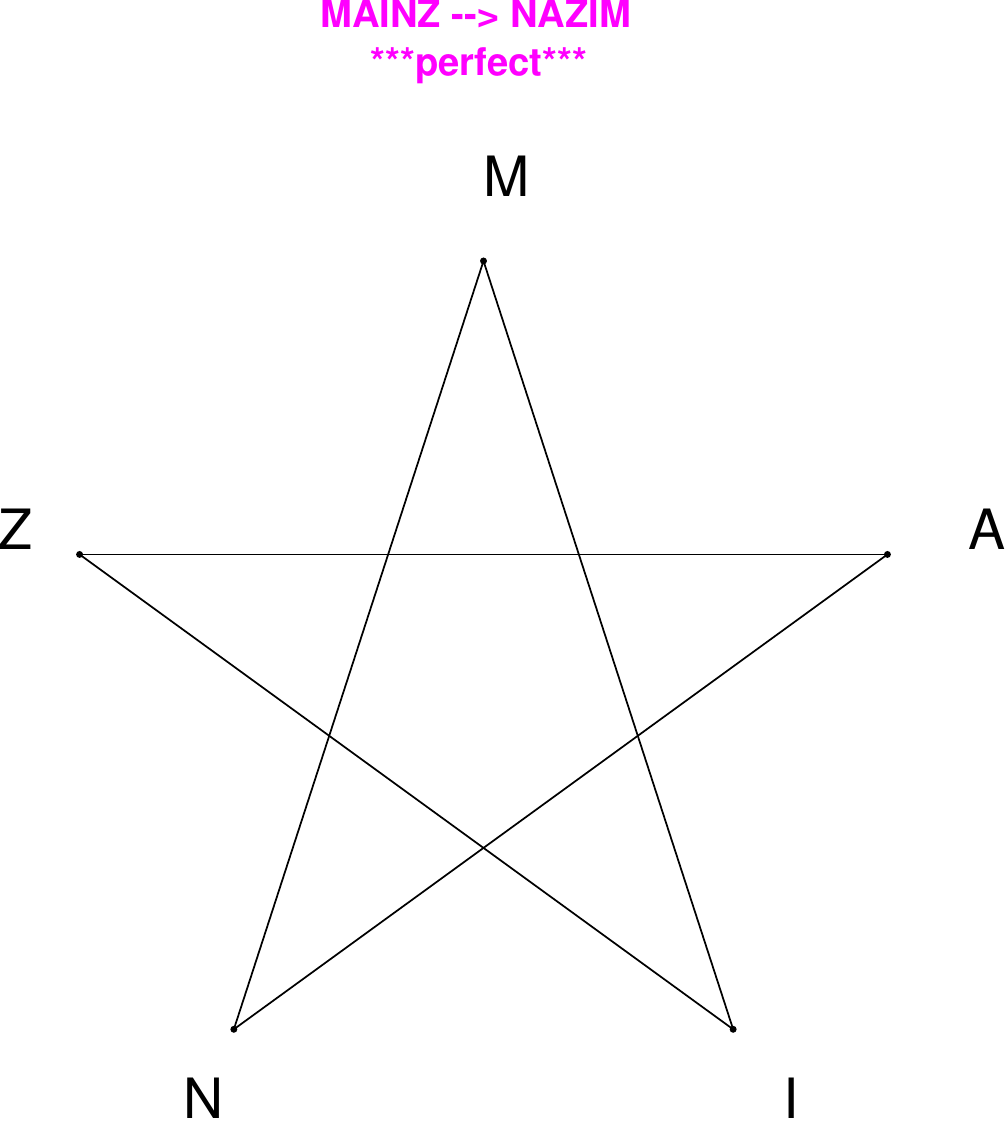}
\end{subfigure}
\hfill
\begin{subfigure}[T]{0.19\textwidth}
\centering
\includegraphics[width=\textwidth]{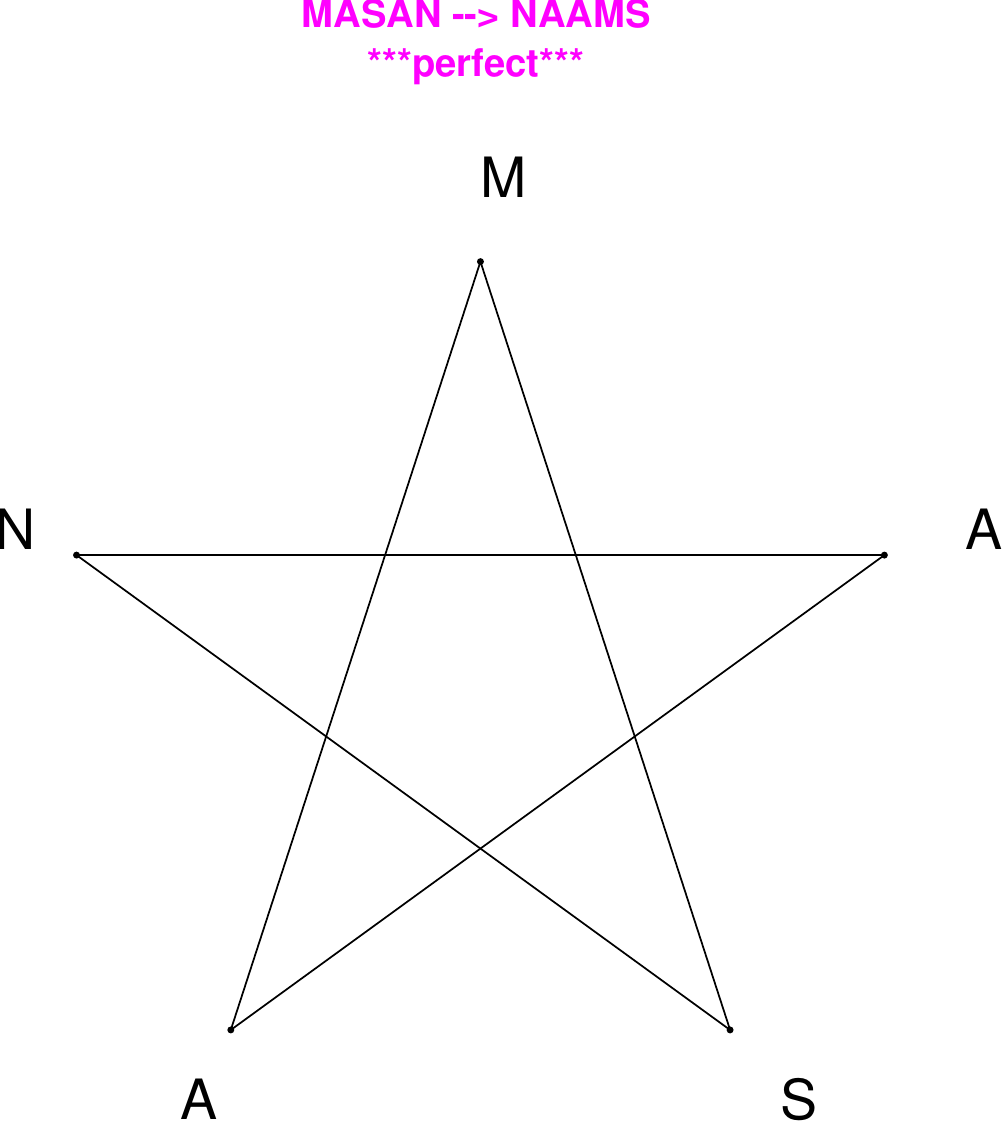}
\end{subfigure}
\hfill
\begin{subfigure}[T]{0.19\textwidth}
\centering
\includegraphics[width=\textwidth]{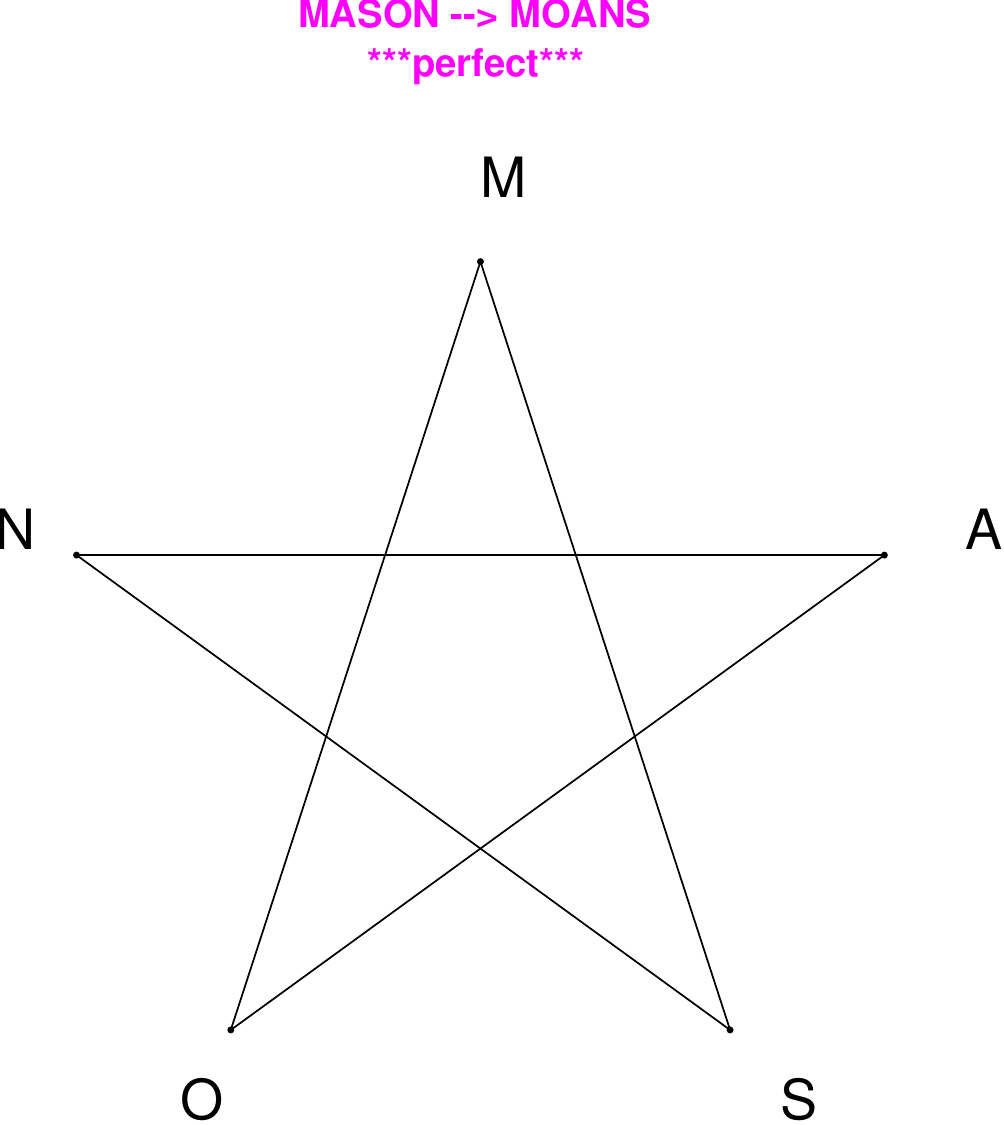}
\end{subfigure}
\hfill
\begin{subfigure}[T]{0.19\textwidth}
\centering
\includegraphics[width=\textwidth]{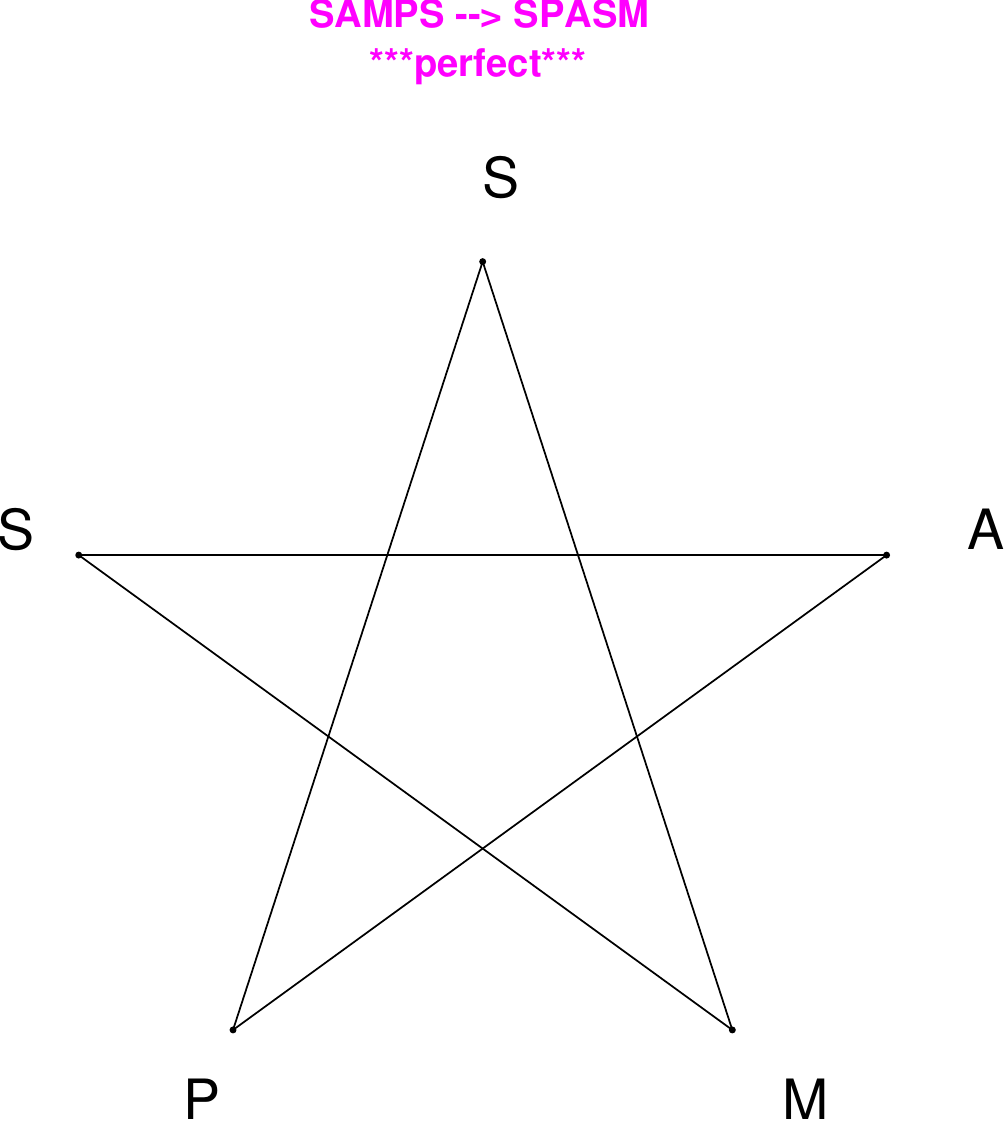}
\end{subfigure}
\hfill
\begin{subfigure}[T]{0.19\textwidth}
\centering
\includegraphics[width=\textwidth]{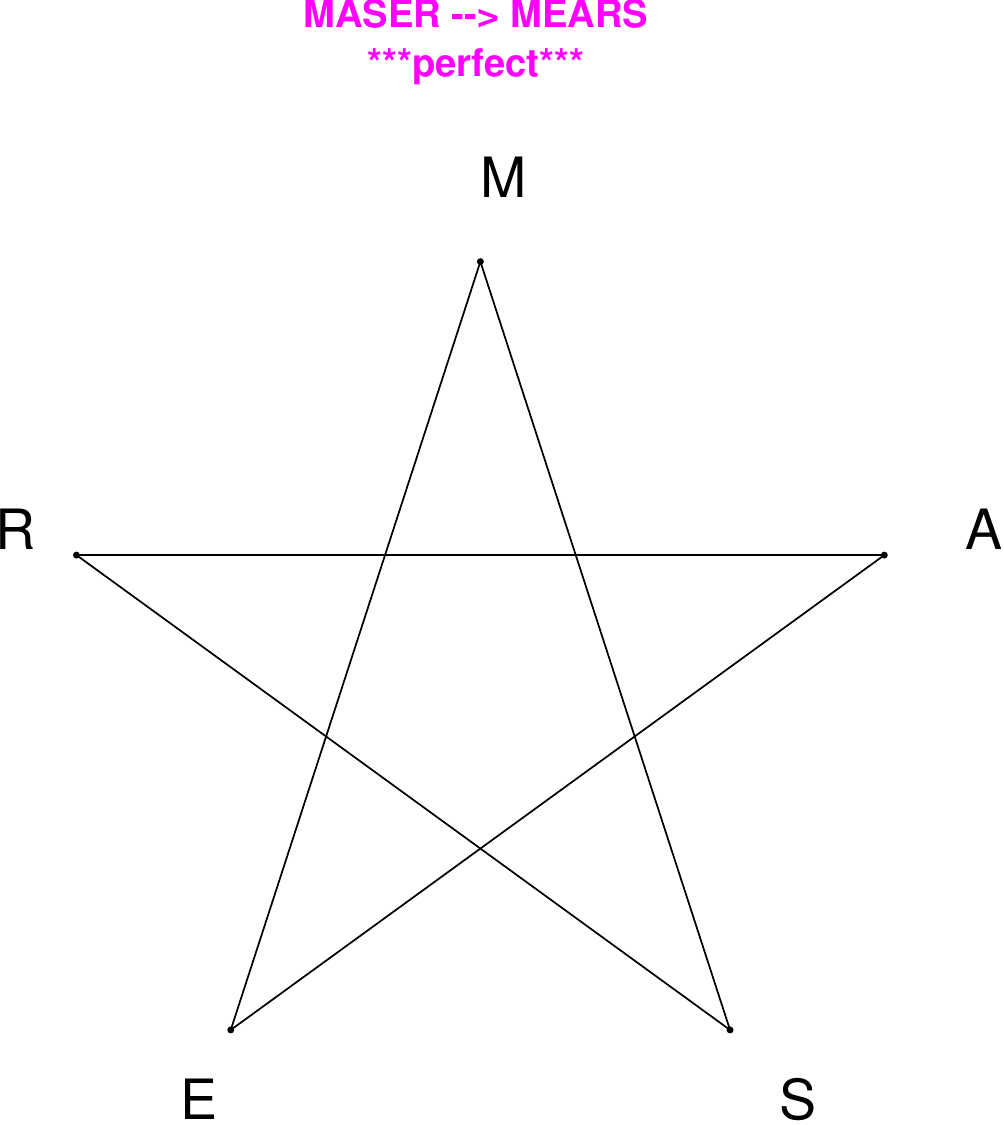}
\end{subfigure}
\end{figure}

\begin{figure}[H]
\centering
\begin{subfigure}[T]{0.19\textwidth}
\centering
\includegraphics[width=\textwidth]{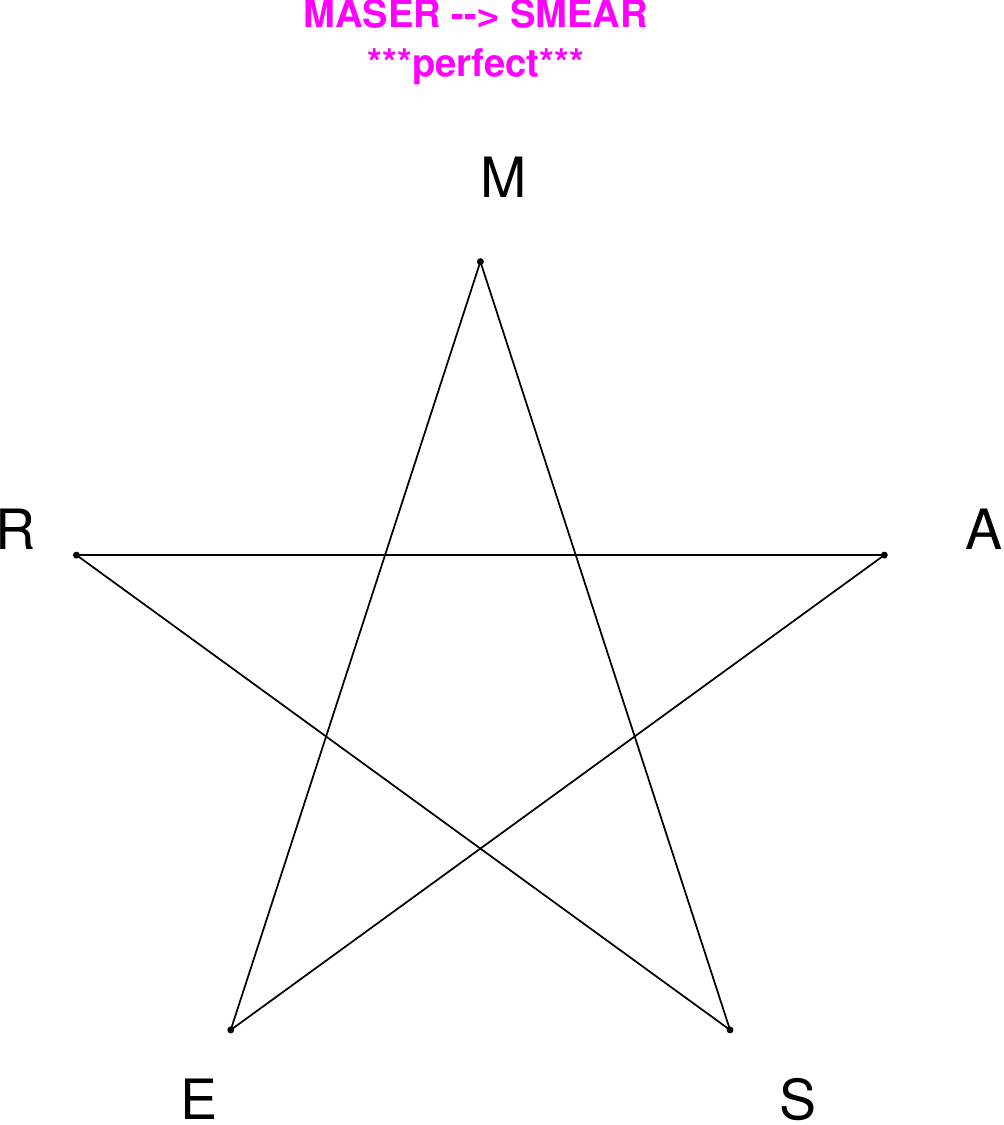}
\end{subfigure}
\hfill
\begin{subfigure}[T]{0.19\textwidth}
\centering
\includegraphics[width=\textwidth]{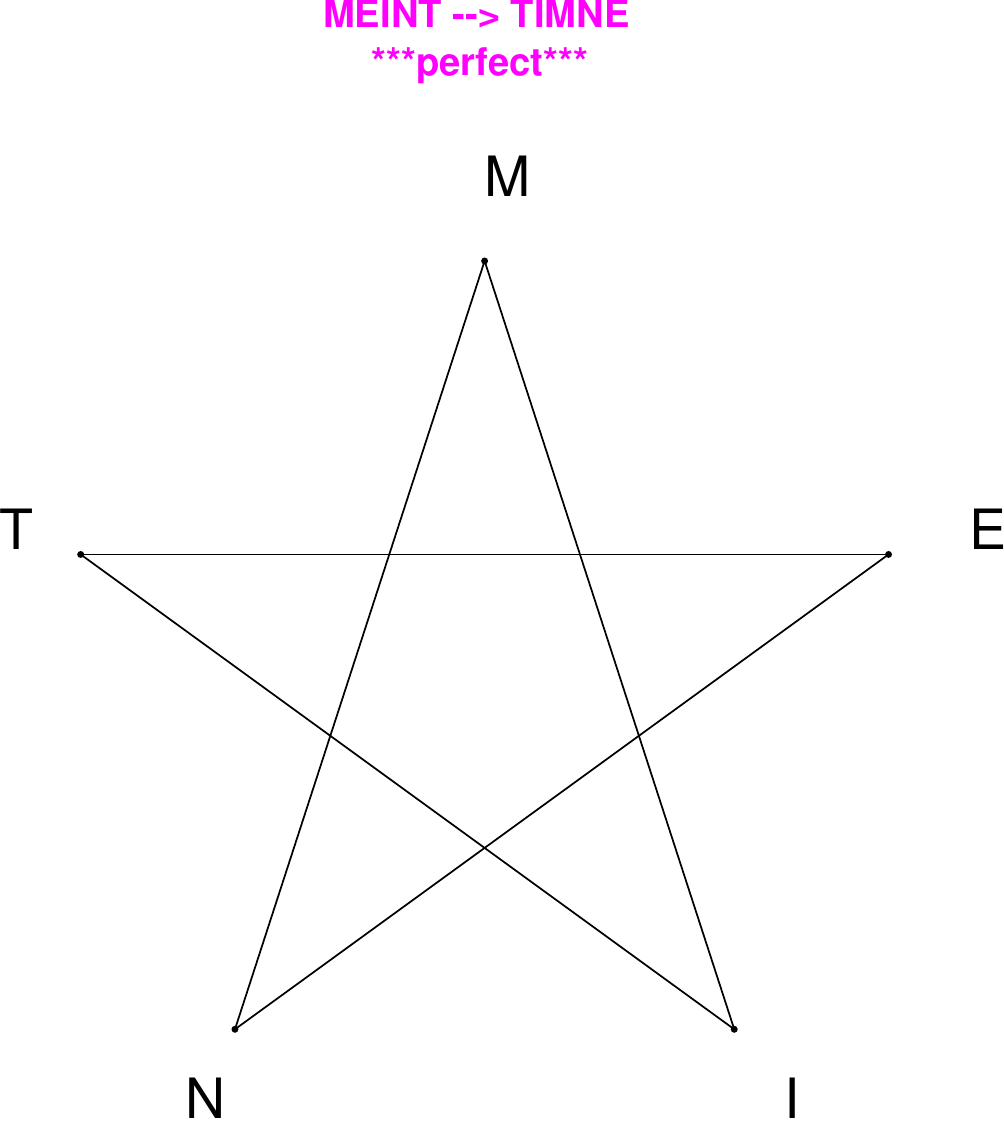}
\end{subfigure}
\hfill
\begin{subfigure}[T]{0.19\textwidth}
\centering
\includegraphics[width=\textwidth]{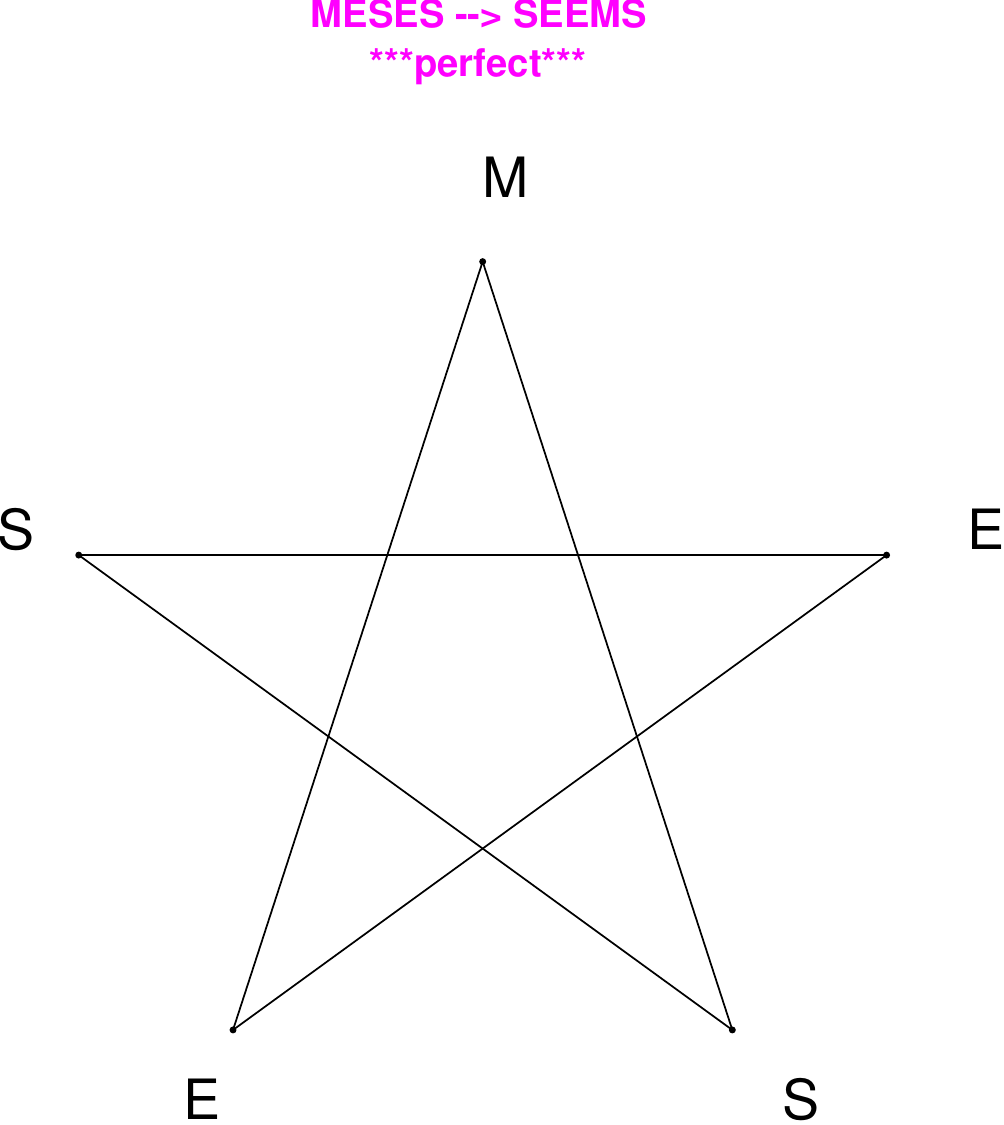}
\end{subfigure}
\hfill
\begin{subfigure}[T]{0.19\textwidth}
\centering
\includegraphics[width=\textwidth]{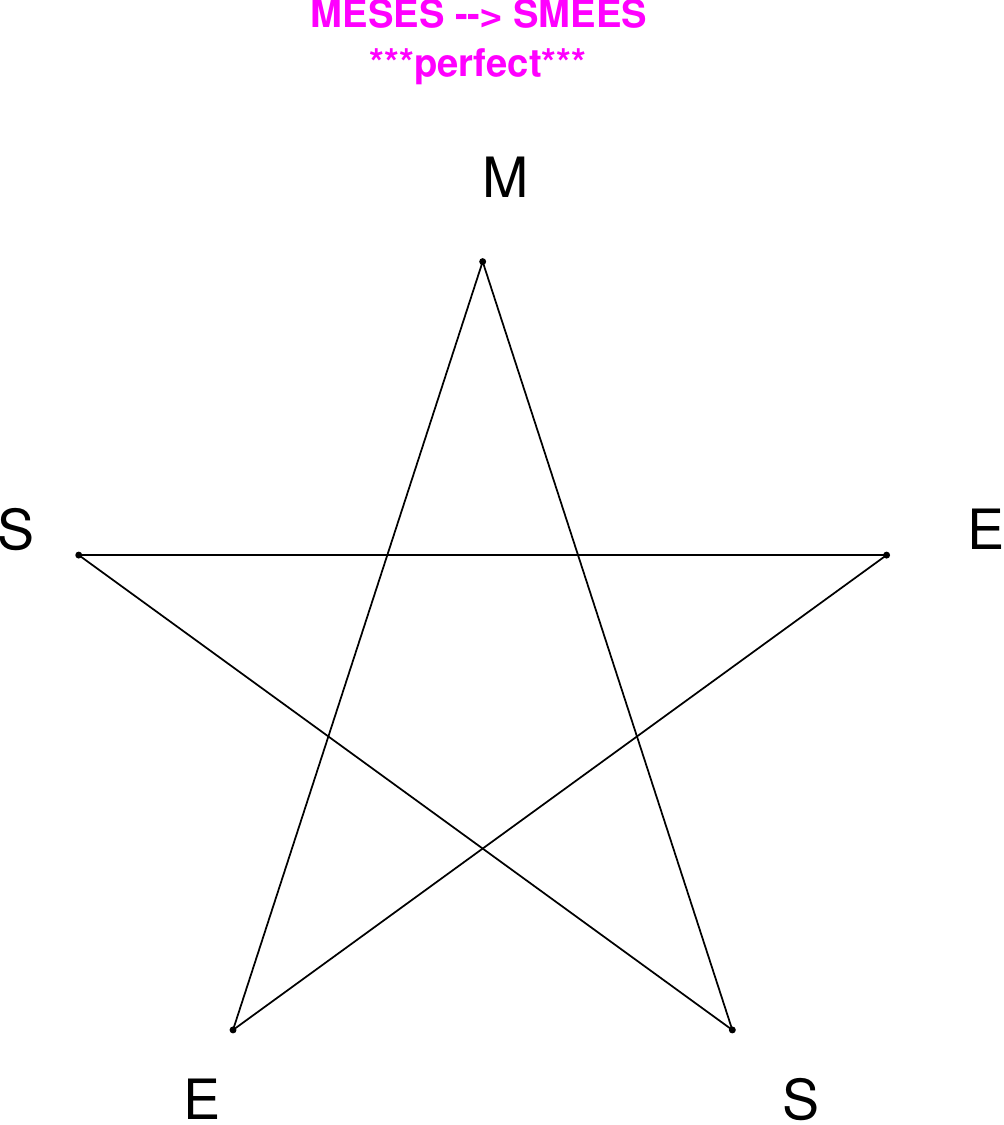}
\end{subfigure}
\hfill
\begin{subfigure}[T]{0.19\textwidth}
\centering
\includegraphics[width=\textwidth]{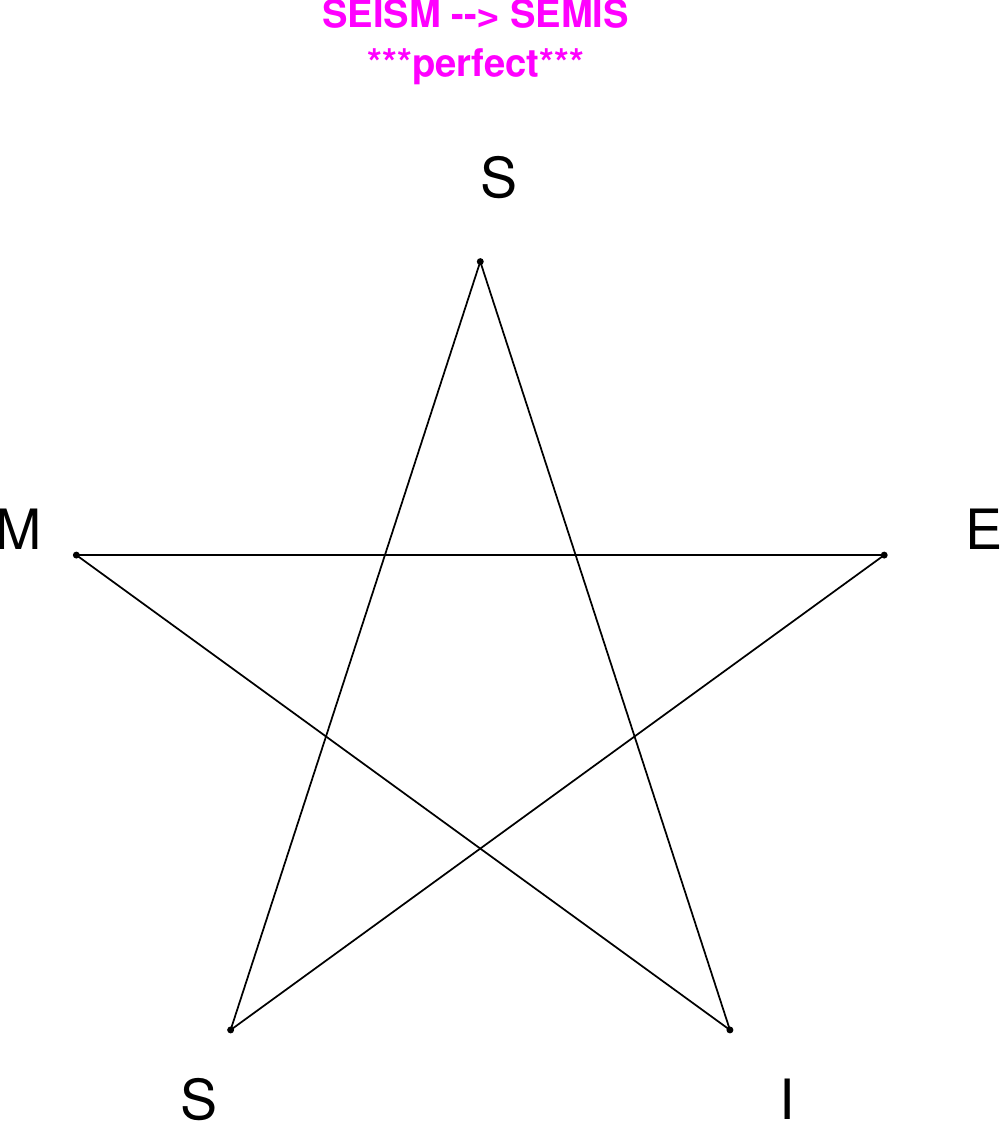}
\end{subfigure}
\end{figure}

\begin{figure}[H]
\centering
\begin{subfigure}[T]{0.19\textwidth}
\centering
\includegraphics[width=\textwidth]{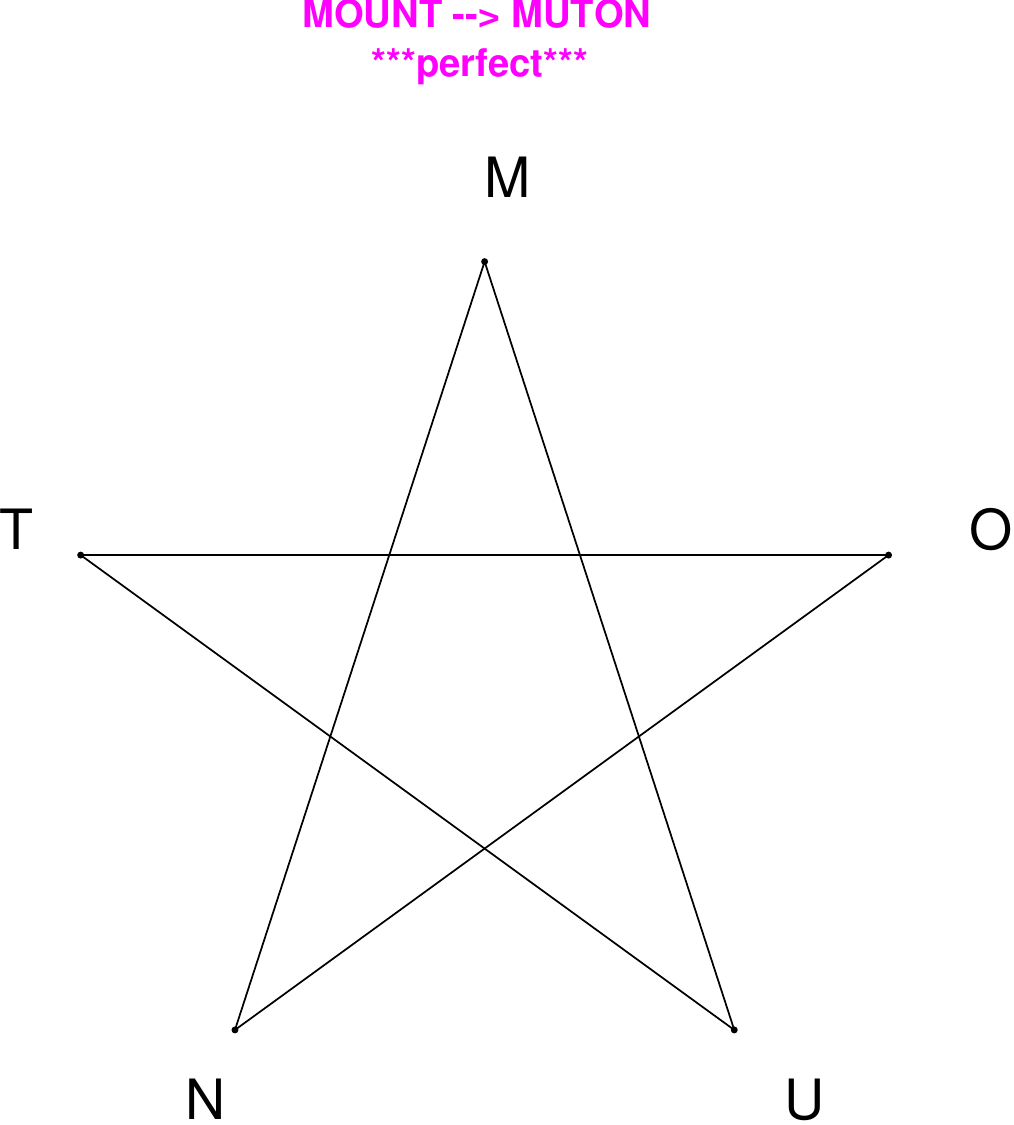}
\end{subfigure}
\hfill
\begin{subfigure}[T]{0.19\textwidth}
\centering
\includegraphics[width=\textwidth]{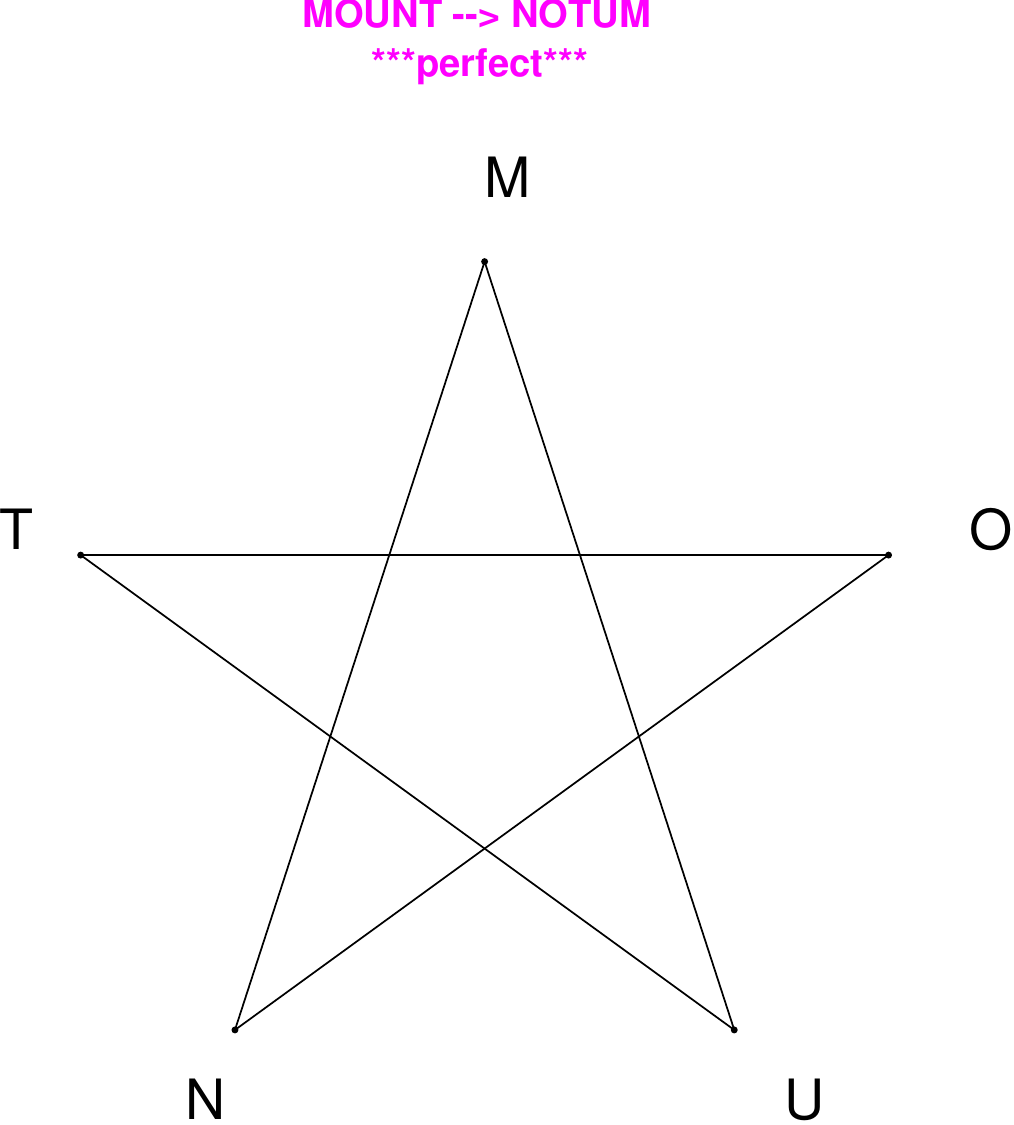}
\end{subfigure}
\hfill
\begin{subfigure}[T]{0.19\textwidth}
\centering
\includegraphics[width=\textwidth]{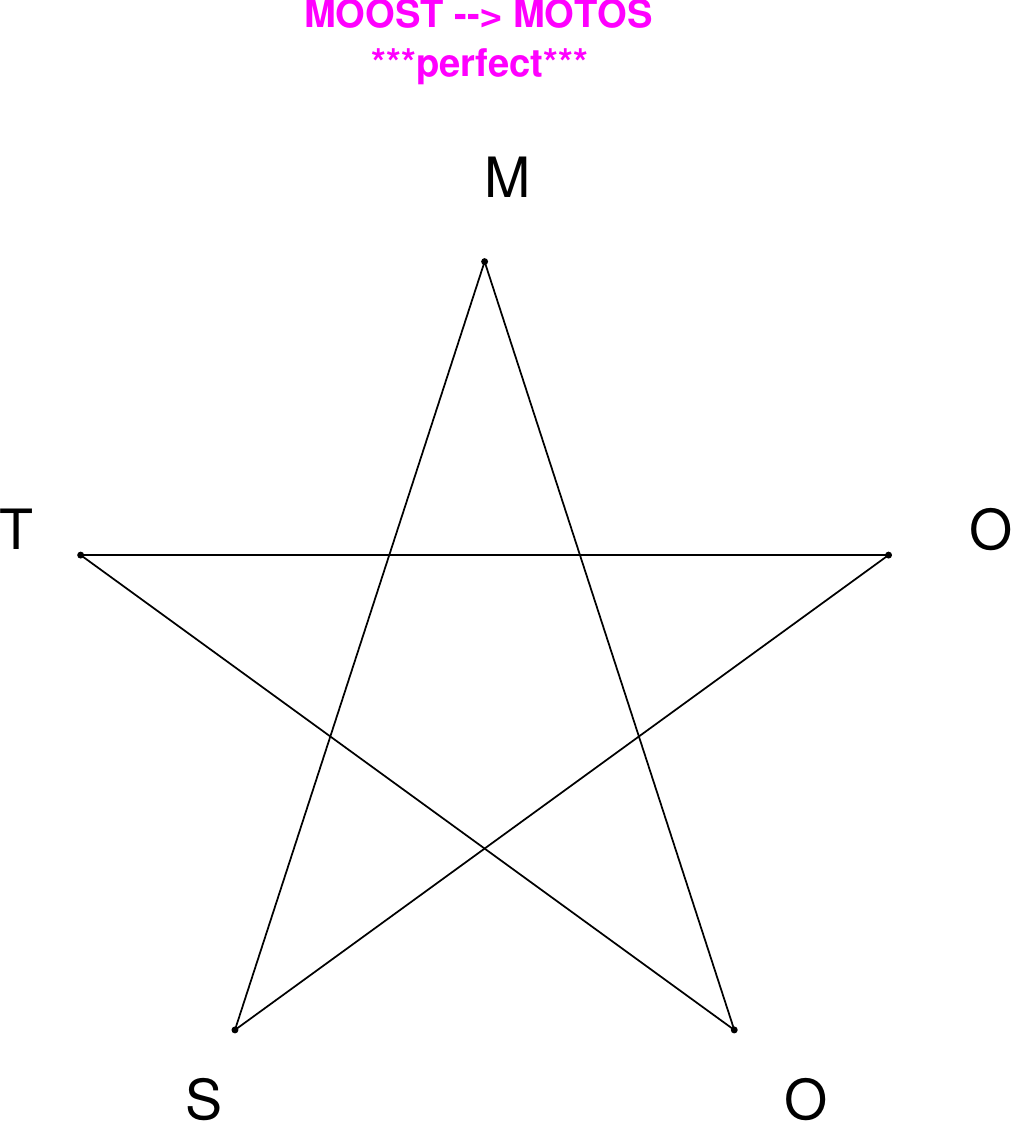}
\end{subfigure}
\hfill
\begin{subfigure}[T]{0.19\textwidth}
\centering
\includegraphics[width=\textwidth]{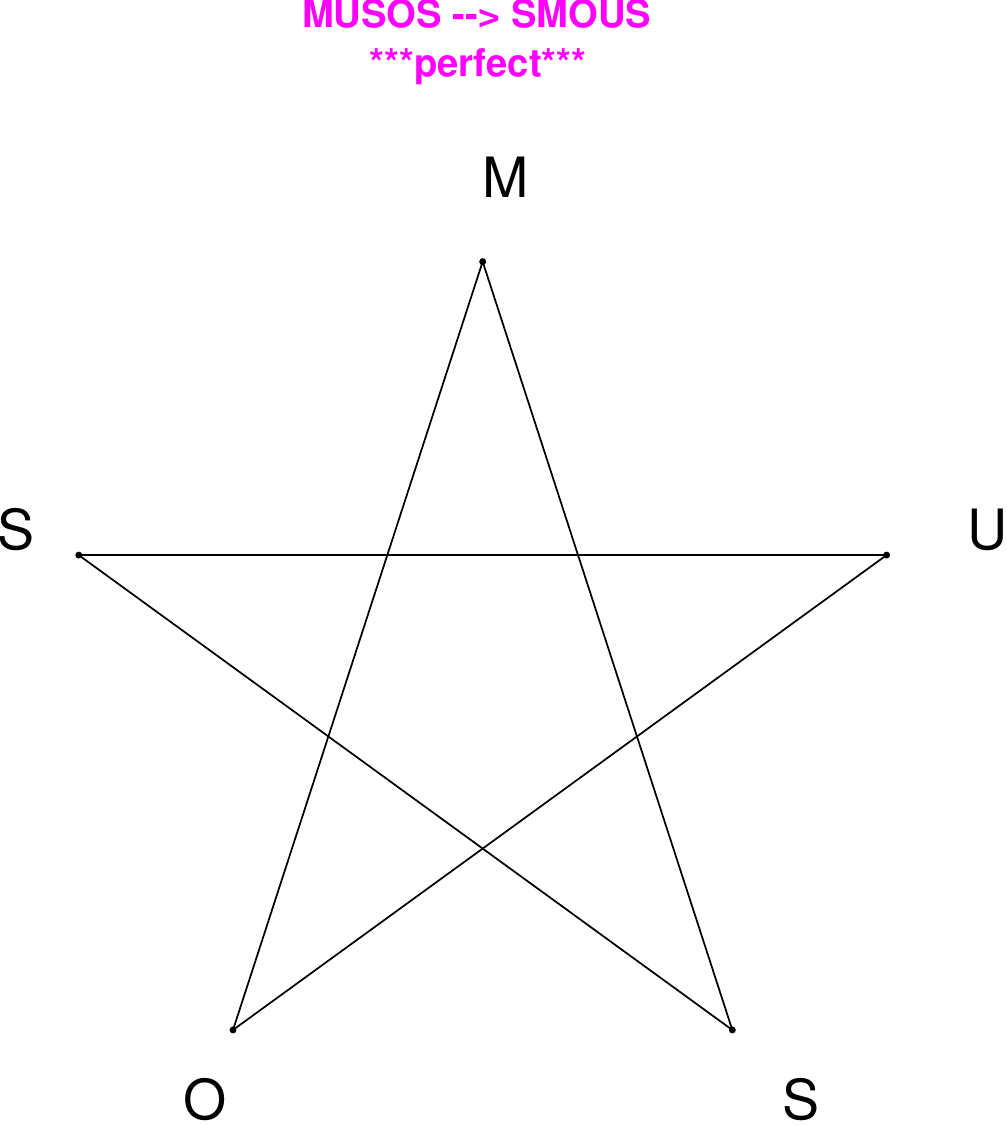}
\end{subfigure}
\hfill
\begin{subfigure}[T]{0.19\textwidth}
\centering
\includegraphics[width=\textwidth]{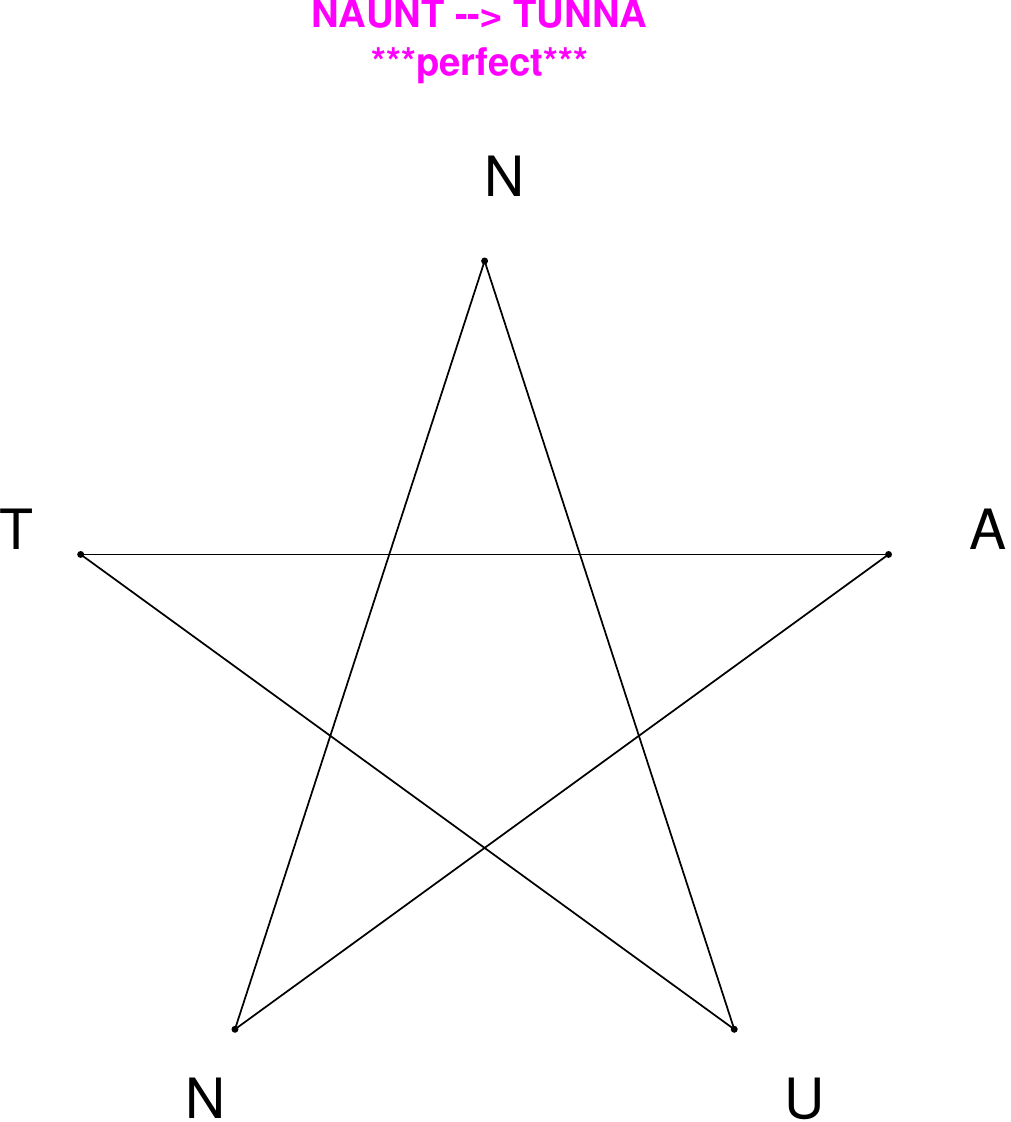}
\end{subfigure}
\end{figure}

\begin{figure}[H]
\centering
\begin{subfigure}[T]{0.19\textwidth}
\centering
\includegraphics[width=\textwidth]{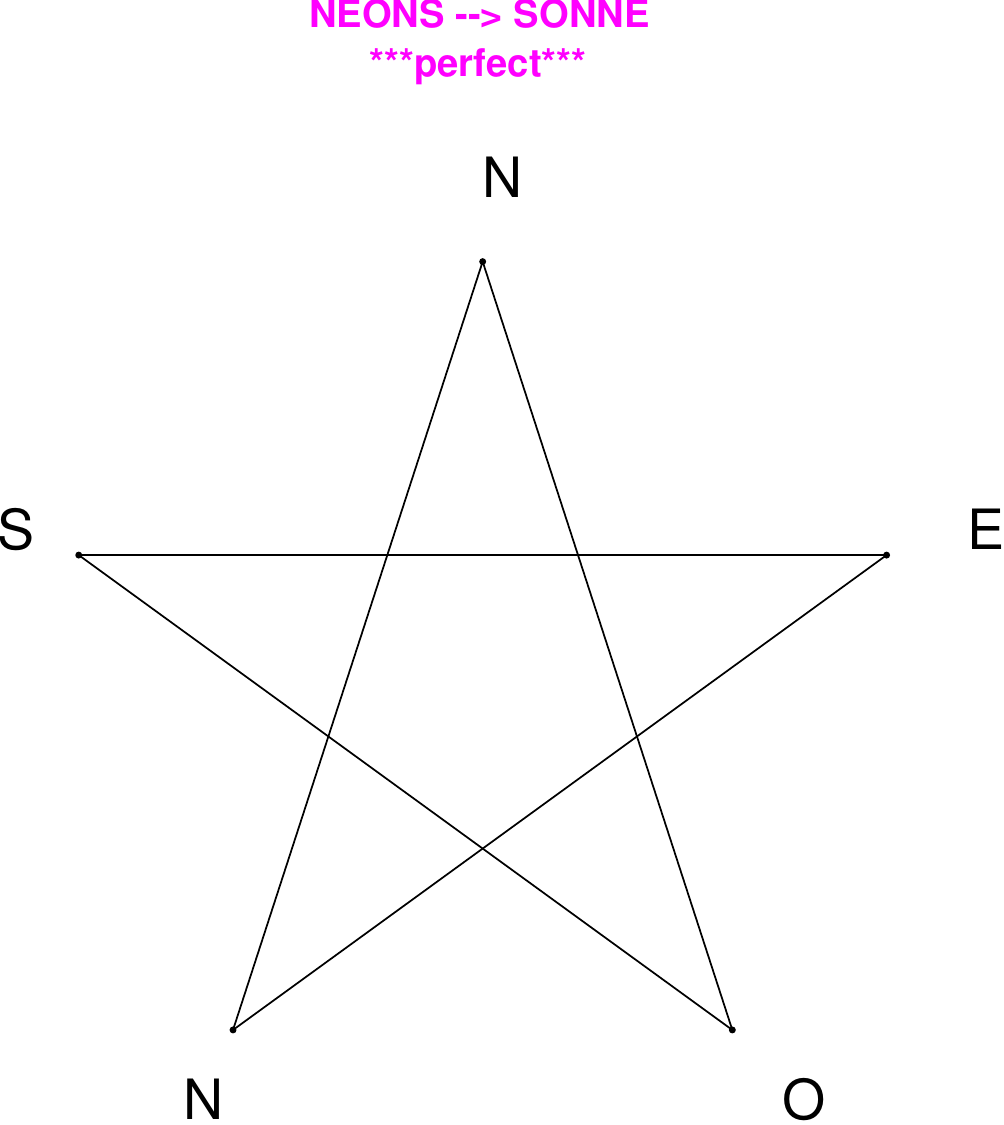}
\end{subfigure}
\hfill
\begin{subfigure}[T]{0.19\textwidth}
\centering
\includegraphics[width=\textwidth]{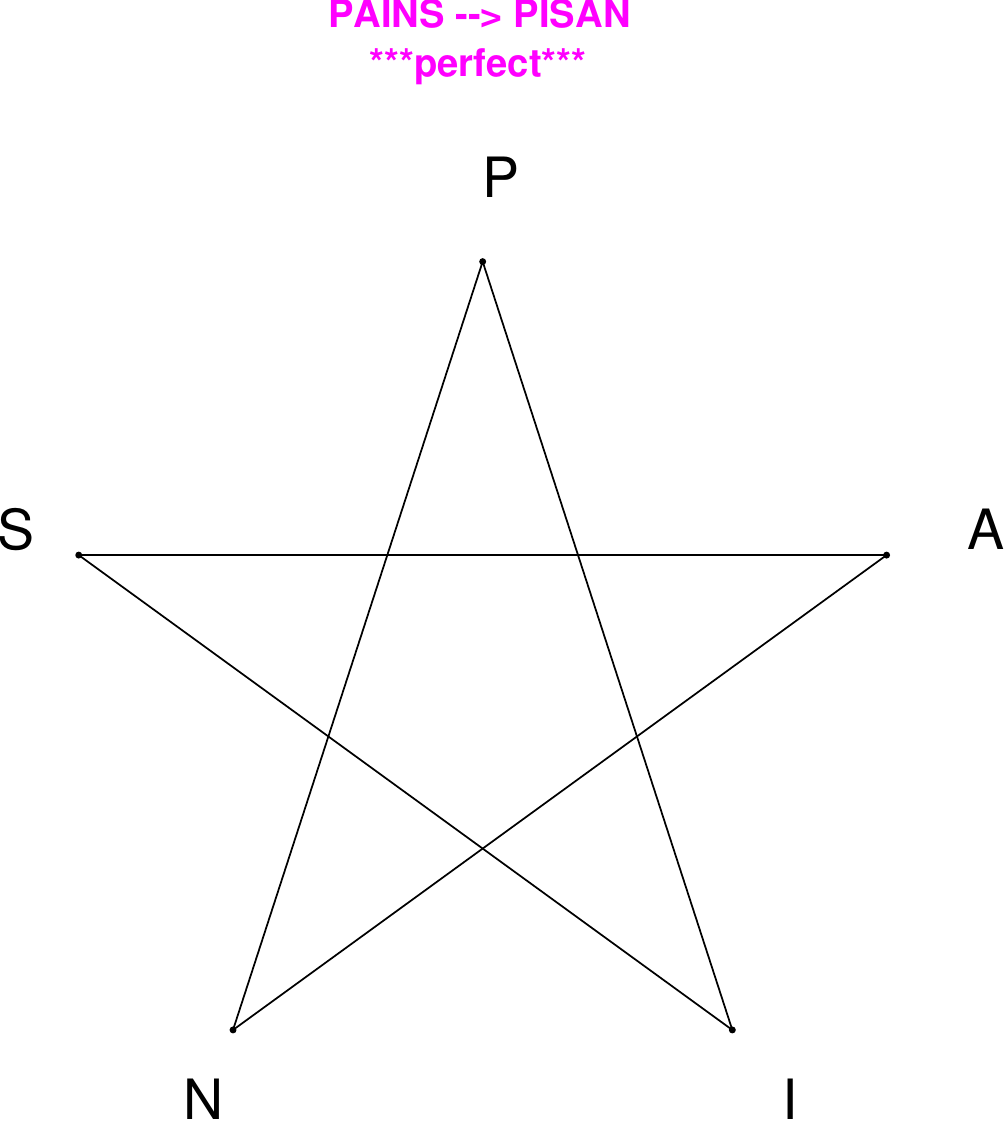}
\end{subfigure}
\hfill
\begin{subfigure}[T]{0.19\textwidth}
\centering
\includegraphics[width=\textwidth]{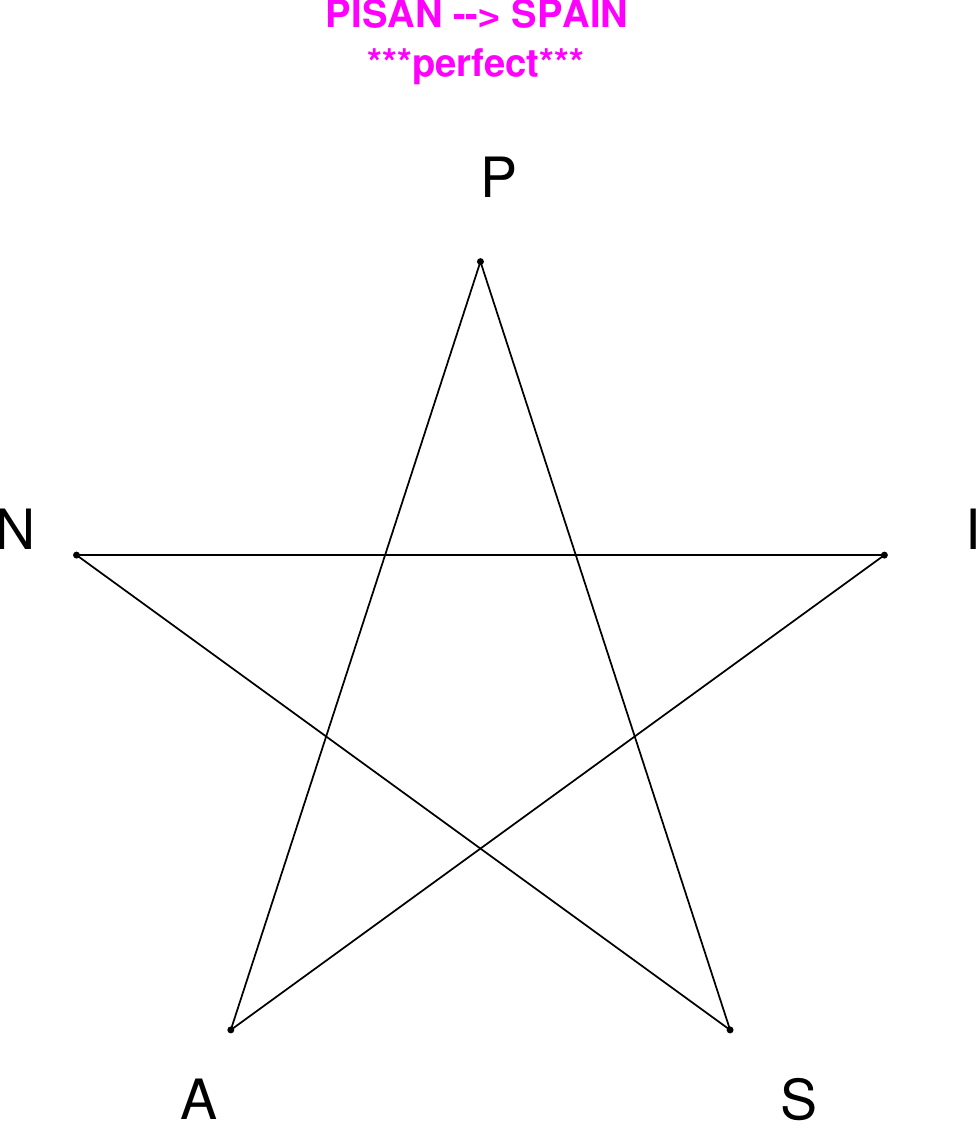}
\end{subfigure}
\hfill
\begin{subfigure}[T]{0.19\textwidth}
\centering
\includegraphics[width=\textwidth]{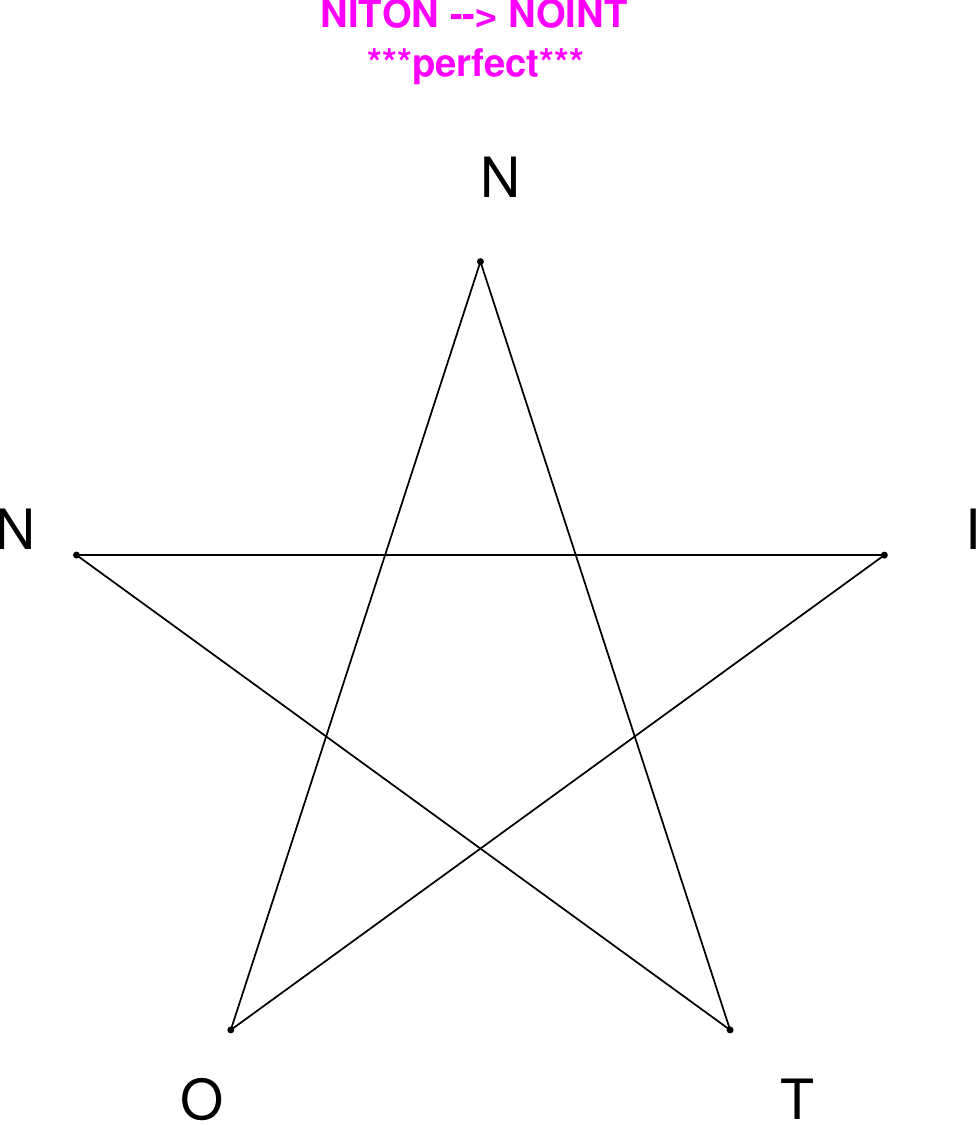}
\end{subfigure}
\hfill
\begin{subfigure}[T]{0.19\textwidth}
\centering
\includegraphics[width=\textwidth]{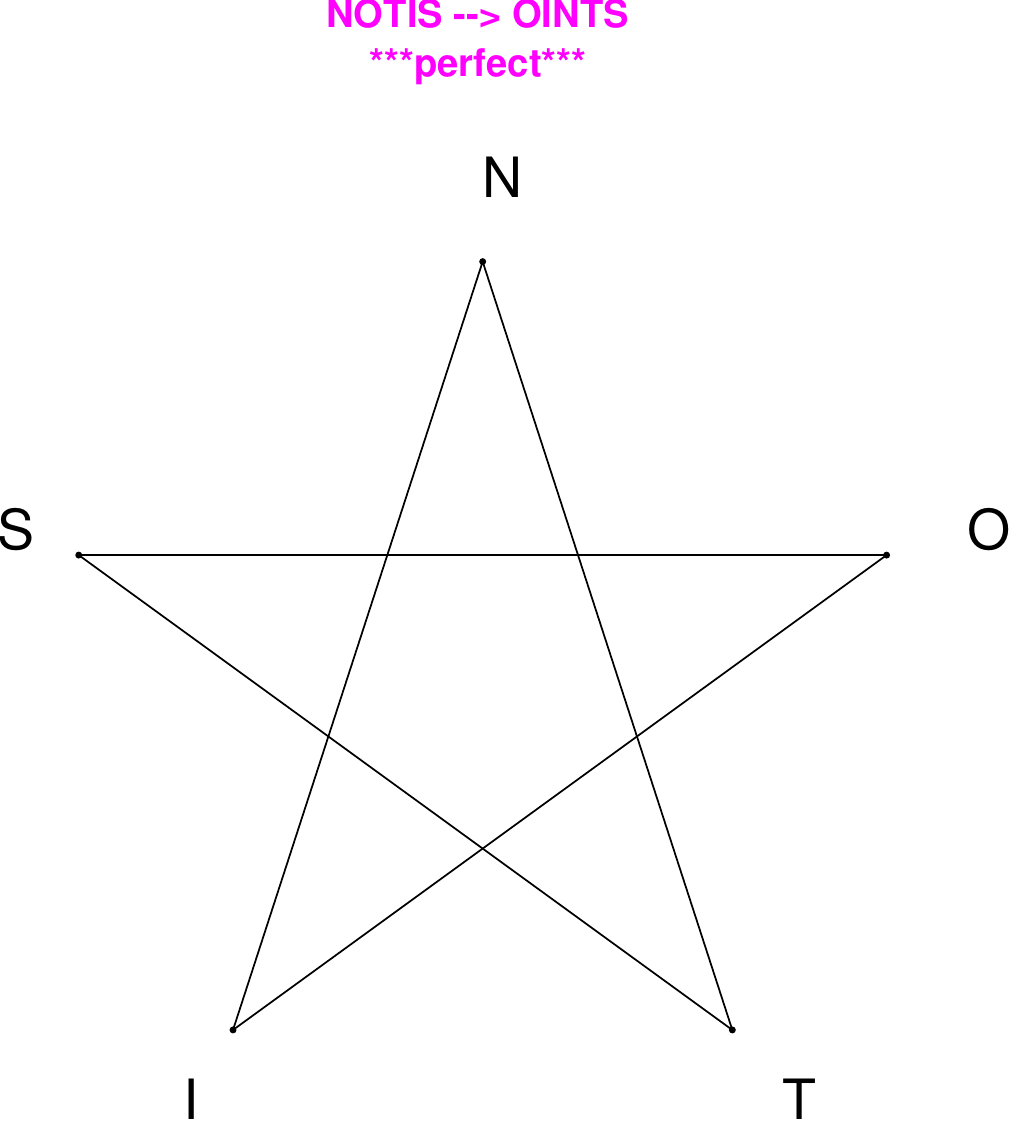}
\end{subfigure}
\end{figure}

\begin{figure}[H]
\centering
\begin{subfigure}[T]{0.19\textwidth}
\centering
\includegraphics[width=\textwidth]{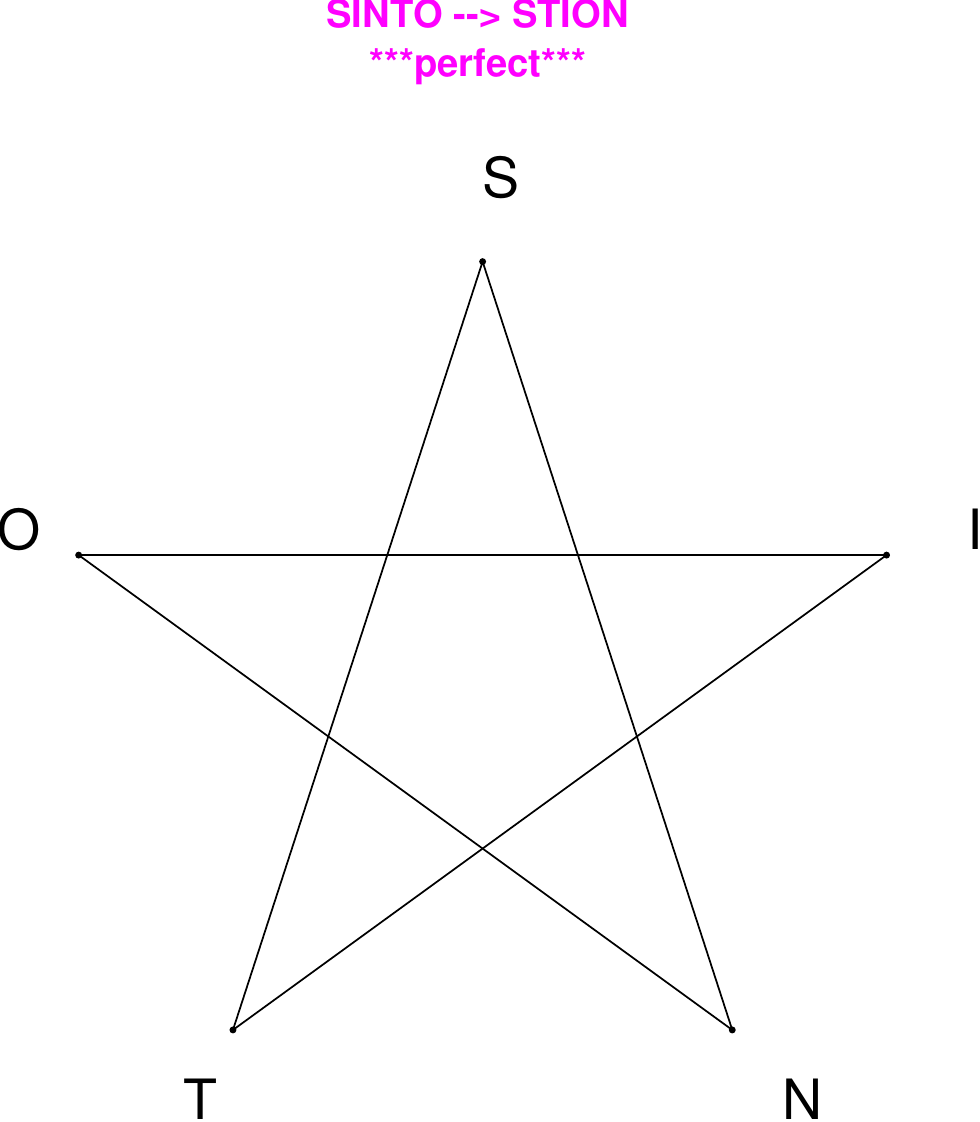}
\end{subfigure}
\hfill
\begin{subfigure}[T]{0.19\textwidth}
\centering
\includegraphics[width=\textwidth]{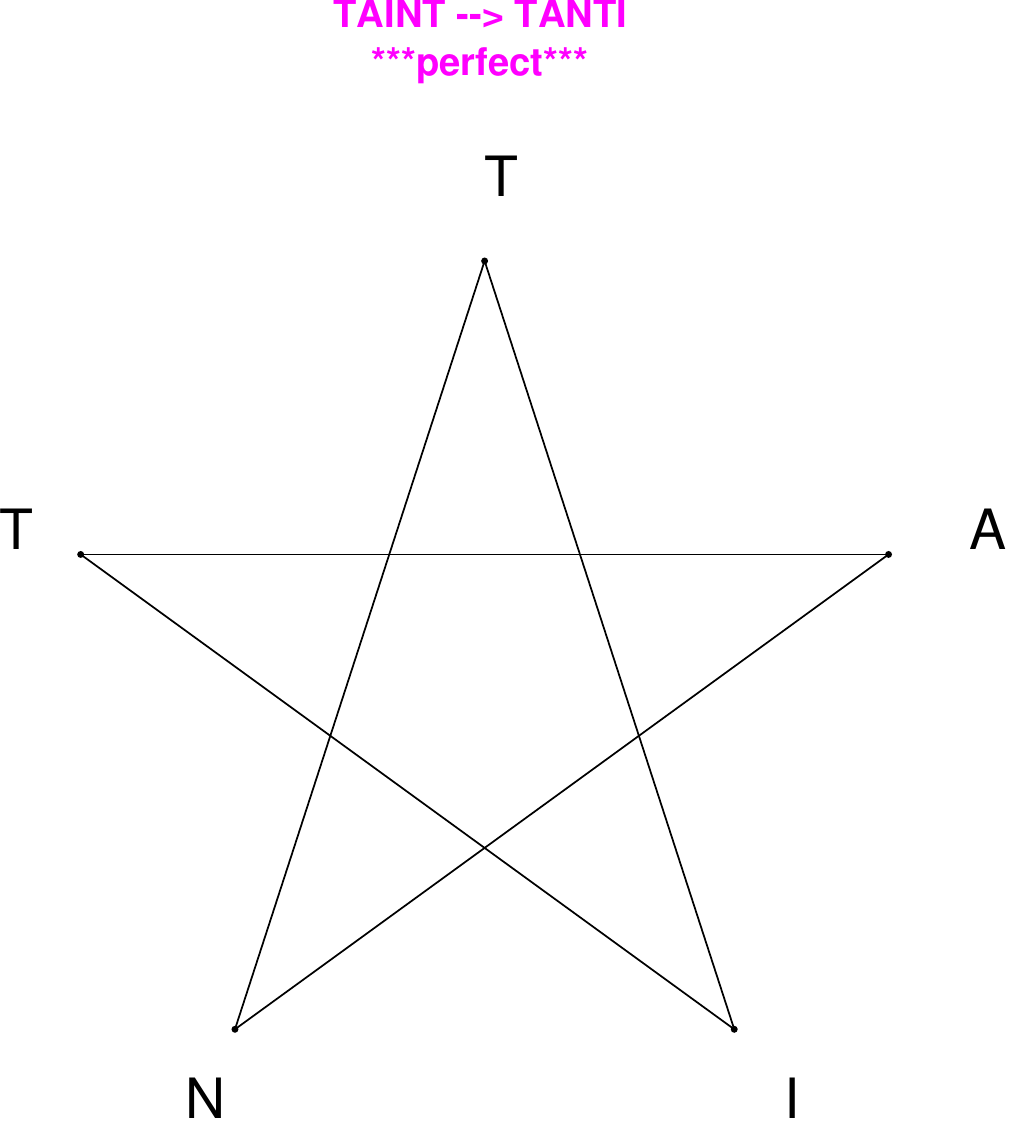}
\end{subfigure}
\hfill
\begin{subfigure}[T]{0.19\textwidth}
\centering
\includegraphics[width=\textwidth]{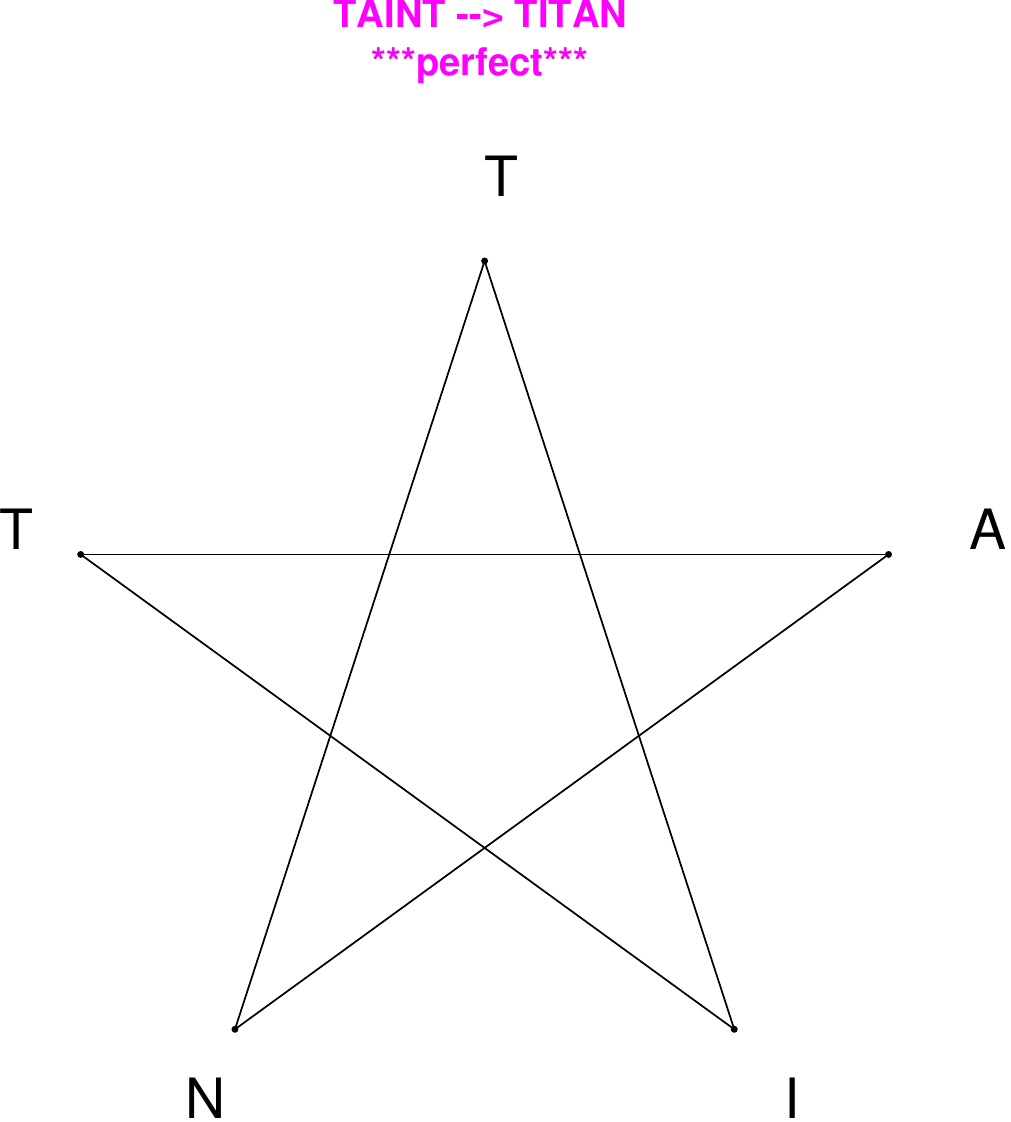}
\end{subfigure}
\hfill
\begin{subfigure}[T]{0.19\textwidth}
\centering
\includegraphics[width=\textwidth]{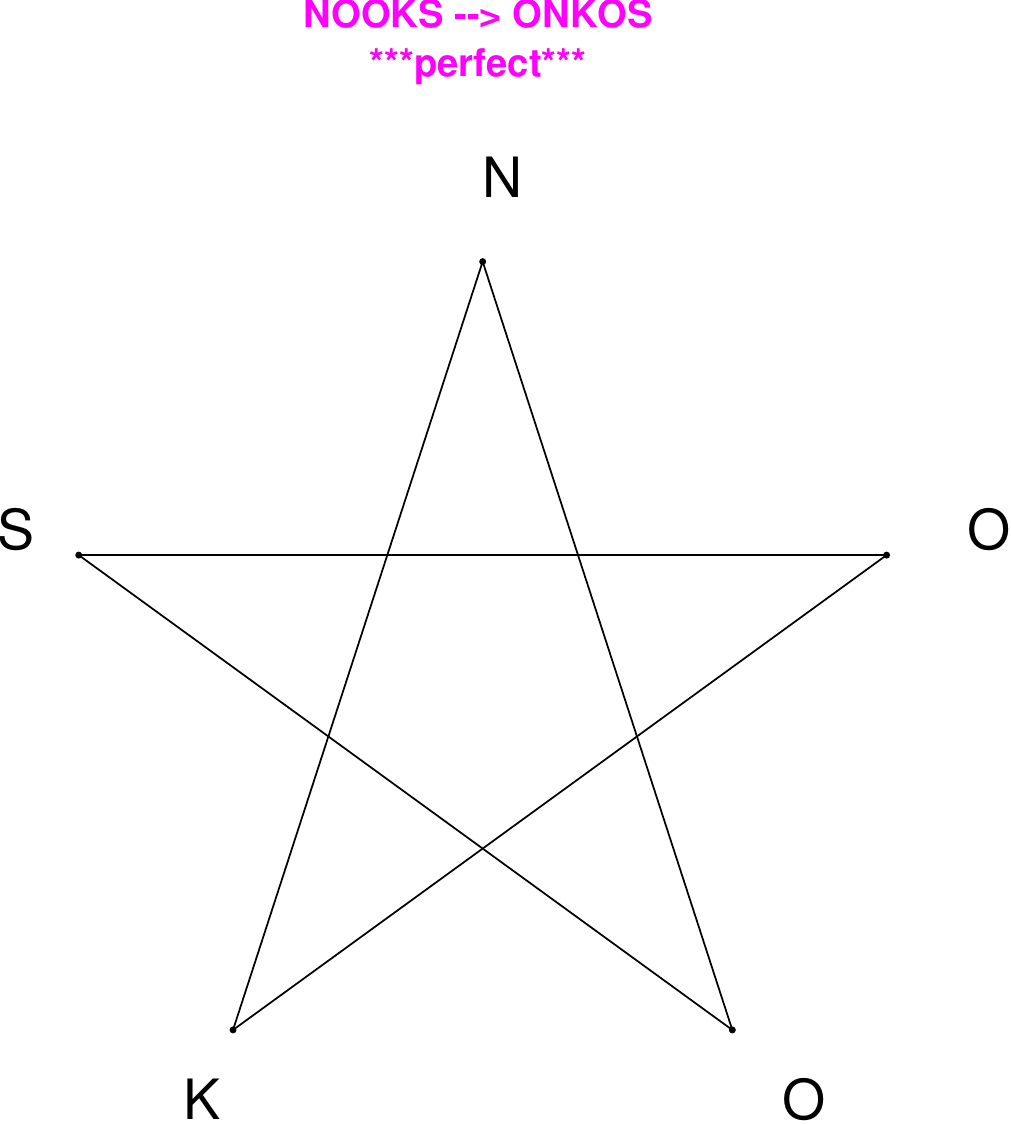}
\end{subfigure}
\hfill
\begin{subfigure}[T]{0.19\textwidth}
\centering
\includegraphics[width=\textwidth]{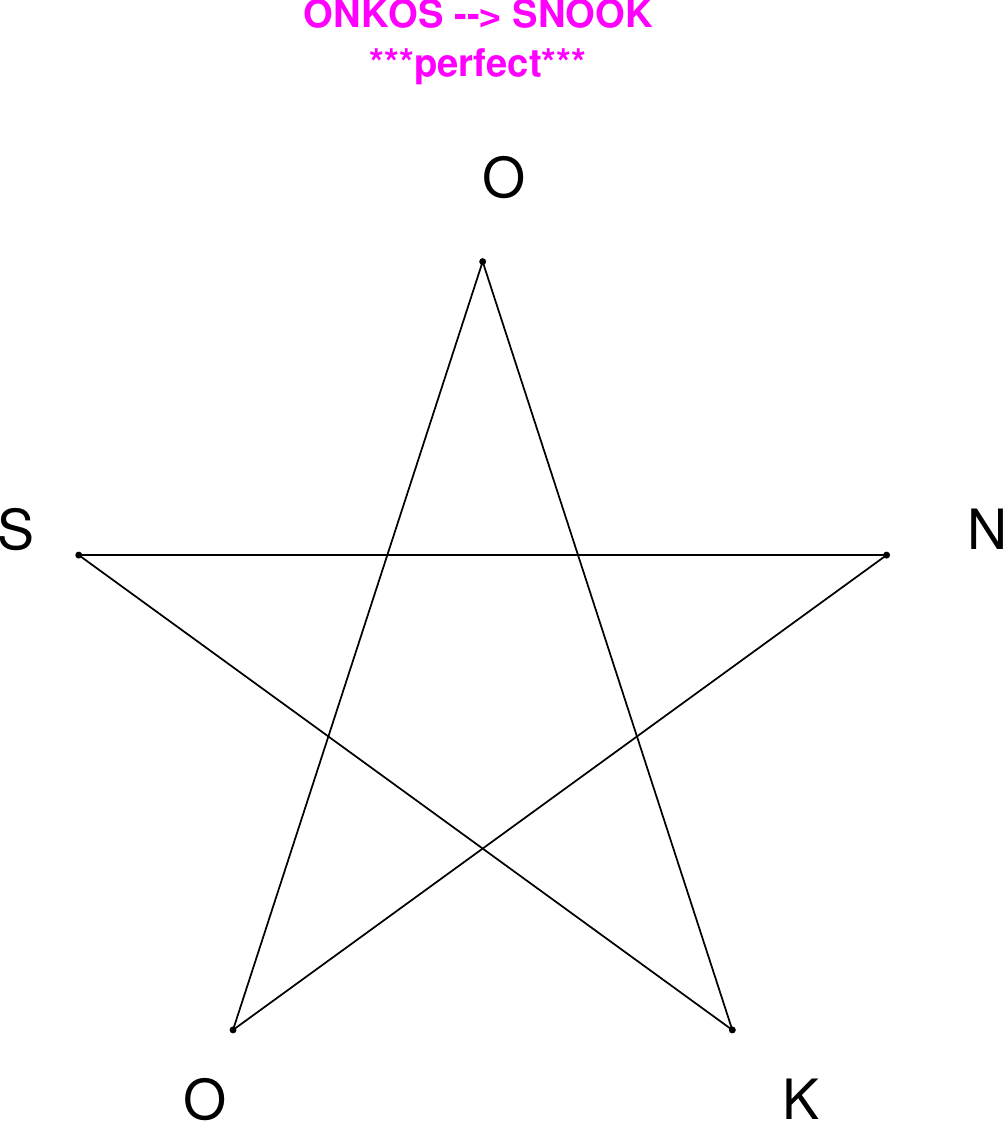}
\end{subfigure}
\end{figure}

\begin{figure}[H]
\centering
\begin{subfigure}[T]{0.19\textwidth}
\centering
\includegraphics[width=\textwidth]{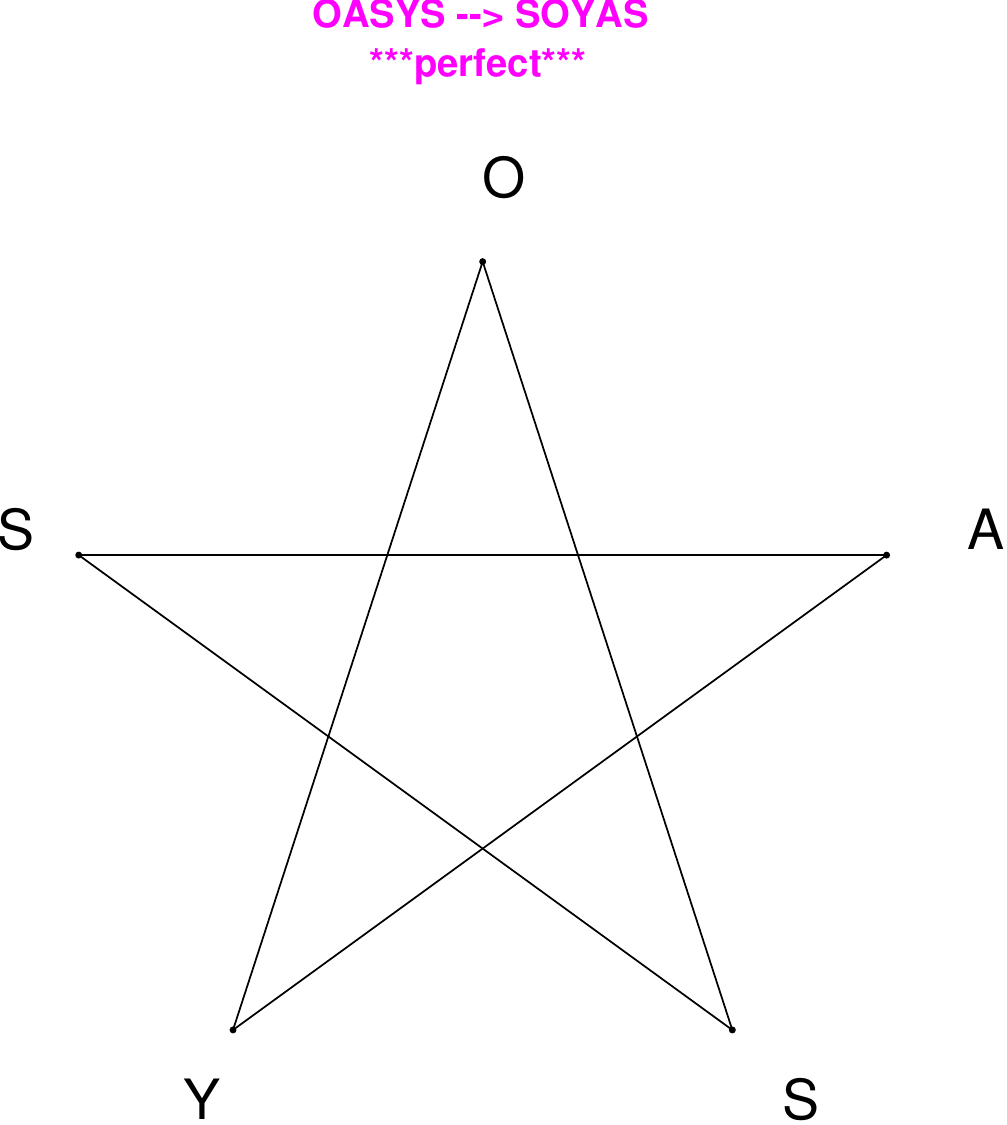}
\end{subfigure}
\hfill
\begin{subfigure}[T]{0.19\textwidth}
\centering
\includegraphics[width=\textwidth]{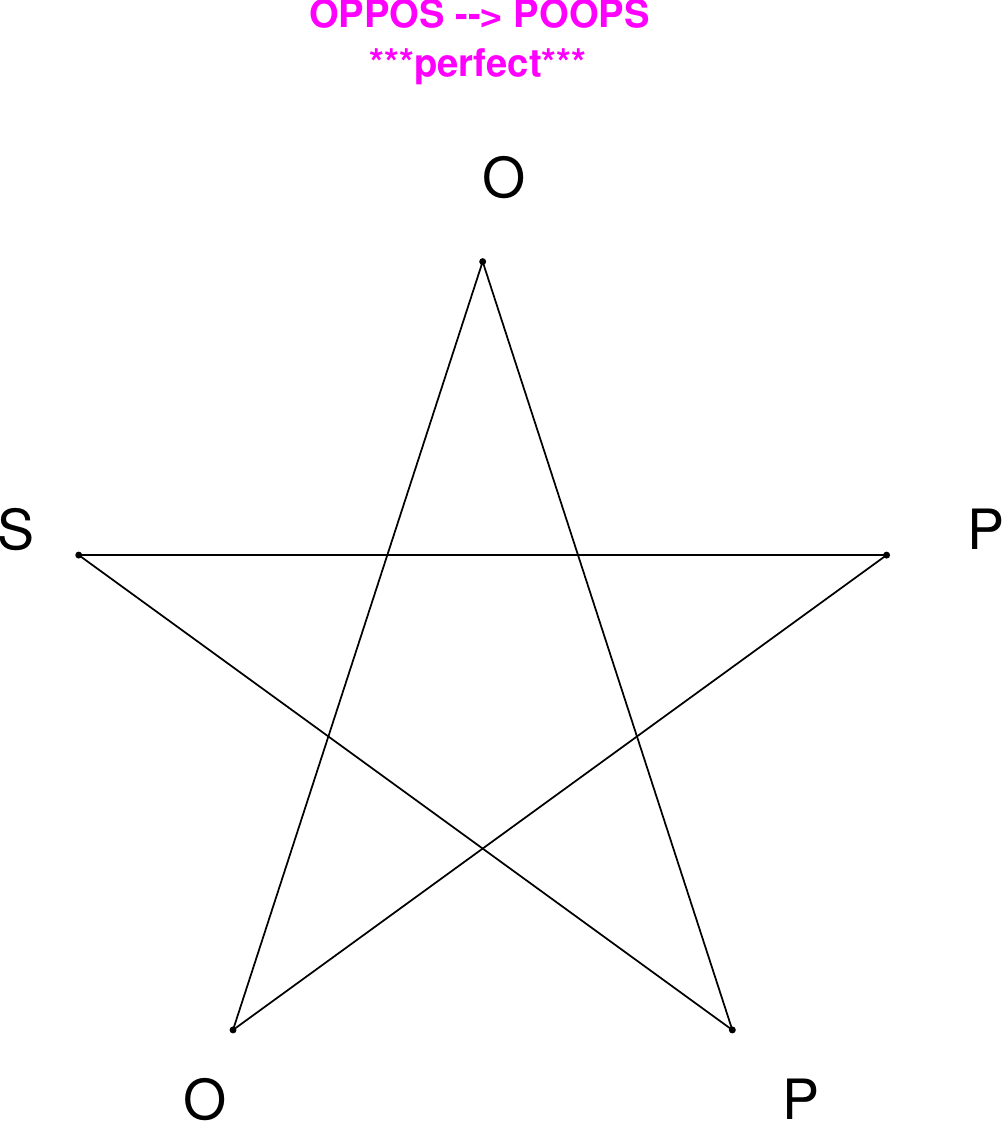}
\end{subfigure}
\hfill
\begin{subfigure}[T]{0.19\textwidth}
\centering
\includegraphics[width=\textwidth]{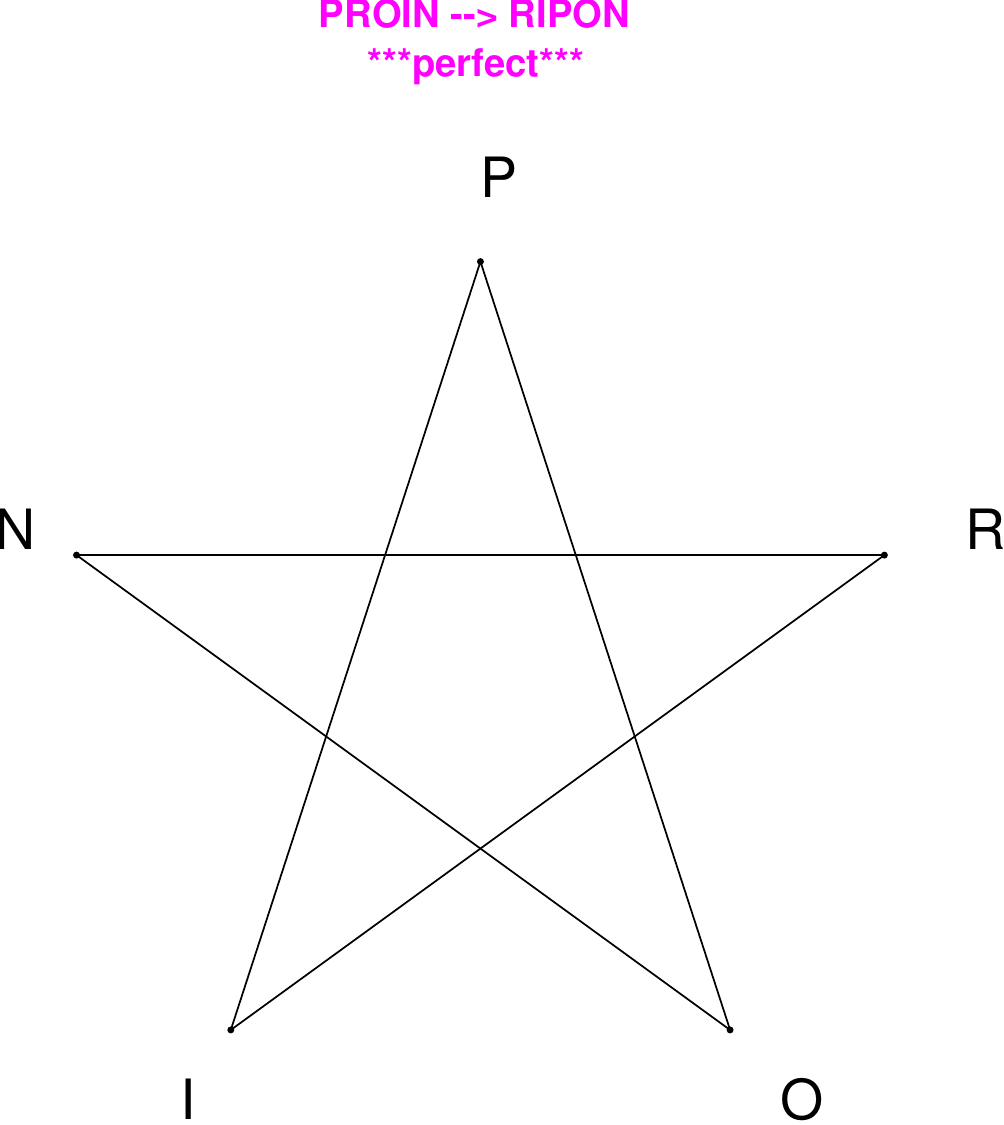}
\end{subfigure}
\hfill
\begin{subfigure}[T]{0.19\textwidth}
\centering
\includegraphics[width=\textwidth]{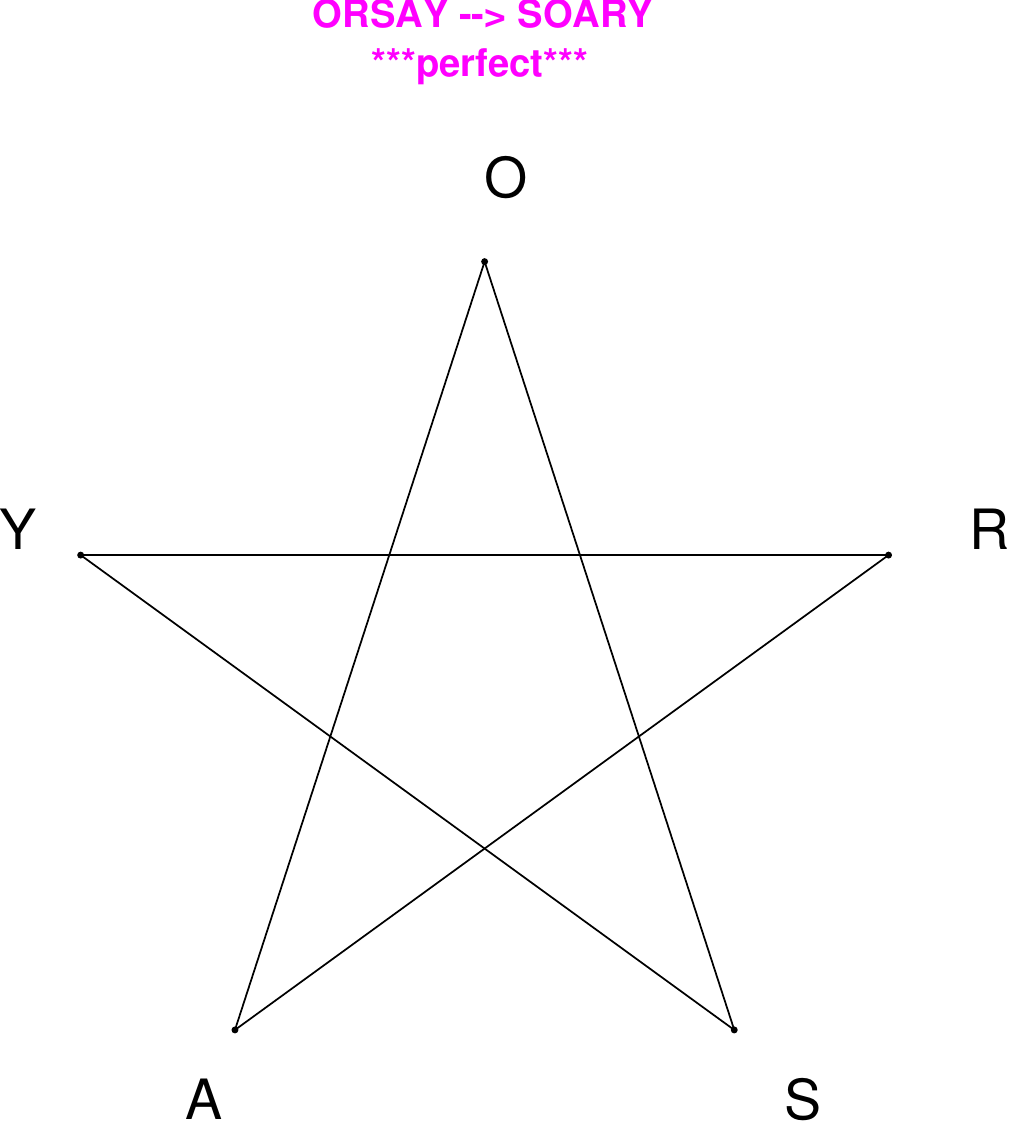}
\end{subfigure}
\hfill
\begin{subfigure}[T]{0.19\textwidth}
\centering
\includegraphics[width=\textwidth]{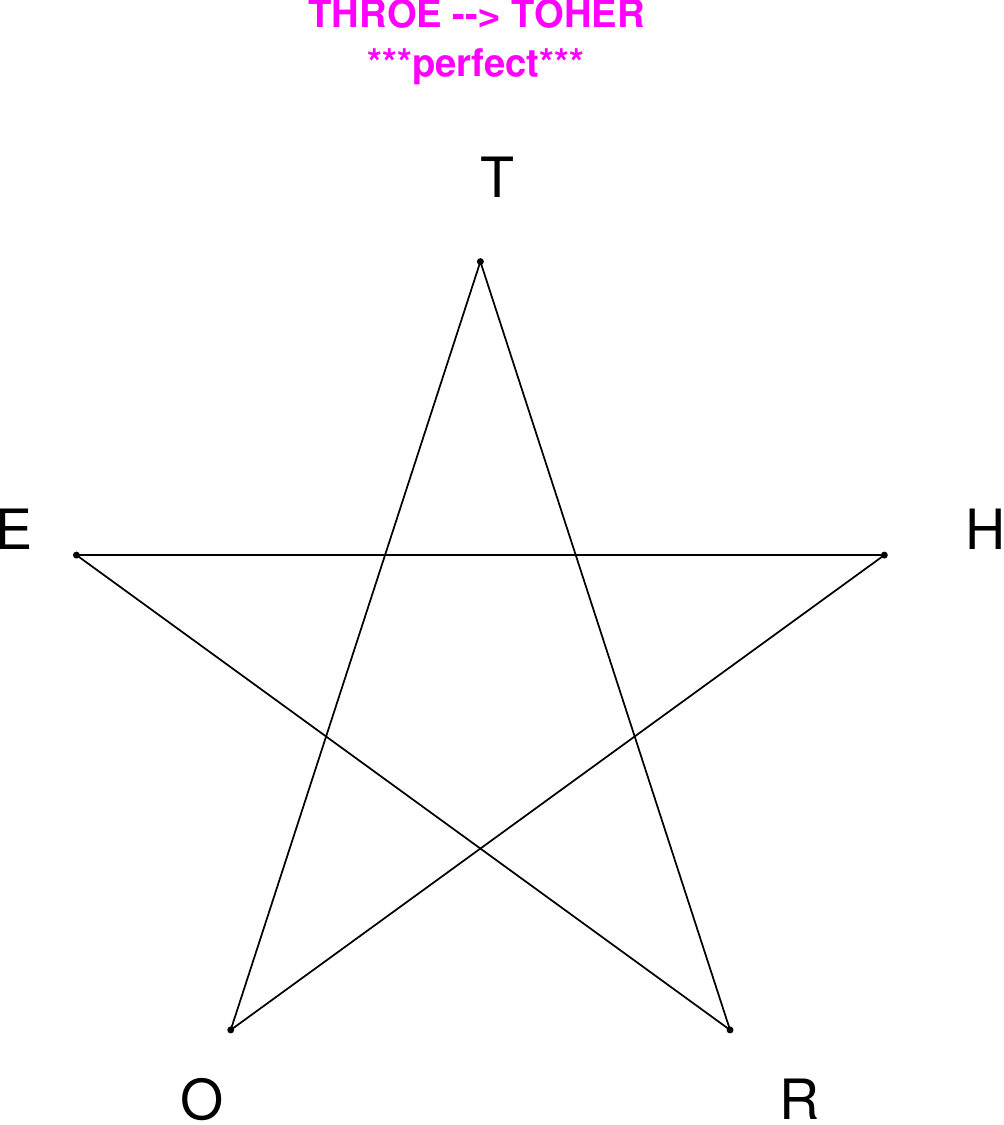}
\end{subfigure}
\end{figure}

\begin{figure}[H]
\centering
\begin{subfigure}[T]{0.19\textwidth}
\centering
\includegraphics[width=\textwidth]{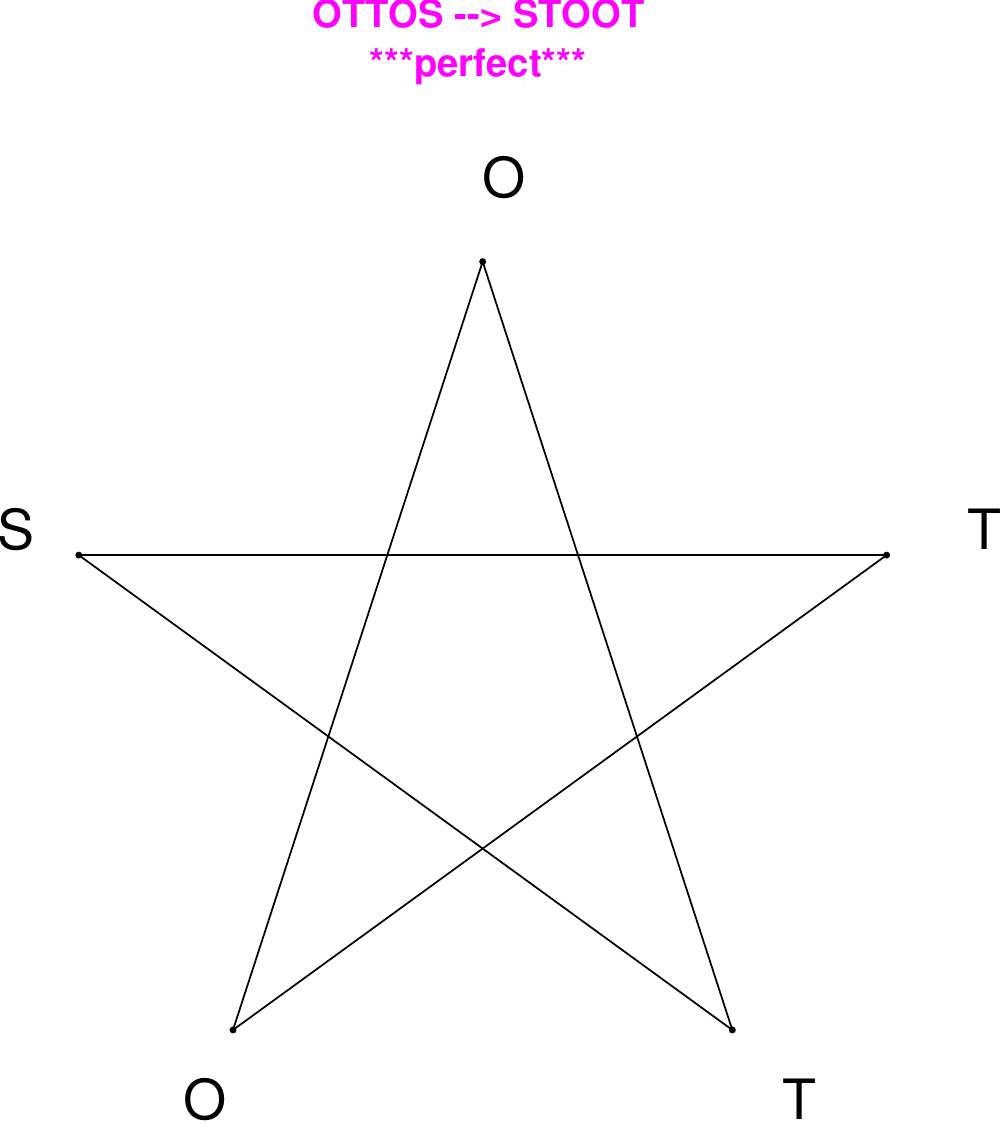}
\end{subfigure}
\hfill
\begin{subfigure}[T]{0.19\textwidth}
\centering
\includegraphics[width=\textwidth]{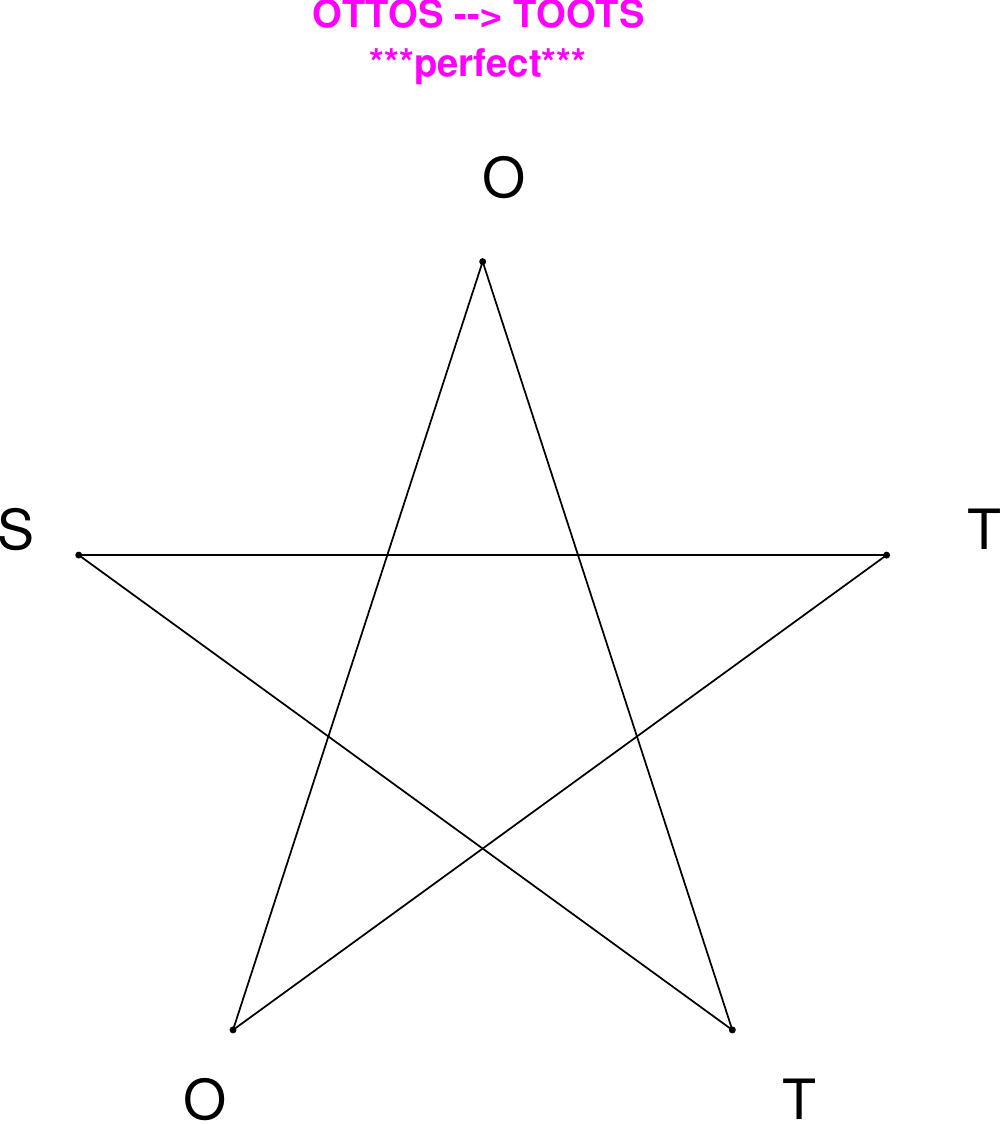}
\end{subfigure}
\hfill
\begin{subfigure}[T]{0.19\textwidth}
\centering
\includegraphics[width=\textwidth]{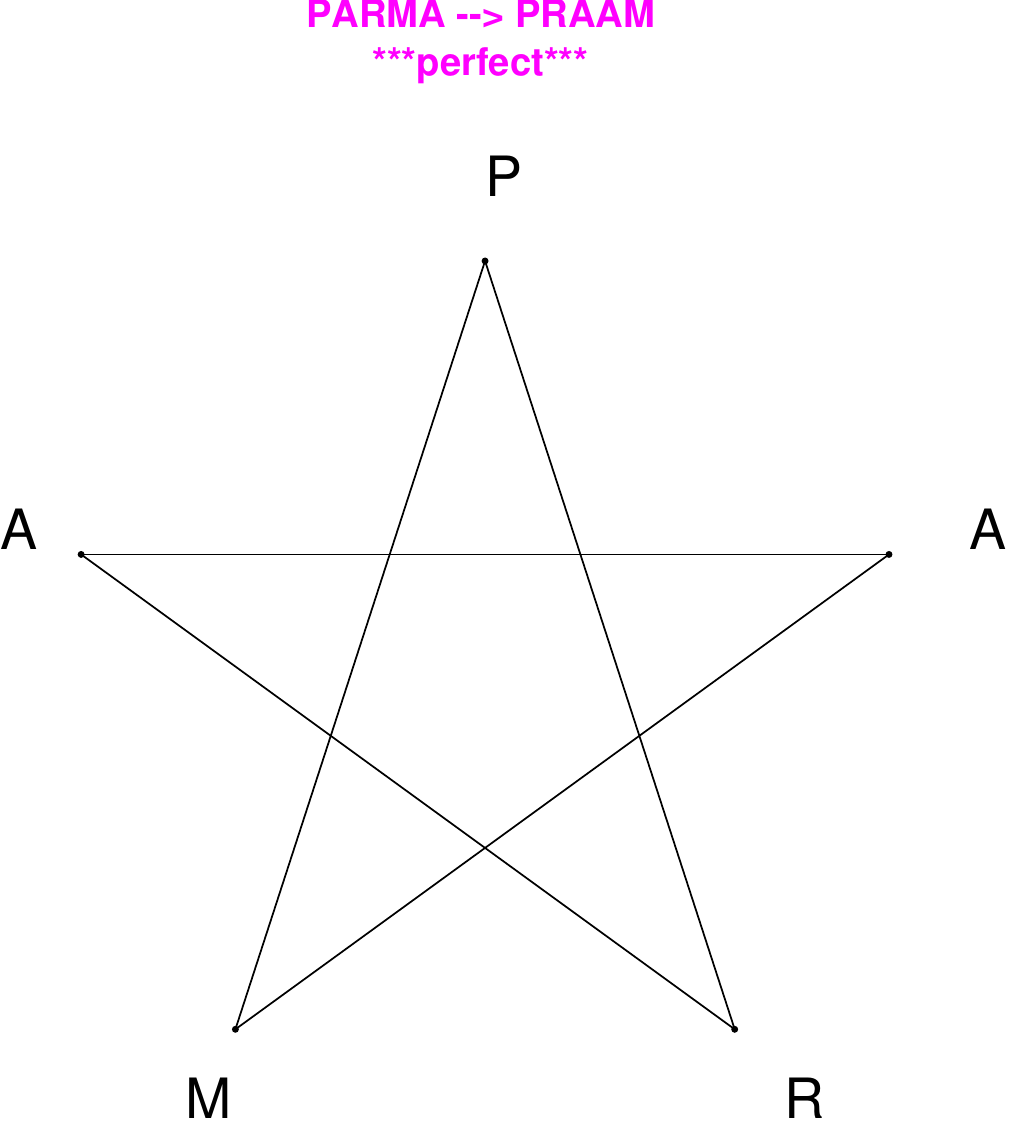}
\end{subfigure}
\hfill
\begin{subfigure}[T]{0.19\textwidth}
\centering
\includegraphics[width=\textwidth]{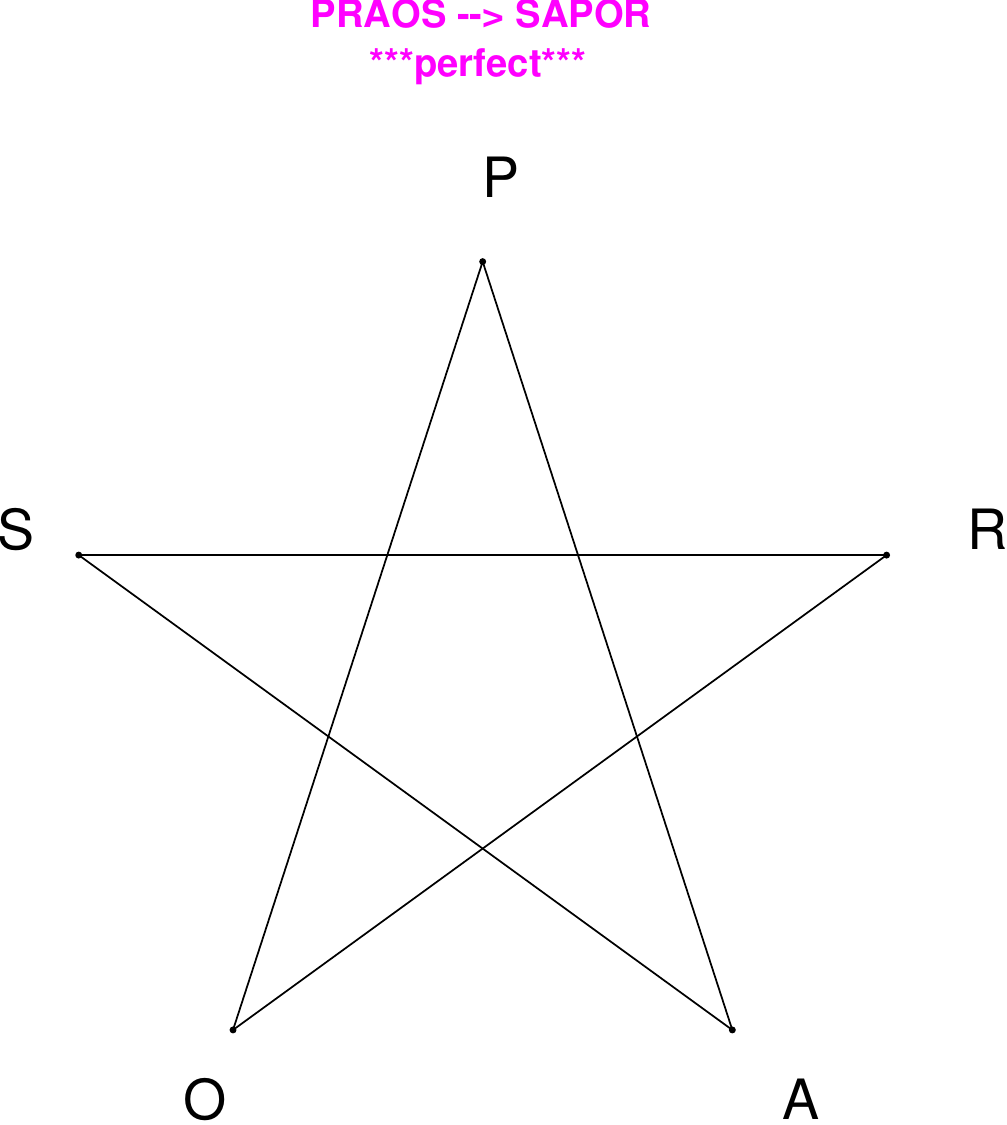}
\end{subfigure}
\hfill
\begin{subfigure}[T]{0.19\textwidth}
\centering
\includegraphics[width=\textwidth]{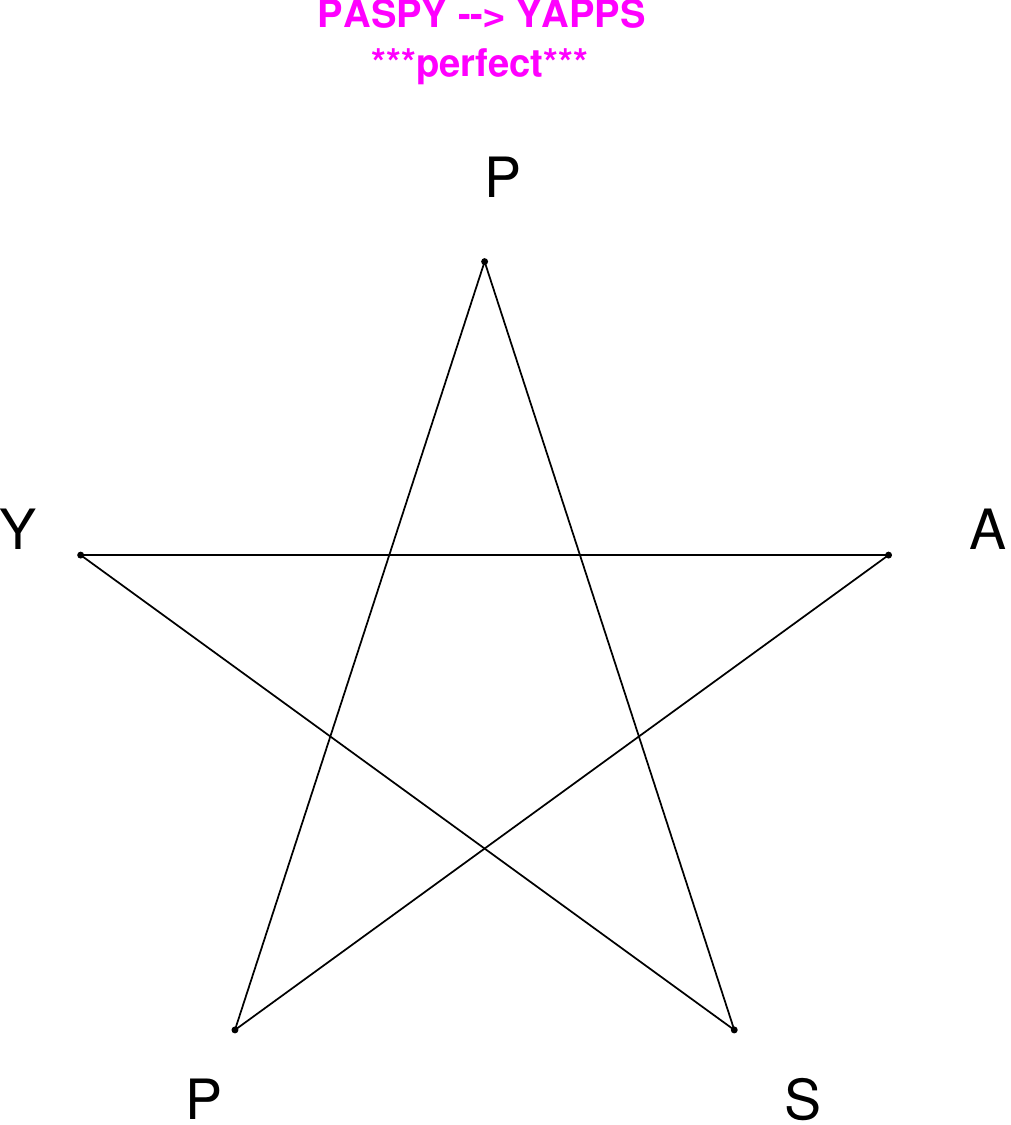}
\end{subfigure}
\end{figure}

\begin{figure}[H]
\centering
\begin{subfigure}[T]{0.19\textwidth}
\centering
\includegraphics[width=\textwidth]{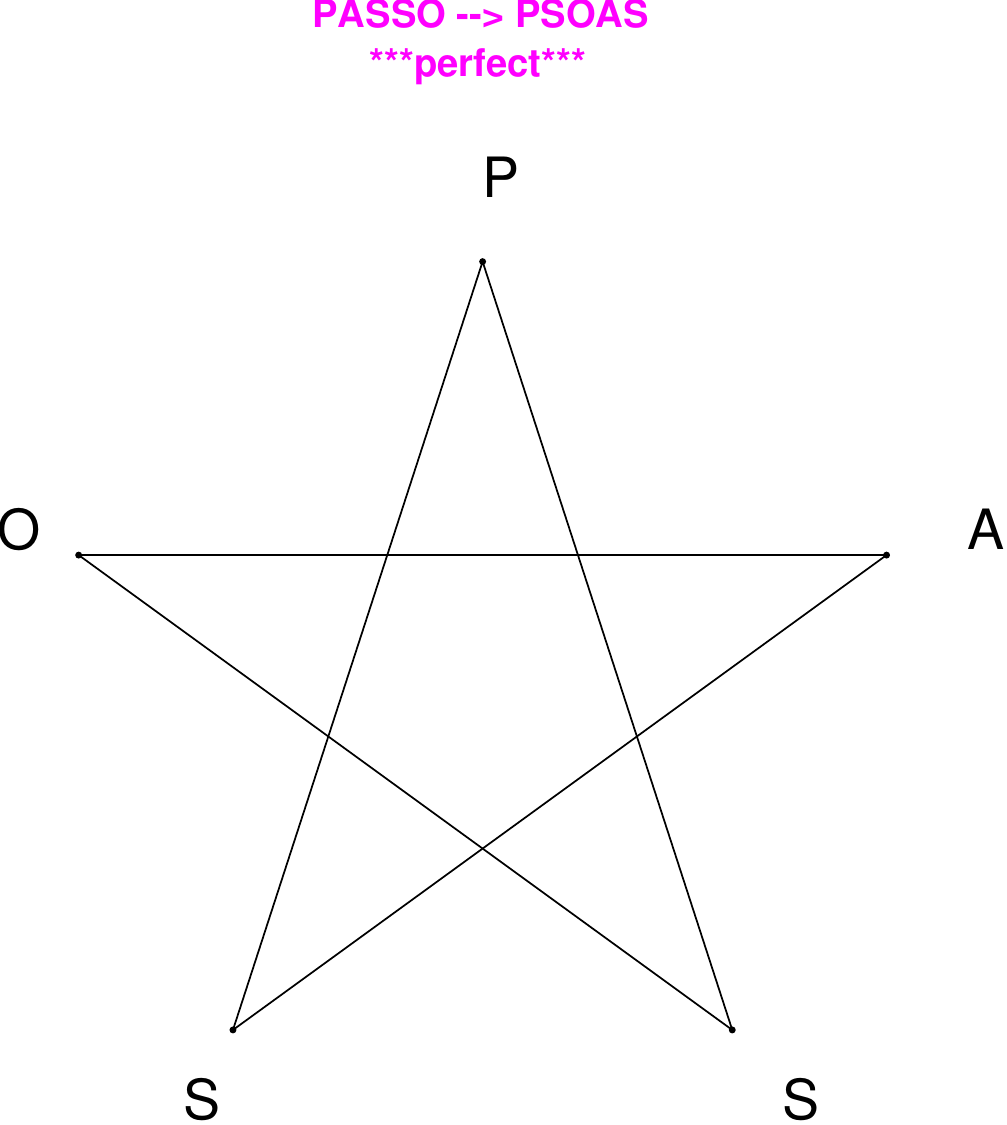}
\end{subfigure}
\hfill
\begin{subfigure}[T]{0.19\textwidth}
\centering
\includegraphics[width=\textwidth]{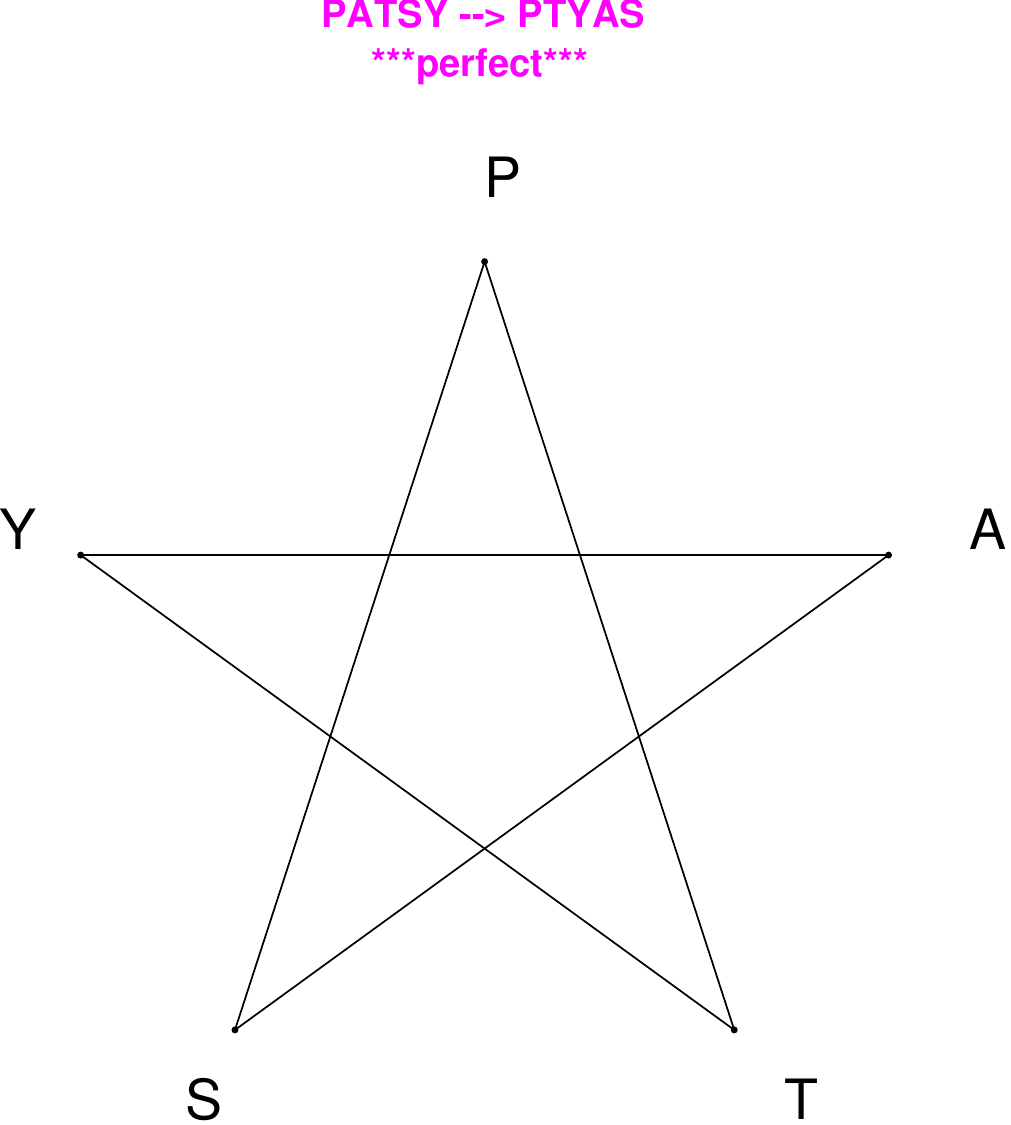}
\end{subfigure}
\hfill
\begin{subfigure}[T]{0.19\textwidth}
\centering
\includegraphics[width=\textwidth]{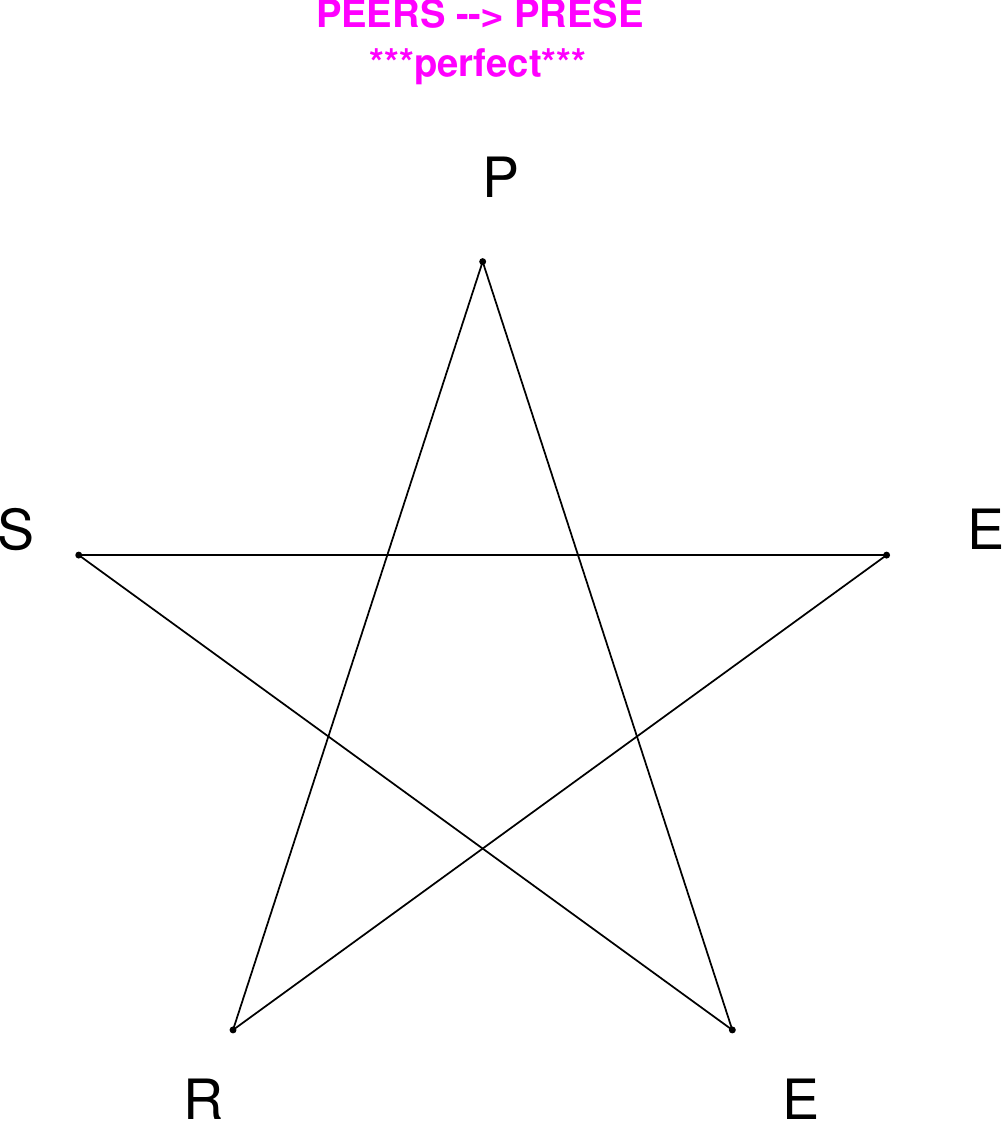}
\end{subfigure}
\hfill
\begin{subfigure}[T]{0.19\textwidth}
\centering
\includegraphics[width=\textwidth]{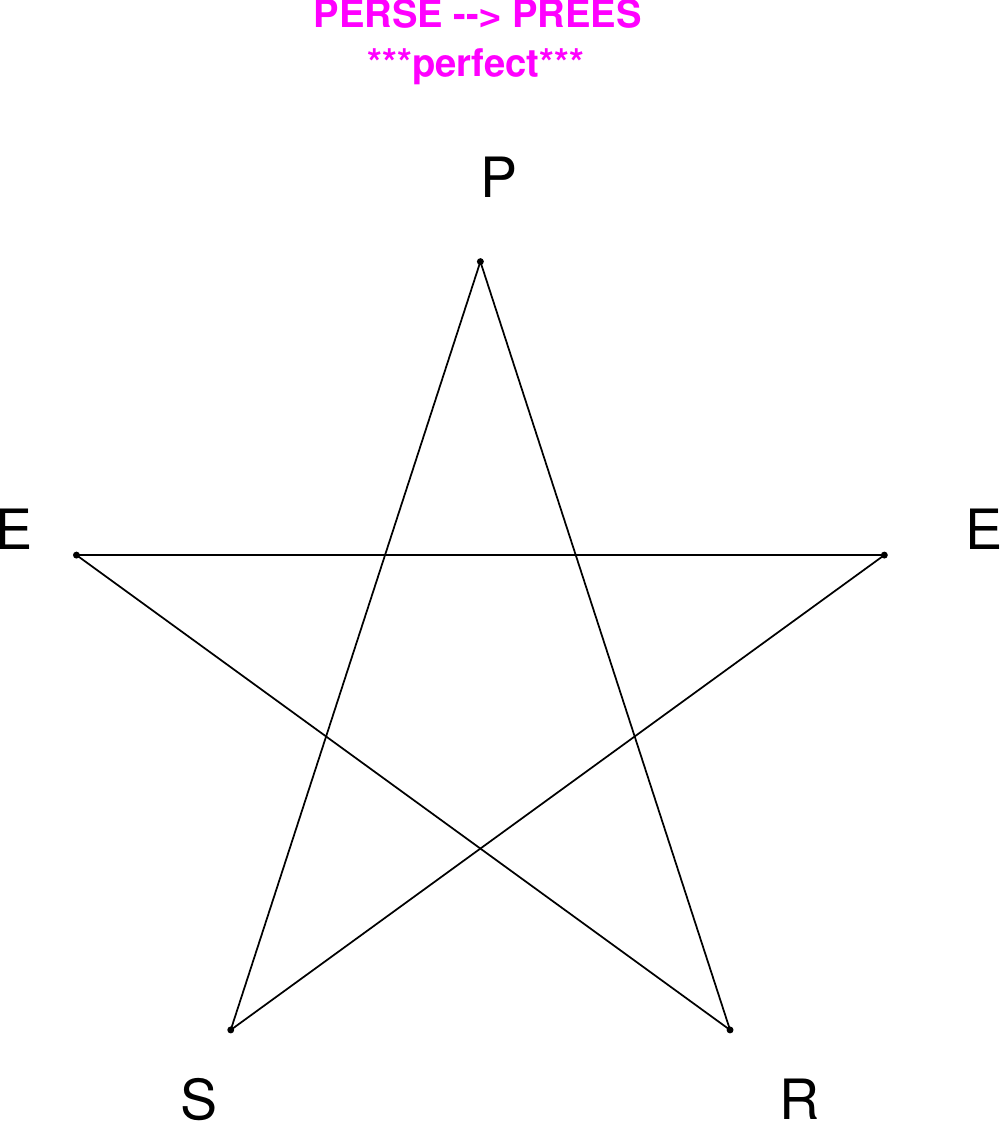}
\end{subfigure}
\hfill
\begin{subfigure}[T]{0.19\textwidth}
\centering
\includegraphics[width=\textwidth]{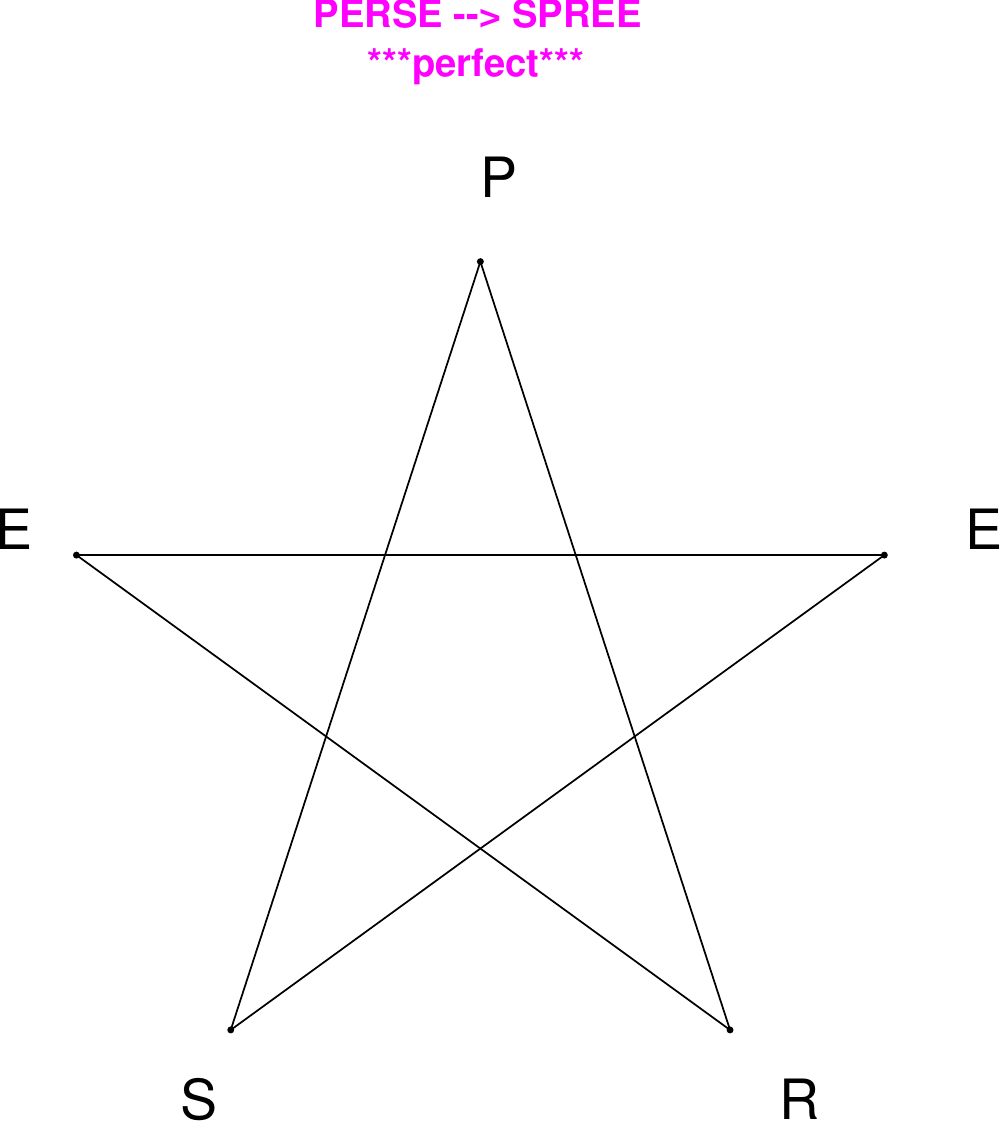}
\end{subfigure}
\end{figure}

\begin{figure}[H]
\centering
\begin{subfigure}[T]{0.19\textwidth}
\centering
\includegraphics[width=\textwidth]{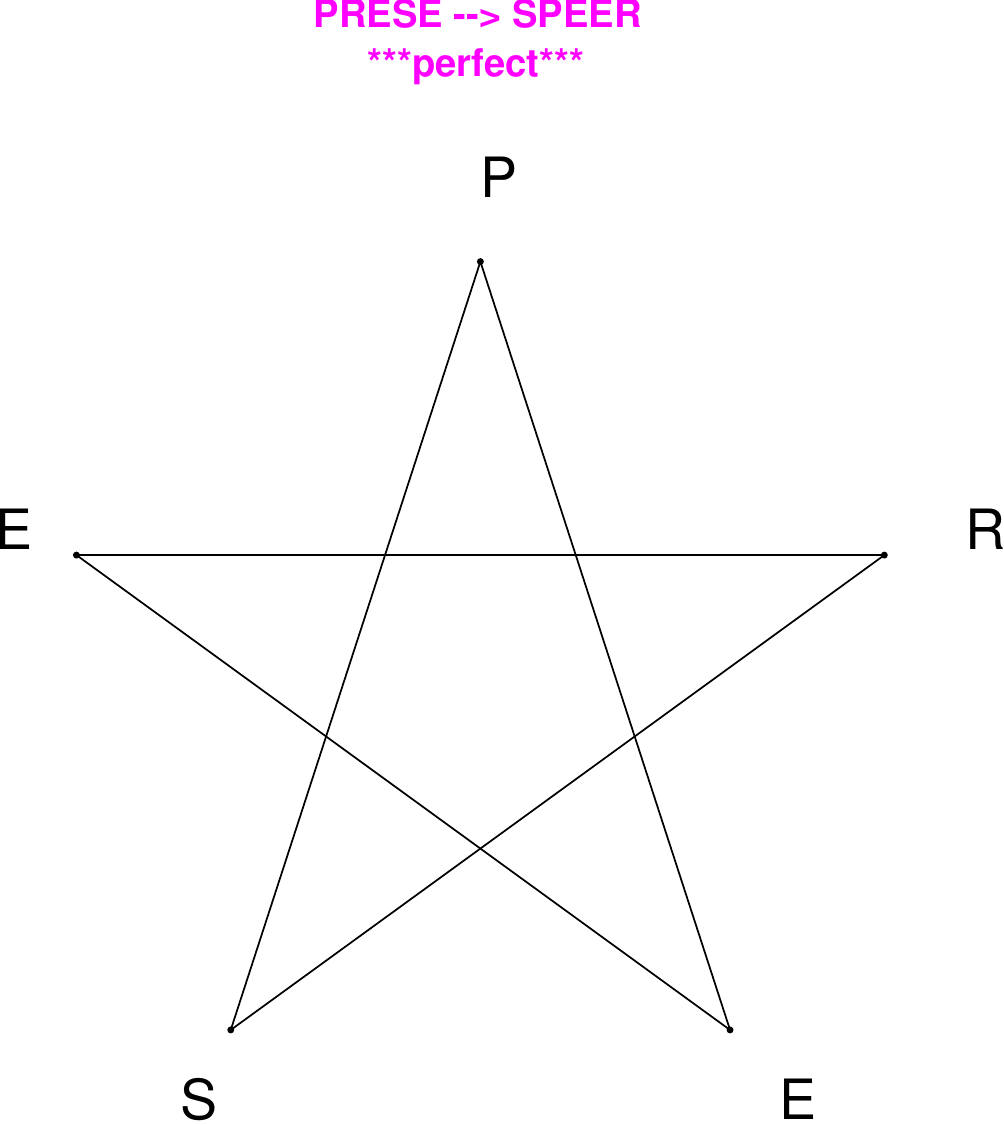}
\end{subfigure}
\hfill
\begin{subfigure}[T]{0.19\textwidth}
\centering
\includegraphics[width=\textwidth]{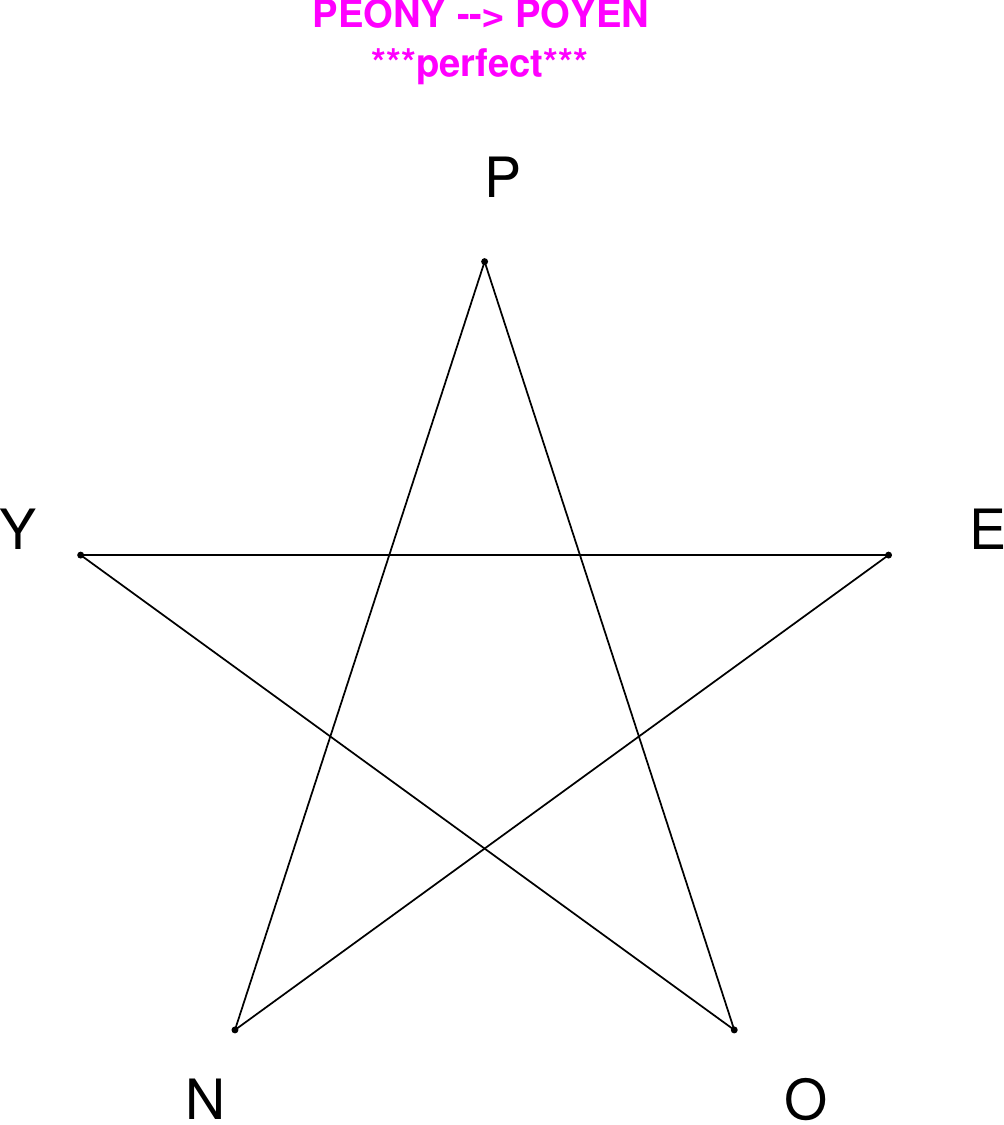}
\end{subfigure}
\hfill
\begin{subfigure}[T]{0.19\textwidth}
\centering
\includegraphics[width=\textwidth]{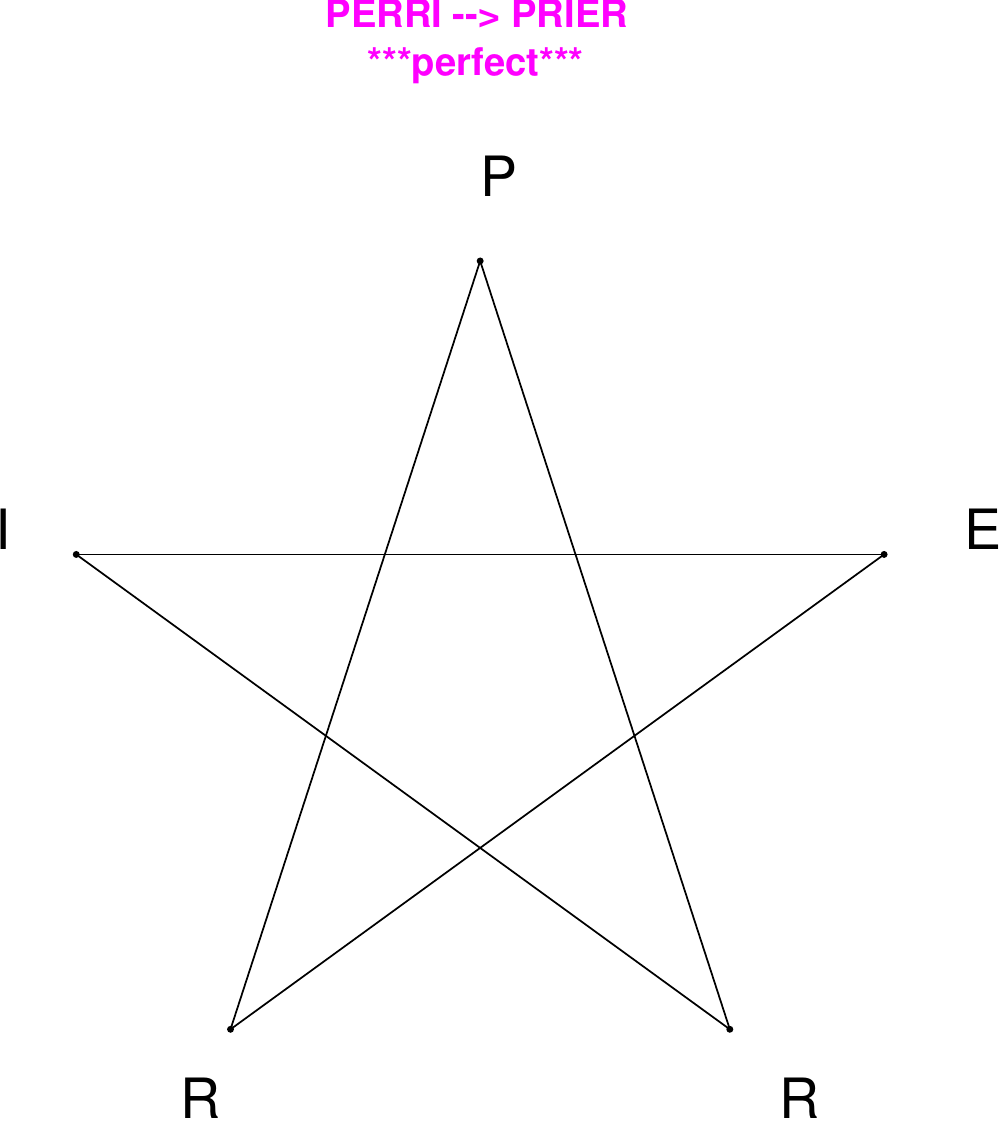}
\end{subfigure}
\hfill
\begin{subfigure}[T]{0.19\textwidth}
\centering
\includegraphics[width=\textwidth]{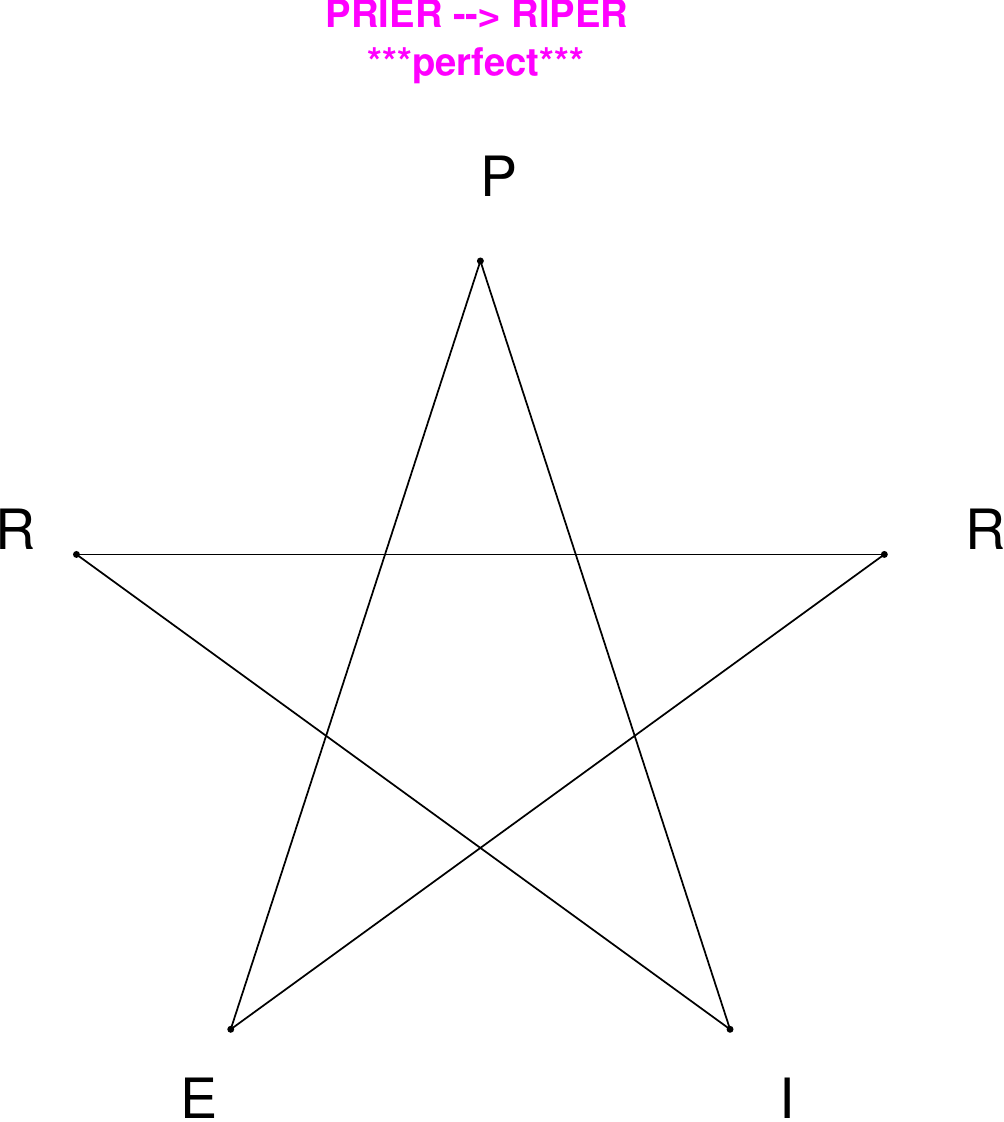}
\end{subfigure}
\hfill
\begin{subfigure}[T]{0.19\textwidth}
\centering
\includegraphics[width=\textwidth]{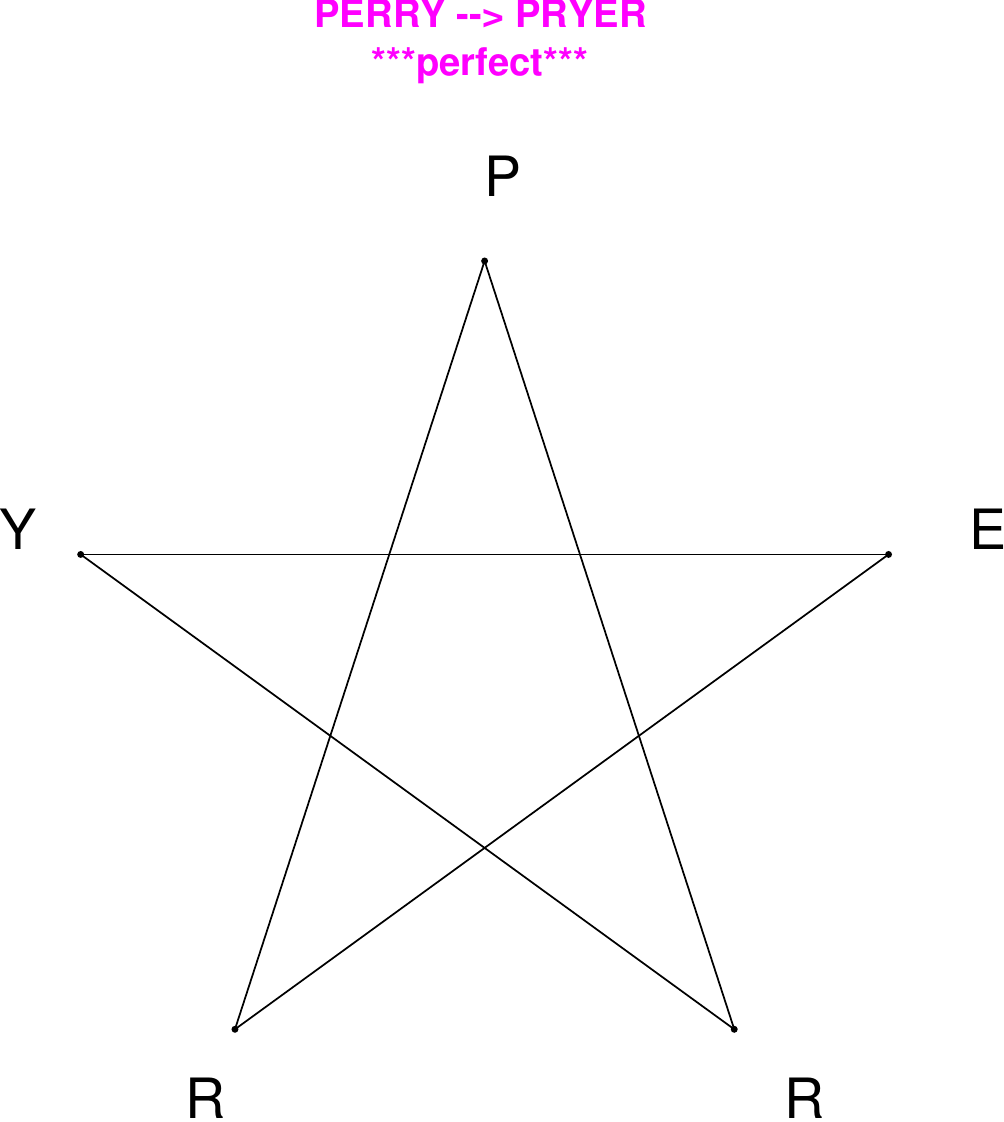}
\end{subfigure}
\end{figure}

\begin{figure}[H]
\centering
\begin{subfigure}[T]{0.19\textwidth}
\centering
\includegraphics[width=\textwidth]{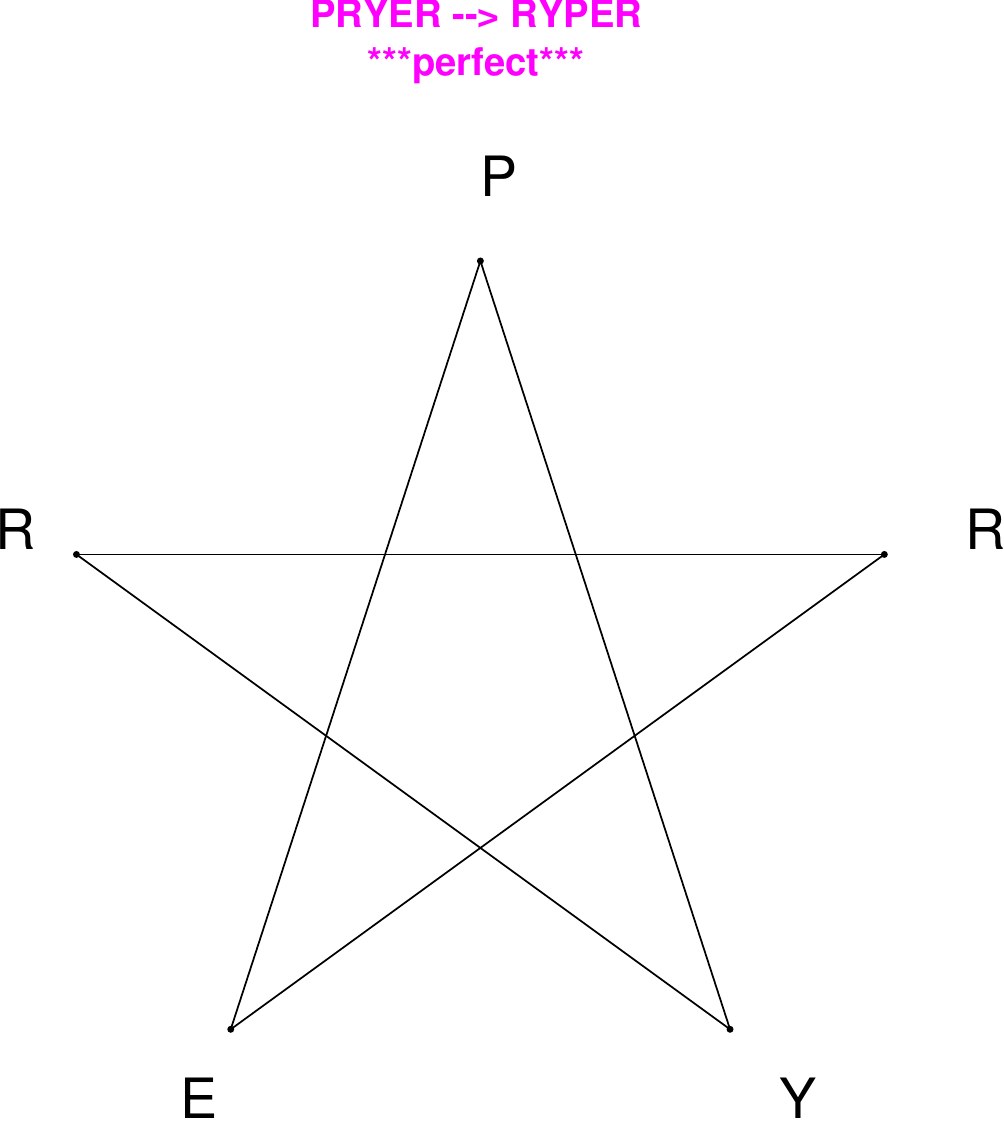}
\end{subfigure}
\hfill
\begin{subfigure}[T]{0.19\textwidth}
\centering
\includegraphics[width=\textwidth]{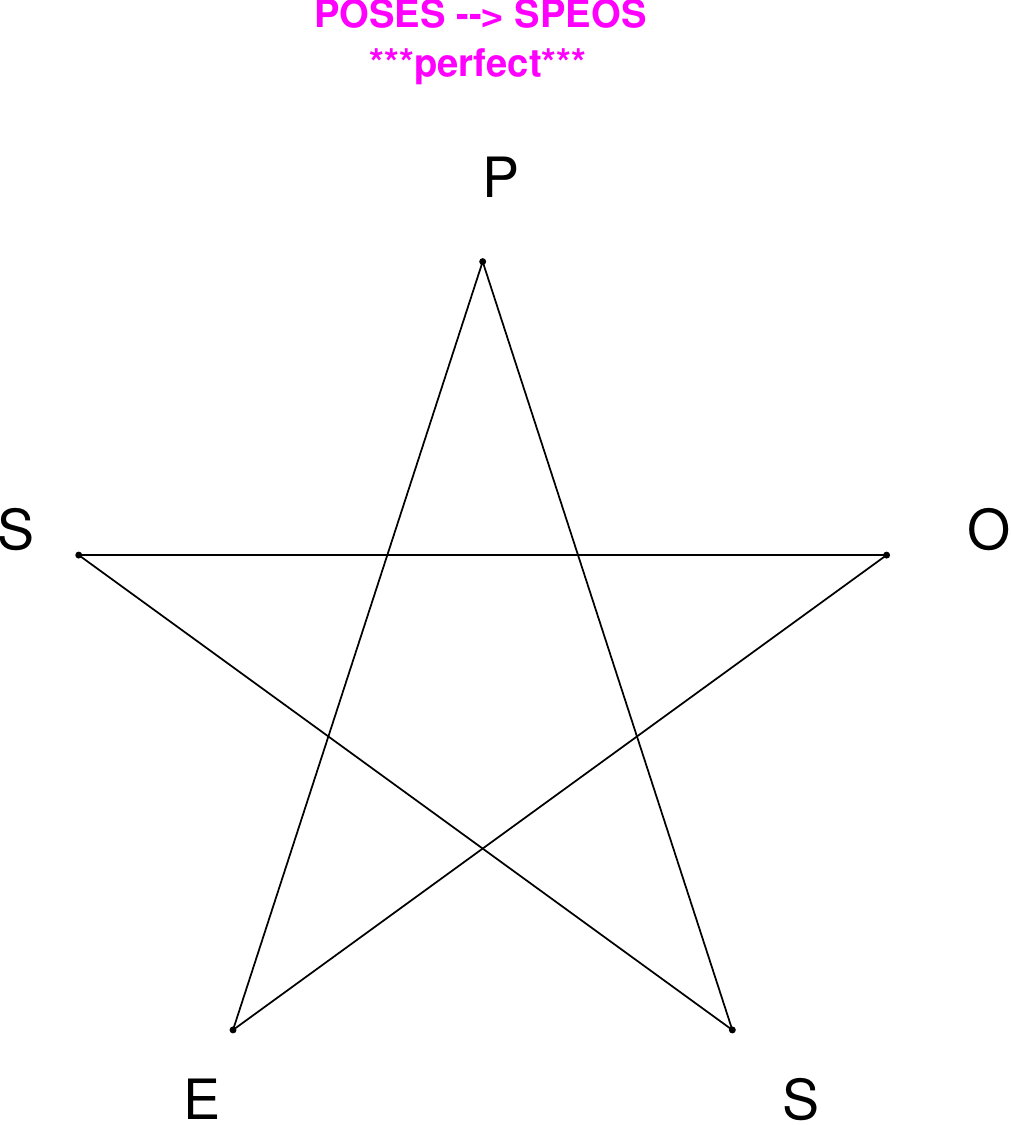}
\end{subfigure}
\hfill
\begin{subfigure}[T]{0.19\textwidth}
\centering
\includegraphics[width=\textwidth]{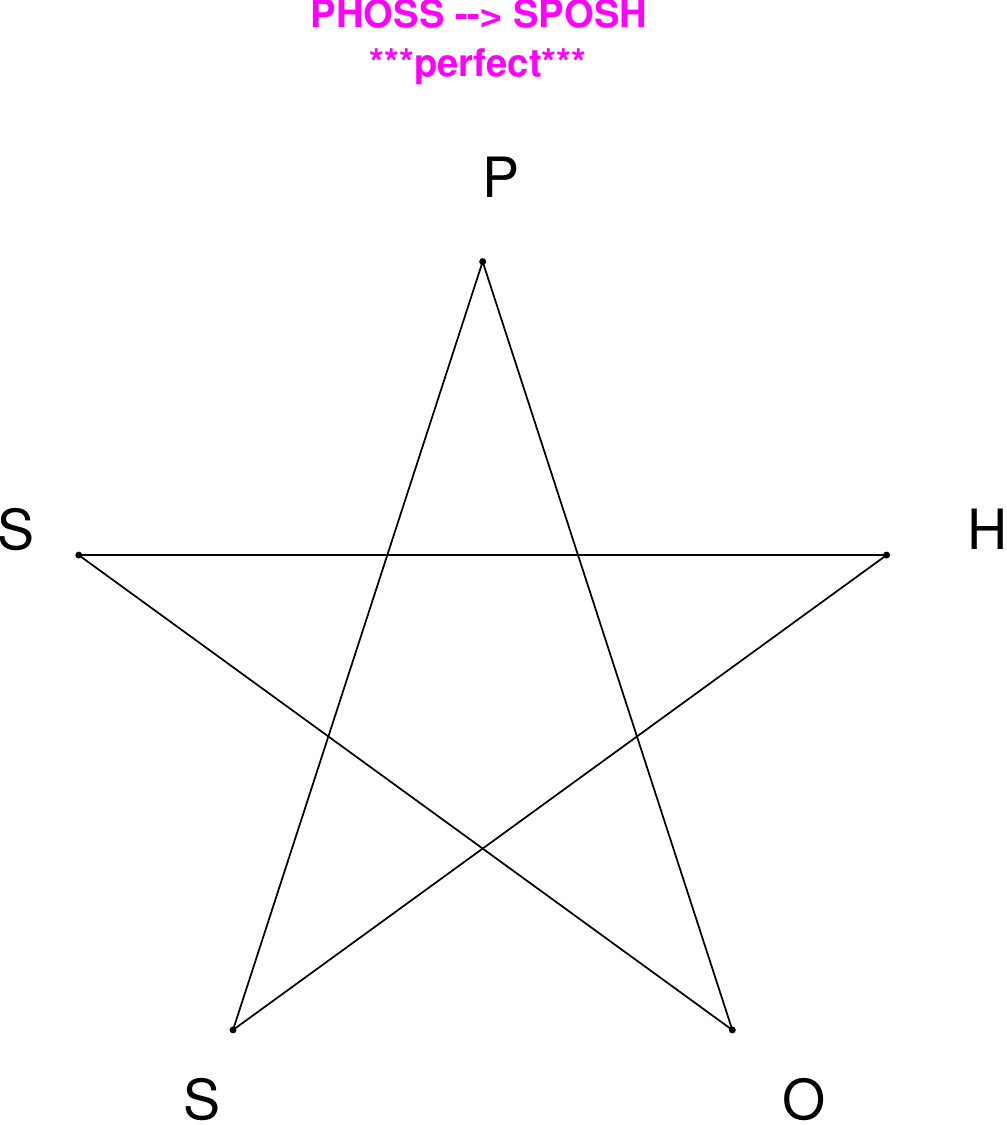}
\end{subfigure}
\hfill
\begin{subfigure}[T]{0.19\textwidth}
\centering
\includegraphics[width=\textwidth]{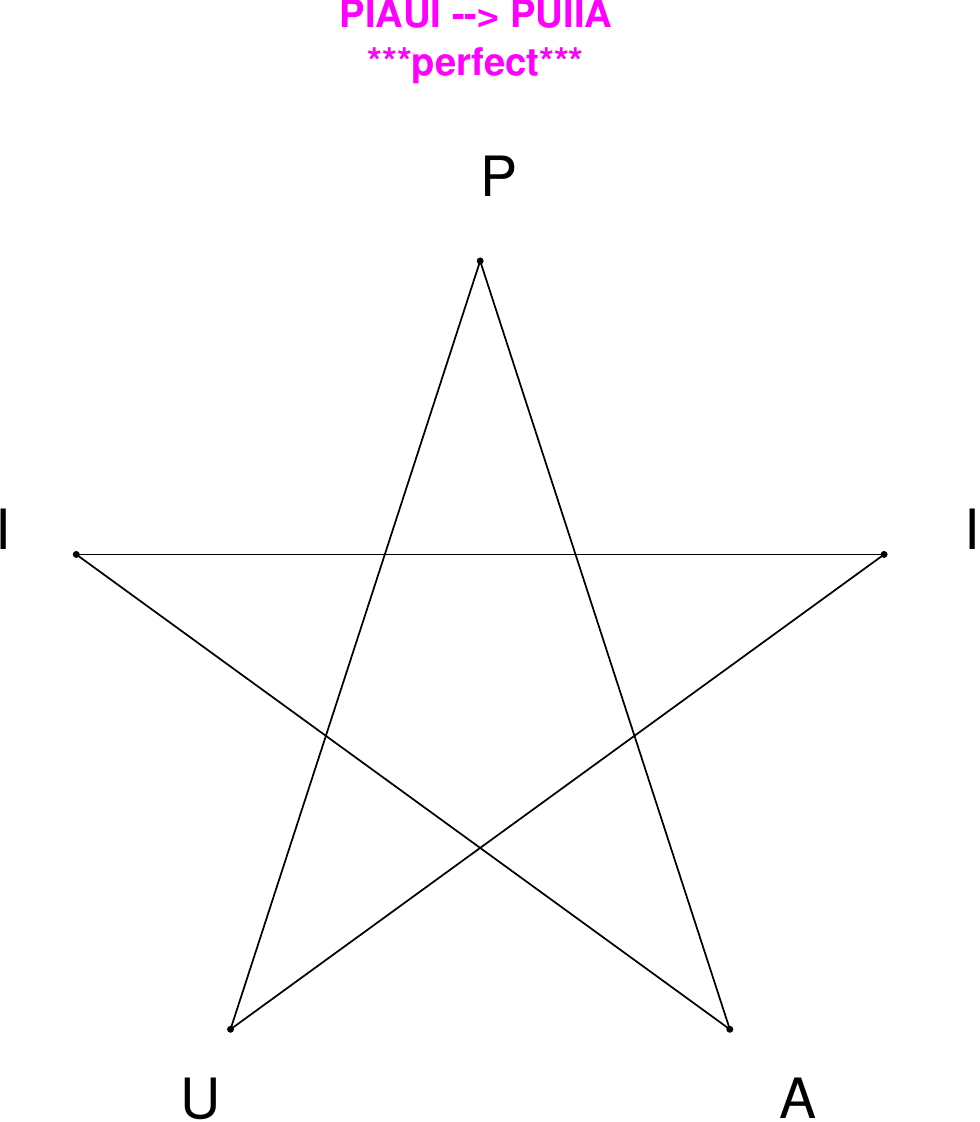}
\end{subfigure}
\hfill
\begin{subfigure}[T]{0.19\textwidth}
\centering
\includegraphics[width=\textwidth]{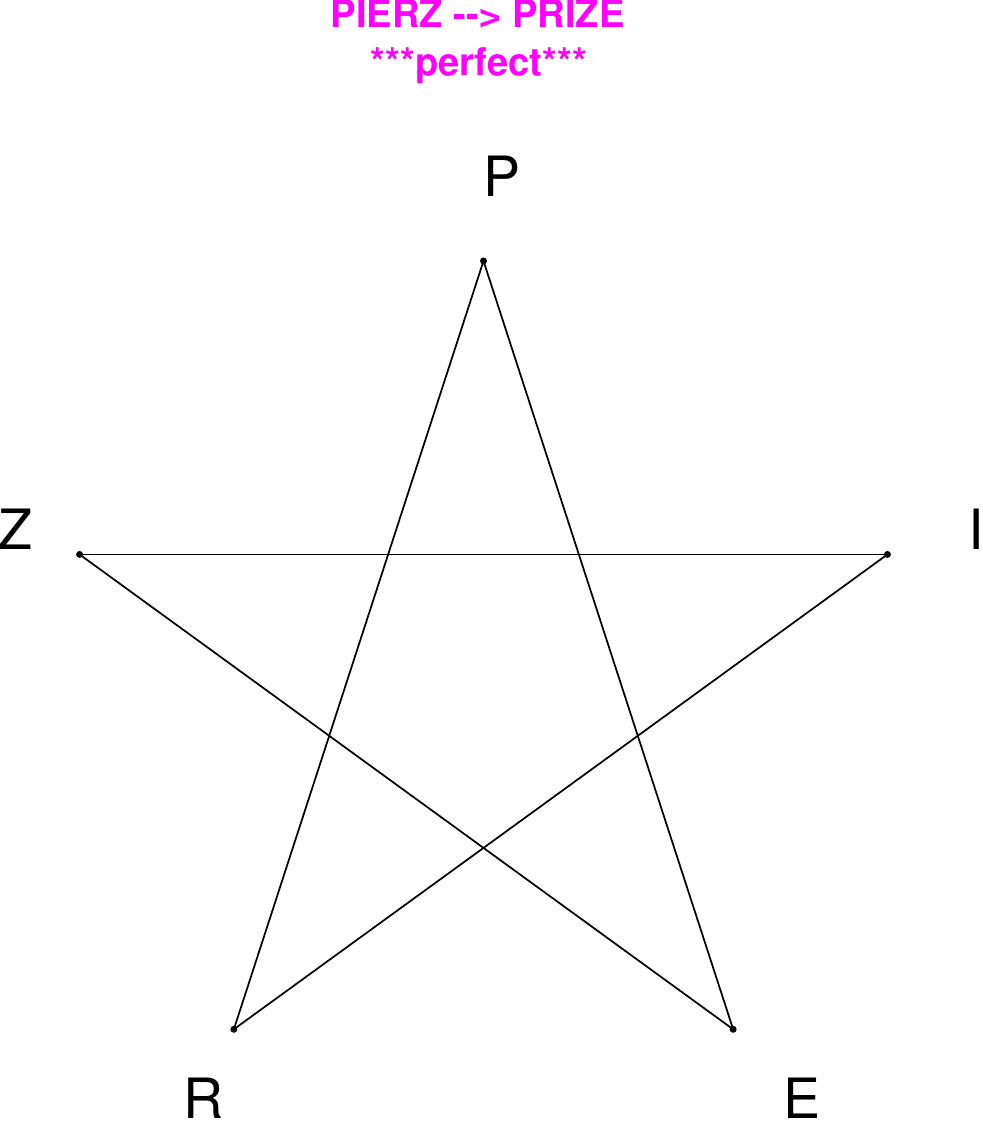}
\end{subfigure}
\end{figure}

\begin{figure}[H]
\centering
\begin{subfigure}[T]{0.19\textwidth}
\centering
\includegraphics[width=\textwidth]{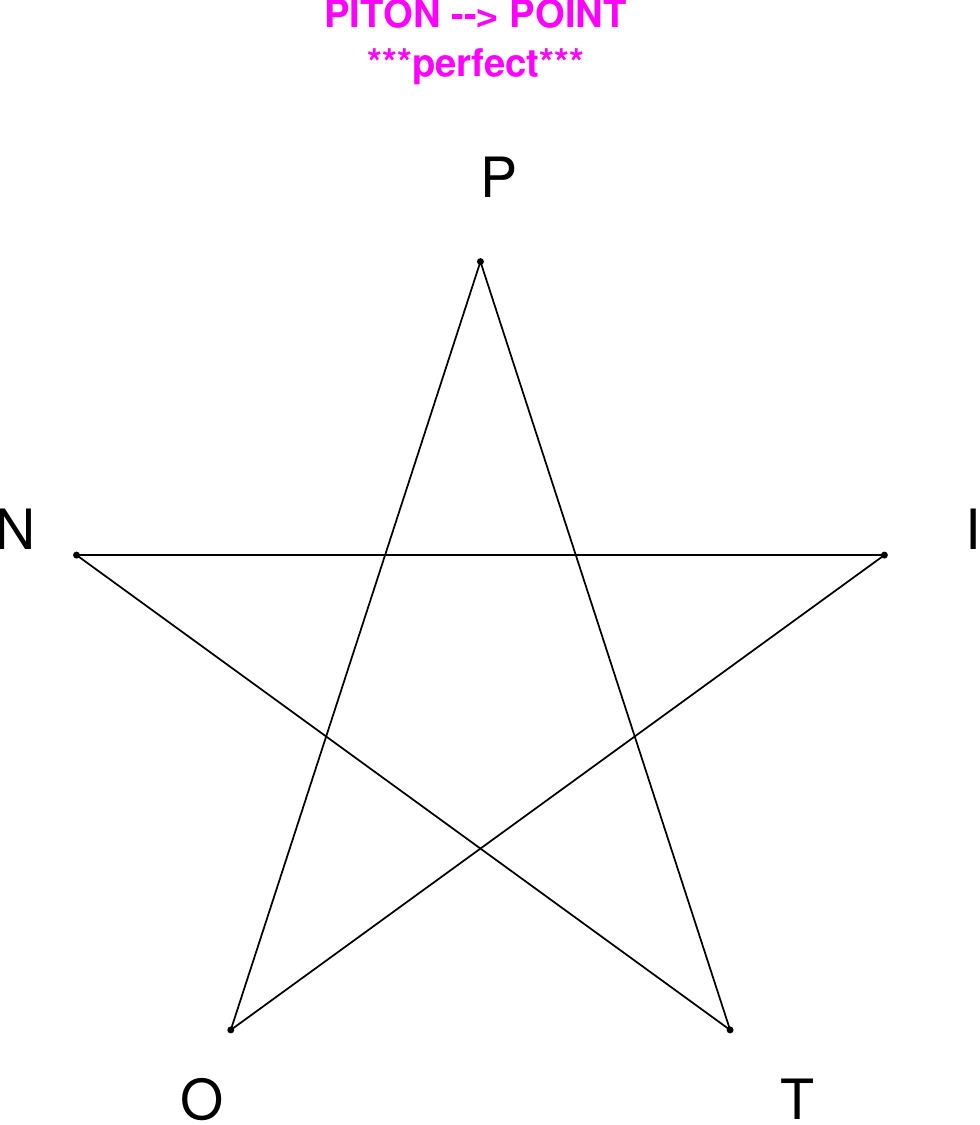}
\end{subfigure}
\hfill
\begin{subfigure}[T]{0.19\textwidth}
\centering
\includegraphics[width=\textwidth]{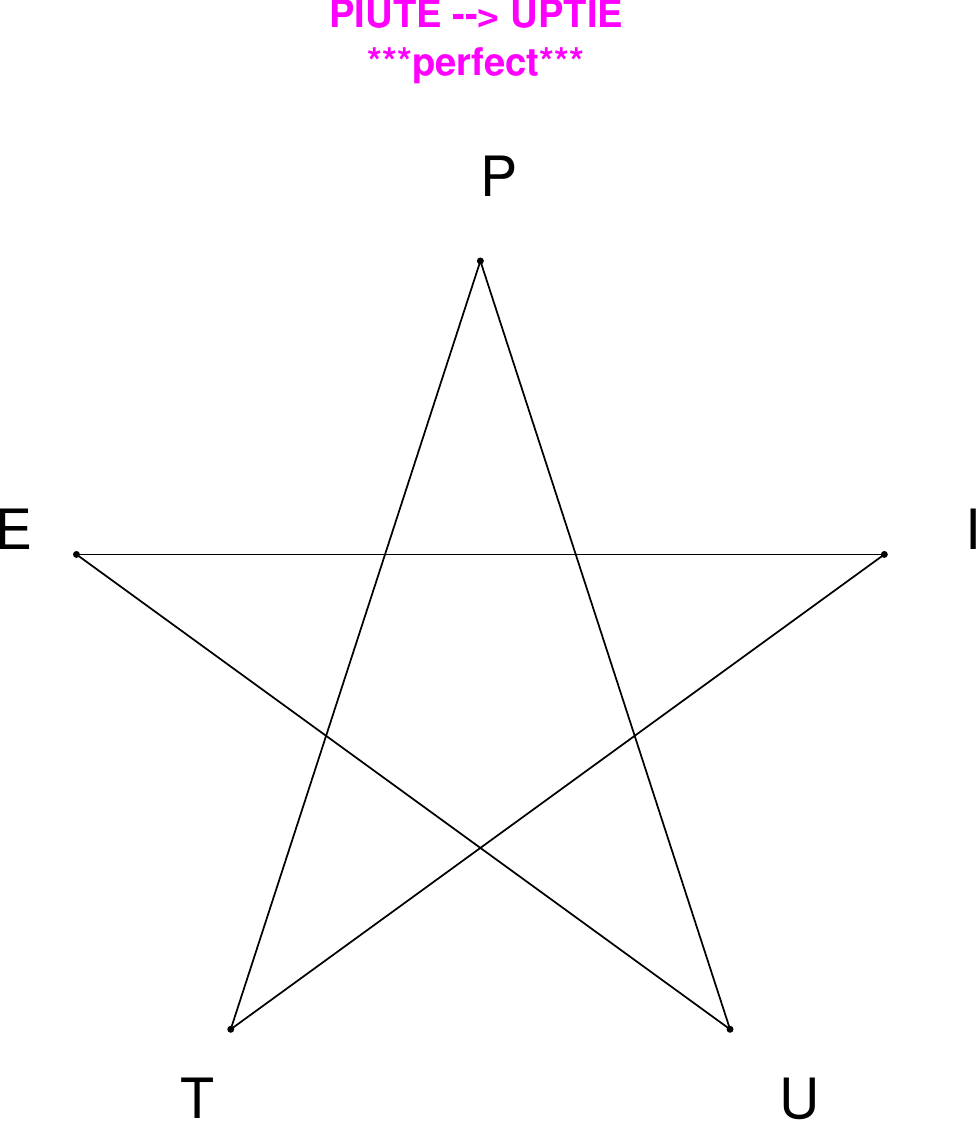}
\end{subfigure}
\hfill
\begin{subfigure}[T]{0.19\textwidth}
\centering
\includegraphics[width=\textwidth]{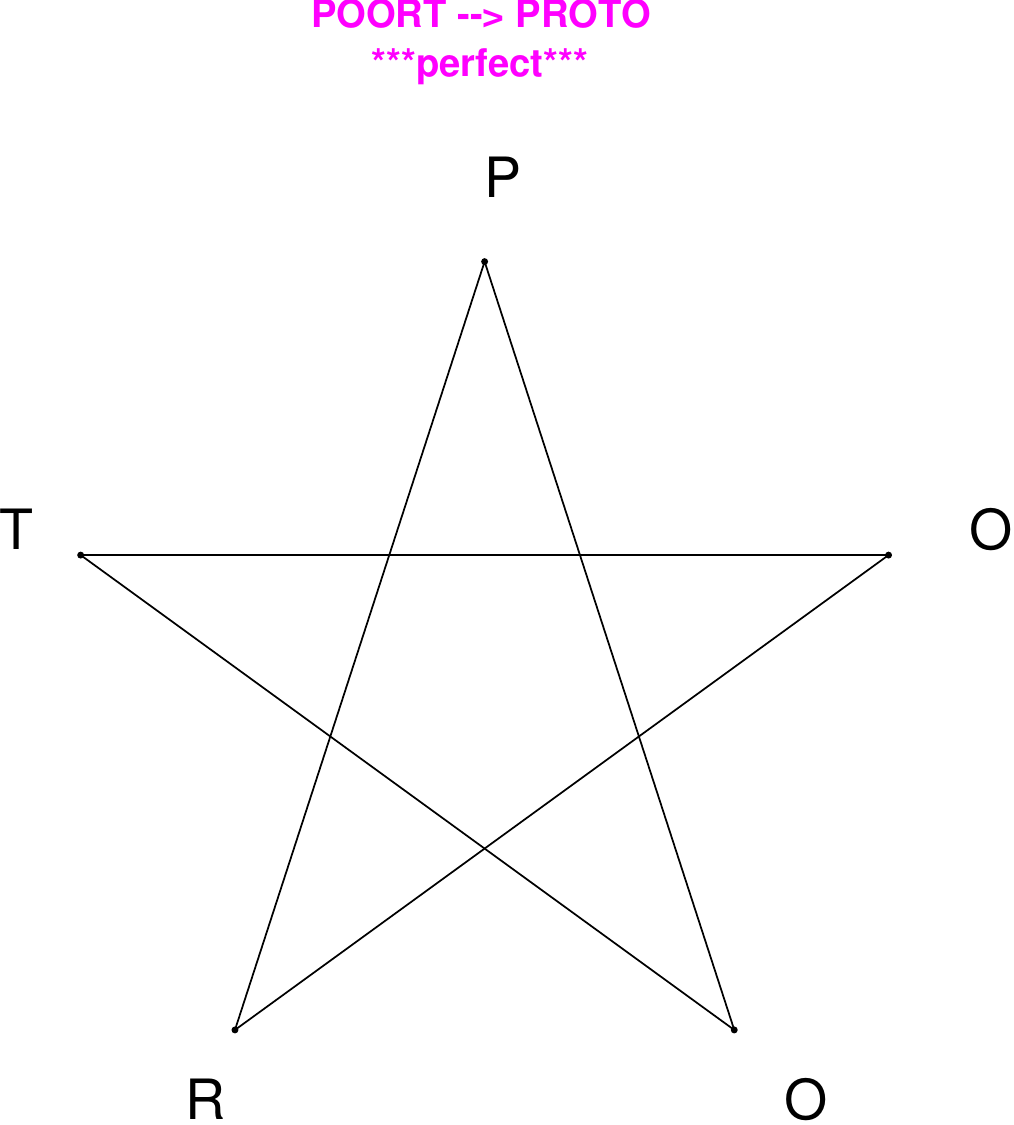}
\end{subfigure}
\hfill
\begin{subfigure}[T]{0.19\textwidth}
\centering
\includegraphics[width=\textwidth]{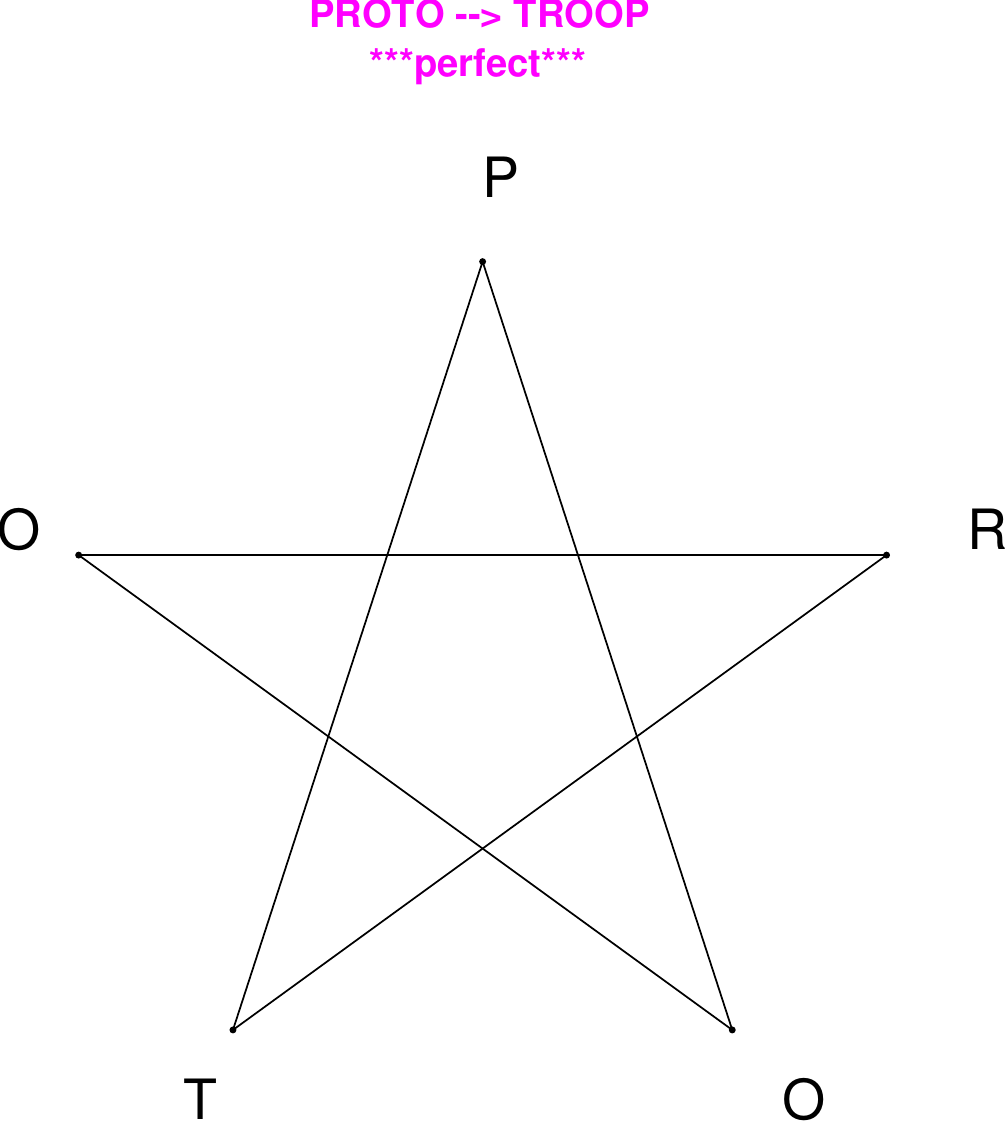}
\end{subfigure}
\hfill
\begin{subfigure}[T]{0.19\textwidth}
\centering
\includegraphics[width=\textwidth]{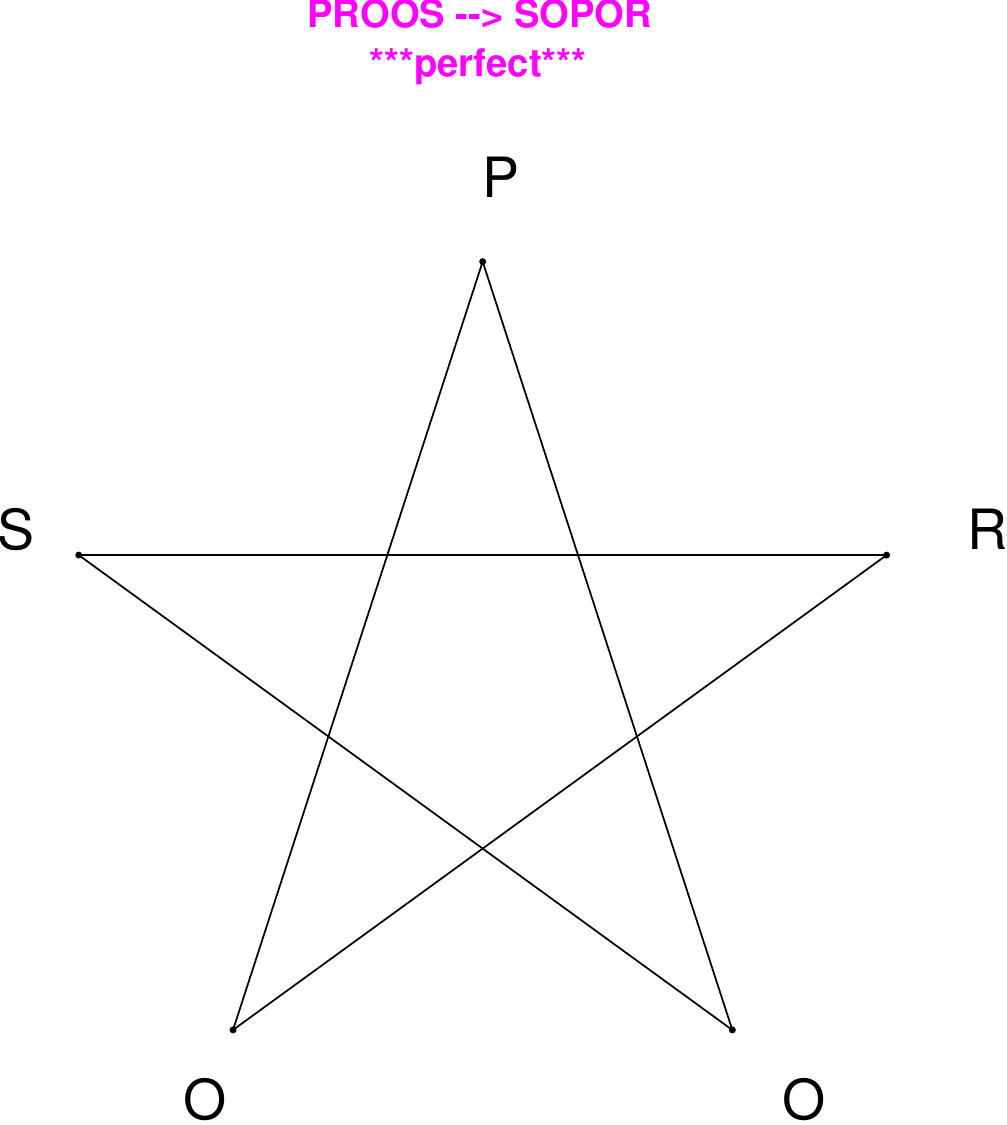}
\end{subfigure}
\end{figure}

\begin{figure}[H]
\centering
\begin{subfigure}[T]{0.19\textwidth}
\centering
\includegraphics[width=\textwidth]{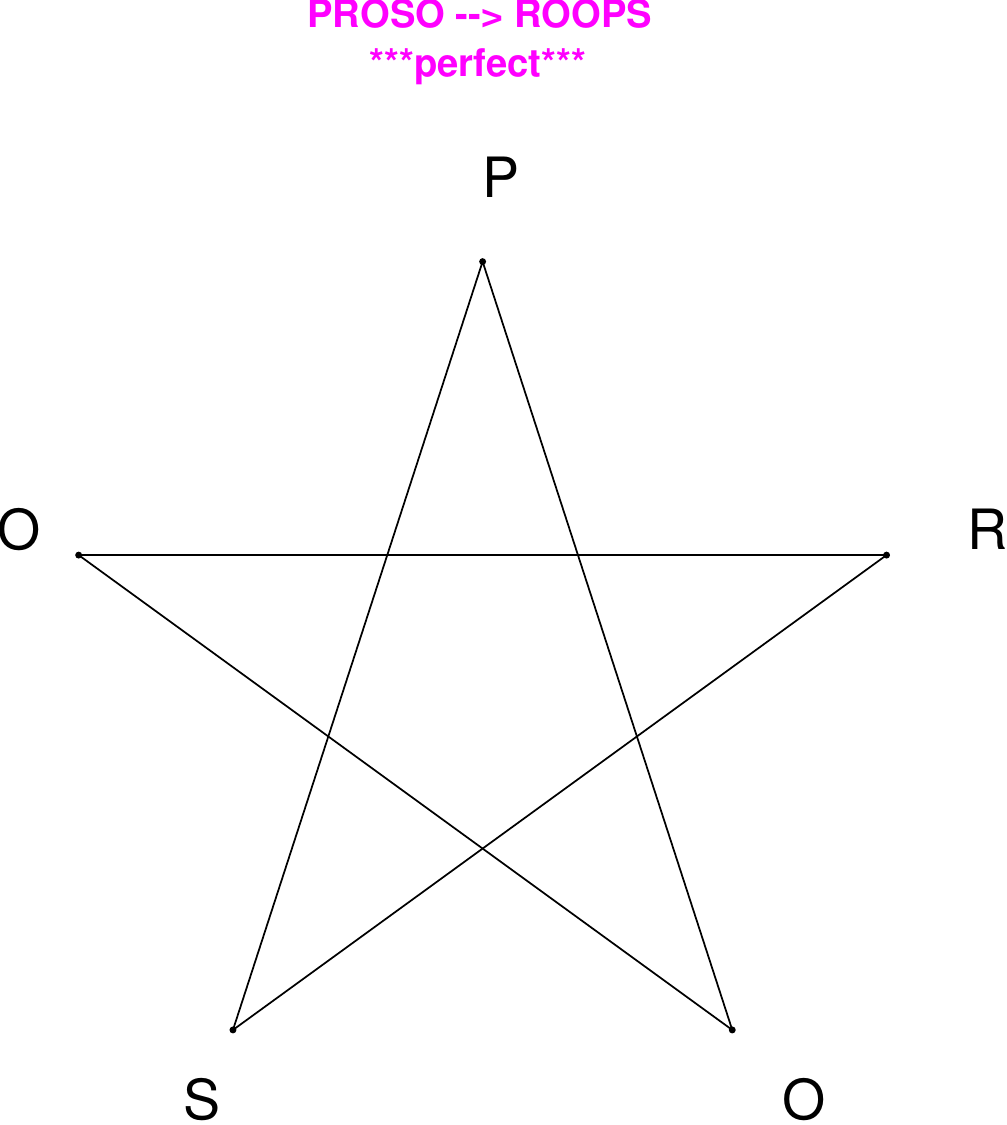}
\end{subfigure}
\hfill
\begin{subfigure}[T]{0.19\textwidth}
\centering
\includegraphics[width=\textwidth]{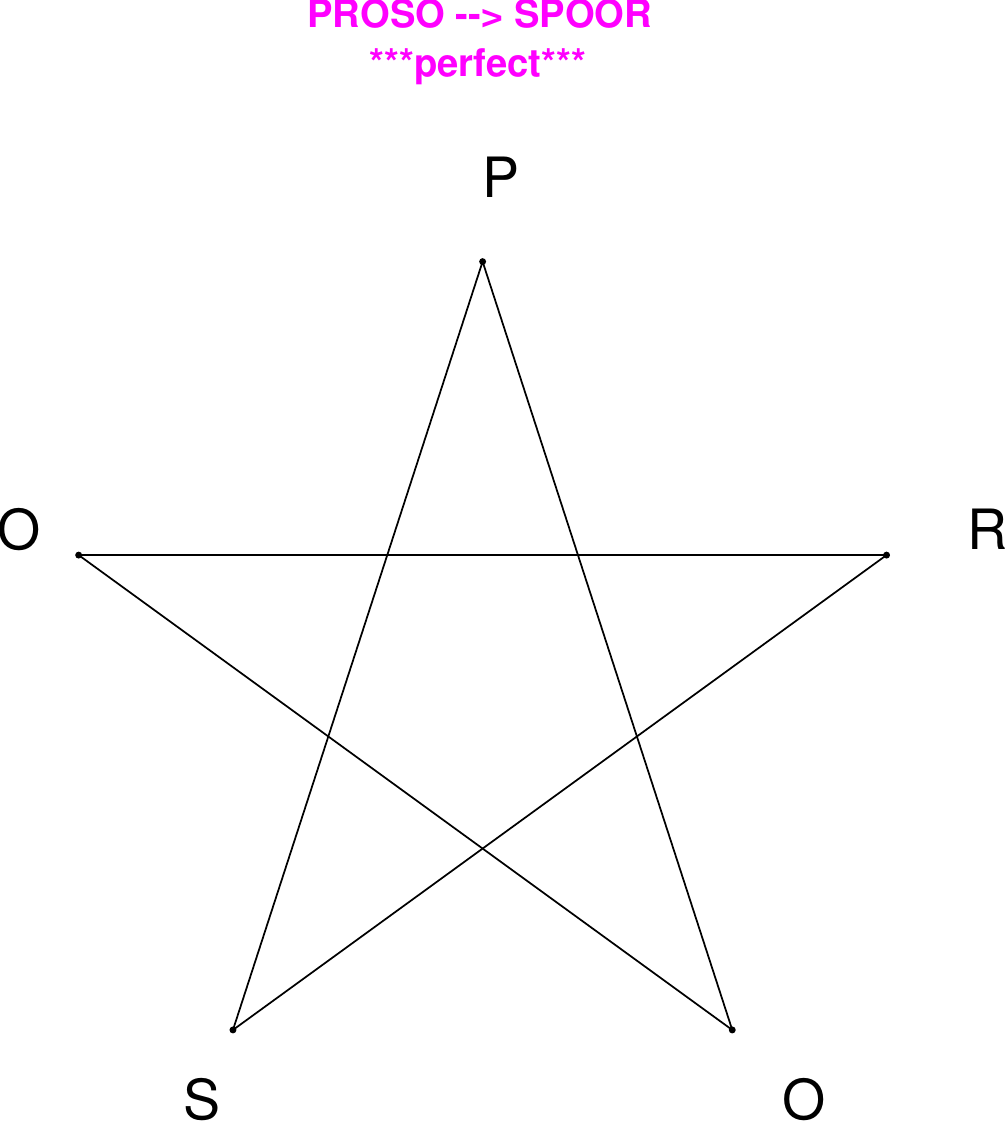}
\end{subfigure}
\hfill
\begin{subfigure}[T]{0.19\textwidth}
\centering
\includegraphics[width=\textwidth]{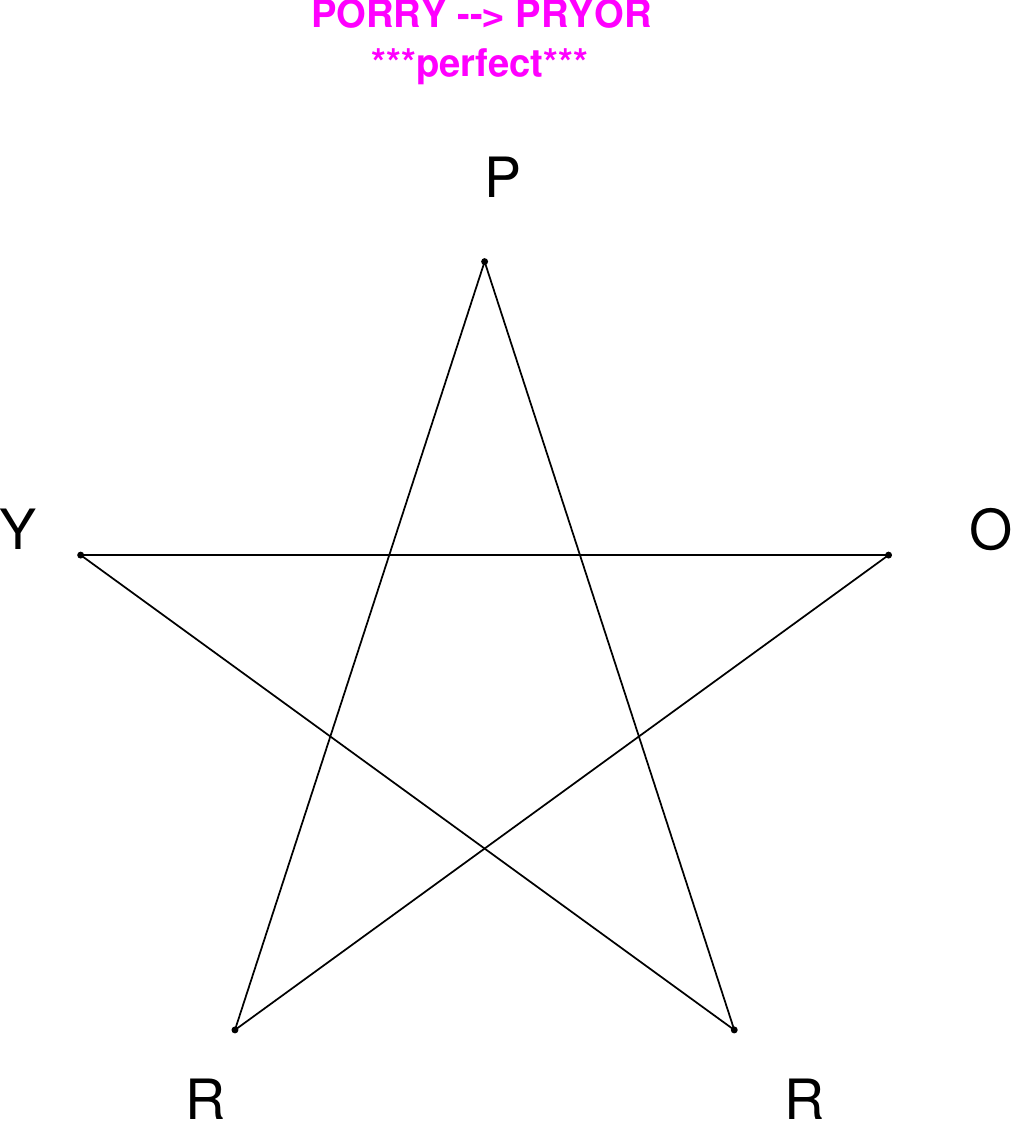}
\end{subfigure}
\hfill
\begin{subfigure}[T]{0.19\textwidth}
\centering
\includegraphics[width=\textwidth]{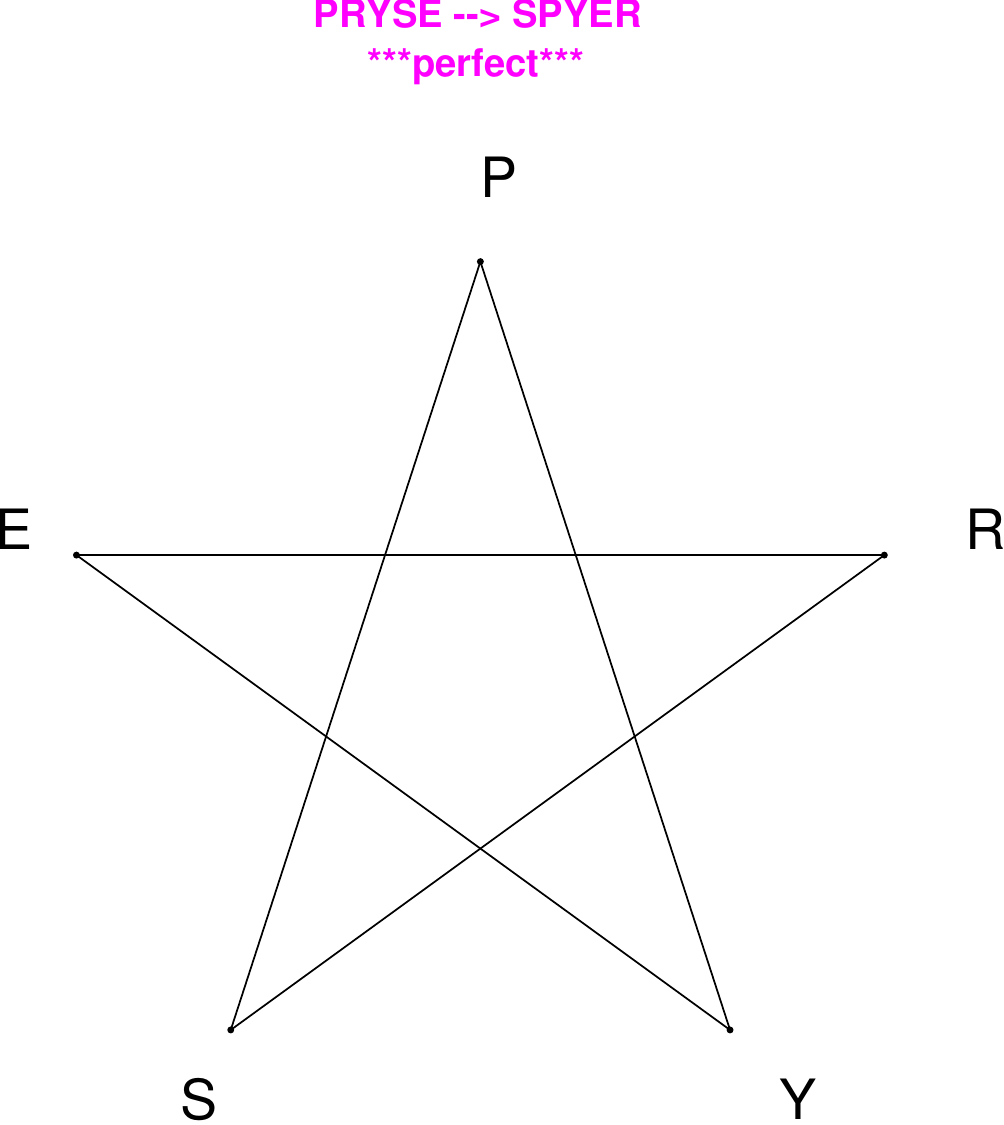}
\end{subfigure}
\hfill
\begin{subfigure}[T]{0.19\textwidth}
\centering
\includegraphics[width=\textwidth]{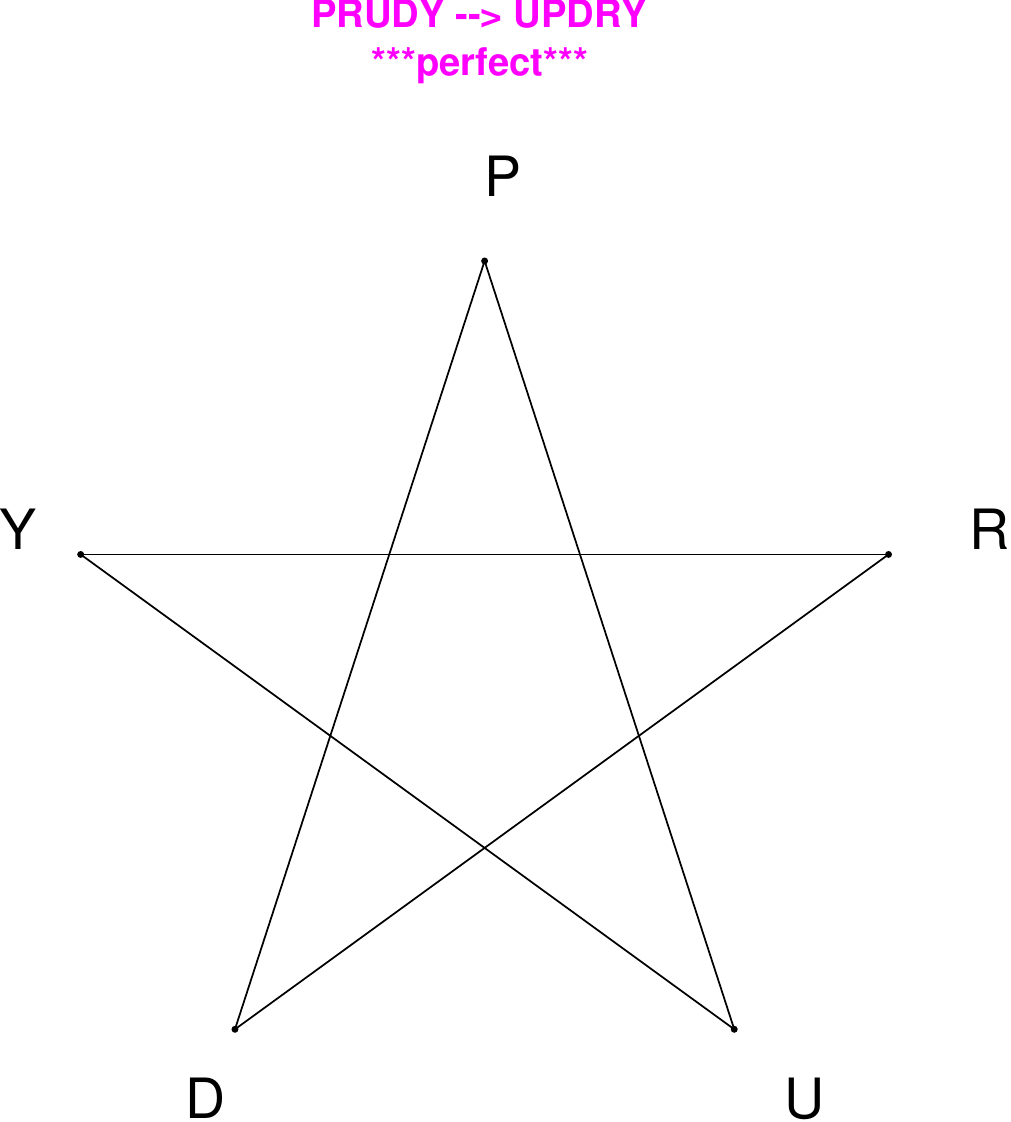}
\end{subfigure}
\end{figure}

\begin{figure}[H]
\centering
\begin{subfigure}[T]{0.19\textwidth}
\centering
\includegraphics[width=\textwidth]{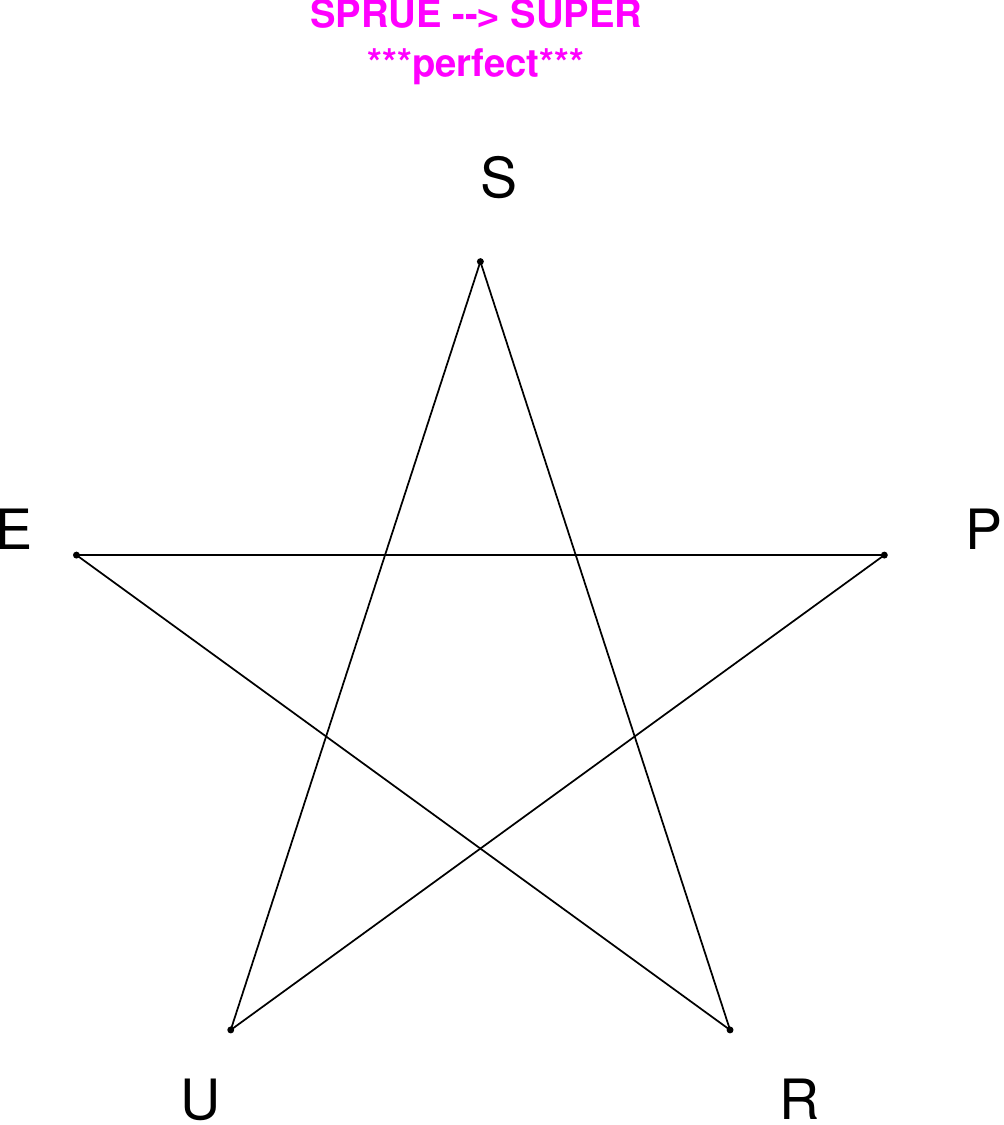}
\end{subfigure}
\hfill
\begin{subfigure}[T]{0.19\textwidth}
\centering
\includegraphics[width=\textwidth]{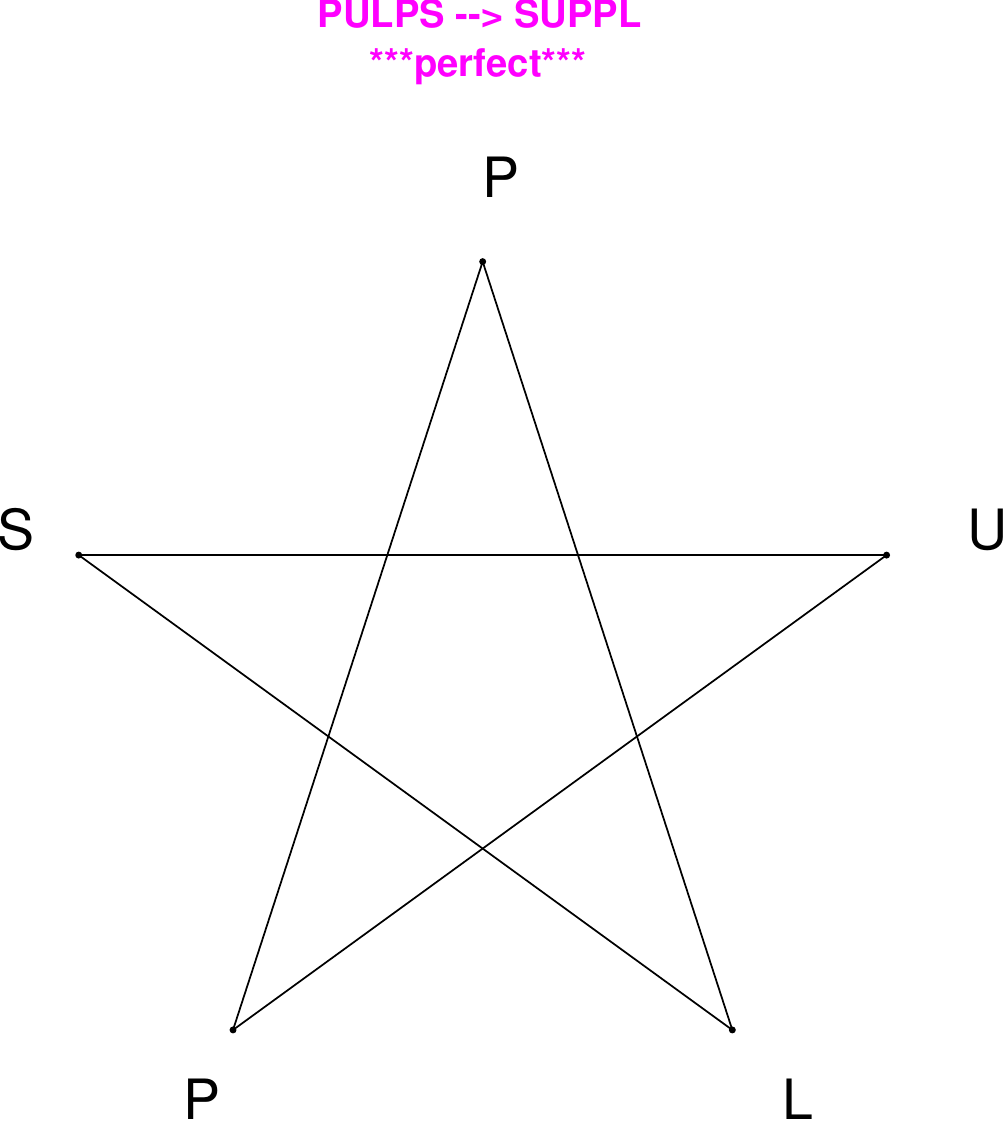}
\end{subfigure}
\hfill
\begin{subfigure}[T]{0.19\textwidth}
\centering
\includegraphics[width=\textwidth]{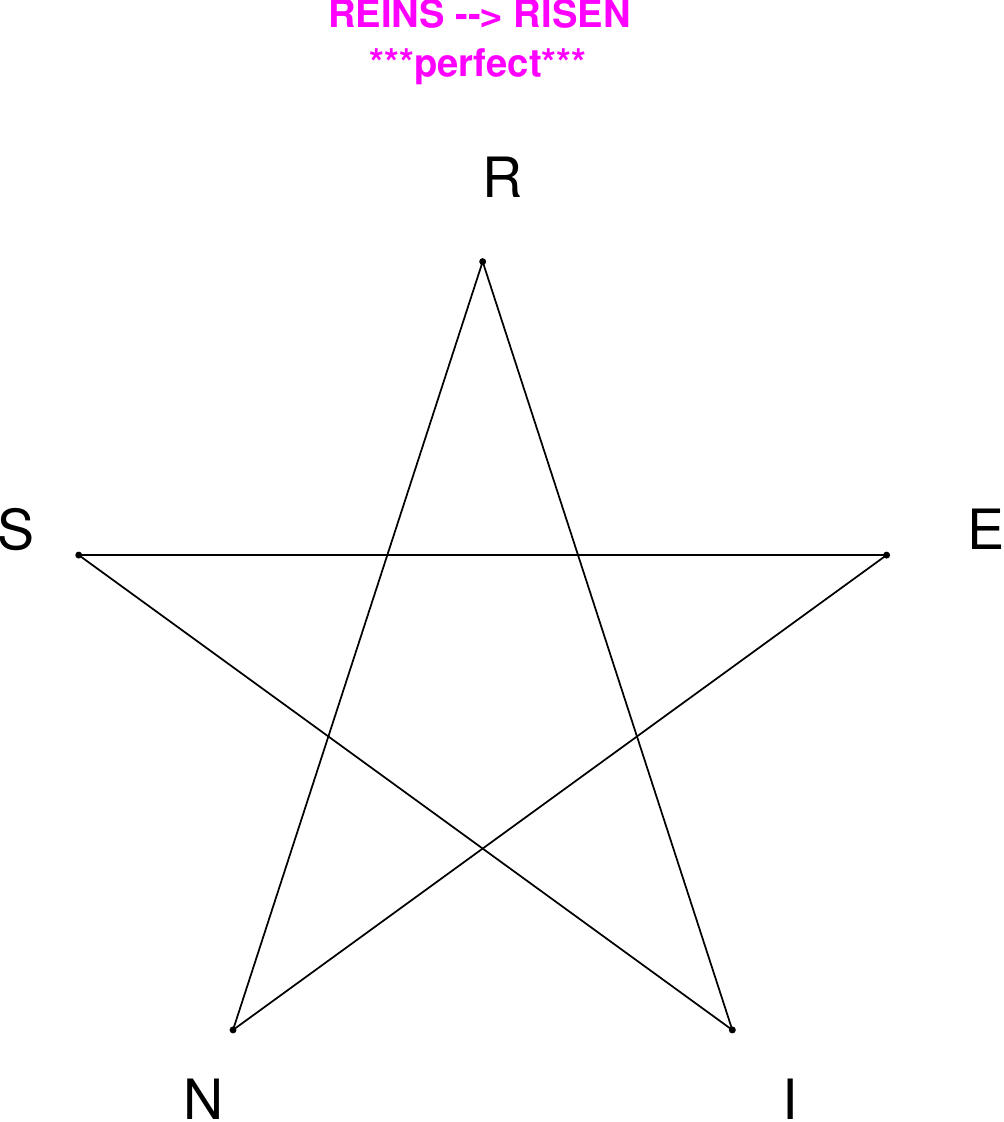}
\end{subfigure}
\hfill
\begin{subfigure}[T]{0.19\textwidth}
\centering
\includegraphics[width=\textwidth]{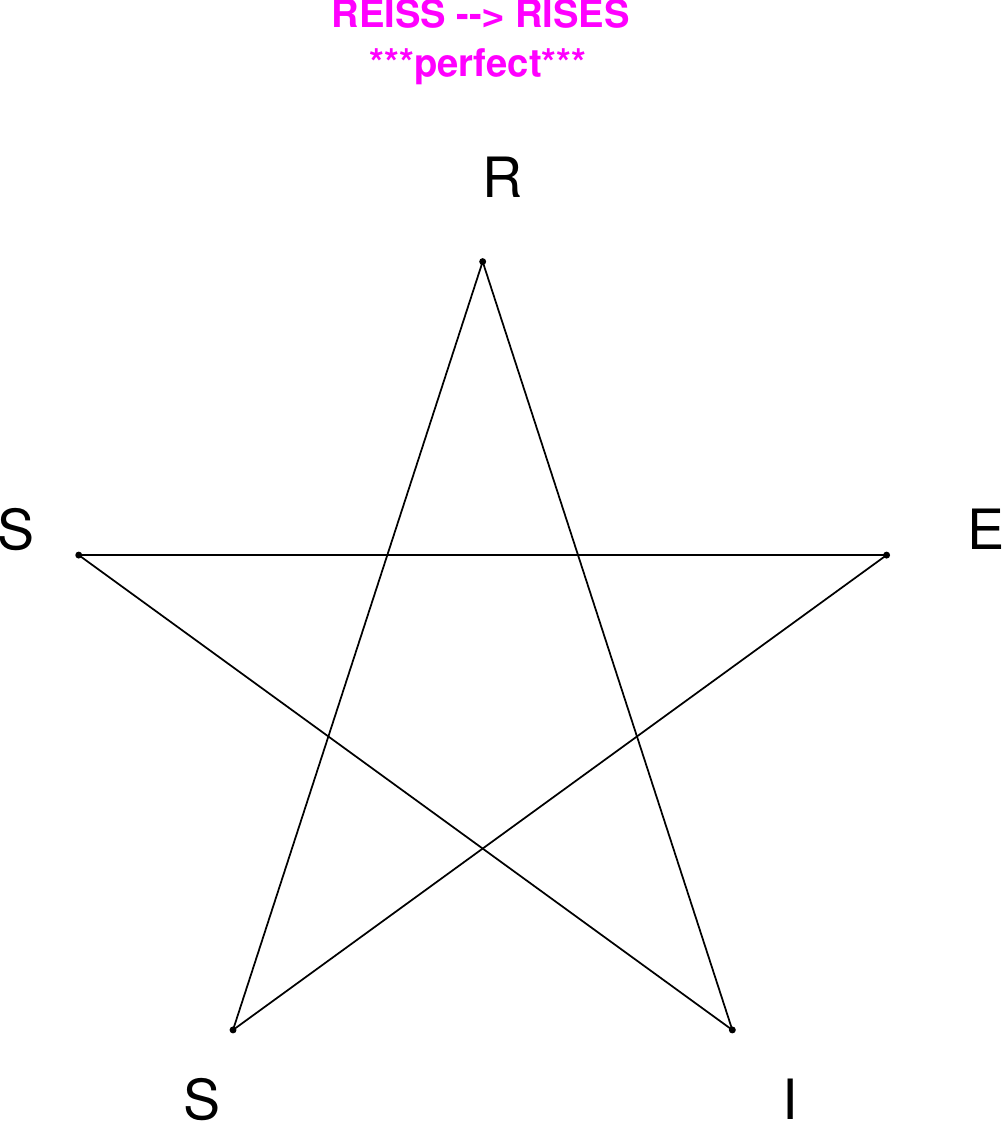}
\end{subfigure}
\hfill
\begin{subfigure}[T]{0.19\textwidth}
\centering
\includegraphics[width=\textwidth]{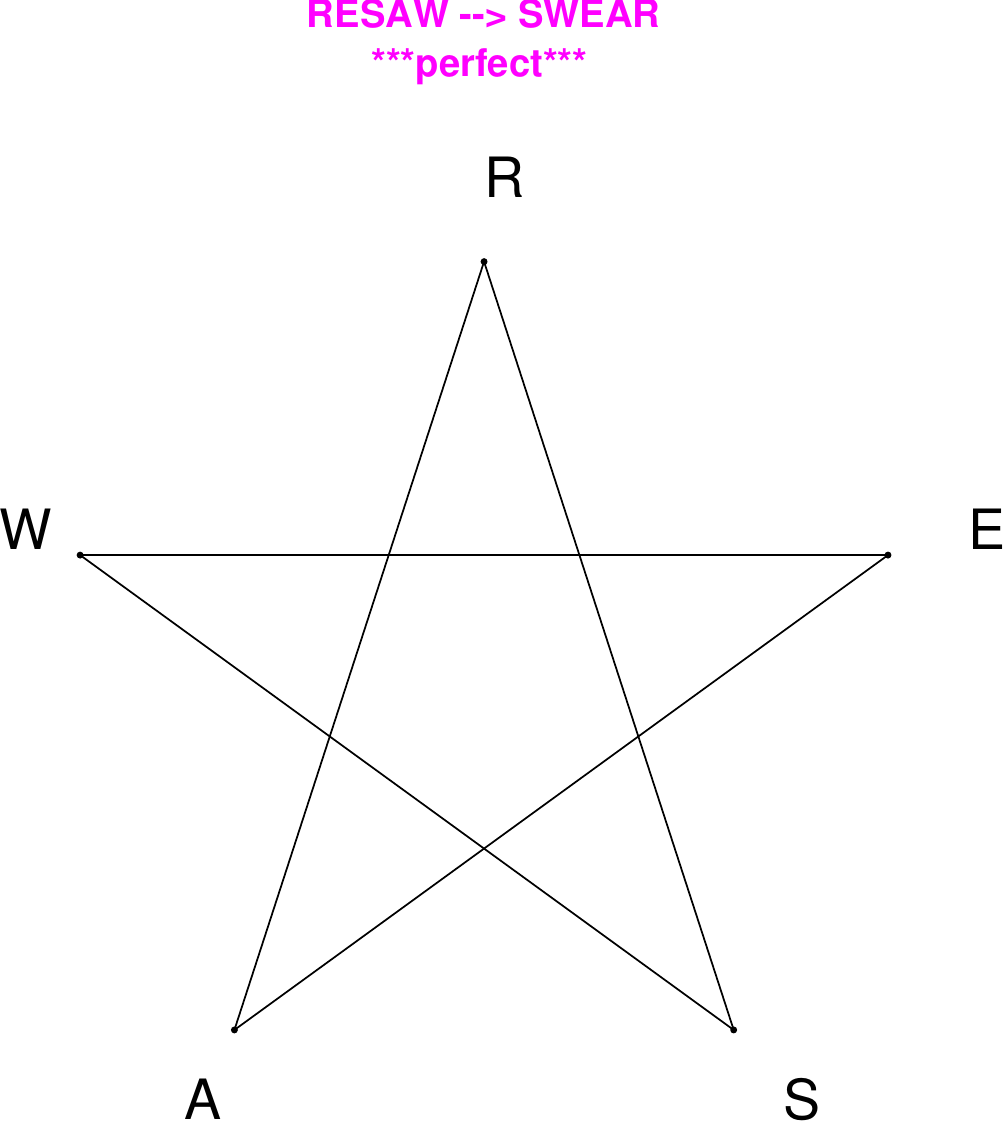}
\end{subfigure}
\end{figure}

\begin{figure}[H]
\centering
\begin{subfigure}[T]{0.19\textwidth}
\centering
\includegraphics[width=\textwidth]{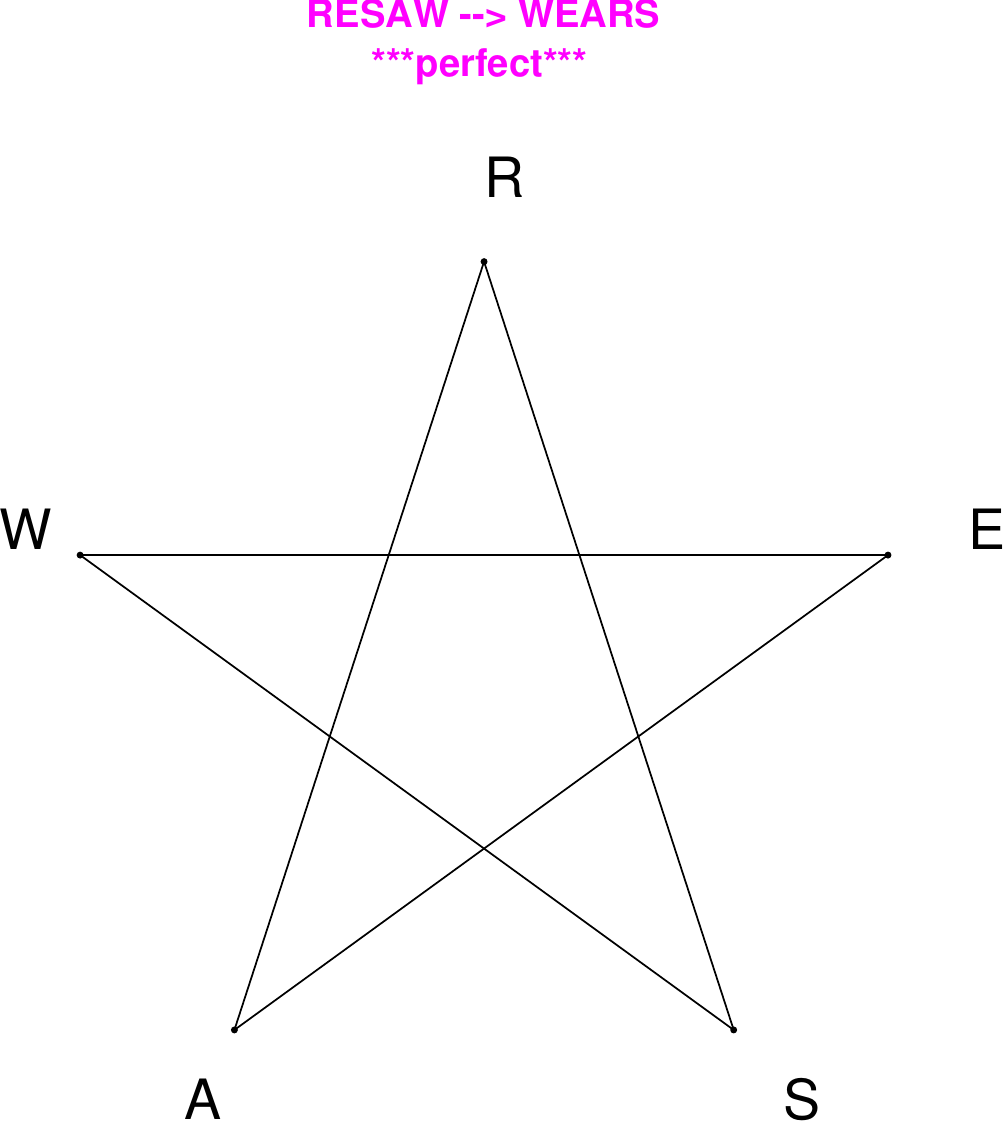}
\end{subfigure}
\hfill
\begin{subfigure}[T]{0.19\textwidth}
\centering
\includegraphics[width=\textwidth]{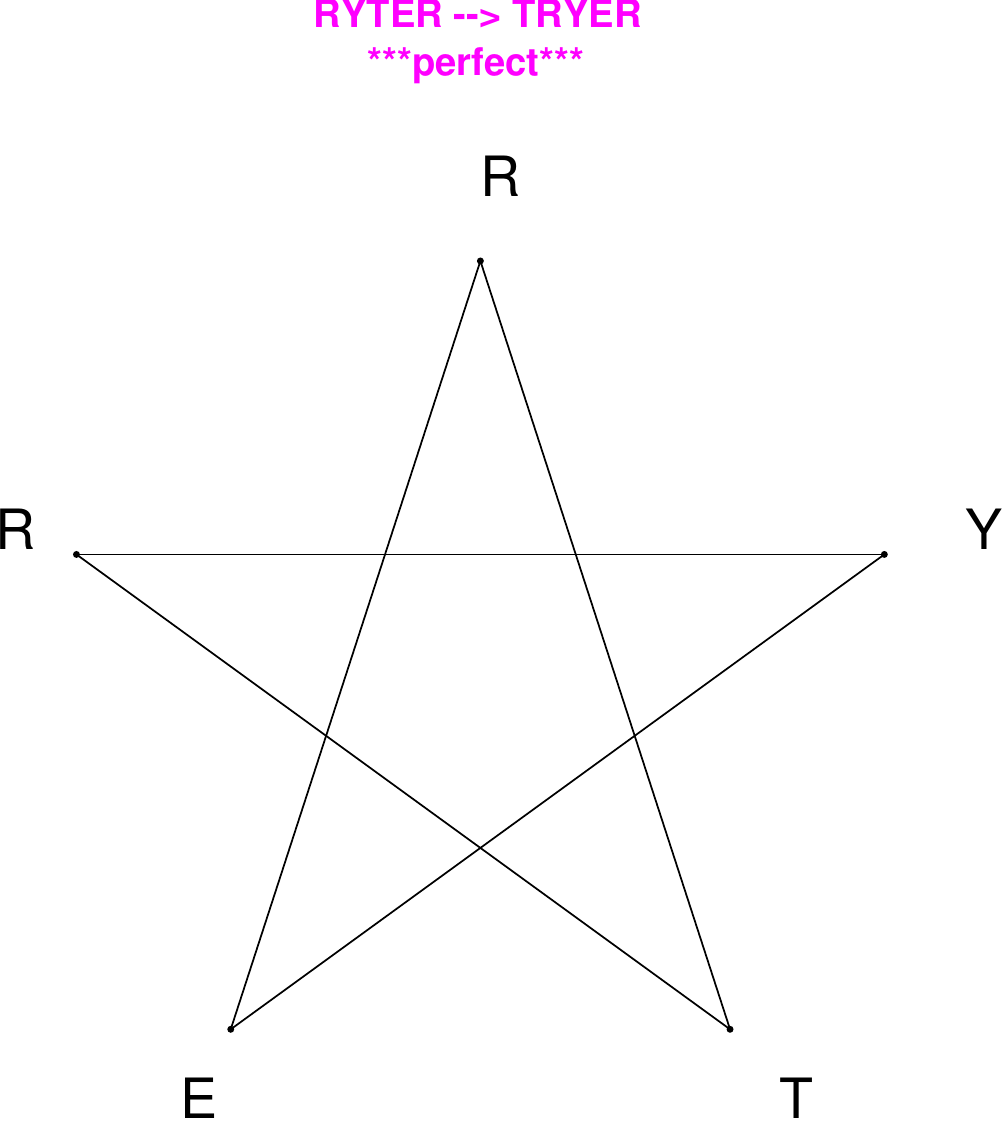}
\end{subfigure}
\hfill
\begin{subfigure}[T]{0.19\textwidth}
\centering
\includegraphics[width=\textwidth]{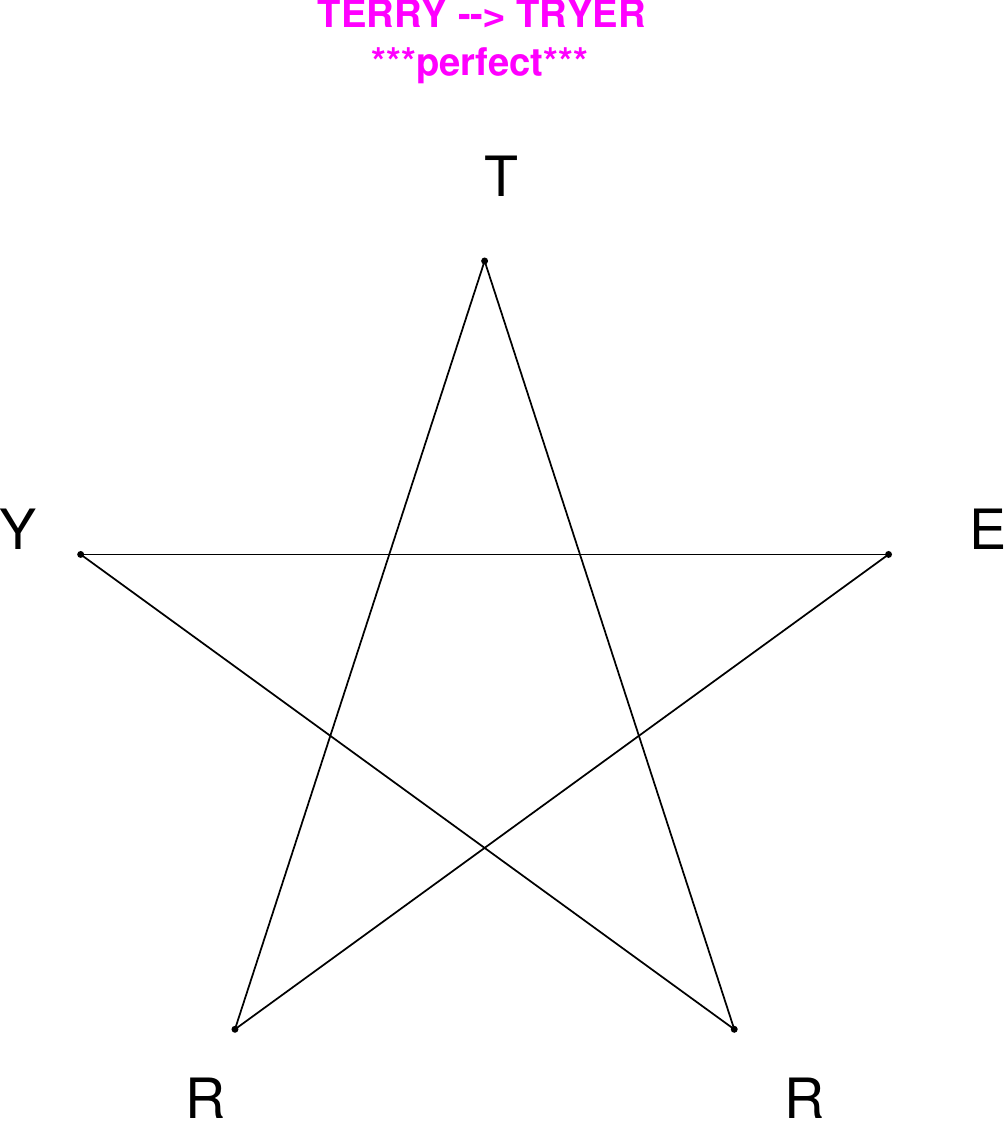}
\end{subfigure}
\hfill
\begin{subfigure}[T]{0.19\textwidth}
\centering
\includegraphics[width=\textwidth]{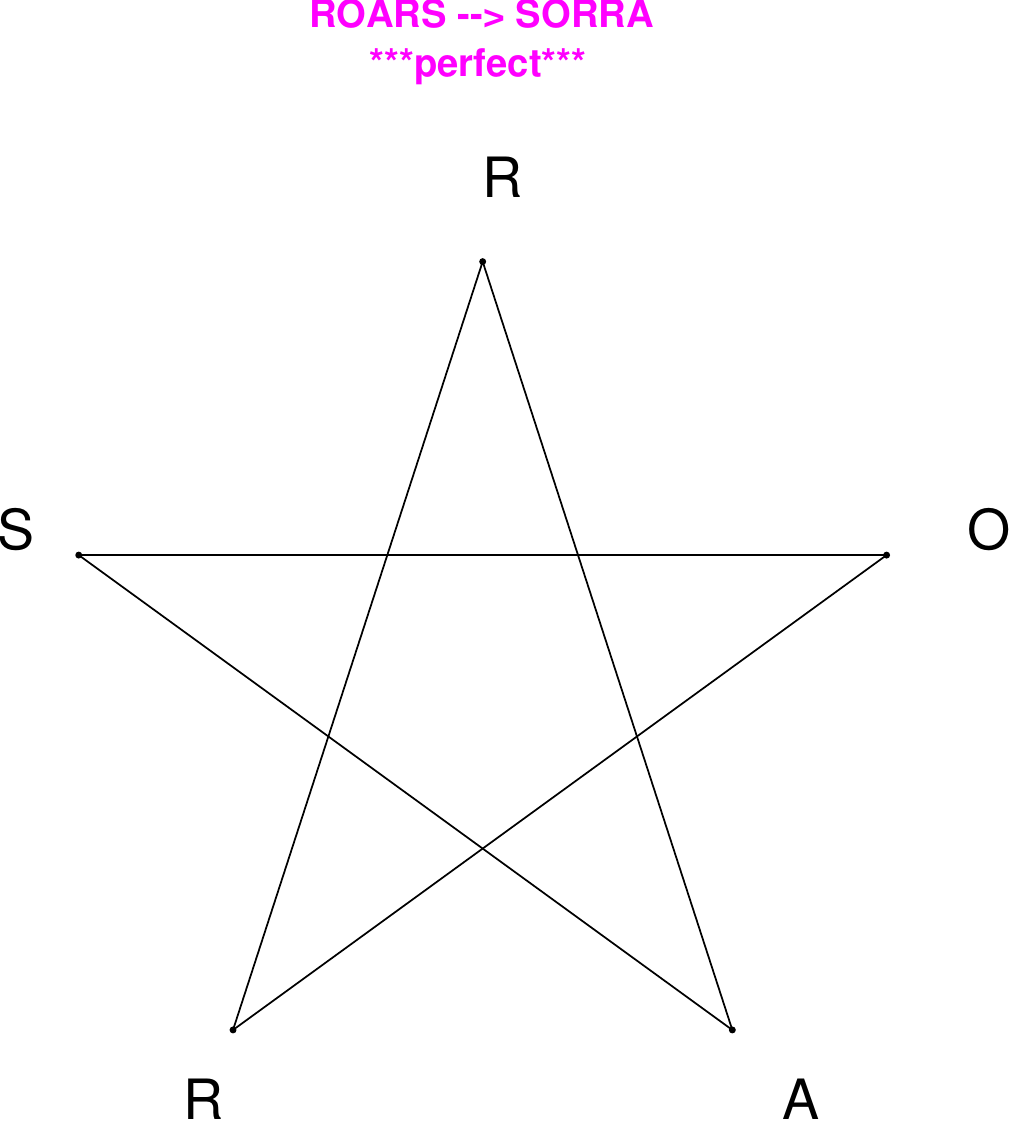}
\end{subfigure}
\hfill
\begin{subfigure}[T]{0.19\textwidth}
\centering
\includegraphics[width=\textwidth]{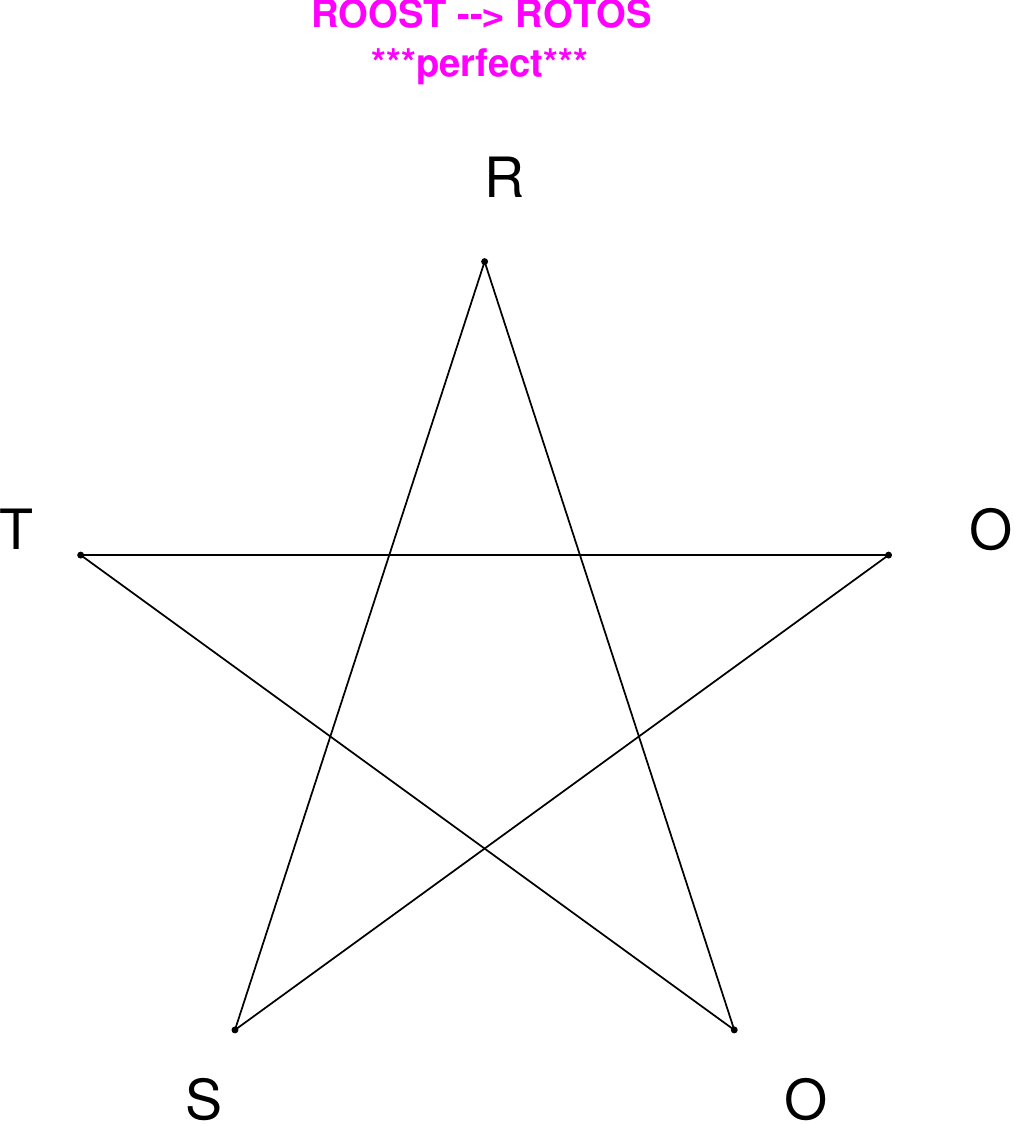}
\end{subfigure}
\end{figure}

\begin{figure}[H]
\centering
\begin{subfigure}[T]{0.19\textwidth}
\centering
\includegraphics[width=\textwidth]{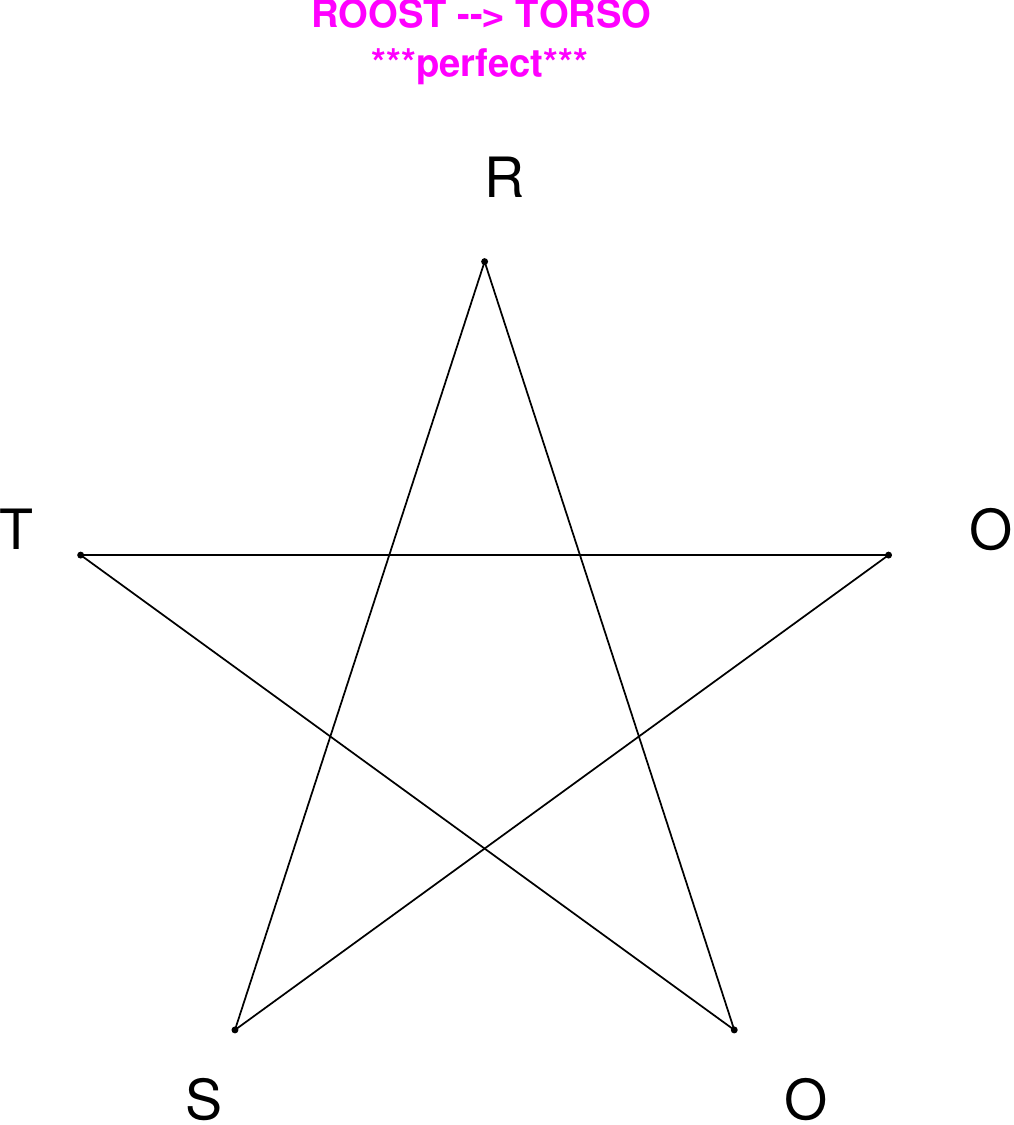}
\end{subfigure}
\hfill
\begin{subfigure}[T]{0.19\textwidth}
\centering
\includegraphics[width=\textwidth]{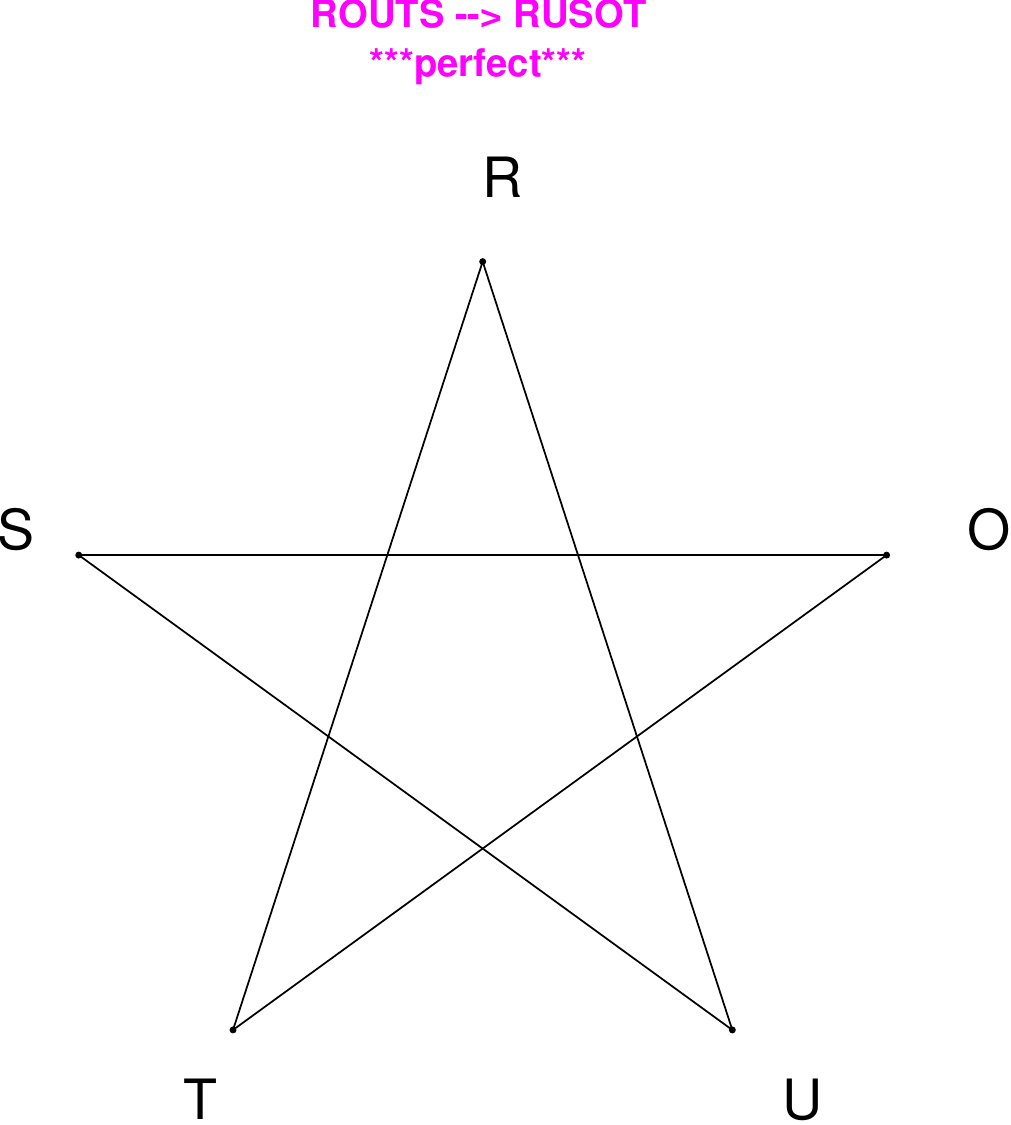}
\end{subfigure}
\hfill
\begin{subfigure}[T]{0.19\textwidth}
\centering
\includegraphics[width=\textwidth]{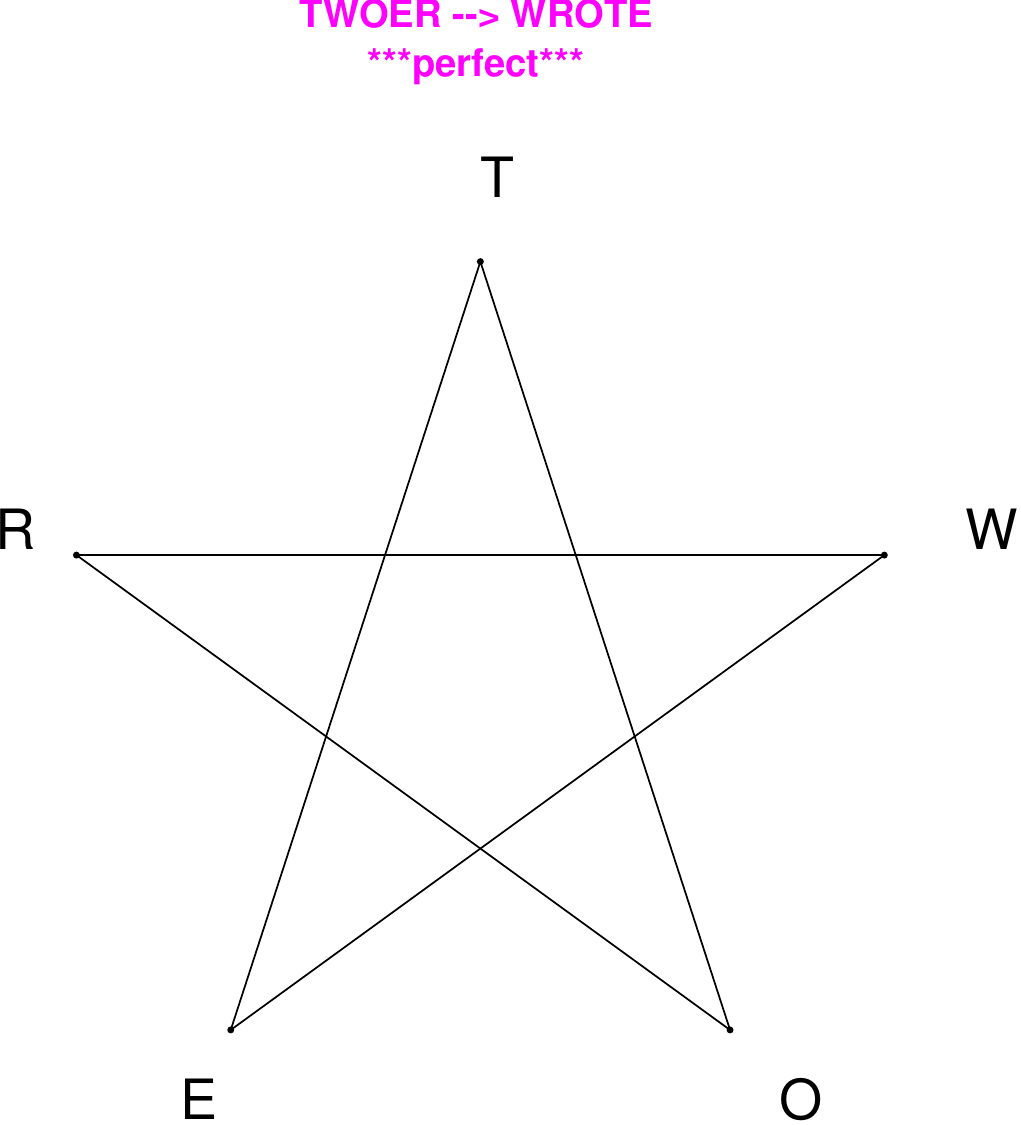}
\end{subfigure}
\hfill
\begin{subfigure}[T]{0.19\textwidth}
\centering
\includegraphics[width=\textwidth]{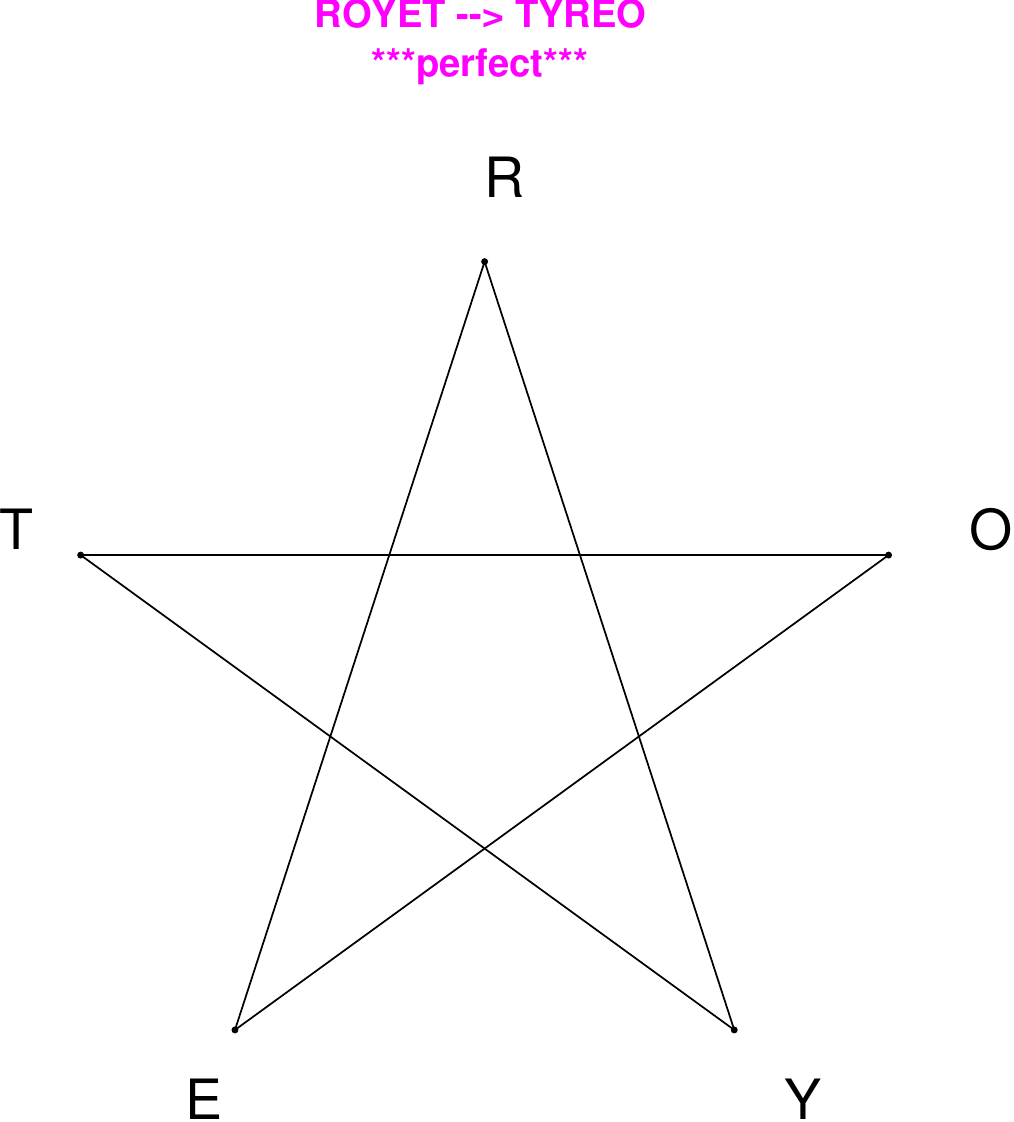}
\end{subfigure}
\hfill
\begin{subfigure}[T]{0.19\textwidth}
\centering
\includegraphics[width=\textwidth]{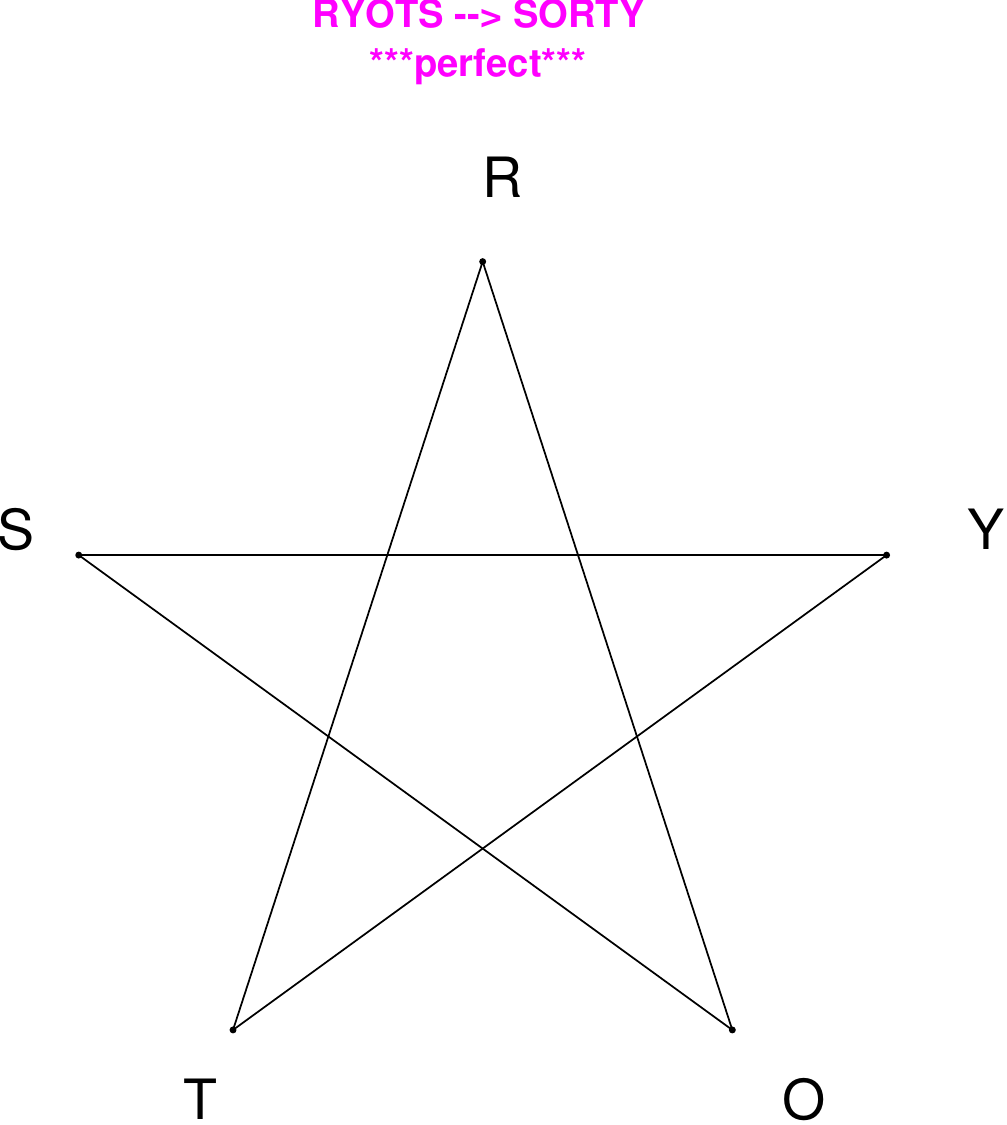}
\end{subfigure}
\end{figure}

\begin{figure}[H]
\centering
\begin{subfigure}[T]{0.19\textwidth}
\centering
\includegraphics[width=\textwidth]{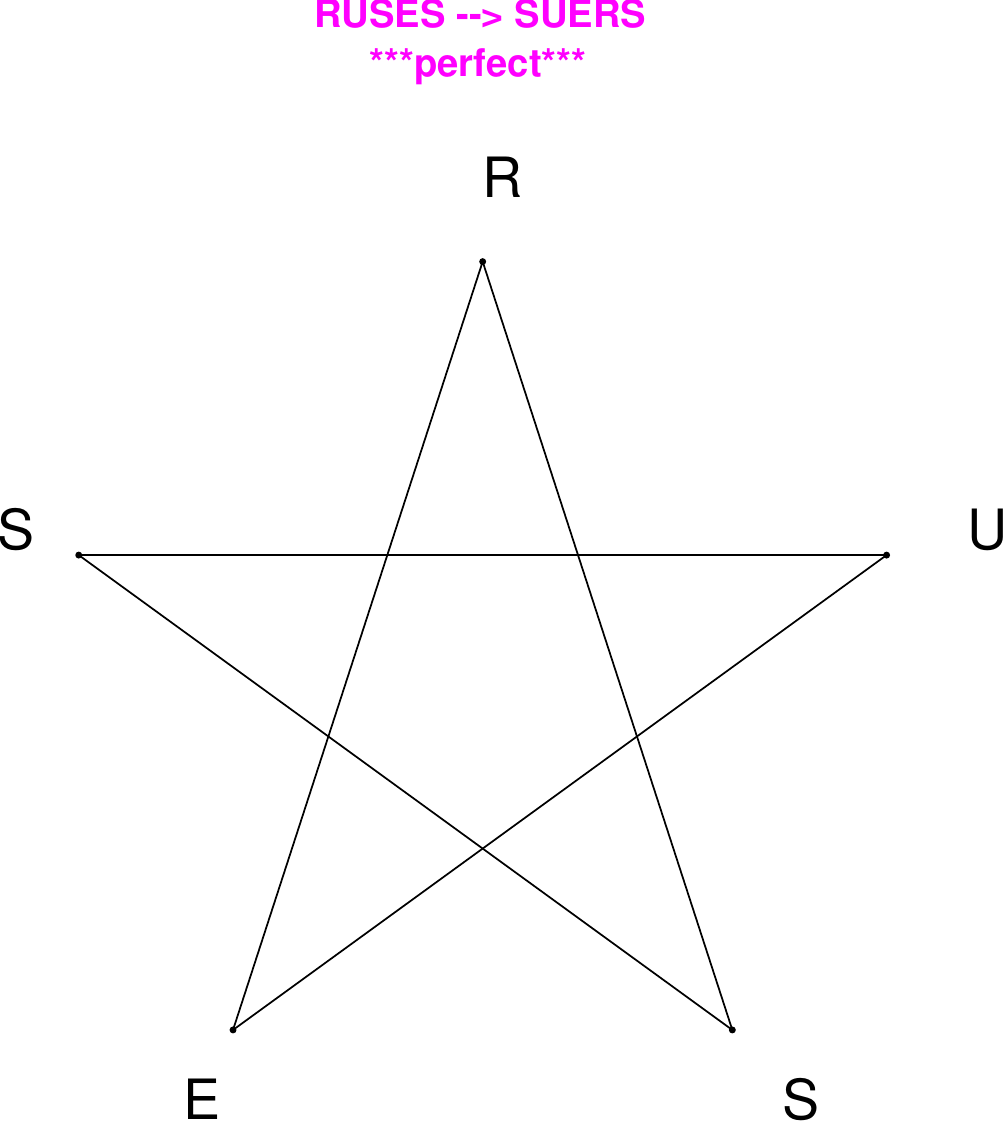}
\end{subfigure}
\hfill
\begin{subfigure}[T]{0.19\textwidth}
\centering
\includegraphics[width=\textwidth]{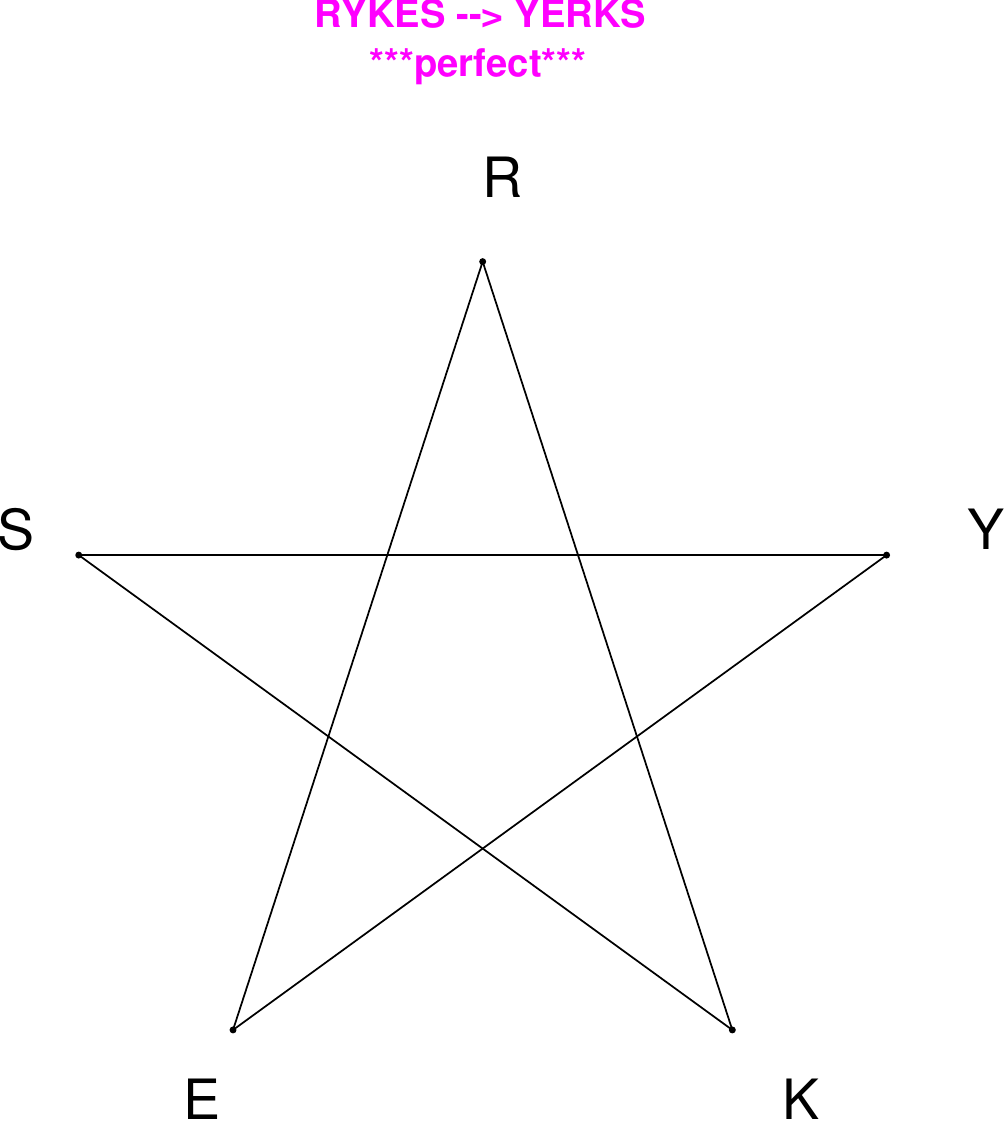}
\end{subfigure}
\hfill
\begin{subfigure}[T]{0.19\textwidth}
\centering
\includegraphics[width=\textwidth]{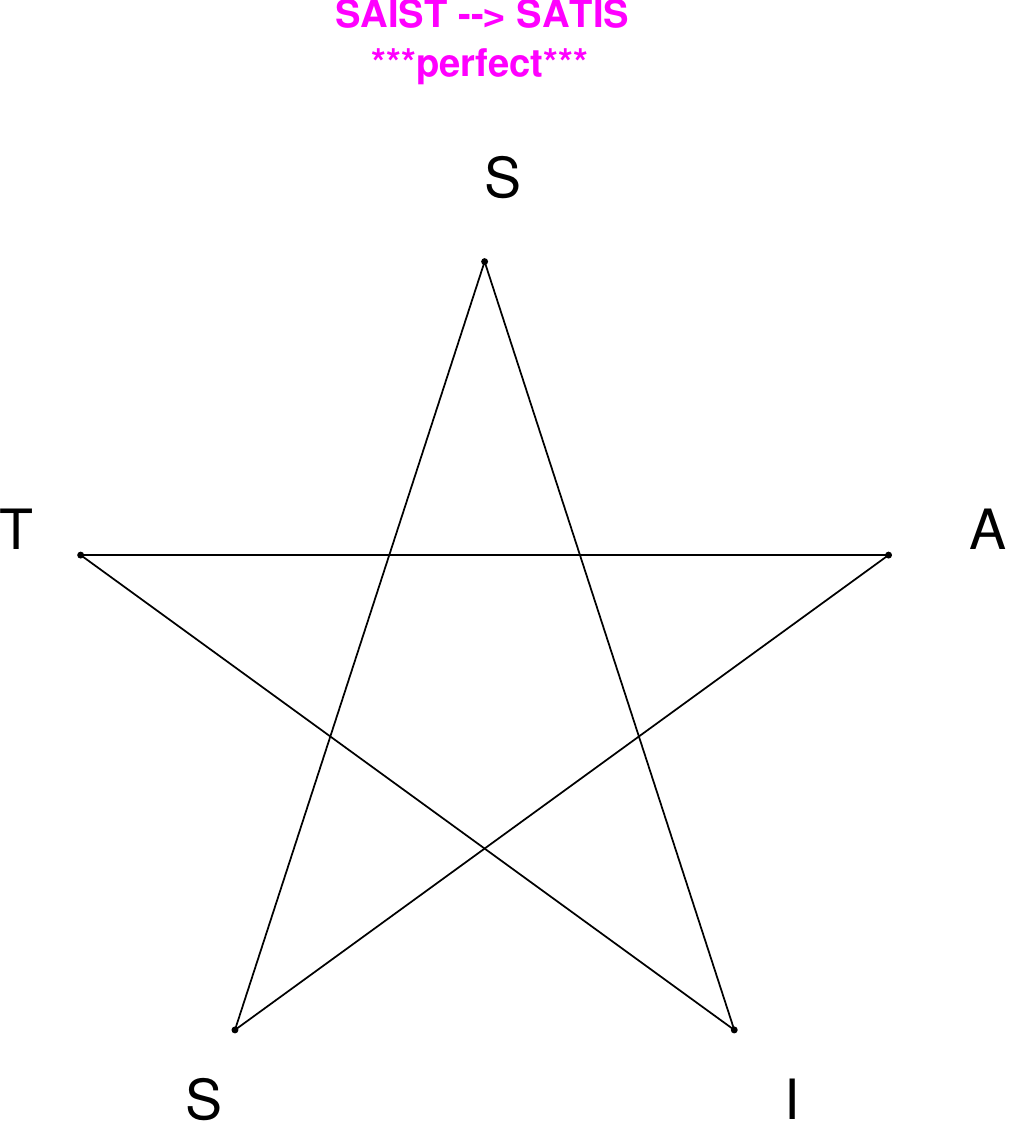}
\end{subfigure}
\hfill
\begin{subfigure}[T]{0.19\textwidth}
\centering
\includegraphics[width=\textwidth]{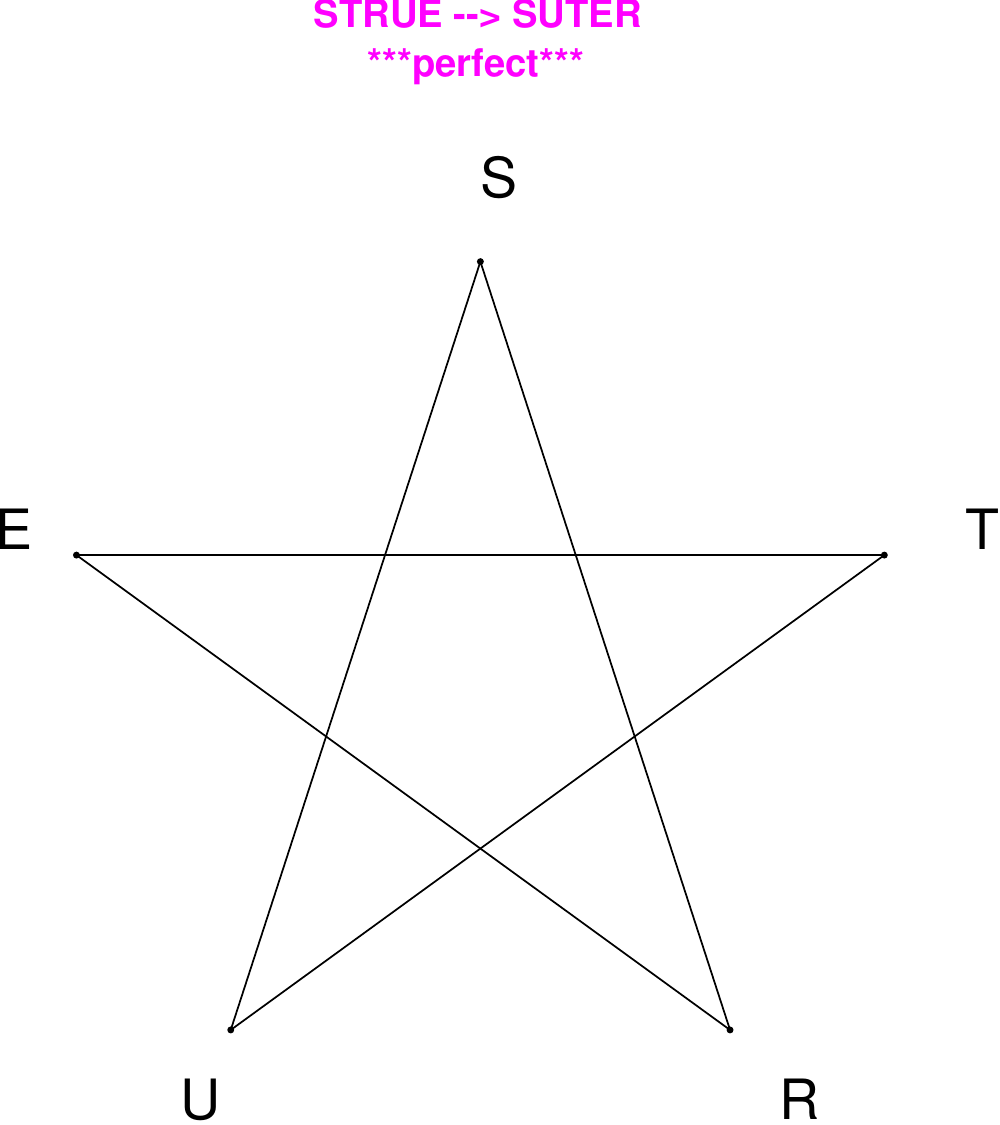}
\end{subfigure}
\hfill
\begin{subfigure}[T]{0.19\textwidth}
\centering
\includegraphics[width=\textwidth]{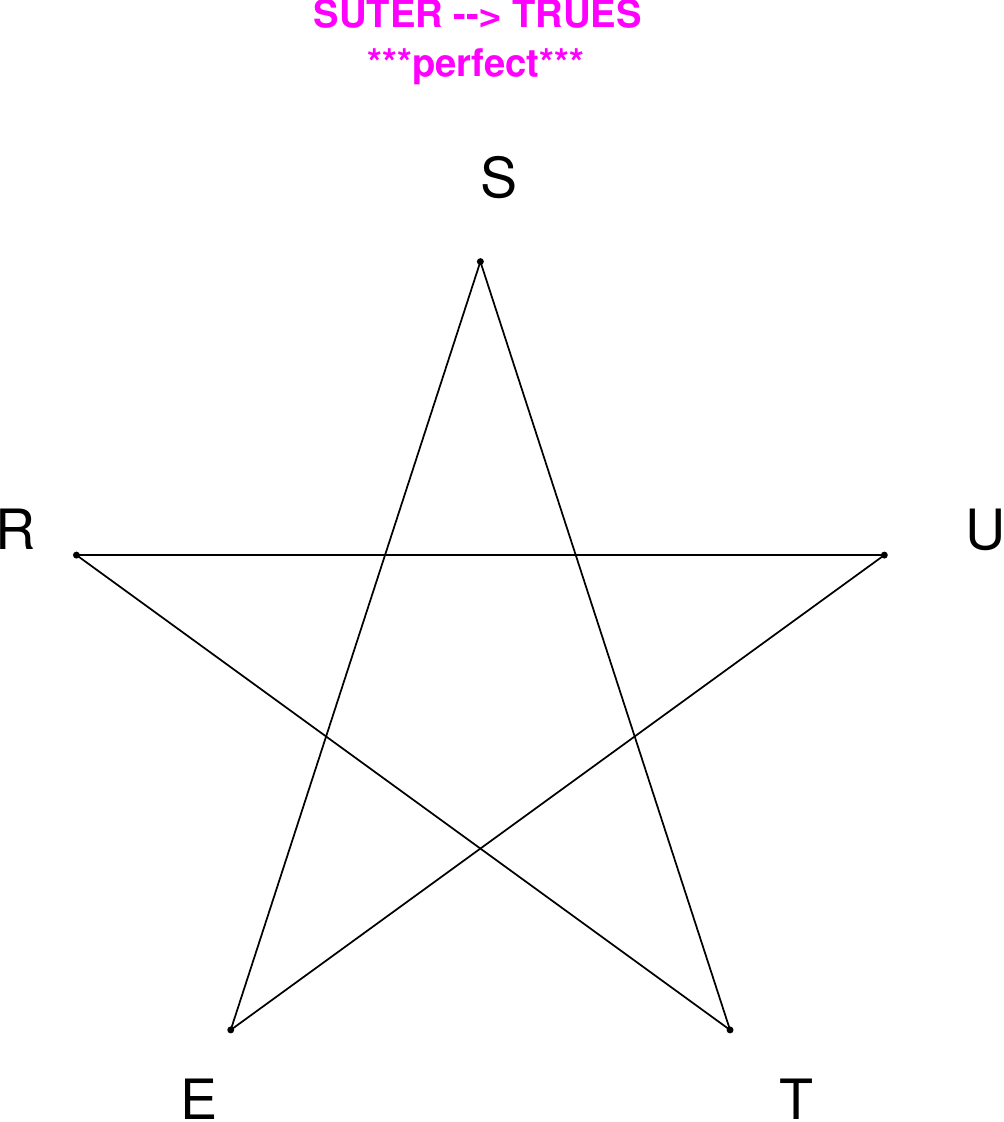}
\end{subfigure}
\end{figure}

\begin{figure}[H]
\centering
\begin{subfigure}[T]{0.19\textwidth}
\centering
\includegraphics[width=\textwidth]{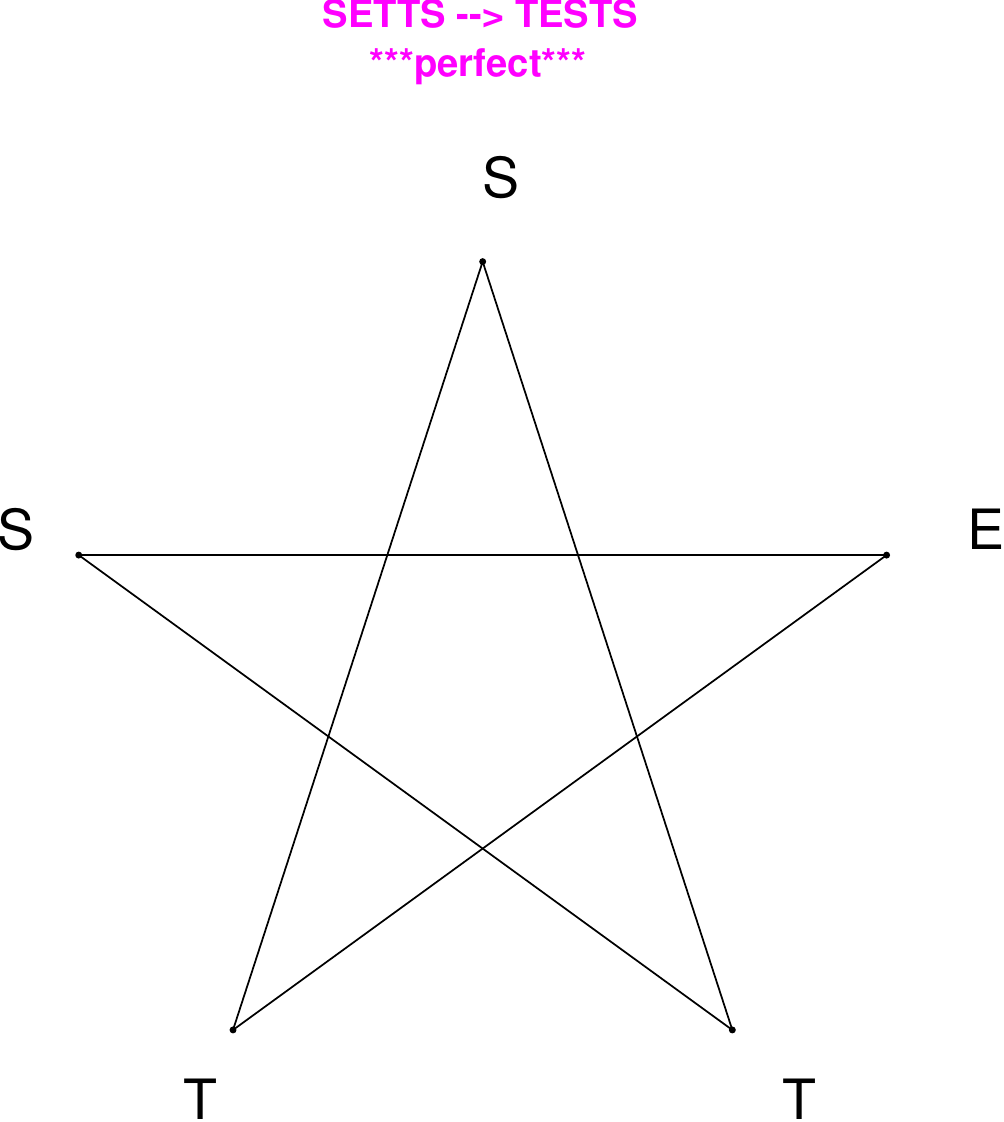}
\end{subfigure}
\hfill
\begin{subfigure}[T]{0.19\textwidth}
\centering
\includegraphics[width=\textwidth]{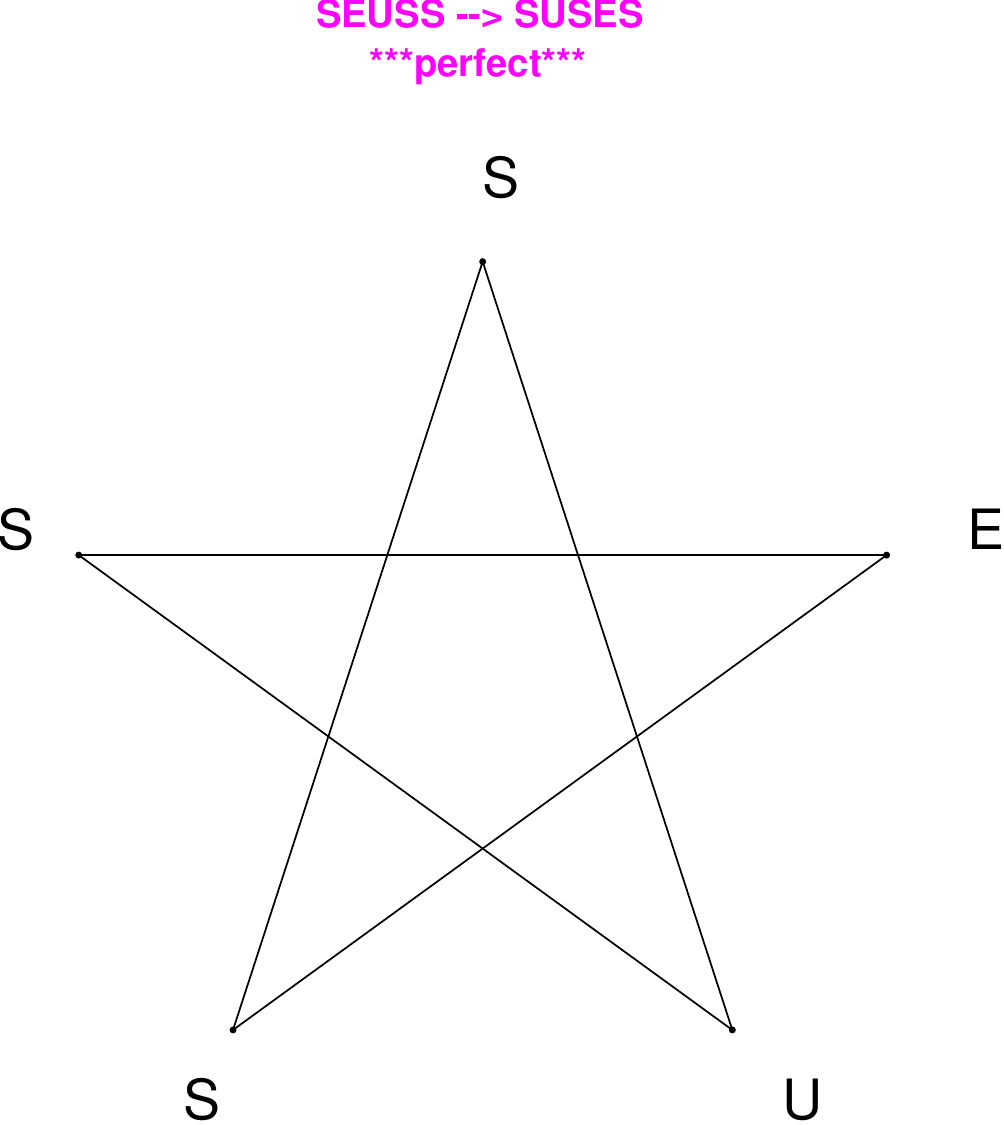}
\end{subfigure}
\hfill
\begin{subfigure}[T]{0.19\textwidth}
\centering
\includegraphics[width=\textwidth]{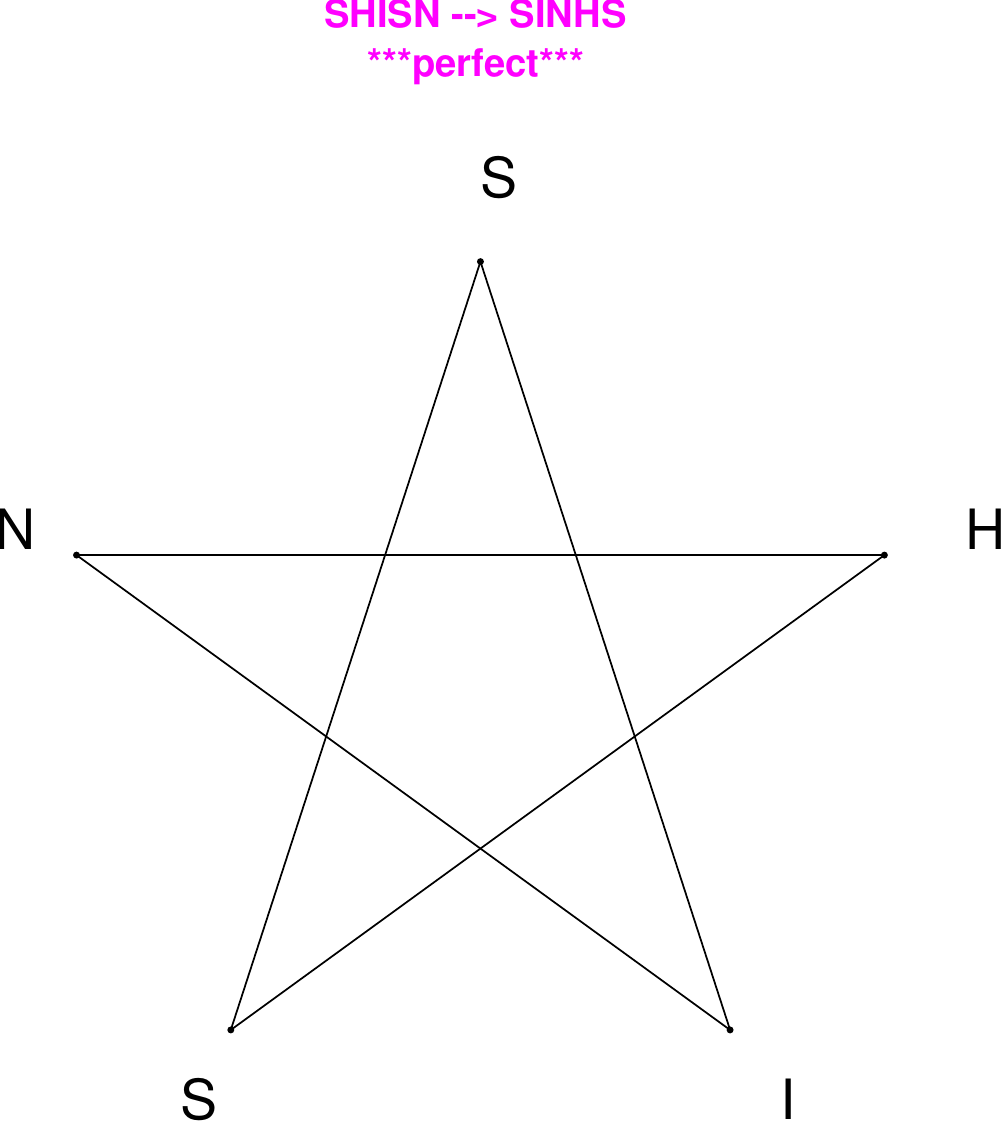}
\end{subfigure}
\hfill
\begin{subfigure}[T]{0.19\textwidth}
\centering
\includegraphics[width=\textwidth]{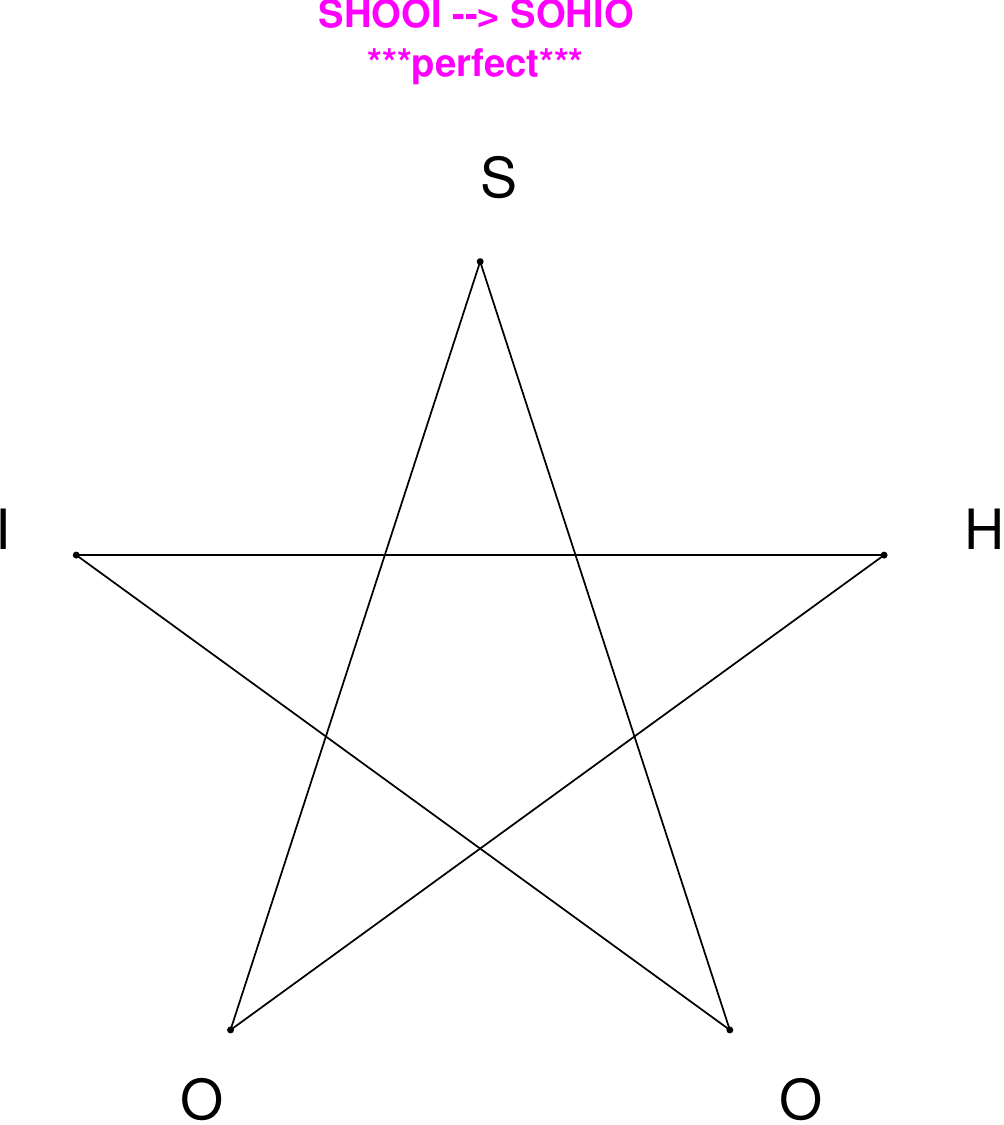}
\end{subfigure}
\hfill
\begin{subfigure}[T]{0.19\textwidth}
\centering
\includegraphics[width=\textwidth]{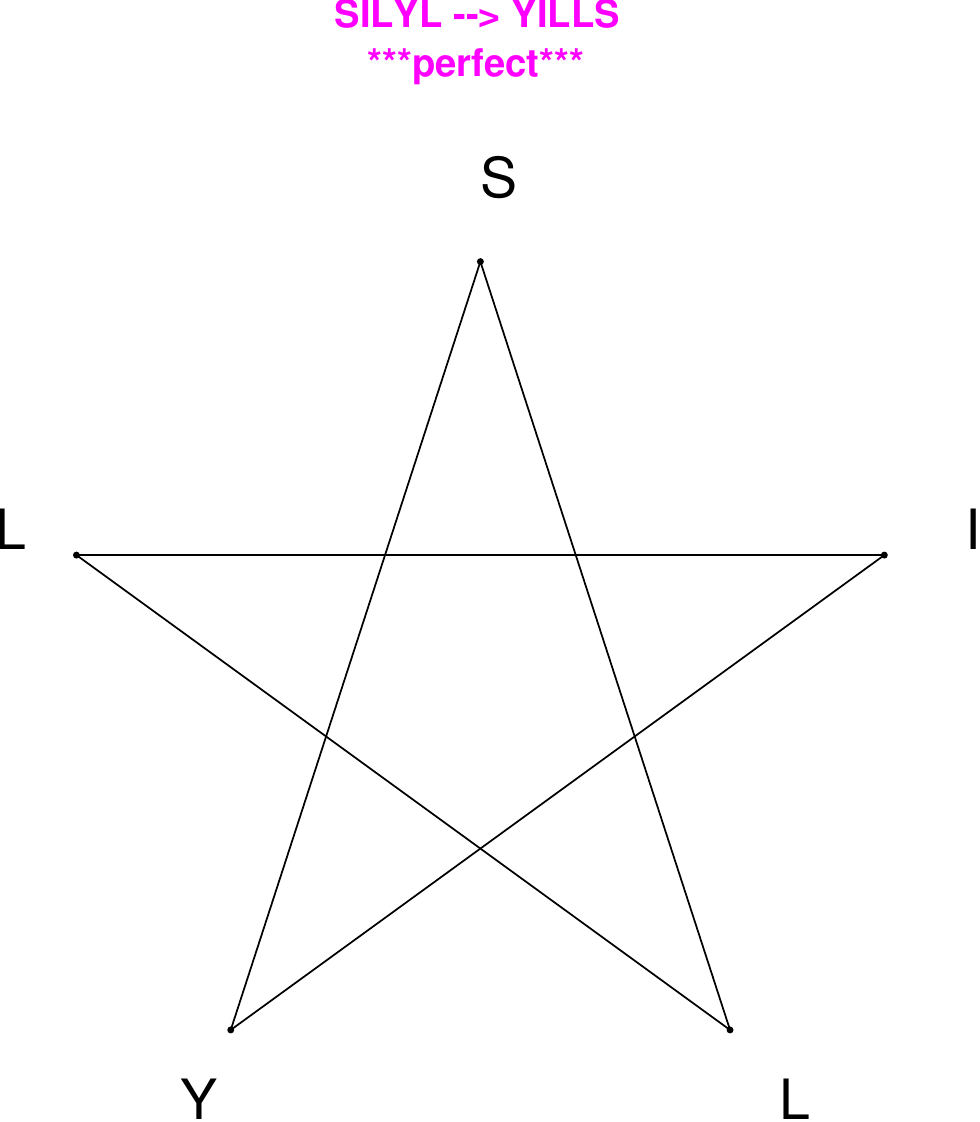}
\end{subfigure}
\end{figure}

\begin{figure}[H]
\centering
\begin{subfigure}[T]{0.19\textwidth}
\centering
\includegraphics[width=\textwidth]{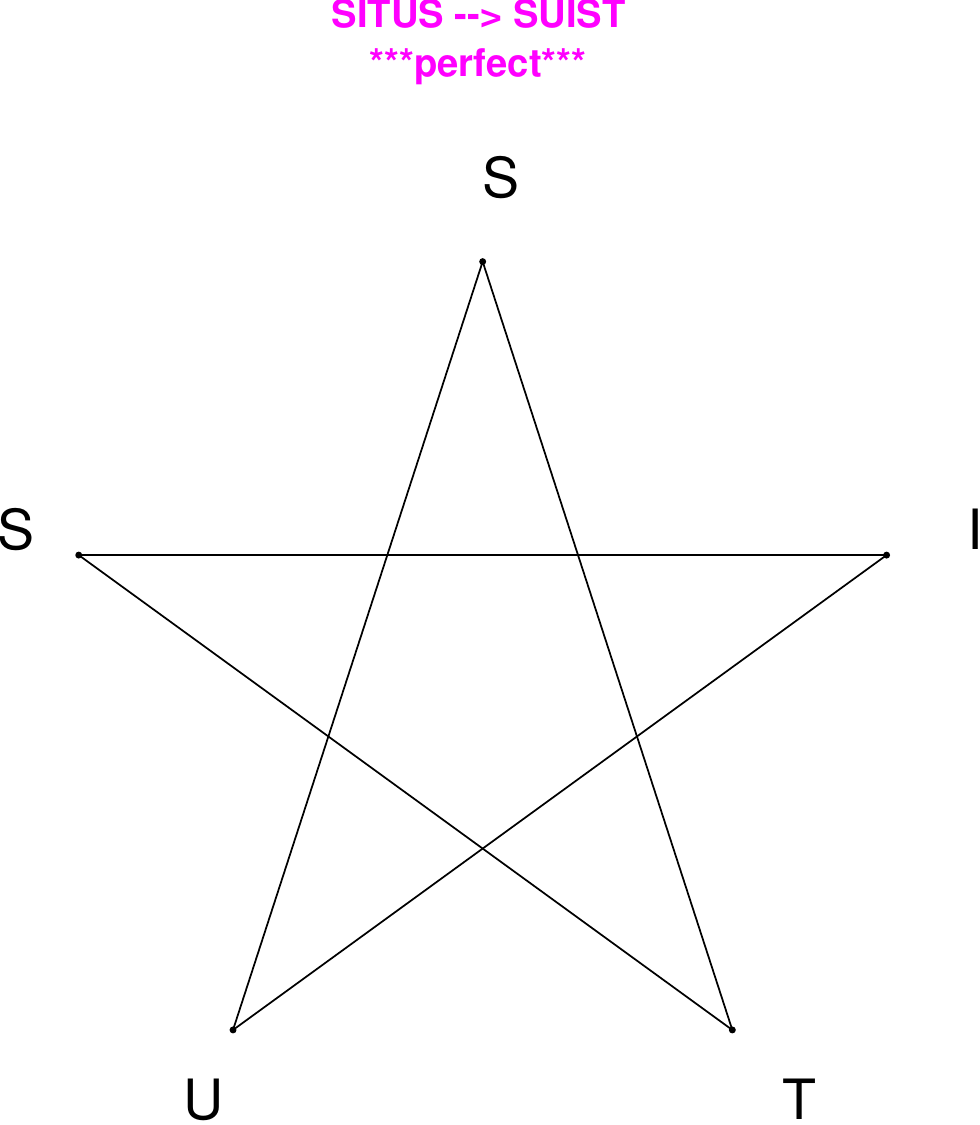}
\end{subfigure}
\hfill
\begin{subfigure}[T]{0.19\textwidth}
\centering
\includegraphics[width=\textwidth]{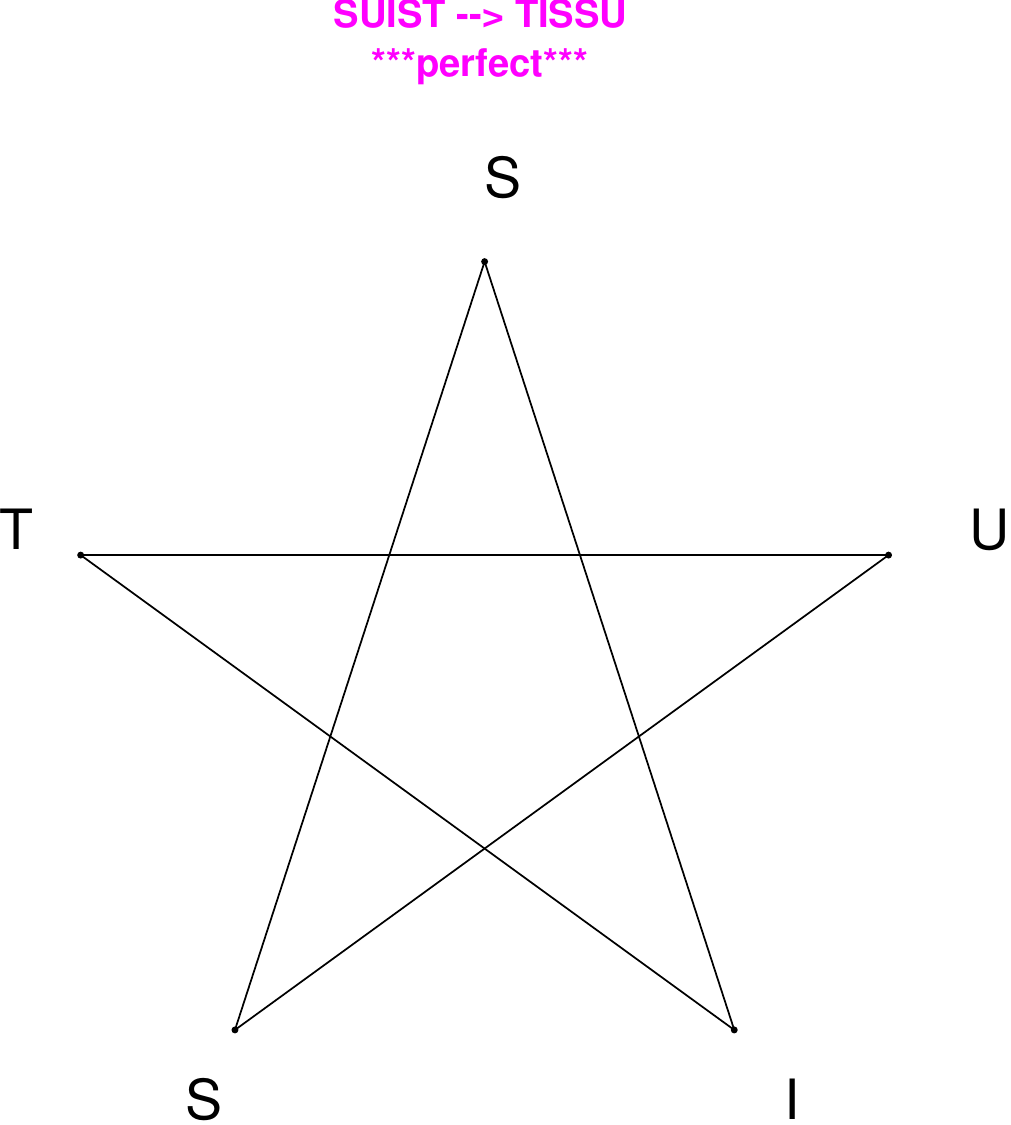}
\end{subfigure}
\hfill
\begin{subfigure}[T]{0.19\textwidth}
\centering
\includegraphics[width=\textwidth]{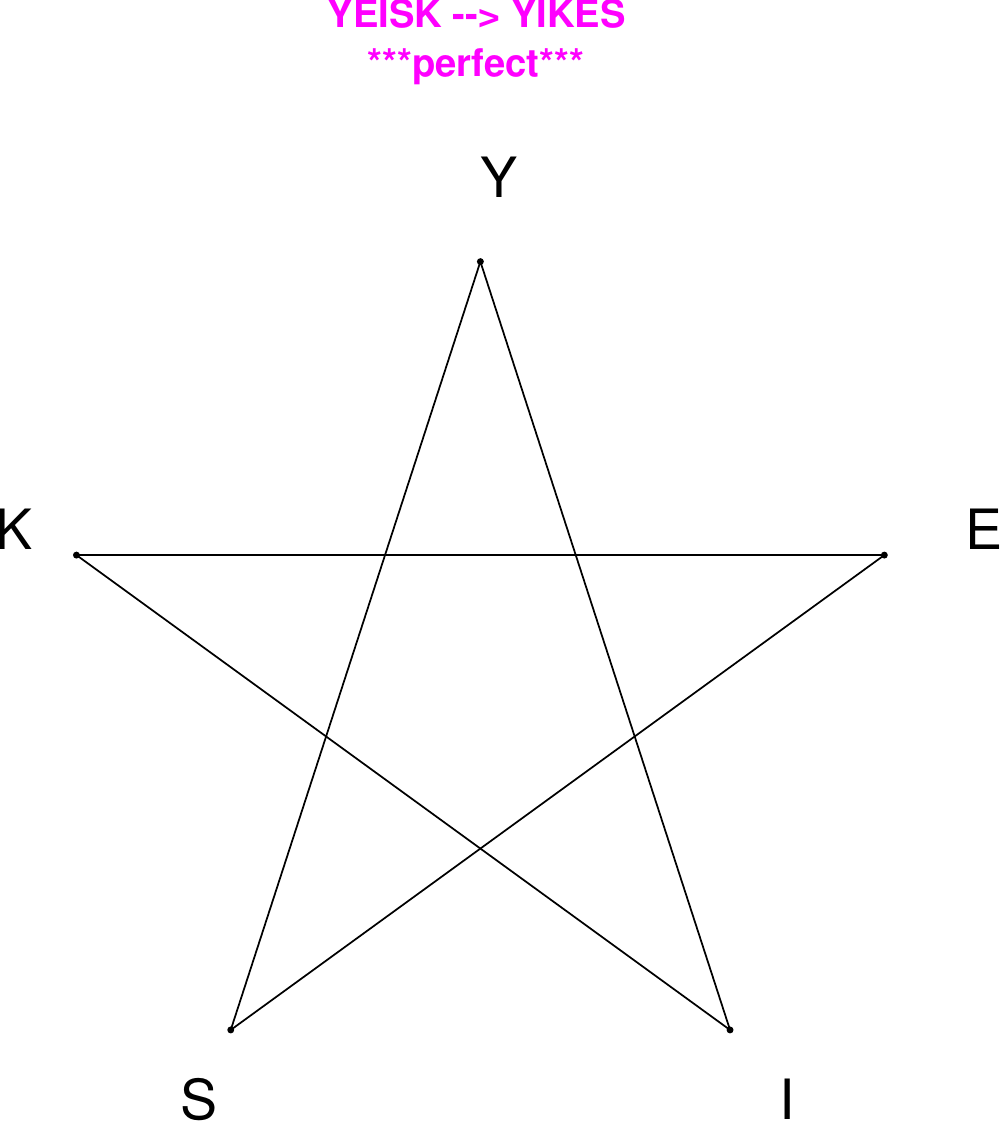}
\end{subfigure}
\hfill
\begin{subfigure}[T]{0.19\textwidth}
\centering
\includegraphics[width=\textwidth]{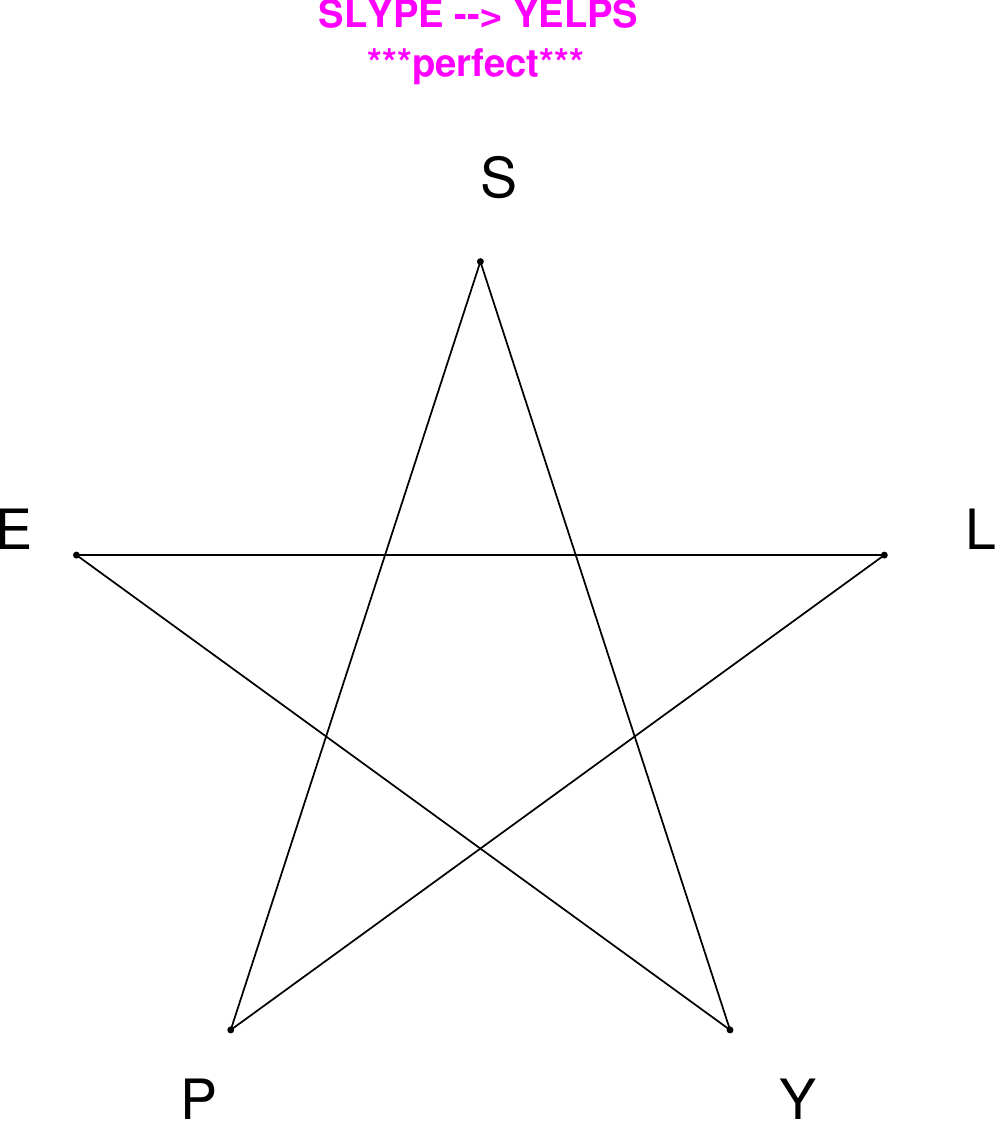}
\end{subfigure}
\hfill
\begin{subfigure}[T]{0.19\textwidth}
\centering
\includegraphics[width=\textwidth]{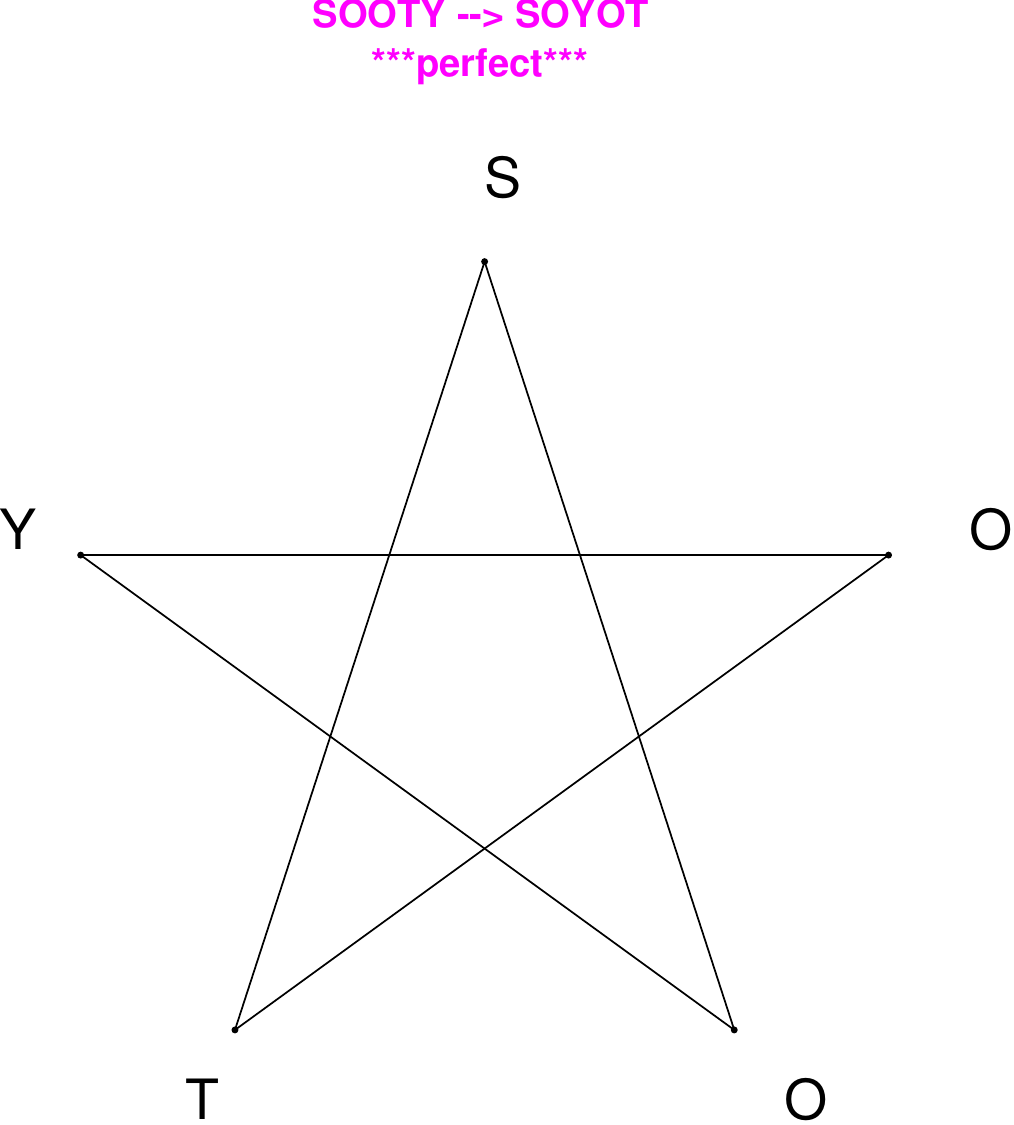}
\end{subfigure}
\end{figure}

\begin{figure}[H]
\centering
\begin{subfigure}[T]{0.19\textwidth}
\centering
\includegraphics[width=\textwidth]{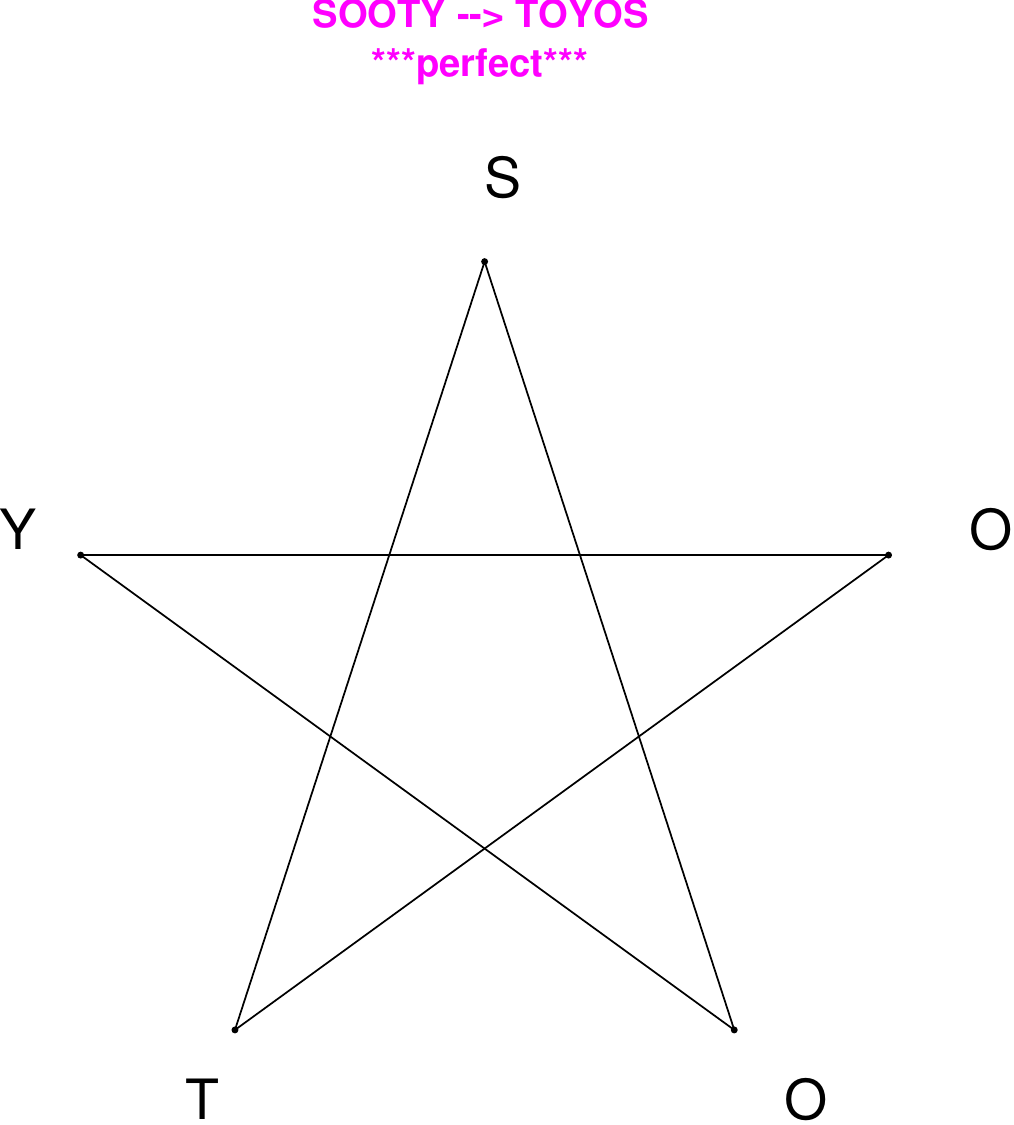}
\end{subfigure}
\hfill
\begin{subfigure}[T]{0.19\textwidth}
\centering
\includegraphics[width=\textwidth]{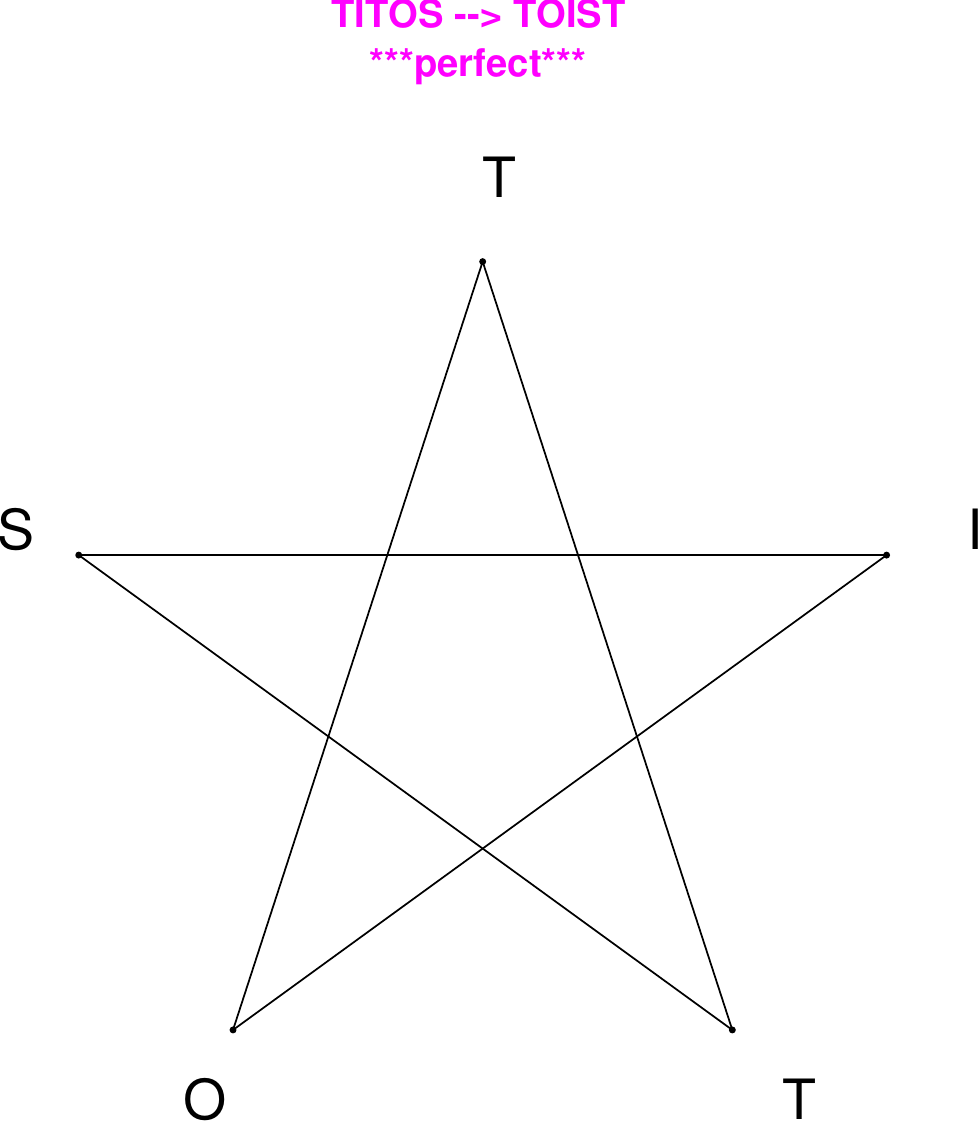}
\end{subfigure}
\hfill
\begin{subfigure}[T]{0.19\textwidth}
\centering
\includegraphics[width=\textwidth]{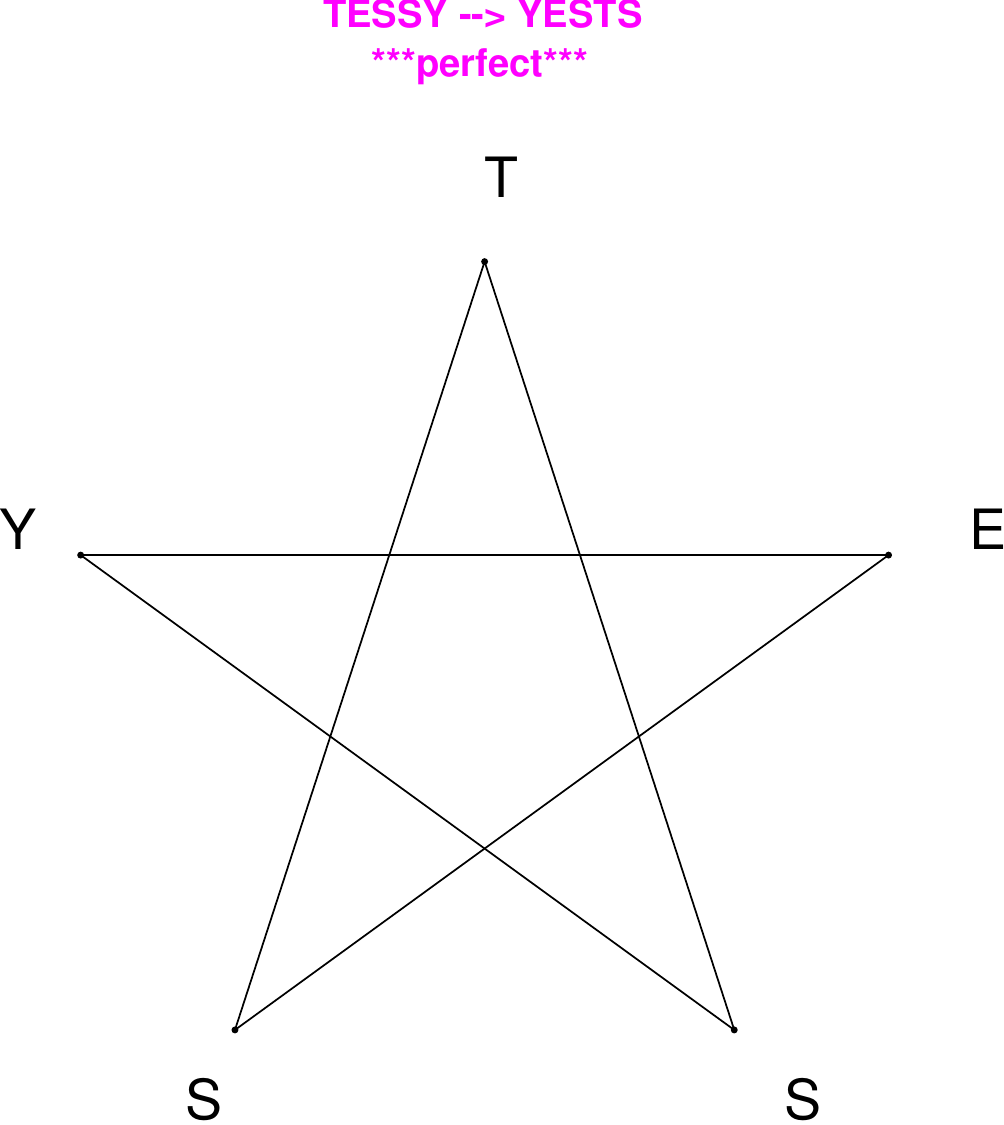}
\end{subfigure}
\hfill
\begin{subfigure}[T]{0.19\textwidth}
\centering
\includegraphics[width=\textwidth]{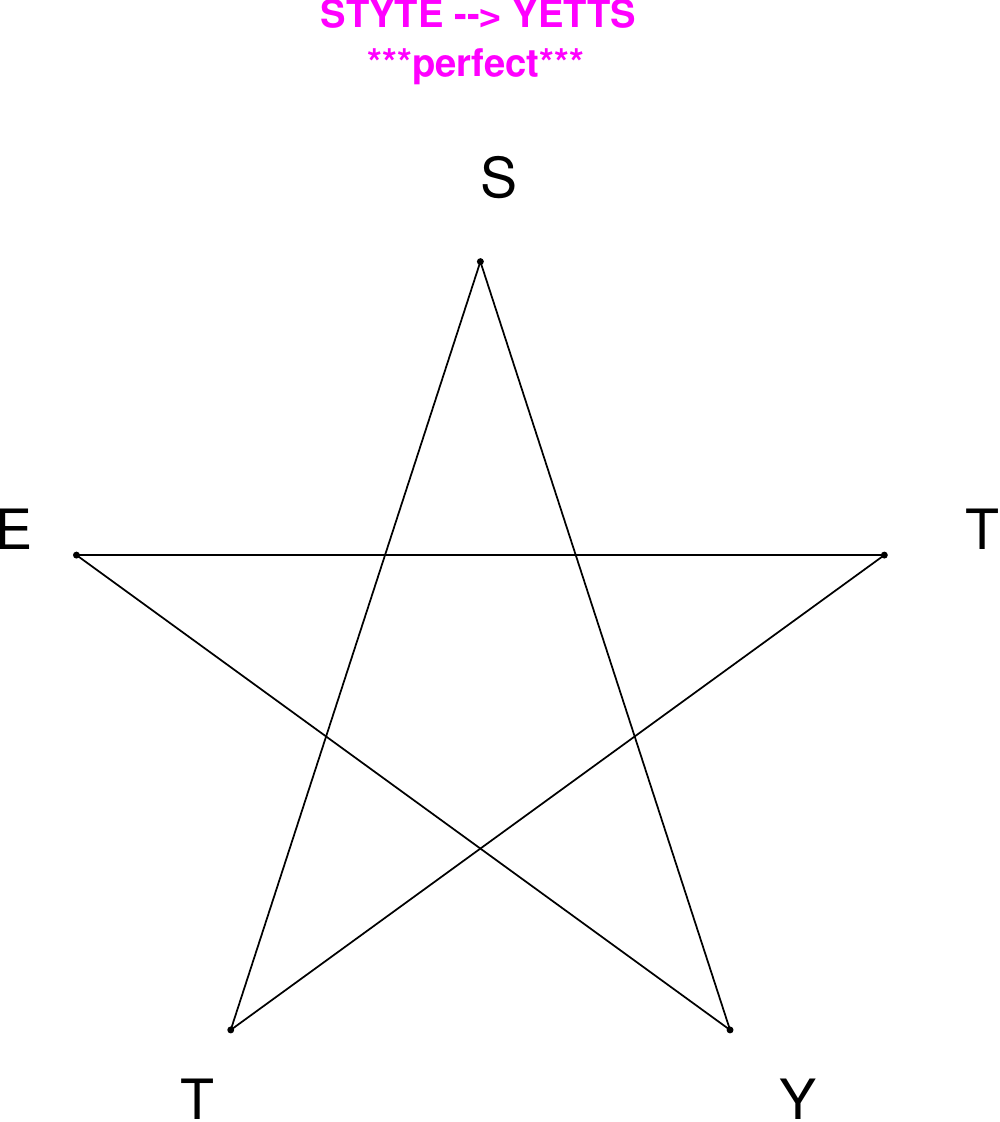}
\end{subfigure}
\hfill
\begin{subfigure}[T]{0.19\textwidth}
\centering
\includegraphics[width=\textwidth]{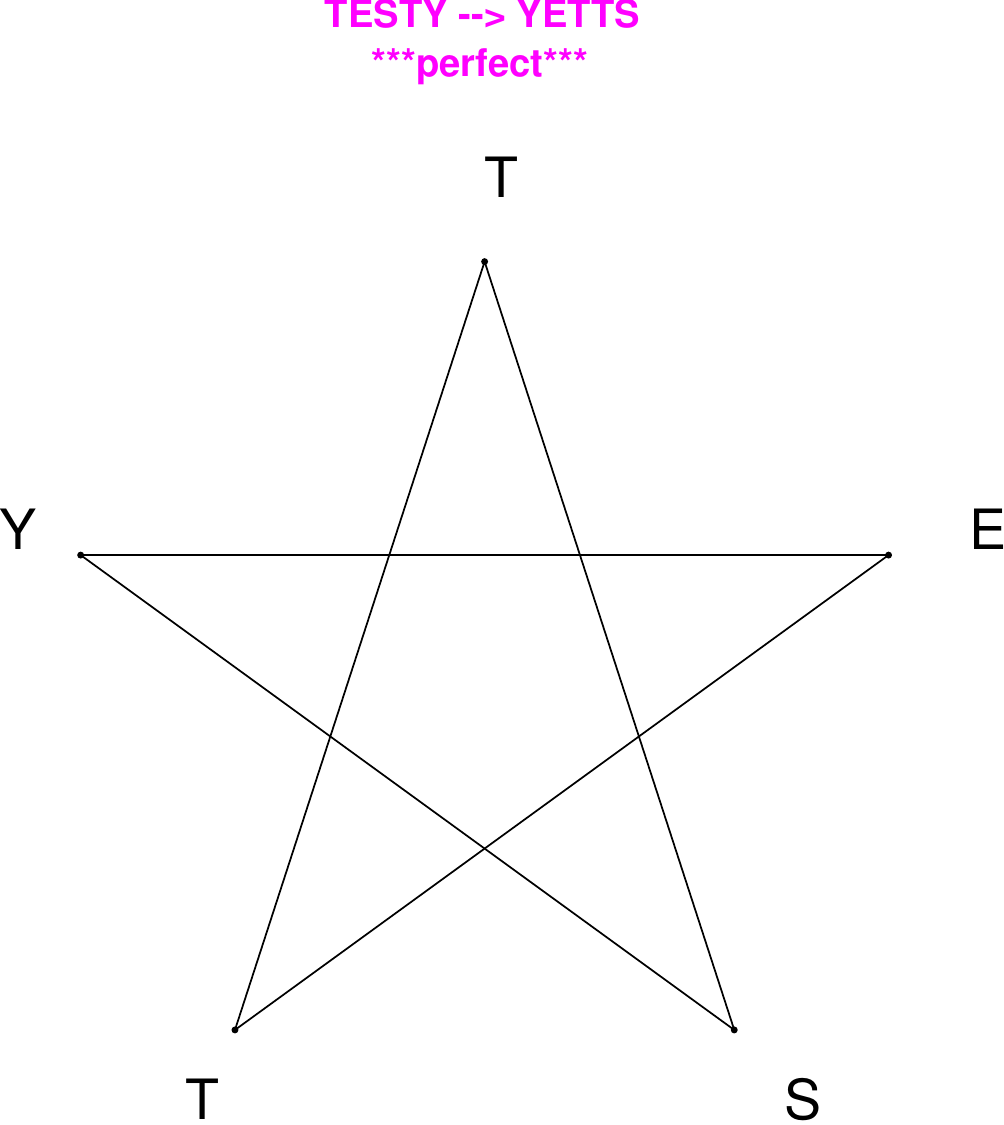}
\end{subfigure}
\end{figure}

\begin{figure}[H]
\centering
\begin{subfigure}[T]{0.19\textwidth}
\centering
\includegraphics[width=\textwidth]{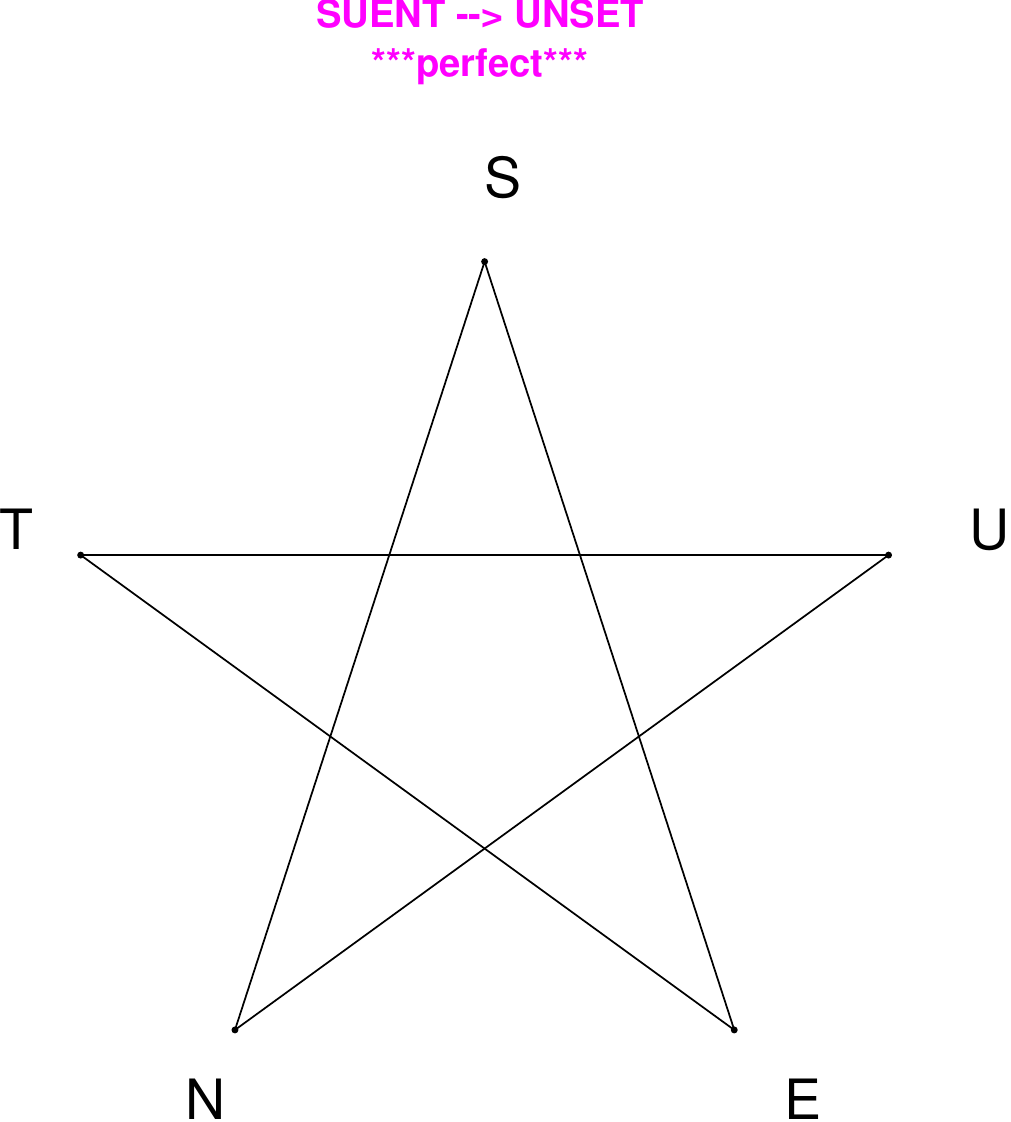}
\end{subfigure}
\hfill
\begin{subfigure}[T]{0.19\textwidth}
\centering
\includegraphics[width=\textwidth]{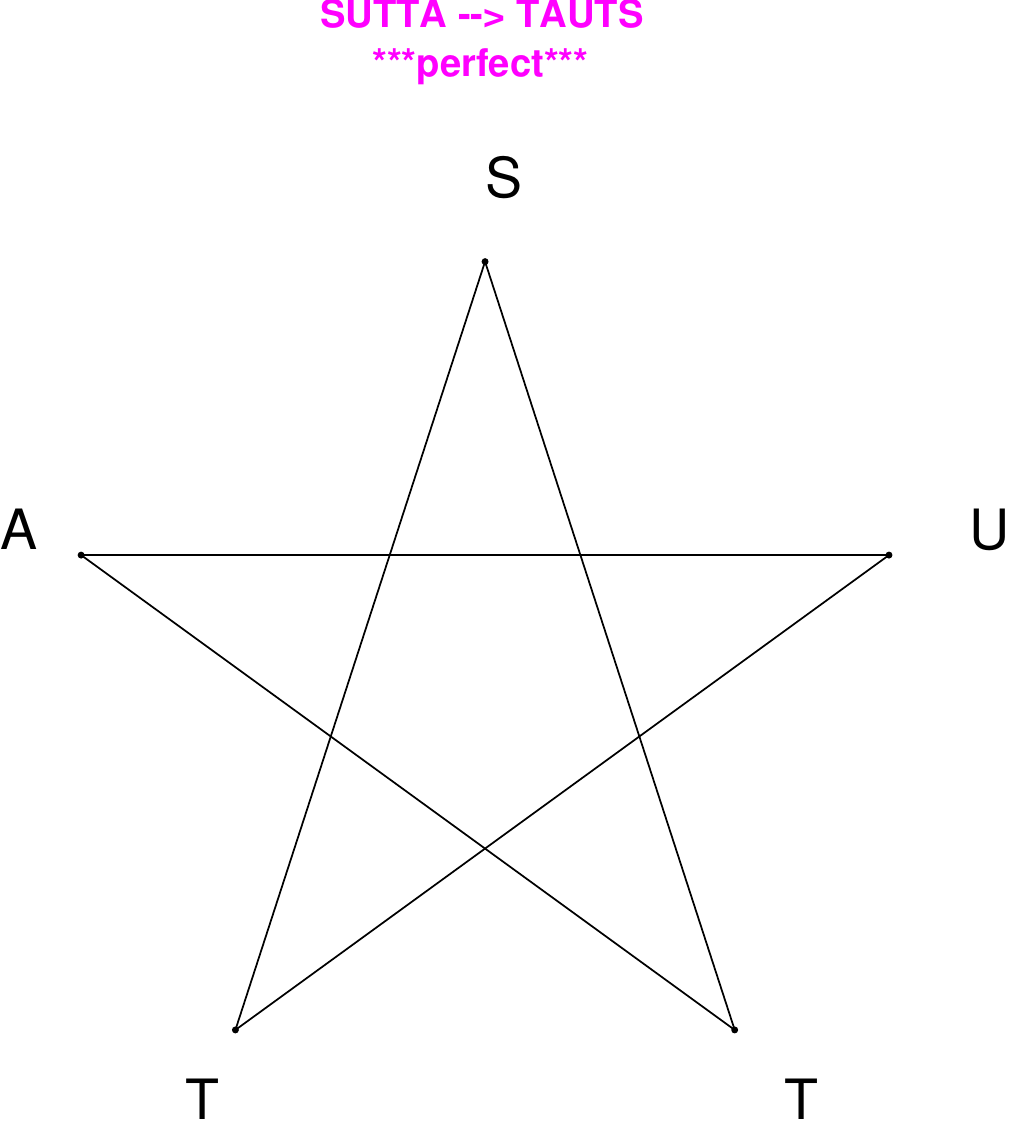}
\end{subfigure}
\hfill
\begin{subfigure}[T]{0.19\textwidth}
\centering
\includegraphics[width=\textwidth]{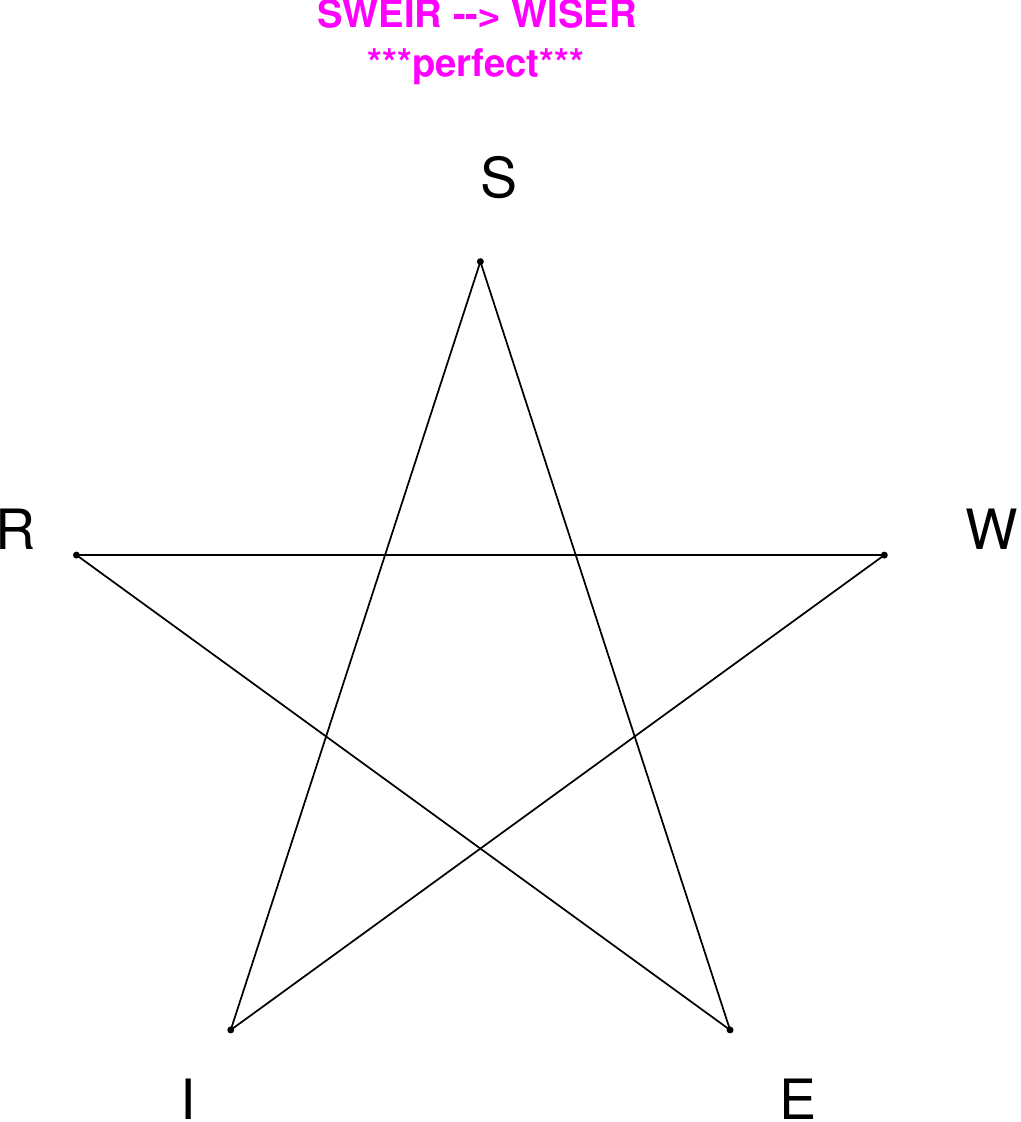}
\end{subfigure}
\hfill
\begin{subfigure}[T]{0.19\textwidth}
\centering
\includegraphics[width=\textwidth]{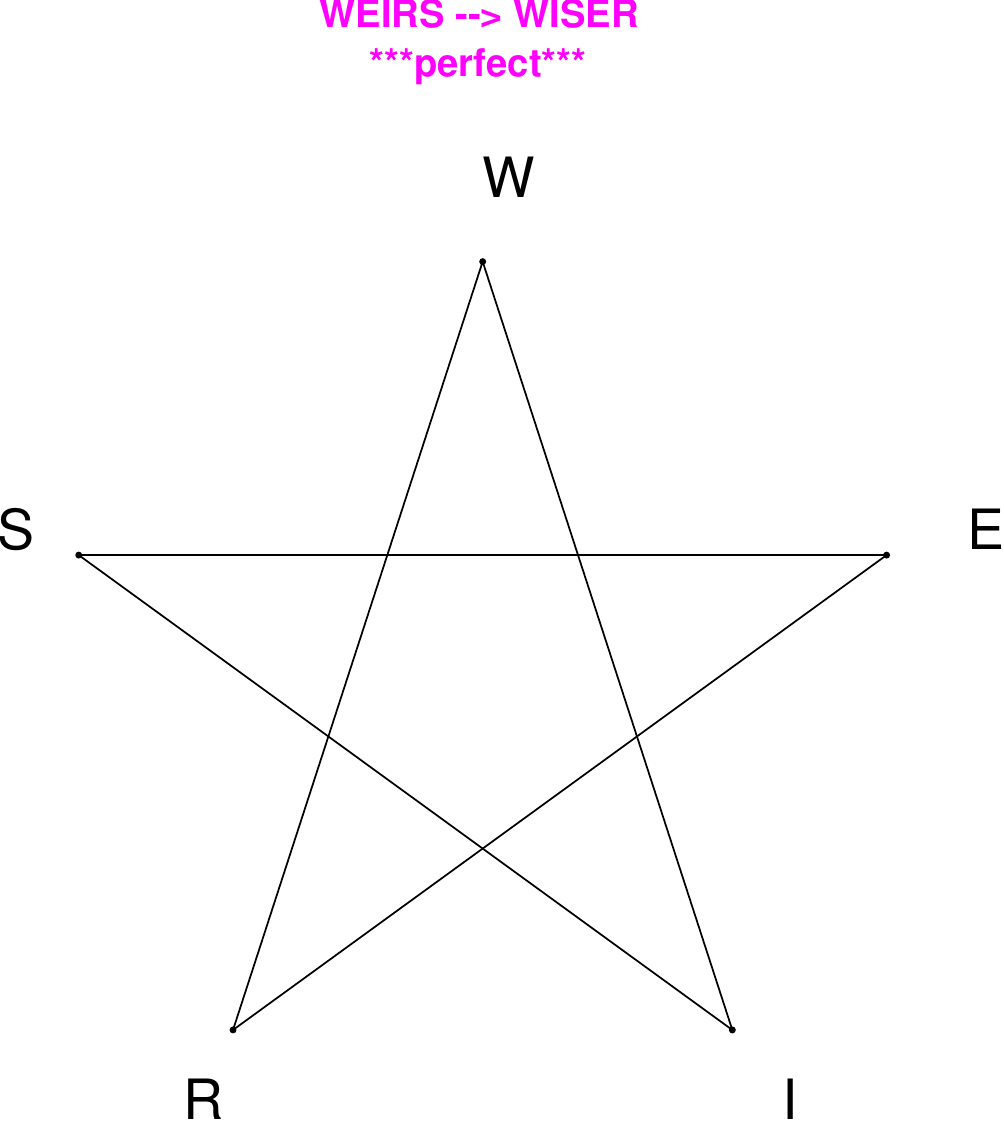}
\end{subfigure}
\hfill
\begin{subfigure}[T]{0.19\textwidth}
\centering
\includegraphics[width=\textwidth]{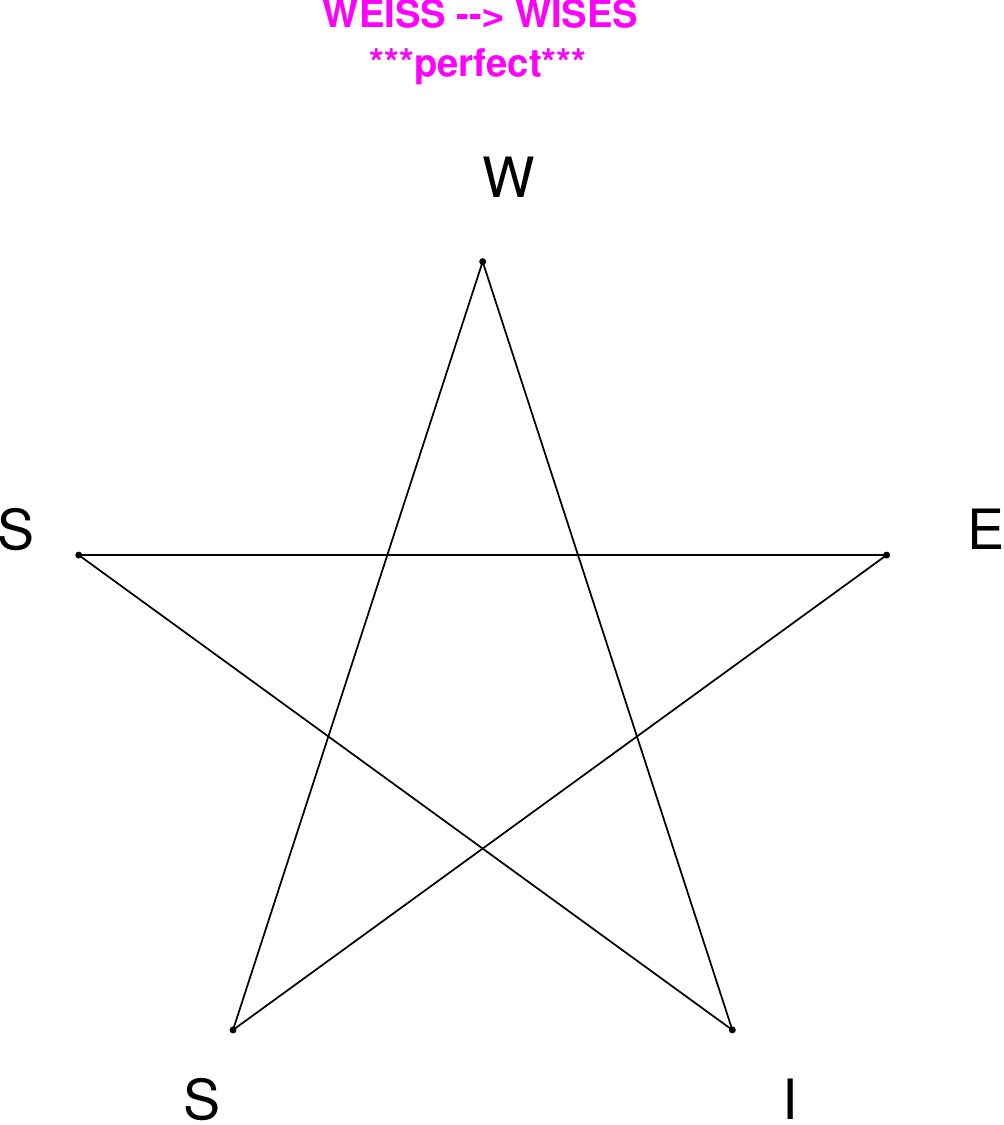}
\end{subfigure}
\end{figure}

\begin{figure}[H]
\centering
\begin{subfigure}[T]{0.19\textwidth}
\centering
\includegraphics[width=\textwidth]{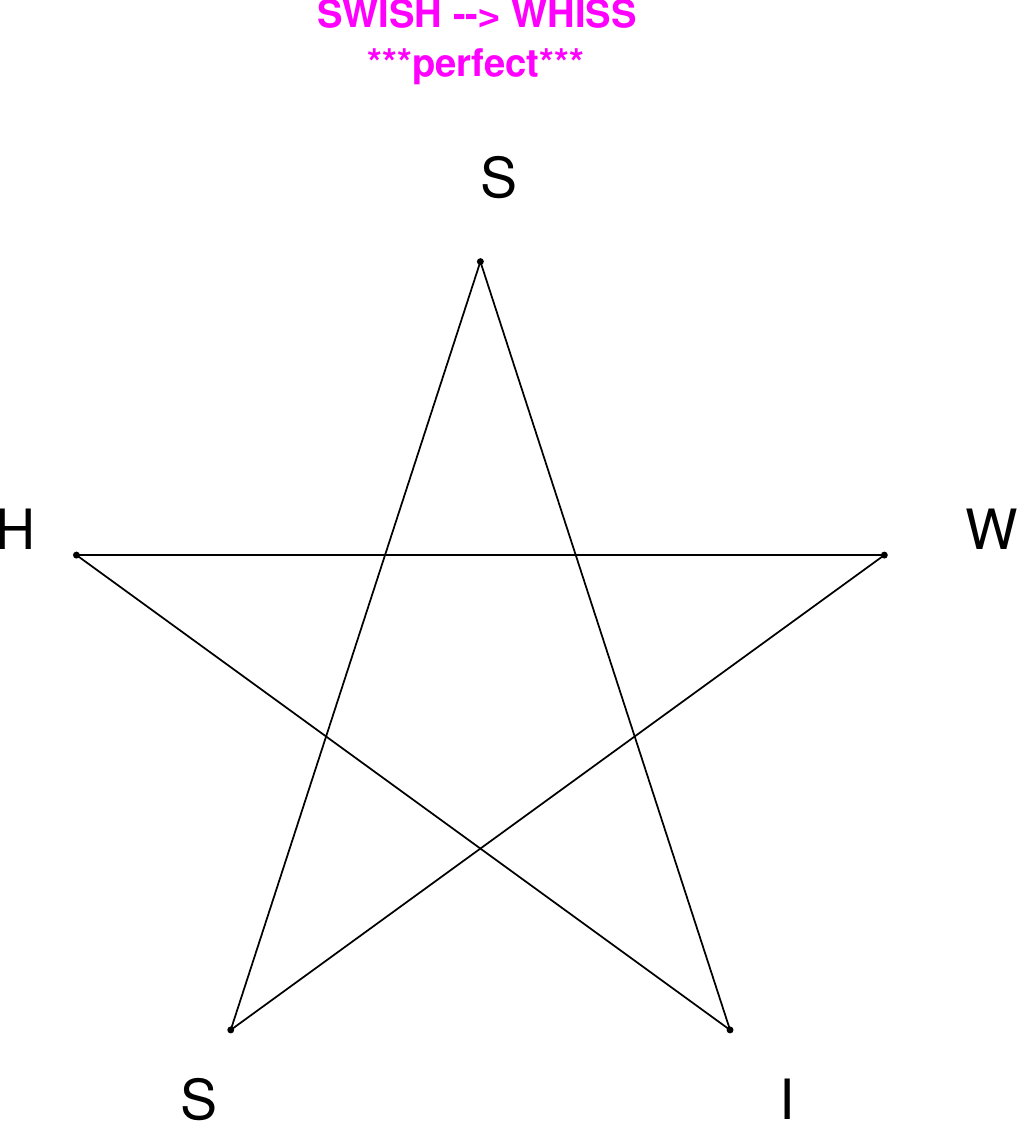}
\end{subfigure}
\hfill
\begin{subfigure}[T]{0.19\textwidth}
\centering
\includegraphics[width=\textwidth]{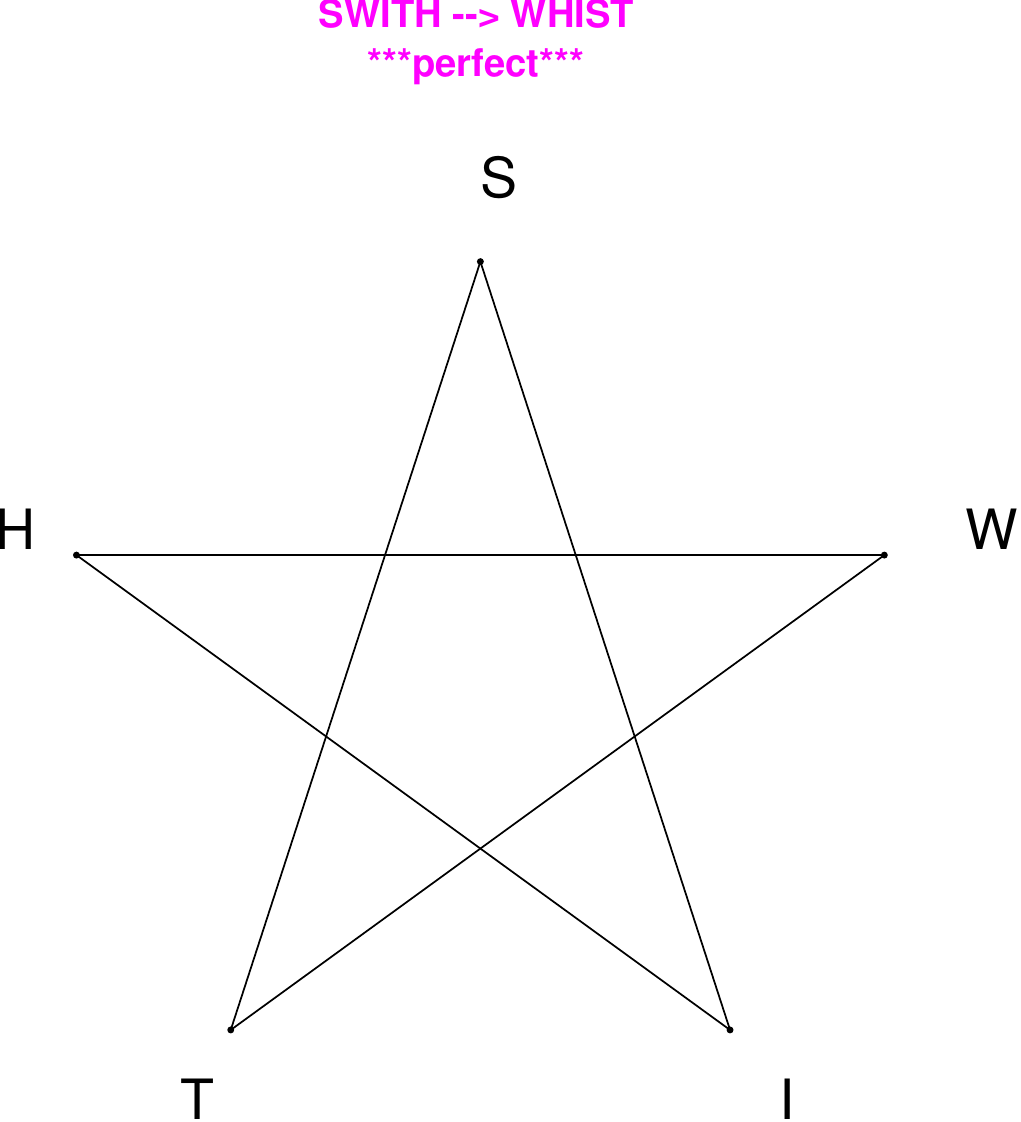}
\end{subfigure}
\hfill
\begin{subfigure}[T]{0.19\textwidth}
\centering
\includegraphics[width=\textwidth]{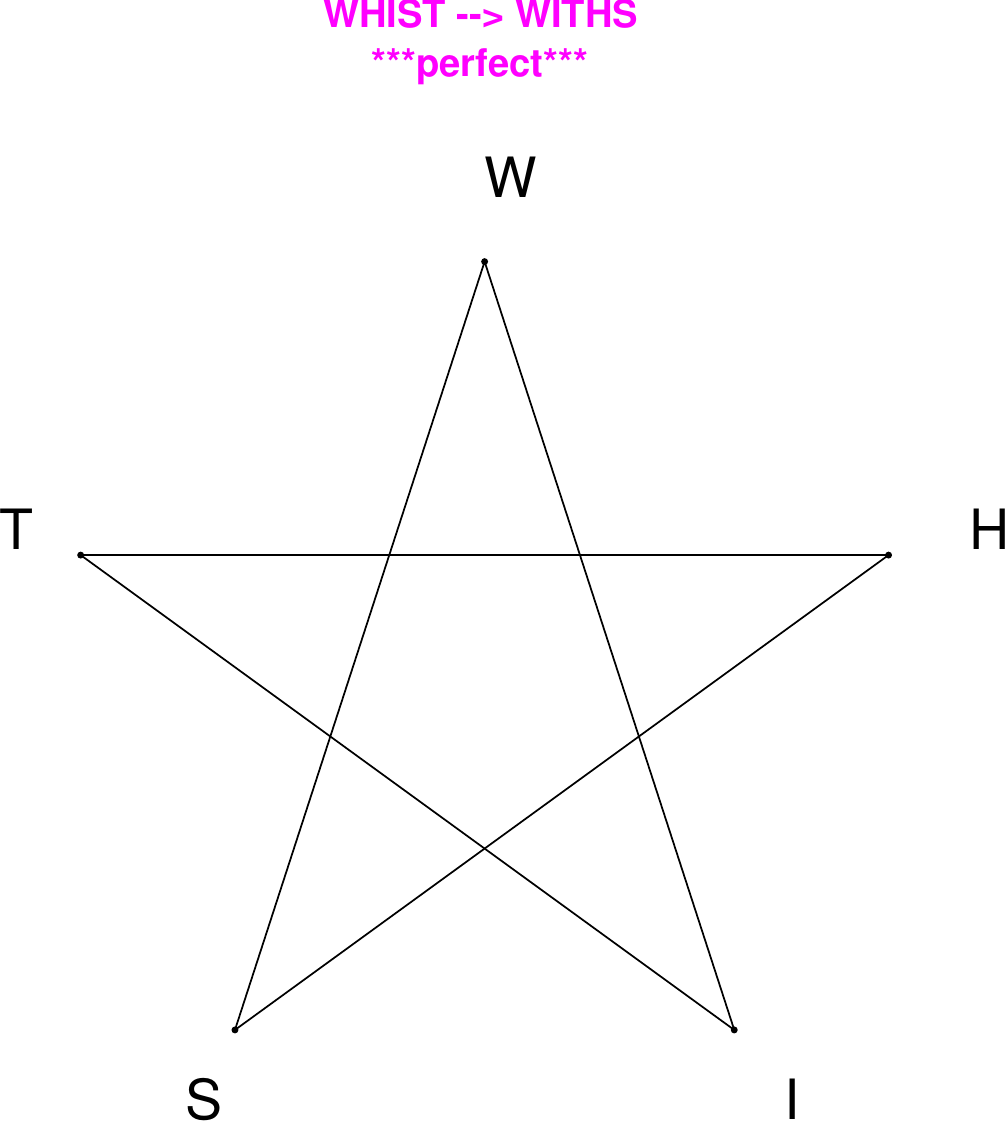}
\end{subfigure}
\hfill
\begin{subfigure}[T]{0.19\textwidth}
\centering
\includegraphics[width=\textwidth]{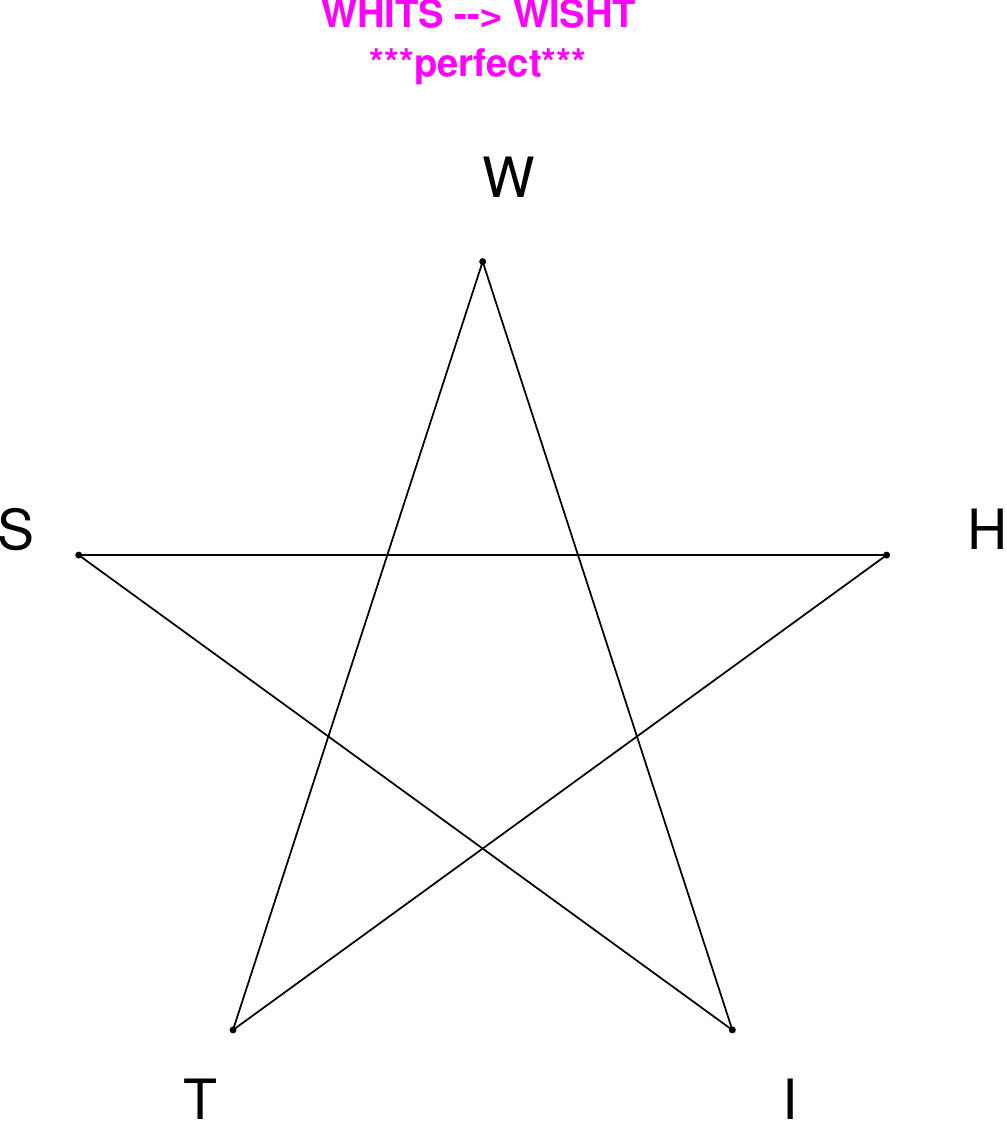}
\end{subfigure}
\hfill
\begin{subfigure}[T]{0.19\textwidth}
\centering
\includegraphics[width=\textwidth]{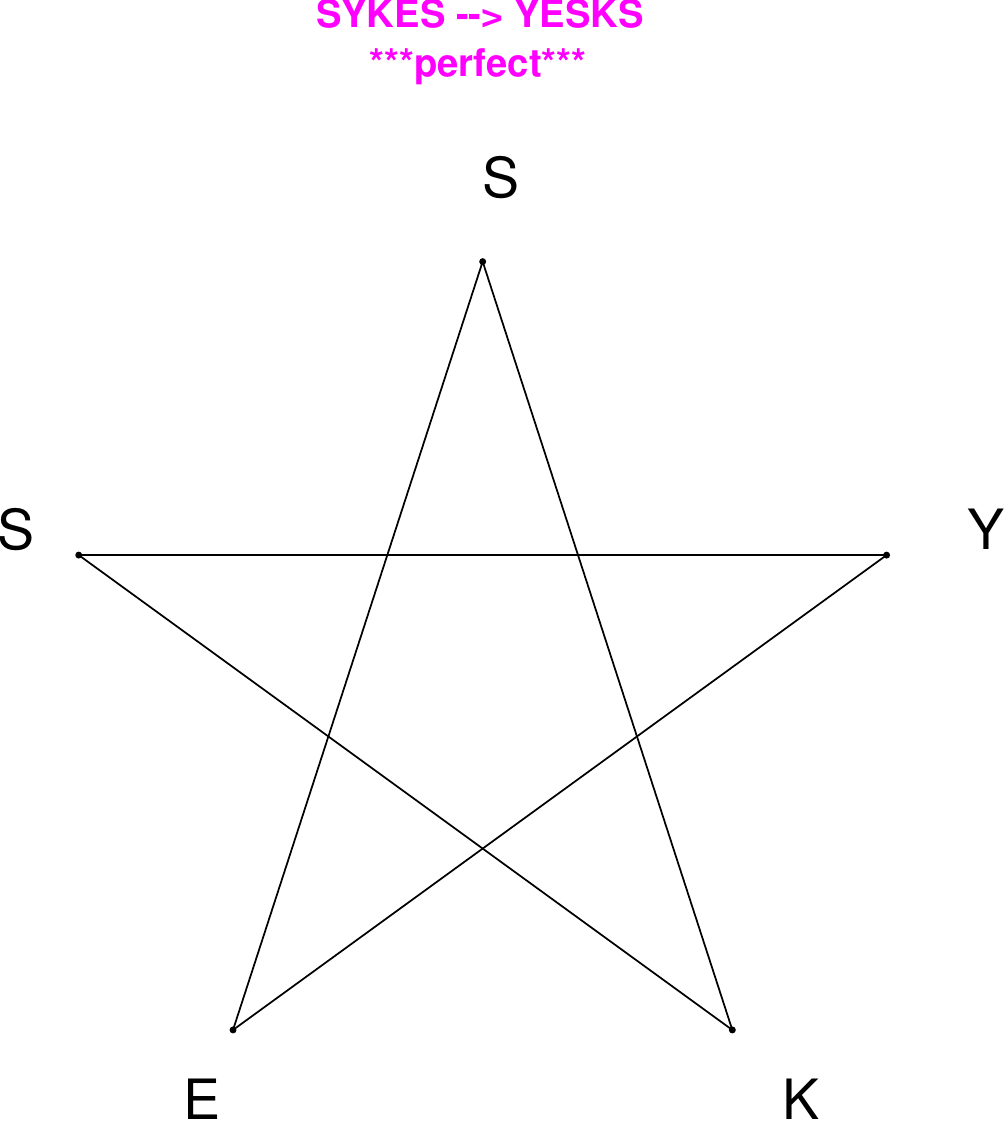}
\end{subfigure}
\end{figure}

\begin{figure}[H]
\centering
\begin{subfigure}[T]{0.19\textwidth}
\centering
\includegraphics[width=\textwidth]{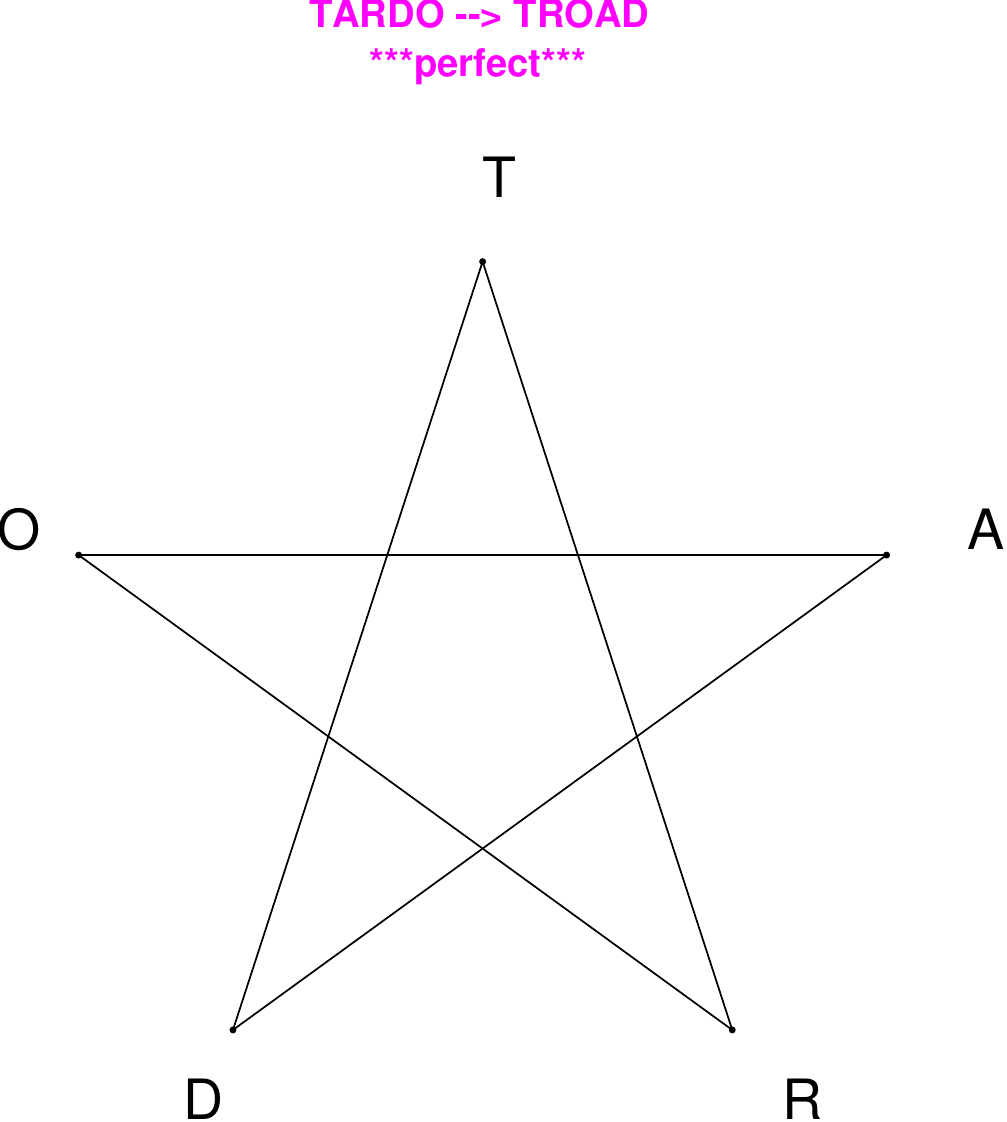}
\end{subfigure}
\hfill
\begin{subfigure}[T]{0.19\textwidth}
\centering
\includegraphics[width=\textwidth]{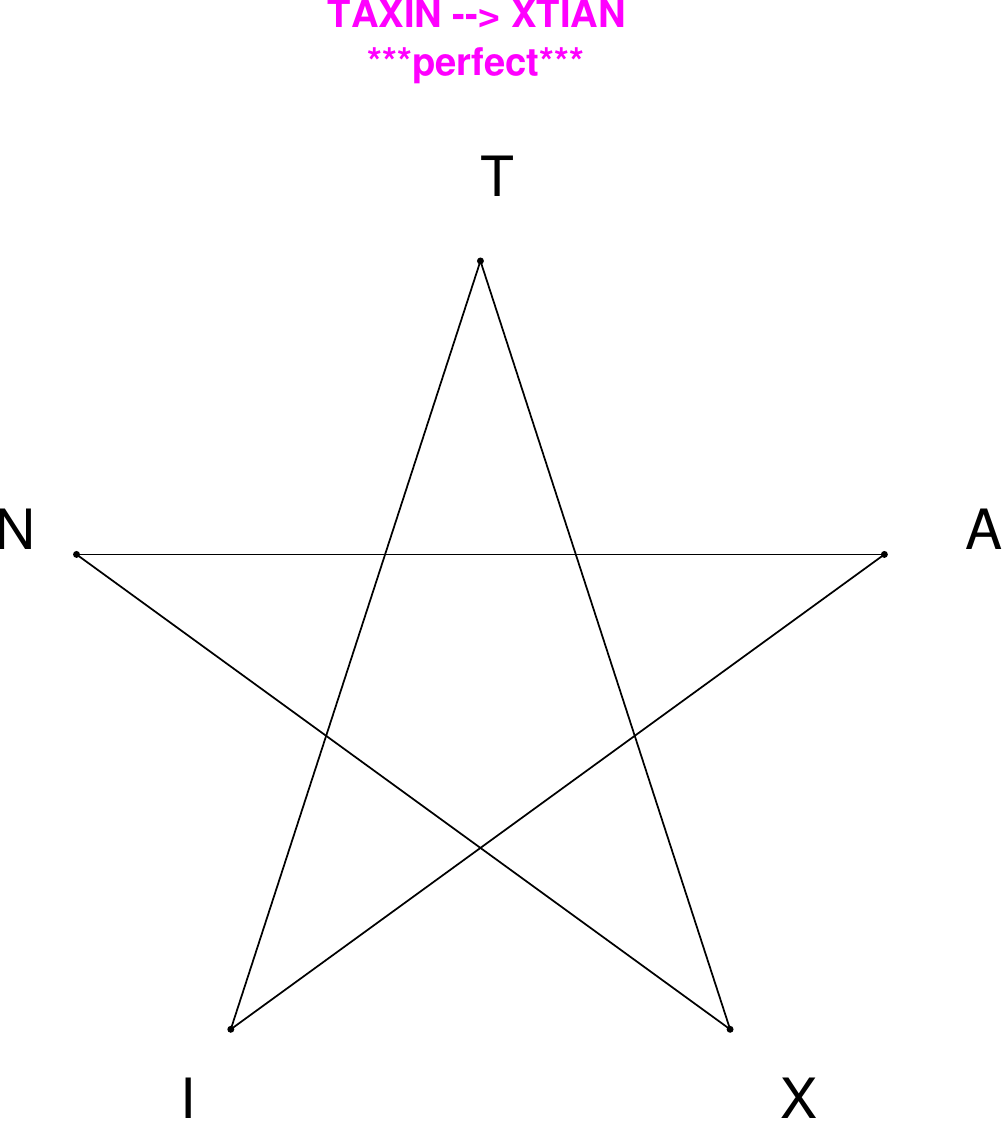}
\end{subfigure}
\hfill
\begin{subfigure}[T]{0.19\textwidth}
\centering
\includegraphics[width=\textwidth]{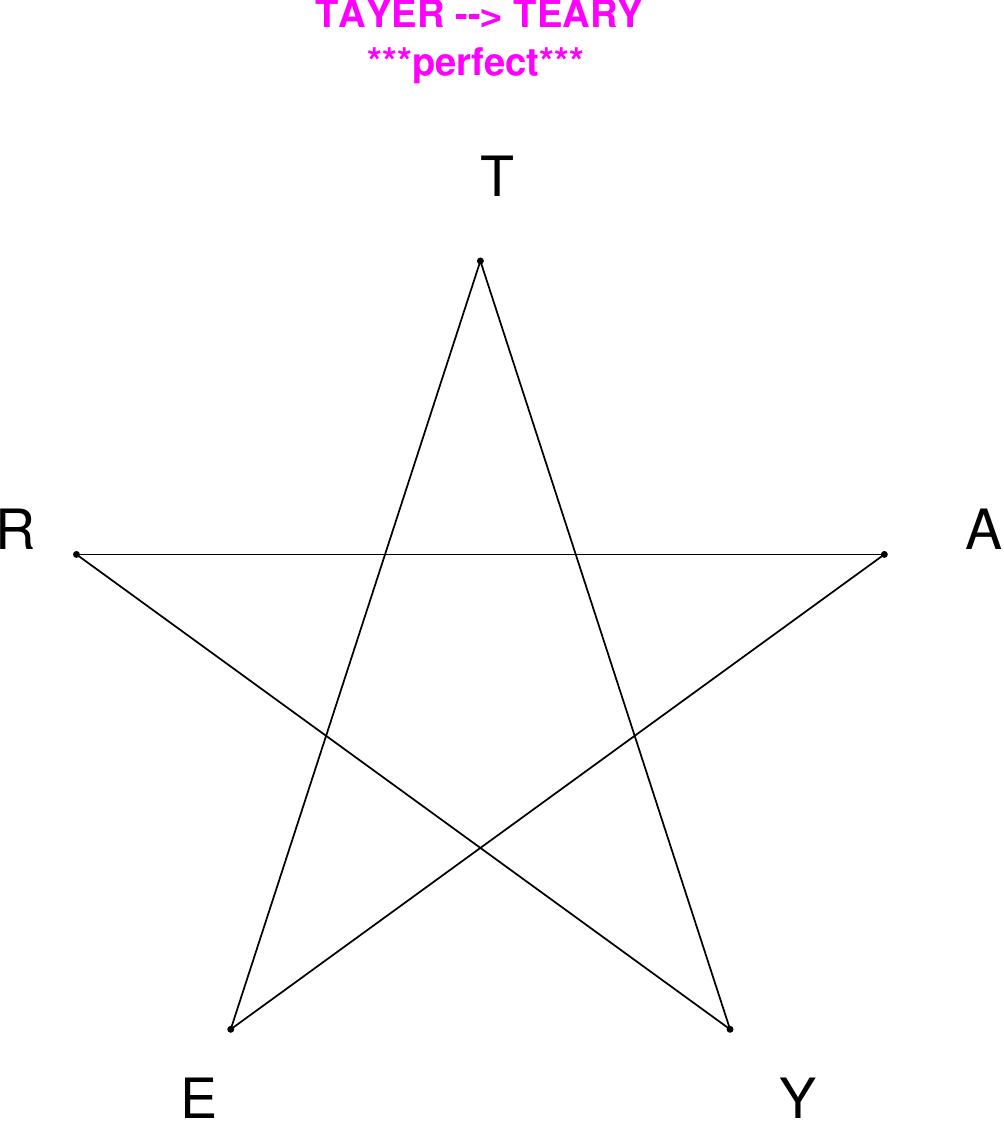}
\end{subfigure}
\hfill
\begin{subfigure}[T]{0.19\textwidth}
\centering
\includegraphics[width=\textwidth]{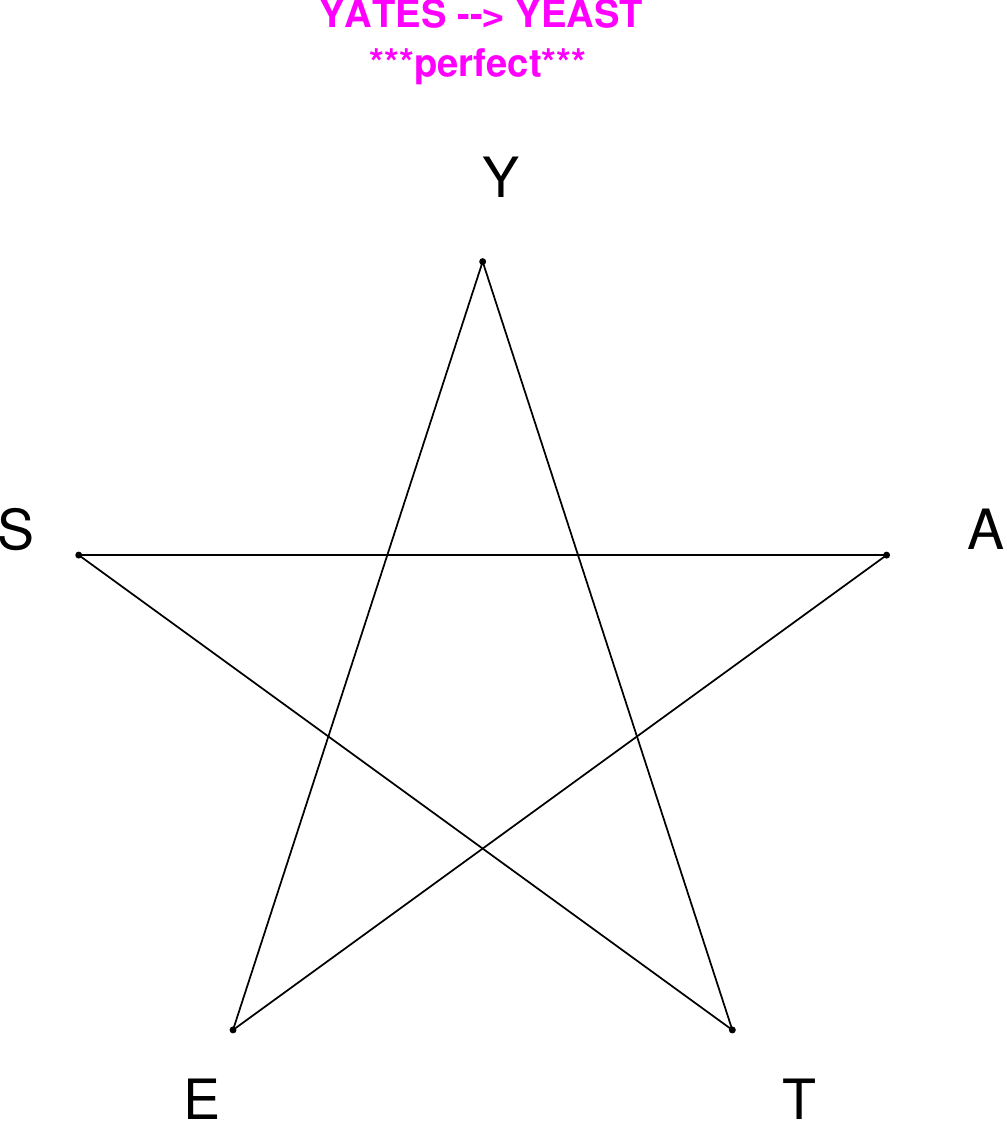}
\end{subfigure}
\hfill
\begin{subfigure}[T]{0.19\textwidth}
\centering
\includegraphics[width=\textwidth]{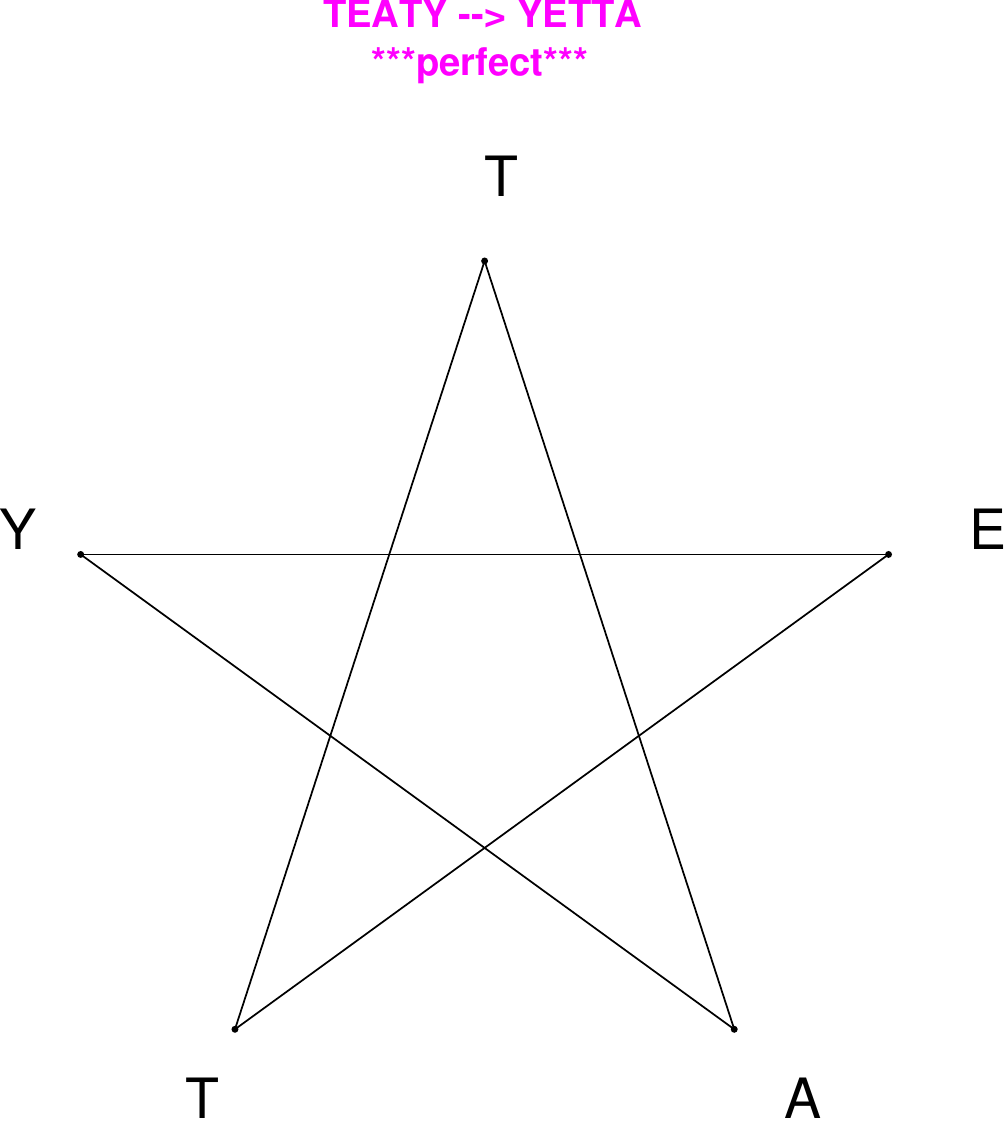}
\end{subfigure}
\end{figure}

\begin{figure}[H]
\centering
\begin{subfigure}[T]{0.19\textwidth}
\centering
\includegraphics[width=\textwidth]{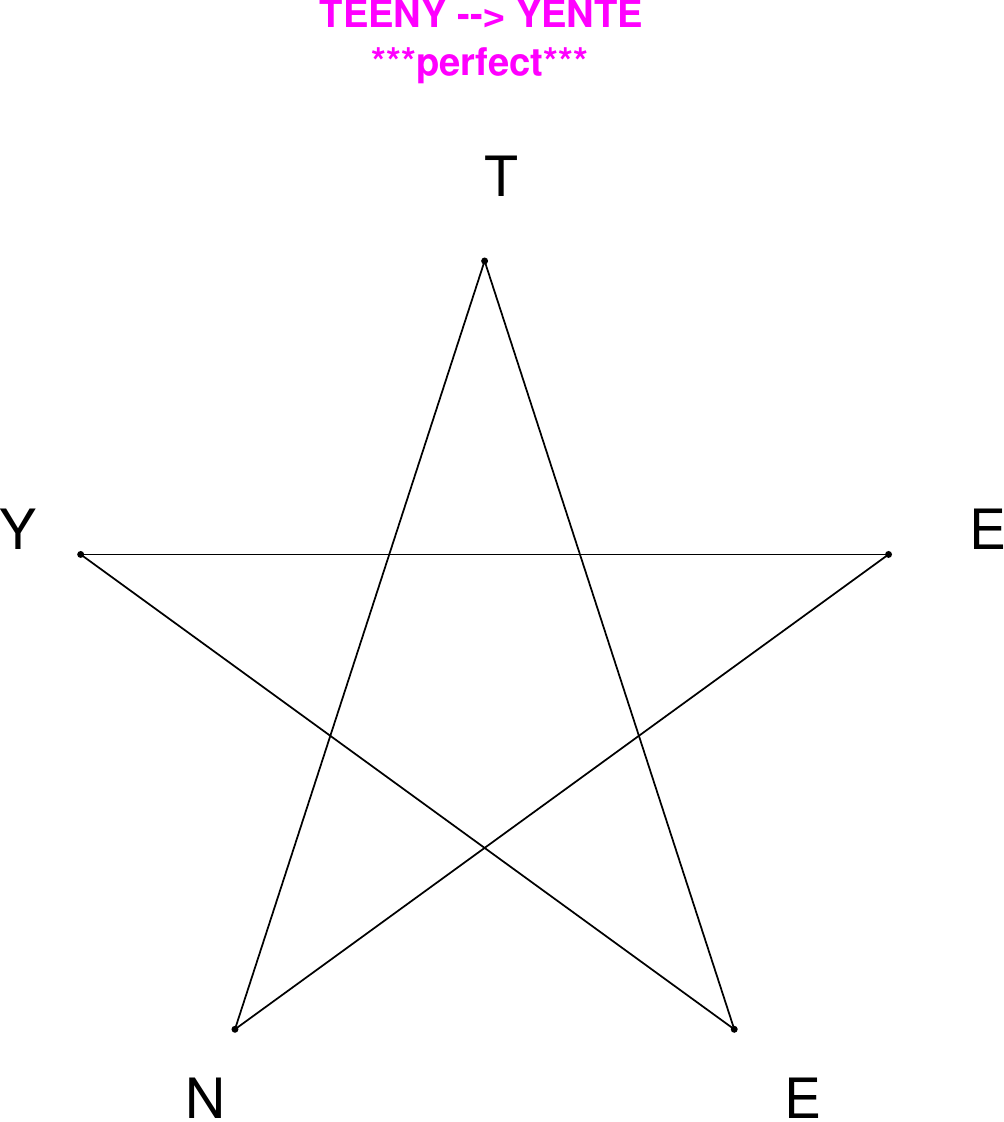}
\end{subfigure}
\hfill
\begin{subfigure}[T]{0.19\textwidth}
\centering
\includegraphics[width=\textwidth]{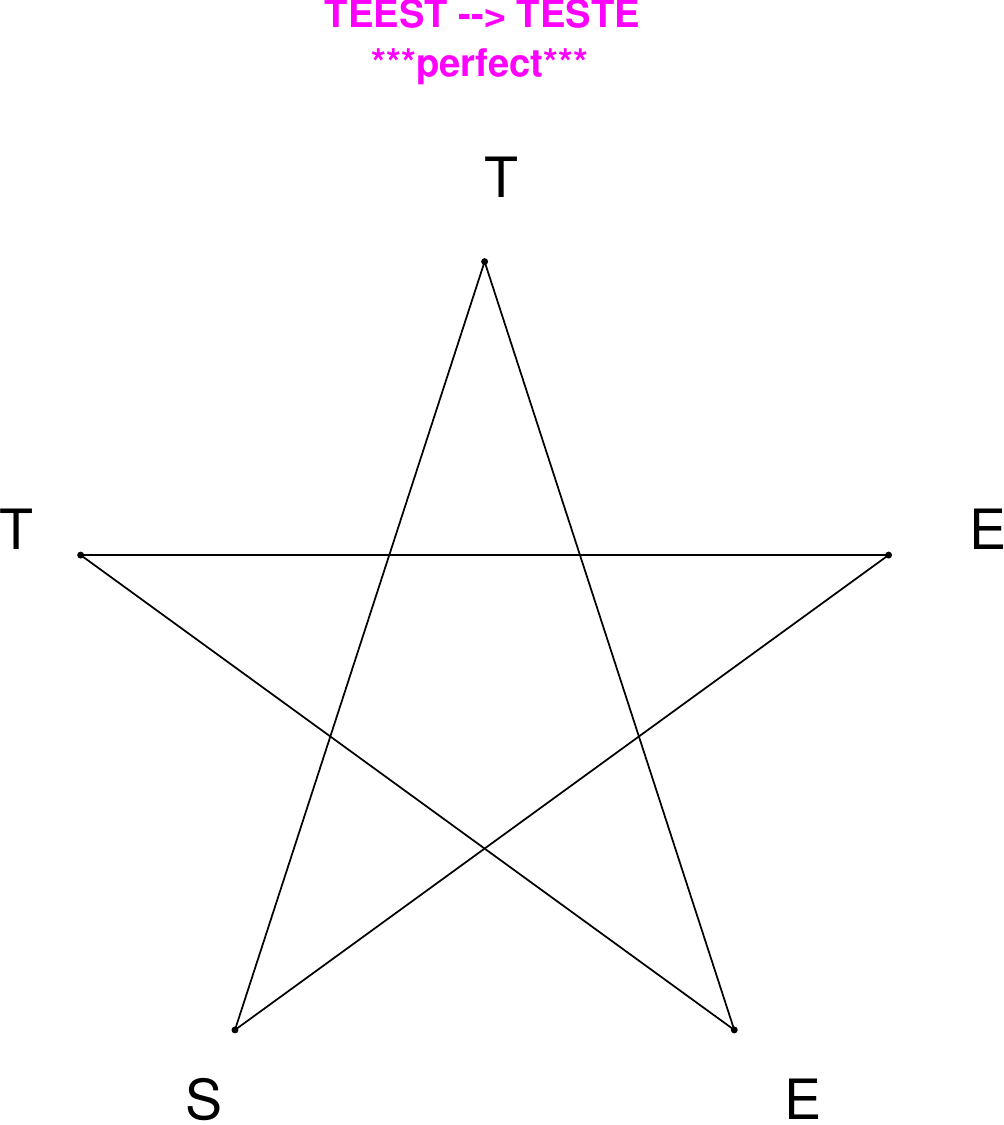}
\end{subfigure}
\hfill
\begin{subfigure}[T]{0.19\textwidth}
\centering
\includegraphics[width=\textwidth]{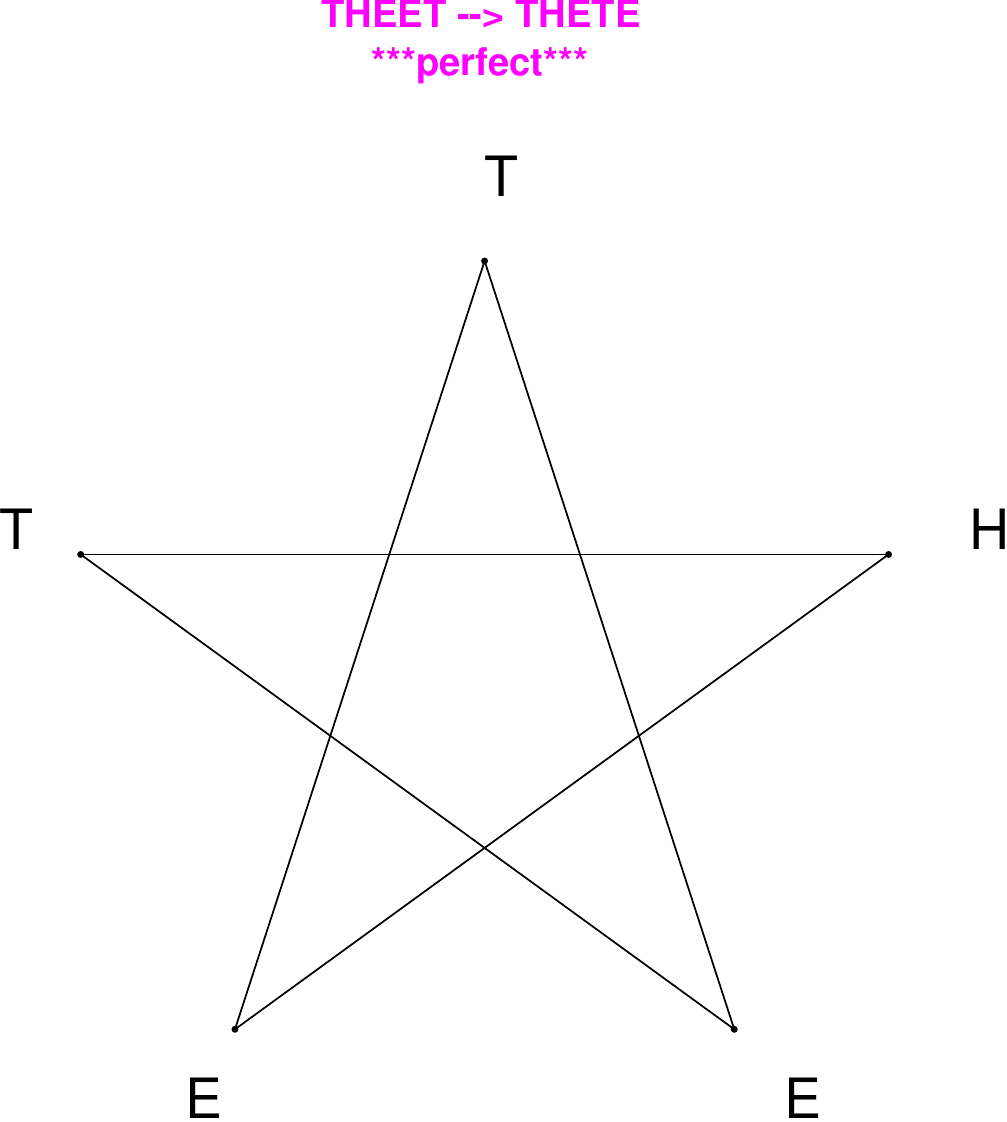}
\end{subfigure}
\hfill
\begin{subfigure}[T]{0.19\textwidth}
\centering
\includegraphics[width=\textwidth]{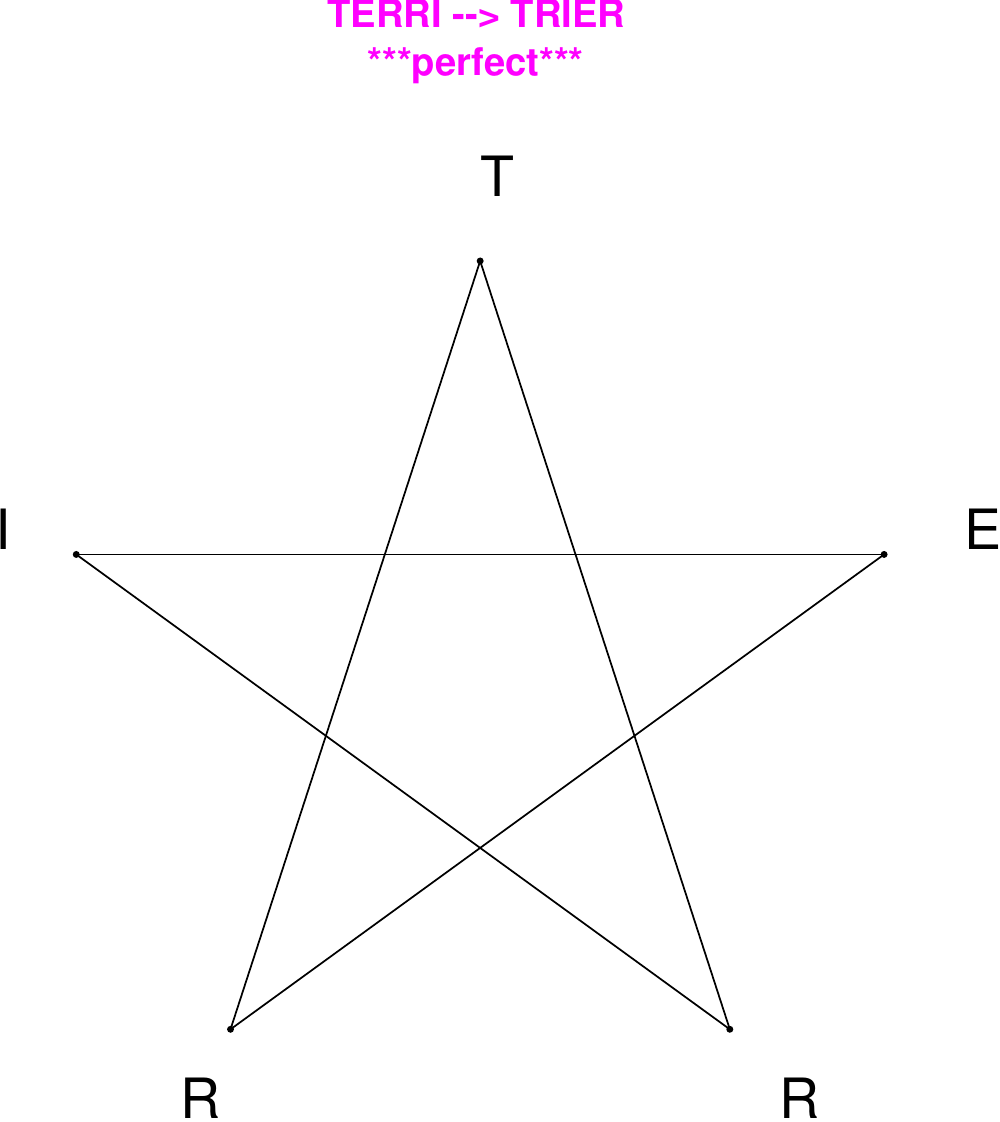}
\end{subfigure}
\hfill
\begin{subfigure}[T]{0.19\textwidth}
\centering
\includegraphics[width=\textwidth]{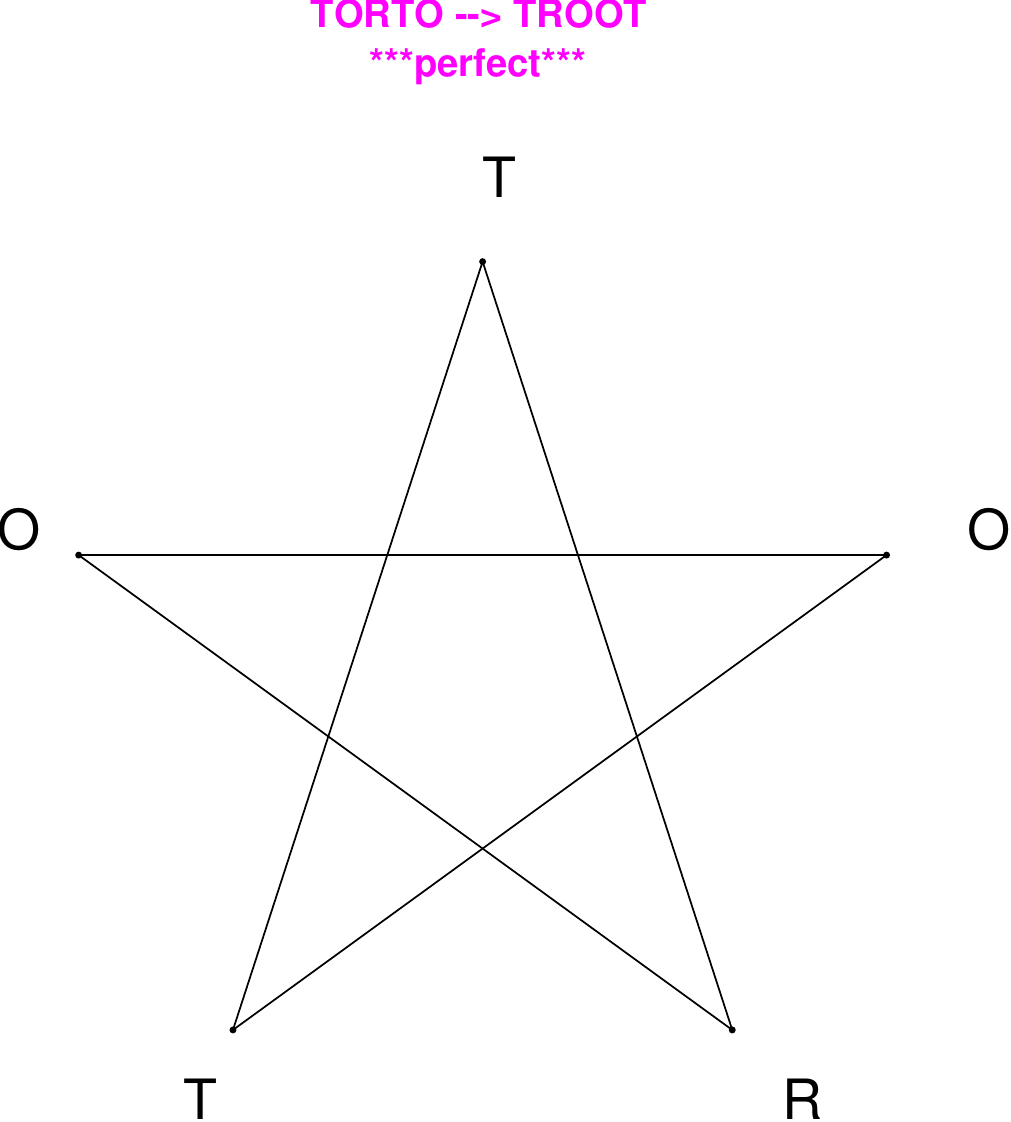}
\end{subfigure}
\end{figure}

\begin{figure}[H]
\centering
\begin{subfigure}[T]{0.19\textwidth}
\centering
\includegraphics[width=\textwidth]{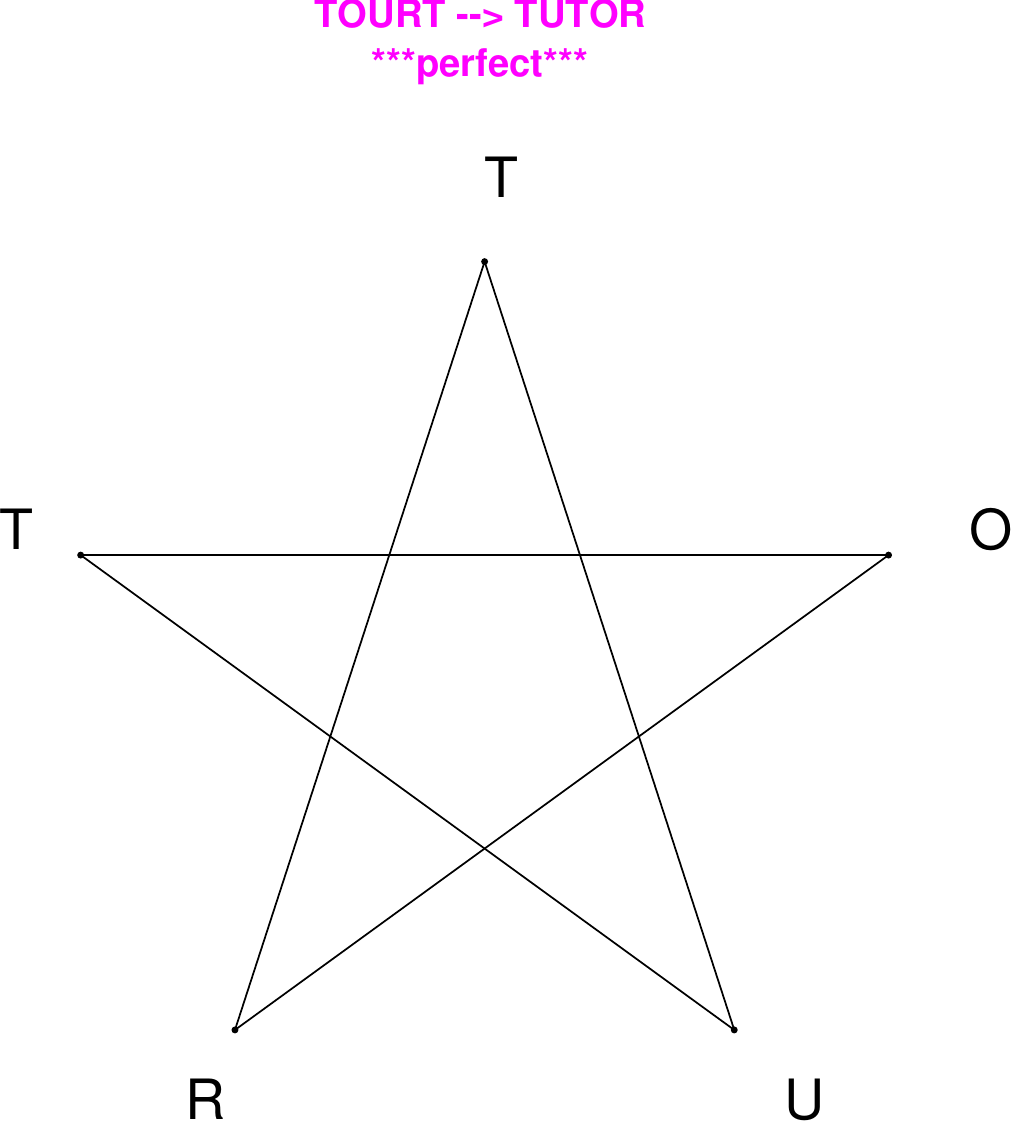}
\end{subfigure}
\hfill
\begin{subfigure}[T]{0.19\textwidth}
\centering
\includegraphics[width=\textwidth]{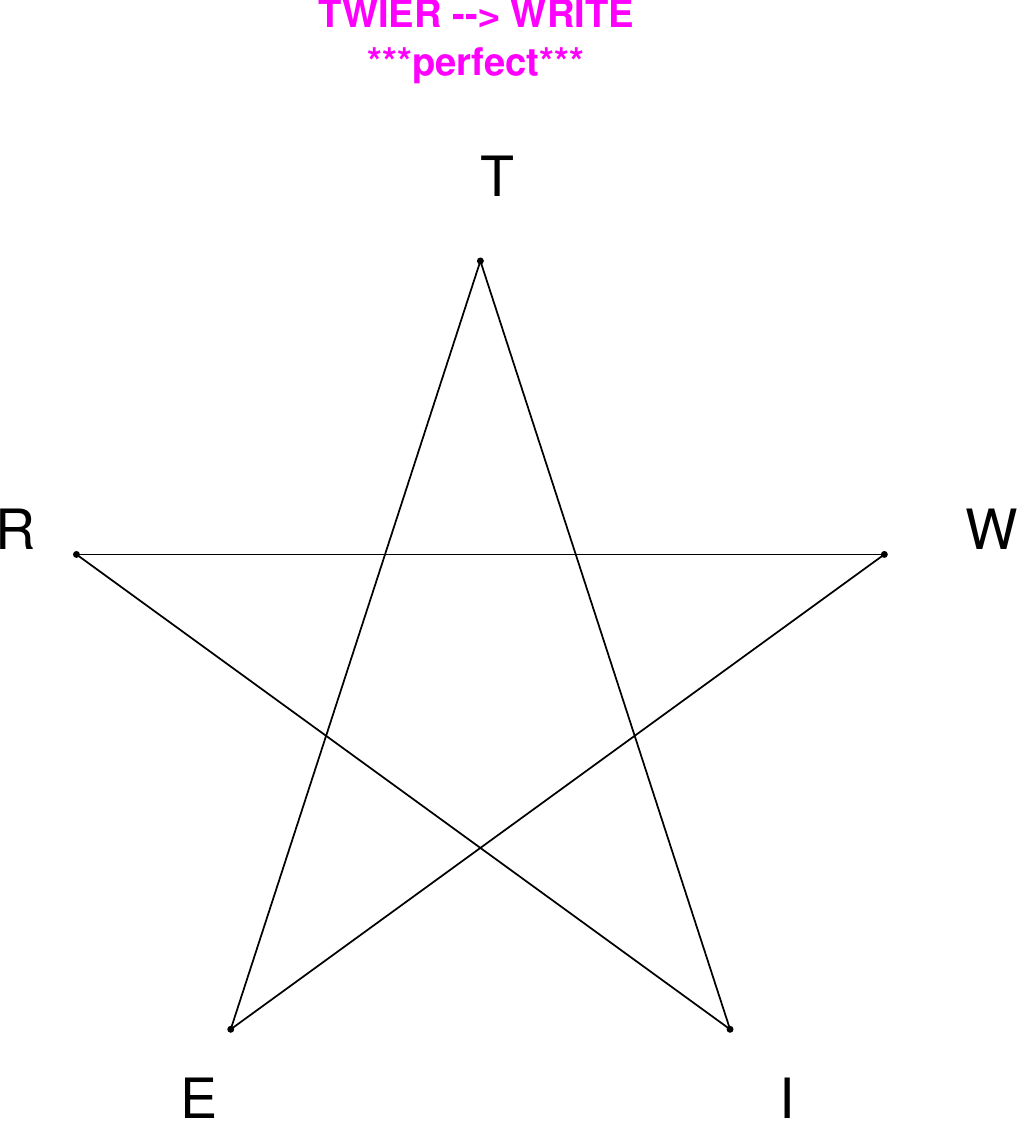}
\end{subfigure}
\hfill
\begin{subfigure}[T]{0.19\textwidth}
\centering
\includegraphics[width=\textwidth]{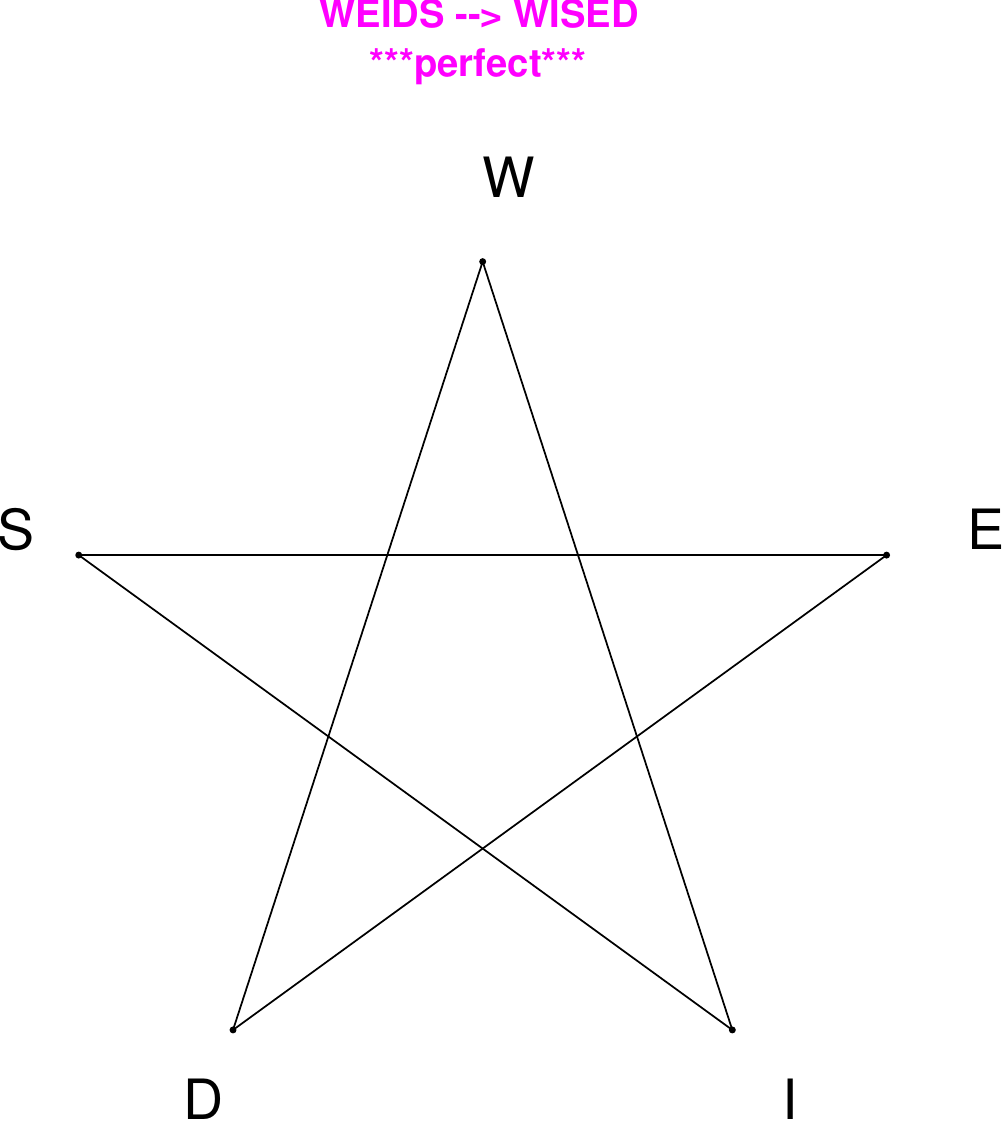}
\end{subfigure}
\hfill
\begin{subfigure}[T]{0.19\textwidth}
\centering
\includegraphics[width=\textwidth]{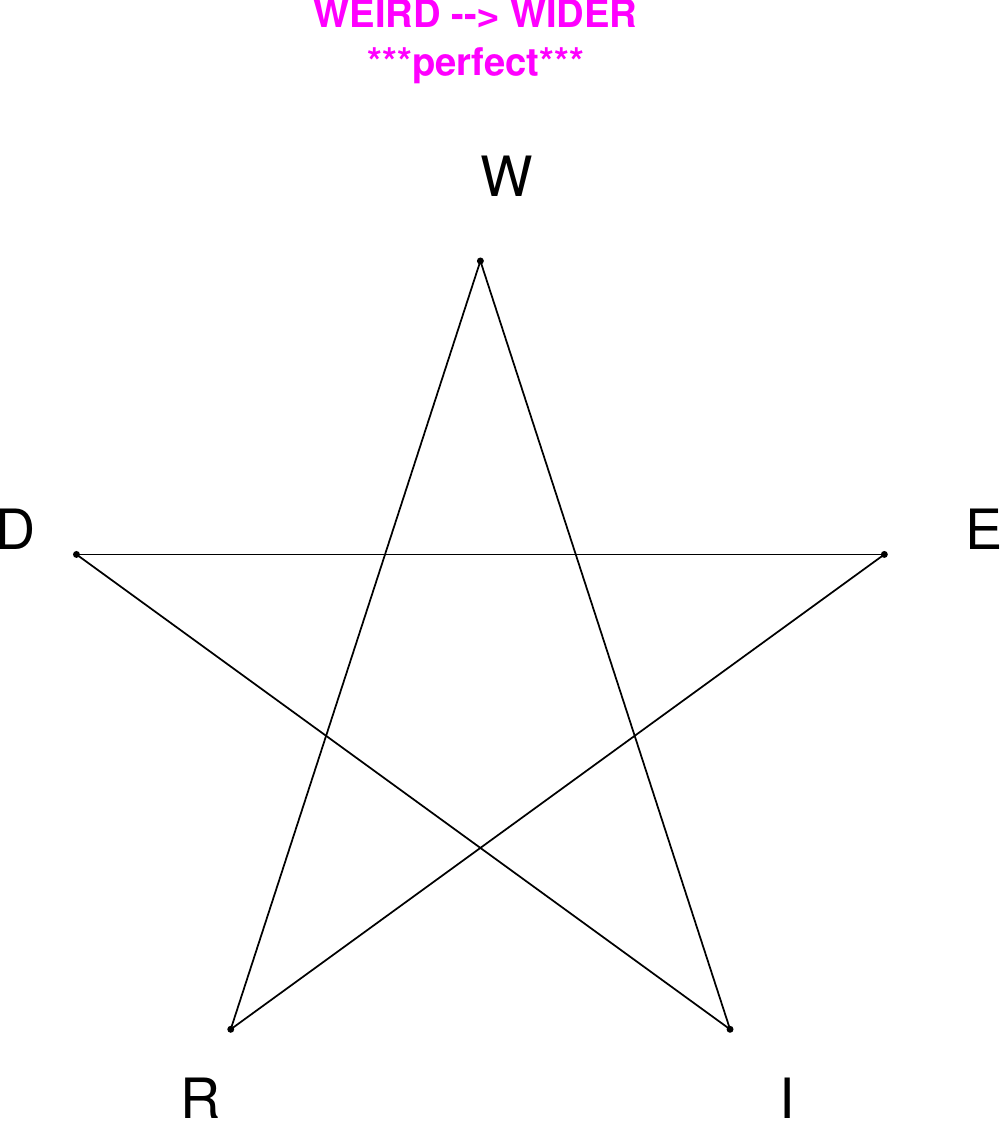}
\end{subfigure}
\hfill
\begin{subfigure}[T]{0.19\textwidth}
\centering
\includegraphics[width=\textwidth]{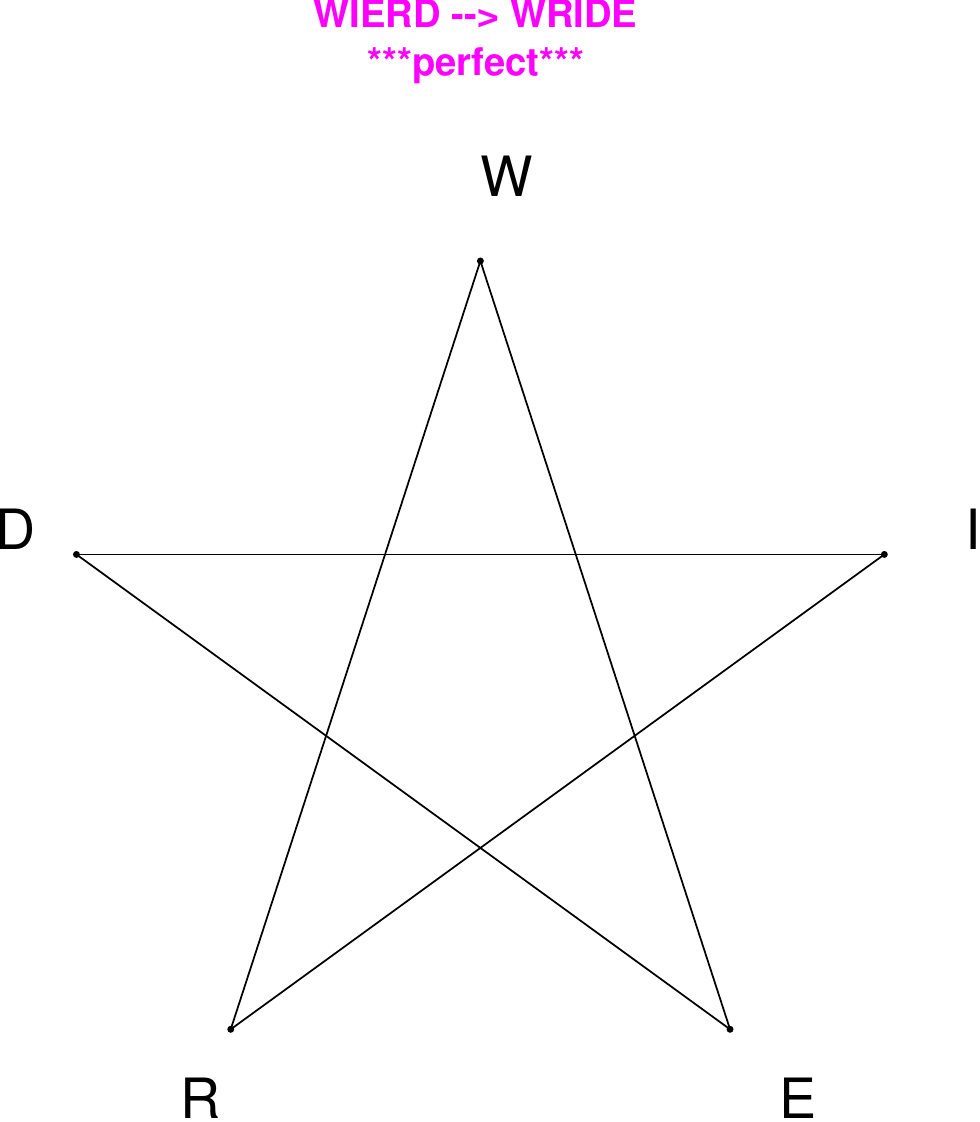}
\end{subfigure}
\end{figure}

\end{document}